\documentclass{raa}
\usepackage{longtable}
\usepackage{threeparttablex}
\usepackage{natbib}
\usepackage{booktabs}
\usepackage{graphicx,times}             
\usepackage{amssymb,amsmath}
\usepackage{epstopdf}
\usepackage{pdflscape}
\bibpunct{(}{)}{;}{a}{}{,}
\usepackage[pagebackref=true]{hyperref}
\usepackage{ulem}
\hypersetup{colorlinks = true, linkcolor = green, anchorcolor = red, citecolor = blue, filecolor = red,  urlcolor = red}
\newcommand\kmps{\mbox{km s$^{-1}$}}
\newcommand{\abs[1]}{\left| #1 \right|}
\def\degree{${}^{\circ}$} 
\newcommand\co[1][12]{\mbox{$^{#1}$CO}}
\newcommand\CO{\mbox{C$^{18}$O}}
\newcommand\vlsr{V$_{LSR}$}
\newcommand{\nodata}{...}
\graphicspath{{./figures-new/}}

\begin{document}

\title{In search for infalling clumps in molecular clouds }
\subtitle{A catalogue of CO blue-profiles}

   \volnopage{Vol.0 (20xx) No.0, 000--000}      
   \setcounter{page}{1}          


      \author{Zhibo Jiang
      \inst{1,2,3}
      \and Shaobo Zhang
      \inst{1}      \and Zhiwei Chen
      \inst{1}
      \and Yang Yang
      \inst{1,2}
      \and Shuling Yu
      \inst{1,2}
      \and Haoran Feng
      \inst{1,2}
      \and Ji Yang
      \inst{1}
      \and the MWISP group
      }

\institute{Purple Mountain Observatory, Chinese Academy of Sciences,
10 Yuanhua Road, 210023
Nanjing, China; {\it zbjiang@pmo.ac.cn} \\
\and  University of Science and Technology of China, Hefei 230026, China\\
\and Center for Astronomy and Space Sciences, Three Gorges University, Yichang 443002, China \\
}

\abstract{ 
We have started a systematic survey of molecular clumps with infall motions to study the very early phase of star formation. Our first step is to utilize the data products by MWISP to make an unbiased survey for blue asymmetric line profiles of CO isotopical molecules. Within a total area of $\sim$ 2400 square degrees nearby the Galactic plane, we have found 3533 candidates showing blue-profiles, in which 3329 are selected from the \co\&\co[13]{} pair and 204 are from the \co[13]\&\CO{} pair. Exploration of the parametric spaces suggests our samples are in the cold phase with relatively high column densities ready for star formation. Analysis of the spatial distribution of our samples suggests that they exist  virtually in all major components of the Galaxy. The vertical distribution suggest that the sources are located mainly in the thick disk of $\sim$ 85 parsec, but still a small part are located far beyond Galactic midplane.   Our follow-up observation indicates that these candidates are a good sample to start a search for infall motions, and to study the condition of very early phase of star formation. 
\keywords{line: profiles --- stars: formation  --- ISM: clouds
--- catalogues --- surveys}
}

   \authorrunning{Z. Jiang et al. }            
   \titlerunning{In Search for Infalls }  
 
   \maketitle

\section{Introduction}           
   \label{sec:intro}
Gravitational collapse of dense molecular cloud cores is a key step in the formation of stars \citep[e.g.,][]{1987ARA&A..25...23S}. Presently due to the limitation of the observational facilities,  it is difficult to obtain a motion picture showing a core in the process of gravitational collapse. However, line profiles can provide signatures of inward collapse motion.  As theoretical modelings show \citep[e.g.,][]{1996ApJ...465L.133M}, after a collapse happens, the center of the core becomes warmer due to the accumulation of gravitational energy while the outside envelope remains relative cool.  In the configuration of an inflowing core surrounded with a static envelope, the radiation transfer along the line of sight can be simplified by a two layer model for both the gas flowing away and toward the observer \citep[see fig.5 in][]{1999ARA&A..37..311E}. For the gas flowing toward the observer, the inner layer with higher excitation is nearer to the observer than the outer layer with lower excitation; for the flow-away gas, emission from the inner layer with higher excitation is absorbed by the outer layer with lower excitation lying closer to the observer. A double-peaked line profile with an absorption dip is produced by the above radiation transfer. The blue peak (gas flowing toward the observer) is stronger than the red peak (gas flowing away from the observer). This effect is especially conspicuous for molecular transition lines with a suitable optical depth and critical density. This kind of profiles are commonly referred to as ``blue-profiles''. They have been discussed by many researchers, and are regarded as an indicator of gas inflow motion \citep[e.g.,][]{1993ApJ...404..232Z,1994ASPC...65..192M,1997ApJ...489..719M,1996ApJ...465L.133M}, or infall signature. To distinguish a blue-profile from  multi-component emissions in the line of sight that may also show double-peaks, one needs an optically thin line, whose profile is single-peaked without  a self-absorption feature.  The peak position of the optically thin line should be close to the self-absorption dip between the blue and red peaks.

   Observational studies of the blue-profiles started in early 1990's \citep{1993ApJ...404..232Z,1994ASPC...65..192M} and are seen accelerated recently. Up to date, the studies can be classified into two categories. One is case study, i.e. detailed observations toward some particular star-forming complexes to study the physical conditions of regions showing infall signature \citep[e.g.,][]{1993ApJ...404..232Z, 2011ApJ...740..114Z, 2012MNRAS.422.1098R, 2014MNRAS.437.3766M,2018ApJ...852...12Y}. The other is the deliberated search, i.e., systematical searches for infall signatures in deliberately selected samples of different properties, such as low-mass protostellar objects \citep{1997ApJ...489..719M}, high-mass star-forming regions \citep{2003ApJ...592L..79W,2007ApJ...669L..37W,2007ApJ...663.1092K,2009MNRAS.392..170S,0067-0049-225-2-21}, high-mass protostellar objects \citep{2018ApJS..235...31Y}, high infrared extinction clouds \citep{2013AA...549A...5R}, massive star-forming cores \citep{2009MNRAS.392..170S,2015MNRAS.450.1926H}, and early dense cores \citep{2018ApJ...862...63C}, extended green object \citep{2010ApJ...710..150C}, etc.

Above studies have presented us with diverse hints on the initial phase of star formation, e.g., when the gravitational collapse starts and ends, whether high- and low-mass stars start their life in similar ways, and the time scale of mass-infall phase. However, a comprehensive study on the infall phase is necessary to understand fully the process how dense molecular cores turn into stars, and to contribute to solve the long-standing problem of the origin of the stellar initial mass function.  Previous studies mainly utilized known targets that have shown star-forming activities as a start to look into. This kind of researches will inevitably introduce bias if a statistical study is carried out upon them. We therefore do the work in an opposite direction, i.e., by merely looking for infall signatures in the Galaxy without any other assumption,  we study the physical properties and star-forming activities where the signatures happen. The ongoing Milky Way Imaging Scroll Painting Project \citep[MWISP,][]{2019ApJS..240....9S}, a large scale survey of CO (J = 1 - 0) lines in the northern Galaxy, provides an excellent opportunity to start up such a kind of work. 

Our project is to set up a sample of CO blue-profiles using the MWISP data by blind search, and refine it by further observations using some more infall-sensitive lines (such as HCO$^+$ lines). The aim of this project is to obtain a comprehensive sample of infall candidates with a high confidence level. This paper is to present the very first result of our project, a preliminary catalogue of CO blue-profiles.    

The structure of this paper is arranged in this way: in Sec. \ref{sec:met} we give a brief introduction to data preparation, strategy of blind search for blue-profiles; in Sec. \ref{sec:cat} we present the main catalogue and the associated line profiles; in Sec. \ref{sec:disc} we discuss the distribution and parameter space of sources; and Sec. \ref{sec:sum} summarizes our results.

\section{Method} \label{sec:met}

\subsection{Data} \label{sec:dat}

All data used in this work are based on the MWISP data products. A detailed description of the data acquisition, quality control and archiving scheme can be found in \citet{2019ApJS..240....9S}. The archive data are in unit of ``cells'', {the main region of} which is an area of 30\arcmin $\times$ 30\arcmin. According to the survey strategy, 
{the outskirts of each cell are observed with the 9-beam Superconducting Spectroscopic Array Receiver \cite[SSAR,][]{2012ITTST...2..593S}} less than the main region. Consequently, the RMS noise levels are higher in the edge part than in the central. Therefore, a combination with the neighboring cells is necessary to achieve a uniform noise level. At this step, we intentionally extend the area of each cell to some extent (generally by 0.5 arcmin on each side) to avoid possible edge effects. We then cut off unnecessary velocity channels ($\abs[V_{LSR}] \ge$ 200 \kmps) to minimize the unnecessary computation, and possible {unreal} detections {due to untrue signals} which might be a problem in the manual check step. The data reduced as such are ready for  machine work.

\begin{figure}[h]
   \begin{minipage}[t]{0.495\linewidth}
   \centering
    \includegraphics[width=60mm]{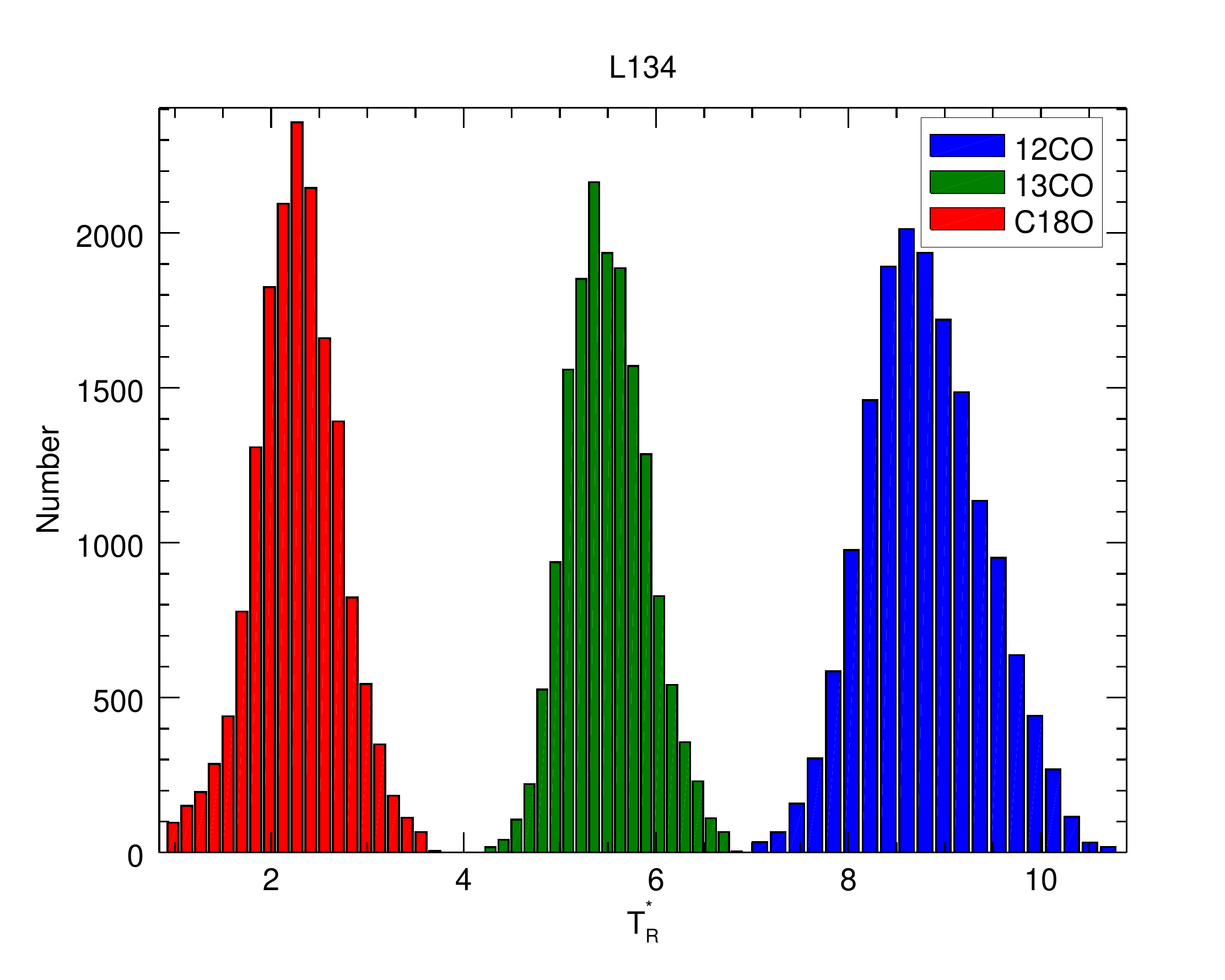}
   \end{minipage}%
   \begin{minipage}[t]{0.495\textwidth}
   \centering
    \includegraphics[width=60mm]{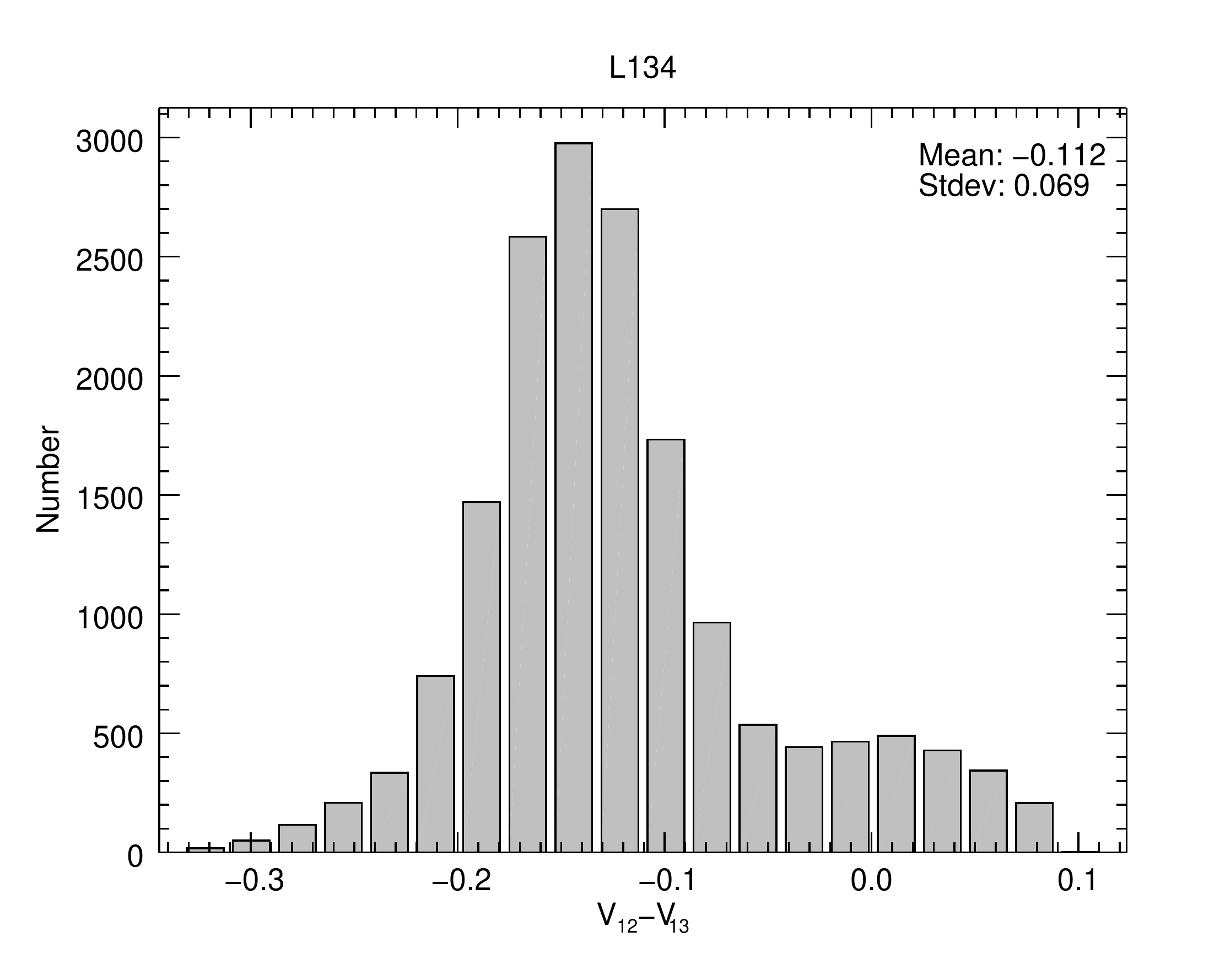}
   \end{minipage}%
   \caption{(left) The distribution of main-beam temperatures of L134 on a time scale of seven years (2011-2017). The blue, green and red histograms represent \co, \co[13]{} and \CO{} peak intensities, respectively. (right) The distribution of velocity difference between \co[12]{} and \co[13]{} in the same period. The figure shows the radial velocities of \co[12] are overall less than that of \co[13] by $\sim$ 0.15 \kmps{} (about one channel). A secondary peak at $\sim$ 0.1 \kmps{} is found, mainly caused in the the 2011 observation season.  The results of all nine beams are included. \label{fig:l134} }
  
 \end{figure}

   Because our work relies on the comparison among different lines, it is an essential demand that the observation facility should be stable on a large time scale. Fortunately, MWISP has a scheme that immediately before and after the observation to each cell,  observations toward a ``standard source''  were  made to monitor the system performance.  As an example, in Fig. \ref{fig:l134} we show the distributions of T$_{MB}$ and V$_{LSR}$ difference of L134, one of the ``standard sources''. We choose this source as an illustration because its line-widths are relative small to allow accurate estimation of the central velocity. The gross data  were collected from 2011 November to 2017 December in all three lines.  The 1$\sigma$  variations of T$_{MB}$ are 0.66 K, 0.47 K, and  0.47 K, or 7.4\%, 8.4\%, and 20\% on relative scale for the three lines, respectively. This suggests the system was reasonably stable within the period of seven years. The radial velocities with respect to local standard of rest (LSR) are also checked within the period. As shown in Fig. \ref{fig:l134},  V$_{LSR}^{12CO}$ is roughly less than  V$_{LSR}^{13CO}$ by $\sim$ 0.15 km $s^{-1}$ (blue shifted). \citet{1978ApJS...36....1M} showed a slight velocity shift between \co{} and \co[13]{} which is consistent with this result. This  difference could be intrinsic or arise in the observations. Considering that the V$_{LSR}$ differences between $^{12}$CO blue peak and $^{13}$CO center in our selected sample are mostly significantly greater than that value, this discrepancy, even if could arise due to the instrument and observation, does not affect our result too much.
   
   \subsection{Search Strategy} \label{sec:ss}
   As stated above, two lines, one being optically thick and the other optically thin, are necessary to discriminate between blue-profiles and multi-components. We use the $^{12}$CO with $^{13}$CO (hereafter Pair-1), and $^{13}$CO with C$^{18}$O (hereafter Pair-2) as two line pairs to do the work, i.e., we arbitrarily assume  $^{13}$CO as an optically thin line and $^{12}$CO as the optically thick to search for blue-profiles, and then do it again with the other line pair. This scheme is reasonable because  the  $^{13}$CO line could be optically thin in some cases while thick elsewhere. We note that in the former case, the optically thin assumption of \co[13] lines, which might be not true, does not affect the judgement of \co{} blue-profiles, but might miss some candidates if \CO{} emissions are not detected. 
   
   Similar to the scheme of the MWISP survey,  {the search regions for machine work are split into cells. Then the automatical search is done pixel by pixel within each cell. } As stated in Sec. \ref{sec:dat}, the area for each cell has a 0.5\arcmin{} overlap with the surrounding cells, redundant detections are inevitable, which are cleaned out at later steps.

    \subsection{Automatic Search}
   Up to date, MWISP has been accomplished the region l = [12\degree, 230\degree], and b = [-5.25\degree, +5.25\degree], with some coverage at l $<$ 12\degree, covering a total area of $\sim$ 2400 square degrees,  and produced $\sim 10^8$ spectra.  It is impossible to deal with such a large amount of data manually. Initial automatic searches  have to be conducted to minimize the manual tasks. 
   At this step for each pixel, we assume \co[13] to be optically thin for Pair-1 and \CO{} for Pair-2. First of all, we decomposed the optically thin lines into components with a 1-dimensional watershed algorithm. Each component is Gaussian-like with a peak exceeding the nearby baseline or dip by over 5 RMS of the whole spectrum. Then the first and second moments were extracted for each component as center velocity and velocity dispersion. They were aligned with the optically thick counterpart to cut out segments of profiles to be selected. Two methods were then adopted to find profiles with double peaks.

   {Method one: we select optically thick profiles fulfilling the following criteria.}
   
   {1. A peak should appear on the blue side of central velocity. Such peak must exceed the nearby dip or baseline by 3 RMS, and away from the center velocity by more than velocity resolution.}

   {2. The peak on the blue side should be higher than the profile on the red side by at least 3 RMS. If a peak appears on the red side, the component is labelled with ``double-peak", or a "shoulder" label will be assigned.}
   
   {Method two: Gaussian fittings with two components were conducted to the optically thick profiles, and then results fulfilling the following criteria are selected.}

   {1. Two components should reside on both sides of the center velocity, and away from it by more than velocity resolution.}

   {2. The blue side Gaussian component must have a higher integrated intensity and peak.}

   \subsection{Manual Check}{\label{sec:man}}
   
   First look at the spectra of the candidates resulted from the automatic search suggests that many of them are not really blue-profiles, or hardly to say that. Therefore, further manual check is  necessary to refine the candidate list. Picking up the candidates of  likely blue-profiles is a hard work. We wrote some codes with graphic user interface to assist task. Then the pick-up work is only a couple of mouse clicks.  
   
   By checking the spectra of the candidates and comparing with the numeric simulations (c.f., Sec. \ref{sec:num}), the  candidates are rejected according to the following general rules:

   \begin{itemize}
   \item[-]Spectra of either optically thick or thin lines are noisy: In the real data, the situation is much more complicated than the simulation due to the introduction of noises.  Further more, some candidates appear at the edge of a certain cell where the noise level is higher than those of central areas. Though this situation has been considered in the automatic search, there are still cases that the machine  would misjudge whether a peak is real or noise-induced, especially for optically thin lines.  
   \item[-]Two peaks of the optically thick line are far away (e.g. V$_{red}$-V$_{blue}$$>$ 10 \kmps): This is just empirical since we seldom find blue-profiles with very large peak separation in the literature. The line width (Full Width at Half Maximum, FWHM) of the optically thin line is too large, i.e., $>$10 \kmps.

   \item[-]Multi-peaks or flattened top in the optically thin line: It is usually very difficult to judge whether a blue-profile of the optically thick line comes from a single or multiple components if there exist multi-peaks or flattened top in the assumed optically thin line. Though there are cases that the peaks of the optically thin lines do not coincident with those of the optically thick lines, we suppose they still could arise from multi-components with very complicated physical conditions. However, the candidates with one situation are retained: for Pair-1 selected candidates, the \co[13]{} line also shows a blue-profile while the \CO{} emission is not detected. The two peaks of the $^{13}$CO line are between those of $^{12}$CO line. In this case it is likely that $^{13}$CO line is also optically thick, but the absorption is fainter than that of the $^{12}$CO line. In such a case self-absorption still exists to show a blue-profile, but the peak separation is smaller than that of the $^{12}$CO line.

   \end{itemize}
   
   \subsection{Cleaning Candidates}
   The manual check is a tiresome work since the output of the automatic search returns hundreds of thousands of candidates. In the manual check step we cautiously relax the above criteria in order to avoid loss of some reliable candidates. After the manual checks, we make the Gaussian fittings to the line profiles. Since the fittings need human interactions in deciding parameters such as fitting ranges, line profiles of the candidates are checked again and some candidates are rejected at this step.  
   
   After the manual check, candidates with similar positions and velocities are grouped together. The similarity of two candidates is defined as {the following two conditions are satisfied}: 
   $$\Delta p = \sqrt{\Delta l^2cos{^2}b+\Delta b^2} \leq \Delta p_{0}; \Delta V_{c} \leq \Delta V_{0}$$ 
   where $\Delta l$, $\Delta b$ and $\Delta V_{c}$ are the differences of the Galactic coordinates and central velocities of two candidates; $\Delta p_0 = 1.0^\prime$ and $\Delta V_0 =0.2\, $\kmps{} in this work being based on the spatial and spectral resolution of the MWISP products.  For each group the sources are checked by eyes and those showing  blue-profile most significantly are  selected. 

   {Lastly, in a few cases blue-profiles are detected in both Pair-1 and Pair-2, we merge them to the Pair-2 group since the physical parameters (c.f., Fig. \ref{fig:specse}), such as column densities and system velocities, could be more correctly estimated due to the fact that the optically thin lines are single-peaked. }
     
   \section{The catalogue}\label{sec:cat}
   
   Finally, a total of 3533 sources are registered as possessing  blue-profiled lines, out of which 3329 sources are selected from Pair-1, and 204 sources are selected from Pair-2.  In Table \ref{tab:tab1} we present a subset of the candidates for illustration, while the whole catalogue is presented in the appendix\footnote{The full table can also be found at https://www.scidb.cn/en/detail?dataSetId=bc1c08c6e7ba426a9a07b8bad69b2ffb}. For each candidate we assign a name code, which is presented in Column 2. Name codes are based on the positions and velocities of the candidates, i.e., the (l, b) codes are reserved to the second decimal places and v to the first decimal places, and decimal points are removed from the final code.  As an example, if (l, b, v) = (123.450000, 1.230000, 123.4500), the code is assigned as ``12345+0123+1235''. Column 3 \& 4 give the Galactic coordinates of sources.  Column 5 gives the spatial association to the objects  from some well known catalogs, such as  IRAS PSC, NGC, the RAFGL catalog, the Sharpless catalog, and Lynds' dark nebular catalog, etc. We also checked some other catalogs such as H$_\textsc{II}$ candidates \citep{2014ApJS..212....1A,2018MNRAS.476.3981Y}, and the MSX dark cloud catalog \citep{2006ApJ...639..227S}.  In Column 6 we indicate whether our candidates show blue-profiles with other tracers in the literature. The associations in Column 5 \& 6 are shown in codes to save the table space, and noted as table footnotes. In Column 7 we give the distances of the sources. The distance of a source is obtained using the online distance calculator\footnote{http://bessel.vlbi-astrometry.org/node/378} which is based on the trigonometric parallax measurements of Galactic maser sources \citep{2016ApJ...823...77R,2019ApJ...885..131R} and assumes all molecular clouds are located on spiral arms. Using a Bayesian approach, this tool is believed to be a major improvement of the kinematic distance estimate of Galactic molecular clouds. For a few sources that are near the tangent direction (l $\sim$ 90\degree), the calculator does not give proper distances.

   The second half of the table presents the derived parameters of the candidates.  Columns 8 \& 9 (V$_c$, V$_b$) are the characteristic velocities of sources.  V$_c$ represents the ``central'' velocities, which are derived from Gaussian fittings to the optically thin lines. V$_b$ represents positions where blue peaks are found, which are largely obtained from double-component Gaussian fittings. However, in some cases when Gaussian fittings do not converge, we use the peak positions by eyes. Column 10 presents the skewness parameter, $\delta$V = (V$_{thick}$-V$_{thin}$)/$\Delta$V$_{thin}$, as suggested by \citet{1997ApJ...489..719M}. 

   \small
\setlength{\tabcolsep}{3pt}
\begin{center}
    \begin{landscape}
        \begin{ThreePartTable}
            \begin{TableNotes}
                \item[] Column Heads: (1): internal serial number; (2): name code; (3)\&(4): Galactic Coordinates; (5): association to some known catalogues (c.f. next foot note); (6): infall signature reported in the literature (c.f. next foot note); (7): distance; (8): central velocity; (9): velocity at the blue peak; (10): skewness parameter (see text); (11): expected Tmb of optically thick line; (12): peak Tmb of optically thin line; (13): excitation temperature; (14): H$_2$ column density; (15): pair from which the blue-profile is obtained.
                \item[$\boldsymbol{\alpha}$] Ref. Codes: (1) IRAS Point Source Catalog; (2) New General Catalog (NGC); (3) \citeauthor{1999yCat.2094....0P} \citeyear{1999yCat.2094....0P}; (4) \citeauthor{1953ApJ...118..362S} \citeyear{1953ApJ...118..362S}; (5) \citeauthor{1996yCat.7007....0L} \citeyear{1996yCat.7007....0L}; (6) \citeauthor{2014ApJS..212....1A} \citeyear{2014ApJS..212....1A}; (7) \citeauthor{2006ApJ...639..227S} \citeyear{2006ApJ...639..227S}; (8) \citeauthor{2018MNRAS.476.3981Y} \citeyear{2018MNRAS.476.3981Y}; (9) \citeauthor{2003ApJS..149..375S} \citeyear{2003ApJS..149..375S}; (10) \citeauthor{2005A&A...442..949F} \citeyear{2005A&A...442..949F}; (11) \citeauthor{2007ApJ...663.1092K} \citeyear{2007ApJ...663.1092K}; (12) \citeauthor{2009MNRAS.392..170S} \citeyear{2009MNRAS.392..170S}; (13) \citeauthor{2011ApJ...740...40R} \citeyear{2011ApJ...740...40R}; (14) \citeauthor{2012A&A...538A.140K} \citeyear{2012A&A...538A.140K}; (15) \citeauthor{2013ApJ...776...29L} \citeyear{2013ApJ...776...29L}; (16) \citeauthor{2016MNRAS.456.2681Q} \citeyear{2016MNRAS.456.2681Q}; (17) \citeauthor{2016MNRAS.461.2288H} \citeyear{2016MNRAS.461.2288H}; (18) \citeauthor{2020RAA....20..115Y} \citeyear{2020RAA....20..115Y}; (19) \citeauthor{2021ApJ...910..112K} \citeyear{2021ApJ...910..112K}
                \item[$\boldsymbol{\beta}$] For a few sources the distances are not given because the distance calculator does not output proper values.
                \item[$\boldsymbol{\gamma}$] For a number of entries the values of N(H$_2$) are not given because the optical depths ($\tau_{thin}$) cannot be estimated due to the complex line profiles.
            \end{TableNotes}
            \begin{longtable}{rrrrllrrrrrrrrr}
                \caption{The catalog}\label{tab:tab1}                                                                                                                                                                                                                                                                                                                                                                                                                                                     \\
                \toprule
                \multicolumn{1}{c}{No}  & \multicolumn{1}{c}{Code} & \multicolumn{1}{c}{L}   & \multicolumn{1}{c}{B}   & \multicolumn{1}{c}{Association$^\alpha$} & \multicolumn{1}{c}{I.S.$^\alpha$} & \multicolumn{1}{c}{D$^\beta$} & \multicolumn{1}{c}{V$_c$}         & \multicolumn{1}{c}{V$_b$}         & \multicolumn{1}{c}{$\delta$V} & \multicolumn{1}{c}{T$_1$} & \multicolumn{1}{c}{T$_2$} & \multicolumn{1}{c}{T$_{ex}$} & \multicolumn{1}{c}{N$_{H2}${$^\gamma$}} & \multicolumn{1}{c}{Pair} \\
                \multicolumn{1}{c}{}    & \multicolumn{1}{c}{}     & \multicolumn{1}{c}{}    & \multicolumn{1}{c}{}    & \multicolumn{1}{c}{}                     & \multicolumn{1}{c}{}              & \multicolumn{1}{c}{(kpc)}     & \multicolumn{1}{c}{(km s$^{-1}$)} & \multicolumn{1}{c}{(km s$^{-1}$)} & \multicolumn{1}{c}{}          & \multicolumn{1}{c}{(K)}   & \multicolumn{1}{c}{(K)}   & \multicolumn{1}{c}{(K)}      & \multicolumn{1}{c}{(cm$^{-2}$)}         & \multicolumn{1}{c}{}     \\
                \multicolumn{1}{c}{(1)} & \multicolumn{1}{c}{(2)}  & \multicolumn{1}{c}{(3)} & \multicolumn{1}{c}{(4)} & \multicolumn{1}{c}{(5)}                  & \multicolumn{1}{c}{(6)}           & \multicolumn{1}{c}{(7)}       & \multicolumn{1}{c}{(8)}           & \multicolumn{1}{c}{(9)}           & \multicolumn{1}{c}{(10)}      & \multicolumn{1}{c}{(11)}  & \multicolumn{1}{c}{(12)}  & \multicolumn{1}{c}{(13)}     & \multicolumn{1}{c}{(14)}                & \multicolumn{1}{c}{(15)} \\

                \midrule
                \endfirsthead

                \multicolumn{15}{c}{Table \ref{tab:tab1}: Continued}                                                                                                                                                                                                                                                                                                                                                                                                                                      \\
                \toprule
                \multicolumn{1}{c}{(1)} & \multicolumn{1}{c}{(2)}  & \multicolumn{1}{c}{(3)} & \multicolumn{1}{c}{(4)} & \multicolumn{1}{c}{(5)}                  & \multicolumn{1}{c}{(6)}           & \multicolumn{1}{c}{(7)}       & \multicolumn{1}{c}{(8)}           & \multicolumn{1}{c}{(9)}           & \multicolumn{1}{c}{(10)}      & \multicolumn{1}{c}{(11)}  & \multicolumn{1}{c}{(12)}  & \multicolumn{1}{c}{(13)}     & \multicolumn{1}{c}{(14)}                & \multicolumn{1}{c}{(15)} \\
                \midrule
                \endhead

                \bottomrule
                \multicolumn{15}{c}{Continued on next page}                                                                                                                                                                                                                                                                                                                                                                                                                                               \\
                \endfoot

                \bottomrule
                \insertTableNotes
                \endlastfoot

                1001                    & 03105$+$0017$+$0118      & 31.0500                 & 0.1667                  & 1,  3,  6                                & \nodata                           & 1.90                          & 11.83                             & 11.20                             & $-$0.65                       & 4.1                       & 1.2                       & 11.2                         & 5.04E$+$20                              & 1                        \\
                1002                    & 03105$+$0076$+$0506      & 31.0500                 & 0.7583                  & 6,  7                                    & \nodata                           & 0.68                          & 50.62                             & 50.07                             & $-$0.44                       & 4.2                       & 1.3                       & 8.8                          & 5.82E$+$21                              & 2                        \\
                1003                    & 03105$+$0407$+$0063      & 31.0500                 & 4.0750                  & 5,  6                                    & \nodata                           & 0.24                          & 6.29                              & 5.68                              & $-$0.59                       & 5.8                       & 1.2                       & 8.4                          & 5.27E$+$20                              & 1                        \\
                1004                    & 03106$-$0067$+$0088      & 31.0583                 & $-$0.6667               & 6,  7                                    & \nodata                           & 0.24                          & 8.77                              & 8.41                              & $-$0.42                       & 4.1                       & 1.8                       & 8.3                          & 7.75E$+$20                              & 1                        \\
                1005                    & 03107$+$0078$+$0512      & 31.0667                 & 0.7833                  & 6,  7                                    & \nodata                           & 0.60                          & 51.22                             & 50.26                             & $-$0.48                       & 6.5                       & 2.9                       & 11.2                         & 3.02E$+$21                              & 1                        \\
                1006                    & 03107$+$0313$+$0103      & 31.0750                 & 3.1333                  & 5,  6                                    & \nodata                           & 0.24                          & 10.31                             & 9.32                              & $-$0.46                       & 6.8                       & 3.4                       & 10.0                         & 4.08E$+$21                              & 1                        \\
                1007                    & 03107$+$0524$+$0084      & 31.0750                 & 5.2417                  & 5,  6                                    & \nodata                           & 0.24                          & 8.41                              & 7.40                              & $-$0.51                       & 2.7                       & 3.4                       & 6.9                          & 6.93E$+$21                              & 1                        \\
                1008                    & 03107$-$0044$+$0085      & 31.0750                 & $-$0.4417               & 6                                        & \nodata                           & 0.24                          & 8.50                              & 8.14                              & $-$0.32                       & 4.3                       & 1.3                       & 10.9                         & 6.56E$+$20                              & 1                        \\
                1009                    & 03108$-$0042$+$0086      & 31.0833                 & $-$0.4167               & 6                                        & \nodata                           & 0.24                          & 8.61                              & 7.96                              & $-$0.68                       & 4.3                       & 1.3                       & 9.1                          & 5.32E$+$20                              & 1                        \\
                1010                    & 03109$-$0339$+$0100      & 31.0917                 & $-$3.3917               & 5,  6                                    & \nodata                           & 0.24                          & 10.00                             & 9.83                              & $-$0.23                       & 5.2                       & 2.1                       & 10.3                         & 7.41E$+$20                              & 1                        \\
            \end{longtable}
        \end{ThreePartTable}
    \end{landscape}
\end{center}

   Columns 11-13 (T$_1$, T$_2$, T$_{ex}$) are related temperatures of the sources. T$_2$ represents the peak main beam temperature of the optically thin line. T$_1$ represents the expected main beam temperature of the optically thick line. Since the line profile shows self-absorbed feature, the ``expected'' peak is derived by Gaussian fitting to the whole profile while excluding the portion that shows self-absorption, i.e., we use the waist and foot portions in the fittings. Though the fitting portions have been carefully adjusted,  {we caution that in some cases the derived values of T$_1$ are  sensitive to the fitting portions adopted, so that the errors could be very large -- they may reach to the values of T$_1$ themselves in some extreme cases depending on the environments of individual sources, e.g. when multi-components coexist with the blue-profiles. However, in most cases, we expect the uncertainties being less than 20\%.} T$_{ex}$ is the excitation temperature calculated from T$_1$\citep{2015PASP..127..266M}: 
    $$T_{ex} = T_0[ln(1+\frac{1}{\frac{T_1}{T_0}+\frac{1}{exp({T_0/T_{bg})}-1}})]^{-1} $$  where $T_0 = h\nu_0/k$ =  5.53 K  for Pair-1, or   5.29 K for Pair-2, T$_{bg}$ (= 2.73 K) is the background temperature of the Universe. {Here we assume the filling factor being close to unity. We use \co[13]{} to estimate the excitation temperatures of Pair-2 sources because in many cases \co{} profiles are very complicated so that we cannot obtain reliable ``expected'' main beam temperatures. In such cases T$_{ex}$ might be underestimated since the \co[13]{} lines are not optically thick enough}.
   
    The column densities of H$_2$ (Column 14) are estimated from the integrated intensities of the optically thin lines assuming local thermal equilibrium {(LTE)}. Under such assumption, the column densities of \co[13]{} or \CO{} are calculated by the following formula \citep{2016PASP..128b9201M}:
    $$ N(\co[13]|\CO) = f_1 \times f_0\times\frac{(T_{ex}+0.88)exp(\frac{T_0}{T_{ex}})}{exp(\frac{T_0}{T_{ex}})-1}\times\frac{\int T_{MB}\,dv \quad(K\,km\,s^{-1})}{f(J_\nu(T_{ex})-J_\nu(T_{bg}))} \qquad(cm^{-2})$$
    where f$_0$ is approximately 2.48 $\times 10^{14}$ for both \co[13]{} and \CO{}, T$_0$ equals 5.29 for \co[13]{} and 5.27 for \CO, f is the beam filling factor which we assume to be 1.0 in this work, and $J_\nu(T) = T_0/(exp(T_0/T)-1)$. The optical depth correction factor $f_1 = \tau/(1-exp(-\tau))$. Where $\tau$ is calculated by:
    $$\tau=-ln[1-\frac{T_2}{f(J_\nu(T_{ex})-J_\nu(T_{bg}))}]$$
    
    The total column densities of molecular hydrogen, $N(H_2)$, are obtained assuming the CO abundance ratios, $ N(H_2) \approx 3.9 \times 10^5 N(^{13}CO)$ or $N(H_2) \approx 3.0 \times 10^6 N(C^{18}O)$\citep{2019PASA...36...49A}, which are derived from the infrared dark cloud and close to those from star-forming regions\citep{2008ApJ...679..481P}. {Several factors affect estimations of N($_{H2}$): 1) Assumption of LTE which may not be true if the sources are actually in the process of infall; 2) T$_{ex}$s estimations; 3) Abundances of \co[13] and \CO{} relative to those of molecular hydrogen change from region to region \citep{2015ARA&A..53..583H};  4) Beam filling factors for the sources are not always close to unity. Therefore, it is difficult to estimate the overall uncertainties of N($_{H2}$).} In the last column (Column 15) we indicate the line pairs adopted in searching for the blue-profiles.

    \begin{figure*}
        \subcaptionbox{05363+0004+0235\label{fig:specsa}}[0.495\textwidth]{\includegraphics[width=8.0cm,angle=0]{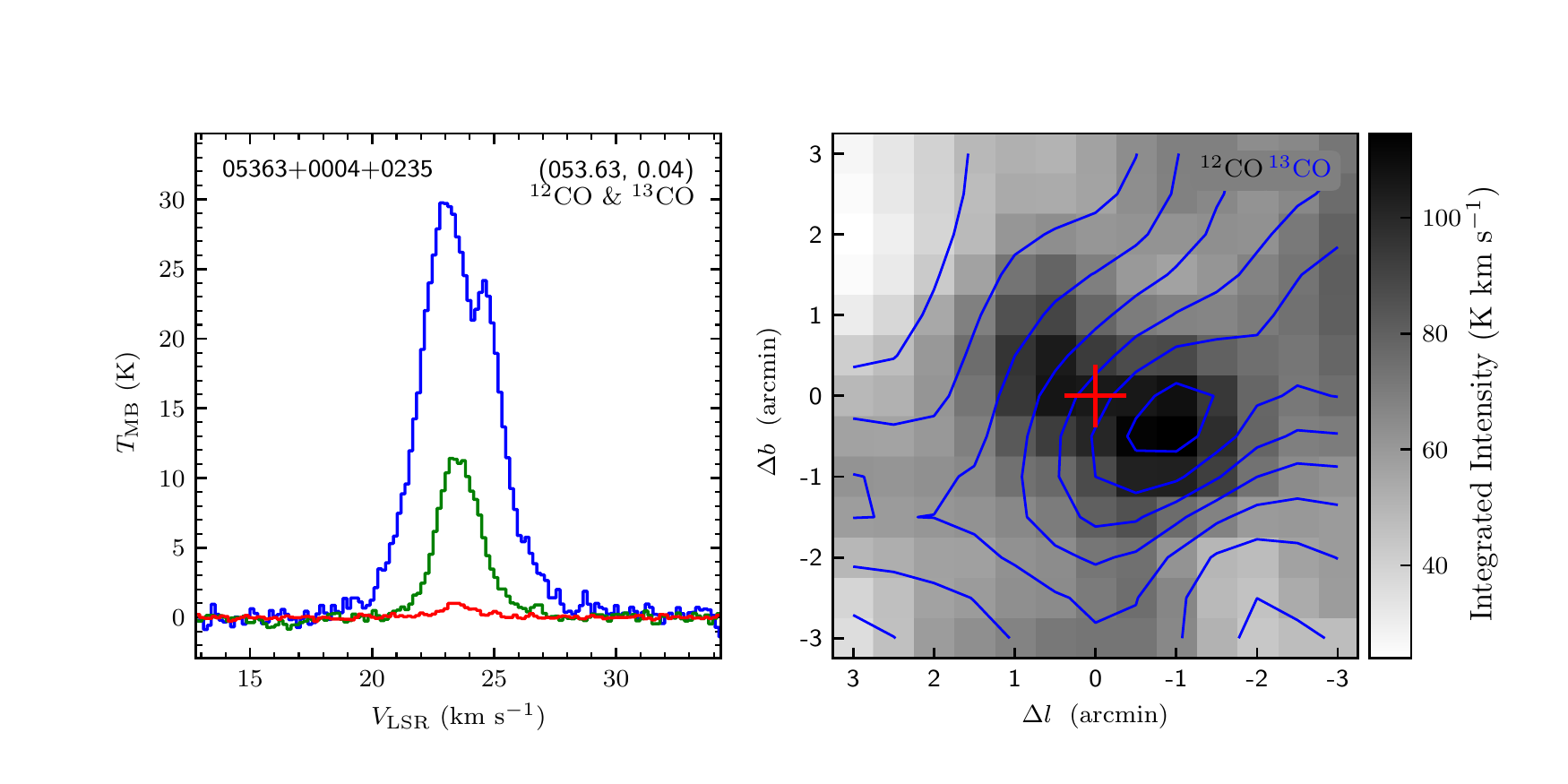}} 
     \subcaptionbox{02869+0377+0073\label{fig:specsb}}[0.495\textwidth]{\includegraphics[width=8.0cm,angle=0]{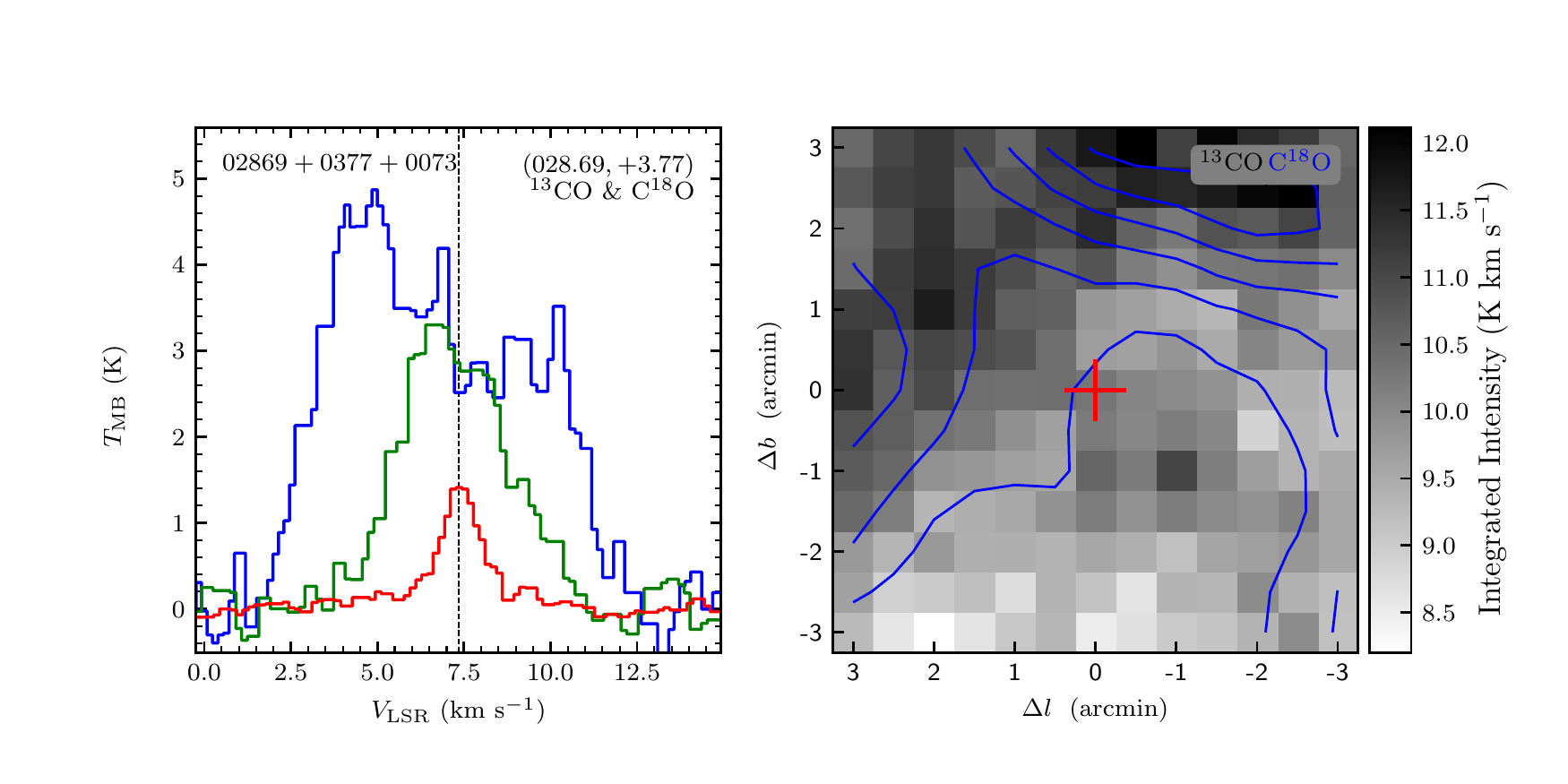}} 
     \subcaptionbox{01272+0069+0173\label{fig:specsc}}[0.495\textwidth]{\includegraphics[width=8.0cm,angle=0]{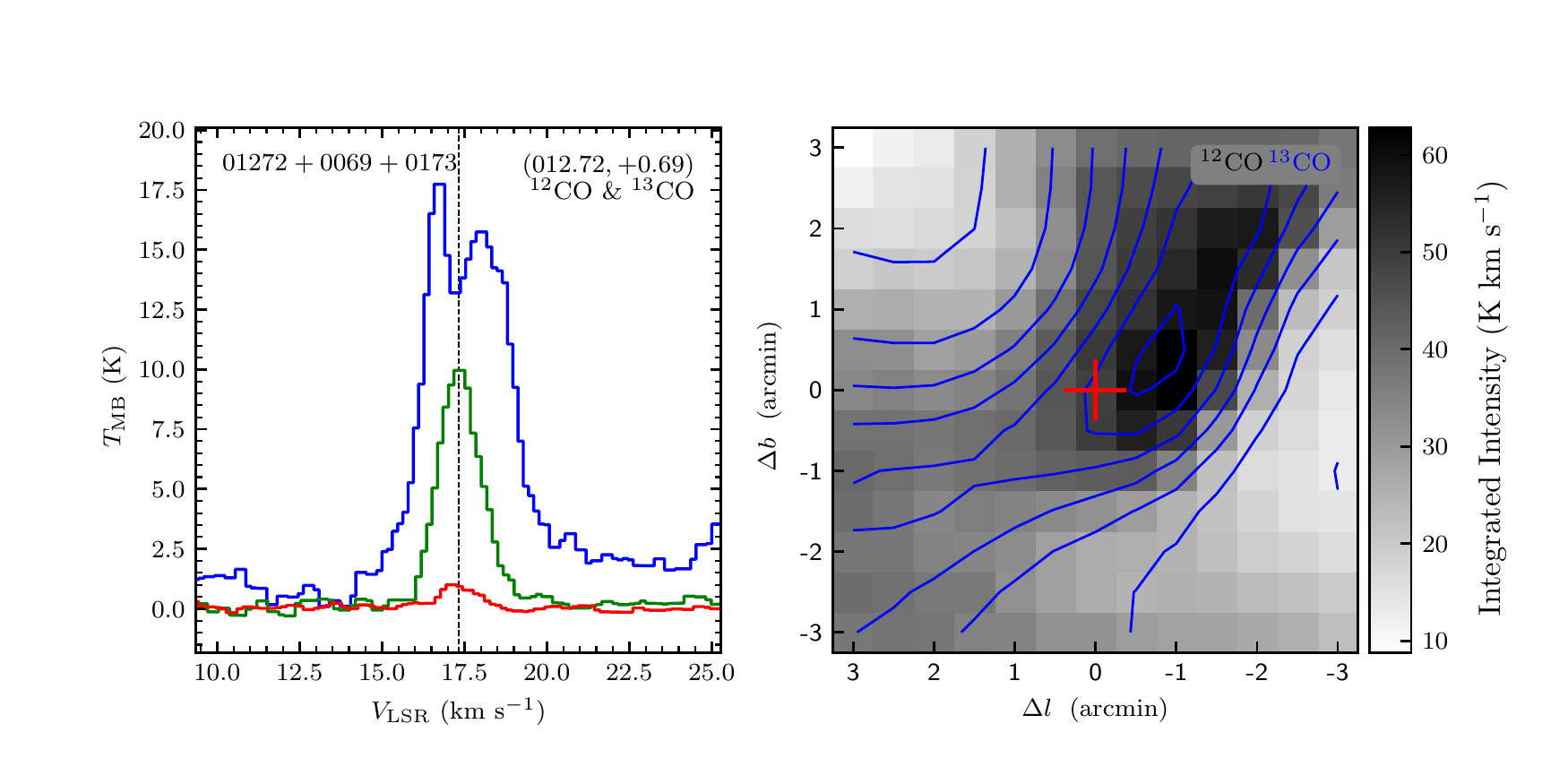}} 
     \subcaptionbox{01282-0019+0349\label{fig:specsd}}[0.495\textwidth]{\includegraphics[width=8.0cm,angle=0]{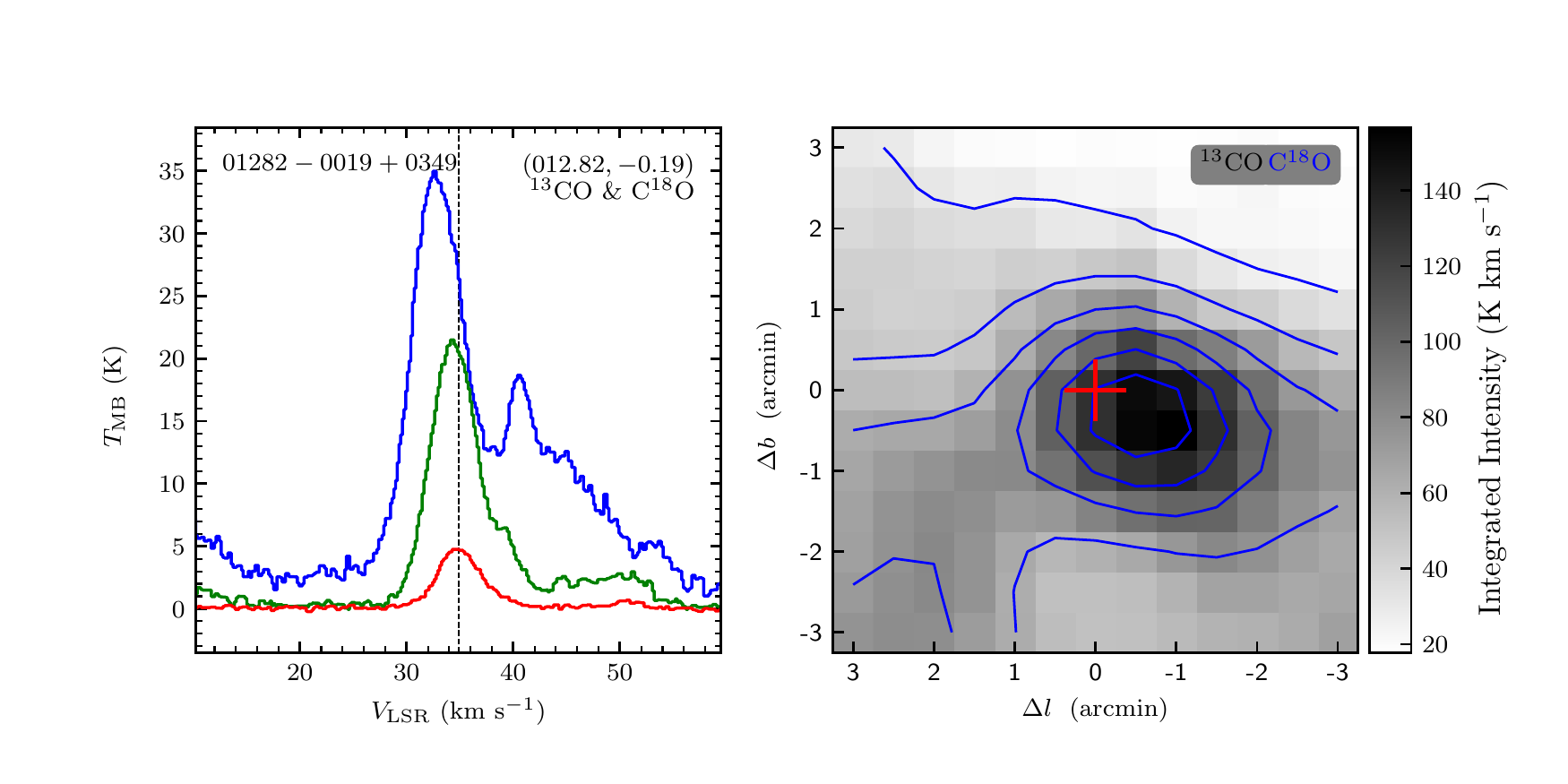}} 
     \subcaptionbox{01432+0268+0156\label{fig:specse}}[0.495\textwidth]{\includegraphics[width=8.0cm,angle=0]{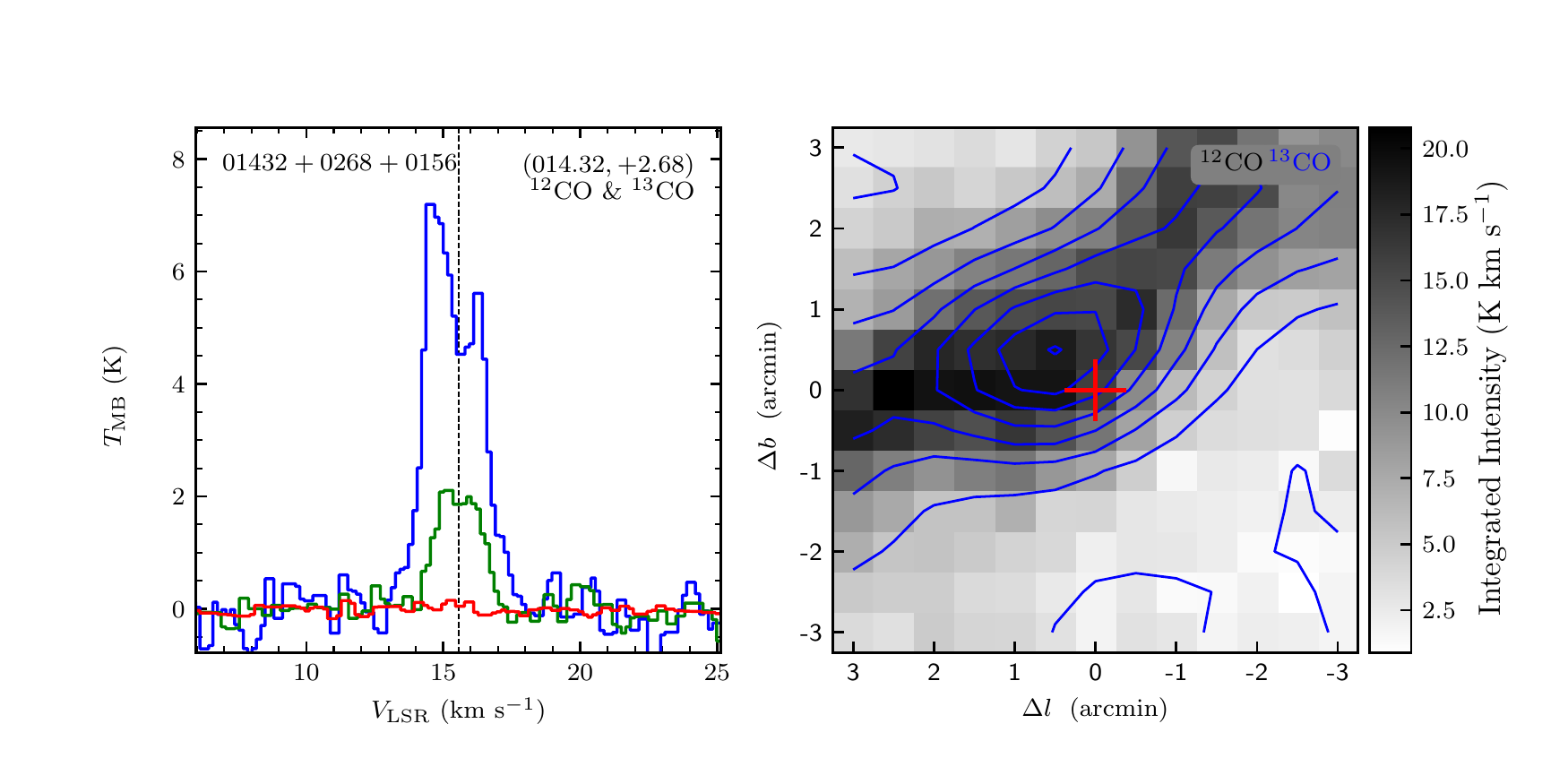}} 
     \subcaptionbox{01288-0024+0355\label{fig:specsf}}[0.495\textwidth]{\includegraphics[width=8.0cm,angle=0]{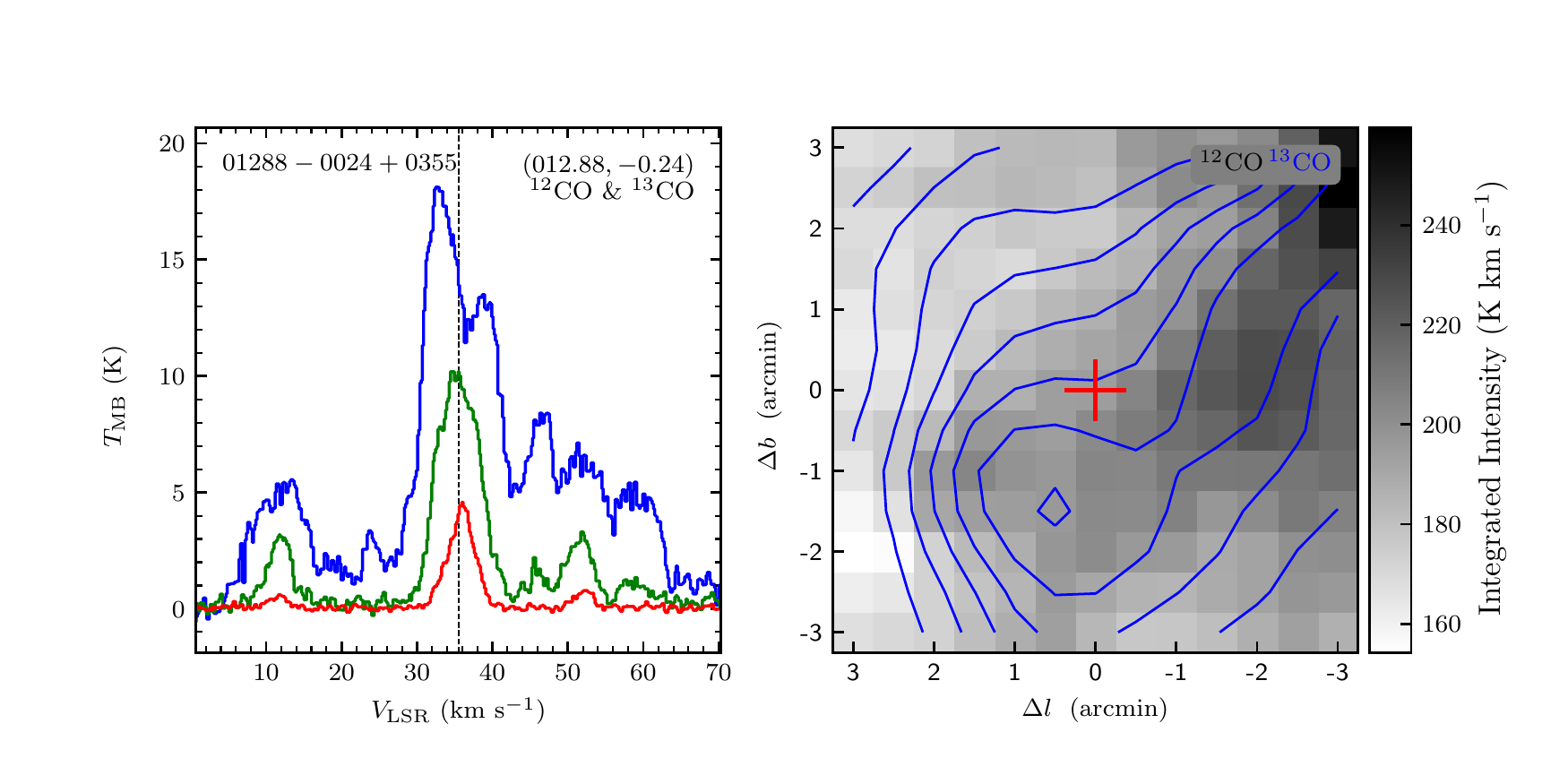}} 
     \subcaptionbox{01116-0014+1503\label{fig:specsg}}[0.495\textwidth]{\includegraphics[width=8.0cm,angle=0]{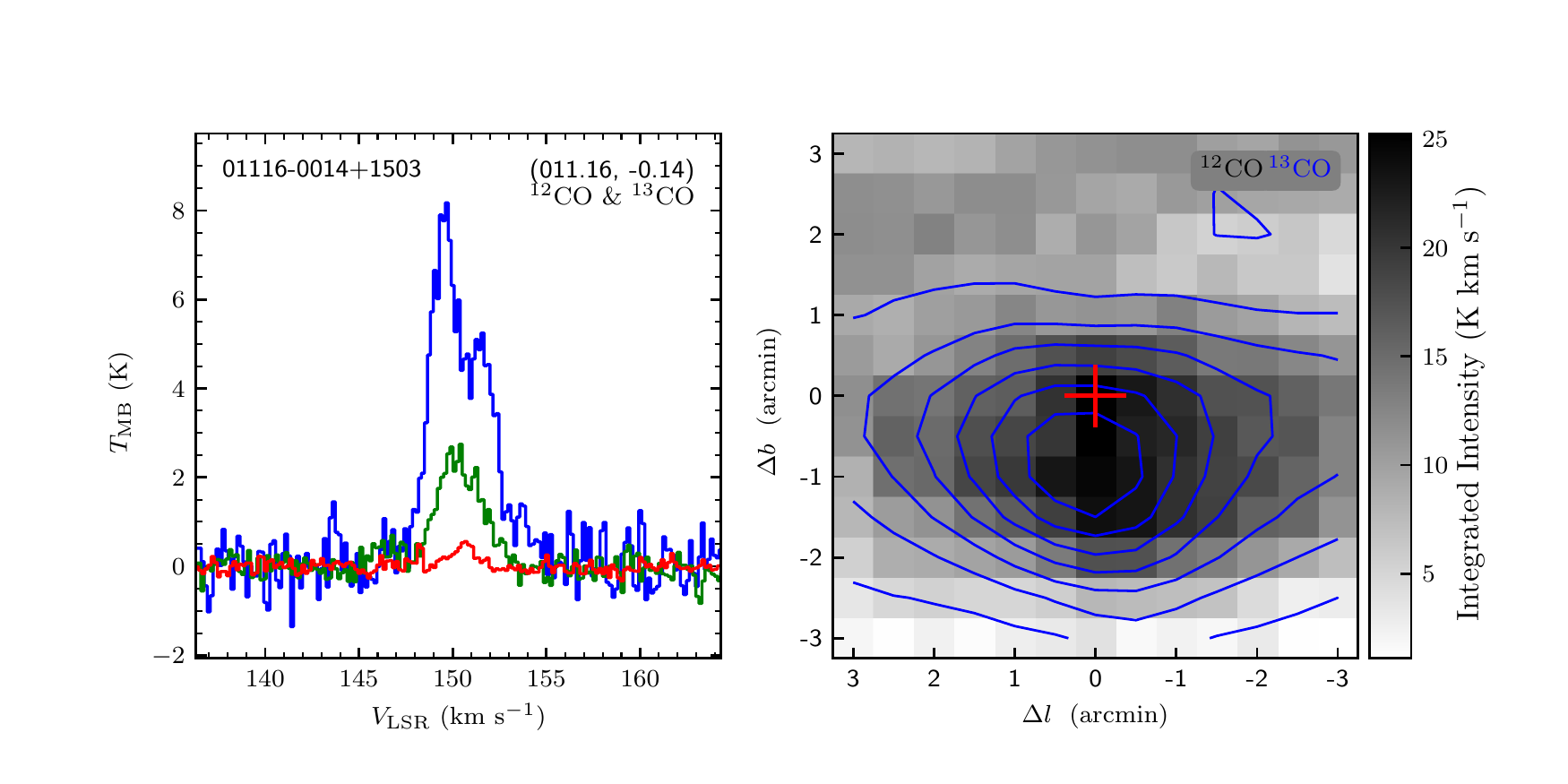}} 
     \subcaptionbox{02849+0399+0064\label{fig:specsh}}[0.495\textwidth]{\includegraphics[width=8.0cm,angle=0]{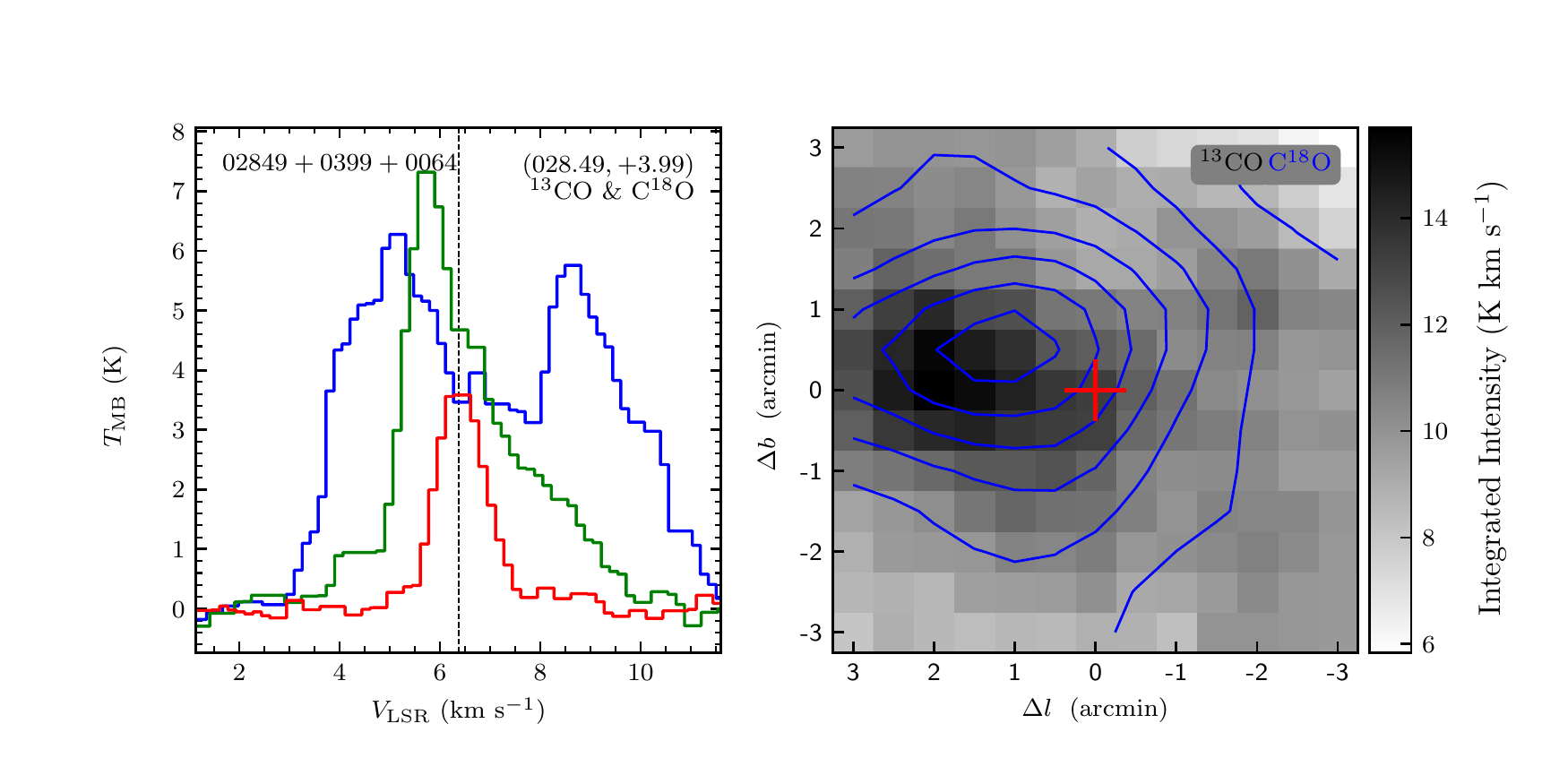}} 
     
        \caption{The spectra (left) and gas distributions (right) of the candidates. On the left panels, the red, green and blue lines represent {\CO, \co[13] and \co} emissions, respectively. To enhance s/n ratios of the lines, the spectra are smoothed with median filter for each 3 velocity channels. For the same reason, the \CO{} spectra are also smoothed with 3$\times$3 spatial pixels,   The  dashed line indicates the system velocity. On the upper-right corners we indicate the Galactic coordinates (l, b), and the line-pairs used in the searching procedure. On the upper-left are the name codes of the candidates. The right panels show the integrated intensity map of the optically thin lines (contours) overlaid on those of the optically thick lines (grey scale). The red pluses indicate the positions where the blue-profiles are detected. \label{fig:specs}}
        \end{figure*}

    Fig. \ref{fig:specs} shows some examples of the spectra (left) and the relative location (right) where blue-profiles are detected. Again, for the limitation of the main body, the whole set of the figures is given in the appendix.  Fig. \ref{fig:specsa} (05363+0004+0235) shows an ideal blue-profile in \co{} emission with virtually symmetric profiles both in \co[13]{} and \CO{} lines, the blue-profile happens at the saddle point between two local maxima (right panel); Fig. \ref{fig:specsb} (02869+0377+0073) shows a candidate selected from Pair-2. The \co{} line of this source shows complicated peaks;      
    Fig. \ref{fig:specsc} (01272+0069+0173) shows a blue-profile selected from Pair-1, but the red side of the \co{} line is probably contaminated by another component. The \CO{} line, weak as it is,  also suggests such a possibility. However, we could recognize this candidate since the contamination is not too serious to affect our judgement;  Fig. \ref{fig:specsd} (01282-0019+0349) shows a candidate that does not show clear blue-profile, but the \CO{} and \co[13]{} peaks are shifted redward with respect to the \co{}  peak. Such a configuration happens when infall velocity is relative large while the optical depth is moderate (c.f., the lower-right panel of Fig. \ref{fig:nsim}). Fig. \ref{fig:specse} (01432+0268+0156) shows an example that both \co{} and \co[13]{} lines exhibit blue-profiles, with the peak separation of \co{} greater than that of the \co[13]{} line, while \CO{} is marginally detectable. The blue-profile happens at virtually the central part of a dense clump.  The examples presented here are typical cases in our candidates.
    
    In Fig. \ref{fig:specs} we also present some extreme cases, which might be of interests to readers. Fig. \ref{fig:specsf} (01288-0024+0355) shows a source with  wide profiles, even for the optically thin lines (\co[13]{} and C$^{18}$O), up to FWHM $\sim$ 6.6 and 4.0 \kmps, respectively. Fig. \ref{fig:specsg} (01116-0014+1503) shows a target that has a very large system velocity, $\sim$ 150 \kmps. The kinematic distance being estimated $\sim$ 8.0 kpc, it is one of the most distant object in the catalog. Interestingly, this source presents blue-profiles both in \co{} and \co[13]{} lines. Fig. \ref{fig:specsh} (02849+0399+0064) shows a source that is rarely seen in the Galactic molecular clouds, i.e., the \co[13]{} line intensity is stronger than that of the \co{} line. Aside of strong self-absorption of \co[12], there could be also a possibility of unusual \co[13]{} abundance and excitation environment.

    In the right panel of each subset, we show the integrated intensity map of individual candidate. An interesting fact that should be noted here is that the infall-signatures do not always coincide with the clump centers (the points with the highest column densities). It would be interesting to ascertain whether this is a real fact or just an observational effect, which might be helpful in our understandings of the initial phase of star formation.

   \section{Discussion}\label{sec:disc}
   \subsection{Numerical Experiments}\label{sec:num}
   
   In order to see what a blue-profiled line appears in different environments, {and to assist us to identify infall candidates from a large number of outputs from the machine work, } we have done a series of numerical experiments. The experiments tried to simulate cores with inflowing gas, but with very simple models, e.g., fixed infall velocities, optical depths and line widths. Temperature and density profile variations are not considered in the experiments. To do this, we generate an emission line of Gaussian profile centered at 0.0 \kmps, and $\sigma$ = 2.0 \kmps, which has no physical meaning other than a scale in the velocity dimension, and upon which many other parameters are assumed.  The peak value is normalized to unity. A number of absorption lines, which are also of Gaussian profiles but with negative values, are put upon the emission line to mimic the absorption. The centers of the absorption lines are shifted a little to the red by fractions of $\sigma$, i.e., V$_{in}$ = f$_{dV}\sigma$. Here we use V$_{in}$ to designate the central velocity shift of the absorption feature because it is somehow related to the infall velocity in the reality. The negative maxima are assumed to be a few percent of unit (dip = f$_{dip}$) because the optical depths of infalling gas are expected to be optically thinner than the background emission. In the reality, the widths of the absorption lines are expected to be narrower than those of the emission lines, and so that are  set to some fractions of $\sigma$ ($\sigma_{abs}$ = f$_\sigma\sigma$). 
   \begin{figure}
   \includegraphics[width=16cm]{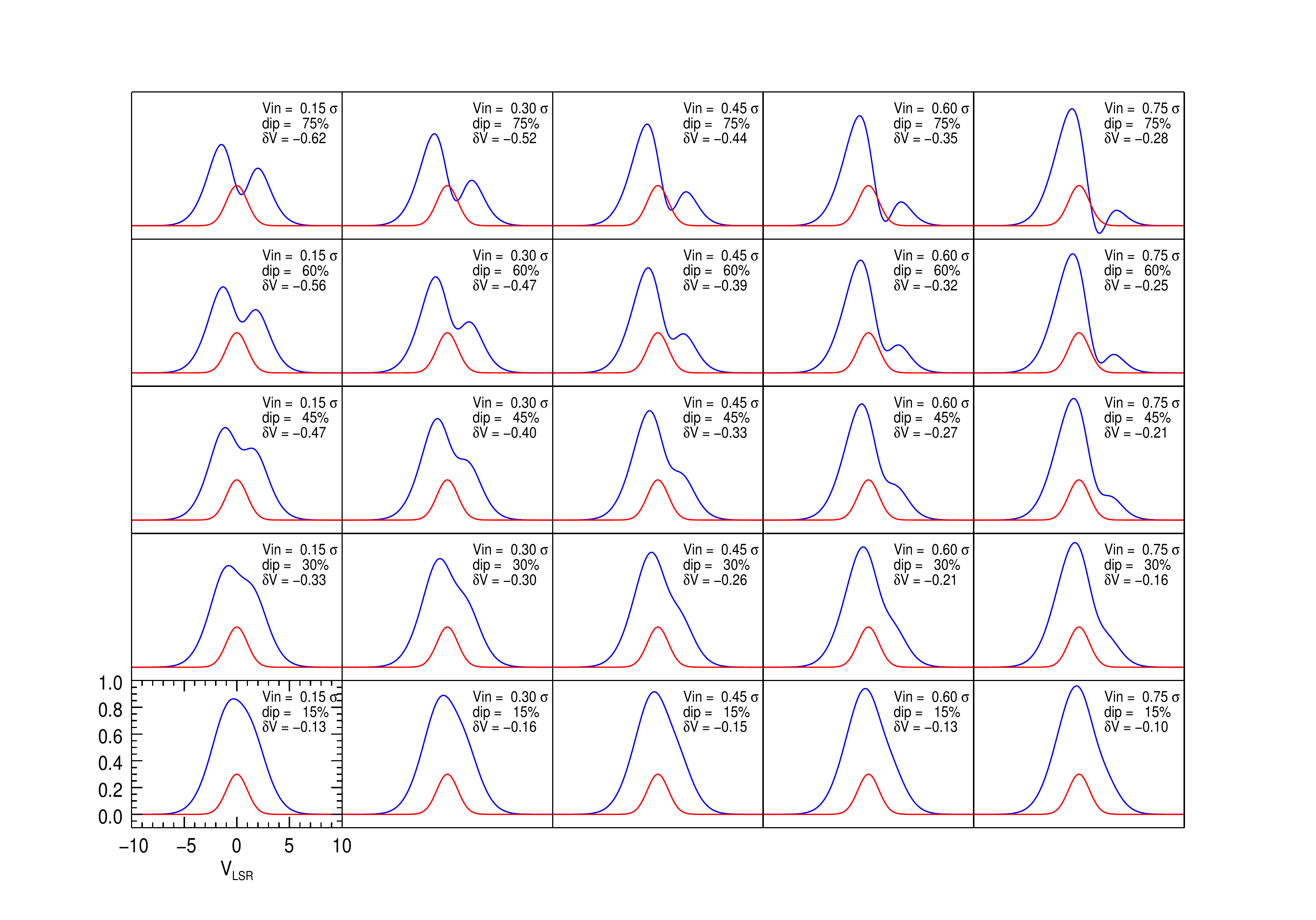}
   \caption{One example of the numerical experiments with f$_\sigma$ = 0.5. The parameters are shown at the upper-right corner of each frame. The blue lines represent the self-absorbed optically thick lines showing  blue-profile lines, and the red lines indicate the un-absorbed optically thin lines for comparison. We also present $\delta V$ suggested by \citet{1997ApJ...489..719M} for different cases. \label{fig:nsim}}
   \end{figure}
   
   In Fig. \ref{fig:nsim} we show some results of the experiments with f$_\sigma$ = 0.5 and various f$_{dV}$'s and f$_{dip}$'s. The parameters are shown at the upper-right corner of each frame.  Though we also have done the experiments by varying f$_\sigma$, the results do not show very much difference and are not shown in this paper. 
   From Fig. \ref{fig:nsim} we see that the observability of a blue-profile line strongly depends on the strength of the absorption (f$_{dip}$), but less sensitive to the velocity shift (V$_{in}$). For f$_{dip}$ $\sim$ 0.15, approximately corresponding to the optical depth $\tau\sim$ 0.15, a blue-profile is hardly seen even in our noise-free experiments and is thus not likely to be detected in real observations. In the cases of f$_{dip}\sim$ 0.3 ($\tau\sim$ 0.36),  red shoulders are seen clearly. In such cases real detections are  mostly missed because our auto-search algorithm searches only for double peaks.   The blue-profiles become observable when f$_{dip}\sim$ 0.45 ($\tau\sim$ 0.6) with a small velocity shift but are still unclear at large velocity shifts. For nearly all velocity shifts, the signatures are clearly observable at f$_{dip}\sim$0.6 ($\tau\sim$ 0.9).

   At the corner of each frame, we also present the dimensionless skewness parameter, $\delta$V,  suggested by \citet{1997ApJ...489..719M} as an indicator of the significance of blue asymmetry.  As can be seen from Fig. \ref{fig:nsim}, $\abs[\delta V]$ generally increase as f$_{dip}$ but decrease as f$_{dV}$. In fact, our experiments suggest the absolute value of $\delta$V decreases also with  f$_\sigma$. 
   
   \subsection{Parameter Distributions}
   Up to date, the MWISP project covers an area of $\sim$ 2400 square degrees toward the northern Galactic plane.  Though the sample is still quite far from complete, due to the fact that the blue-profiles are recognized without any other presumption, the blind search strategy enables us to avoid  bias toward known star-forming regions. This kind of work will provide a better approach to explore parametric space of the infalling molecular clumps. In the following we discuss the parametric distributions of our samples.   
   
\begin{figure}
      \subcaptionbox{Vc\label{fig:vc_distr}}[0.495\textwidth]{\includegraphics[width=8cm]{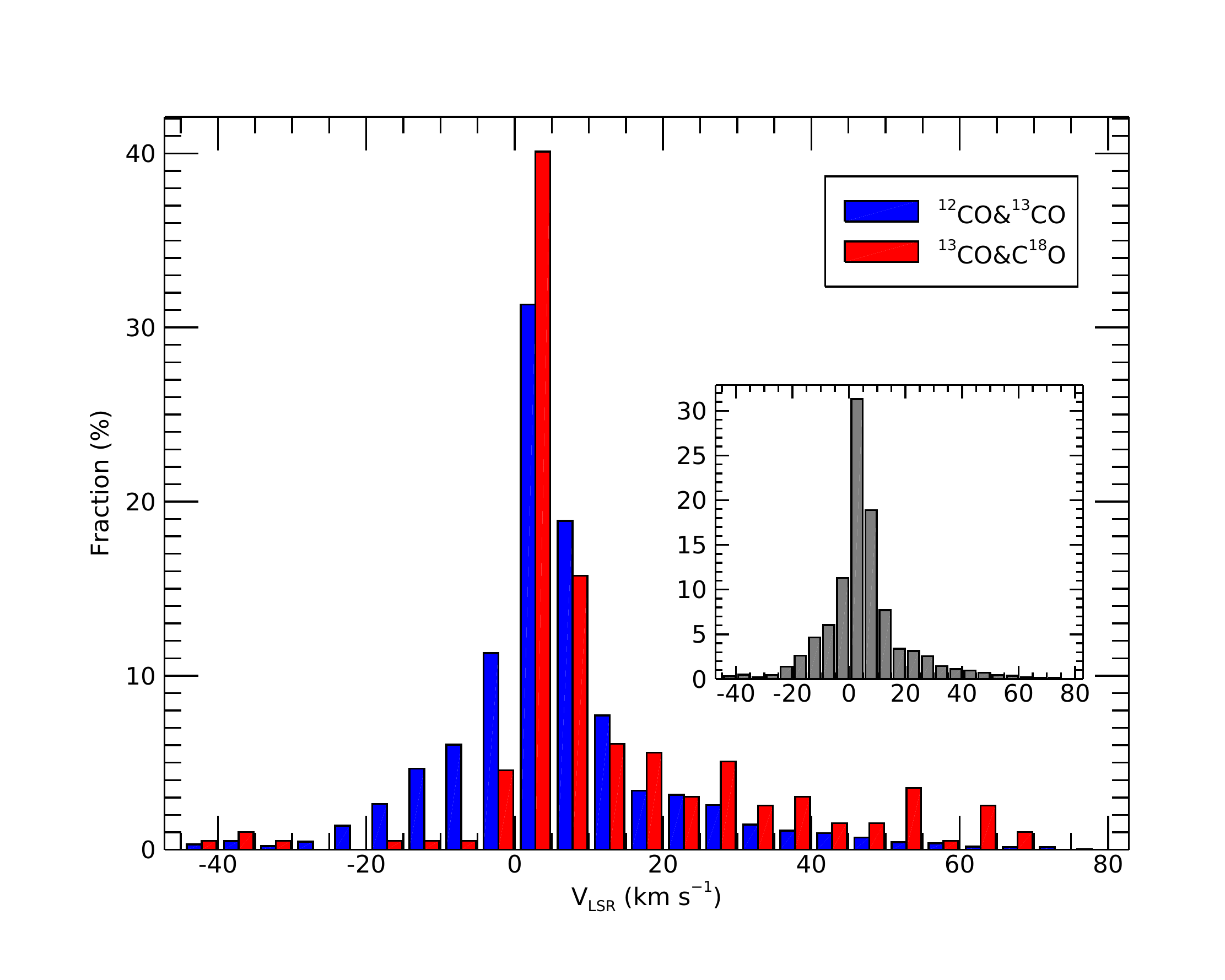}}
      \subcaptionbox{T$_{ex}$\label{fig:tex_distr}}[0.495\textwidth]{\includegraphics[width=8cm]{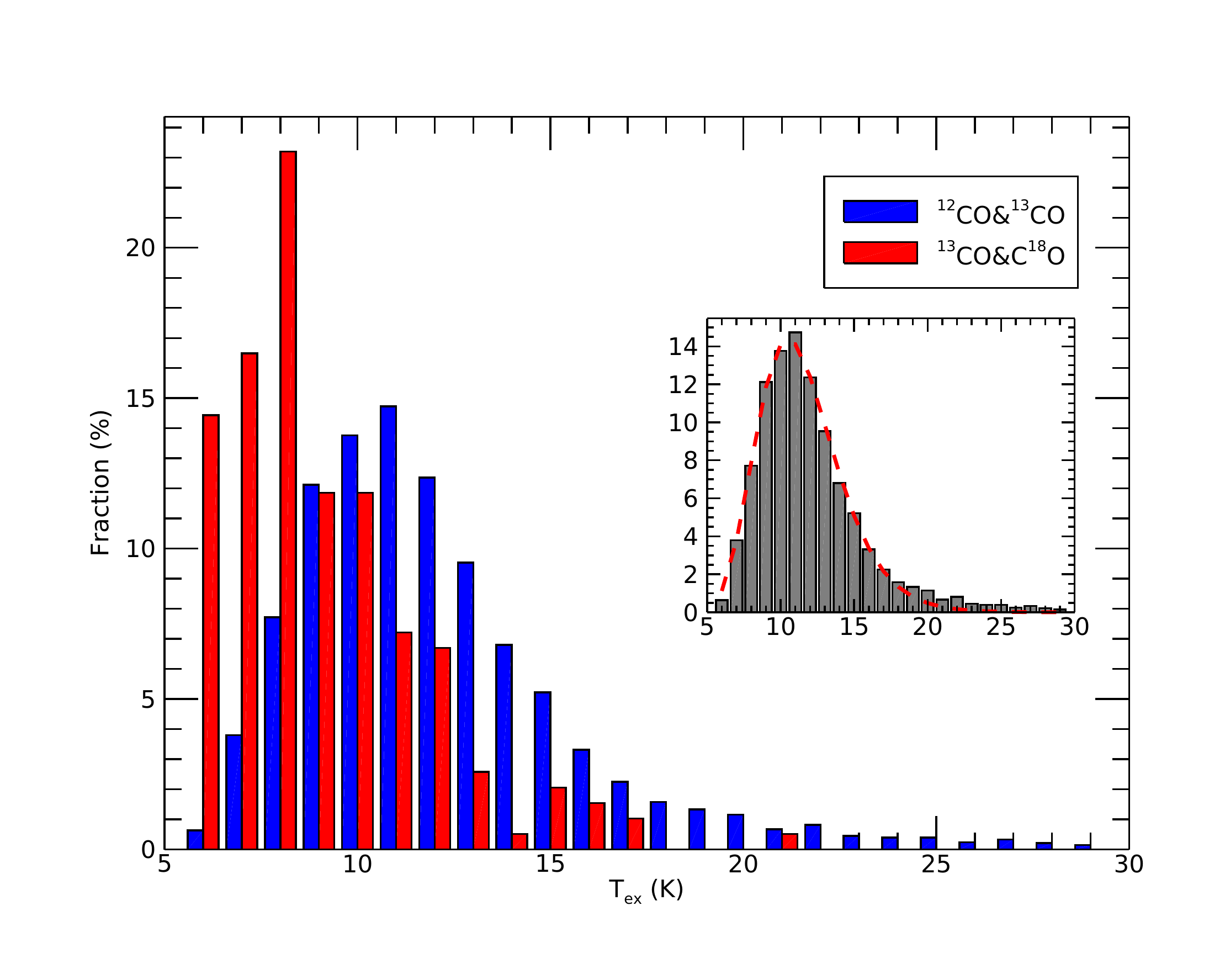}} 
      \subcaptionbox{N(H$_2$)\label{fig:nh_distr}}[0.495\textwidth]{\includegraphics[width=8cm]{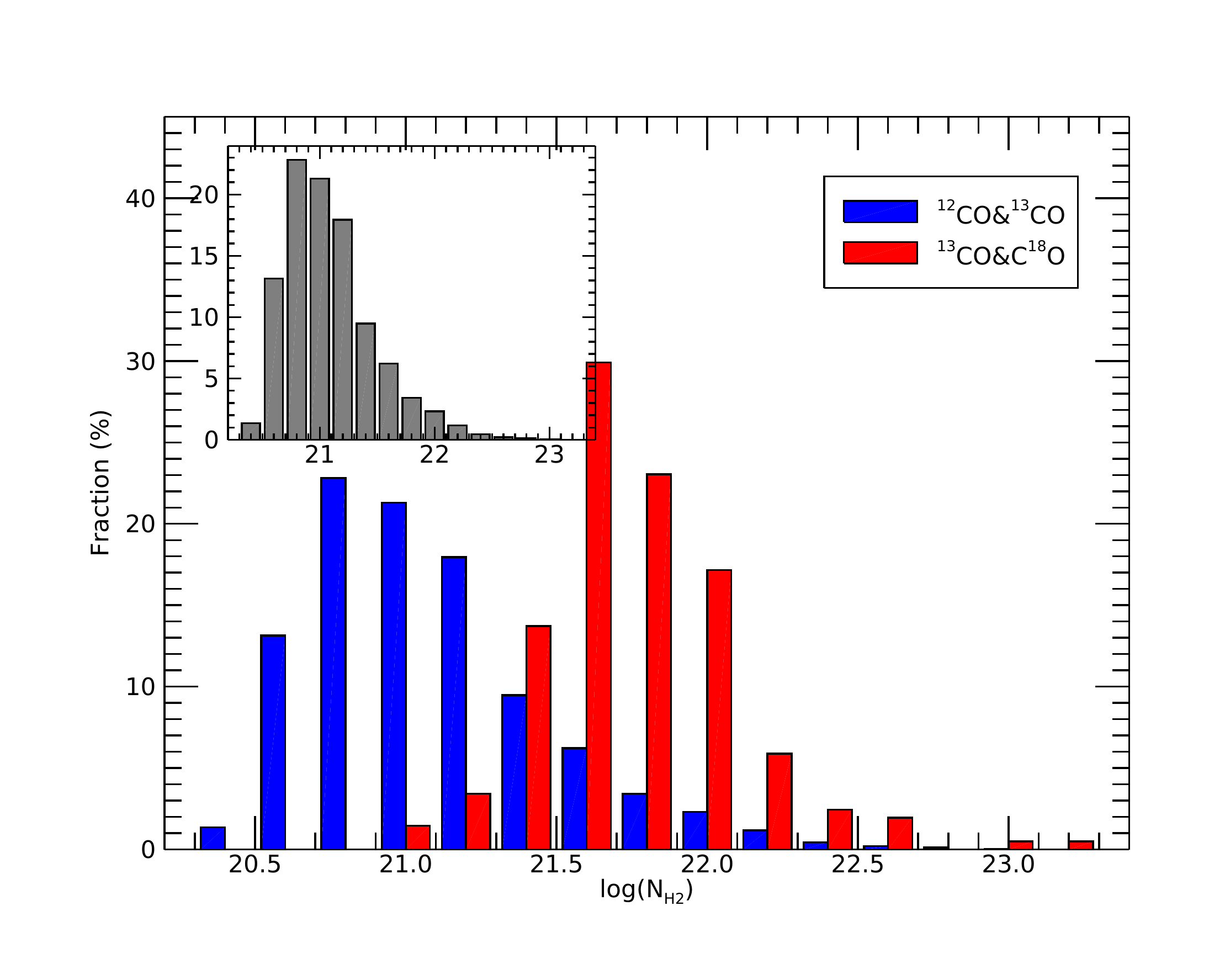}} 
      \subcaptionbox{FWHM\label{fig:fwhm_distr}}[0.495\textwidth]{\includegraphics[width=8cm]{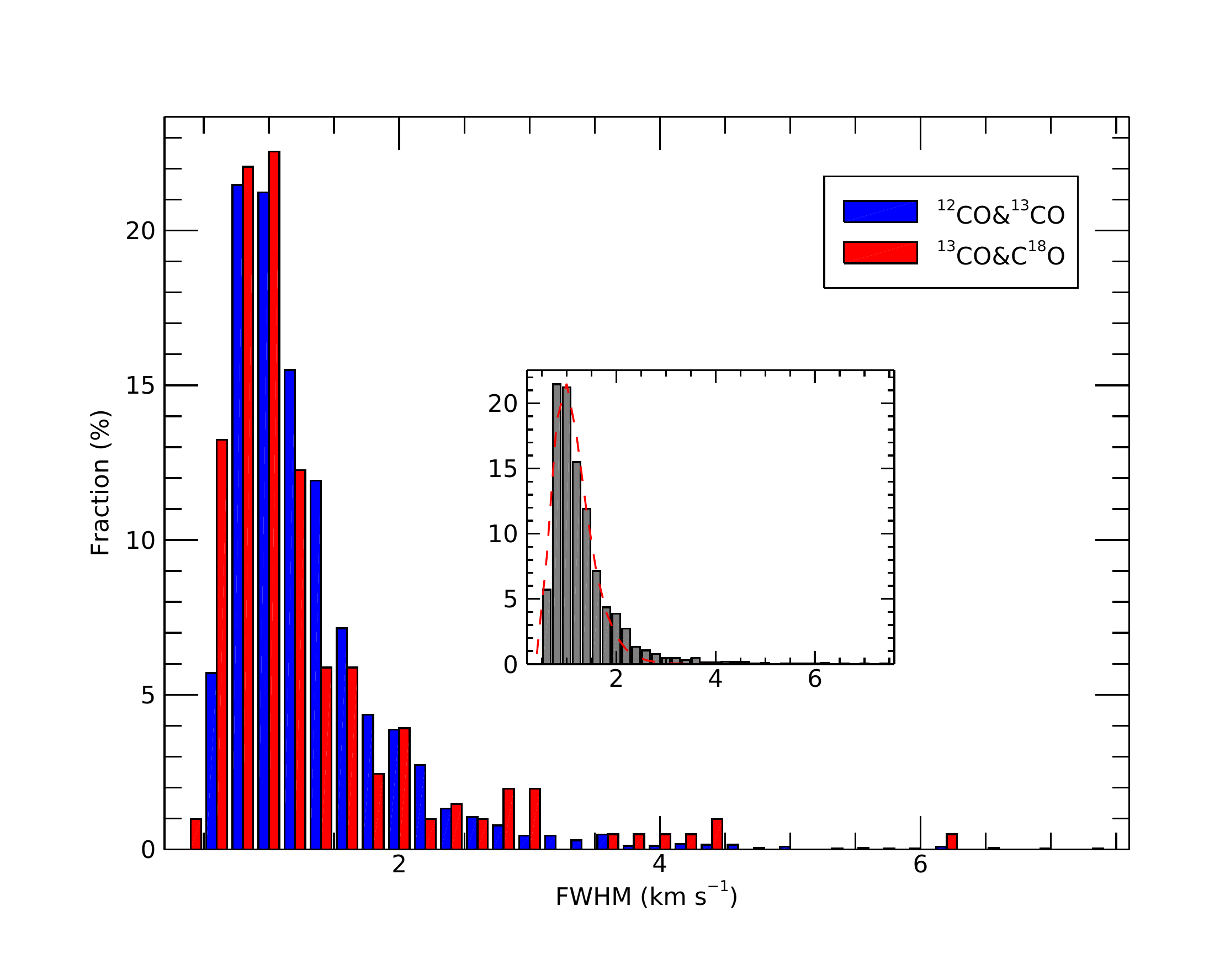}} 

   \caption{Frequency distributions of (a) central velocities; (b) excitation temperatures;  (c) derived H$_2$ column densities (cm$^{-2}$ on log scale); (d) Line widths (FWHM) of the optically thin lines. In the inset of (b), the  dashed curve is a log-normal fit with ($\sigma, \mu$) = (2.41, 0.25) to overall T$_{ex}$ distribution. In the inset of (d), the dashed curve is the best log-normal fit with ($\sigma, \mu$) = (0.34, 0.076), to the overall distribution of FWHMs. For each parameter, the bins for both Pair-1 and Pair-2 as well as for the overall are set to same.  \label{fig:p_dist}}
   \end{figure}

   In Fig. \ref{fig:vc_distr} we present the distributions of central velocities of the optically thin lines. In the main frame we show the pair-separated distributions while the overall distribution is shown in the inset, and the rest figures are treated in the same way. The distribution is generally in Gaussian style peaked around 5 \kmps, and shows excess in both wings.  A large amount of sources gather within  $\abs[V_{LSR}] \le$   10 \,\kmps, suggesting that our sources are mainly located  close to the Sun. We notice here, though not shown in the figure, the smallest and largest central velocity are -75.5 and 150.3 \kmps, respectively. An interesting feature seen from Fig. \ref{fig:vc_distr} is that Pair-1 selected sources have lower radial velocities (with a median value of $\sim$ 6.8 \kmps) than  Pair-2 selected sources (median $\sim$ 7.9 \kmps).
   
   Fig. \ref{fig:tex_distr} shows the distributions of excitation temperatures T$_{ex}$. The excitation temperatures are generally low, all less than 50 K, with a minimum value $\lesssim$ 4.5 K. A vast majority of sources ($\sim$ 94\%) show T$_{ex}$ less than 20 K. The median value is $\sim$ 11.7 K, which is a little bit less than that of all infall sources in the literature \citep{2022RAA....22i5014Y}. The distribution of Pair-2 does not seem to follow a regular function, probably because of the paucity of samples. On the other hand, the profiles of Pair-1 as well as of the overall are quite smooth. The best log-normal fitting to the overall distribution gives a parameter-pair ($\sigma,\mu$) = (2.41, 0.25) with $\chi^2$ = 3.2. This results is very close to that derived from Planck cold clumps \citep{2012ApJ...756...76W}, indicating that our sources may be of similar properties to those cold clumps. {We note however, as stated in Section \ref{sec:cat}, the excitation temperatures derived from this work are subject to rather large uncertainties.}

   Fig. \ref{fig:nh_distr} presents the distributions of column densities of molecular hydrogen, N(H$_2$). It can be clearly seen the column densities traced by \co[13]{} (Pair-1) and \CO{} (Pair-2) are obviously separated. The values for the former group range from $2.6\times10^{20}$ to $1.1\times10^{23}$ cm$^{-2}$ with a median value of $1.3\times10^{21}$ cm$^{-2}$, while  those for the latter range between $1.1\times10^{21}$ and $1.8\times10^{23}$ cm$^{-2}$ with a median value of $6.4\times10^{21}$ cm$^{-2}$. The column densities of our samples are distributed narrower, and have smaller median value than those infall sources in the literature \citep{2022RAA....22i5014Y}. This may be because the latter utilized a large variety of molecular lines which include high density tracers, and used different observing facilities including ones with high angular resolution. 
   
   Our result means that a major part of the Pair-2  and a significant part of Pair-1 sources have higher column densities than the possible star formation threshold, i.e. 6.3$\times10^{21}$ cm$^{-2}$ \citep{2014Sci...344..183K,2010ApJ...724..687L,2021ApJS..254....3M}. All of our candidates have column densities less than the threshold of massive star formation proposed by \citet{2008Natur.451.1082K}, i.e., N(H$_2$) $\sim$2.1$\times10^{23}$ cm$^{-2}$. Three factors may reduce the estimated value of N(H$_2$): 1) large beam size of the telescope (hence low filling factor) may dilute the peak values; 2) our candidates are frequently seen outside of the central part of certain  clumps; and 3) T$_{ex}$'s of Pair-2 selected candidates are often underestimated hence the derived H$_2$ column densities. Therefore, whether our sources are massive star-forming candidates should await further deliberate studies.  
   
   The distribution of the column densities from \CO{} is roughly symmetric with respect to the median value, suggesting a log-normal profile (note the abscissa is in log-scale). On the other hand, that from \co[13]{} shows a log-normal style only in the first several bins, up to $\sim 4.0\times10^{21}$ cm$^{-2}$, and deviate from the log-normal distribution at higher column densities. Since Pair-1 objects account for a vast majority of our sample, the overall distribution generally resembles that from \co[13]{} (inset of Fig. \ref{fig:nh_distr}).

   Fig. \ref{fig:fwhm_distr} shows the distributions of the FWHMs of optically thin lines.  The distributions of the sources selected from two pairs do not show significant difference, except for a slightly shift of median values (1.21 vs 1.10 \kmps{} for Pair-1 and Pair-2, respectively). A Kolmogorov–Smirnov (K-S) test with a +0.10 \kmps{} shift of Pair-2 sources results in P-value of $\sim$0.47, rejecting the hypothesis that the two profiles are different. A log-normal fit to the overall distribution (inset) results in a $\sigma$ = 0.35 $\pm$ 0.013 and $\mu$ =0.076 $\pm$ 0.014. A K-S test results in P-value of 0.23, thus accepting the log-normal distribution hypothesis.

\begin{figure}
      \centering
      \includegraphics[width=12cm,angle=0]{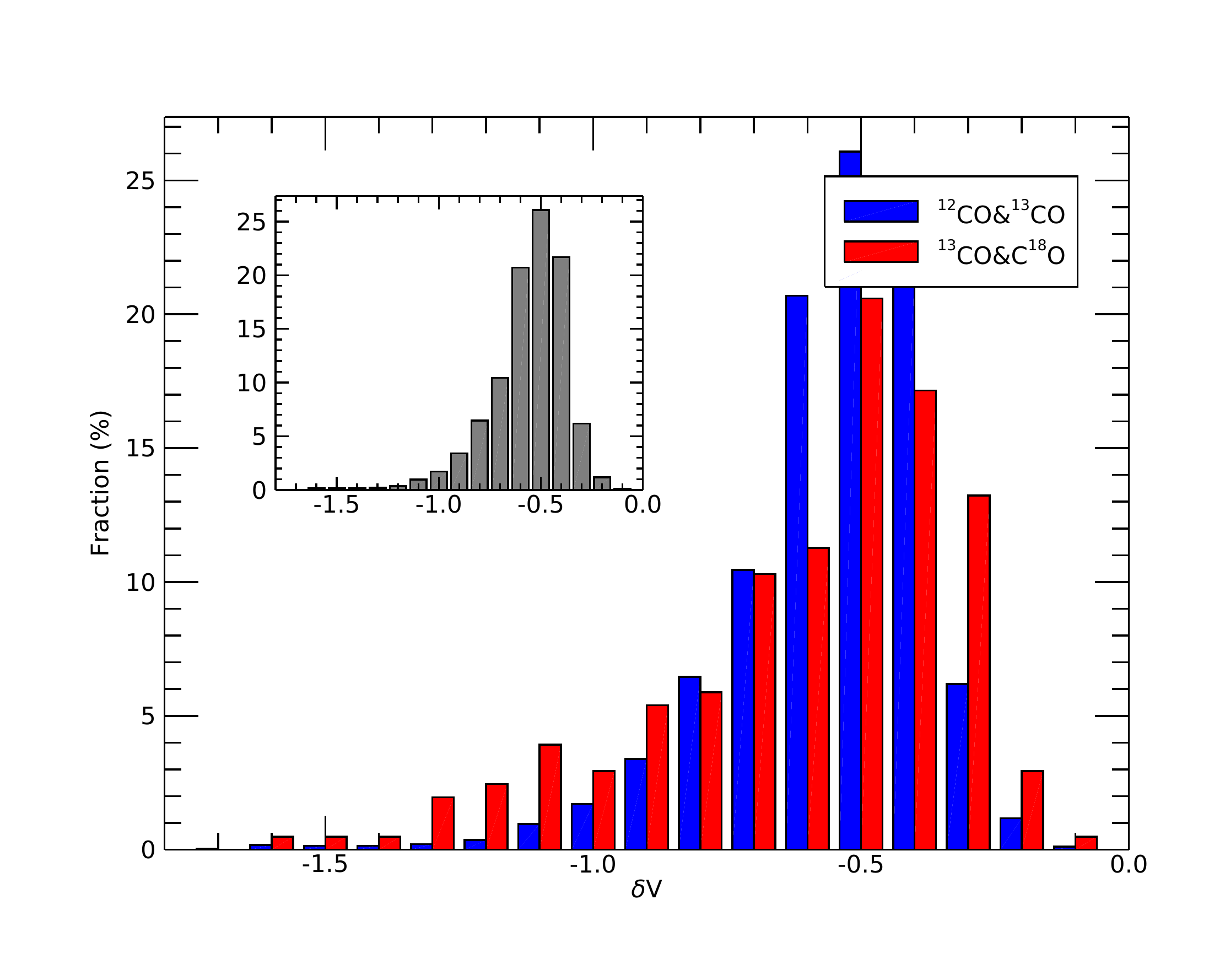}
      \caption{The distributions of the skewness parameters, $\delta$V, of the sources selected from Pair-1 (blue), Pair-2 (red) and overall (inset). \label{fig:dv_distr}}
      \end{figure}

   In Fig. \ref{fig:dv_distr} we show the distribution of the skewness parameters, $\delta$V, defined by \citet{1997ApJ...489..719M}.  As can be seen in the figure, the values range between -1.6 and 0.0 with a mean value $\sim$ -0.5. Since our samples are selected in favor of blue-profiles, it is not surprising that the values are mostly less than -0.25, a criterion as significant infall signature suggested by \citet{1997ApJ...489..719M}.

\begin{figure}
    \centering
   \includegraphics[width=10cm,angle=0]{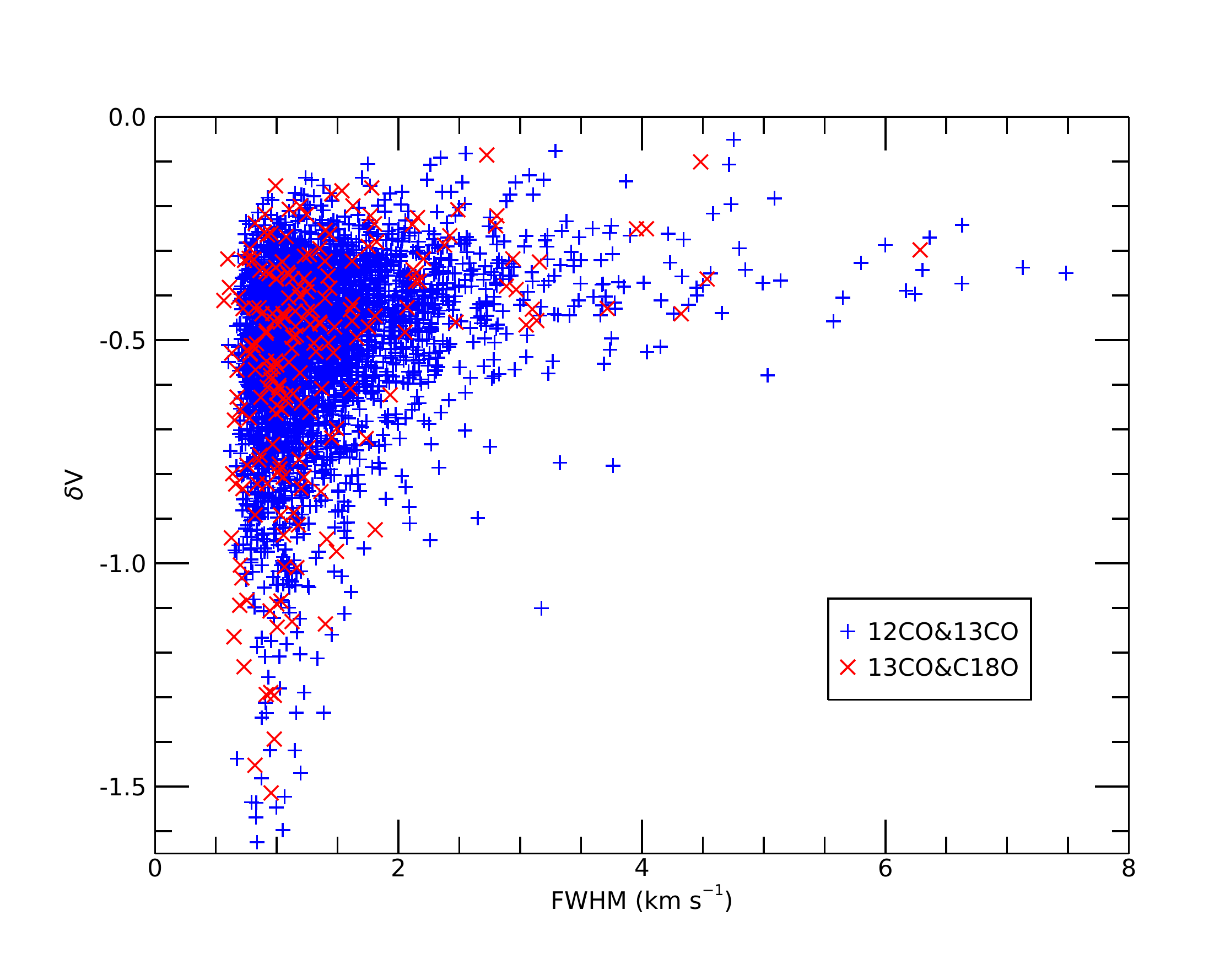}
   \caption{A plot of $\delta$V and line width (FWHM) of all candidates. The blue pluses and red crosses represent Pair-1 and Pair-2 selected sources, respectively.\label{fig:dv_fwhm}}
   \end{figure}

   Though  $\delta$V is defined as inversely correlated  to the line widths of the optically thin line, as shown in Fig. \ref{fig:dv_fwhm}, we have not found any clear trend of such correlation.

   \subsection{Spatial Distribution}

\begin{figure}
      \centering
      \includegraphics[width=12cm,angle=0]{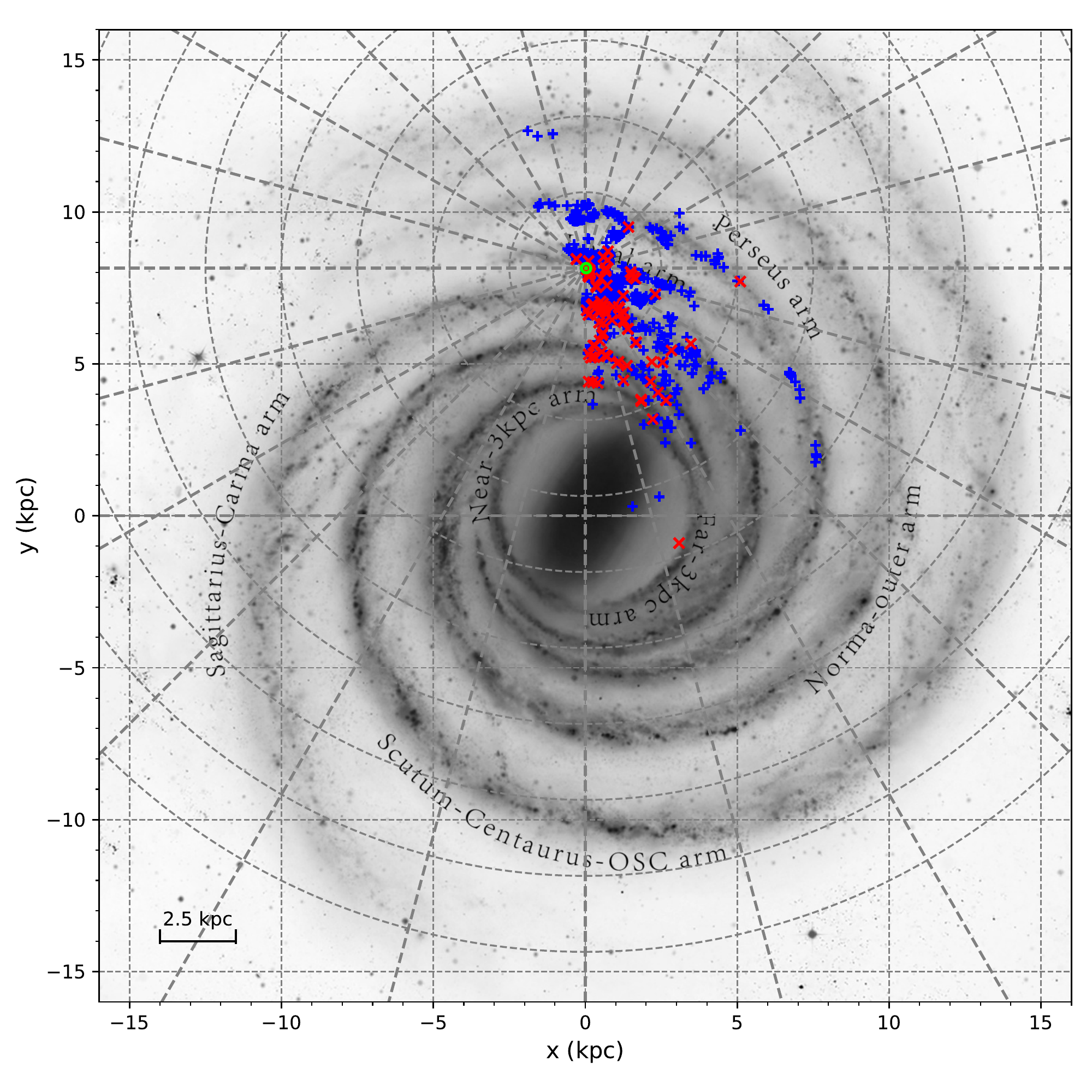}
      \caption{Spatial distribution of the infall candidates overlaid on the imaginary face-on view of the Milky Way (credit: Xing-Wu Zheng \& Mark Reid  BeSSeL/NJU/CFA). The green $\odot$ sign indicates the location of the Sun. The Galactic center is at the origin. Our candidates are designated by blue pluses (Pair-1) and red crosses(Pair-2). \label{fig:s_distr}}
   \end{figure}

   Fig. \ref{fig:s_distr} shows the spatial distribution of our candidates, overlaid on the imaginary face-on view of the Milky Way (credit: Xing-Wu Zheng \& Mark Reid  BeSSeL/NJU/CFA\footnote{https://astronomy.nju.edu.cn/xtzl/EN/index.html}). It can be seen clearly that a vast majority of our candidates are  located  in the first quadrant, especially those selected from Pair-2, and outside of the 3-kpc molecular ring. This is not surprising because of the facility limitation and the confusion in the inner Galactic molecular clouds. It should be noted here that most of the sources are located in the spiral arms. This may be due to the distance estimator \citep{2016ApJ...823...77R,2019ApJ...885..131R}, which assumes that molecular clouds are presumably located in the spiral arms. Though this could be true  for a majority of the our sources, we advice the  researchers to interpret this result with caution.

\begin{figure}
      \centering
      \includegraphics[width=12cm,angle=0]{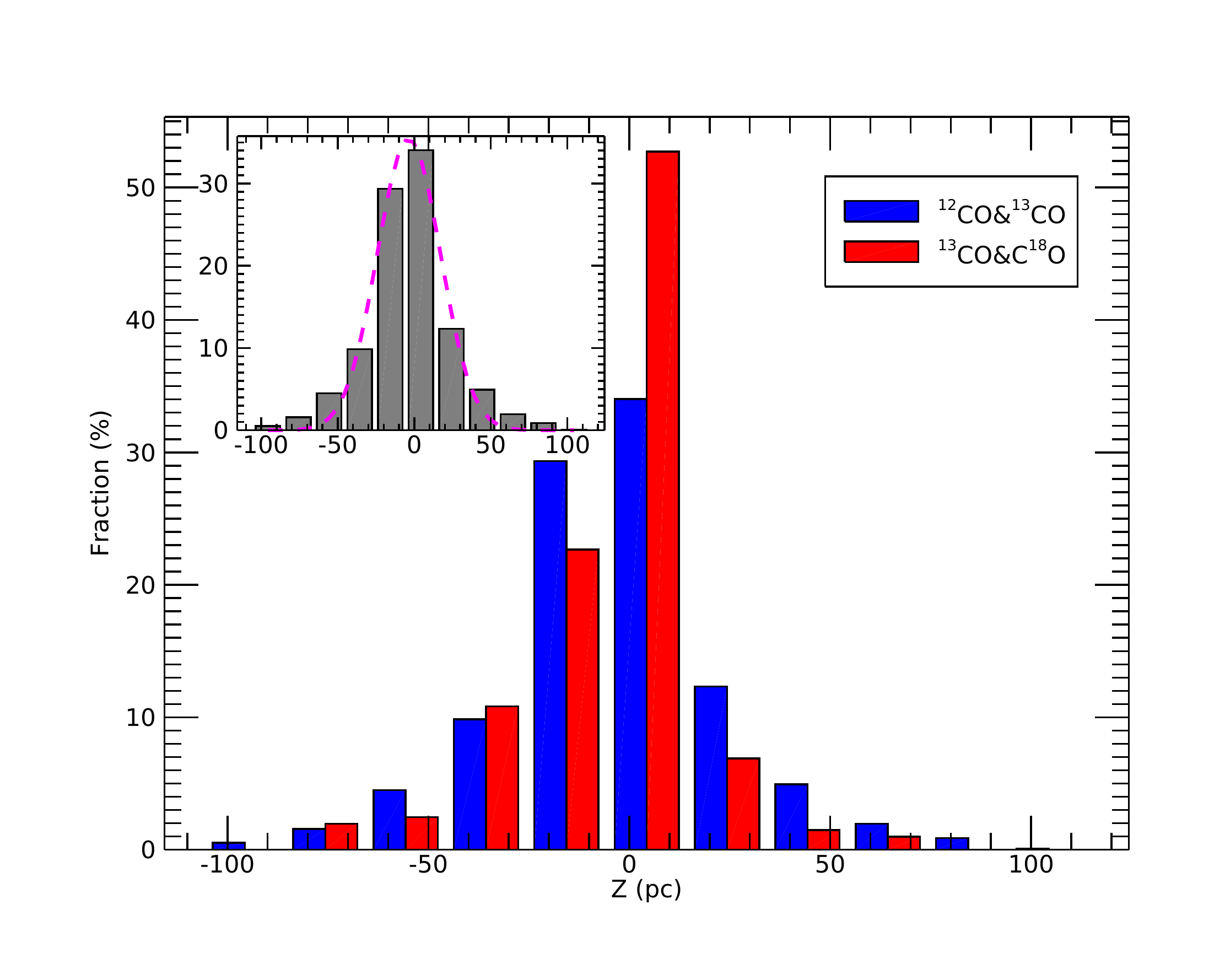}
      \caption{The vertical distribution. The blue and red bars represents the Pair-1 and Pair-2 selected sources, respectively. The pink dashed curve in the inset is a Gaussian fit to the overall distribution with $\sigma$ = 36 pc (FWHM = 85 pc).    \label{fig:z_distr}}
   \end{figure}

   Fig. \ref{fig:z_distr} is the vertical distribution (source offset from the Galactic midplane) of our candidates. The main frame shows the pair-separated distribution while the inset is that of overall.  Gaussian fitting of the overall distribution shows the FWHM thickness is $\sim$ 85 pc, which is reminiscent of the thin molecular disk suggested  by \citet{2021ApJ...910..131S}. A majority (3034, $\sim$85.8 \%) of the sources are located within this range (i.e., $\abs[z]{} \le$ 42.5 pc). The profile of Pair-1 selected candidates is wider than that of the Pair-2 ones (85 pc vs 54 pc in FWHM). Interestingly, we notice the presence of excess of sources at the foot of the Gaussian fit (inset of Fig. \ref{fig:z_distr}), similar to that of the vertical distribution of molecular clouds by \citet{2021ApJ...910..131S}, and suggesting that star formation may also take place in the thick disk component. It is also interesting to mention that a number (16) of sources are located far beyond the Galactic midplane (i.e., $\abs[z]{} \ge$ 200 pc). These sources are worthy to investigate to show if there are star-forming activities even far from the Galactic plane.

\begin{figure}
      \centering
      \includegraphics[width=12cm,angle=0]{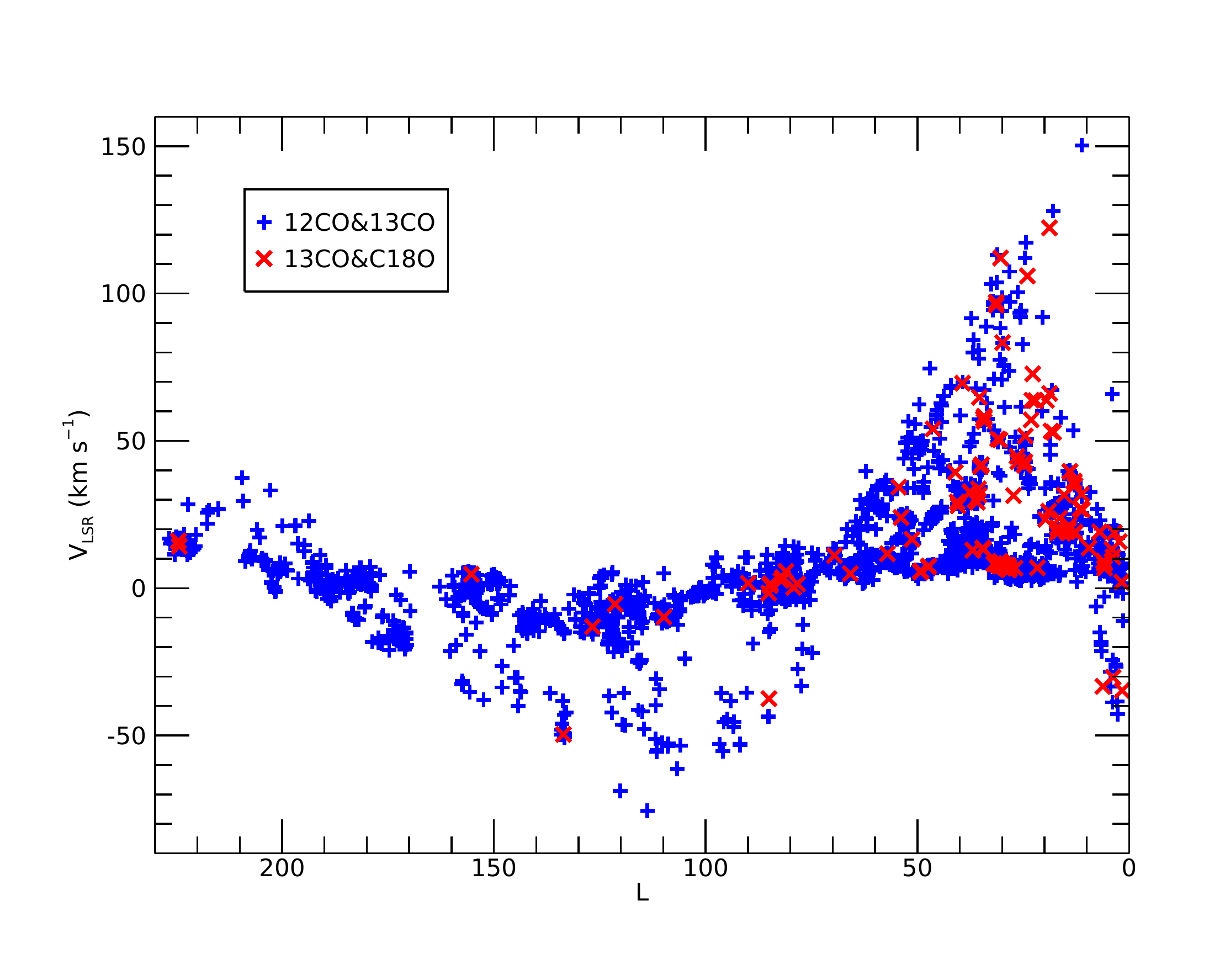}
      \caption{The source distribution in L - \vlsr{} space. As in Fig. \ref{fig:s_distr}, our sources are designated as the blue pluses (Pair-1) and red crosses (Pair-2). \label{fig:l_v}}
      \end{figure}

   In Fig. \ref{fig:l_v} we show source distribution in L - \vlsr{} space. Apart from the fact that sources selected from Pair-2 are more concentrated to the first quadrant, the overall distribution is reminiscent of the result by \citet[Fig. 3]{2001ApJ...547..792D}. Our sources exist in virtually all large-scale components in the northern Galaxy, such as  the Sagittarius-Scutum arm, the 3-kpc molecular ring, Lindblad Ring \& the local arm, the Perseus arm, and even the outer arm. If our sources represent quite a part of the infalling gas motion, the overall distribution indicates that there exist star-forming activities in nearly all large components in the Galaxy at present epoch. 

\subsection{Confirmation Rates}
 As known to researchers, \co{} and its isotopic molecules are not ideal tracers to the infall motions, and HCO$^+$ would be much more efficient. Therefore,  shortly after we initiated this project, follow-up observations were made with HCO$^+$ (J = 1 - 0) and HCN (J = 1 - 0) lines. \citet{2020RAA....20..115Y} selected 133 candidates from our sample with a restriction that T$_{MB}$(\CO) $\geqq$ 1 K. They confirmed 56 sources to have infall signatures in HCO$^+$ and/or HCN lines. The overall confirmation rate is $\sim$ 42\%, the respective  rate being 40\% and 49\% for Pair-1 and Pair-2. Further, \citet{2021ApJ...922..144Y} used IRAM to carry out mapping observations towards 24 out of the 56 confirmed sources and finally confirmed 9 blue asymmetry profiles in HCO$^+$ (J = 1 - 0) line. These 9 sources are regarded as infall sources with high confidence. 
 
 On the other hand, Yu et al. (2023, in prep.) selected a sample of 210 sources with  more relaxed conditions to study the infall signatures. They found a total of 40 sources showing blue-profiles in HCO$^+$ emission line. The gross detection rate is $\sim$ 19\%, with 18\% for Pair-1 and 28\% for Pair-2 objects. In general, Pair-2 selected candidates have higher confirmation rate than Pair-1 selected ones. This is not surprising in light of that \CO{} traces higher column density than \co[13]{} does. If the sub-sample represents our overall sample adequately, we would expect the confirmation rate being no less $\sim$ 20\%. 

 In conclusion, starting from the \co[12], \co[13] and \CO{} (J = 1 - 0) lines, our approach to search for infall sources is effective. 

\section{Summary}\label{sec:sum}
Based on the MWISP data, we conducted a survey of infalling clumps in the Galaxy, which presumably show blue asymmetric profiles in the optically thick lines and single peaks in the  optically thin lines. Both \co\&\co[13]{} and \co[13]\&\CO{} pairs are utilized to carry out the work. The automatic search and manual check are conducted to select candidates out of $\sim$ 10$^8$ spectra. The main results are summarized as following:
\begin{enumerate}
\item A total of 3533 sources are finalized as candidates in the infalling phase of star formation, in which 3329 candidates are selected from the \co\&\co[13]{} pair and 204 are from the \co[13]\&\CO{} pair. 
\item Though being manually checked and filtered, the candidates show a wide range of spectra with complicated profiles. The locations that show blue-profiles do not always coincide with central parts of molecular clumps where the highest column densities are detected, but a significant part are rather located at the edges. 
\item The analysis of physical parameters of the sources suggests the Pair-2 candidates are colder, and have higher column densities, than Pair-1 ones. The overall distribution of T$_{ex}$ follows a log-normal style and is consistent to that by \citet{2012ApJ...756...76W}, suggesting the properties of our sources are quite similar to those of Planck cold  clumps. 
\item The line widths of the optically thin lines for both Pair-1 and Pair-2 sources are in log-normal style. The mean values of the two groups are different by < 0.1 \kmps.  A K-S test indicates that the distribution are the same when line widths of Pair-2 sources are added by 0.1 \kmps. 
\item Most of the sources are located within the first quadrant, especially those selected from the \co[13]\&\CO{} pair. The system velocities of the candidates ranges from $\sim$ -70 to $\sim$ 150 \kmps, and are present in virtually all large scale components in the northern Galaxy. The vertical distribution suggests that our sources are located primarily within the thin disk, but still present in the extended thick disk.
\item A sketchy estimation suggests that the confirmation rate of our sample could be no less than $\sim$ 20\%, indicating our strategy is a good start to study the very early phase of star formation systematically. 
\end{enumerate}
   \begin{acknowledgements}
      We are deeply obliged to the MWISP working group, Yang Su, Xin Zhou, Yan Sun, Jixian Sun, Dengrong Lu and Binggang Ju, and the observation assistants of the project, who have been working hard in instrument maintenance, taking and reducing the data, without which this work cannot be done. This work is supported by the National Key R\&D Program of China (Grant No. 2017YFA0402702), and the National Natural Science Foundation of China (NSFC, Grant Nos., 11873093, U2031202,11903083). MWISP is sponsored by National Key R\&D Program of China with grant 2017YFA0402701 and by CAS Key Research Program of Frontier Sciences with grant QYZDJ-SSW-SLH047.  We would like to express our thanks to the anonymous reviewer for constructive suggestions that enable improvement of the manuscript.  
   \end{acknowledgements}

   \appendix  

   \section{The full catalogue}
\input{ms_2022-0304_table_appendix.tex}

   \section{Spectra and gas distributions of the sources}

\begin{figure}
\includegraphics[width=9.0cm,angle=0]{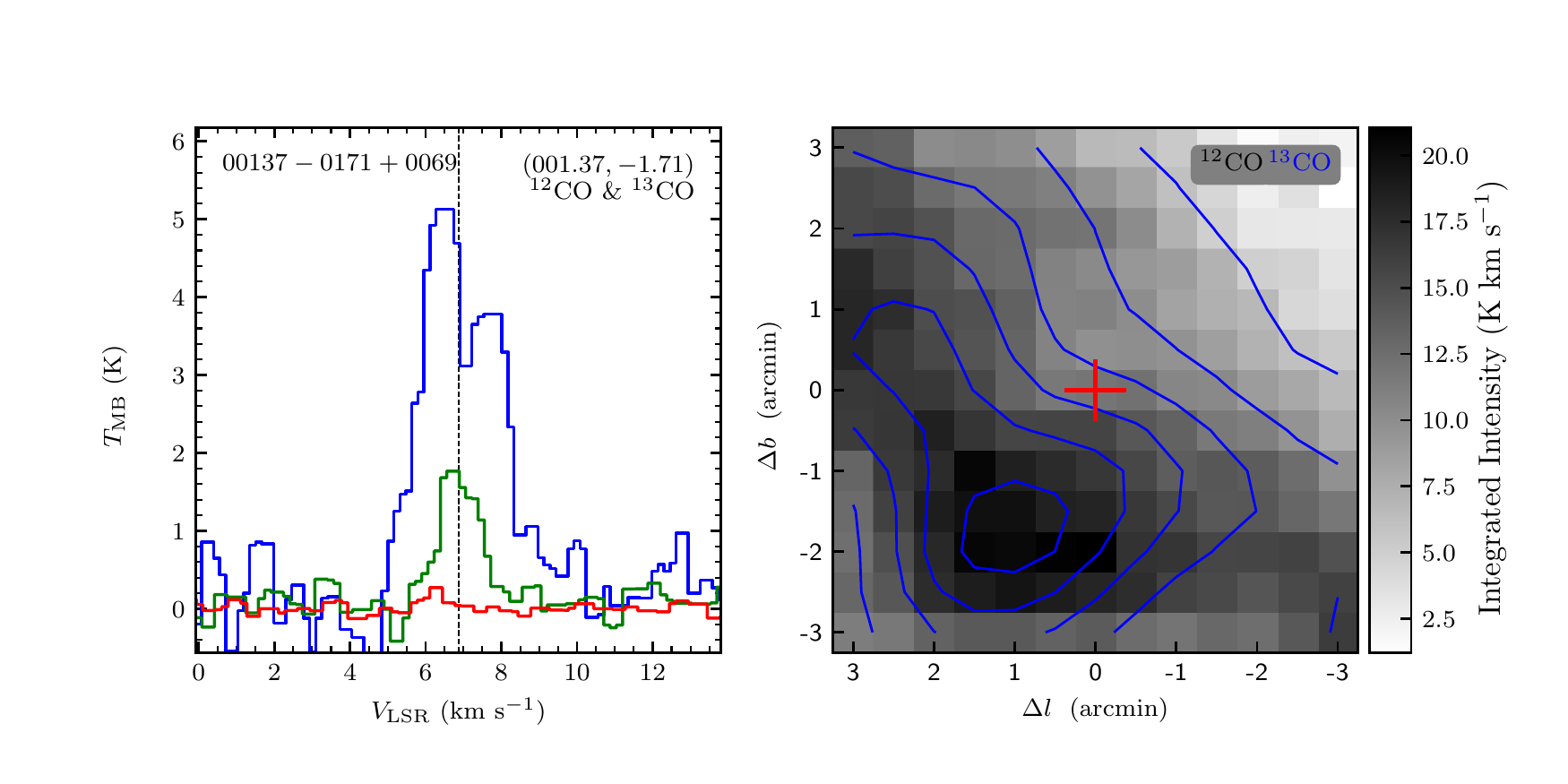}
\includegraphics[width=9.0cm,angle=0]{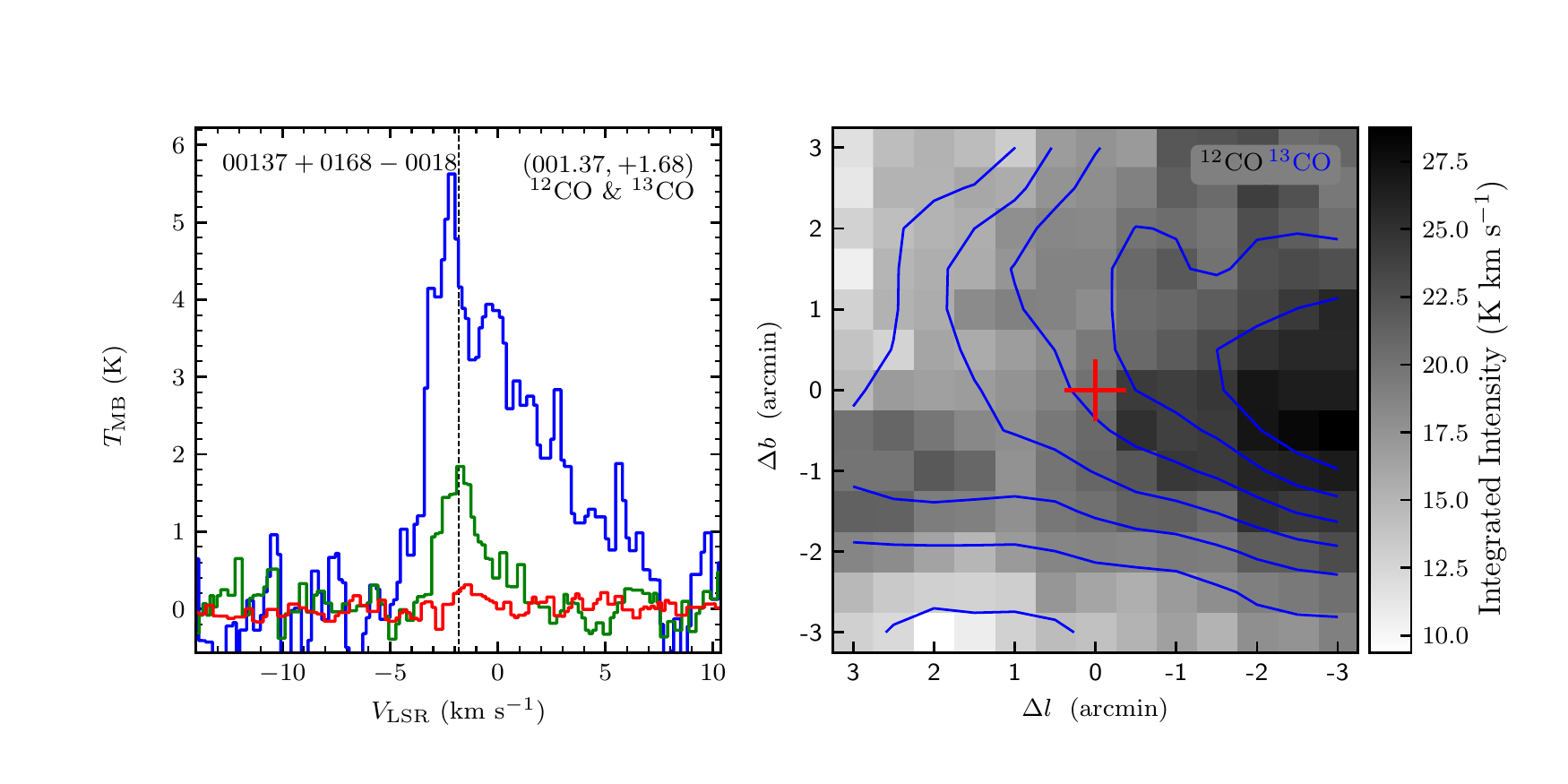}
\vspace{-0.5cm}

\includegraphics[width=9.0cm,angle=0]{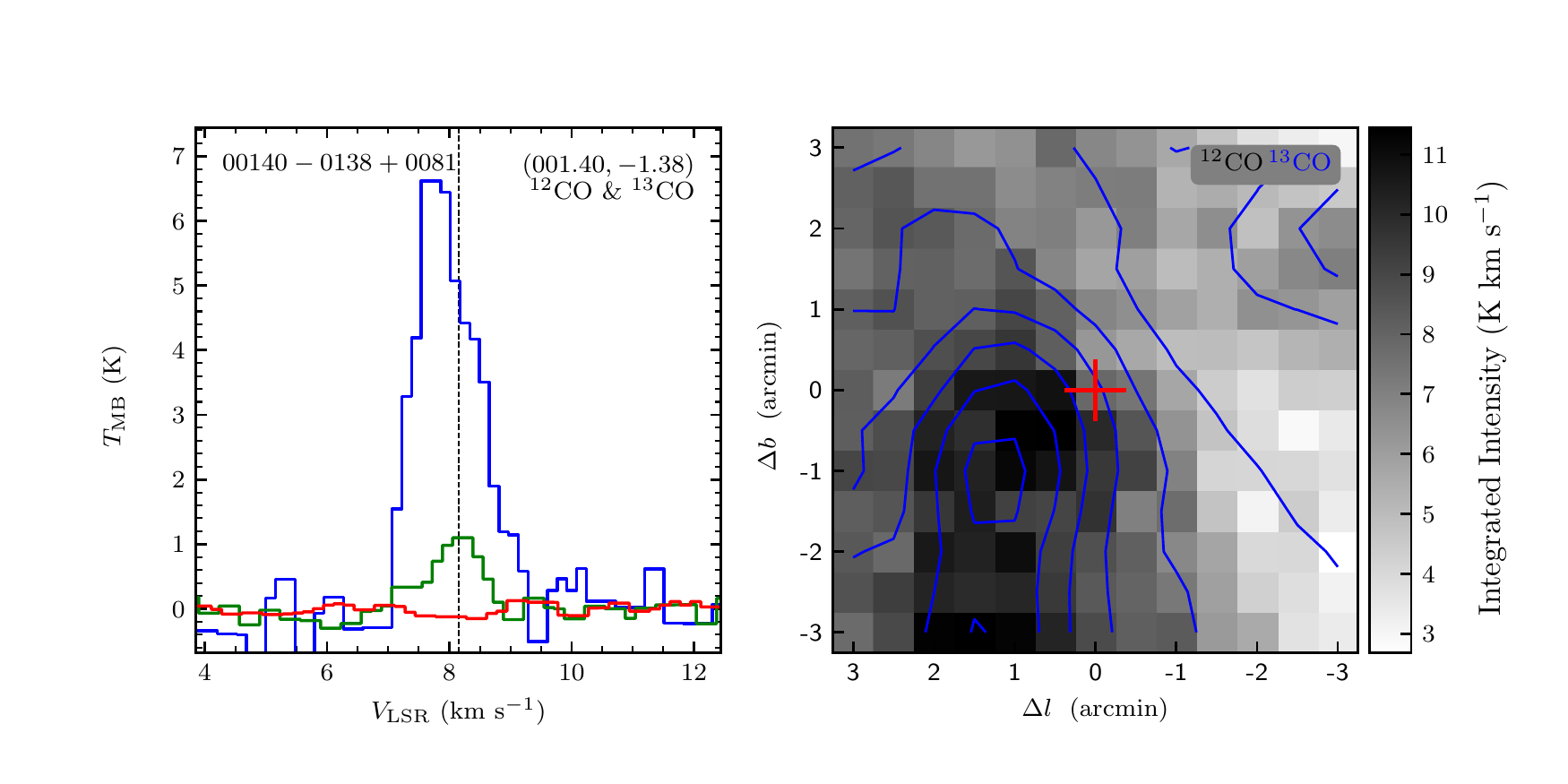}
\includegraphics[width=9.0cm,angle=0]{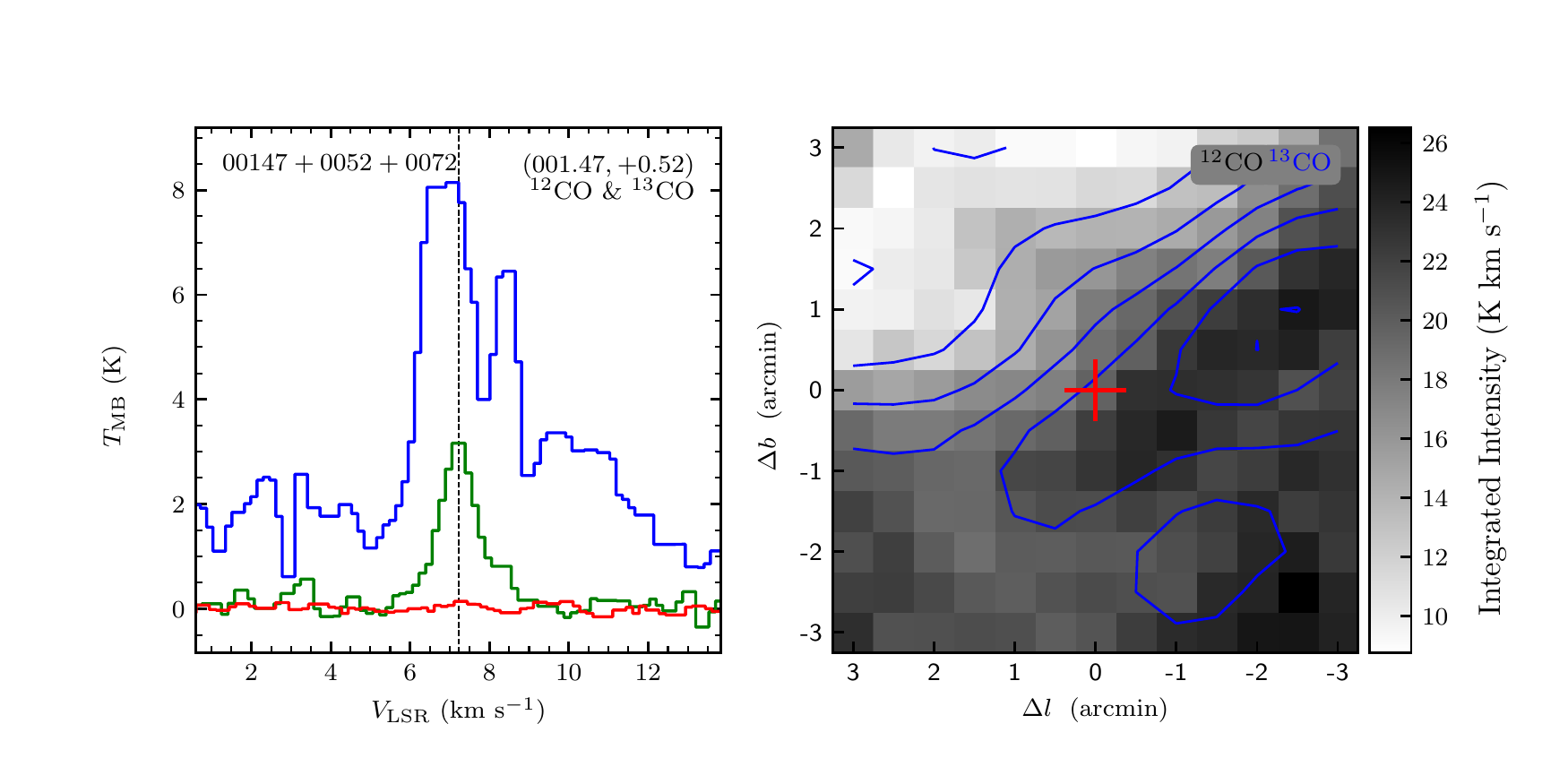}
\vspace{-0.5cm}

\includegraphics[width=9.0cm,angle=0]{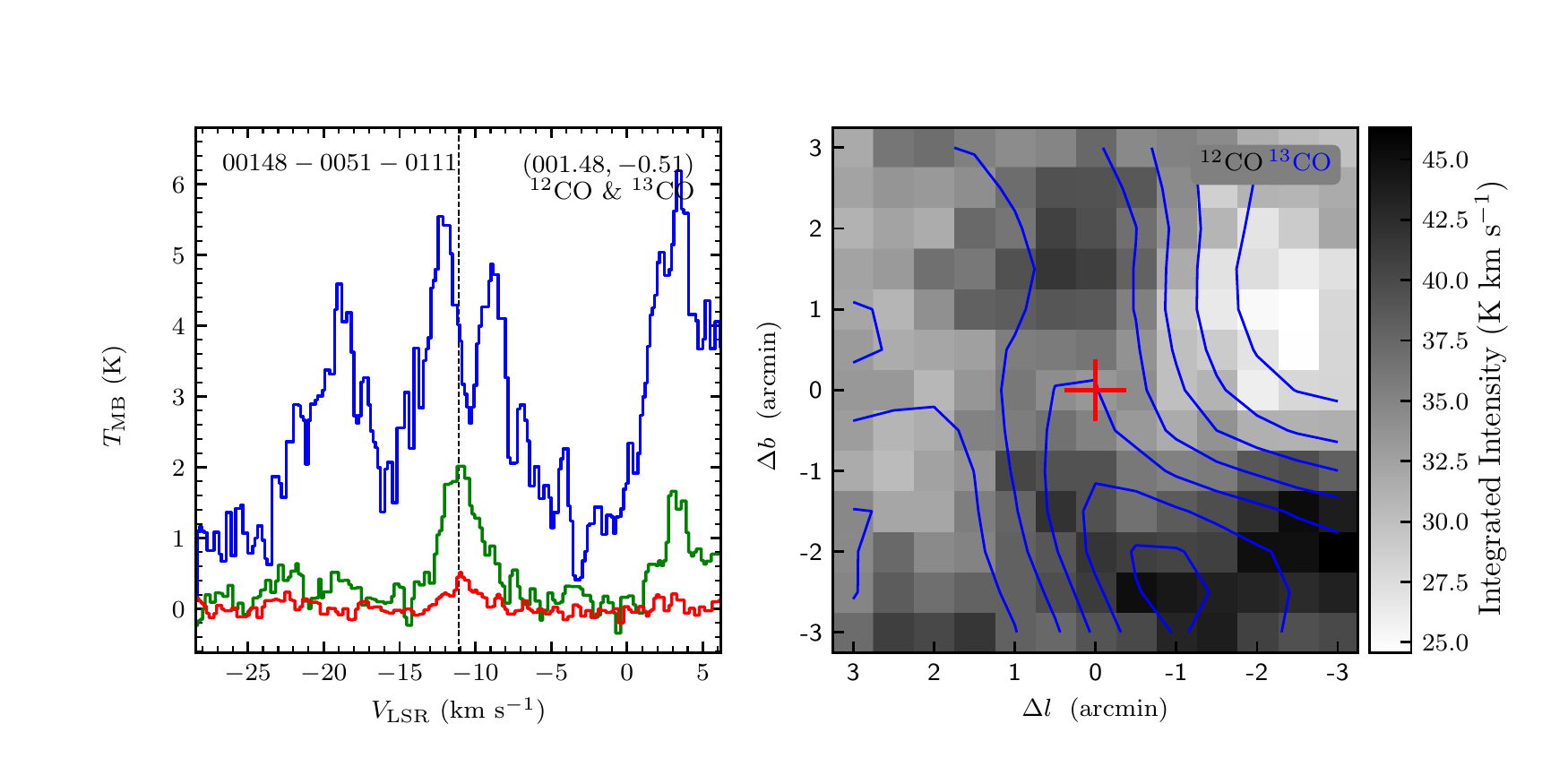}
\includegraphics[width=9.0cm,angle=0]{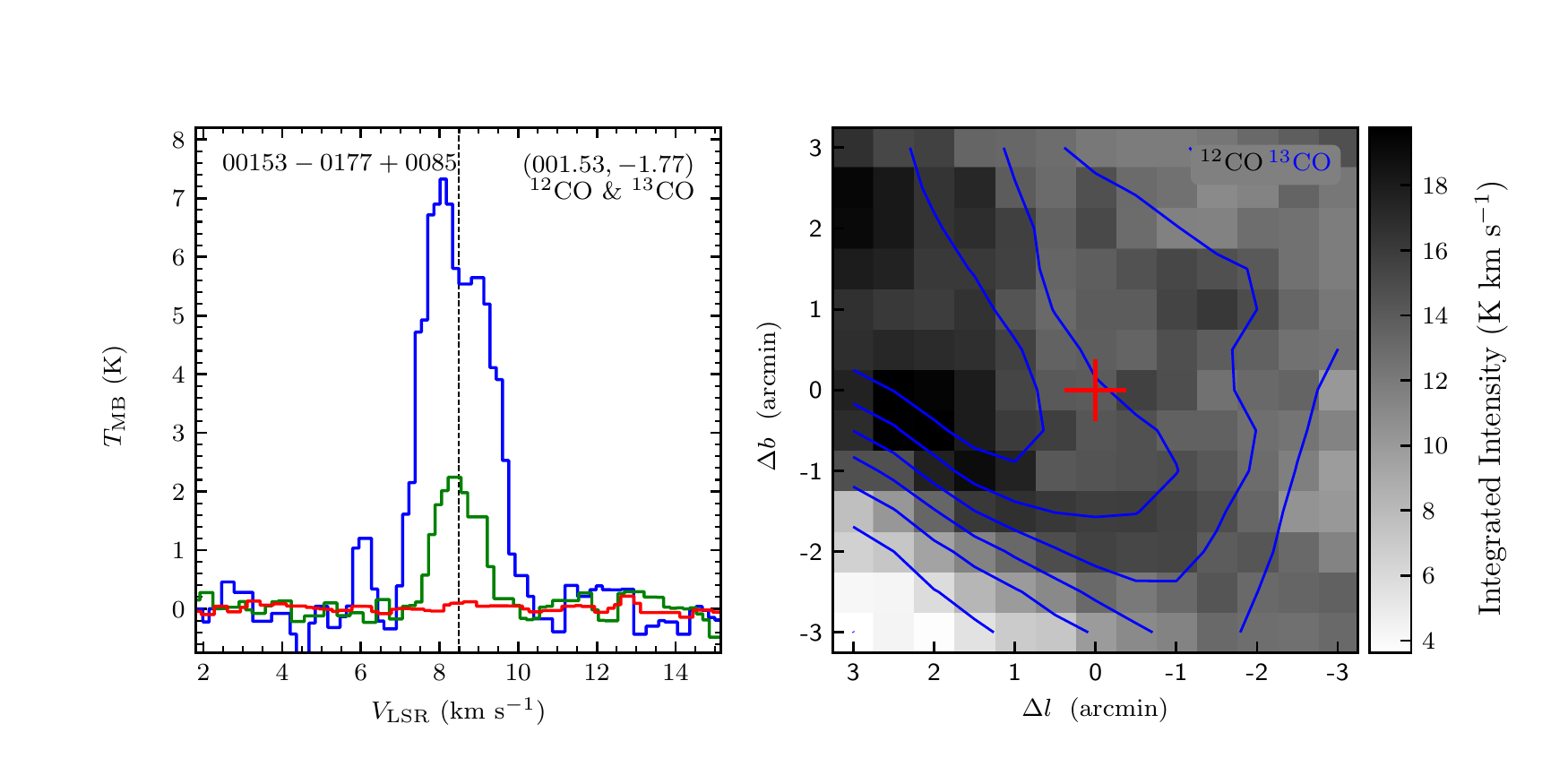}
\vspace{-0.5cm}

\includegraphics[width=9.0cm,angle=0]{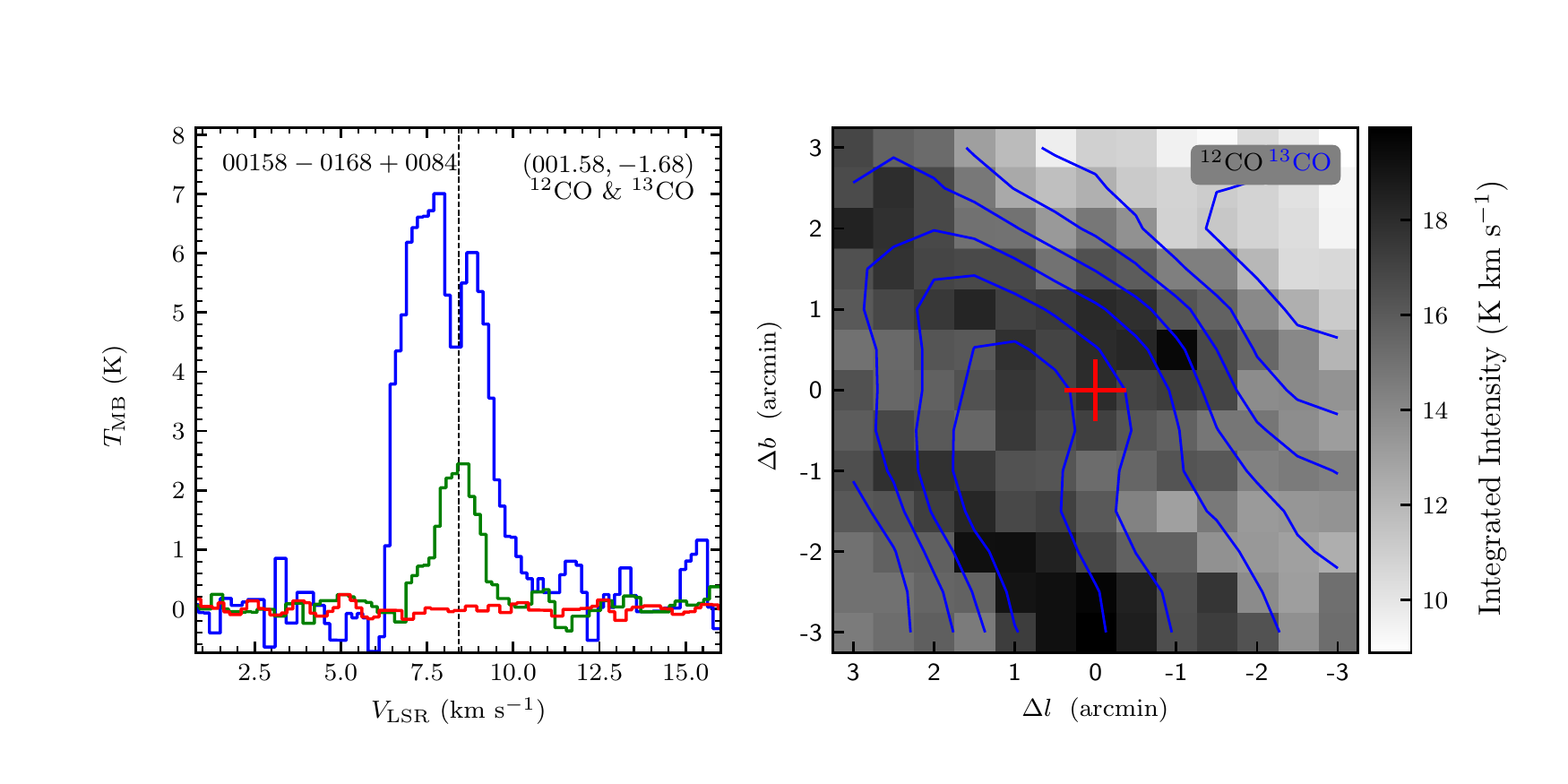}
\includegraphics[width=9.0cm,angle=0]{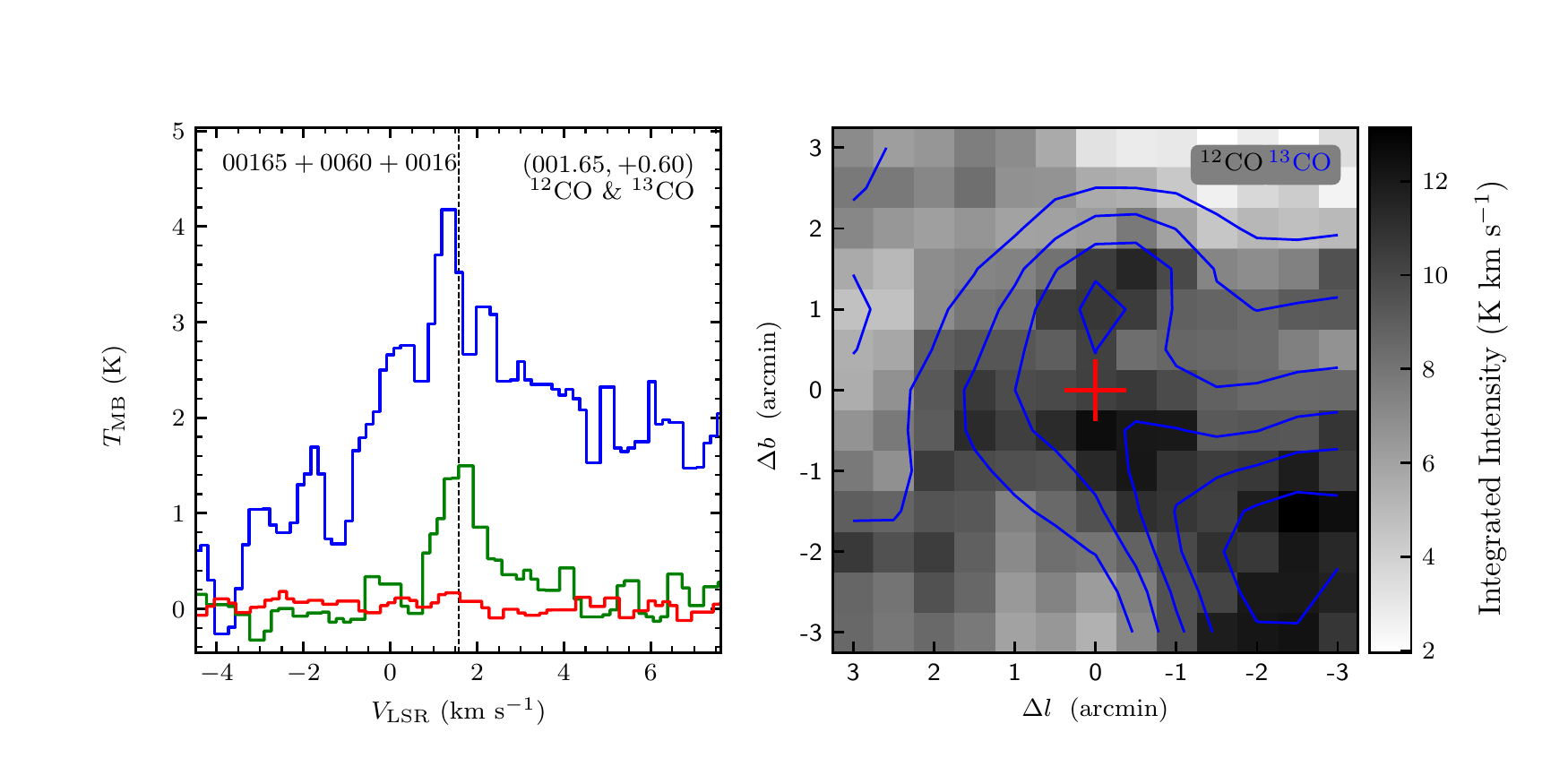}
\vspace{-0.5cm}

\includegraphics[width=9.0cm,angle=0]{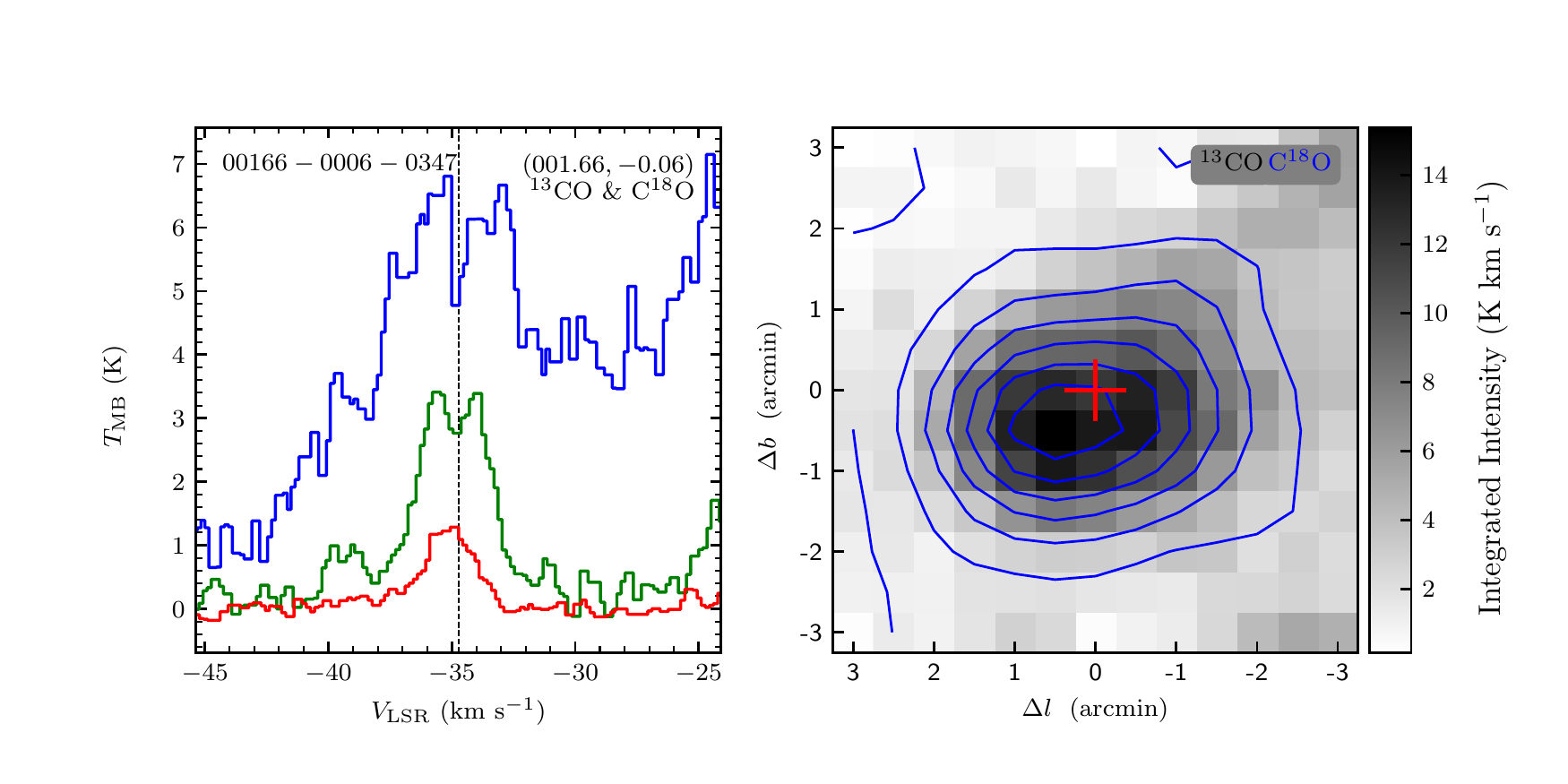}
\includegraphics[width=9.0cm,angle=0]{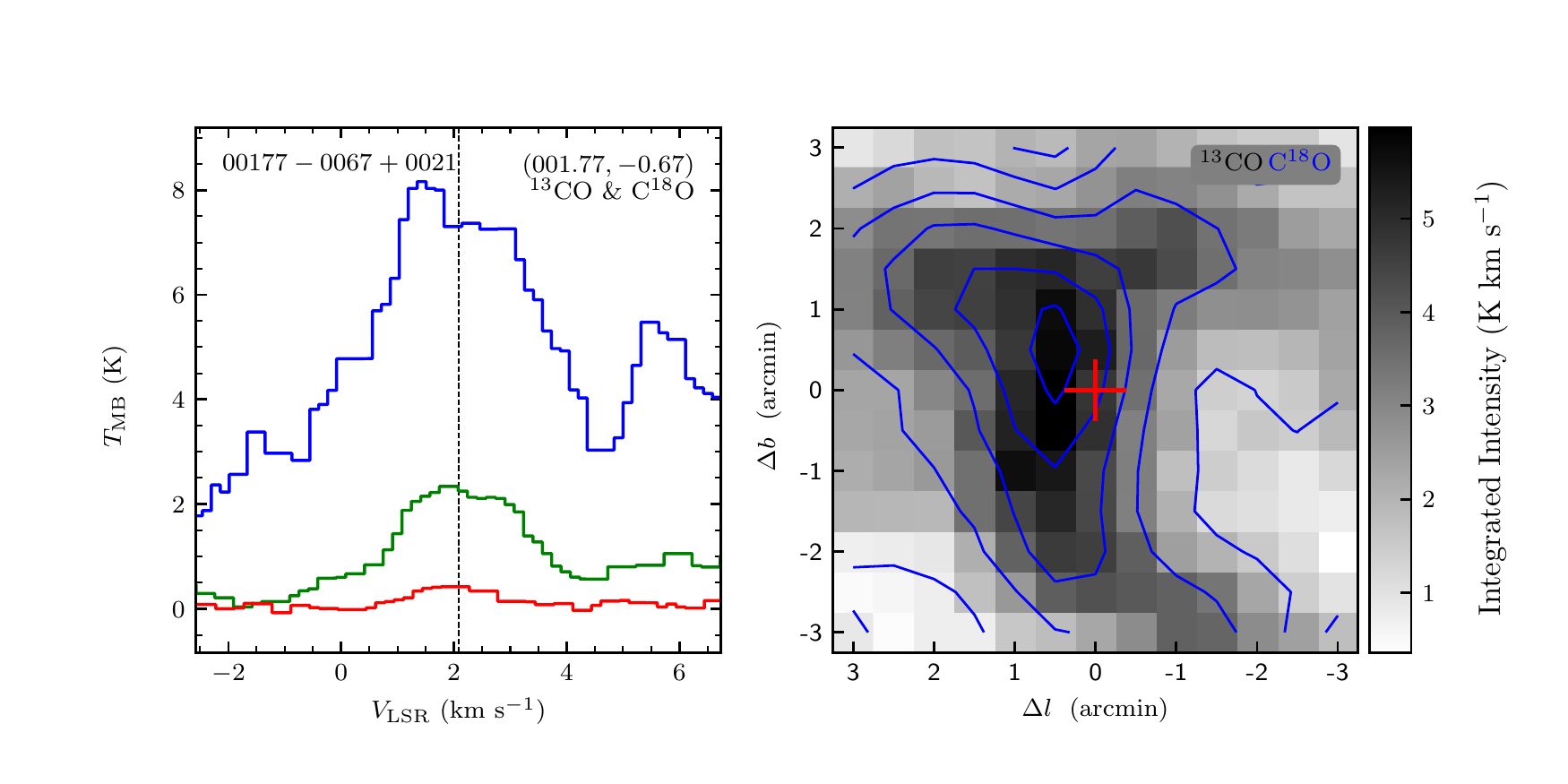}
\end{figure}
\clearpage

\begin{figure}
\includegraphics[width=9.0cm,angle=0]{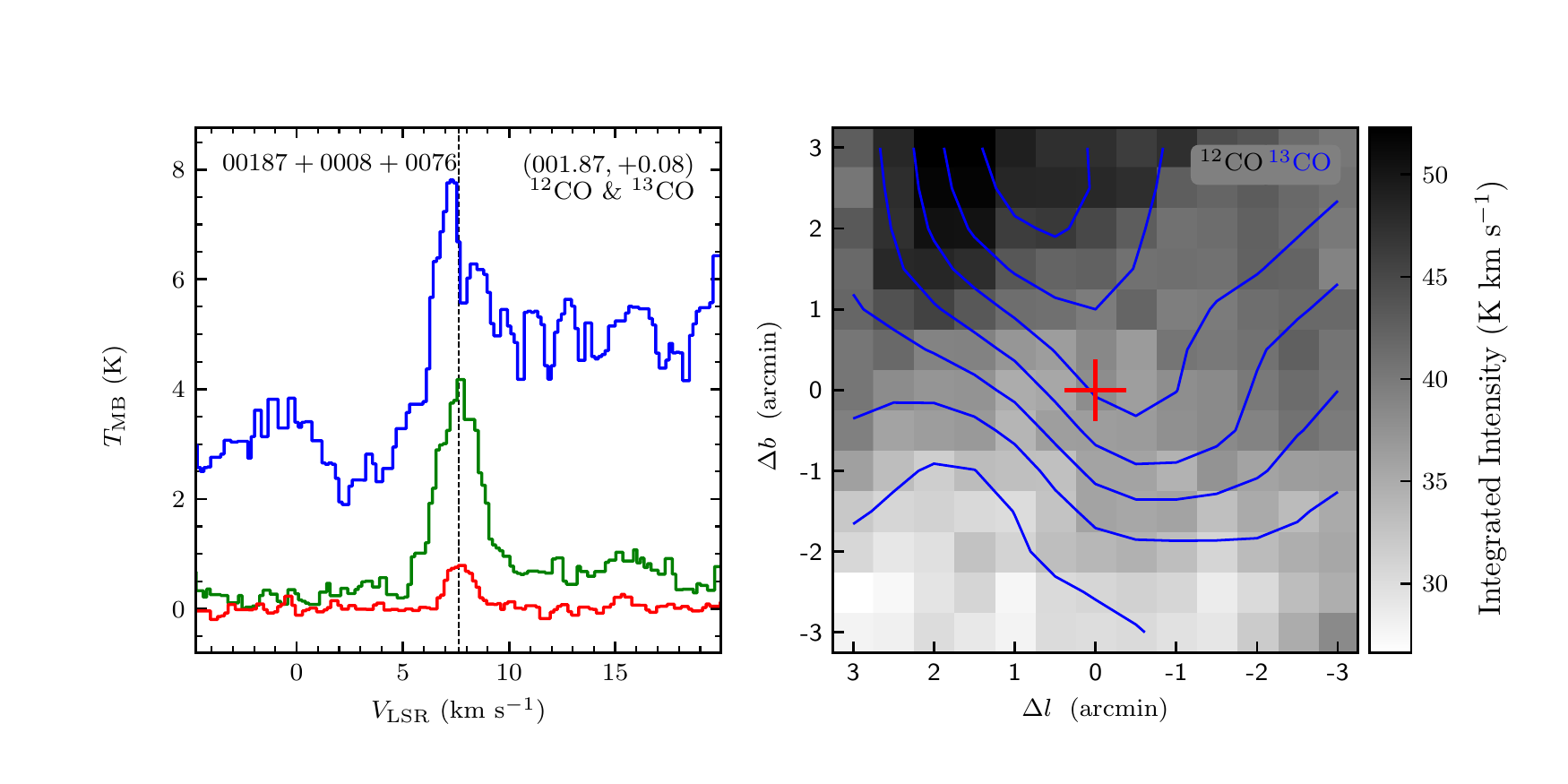}
\includegraphics[width=9.0cm,angle=0]{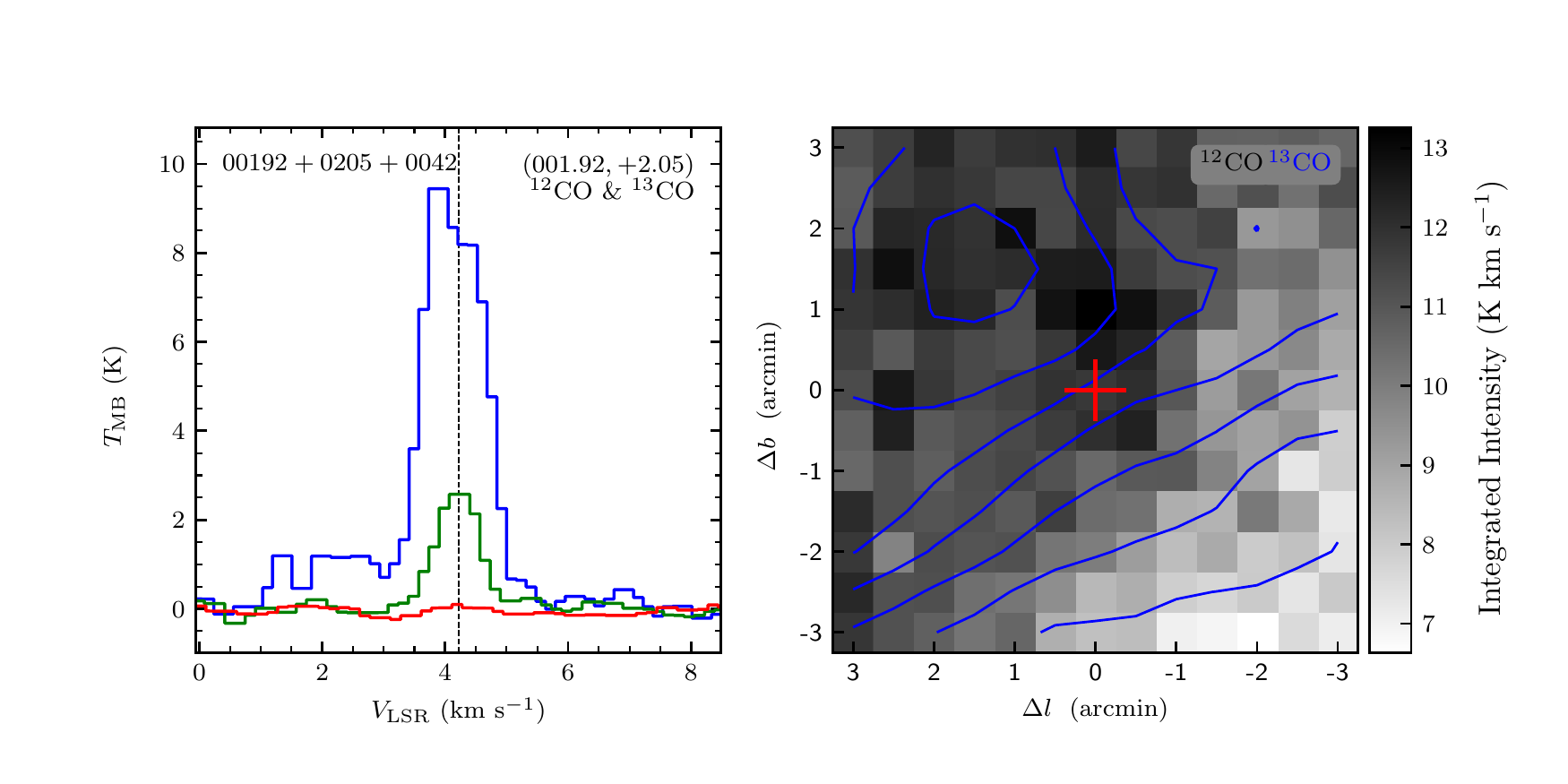}
\vspace{-0.5cm}

\includegraphics[width=9.0cm,angle=0]{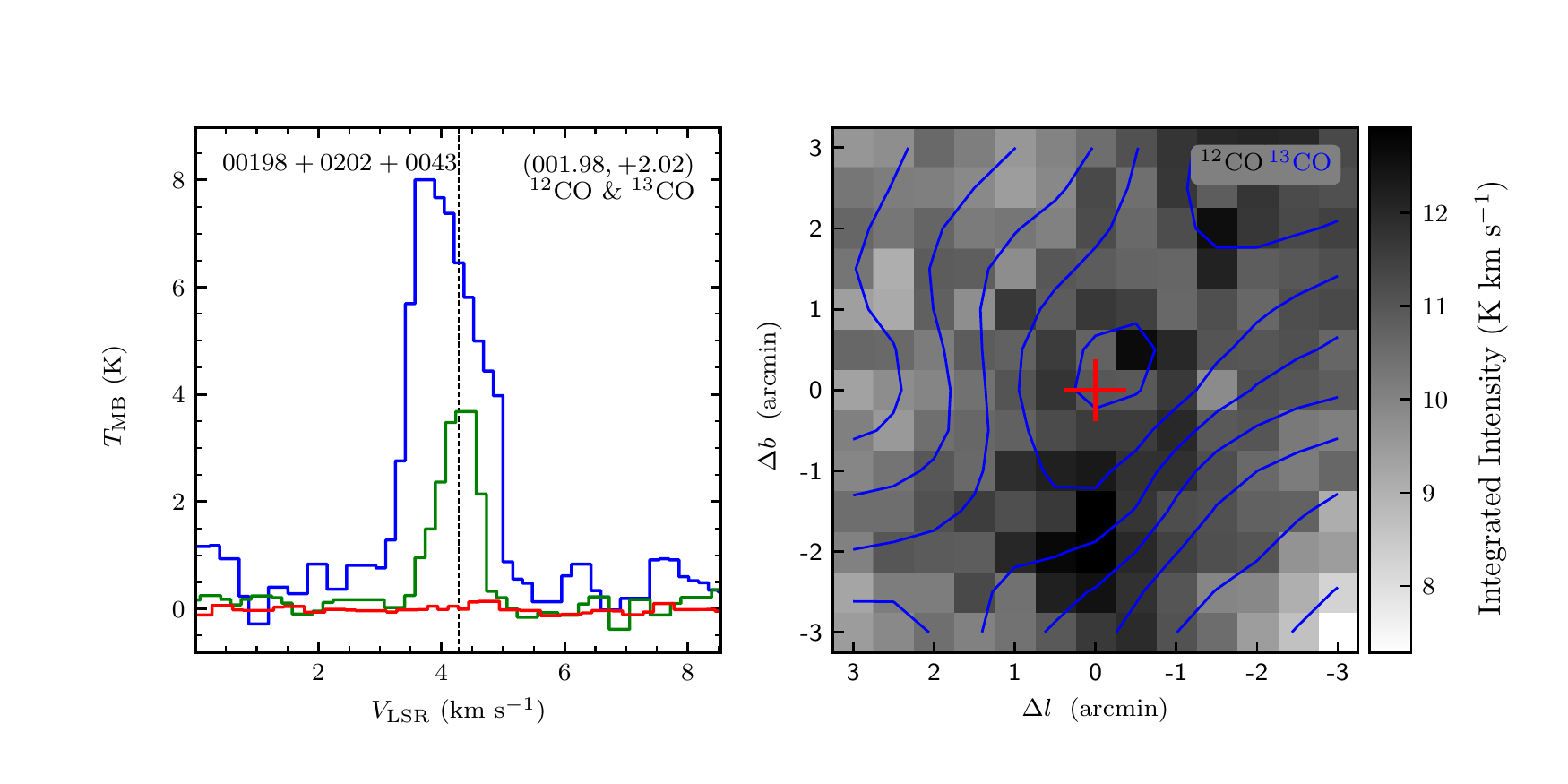}
\includegraphics[width=9.0cm,angle=0]{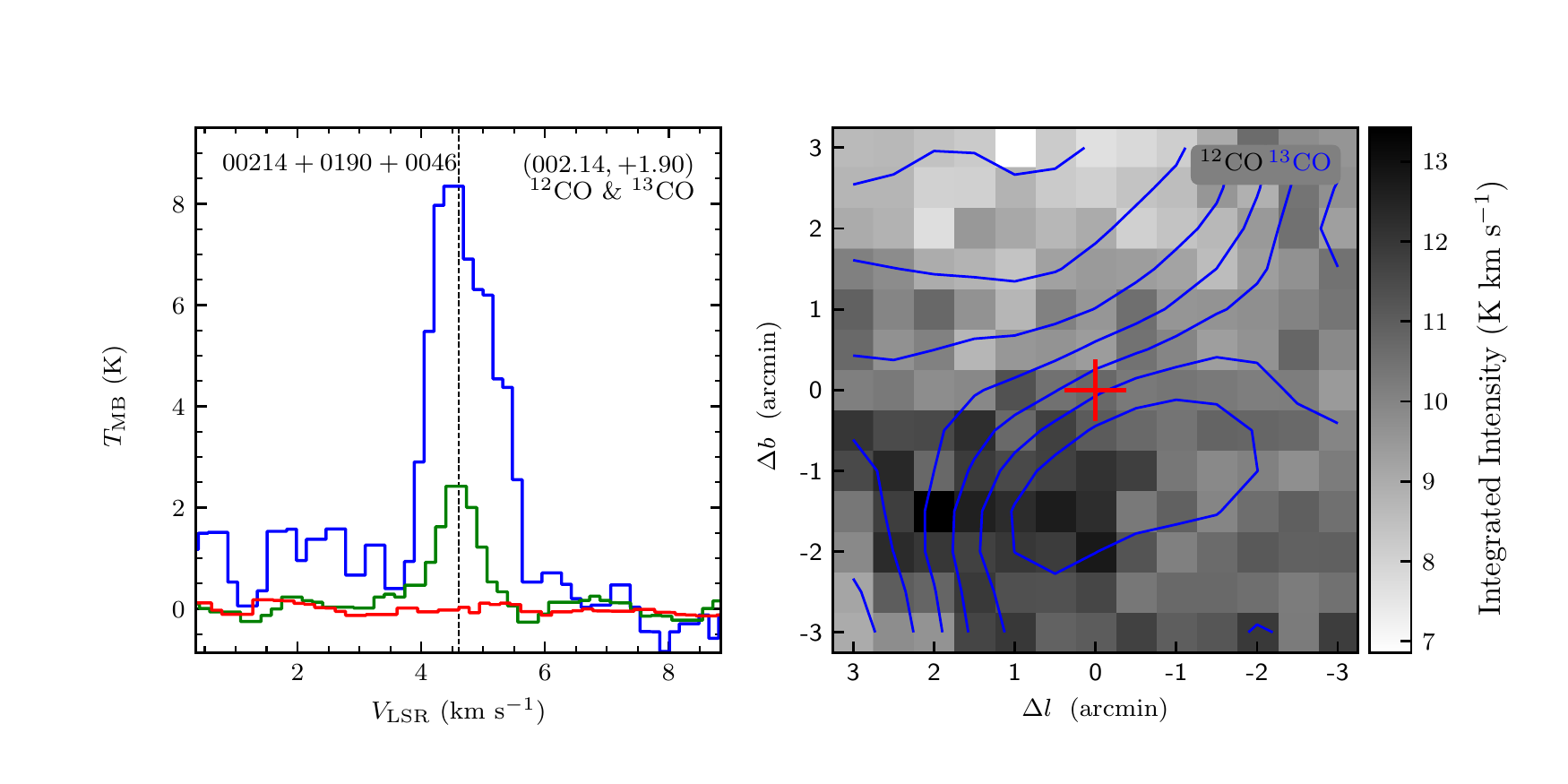}
\vspace{-0.5cm}

\includegraphics[width=9.0cm,angle=0]{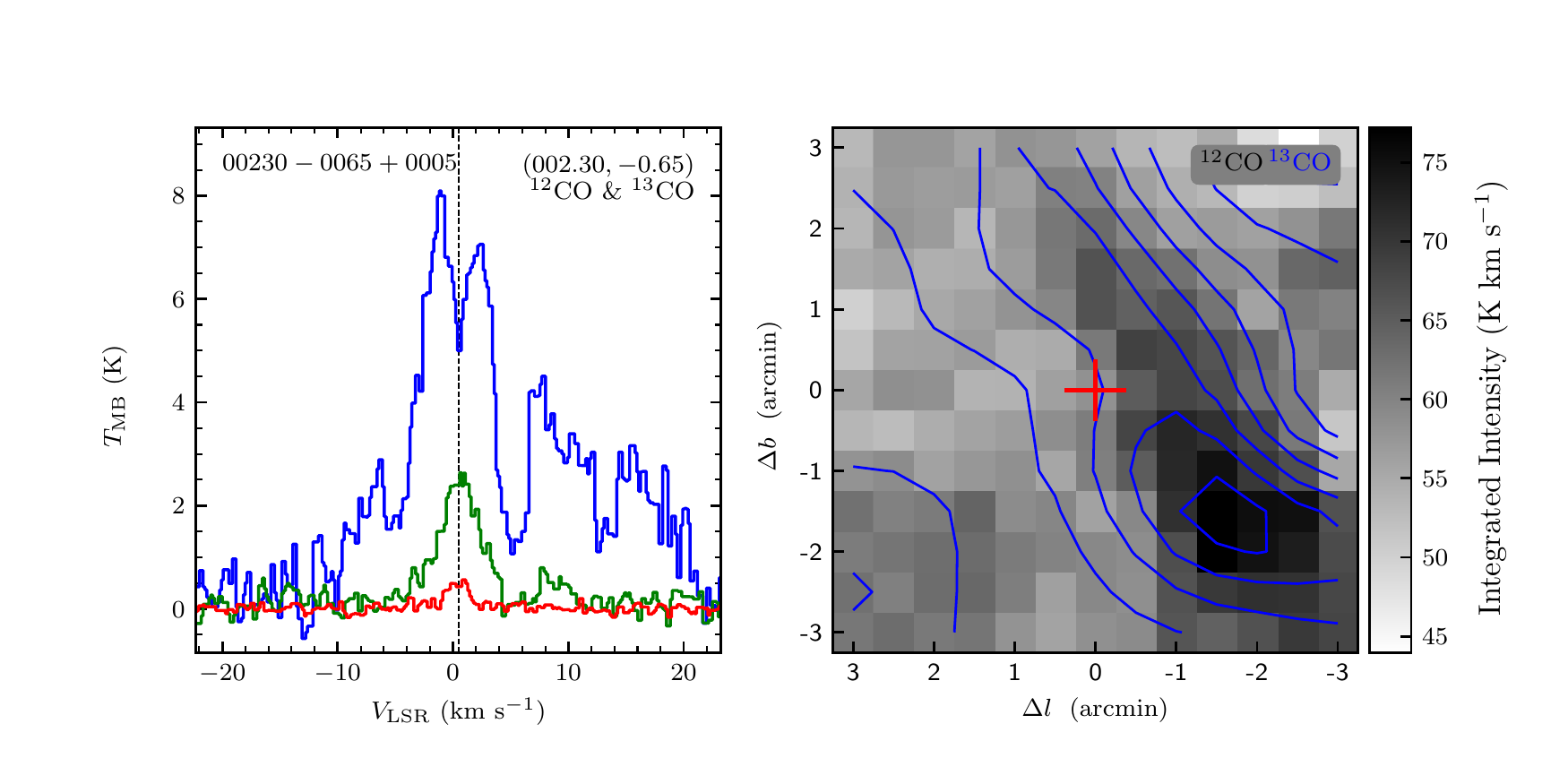}
\includegraphics[width=9.0cm,angle=0]{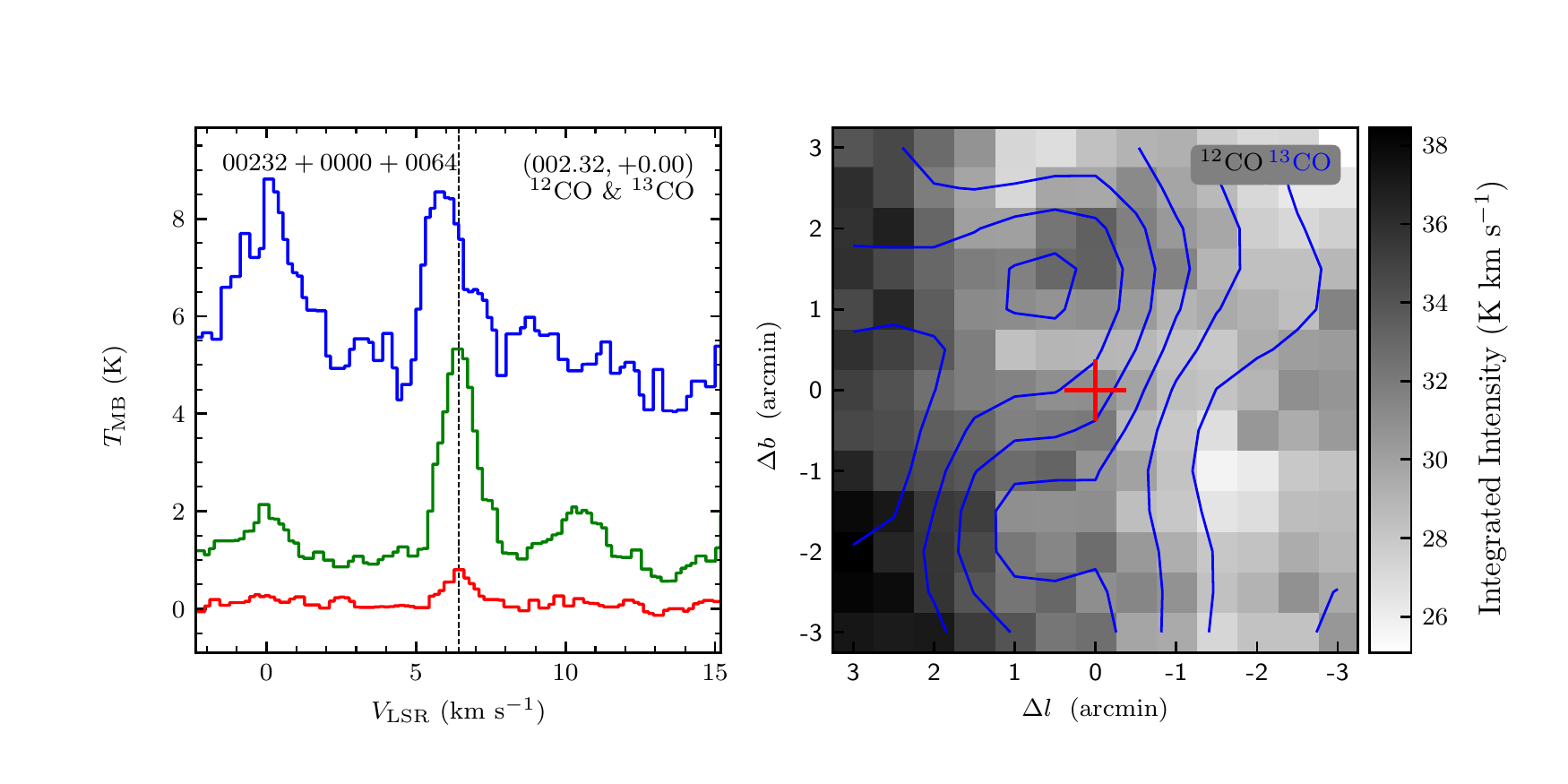}
\vspace{-0.5cm}

\includegraphics[width=9.0cm,angle=0]{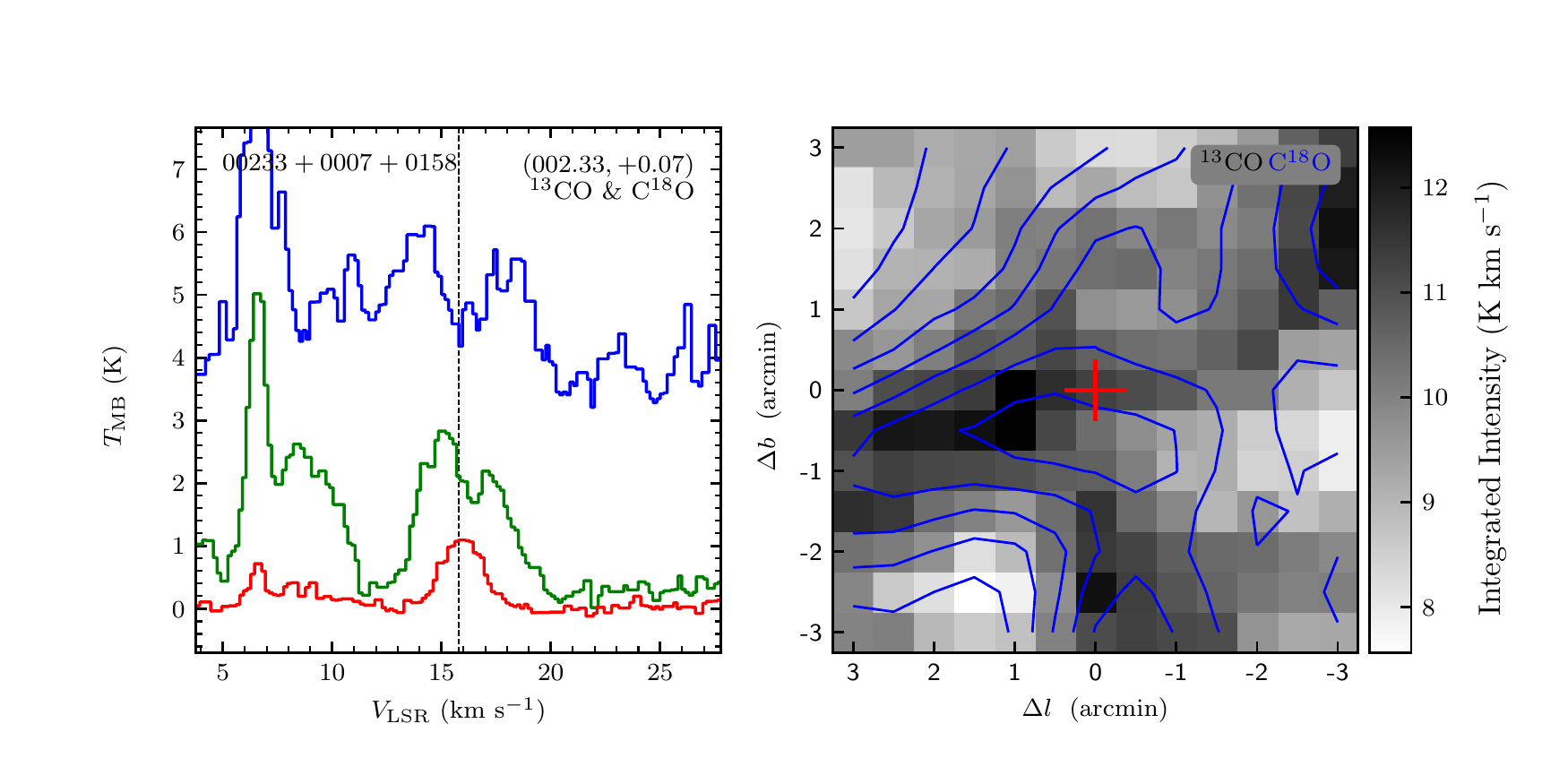}
\includegraphics[width=9.0cm,angle=0]{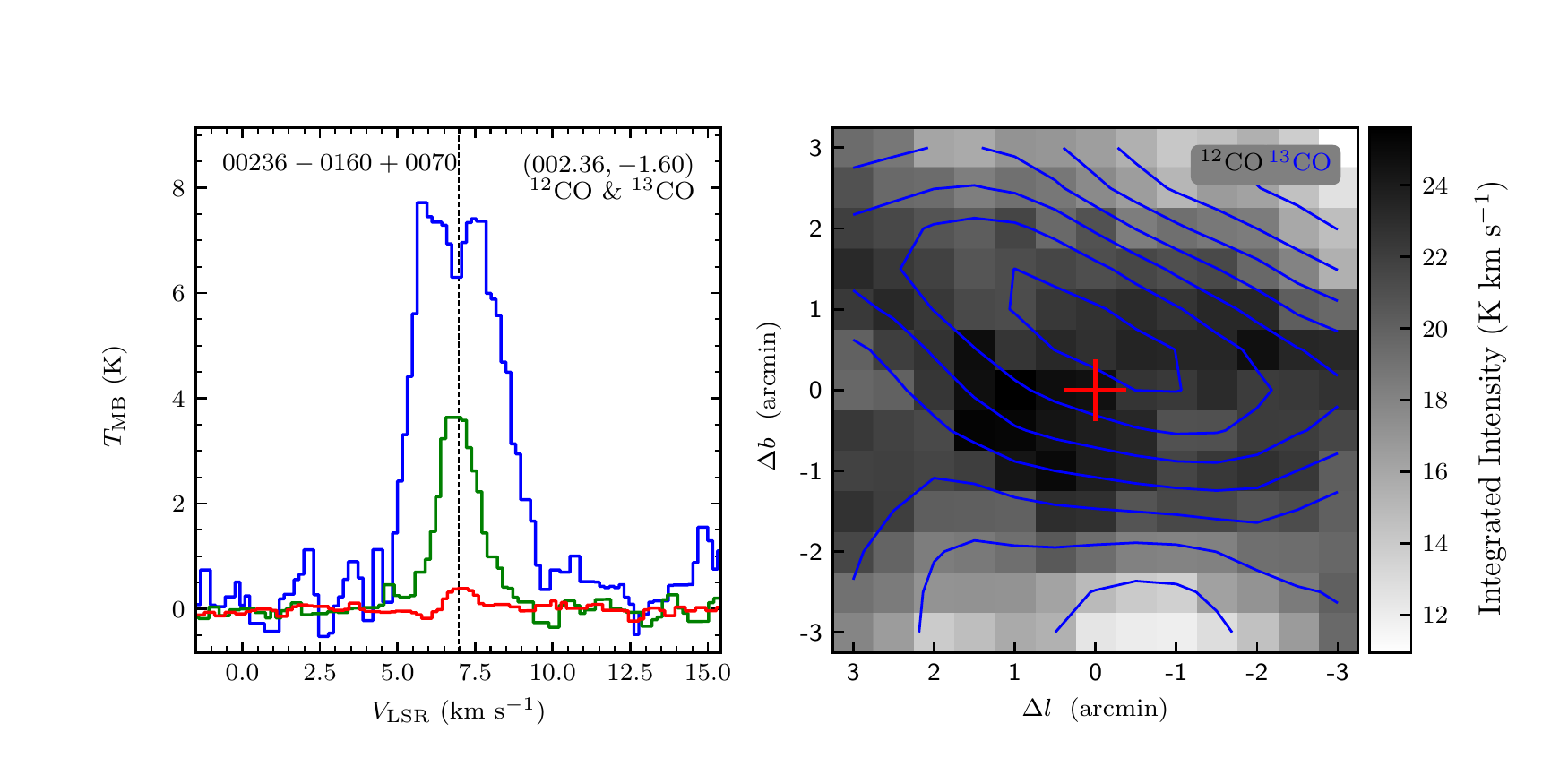}
\vspace{-0.5cm}

\includegraphics[width=9.0cm,angle=0]{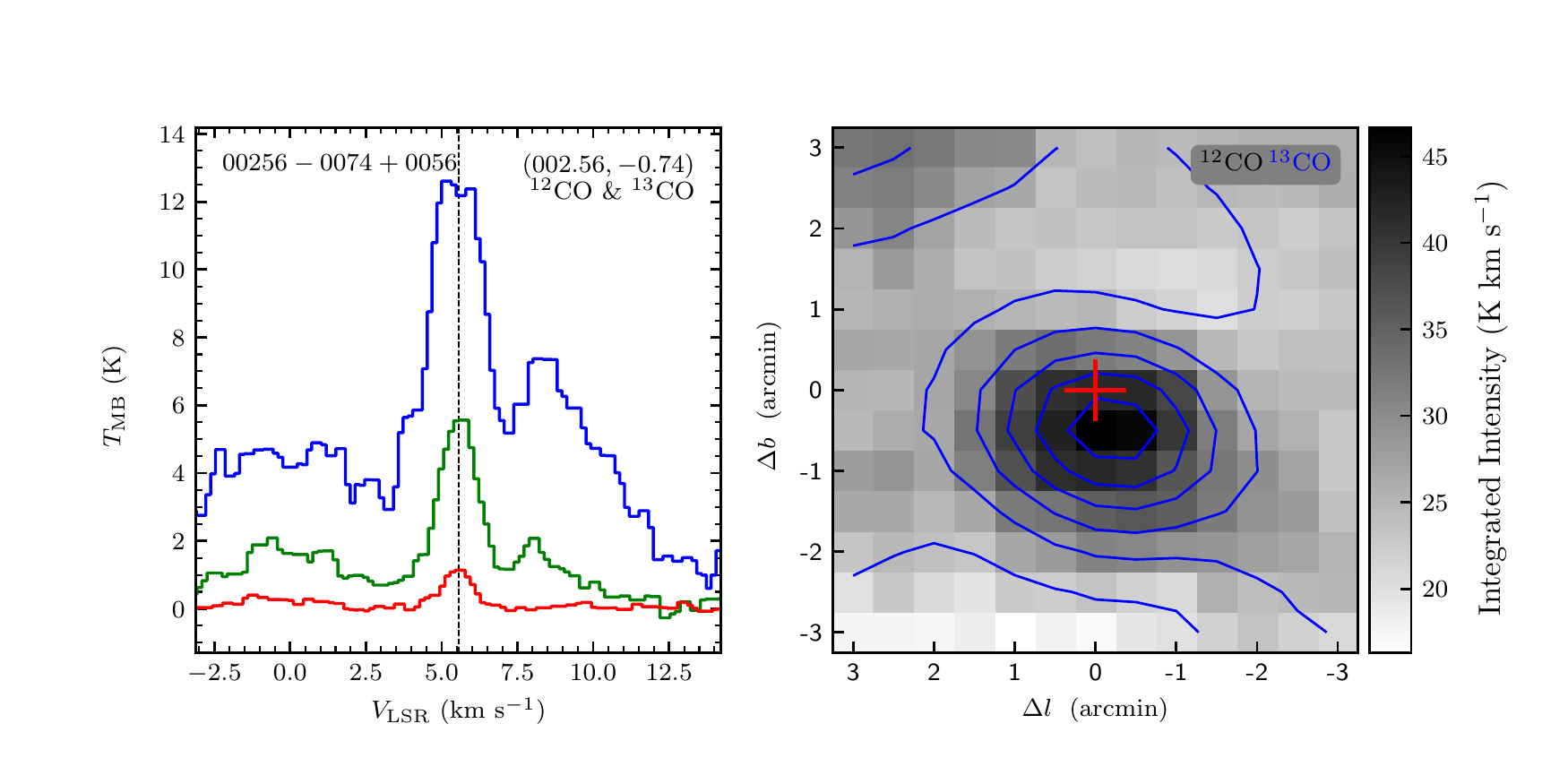}
\includegraphics[width=9.0cm,angle=0]{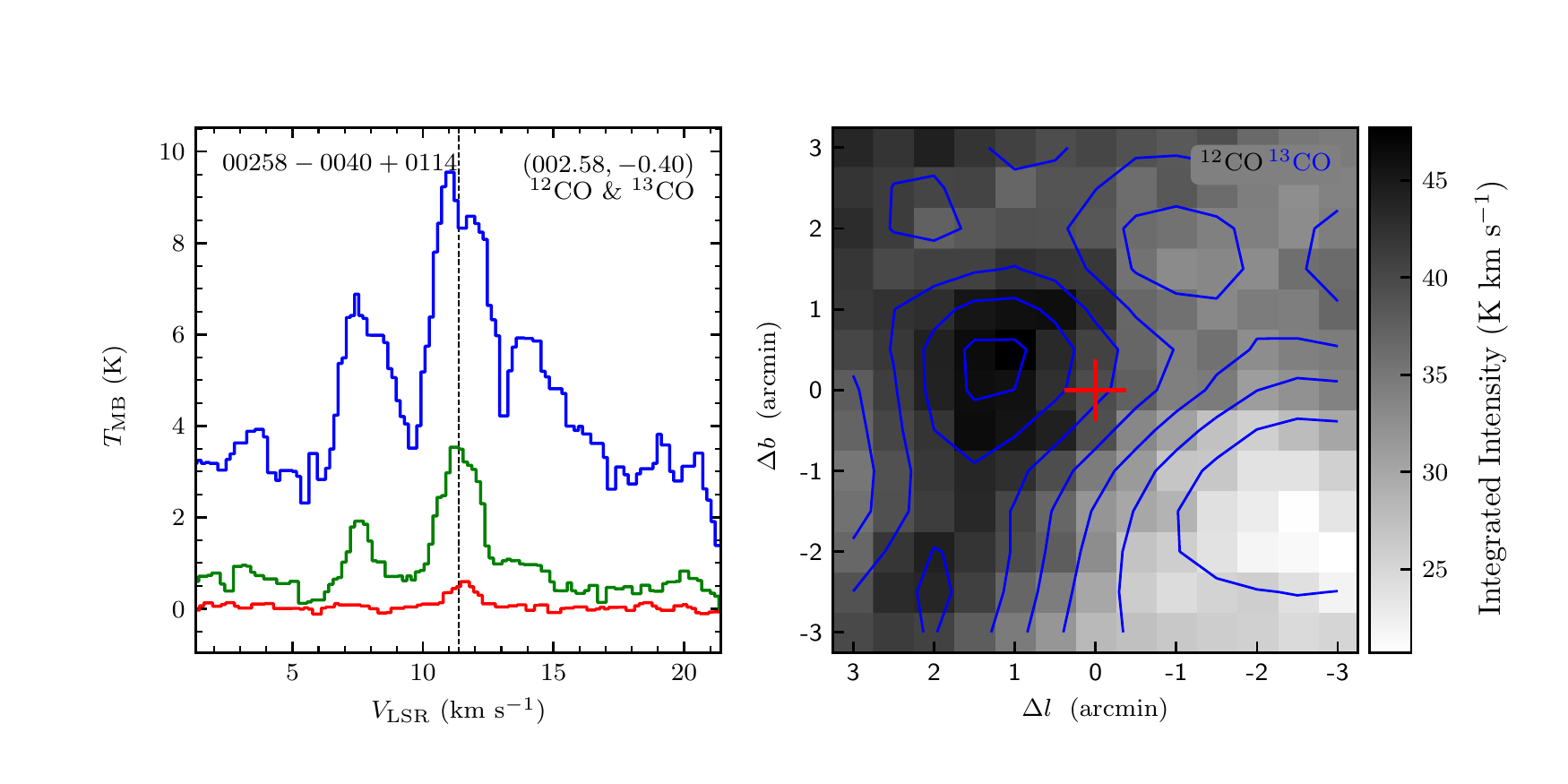}
\end{figure}
\clearpage

\begin{figure}
\includegraphics[width=9.0cm,angle=0]{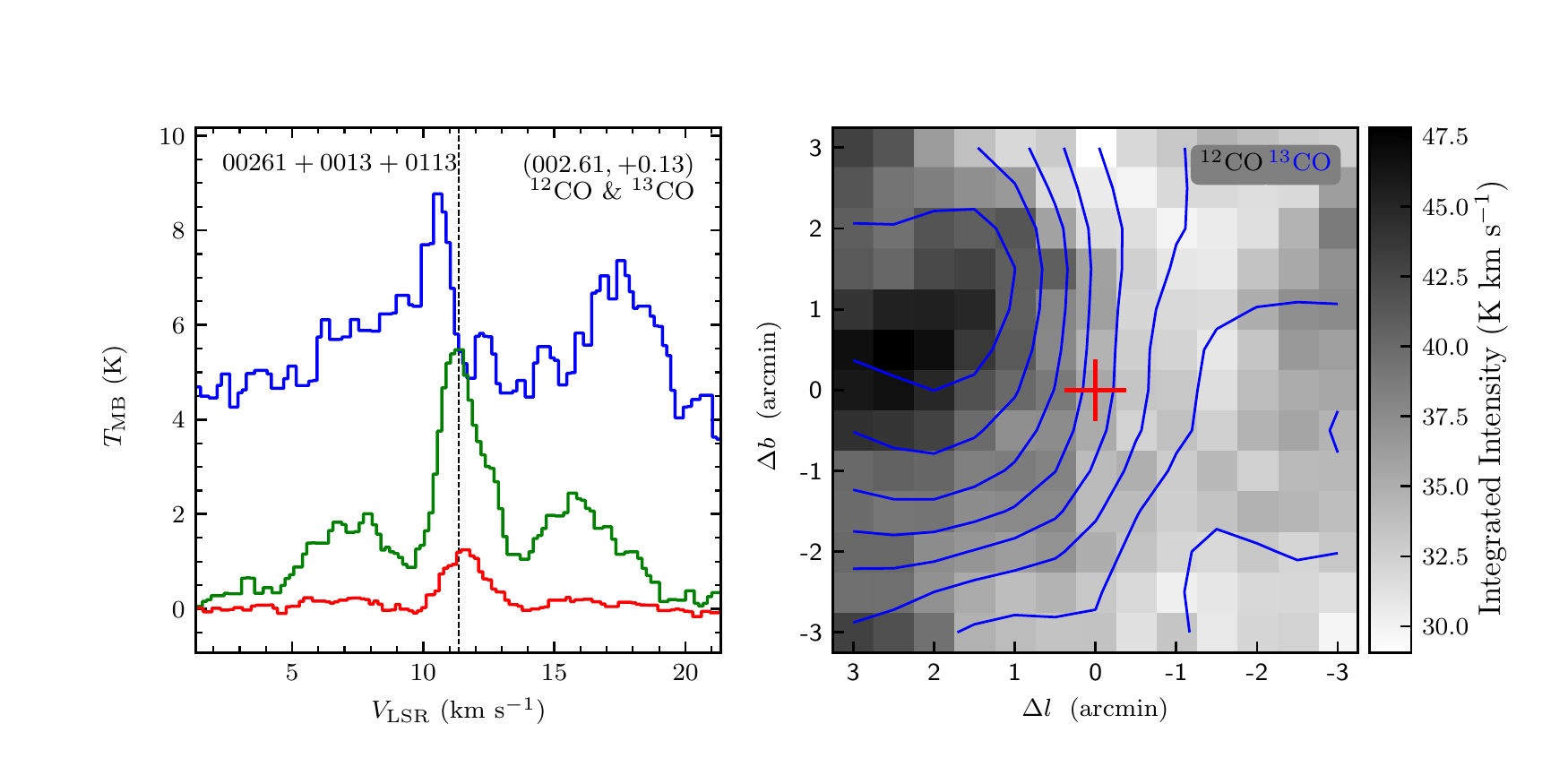}
\includegraphics[width=9.0cm,angle=0]{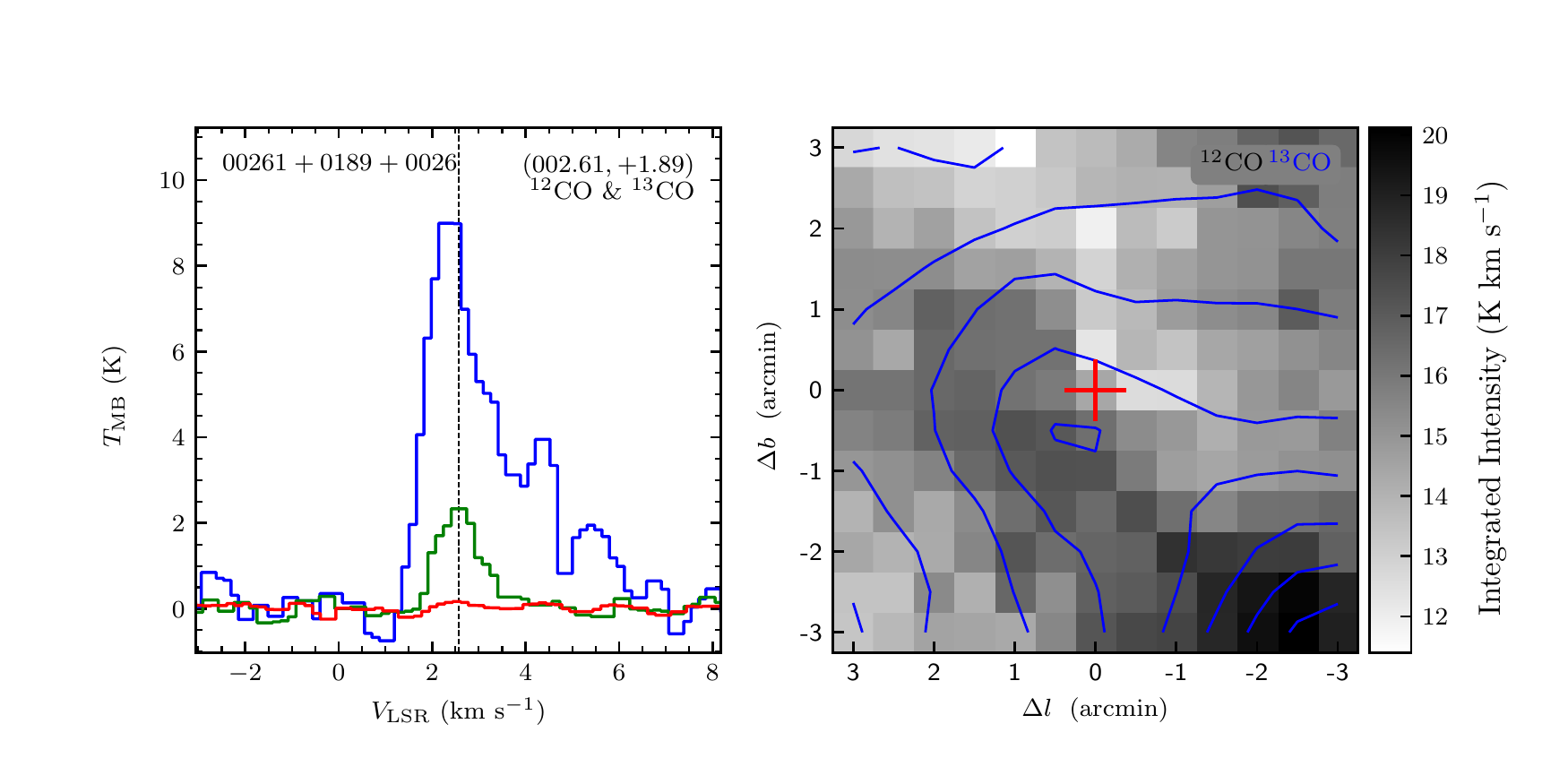}
\vspace{-0.5cm}

\includegraphics[width=9.0cm,angle=0]{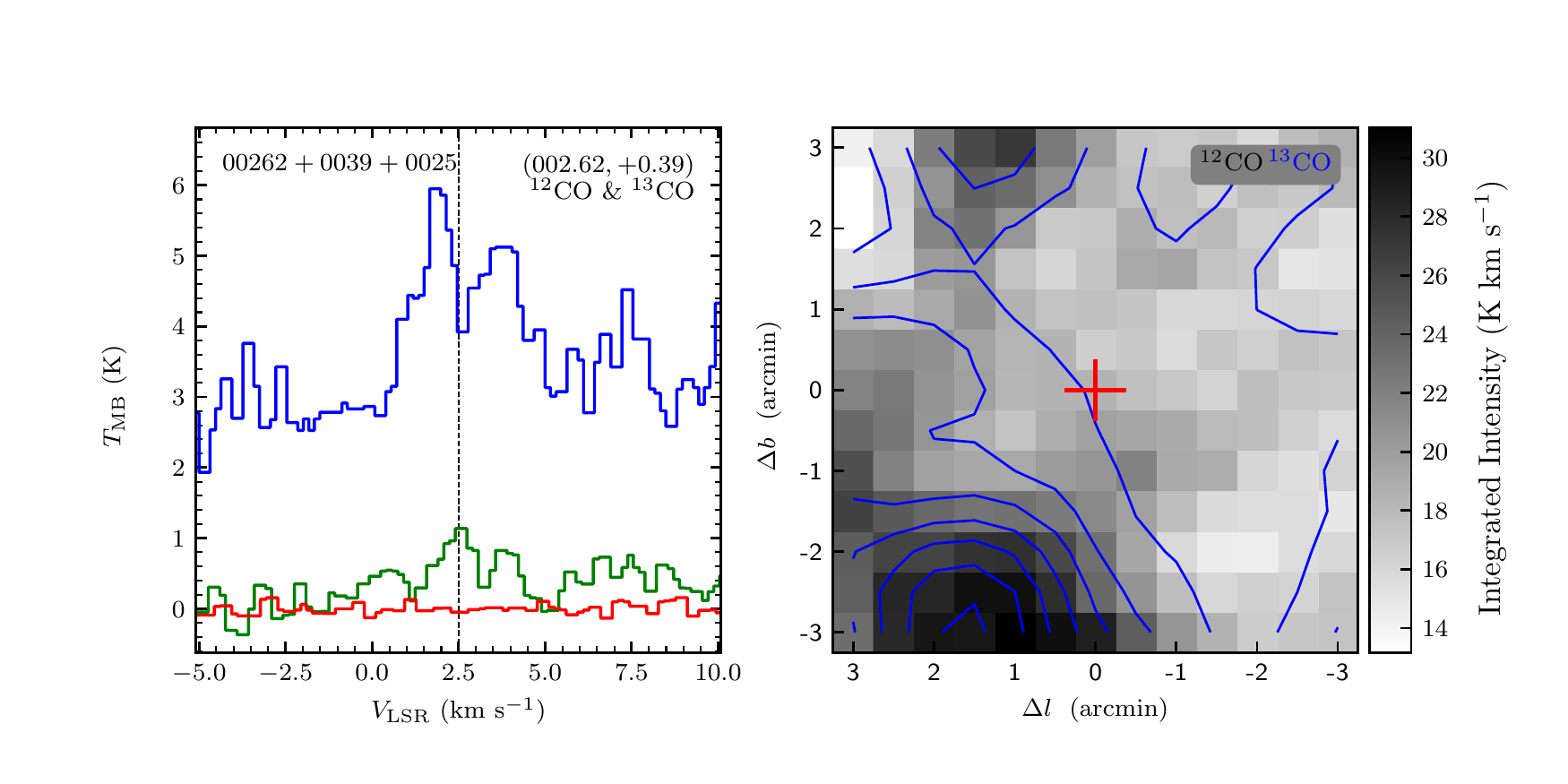}
\includegraphics[width=9.0cm,angle=0]{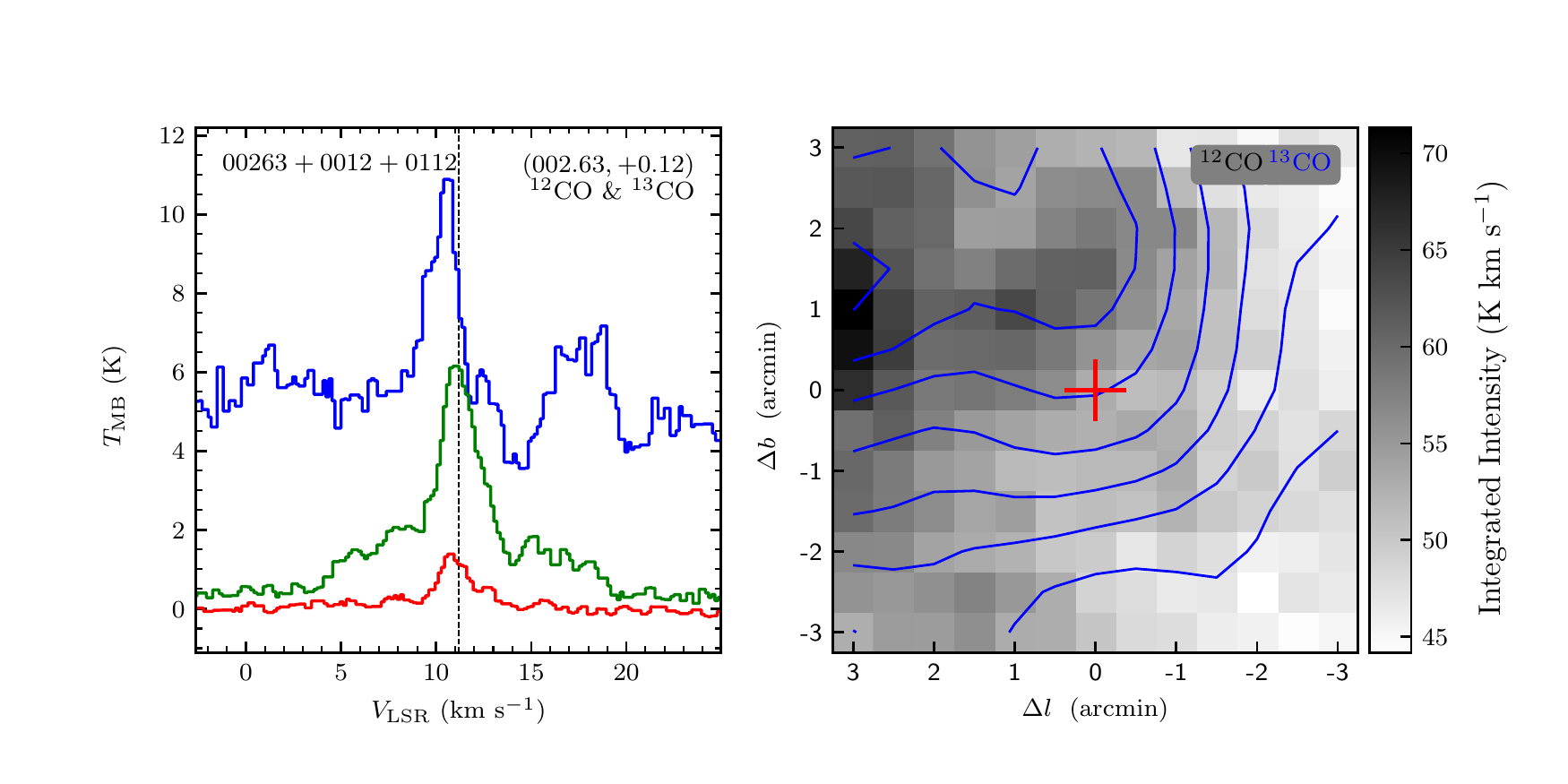}
\vspace{-0.5cm}

\includegraphics[width=9.0cm,angle=0]{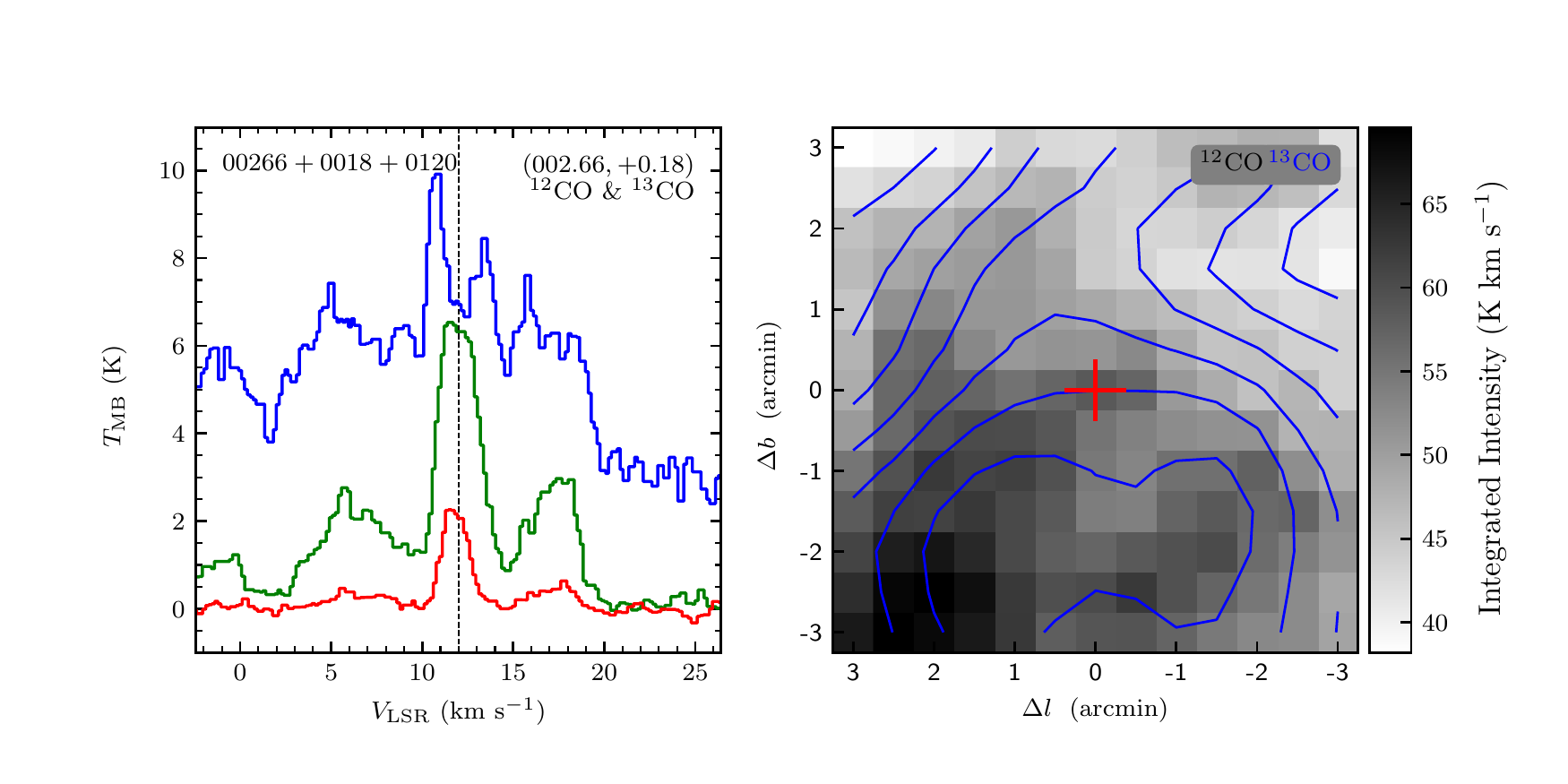}
\includegraphics[width=9.0cm,angle=0]{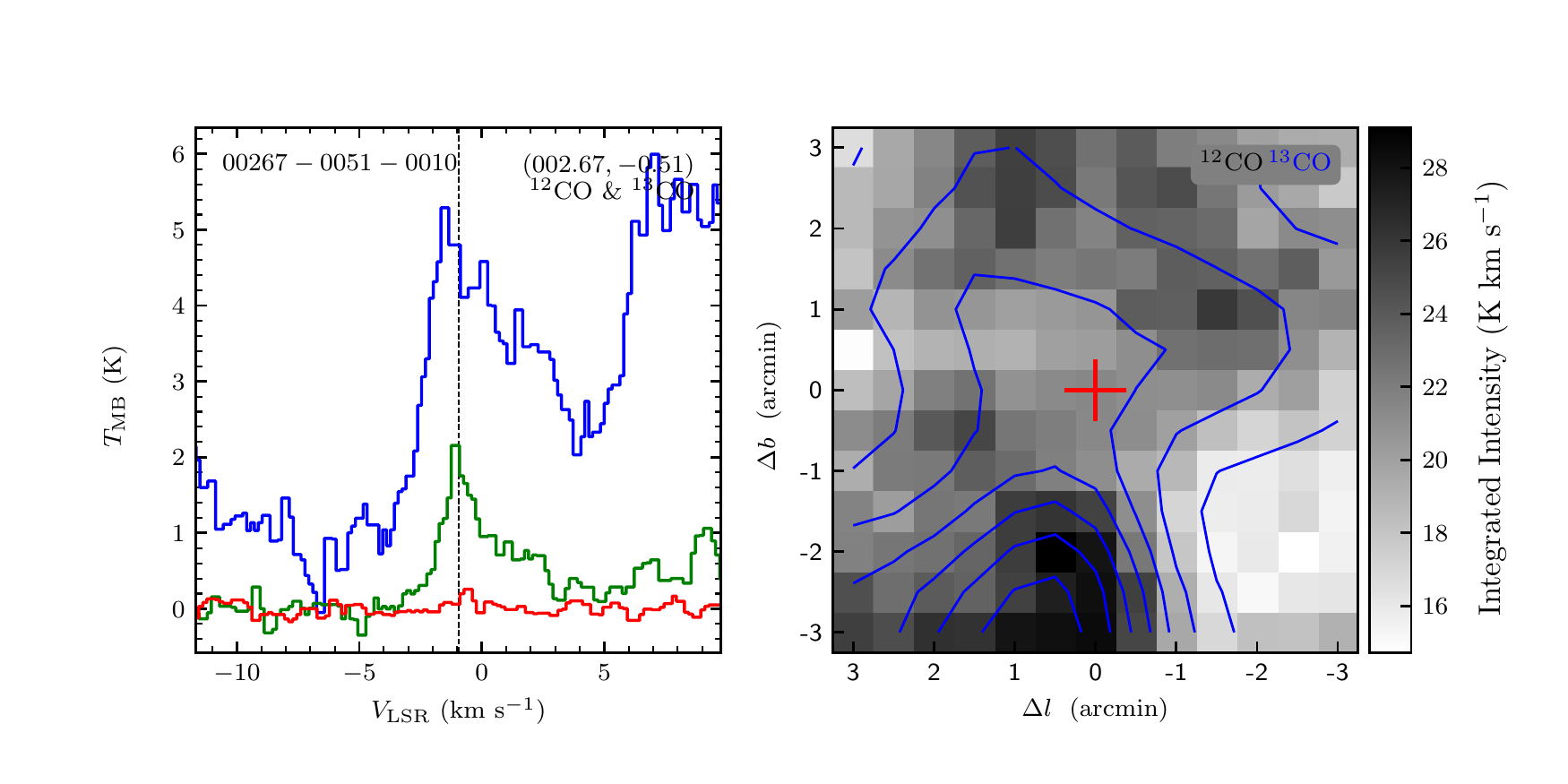}
\vspace{-0.5cm}

\includegraphics[width=9.0cm,angle=0]{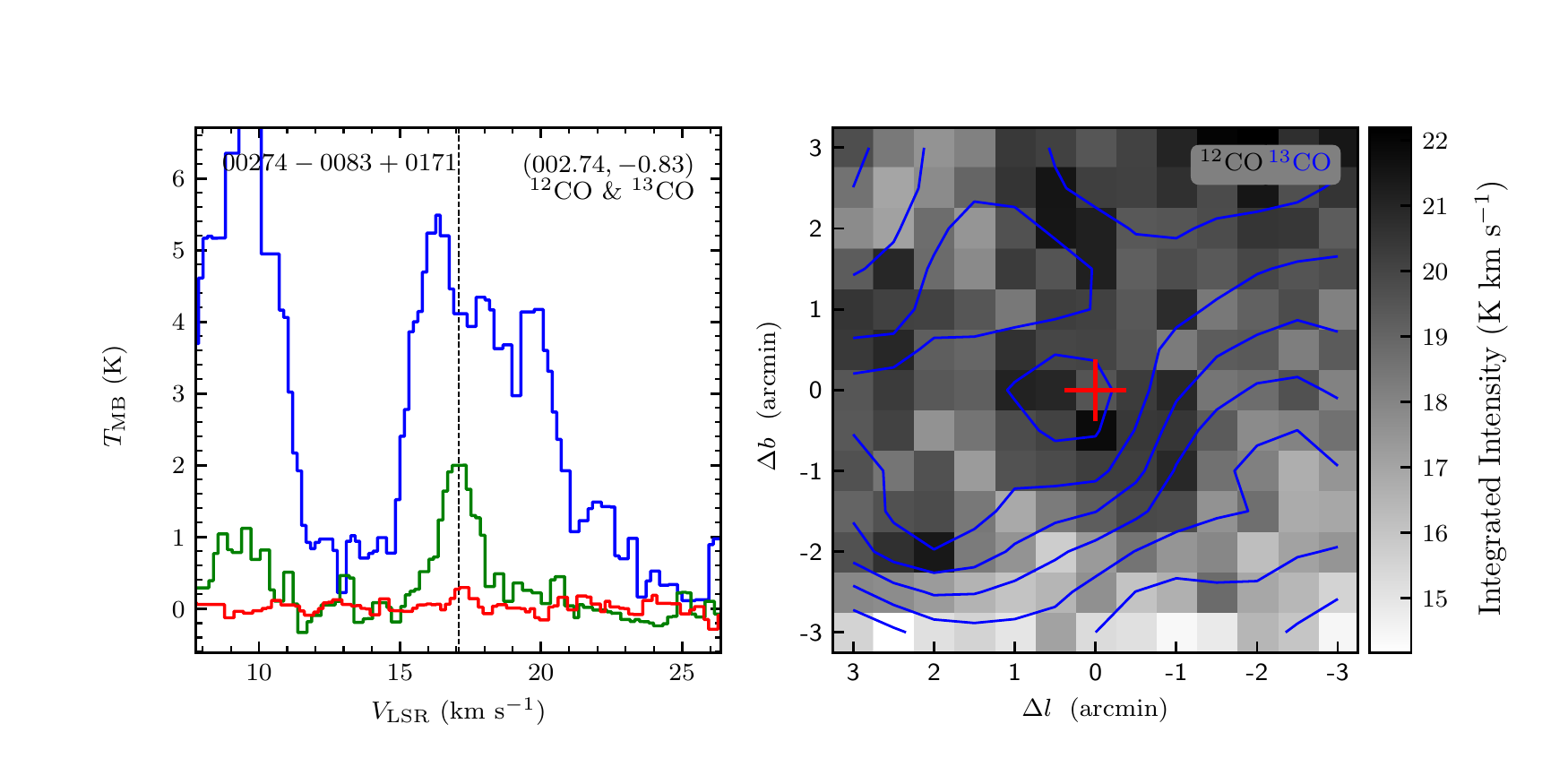}
\includegraphics[width=9.0cm,angle=0]{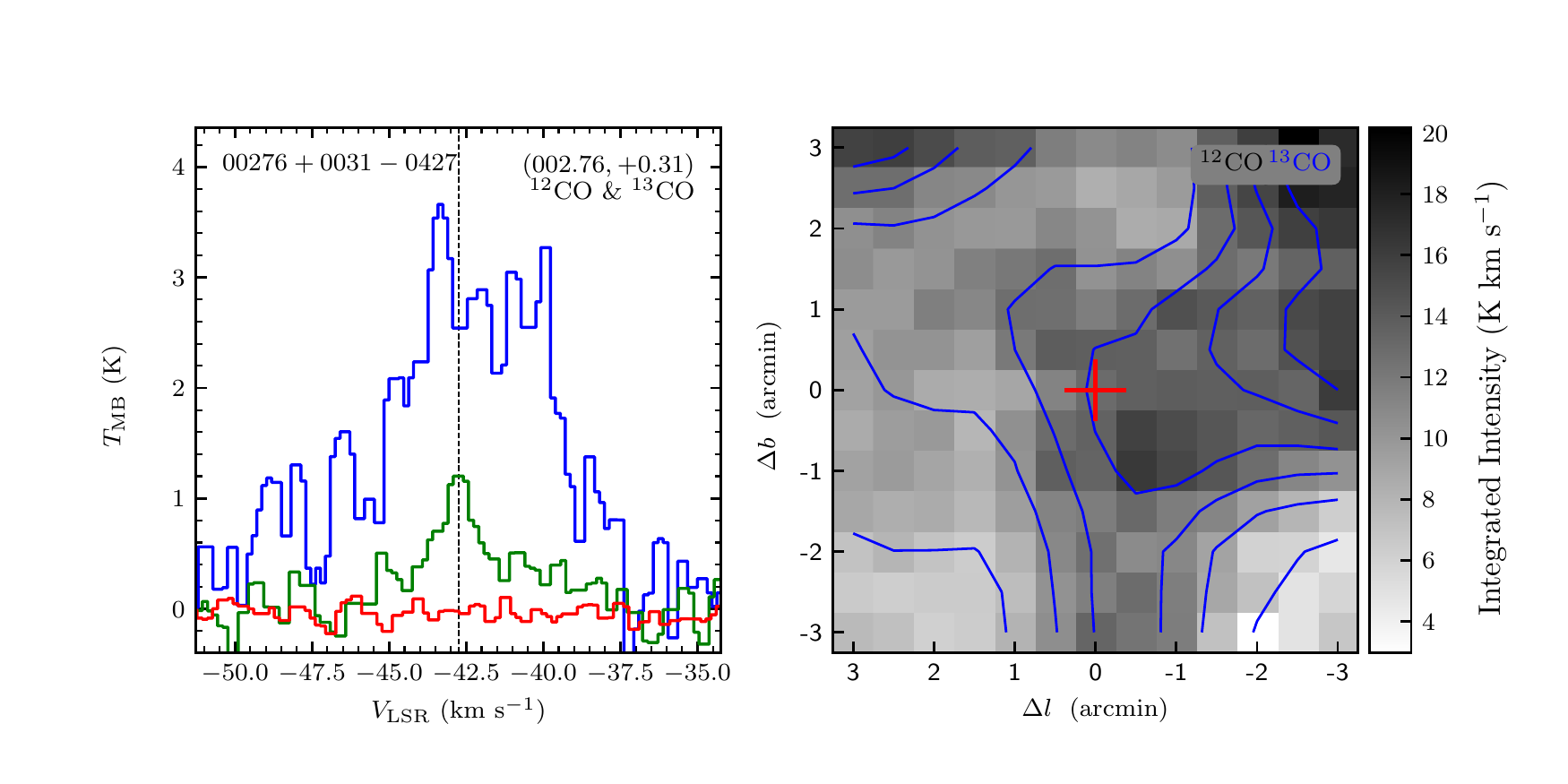}
\vspace{-0.5cm}

\includegraphics[width=9.0cm,angle=0]{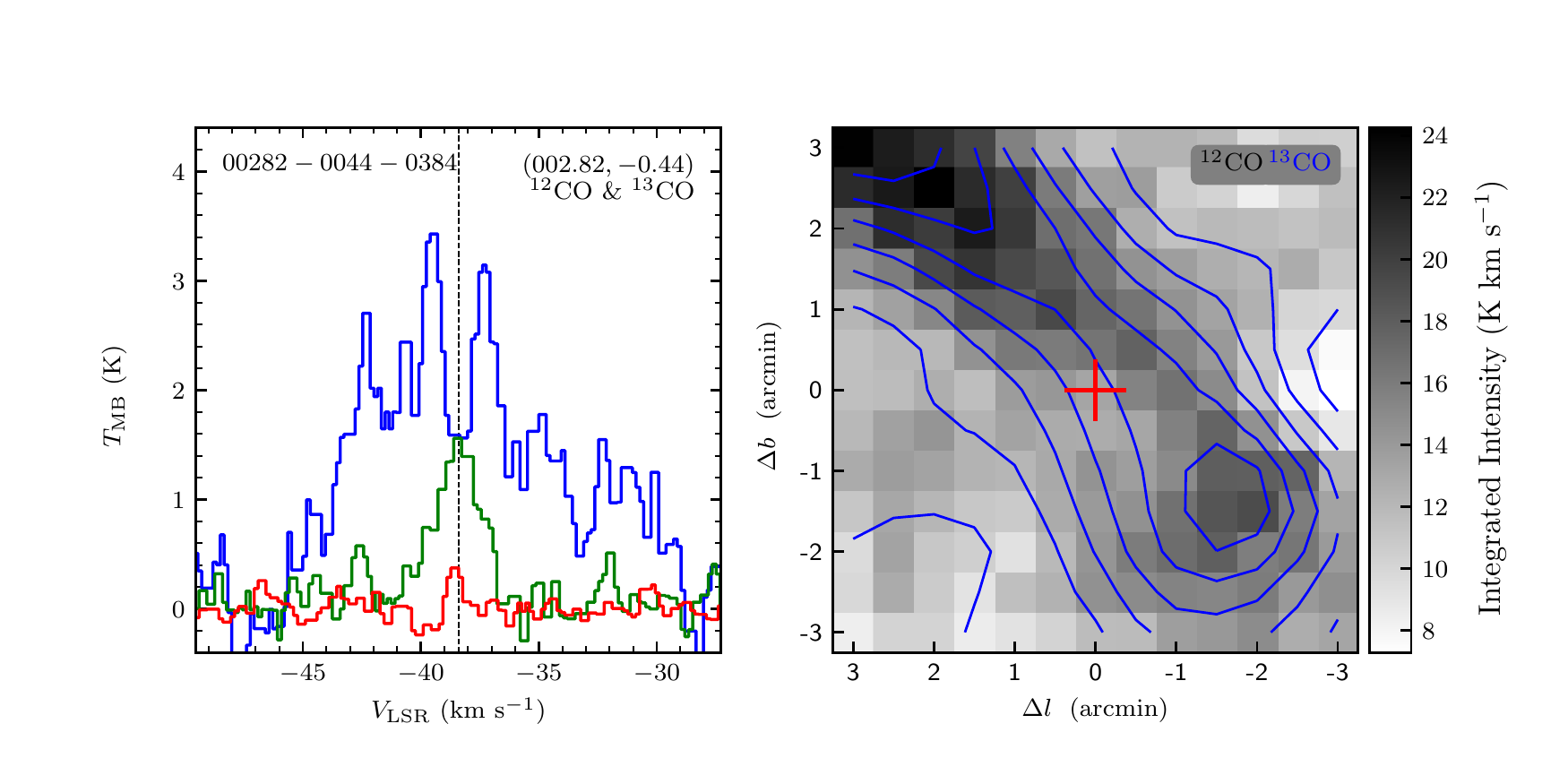}
\includegraphics[width=9.0cm,angle=0]{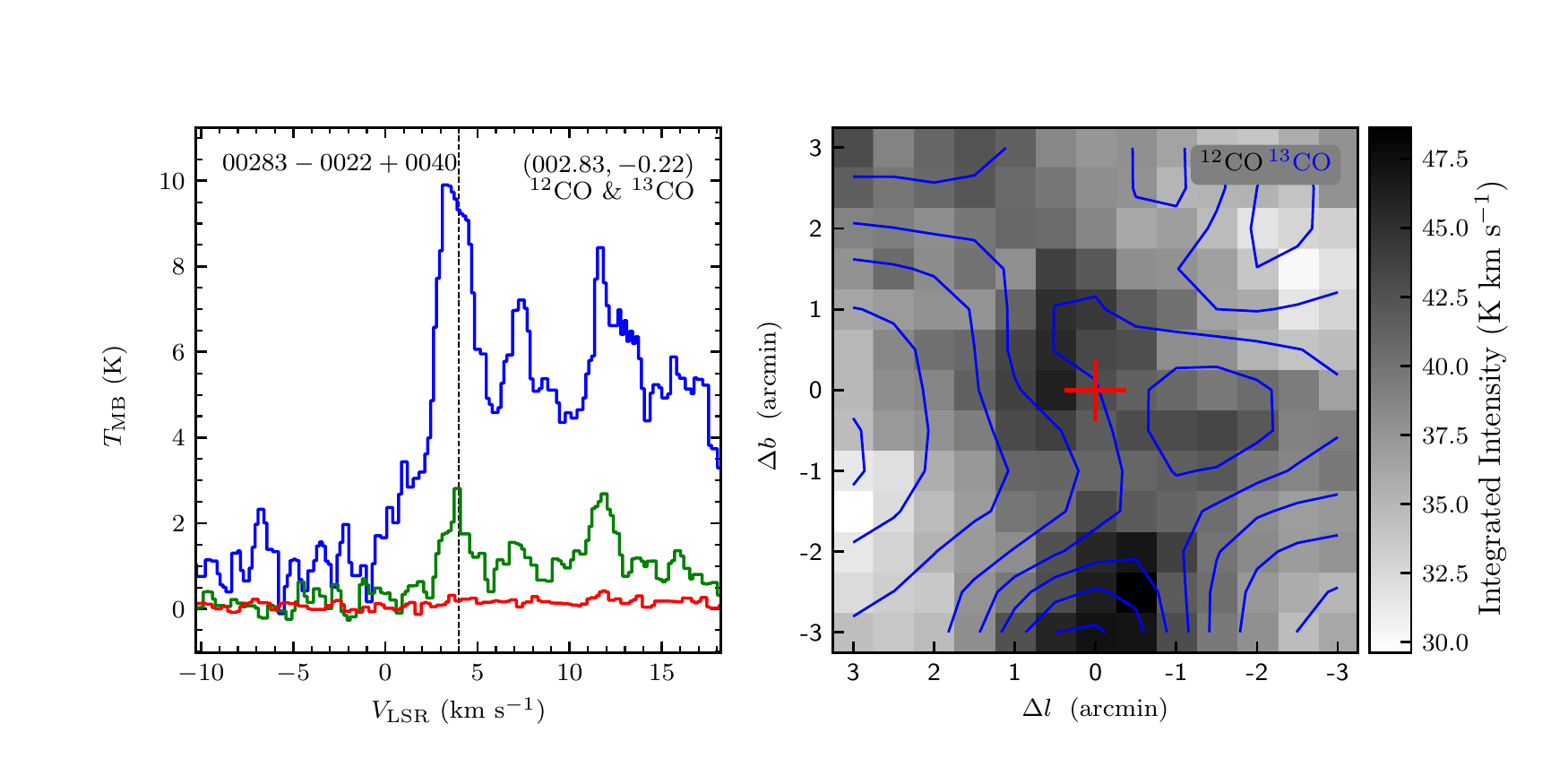}
\end{figure}
\clearpage

\begin{figure}
\includegraphics[width=9.0cm,angle=0]{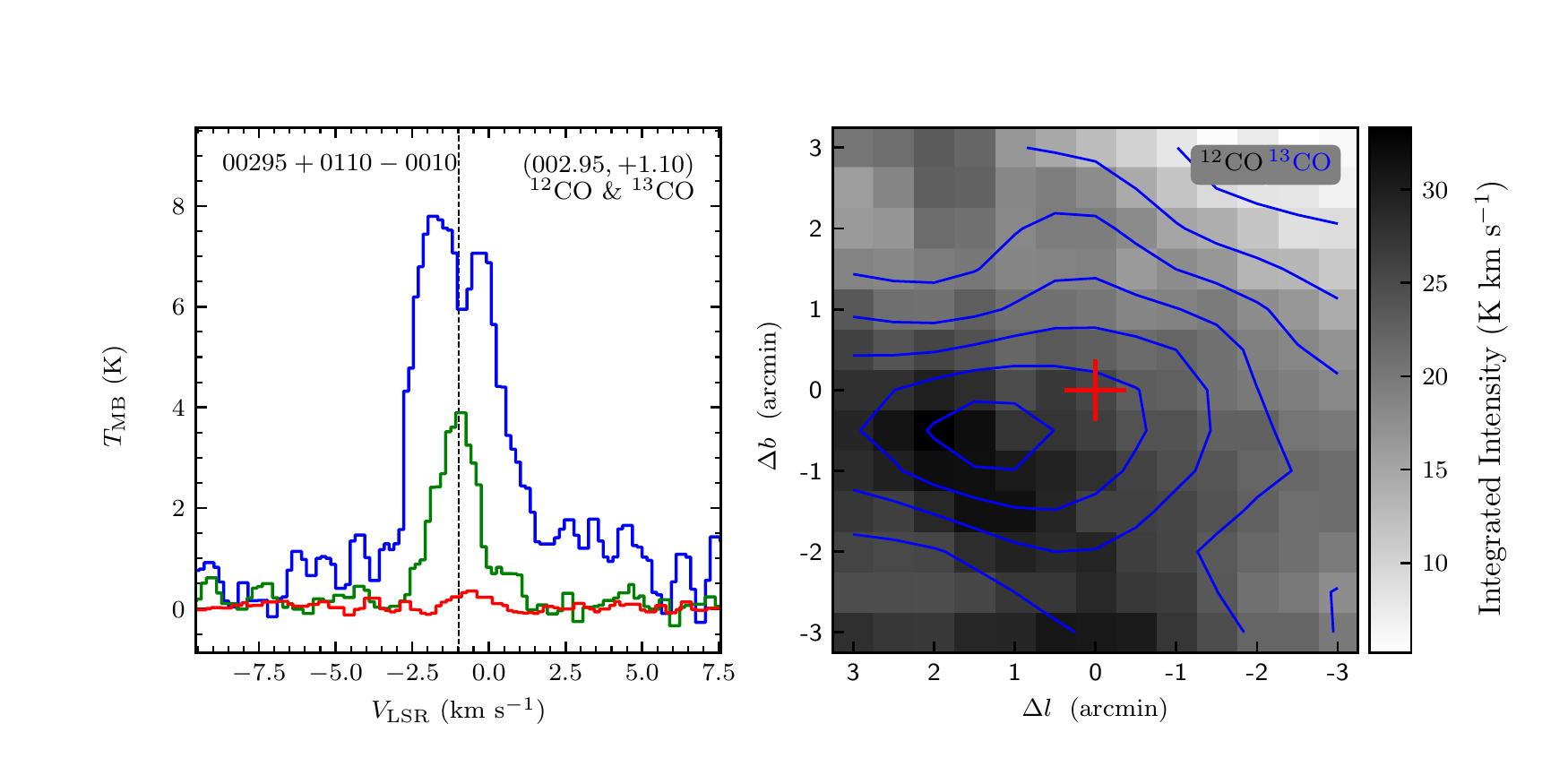}
\includegraphics[width=9.0cm,angle=0]{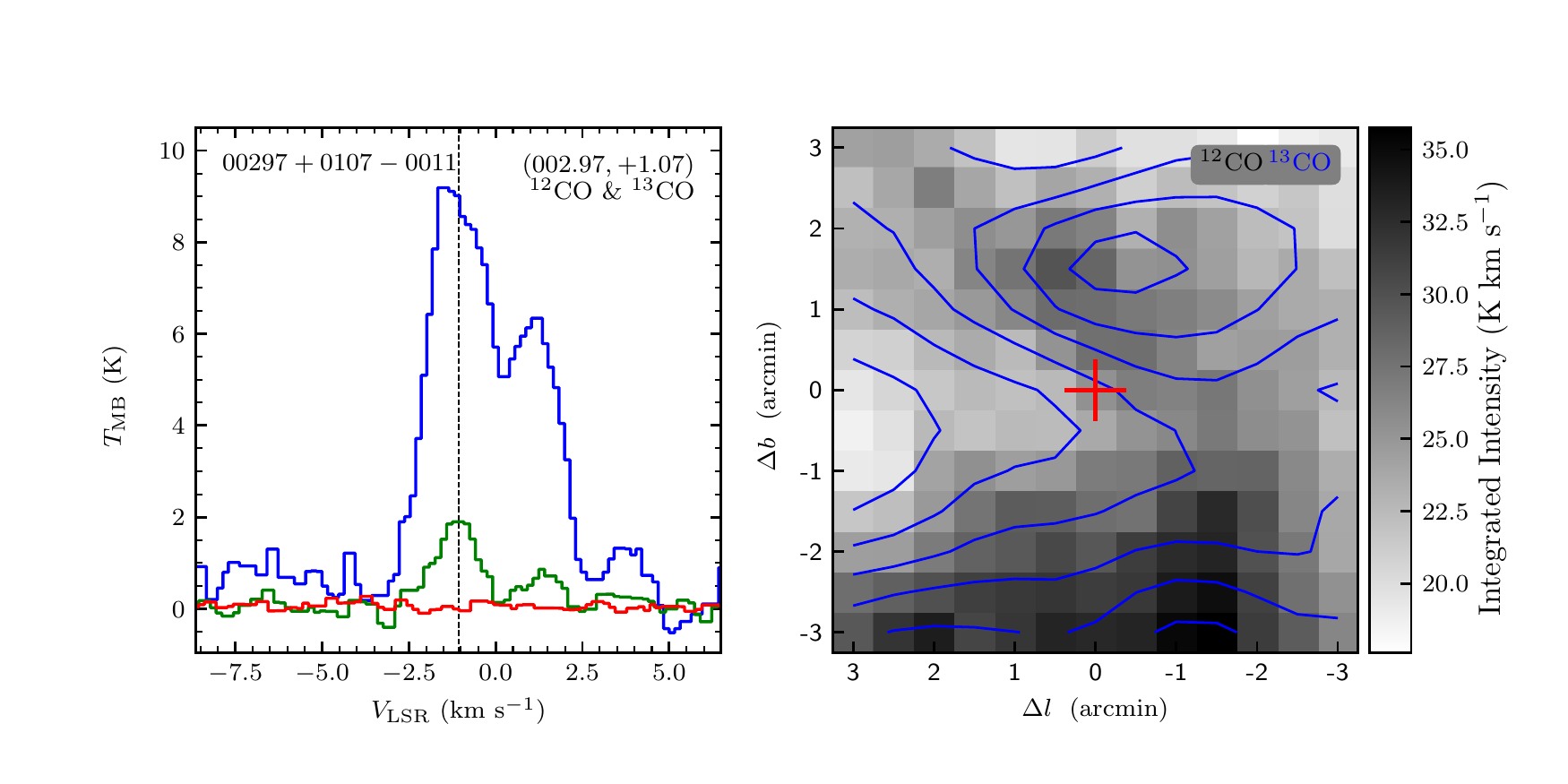}
\vspace{-0.5cm}

\includegraphics[width=9.0cm,angle=0]{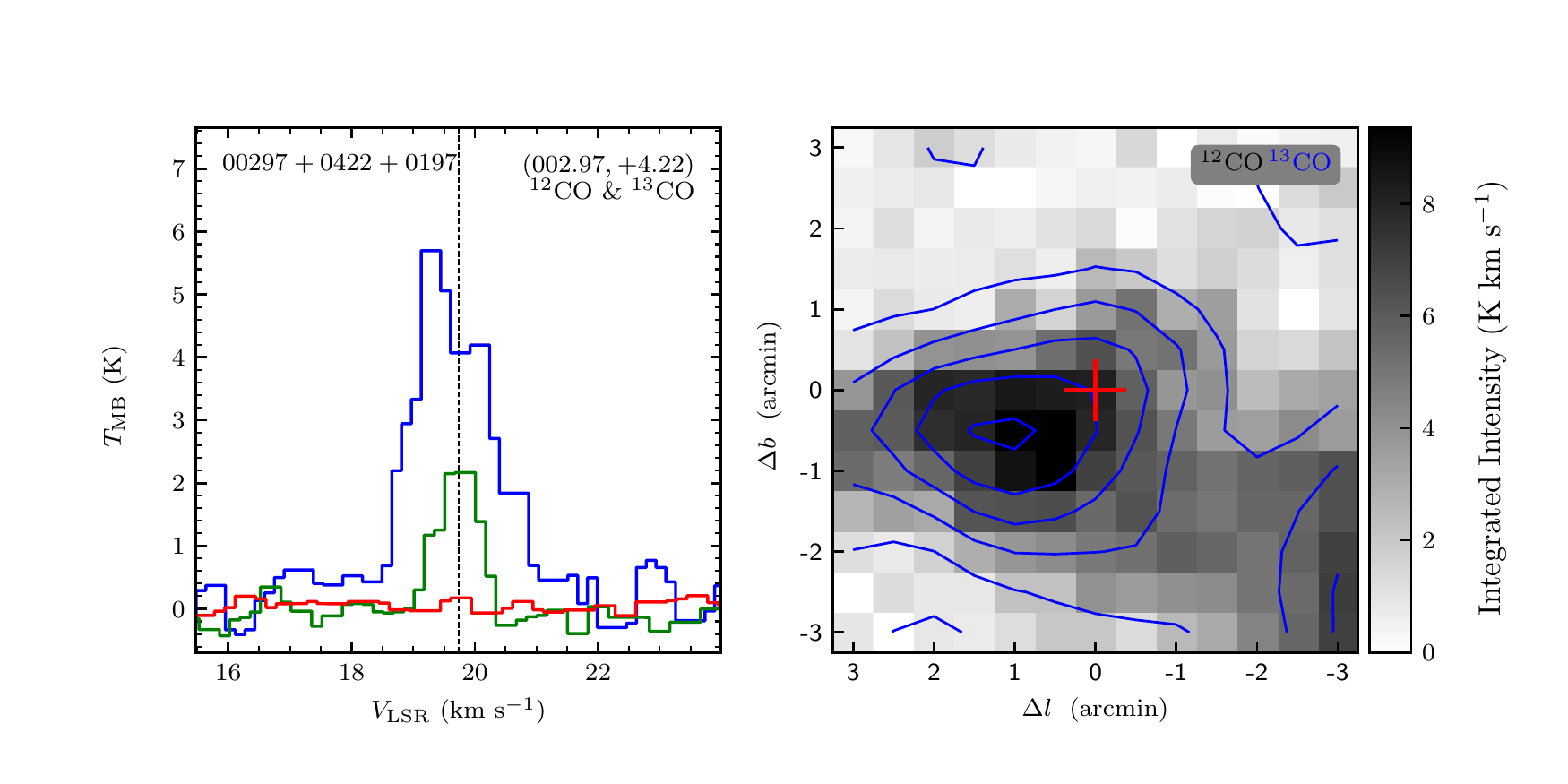}
\includegraphics[width=9.0cm,angle=0]{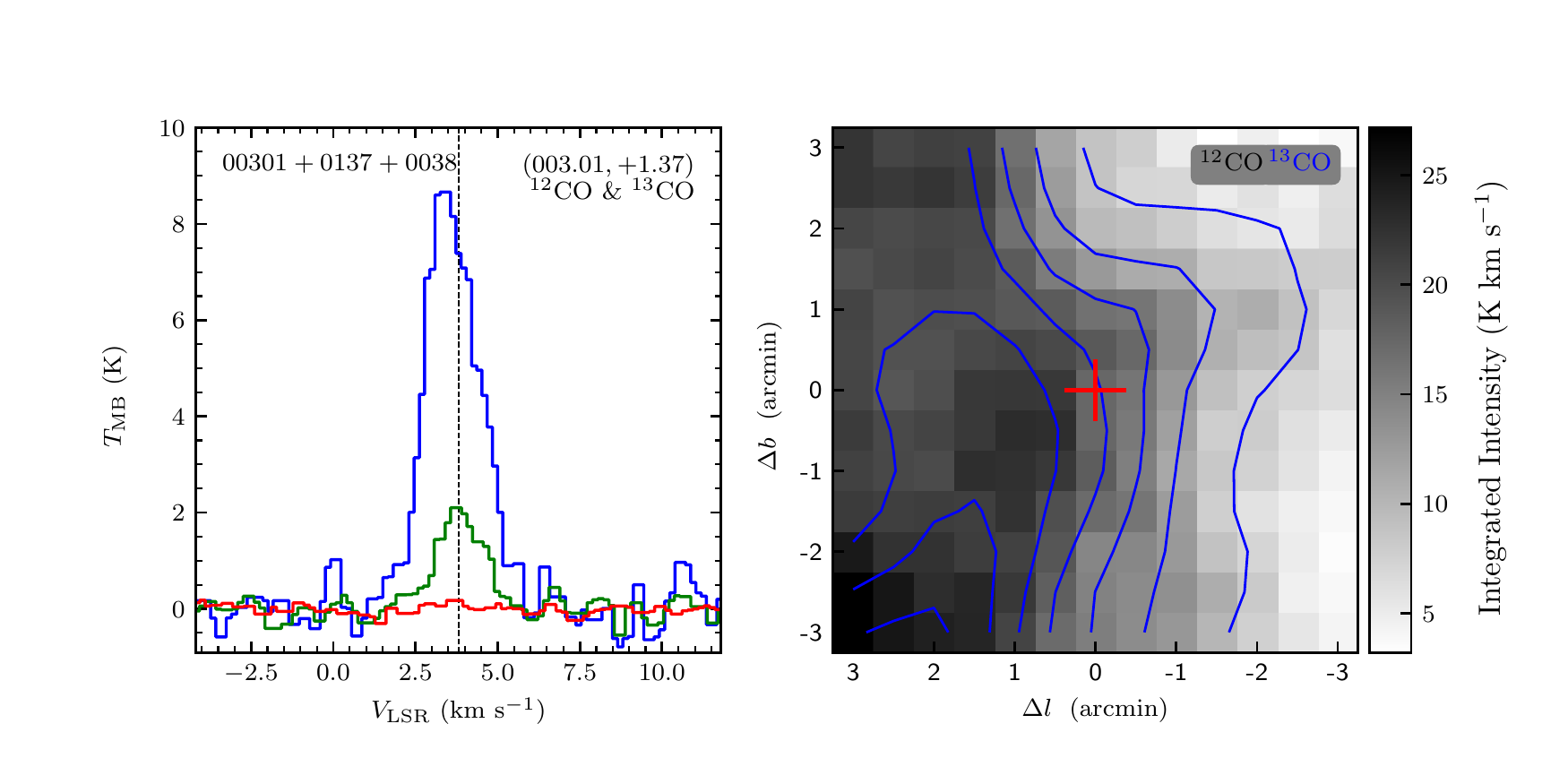}
\vspace{-0.5cm}

\includegraphics[width=9.0cm,angle=0]{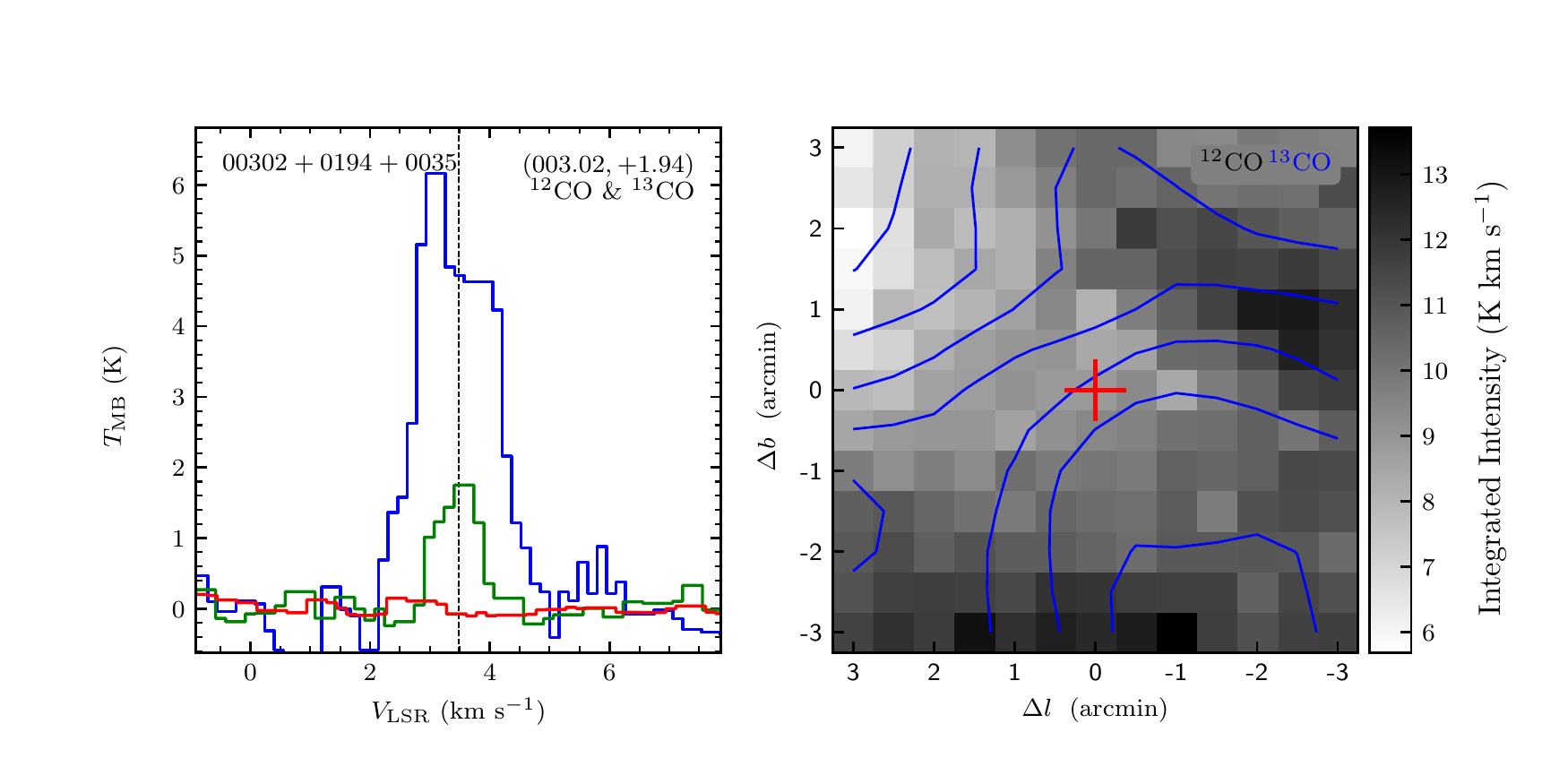}
\includegraphics[width=9.0cm,angle=0]{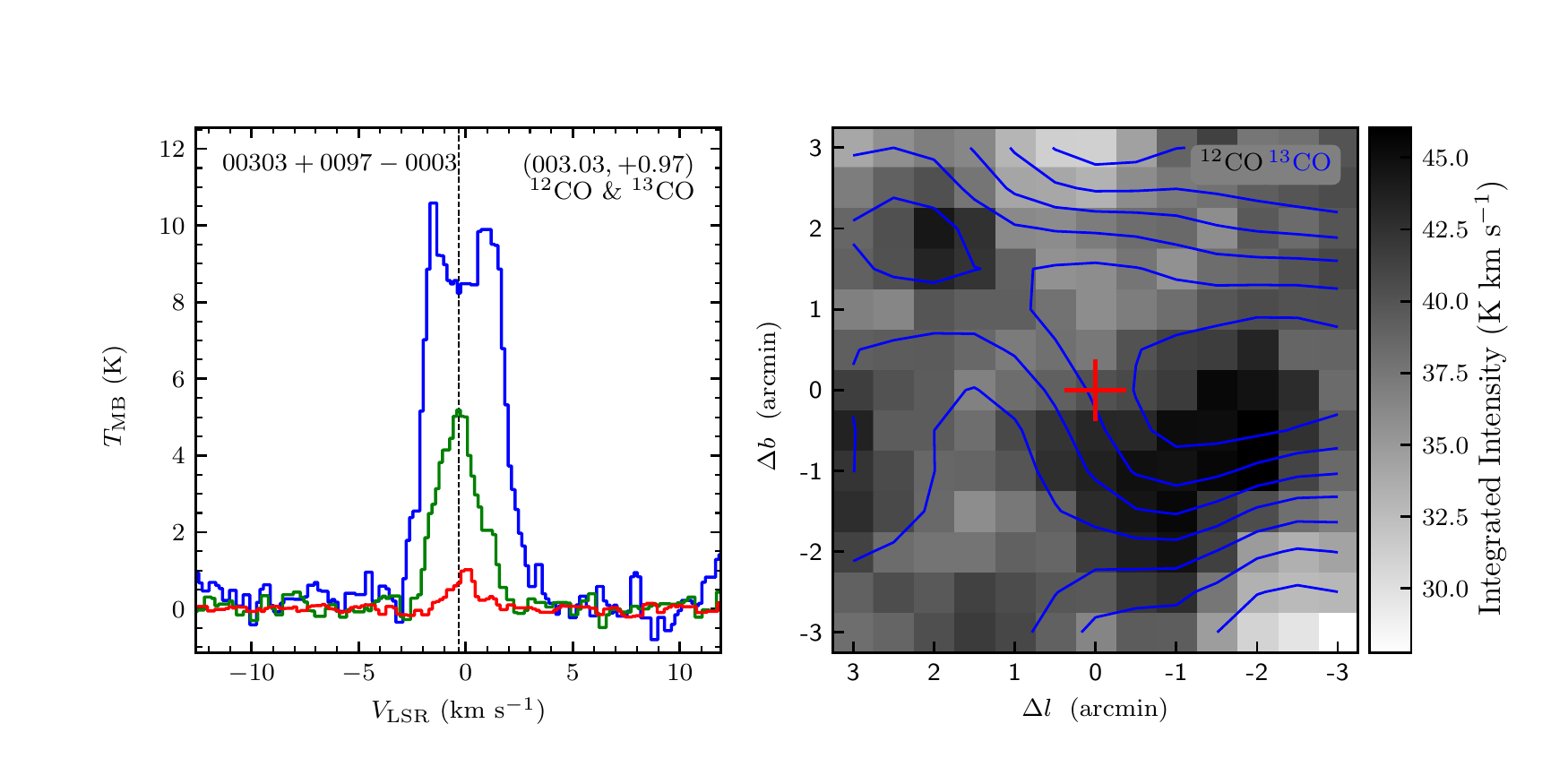}
\vspace{-0.5cm}

\includegraphics[width=9.0cm,angle=0]{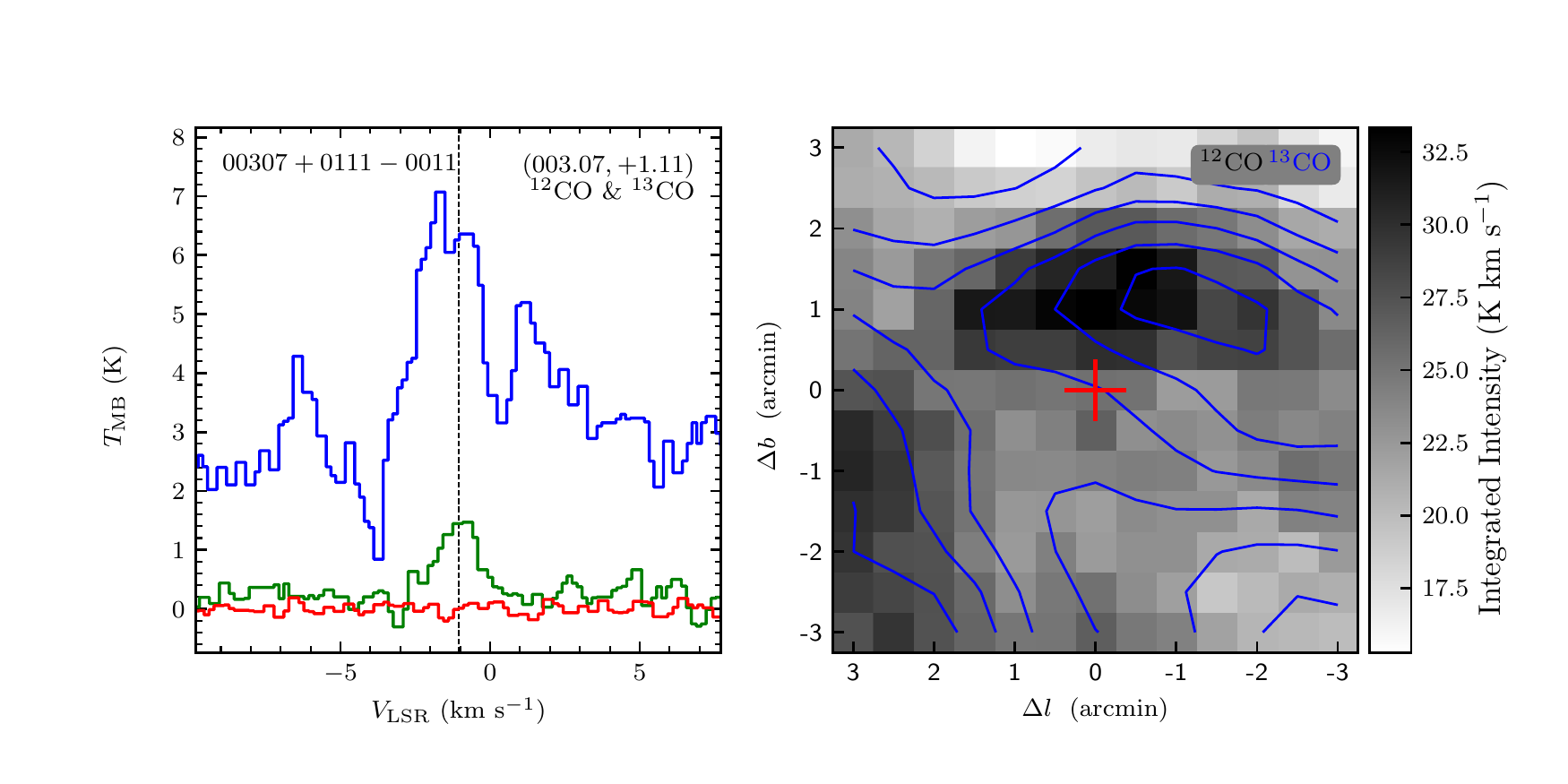}
\includegraphics[width=9.0cm,angle=0]{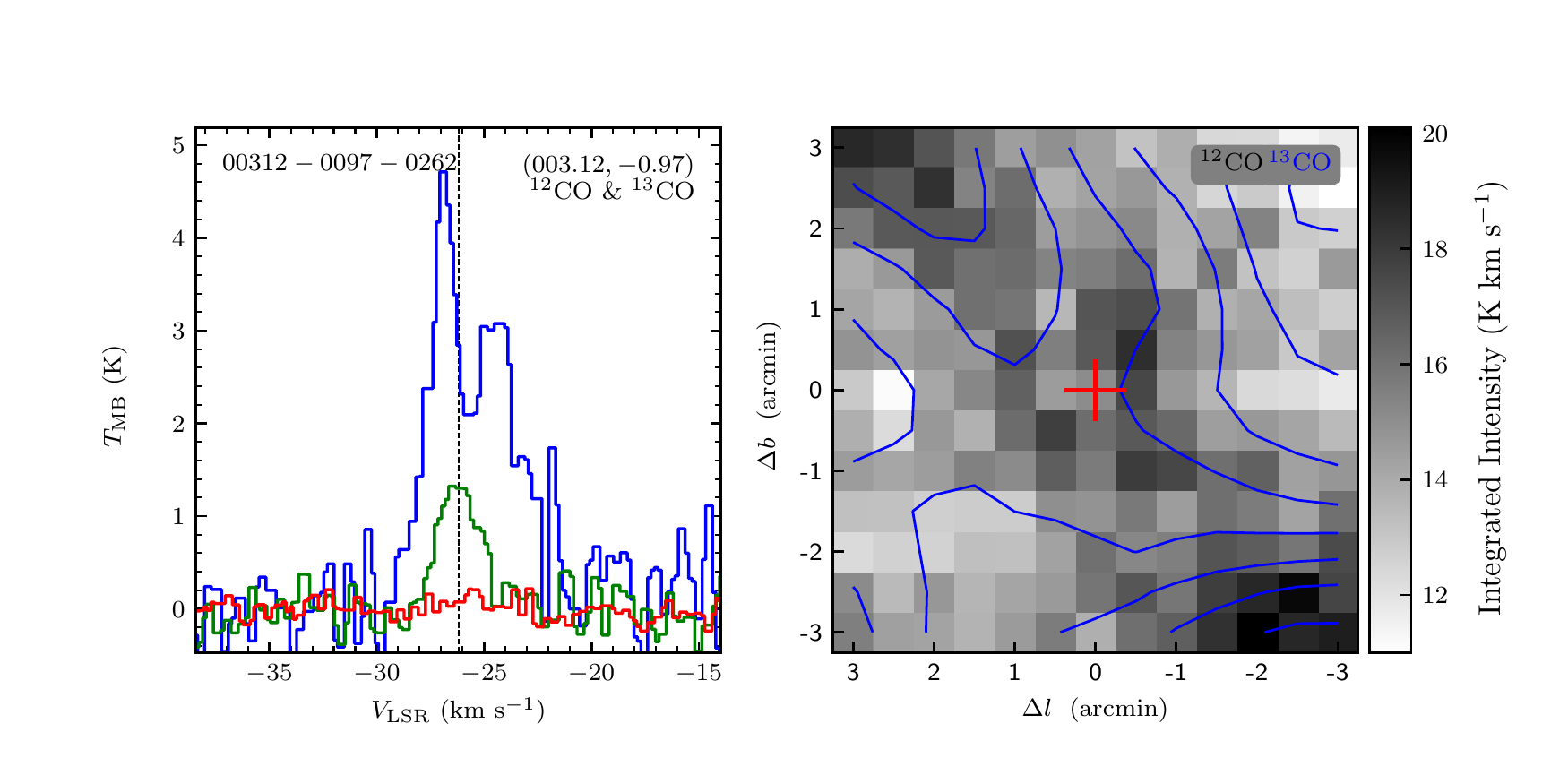}
\vspace{-0.5cm}

\includegraphics[width=9.0cm,angle=0]{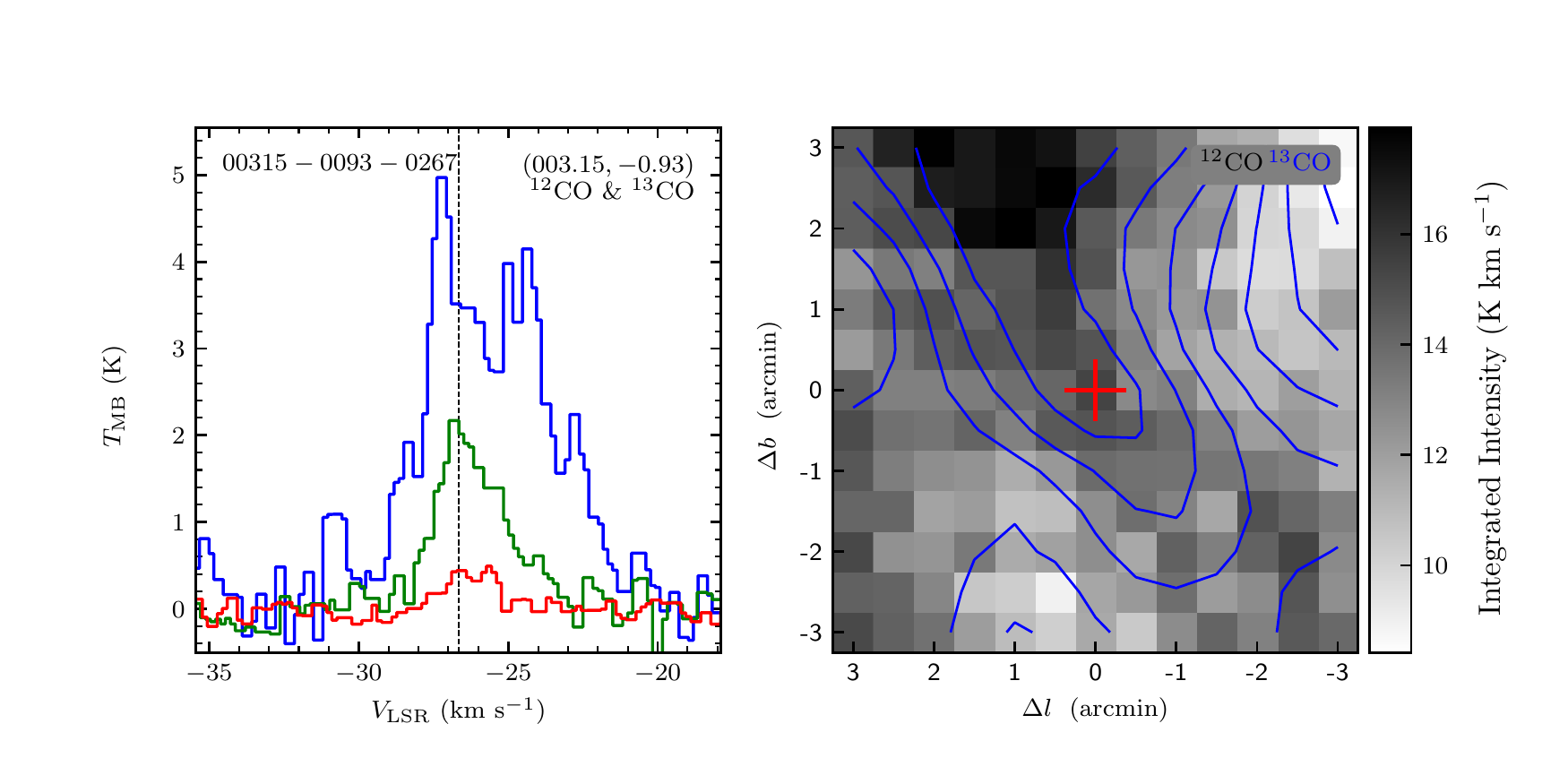}
\includegraphics[width=9.0cm,angle=0]{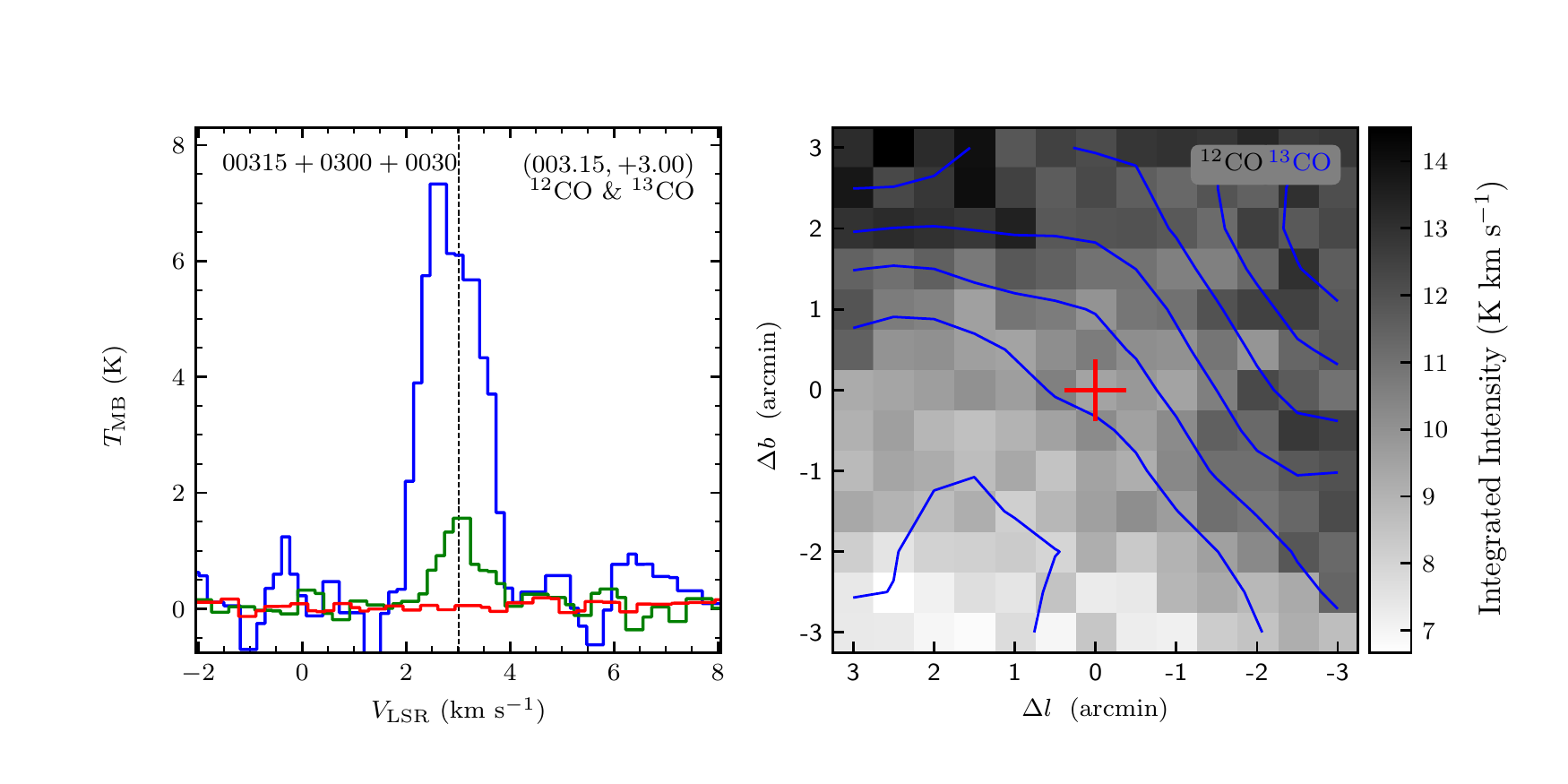}
\end{figure}
\clearpage

\begin{figure}
\includegraphics[width=9.0cm,angle=0]{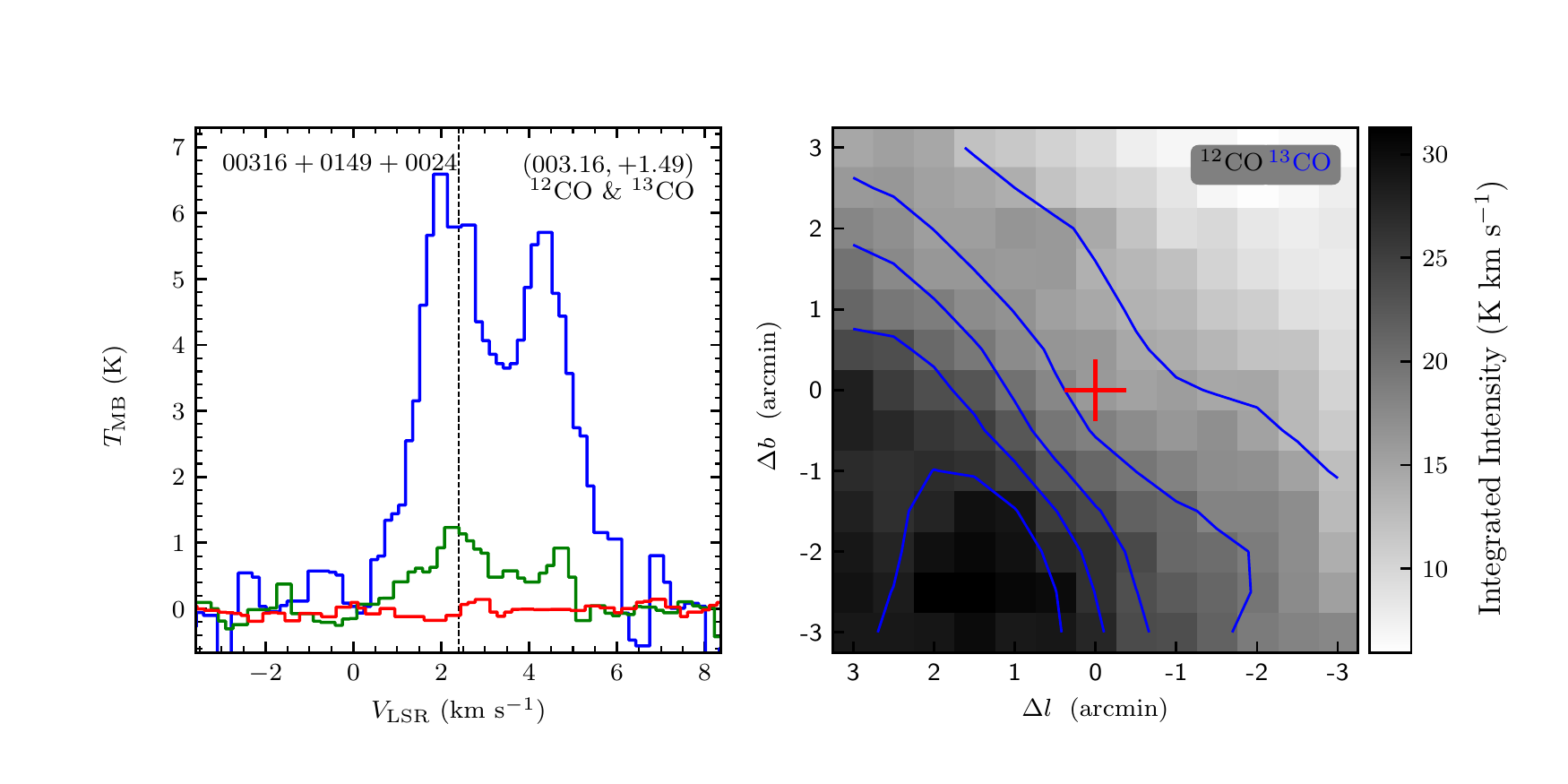}
\includegraphics[width=9.0cm,angle=0]{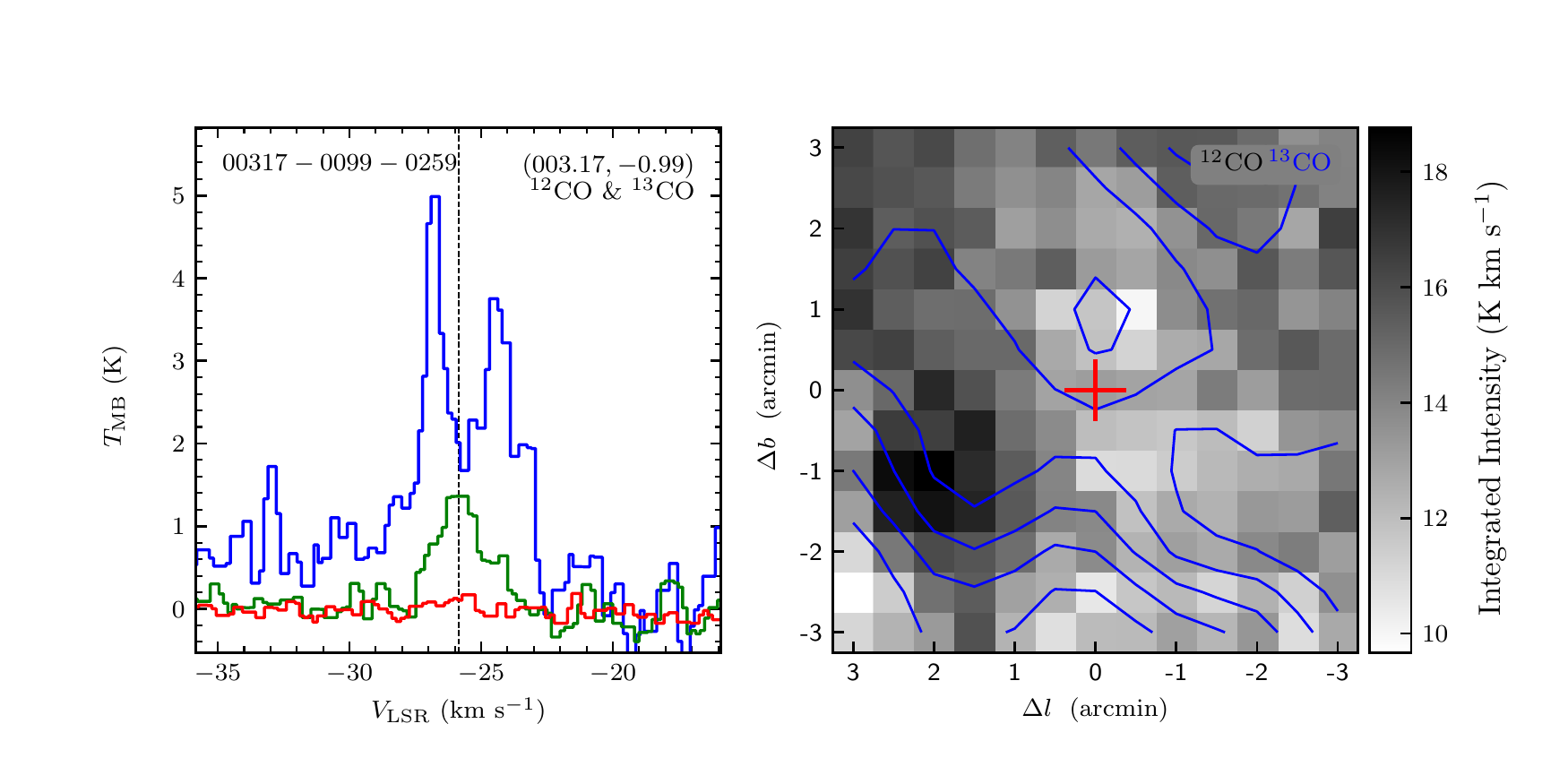}
\vspace{-0.5cm}

\includegraphics[width=9.0cm,angle=0]{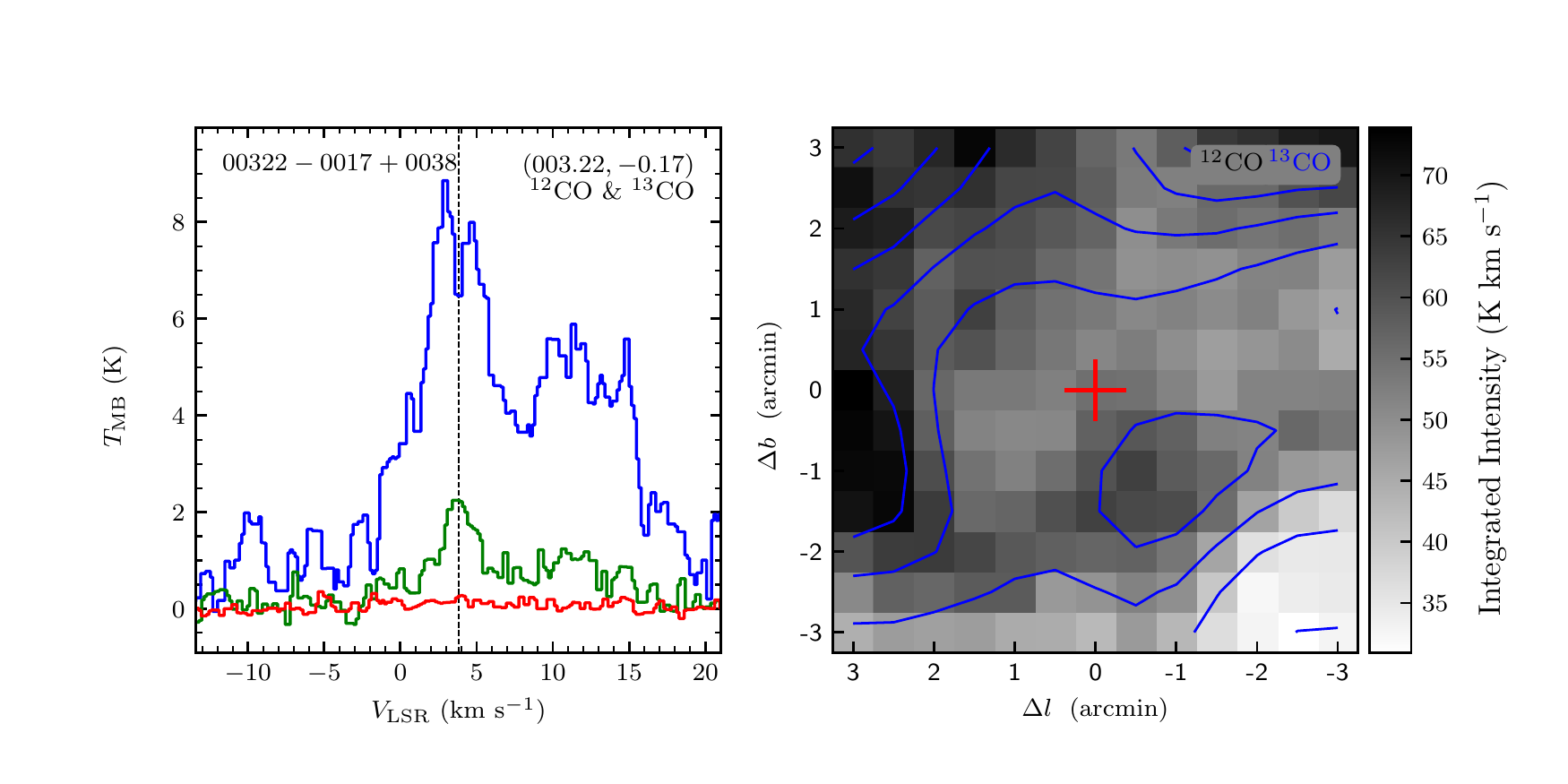}
\includegraphics[width=9.0cm,angle=0]{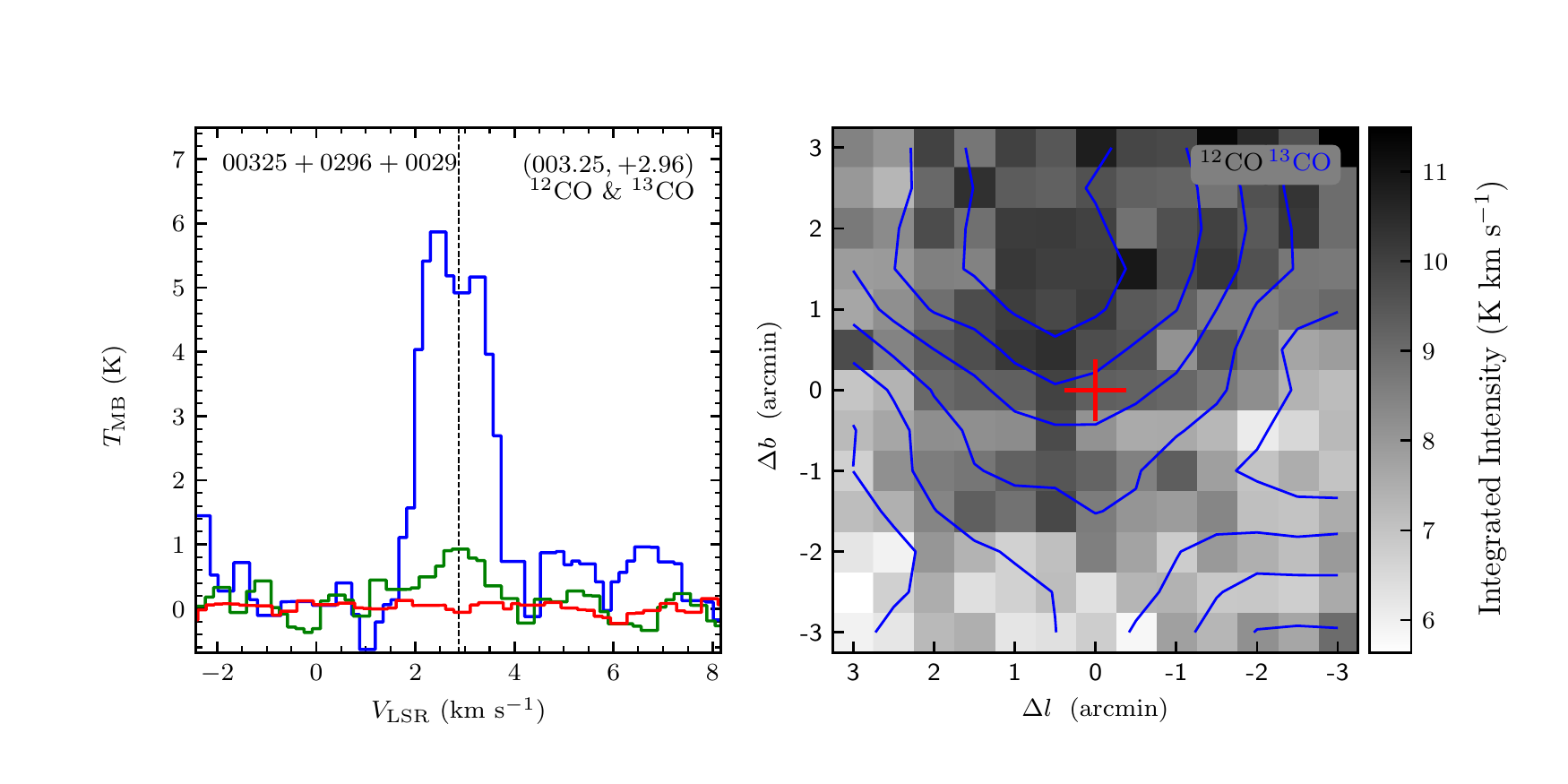}
\vspace{-0.5cm}

\includegraphics[width=9.0cm,angle=0]{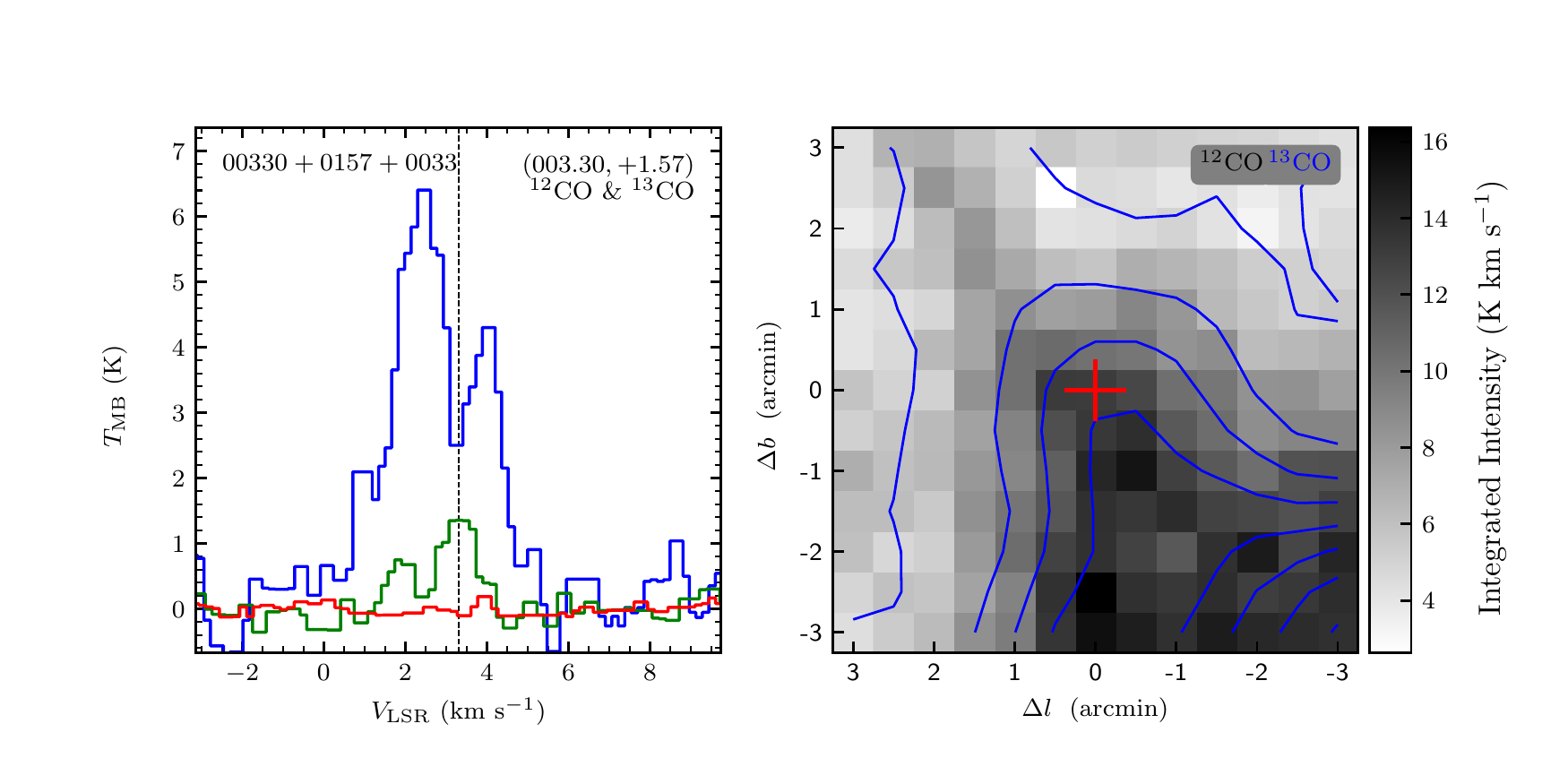}
\includegraphics[width=9.0cm,angle=0]{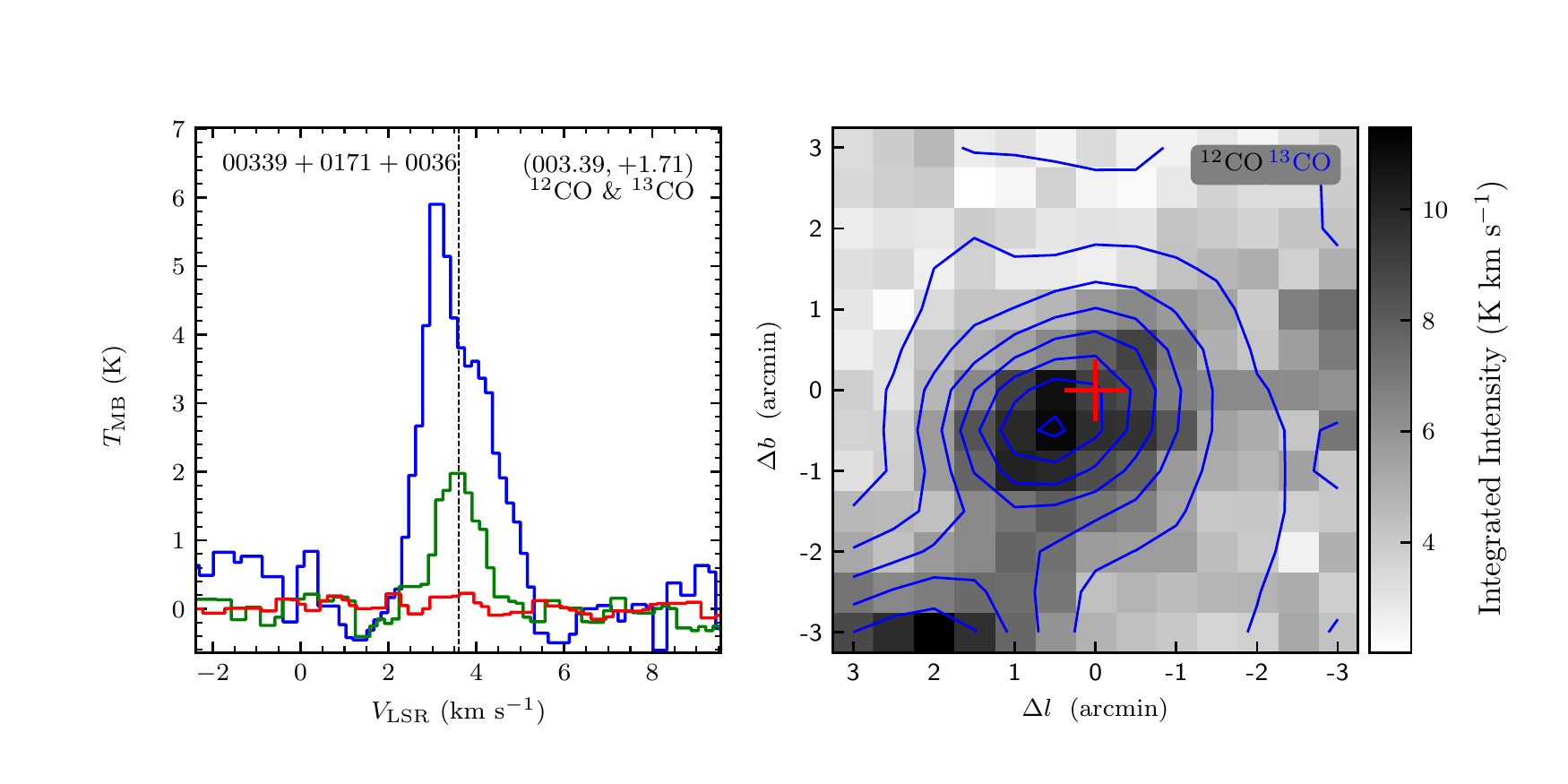}
\vspace{-0.5cm}

\includegraphics[width=9.0cm,angle=0]{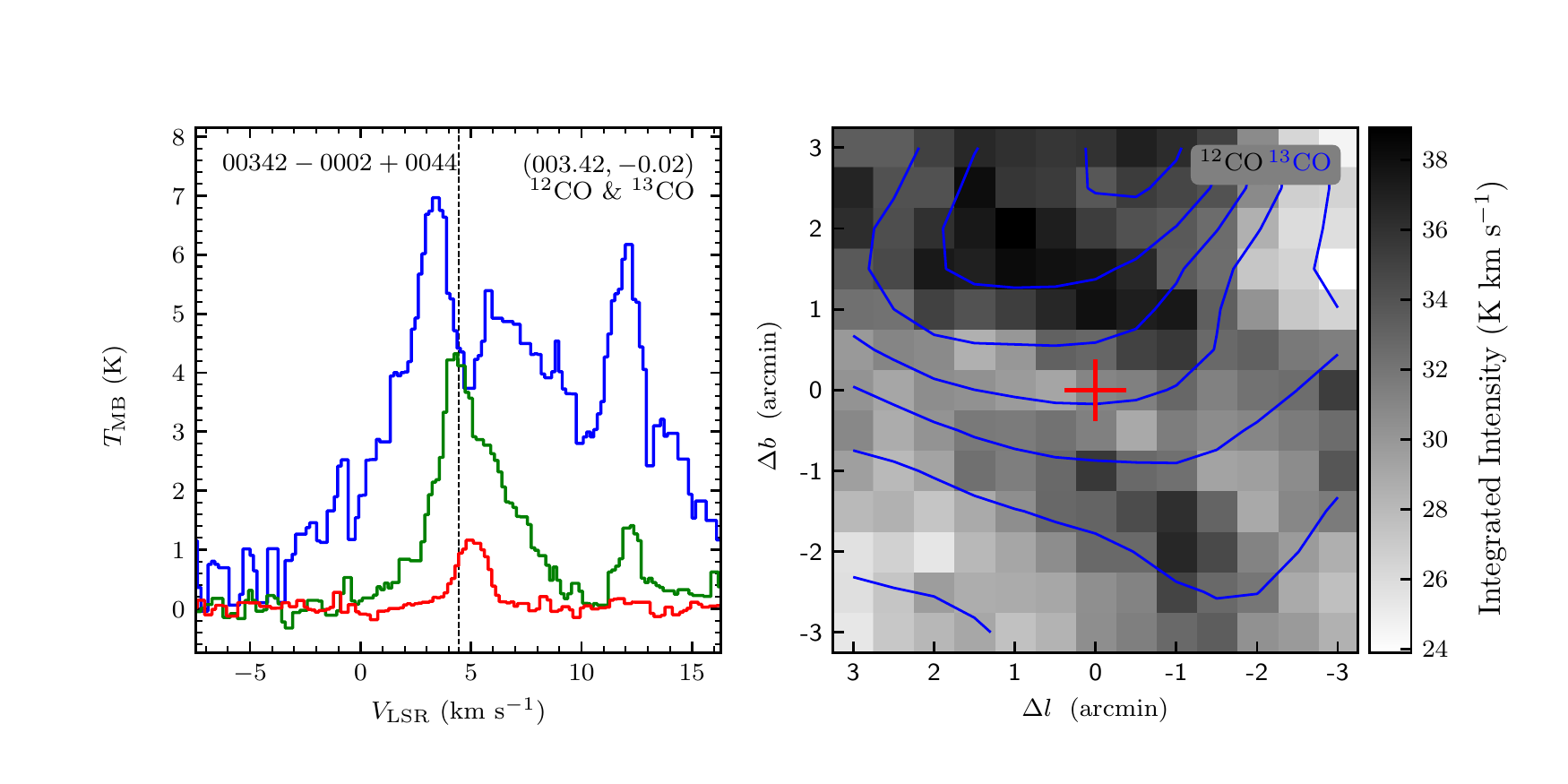}
\includegraphics[width=9.0cm,angle=0]{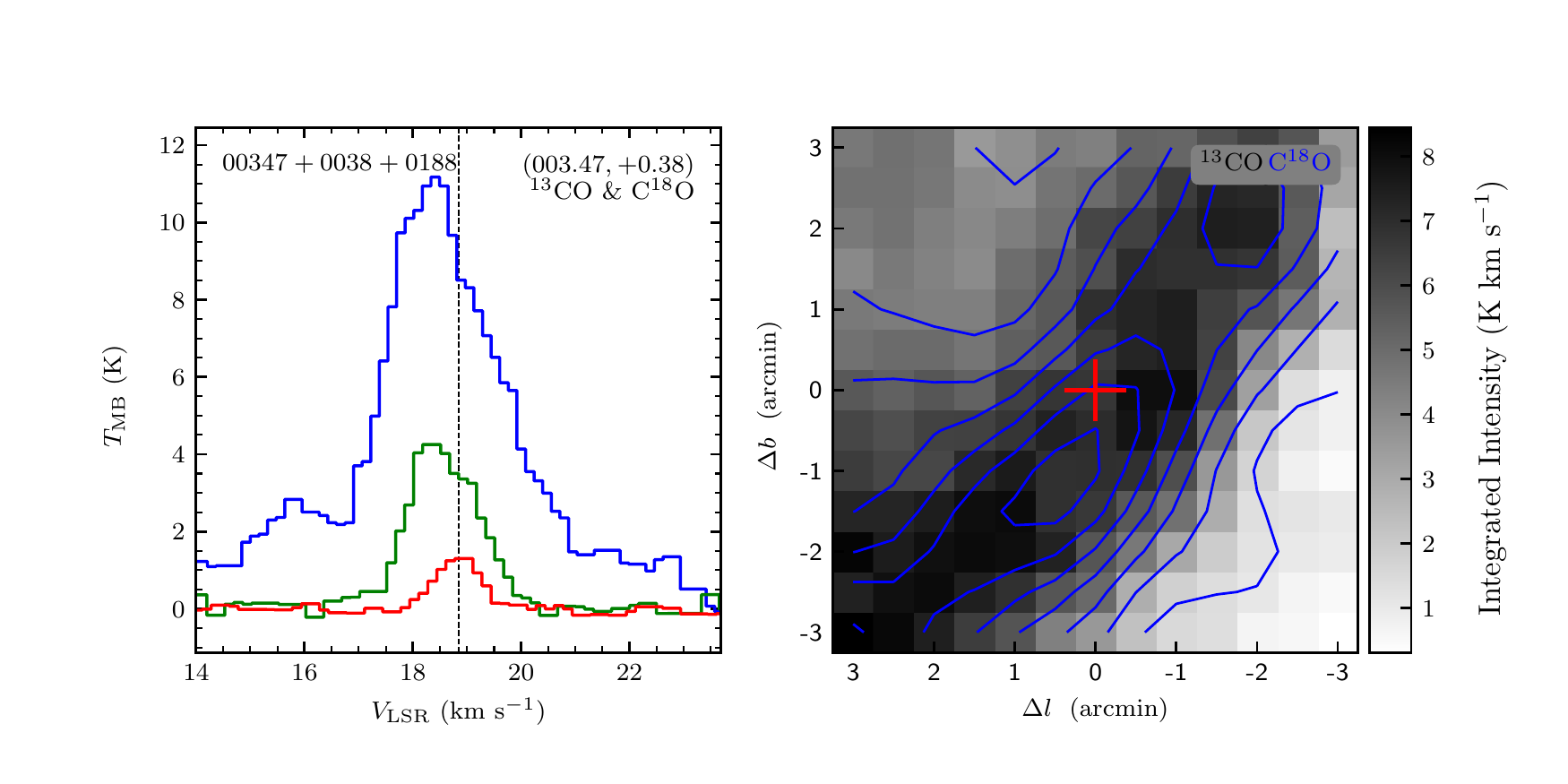}
\vspace{-0.5cm}

\includegraphics[width=9.0cm,angle=0]{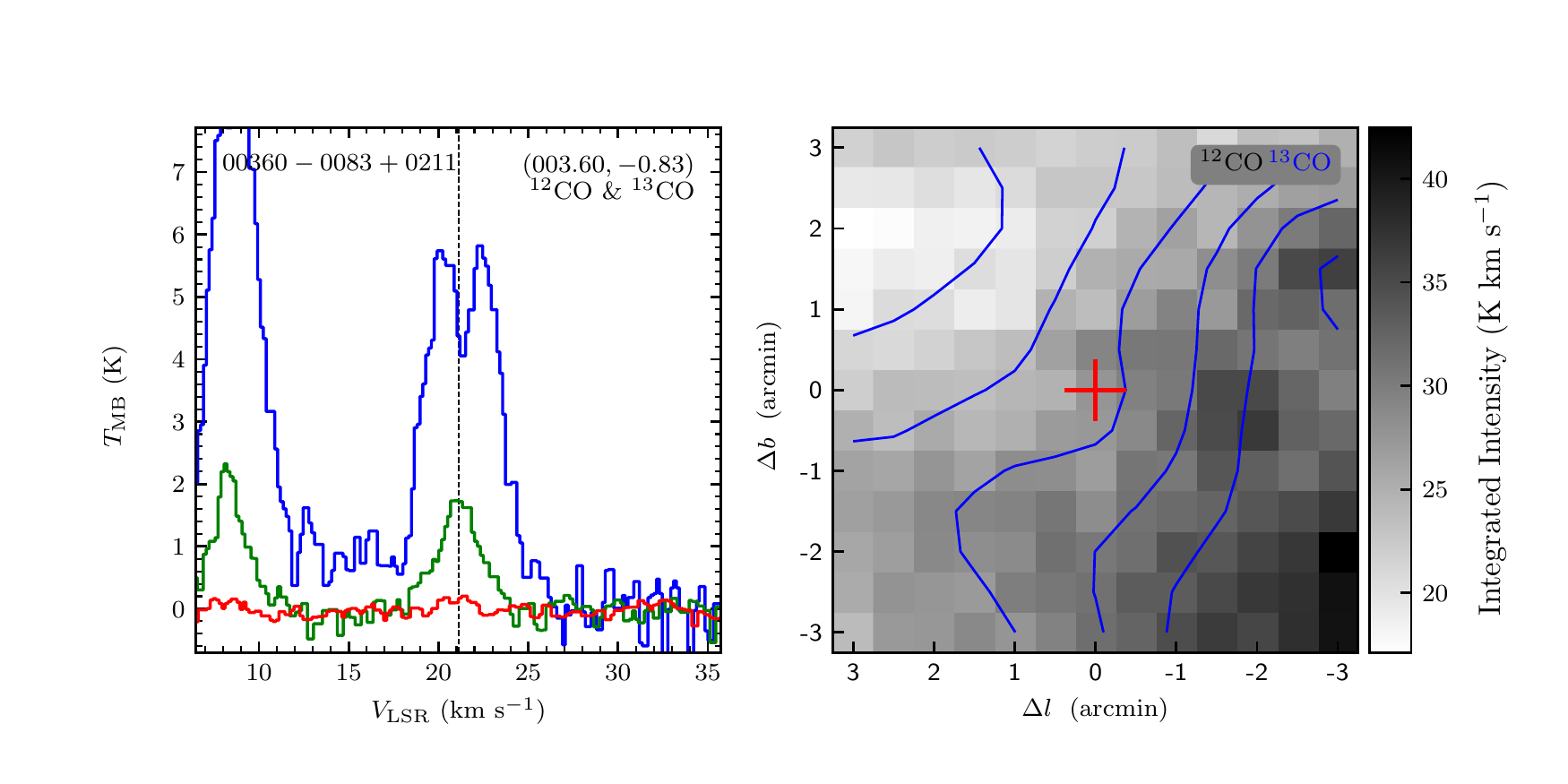}
\includegraphics[width=9.0cm,angle=0]{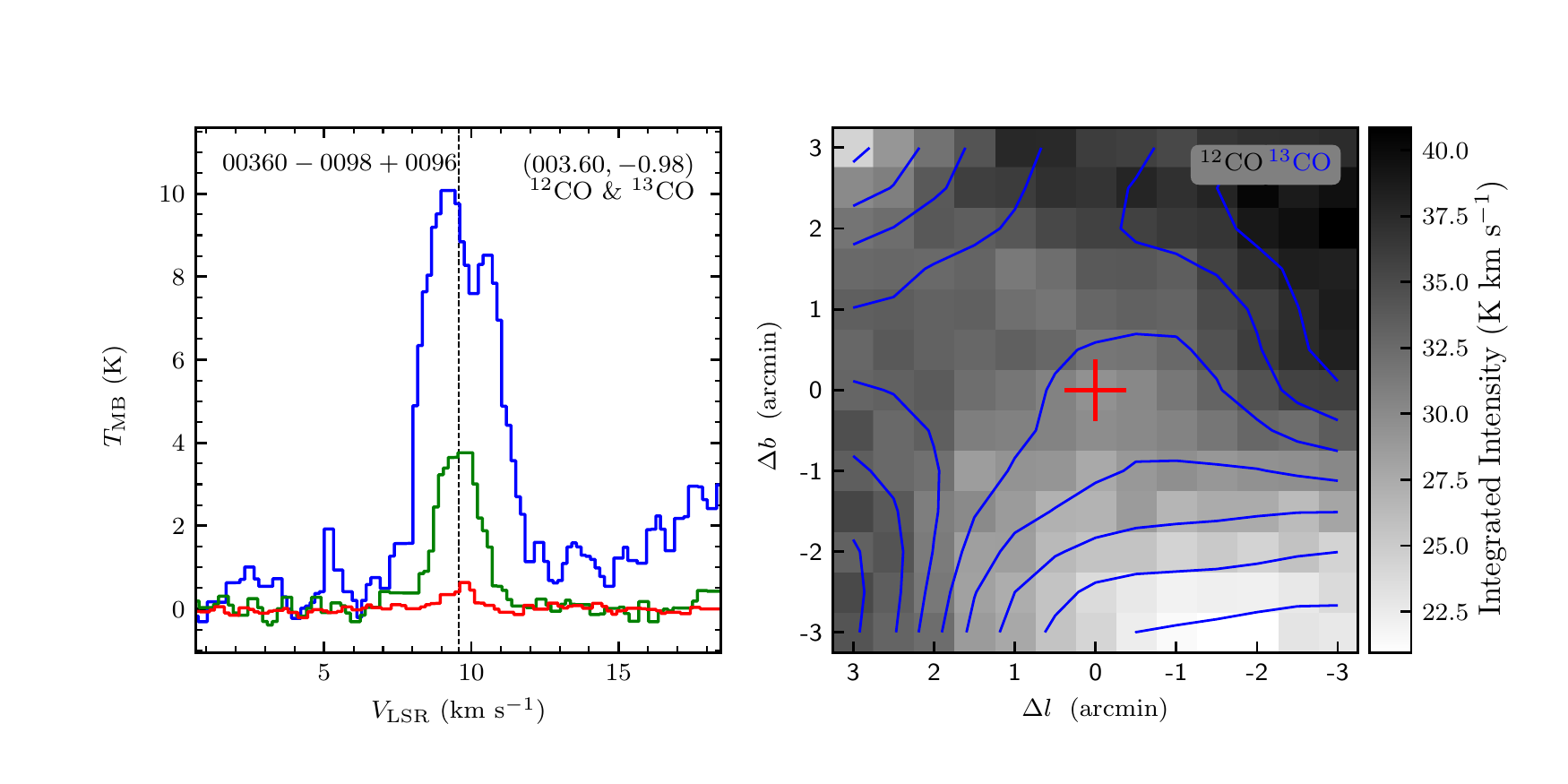}
\end{figure}
\clearpage

\begin{figure}
\includegraphics[width=9.0cm,angle=0]{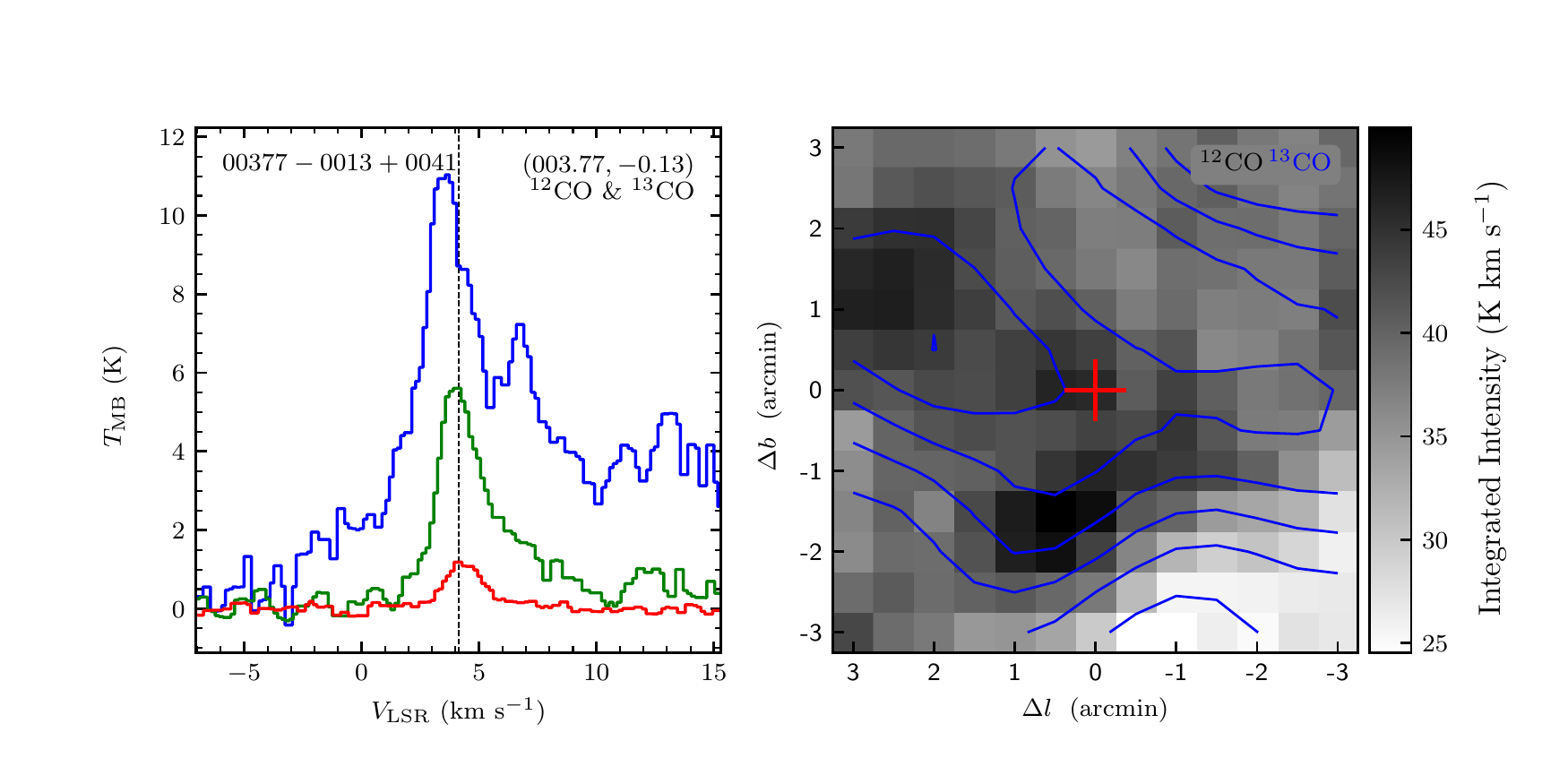}
\includegraphics[width=9.0cm,angle=0]{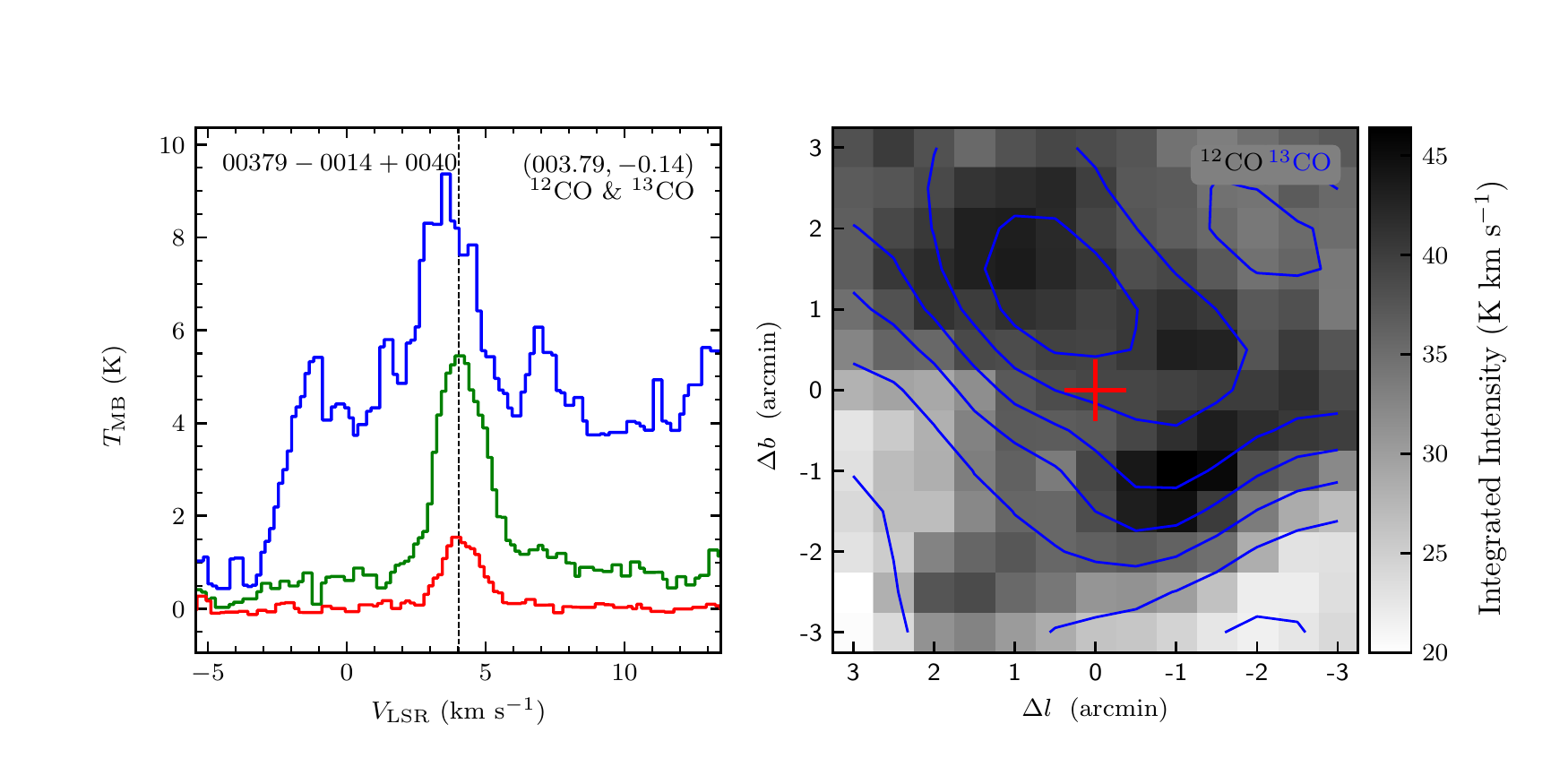}
\vspace{-0.5cm}

\includegraphics[width=9.0cm,angle=0]{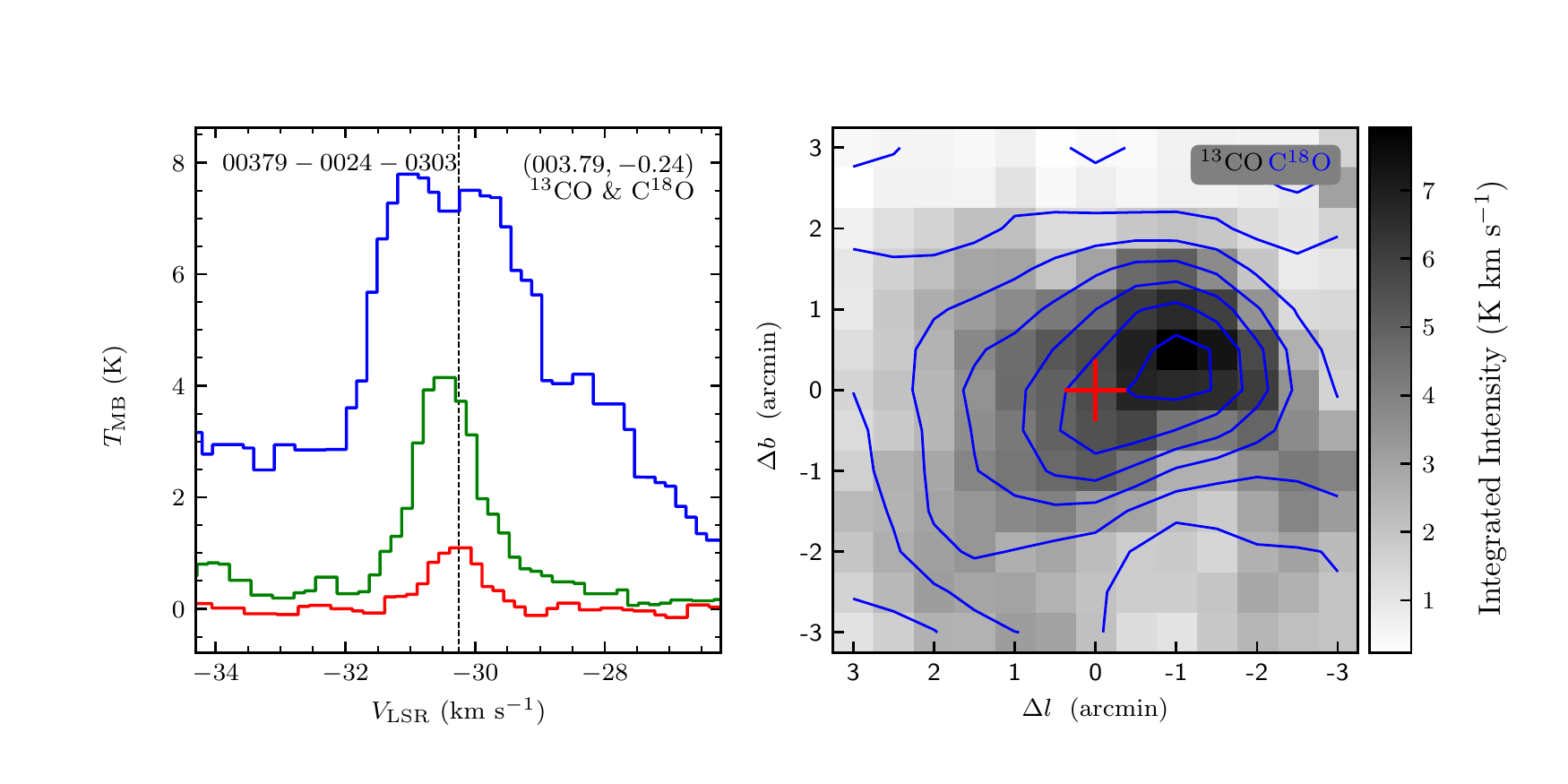}
\includegraphics[width=9.0cm,angle=0]{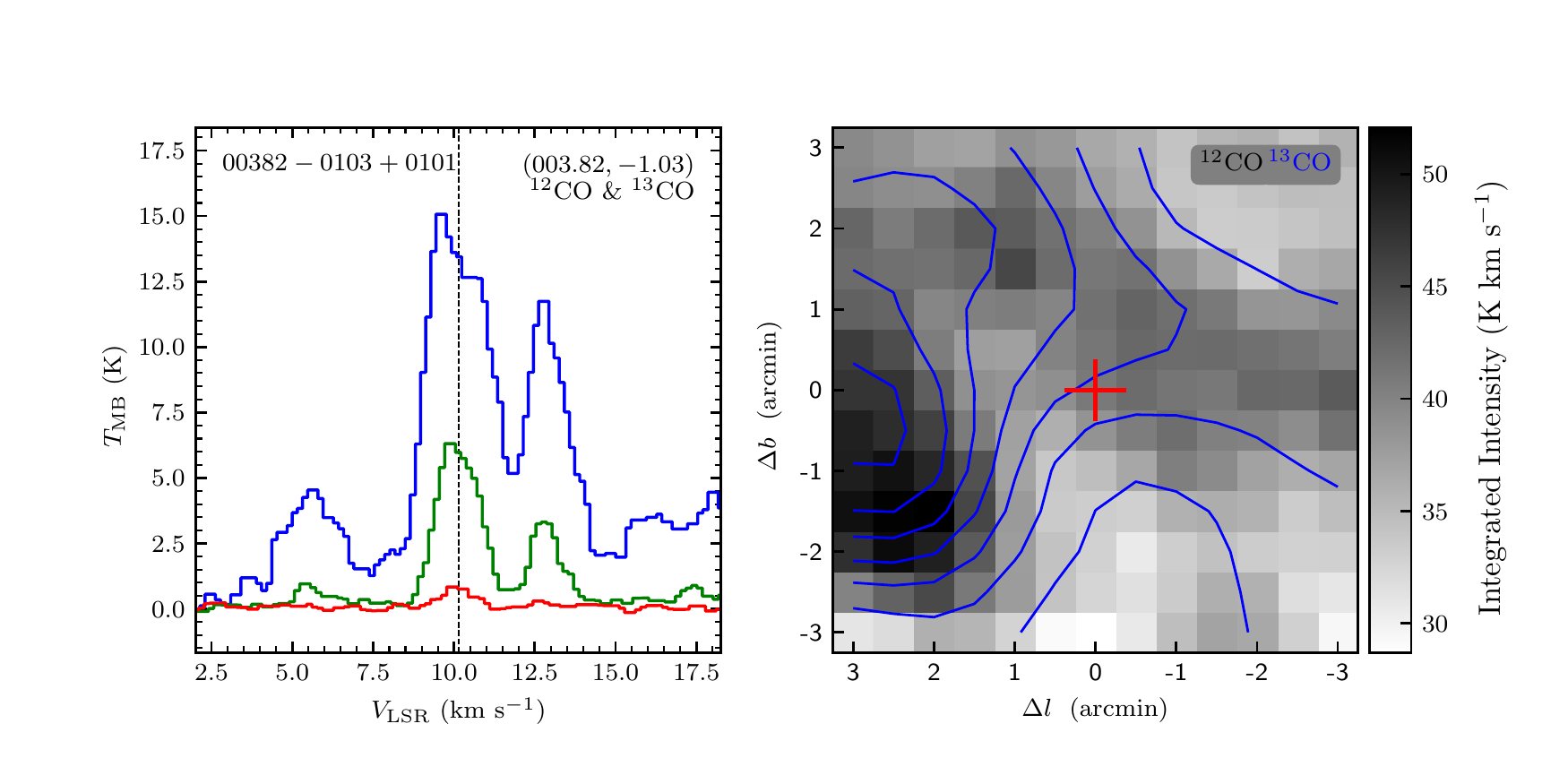}
\vspace{-0.5cm}

\includegraphics[width=9.0cm,angle=0]{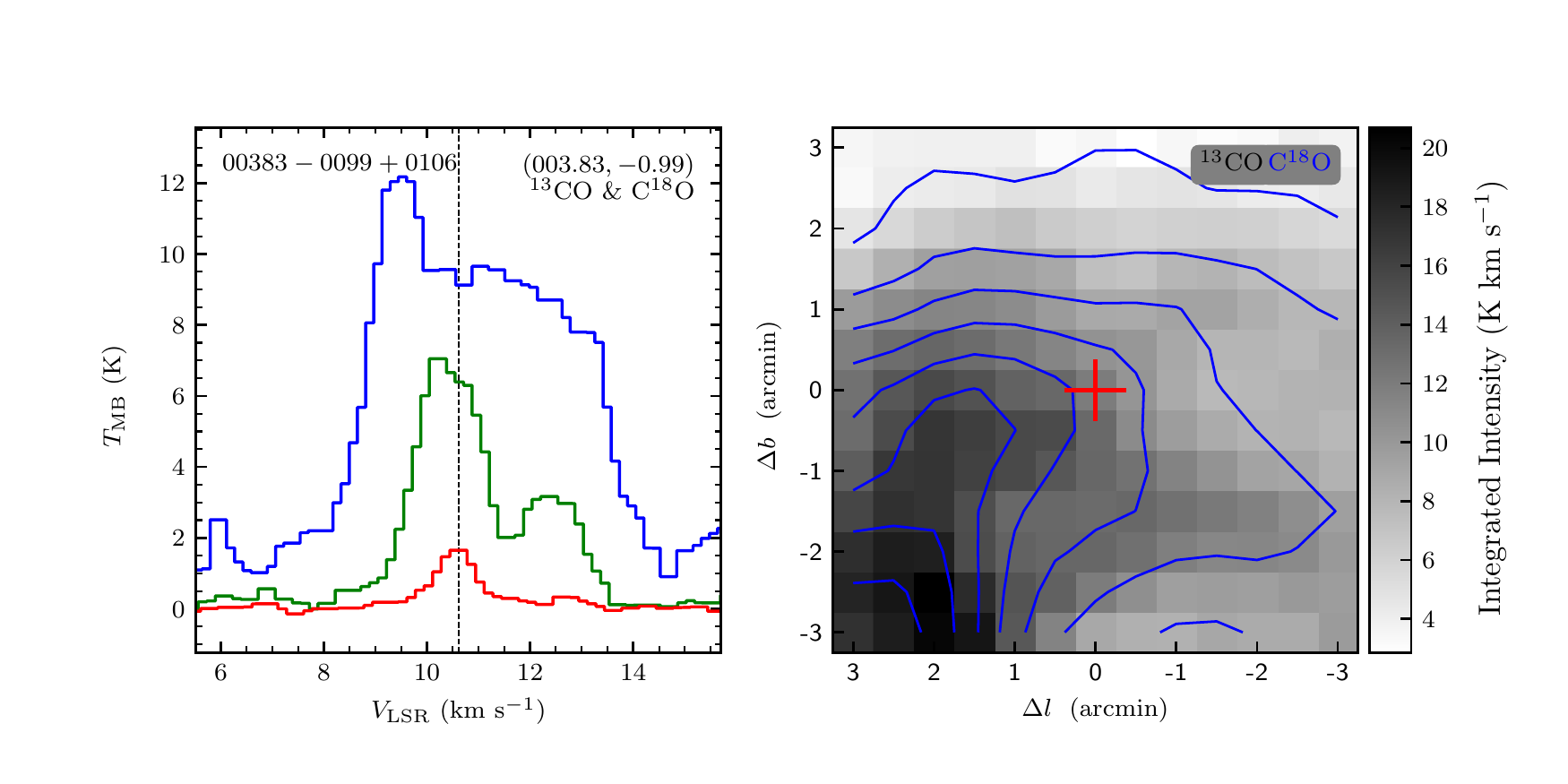}
\includegraphics[width=9.0cm,angle=0]{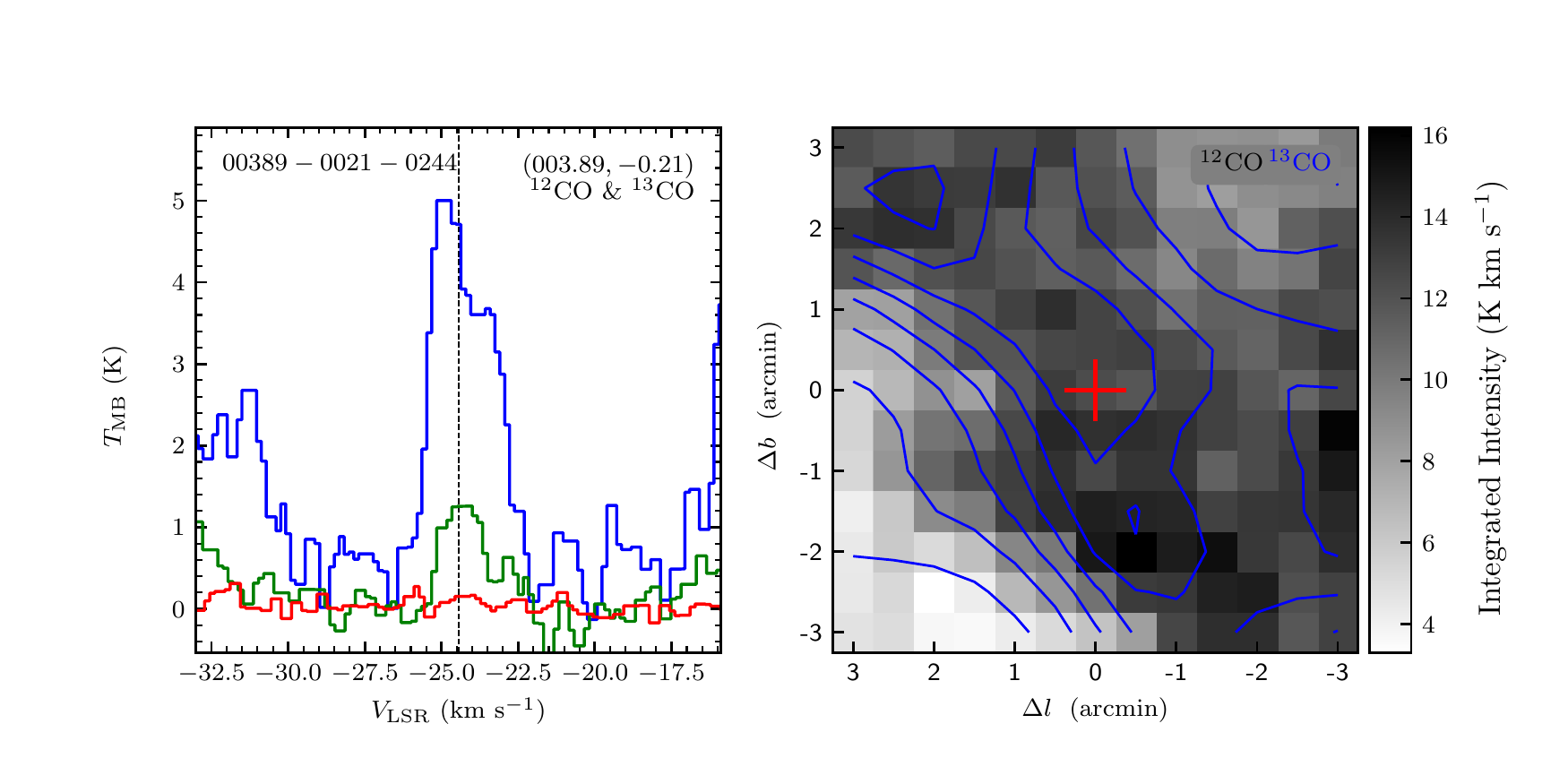}
\vspace{-0.5cm}

\includegraphics[width=9.0cm,angle=0]{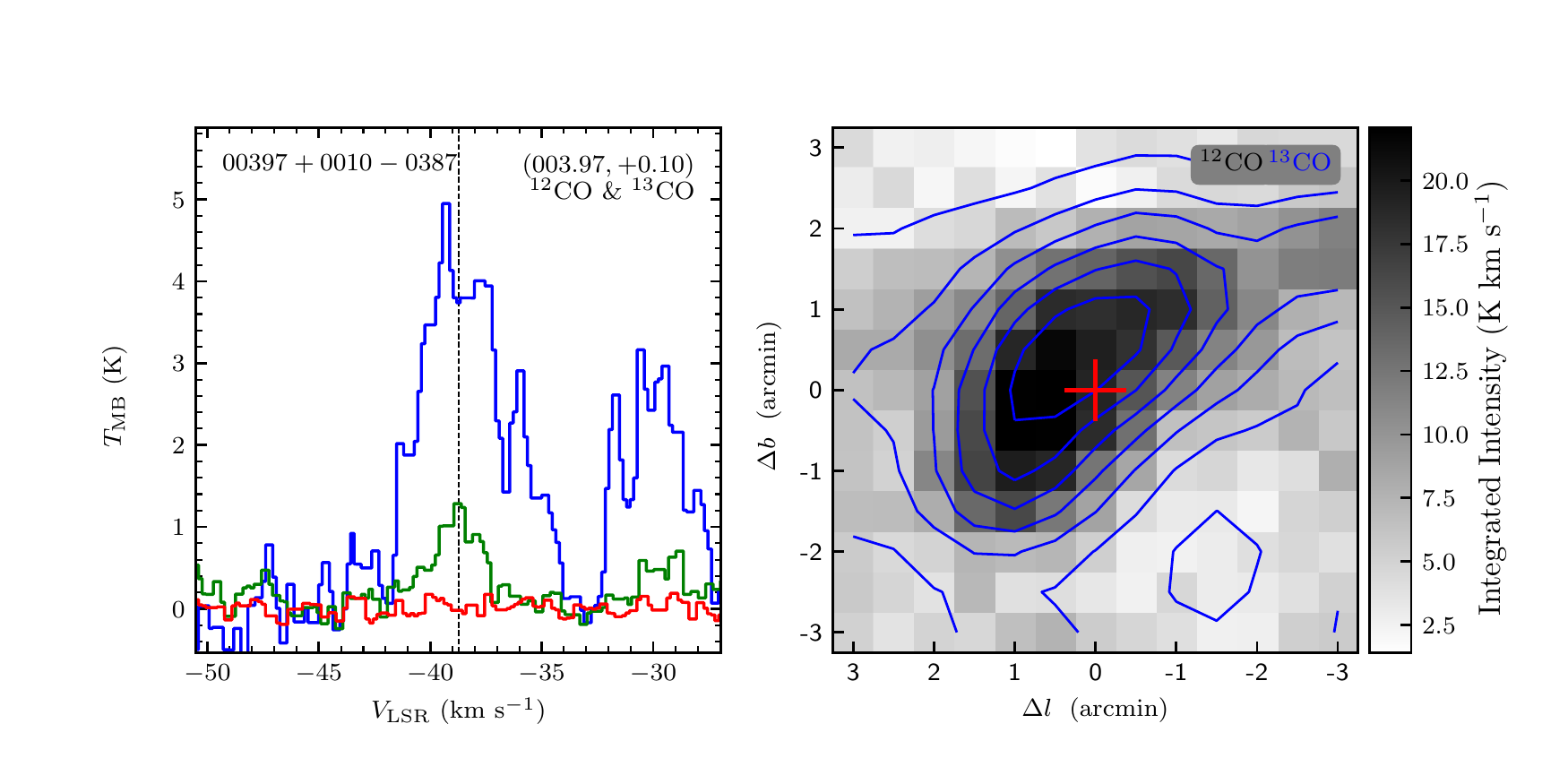}
\includegraphics[width=9.0cm,angle=0]{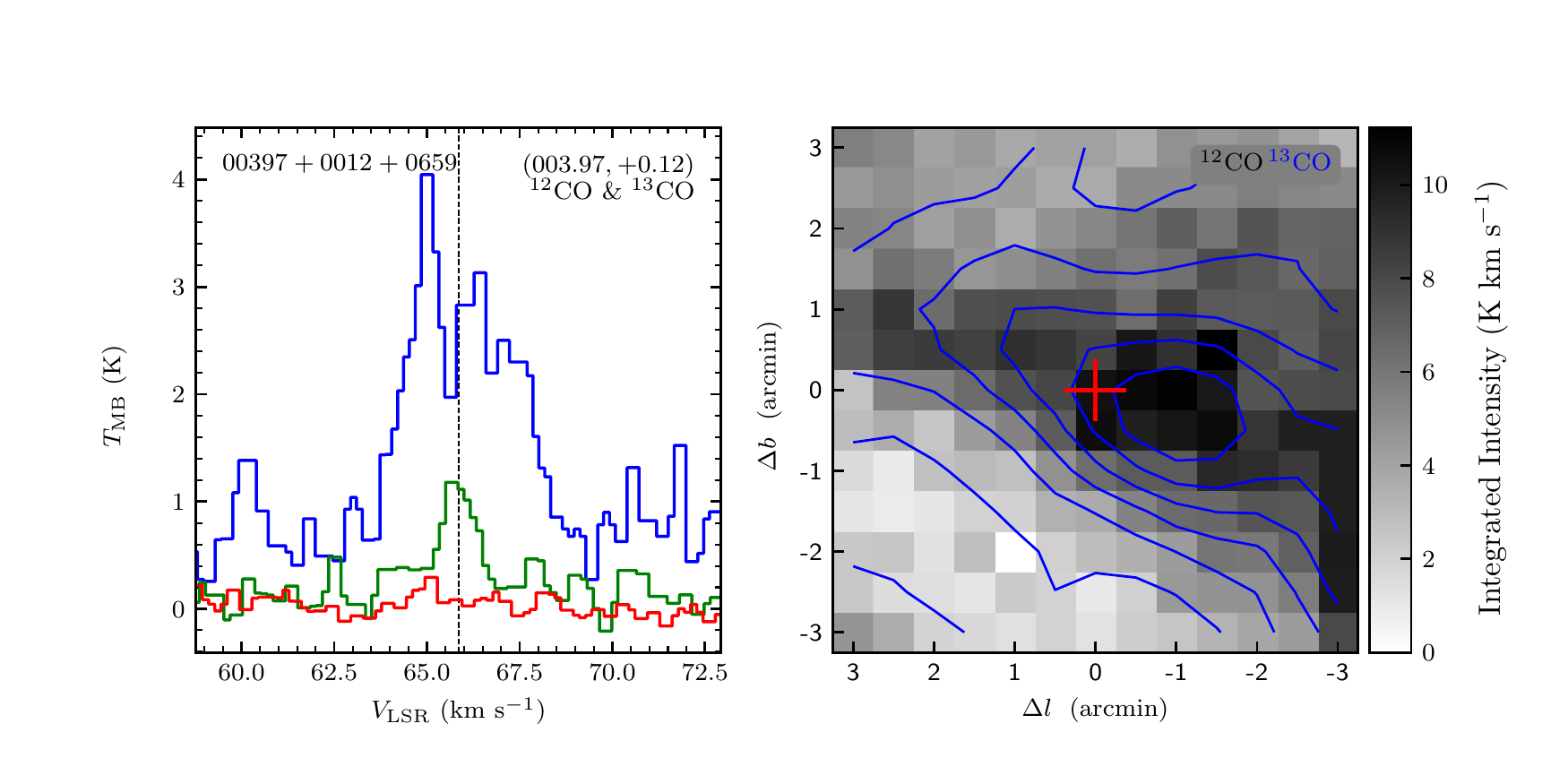}
\vspace{-0.5cm}

\includegraphics[width=9.0cm,angle=0]{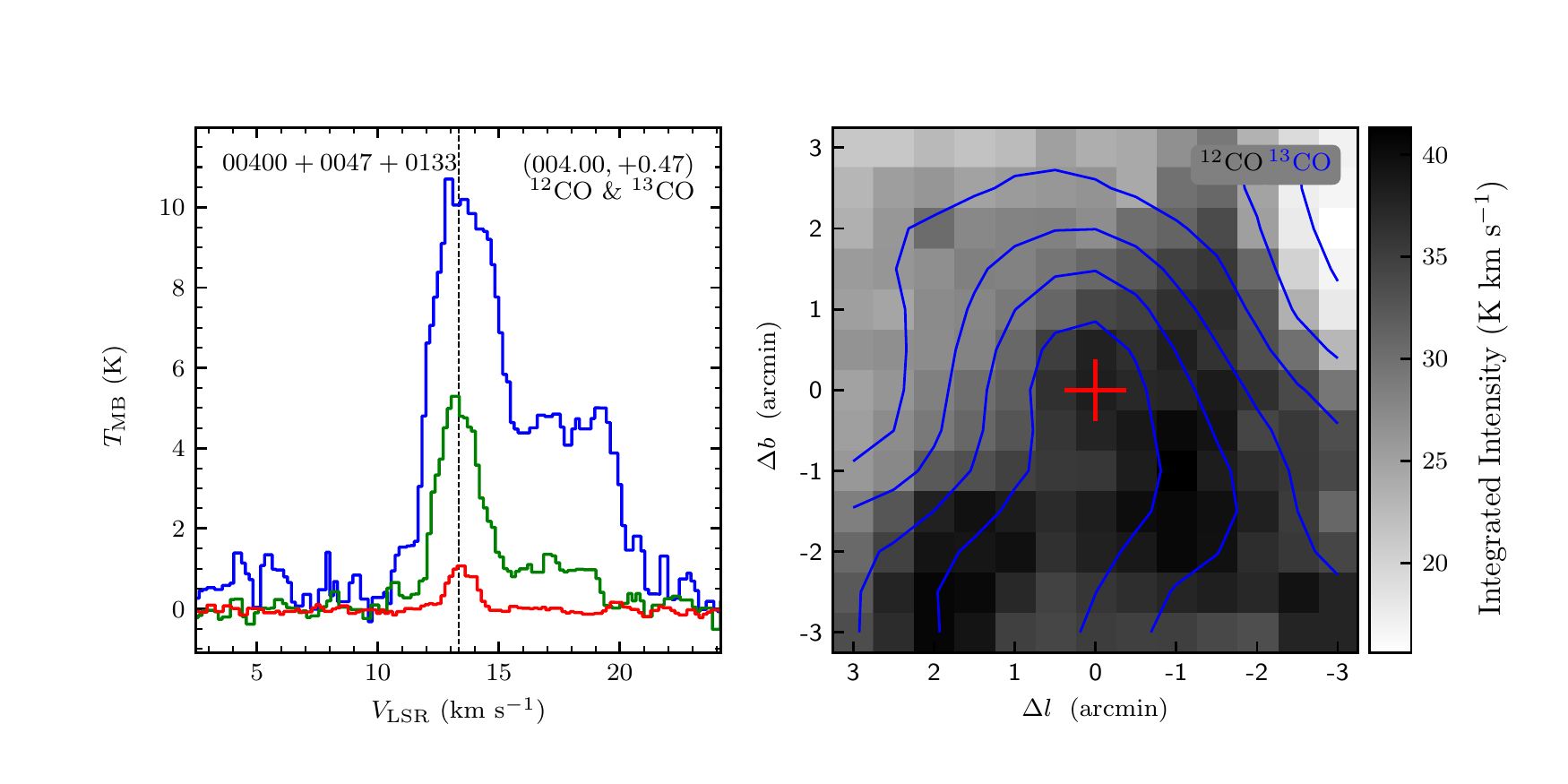}
\includegraphics[width=9.0cm,angle=0]{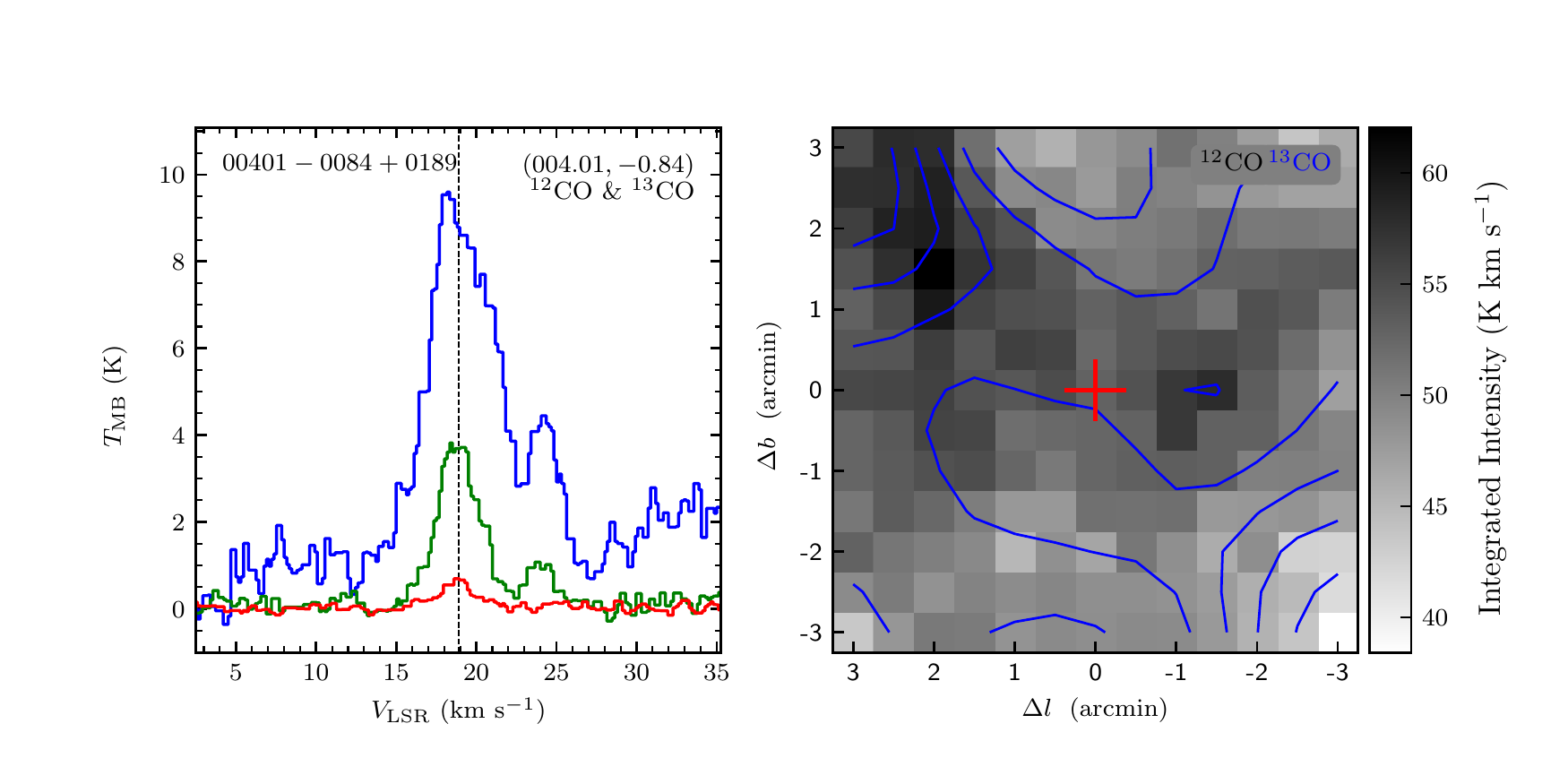}
\end{figure}
\clearpage

\begin{figure}
\includegraphics[width=9.0cm,angle=0]{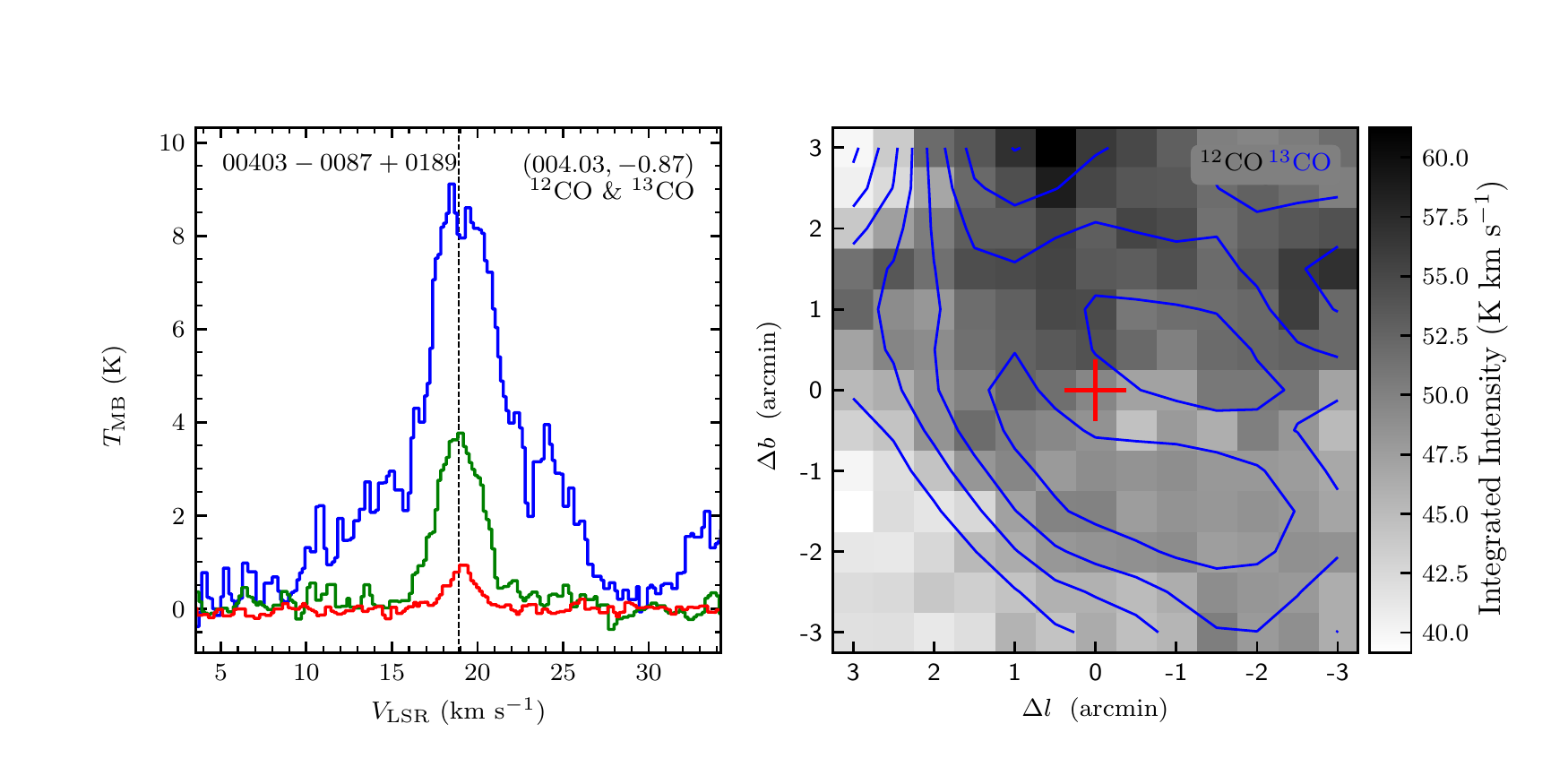}
\includegraphics[width=9.0cm,angle=0]{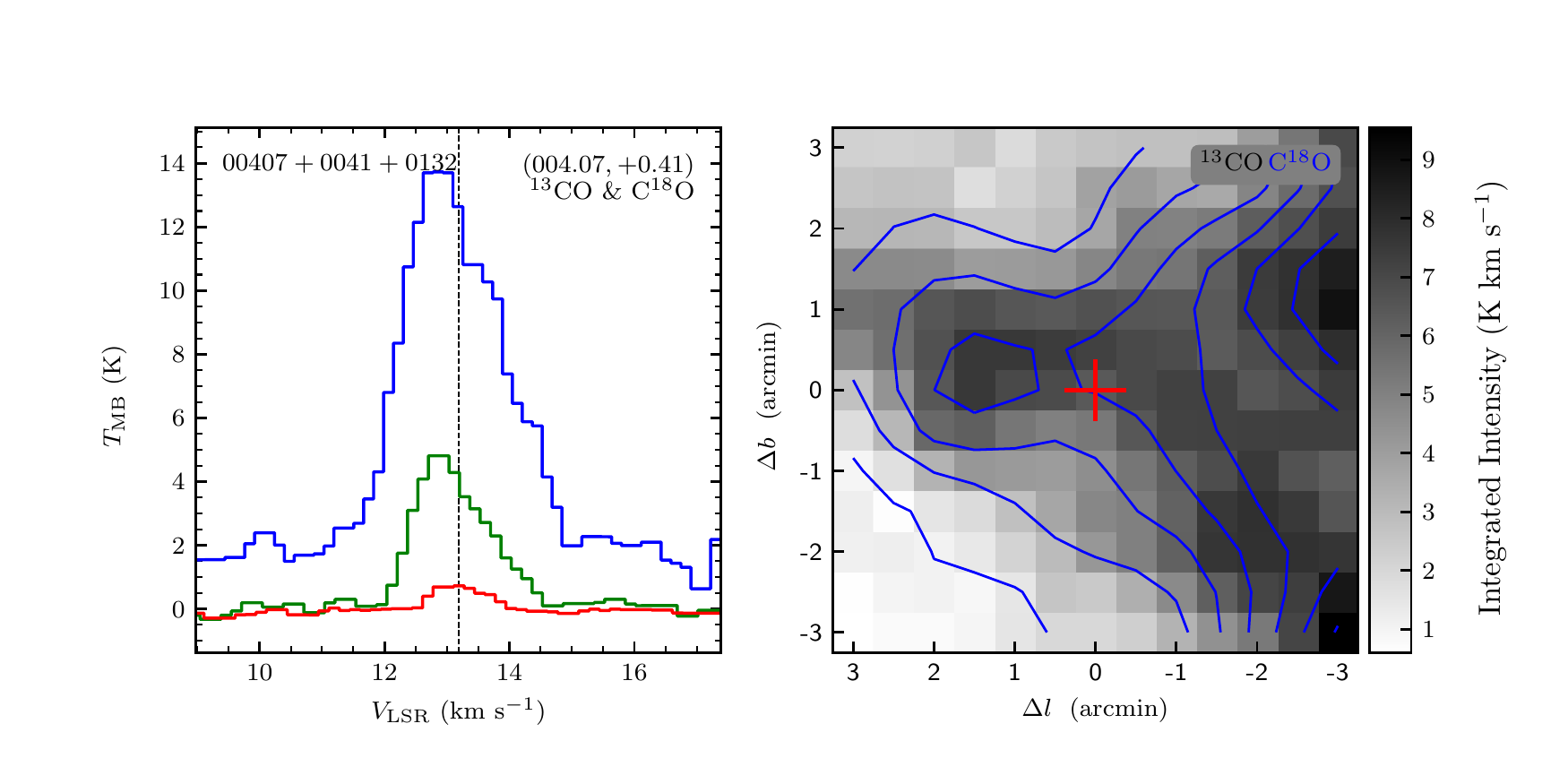}
\vspace{-0.5cm}

\includegraphics[width=9.0cm,angle=0]{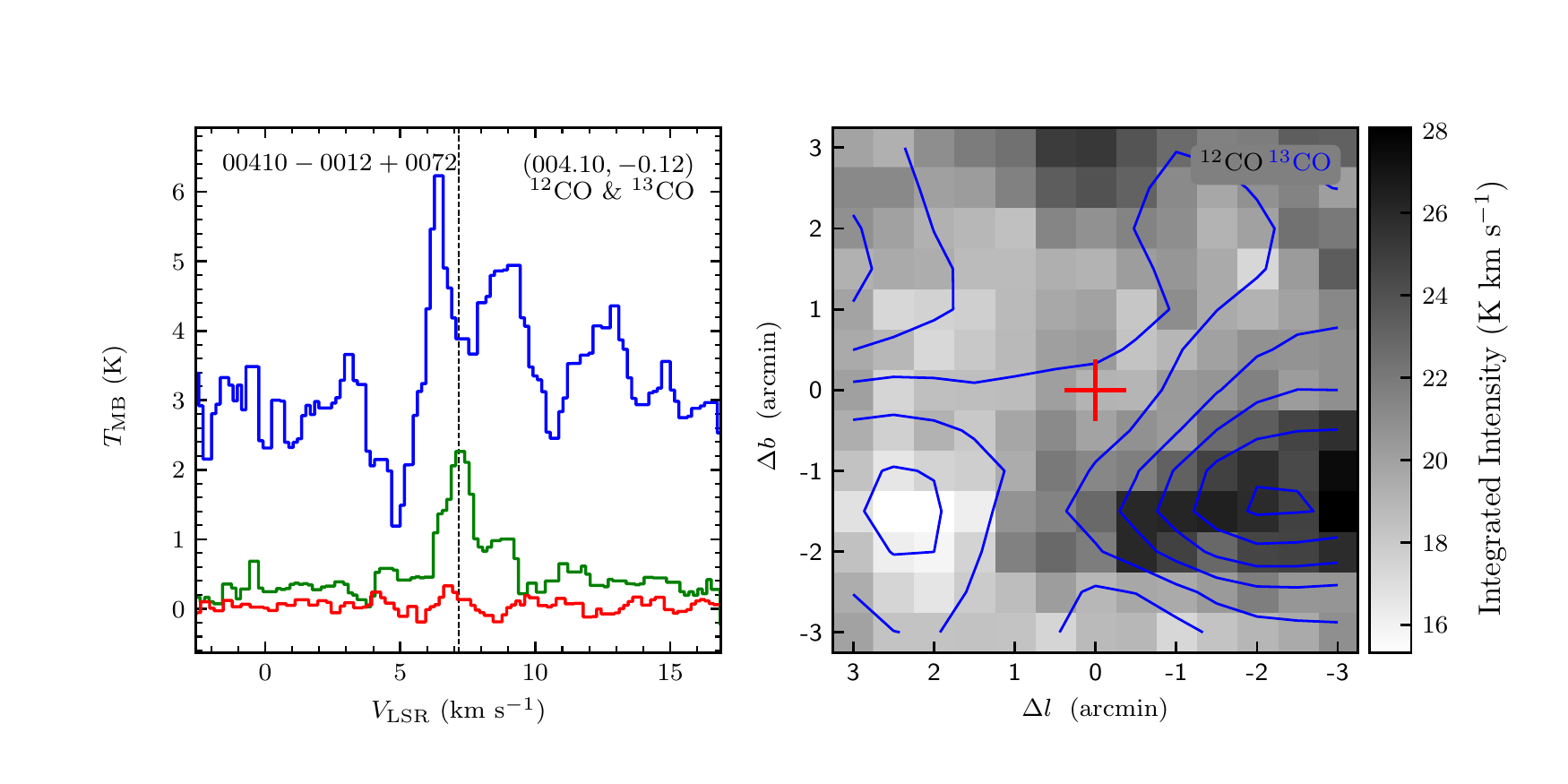}
\includegraphics[width=9.0cm,angle=0]{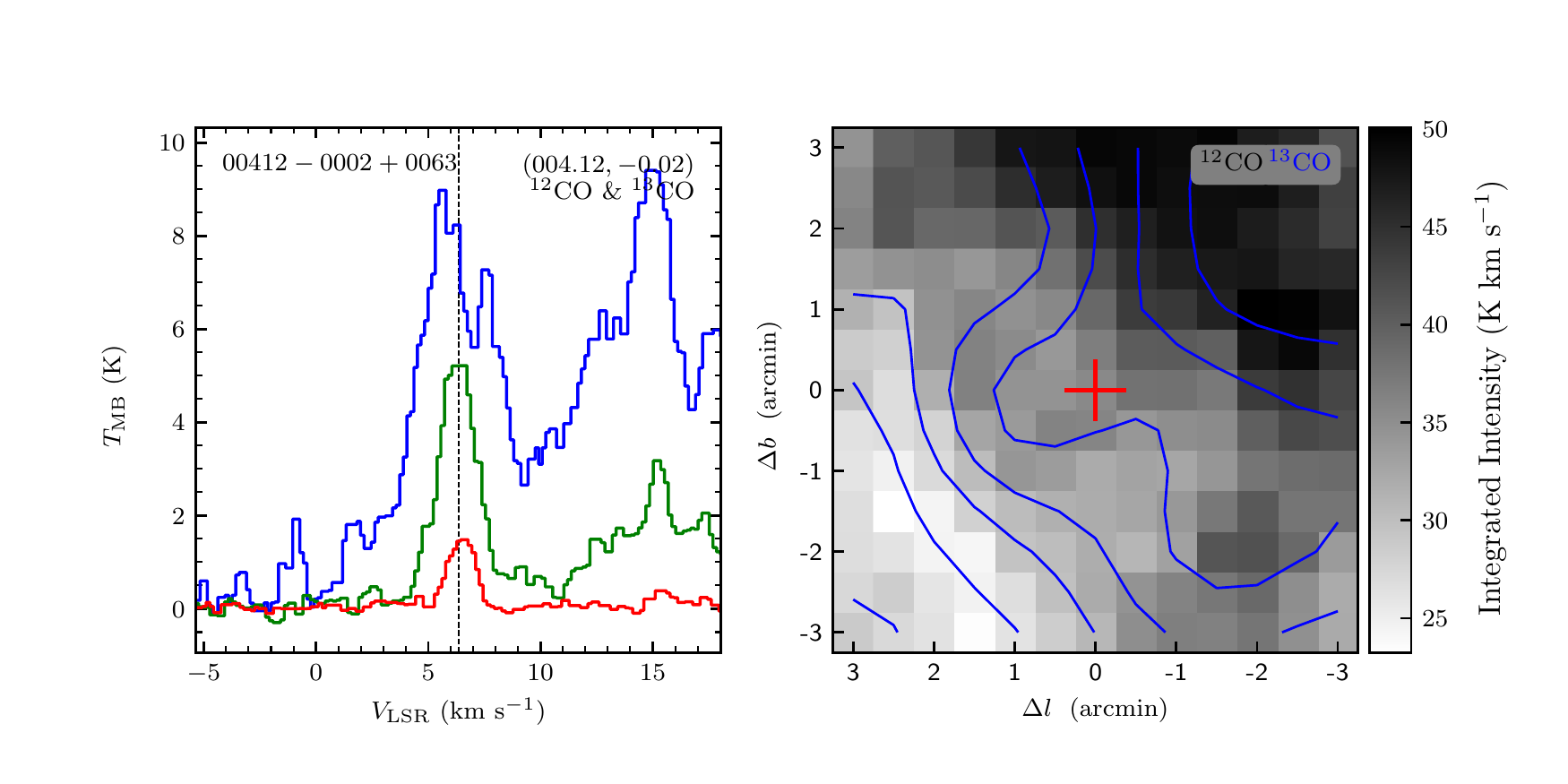}
\vspace{-0.5cm}

\includegraphics[width=9.0cm,angle=0]{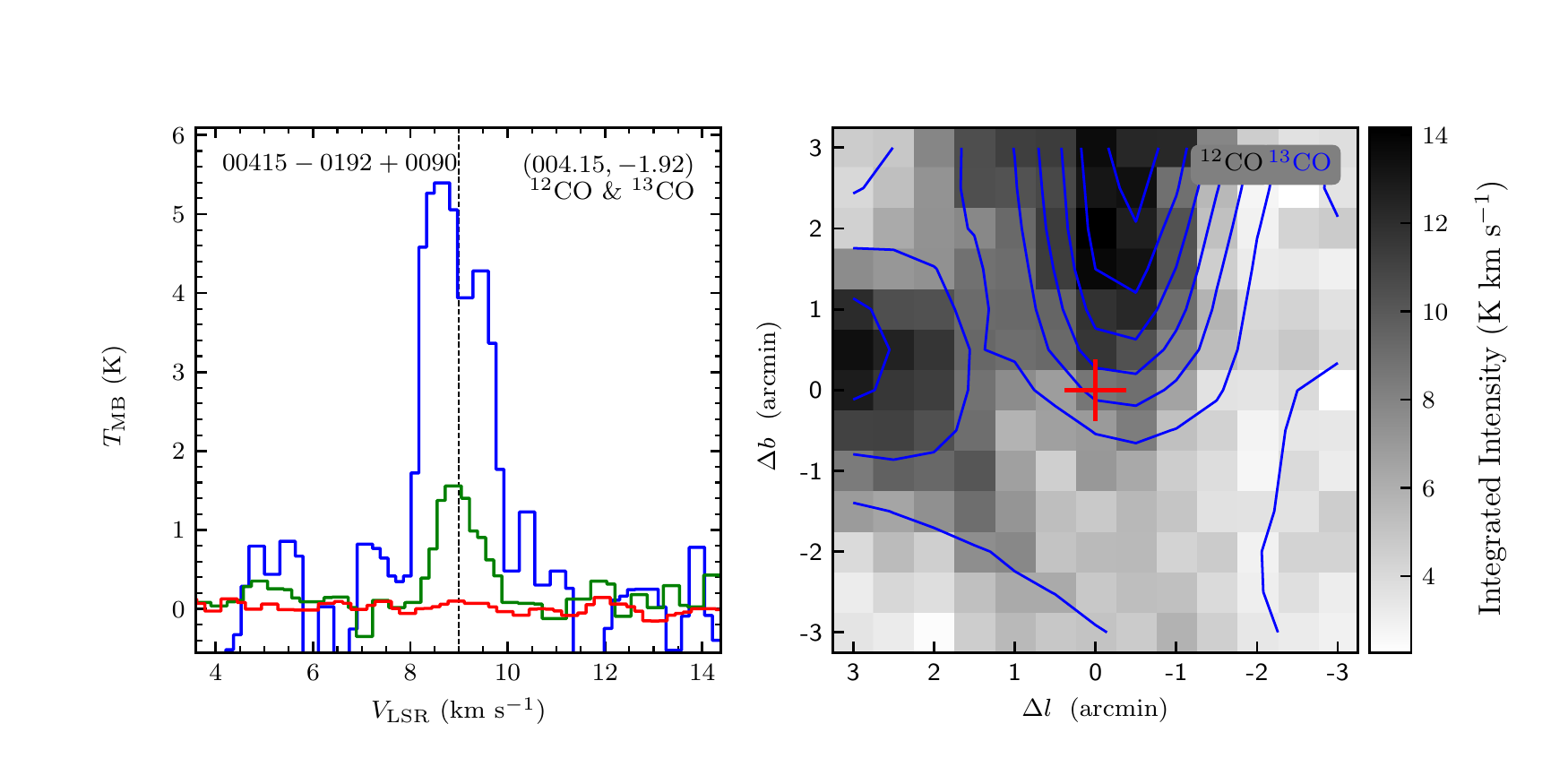}
\includegraphics[width=9.0cm,angle=0]{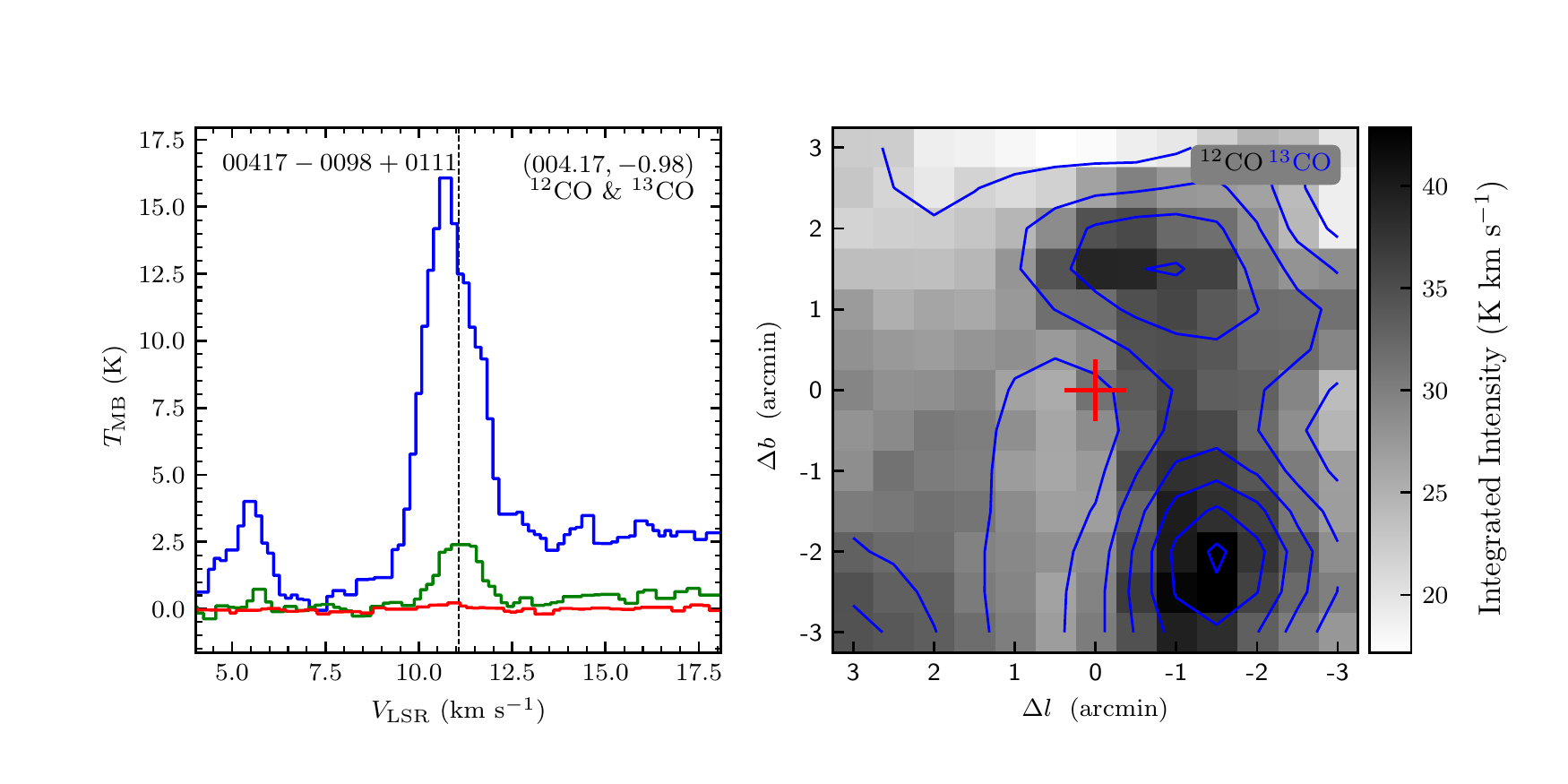}
\vspace{-0.5cm}

\includegraphics[width=9.0cm,angle=0]{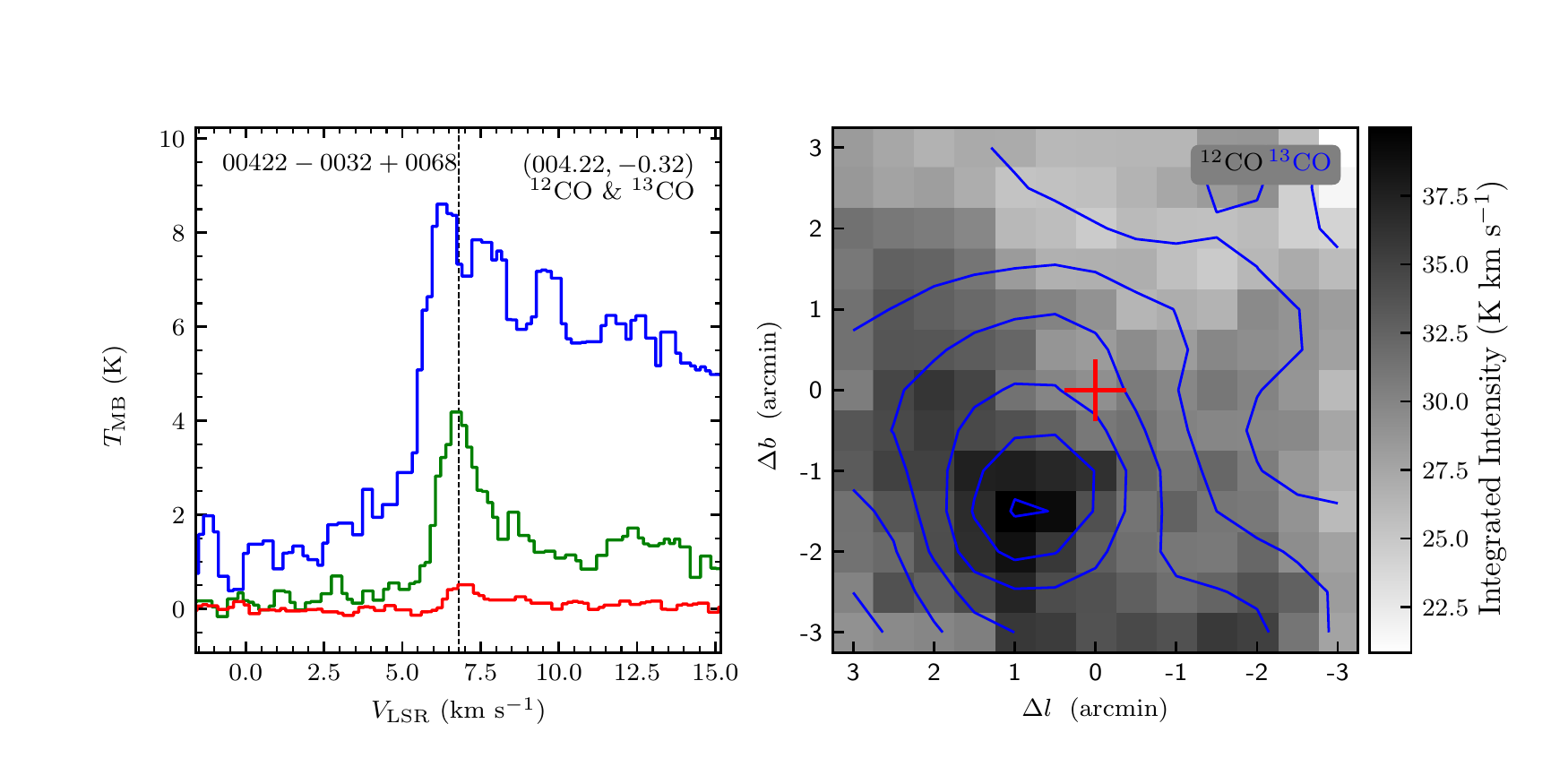}
\includegraphics[width=9.0cm,angle=0]{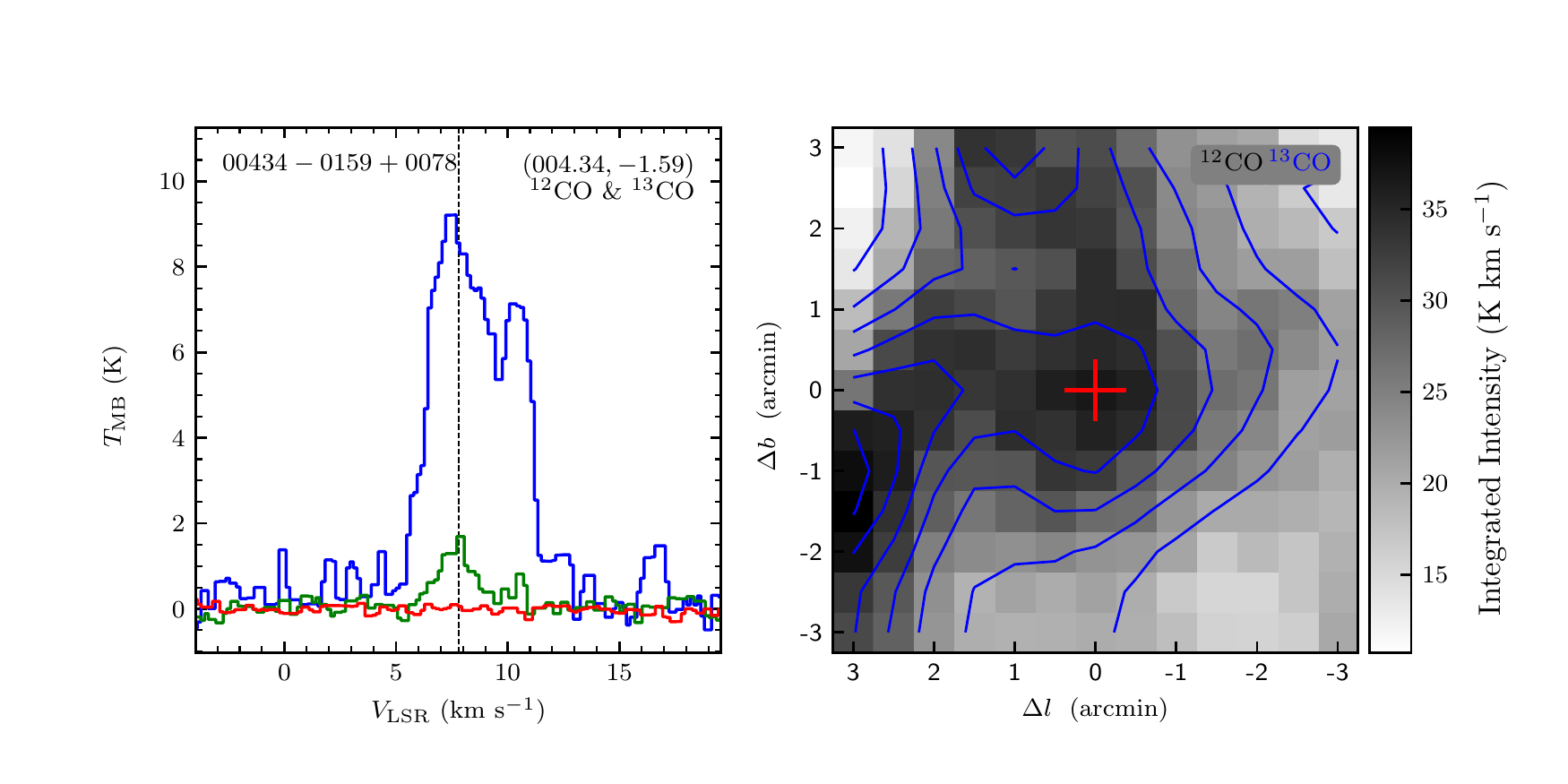}
\vspace{-0.5cm}

\includegraphics[width=9.0cm,angle=0]{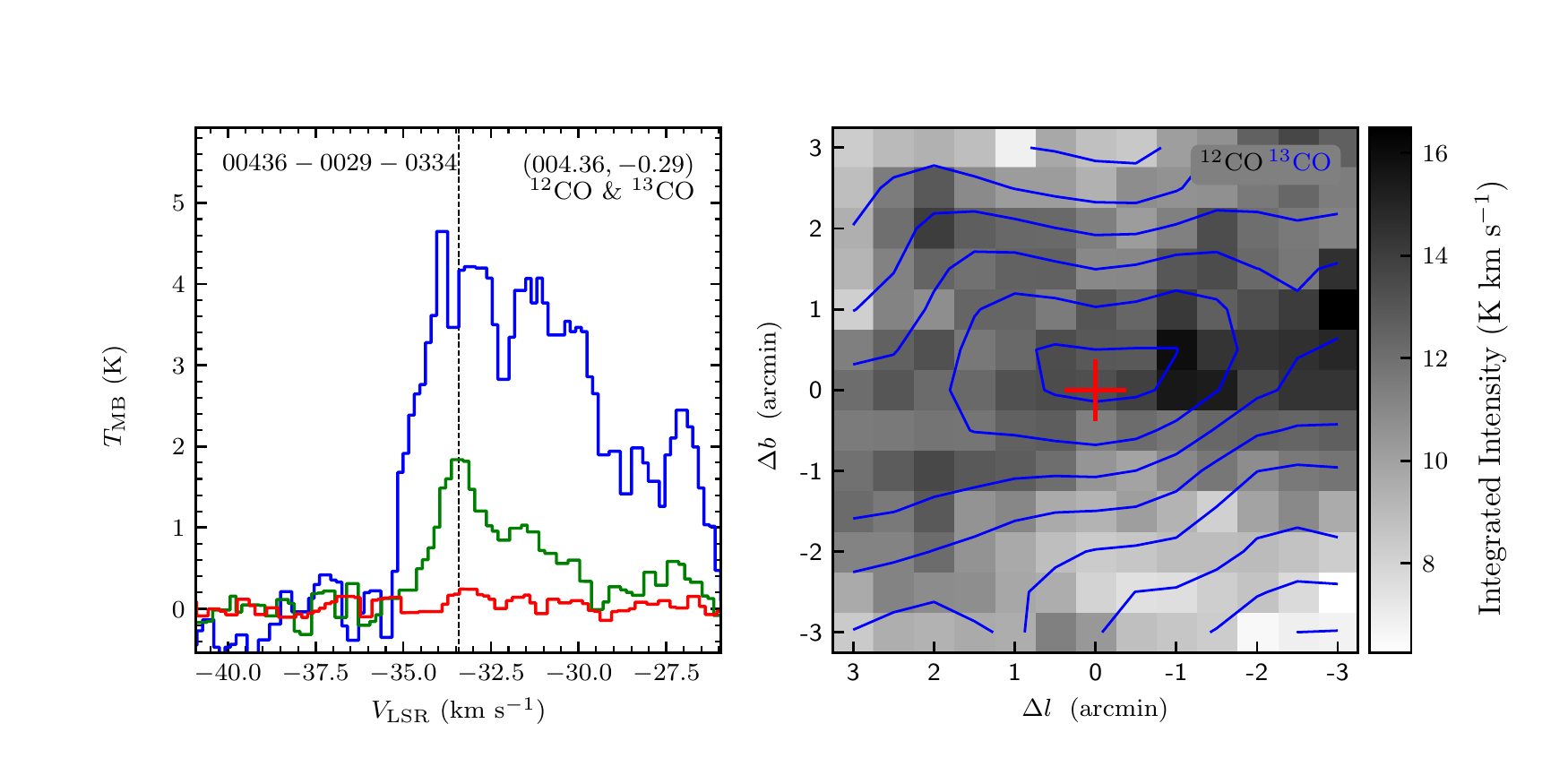}
\includegraphics[width=9.0cm,angle=0]{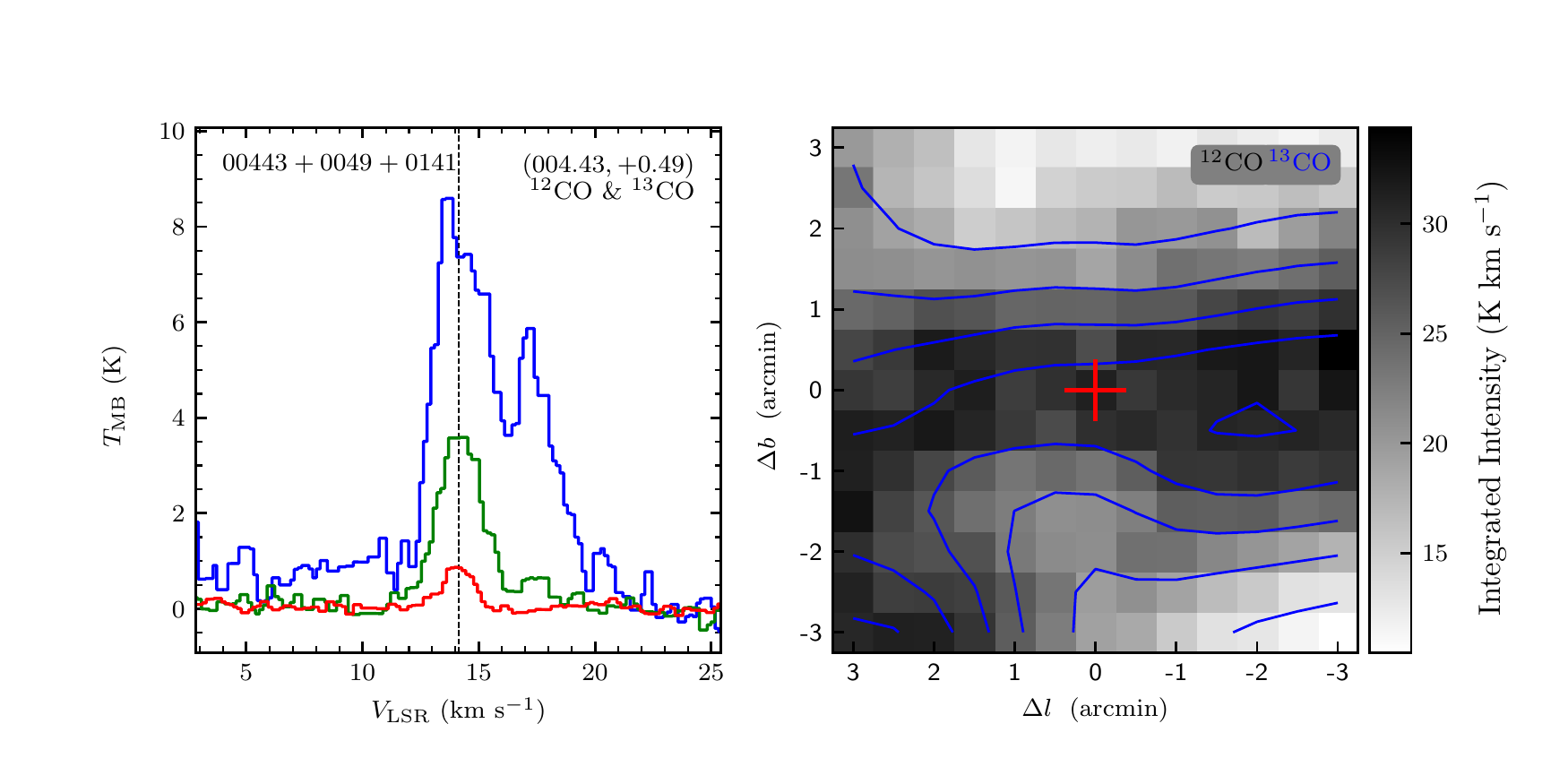}
\end{figure}
\clearpage

\begin{figure}
\includegraphics[width=9.0cm,angle=0]{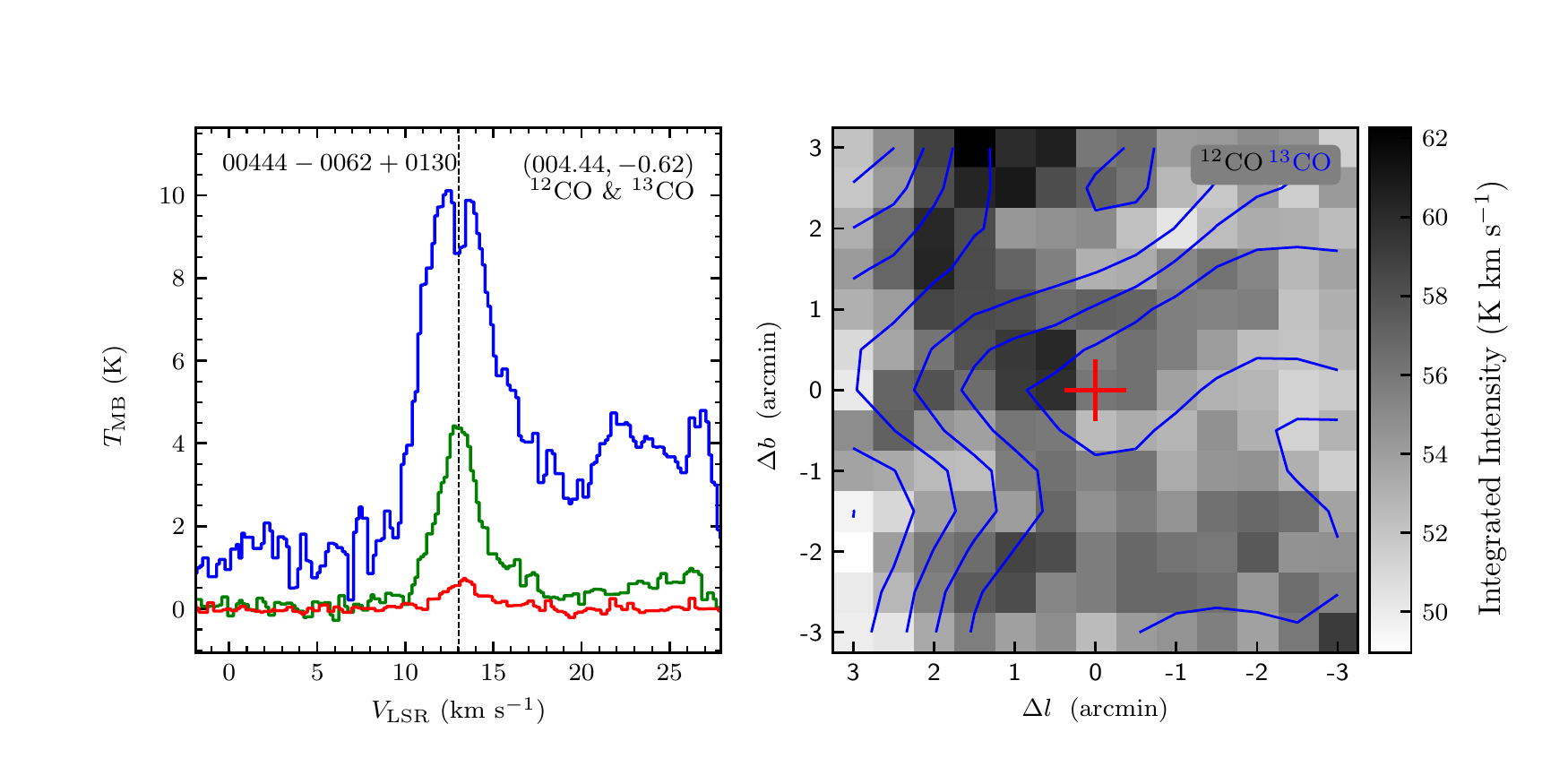}
\includegraphics[width=9.0cm,angle=0]{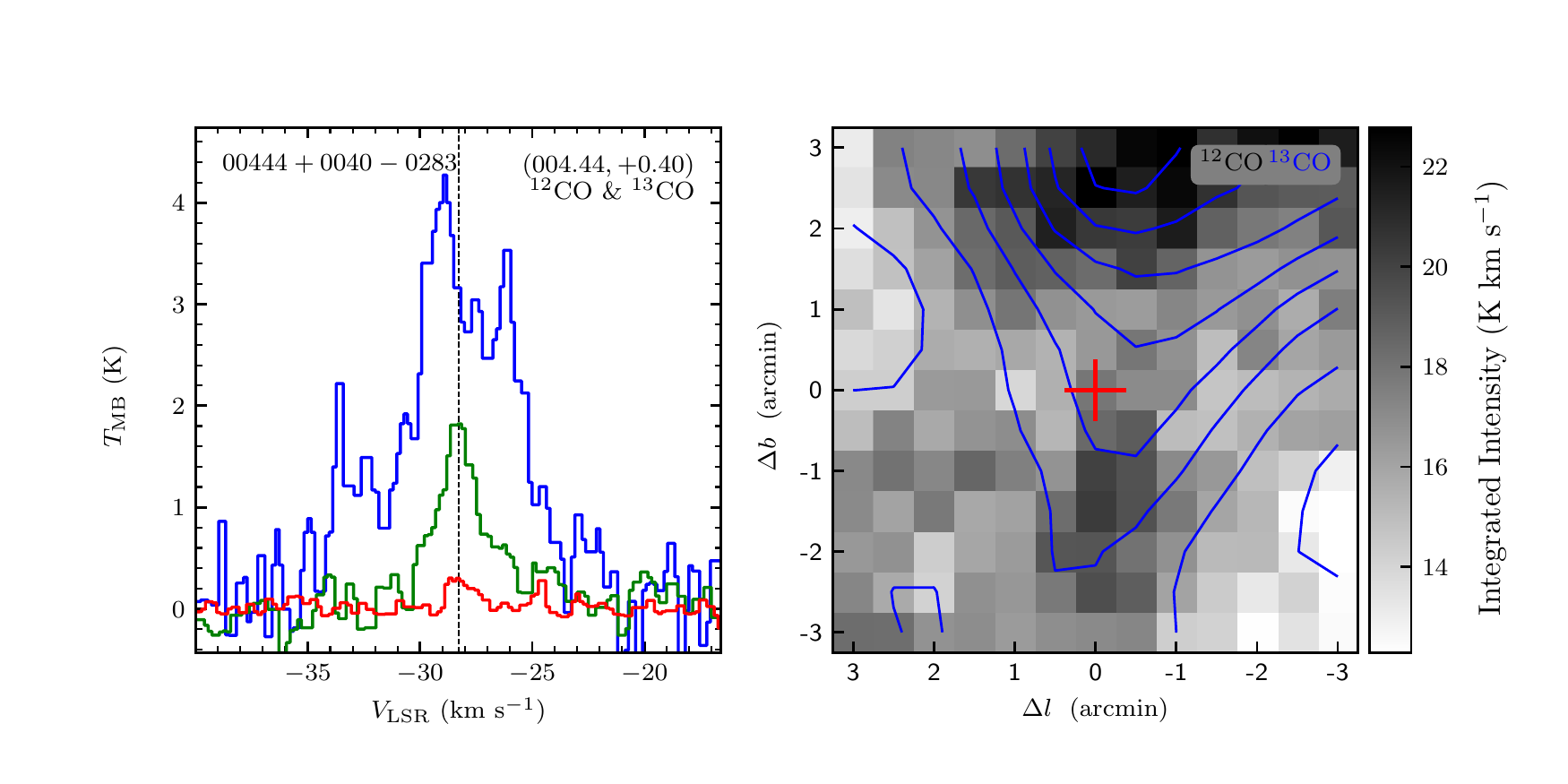}
\vspace{-0.5cm}

\includegraphics[width=9.0cm,angle=0]{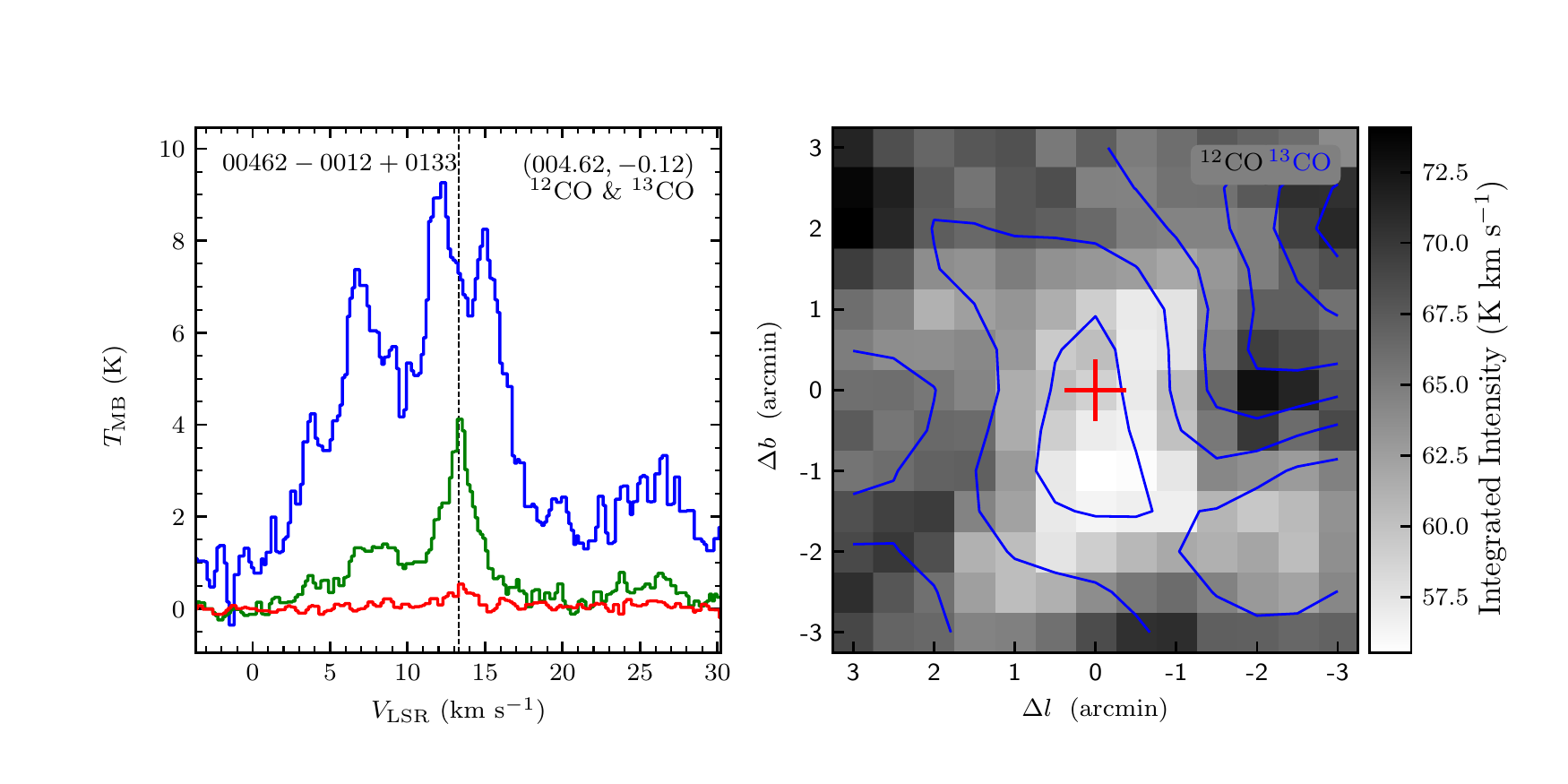}
\includegraphics[width=9.0cm,angle=0]{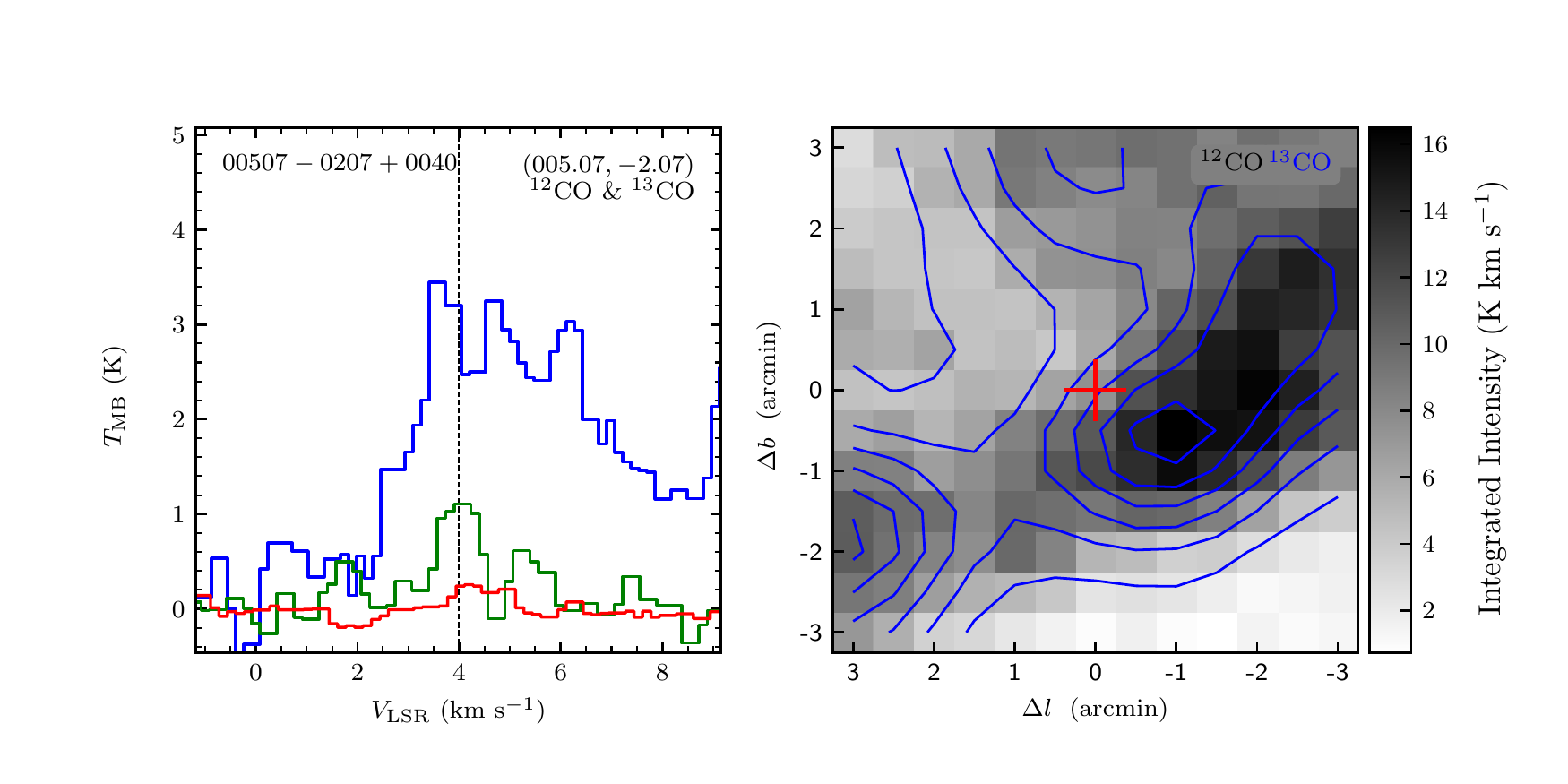}
\vspace{-0.5cm}

\includegraphics[width=9.0cm,angle=0]{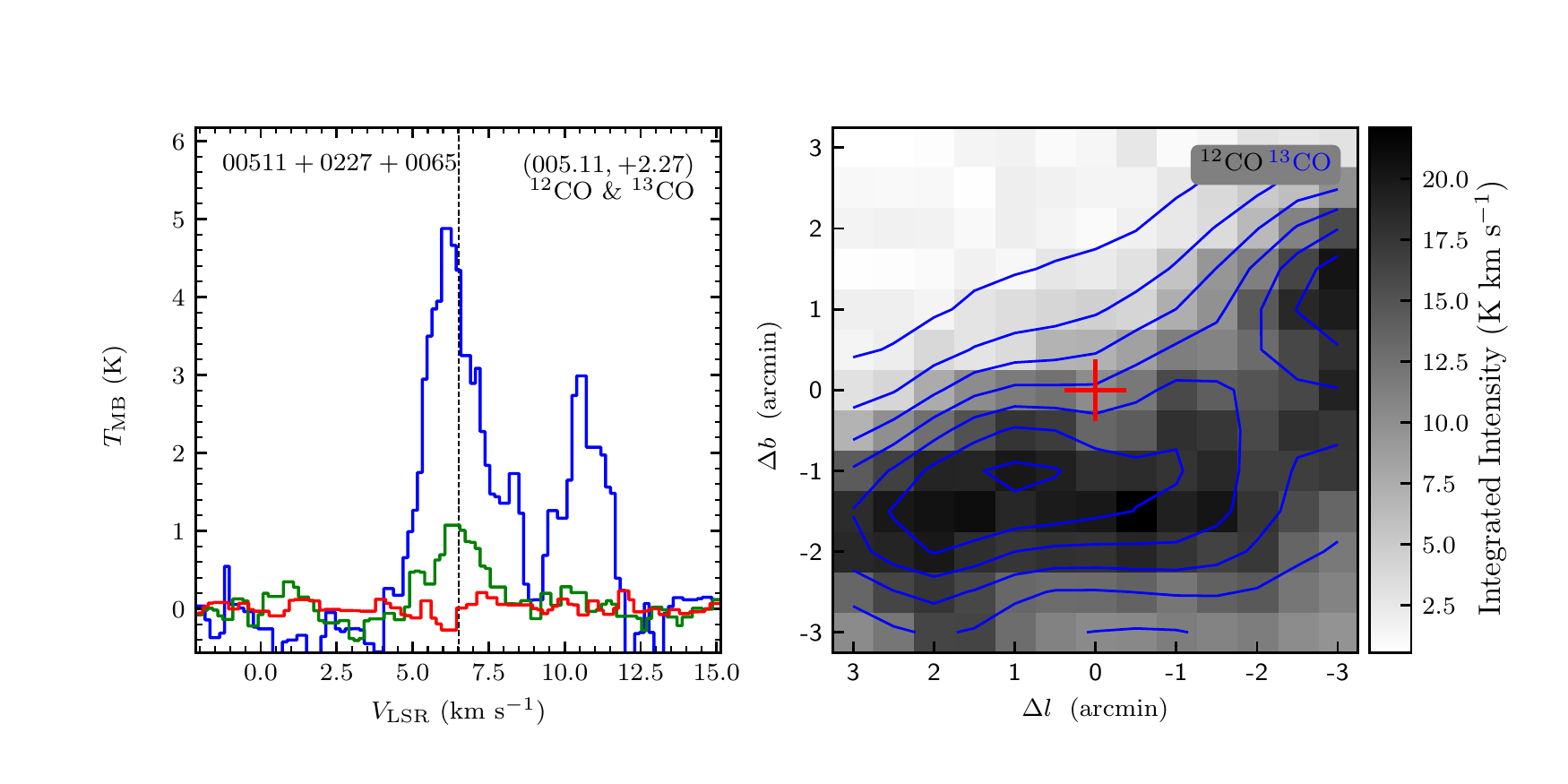}
\includegraphics[width=9.0cm,angle=0]{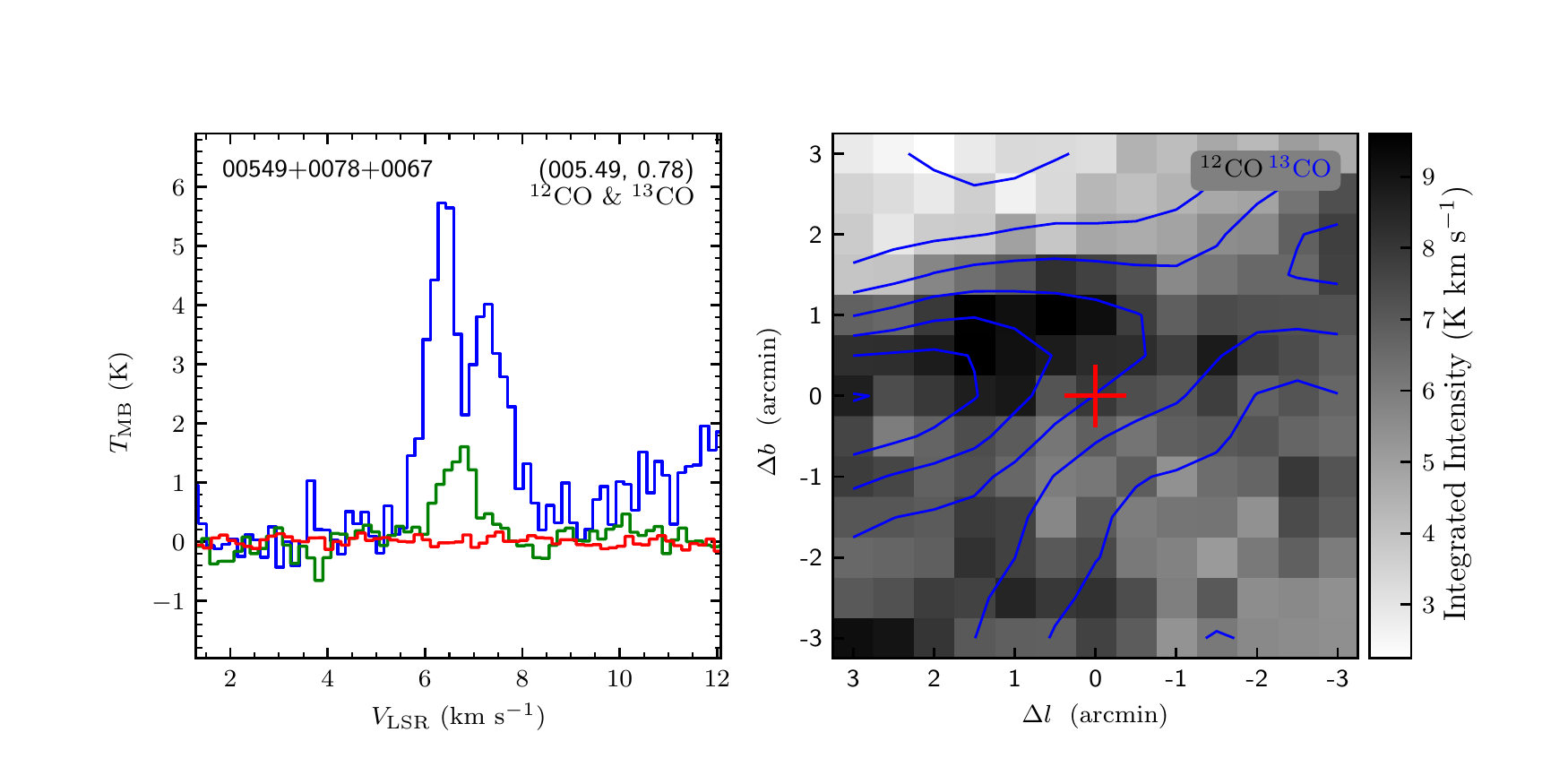}
\vspace{-0.5cm}

\includegraphics[width=9.0cm,angle=0]{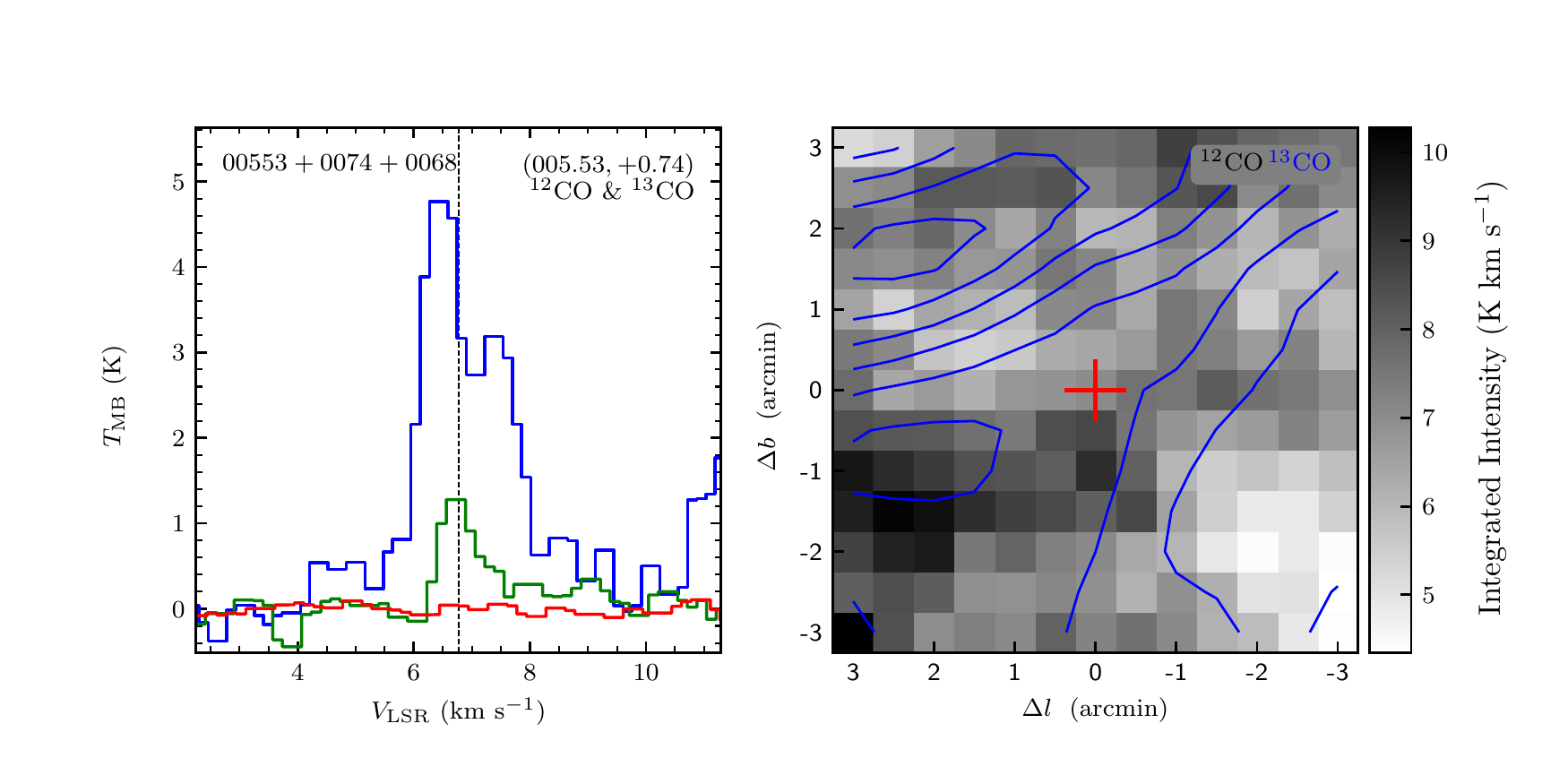}
\includegraphics[width=9.0cm,angle=0]{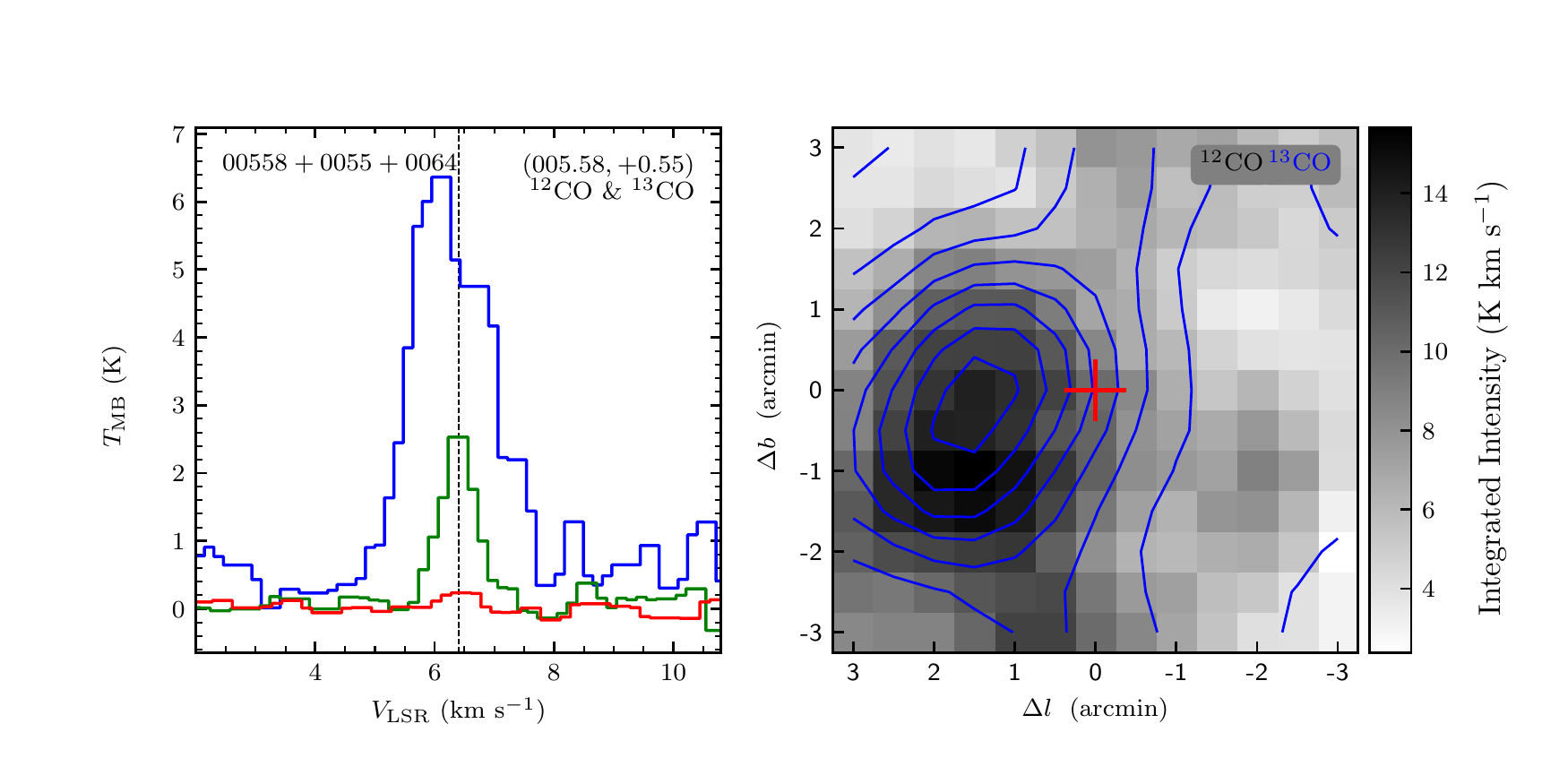}
\vspace{-0.5cm}

\includegraphics[width=9.0cm,angle=0]{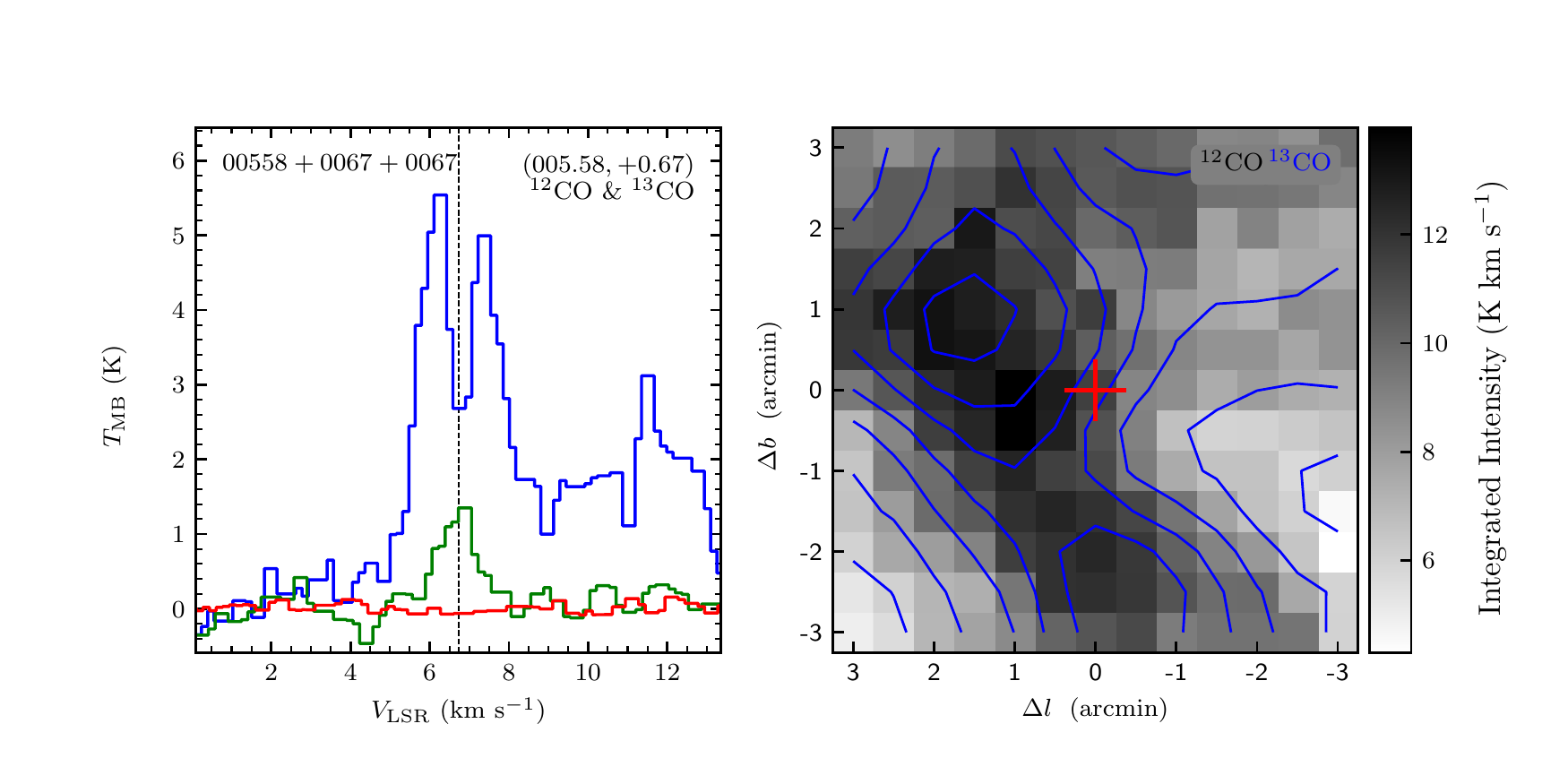}
\includegraphics[width=9.0cm,angle=0]{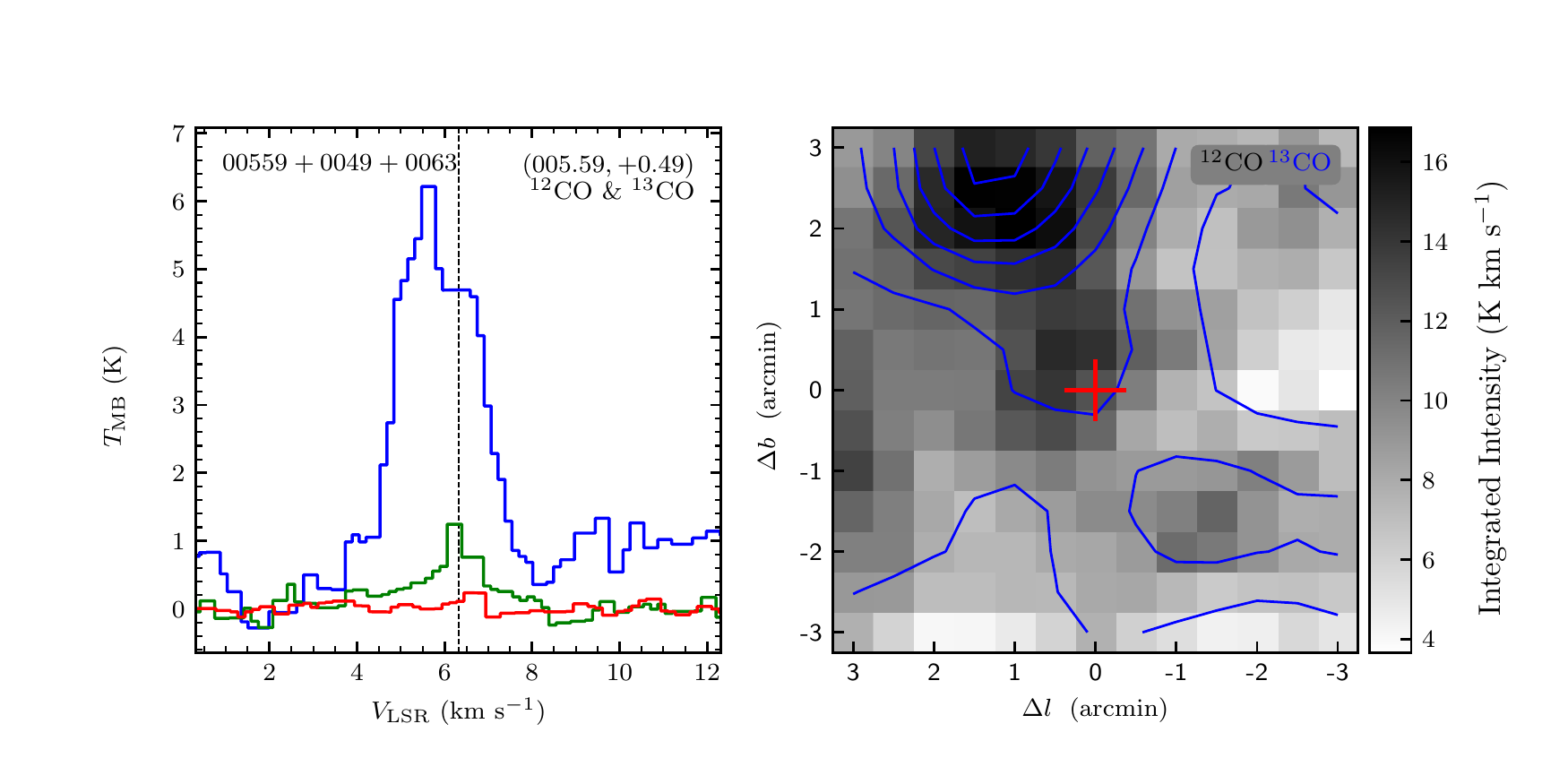}
\end{figure}
\clearpage

\begin{figure}
\includegraphics[width=9.0cm,angle=0]{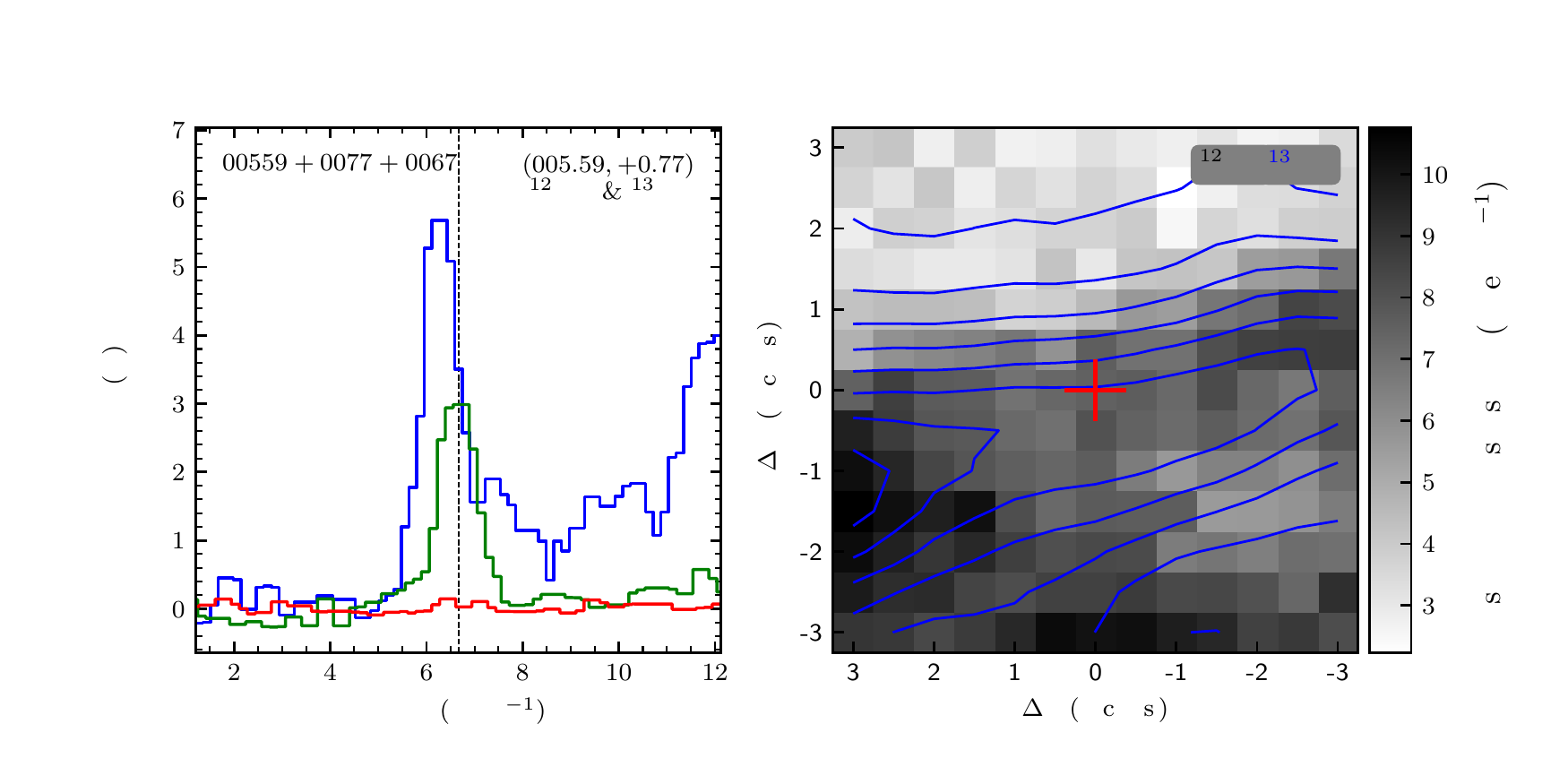}
\includegraphics[width=9.0cm,angle=0]{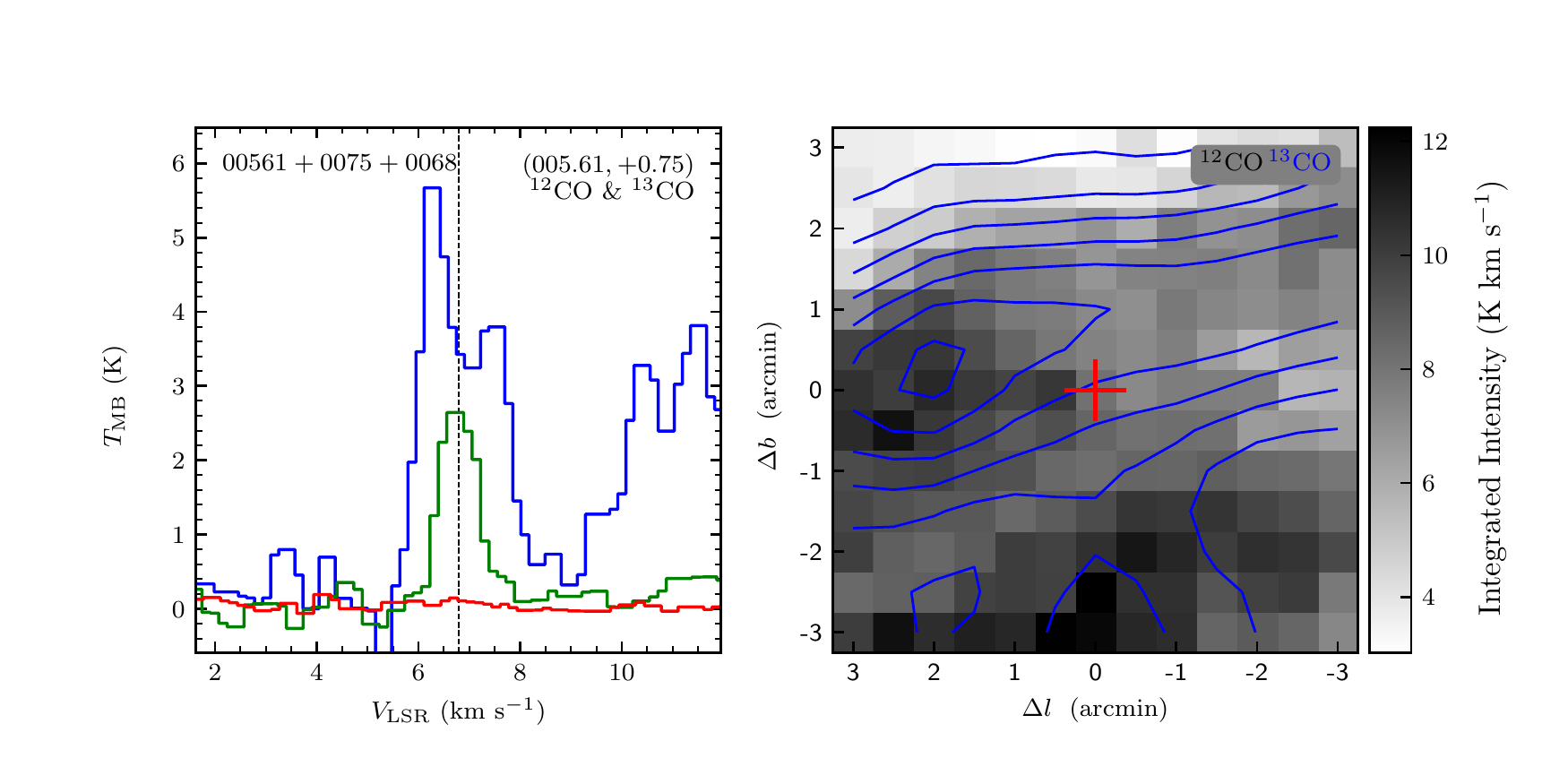}
\vspace{-0.5cm}

\includegraphics[width=9.0cm,angle=0]{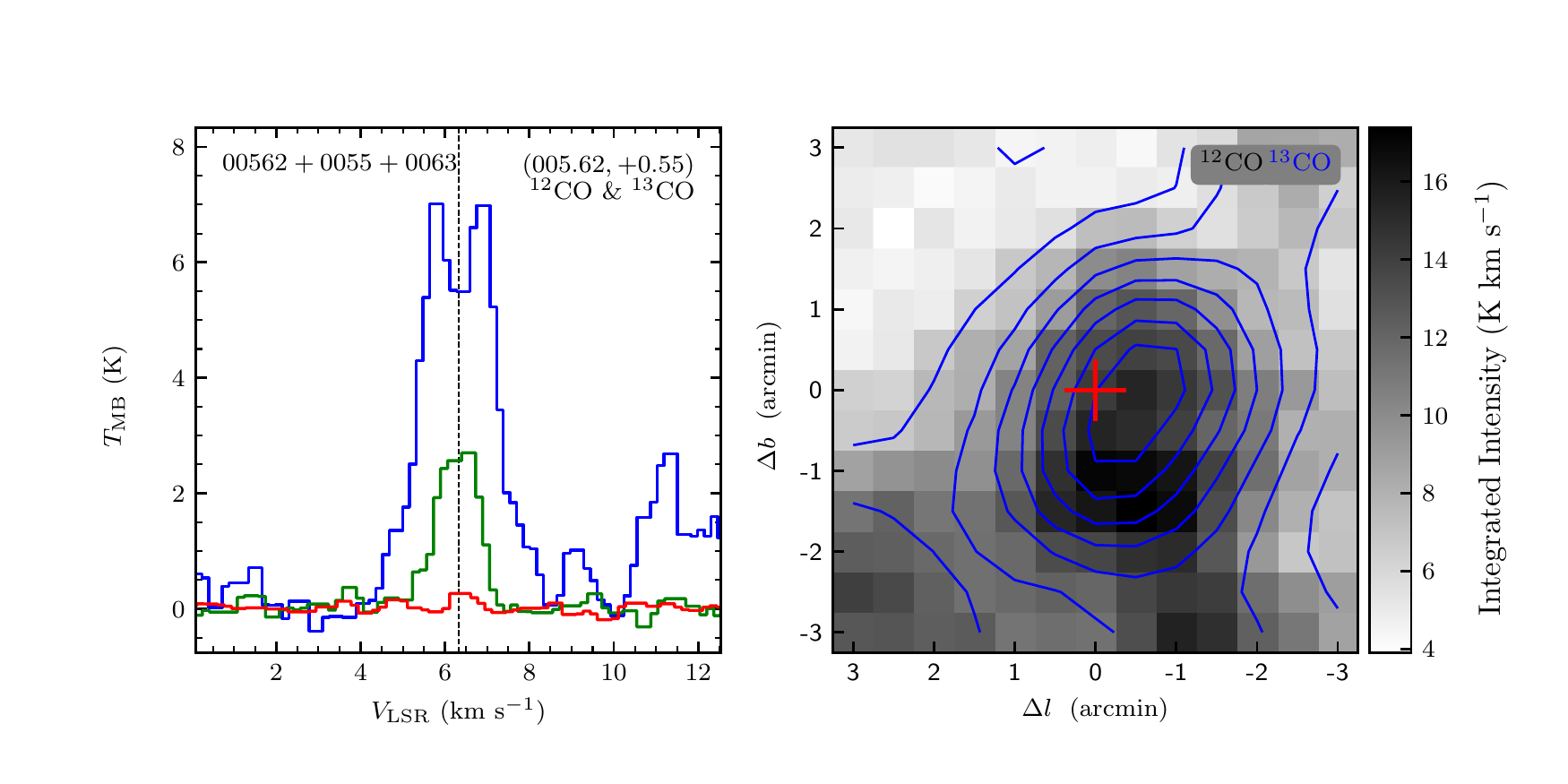}
\includegraphics[width=9.0cm,angle=0]{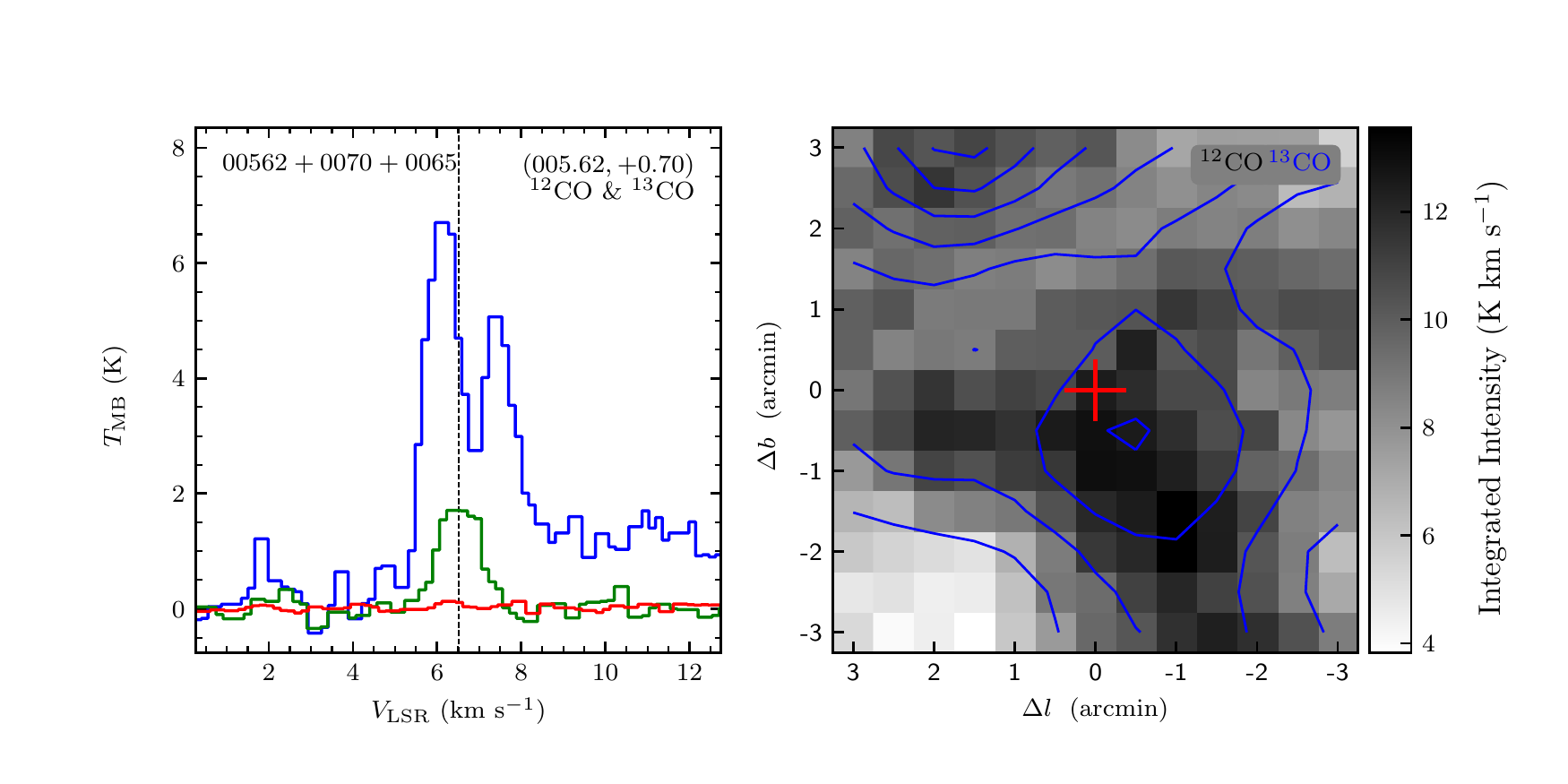}
\vspace{-0.5cm}

\includegraphics[width=9.0cm,angle=0]{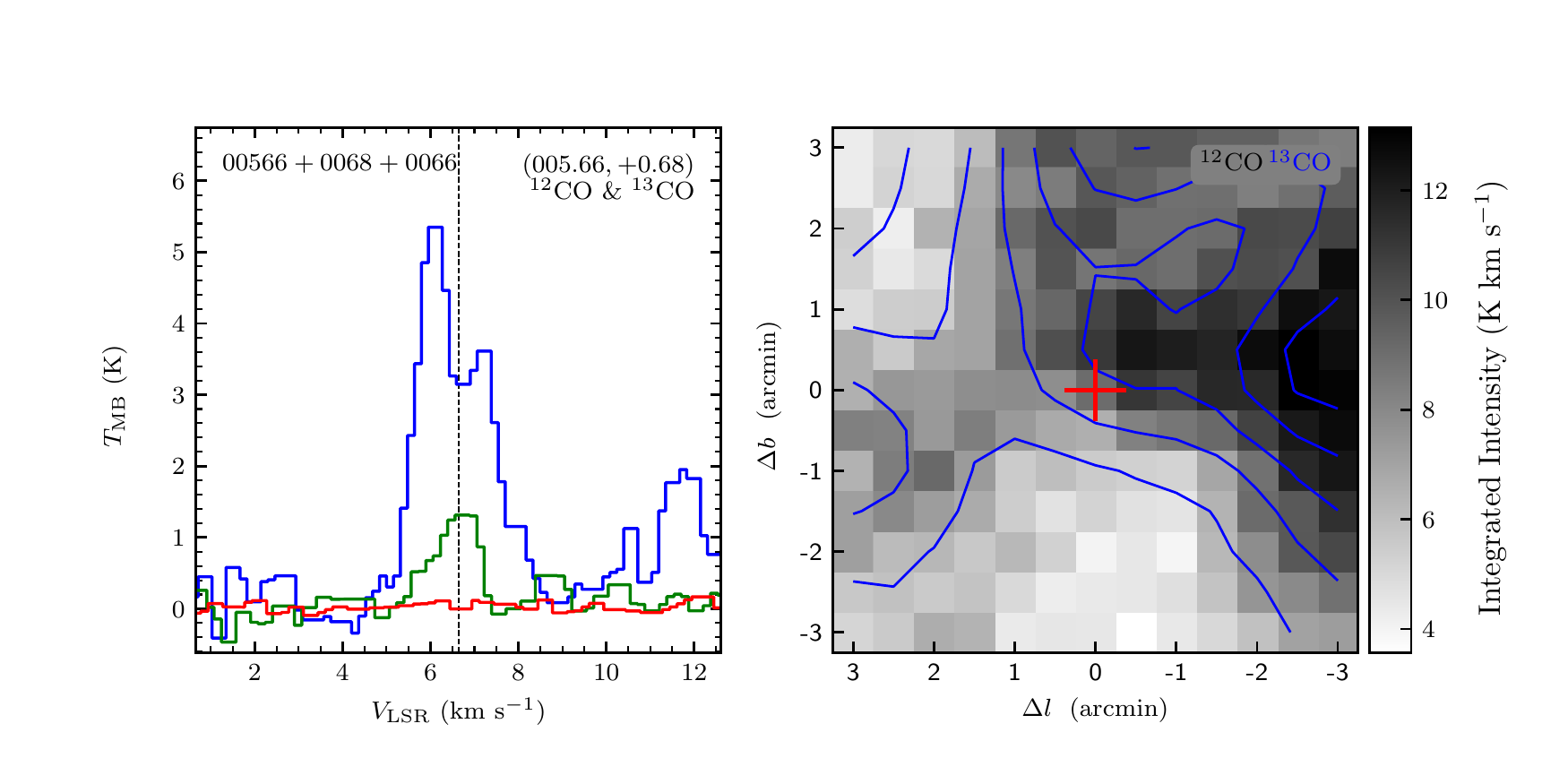}
\includegraphics[width=9.0cm,angle=0]{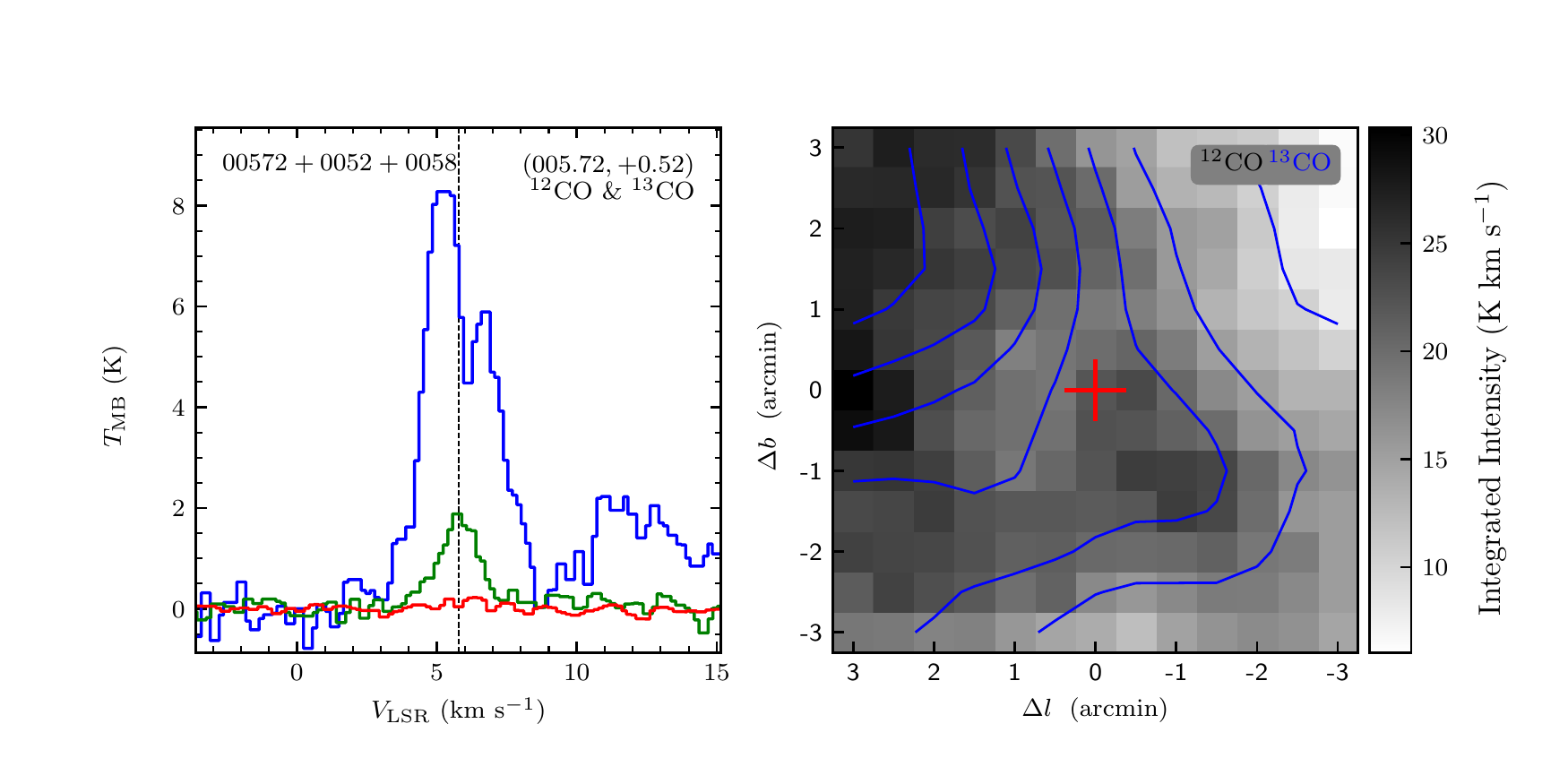}
\vspace{-0.5cm}

\includegraphics[width=9.0cm,angle=0]{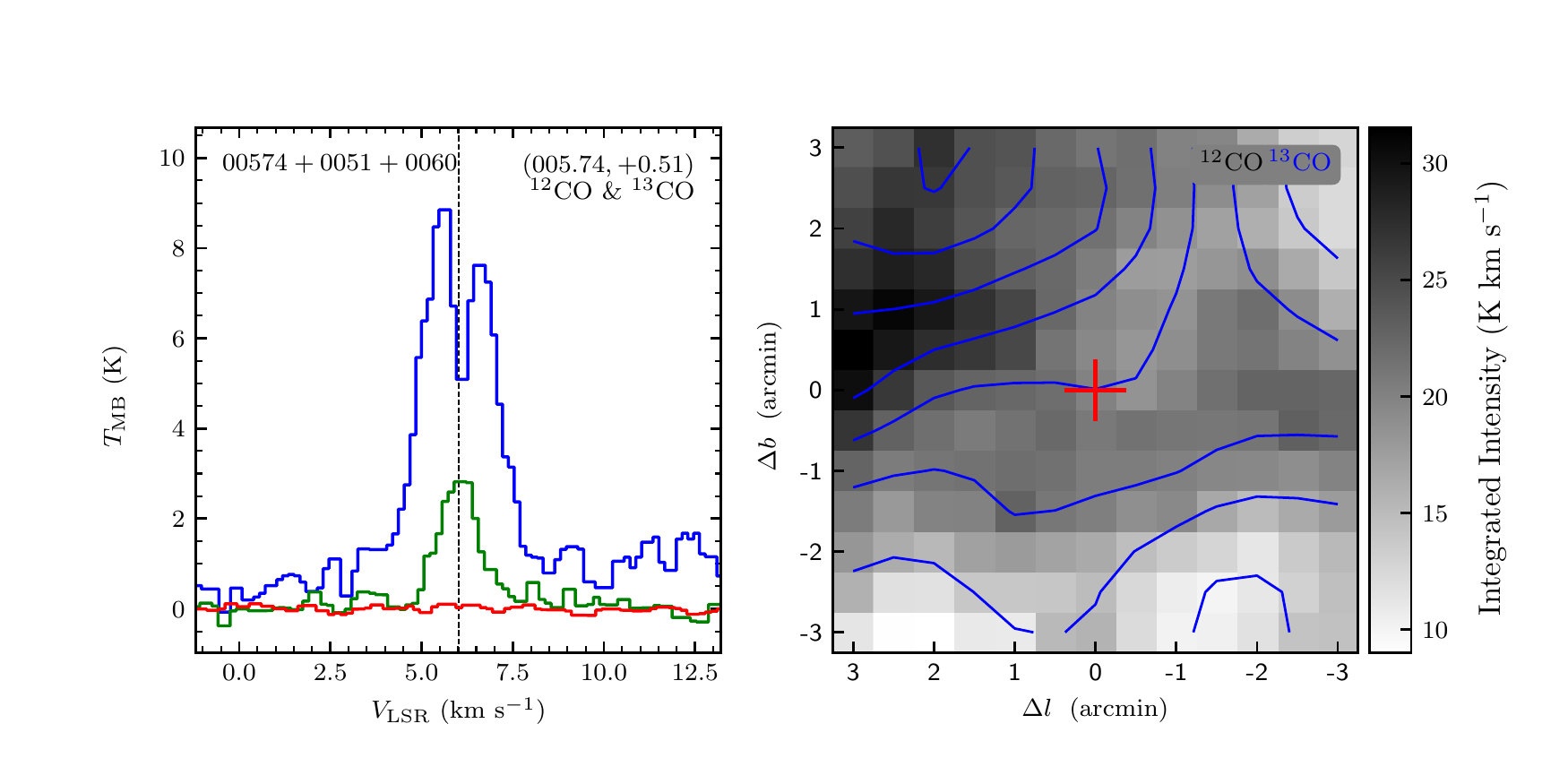}
\includegraphics[width=9.0cm,angle=0]{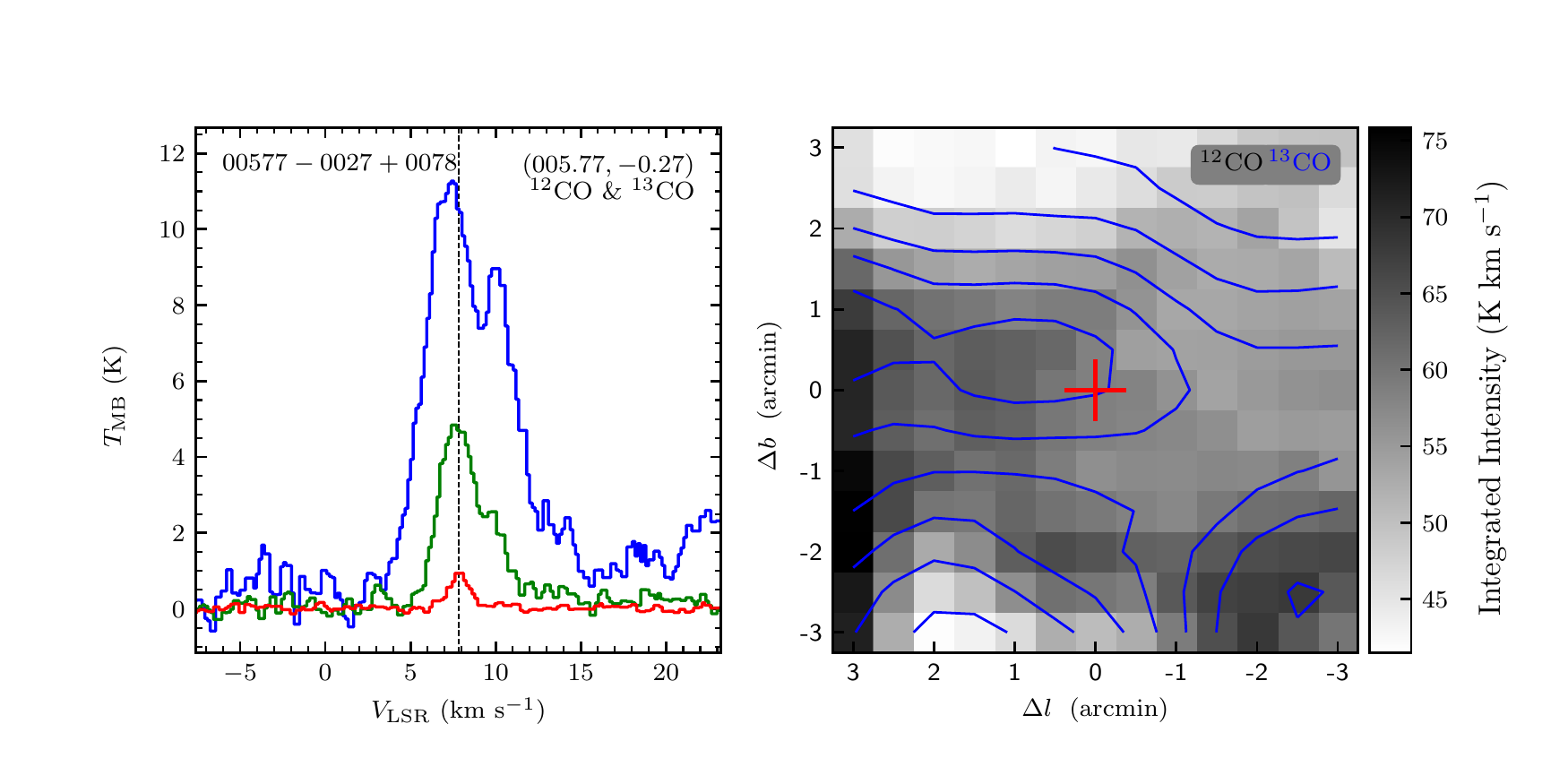}
\vspace{-0.5cm}

\includegraphics[width=9.0cm,angle=0]{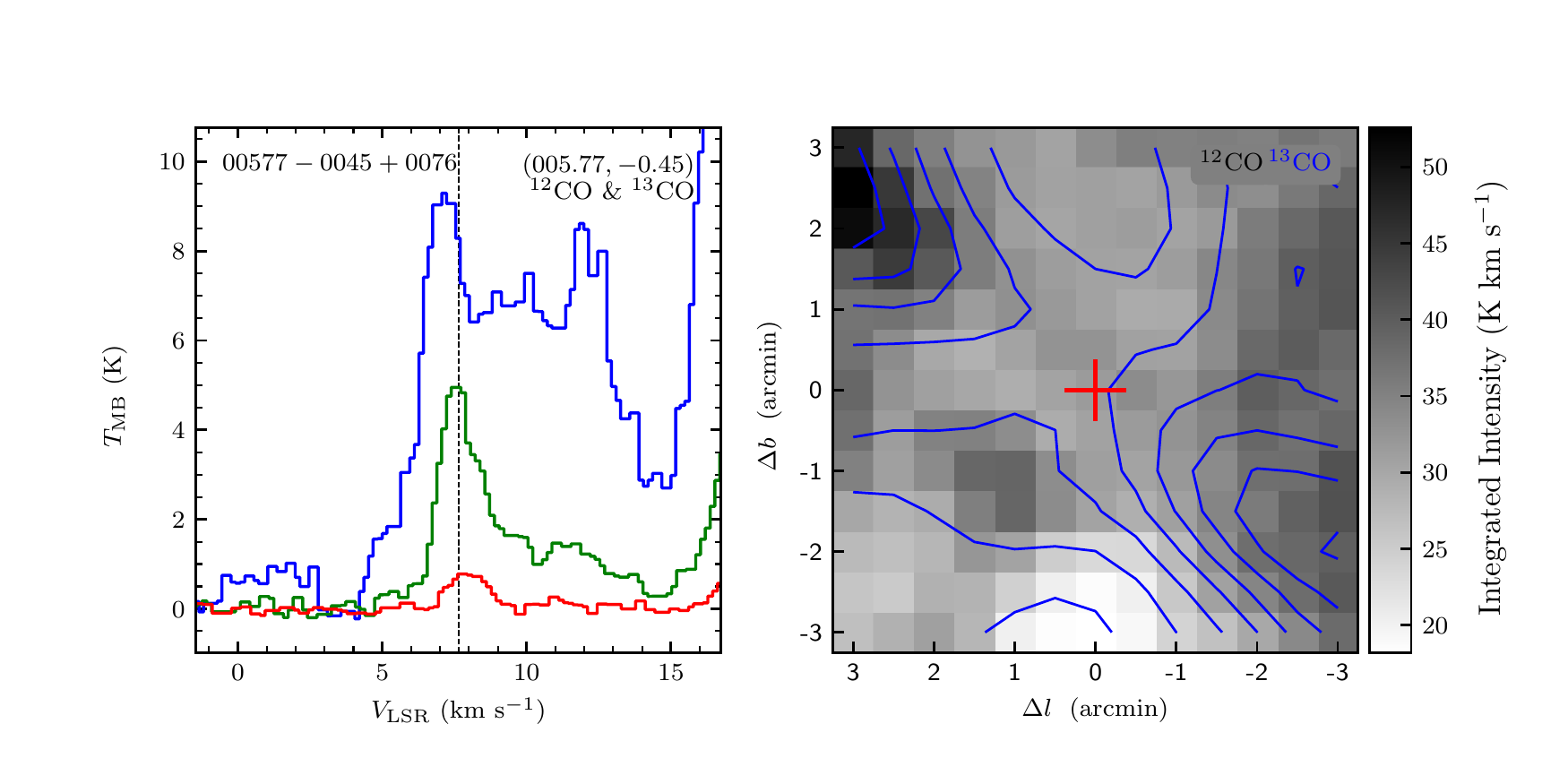}
\includegraphics[width=9.0cm,angle=0]{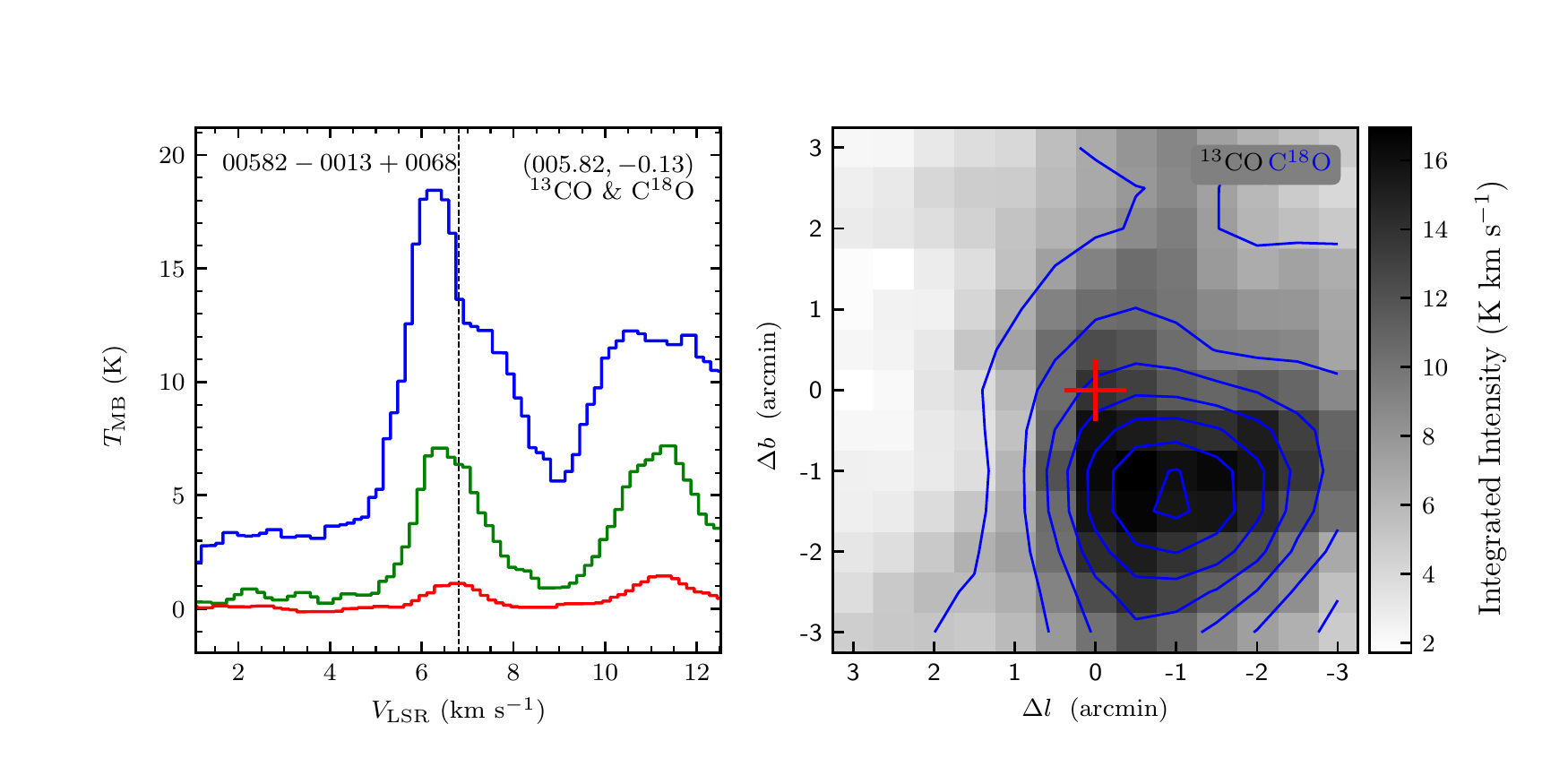}
\end{figure}
\clearpage

\begin{figure}
\includegraphics[width=9.0cm,angle=0]{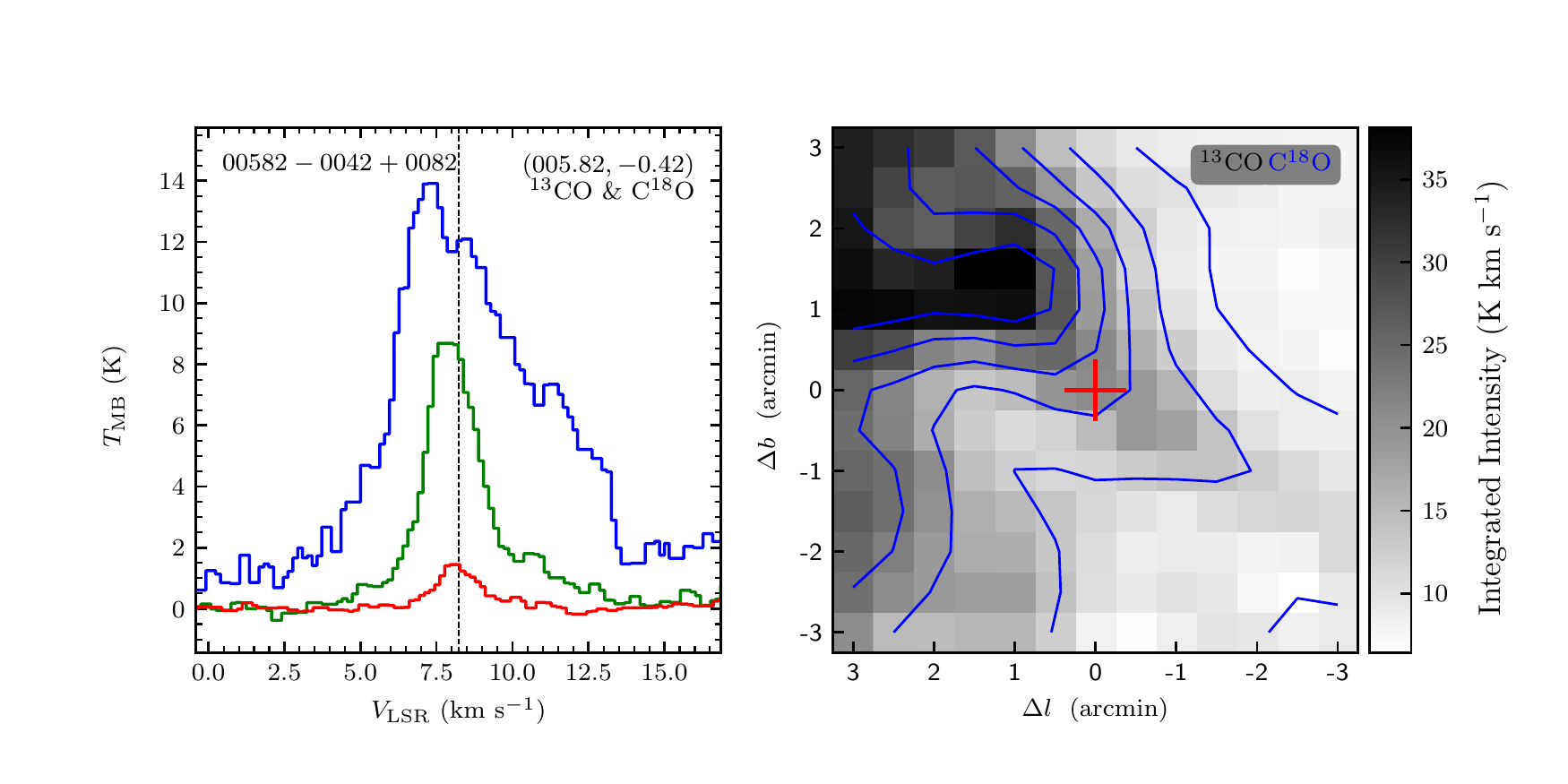}
\includegraphics[width=9.0cm,angle=0]{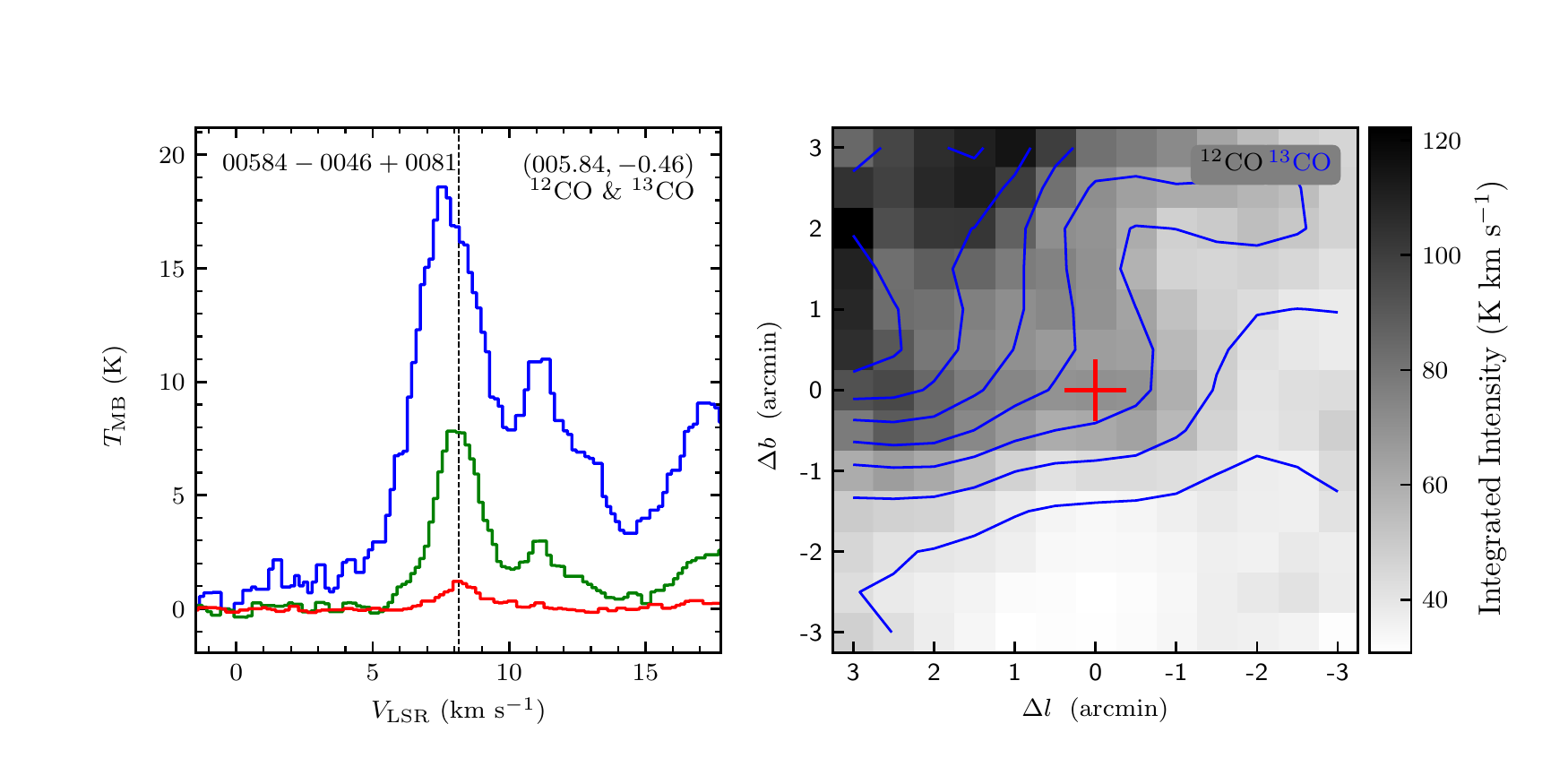}
\vspace{-0.5cm}

\includegraphics[width=9.0cm,angle=0]{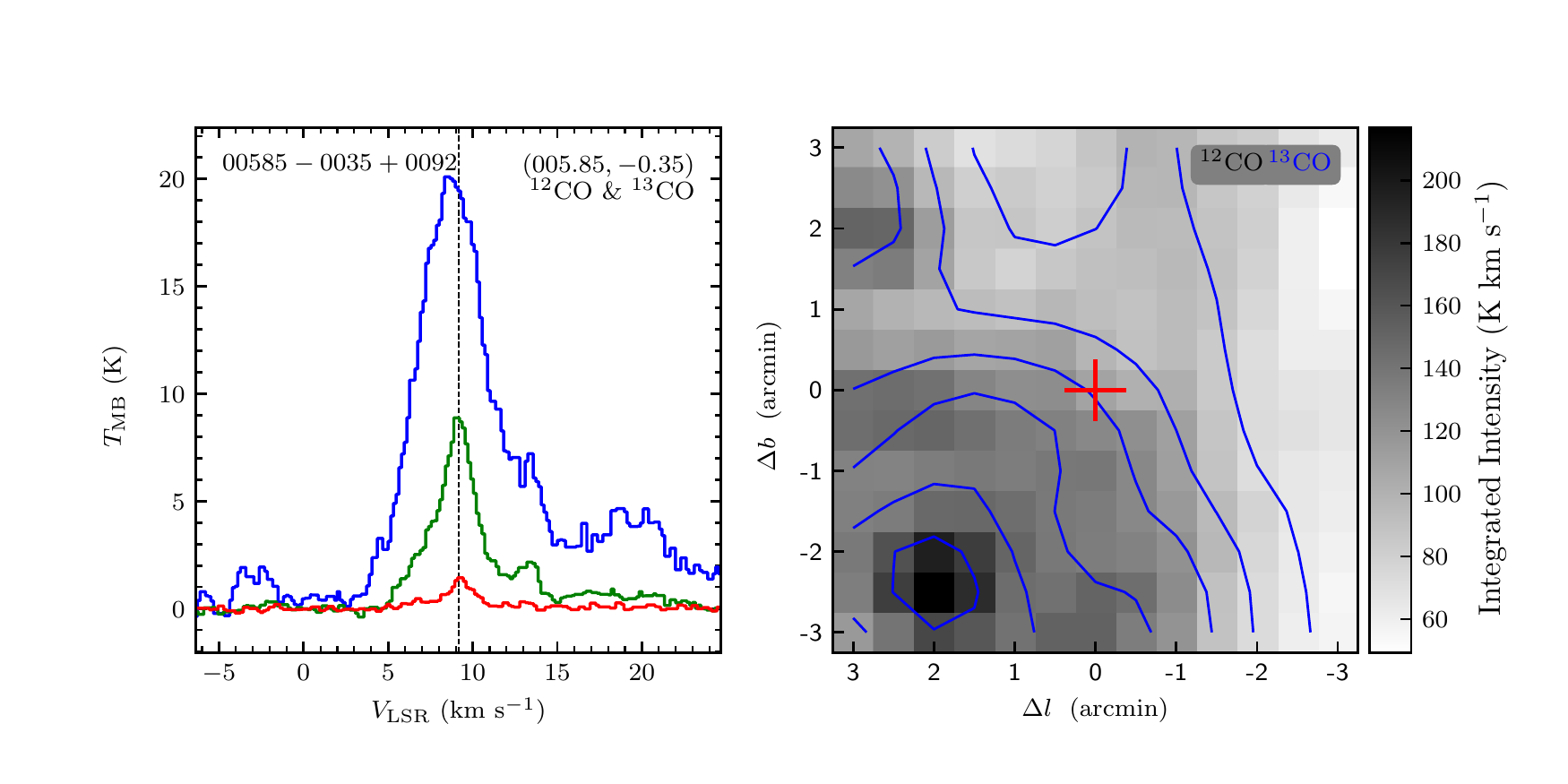}
\includegraphics[width=9.0cm,angle=0]{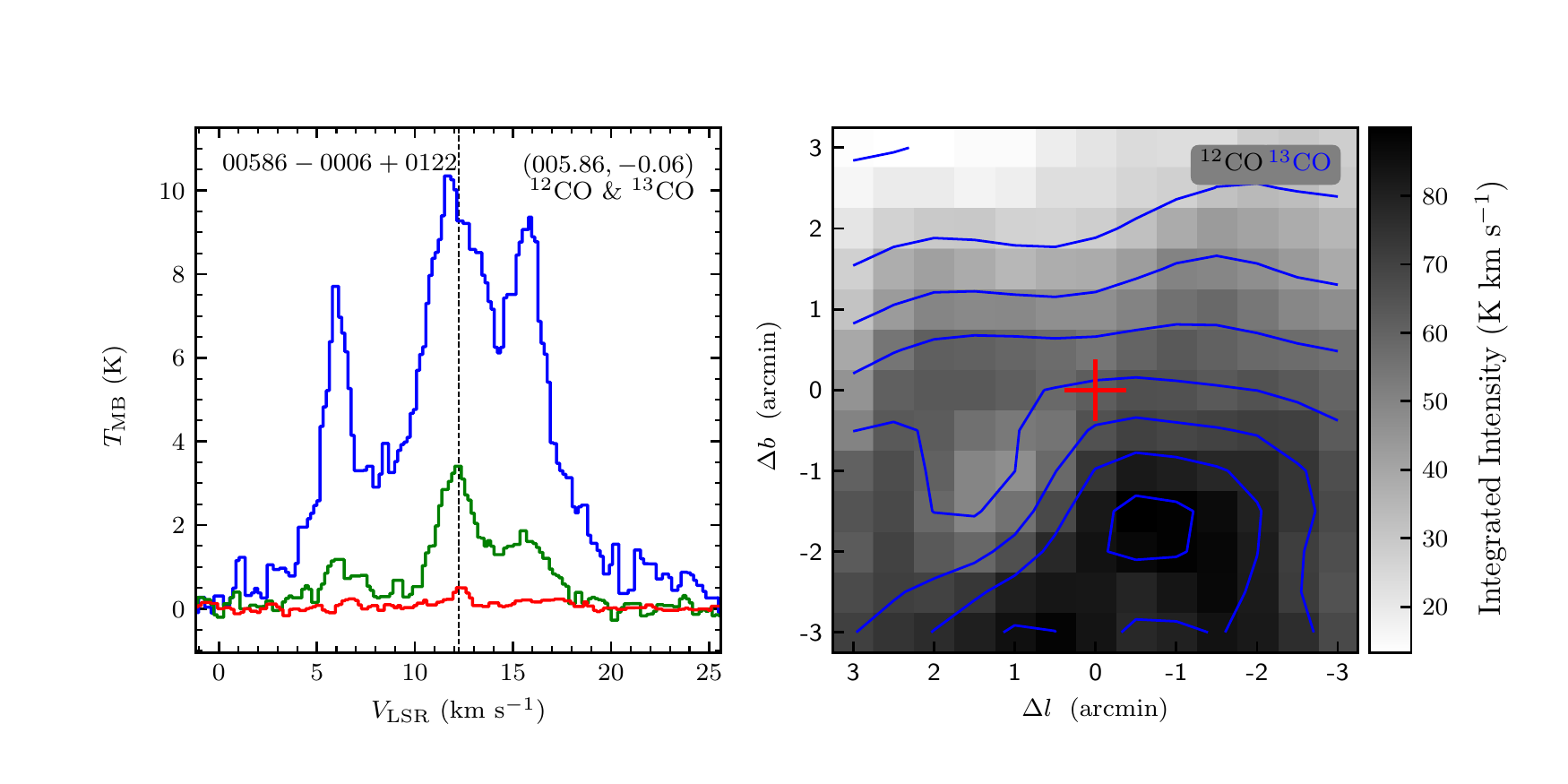}
\vspace{-0.5cm}

\includegraphics[width=9.0cm,angle=0]{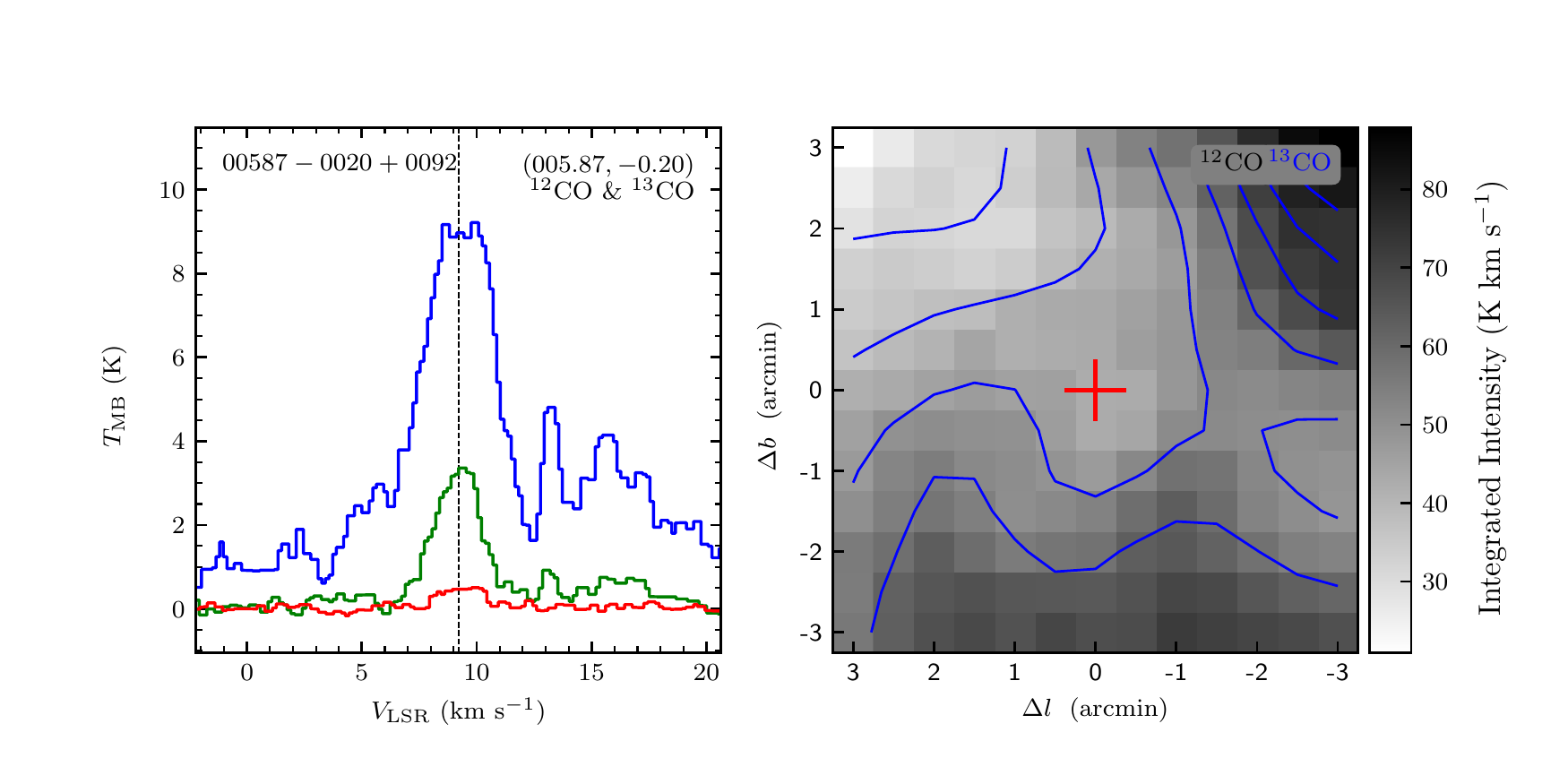}
\includegraphics[width=9.0cm,angle=0]{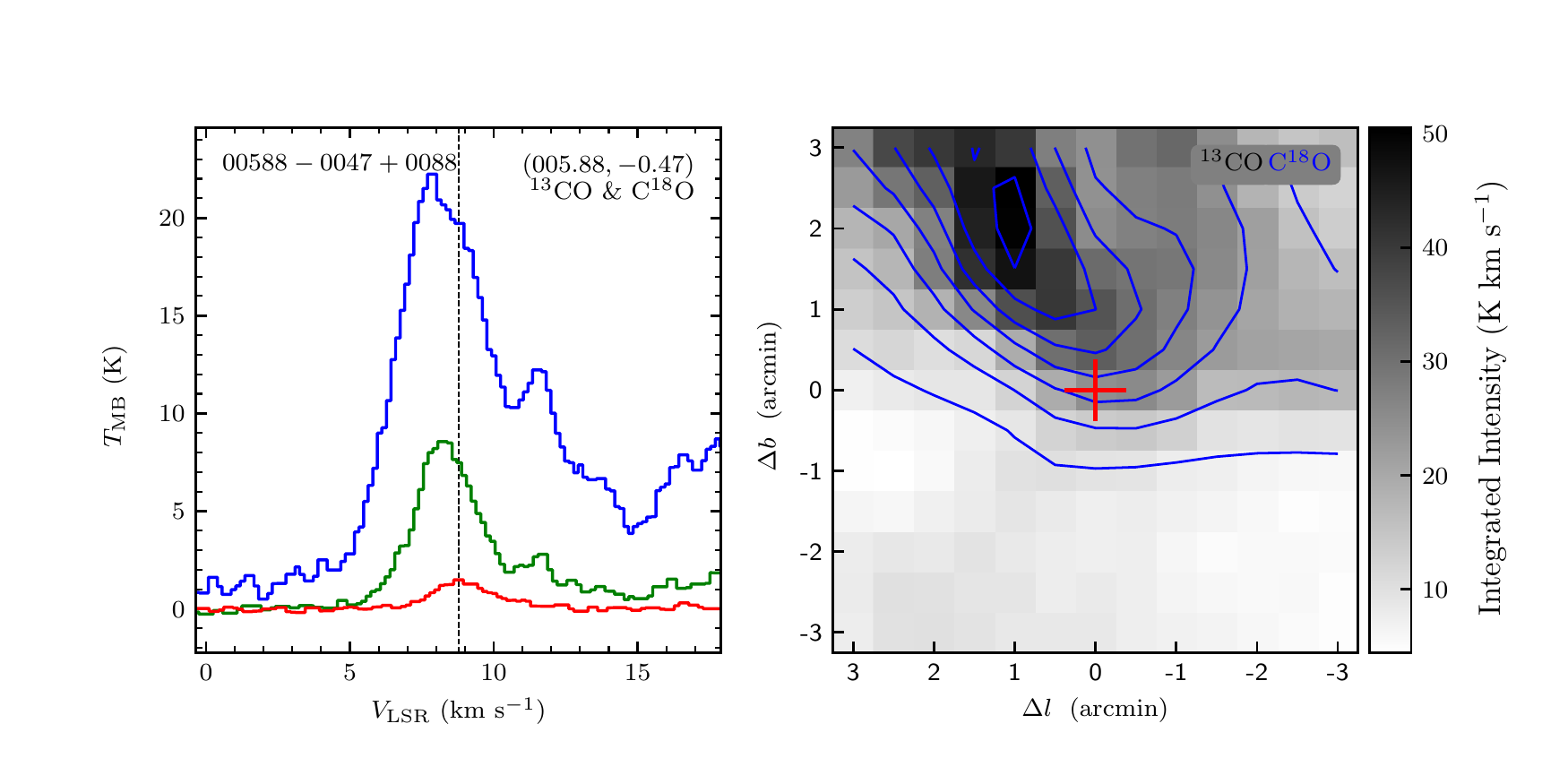}
\vspace{-0.5cm}

\includegraphics[width=9.0cm,angle=0]{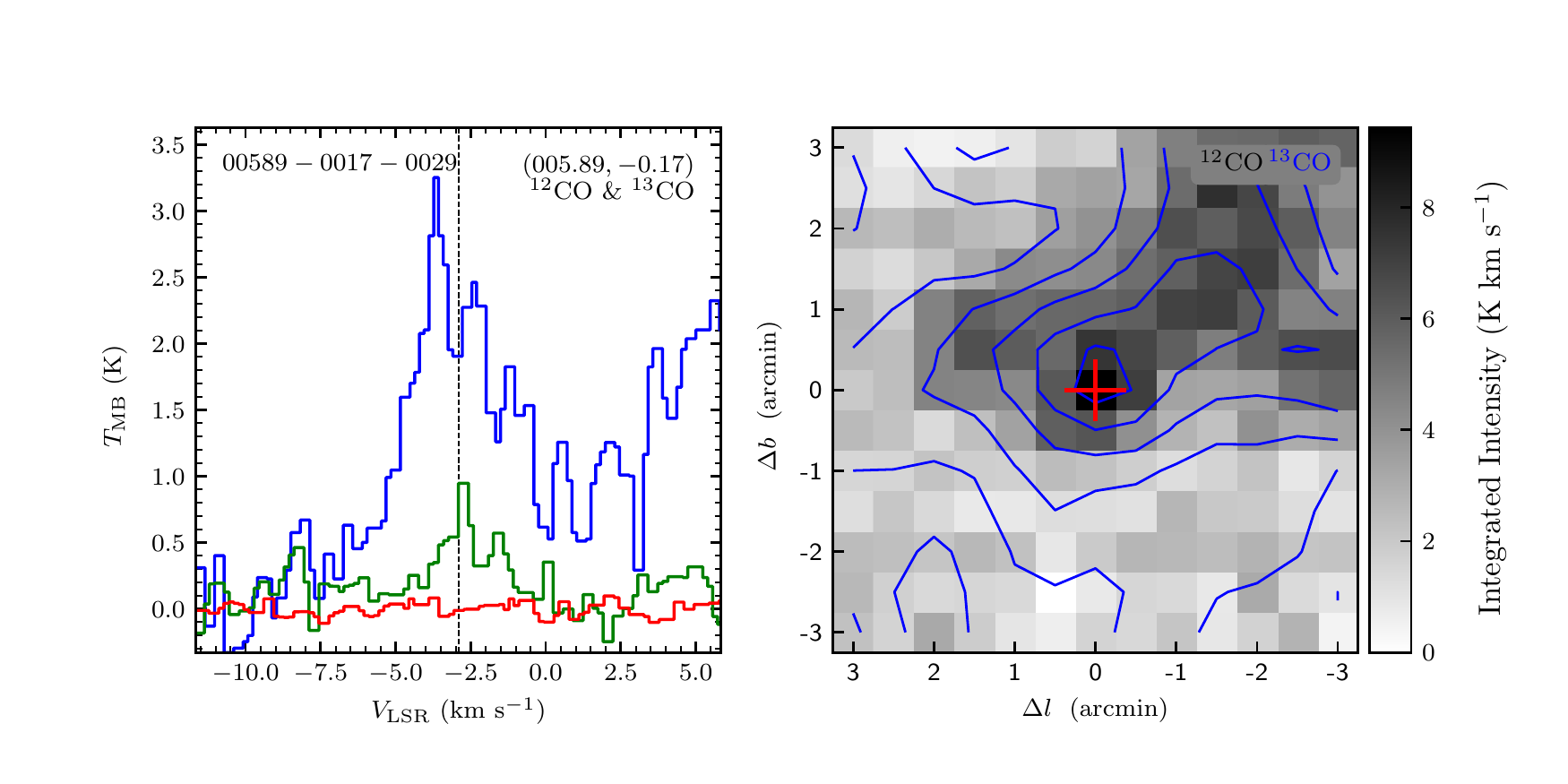}
\includegraphics[width=9.0cm,angle=0]{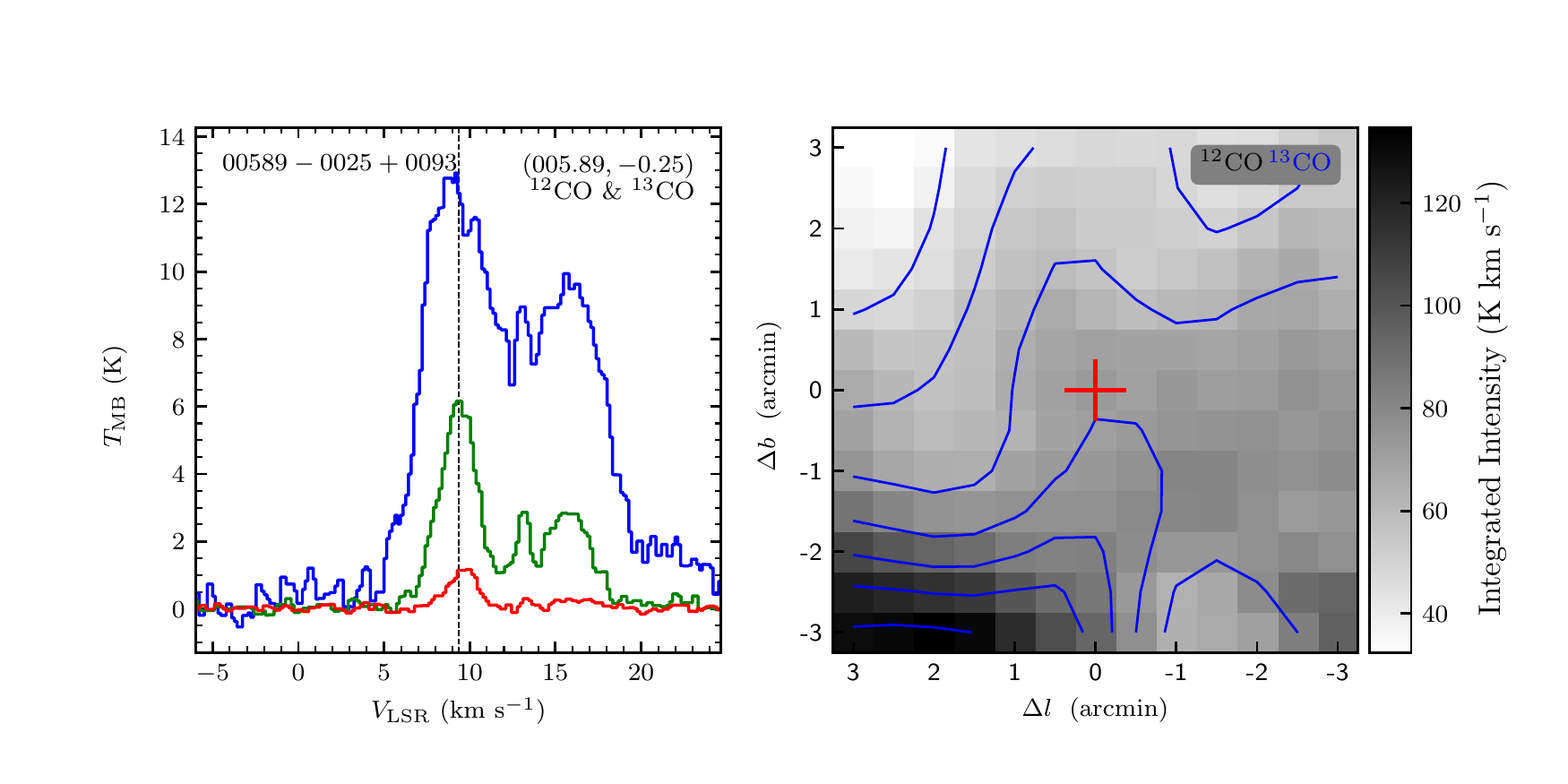}
\vspace{-0.5cm}

\includegraphics[width=9.0cm,angle=0]{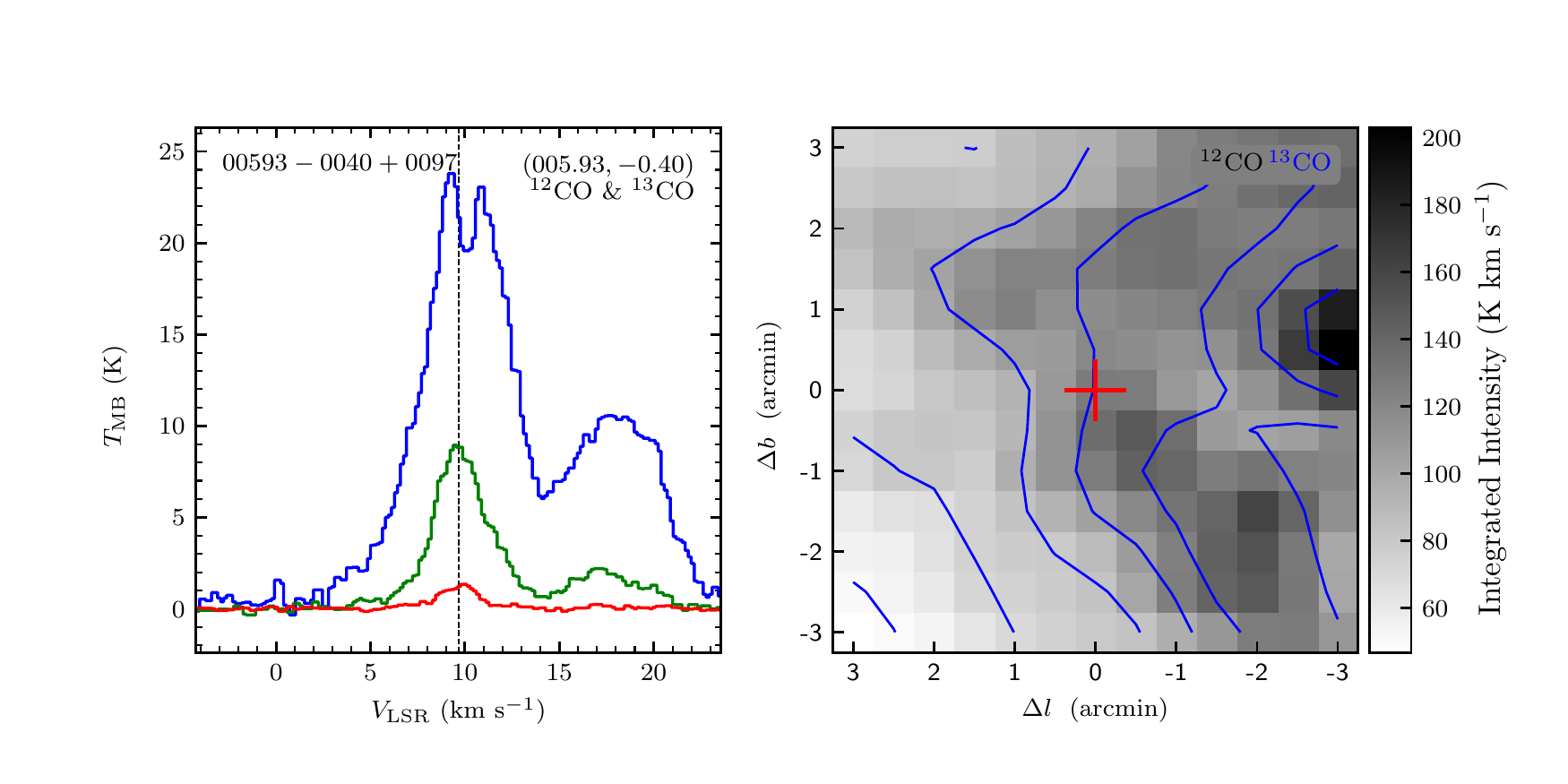}
\includegraphics[width=9.0cm,angle=0]{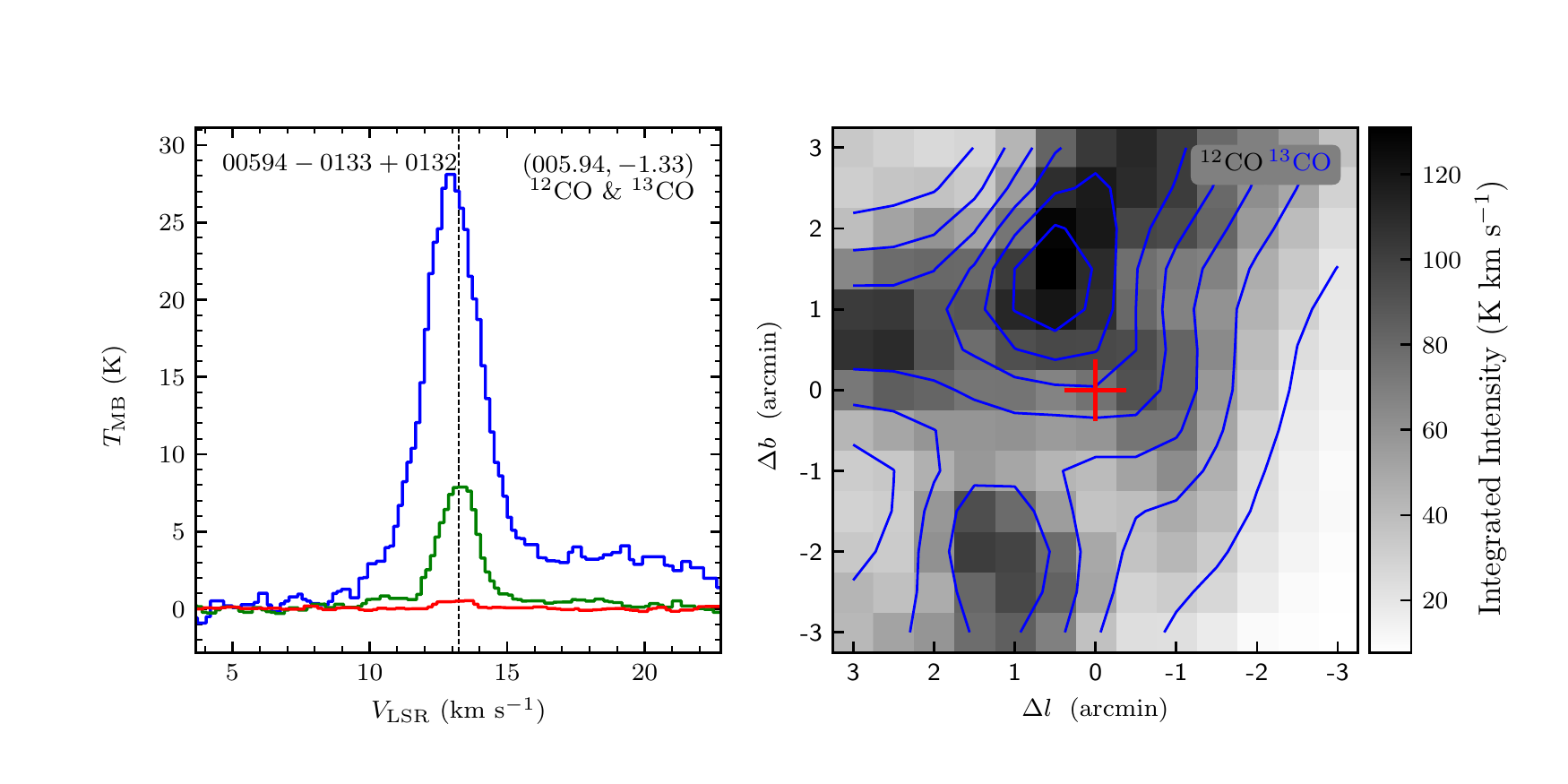}
\end{figure}
\clearpage

\begin{figure}
\includegraphics[width=9.0cm,angle=0]{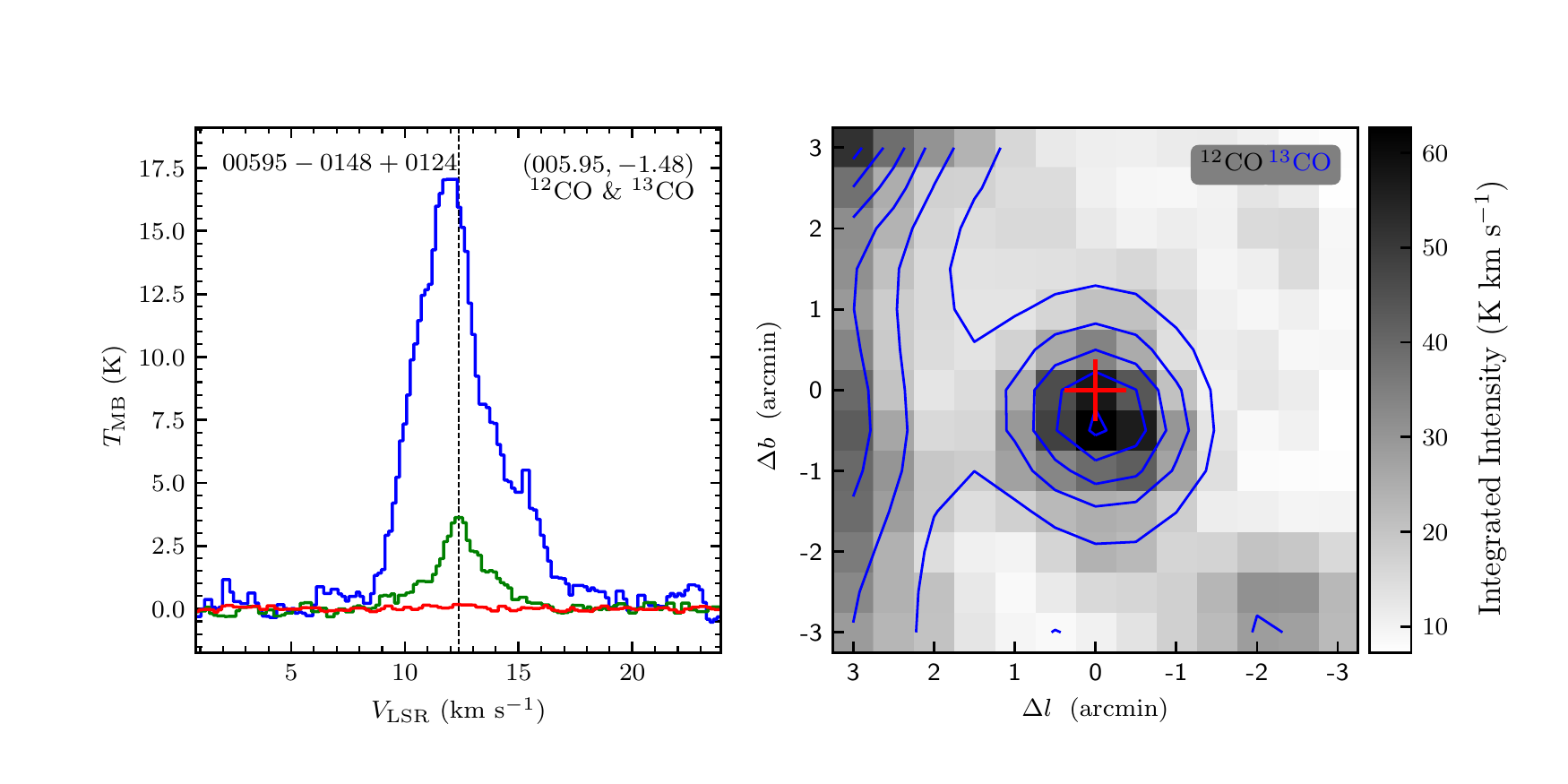}
\includegraphics[width=9.0cm,angle=0]{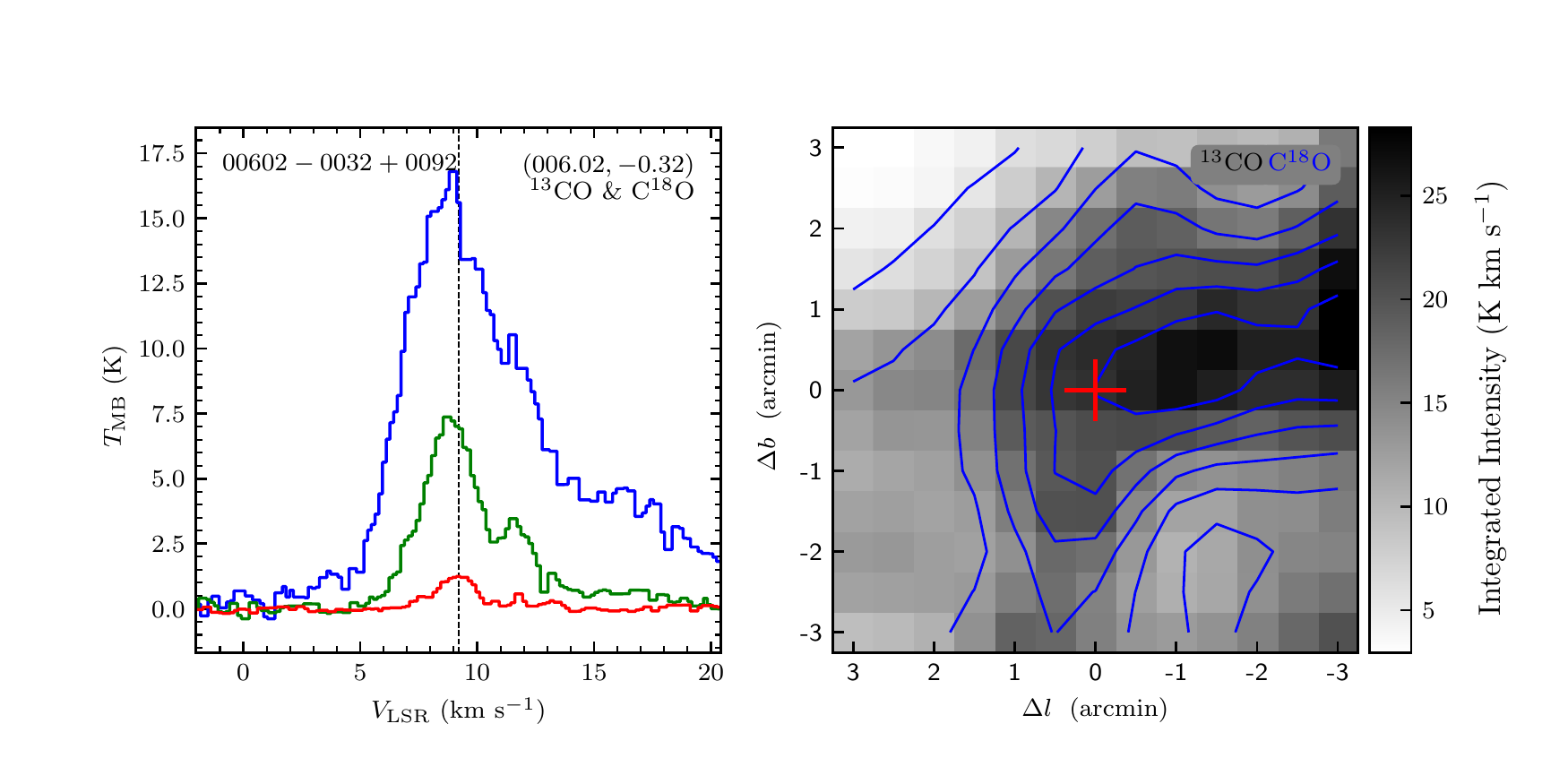}
\vspace{-0.5cm}

\includegraphics[width=9.0cm,angle=0]{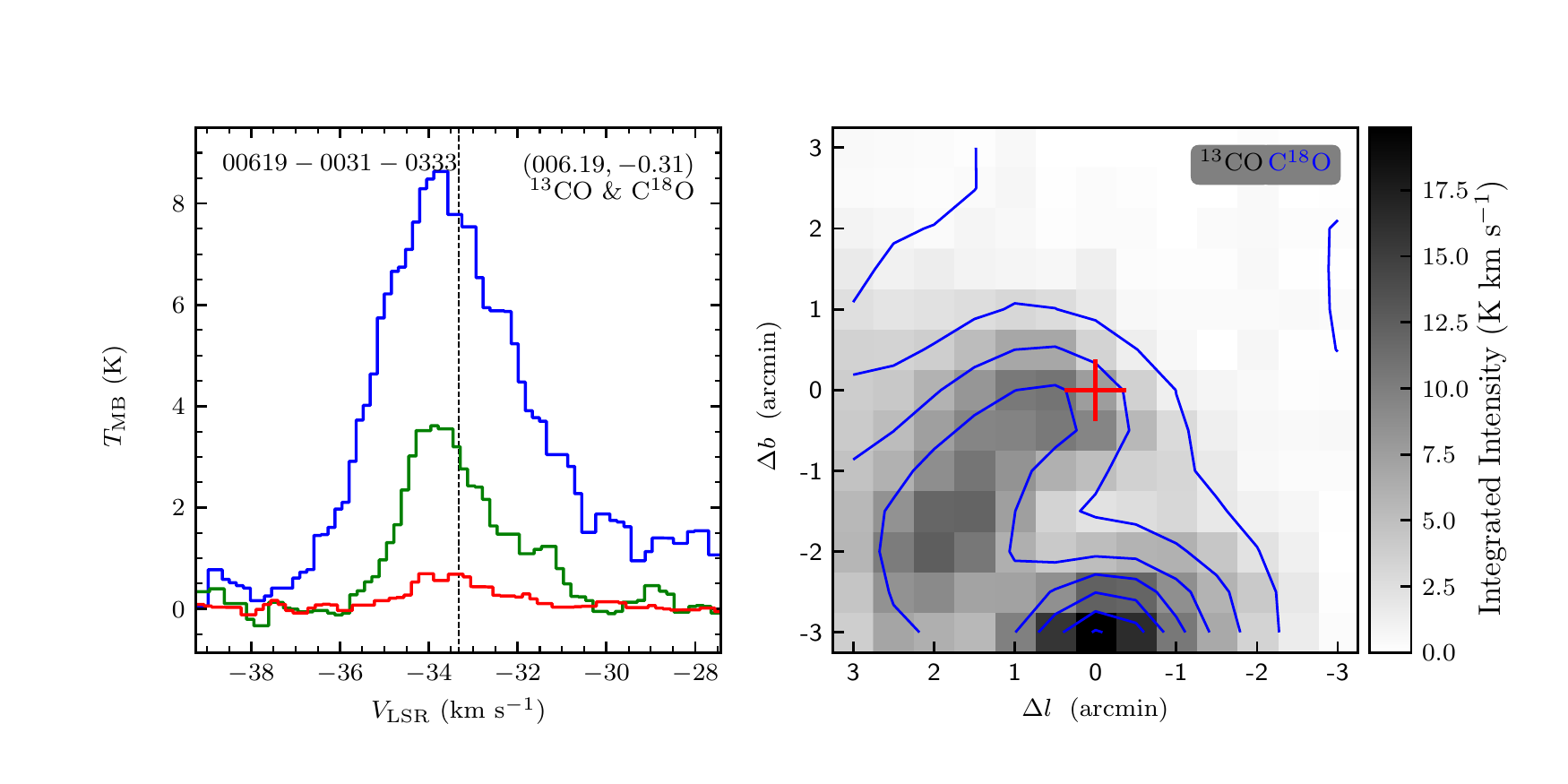}
\includegraphics[width=9.0cm,angle=0]{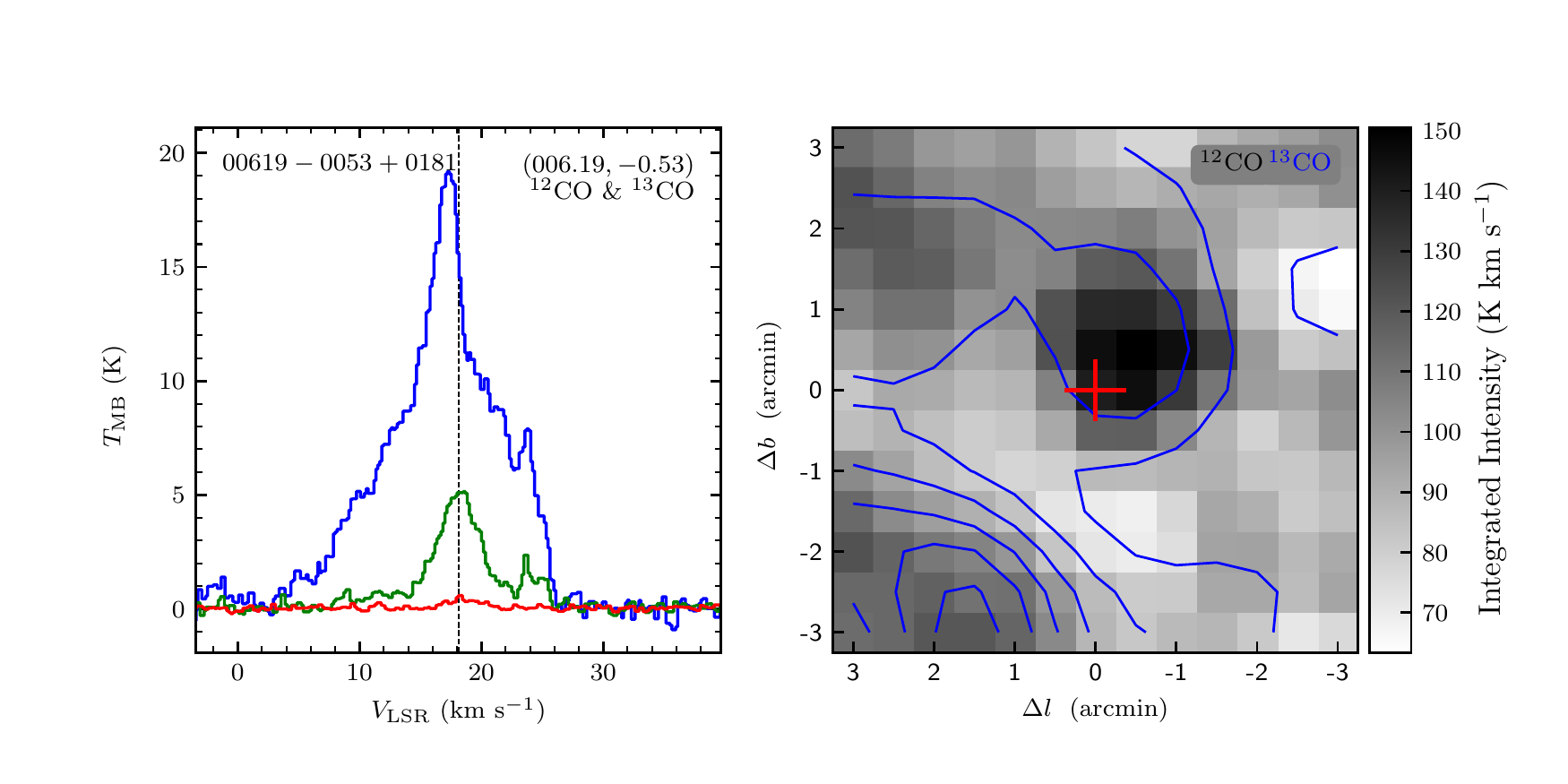}
\vspace{-0.5cm}

\includegraphics[width=9.0cm,angle=0]{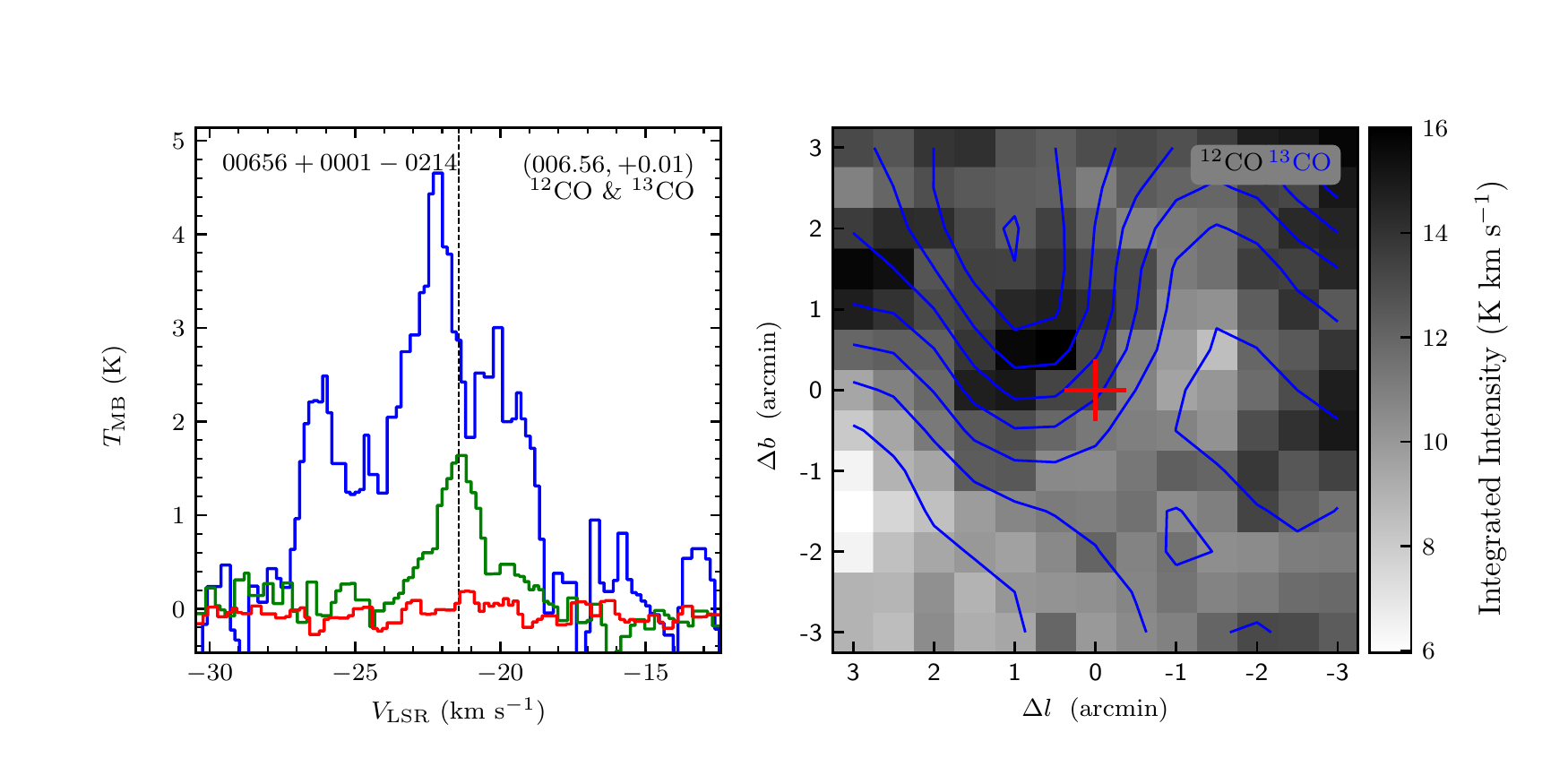}
\includegraphics[width=9.0cm,angle=0]{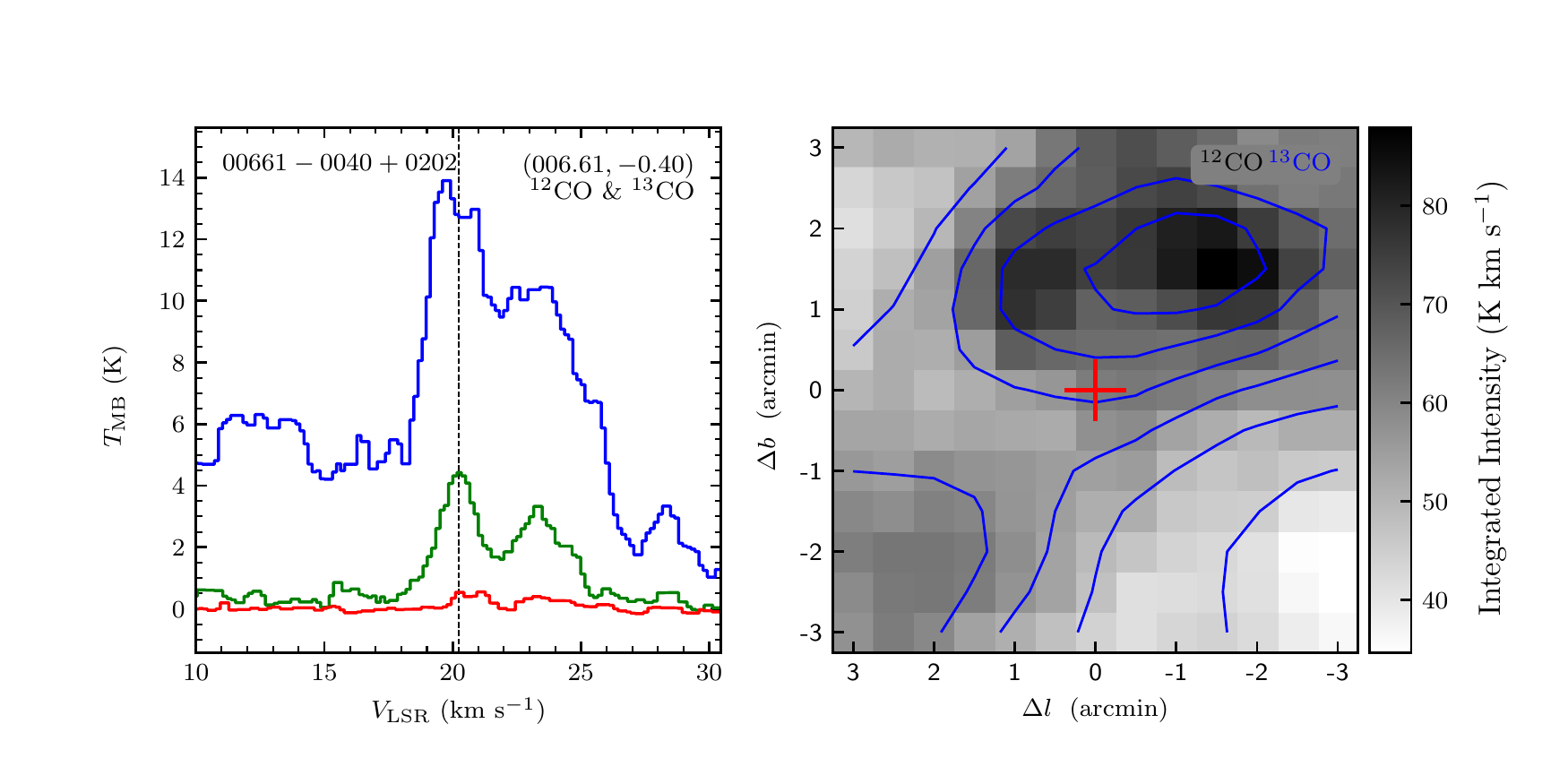}
\vspace{-0.5cm}

\includegraphics[width=9.0cm,angle=0]{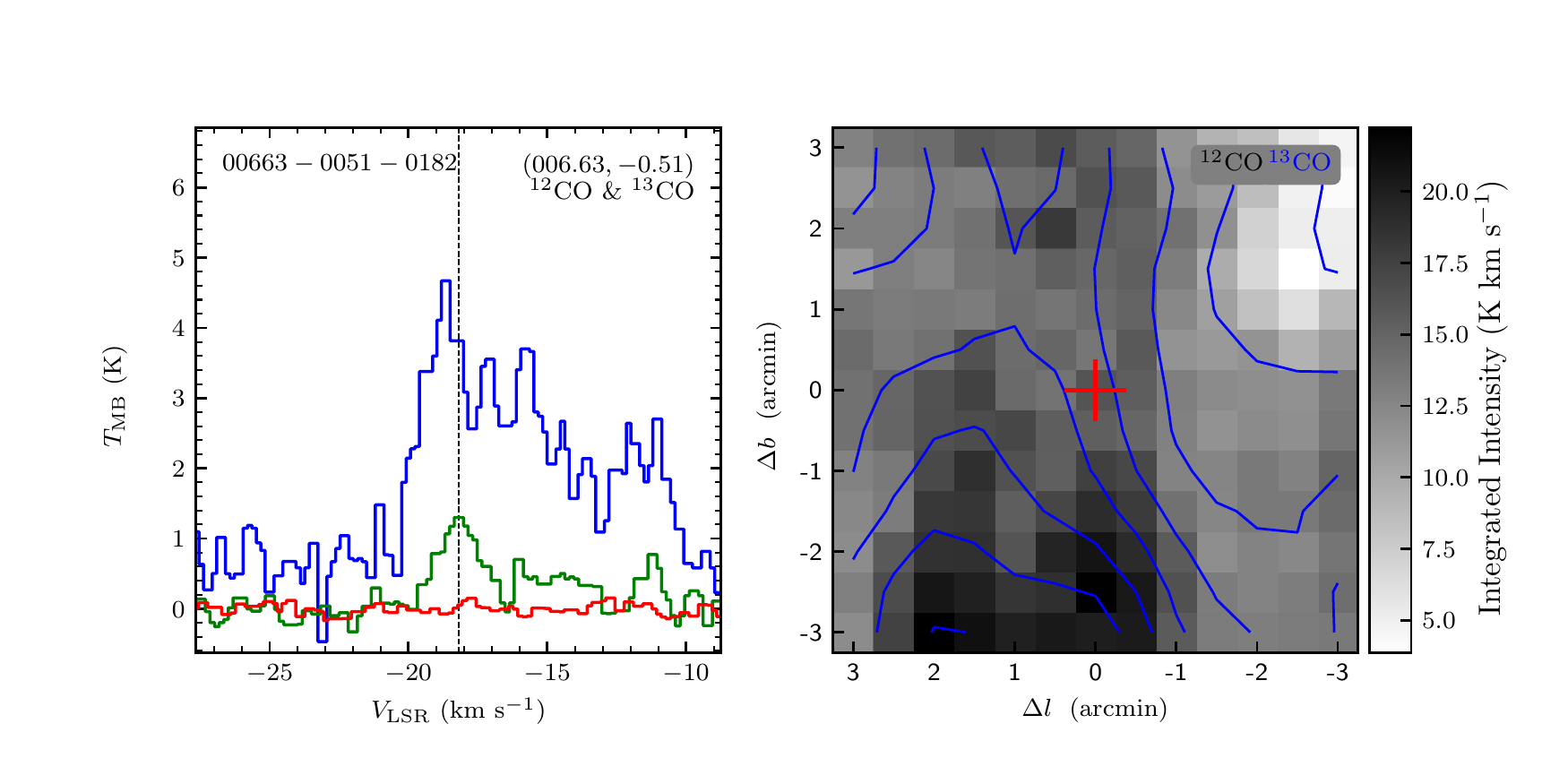}
\includegraphics[width=9.0cm,angle=0]{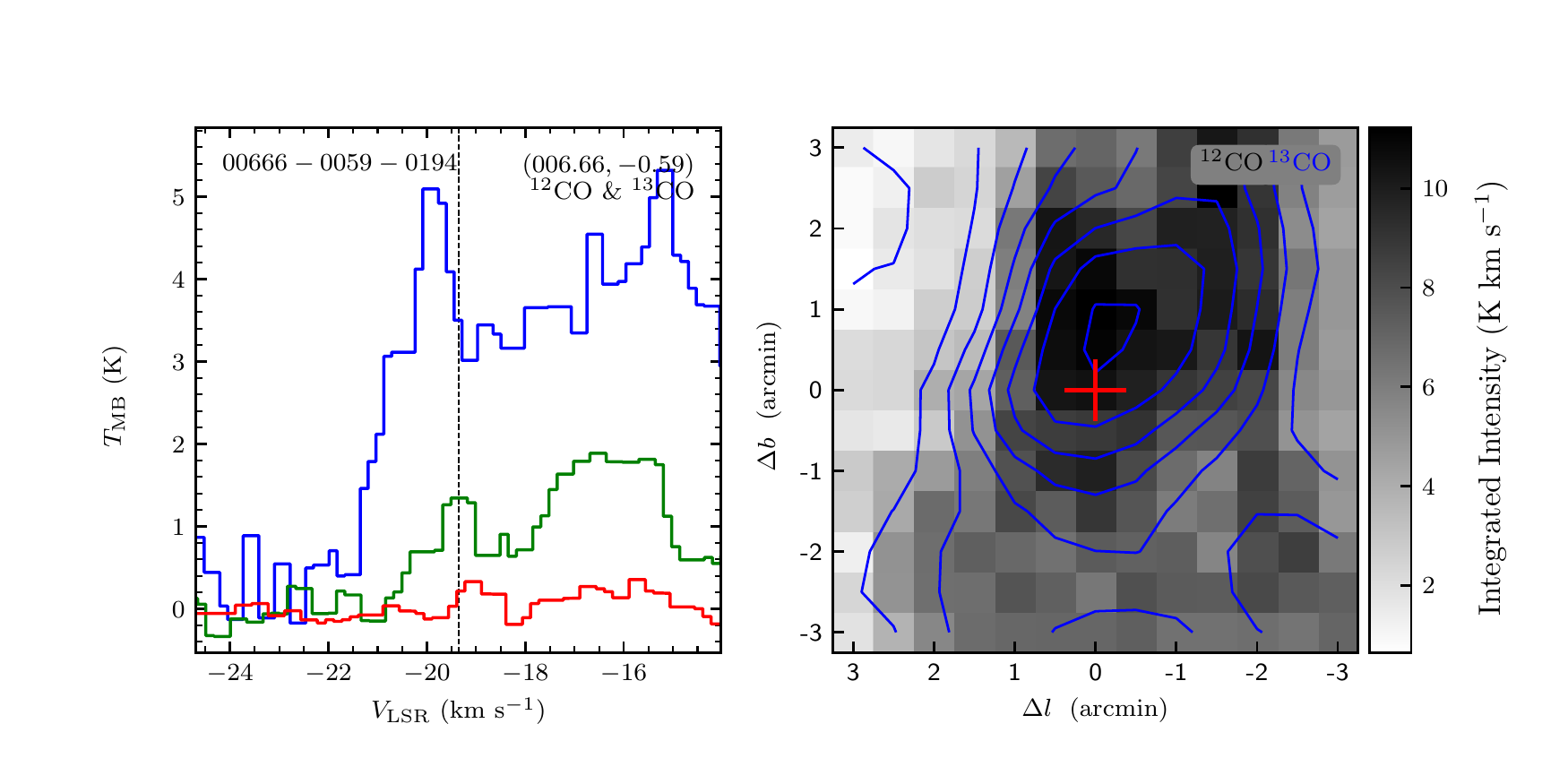}
\vspace{-0.5cm}

\includegraphics[width=9.0cm,angle=0]{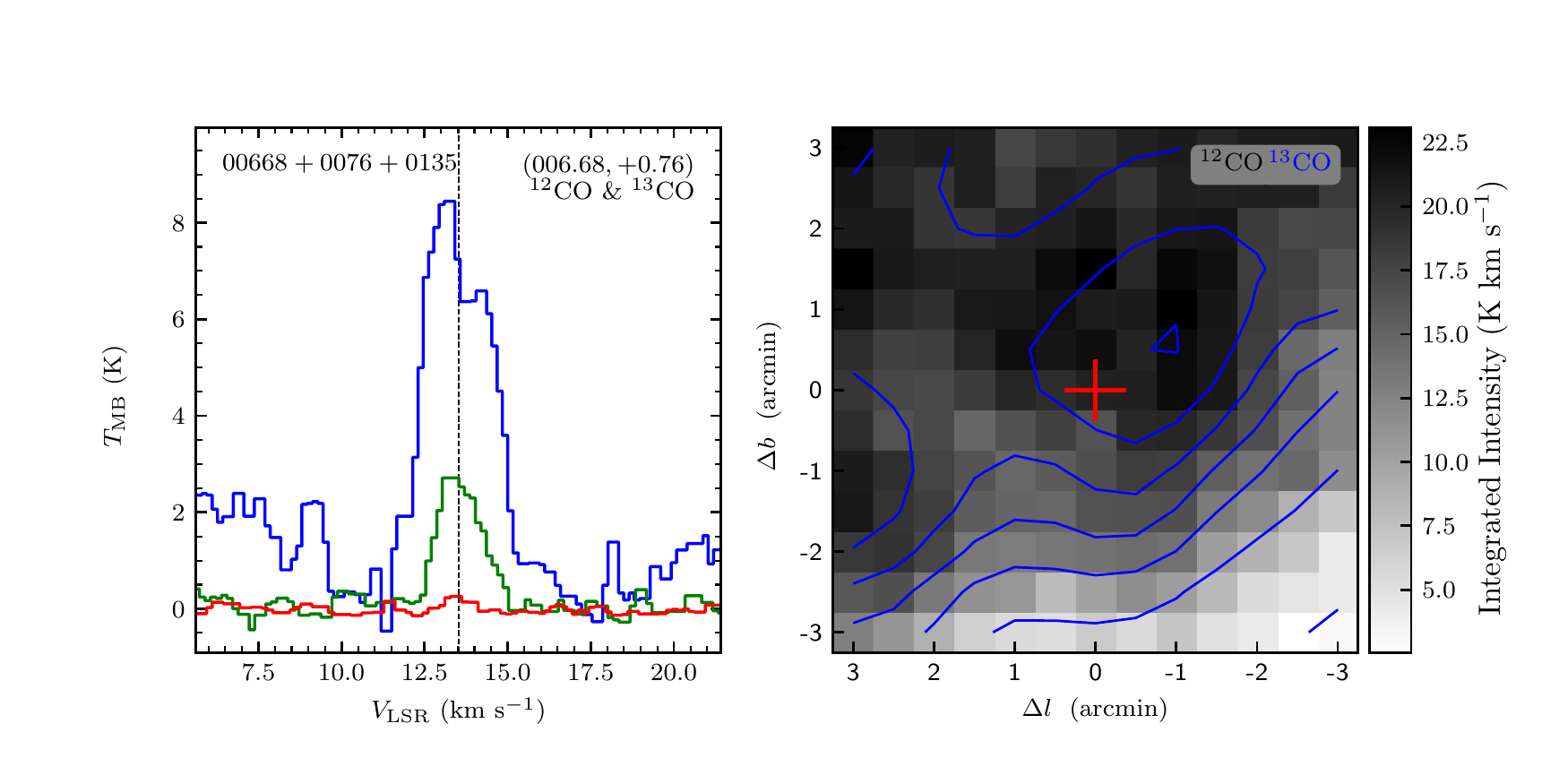}
\includegraphics[width=9.0cm,angle=0]{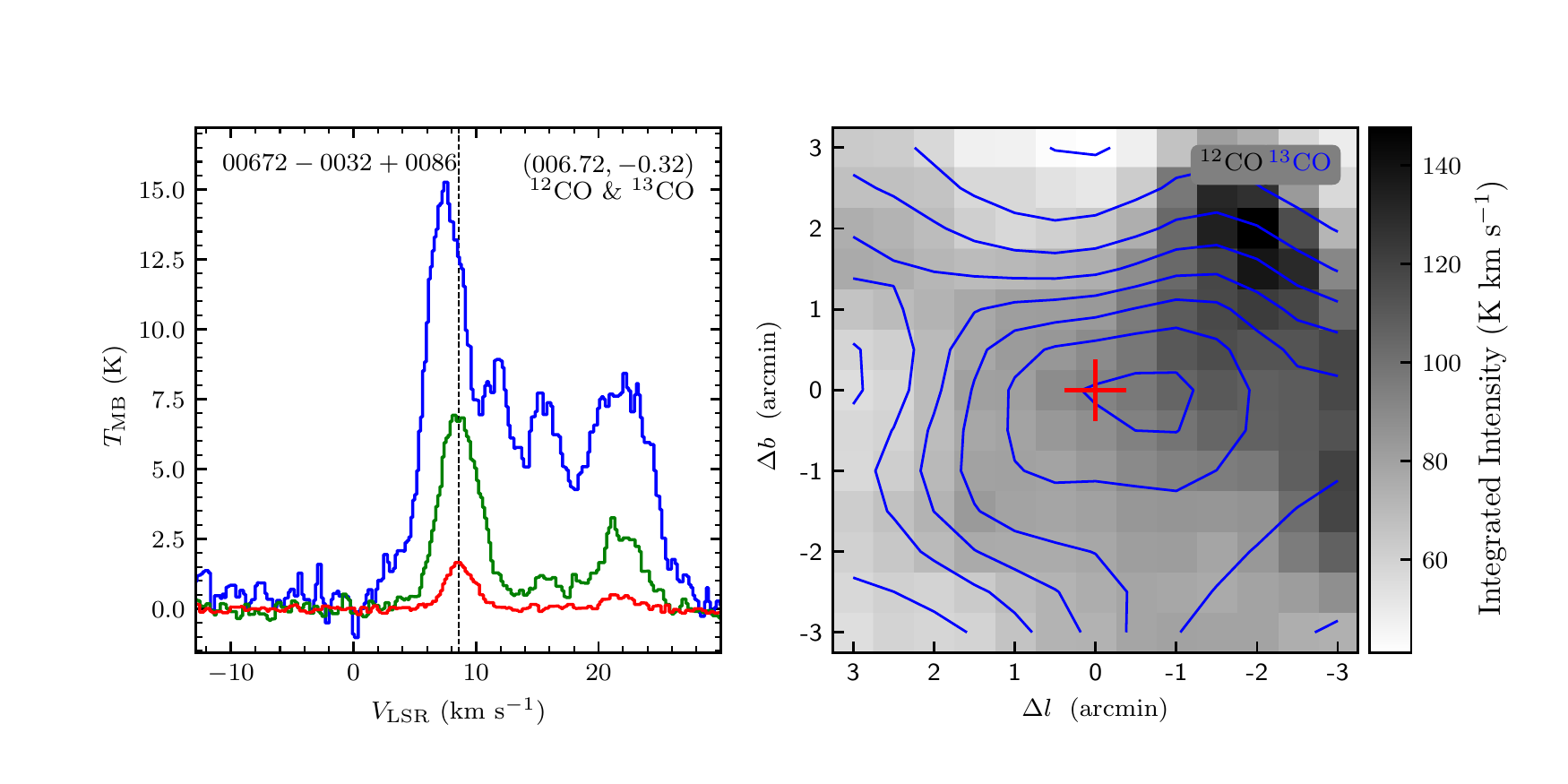}
\end{figure}
\clearpage

\begin{figure}
\includegraphics[width=9.0cm,angle=0]{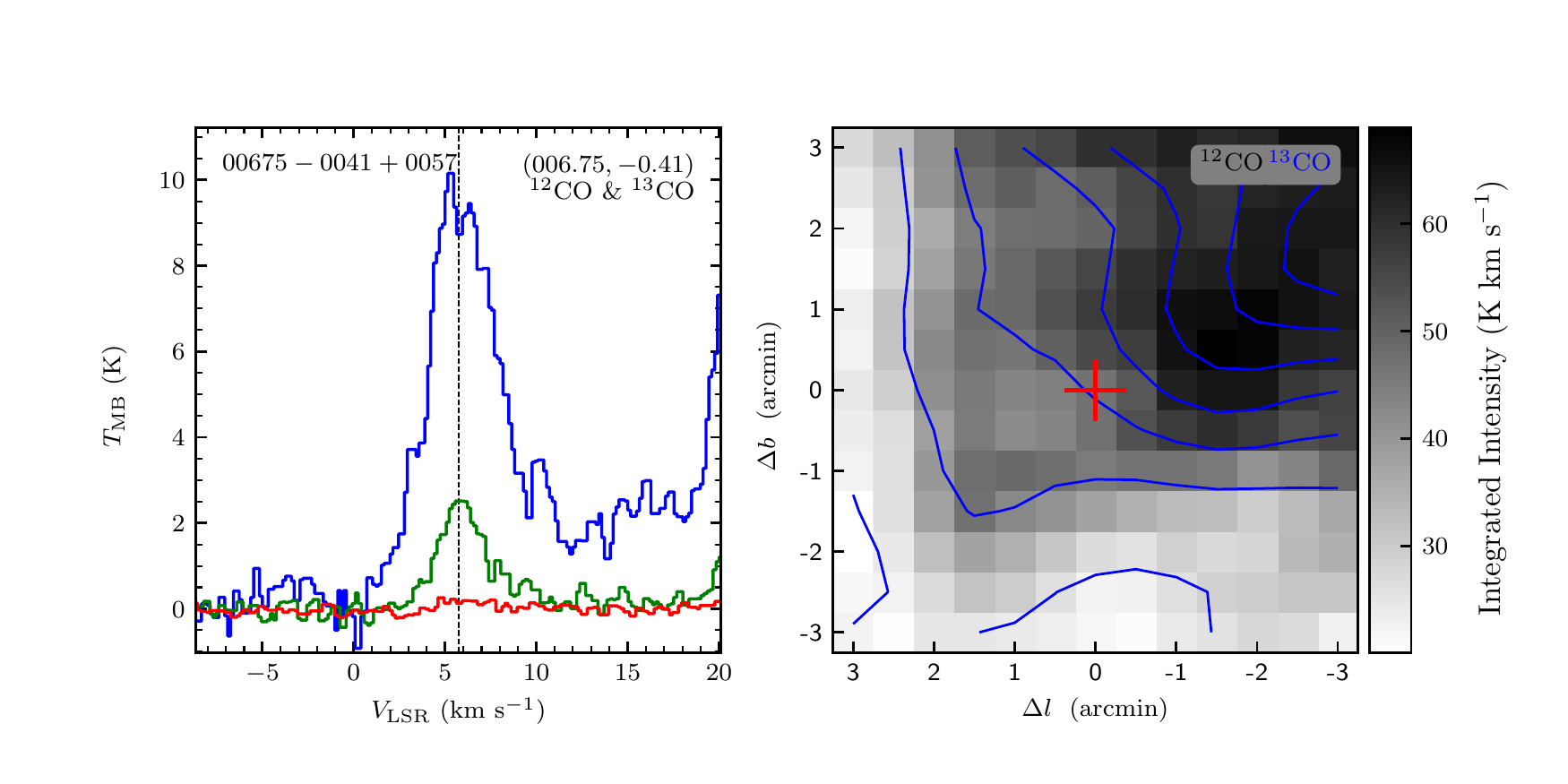}
\includegraphics[width=9.0cm,angle=0]{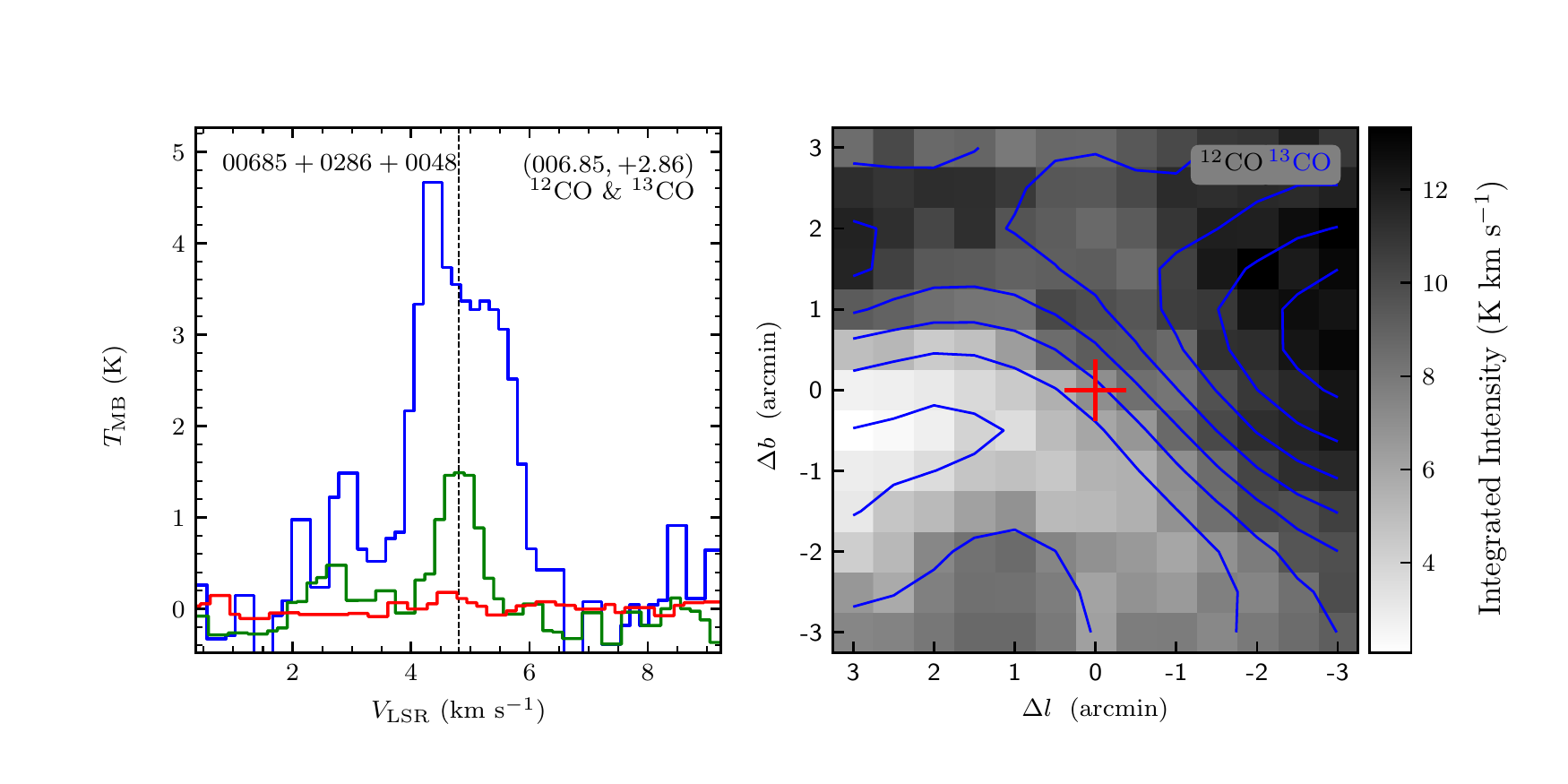}
\vspace{-0.5cm}

\includegraphics[width=9.0cm,angle=0]{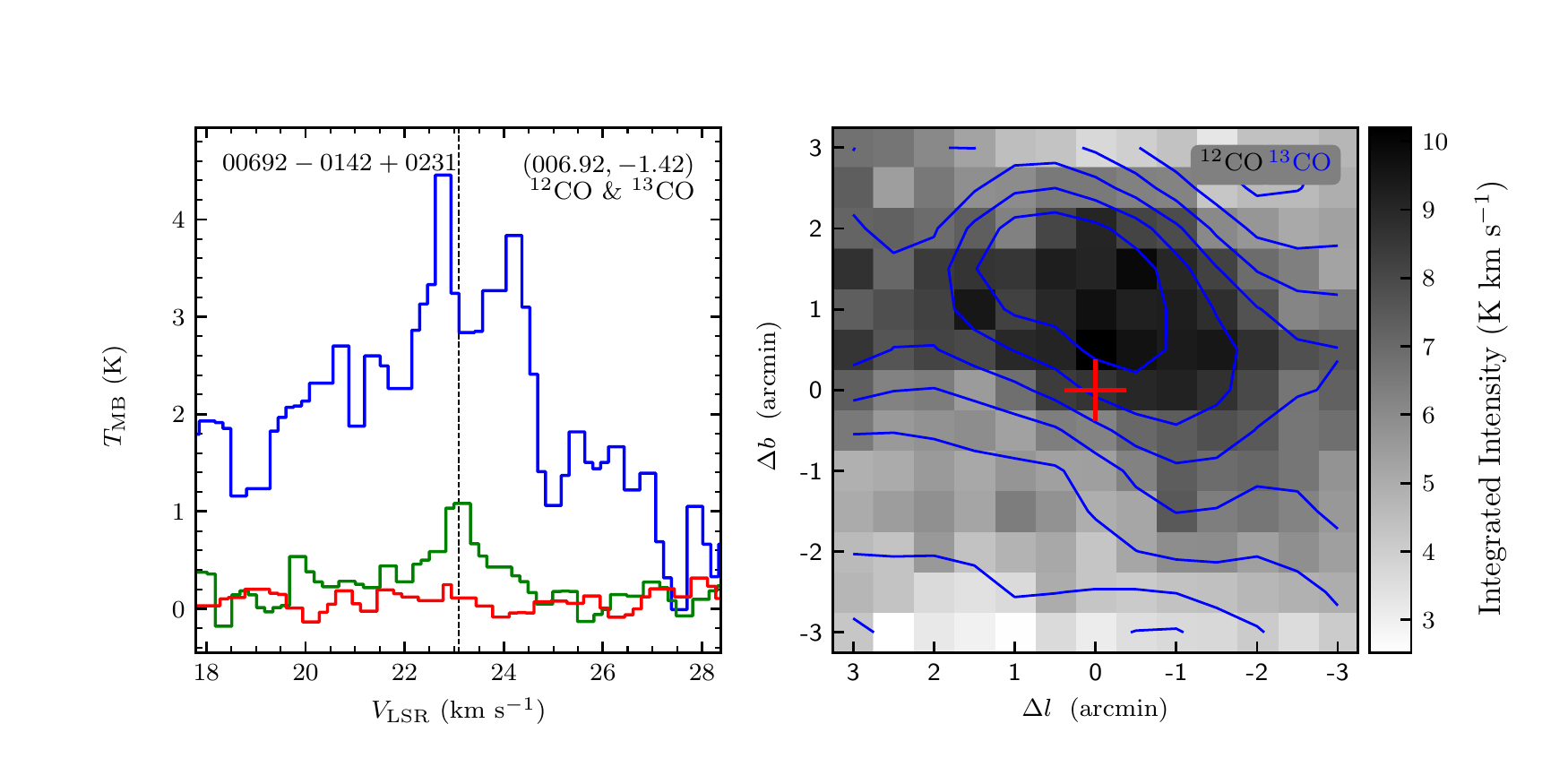}
\includegraphics[width=9.0cm,angle=0]{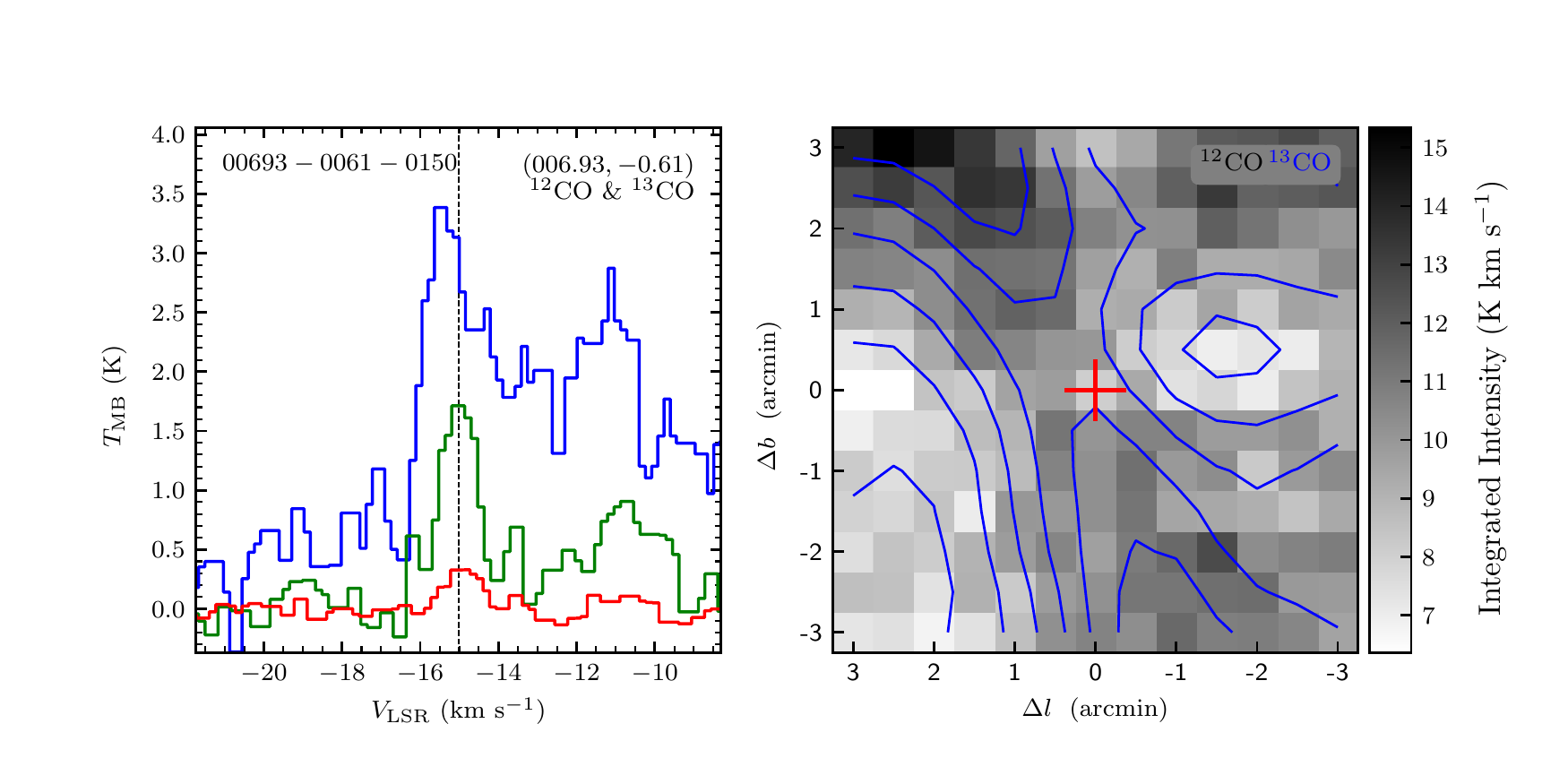}
\vspace{-0.5cm}

\includegraphics[width=9.0cm,angle=0]{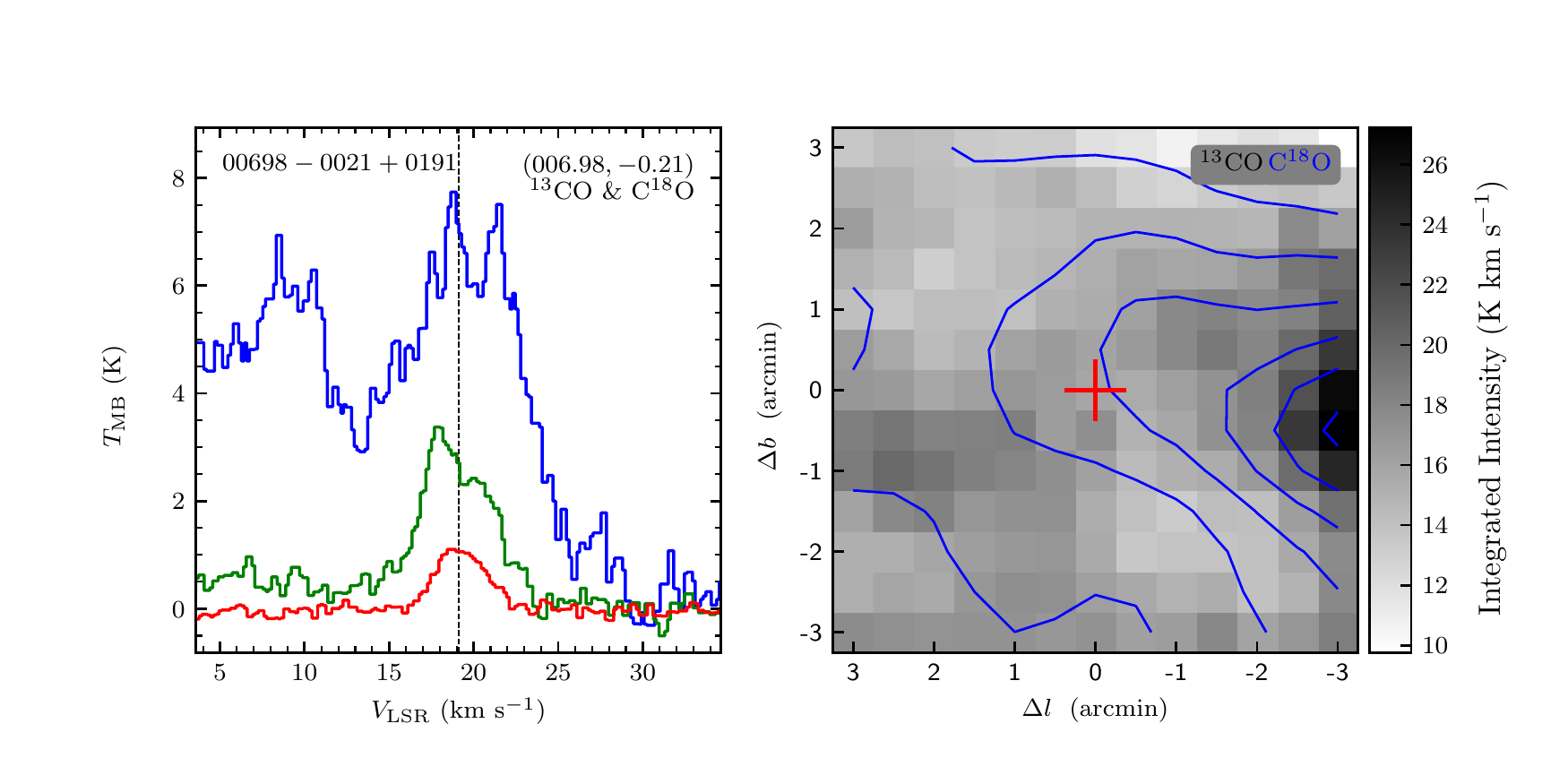}
\includegraphics[width=9.0cm,angle=0]{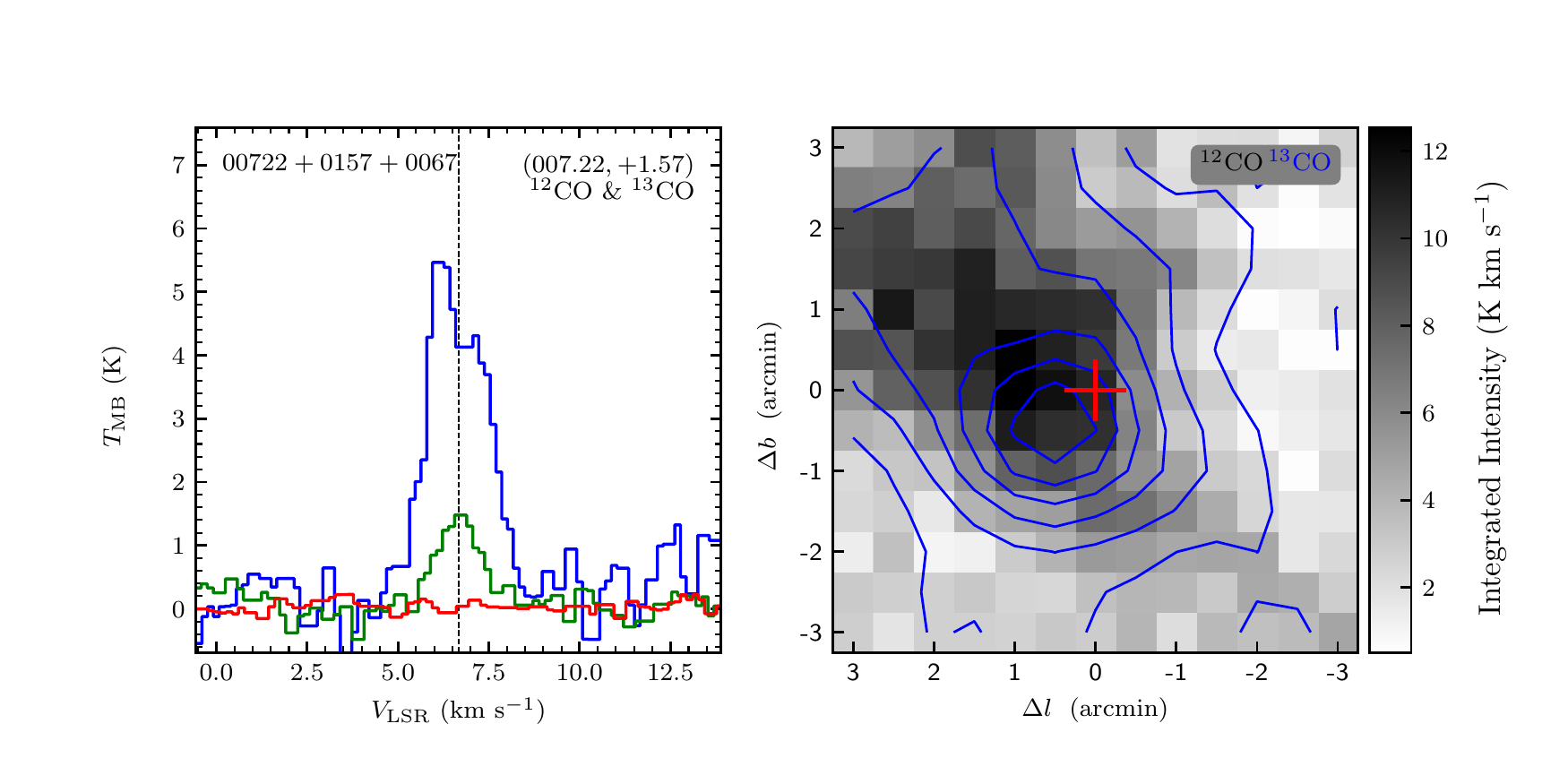}
\vspace{-0.5cm}

\includegraphics[width=9.0cm,angle=0]{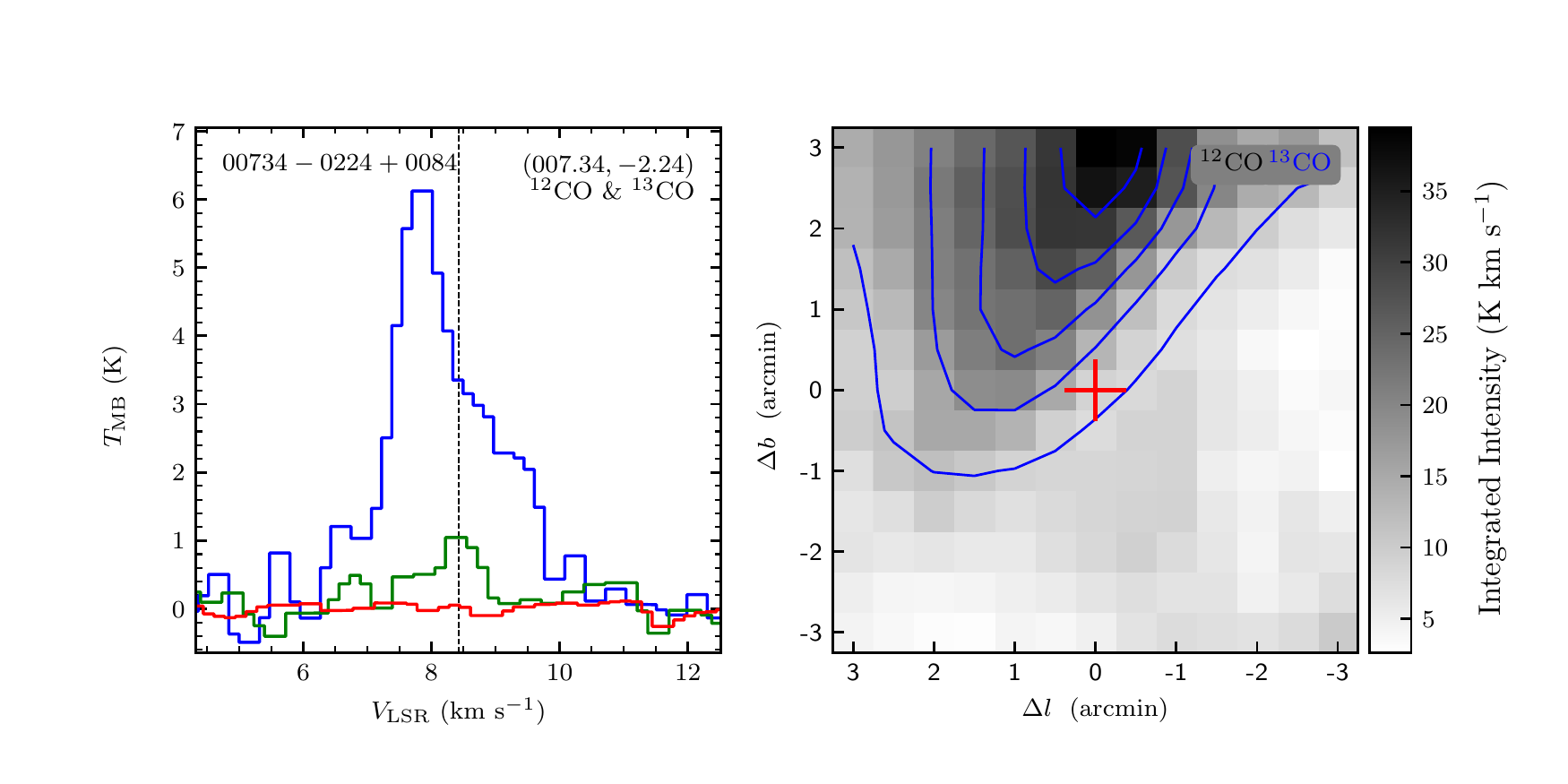}
\includegraphics[width=9.0cm,angle=0]{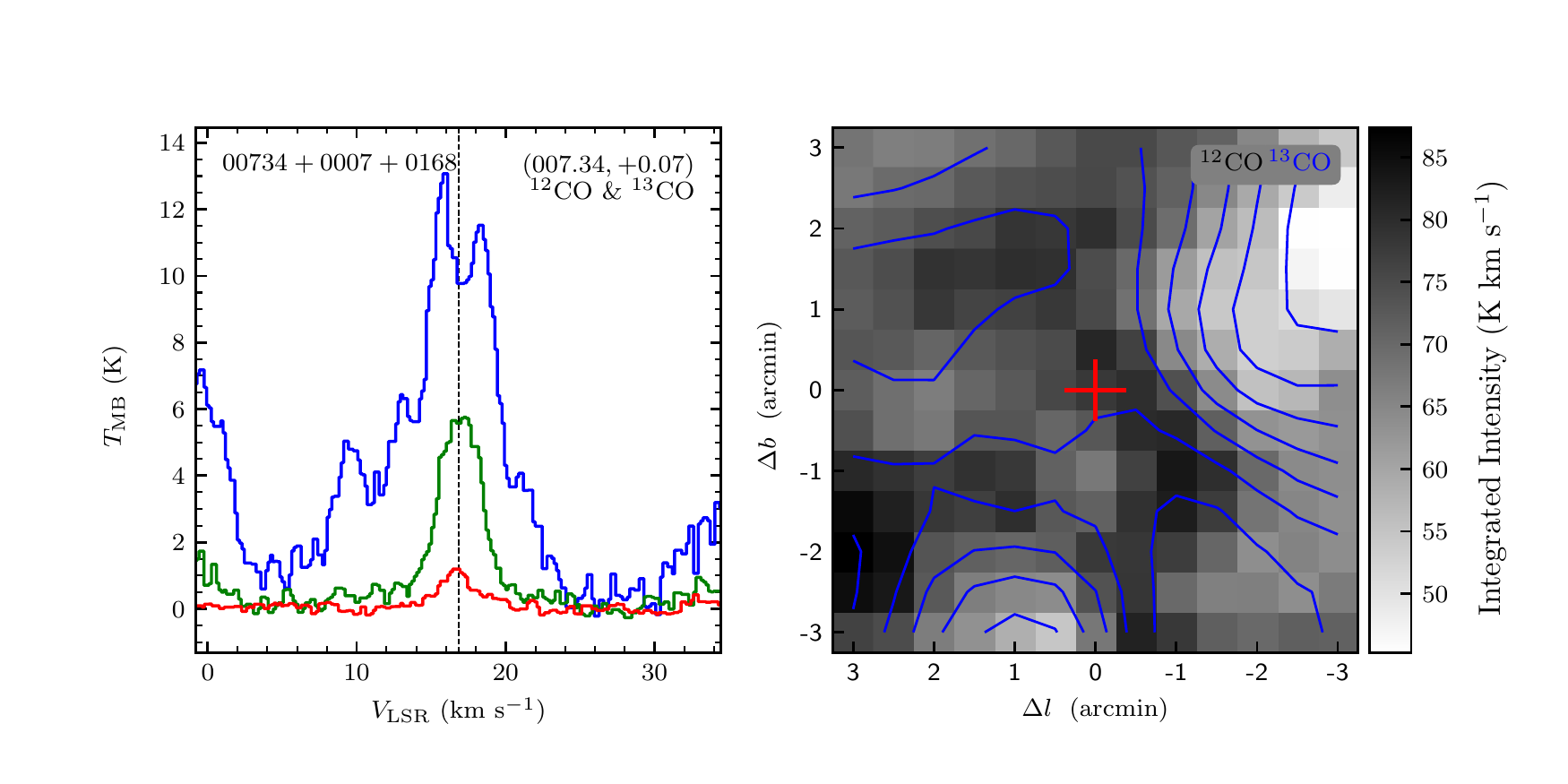}
\vspace{-0.5cm}

\includegraphics[width=9.0cm,angle=0]{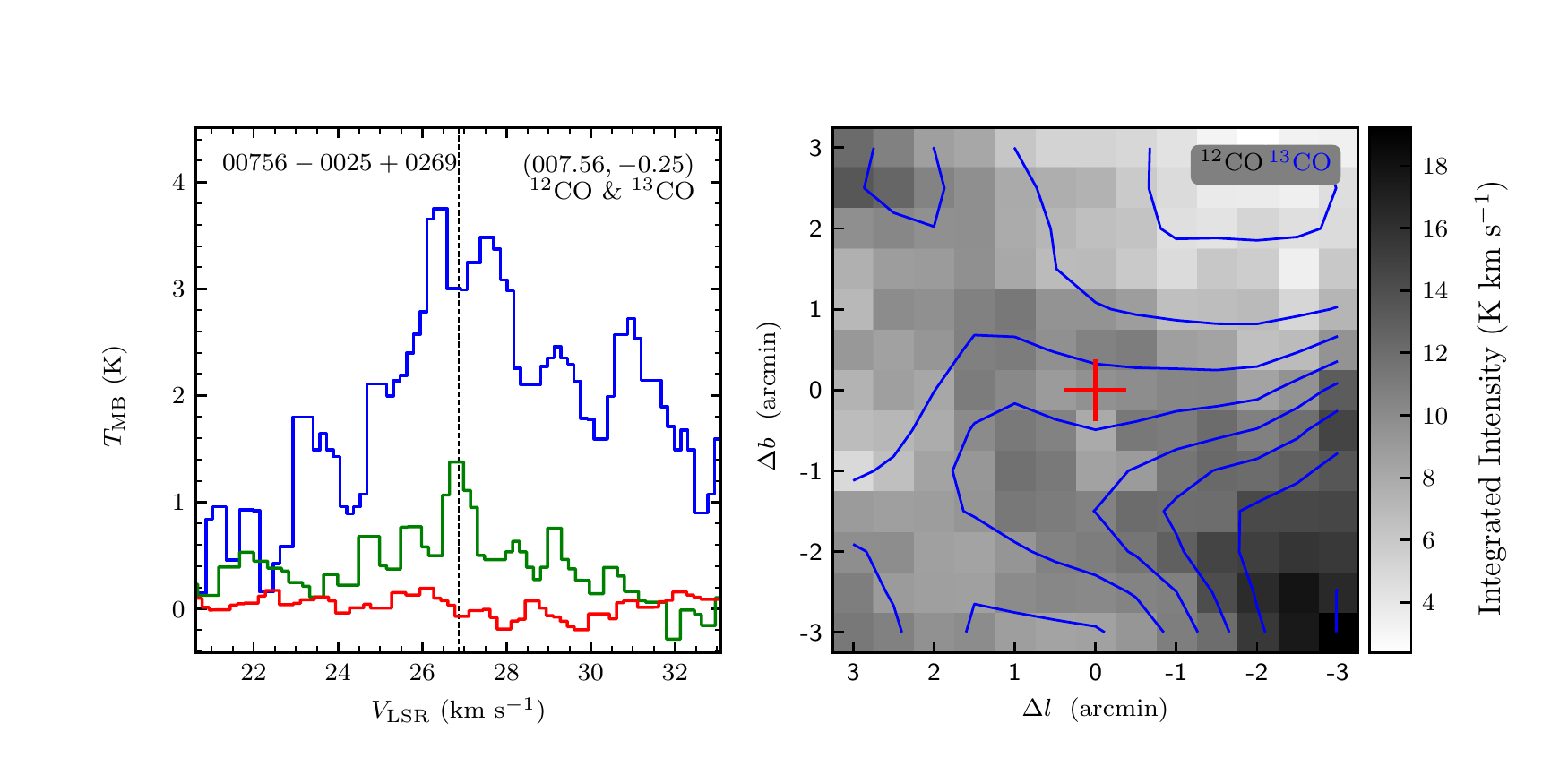}
\includegraphics[width=9.0cm,angle=0]{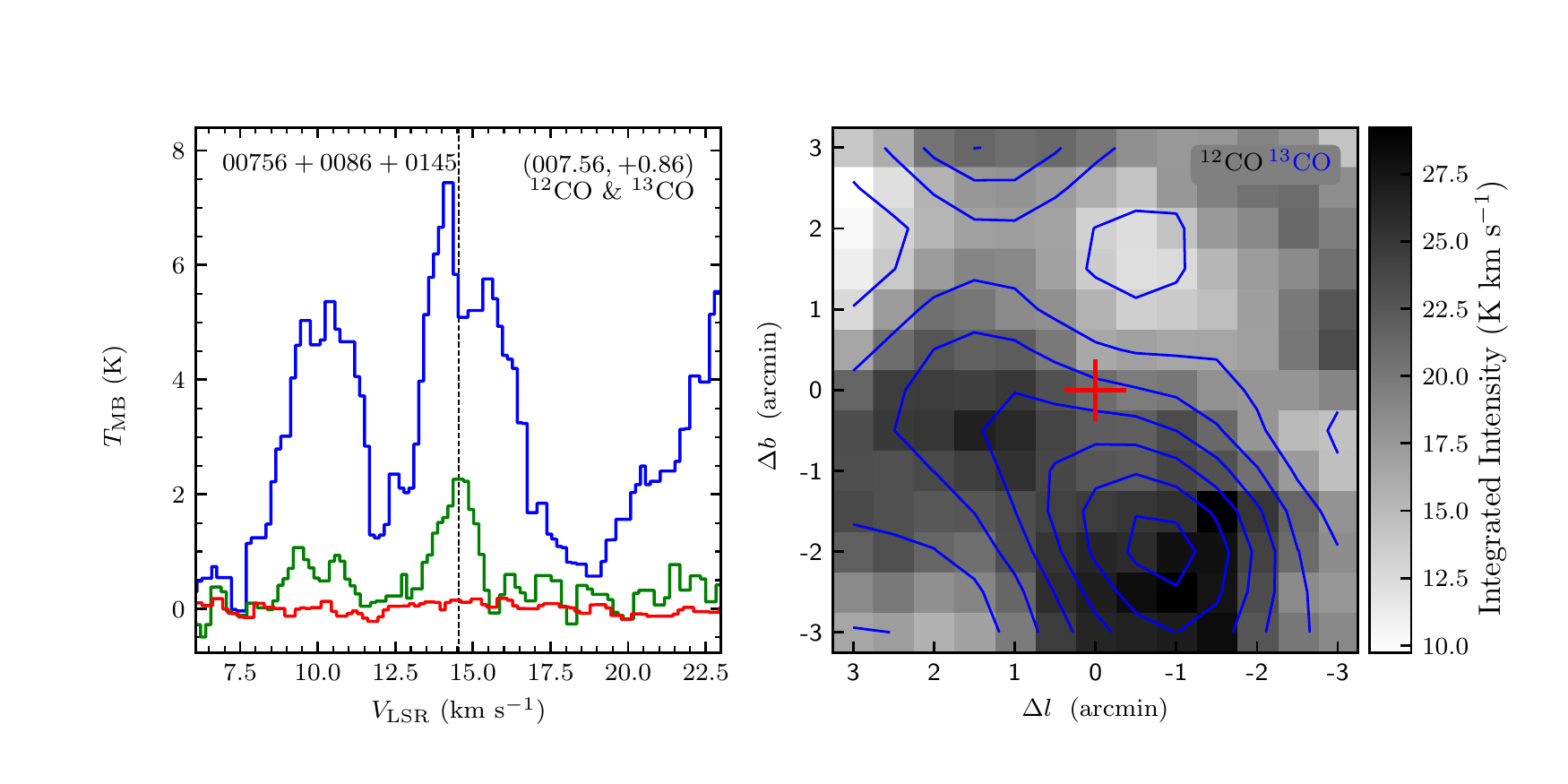}
\end{figure}
\clearpage

\begin{figure}
\includegraphics[width=9.0cm,angle=0]{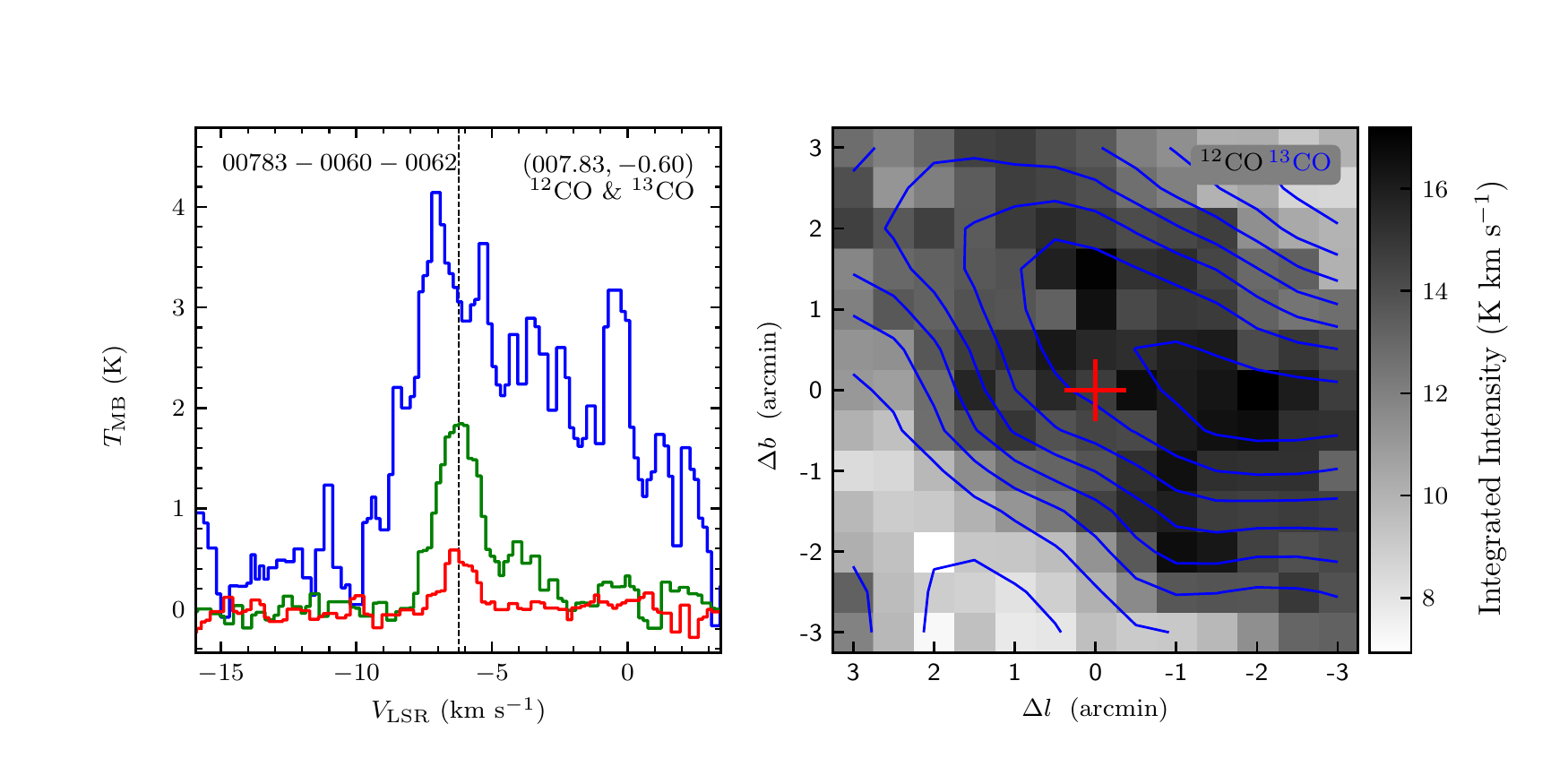}
\includegraphics[width=9.0cm,angle=0]{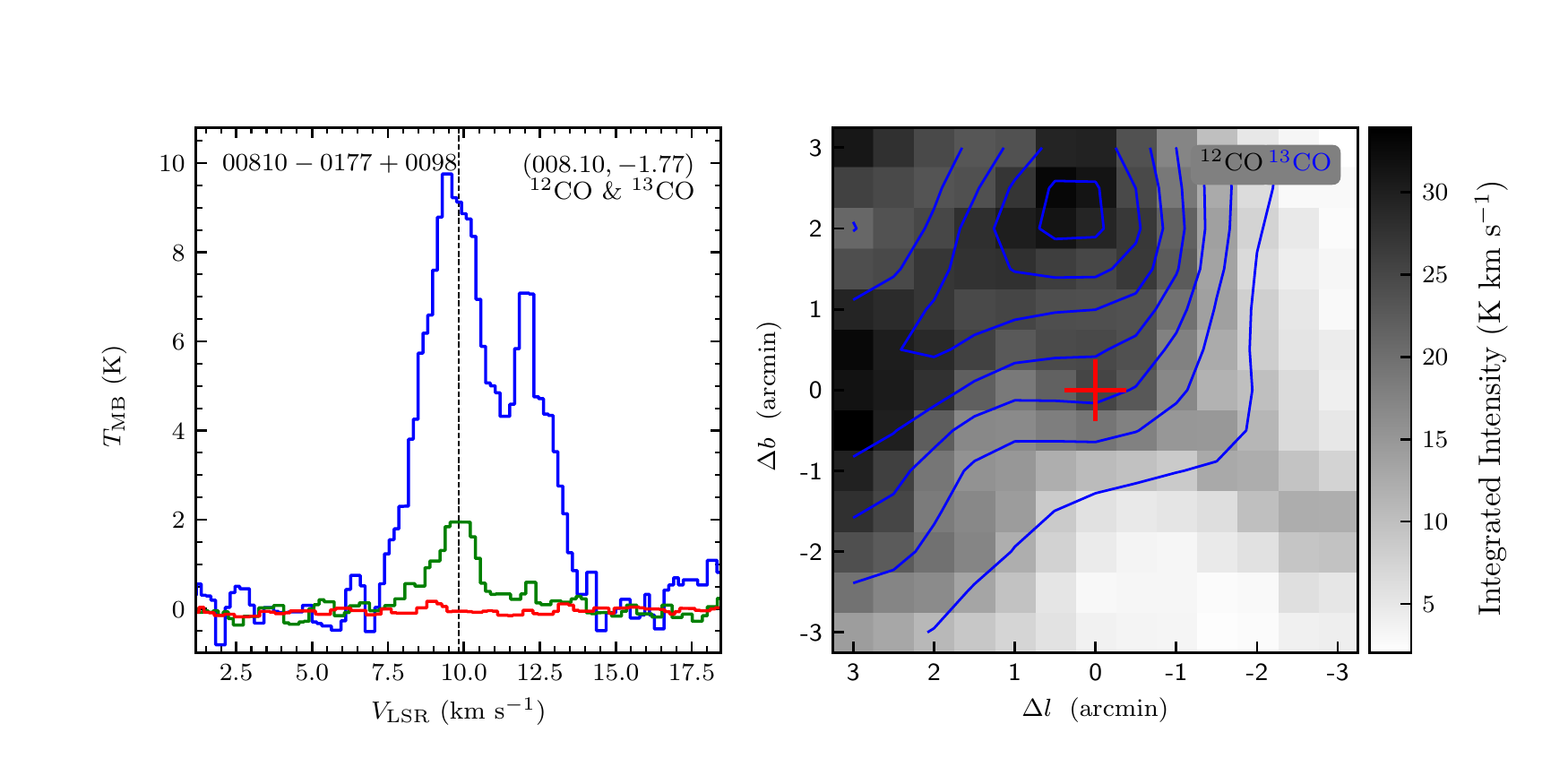}
\vspace{-0.5cm}

\includegraphics[width=9.0cm,angle=0]{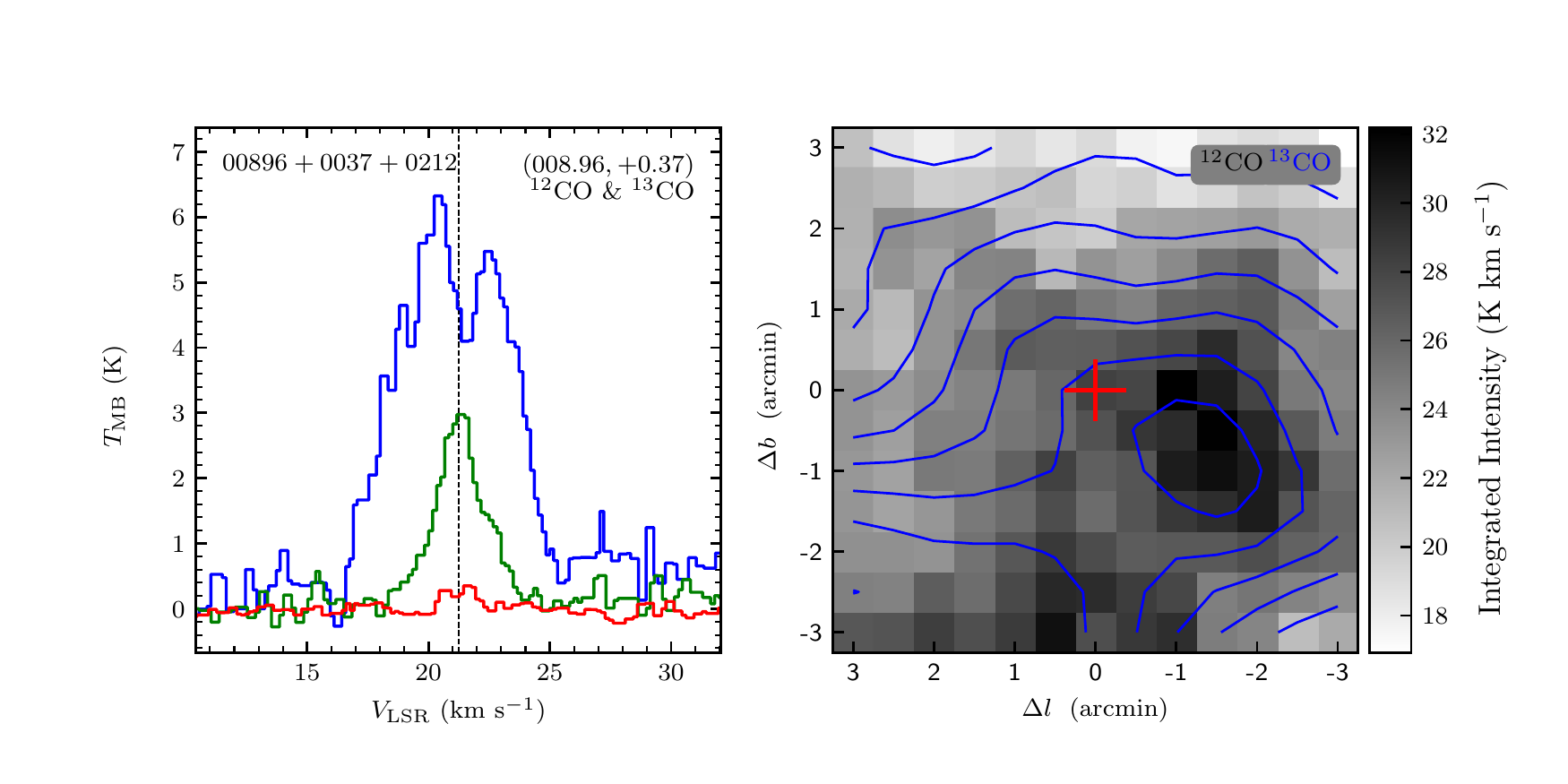}
\includegraphics[width=9.0cm,angle=0]{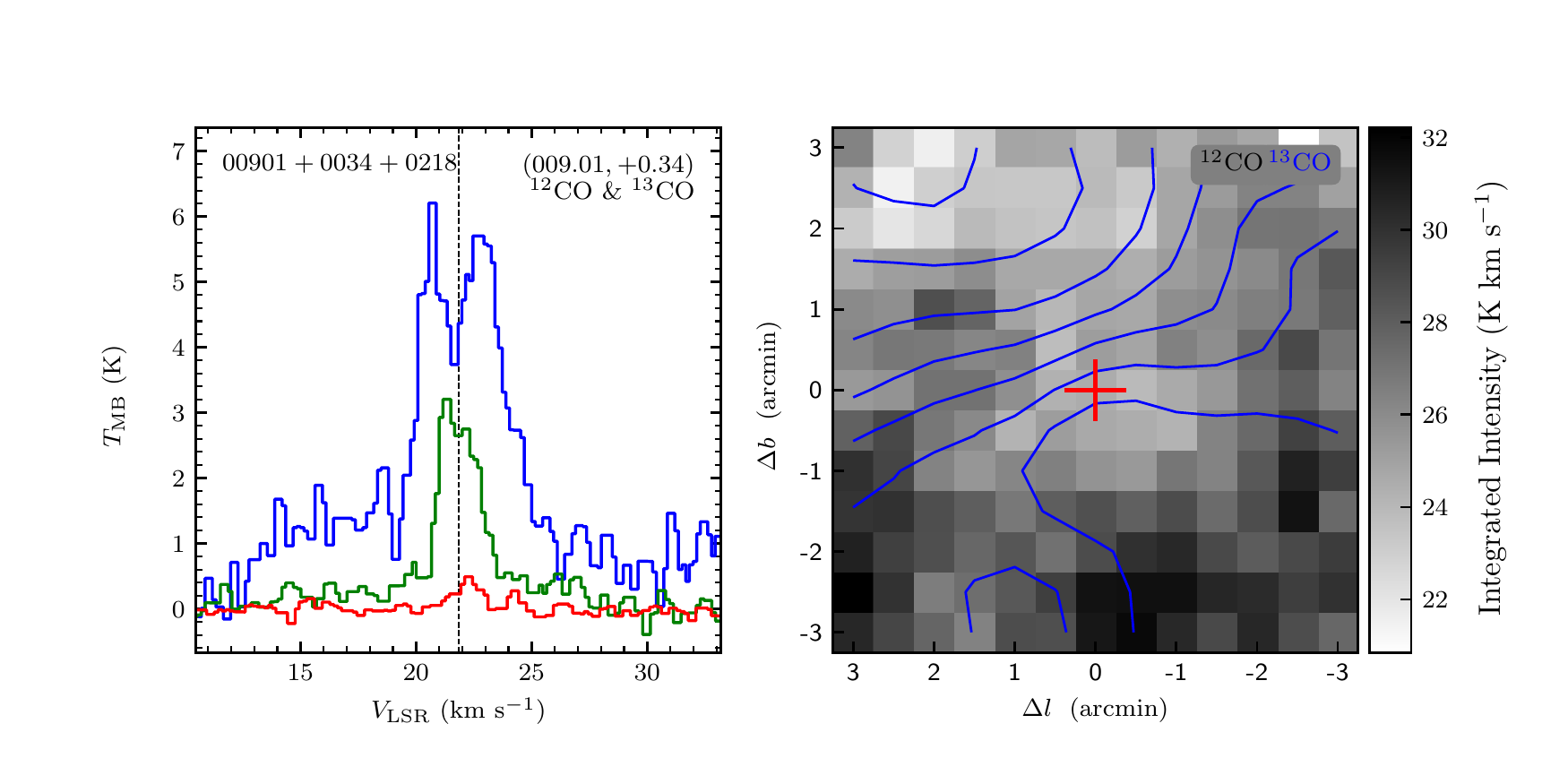}
\vspace{-0.5cm}

\includegraphics[width=9.0cm,angle=0]{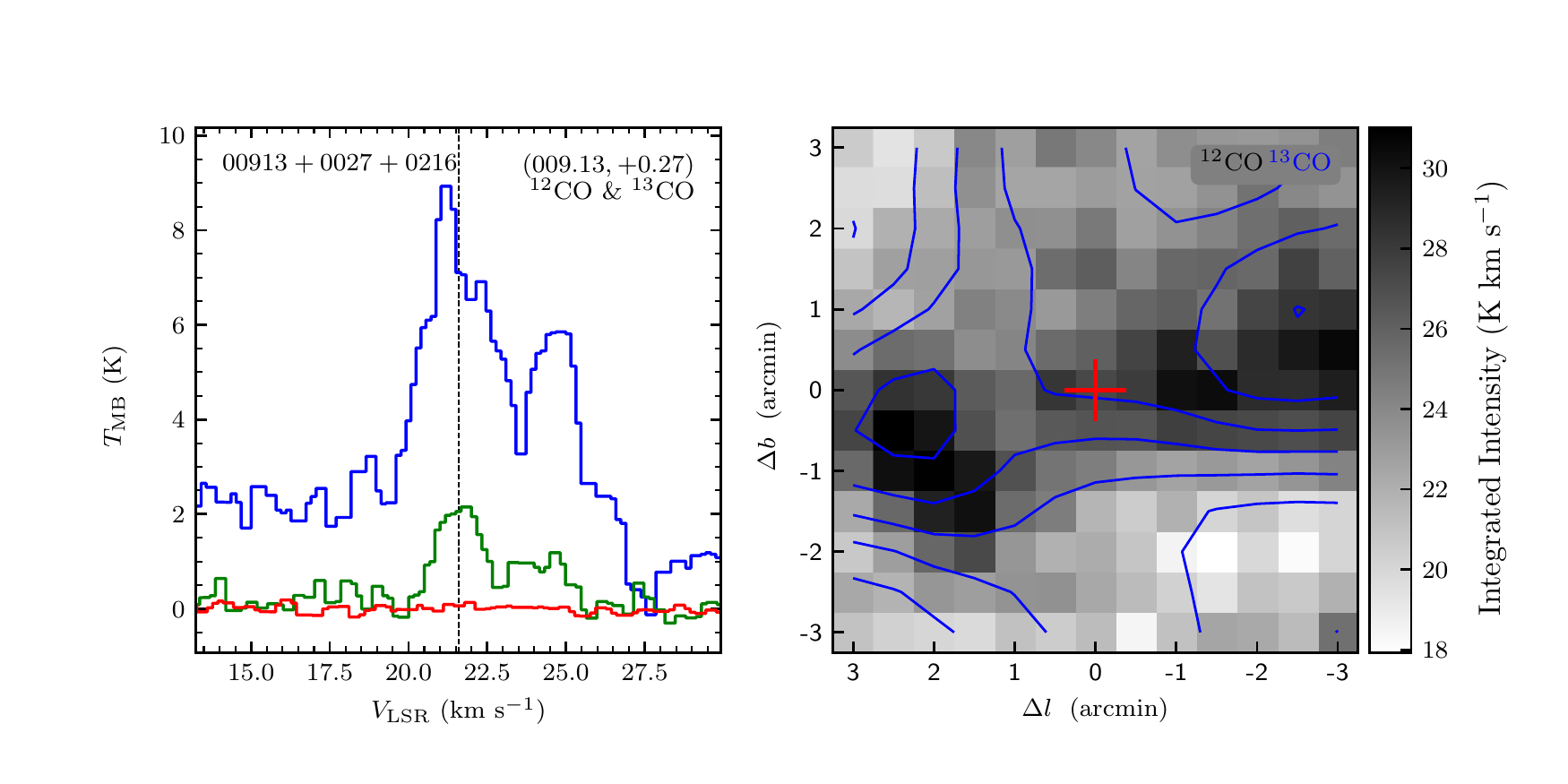}
\includegraphics[width=9.0cm,angle=0]{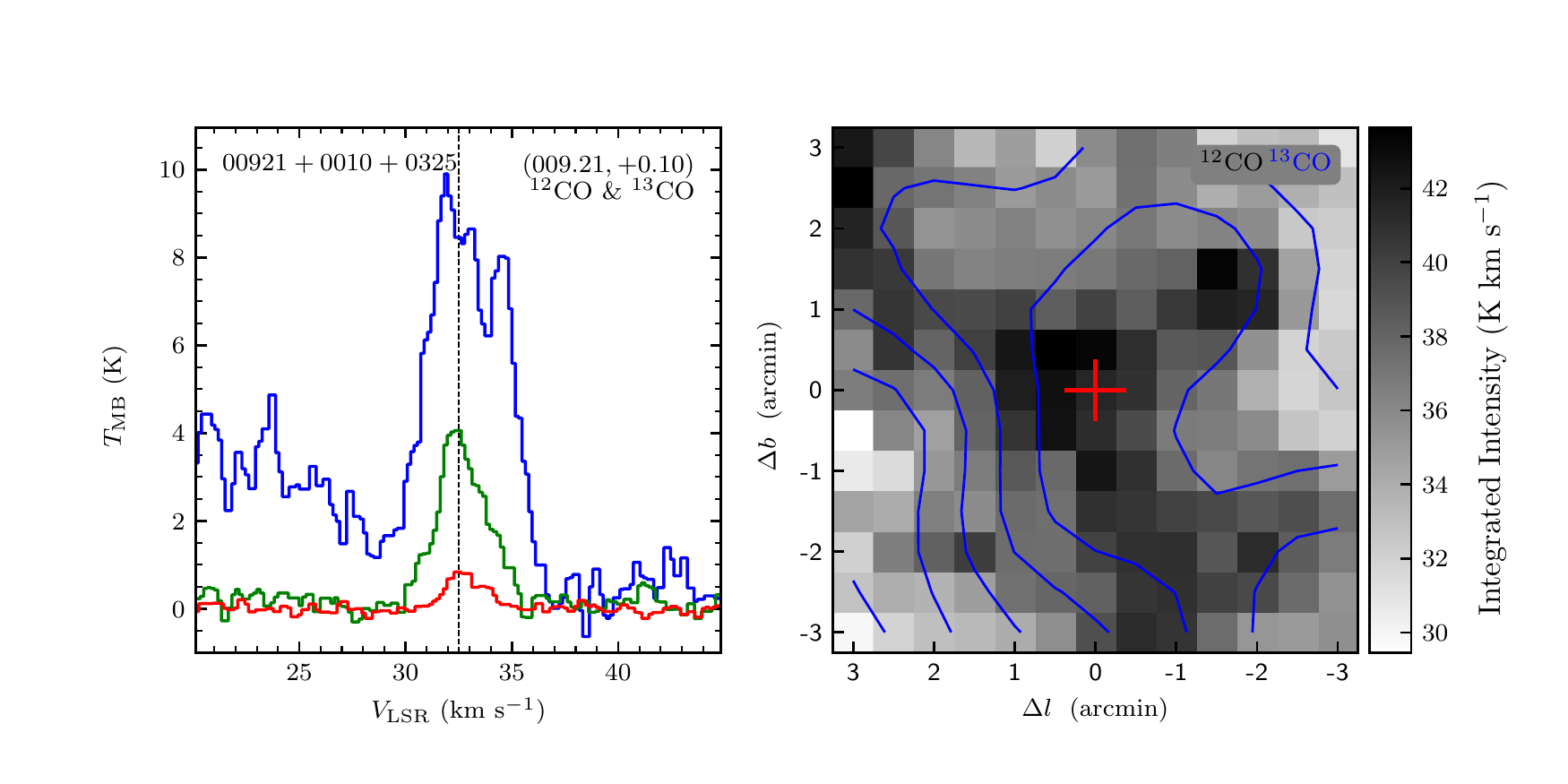}
\vspace{-0.5cm}

\includegraphics[width=9.0cm,angle=0]{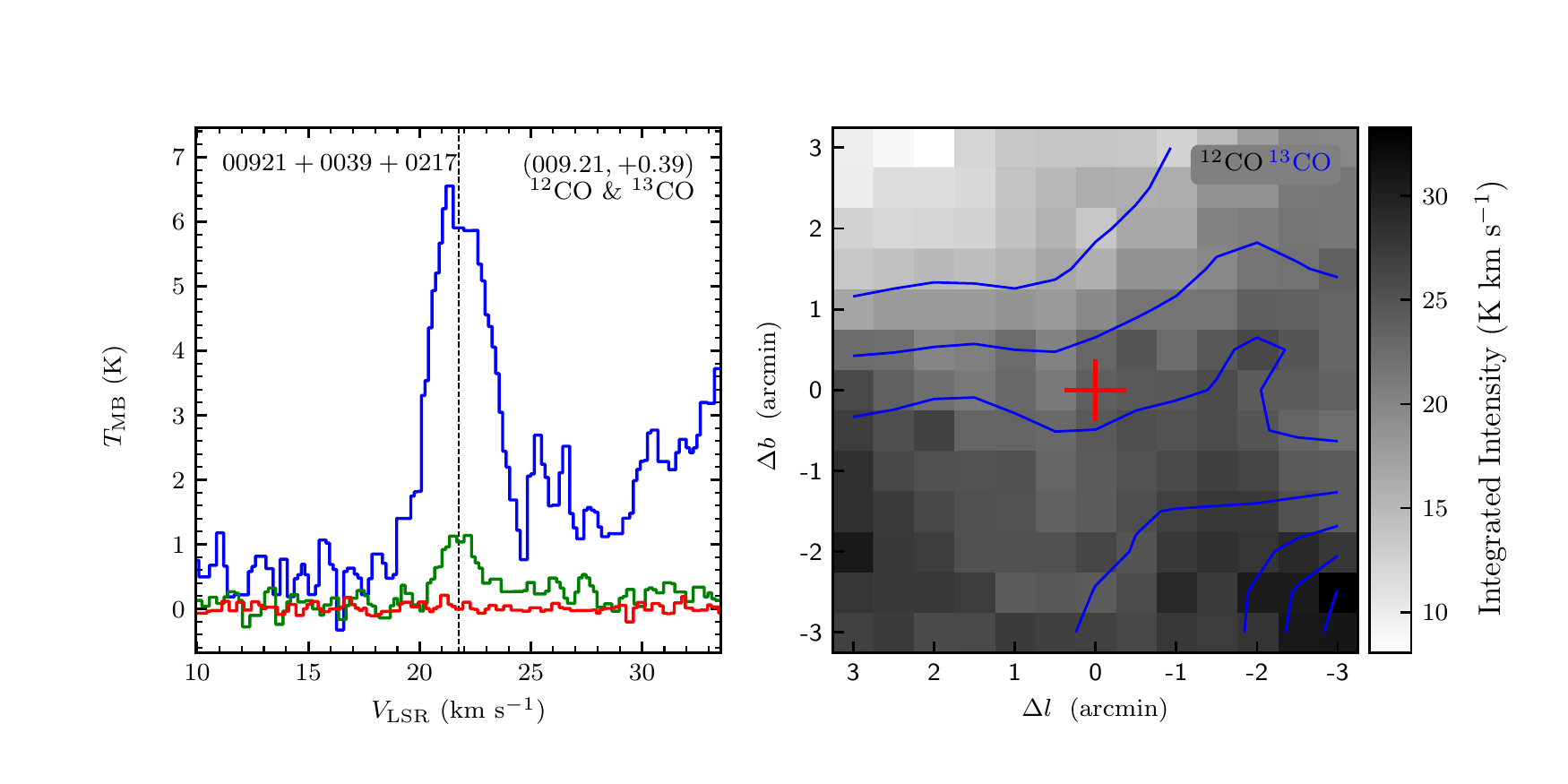}
\includegraphics[width=9.0cm,angle=0]{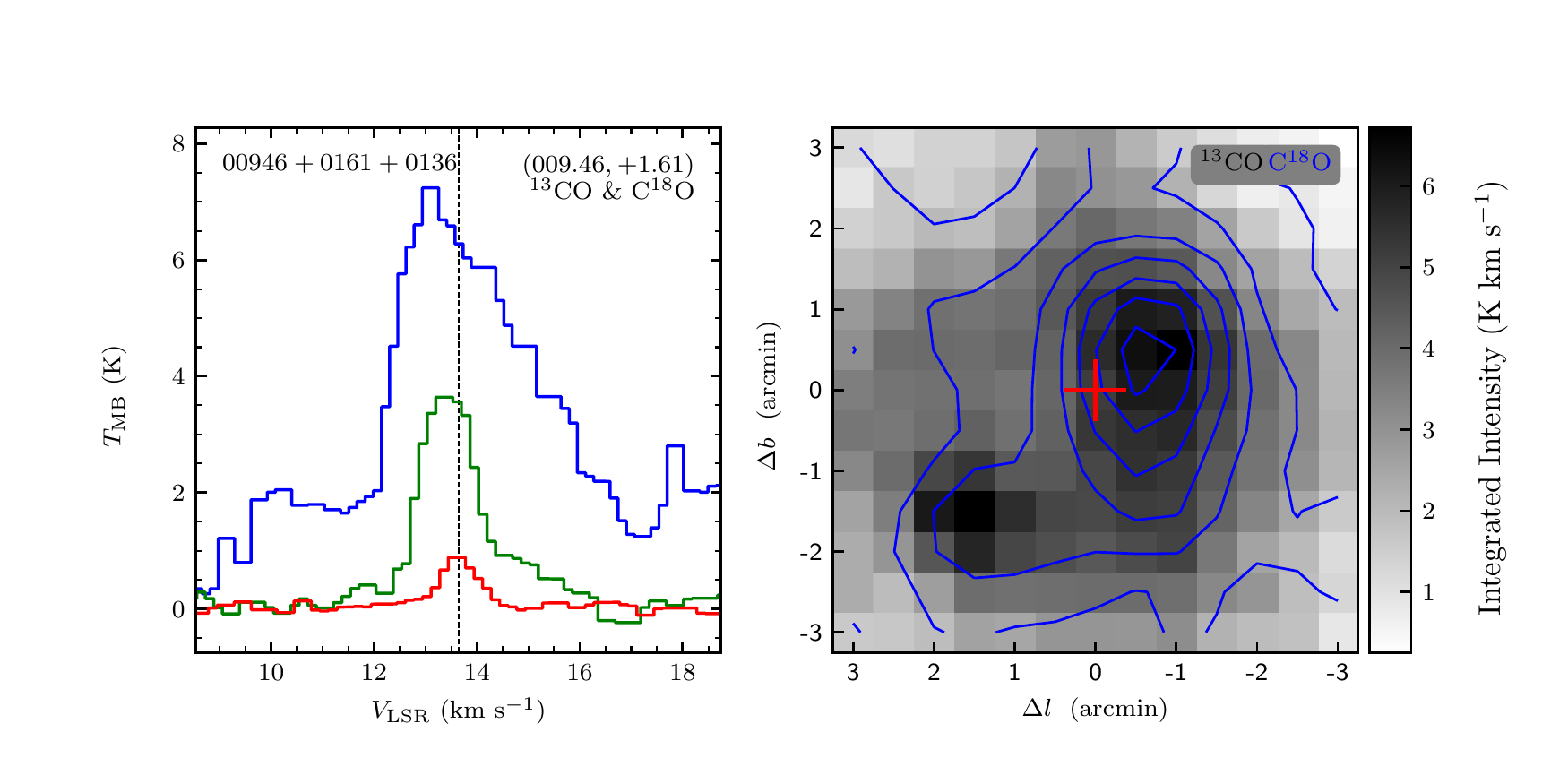}
\vspace{-0.5cm}

\includegraphics[width=9.0cm,angle=0]{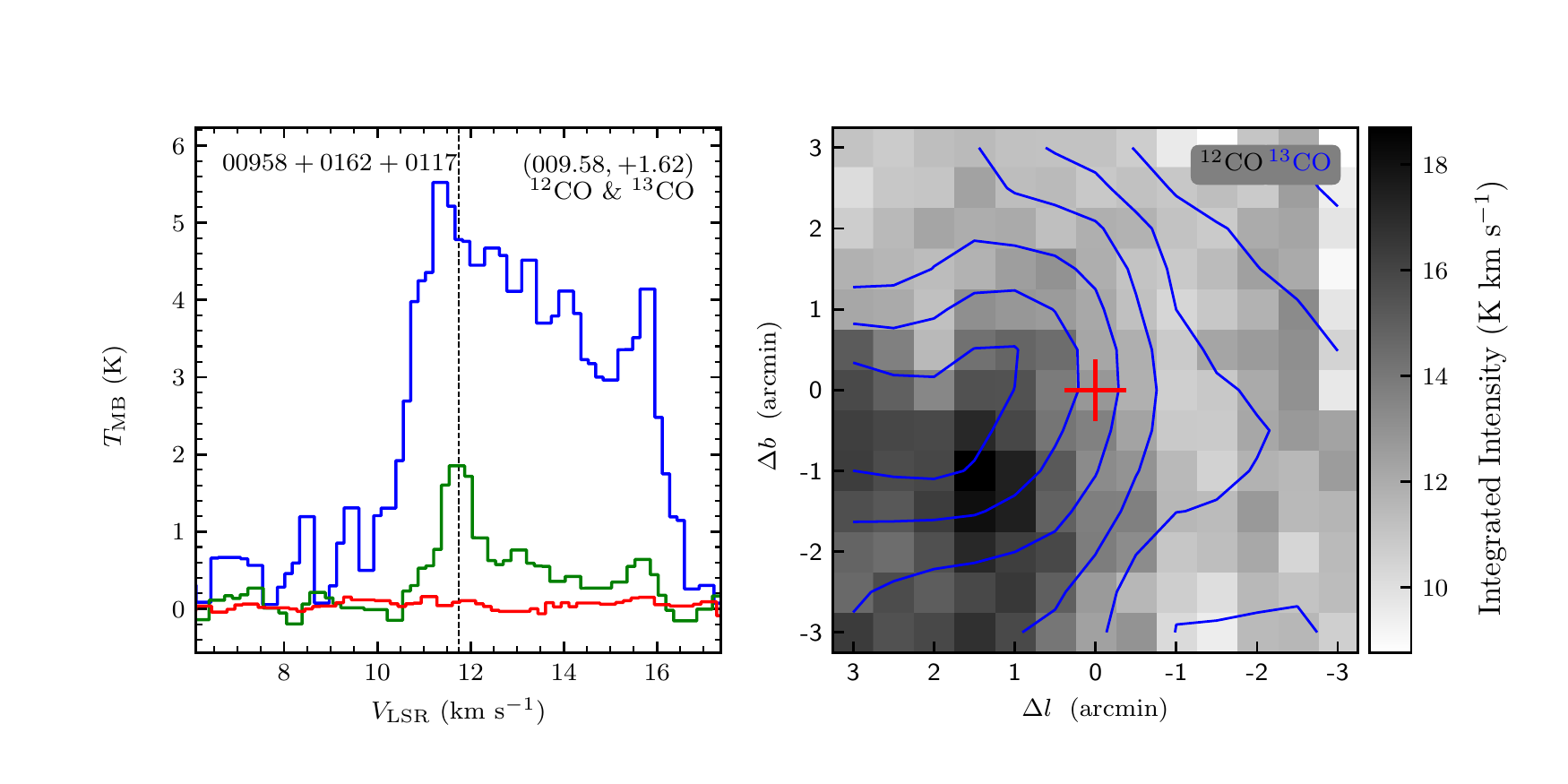}
\includegraphics[width=9.0cm,angle=0]{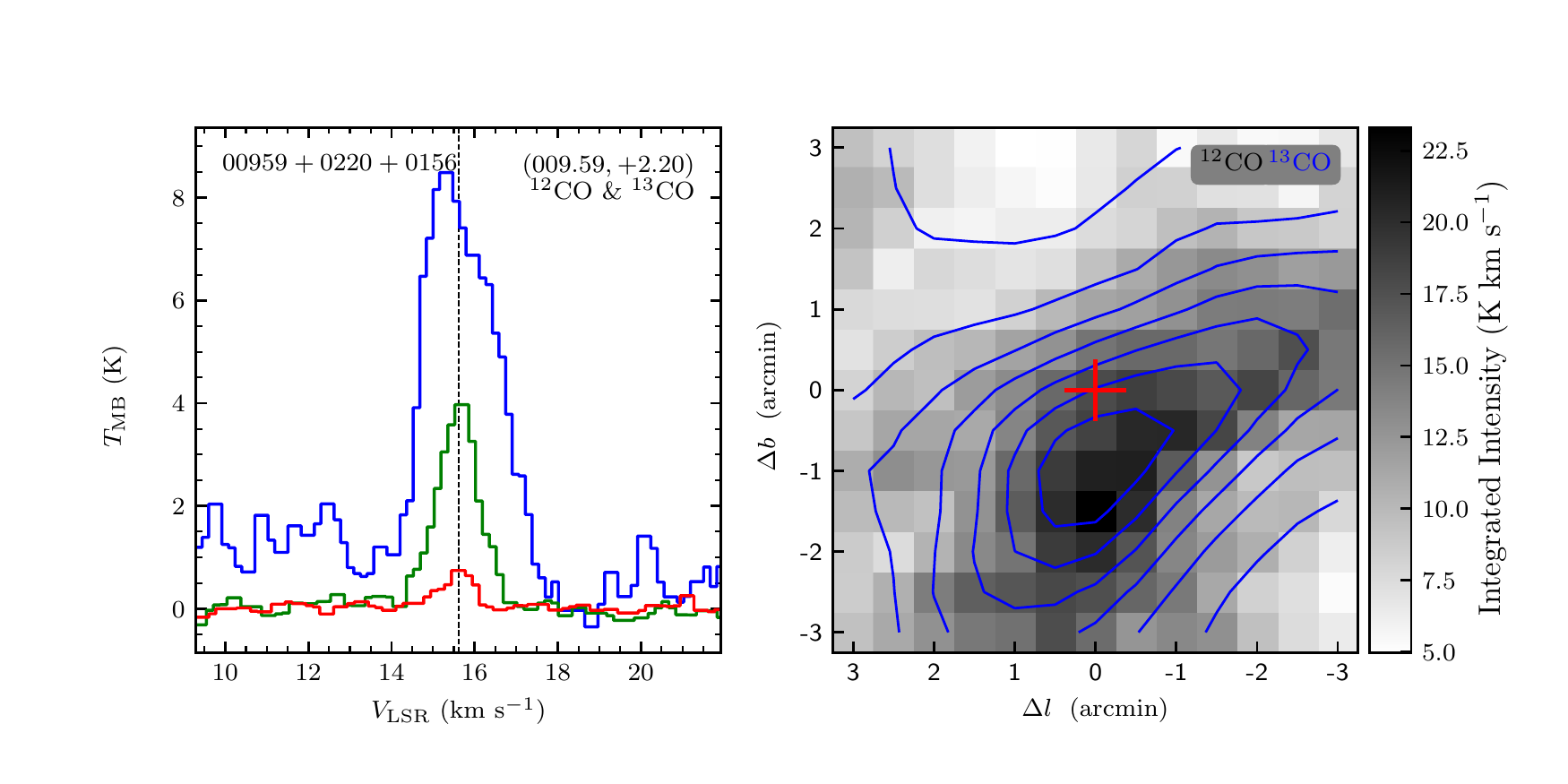}
\end{figure}
\clearpage

\begin{figure}
\includegraphics[width=9.0cm,angle=0]{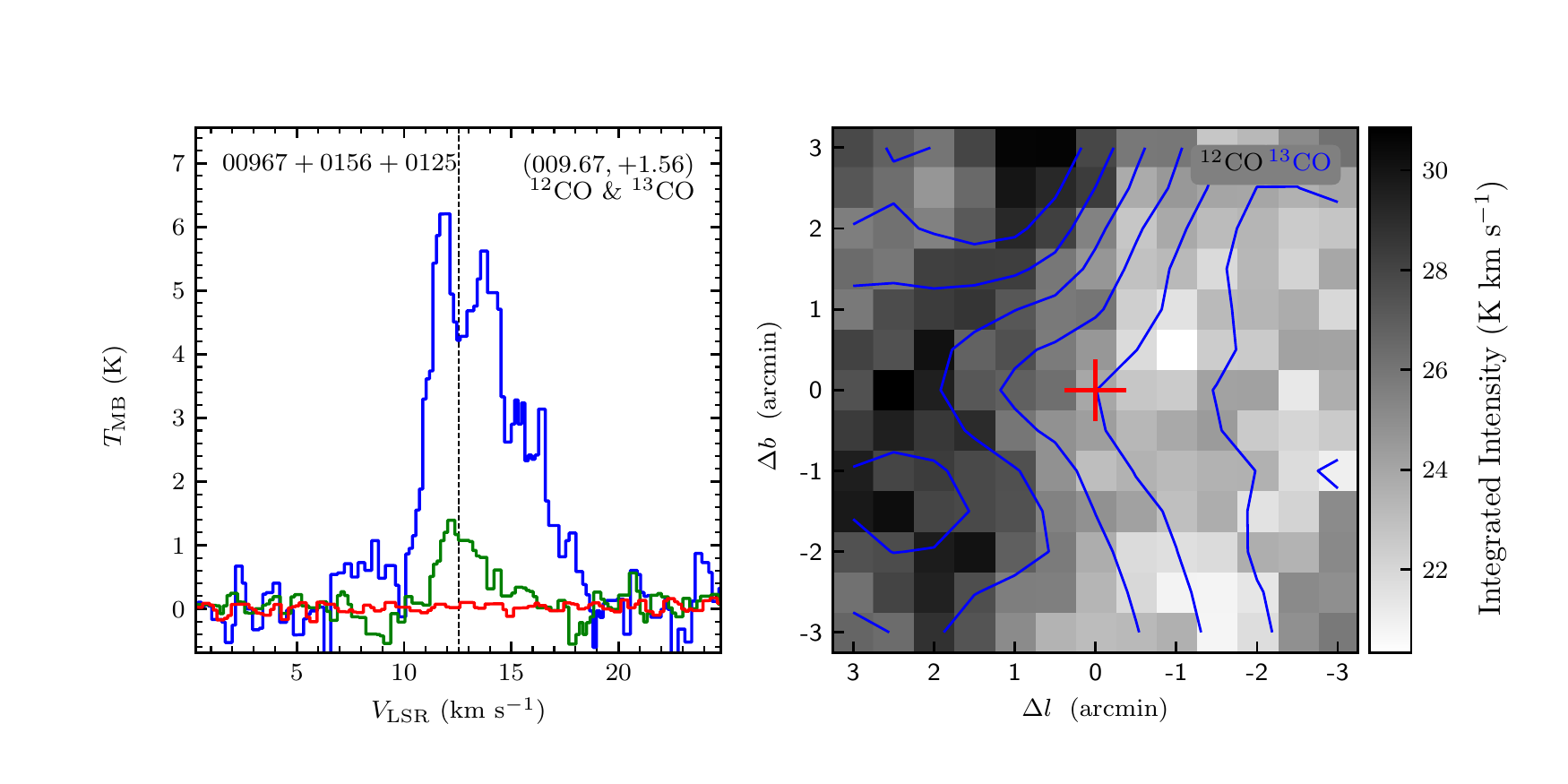}
\includegraphics[width=9.0cm,angle=0]{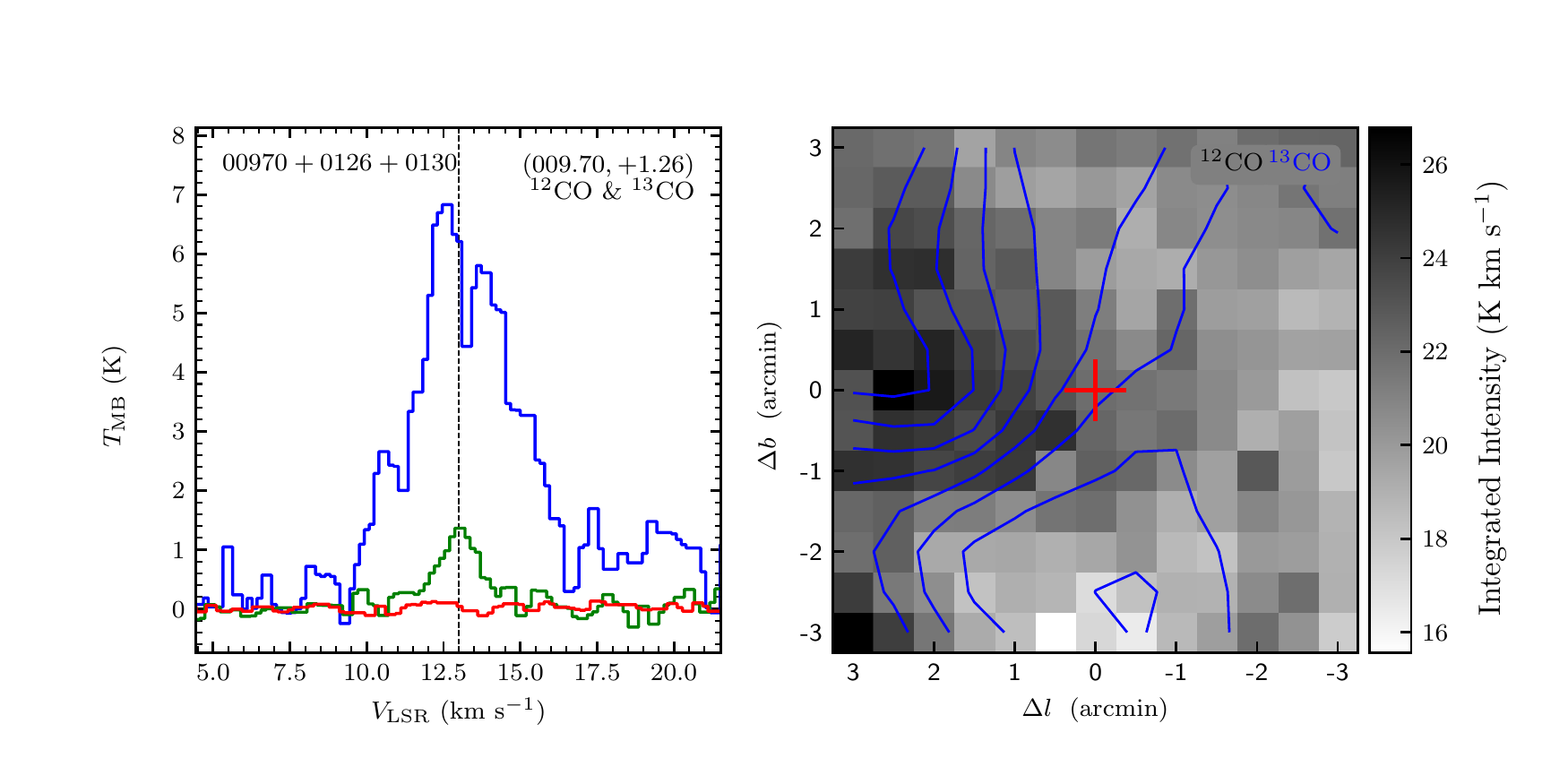}
\vspace{-0.5cm}

\includegraphics[width=9.0cm,angle=0]{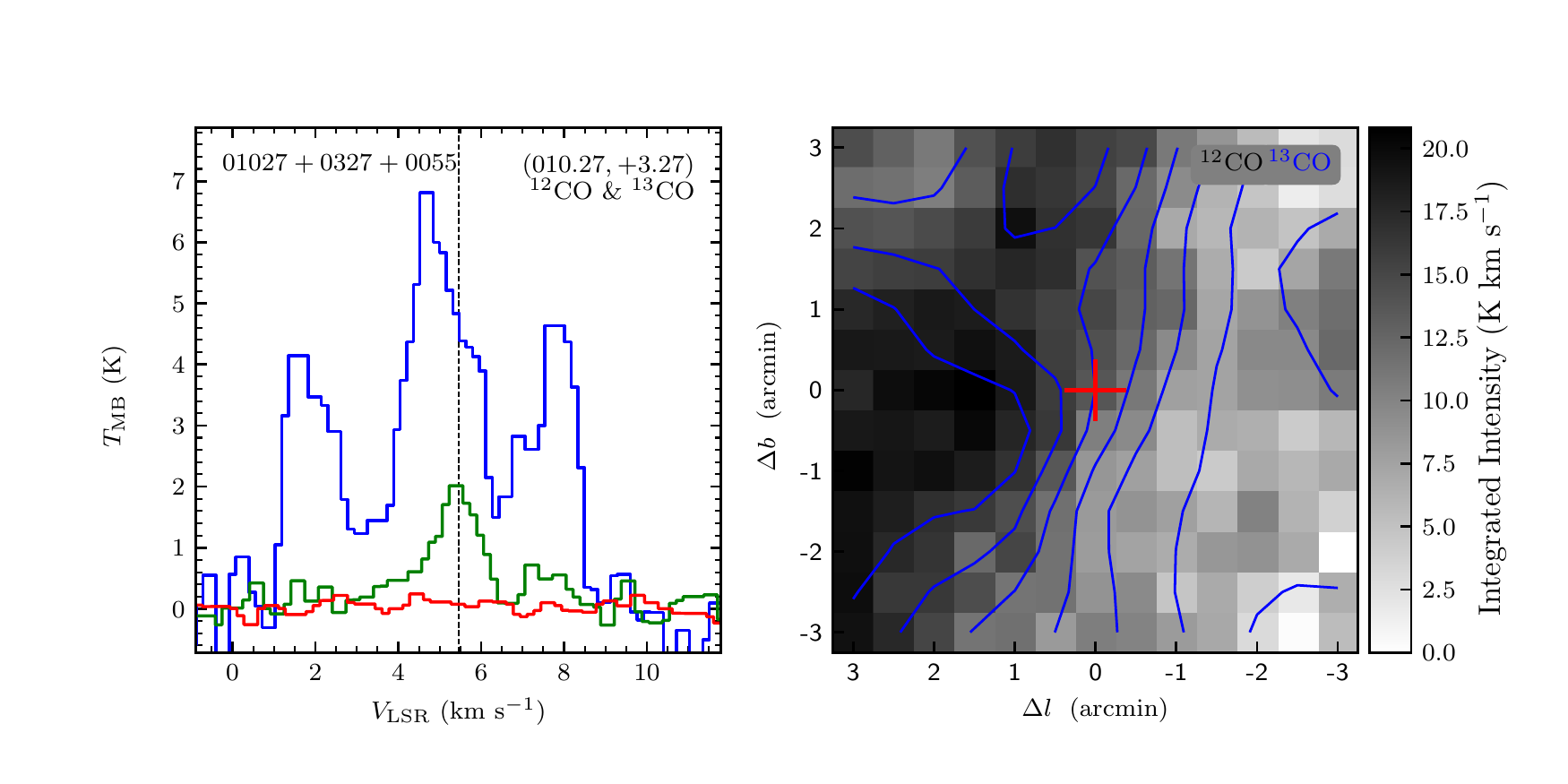}
\includegraphics[width=9.0cm,angle=0]{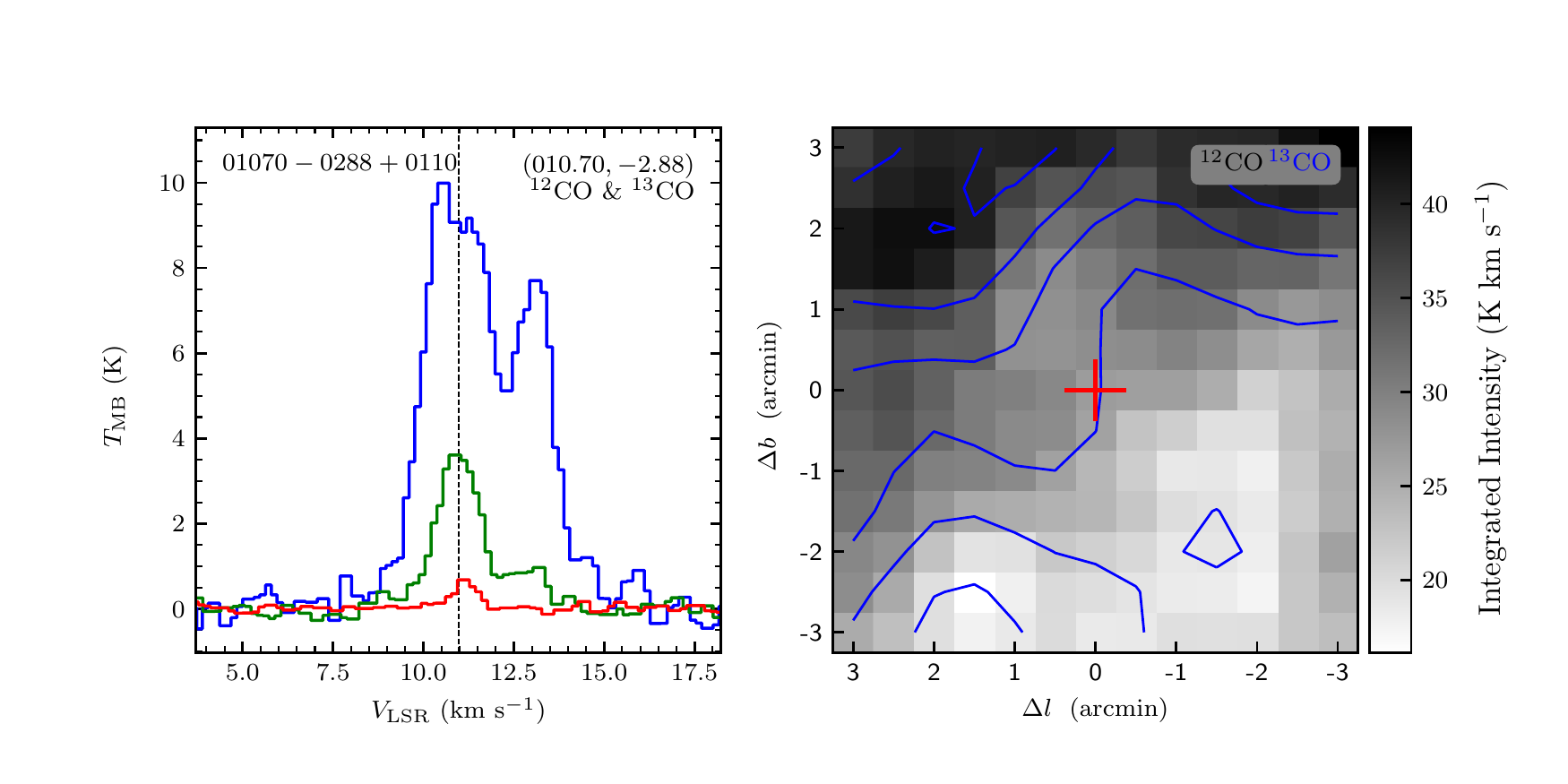}
\vspace{-0.5cm}

\includegraphics[width=9.0cm,angle=0]{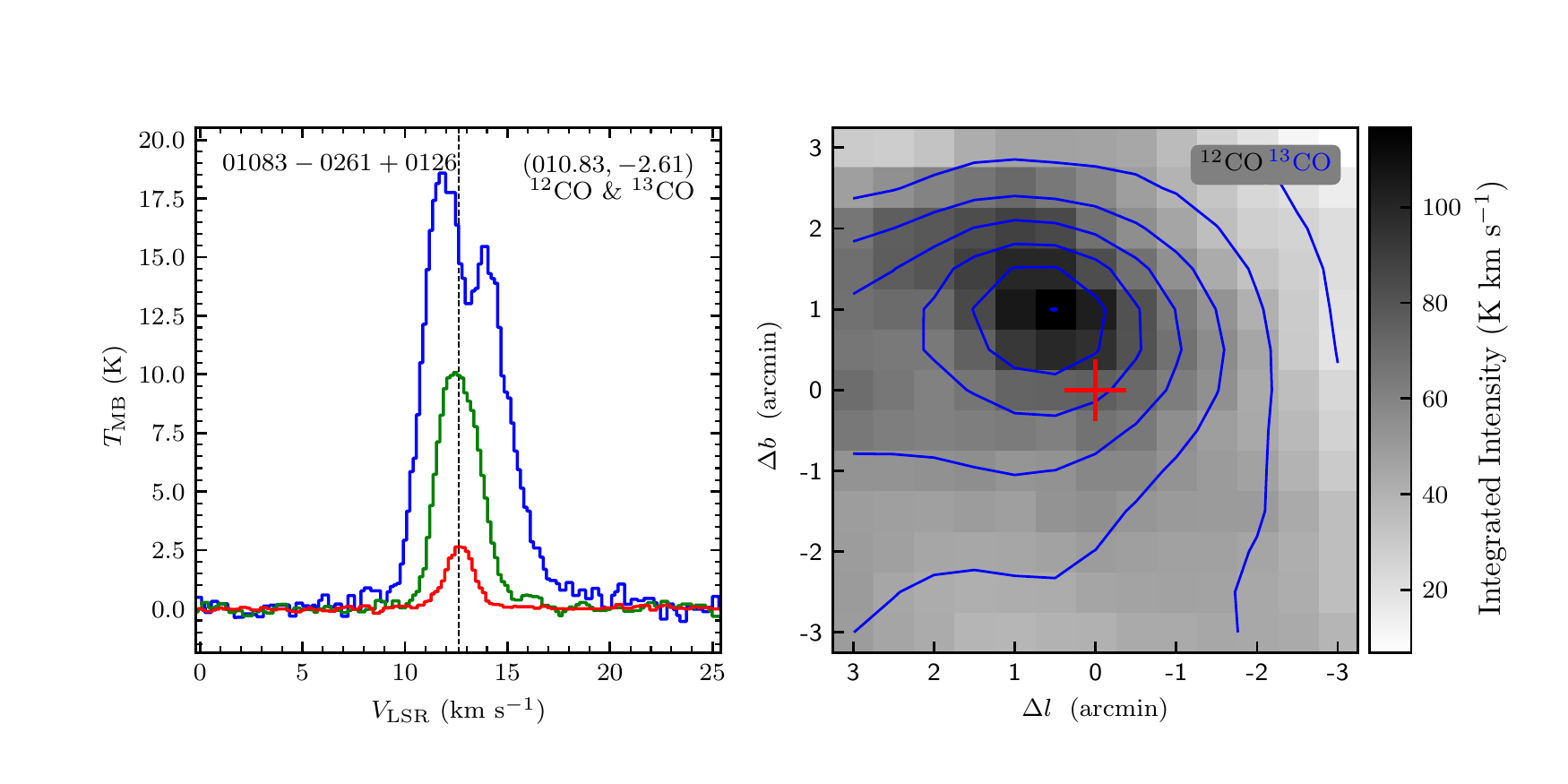}
\includegraphics[width=9.0cm,angle=0]{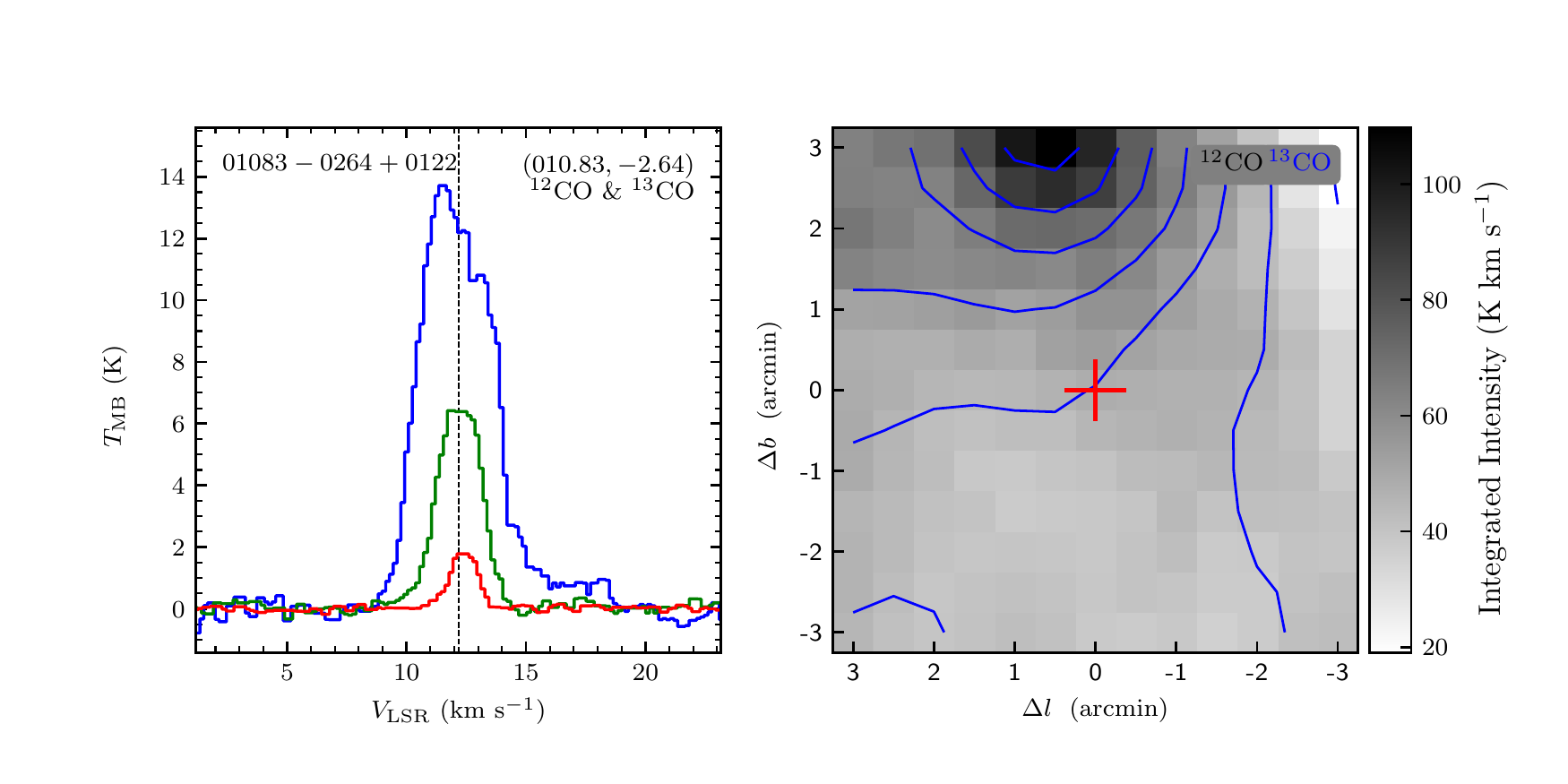}
\vspace{-0.5cm}

\includegraphics[width=9.0cm,angle=0]{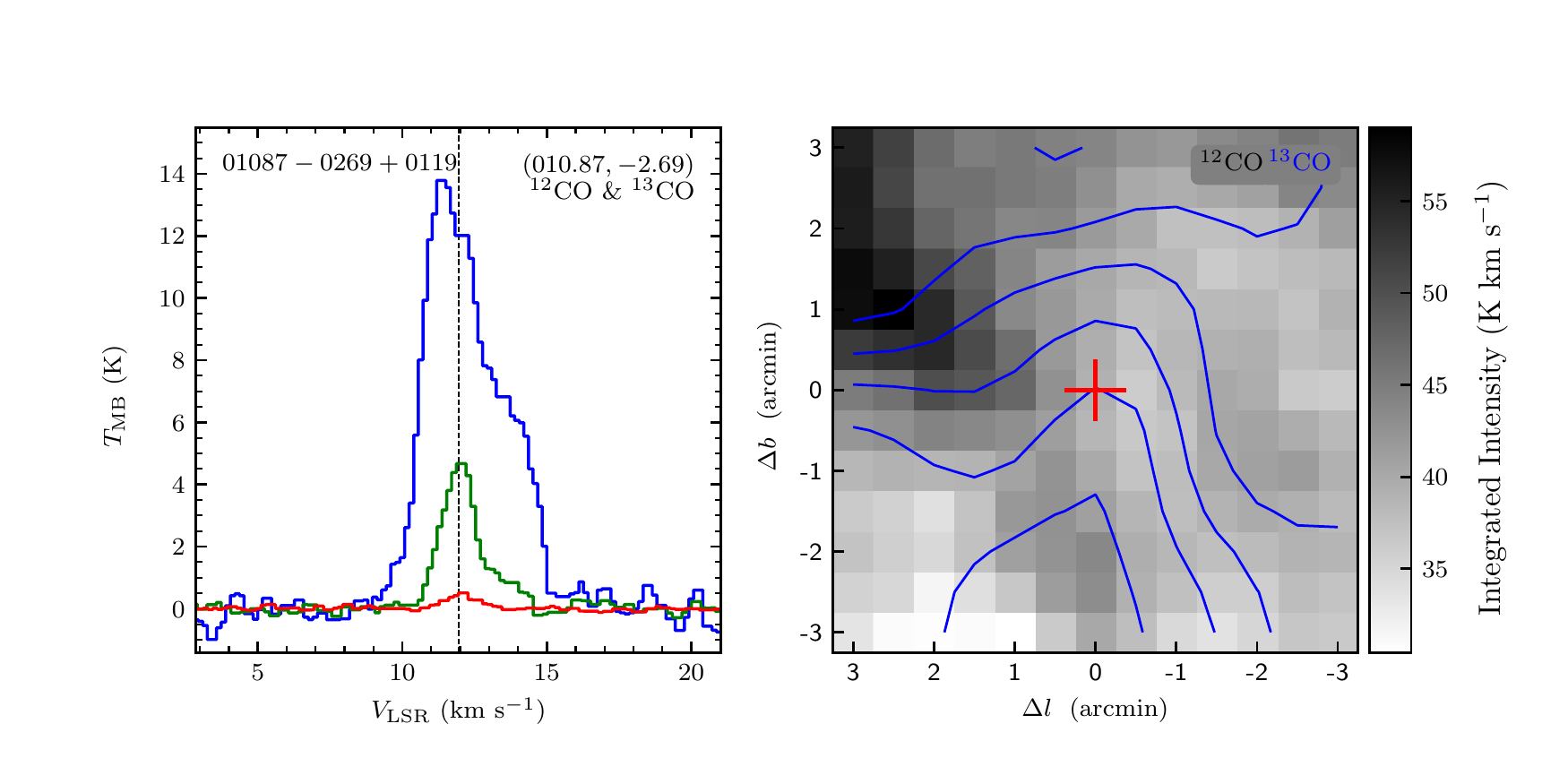}
\includegraphics[width=9.0cm,angle=0]{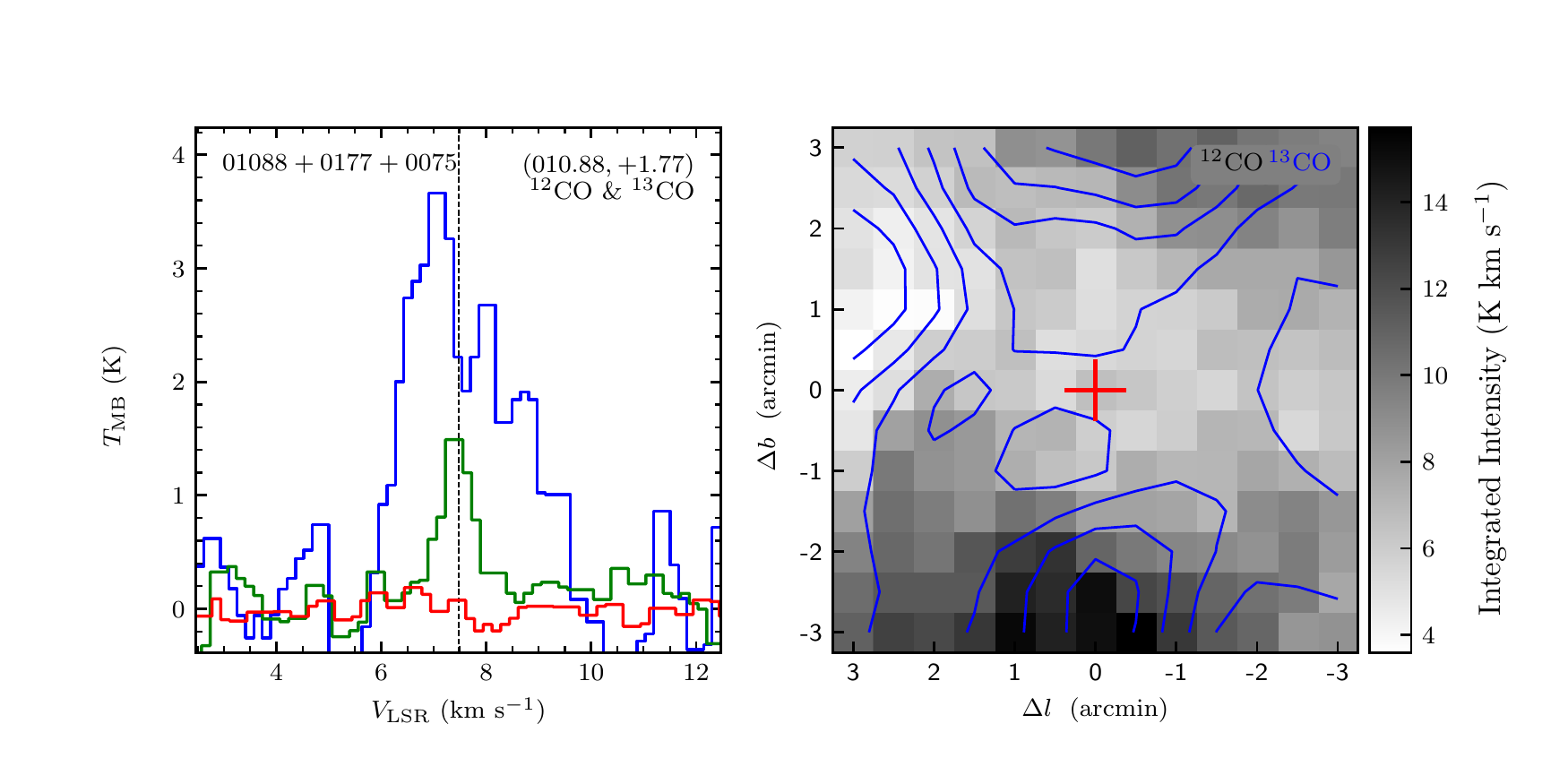}
\vspace{-0.5cm}

\includegraphics[width=9.0cm,angle=0]{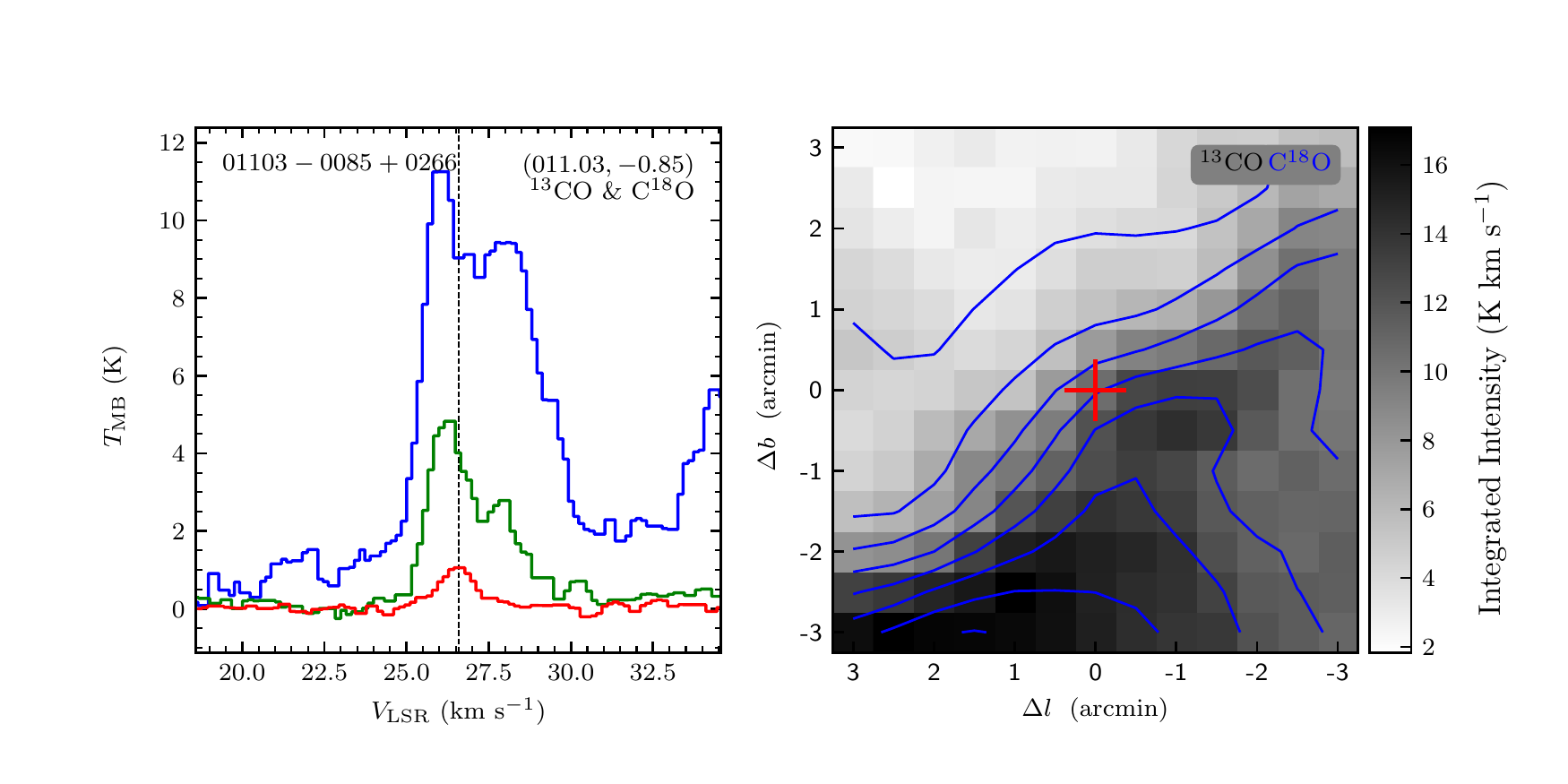}
\includegraphics[width=9.0cm,angle=0]{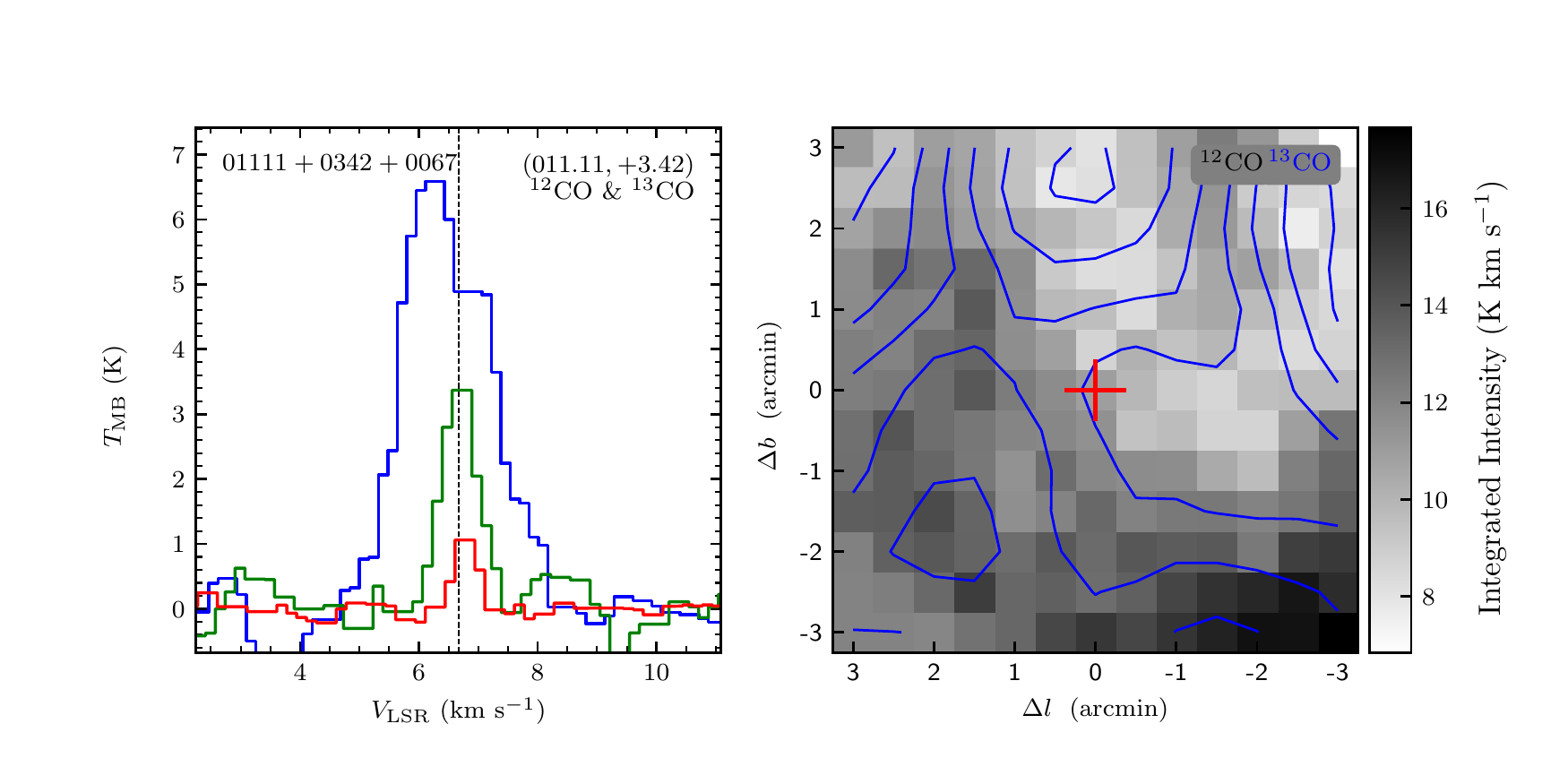}
\end{figure}
\clearpage

\begin{figure}
\includegraphics[width=9.0cm,angle=0]{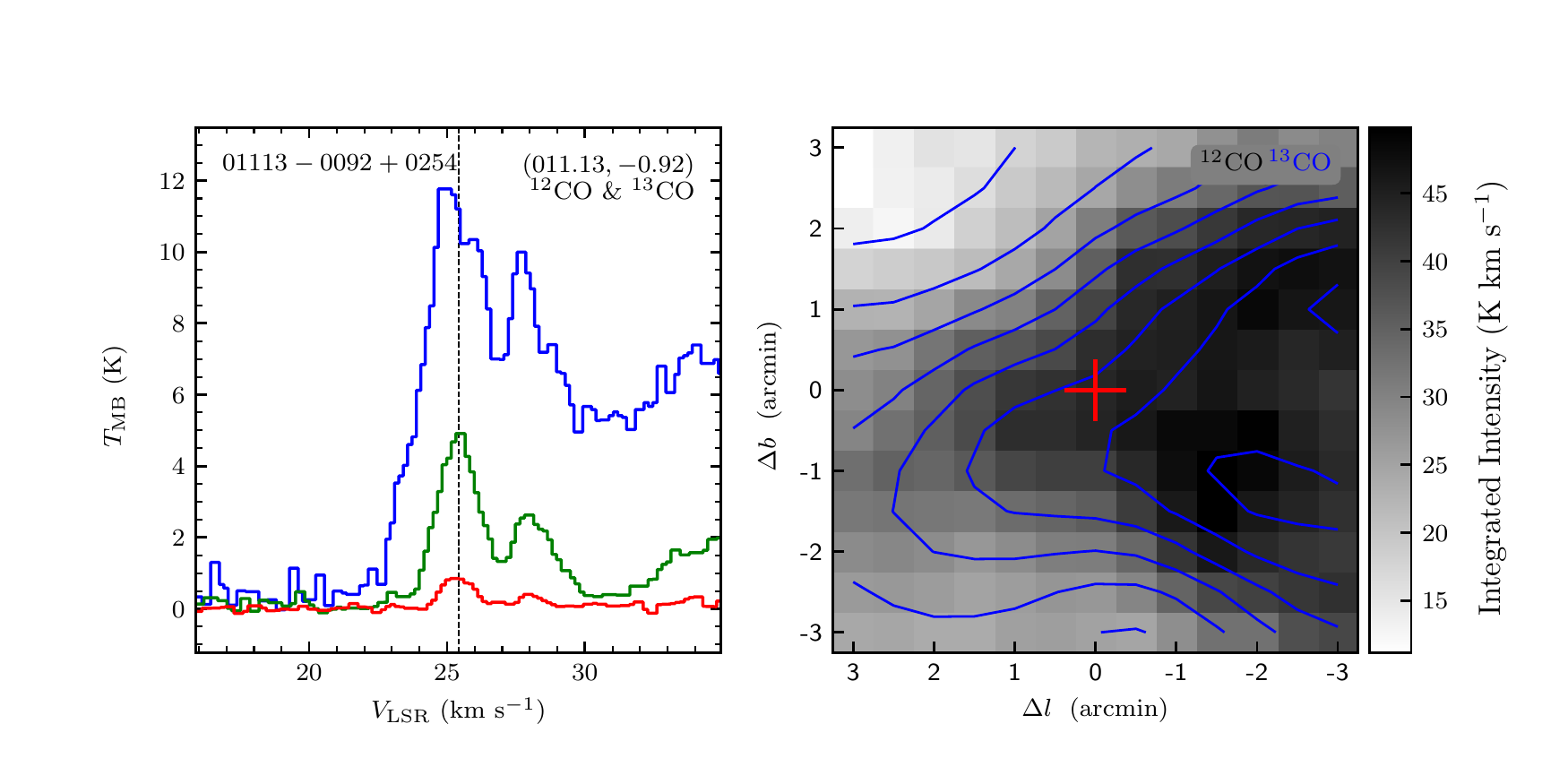}
\includegraphics[width=9.0cm,angle=0]{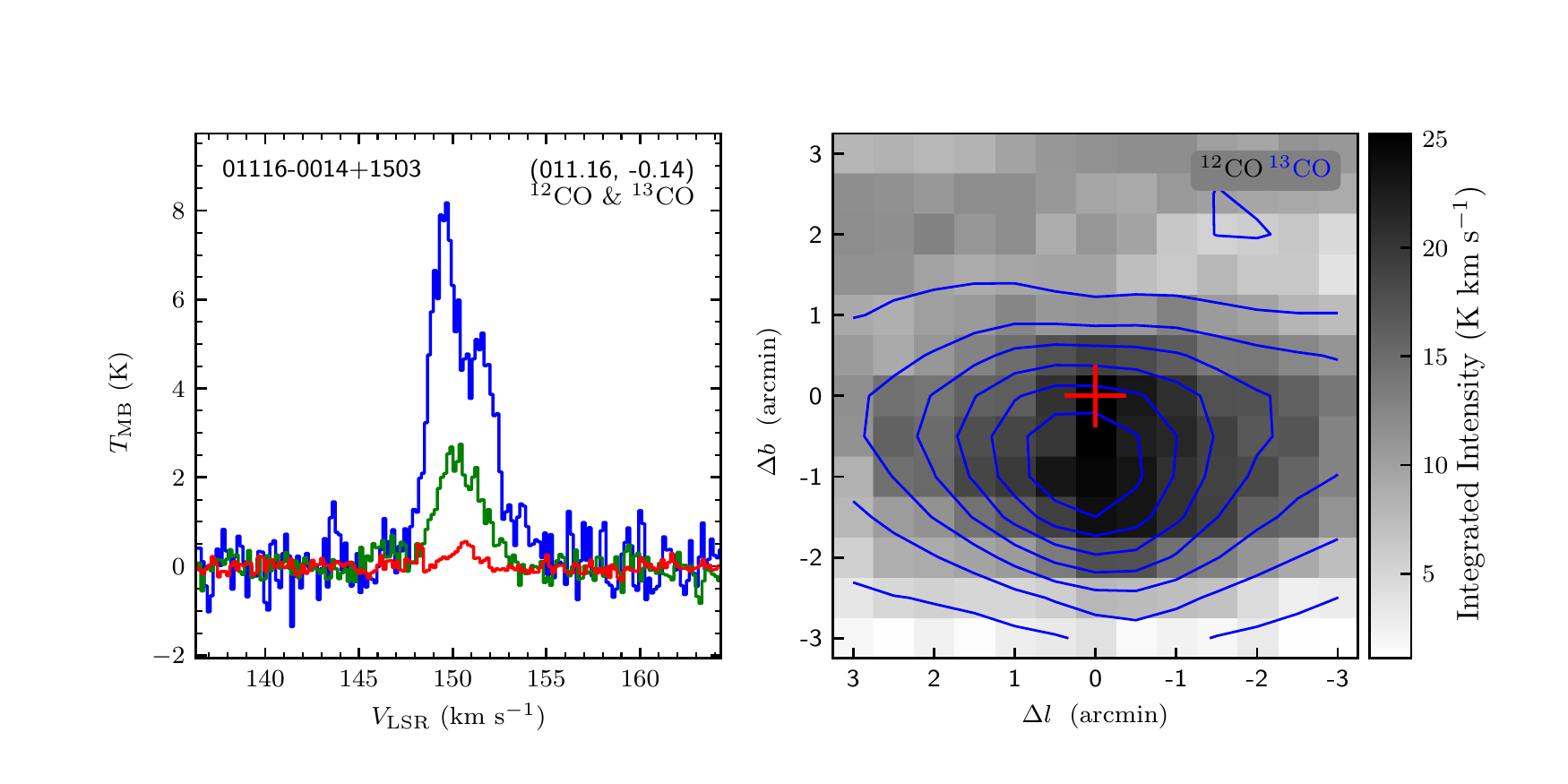}
\vspace{-0.5cm}

\includegraphics[width=9.0cm,angle=0]{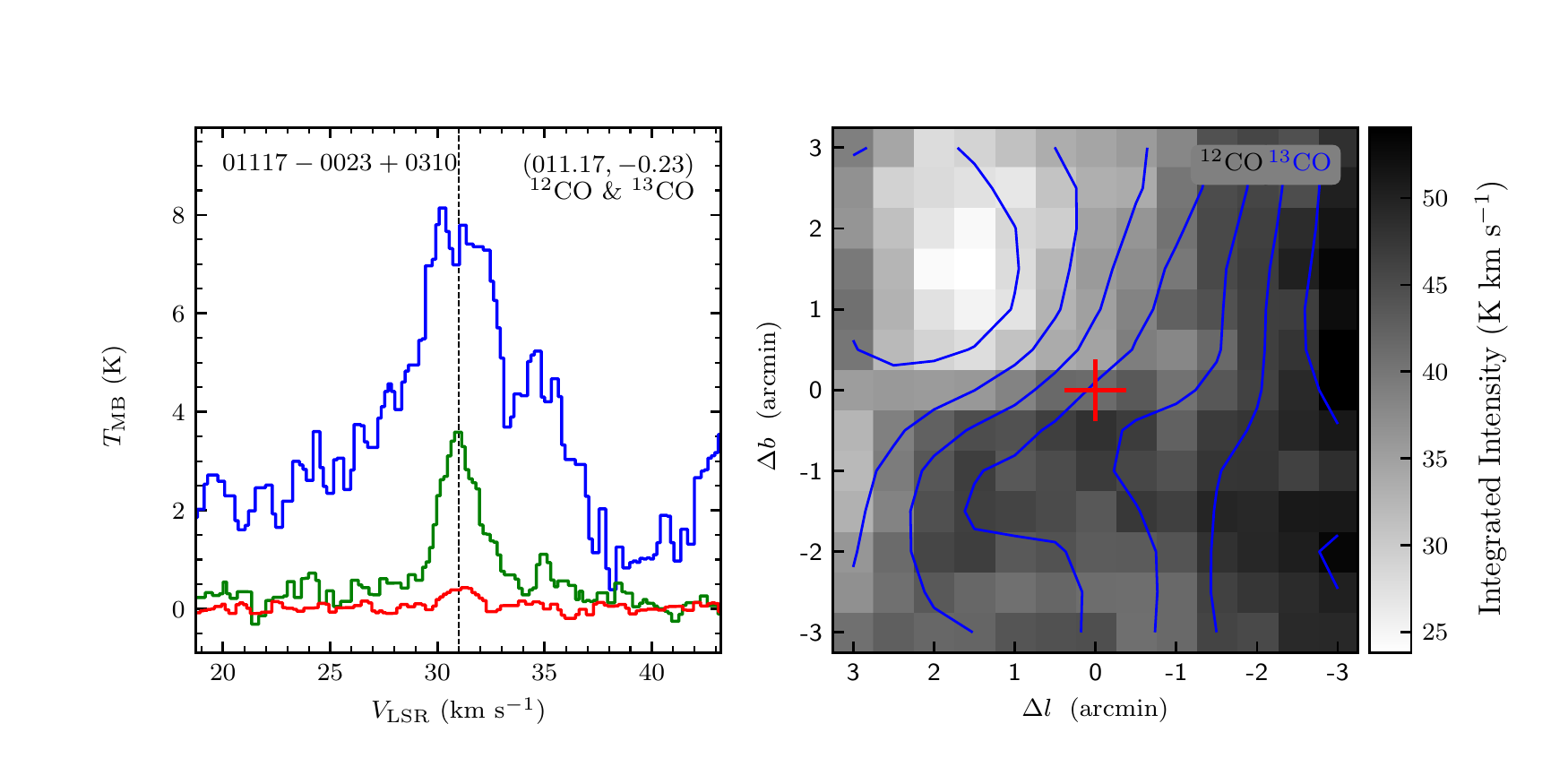}
\includegraphics[width=9.0cm,angle=0]{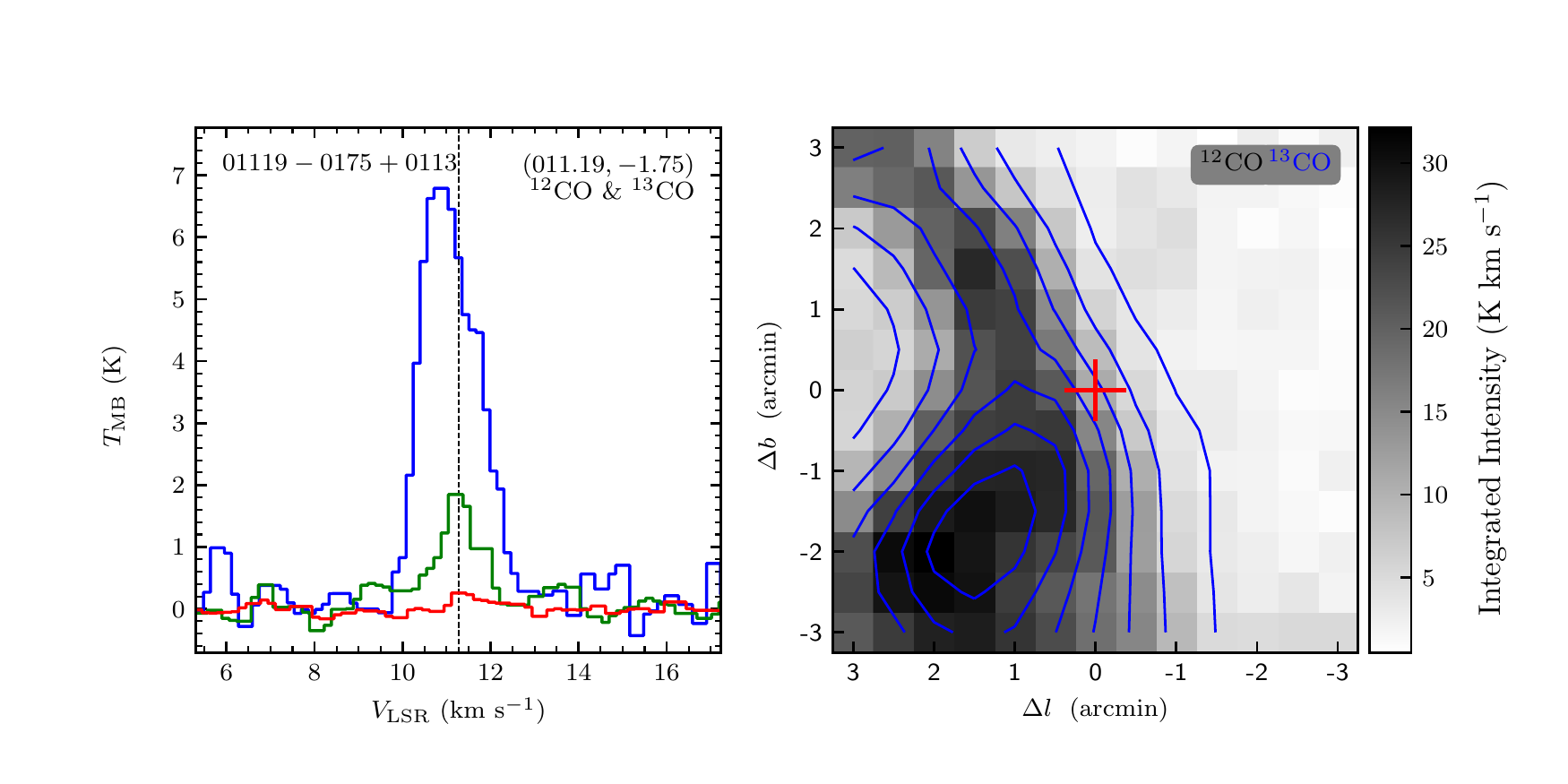}
\vspace{-0.5cm}

\includegraphics[width=9.0cm,angle=0]{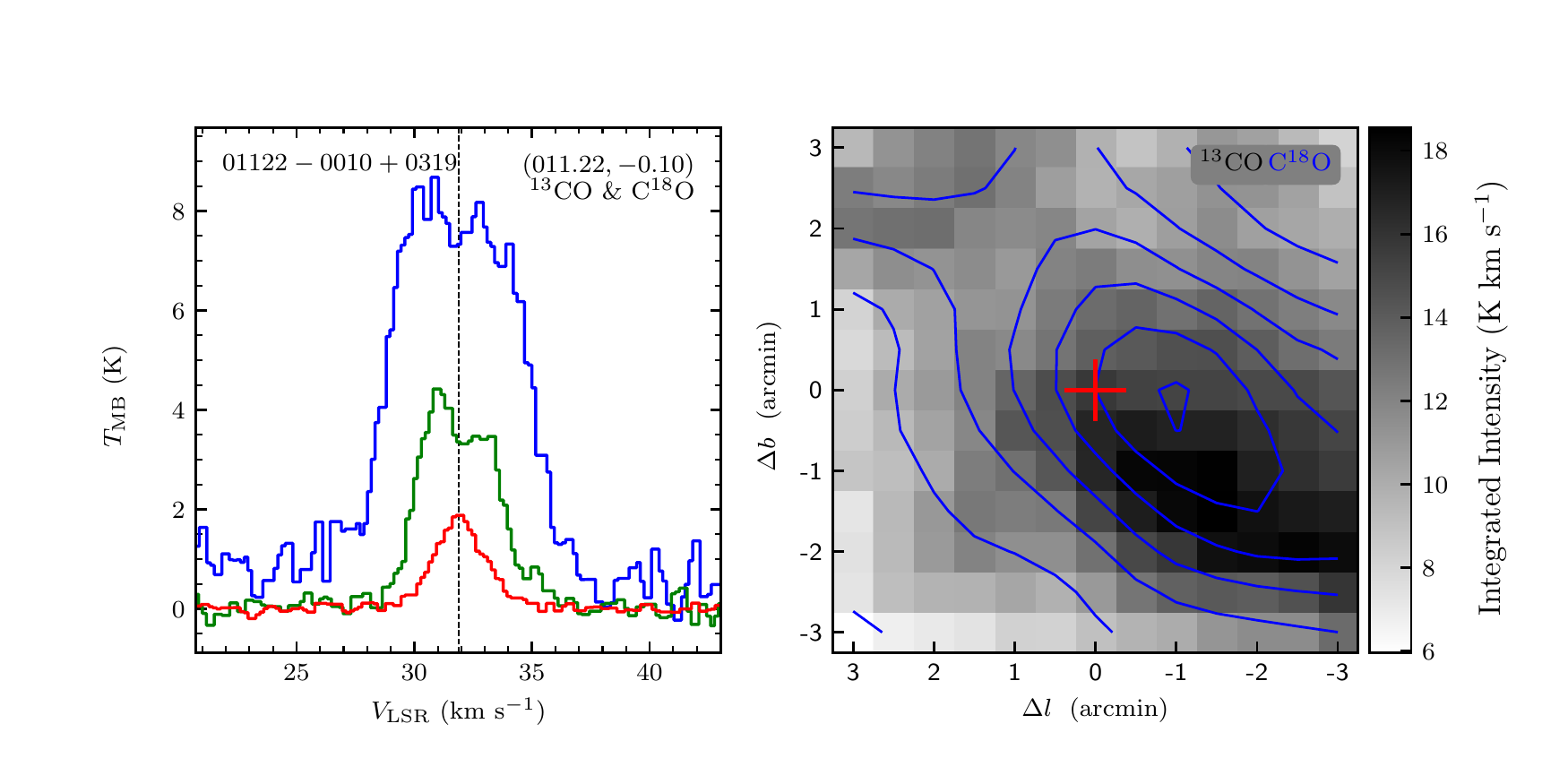}
\includegraphics[width=9.0cm,angle=0]{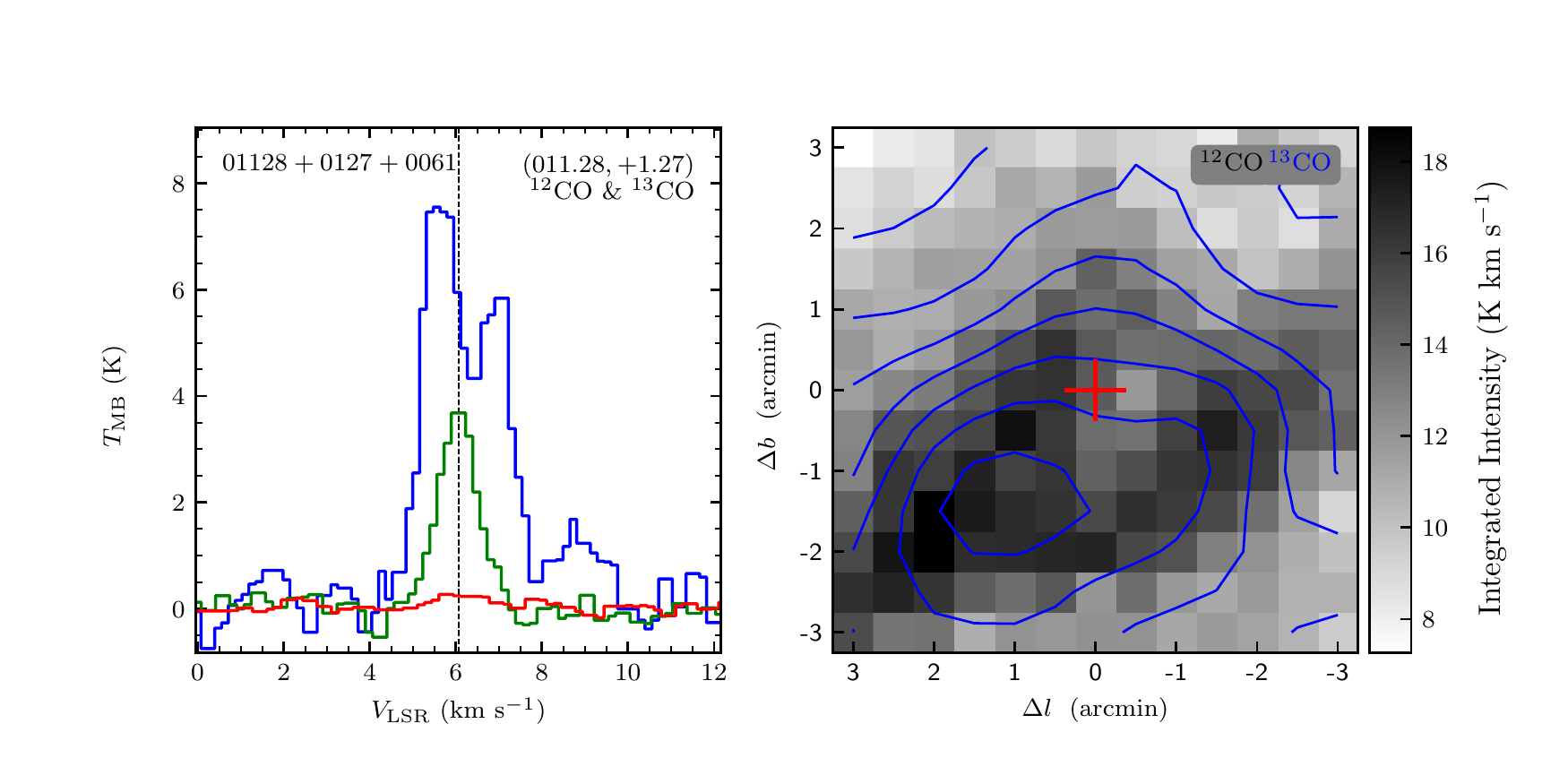}
\vspace{-0.5cm}

\includegraphics[width=9.0cm,angle=0]{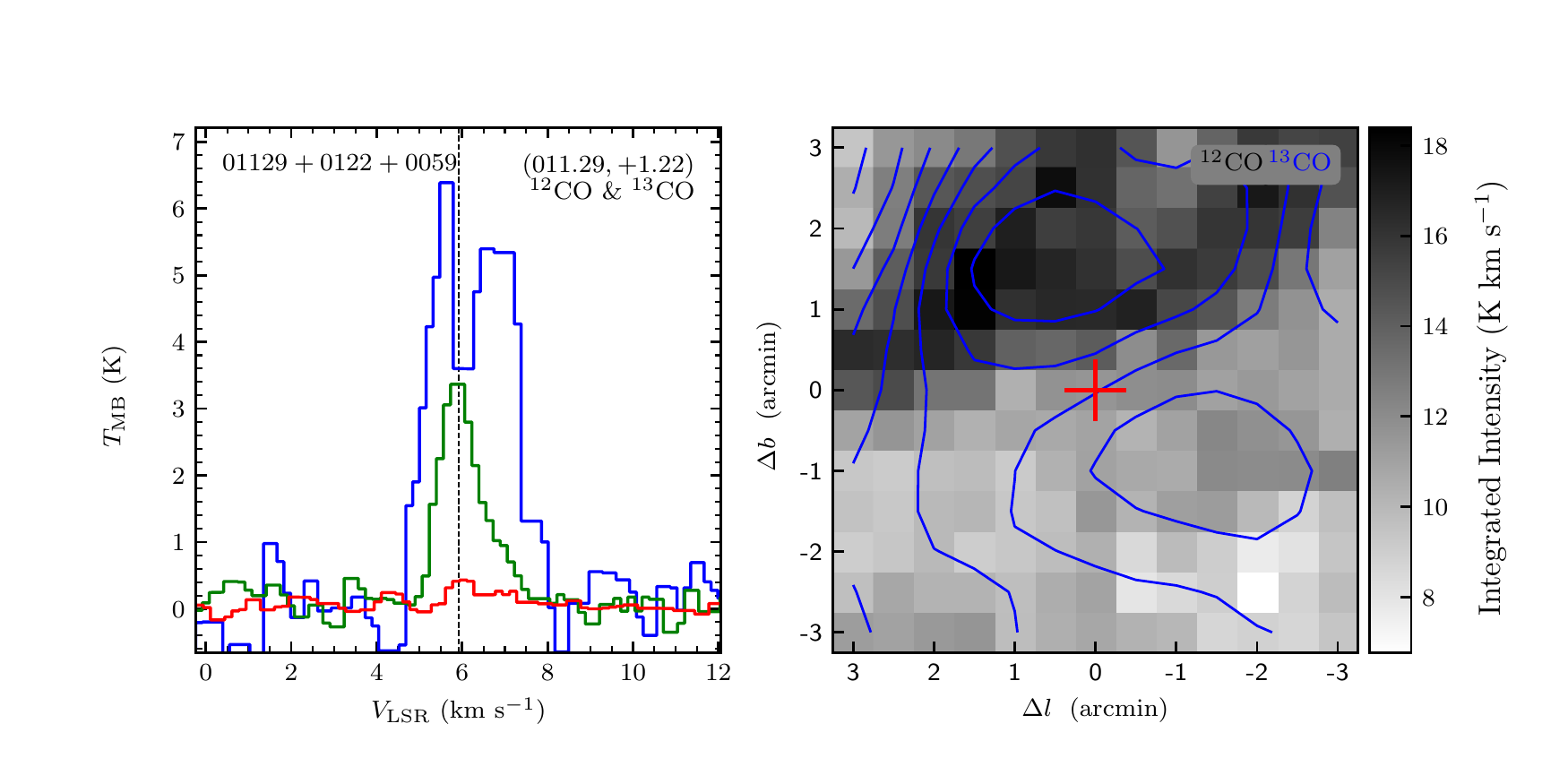}
\includegraphics[width=9.0cm,angle=0]{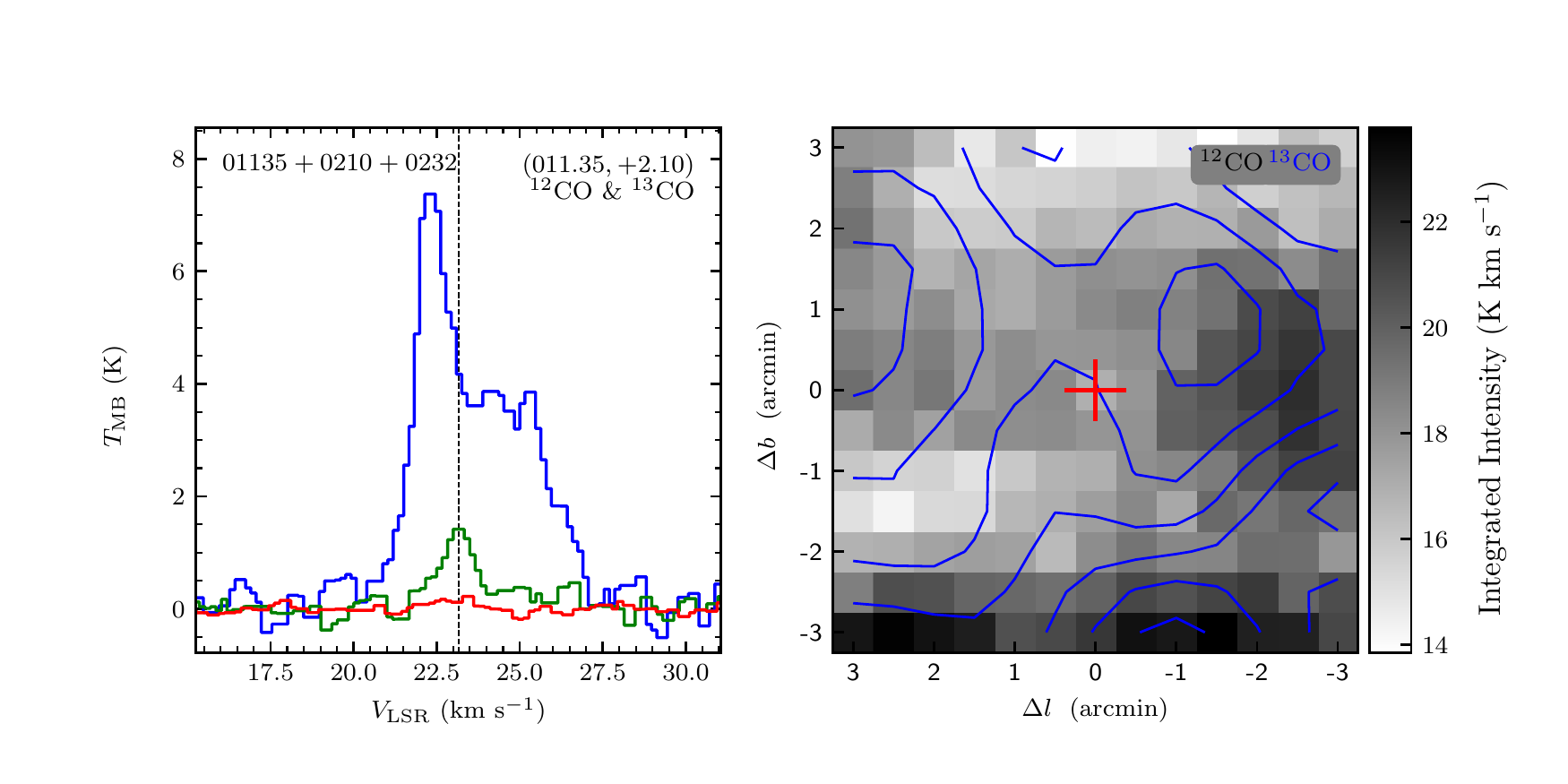}
\vspace{-0.5cm}

\includegraphics[width=9.0cm,angle=0]{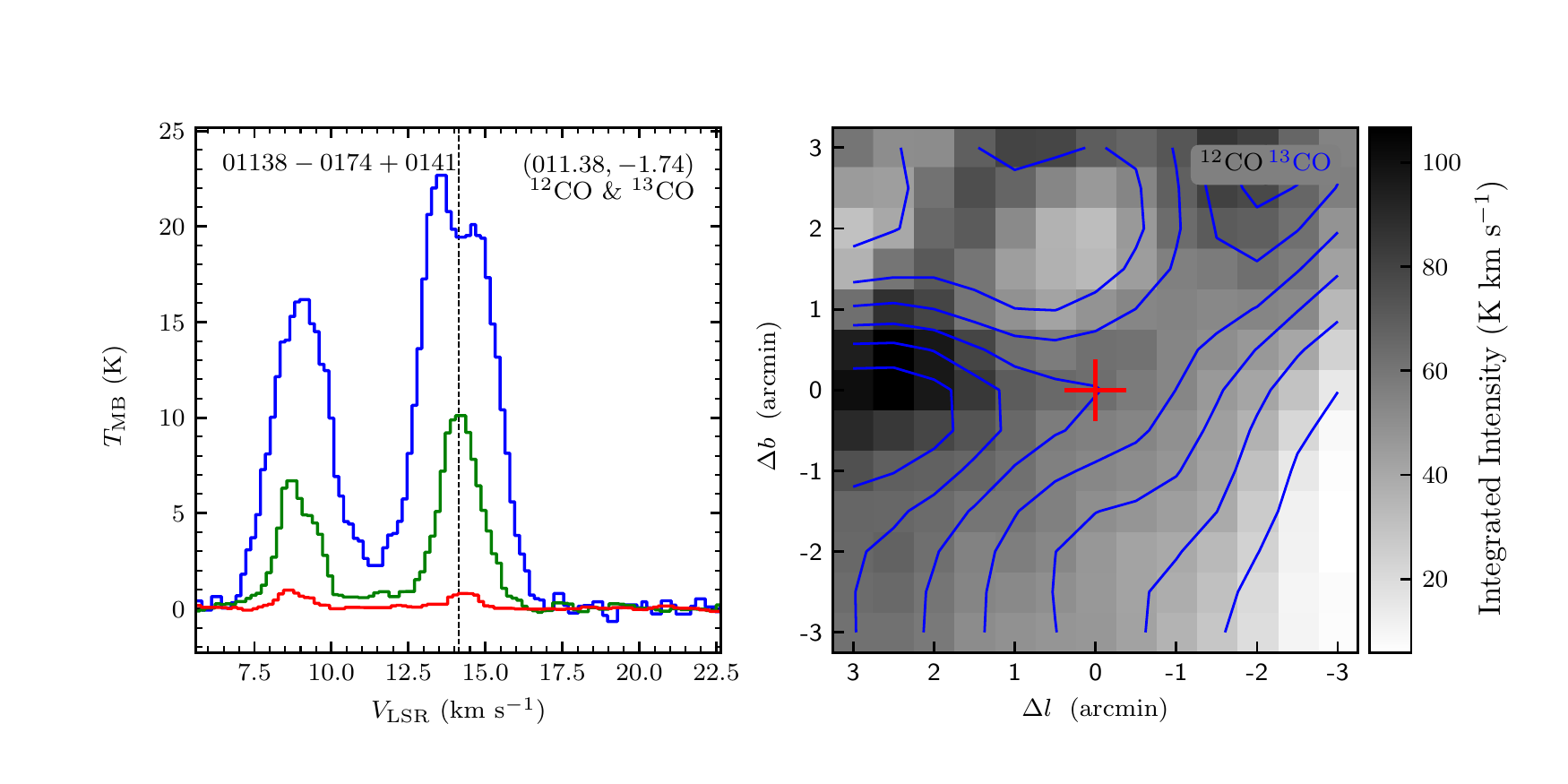}
\includegraphics[width=9.0cm,angle=0]{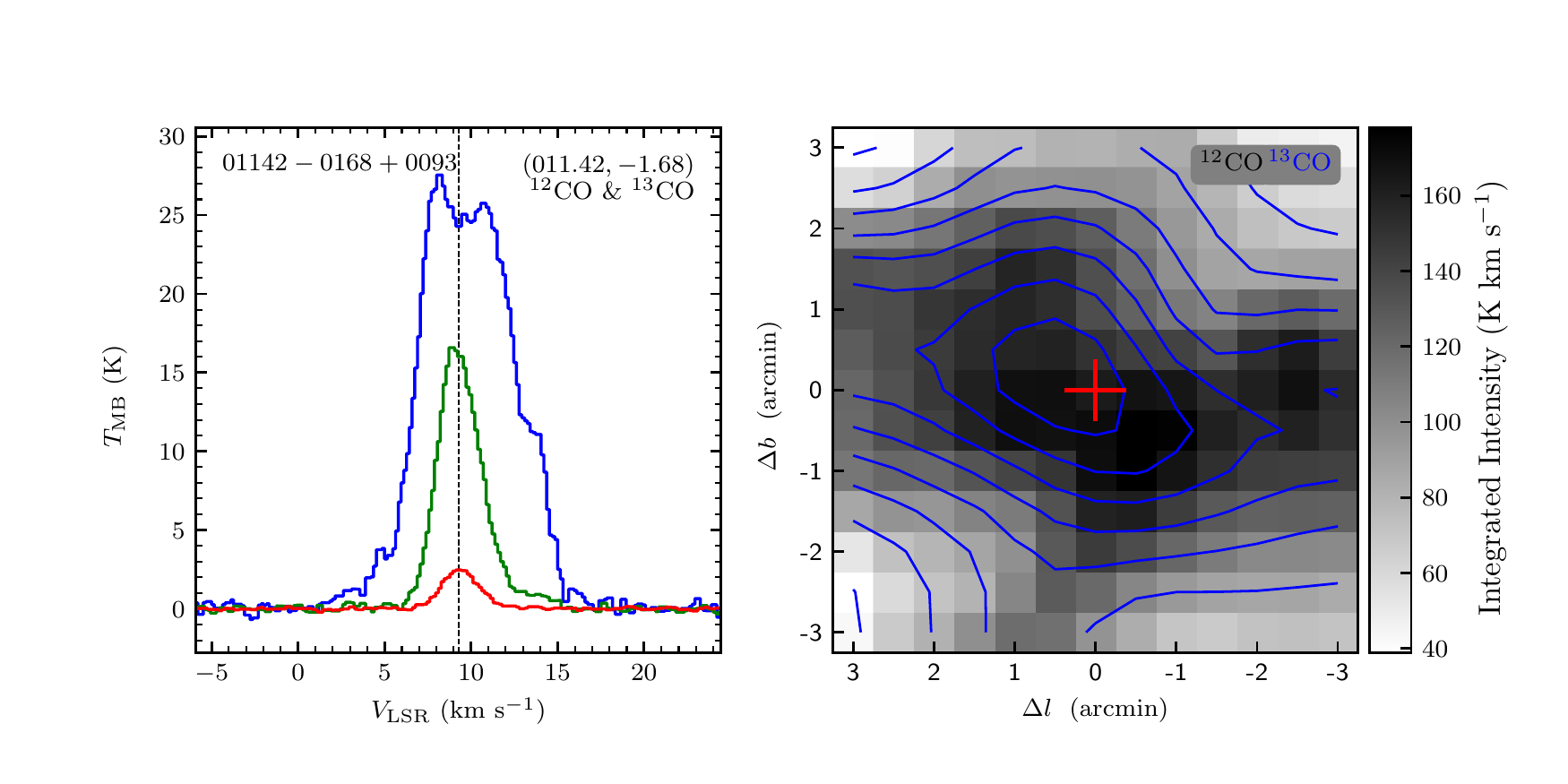}
\end{figure}
\clearpage

\begin{figure}
\includegraphics[width=9.0cm,angle=0]{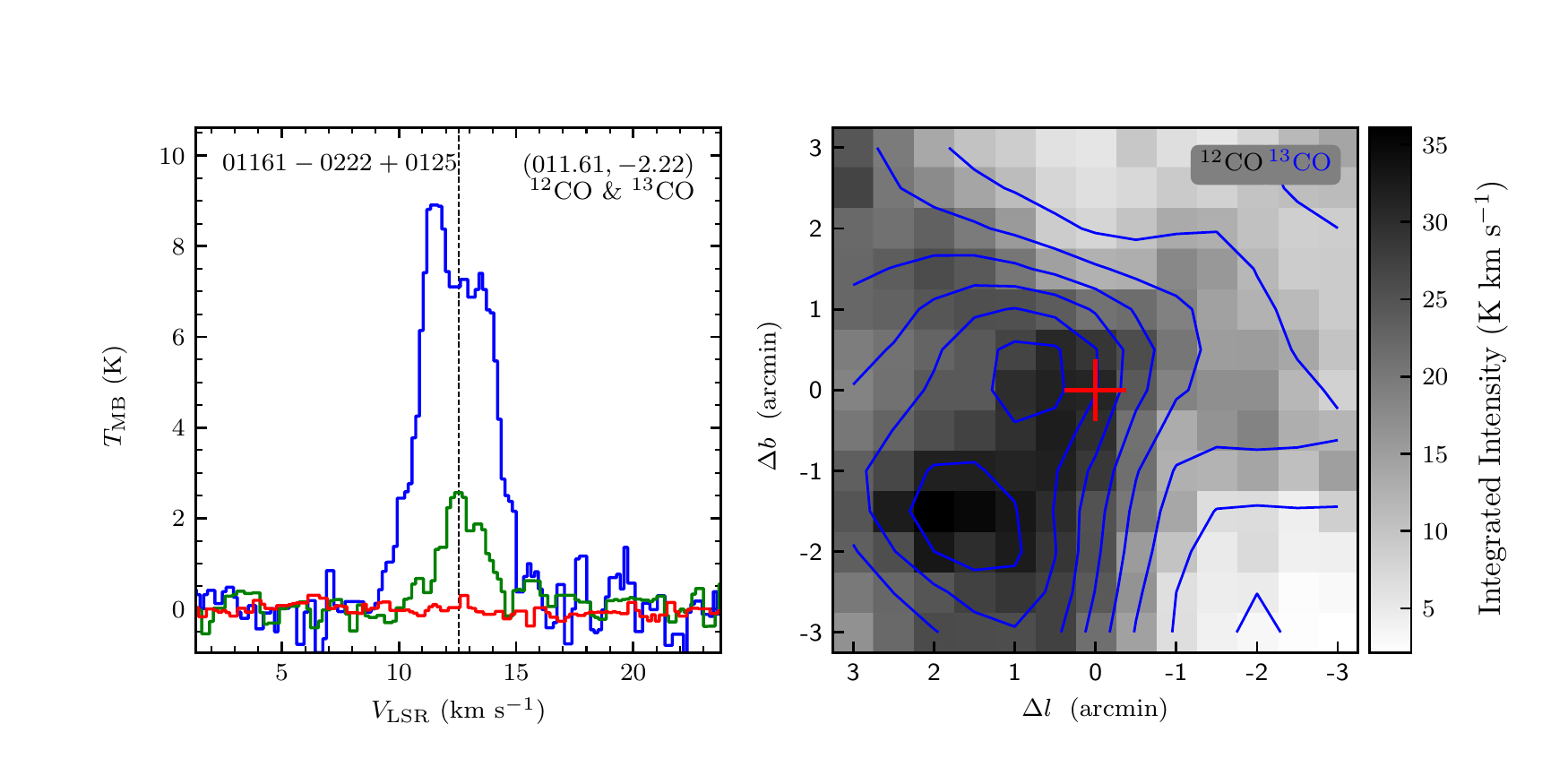}
\includegraphics[width=9.0cm,angle=0]{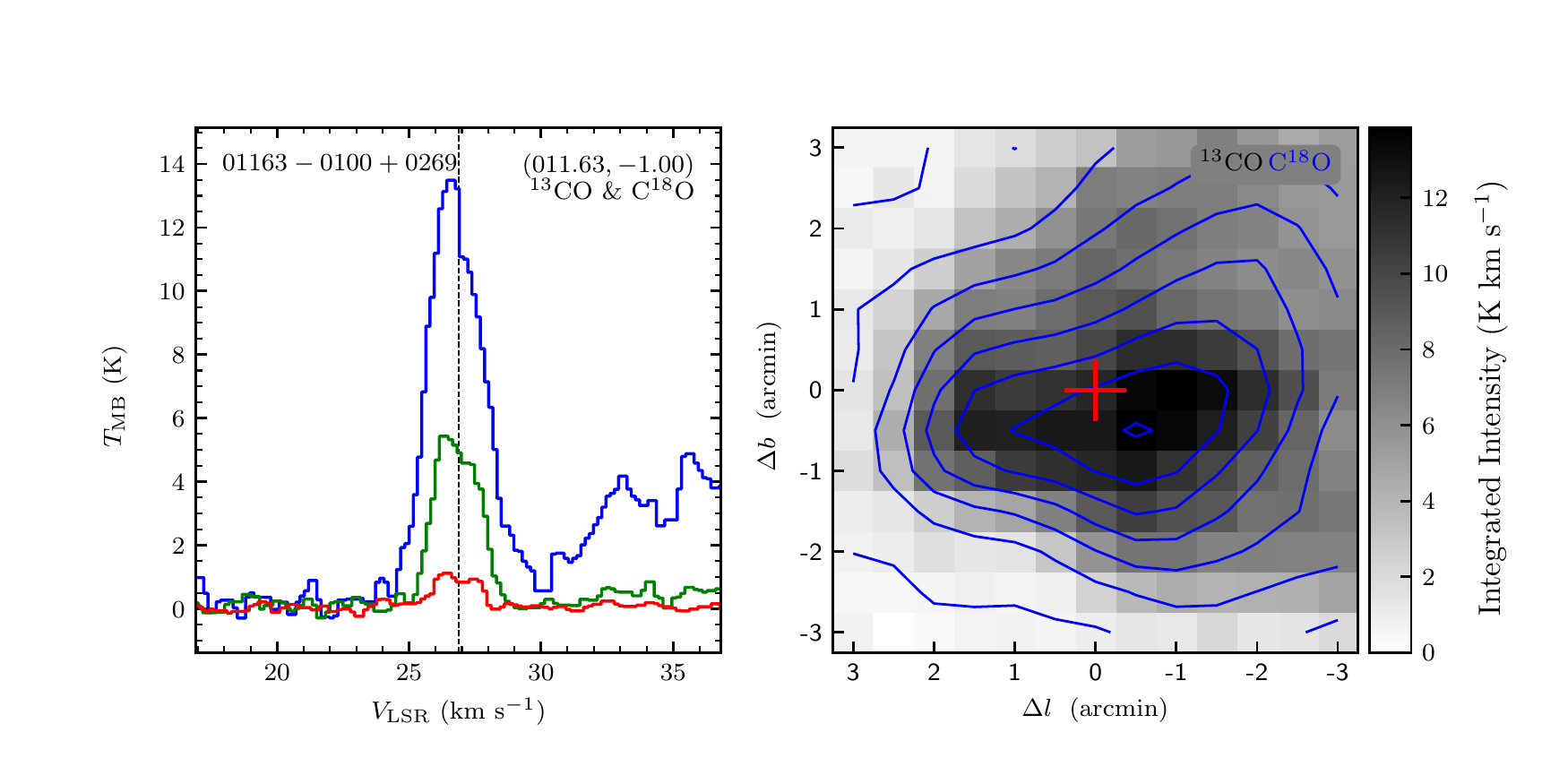}
\vspace{-0.5cm}

\includegraphics[width=9.0cm,angle=0]{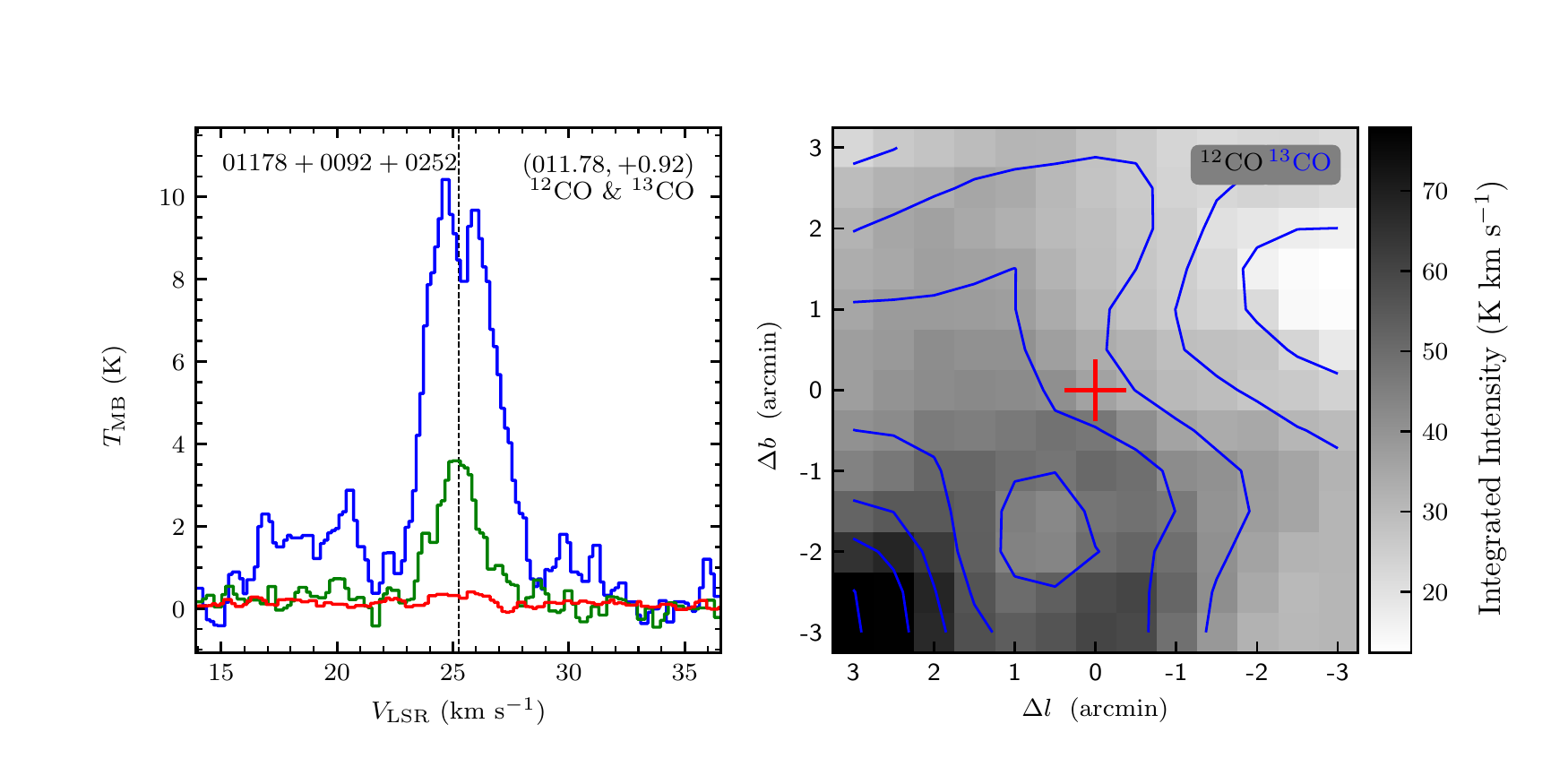}
\includegraphics[width=9.0cm,angle=0]{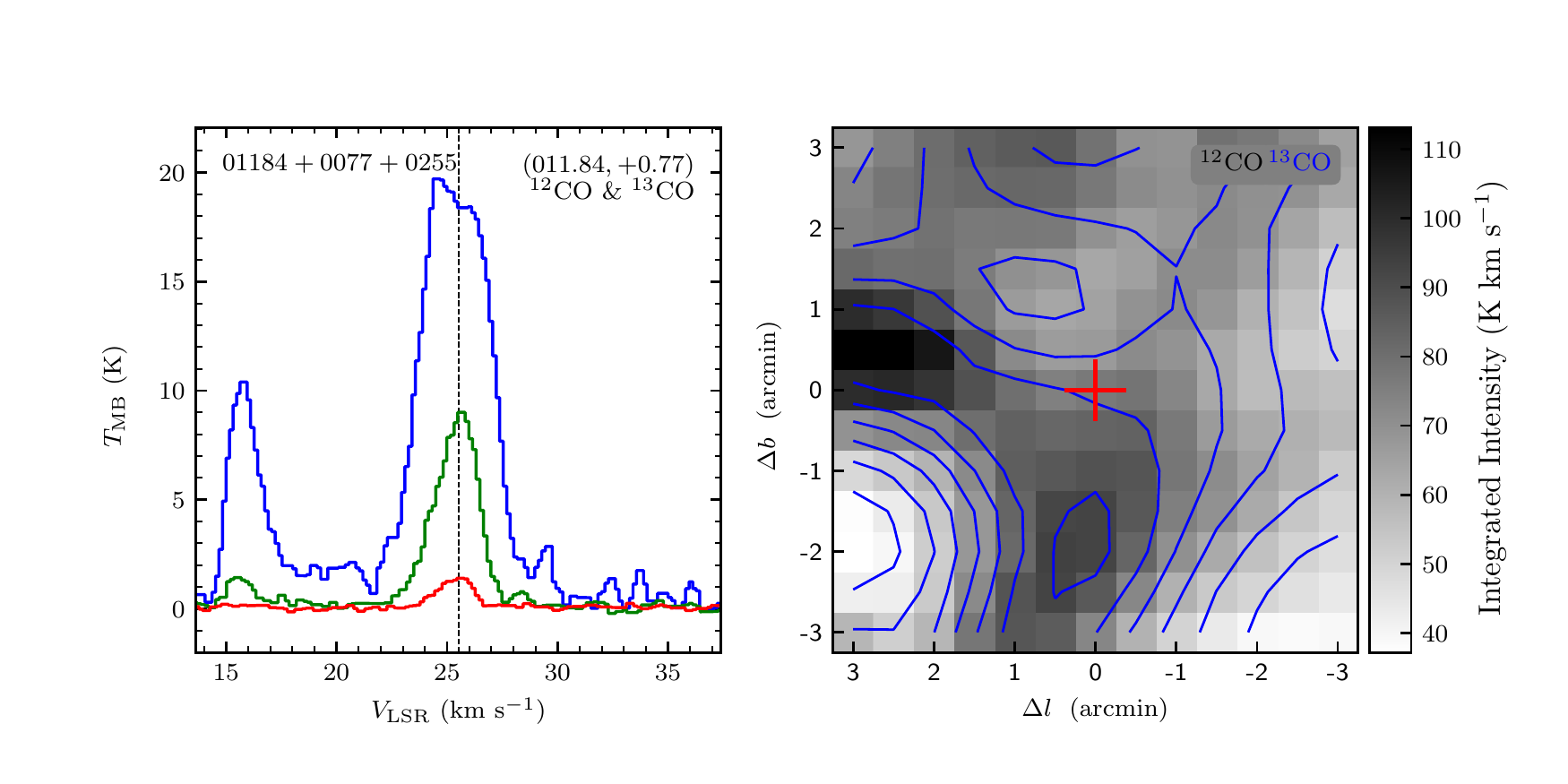}
\vspace{-0.5cm}

\includegraphics[width=9.0cm,angle=0]{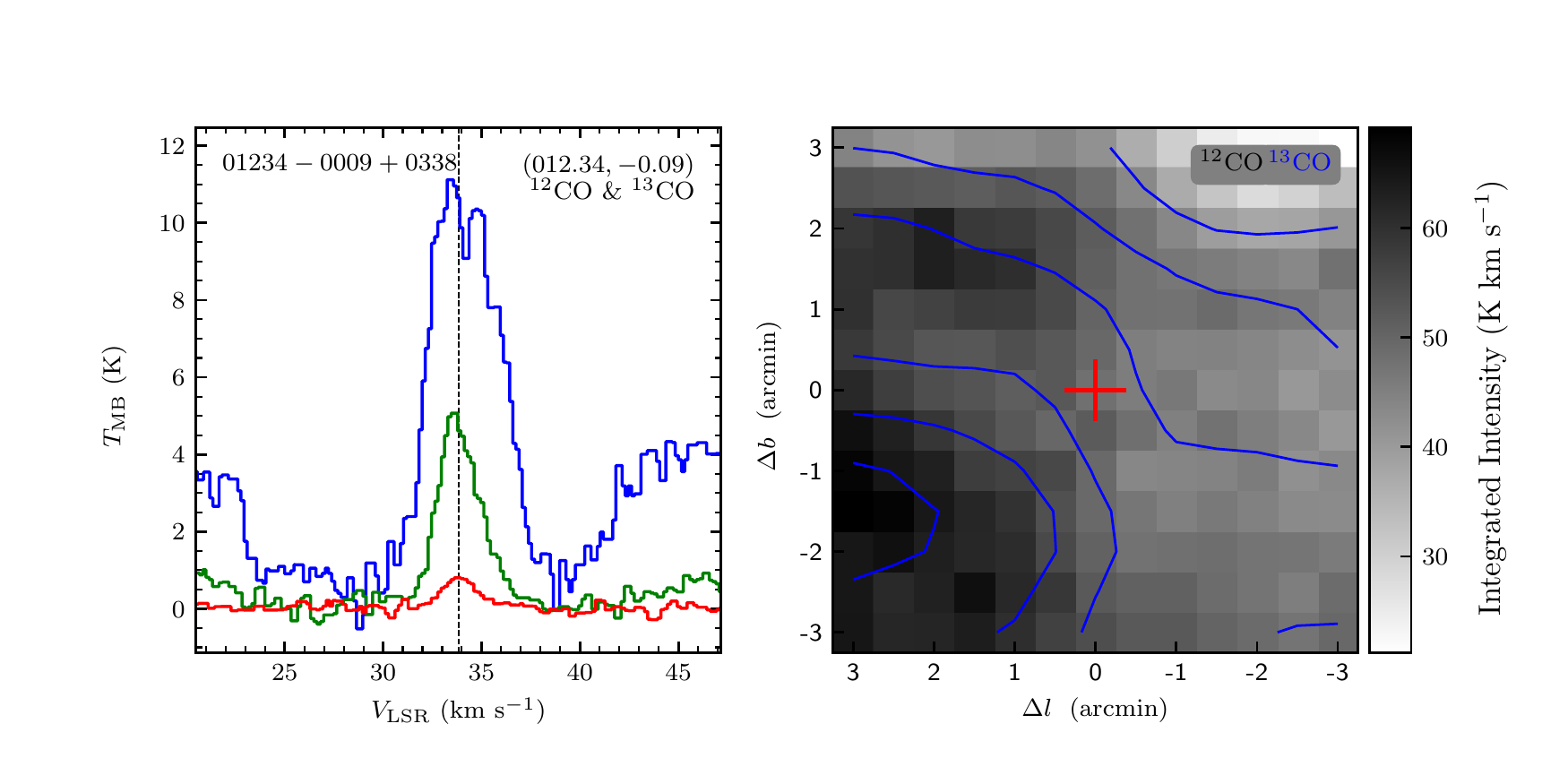}
\includegraphics[width=9.0cm,angle=0]{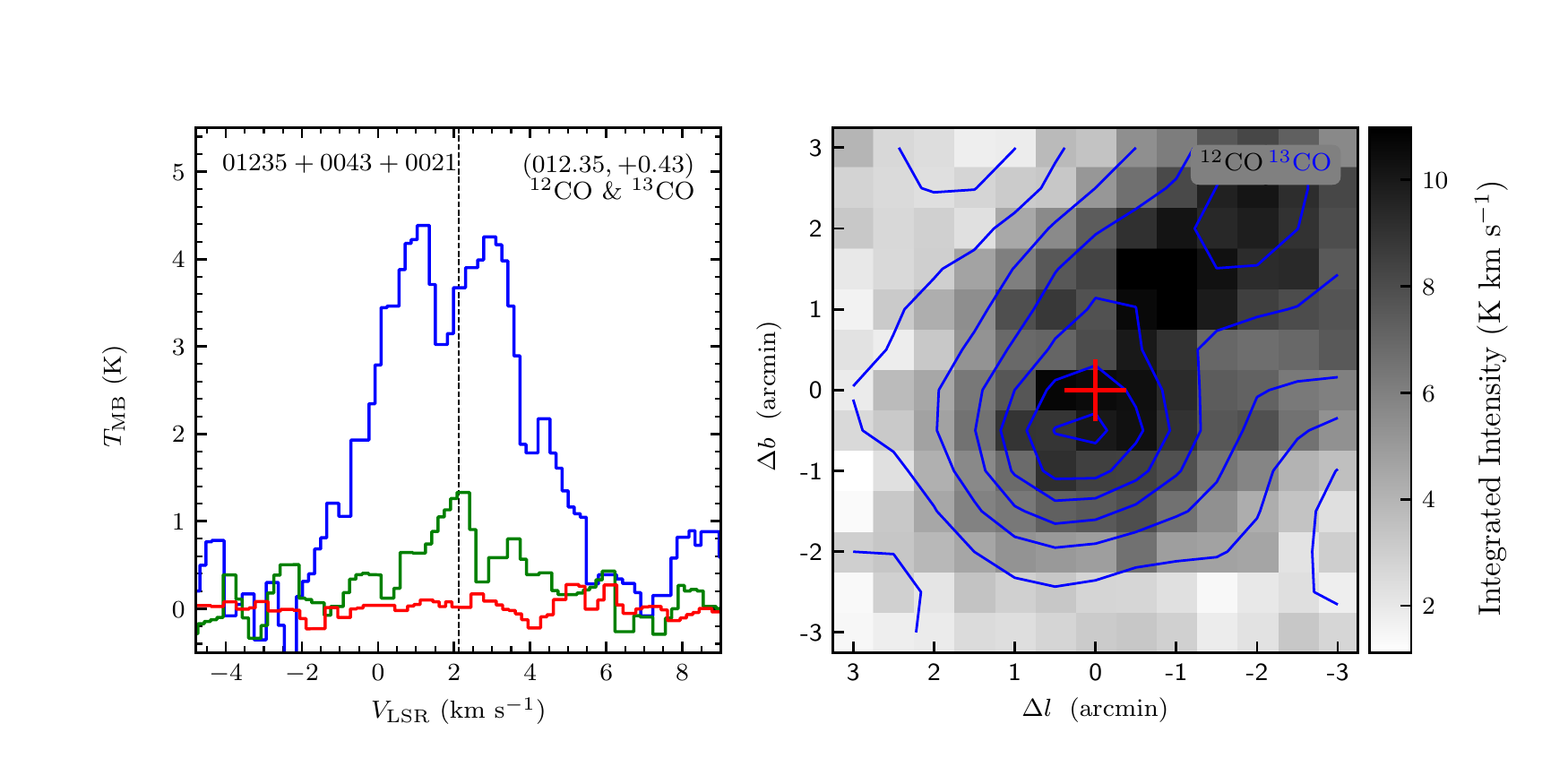}
\vspace{-0.5cm}

\includegraphics[width=9.0cm,angle=0]{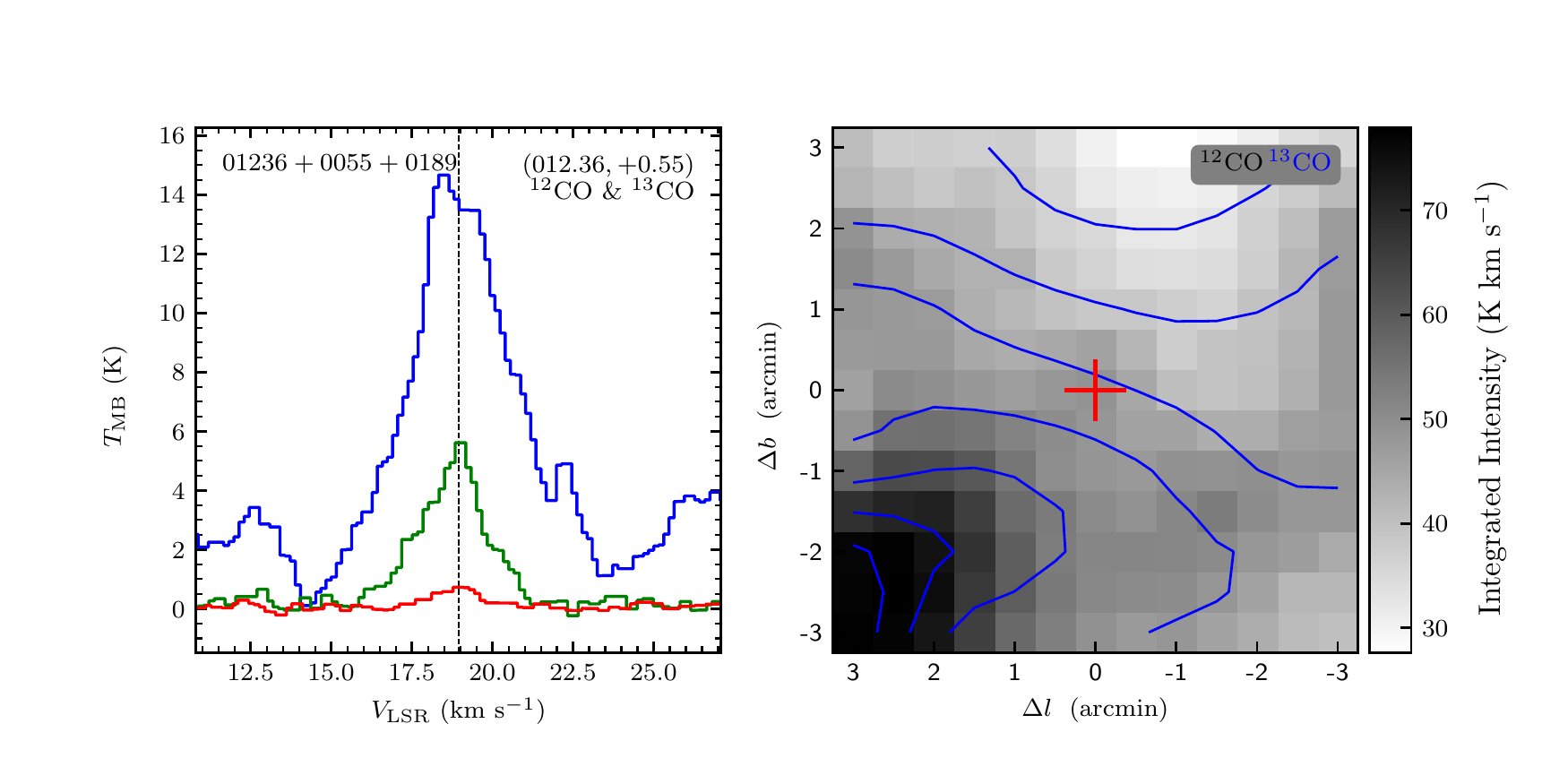}
\includegraphics[width=9.0cm,angle=0]{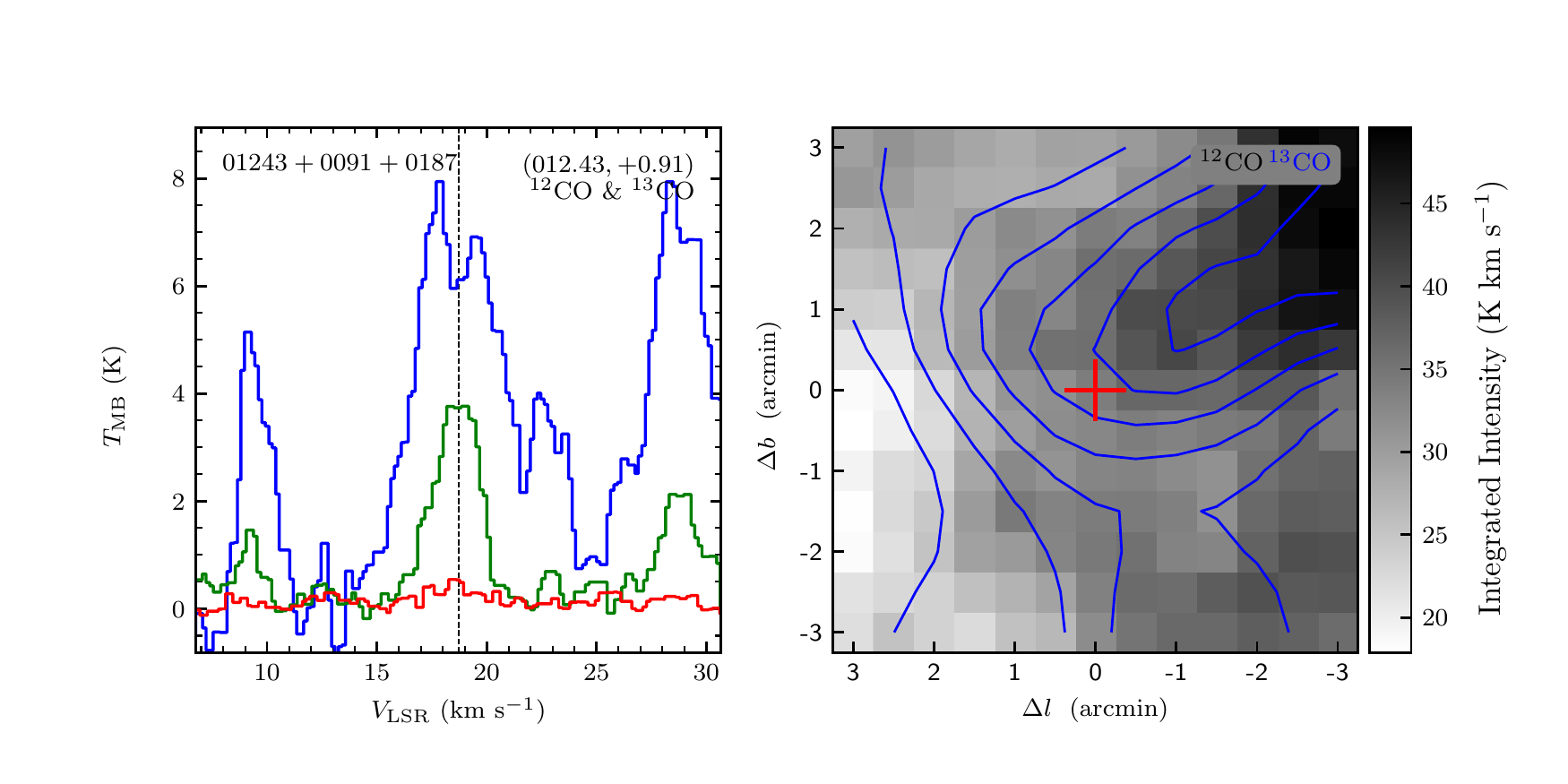}
\vspace{-0.5cm}

\includegraphics[width=9.0cm,angle=0]{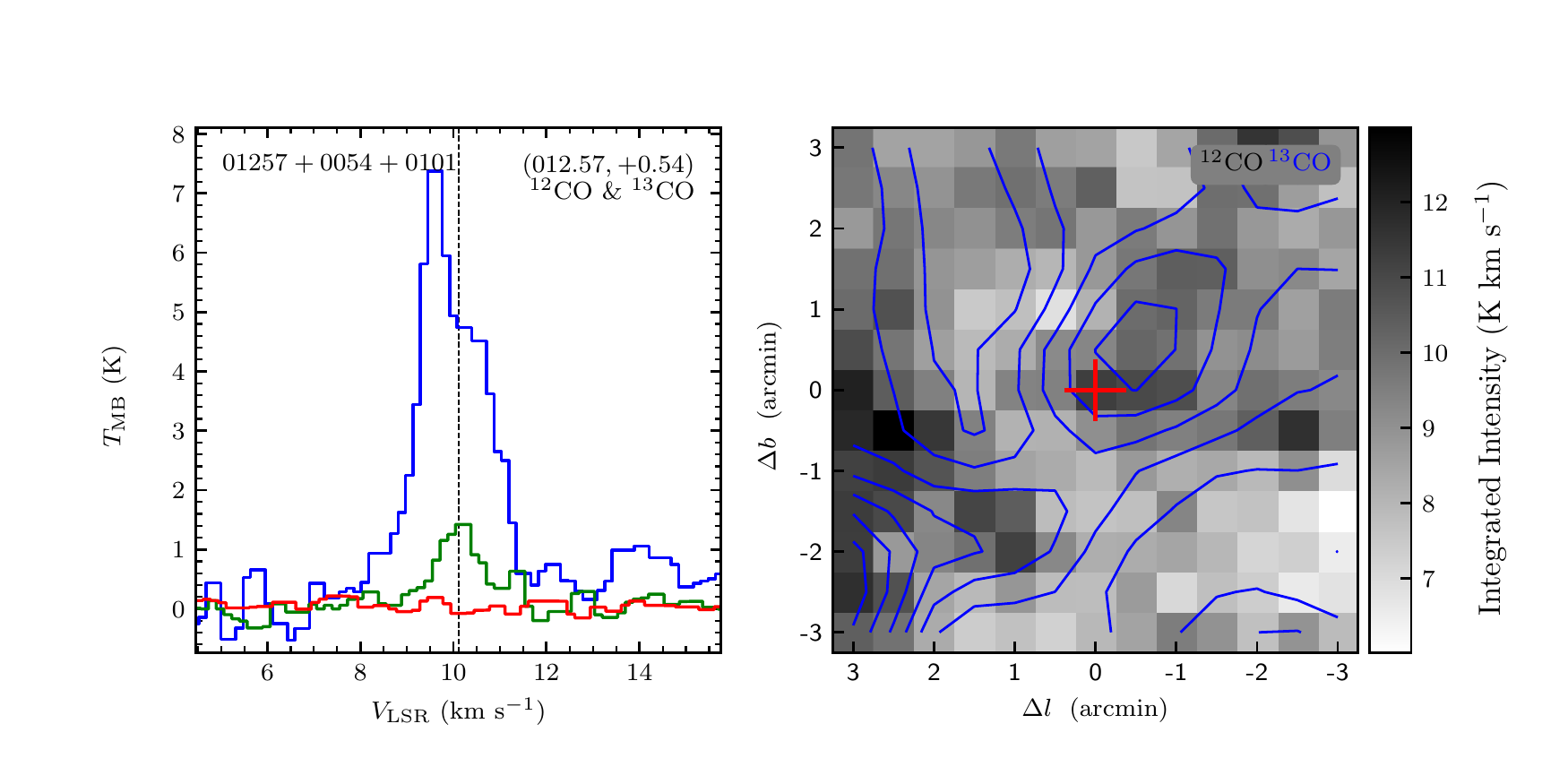}
\includegraphics[width=9.0cm,angle=0]{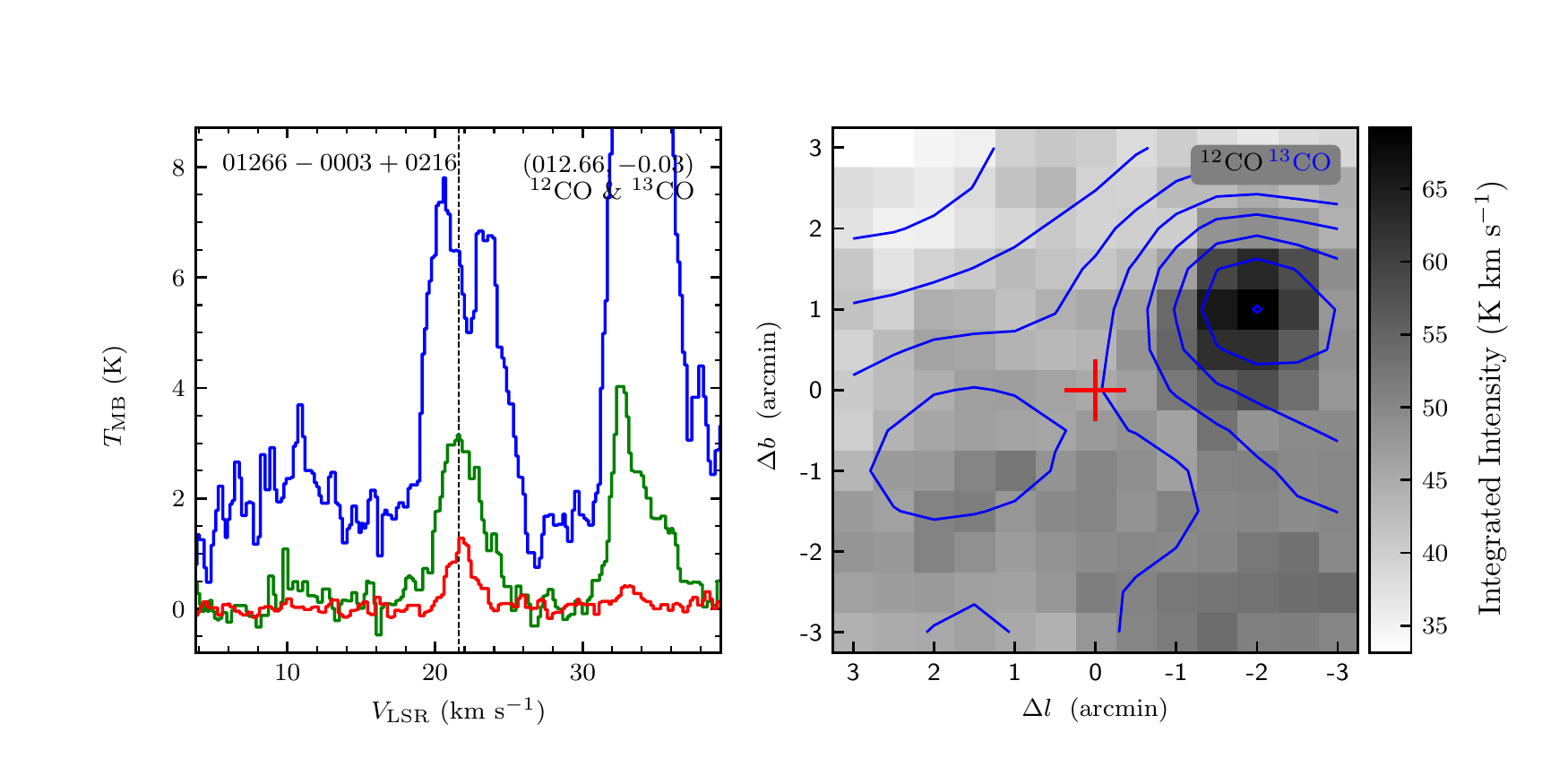}
\end{figure}
\clearpage

\begin{figure}
\includegraphics[width=9.0cm,angle=0]{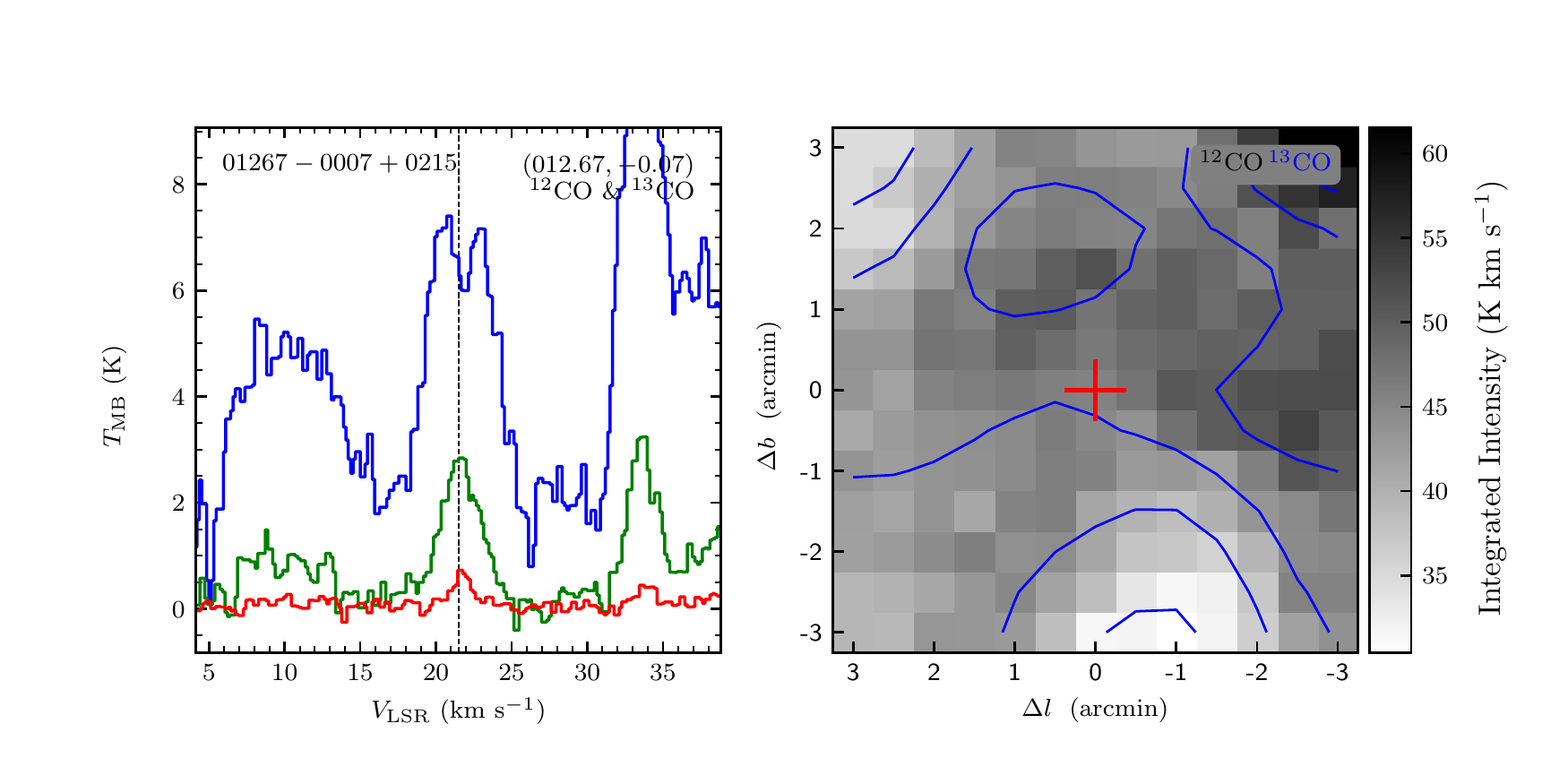}
\includegraphics[width=9.0cm,angle=0]{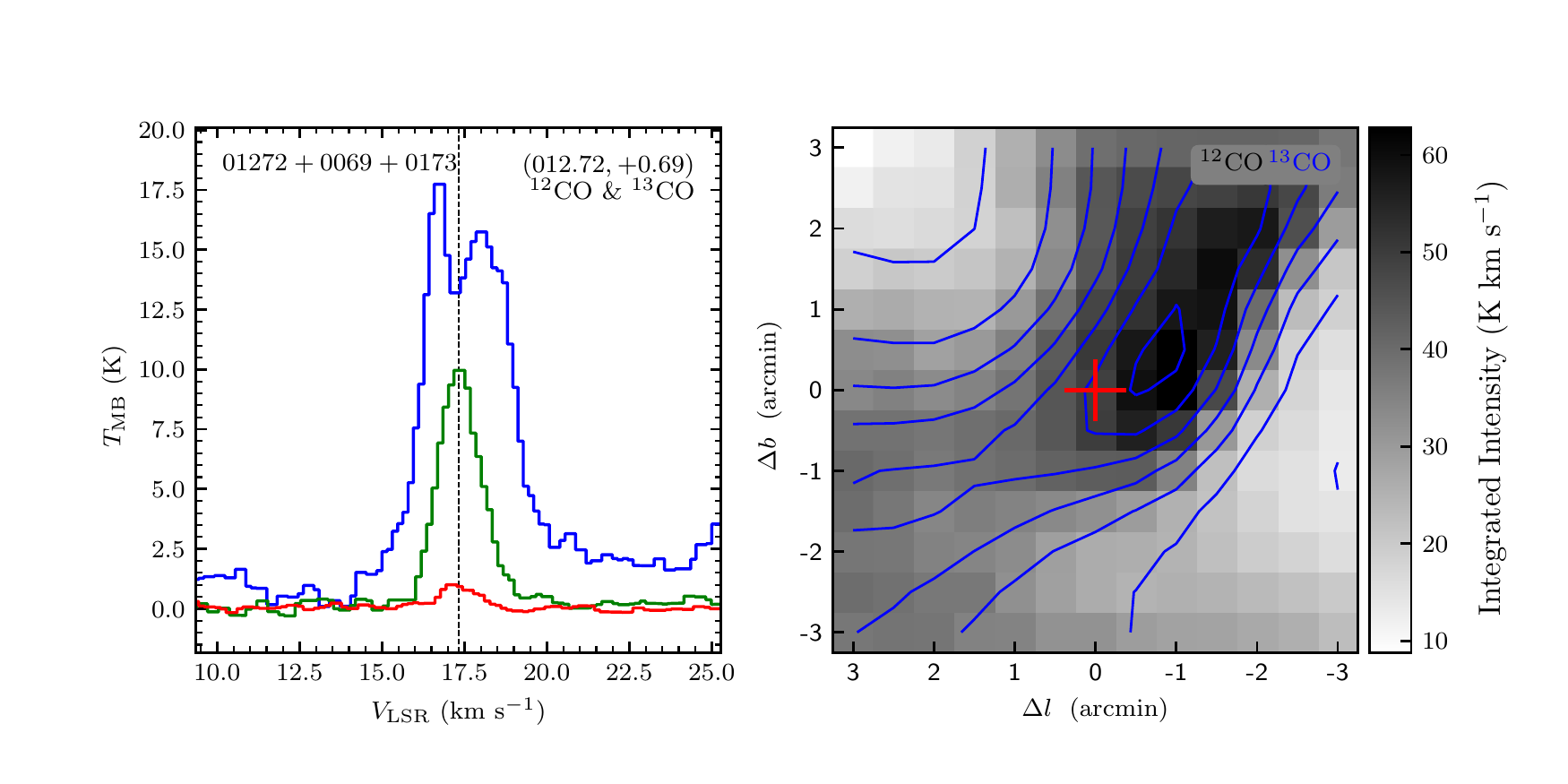}
\vspace{-0.5cm}

\includegraphics[width=9.0cm,angle=0]{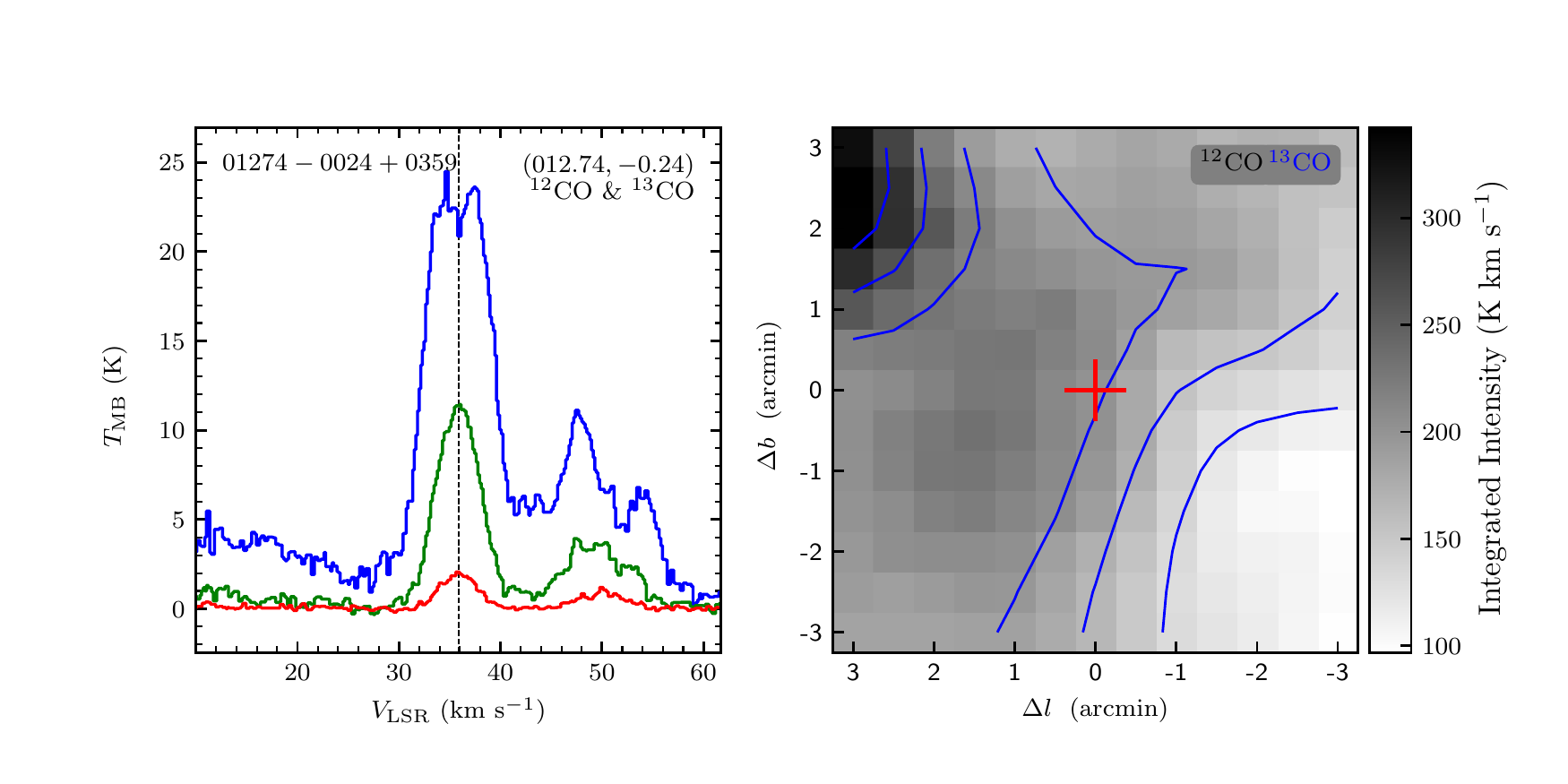}
\includegraphics[width=9.0cm,angle=0]{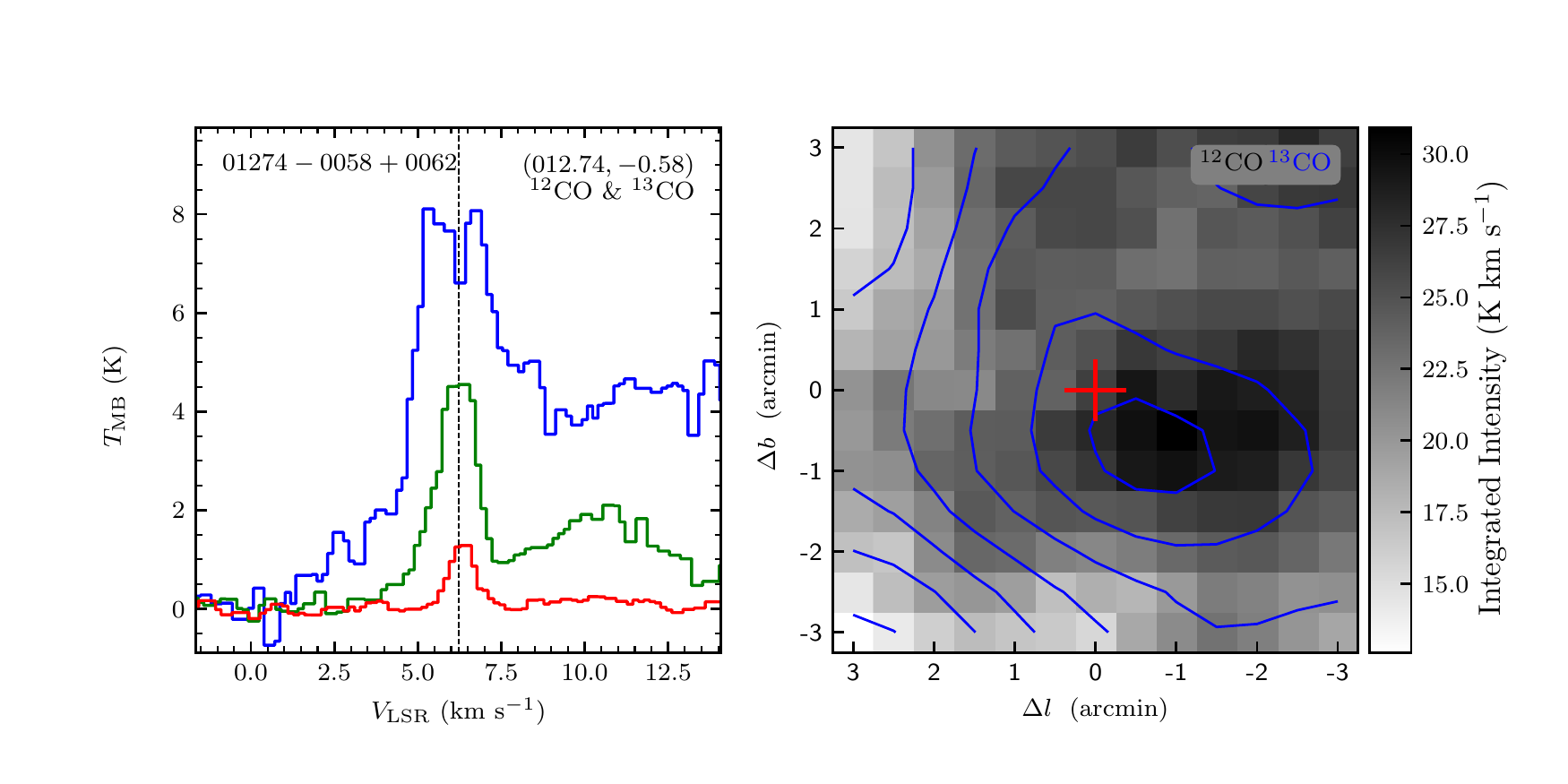}
\vspace{-0.5cm}

\includegraphics[width=9.0cm,angle=0]{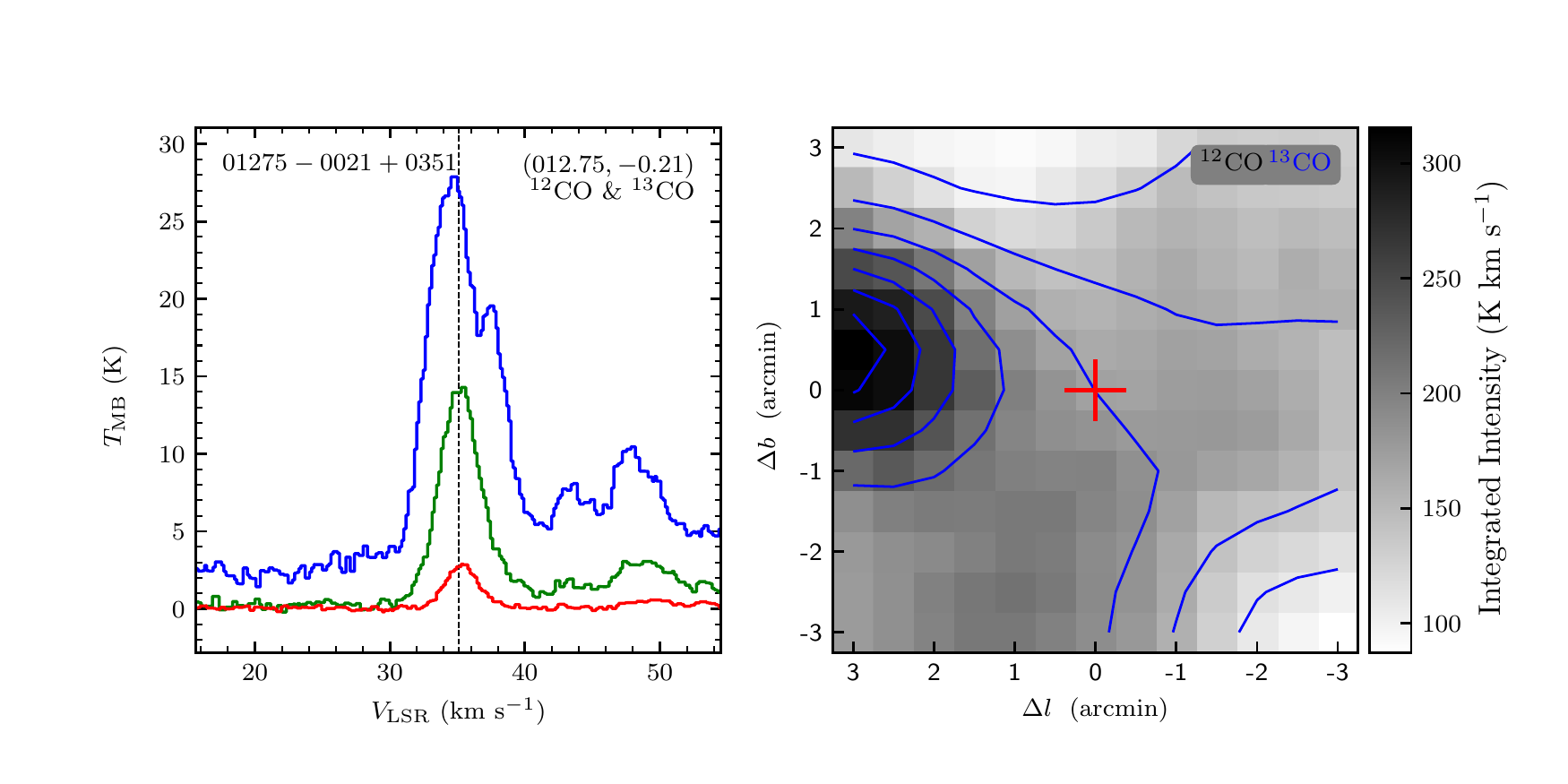}
\includegraphics[width=9.0cm,angle=0]{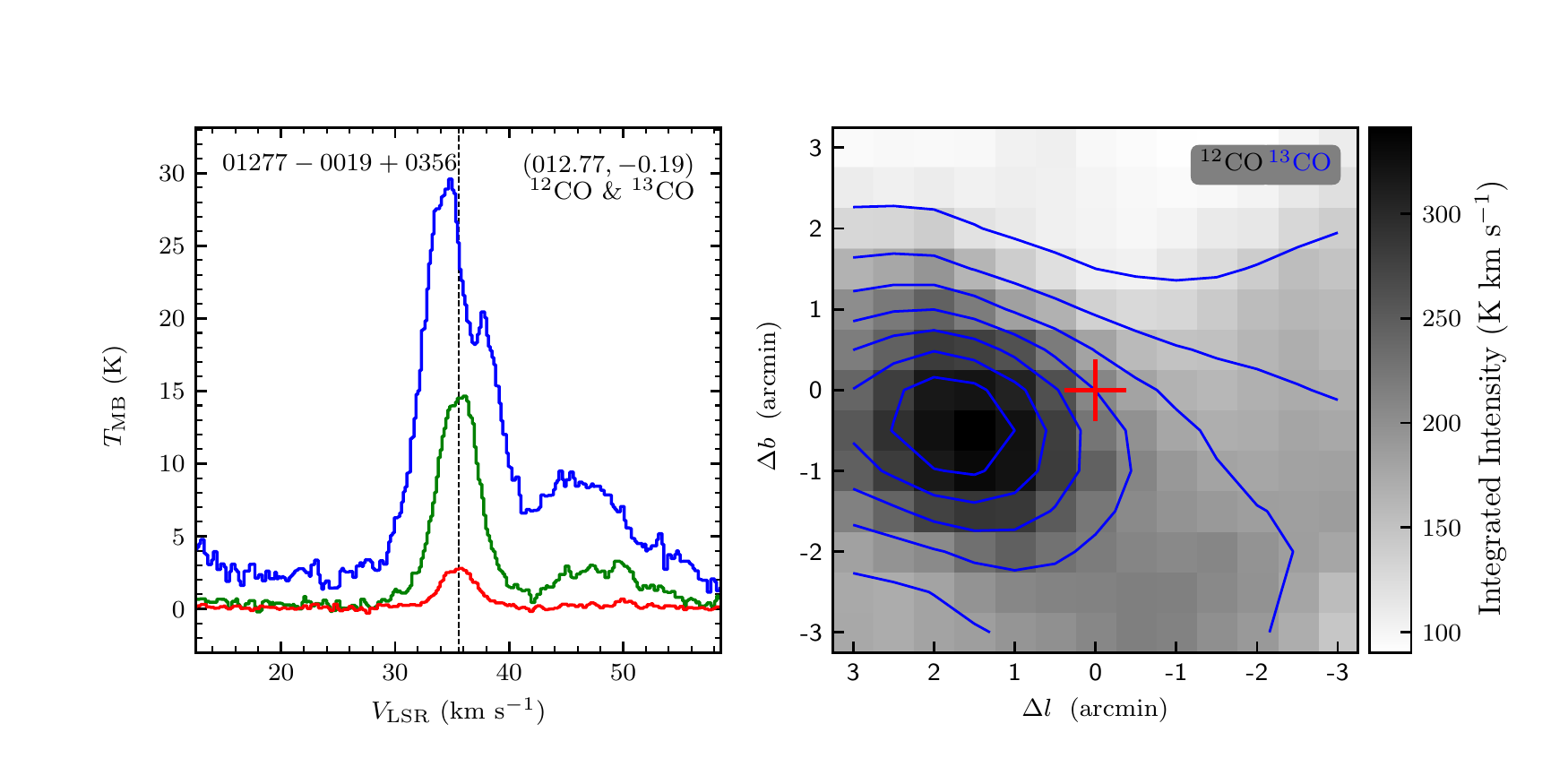}
\vspace{-0.5cm}

\includegraphics[width=9.0cm,angle=0]{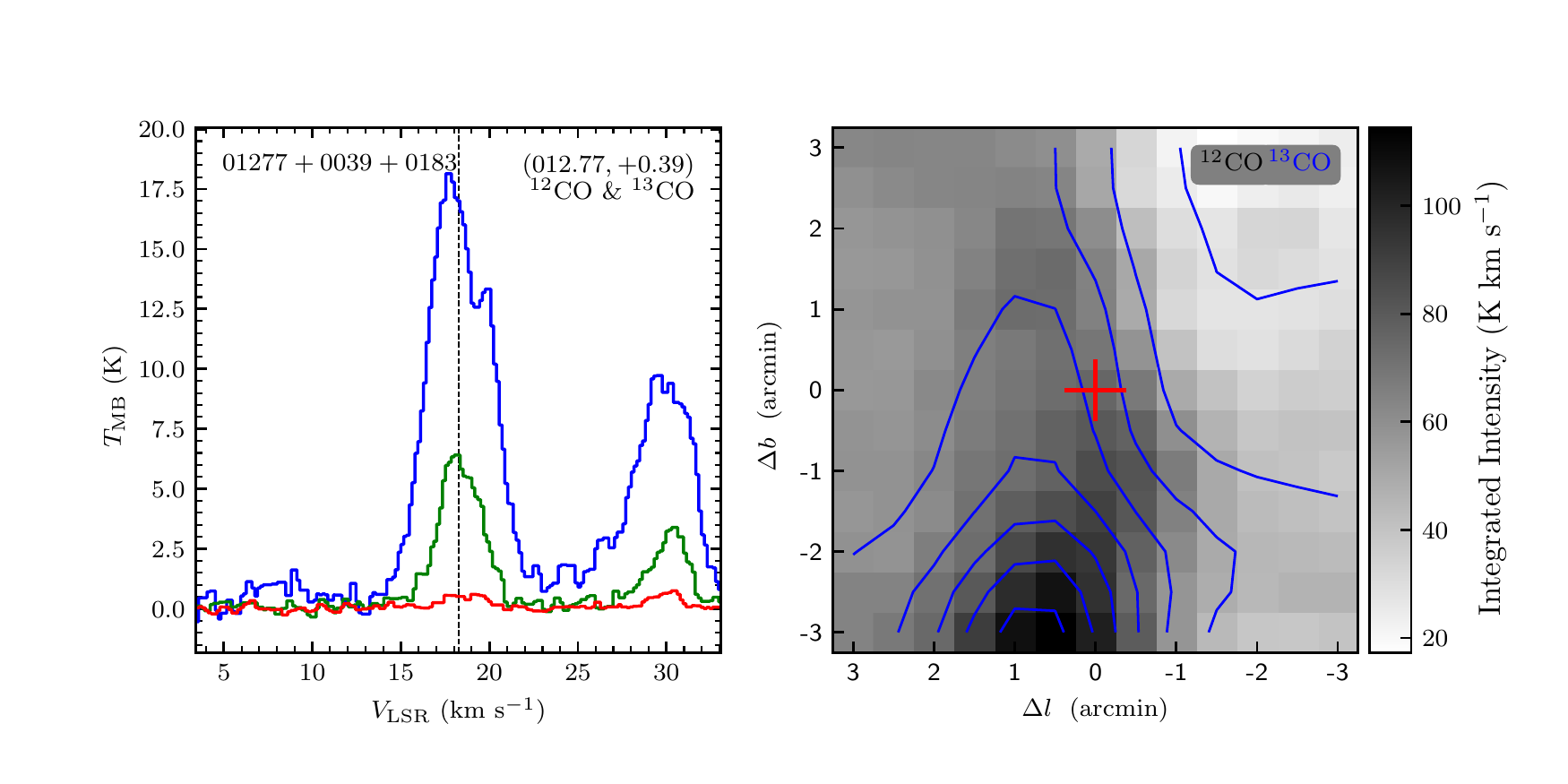}
\includegraphics[width=9.0cm,angle=0]{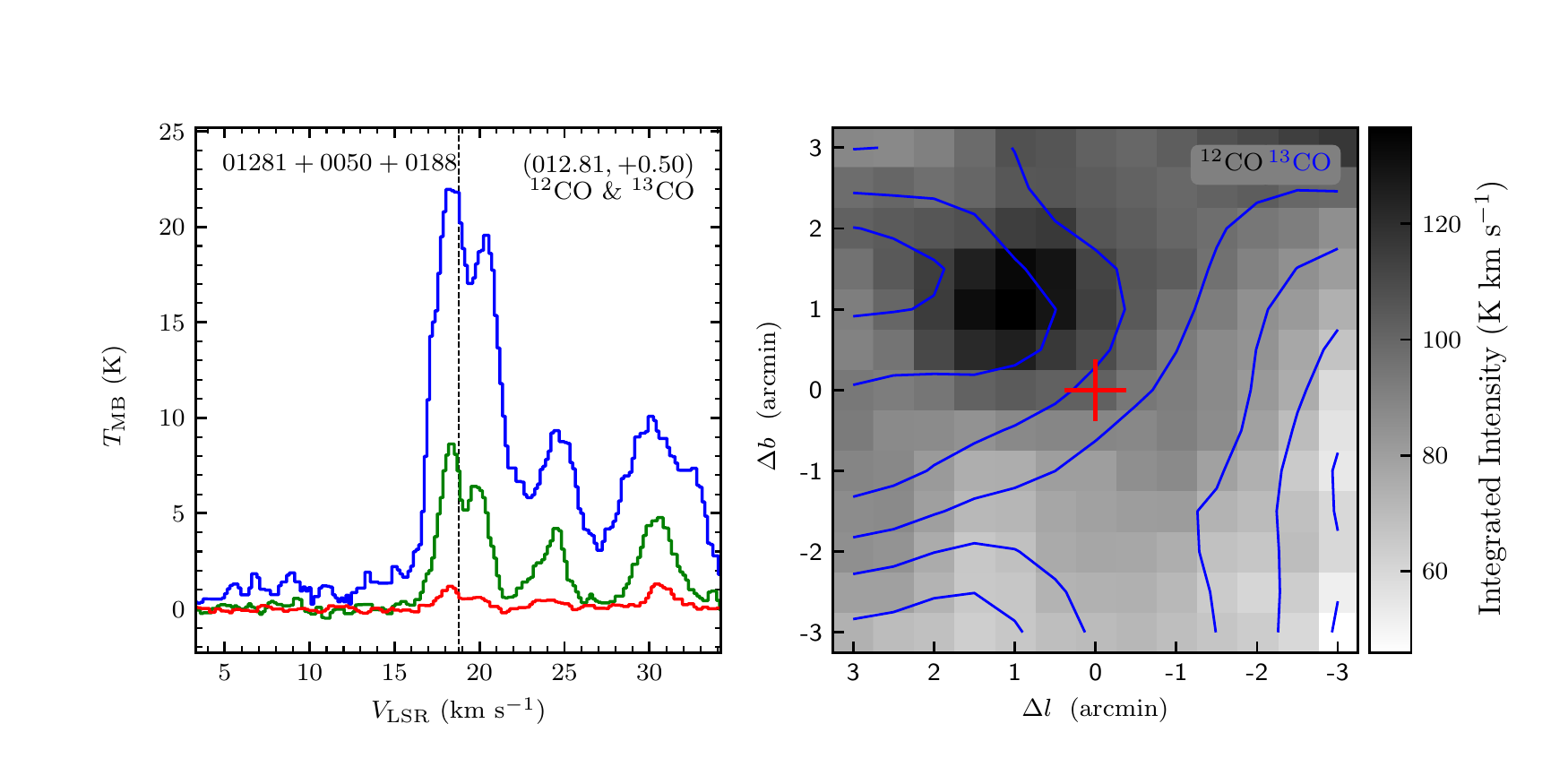}
\vspace{-0.5cm}

\includegraphics[width=9.0cm,angle=0]{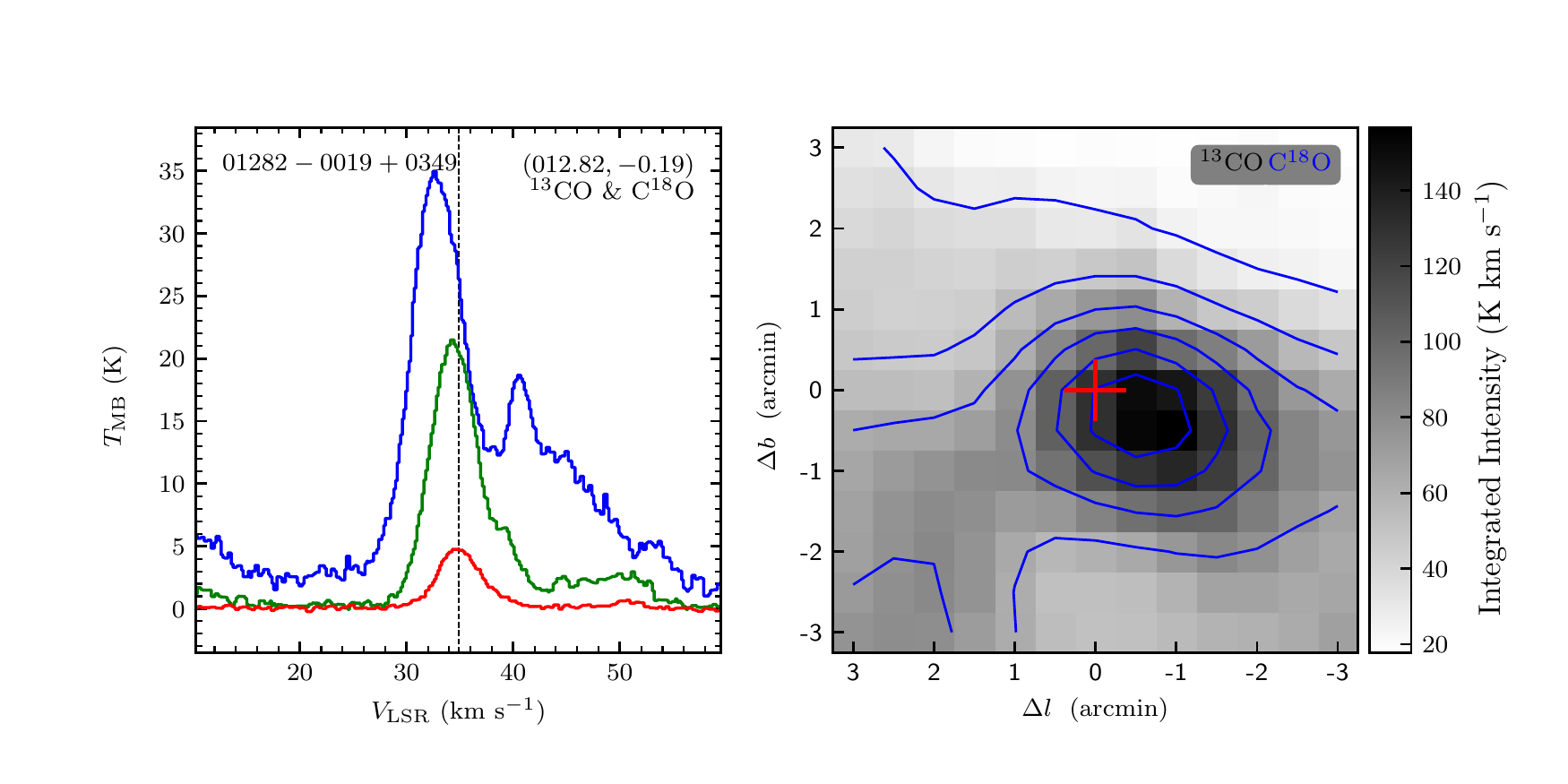}
\includegraphics[width=9.0cm,angle=0]{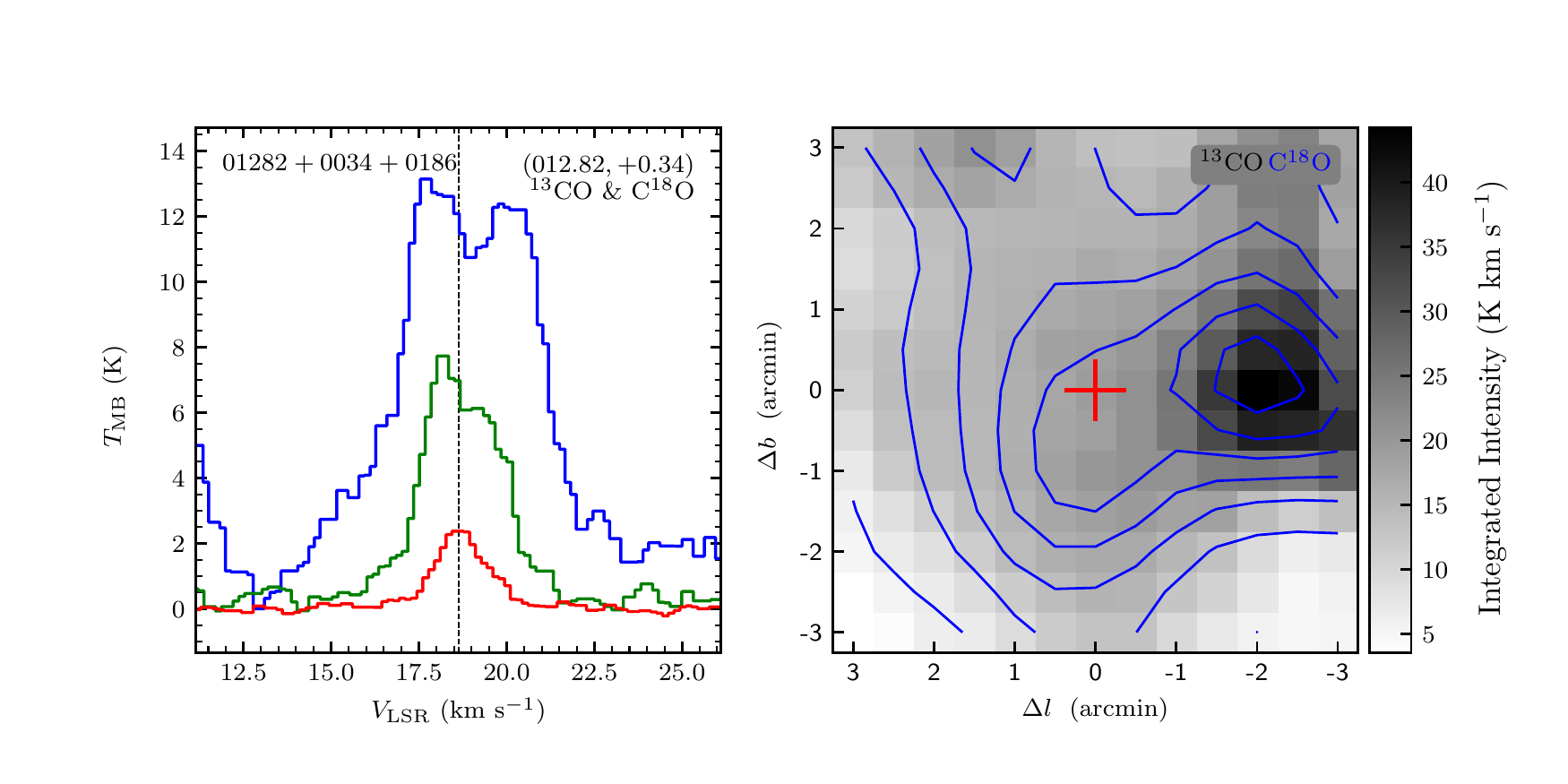}
\end{figure}
\clearpage

\begin{figure}
\includegraphics[width=9.0cm,angle=0]{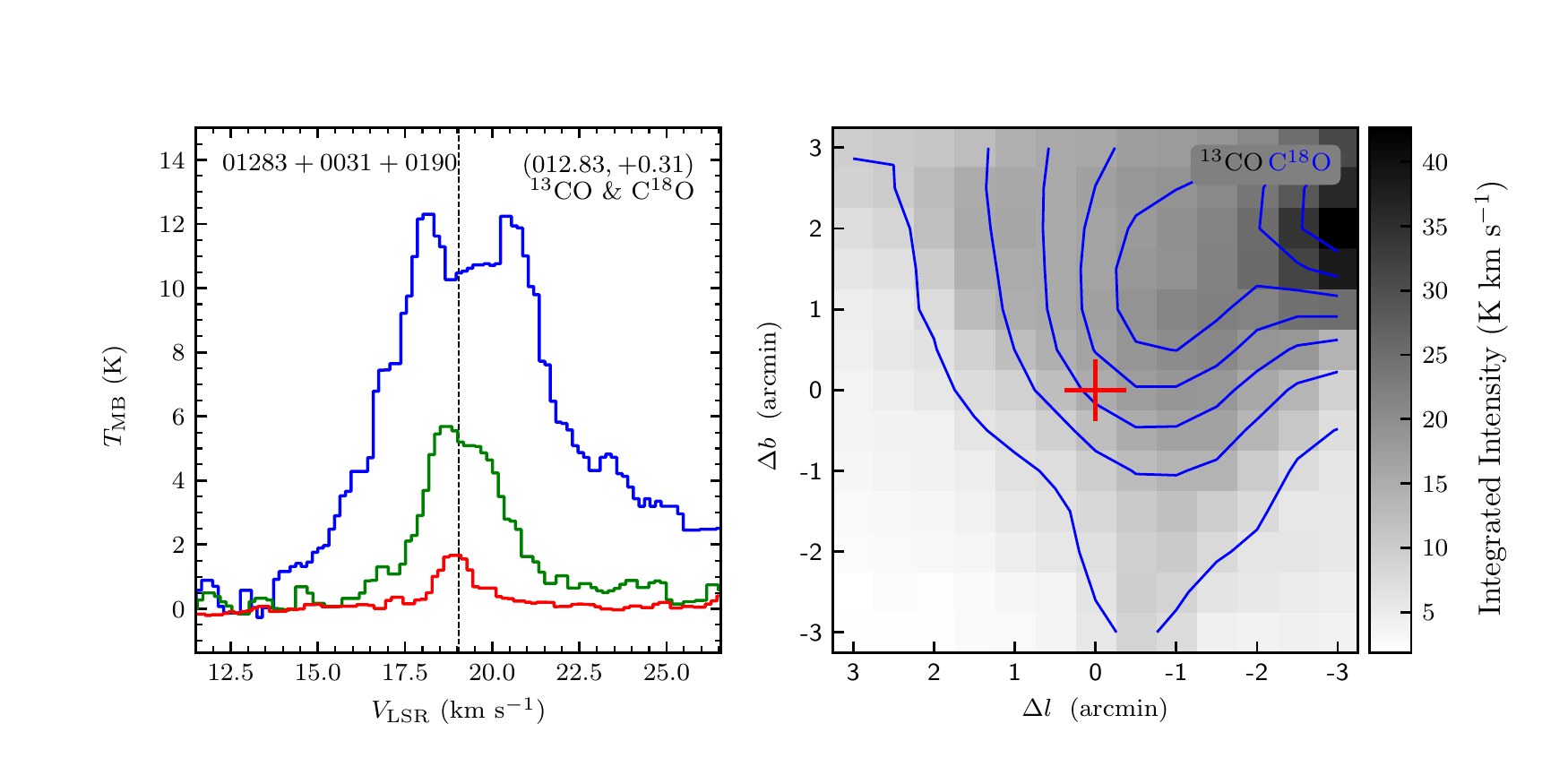}
\includegraphics[width=9.0cm,angle=0]{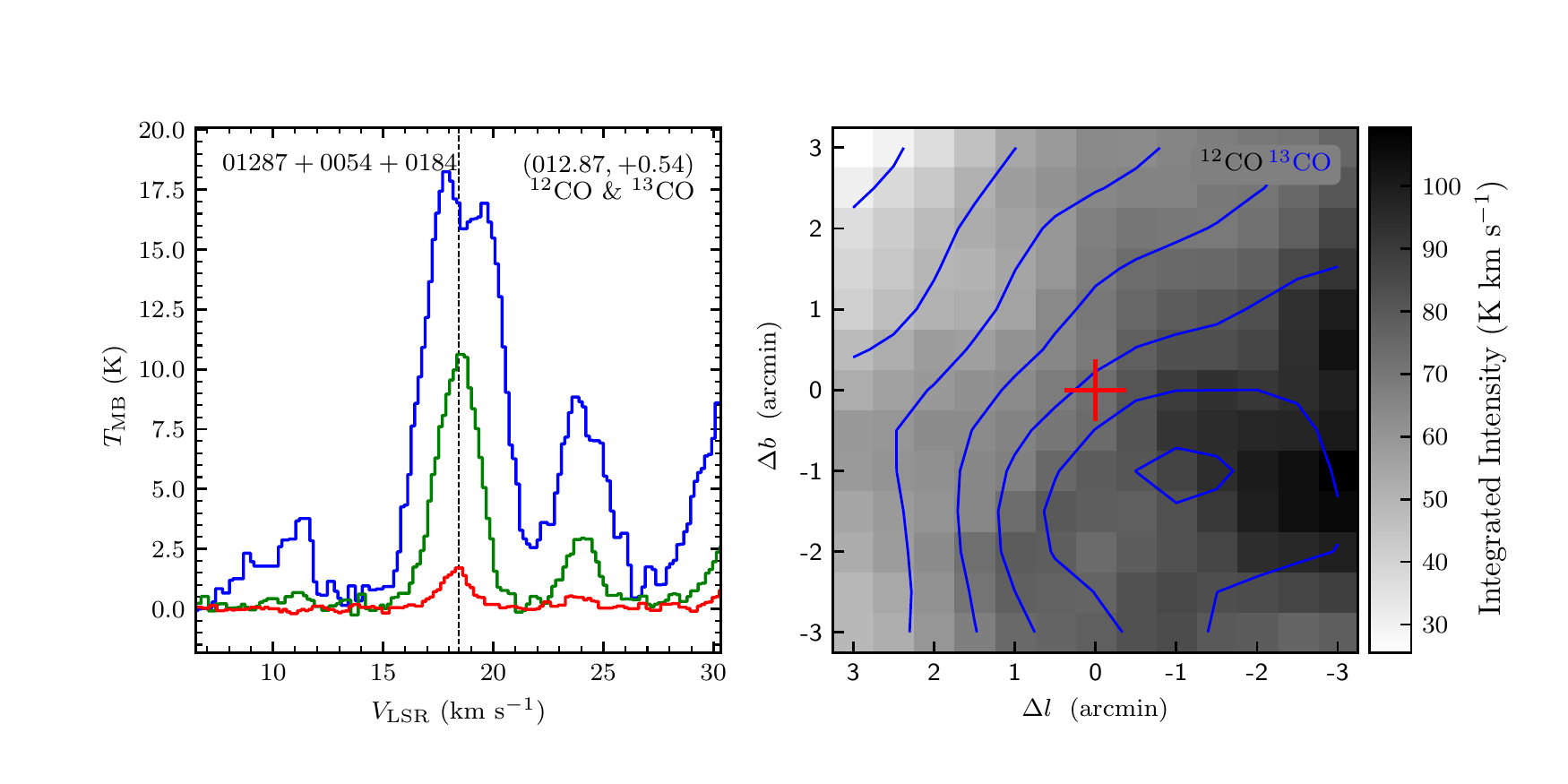}
\vspace{-0.5cm}

\includegraphics[width=9.0cm,angle=0]{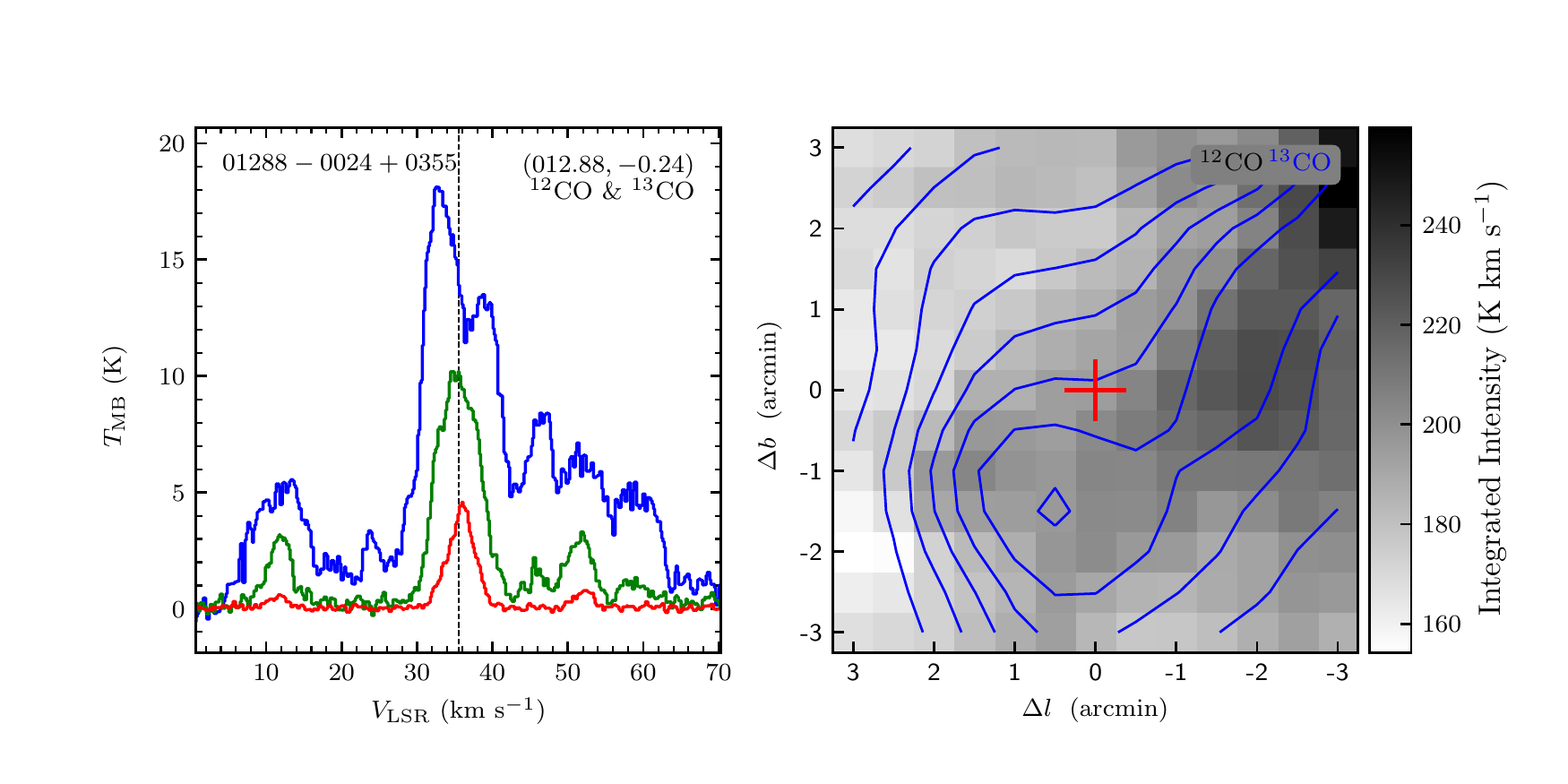}
\includegraphics[width=9.0cm,angle=0]{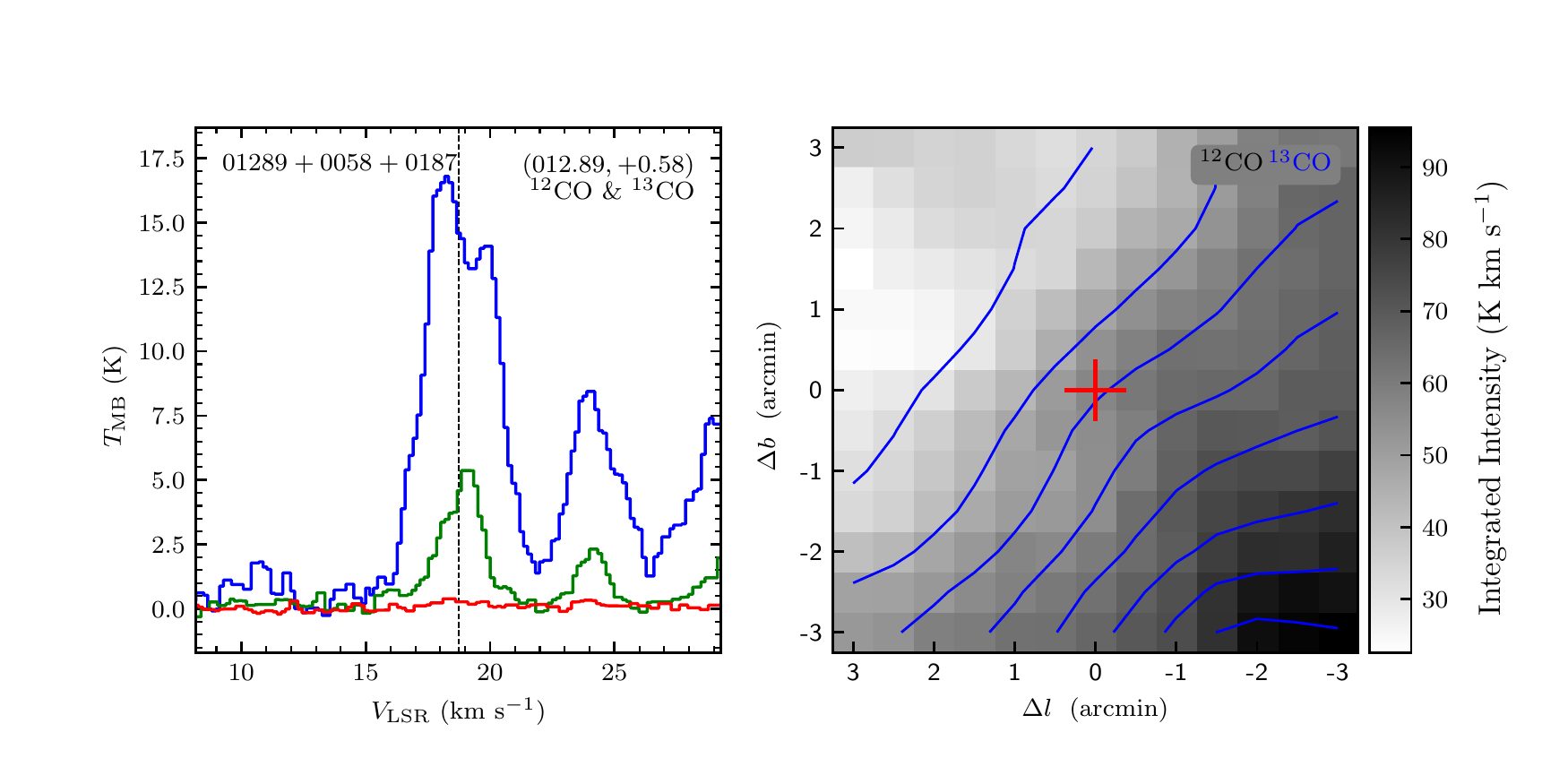}
\vspace{-0.5cm}

\includegraphics[width=9.0cm,angle=0]{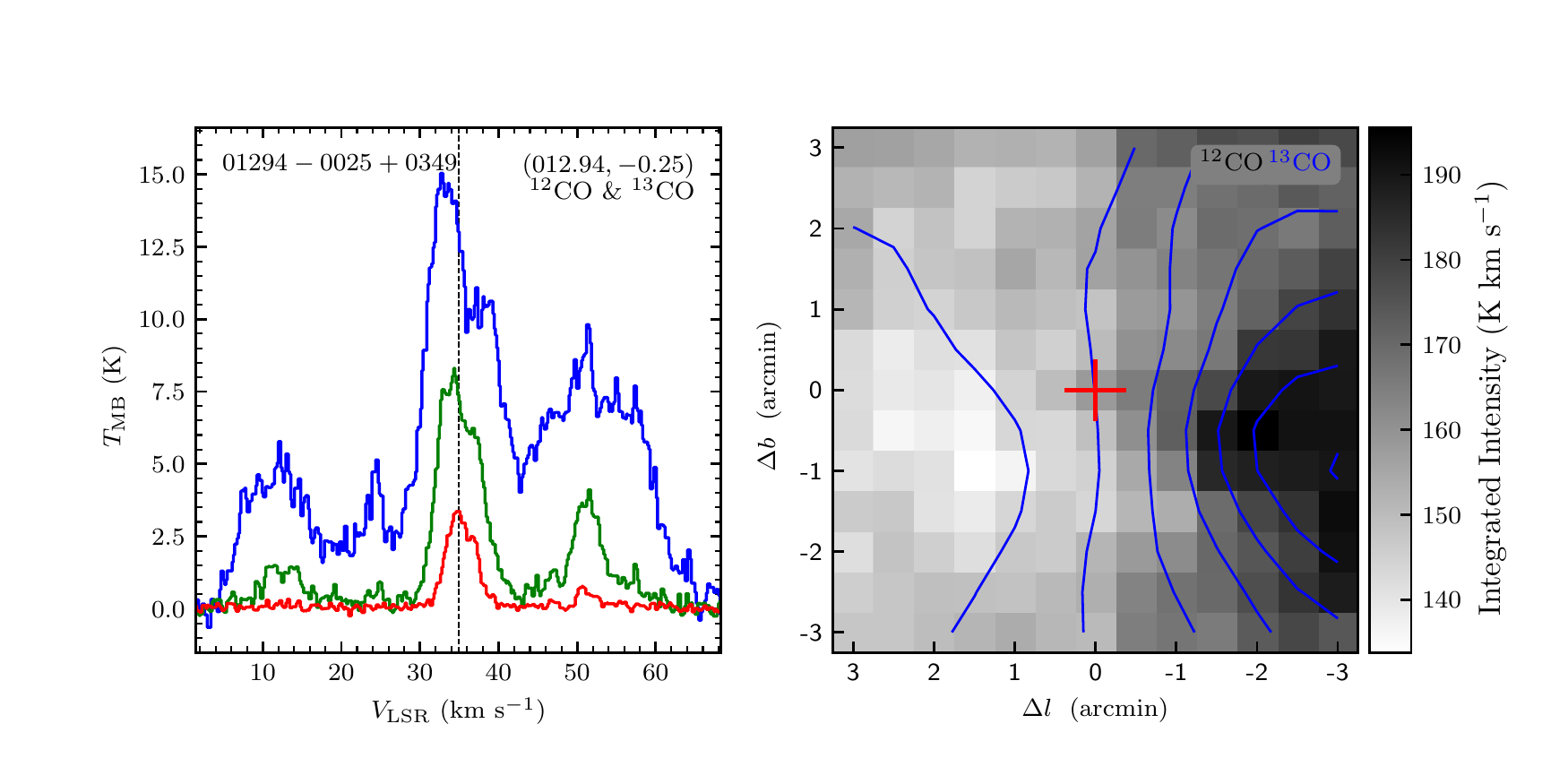}
\includegraphics[width=9.0cm,angle=0]{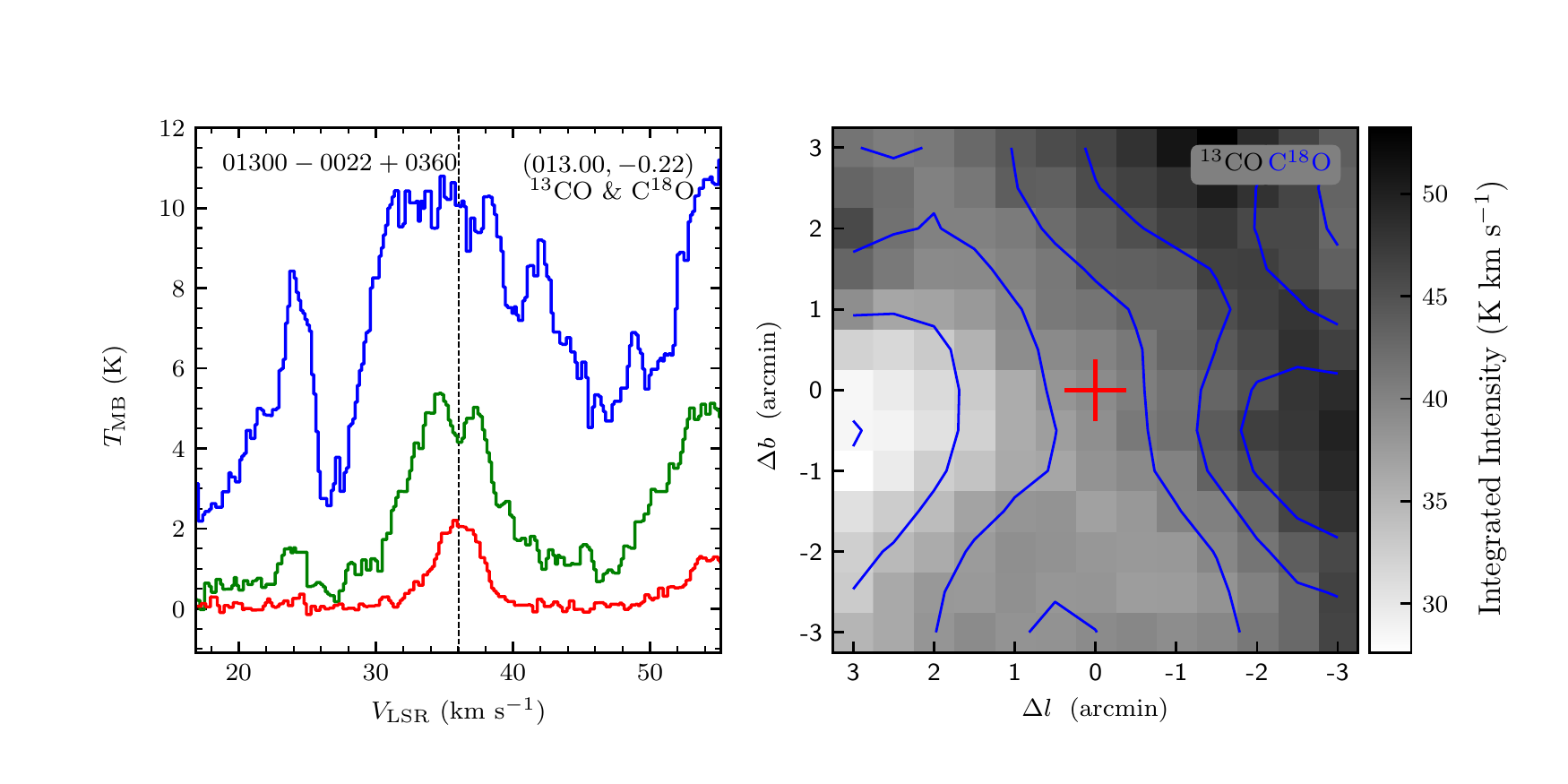}
\vspace{-0.5cm}

\includegraphics[width=9.0cm,angle=0]{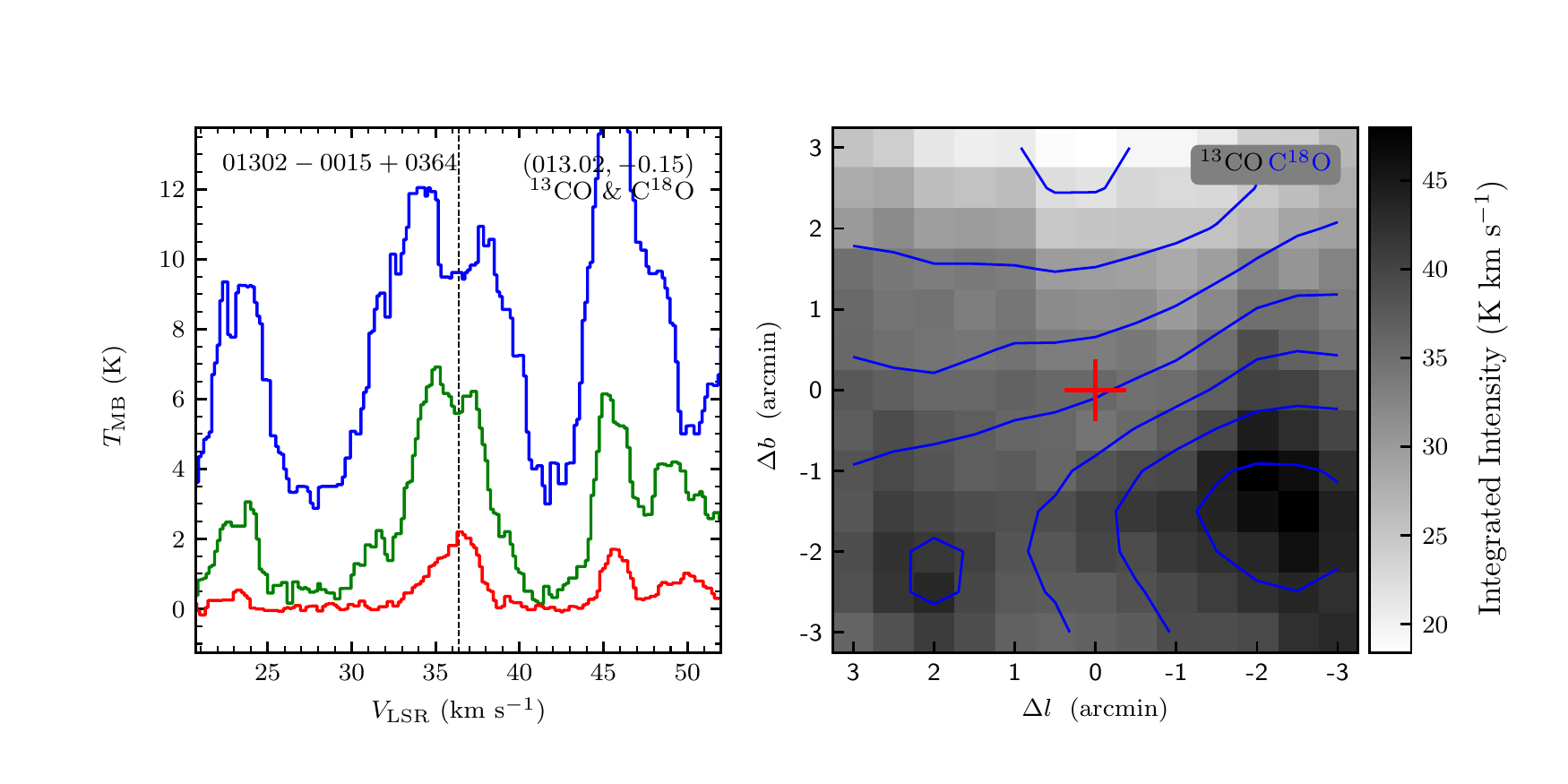}
\includegraphics[width=9.0cm,angle=0]{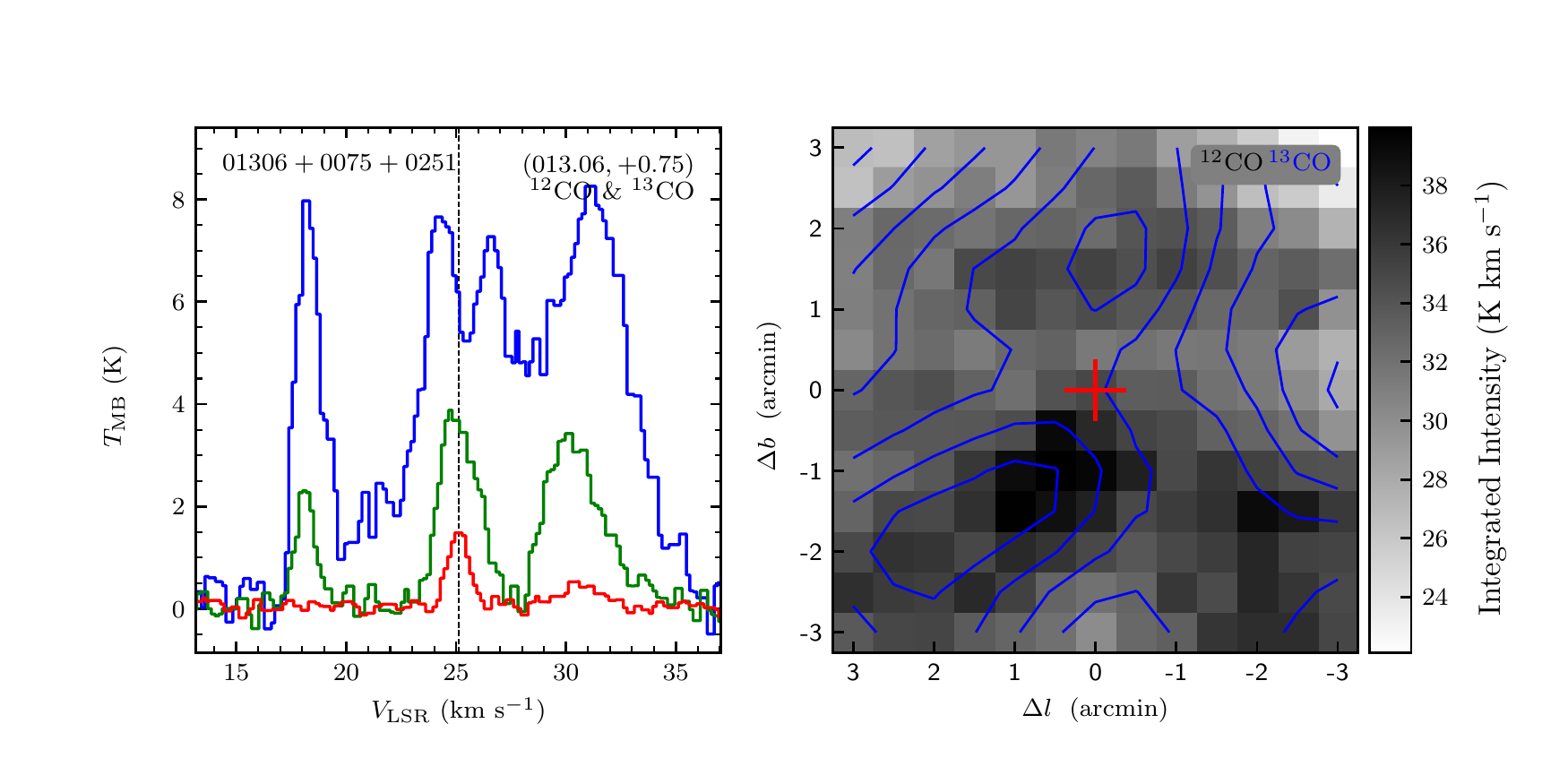}
\vspace{-0.5cm}

\includegraphics[width=9.0cm,angle=0]{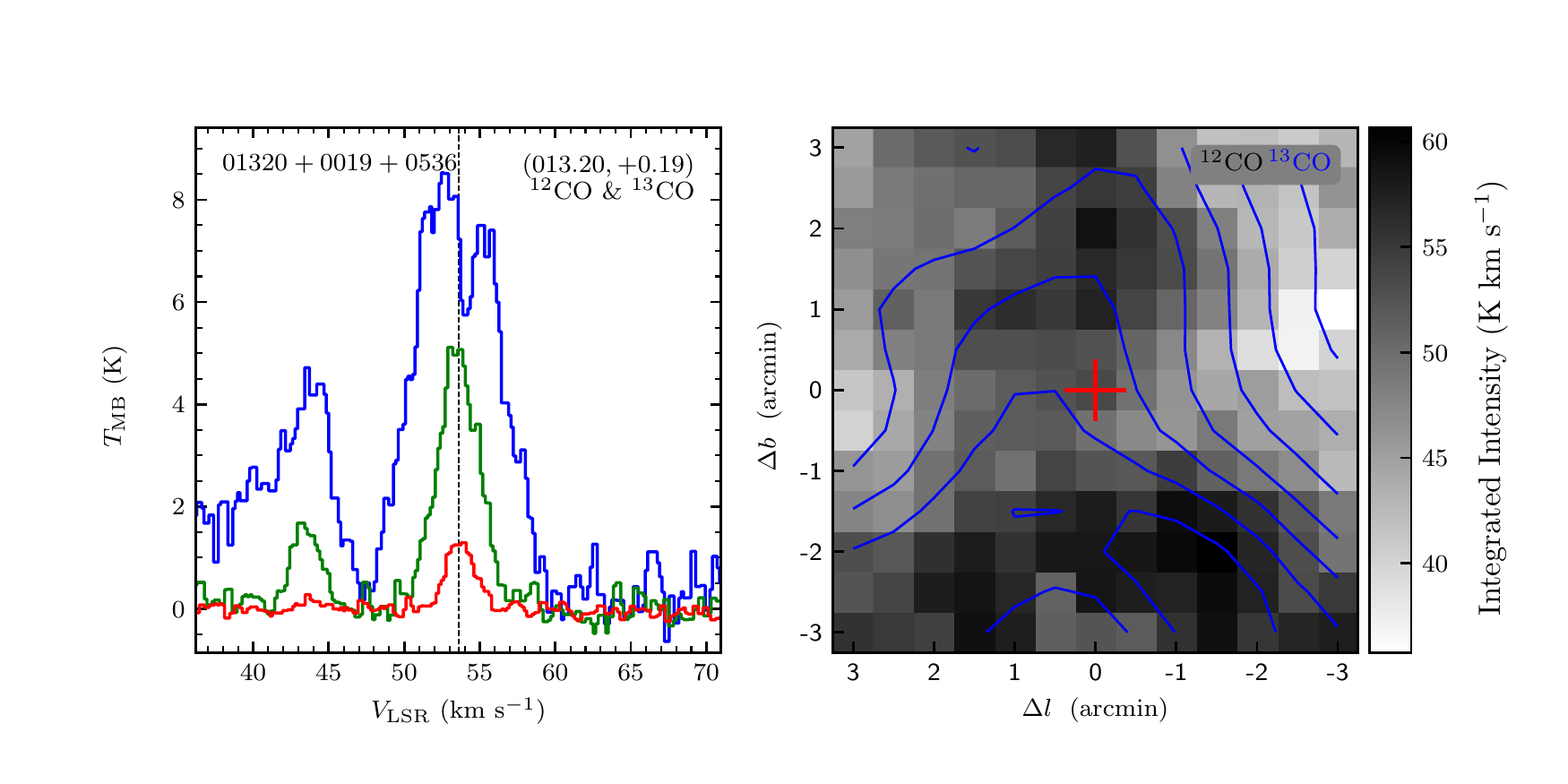}
\includegraphics[width=9.0cm,angle=0]{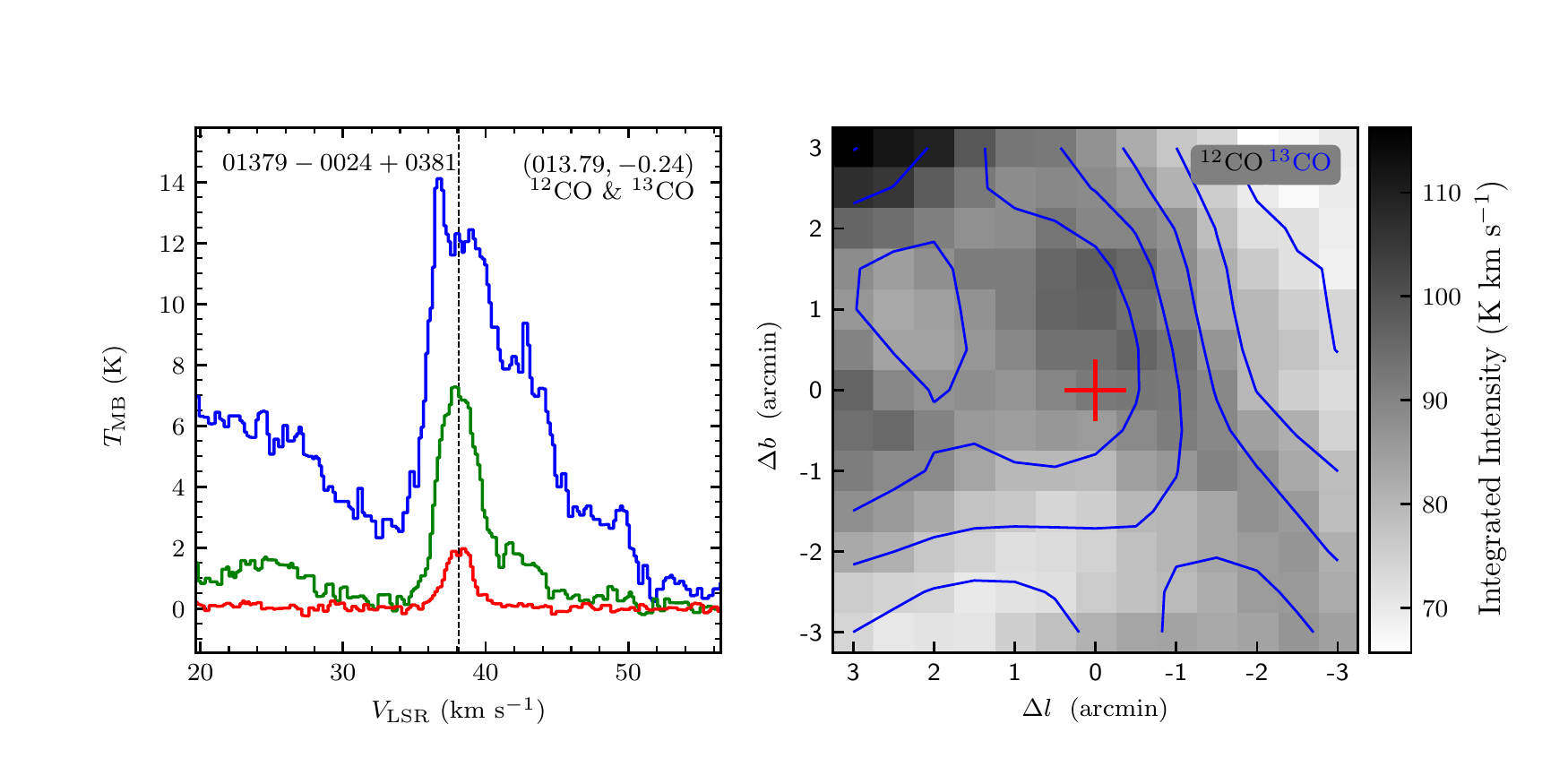}
\end{figure}
\clearpage

\begin{figure}
\includegraphics[width=9.0cm,angle=0]{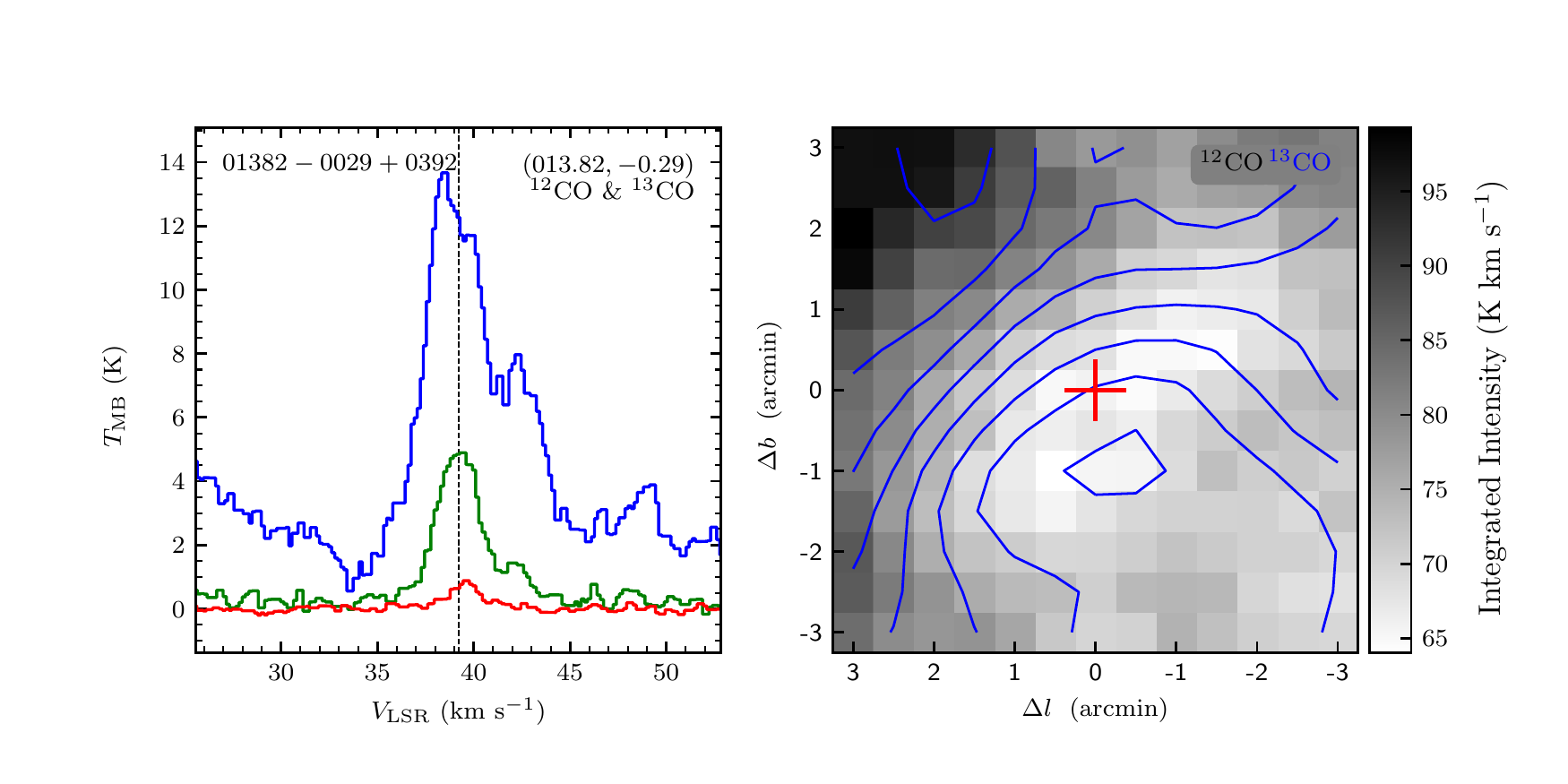}
\includegraphics[width=9.0cm,angle=0]{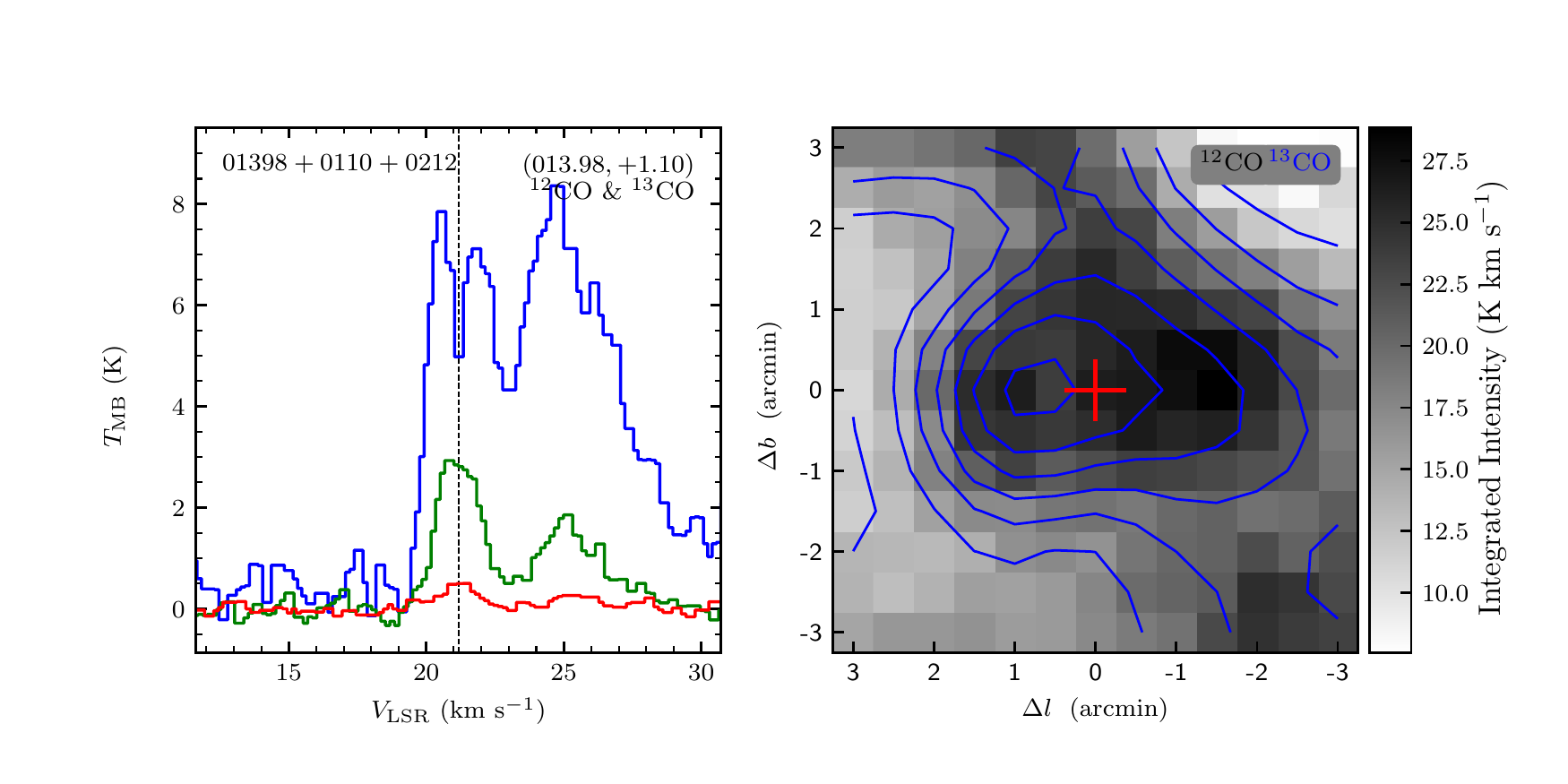}
\vspace{-0.5cm}

\includegraphics[width=9.0cm,angle=0]{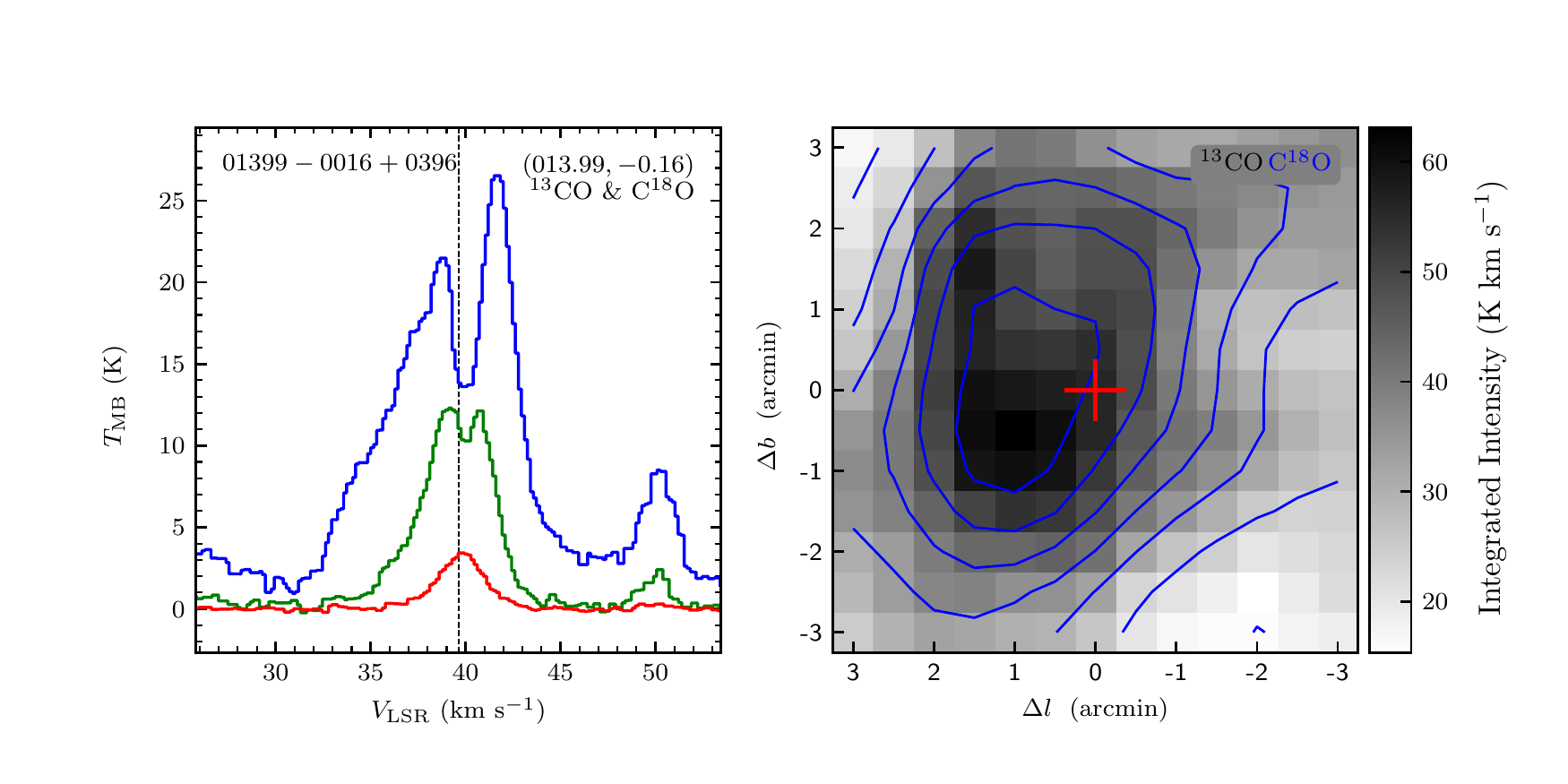}
\includegraphics[width=9.0cm,angle=0]{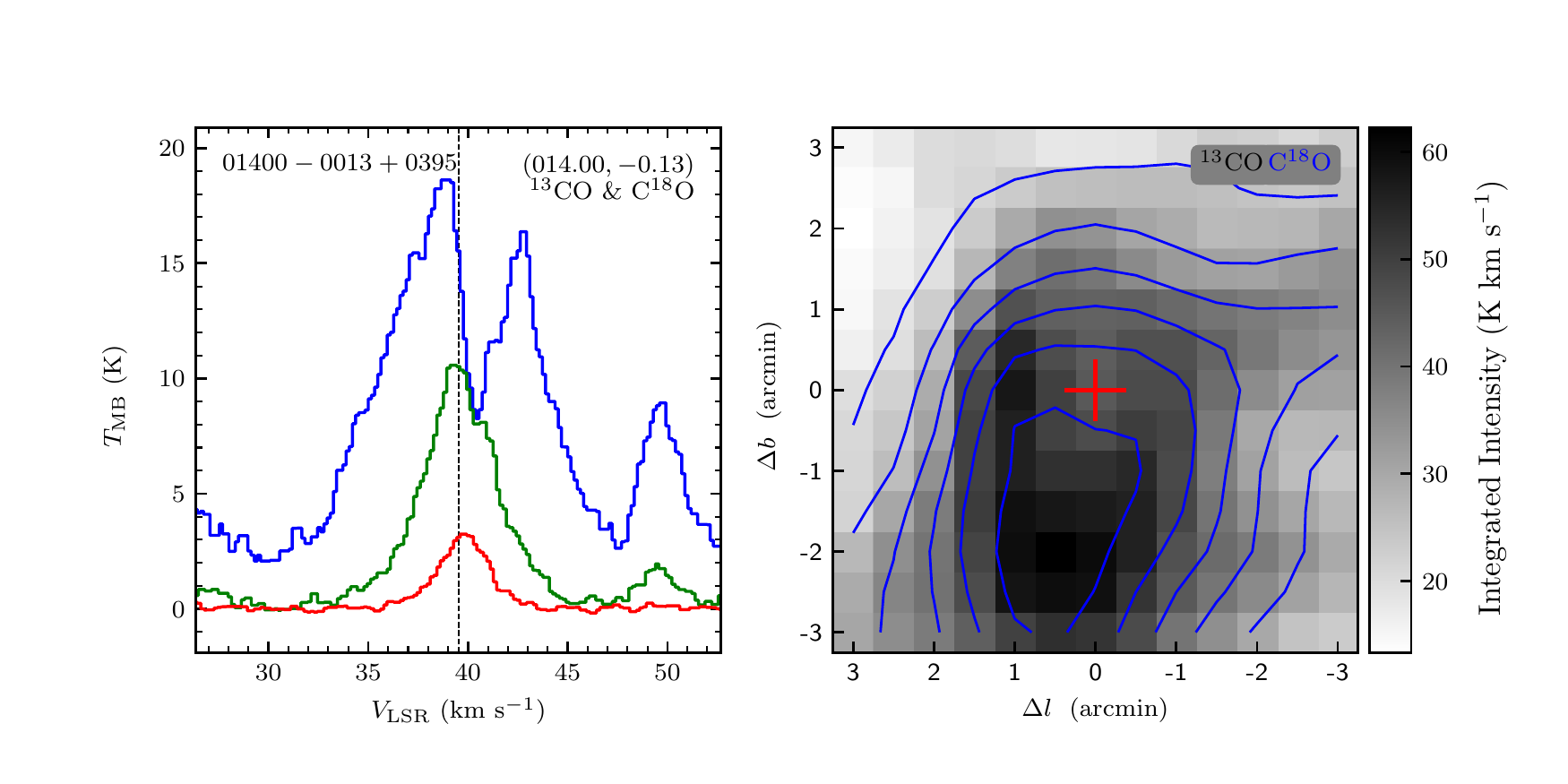}
\vspace{-0.5cm}

\includegraphics[width=9.0cm,angle=0]{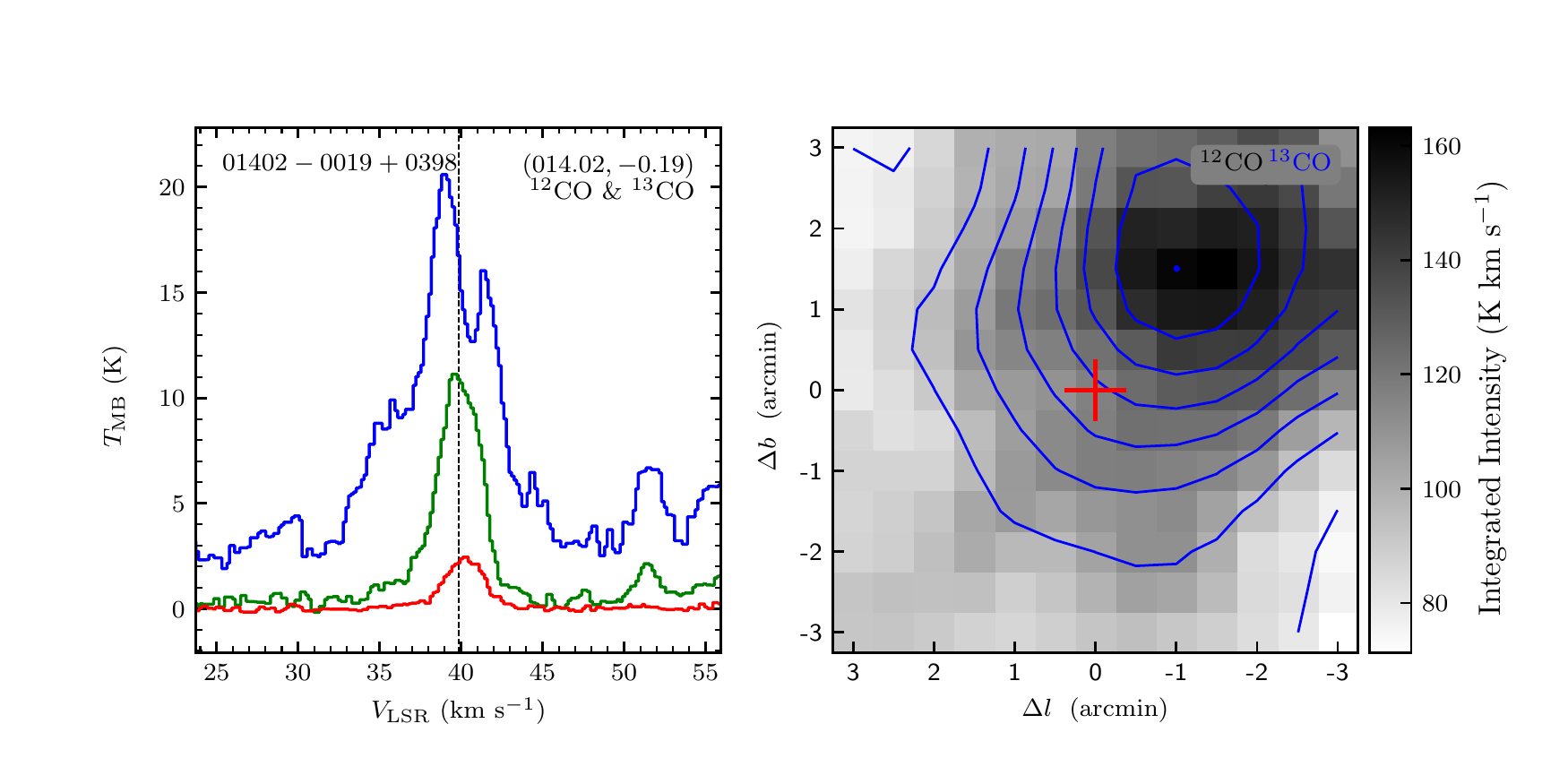}
\includegraphics[width=9.0cm,angle=0]{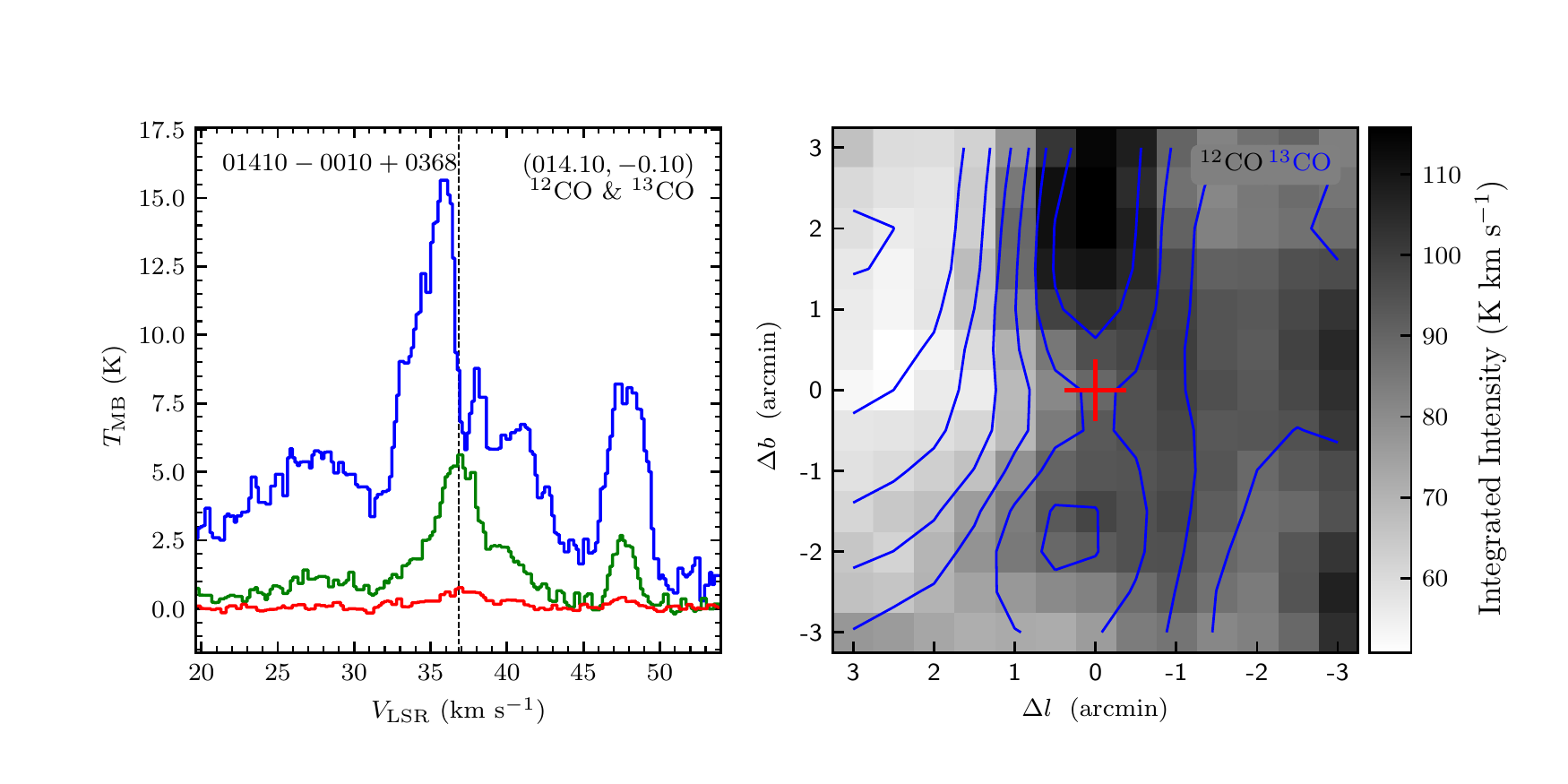}
\vspace{-0.5cm}

\includegraphics[width=9.0cm,angle=0]{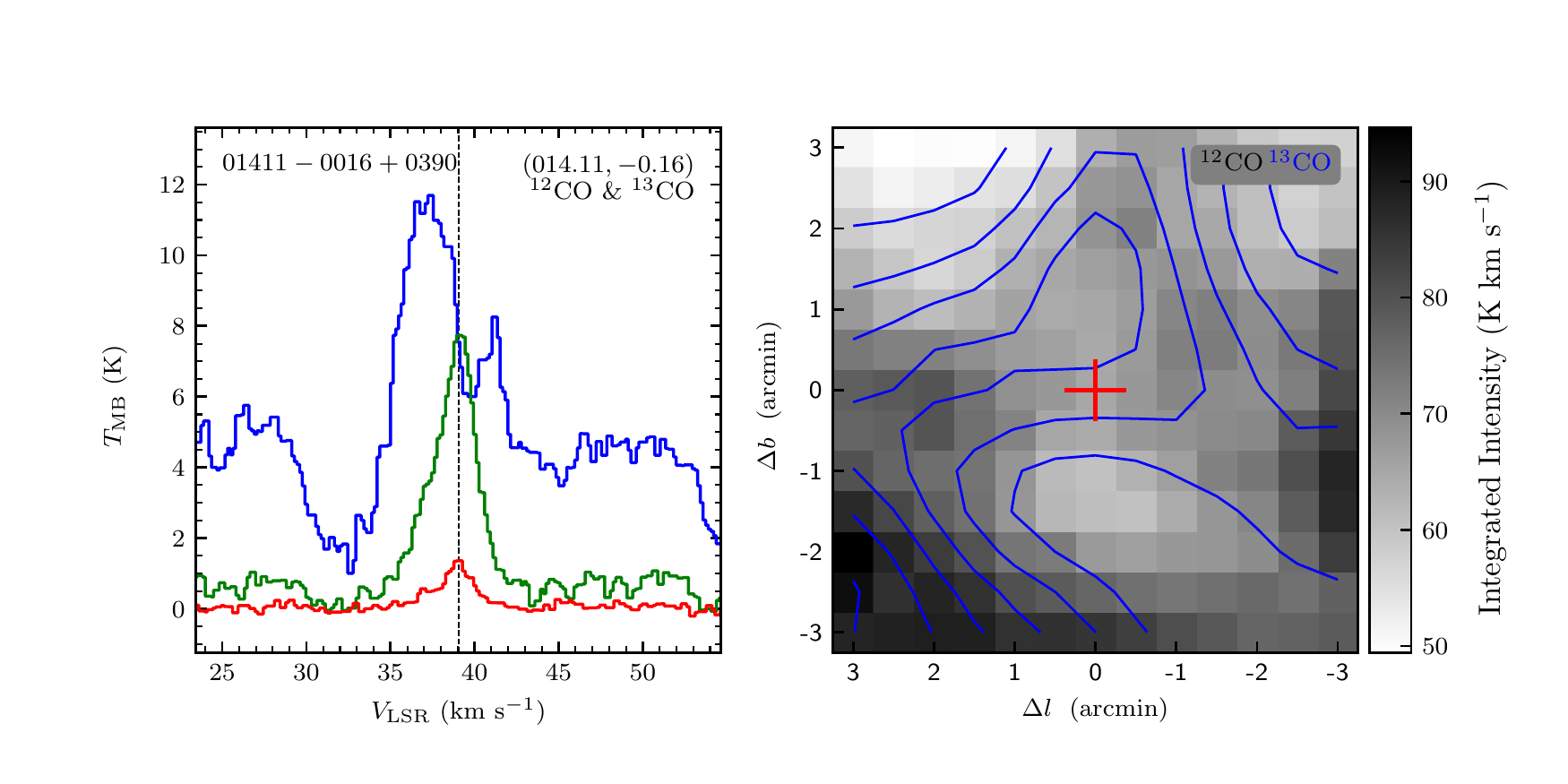}
\includegraphics[width=9.0cm,angle=0]{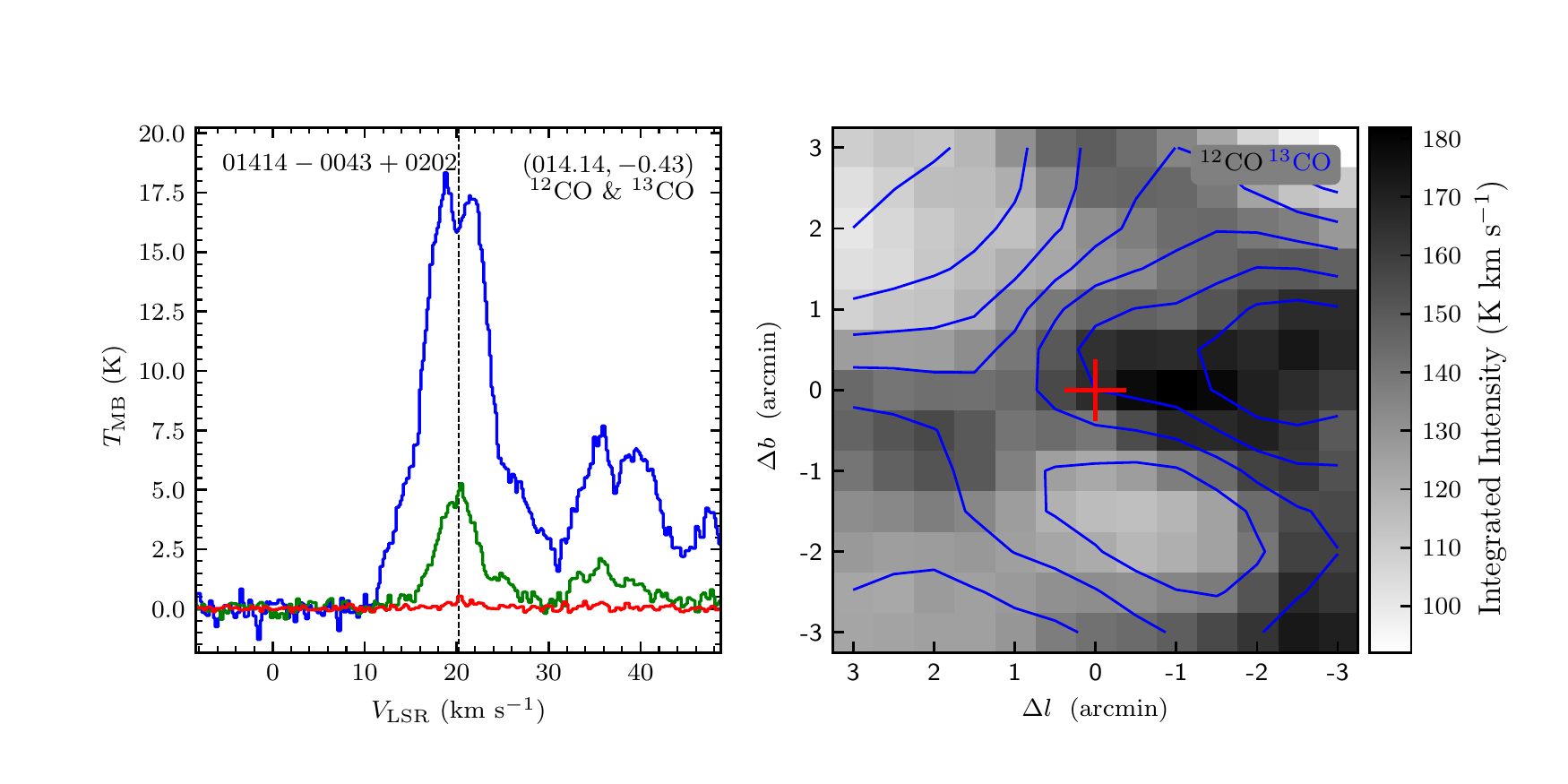}
\vspace{-0.5cm}

\includegraphics[width=9.0cm,angle=0]{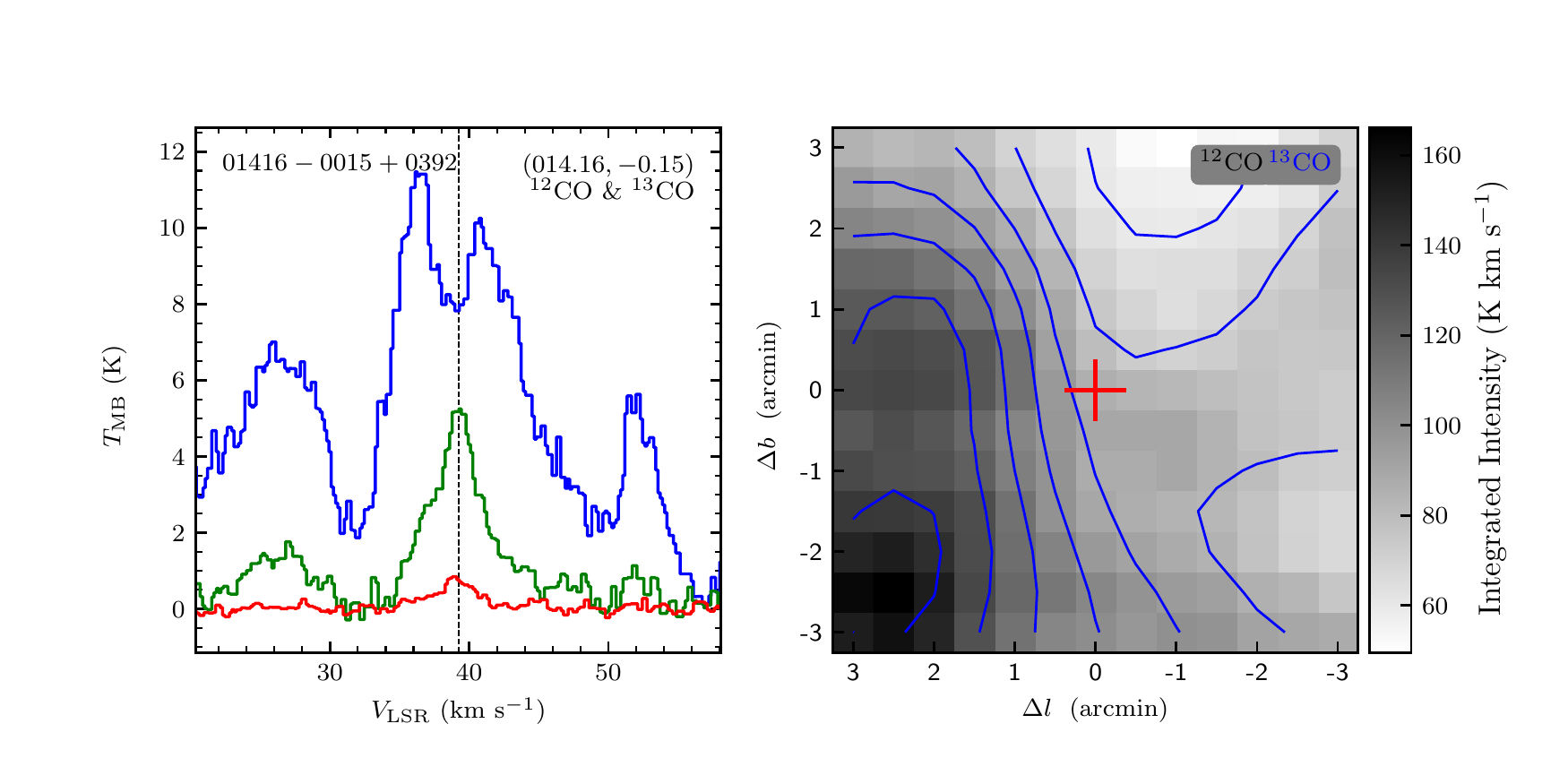}
\includegraphics[width=9.0cm,angle=0]{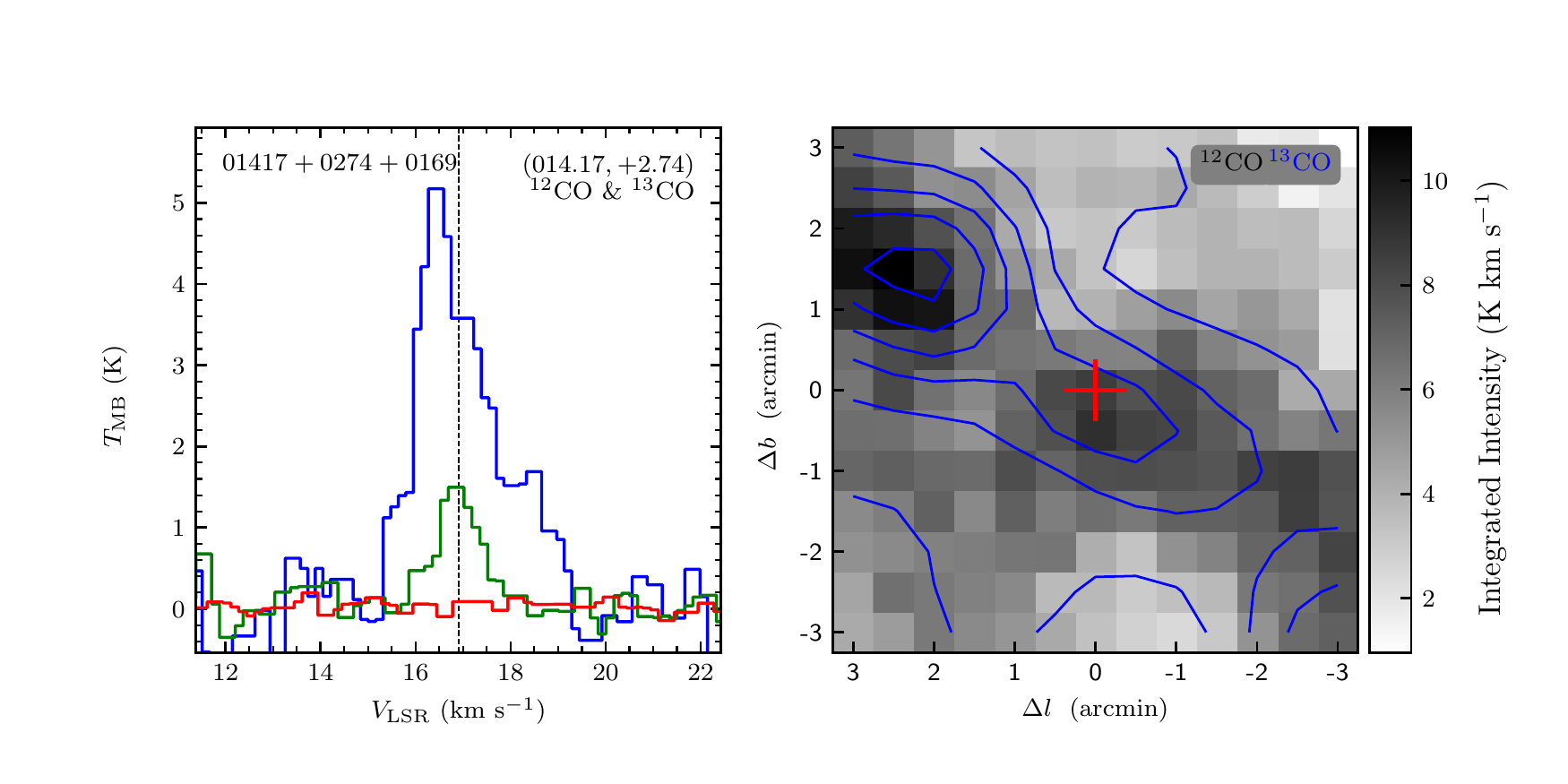}
\end{figure}
\clearpage

\begin{figure}
\includegraphics[width=9.0cm,angle=0]{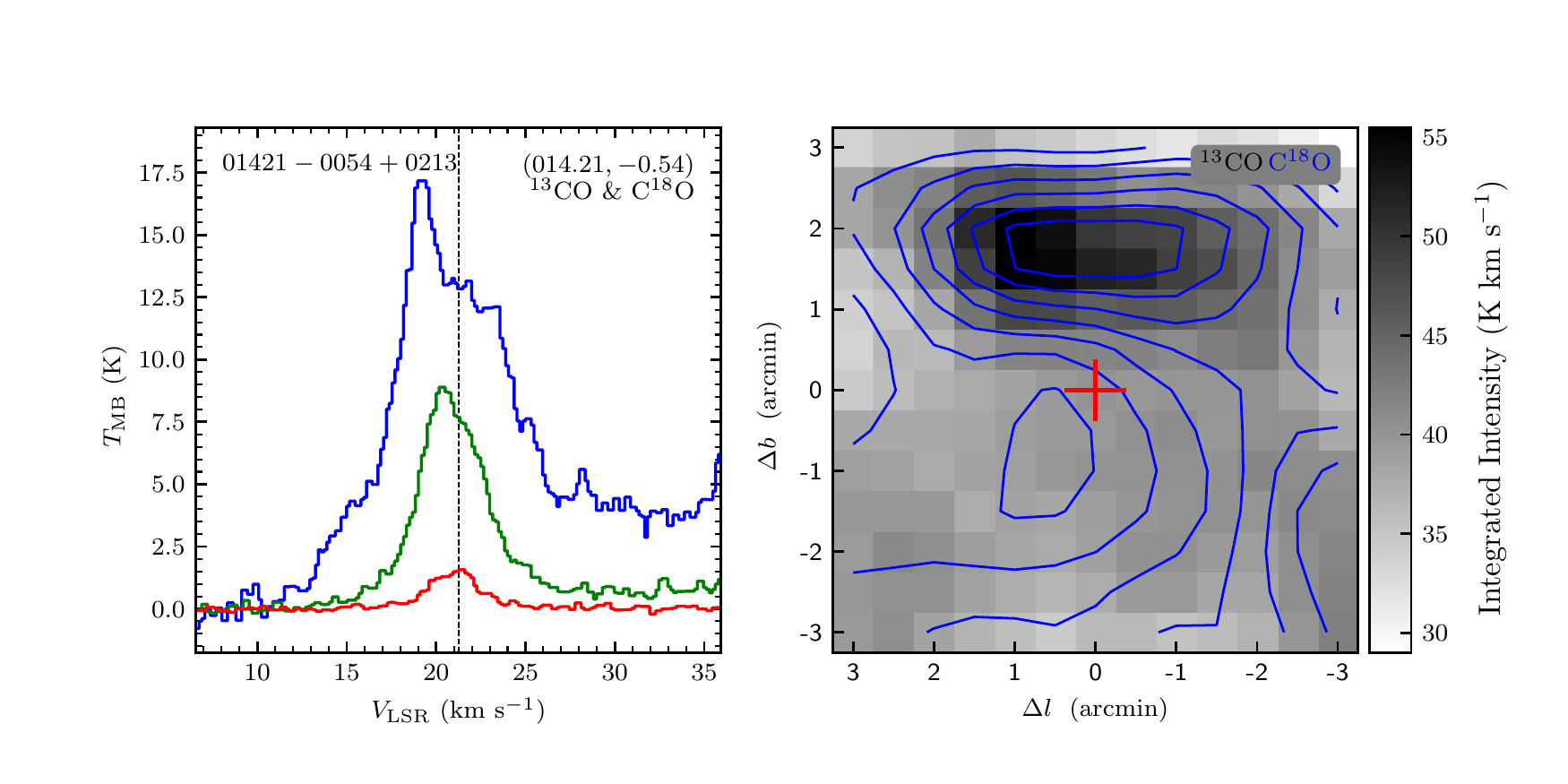}
\includegraphics[width=9.0cm,angle=0]{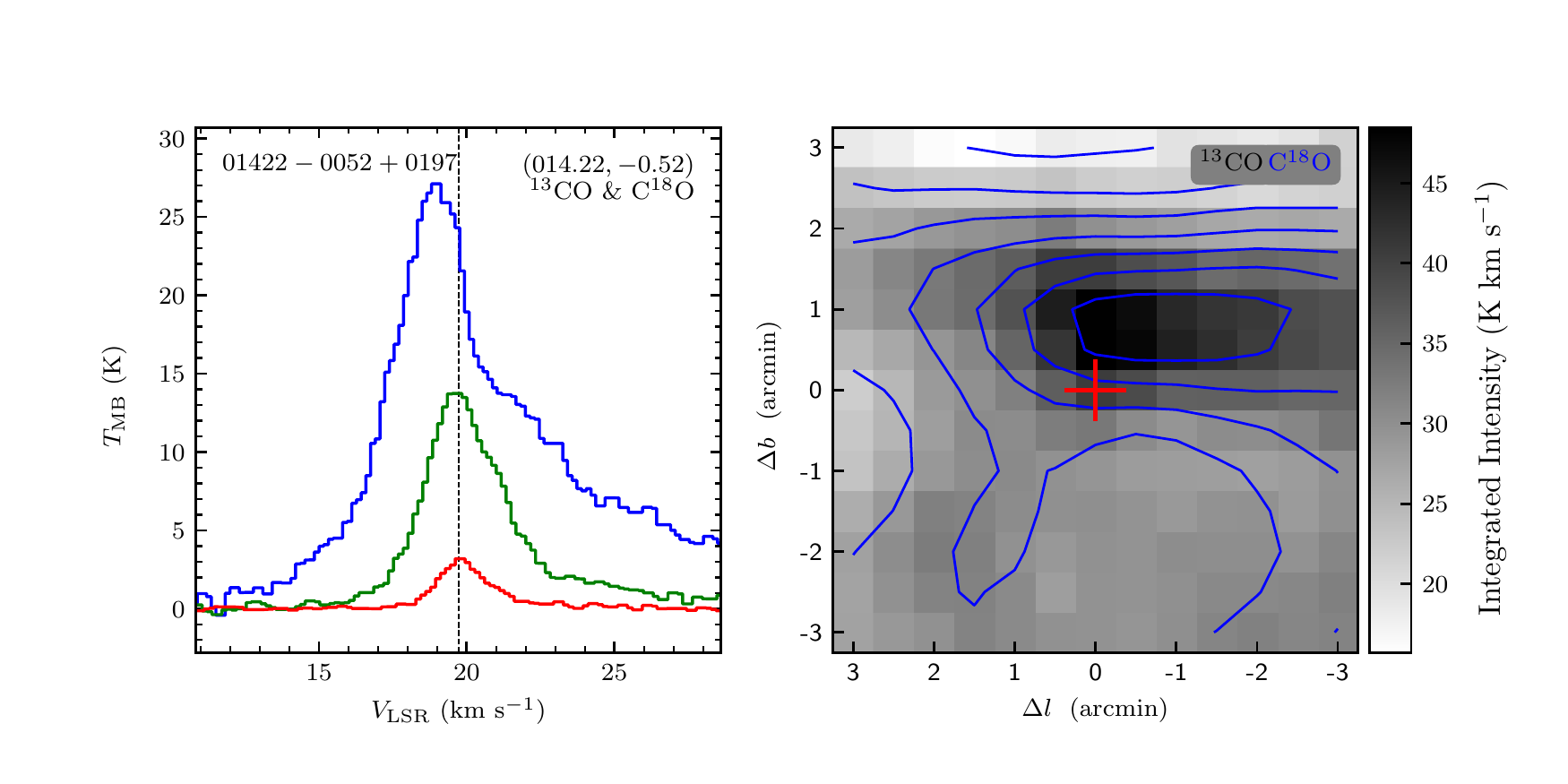}
\vspace{-0.5cm}

\includegraphics[width=9.0cm,angle=0]{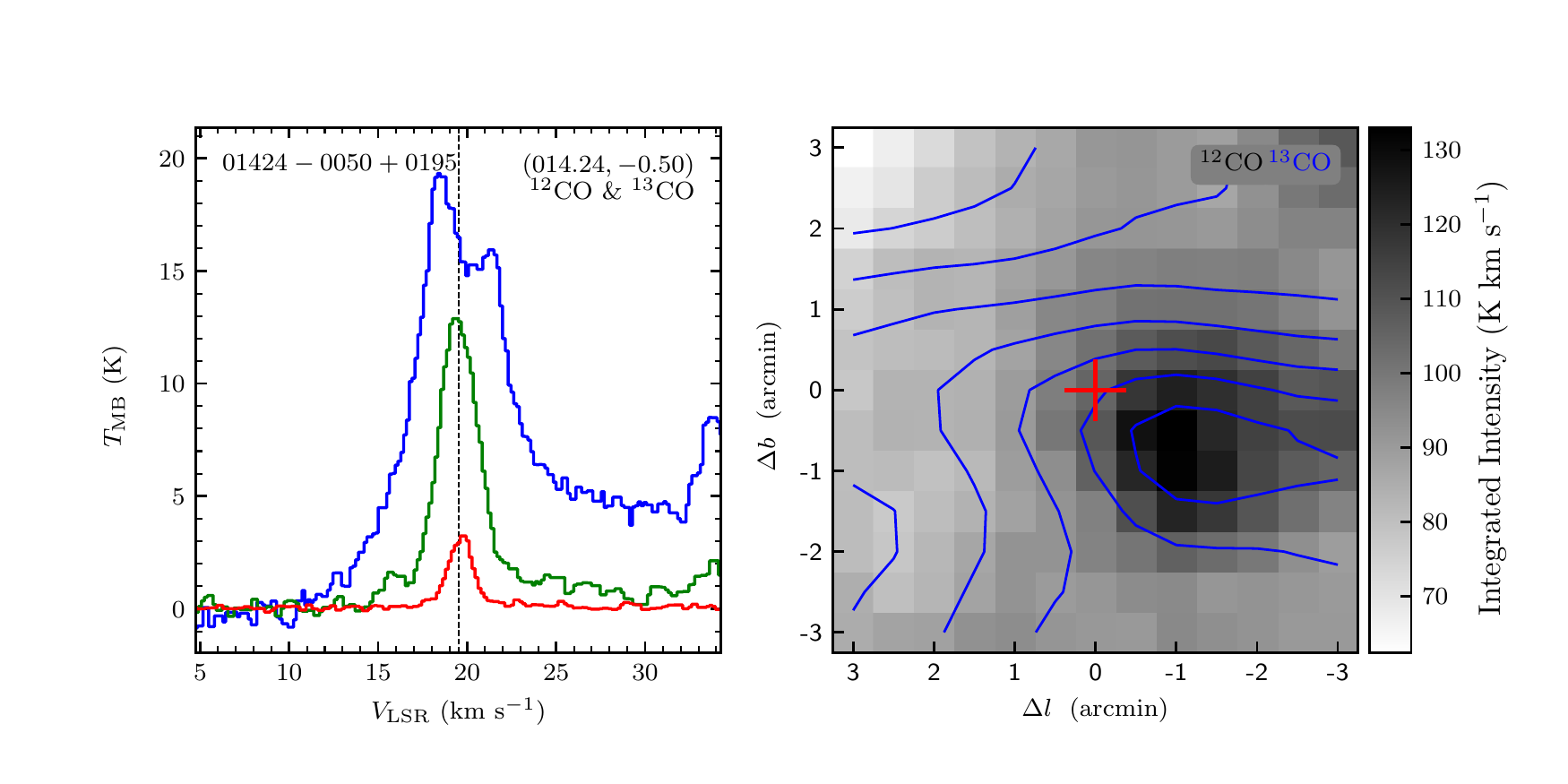}
\includegraphics[width=9.0cm,angle=0]{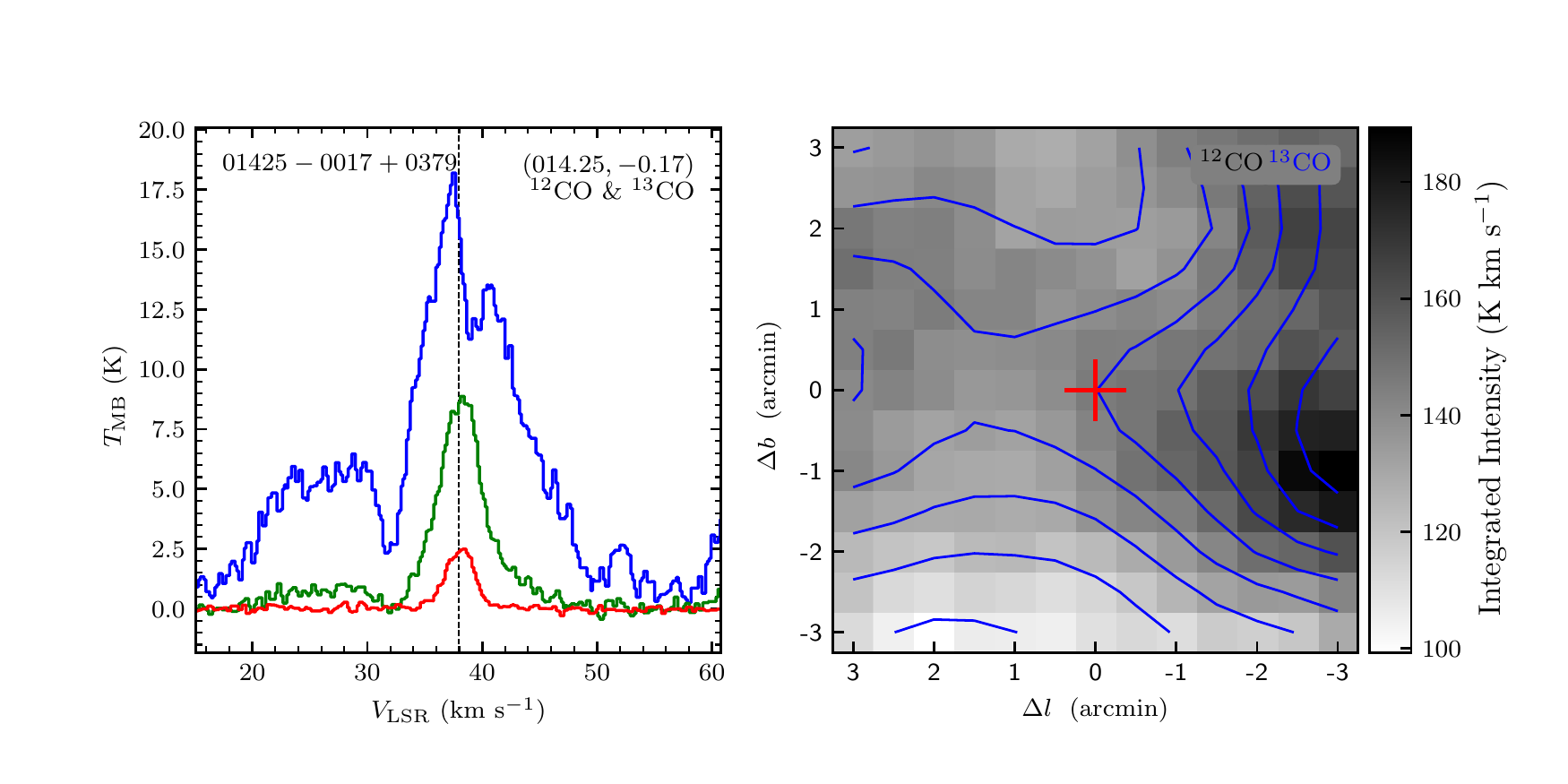}
\vspace{-0.5cm}

\includegraphics[width=9.0cm,angle=0]{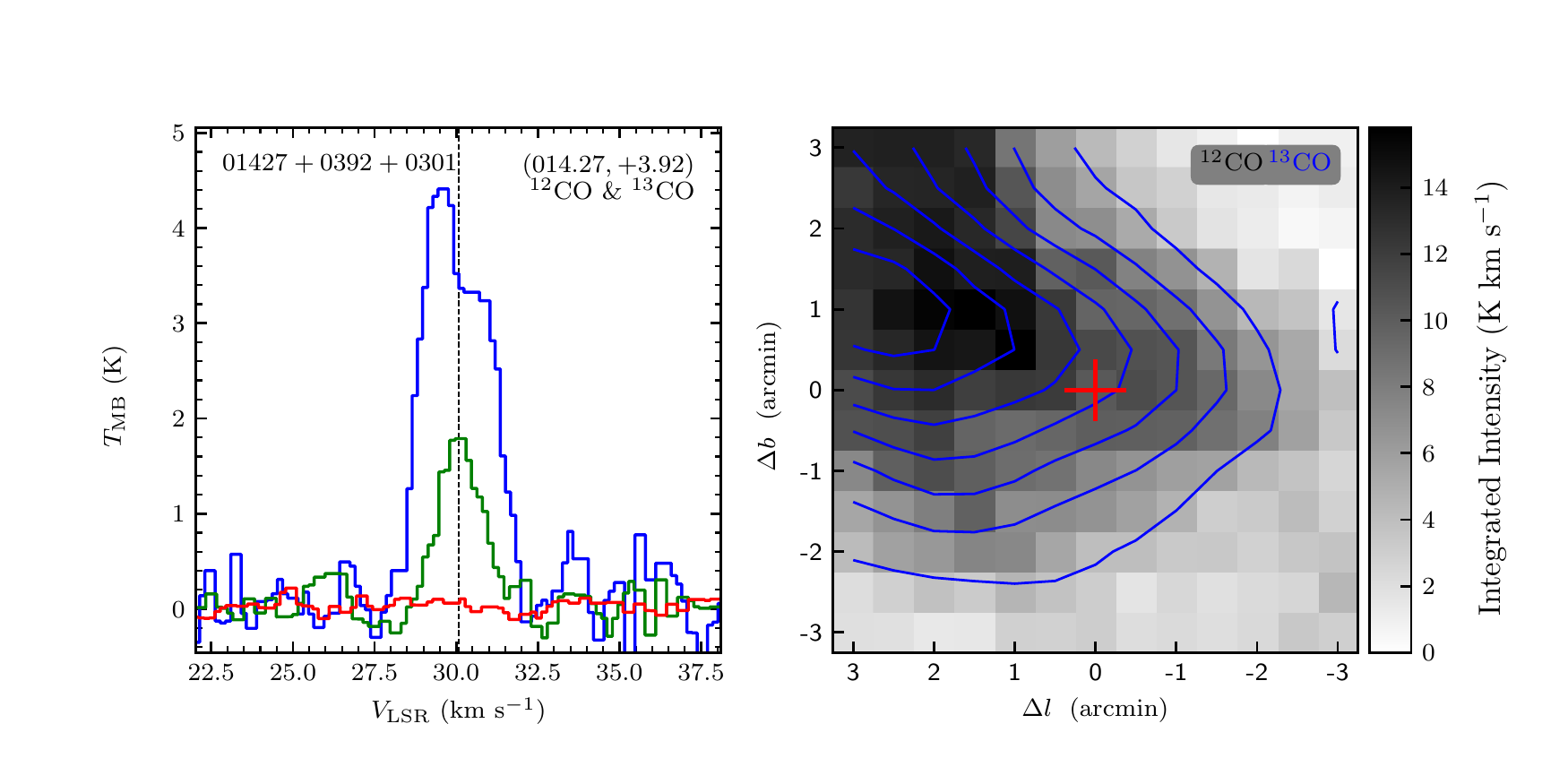}
\includegraphics[width=9.0cm,angle=0]{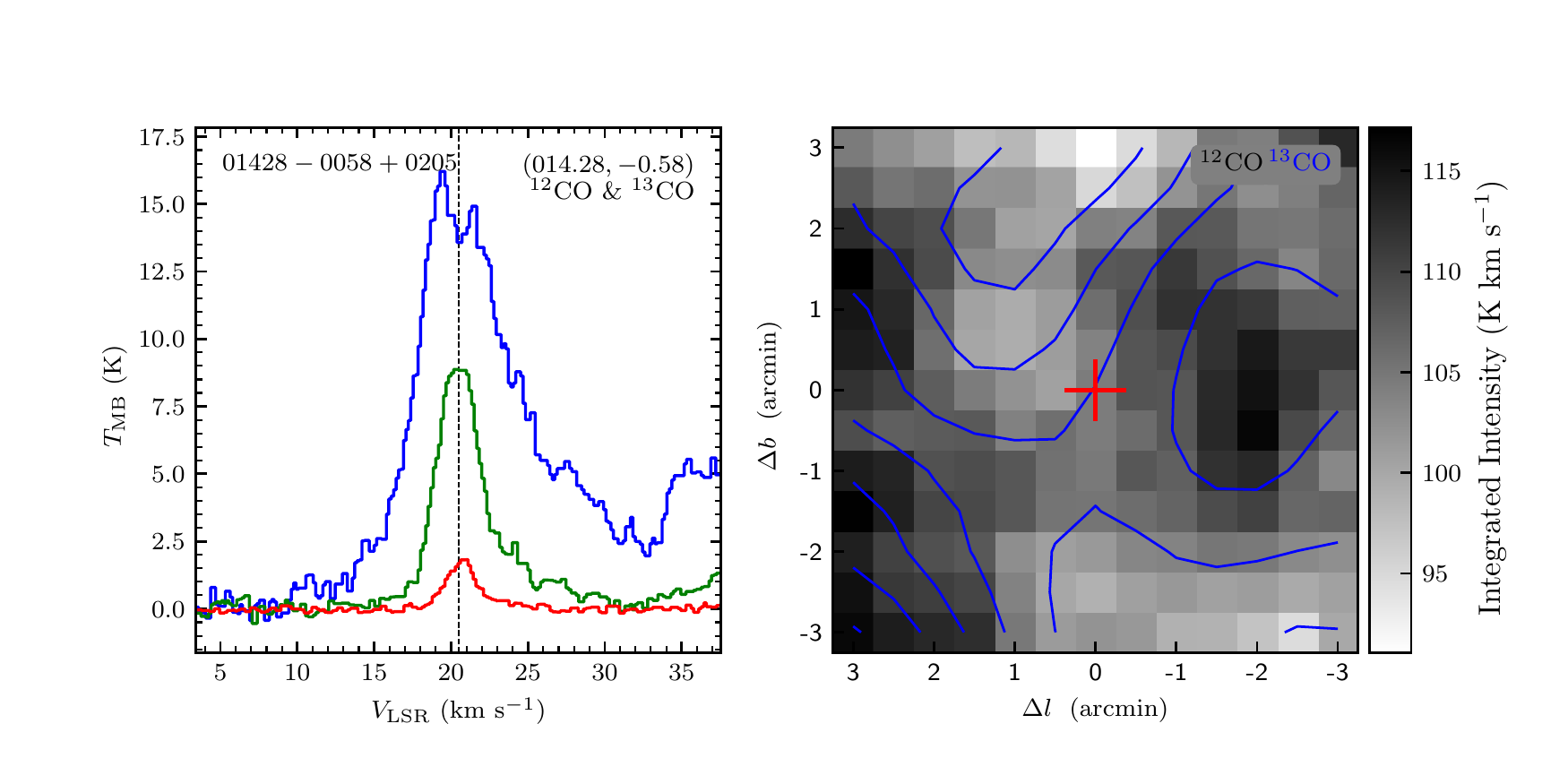}
\vspace{-0.5cm}

\includegraphics[width=9.0cm,angle=0]{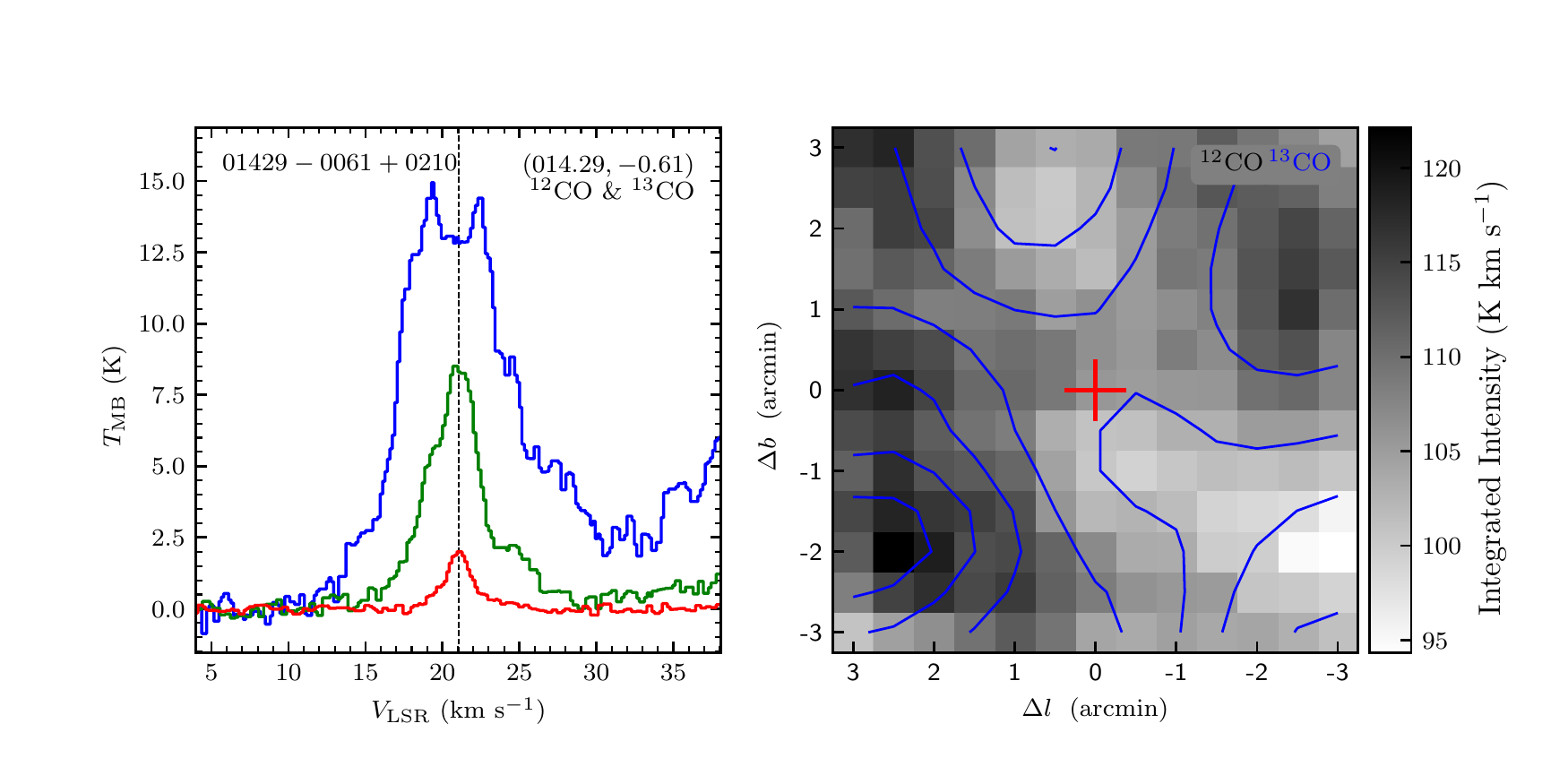}
\includegraphics[width=9.0cm,angle=0]{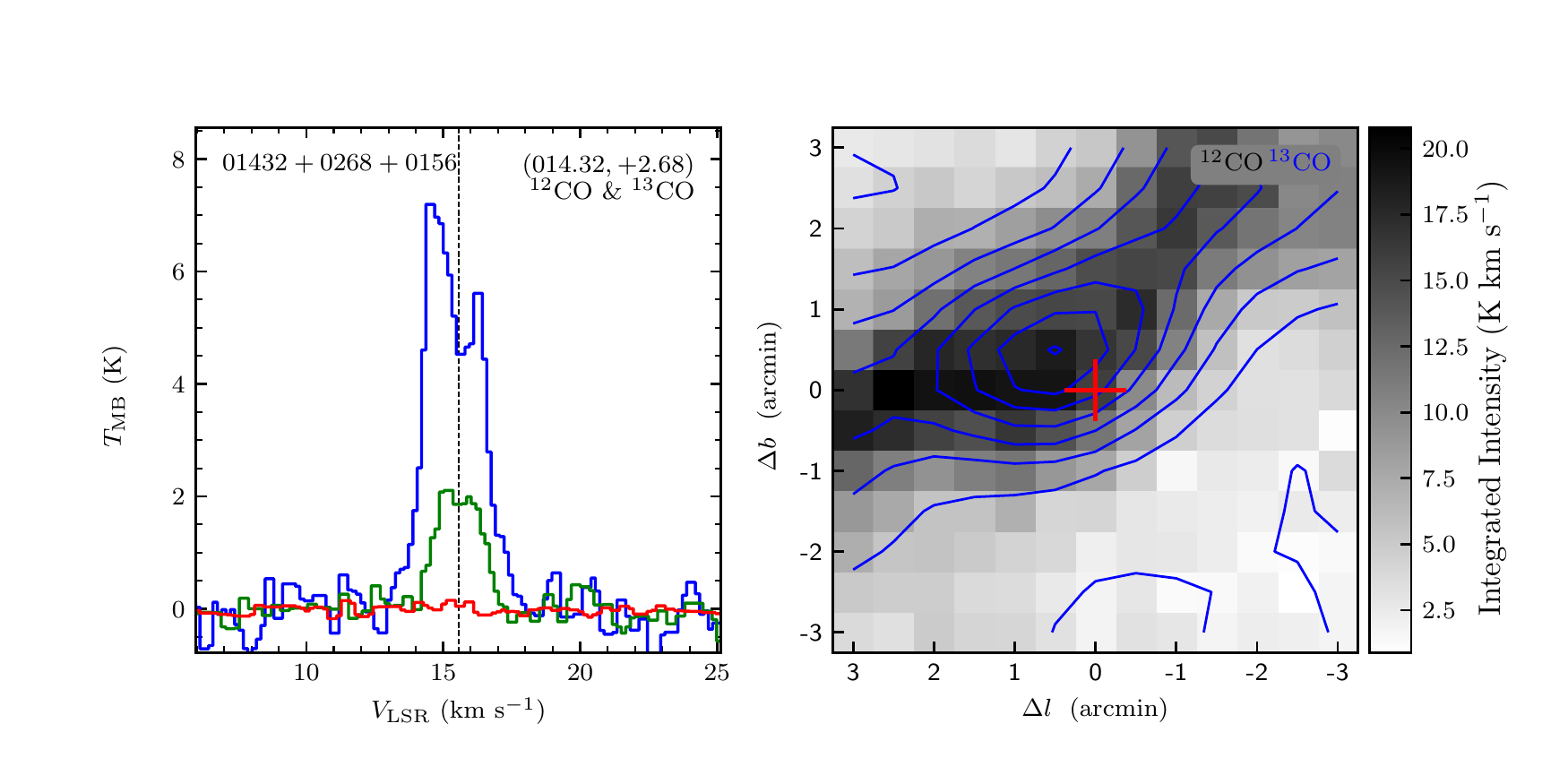}
\vspace{-0.5cm}

\includegraphics[width=9.0cm,angle=0]{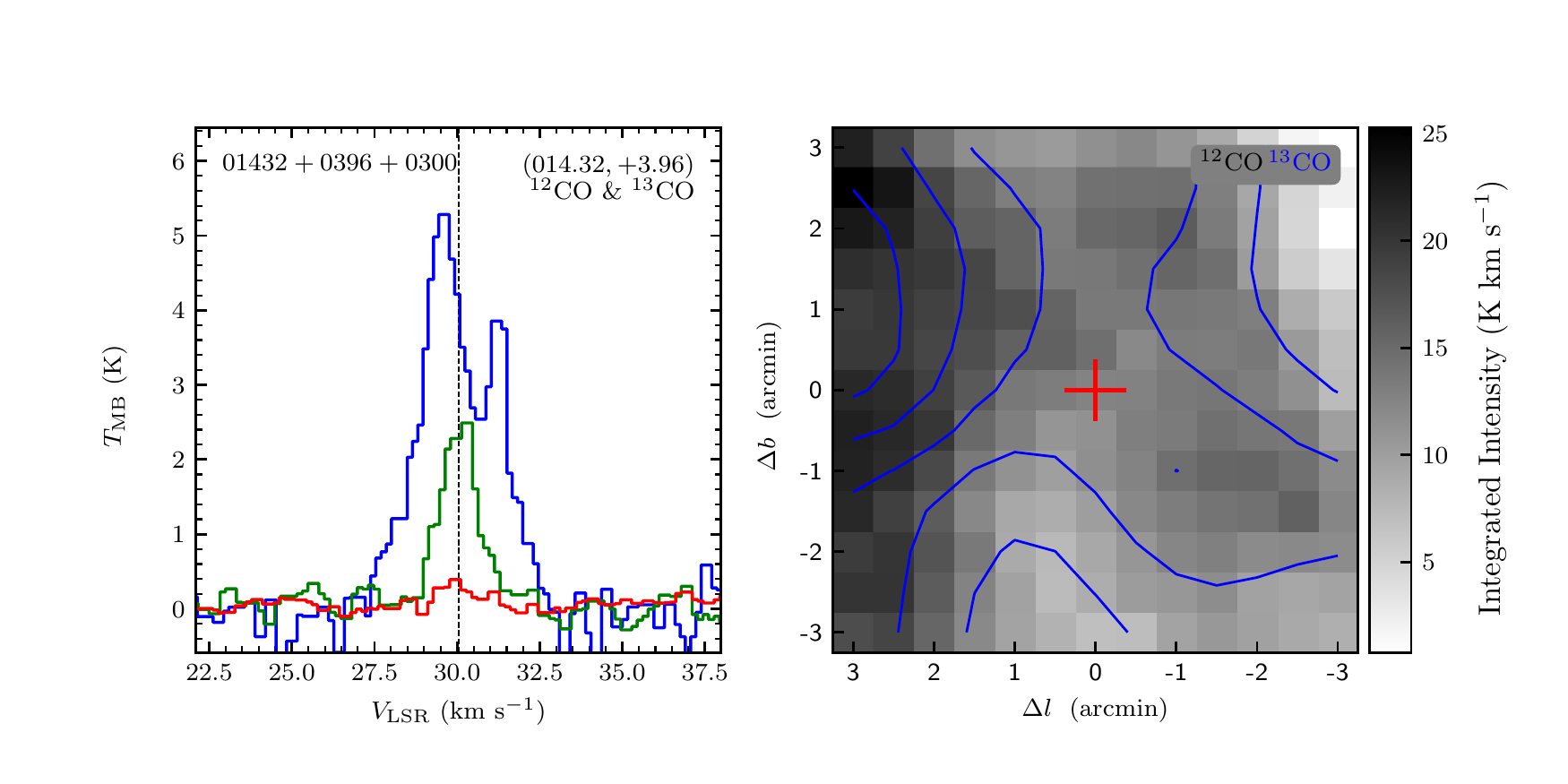}
\includegraphics[width=9.0cm,angle=0]{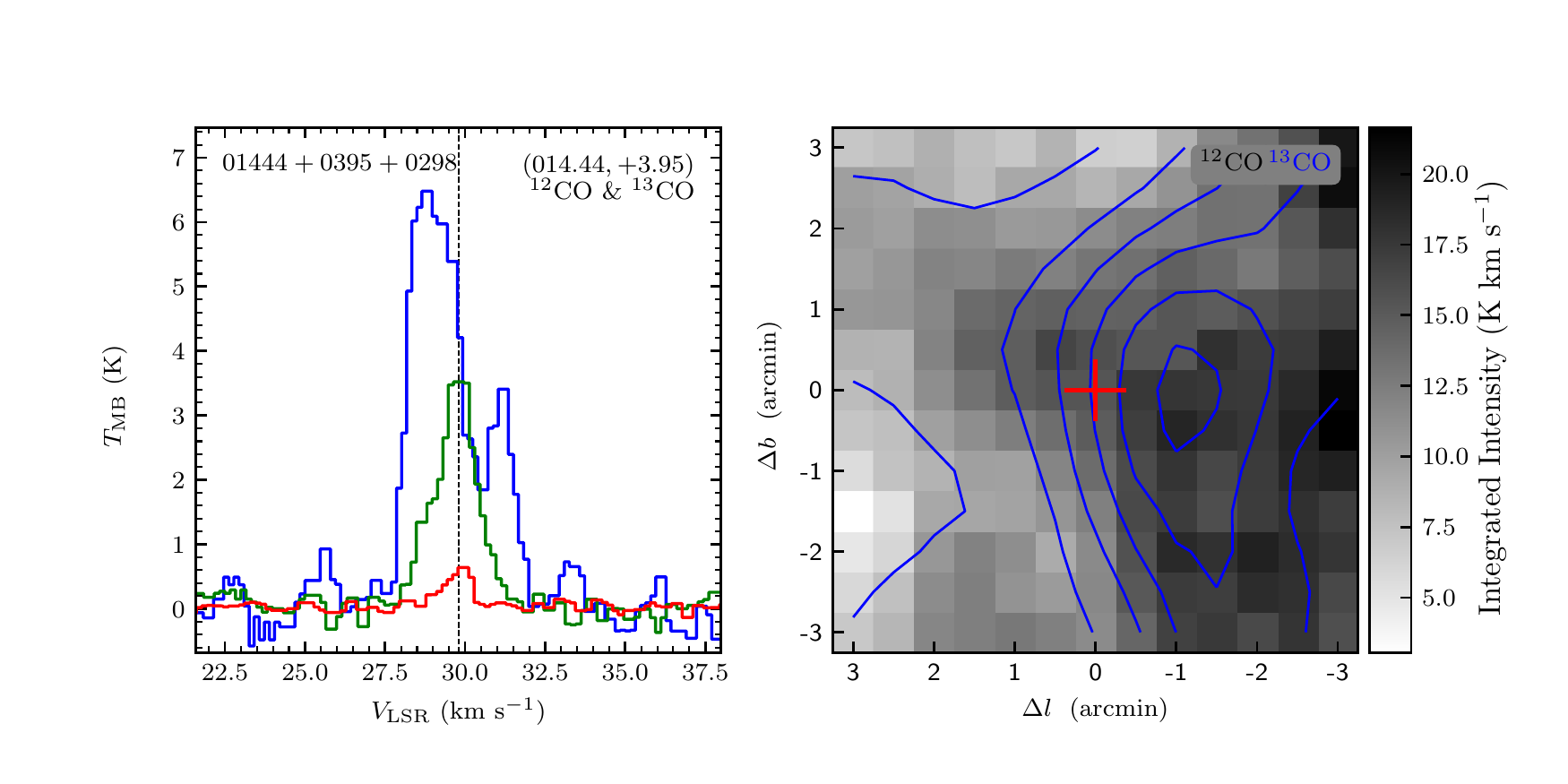}
\end{figure}
\clearpage

\begin{figure}
\includegraphics[width=9.0cm,angle=0]{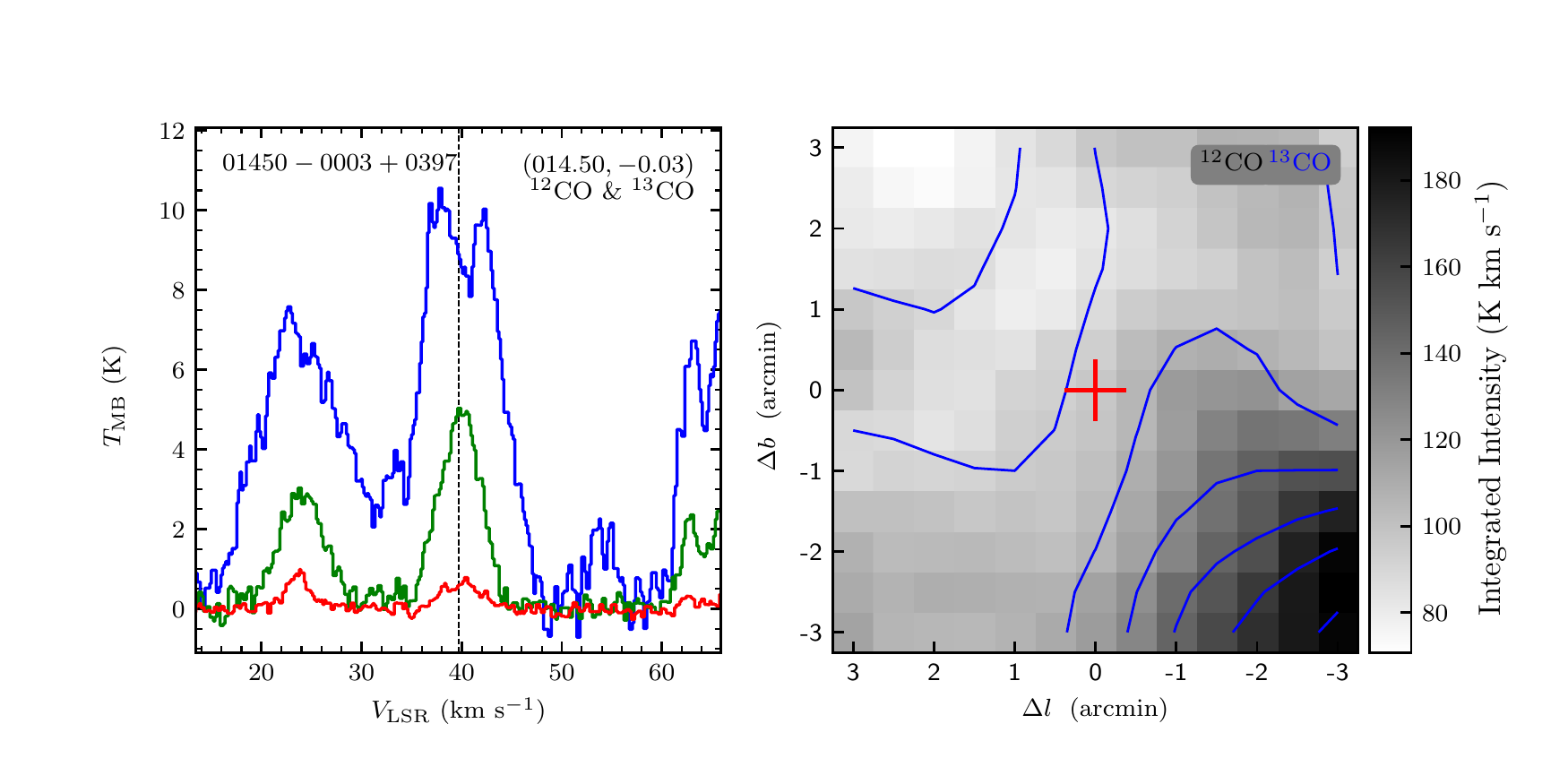}
\includegraphics[width=9.0cm,angle=0]{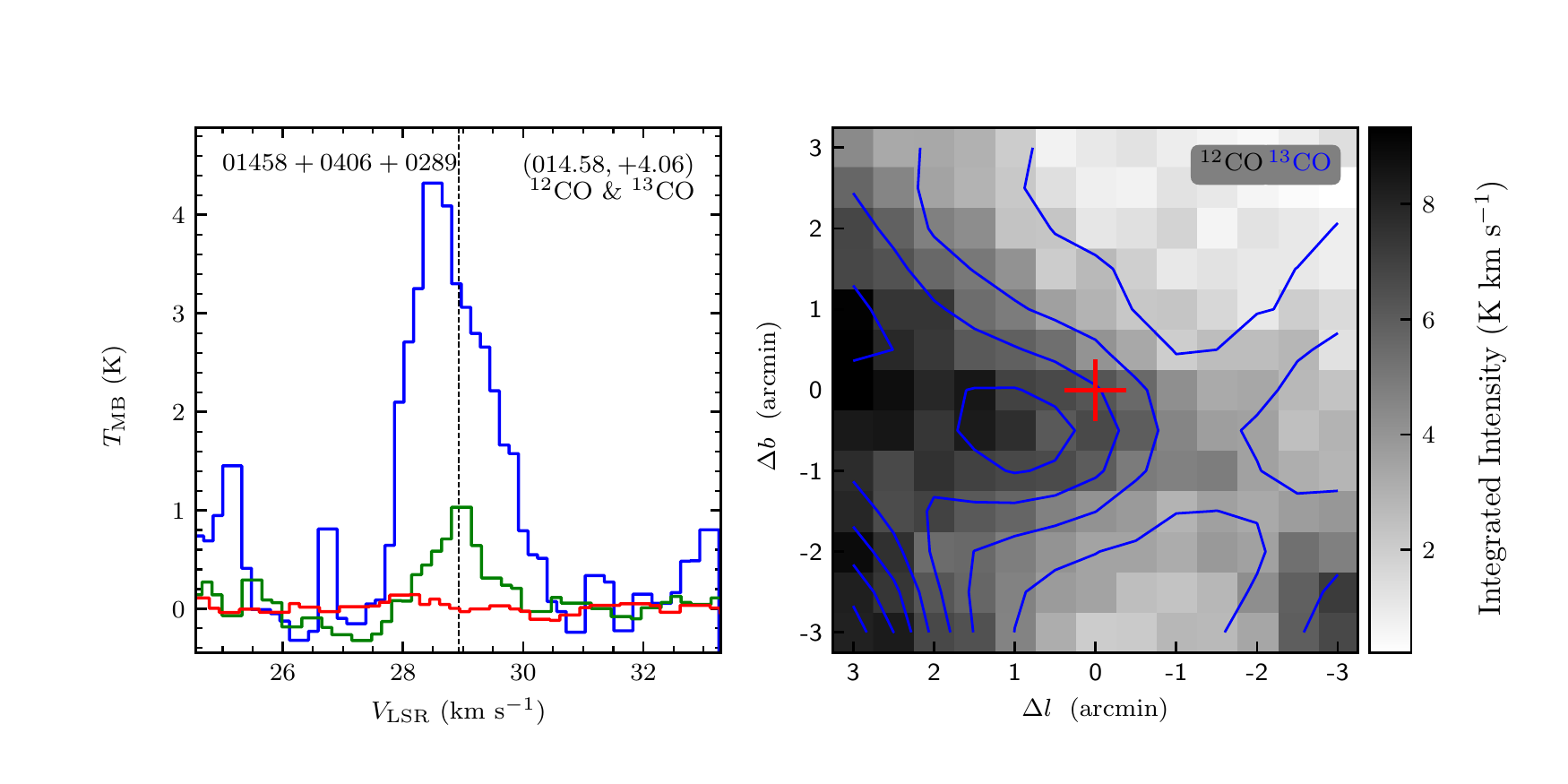}
\vspace{-0.5cm}

\includegraphics[width=9.0cm,angle=0]{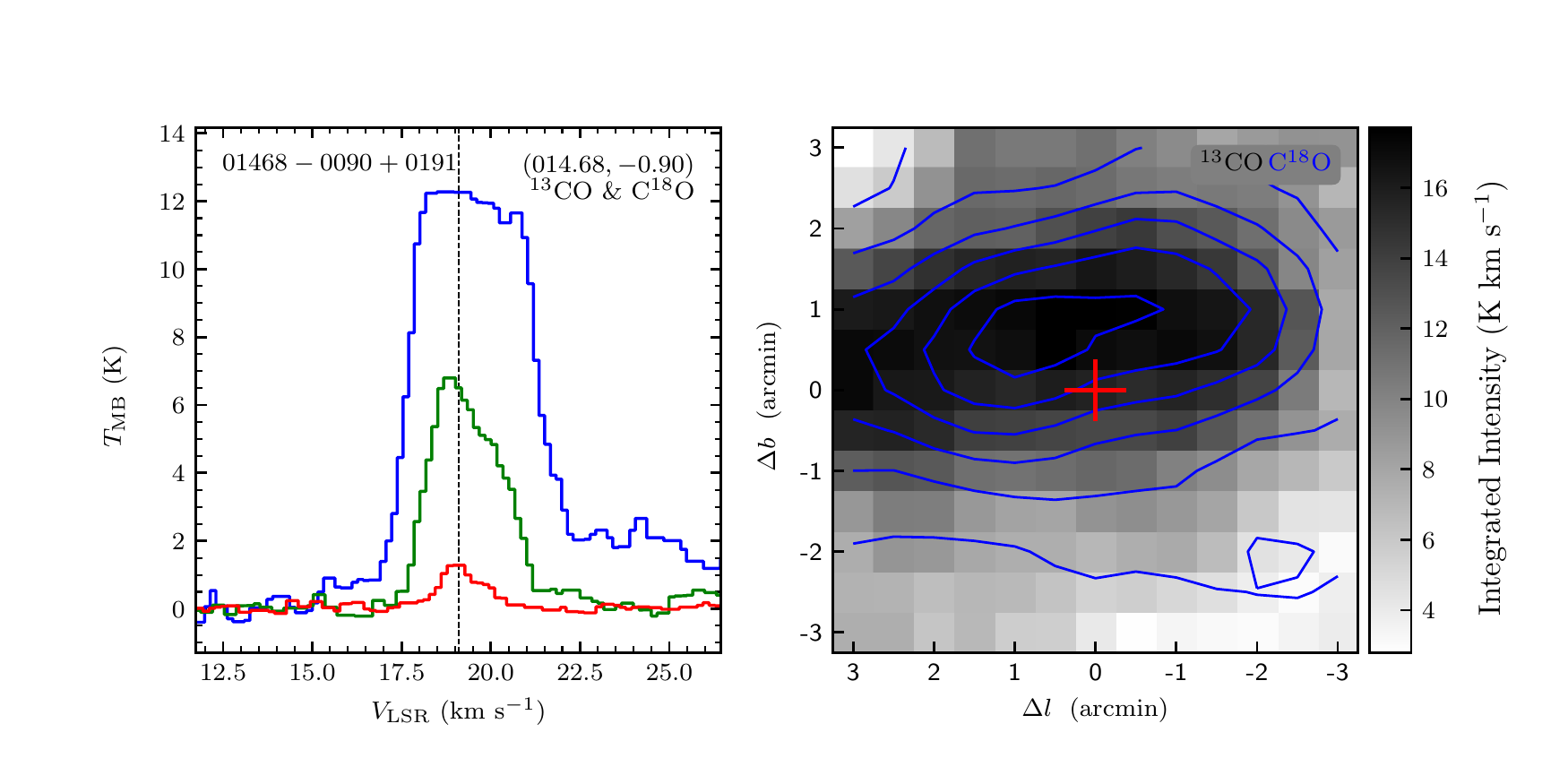}
\includegraphics[width=9.0cm,angle=0]{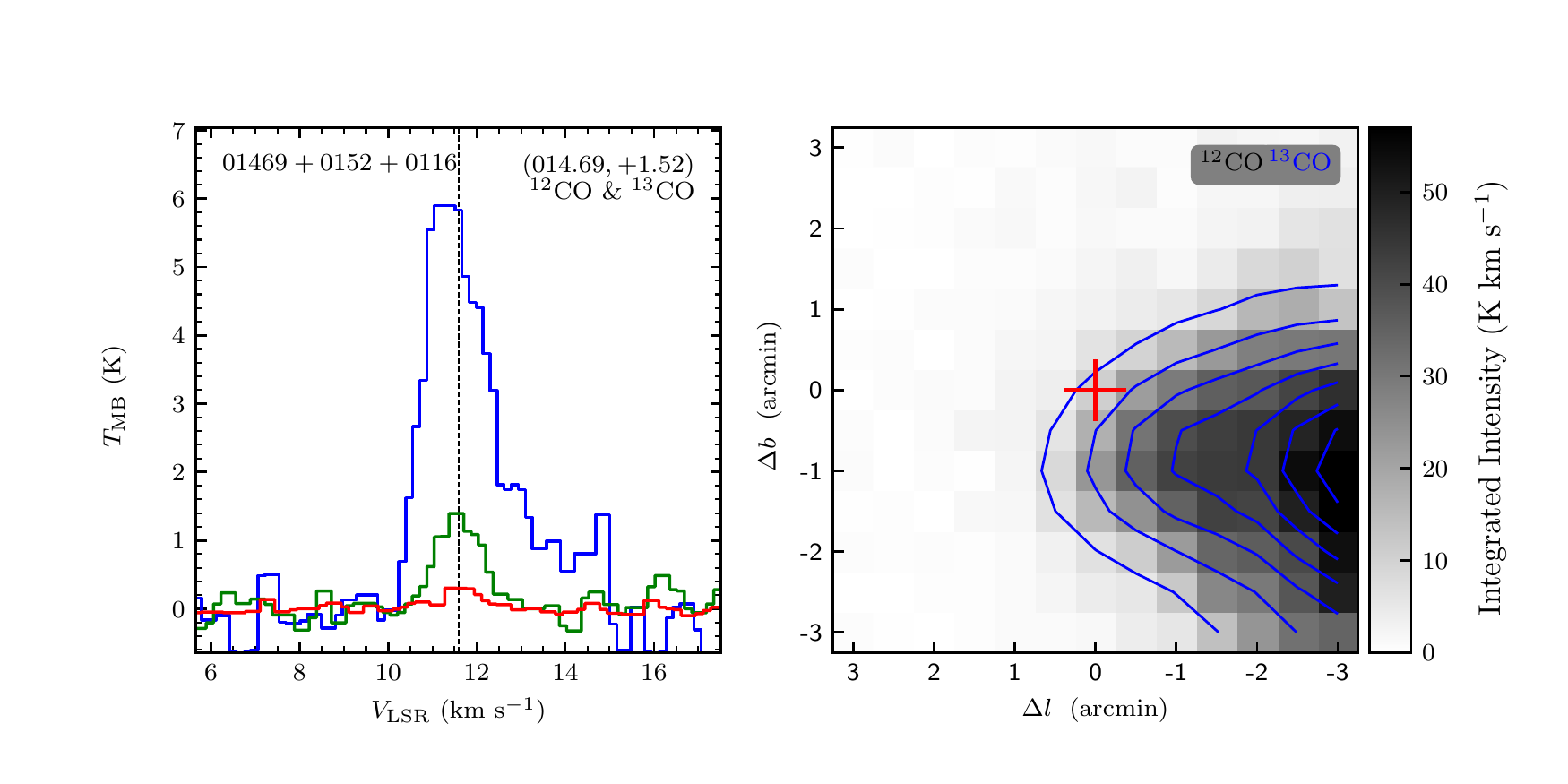}
\vspace{-0.5cm}

\includegraphics[width=9.0cm,angle=0]{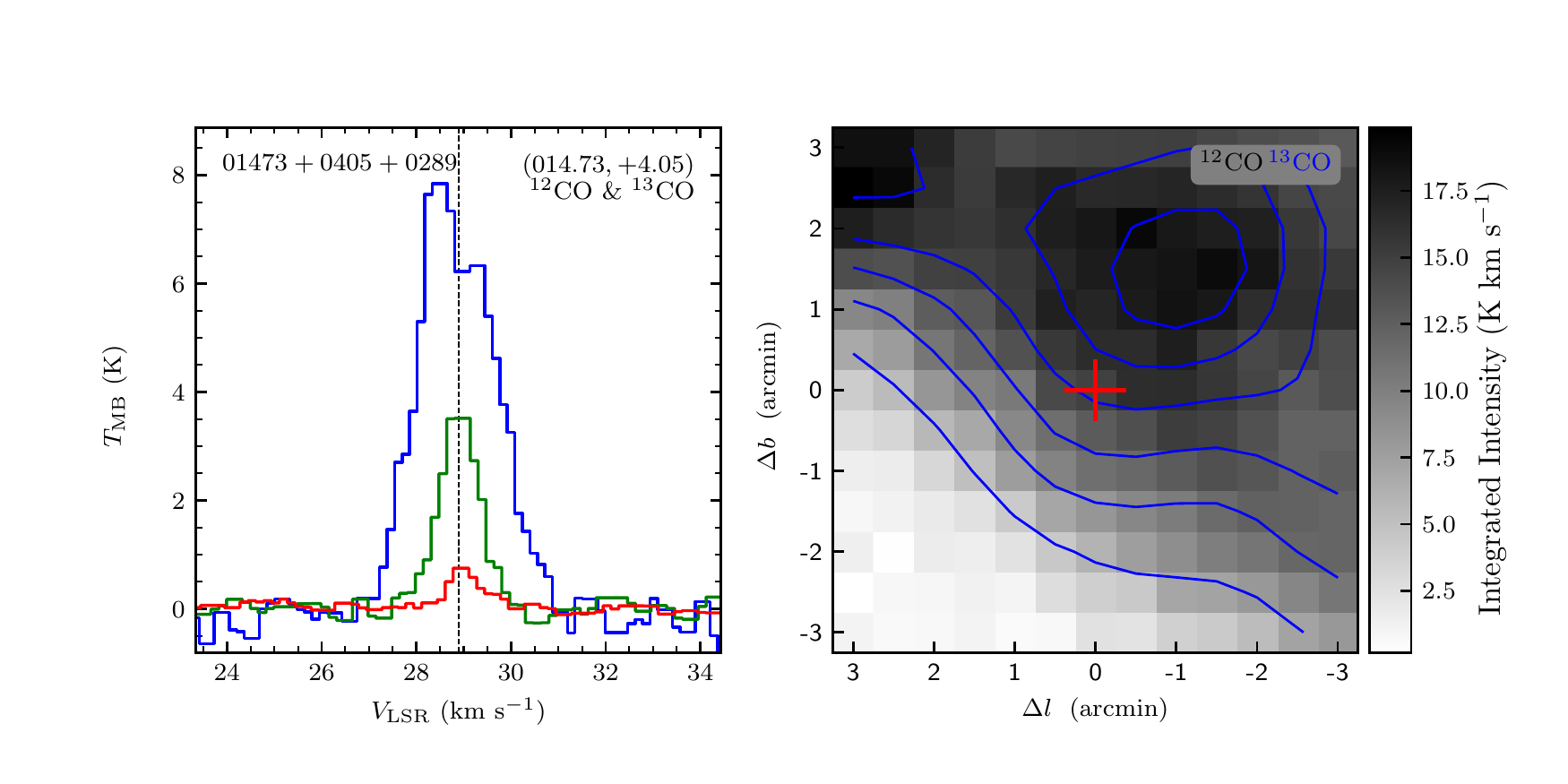}
\includegraphics[width=9.0cm,angle=0]{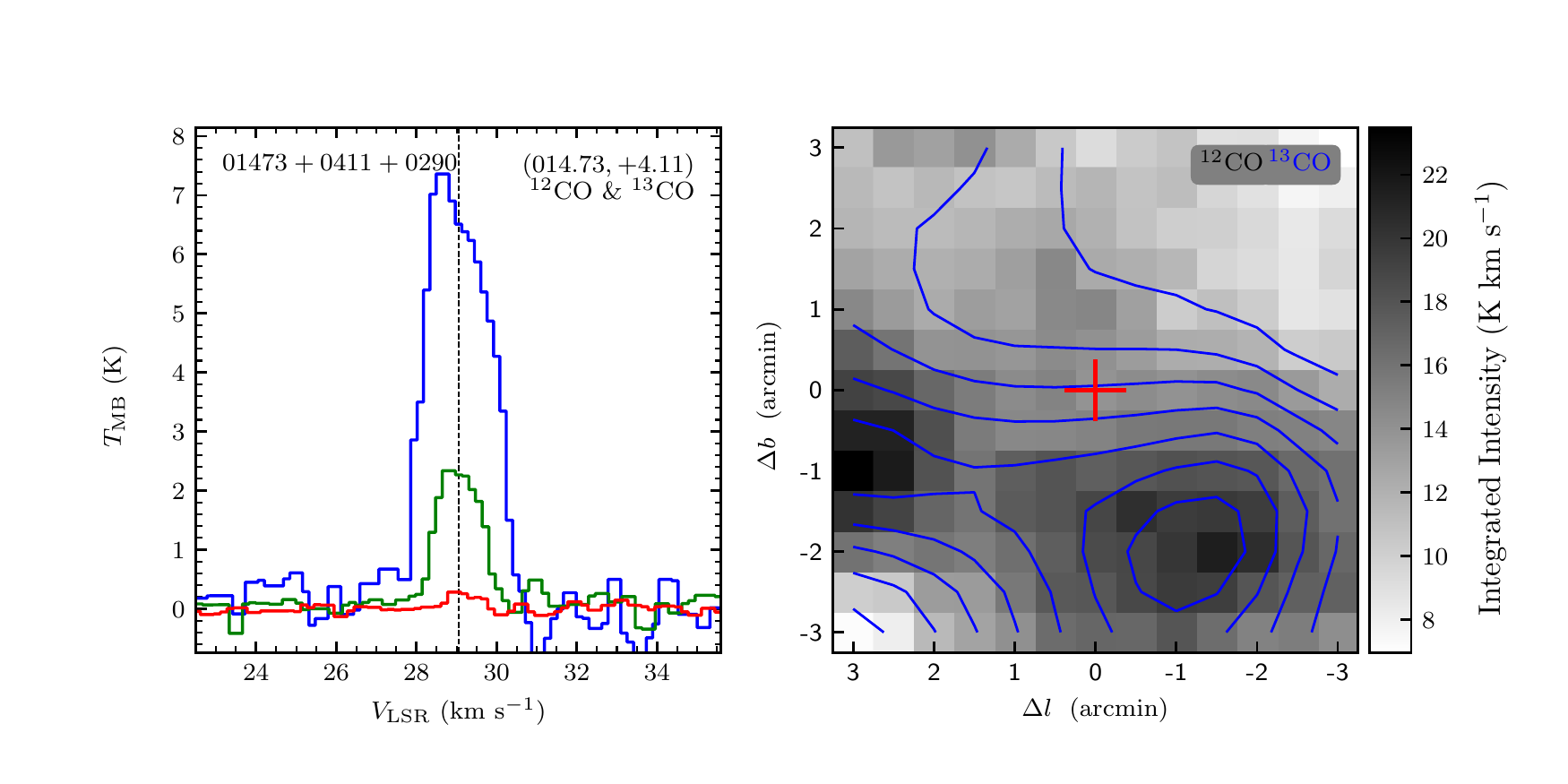}
\vspace{-0.5cm}

\includegraphics[width=9.0cm,angle=0]{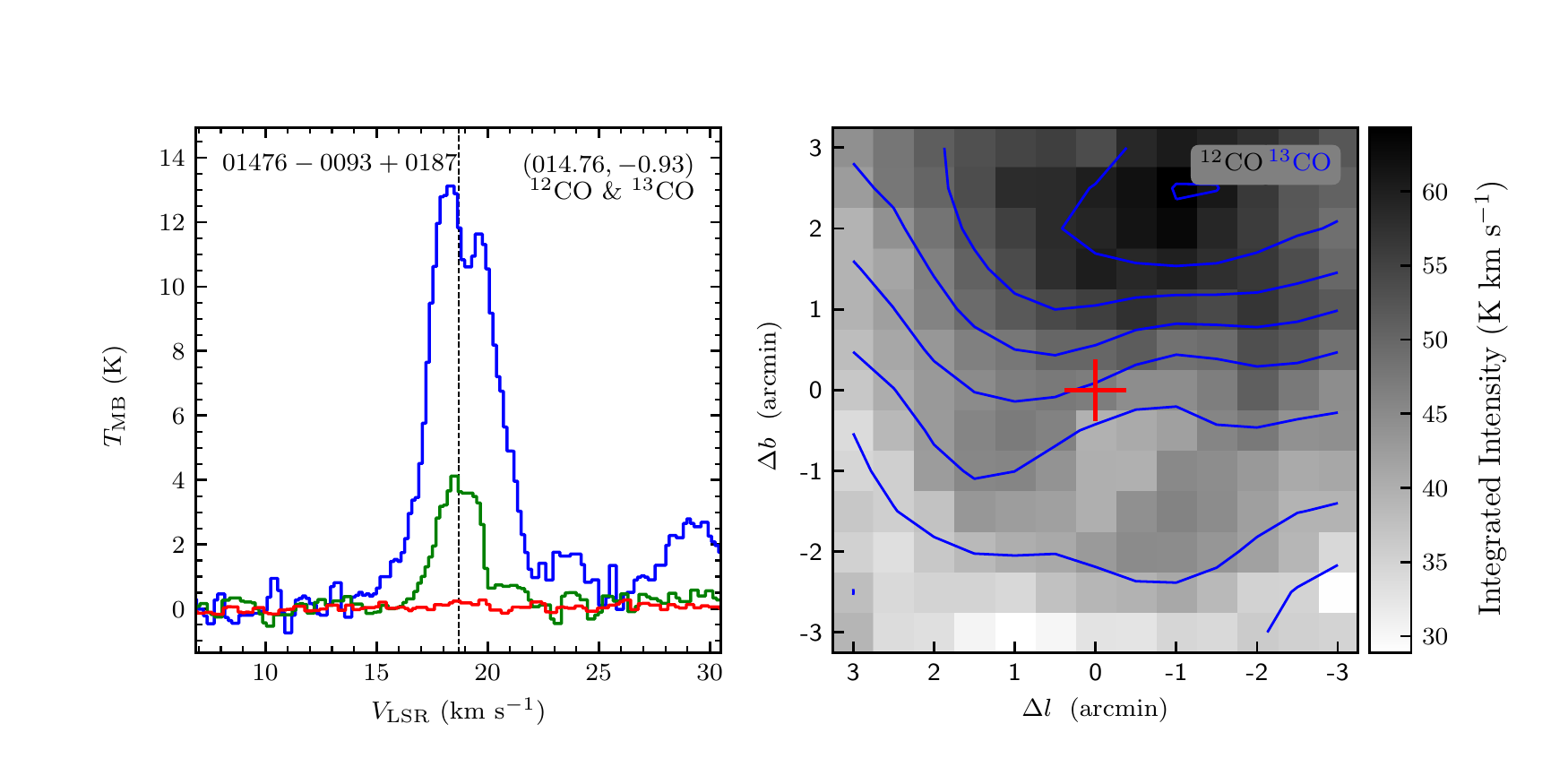}
\includegraphics[width=9.0cm,angle=0]{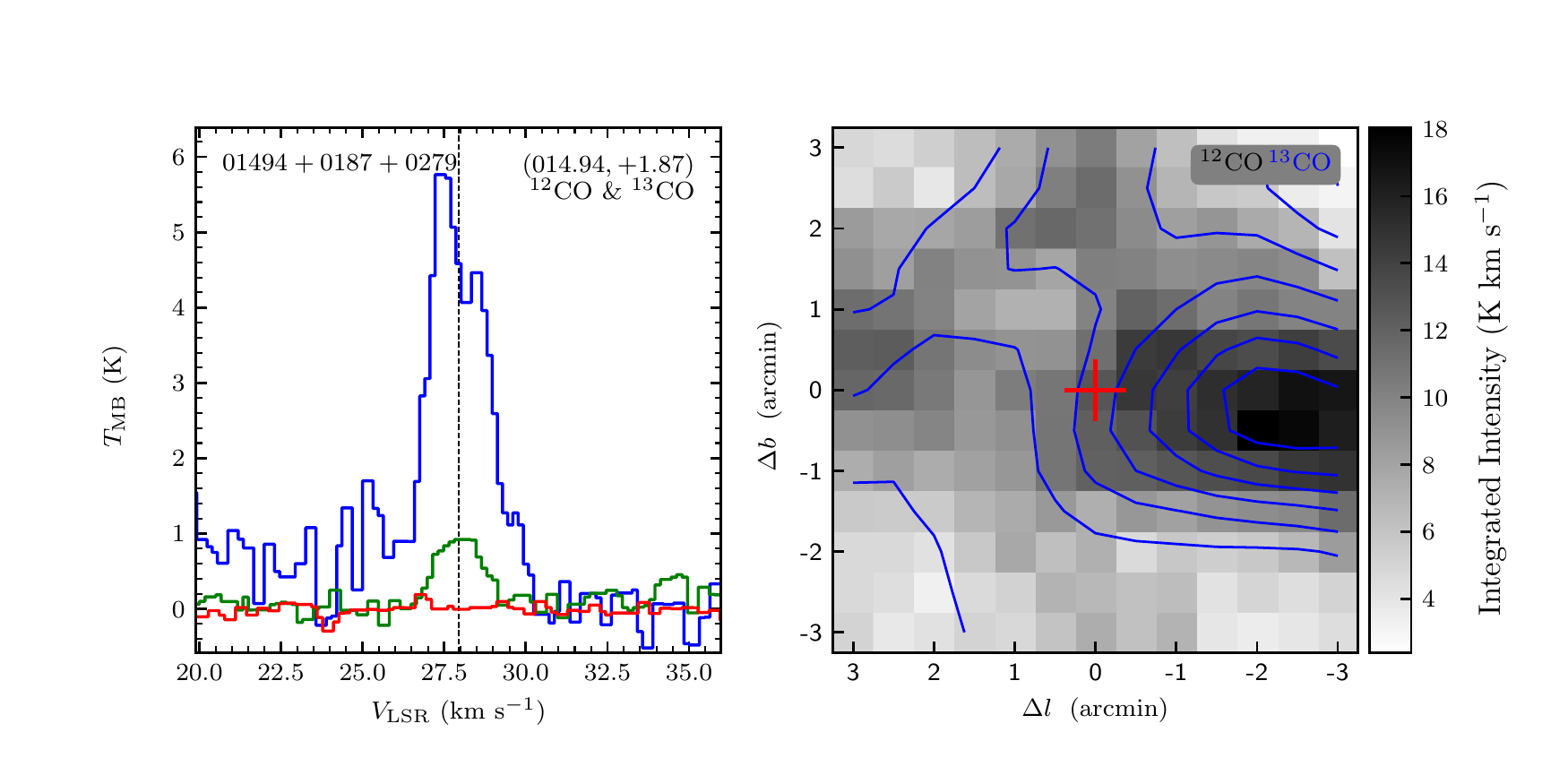}
\vspace{-0.5cm}

\includegraphics[width=9.0cm,angle=0]{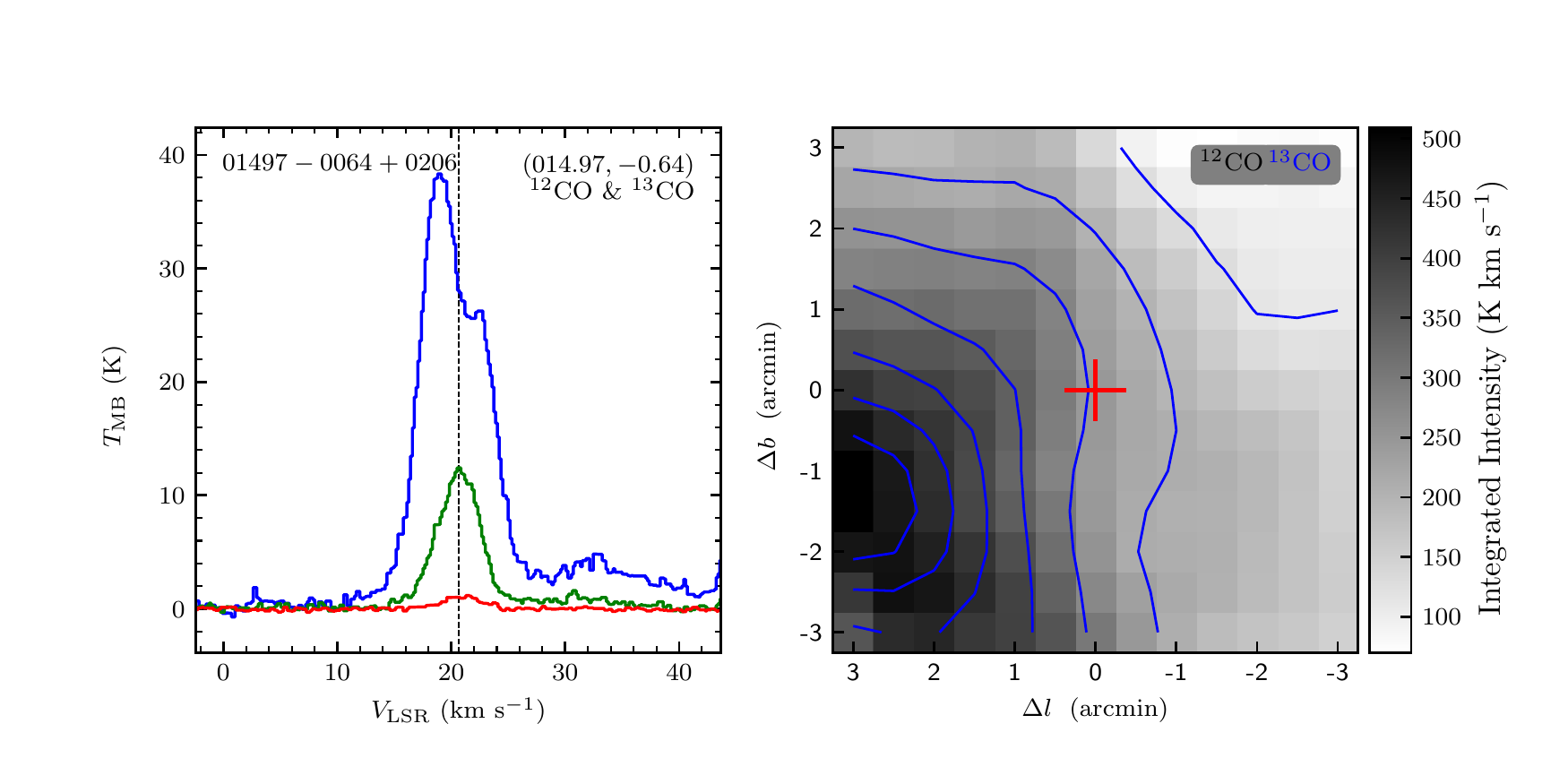}
\includegraphics[width=9.0cm,angle=0]{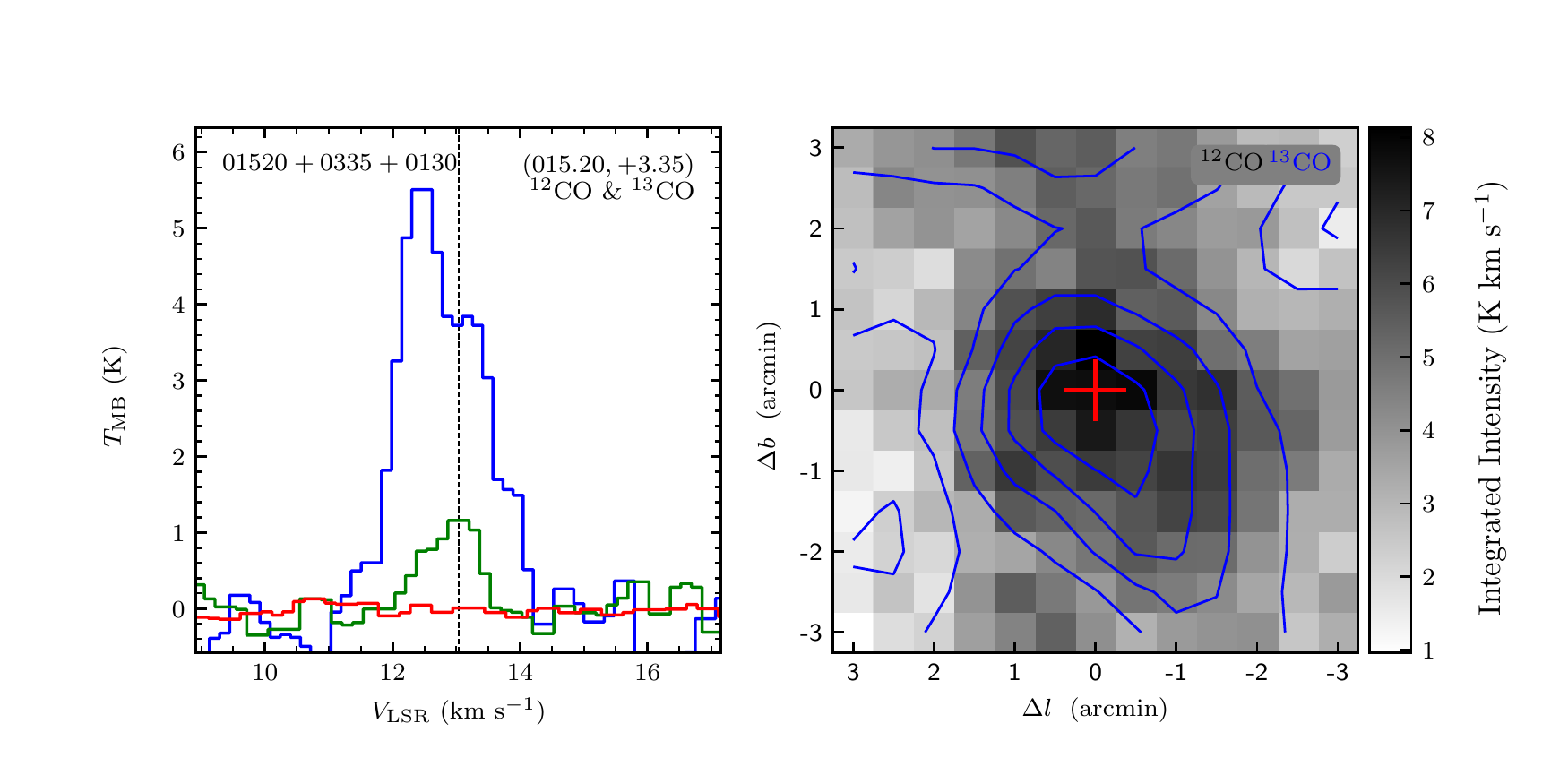}
\end{figure}
\clearpage

\begin{figure}
\includegraphics[width=9.0cm,angle=0]{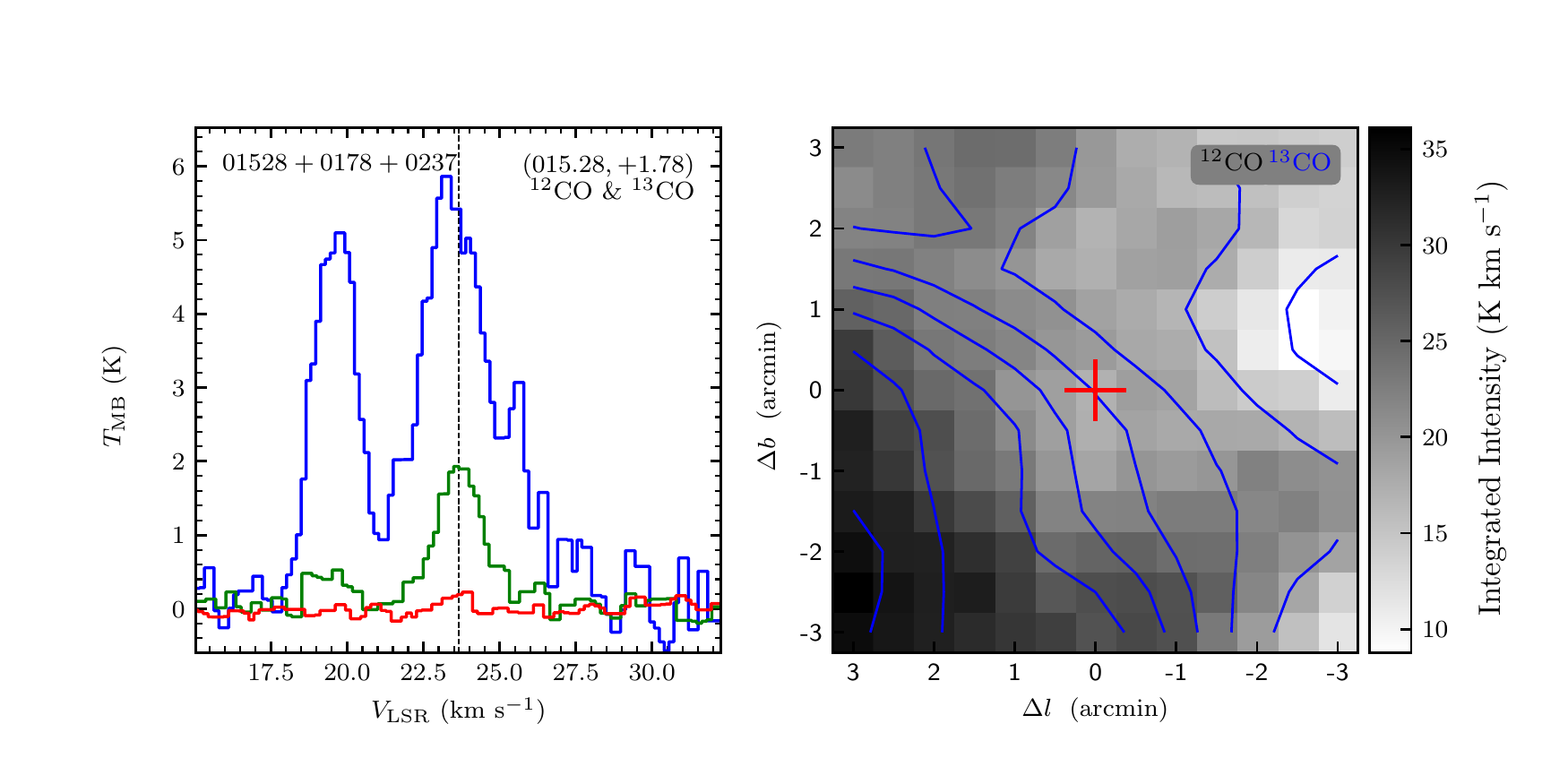}
\includegraphics[width=9.0cm,angle=0]{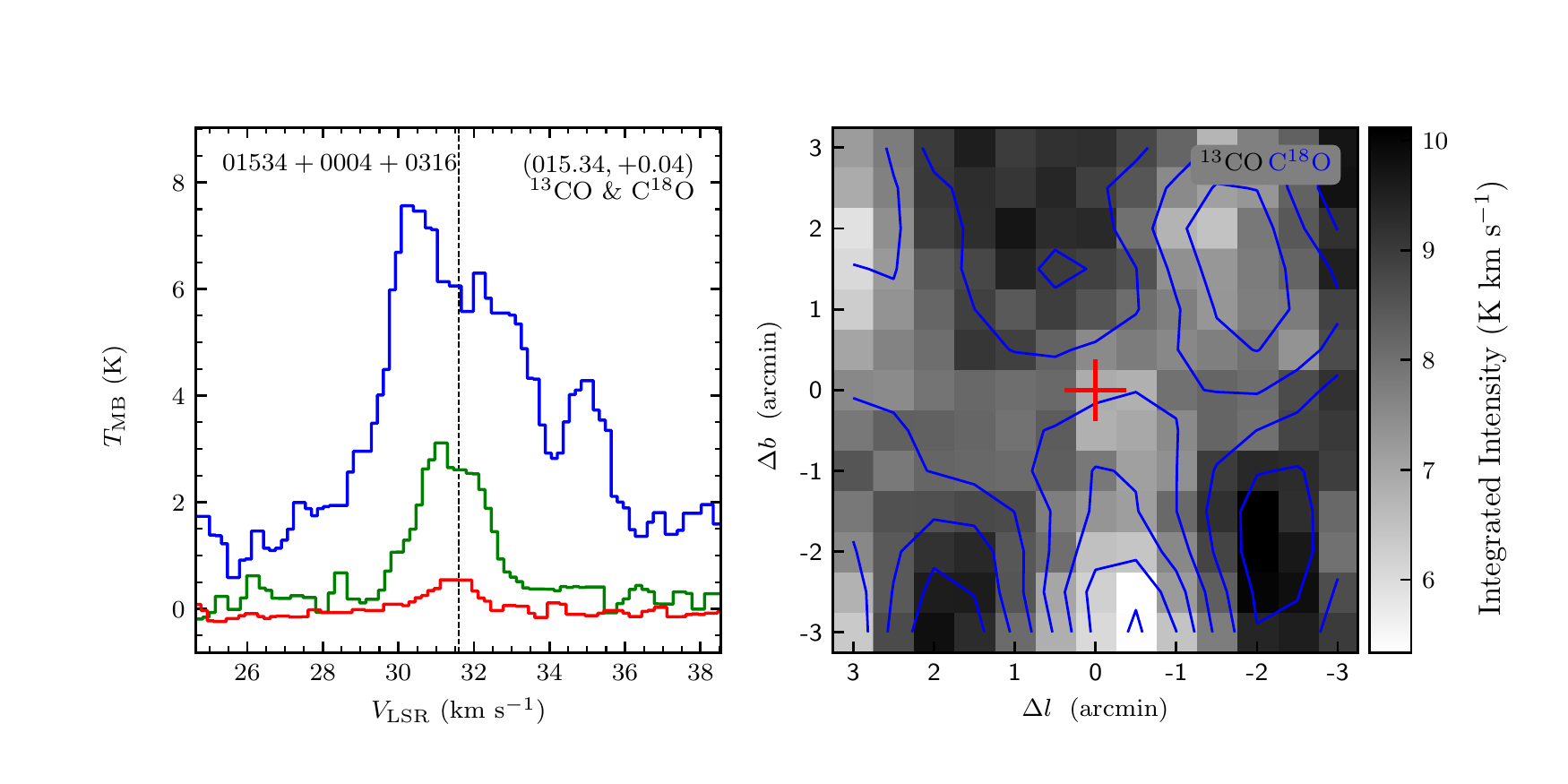}
\vspace{-0.5cm}

\includegraphics[width=9.0cm,angle=0]{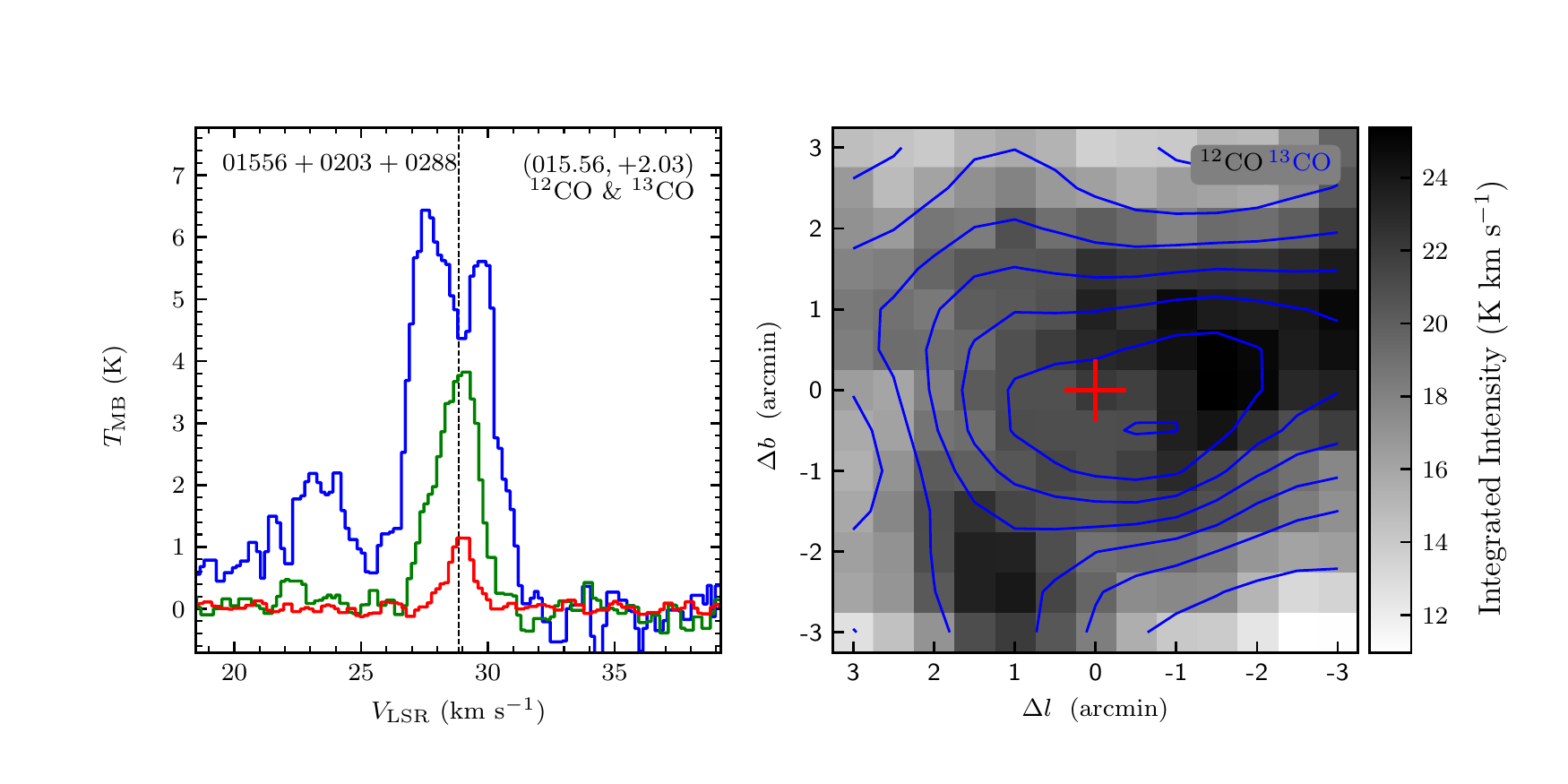}
\includegraphics[width=9.0cm,angle=0]{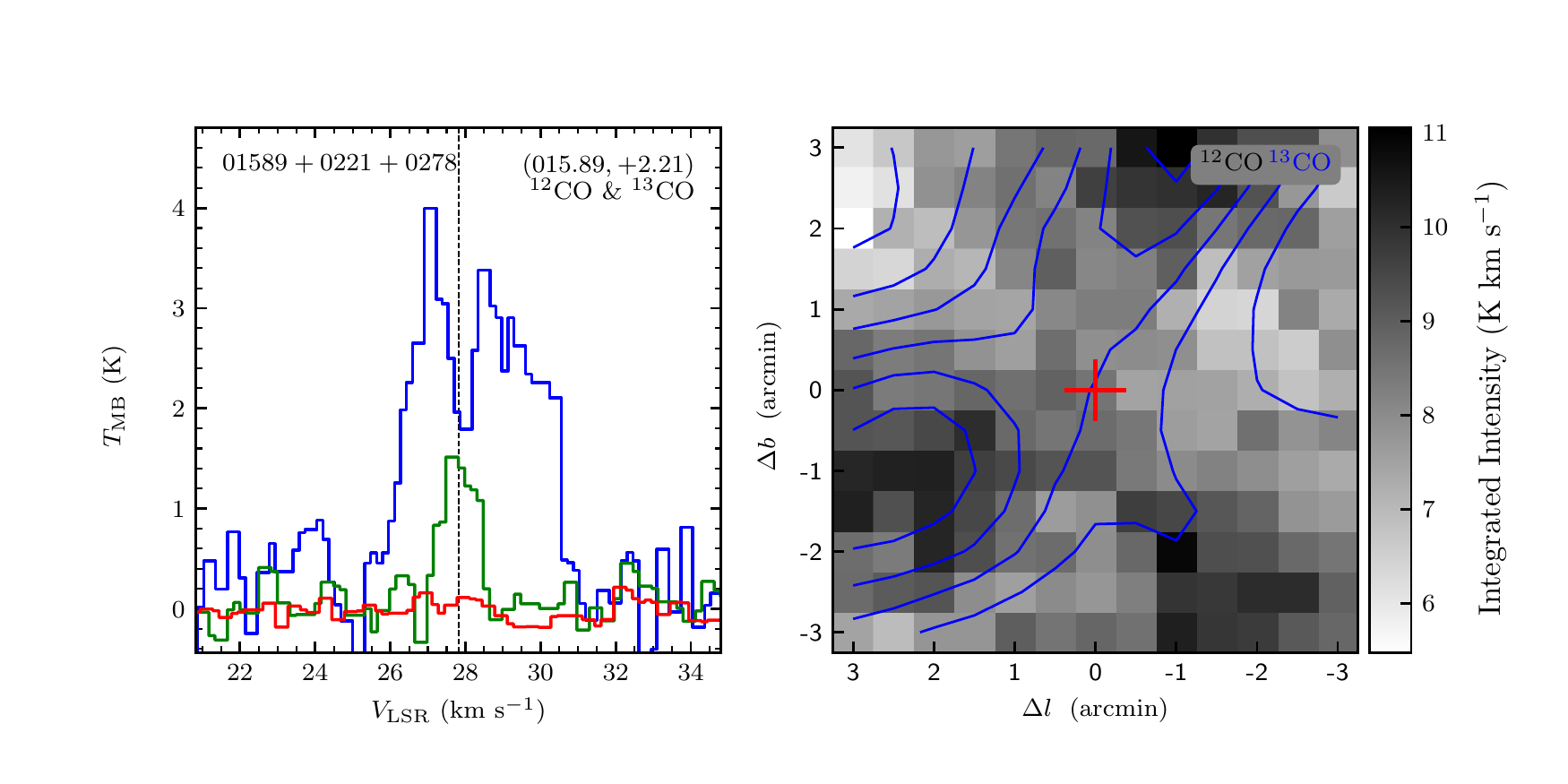}
\vspace{-0.5cm}

\includegraphics[width=9.0cm,angle=0]{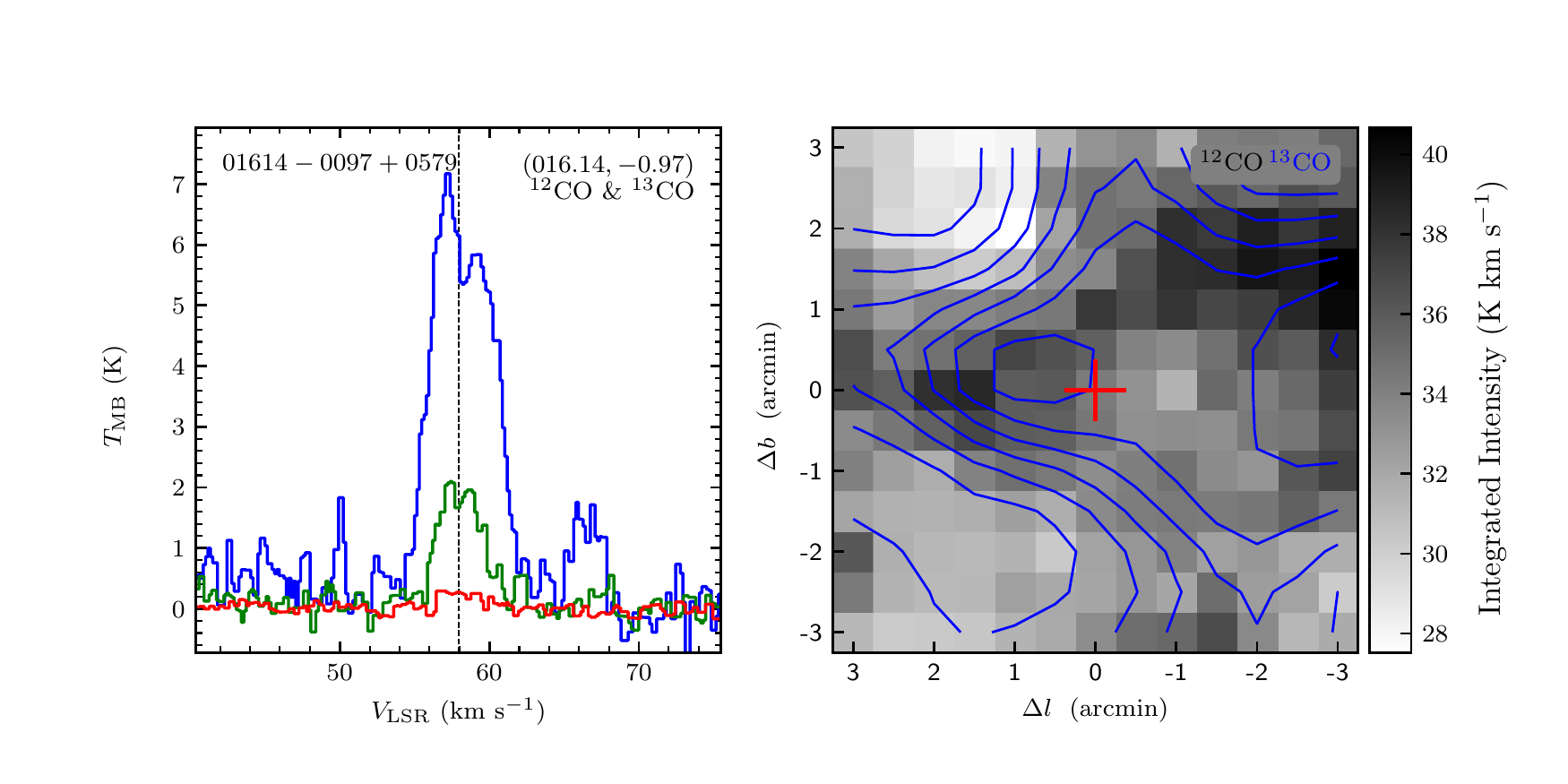}
\includegraphics[width=9.0cm,angle=0]{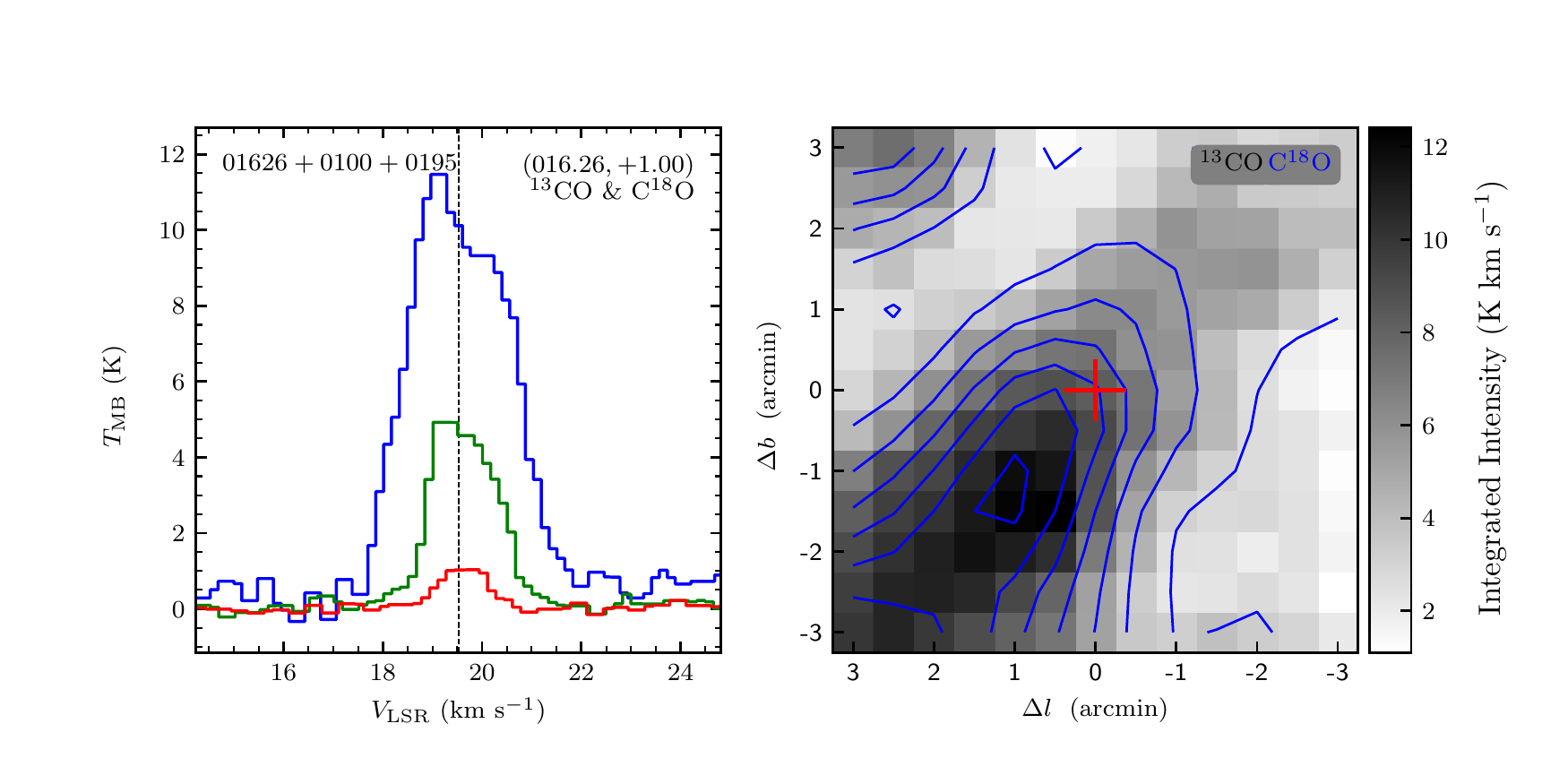}
\vspace{-0.5cm}

\includegraphics[width=9.0cm,angle=0]{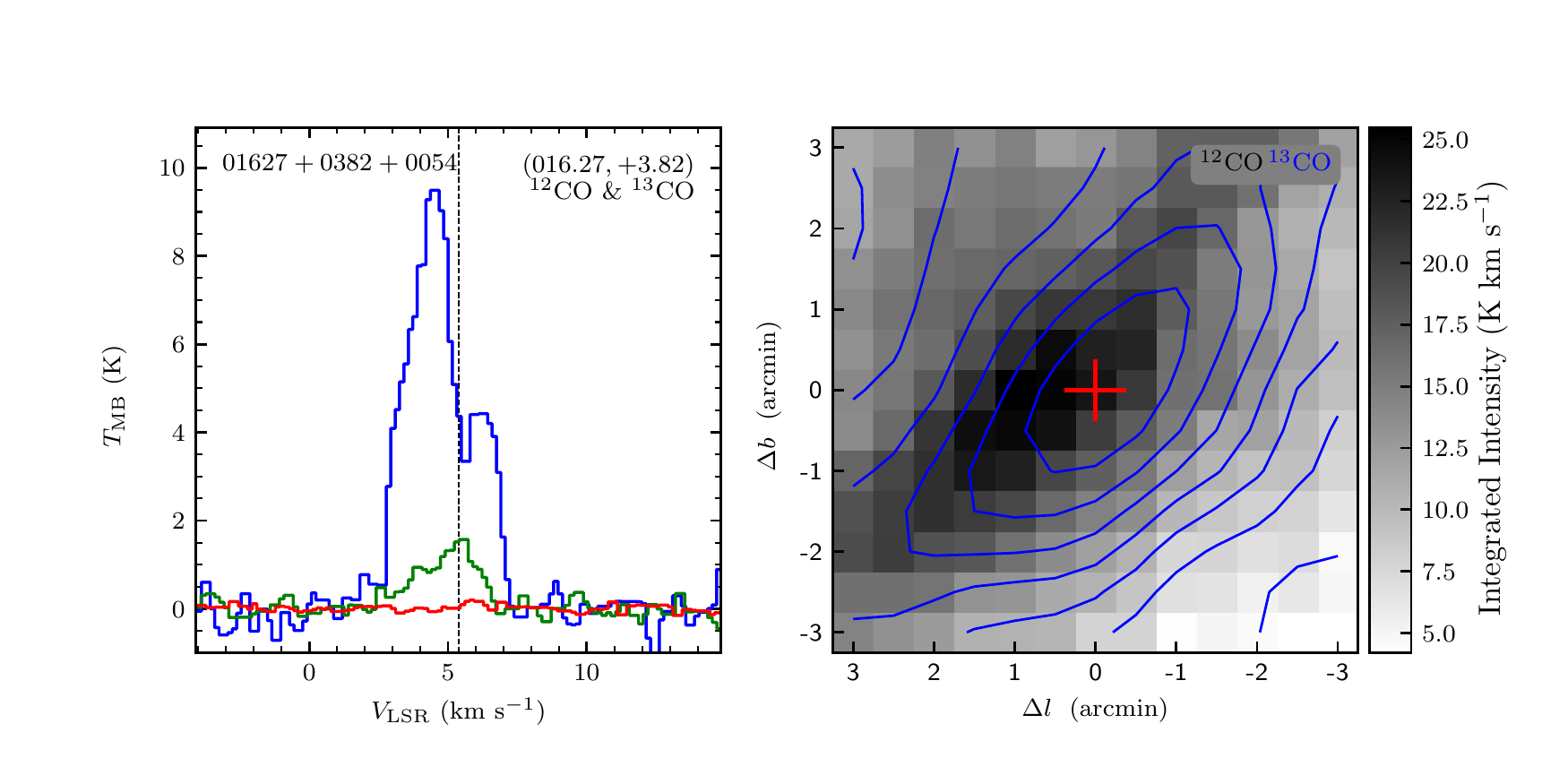}
\includegraphics[width=9.0cm,angle=0]{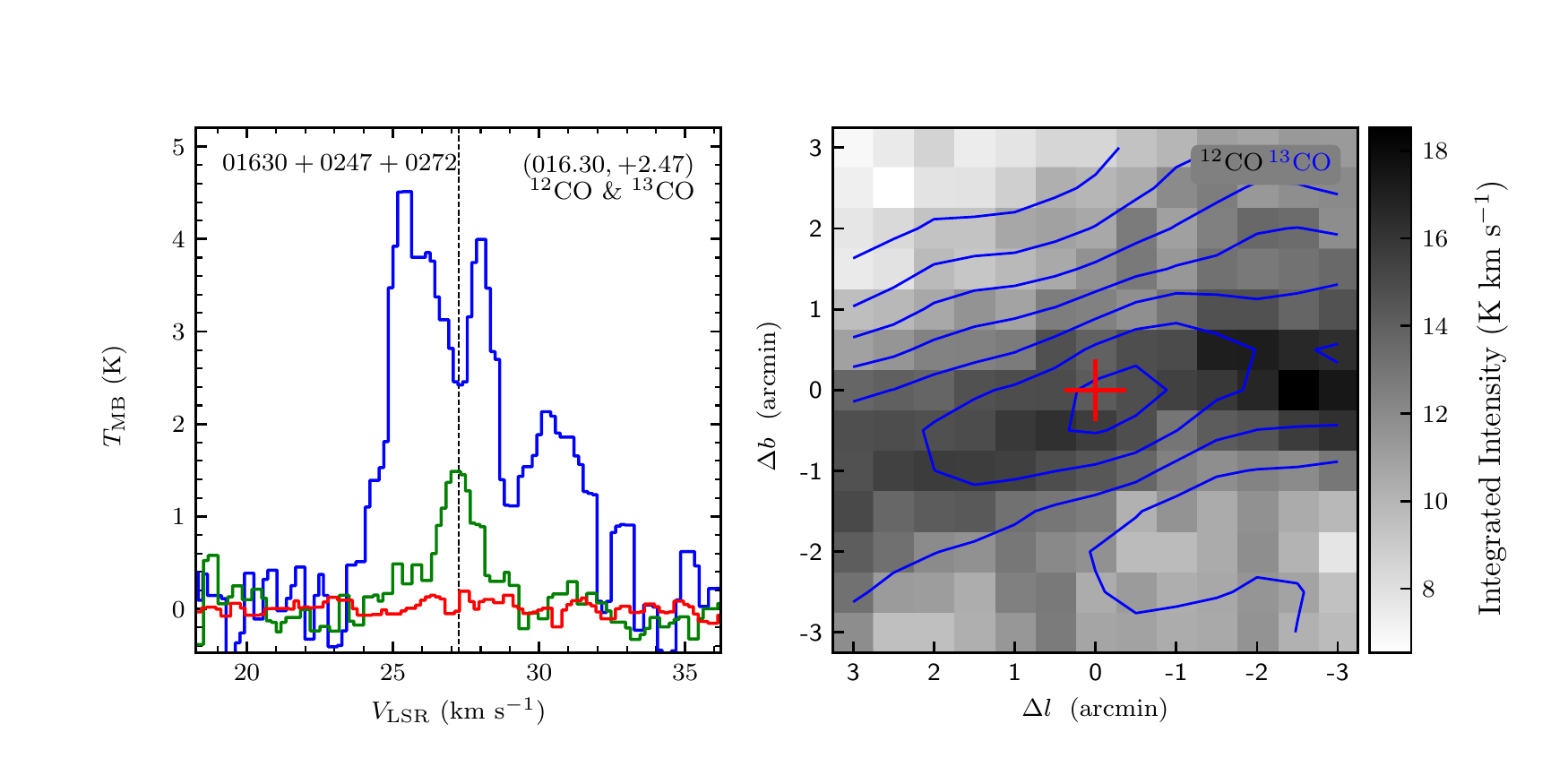}
\vspace{-0.5cm}

\includegraphics[width=9.0cm,angle=0]{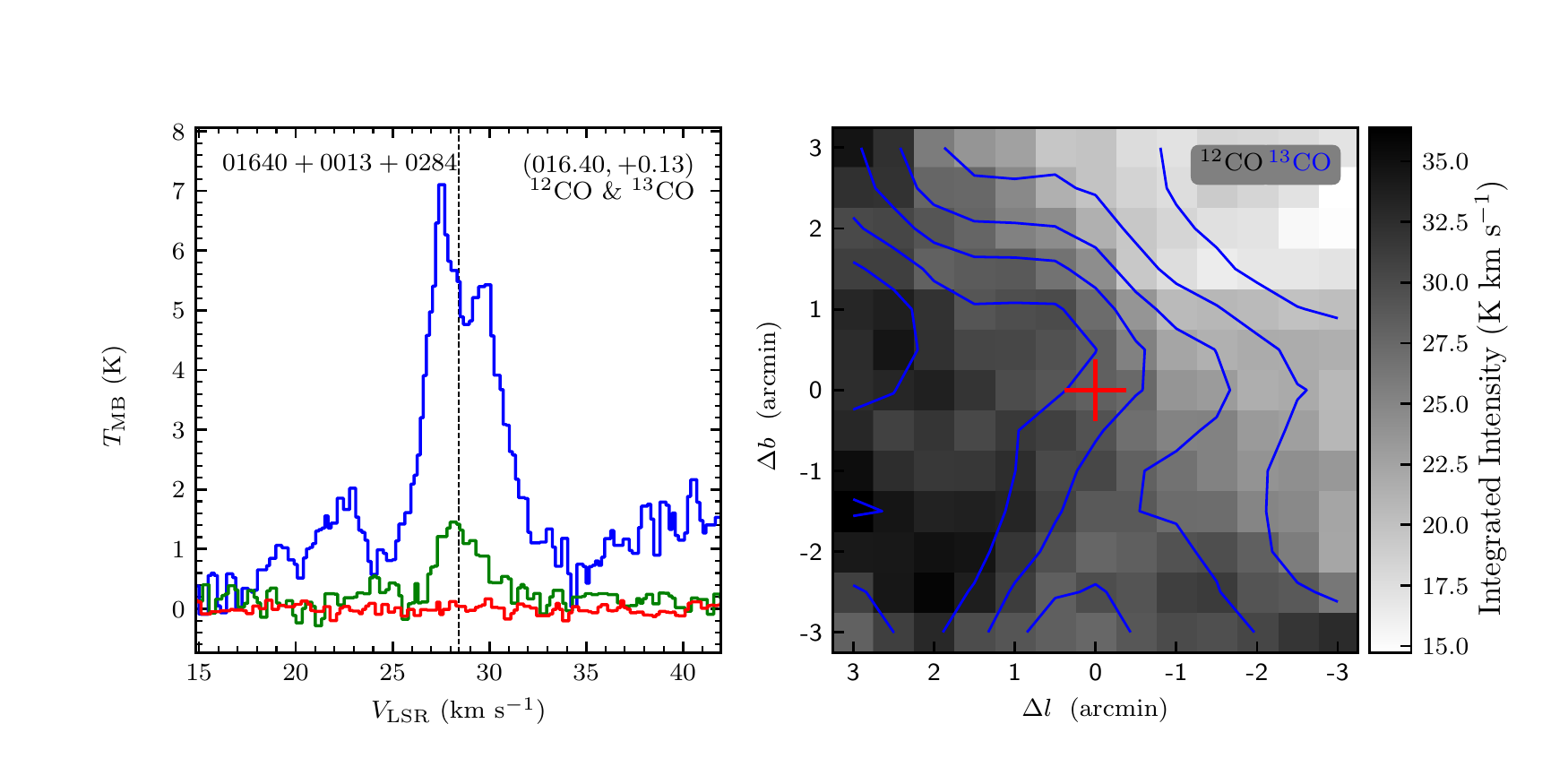}
\includegraphics[width=9.0cm,angle=0]{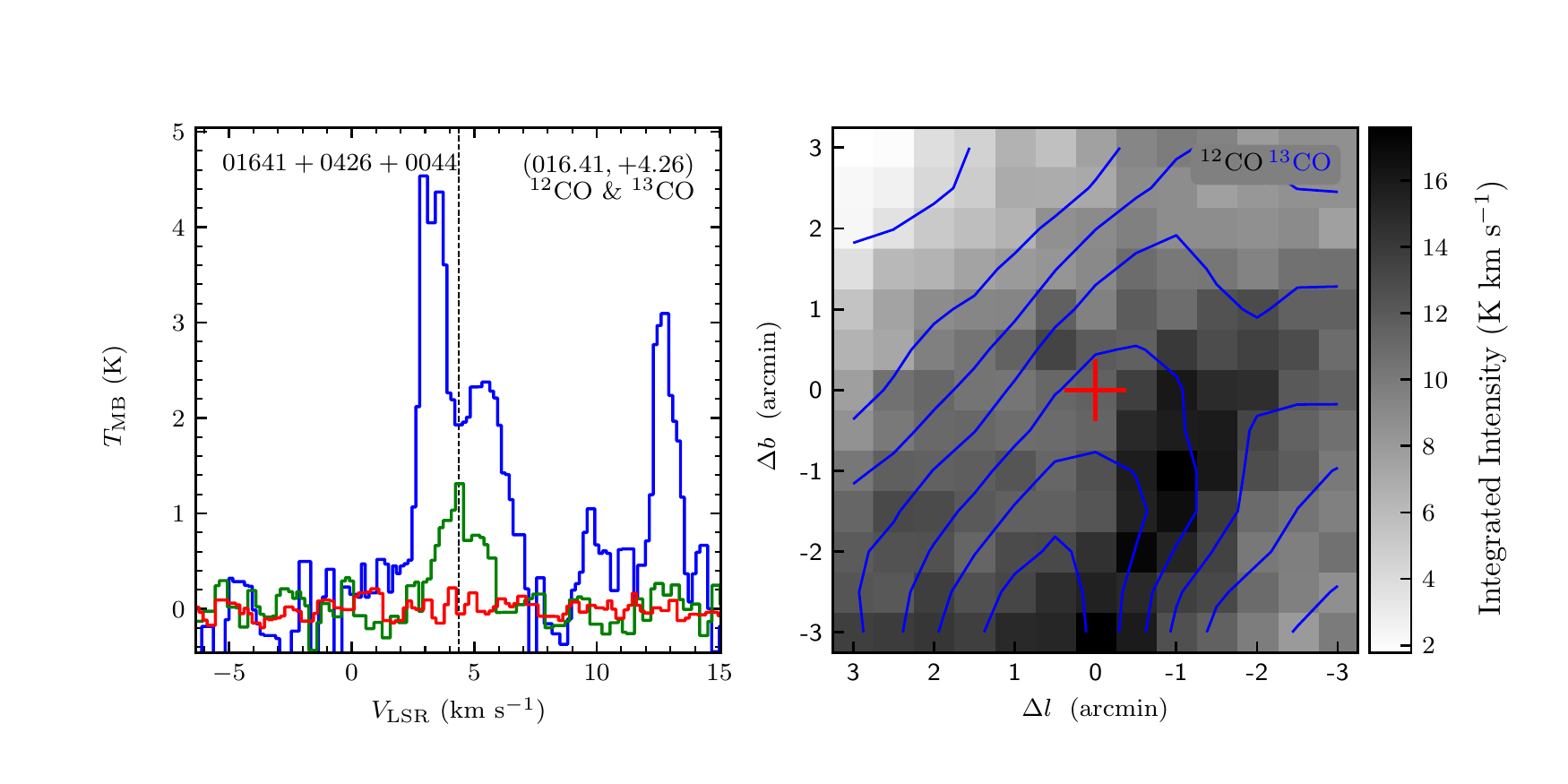}
\end{figure}
\clearpage

\begin{figure}
\includegraphics[width=9.0cm,angle=0]{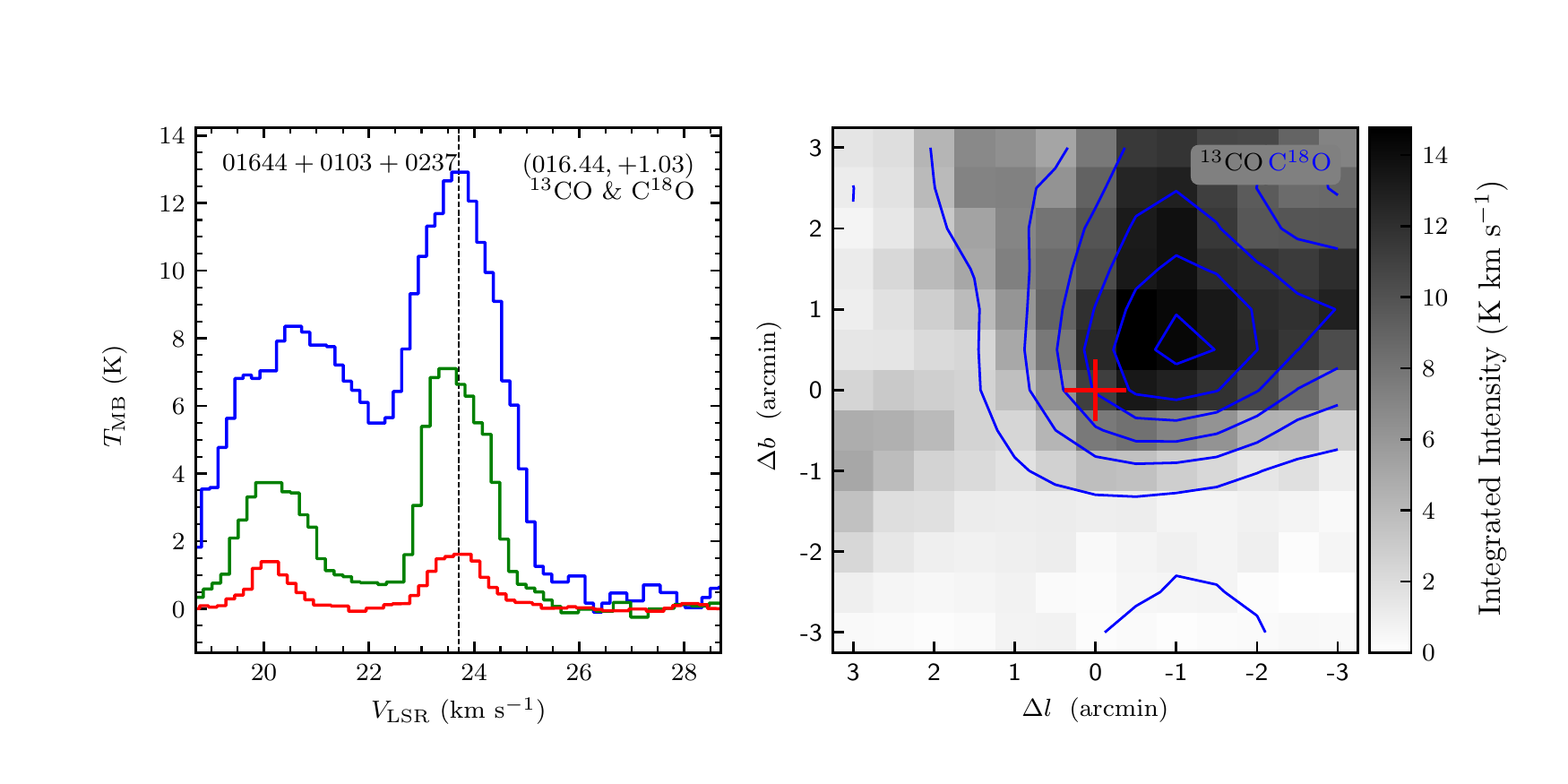}
\includegraphics[width=9.0cm,angle=0]{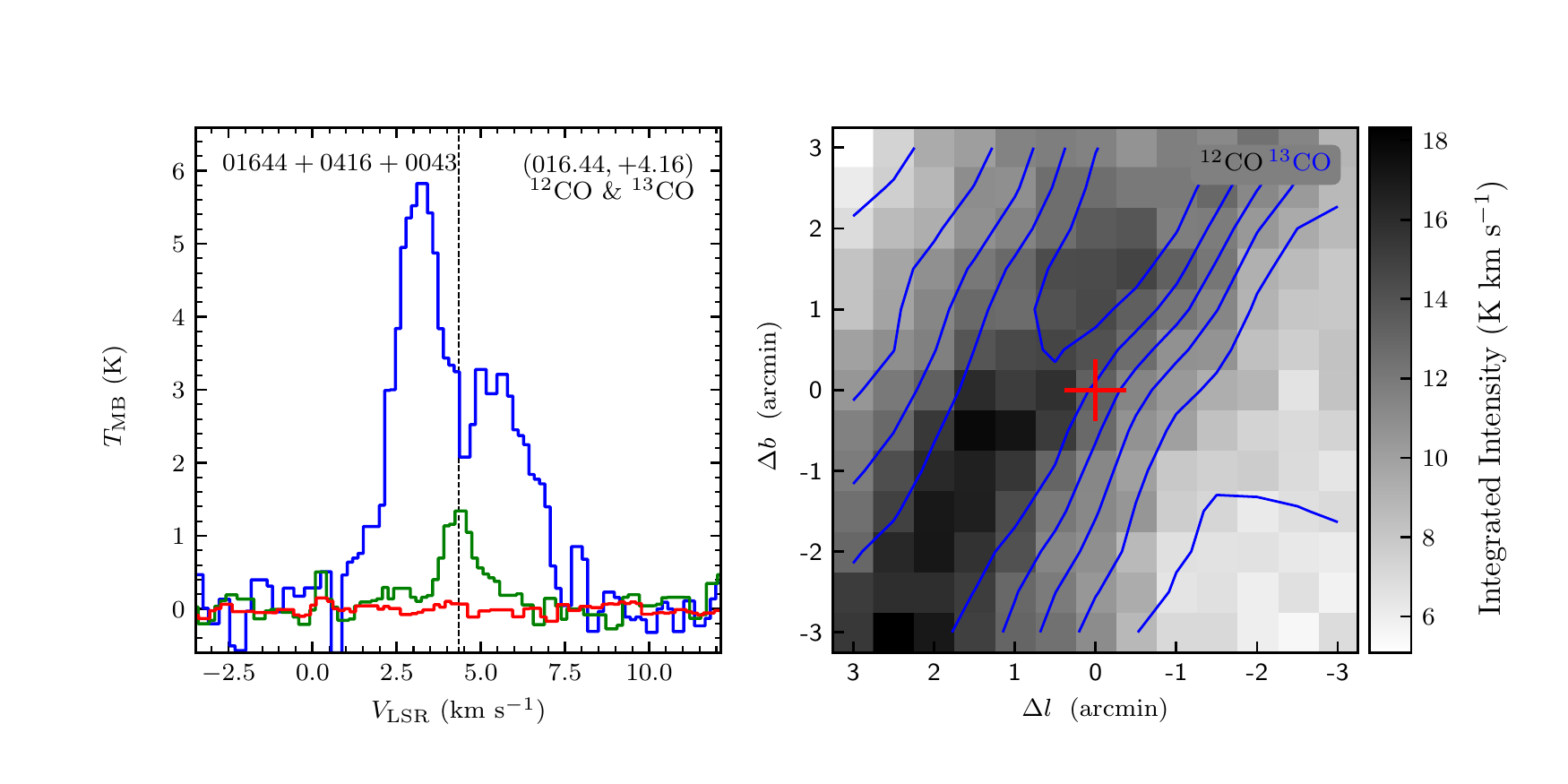}
\vspace{-0.5cm}

\includegraphics[width=9.0cm,angle=0]{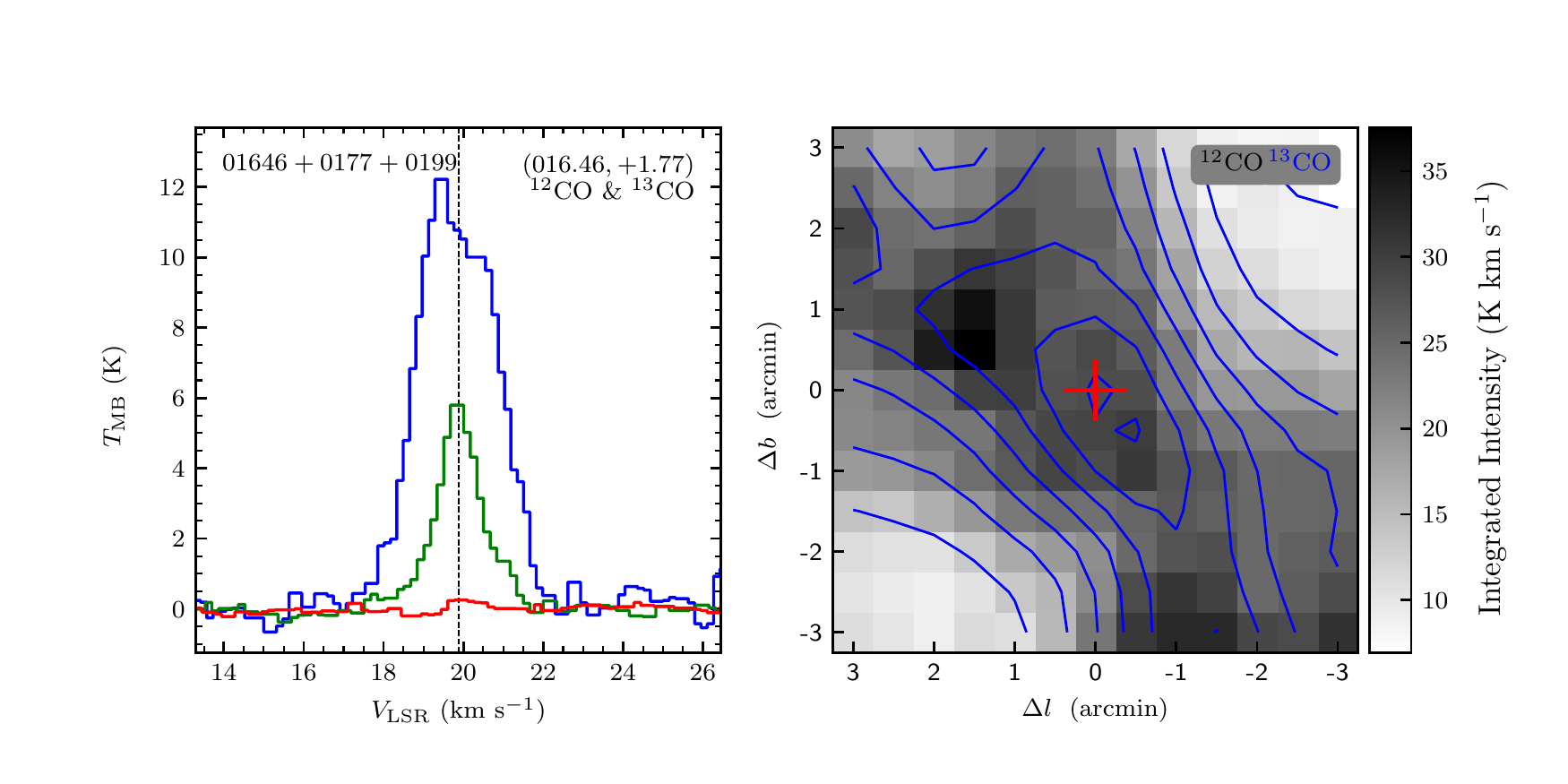}
\includegraphics[width=9.0cm,angle=0]{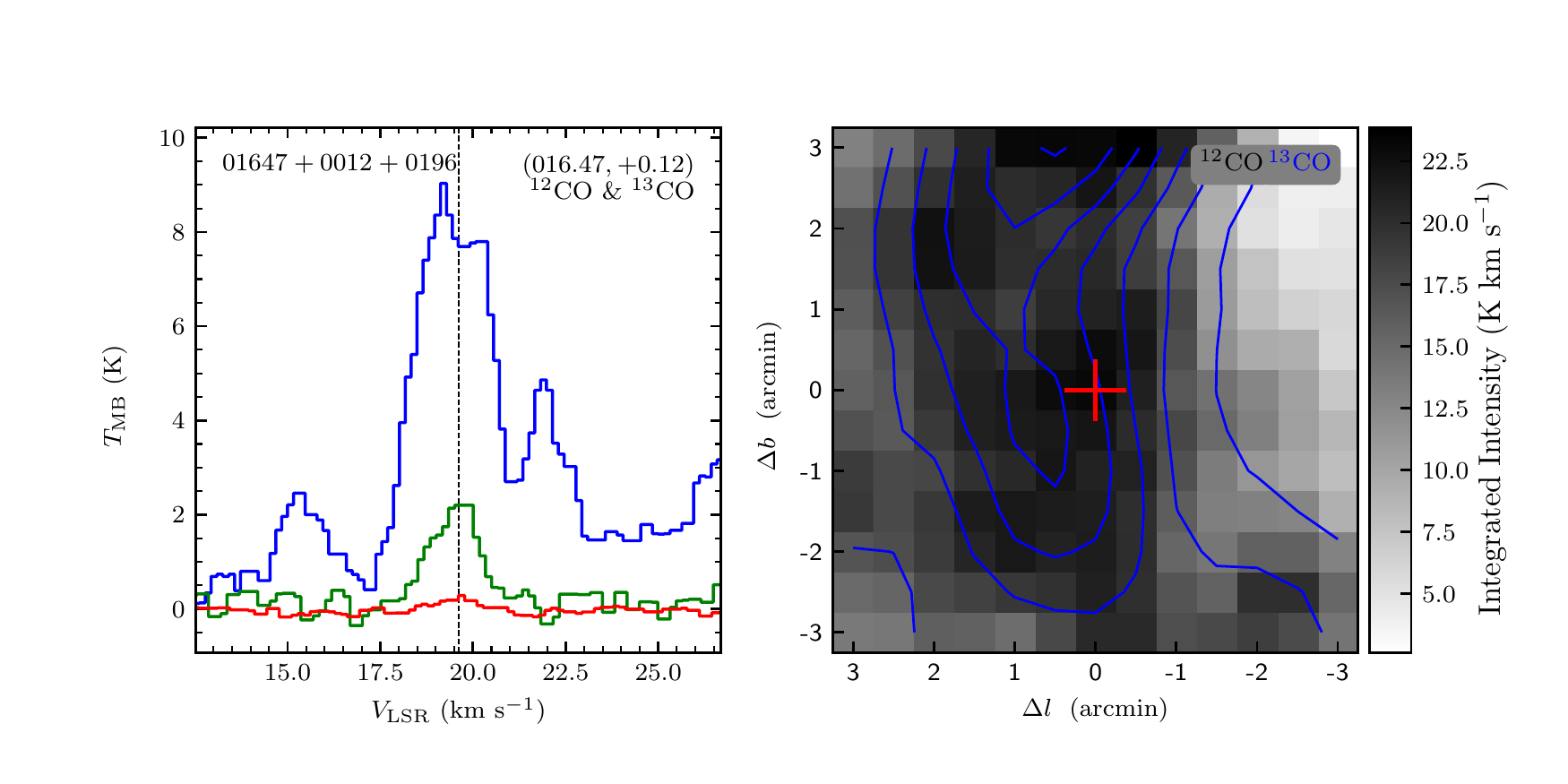}
\vspace{-0.5cm}

\includegraphics[width=9.0cm,angle=0]{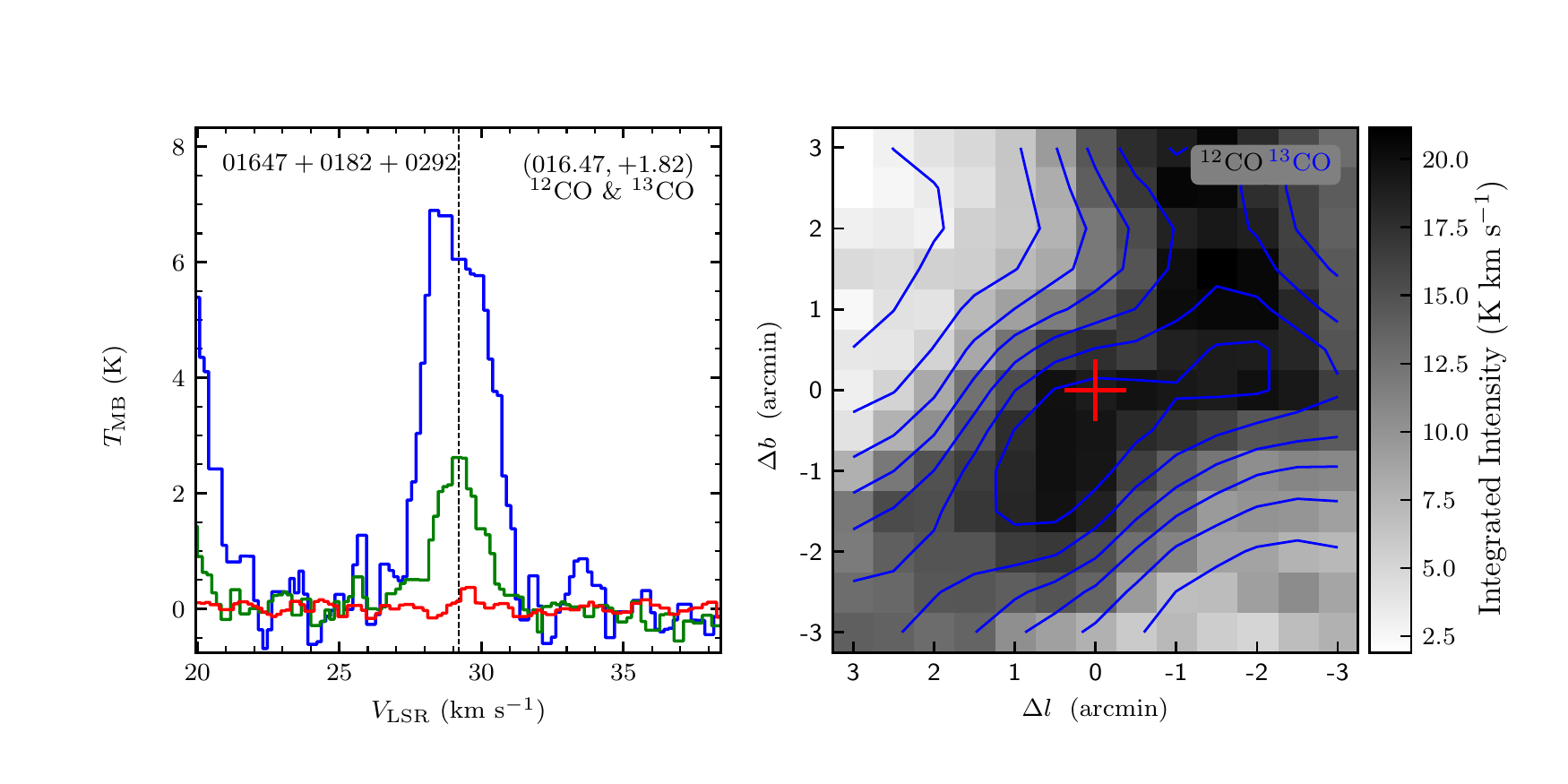}
\includegraphics[width=9.0cm,angle=0]{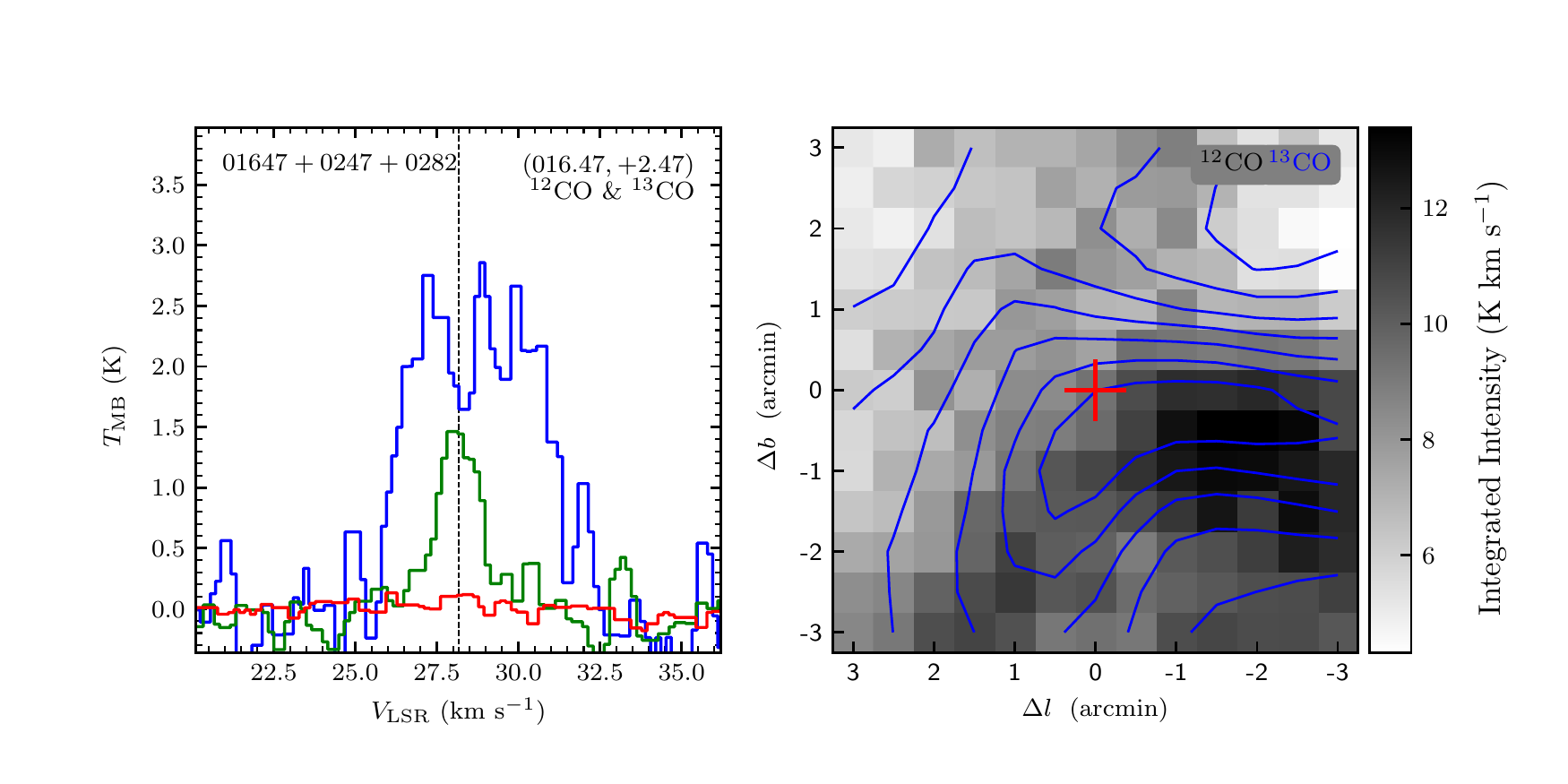}
\vspace{-0.5cm}

\includegraphics[width=9.0cm,angle=0]{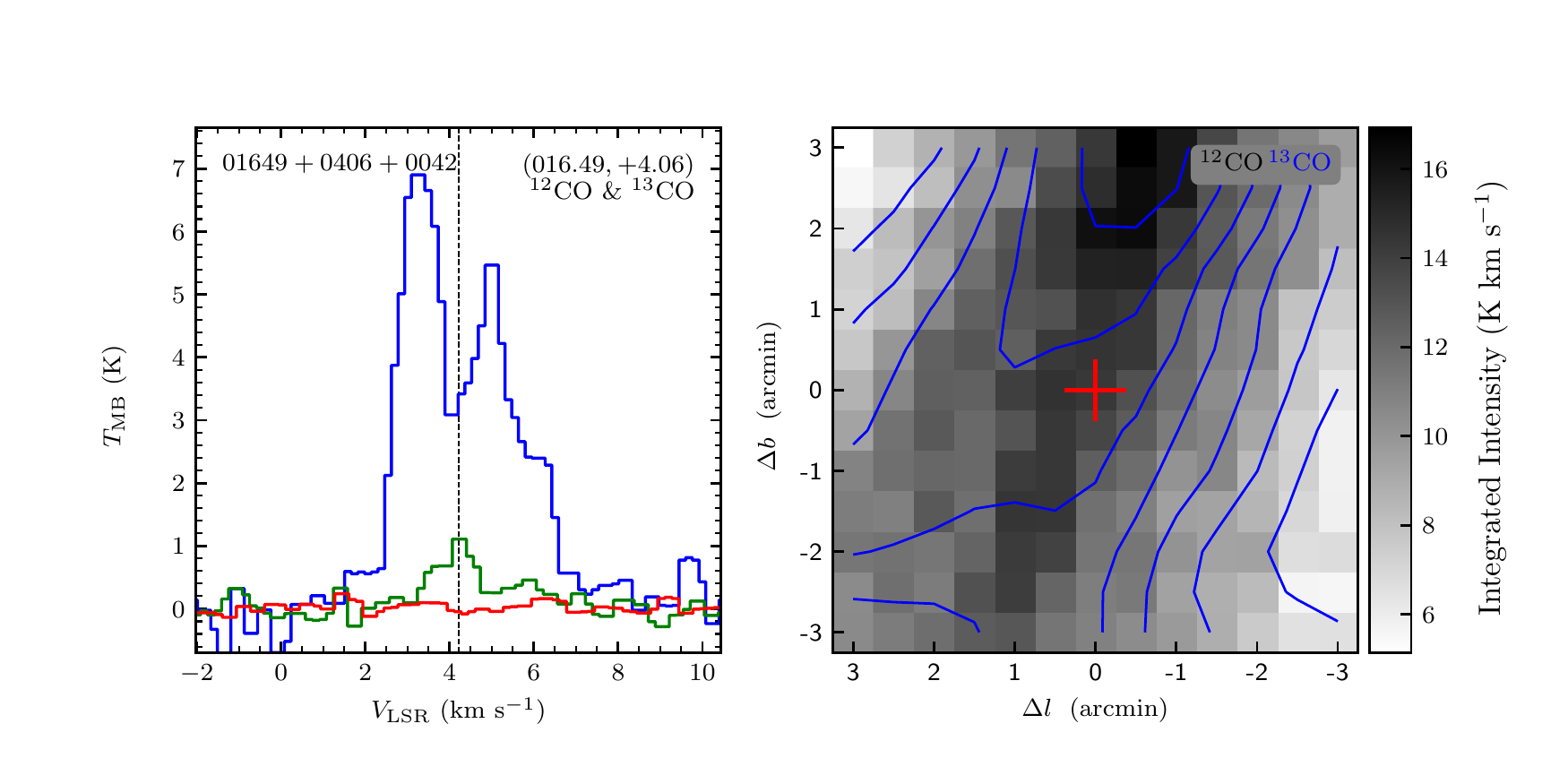}
\includegraphics[width=9.0cm,angle=0]{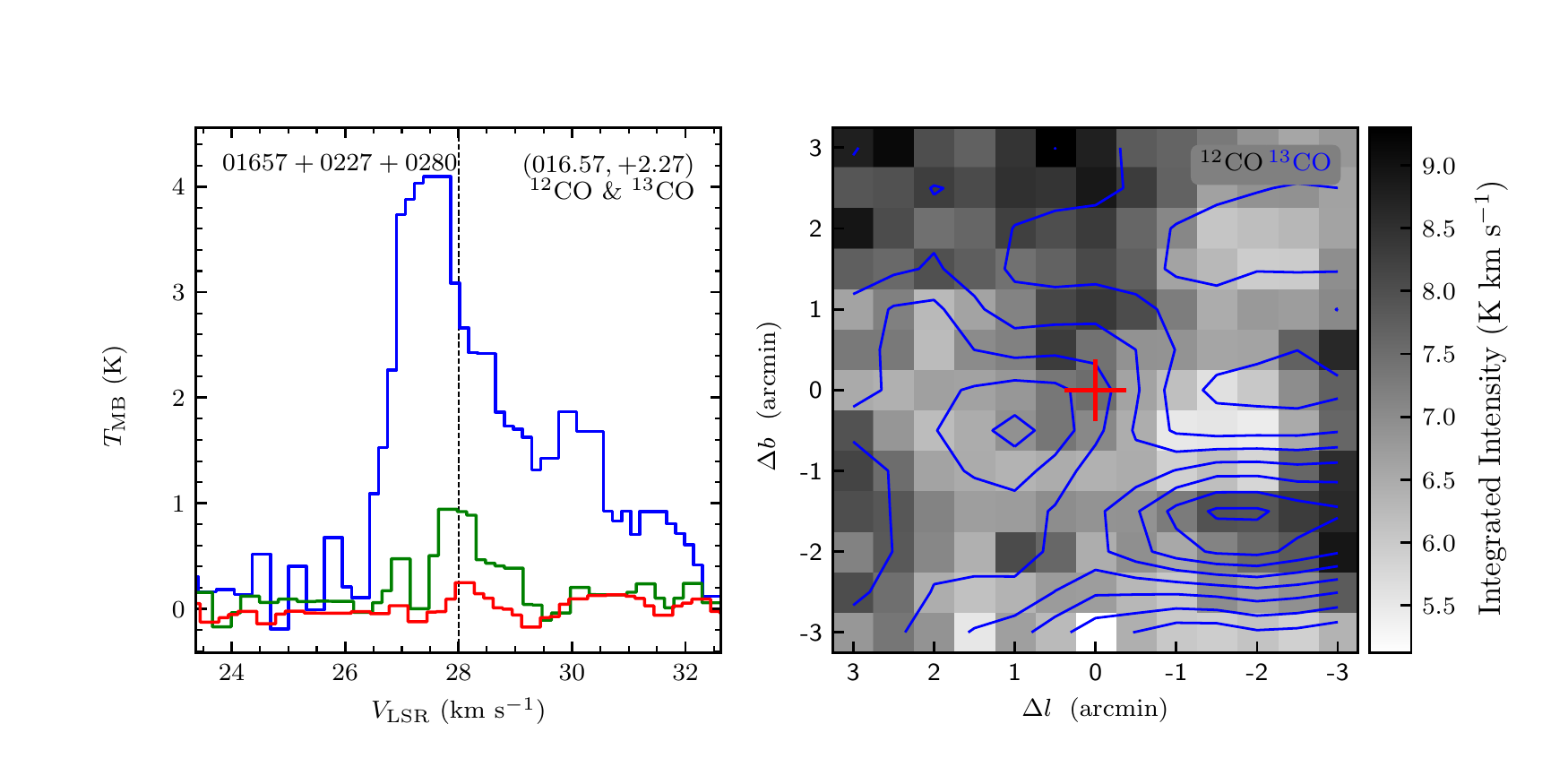}
\vspace{-0.5cm}

\includegraphics[width=9.0cm,angle=0]{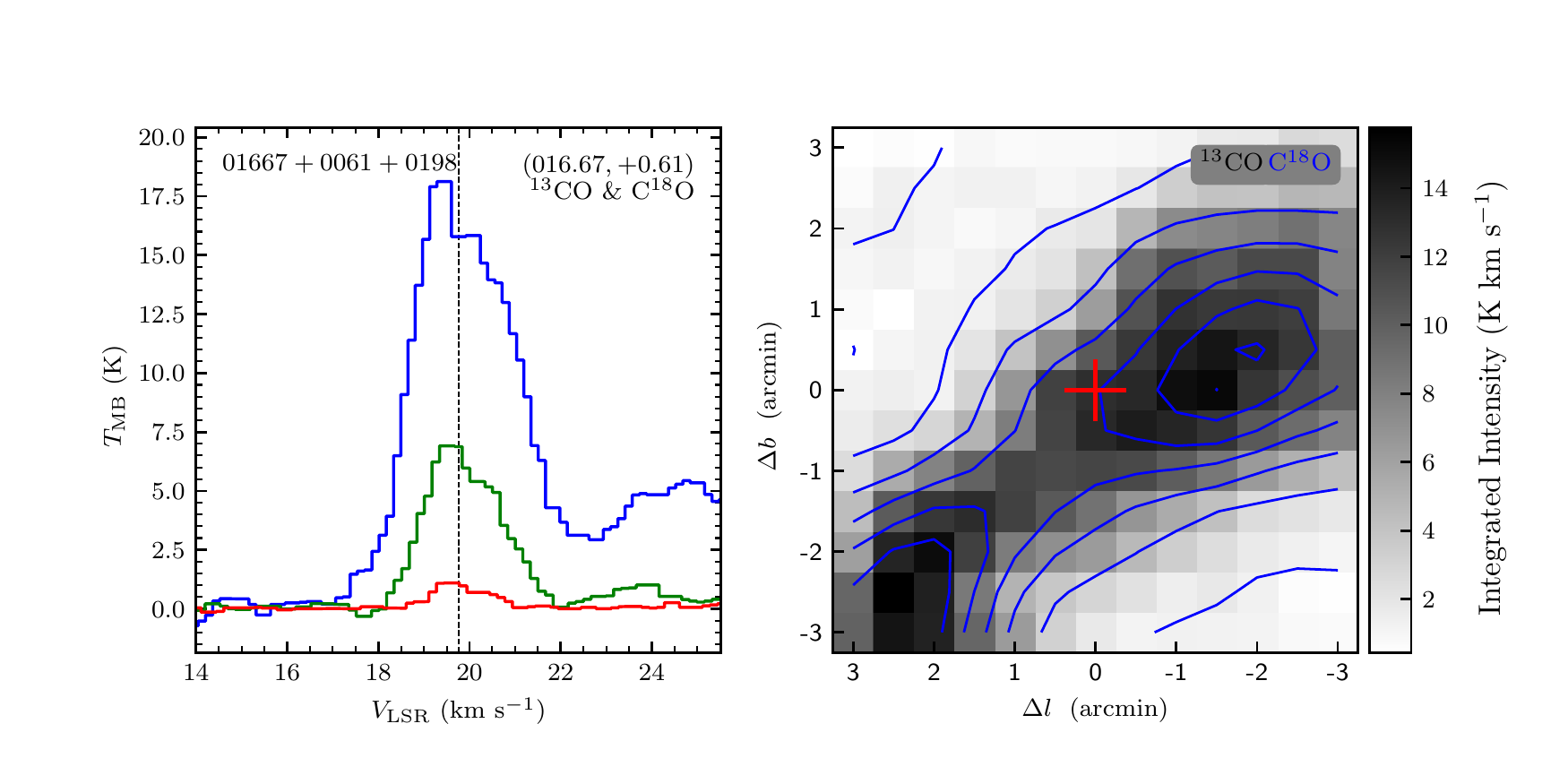}
\includegraphics[width=9.0cm,angle=0]{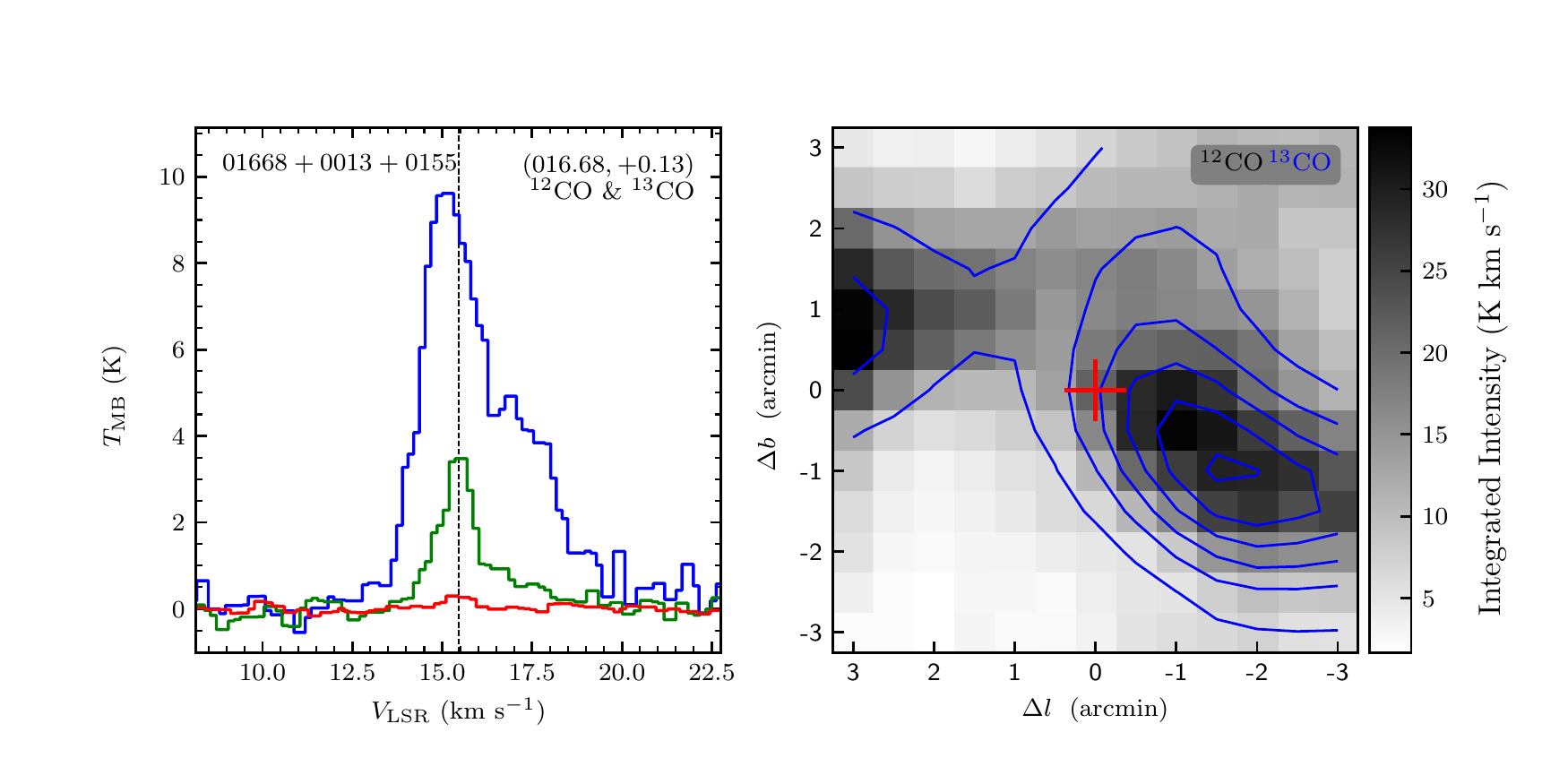}
\end{figure}
\clearpage

\begin{figure}
\includegraphics[width=9.0cm,angle=0]{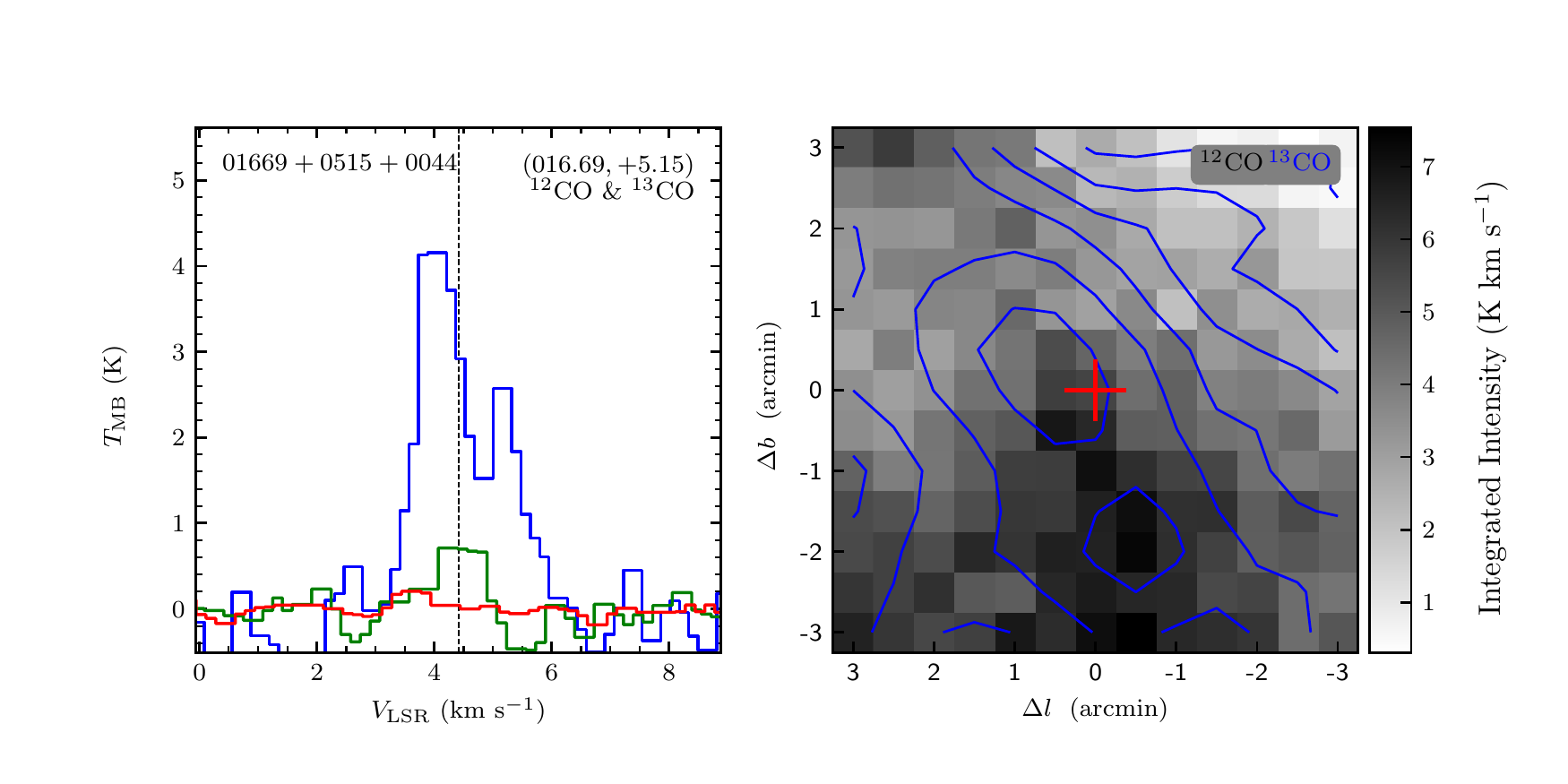}
\includegraphics[width=9.0cm,angle=0]{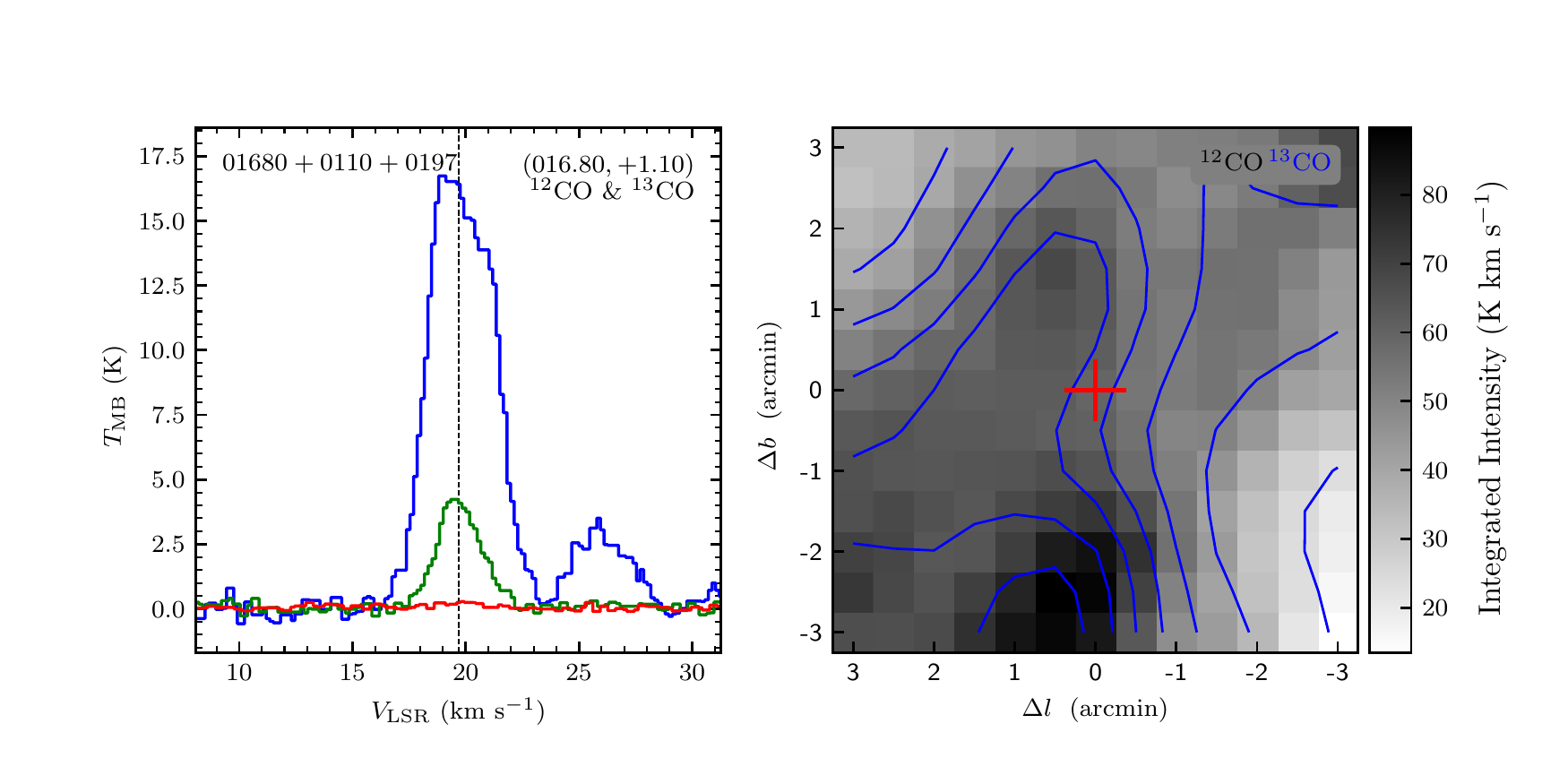}
\vspace{-0.5cm}

\includegraphics[width=9.0cm,angle=0]{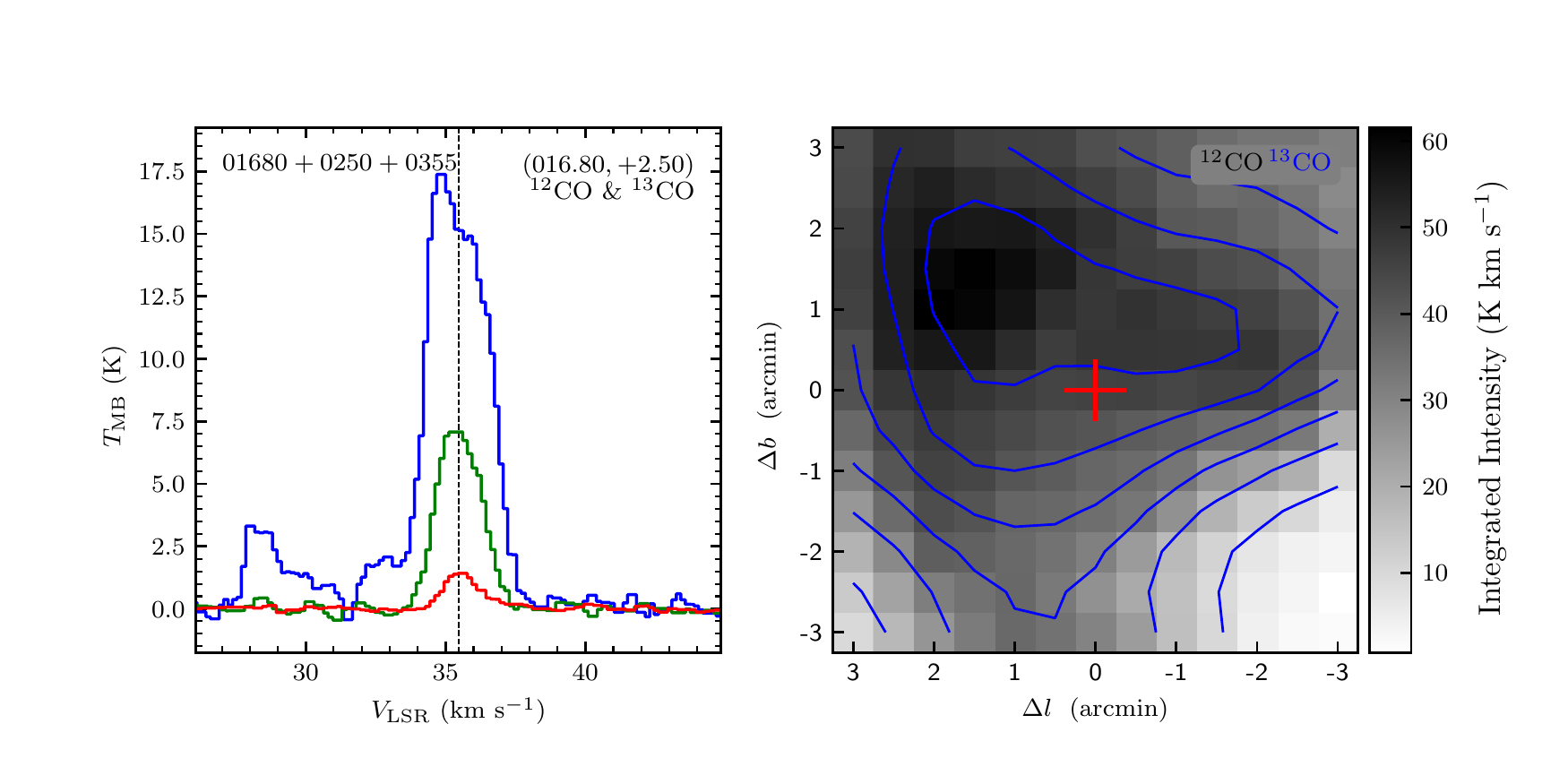}
\includegraphics[width=9.0cm,angle=0]{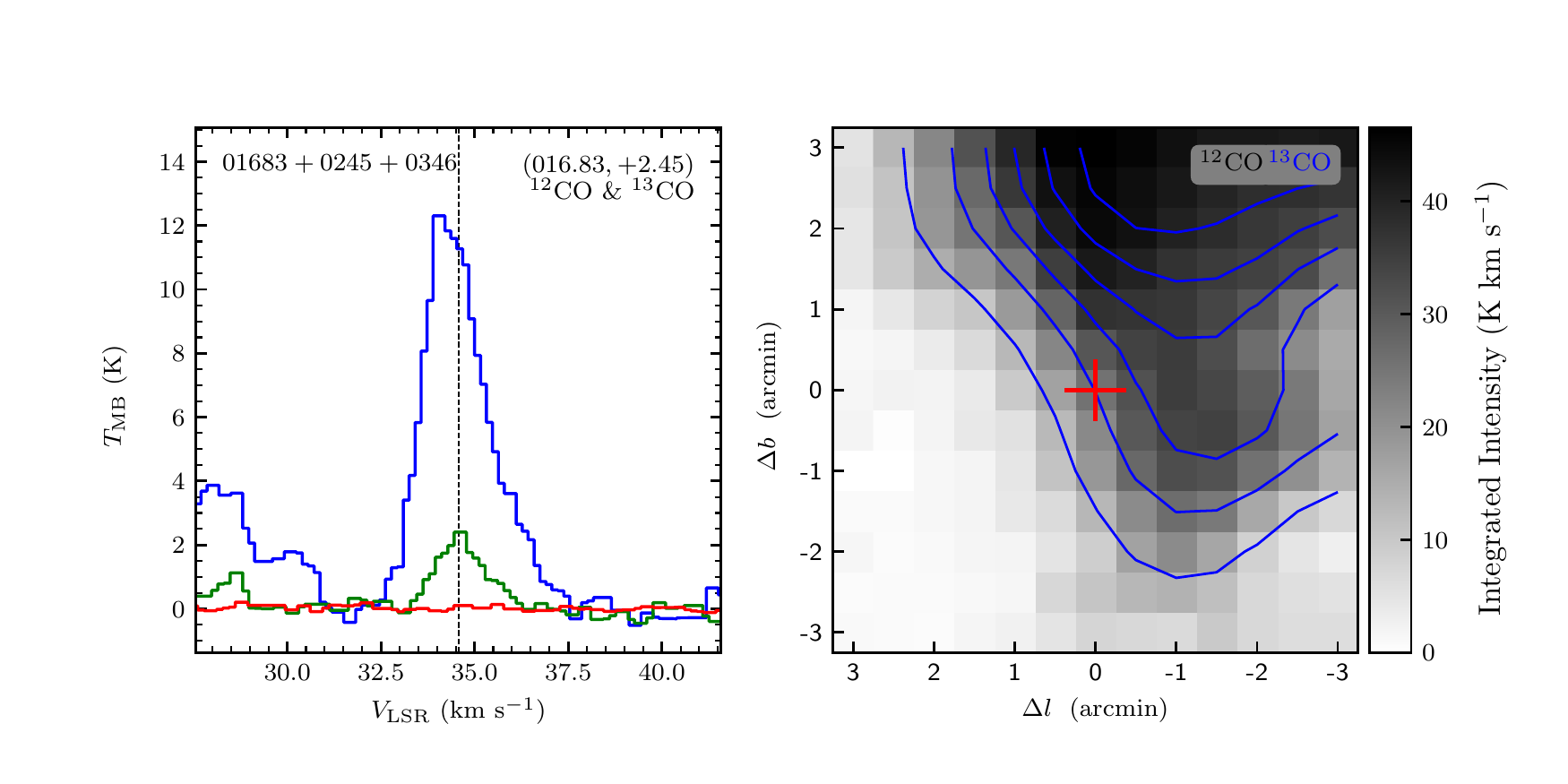}
\vspace{-0.5cm}

\includegraphics[width=9.0cm,angle=0]{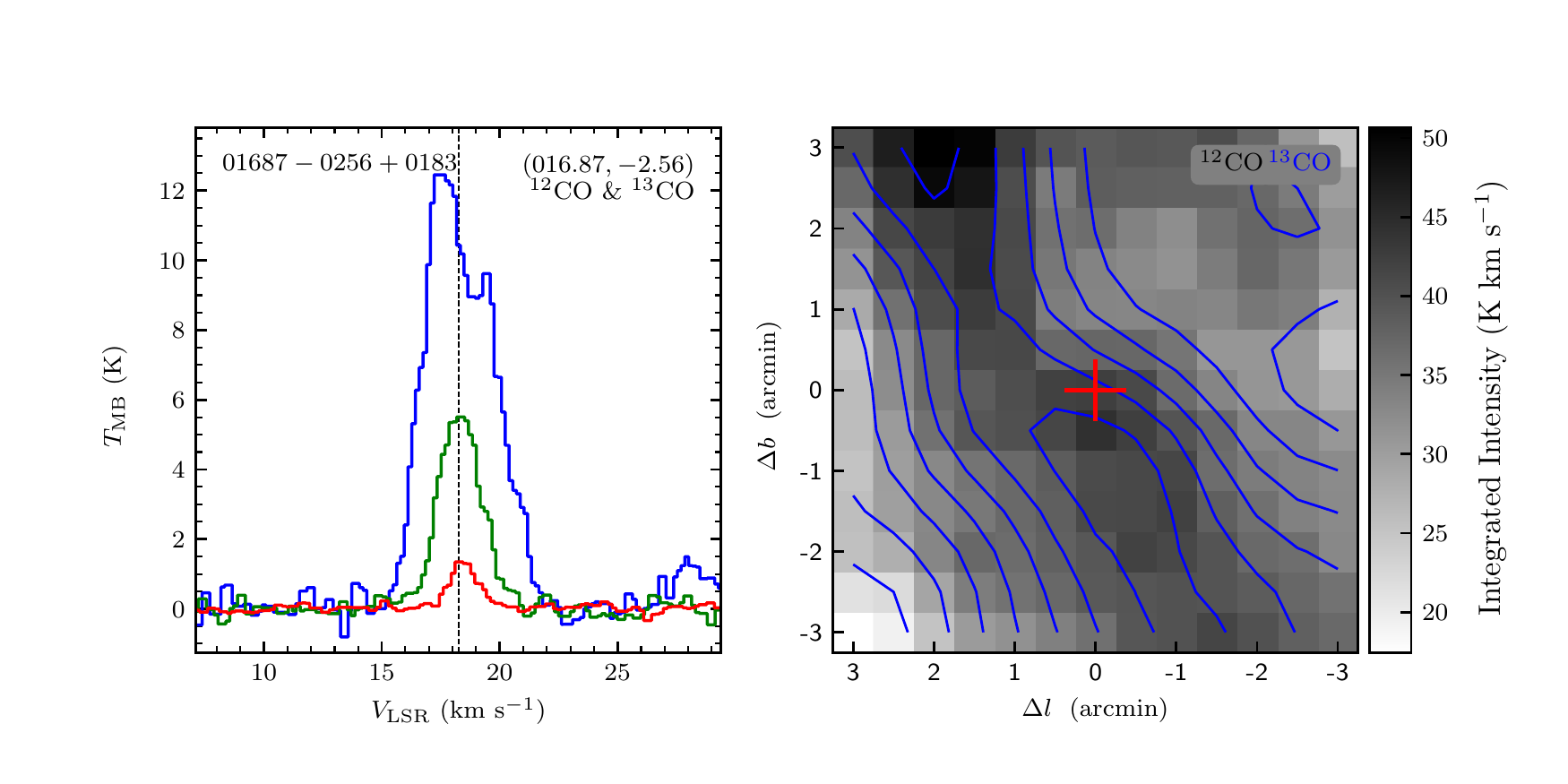}
\includegraphics[width=9.0cm,angle=0]{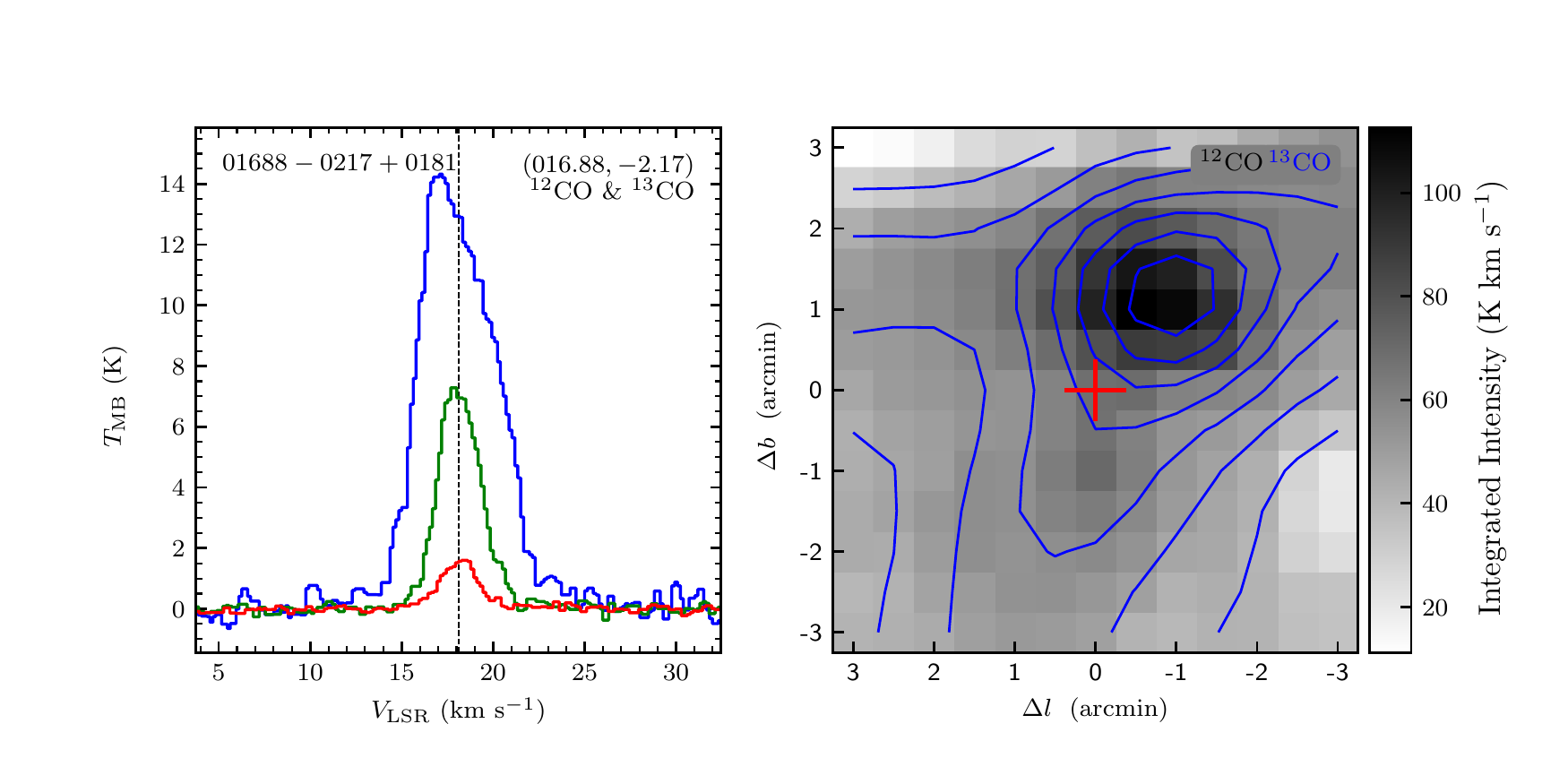}
\vspace{-0.5cm}

\includegraphics[width=9.0cm,angle=0]{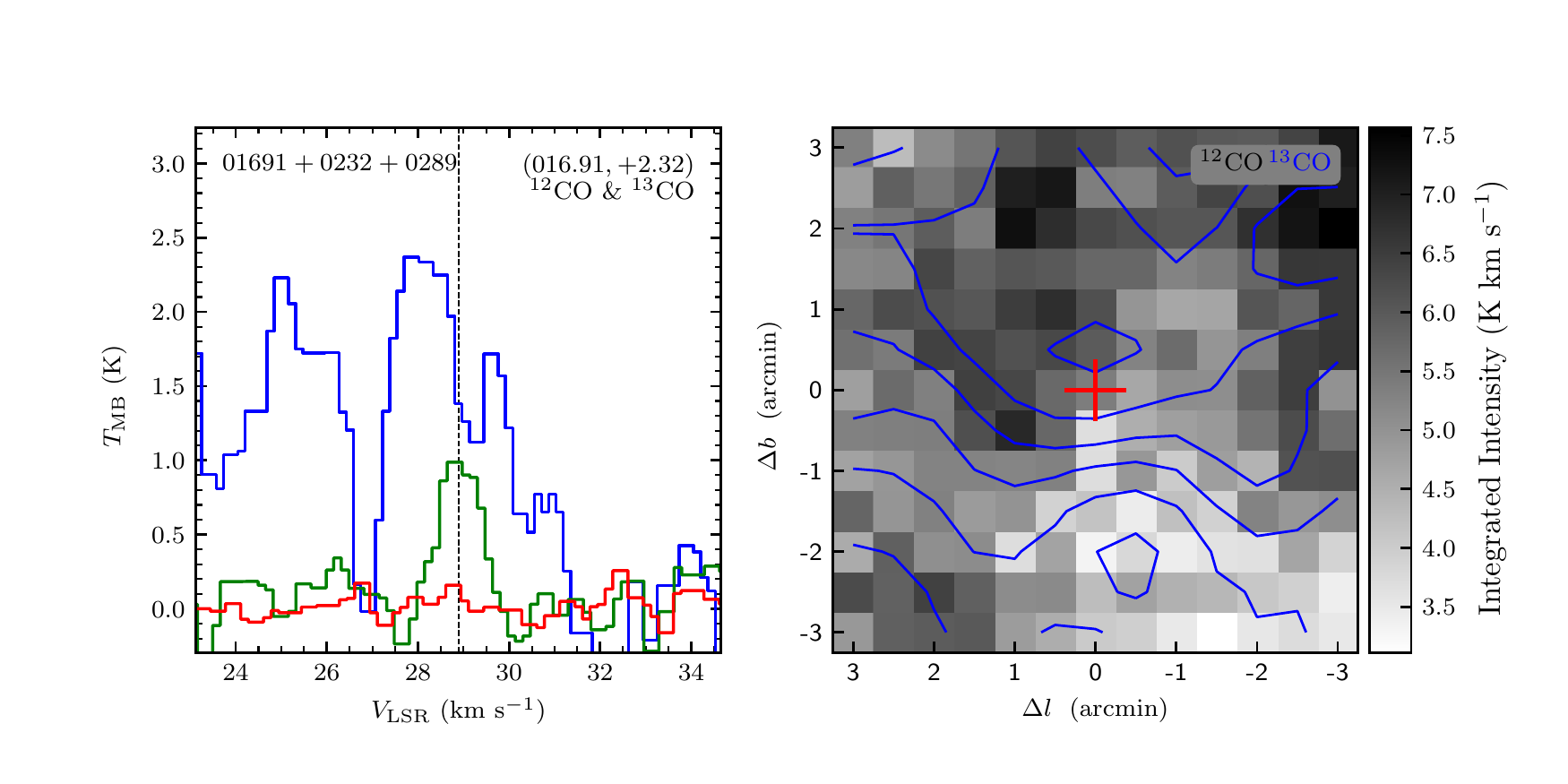}
\includegraphics[width=9.0cm,angle=0]{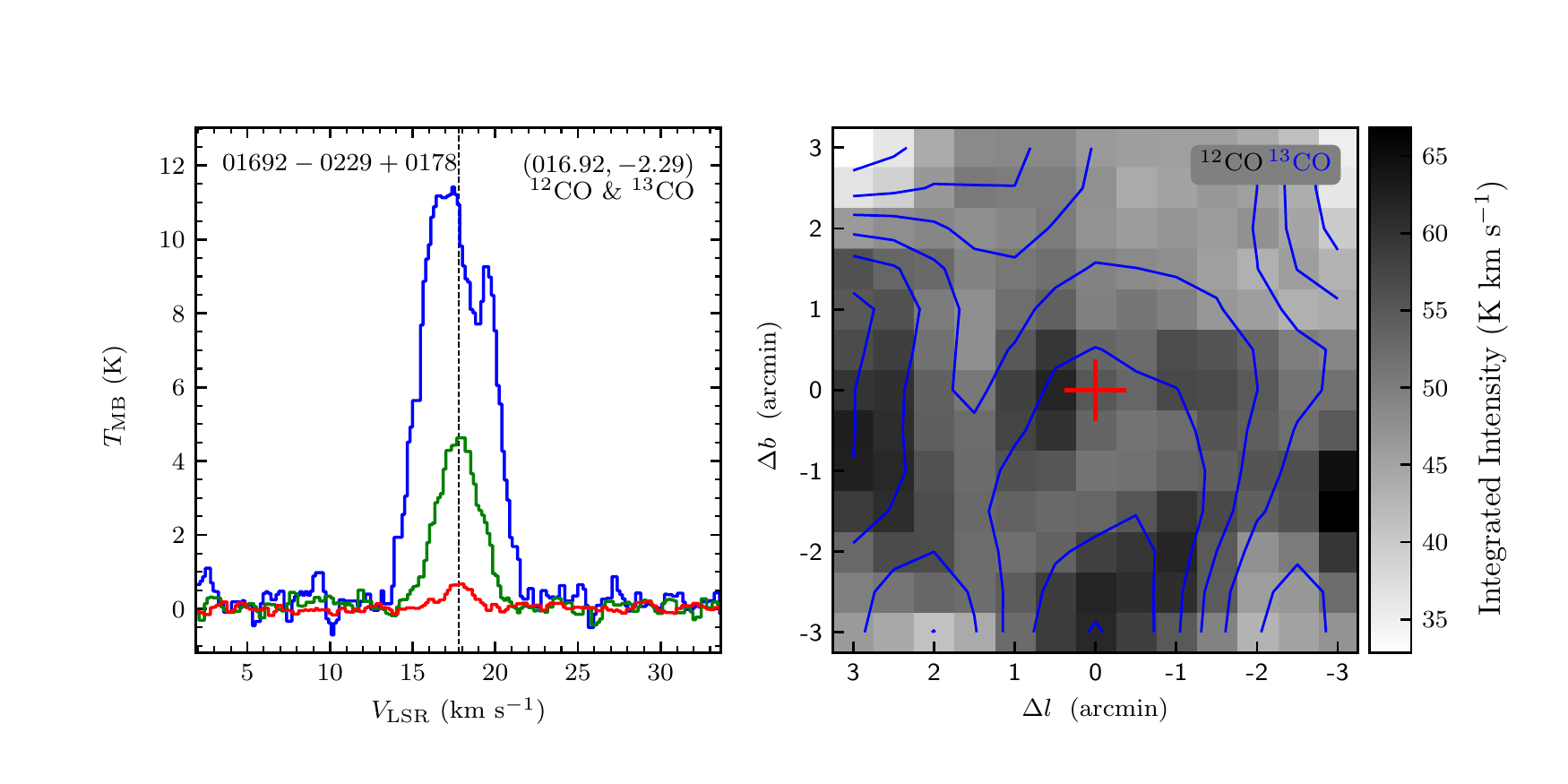}
\vspace{-0.5cm}

\includegraphics[width=9.0cm,angle=0]{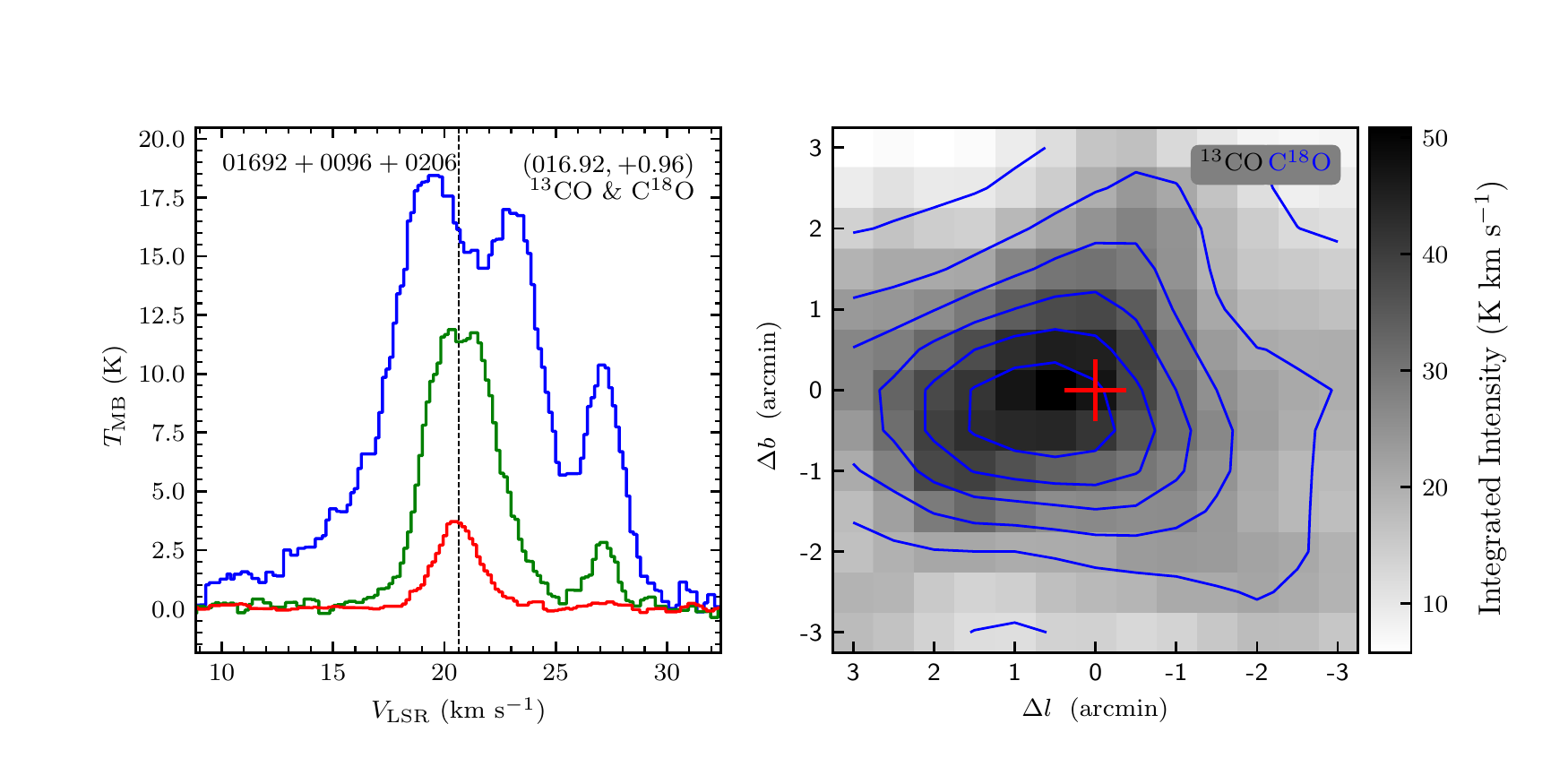}
\includegraphics[width=9.0cm,angle=0]{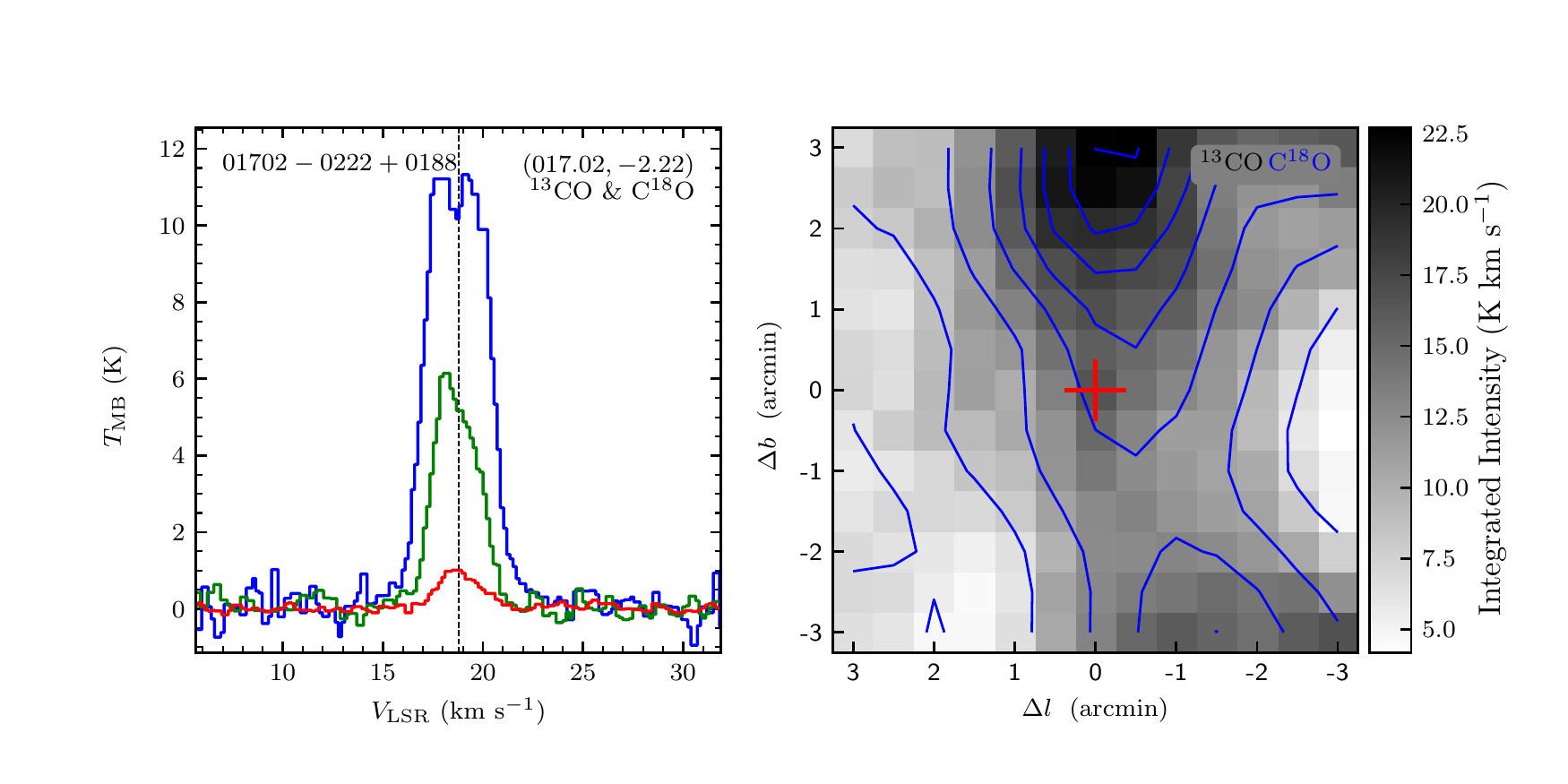}
\end{figure}
\clearpage

\begin{figure}
\includegraphics[width=9.0cm,angle=0]{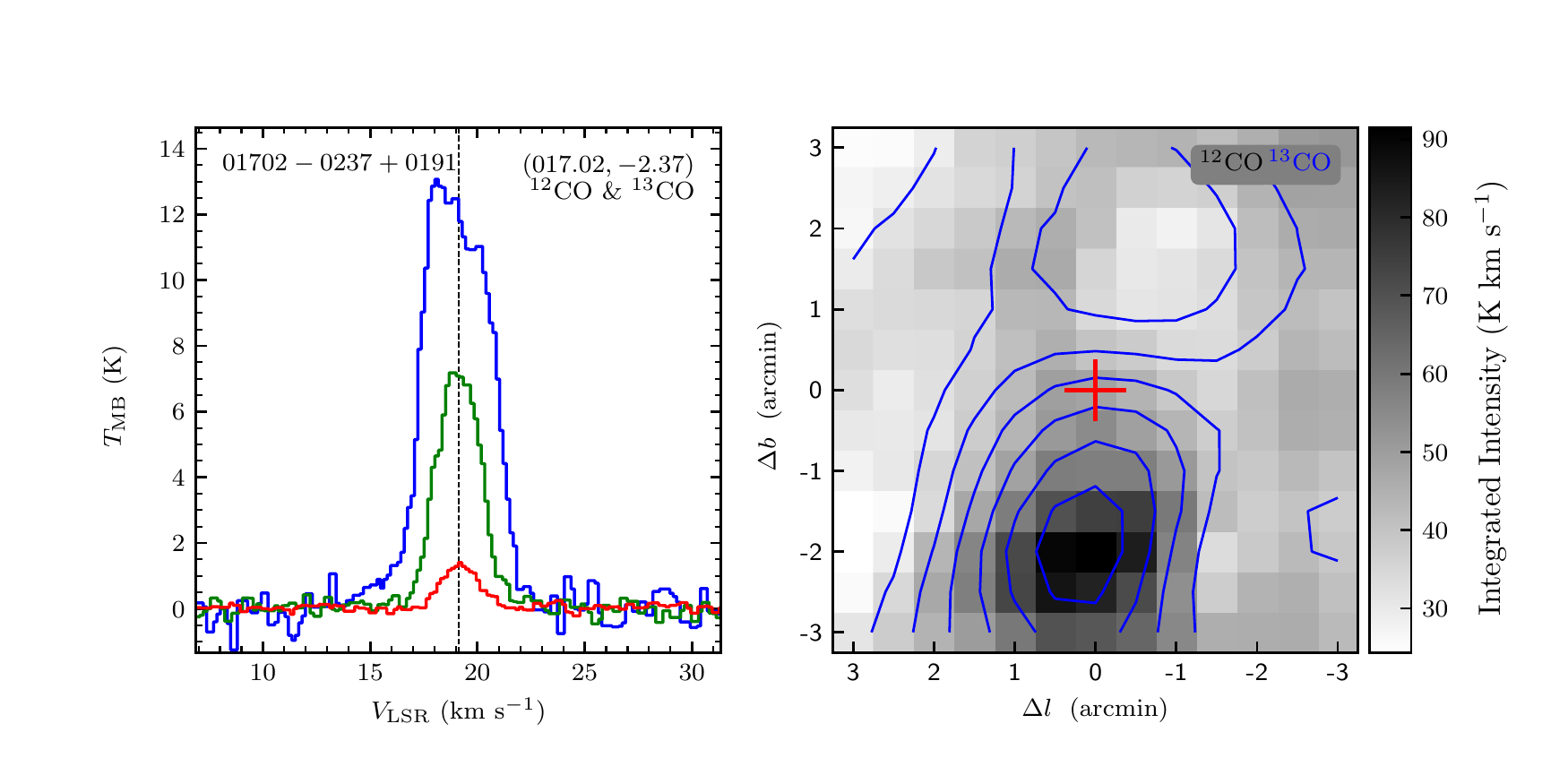}
\includegraphics[width=9.0cm,angle=0]{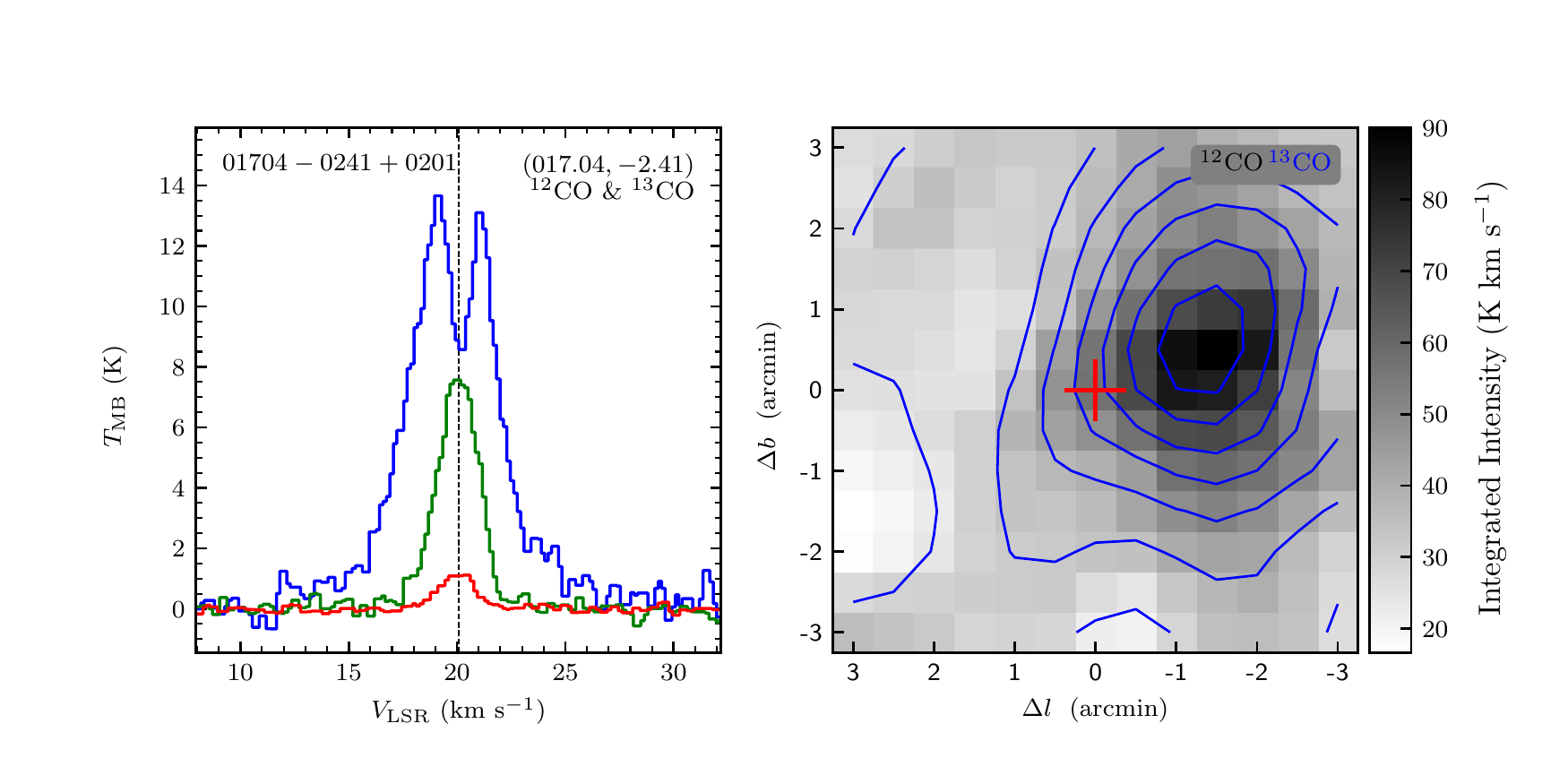}
\vspace{-0.5cm}

\includegraphics[width=9.0cm,angle=0]{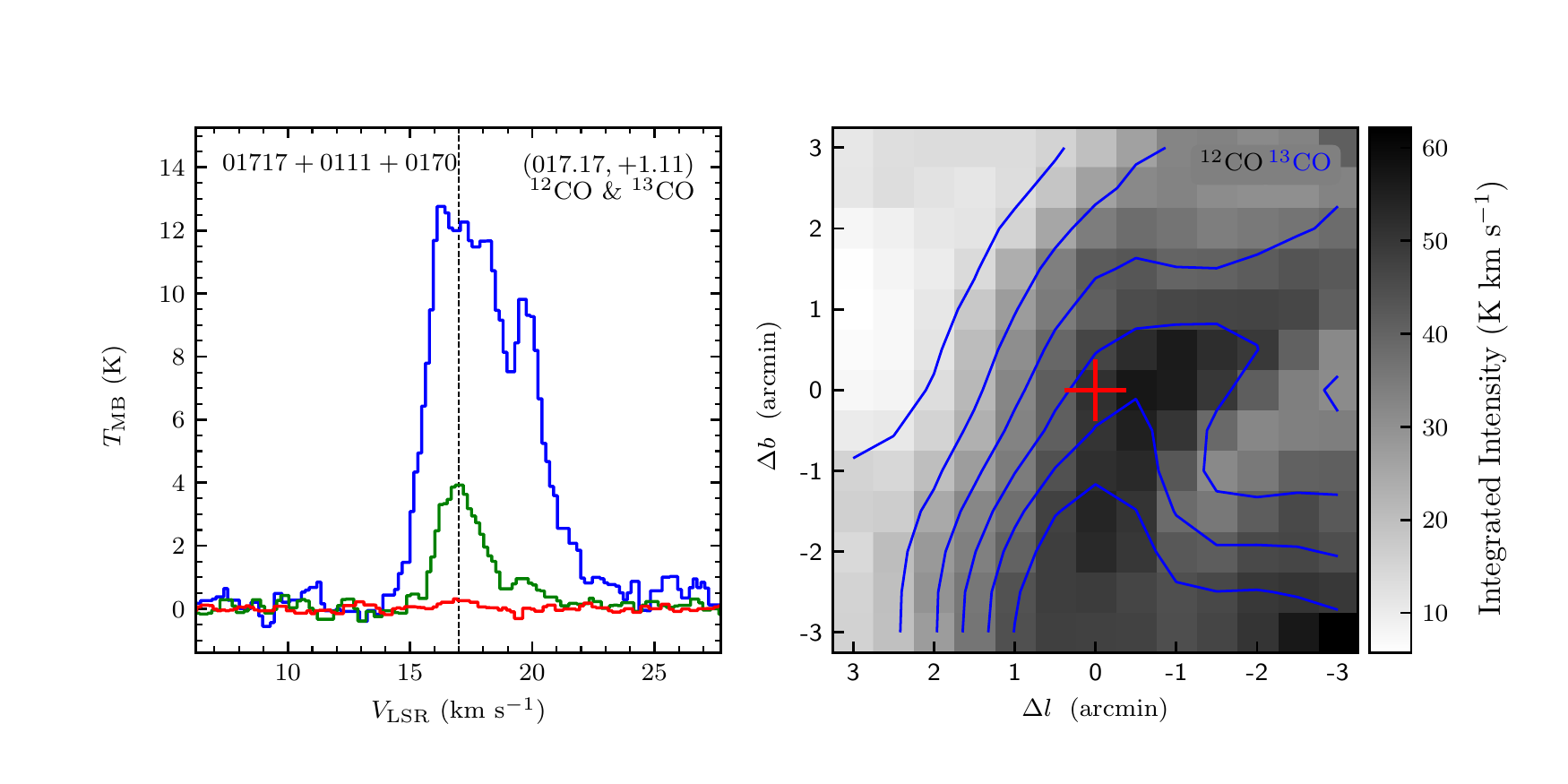}
\includegraphics[width=9.0cm,angle=0]{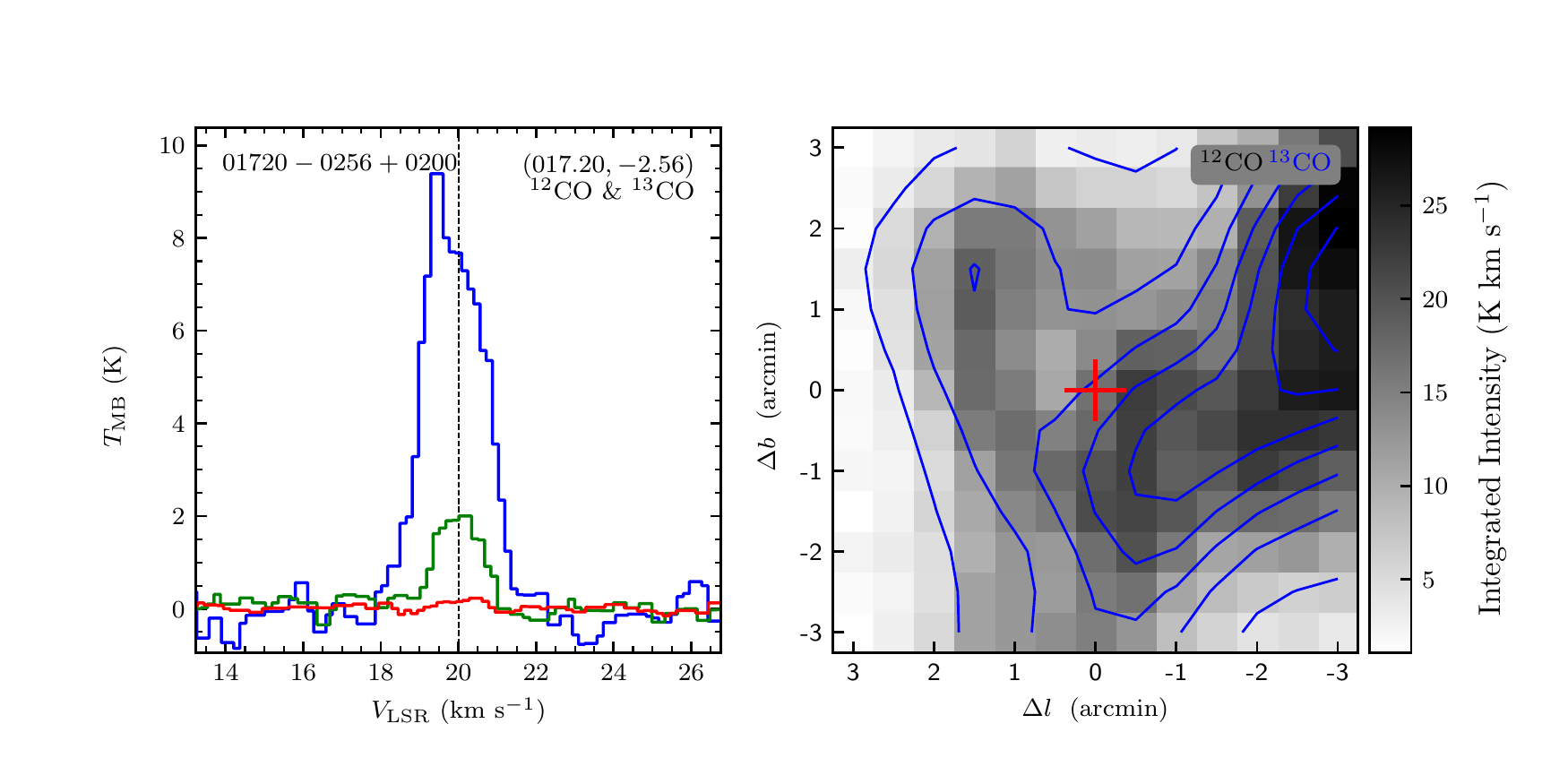}
\vspace{-0.5cm}

\includegraphics[width=9.0cm,angle=0]{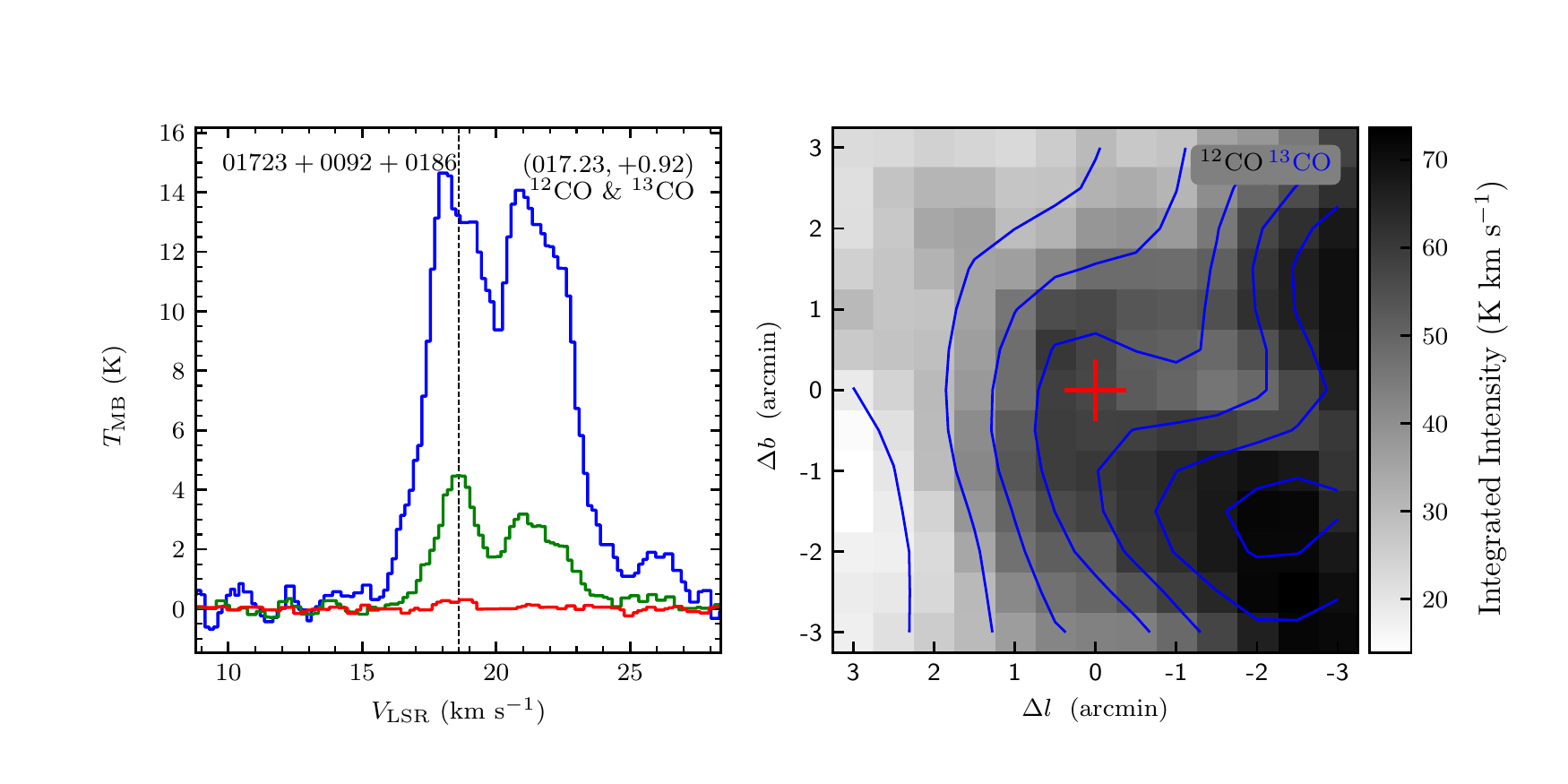}
\includegraphics[width=9.0cm,angle=0]{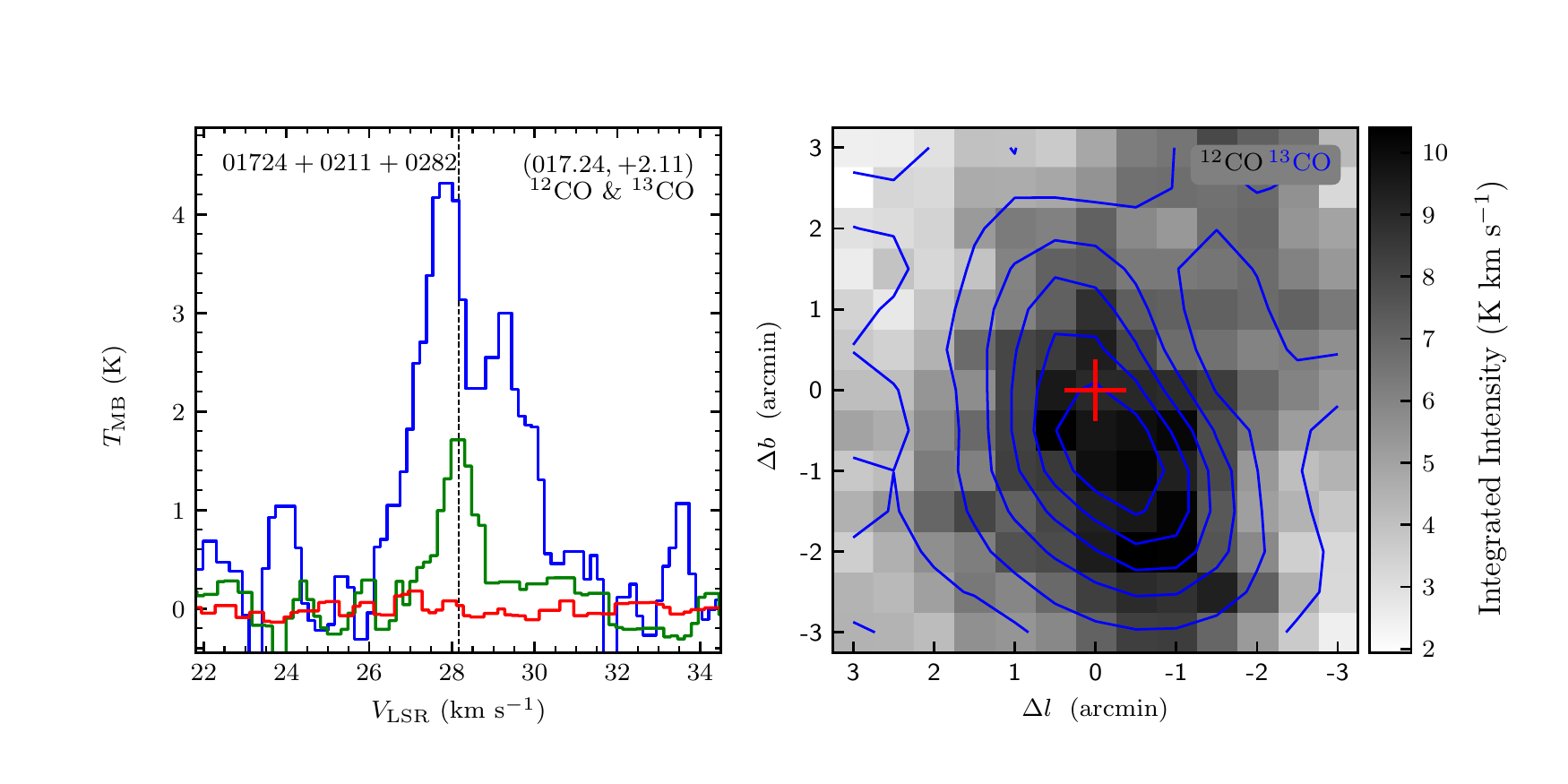}
\vspace{-0.5cm}

\includegraphics[width=9.0cm,angle=0]{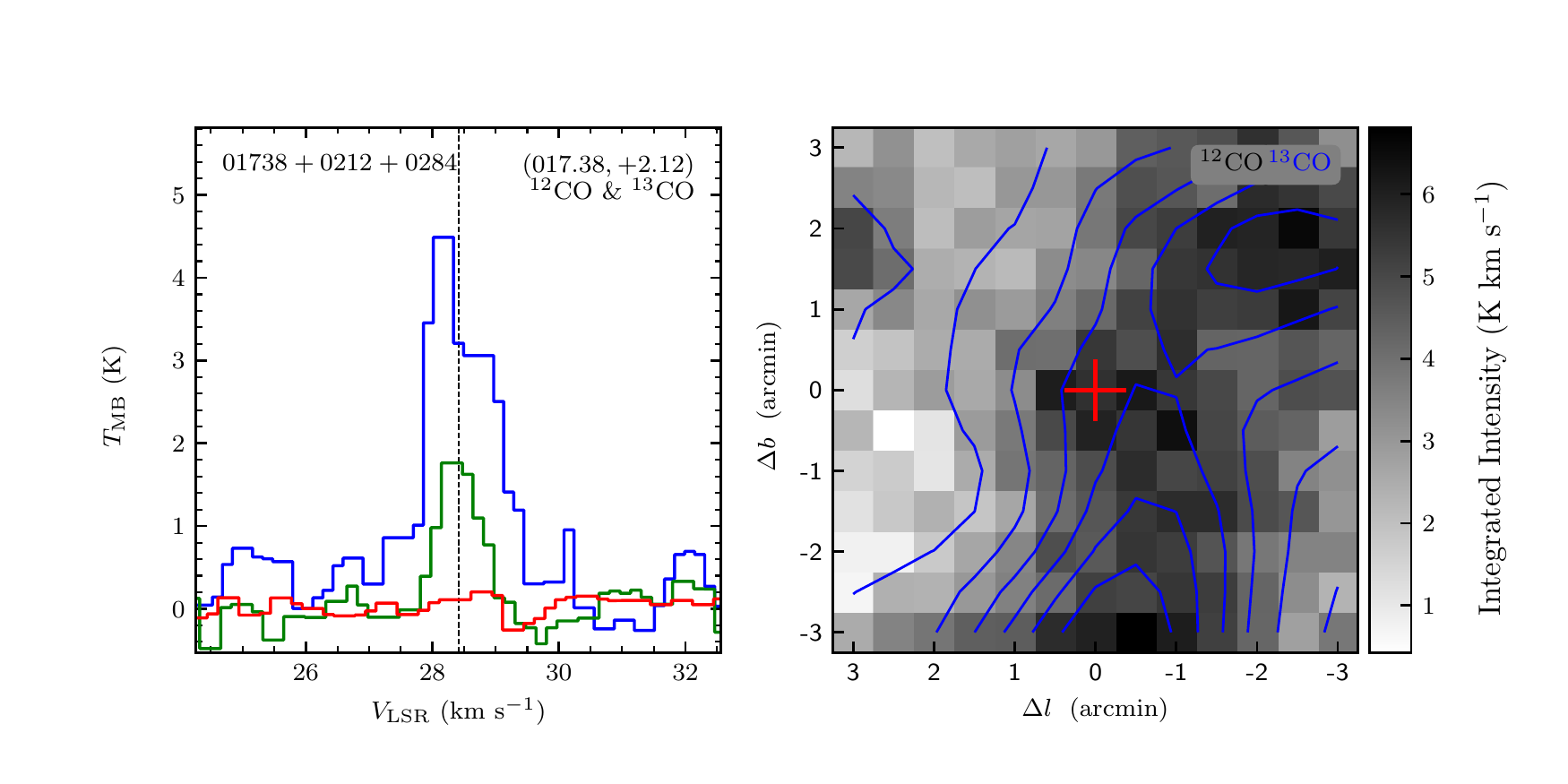}
\includegraphics[width=9.0cm,angle=0]{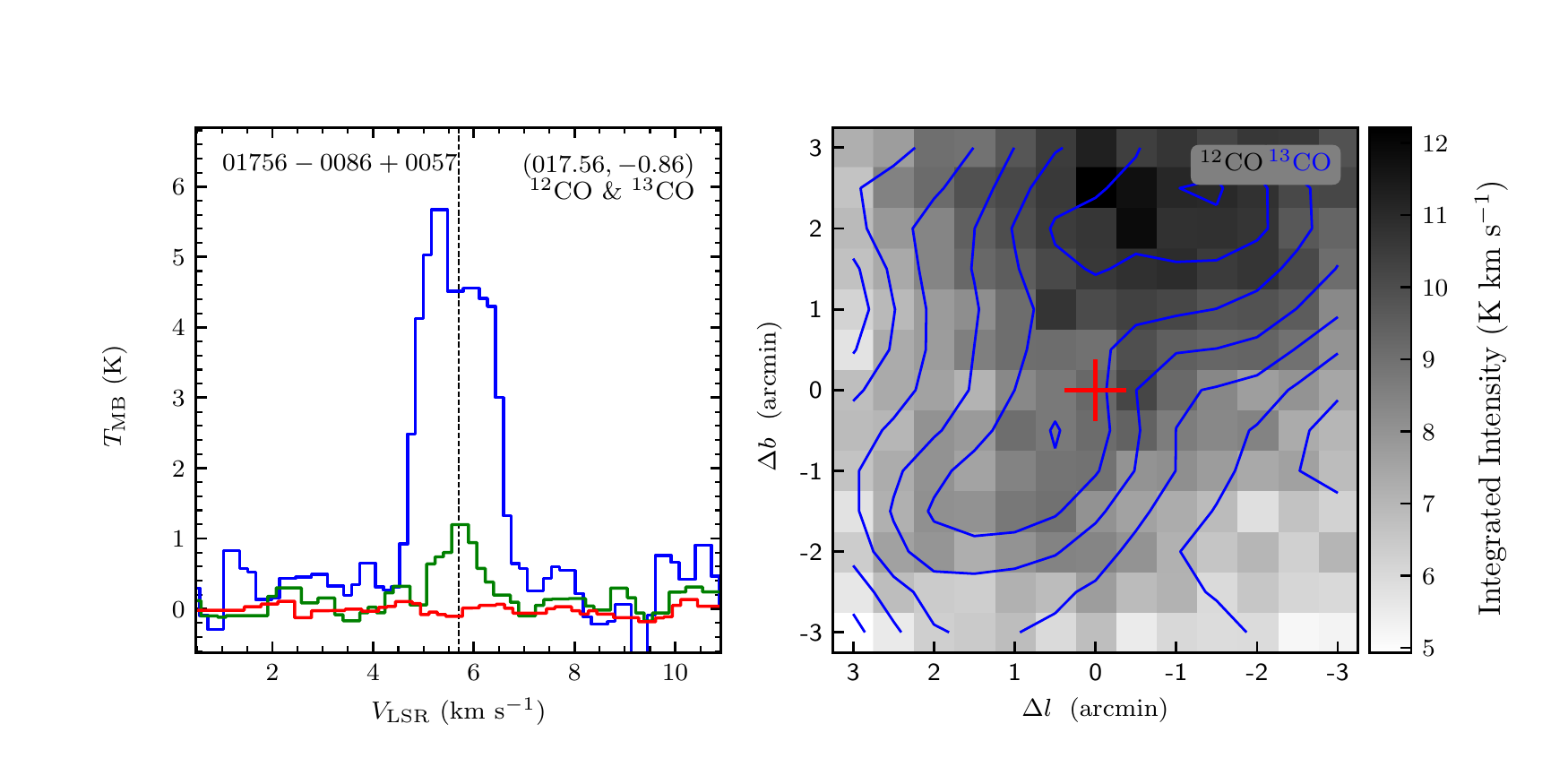}
\vspace{-0.5cm}

\includegraphics[width=9.0cm,angle=0]{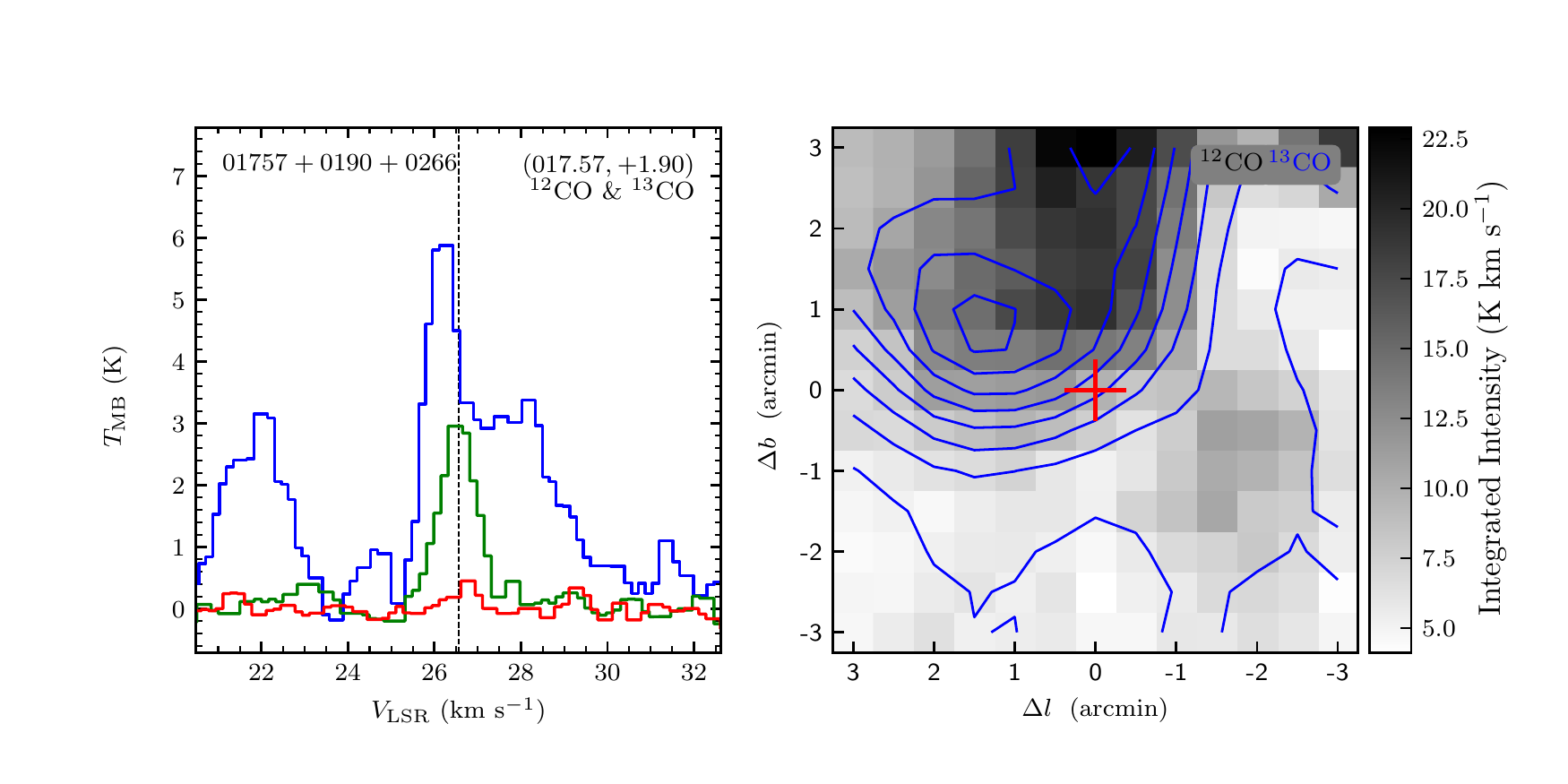}
\includegraphics[width=9.0cm,angle=0]{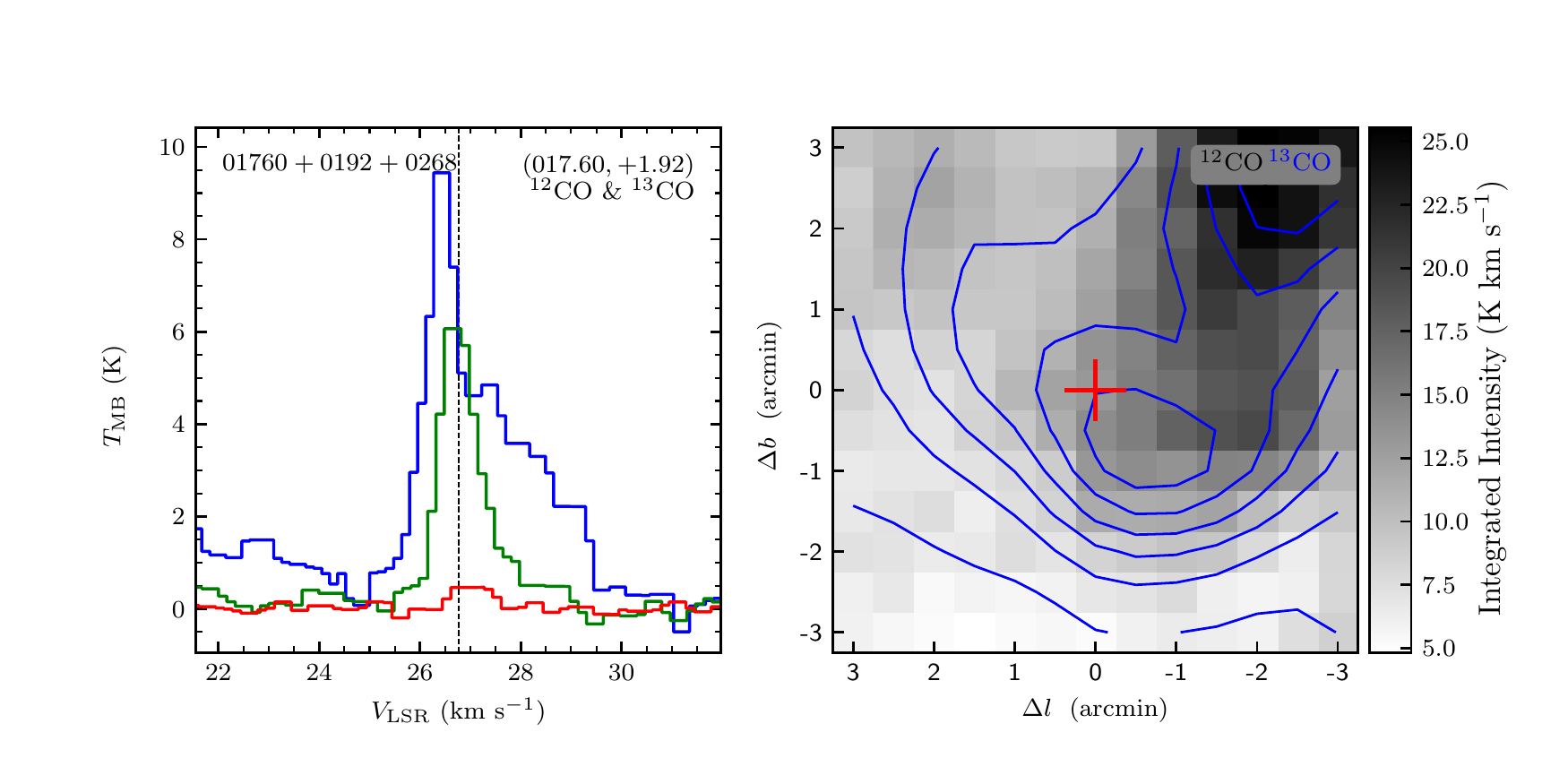}
\end{figure}
\clearpage

\begin{figure}
\includegraphics[width=9.0cm,angle=0]{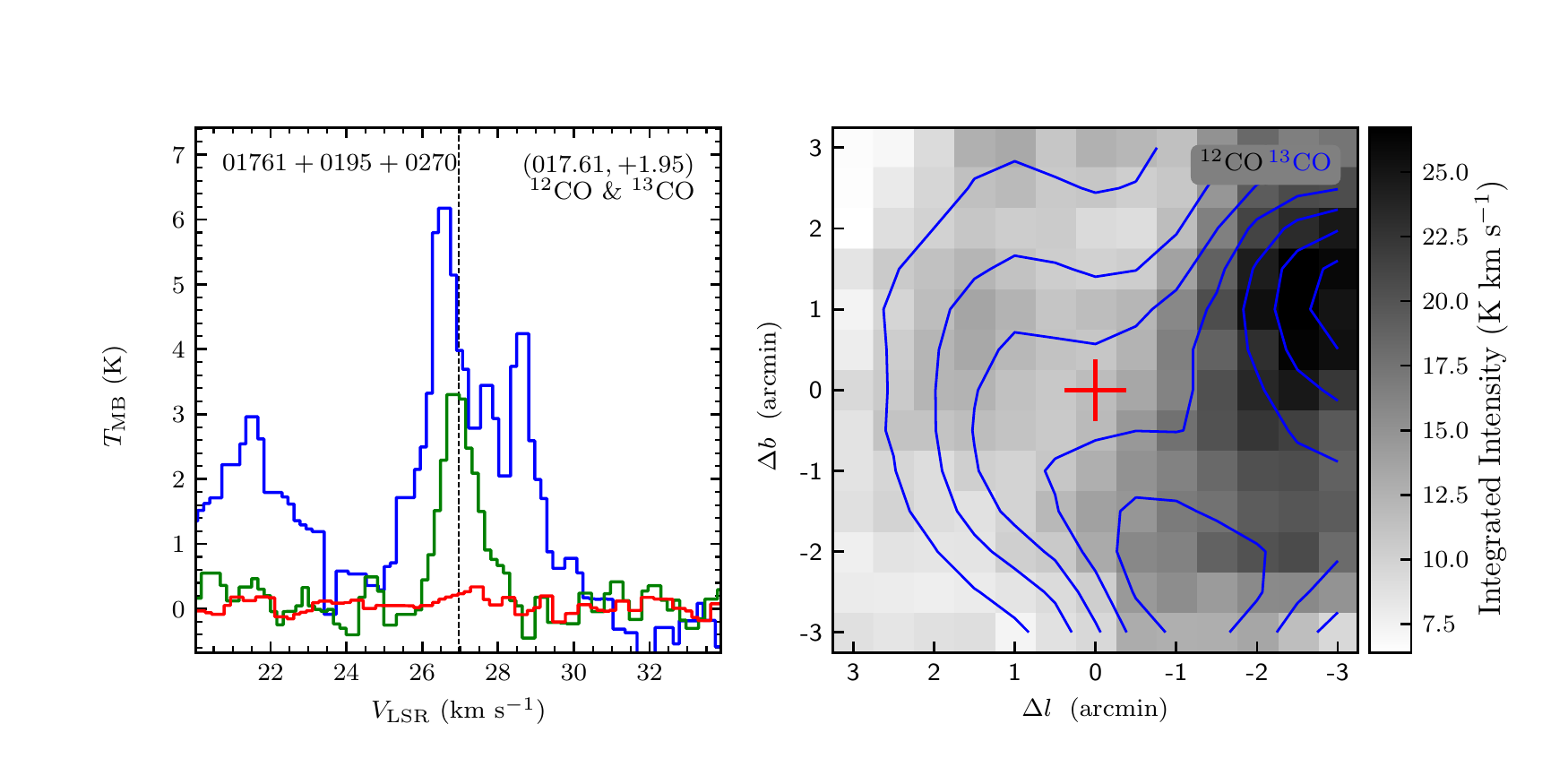}
\includegraphics[width=9.0cm,angle=0]{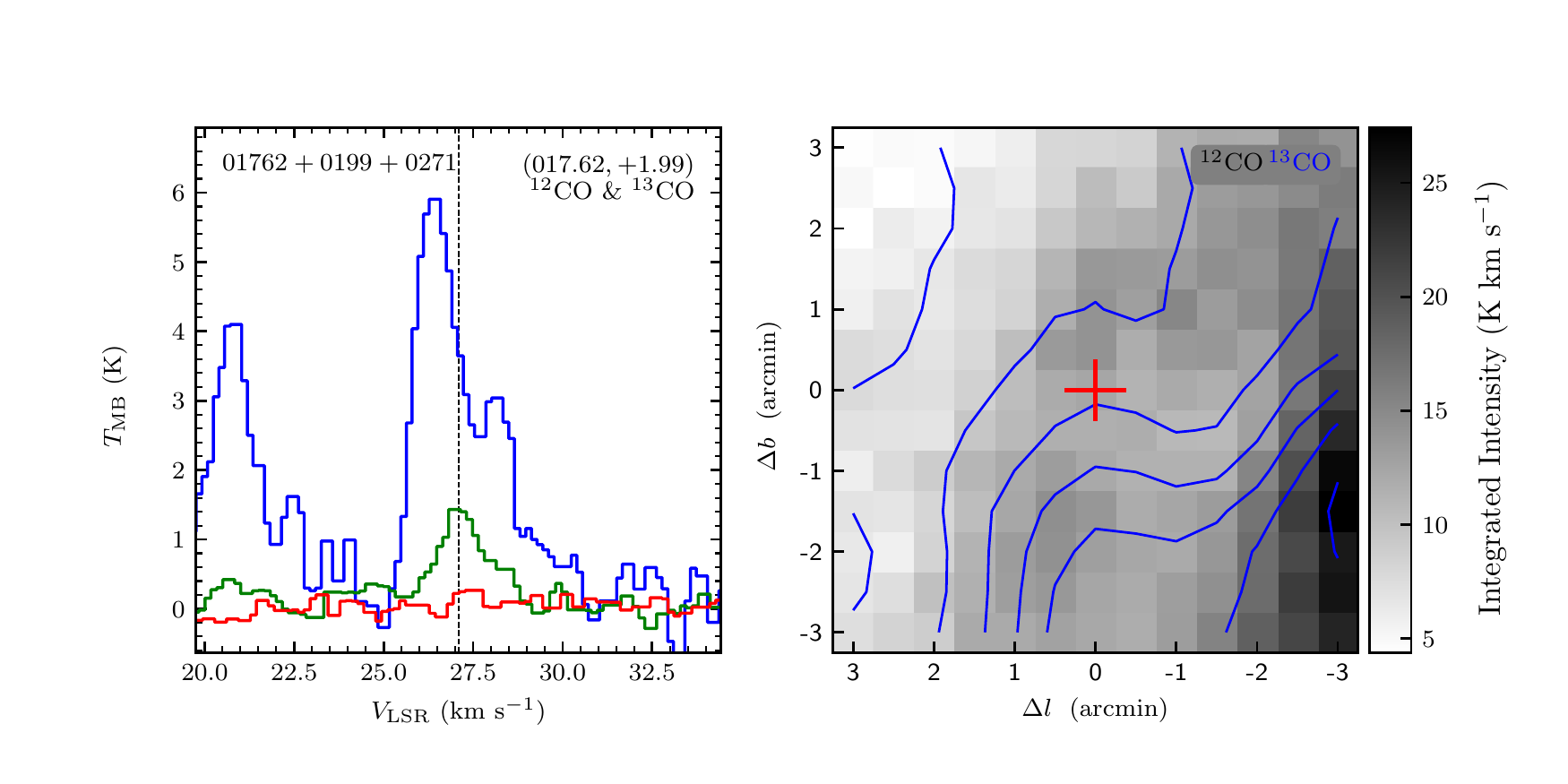}
\vspace{-0.5cm}

\includegraphics[width=9.0cm,angle=0]{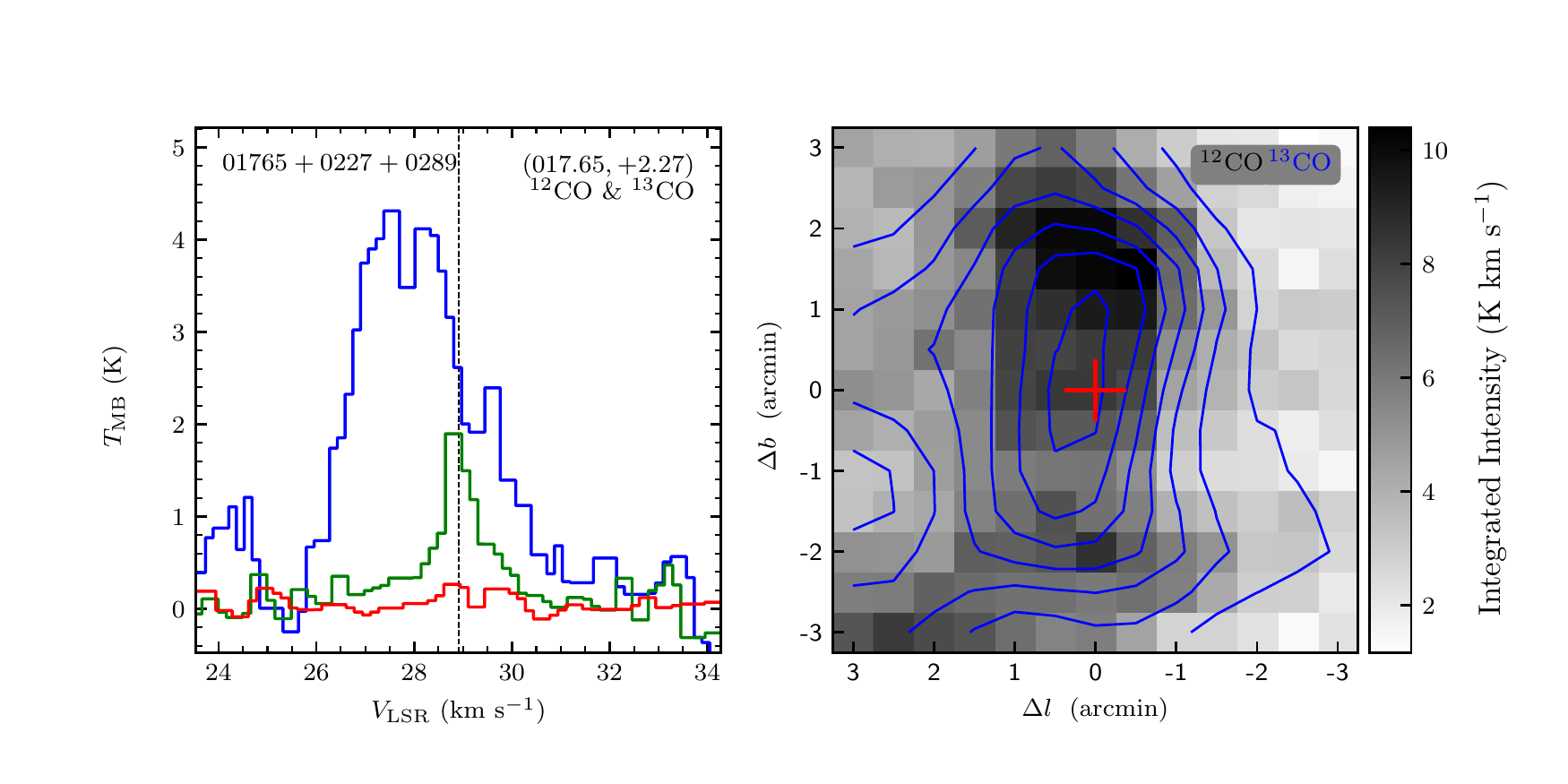}
\includegraphics[width=9.0cm,angle=0]{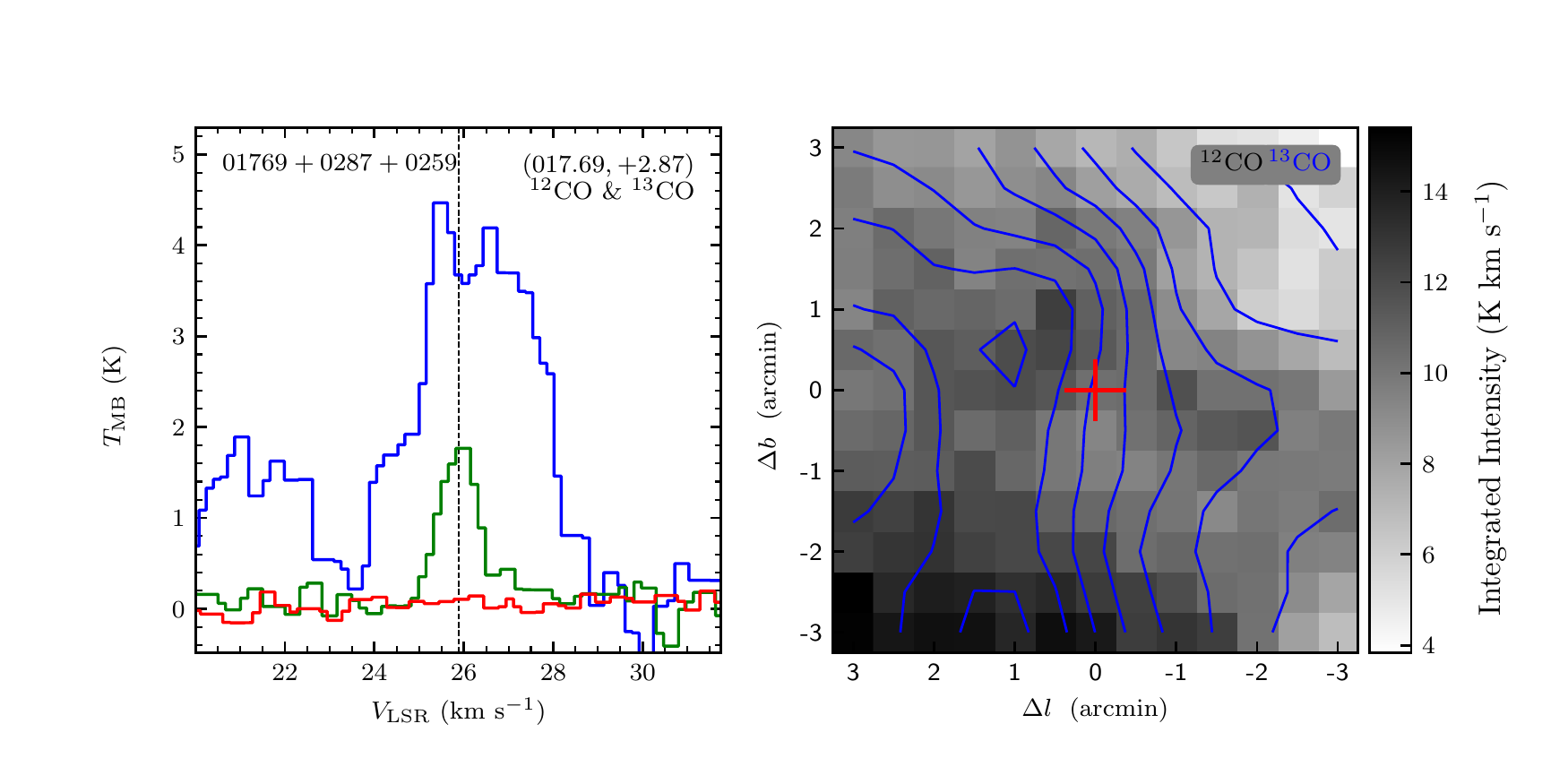}
\vspace{-0.5cm}

\includegraphics[width=9.0cm,angle=0]{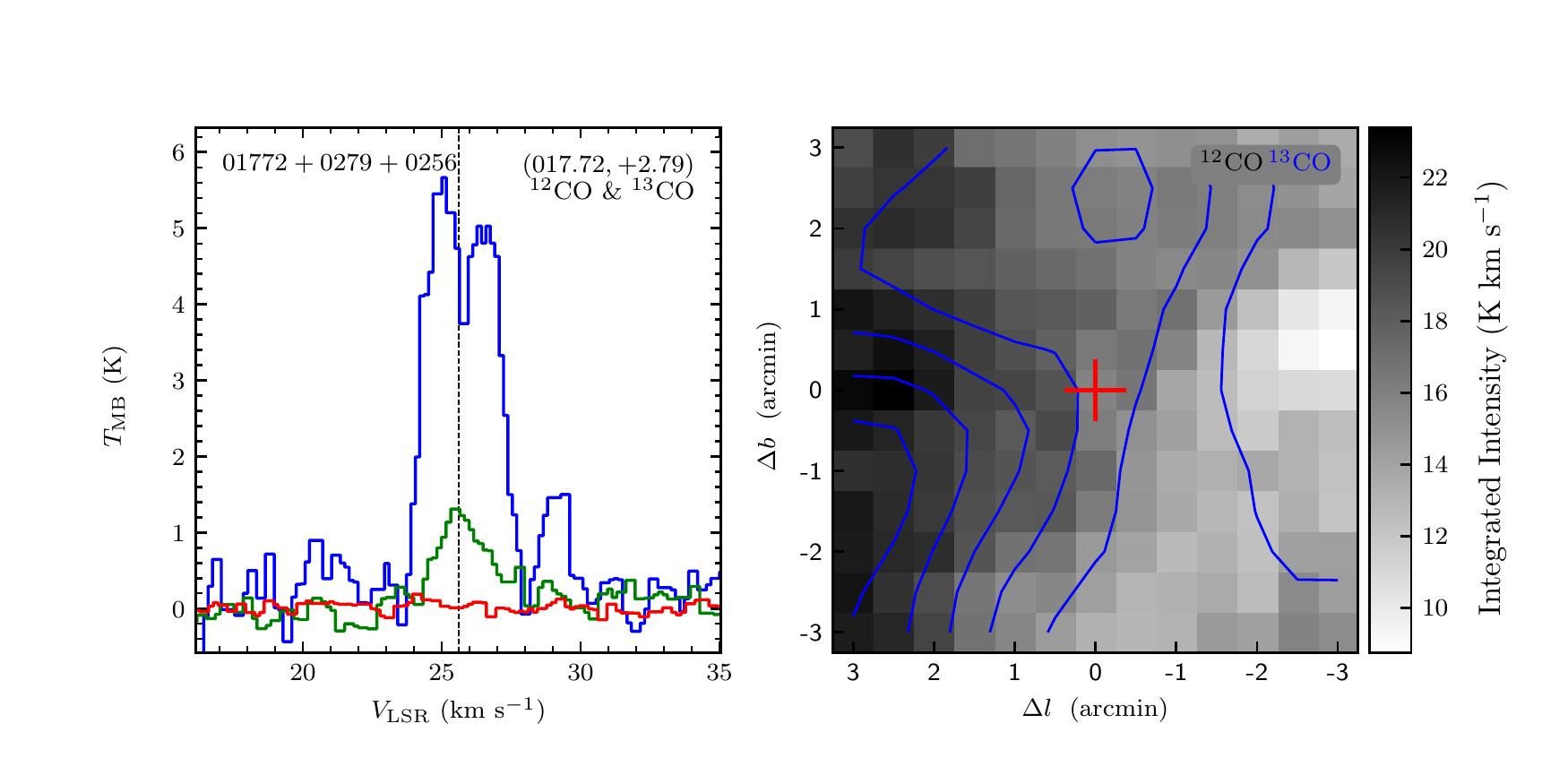}
\includegraphics[width=9.0cm,angle=0]{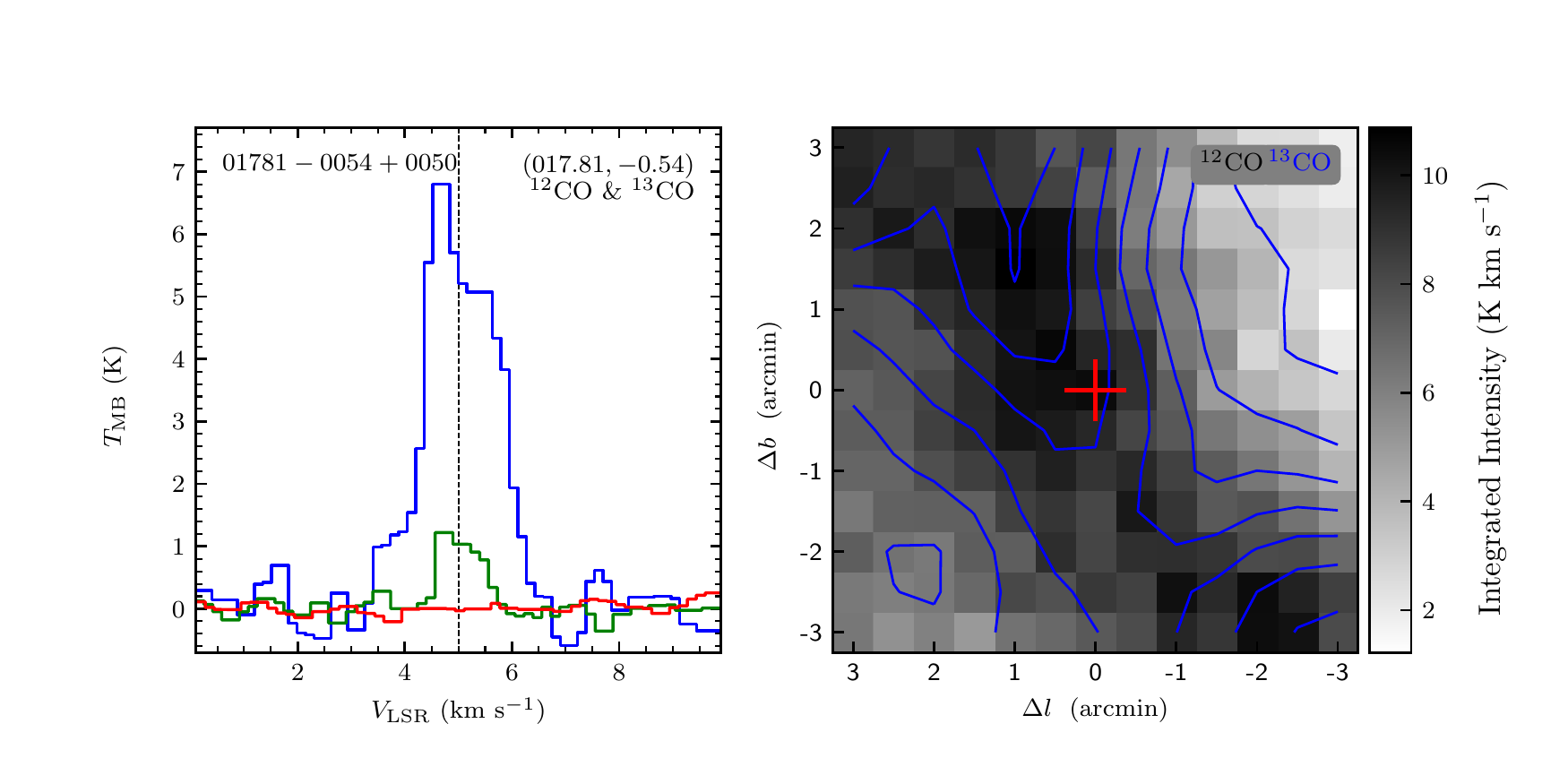}
\vspace{-0.5cm}

\includegraphics[width=9.0cm,angle=0]{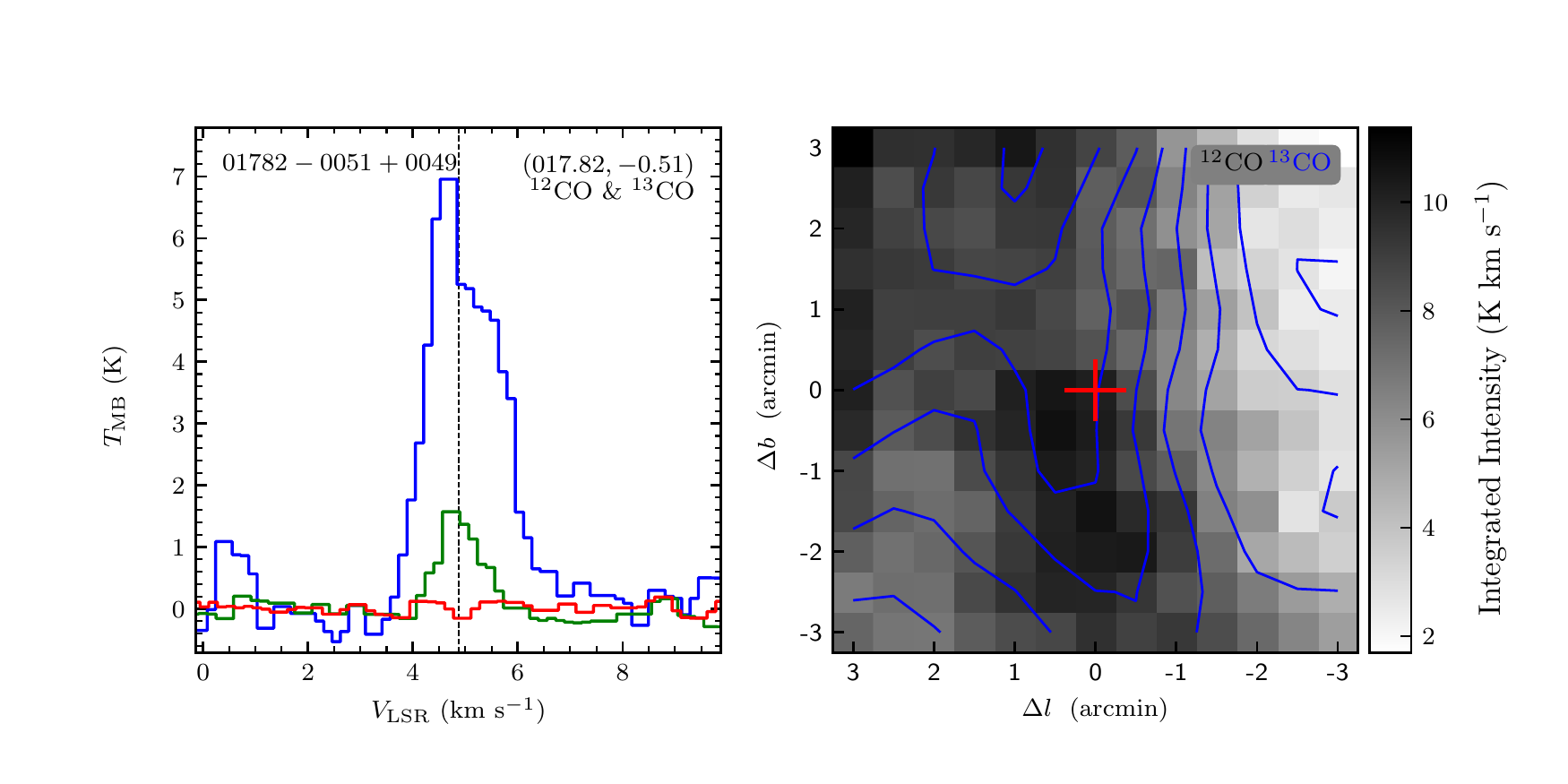}
\includegraphics[width=9.0cm,angle=0]{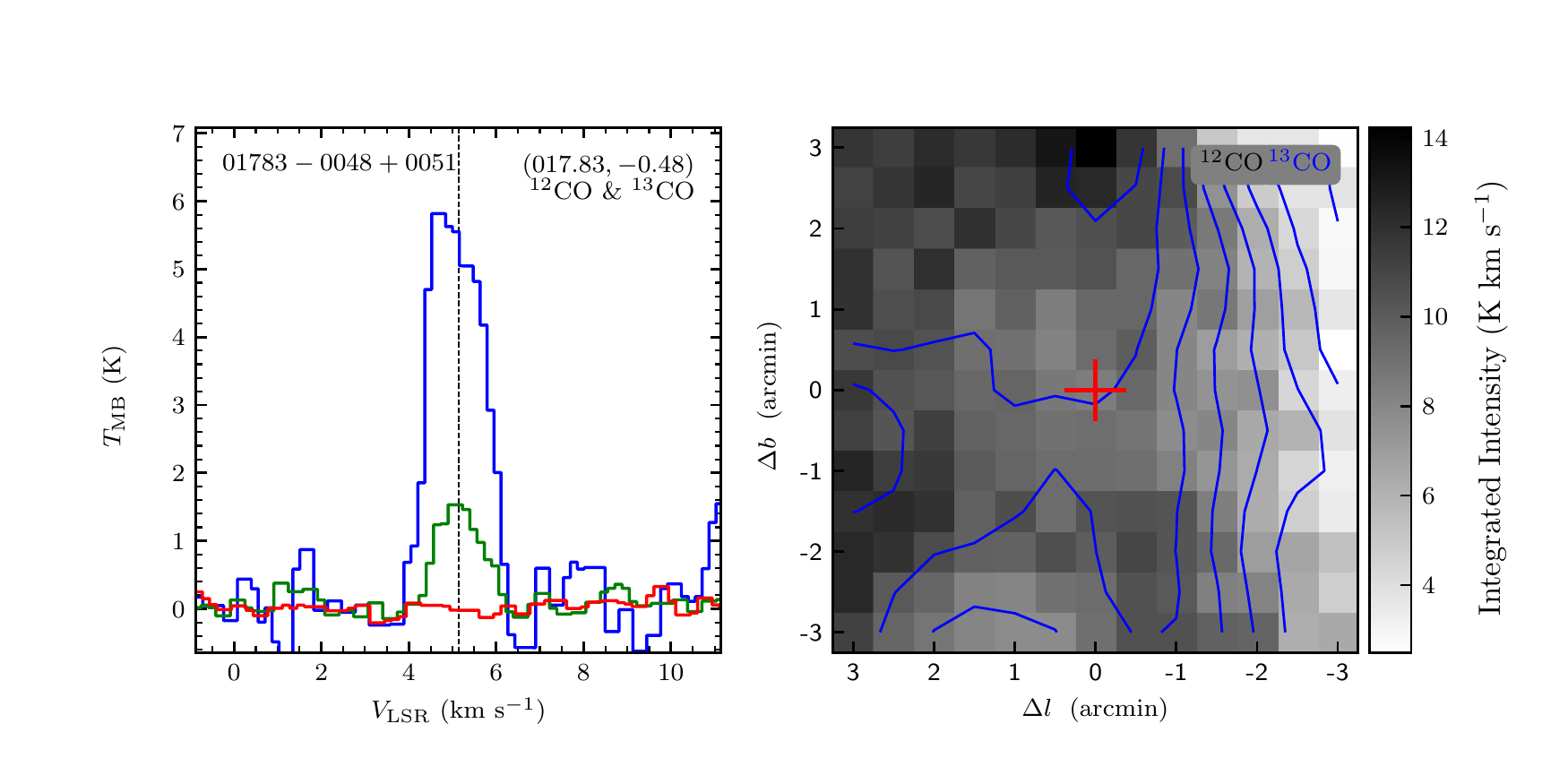}
\vspace{-0.5cm}

\includegraphics[width=9.0cm,angle=0]{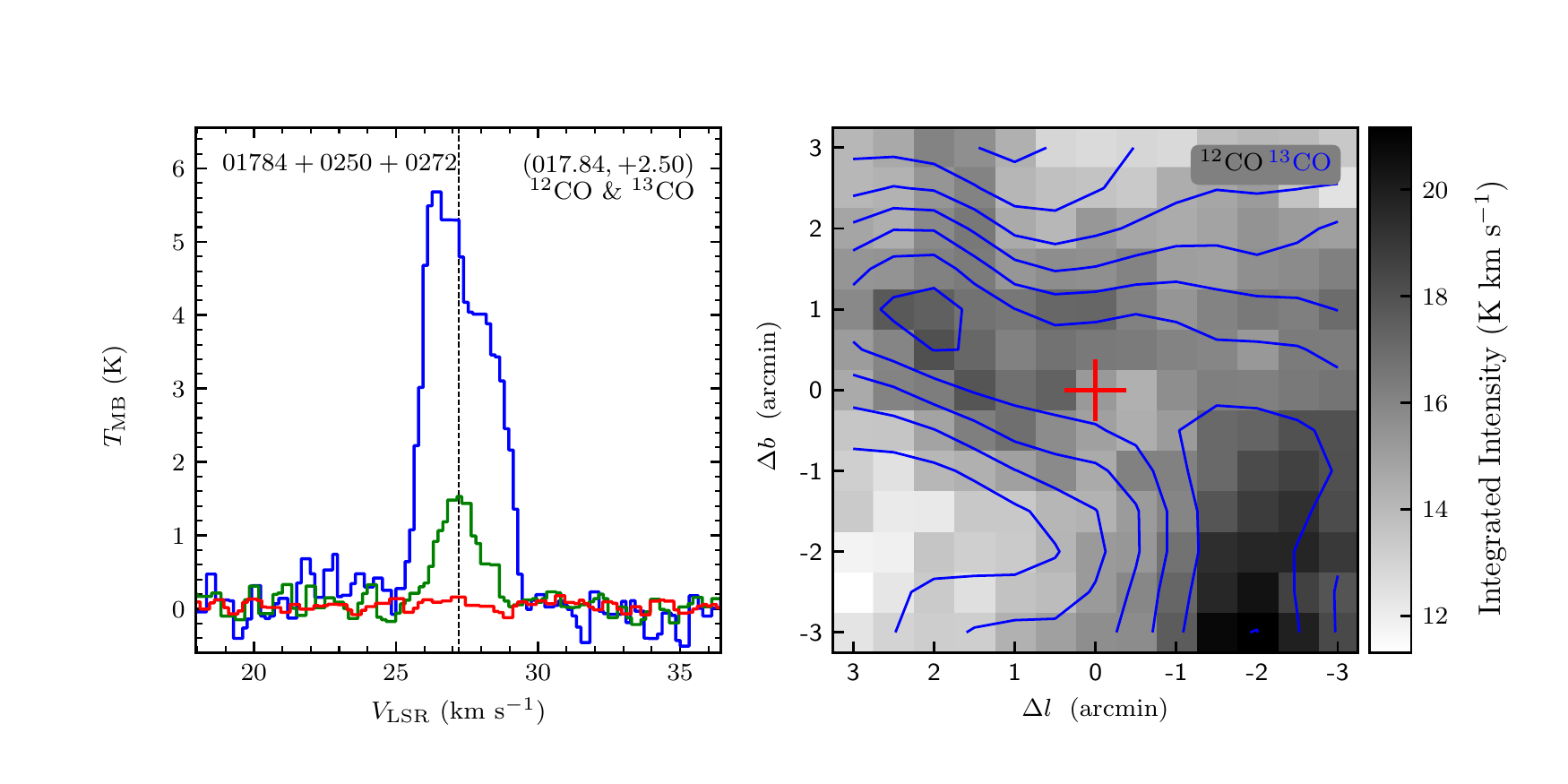}
\includegraphics[width=9.0cm,angle=0]{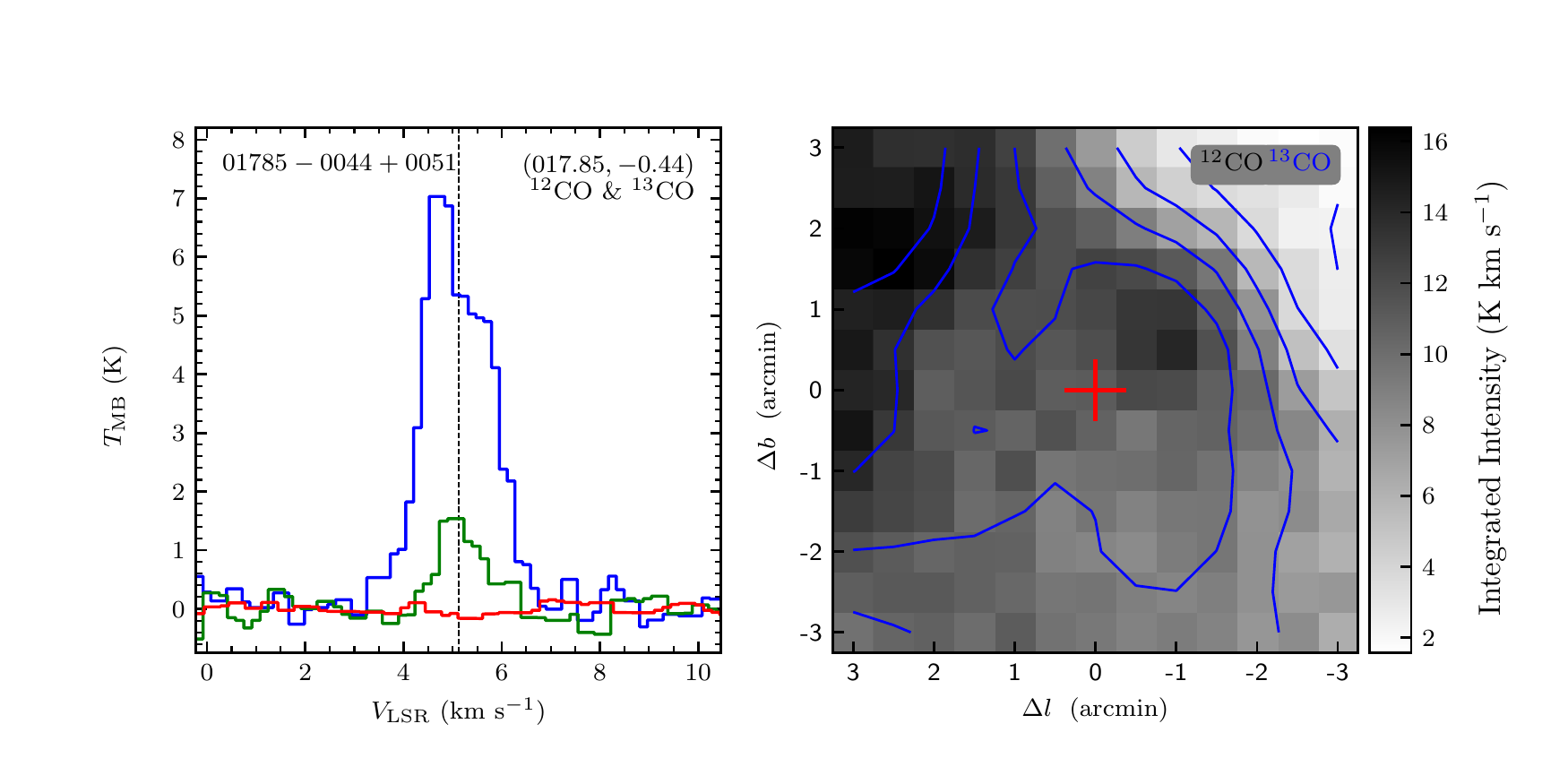}
\end{figure}
\clearpage

\begin{figure}
\includegraphics[width=9.0cm,angle=0]{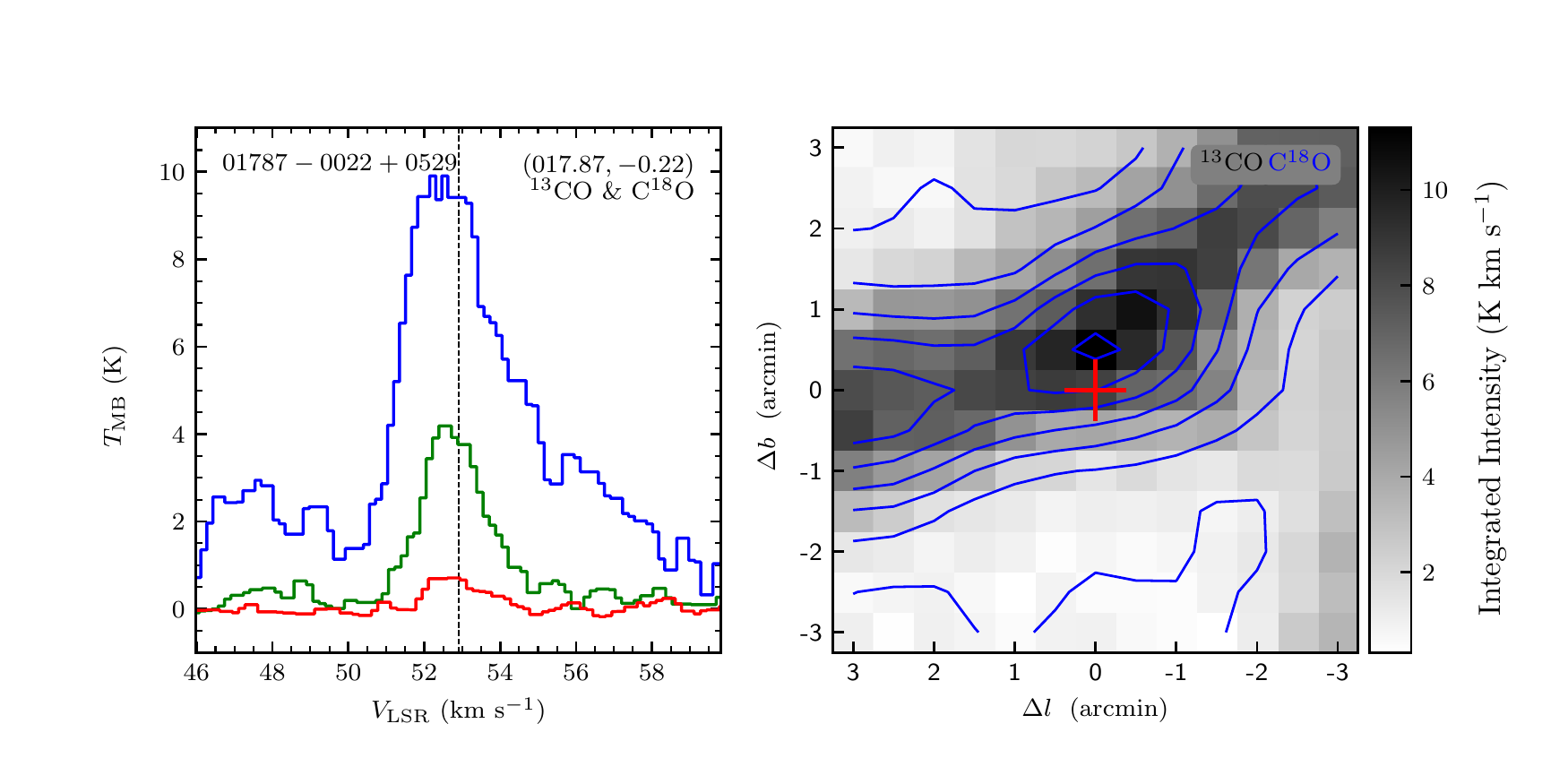}
\includegraphics[width=9.0cm,angle=0]{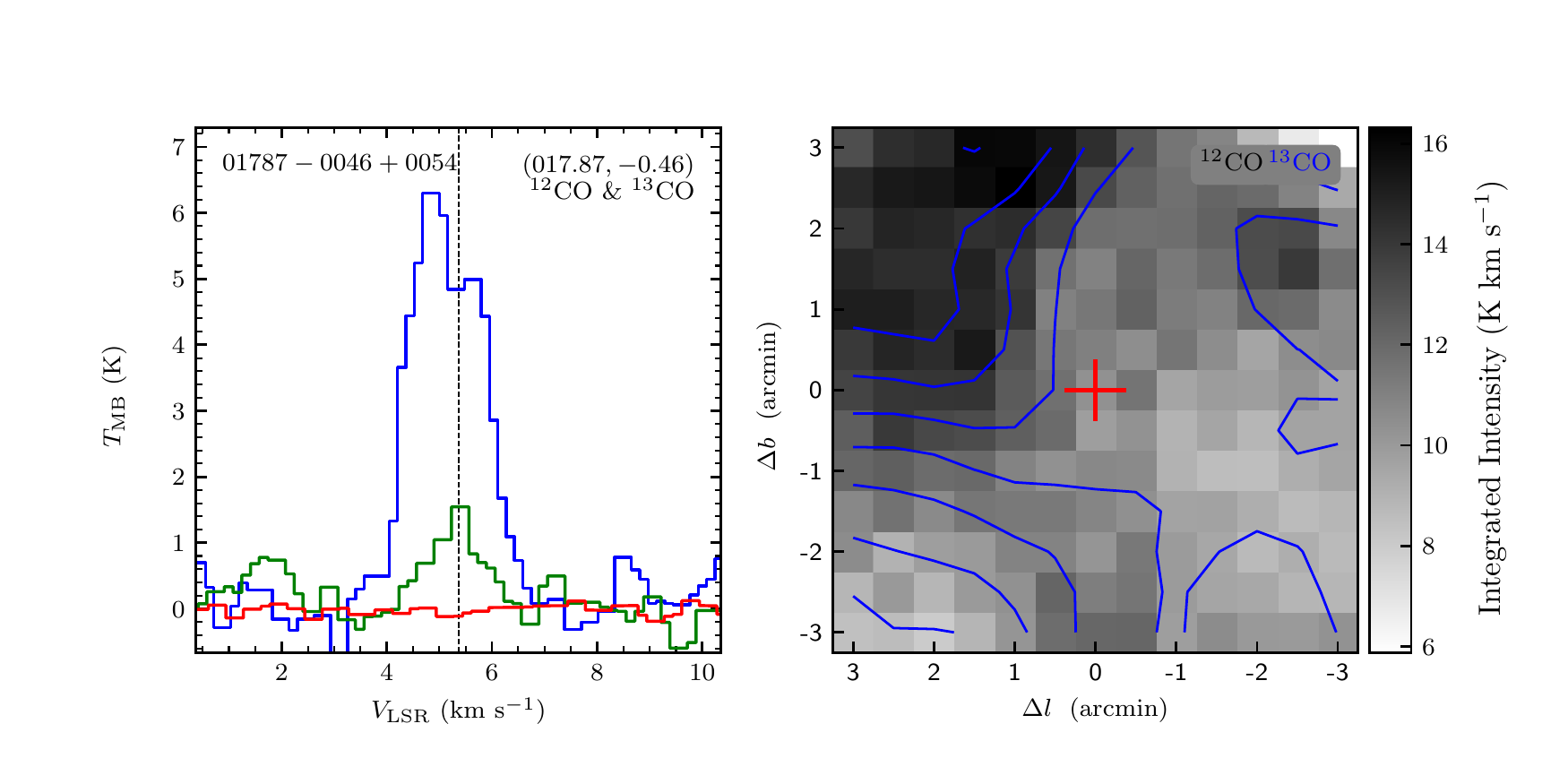}
\vspace{-0.5cm}

\includegraphics[width=9.0cm,angle=0]{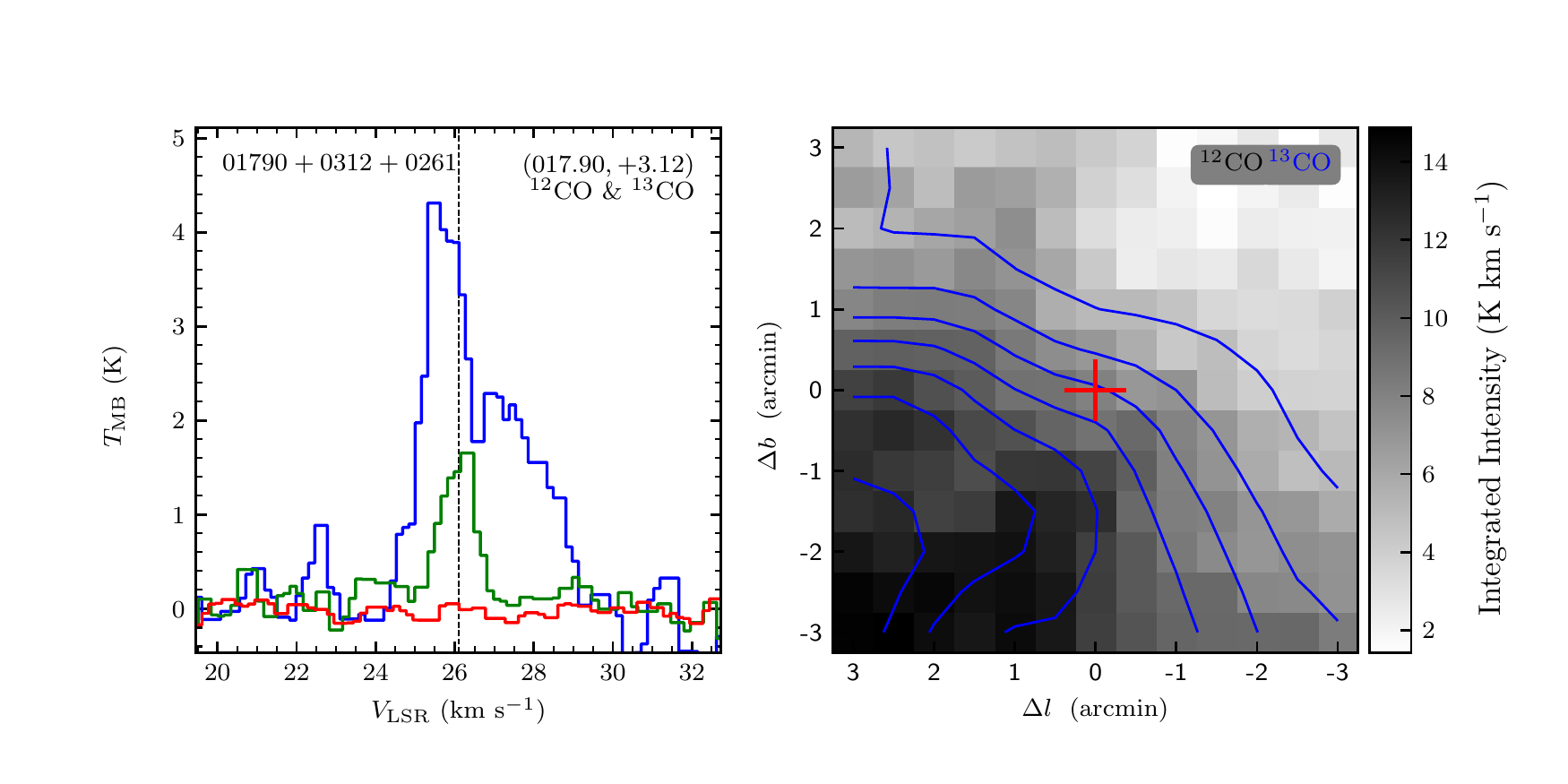}
\includegraphics[width=9.0cm,angle=0]{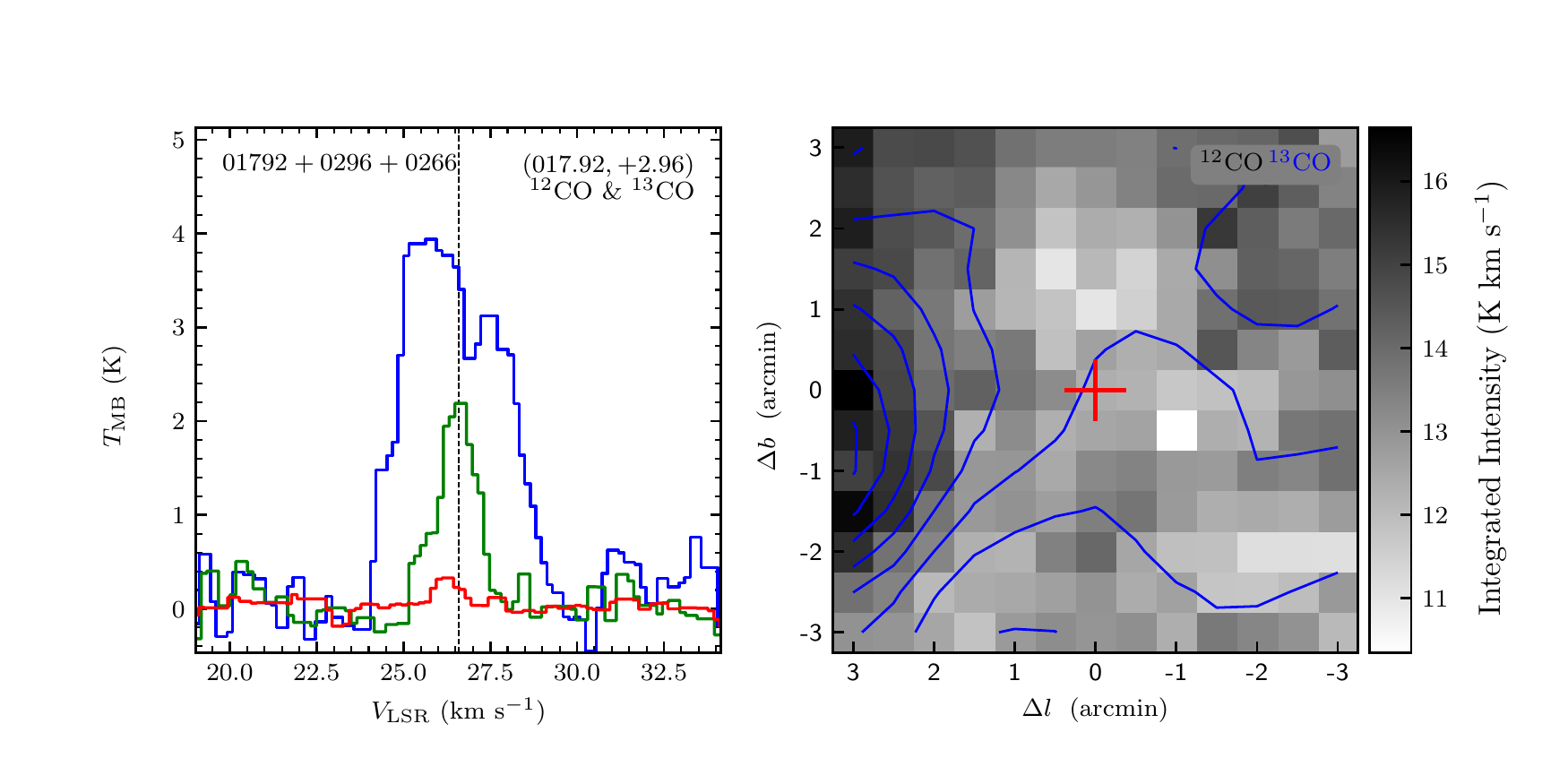}
\vspace{-0.5cm}

\includegraphics[width=9.0cm,angle=0]{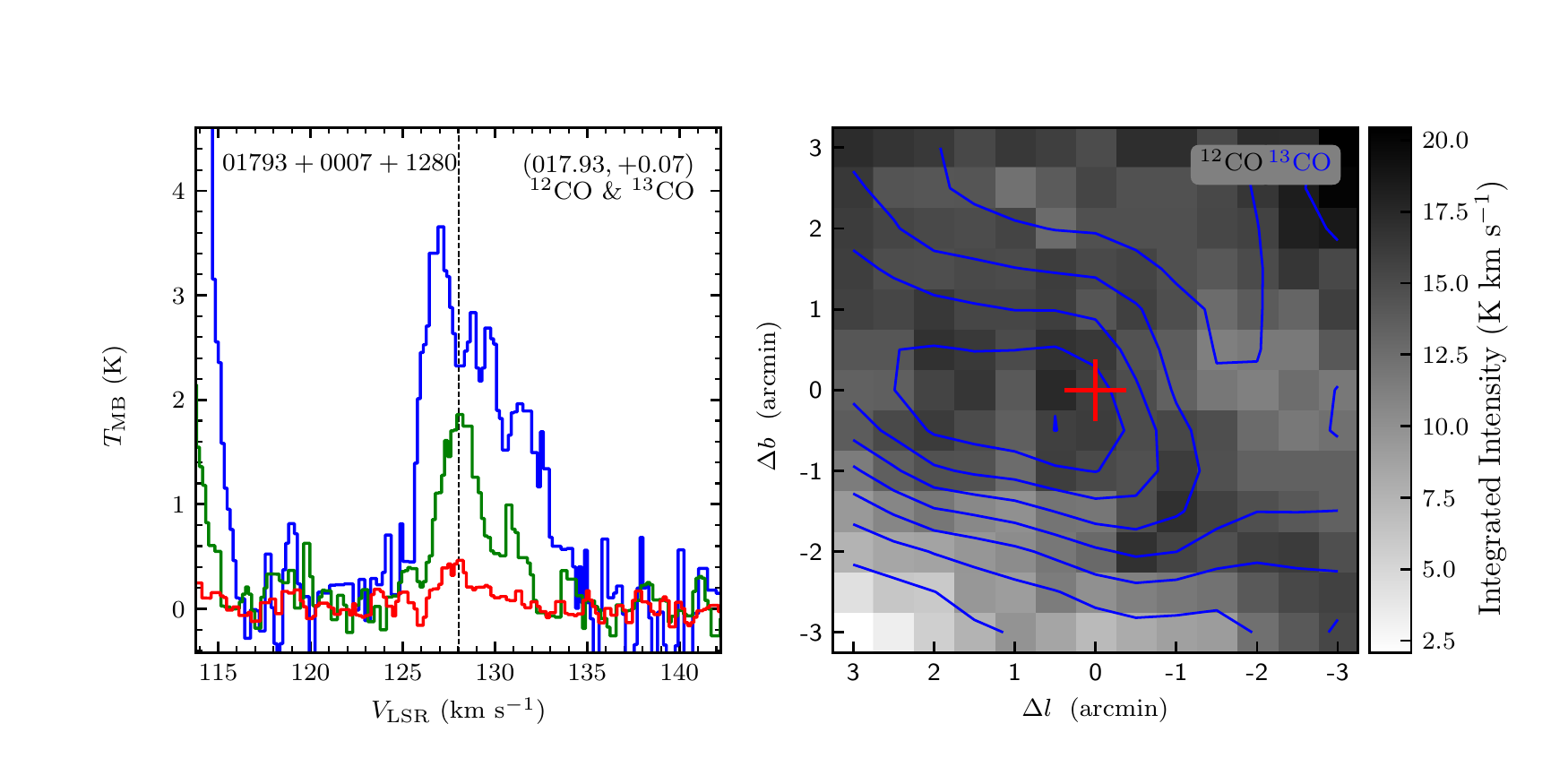}
\includegraphics[width=9.0cm,angle=0]{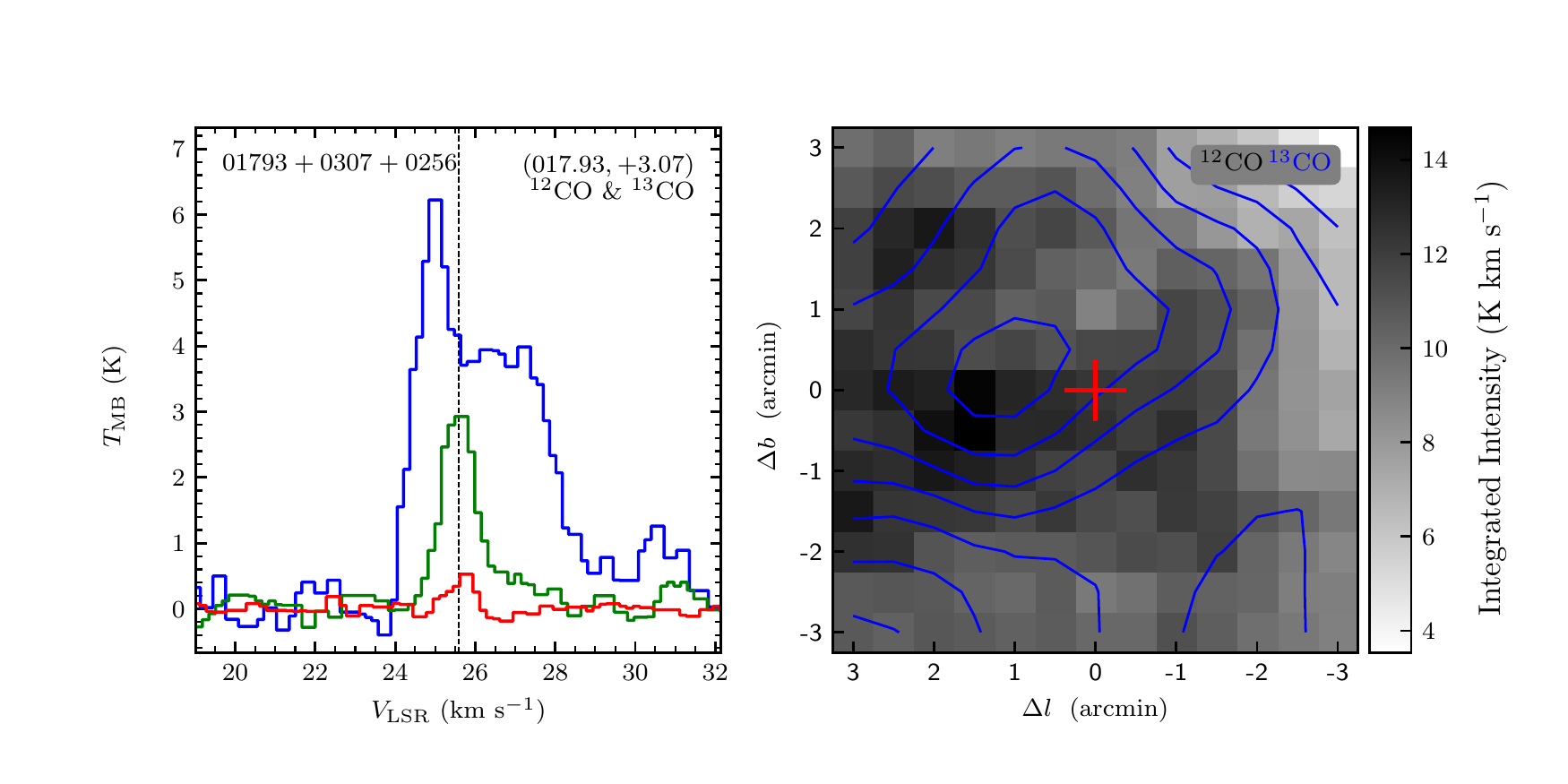}
\vspace{-0.5cm}

\includegraphics[width=9.0cm,angle=0]{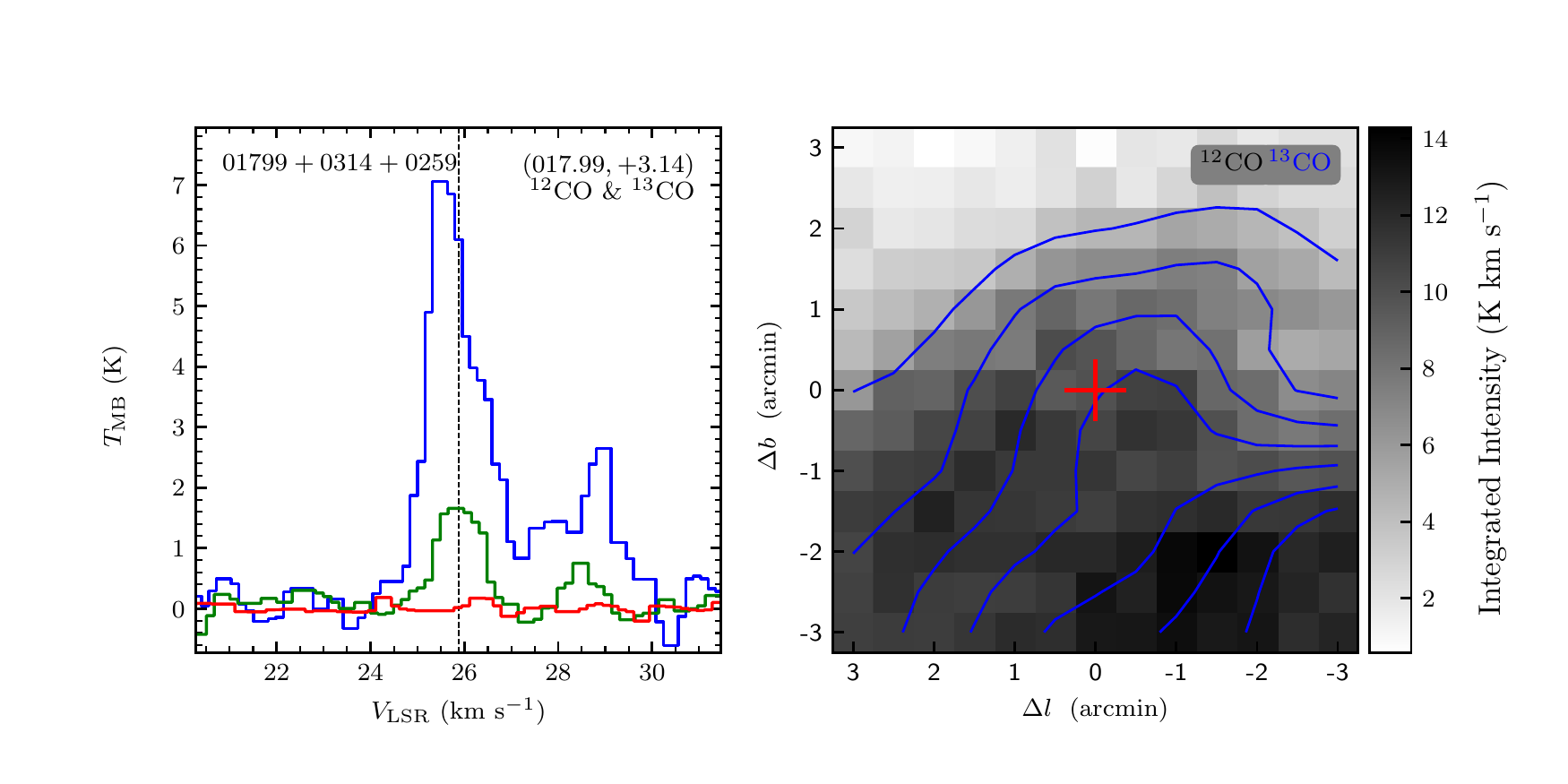}
\includegraphics[width=9.0cm,angle=0]{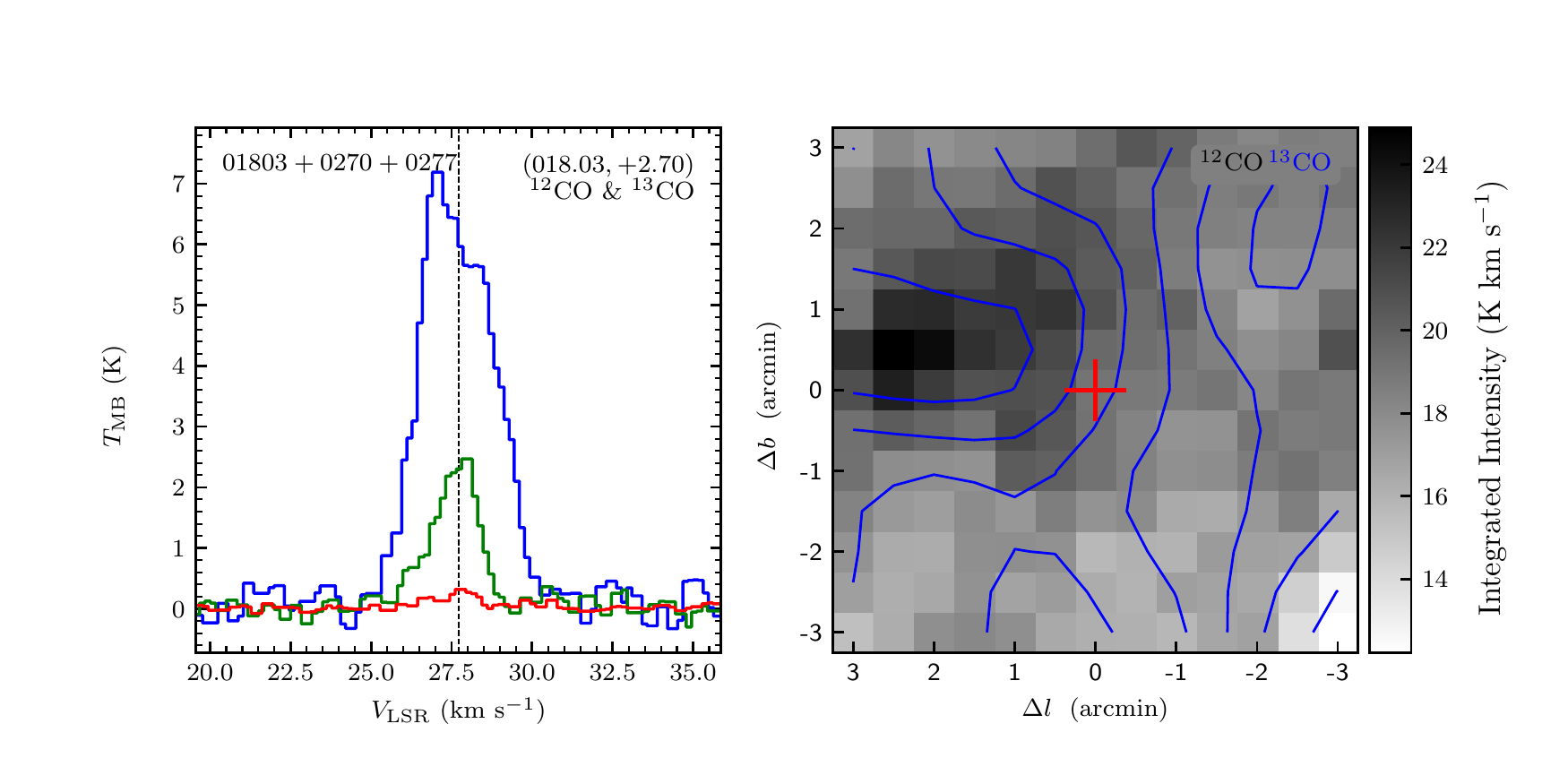}
\vspace{-0.5cm}

\includegraphics[width=9.0cm,angle=0]{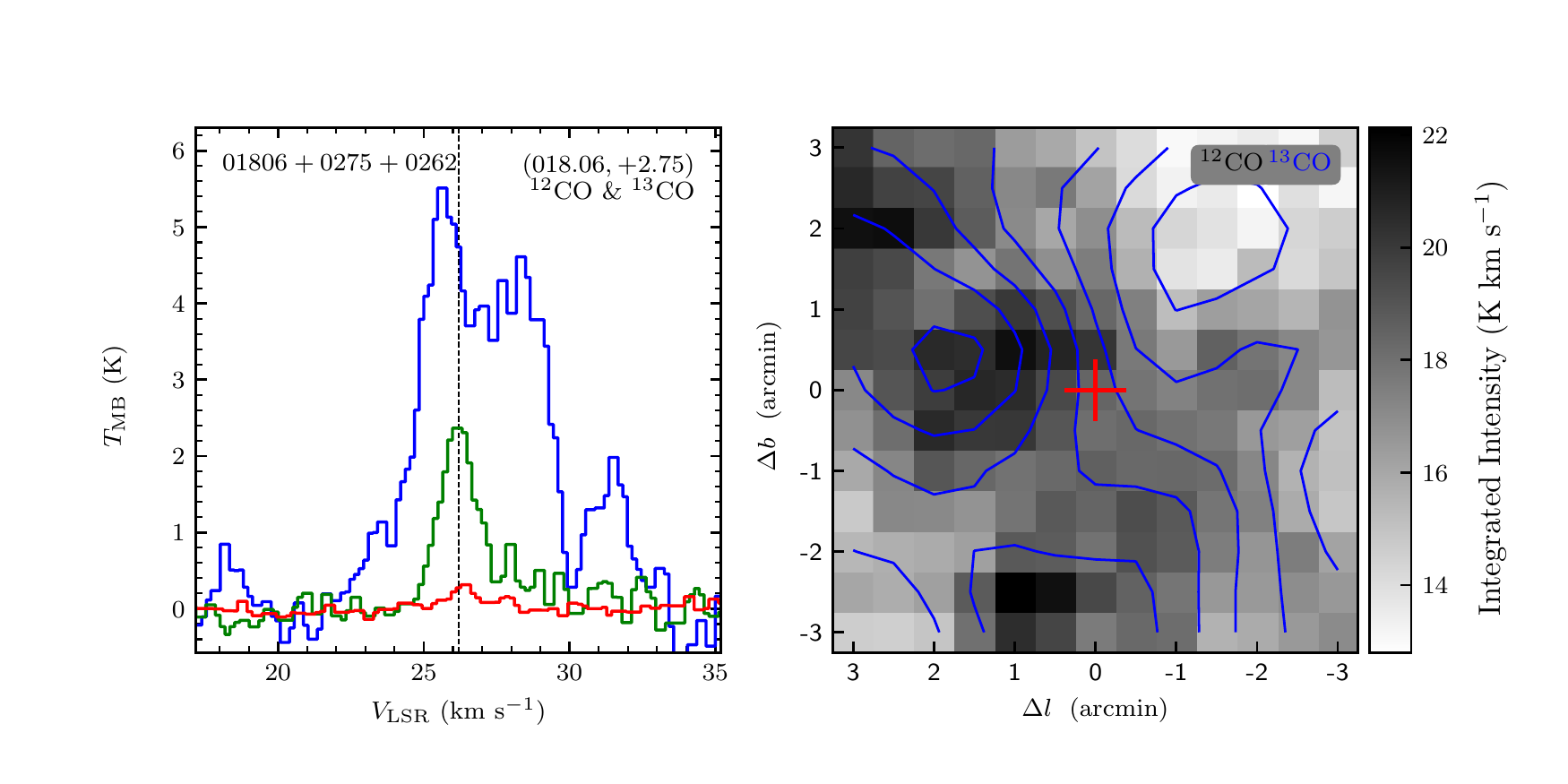}
\includegraphics[width=9.0cm,angle=0]{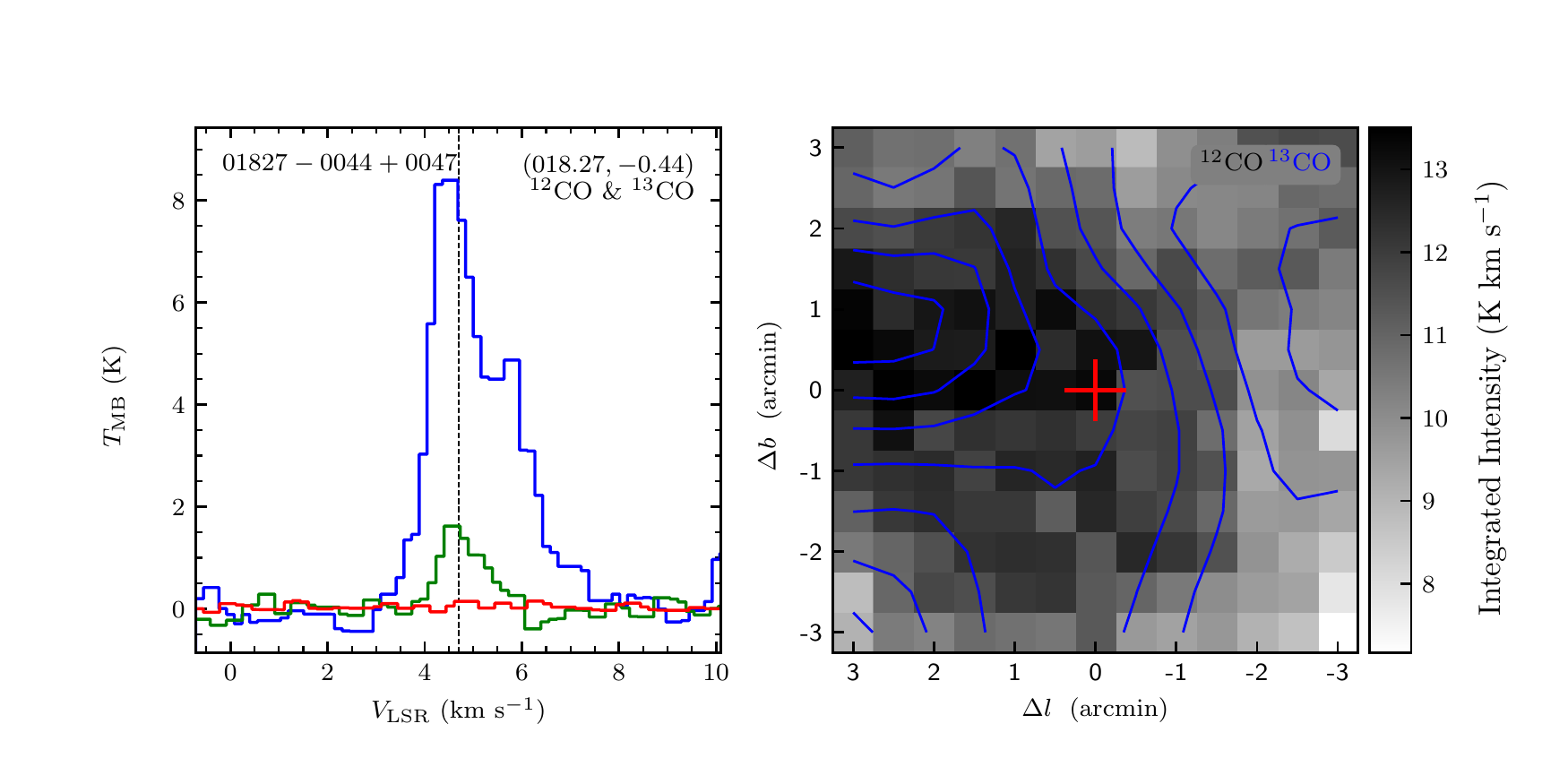}
\end{figure}
\clearpage

\begin{figure}
\includegraphics[width=9.0cm,angle=0]{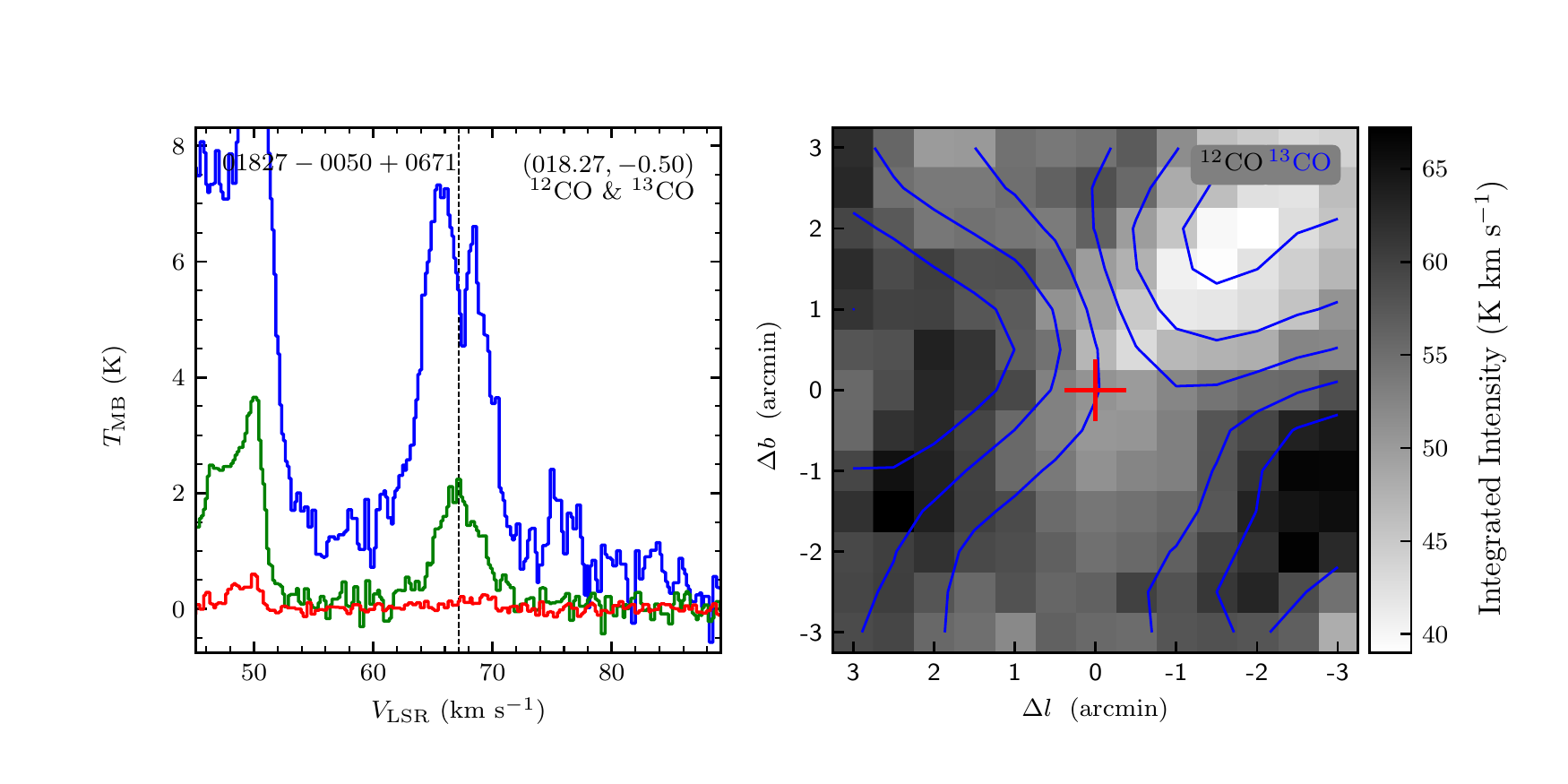}
\includegraphics[width=9.0cm,angle=0]{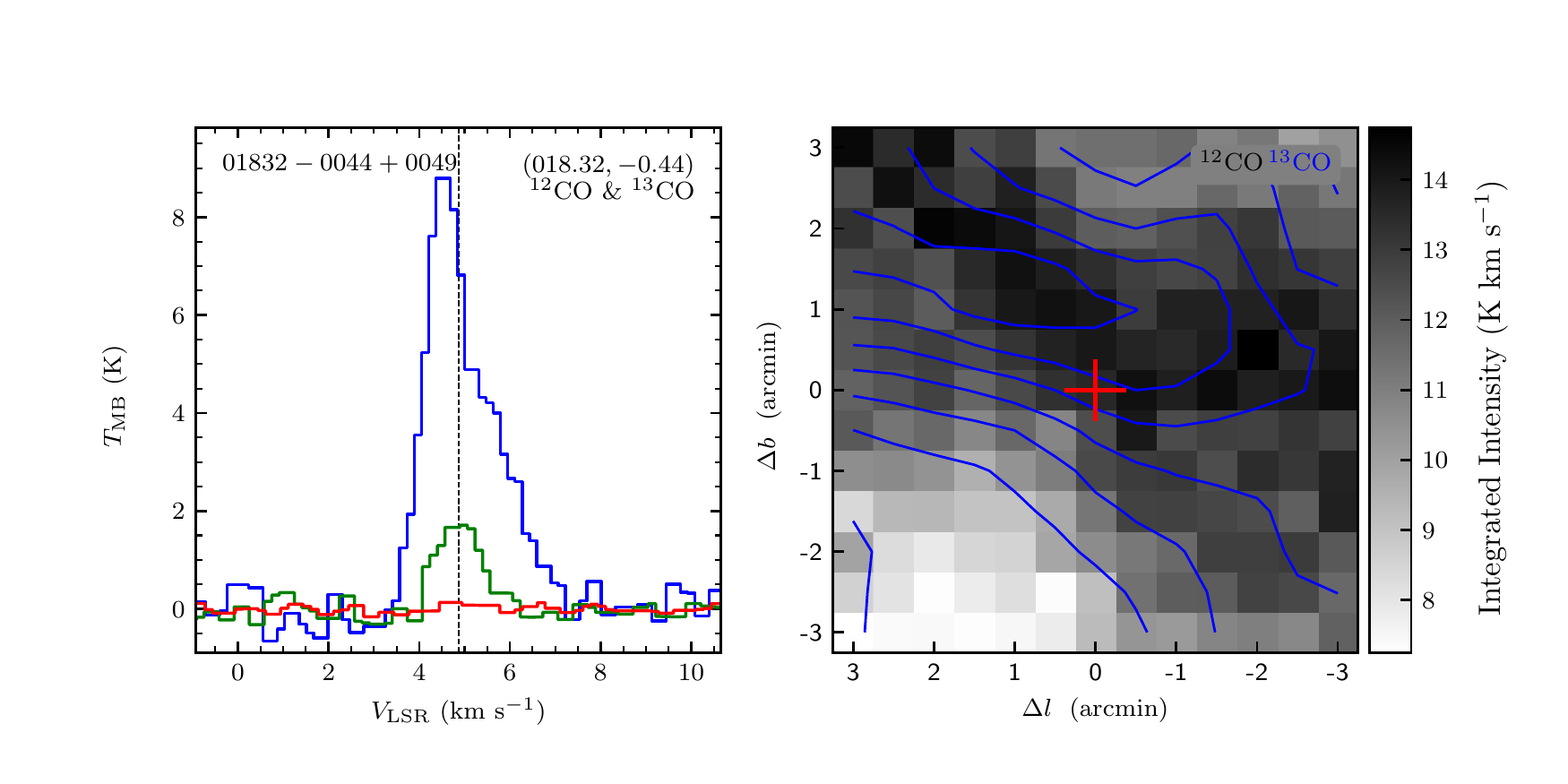}
\vspace{-0.5cm}

\includegraphics[width=9.0cm,angle=0]{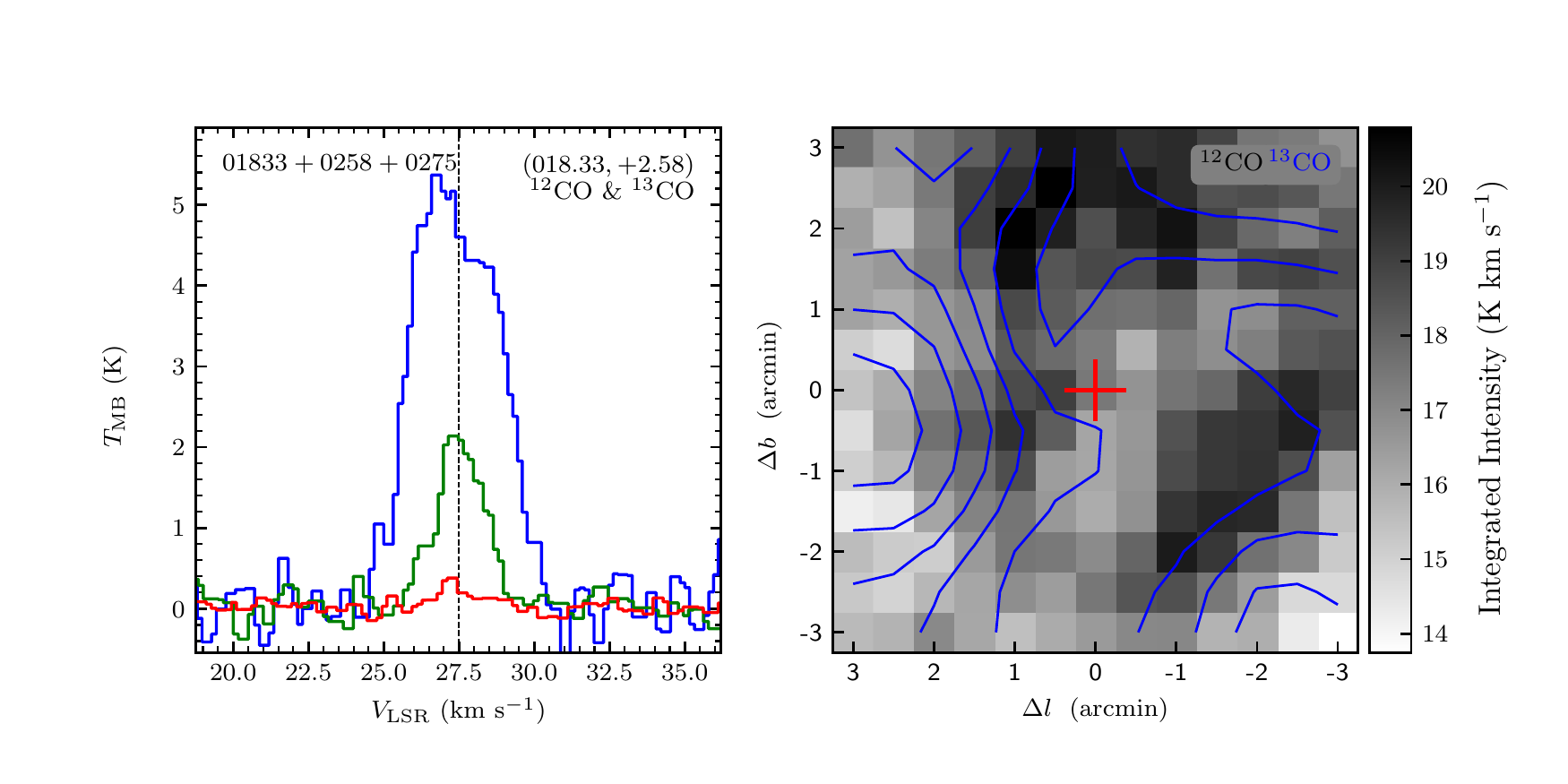}
\includegraphics[width=9.0cm,angle=0]{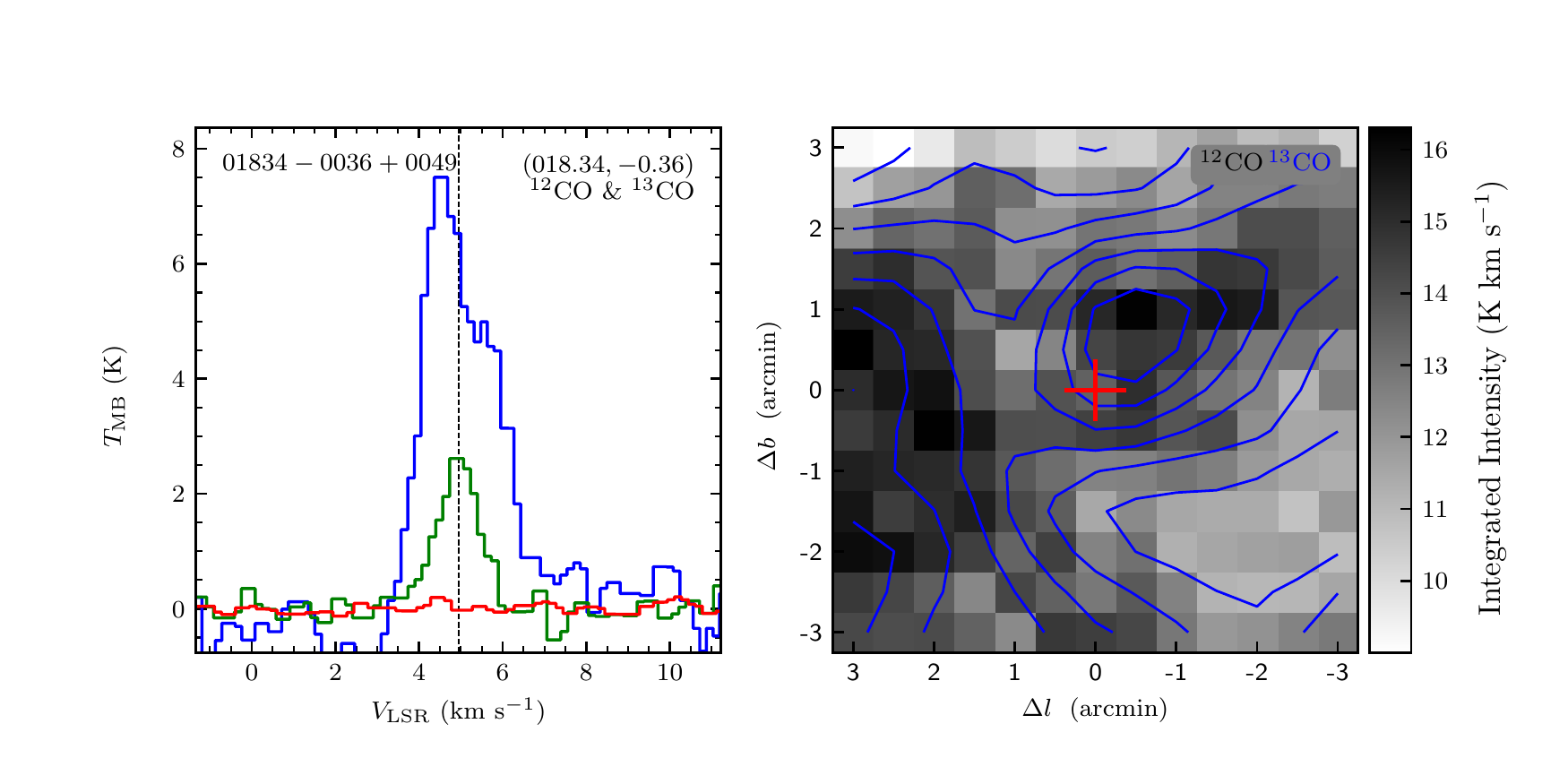}
\vspace{-0.5cm}

\includegraphics[width=9.0cm,angle=0]{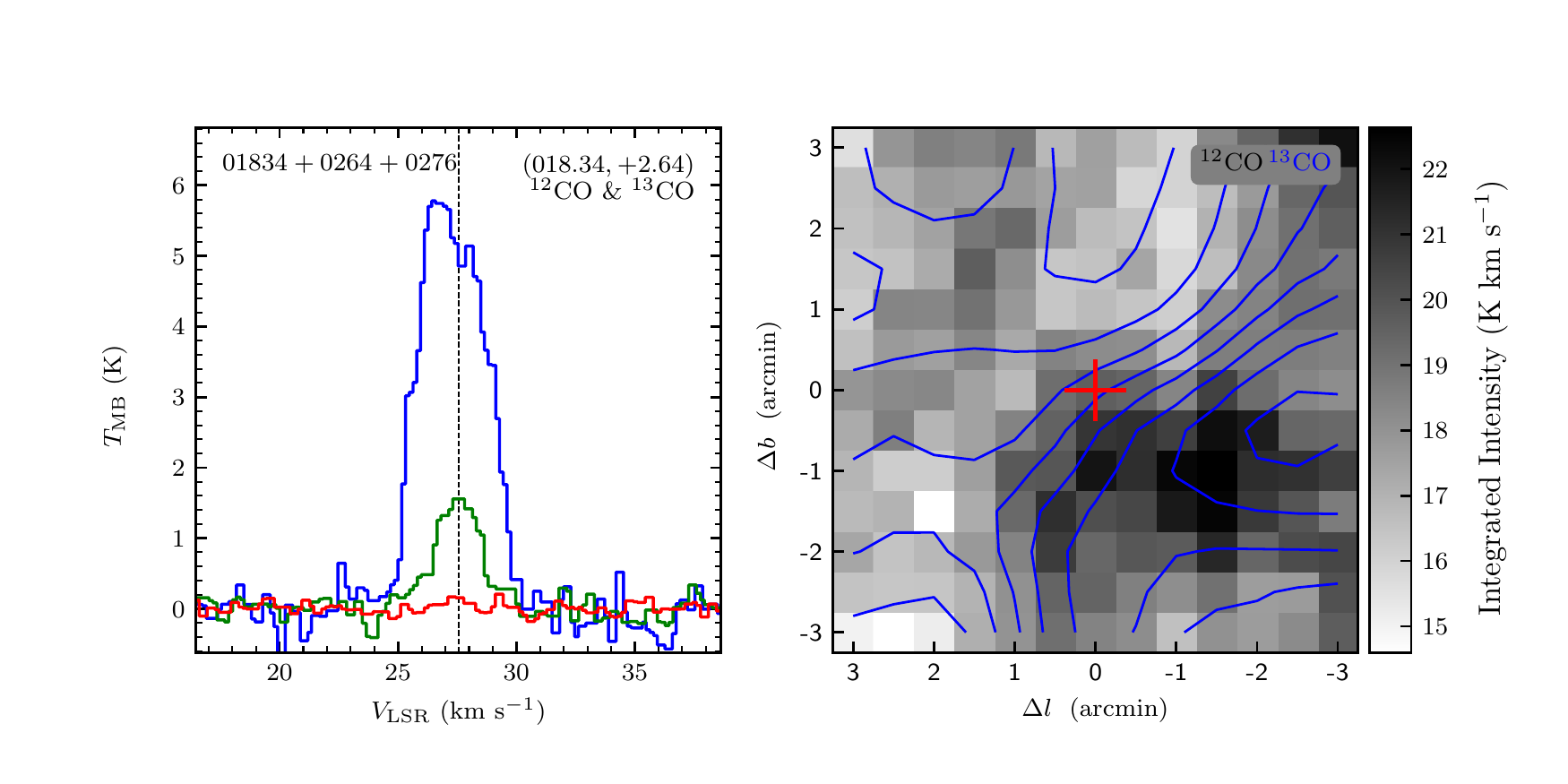}
\includegraphics[width=9.0cm,angle=0]{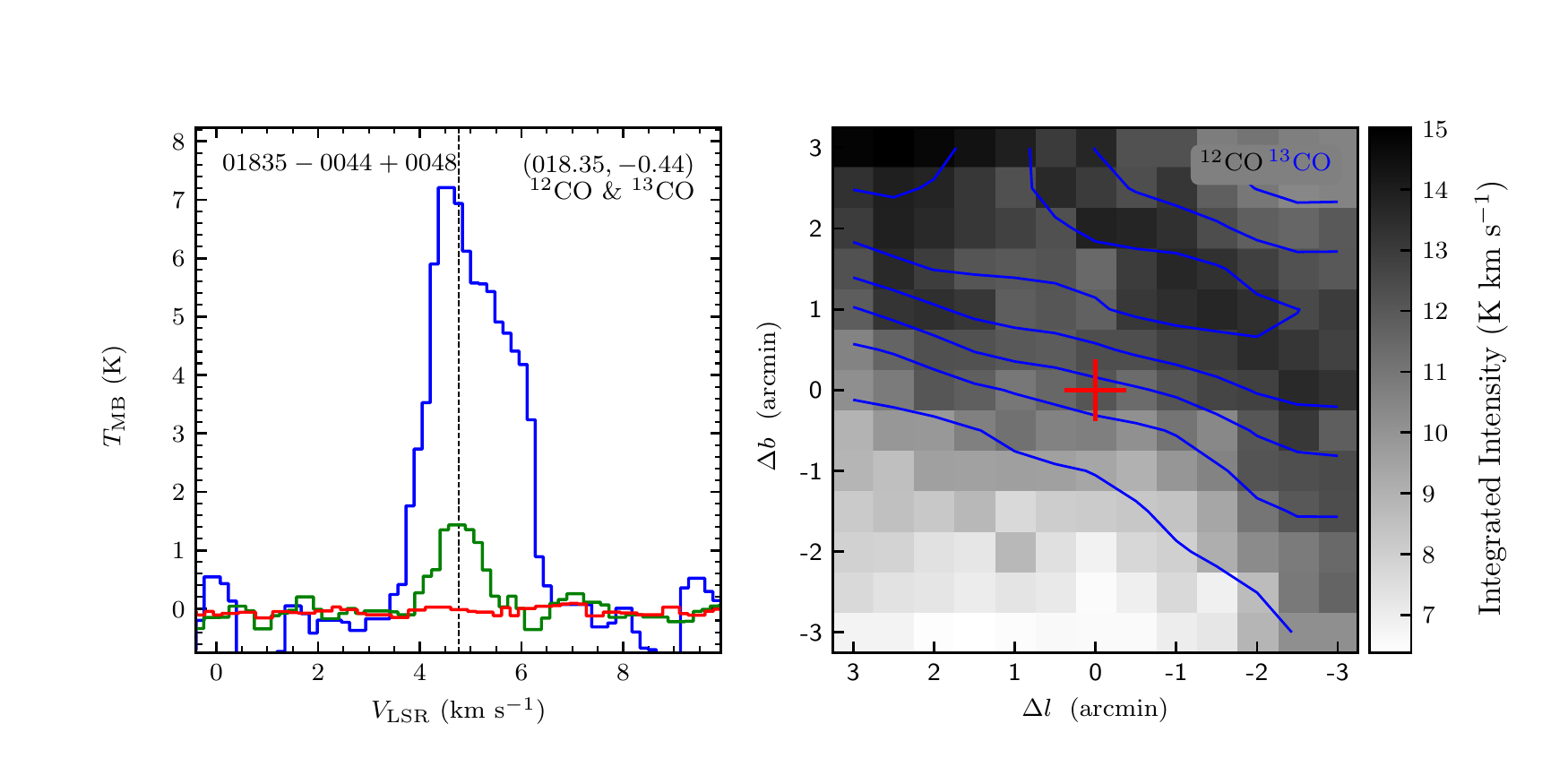}
\vspace{-0.5cm}

\includegraphics[width=9.0cm,angle=0]{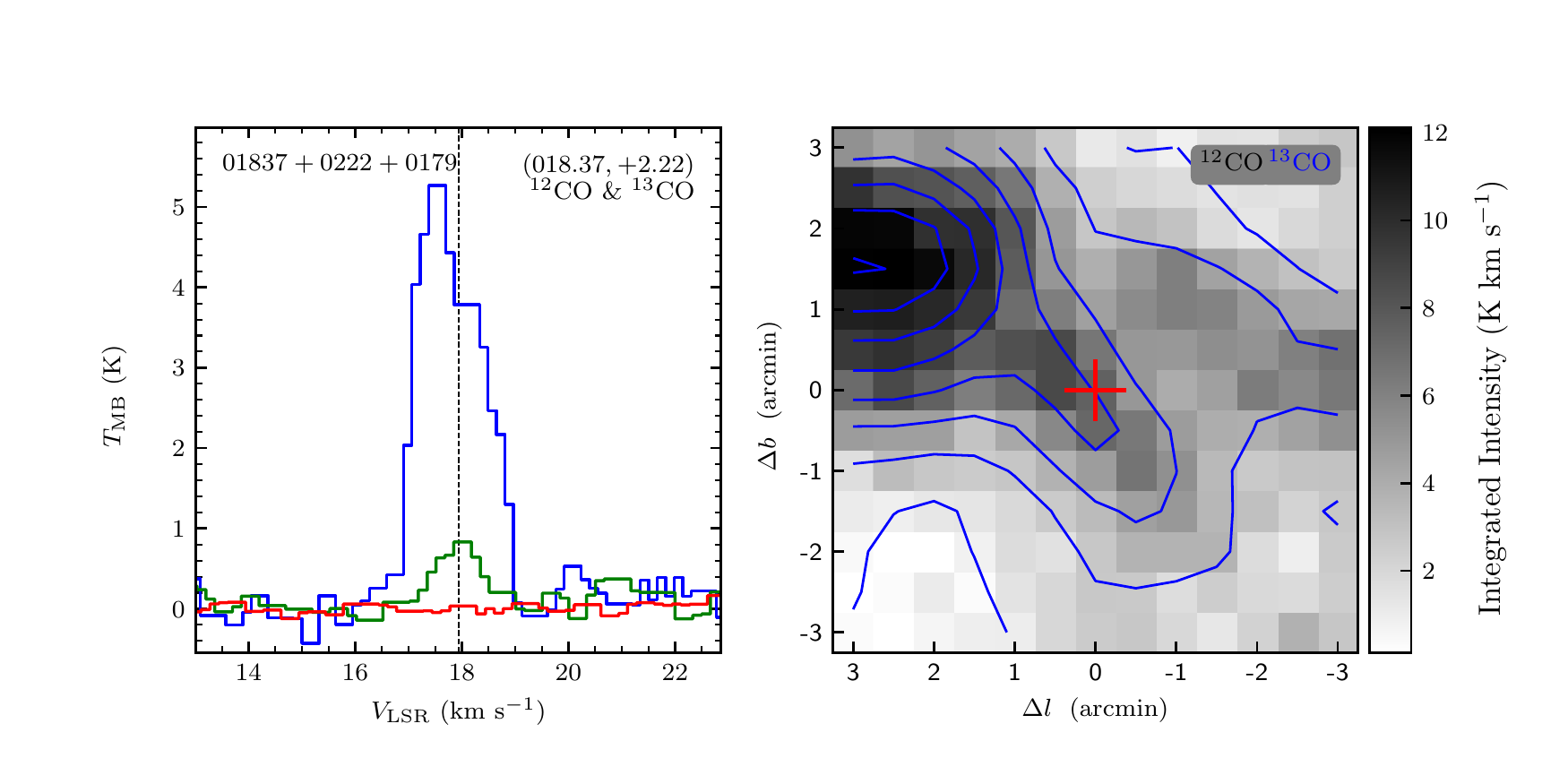}
\includegraphics[width=9.0cm,angle=0]{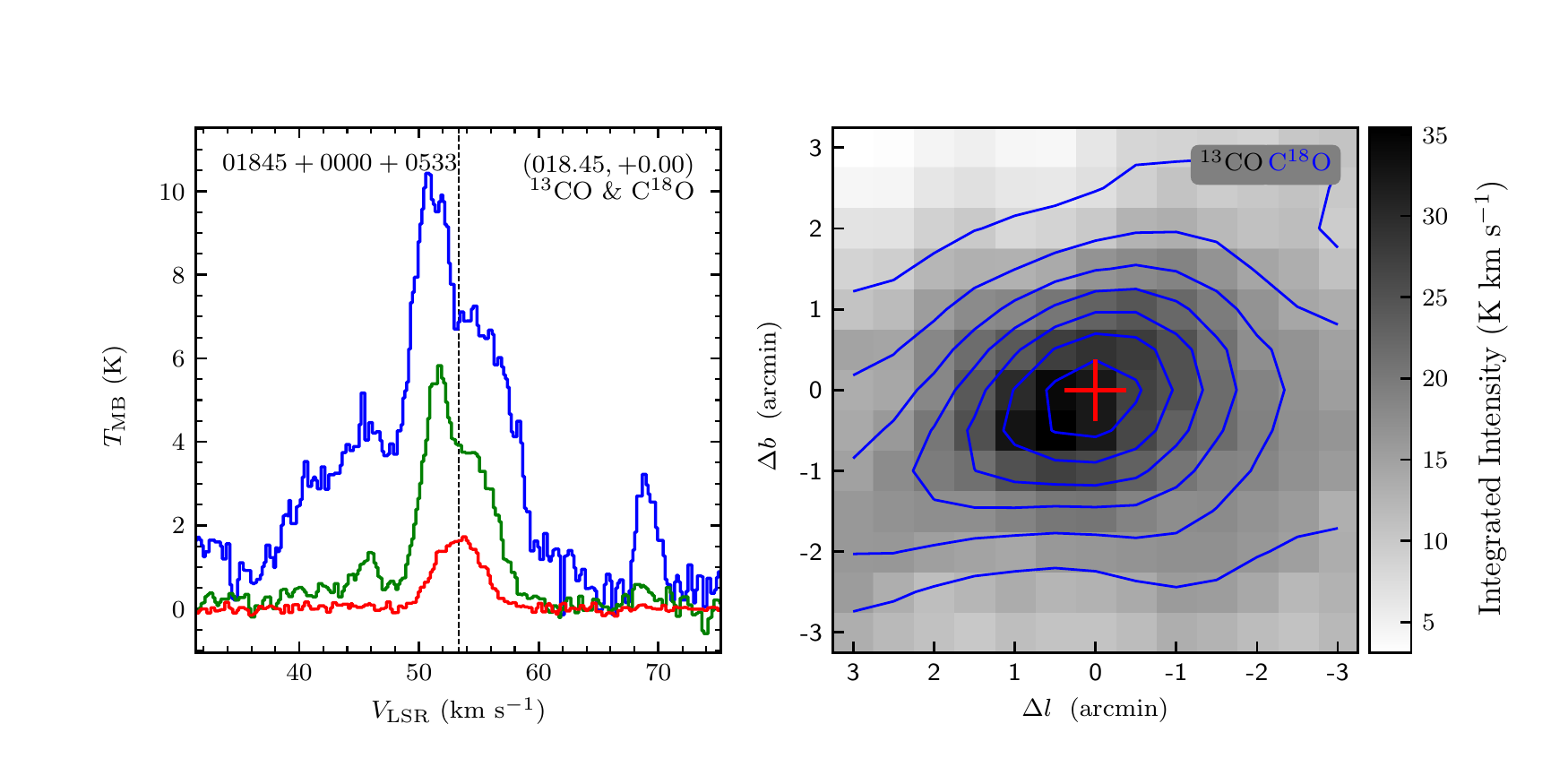}
\vspace{-0.5cm}

\includegraphics[width=9.0cm,angle=0]{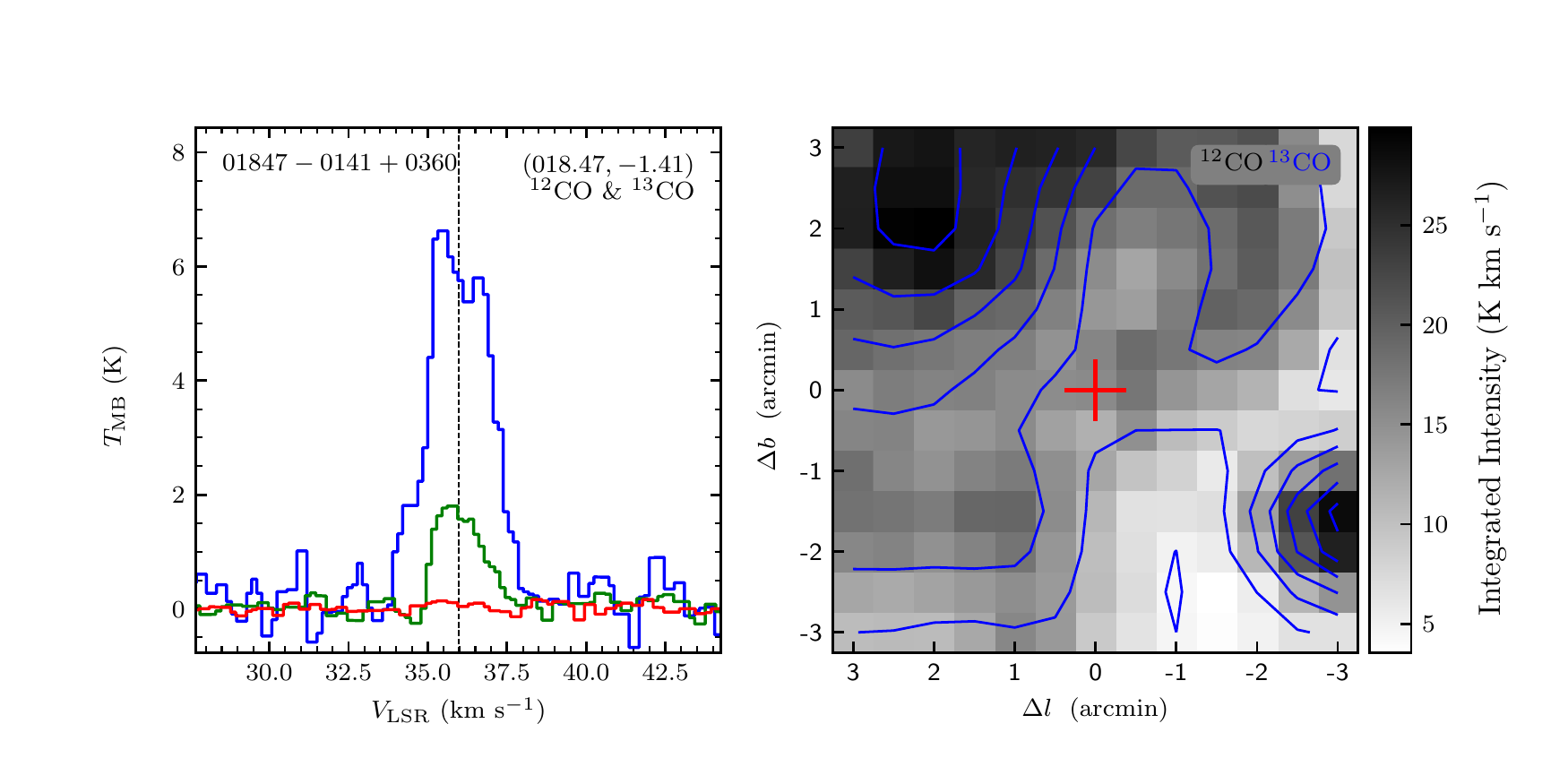}
\includegraphics[width=9.0cm,angle=0]{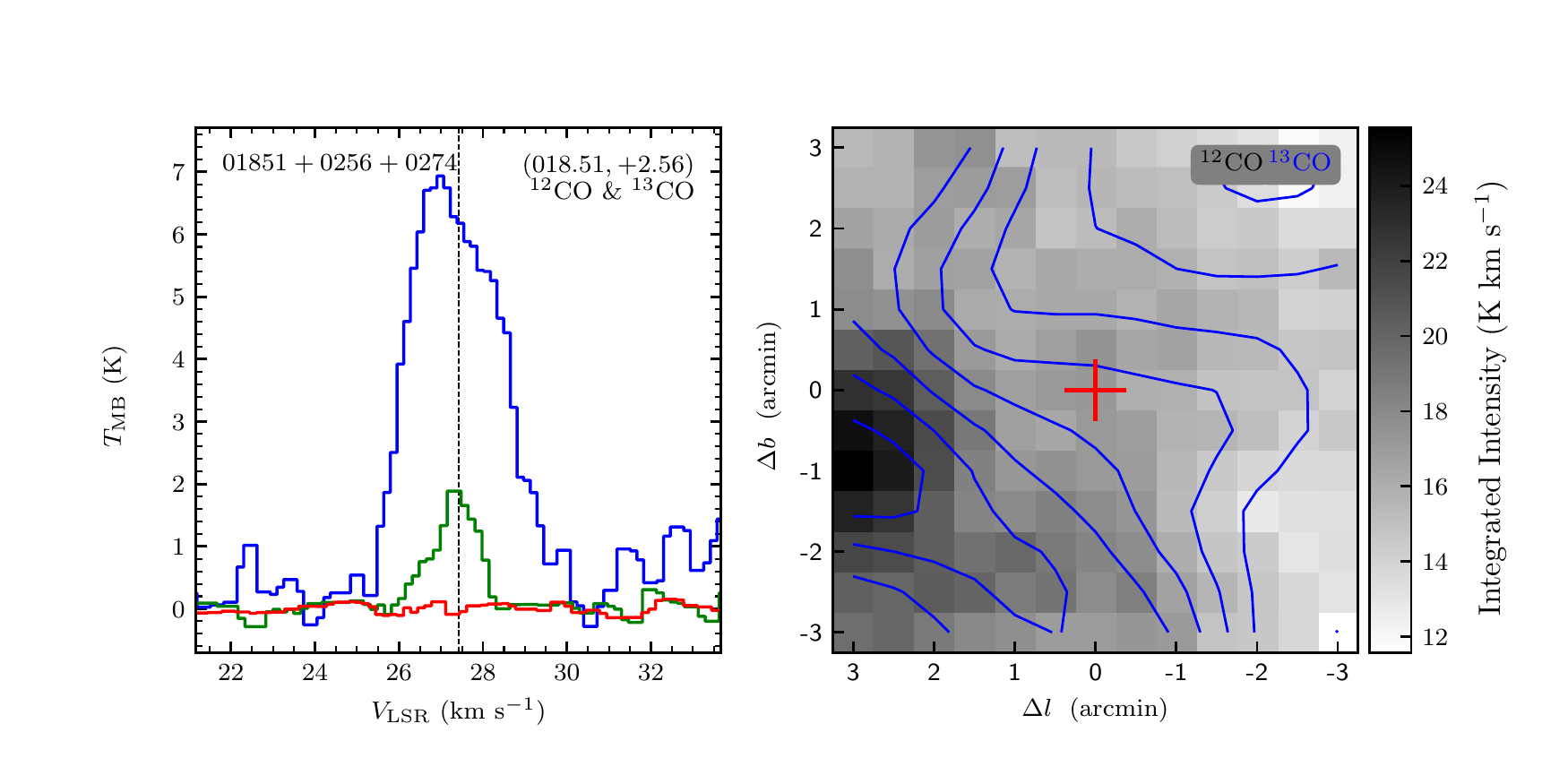}
\end{figure}
\clearpage

\begin{figure}
\includegraphics[width=9.0cm,angle=0]{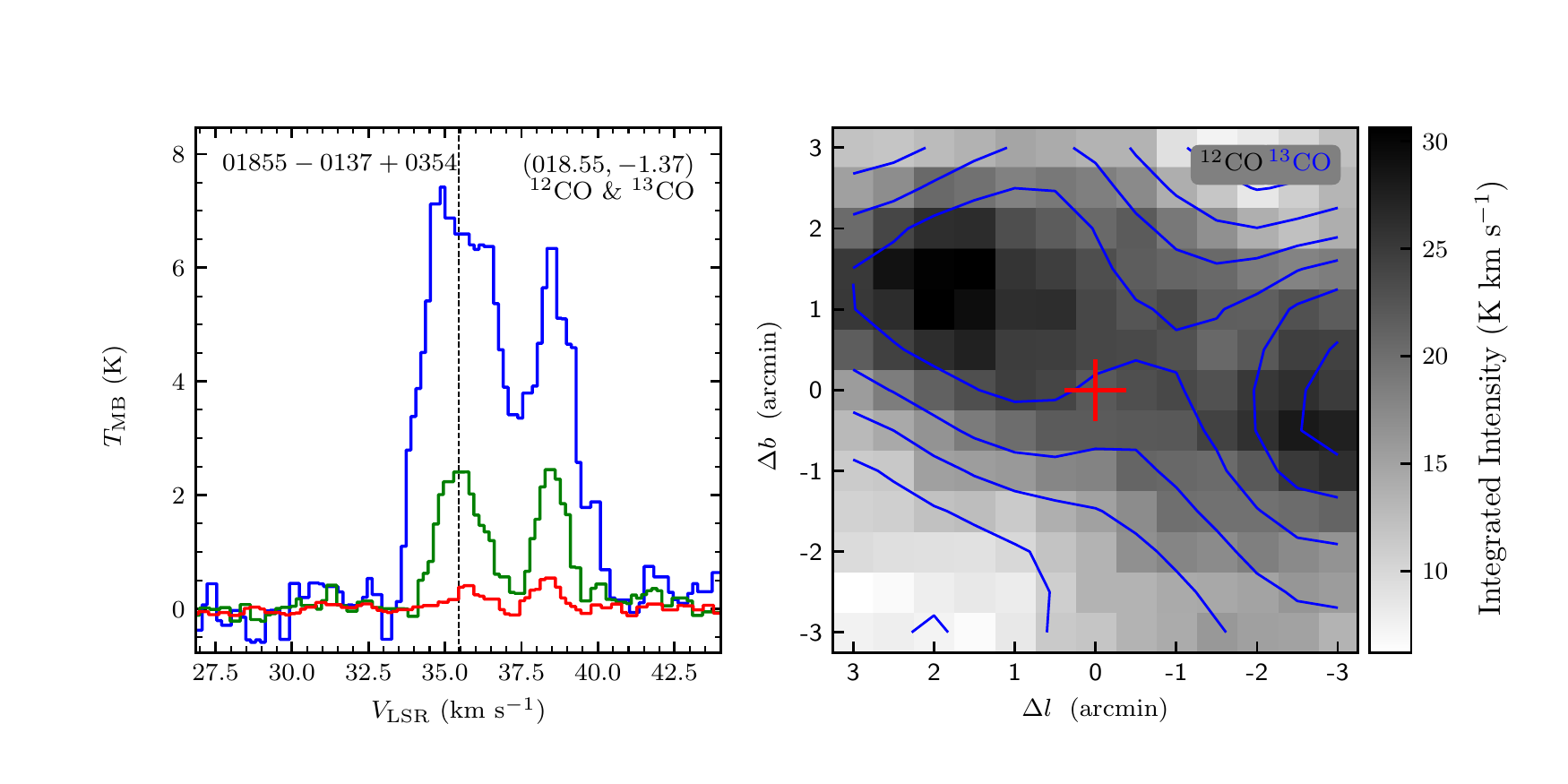}
\includegraphics[width=9.0cm,angle=0]{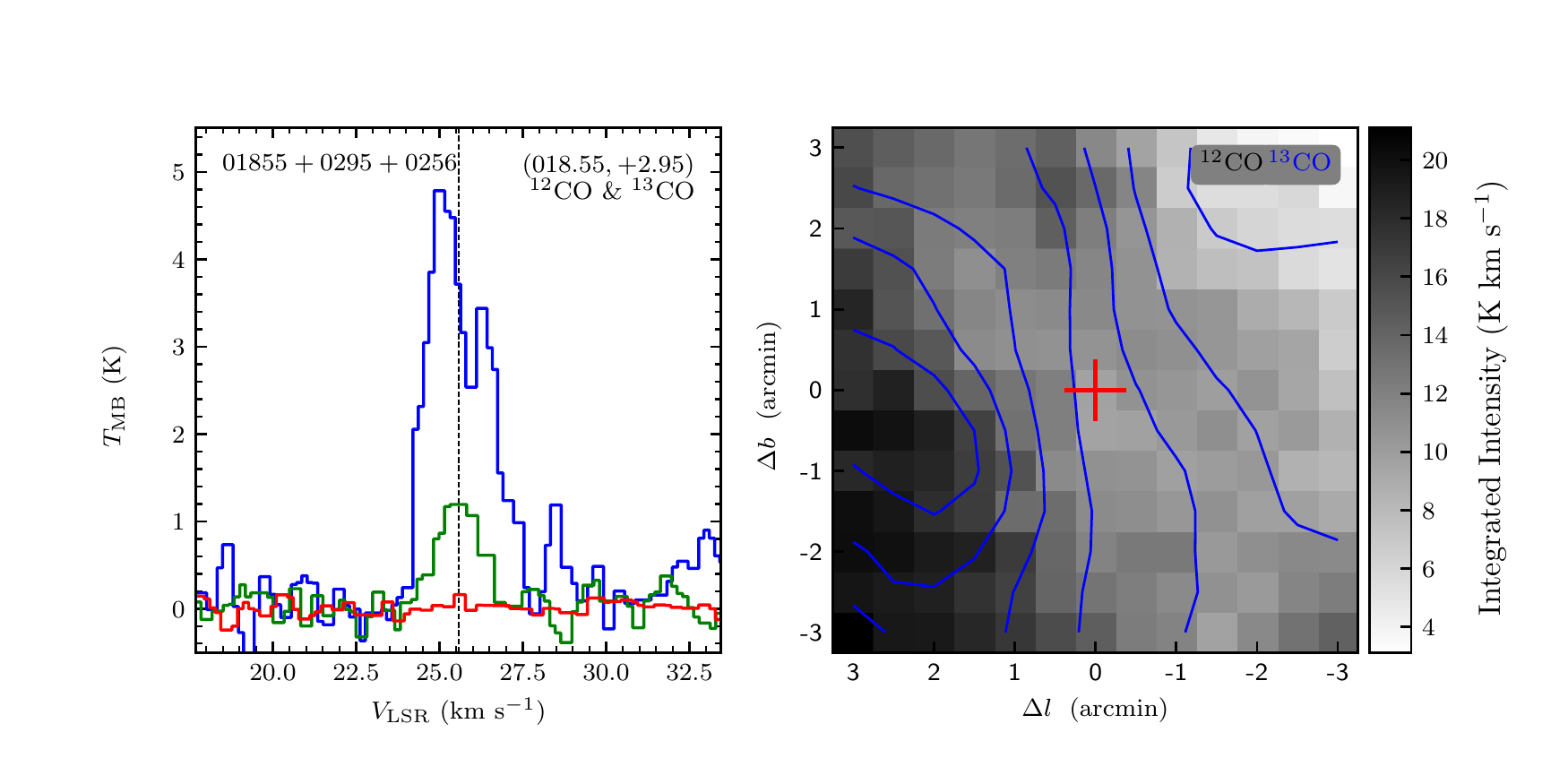}
\vspace{-0.5cm}

\includegraphics[width=9.0cm,angle=0]{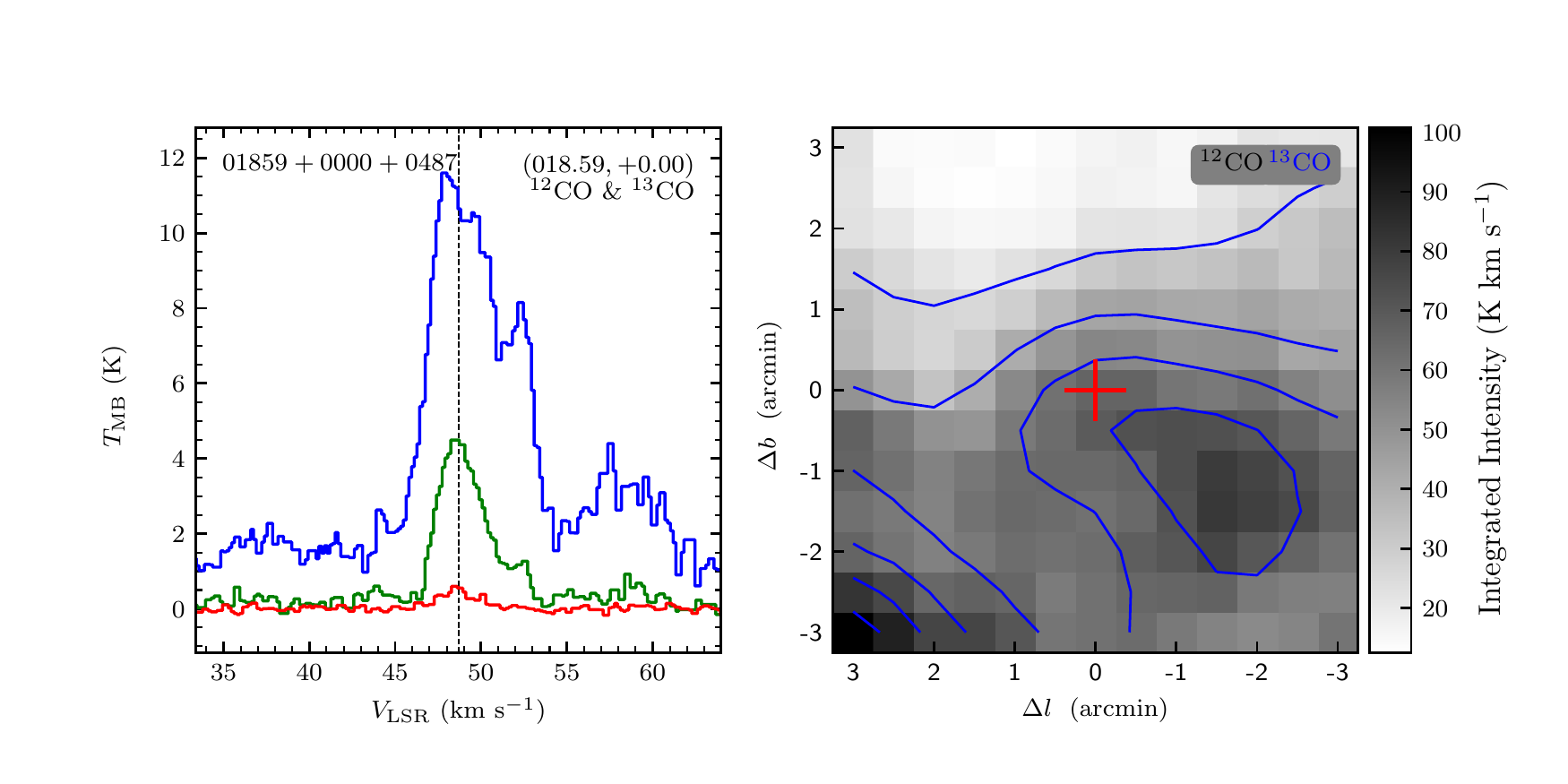}
\includegraphics[width=9.0cm,angle=0]{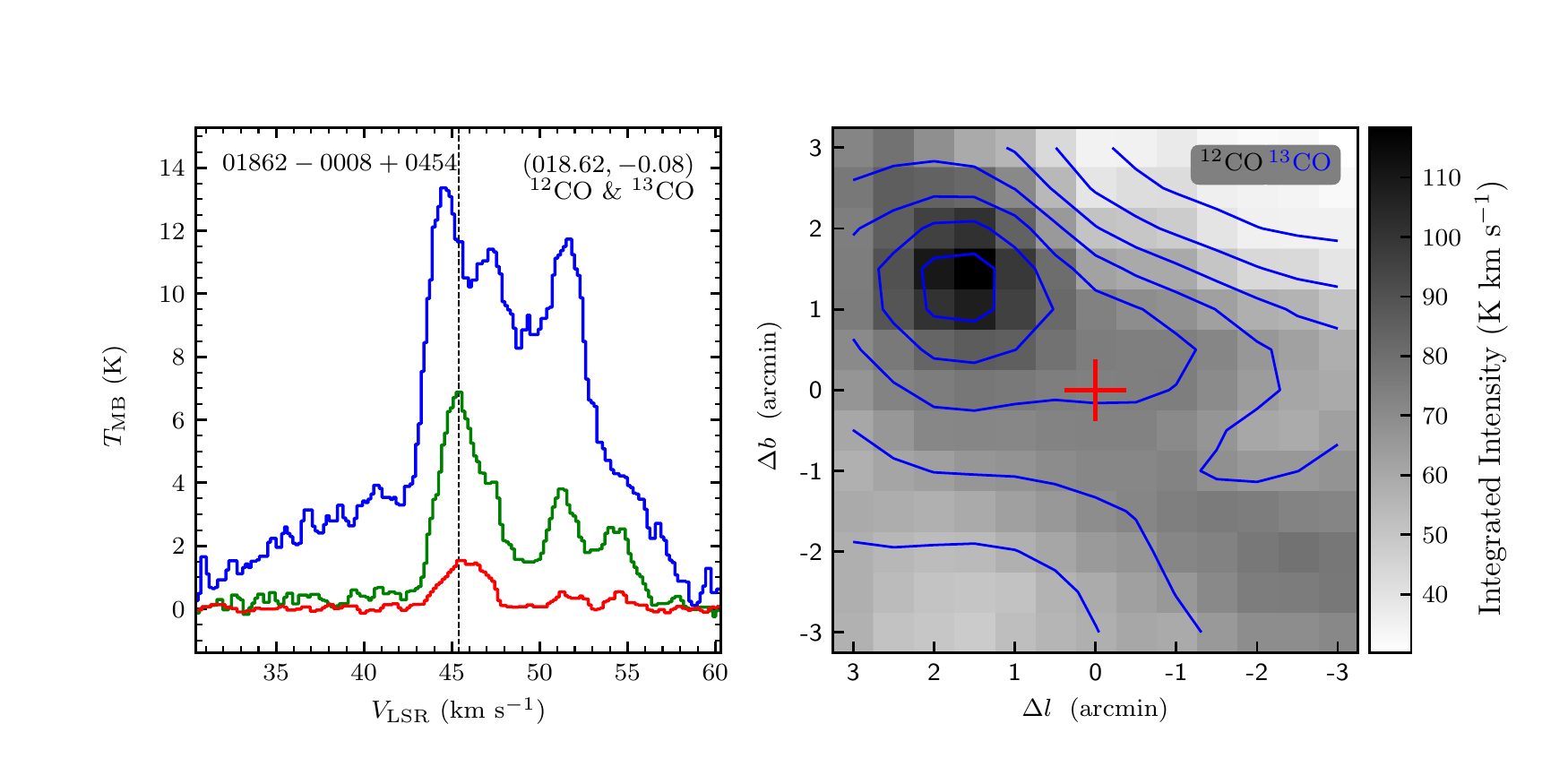}
\vspace{-0.5cm}

\includegraphics[width=9.0cm,angle=0]{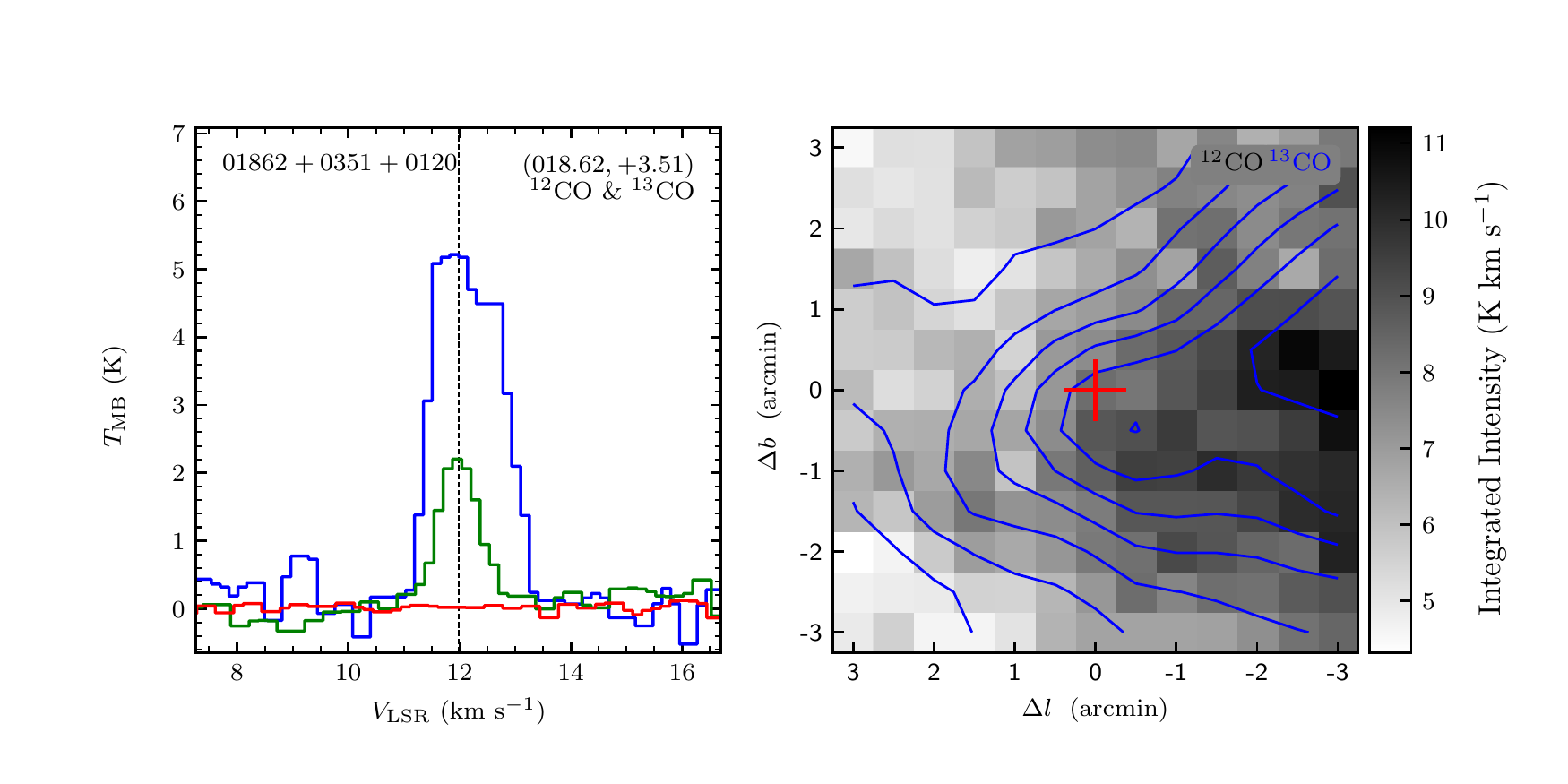}
\includegraphics[width=9.0cm,angle=0]{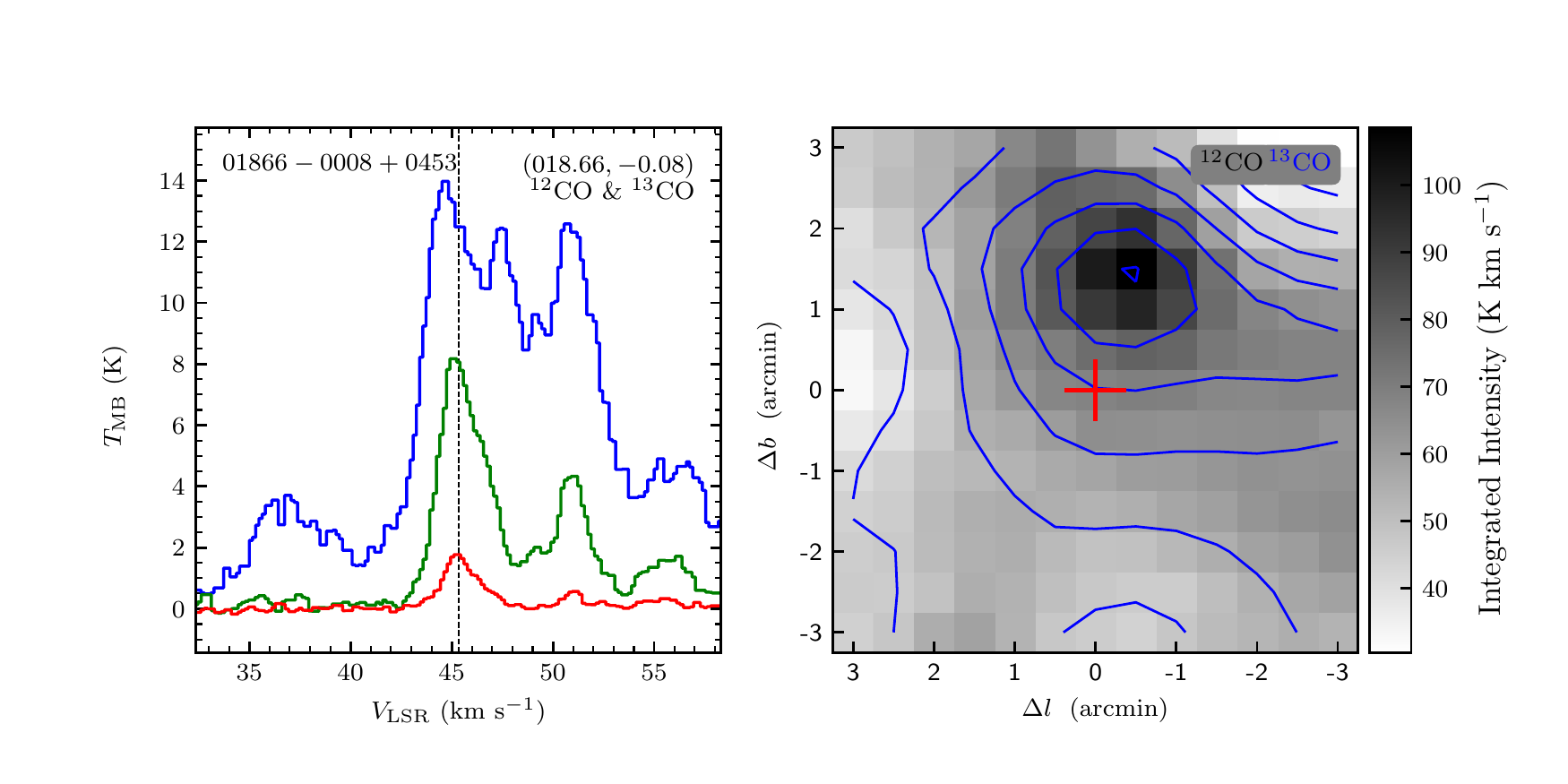}
\vspace{-0.5cm}

\includegraphics[width=9.0cm,angle=0]{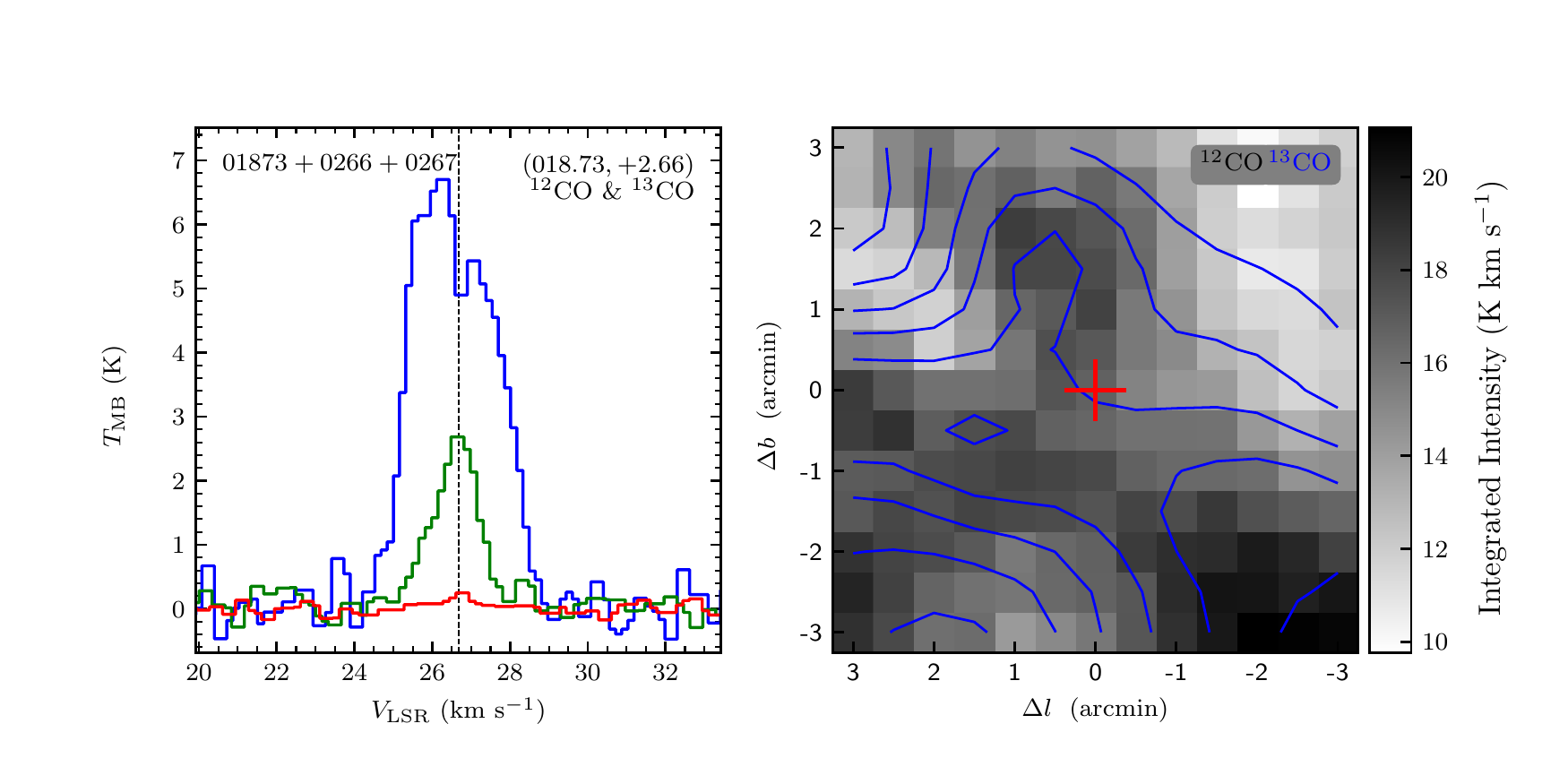}
\includegraphics[width=9.0cm,angle=0]{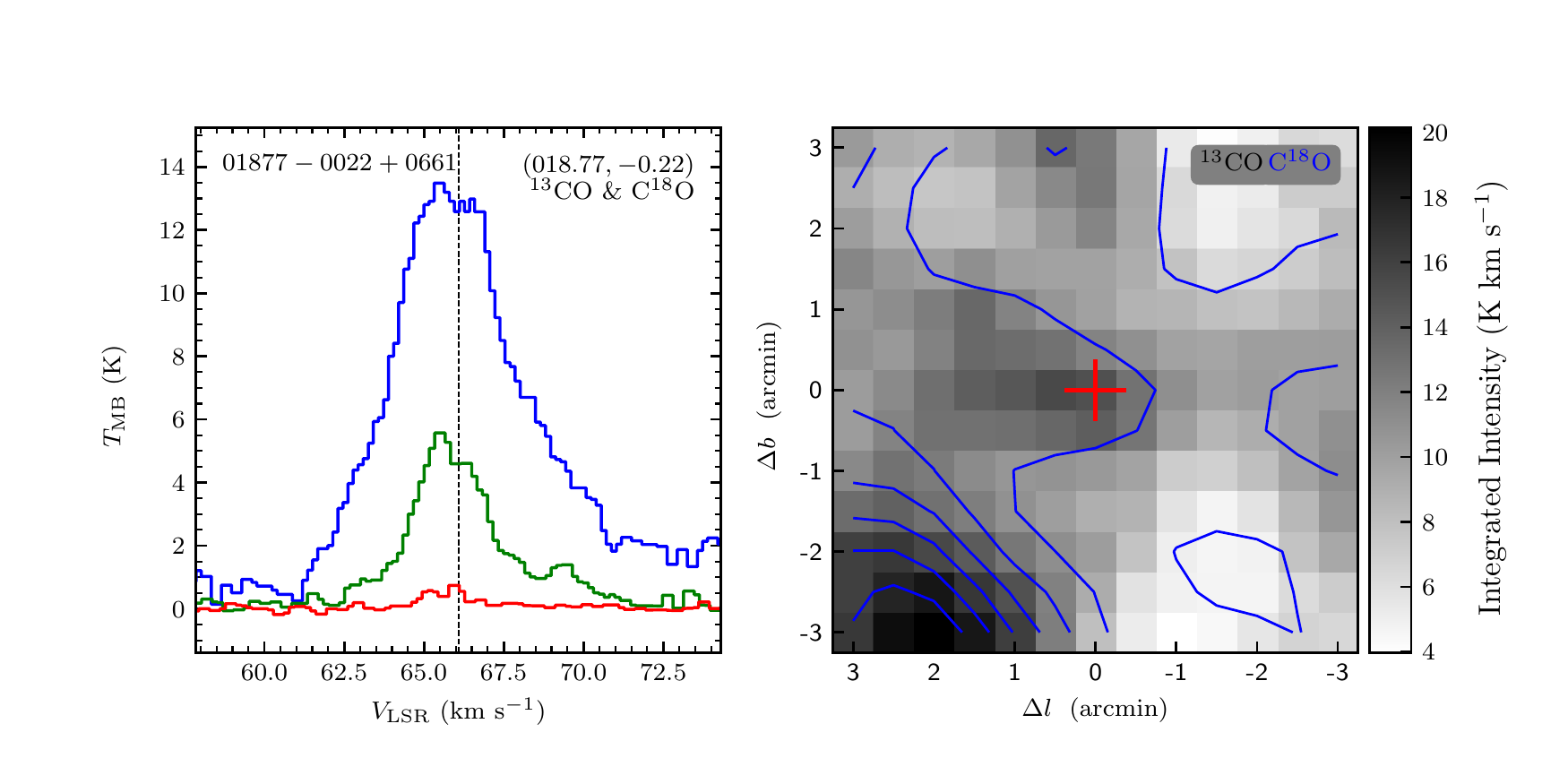}
\vspace{-0.5cm}

\includegraphics[width=9.0cm,angle=0]{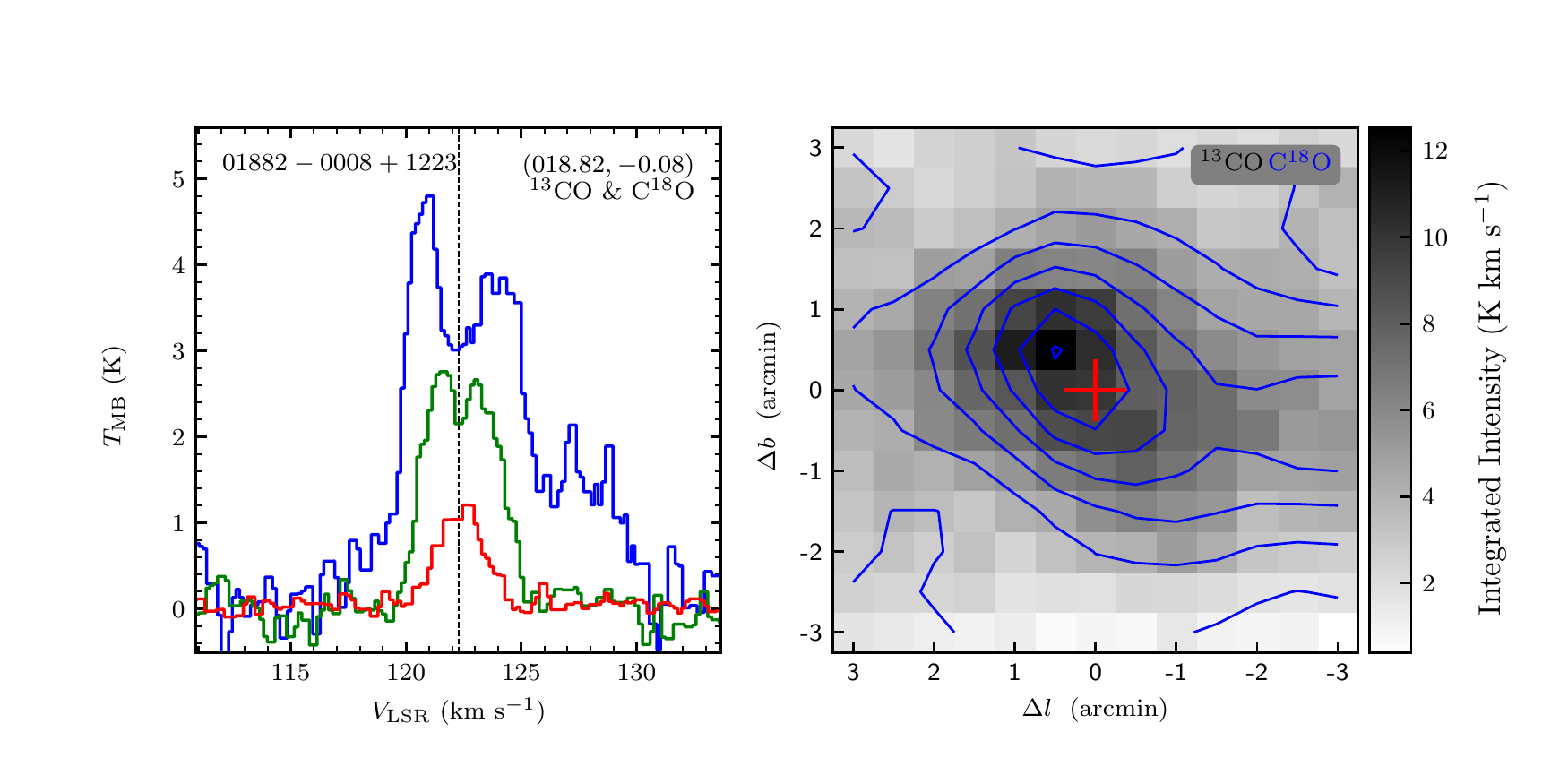}
\includegraphics[width=9.0cm,angle=0]{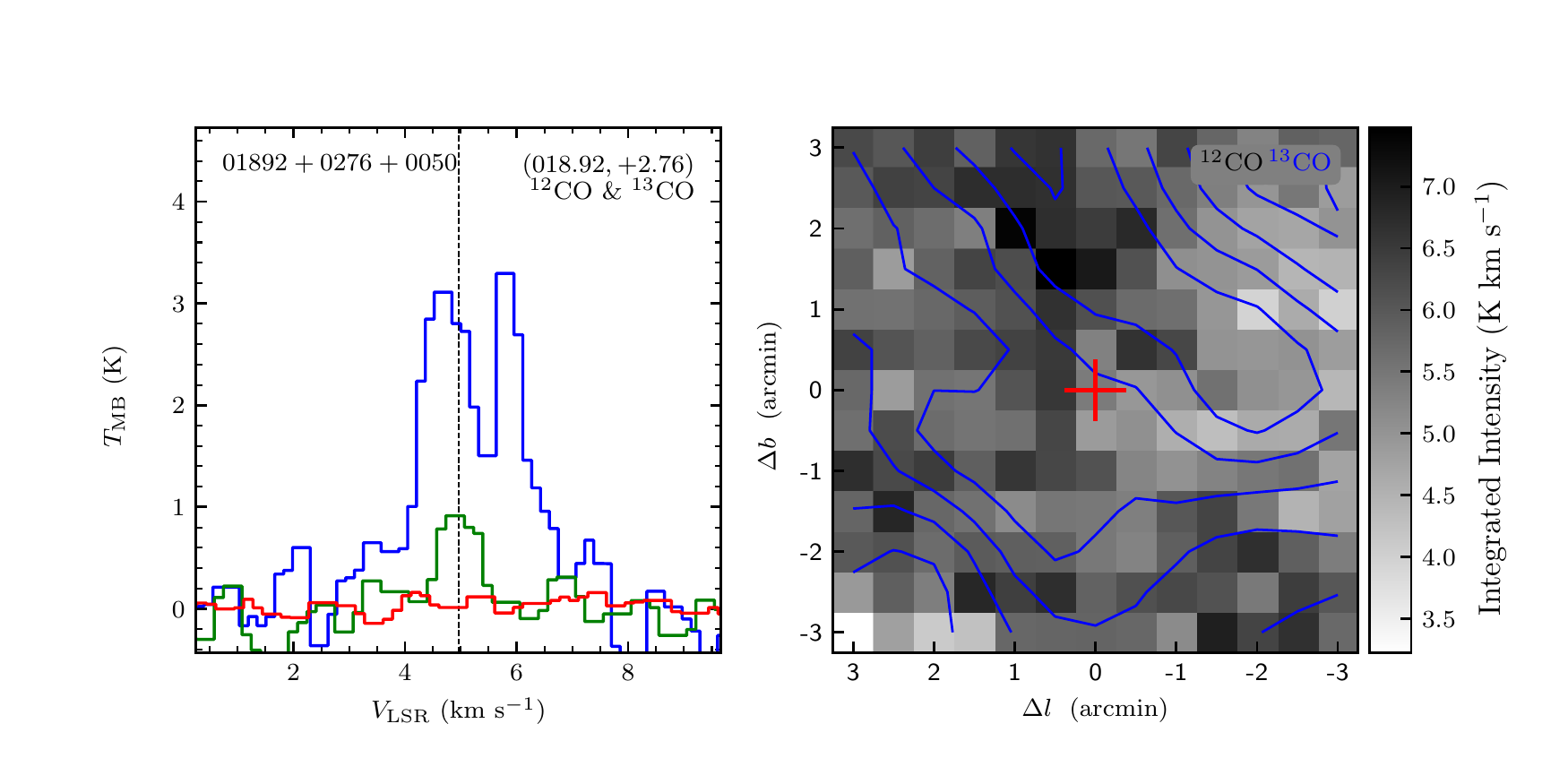}
\end{figure}
\clearpage

\begin{figure}
\includegraphics[width=9.0cm,angle=0]{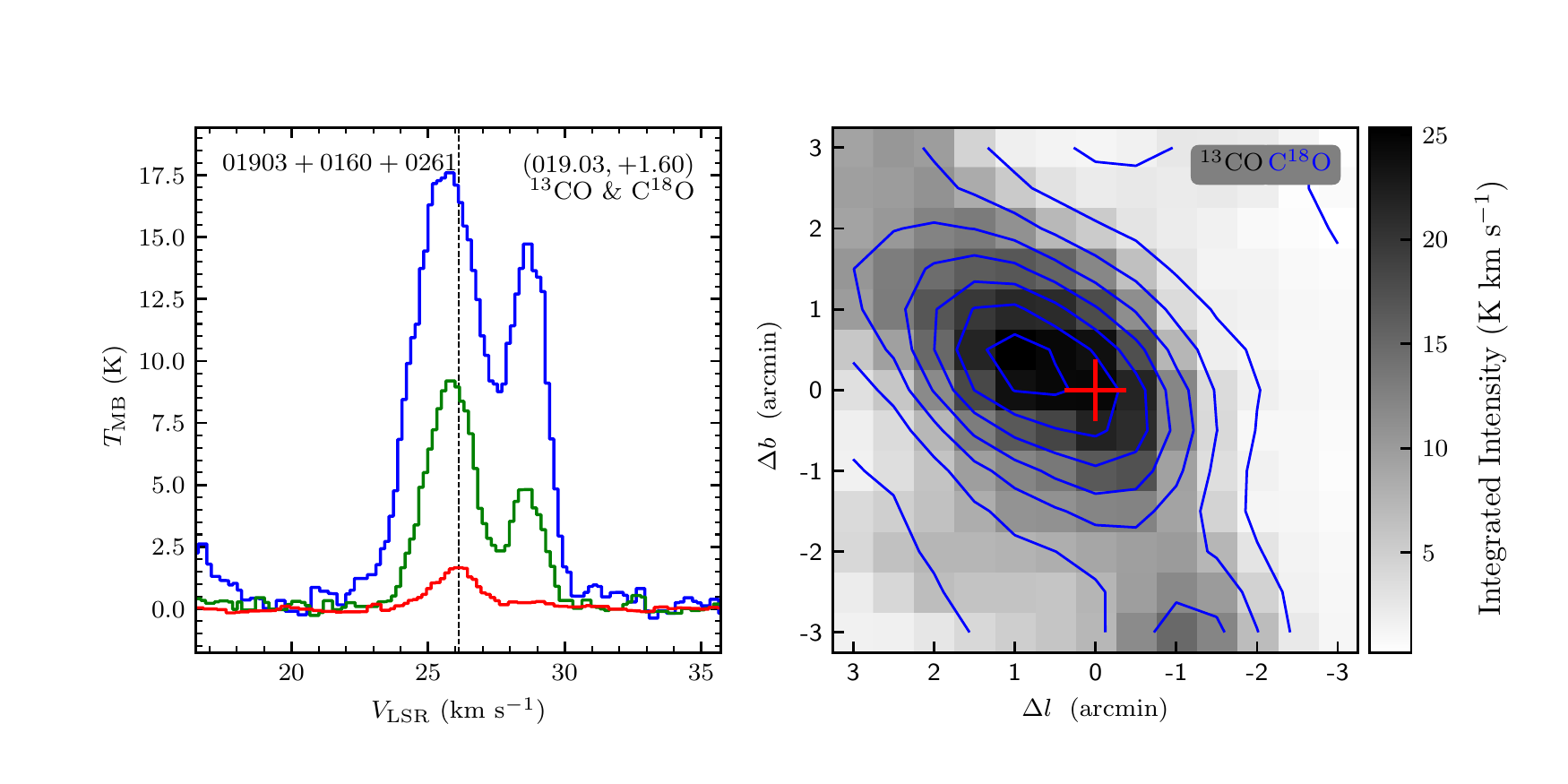}
\includegraphics[width=9.0cm,angle=0]{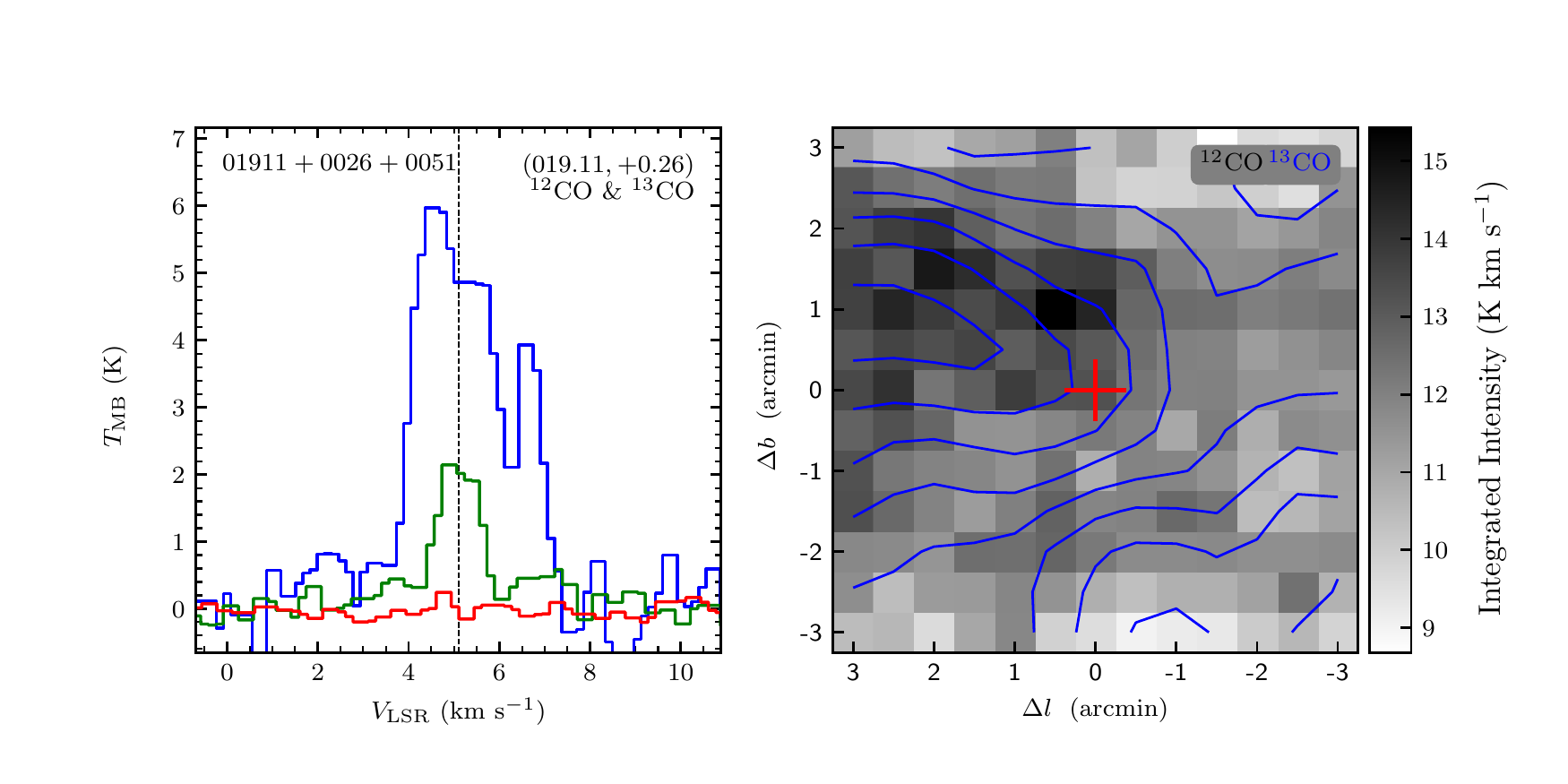}
\vspace{-0.5cm}

\includegraphics[width=9.0cm,angle=0]{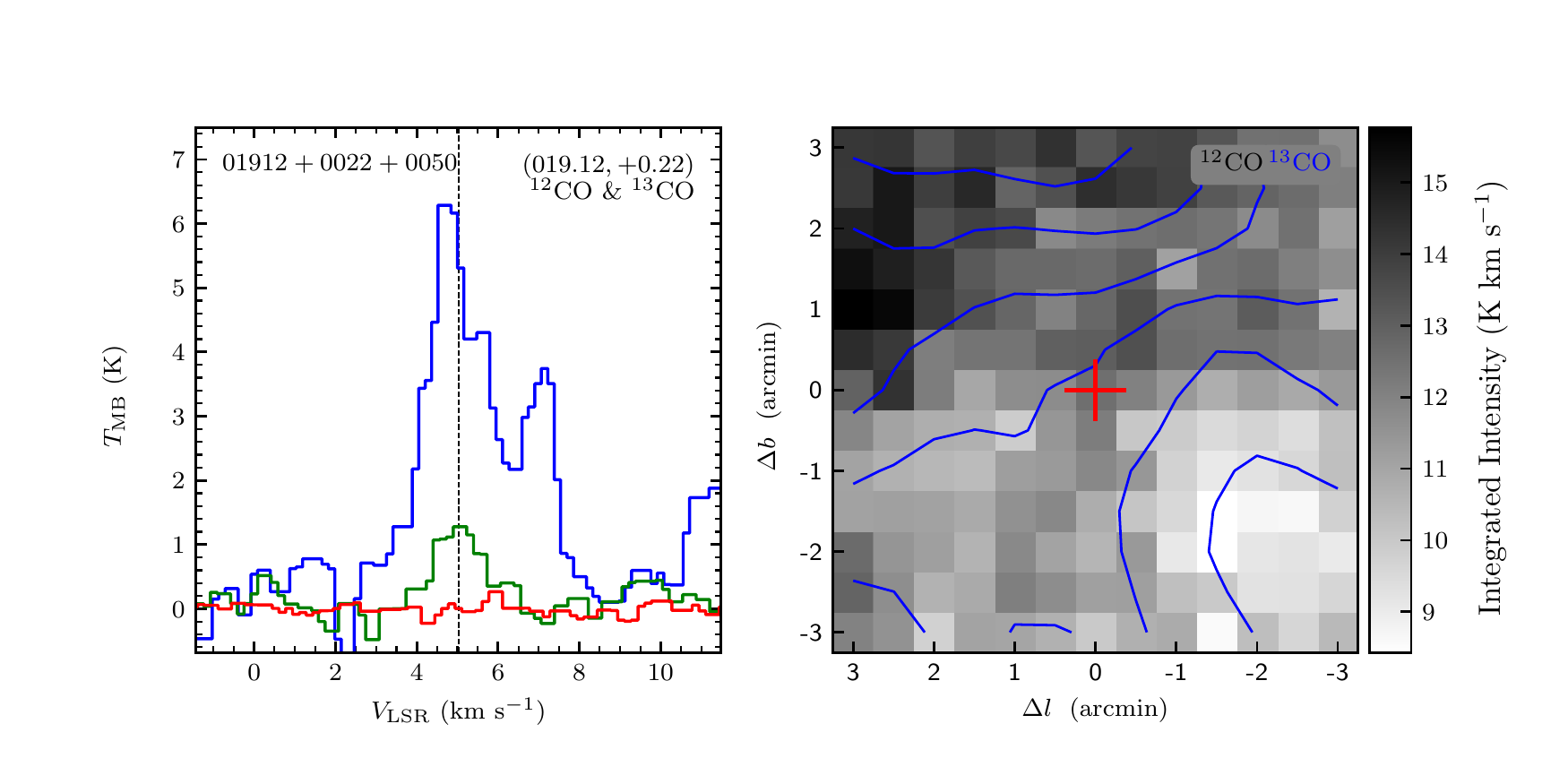}
\includegraphics[width=9.0cm,angle=0]{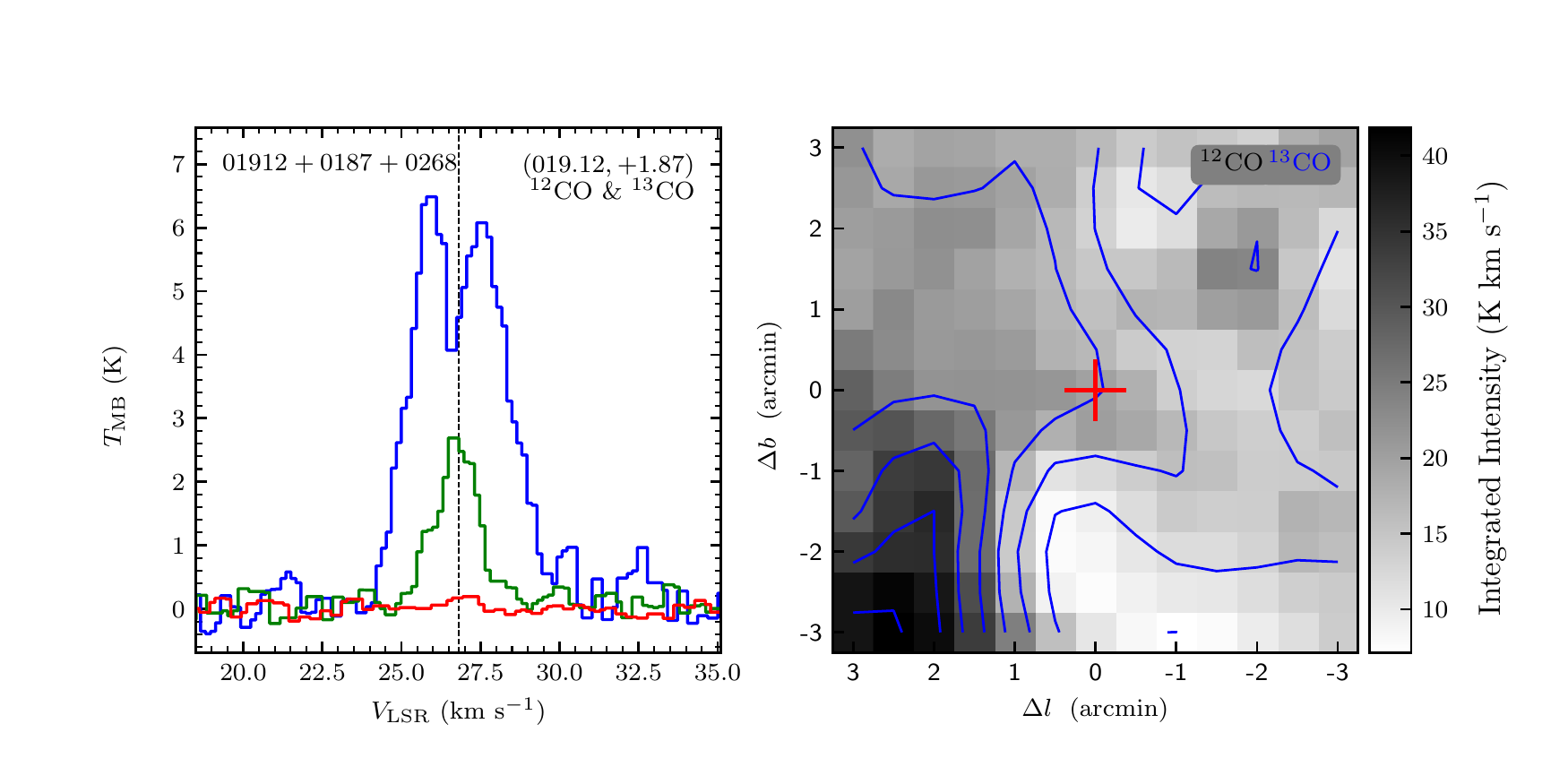}
\vspace{-0.5cm}

\includegraphics[width=9.0cm,angle=0]{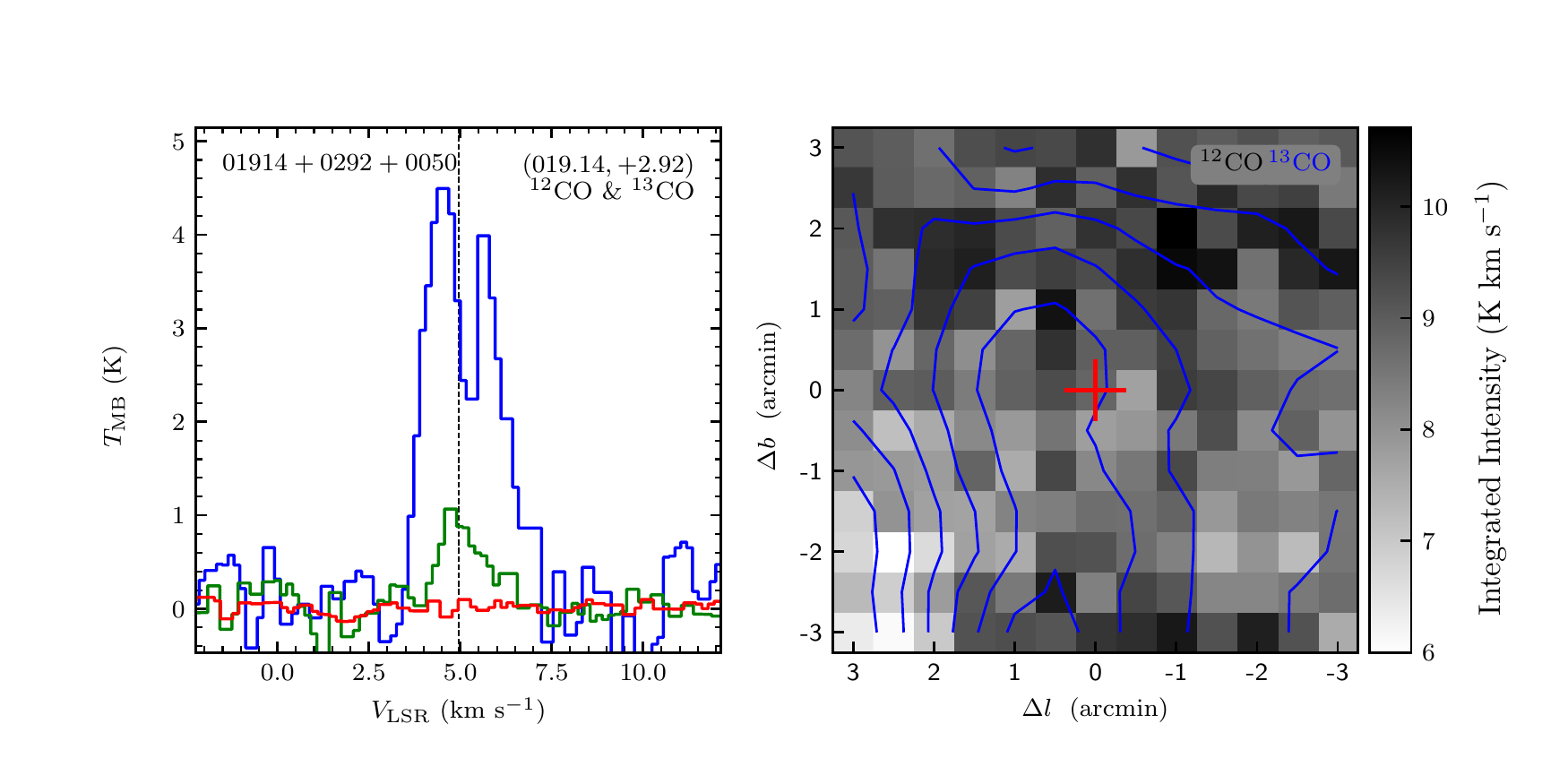}
\includegraphics[width=9.0cm,angle=0]{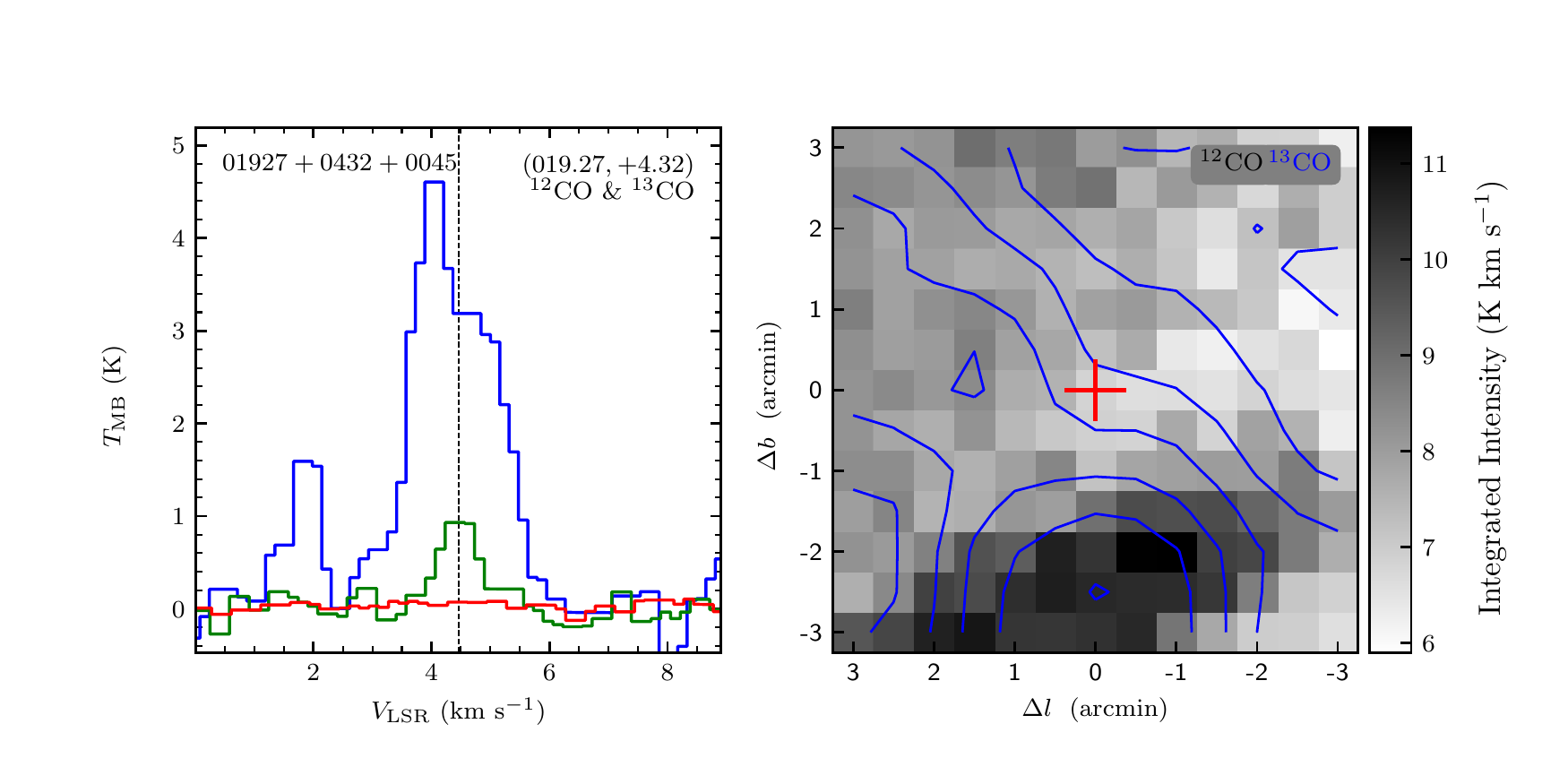}
\vspace{-0.5cm}

\includegraphics[width=9.0cm,angle=0]{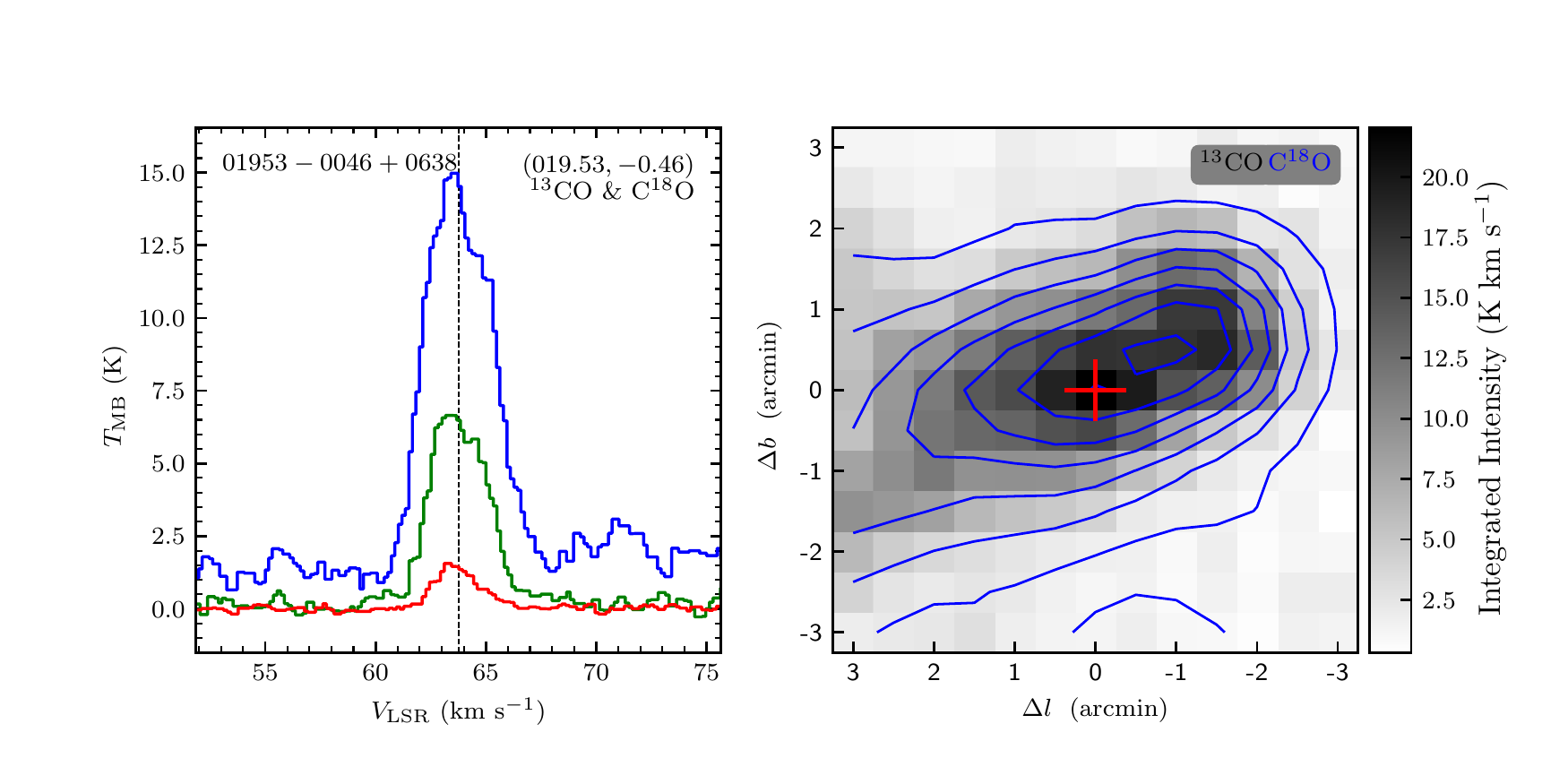}
\includegraphics[width=9.0cm,angle=0]{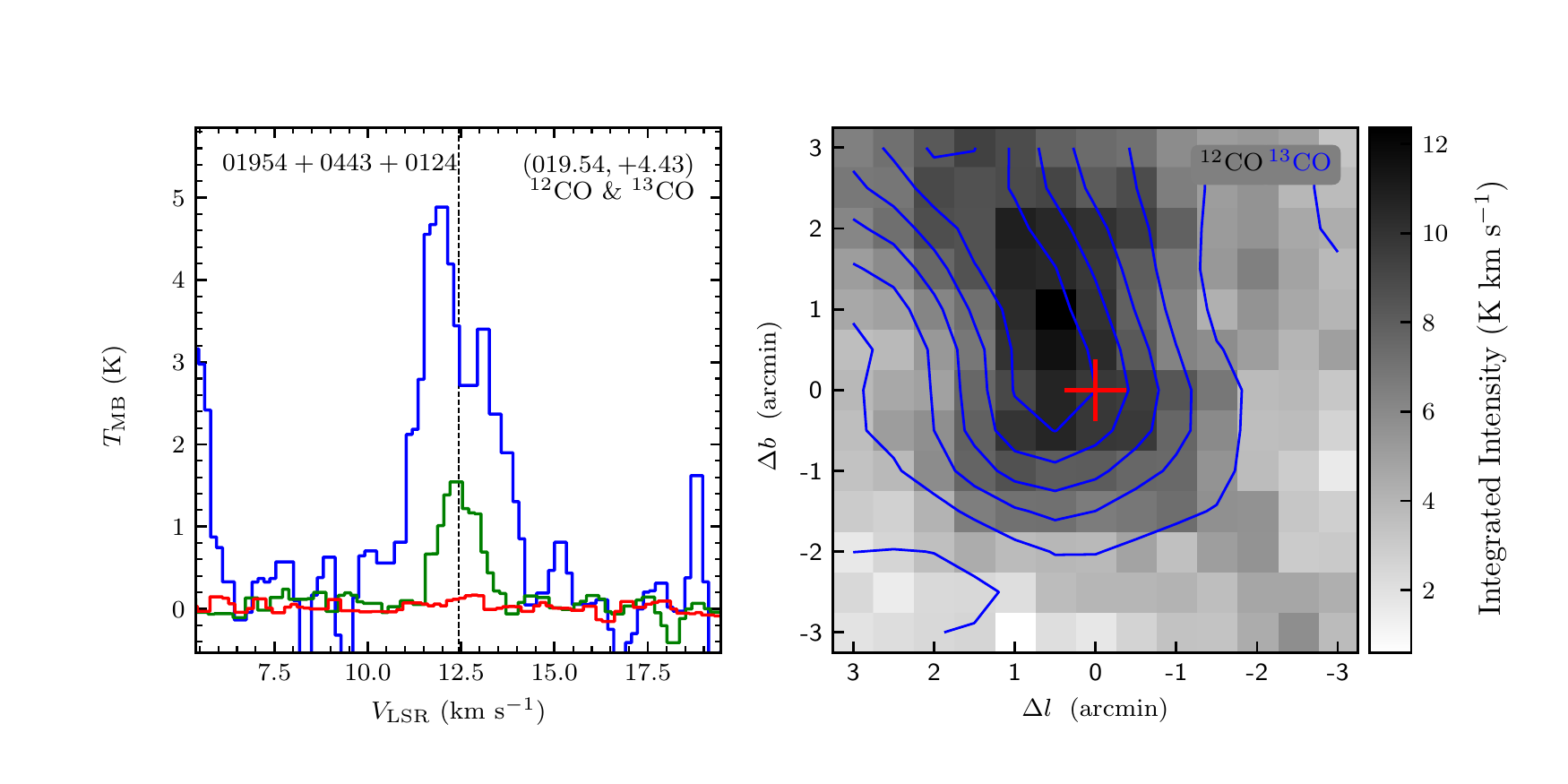}
\vspace{-0.5cm}

\includegraphics[width=9.0cm,angle=0]{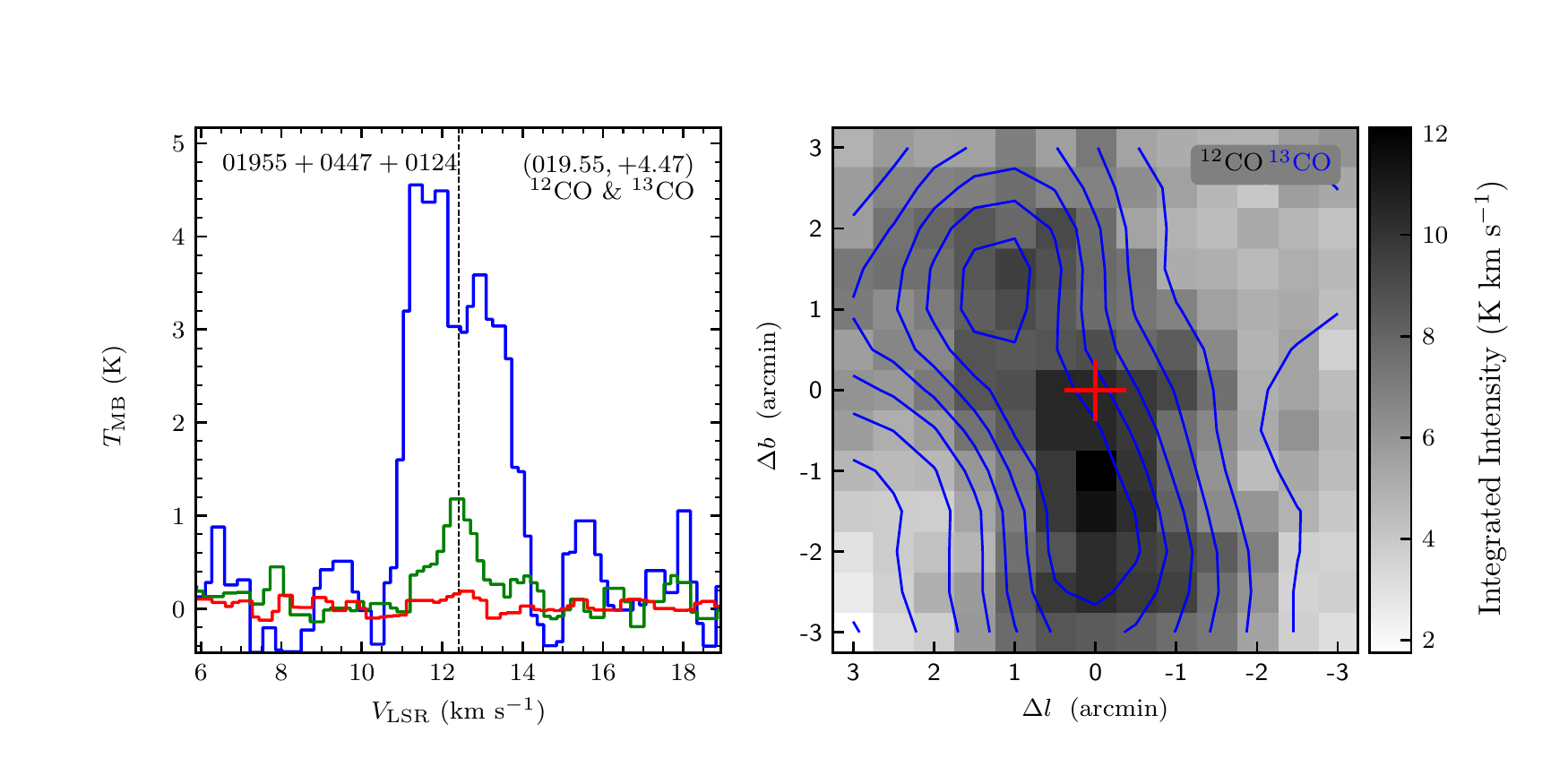}
\includegraphics[width=9.0cm,angle=0]{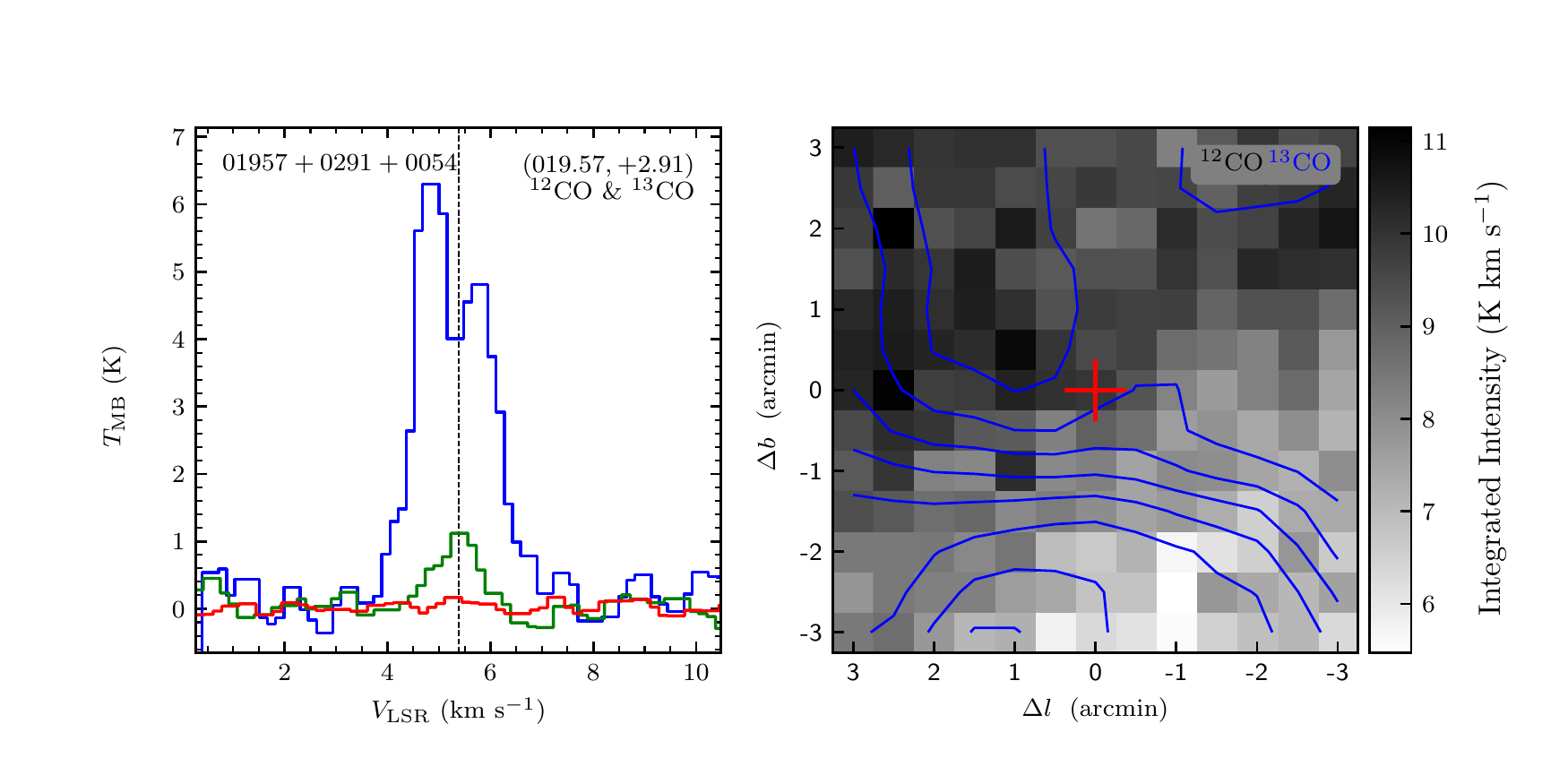}
\end{figure}
\clearpage

\begin{figure}
\includegraphics[width=9.0cm,angle=0]{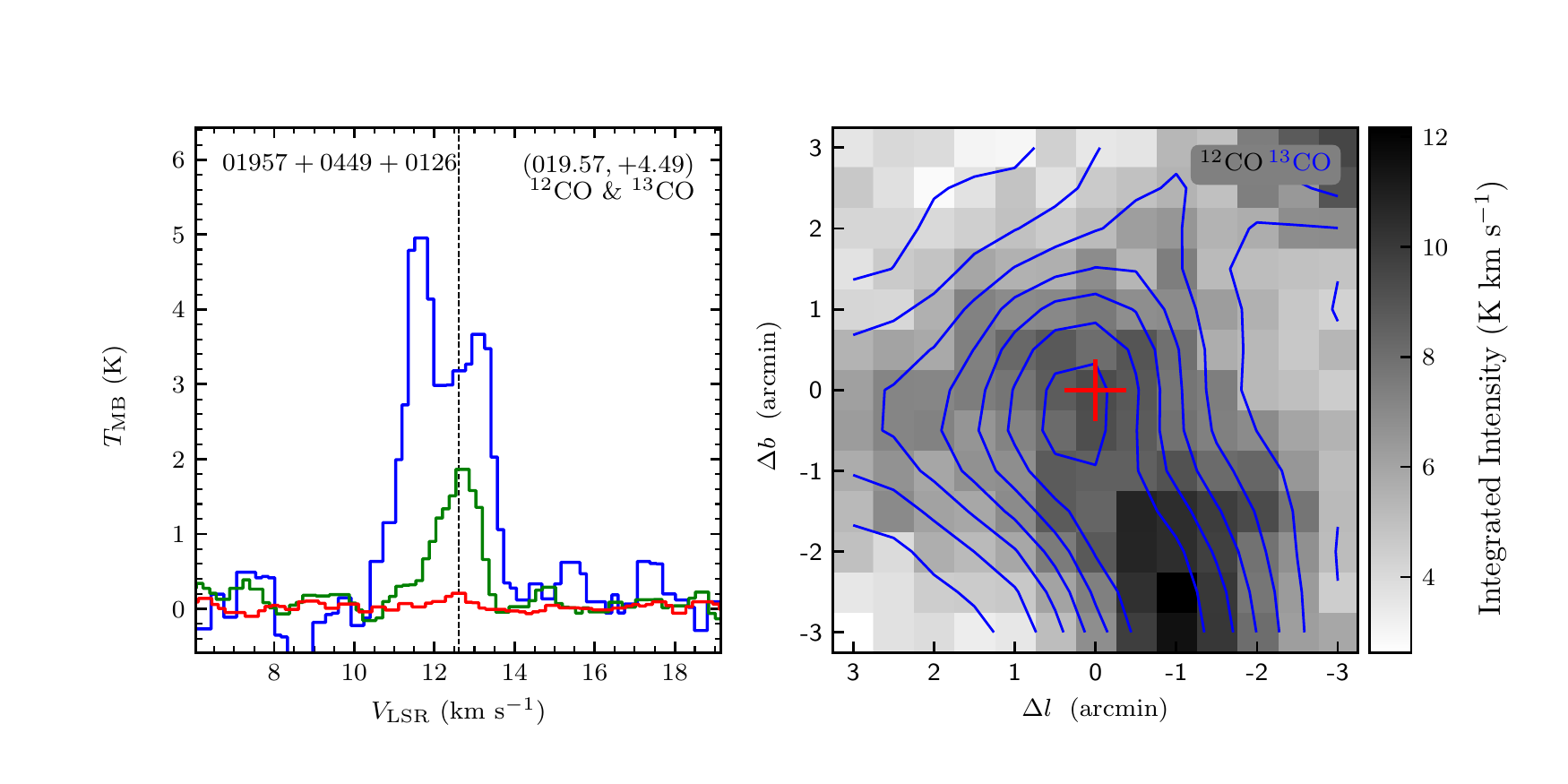}
\includegraphics[width=9.0cm,angle=0]{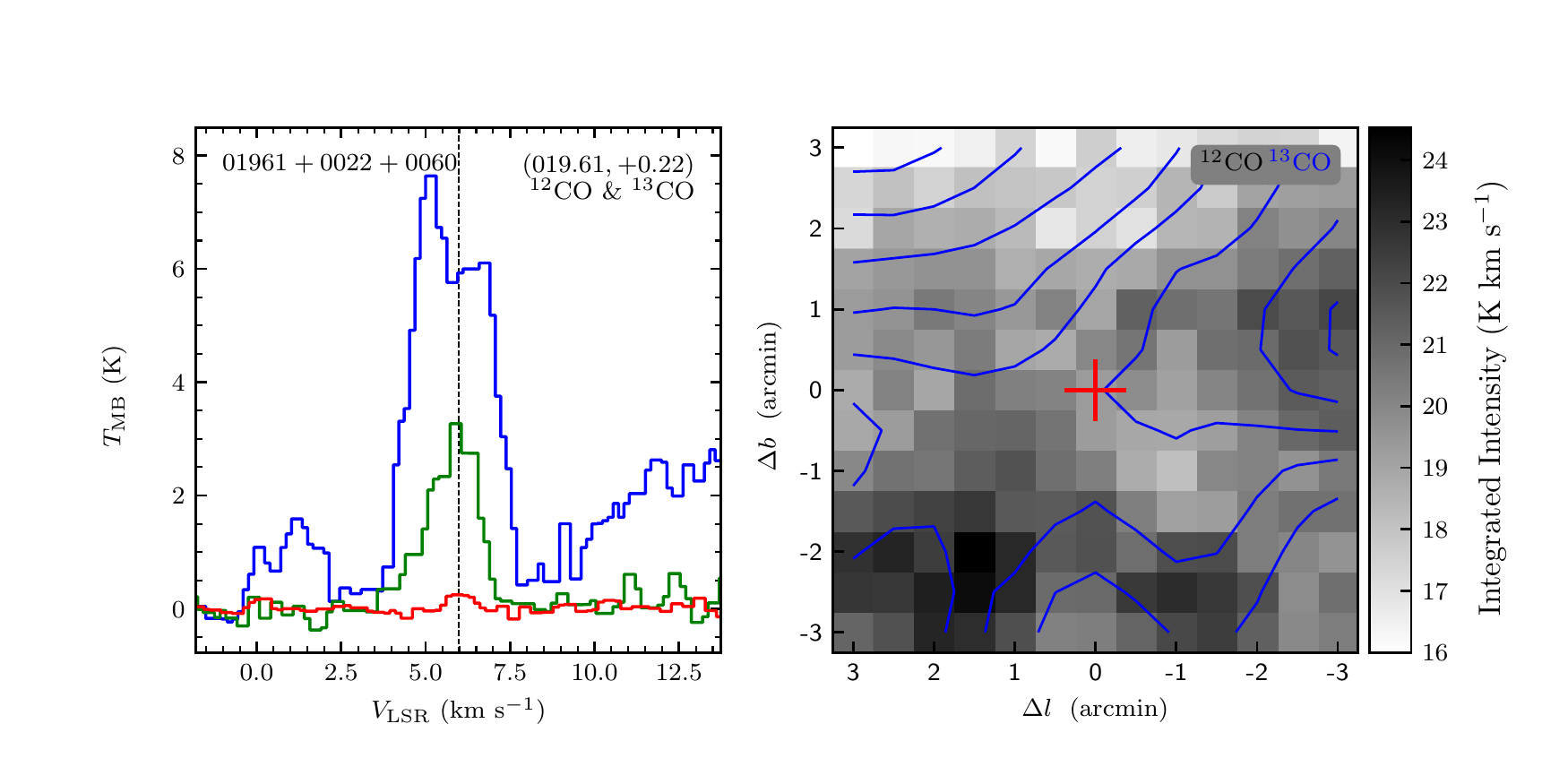}
\vspace{-0.5cm}

\includegraphics[width=9.0cm,angle=0]{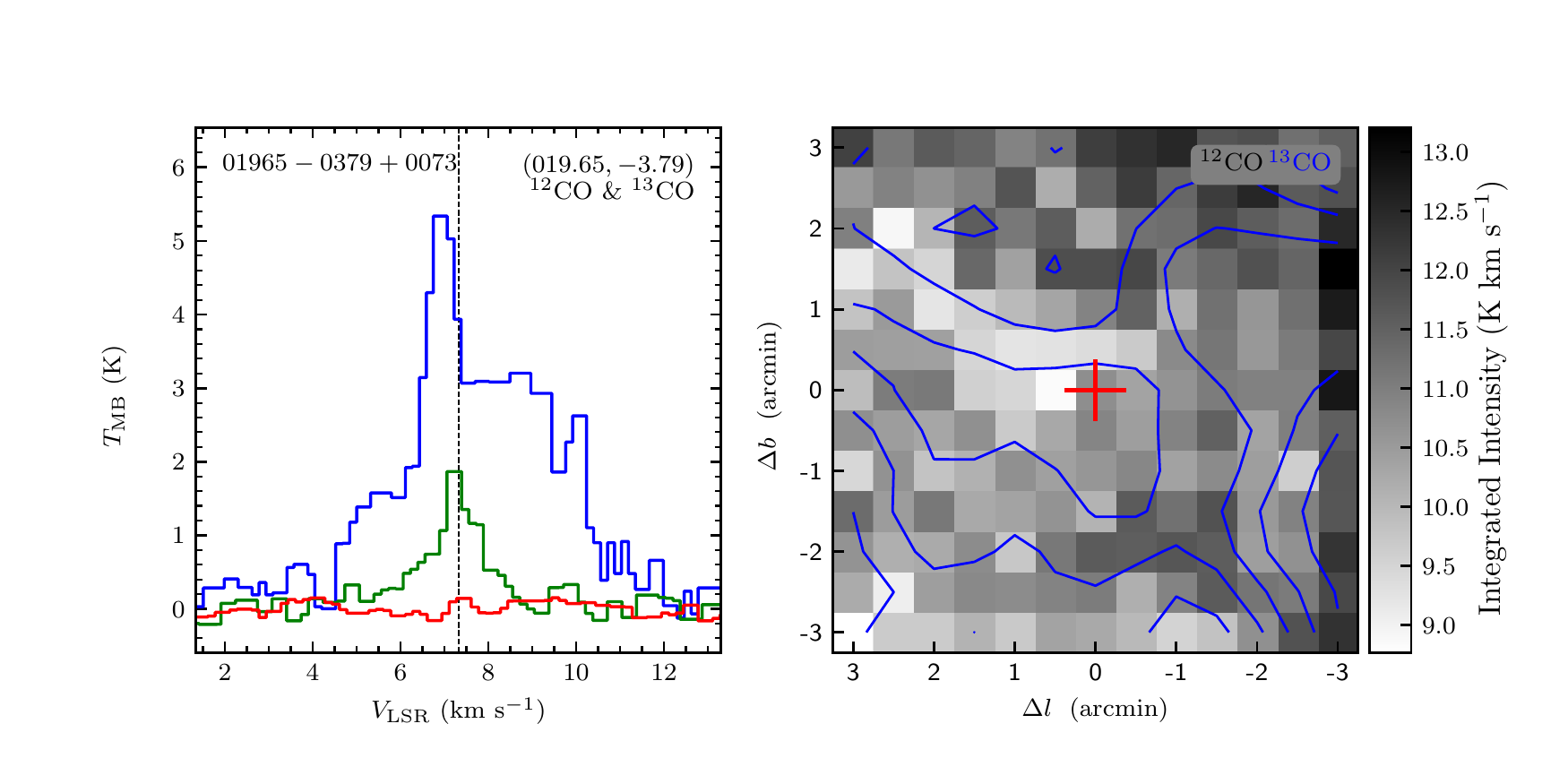}
\includegraphics[width=9.0cm,angle=0]{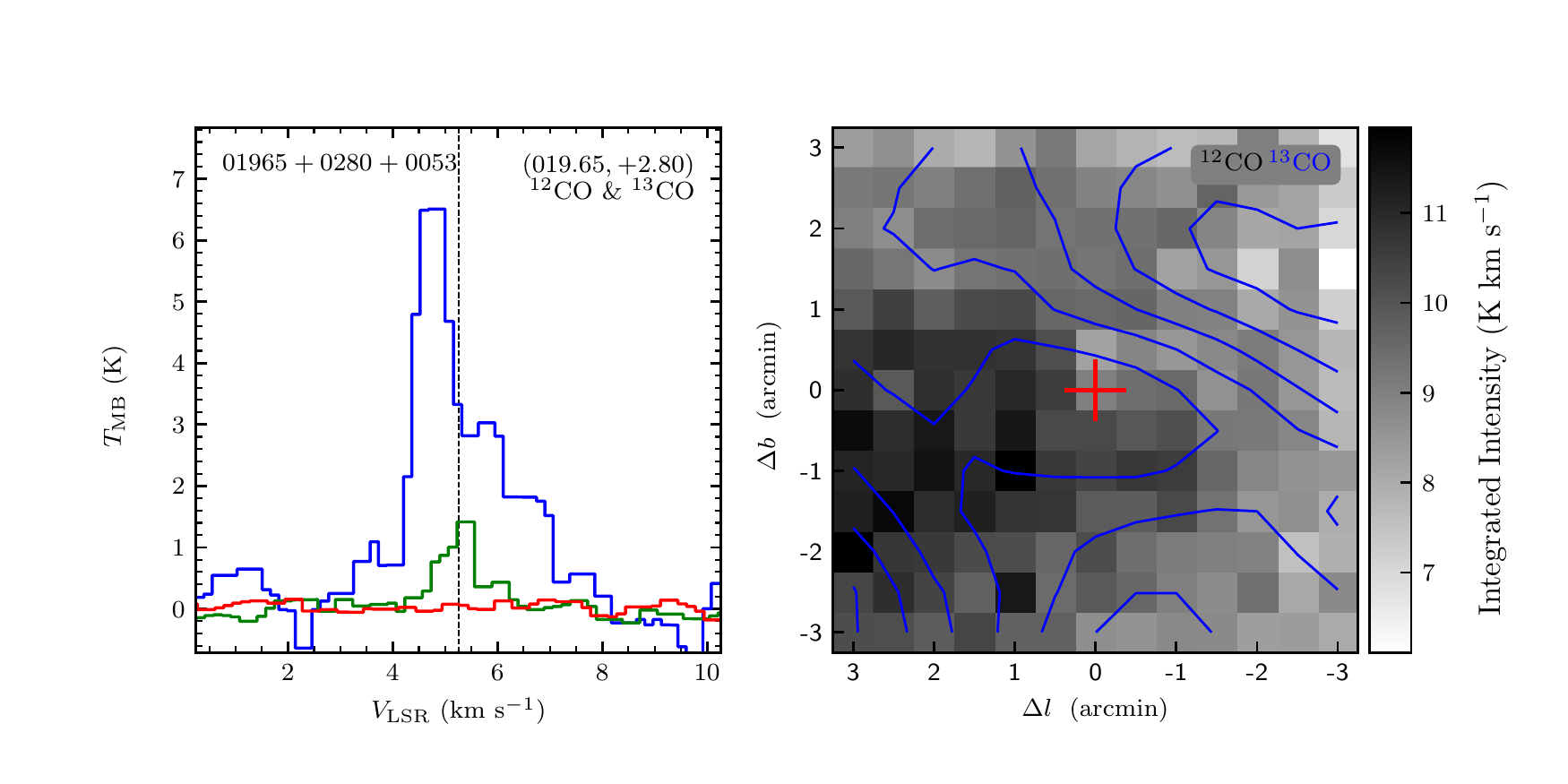}
\vspace{-0.5cm}

\includegraphics[width=9.0cm,angle=0]{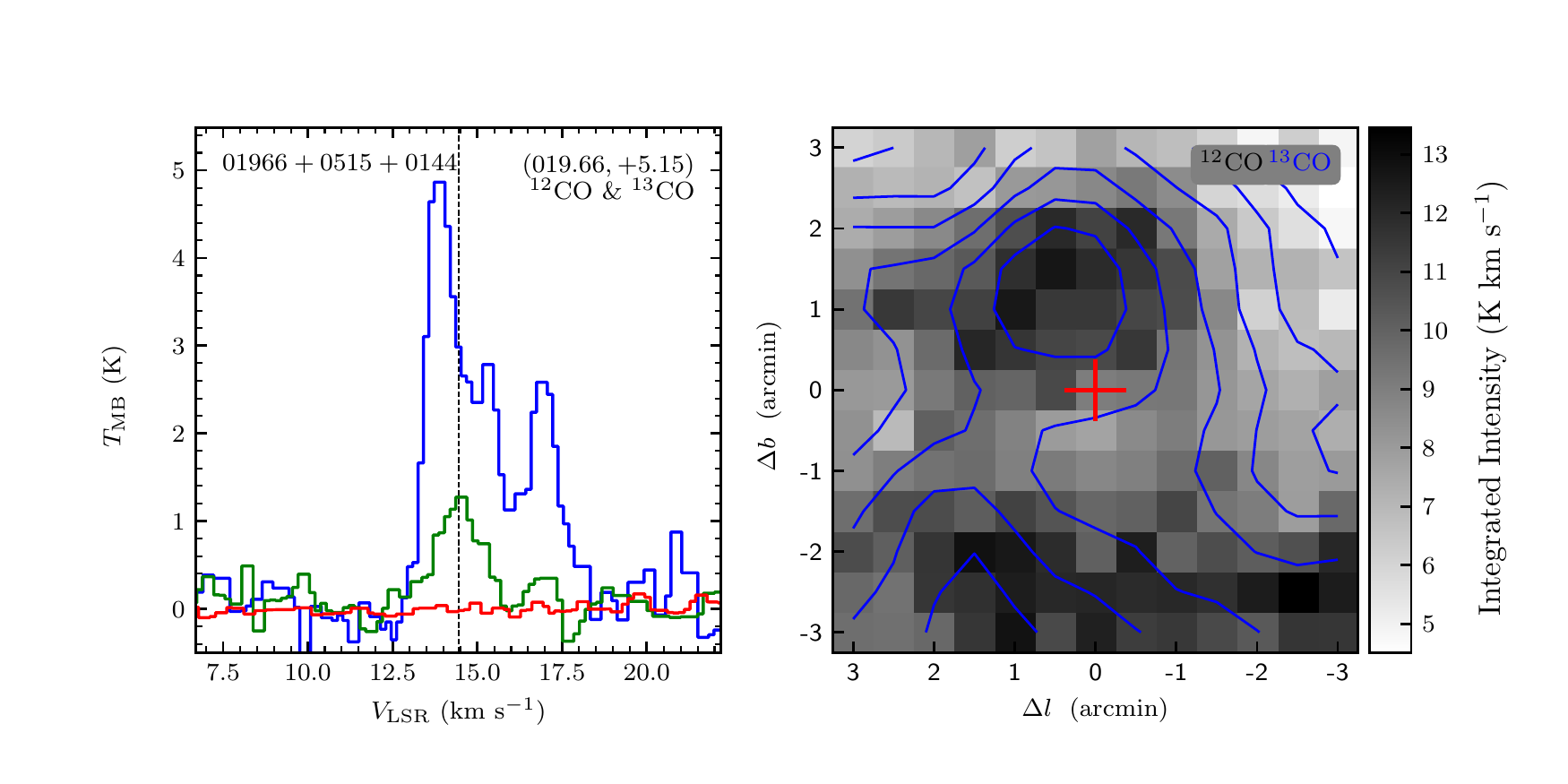}
\includegraphics[width=9.0cm,angle=0]{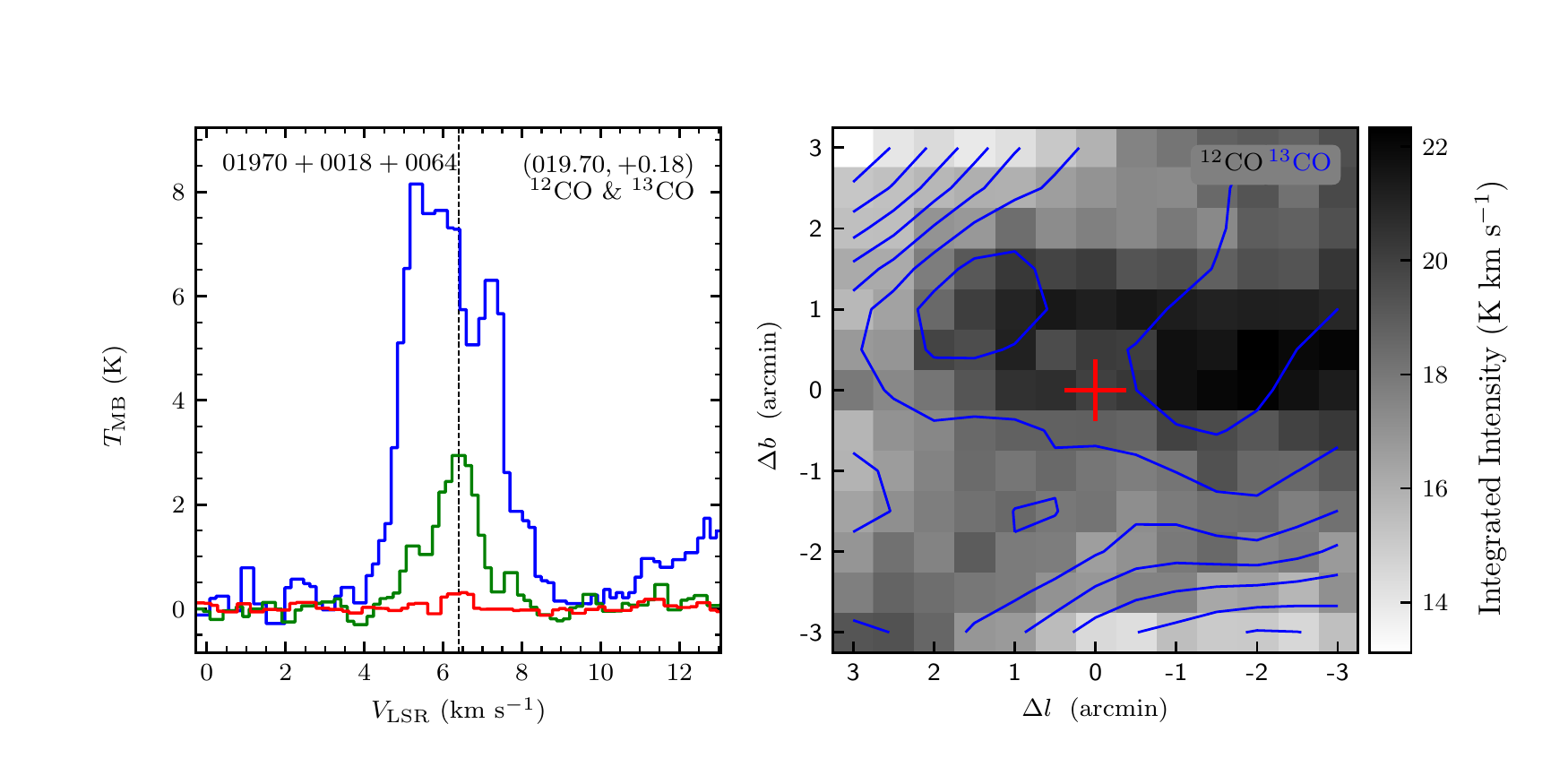}
\vspace{-0.5cm}

\includegraphics[width=9.0cm,angle=0]{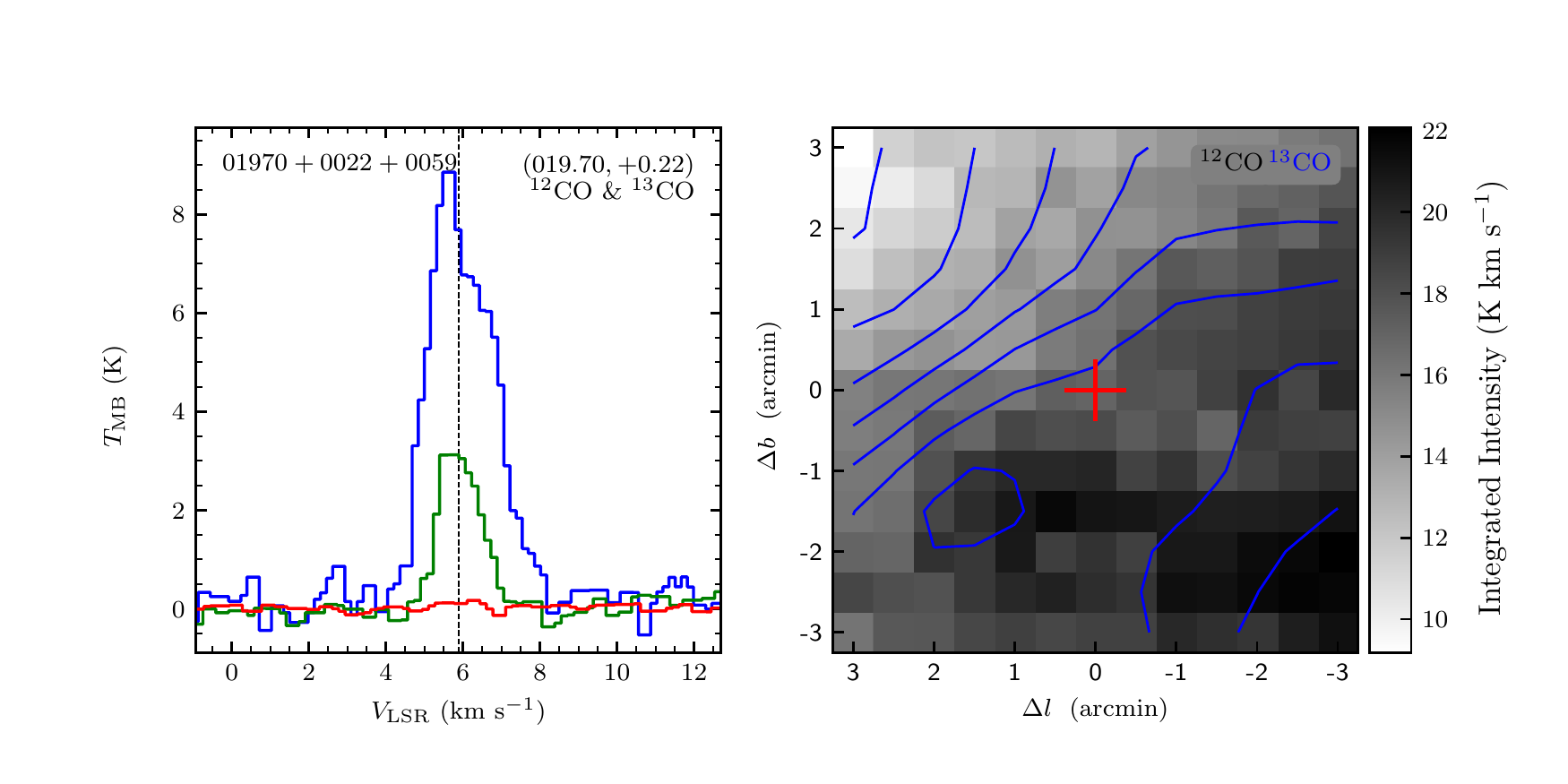}
\includegraphics[width=9.0cm,angle=0]{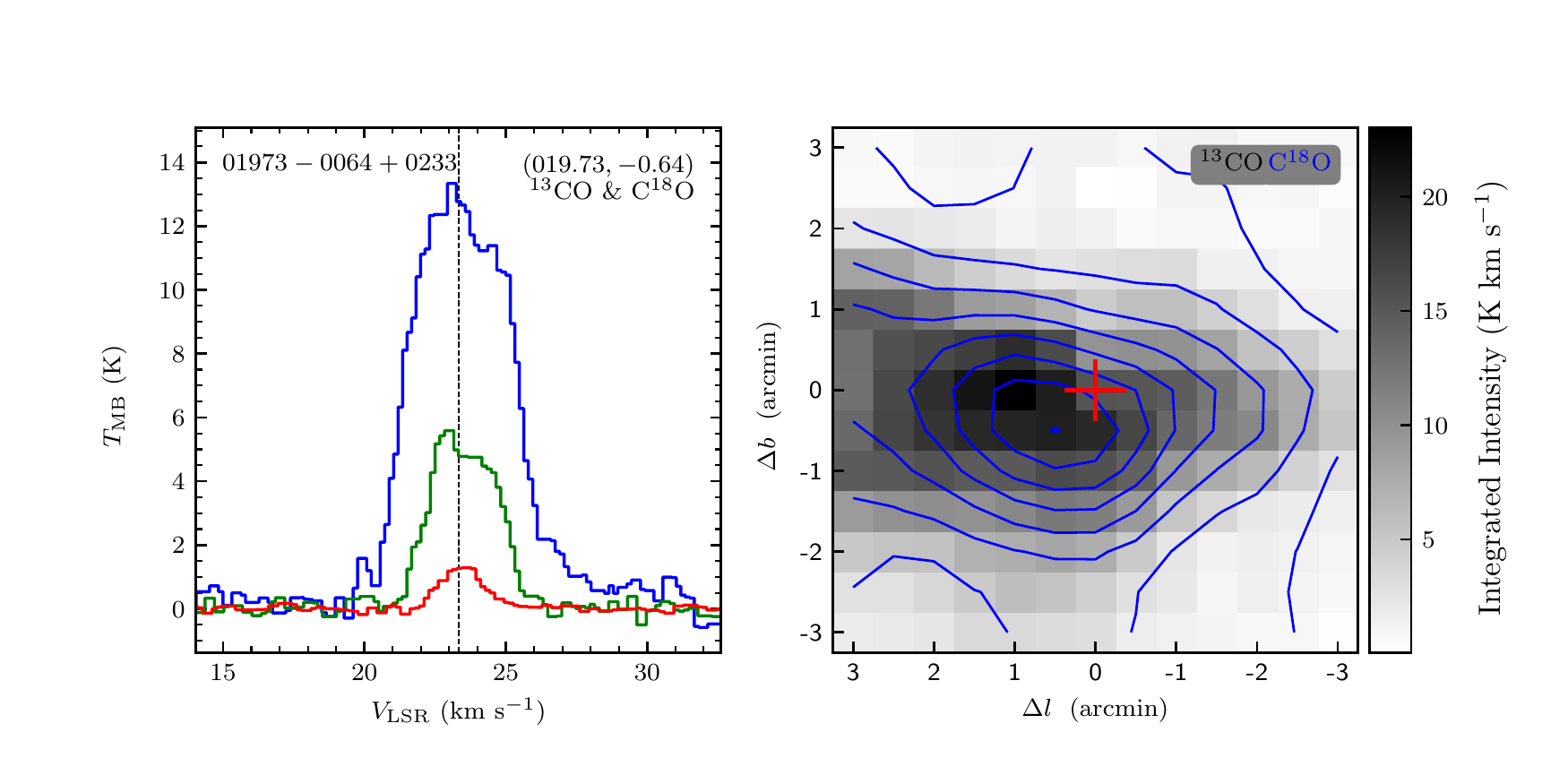}
\vspace{-0.5cm}

\includegraphics[width=9.0cm,angle=0]{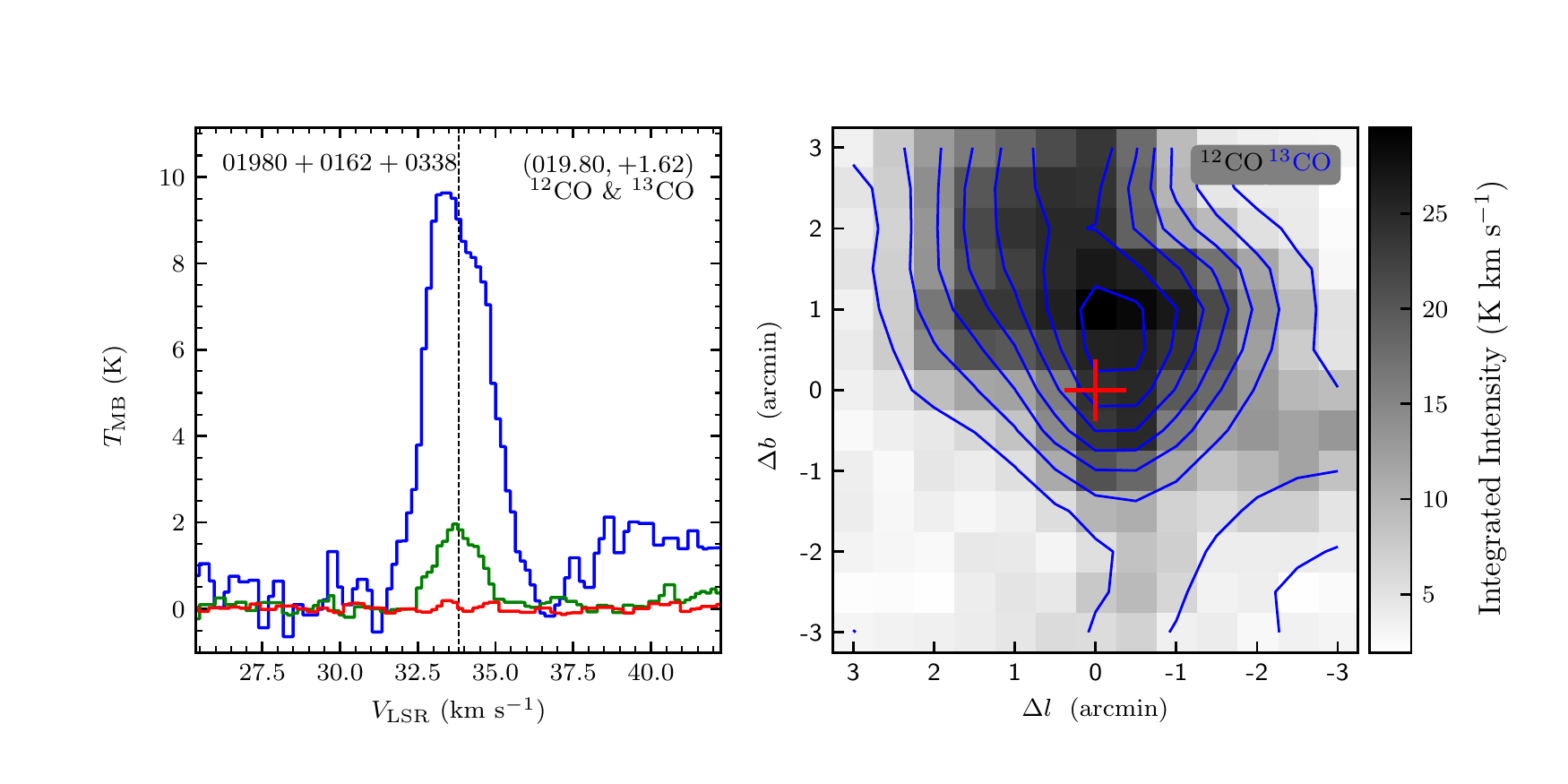}
\includegraphics[width=9.0cm,angle=0]{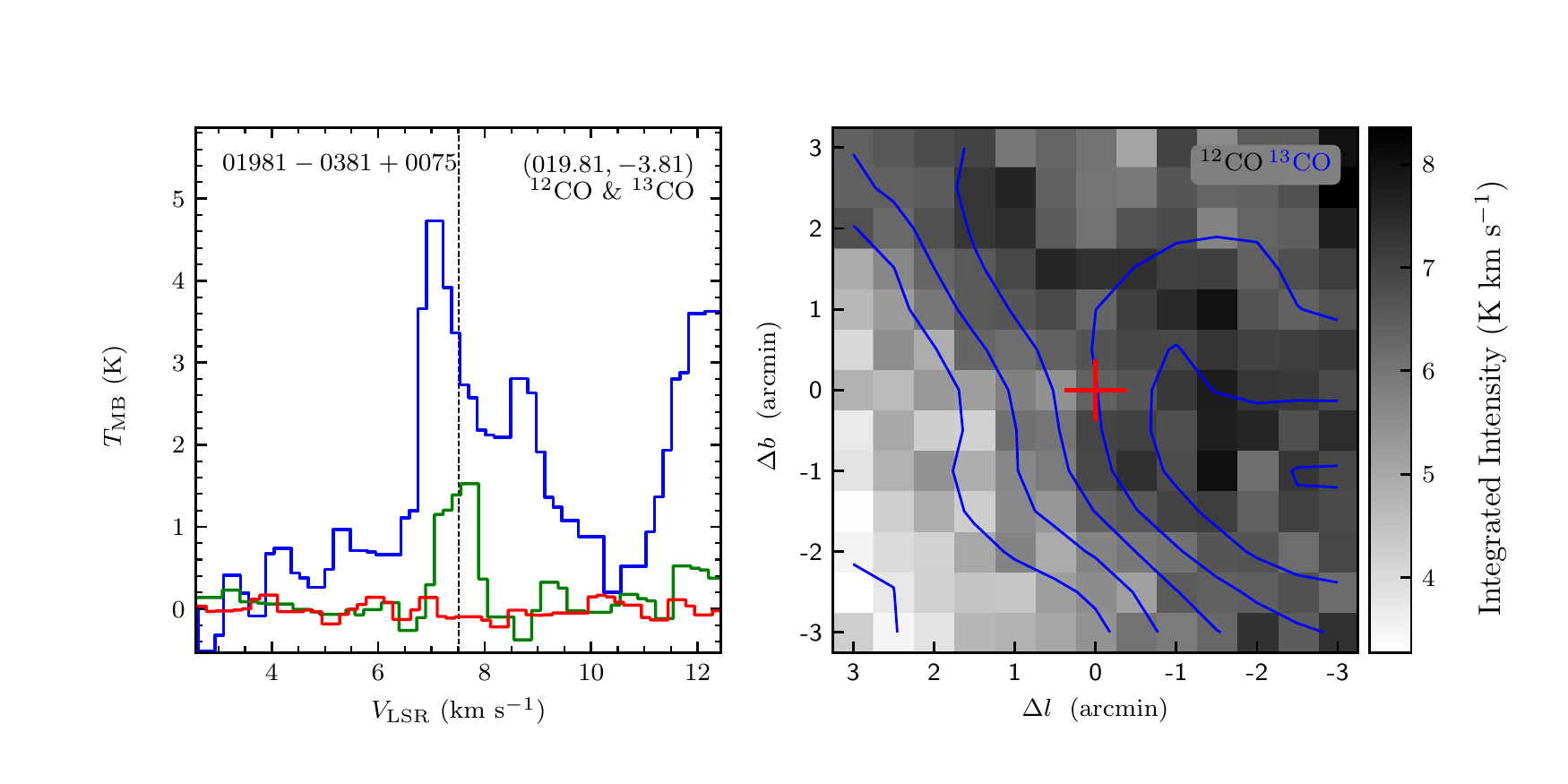}
\end{figure}
\clearpage

\begin{figure}
\includegraphics[width=9.0cm,angle=0]{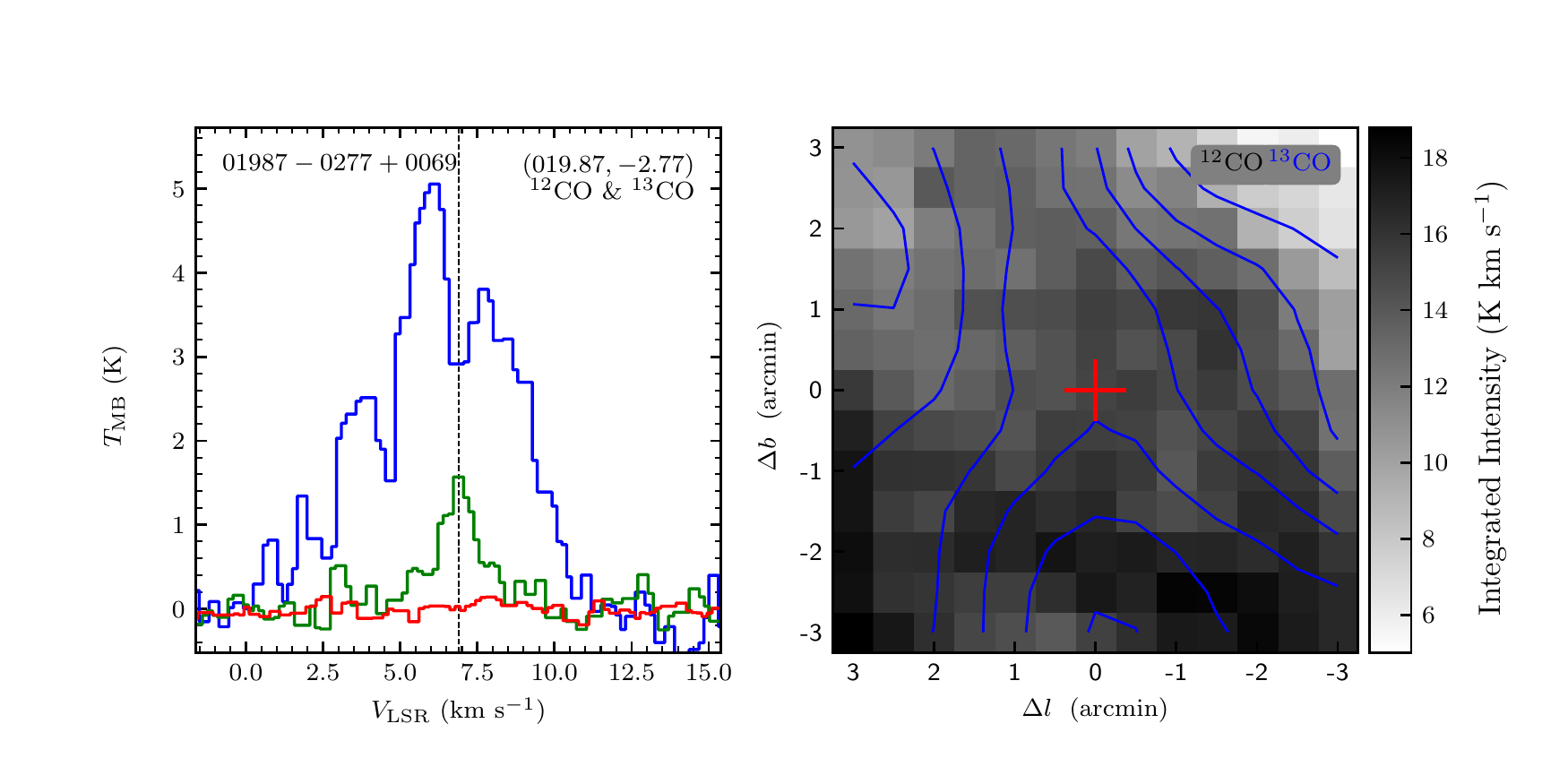}
\includegraphics[width=9.0cm,angle=0]{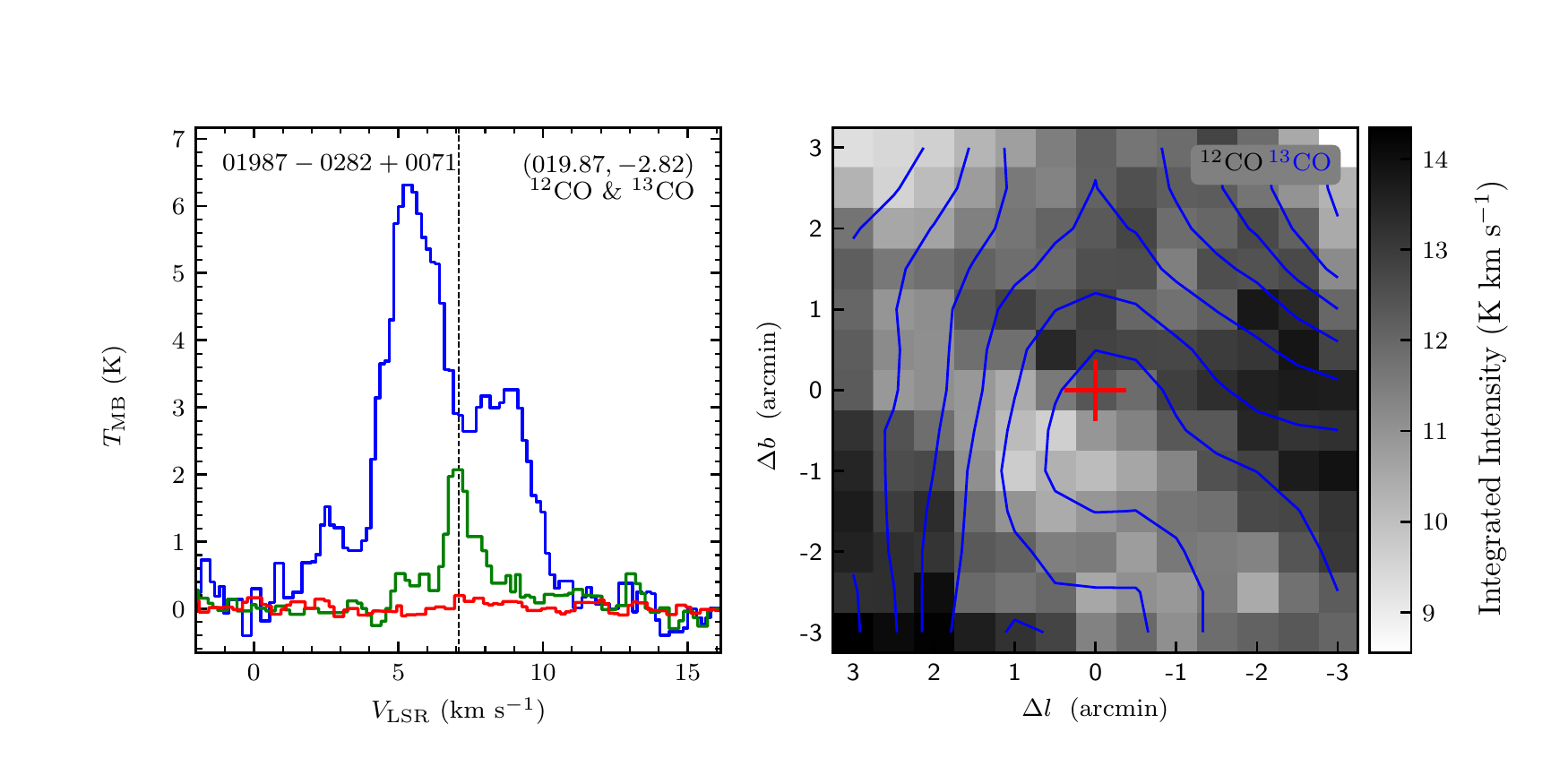}
\vspace{-0.5cm}

\includegraphics[width=9.0cm,angle=0]{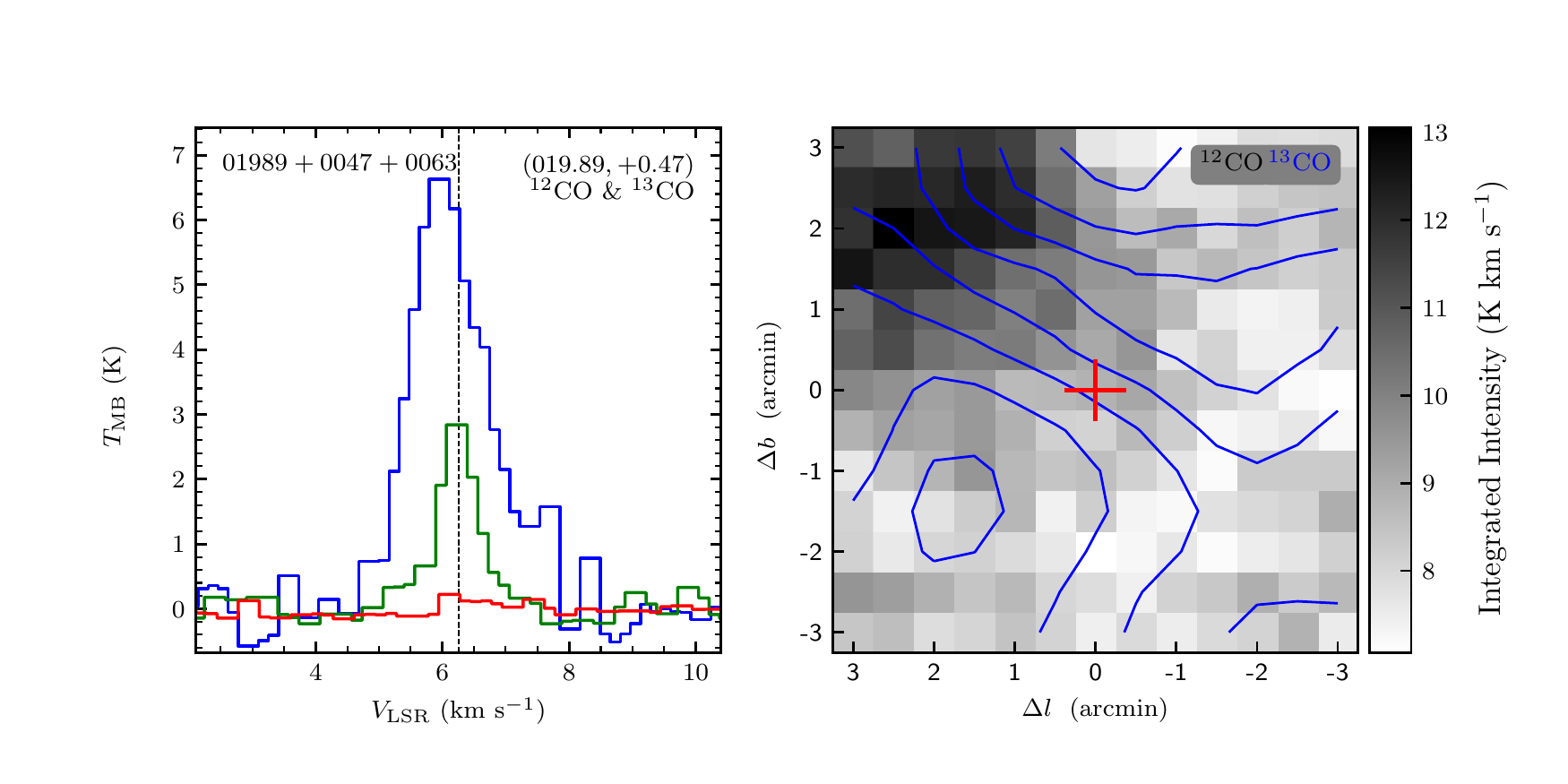}
\includegraphics[width=9.0cm,angle=0]{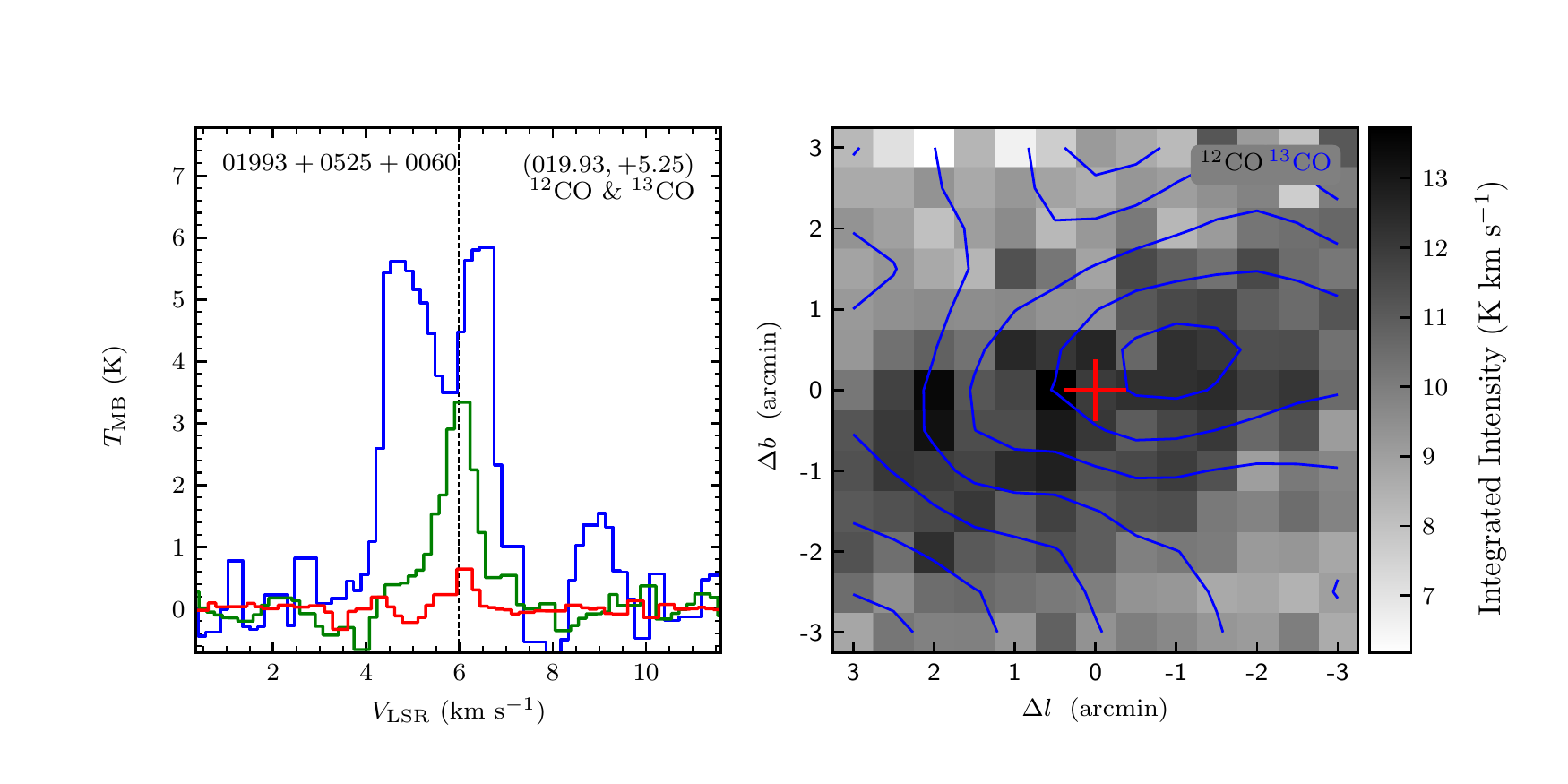}
\vspace{-0.5cm}

\includegraphics[width=9.0cm,angle=0]{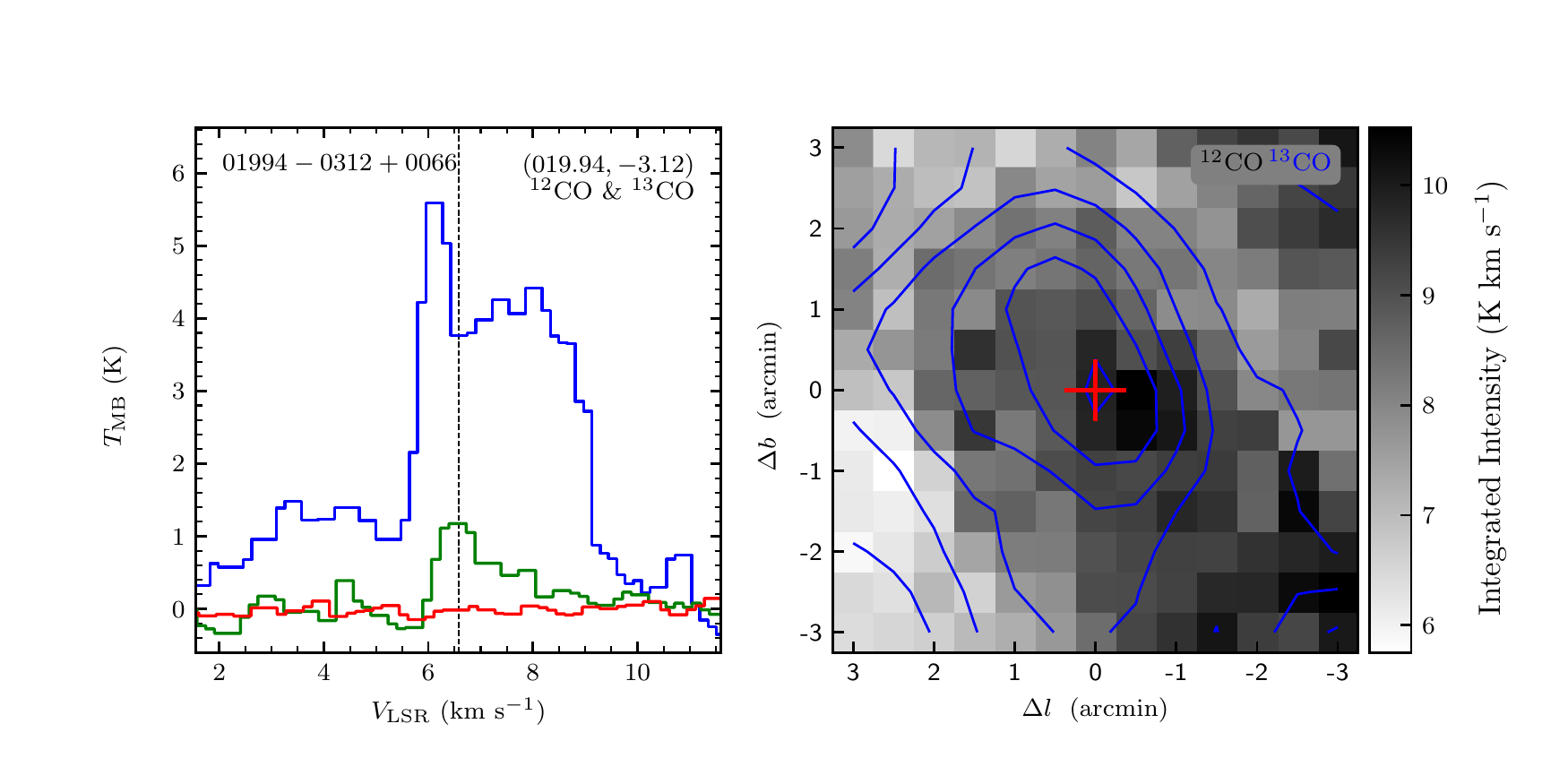}
\includegraphics[width=9.0cm,angle=0]{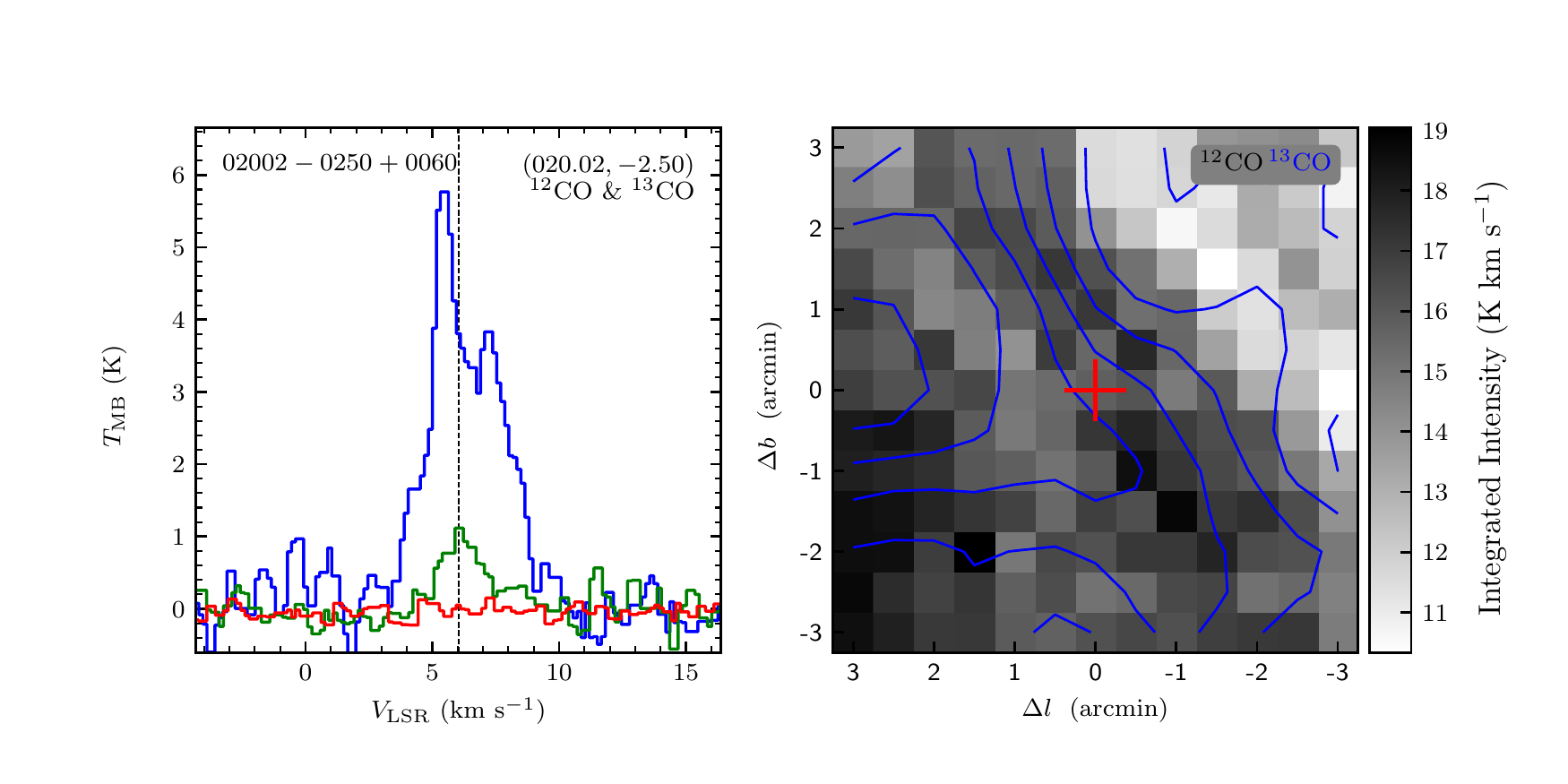}
\vspace{-0.5cm}

\includegraphics[width=9.0cm,angle=0]{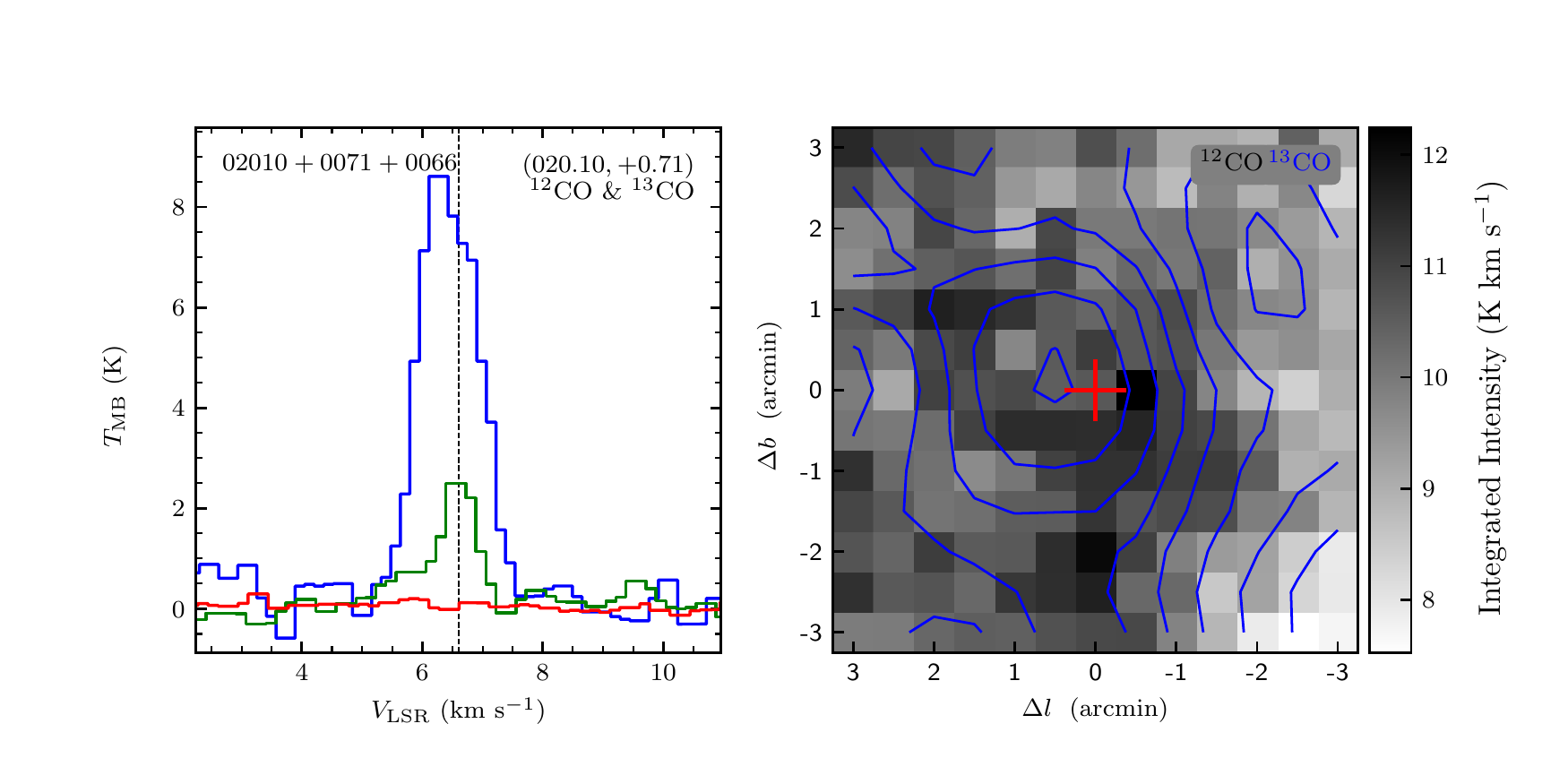}
\includegraphics[width=9.0cm,angle=0]{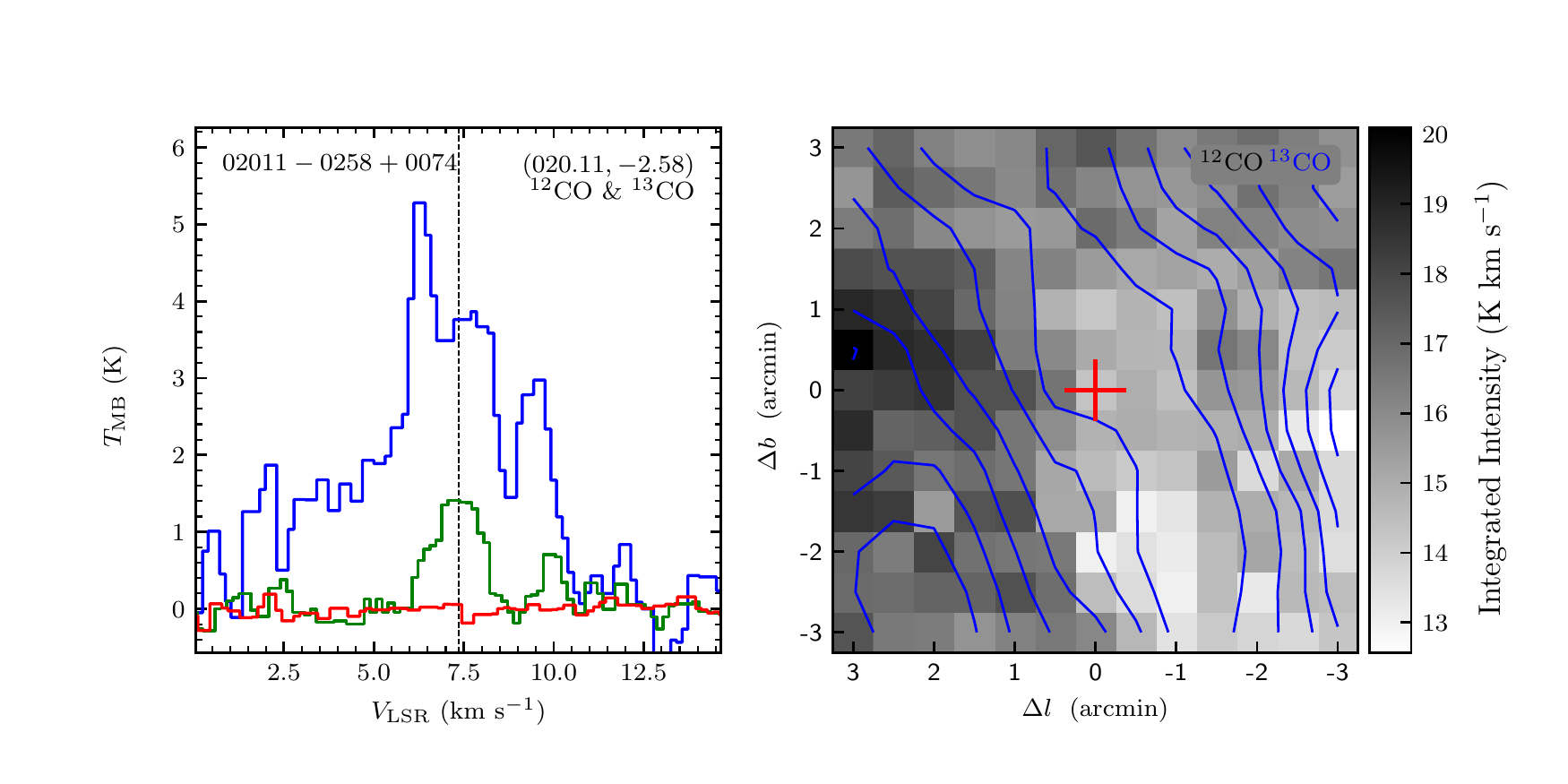}
\vspace{-0.5cm}

\includegraphics[width=9.0cm,angle=0]{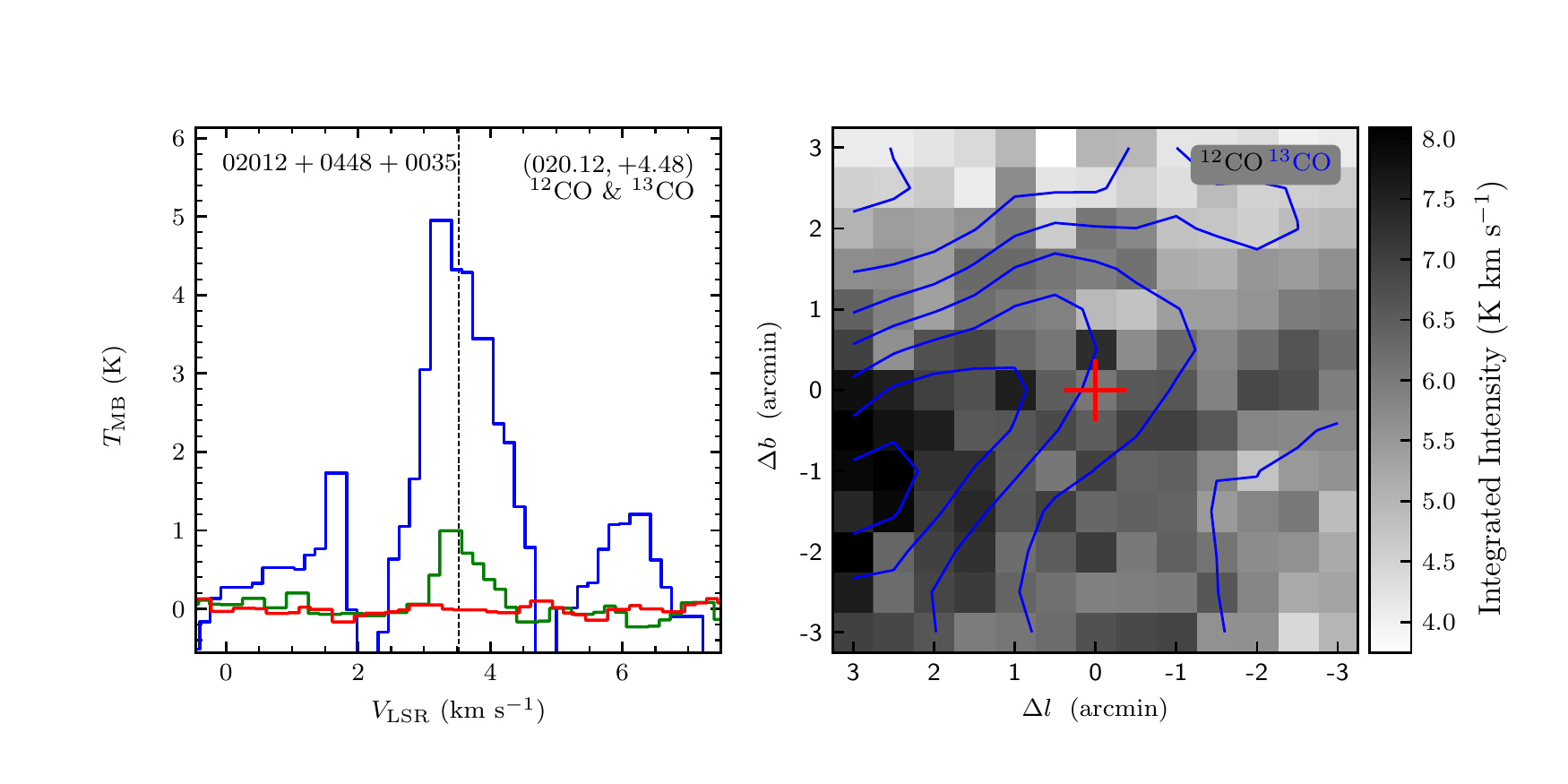}
\includegraphics[width=9.0cm,angle=0]{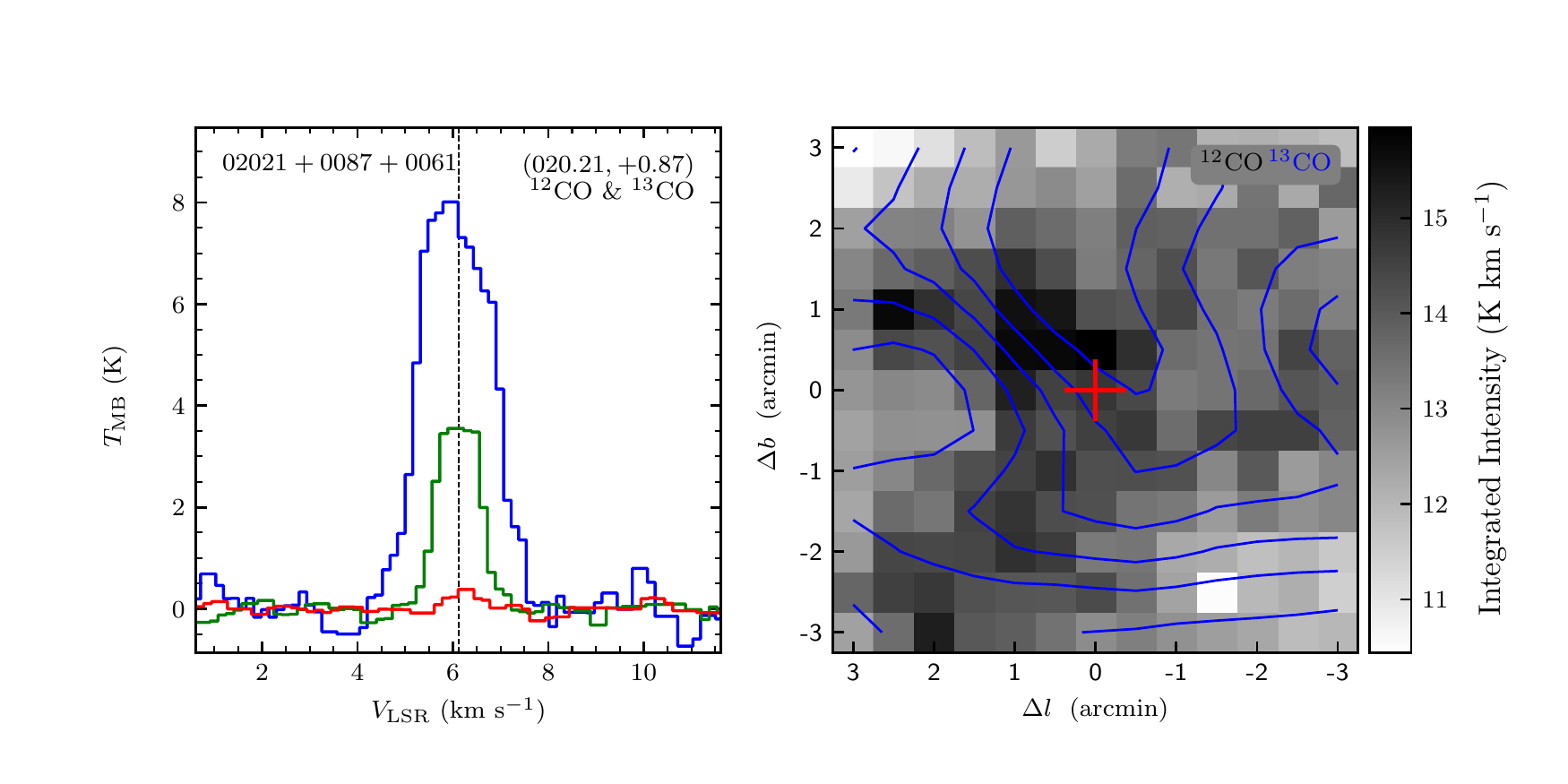}
\end{figure}
\clearpage

\begin{figure}
\includegraphics[width=9.0cm,angle=0]{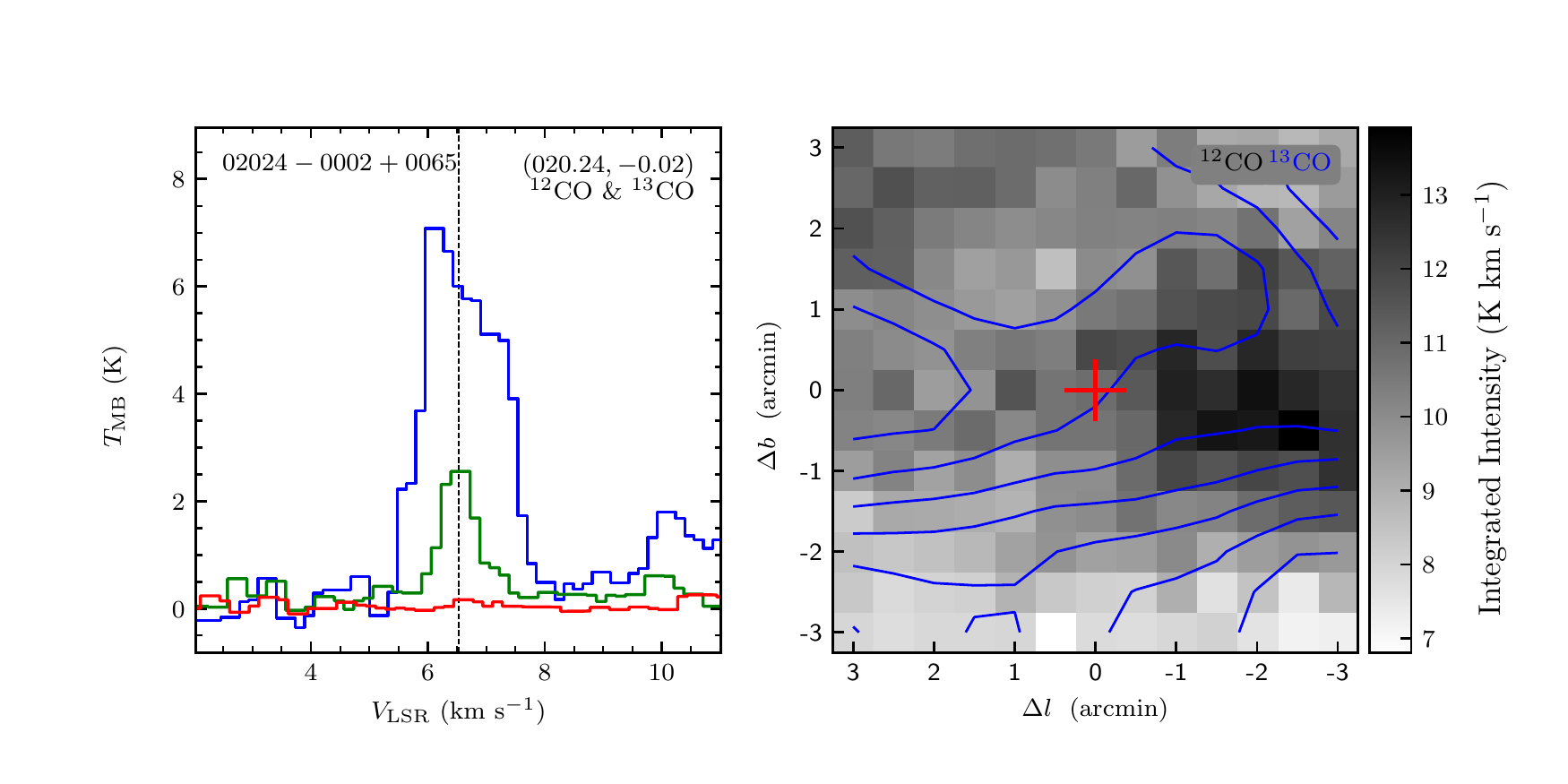}
\includegraphics[width=9.0cm,angle=0]{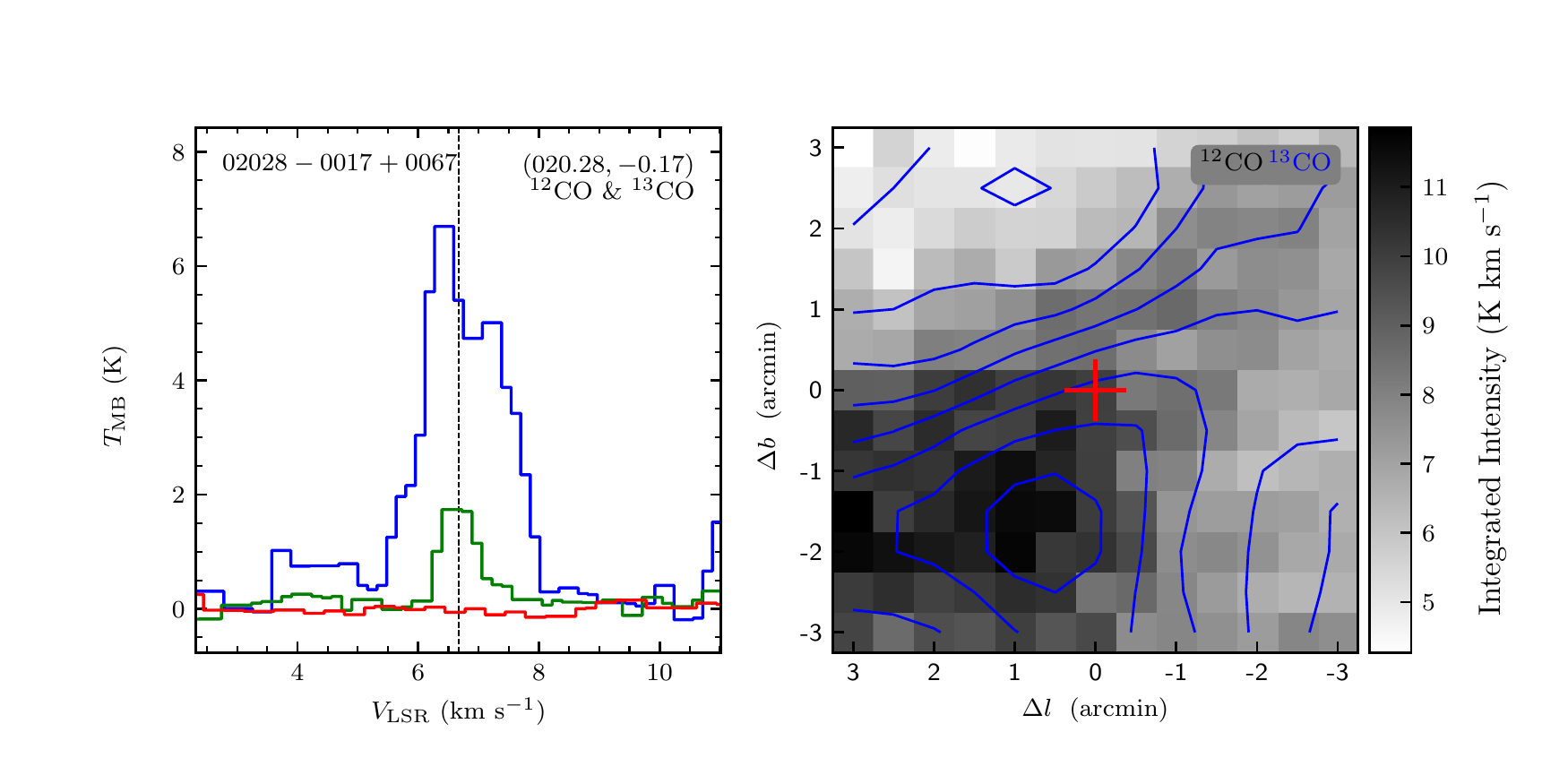}
\vspace{-0.5cm}

\includegraphics[width=9.0cm,angle=0]{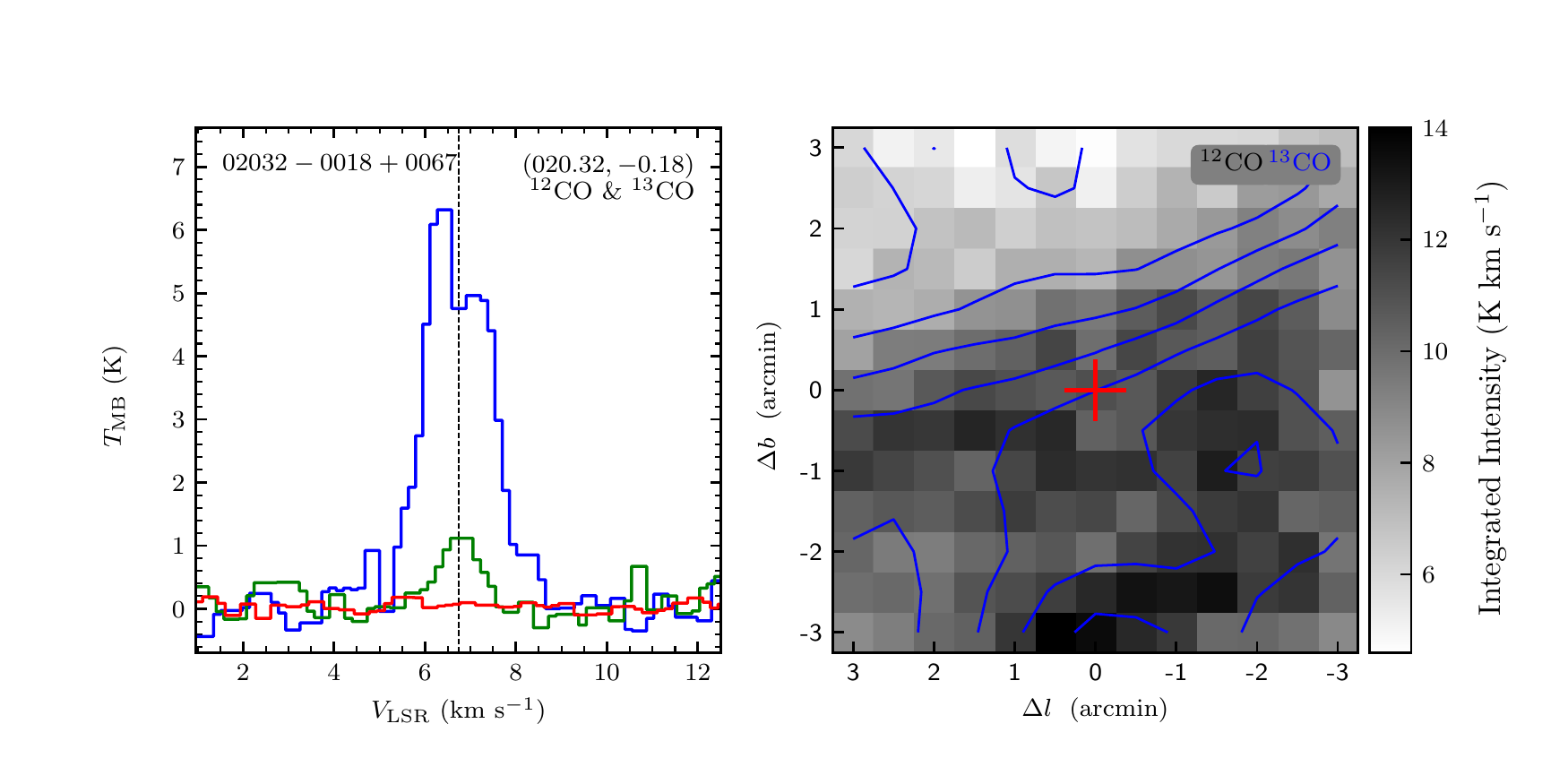}
\includegraphics[width=9.0cm,angle=0]{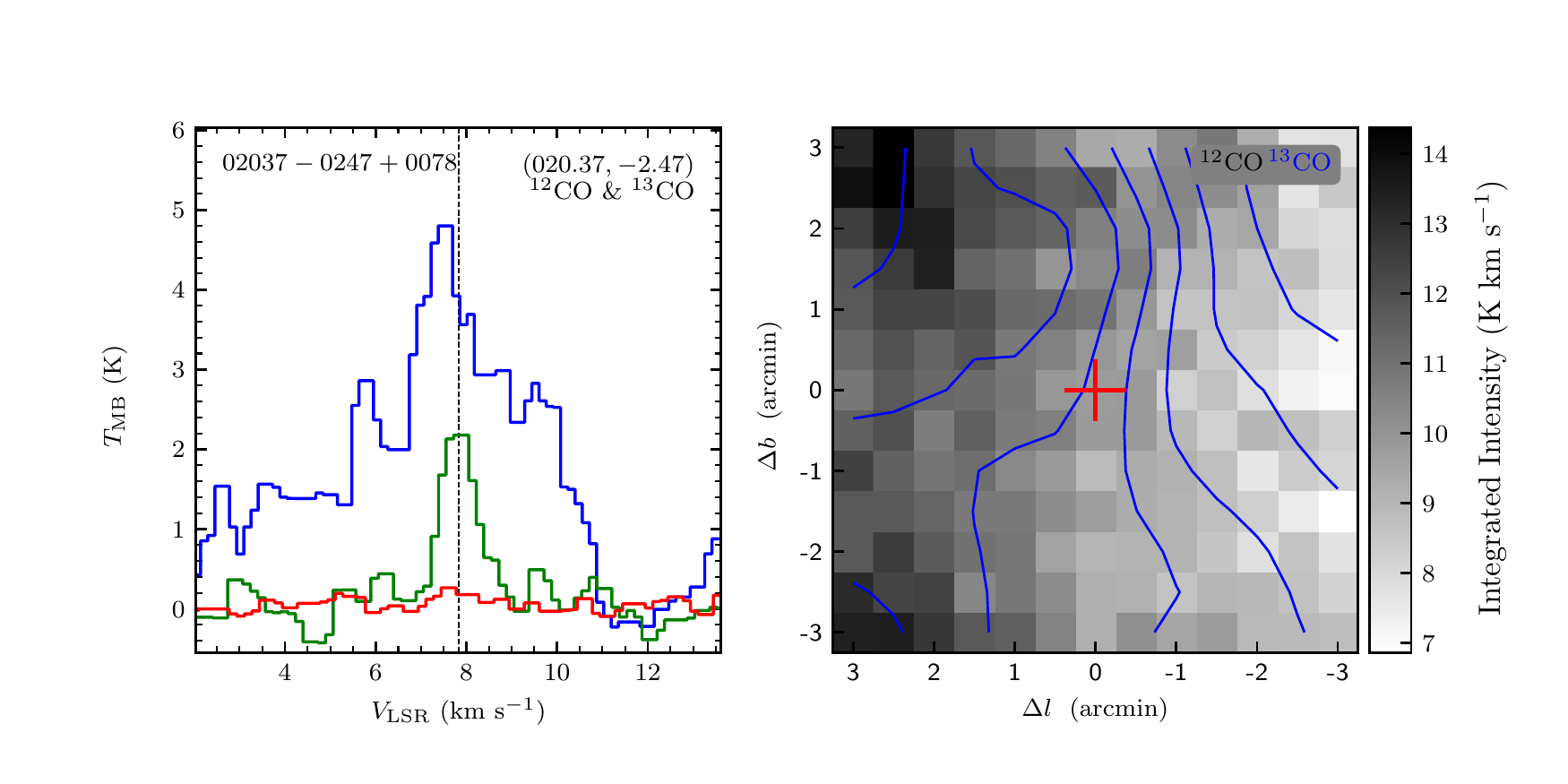}
\vspace{-0.5cm}

\includegraphics[width=9.0cm,angle=0]{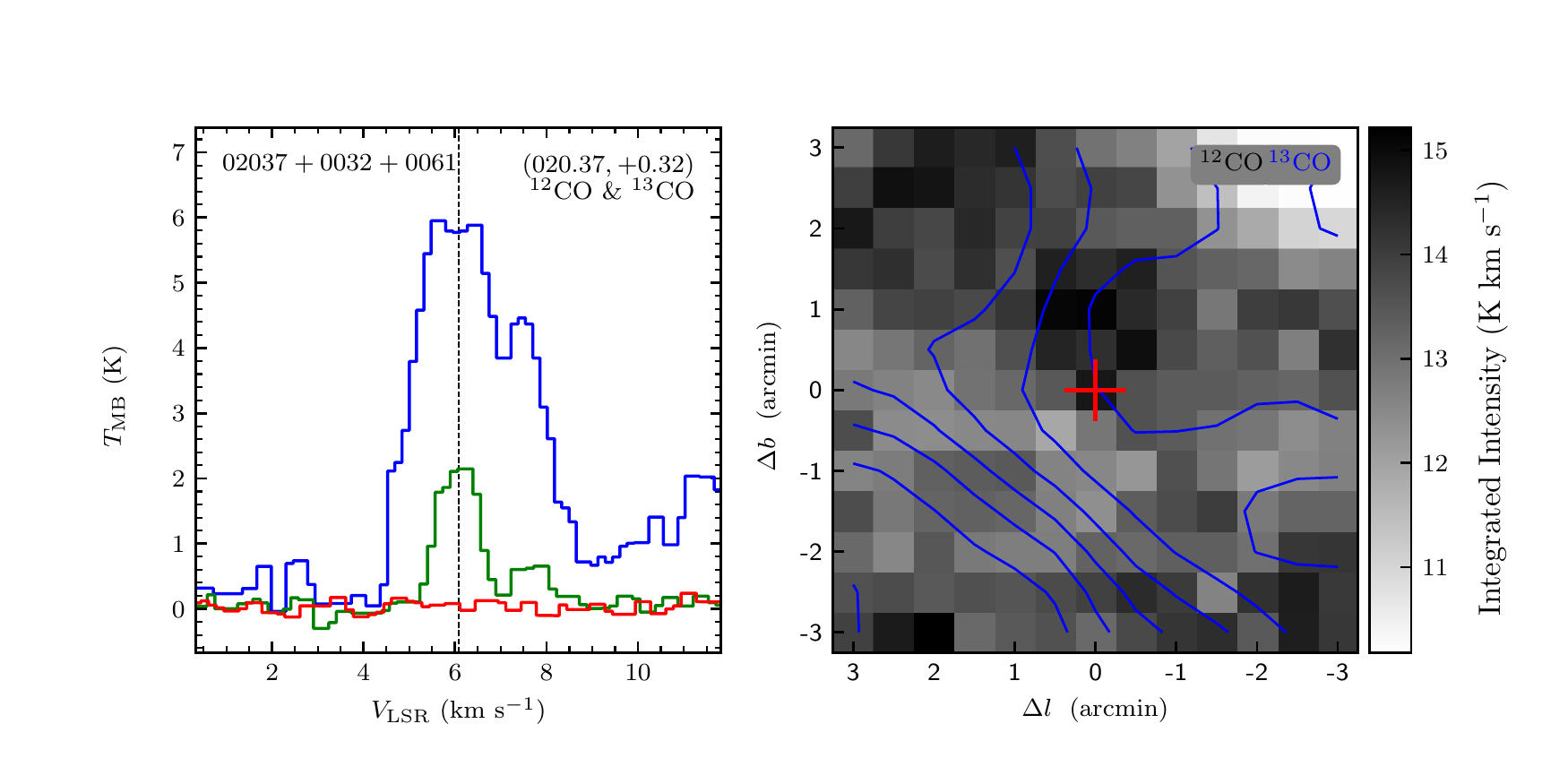}
\includegraphics[width=9.0cm,angle=0]{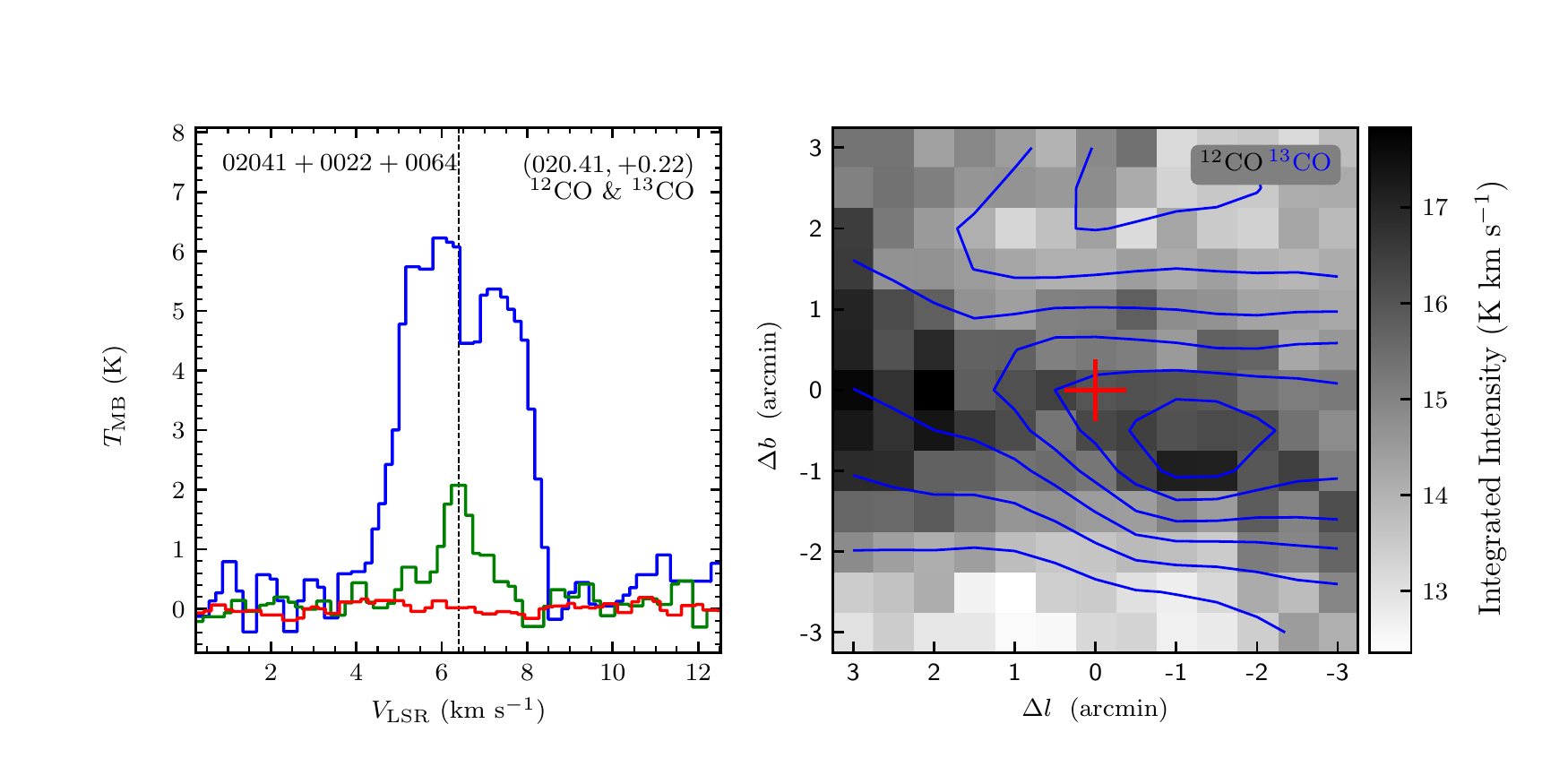}
\vspace{-0.5cm}

\includegraphics[width=9.0cm,angle=0]{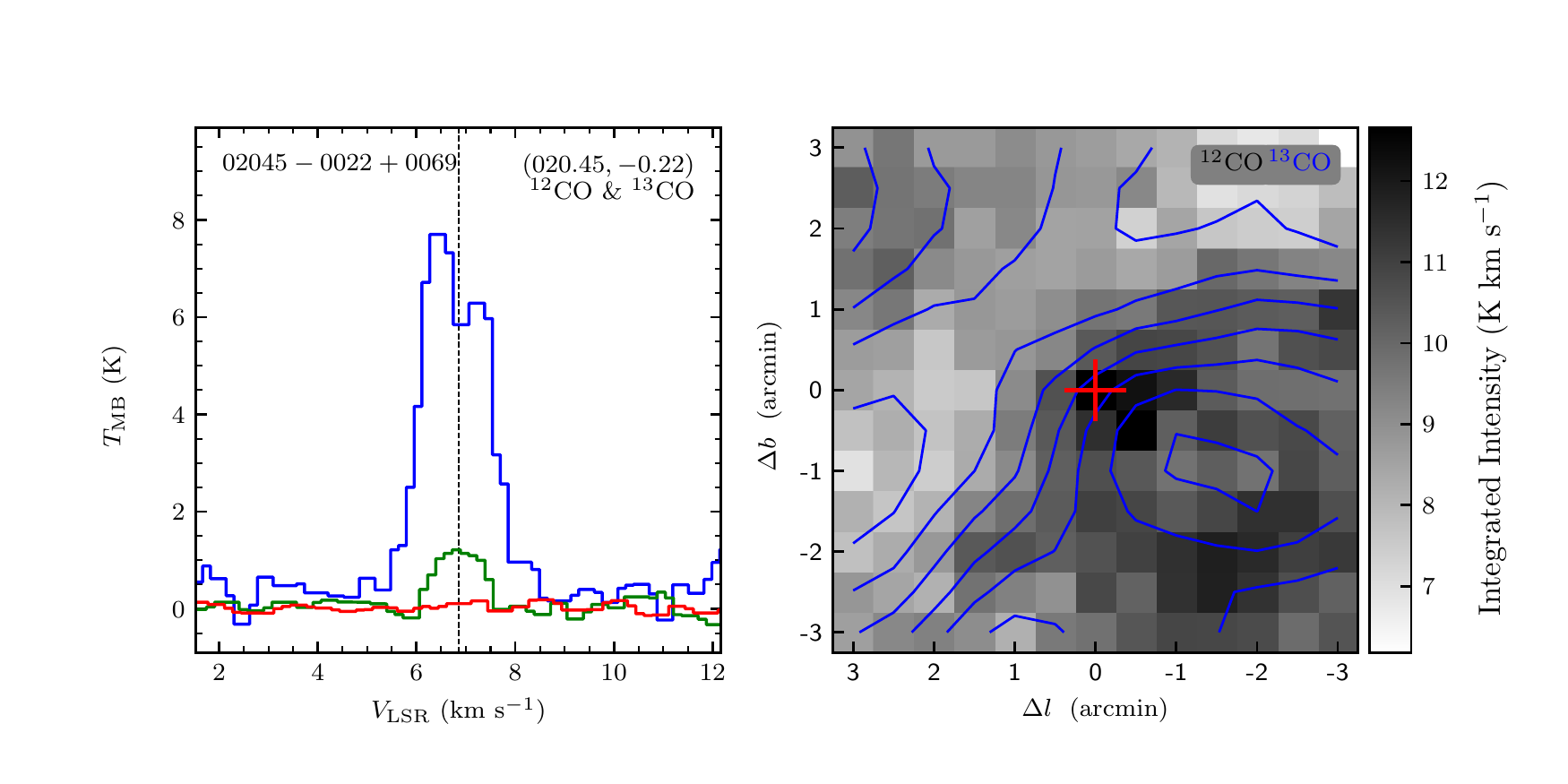}
\includegraphics[width=9.0cm,angle=0]{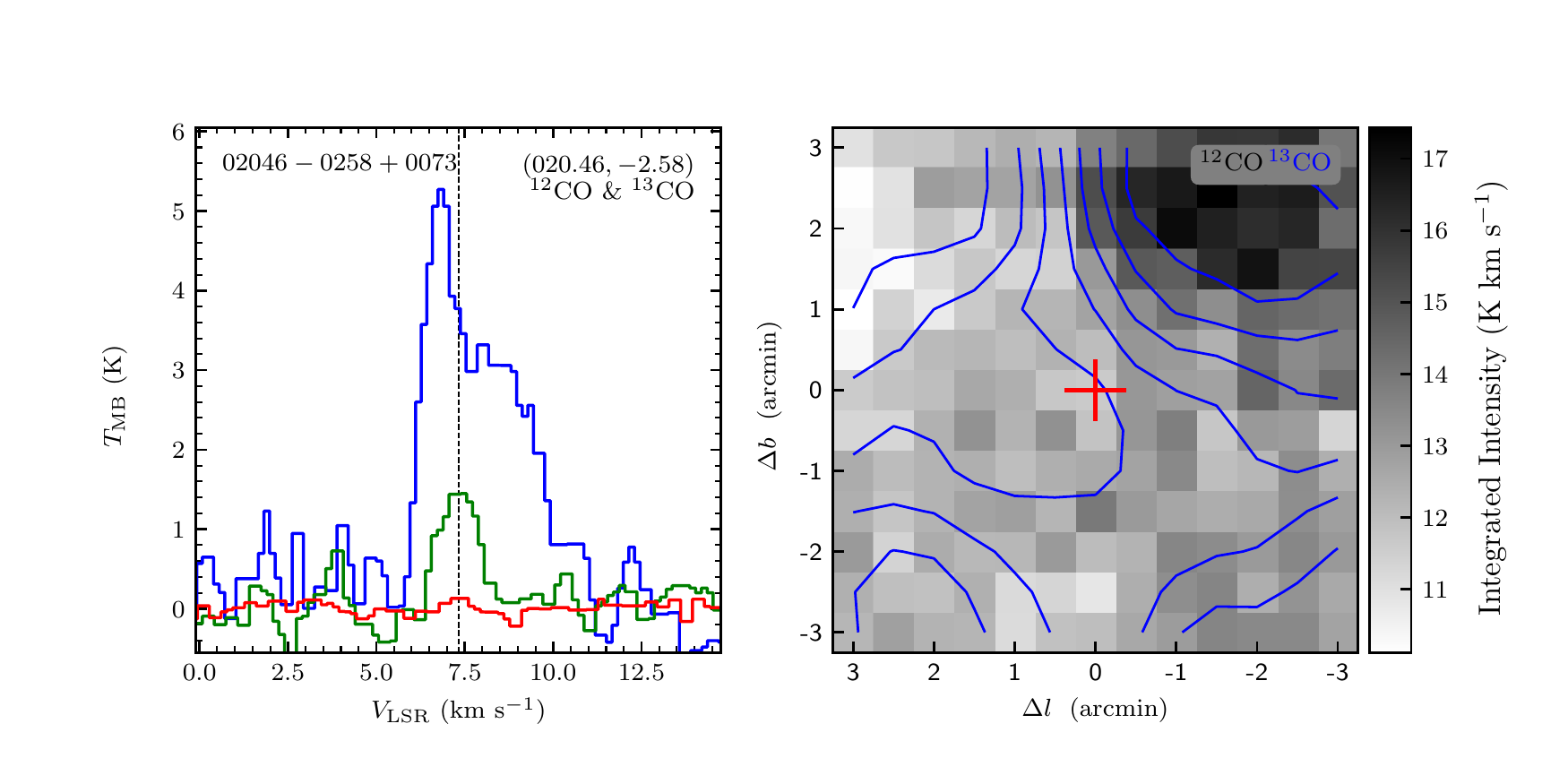}
\vspace{-0.5cm}

\includegraphics[width=9.0cm,angle=0]{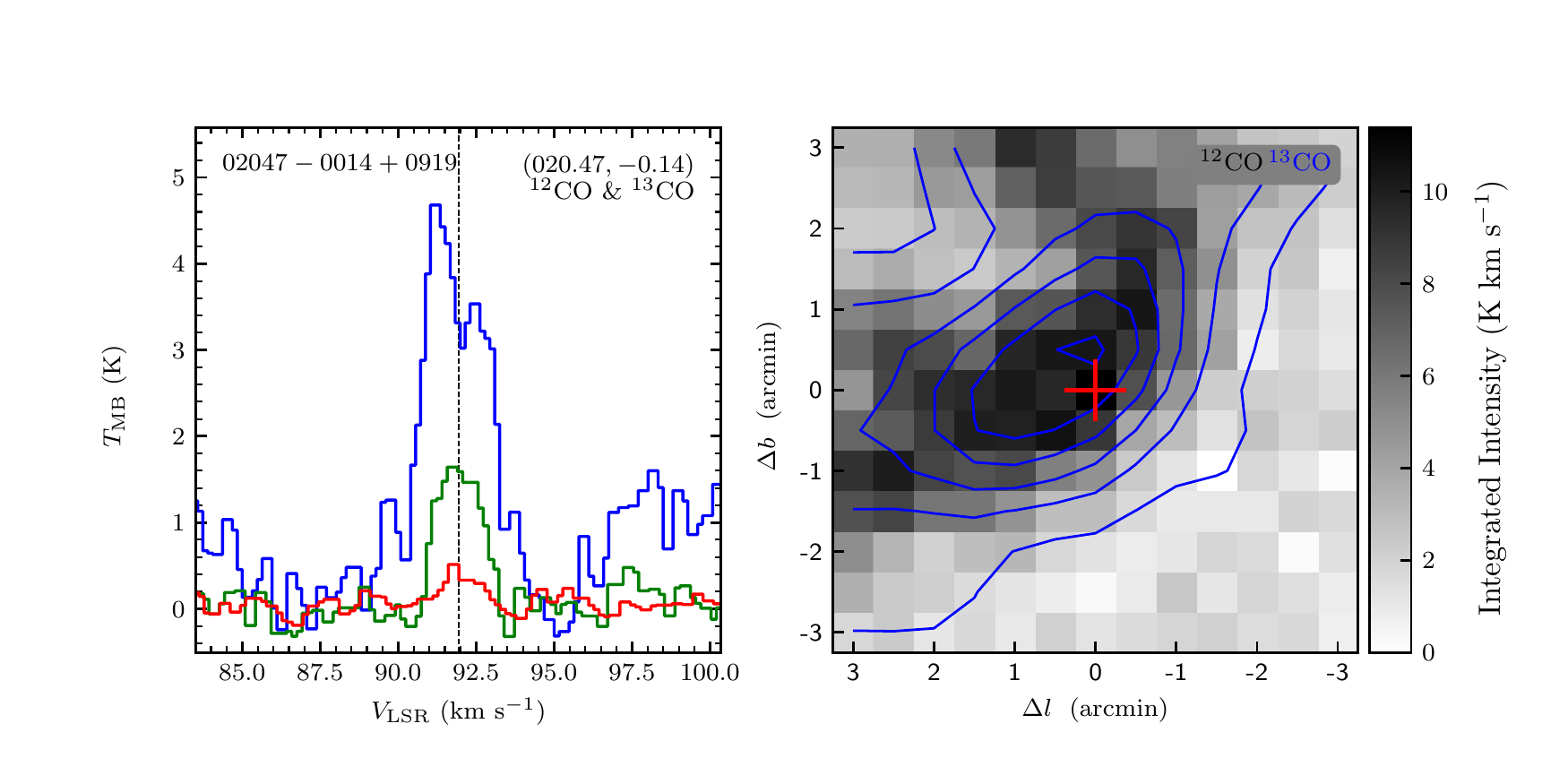}
\includegraphics[width=9.0cm,angle=0]{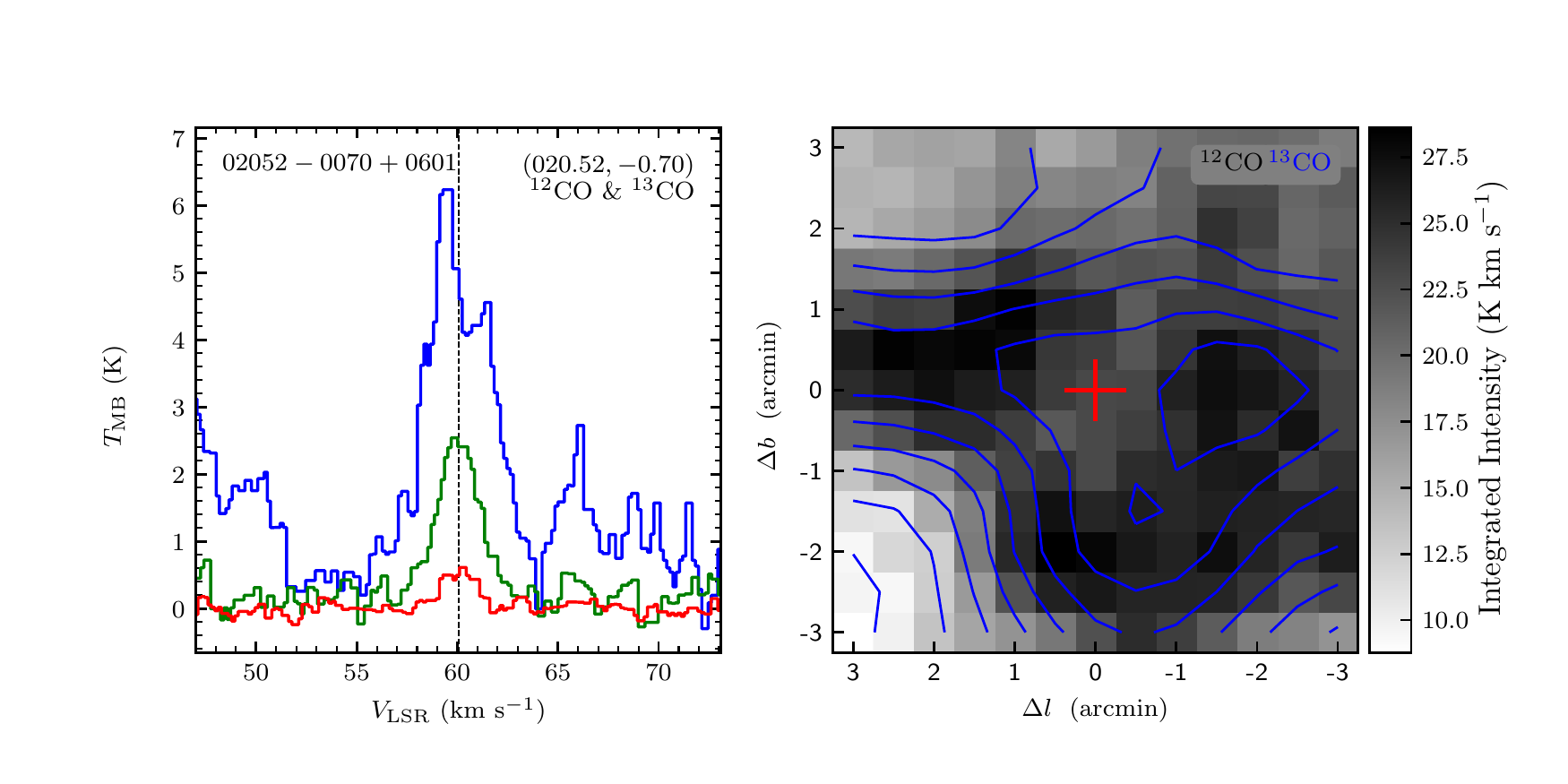}
\end{figure}
\clearpage

\begin{figure}
\includegraphics[width=9.0cm,angle=0]{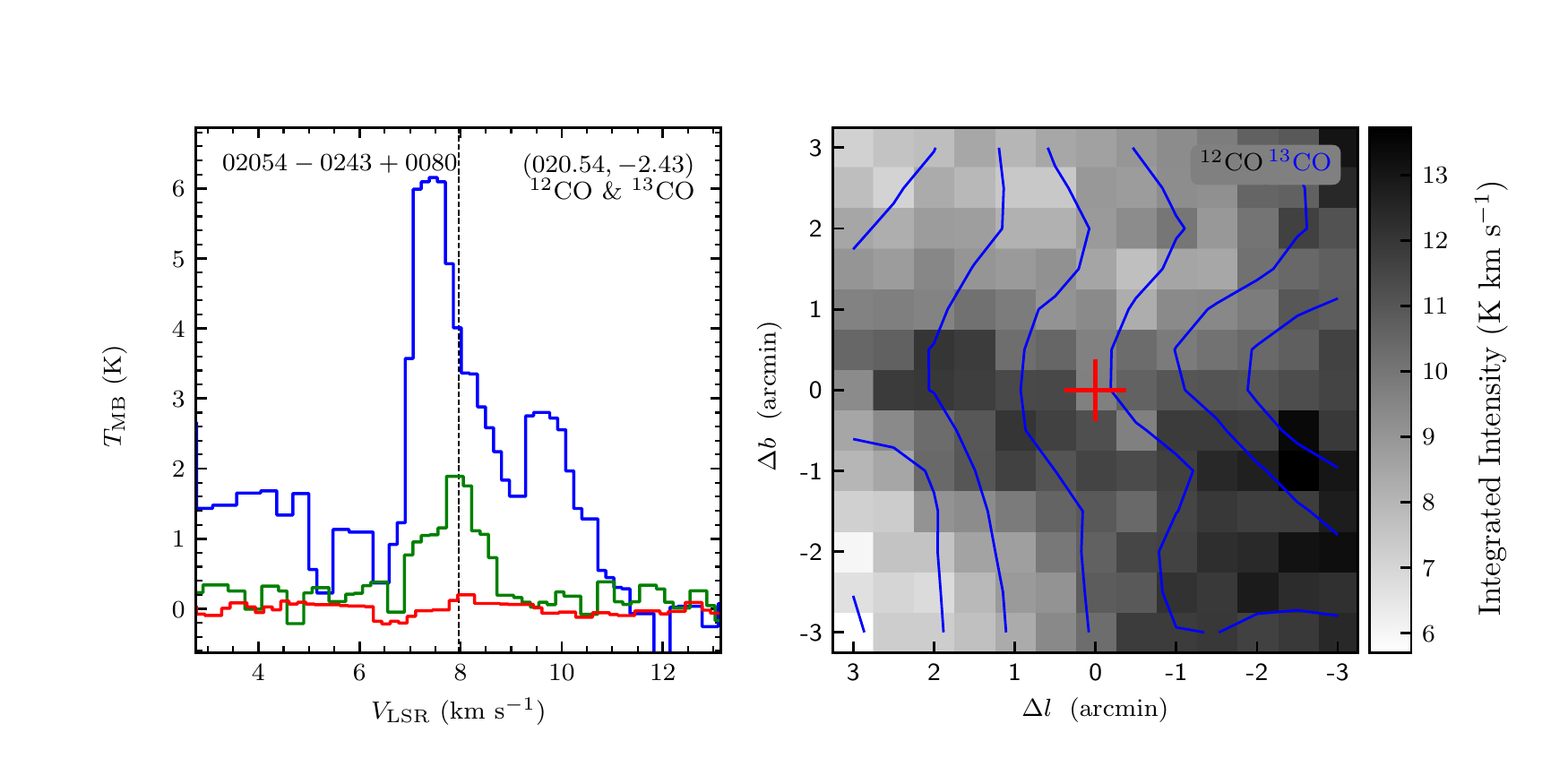}
\includegraphics[width=9.0cm,angle=0]{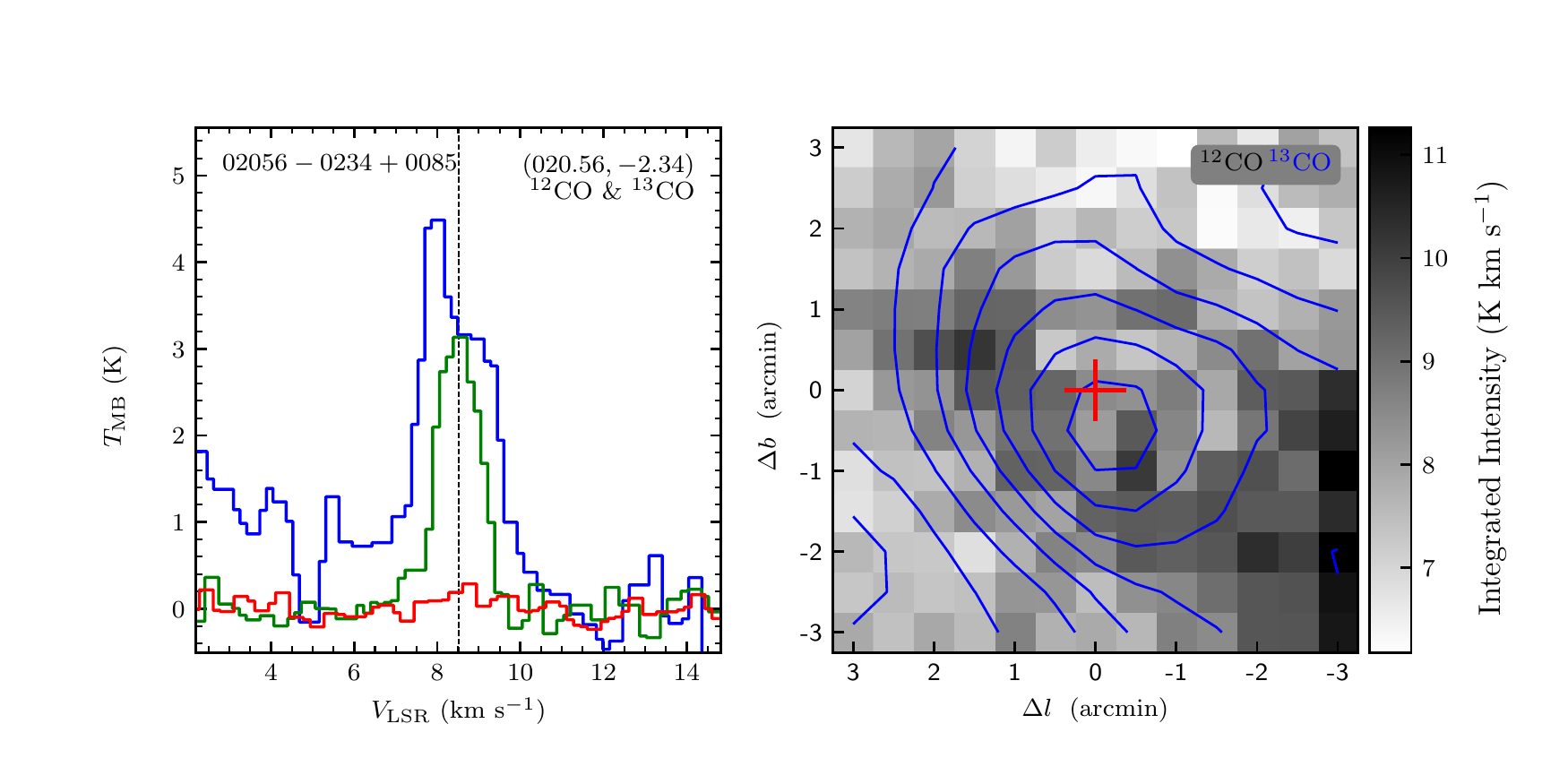}
\vspace{-0.5cm}

\includegraphics[width=9.0cm,angle=0]{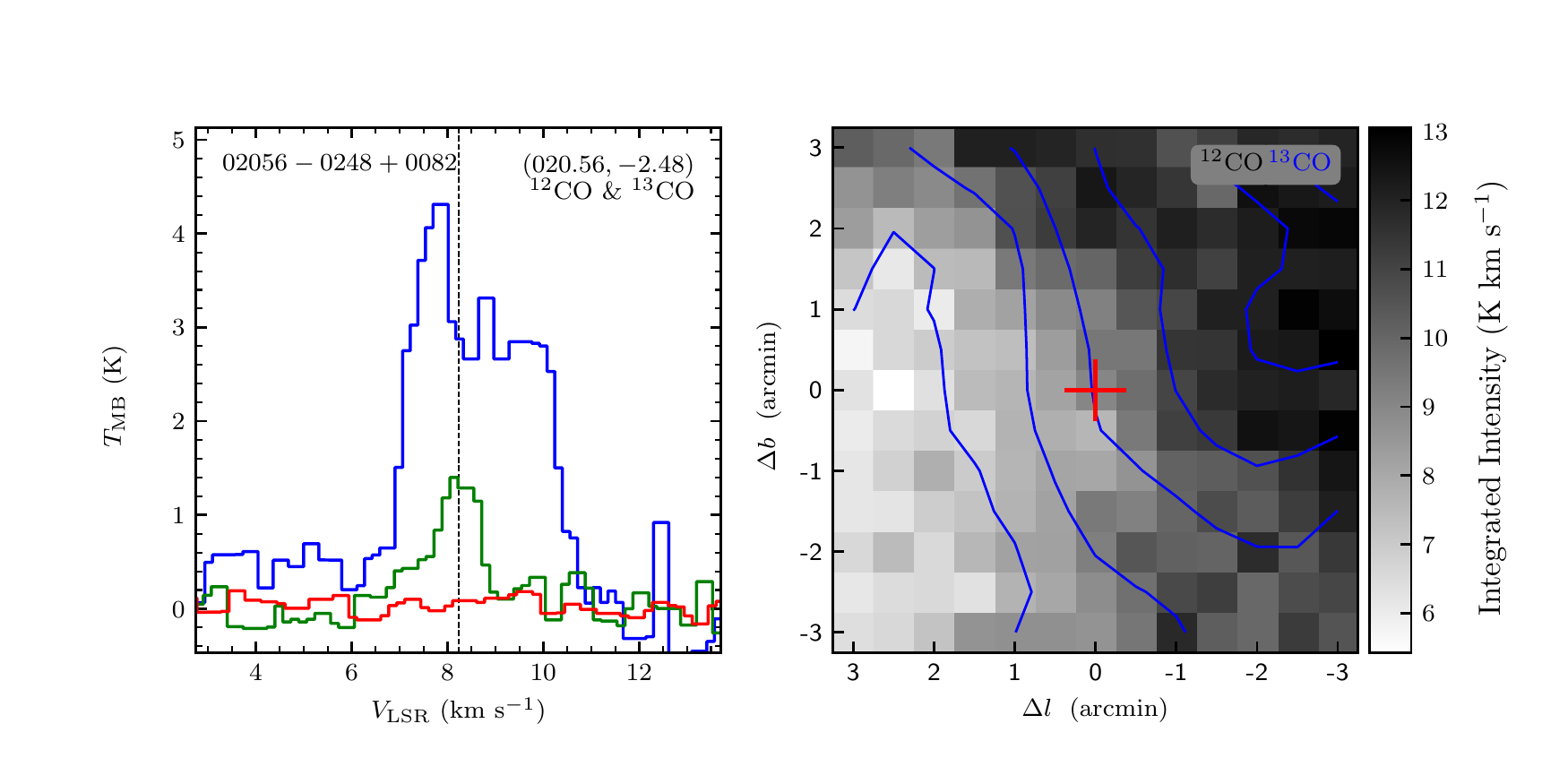}
\includegraphics[width=9.0cm,angle=0]{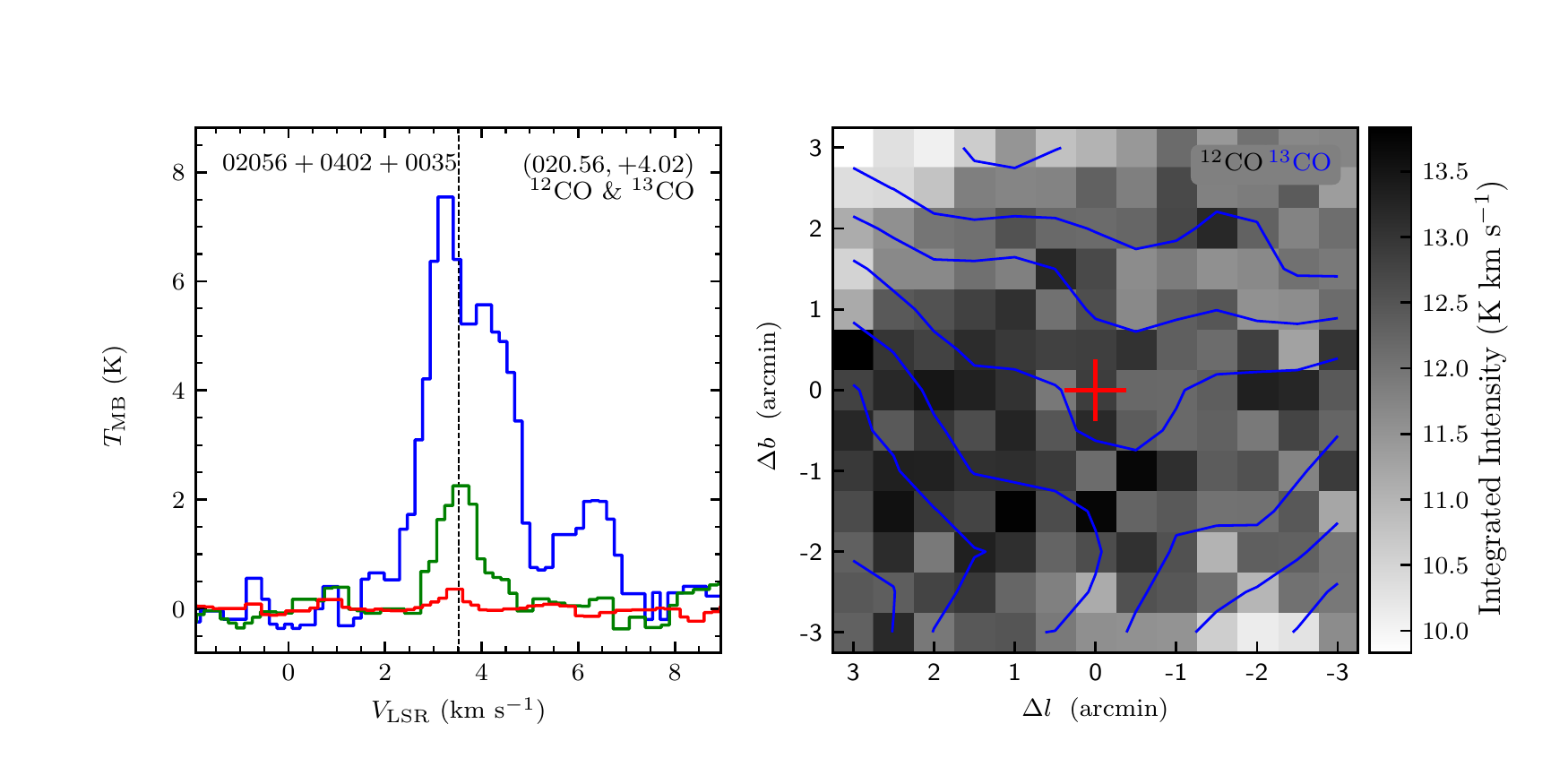}
\vspace{-0.5cm}

\includegraphics[width=9.0cm,angle=0]{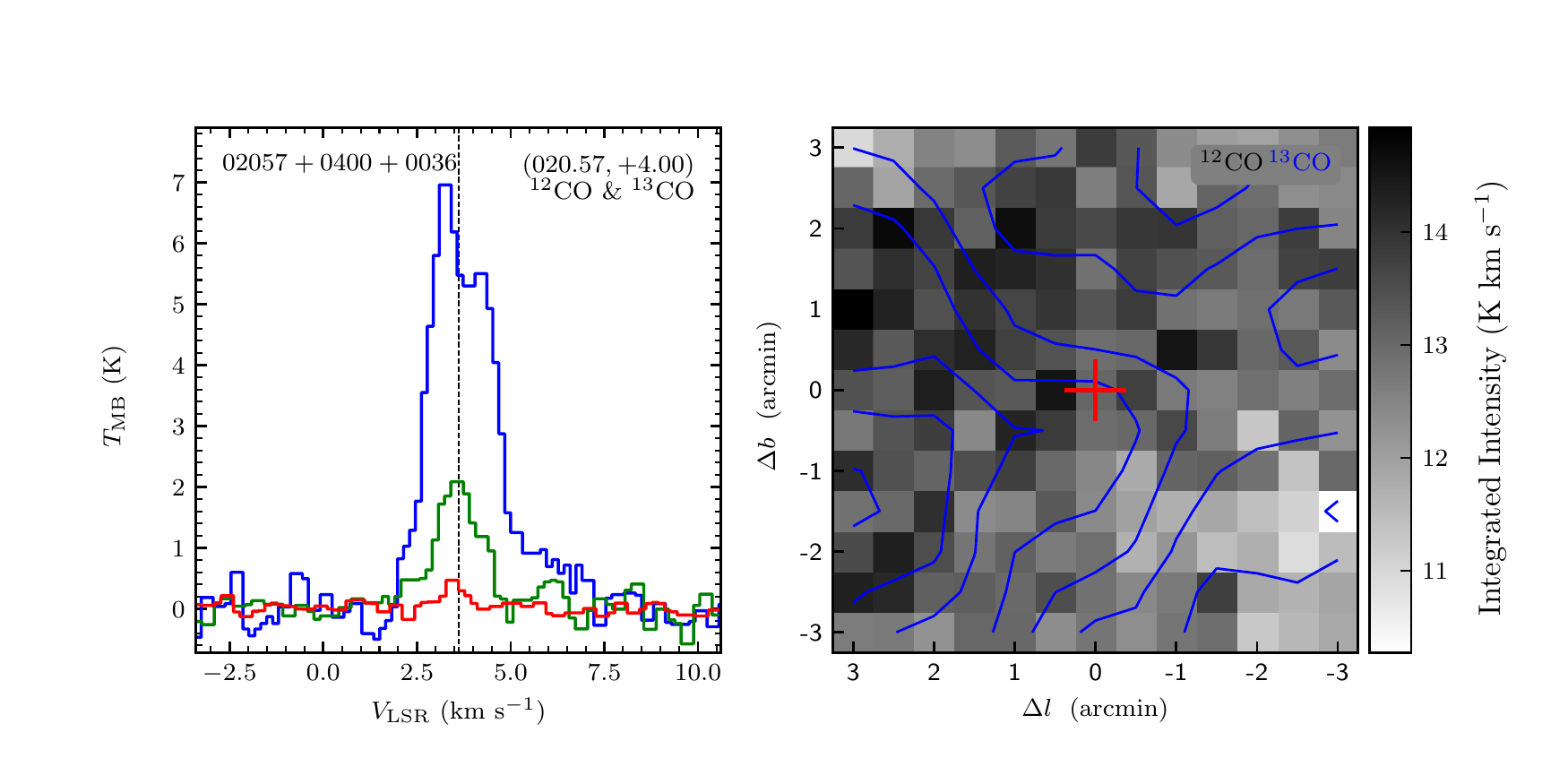}
\includegraphics[width=9.0cm,angle=0]{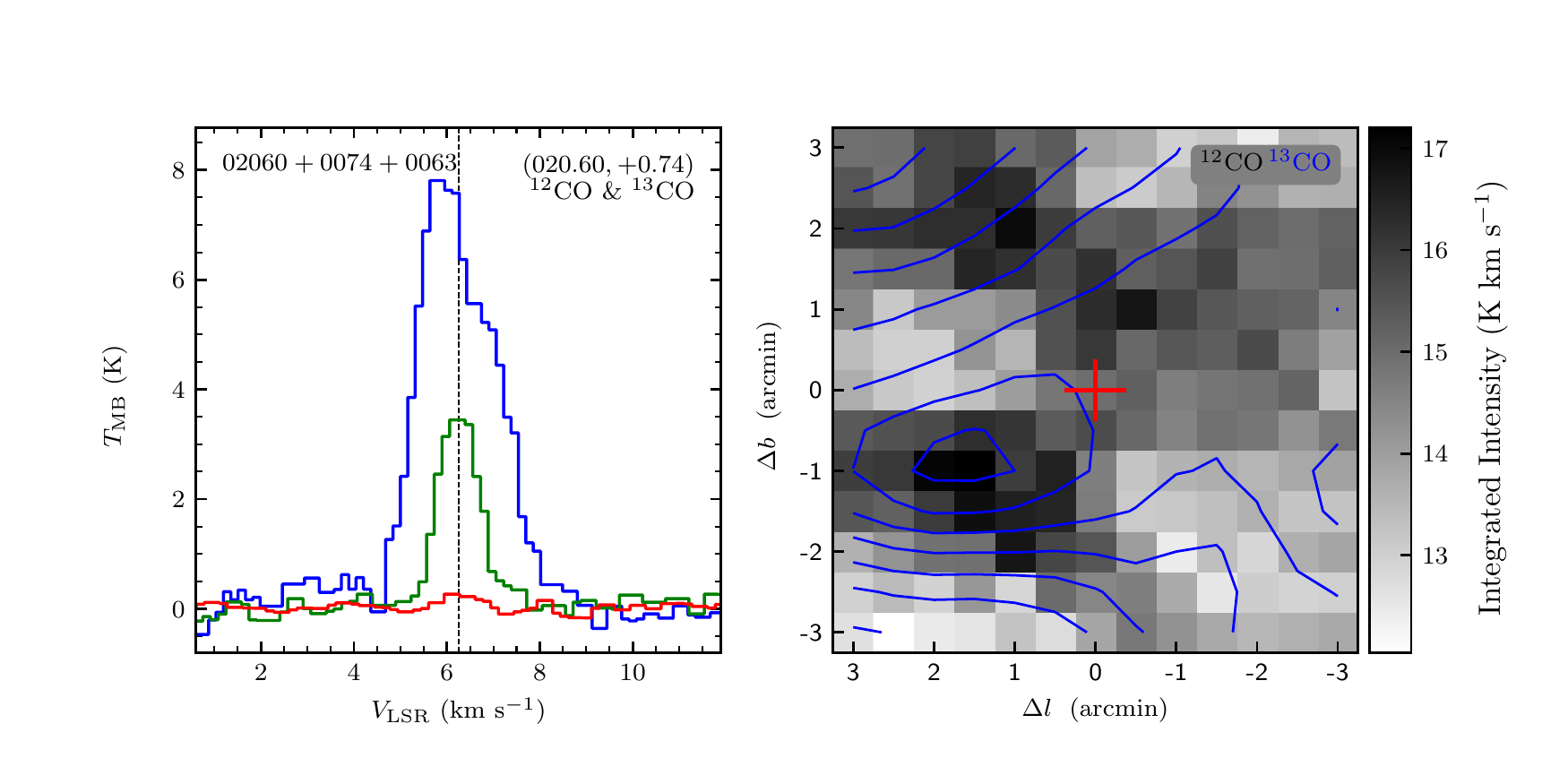}
\vspace{-0.5cm}

\includegraphics[width=9.0cm,angle=0]{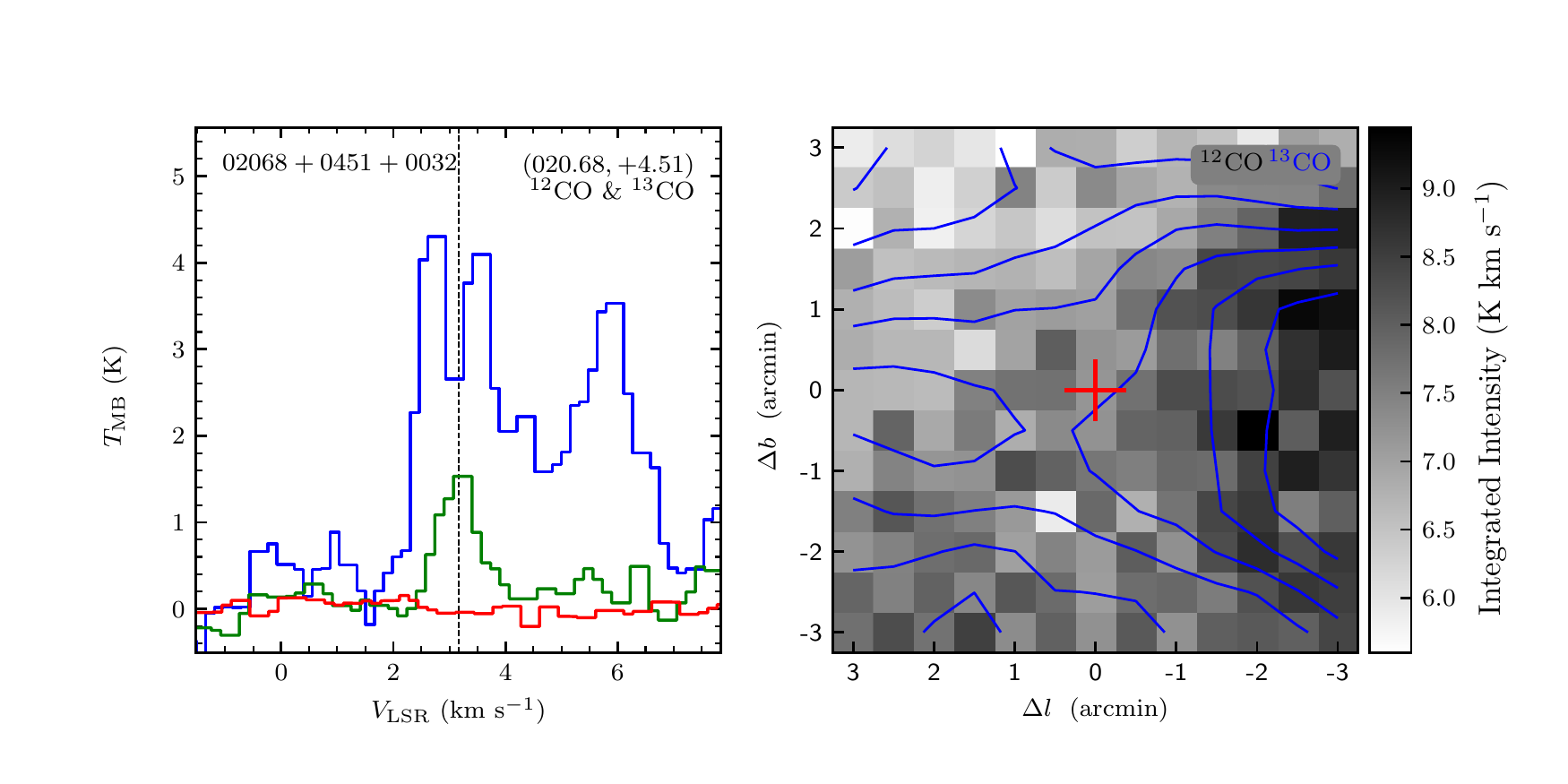}
\includegraphics[width=9.0cm,angle=0]{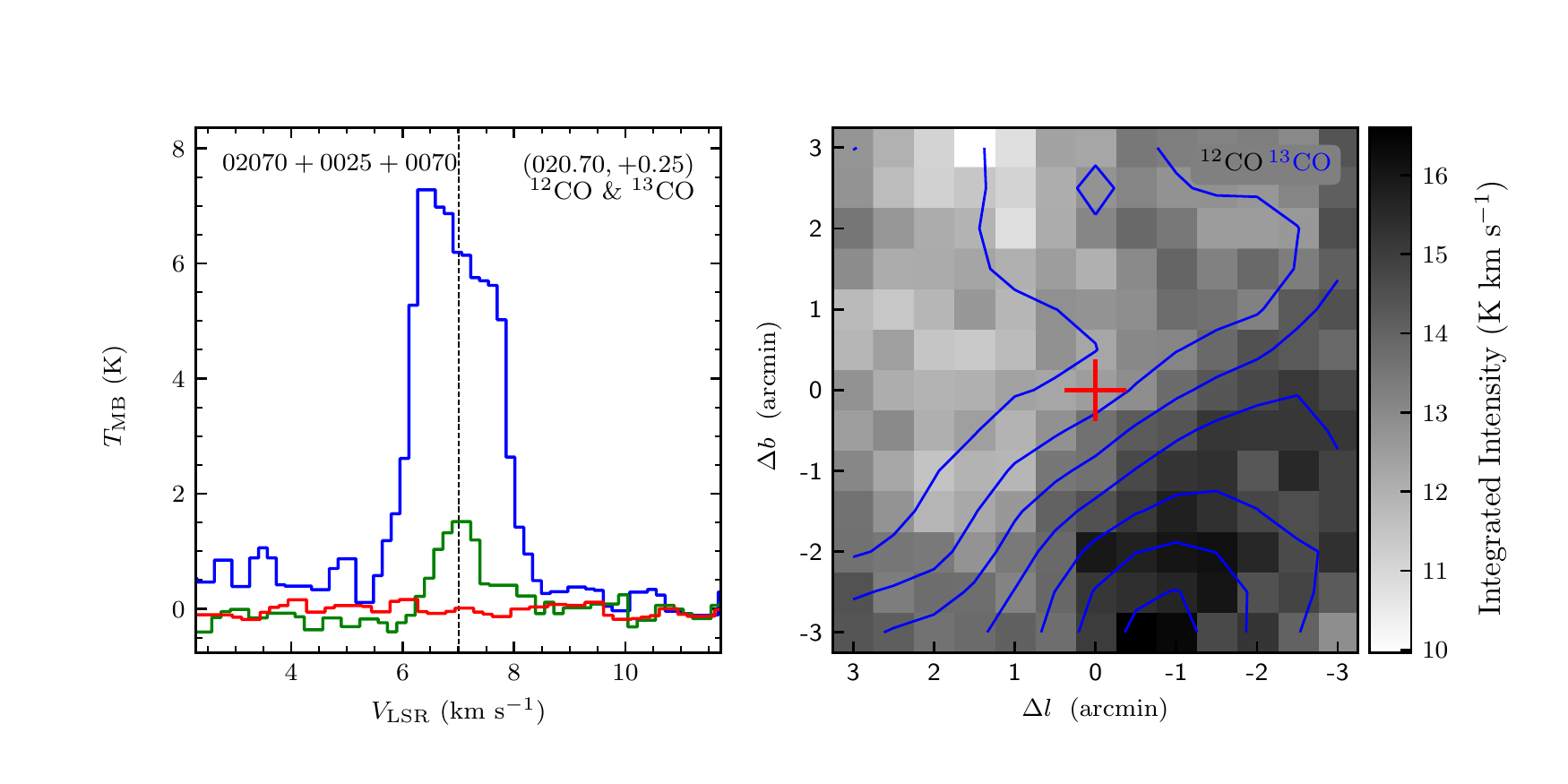}
\vspace{-0.5cm}

\includegraphics[width=9.0cm,angle=0]{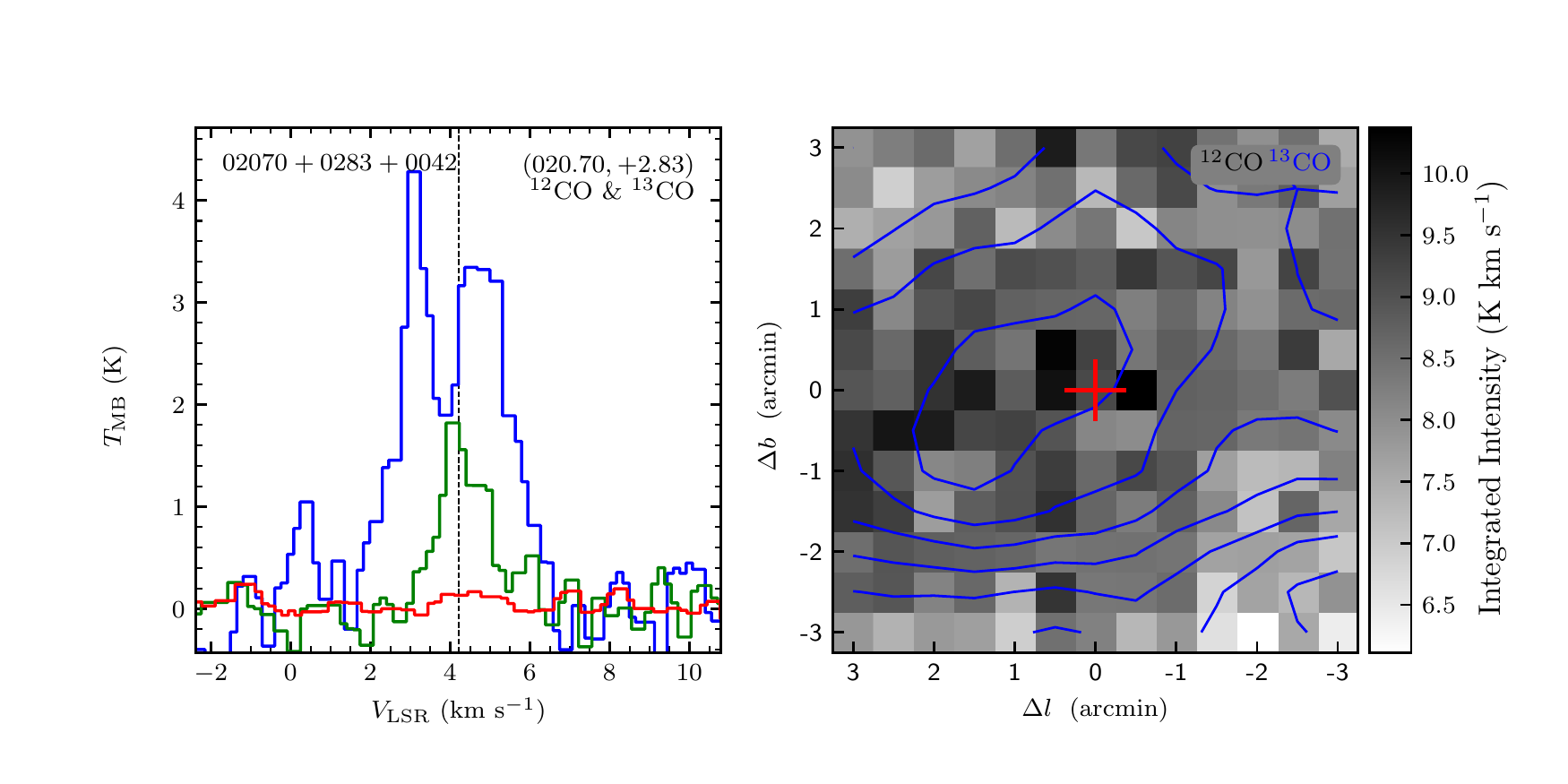}
\includegraphics[width=9.0cm,angle=0]{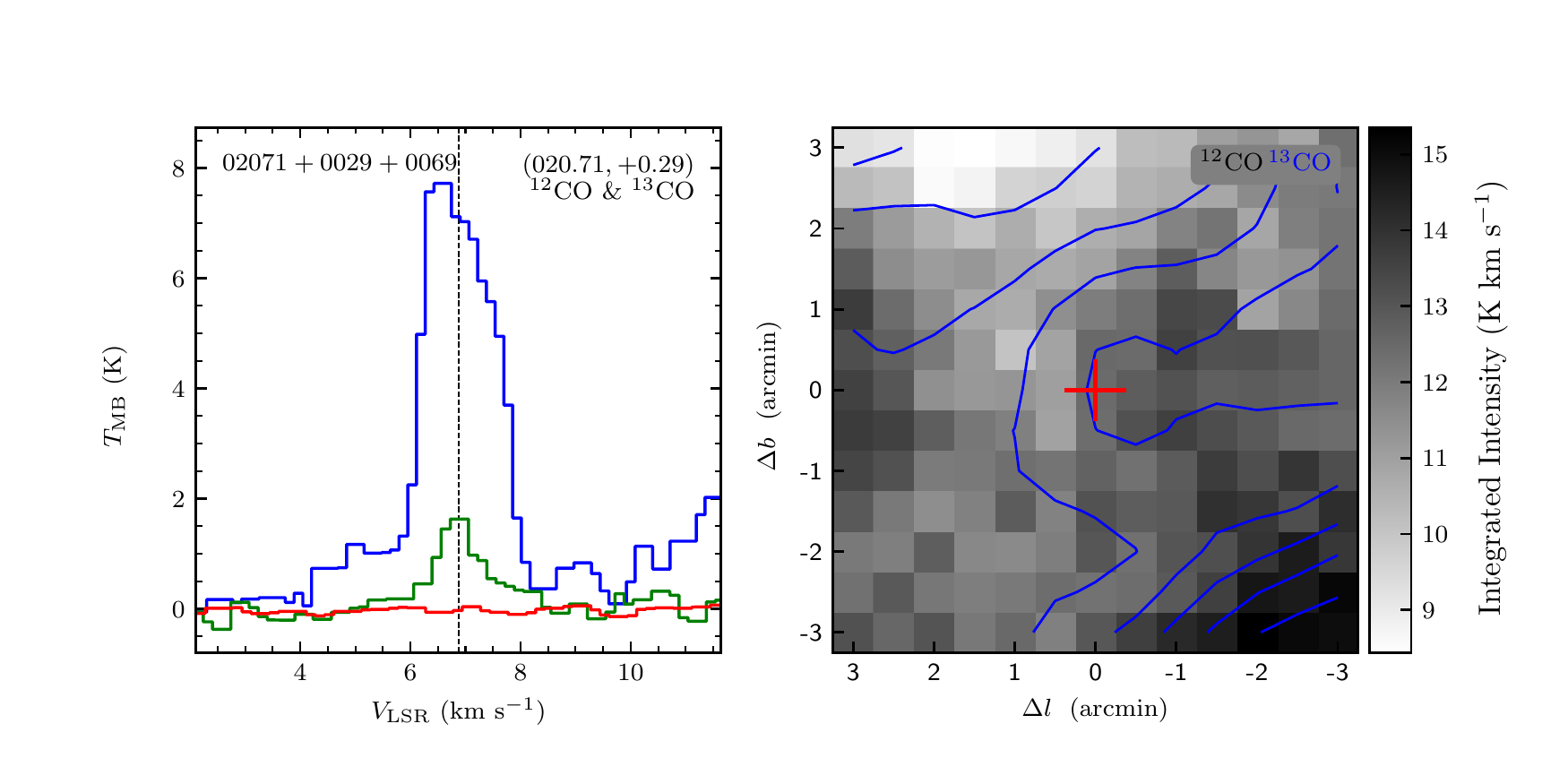}
\end{figure}
\clearpage

\begin{figure}
\includegraphics[width=9.0cm,angle=0]{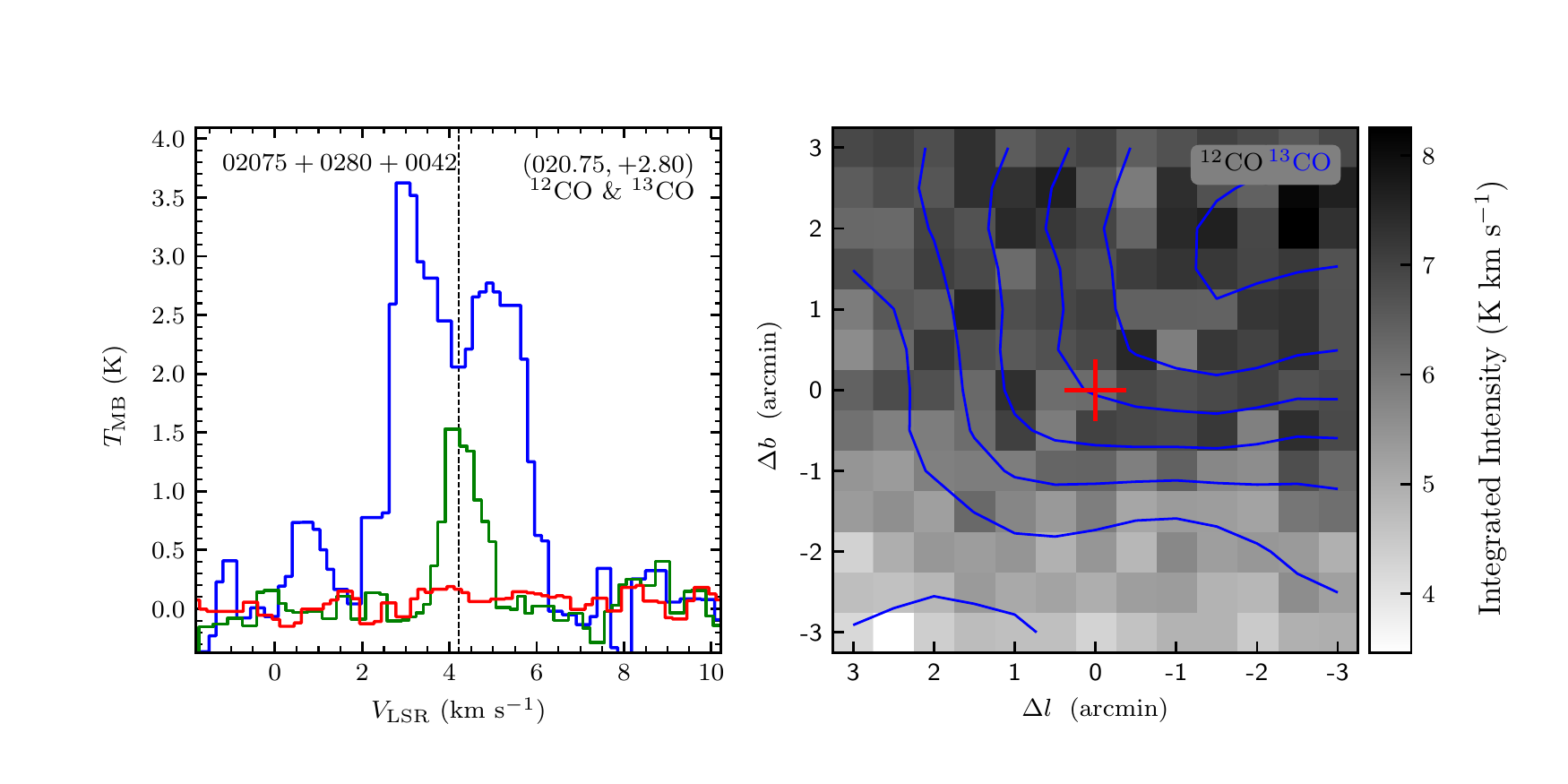}
\includegraphics[width=9.0cm,angle=0]{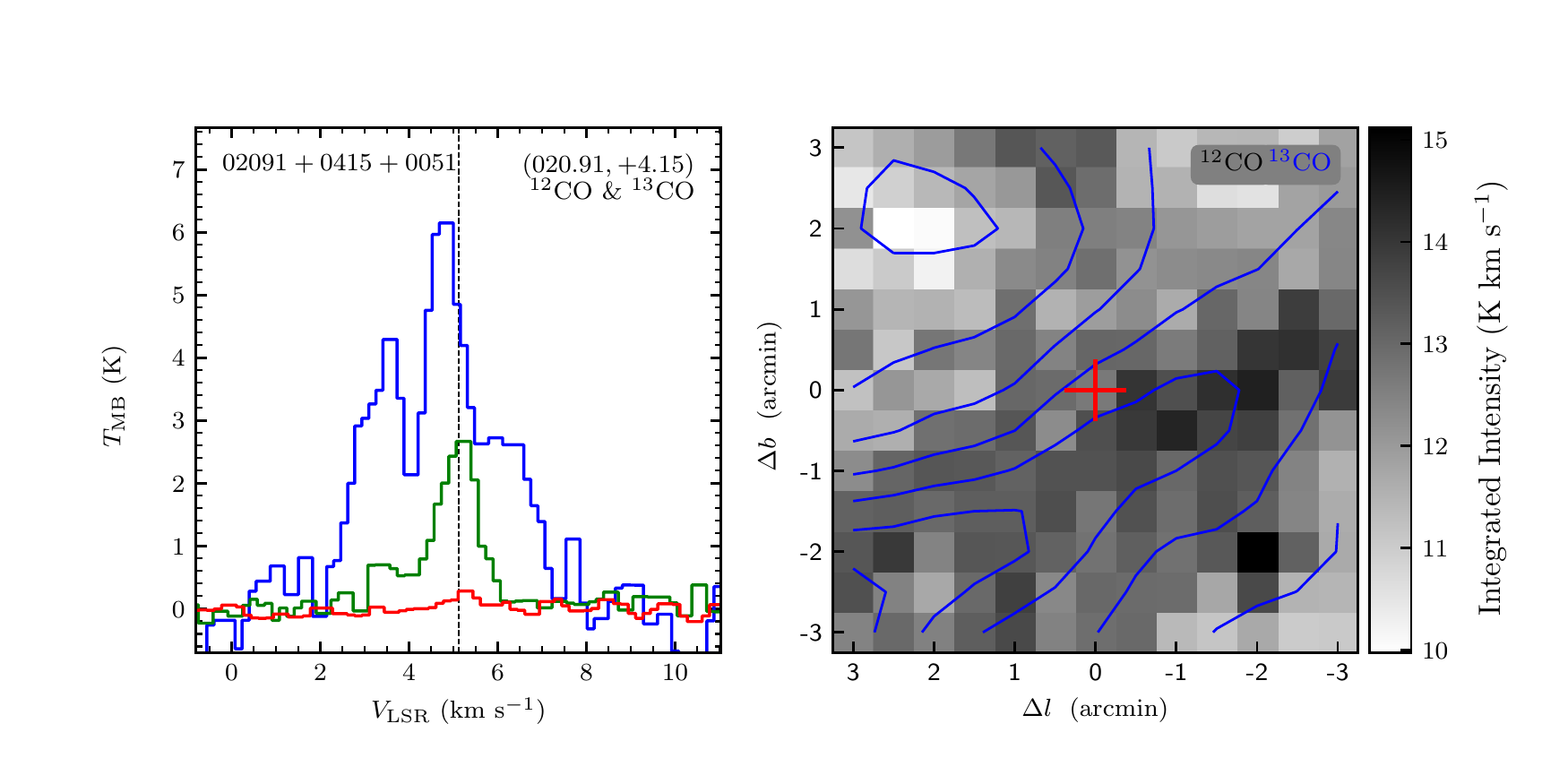}
\vspace{-0.5cm}

\includegraphics[width=9.0cm,angle=0]{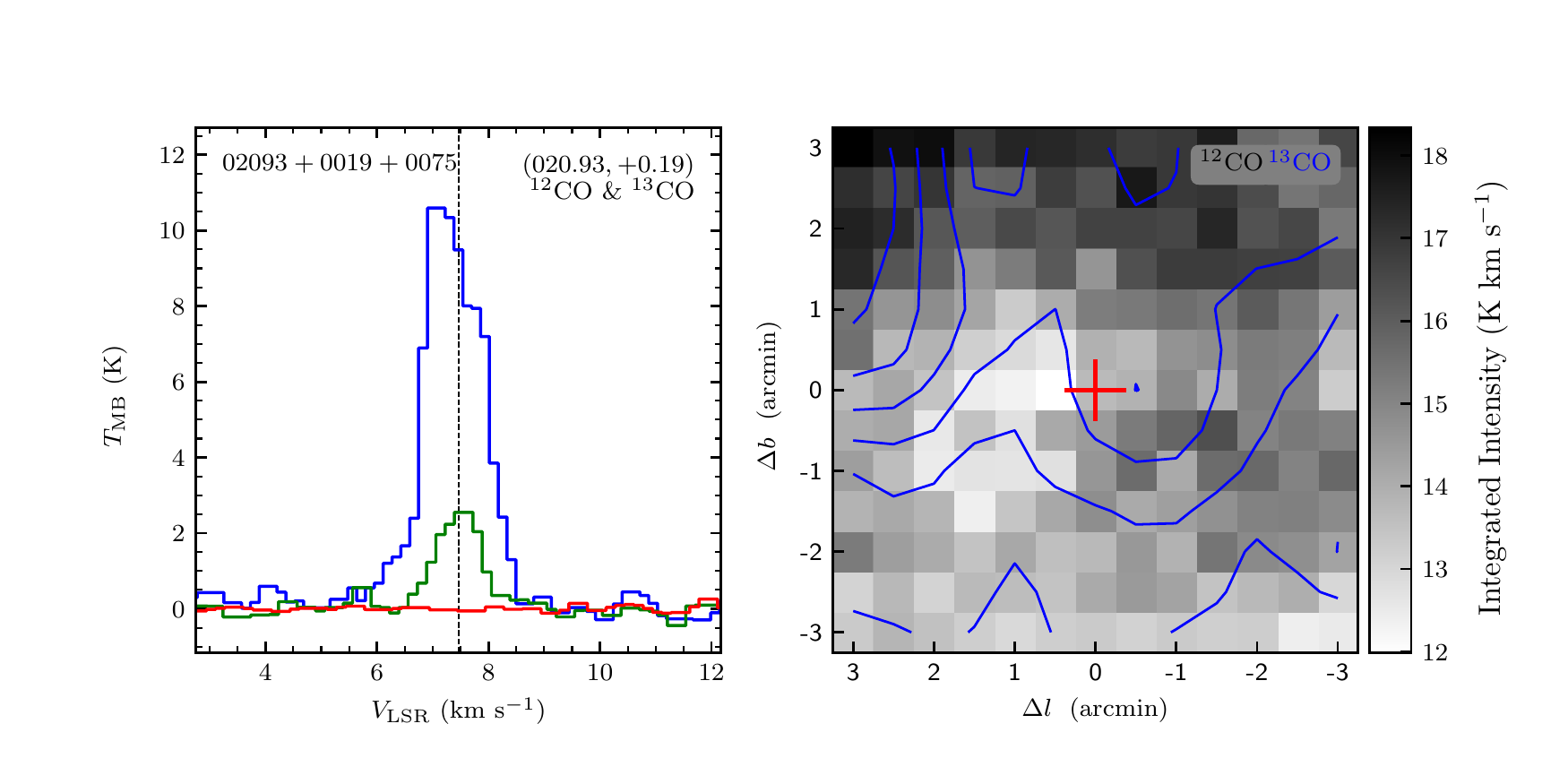}
\includegraphics[width=9.0cm,angle=0]{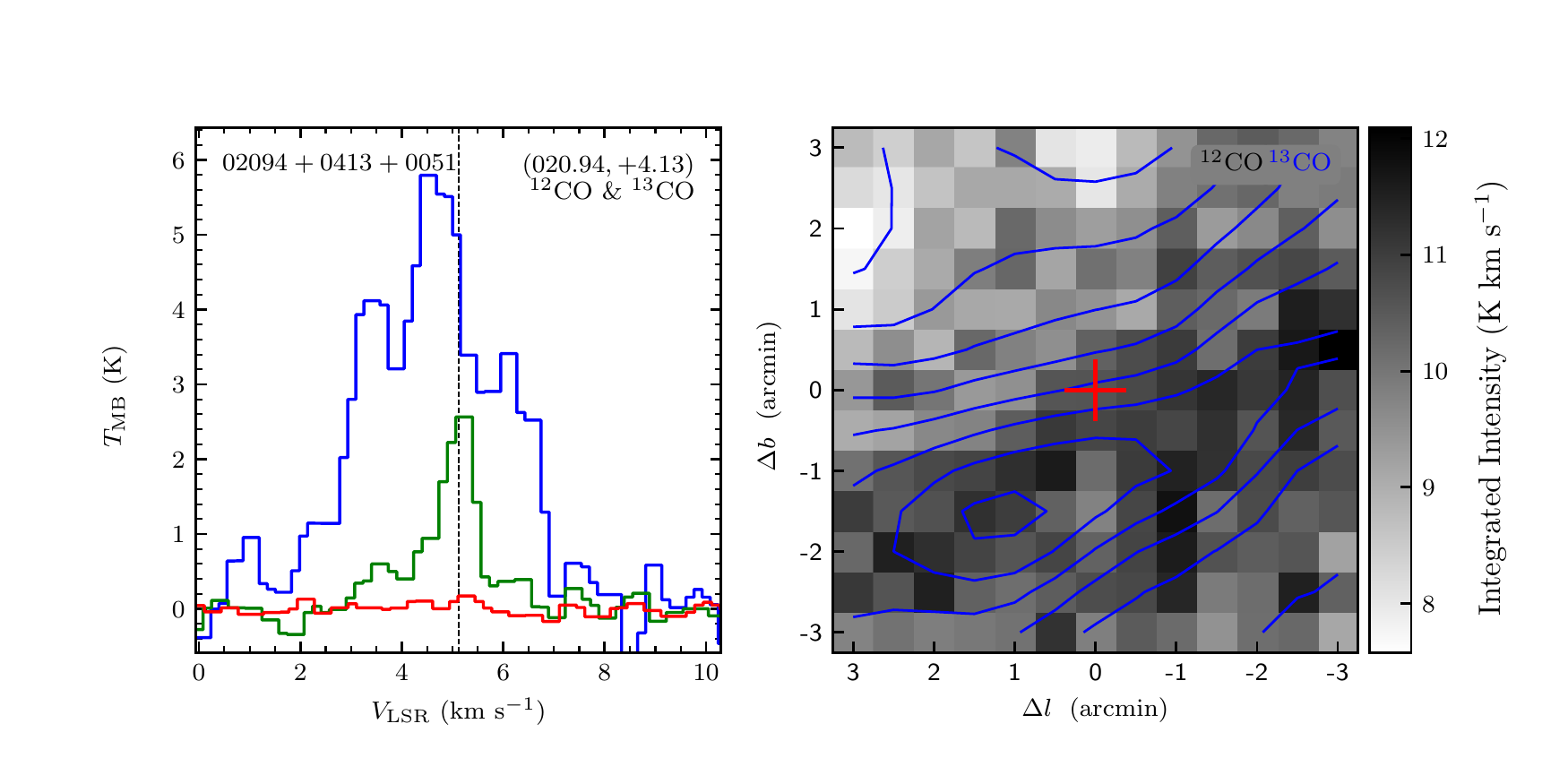}
\vspace{-0.5cm}

\includegraphics[width=9.0cm,angle=0]{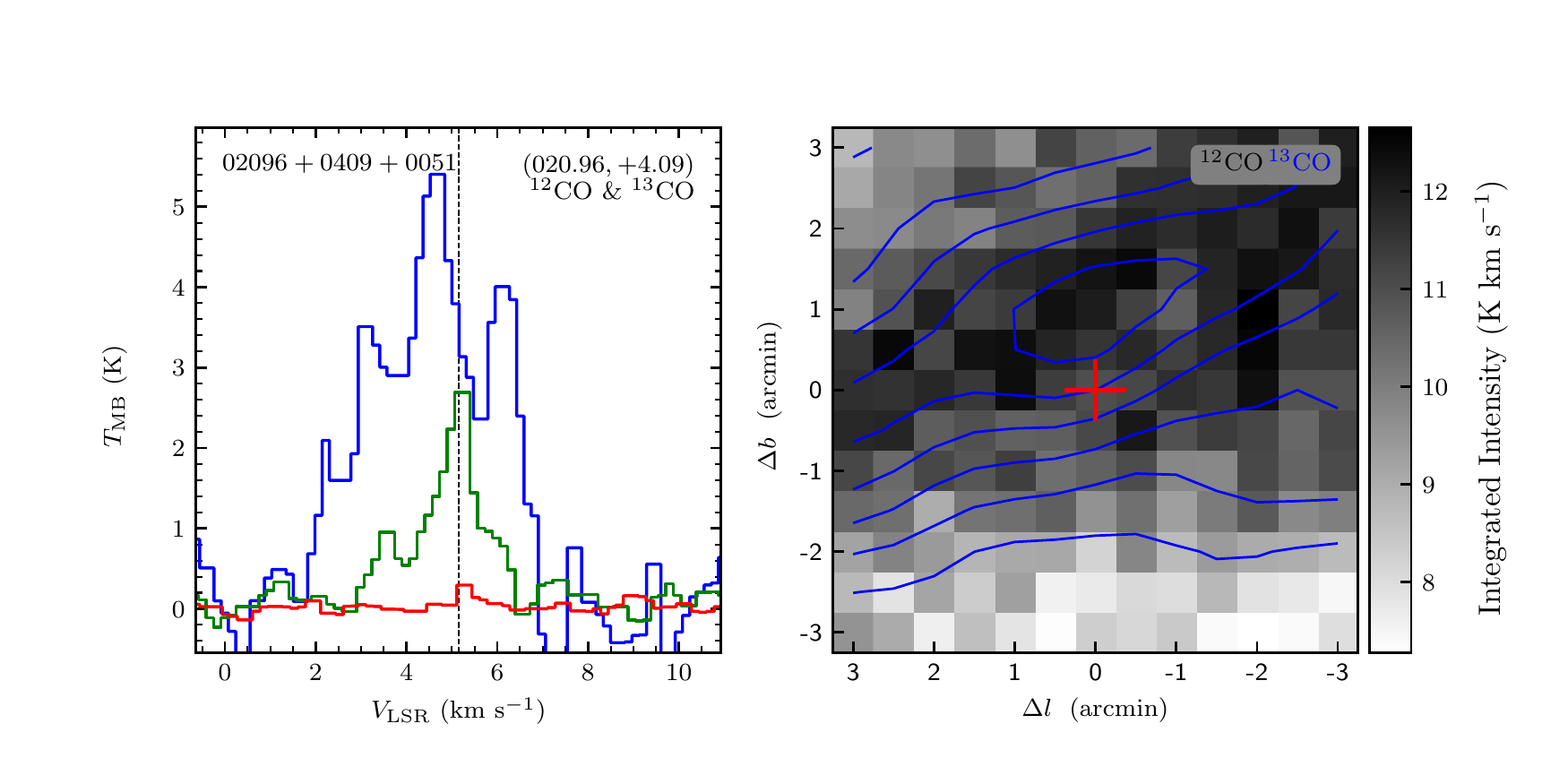}
\includegraphics[width=9.0cm,angle=0]{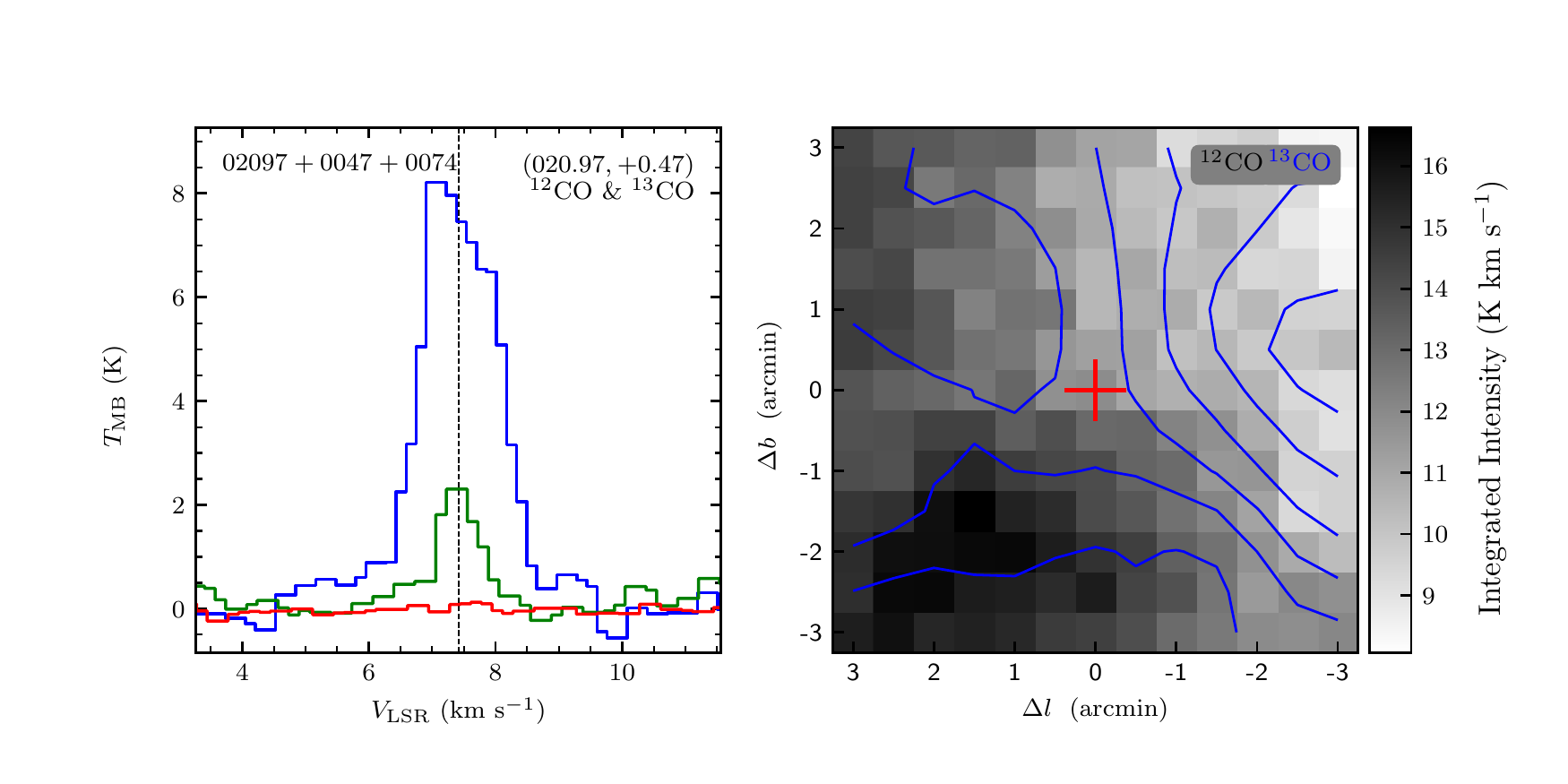}
\vspace{-0.5cm}

\includegraphics[width=9.0cm,angle=0]{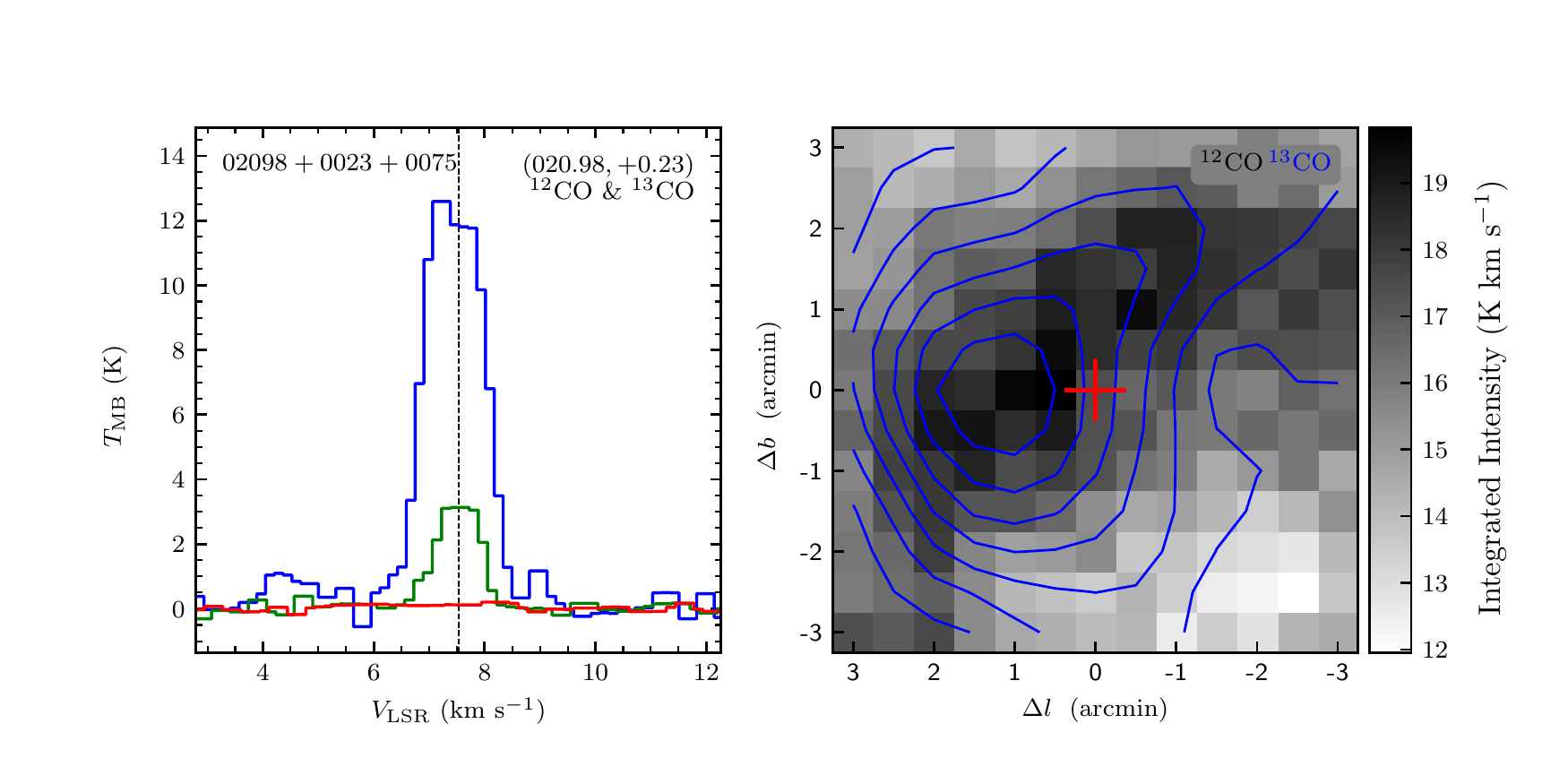}
\includegraphics[width=9.0cm,angle=0]{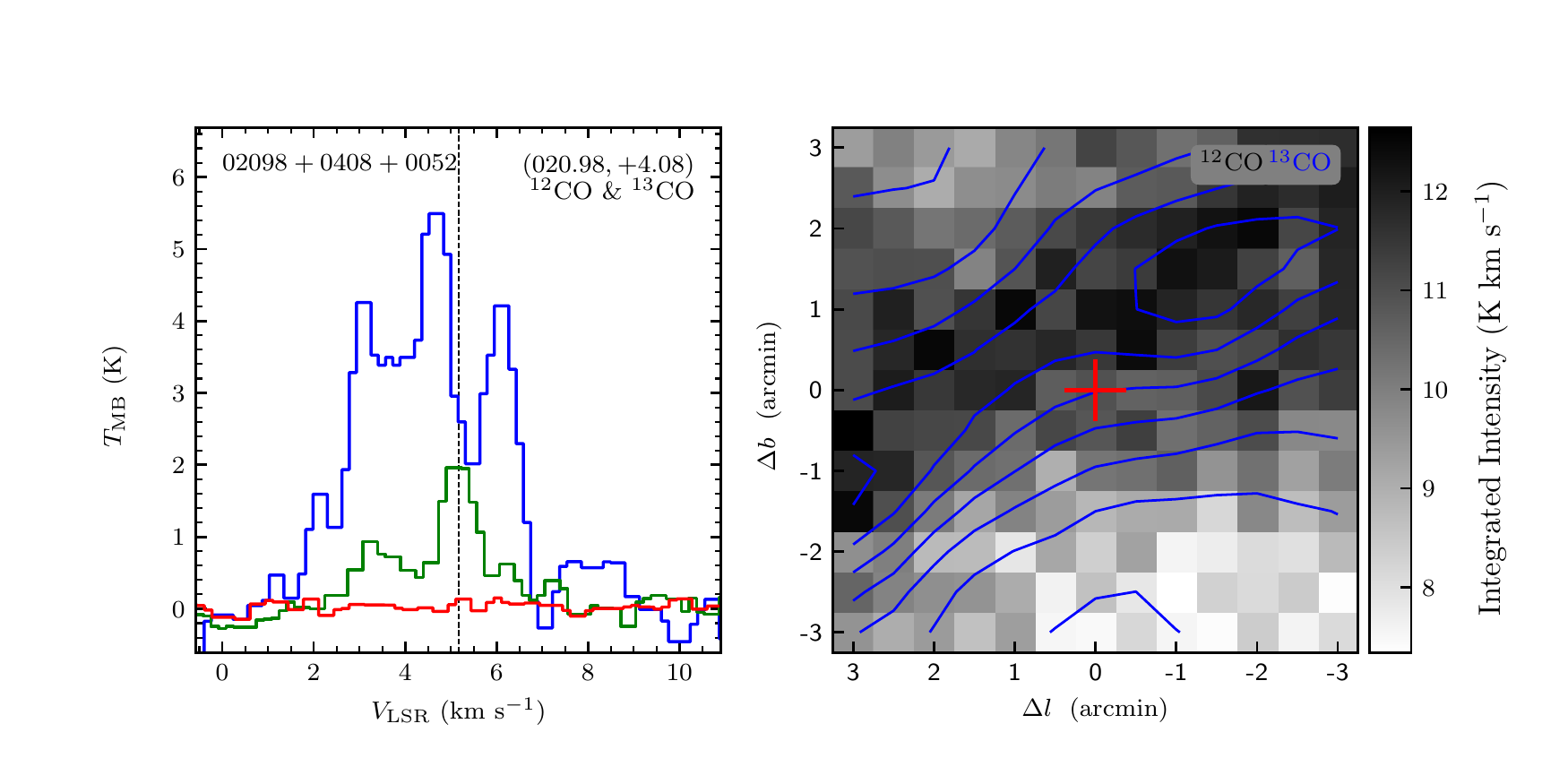}
\vspace{-0.5cm}

\includegraphics[width=9.0cm,angle=0]{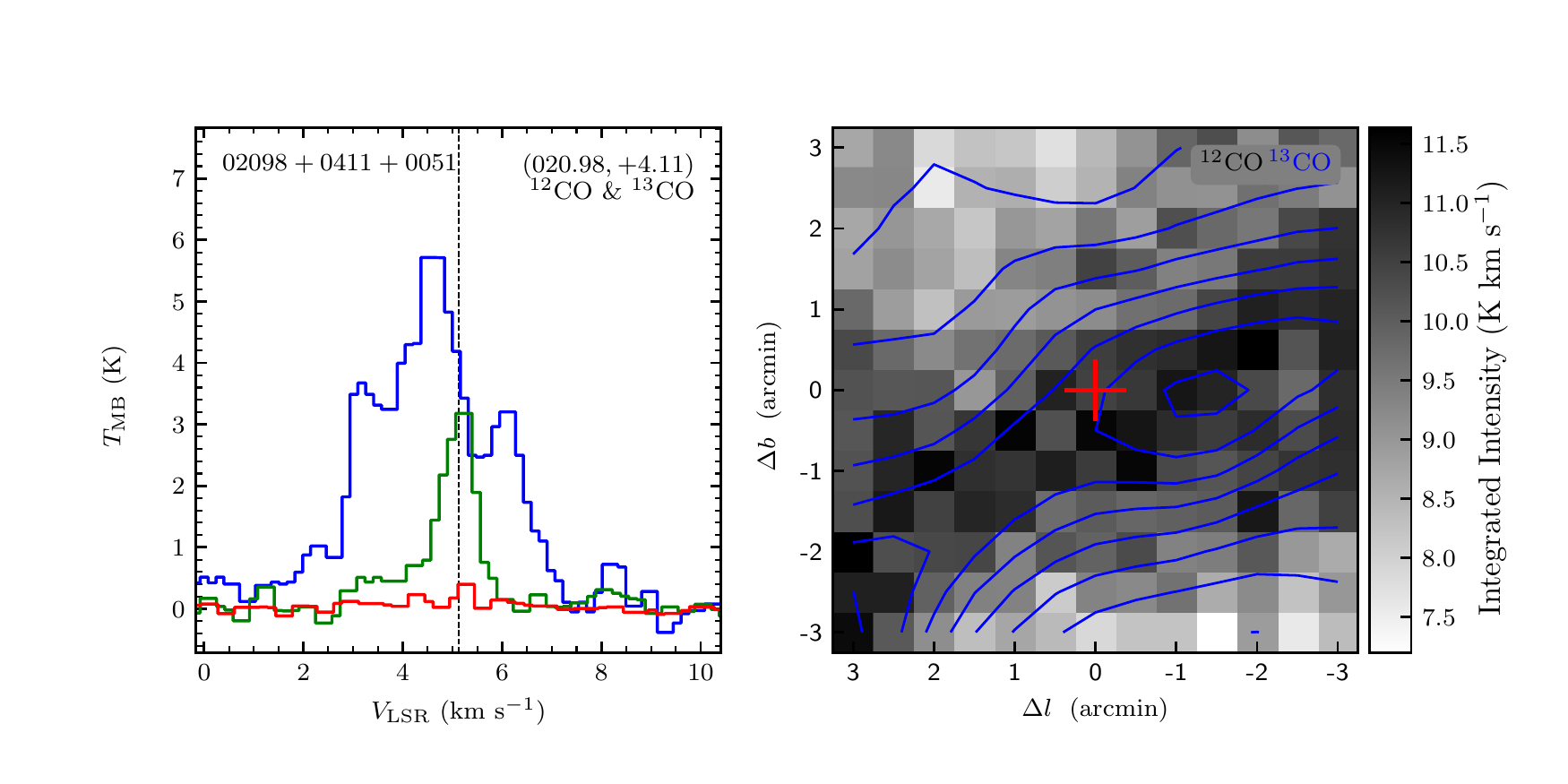}
\includegraphics[width=9.0cm,angle=0]{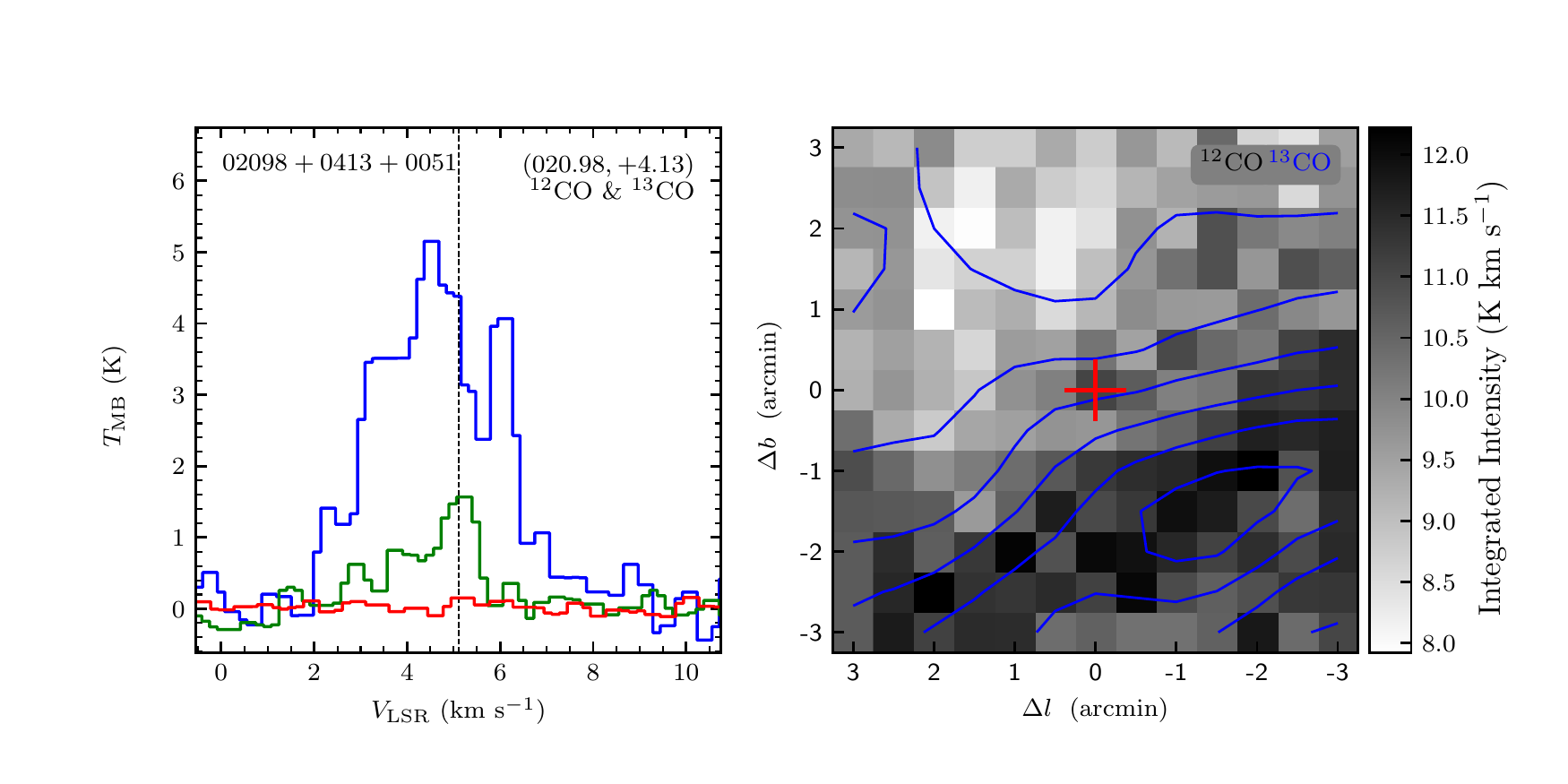}
\end{figure}
\clearpage

\begin{figure}
\includegraphics[width=9.0cm,angle=0]{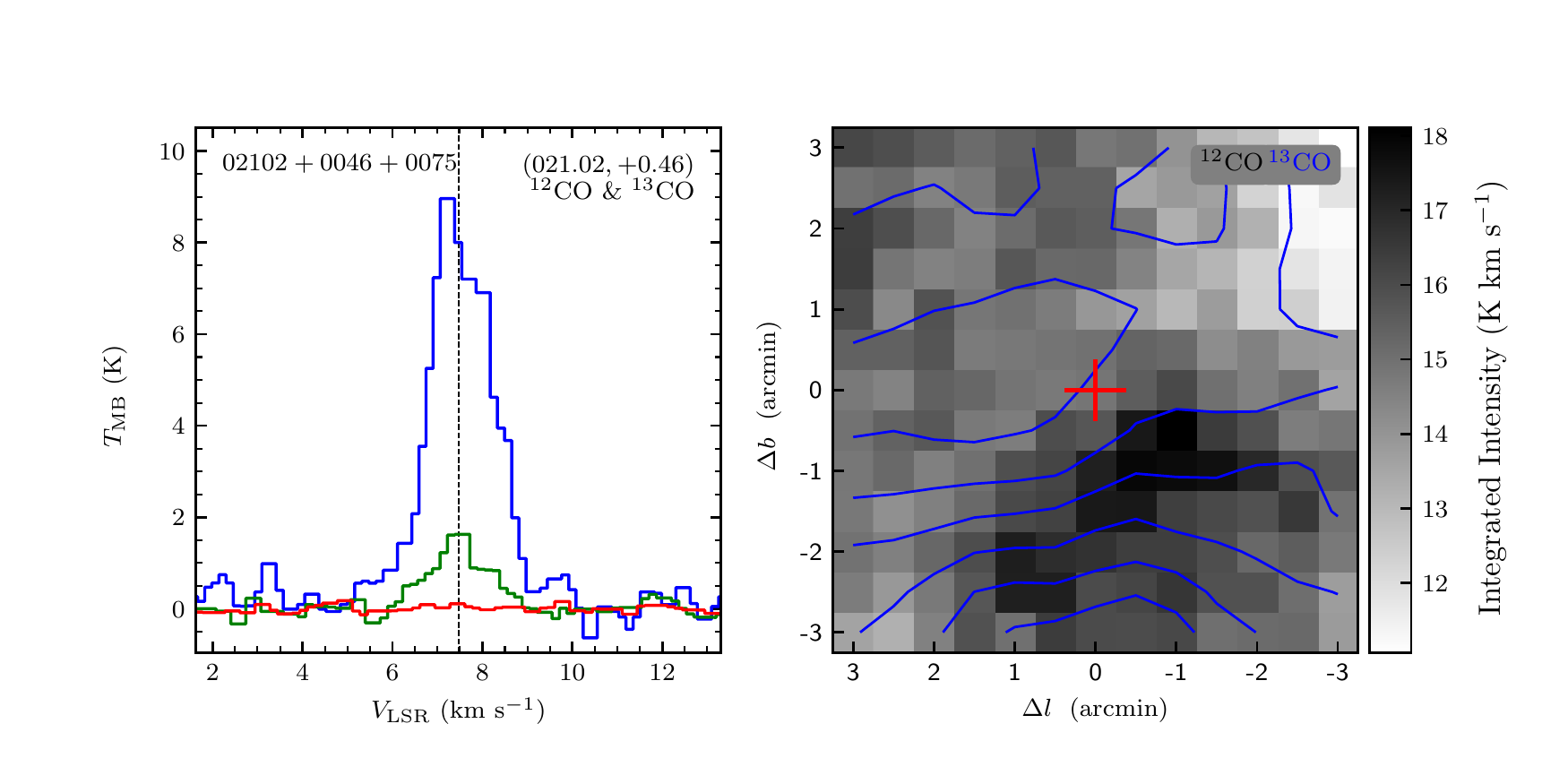}
\includegraphics[width=9.0cm,angle=0]{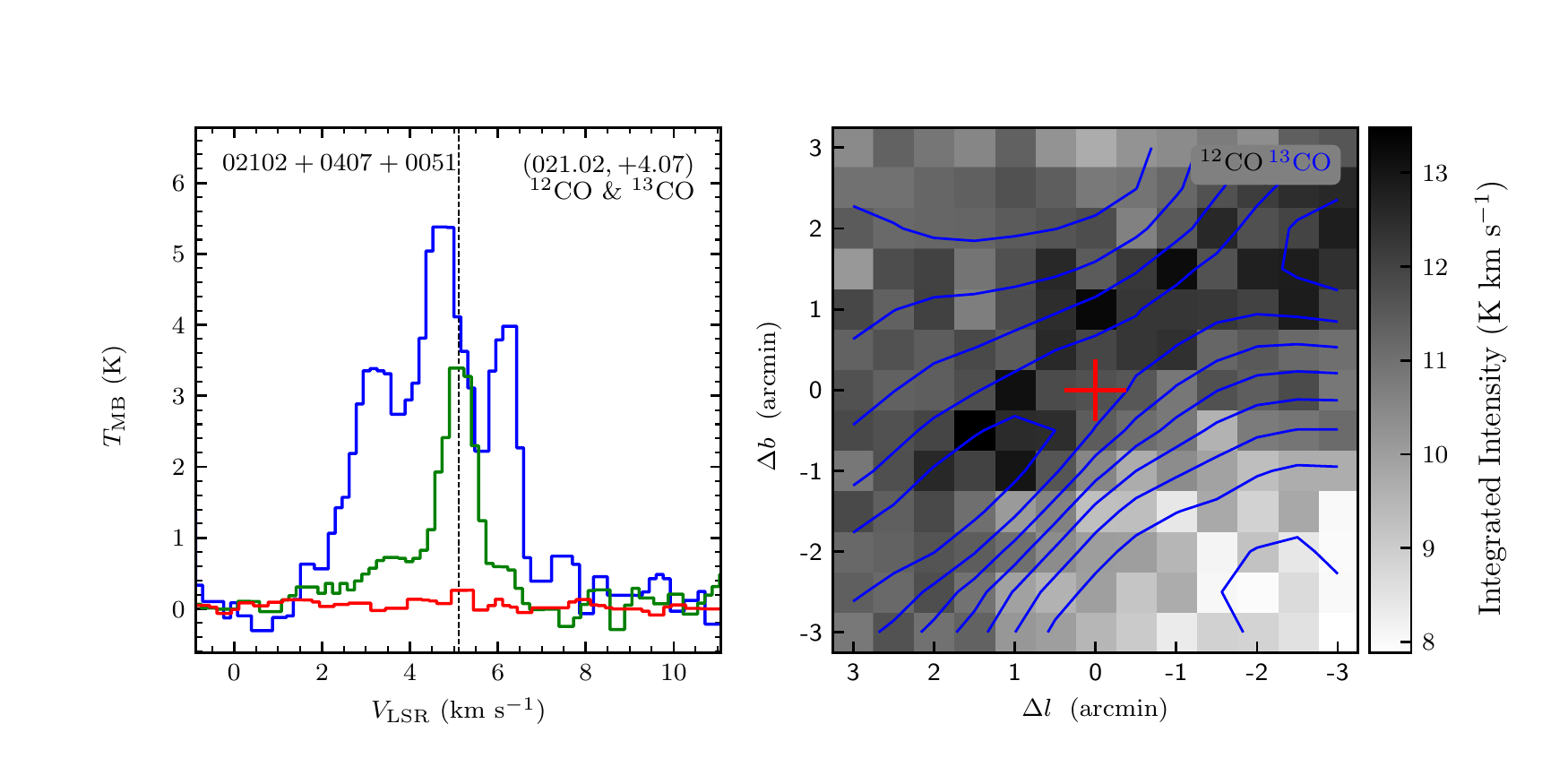}
\vspace{-0.5cm}

\includegraphics[width=9.0cm,angle=0]{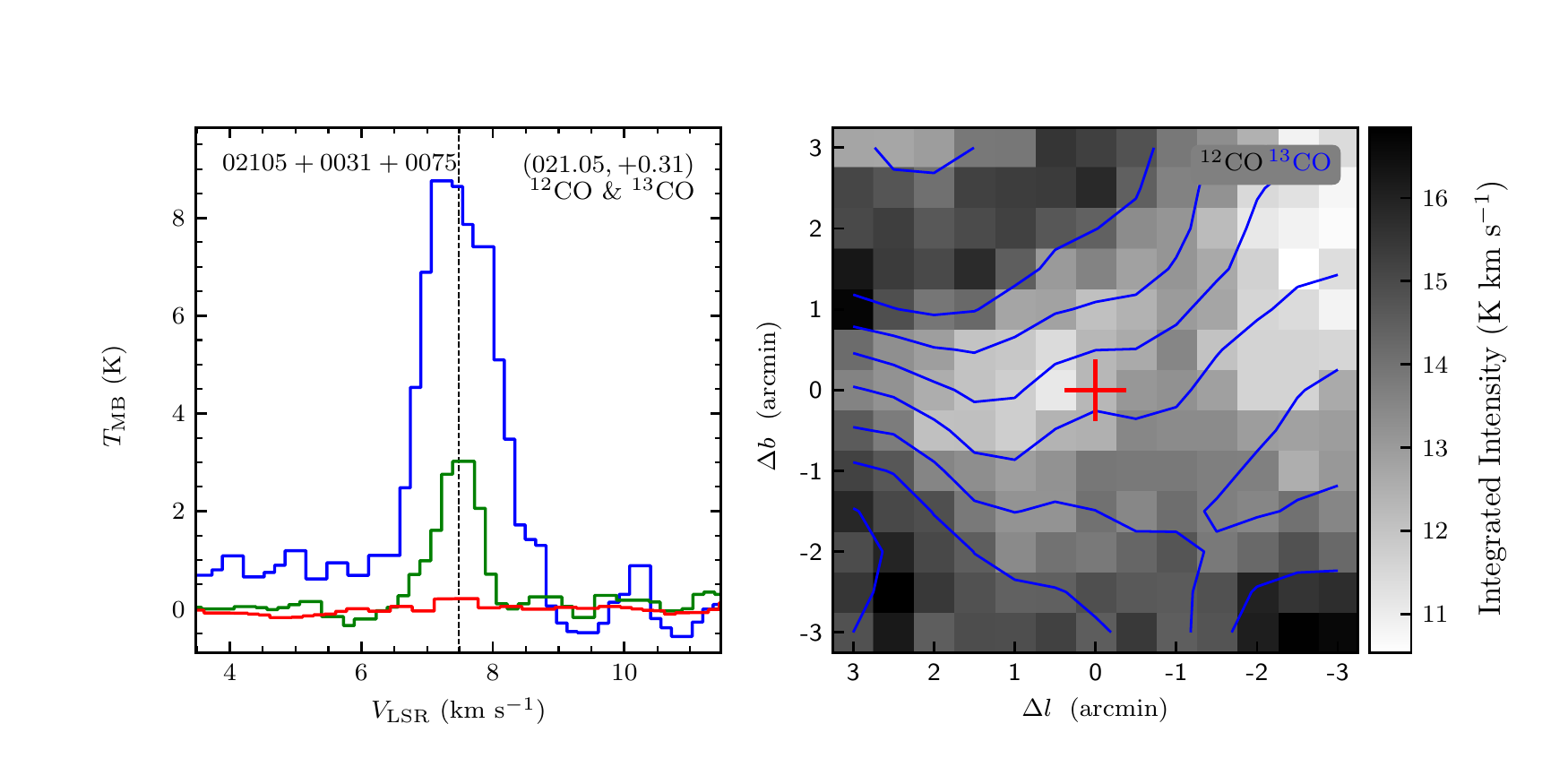}
\includegraphics[width=9.0cm,angle=0]{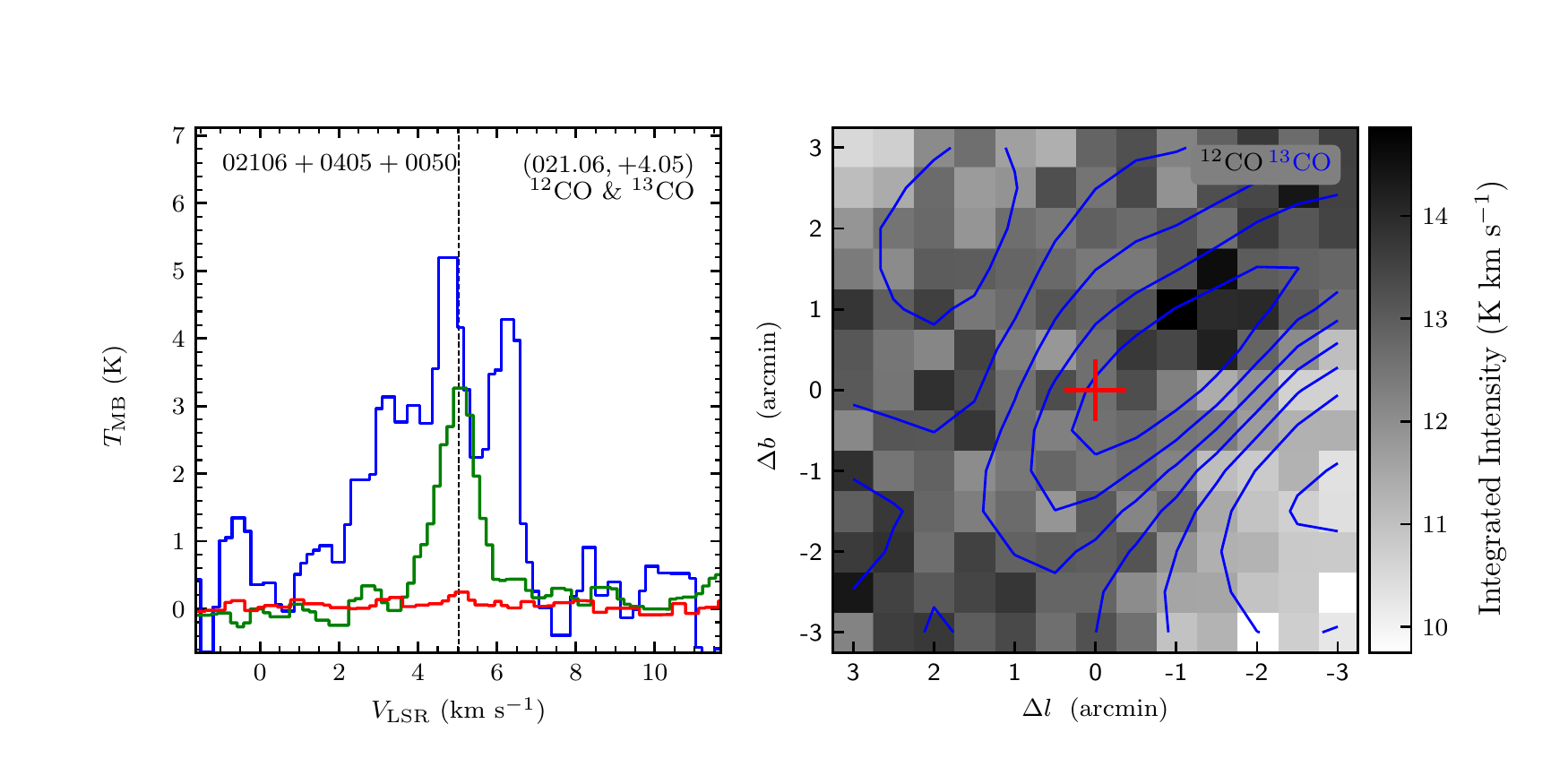}
\vspace{-0.5cm}

\includegraphics[width=9.0cm,angle=0]{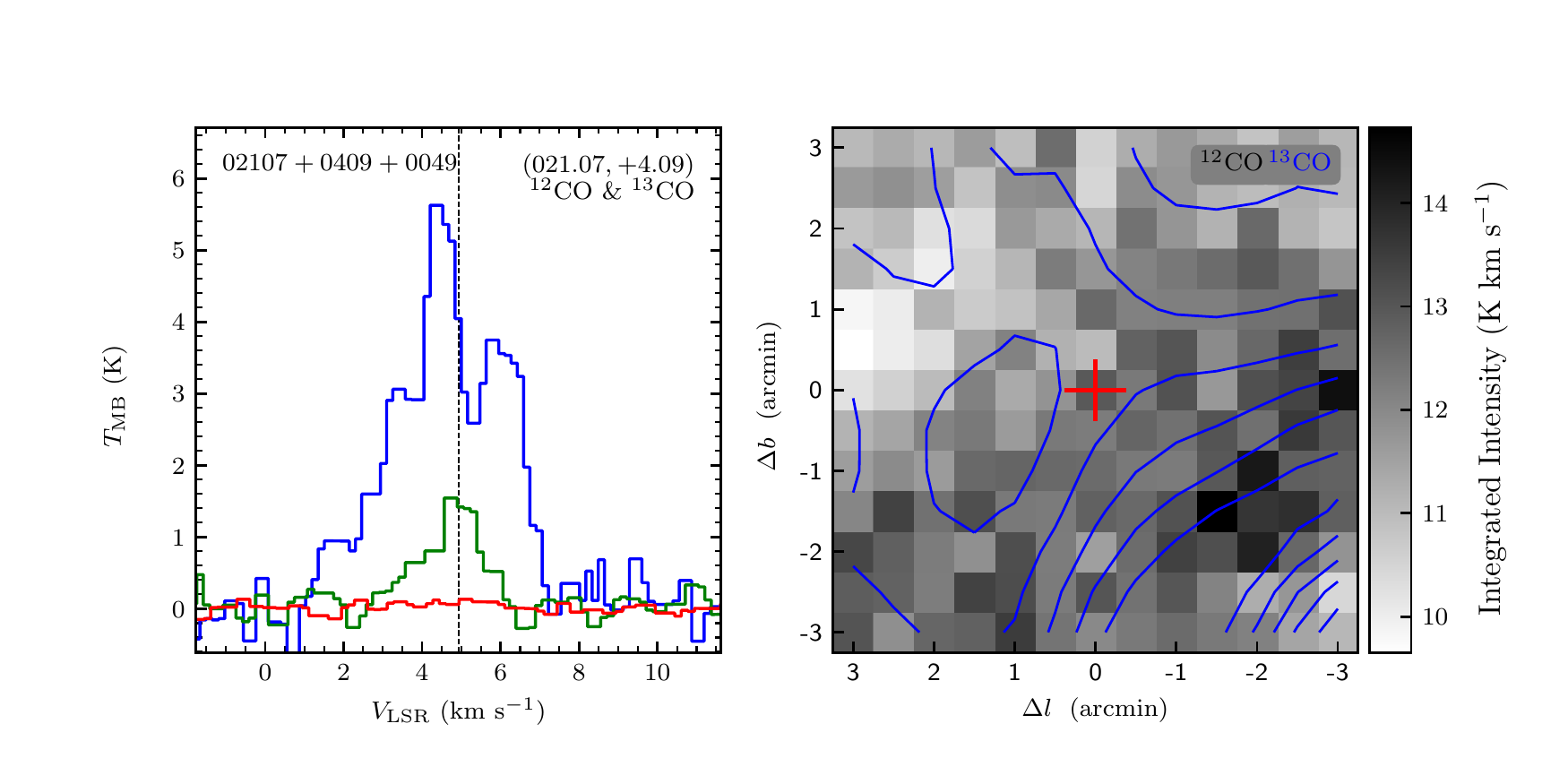}
\includegraphics[width=9.0cm,angle=0]{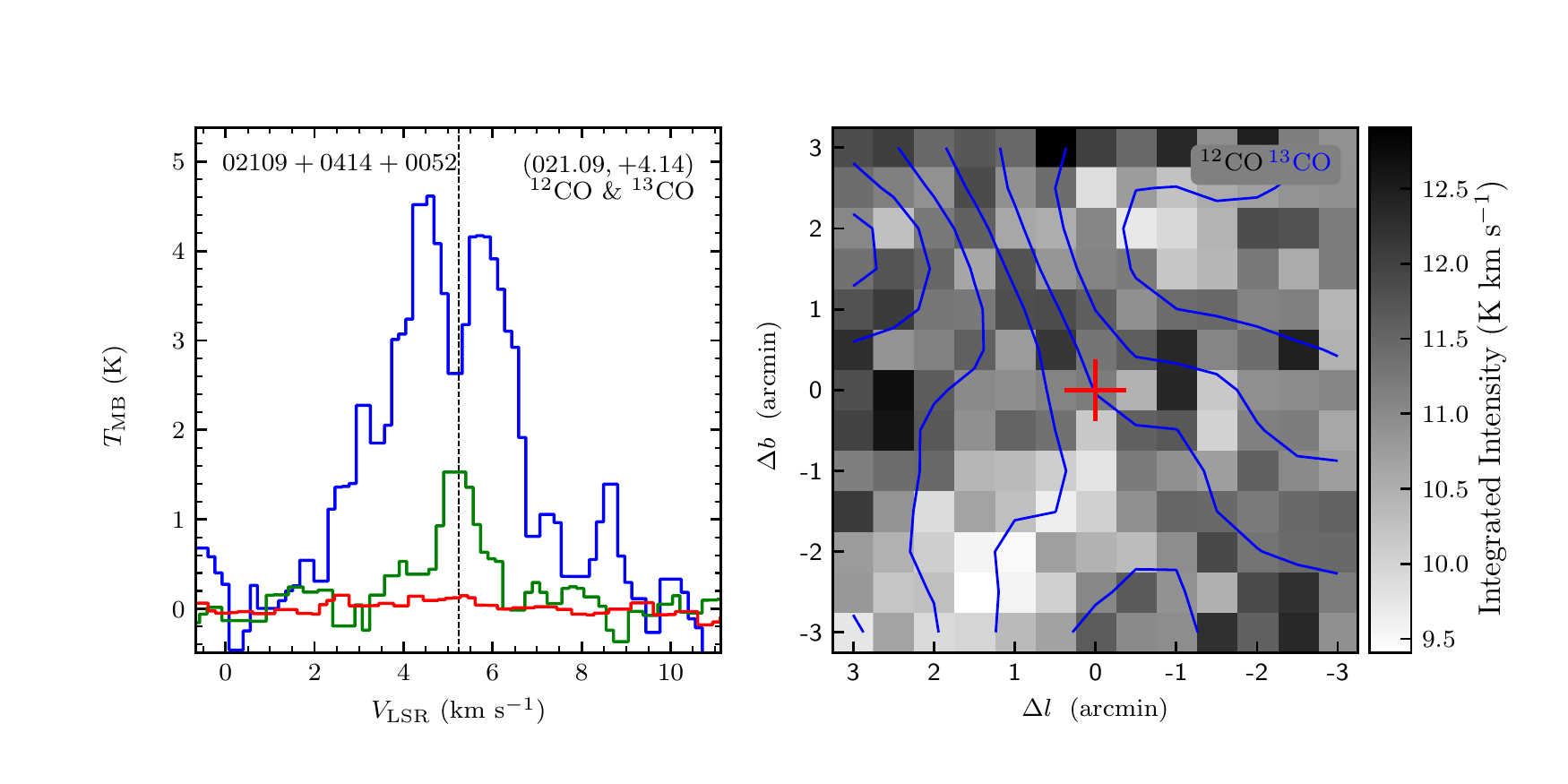}
\vspace{-0.5cm}

\includegraphics[width=9.0cm,angle=0]{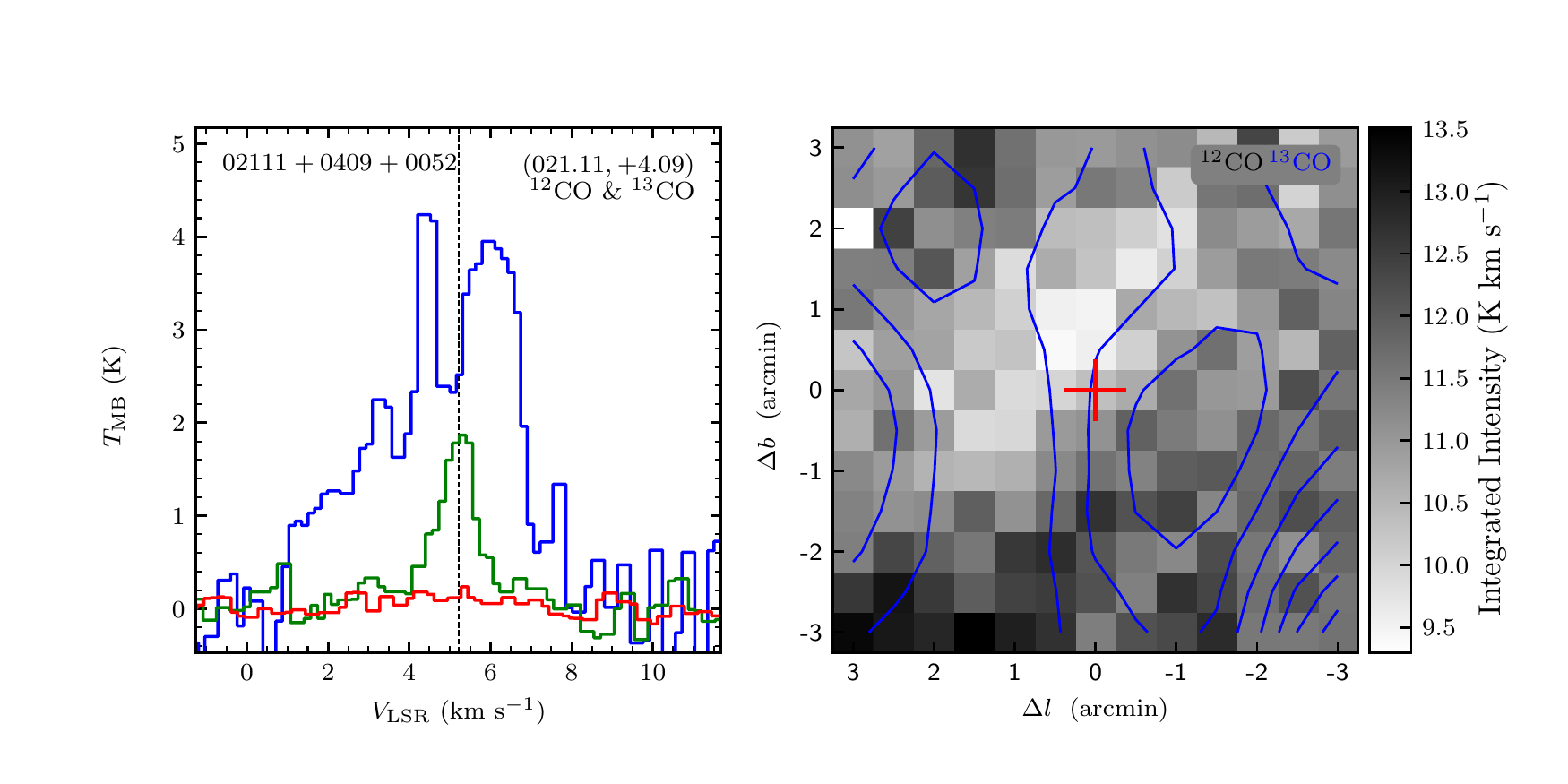}
\includegraphics[width=9.0cm,angle=0]{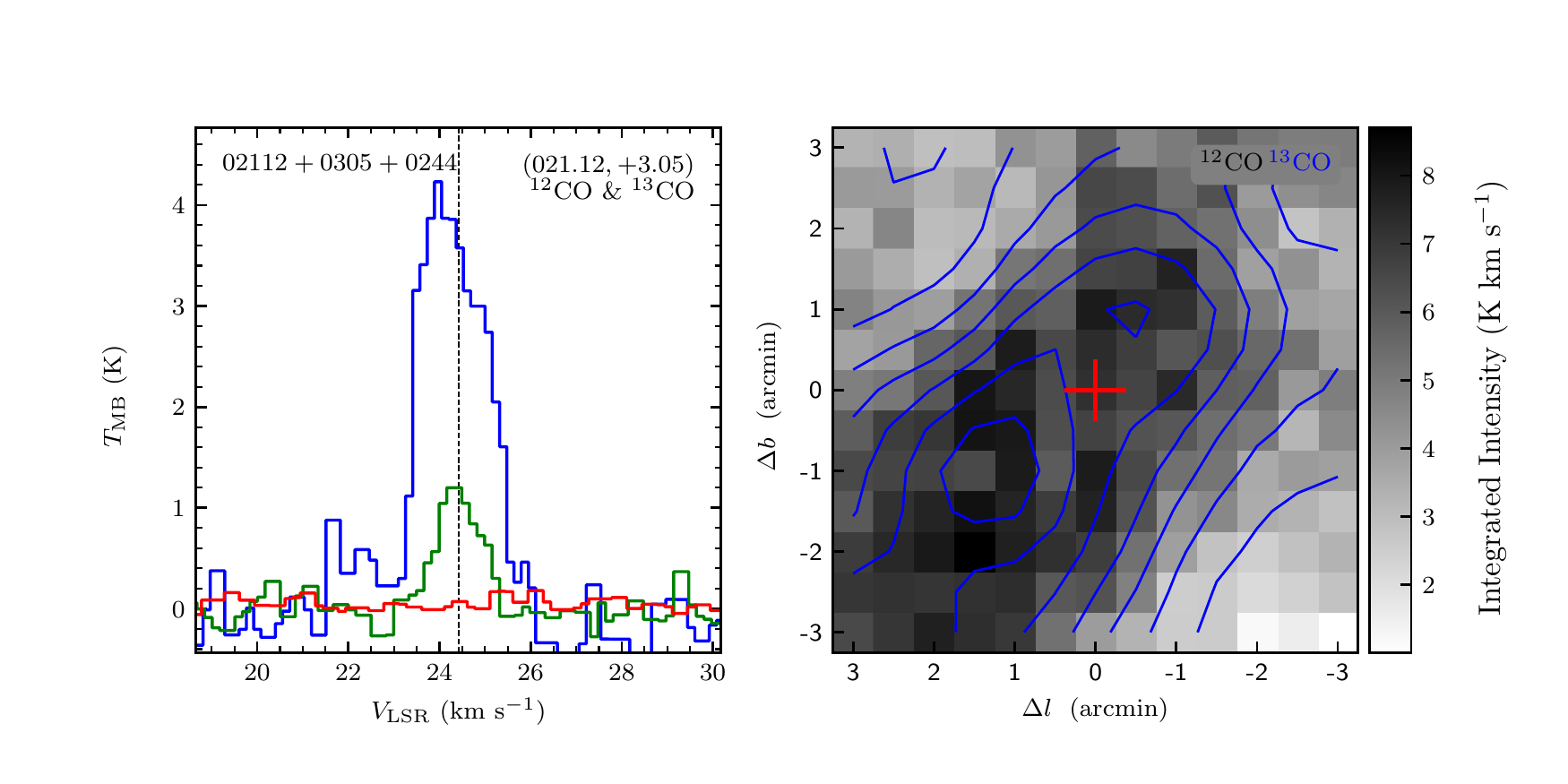}
\vspace{-0.5cm}

\includegraphics[width=9.0cm,angle=0]{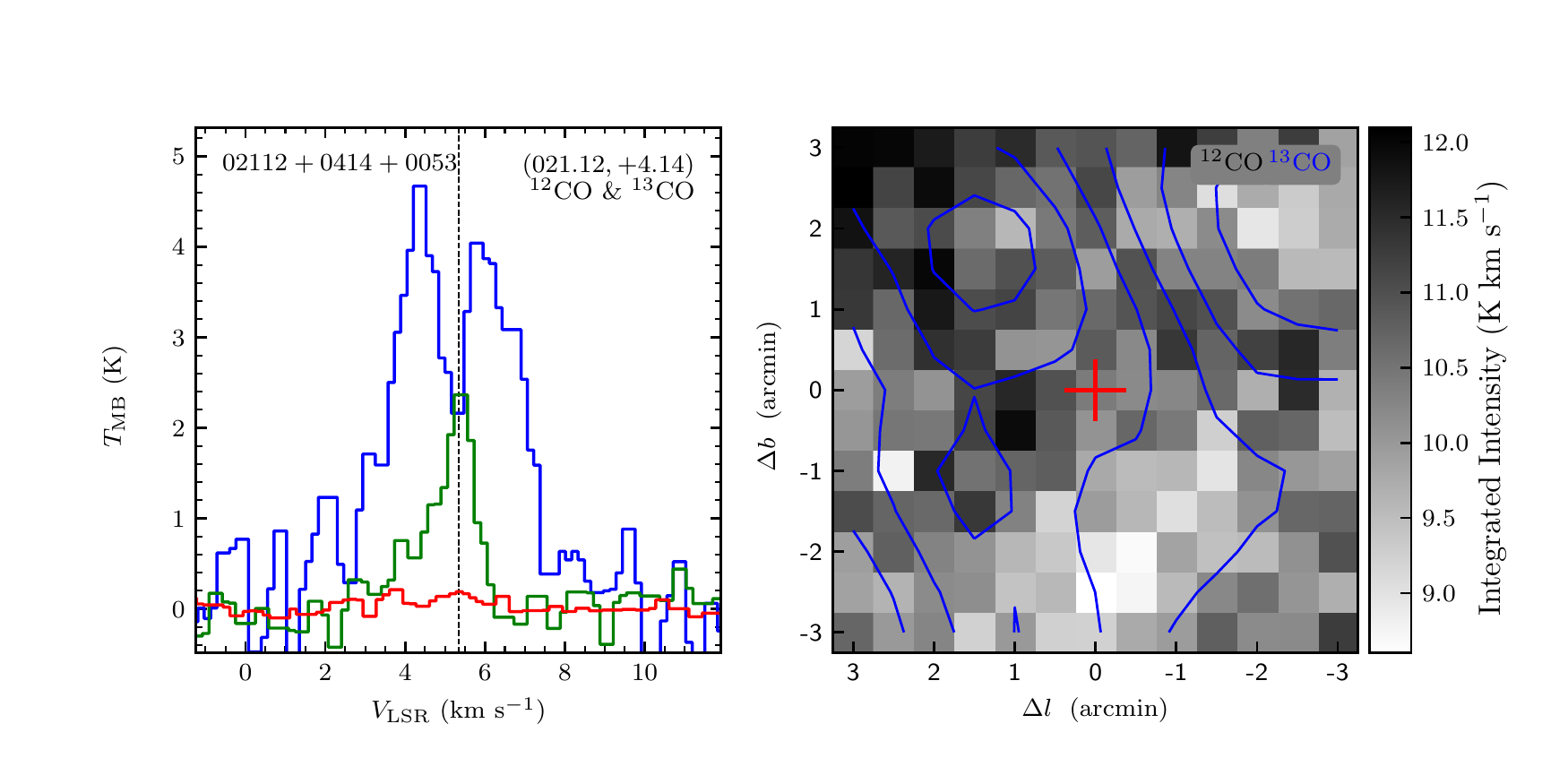}
\includegraphics[width=9.0cm,angle=0]{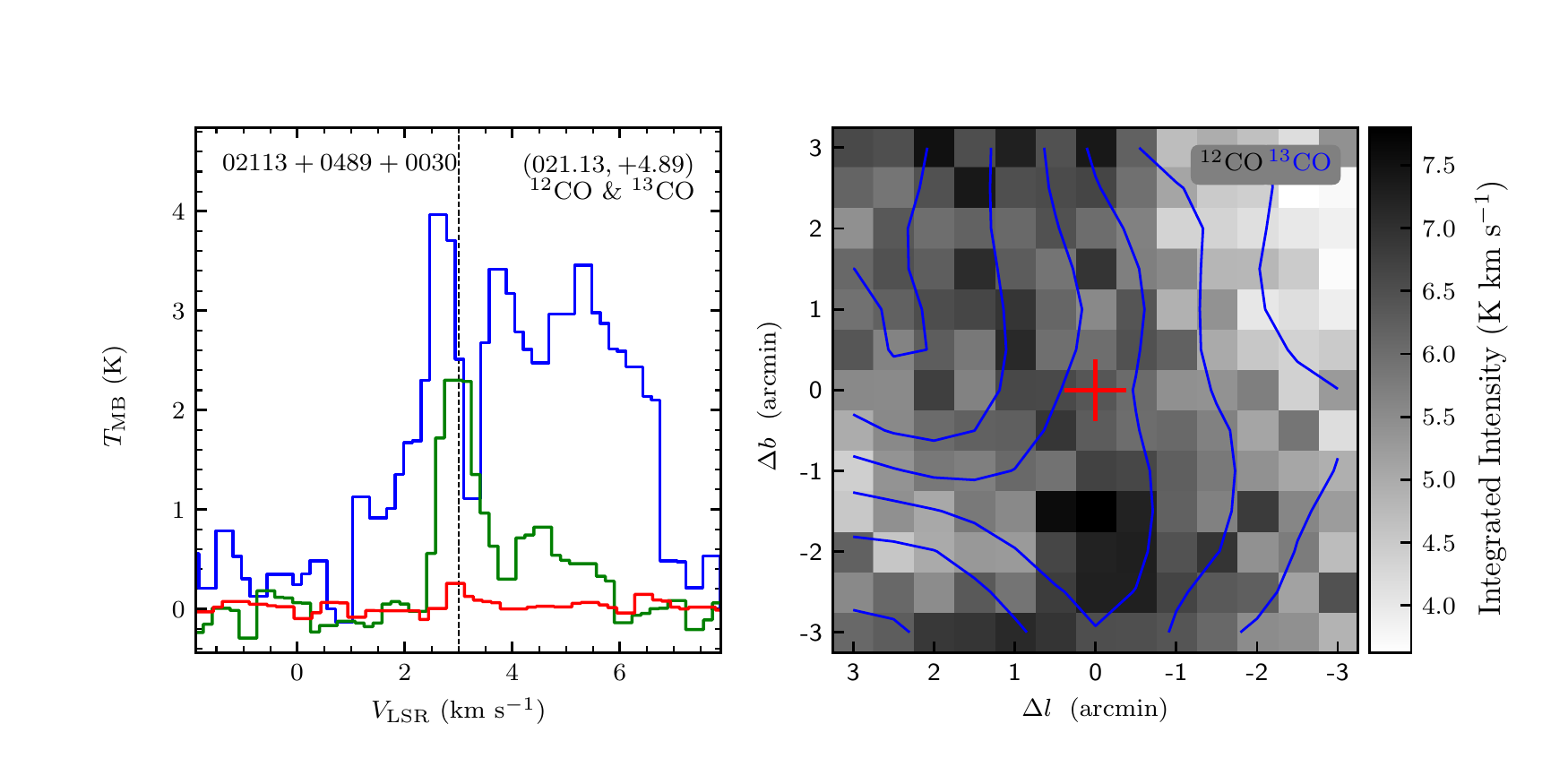}
\end{figure}
\clearpage

\begin{figure}
\includegraphics[width=9.0cm,angle=0]{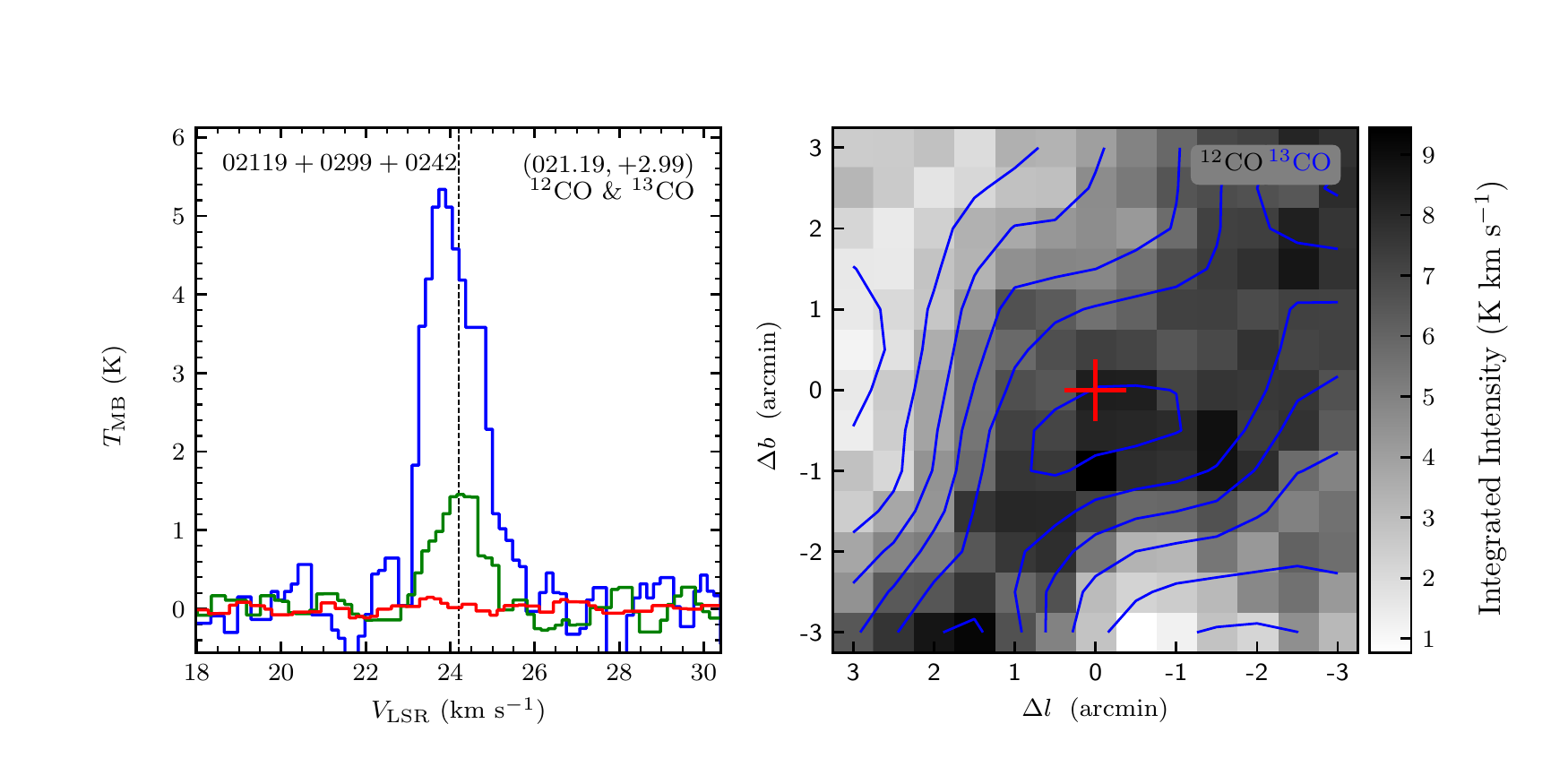}
\includegraphics[width=9.0cm,angle=0]{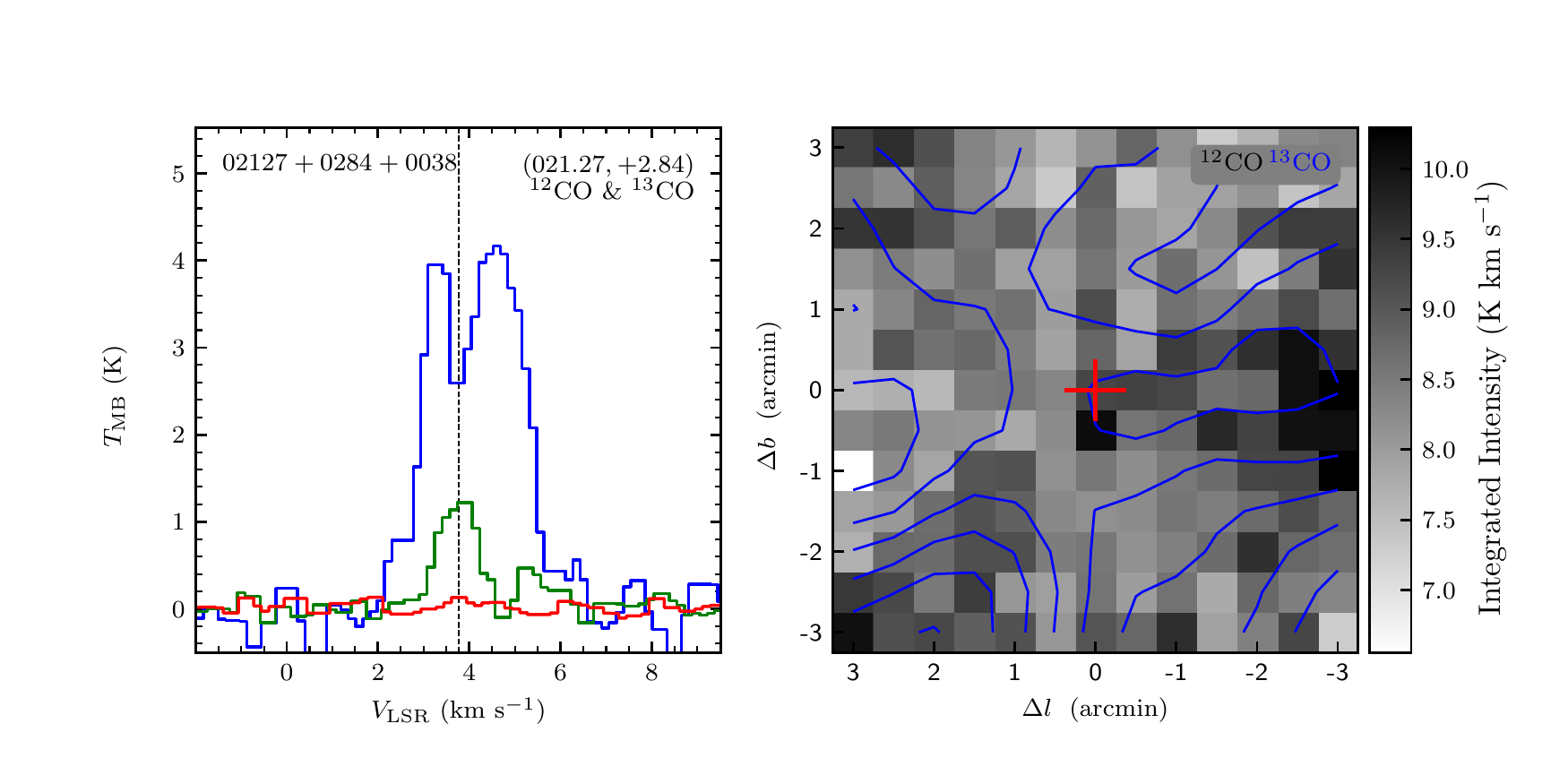}
\vspace{-0.5cm}

\includegraphics[width=9.0cm,angle=0]{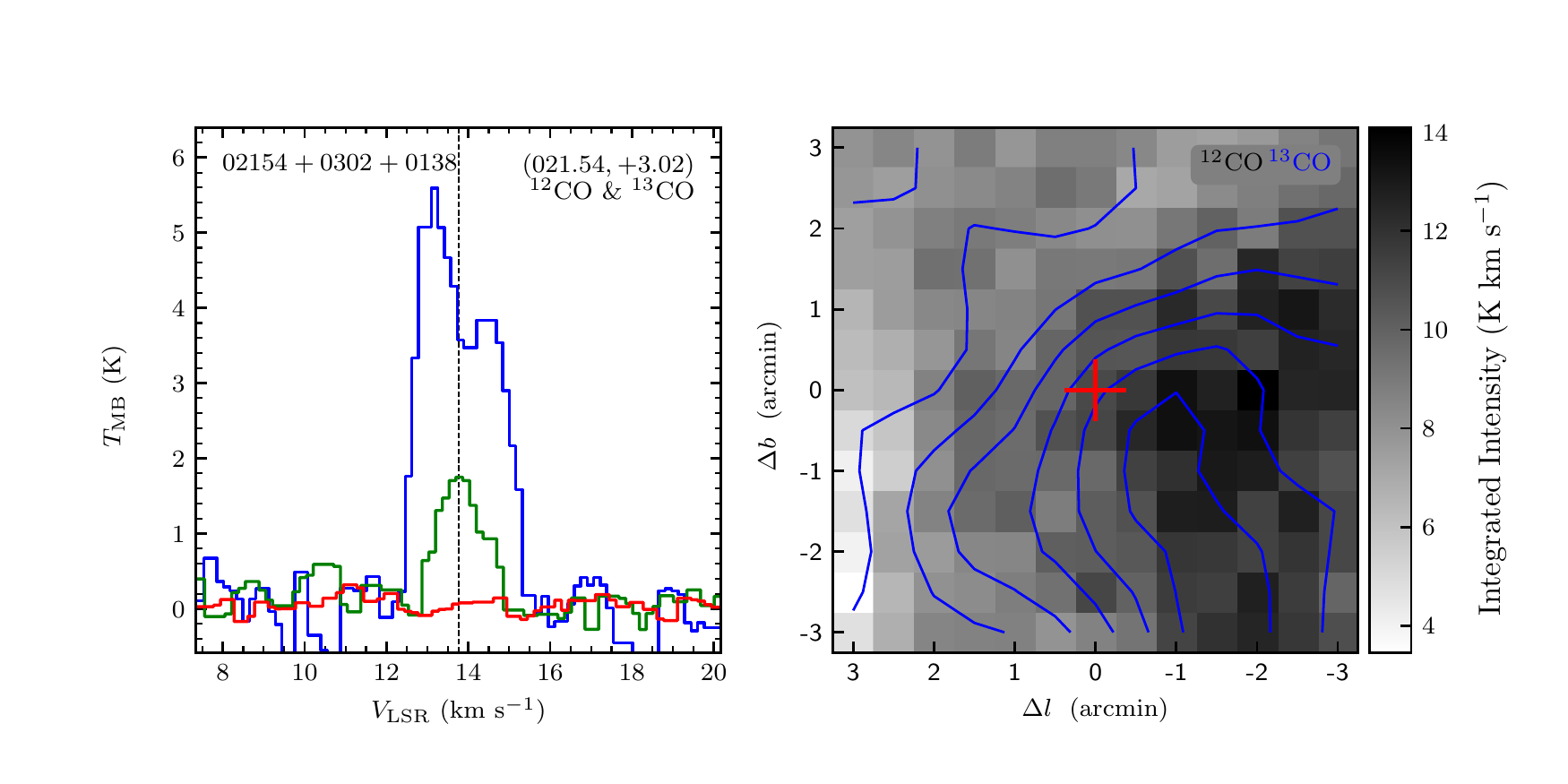}
\includegraphics[width=9.0cm,angle=0]{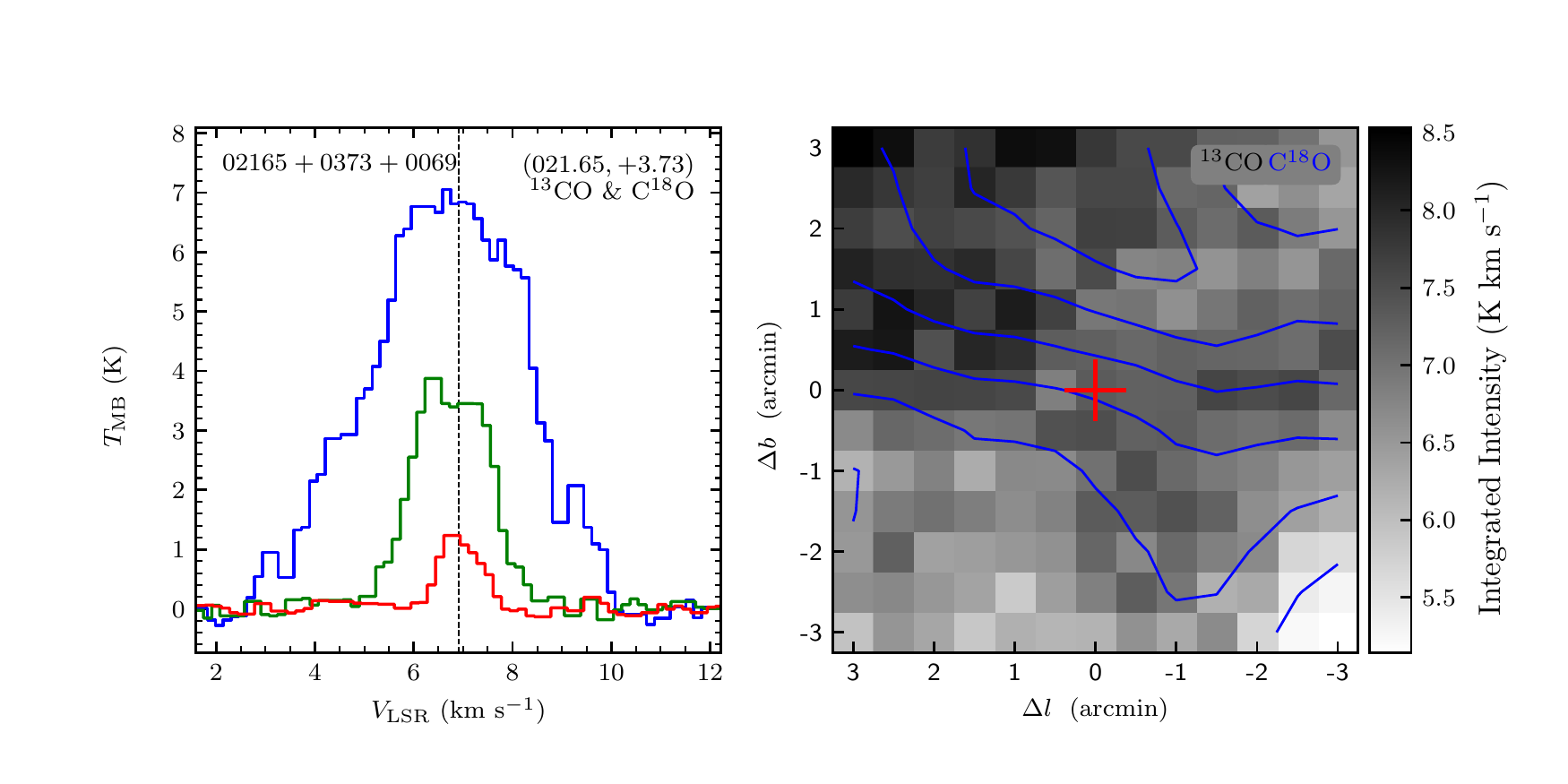}
\vspace{-0.5cm}

\includegraphics[width=9.0cm,angle=0]{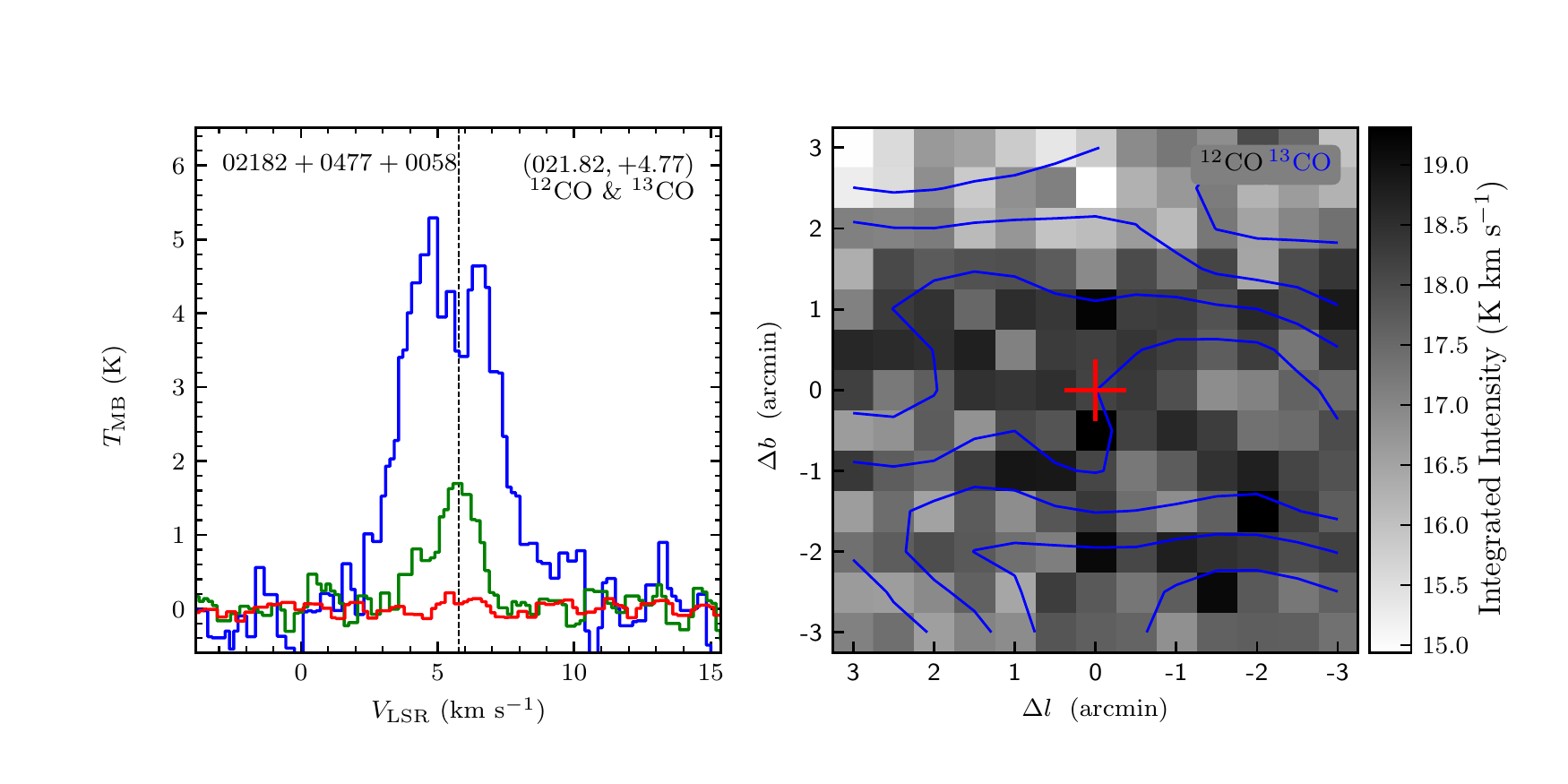}
\includegraphics[width=9.0cm,angle=0]{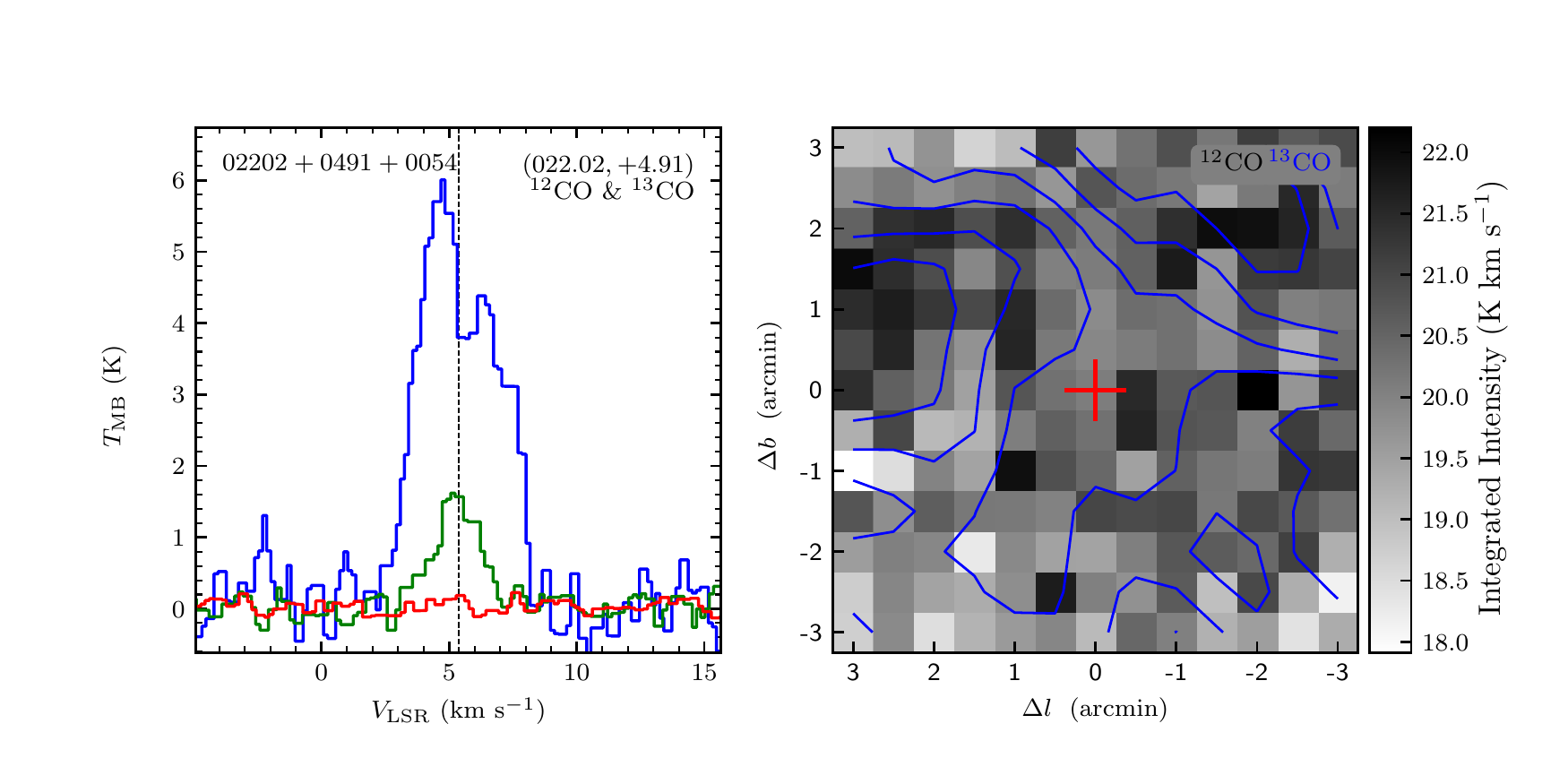}
\vspace{-0.5cm}

\includegraphics[width=9.0cm,angle=0]{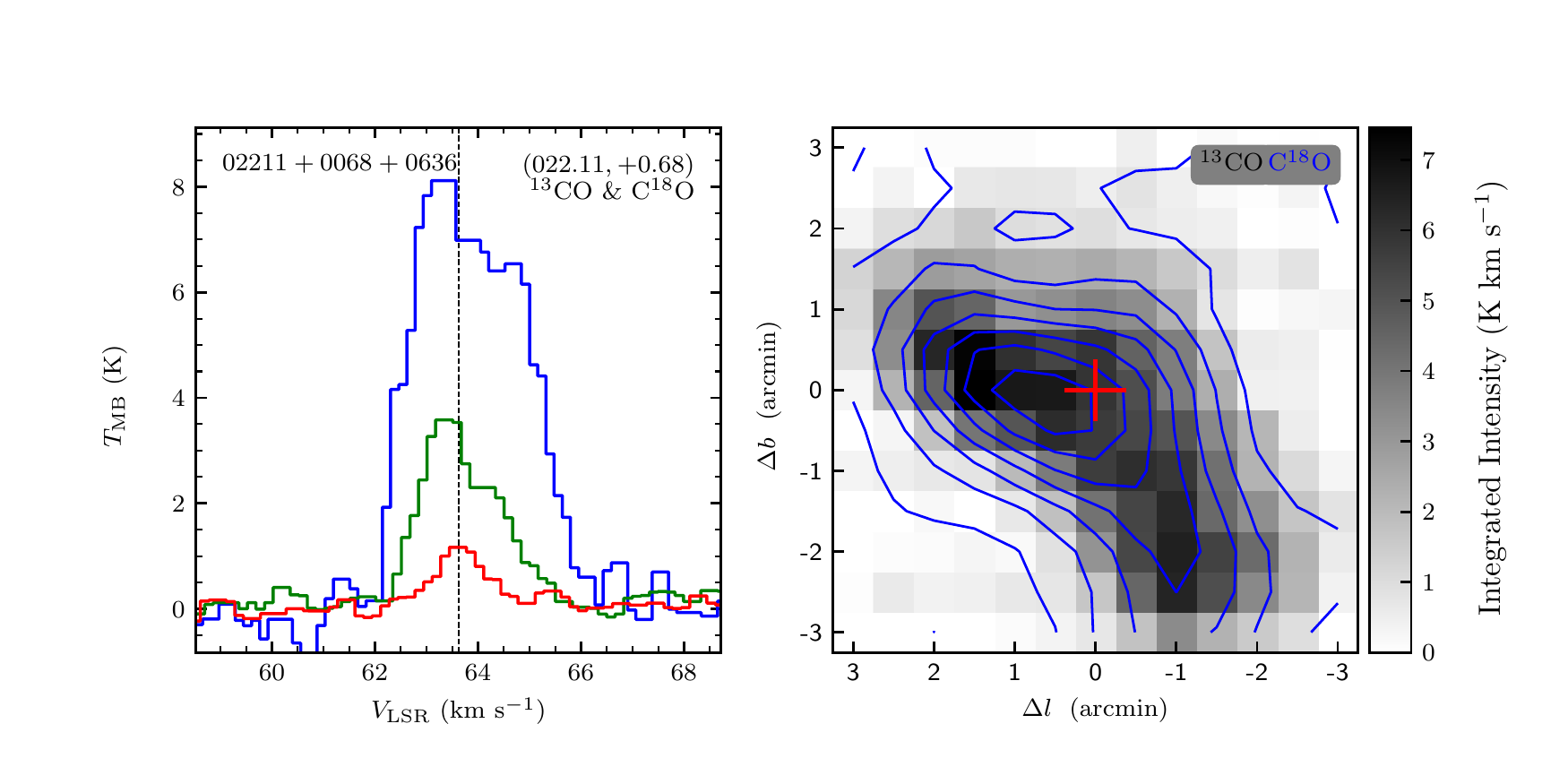}
\includegraphics[width=9.0cm,angle=0]{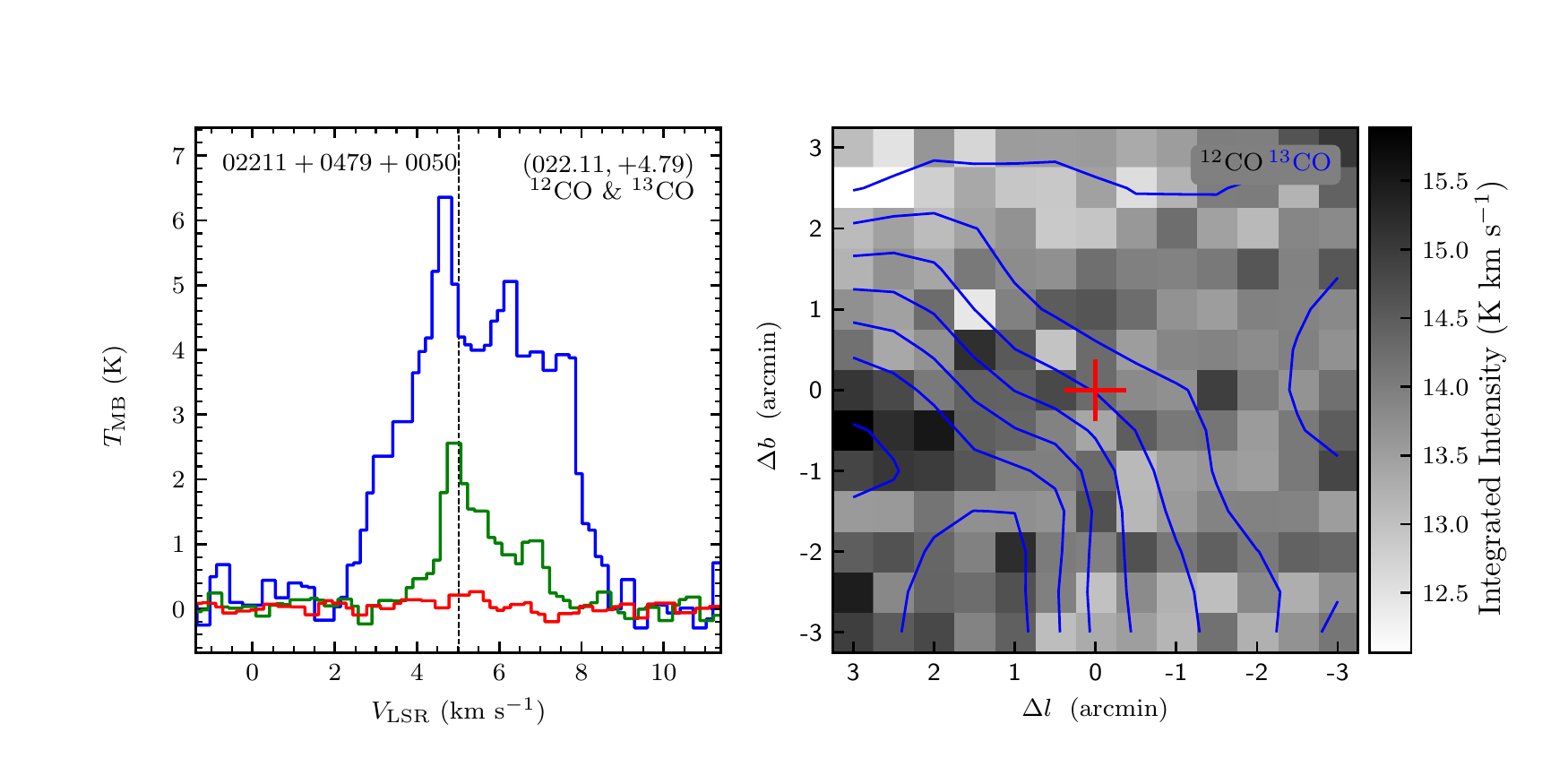}
\vspace{-0.5cm}

\includegraphics[width=9.0cm,angle=0]{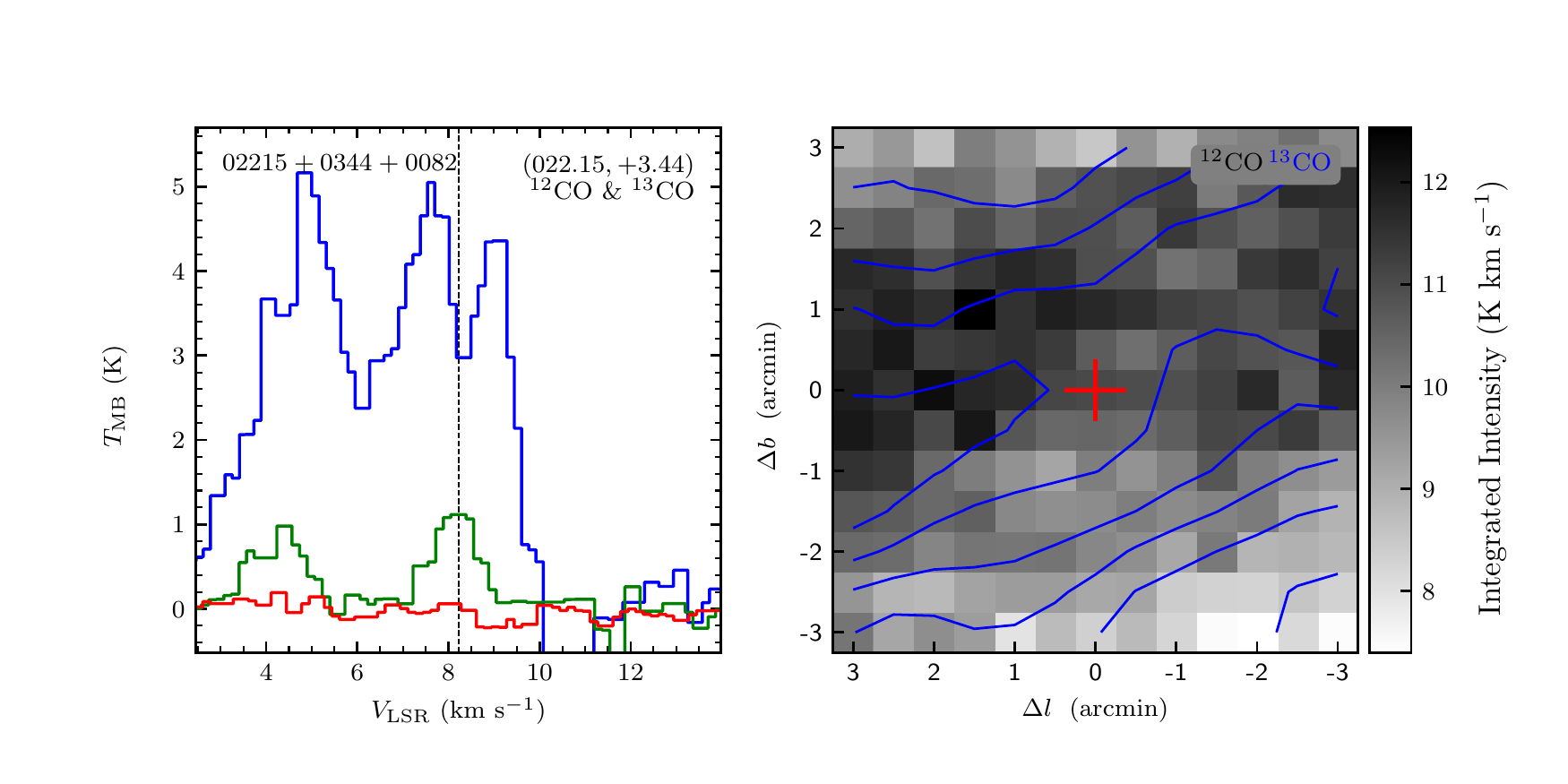}
\includegraphics[width=9.0cm,angle=0]{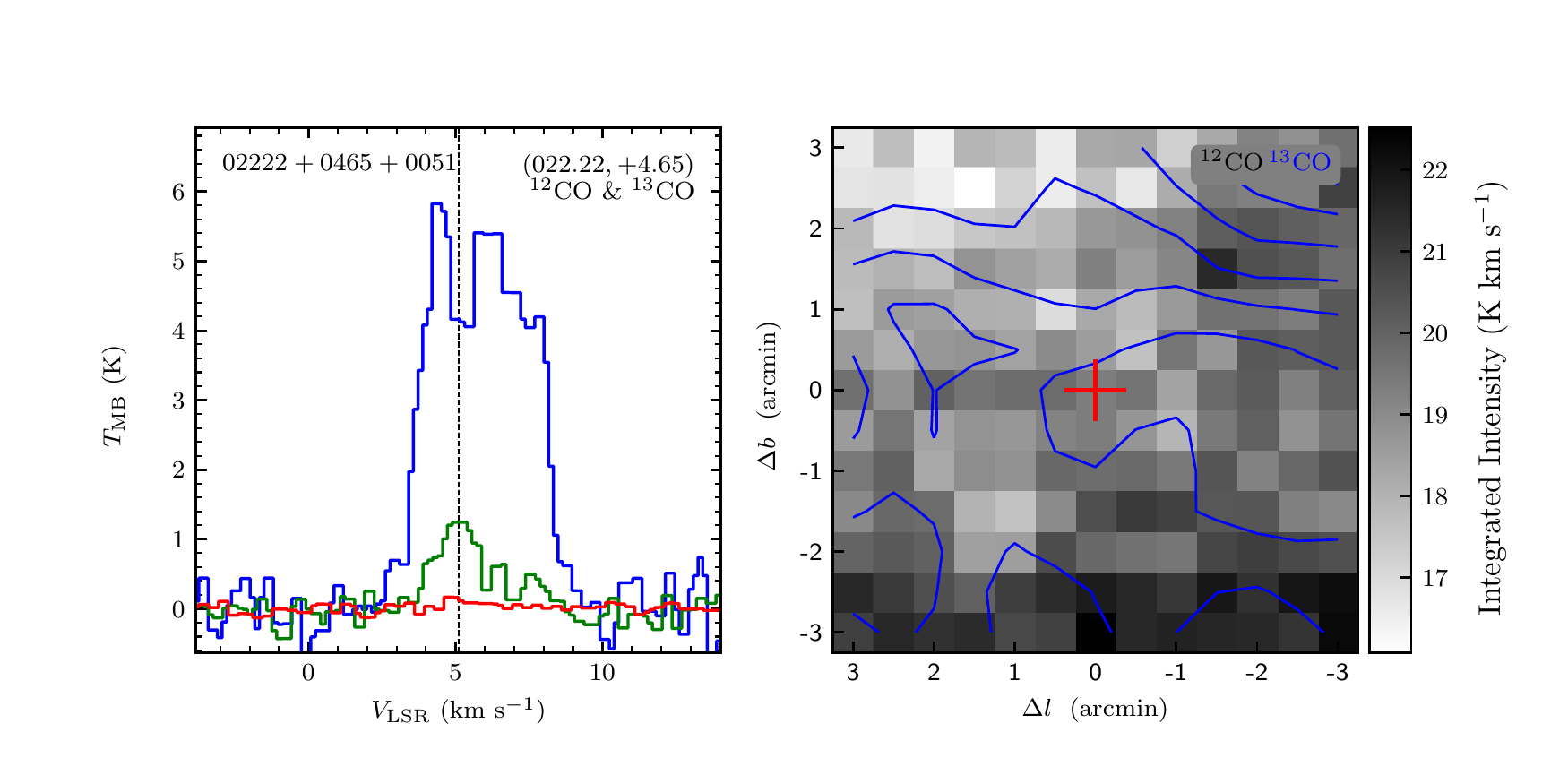}
\end{figure}
\clearpage

\begin{figure}
\includegraphics[width=9.0cm,angle=0]{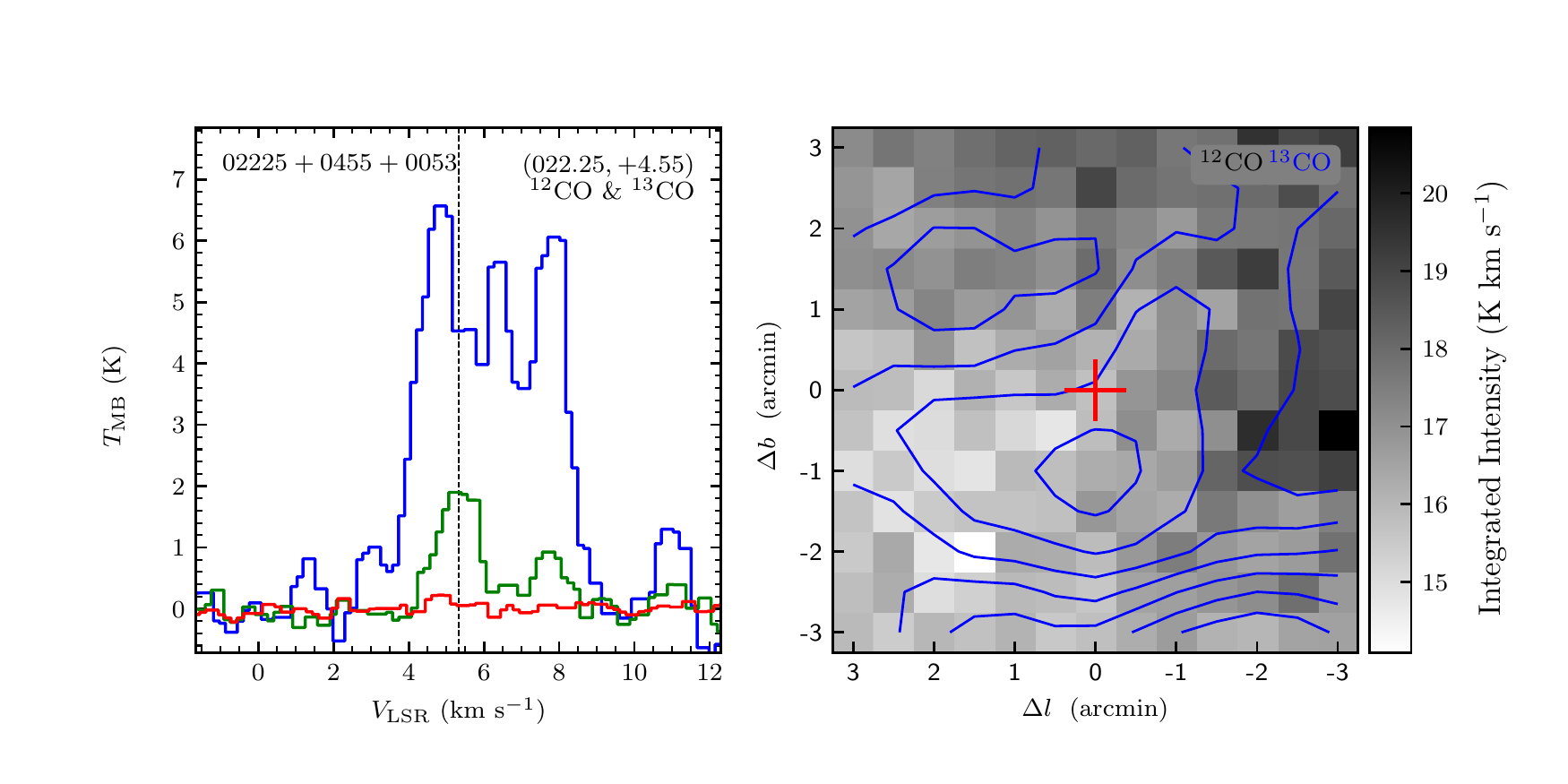}
\includegraphics[width=9.0cm,angle=0]{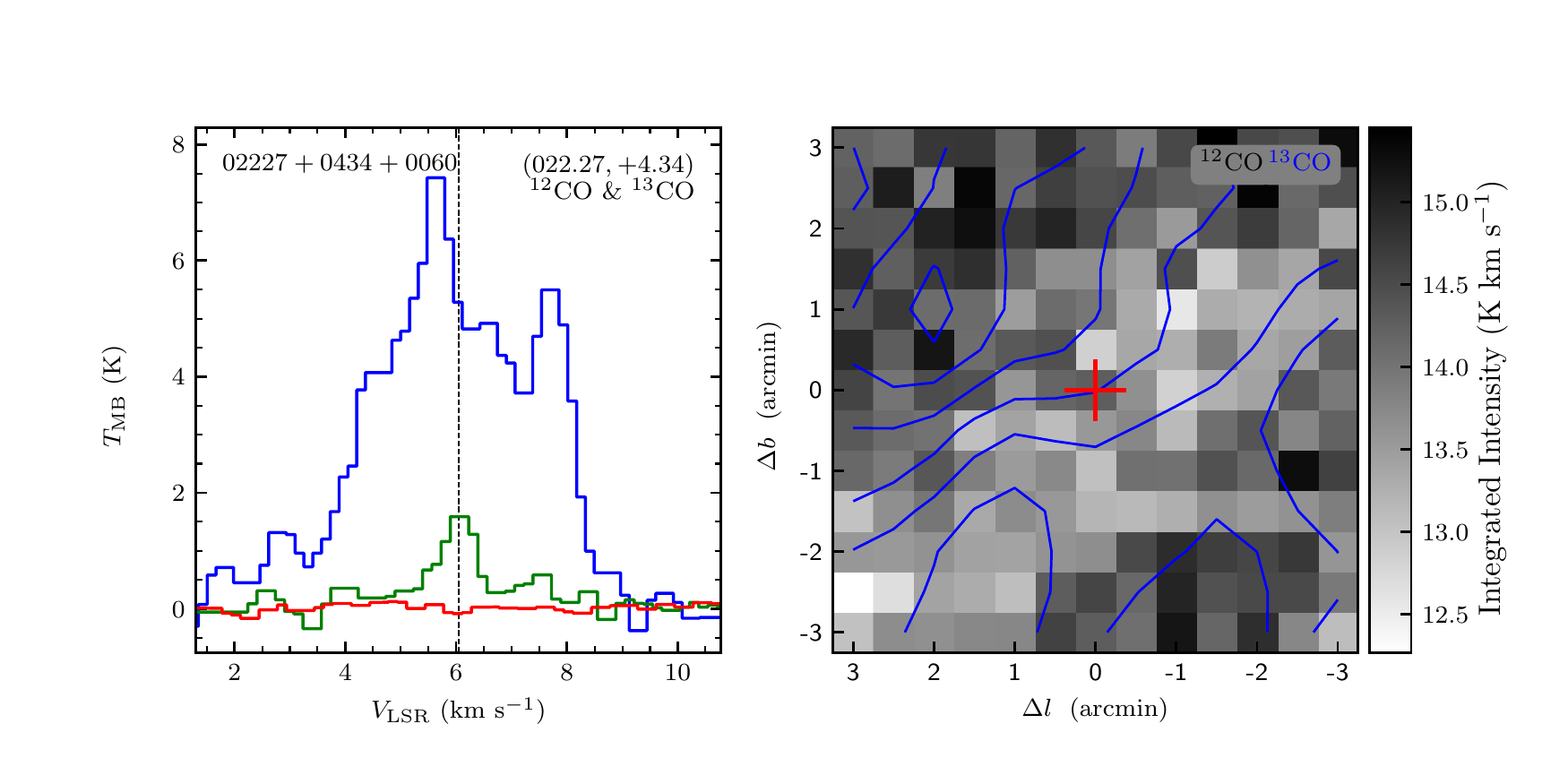}
\vspace{-0.5cm}

\includegraphics[width=9.0cm,angle=0]{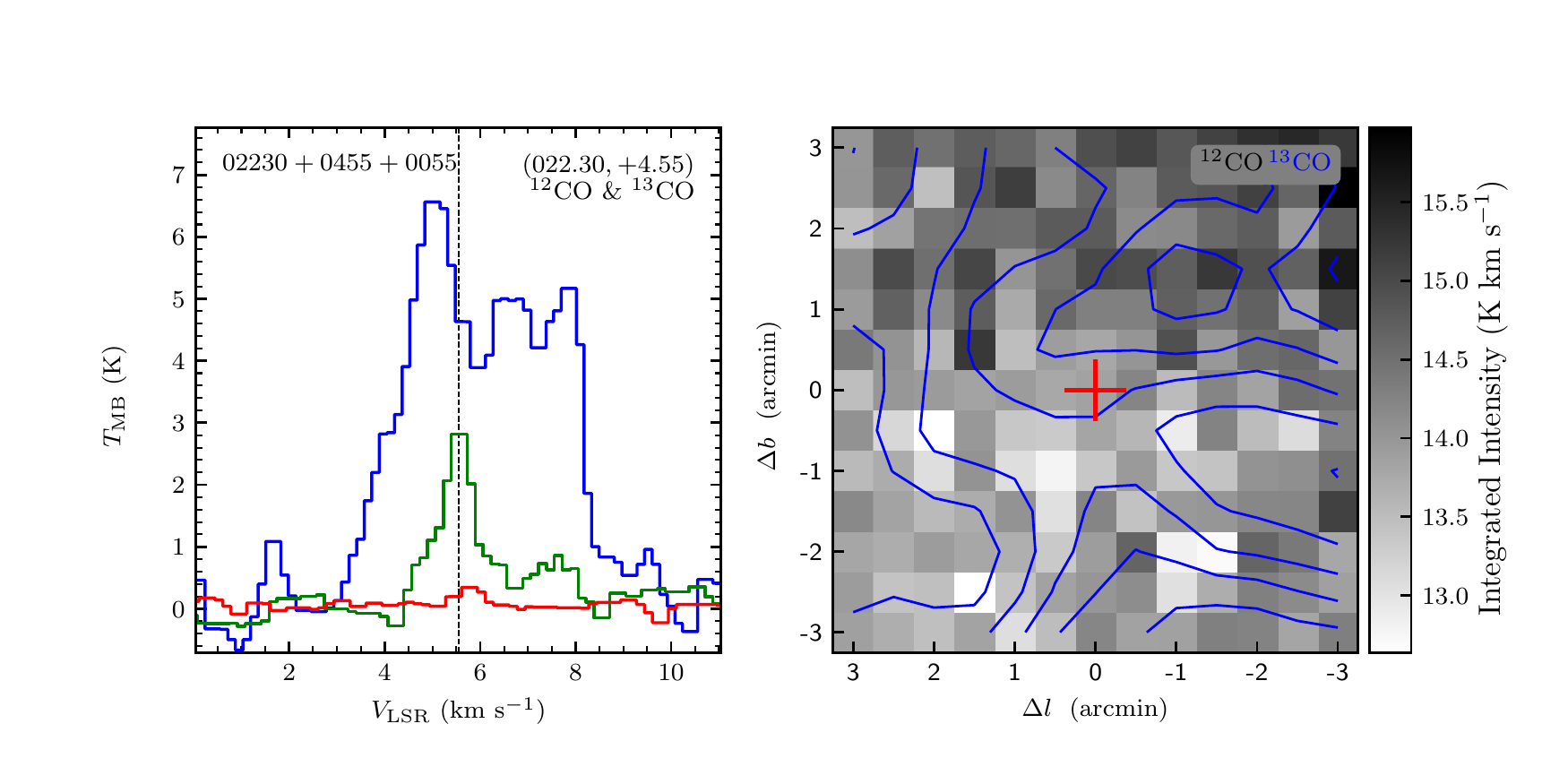}
\includegraphics[width=9.0cm,angle=0]{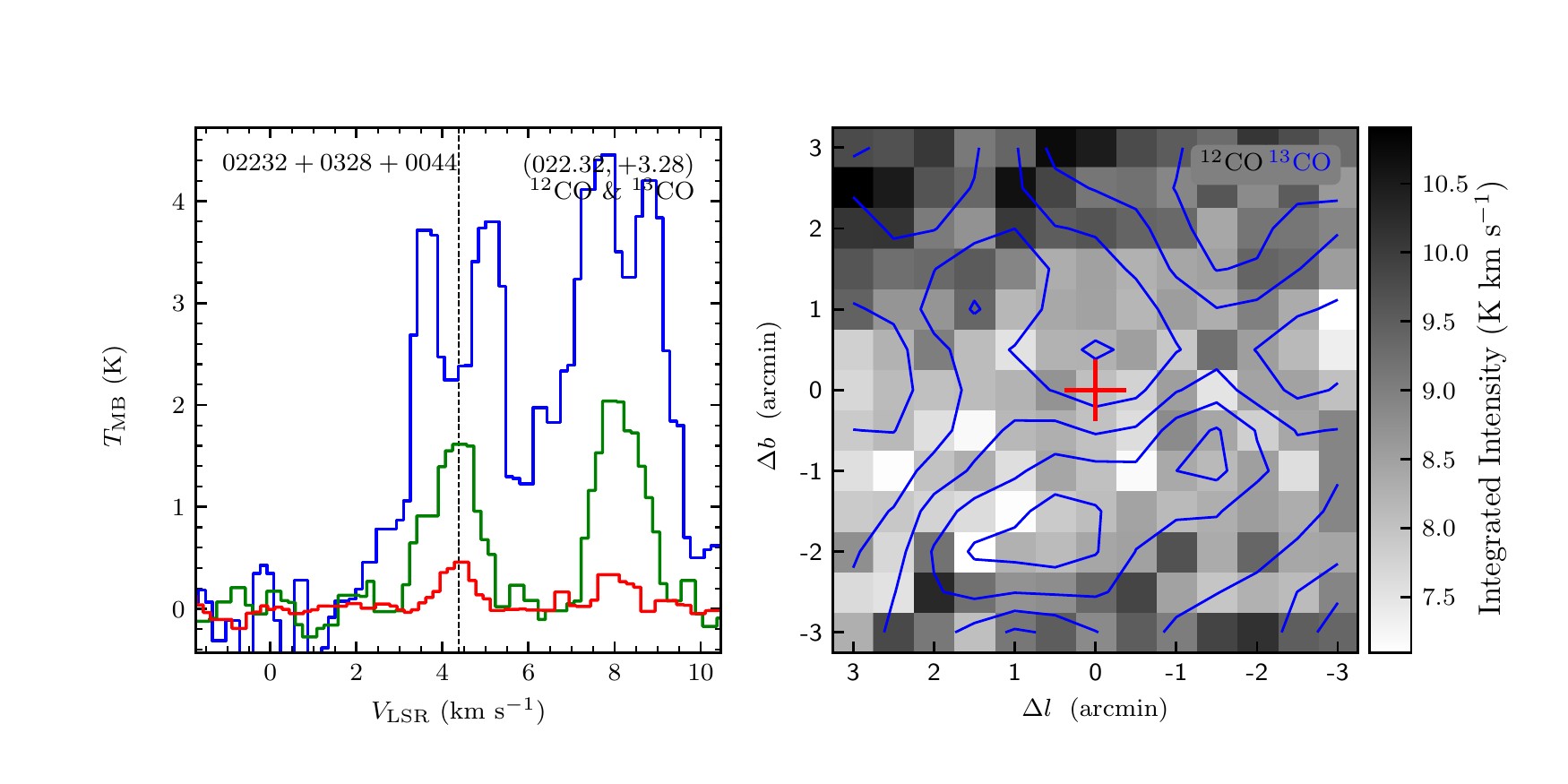}
\vspace{-0.5cm}

\includegraphics[width=9.0cm,angle=0]{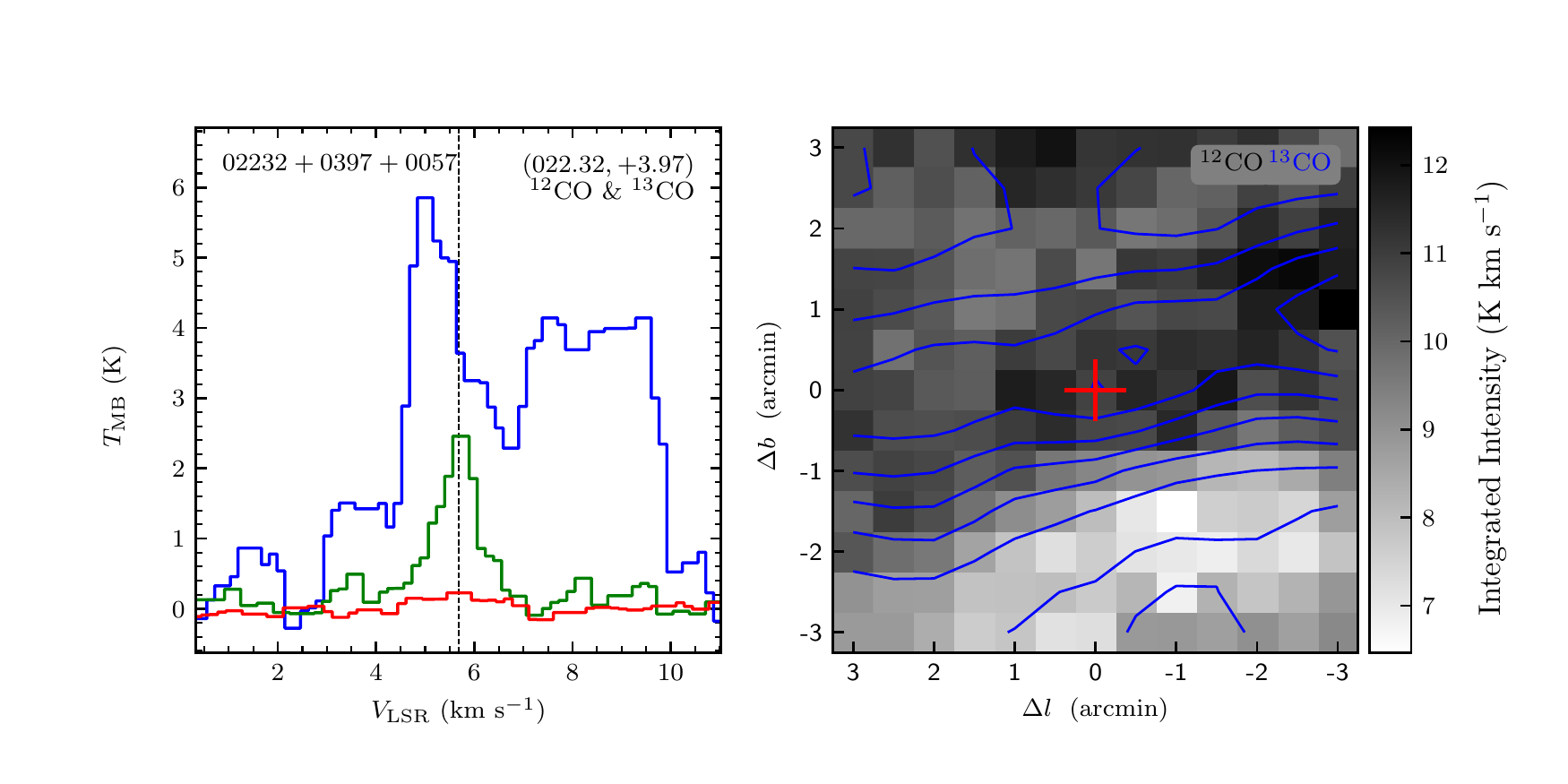}
\includegraphics[width=9.0cm,angle=0]{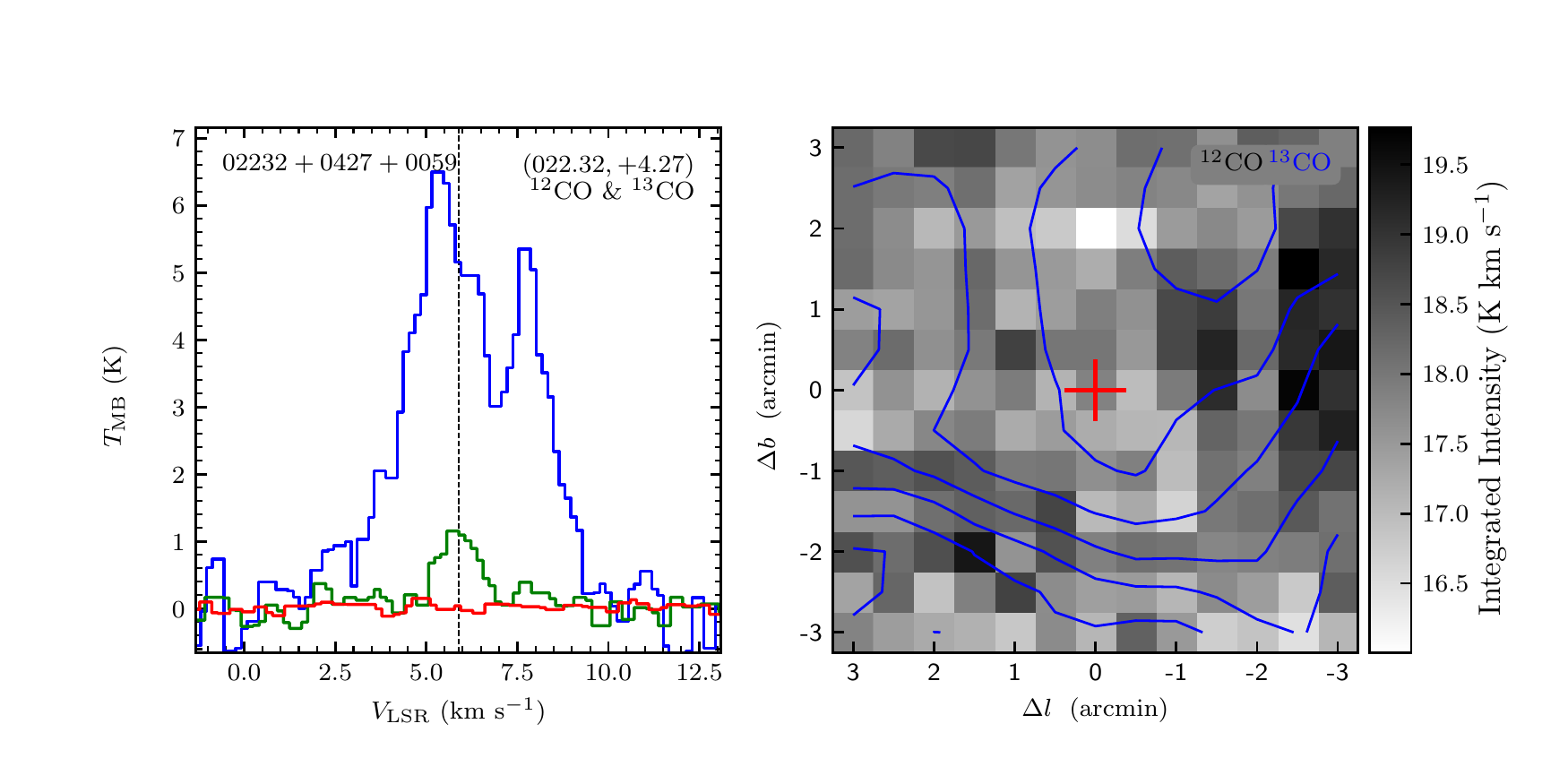}
\vspace{-0.5cm}

\includegraphics[width=9.0cm,angle=0]{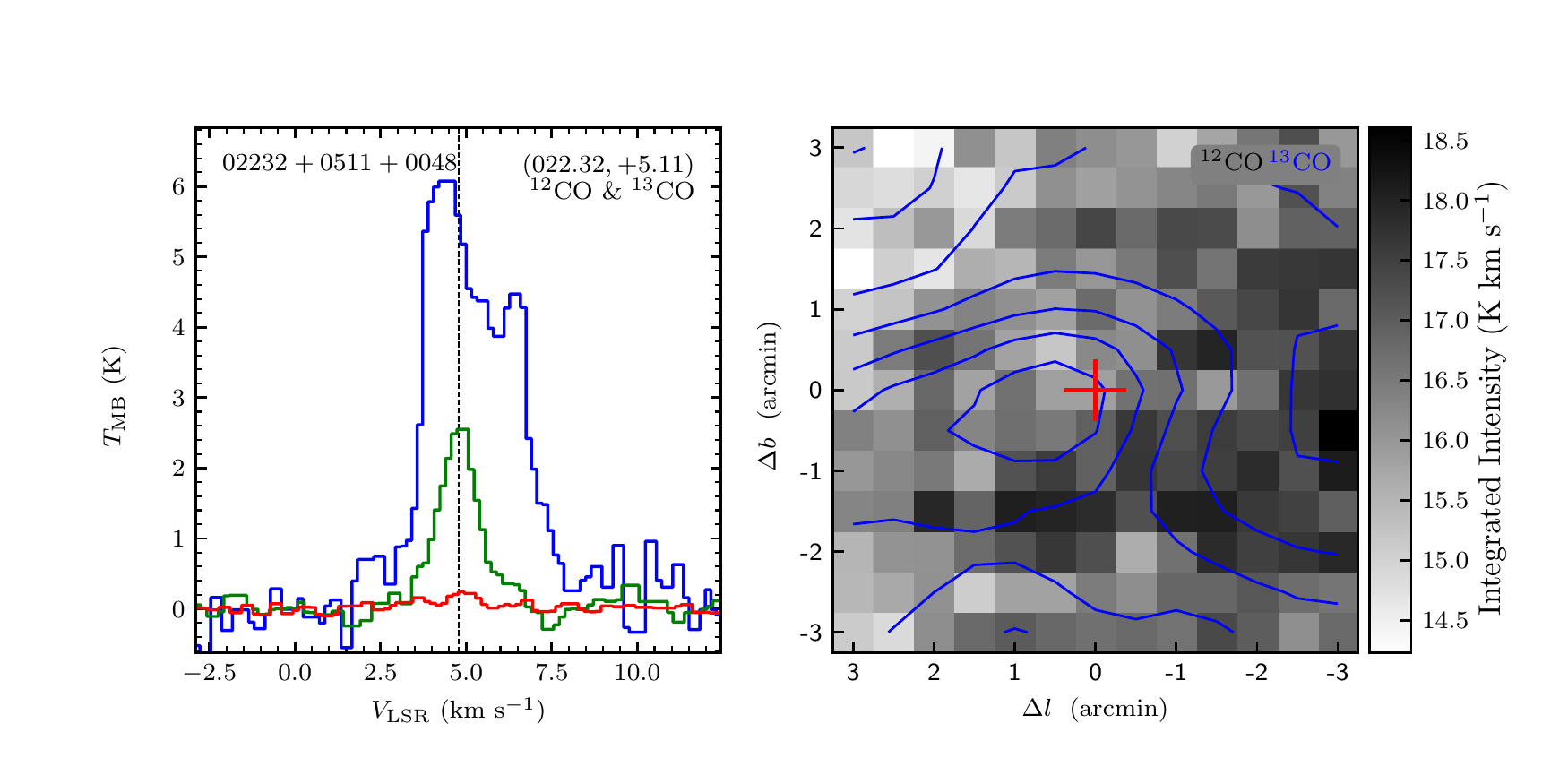}
\includegraphics[width=9.0cm,angle=0]{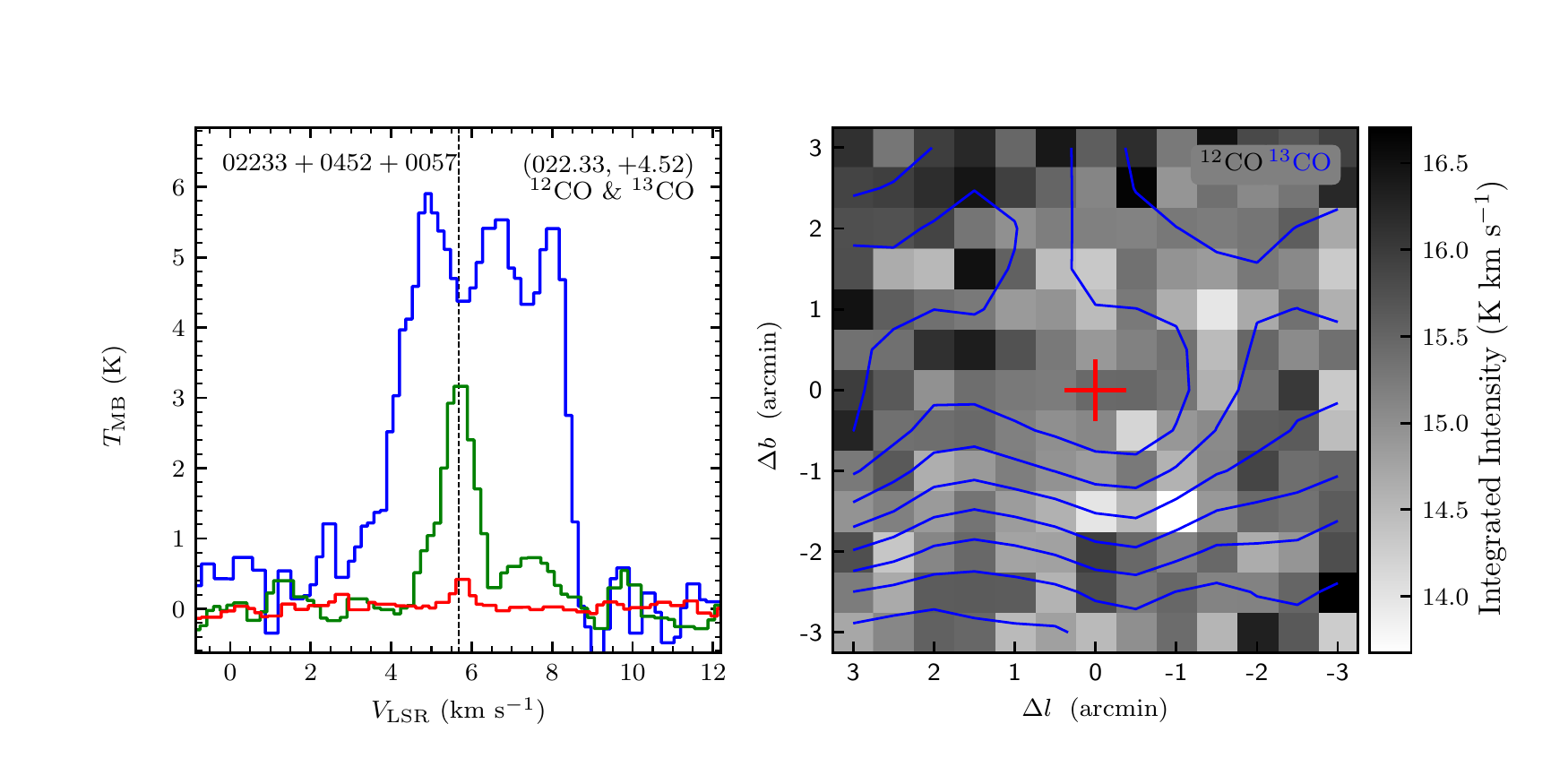}
\vspace{-0.5cm}

\includegraphics[width=9.0cm,angle=0]{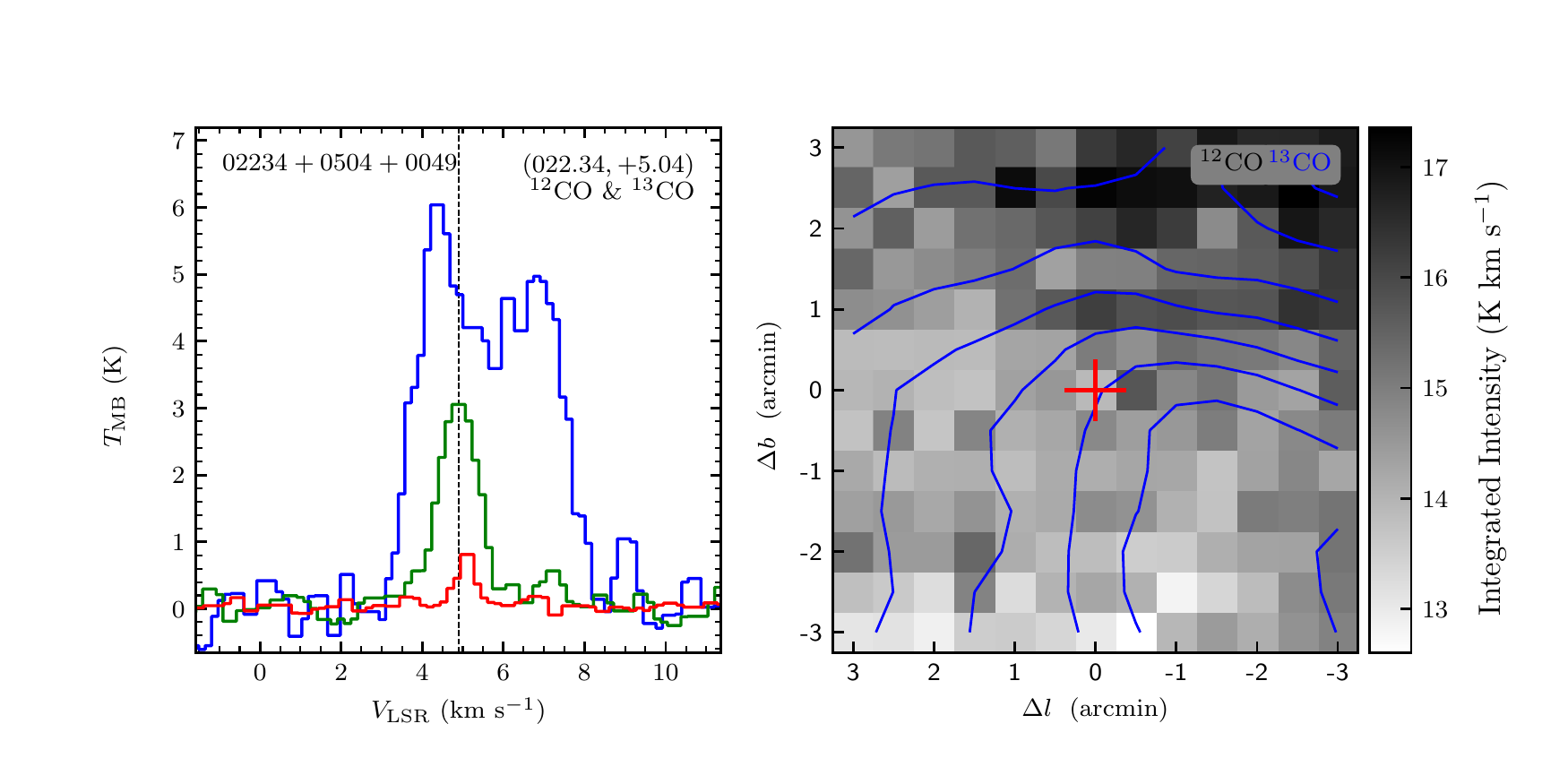}
\includegraphics[width=9.0cm,angle=0]{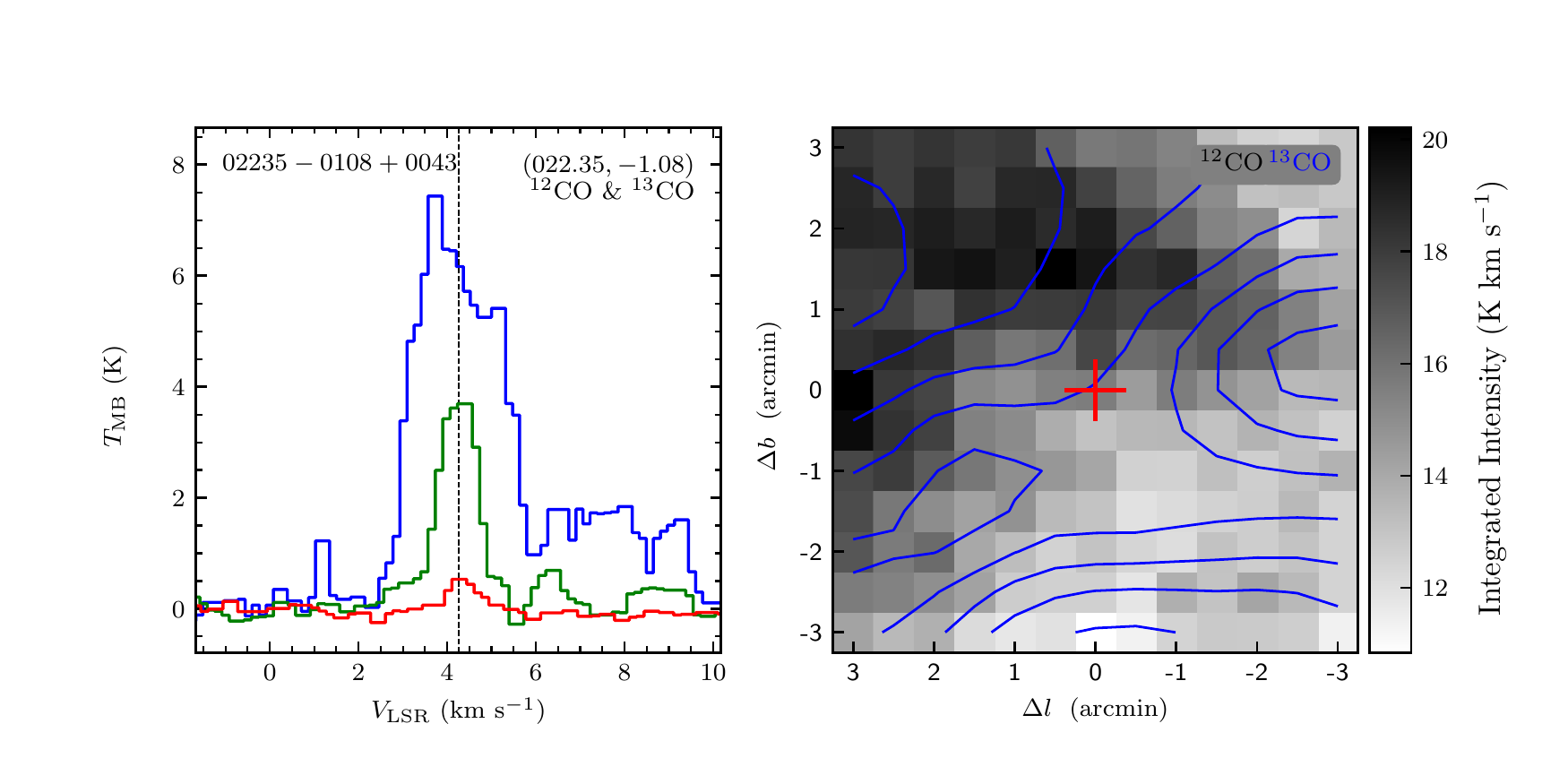}
\end{figure}
\clearpage

\begin{figure}
\includegraphics[width=9.0cm,angle=0]{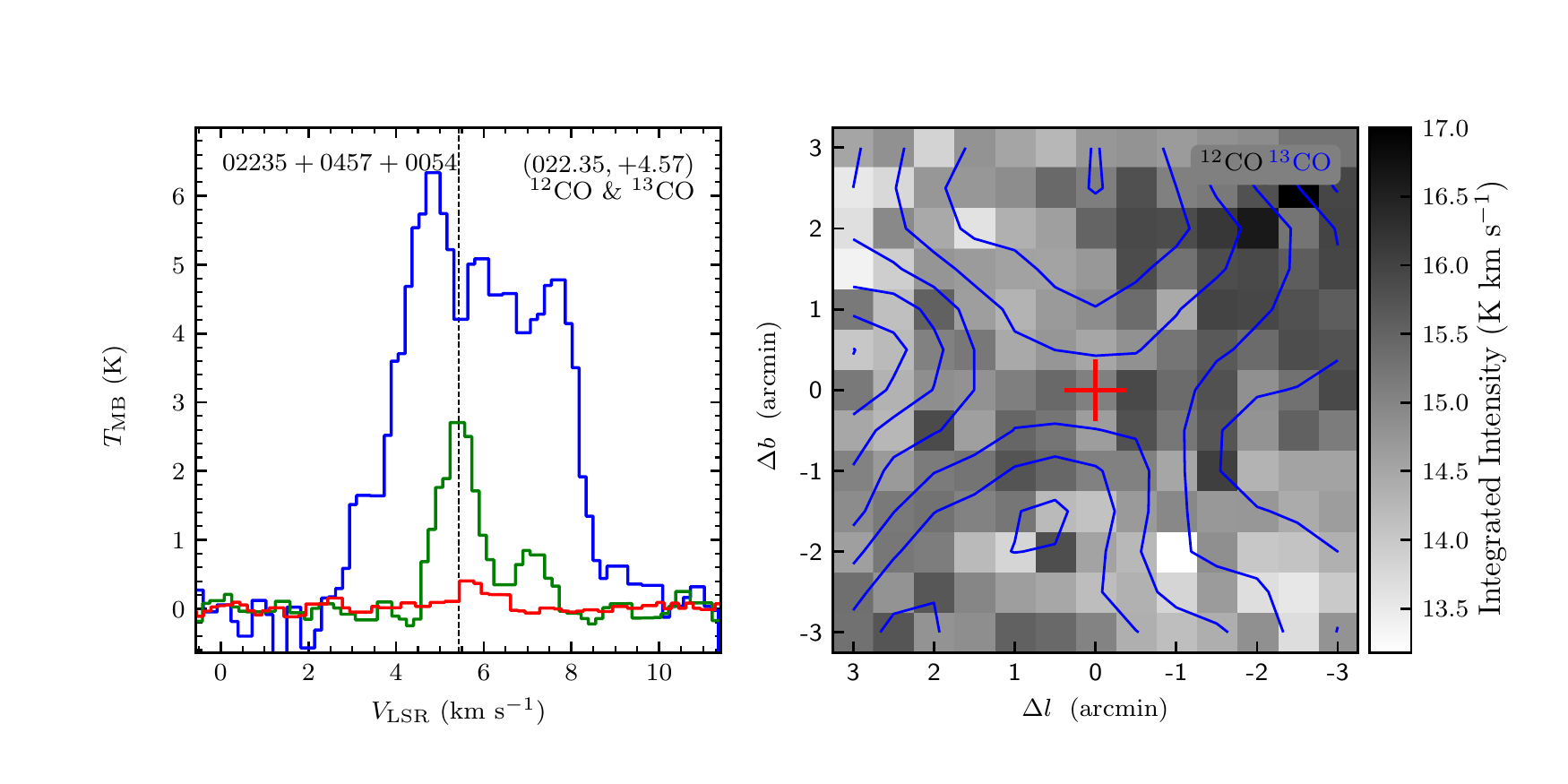}
\includegraphics[width=9.0cm,angle=0]{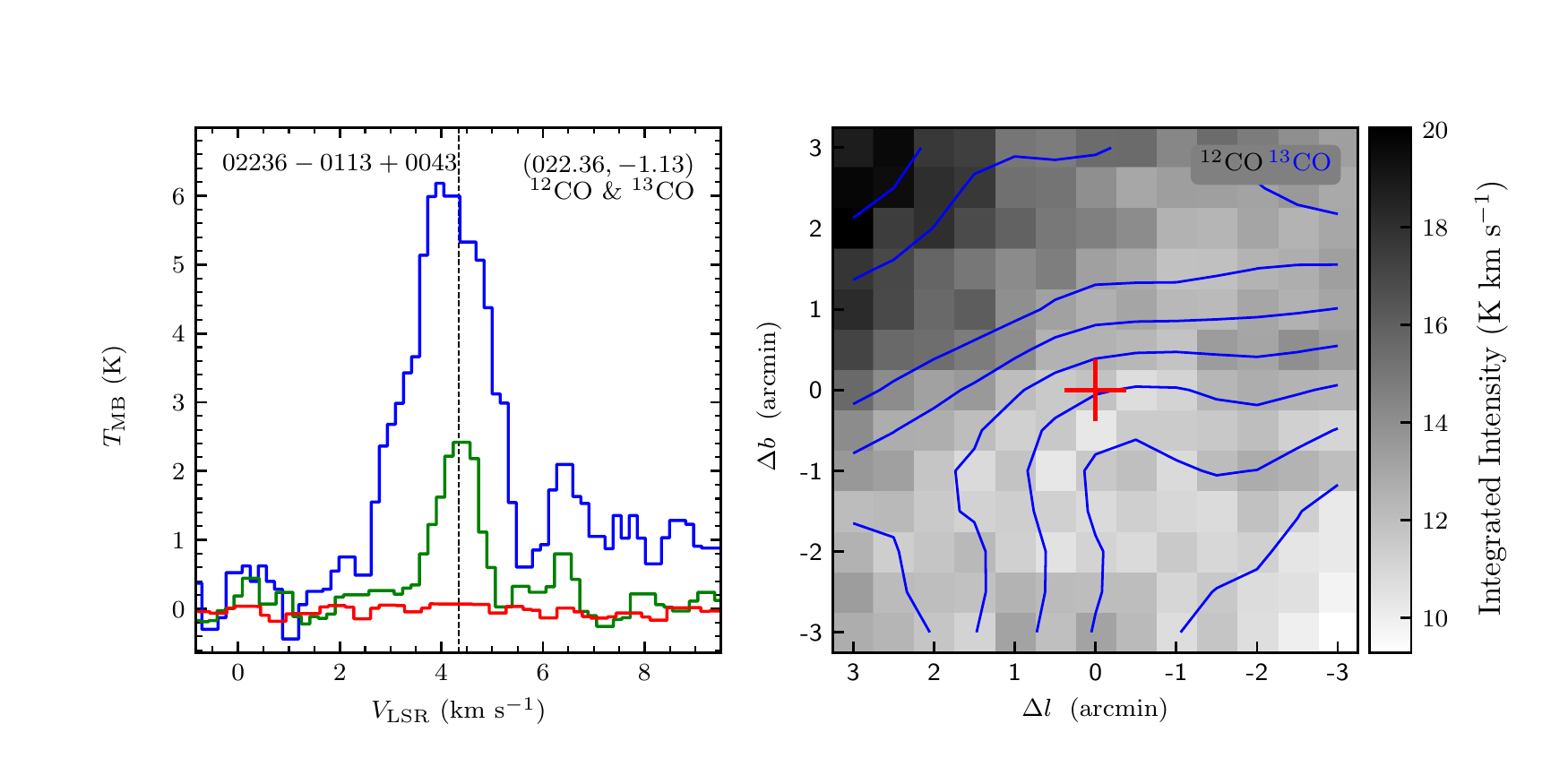}
\vspace{-0.5cm}

\includegraphics[width=9.0cm,angle=0]{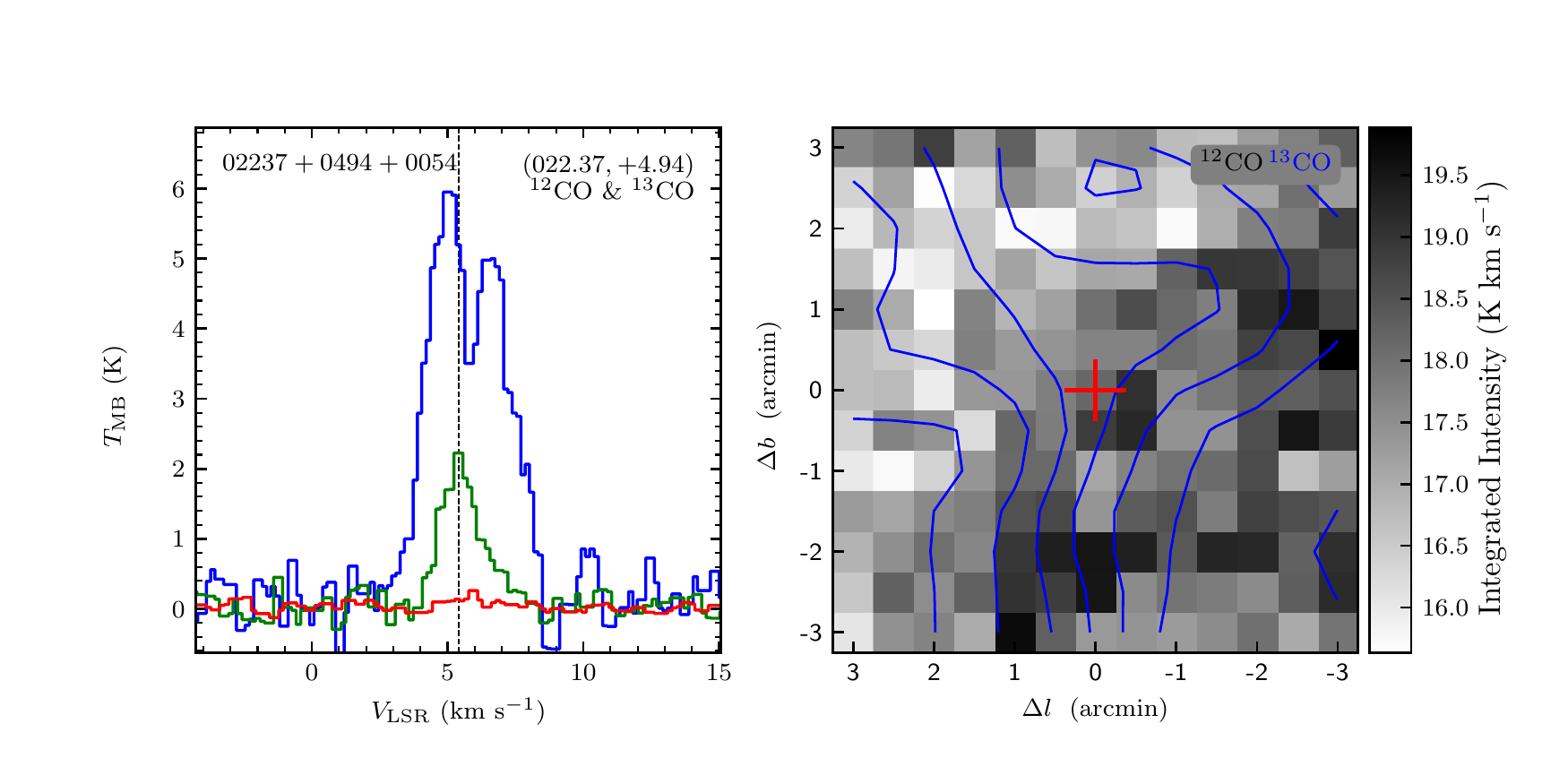}
\includegraphics[width=9.0cm,angle=0]{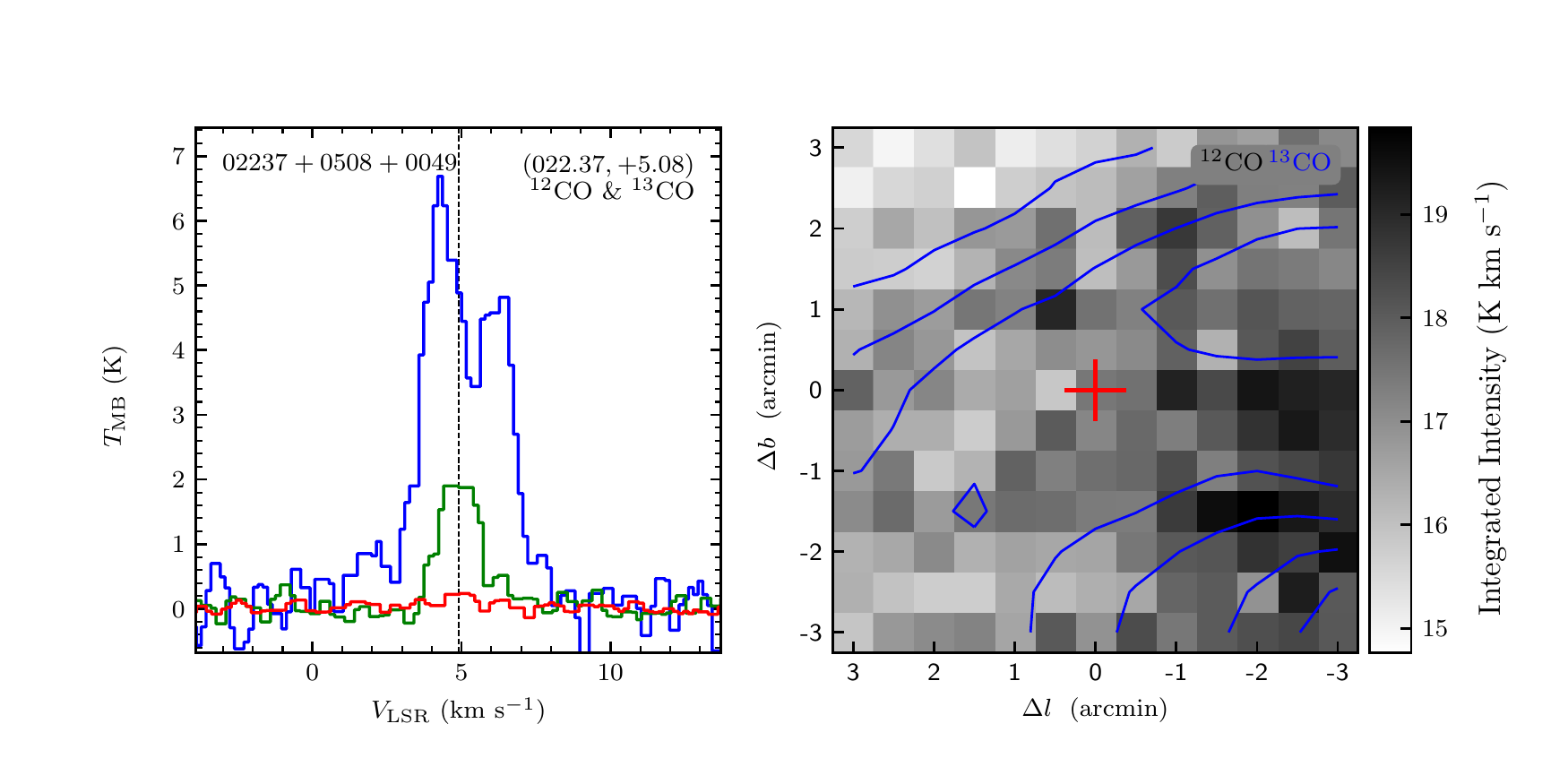}
\vspace{-0.5cm}

\includegraphics[width=9.0cm,angle=0]{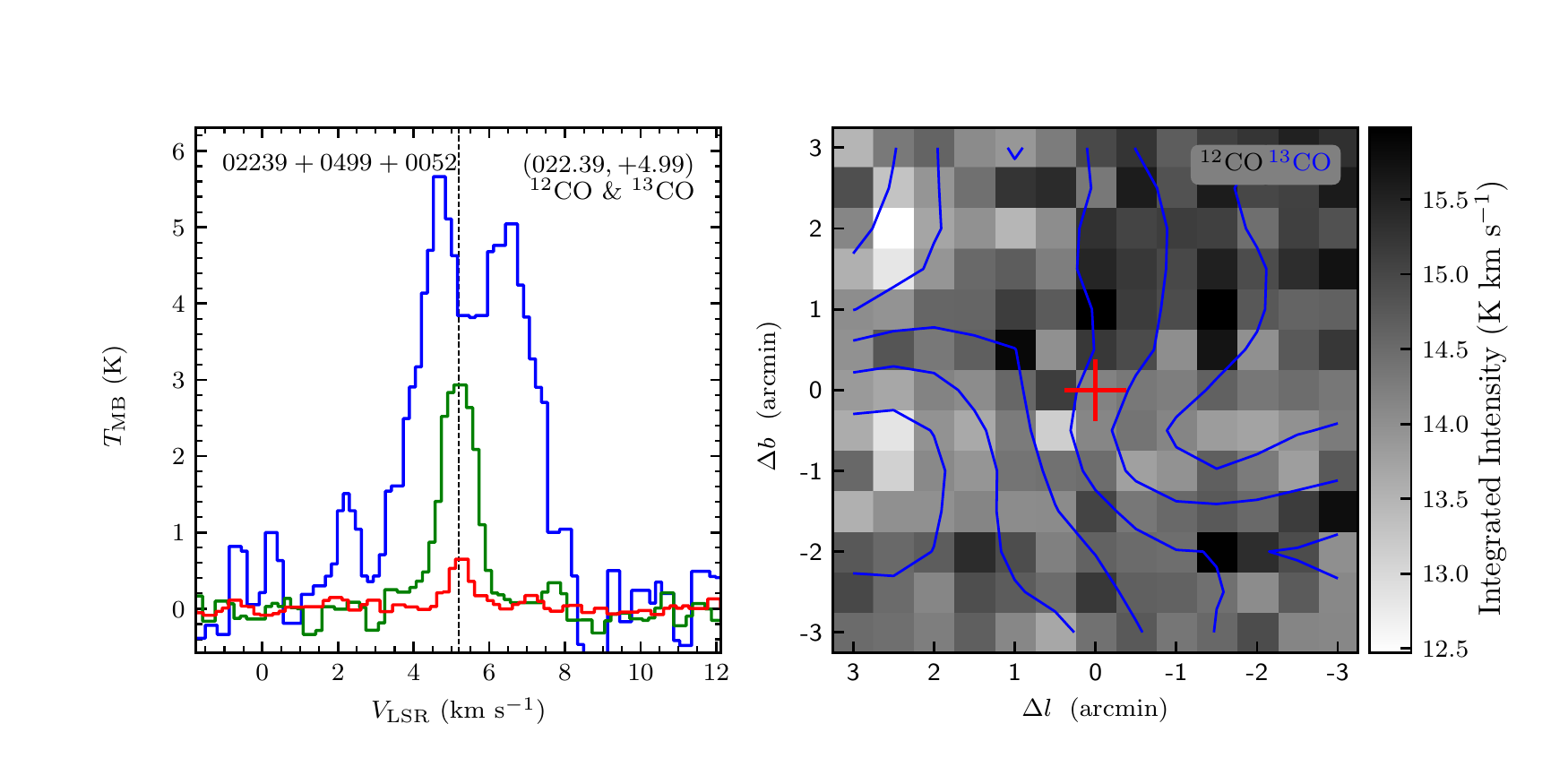}
\includegraphics[width=9.0cm,angle=0]{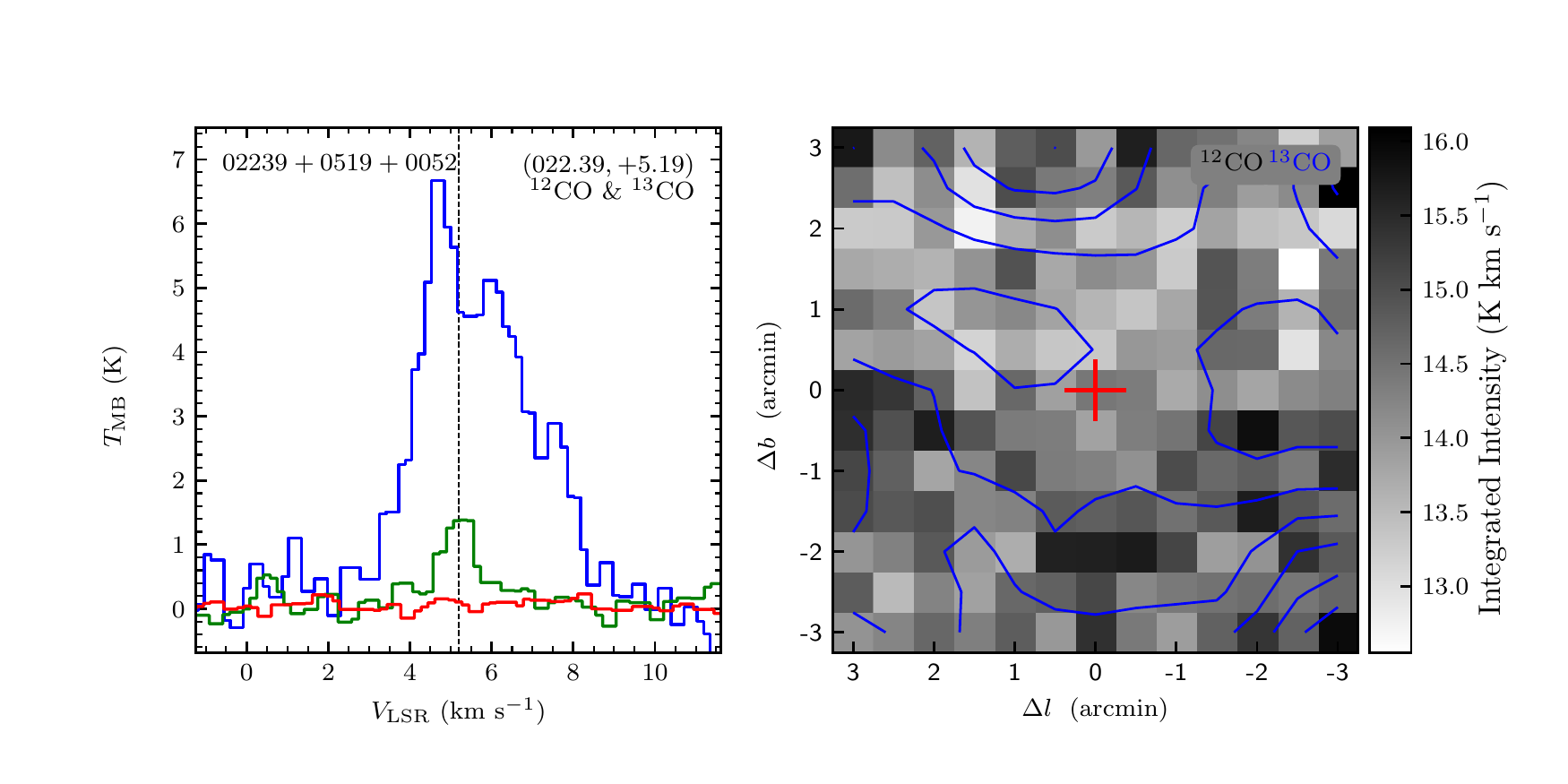}
\vspace{-0.5cm}

\includegraphics[width=9.0cm,angle=0]{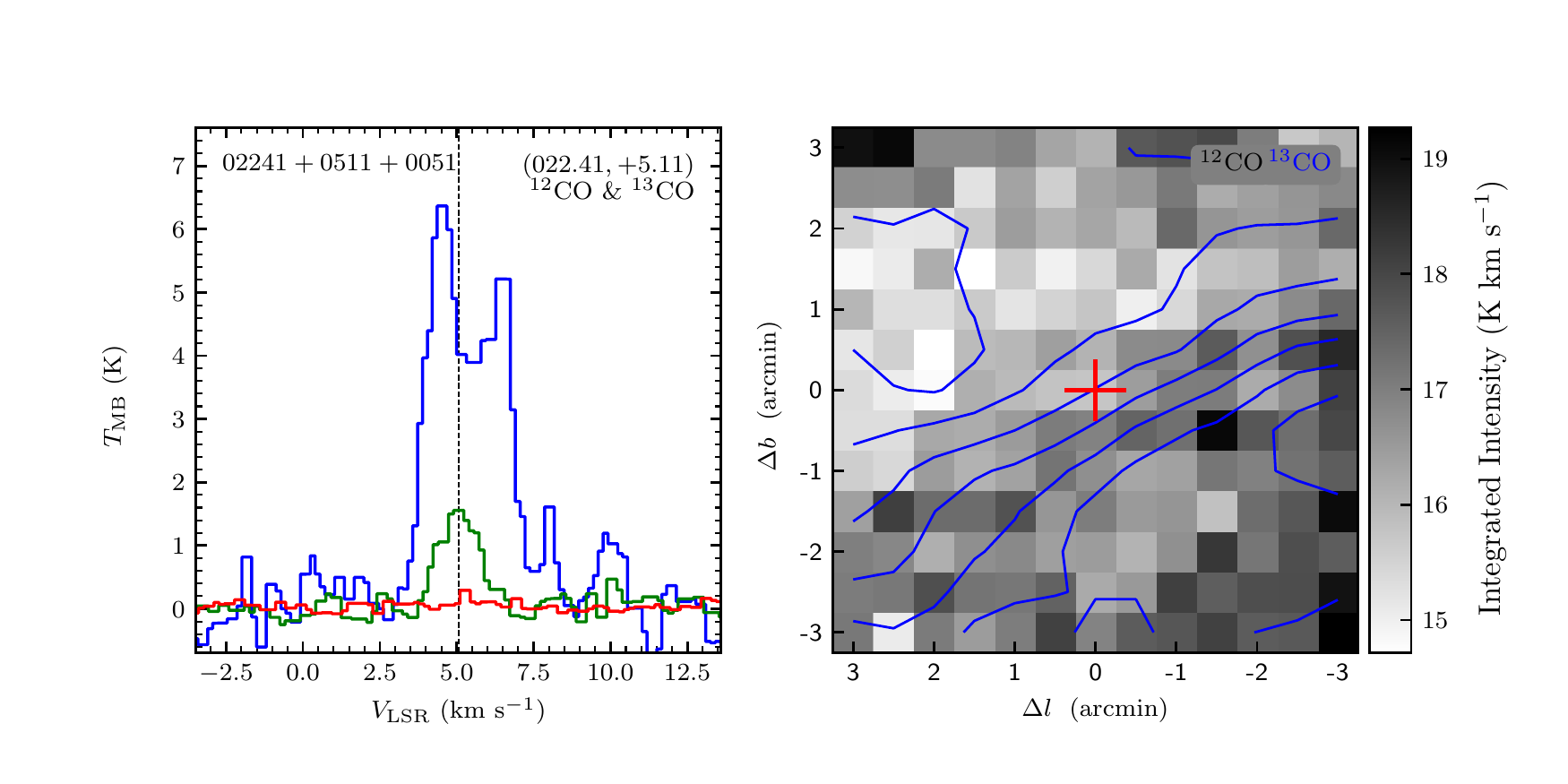}
\includegraphics[width=9.0cm,angle=0]{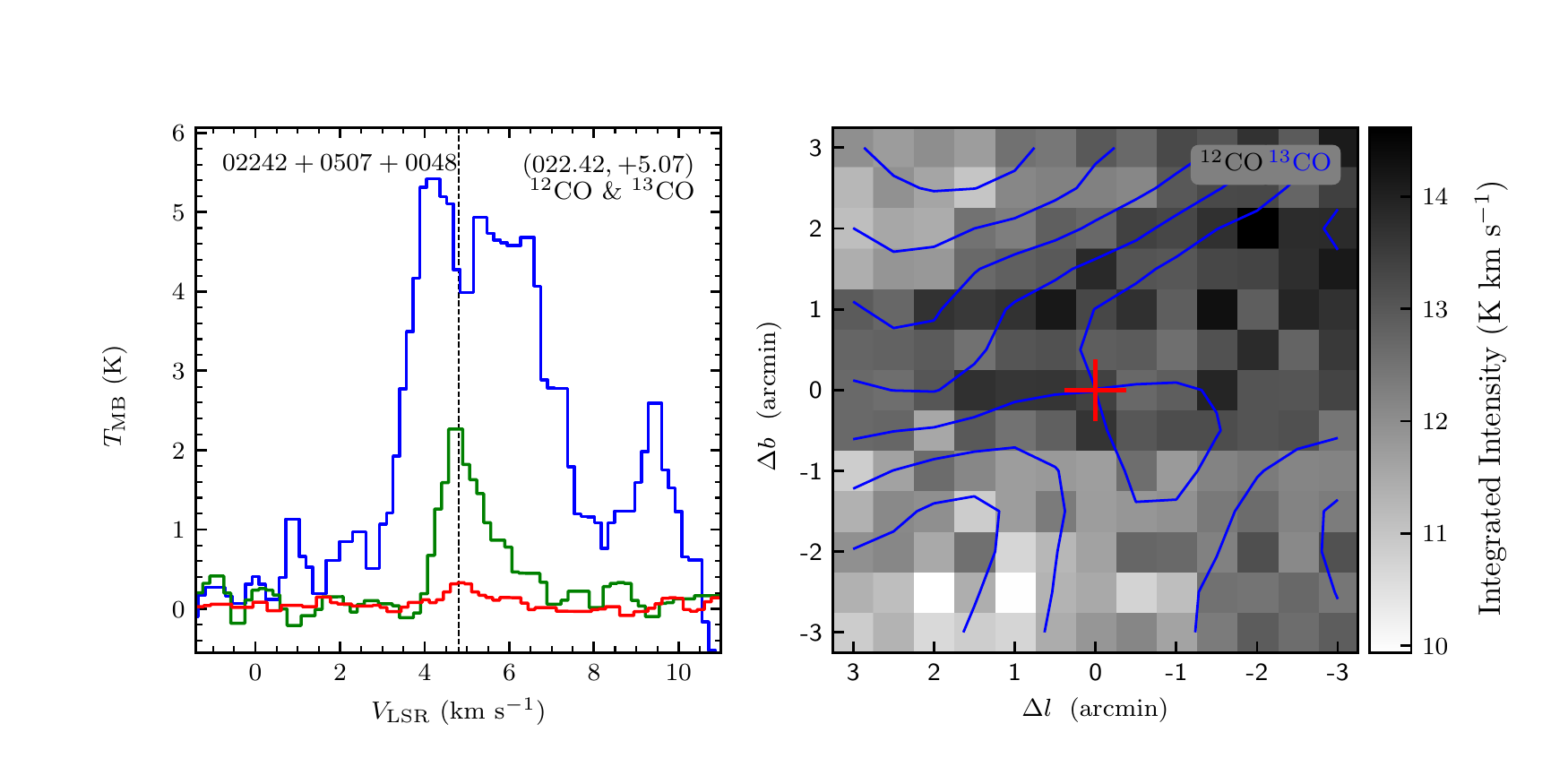}
\vspace{-0.5cm}

\includegraphics[width=9.0cm,angle=0]{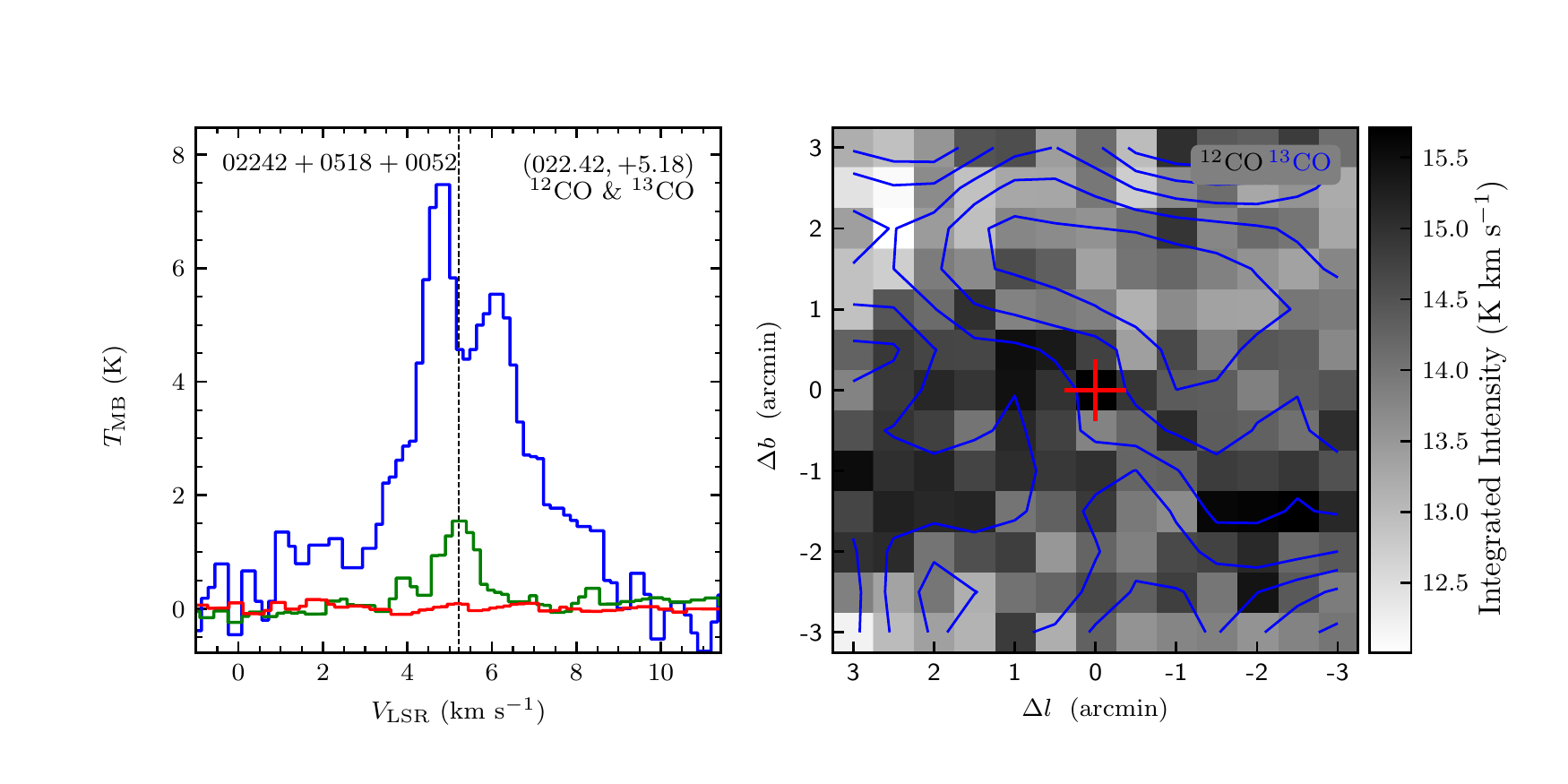}
\includegraphics[width=9.0cm,angle=0]{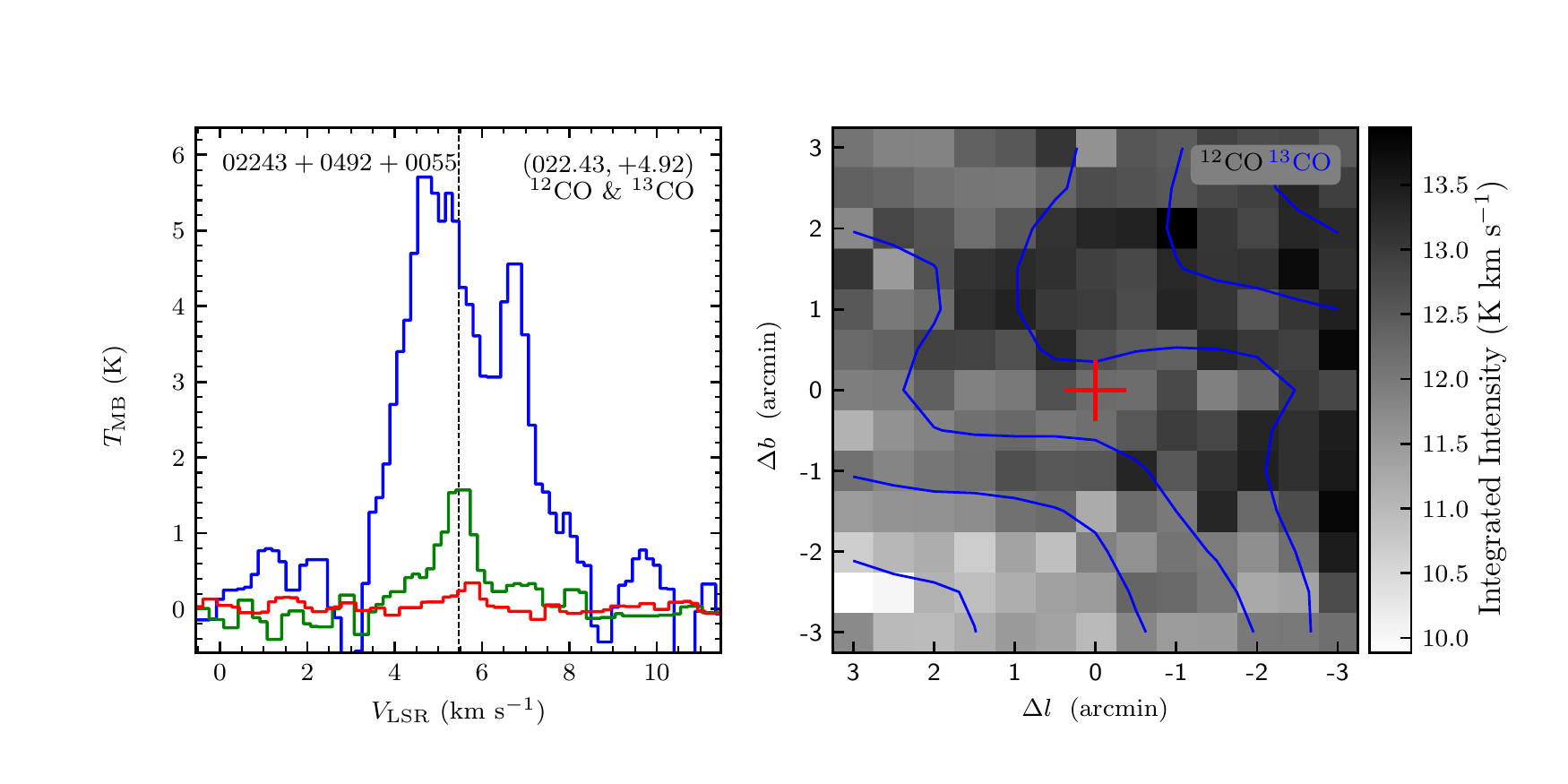}
\end{figure}
\clearpage

\begin{figure}
\includegraphics[width=9.0cm,angle=0]{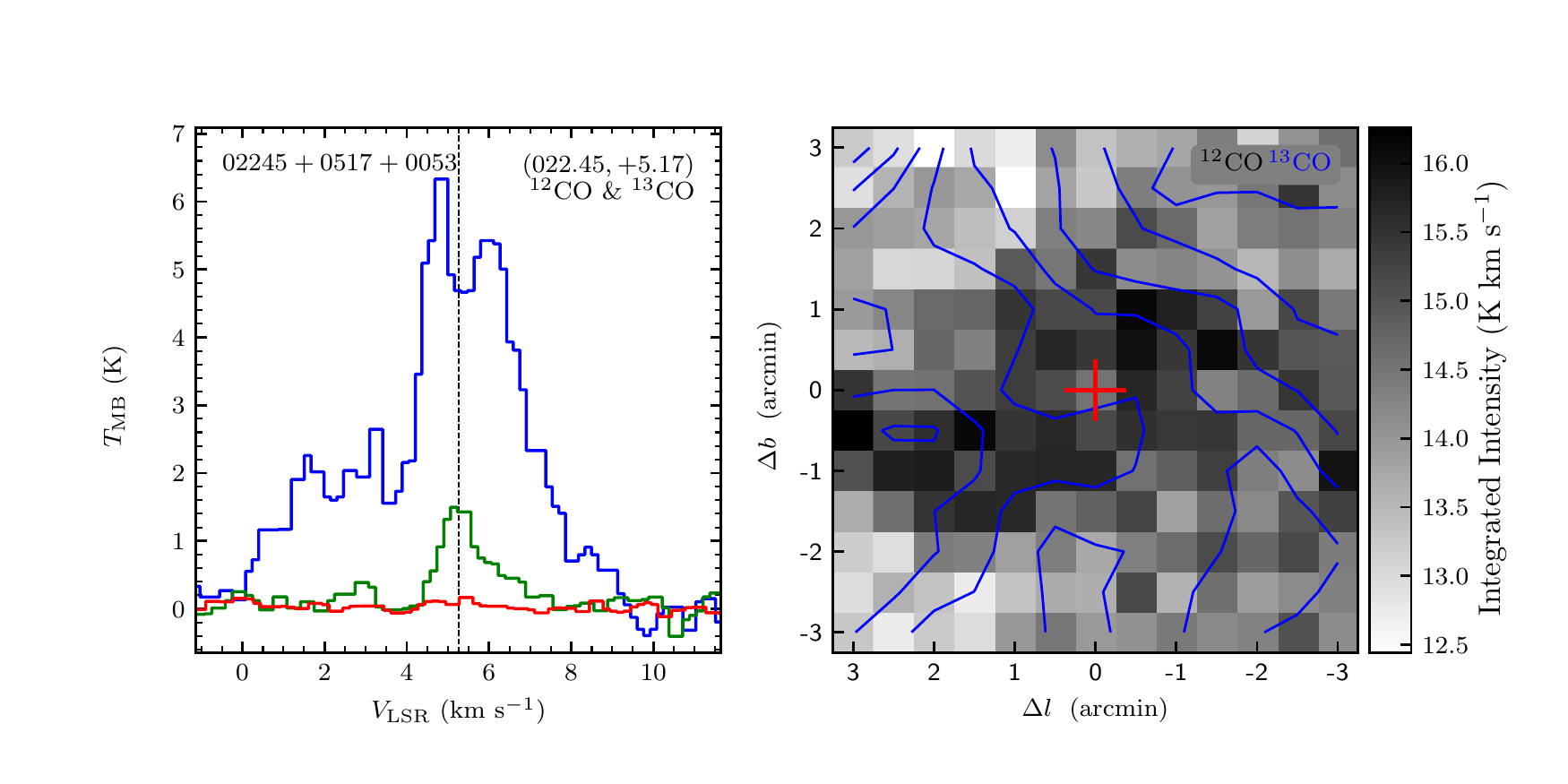}
\includegraphics[width=9.0cm,angle=0]{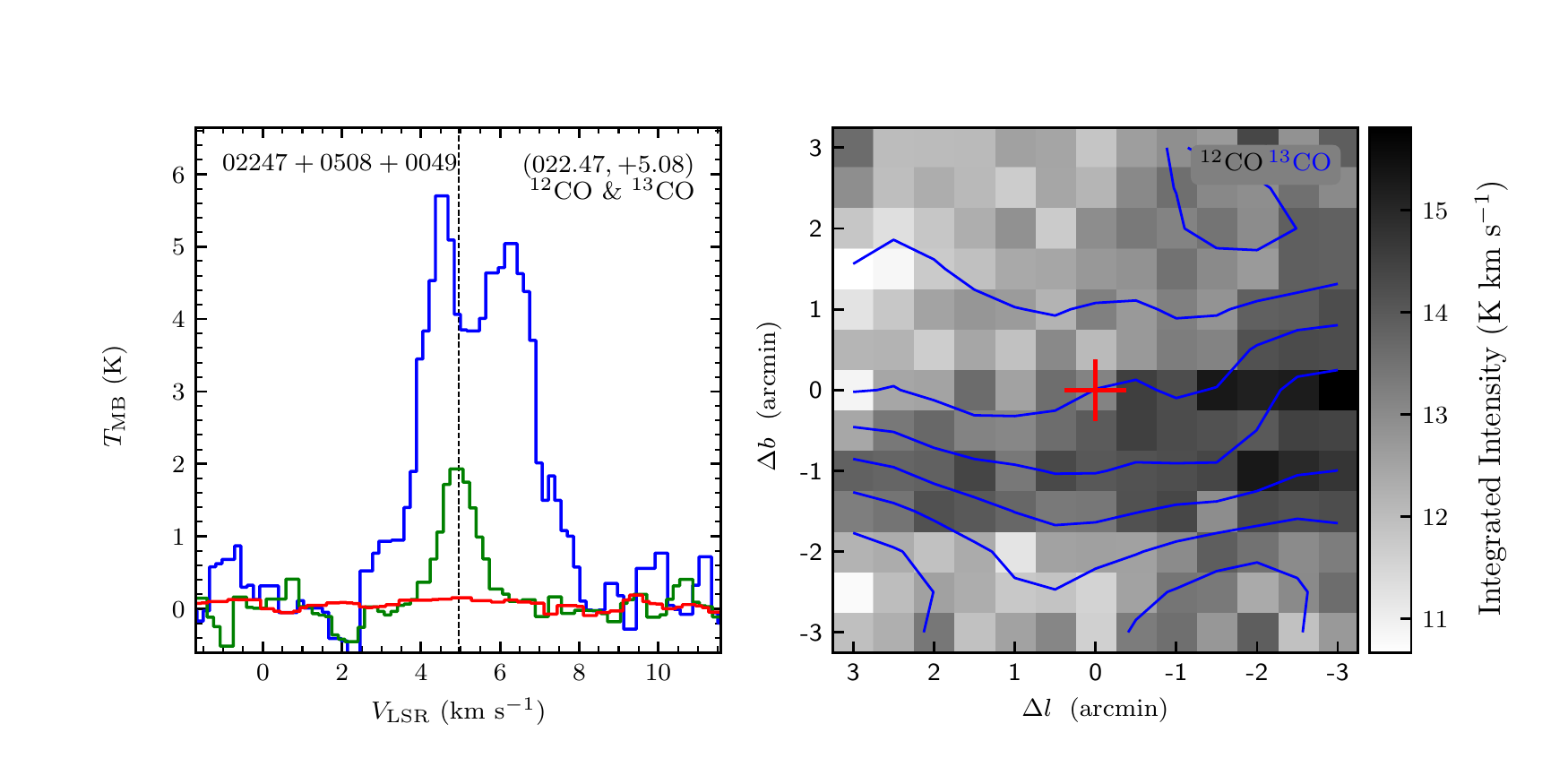}
\vspace{-0.5cm}

\includegraphics[width=9.0cm,angle=0]{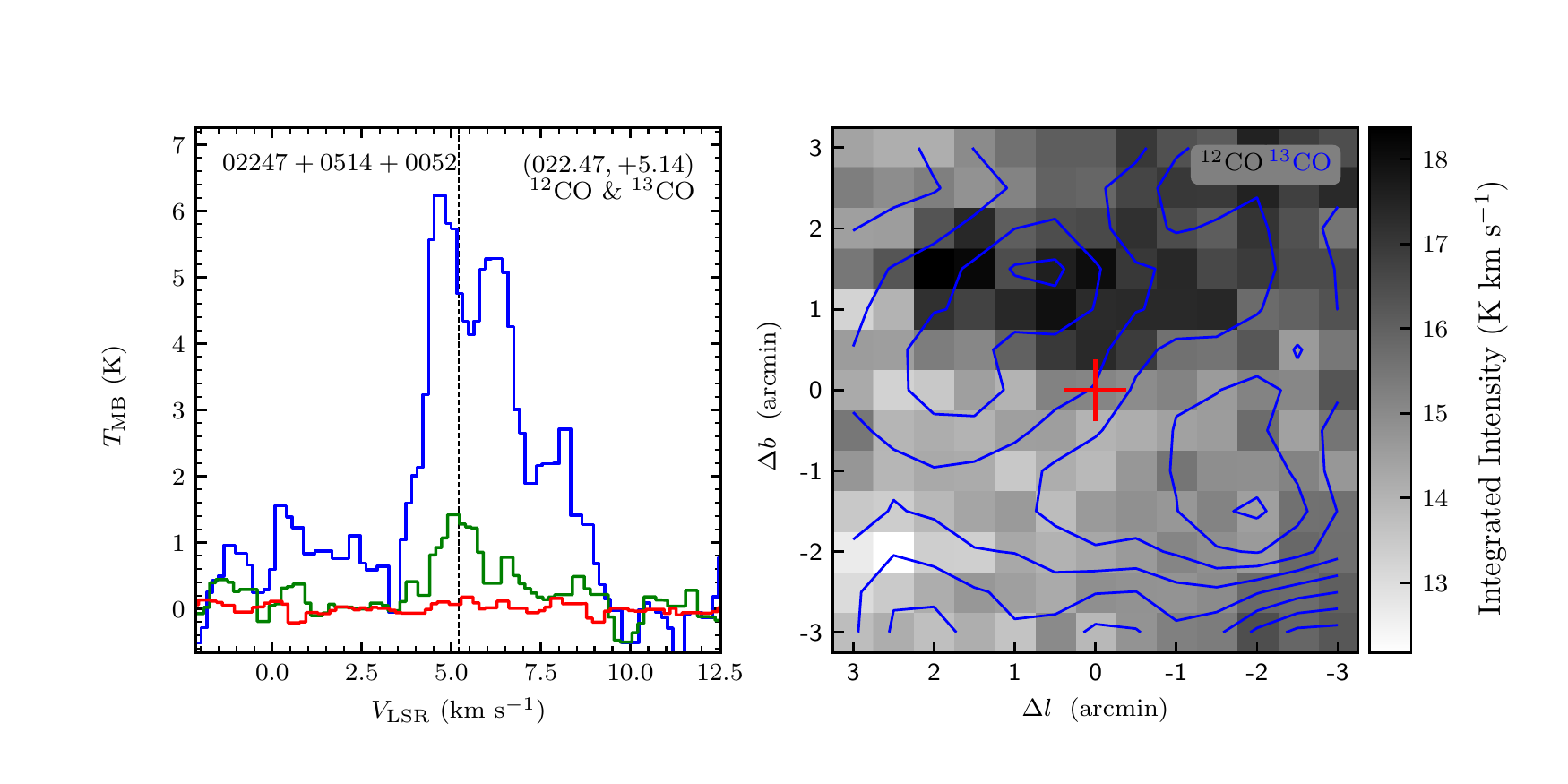}
\includegraphics[width=9.0cm,angle=0]{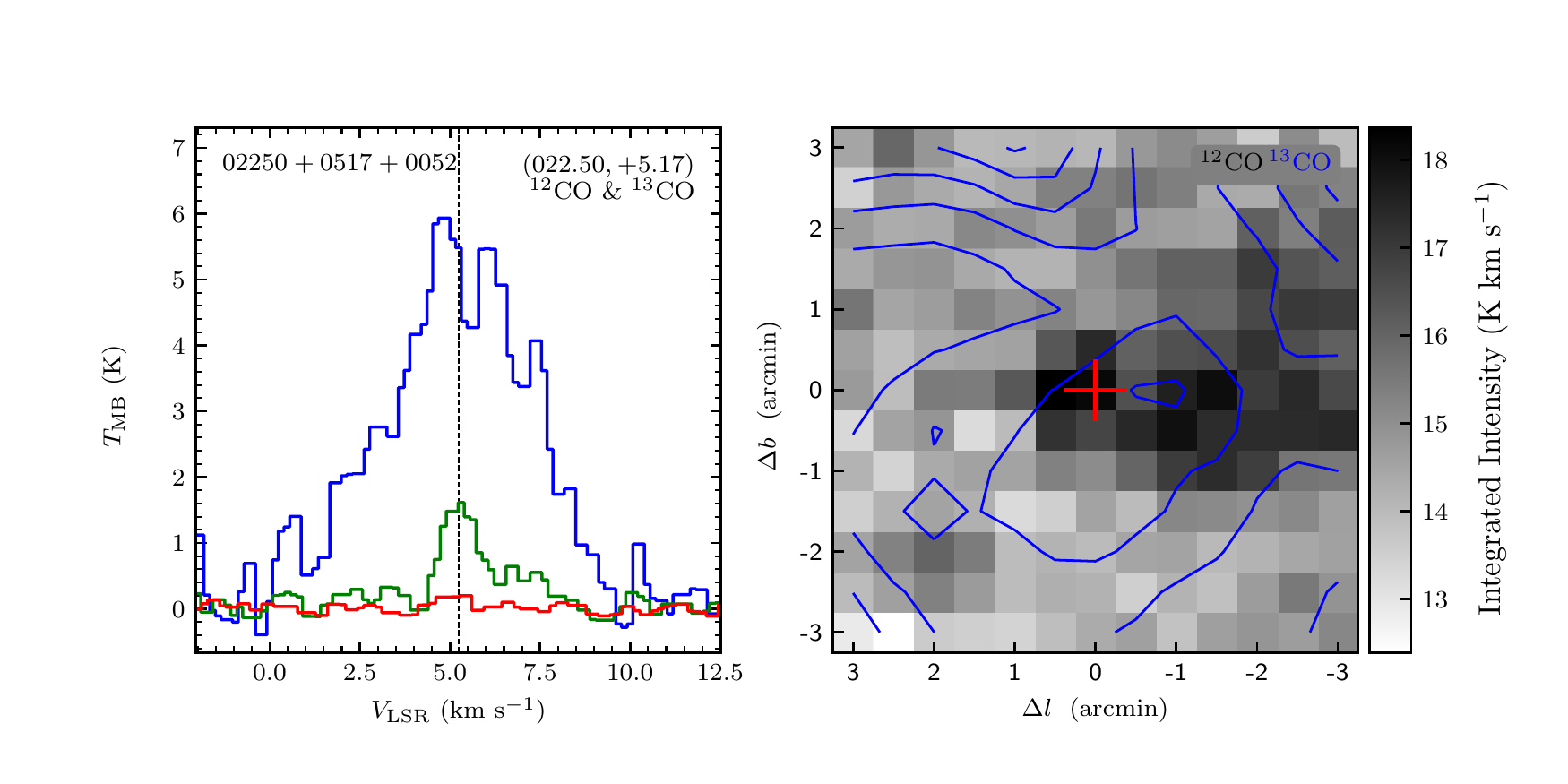}
\vspace{-0.5cm}

\includegraphics[width=9.0cm,angle=0]{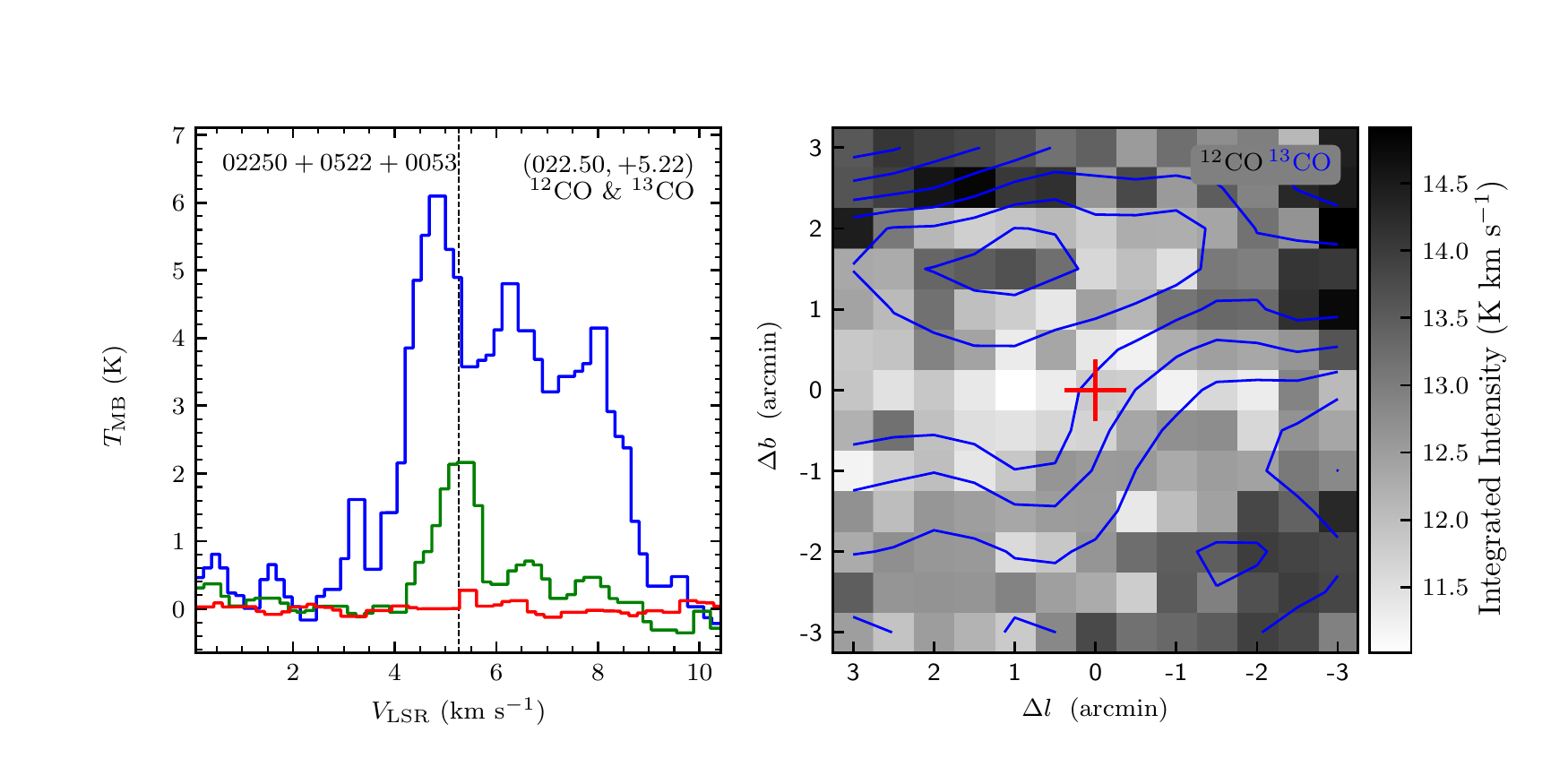}
\includegraphics[width=9.0cm,angle=0]{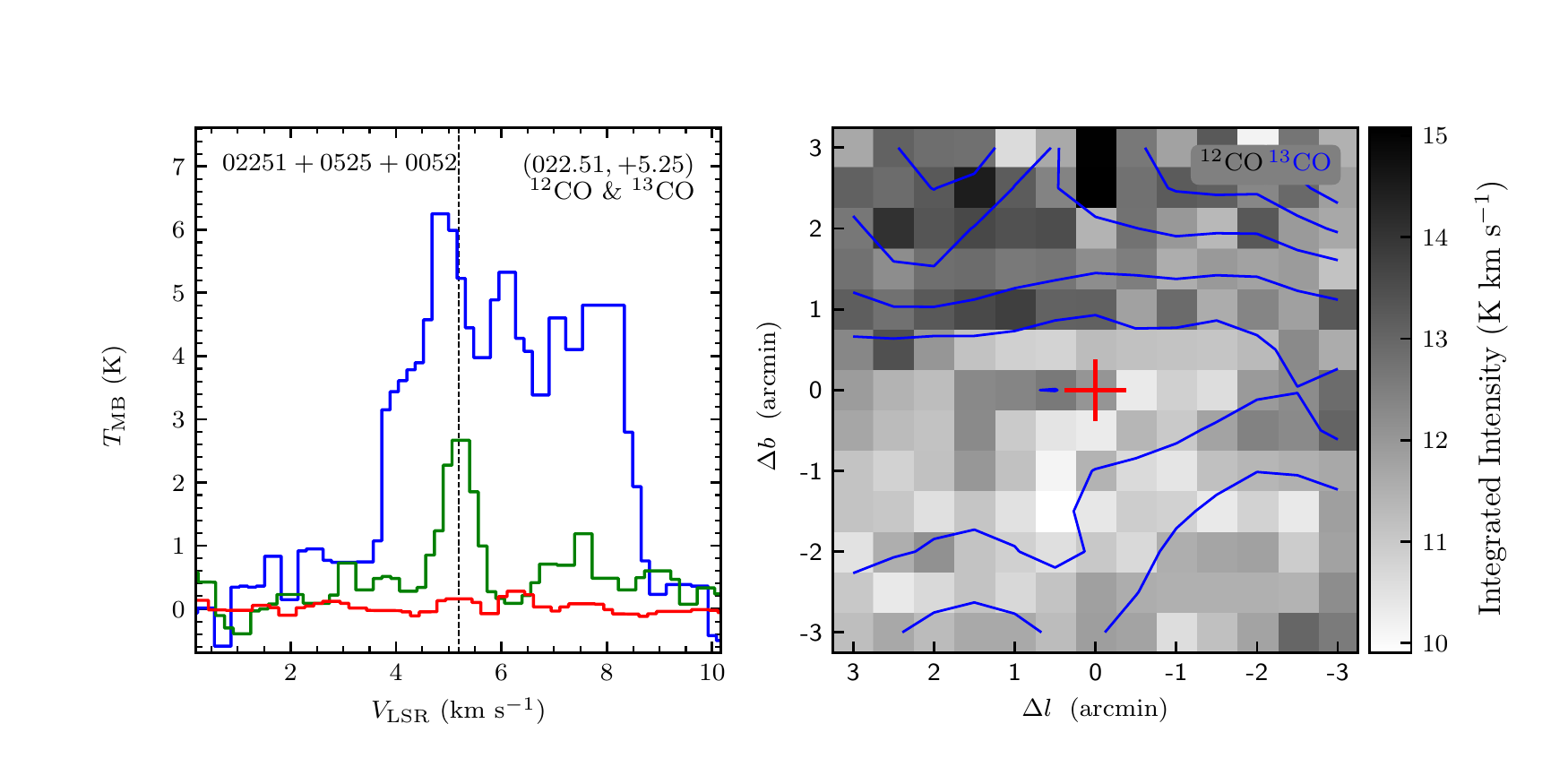}
\vspace{-0.5cm}

\includegraphics[width=9.0cm,angle=0]{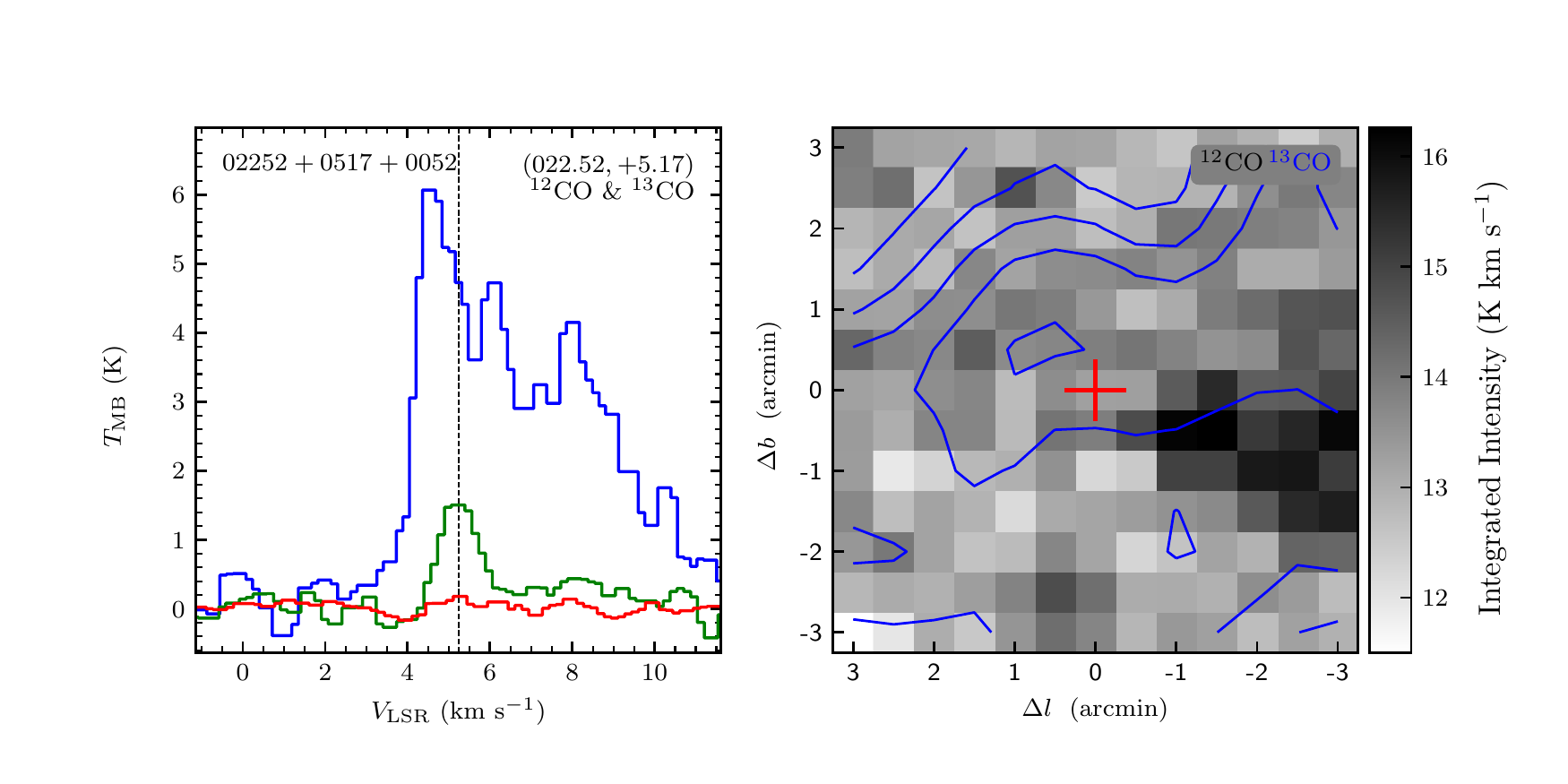}
\includegraphics[width=9.0cm,angle=0]{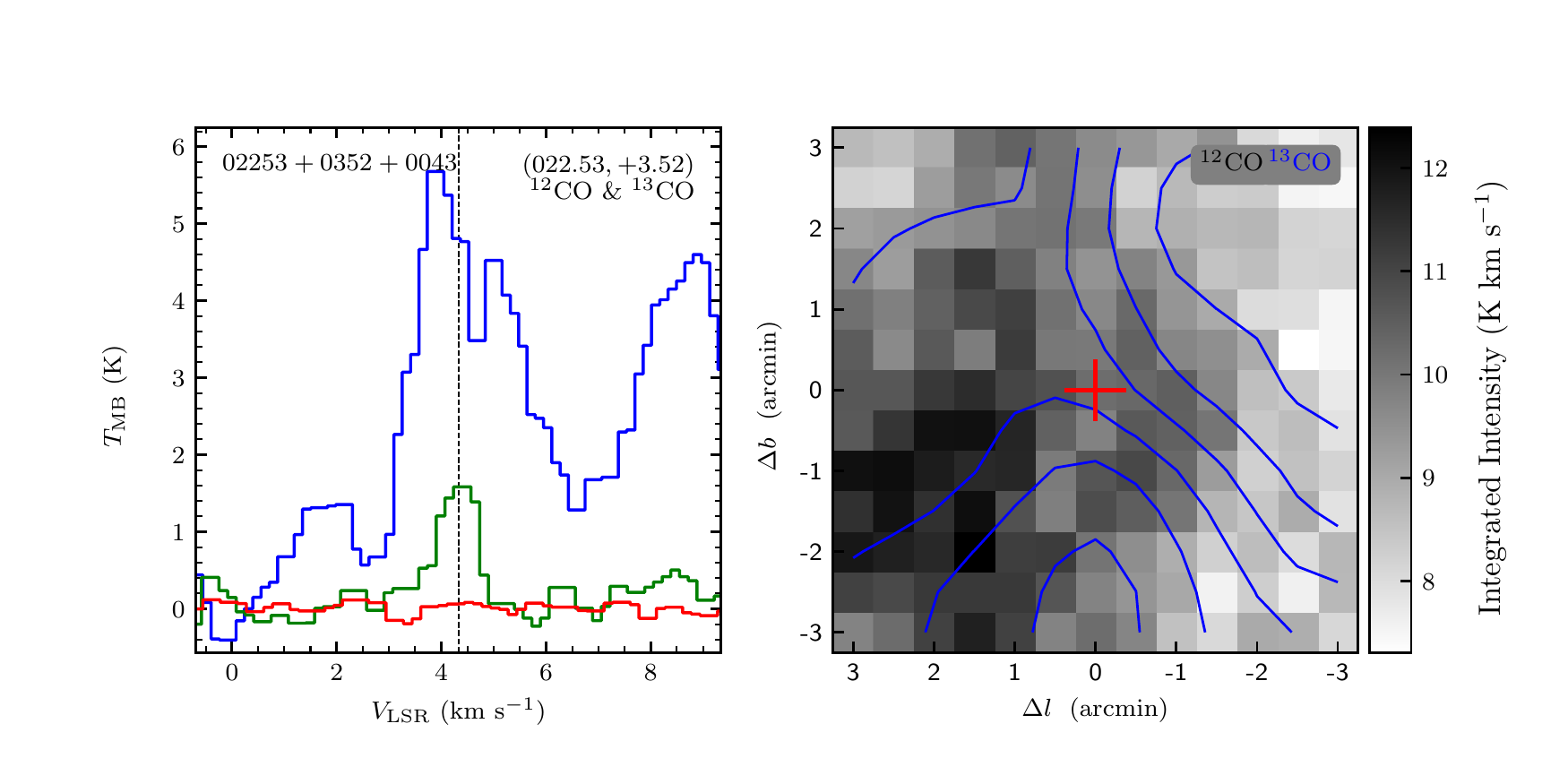}
\vspace{-0.5cm}

\includegraphics[width=9.0cm,angle=0]{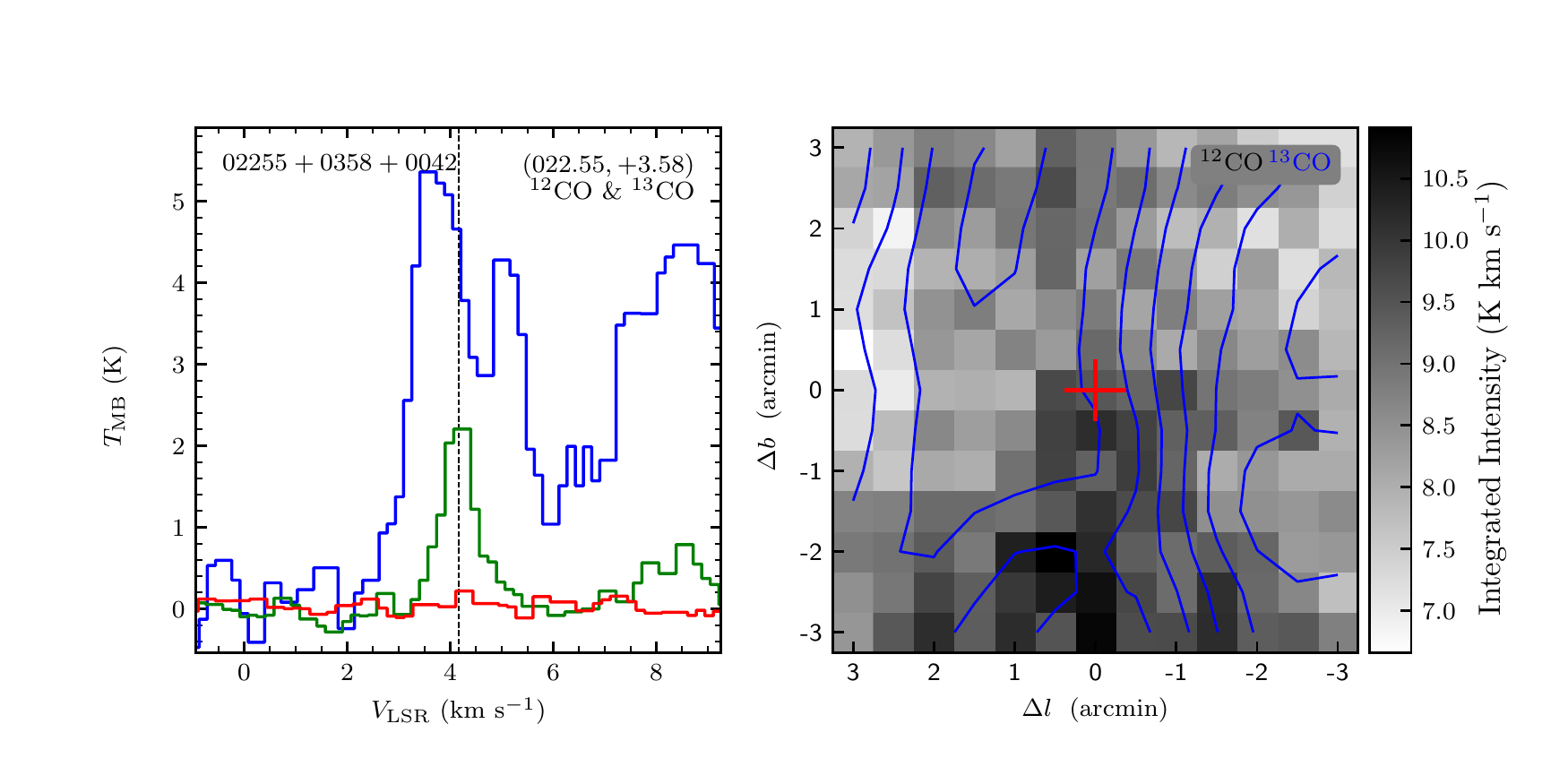}
\includegraphics[width=9.0cm,angle=0]{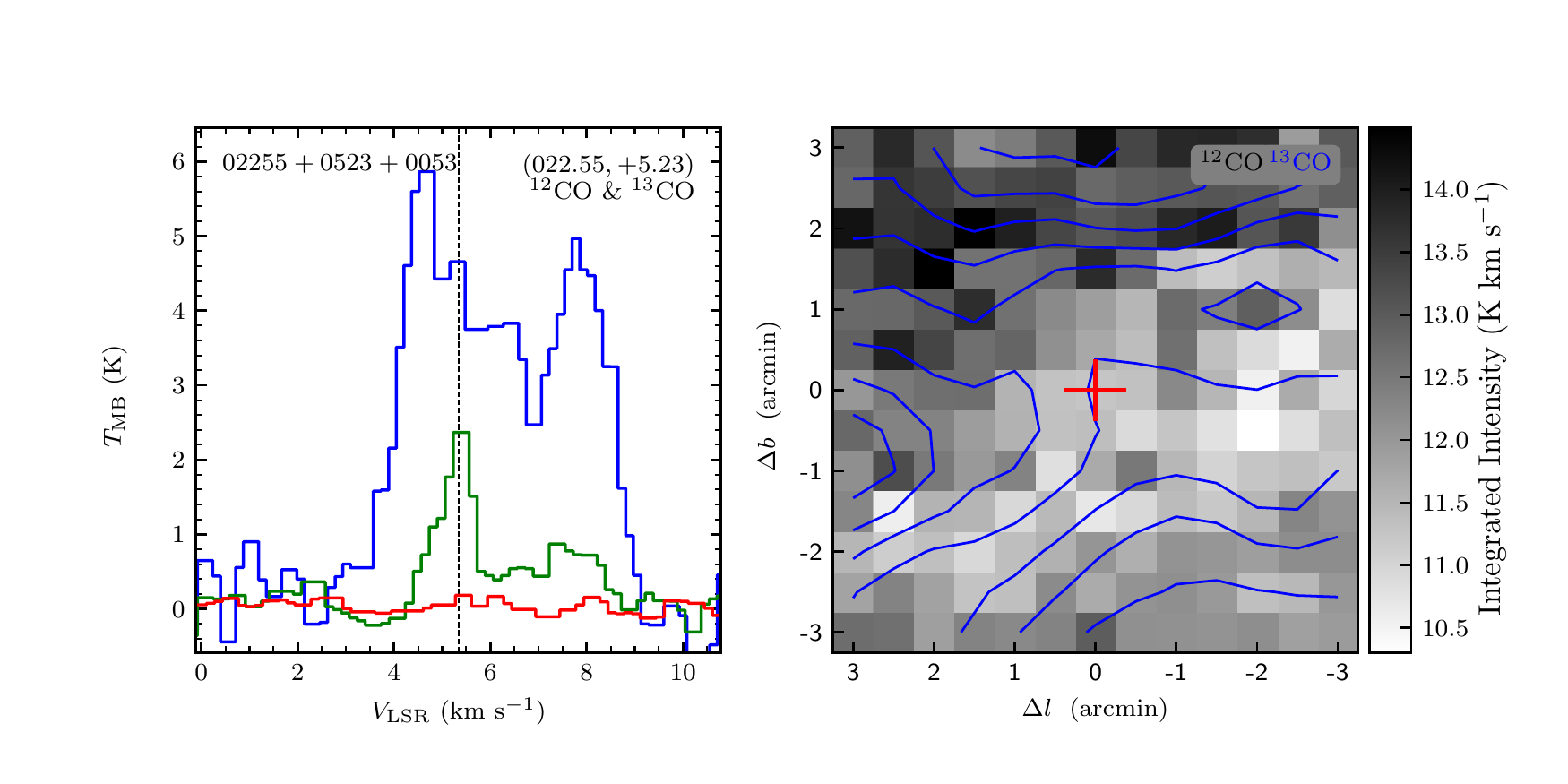}
\end{figure}
\clearpage

\begin{figure}
\includegraphics[width=9.0cm,angle=0]{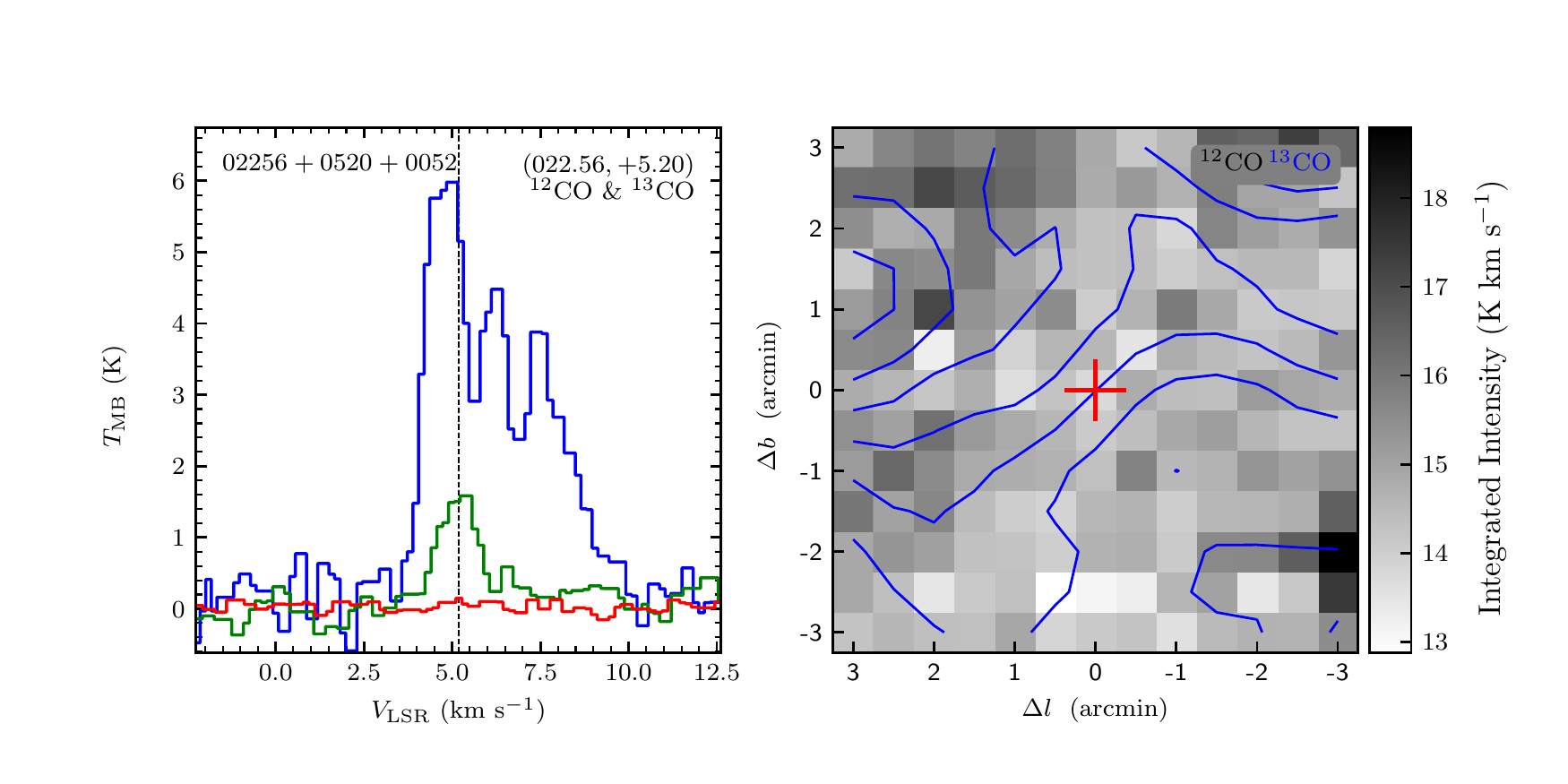}
\includegraphics[width=9.0cm,angle=0]{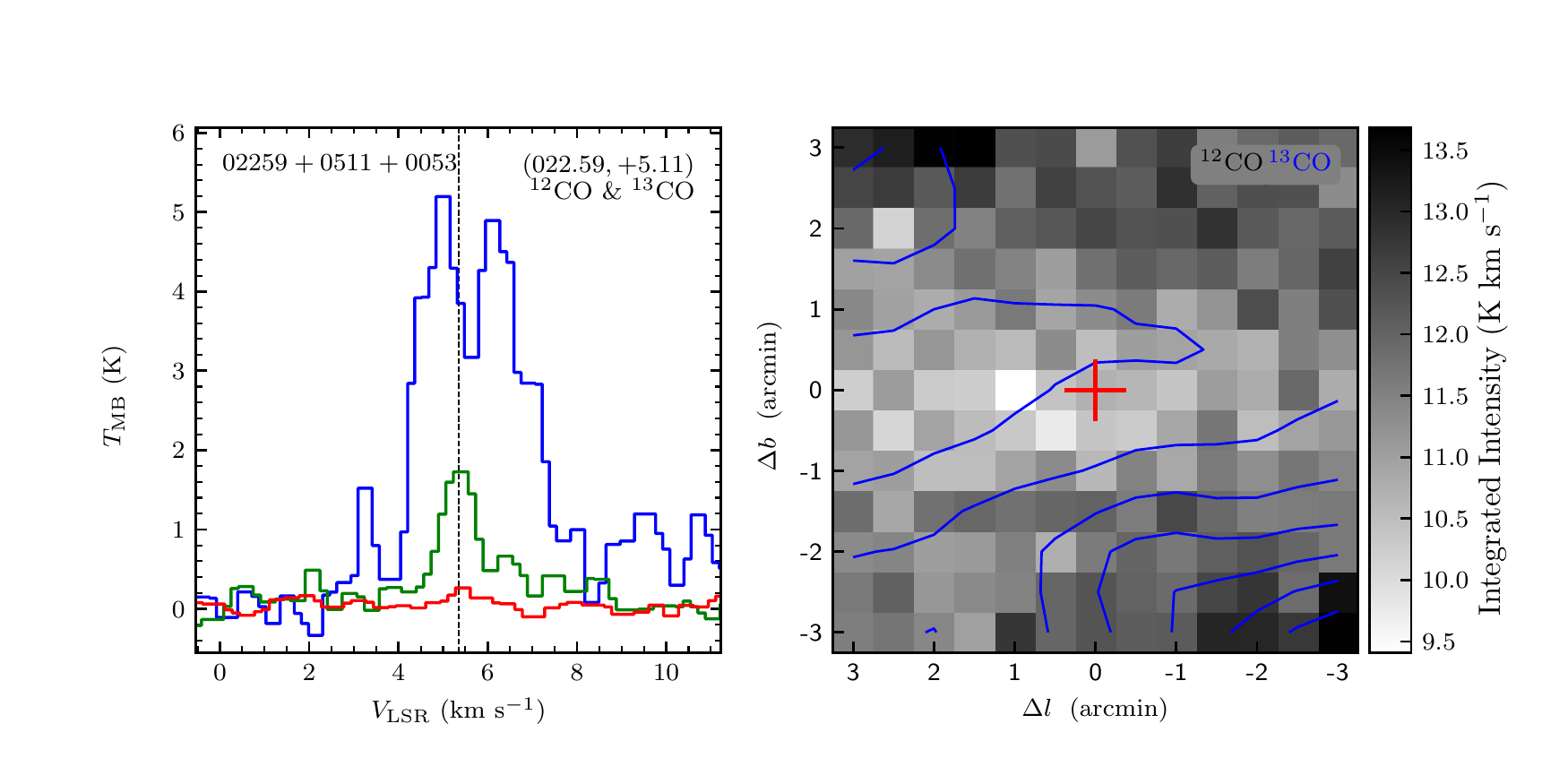}
\vspace{-0.5cm}

\includegraphics[width=9.0cm,angle=0]{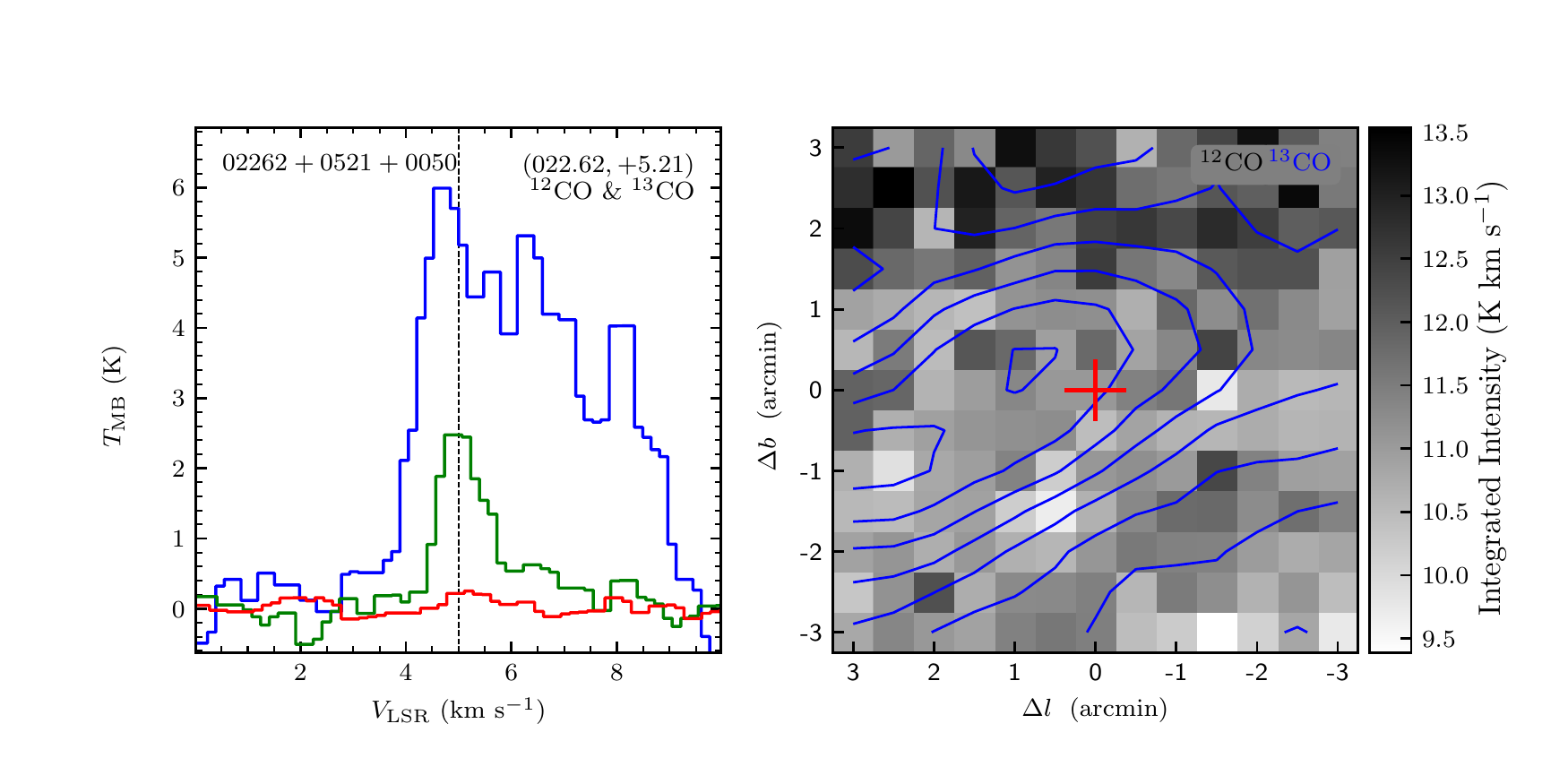}
\includegraphics[width=9.0cm,angle=0]{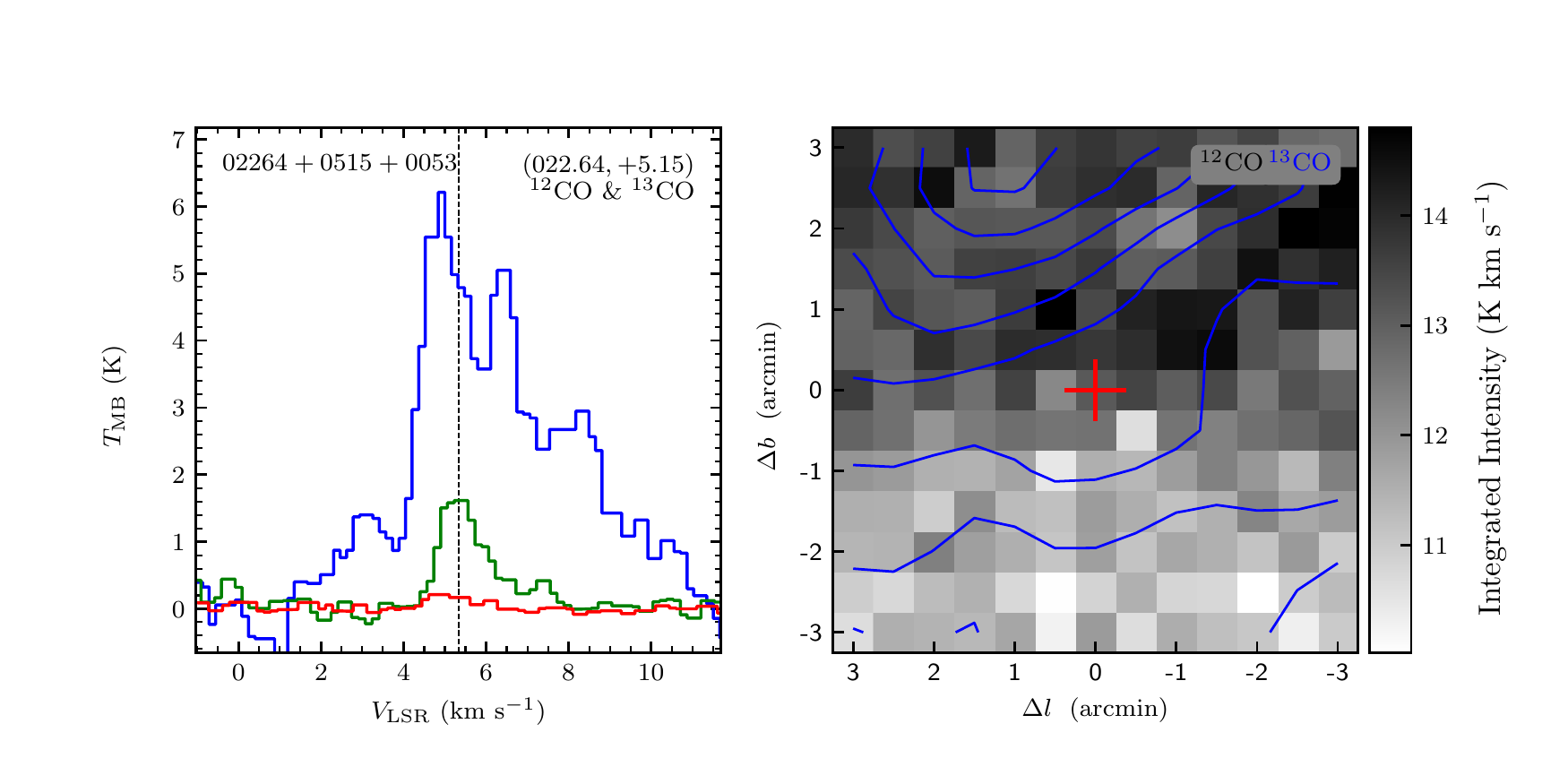}
\vspace{-0.5cm}

\includegraphics[width=9.0cm,angle=0]{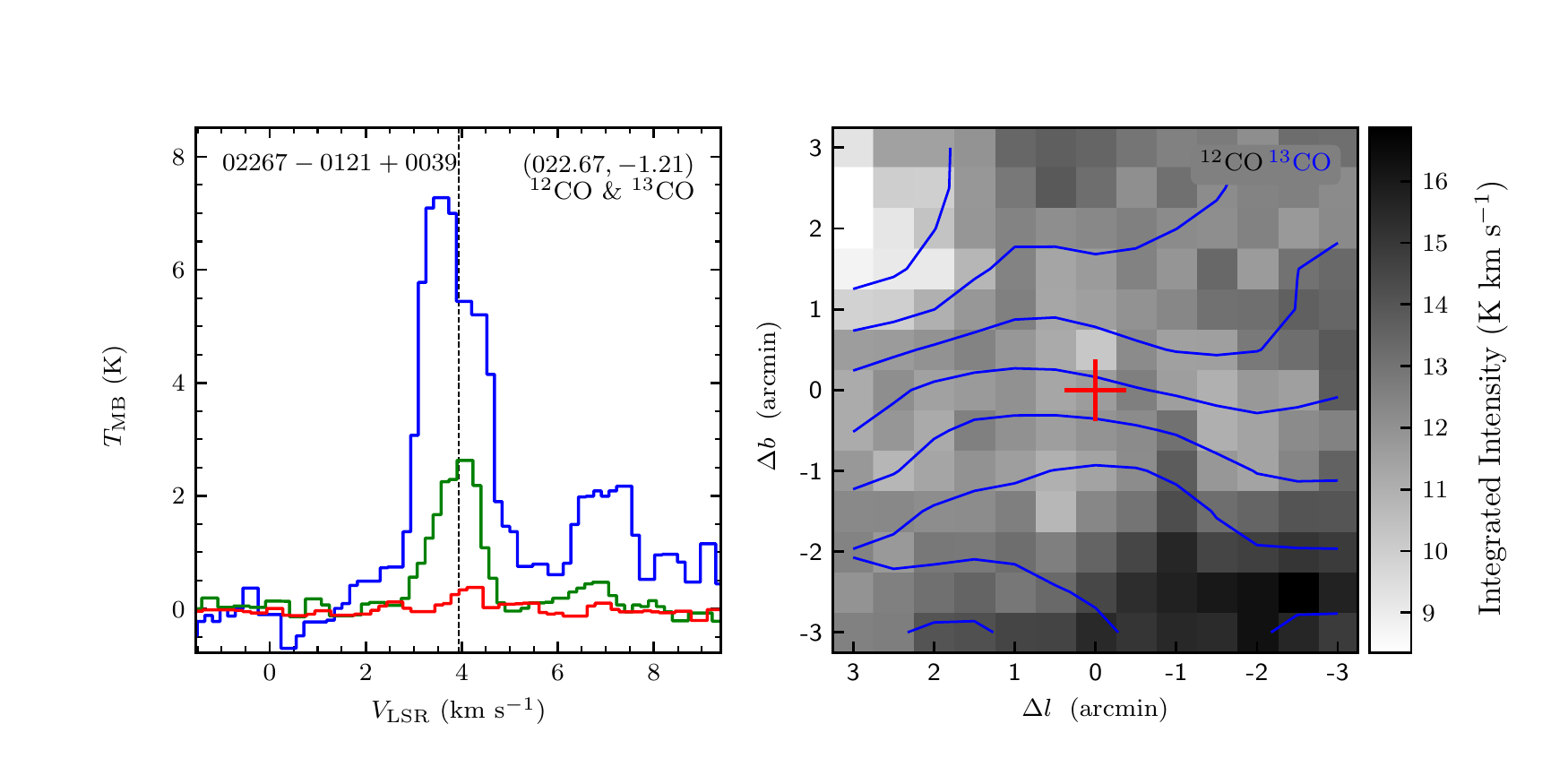}
\includegraphics[width=9.0cm,angle=0]{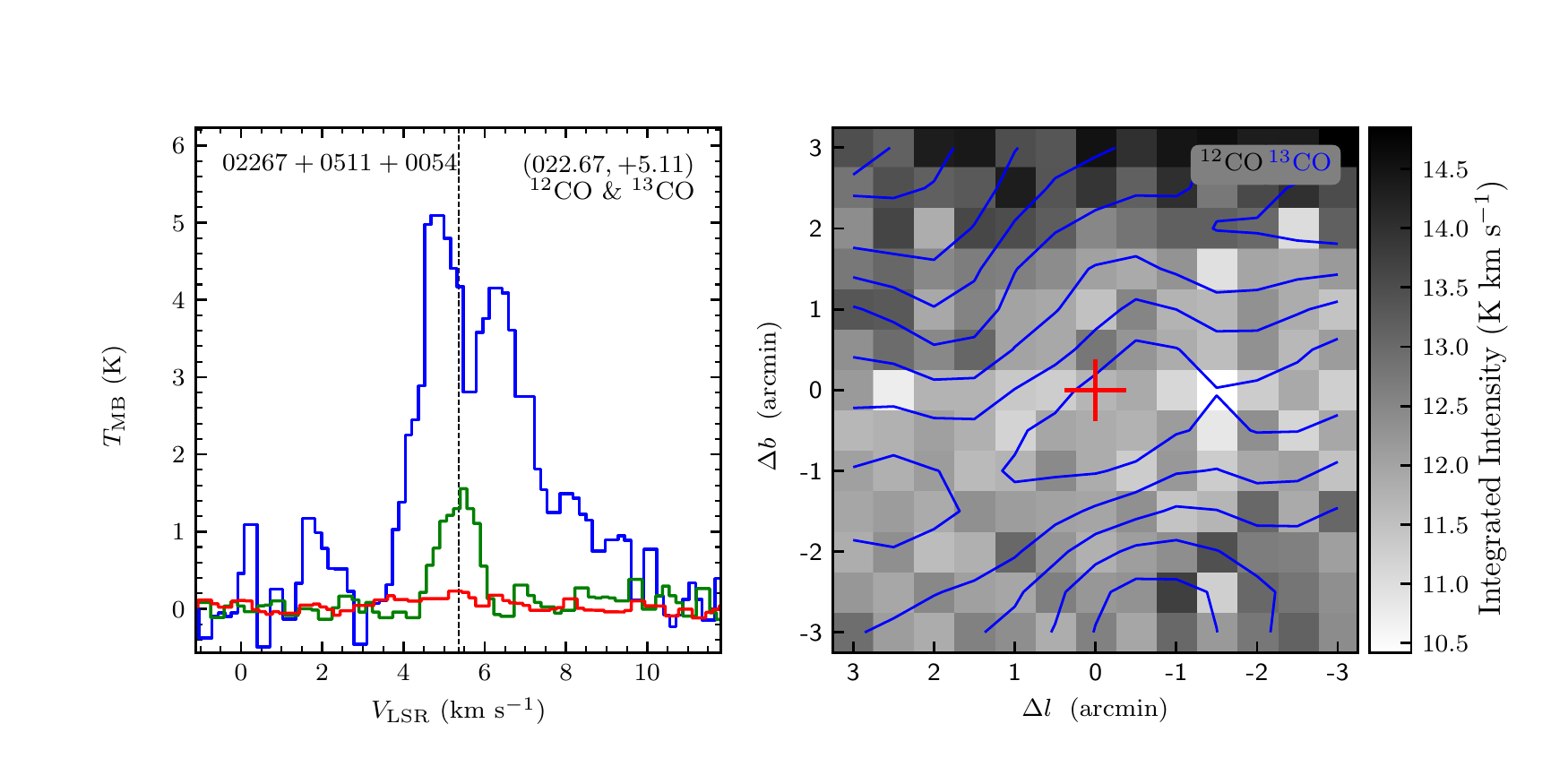}
\vspace{-0.5cm}

\includegraphics[width=9.0cm,angle=0]{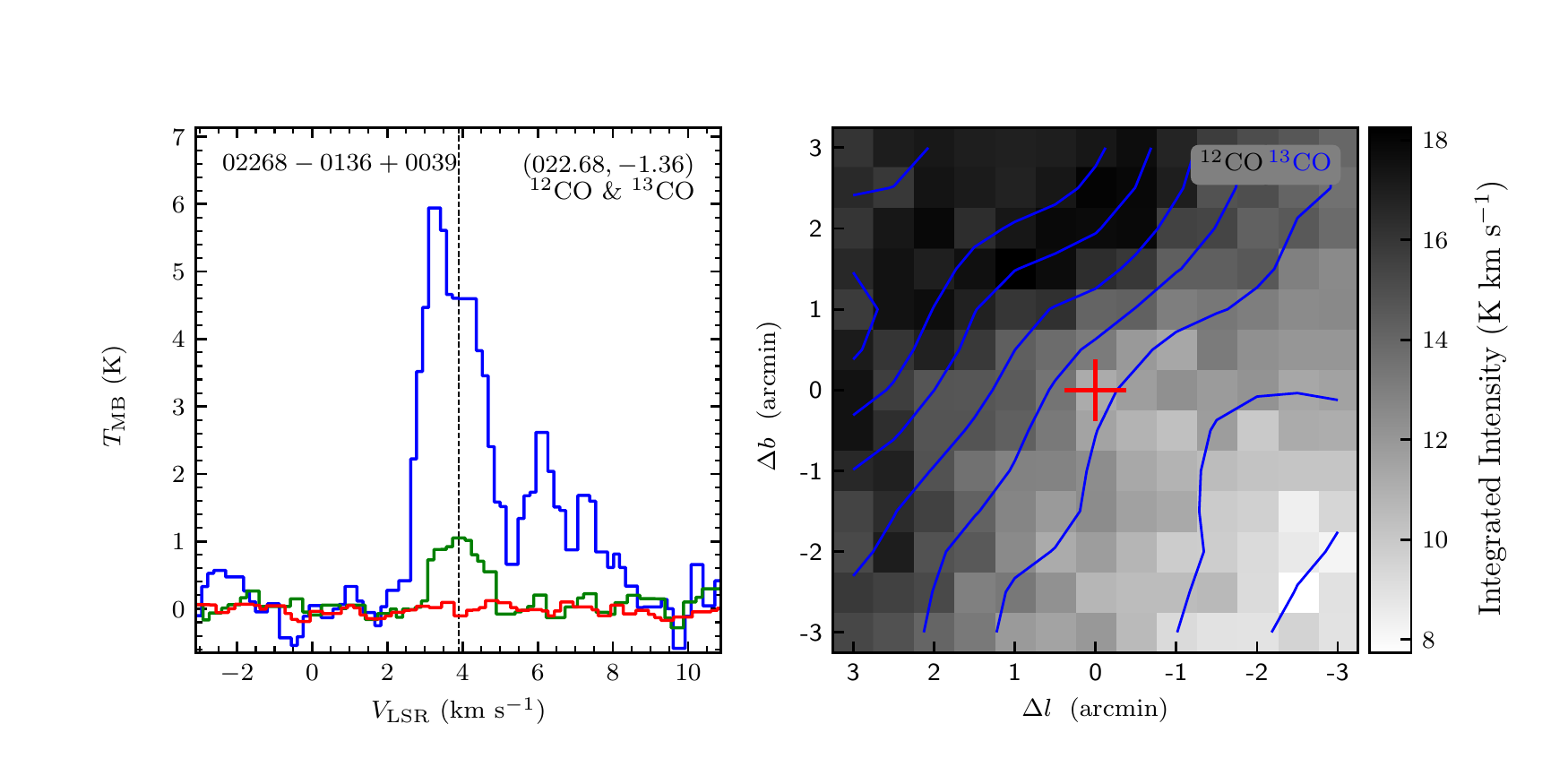}
\includegraphics[width=9.0cm,angle=0]{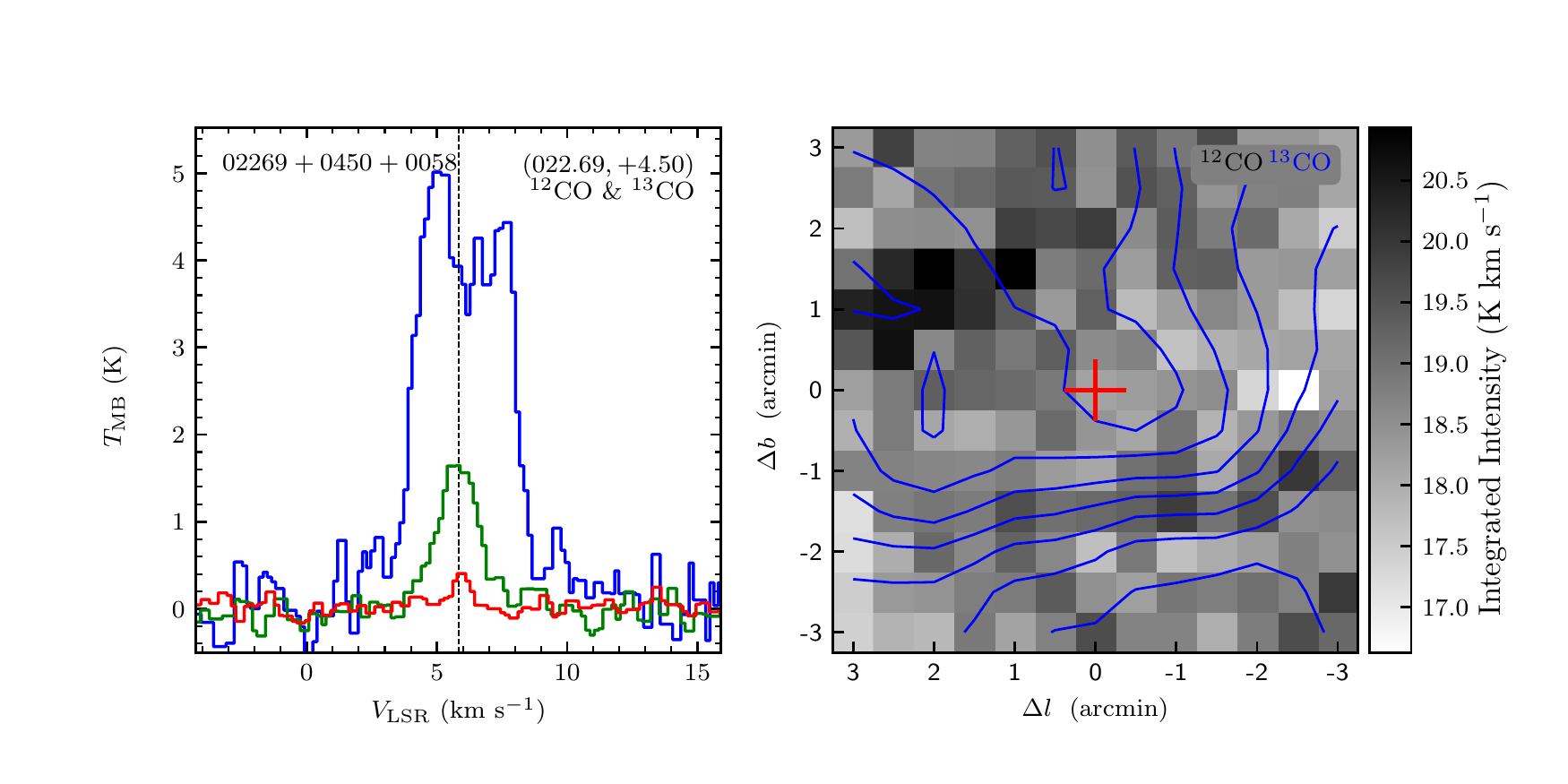}
\vspace{-0.5cm}

\includegraphics[width=9.0cm,angle=0]{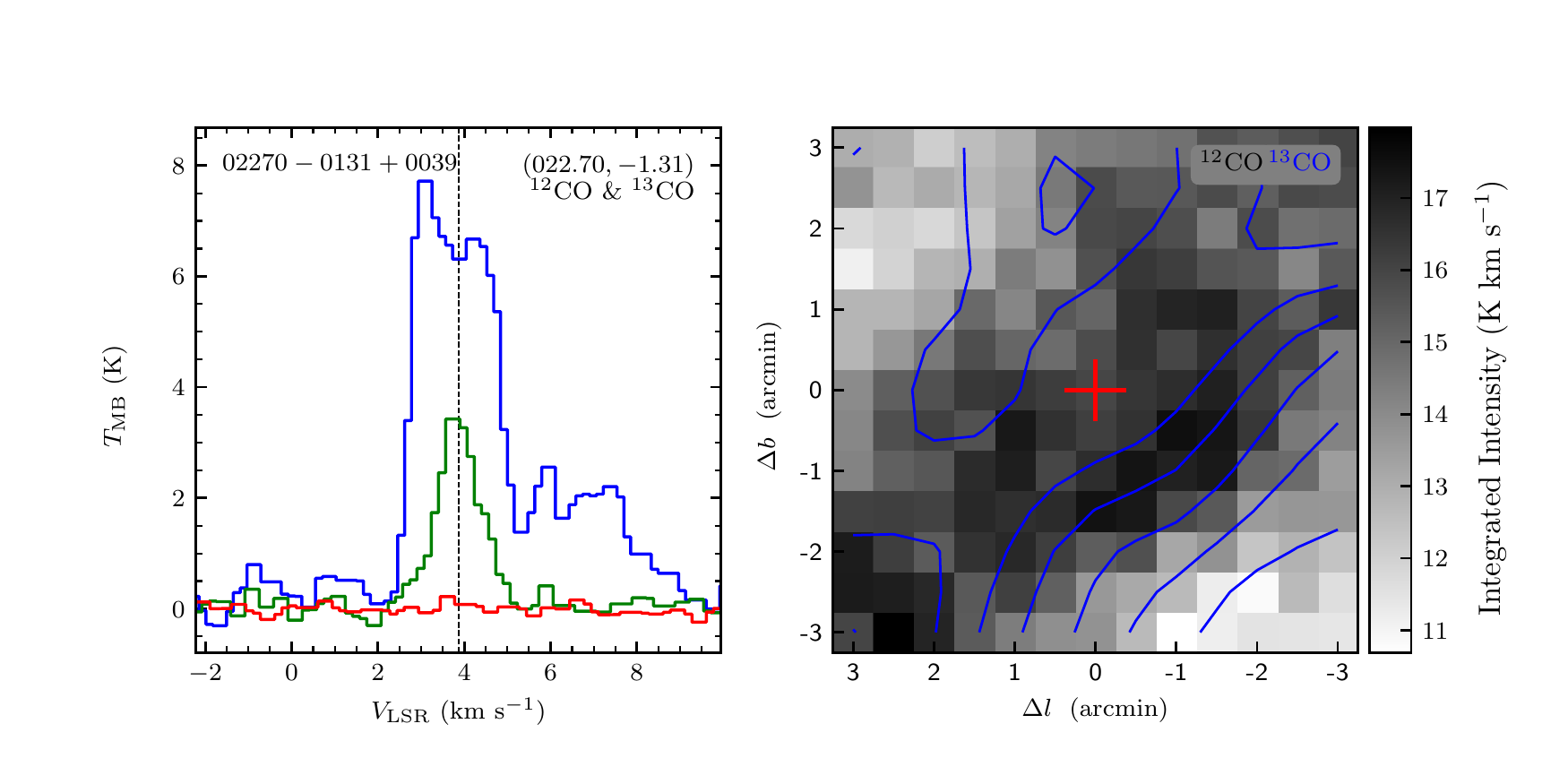}
\includegraphics[width=9.0cm,angle=0]{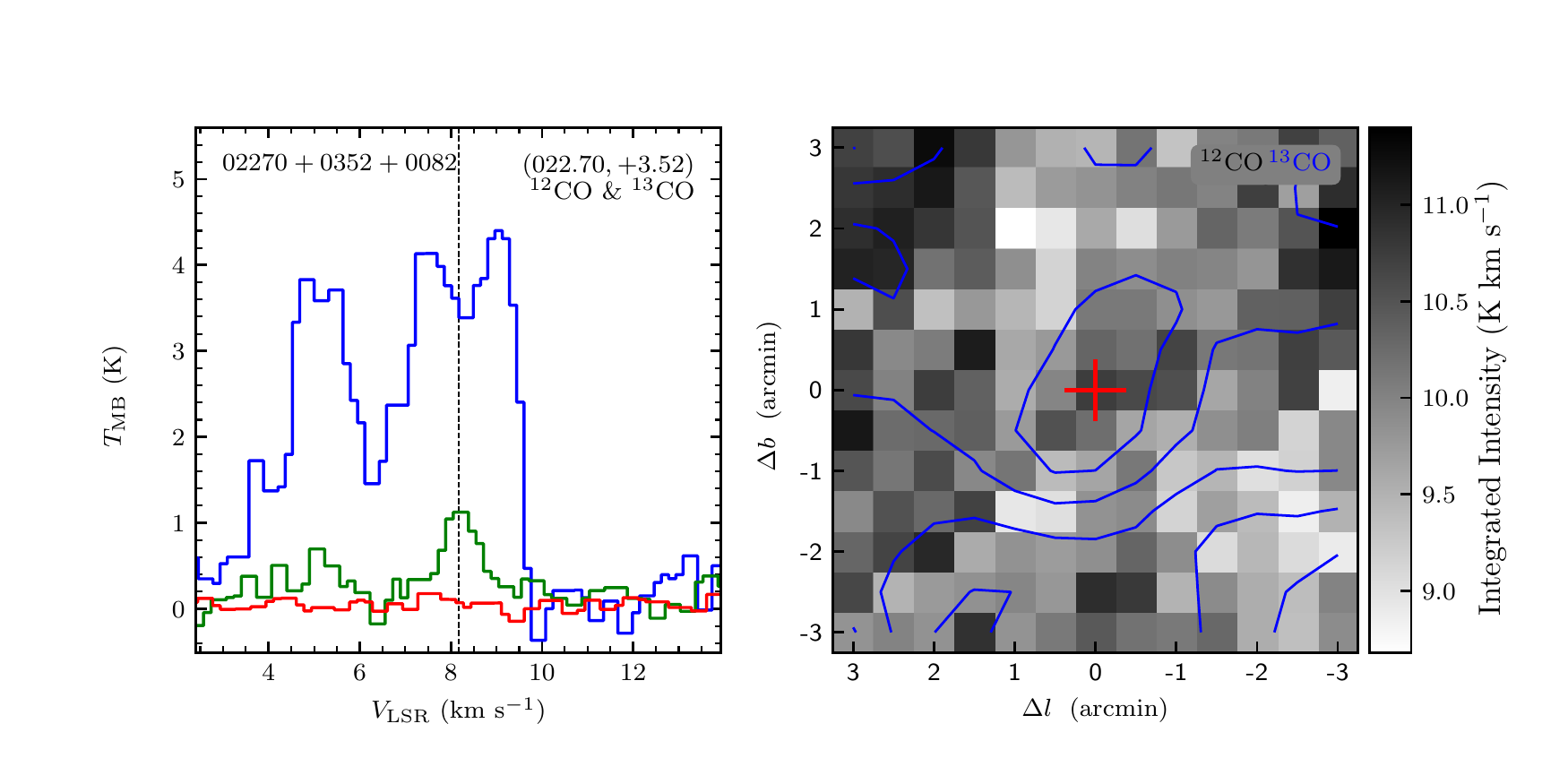}
\end{figure}
\clearpage

\begin{figure}
\includegraphics[width=9.0cm,angle=0]{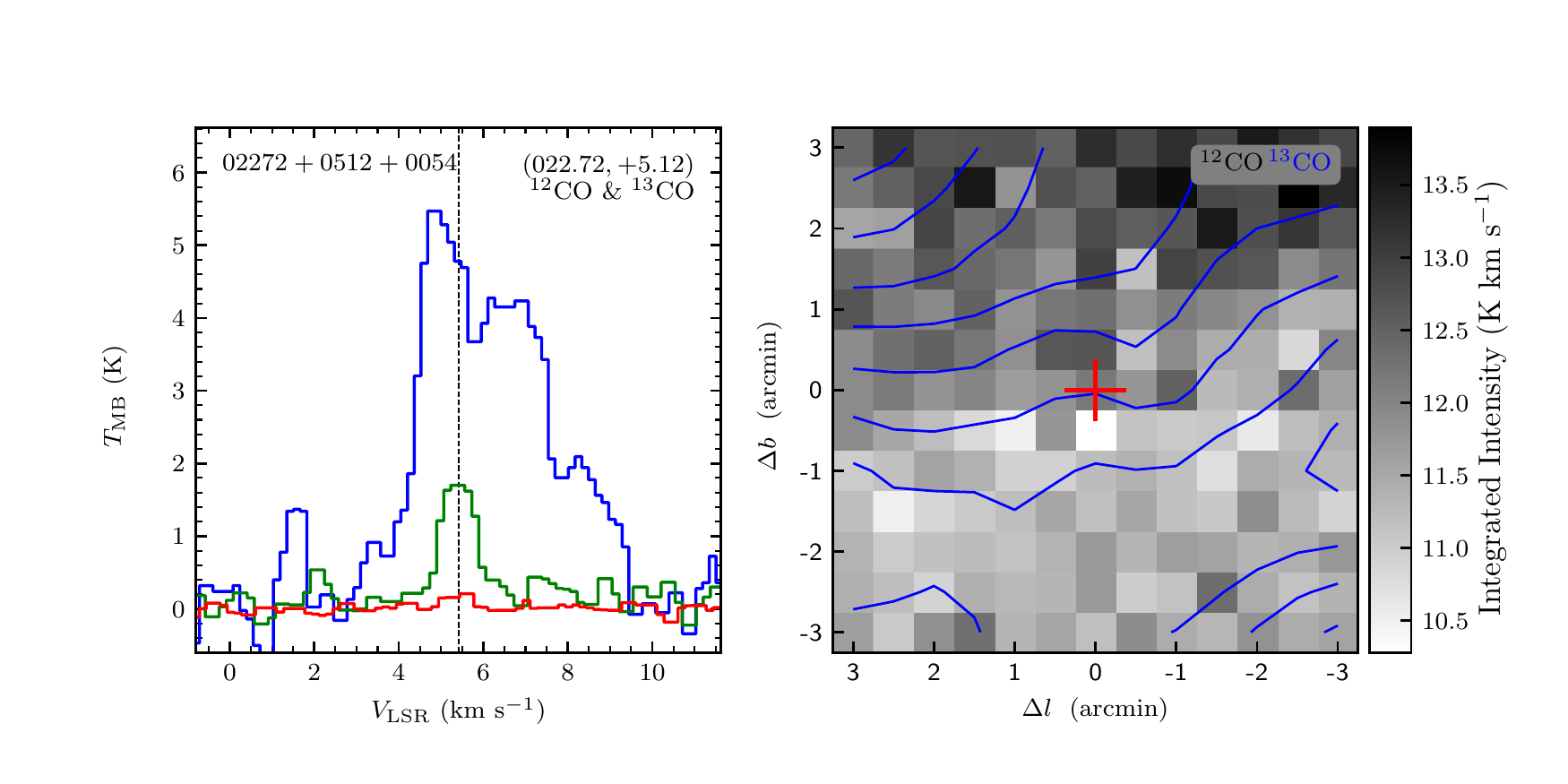}
\includegraphics[width=9.0cm,angle=0]{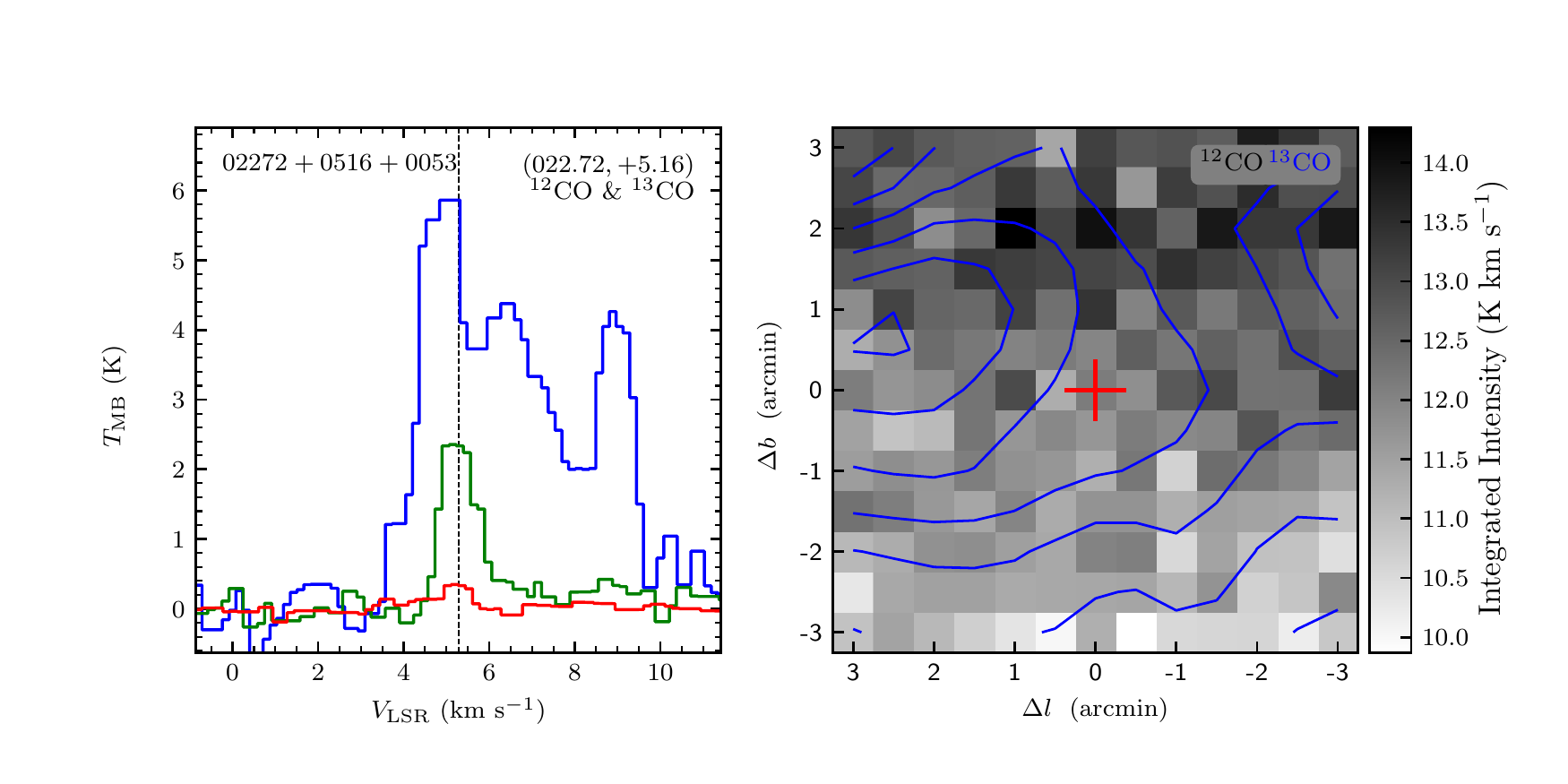}
\vspace{-0.5cm}

\includegraphics[width=9.0cm,angle=0]{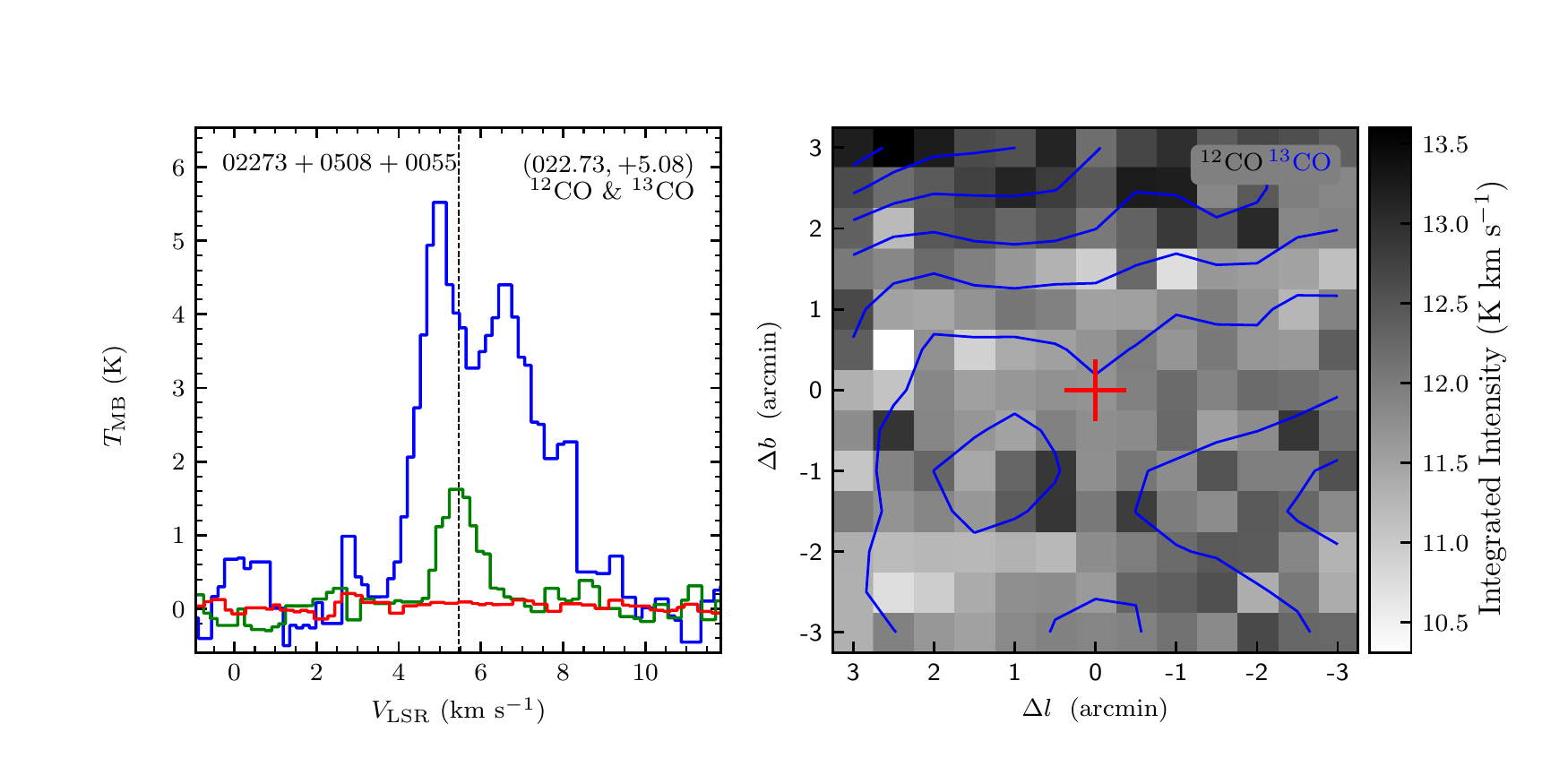}
\includegraphics[width=9.0cm,angle=0]{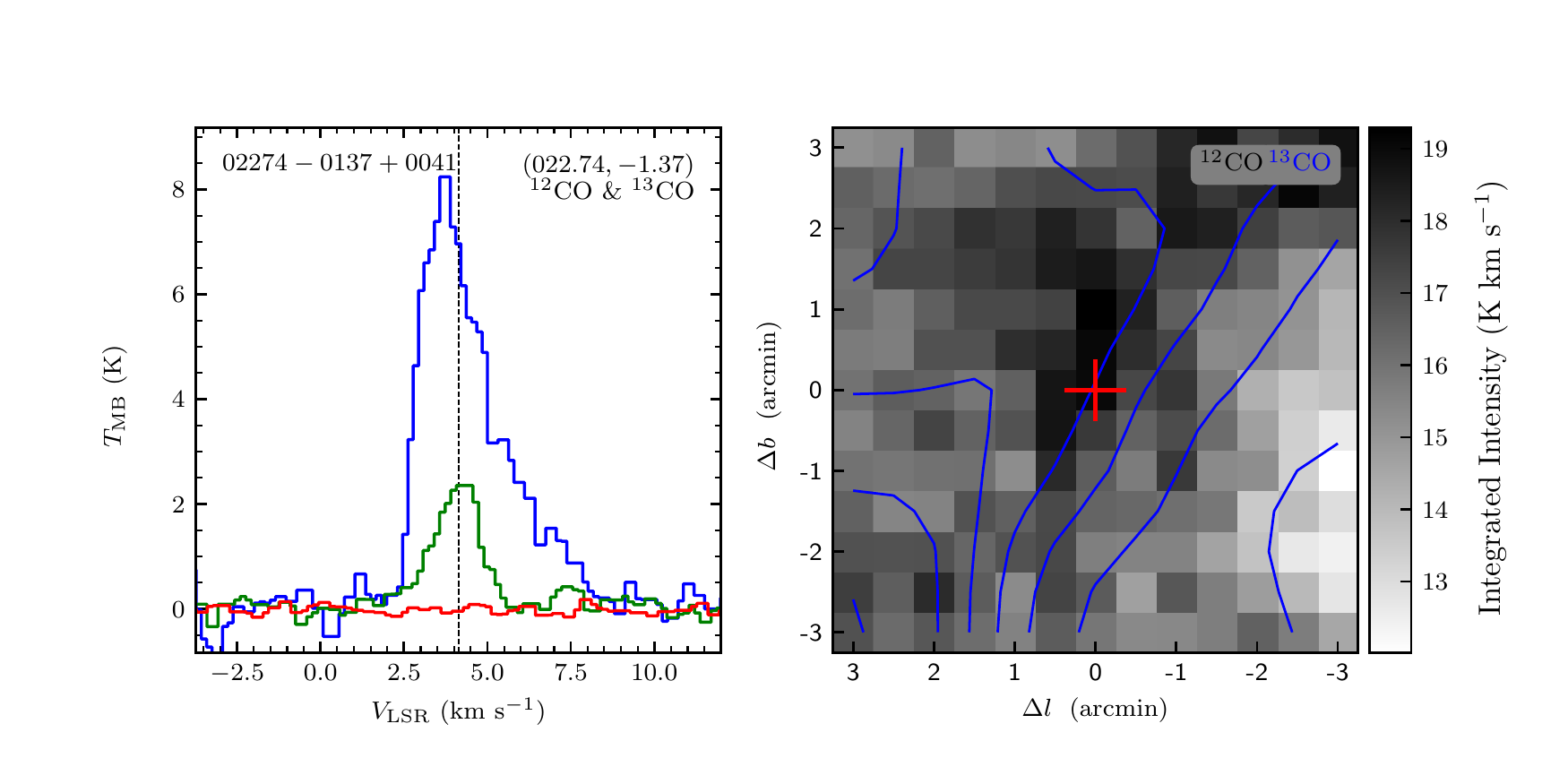}
\vspace{-0.5cm}

\includegraphics[width=9.0cm,angle=0]{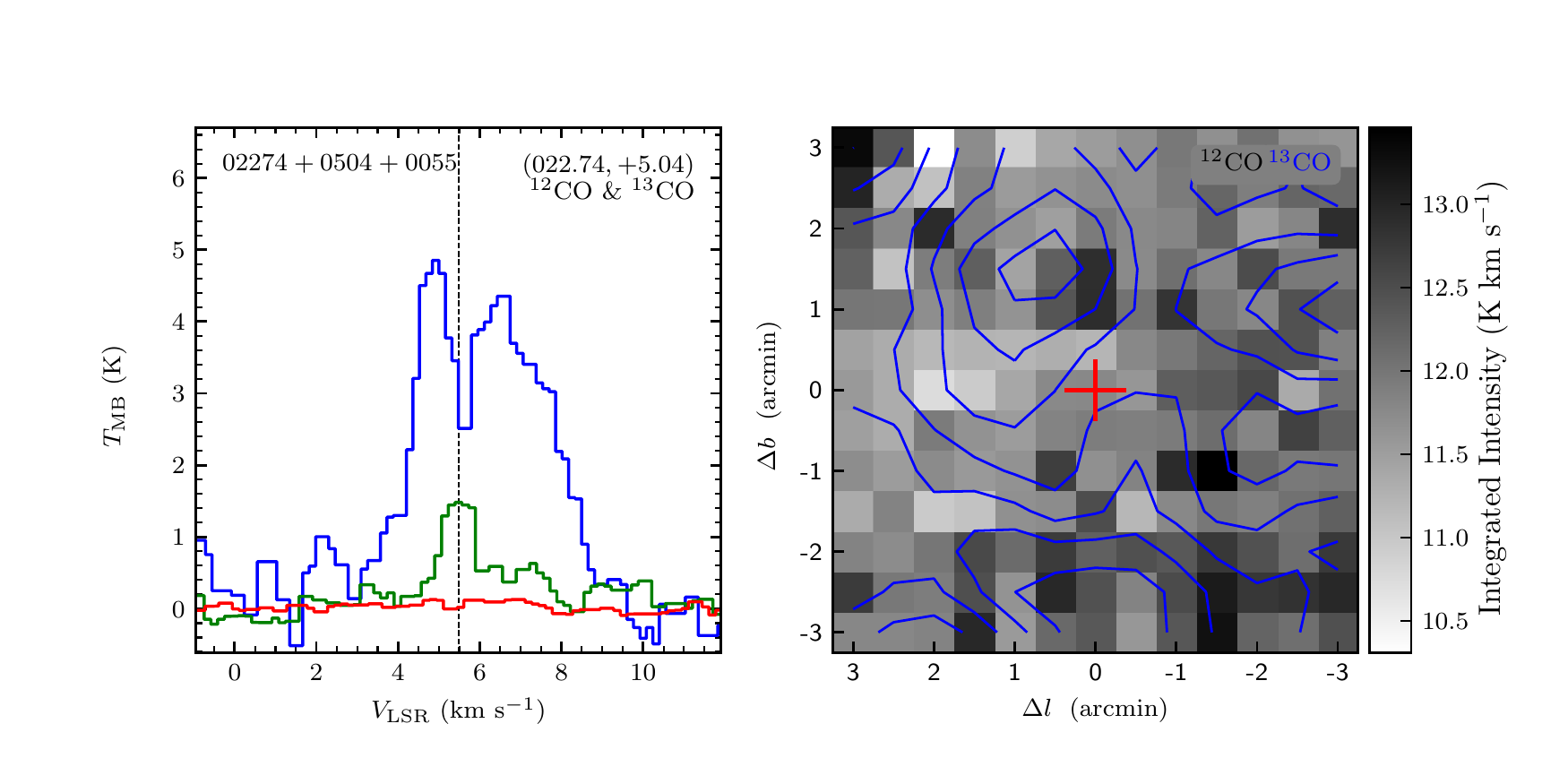}
\includegraphics[width=9.0cm,angle=0]{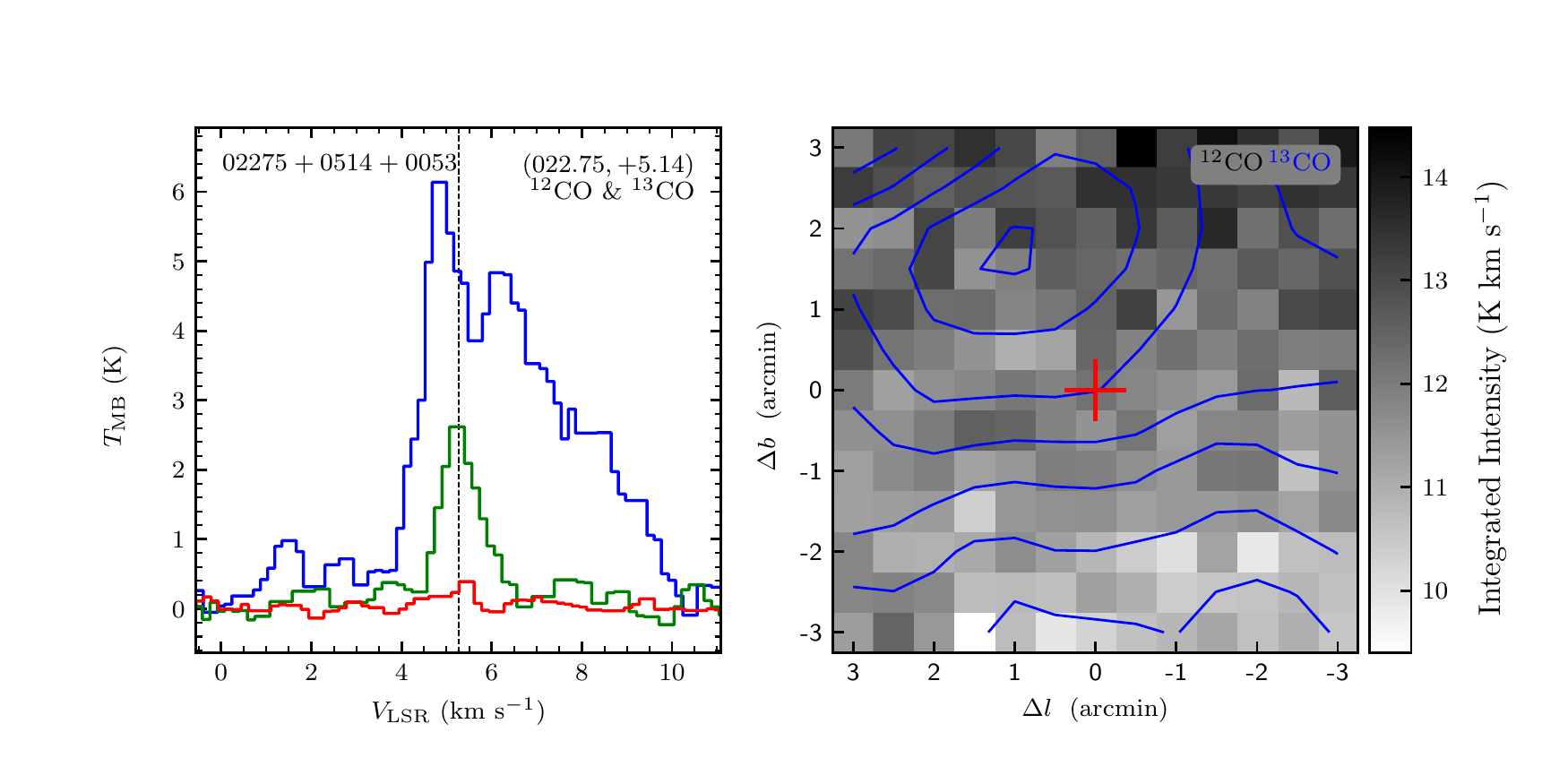}
\vspace{-0.5cm}

\includegraphics[width=9.0cm,angle=0]{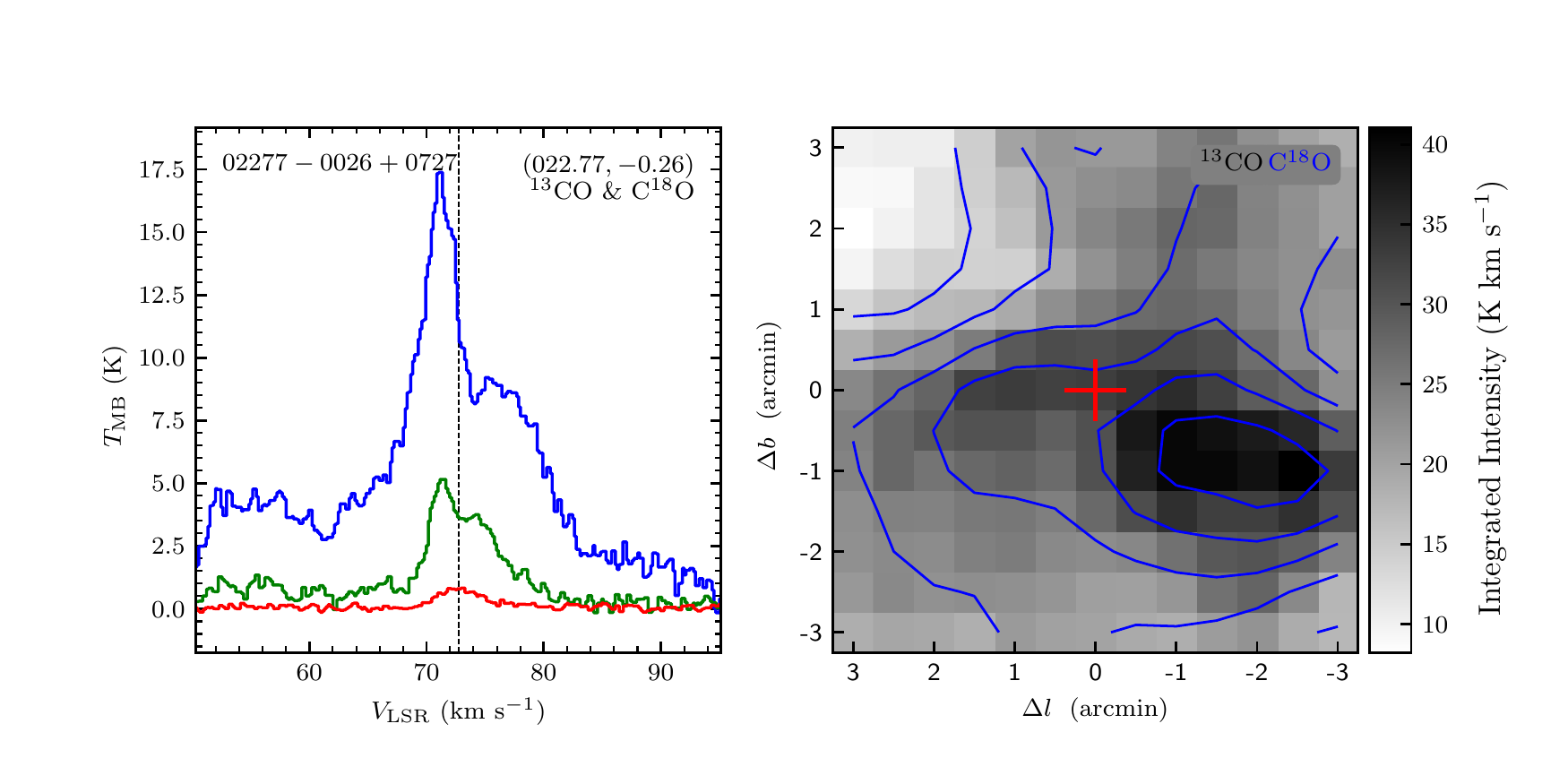}
\includegraphics[width=9.0cm,angle=0]{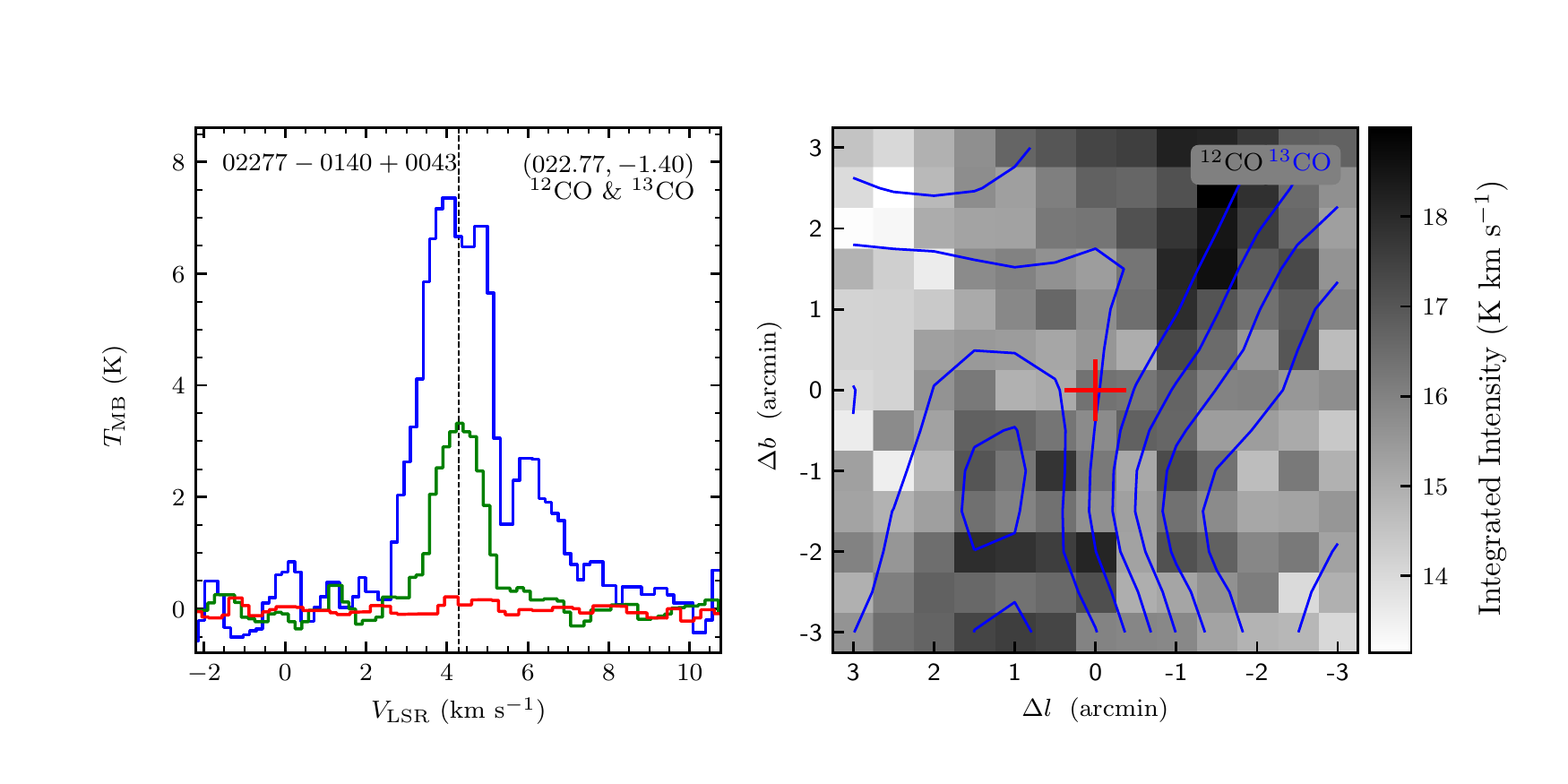}
\vspace{-0.5cm}

\includegraphics[width=9.0cm,angle=0]{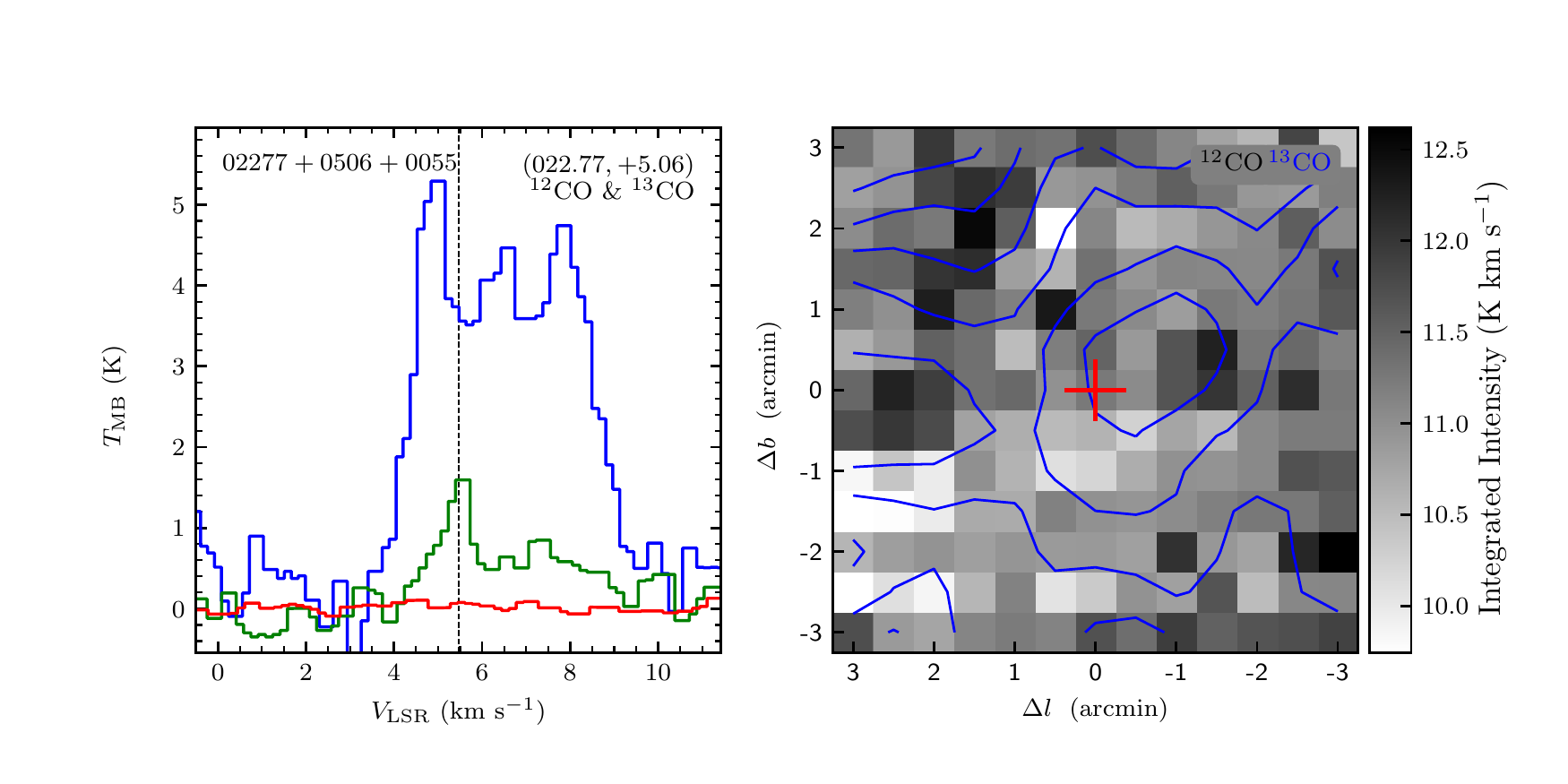}
\includegraphics[width=9.0cm,angle=0]{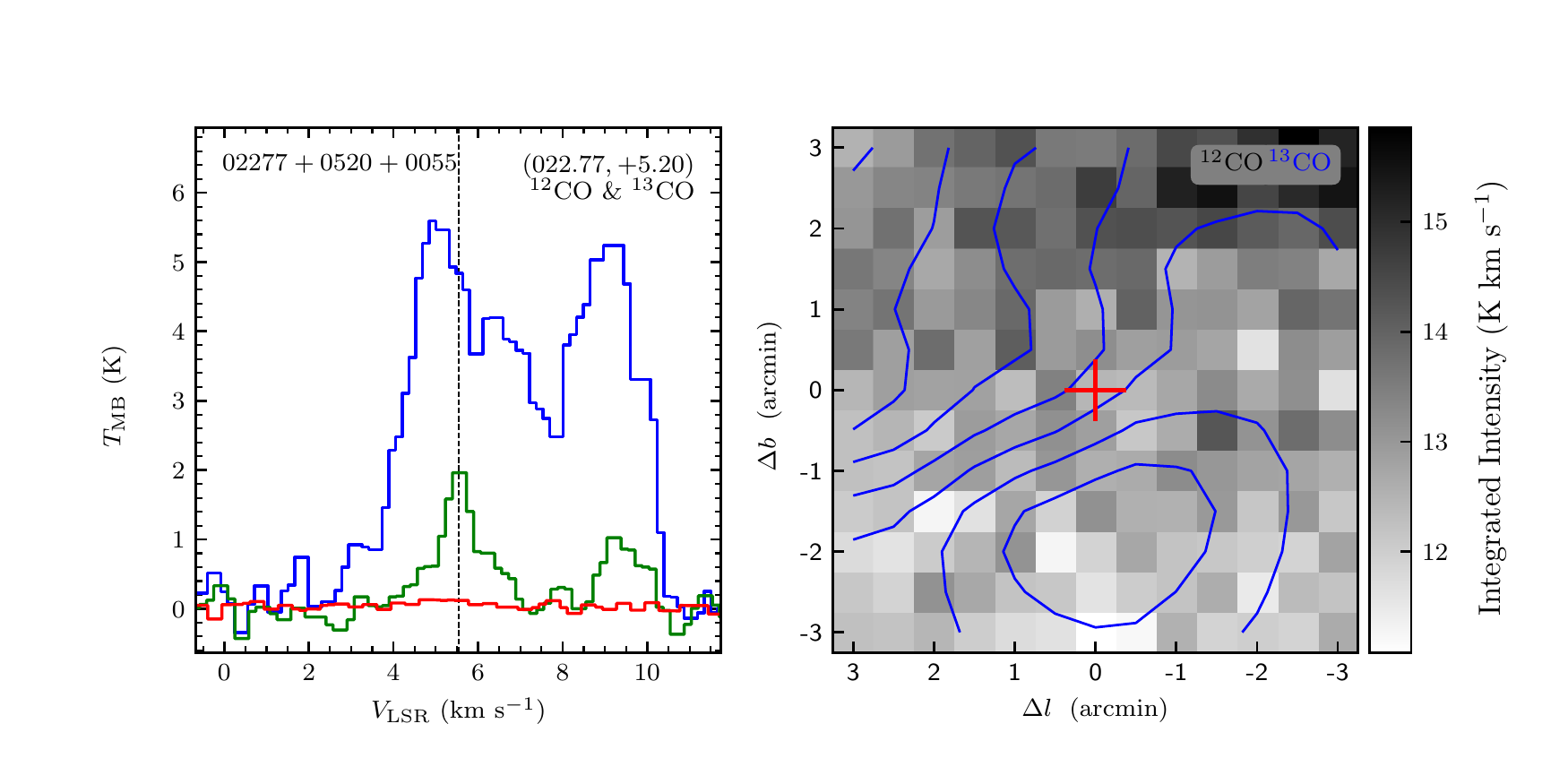}
\end{figure}
\clearpage

\begin{figure}
\includegraphics[width=9.0cm,angle=0]{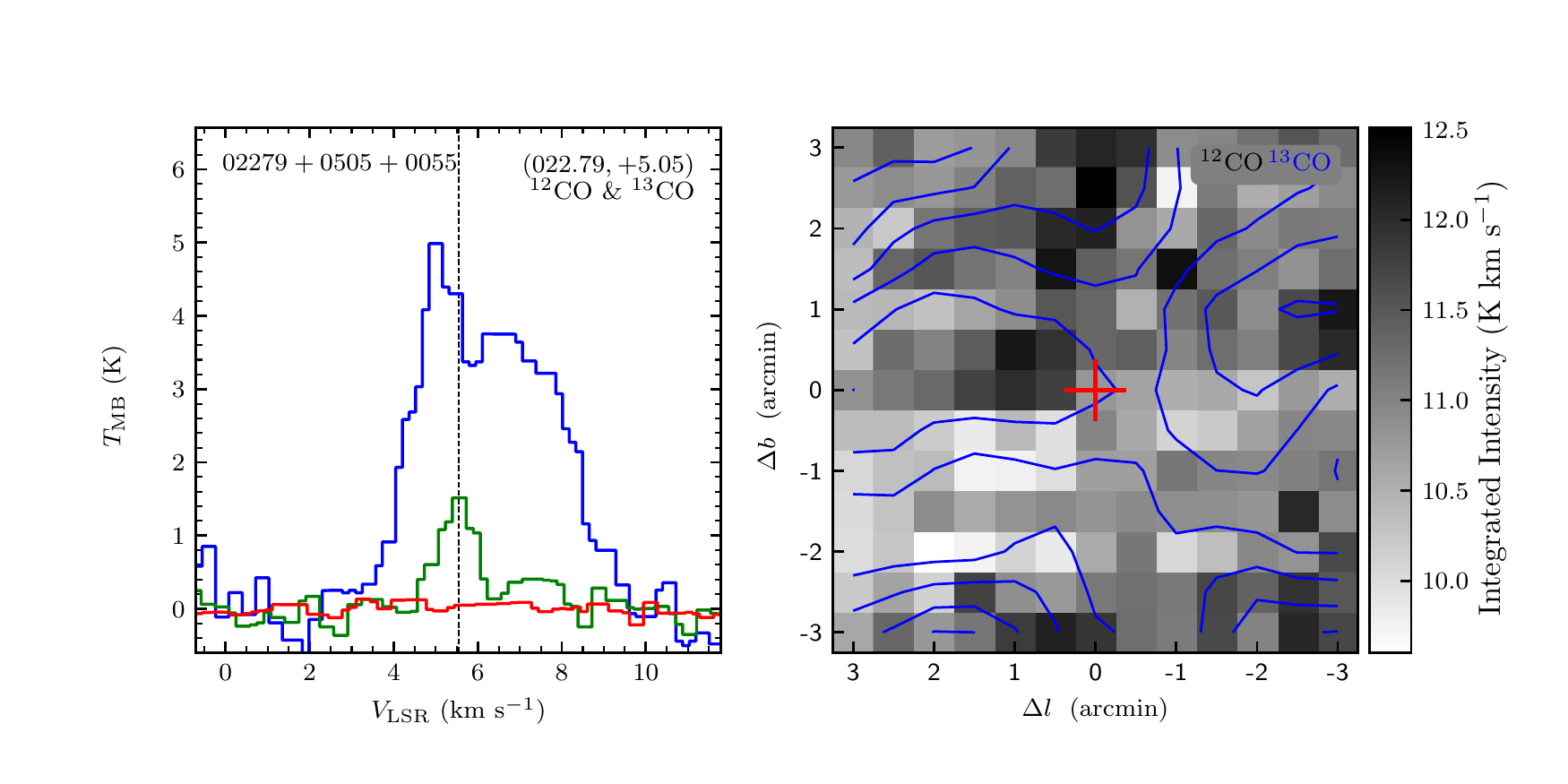}
\includegraphics[width=9.0cm,angle=0]{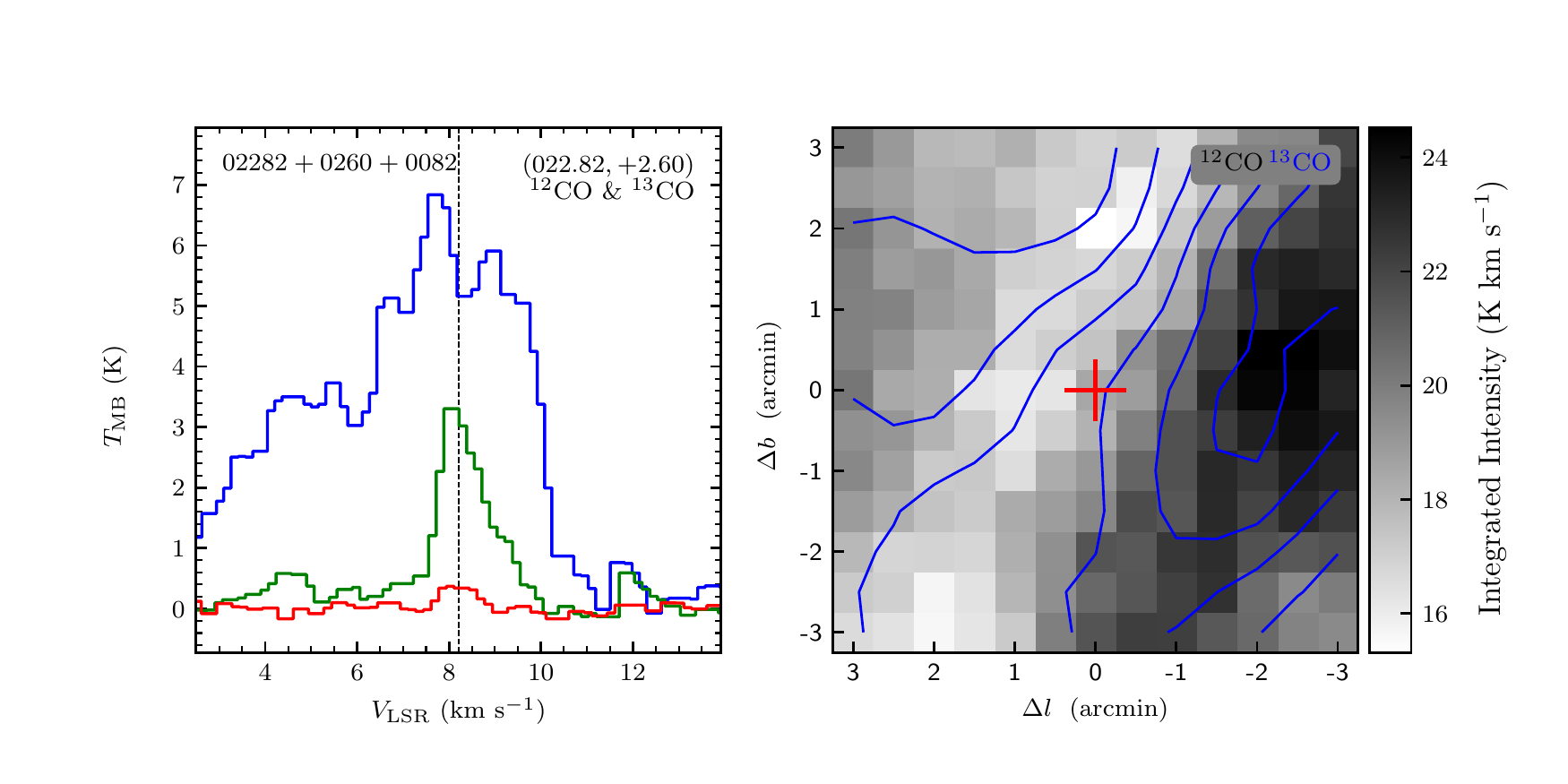}
\vspace{-0.5cm}

\includegraphics[width=9.0cm,angle=0]{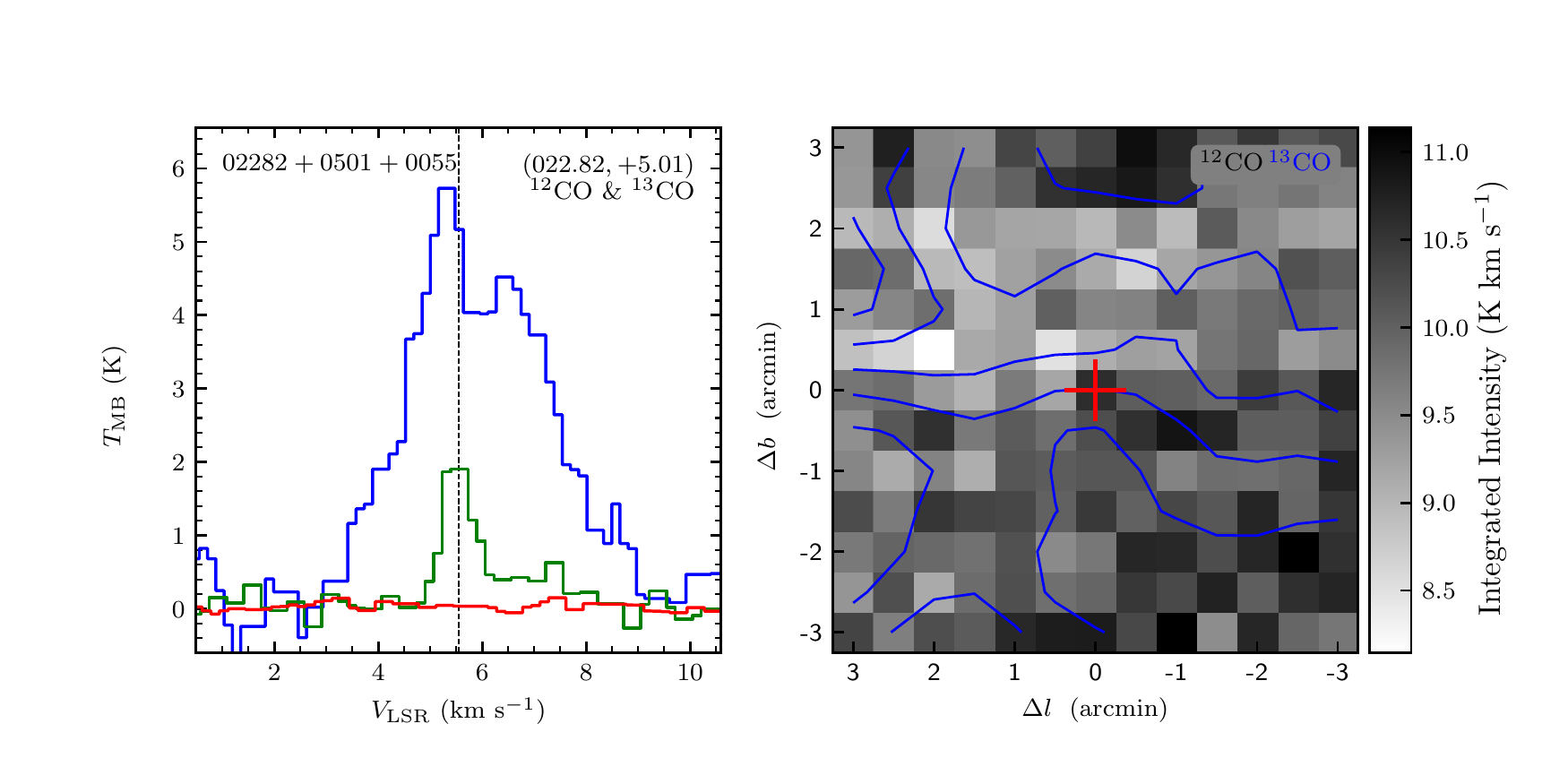}
\includegraphics[width=9.0cm,angle=0]{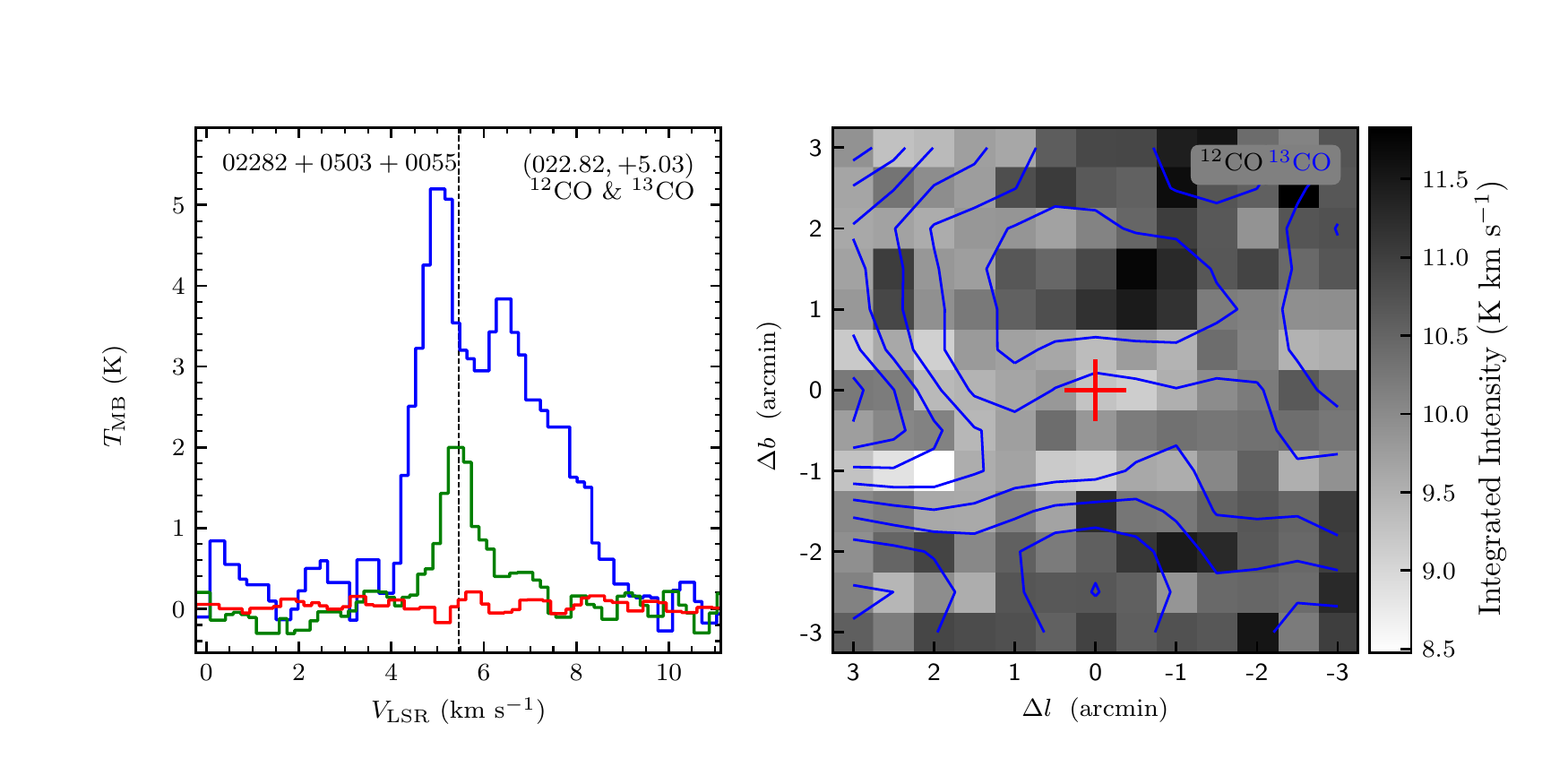}
\vspace{-0.5cm}

\includegraphics[width=9.0cm,angle=0]{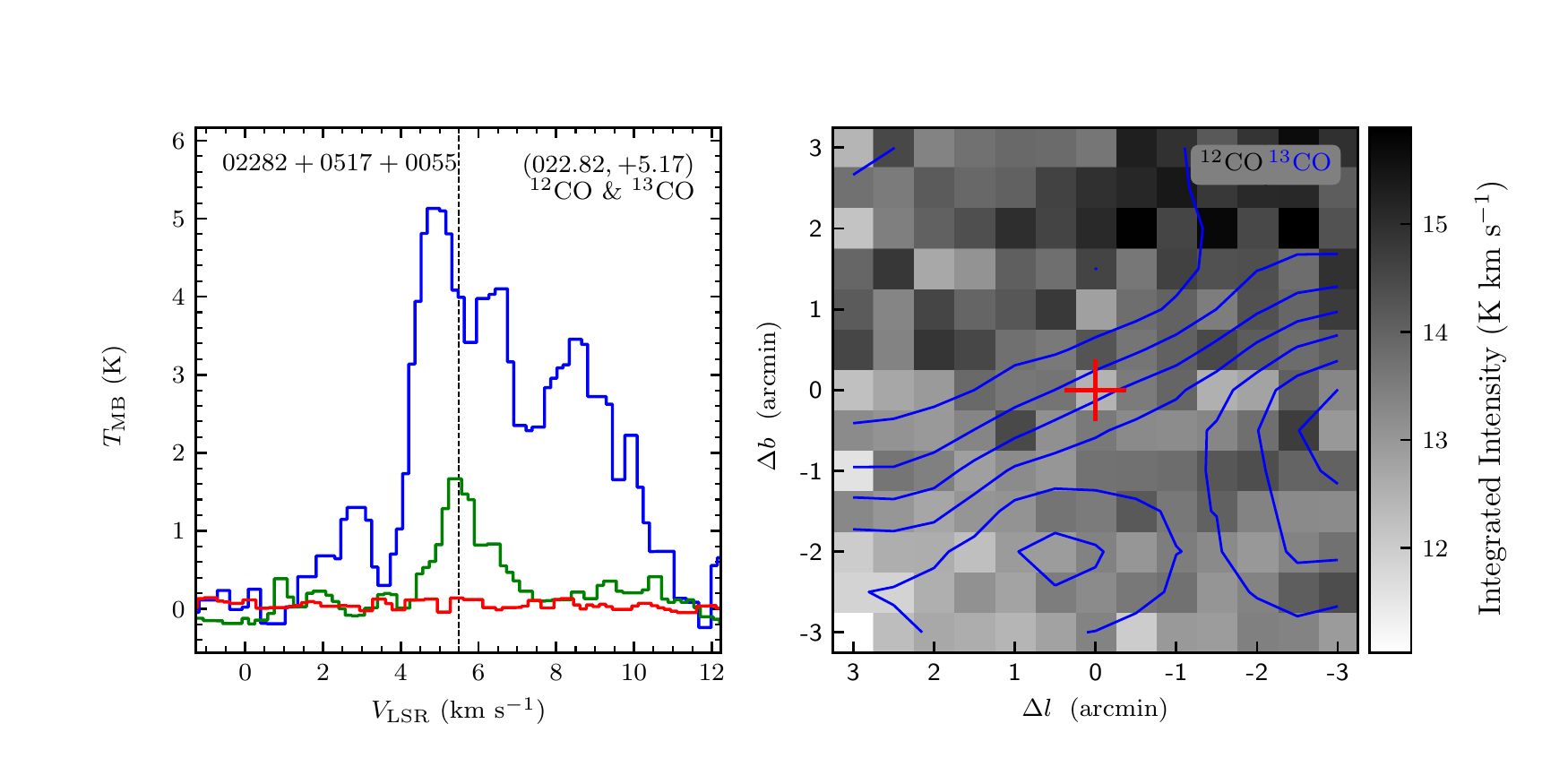}
\includegraphics[width=9.0cm,angle=0]{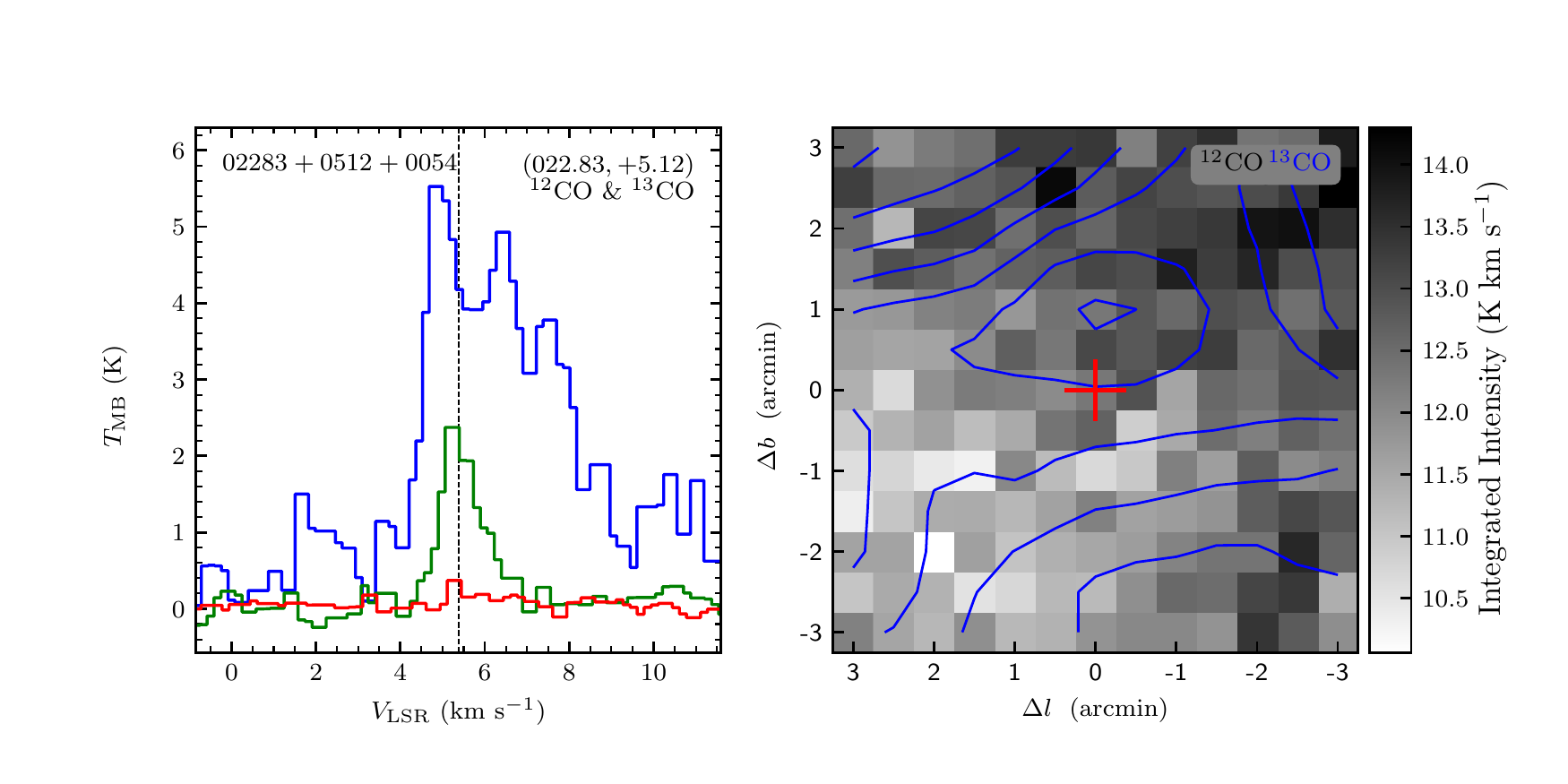}
\vspace{-0.5cm}

\includegraphics[width=9.0cm,angle=0]{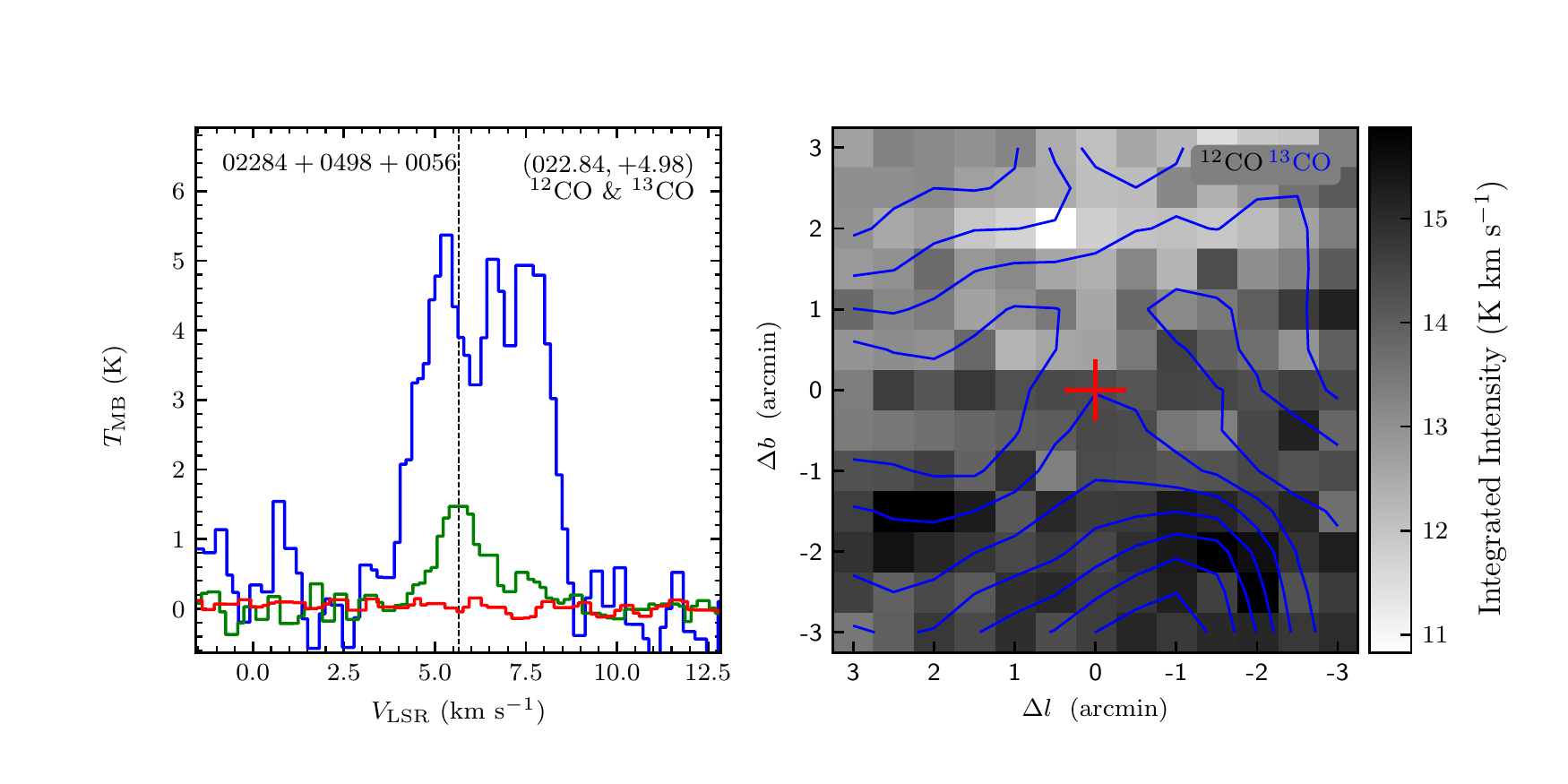}
\includegraphics[width=9.0cm,angle=0]{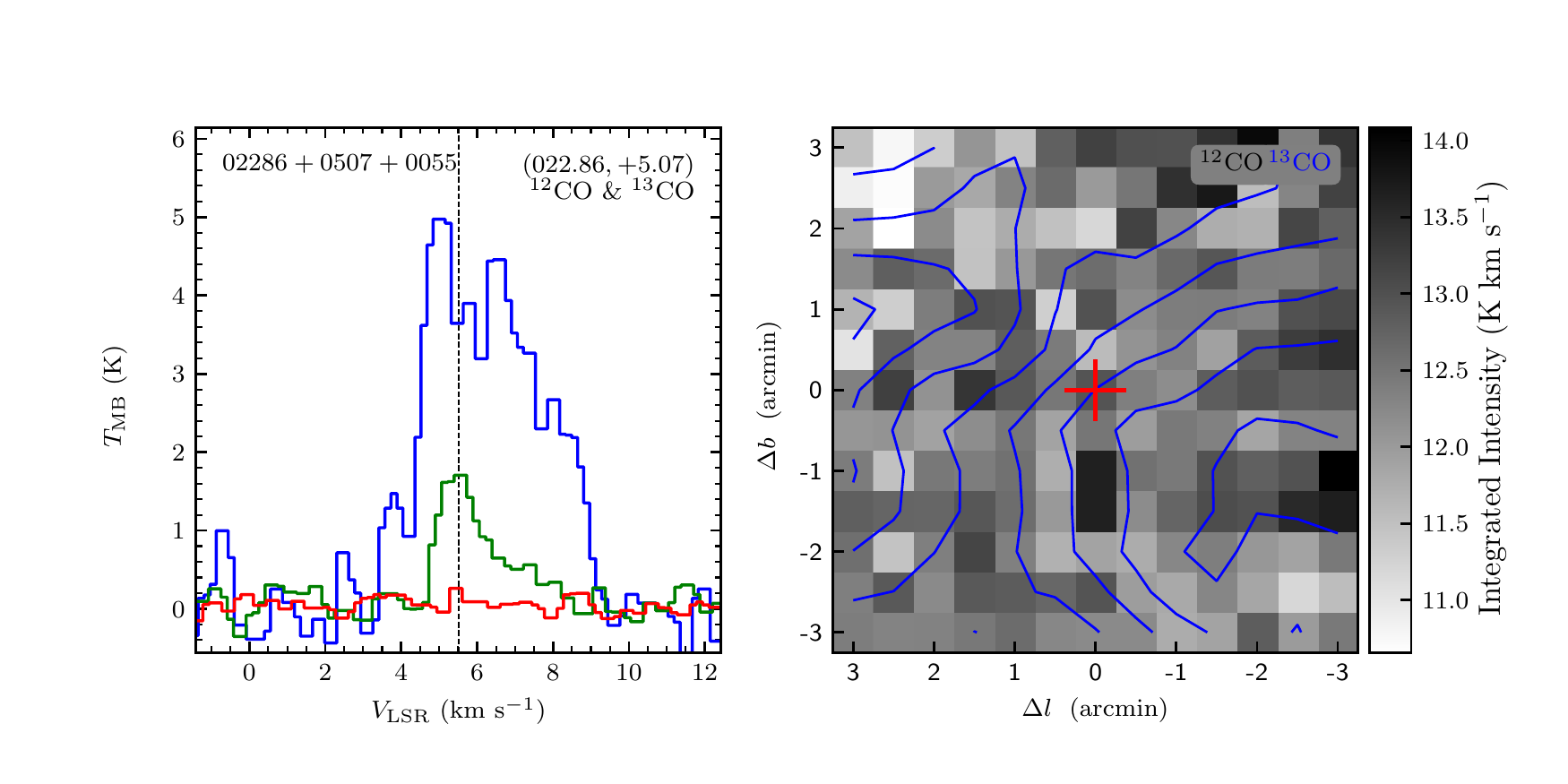}
\vspace{-0.5cm}

\includegraphics[width=9.0cm,angle=0]{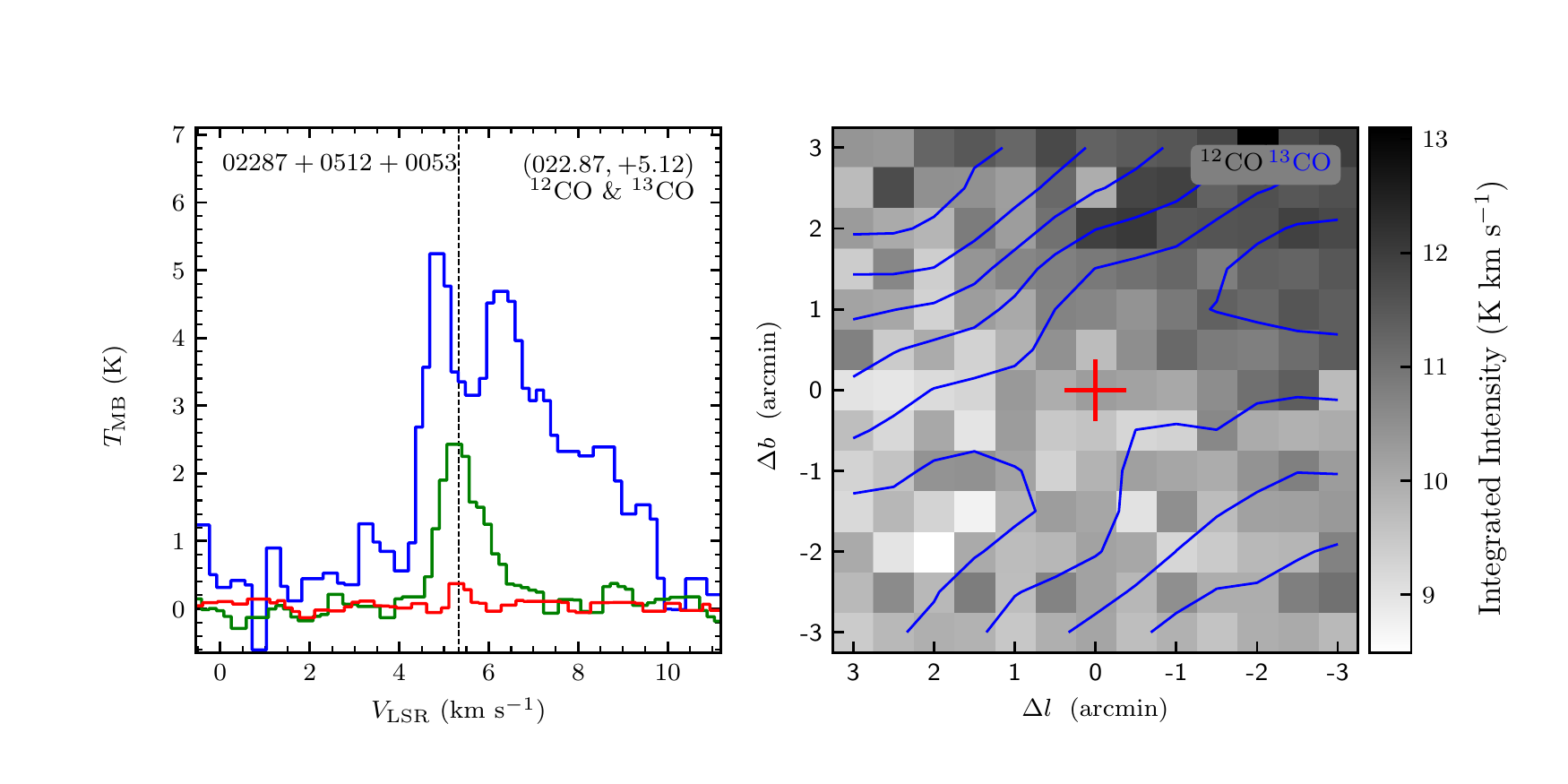}
\includegraphics[width=9.0cm,angle=0]{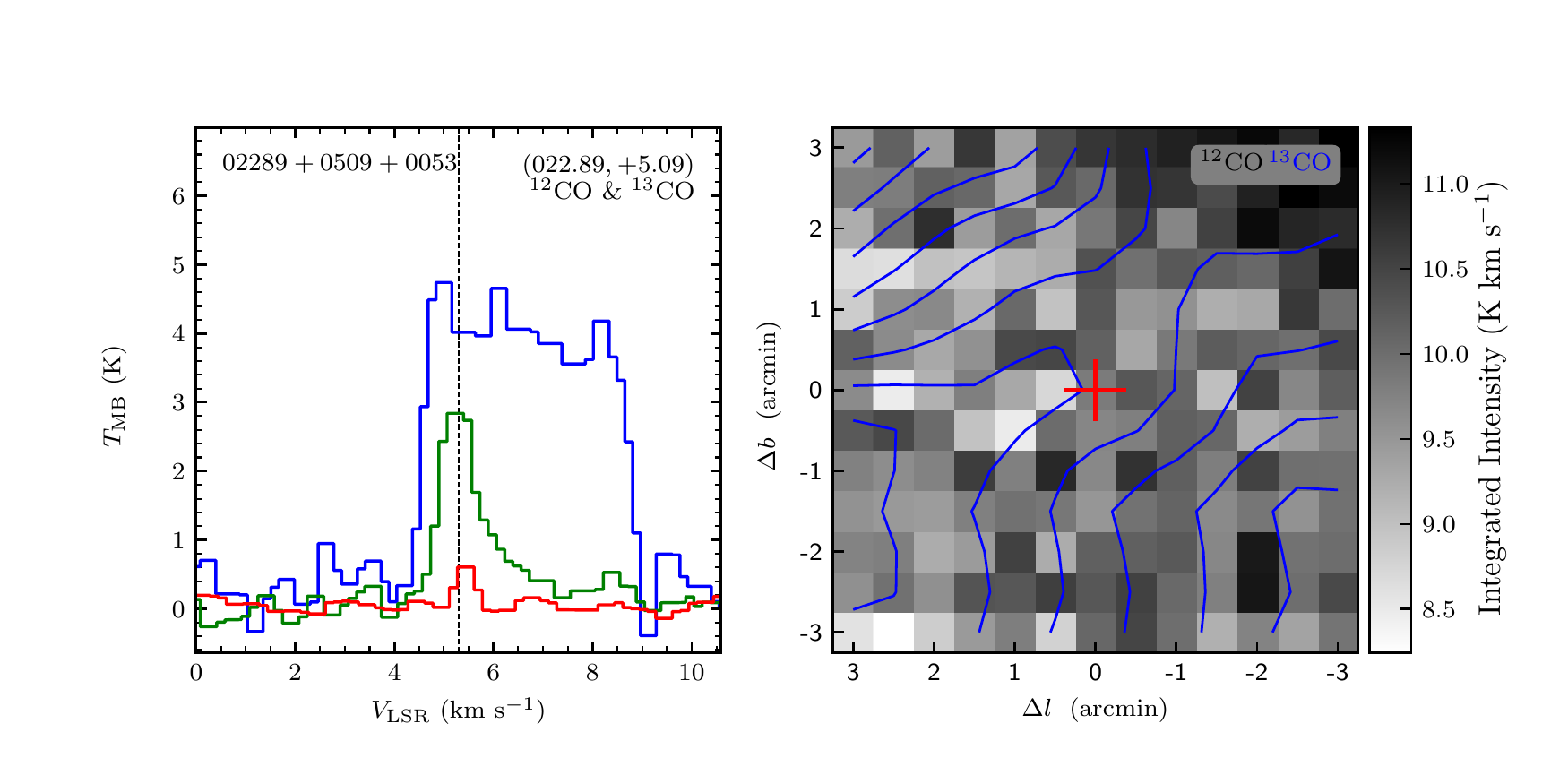}
\end{figure}
\clearpage

\begin{figure}
\includegraphics[width=9.0cm,angle=0]{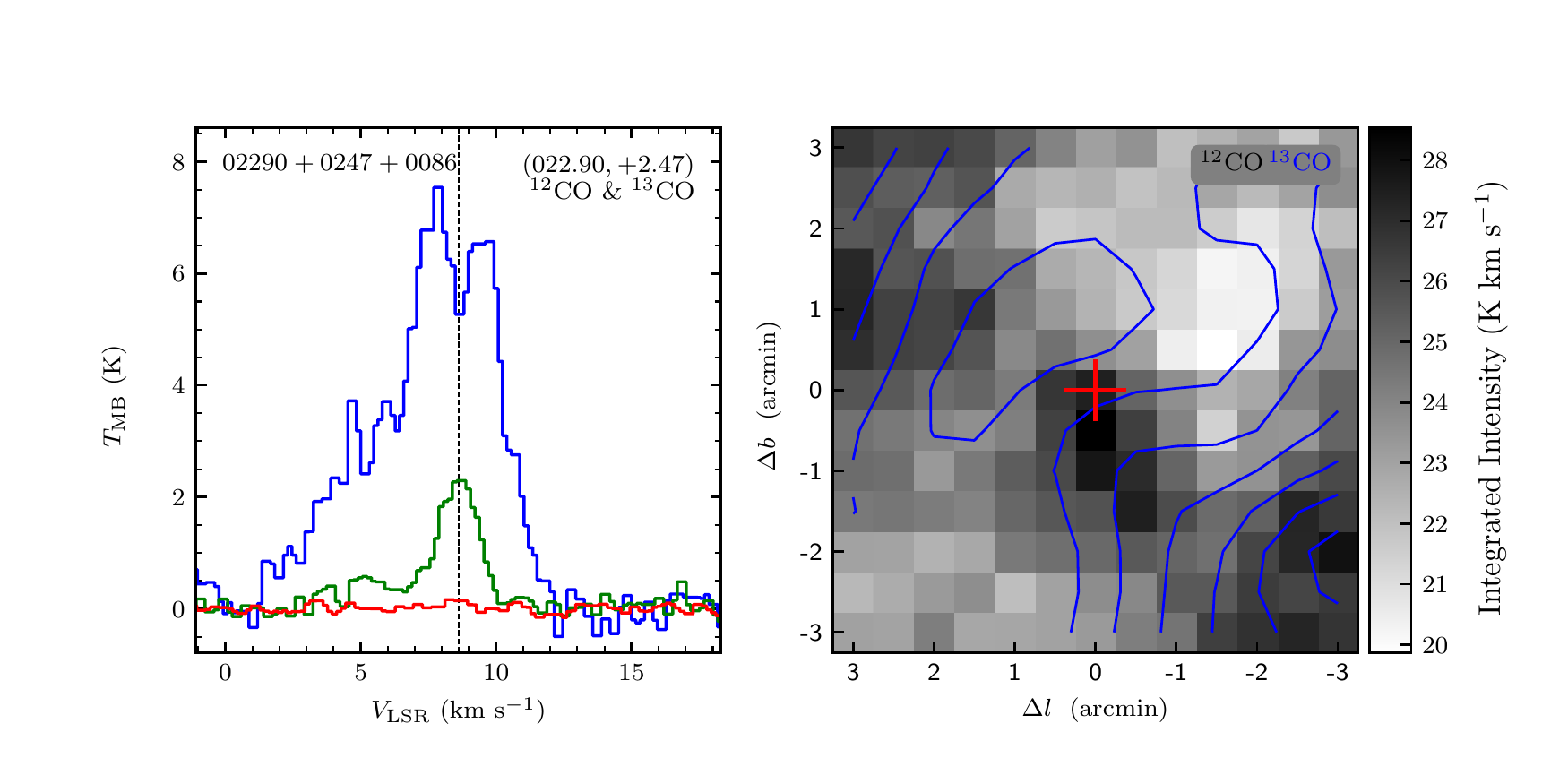}
\includegraphics[width=9.0cm,angle=0]{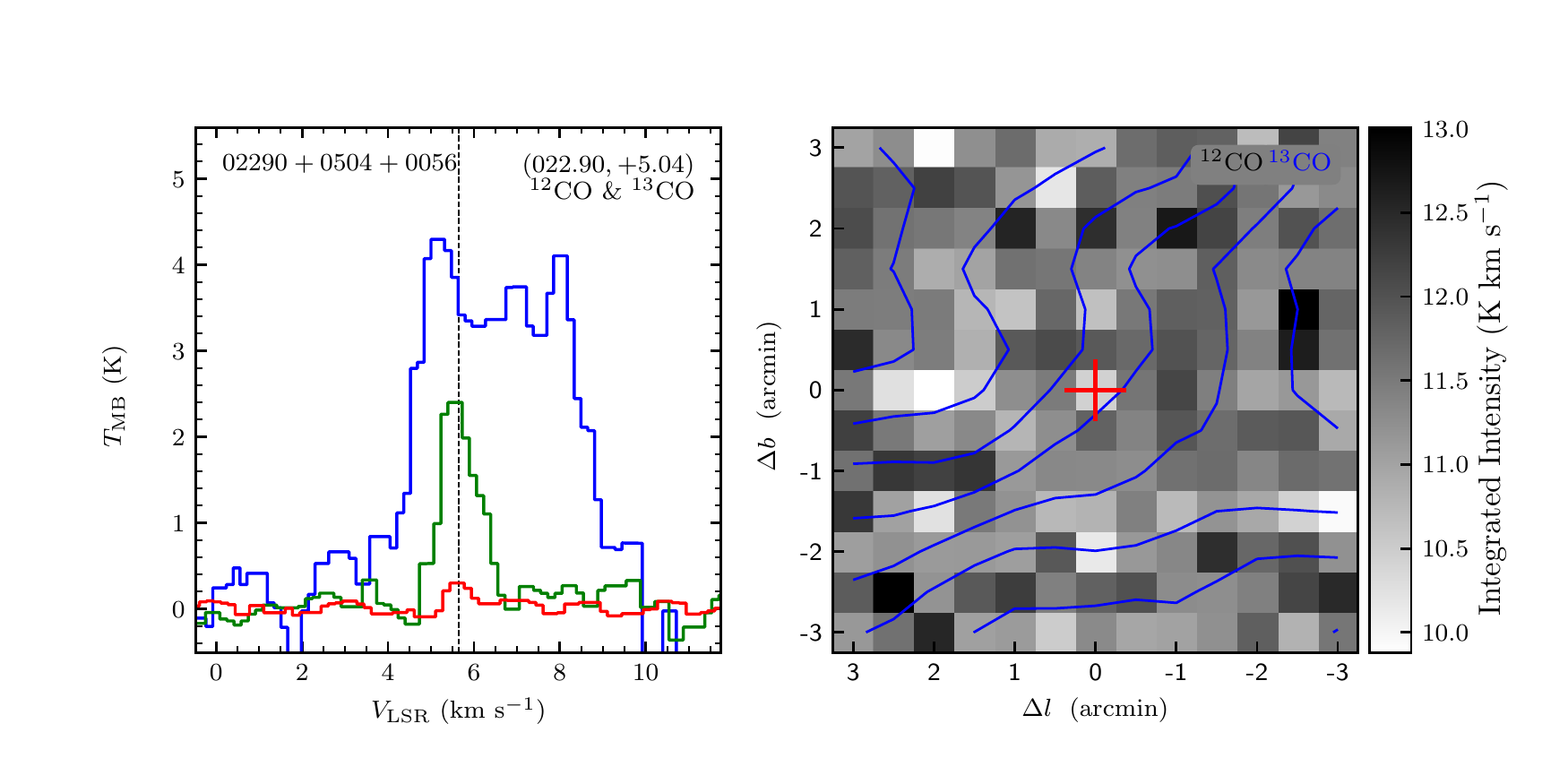}
\vspace{-0.5cm}

\includegraphics[width=9.0cm,angle=0]{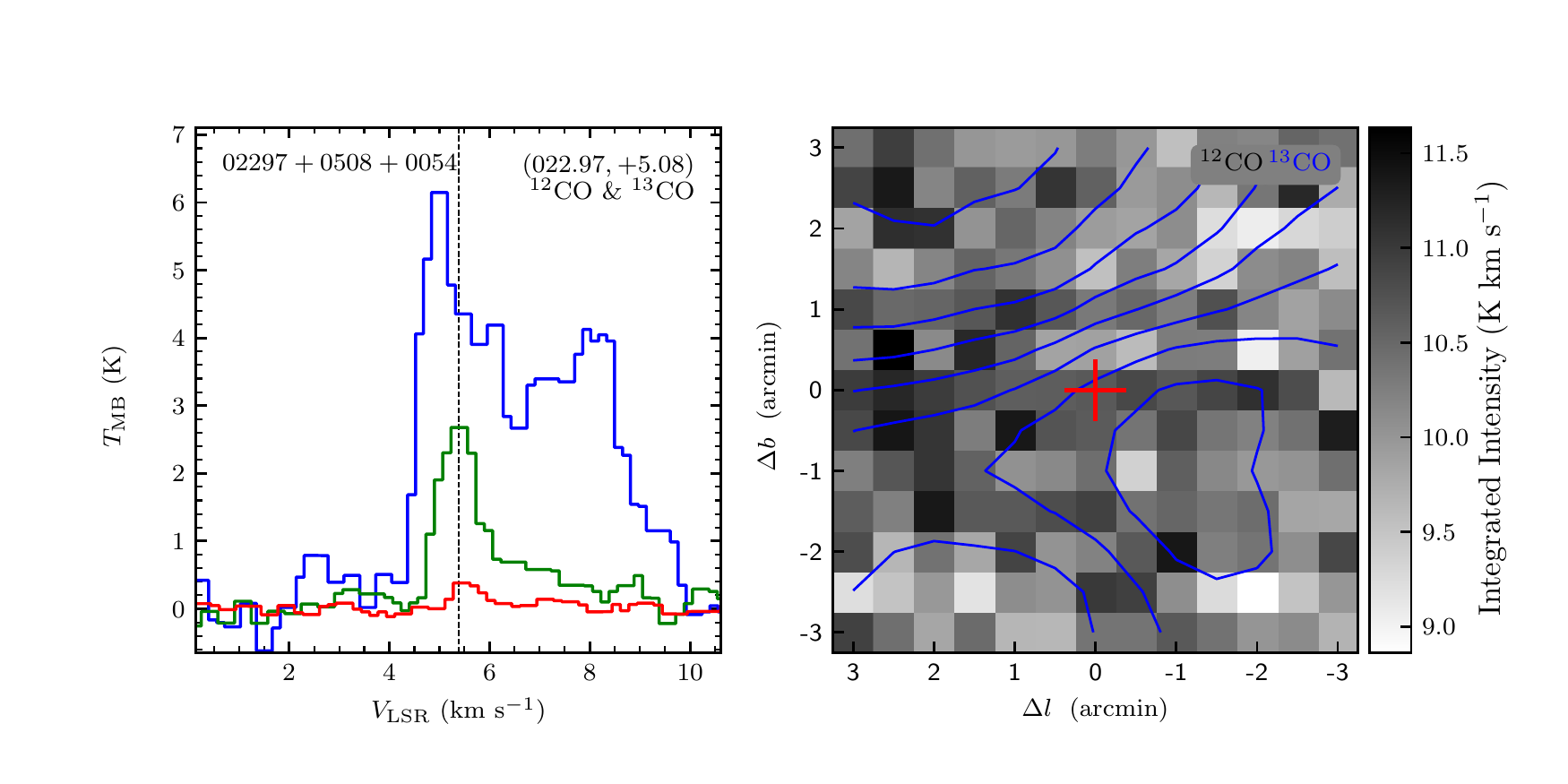}
\includegraphics[width=9.0cm,angle=0]{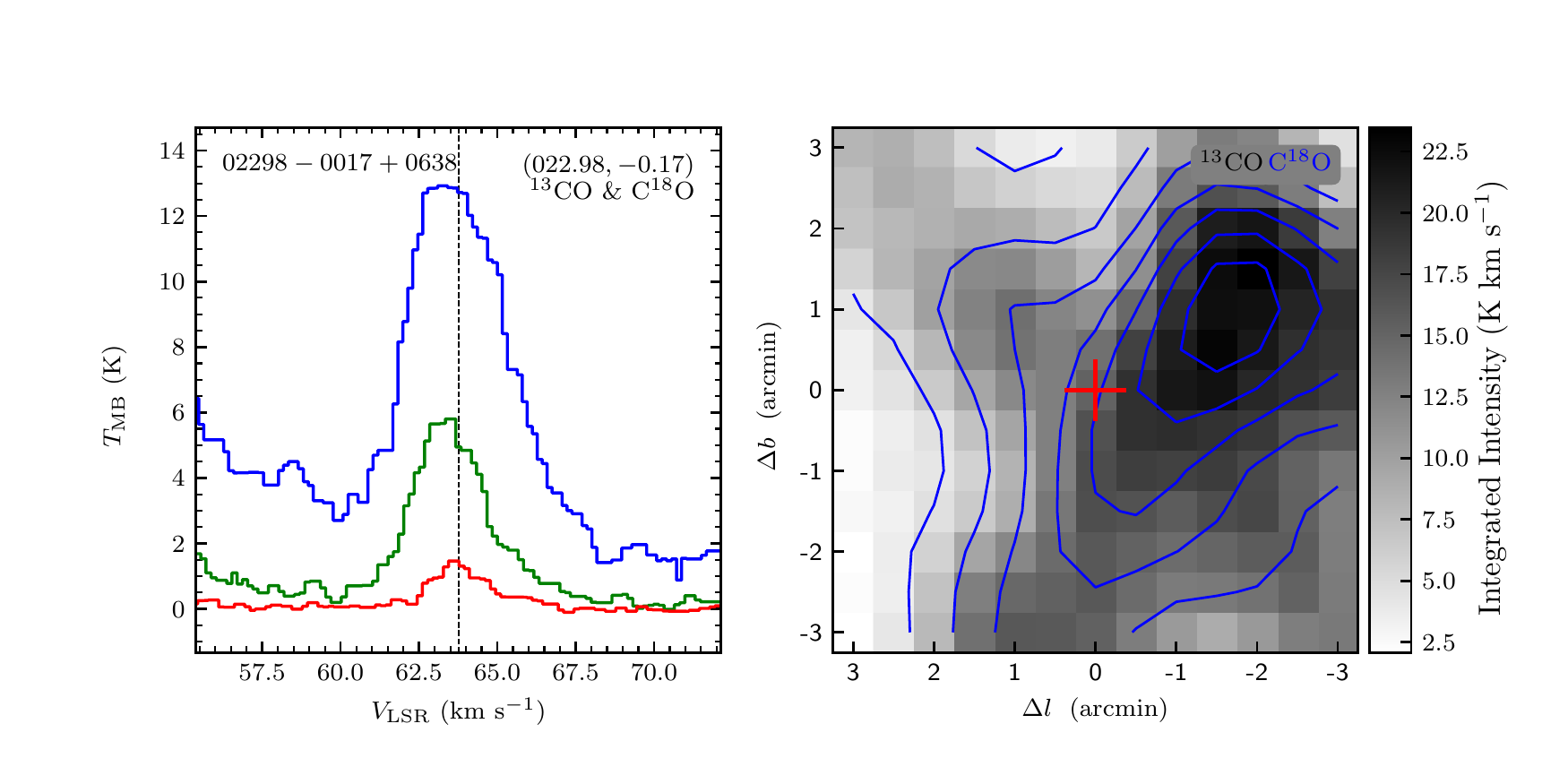}
\vspace{-0.5cm}

\includegraphics[width=9.0cm,angle=0]{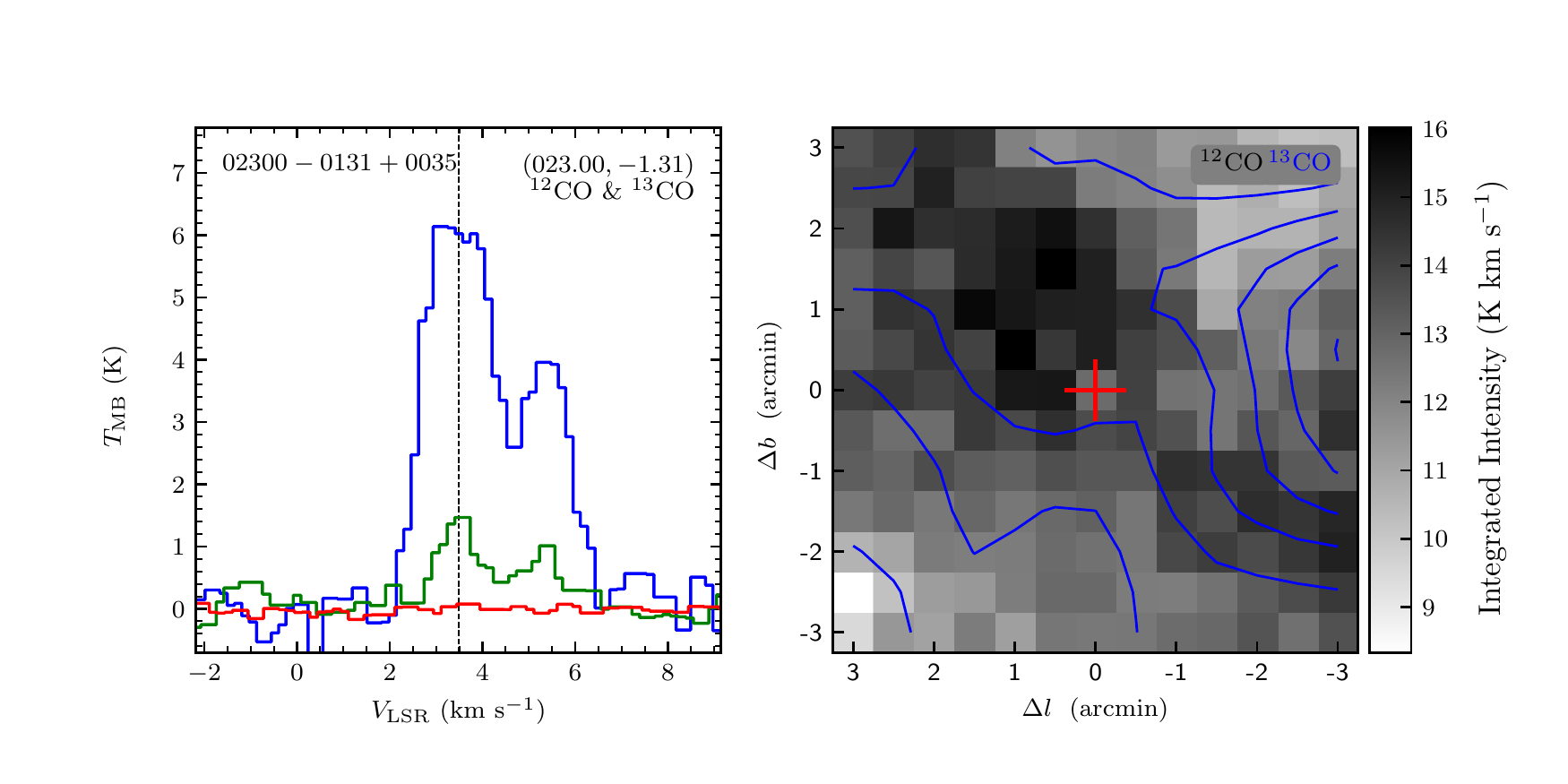}
\includegraphics[width=9.0cm,angle=0]{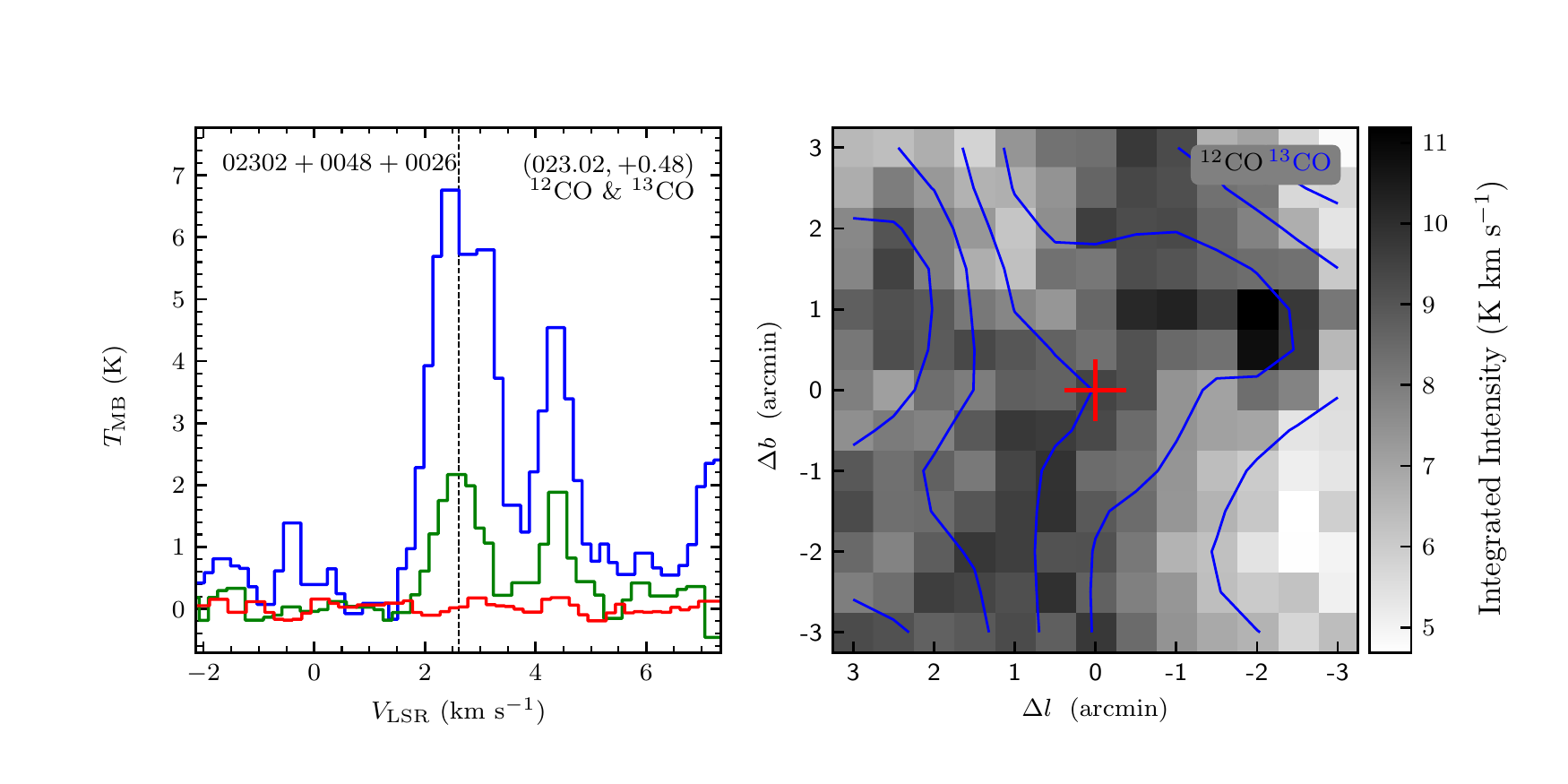}
\vspace{-0.5cm}

\includegraphics[width=9.0cm,angle=0]{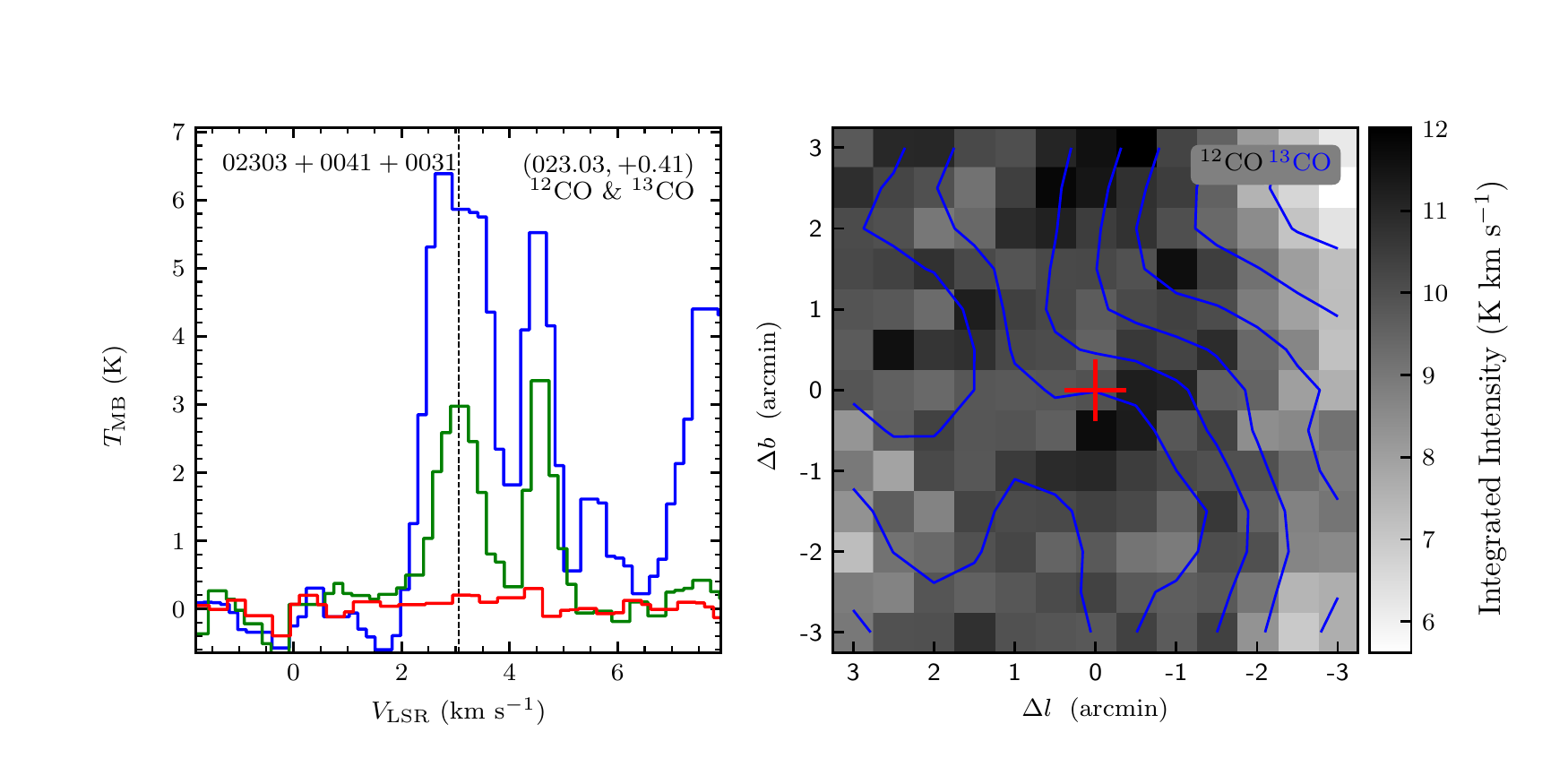}
\includegraphics[width=9.0cm,angle=0]{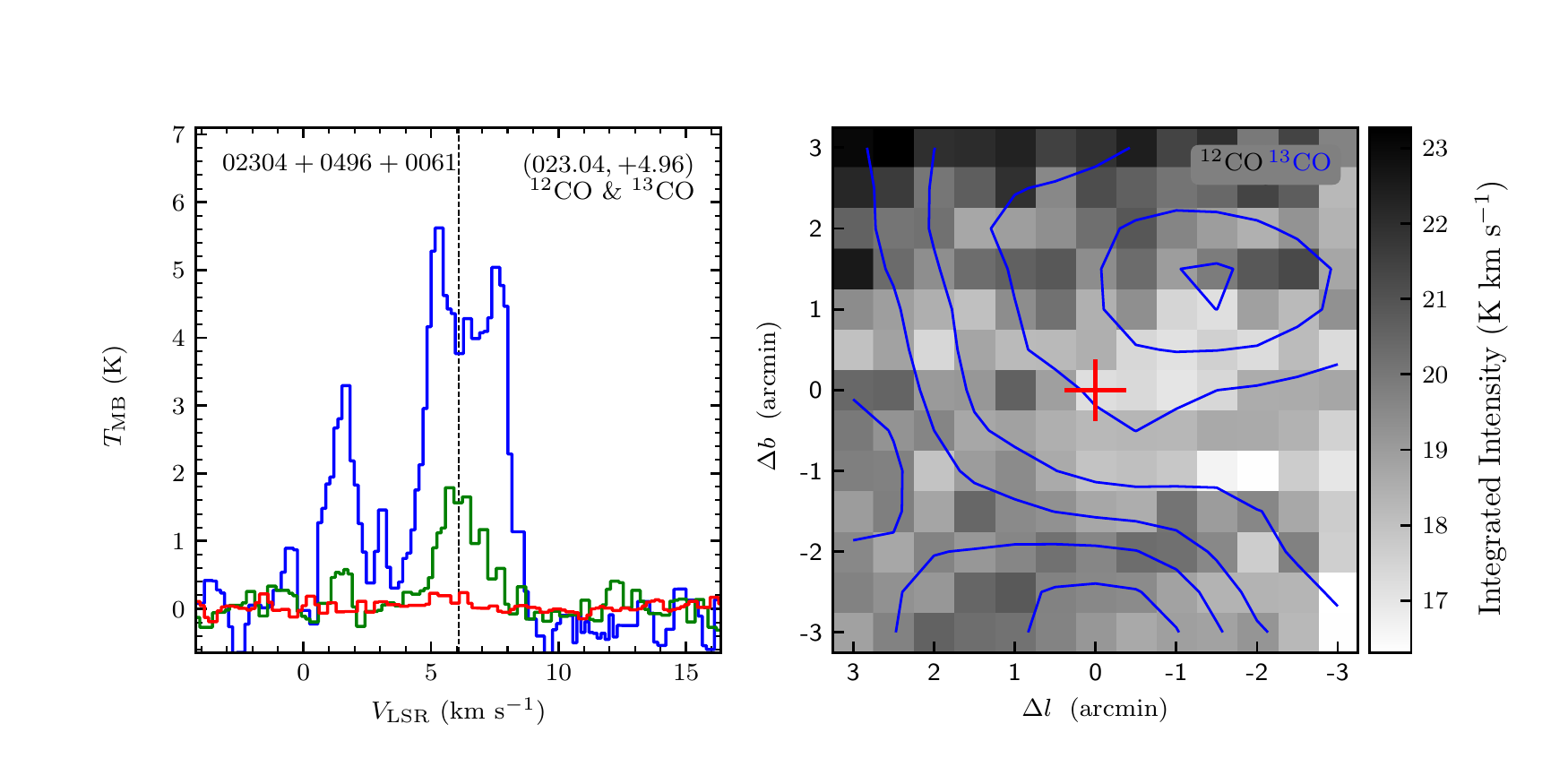}
\vspace{-0.5cm}

\includegraphics[width=9.0cm,angle=0]{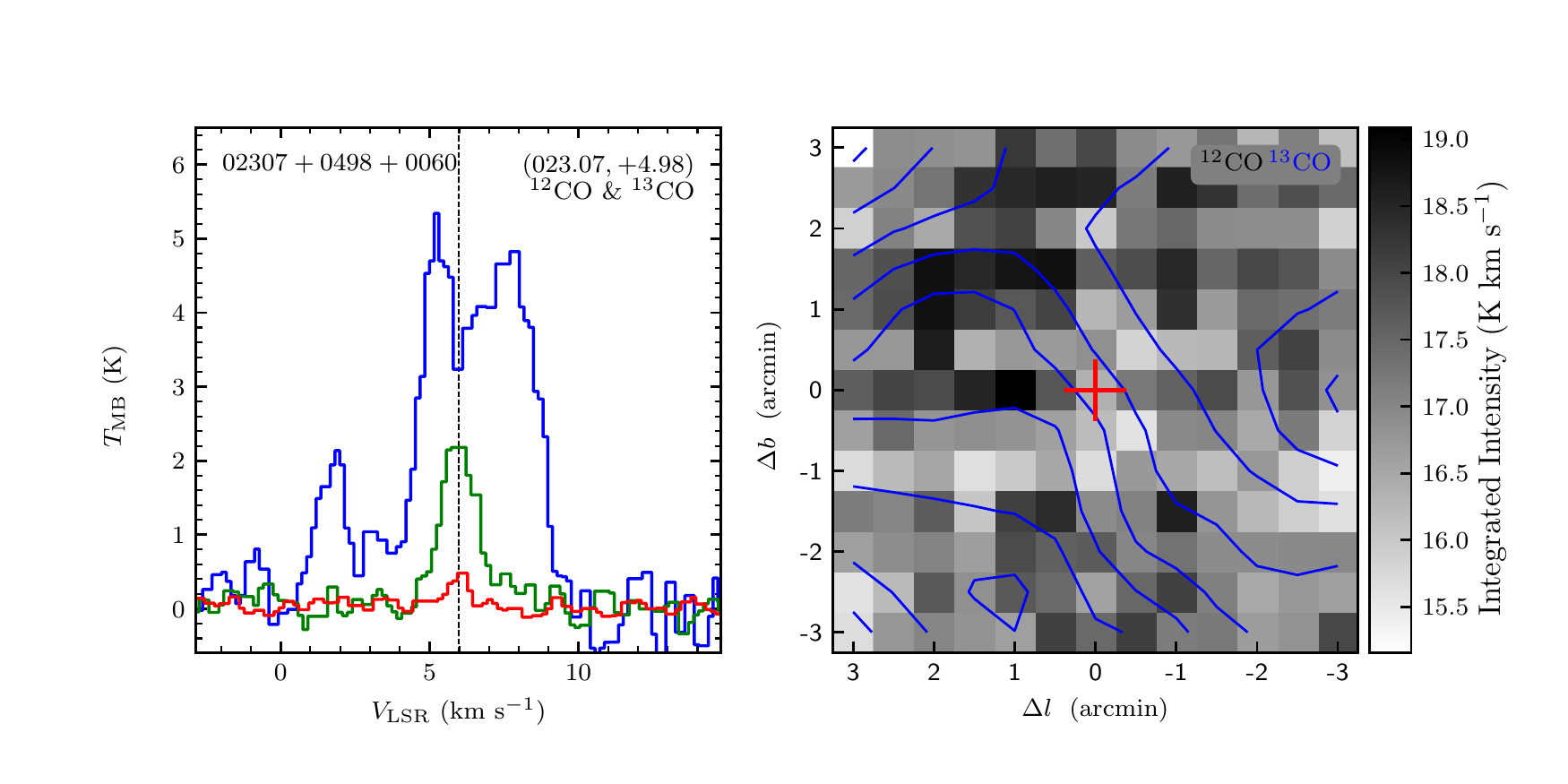}
\includegraphics[width=9.0cm,angle=0]{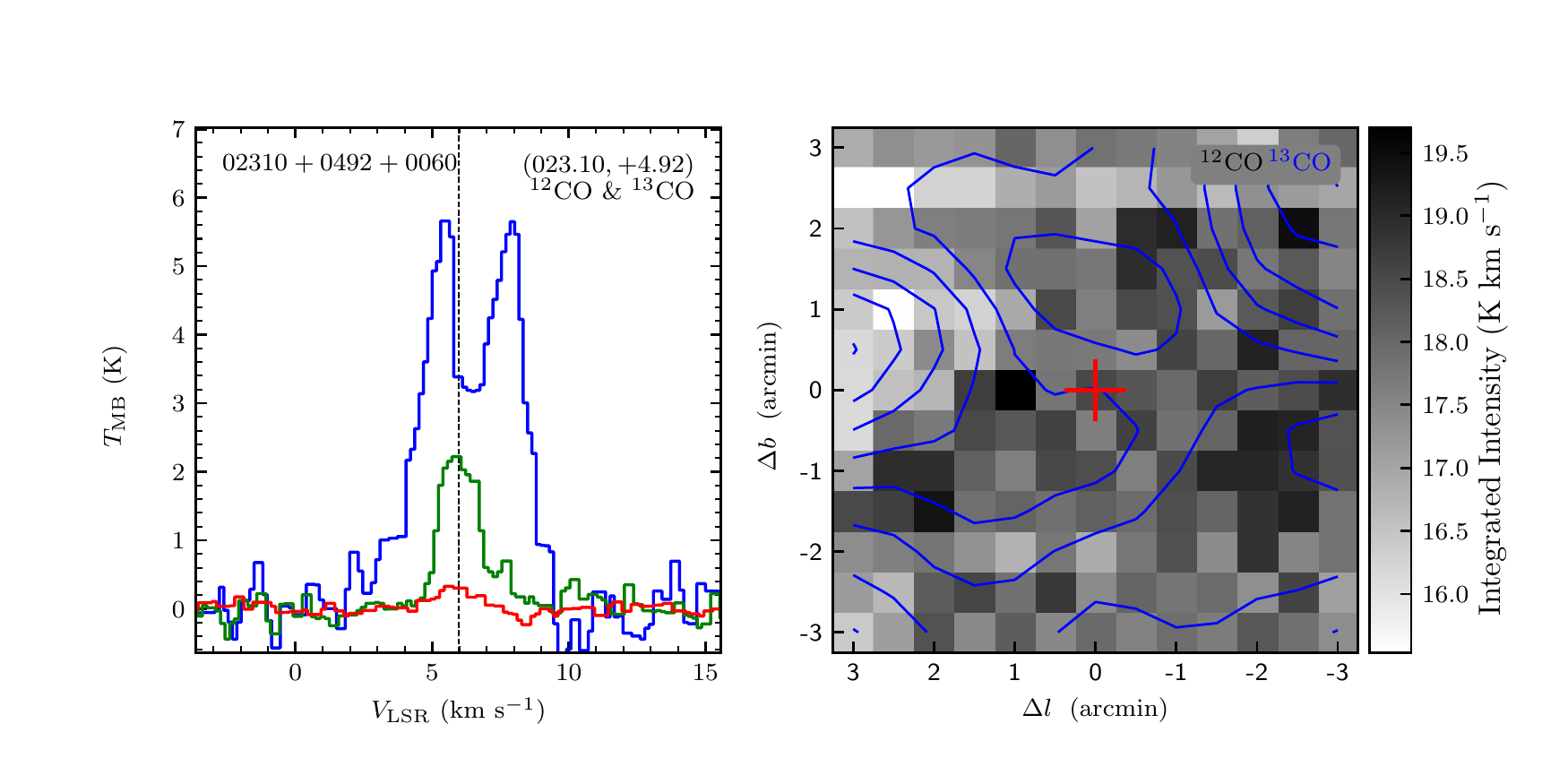}
\end{figure}
\clearpage

\begin{figure}
\includegraphics[width=9.0cm,angle=0]{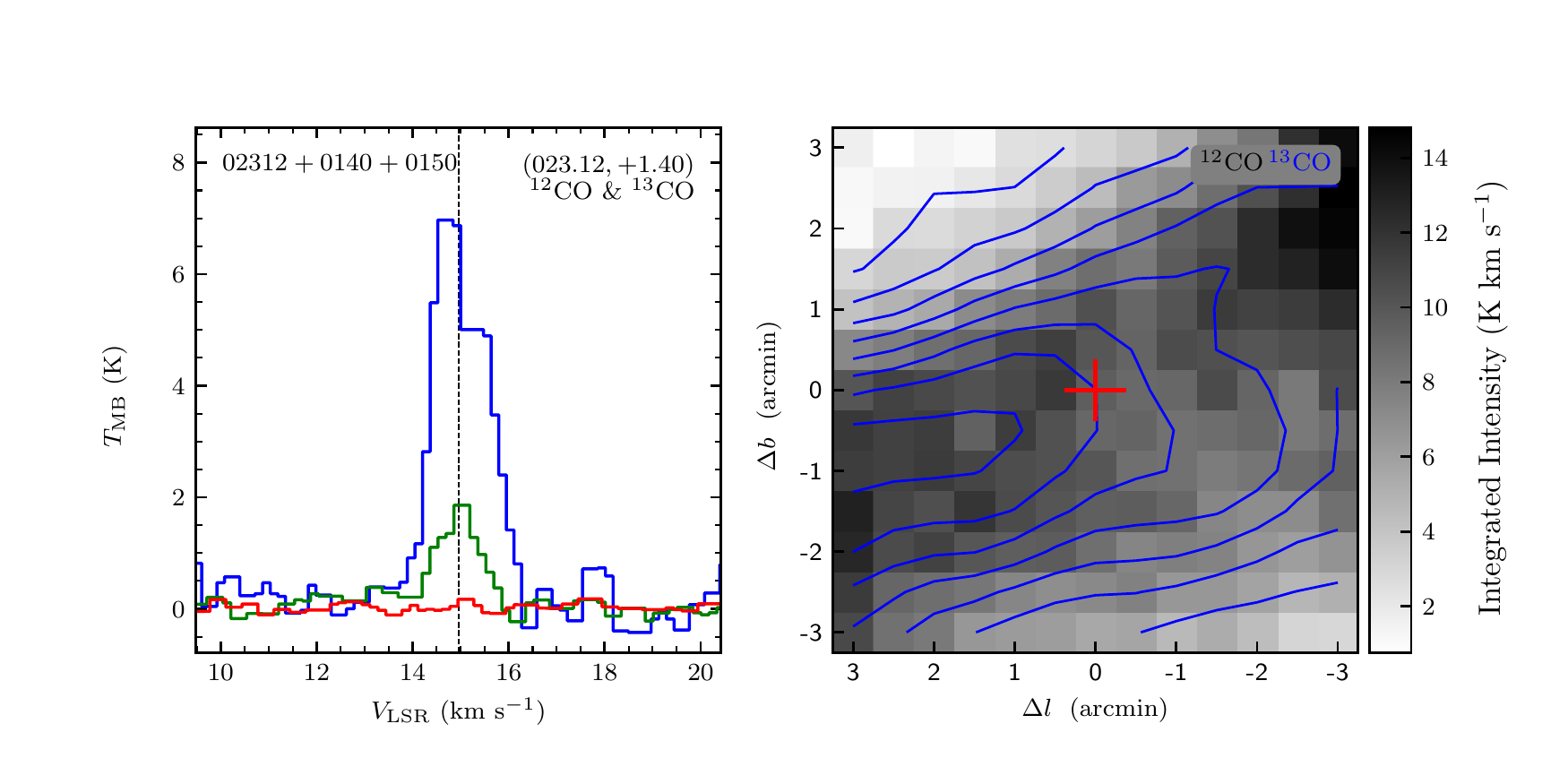}
\includegraphics[width=9.0cm,angle=0]{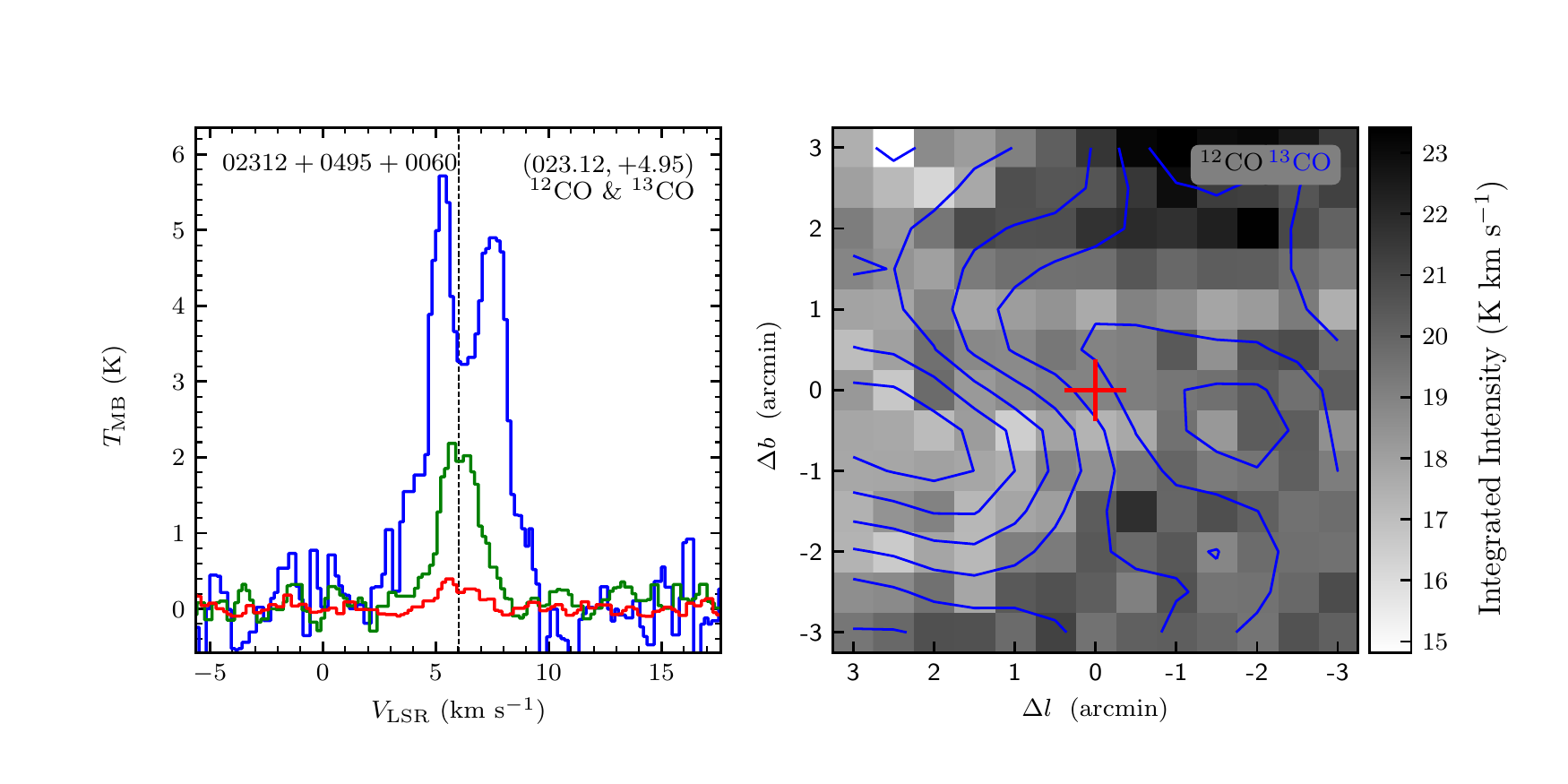}
\vspace{-0.5cm}

\includegraphics[width=9.0cm,angle=0]{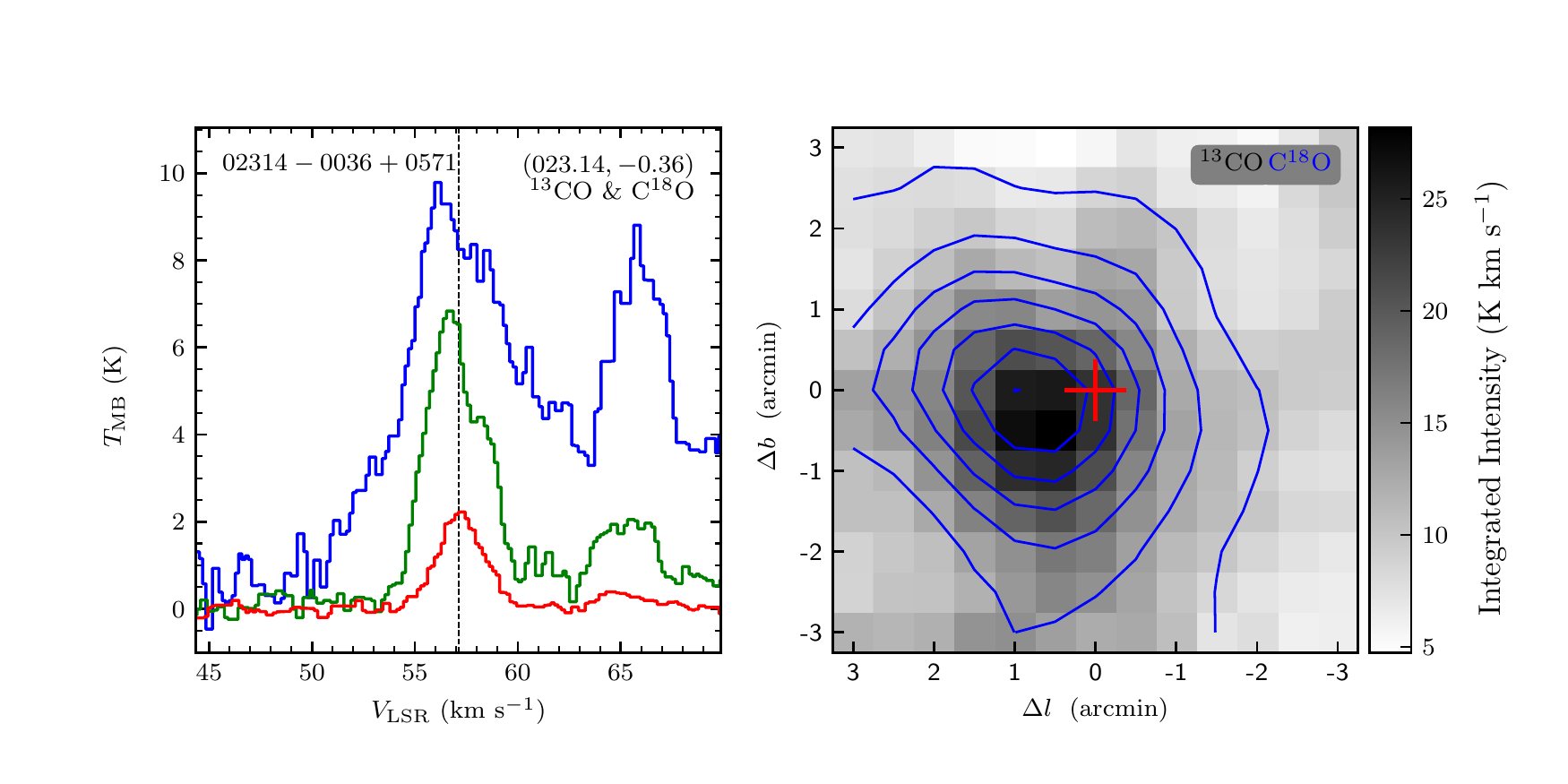}
\includegraphics[width=9.0cm,angle=0]{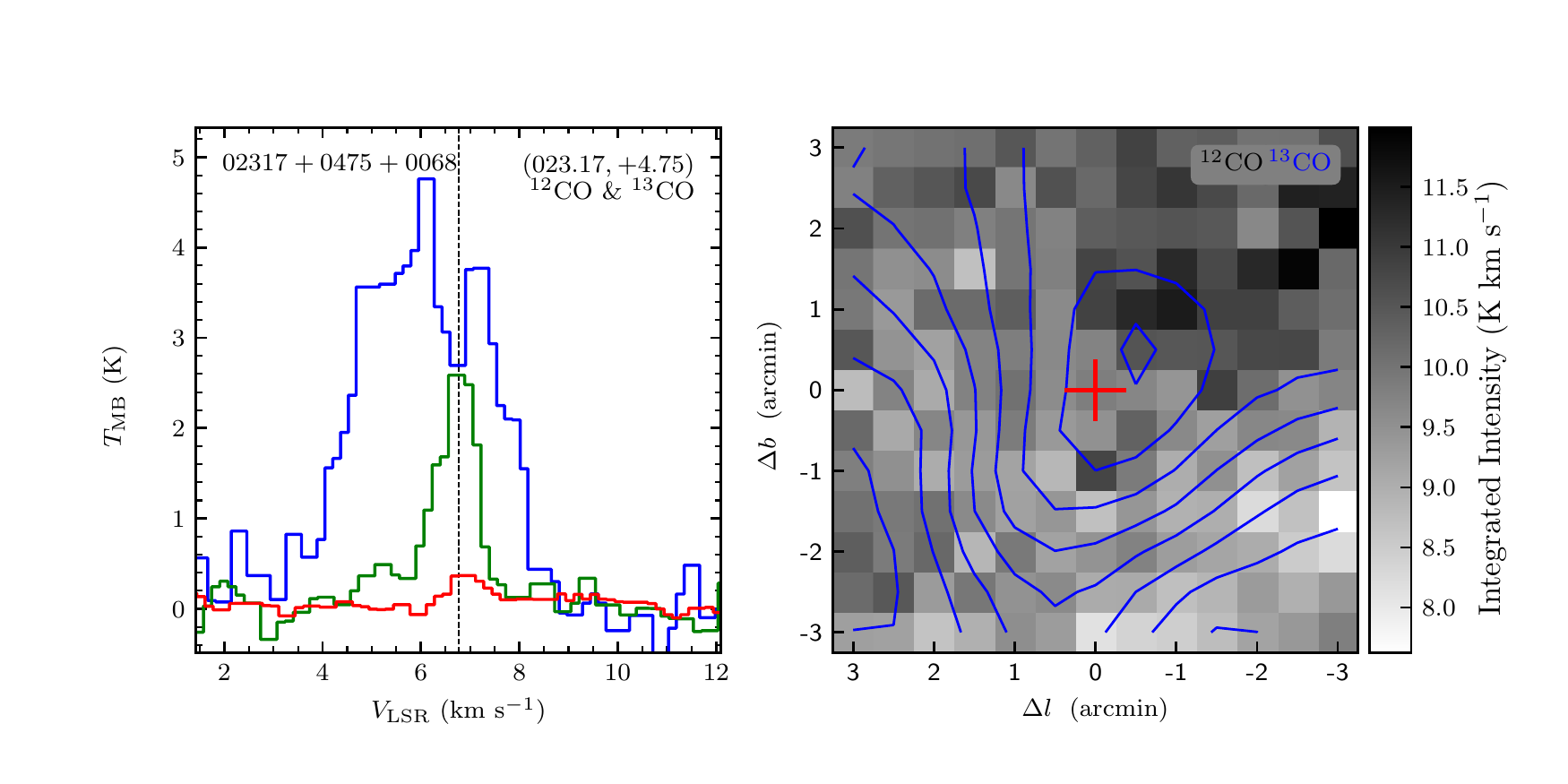}
\vspace{-0.5cm}

\includegraphics[width=9.0cm,angle=0]{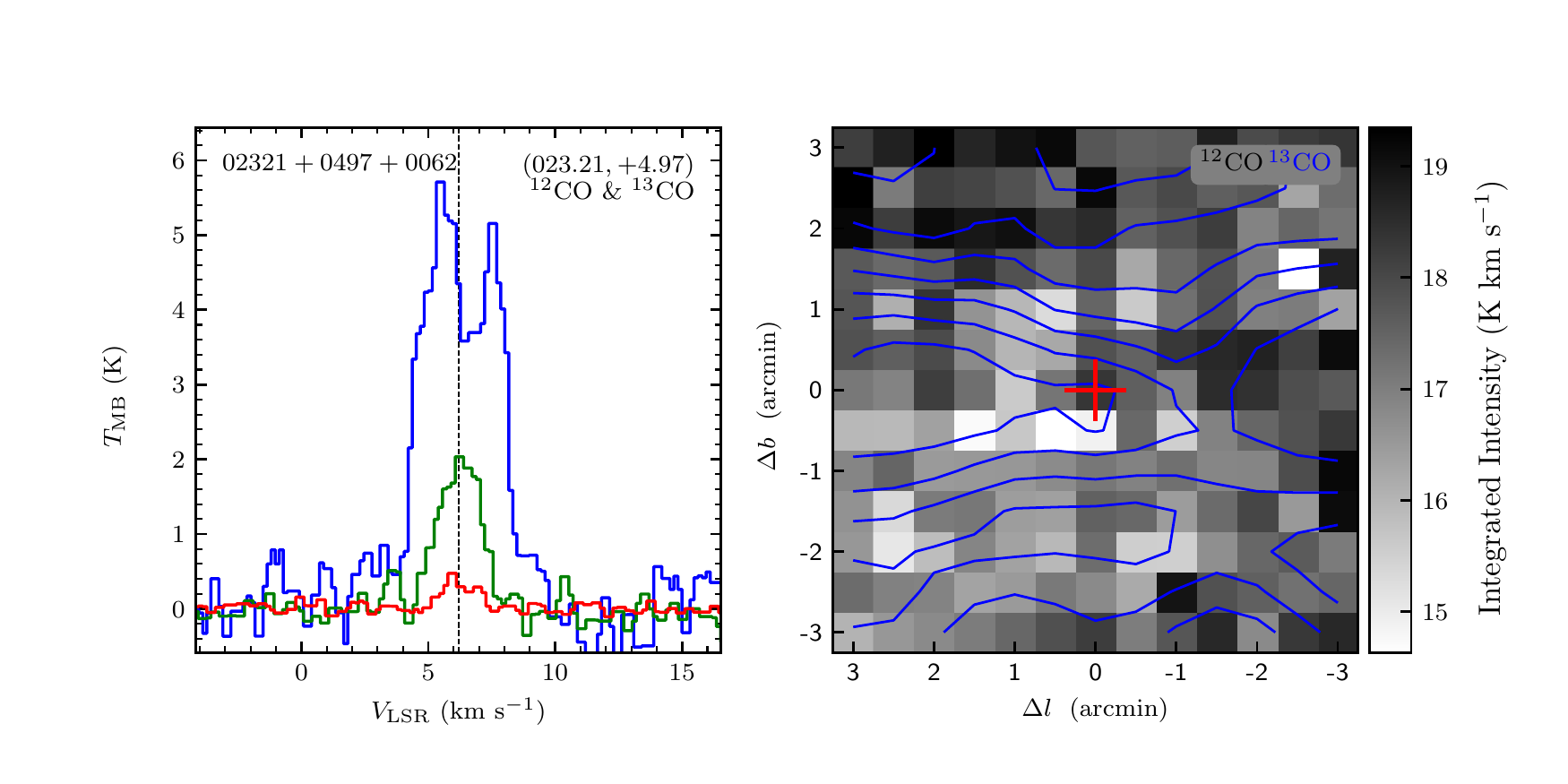}
\includegraphics[width=9.0cm,angle=0]{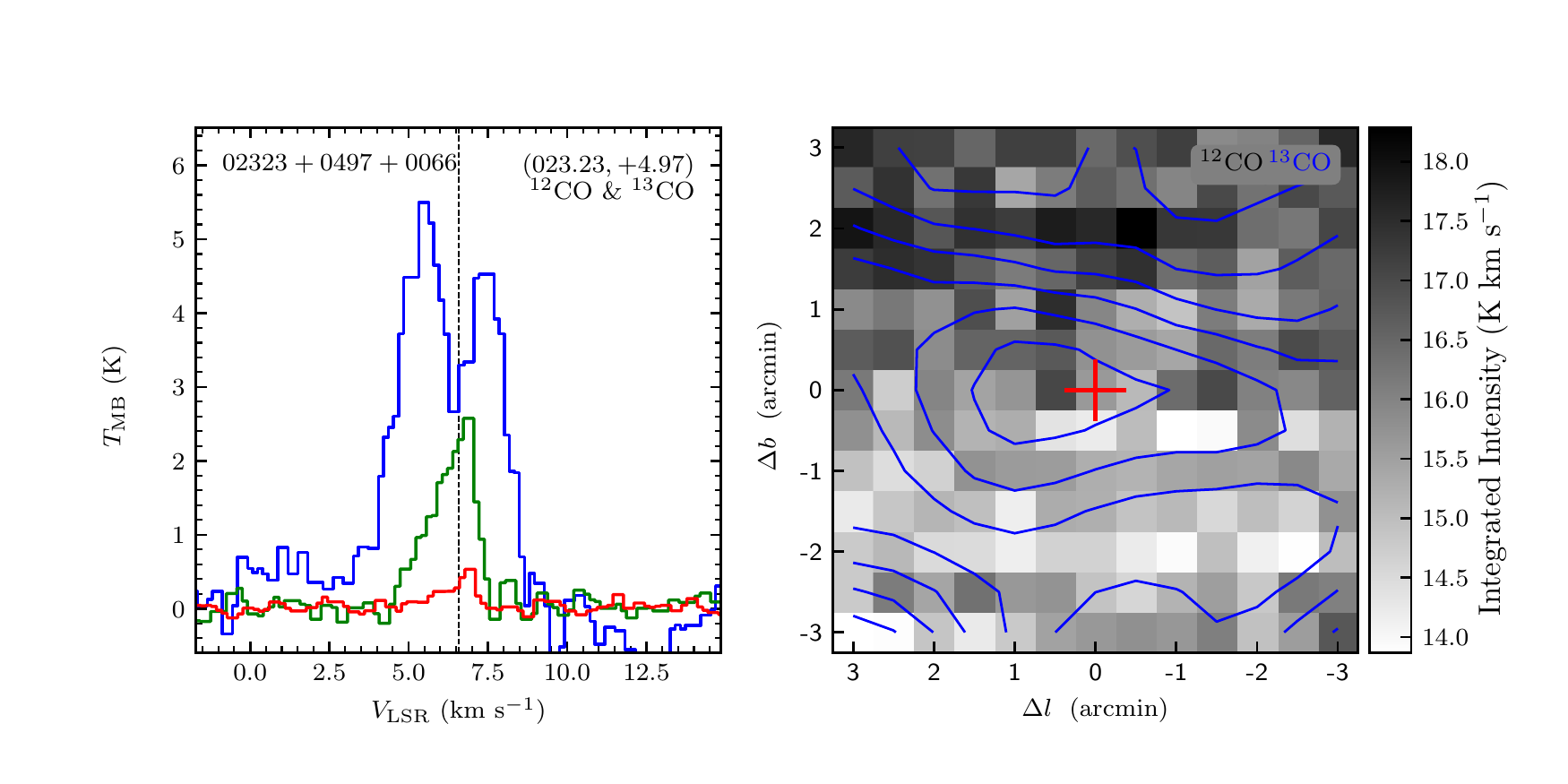}
\vspace{-0.5cm}

\includegraphics[width=9.0cm,angle=0]{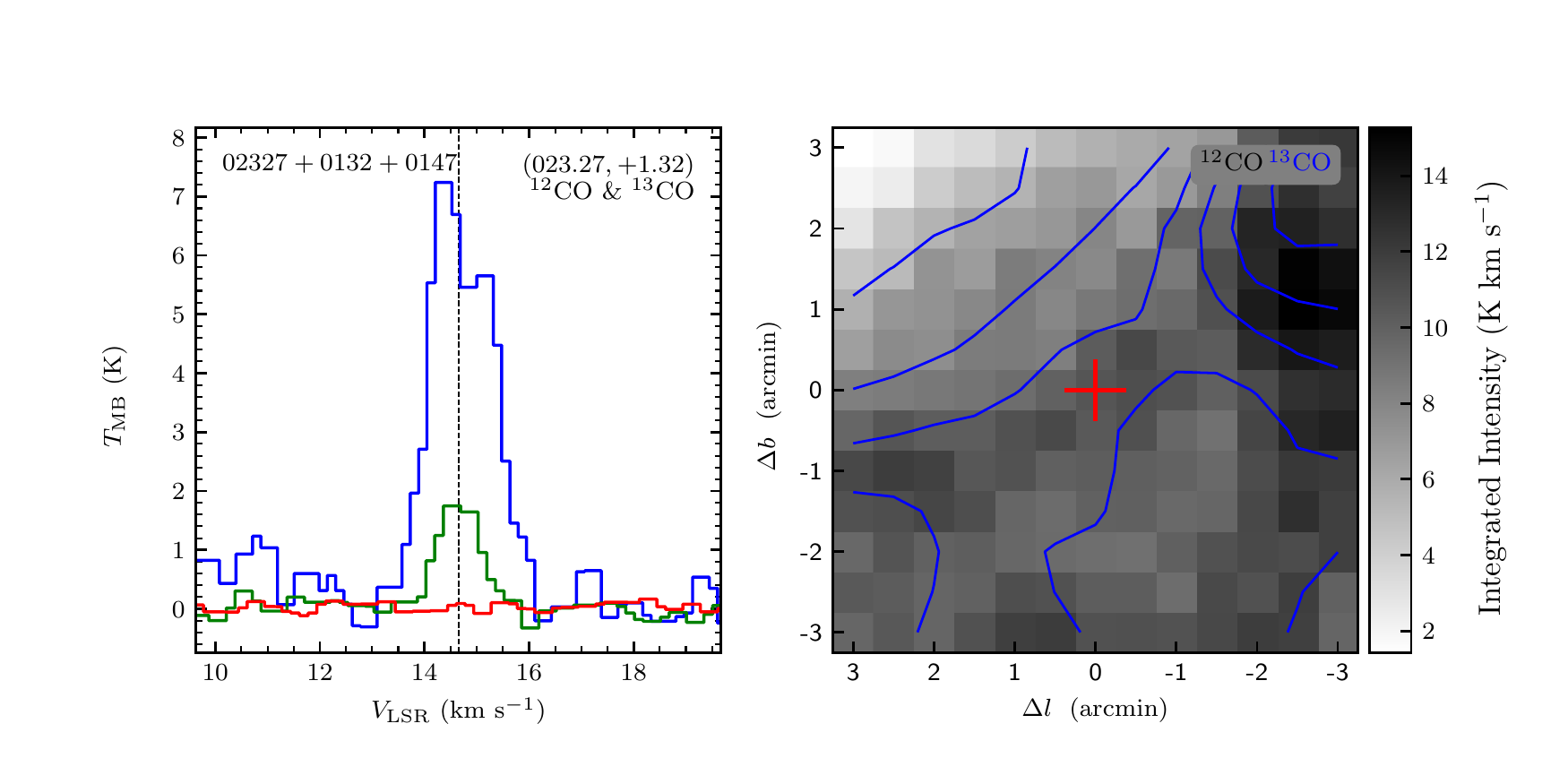}
\includegraphics[width=9.0cm,angle=0]{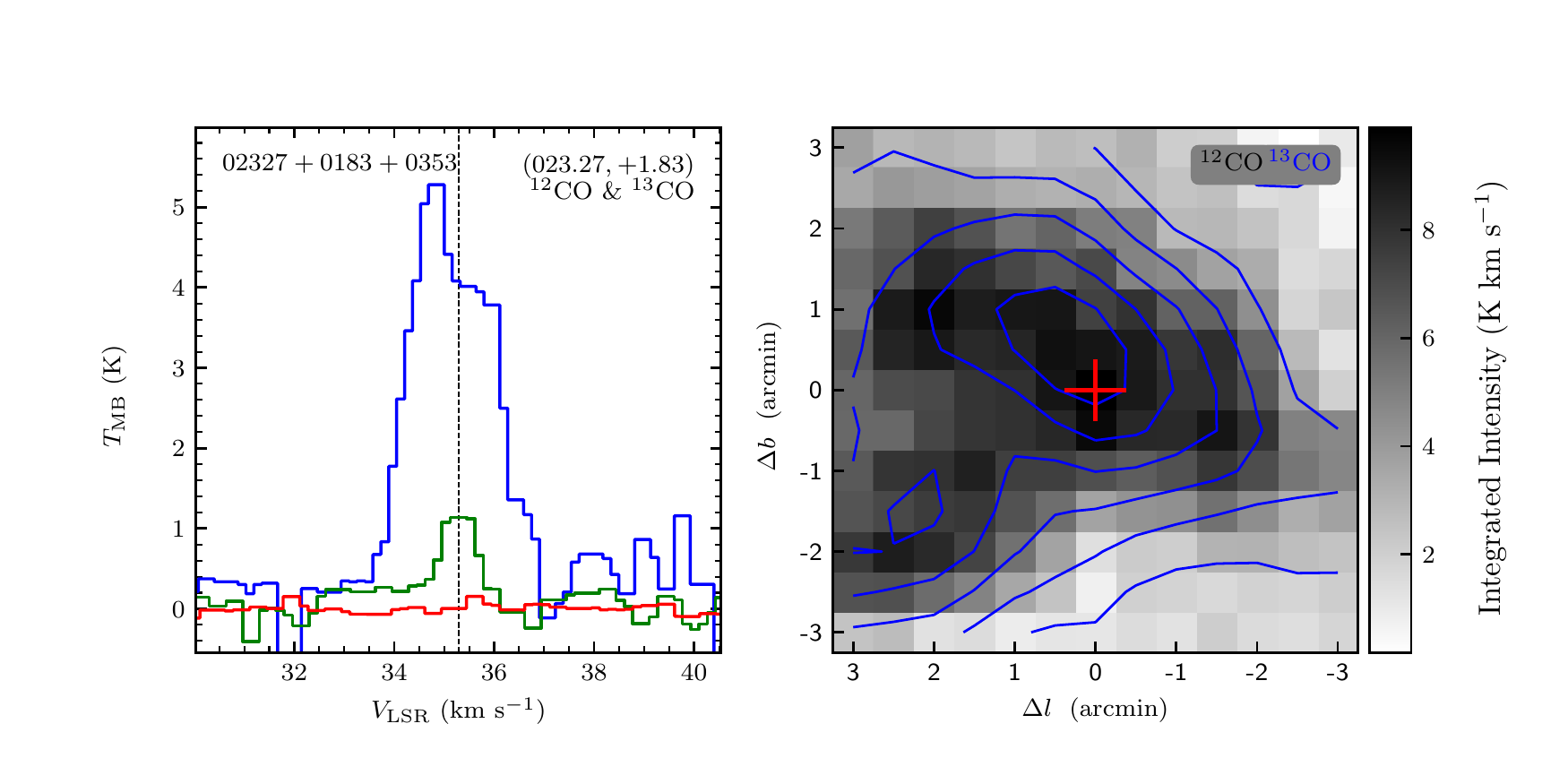}
\vspace{-0.5cm}

\includegraphics[width=9.0cm,angle=0]{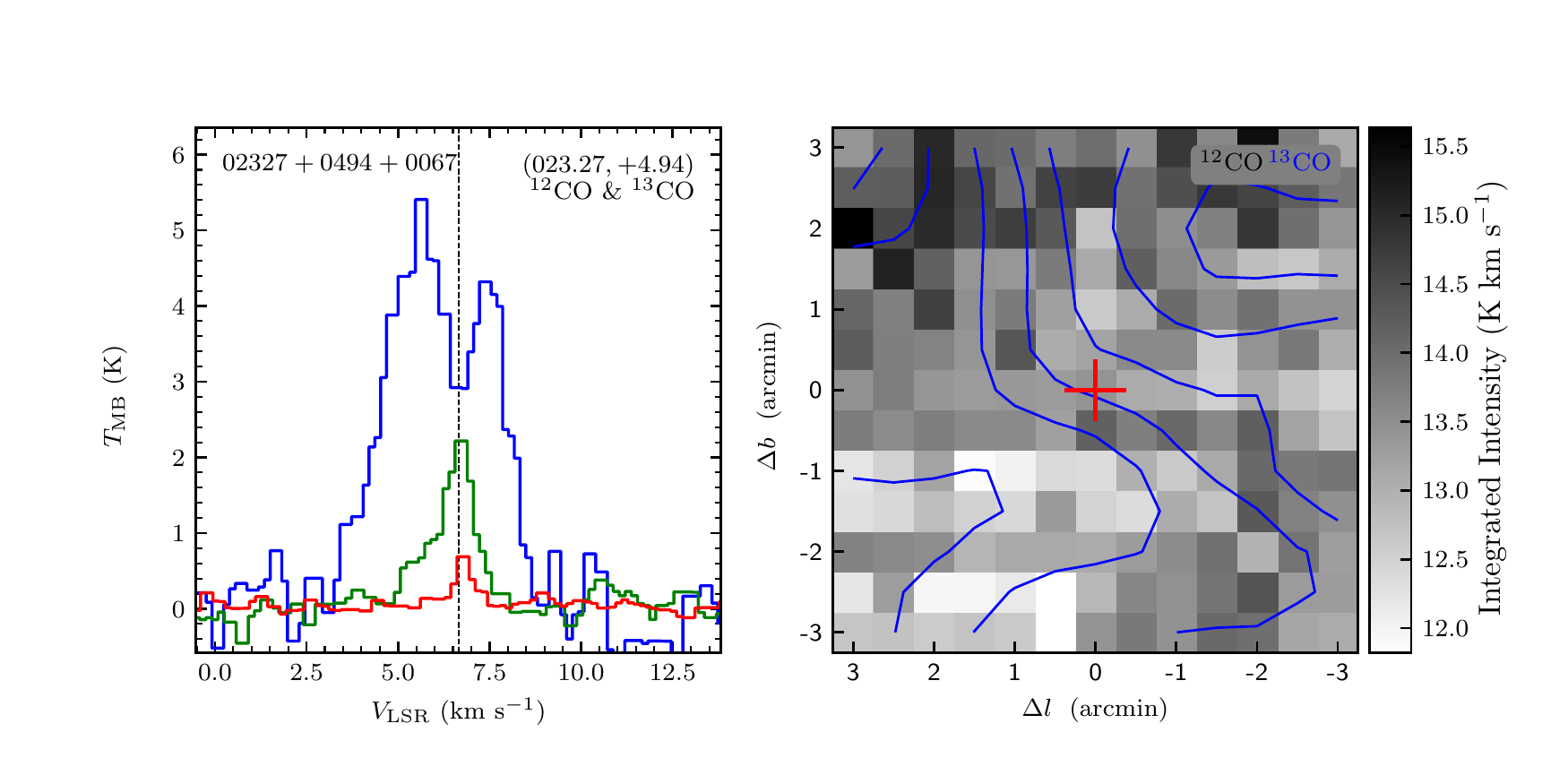}
\includegraphics[width=9.0cm,angle=0]{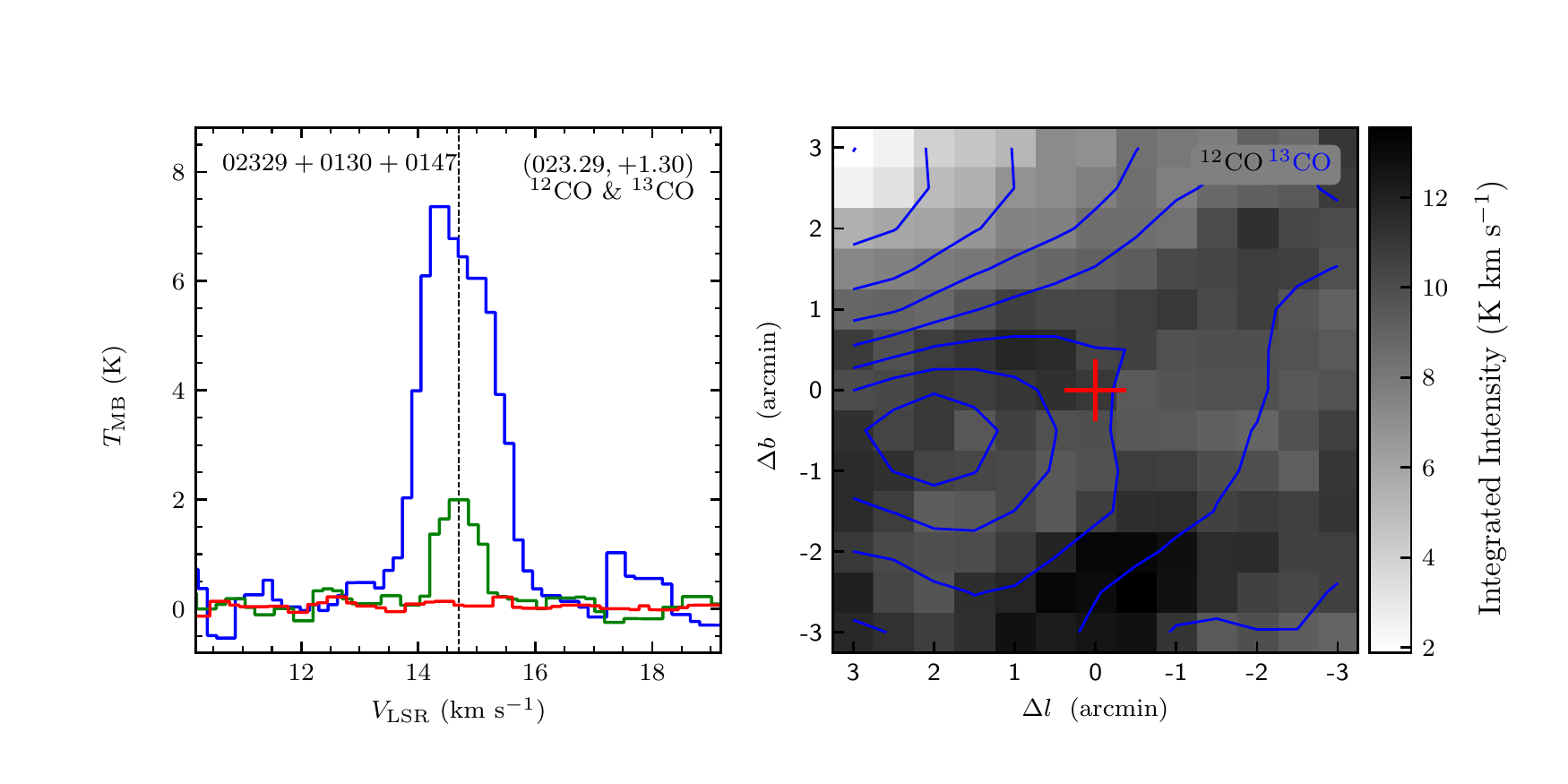}
\end{figure}
\clearpage

\begin{figure}
\includegraphics[width=9.0cm,angle=0]{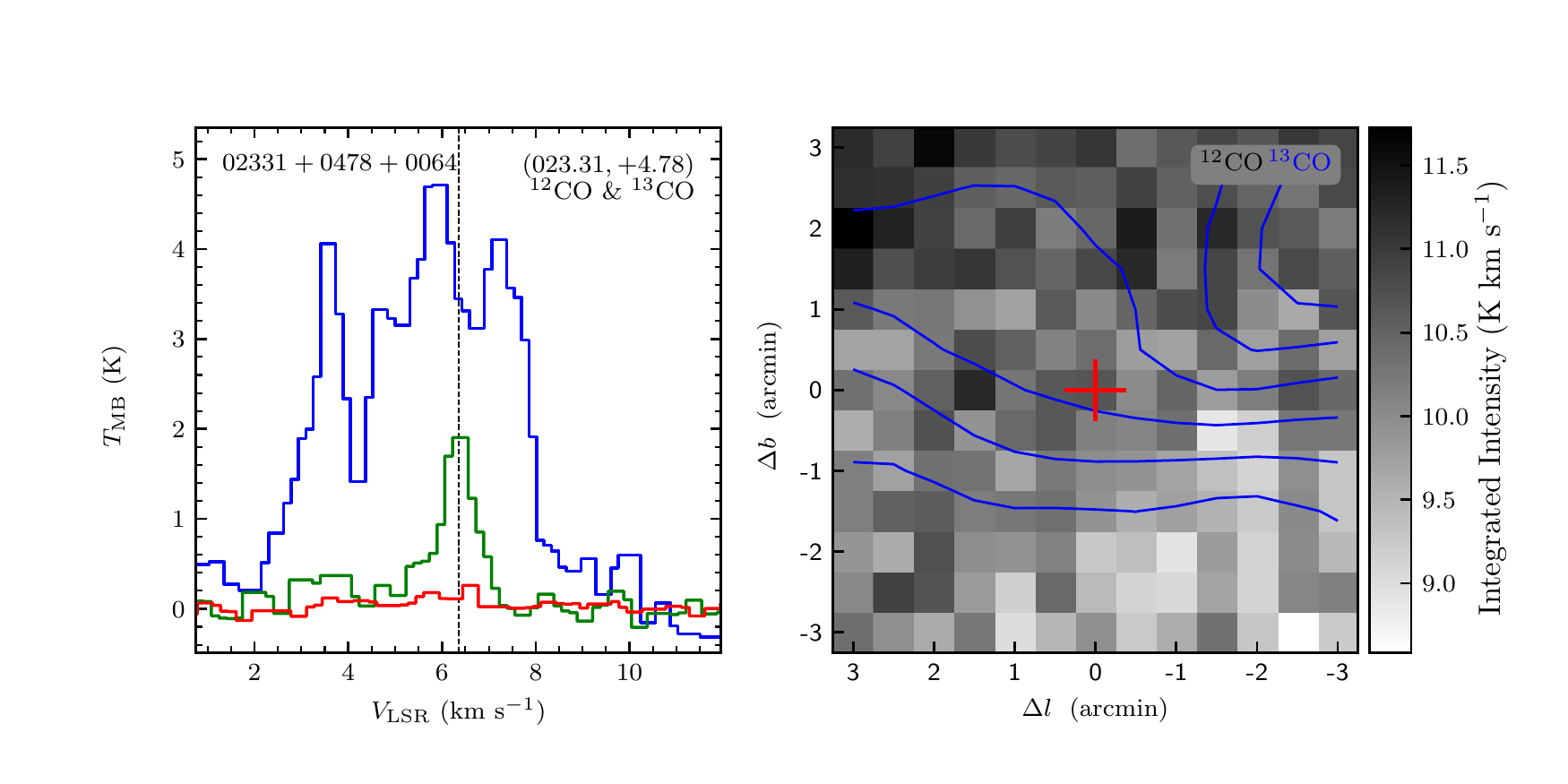}
\includegraphics[width=9.0cm,angle=0]{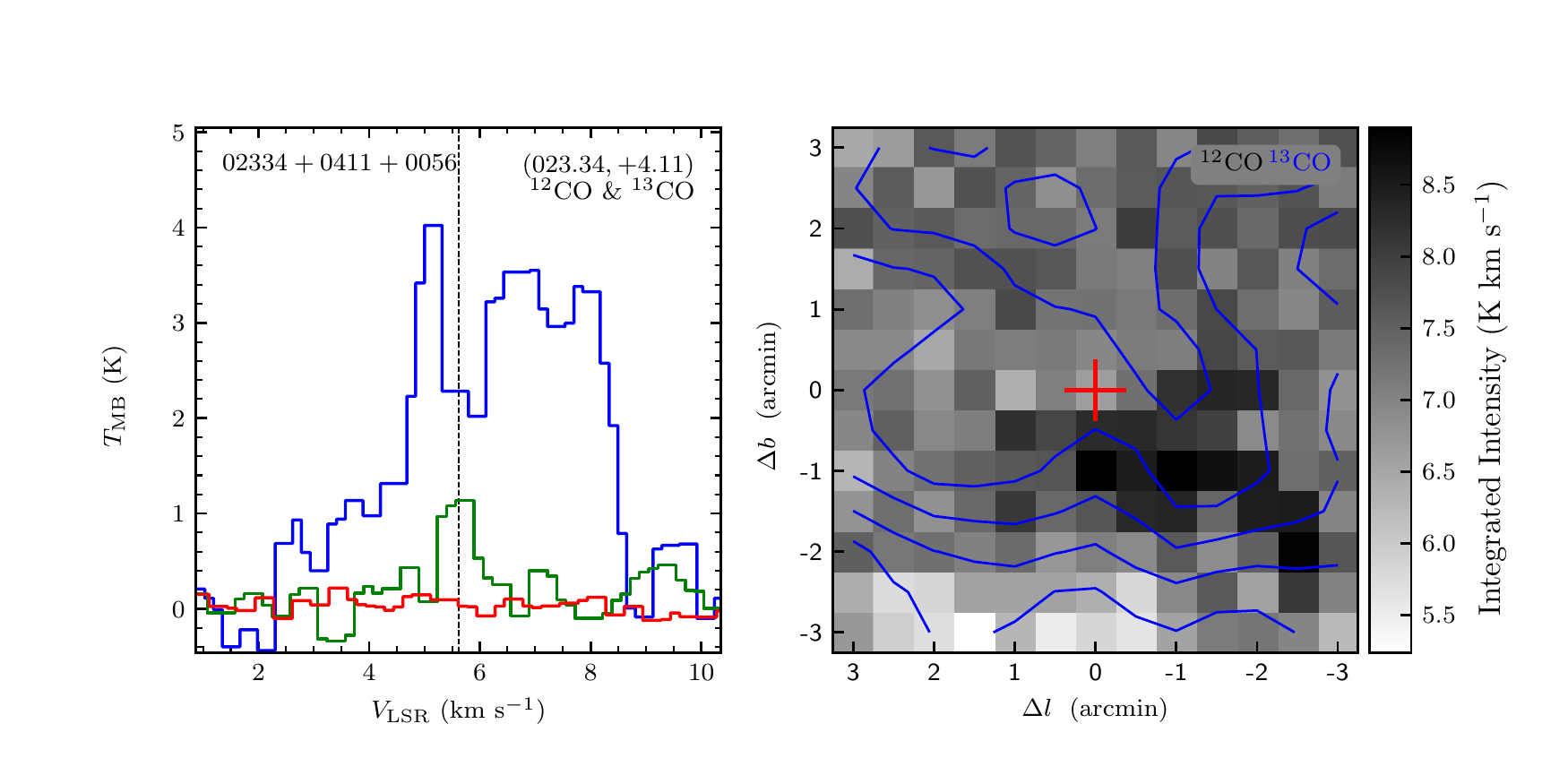}
\vspace{-0.5cm}

\includegraphics[width=9.0cm,angle=0]{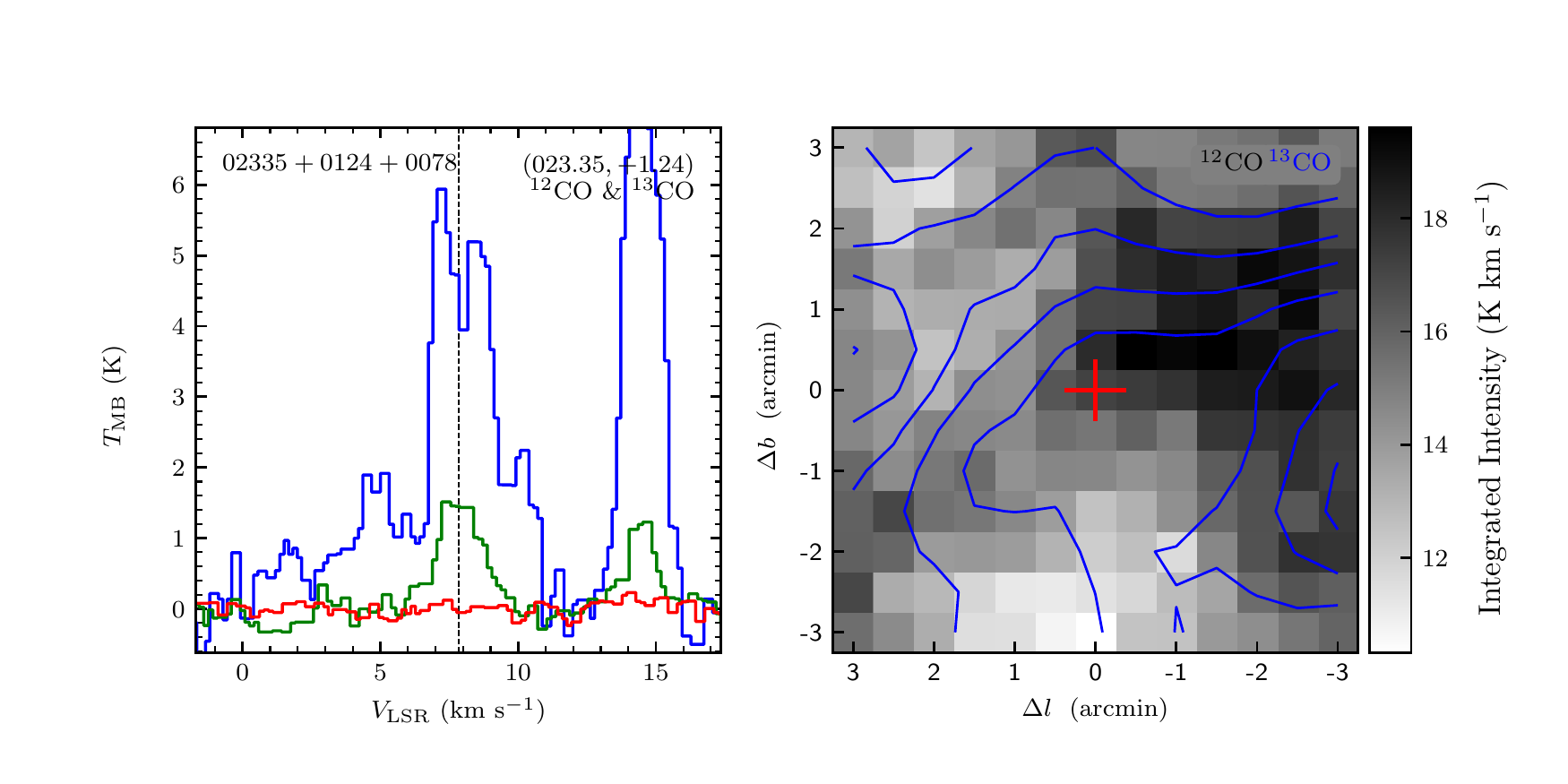}
\includegraphics[width=9.0cm,angle=0]{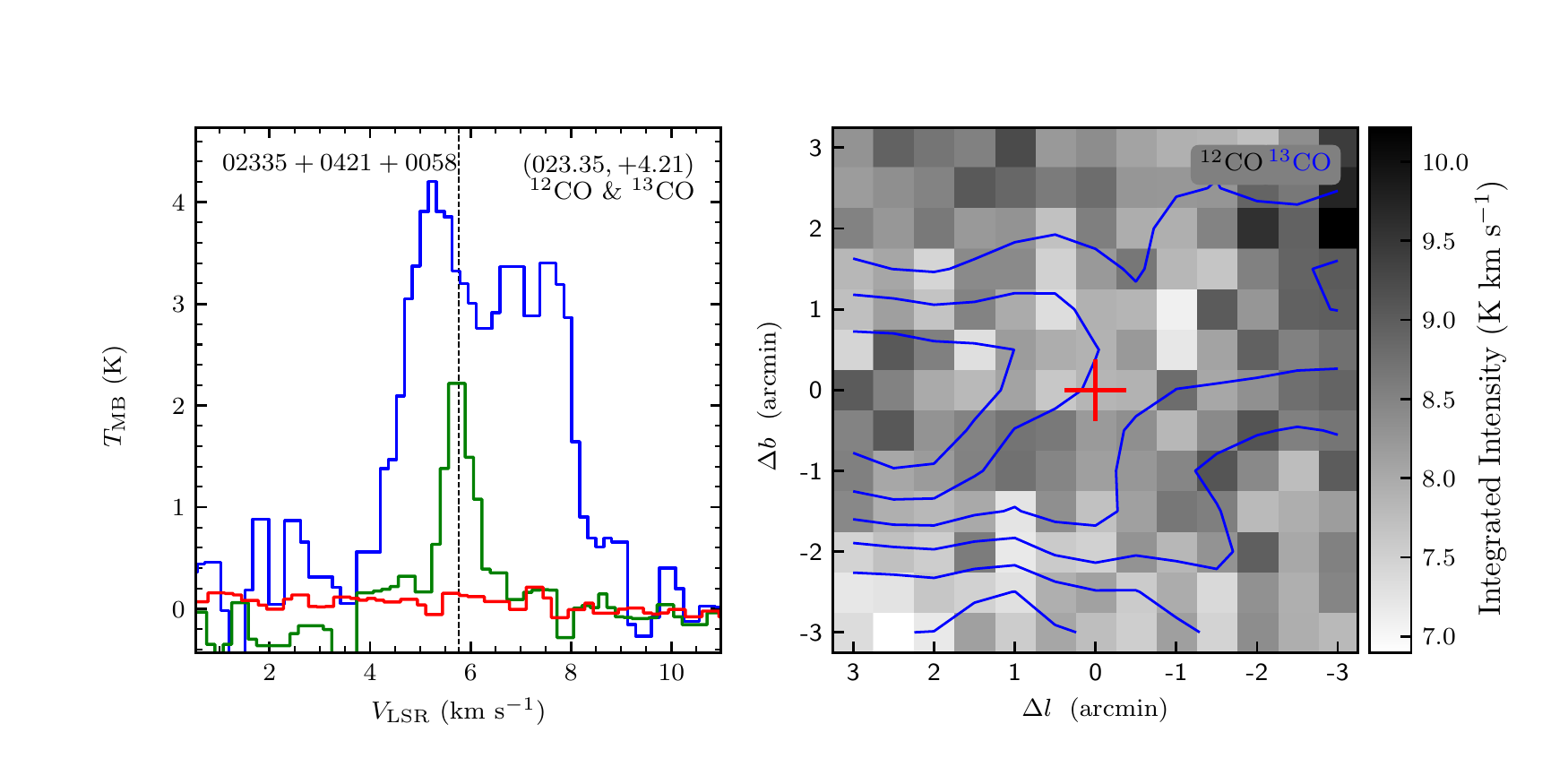}
\vspace{-0.5cm}

\includegraphics[width=9.0cm,angle=0]{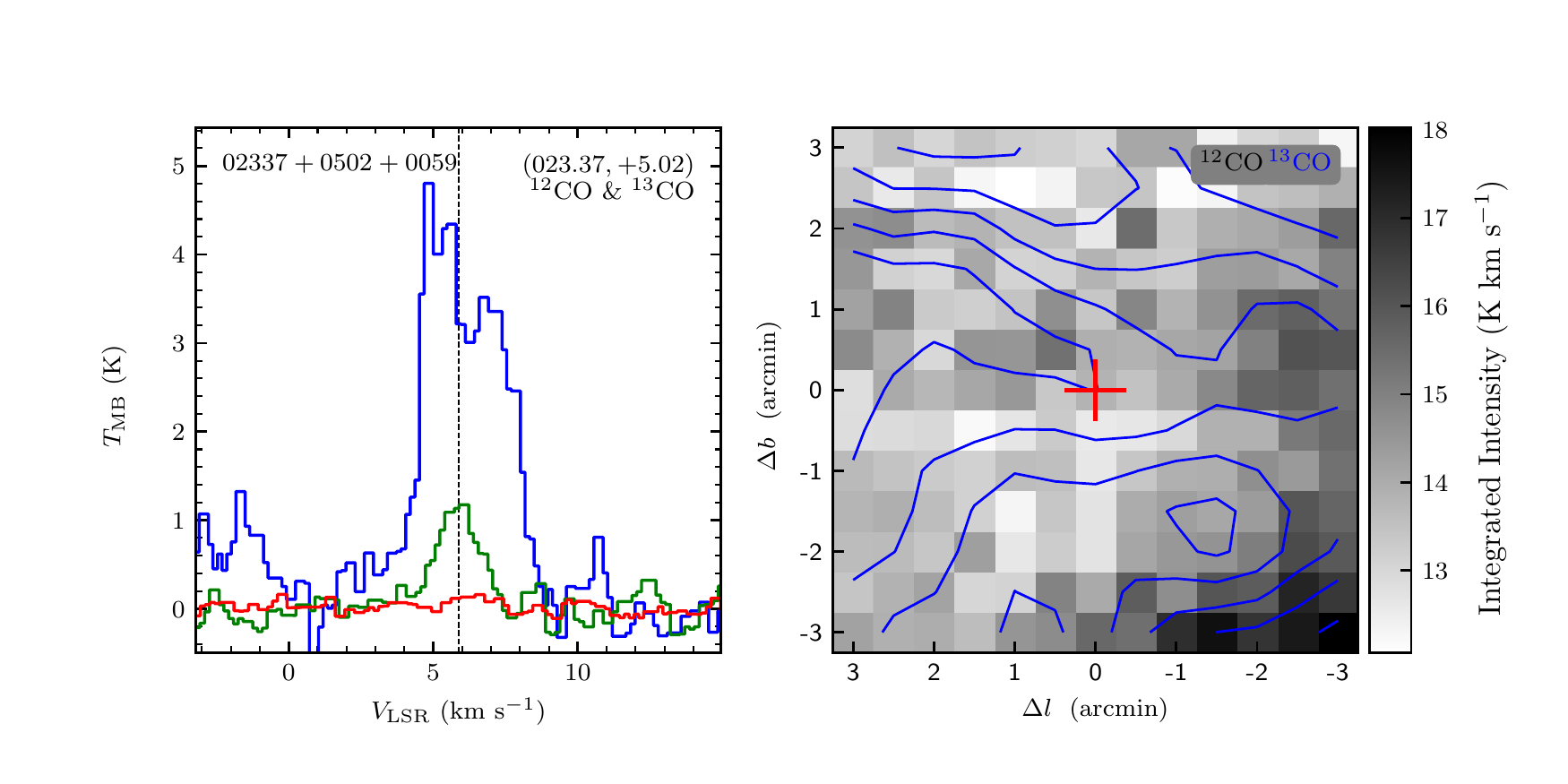}
\includegraphics[width=9.0cm,angle=0]{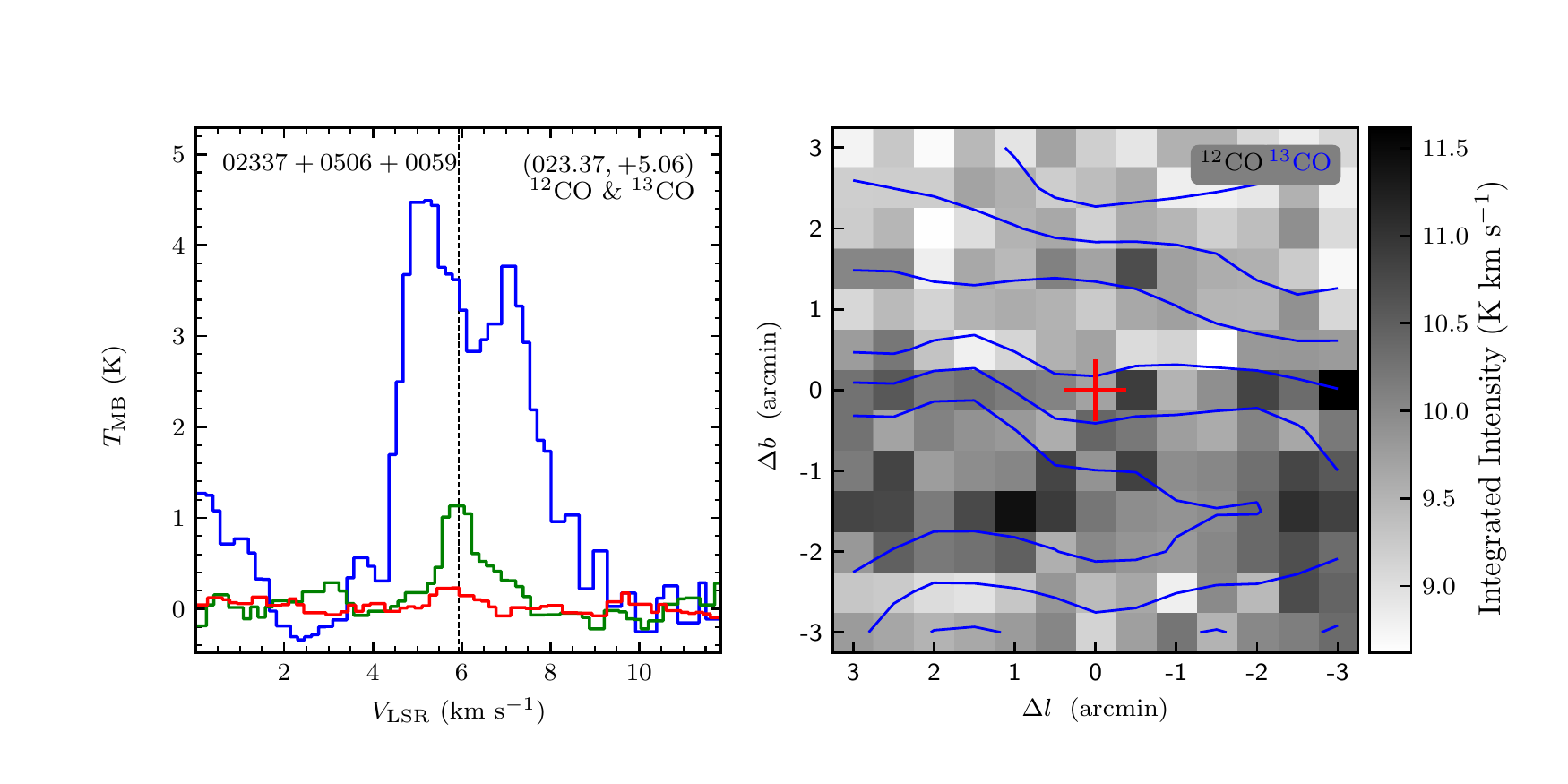}
\vspace{-0.5cm}

\includegraphics[width=9.0cm,angle=0]{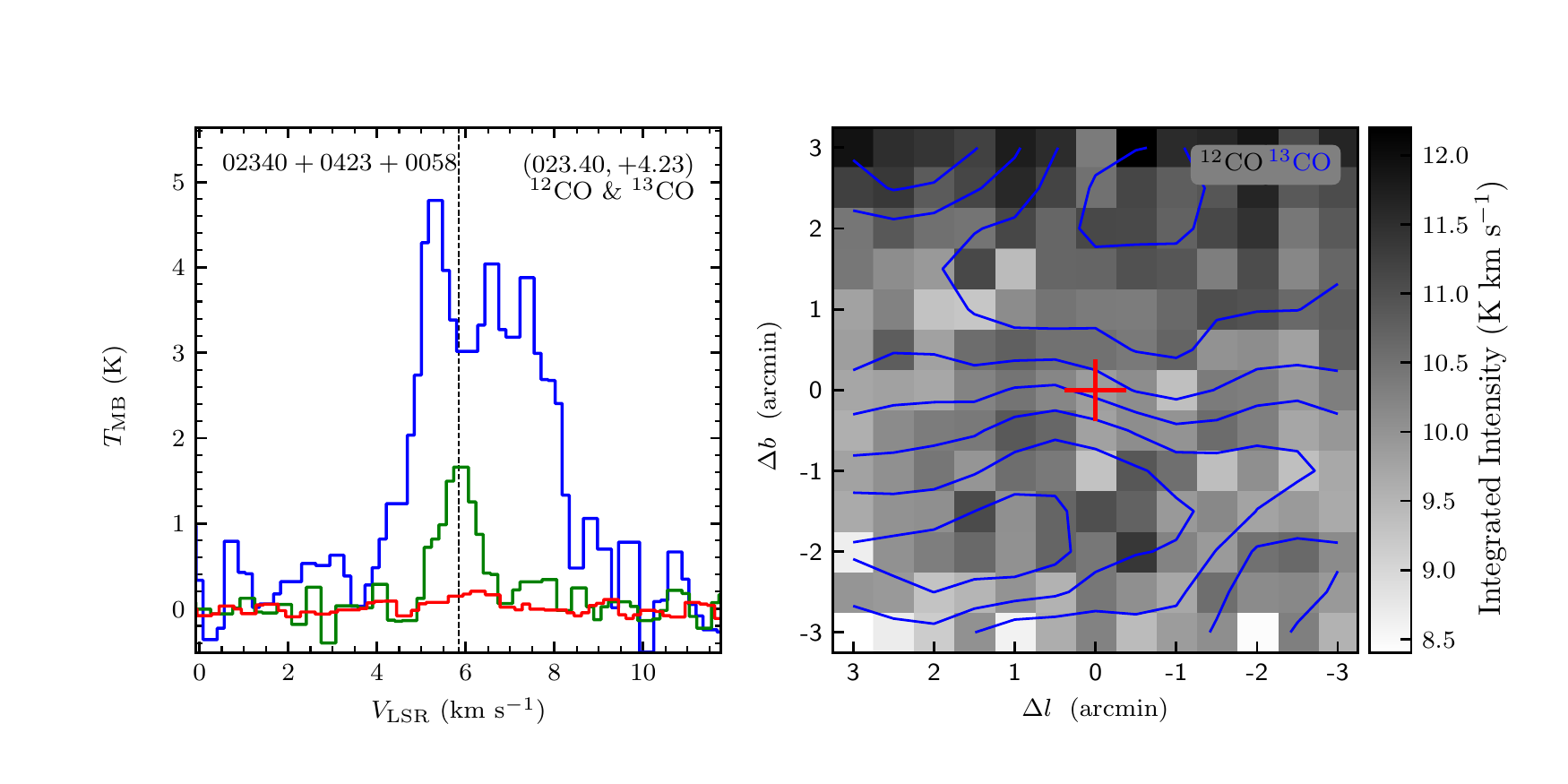}
\includegraphics[width=9.0cm,angle=0]{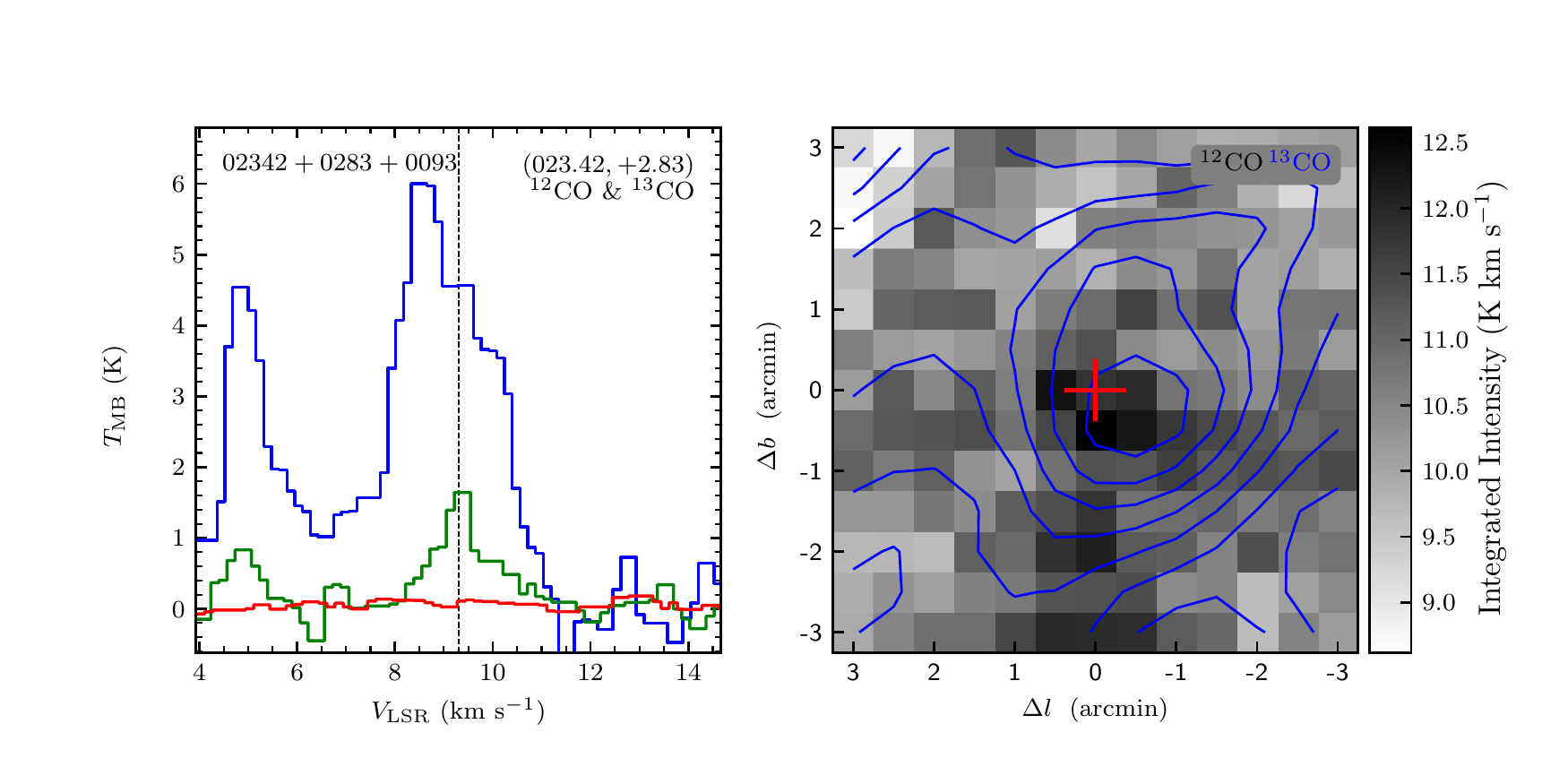}
\vspace{-0.5cm}

\includegraphics[width=9.0cm,angle=0]{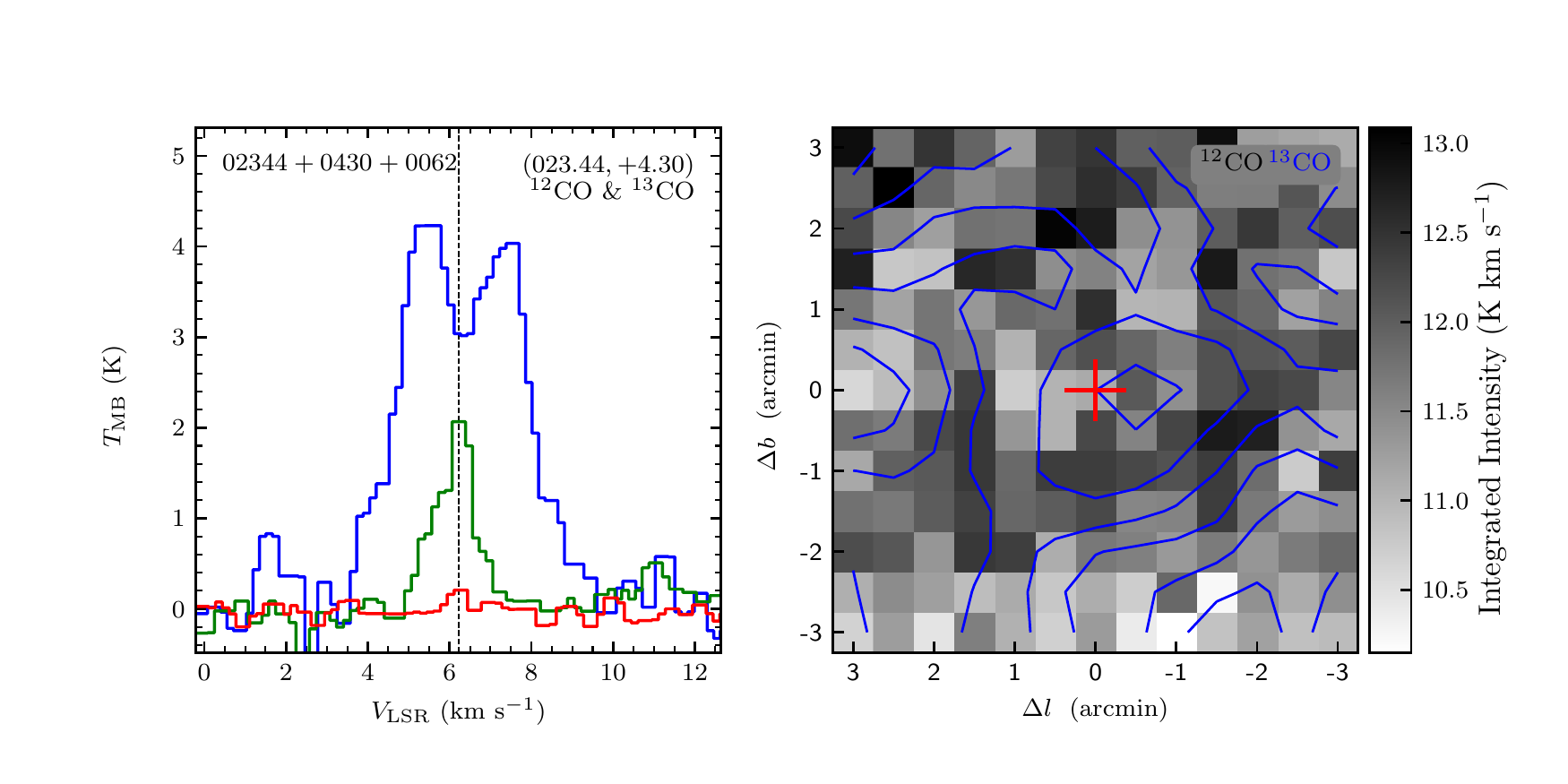}
\includegraphics[width=9.0cm,angle=0]{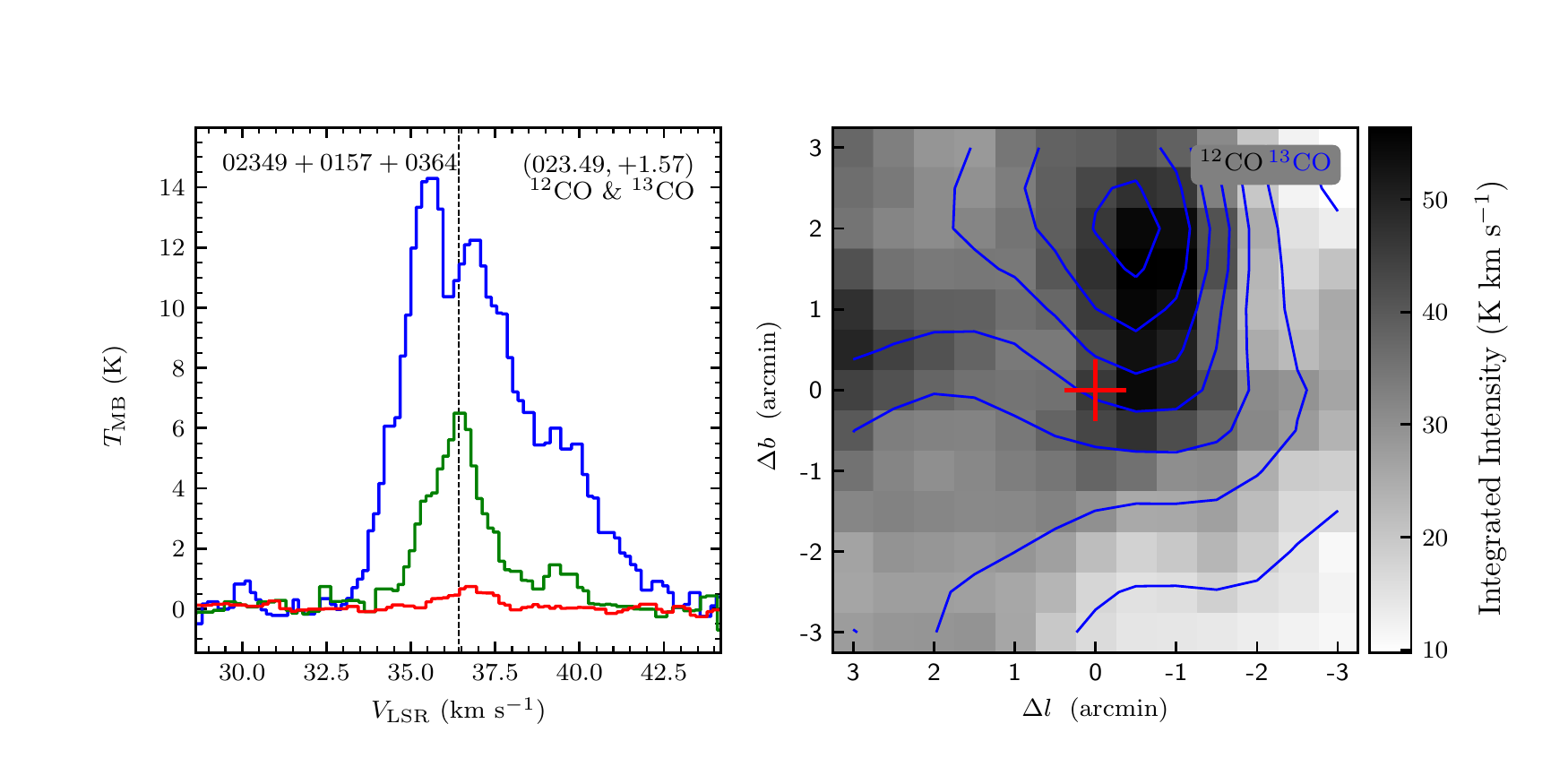}
\end{figure}
\clearpage

\begin{figure}
\includegraphics[width=9.0cm,angle=0]{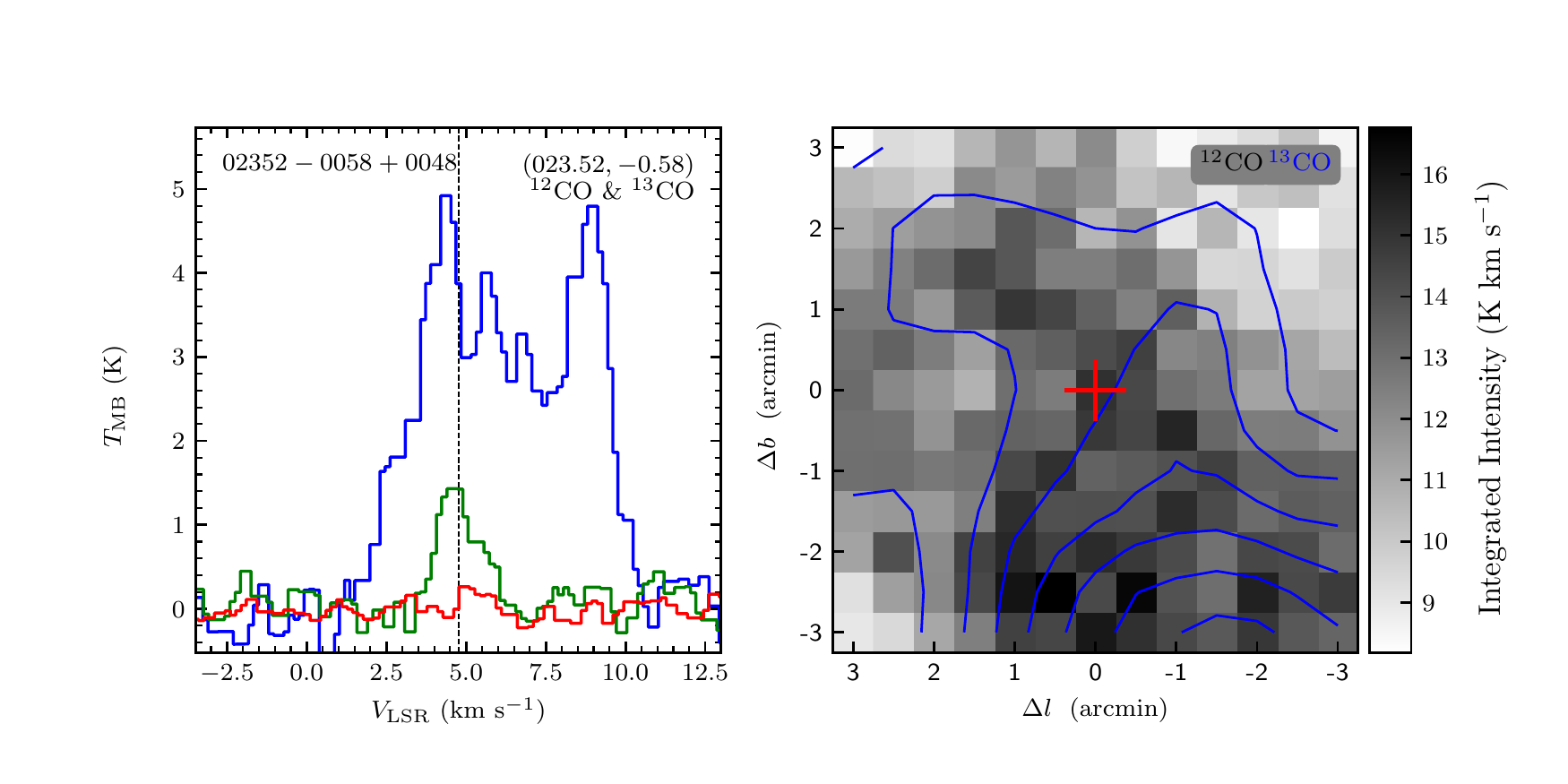}
\includegraphics[width=9.0cm,angle=0]{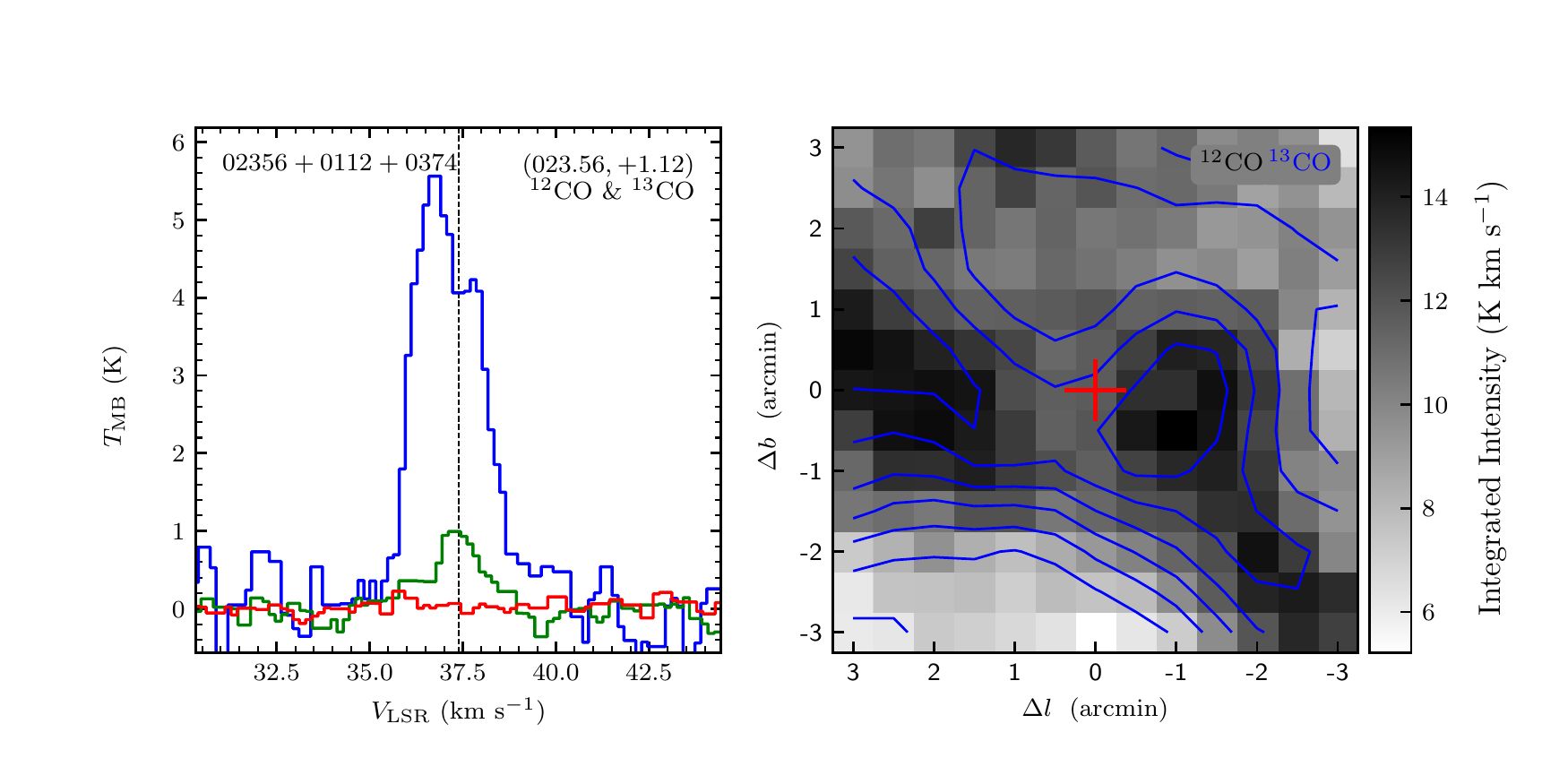}
\vspace{-0.5cm}

\includegraphics[width=9.0cm,angle=0]{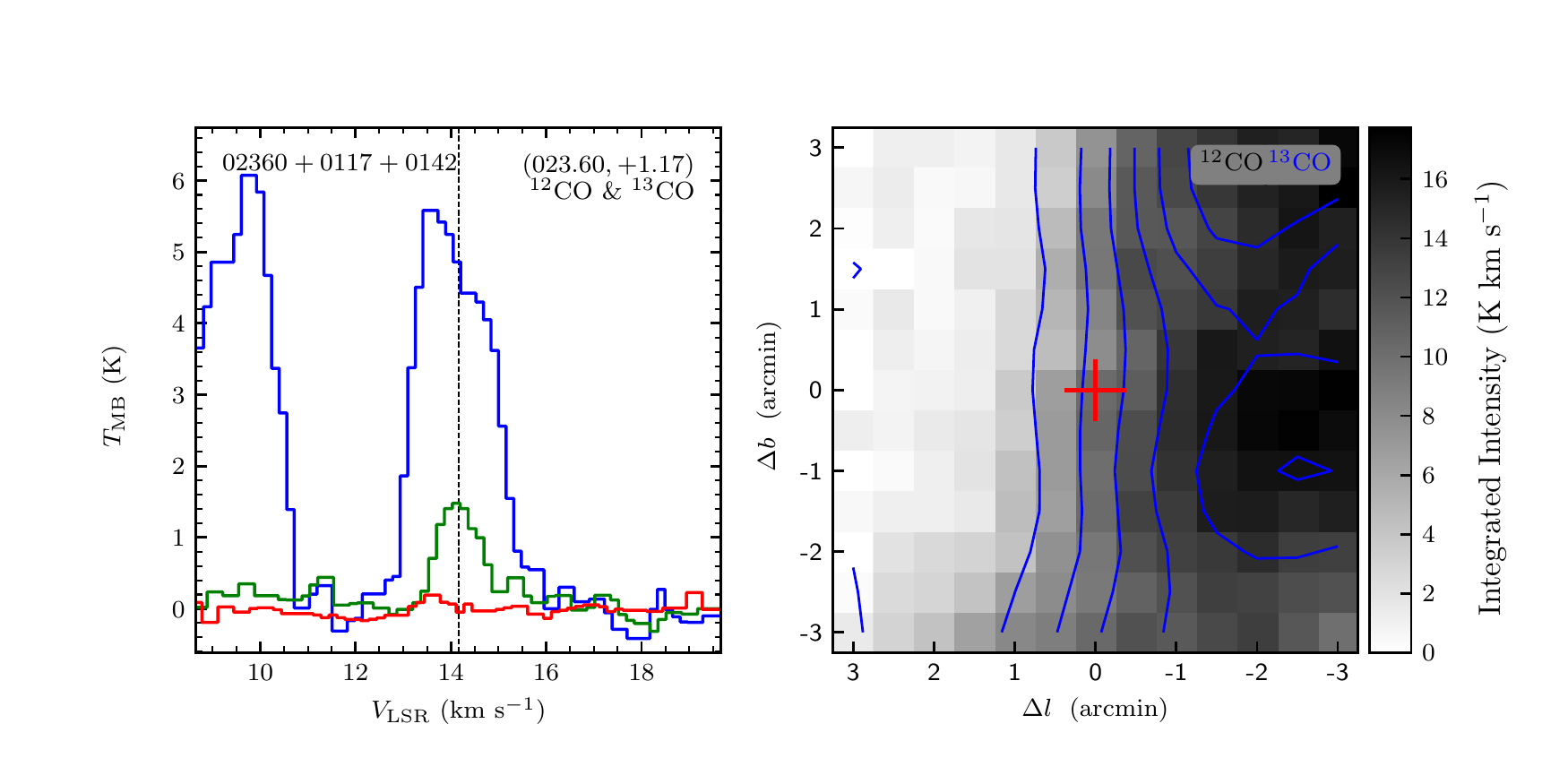}
\includegraphics[width=9.0cm,angle=0]{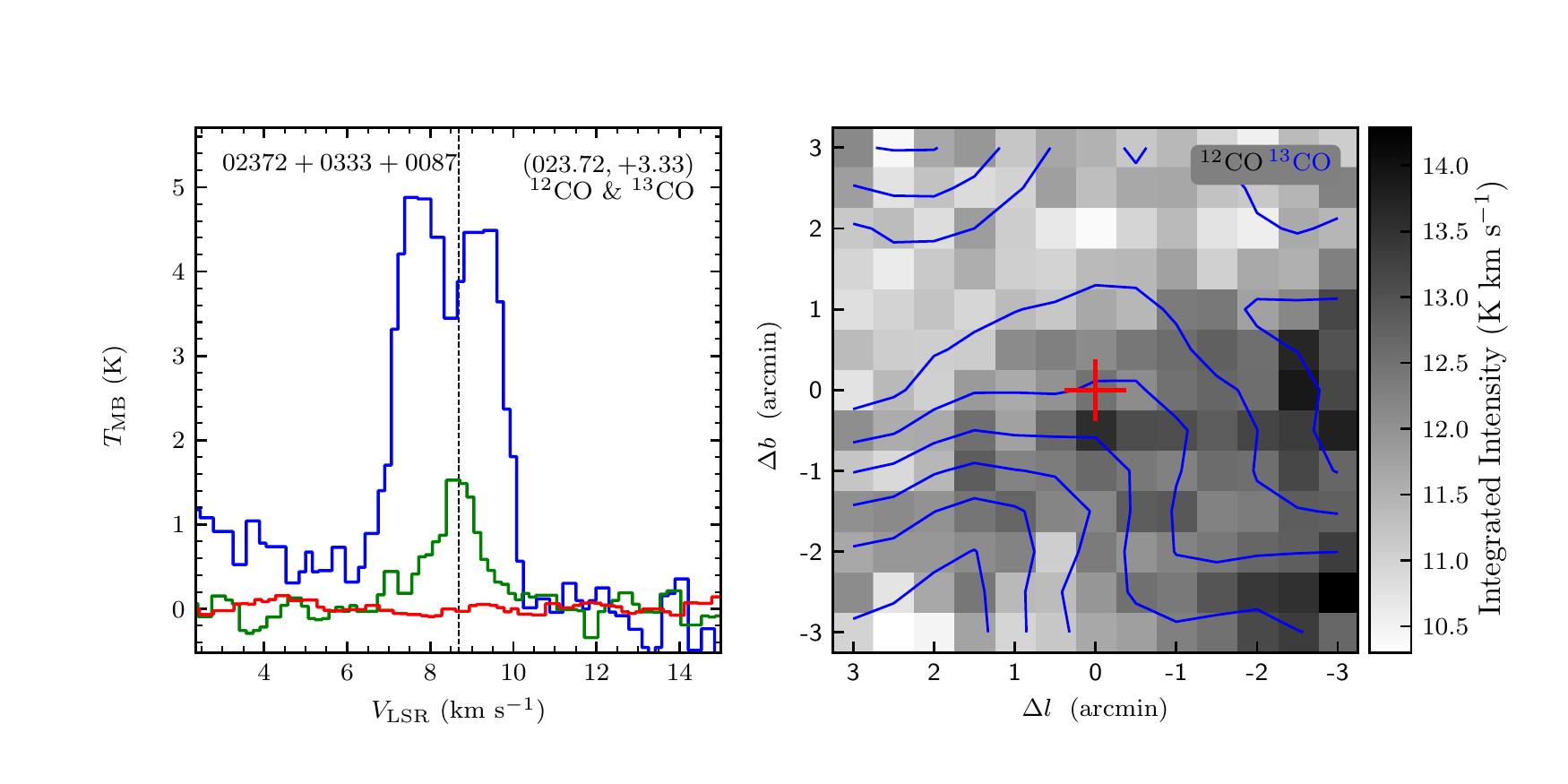}
\vspace{-0.5cm}

\includegraphics[width=9.0cm,angle=0]{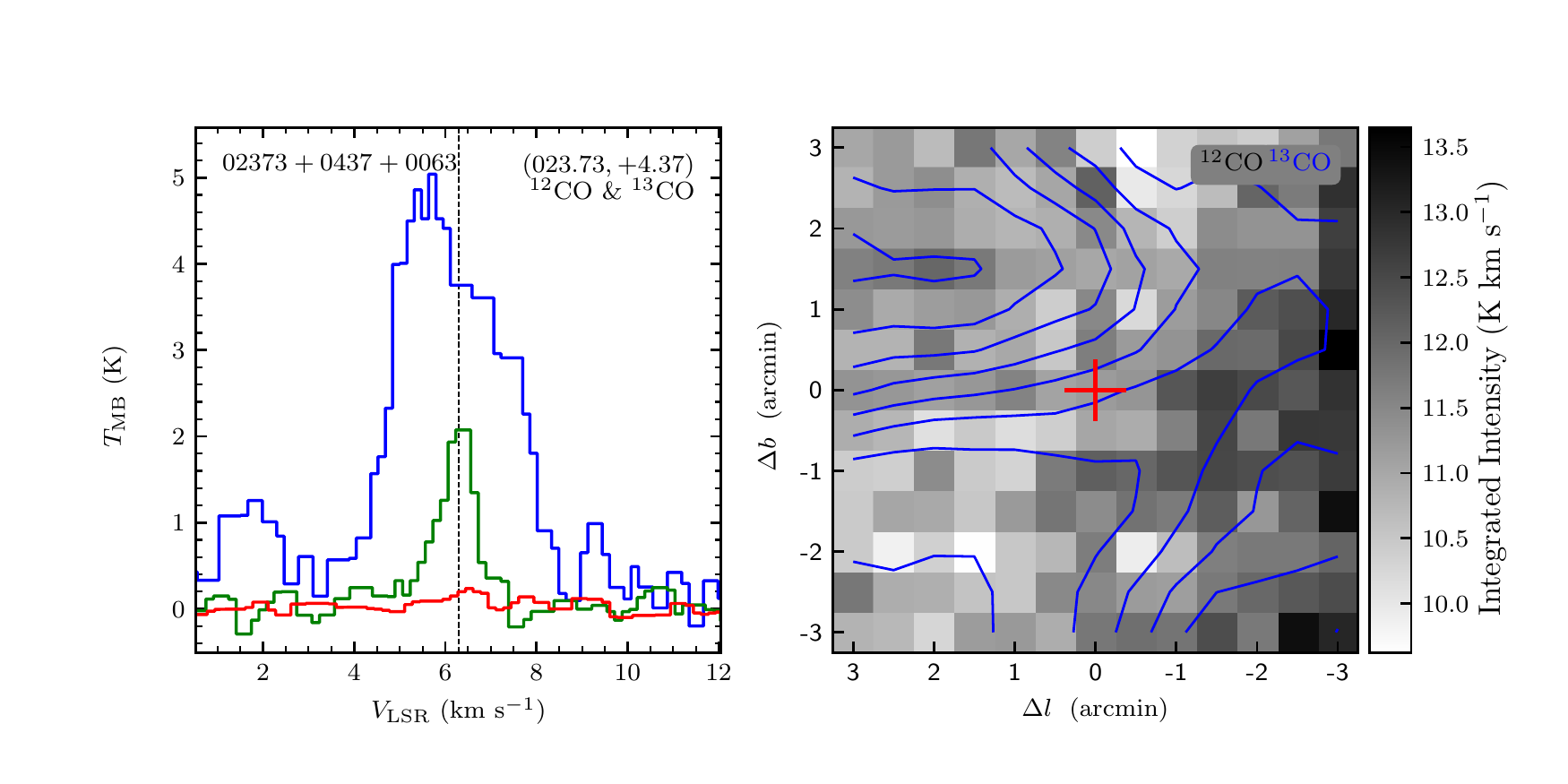}
\includegraphics[width=9.0cm,angle=0]{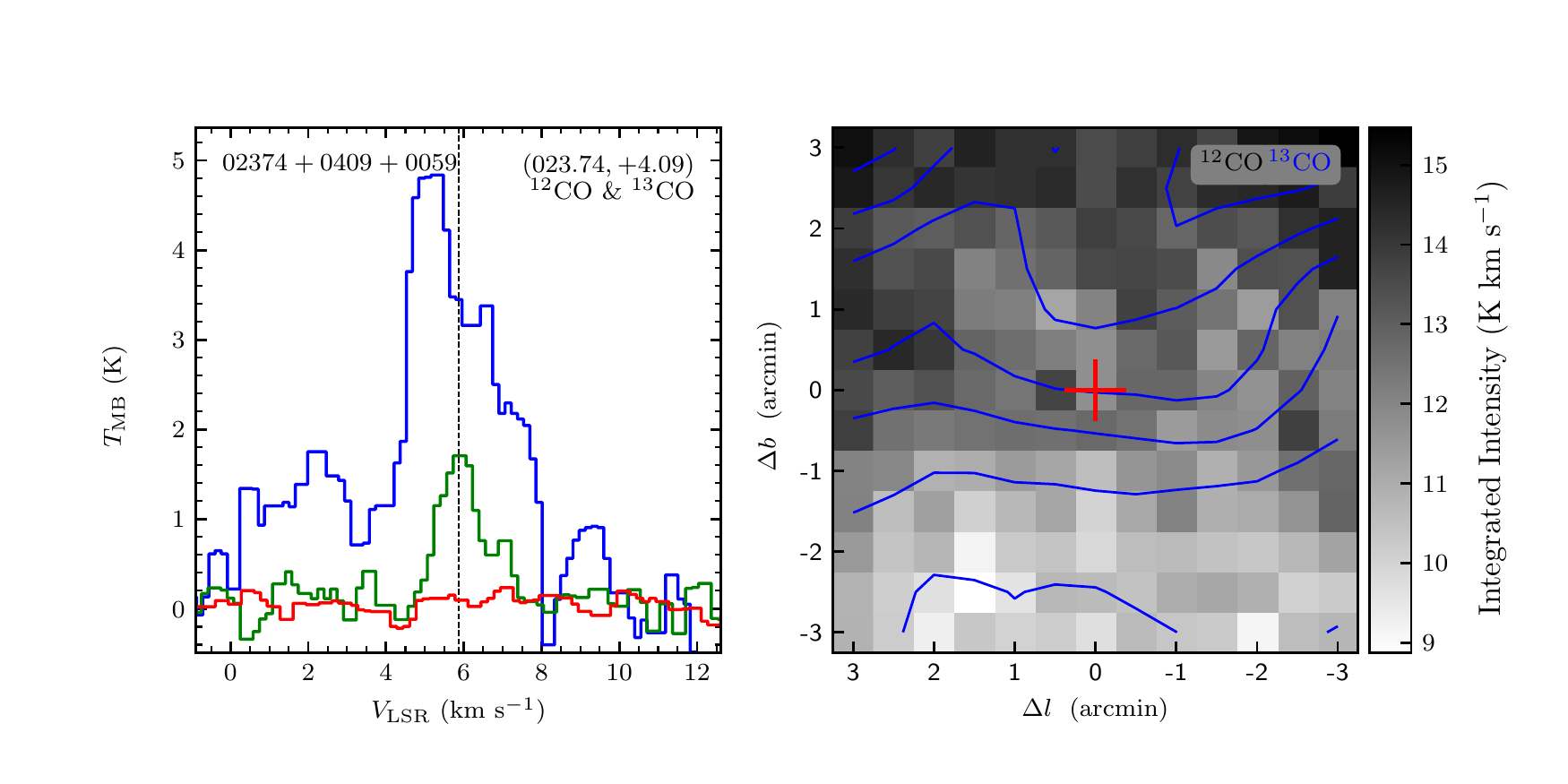}
\vspace{-0.5cm}

\includegraphics[width=9.0cm,angle=0]{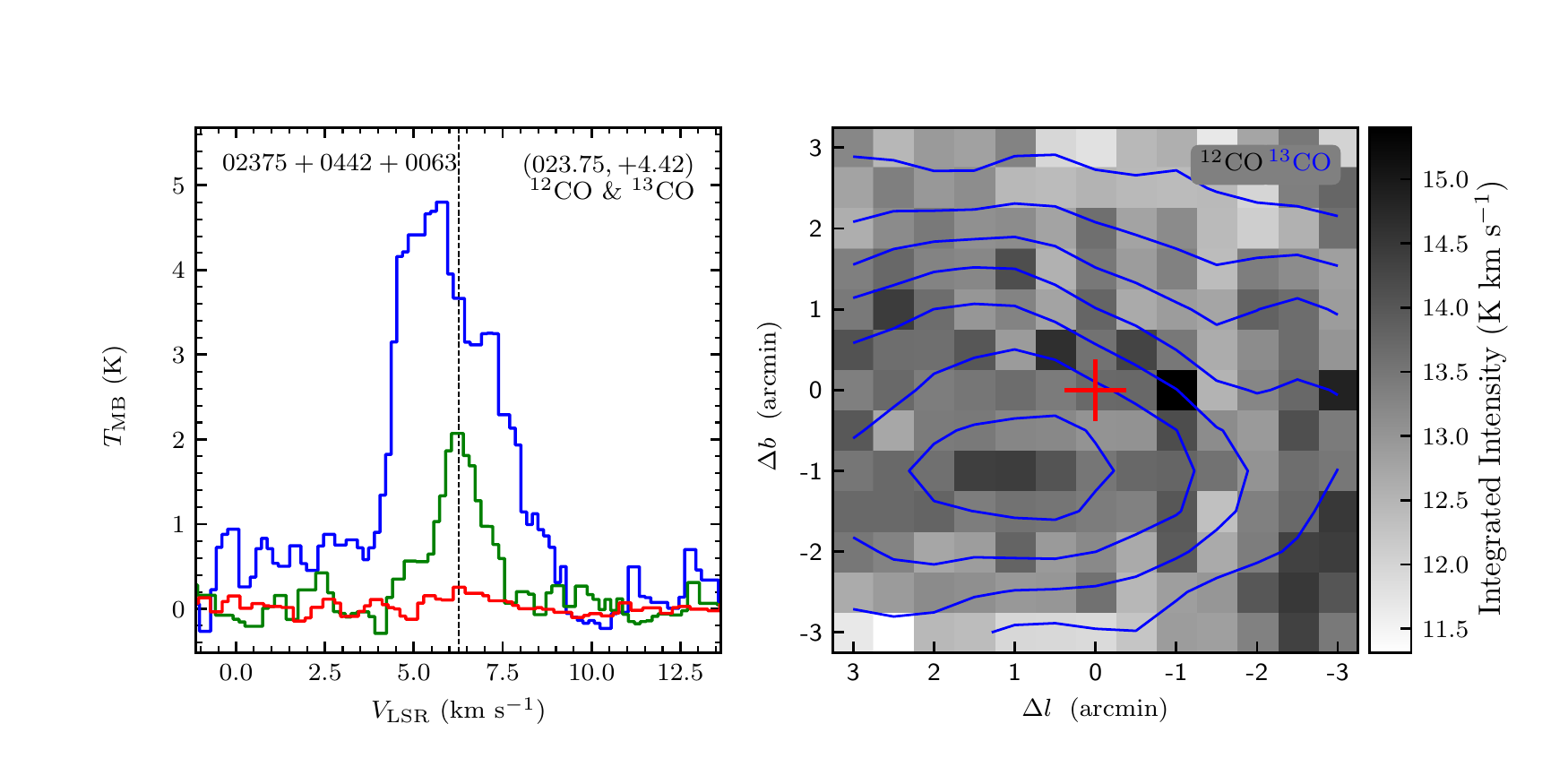}
\includegraphics[width=9.0cm,angle=0]{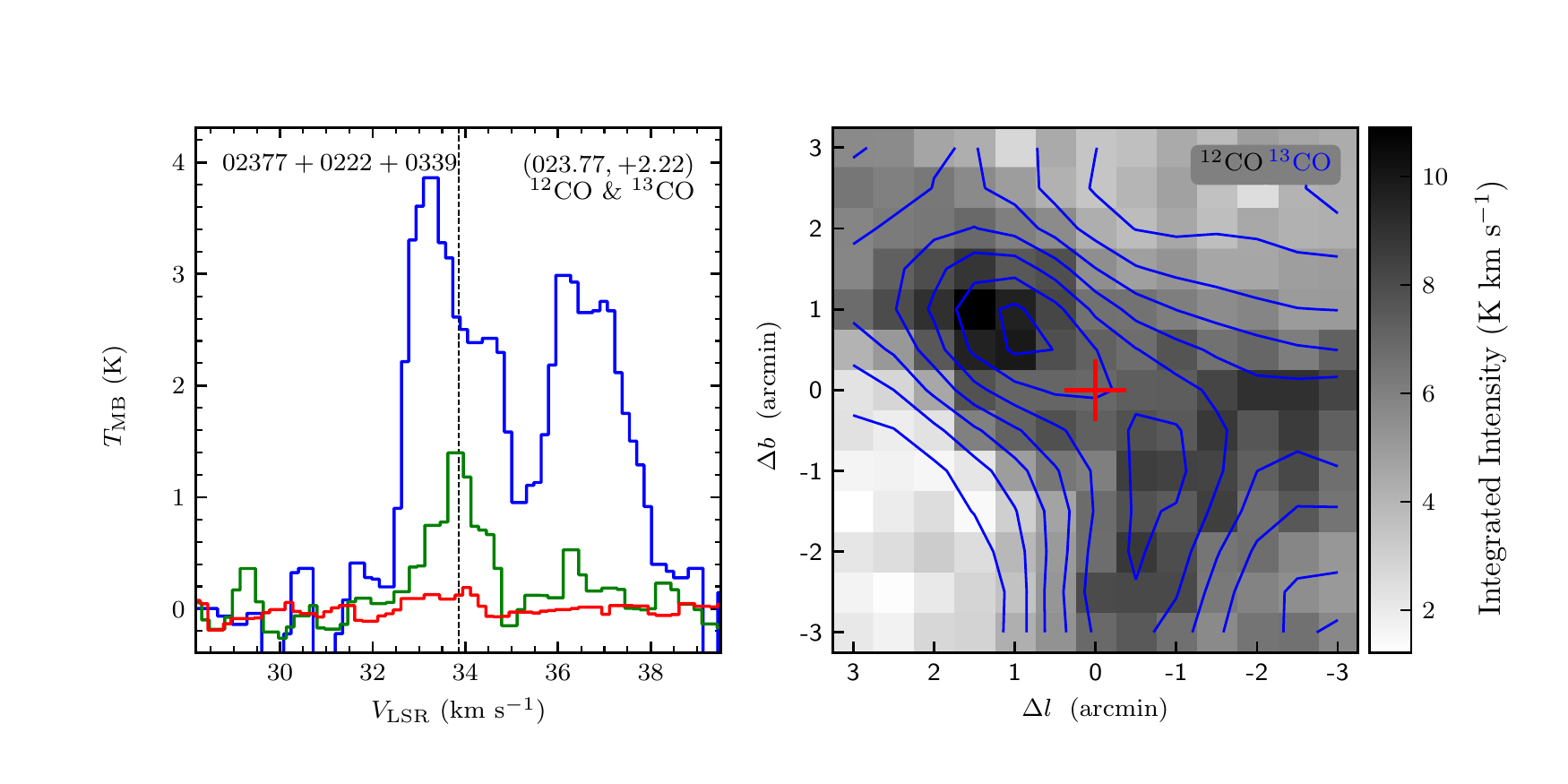}
\vspace{-0.5cm}

\includegraphics[width=9.0cm,angle=0]{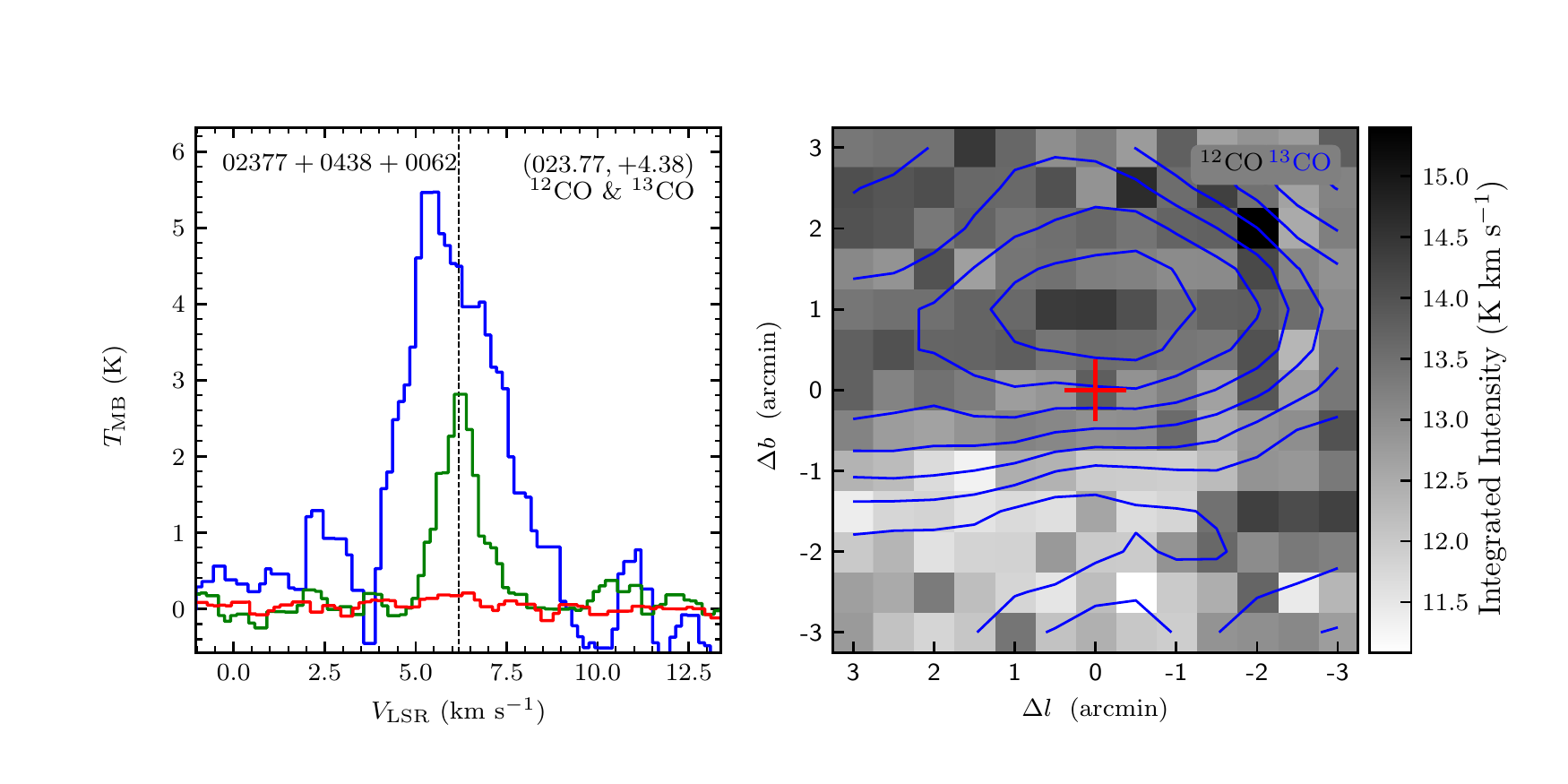}
\includegraphics[width=9.0cm,angle=0]{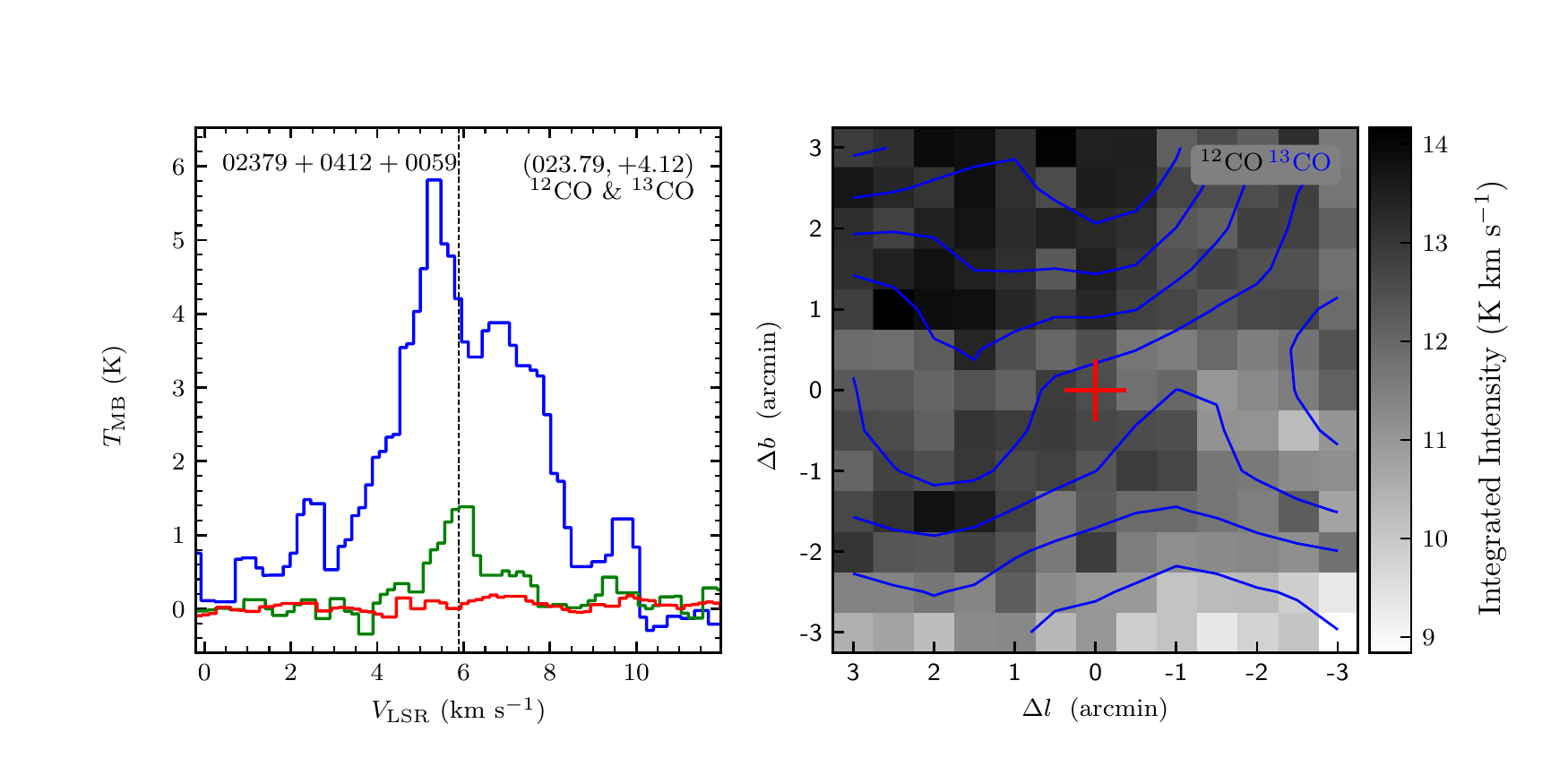}
\end{figure}
\clearpage

\begin{figure}
\includegraphics[width=9.0cm,angle=0]{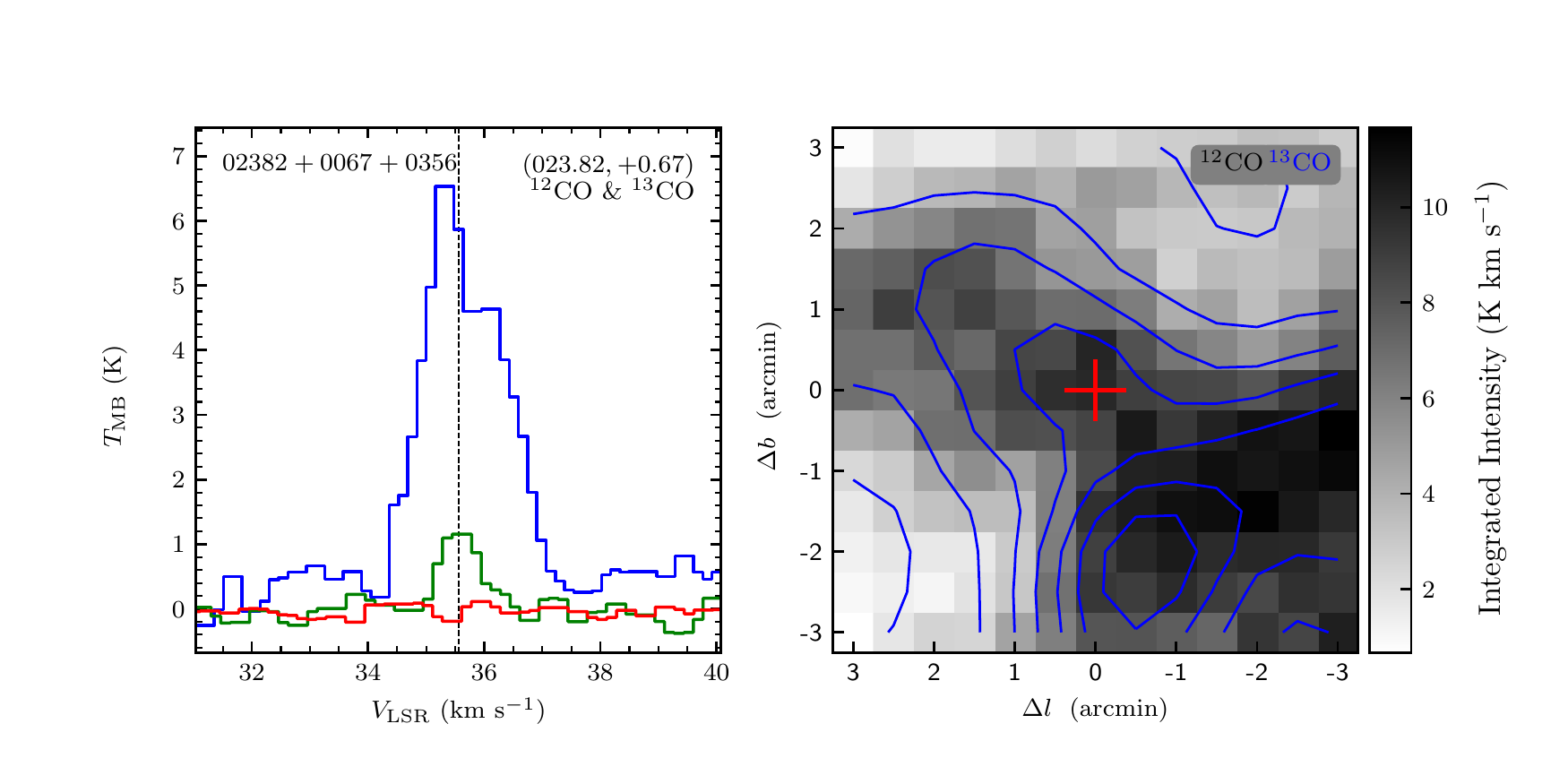}
\includegraphics[width=9.0cm,angle=0]{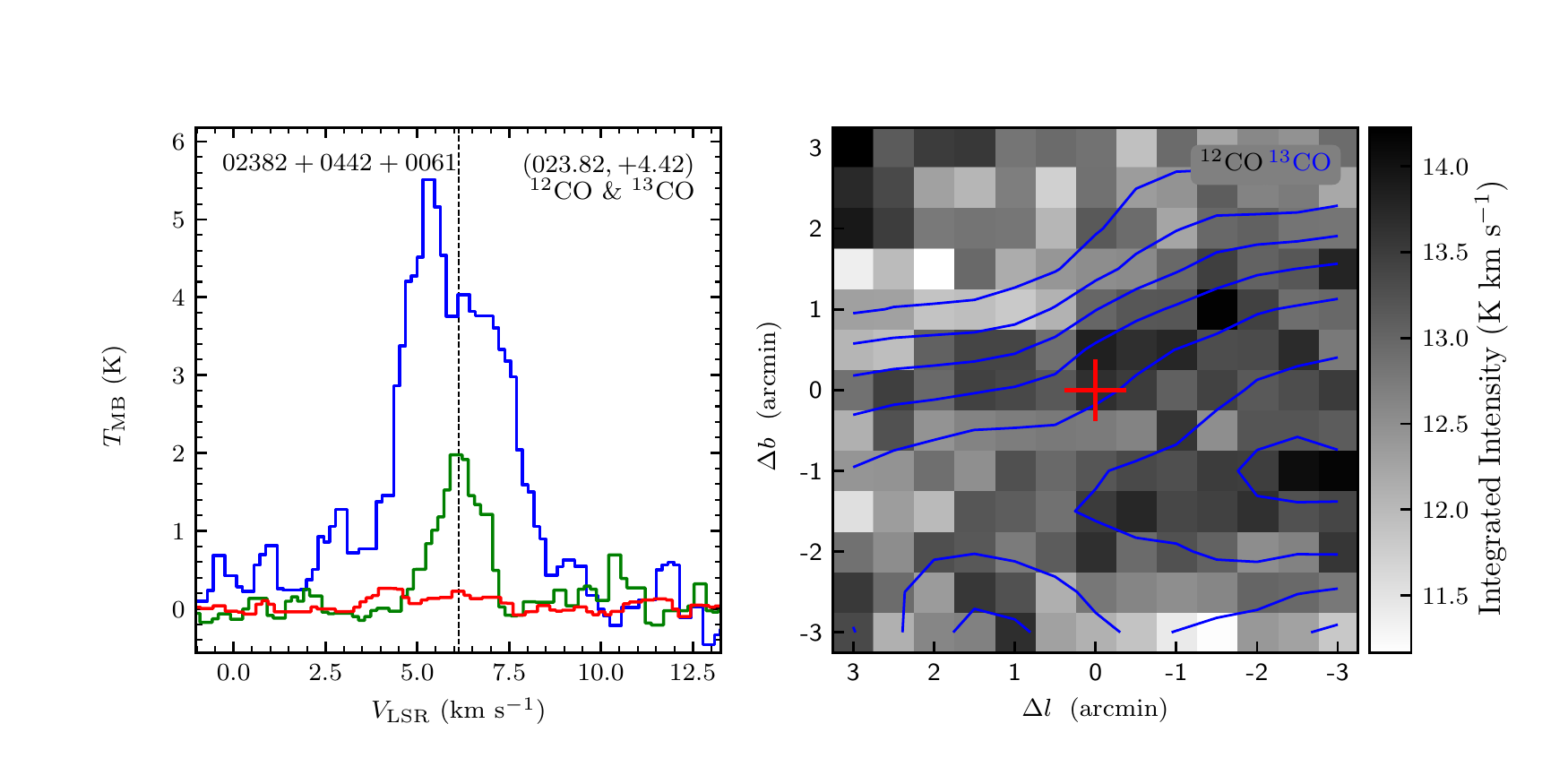}
\vspace{-0.5cm}

\includegraphics[width=9.0cm,angle=0]{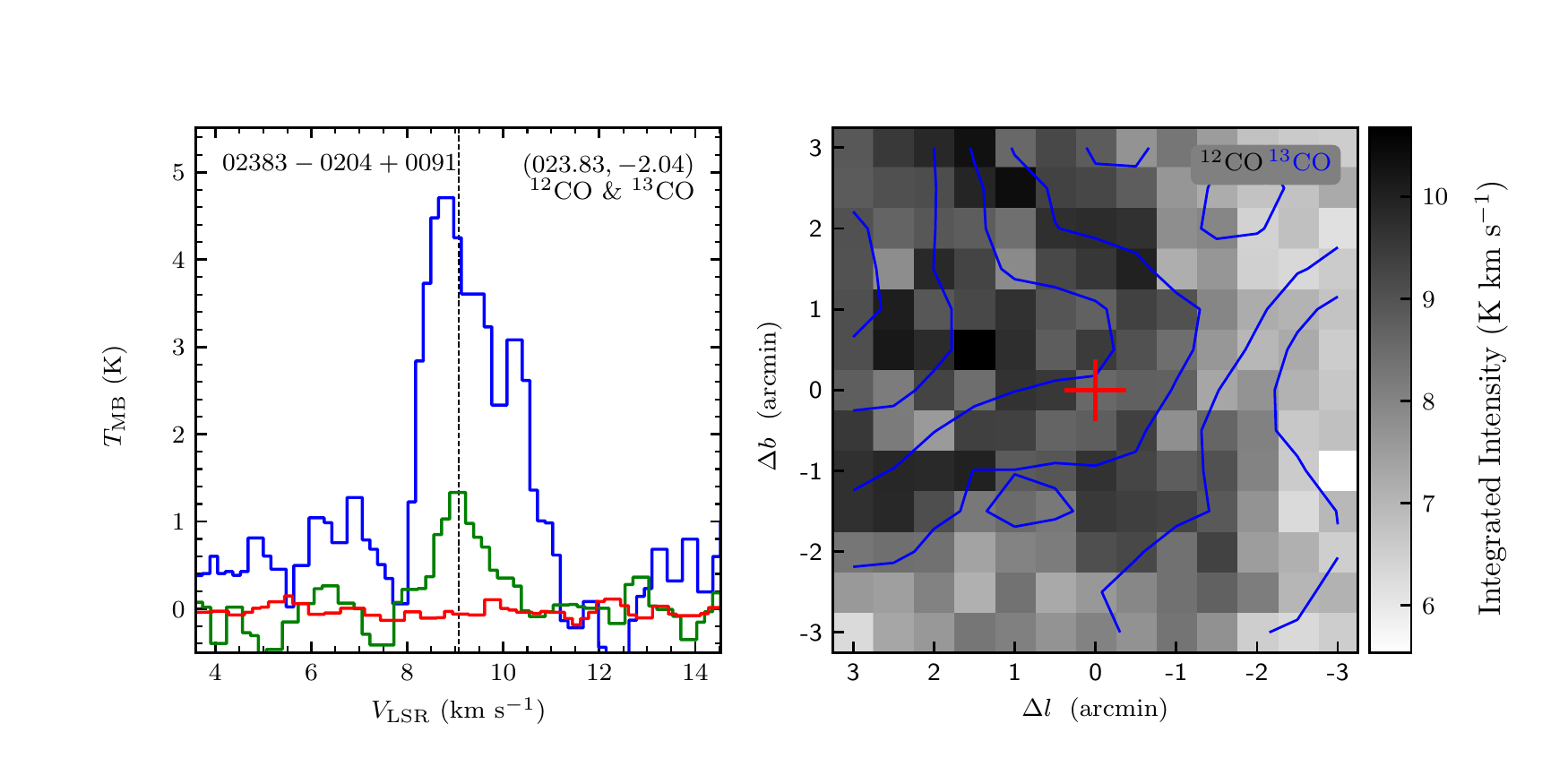}
\includegraphics[width=9.0cm,angle=0]{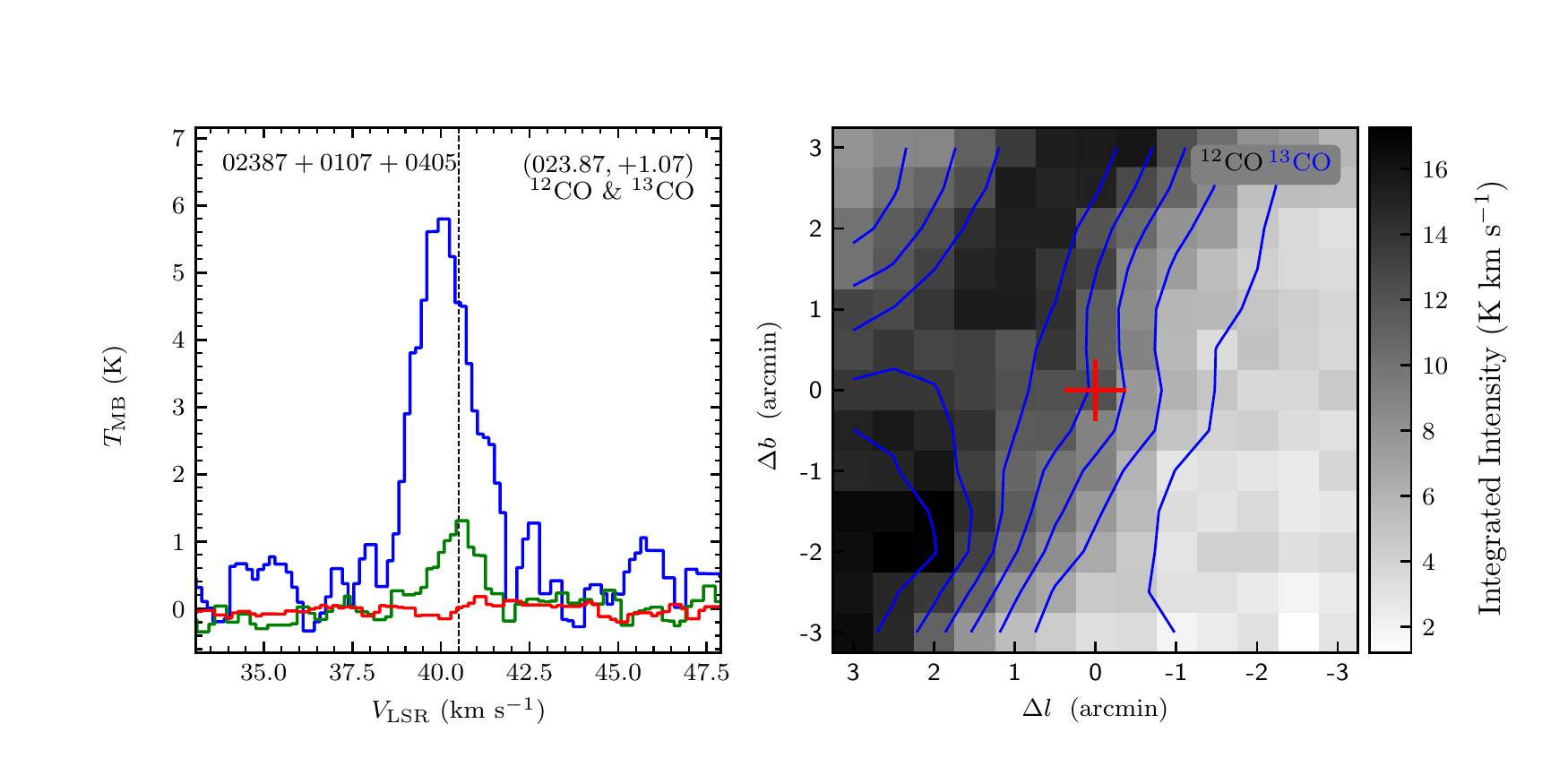}
\vspace{-0.5cm}

\includegraphics[width=9.0cm,angle=0]{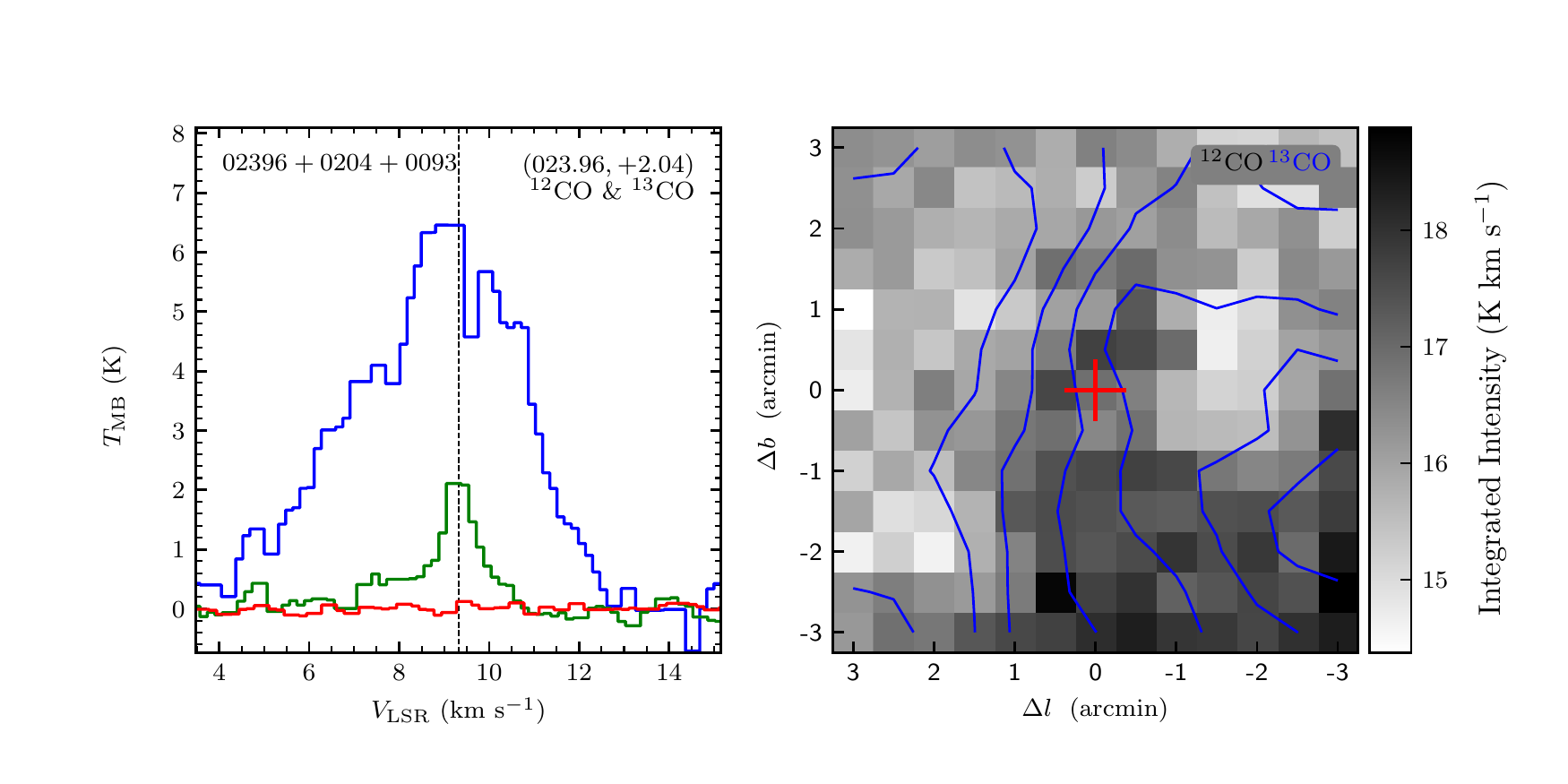}
\includegraphics[width=9.0cm,angle=0]{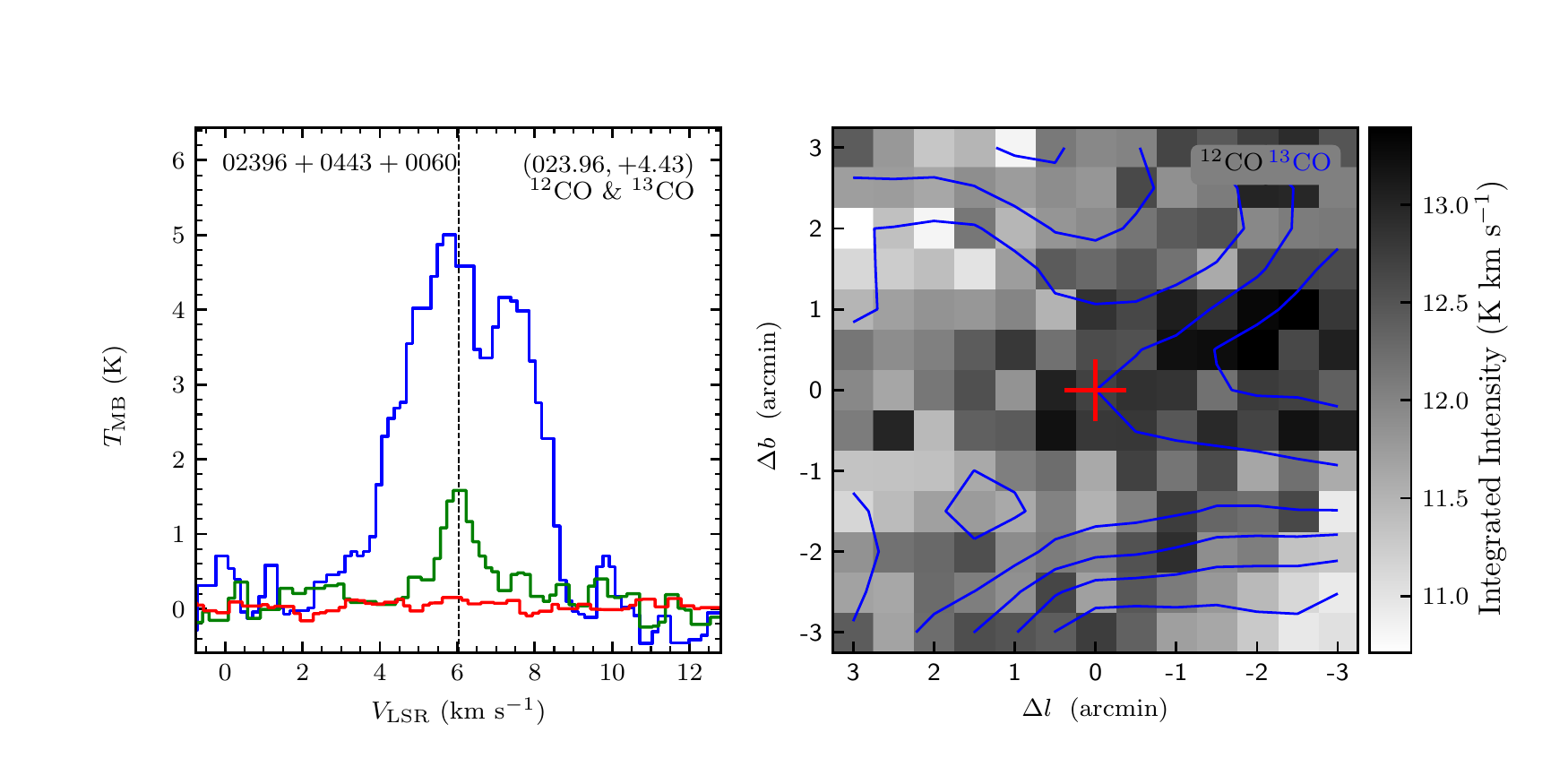}
\vspace{-0.5cm}

\includegraphics[width=9.0cm,angle=0]{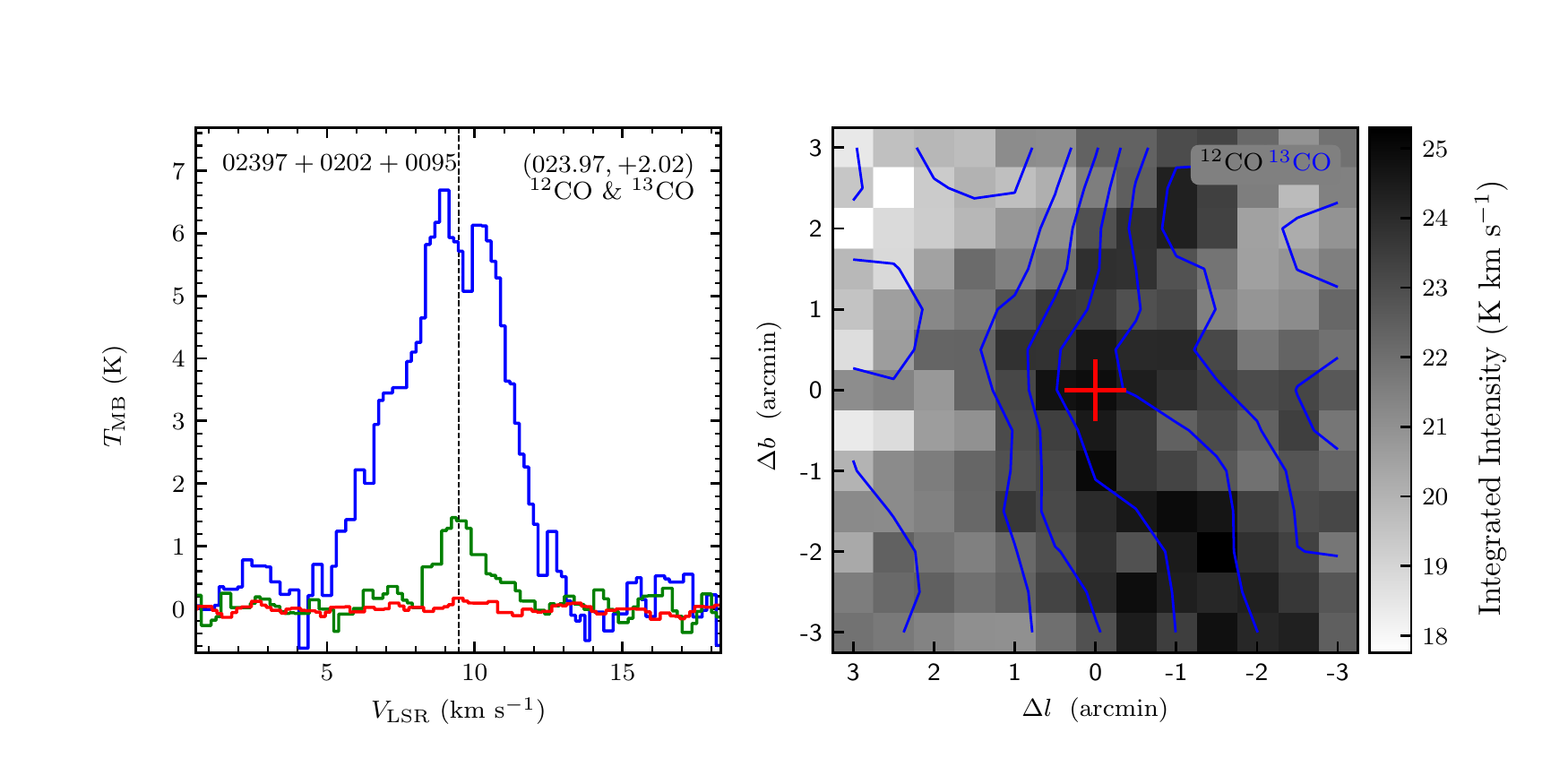}
\includegraphics[width=9.0cm,angle=0]{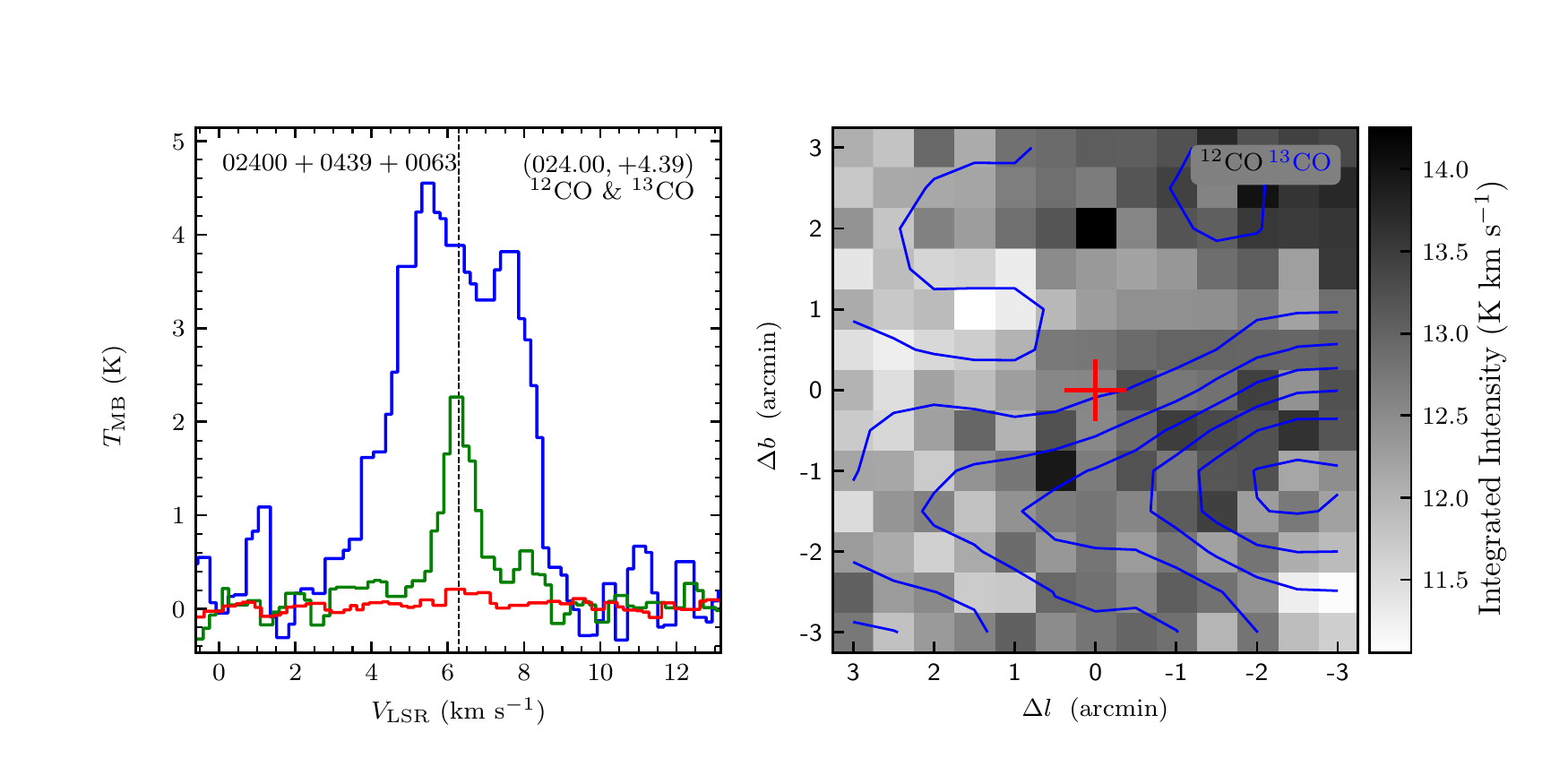}
\vspace{-0.5cm}

\includegraphics[width=9.0cm,angle=0]{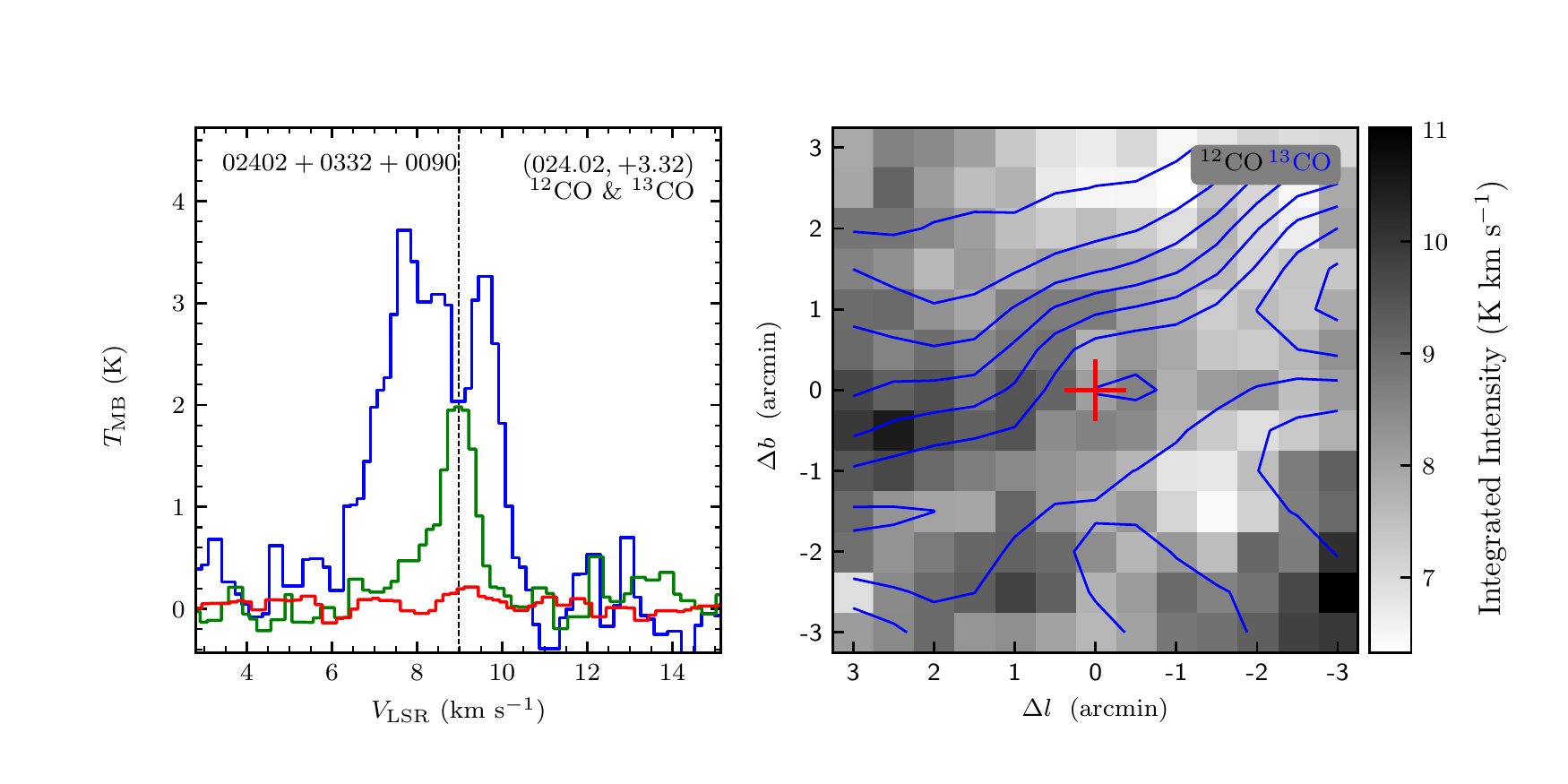}
\includegraphics[width=9.0cm,angle=0]{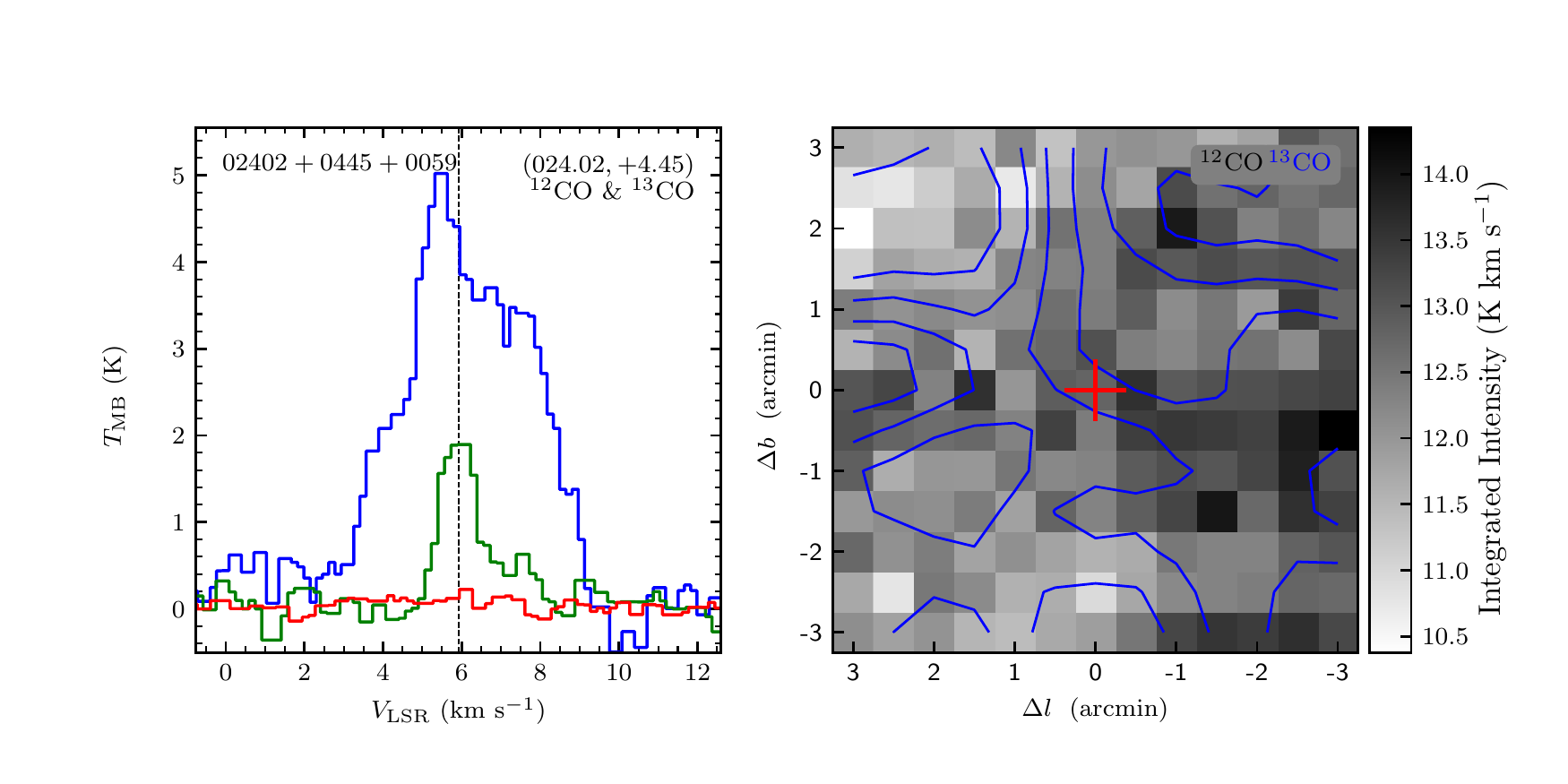}
\end{figure}
\clearpage

\begin{figure}
\includegraphics[width=9.0cm,angle=0]{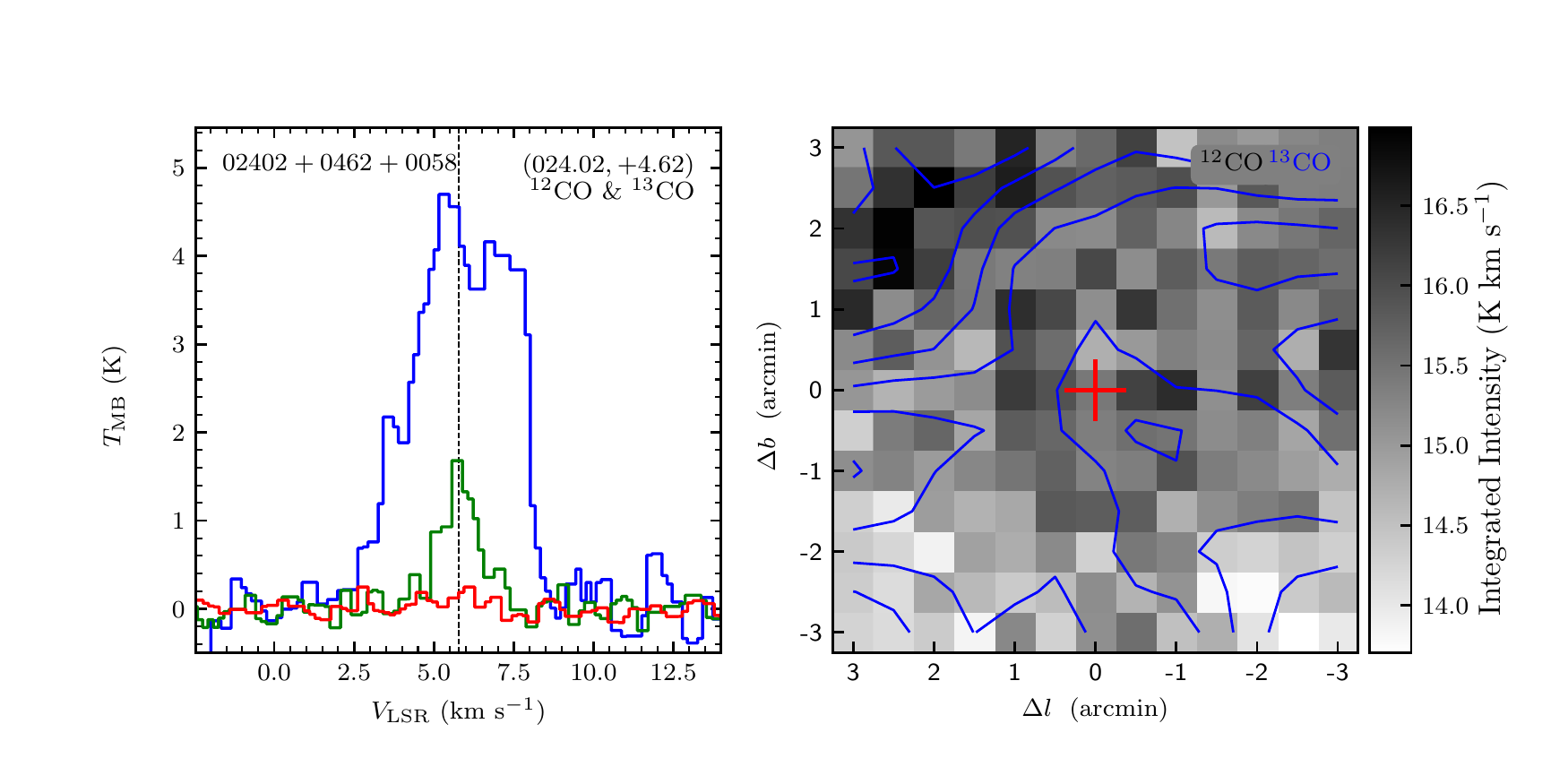}
\includegraphics[width=9.0cm,angle=0]{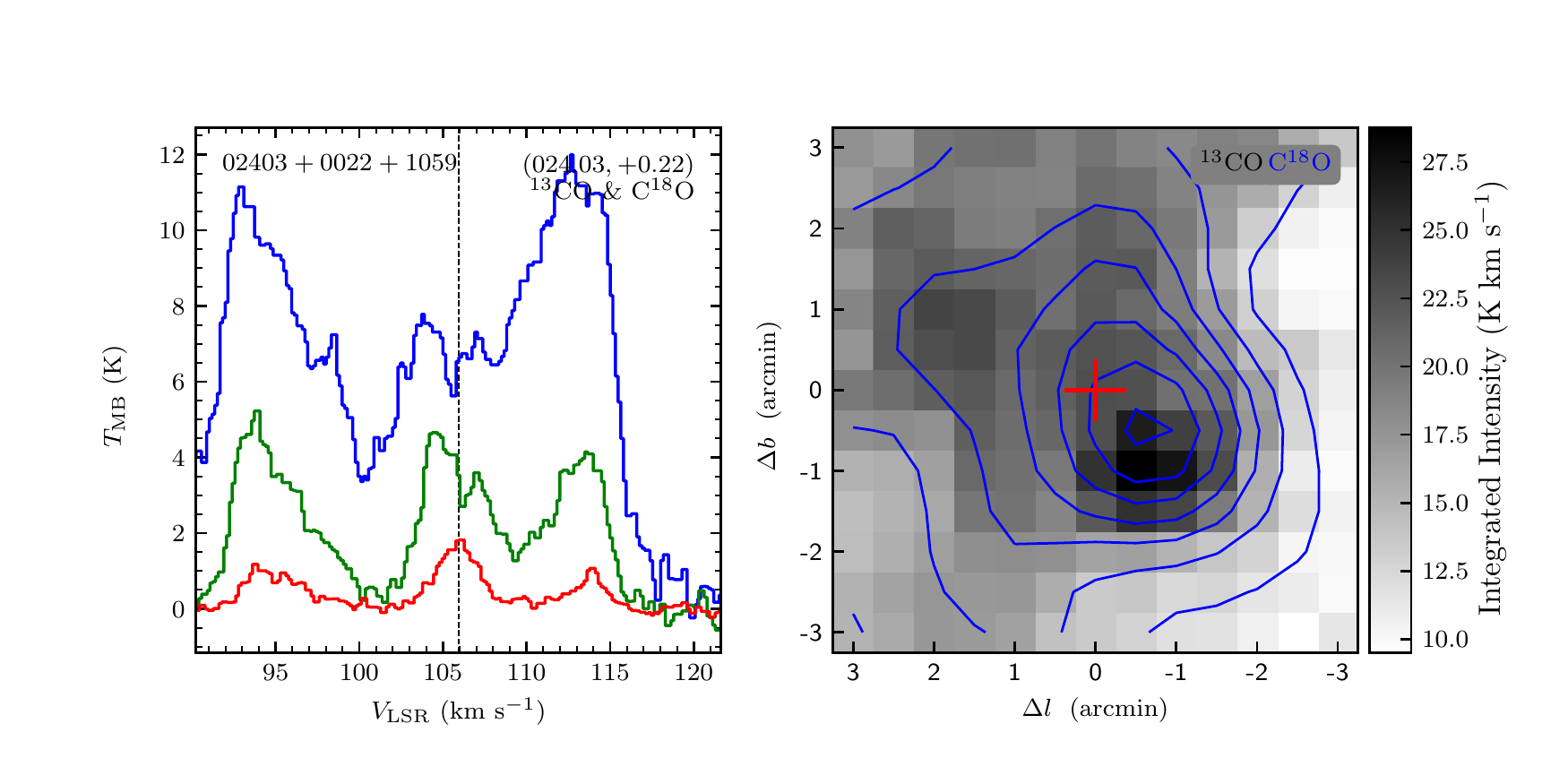}
\vspace{-0.5cm}

\includegraphics[width=9.0cm,angle=0]{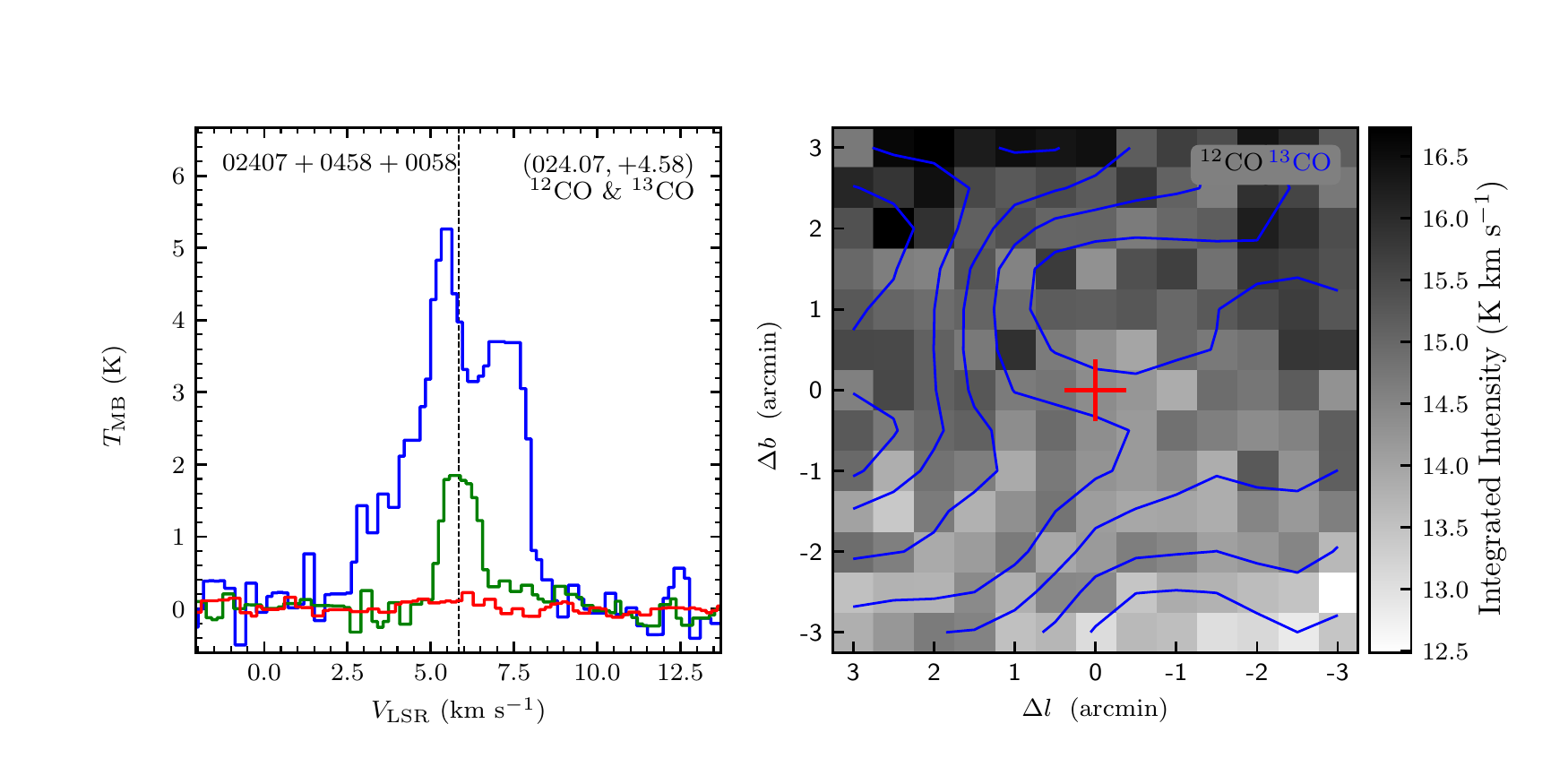}
\includegraphics[width=9.0cm,angle=0]{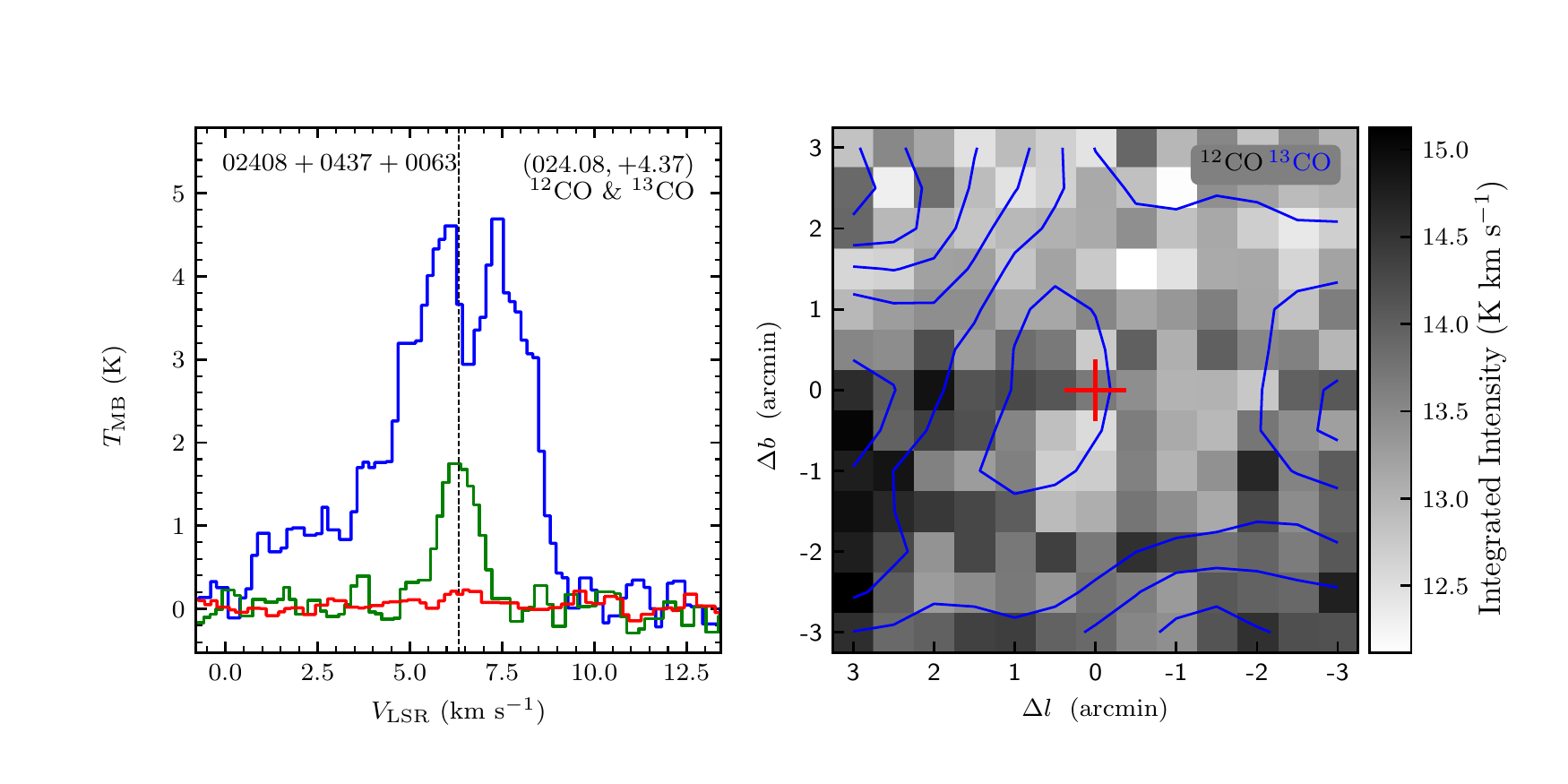}
\vspace{-0.5cm}

\includegraphics[width=9.0cm,angle=0]{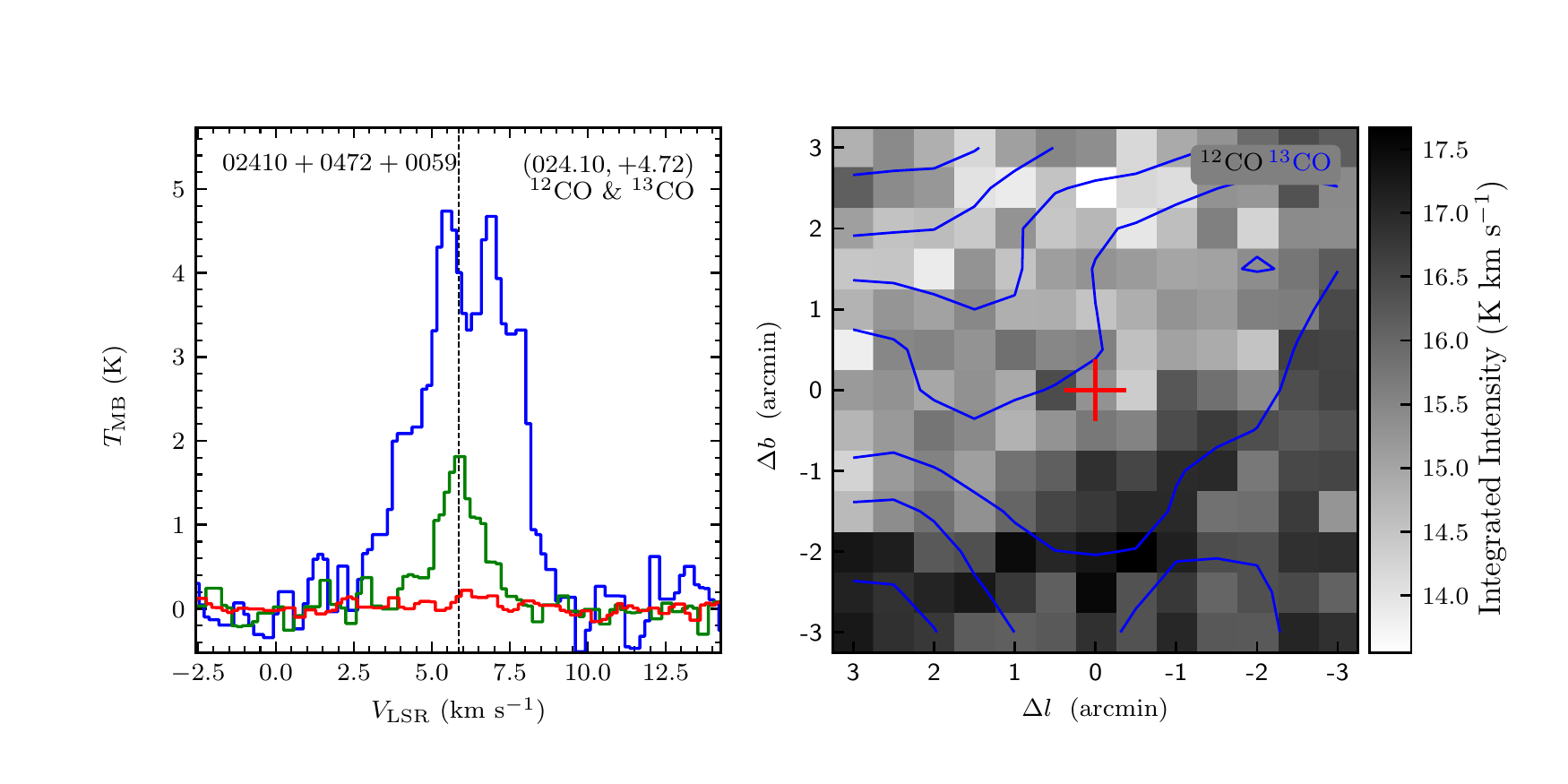}
\includegraphics[width=9.0cm,angle=0]{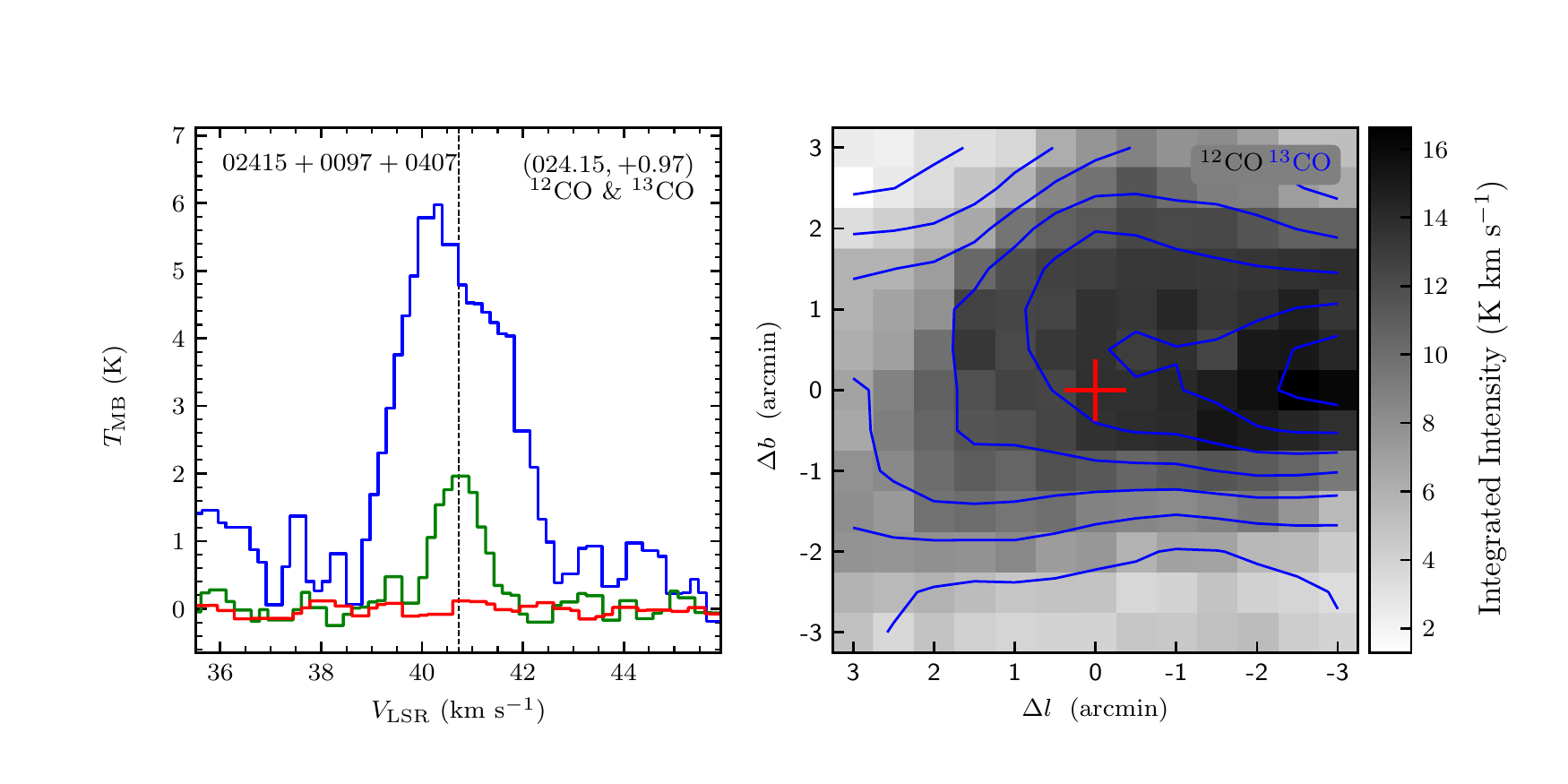}
\vspace{-0.5cm}

\includegraphics[width=9.0cm,angle=0]{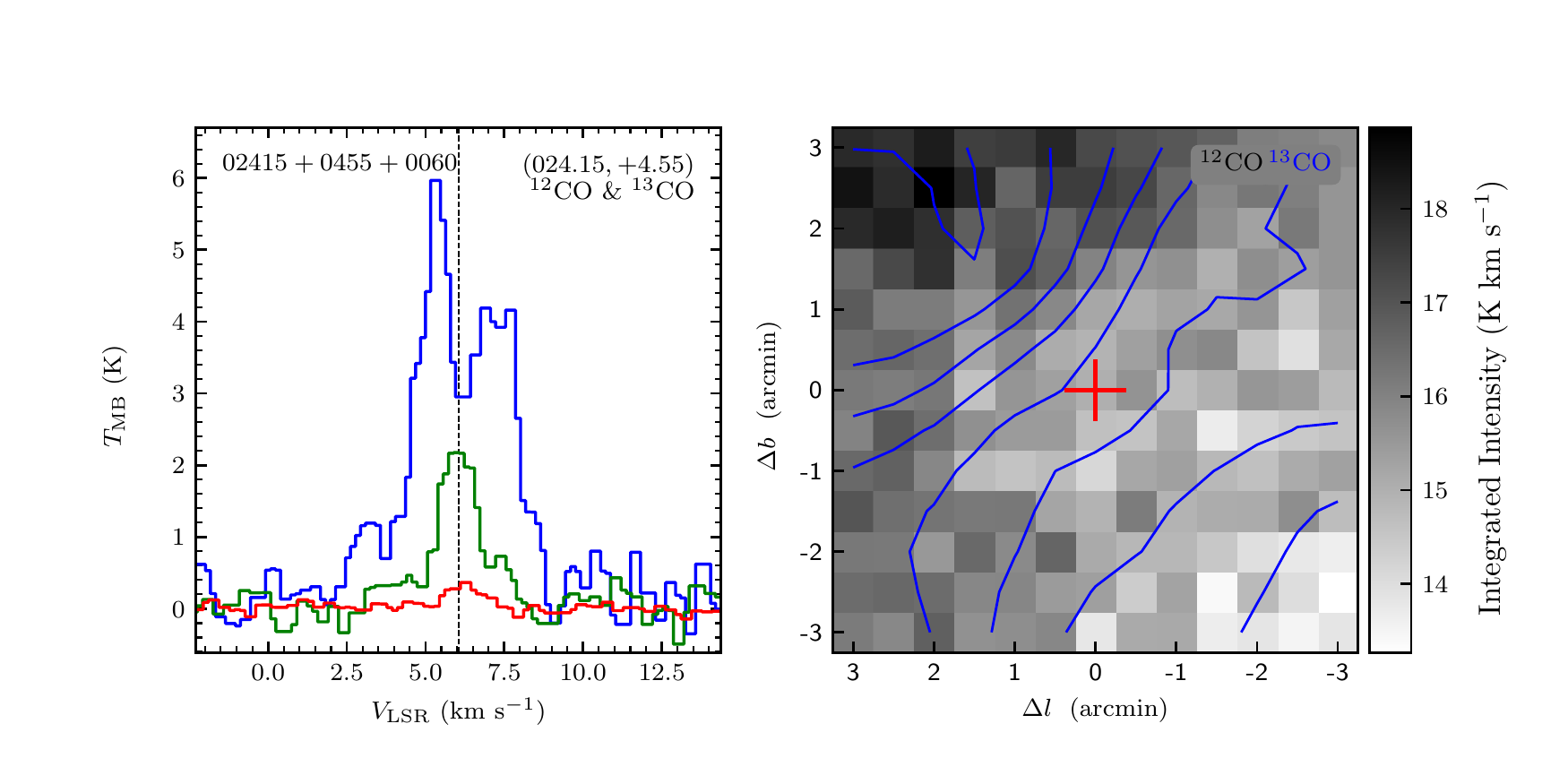}
\includegraphics[width=9.0cm,angle=0]{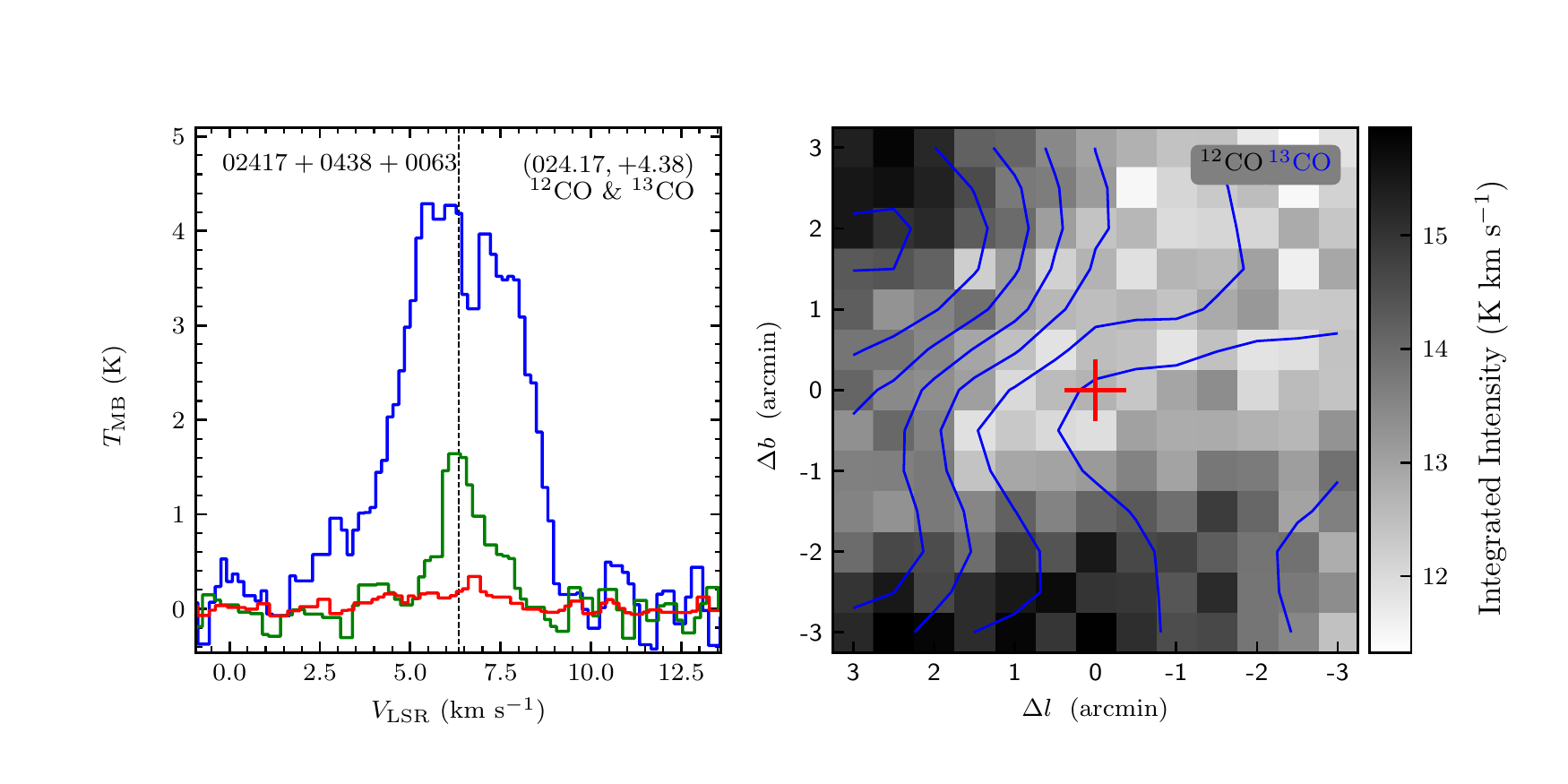}
\vspace{-0.5cm}

\includegraphics[width=9.0cm,angle=0]{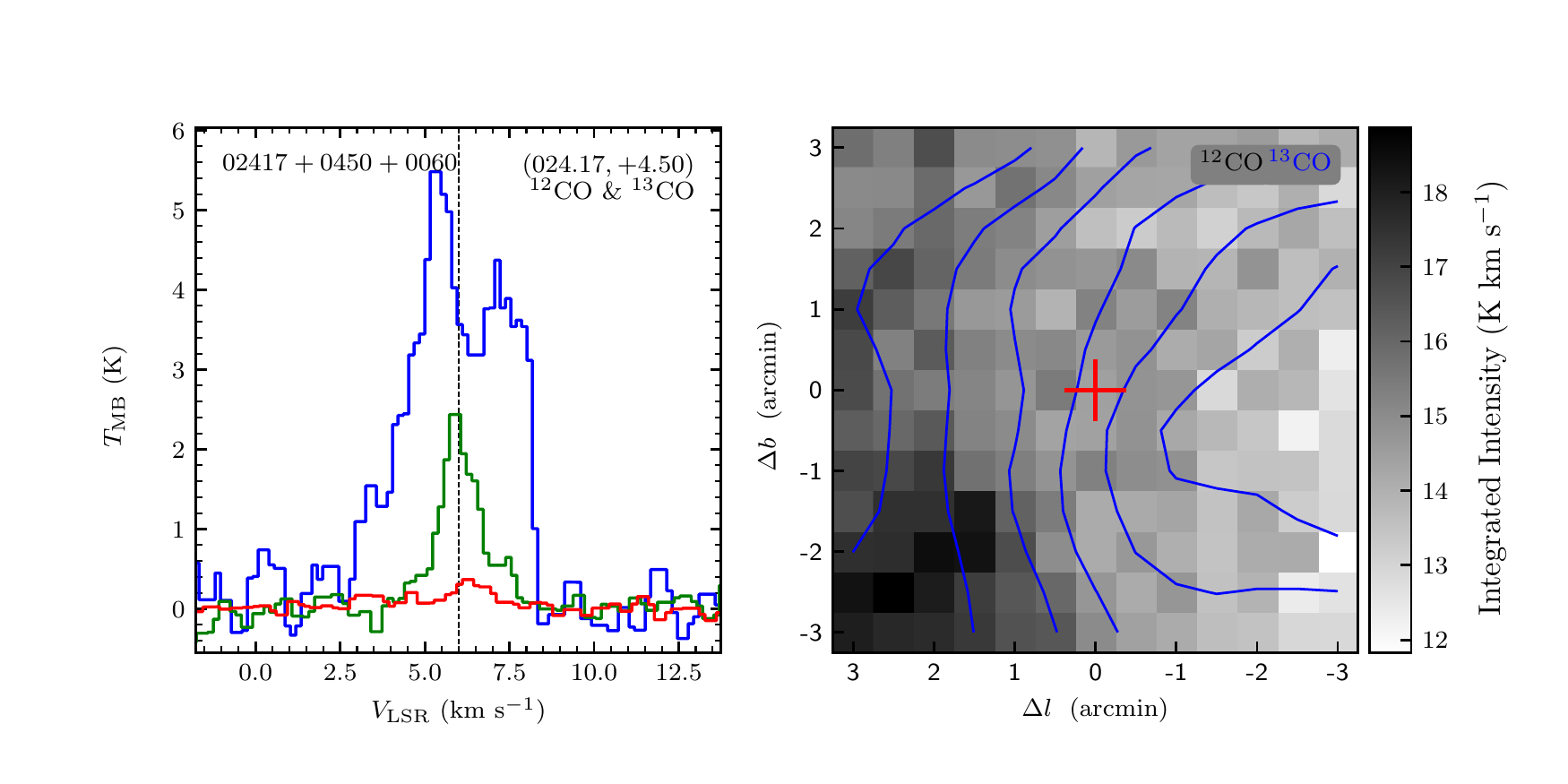}
\includegraphics[width=9.0cm,angle=0]{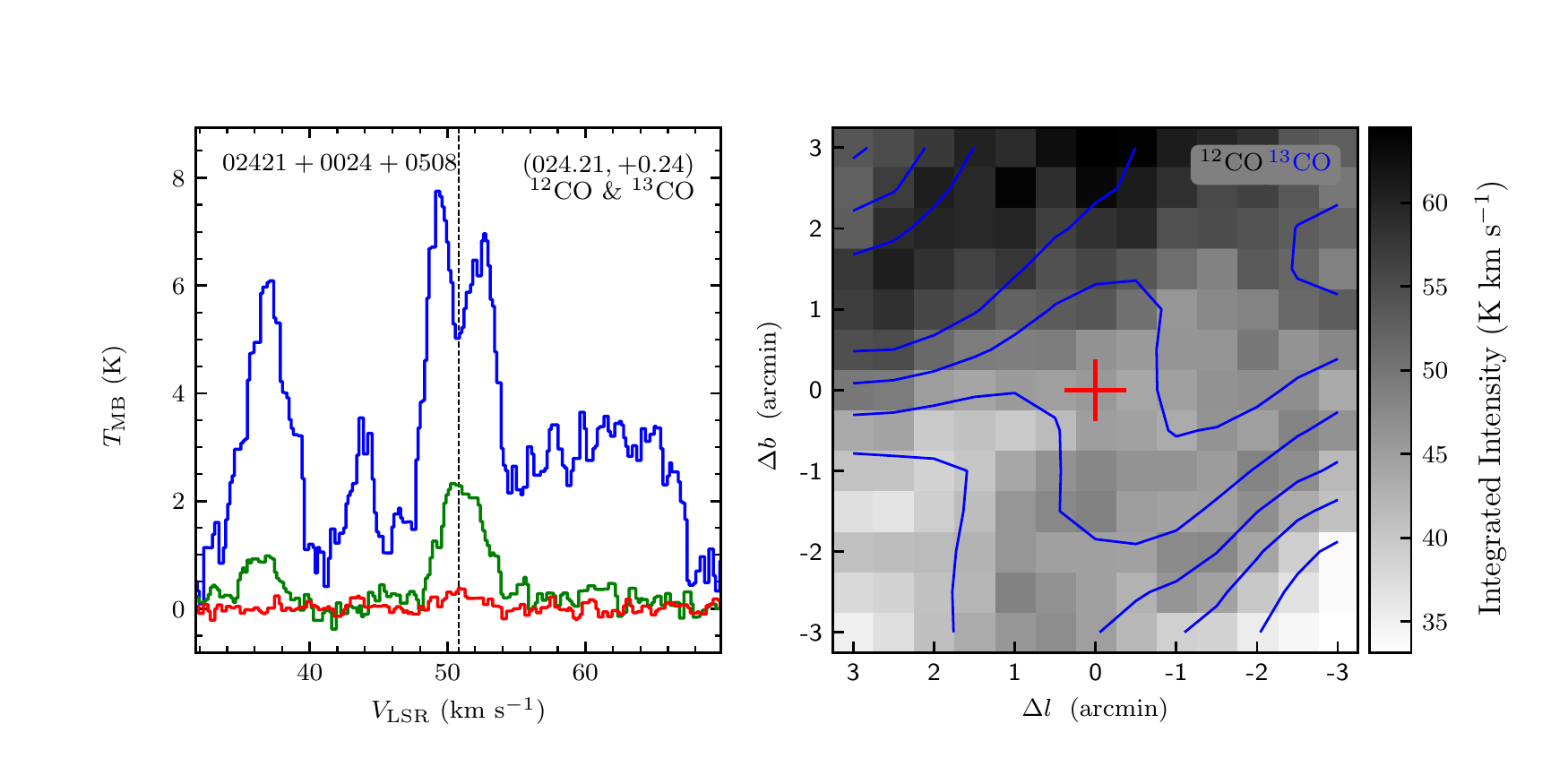}
\end{figure}
\clearpage

\begin{figure}
\includegraphics[width=9.0cm,angle=0]{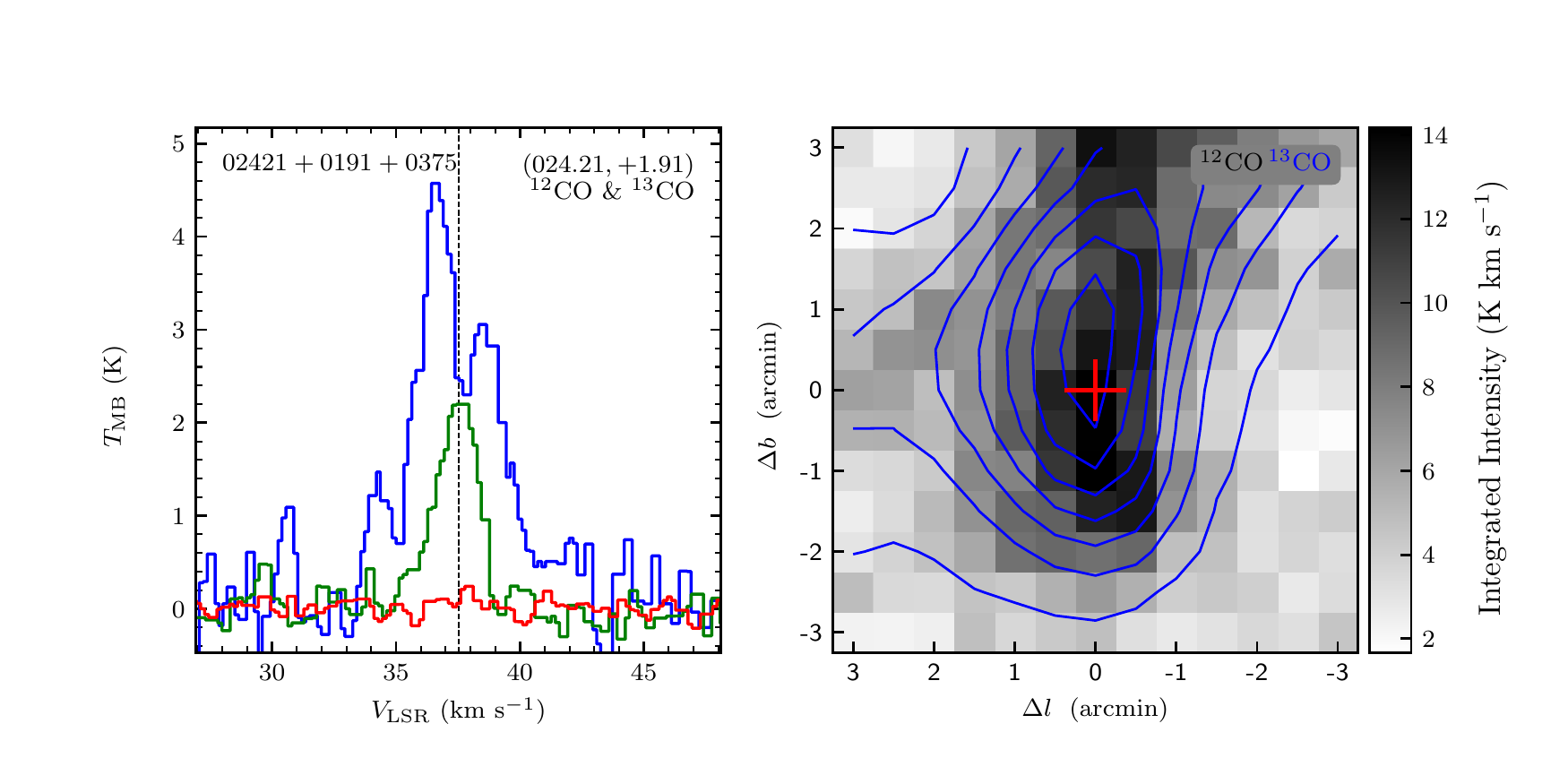}
\includegraphics[width=9.0cm,angle=0]{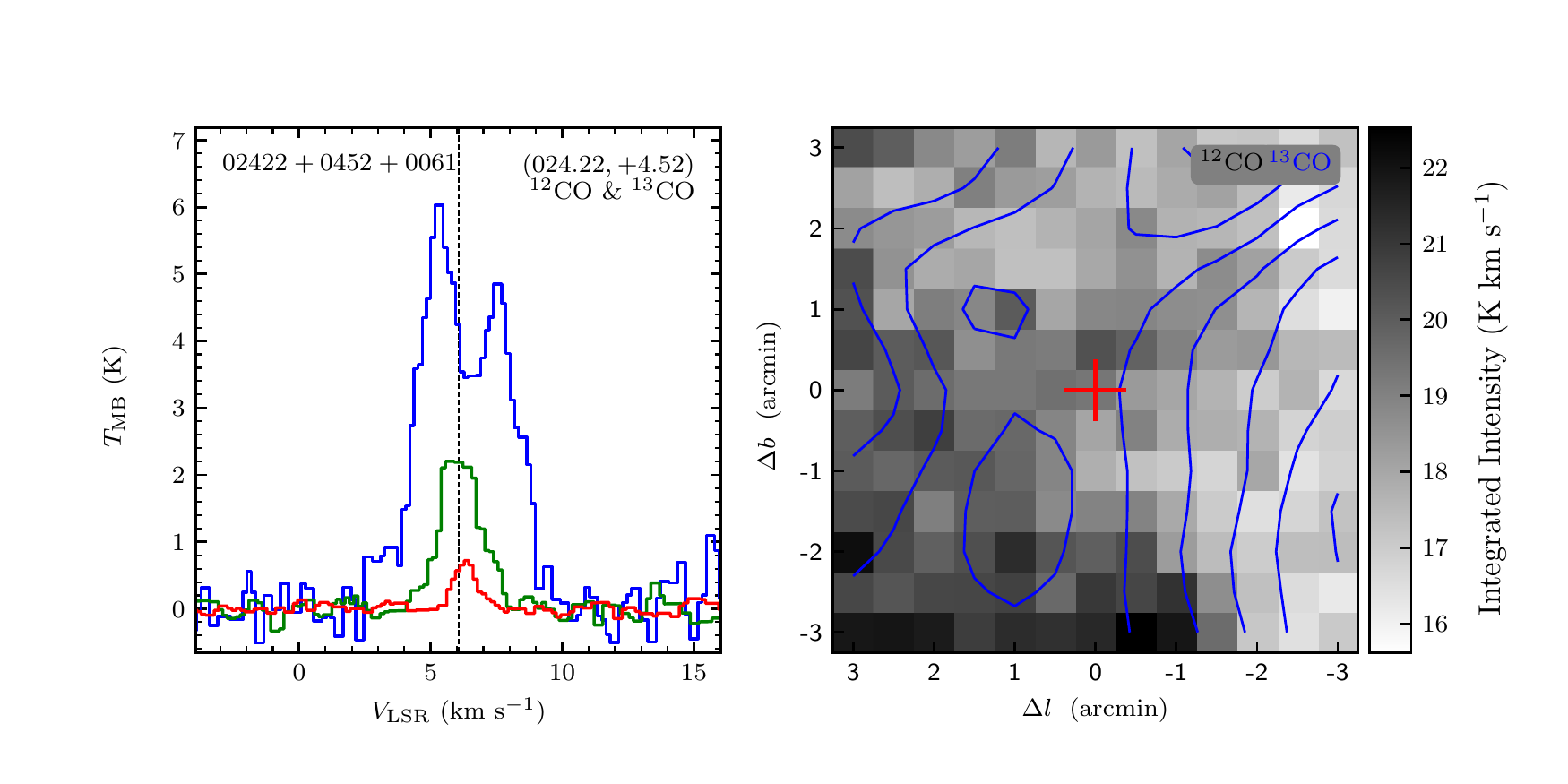}
\vspace{-0.5cm}

\includegraphics[width=9.0cm,angle=0]{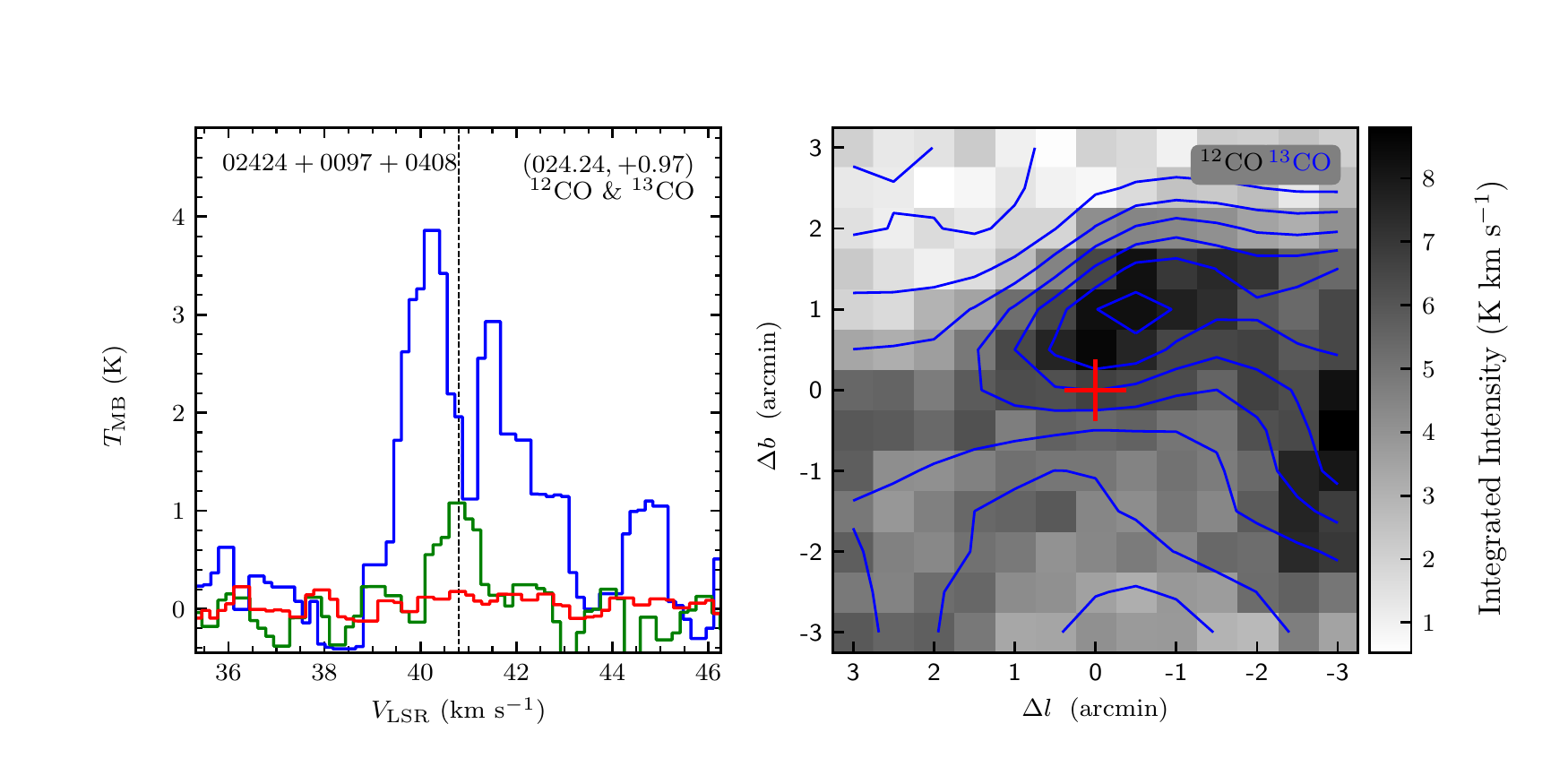}
\includegraphics[width=9.0cm,angle=0]{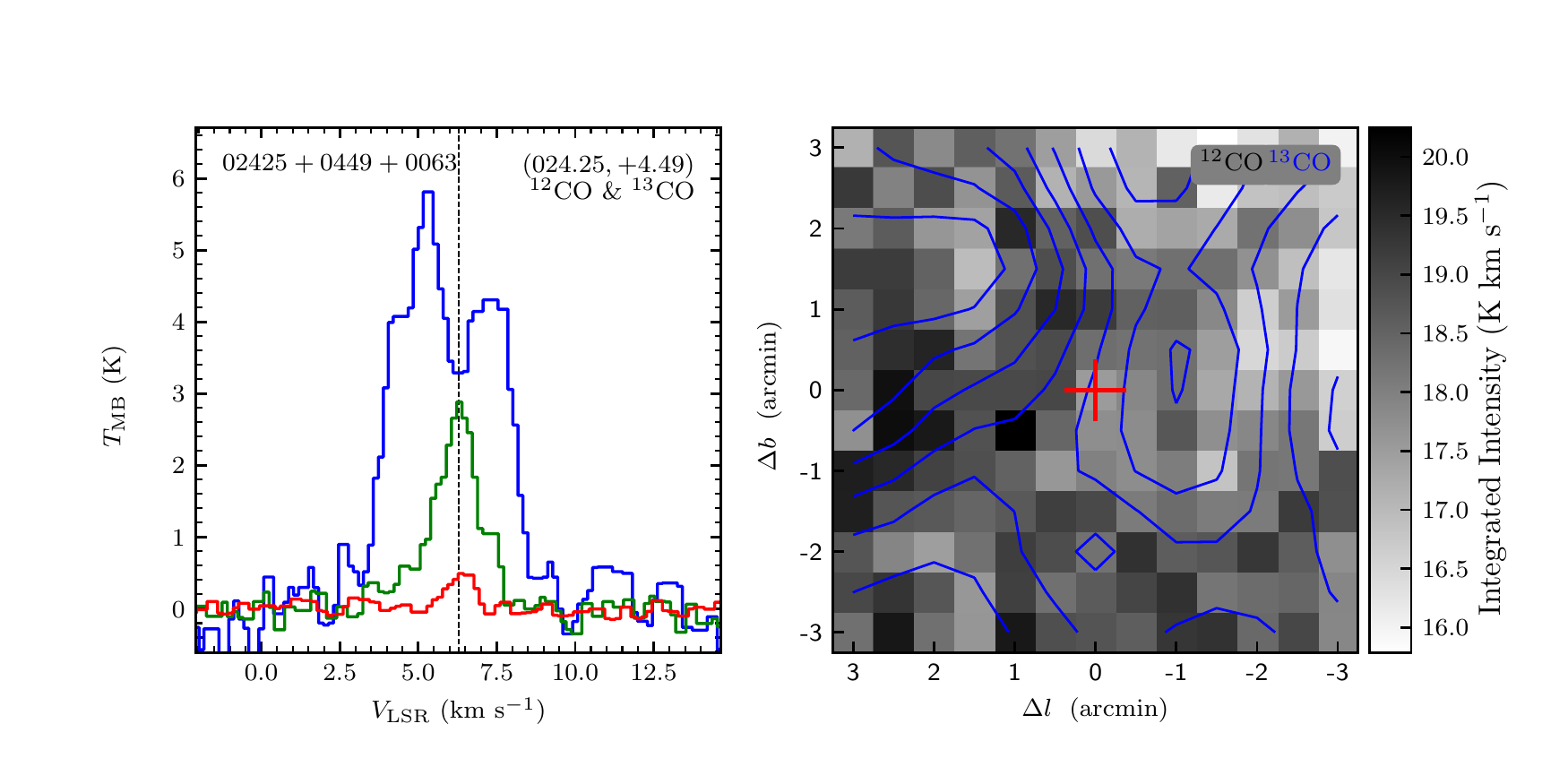}
\vspace{-0.5cm}

\includegraphics[width=9.0cm,angle=0]{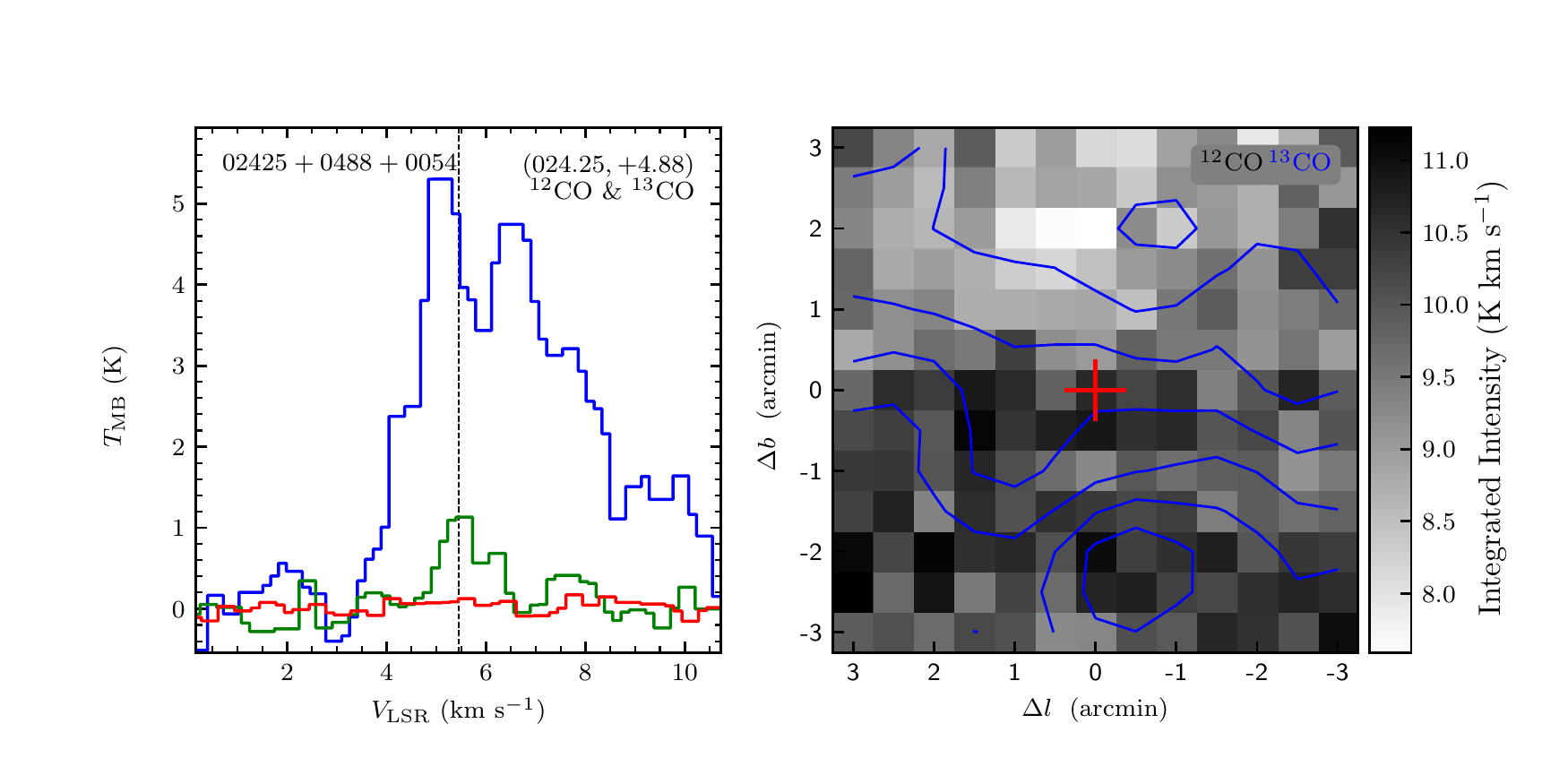}
\includegraphics[width=9.0cm,angle=0]{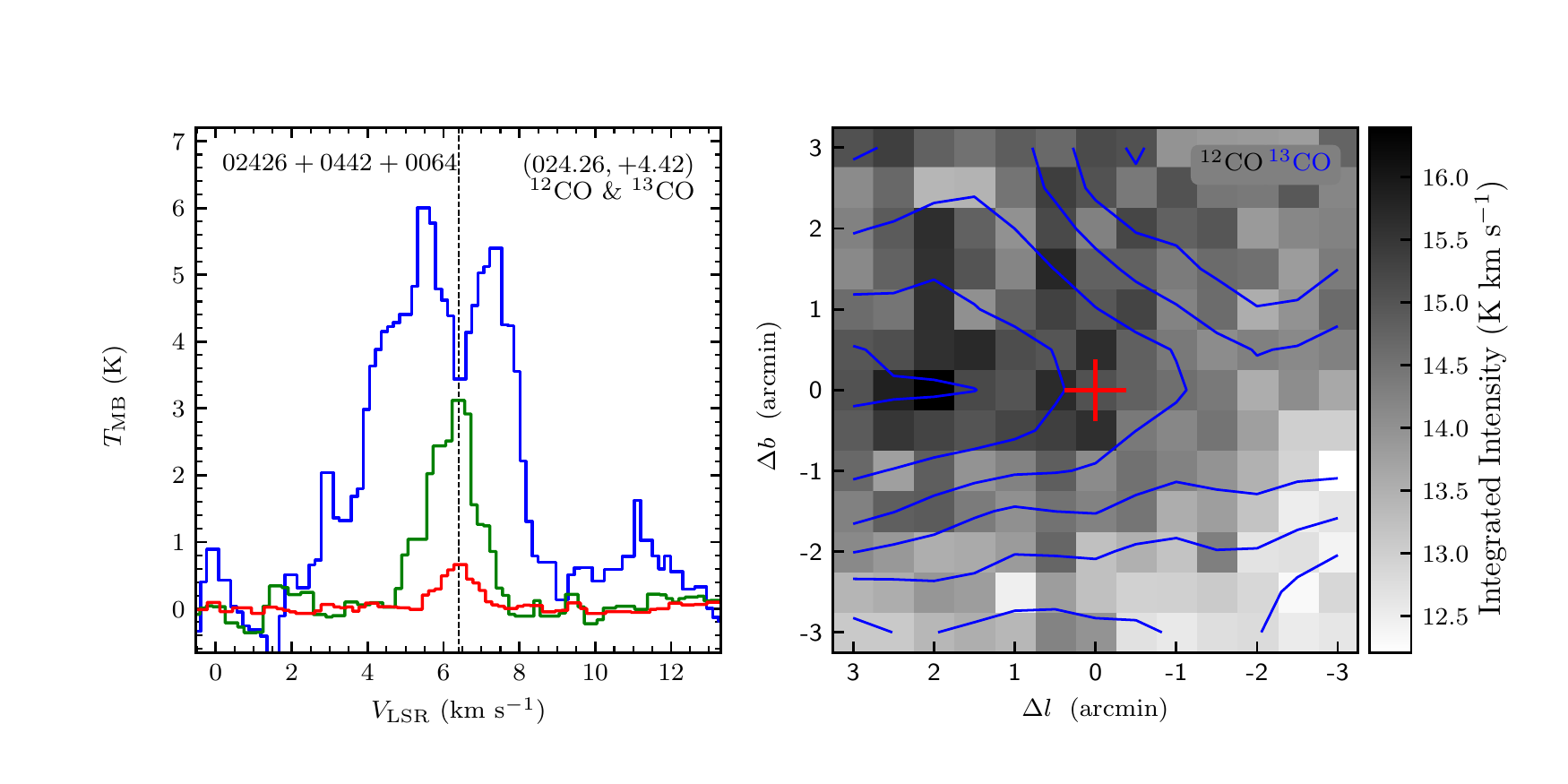}
\vspace{-0.5cm}

\includegraphics[width=9.0cm,angle=0]{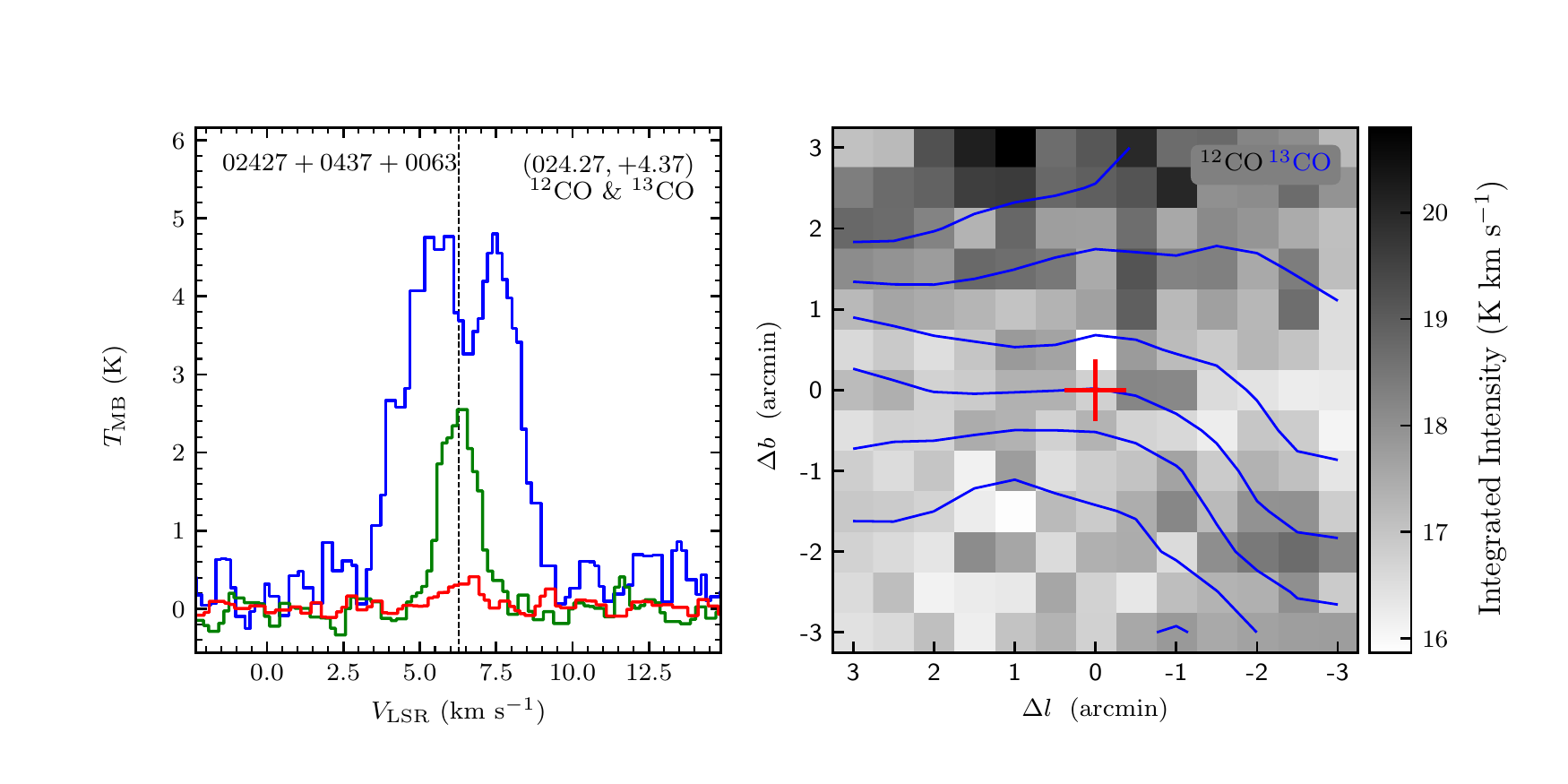}
\includegraphics[width=9.0cm,angle=0]{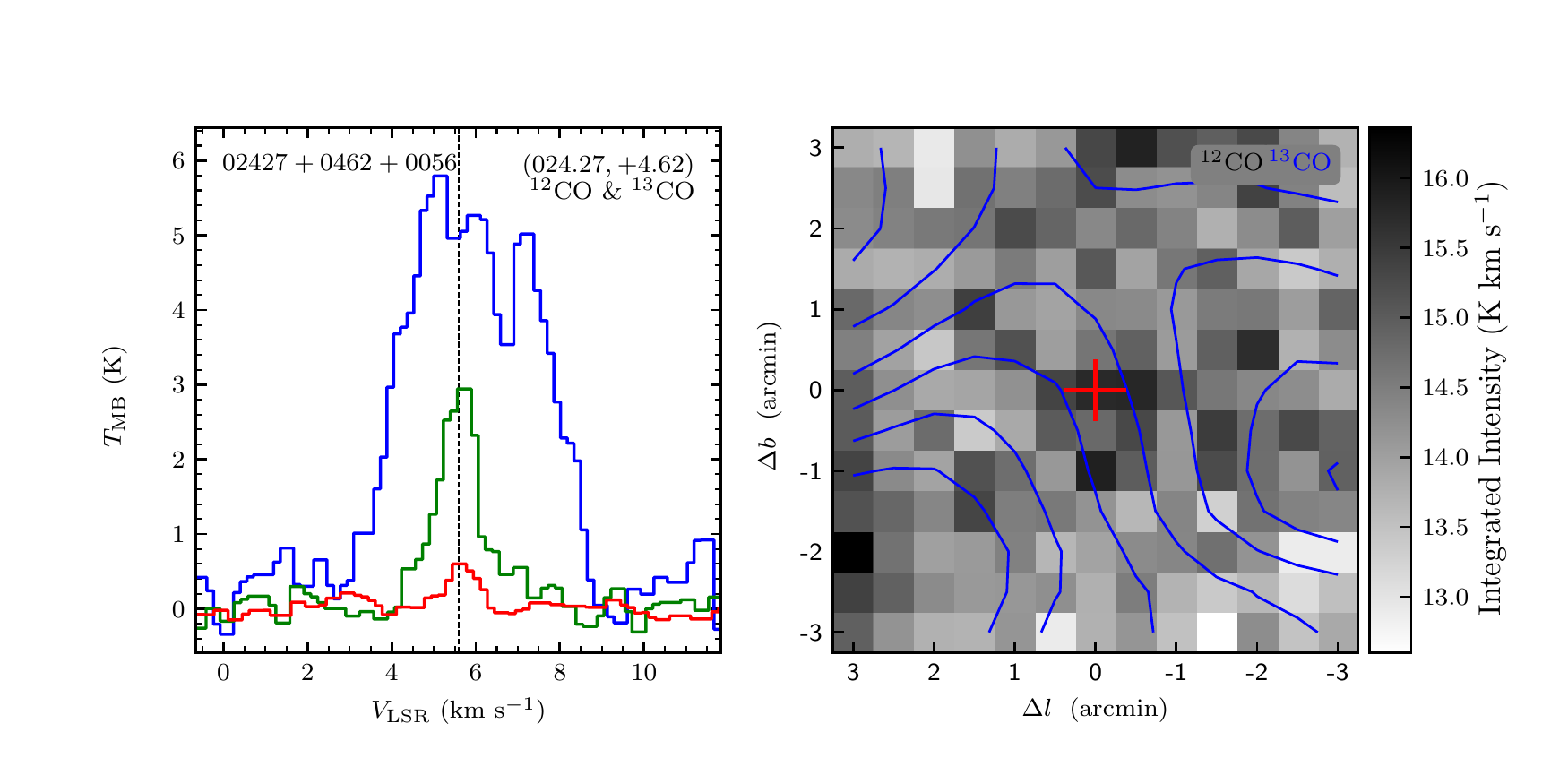}
\vspace{-0.5cm}

\includegraphics[width=9.0cm,angle=0]{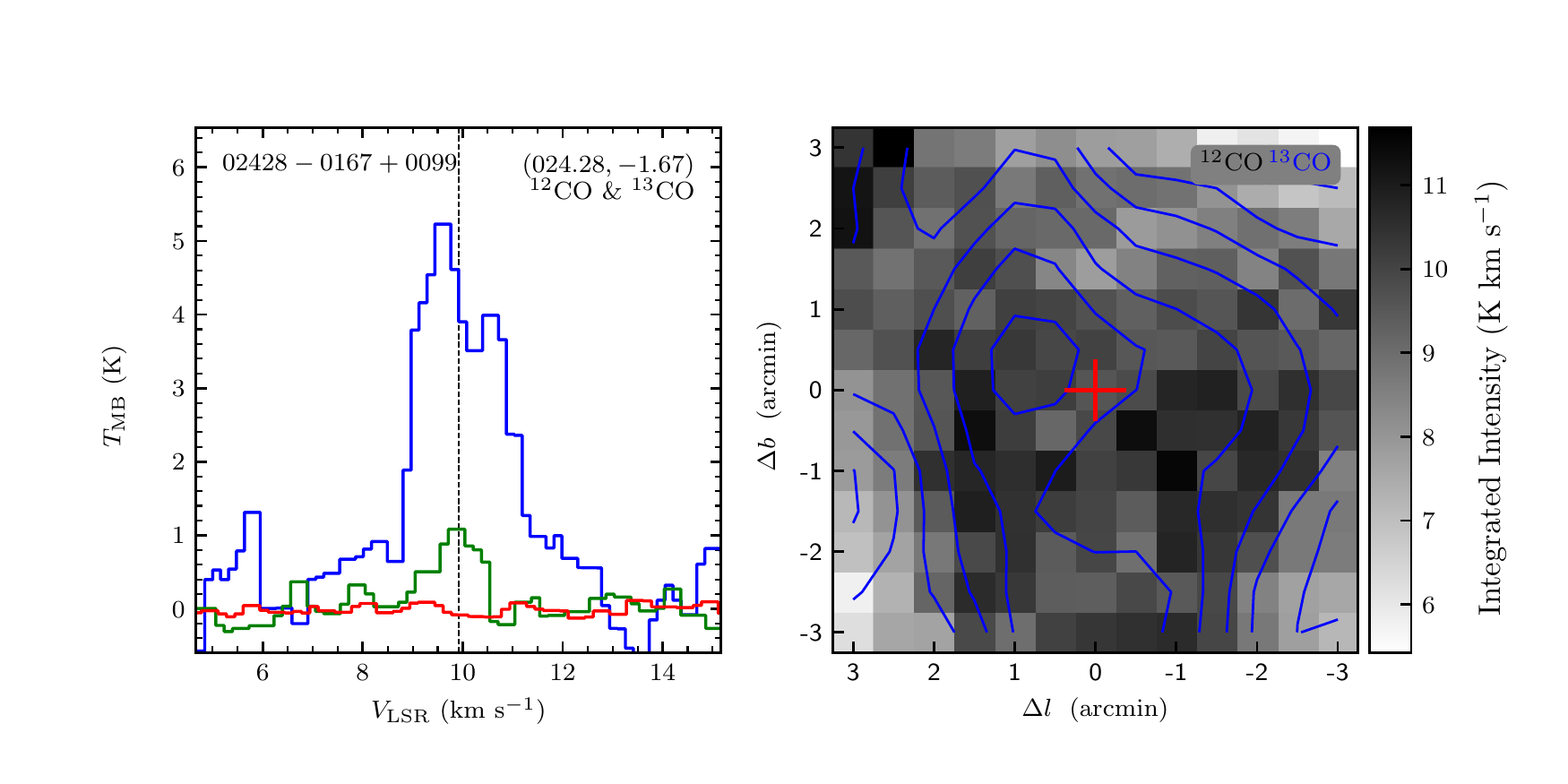}
\includegraphics[width=9.0cm,angle=0]{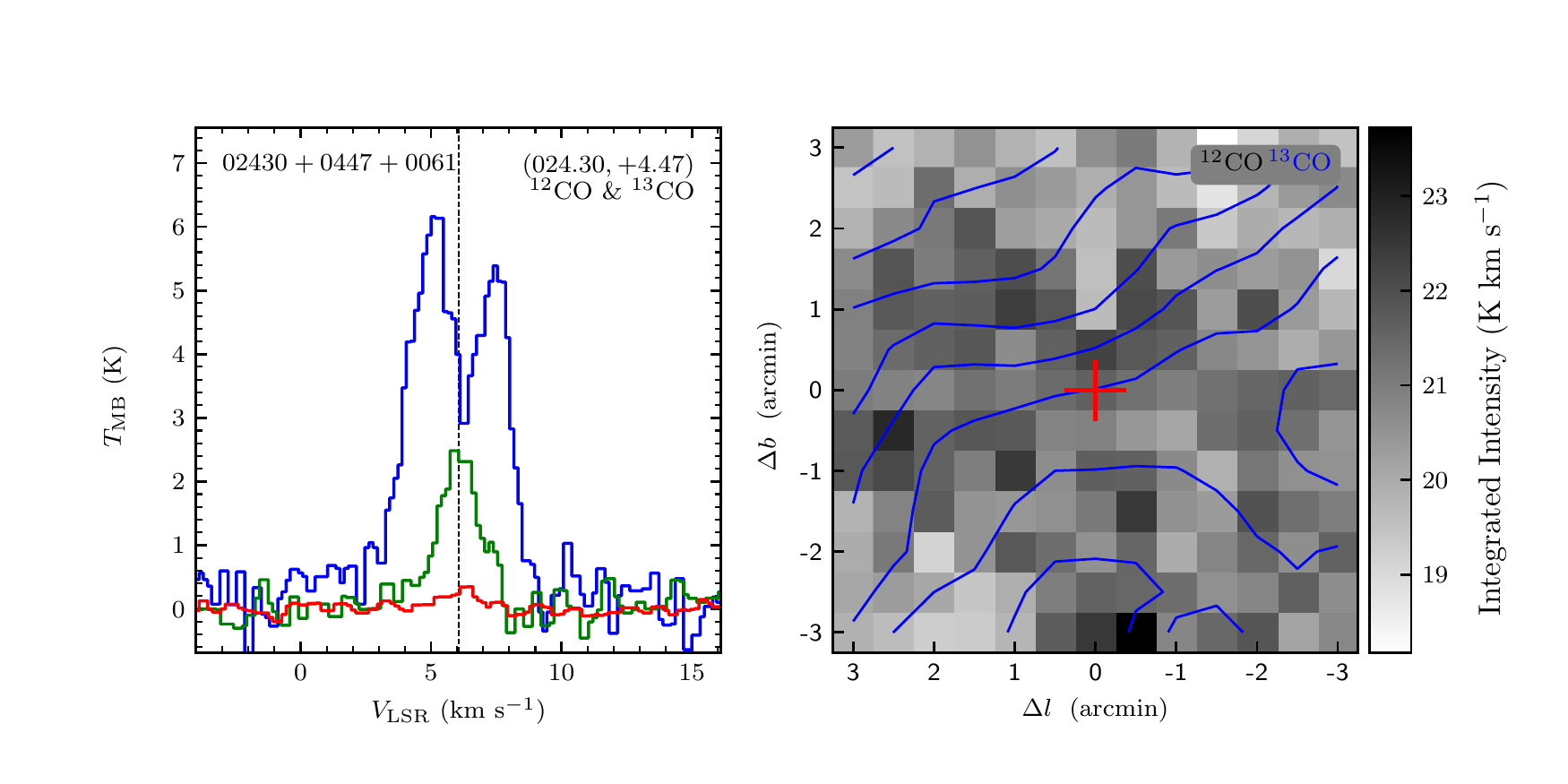}
\end{figure}
\clearpage

\begin{figure}
\includegraphics[width=9.0cm,angle=0]{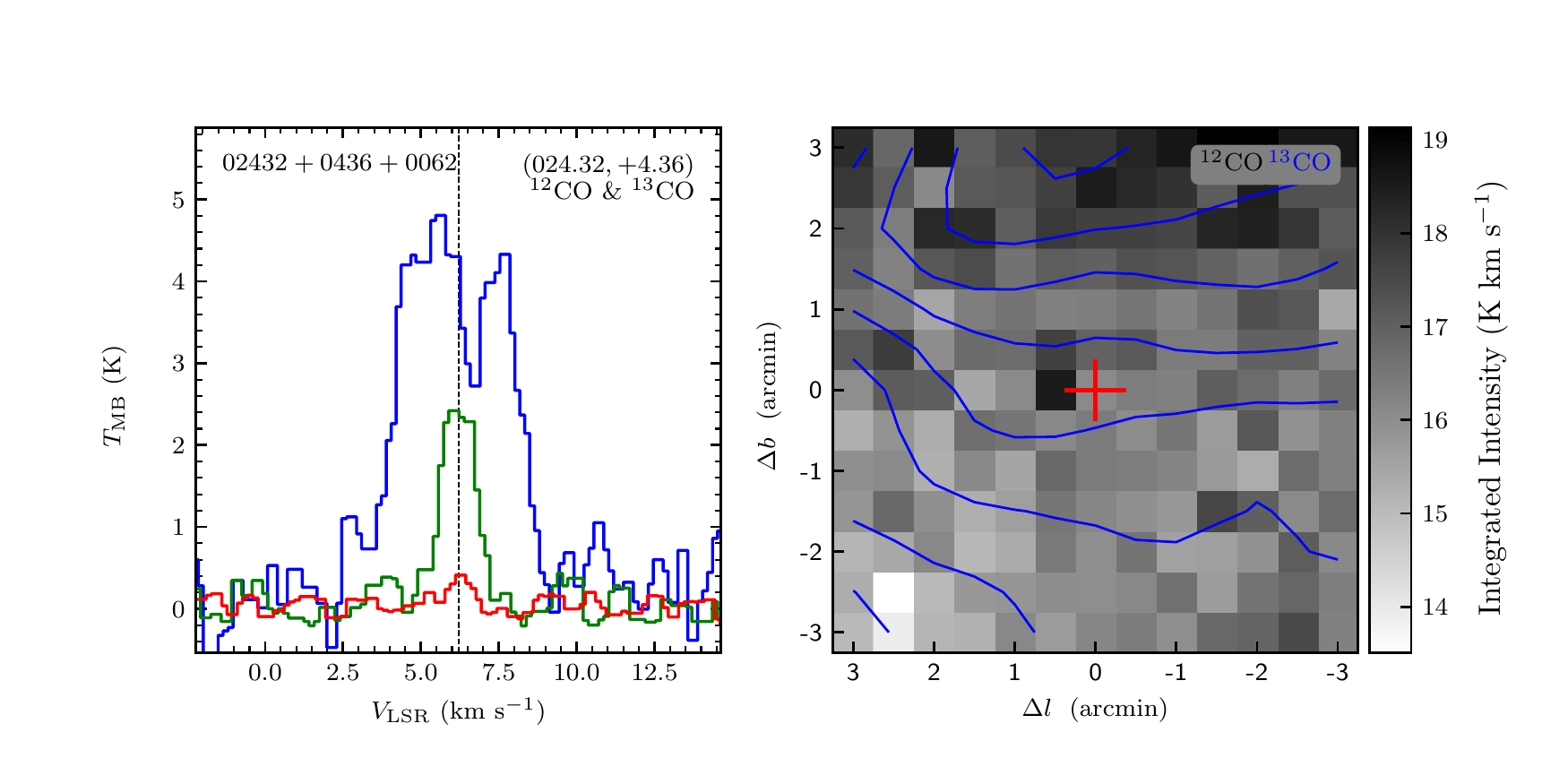}
\includegraphics[width=9.0cm,angle=0]{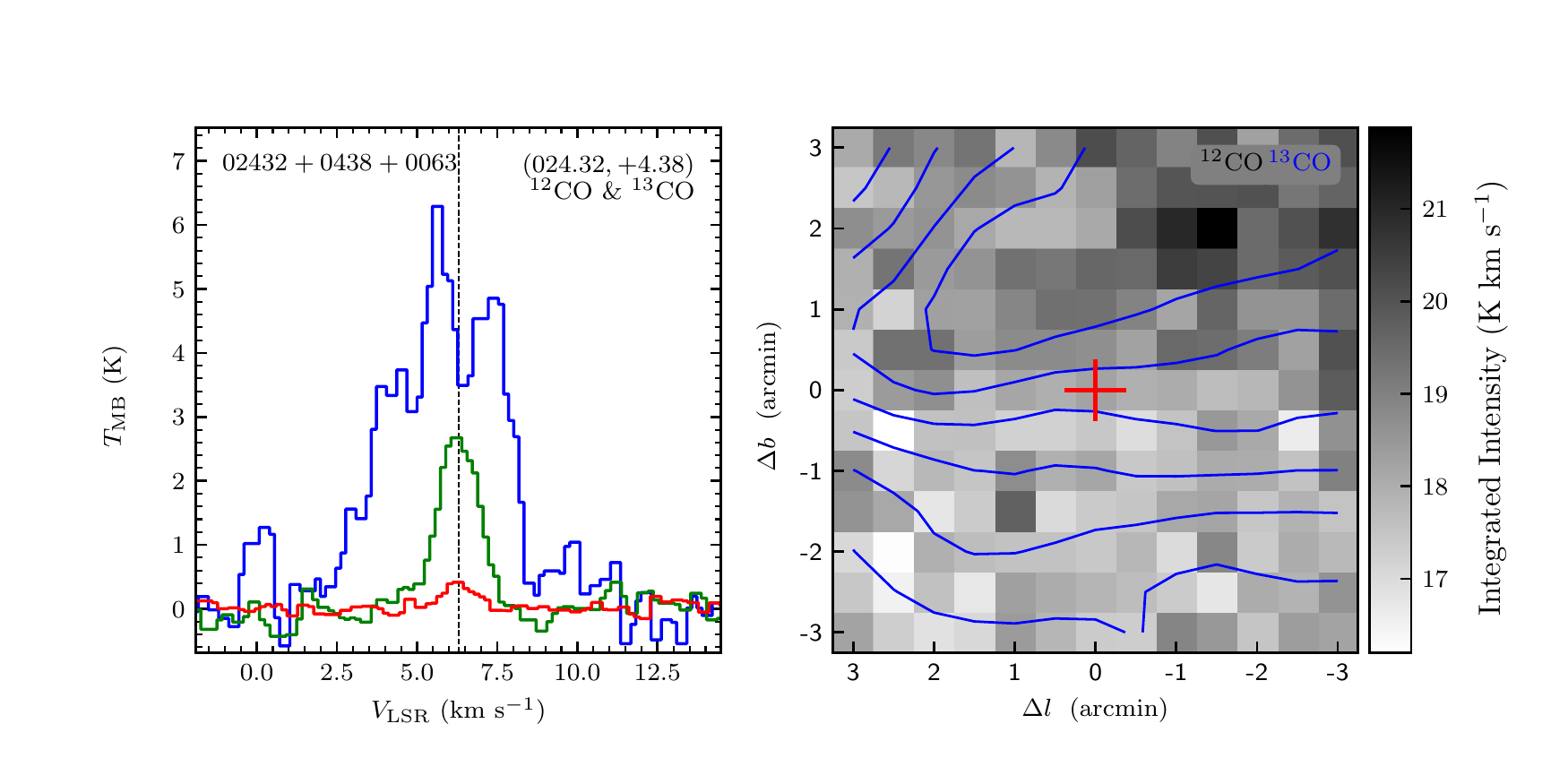}
\vspace{-0.5cm}

\includegraphics[width=9.0cm,angle=0]{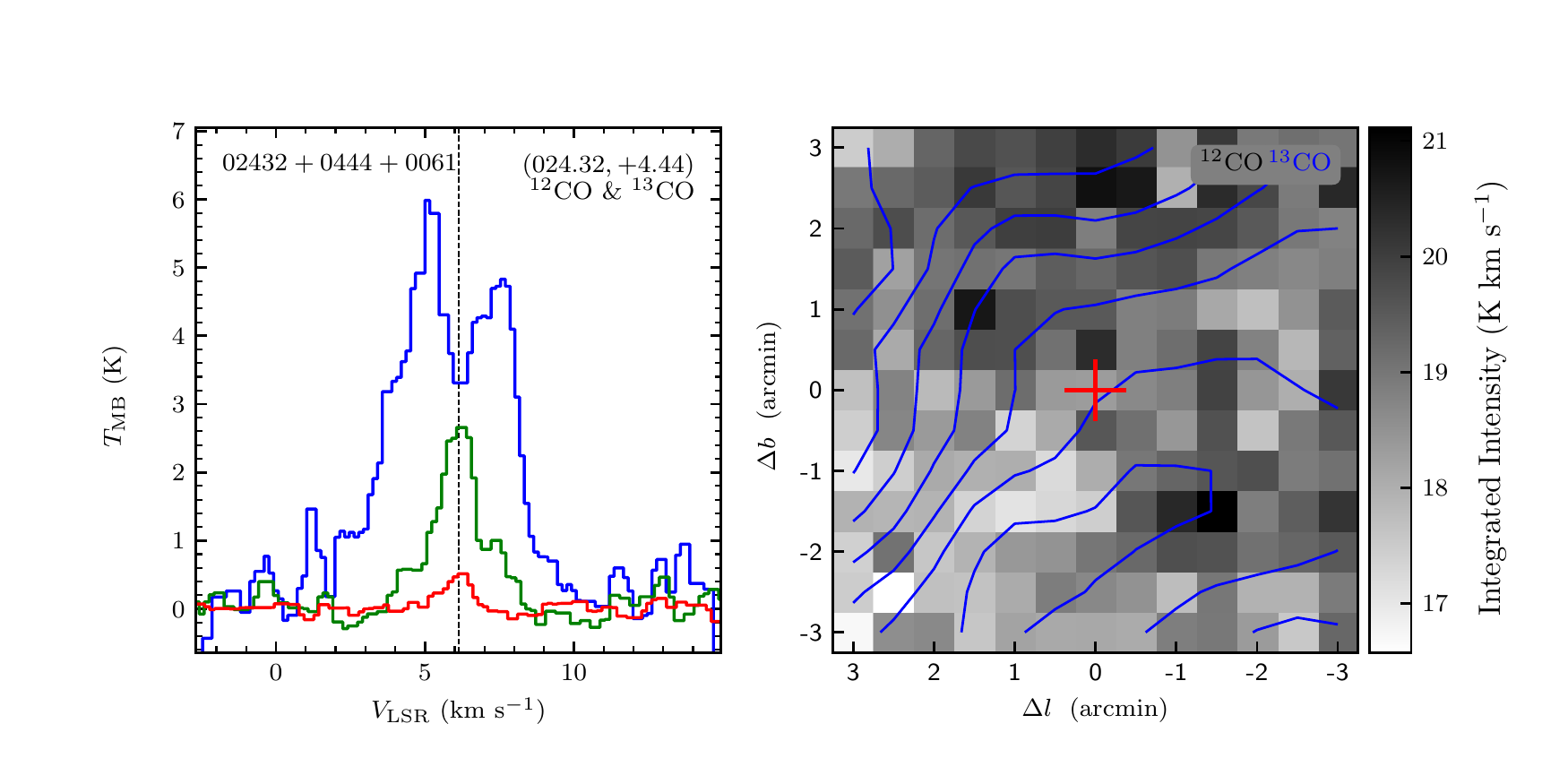}
\includegraphics[width=9.0cm,angle=0]{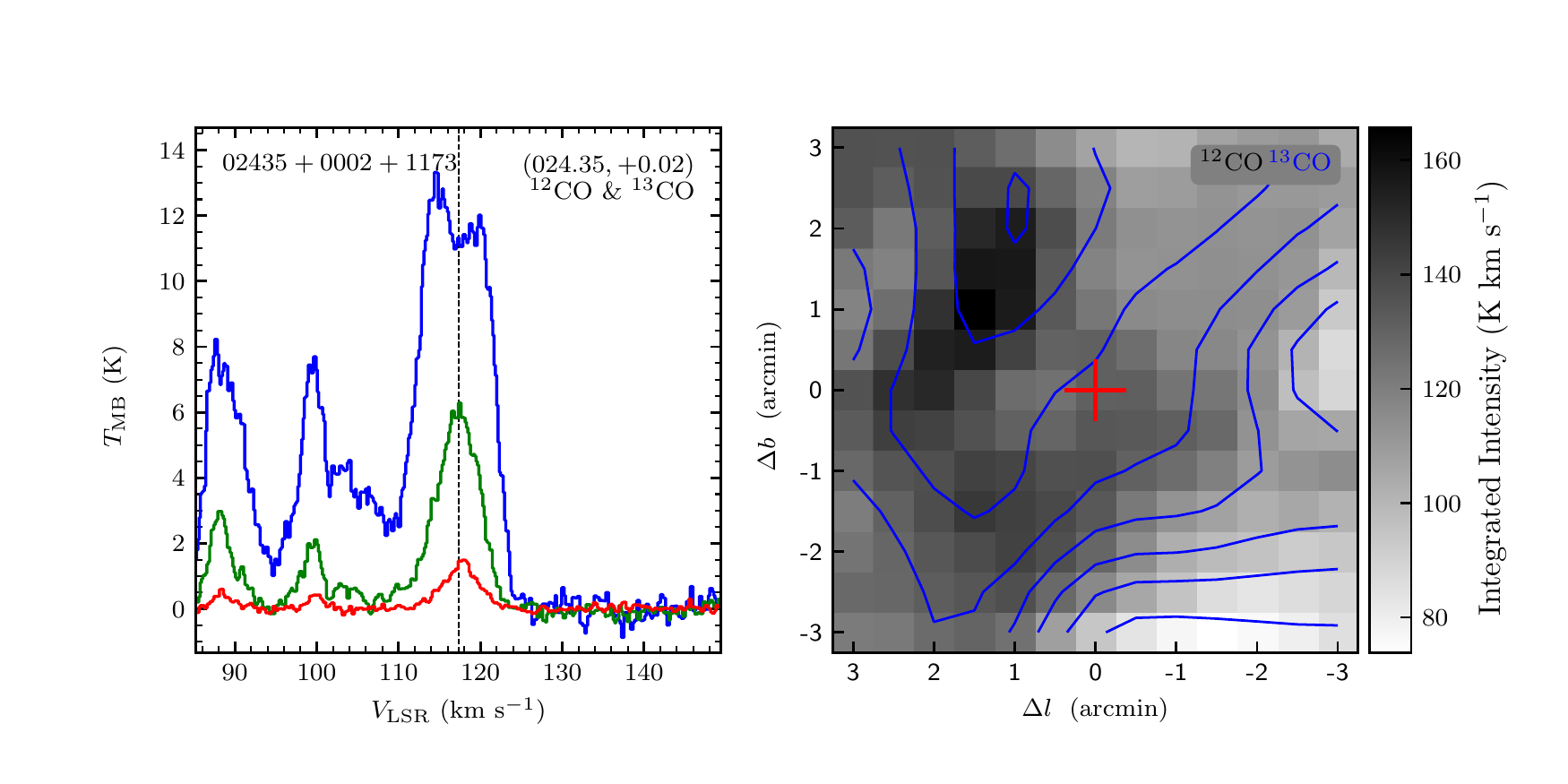}
\vspace{-0.5cm}

\includegraphics[width=9.0cm,angle=0]{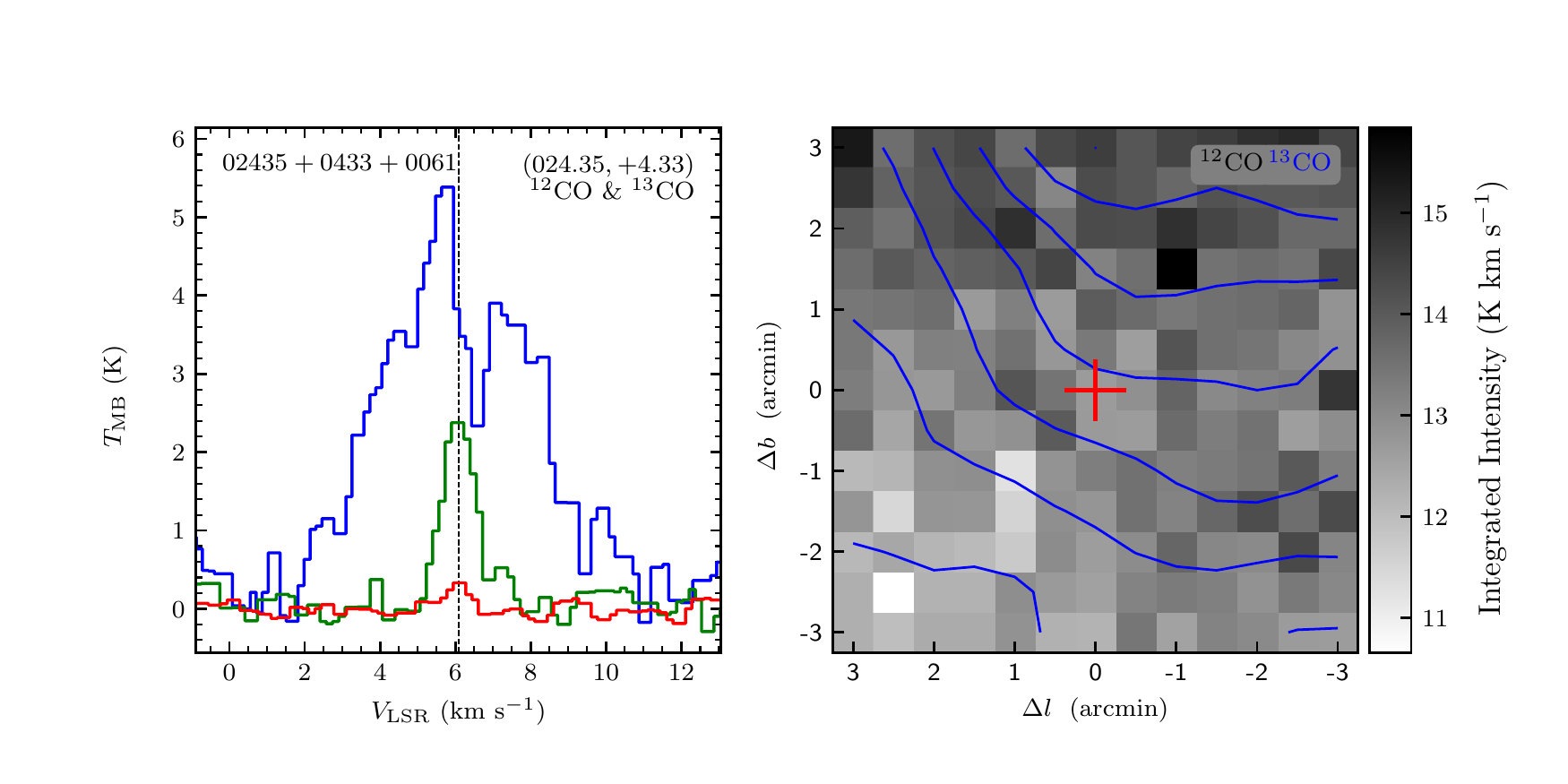}
\includegraphics[width=9.0cm,angle=0]{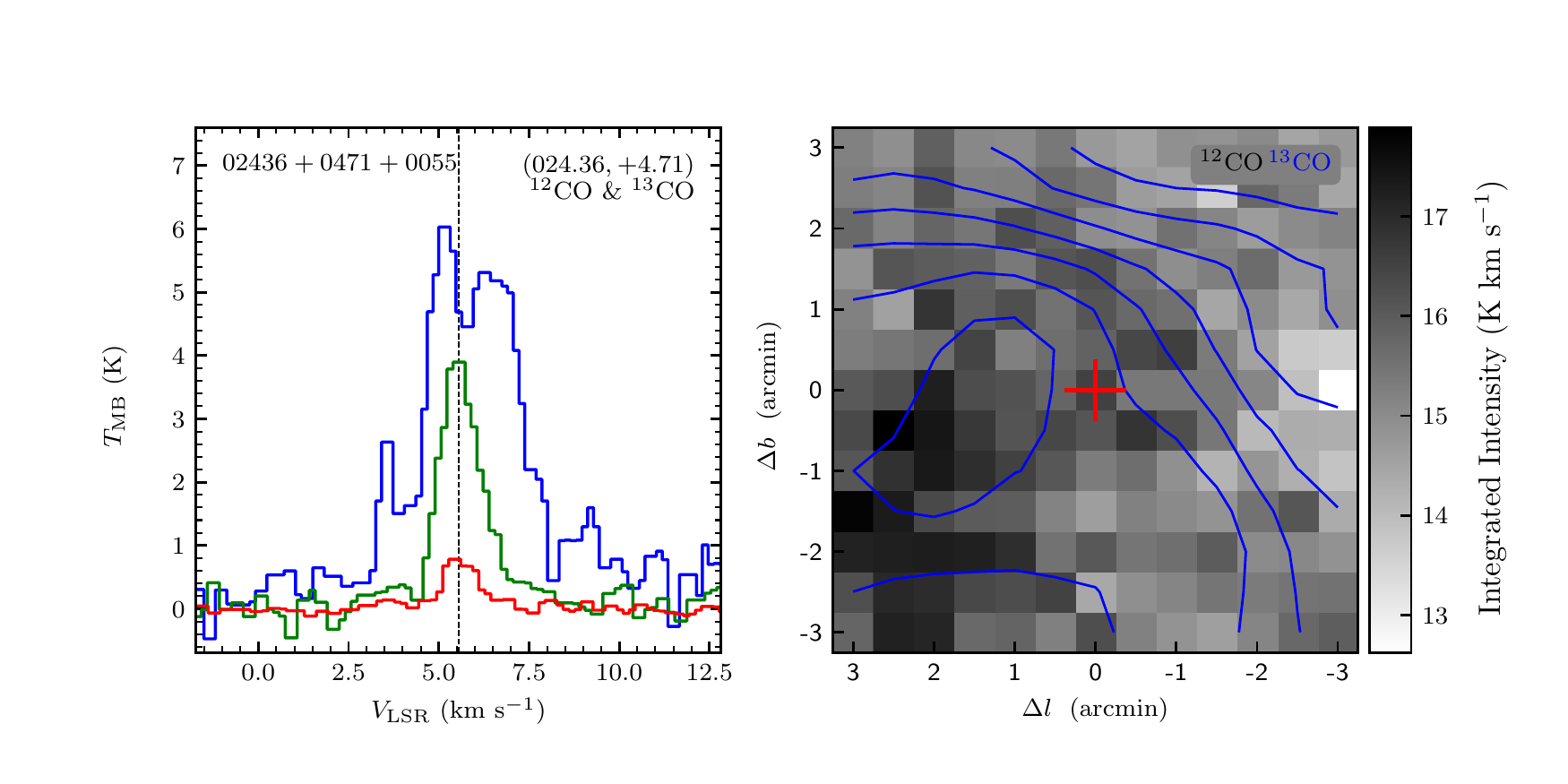}
\vspace{-0.5cm}

\includegraphics[width=9.0cm,angle=0]{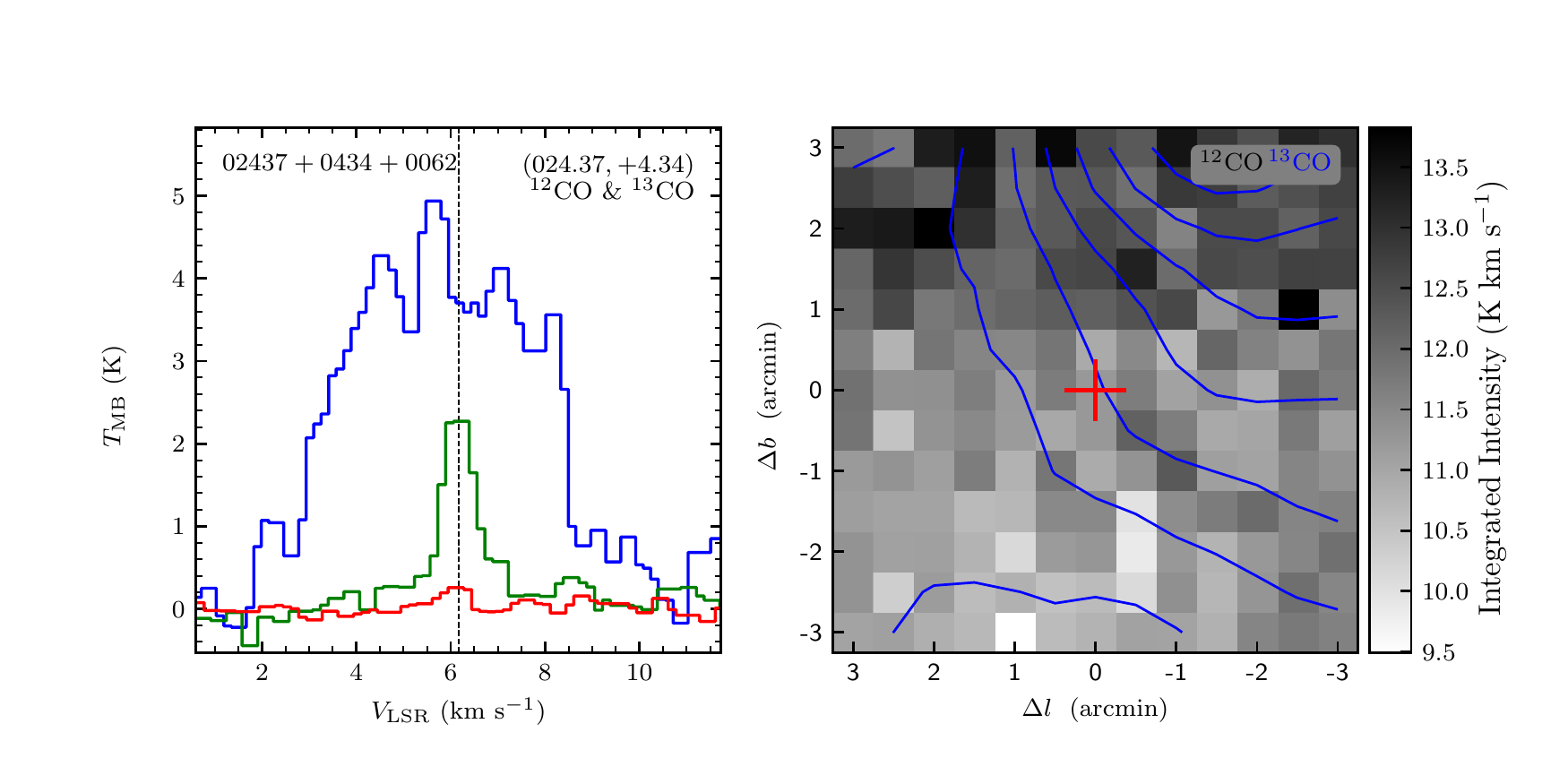}
\includegraphics[width=9.0cm,angle=0]{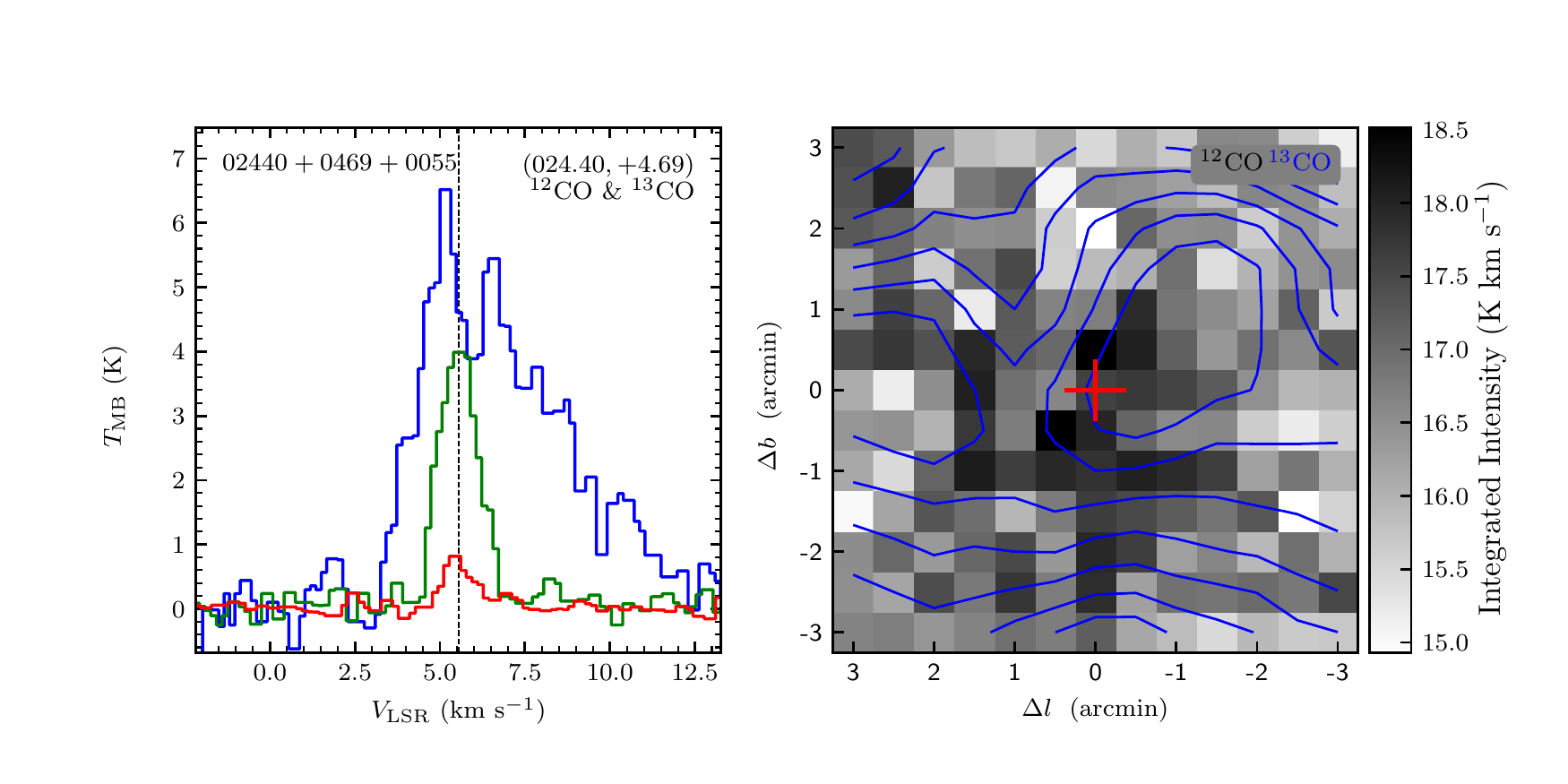}
\vspace{-0.5cm}

\includegraphics[width=9.0cm,angle=0]{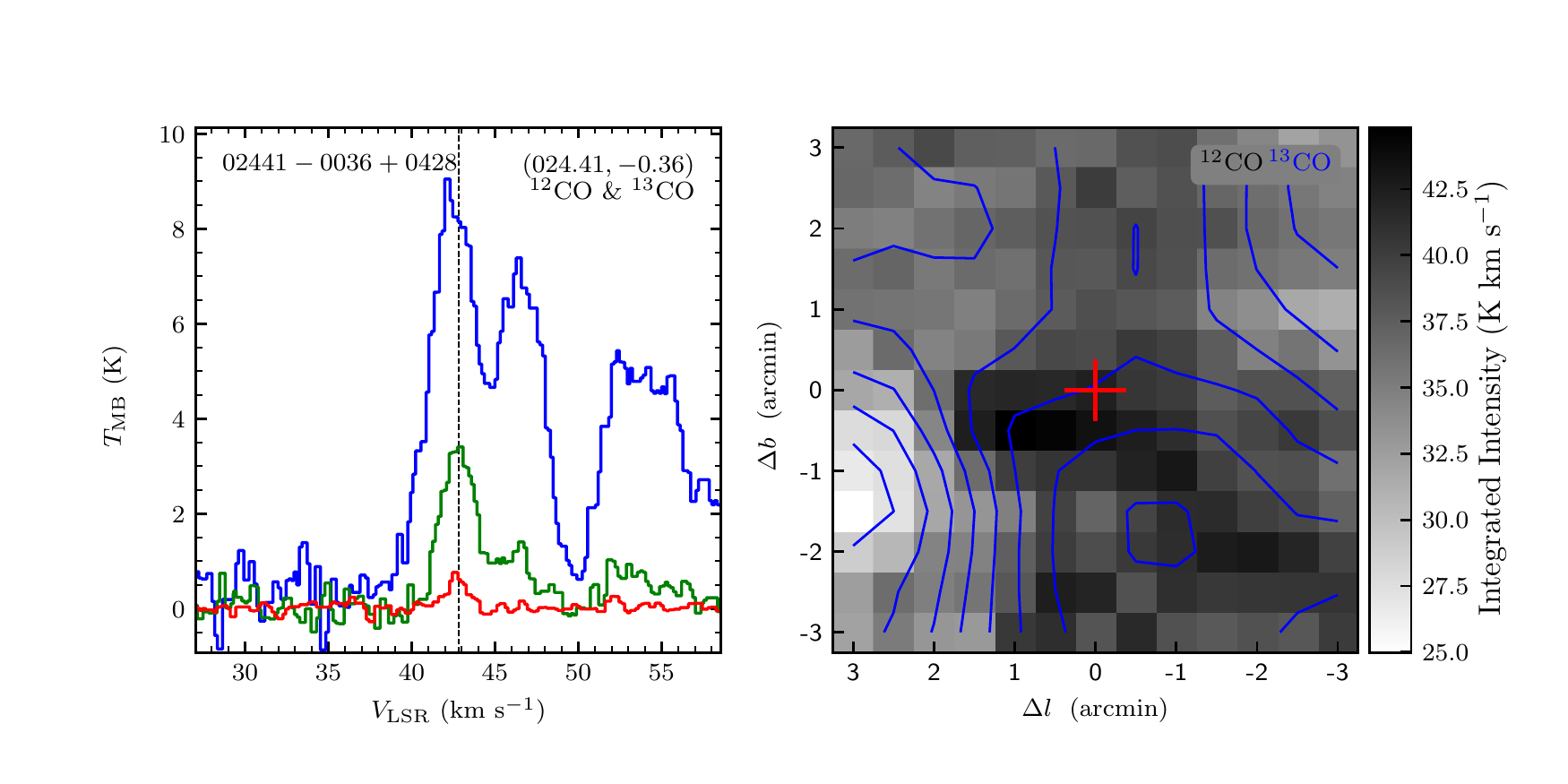}
\includegraphics[width=9.0cm,angle=0]{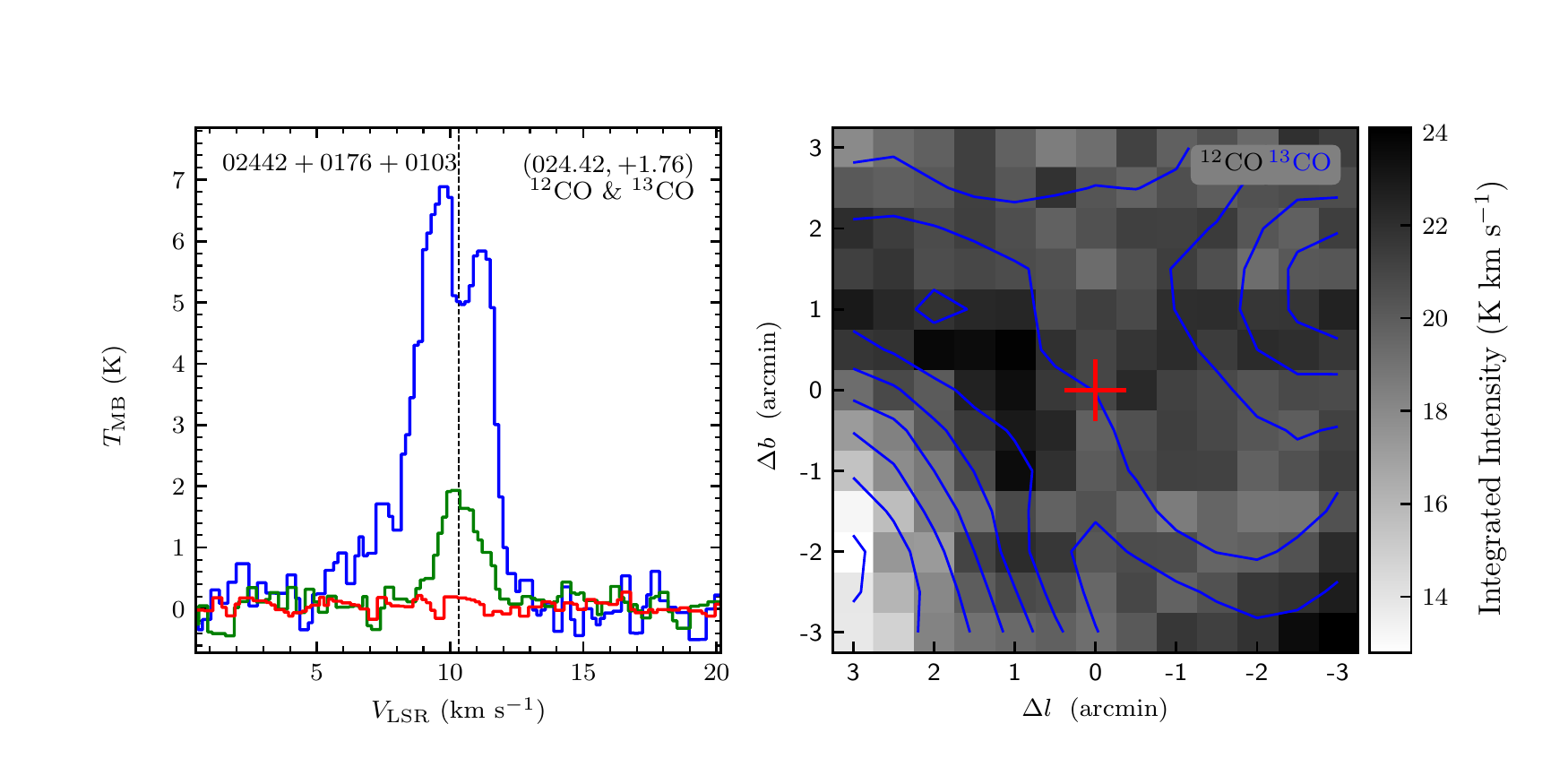}
\end{figure}
\clearpage

\begin{figure}
\includegraphics[width=9.0cm,angle=0]{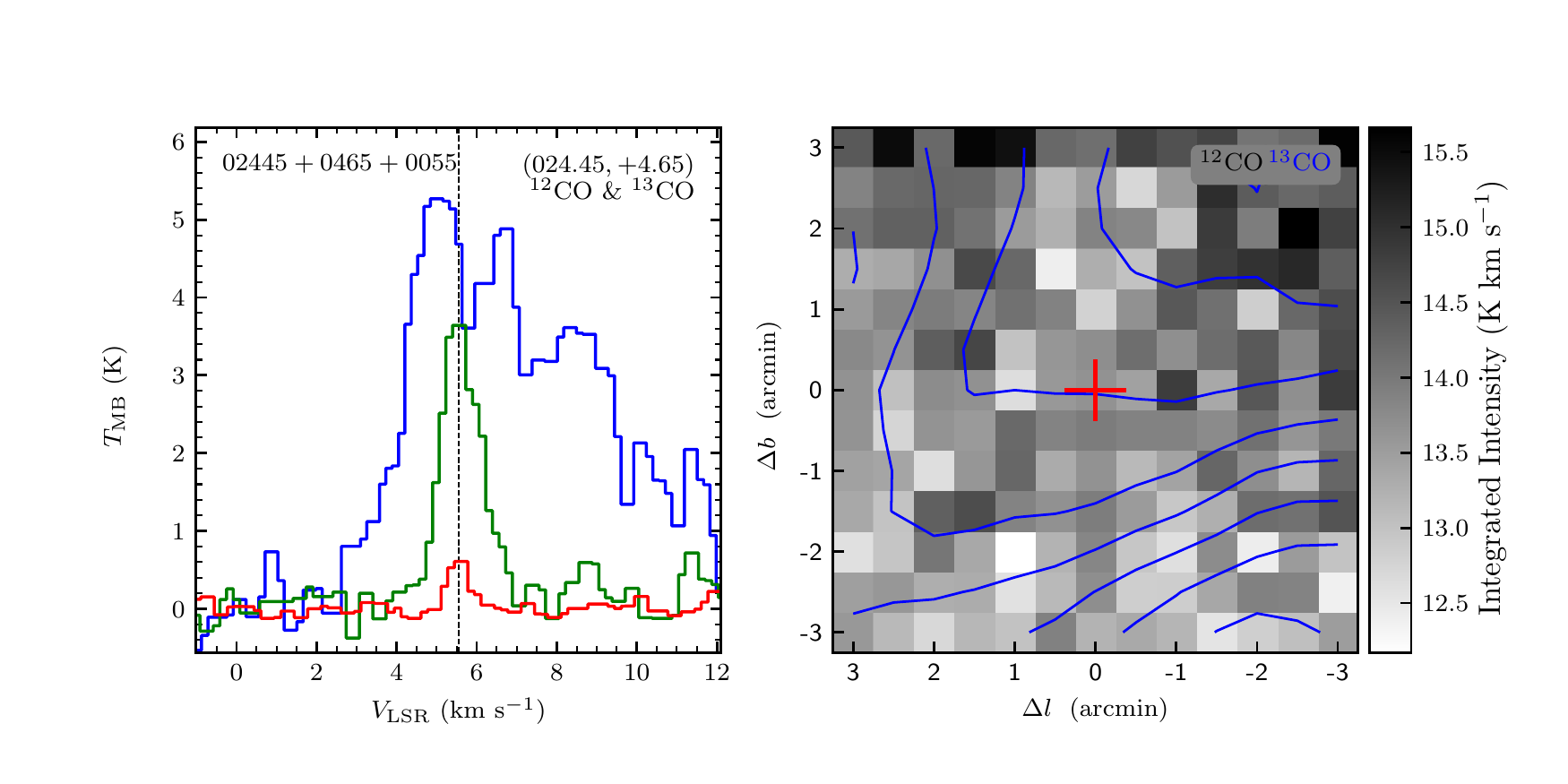}
\includegraphics[width=9.0cm,angle=0]{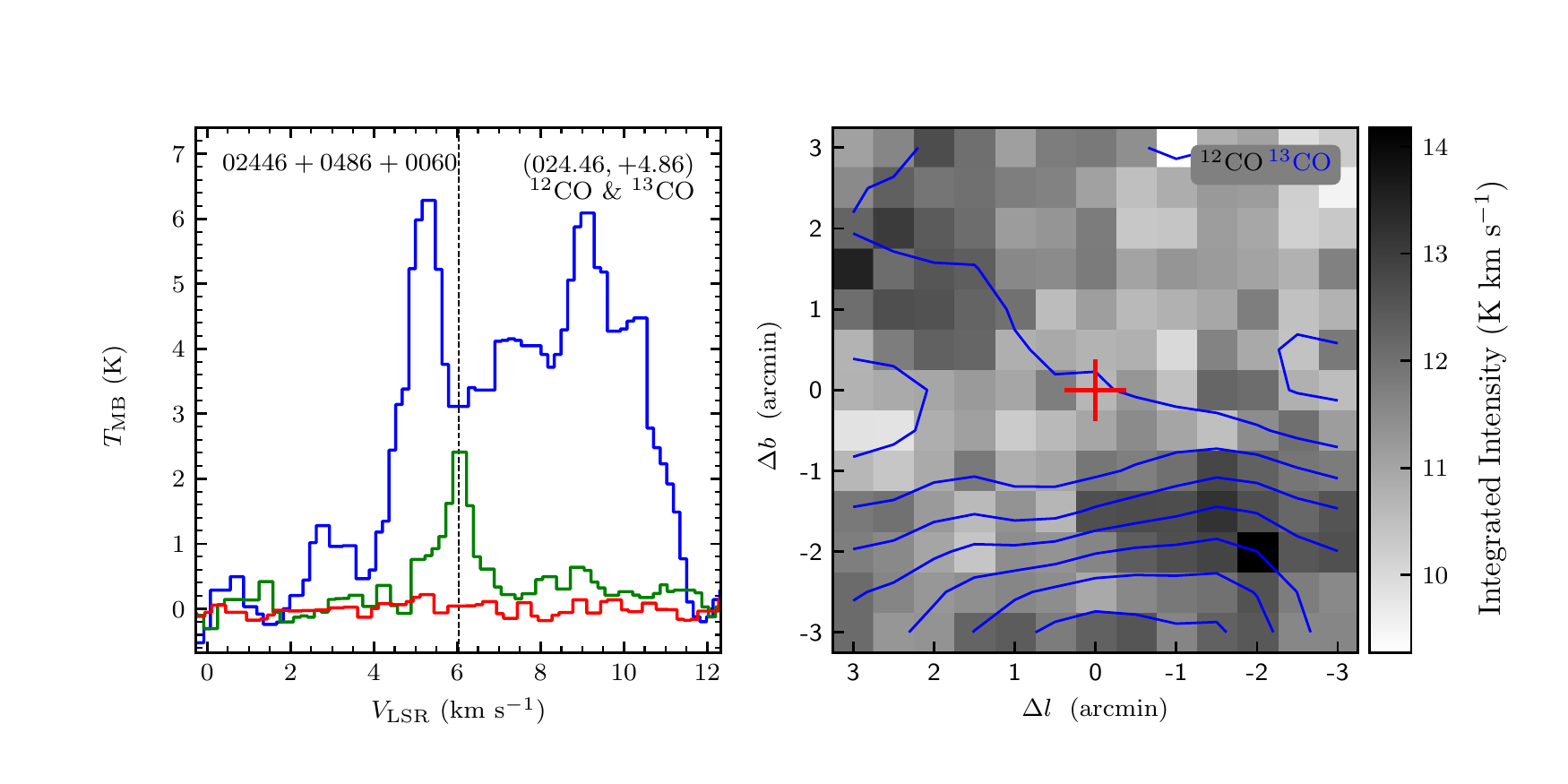}
\vspace{-0.5cm}

\includegraphics[width=9.0cm,angle=0]{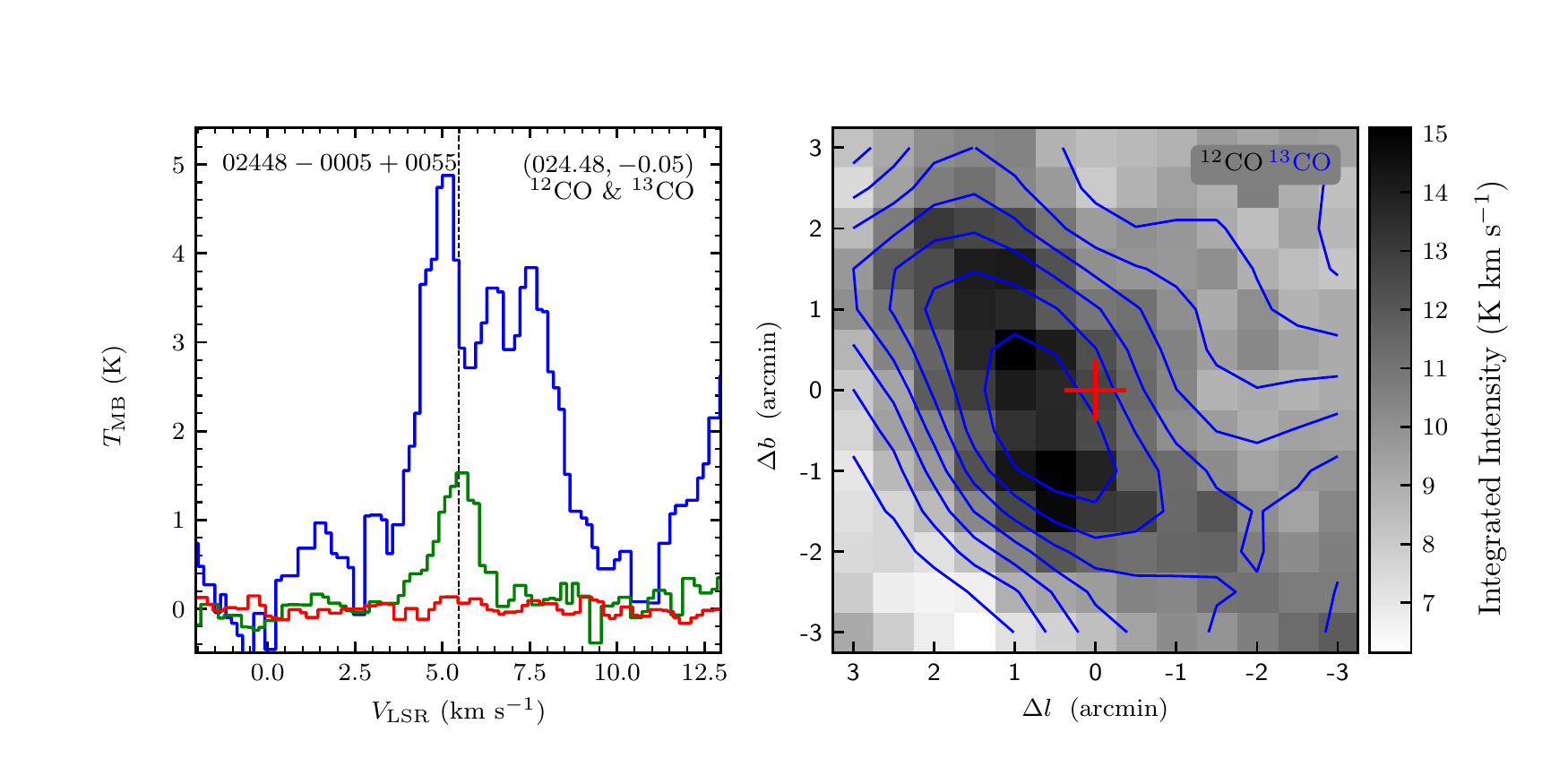}
\includegraphics[width=9.0cm,angle=0]{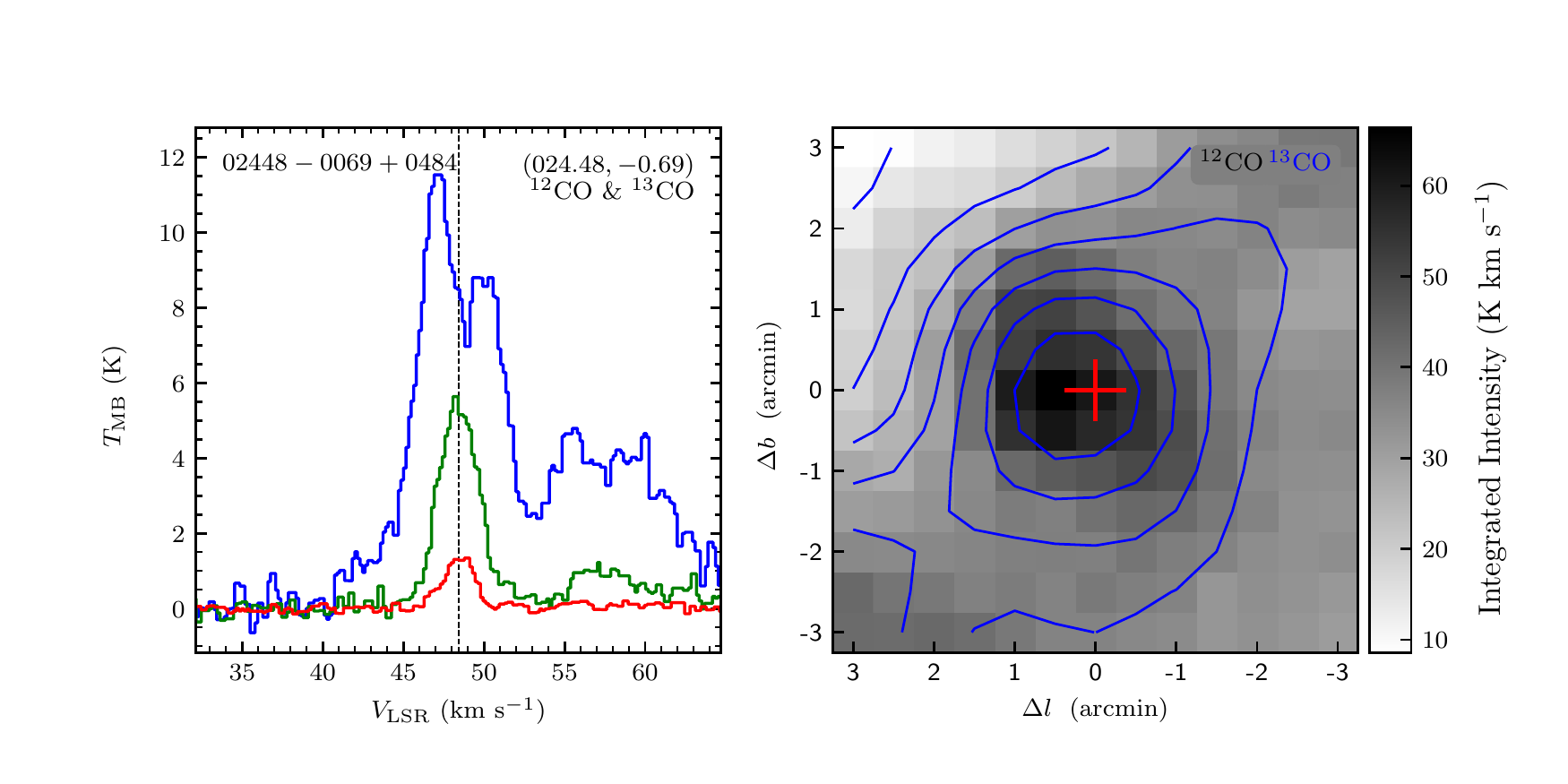}
\vspace{-0.5cm}

\includegraphics[width=9.0cm,angle=0]{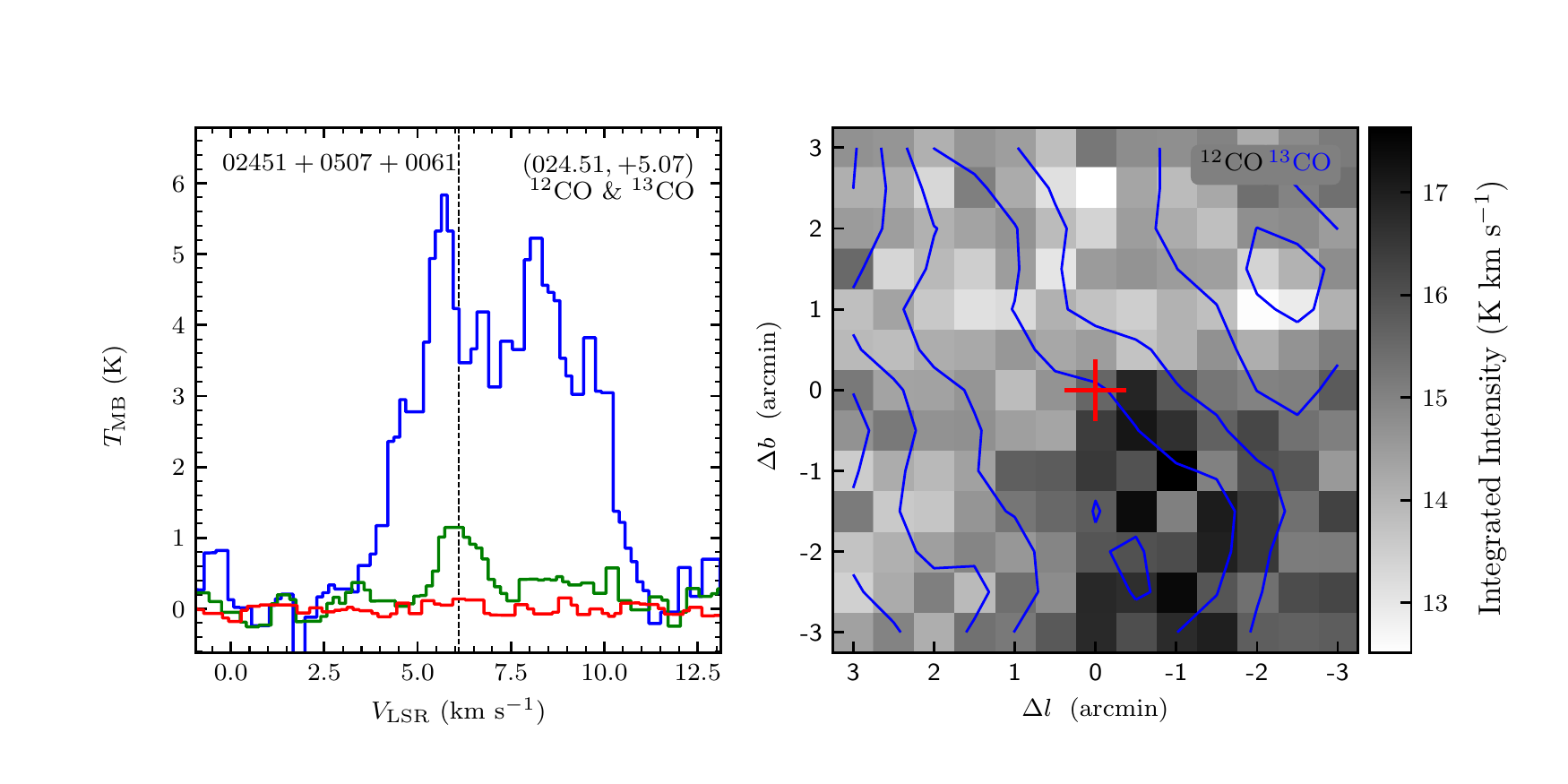}
\includegraphics[width=9.0cm,angle=0]{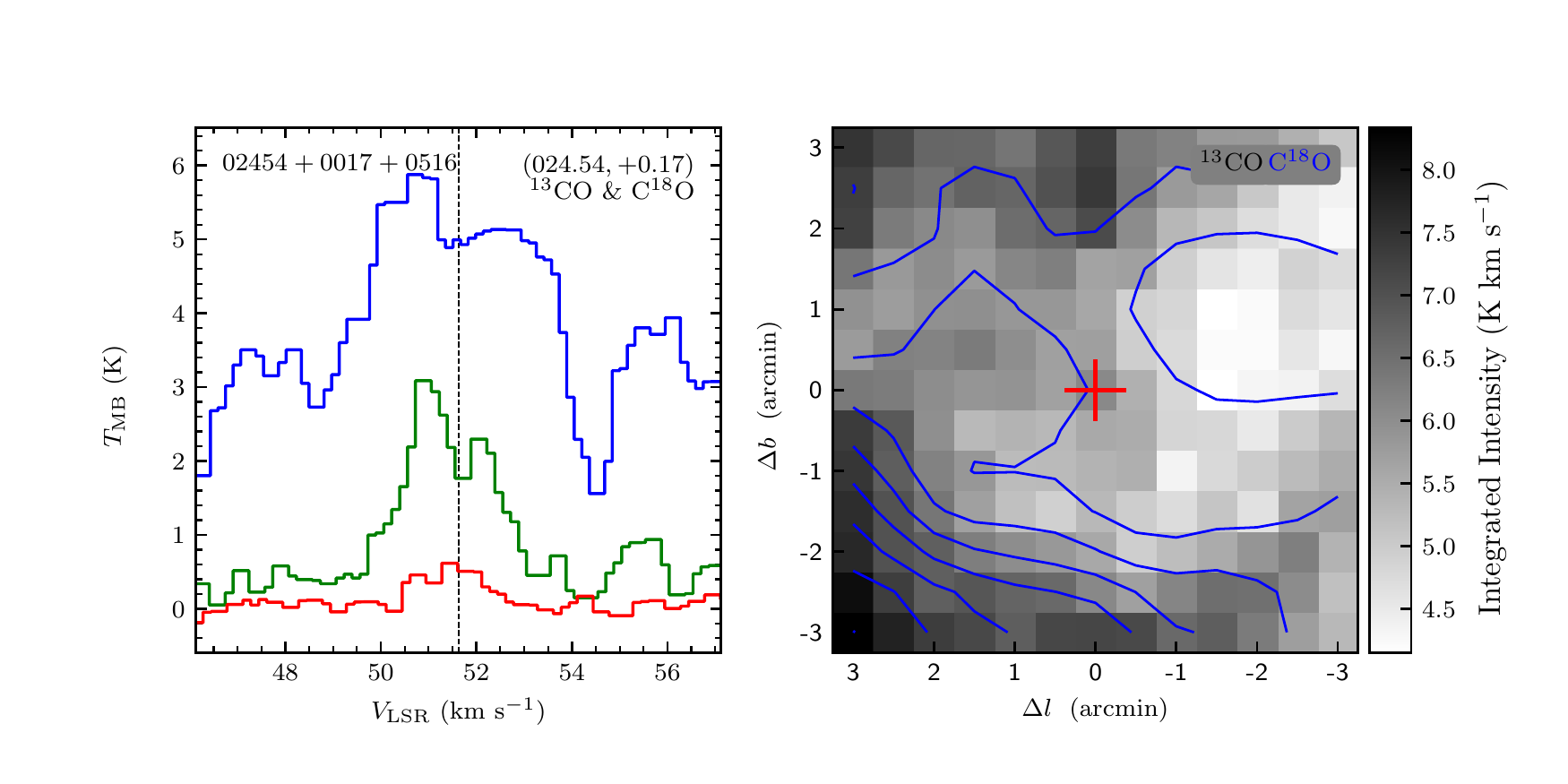}
\vspace{-0.5cm}

\includegraphics[width=9.0cm,angle=0]{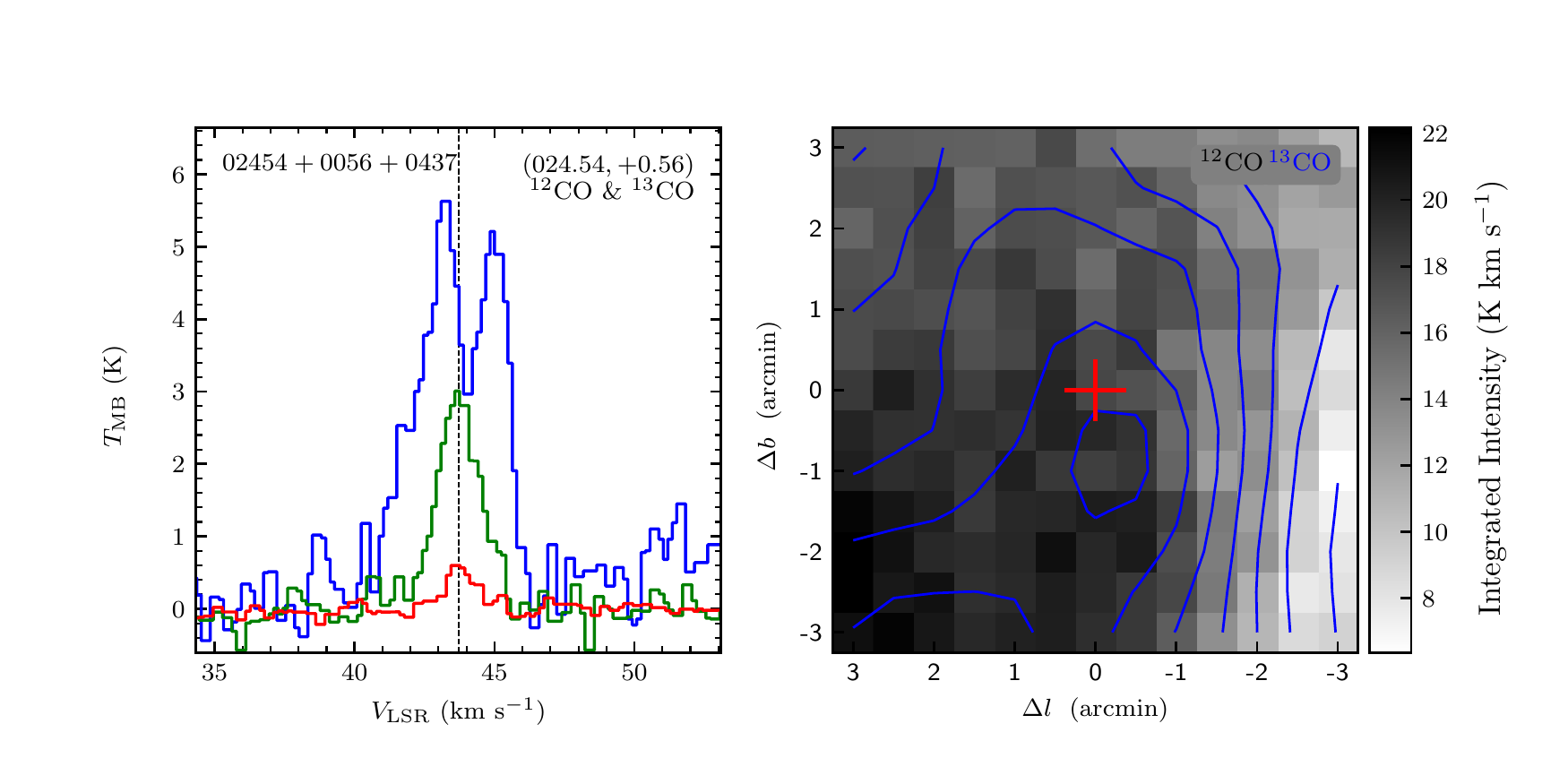}
\includegraphics[width=9.0cm,angle=0]{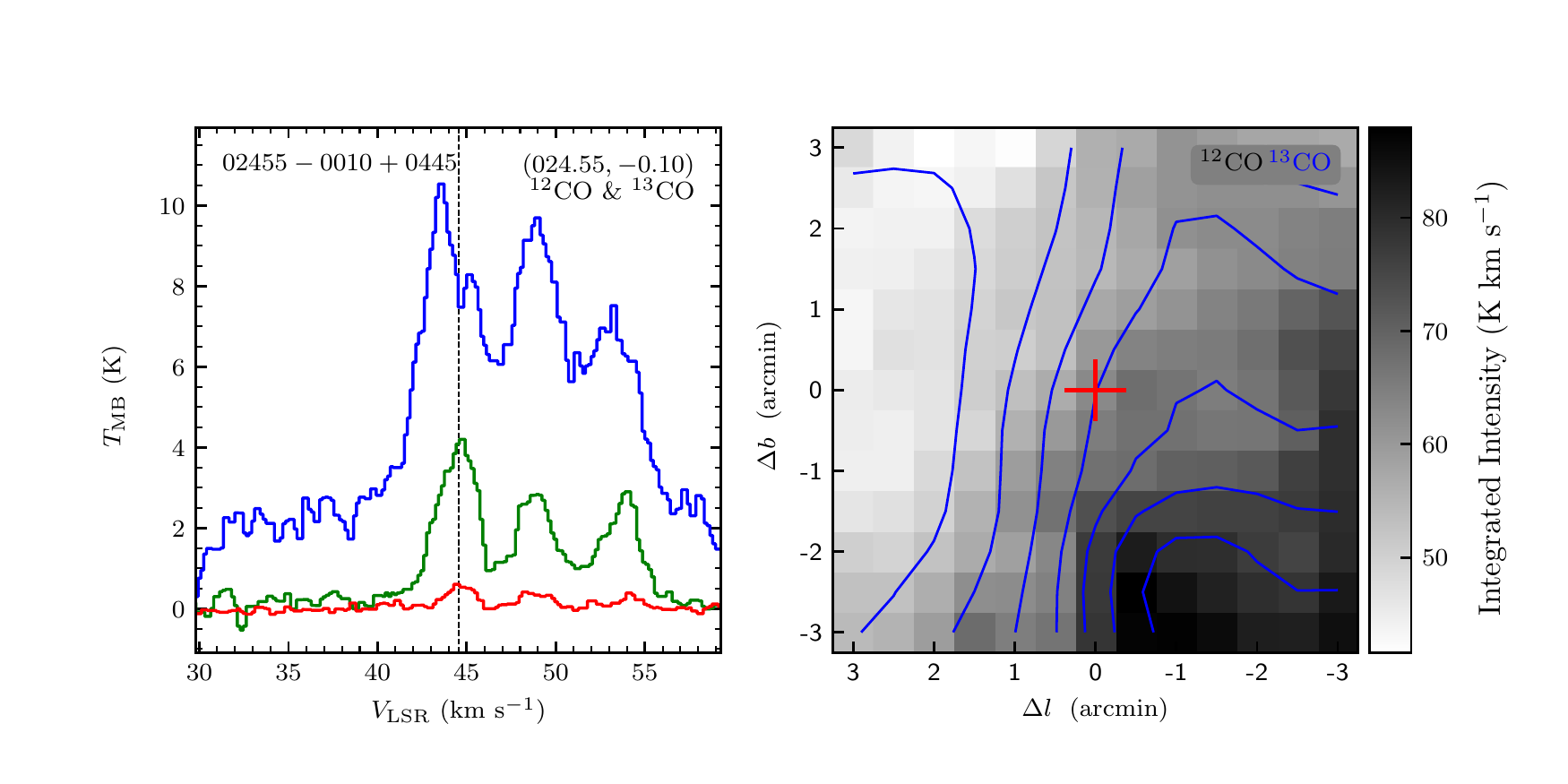}
\vspace{-0.5cm}

\includegraphics[width=9.0cm,angle=0]{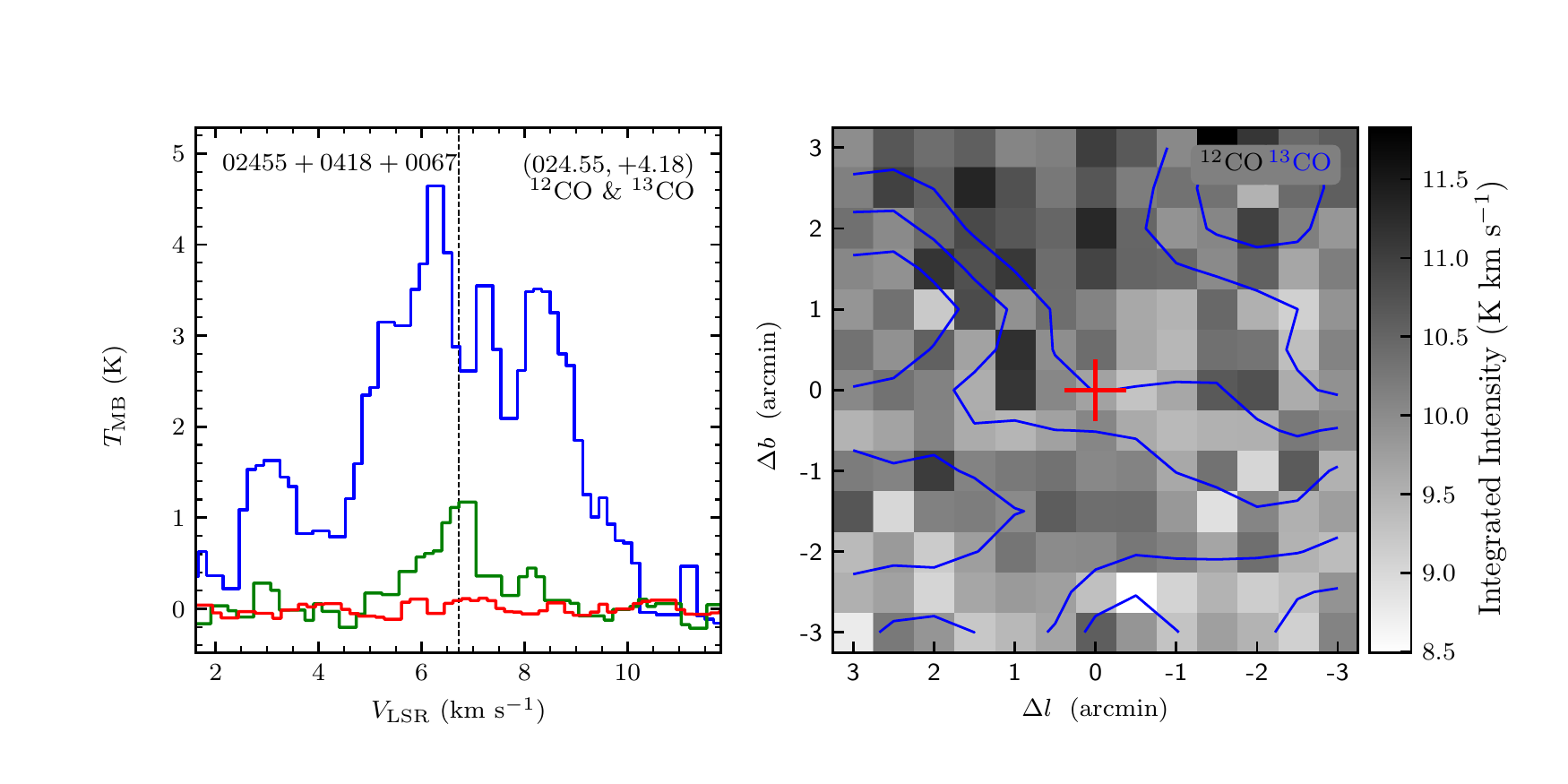}
\includegraphics[width=9.0cm,angle=0]{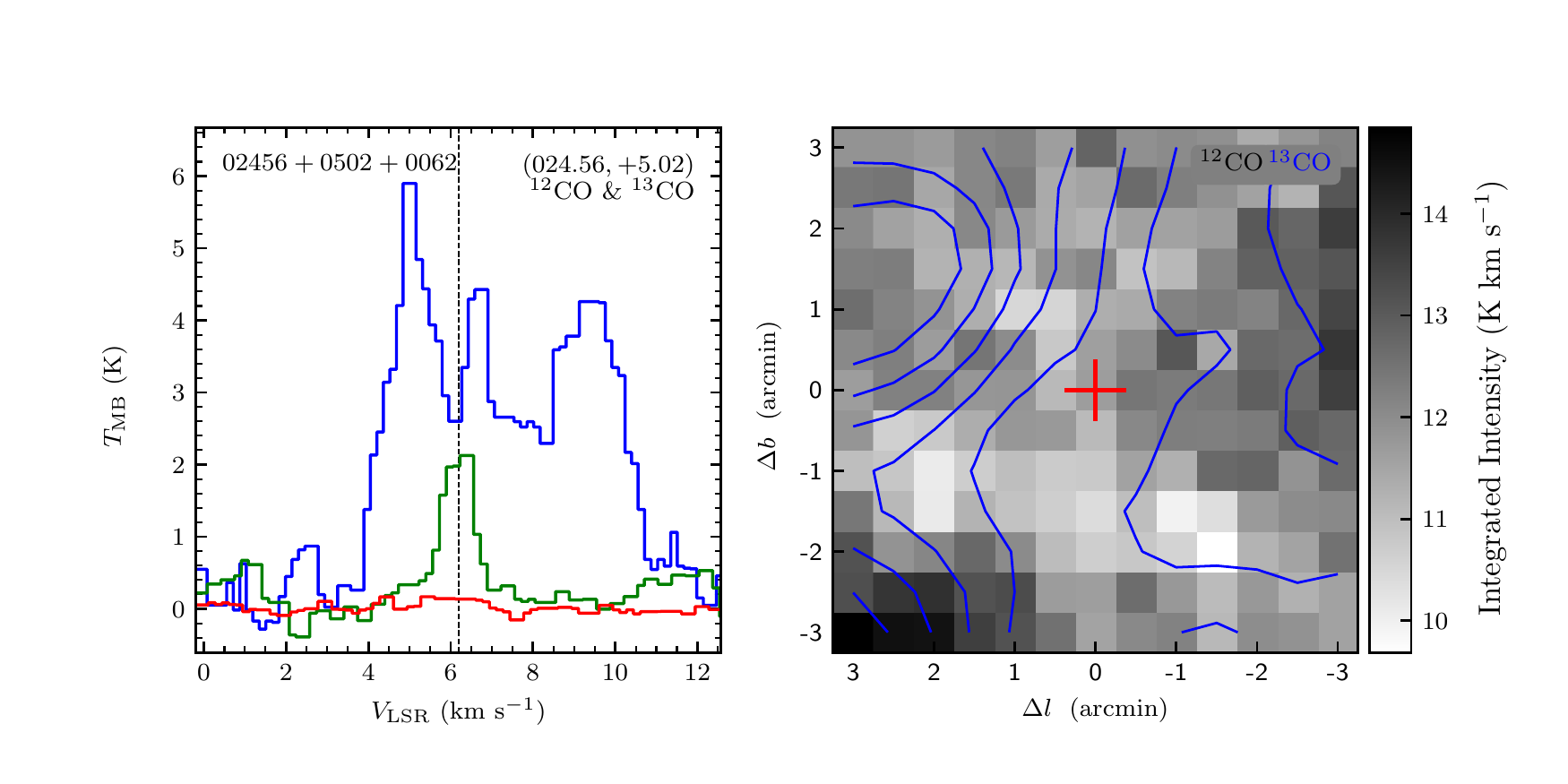}
\end{figure}
\clearpage

\begin{figure}
\includegraphics[width=9.0cm,angle=0]{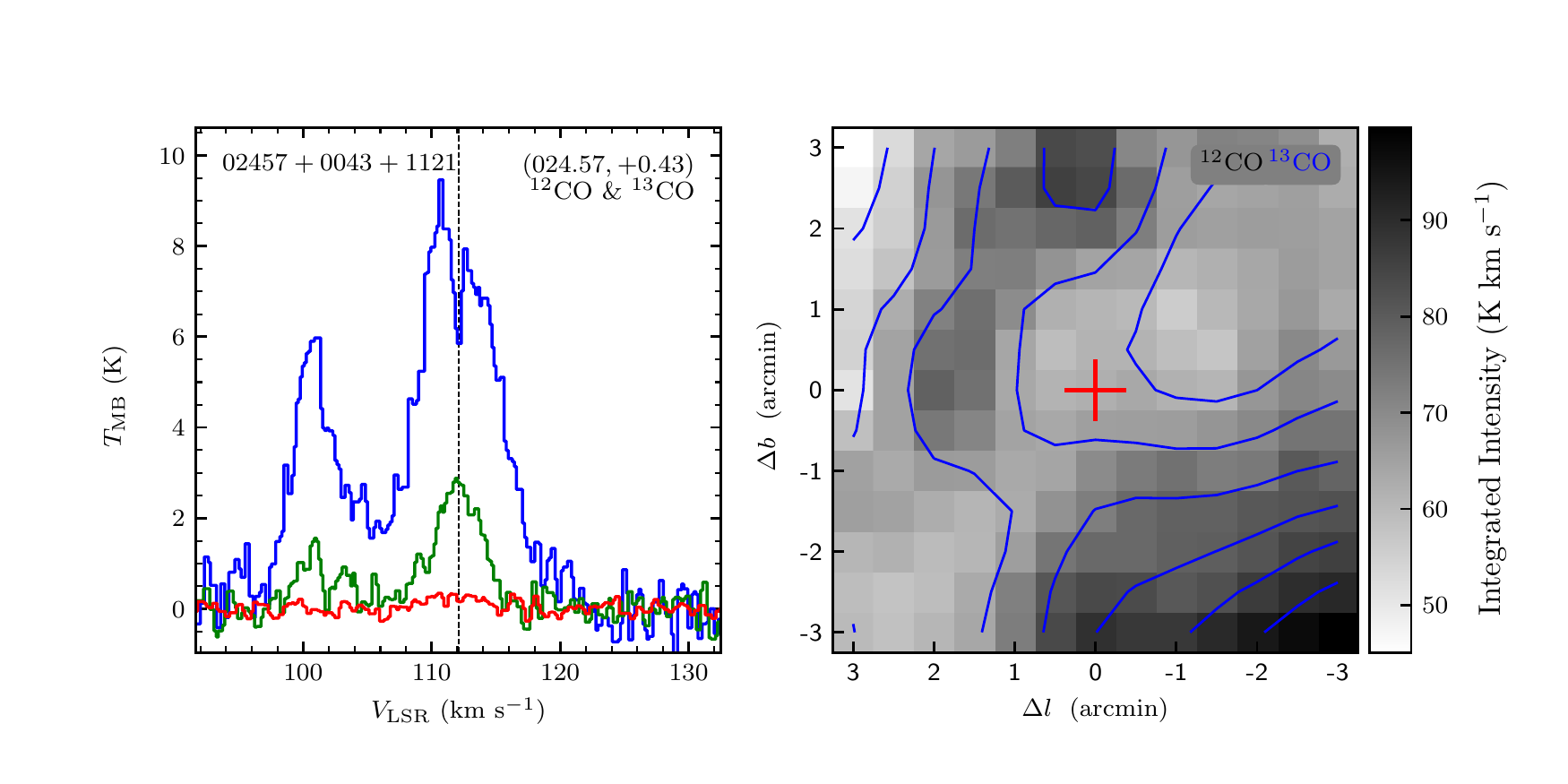}
\includegraphics[width=9.0cm,angle=0]{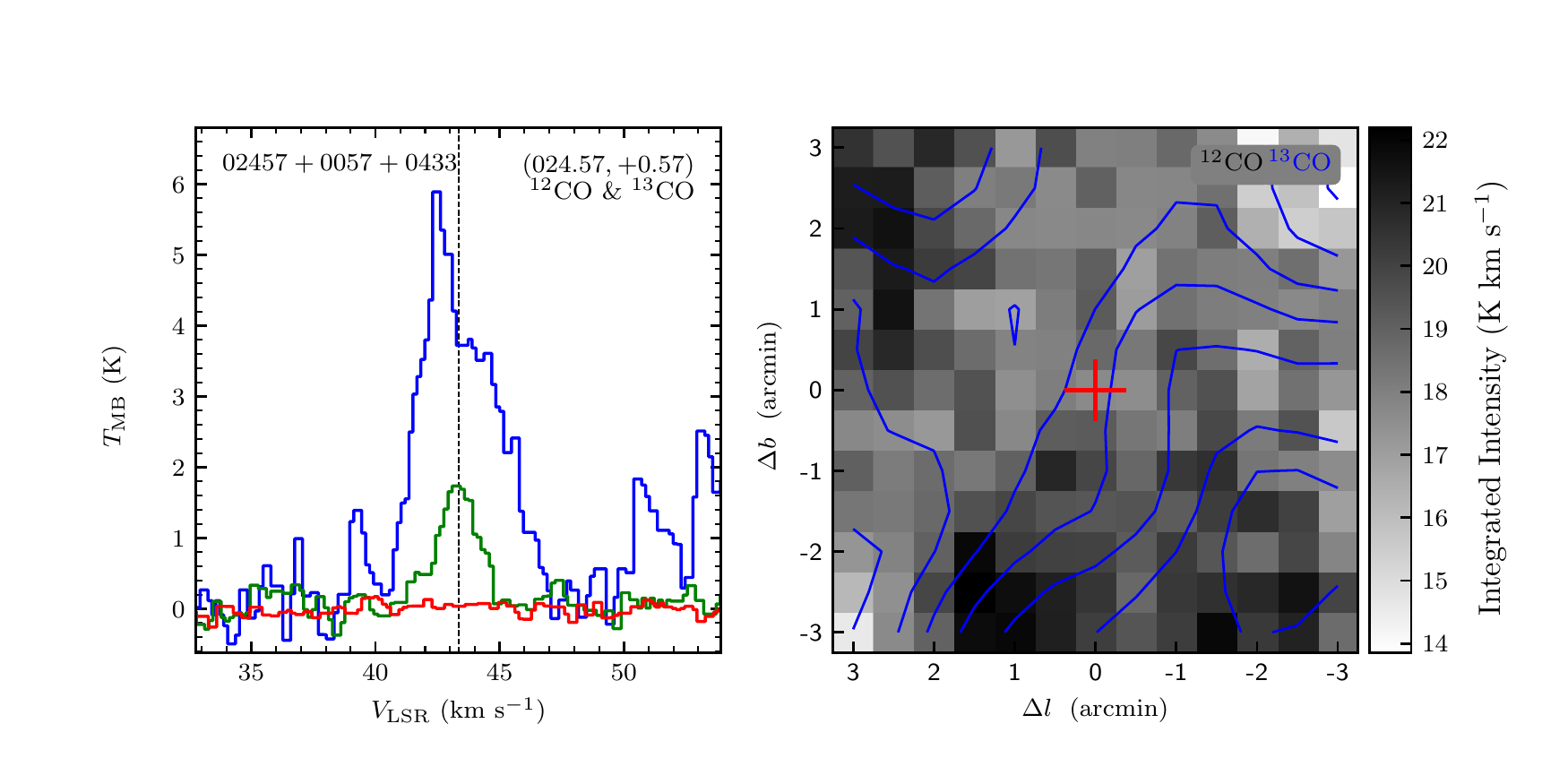}
\vspace{-0.5cm}

\includegraphics[width=9.0cm,angle=0]{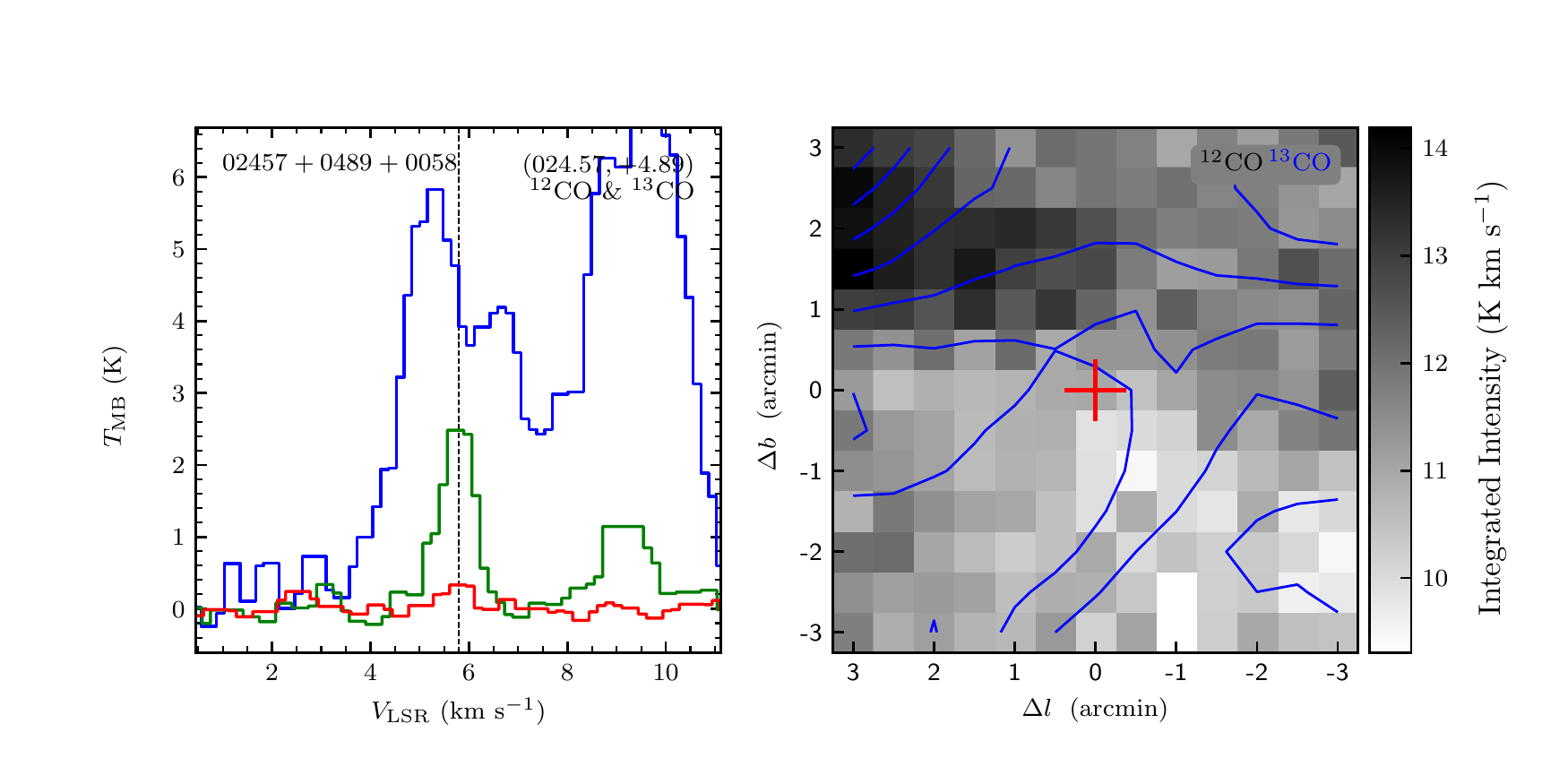}
\includegraphics[width=9.0cm,angle=0]{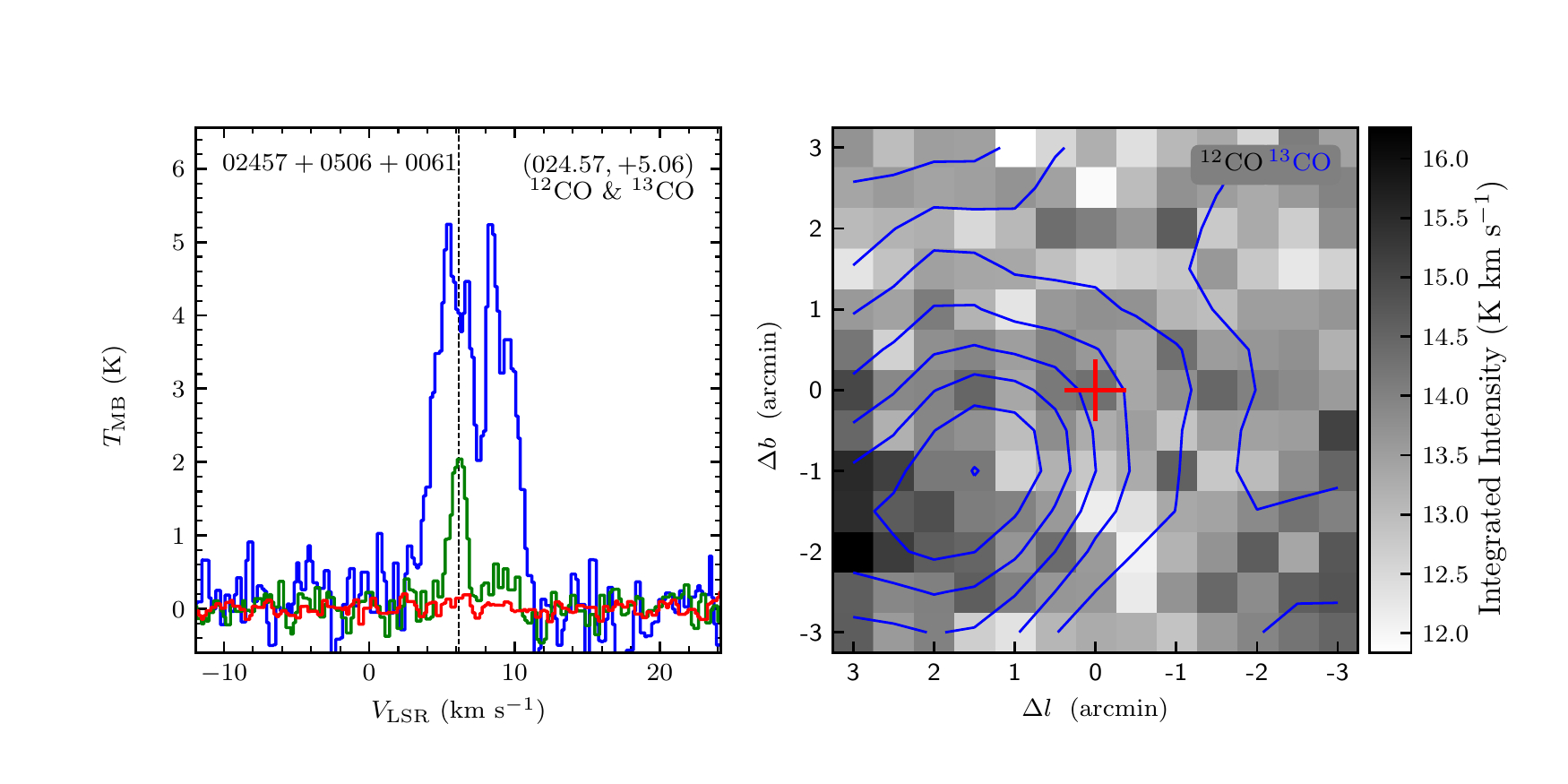}
\vspace{-0.5cm}

\includegraphics[width=9.0cm,angle=0]{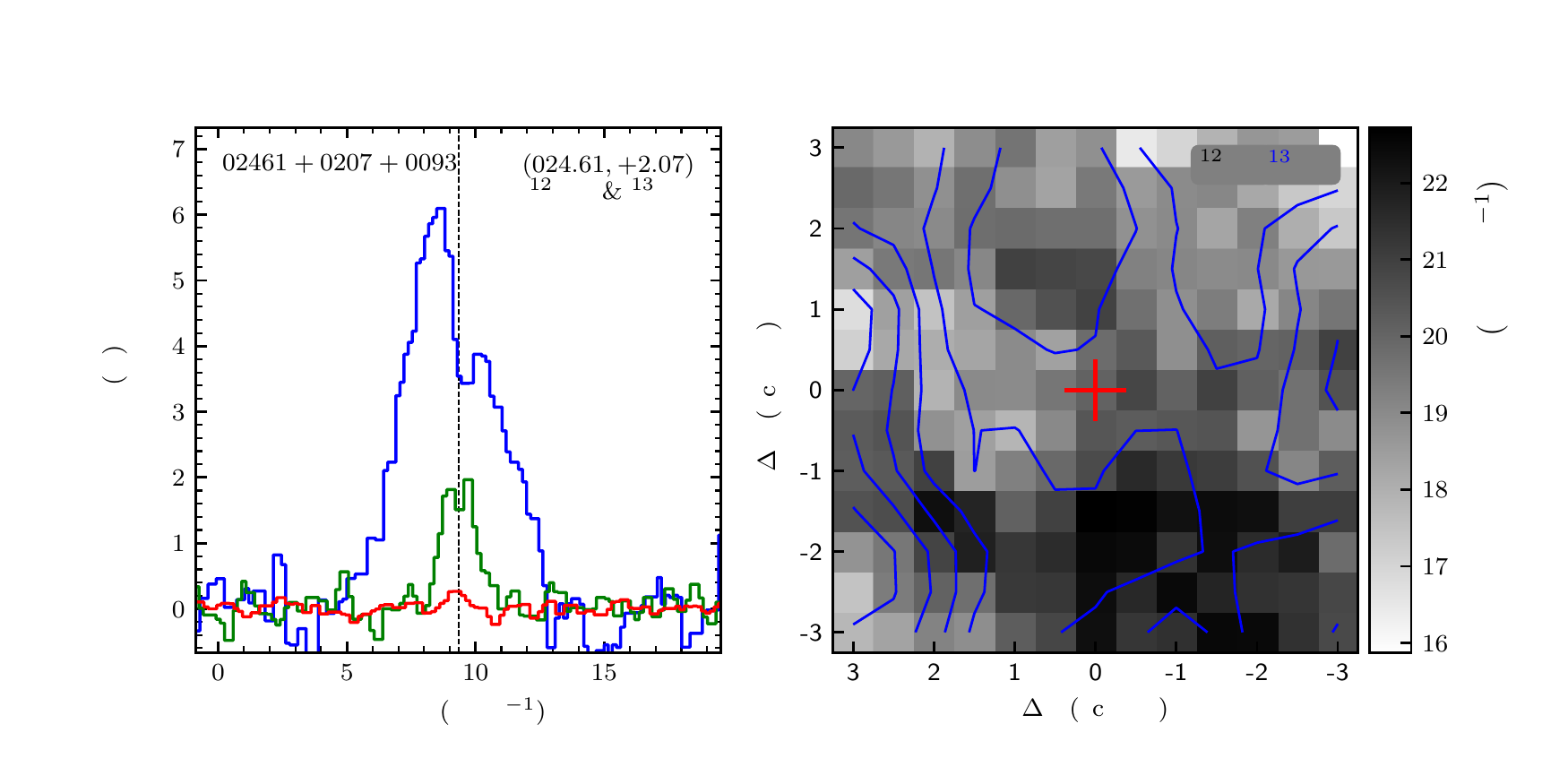}
\includegraphics[width=9.0cm,angle=0]{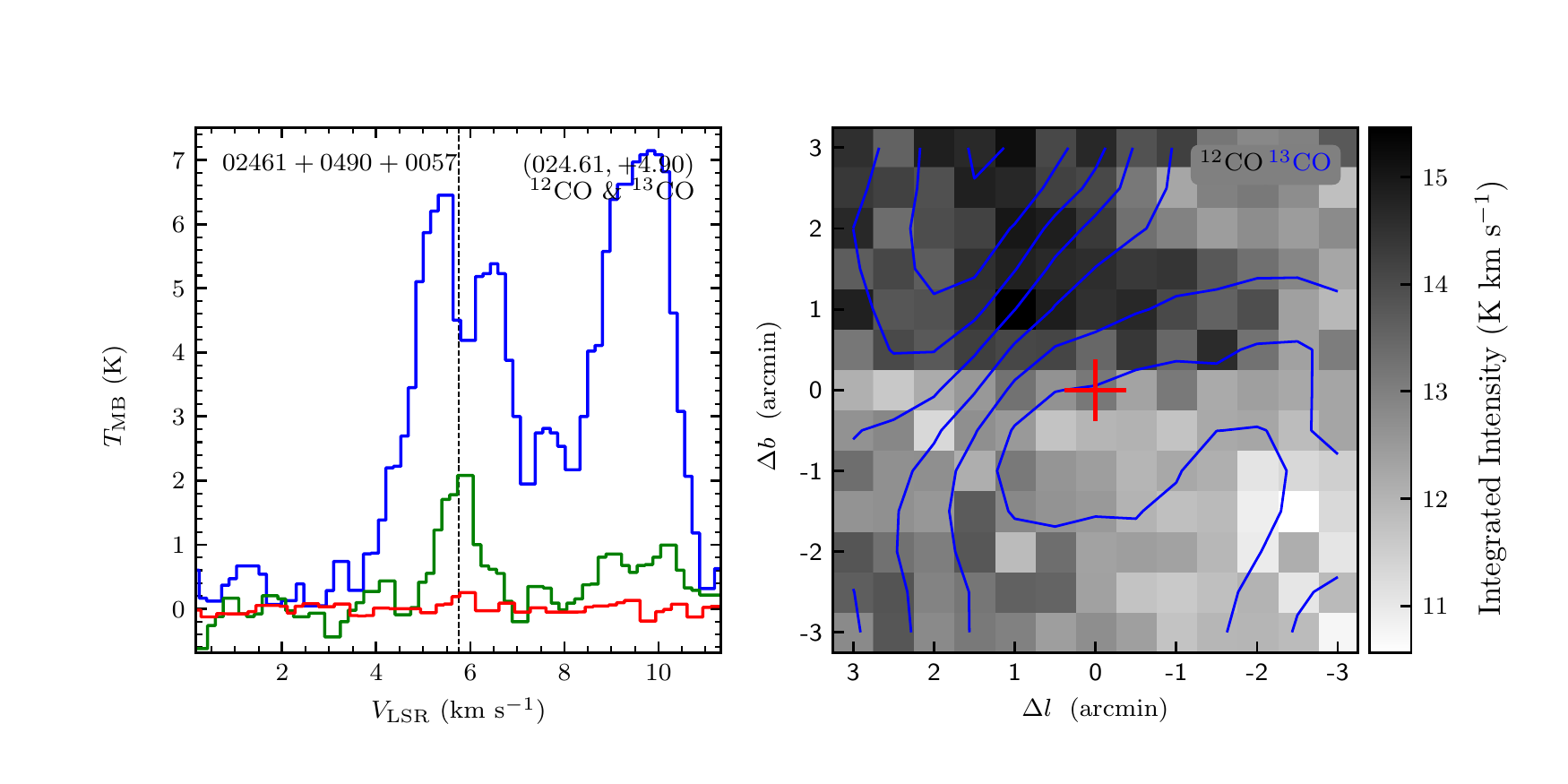}
\vspace{-0.5cm}

\includegraphics[width=9.0cm,angle=0]{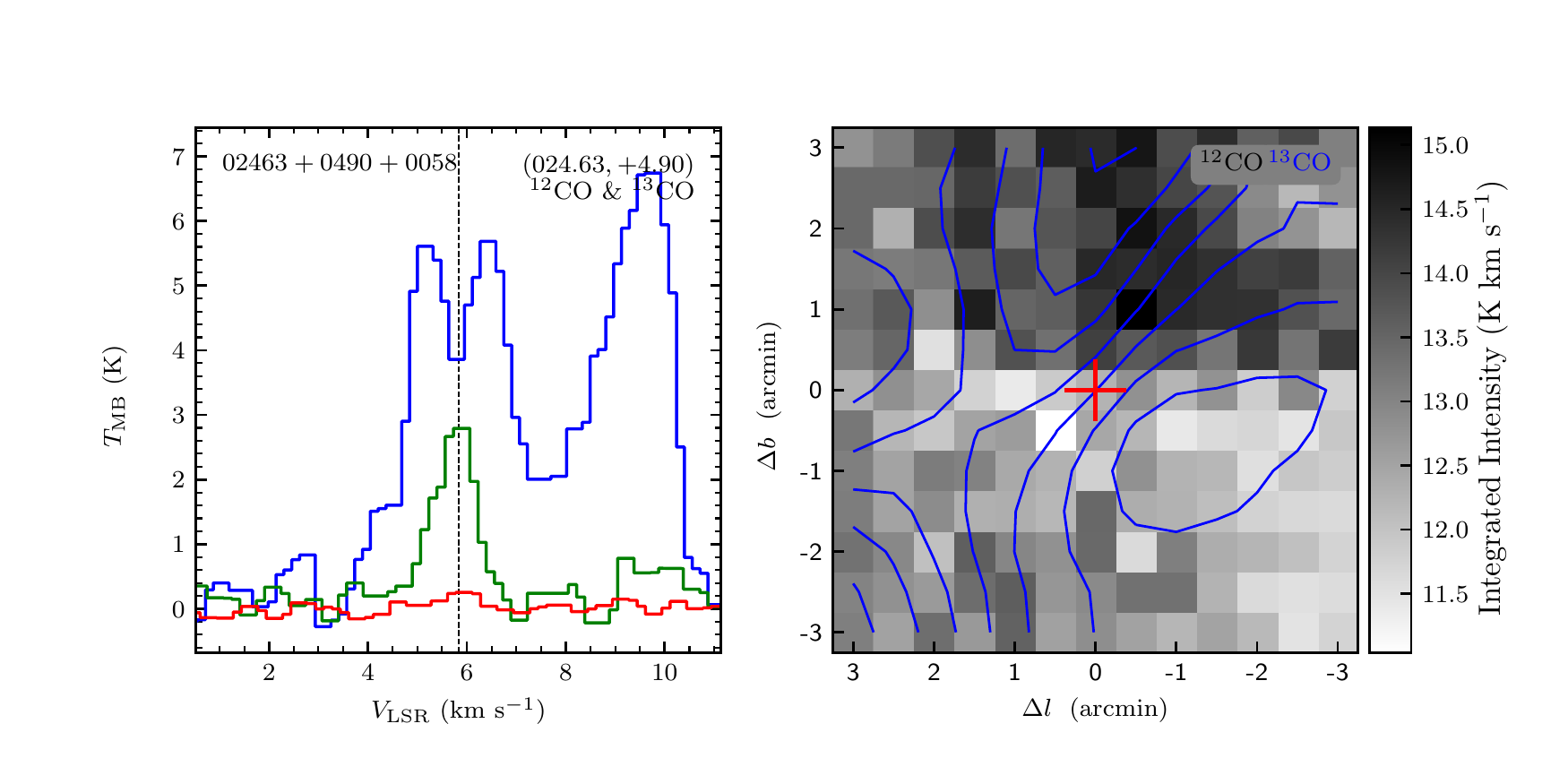}
\includegraphics[width=9.0cm,angle=0]{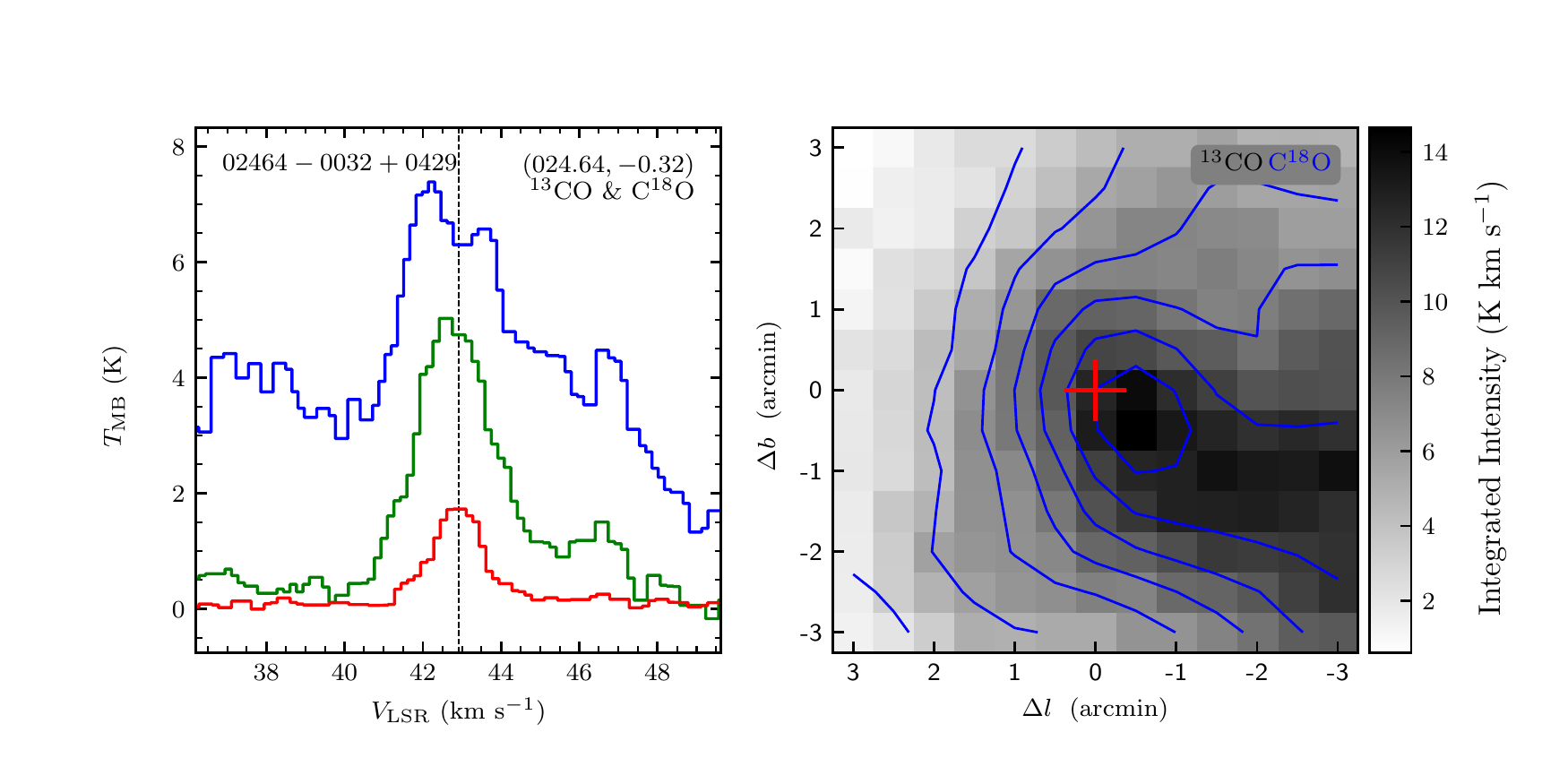}
\vspace{-0.5cm}

\includegraphics[width=9.0cm,angle=0]{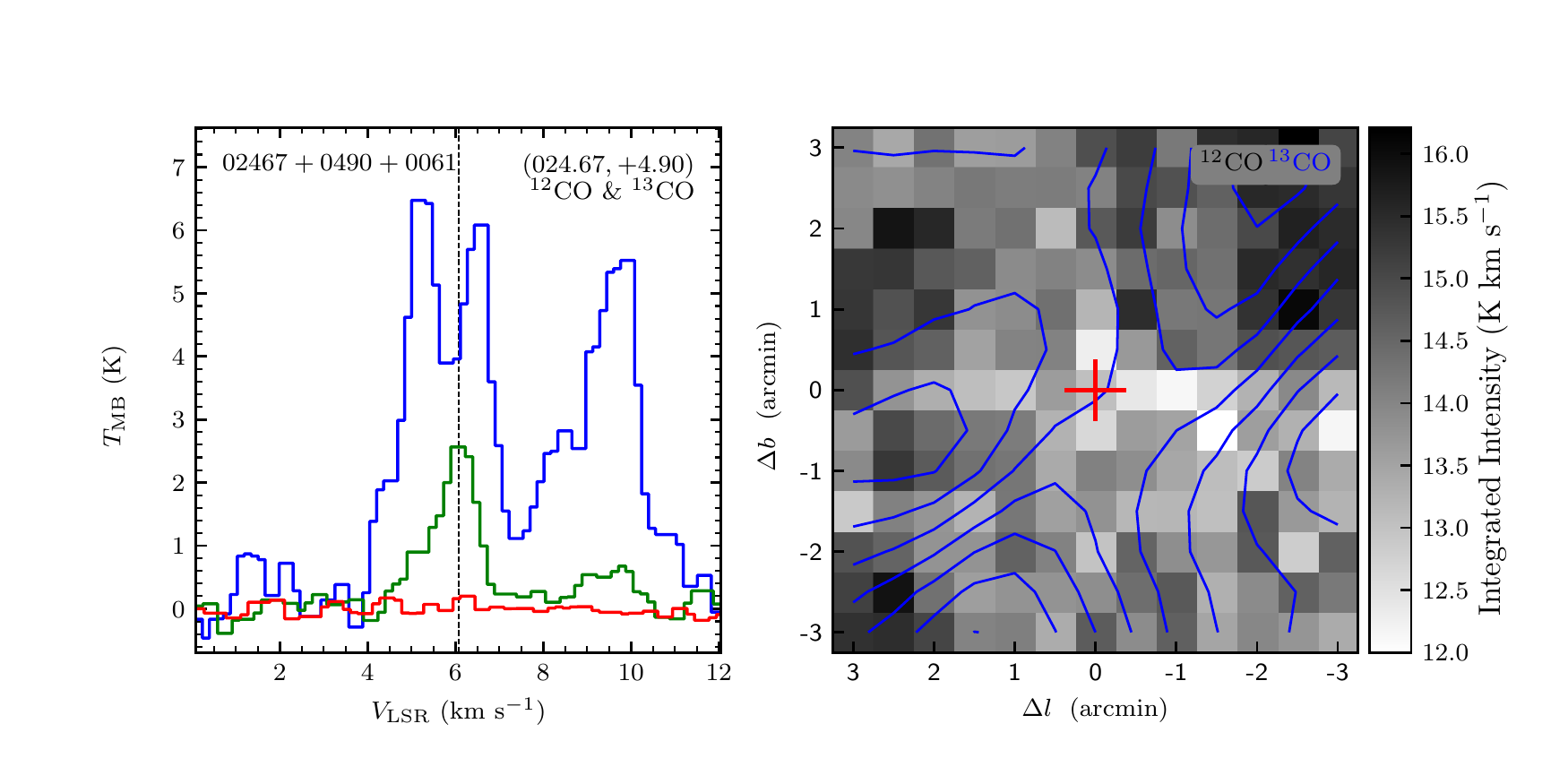}
\includegraphics[width=9.0cm,angle=0]{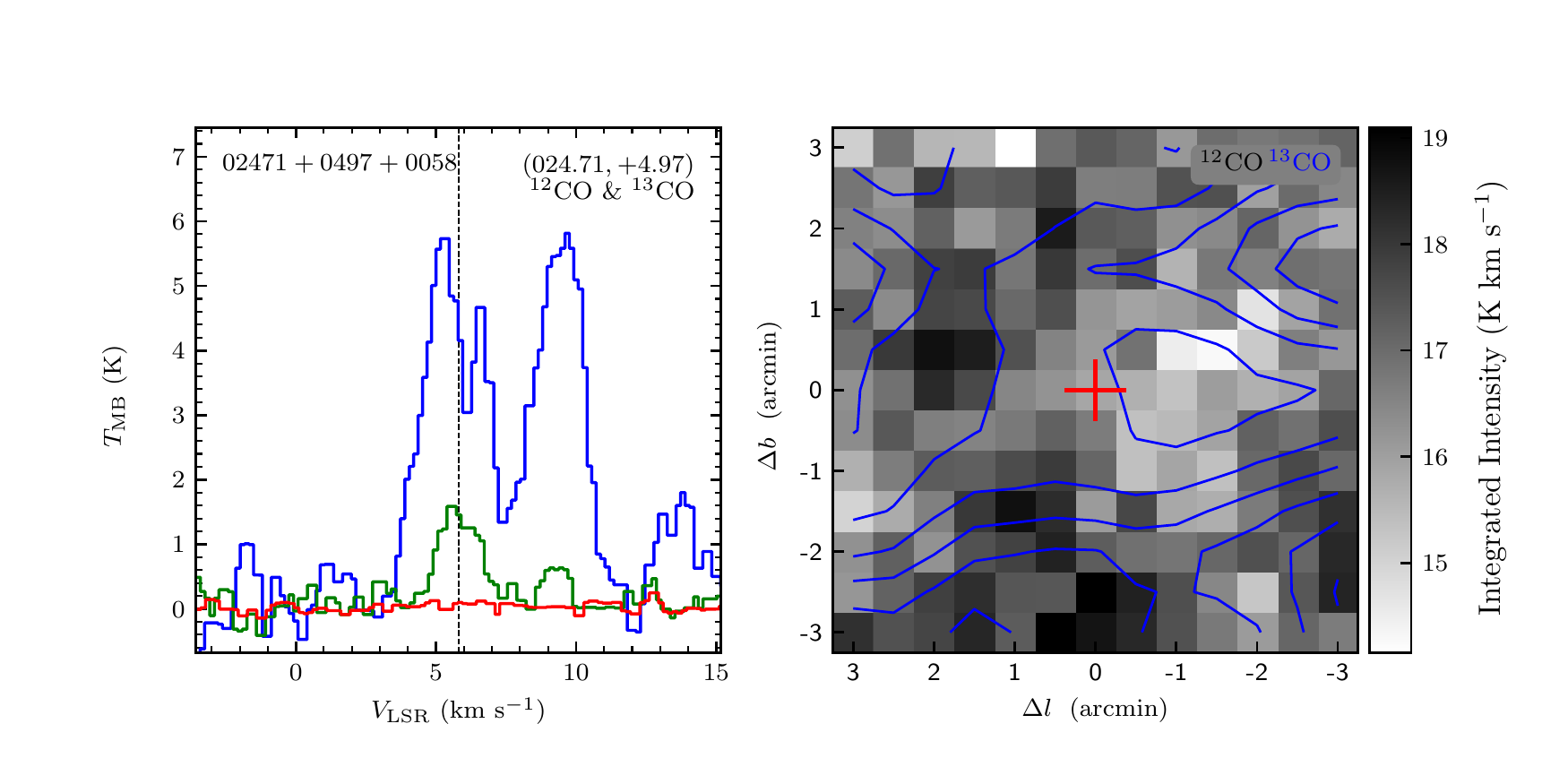}
\end{figure}
\clearpage

\begin{figure}
\includegraphics[width=9.0cm,angle=0]{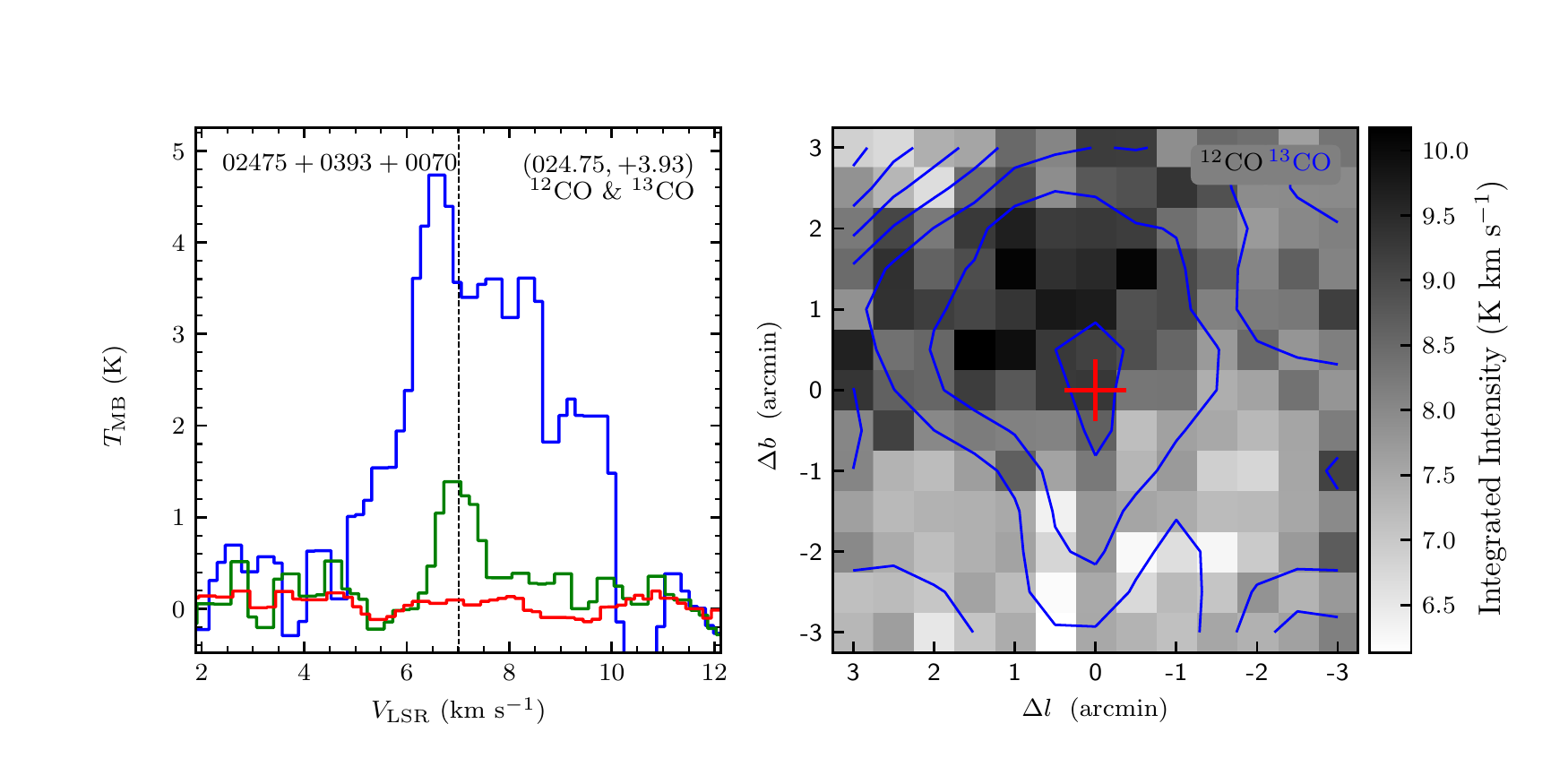}
\includegraphics[width=9.0cm,angle=0]{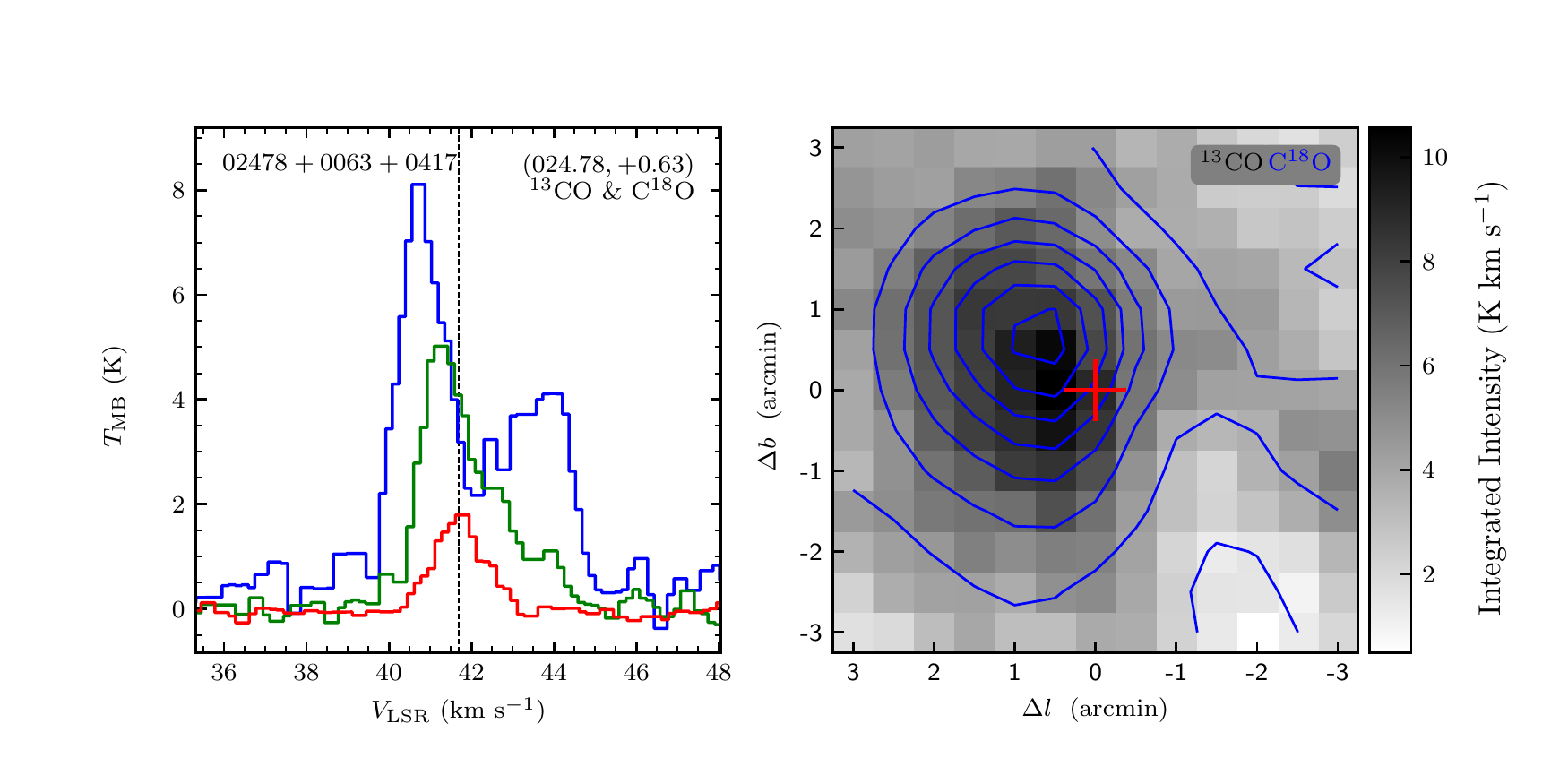}
\vspace{-0.5cm}

\includegraphics[width=9.0cm,angle=0]{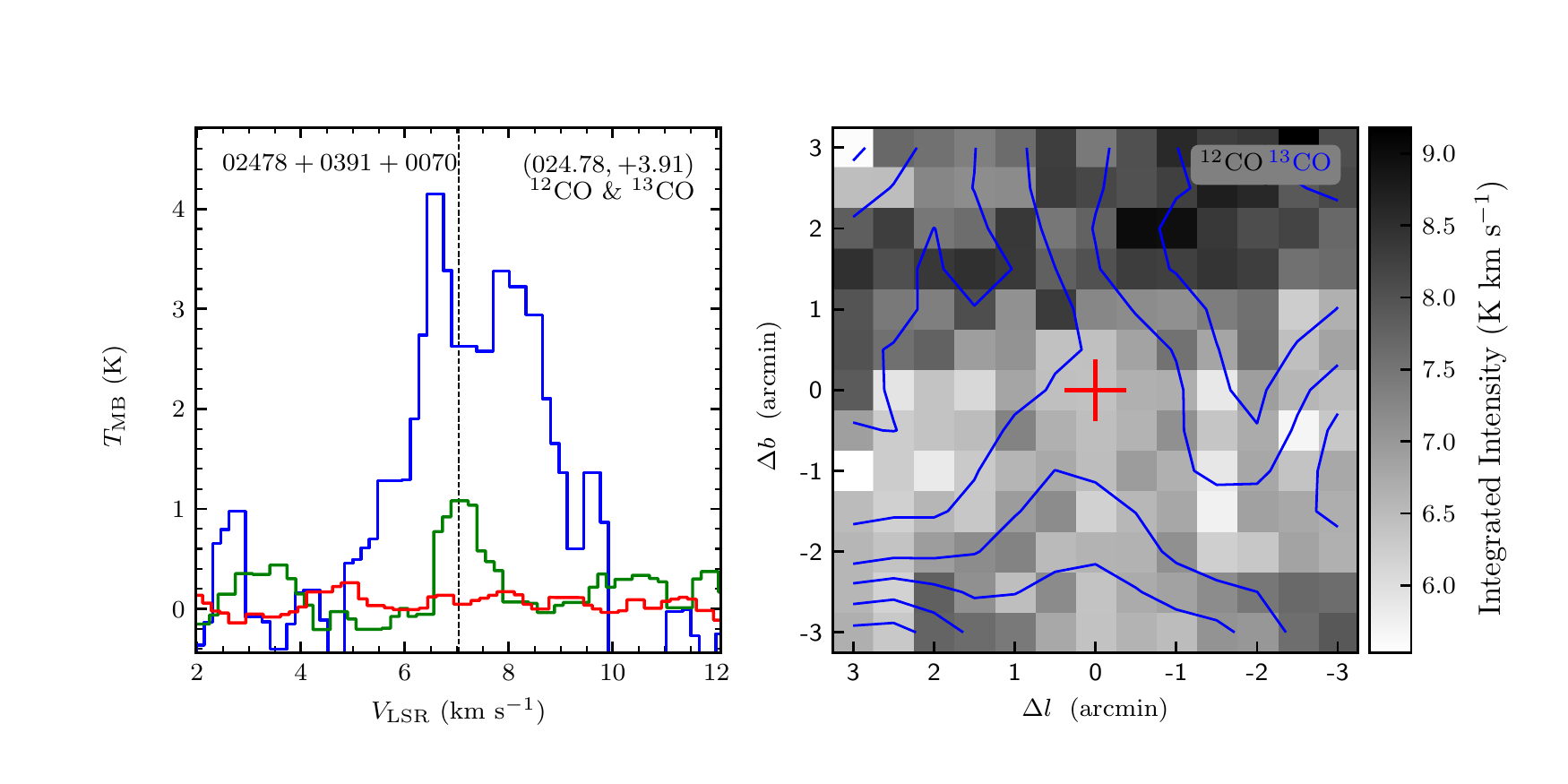}
\includegraphics[width=9.0cm,angle=0]{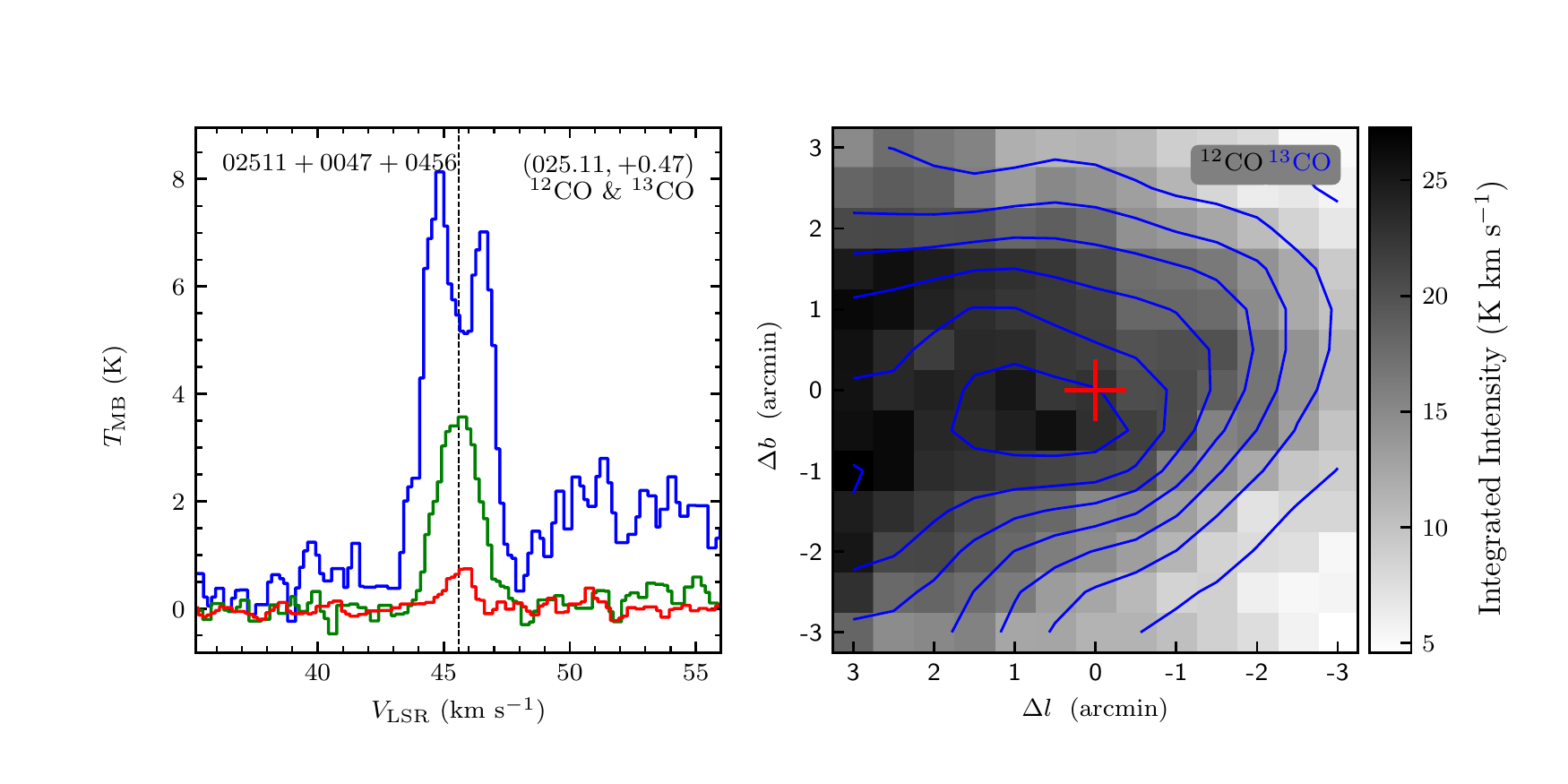}
\vspace{-0.5cm}

\includegraphics[width=9.0cm,angle=0]{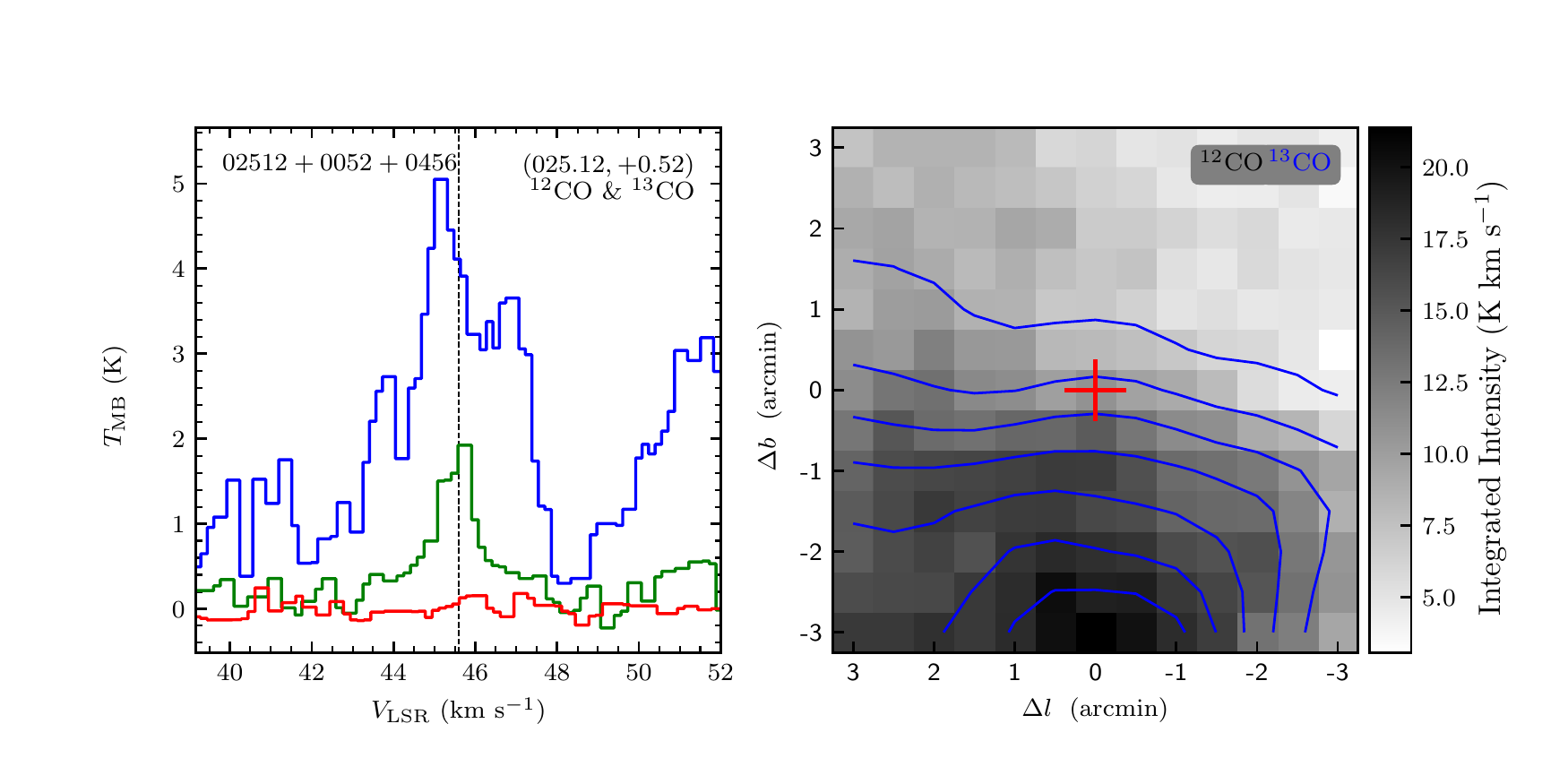}
\includegraphics[width=9.0cm,angle=0]{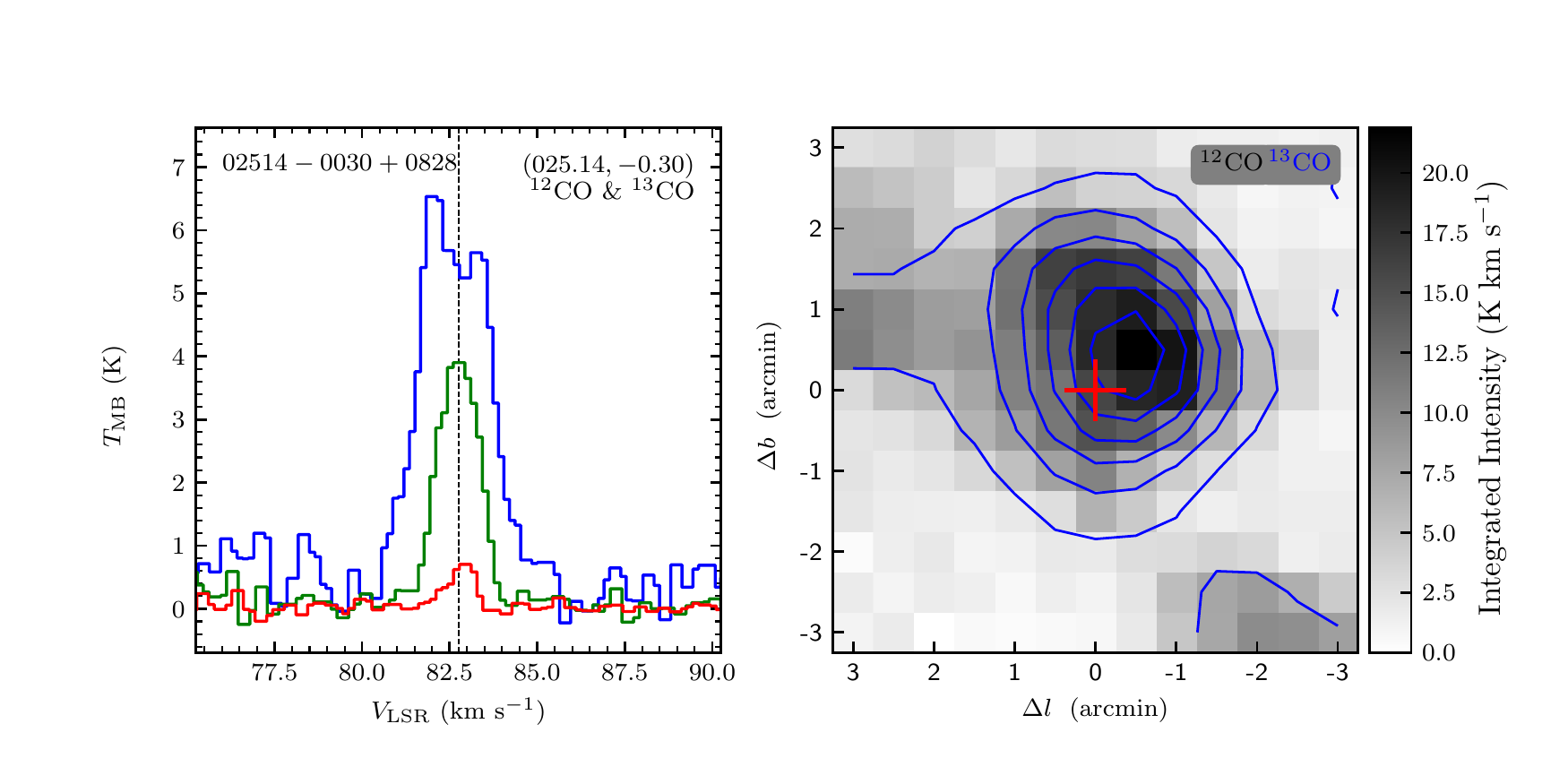}
\vspace{-0.5cm}

\includegraphics[width=9.0cm,angle=0]{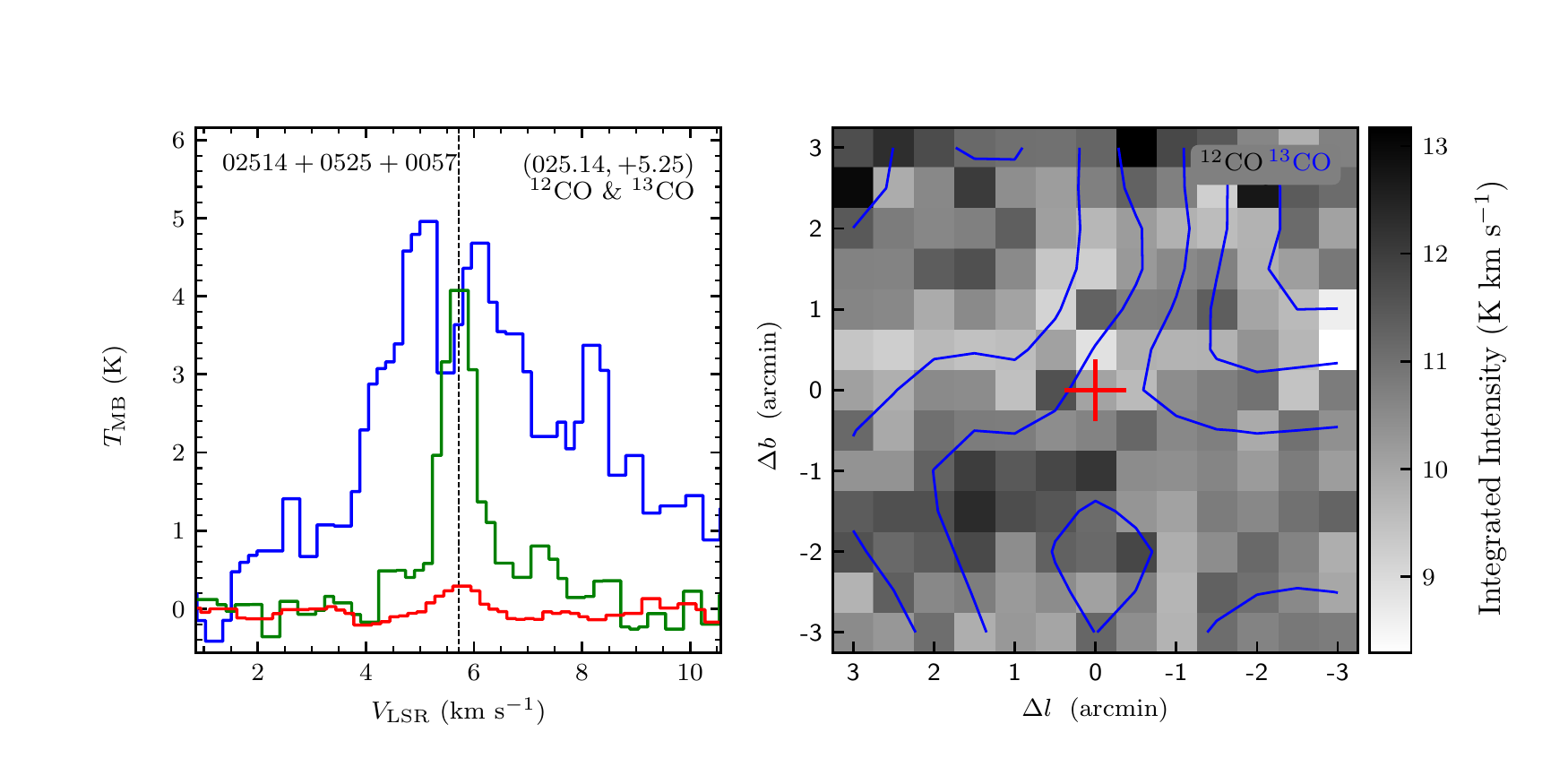}
\includegraphics[width=9.0cm,angle=0]{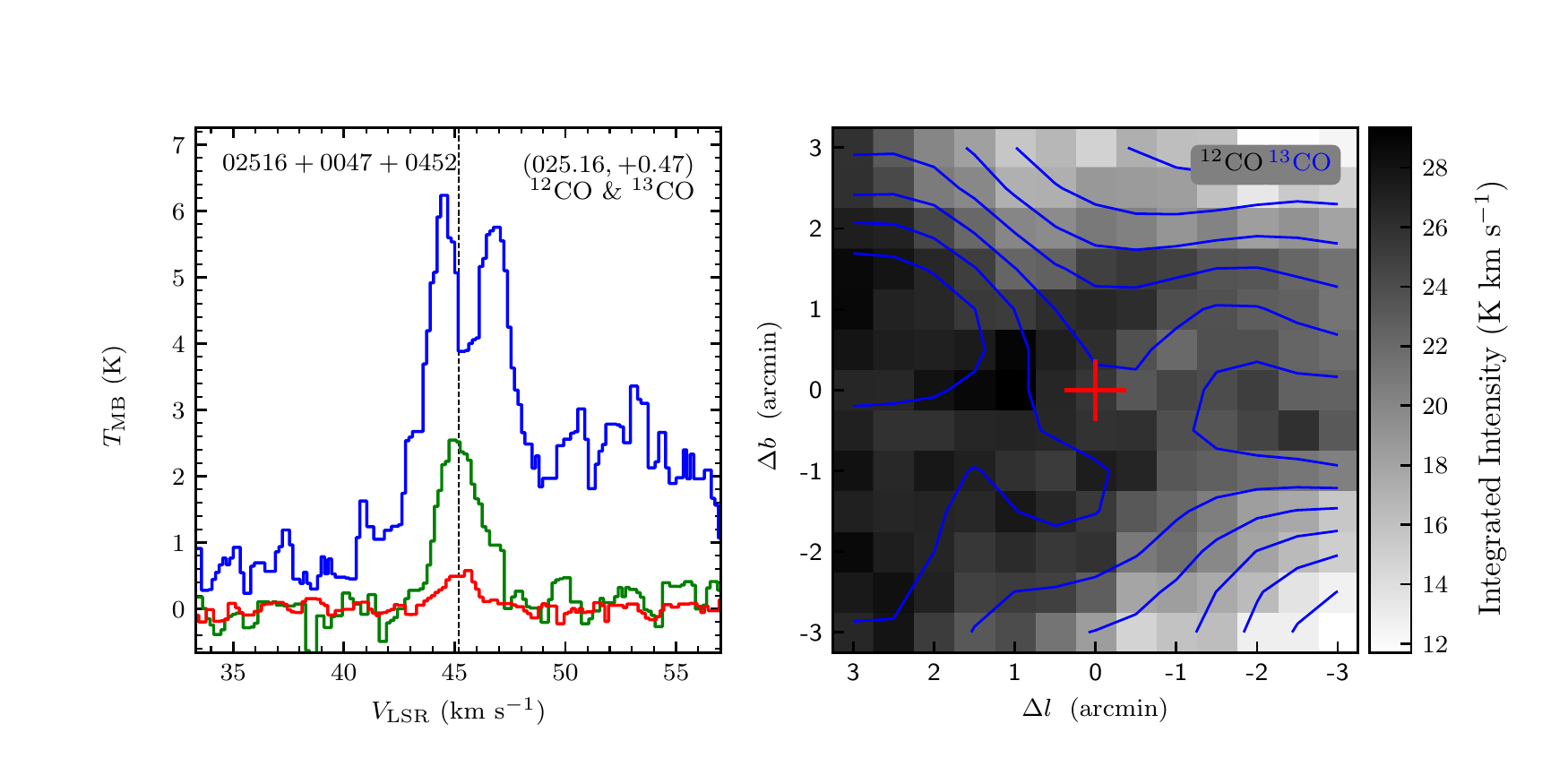}
\vspace{-0.5cm}

\includegraphics[width=9.0cm,angle=0]{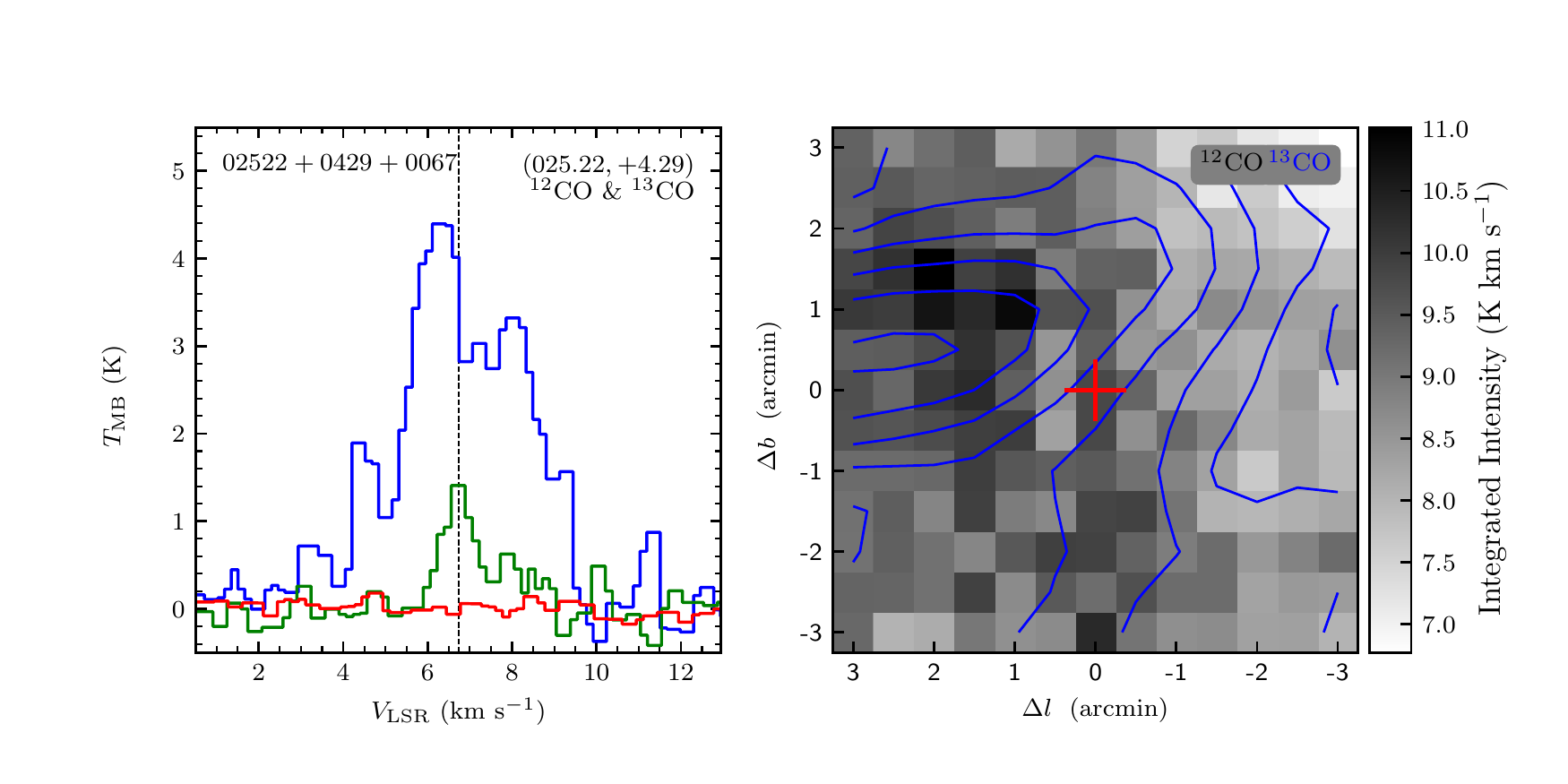}
\includegraphics[width=9.0cm,angle=0]{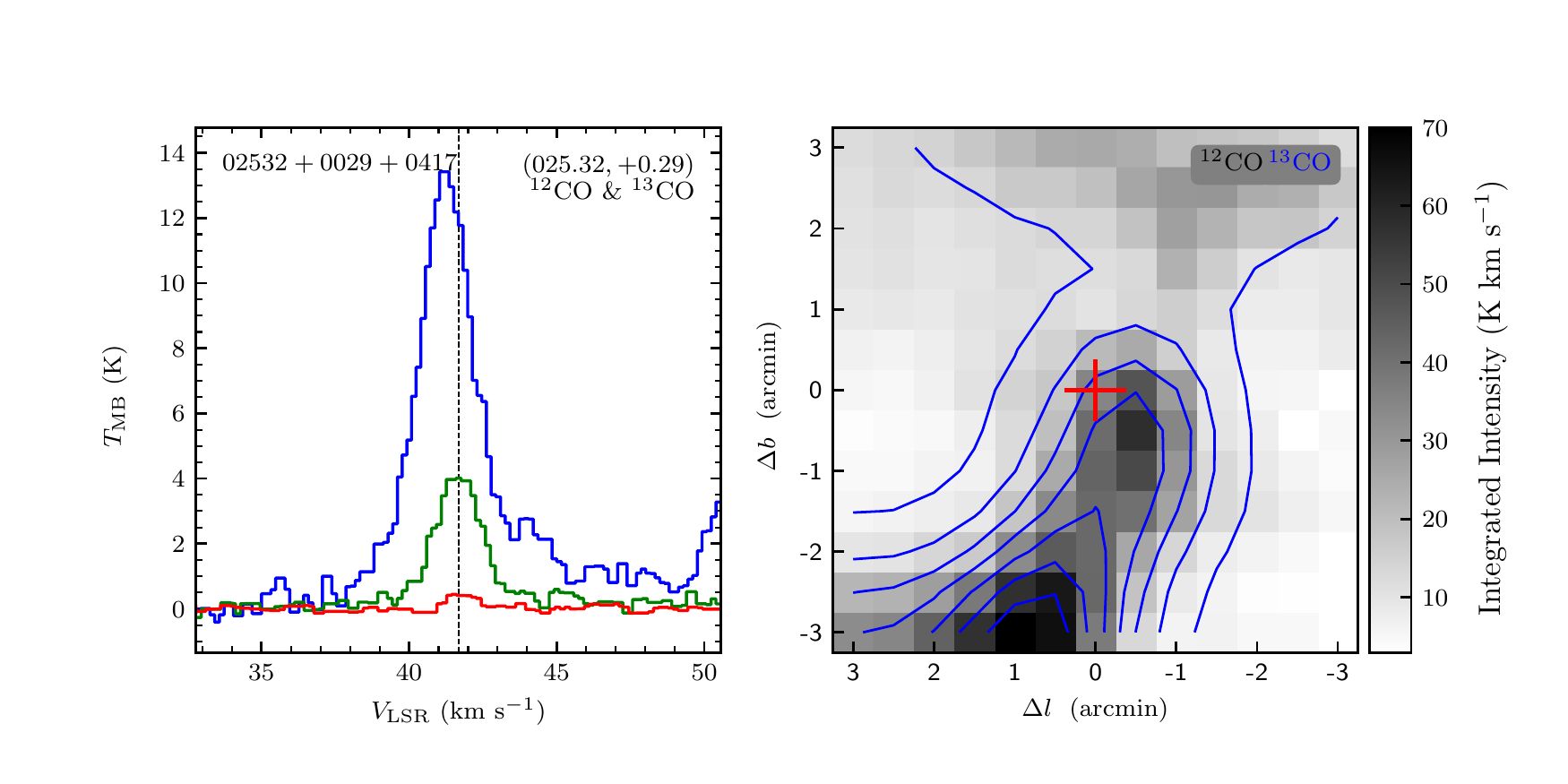}
\end{figure}
\clearpage

\begin{figure}
\includegraphics[width=9.0cm,angle=0]{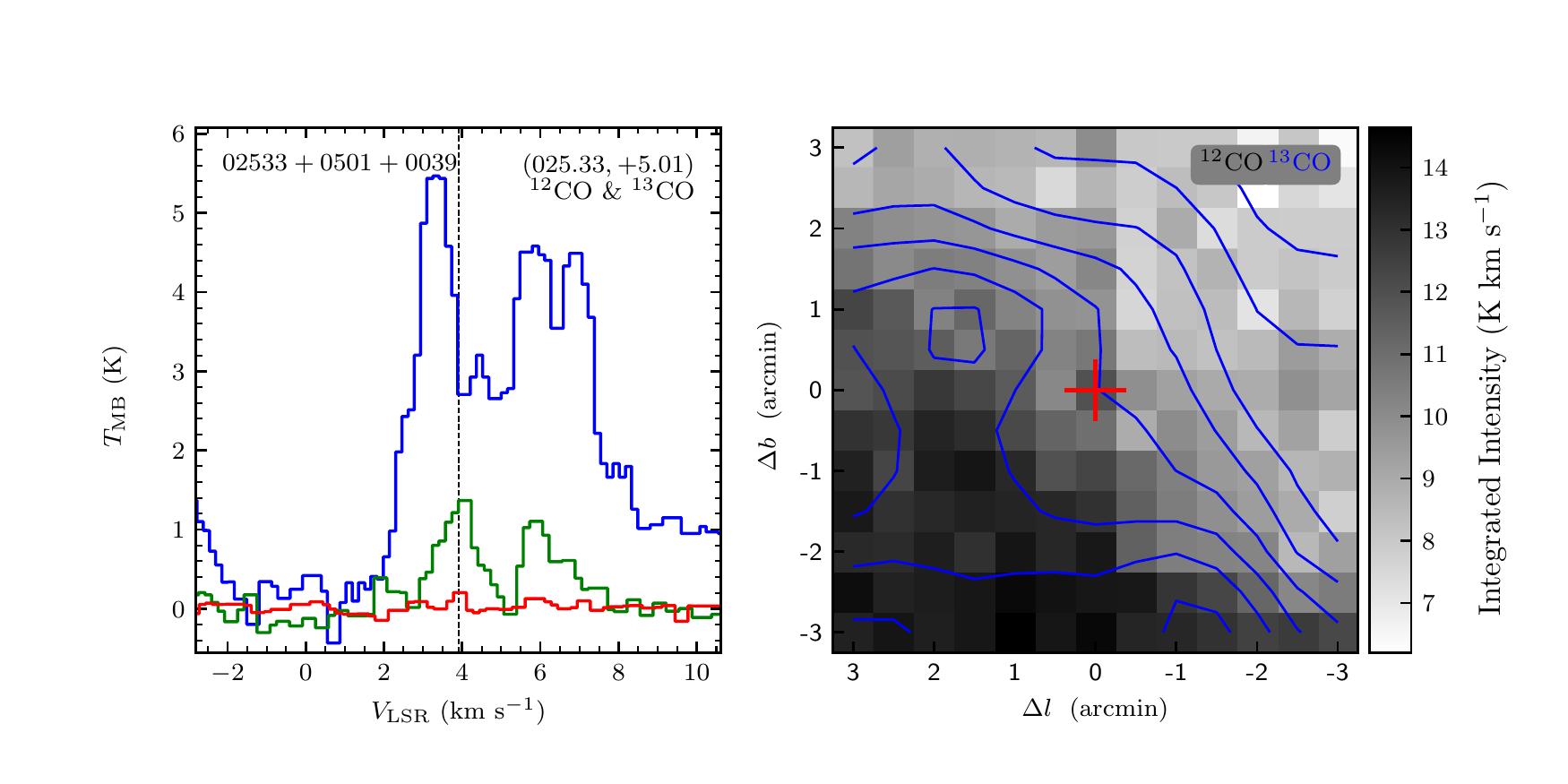}
\includegraphics[width=9.0cm,angle=0]{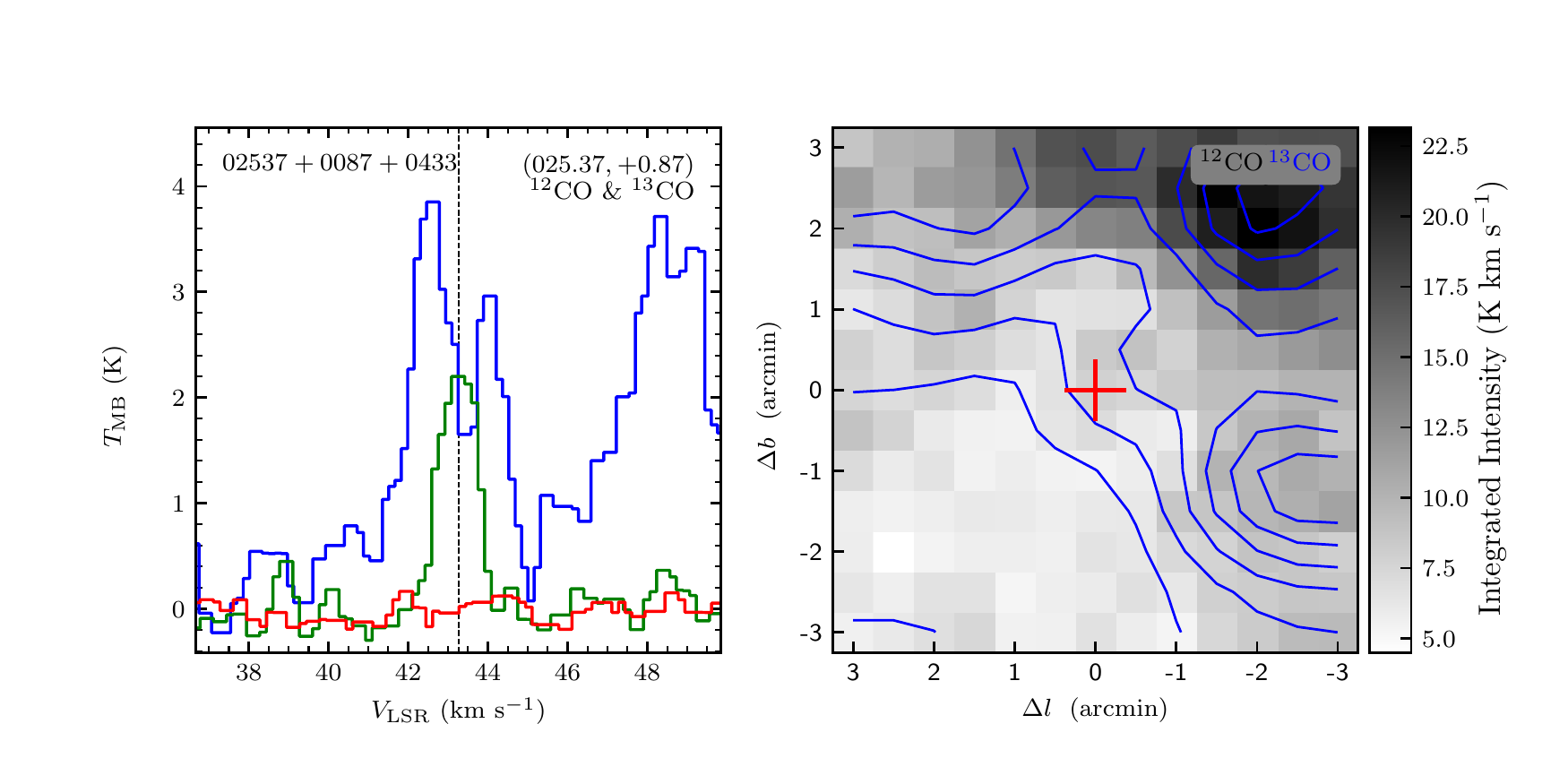}
\vspace{-0.5cm}

\includegraphics[width=9.0cm,angle=0]{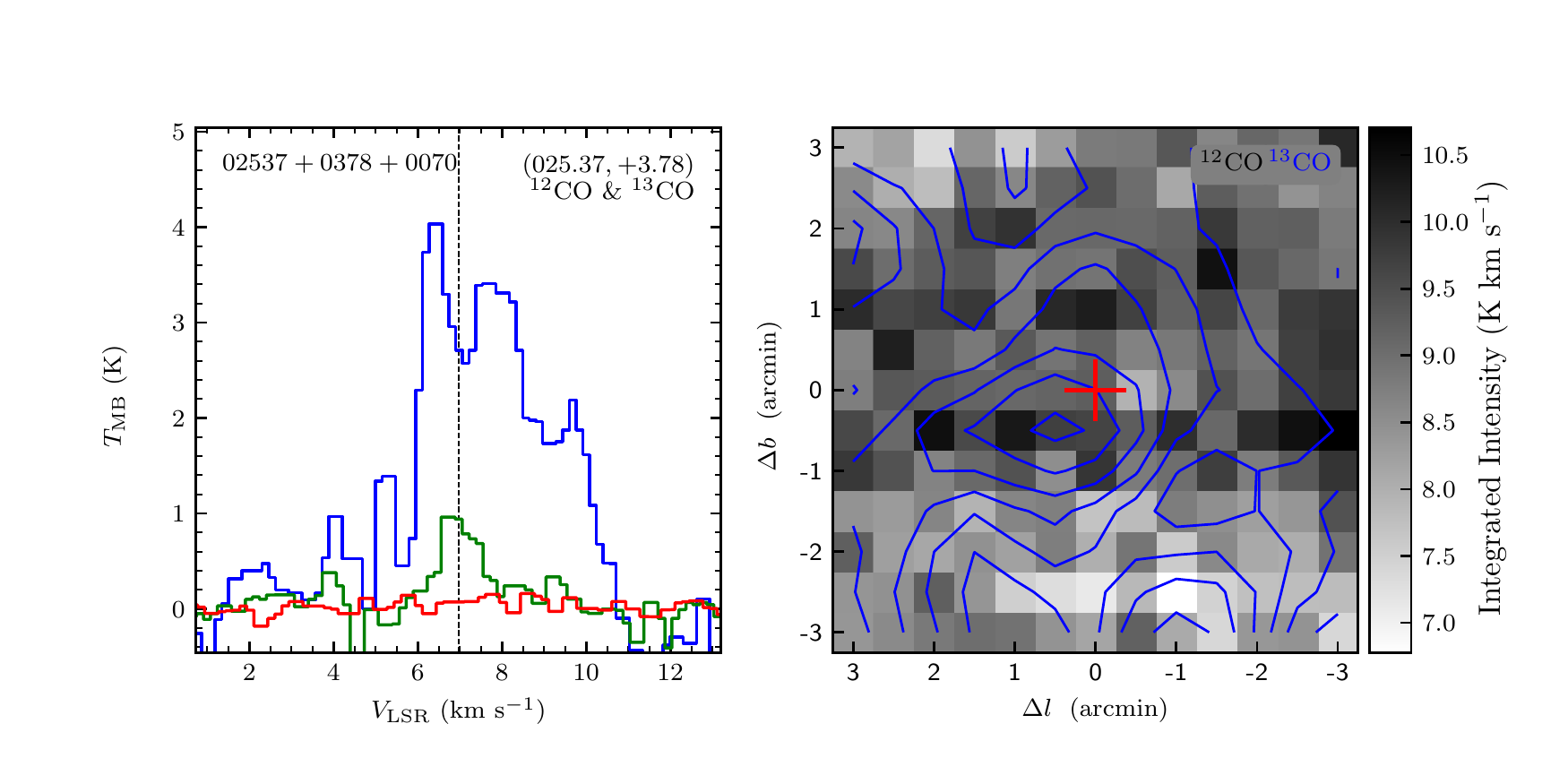}
\includegraphics[width=9.0cm,angle=0]{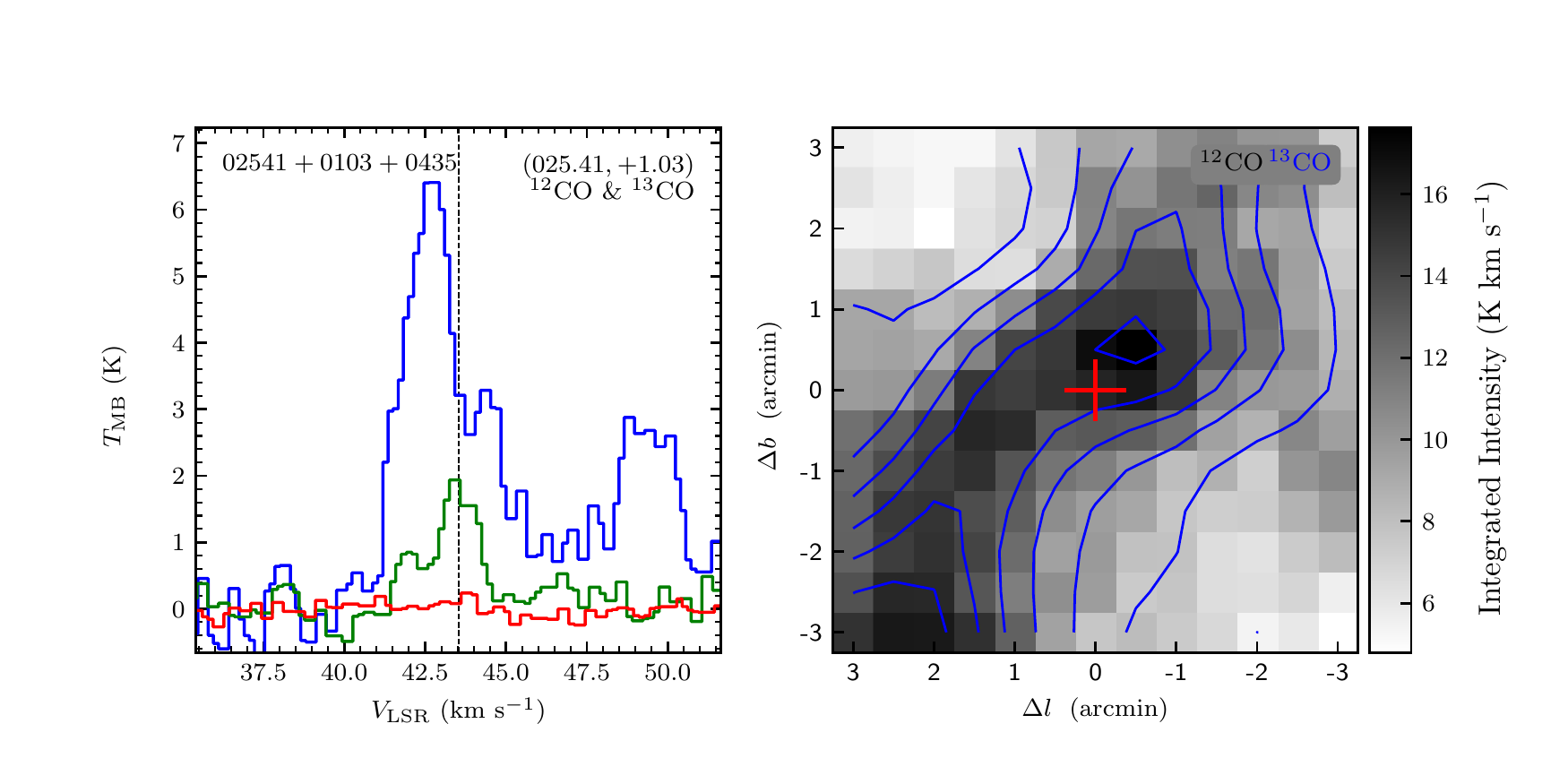}
\vspace{-0.5cm}

\includegraphics[width=9.0cm,angle=0]{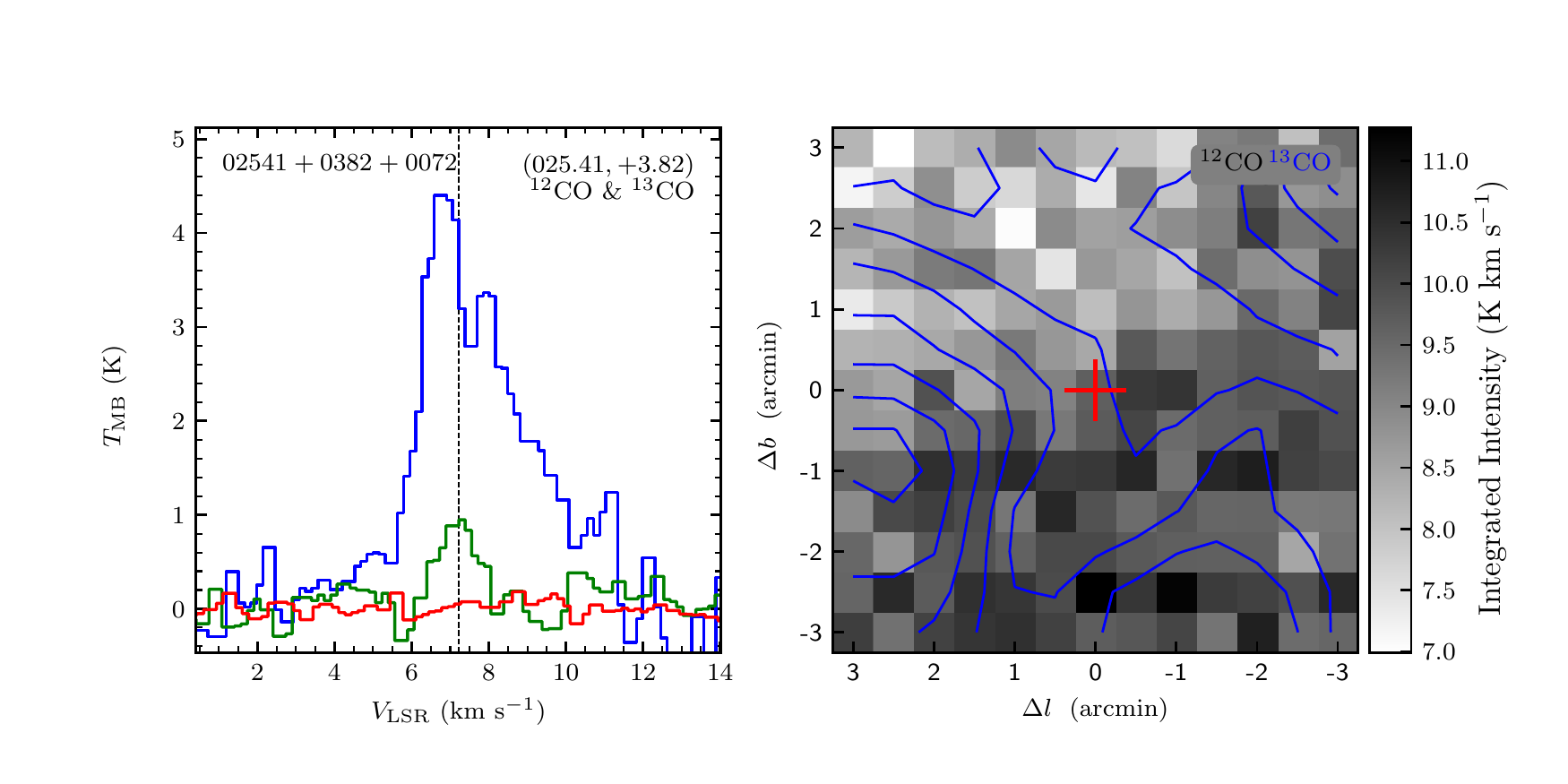}
\includegraphics[width=9.0cm,angle=0]{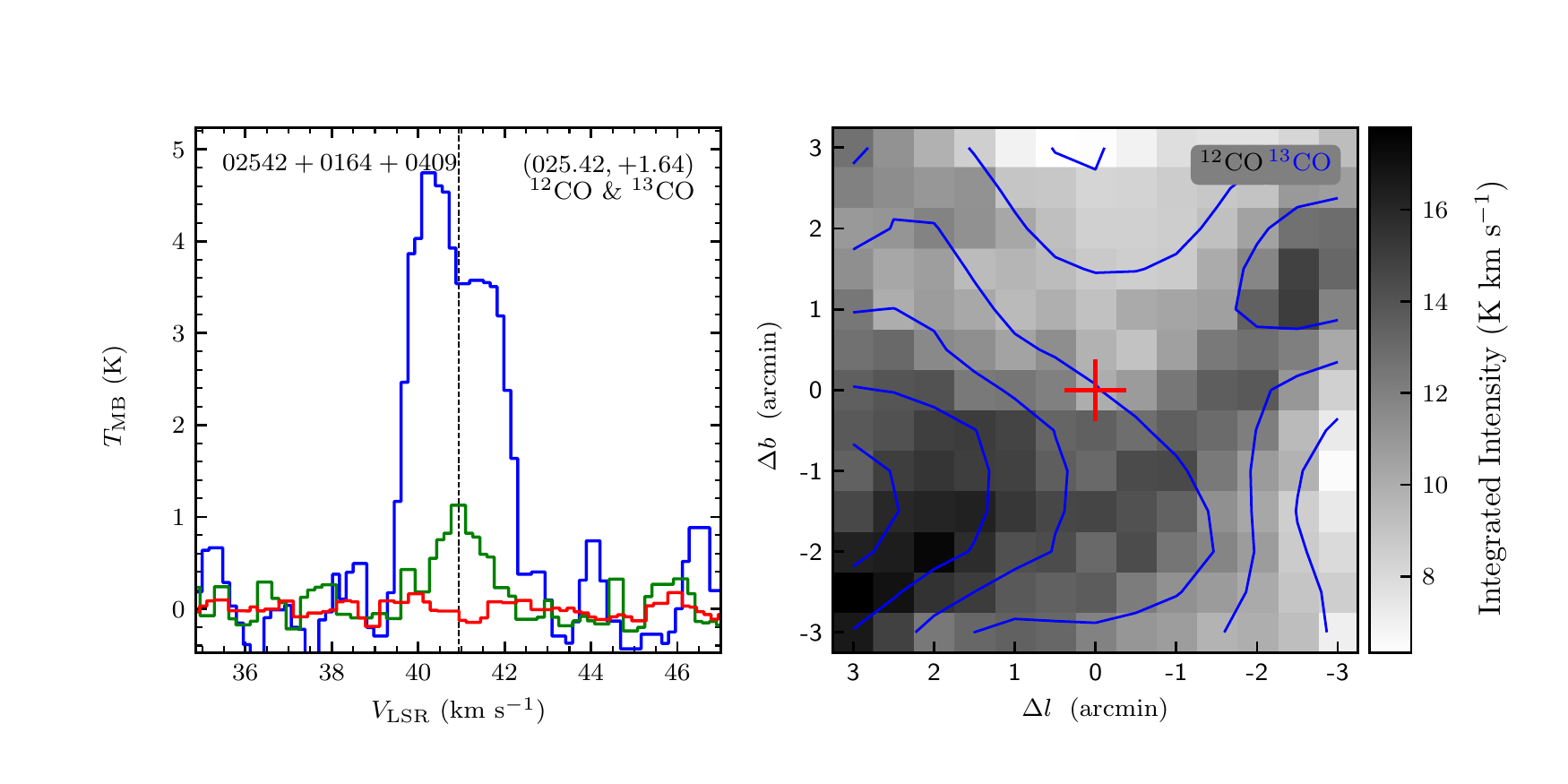}
\vspace{-0.5cm}

\includegraphics[width=9.0cm,angle=0]{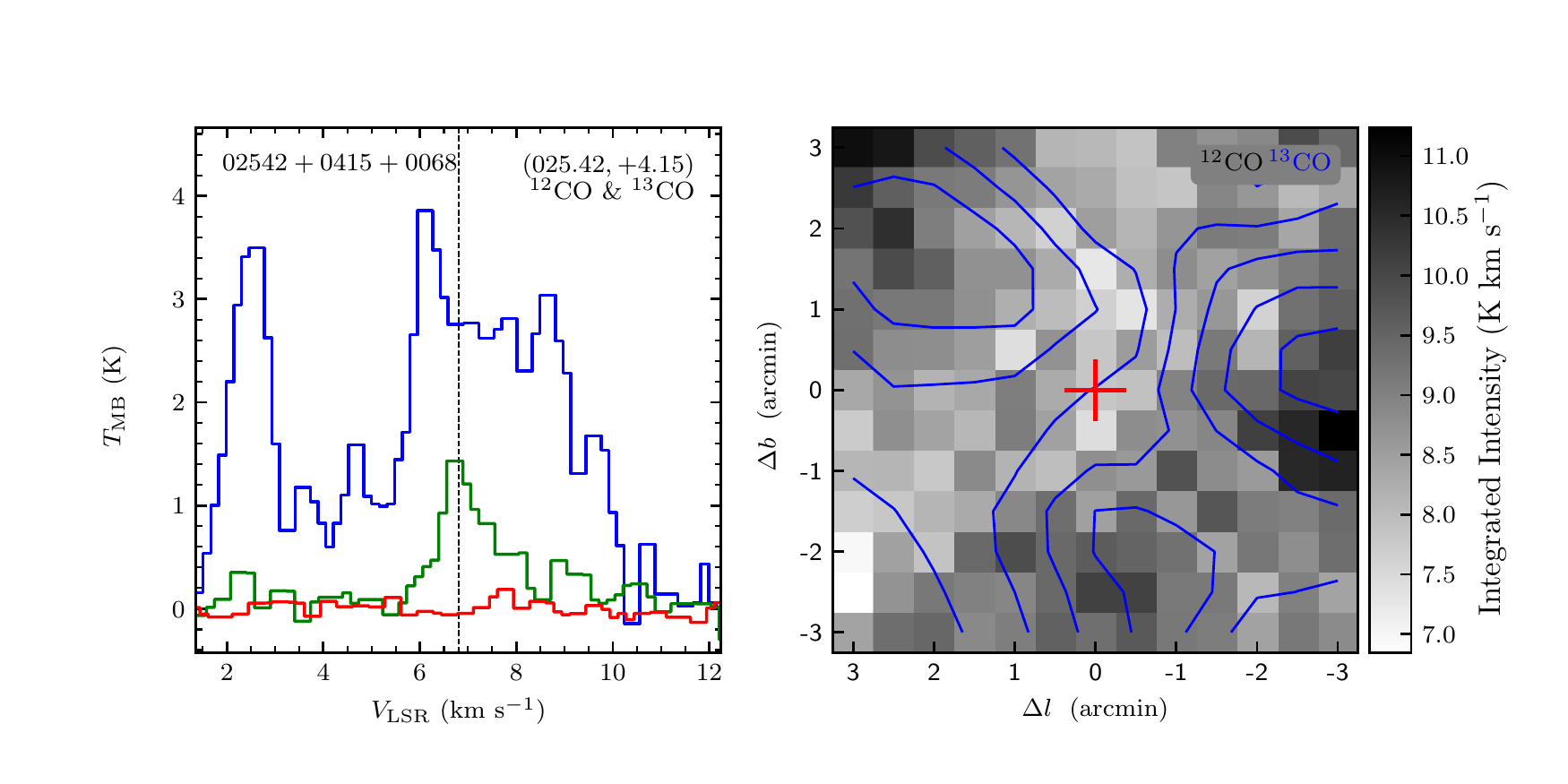}
\includegraphics[width=9.0cm,angle=0]{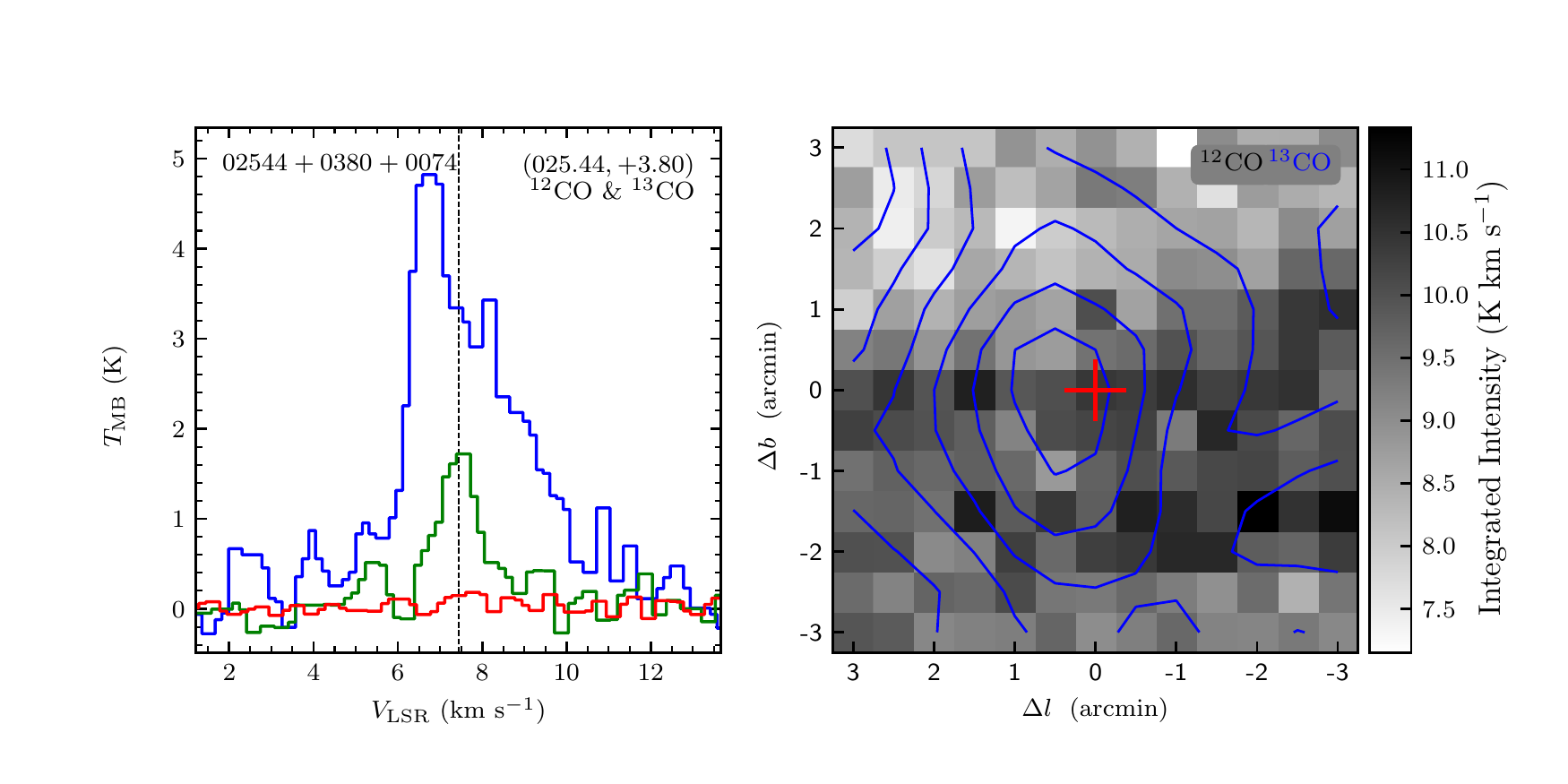}
\vspace{-0.5cm}

\includegraphics[width=9.0cm,angle=0]{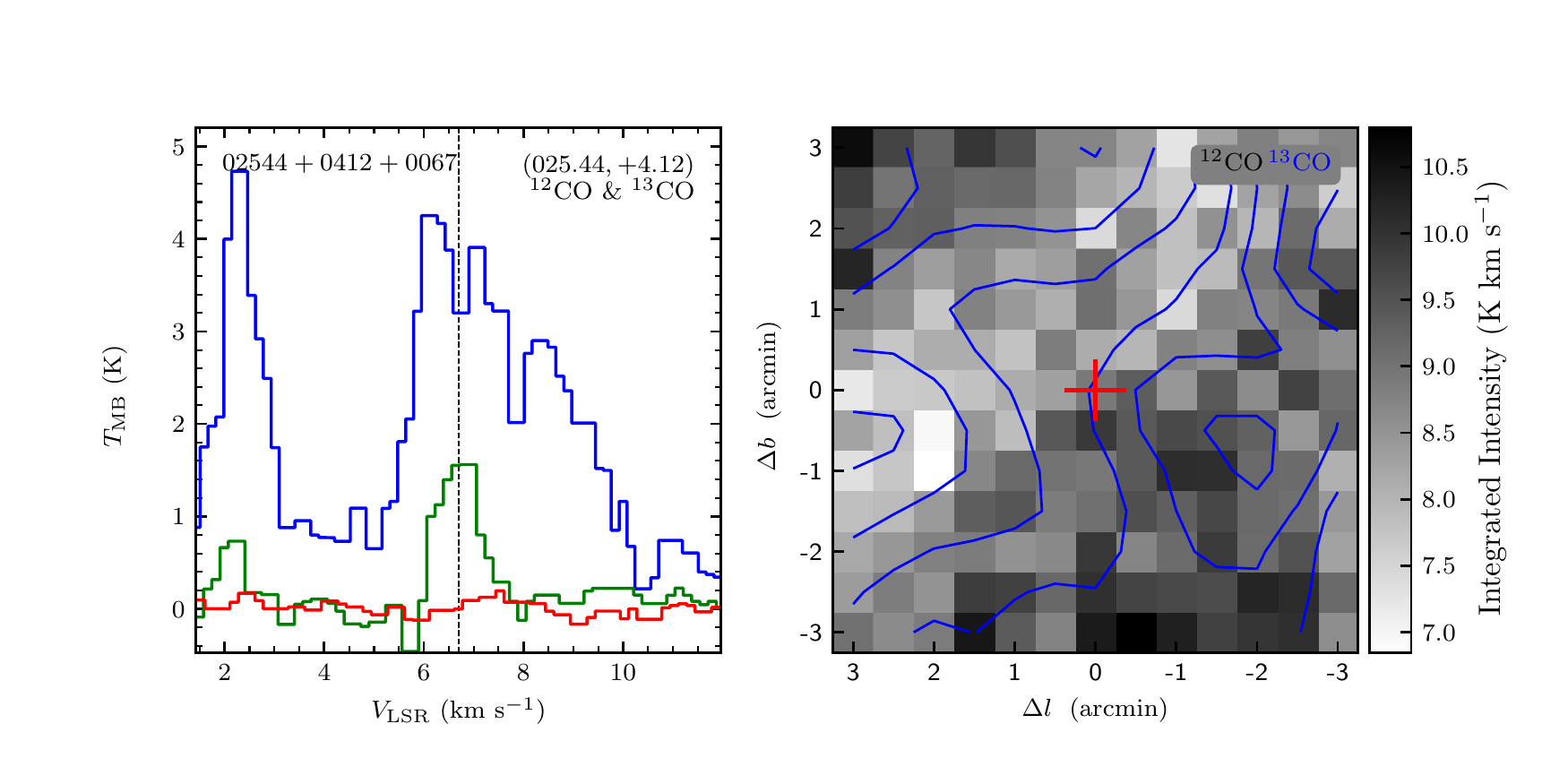}
\includegraphics[width=9.0cm,angle=0]{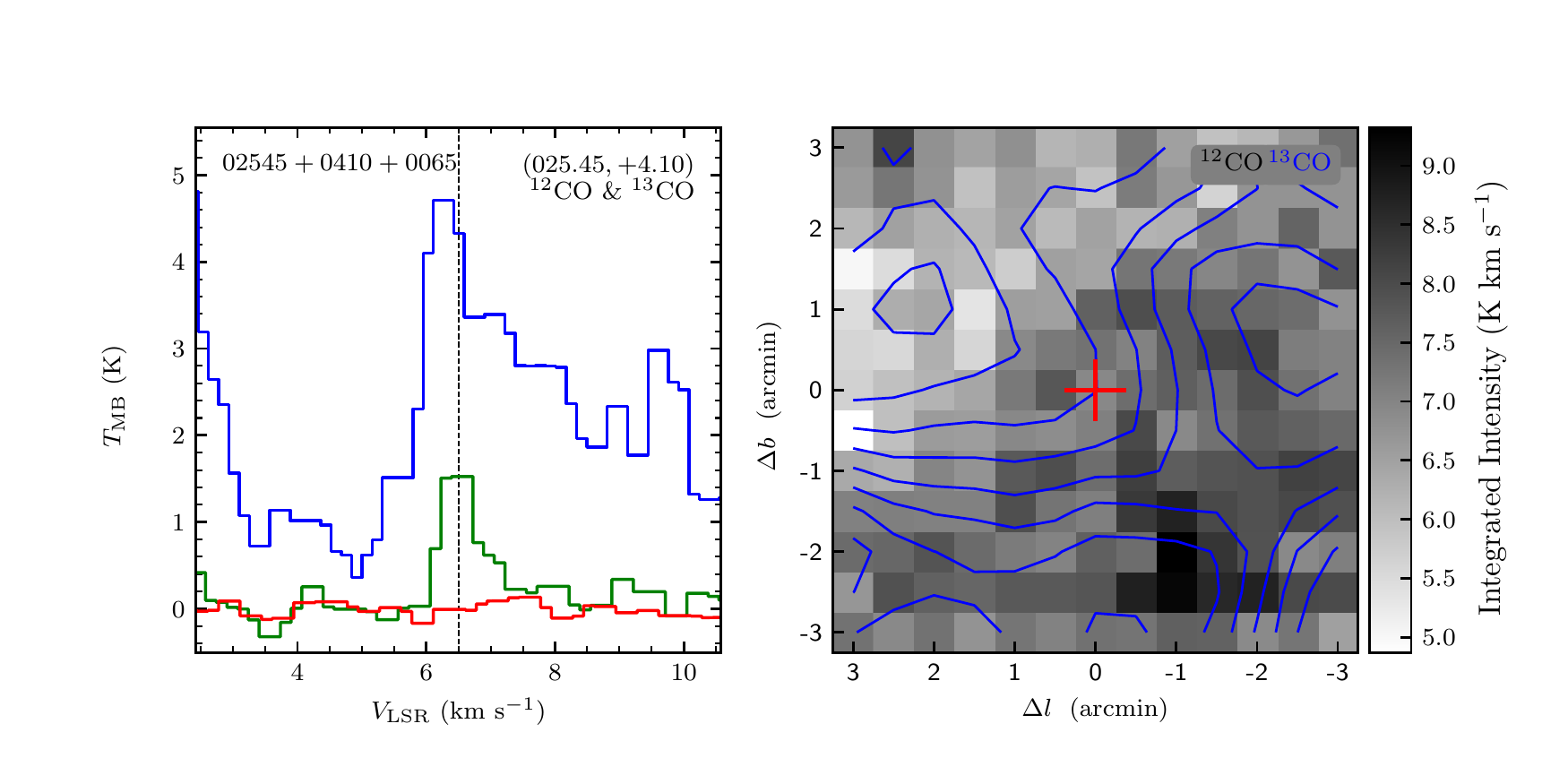}
\end{figure}
\clearpage

\begin{figure}
\includegraphics[width=9.0cm,angle=0]{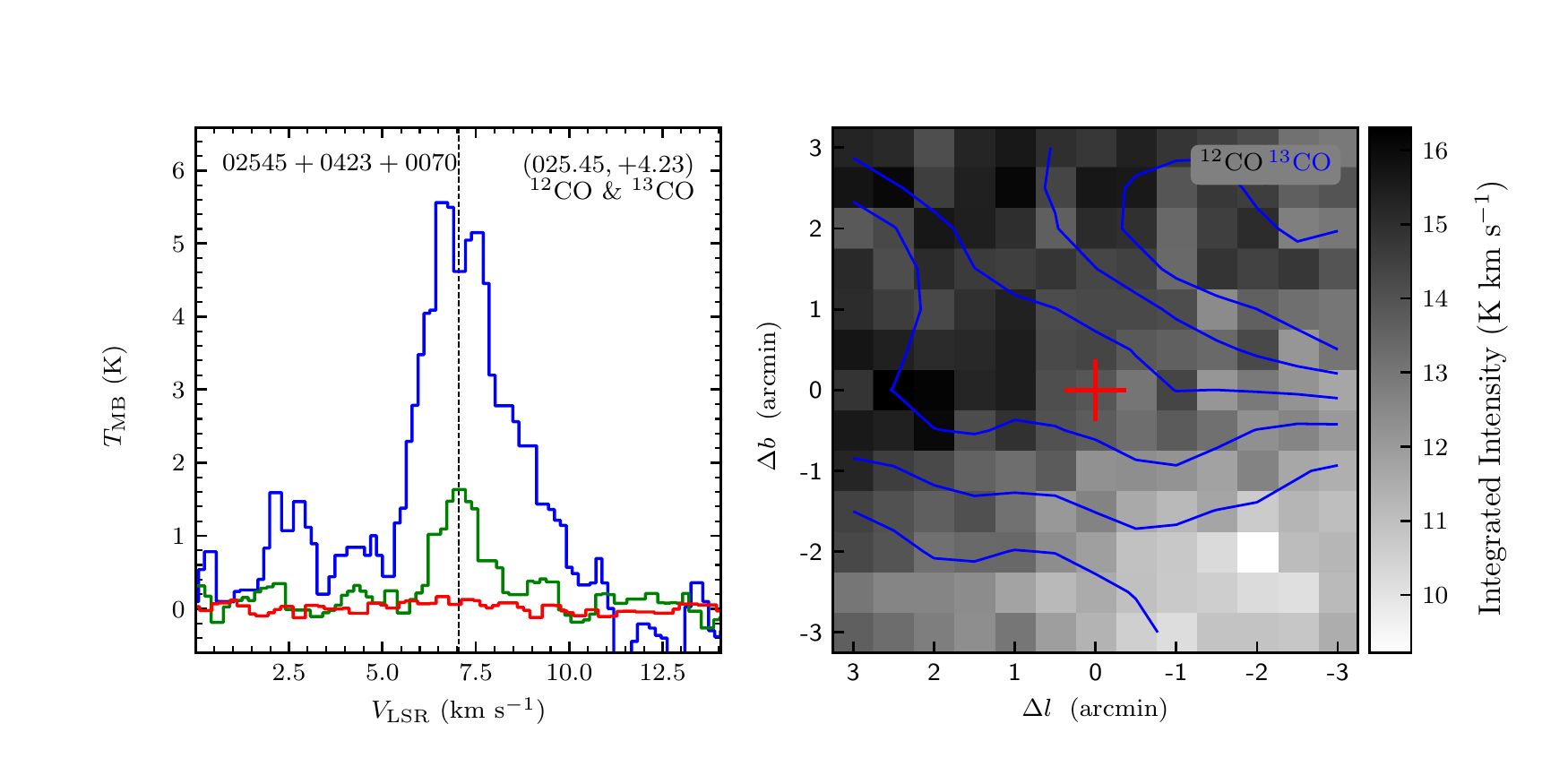}
\includegraphics[width=9.0cm,angle=0]{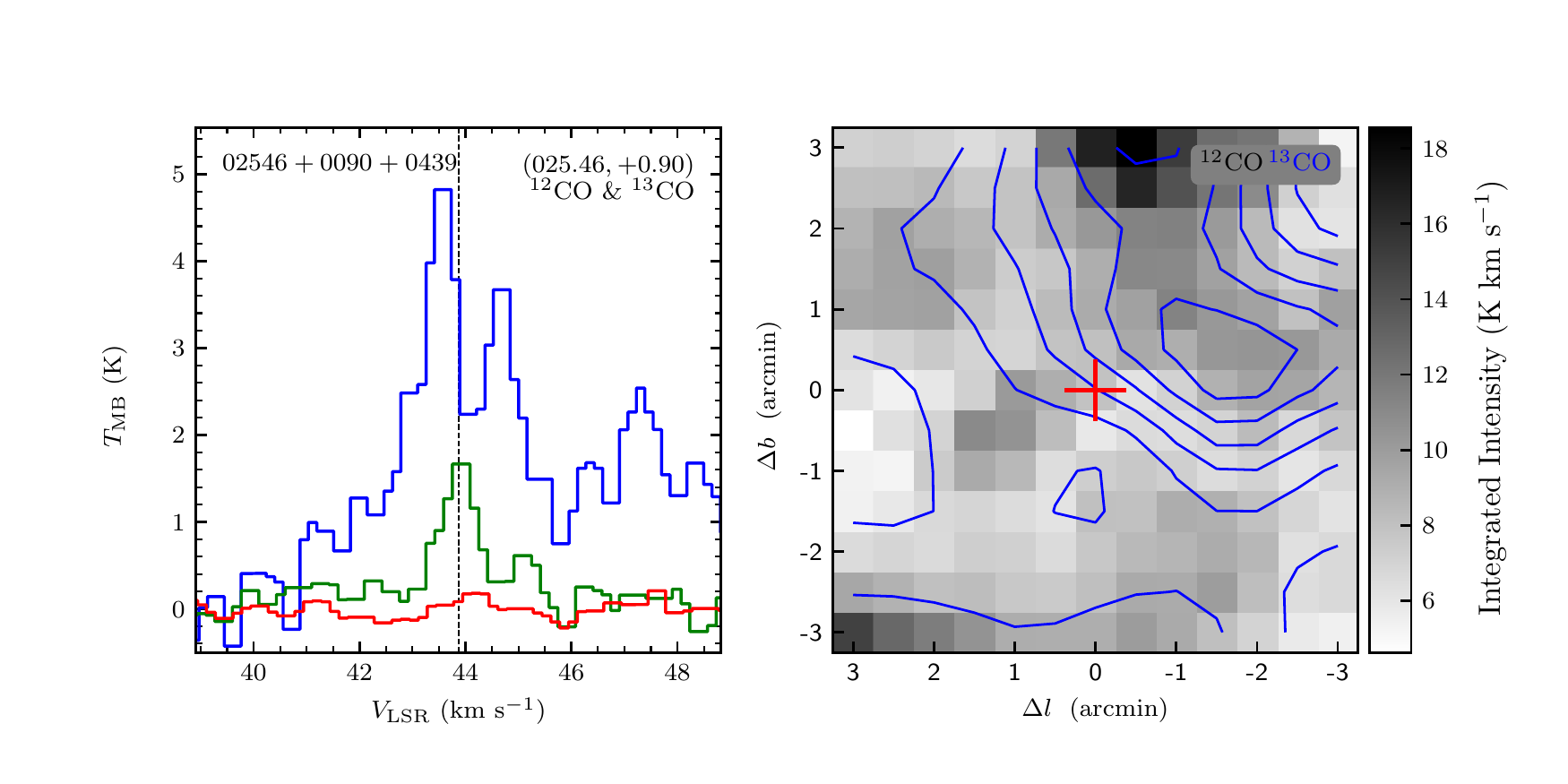}
\vspace{-0.5cm}

\includegraphics[width=9.0cm,angle=0]{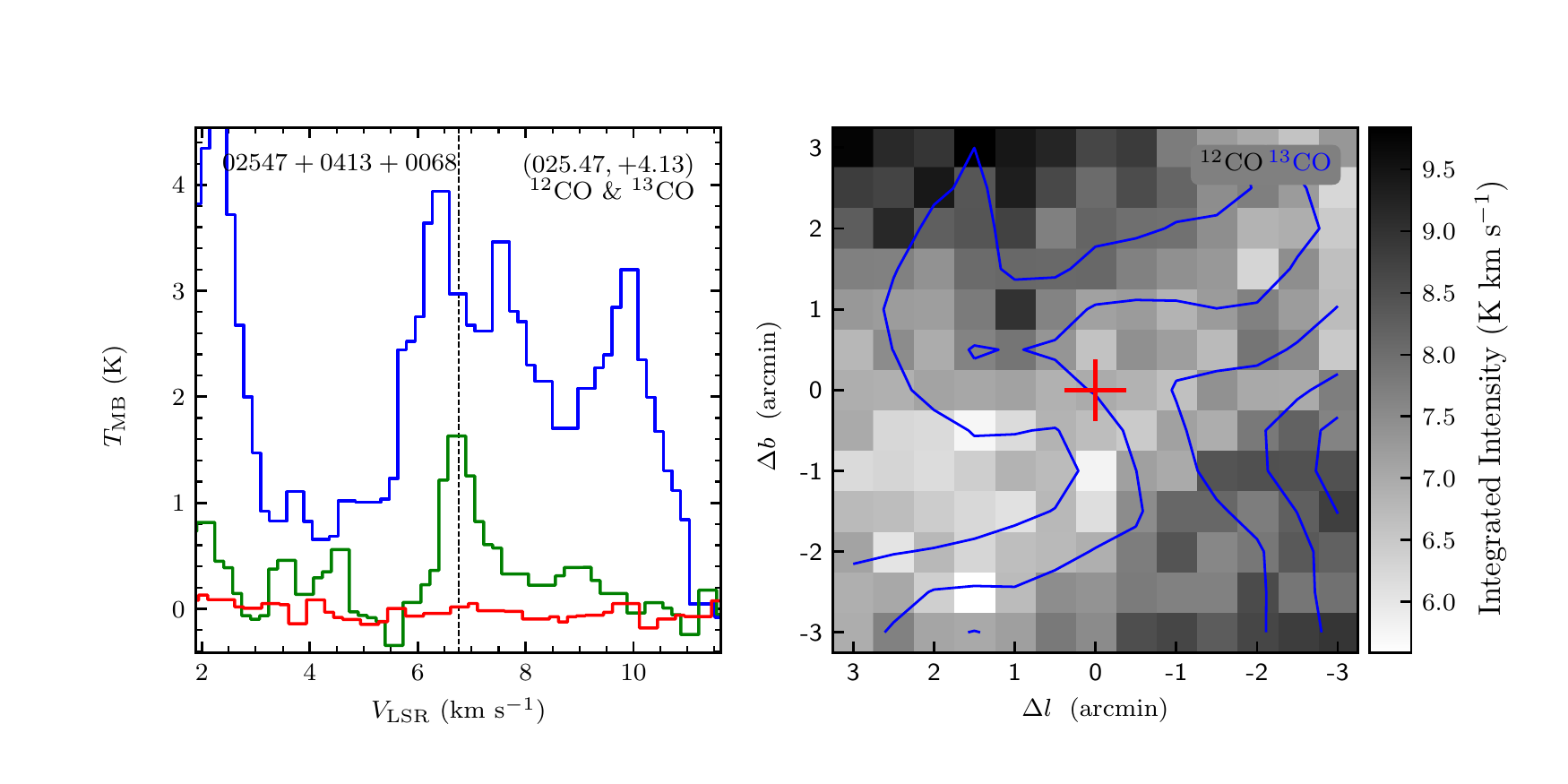}
\includegraphics[width=9.0cm,angle=0]{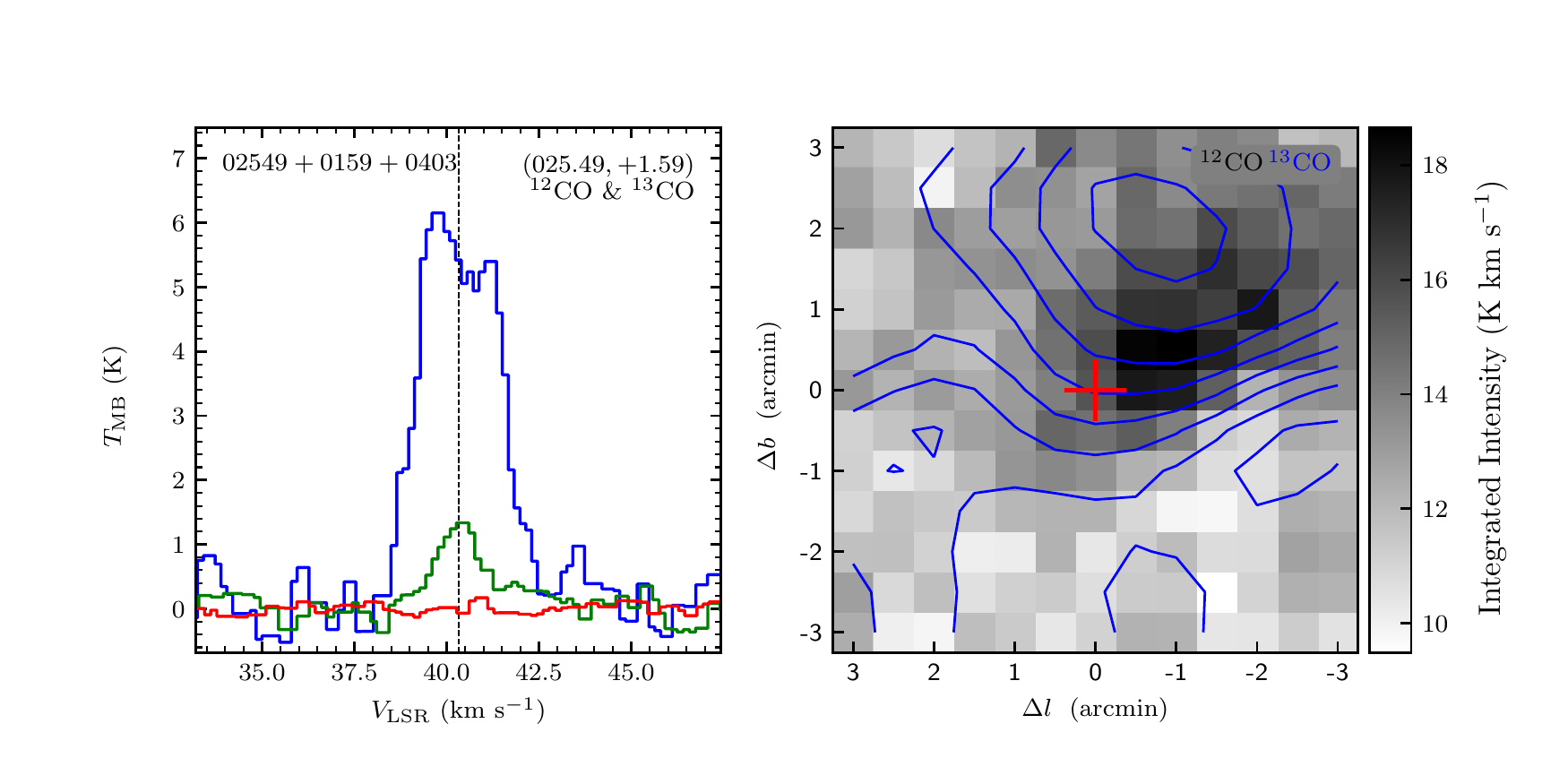}
\vspace{-0.5cm}

\includegraphics[width=9.0cm,angle=0]{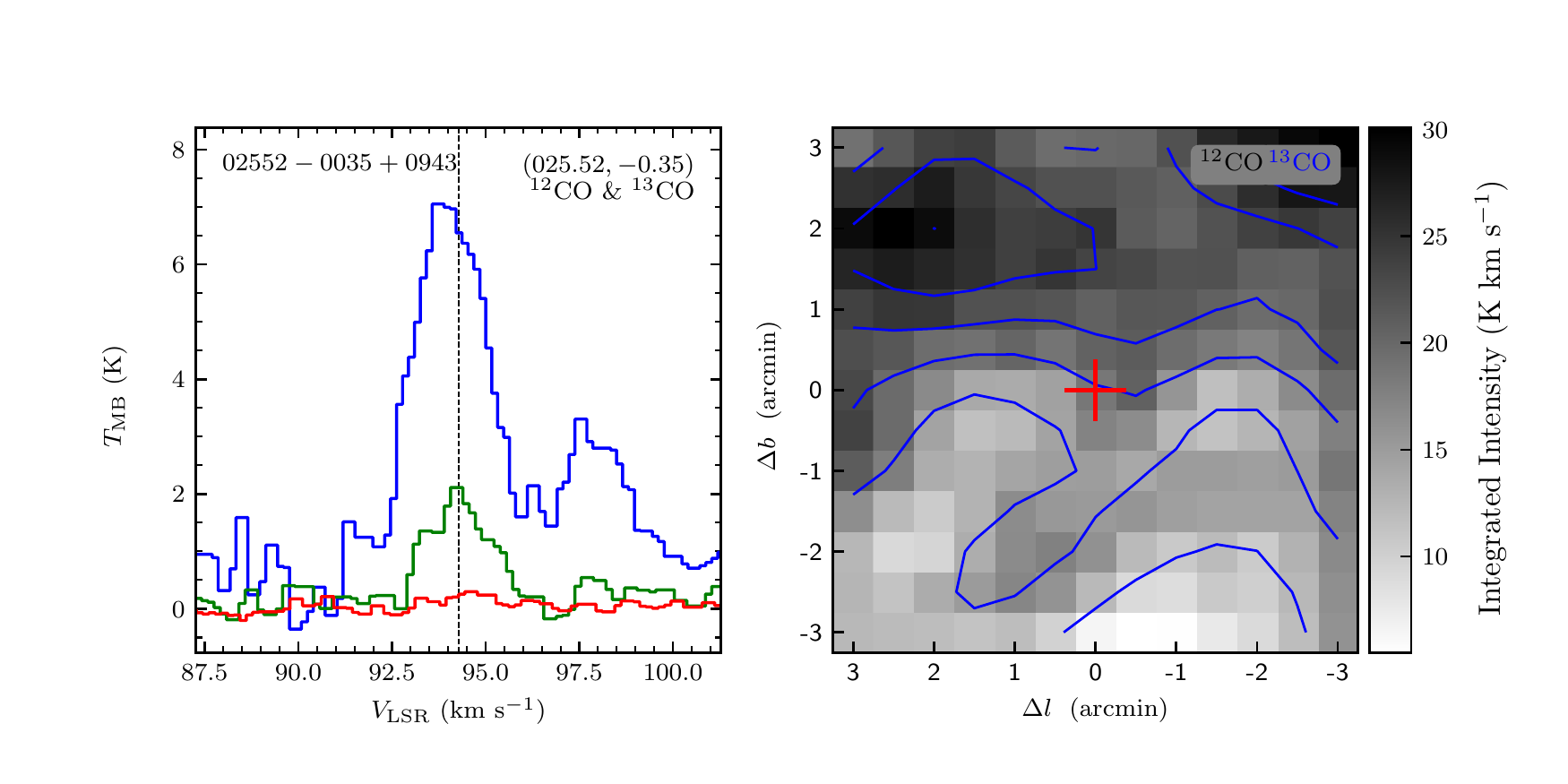}
\includegraphics[width=9.0cm,angle=0]{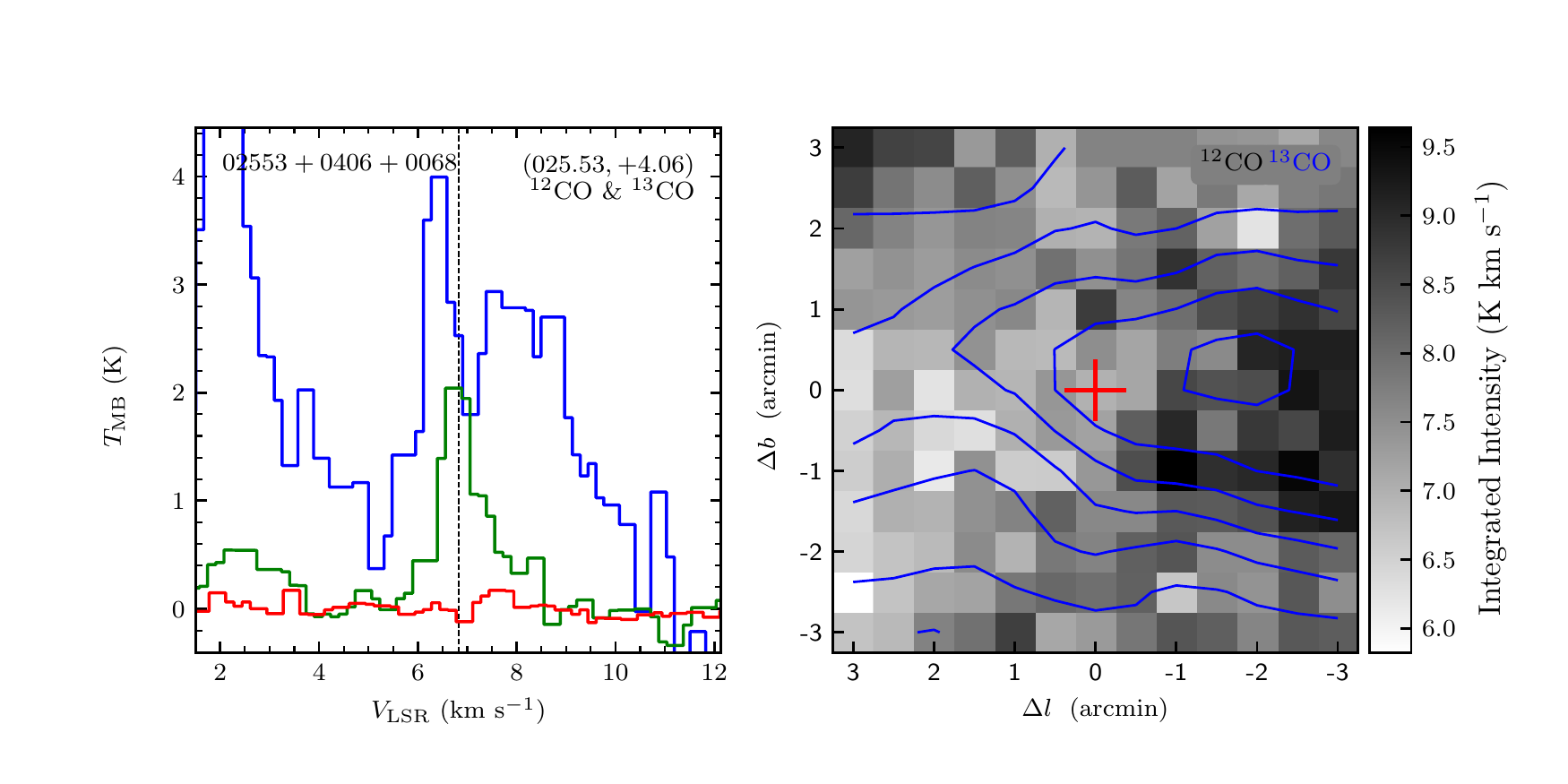}
\vspace{-0.5cm}

\includegraphics[width=9.0cm,angle=0]{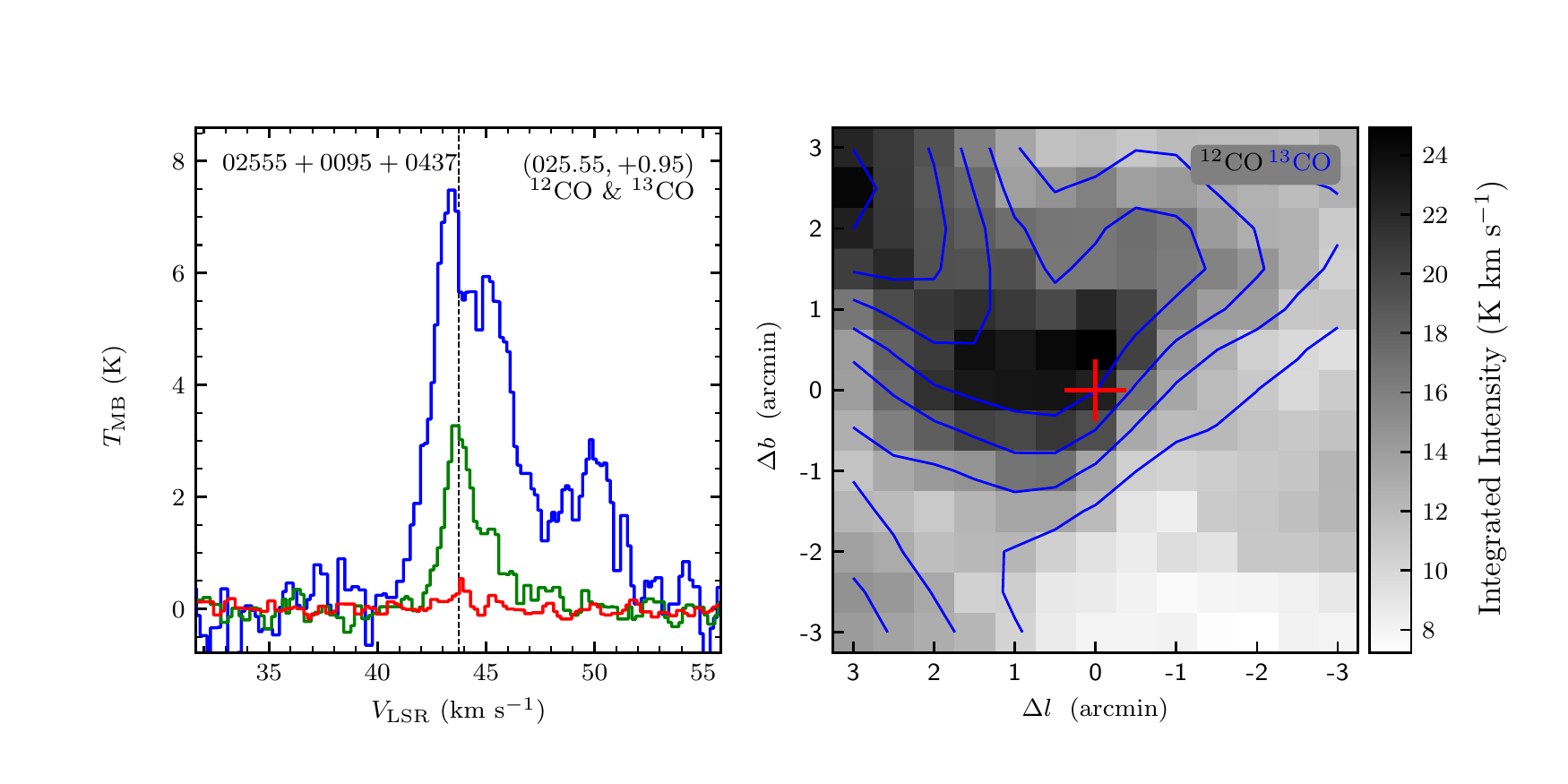}
\includegraphics[width=9.0cm,angle=0]{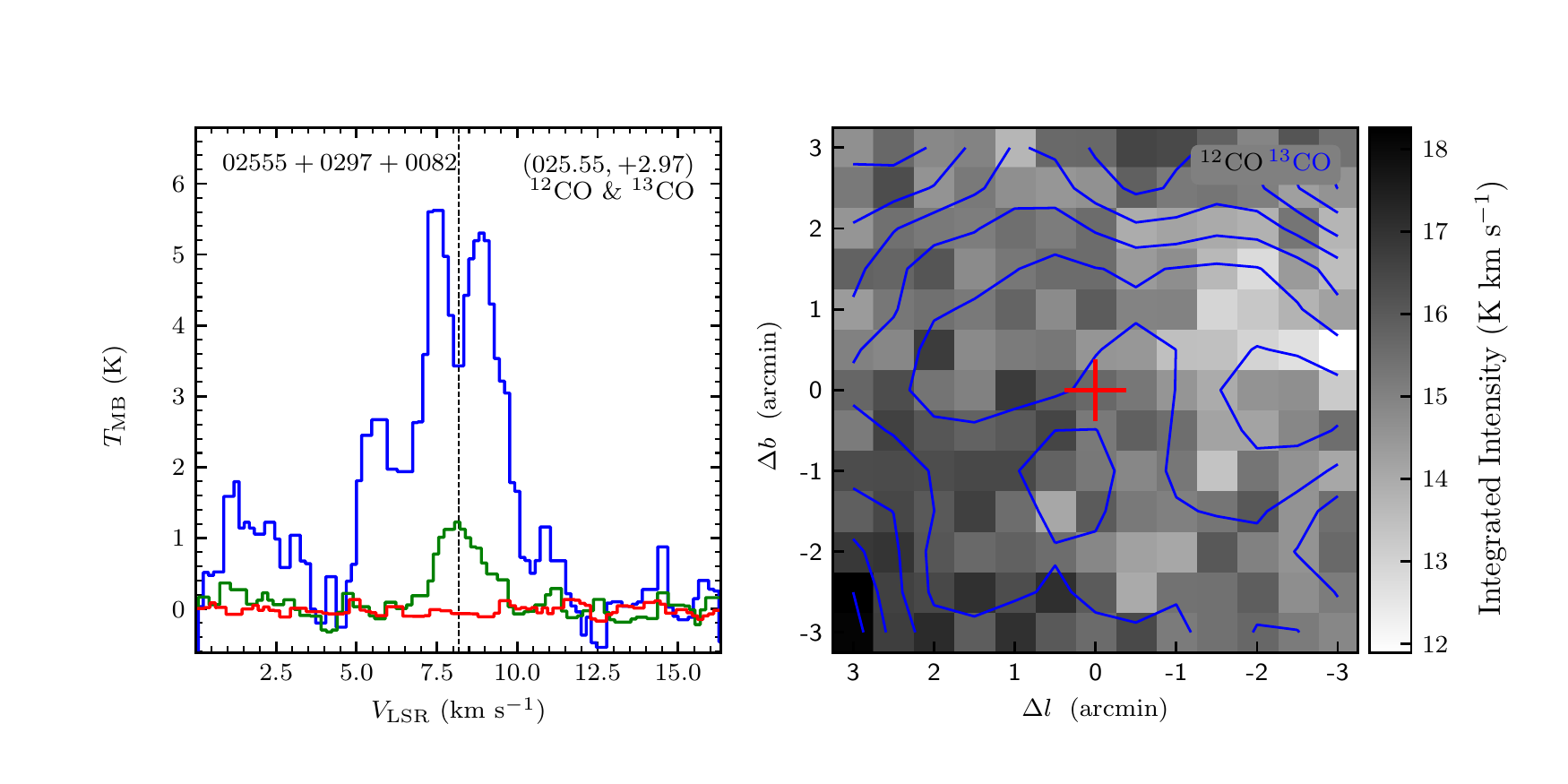}
\vspace{-0.5cm}

\includegraphics[width=9.0cm,angle=0]{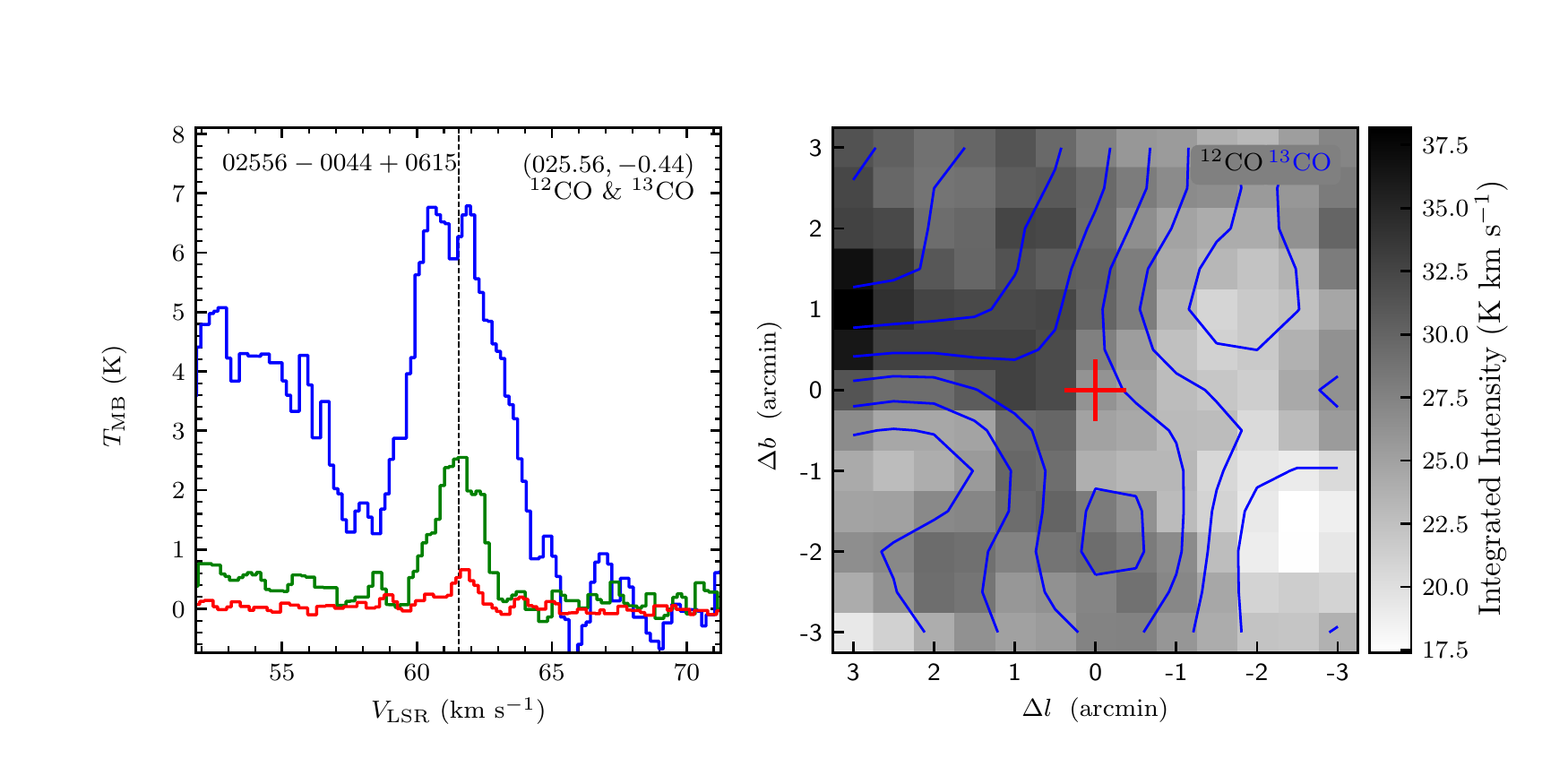}
\includegraphics[width=9.0cm,angle=0]{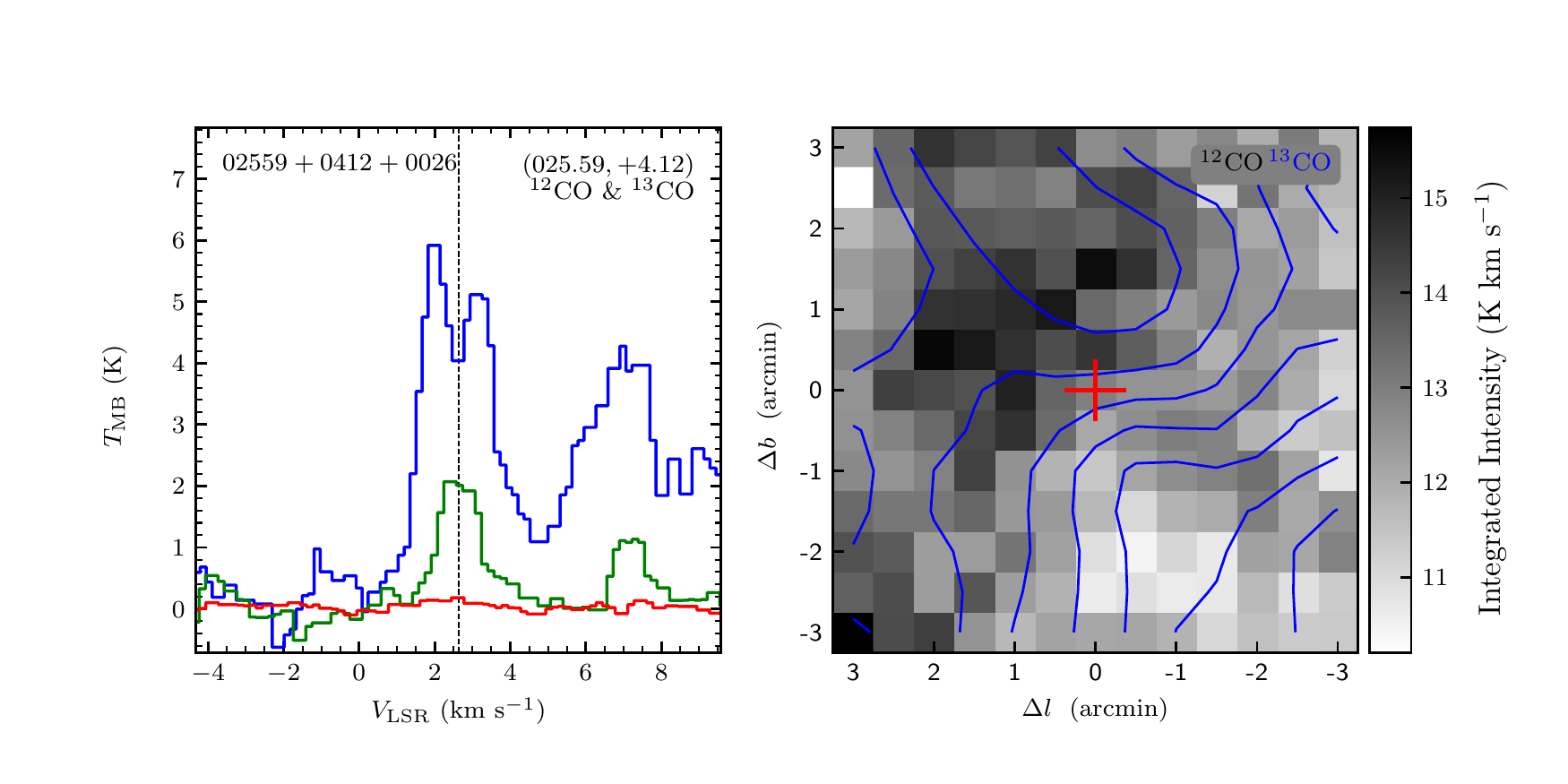}
\end{figure}
\clearpage

\begin{figure}
\includegraphics[width=9.0cm,angle=0]{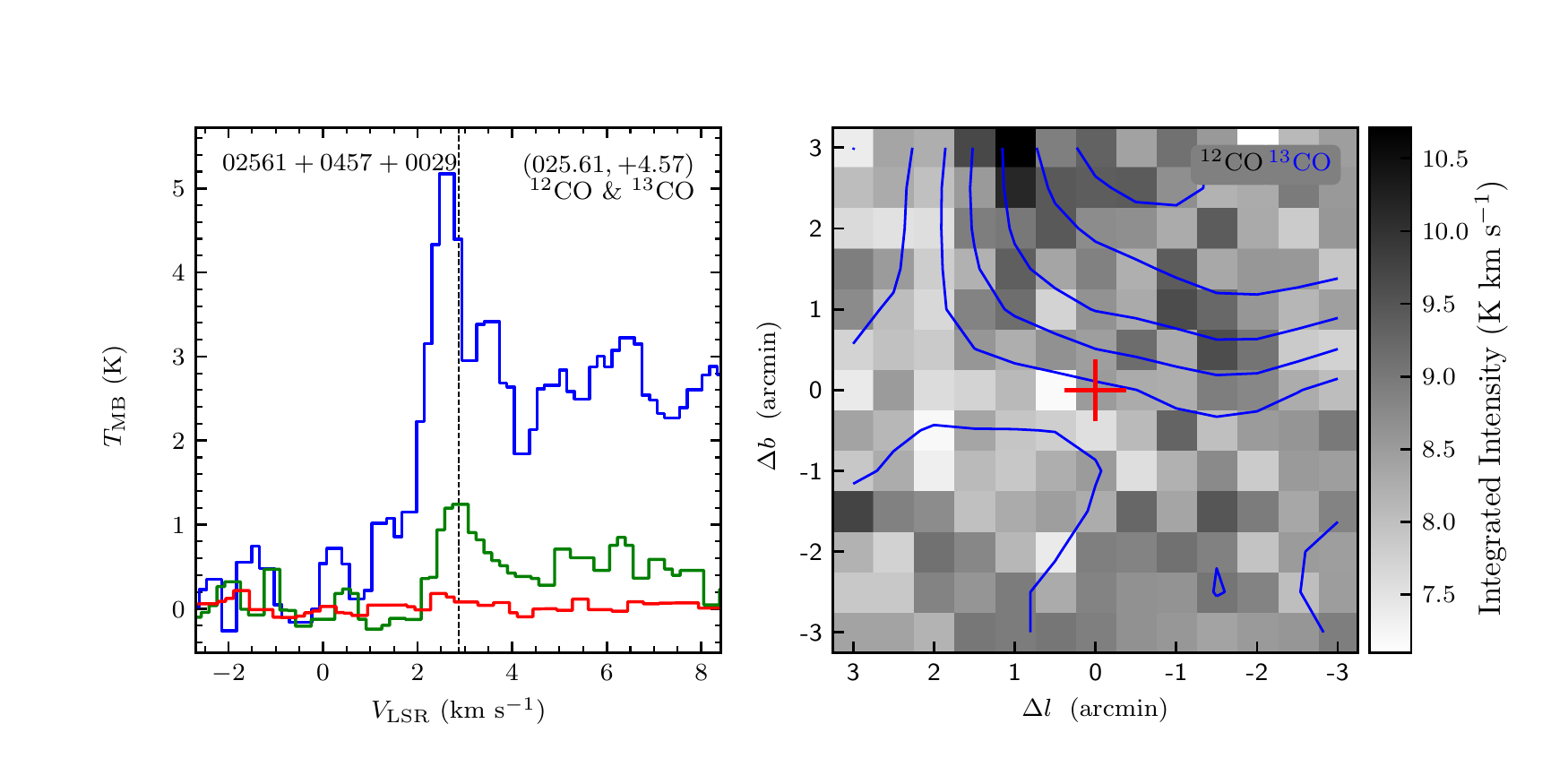}
\includegraphics[width=9.0cm,angle=0]{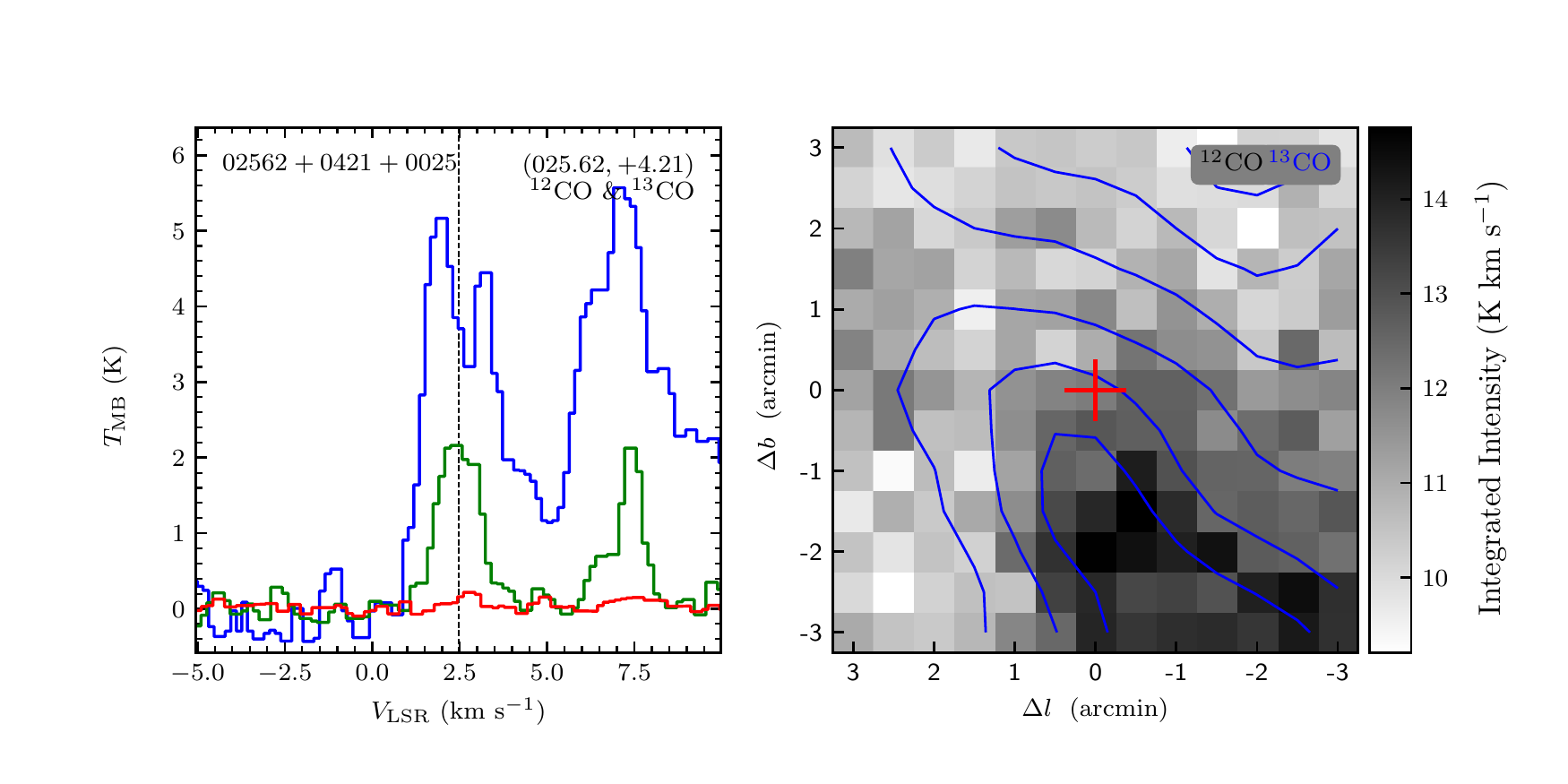}
\vspace{-0.5cm}

\includegraphics[width=9.0cm,angle=0]{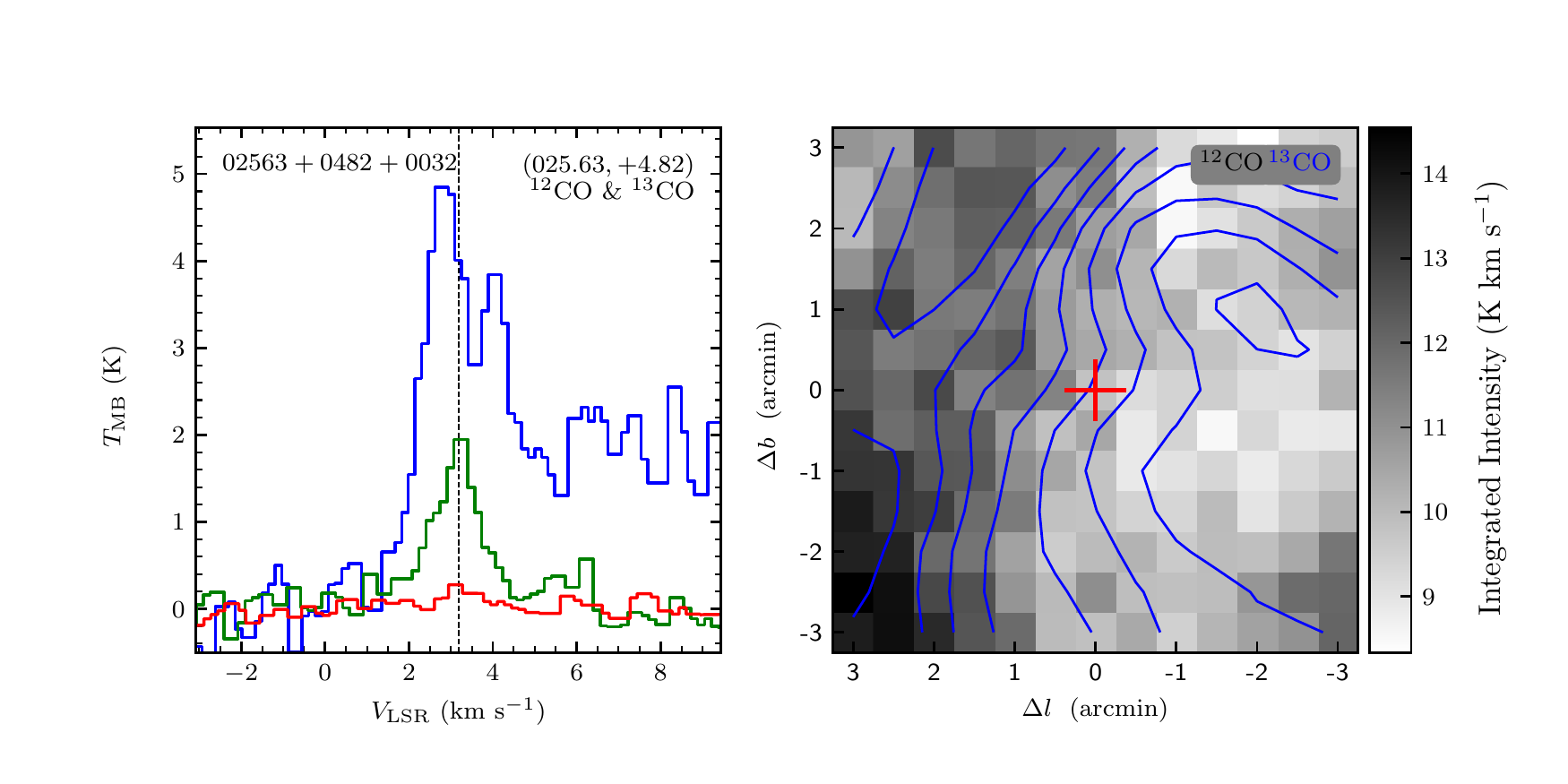}
\includegraphics[width=9.0cm,angle=0]{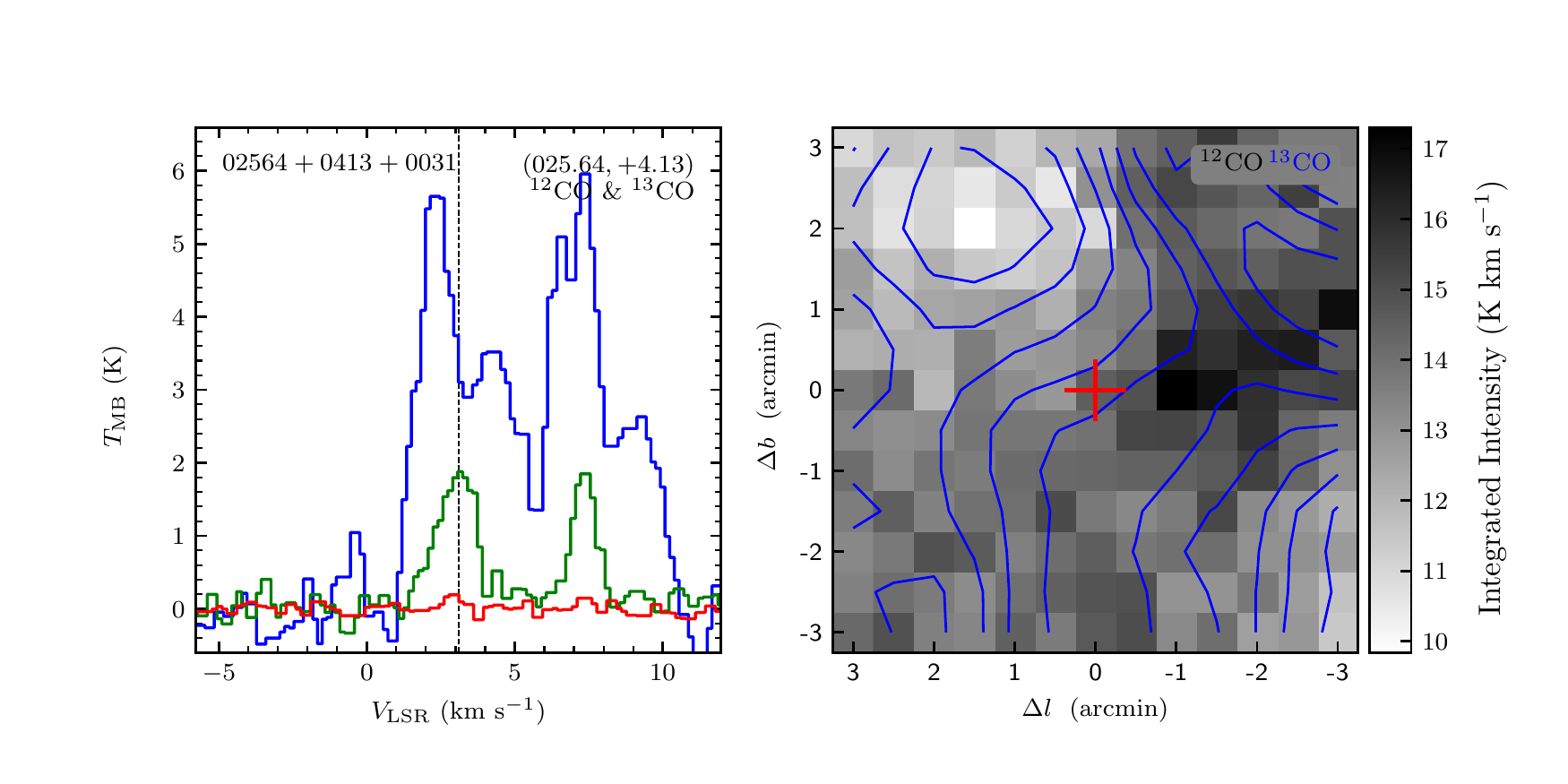}
\vspace{-0.5cm}

\includegraphics[width=9.0cm,angle=0]{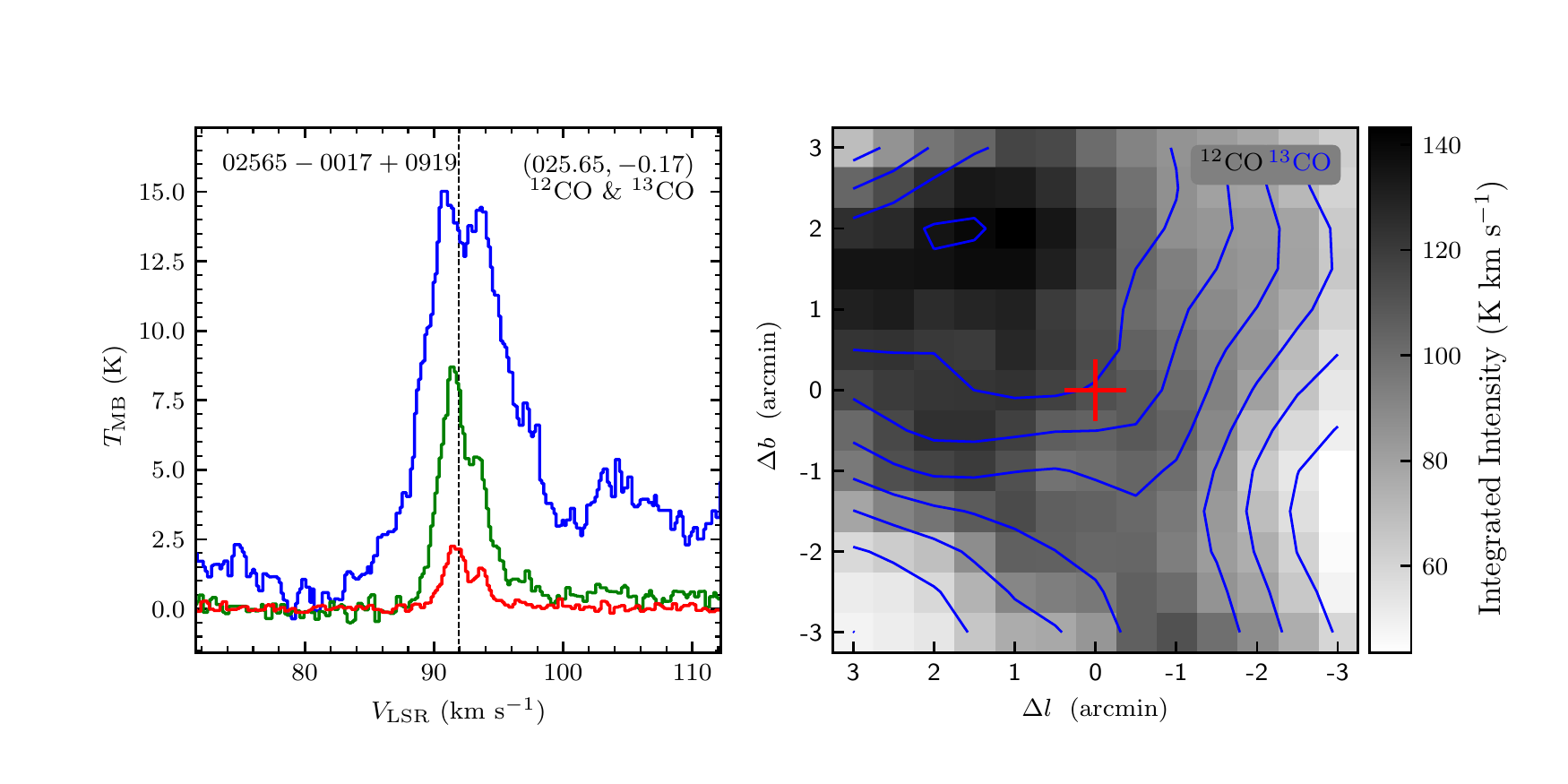}
\includegraphics[width=9.0cm,angle=0]{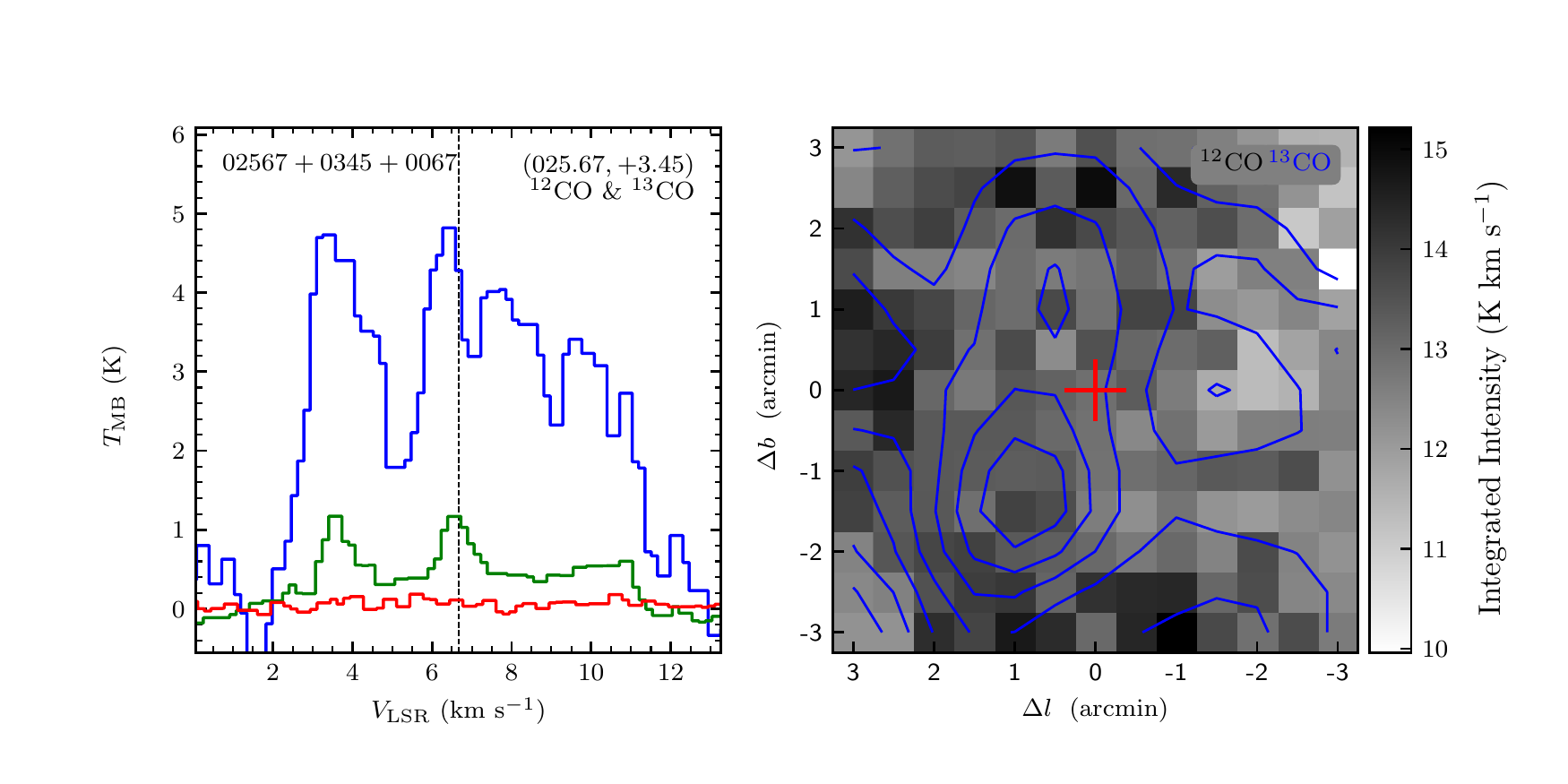}
\vspace{-0.5cm}

\includegraphics[width=9.0cm,angle=0]{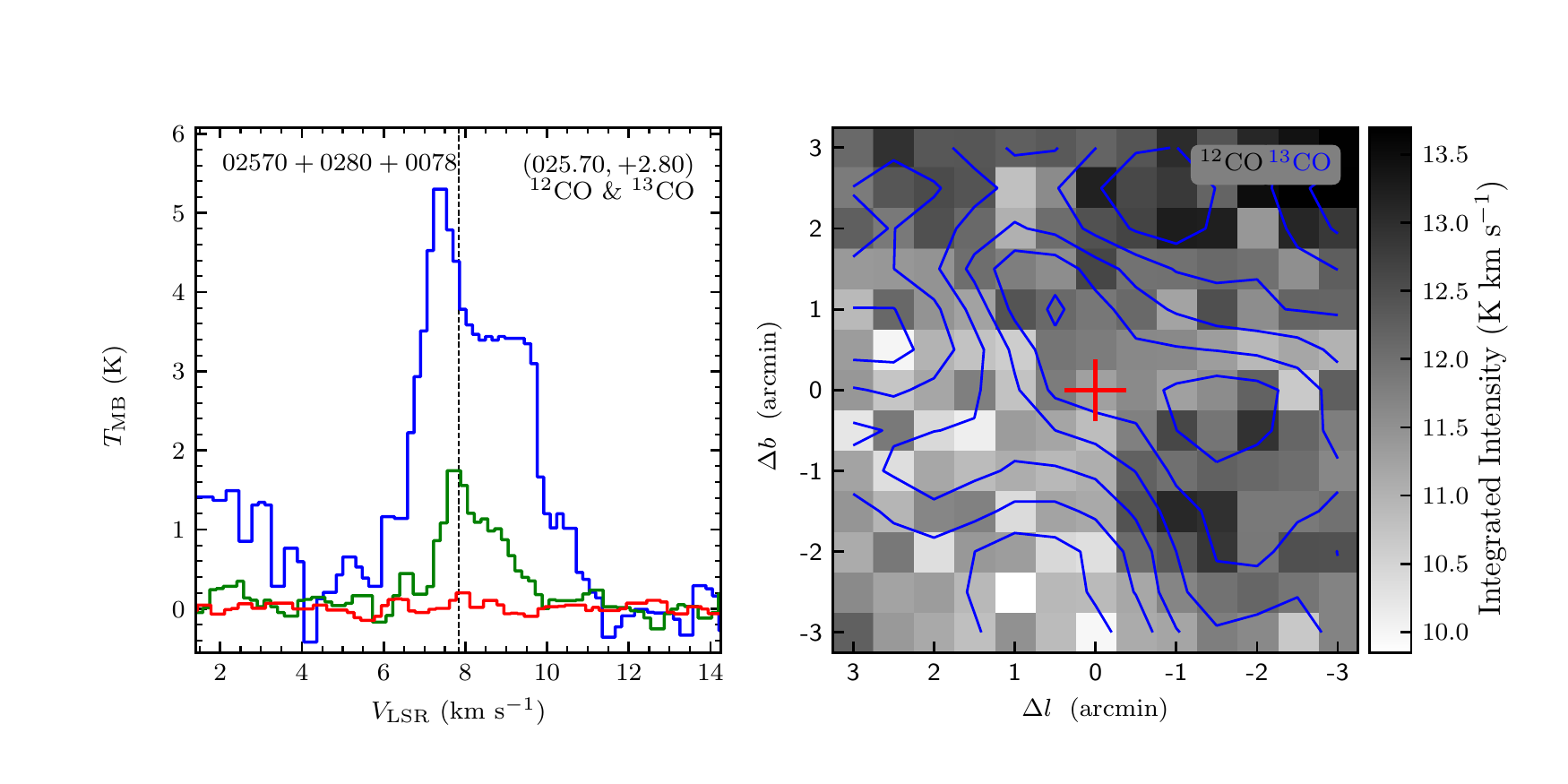}
\includegraphics[width=9.0cm,angle=0]{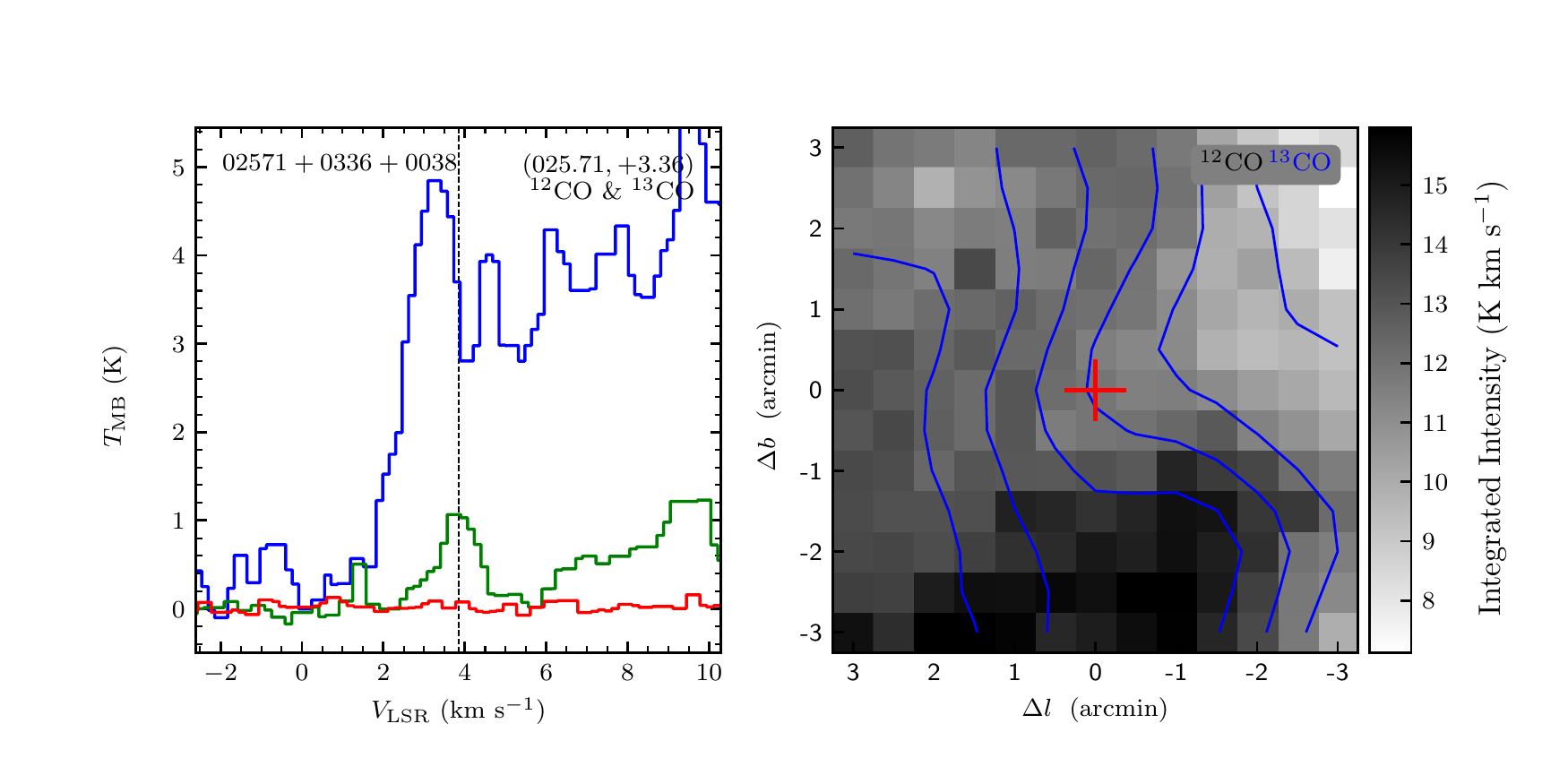}
\vspace{-0.5cm}

\includegraphics[width=9.0cm,angle=0]{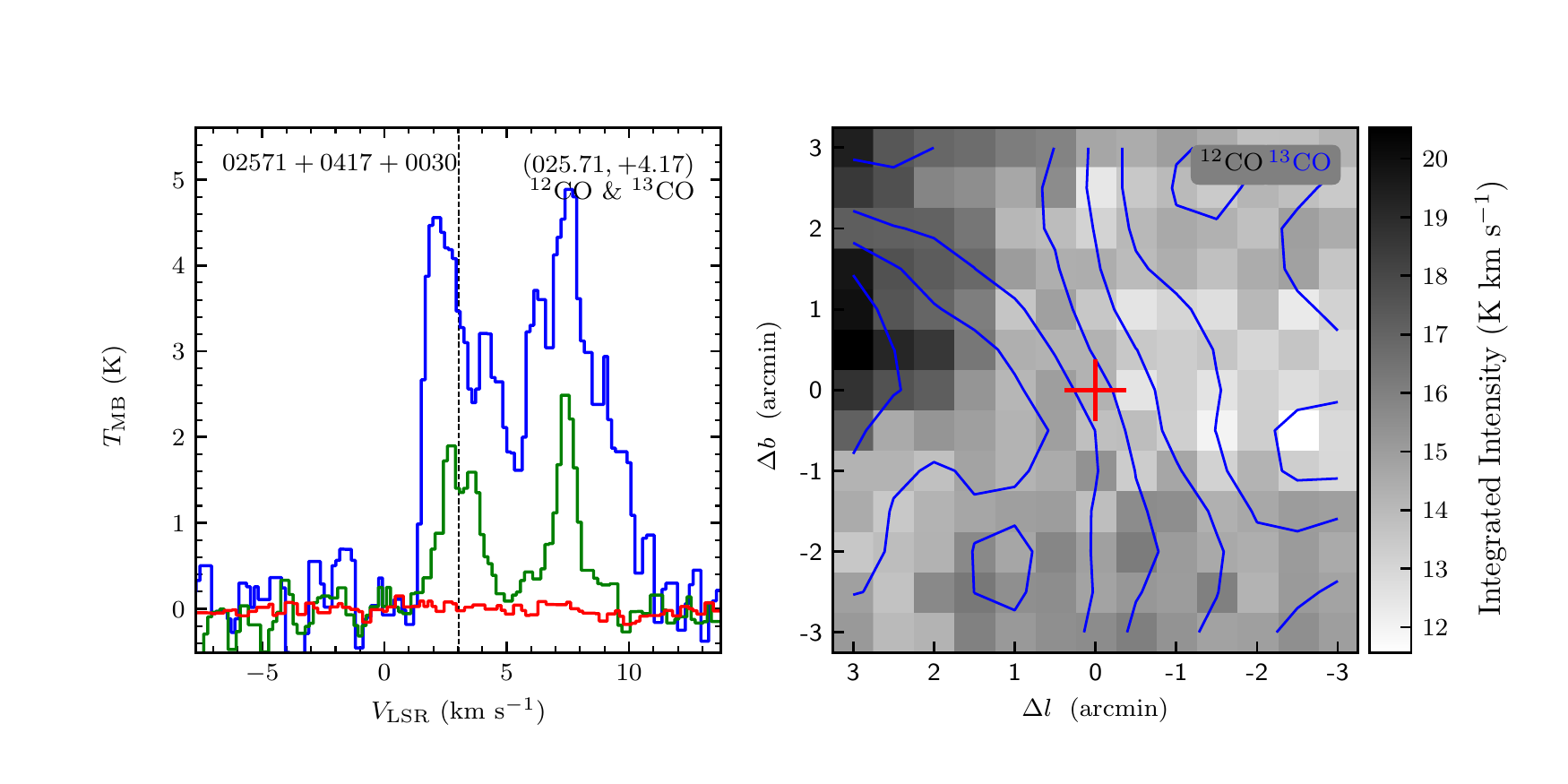}
\includegraphics[width=9.0cm,angle=0]{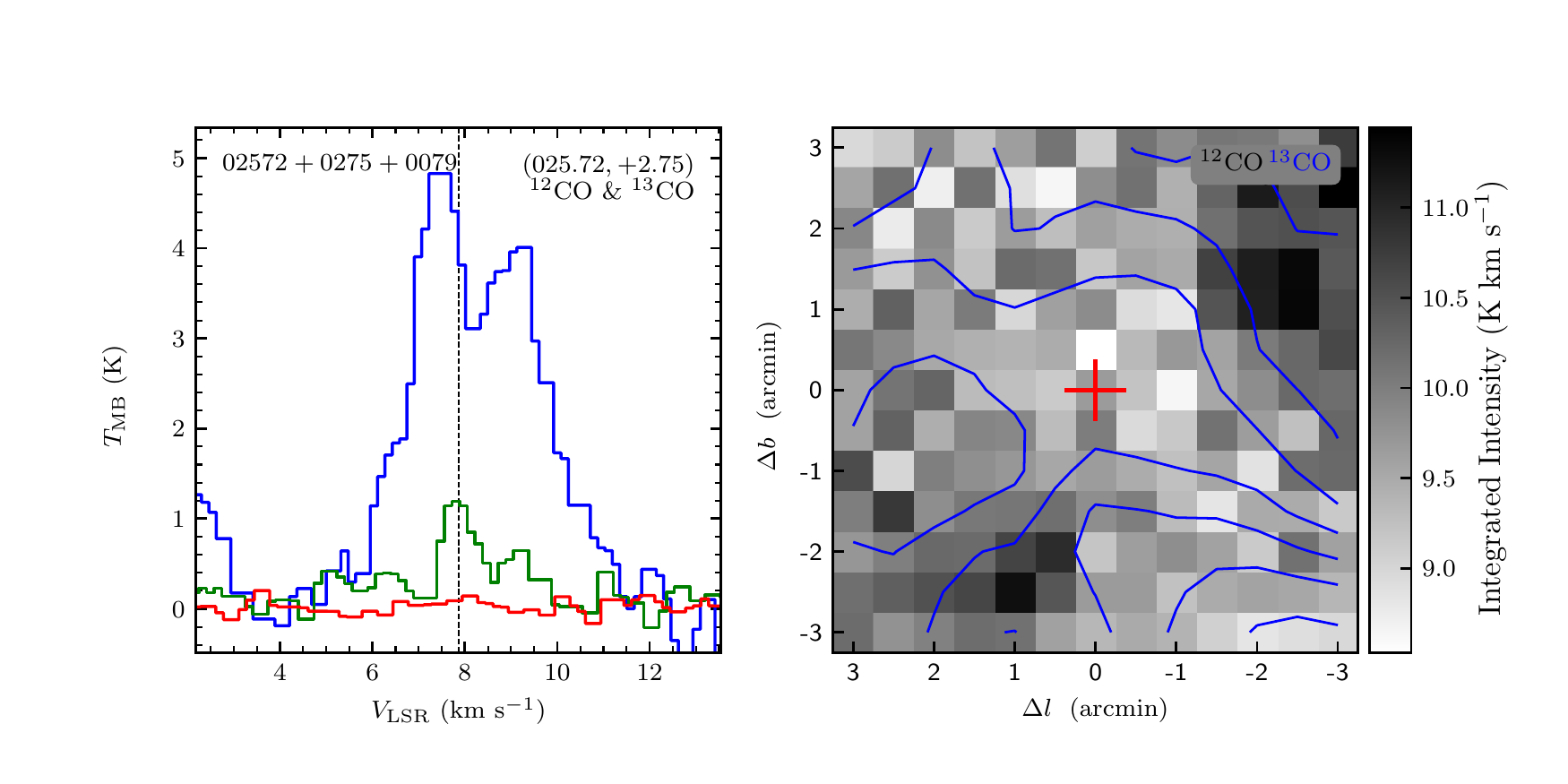}
\end{figure}
\clearpage

\begin{figure}
\includegraphics[width=9.0cm,angle=0]{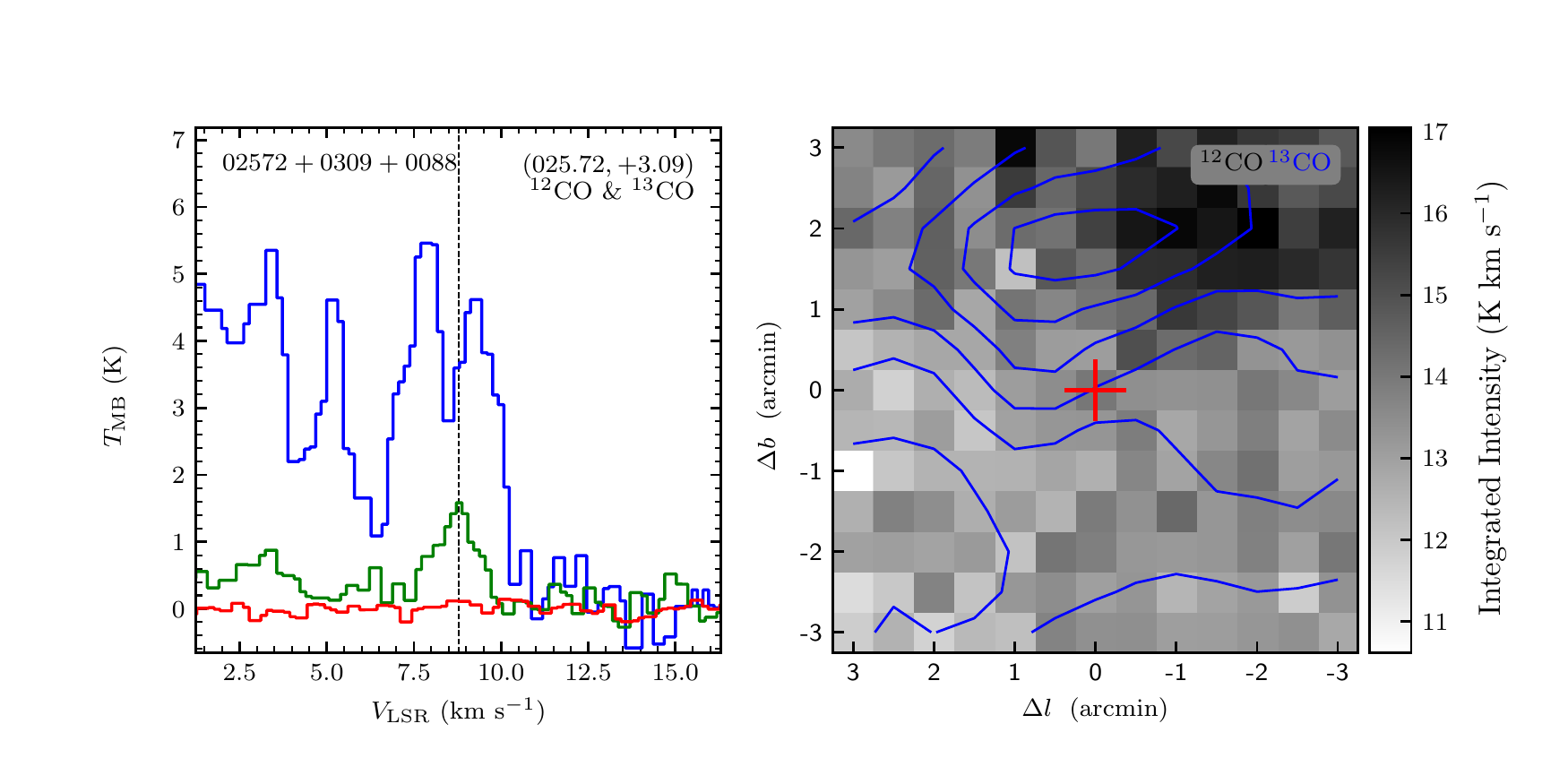}
\includegraphics[width=9.0cm,angle=0]{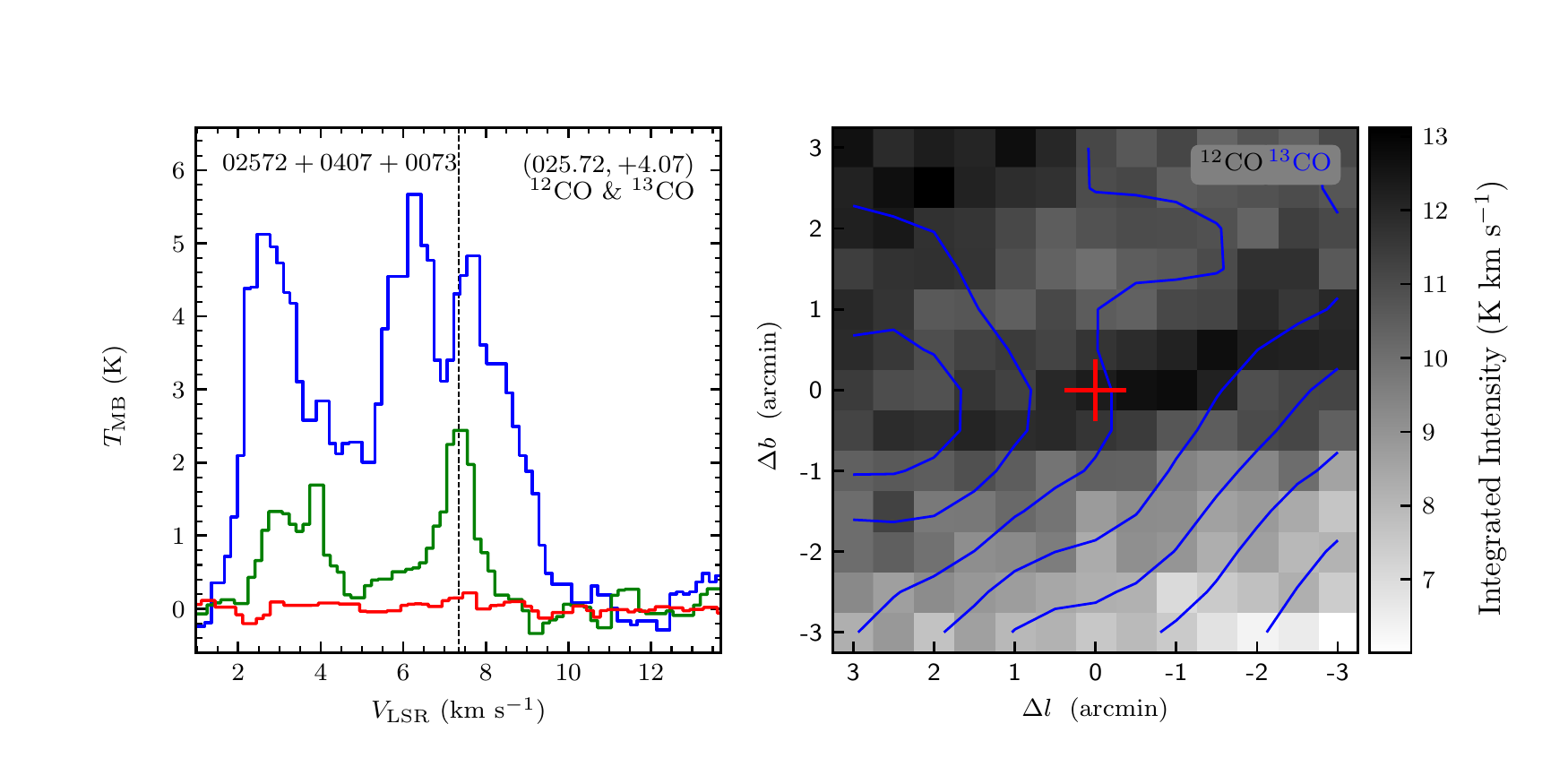}
\vspace{-0.5cm}

\includegraphics[width=9.0cm,angle=0]{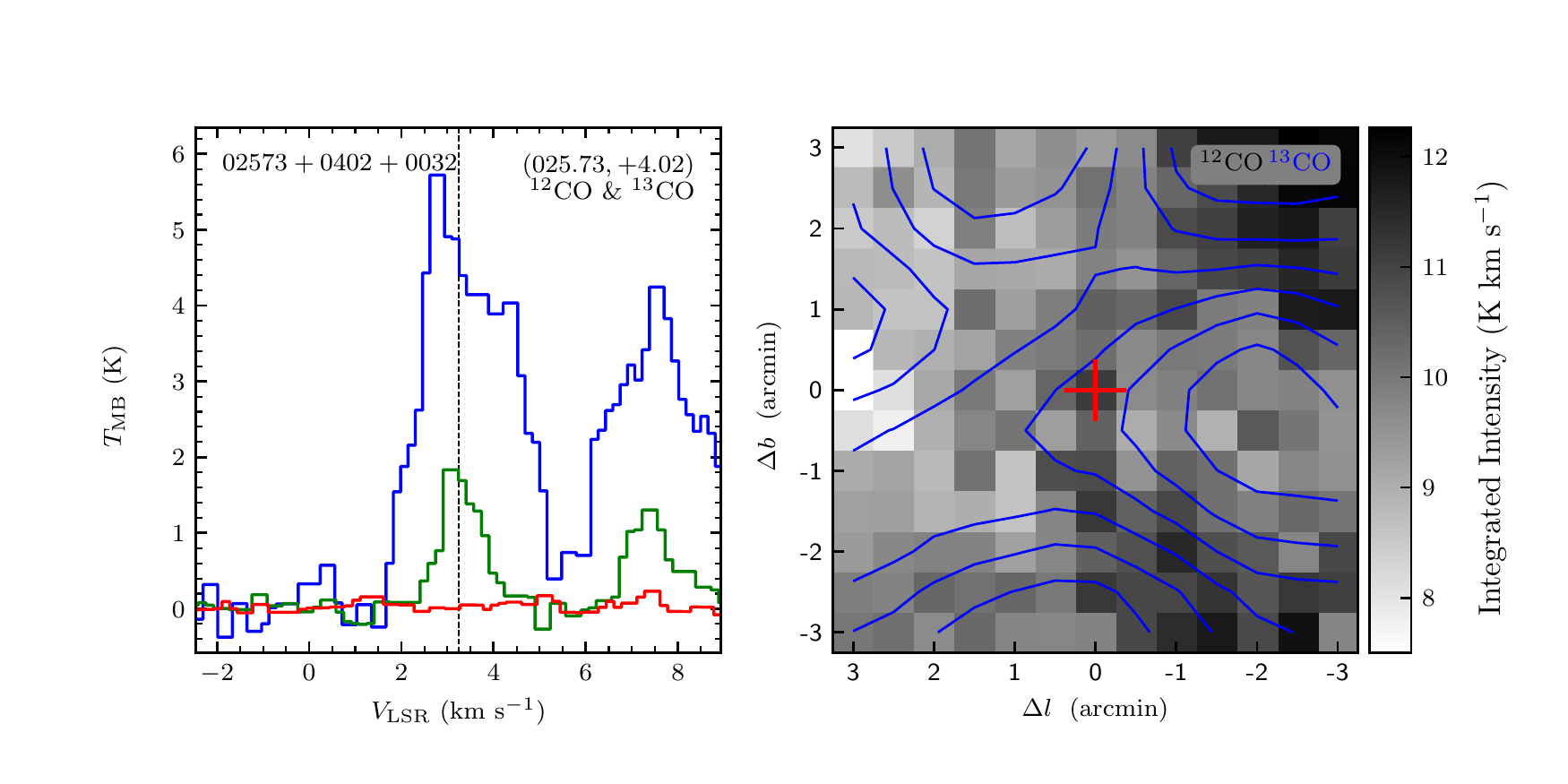}
\includegraphics[width=9.0cm,angle=0]{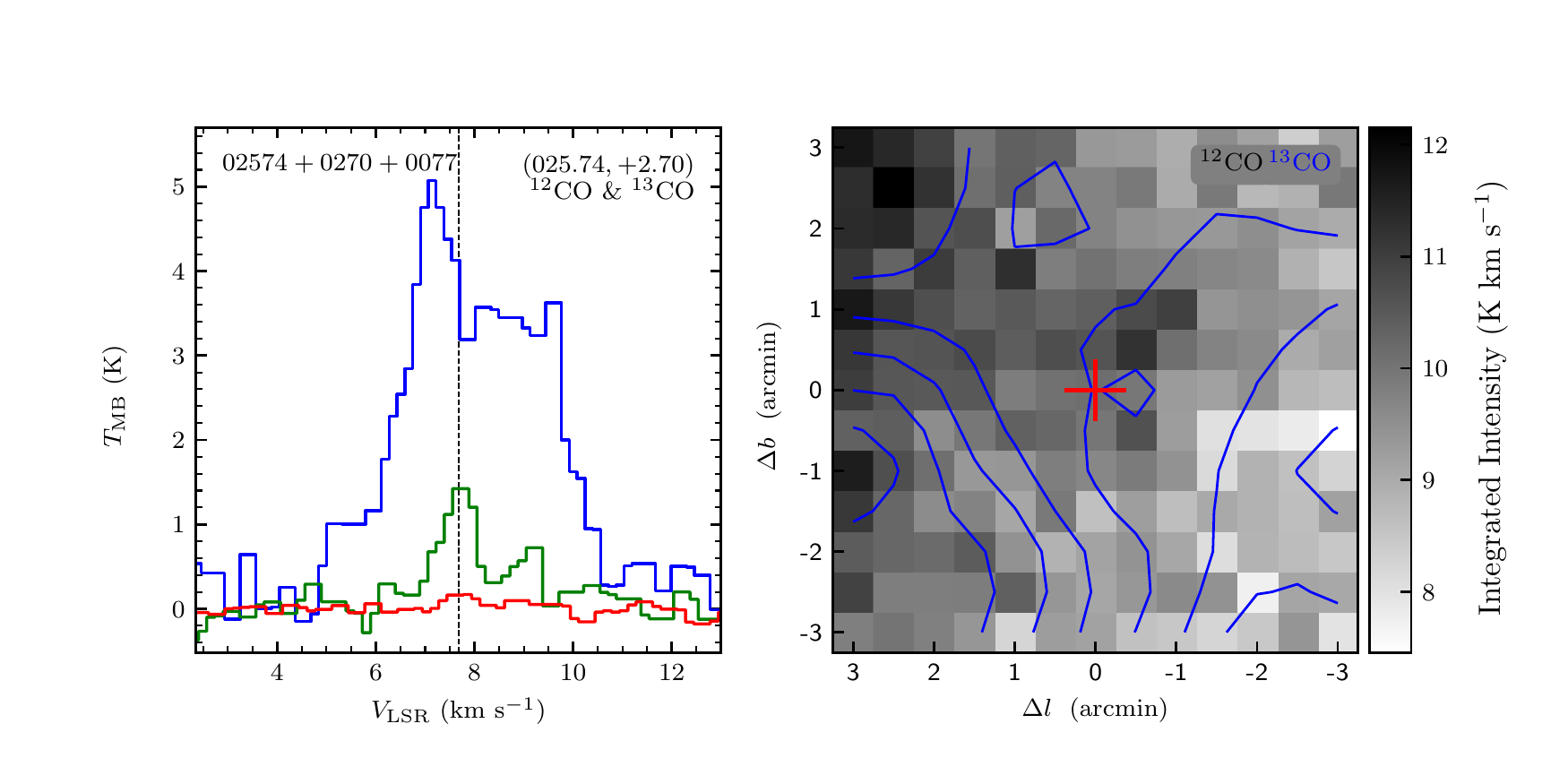}
\vspace{-0.5cm}

\includegraphics[width=9.0cm,angle=0]{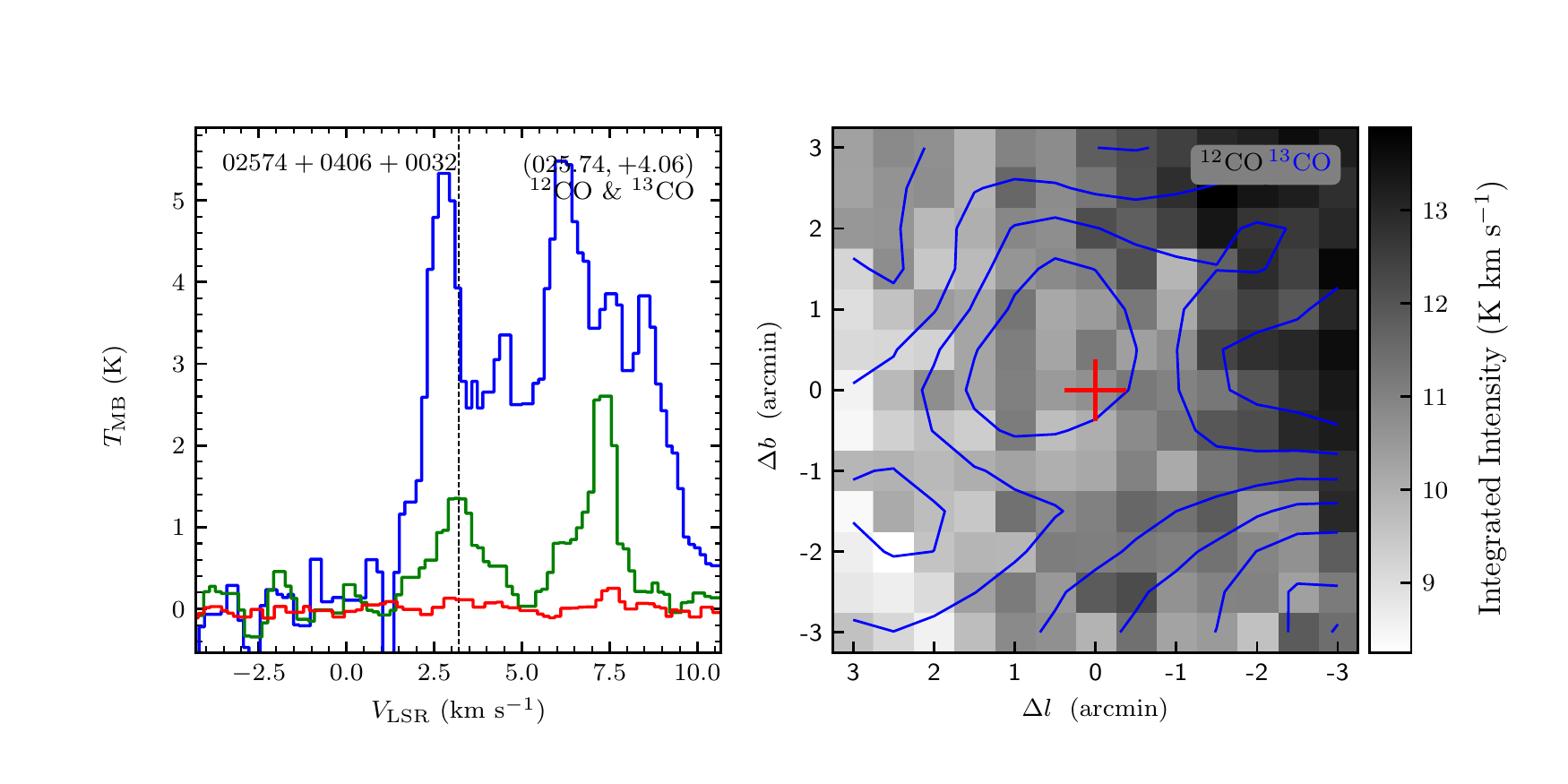}
\includegraphics[width=9.0cm,angle=0]{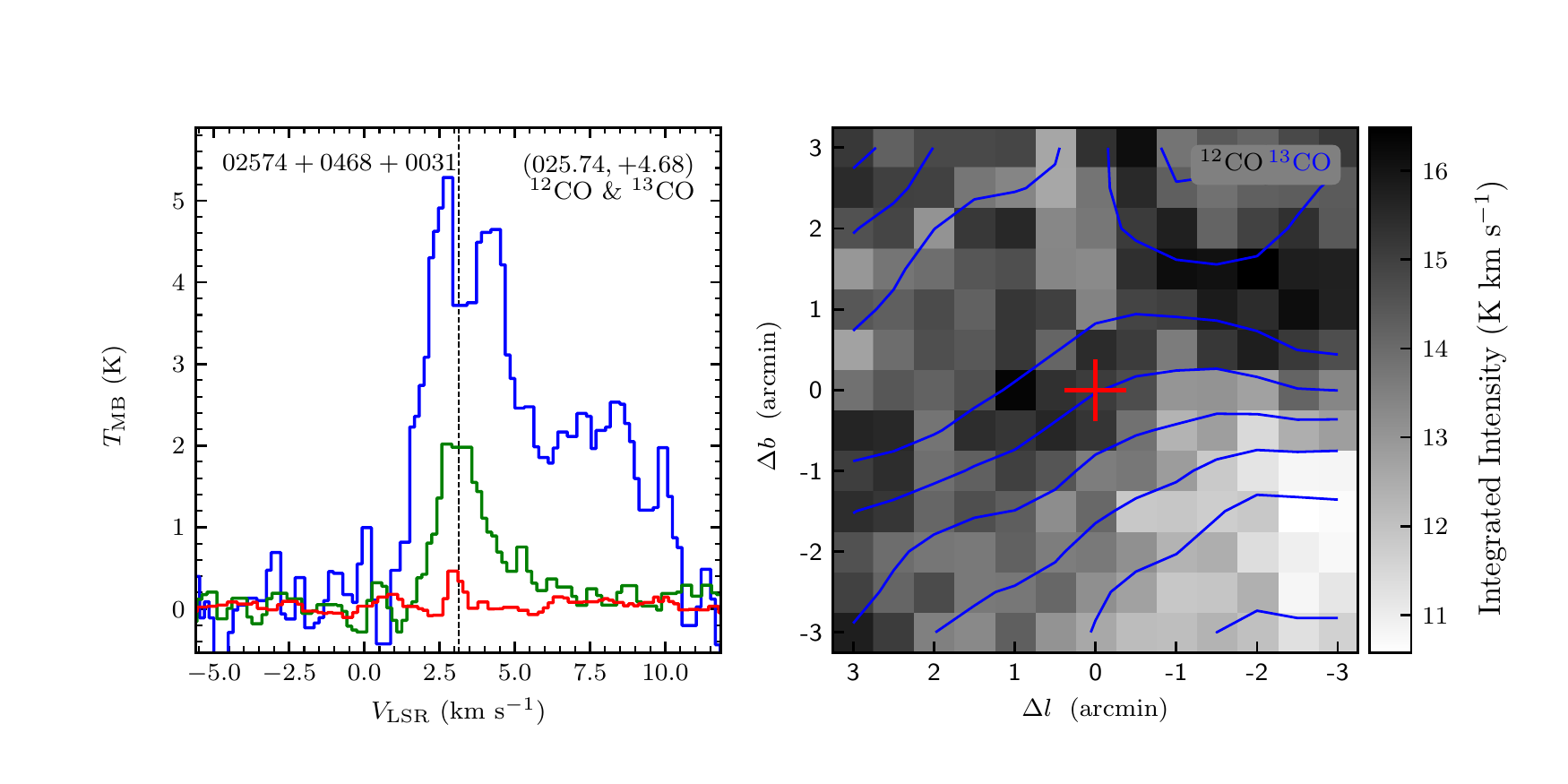}
\vspace{-0.5cm}

\includegraphics[width=9.0cm,angle=0]{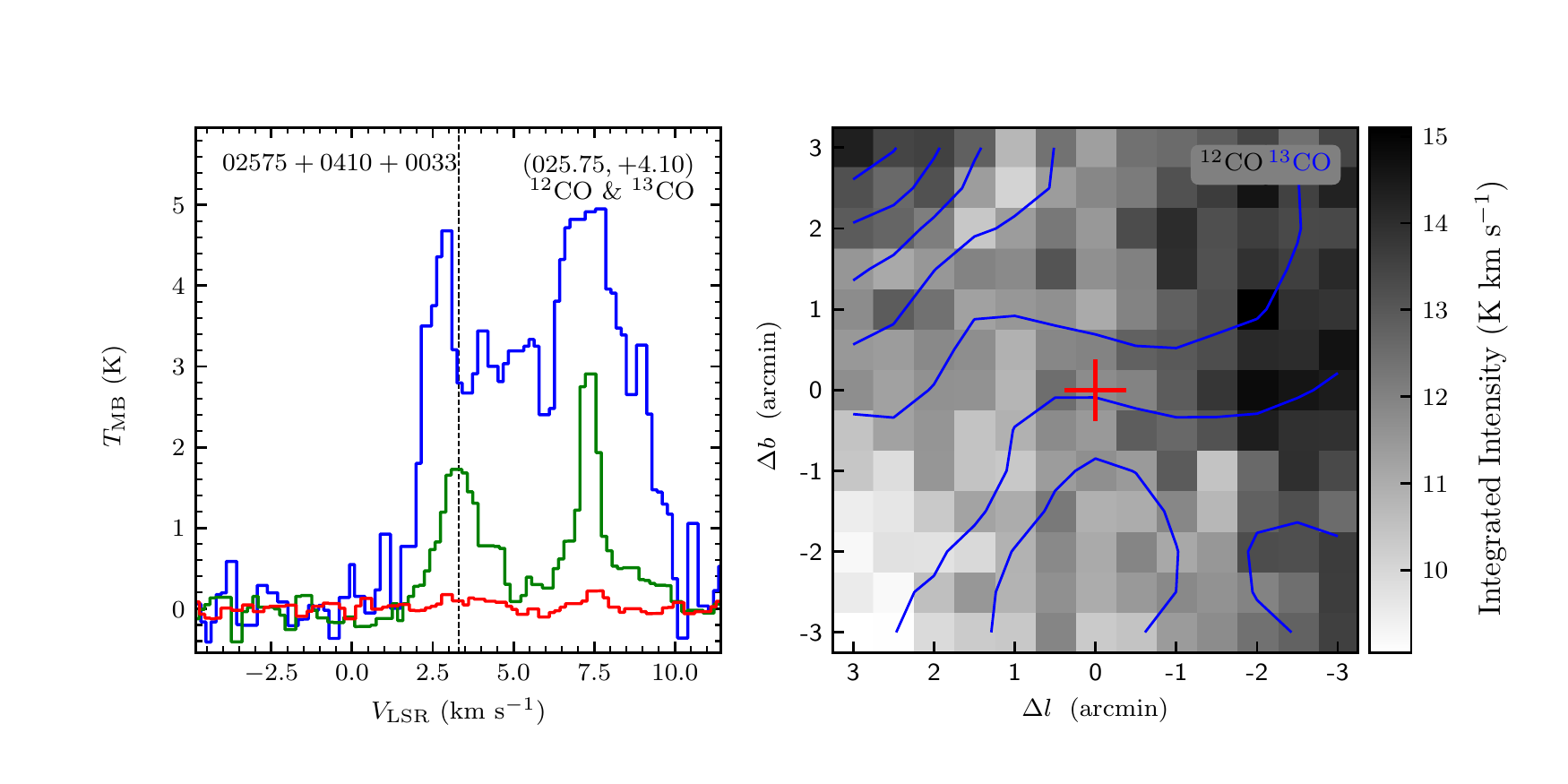}
\includegraphics[width=9.0cm,angle=0]{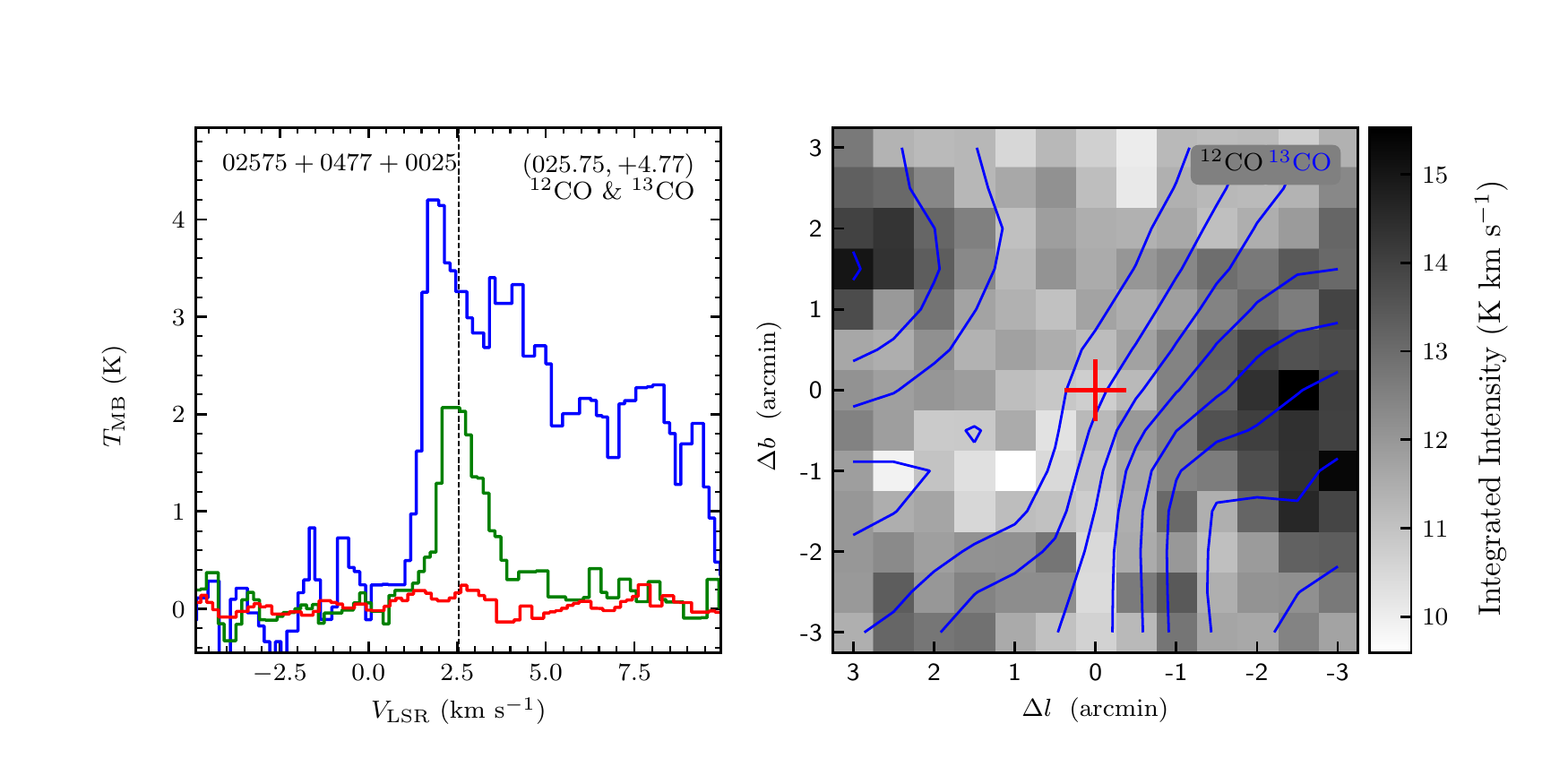}
\vspace{-0.5cm}

\includegraphics[width=9.0cm,angle=0]{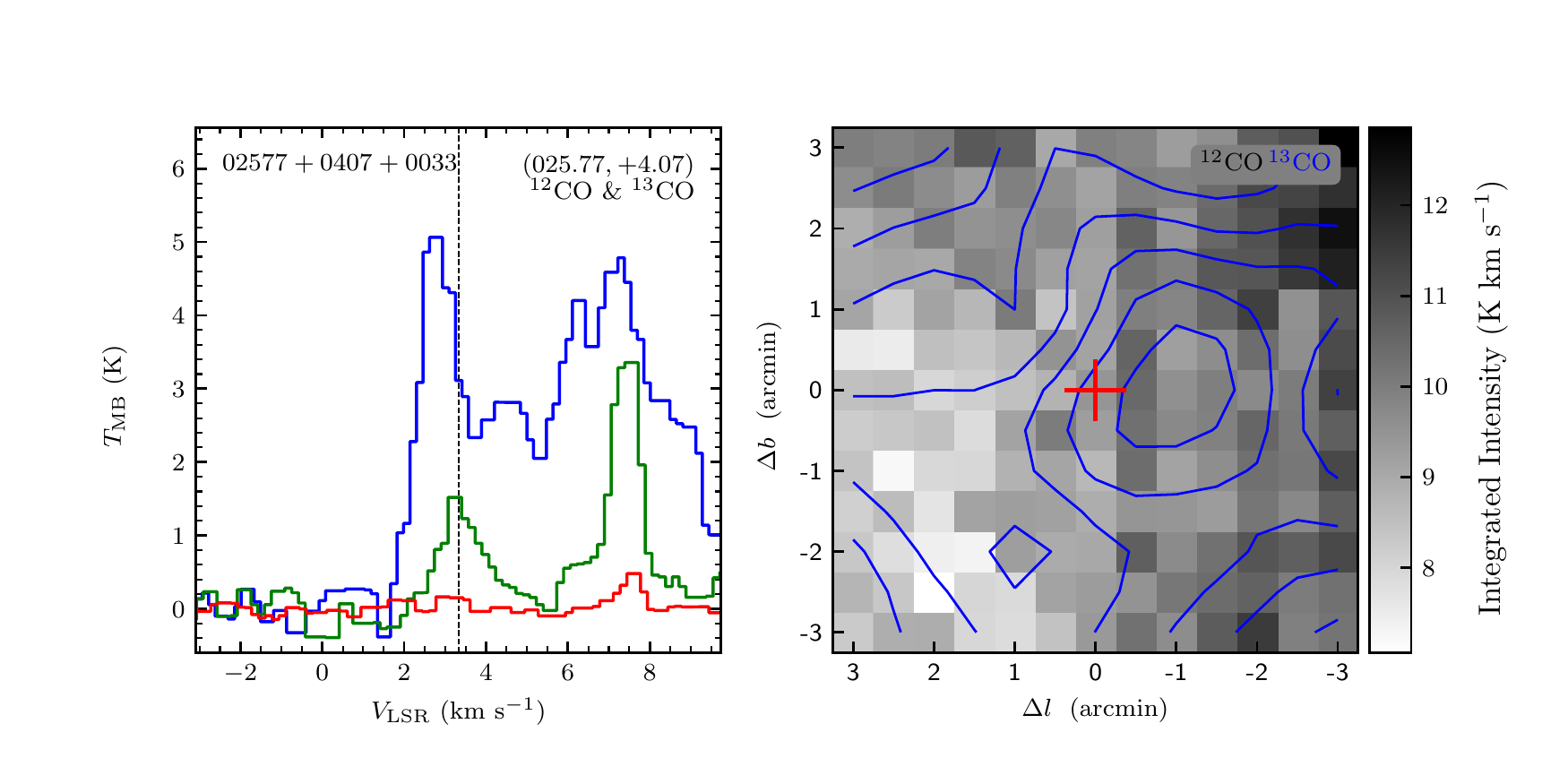}
\includegraphics[width=9.0cm,angle=0]{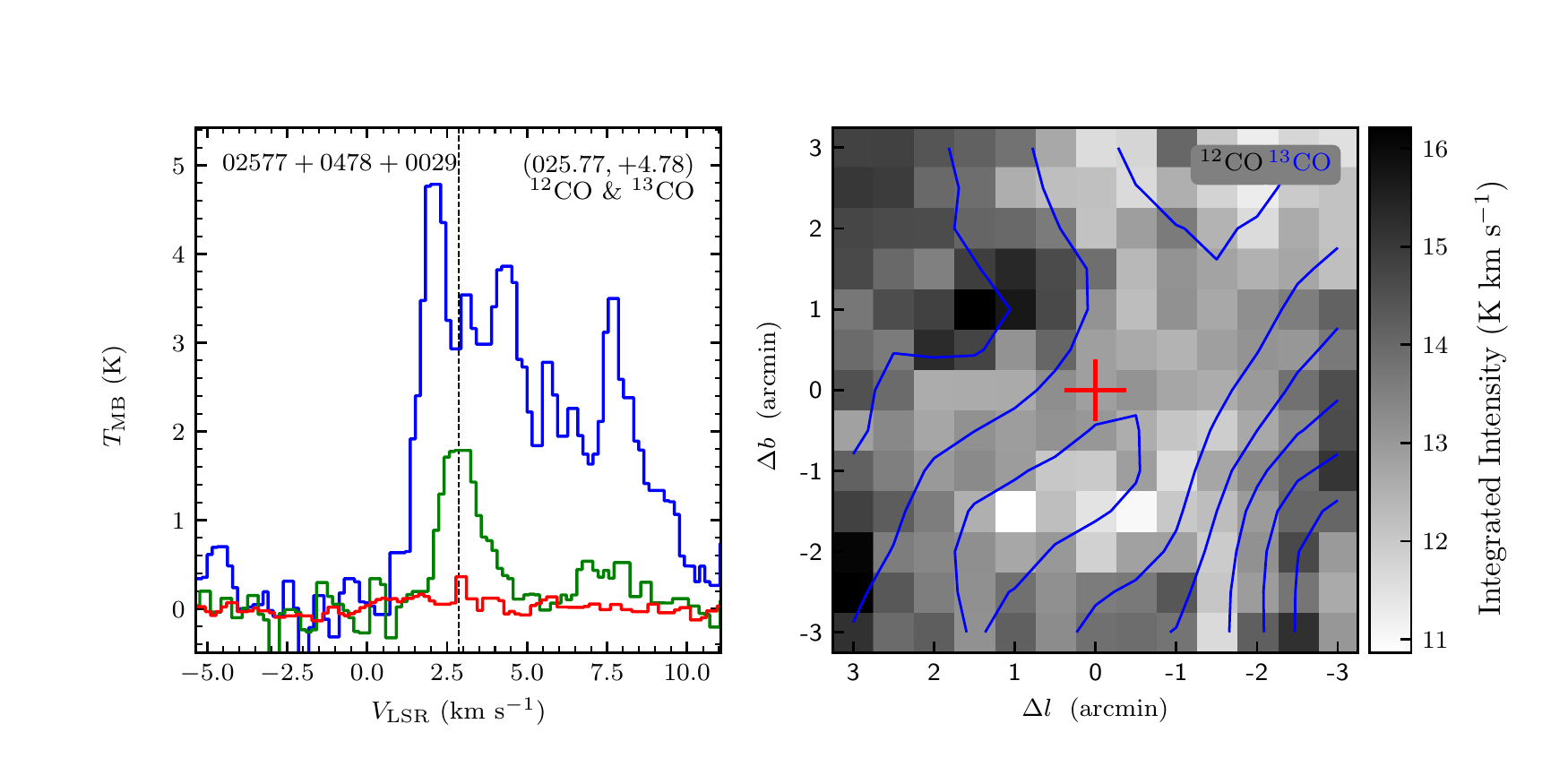}
\end{figure}
\clearpage

\begin{figure}
\includegraphics[width=9.0cm,angle=0]{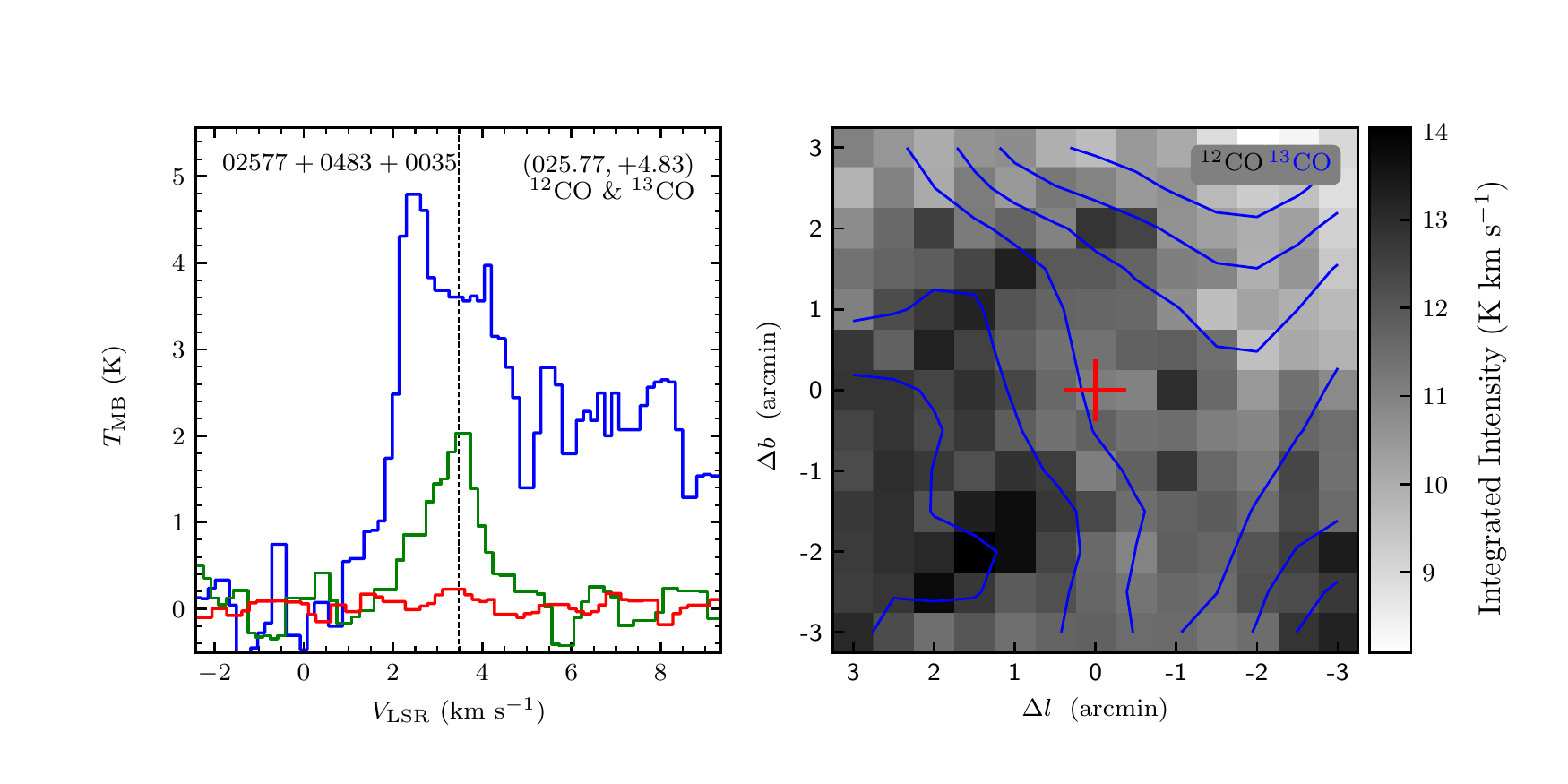}
\includegraphics[width=9.0cm,angle=0]{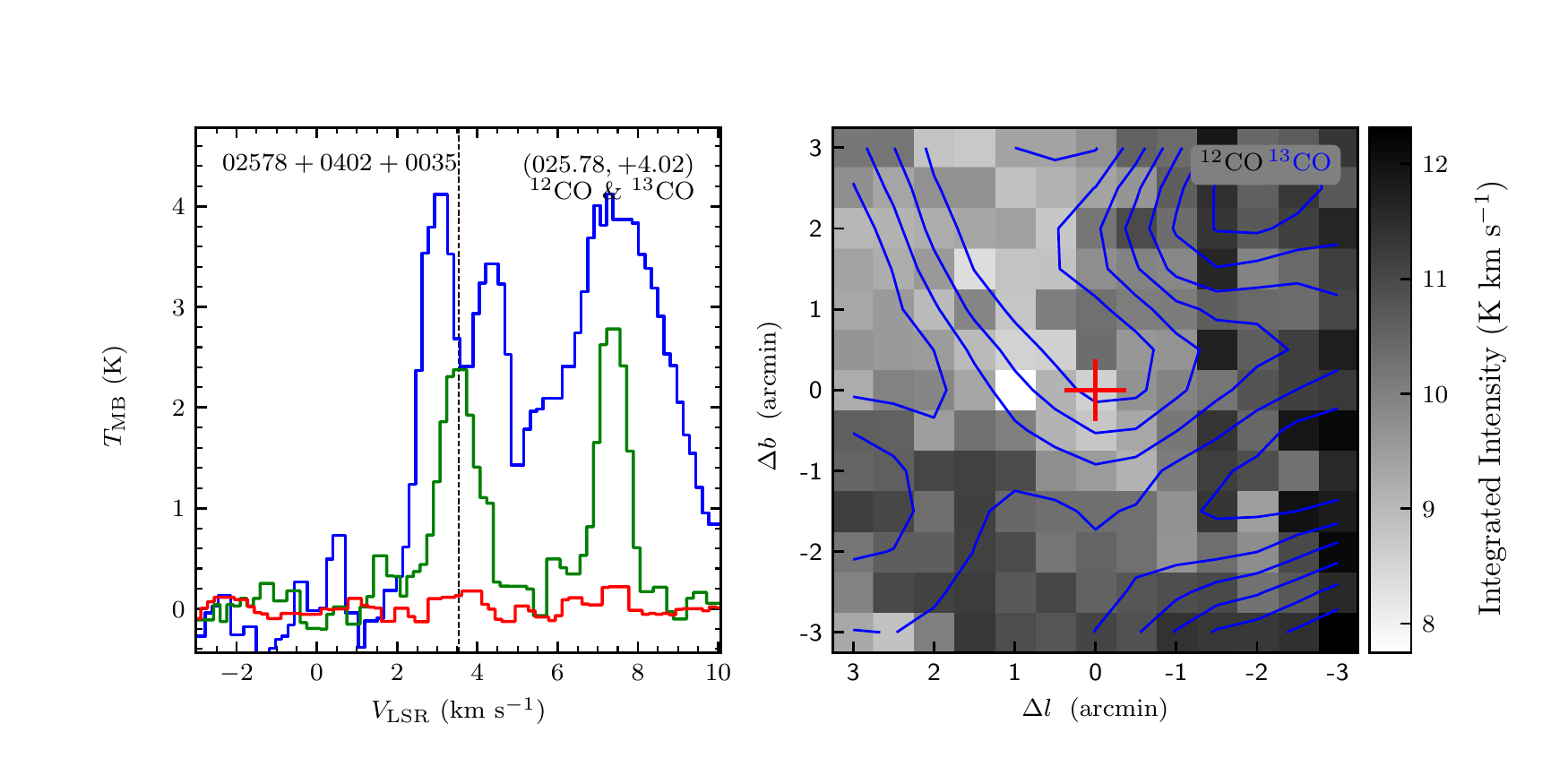}
\vspace{-0.5cm}

\includegraphics[width=9.0cm,angle=0]{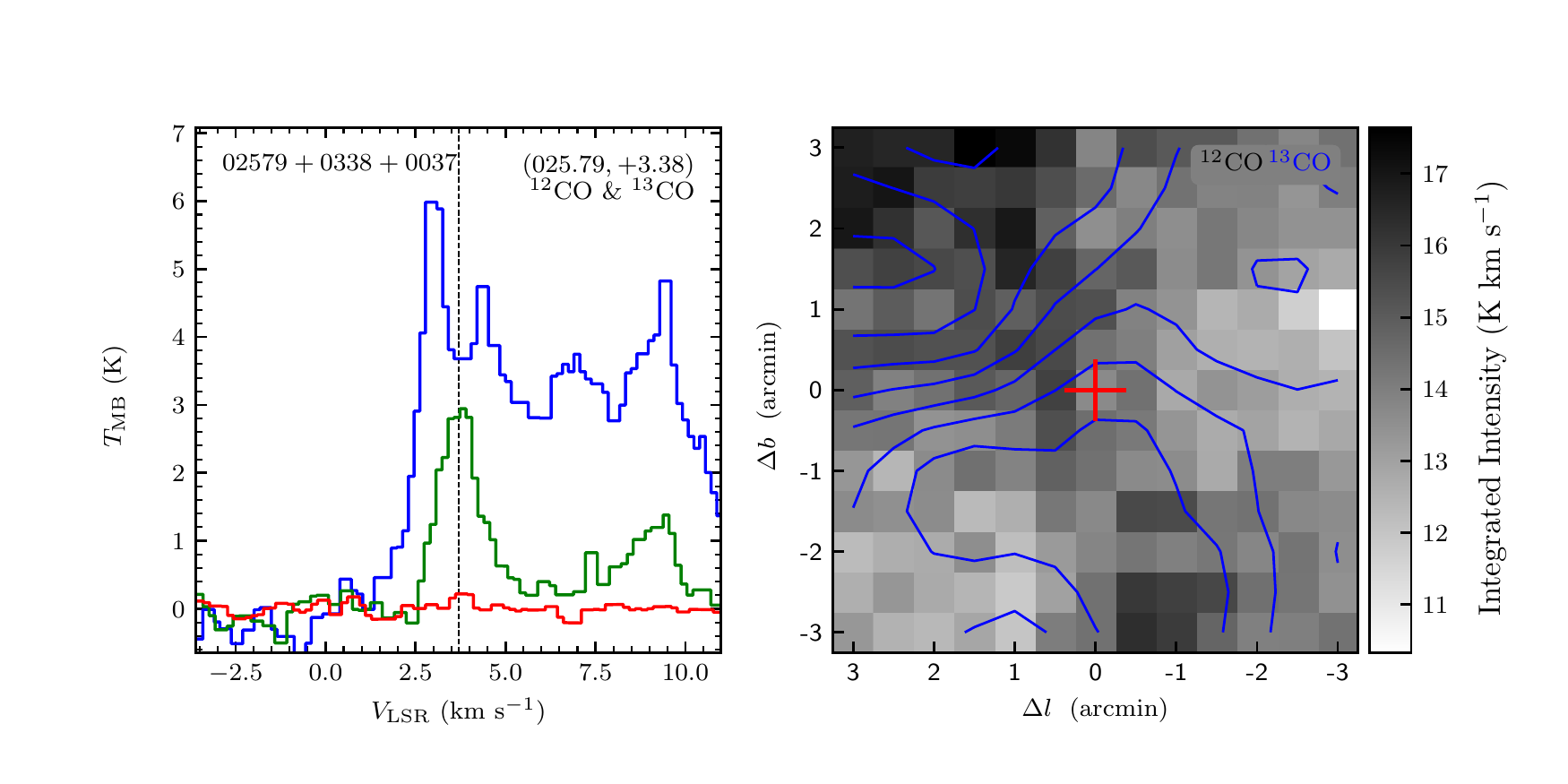}
\includegraphics[width=9.0cm,angle=0]{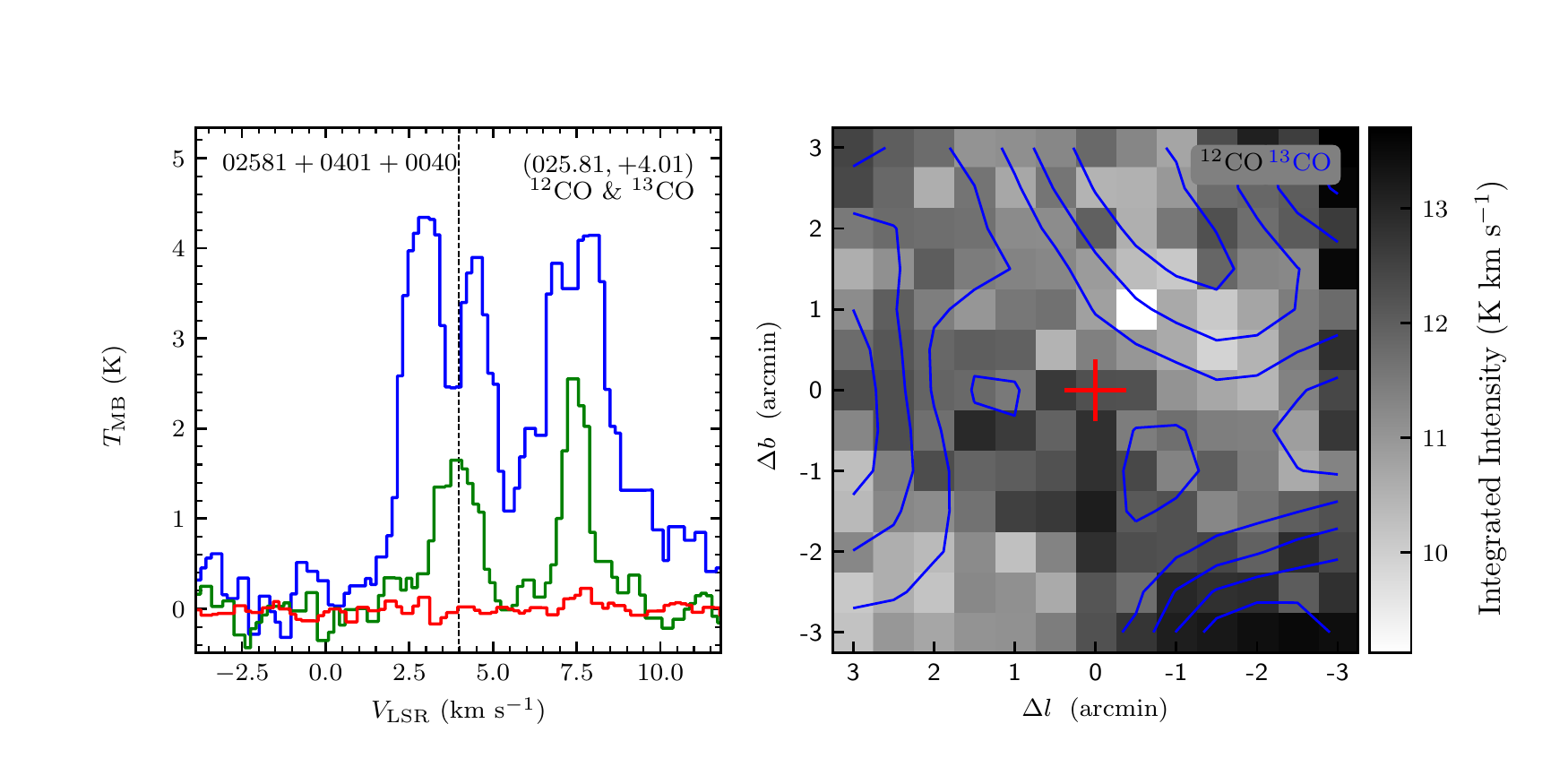}
\vspace{-0.5cm}

\includegraphics[width=9.0cm,angle=0]{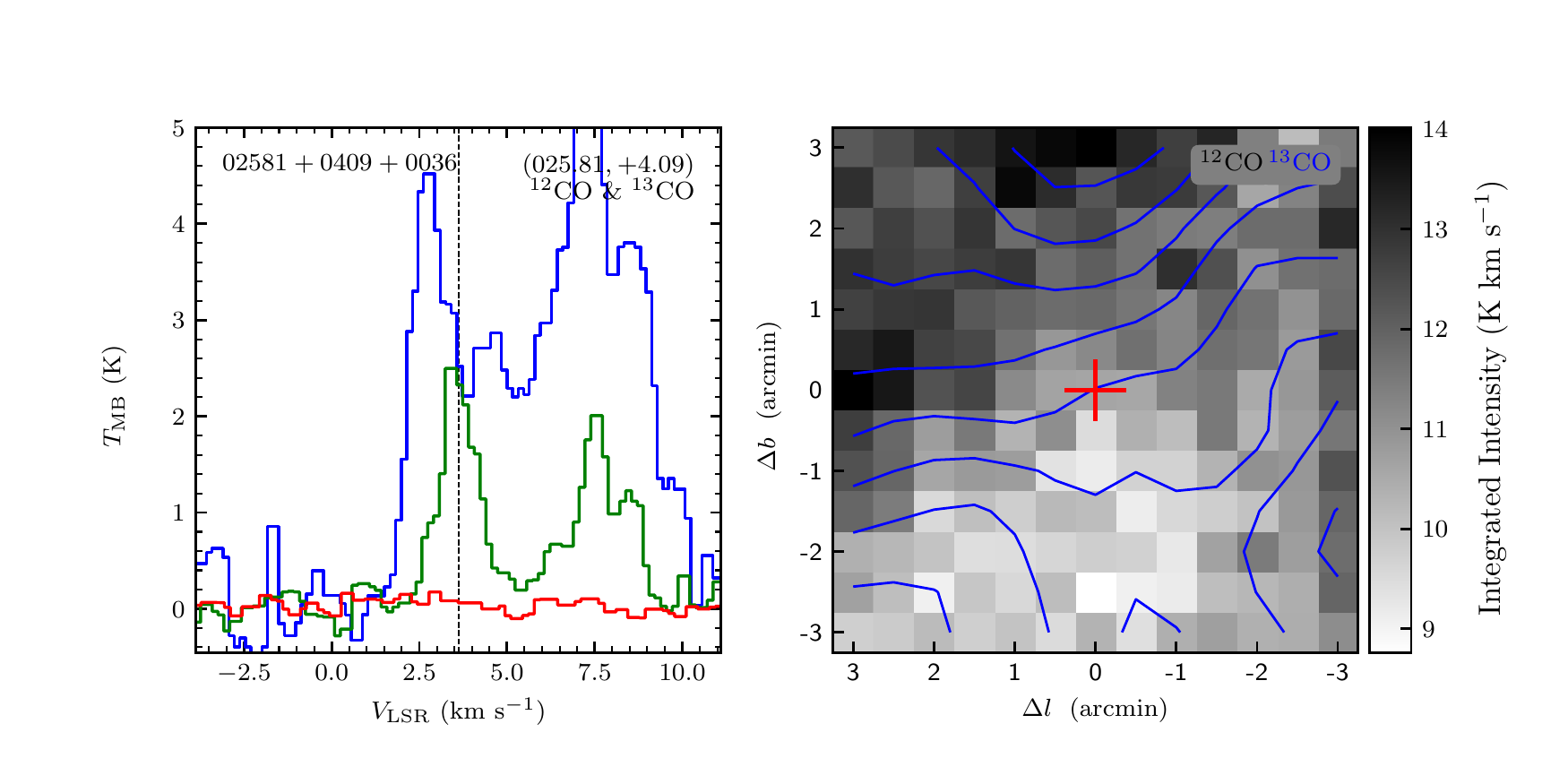}
\includegraphics[width=9.0cm,angle=0]{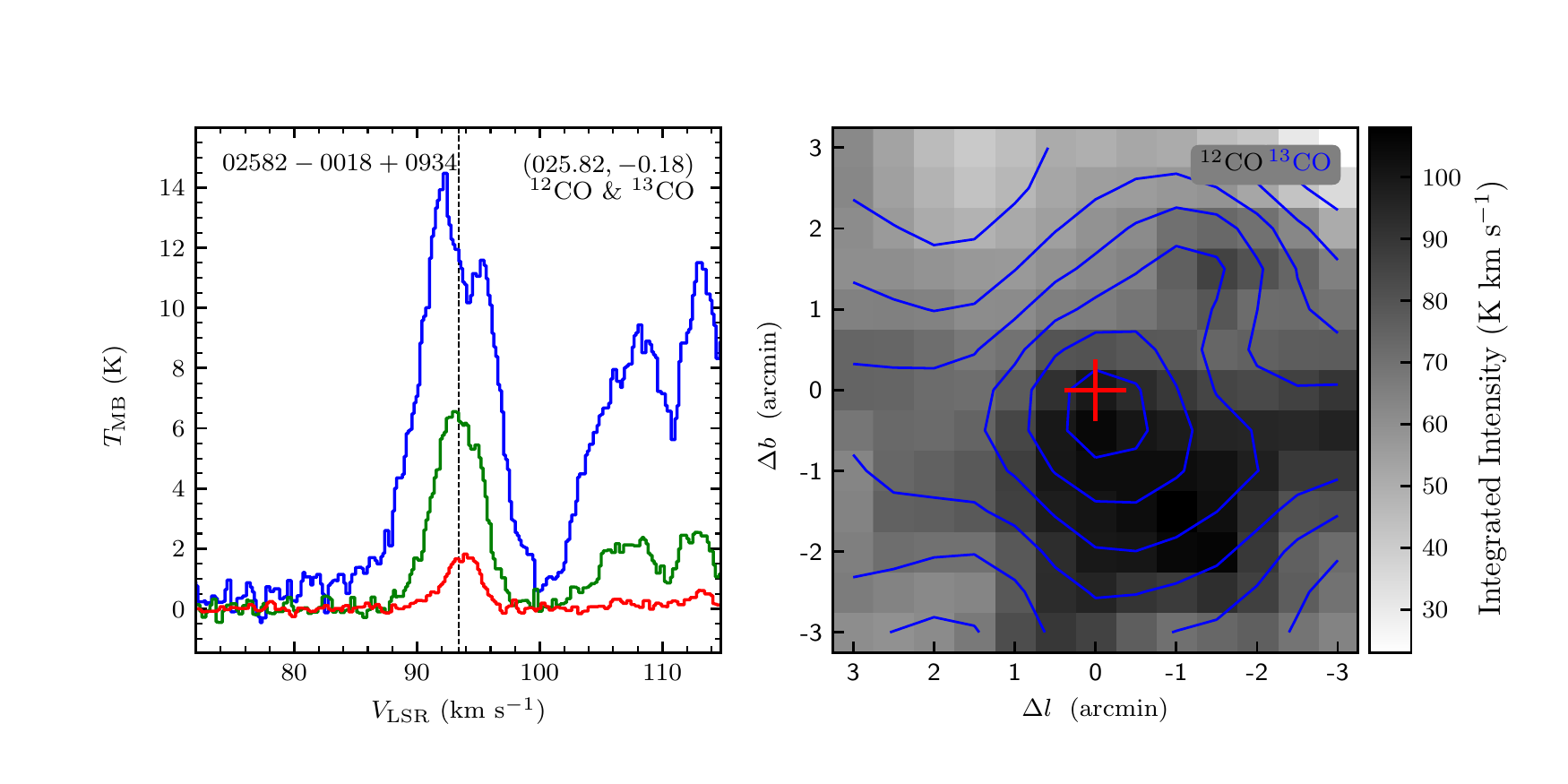}
\vspace{-0.5cm}

\includegraphics[width=9.0cm,angle=0]{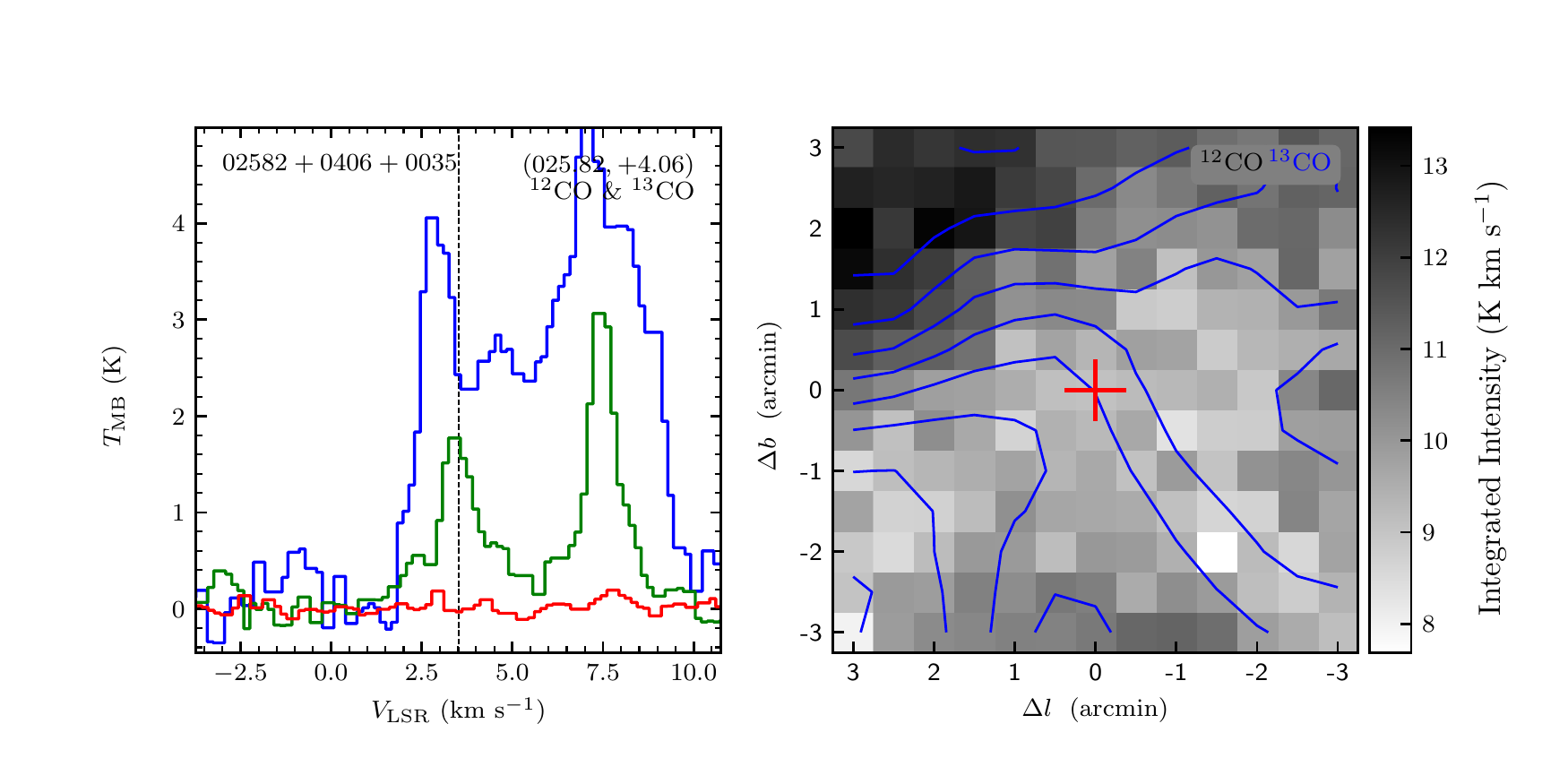}
\includegraphics[width=9.0cm,angle=0]{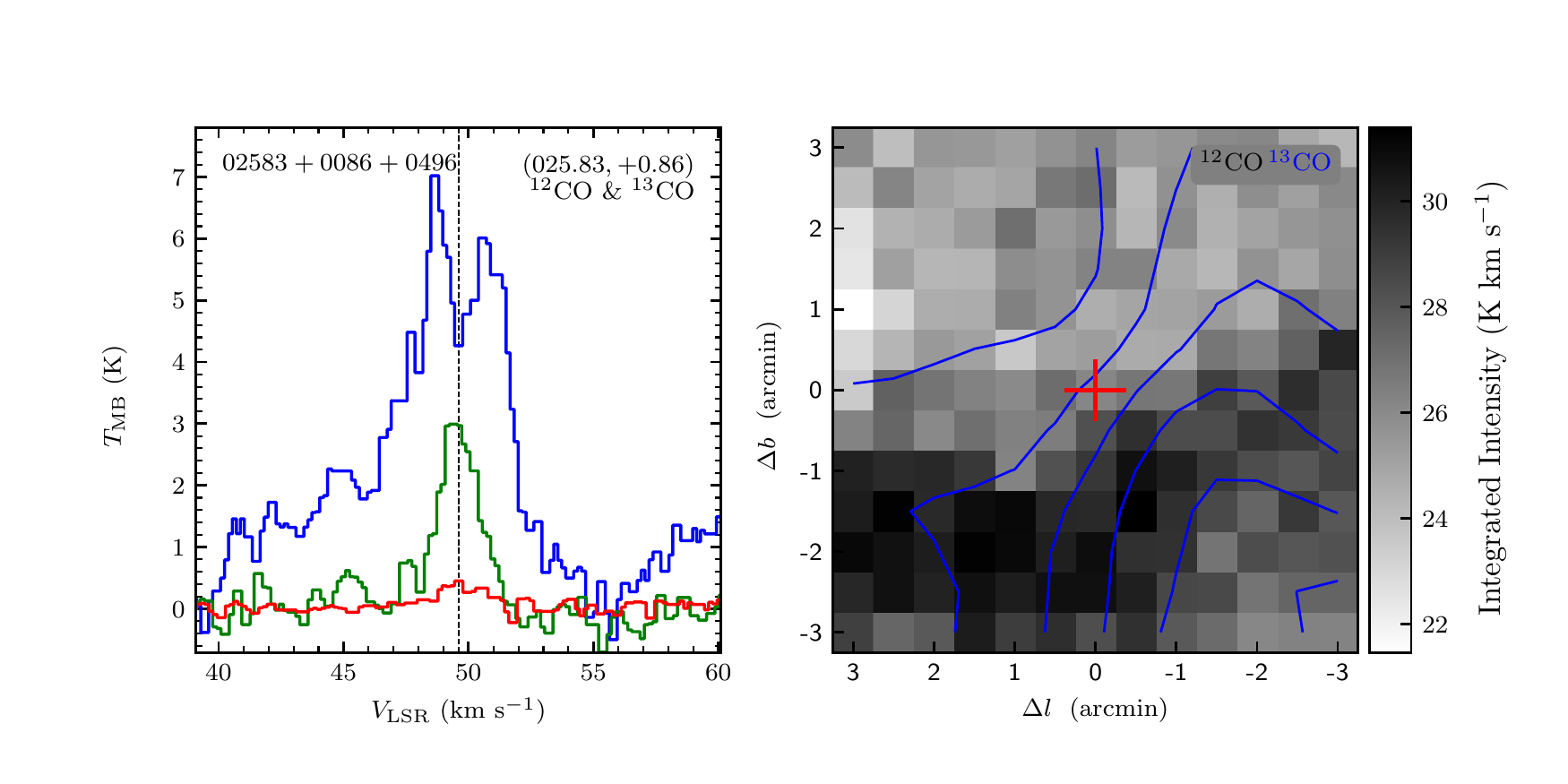}
\vspace{-0.5cm}

\includegraphics[width=9.0cm,angle=0]{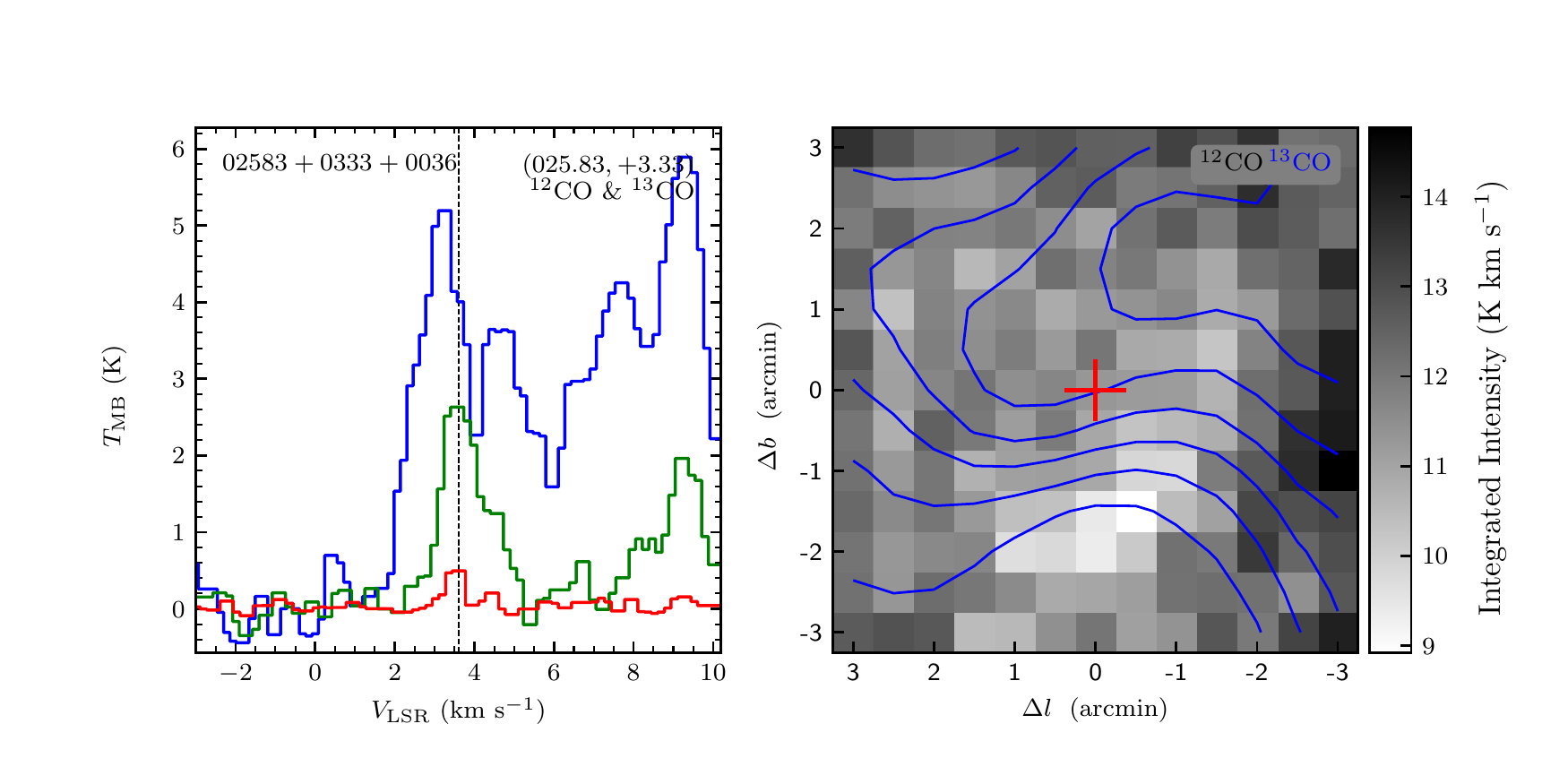}
\includegraphics[width=9.0cm,angle=0]{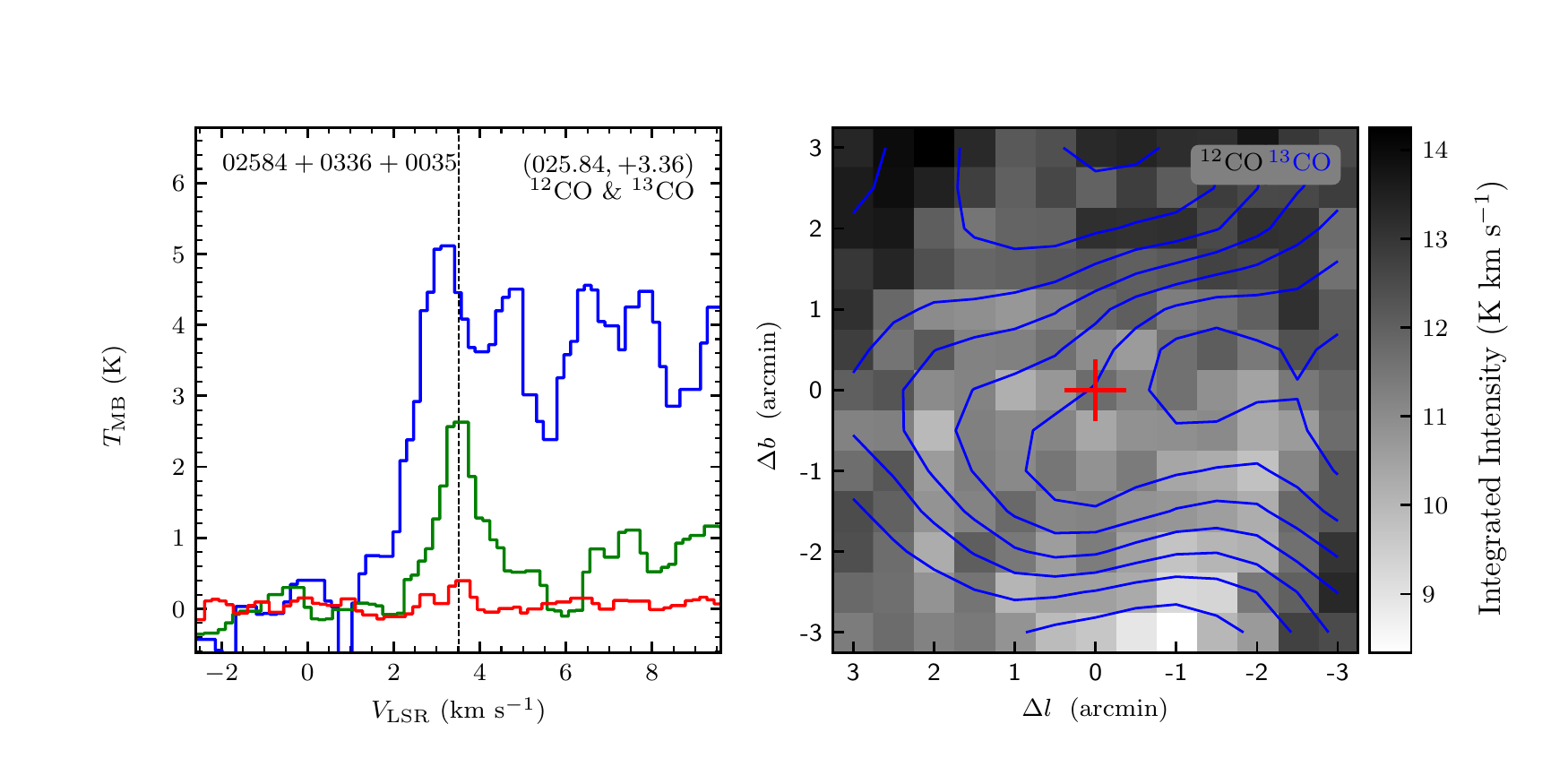}
\end{figure}
\clearpage

\begin{figure}
\includegraphics[width=9.0cm,angle=0]{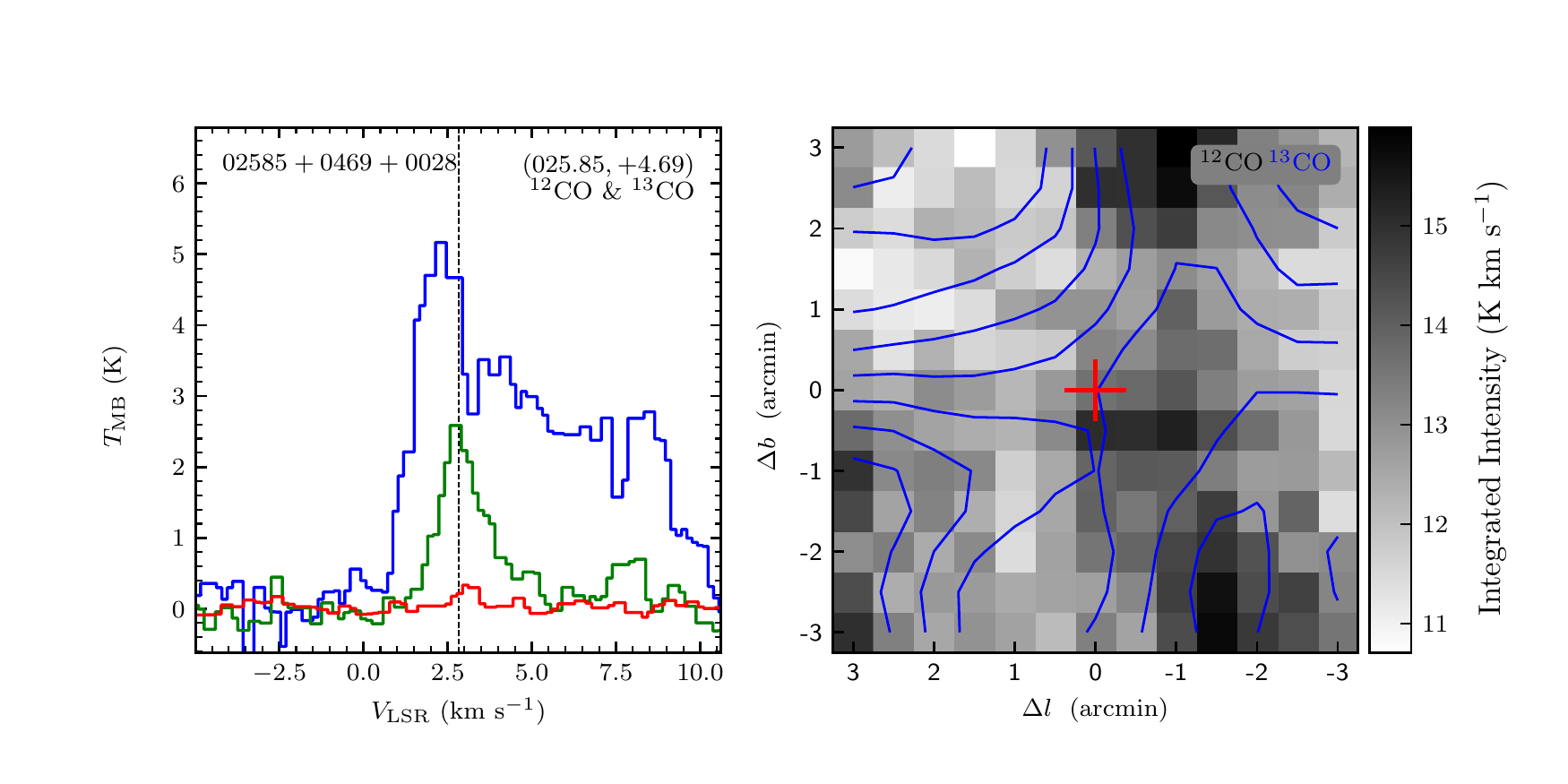}
\includegraphics[width=9.0cm,angle=0]{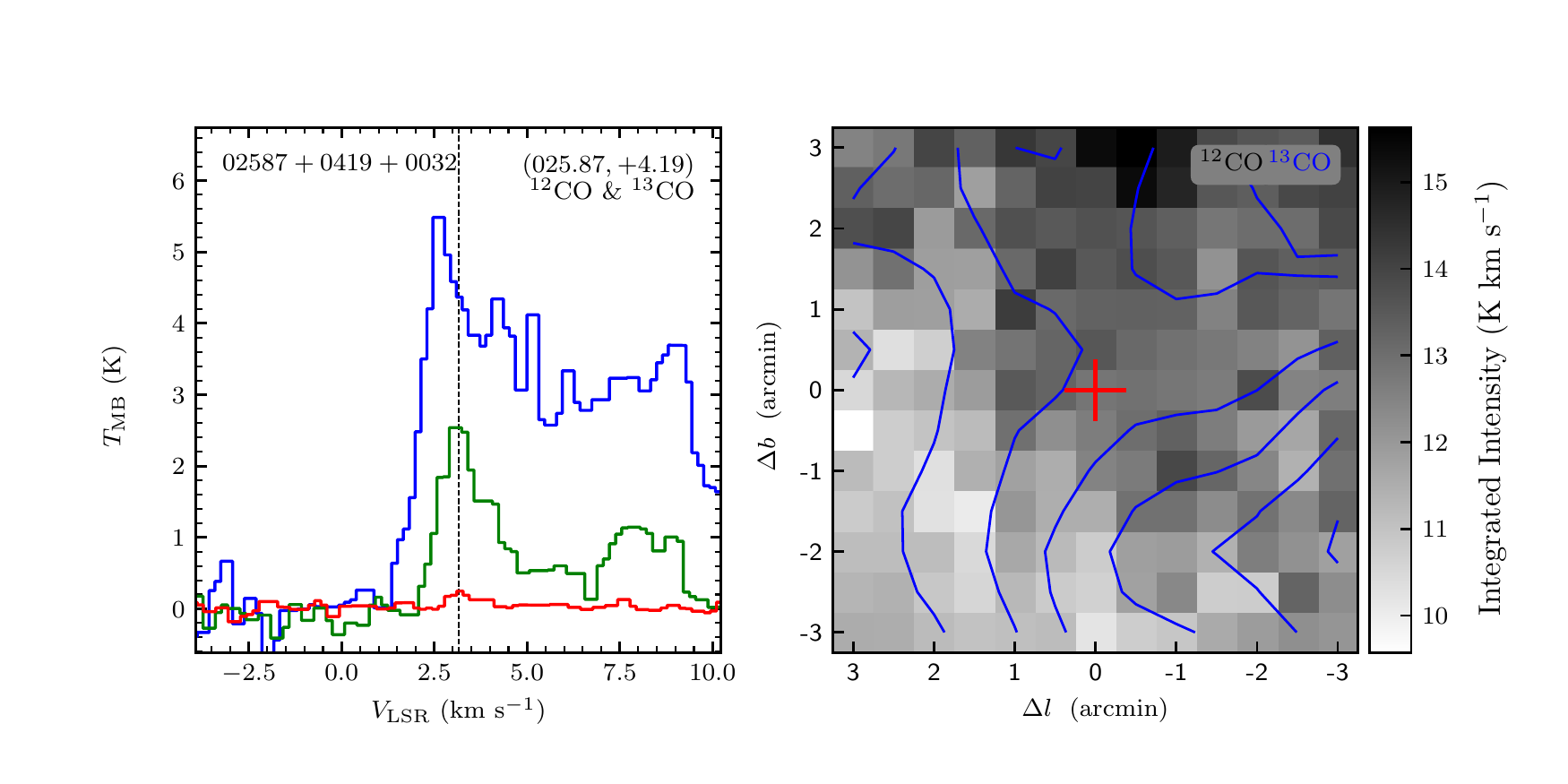}
\vspace{-0.5cm}

\includegraphics[width=9.0cm,angle=0]{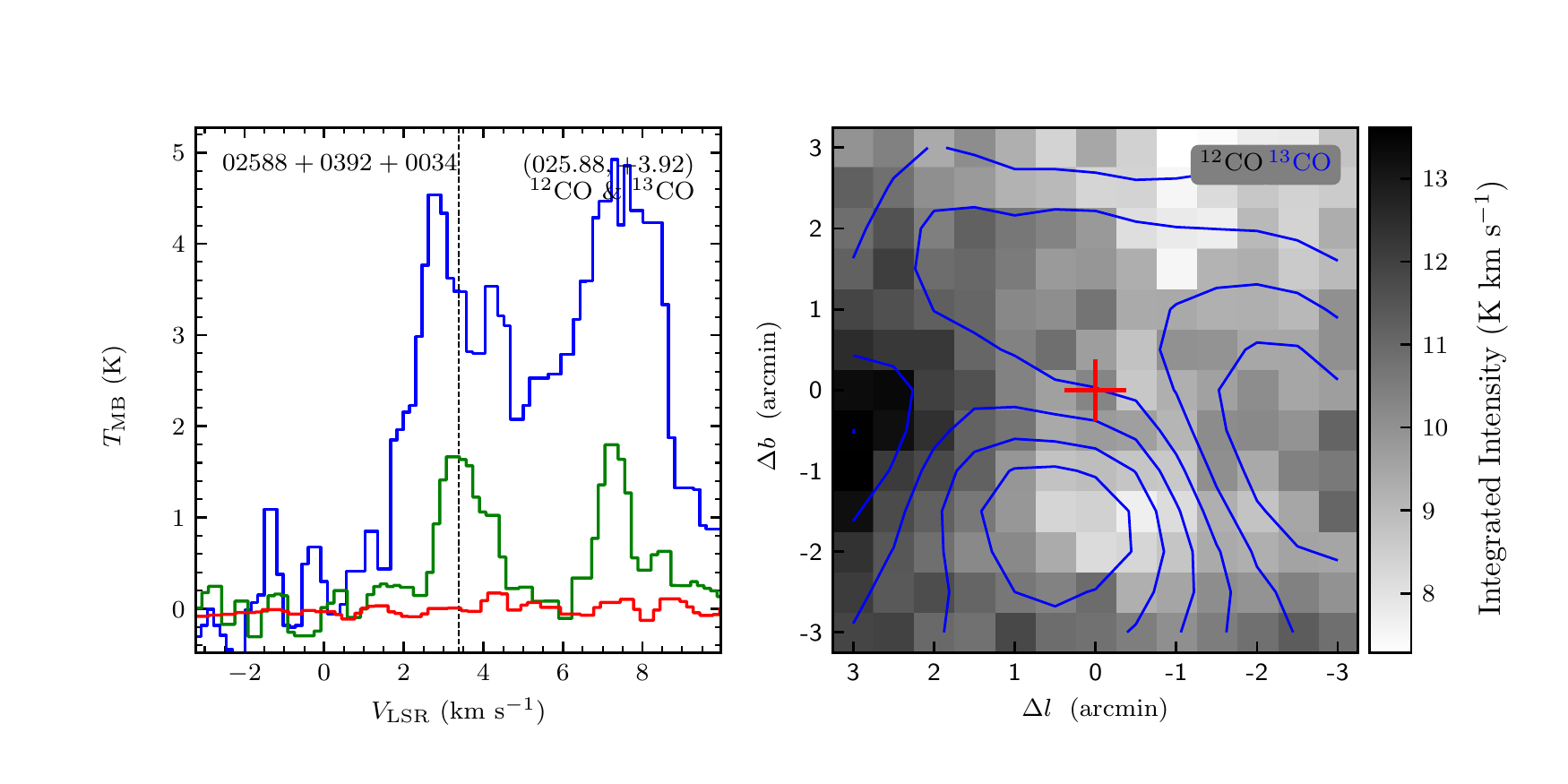}
\includegraphics[width=9.0cm,angle=0]{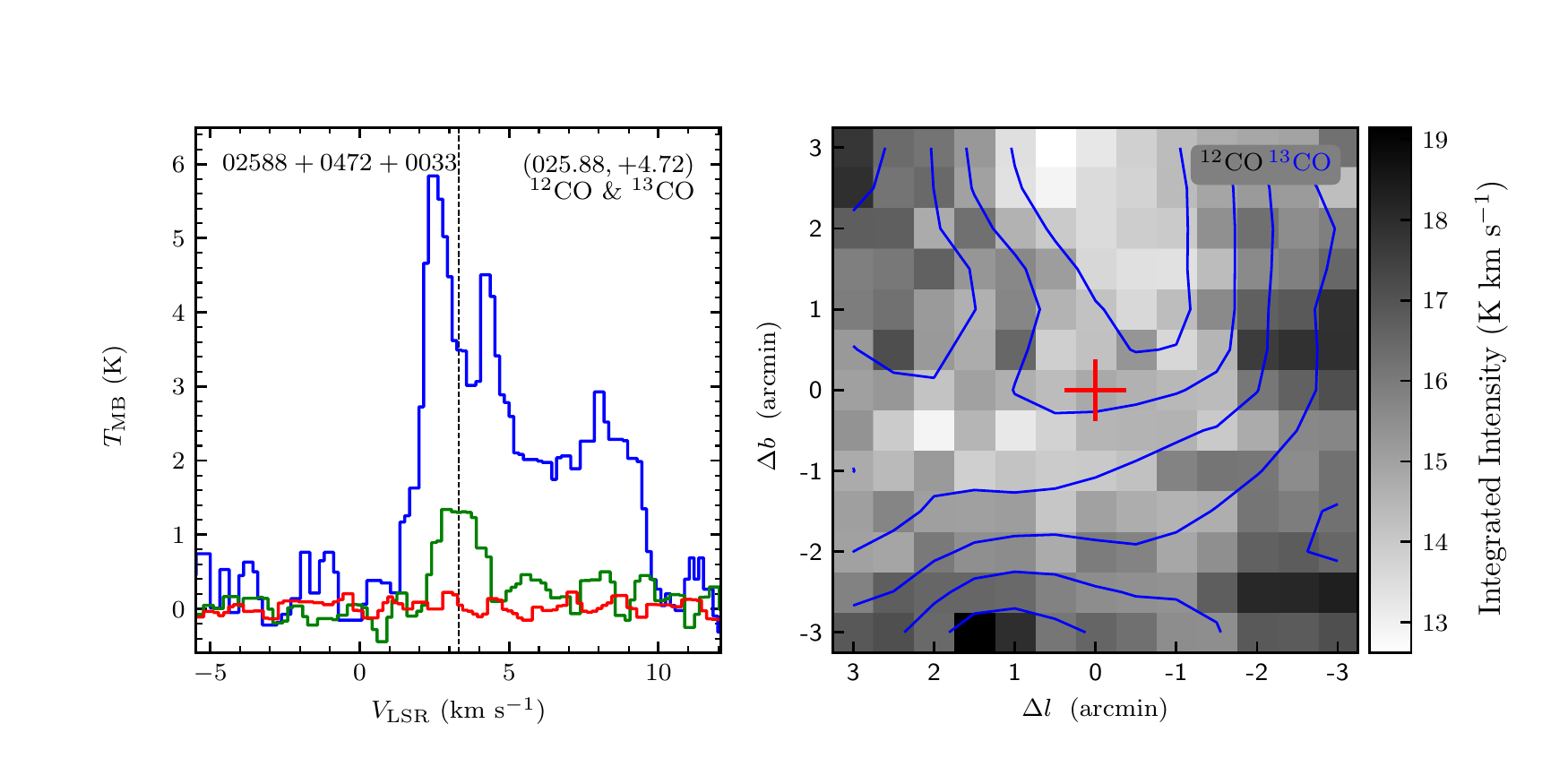}
\vspace{-0.5cm}

\includegraphics[width=9.0cm,angle=0]{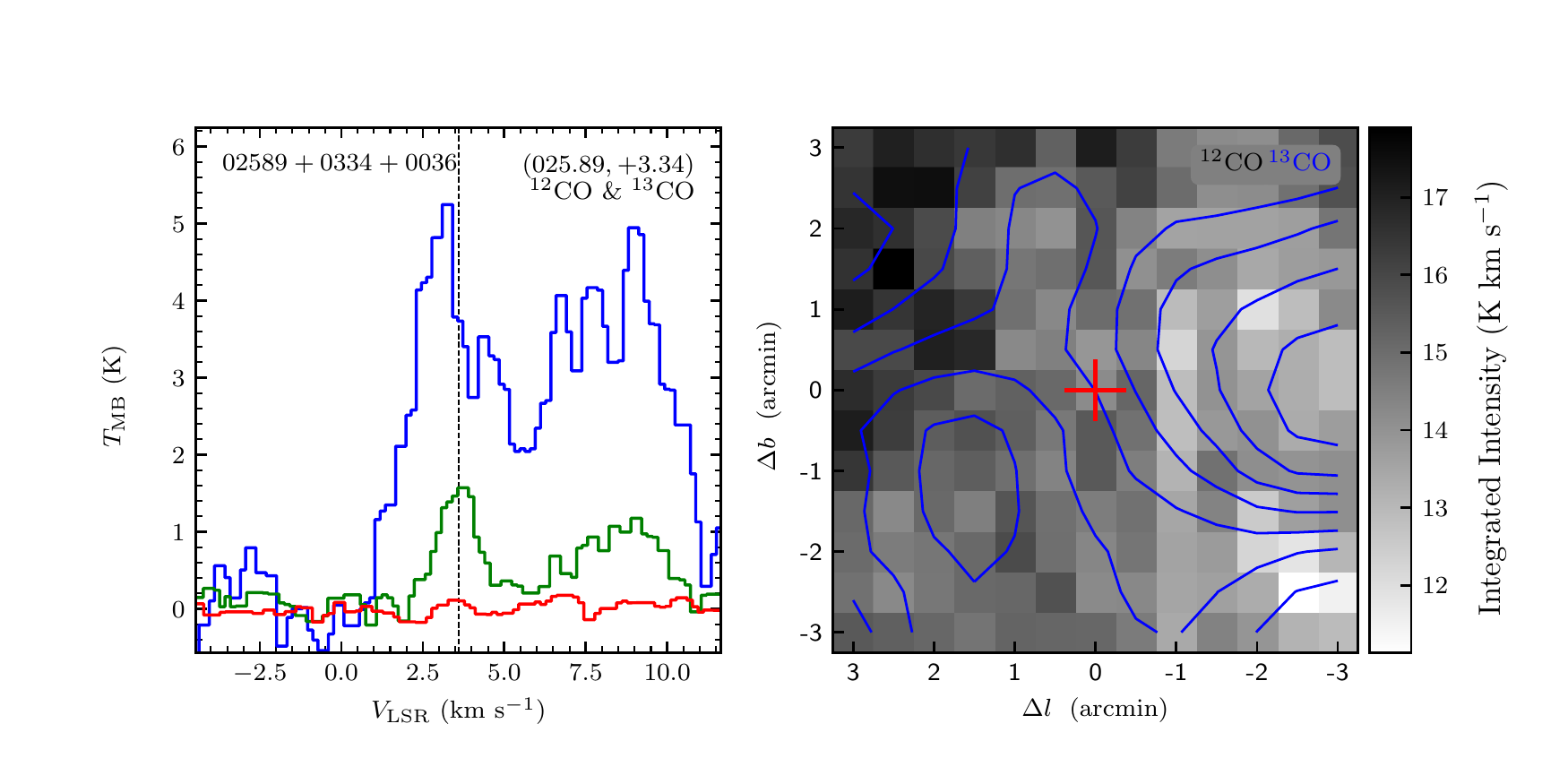}
\includegraphics[width=9.0cm,angle=0]{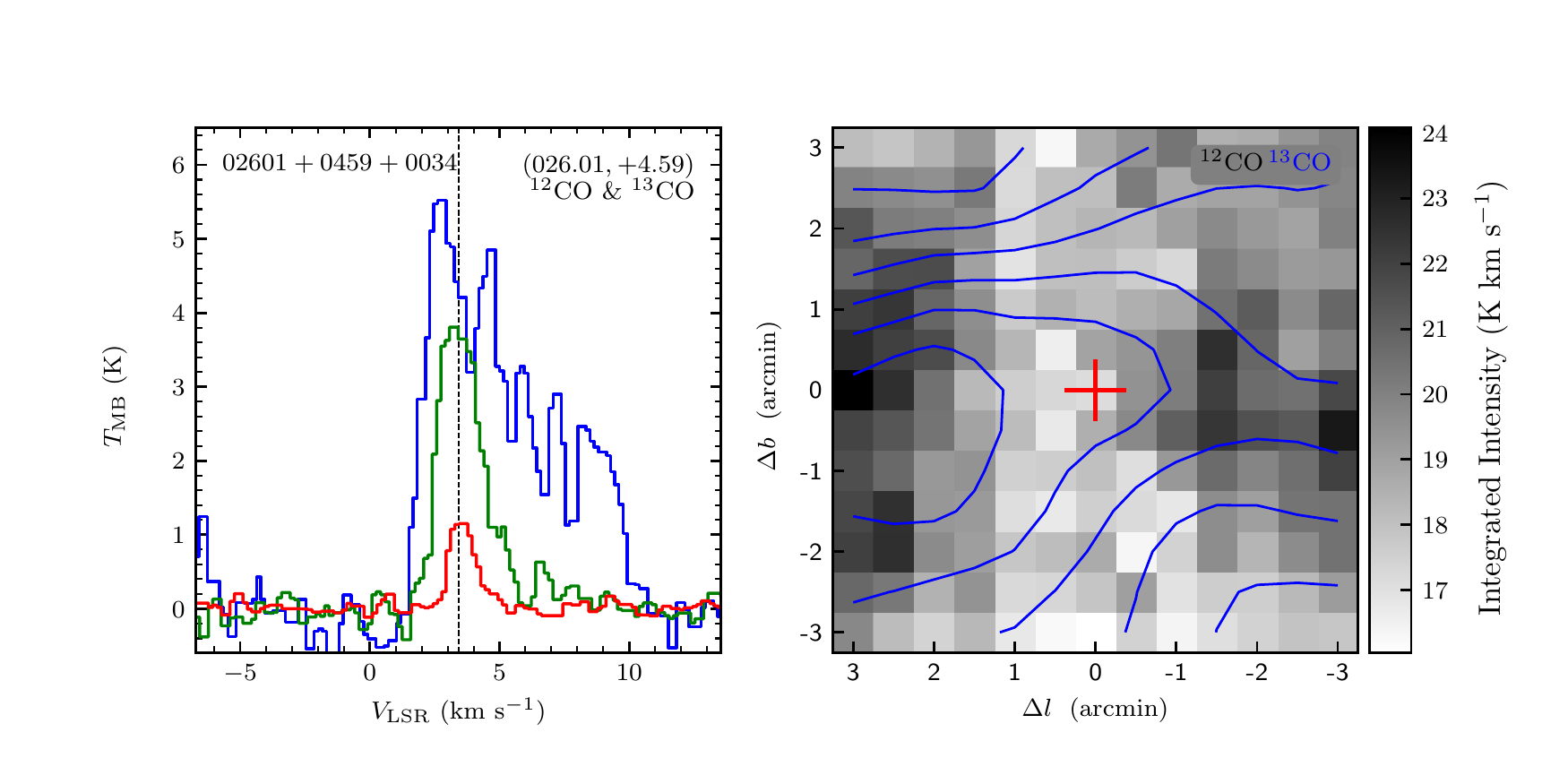}
\vspace{-0.5cm}

\includegraphics[width=9.0cm,angle=0]{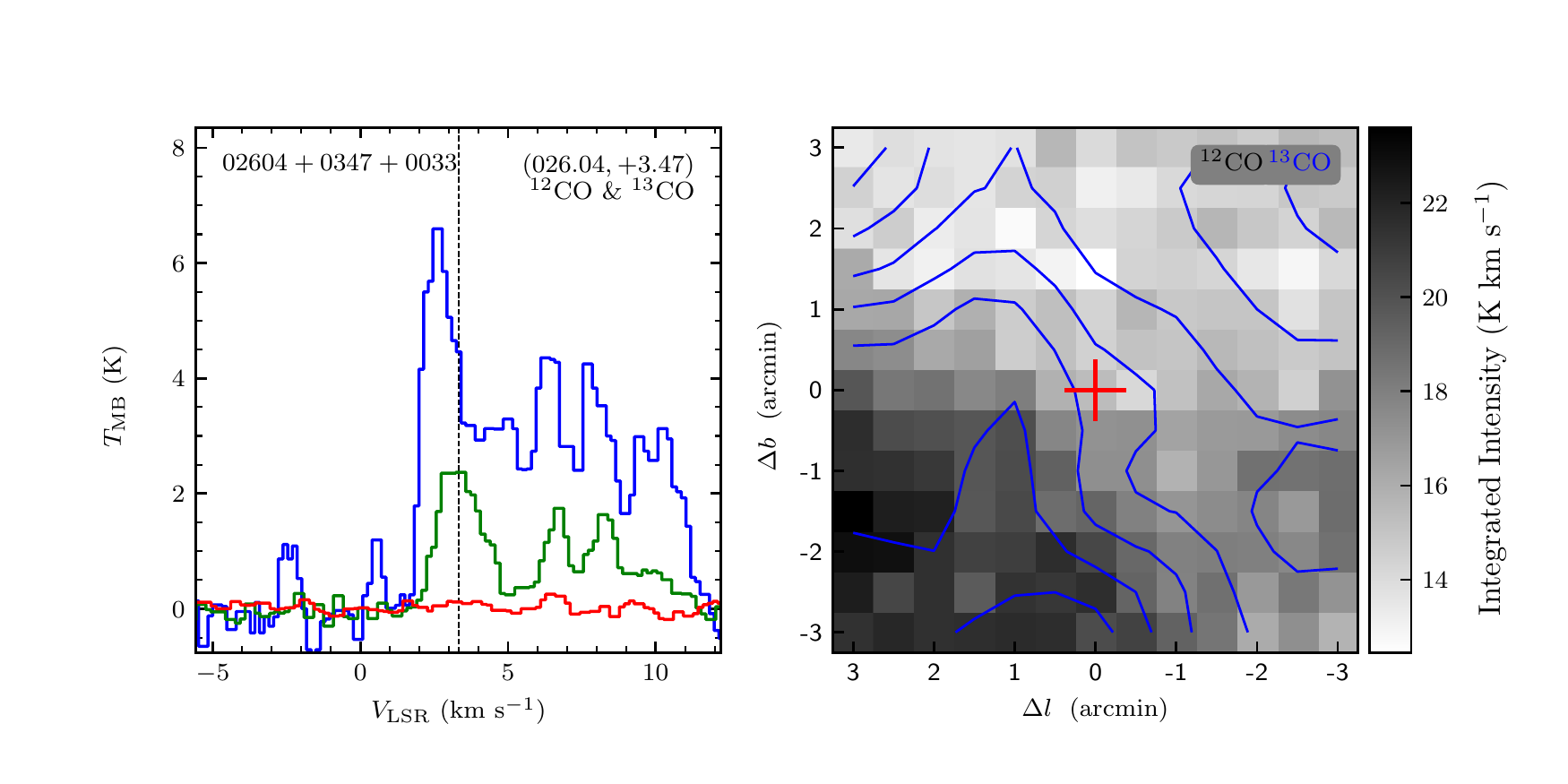}
\includegraphics[width=9.0cm,angle=0]{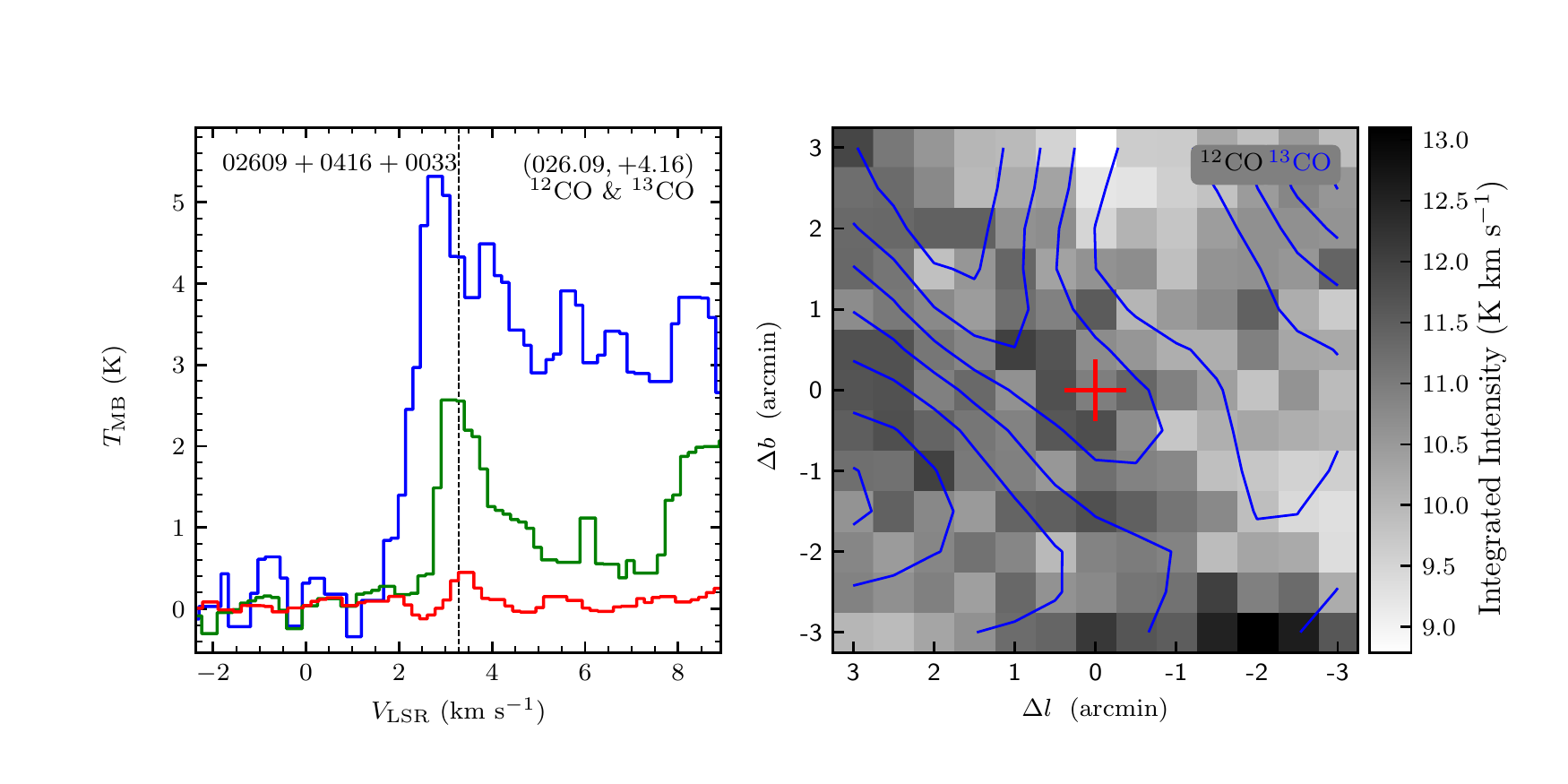}
\vspace{-0.5cm}

\includegraphics[width=9.0cm,angle=0]{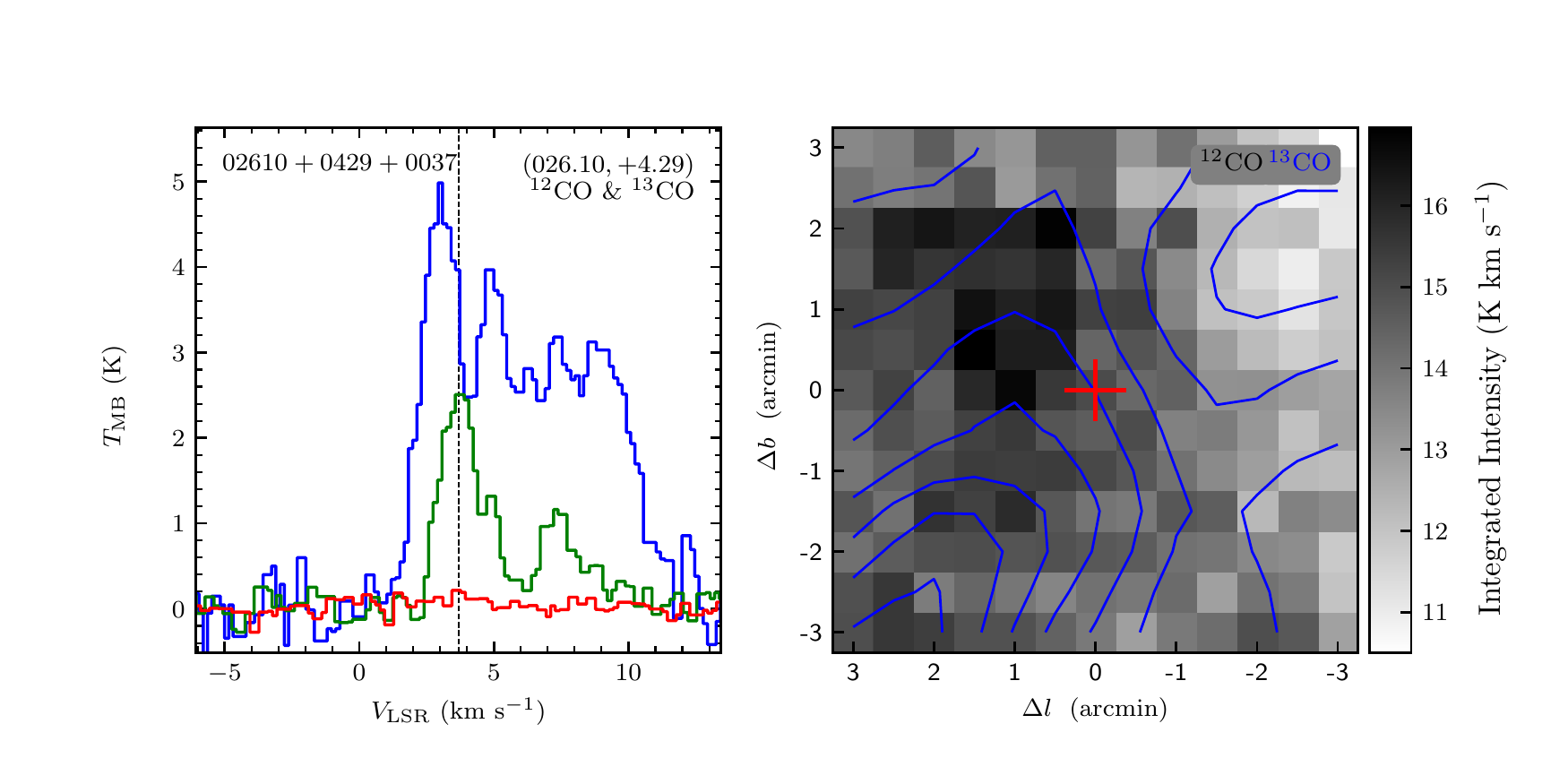}
\includegraphics[width=9.0cm,angle=0]{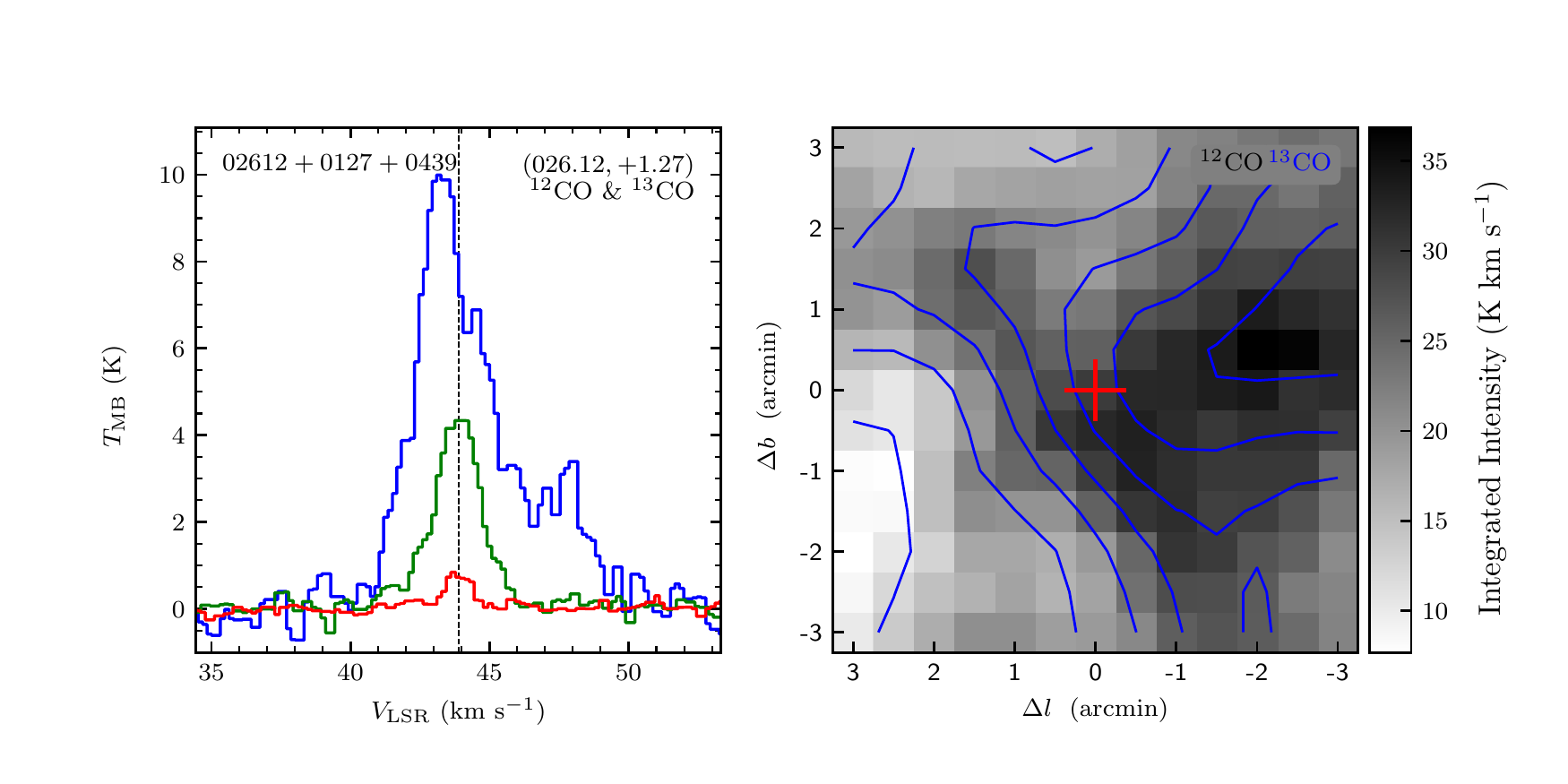}
\end{figure}
\clearpage

\begin{figure}
\includegraphics[width=9.0cm,angle=0]{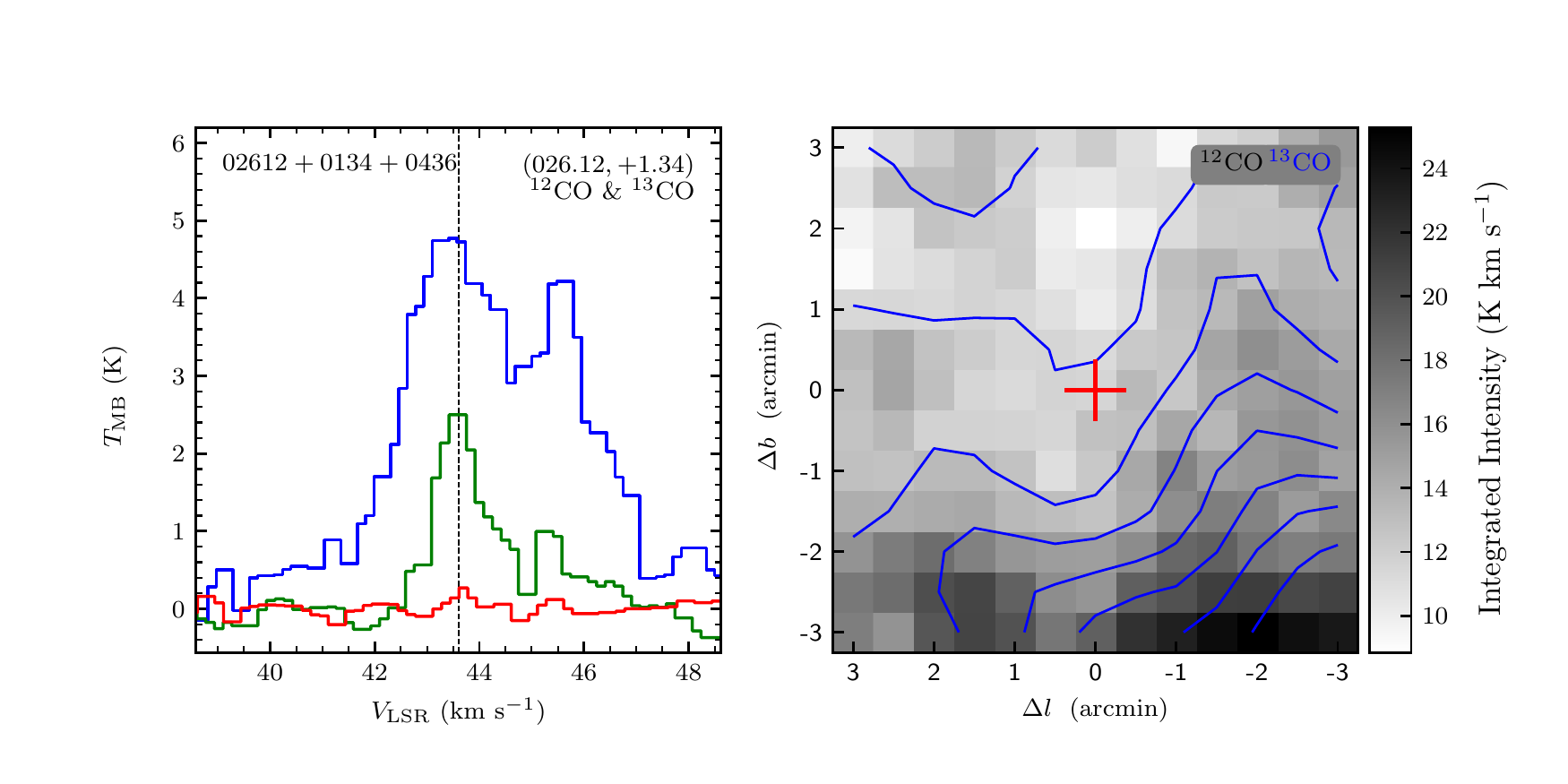}
\includegraphics[width=9.0cm,angle=0]{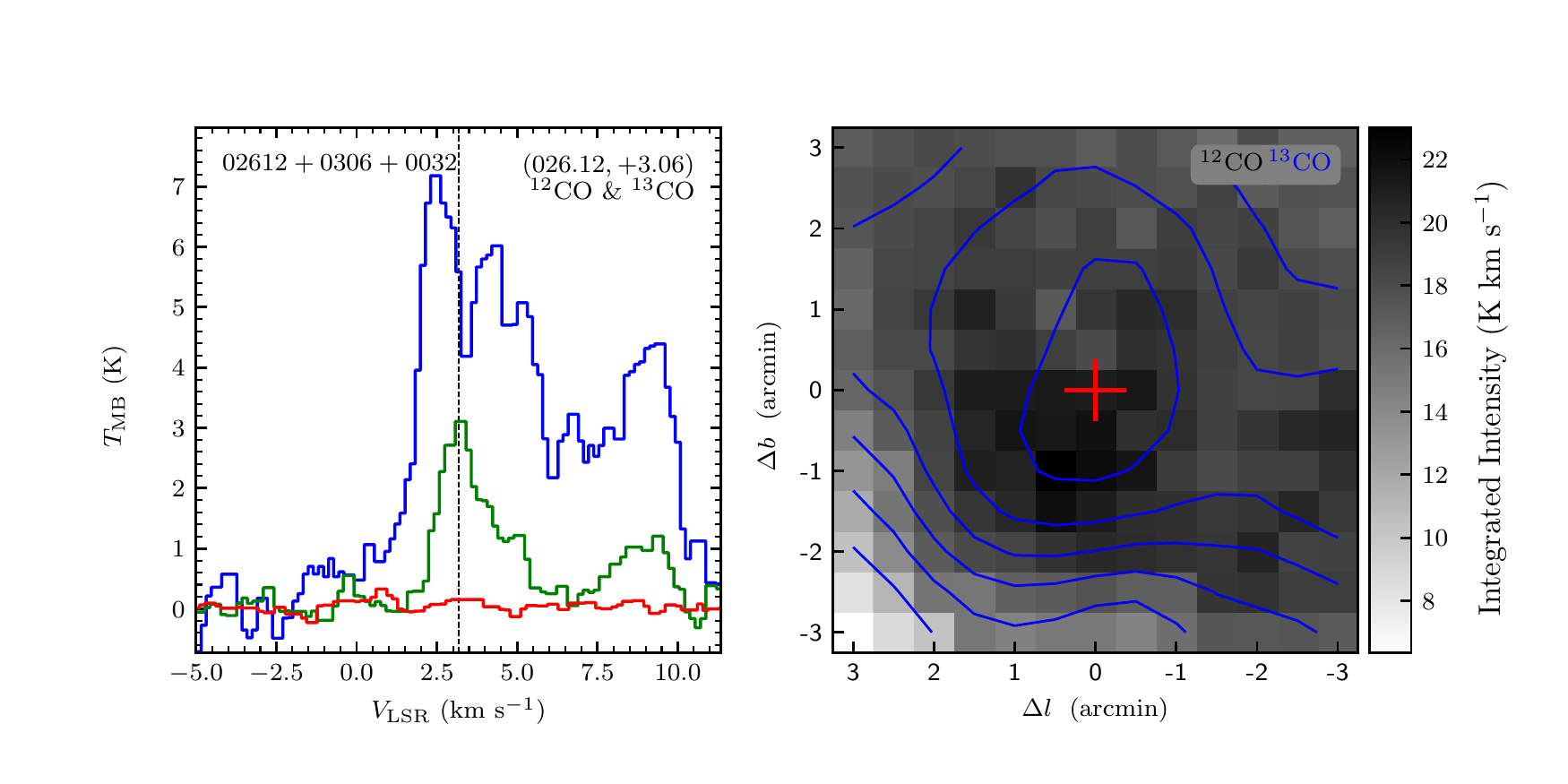}
\vspace{-0.5cm}

\includegraphics[width=9.0cm,angle=0]{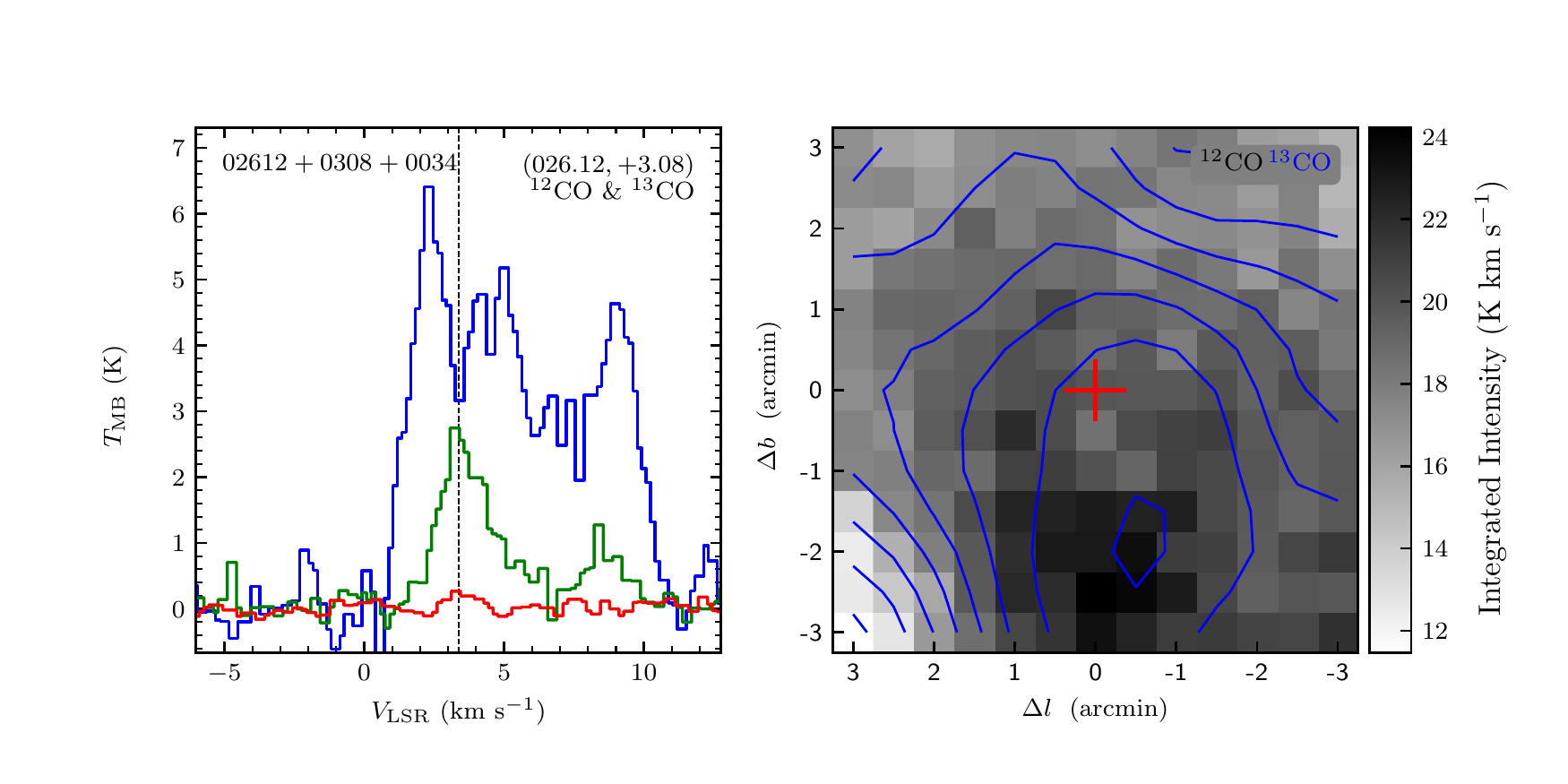}
\includegraphics[width=9.0cm,angle=0]{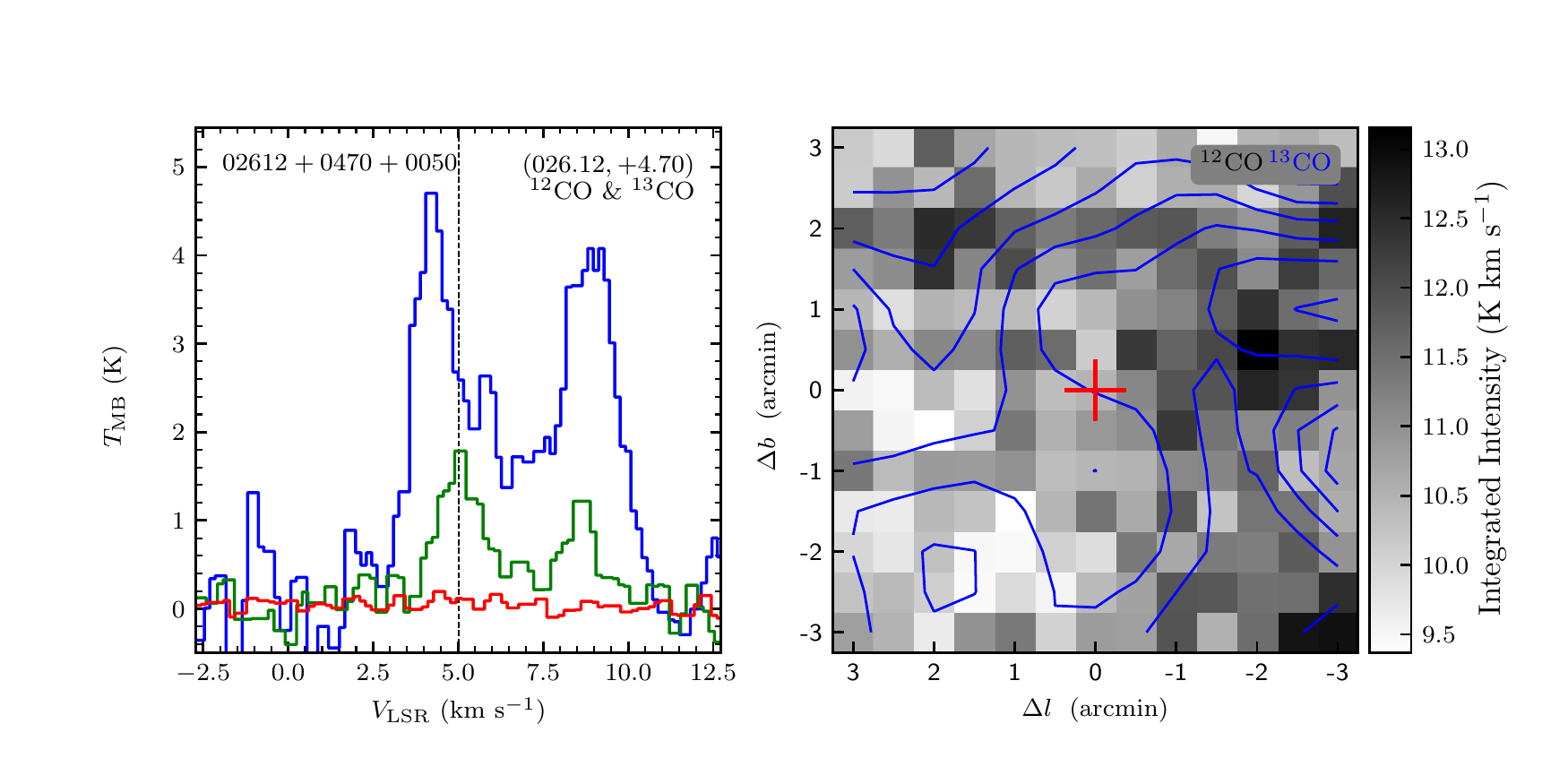}
\vspace{-0.5cm}

\includegraphics[width=9.0cm,angle=0]{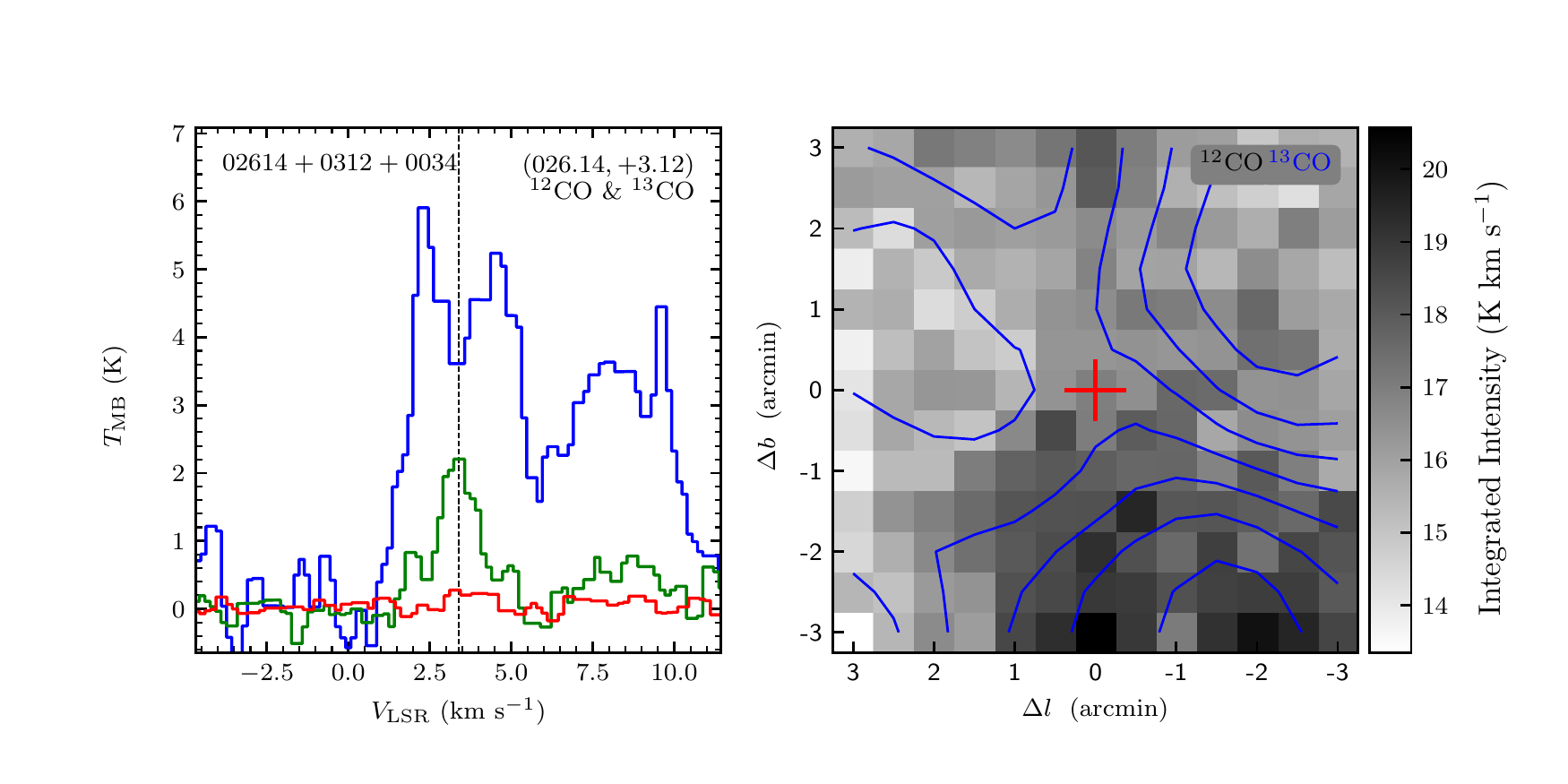}
\includegraphics[width=9.0cm,angle=0]{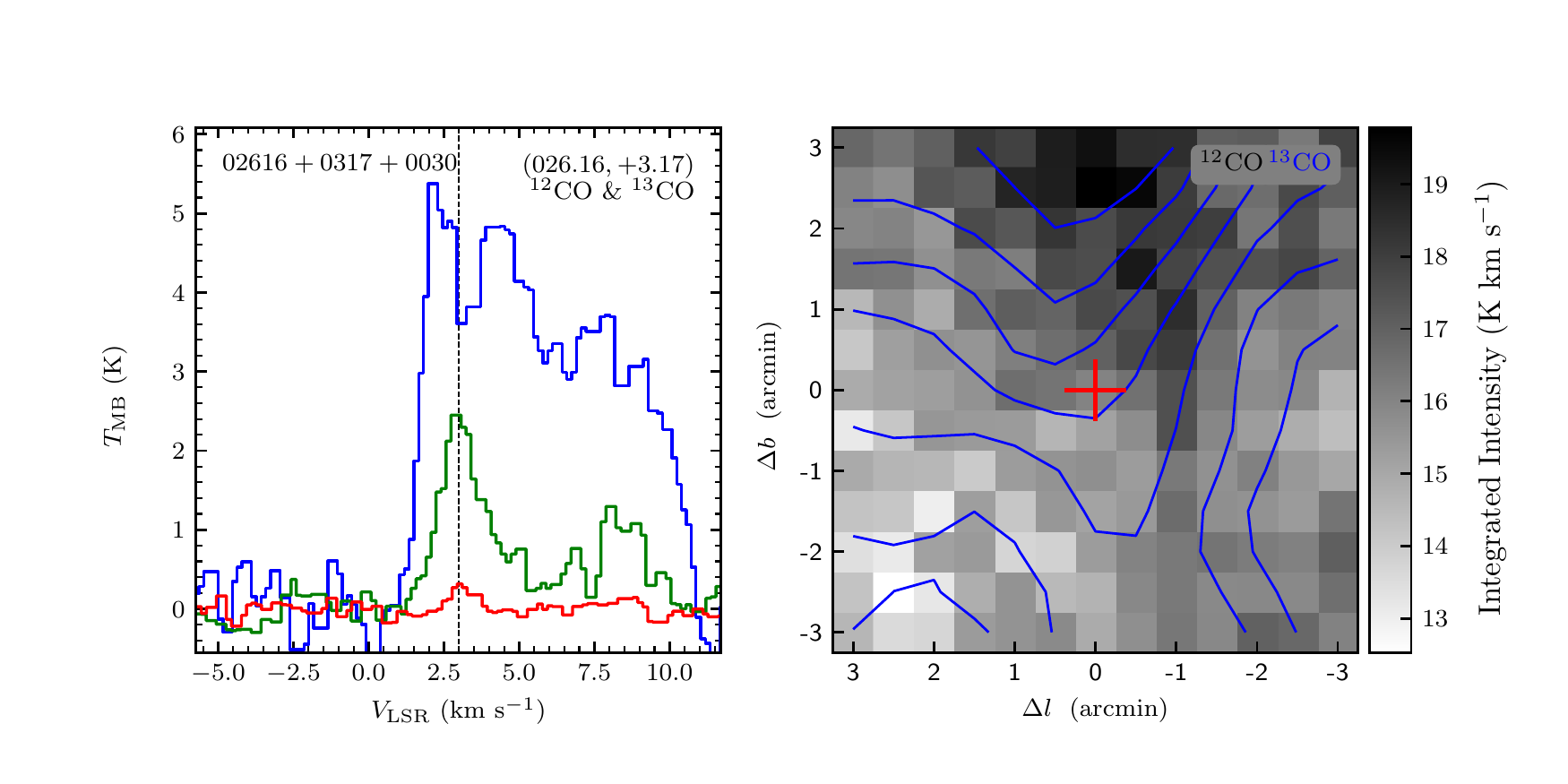}
\vspace{-0.5cm}

\includegraphics[width=9.0cm,angle=0]{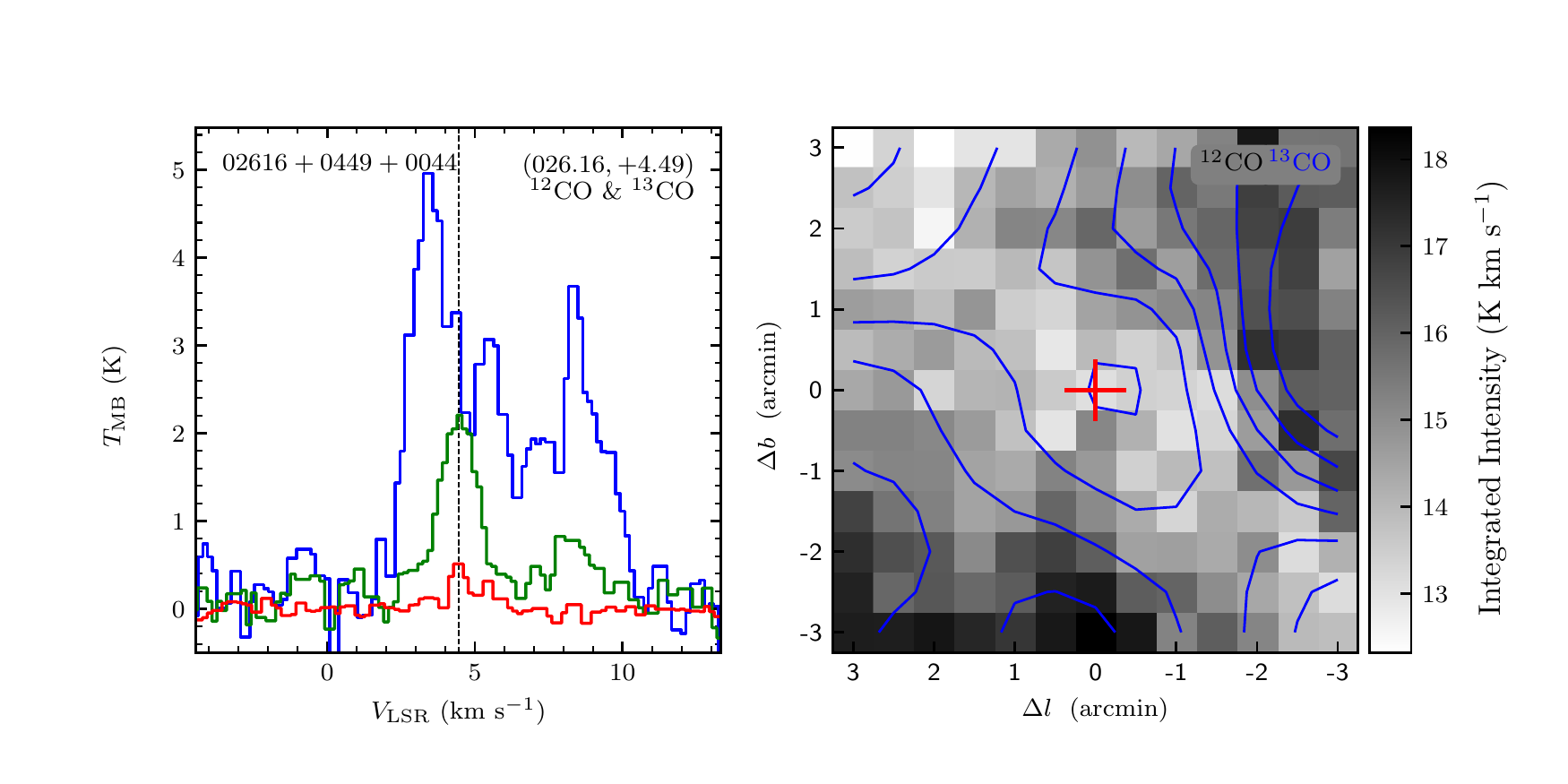}
\includegraphics[width=9.0cm,angle=0]{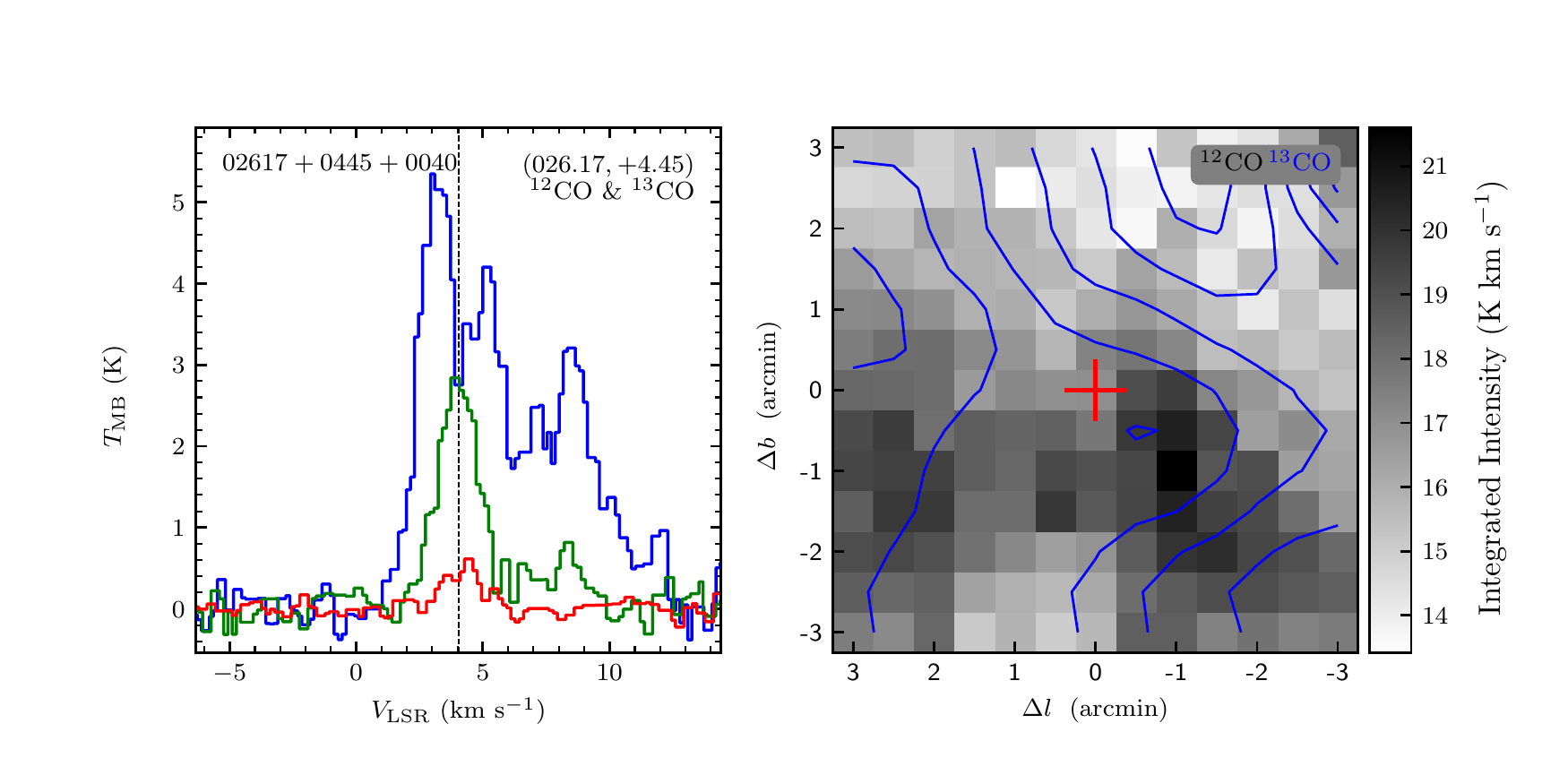}
\vspace{-0.5cm}

\includegraphics[width=9.0cm,angle=0]{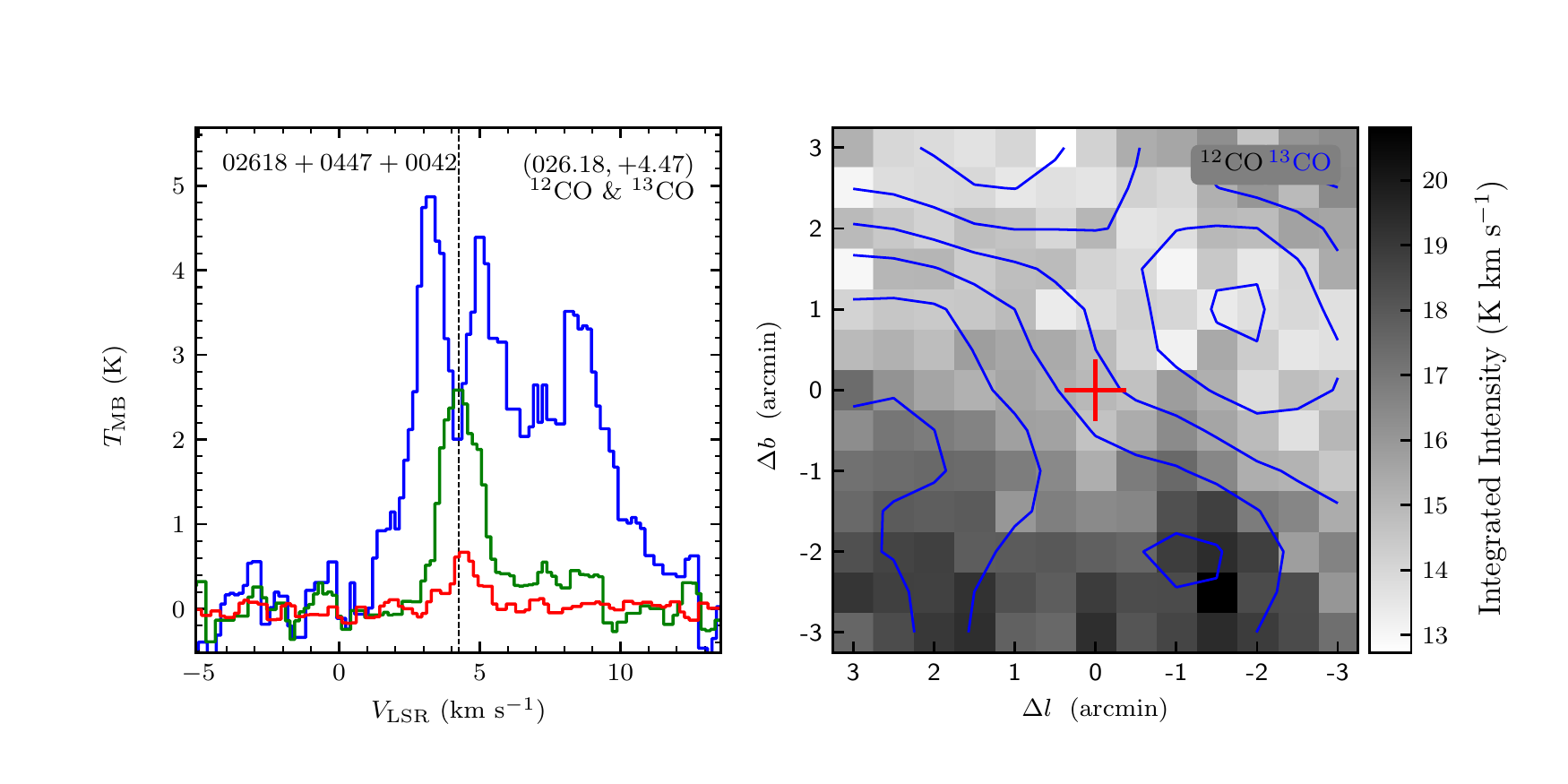}
\includegraphics[width=9.0cm,angle=0]{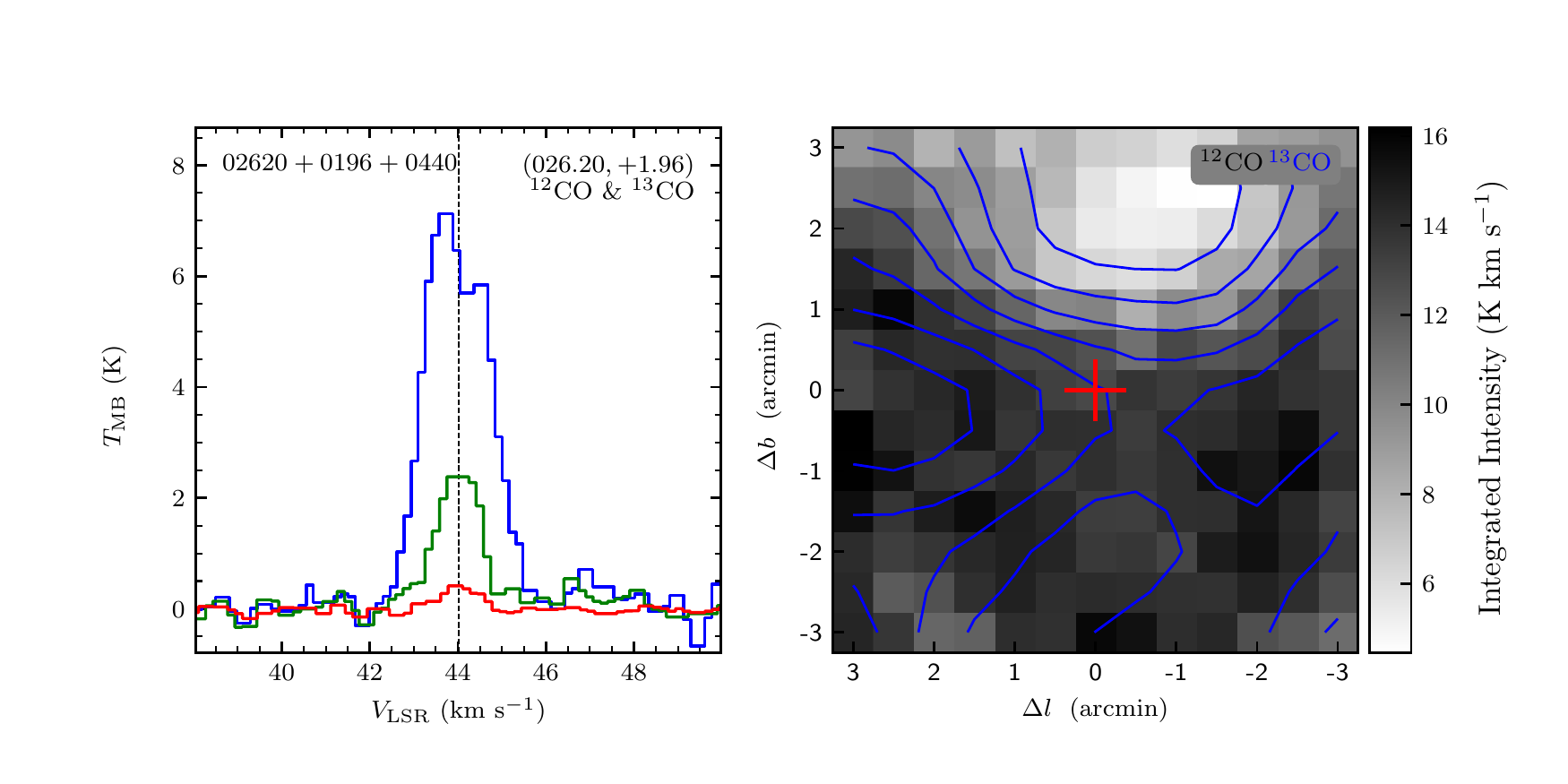}
\end{figure}
\clearpage

\begin{figure}
\includegraphics[width=9.0cm,angle=0]{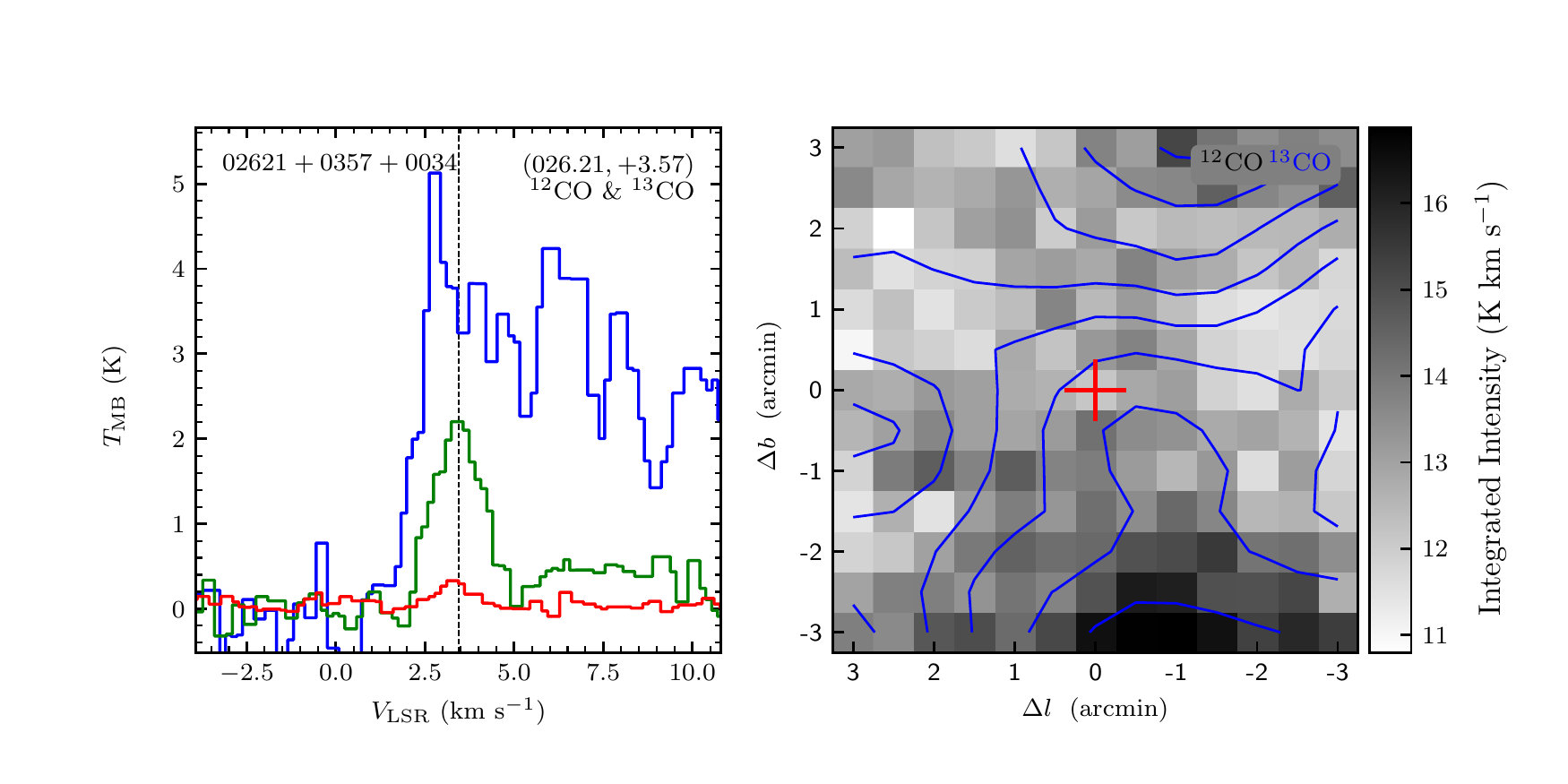}
\includegraphics[width=9.0cm,angle=0]{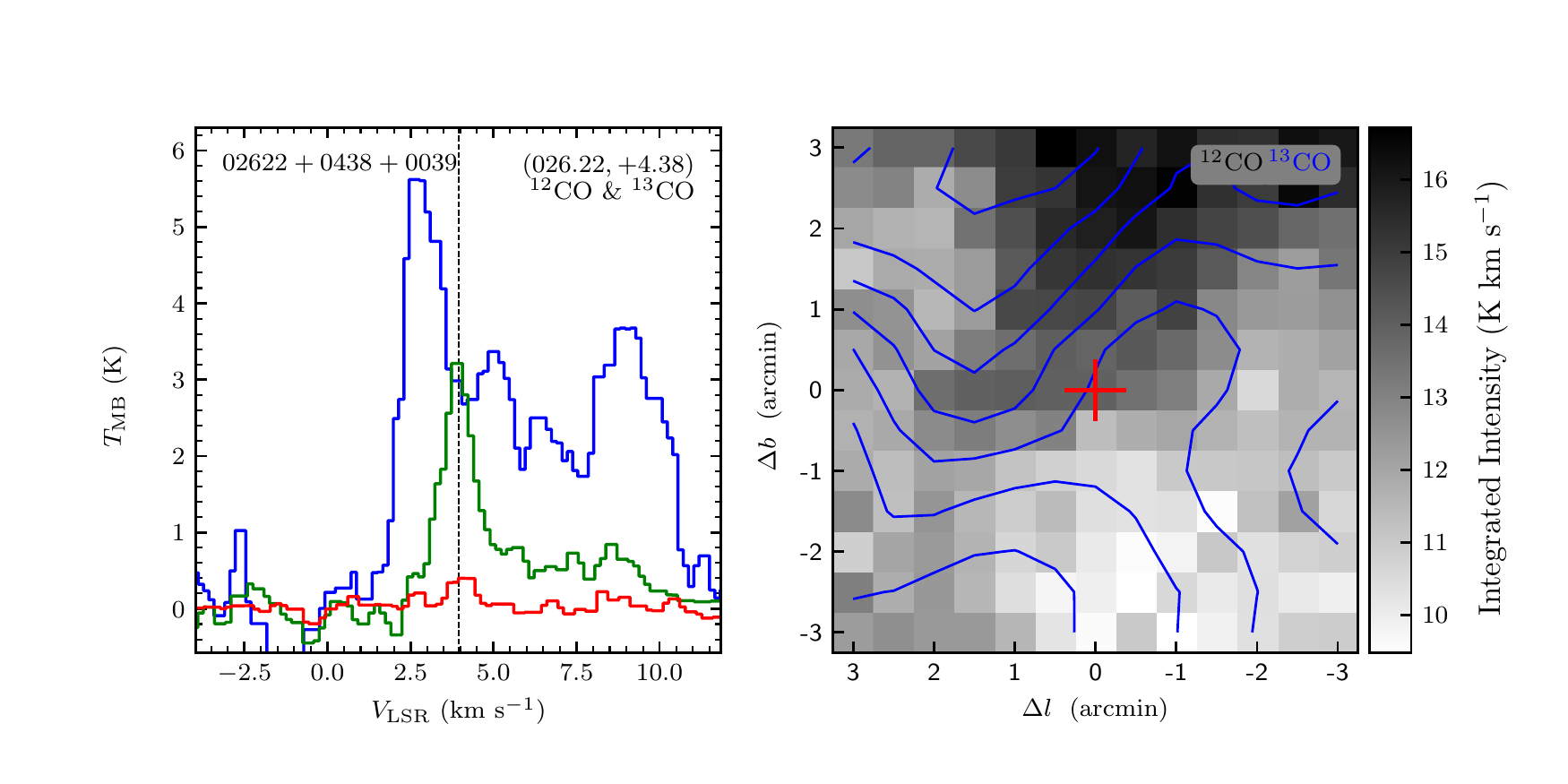}
\vspace{-0.5cm}

\includegraphics[width=9.0cm,angle=0]{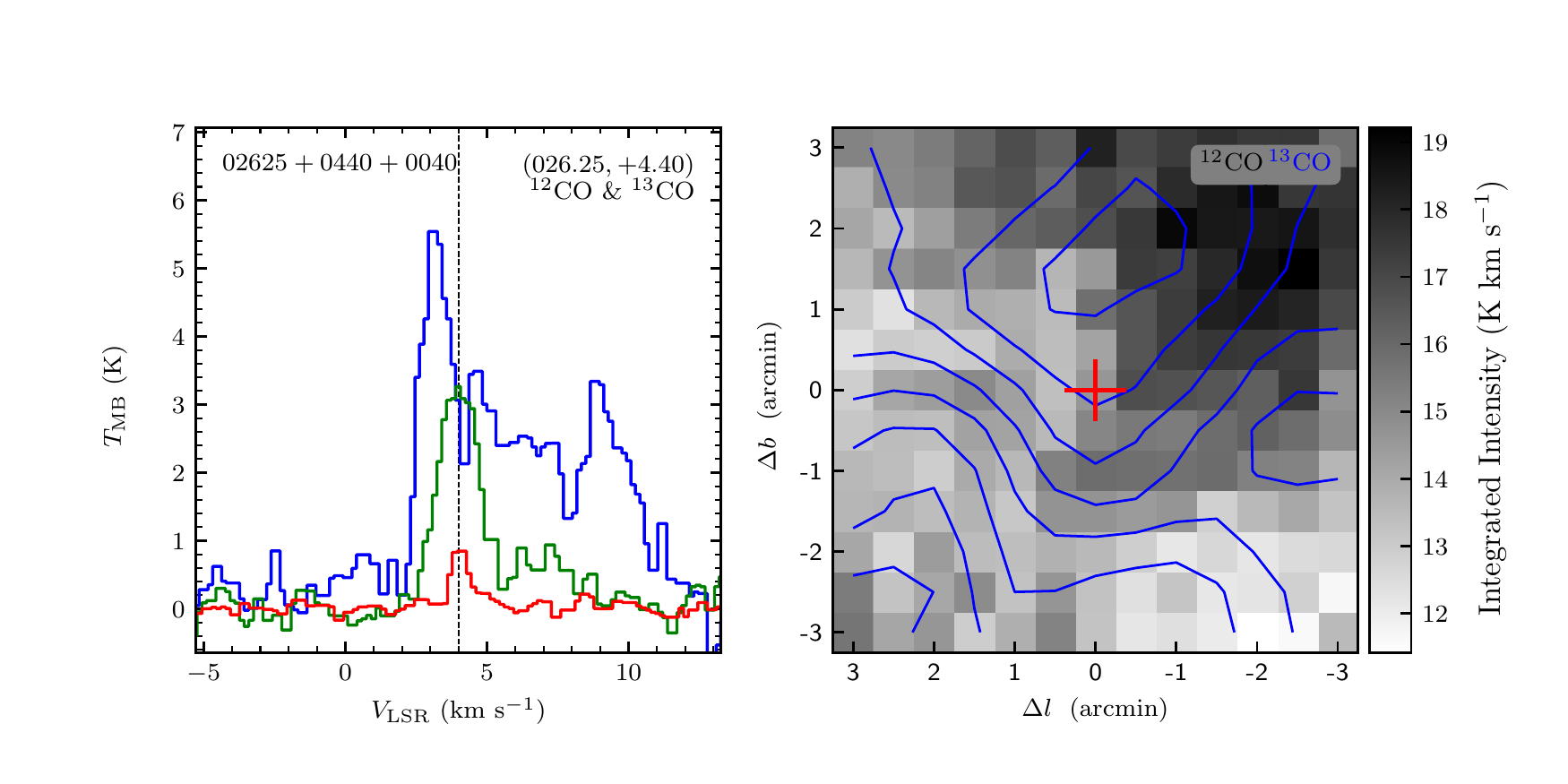}
\includegraphics[width=9.0cm,angle=0]{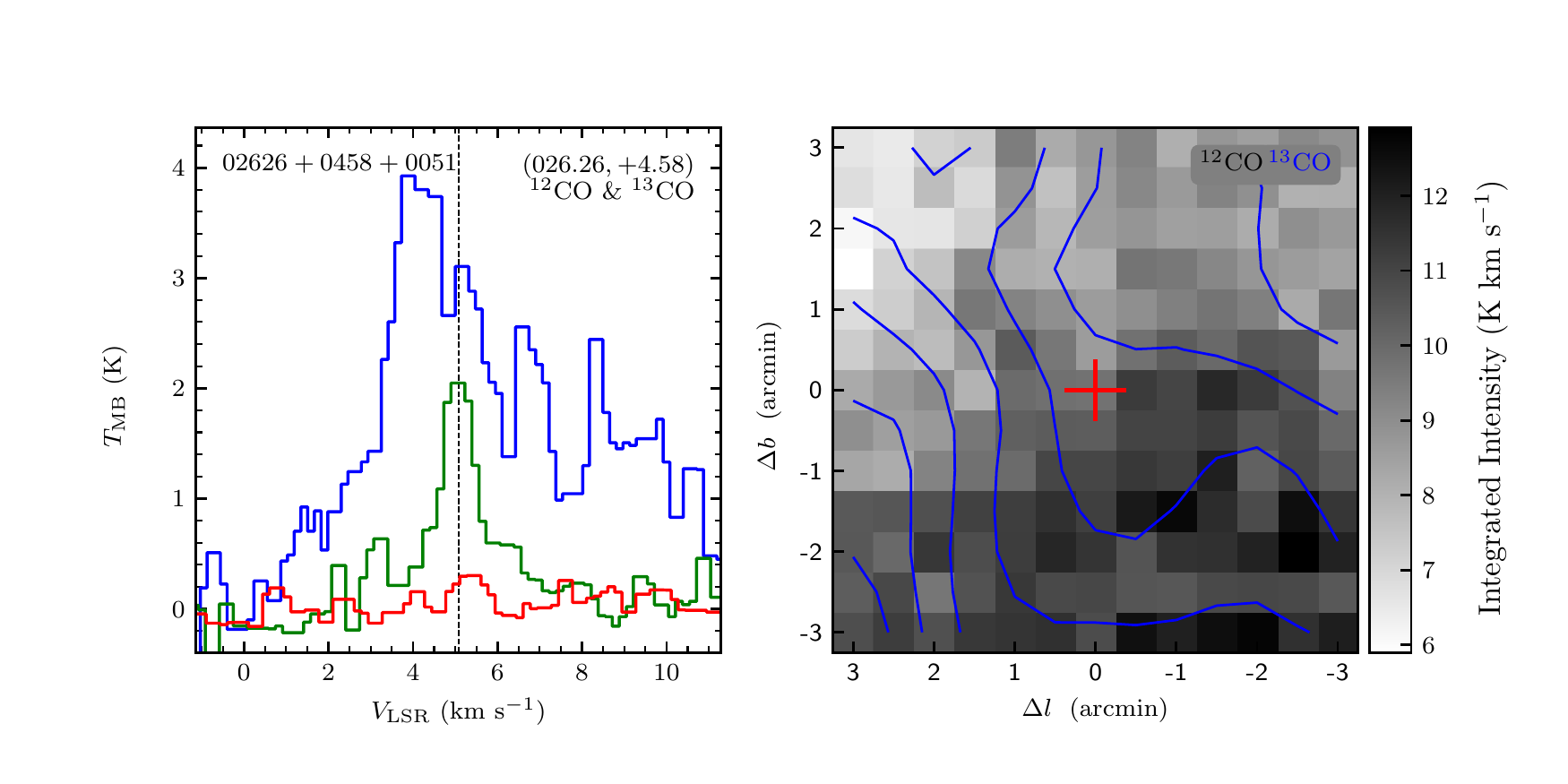}
\vspace{-0.5cm}

\includegraphics[width=9.0cm,angle=0]{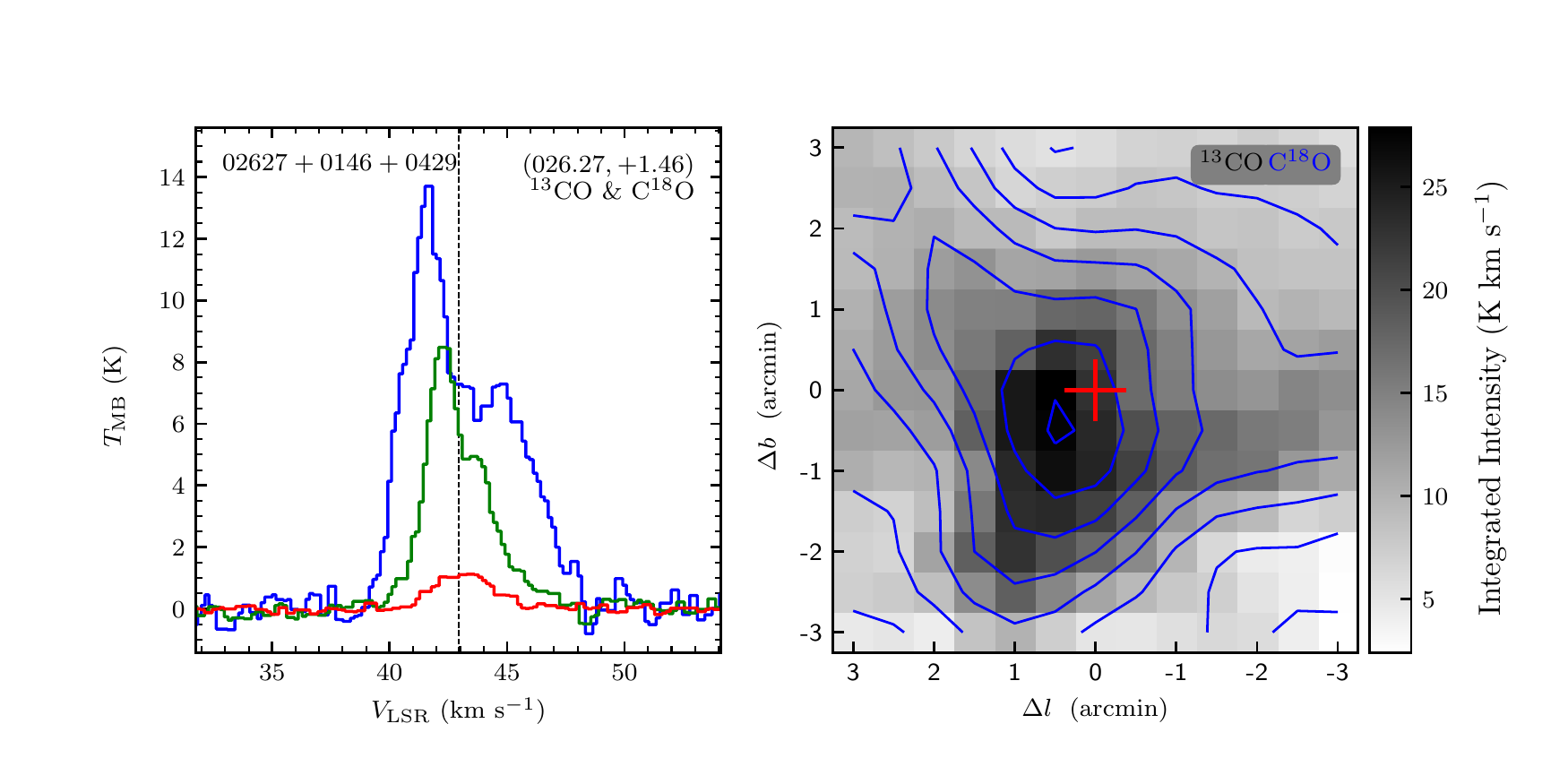}
\includegraphics[width=9.0cm,angle=0]{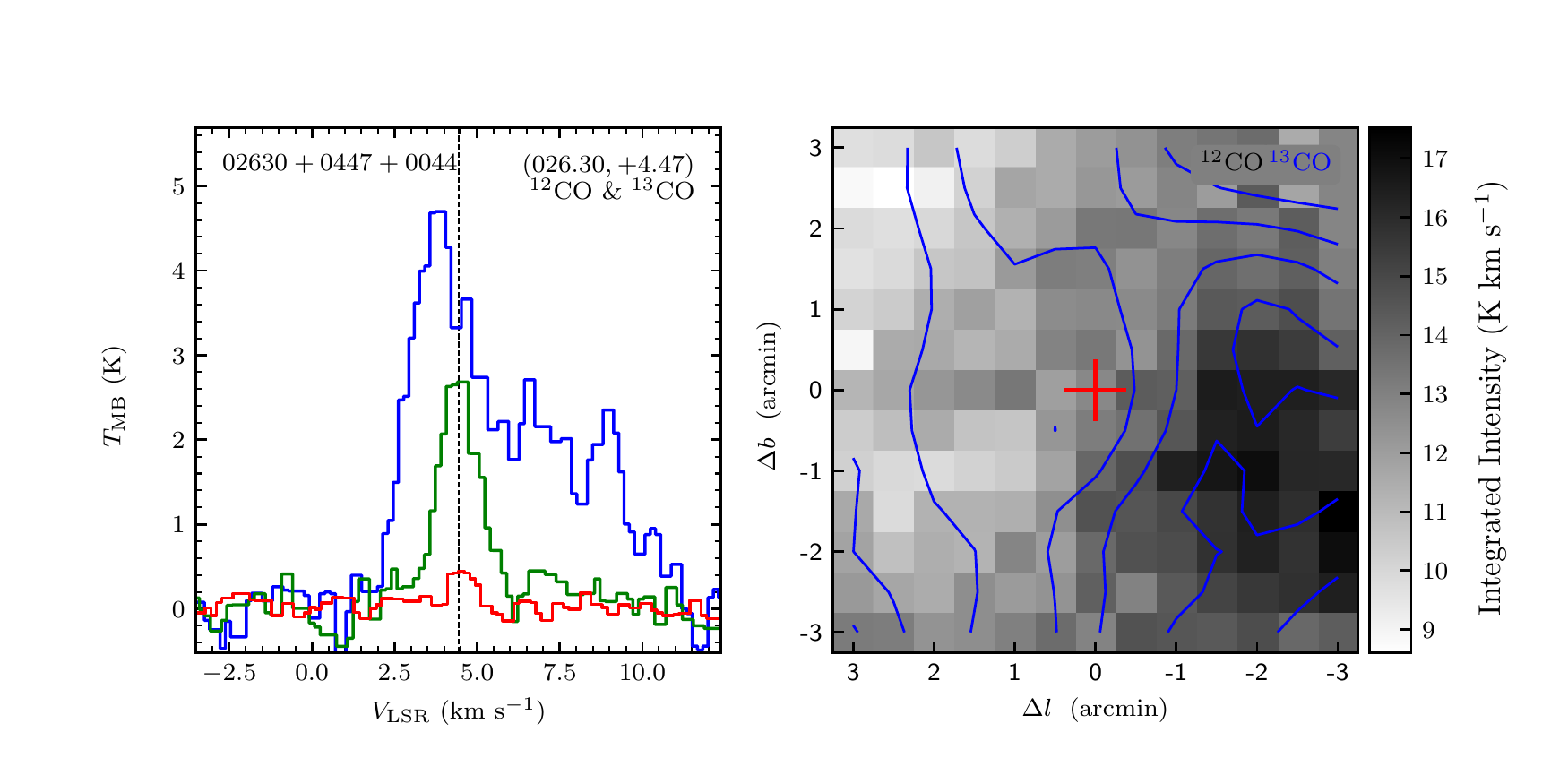}
\vspace{-0.5cm}

\includegraphics[width=9.0cm,angle=0]{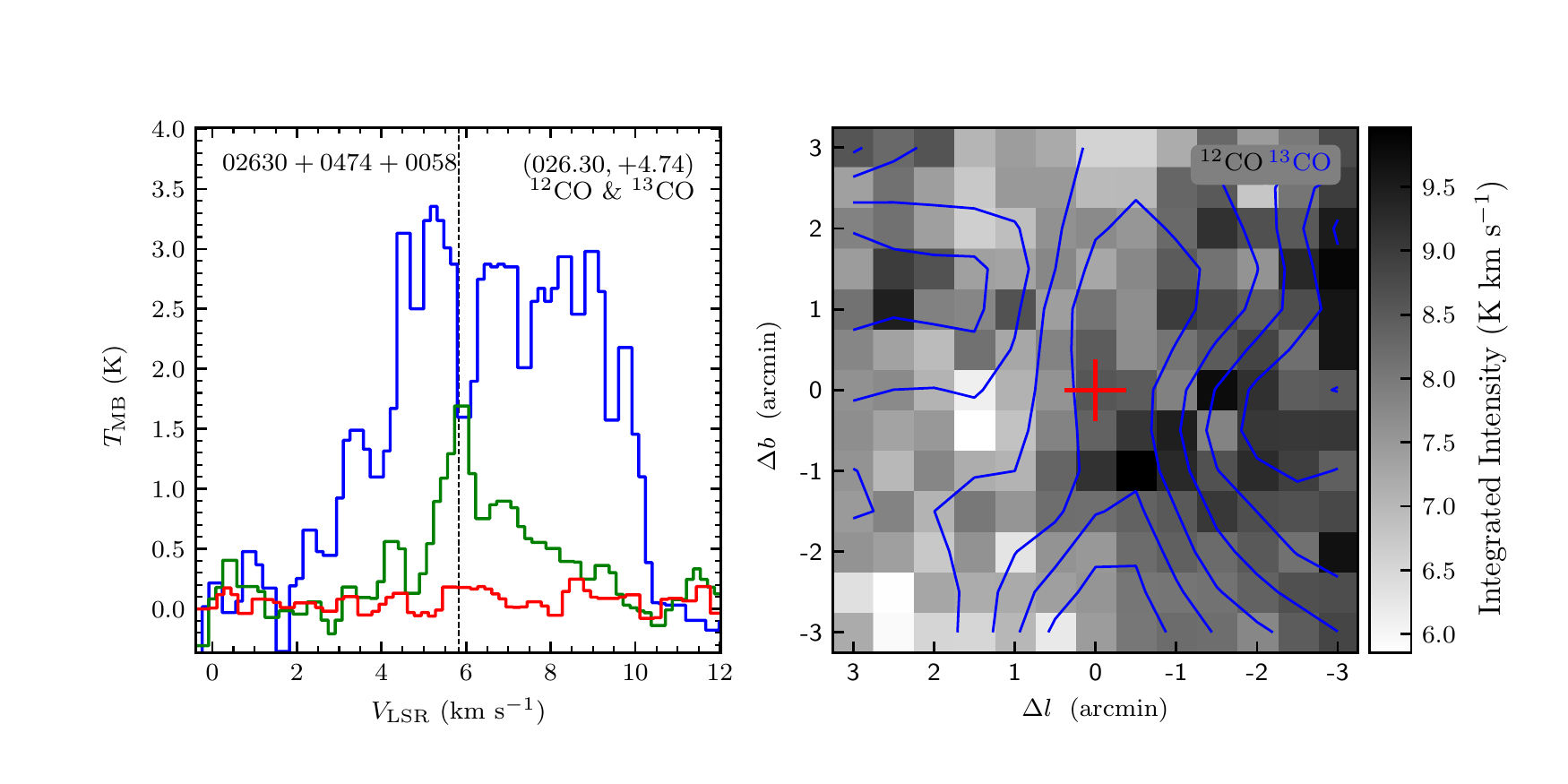}
\includegraphics[width=9.0cm,angle=0]{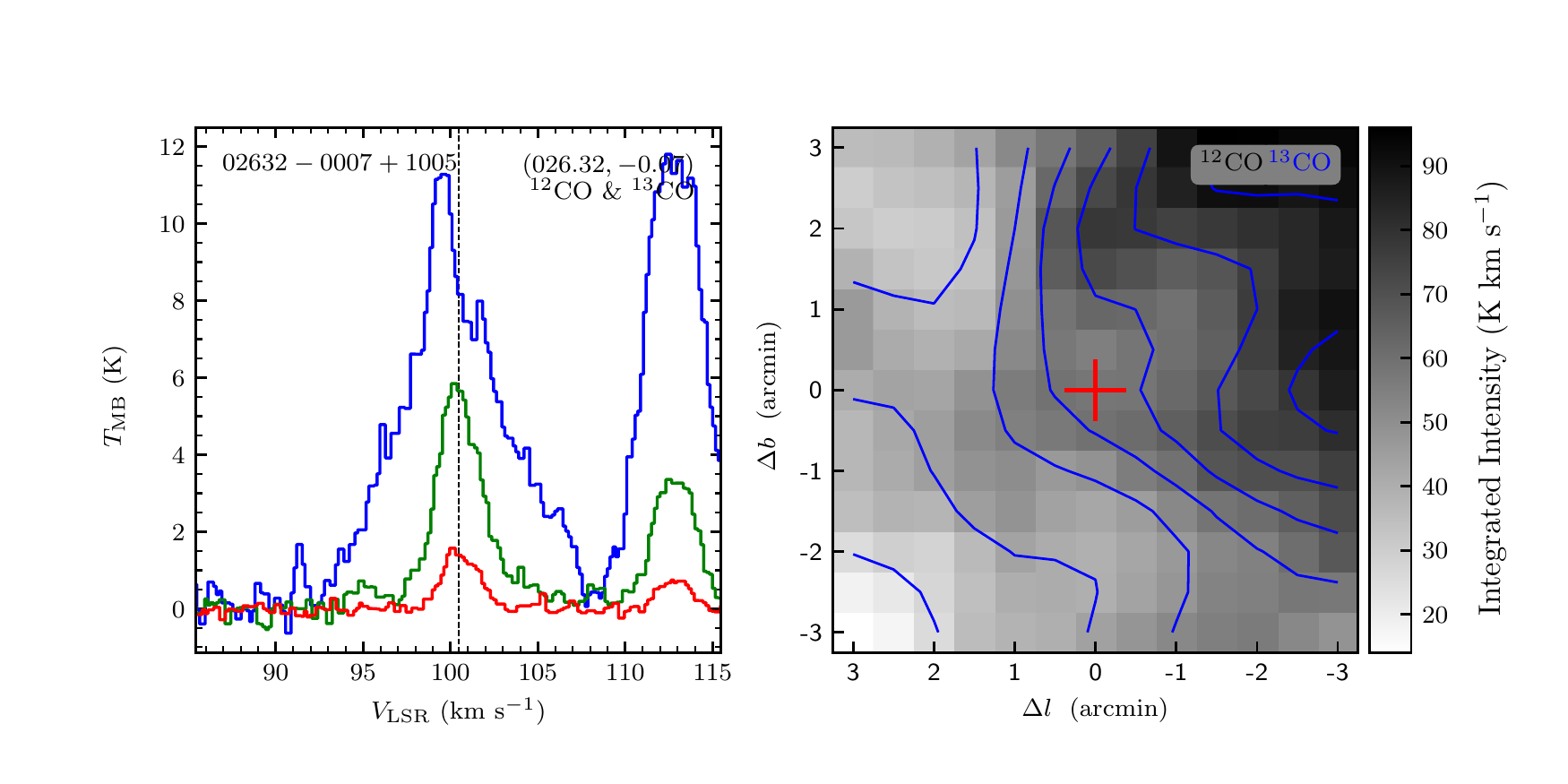}
\vspace{-0.5cm}

\includegraphics[width=9.0cm,angle=0]{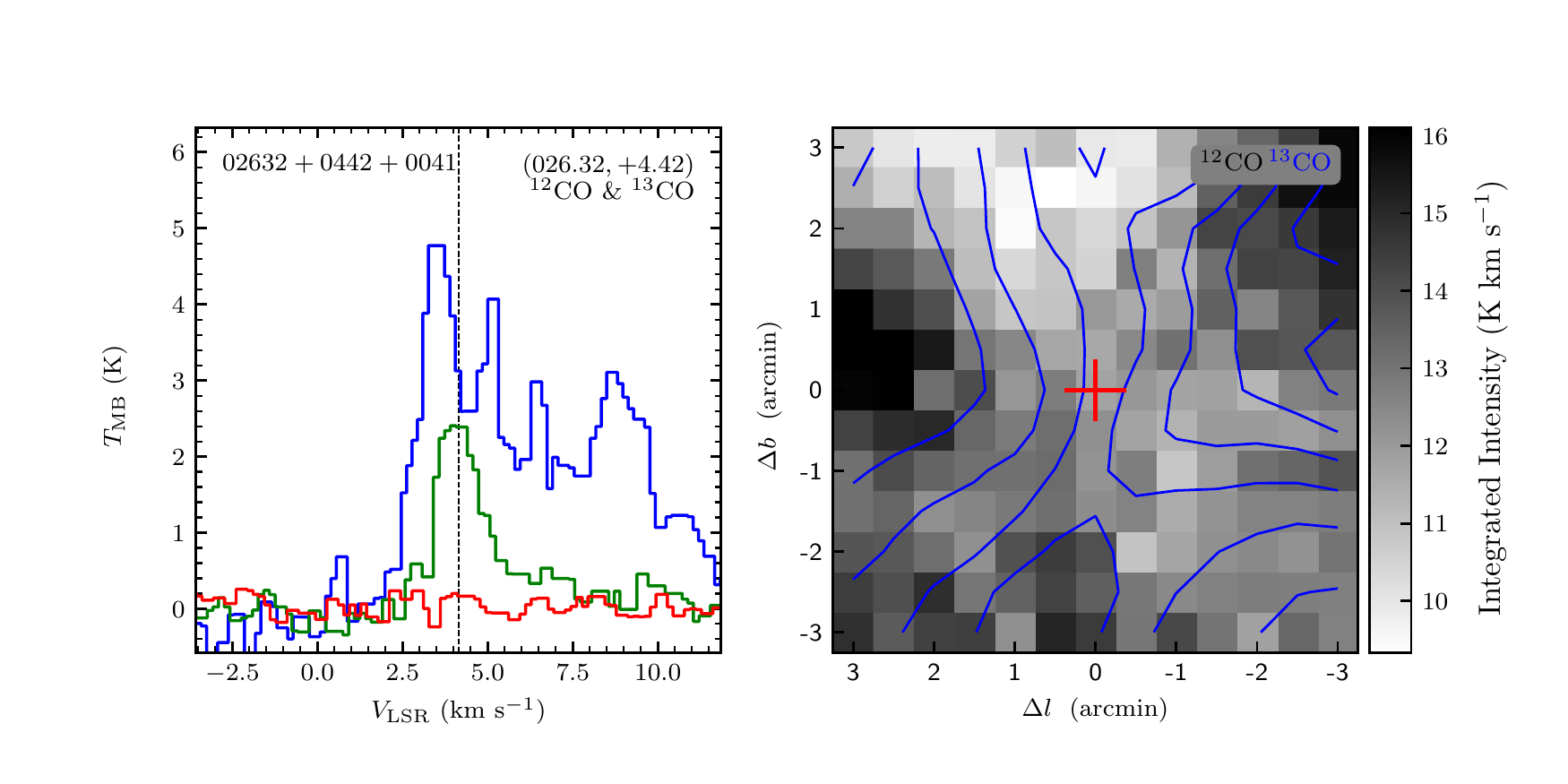}
\includegraphics[width=9.0cm,angle=0]{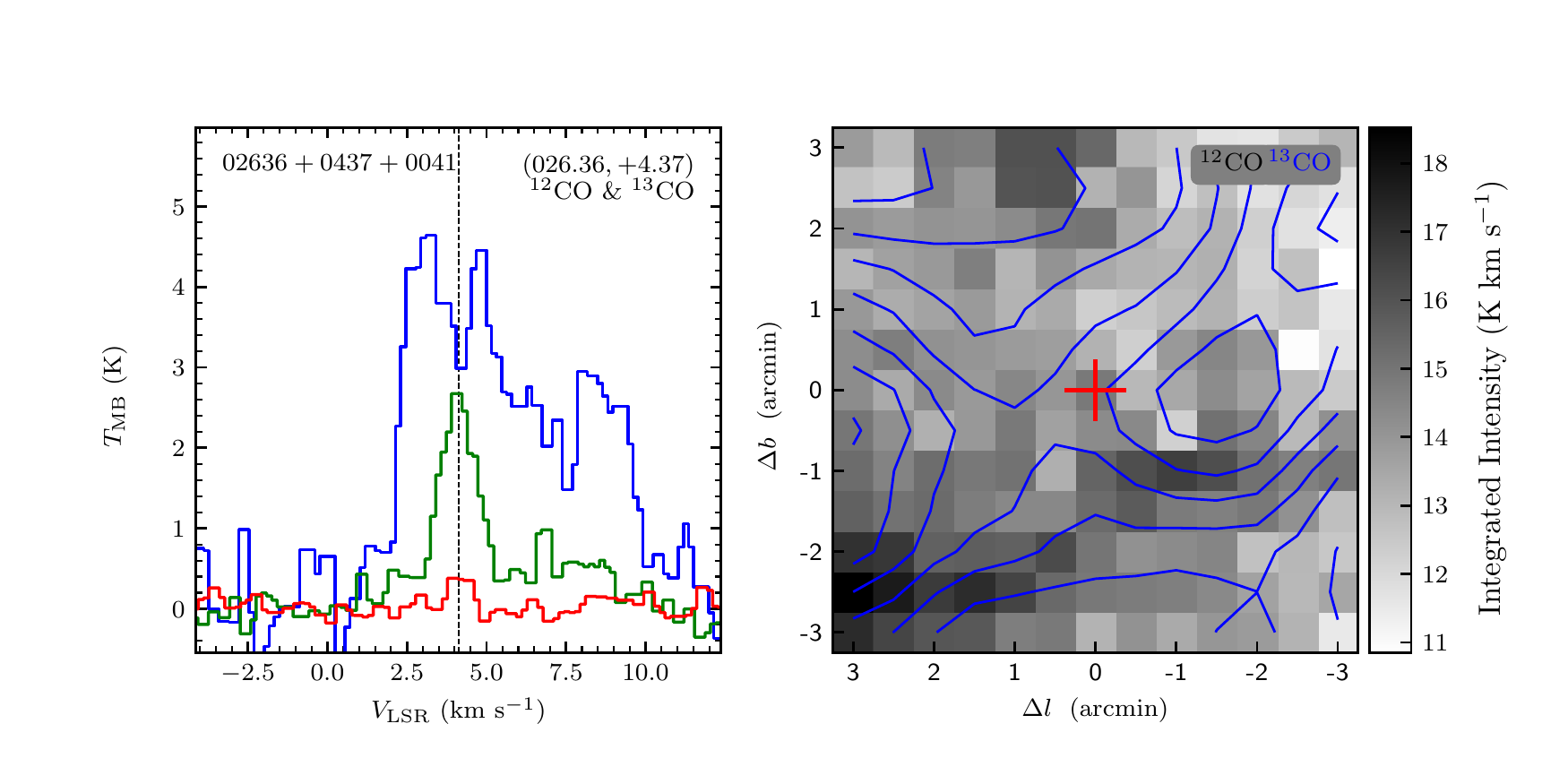}
\end{figure}
\clearpage

\begin{figure}
\includegraphics[width=9.0cm,angle=0]{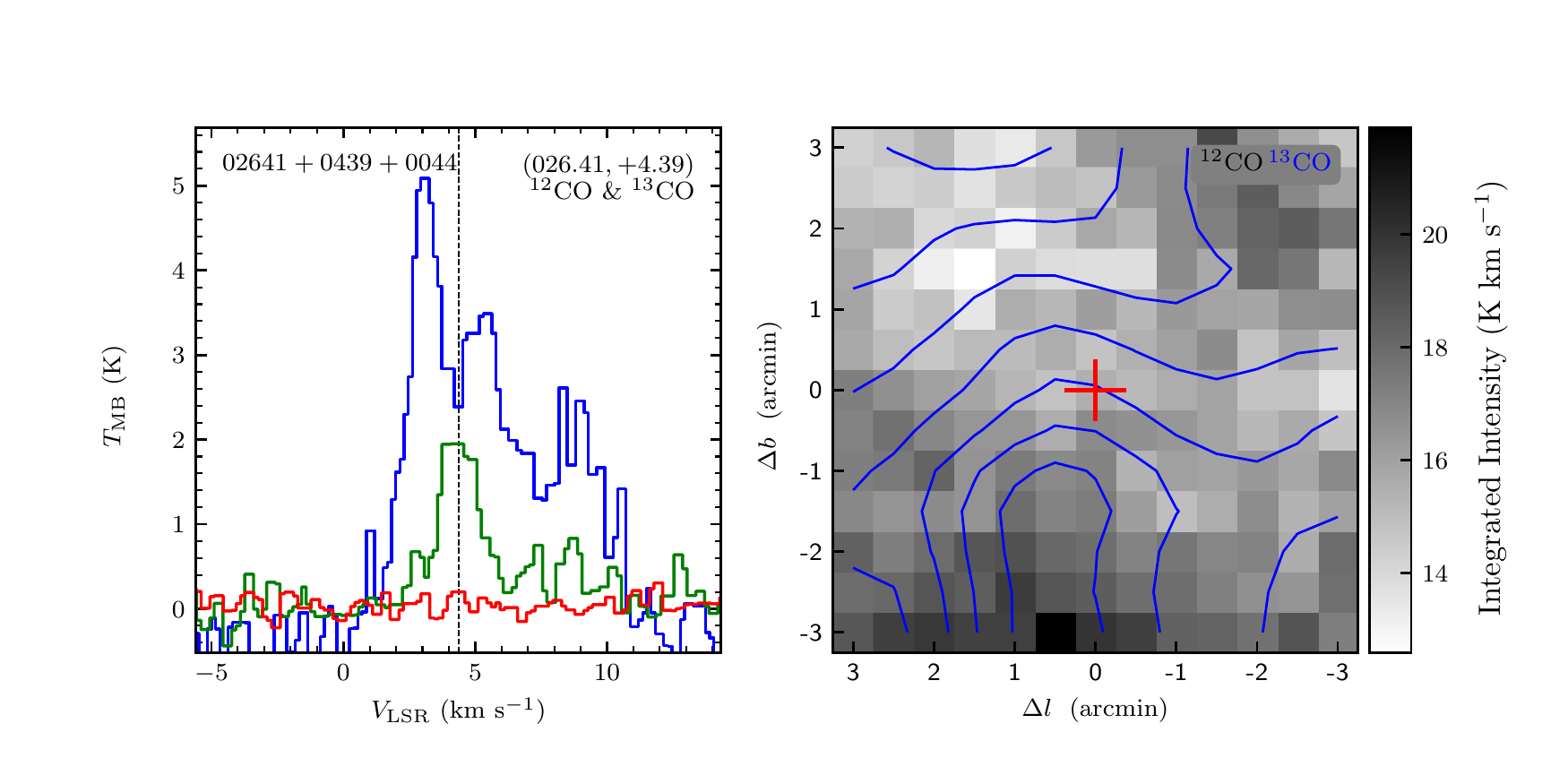}
\includegraphics[width=9.0cm,angle=0]{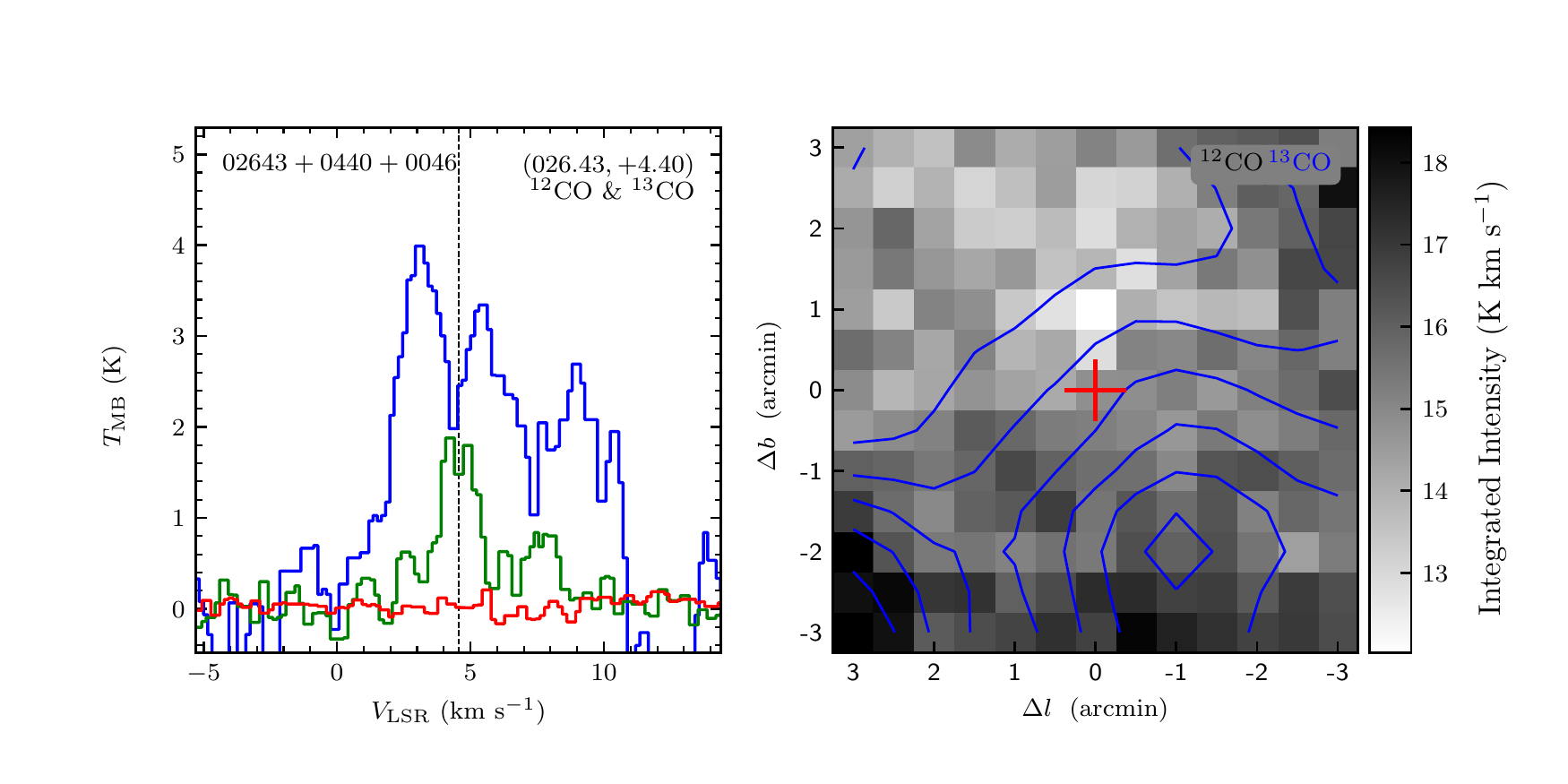}
\vspace{-0.5cm}

\includegraphics[width=9.0cm,angle=0]{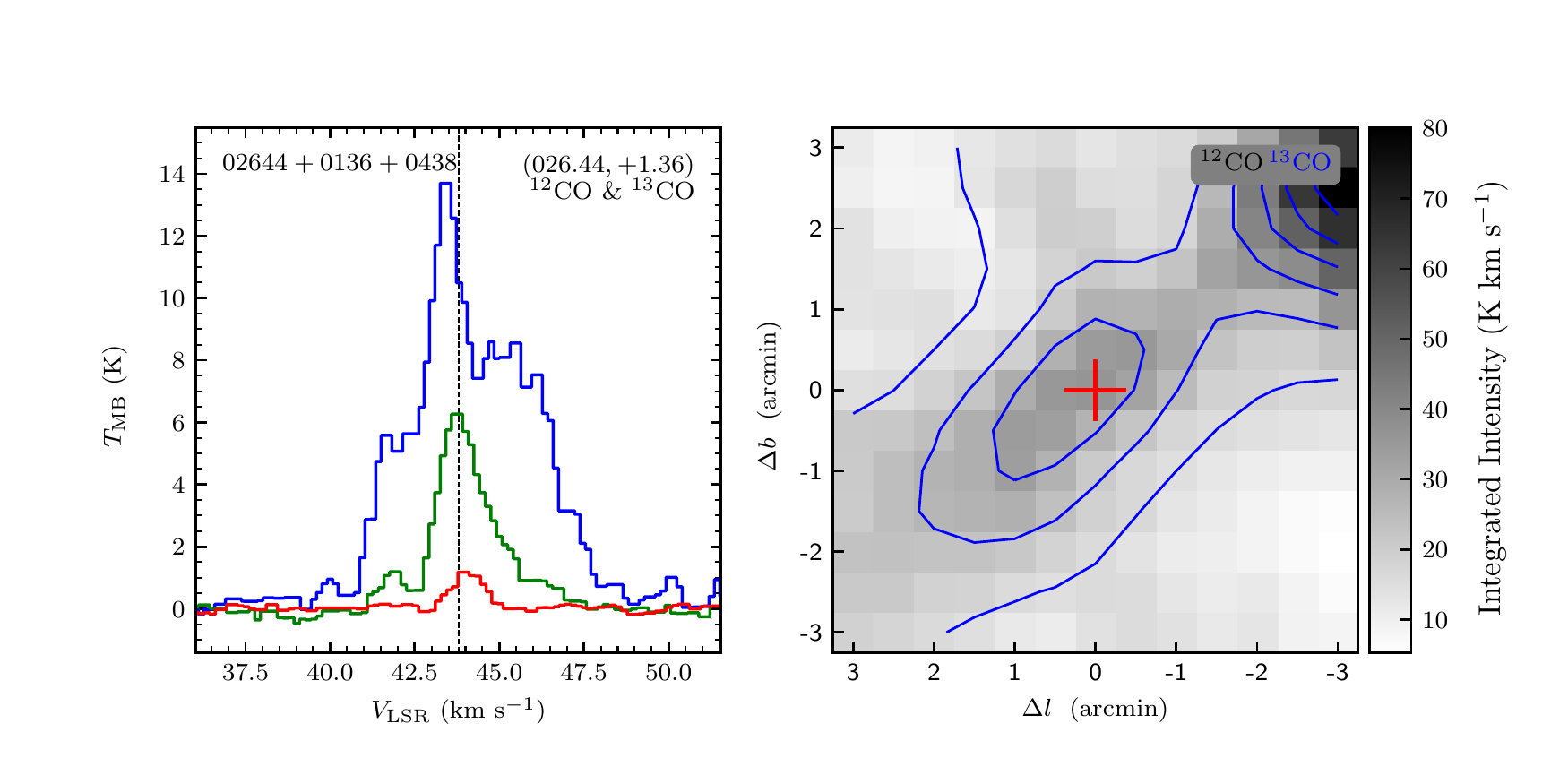}
\includegraphics[width=9.0cm,angle=0]{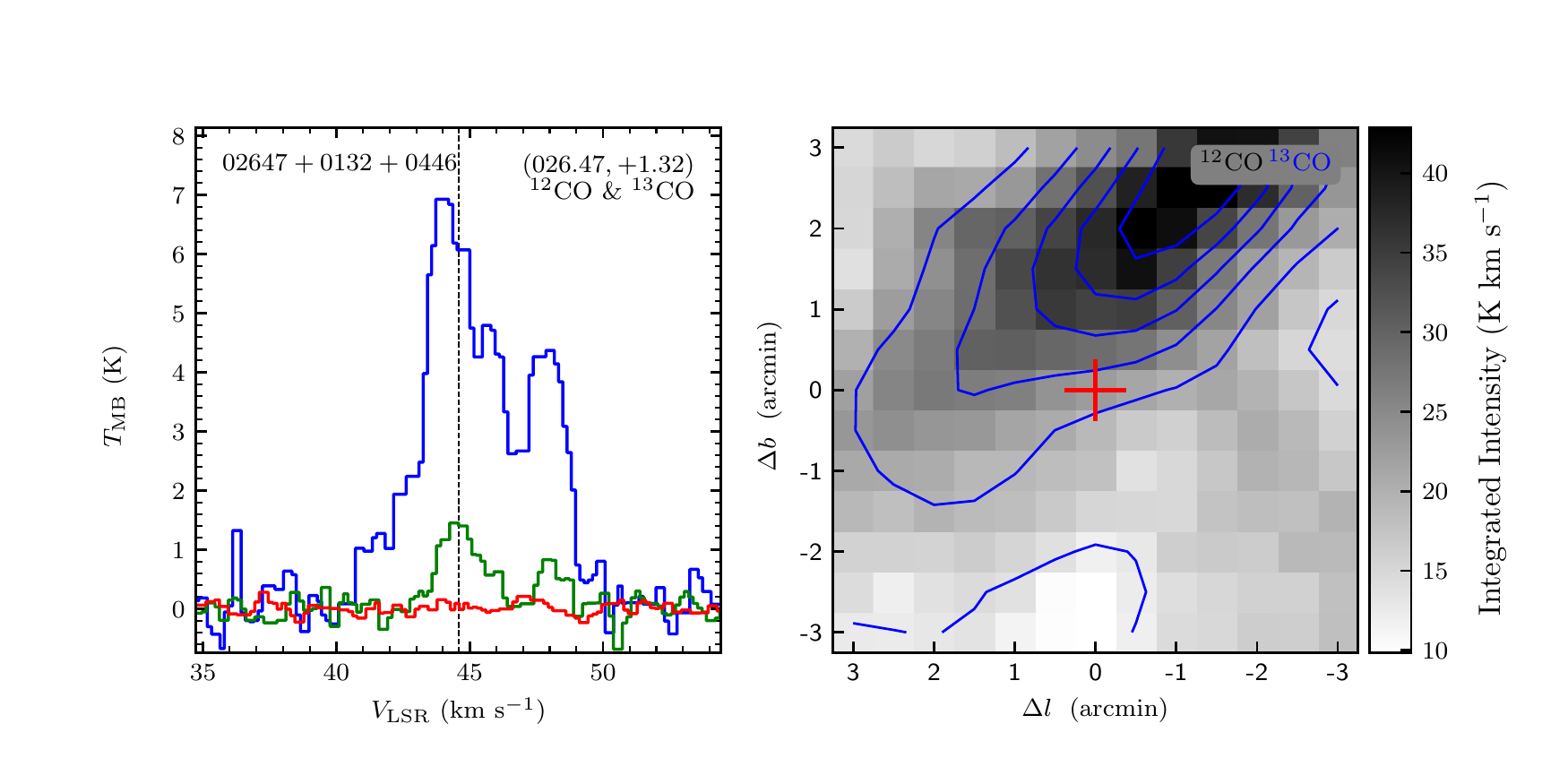}
\vspace{-0.5cm}

\includegraphics[width=9.0cm,angle=0]{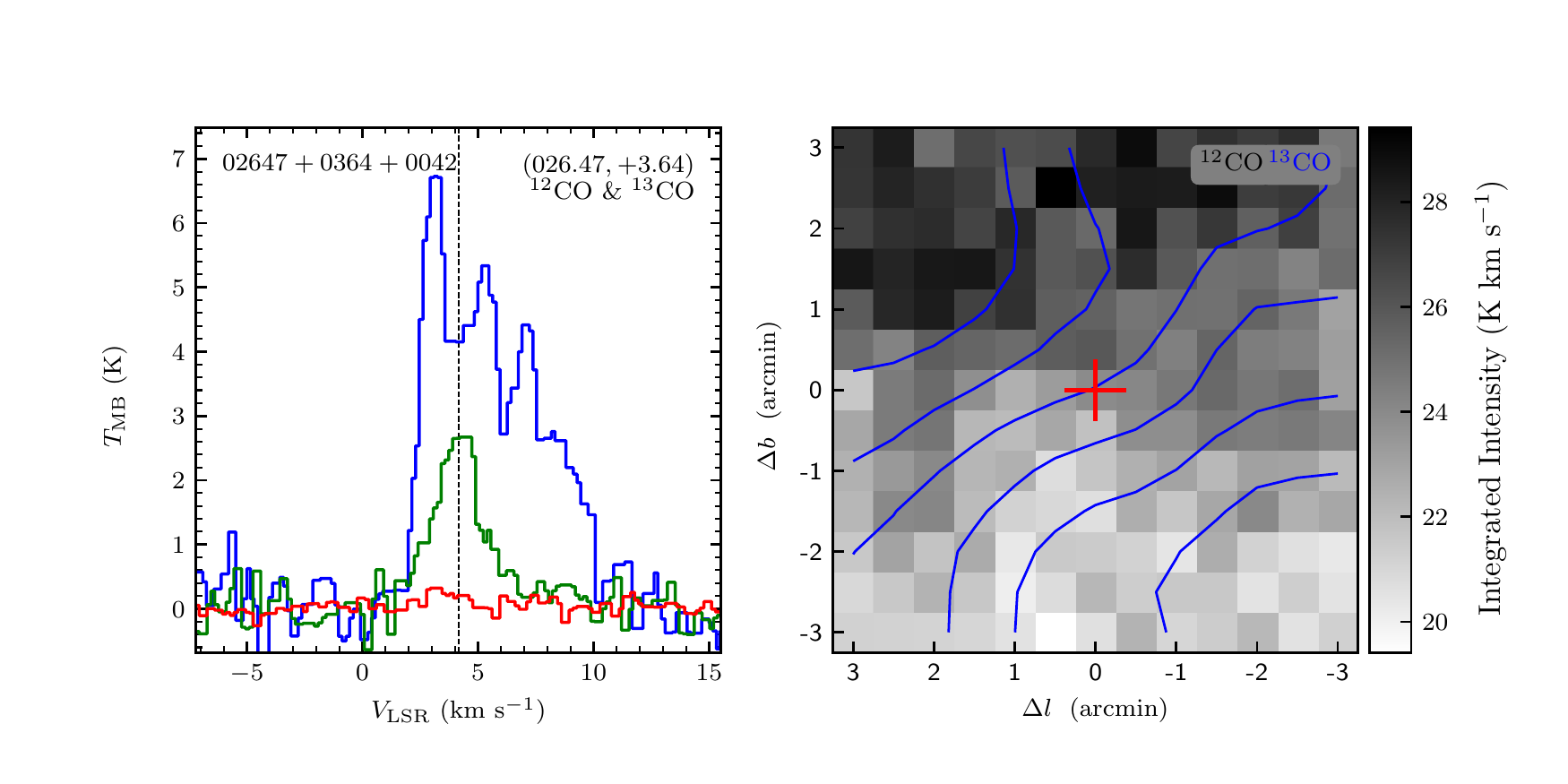}
\includegraphics[width=9.0cm,angle=0]{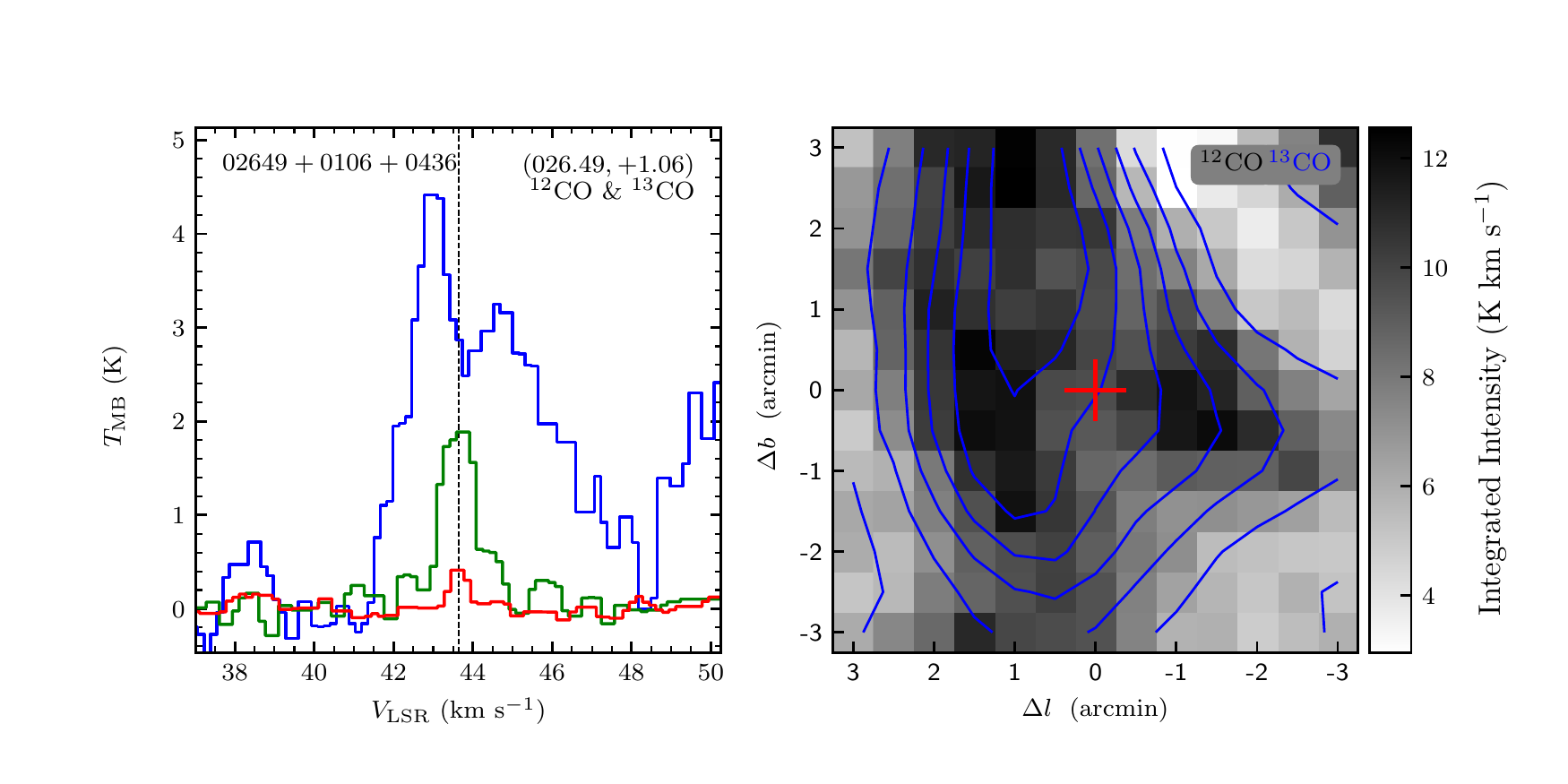}
\vspace{-0.5cm}

\includegraphics[width=9.0cm,angle=0]{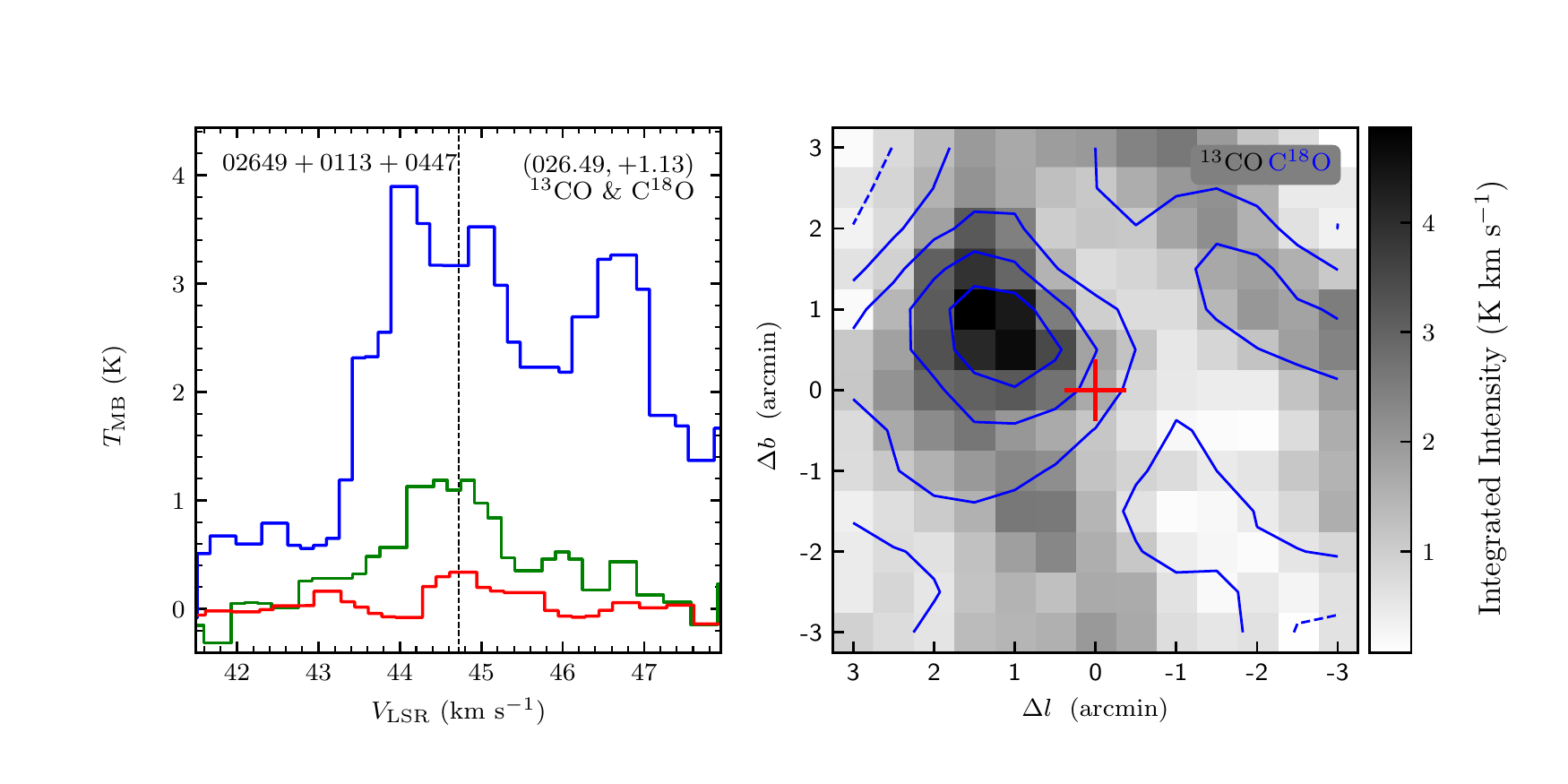}
\includegraphics[width=9.0cm,angle=0]{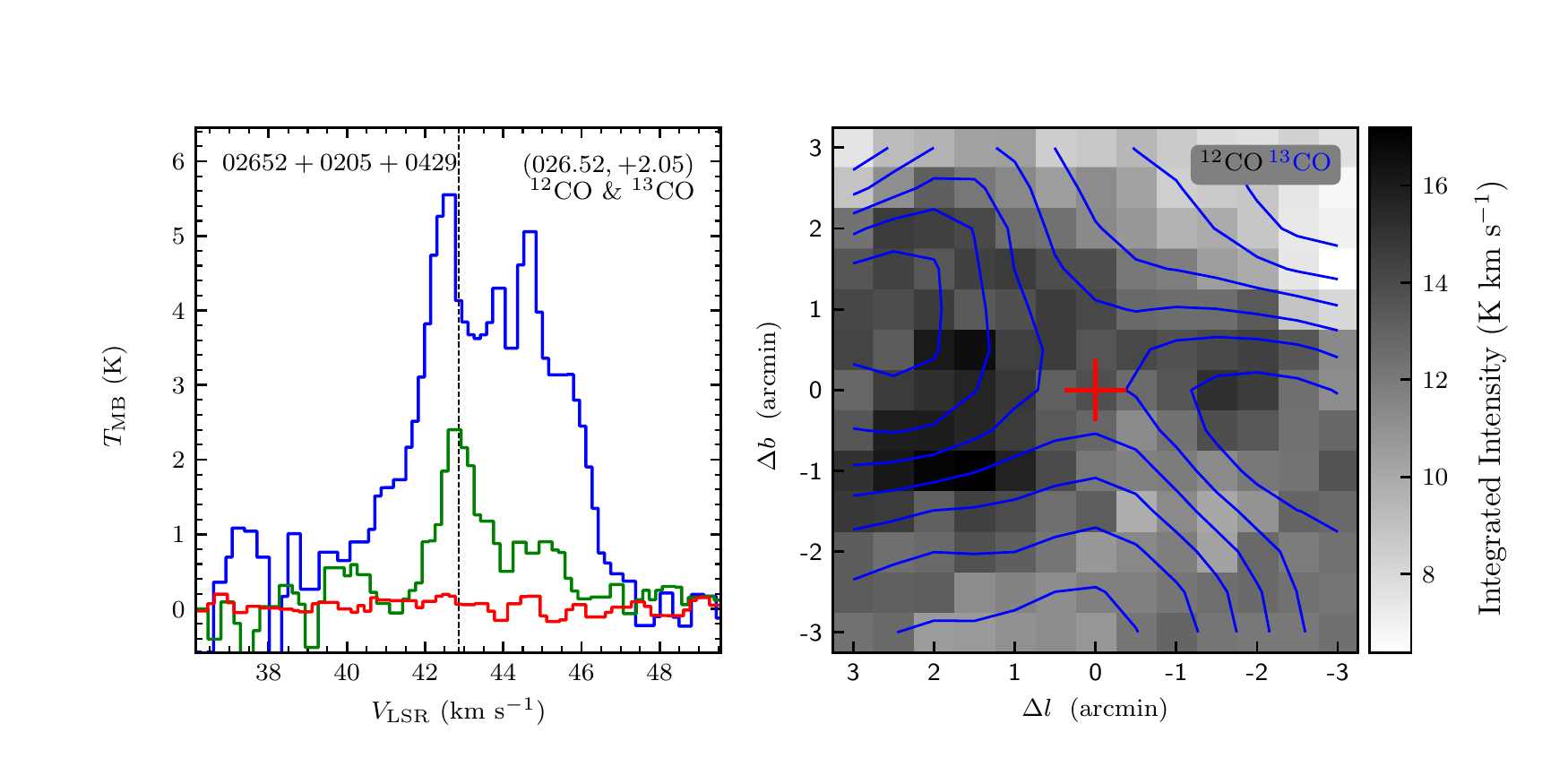}
\vspace{-0.5cm}

\includegraphics[width=9.0cm,angle=0]{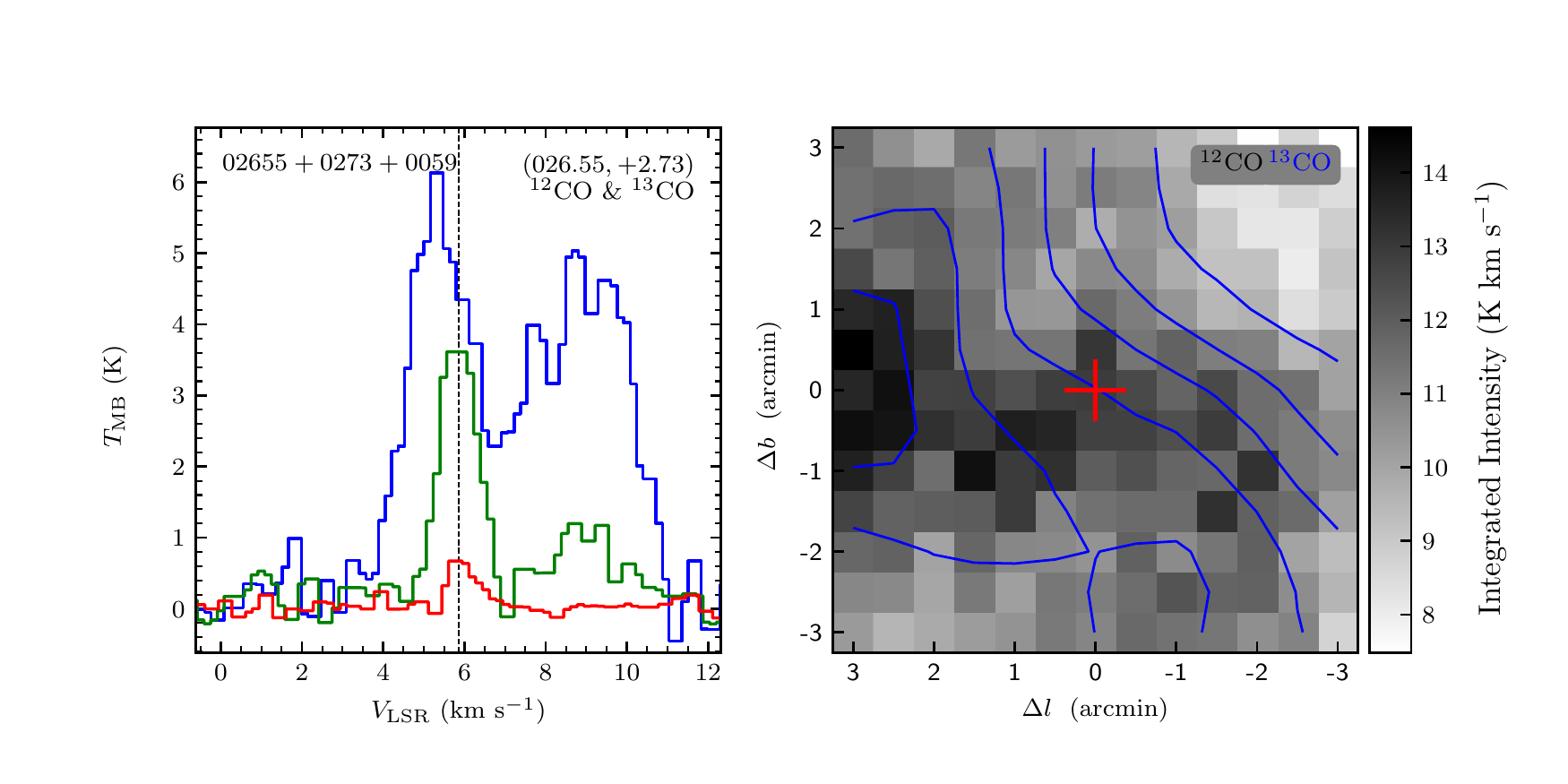}
\includegraphics[width=9.0cm,angle=0]{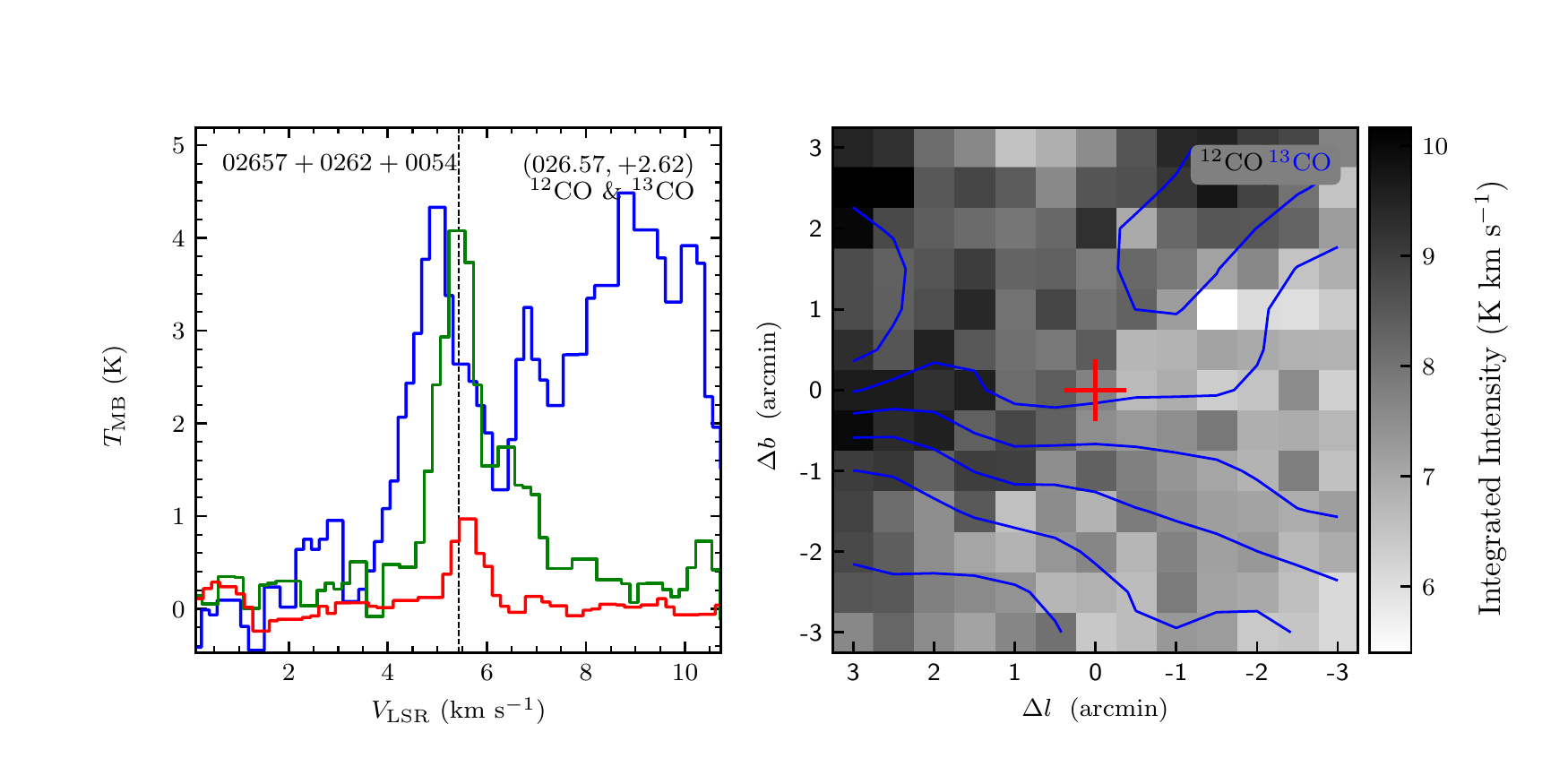}
\end{figure}
\clearpage

\begin{figure}
\includegraphics[width=9.0cm,angle=0]{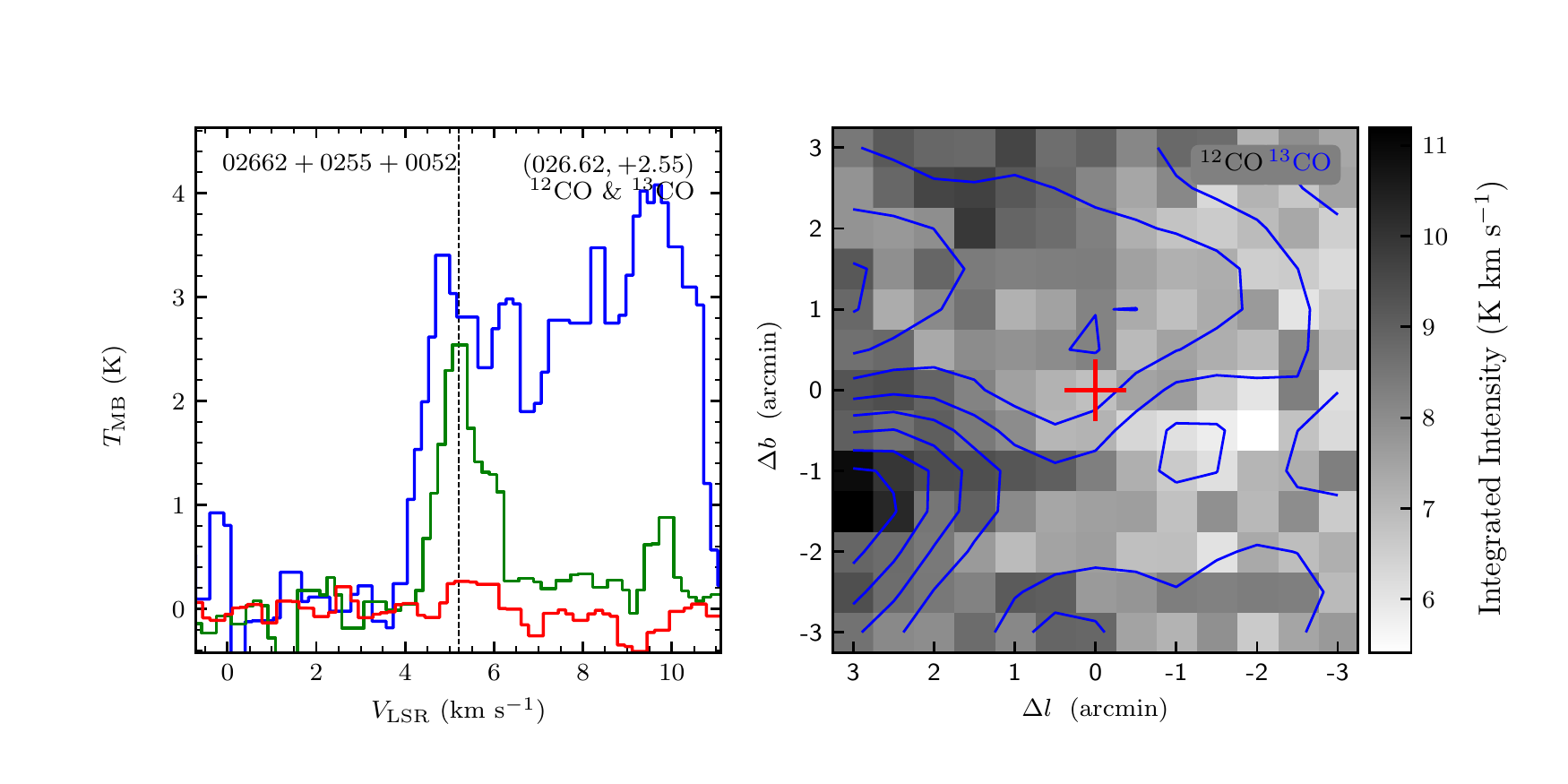}
\includegraphics[width=9.0cm,angle=0]{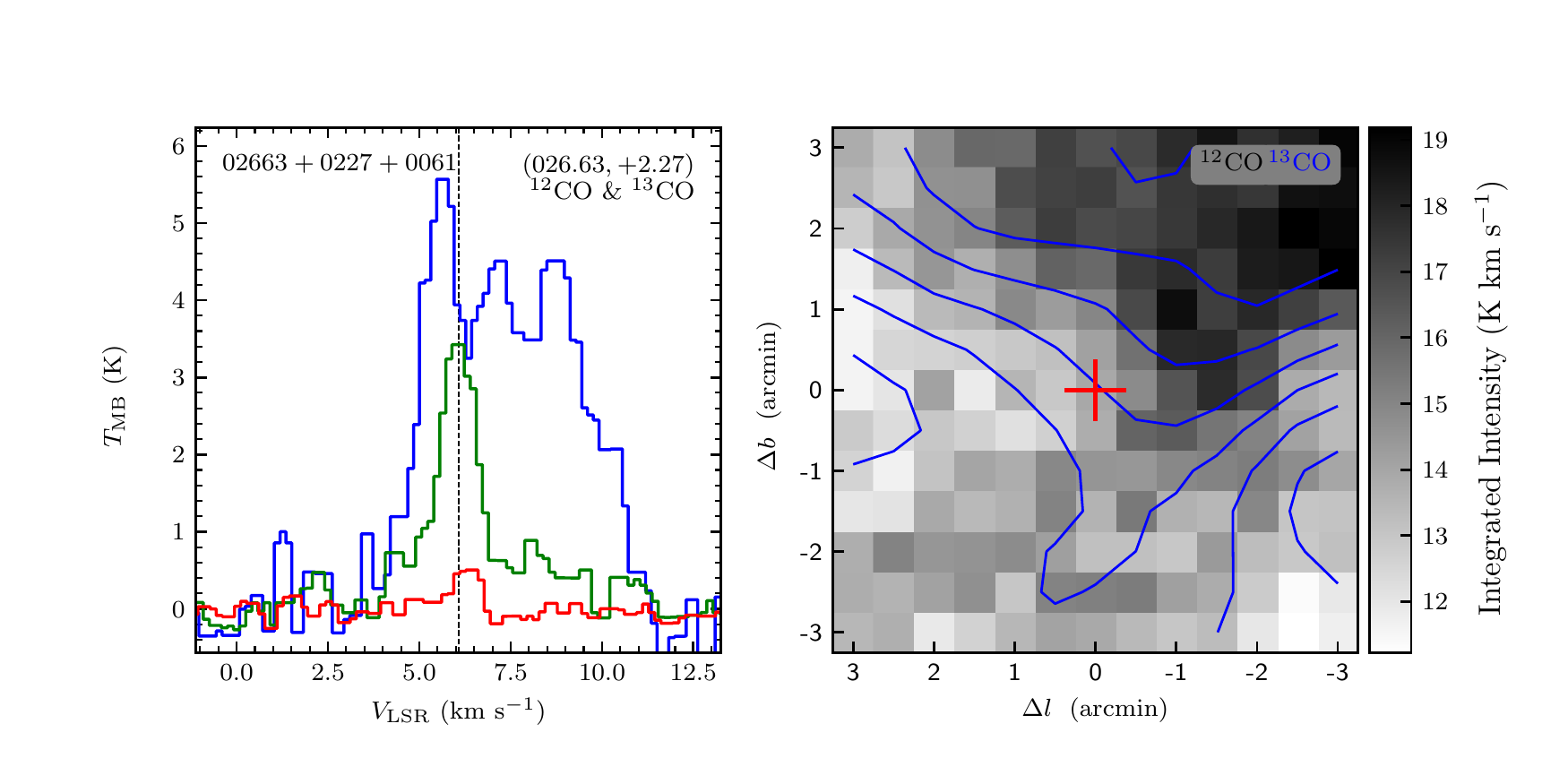}
\vspace{-0.5cm}

\includegraphics[width=9.0cm,angle=0]{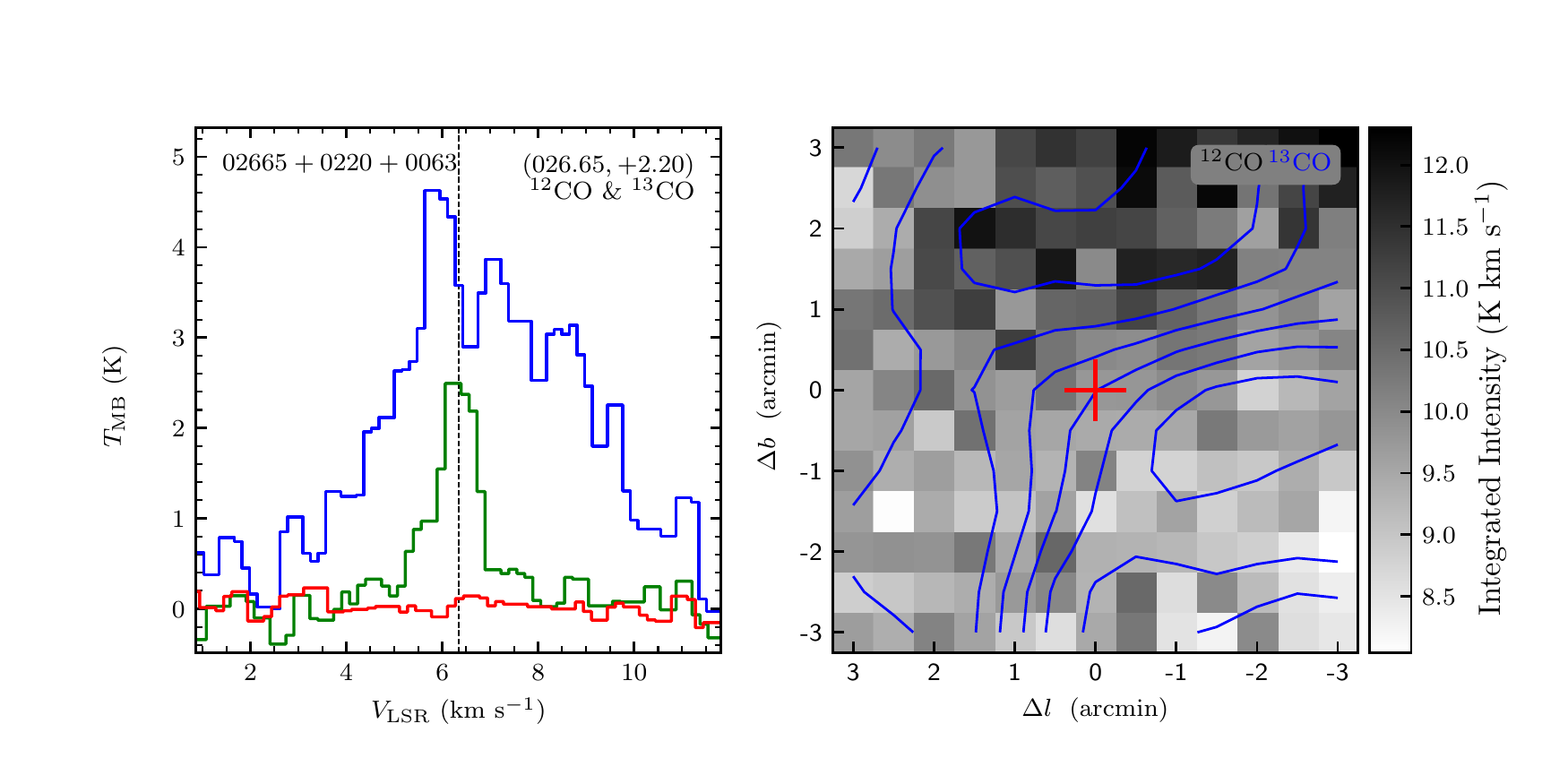}
\includegraphics[width=9.0cm,angle=0]{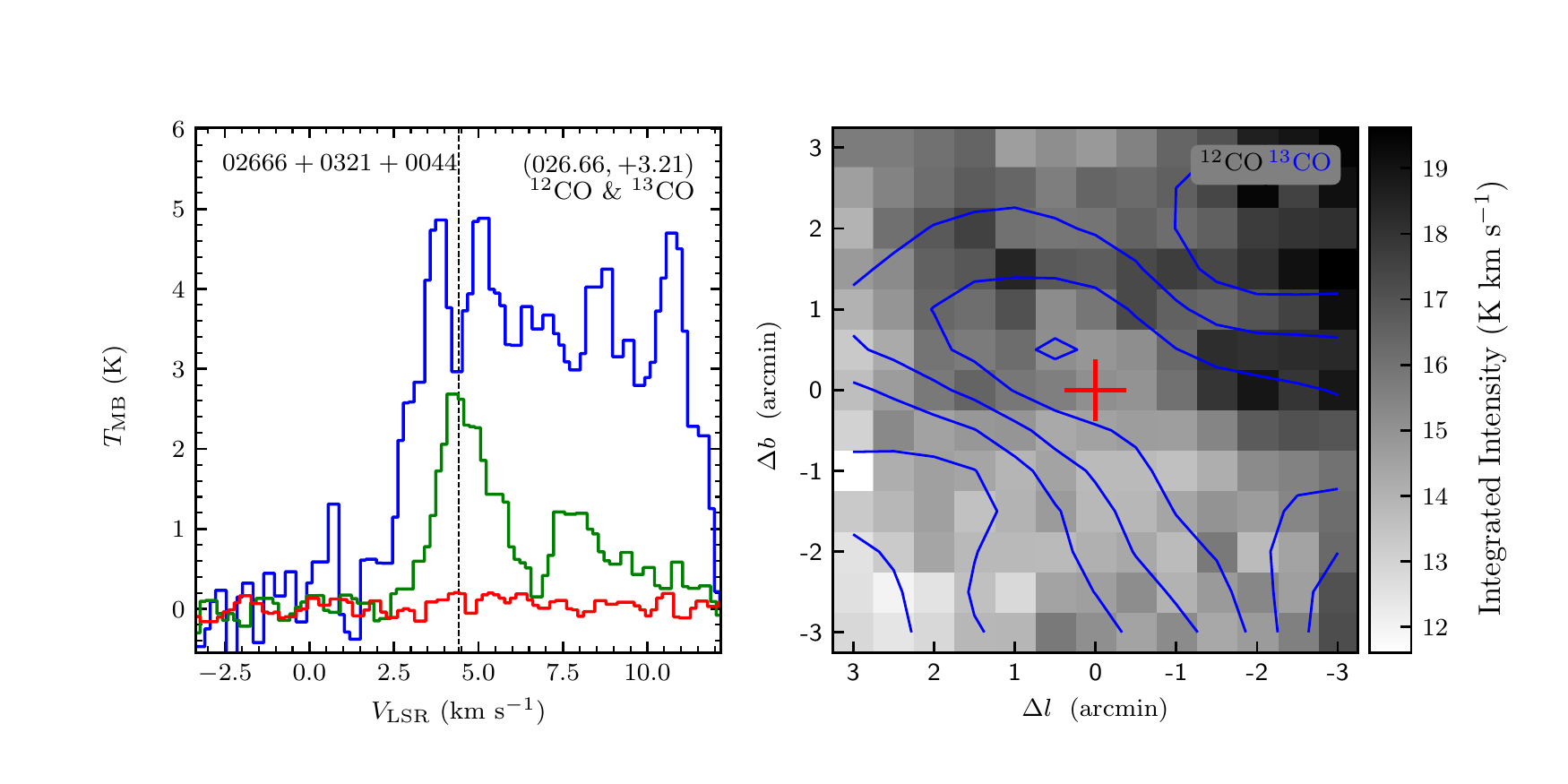}
\vspace{-0.5cm}

\includegraphics[width=9.0cm,angle=0]{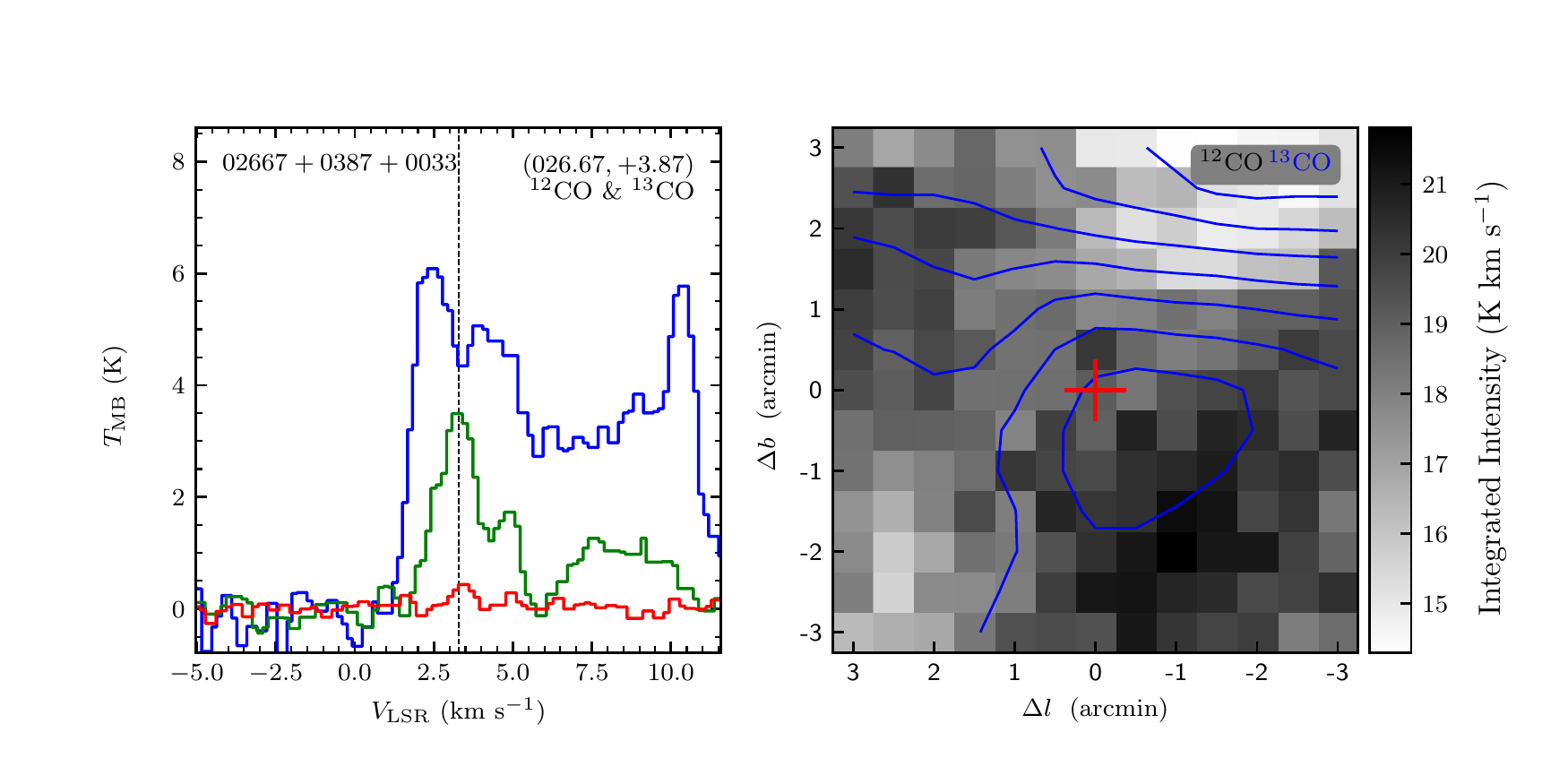}
\includegraphics[width=9.0cm,angle=0]{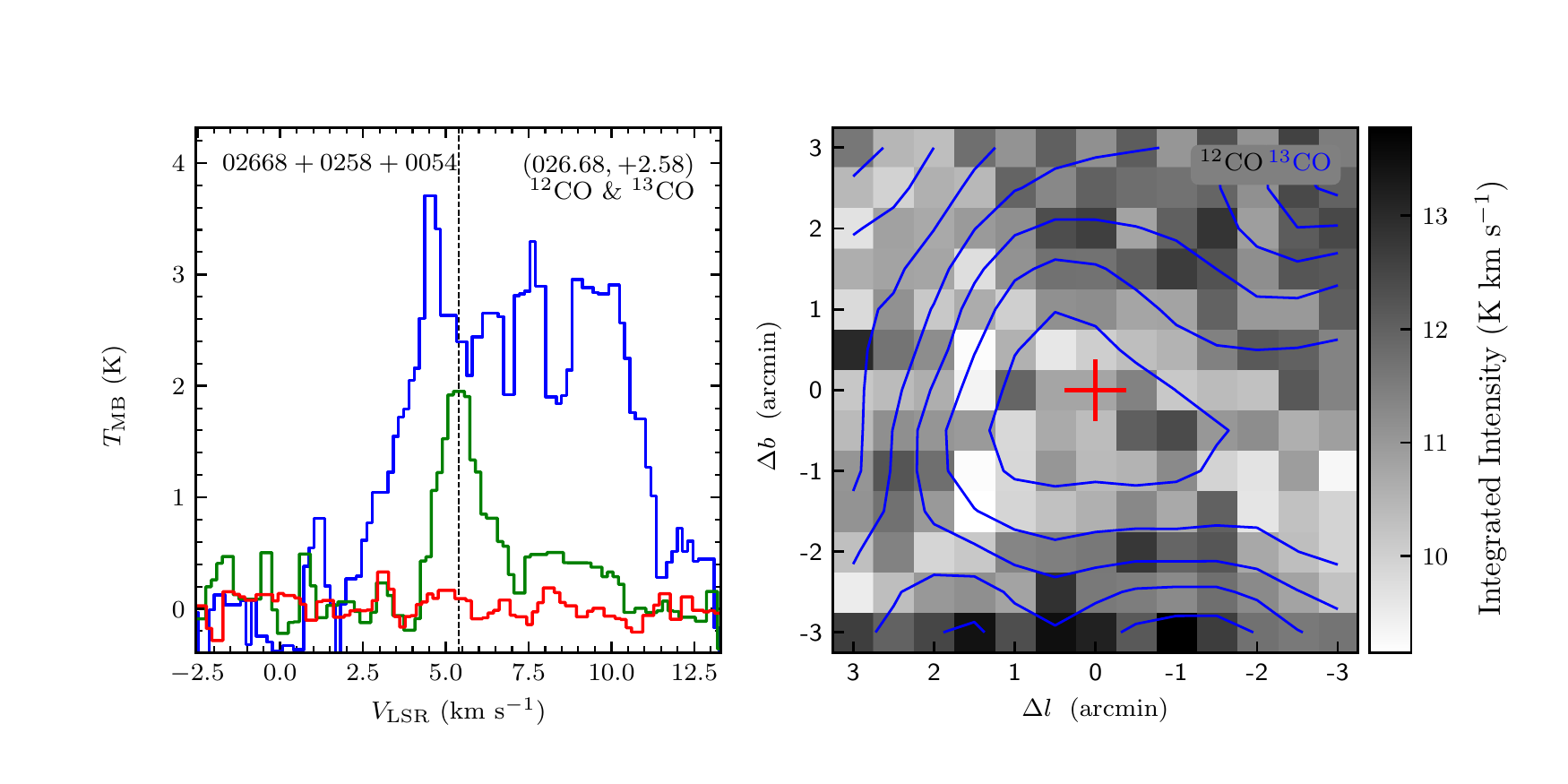}
\vspace{-0.5cm}

\includegraphics[width=9.0cm,angle=0]{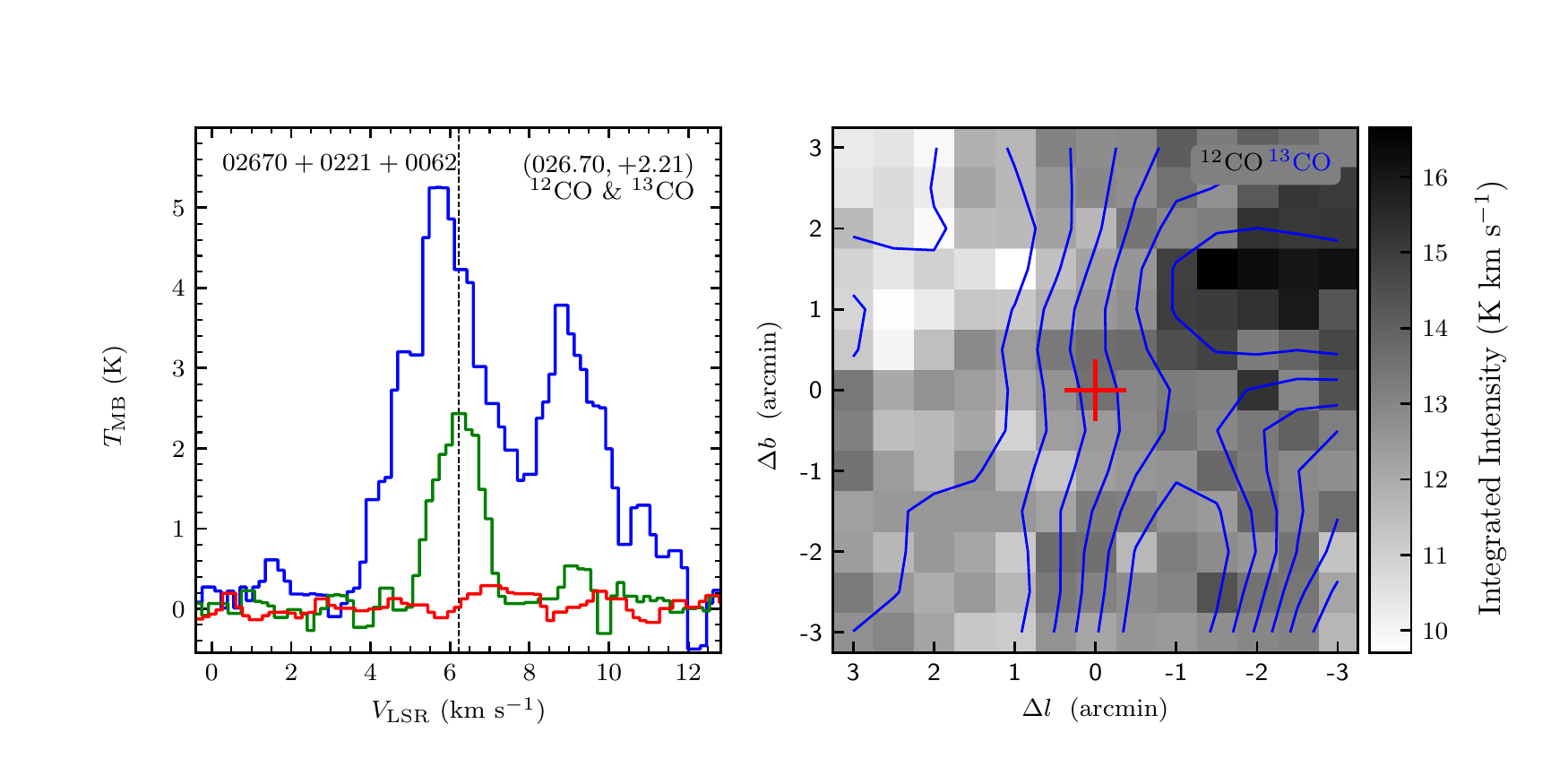}
\includegraphics[width=9.0cm,angle=0]{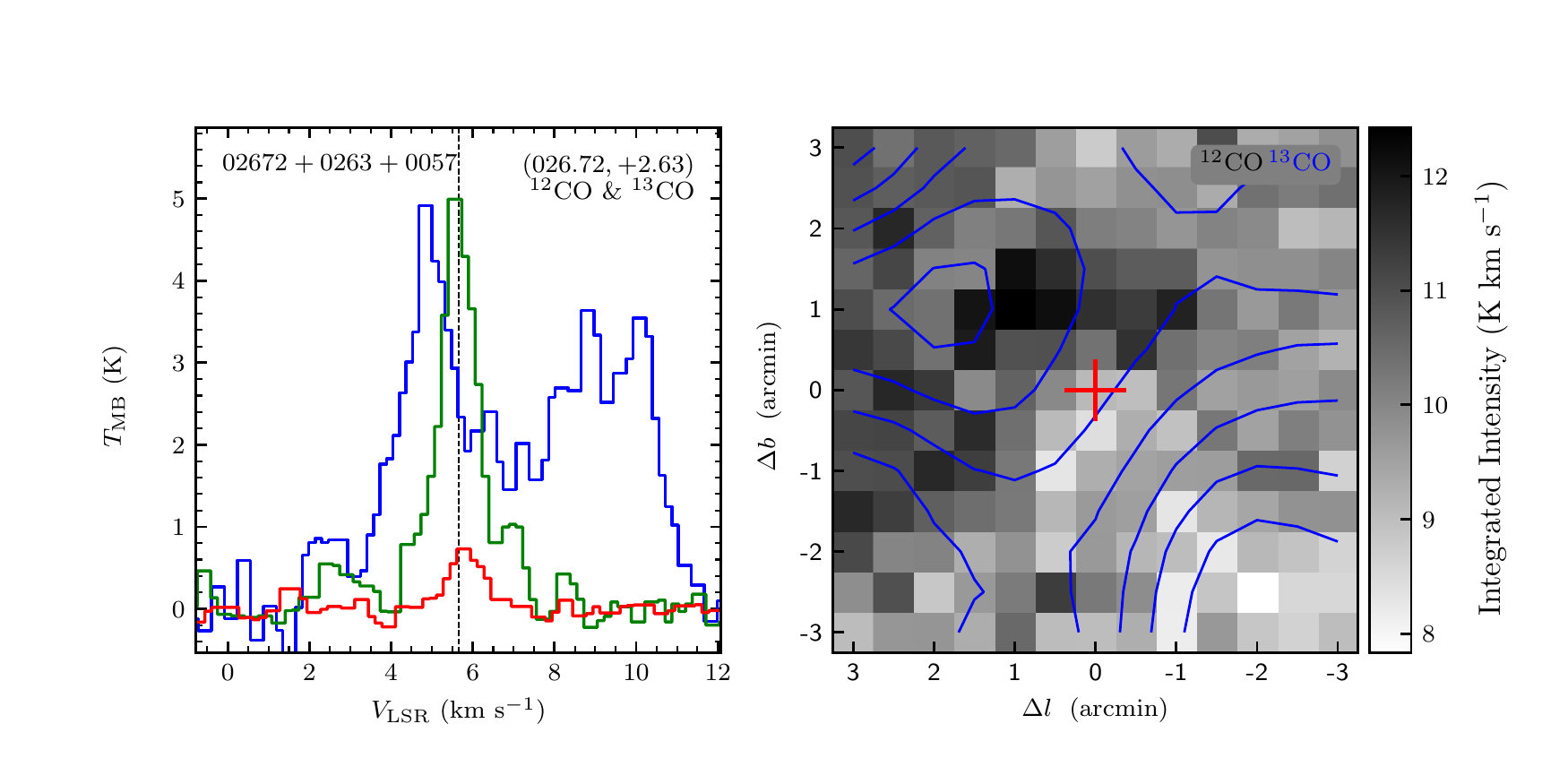}
\vspace{-0.5cm}

\includegraphics[width=9.0cm,angle=0]{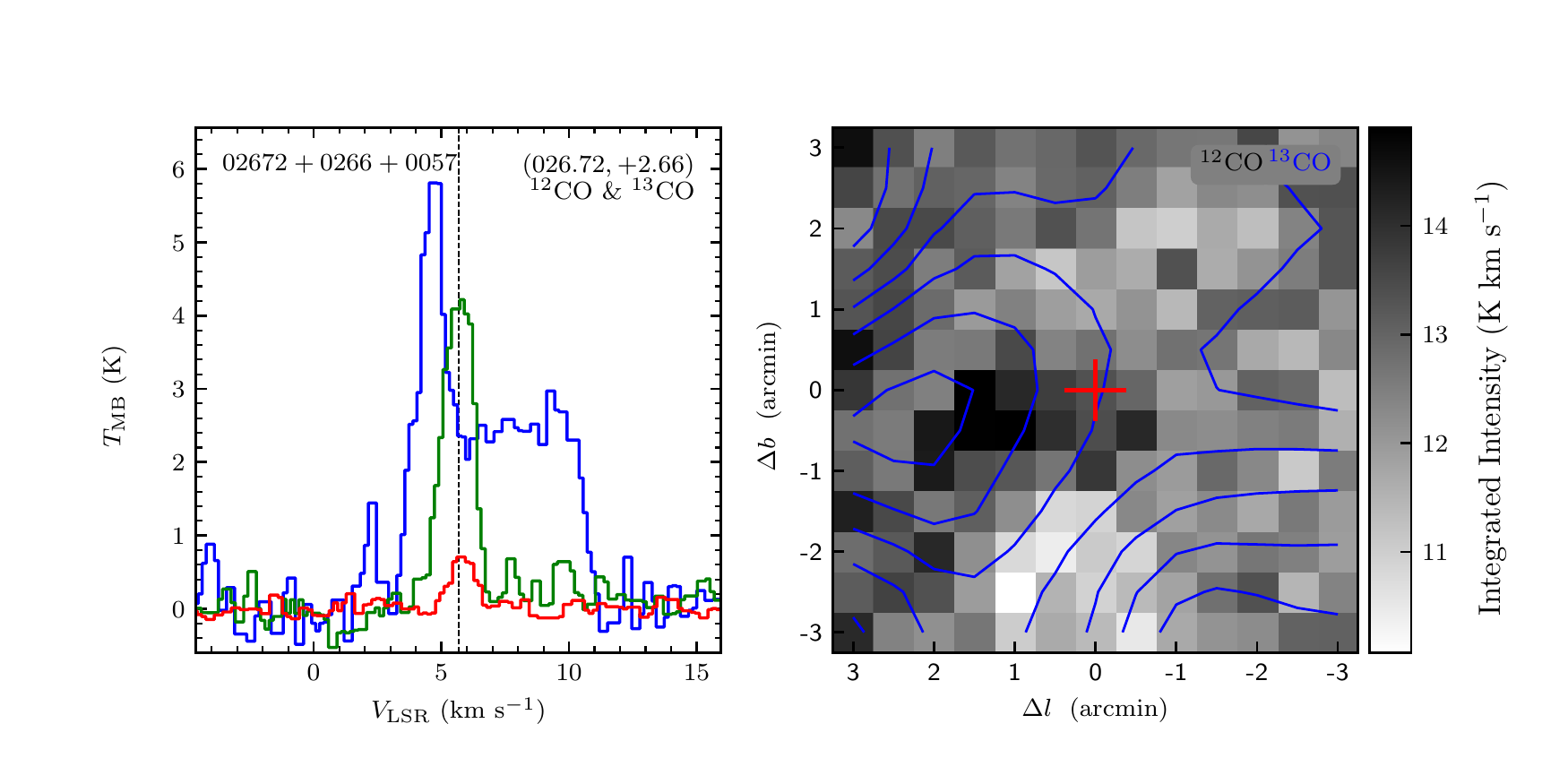}
\includegraphics[width=9.0cm,angle=0]{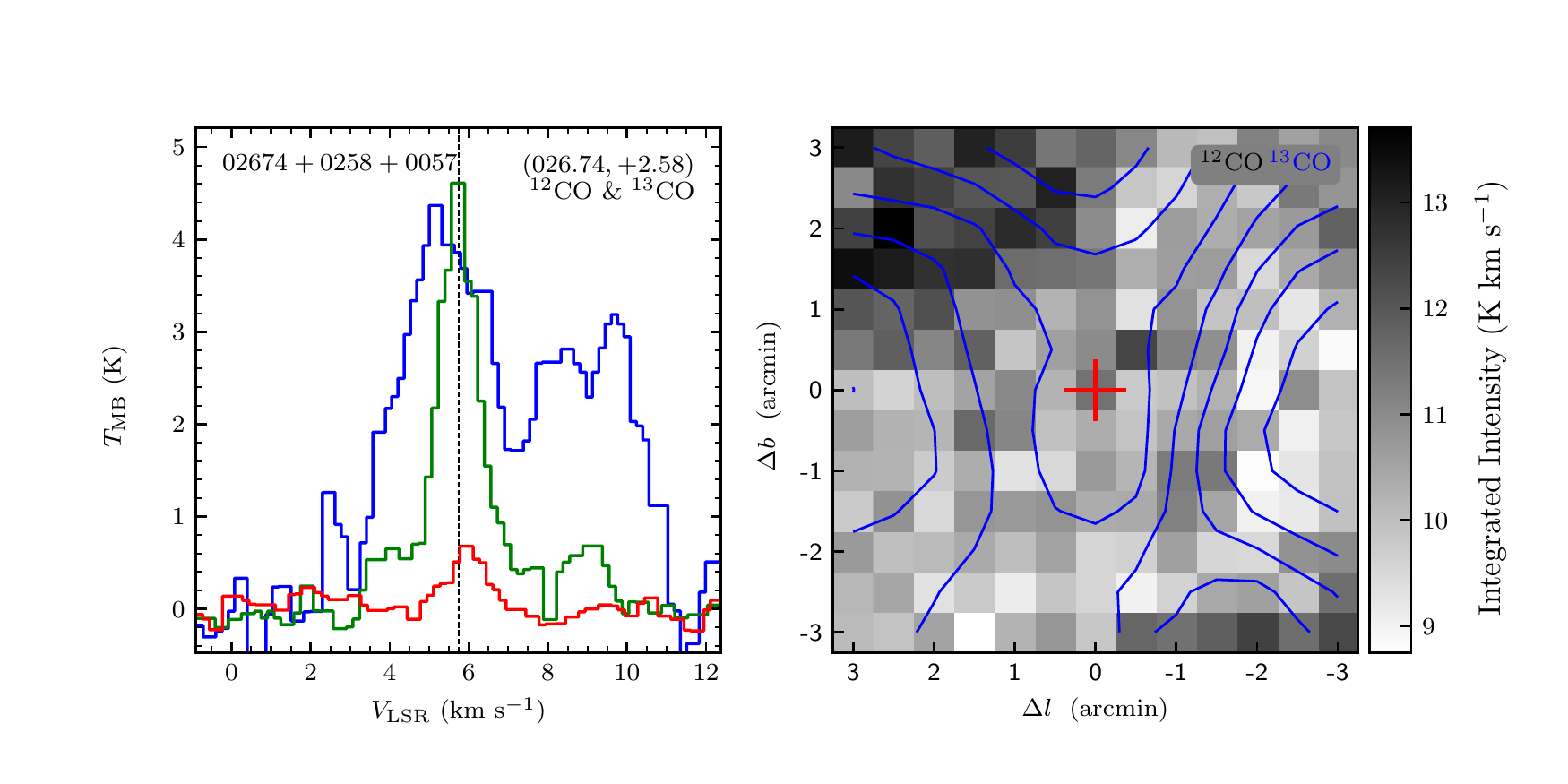}
\end{figure}
\clearpage

\begin{figure}
\includegraphics[width=9.0cm,angle=0]{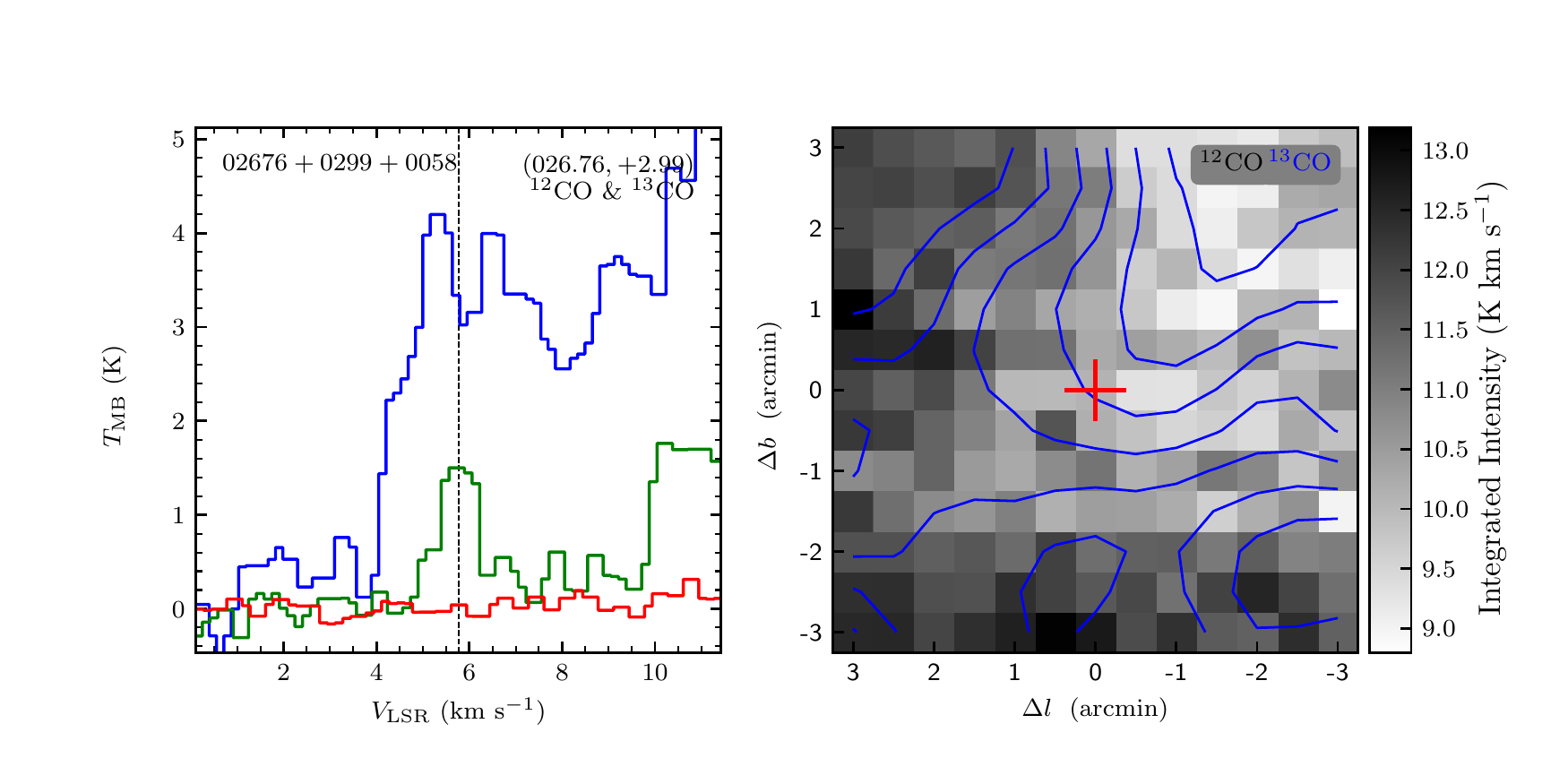}
\includegraphics[width=9.0cm,angle=0]{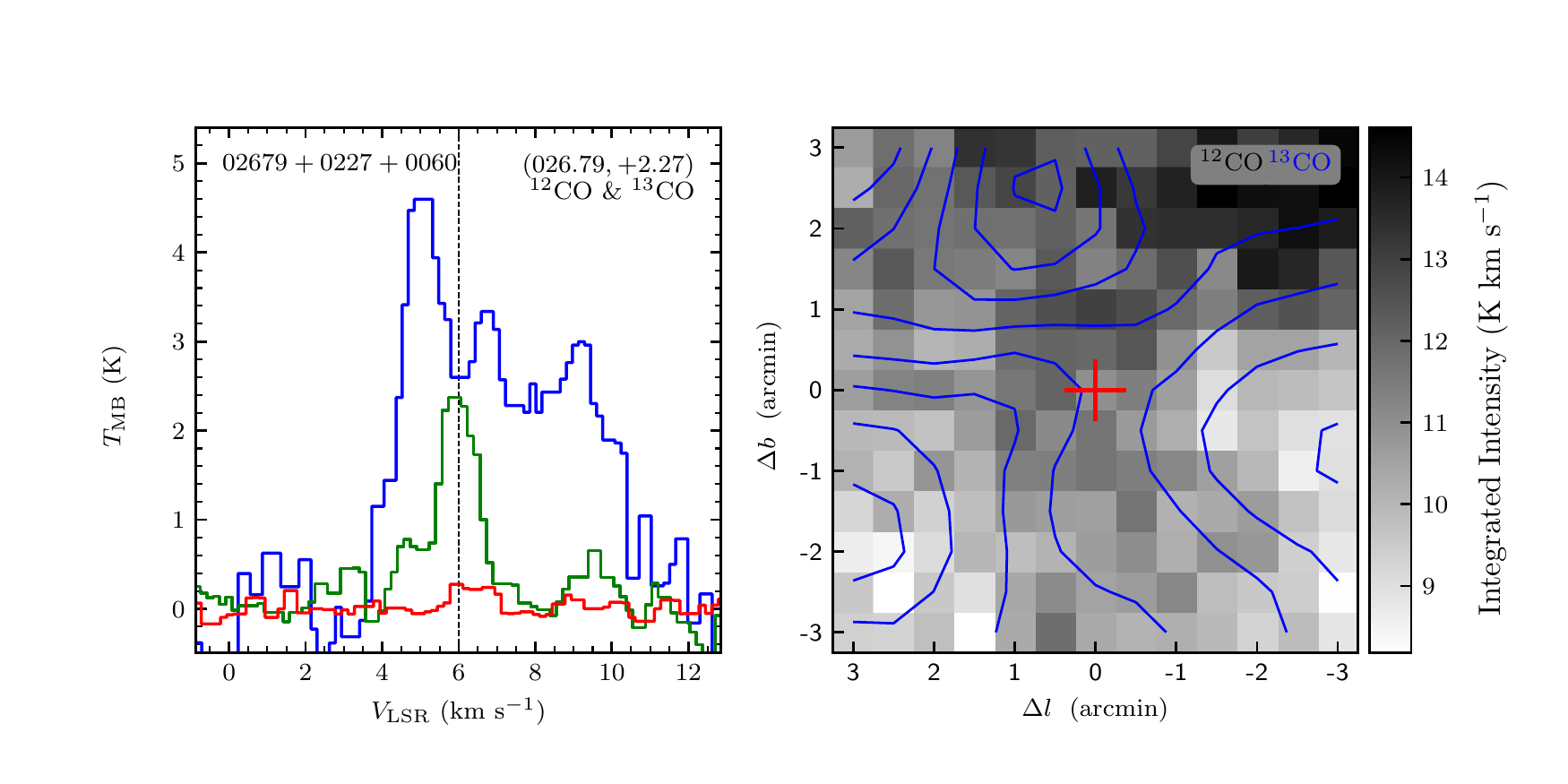}
\vspace{-0.5cm}

\includegraphics[width=9.0cm,angle=0]{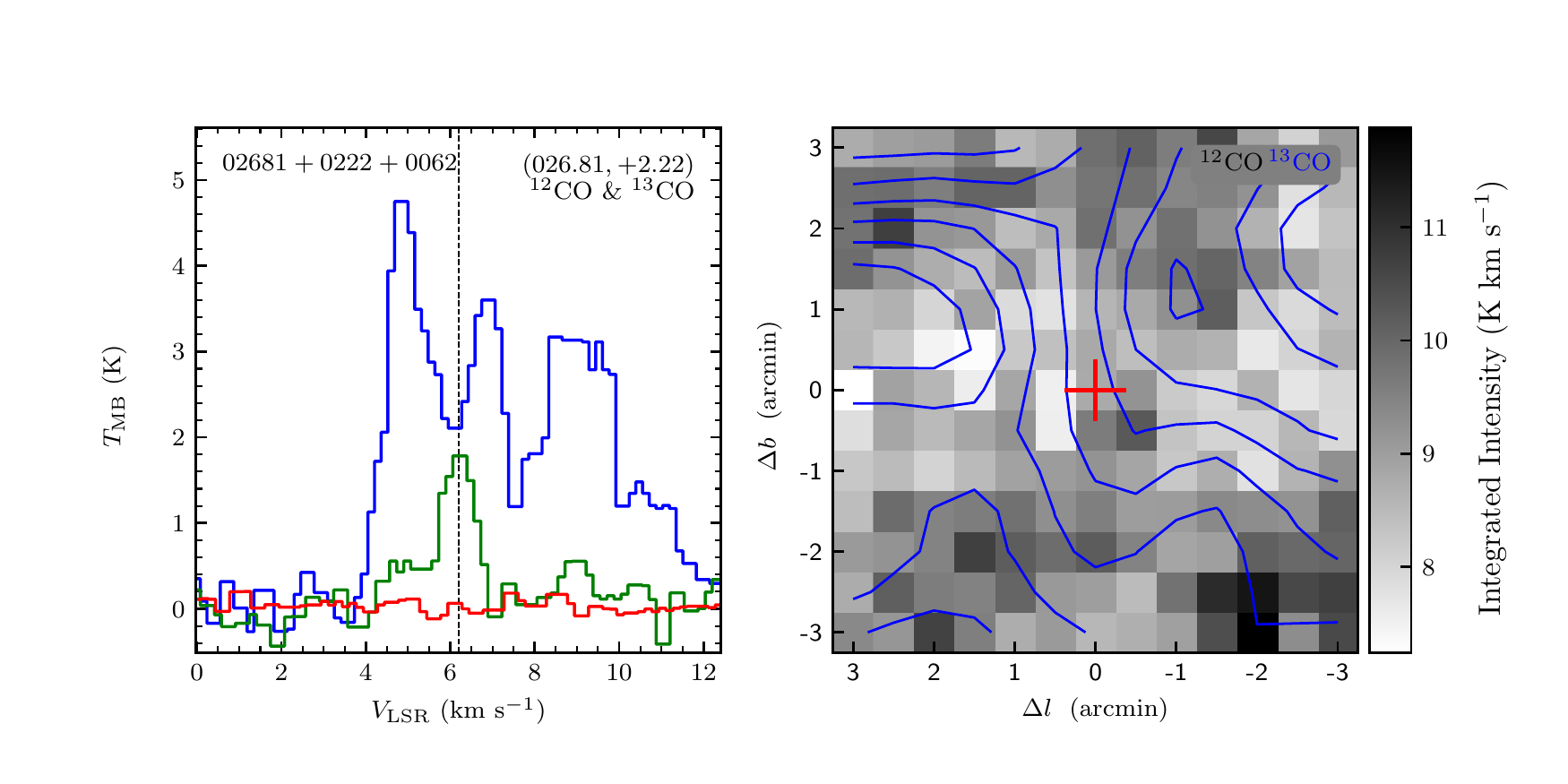}
\includegraphics[width=9.0cm,angle=0]{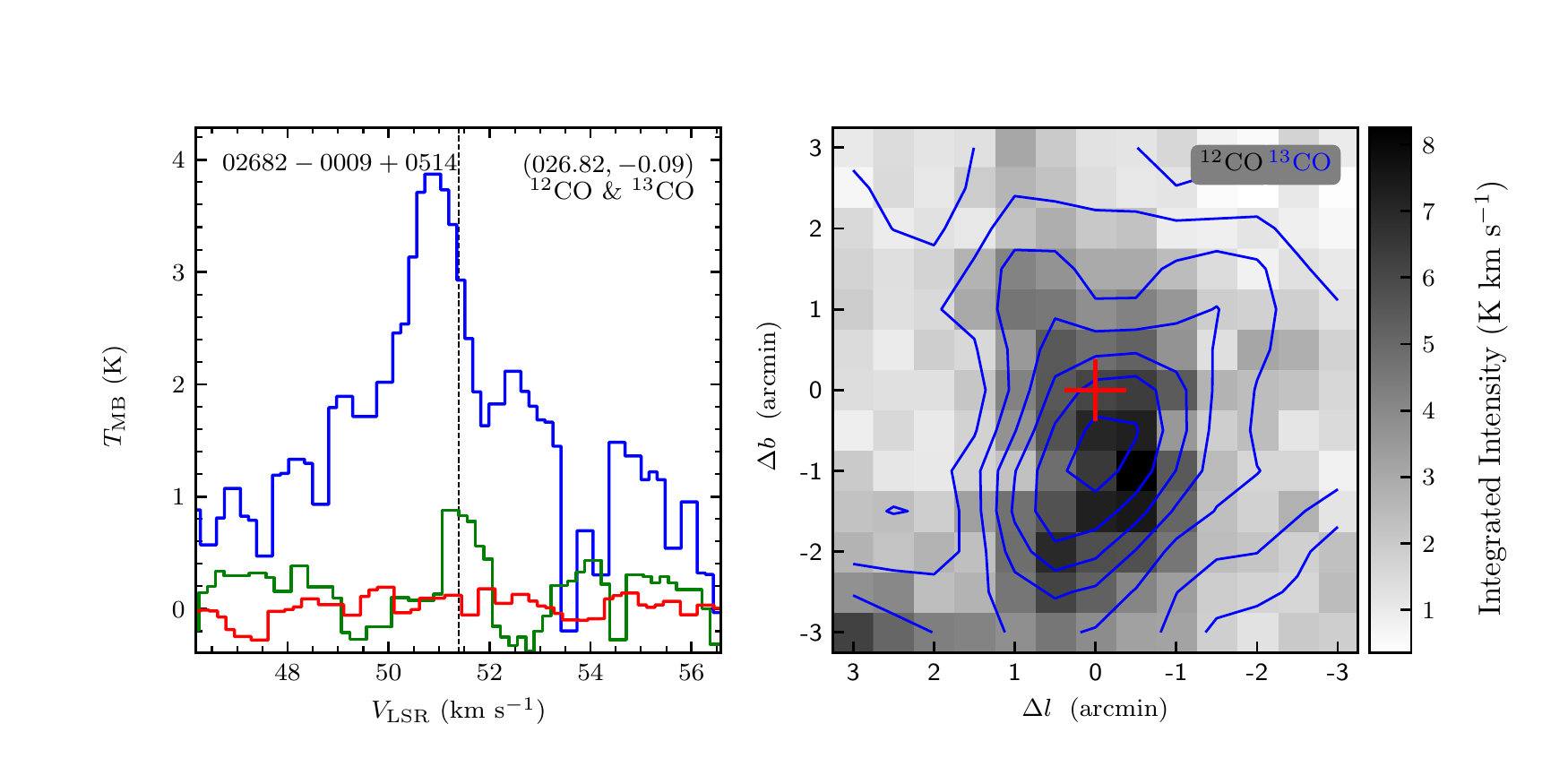}
\vspace{-0.5cm}

\includegraphics[width=9.0cm,angle=0]{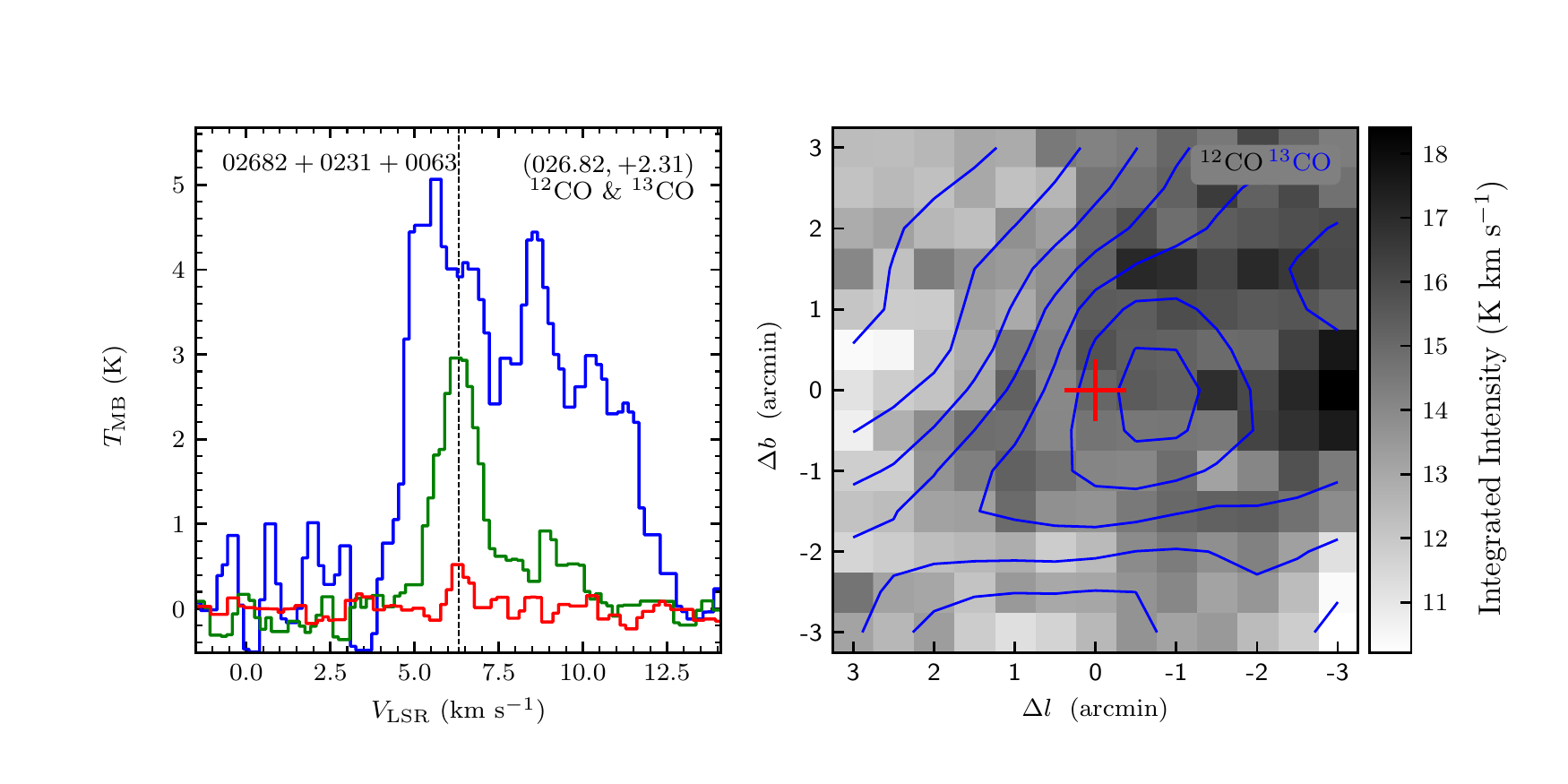}
\includegraphics[width=9.0cm,angle=0]{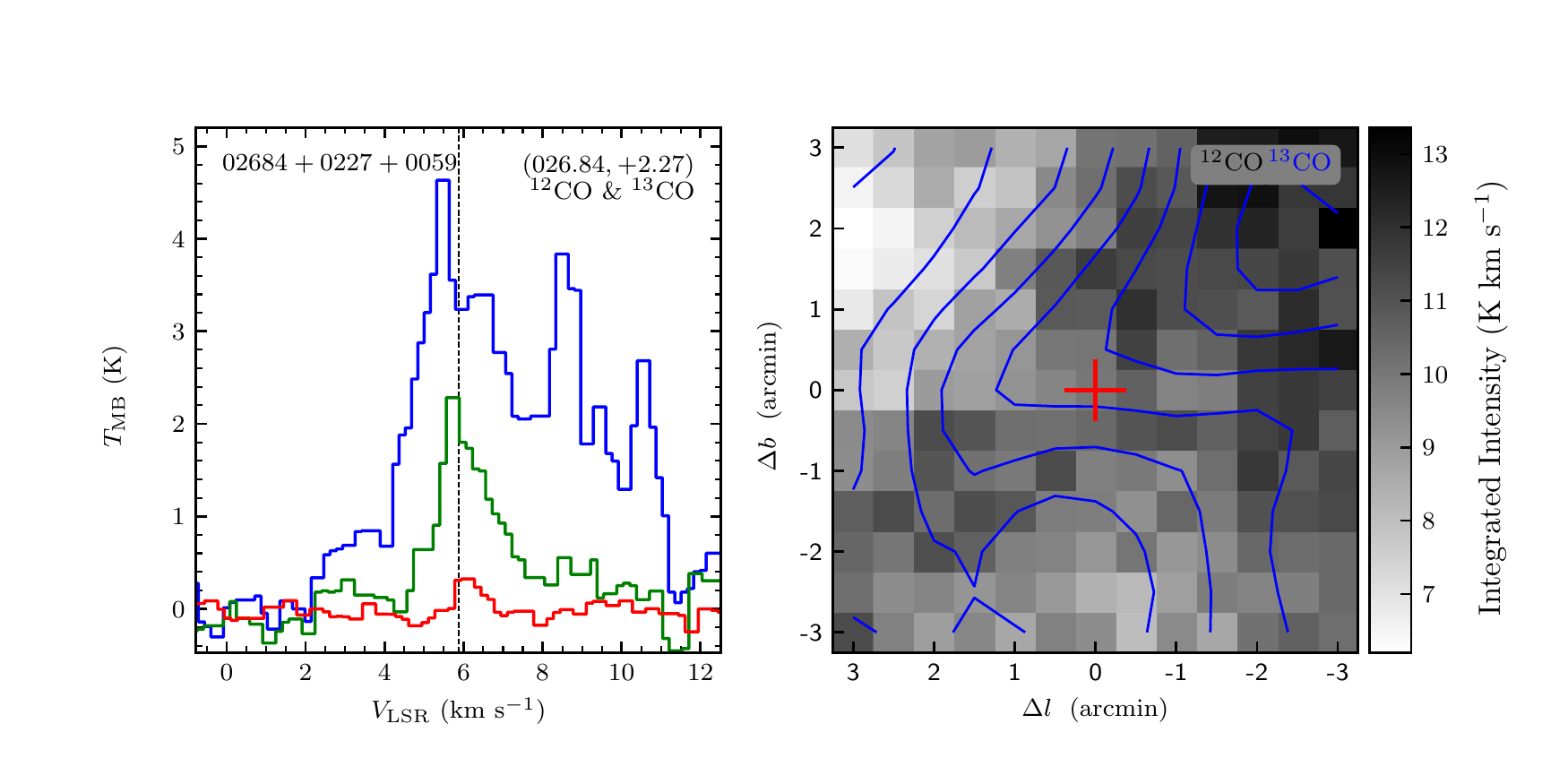}
\vspace{-0.5cm}

\includegraphics[width=9.0cm,angle=0]{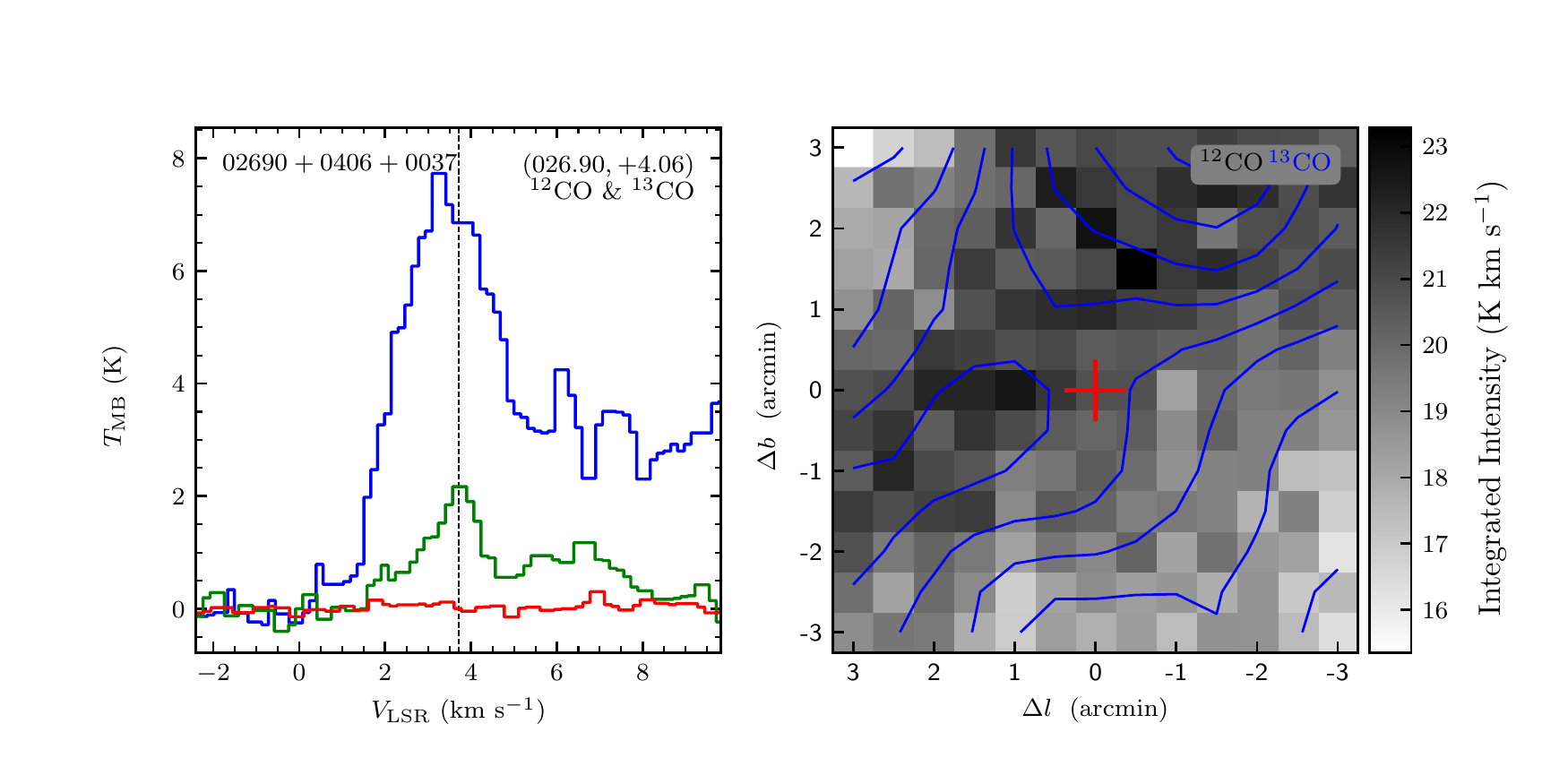}
\includegraphics[width=9.0cm,angle=0]{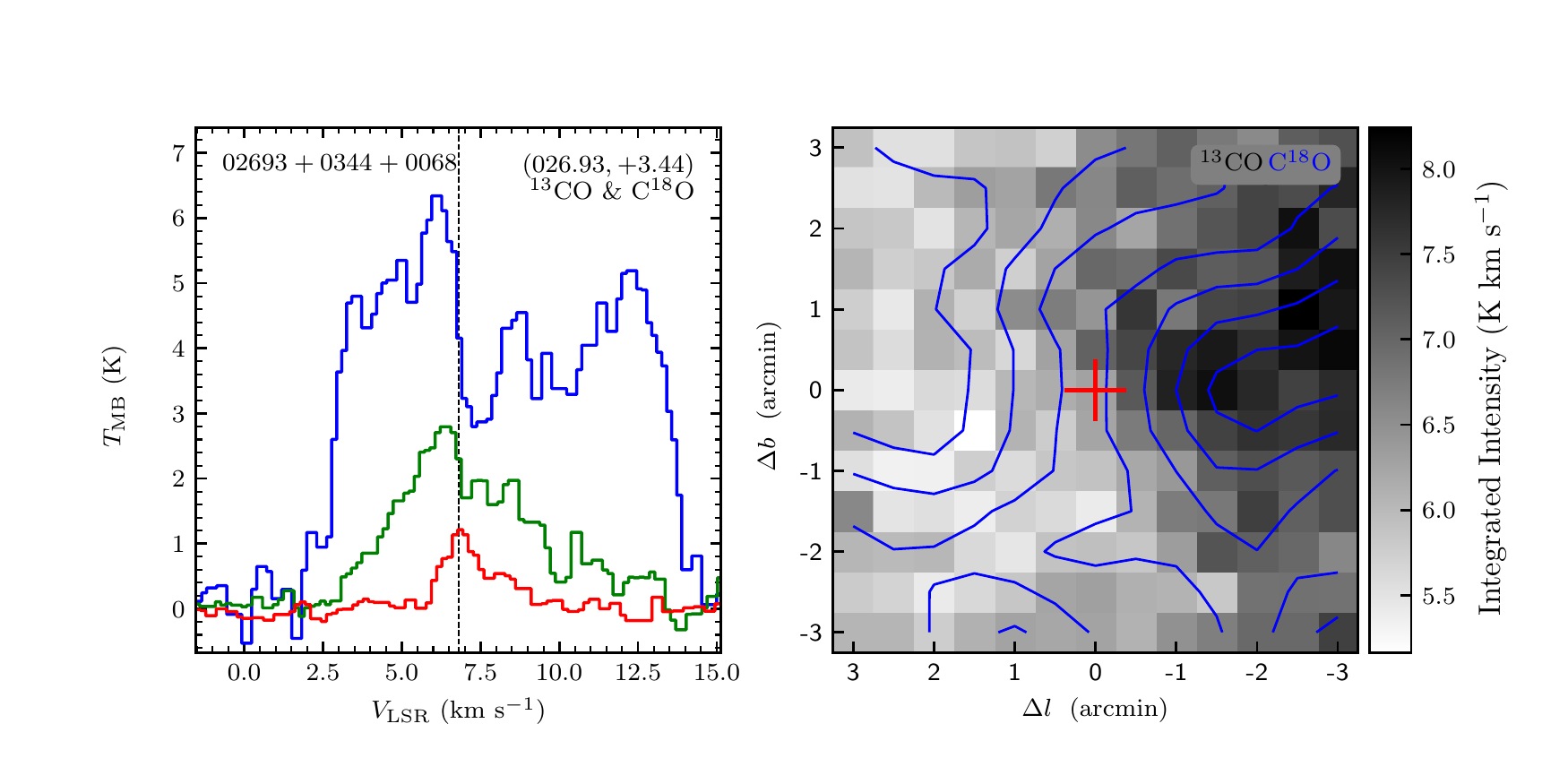}
\vspace{-0.5cm}

\includegraphics[width=9.0cm,angle=0]{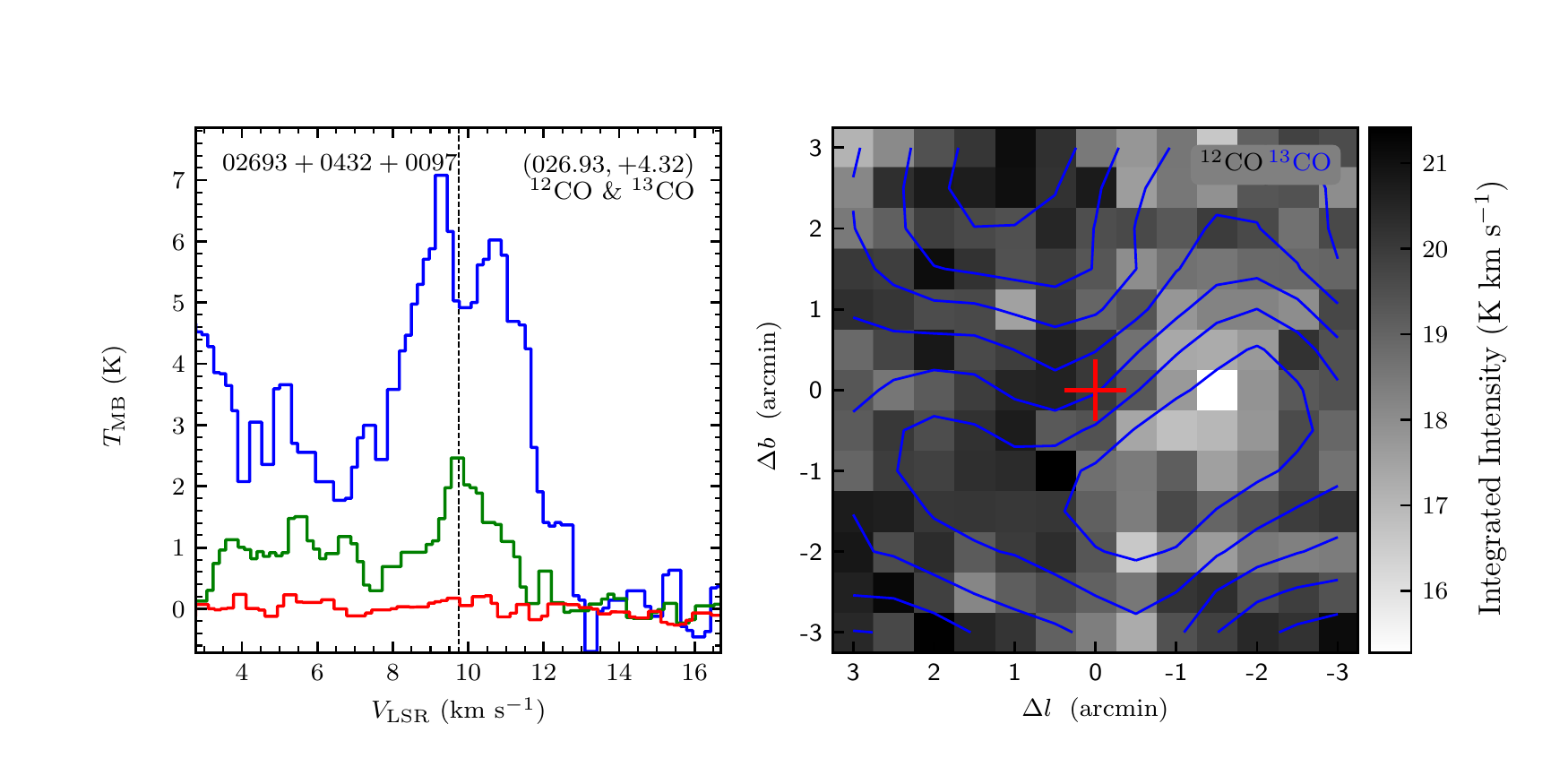}
\includegraphics[width=9.0cm,angle=0]{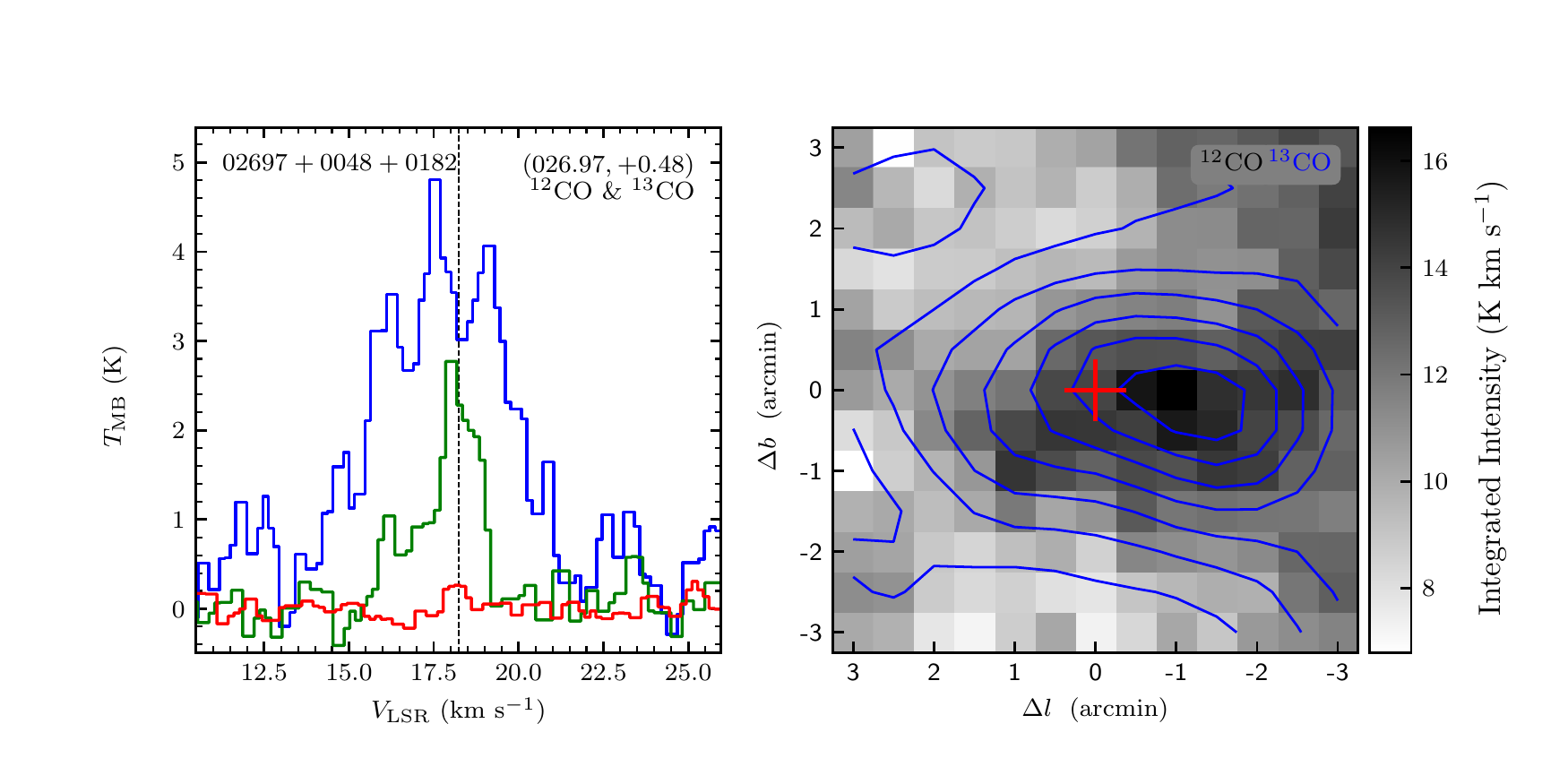}
\end{figure}
\clearpage

\begin{figure}
\includegraphics[width=9.0cm,angle=0]{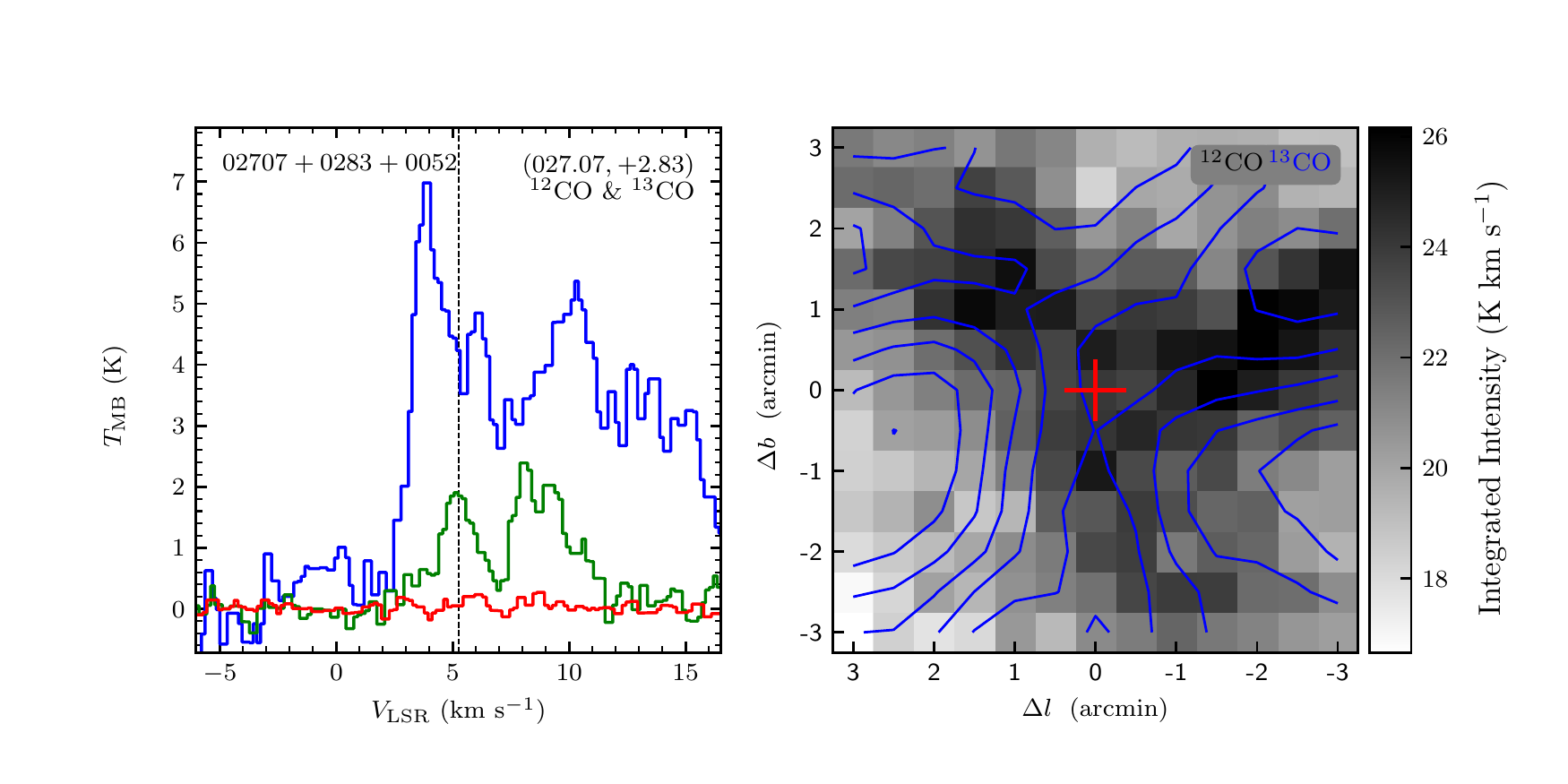}
\includegraphics[width=9.0cm,angle=0]{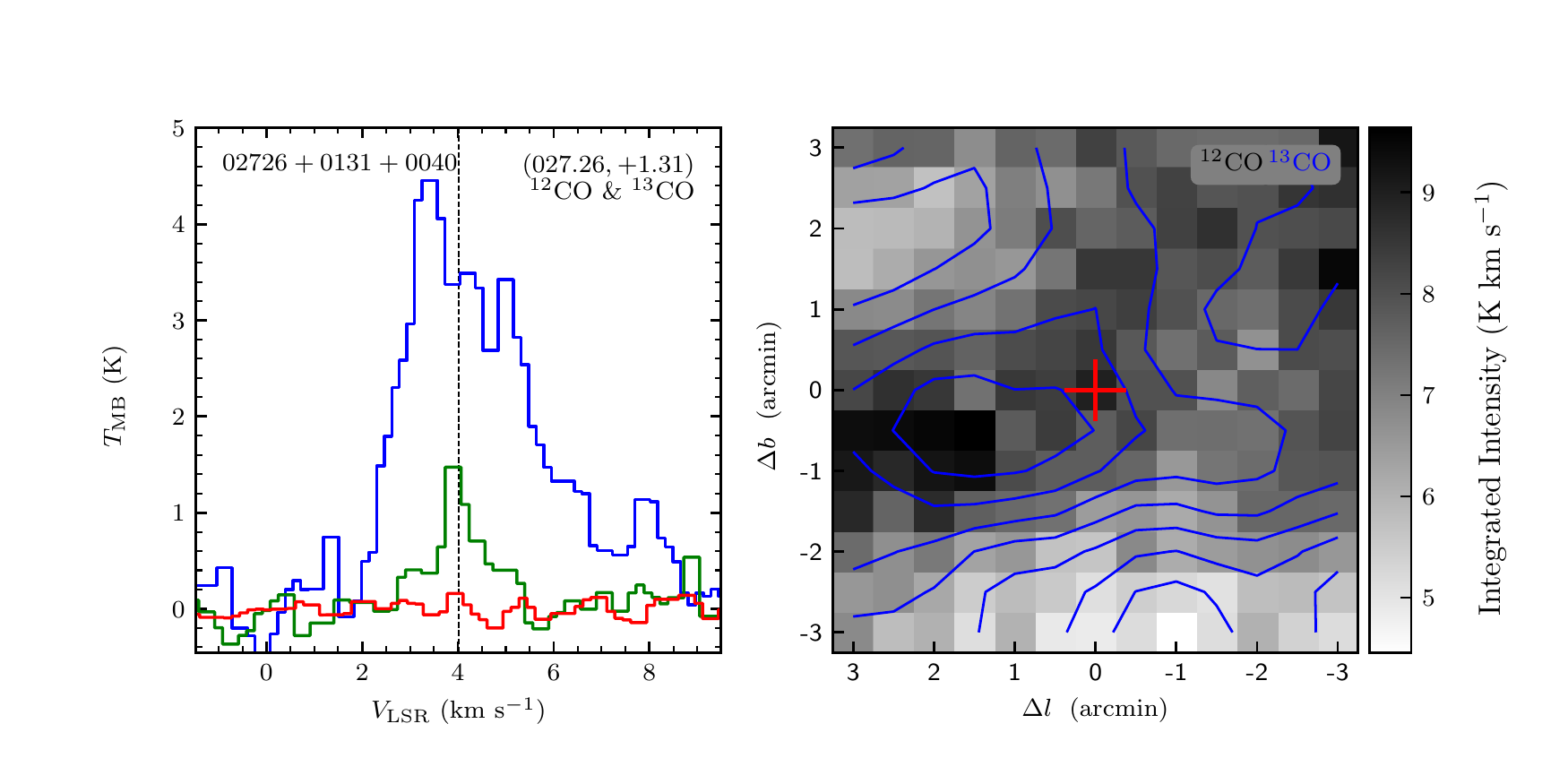}
\vspace{-0.5cm}

\includegraphics[width=9.0cm,angle=0]{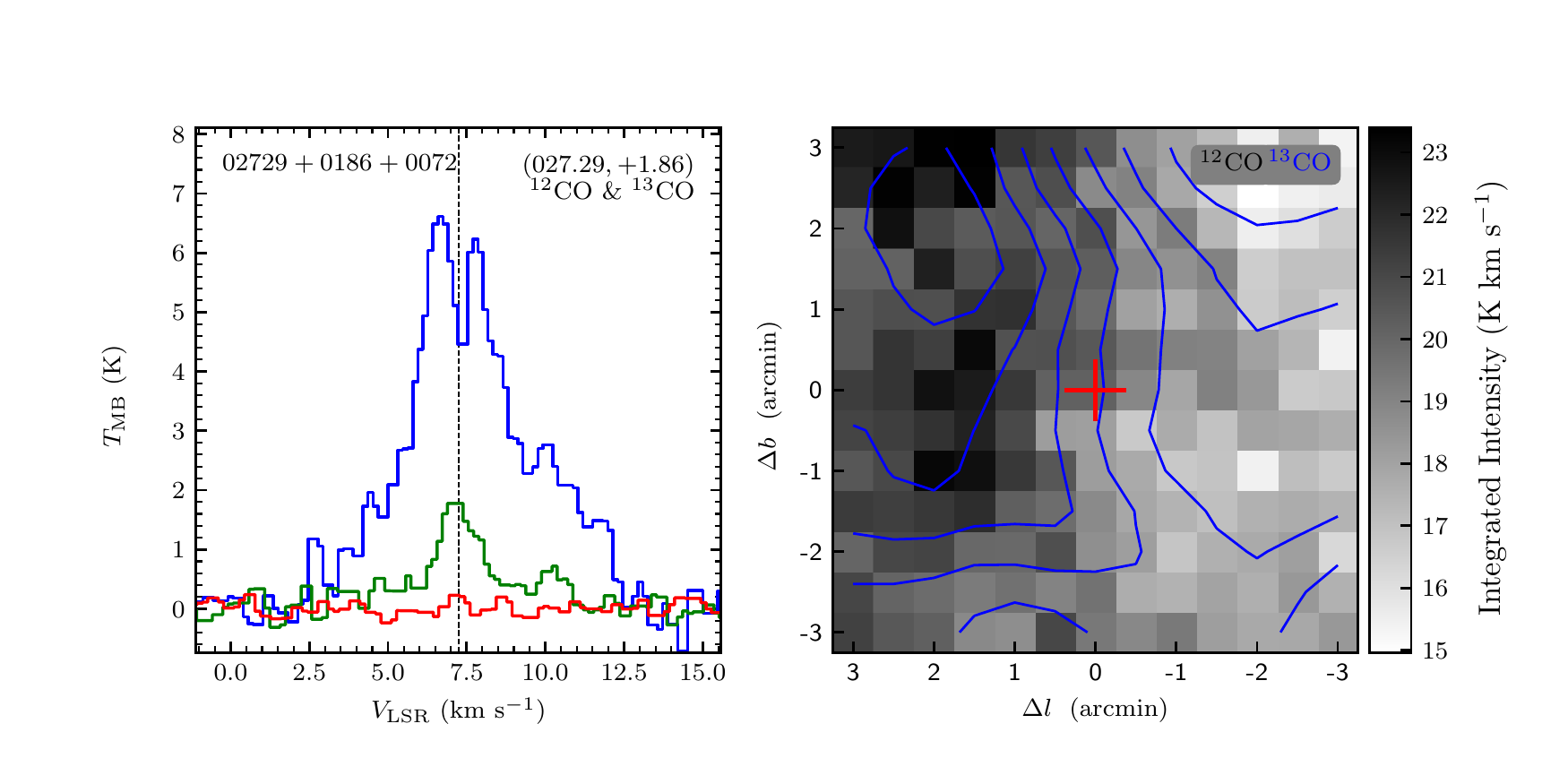}
\includegraphics[width=9.0cm,angle=0]{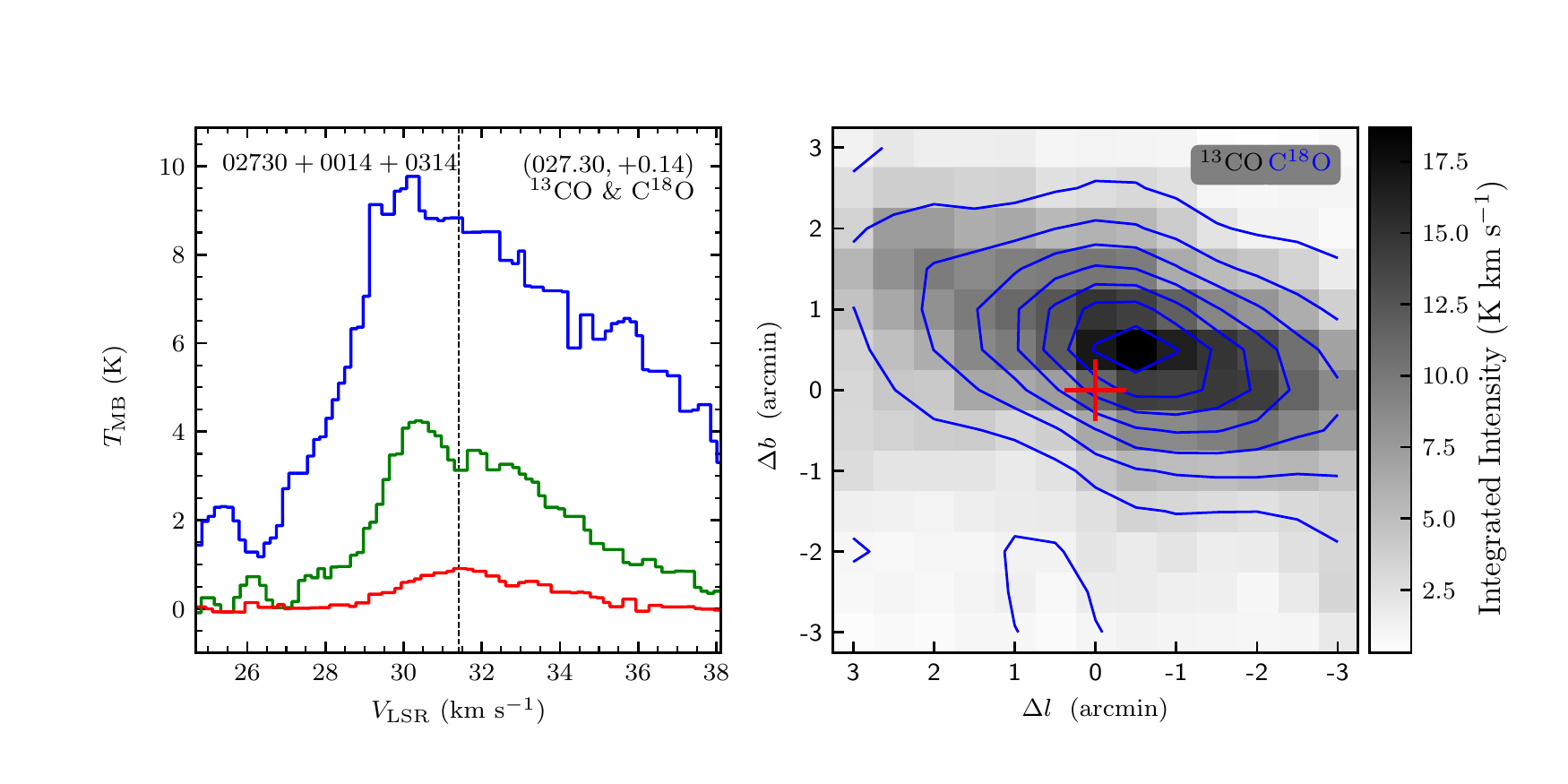}
\vspace{-0.5cm}

\includegraphics[width=9.0cm,angle=0]{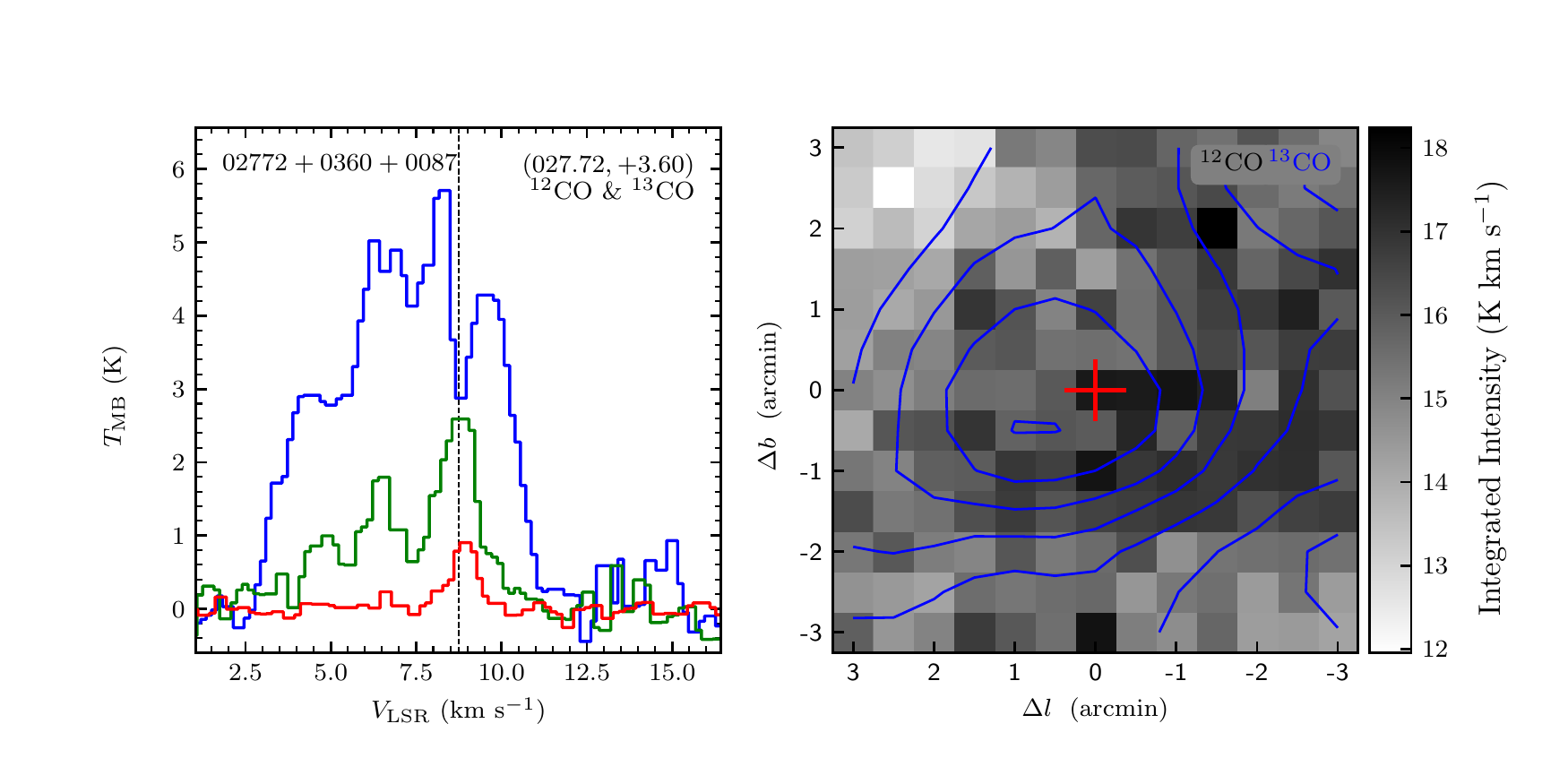}
\includegraphics[width=9.0cm,angle=0]{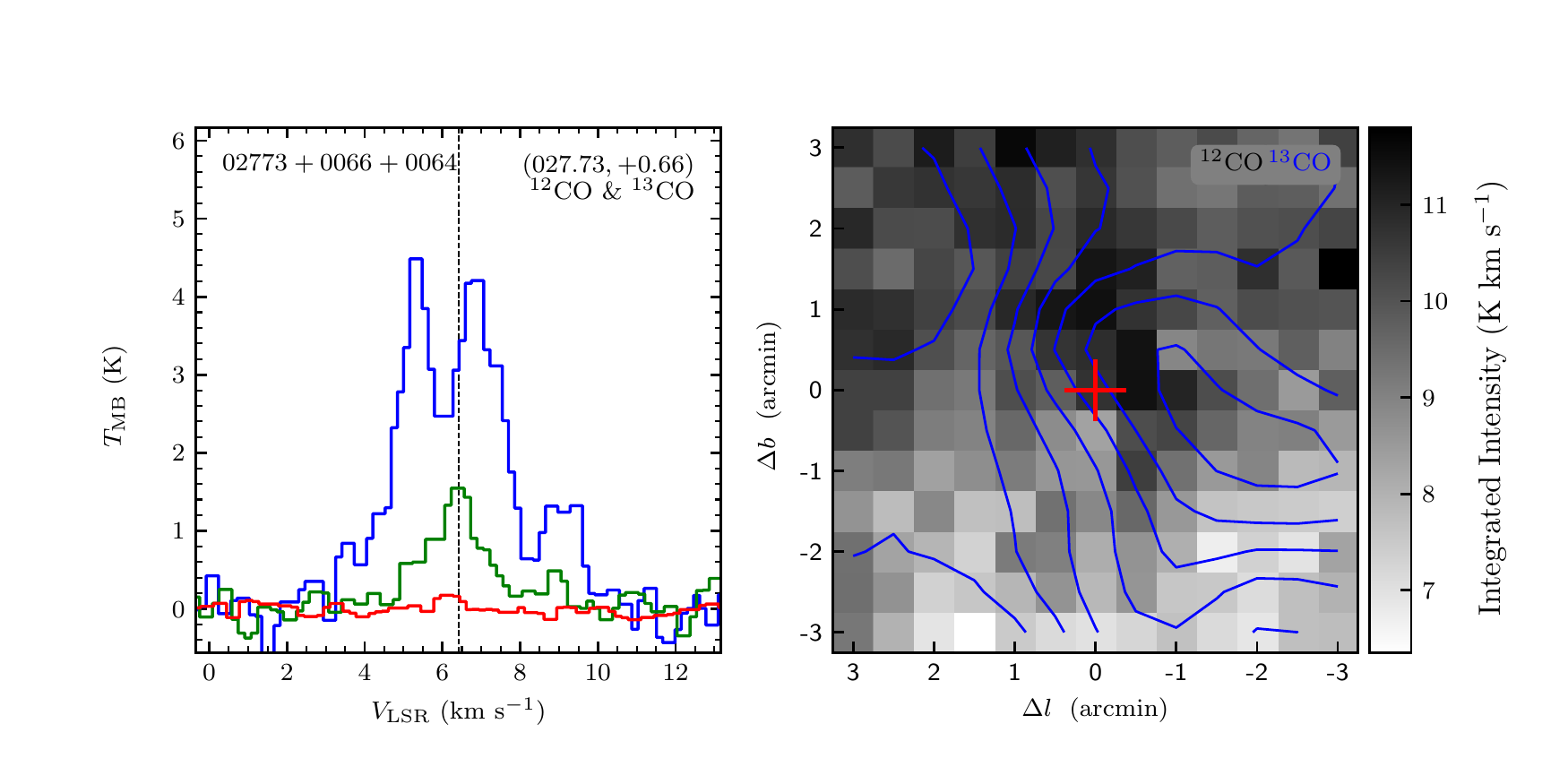}
\vspace{-0.5cm}

\includegraphics[width=9.0cm,angle=0]{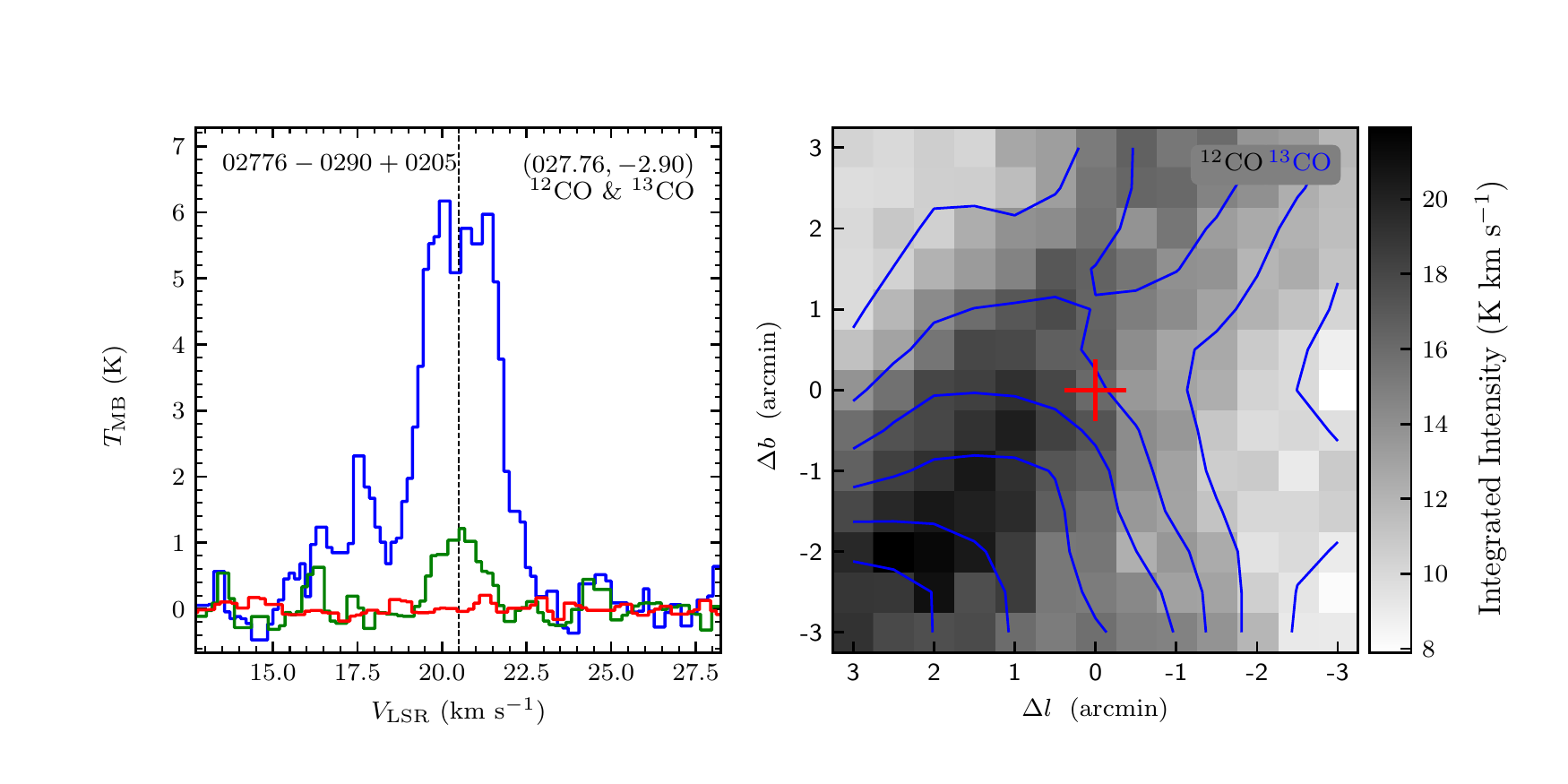}
\includegraphics[width=9.0cm,angle=0]{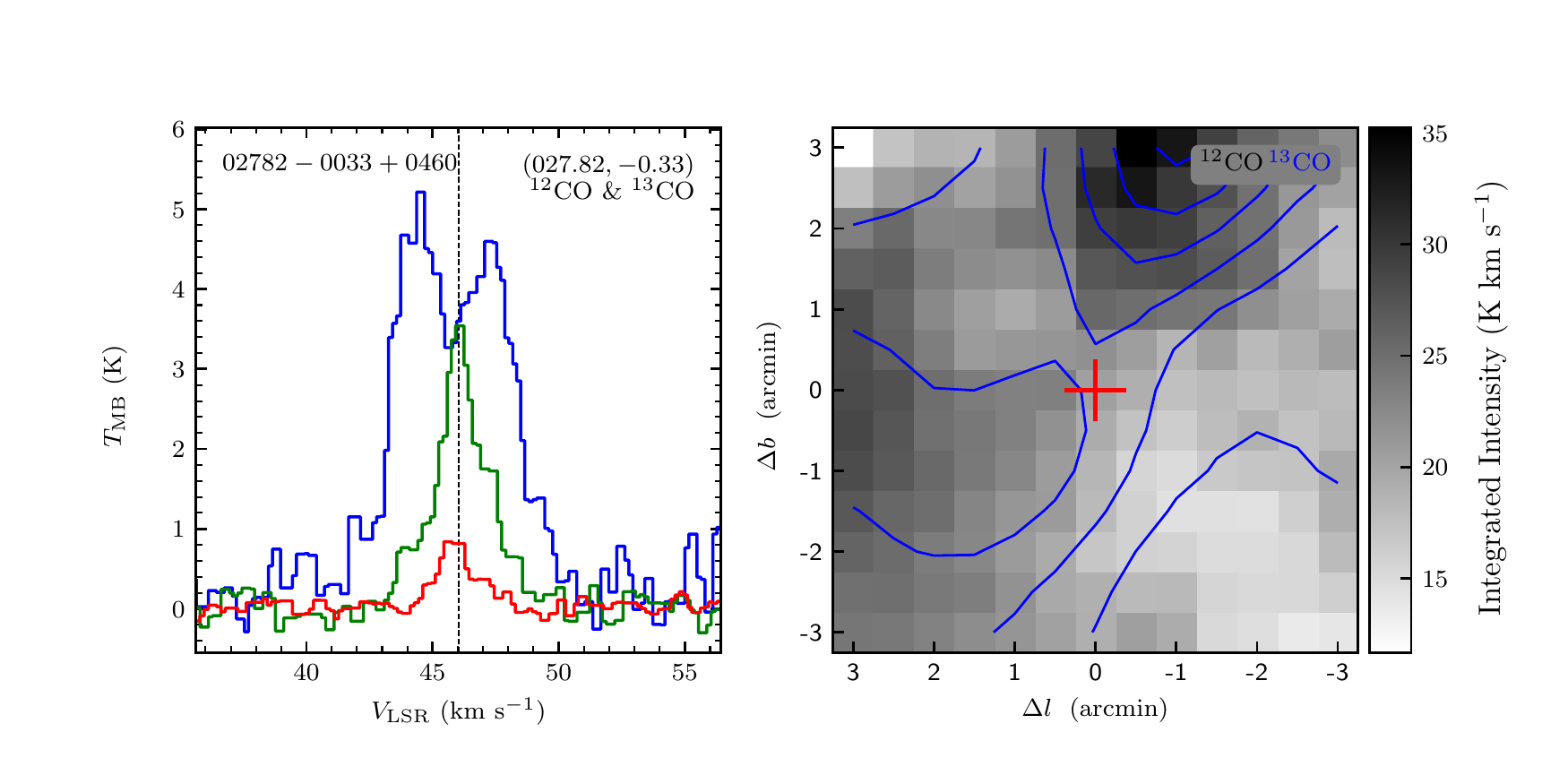}
\vspace{-0.5cm}

\includegraphics[width=9.0cm,angle=0]{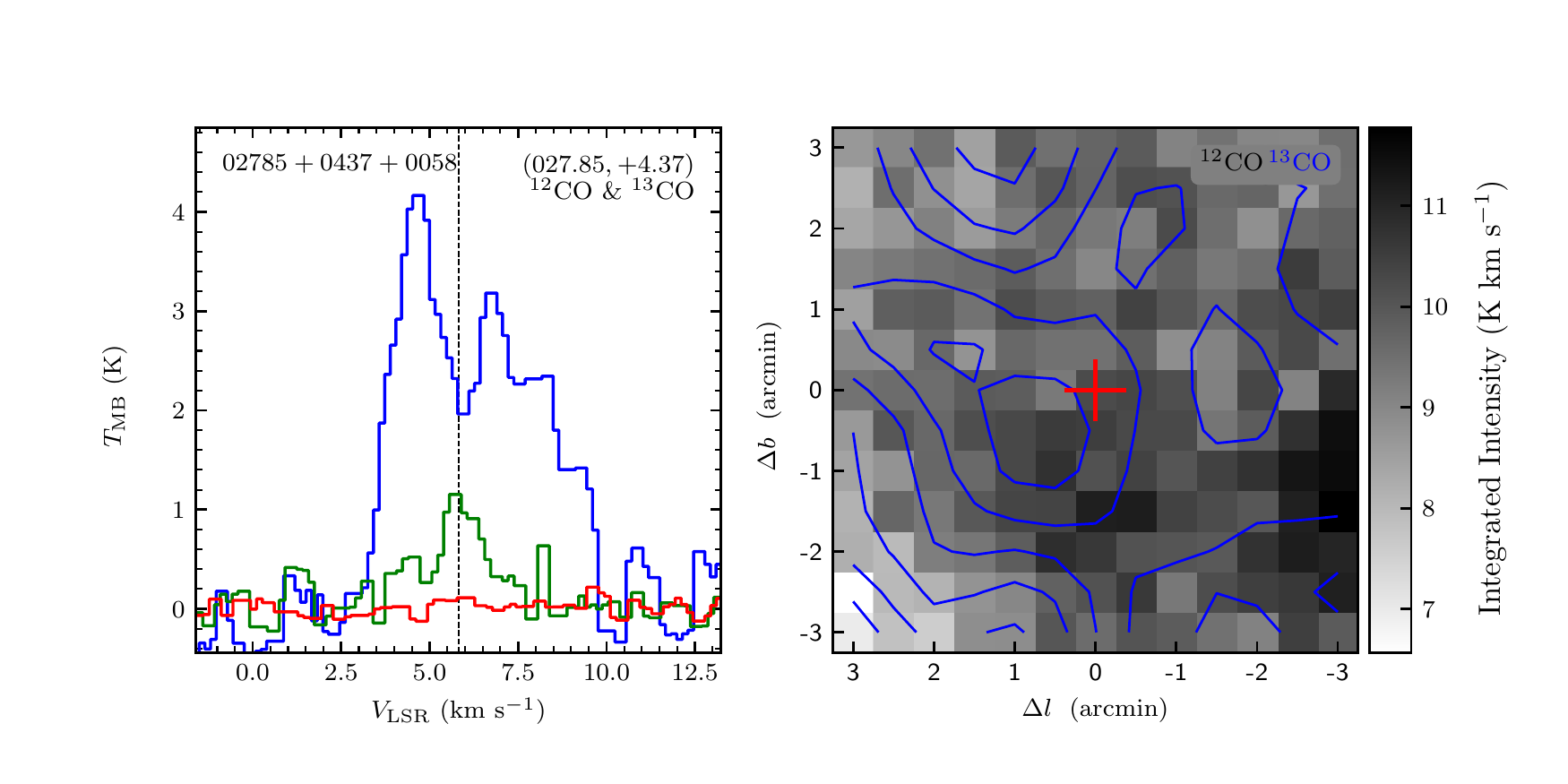}
\includegraphics[width=9.0cm,angle=0]{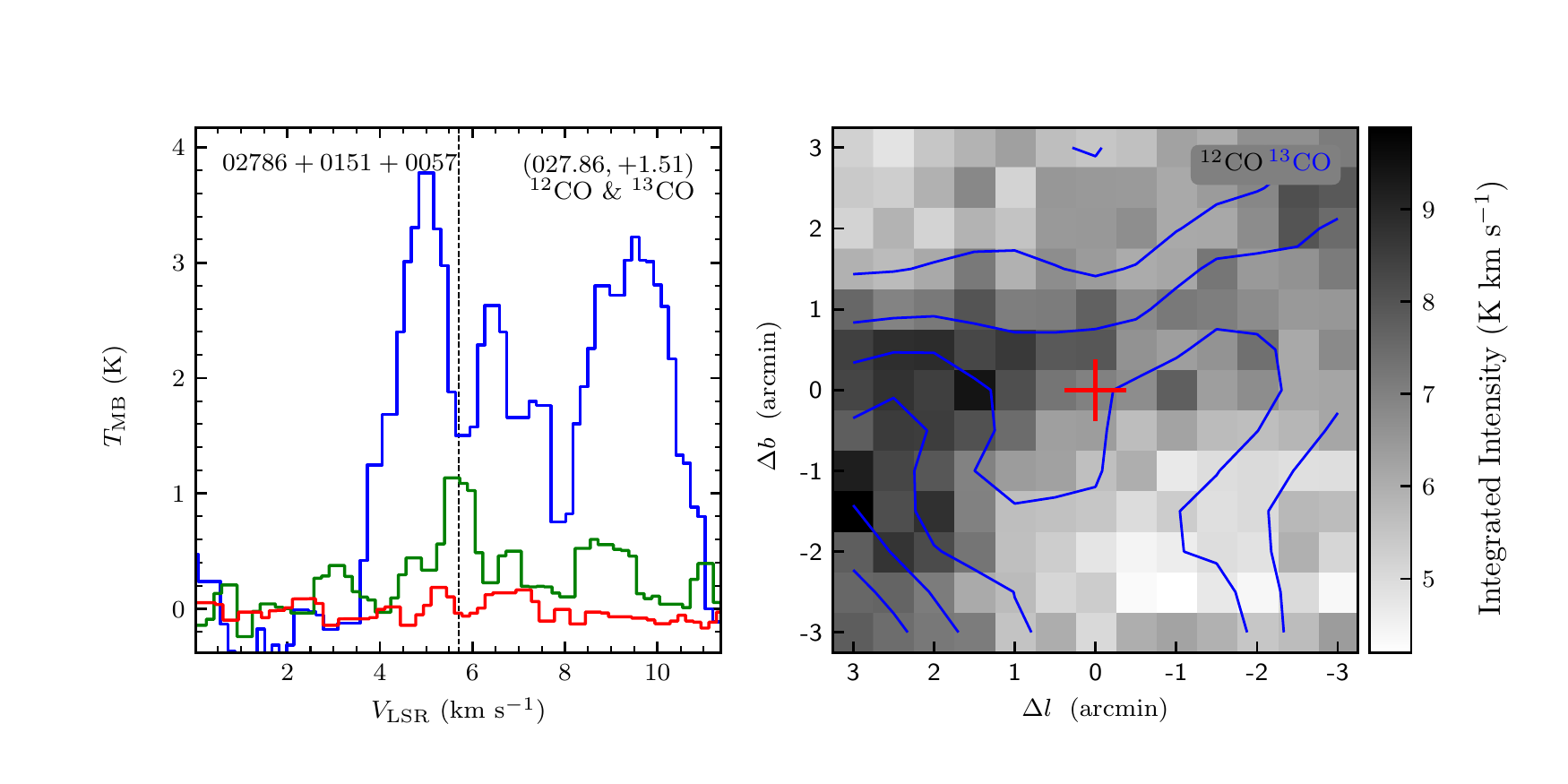}
\end{figure}
\clearpage

\begin{figure}
\includegraphics[width=9.0cm,angle=0]{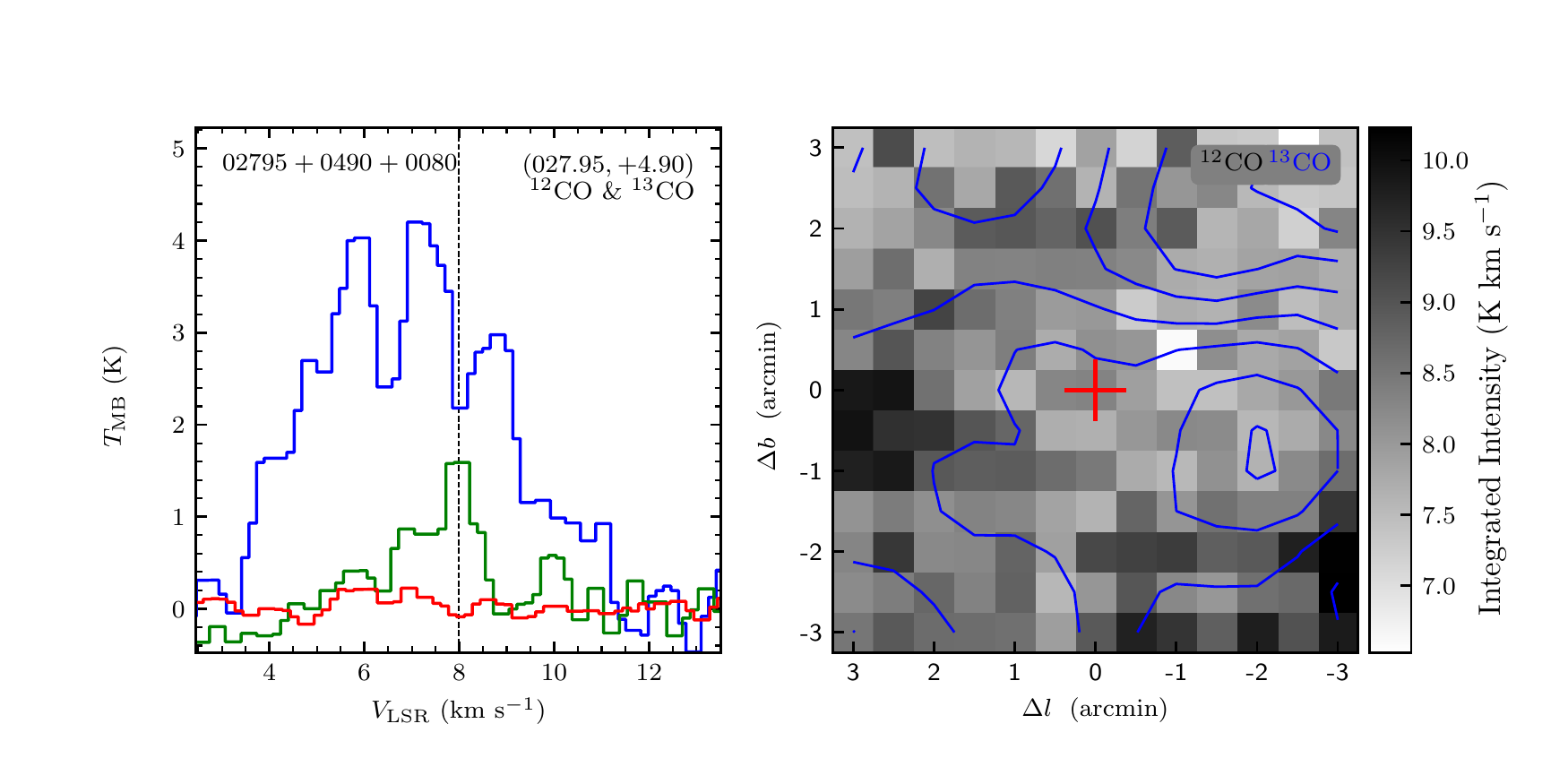}
\includegraphics[width=9.0cm,angle=0]{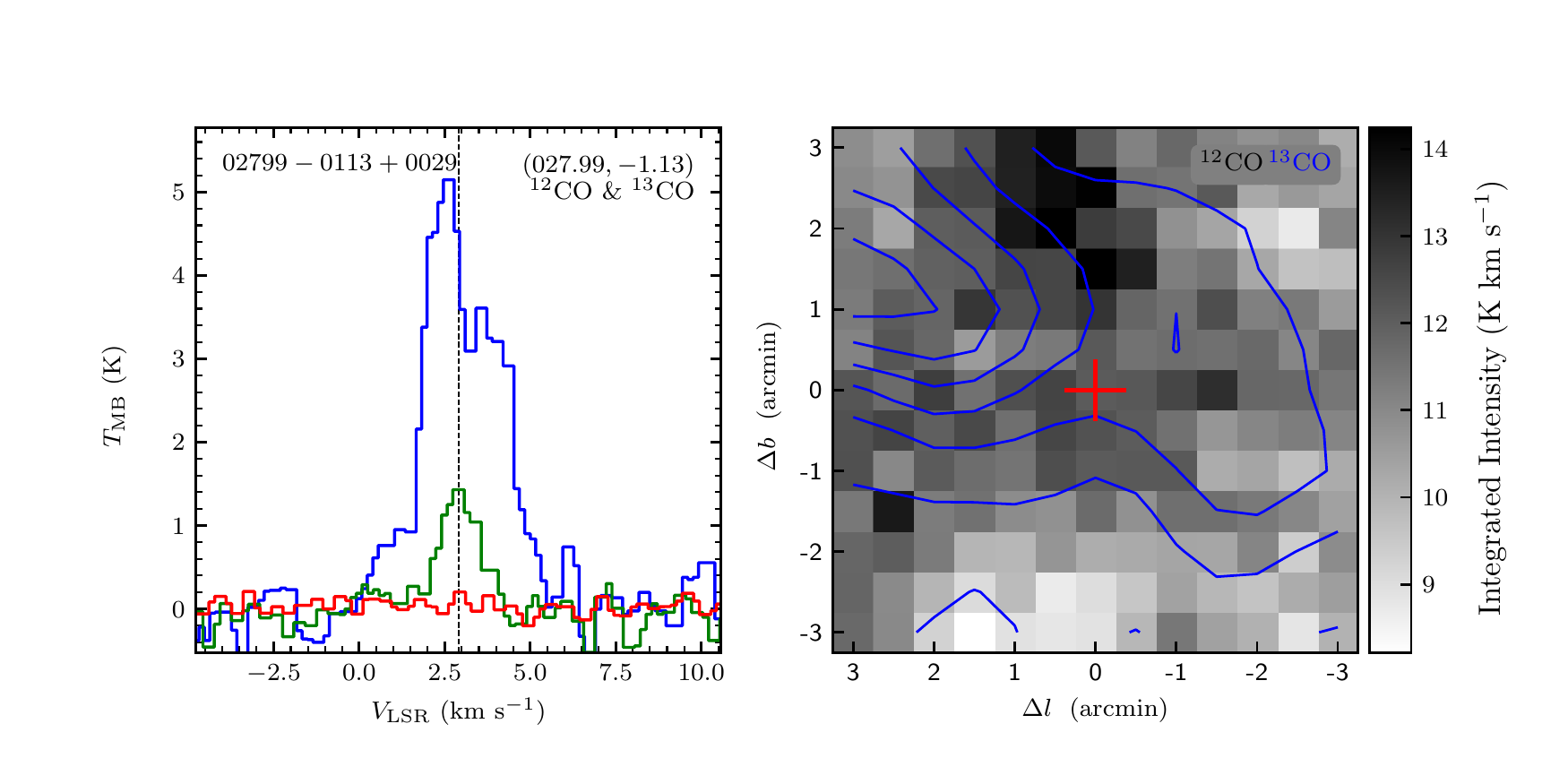}
\vspace{-0.5cm}

\includegraphics[width=9.0cm,angle=0]{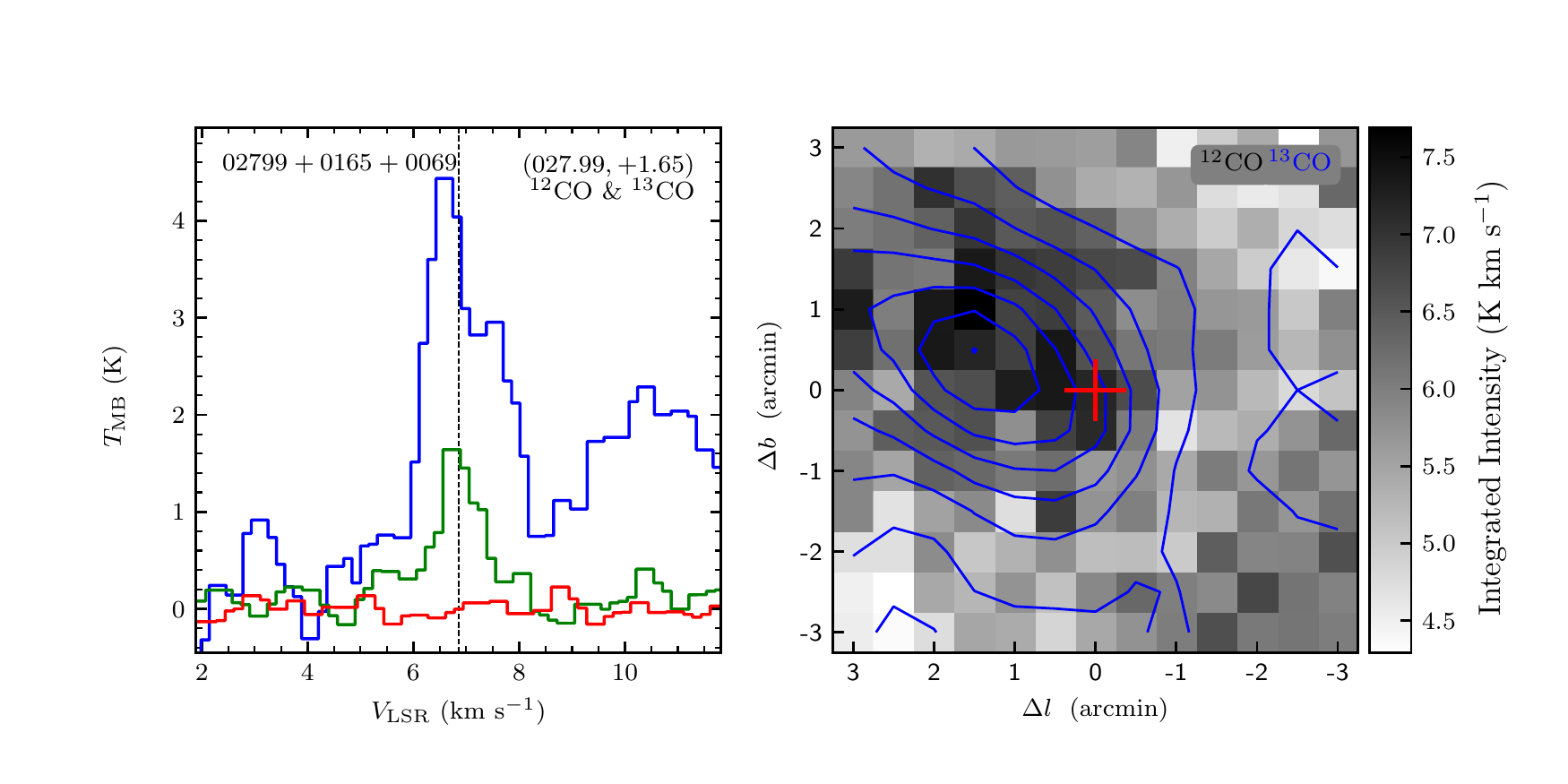}
\includegraphics[width=9.0cm,angle=0]{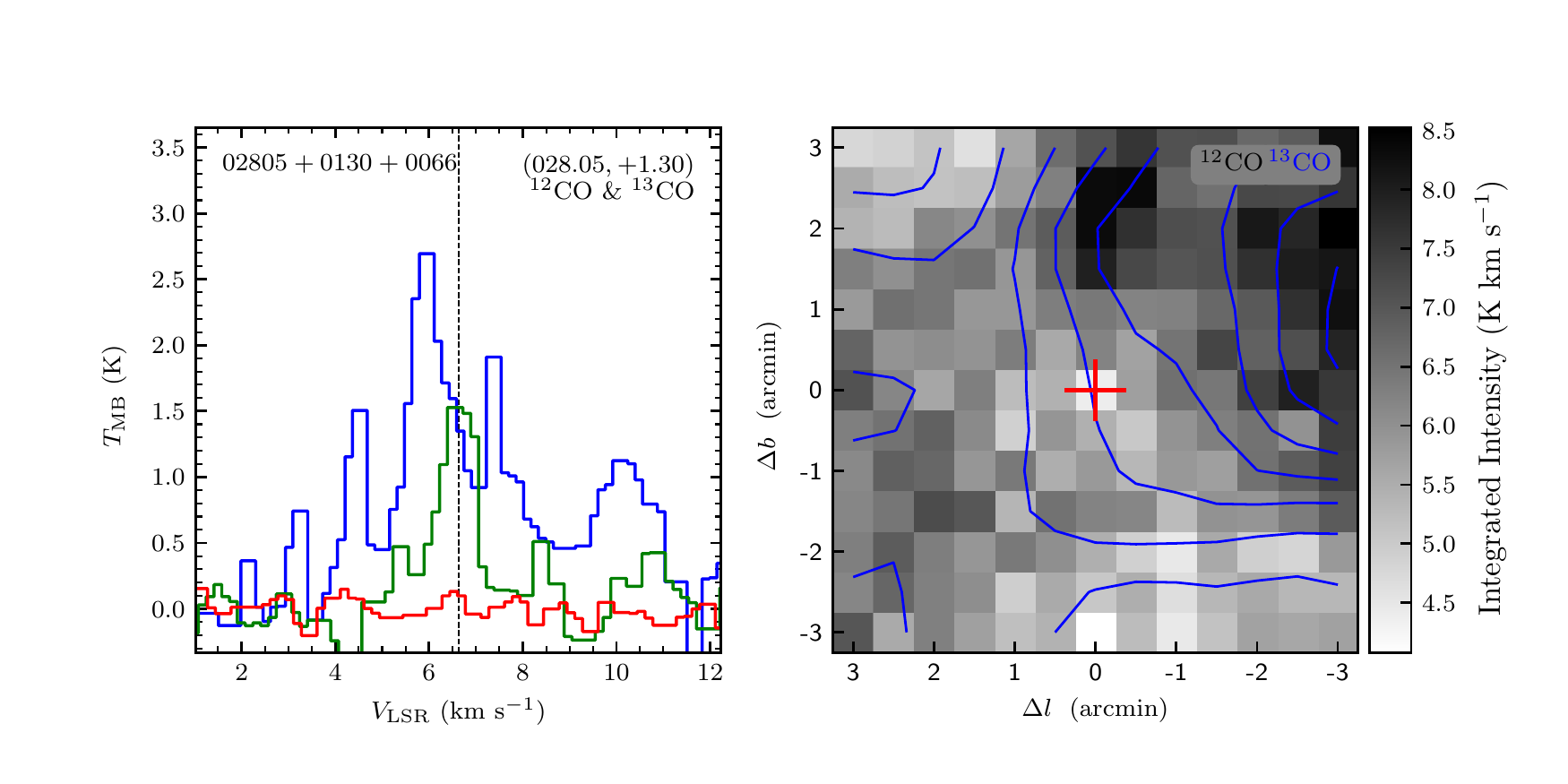}
\vspace{-0.5cm}

\includegraphics[width=9.0cm,angle=0]{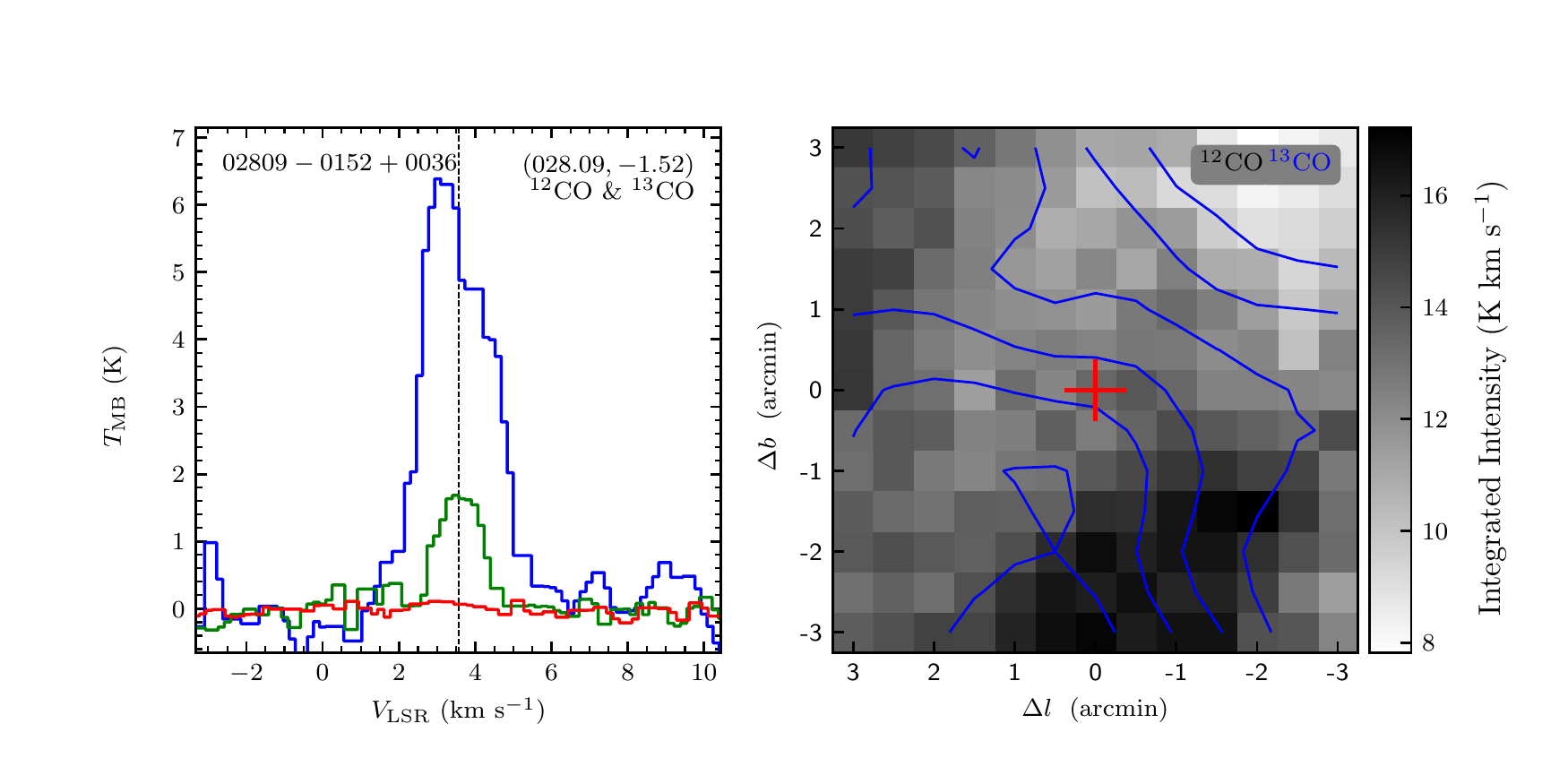}
\includegraphics[width=9.0cm,angle=0]{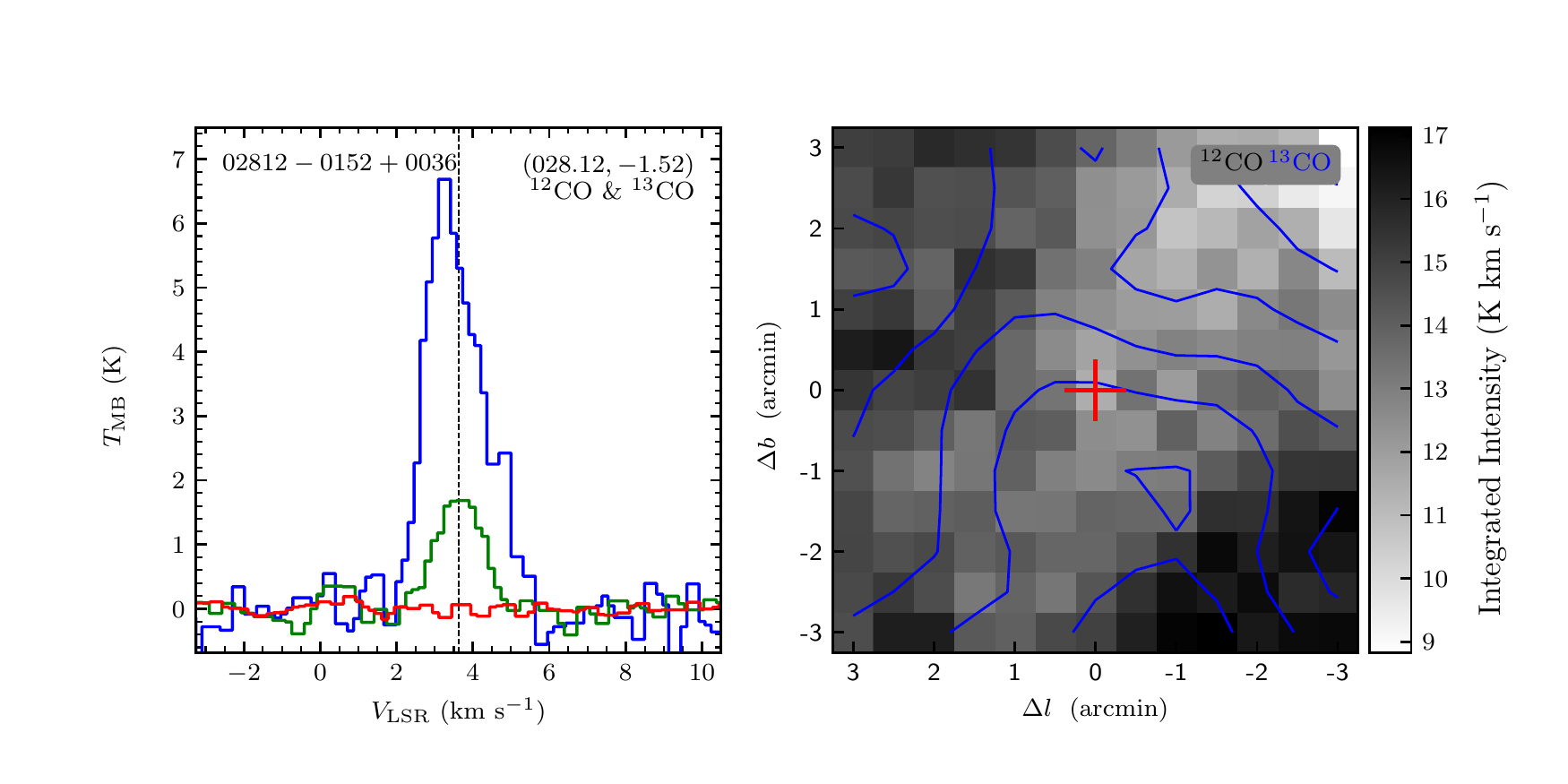}
\vspace{-0.5cm}

\includegraphics[width=9.0cm,angle=0]{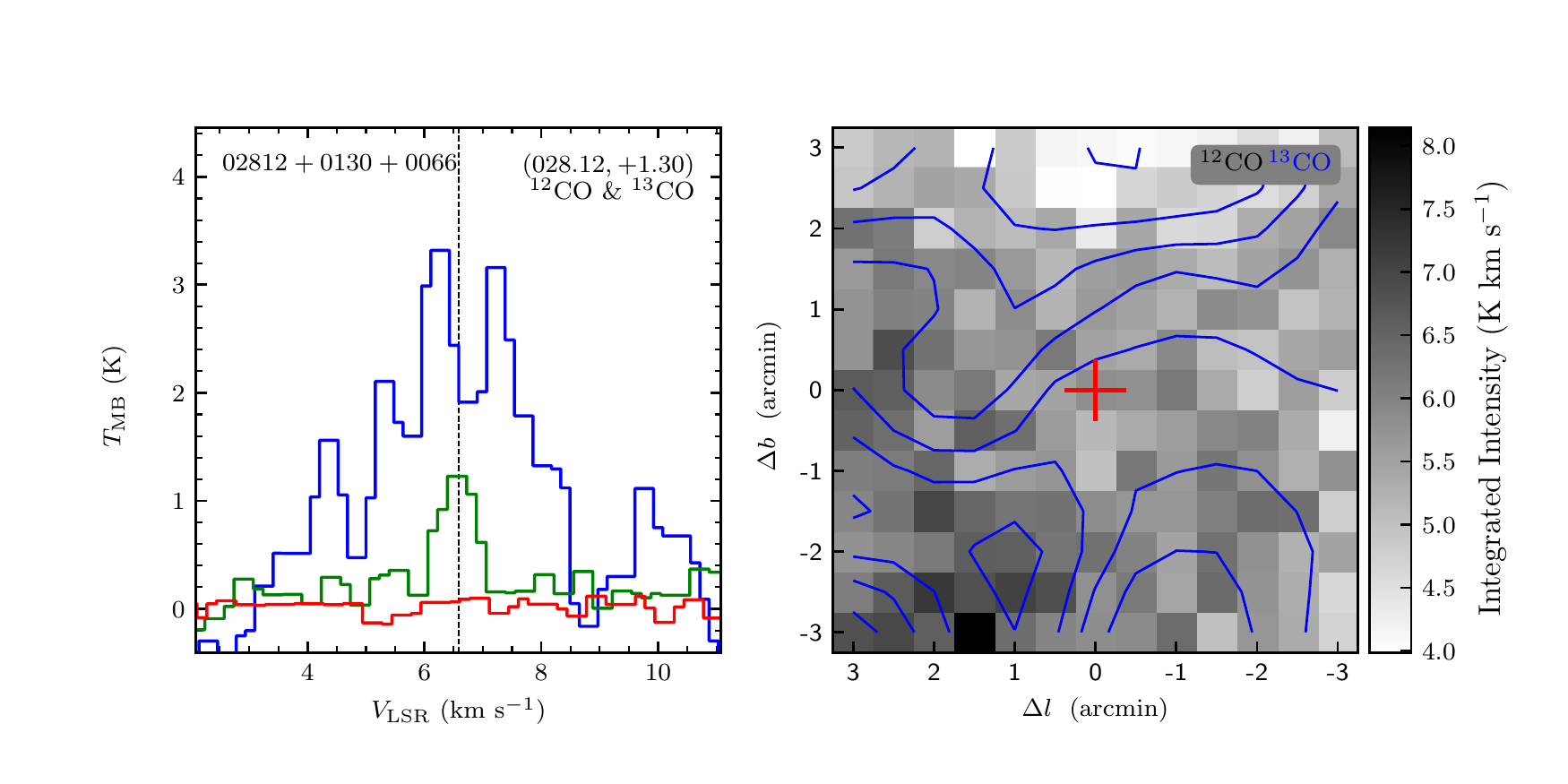}
\includegraphics[width=9.0cm,angle=0]{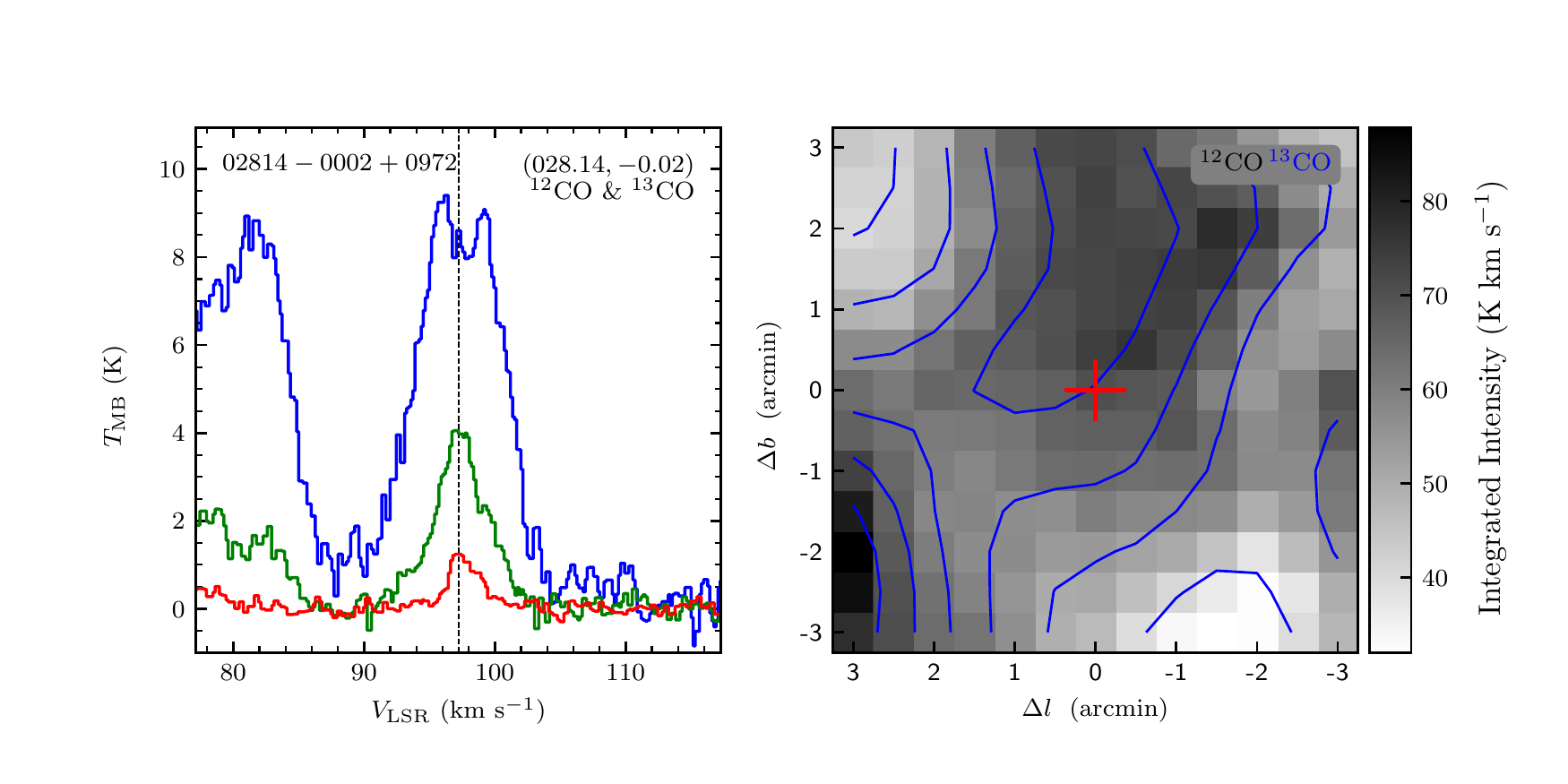}
\vspace{-0.5cm}

\includegraphics[width=9.0cm,angle=0]{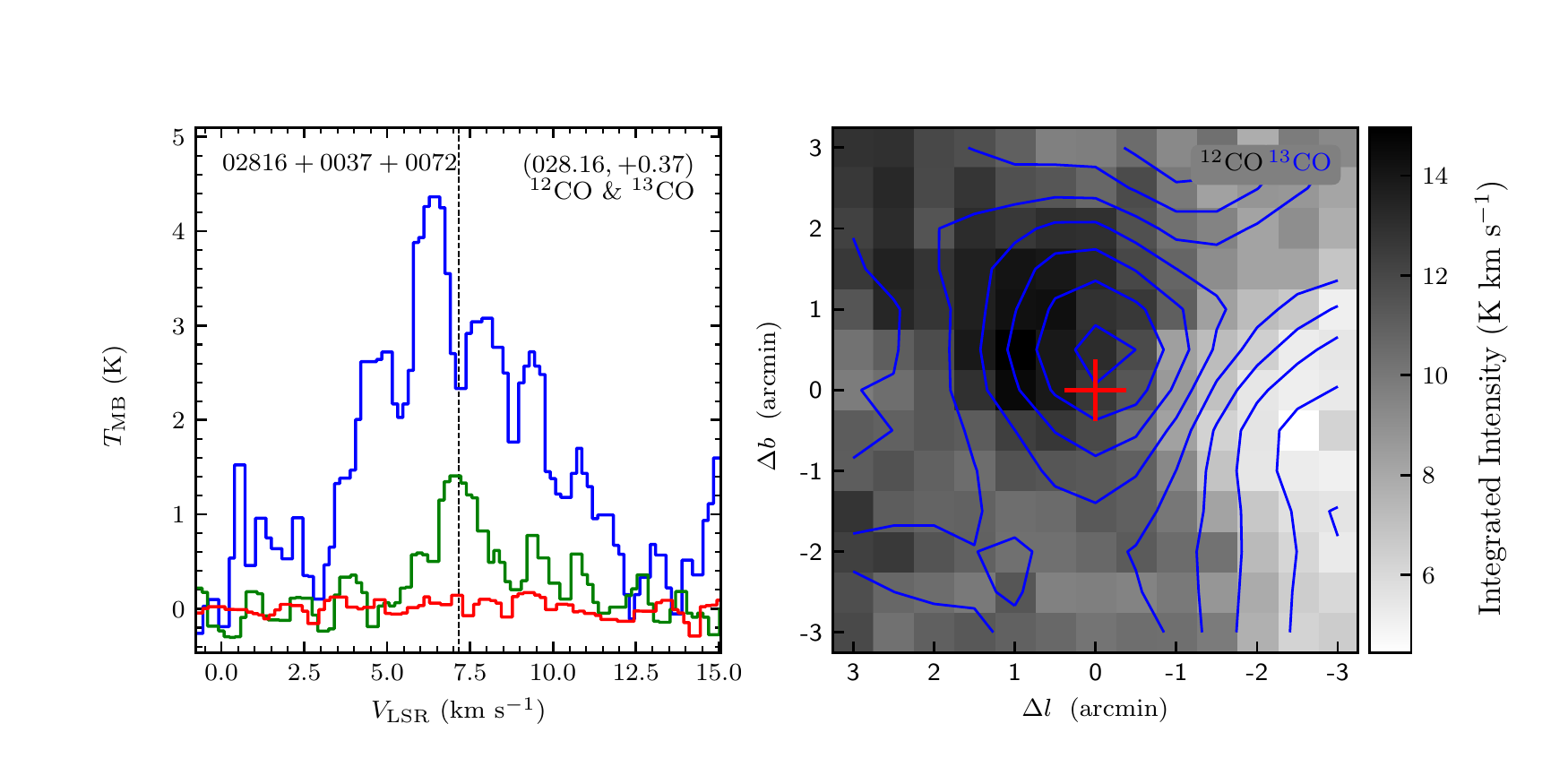}
\includegraphics[width=9.0cm,angle=0]{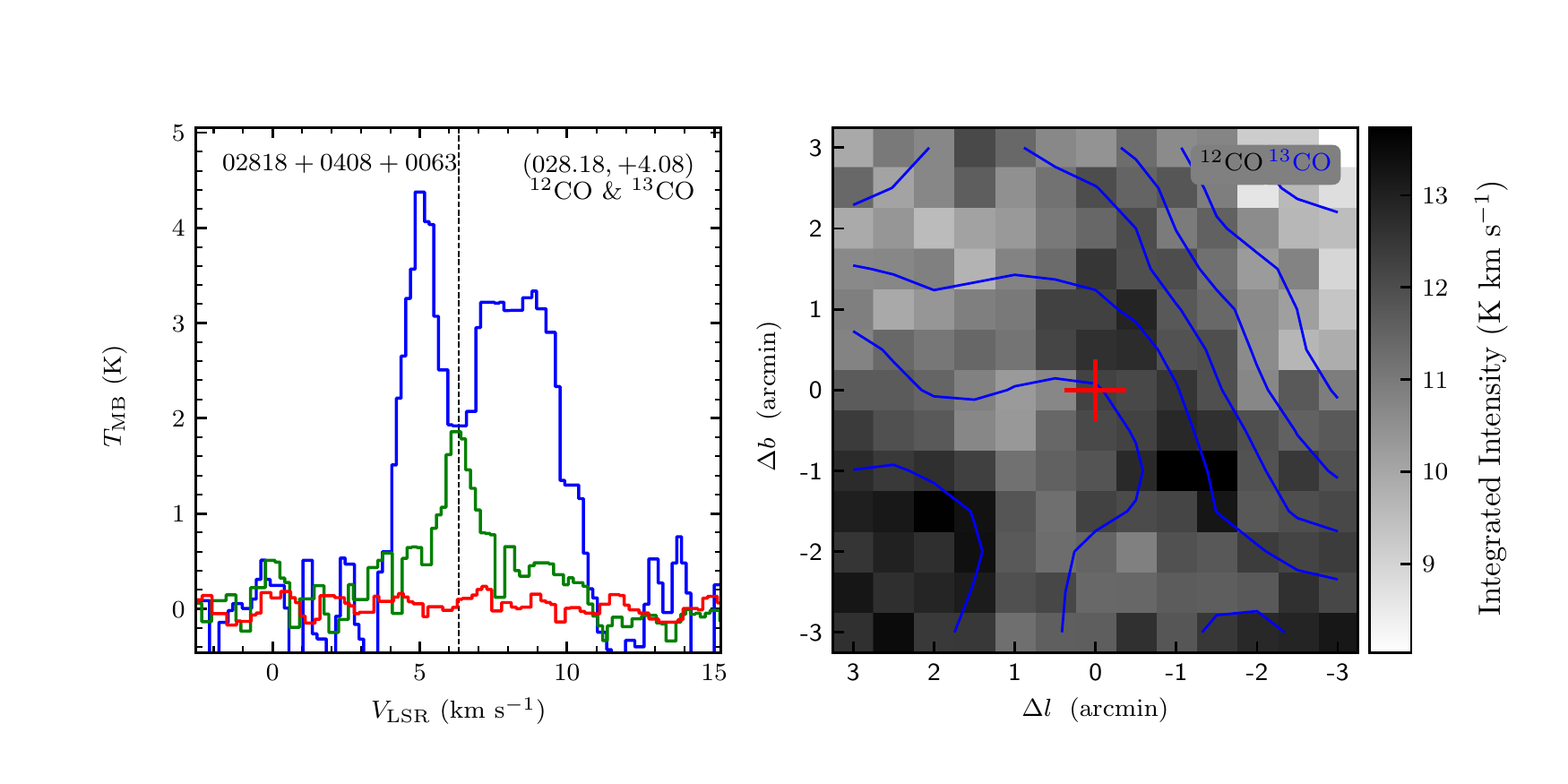}
\end{figure}
\clearpage

\begin{figure}
\includegraphics[width=9.0cm,angle=0]{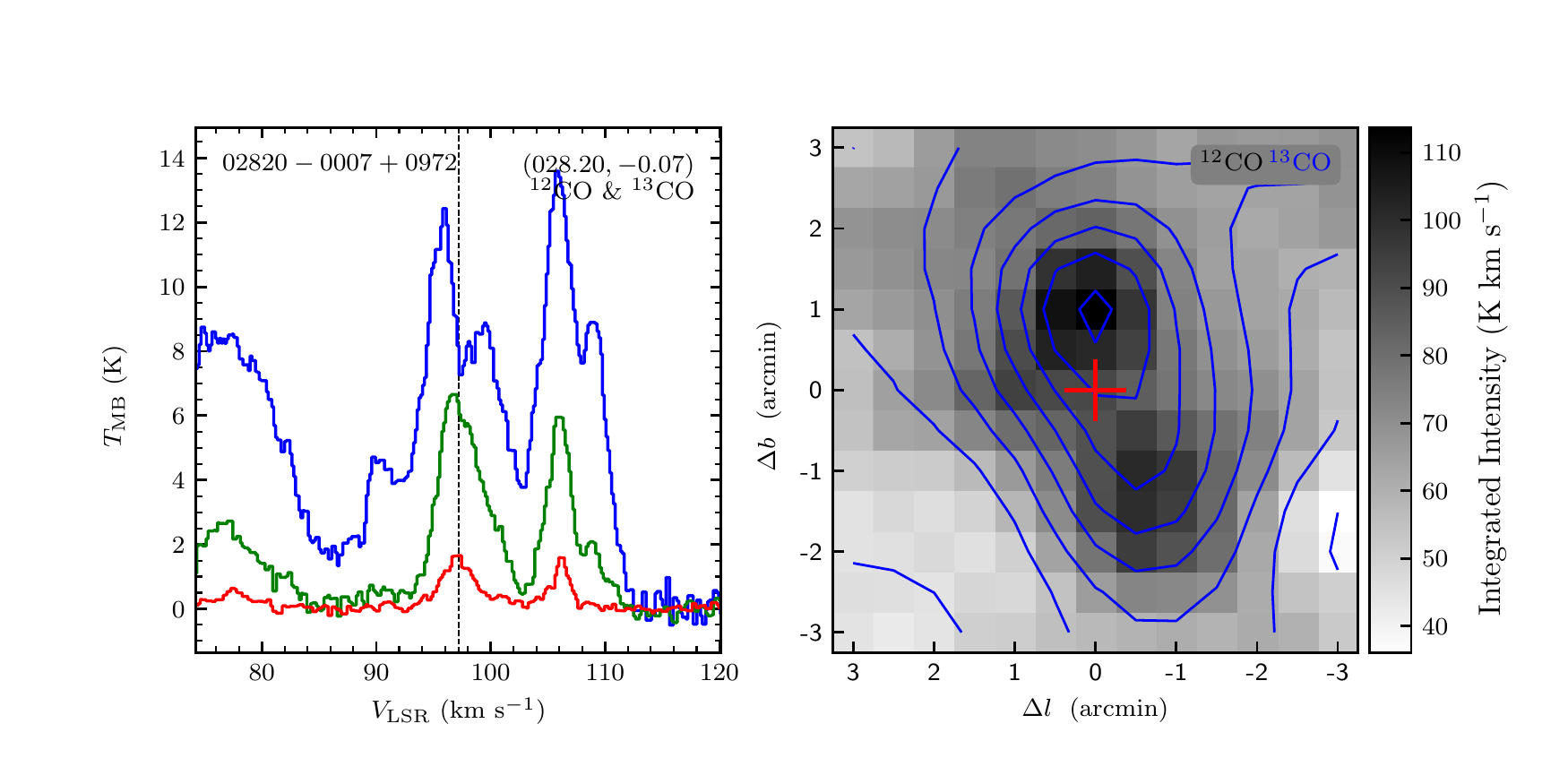}
\includegraphics[width=9.0cm,angle=0]{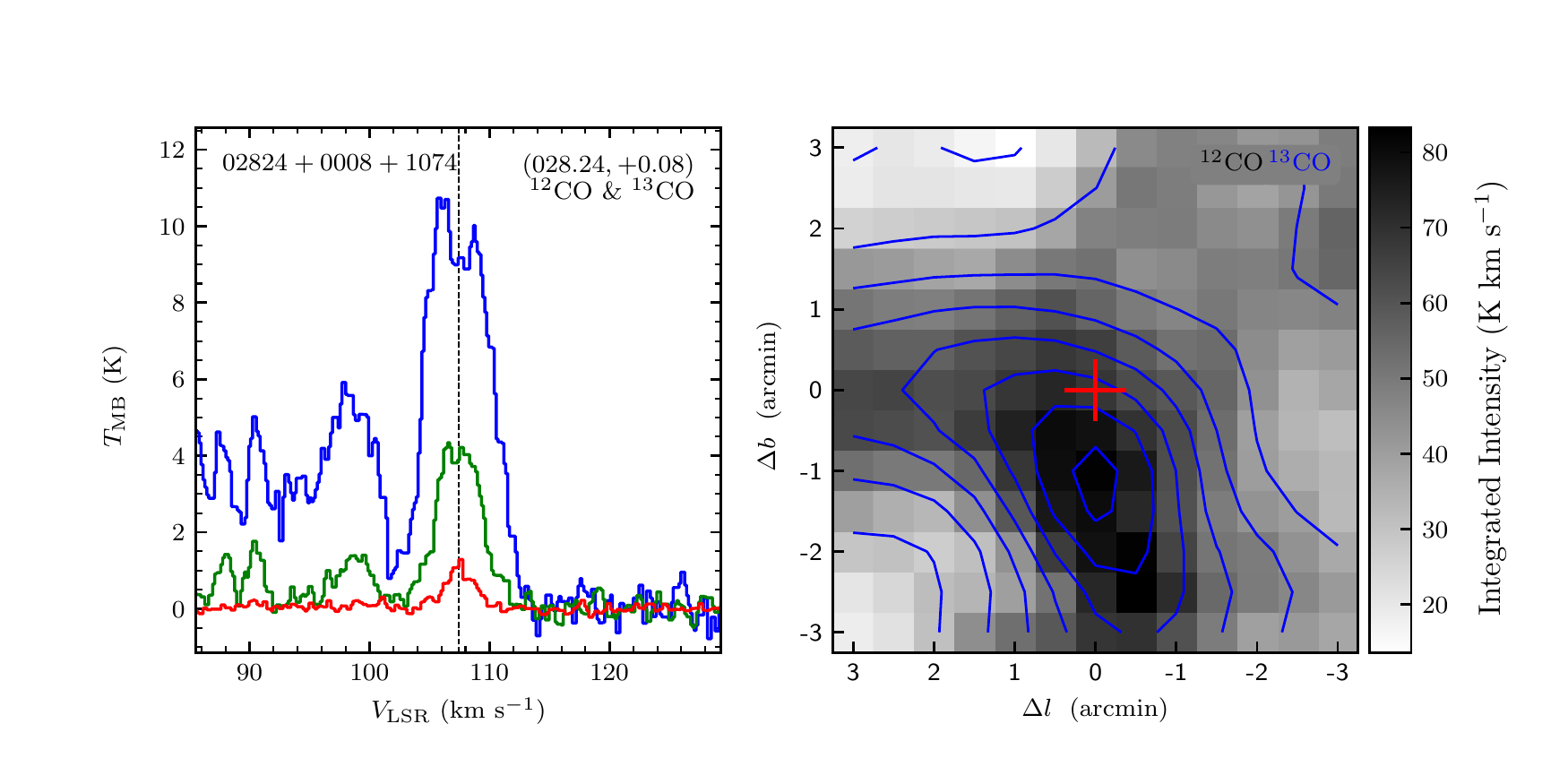}
\vspace{-0.5cm}

\includegraphics[width=9.0cm,angle=0]{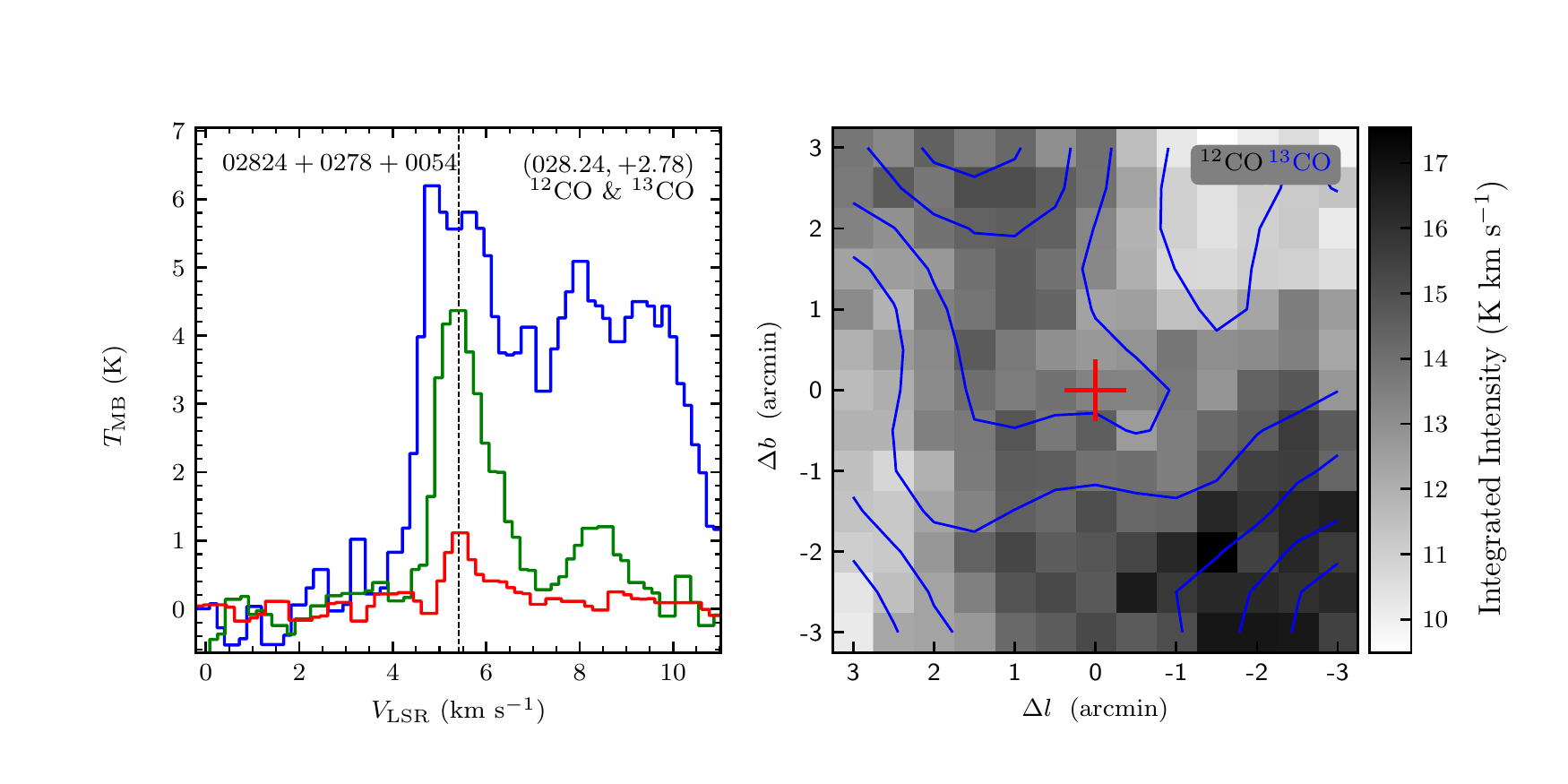}
\includegraphics[width=9.0cm,angle=0]{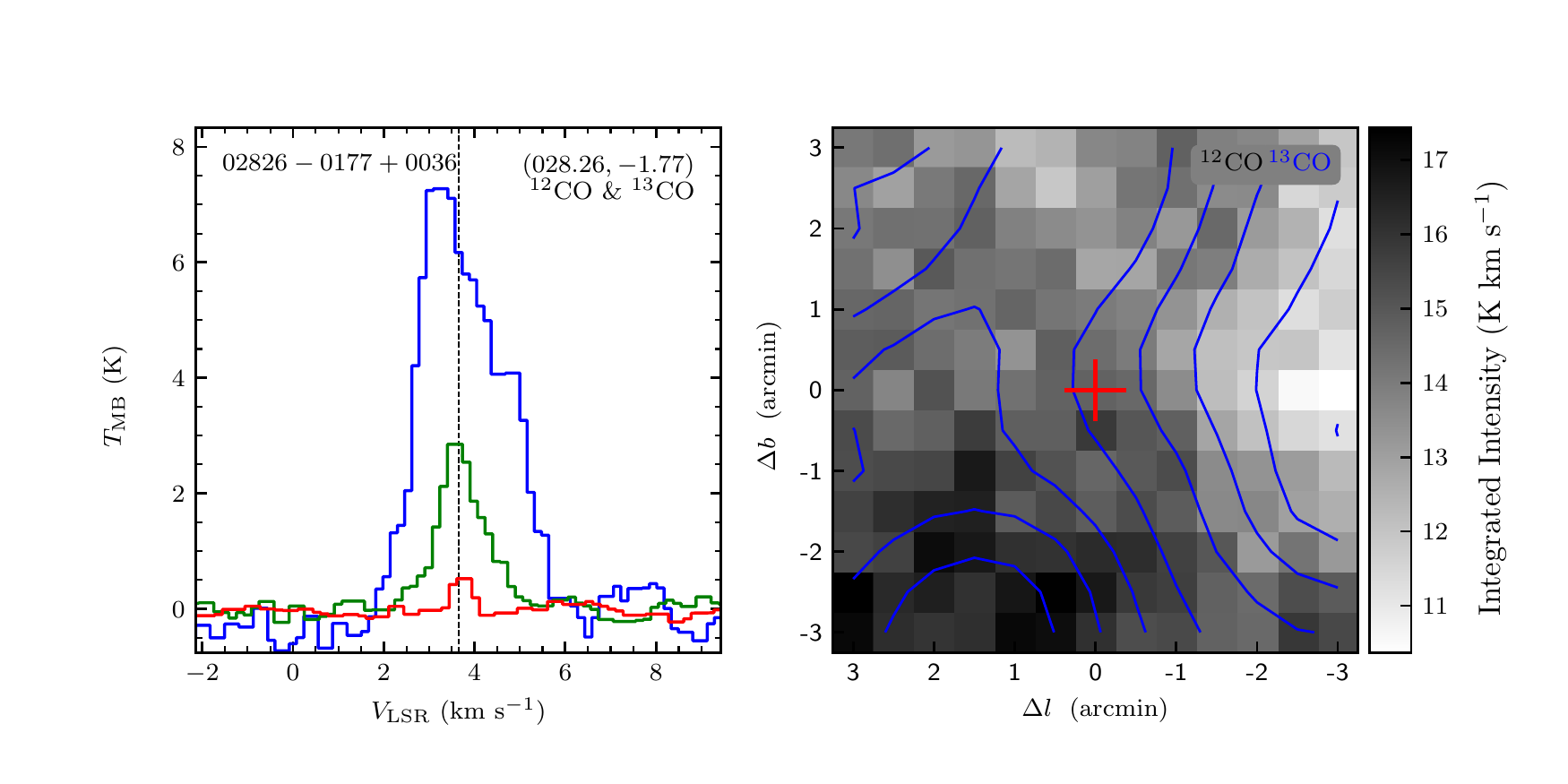}
\vspace{-0.5cm}

\includegraphics[width=9.0cm,angle=0]{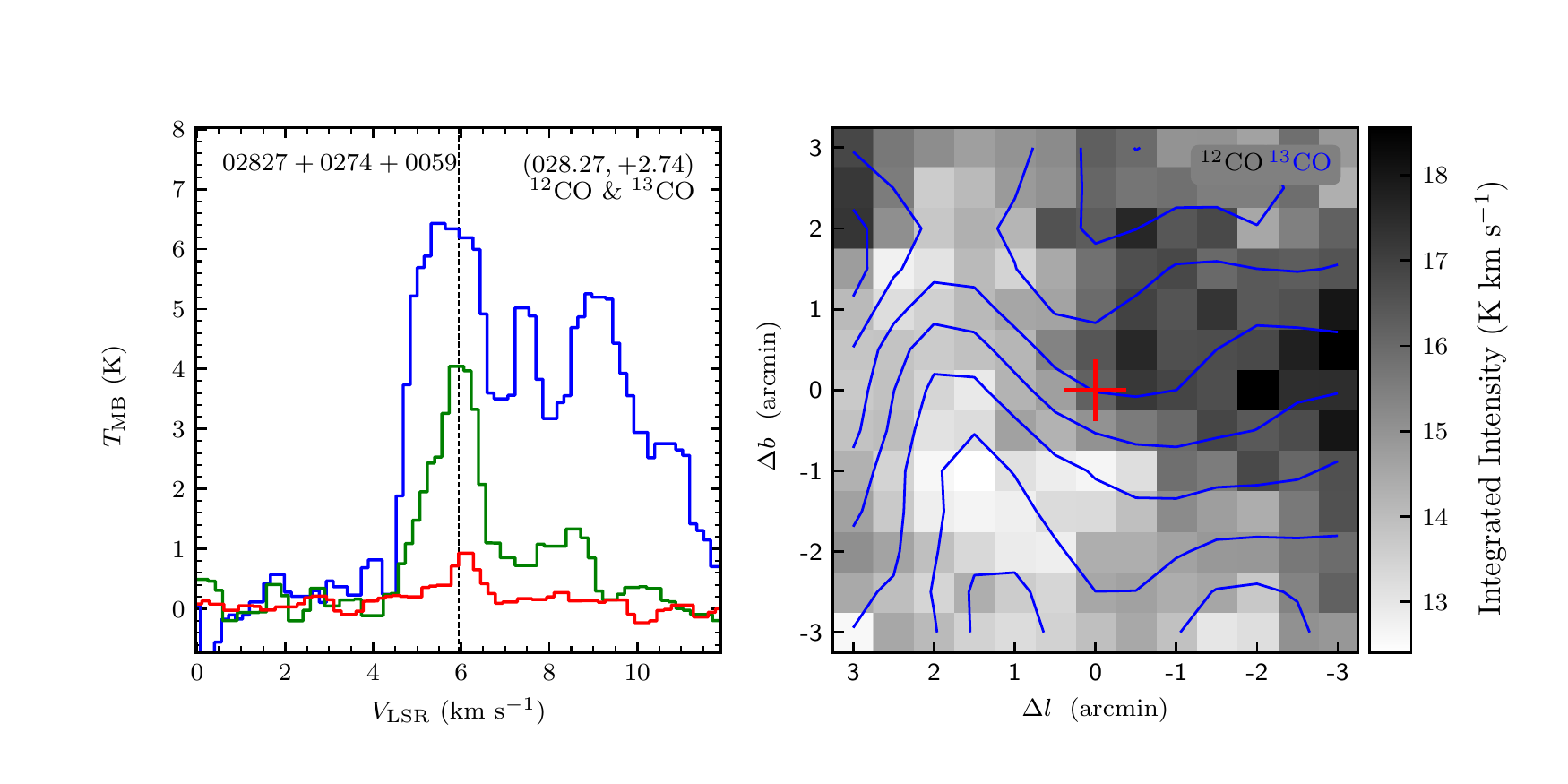}
\includegraphics[width=9.0cm,angle=0]{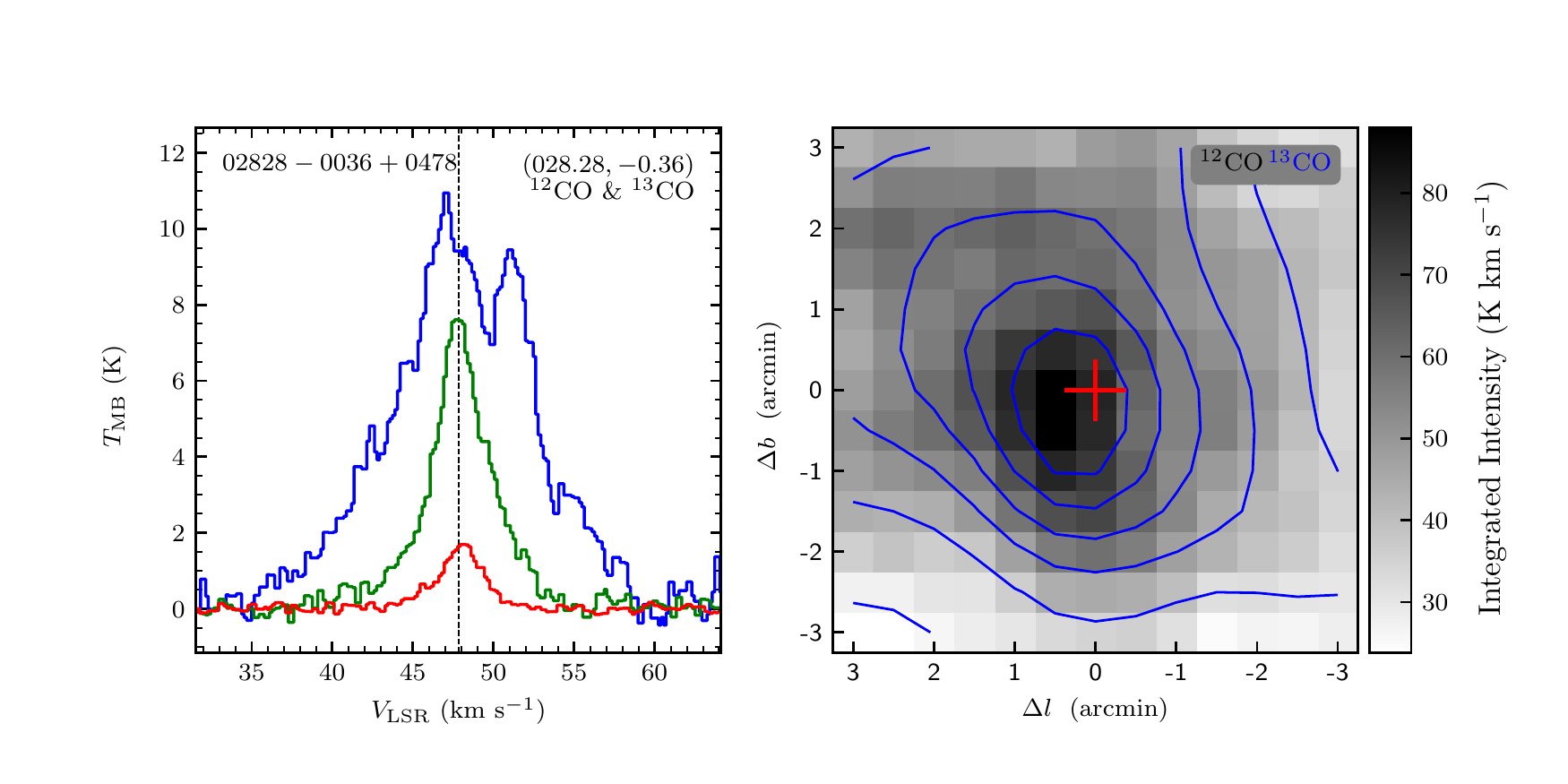}
\vspace{-0.5cm}

\includegraphics[width=9.0cm,angle=0]{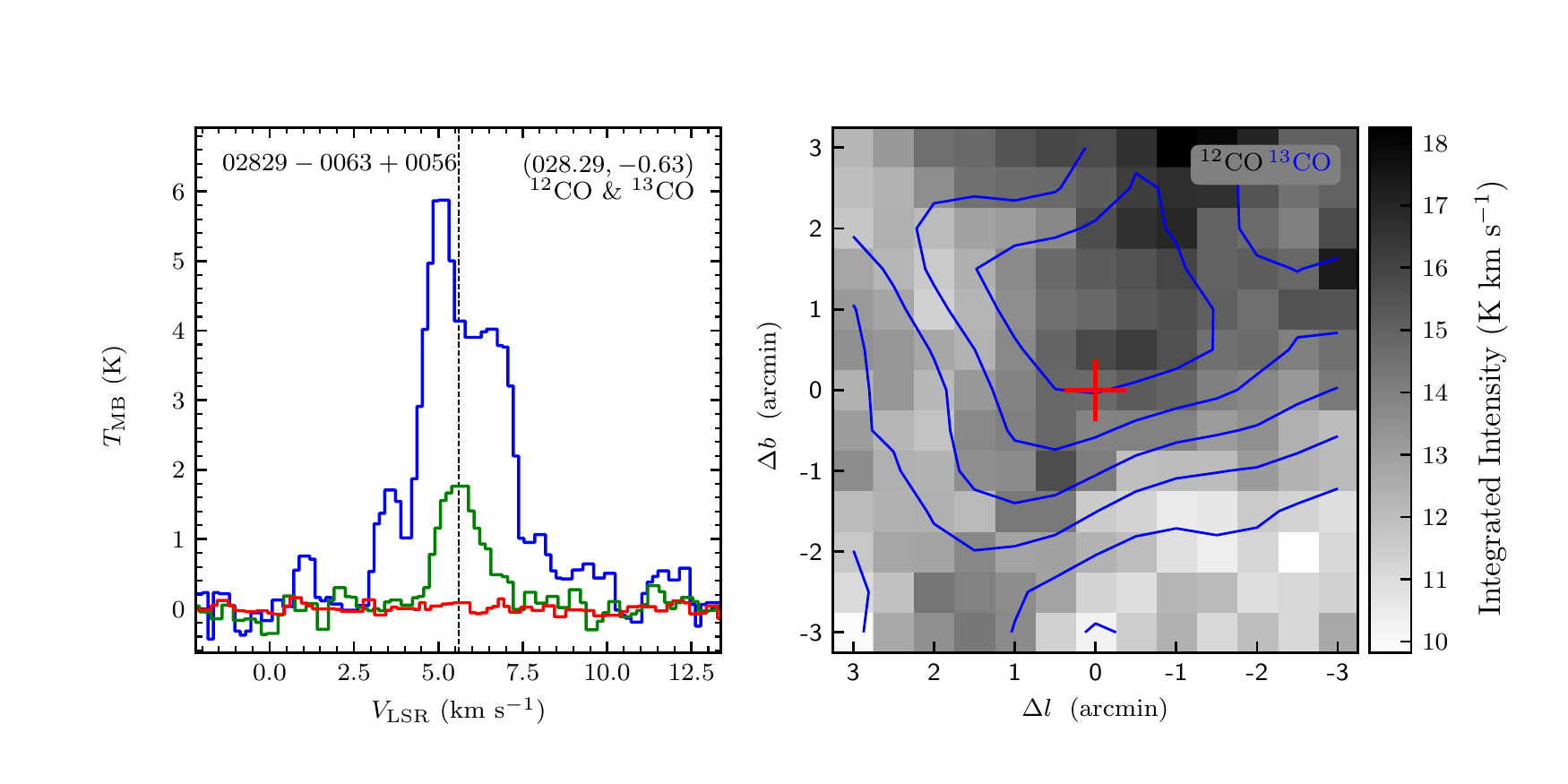}
\includegraphics[width=9.0cm,angle=0]{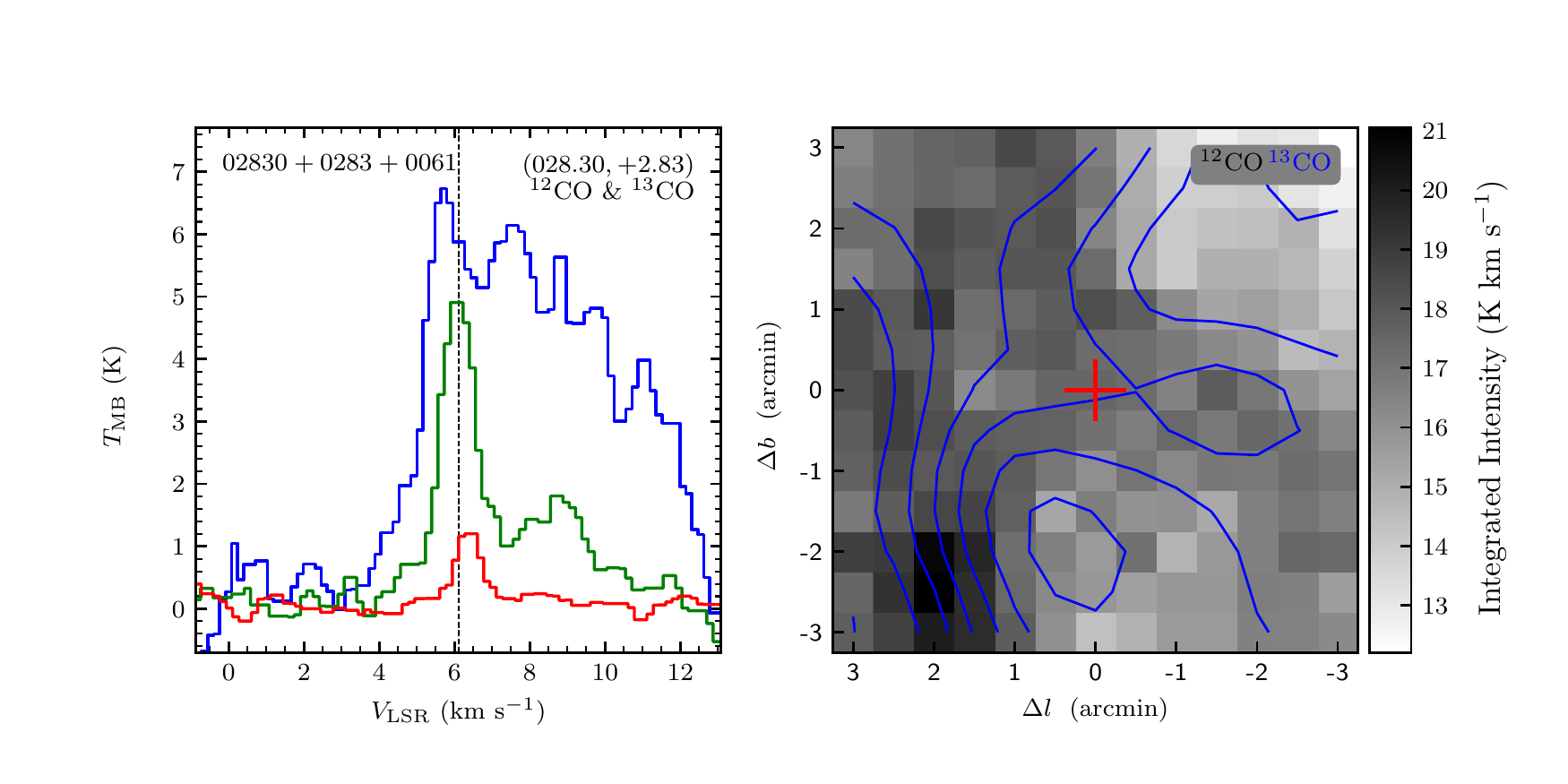}
\vspace{-0.5cm}

\includegraphics[width=9.0cm,angle=0]{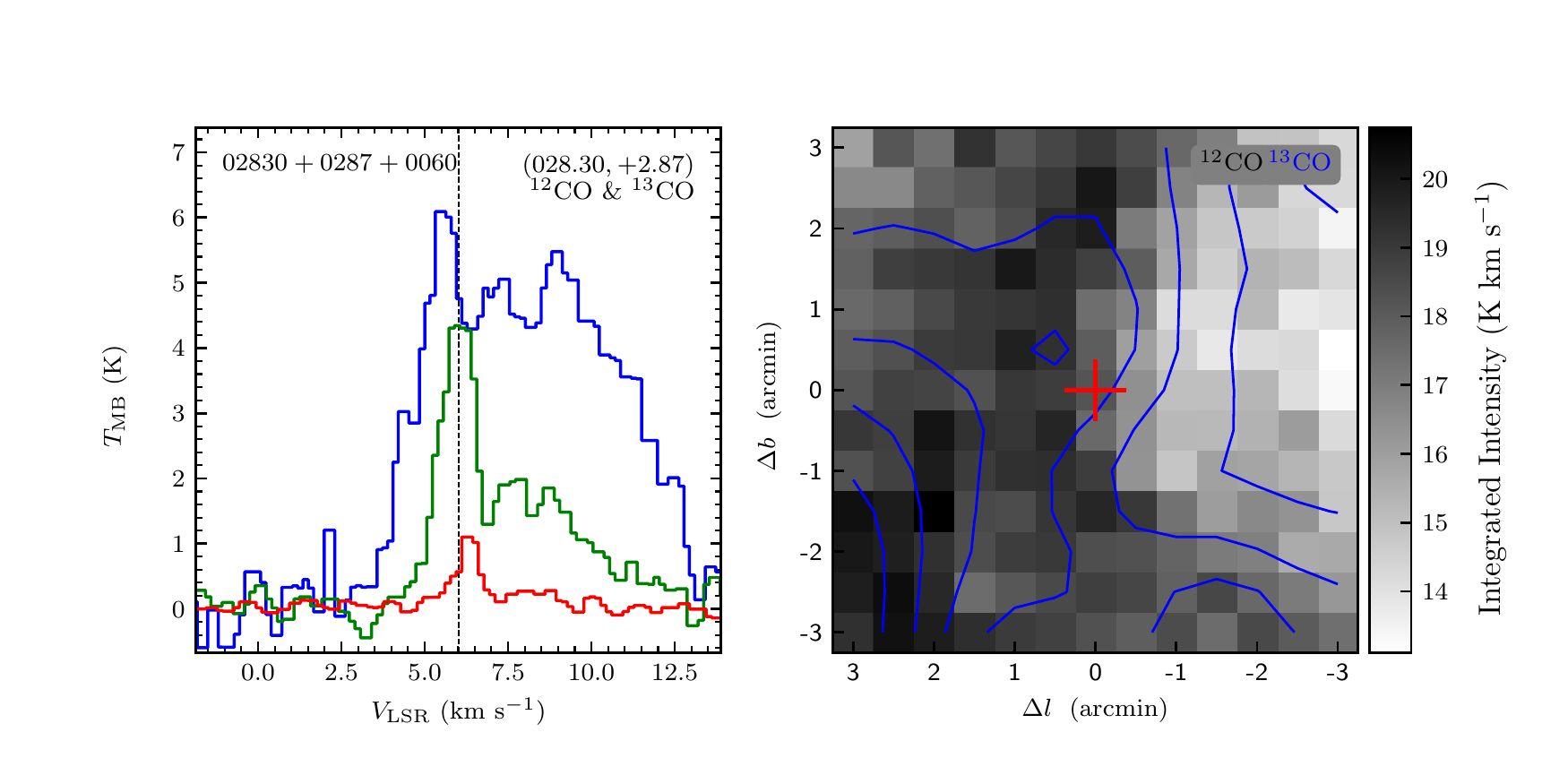}
\includegraphics[width=9.0cm,angle=0]{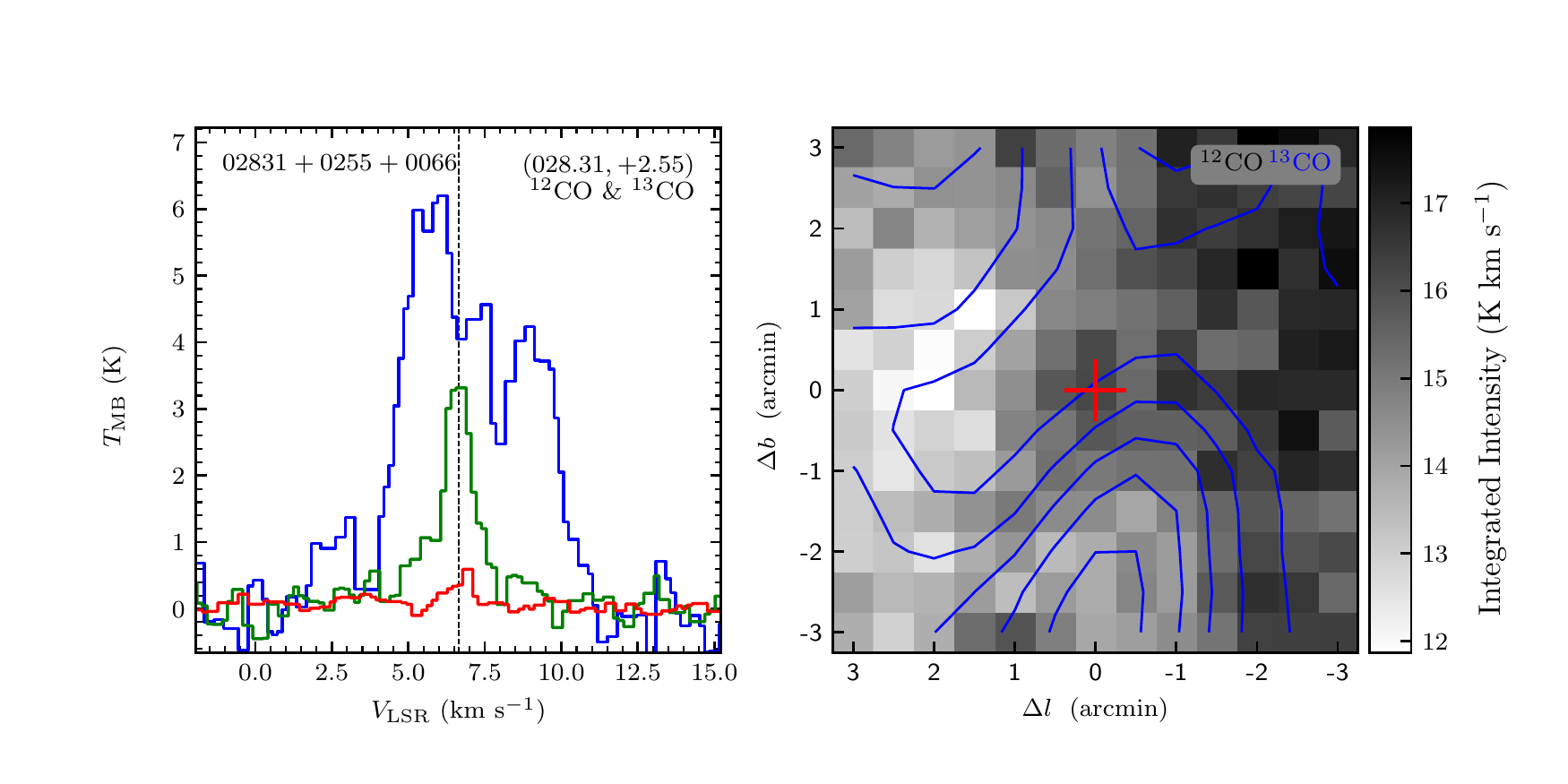}
\end{figure}
\clearpage

\begin{figure}
\includegraphics[width=9.0cm,angle=0]{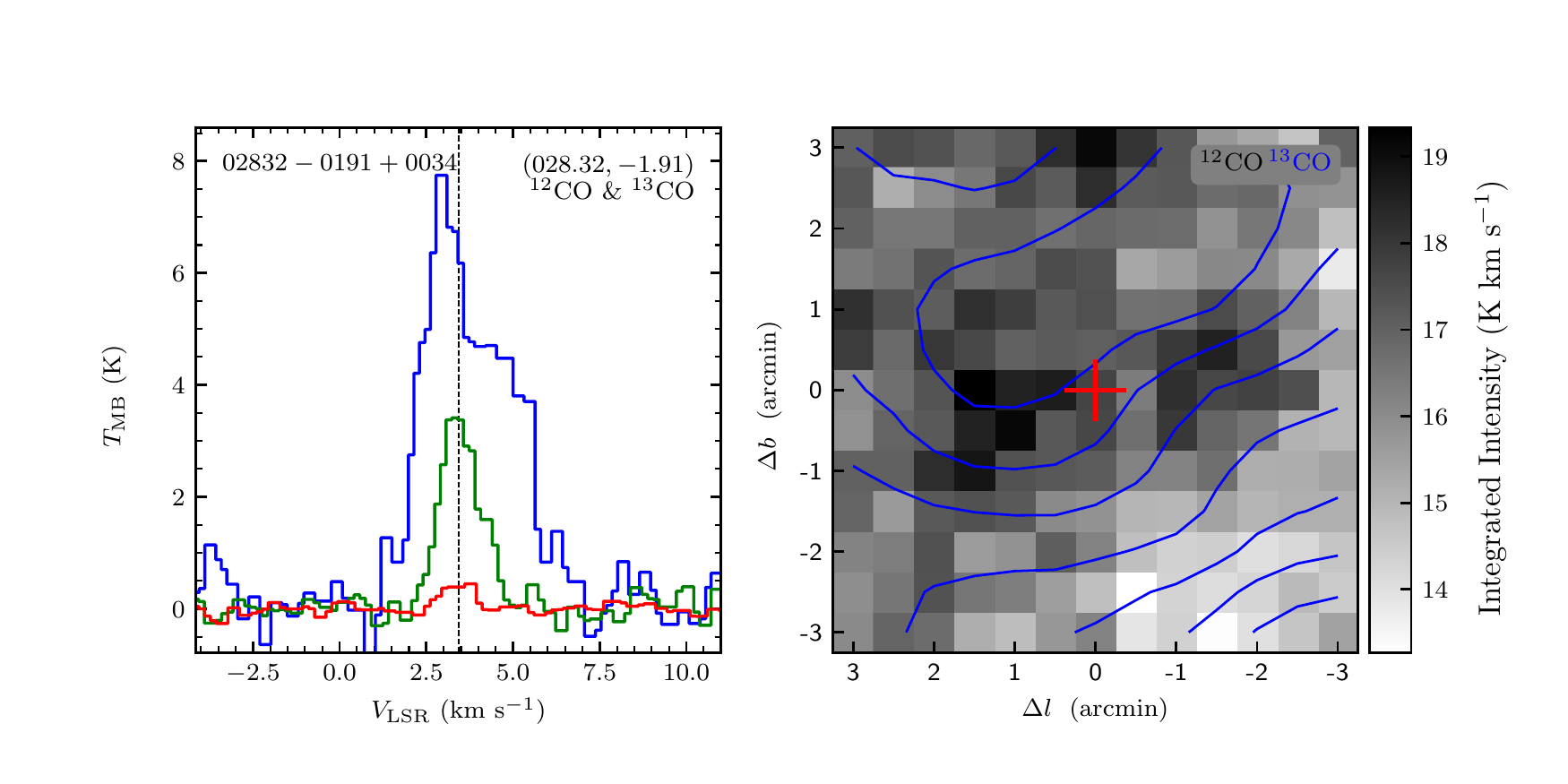}
\includegraphics[width=9.0cm,angle=0]{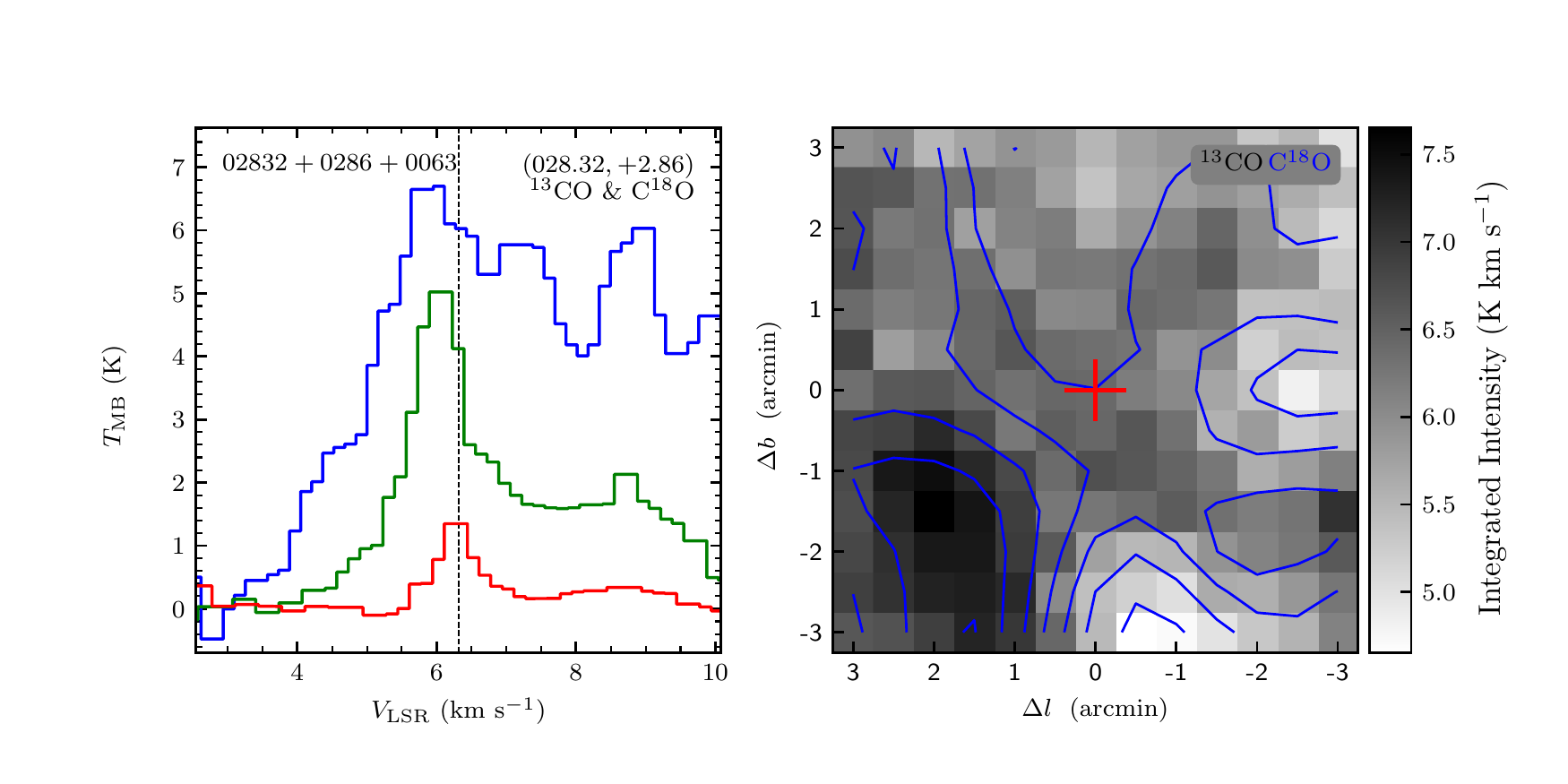}
\vspace{-0.5cm}

\includegraphics[width=9.0cm,angle=0]{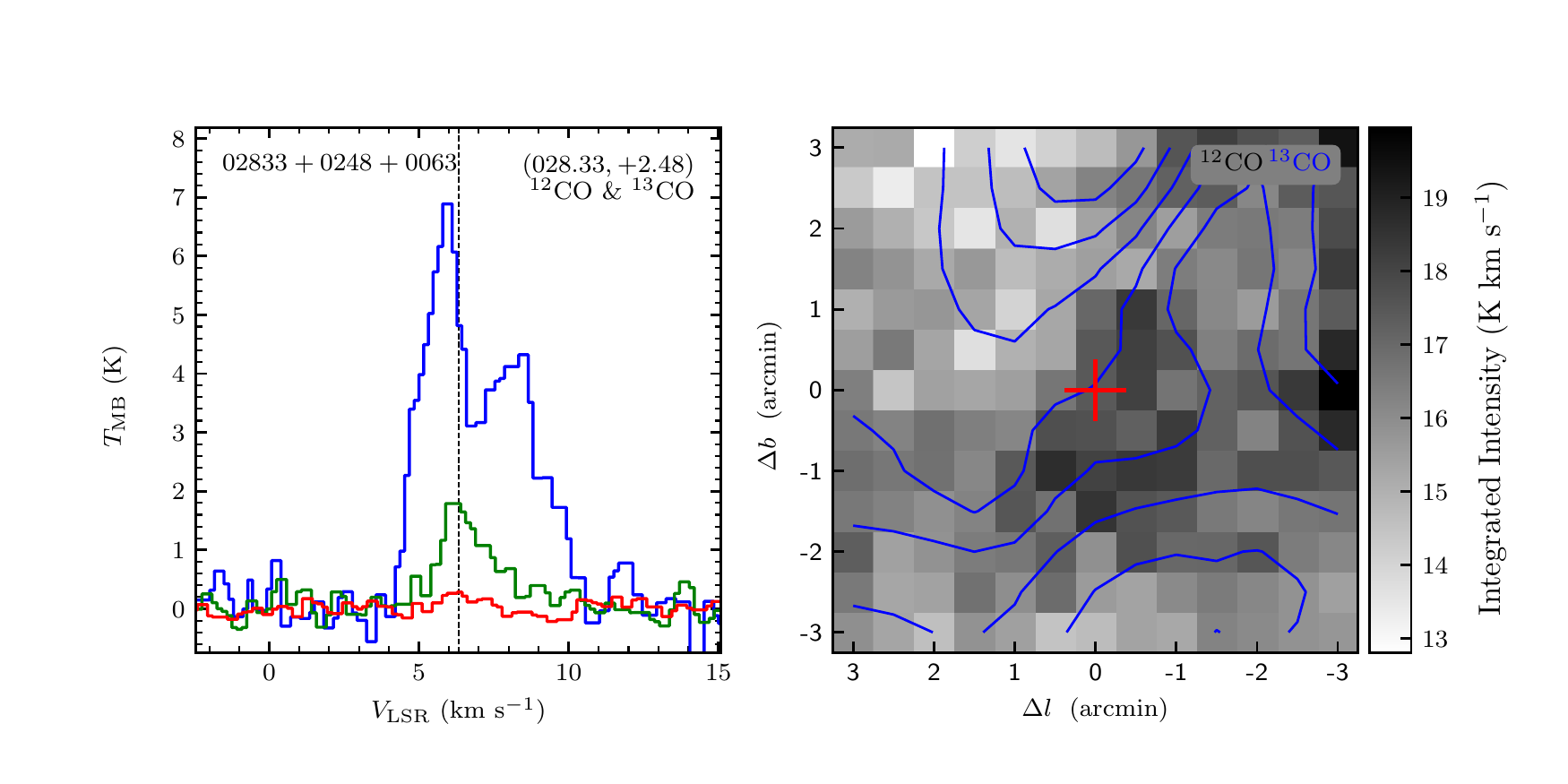}
\includegraphics[width=9.0cm,angle=0]{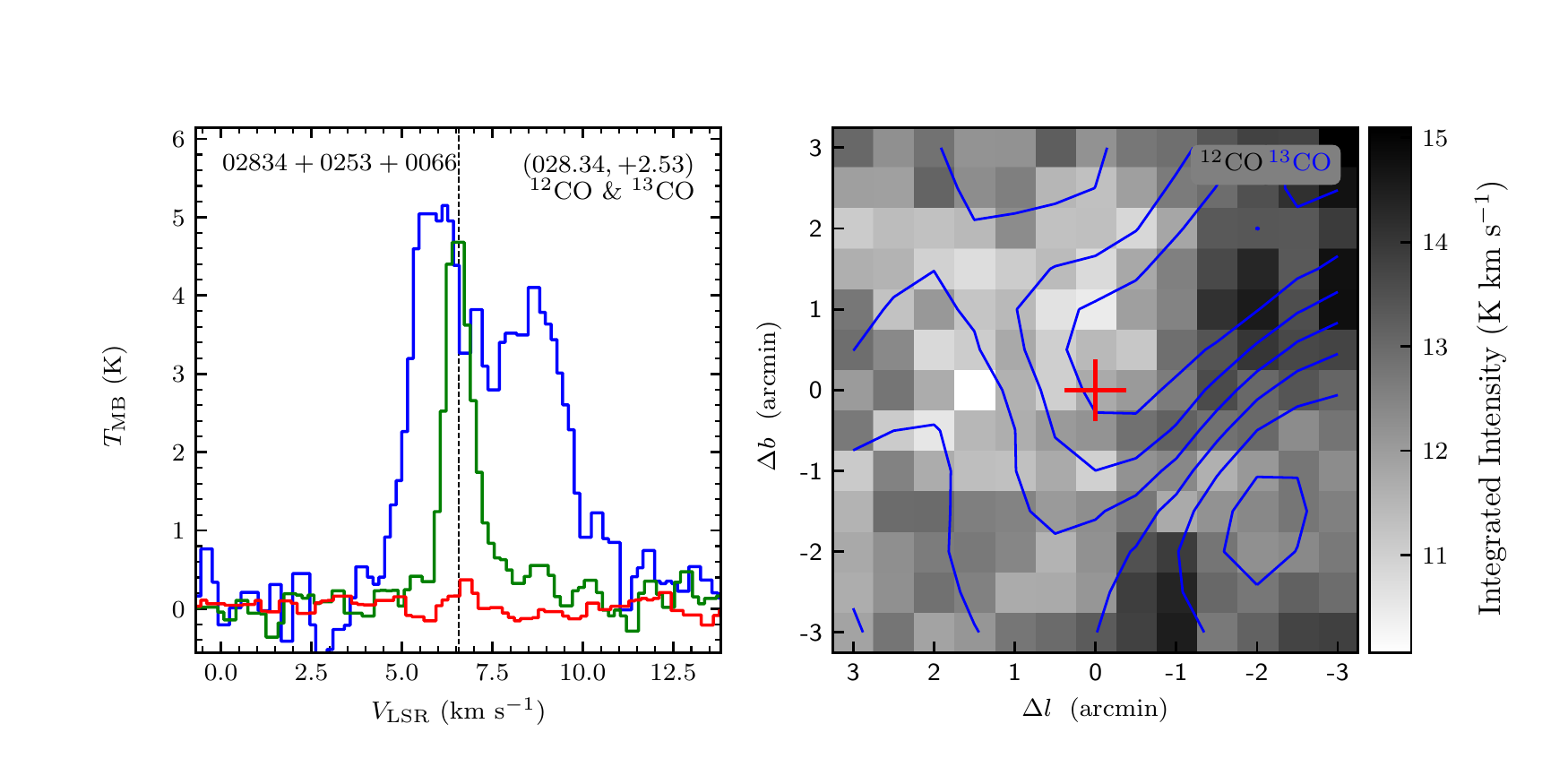}
\vspace{-0.5cm}

\includegraphics[width=9.0cm,angle=0]{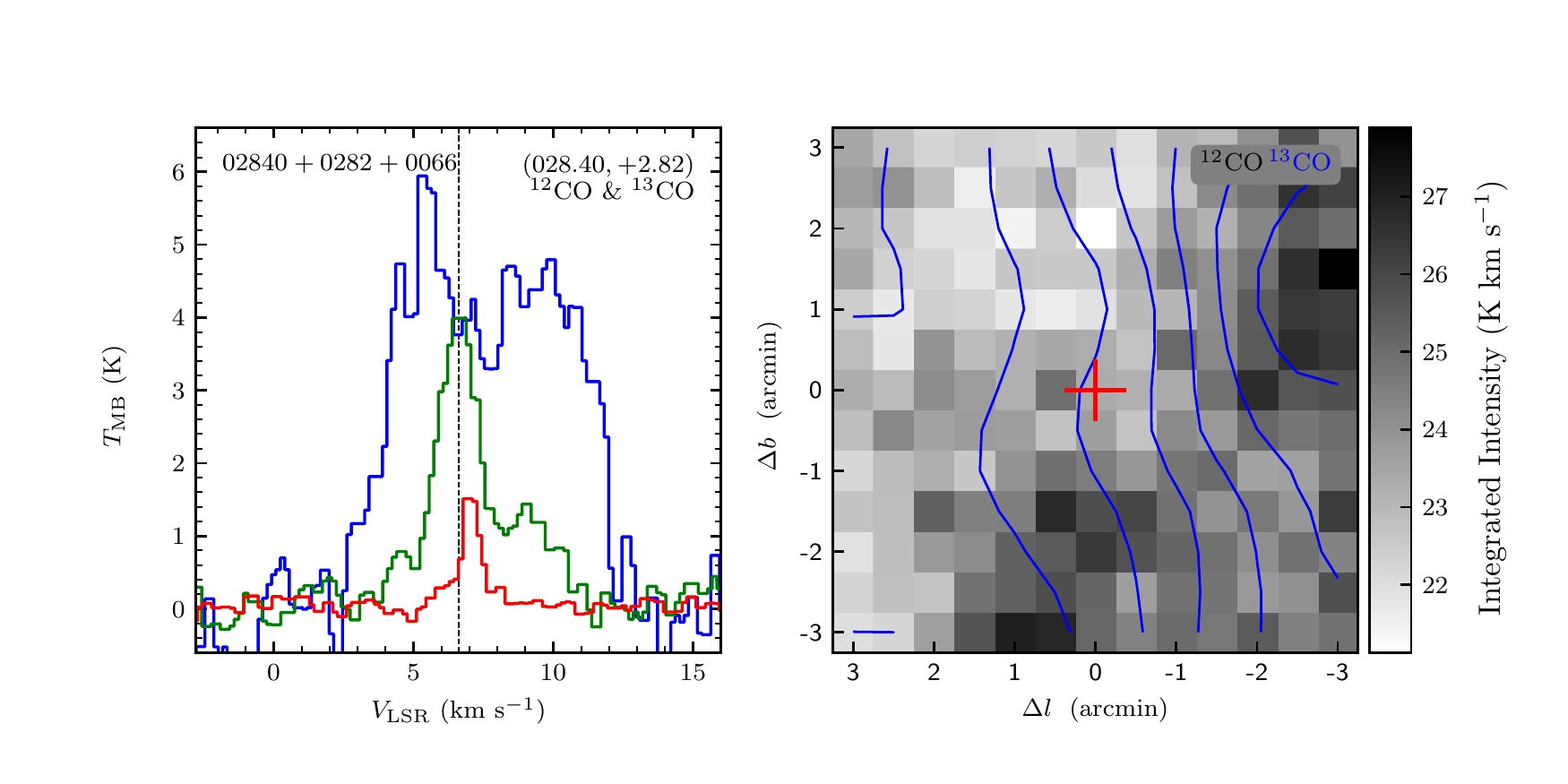}
\includegraphics[width=9.0cm,angle=0]{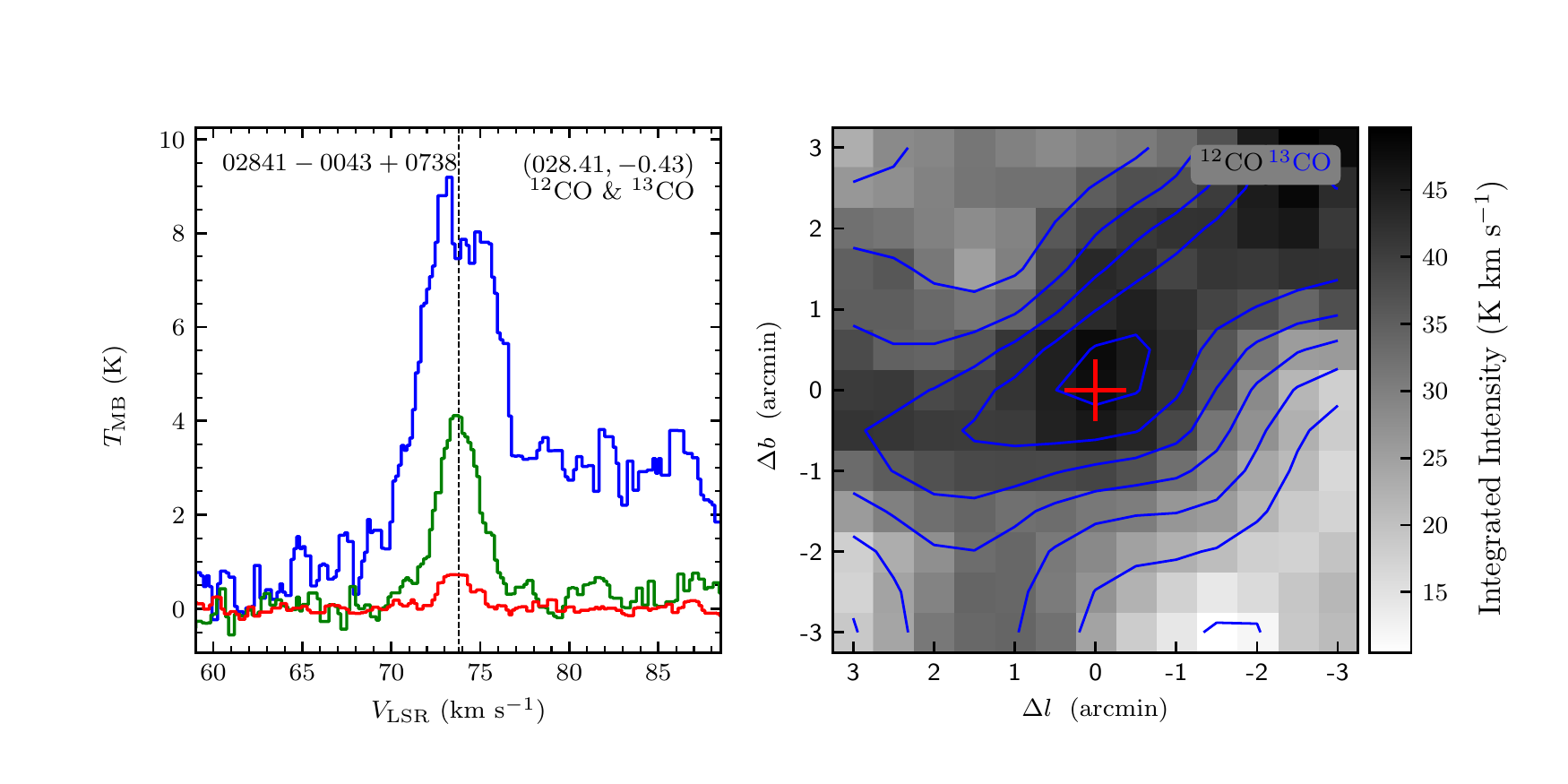}
\vspace{-0.5cm}

\includegraphics[width=9.0cm,angle=0]{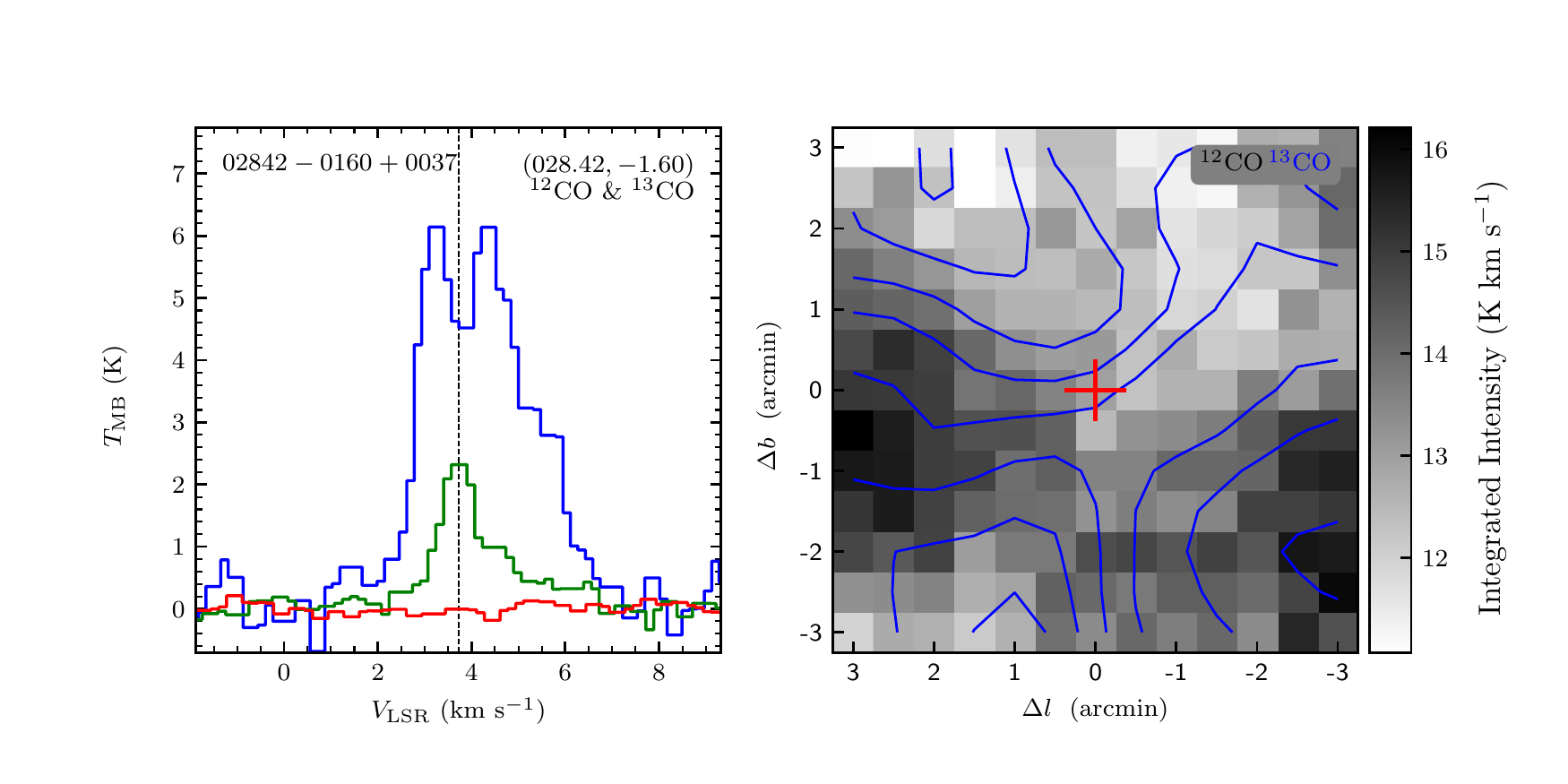}
\includegraphics[width=9.0cm,angle=0]{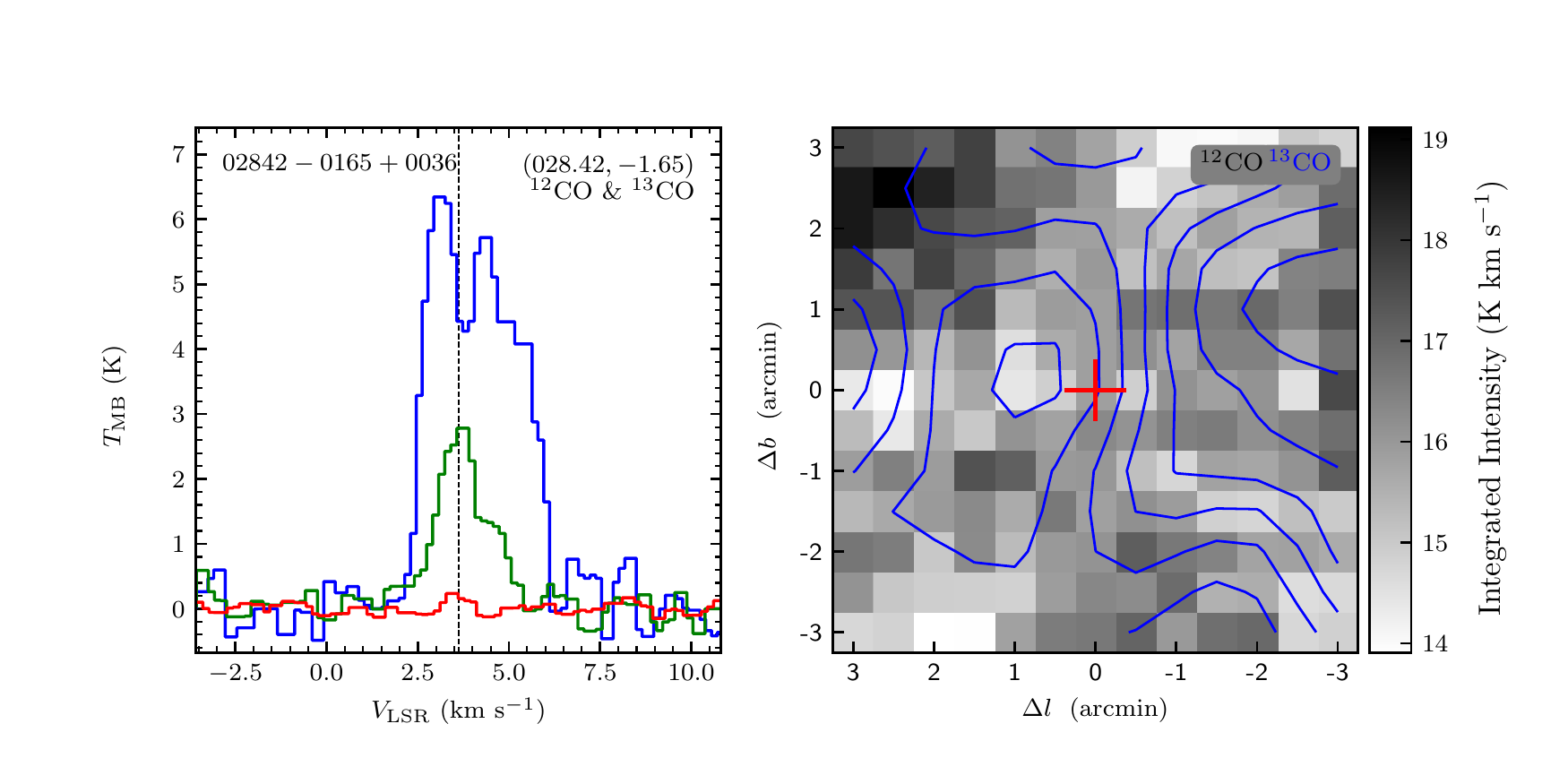}
\vspace{-0.5cm}

\includegraphics[width=9.0cm,angle=0]{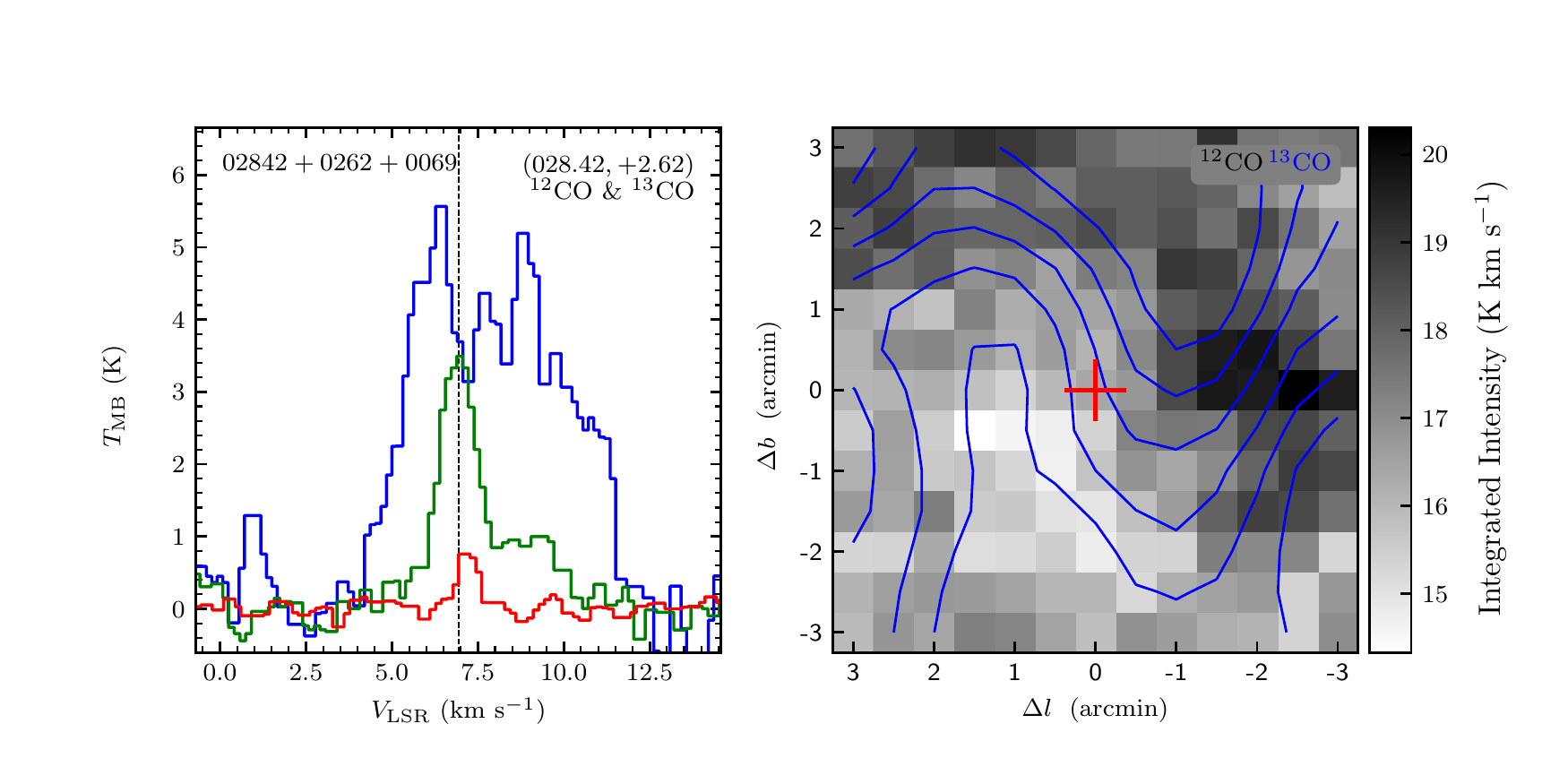}
\includegraphics[width=9.0cm,angle=0]{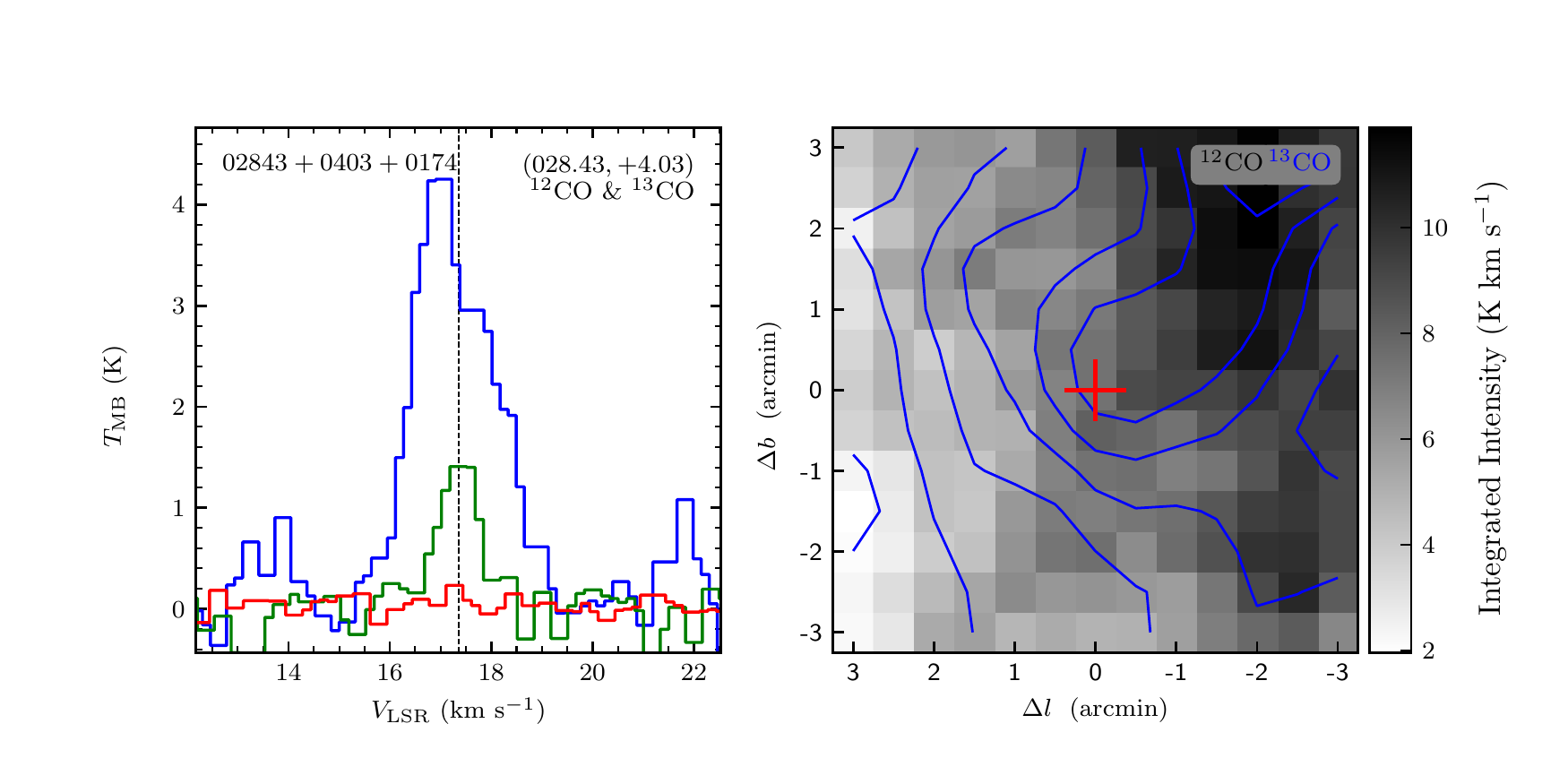}
\end{figure}
\clearpage

\begin{figure}
\includegraphics[width=9.0cm,angle=0]{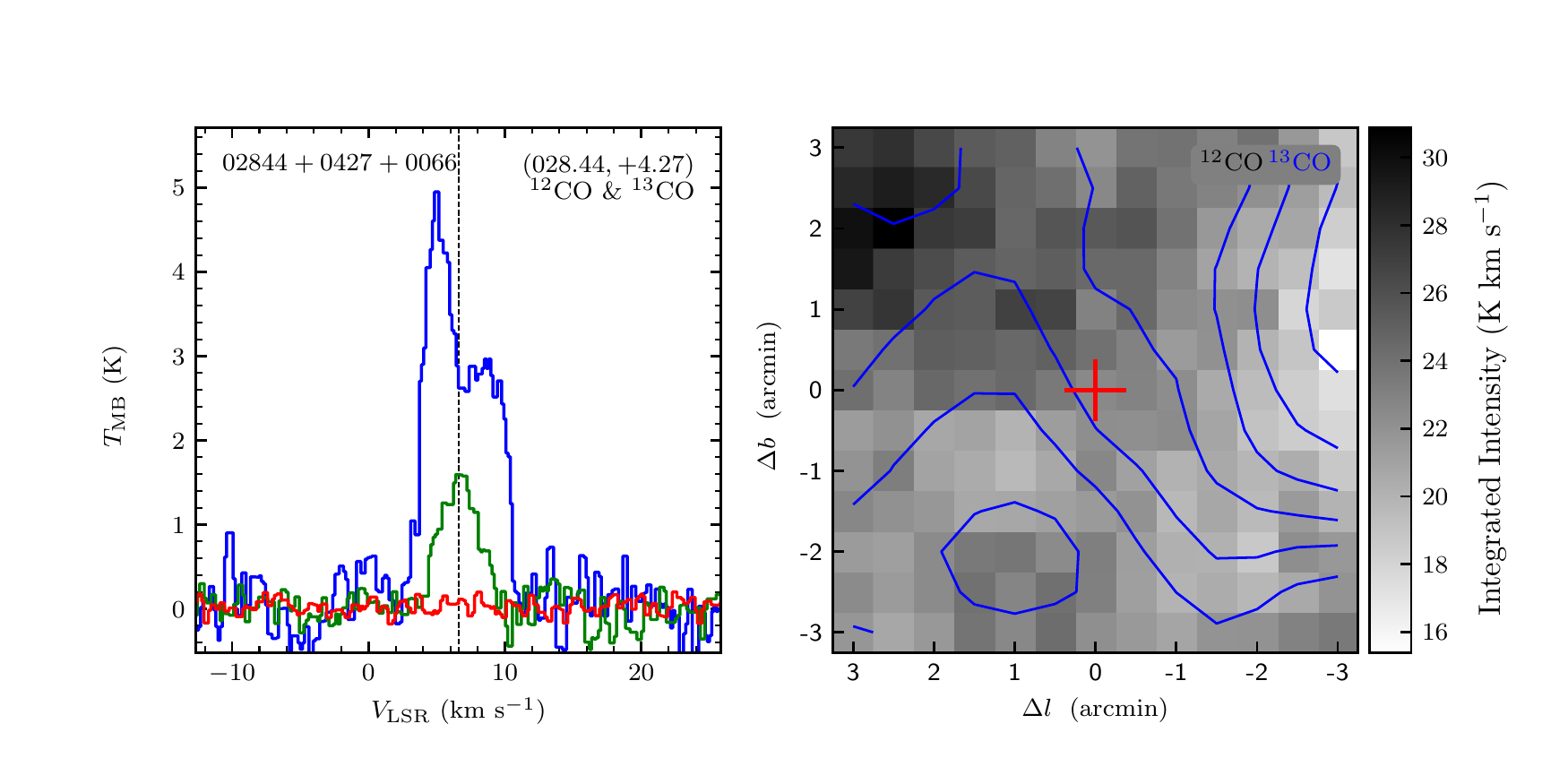}
\includegraphics[width=9.0cm,angle=0]{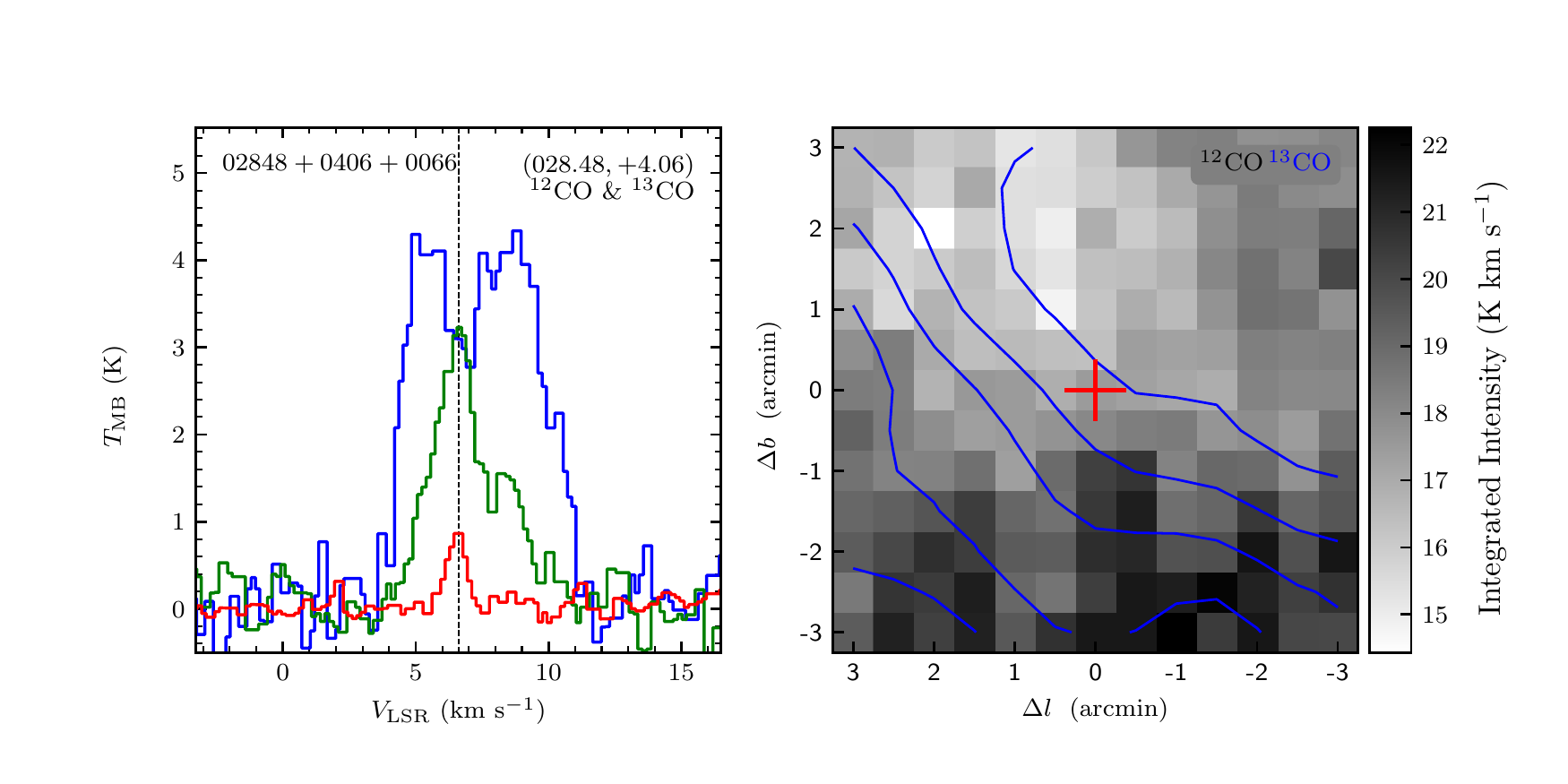}
\vspace{-0.5cm}

\includegraphics[width=9.0cm,angle=0]{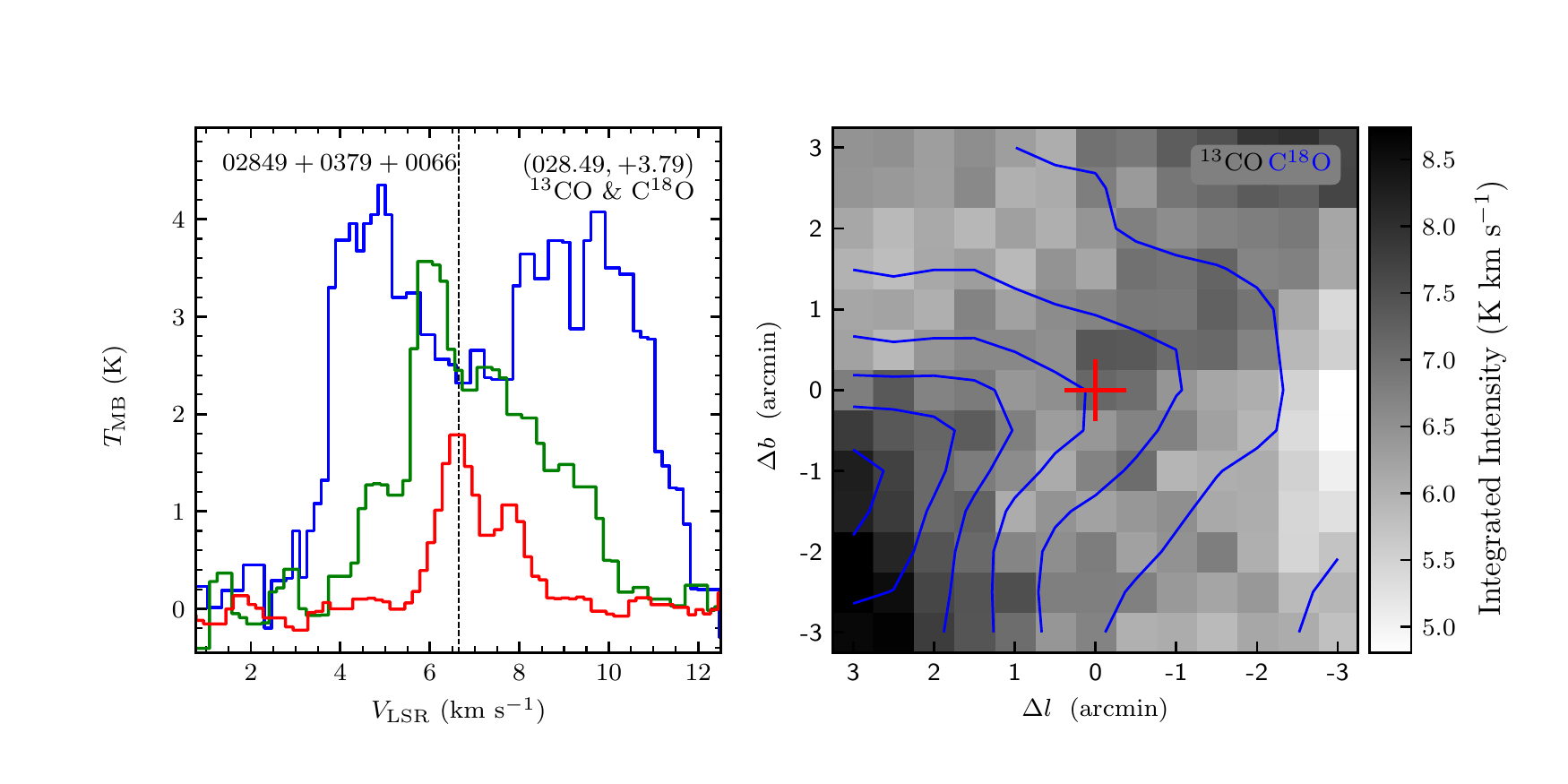}
\includegraphics[width=9.0cm,angle=0]{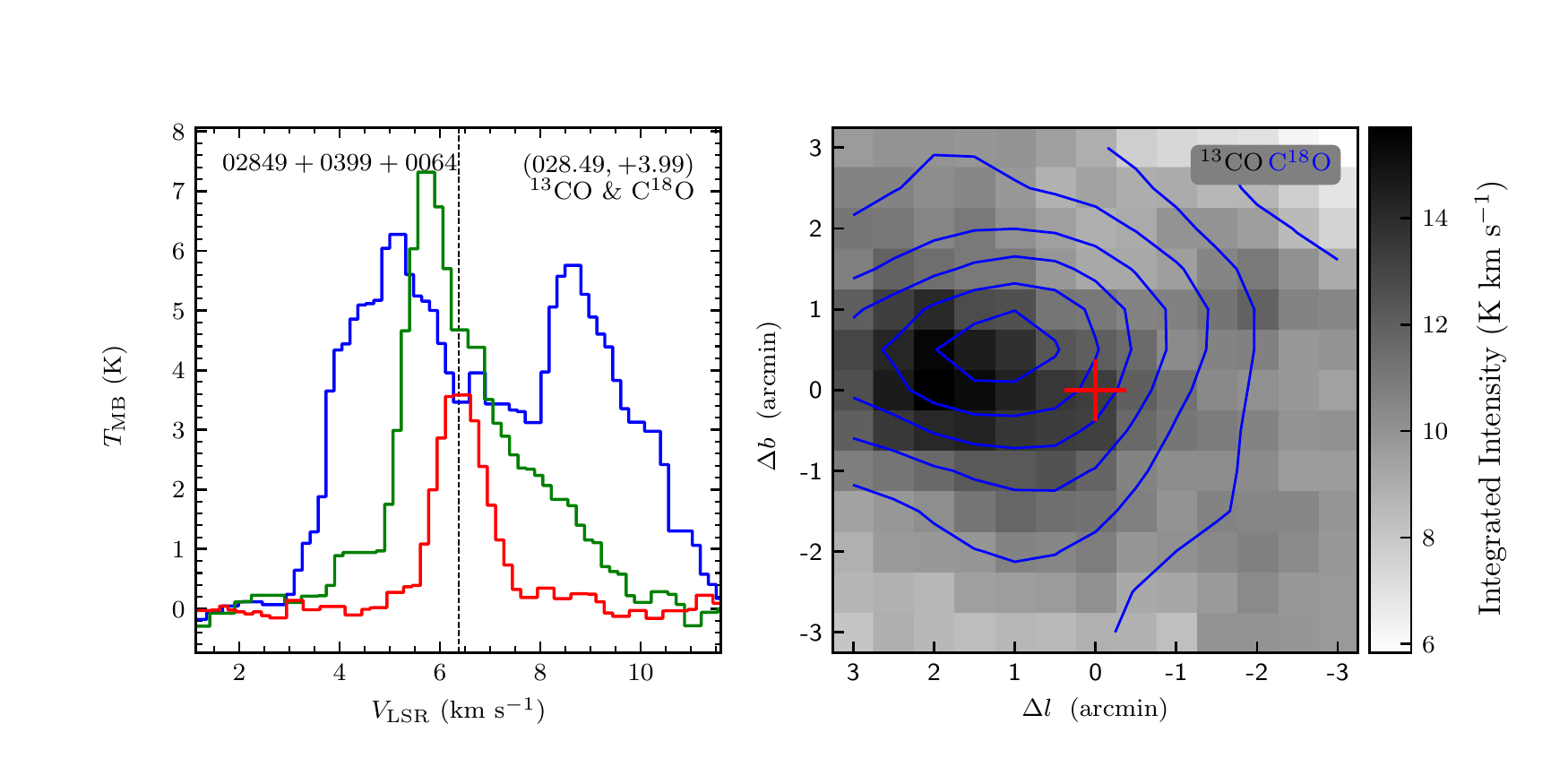}
\vspace{-0.5cm}

\includegraphics[width=9.0cm,angle=0]{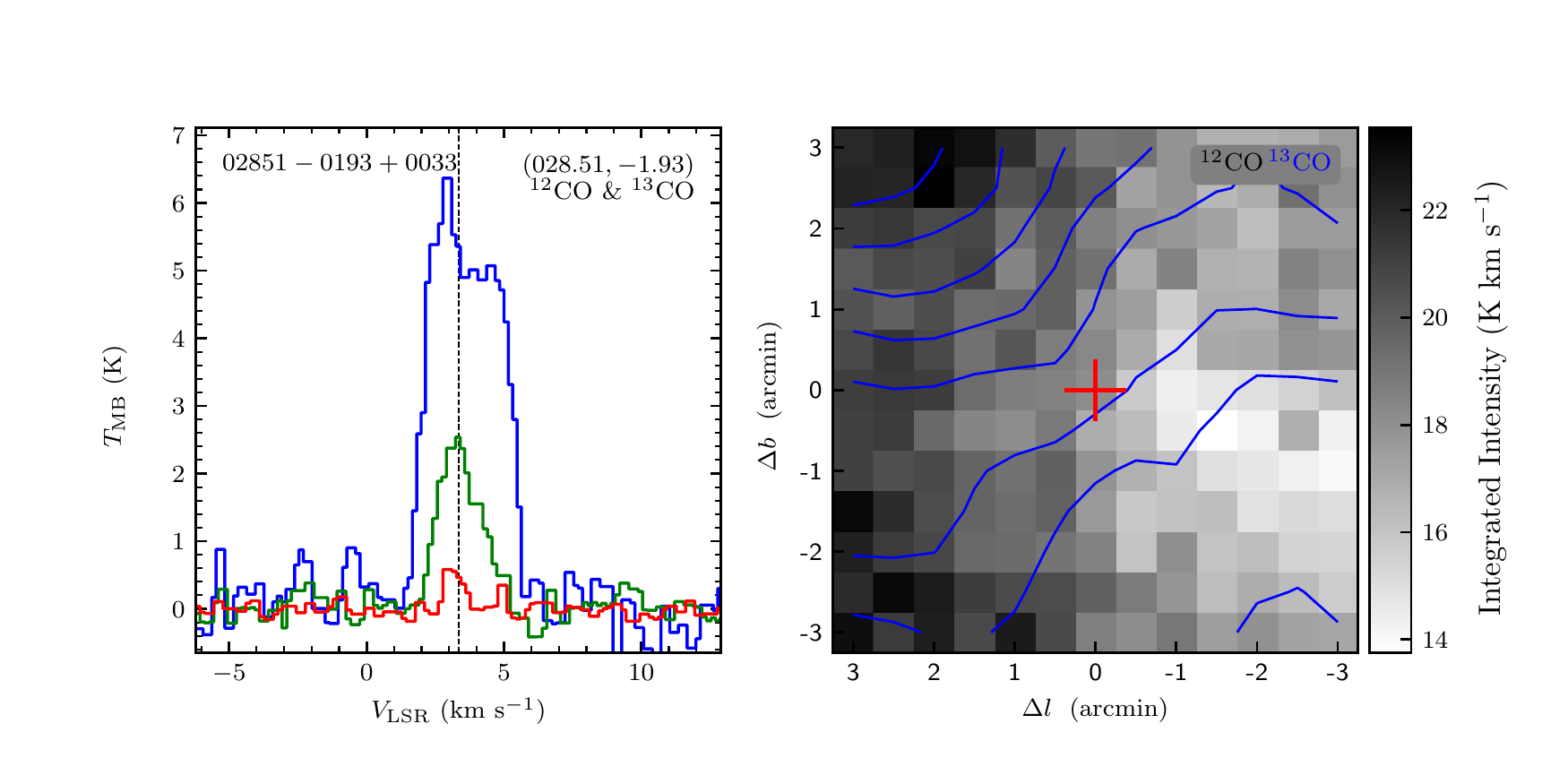}
\includegraphics[width=9.0cm,angle=0]{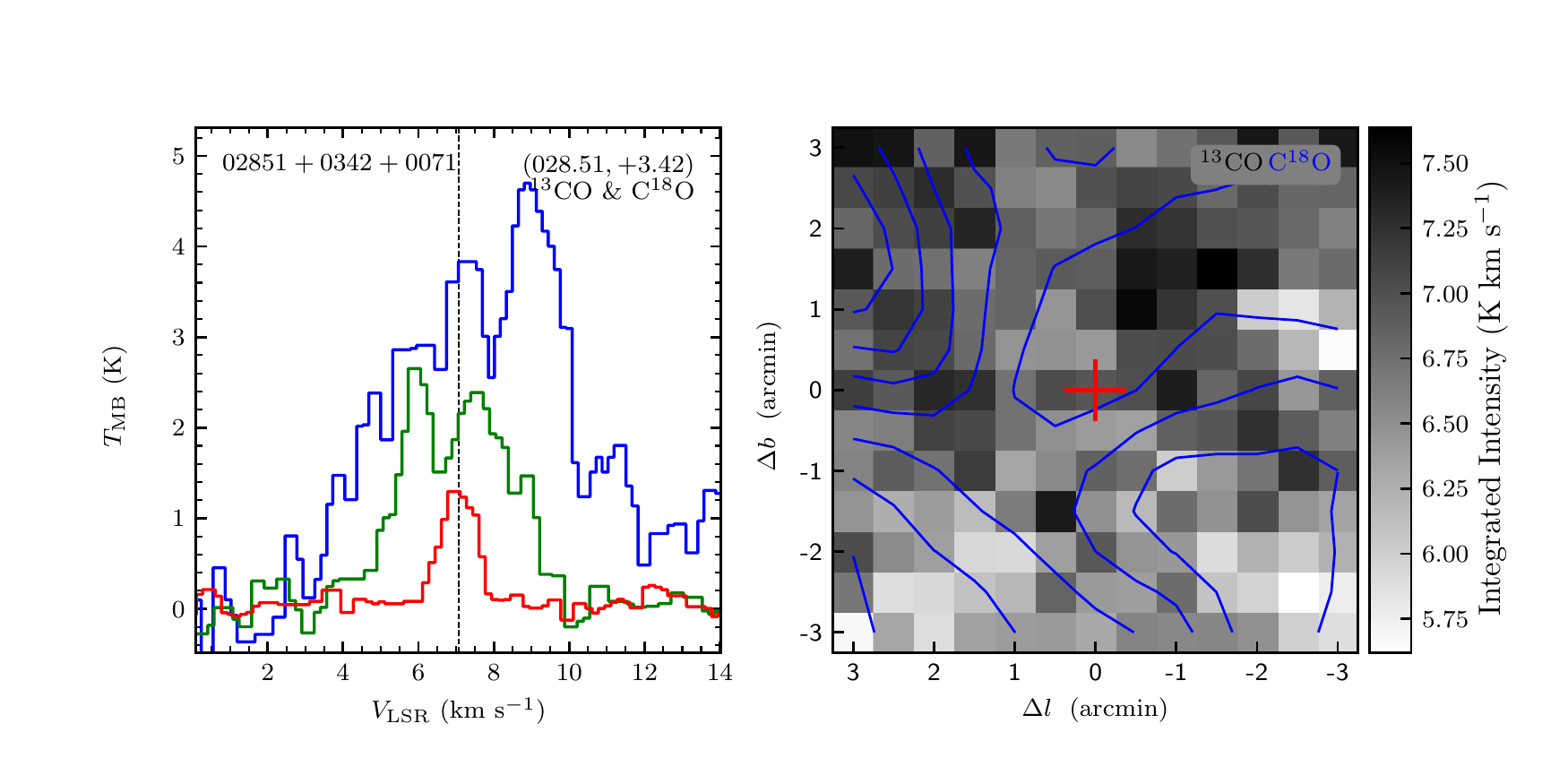}
\vspace{-0.5cm}

\includegraphics[width=9.0cm,angle=0]{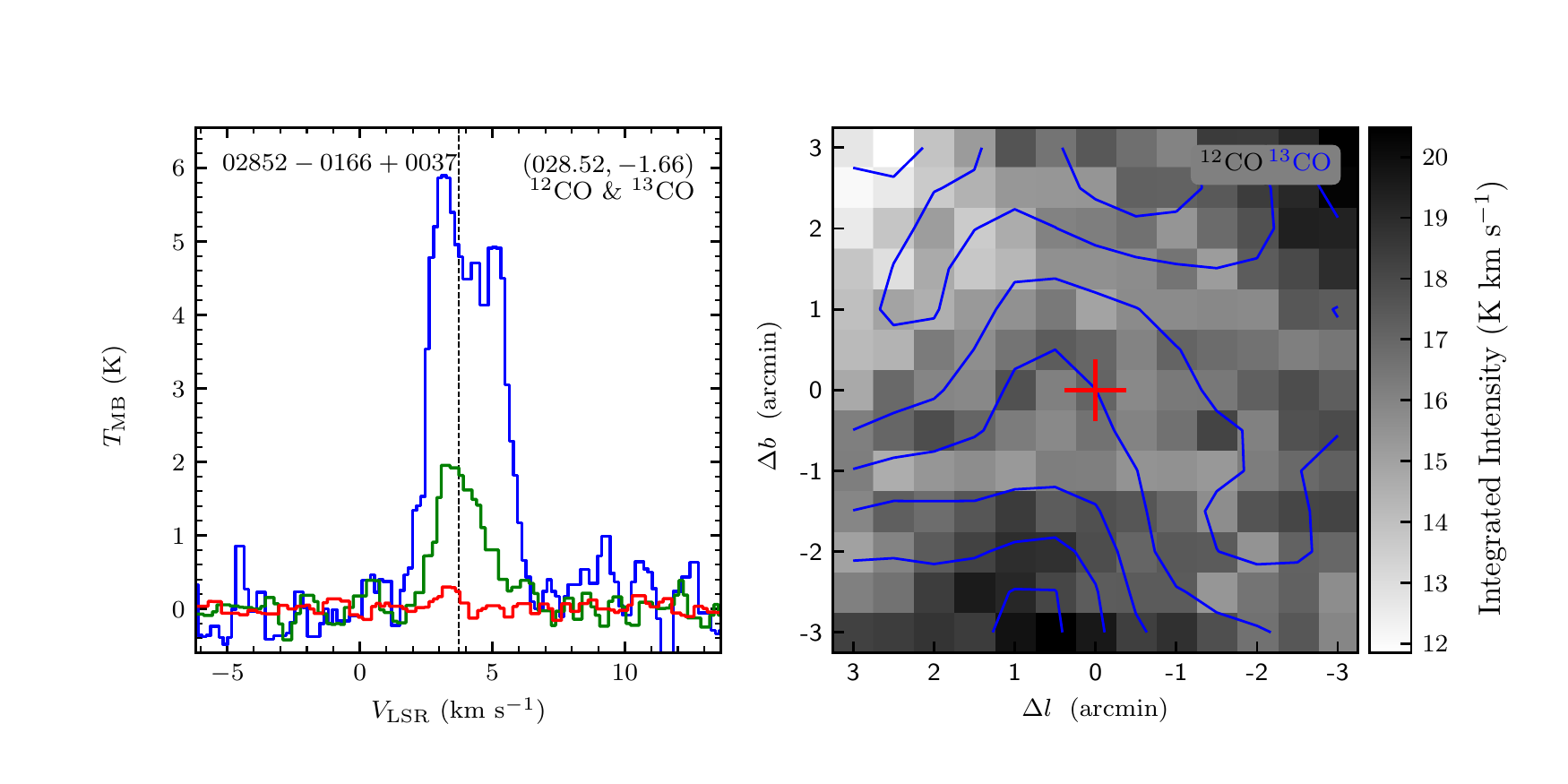}
\includegraphics[width=9.0cm,angle=0]{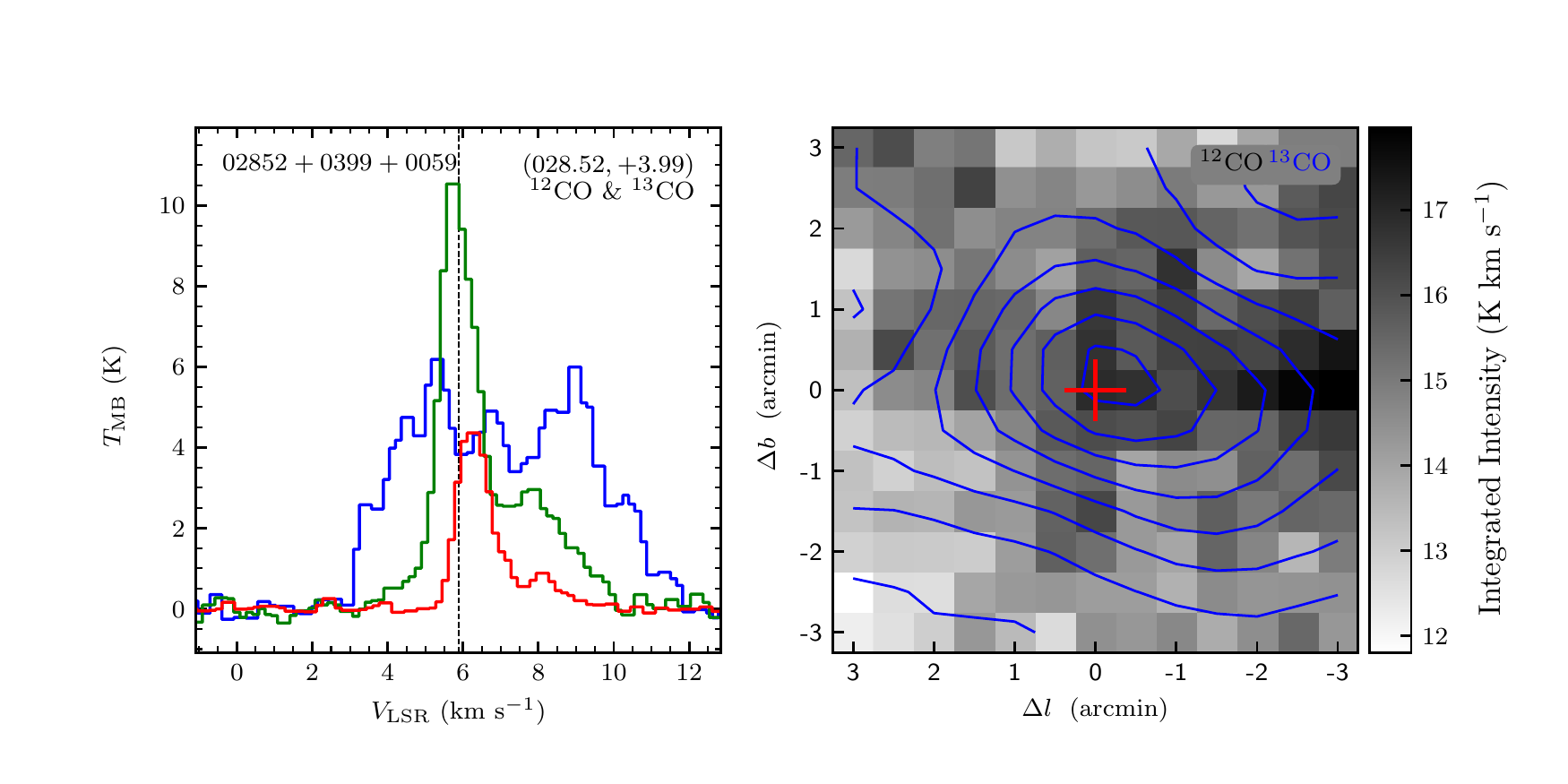}
\vspace{-0.5cm}

\includegraphics[width=9.0cm,angle=0]{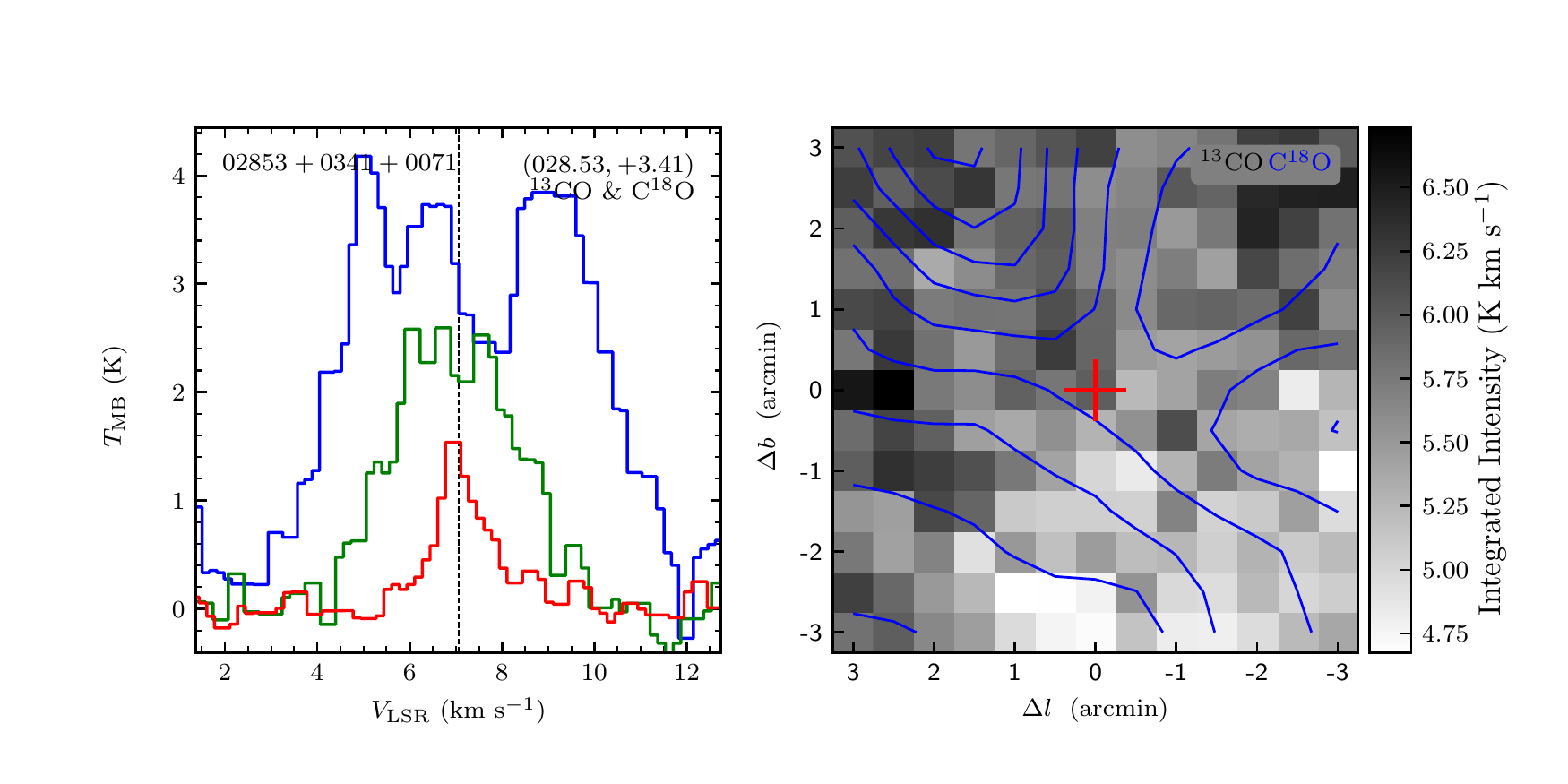}
\includegraphics[width=9.0cm,angle=0]{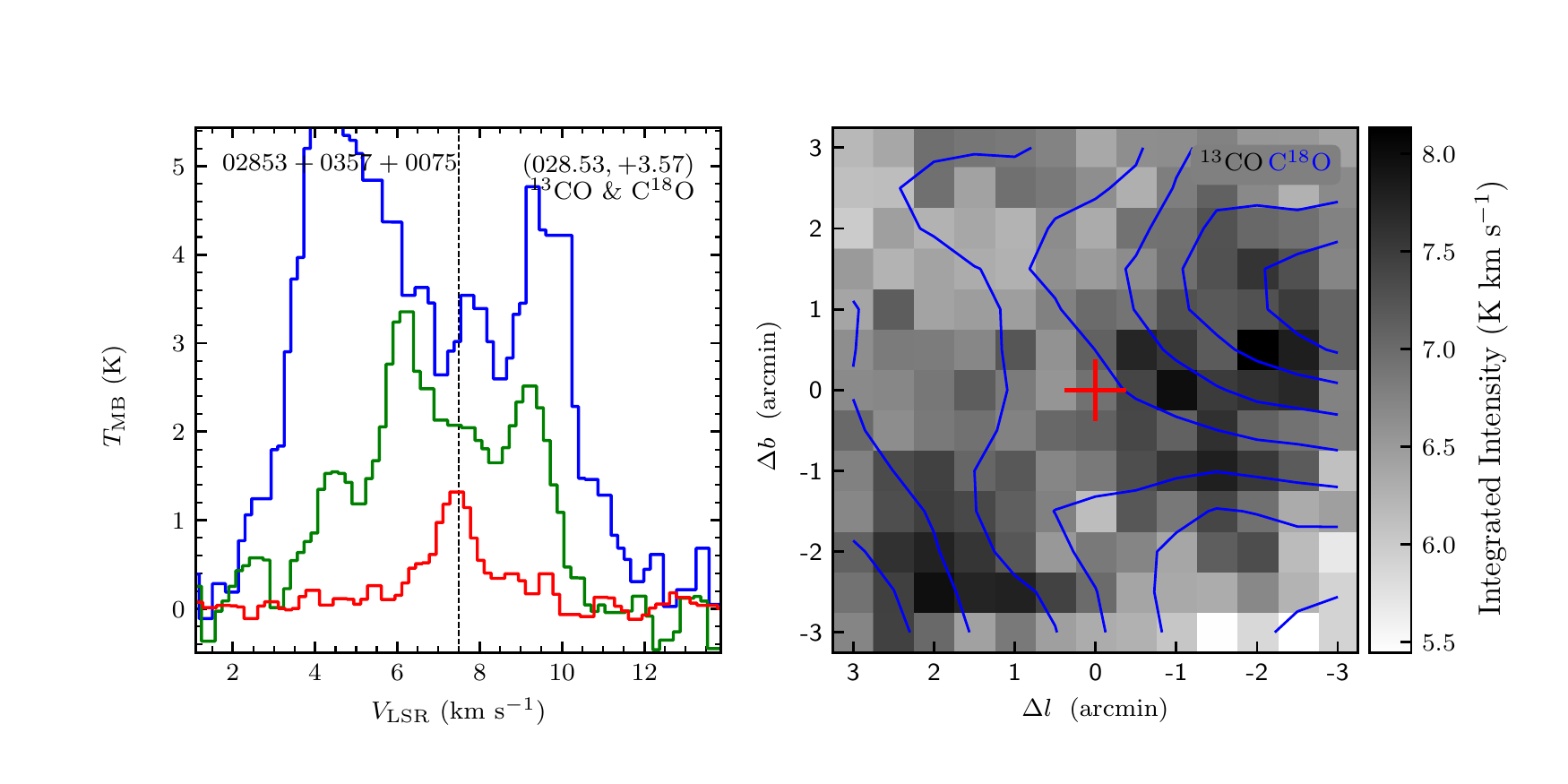}
\end{figure}
\clearpage

\begin{figure}
\includegraphics[width=9.0cm,angle=0]{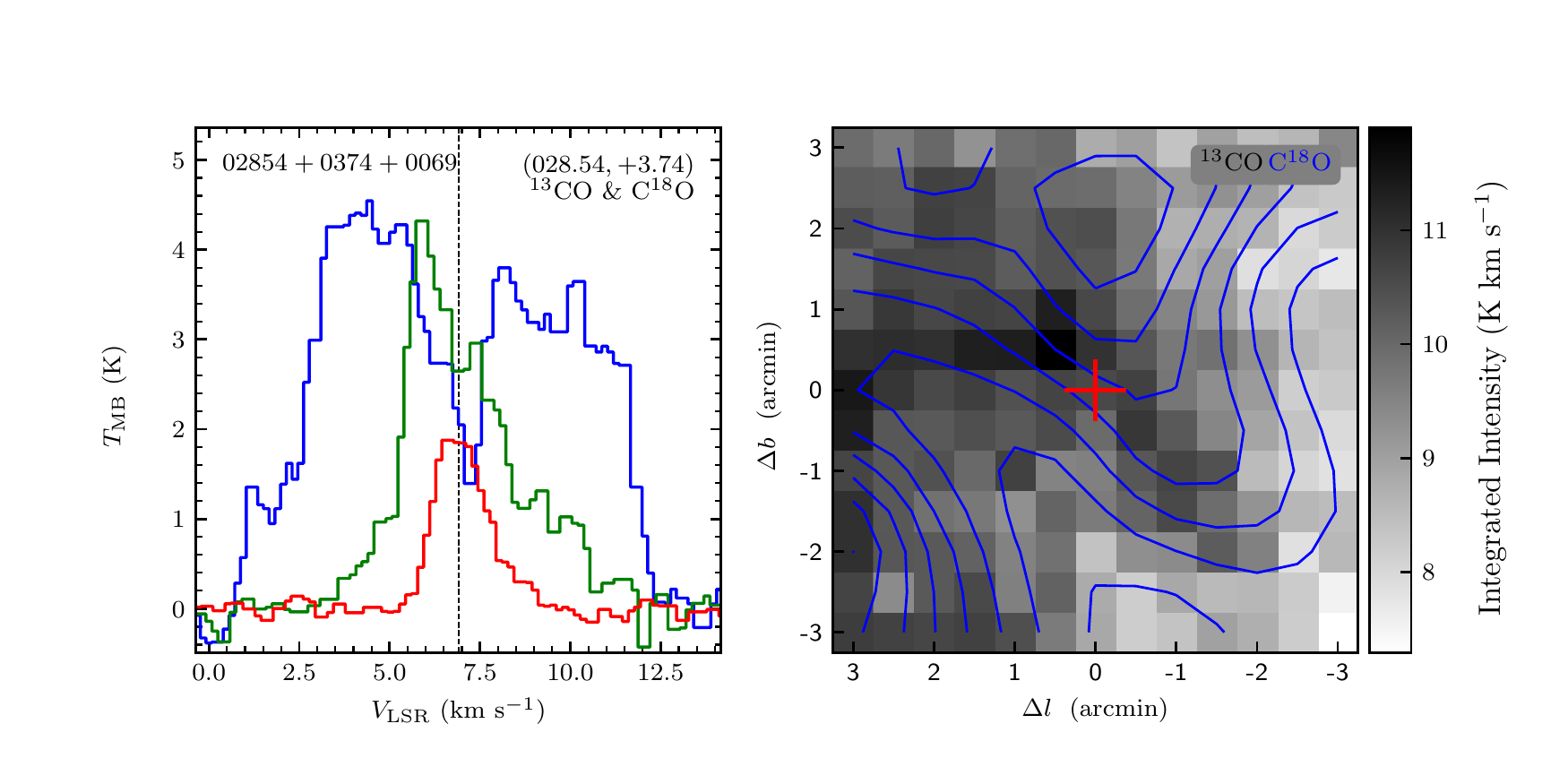}
\includegraphics[width=9.0cm,angle=0]{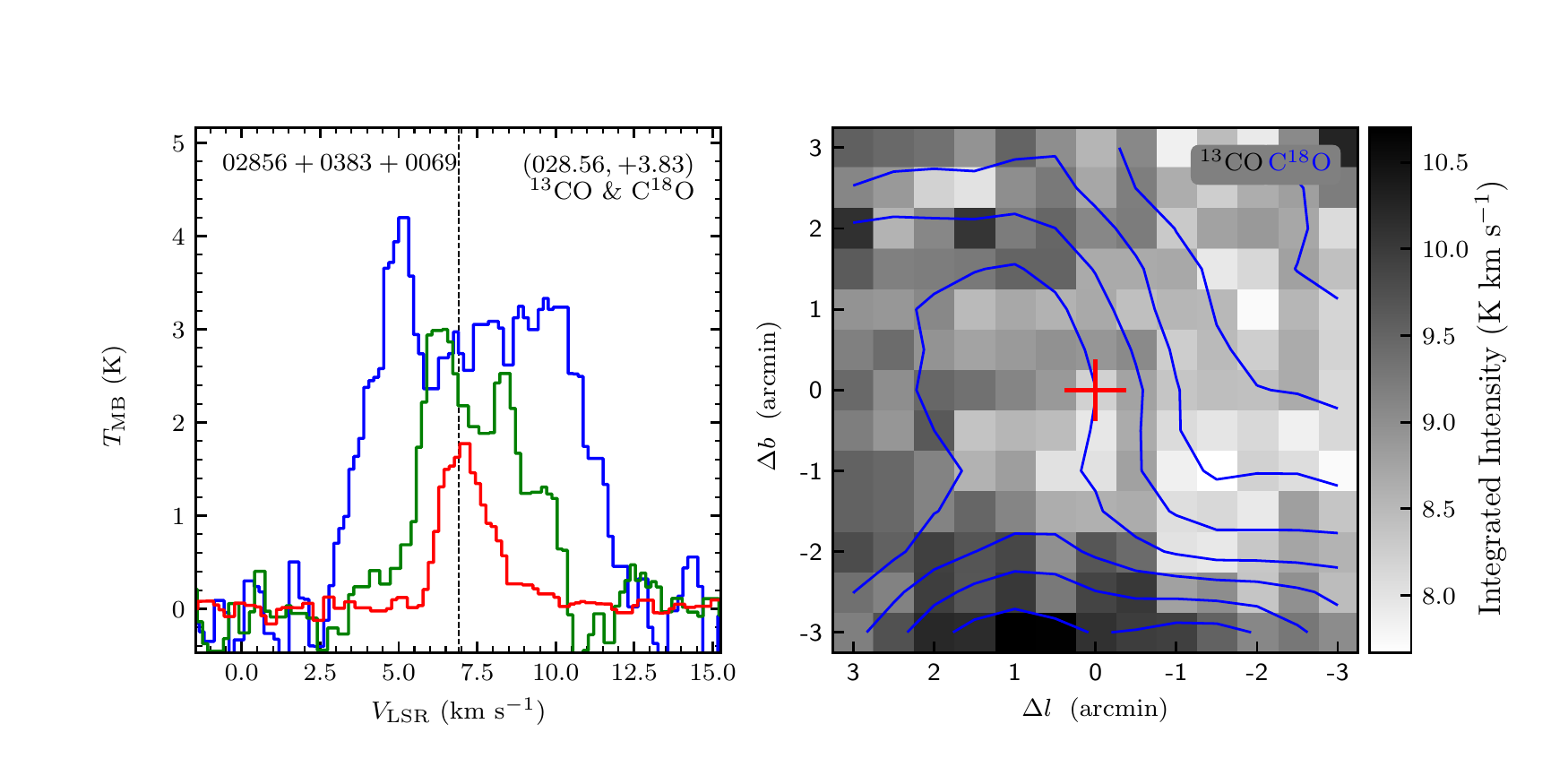}
\vspace{-0.5cm}

\includegraphics[width=9.0cm,angle=0]{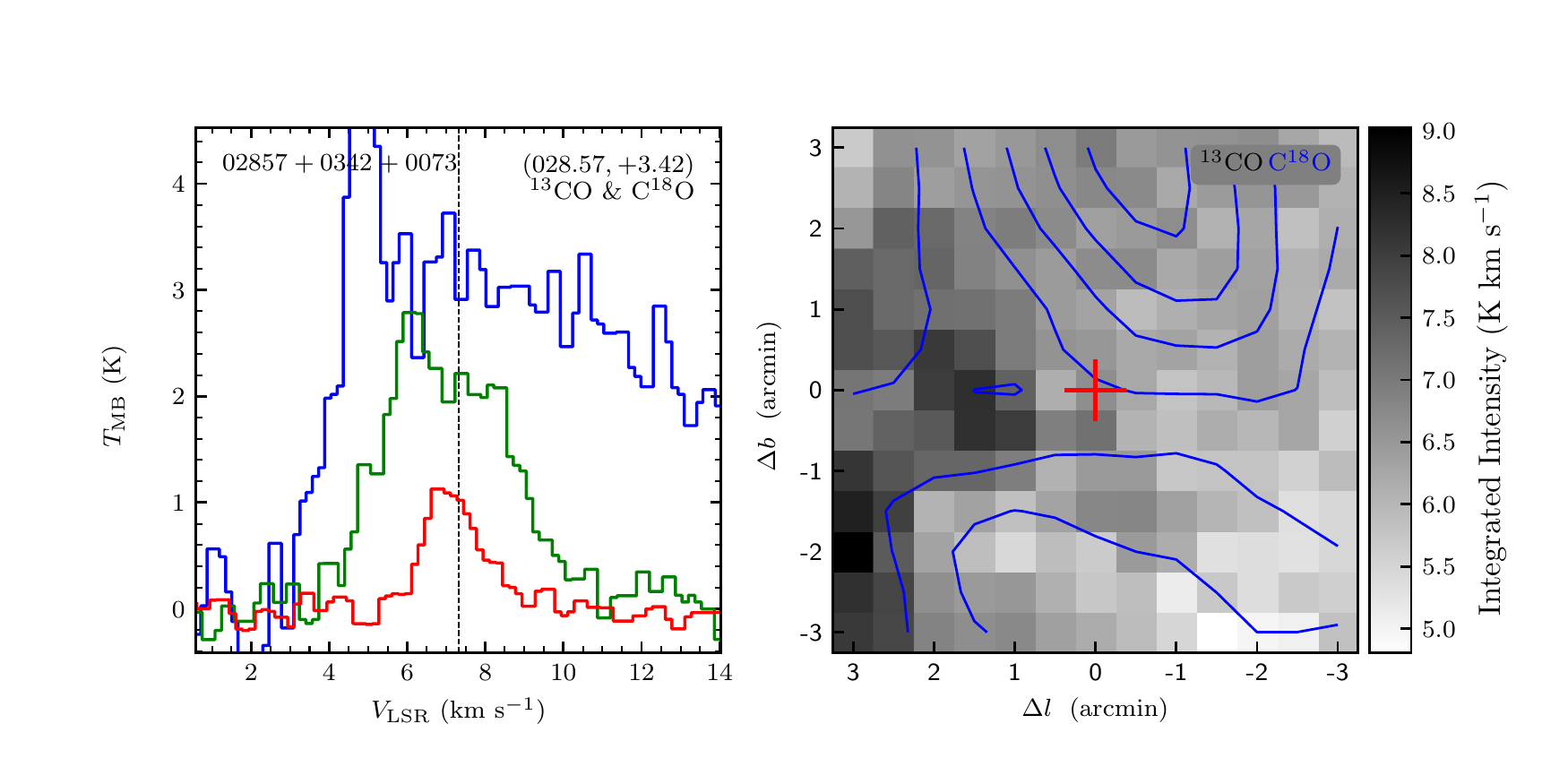}
\includegraphics[width=9.0cm,angle=0]{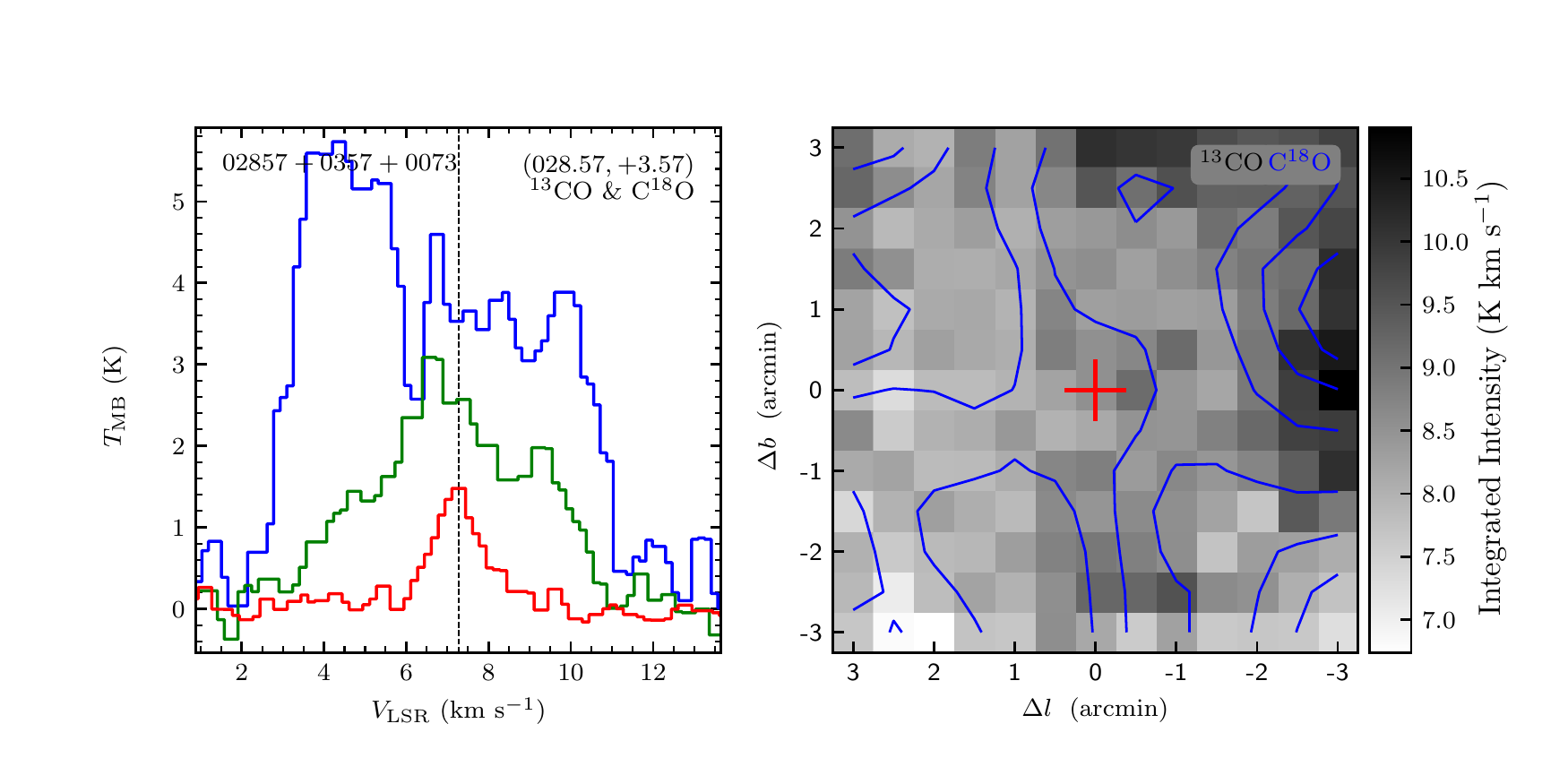}
\vspace{-0.5cm}

\includegraphics[width=9.0cm,angle=0]{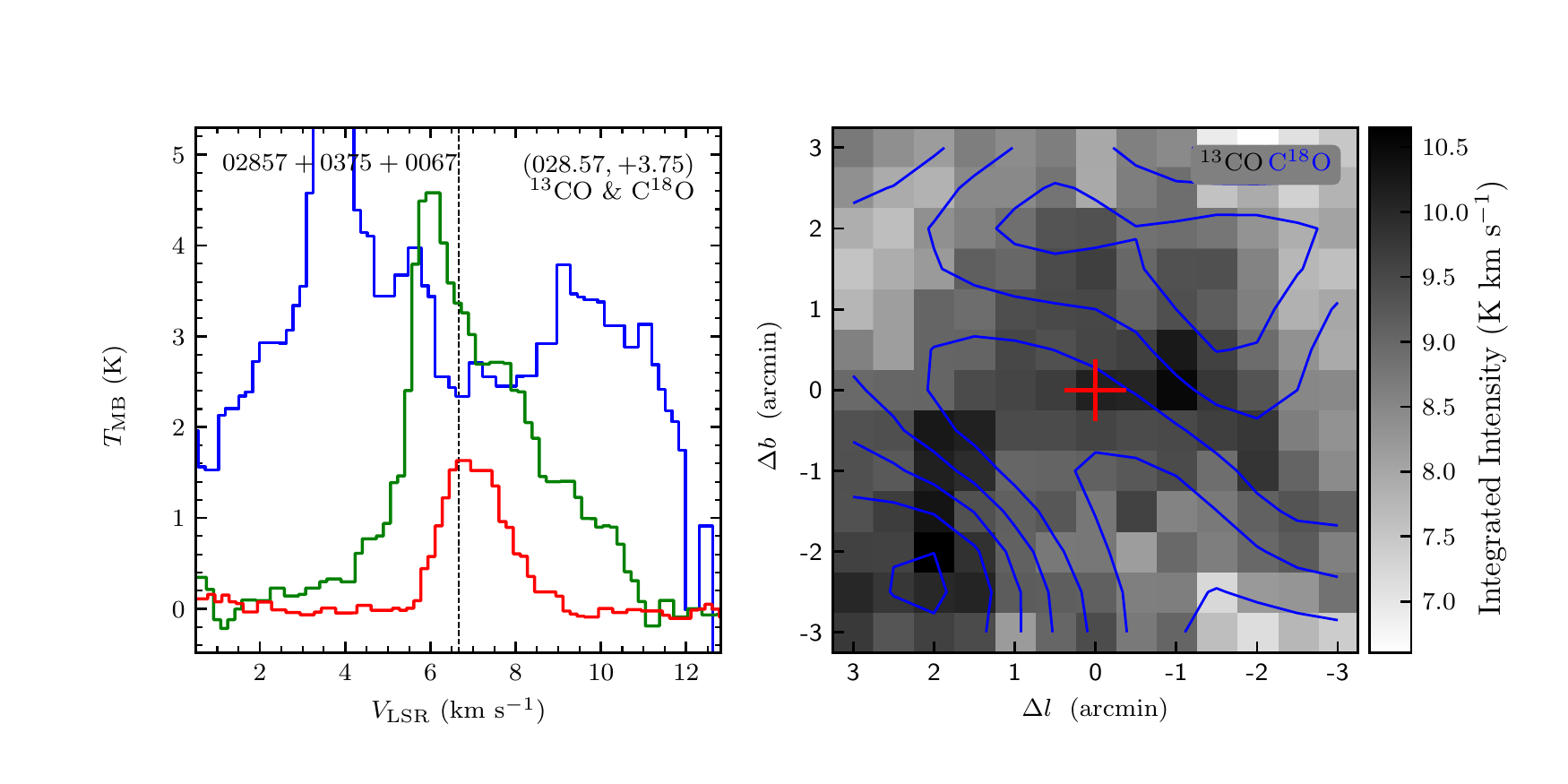}
\includegraphics[width=9.0cm,angle=0]{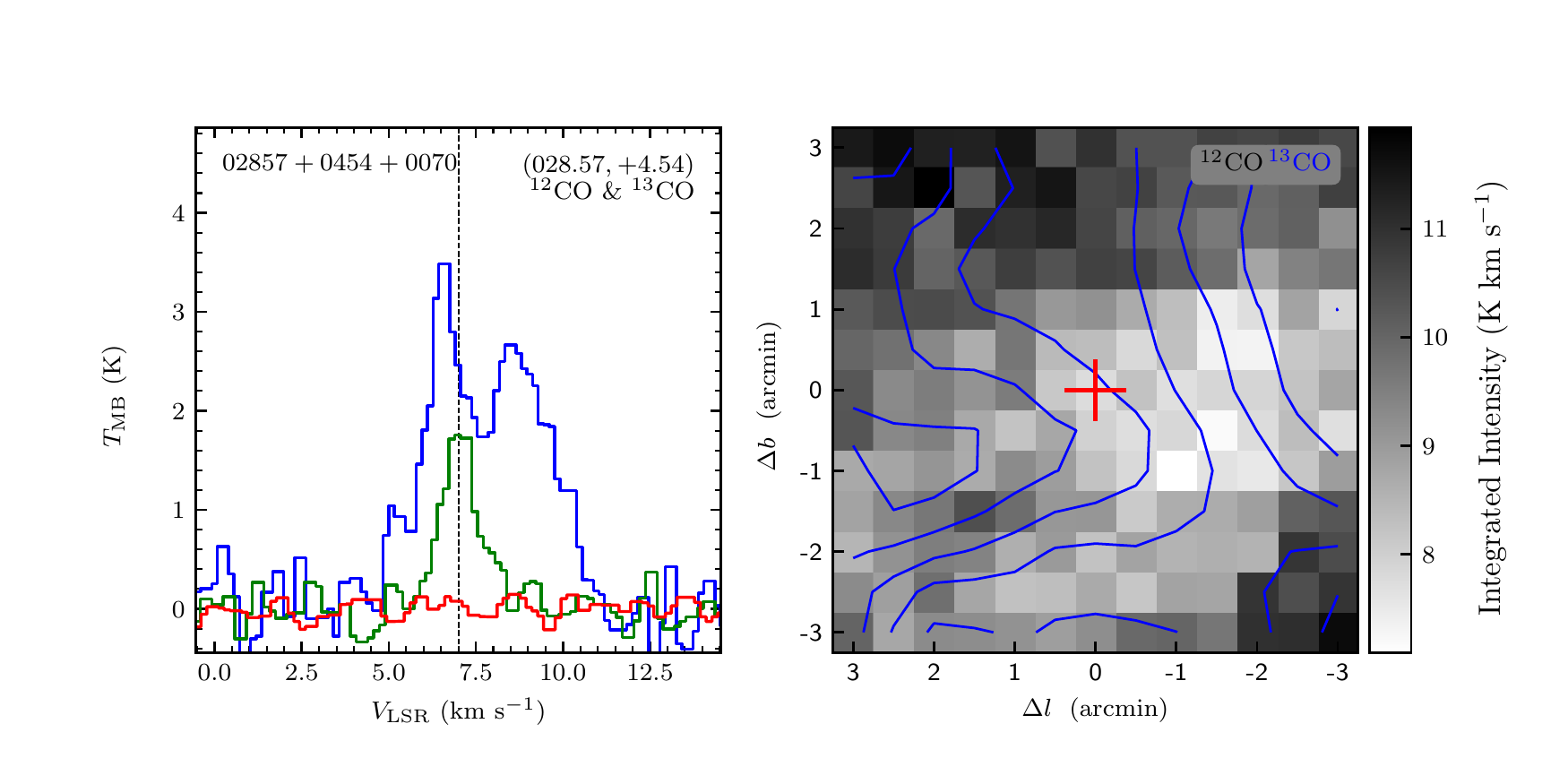}
\vspace{-0.5cm}

\includegraphics[width=9.0cm,angle=0]{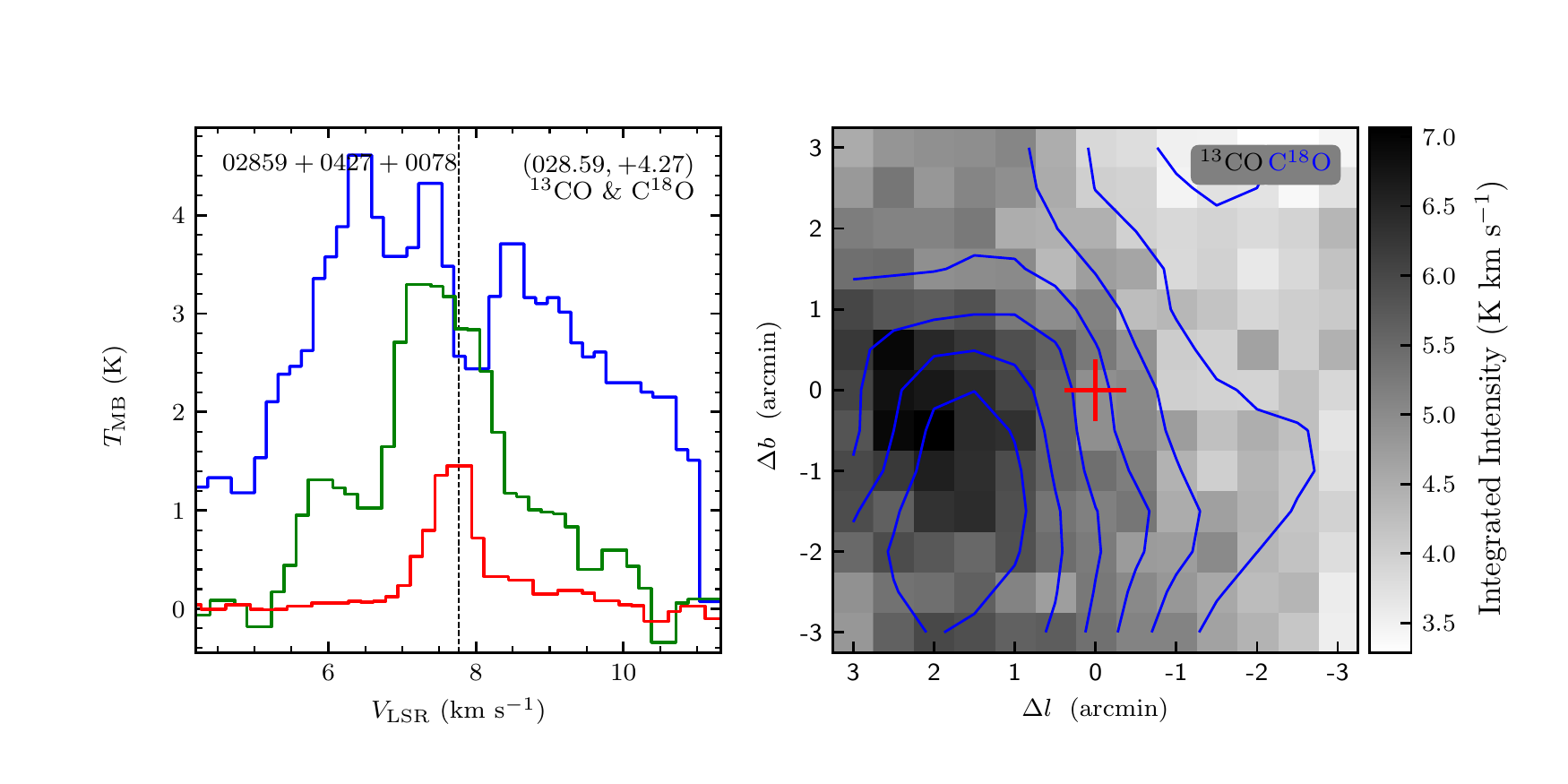}
\includegraphics[width=9.0cm,angle=0]{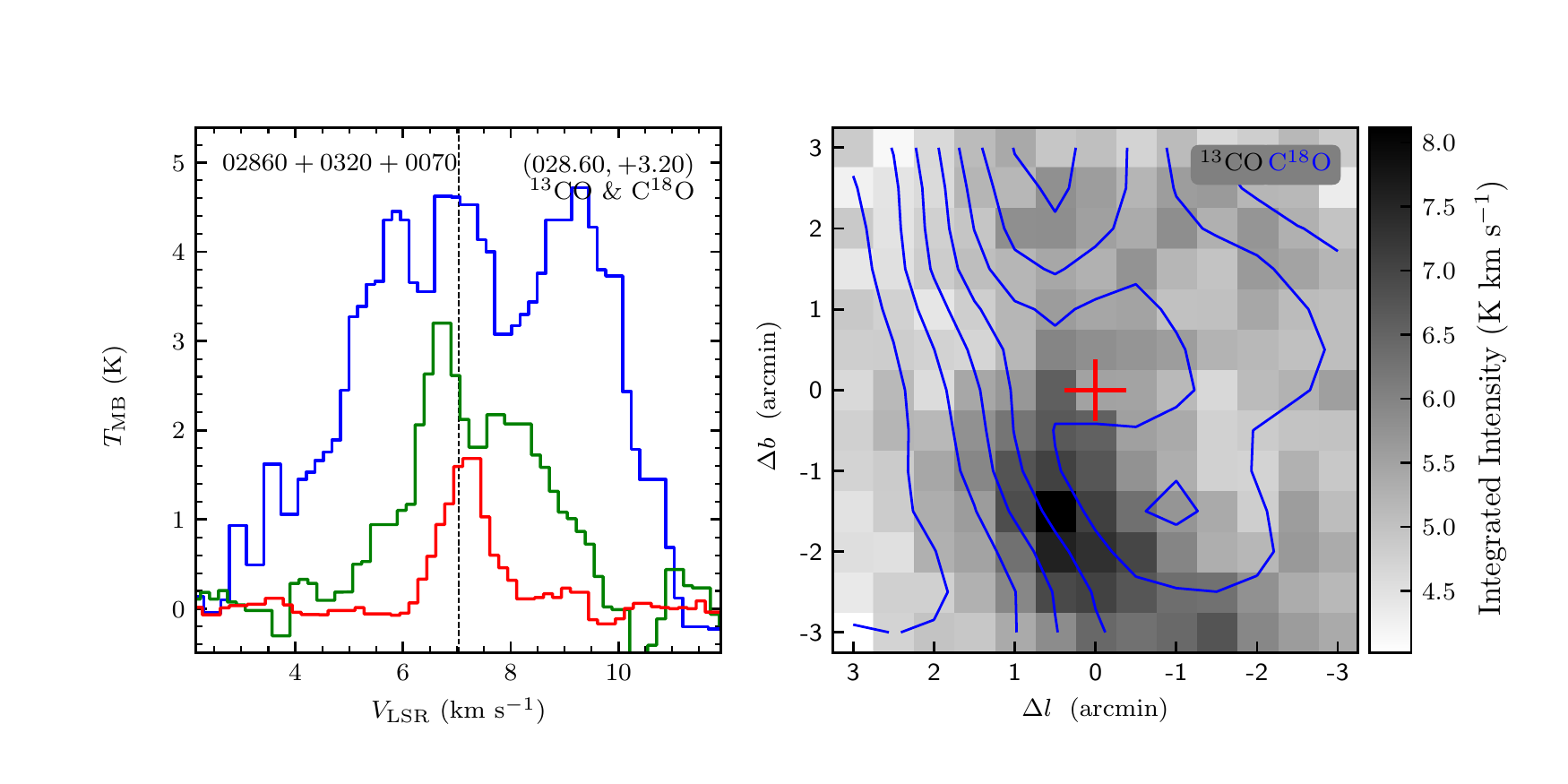}
\vspace{-0.5cm}

\includegraphics[width=9.0cm,angle=0]{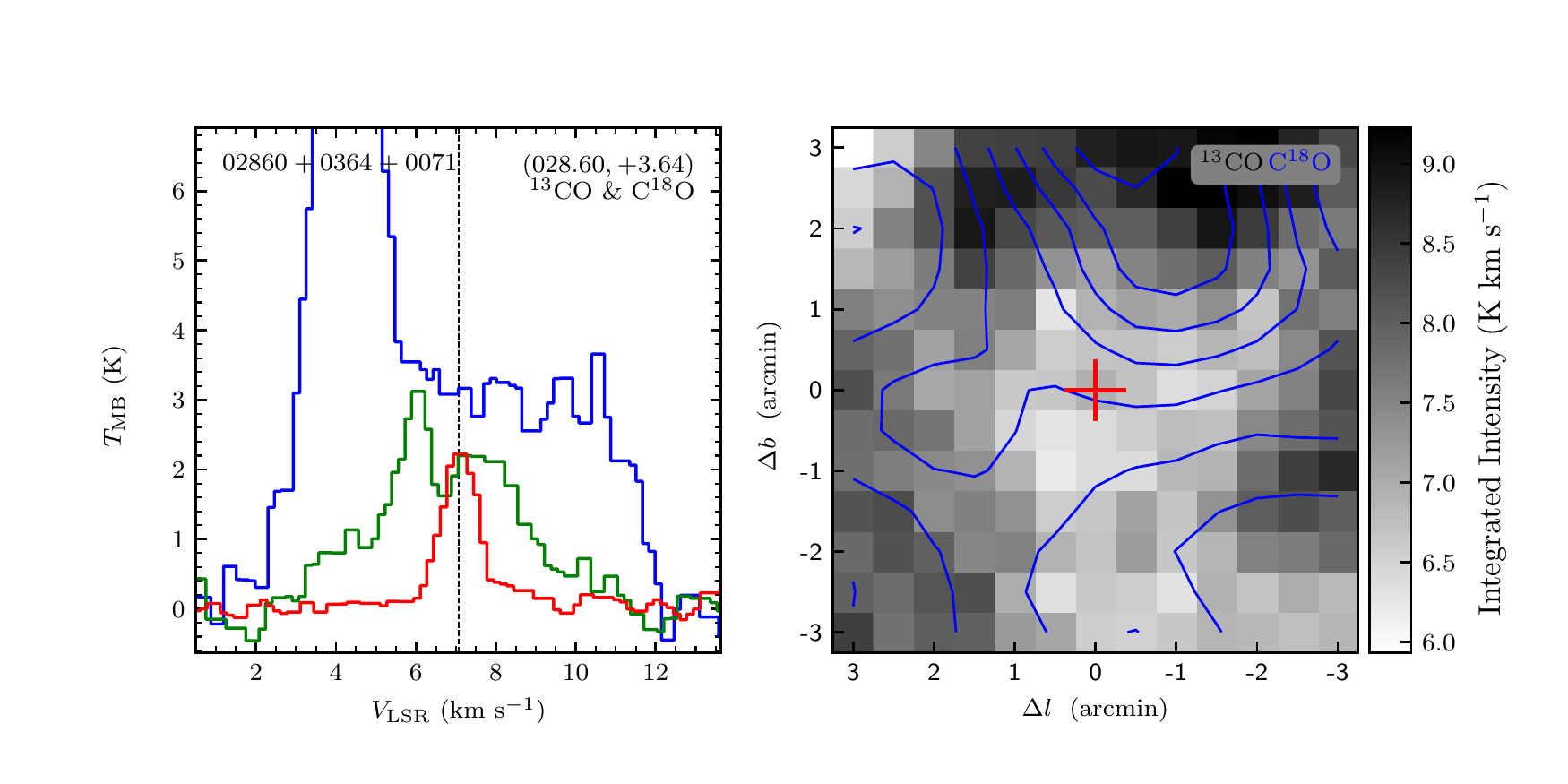}
\includegraphics[width=9.0cm,angle=0]{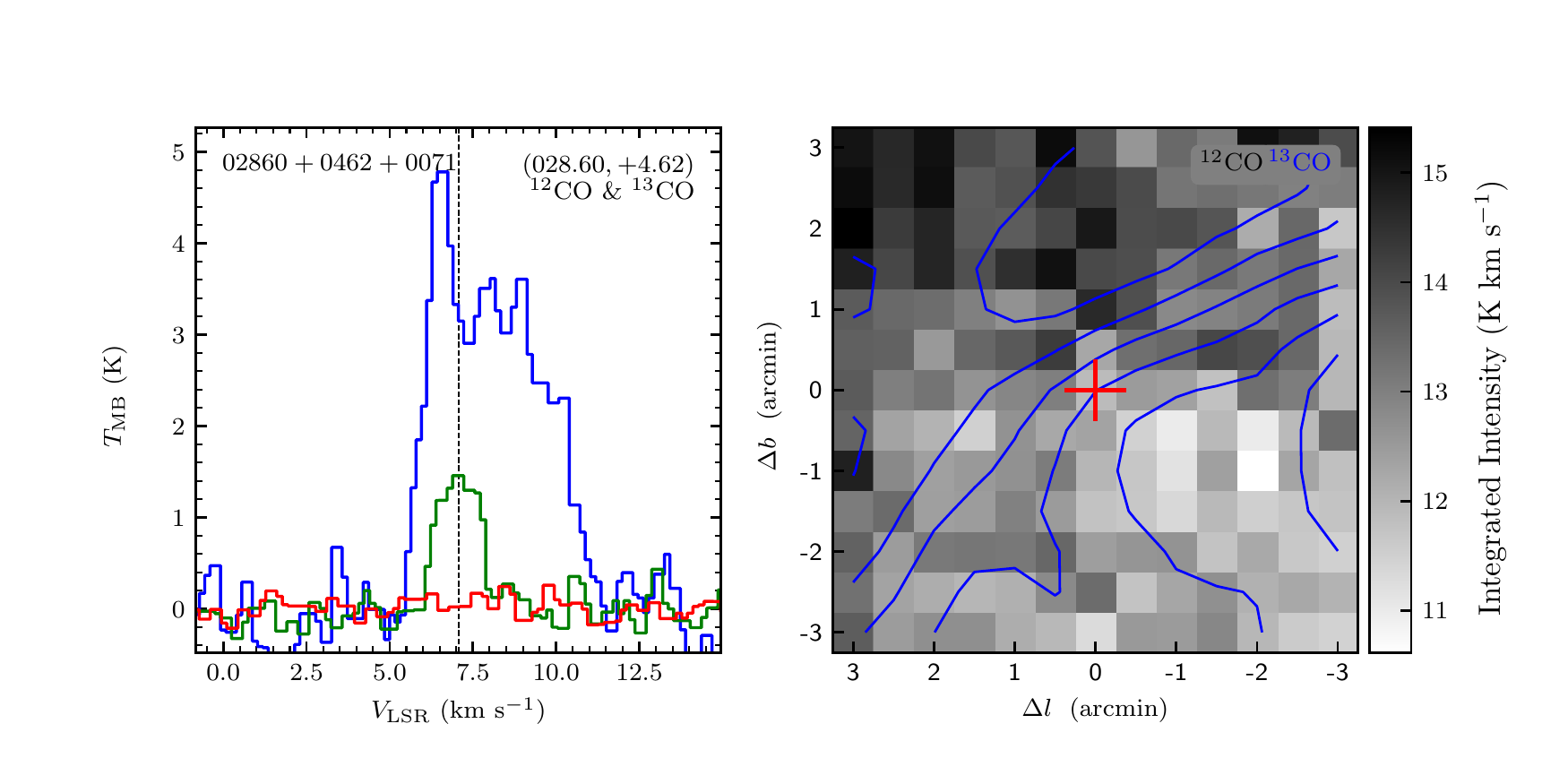}
\end{figure}
\clearpage

\begin{figure}
\includegraphics[width=9.0cm,angle=0]{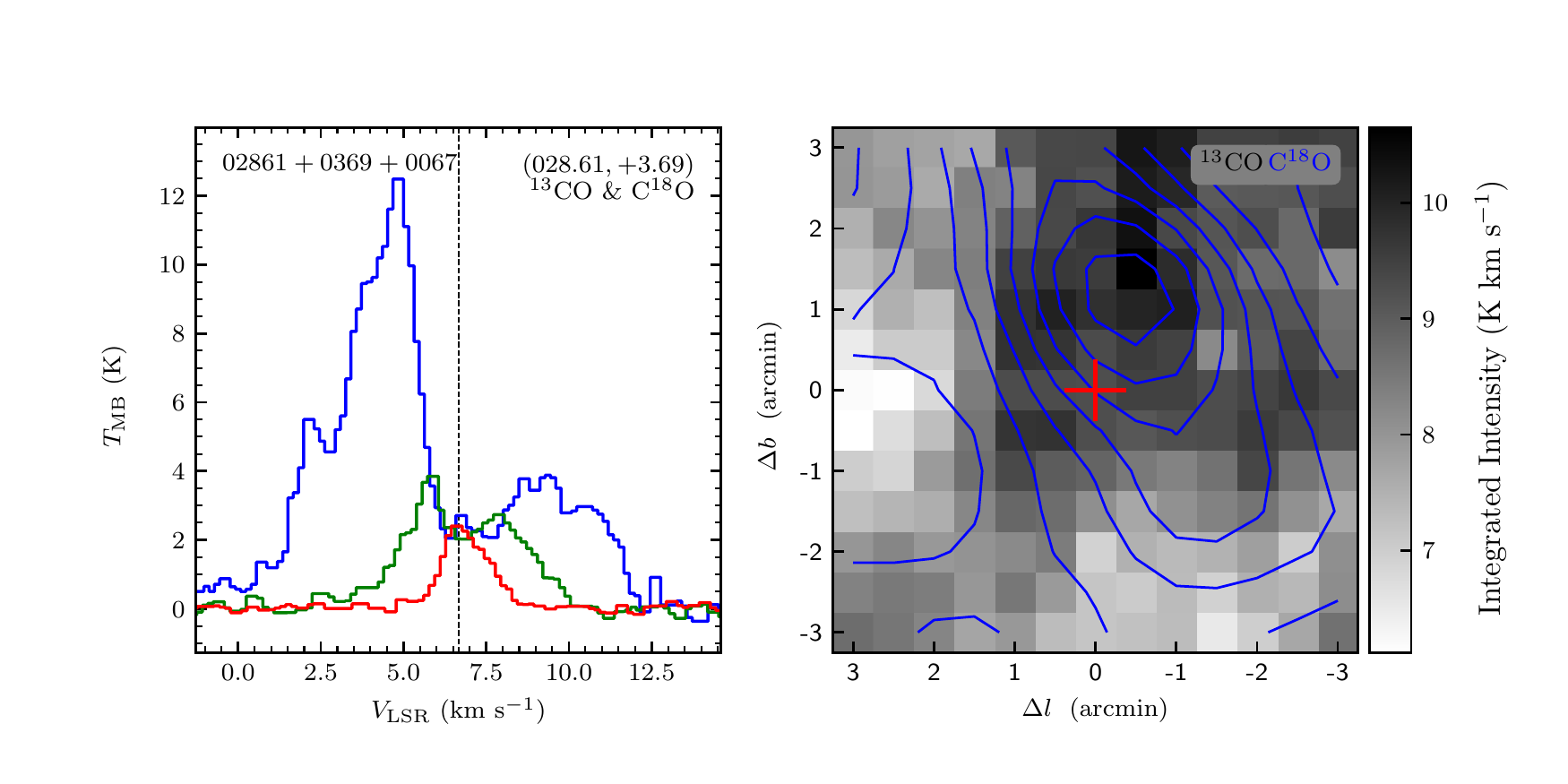}
\includegraphics[width=9.0cm,angle=0]{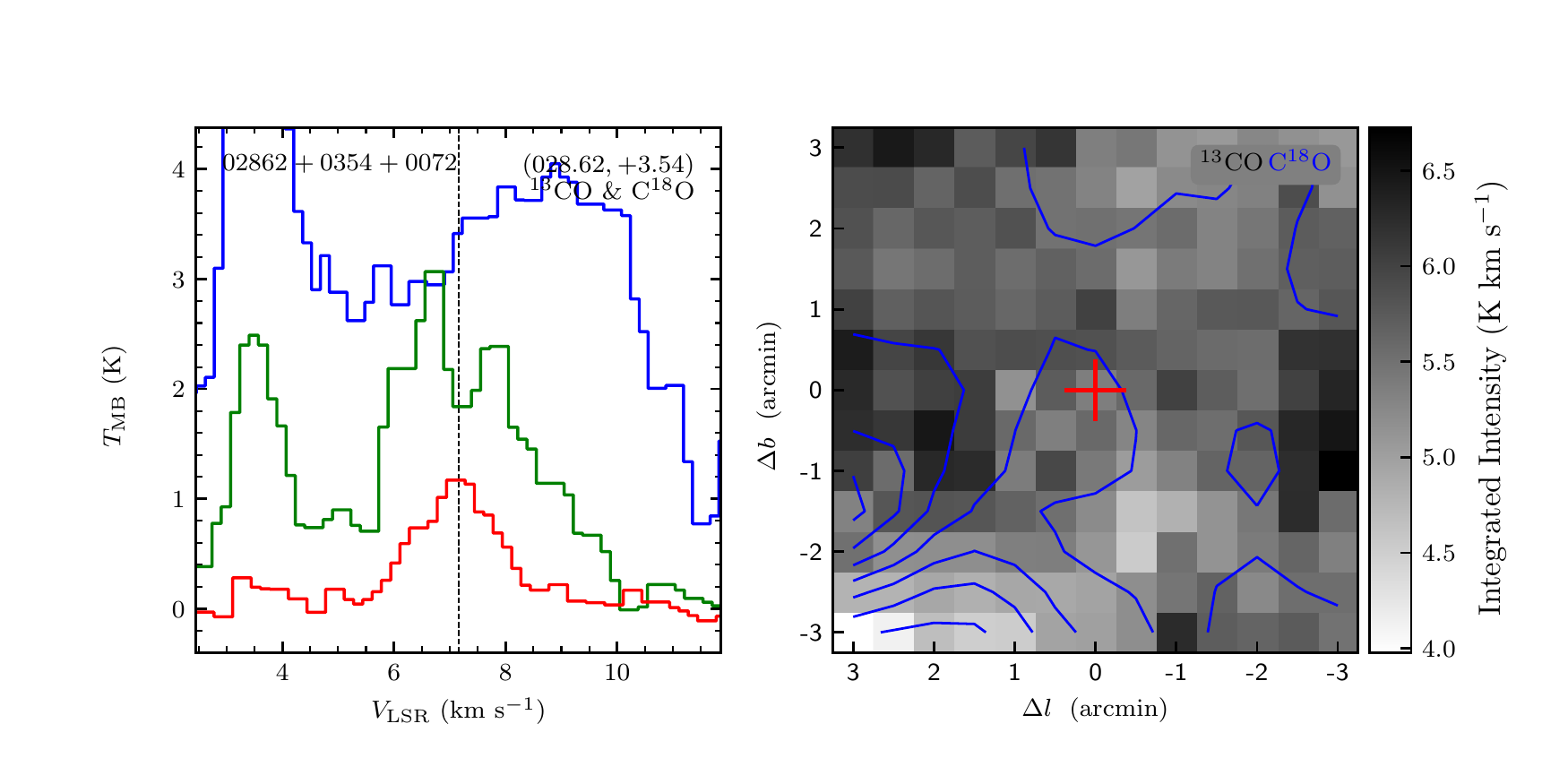}
\vspace{-0.5cm}

\includegraphics[width=9.0cm,angle=0]{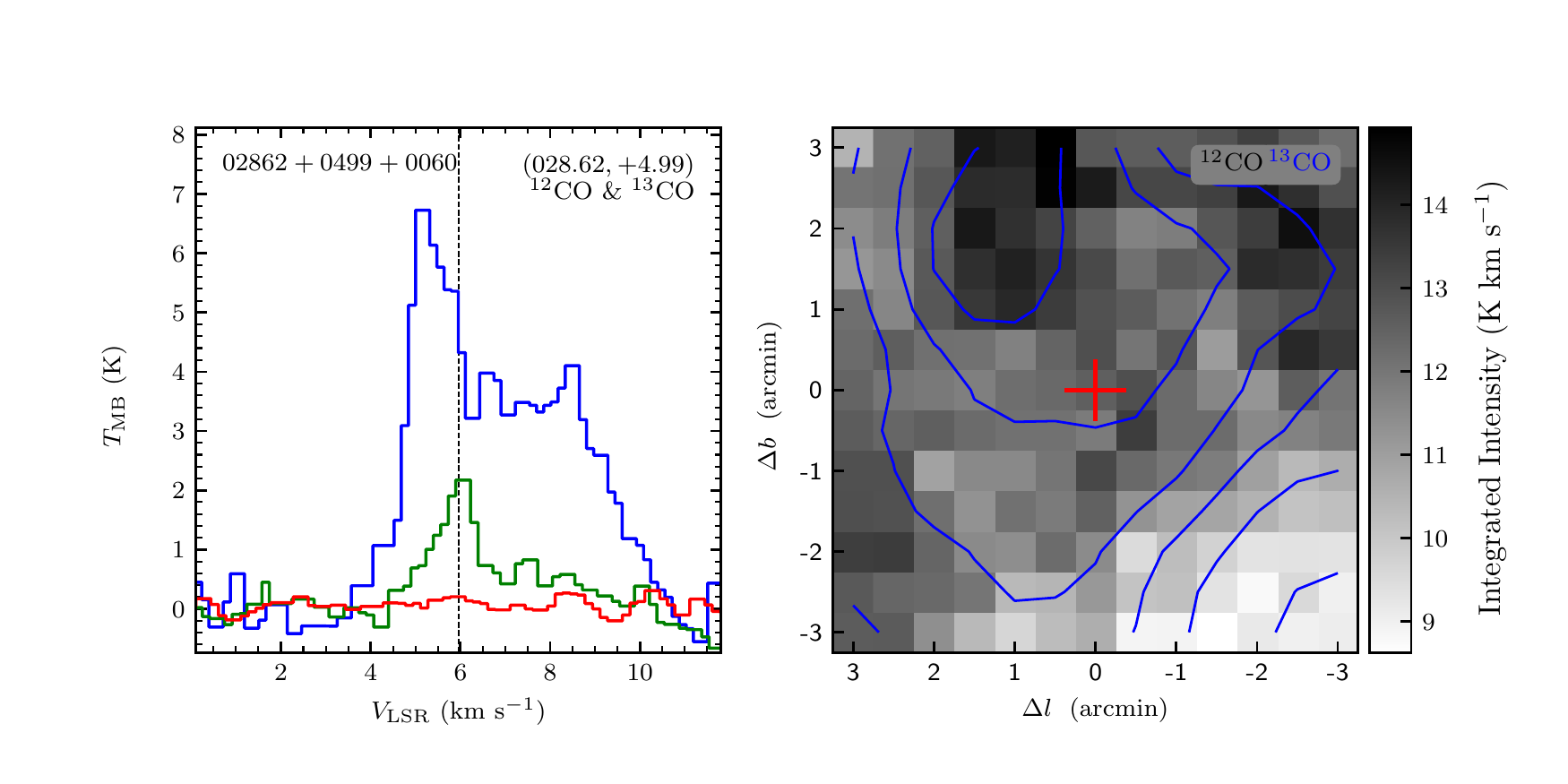}
\includegraphics[width=9.0cm,angle=0]{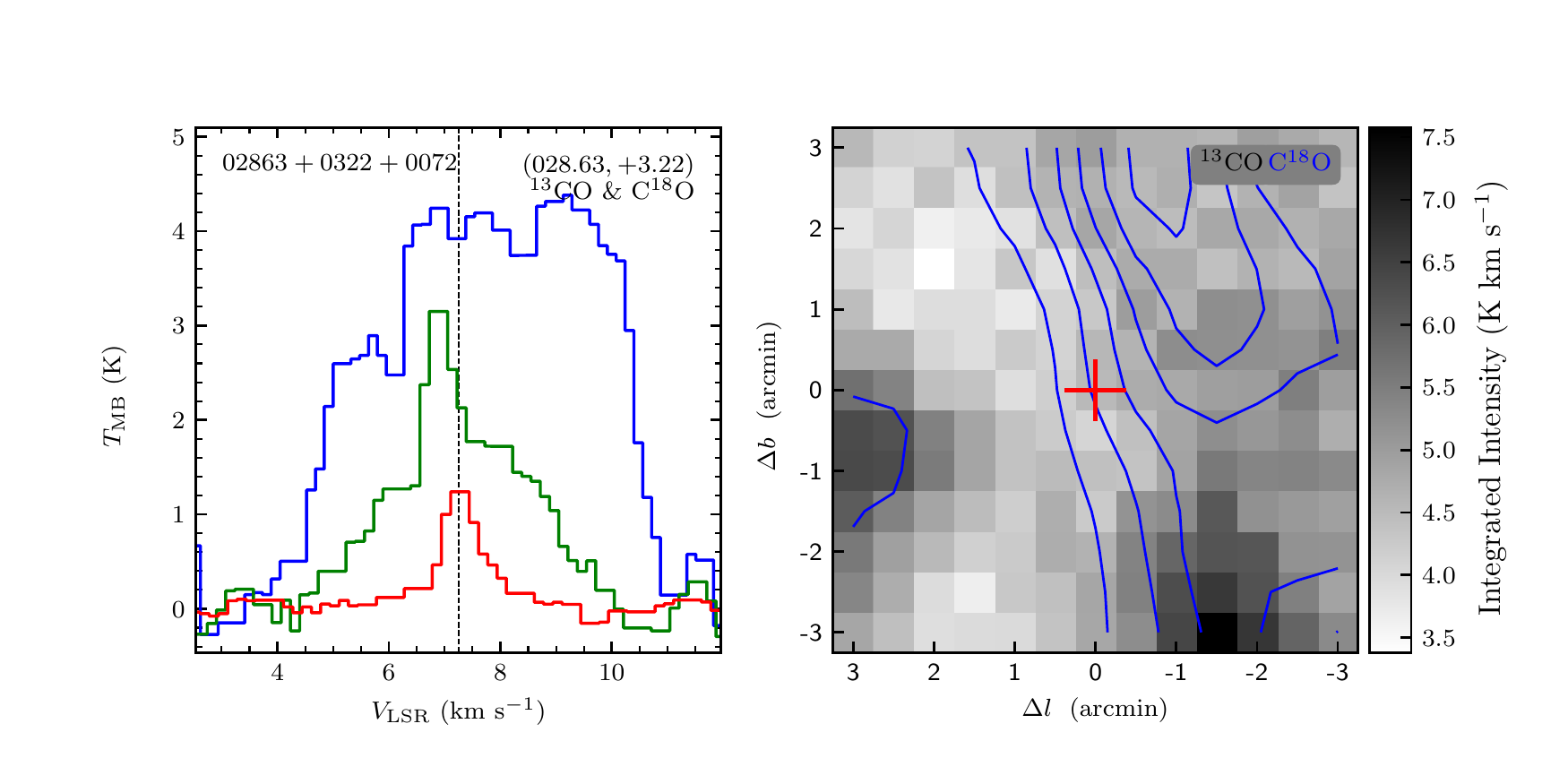}
\vspace{-0.5cm}

\includegraphics[width=9.0cm,angle=0]{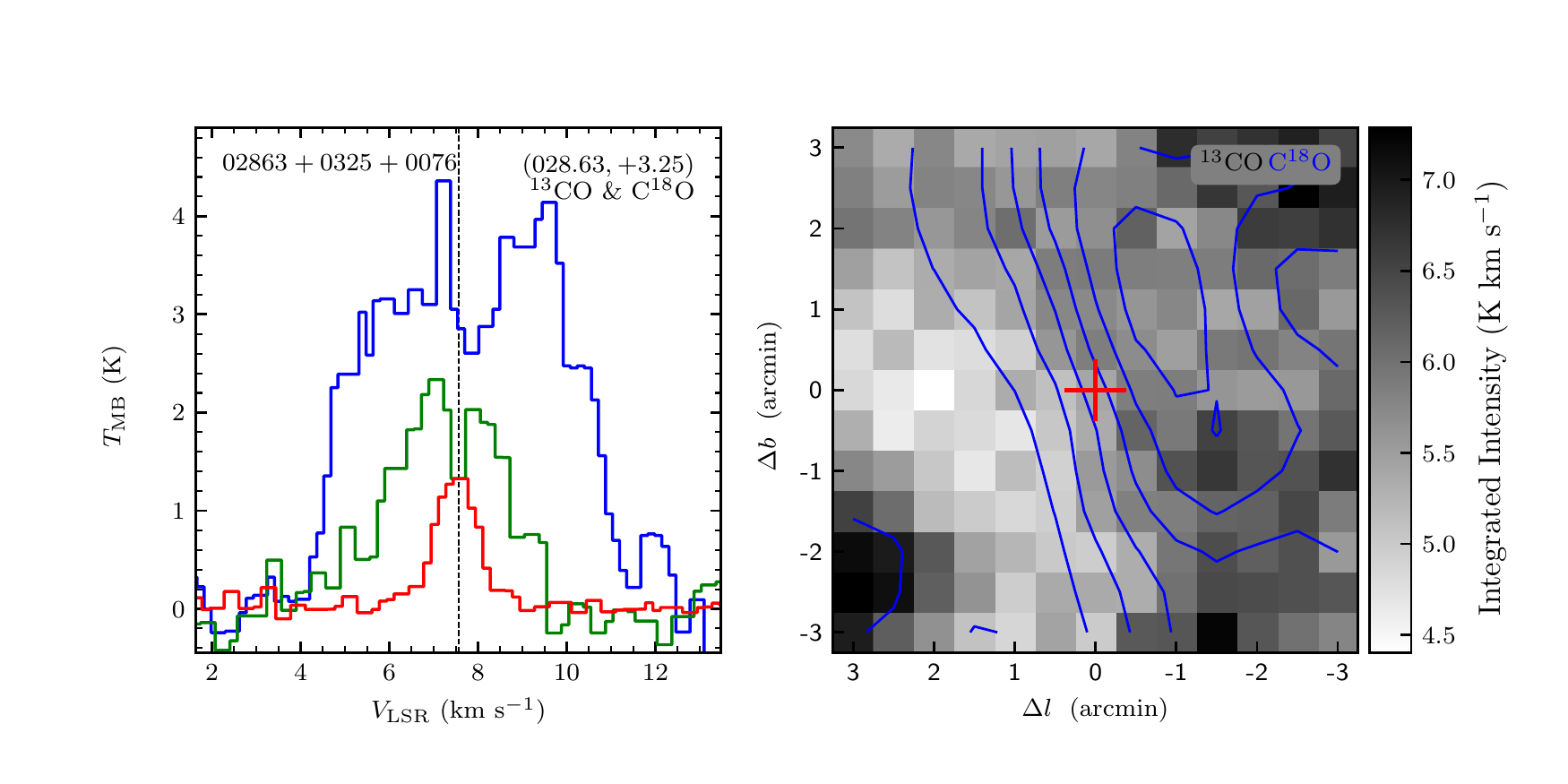}
\includegraphics[width=9.0cm,angle=0]{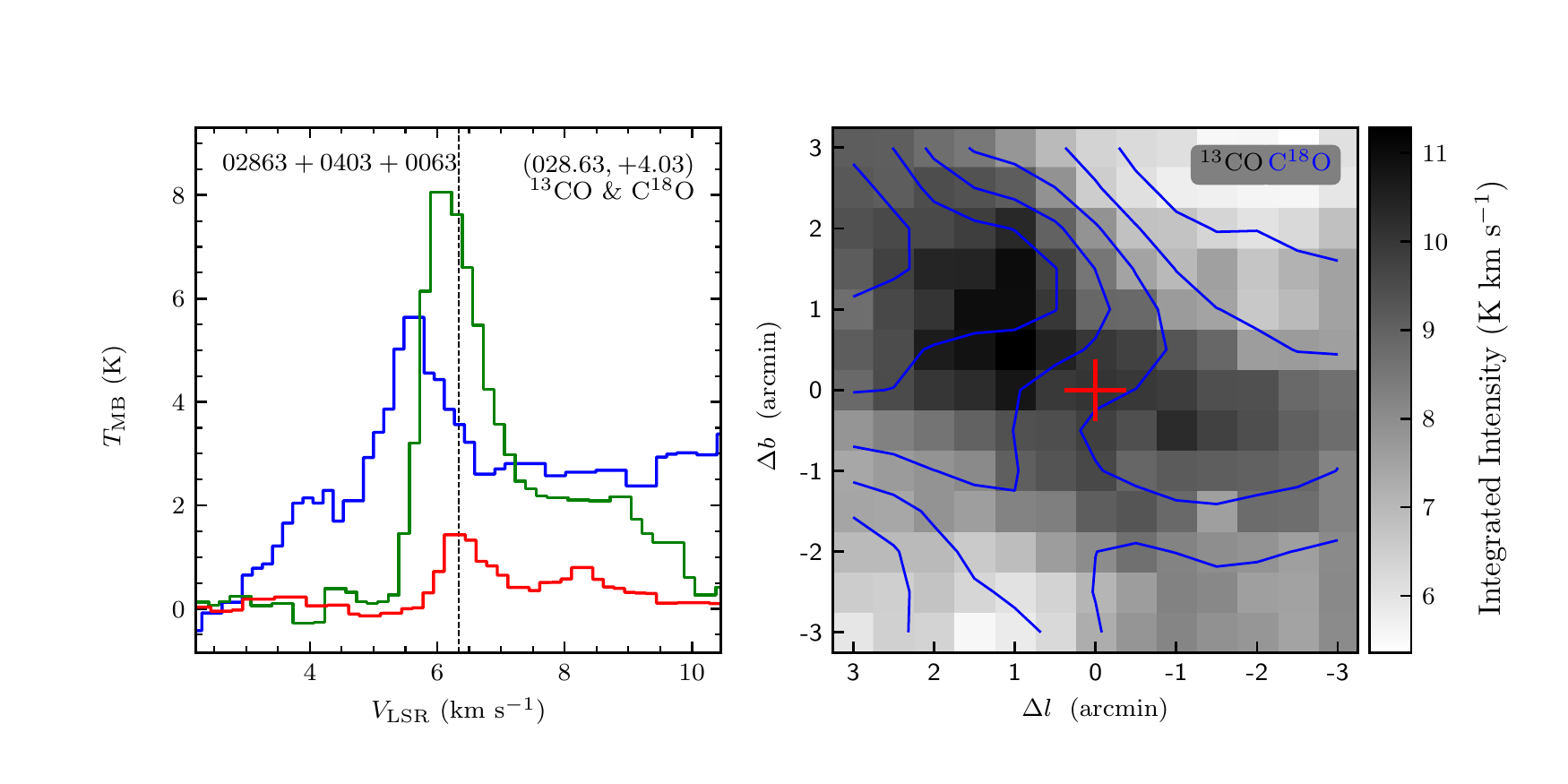}
\vspace{-0.5cm}

\includegraphics[width=9.0cm,angle=0]{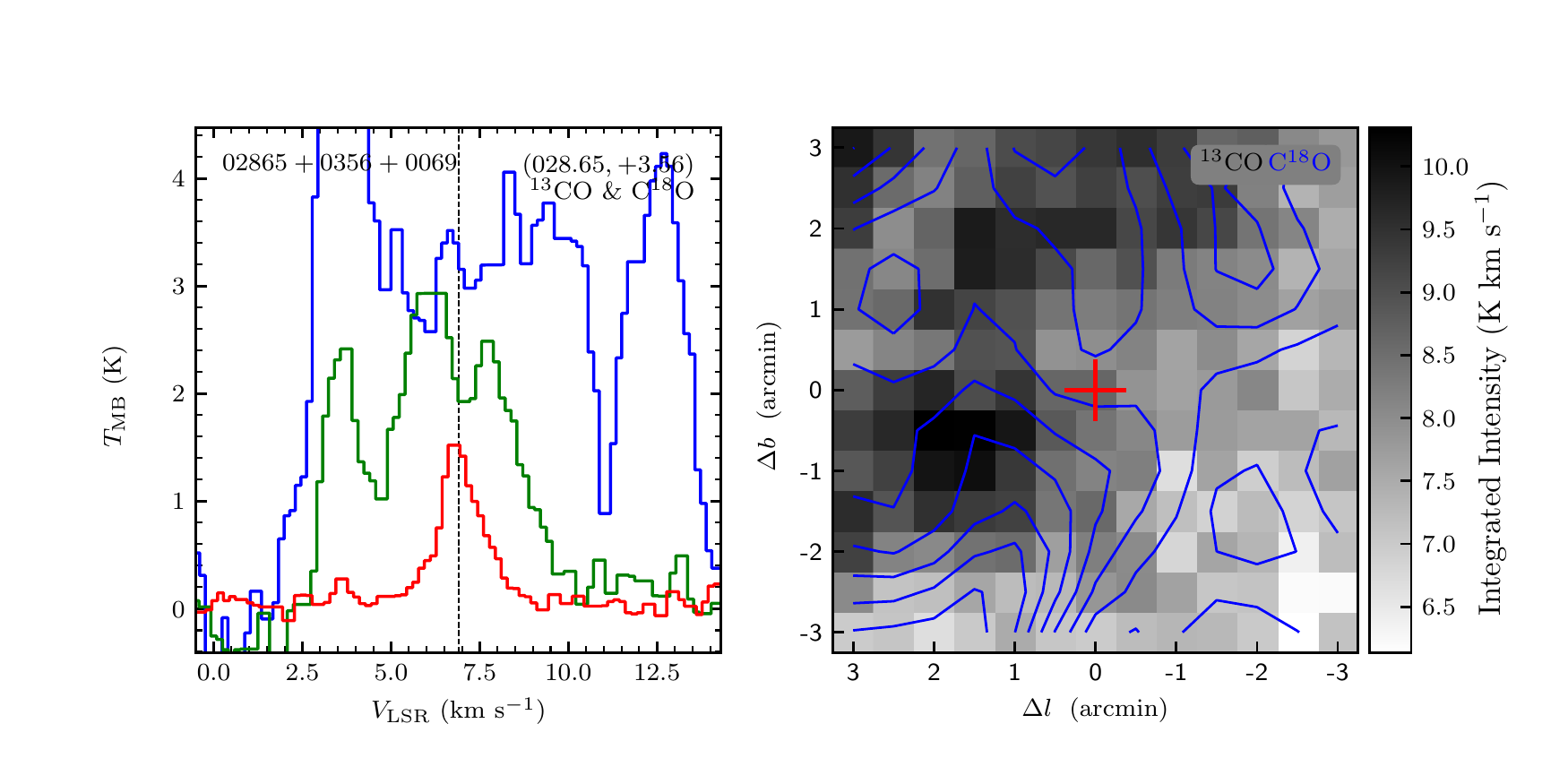}
\includegraphics[width=9.0cm,angle=0]{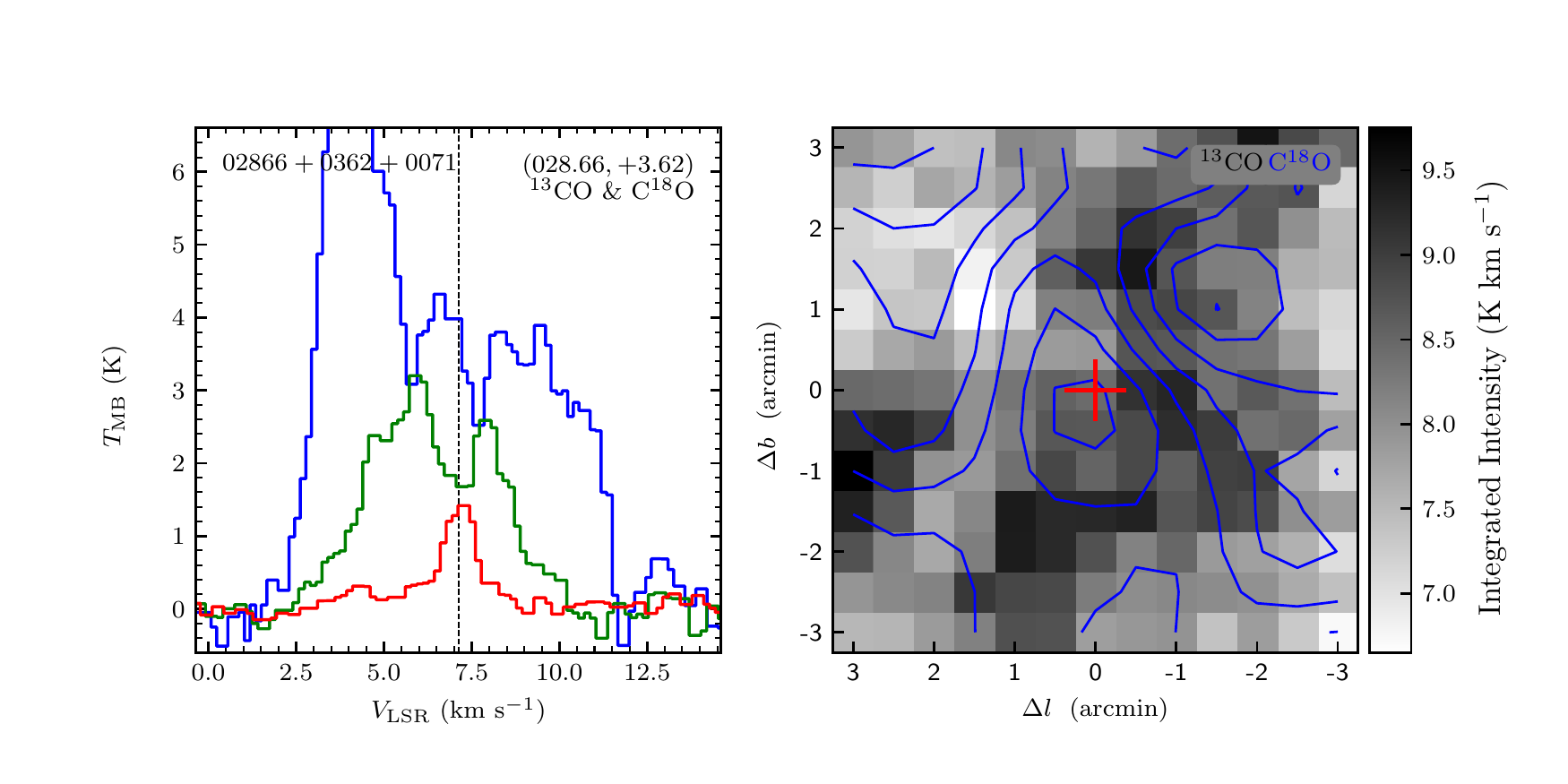}
\vspace{-0.5cm}

\includegraphics[width=9.0cm,angle=0]{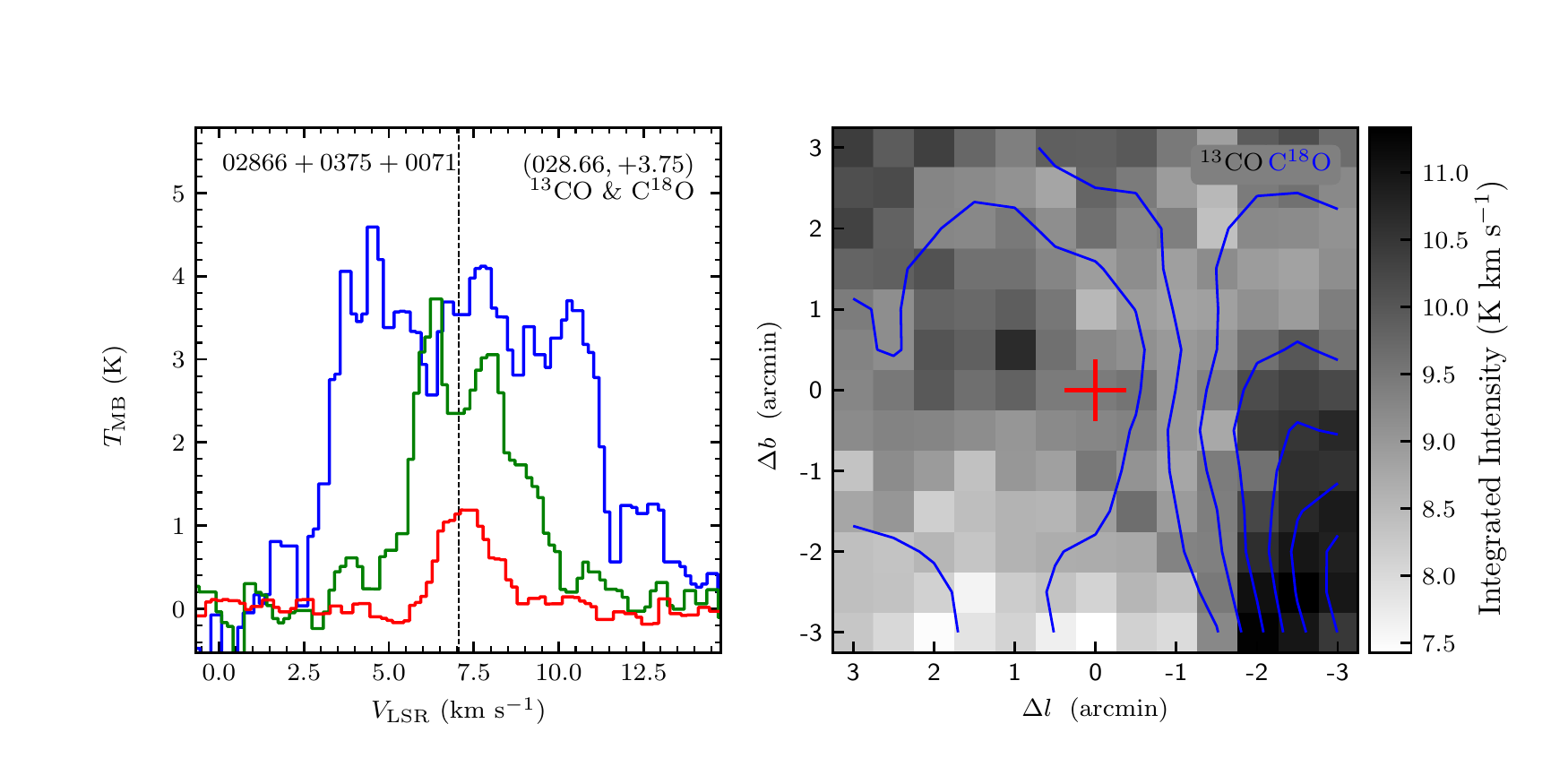}
\includegraphics[width=9.0cm,angle=0]{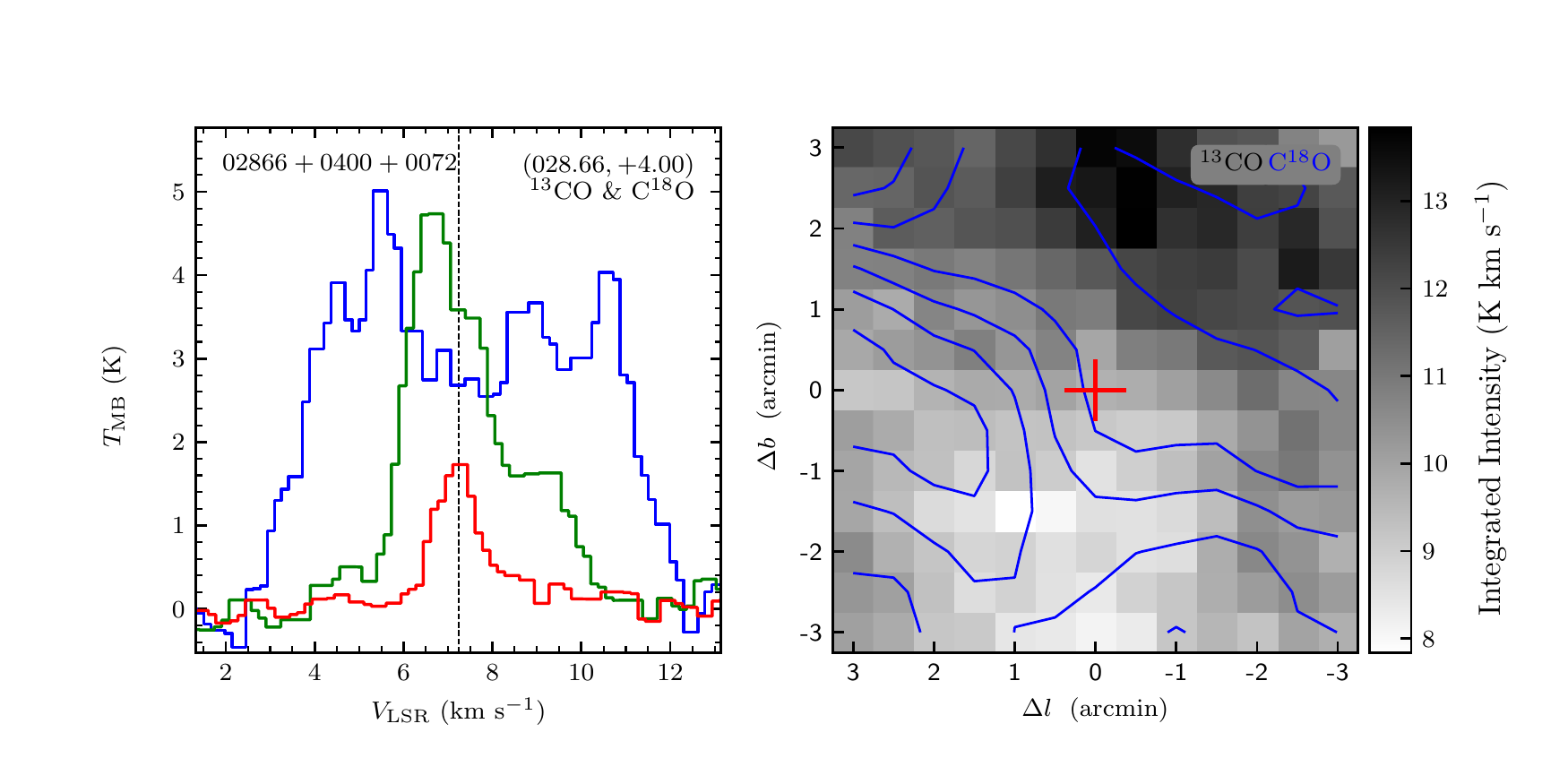}
\end{figure}
\clearpage

\begin{figure}
\includegraphics[width=9.0cm,angle=0]{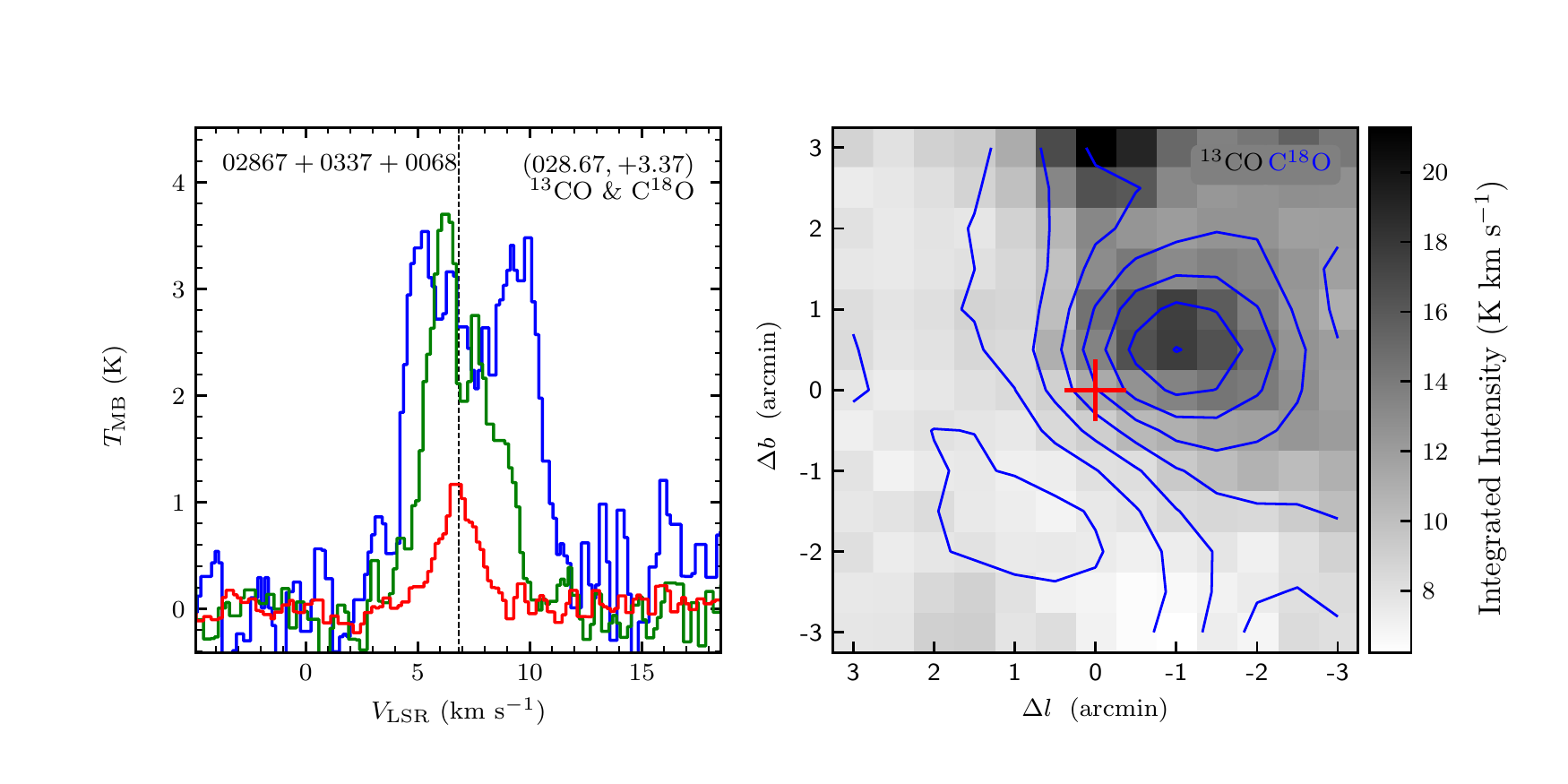}
\includegraphics[width=9.0cm,angle=0]{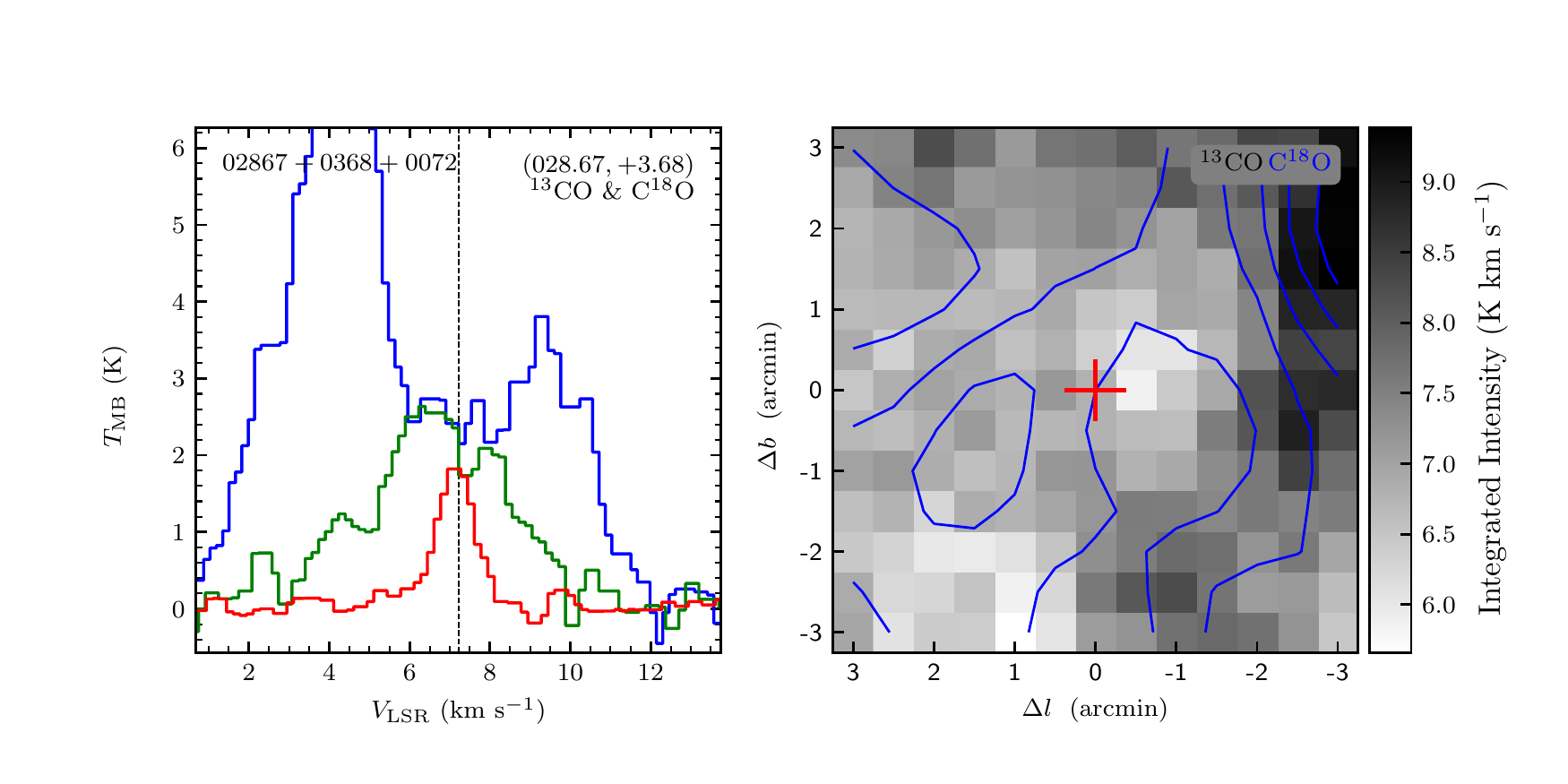}
\vspace{-0.5cm}

\includegraphics[width=9.0cm,angle=0]{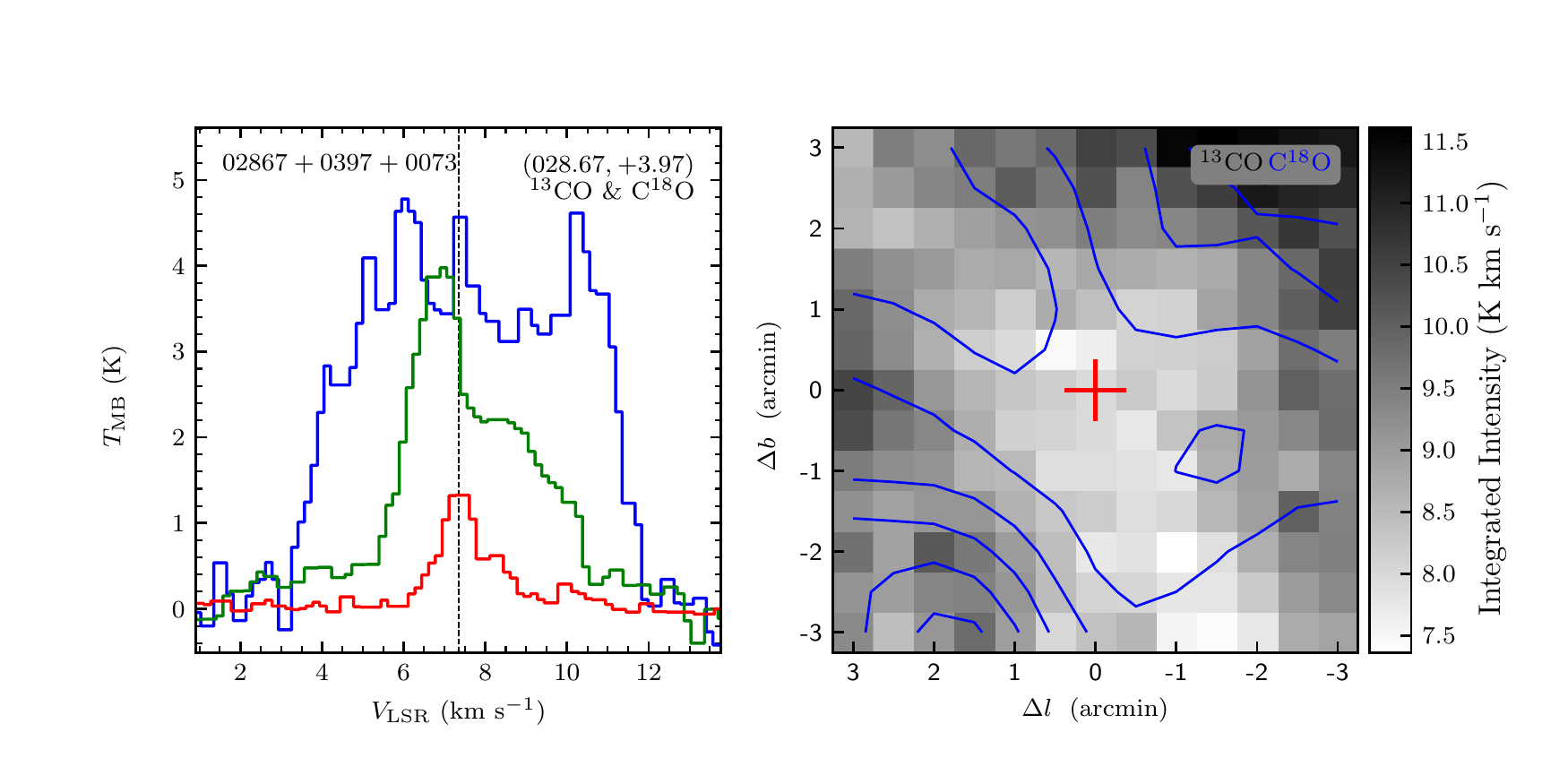}
\includegraphics[width=9.0cm,angle=0]{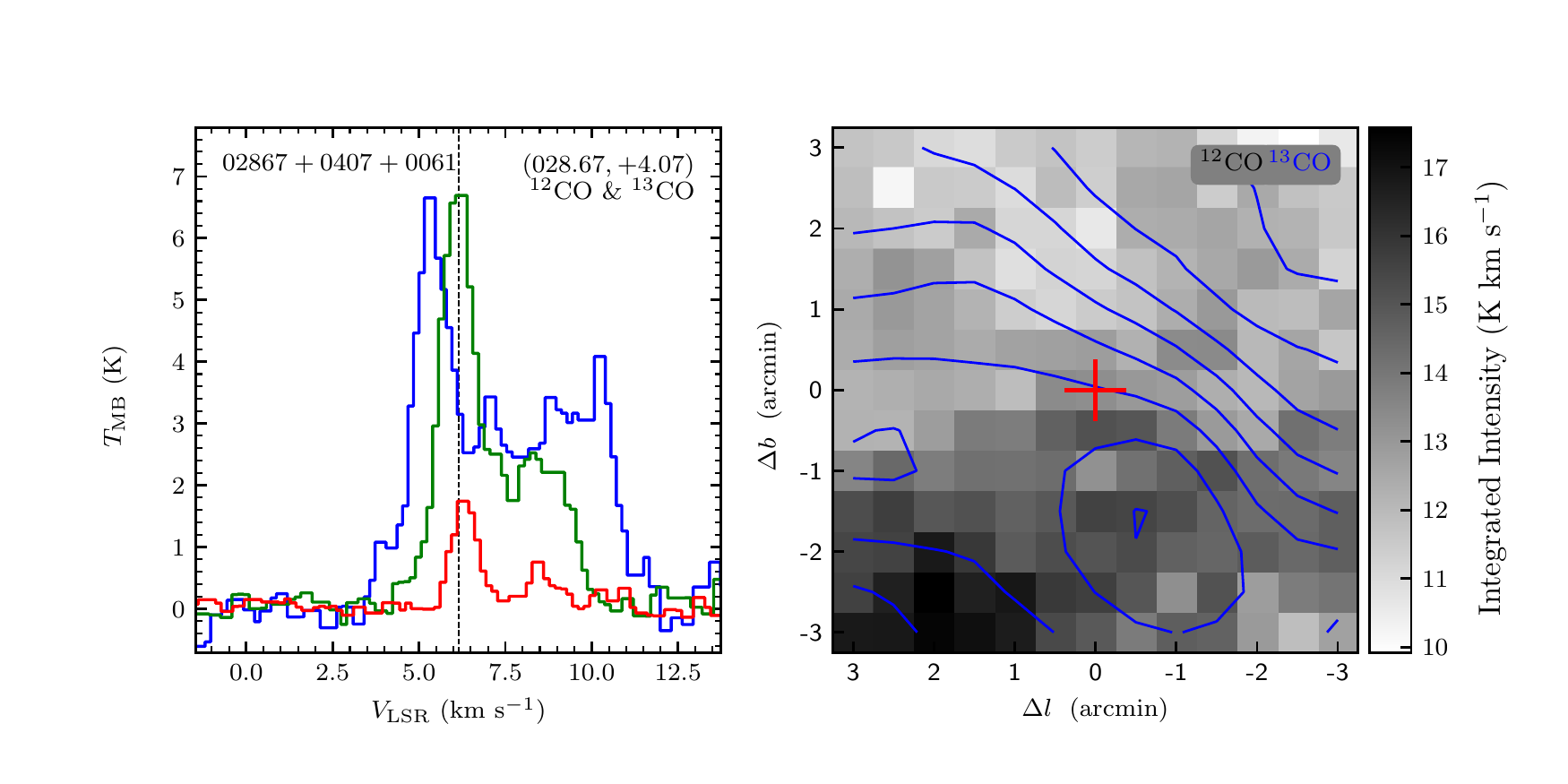}
\vspace{-0.5cm}

\includegraphics[width=9.0cm,angle=0]{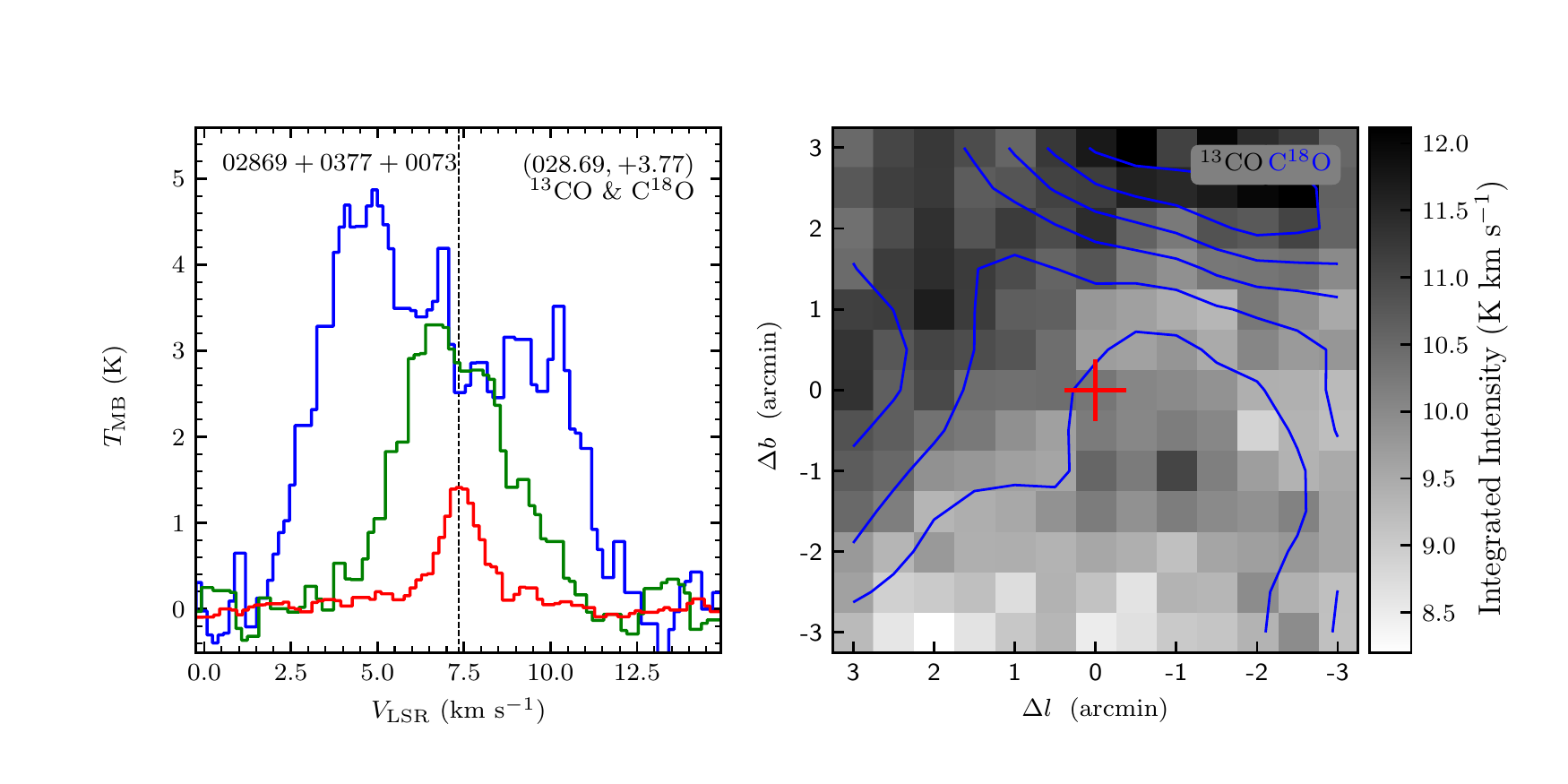}
\includegraphics[width=9.0cm,angle=0]{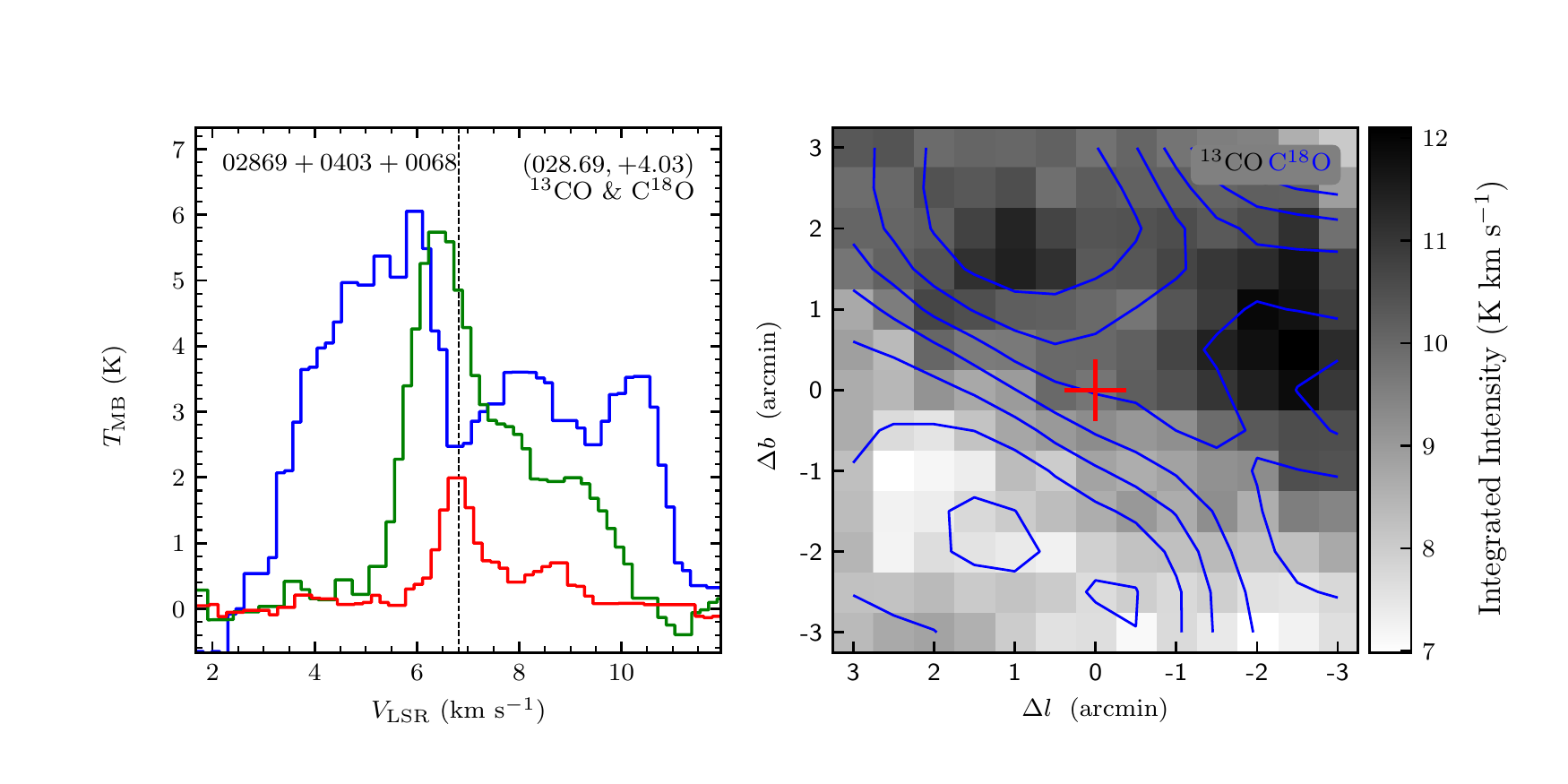}
\vspace{-0.5cm}

\includegraphics[width=9.0cm,angle=0]{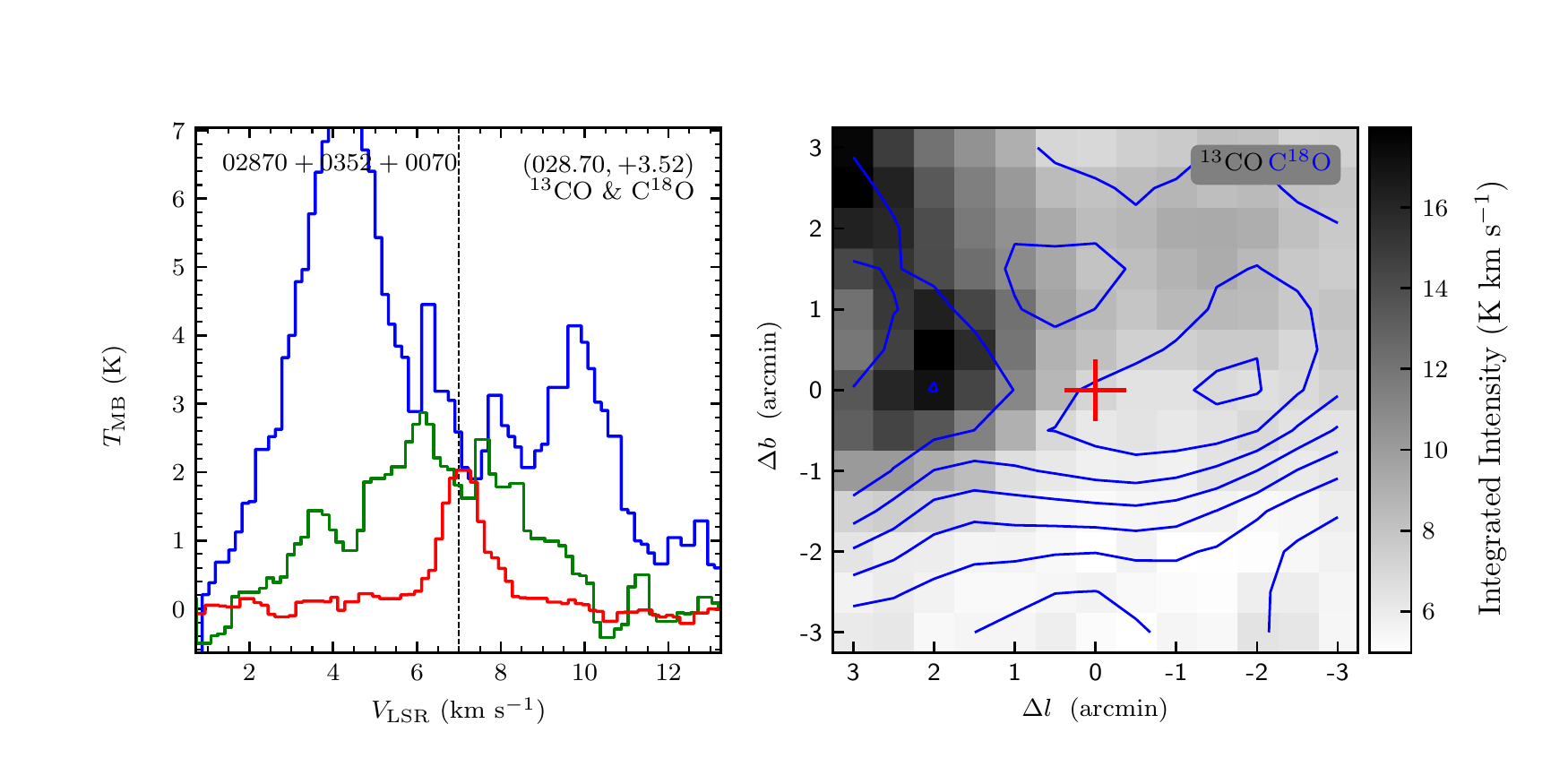}
\includegraphics[width=9.0cm,angle=0]{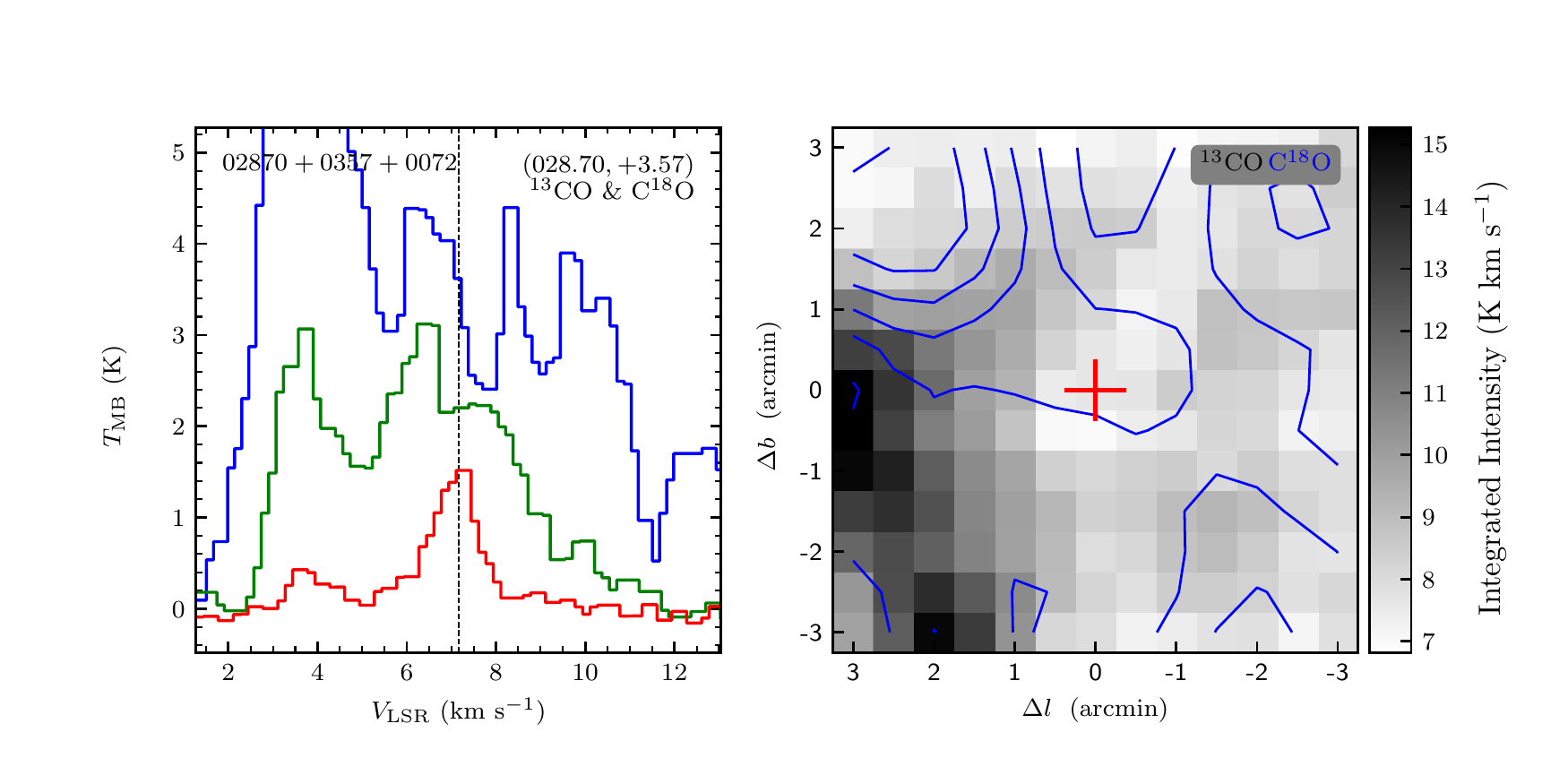}
\vspace{-0.5cm}

\includegraphics[width=9.0cm,angle=0]{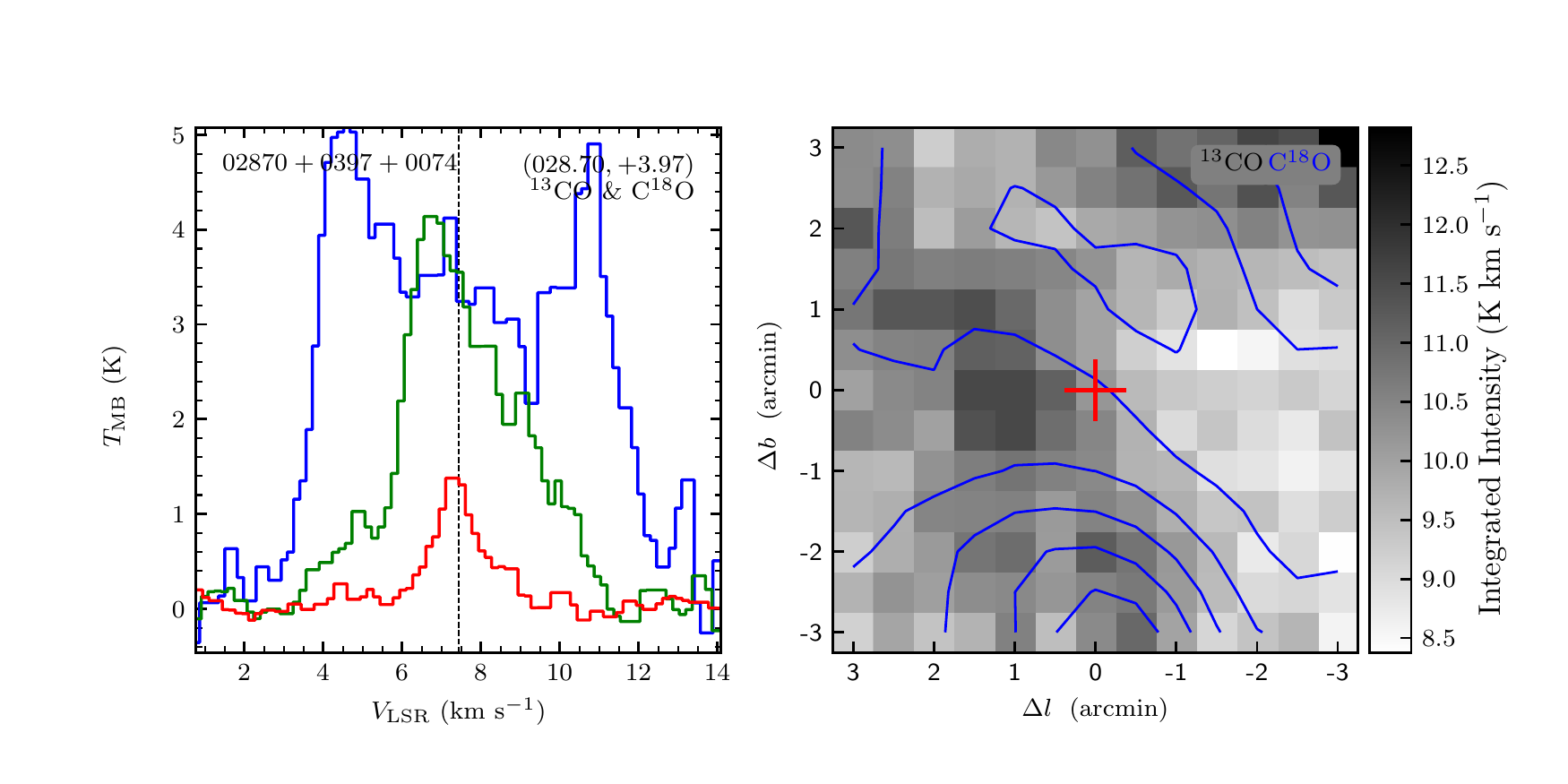}
\includegraphics[width=9.0cm,angle=0]{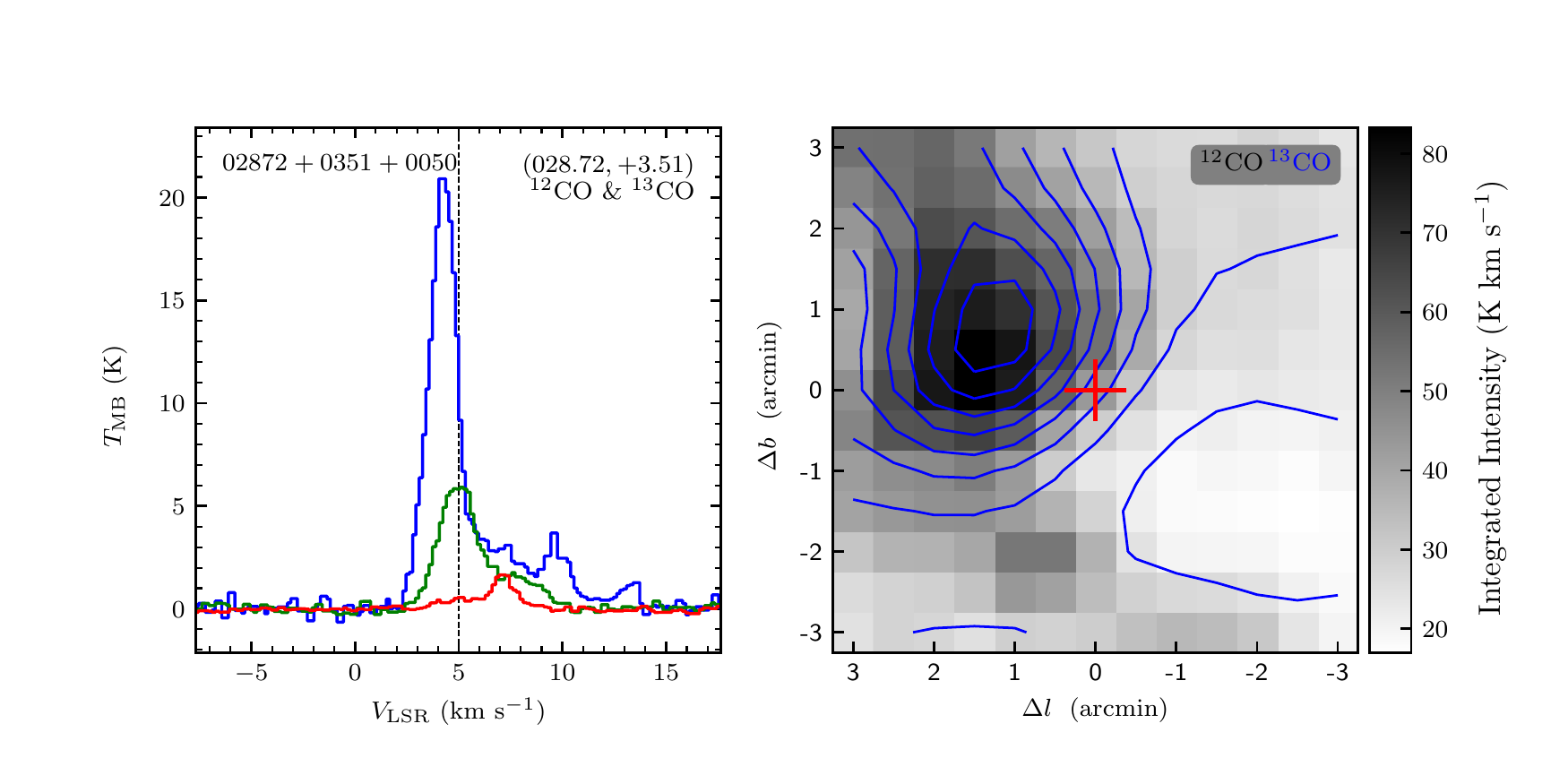}
\end{figure}
\clearpage

\begin{figure}
\includegraphics[width=9.0cm,angle=0]{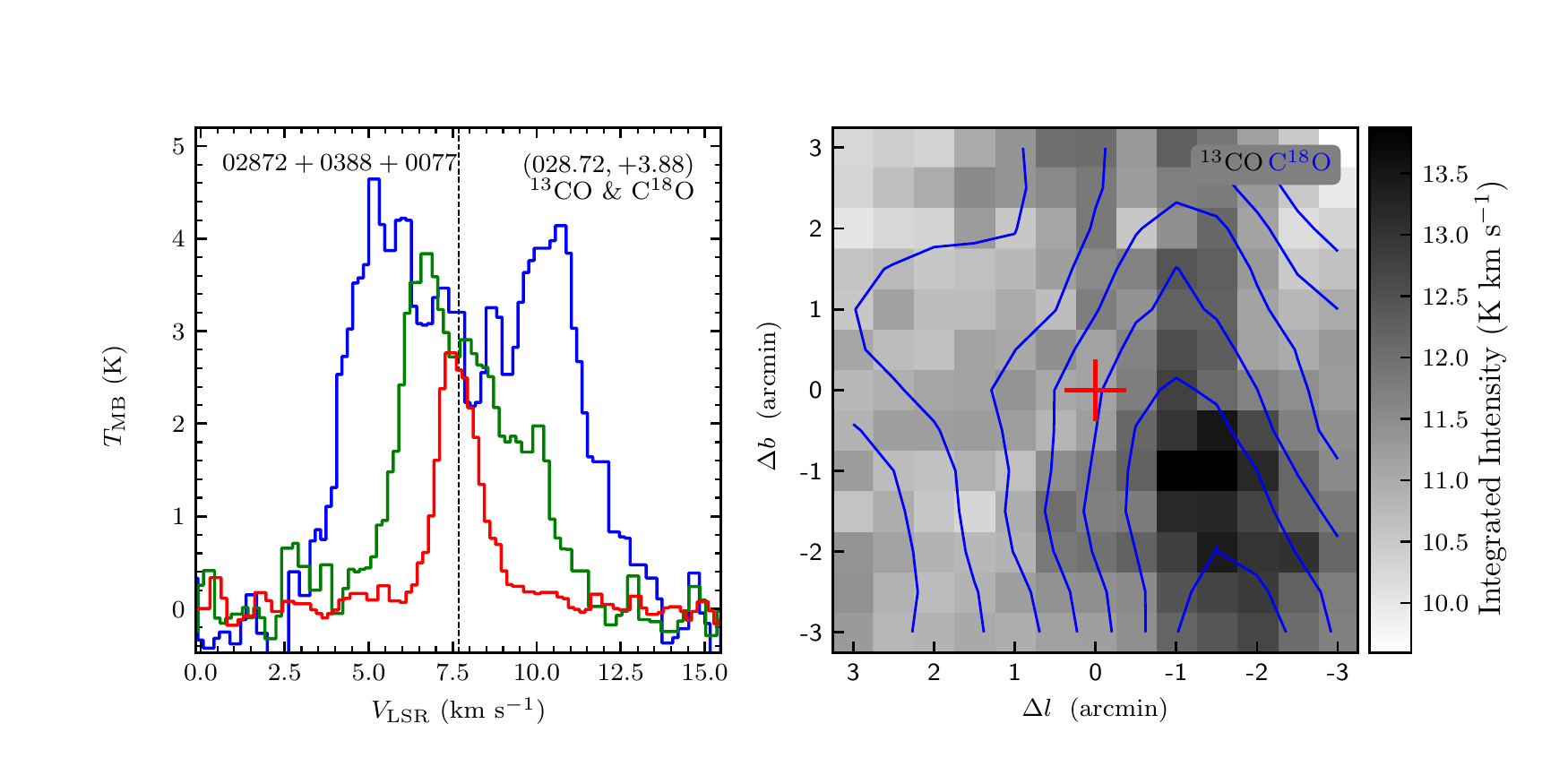}
\includegraphics[width=9.0cm,angle=0]{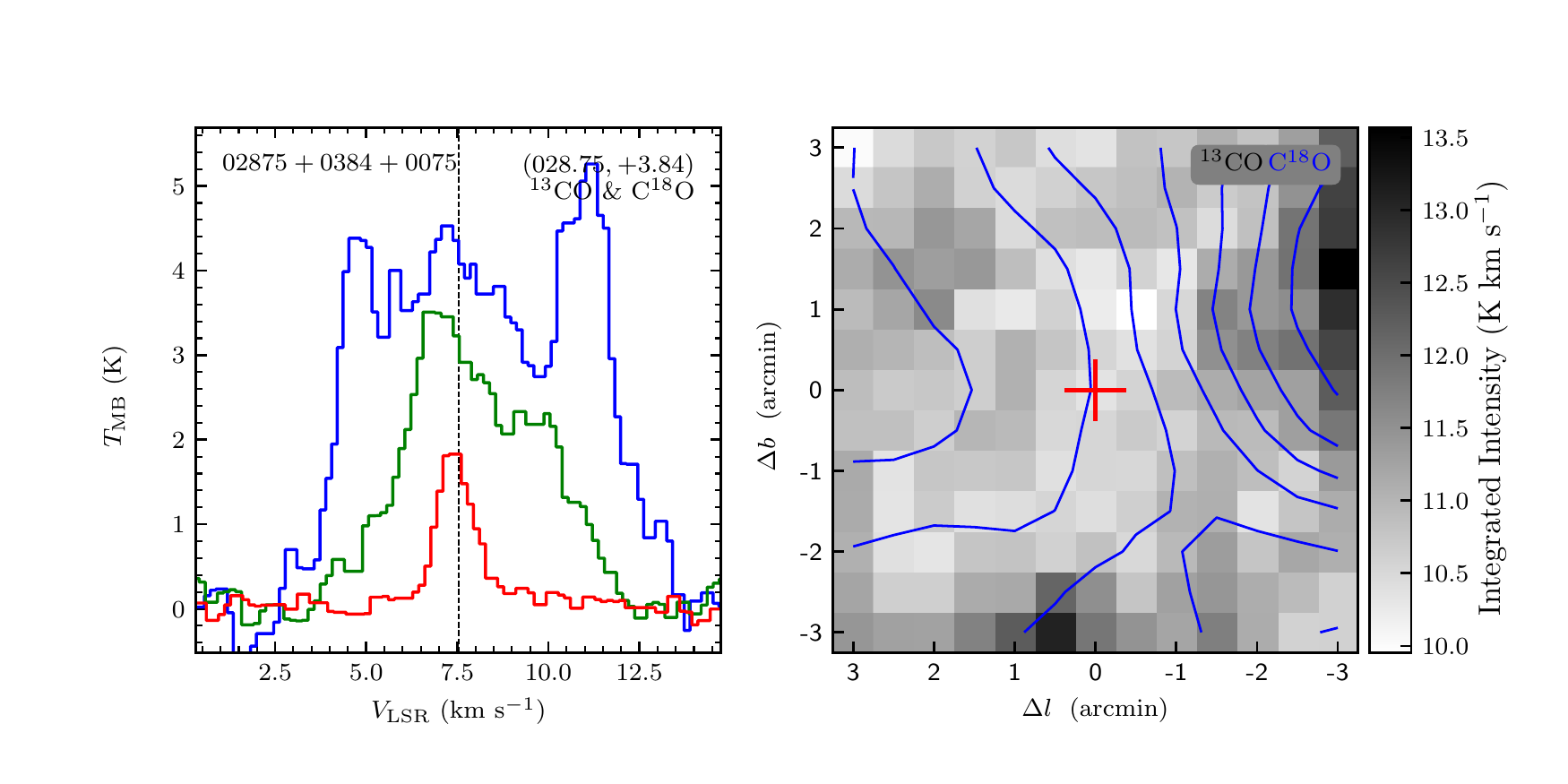}
\vspace{-0.5cm}

\includegraphics[width=9.0cm,angle=0]{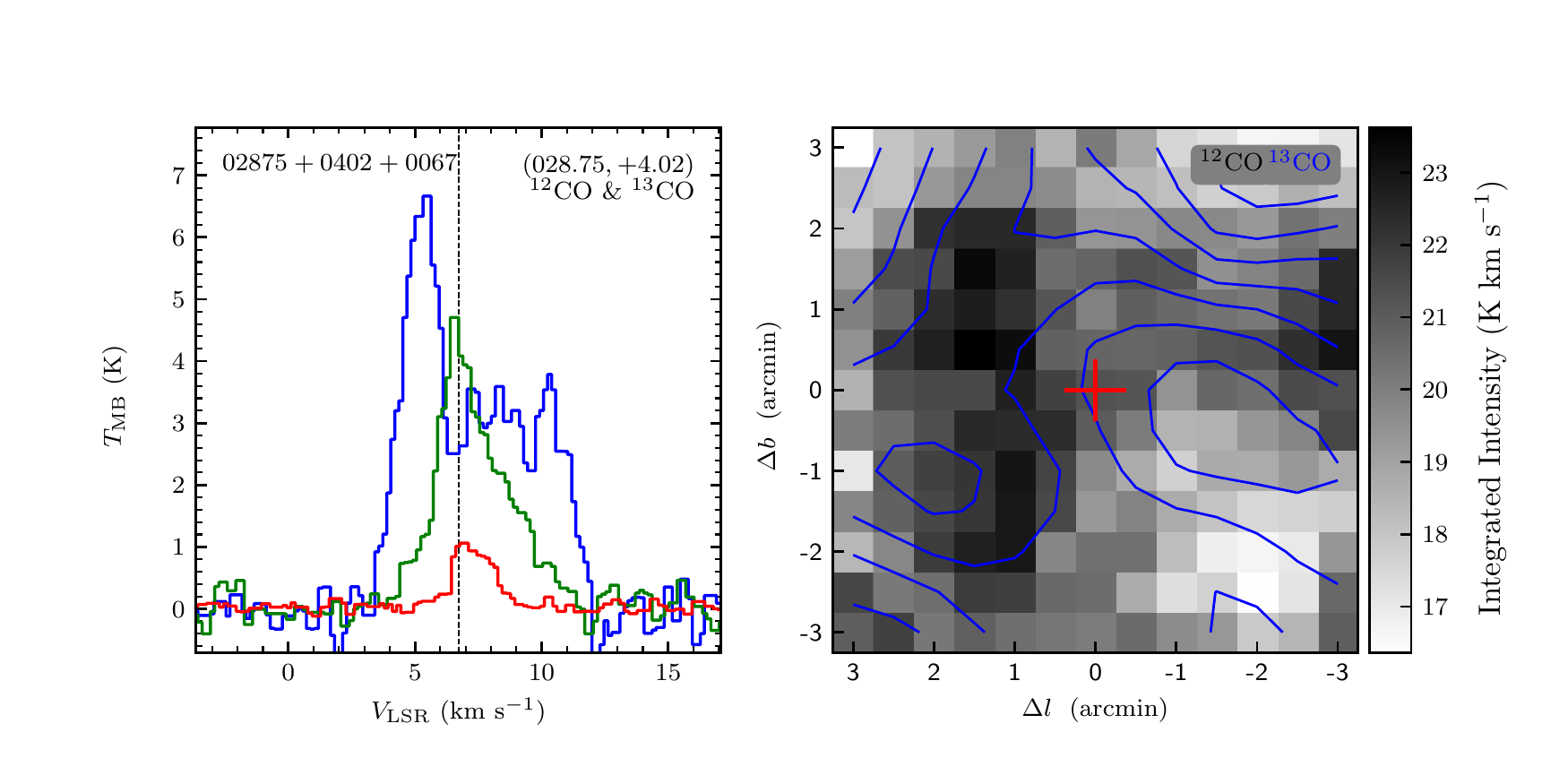}
\includegraphics[width=9.0cm,angle=0]{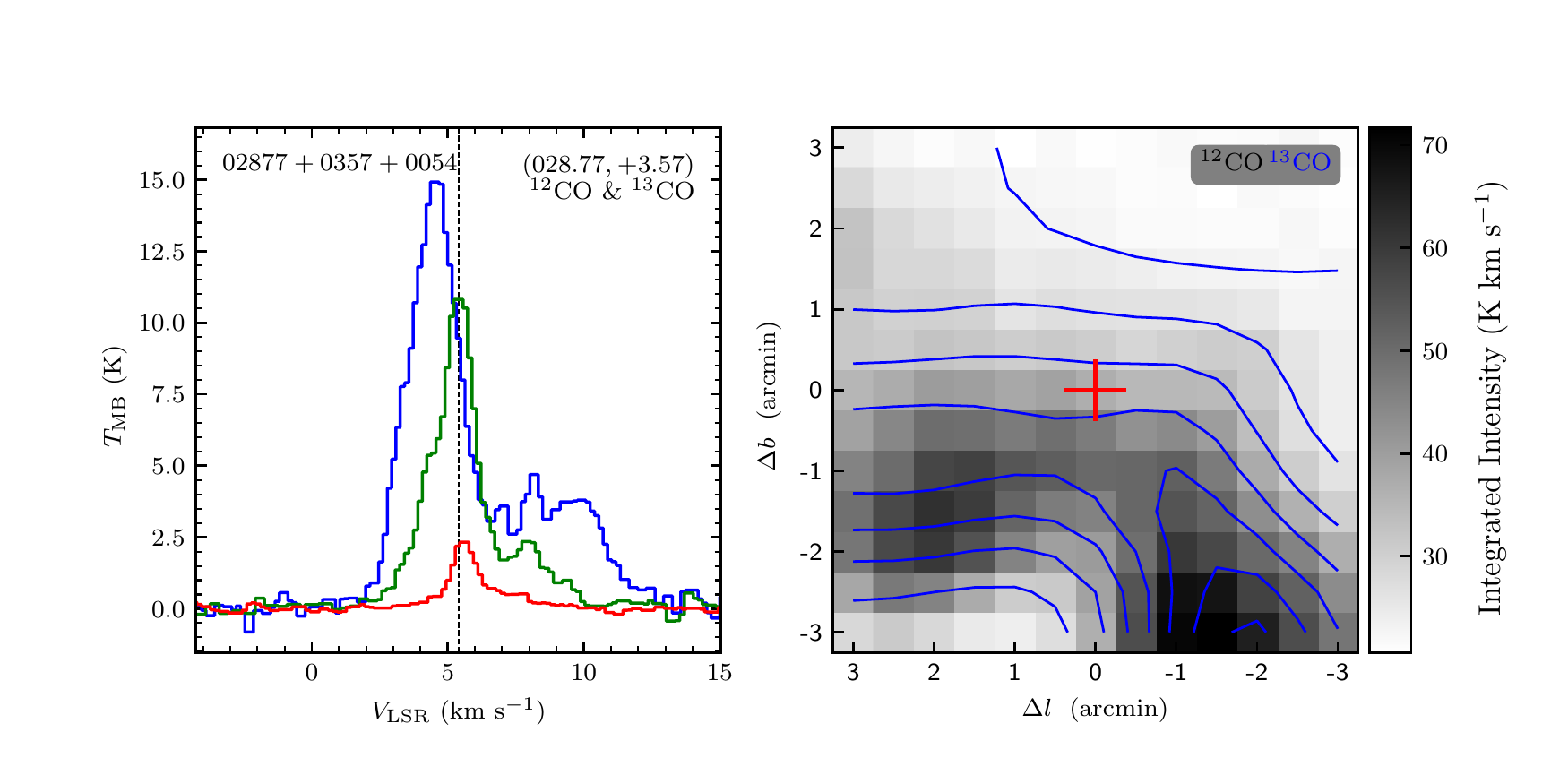}
\vspace{-0.5cm}

\includegraphics[width=9.0cm,angle=0]{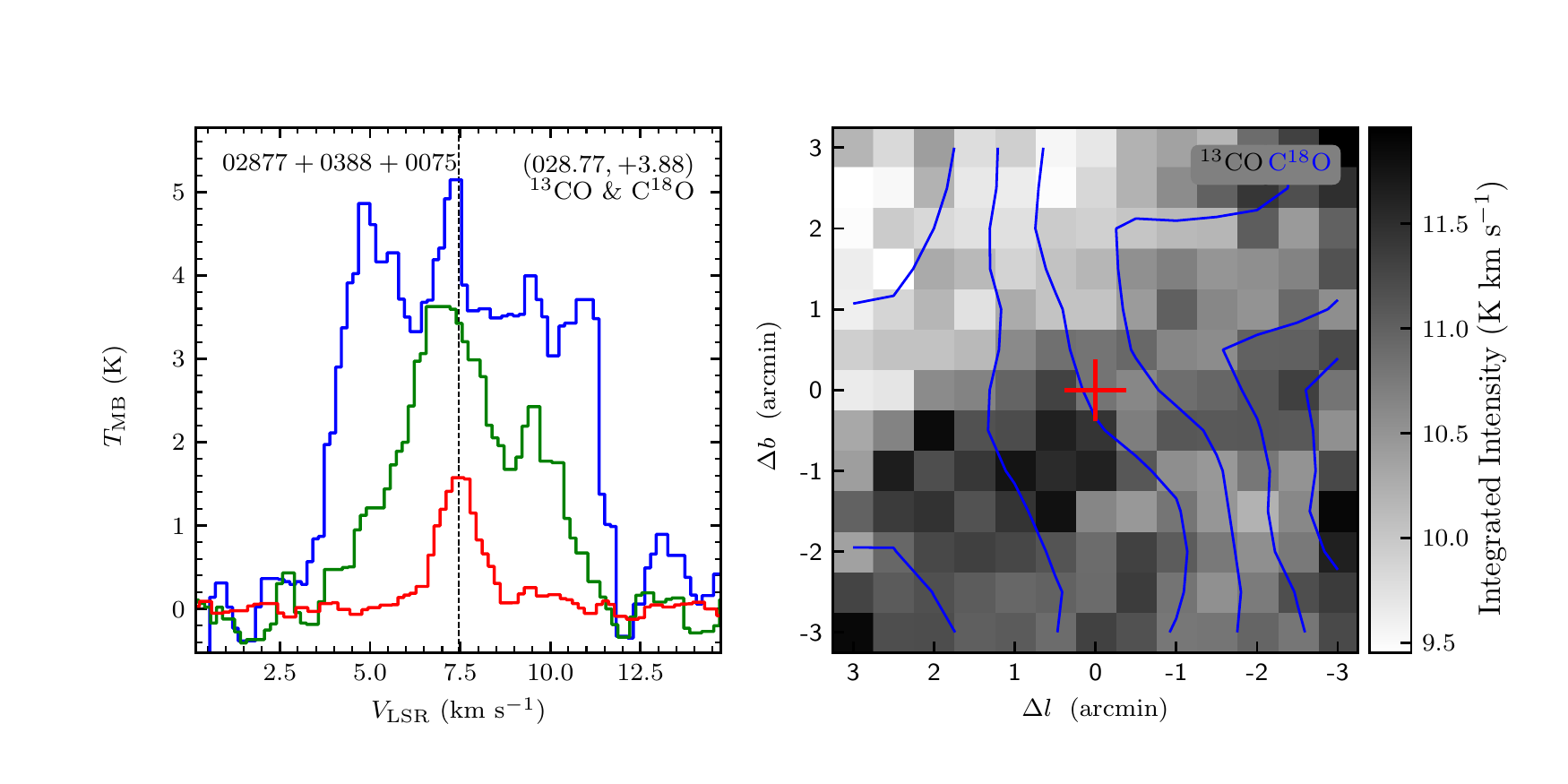}
\includegraphics[width=9.0cm,angle=0]{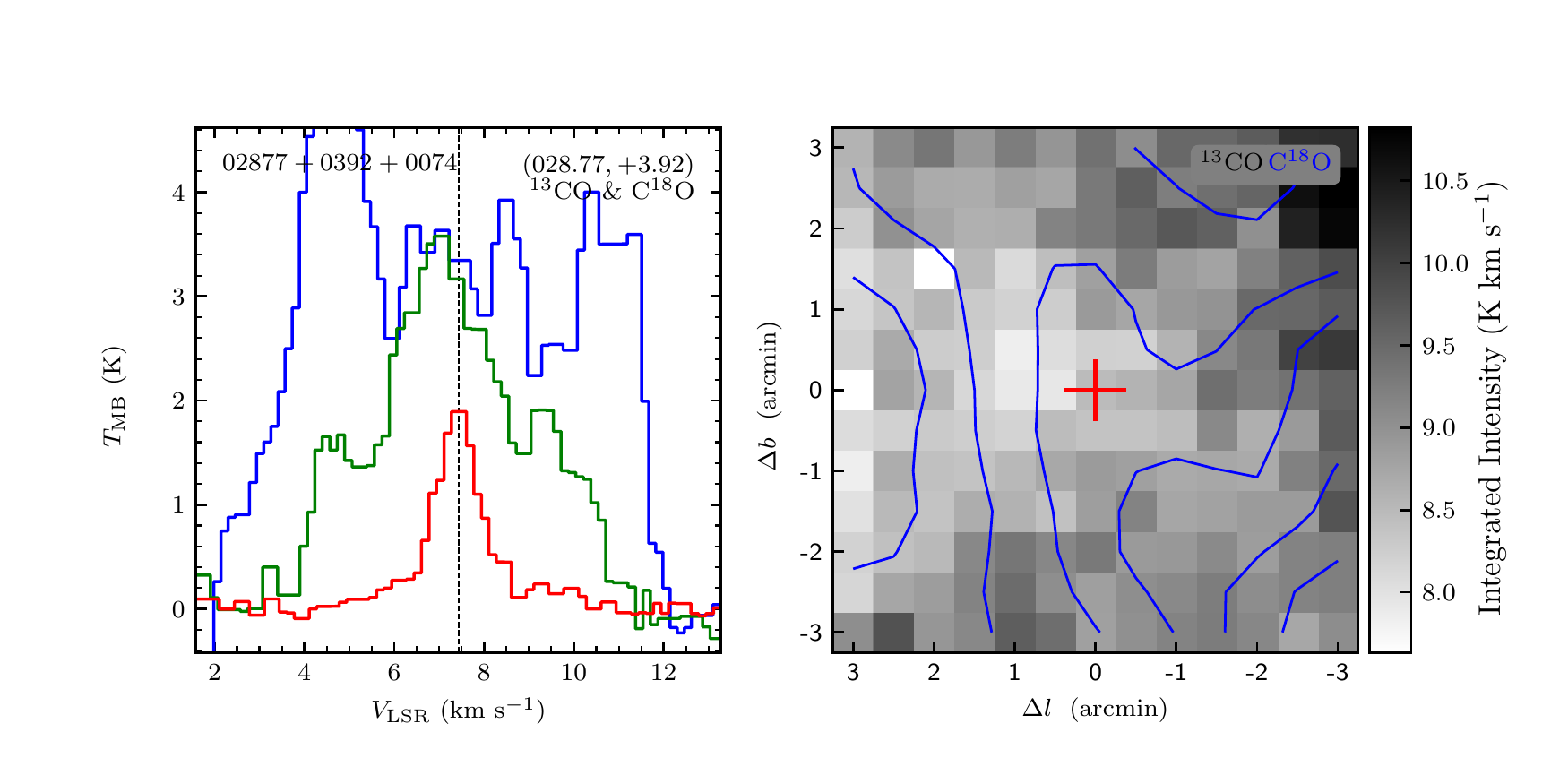}
\vspace{-0.5cm}

\includegraphics[width=9.0cm,angle=0]{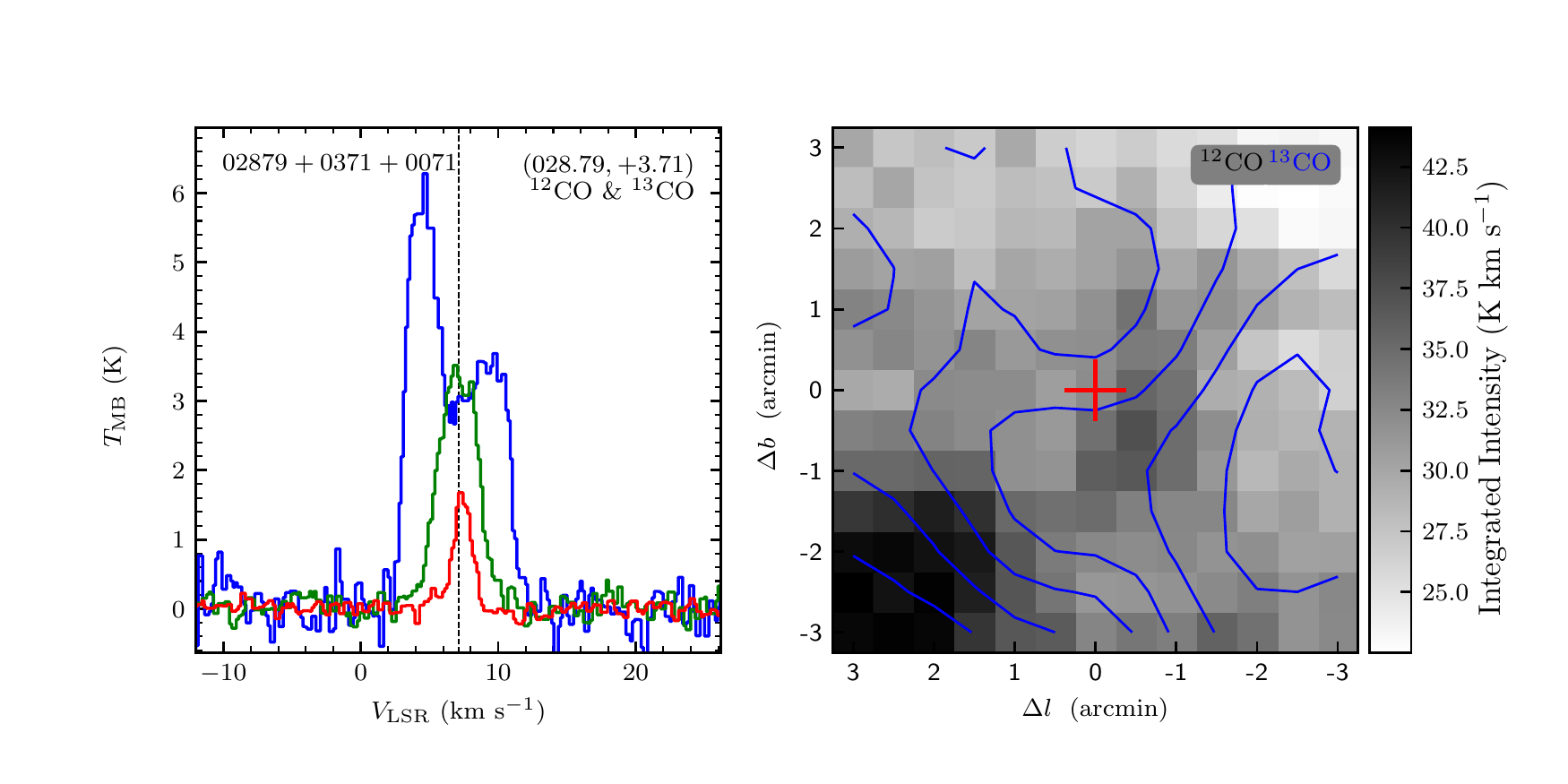}
\includegraphics[width=9.0cm,angle=0]{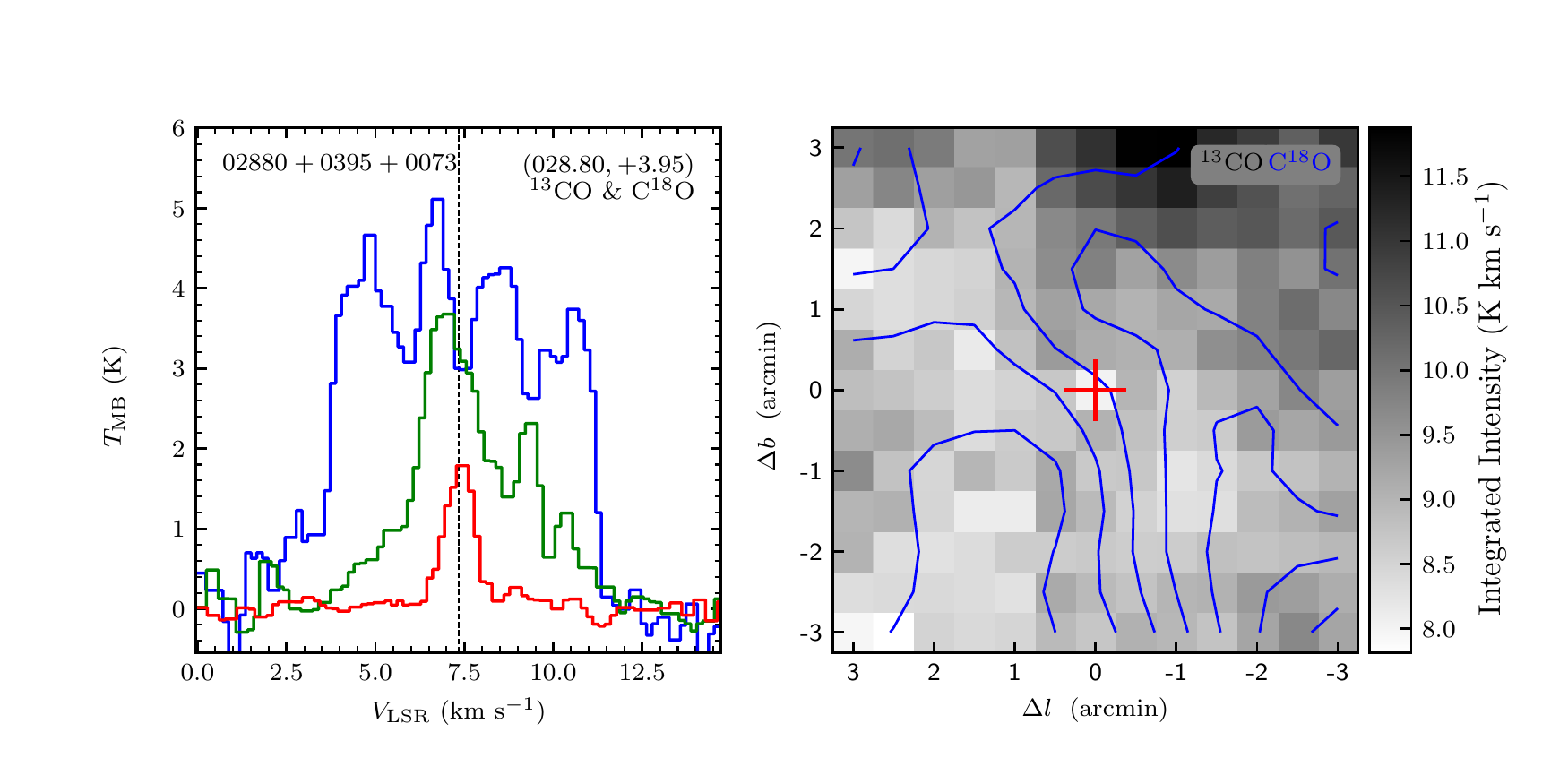}
\vspace{-0.5cm}

\includegraphics[width=9.0cm,angle=0]{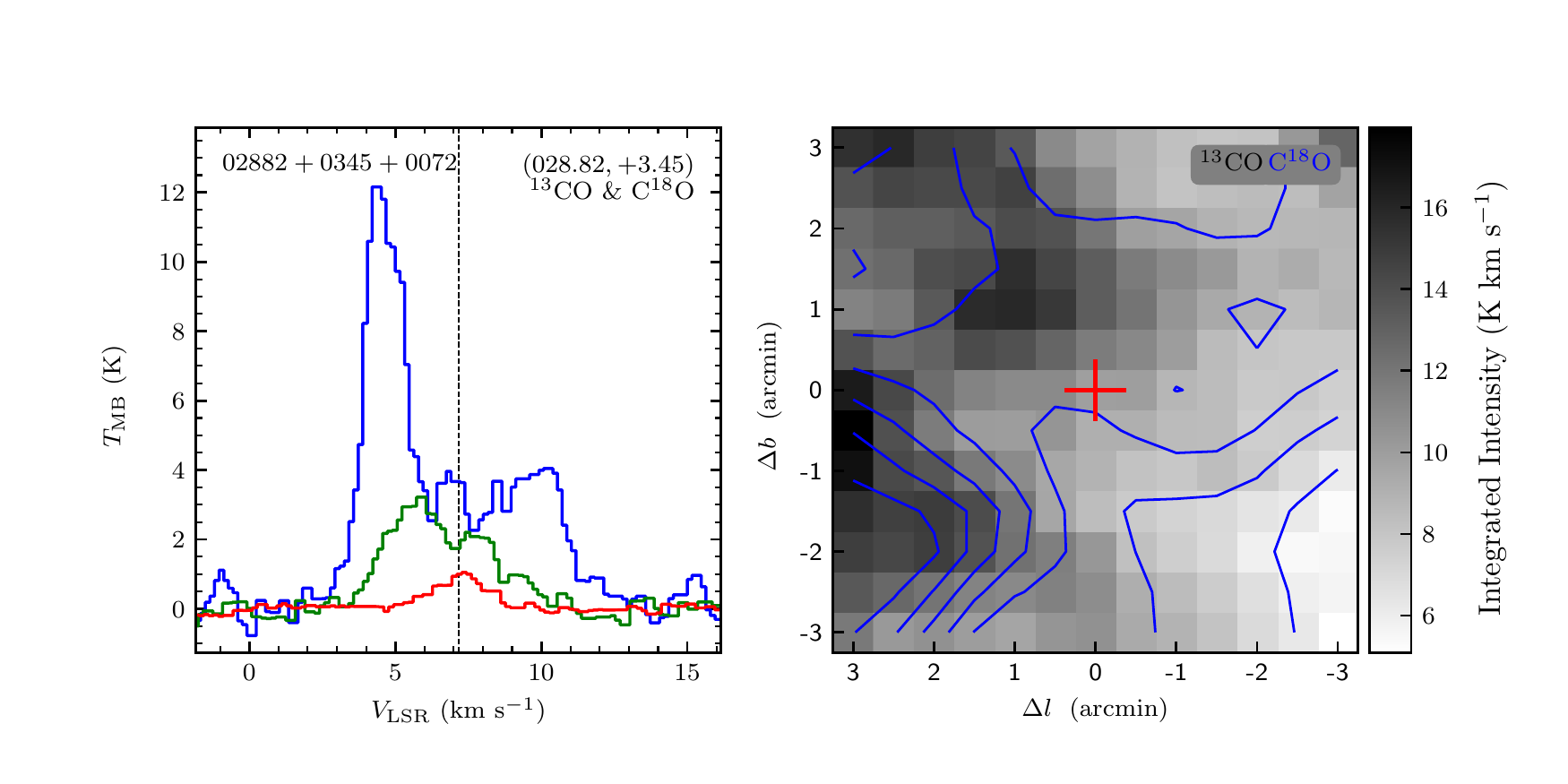}
\includegraphics[width=9.0cm,angle=0]{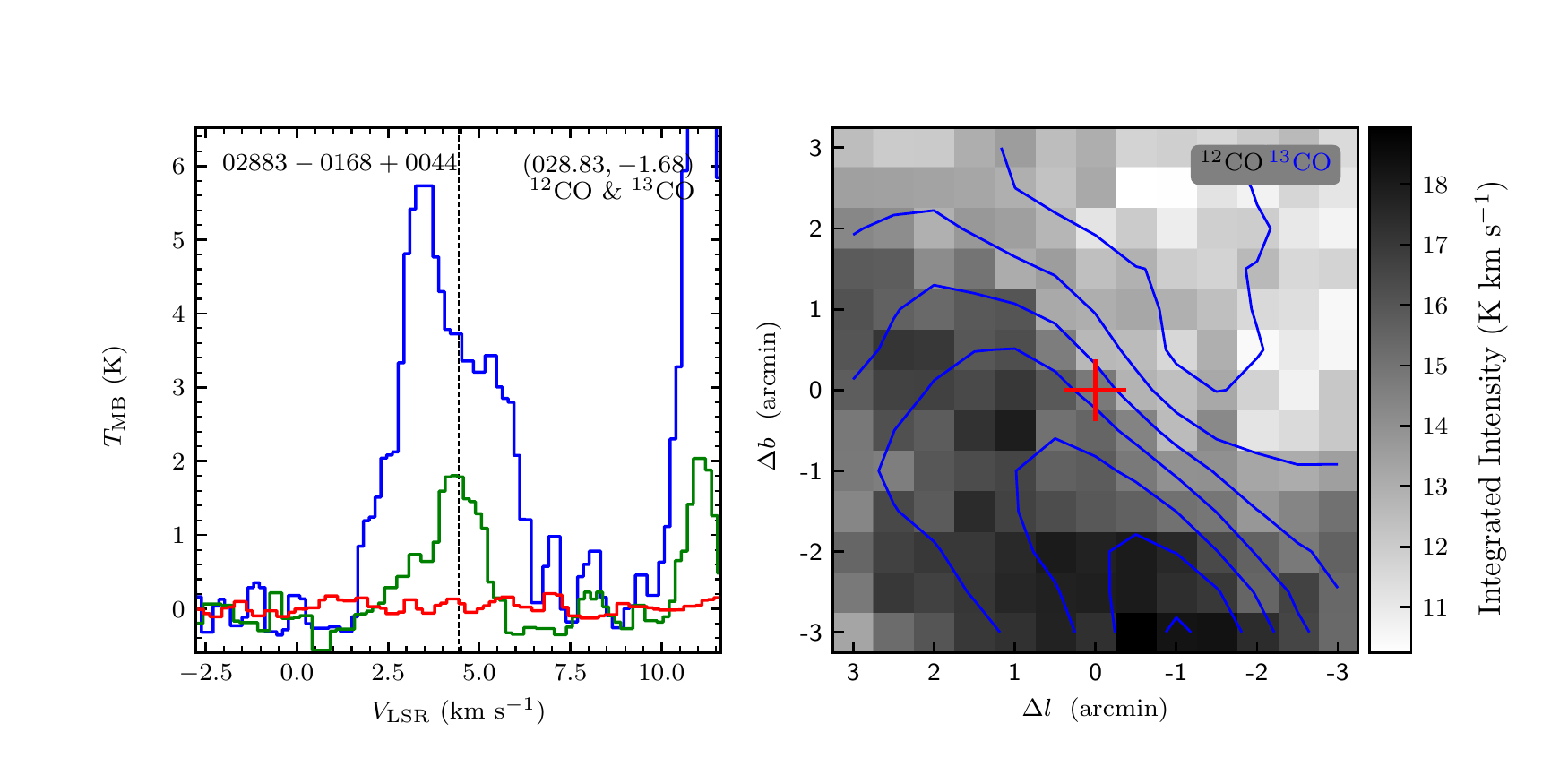}
\end{figure}
\clearpage

\begin{figure}
\includegraphics[width=9.0cm,angle=0]{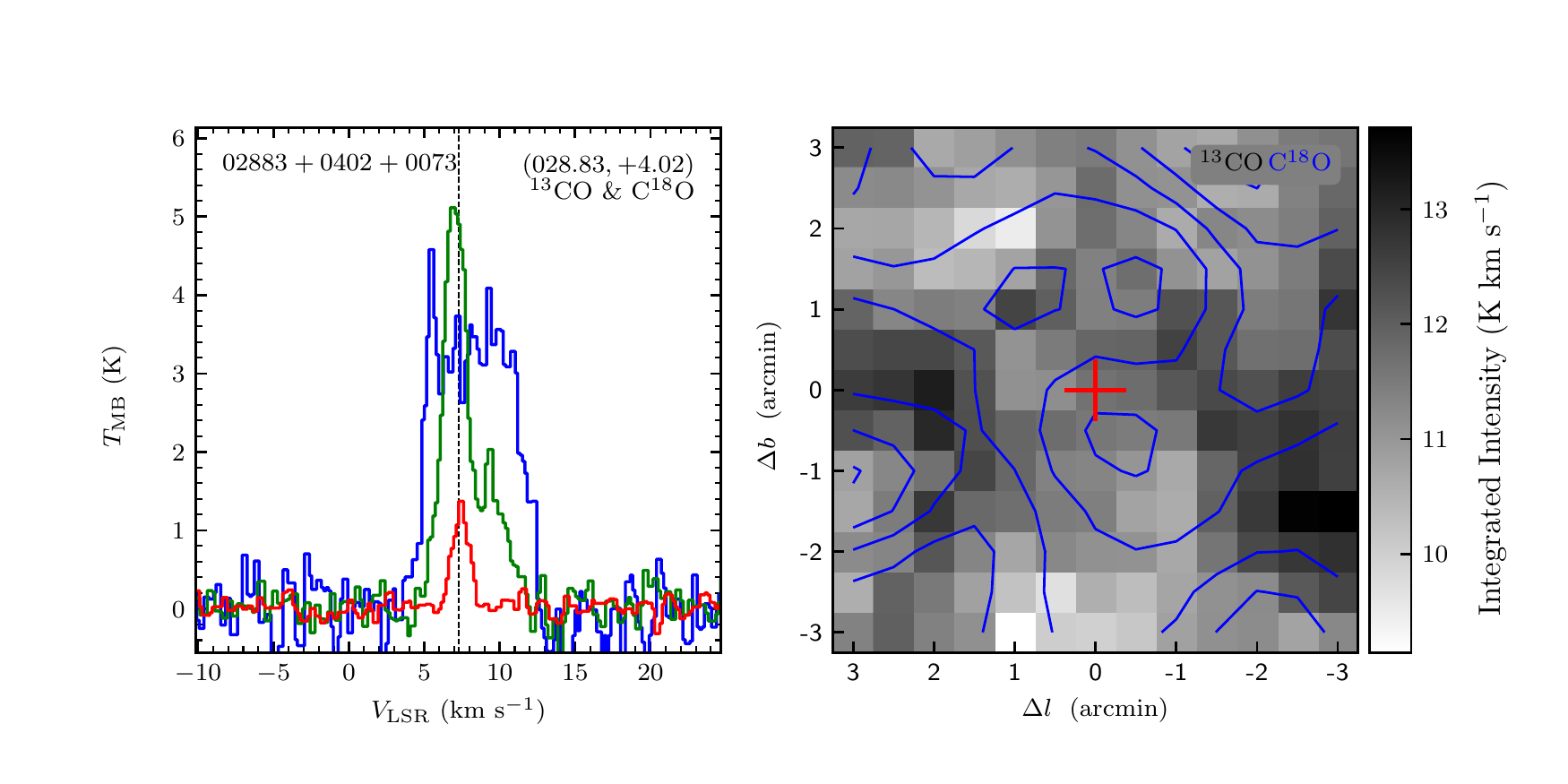}
\includegraphics[width=9.0cm,angle=0]{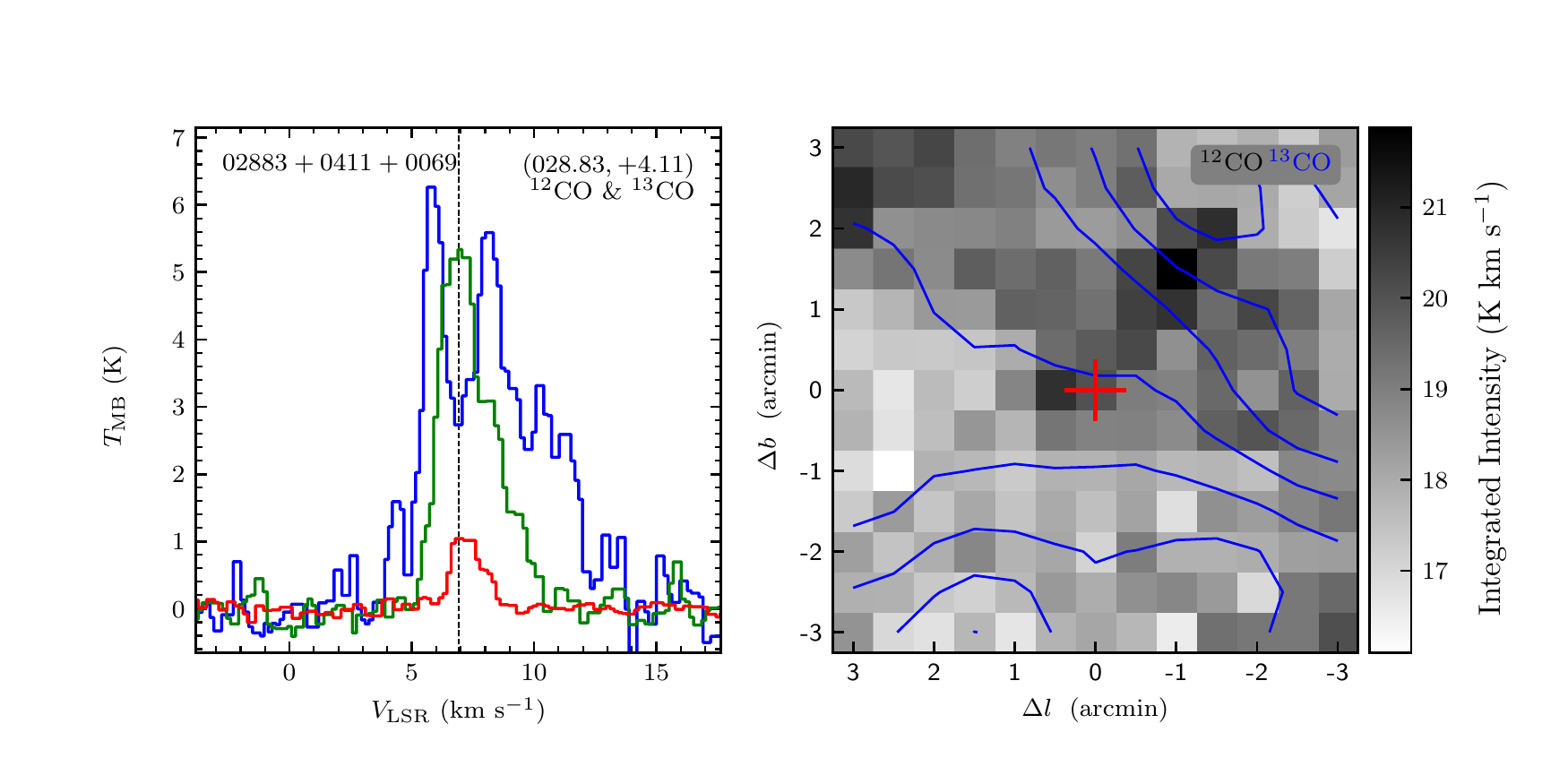}
\vspace{-0.5cm}

\includegraphics[width=9.0cm,angle=0]{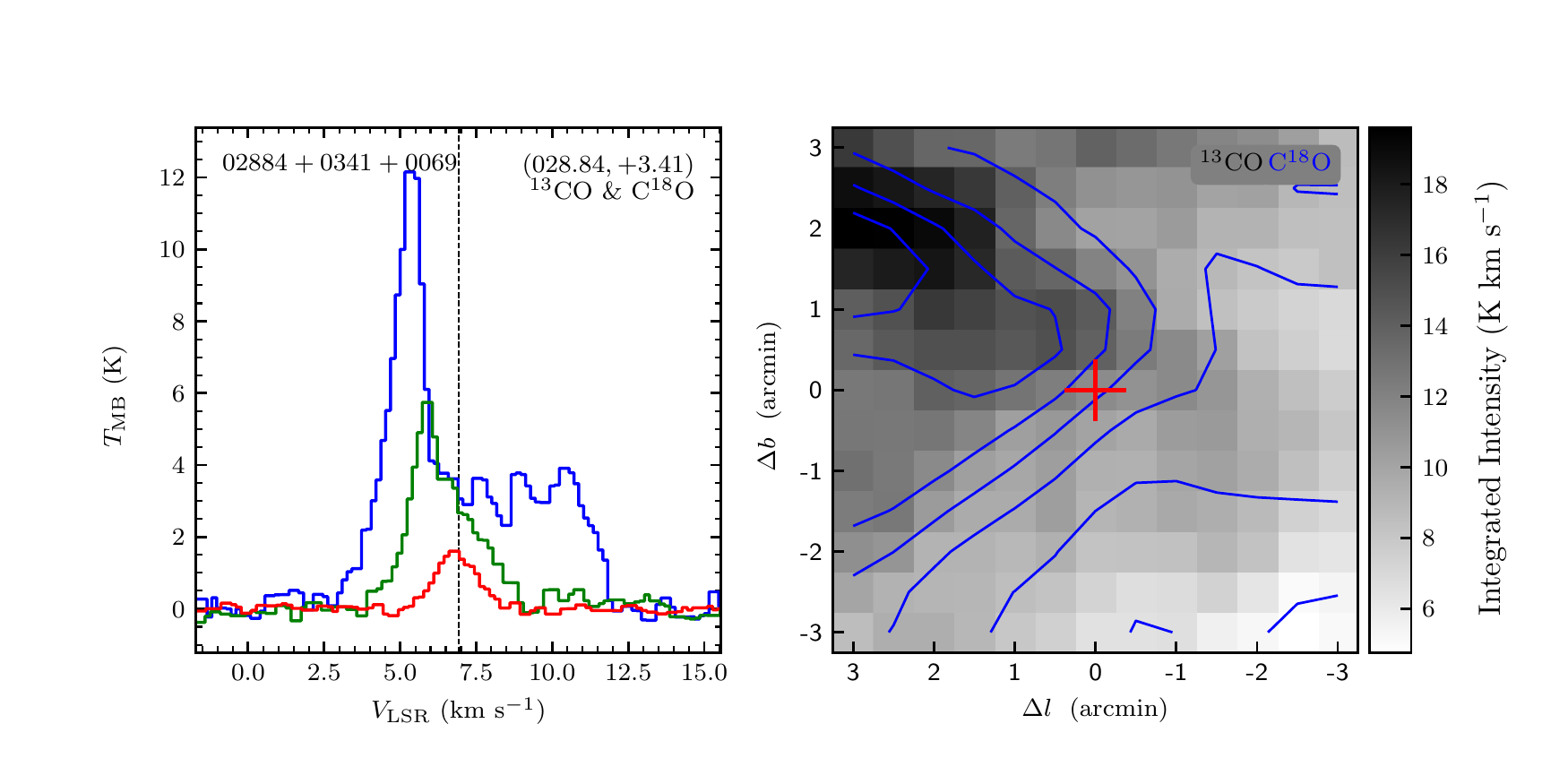}
\includegraphics[width=9.0cm,angle=0]{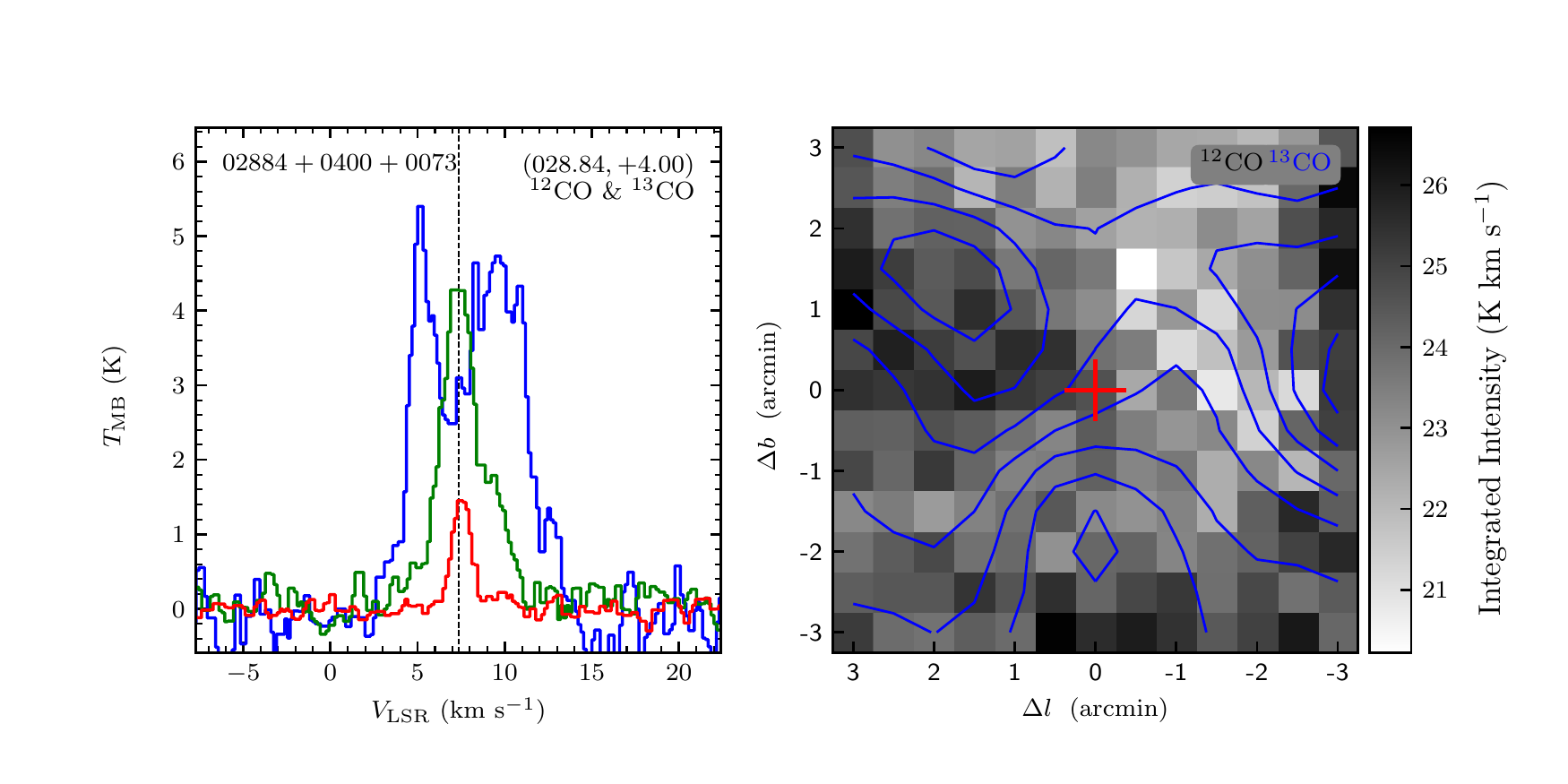}
\vspace{-0.5cm}

\includegraphics[width=9.0cm,angle=0]{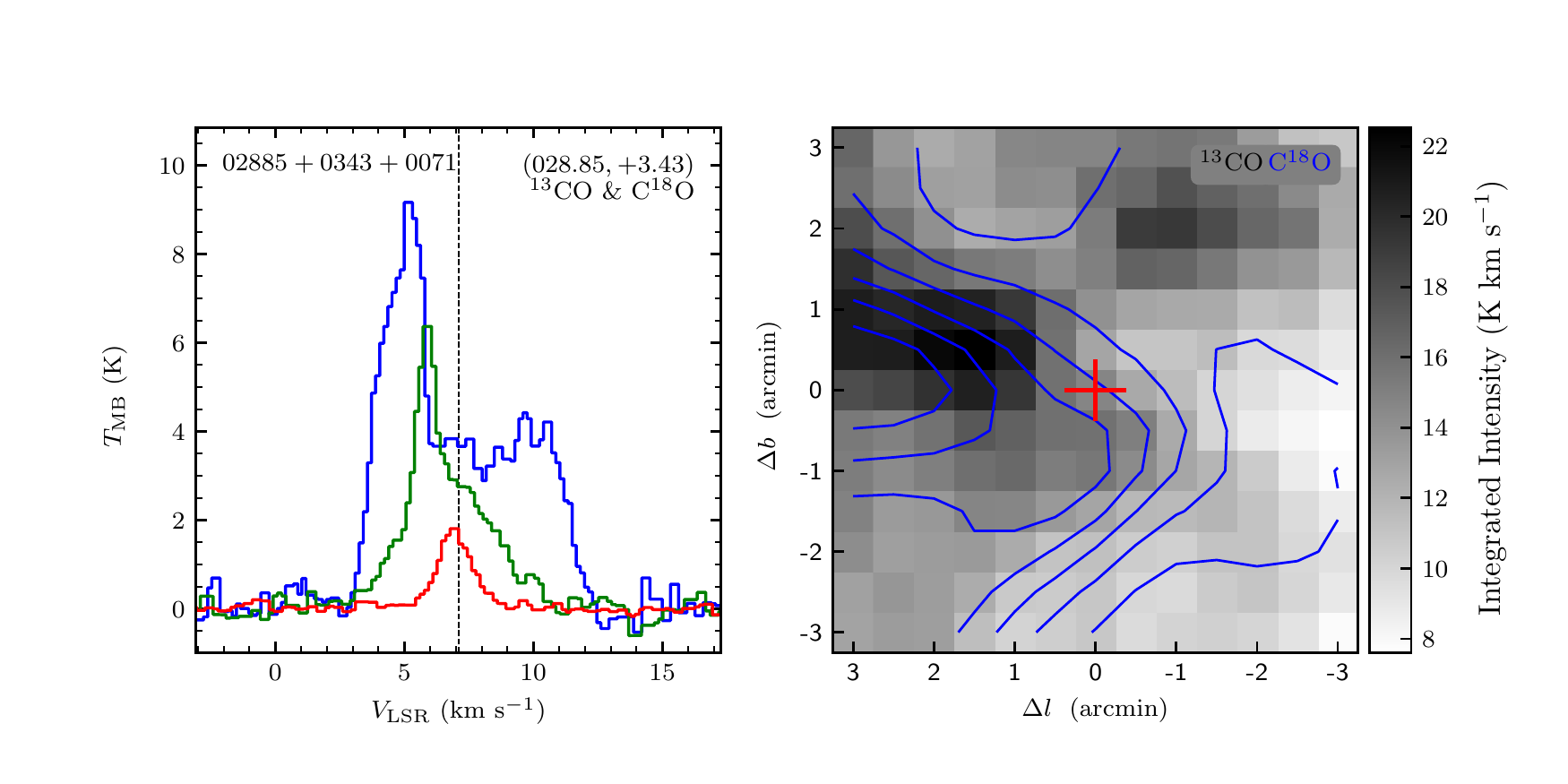}
\includegraphics[width=9.0cm,angle=0]{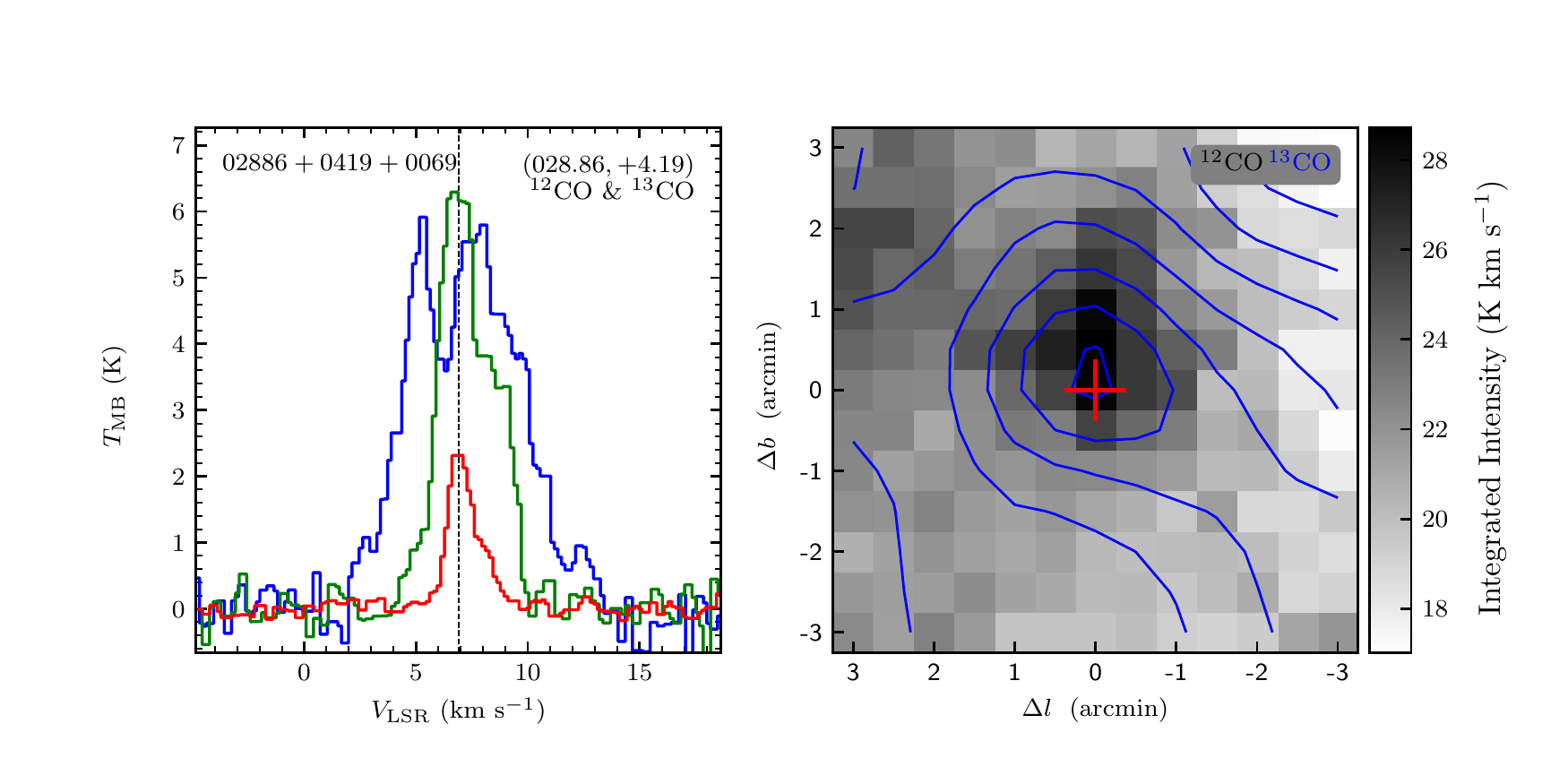}
\vspace{-0.5cm}

\includegraphics[width=9.0cm,angle=0]{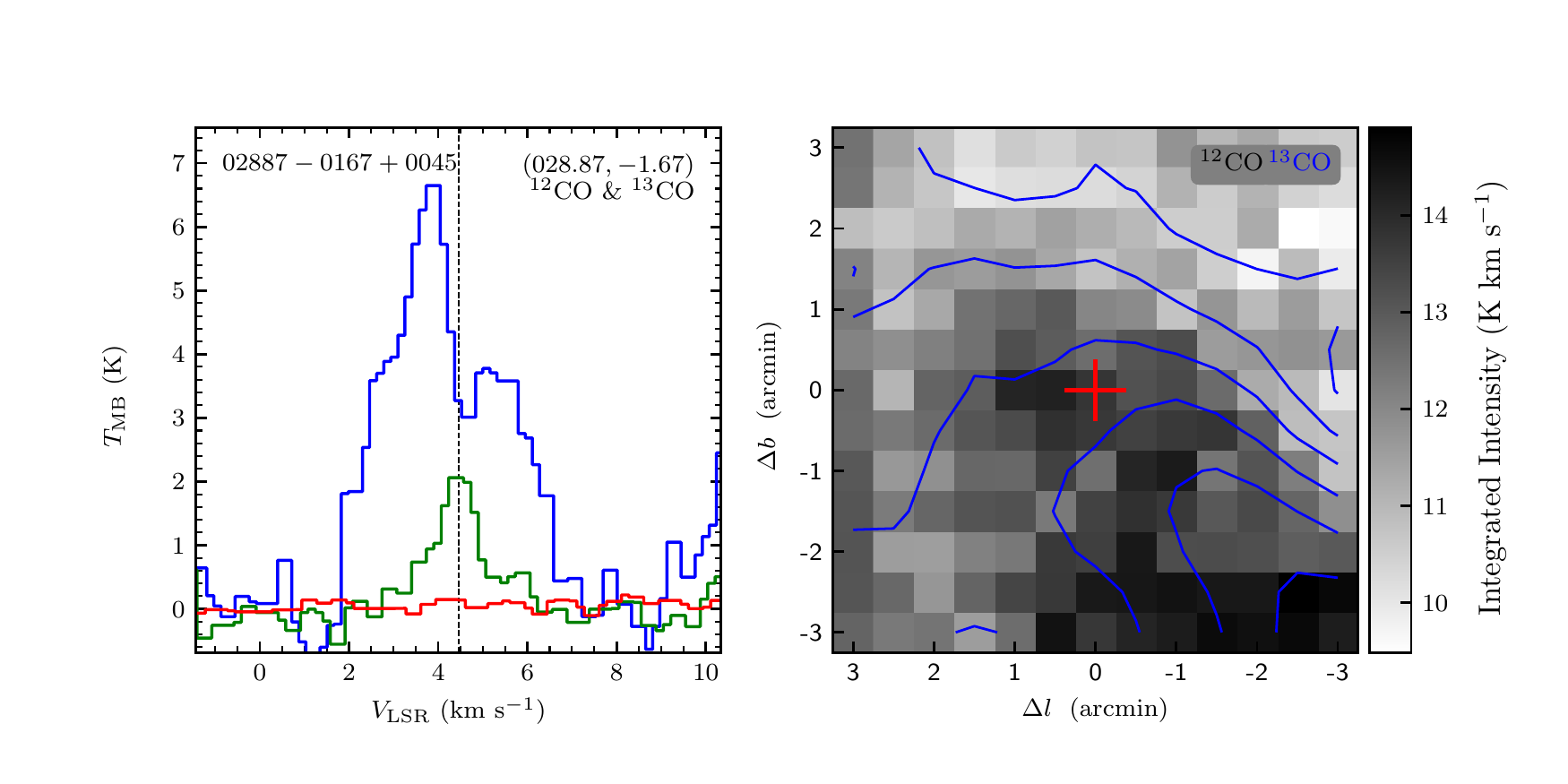}
\includegraphics[width=9.0cm,angle=0]{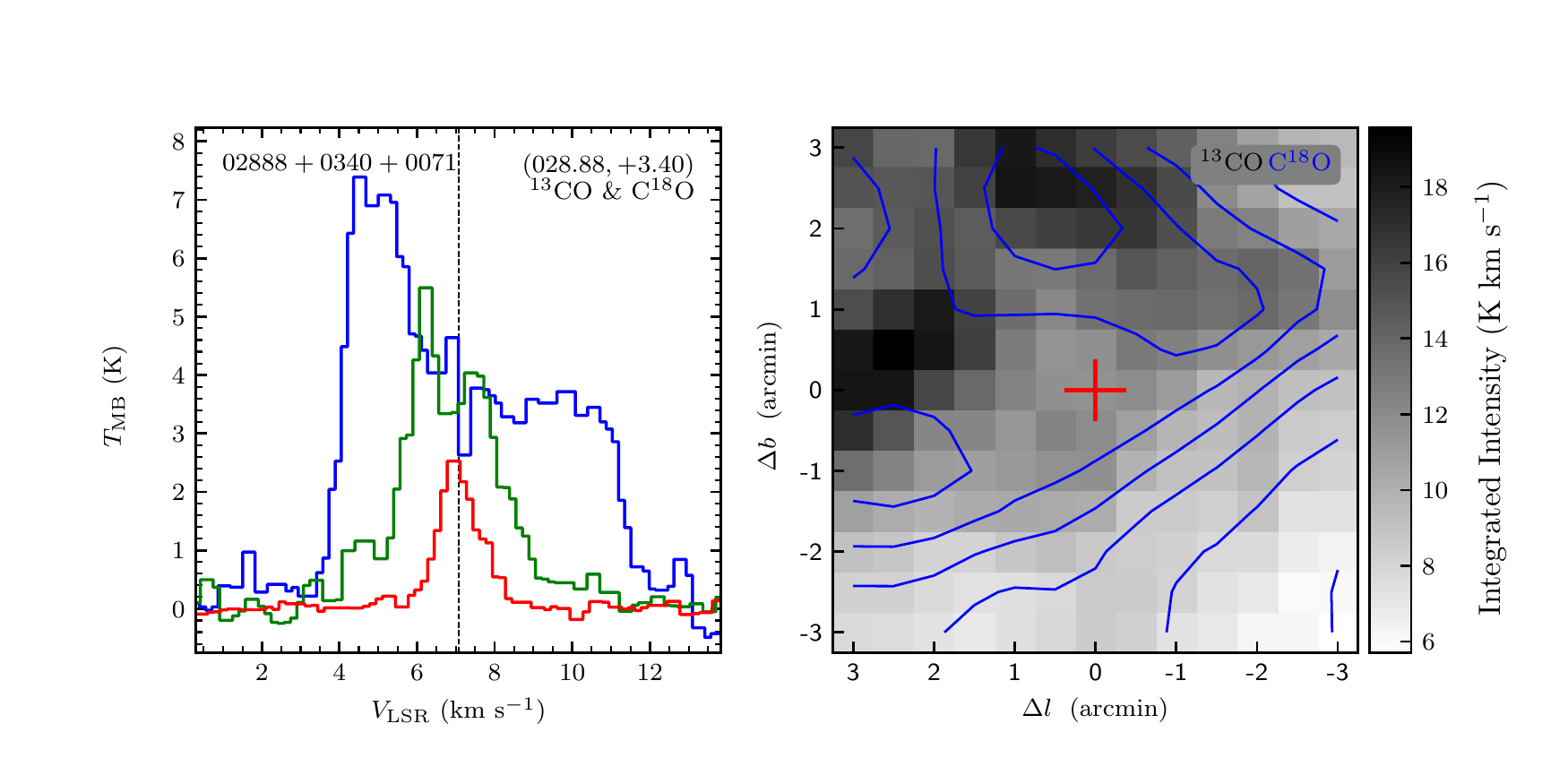}
\vspace{-0.5cm}

\includegraphics[width=9.0cm,angle=0]{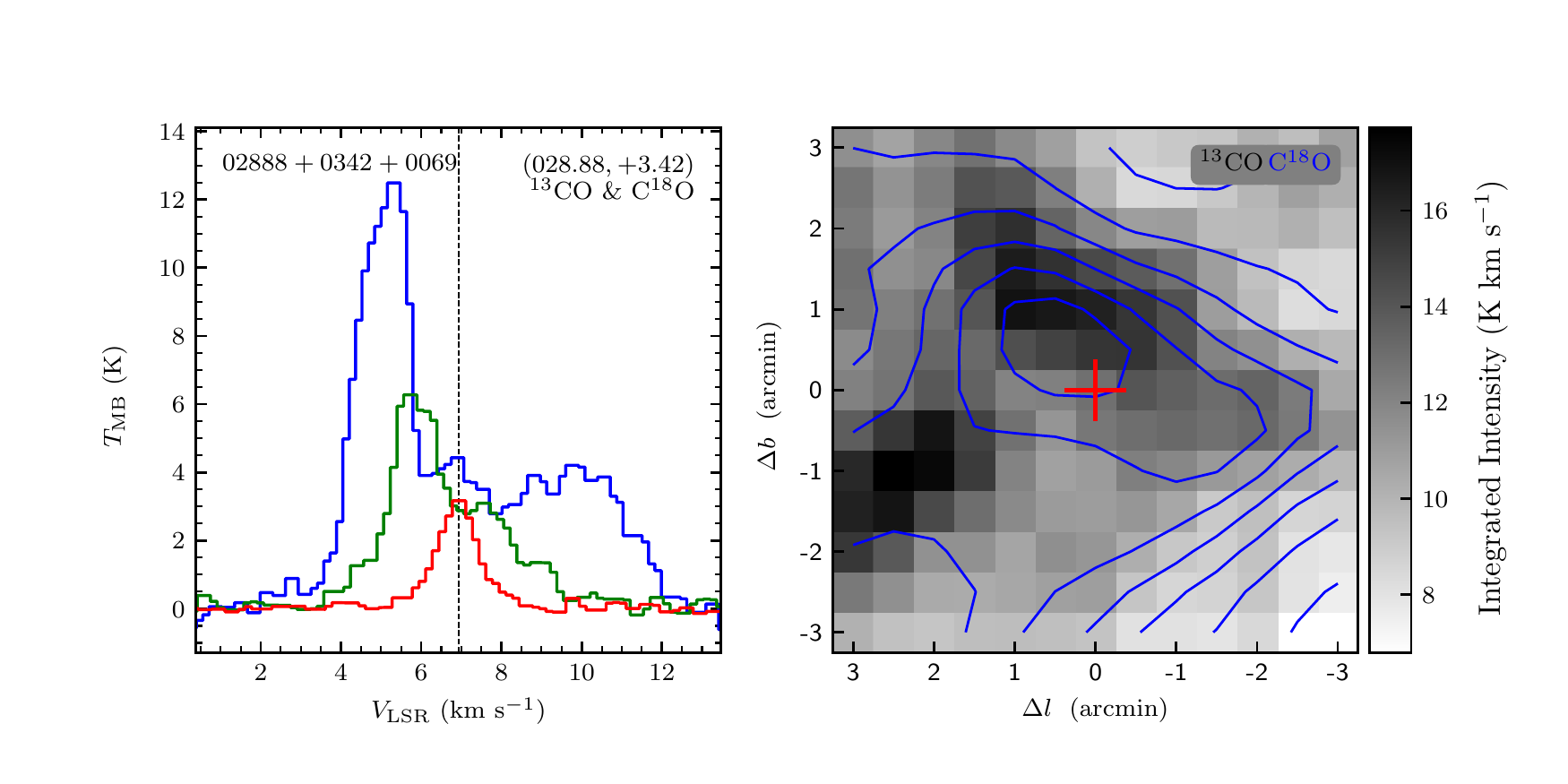}
\includegraphics[width=9.0cm,angle=0]{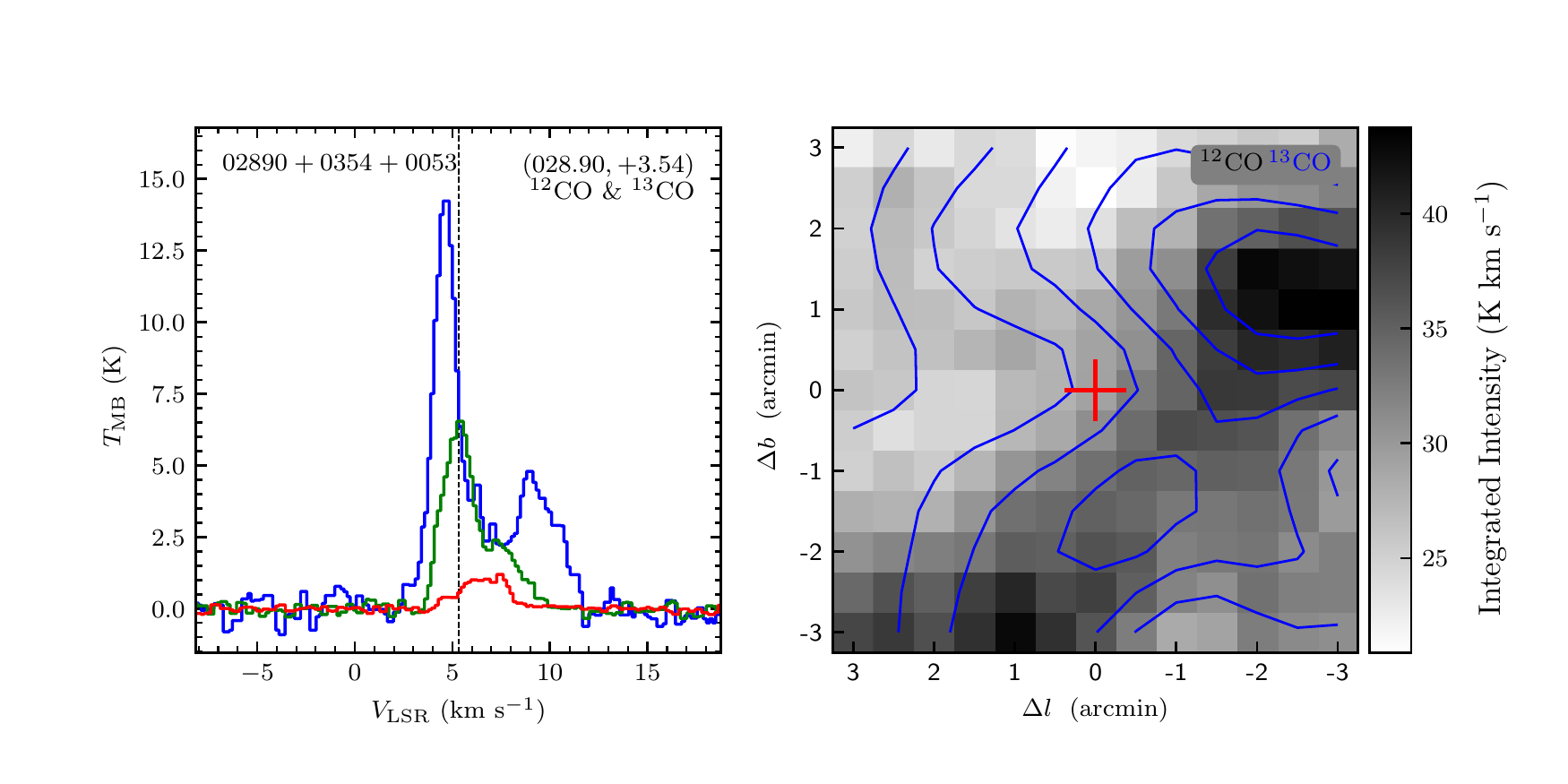}
\end{figure}
\clearpage

\begin{figure}
\includegraphics[width=9.0cm,angle=0]{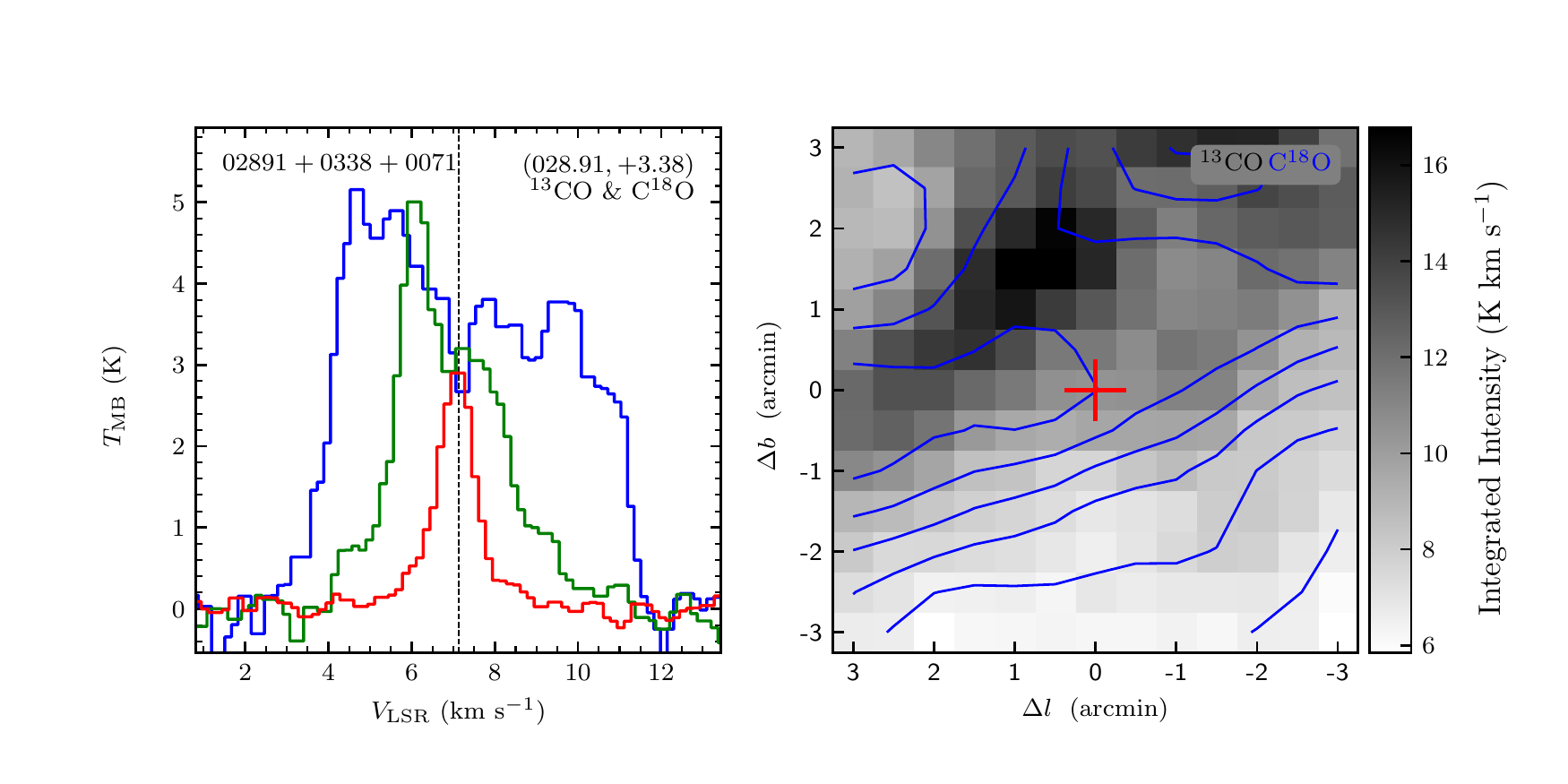}
\includegraphics[width=9.0cm,angle=0]{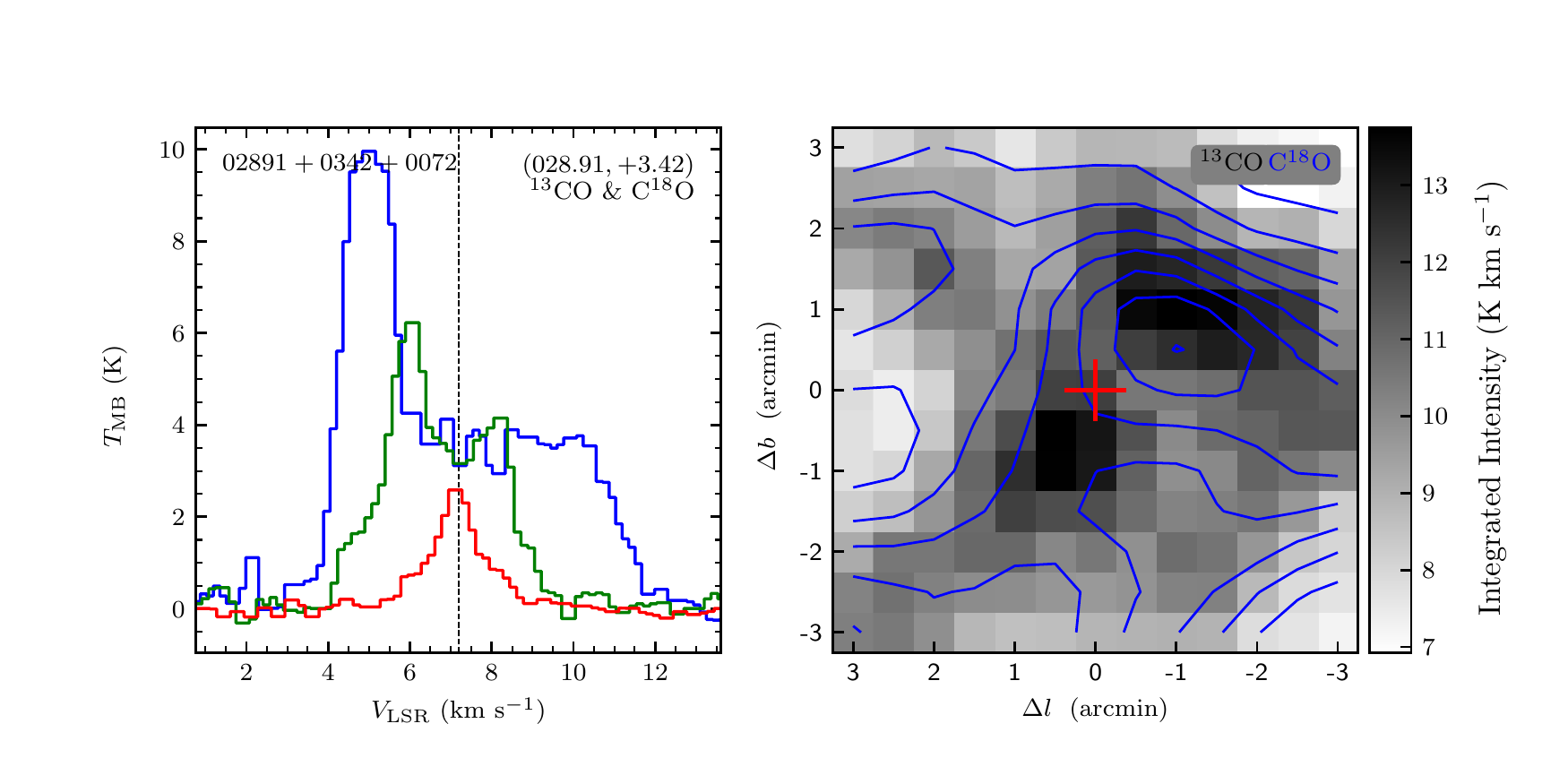}
\vspace{-0.5cm}

\includegraphics[width=9.0cm,angle=0]{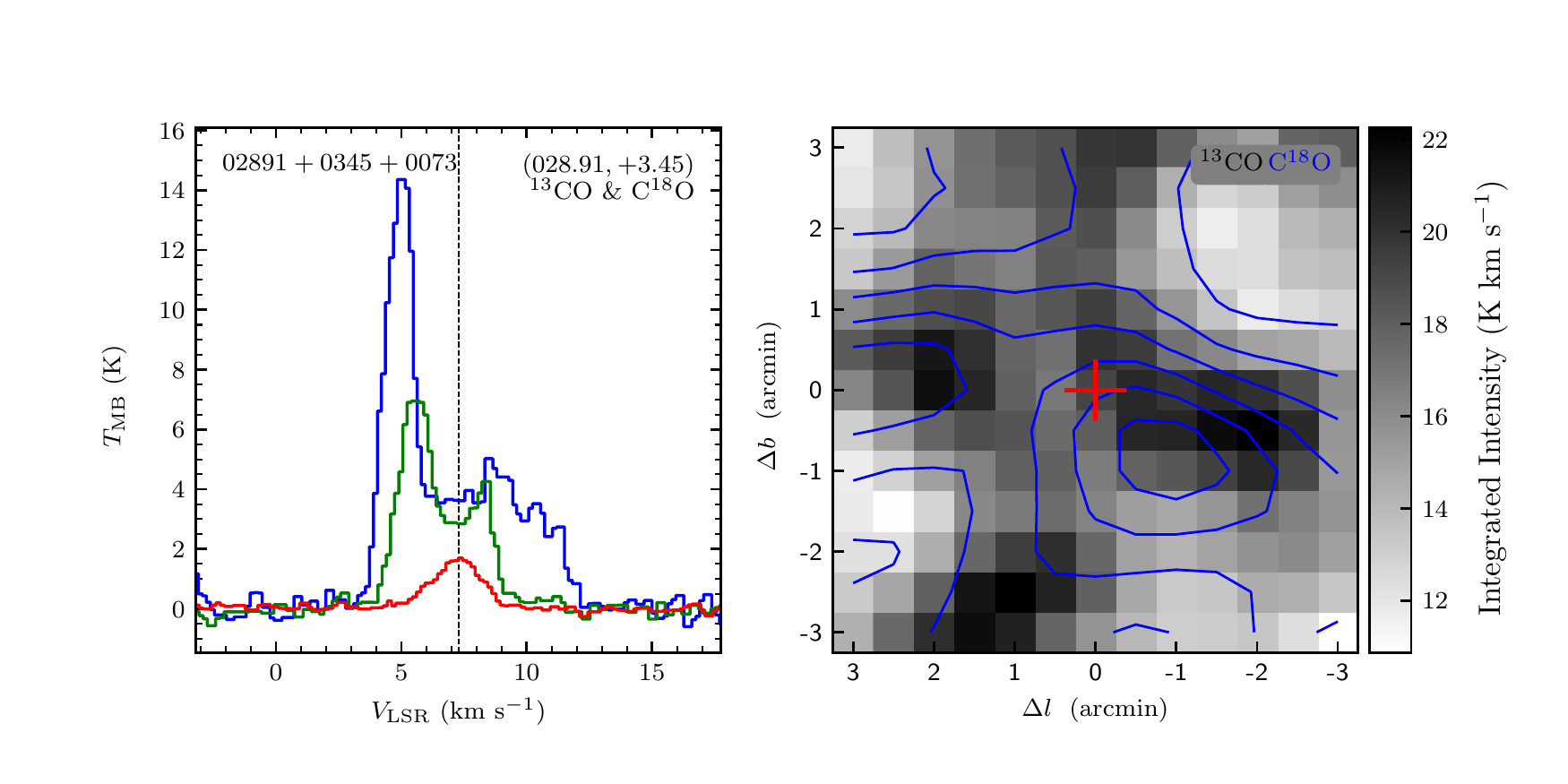}
\includegraphics[width=9.0cm,angle=0]{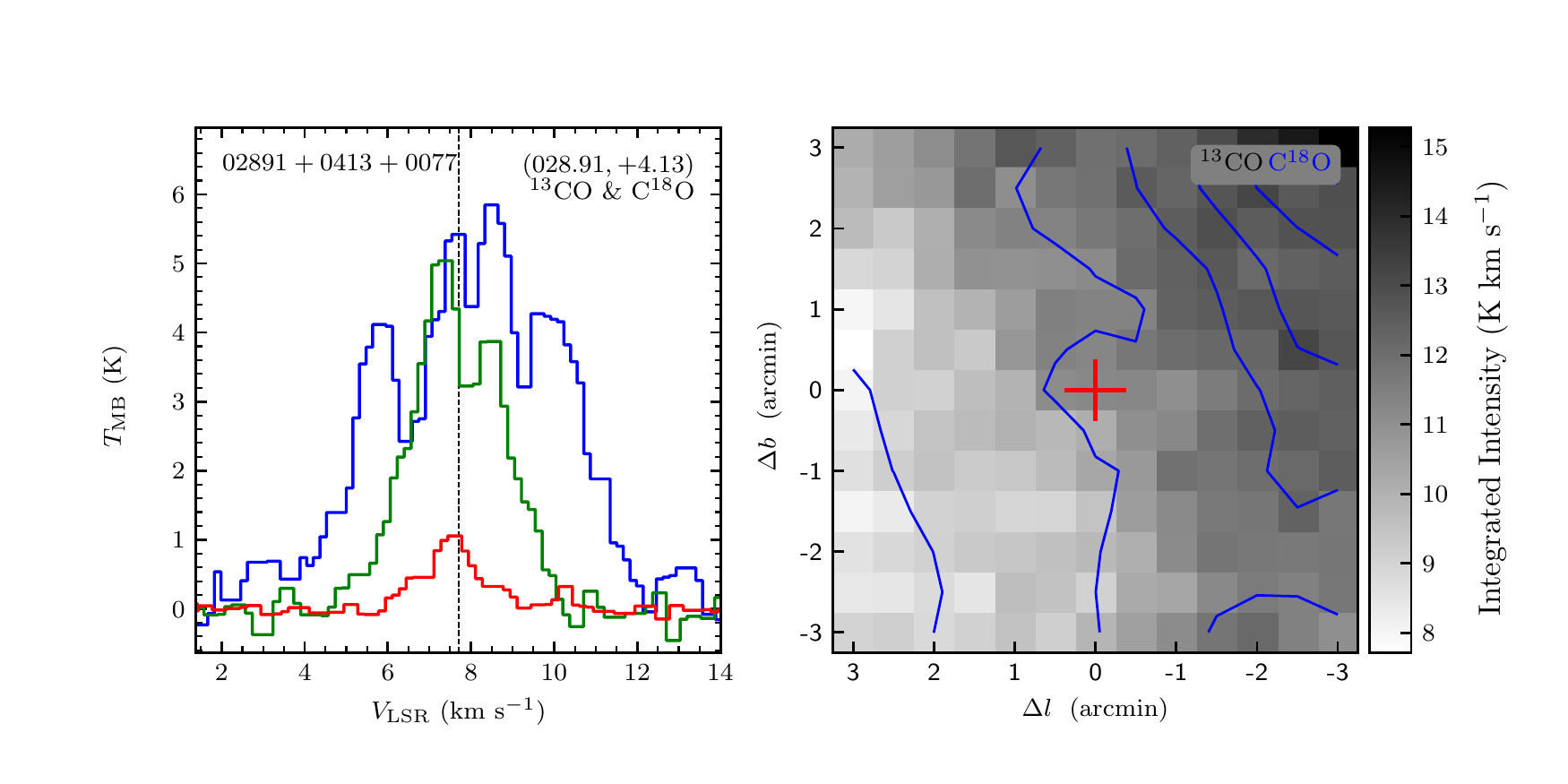}
\vspace{-0.5cm}

\includegraphics[width=9.0cm,angle=0]{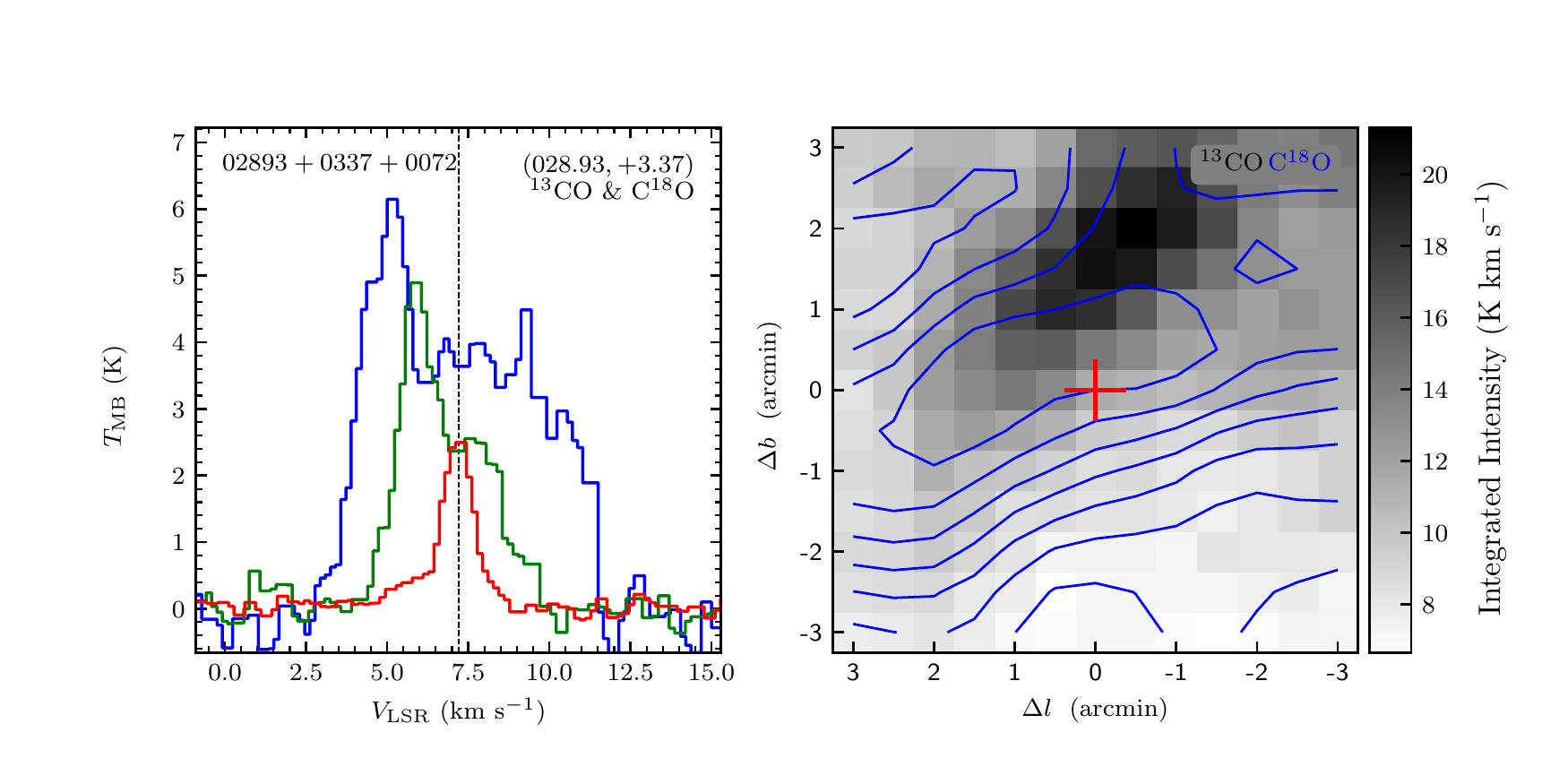}
\includegraphics[width=9.0cm,angle=0]{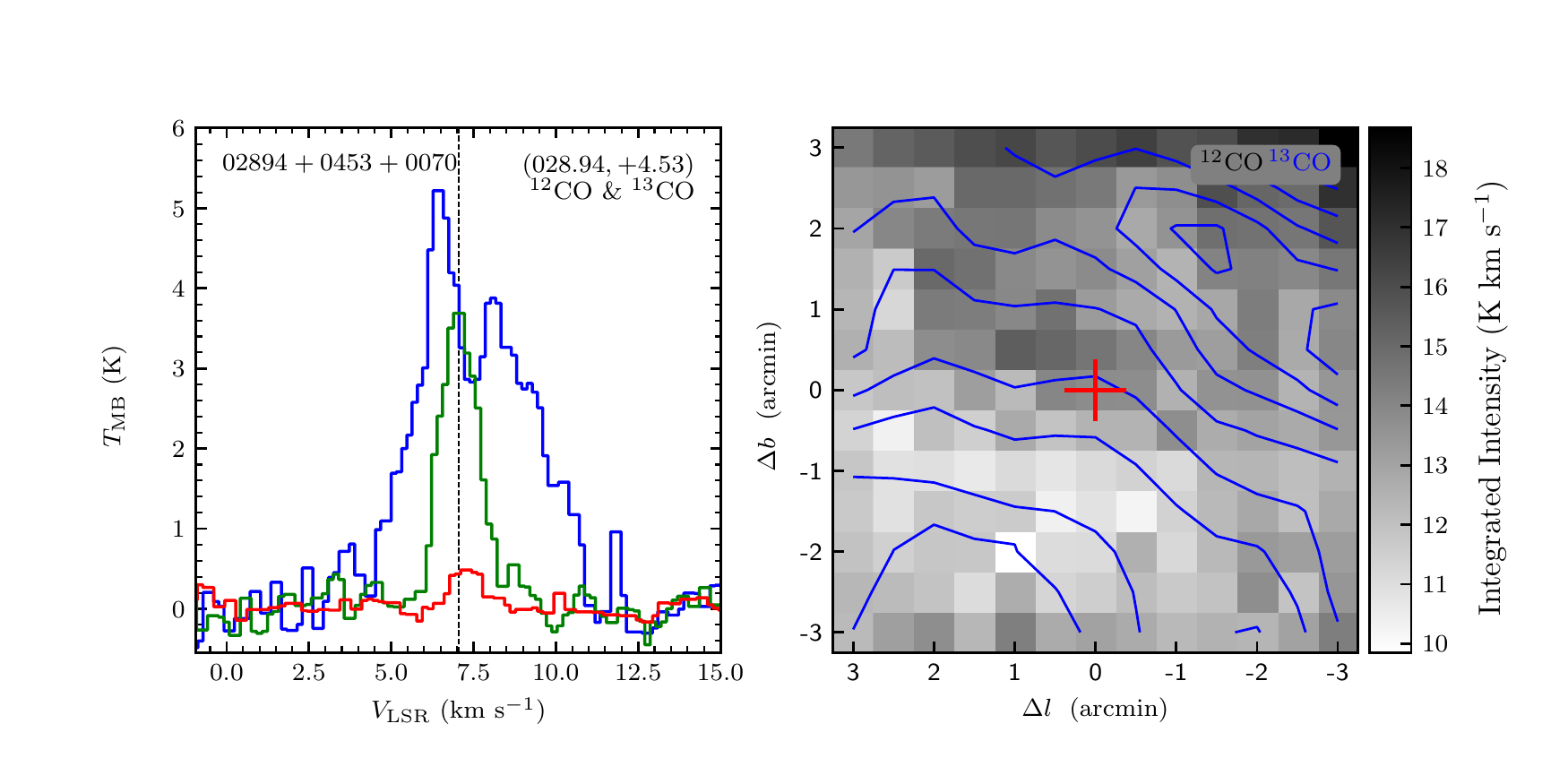}
\vspace{-0.5cm}

\includegraphics[width=9.0cm,angle=0]{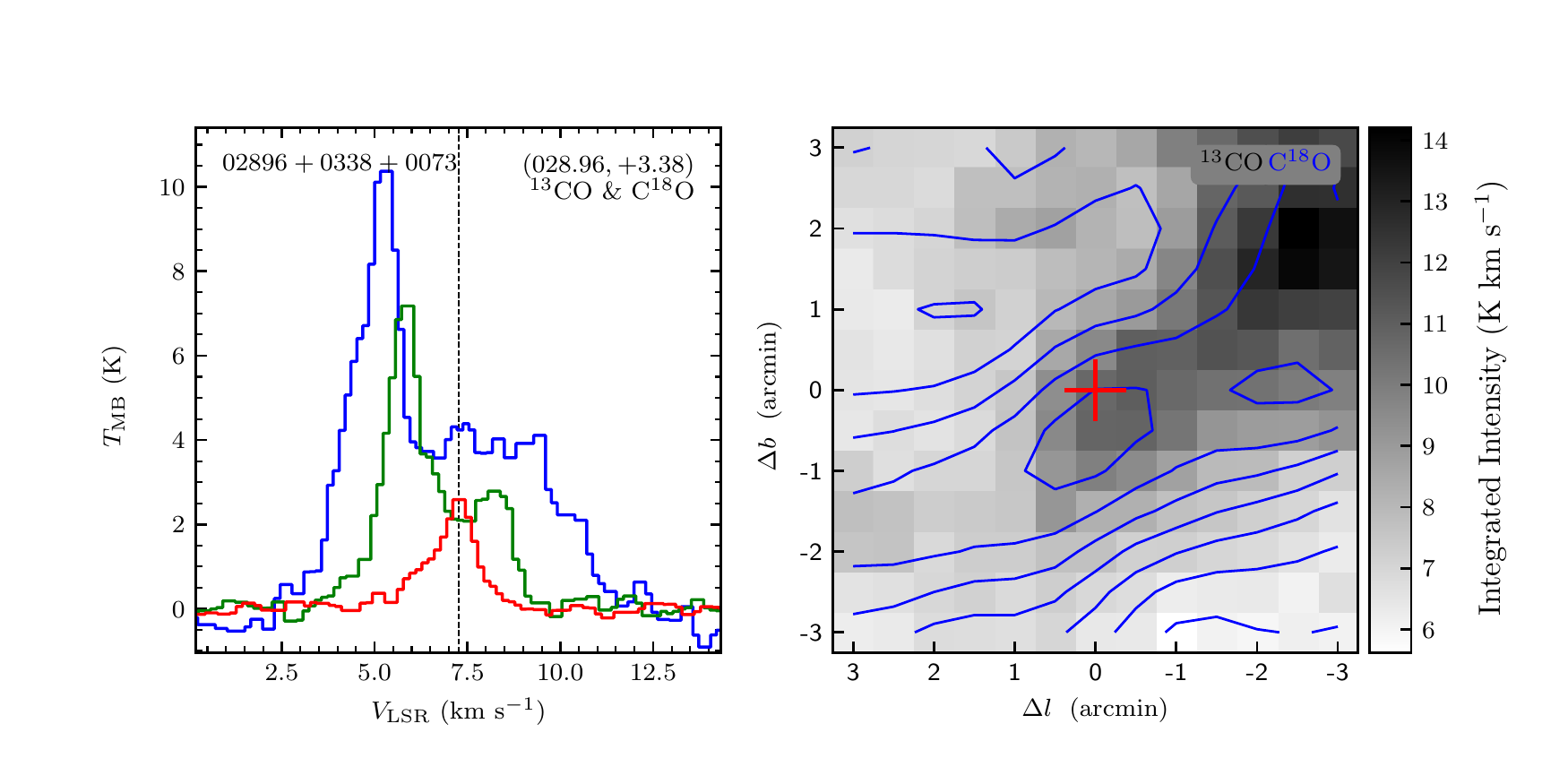}
\includegraphics[width=9.0cm,angle=0]{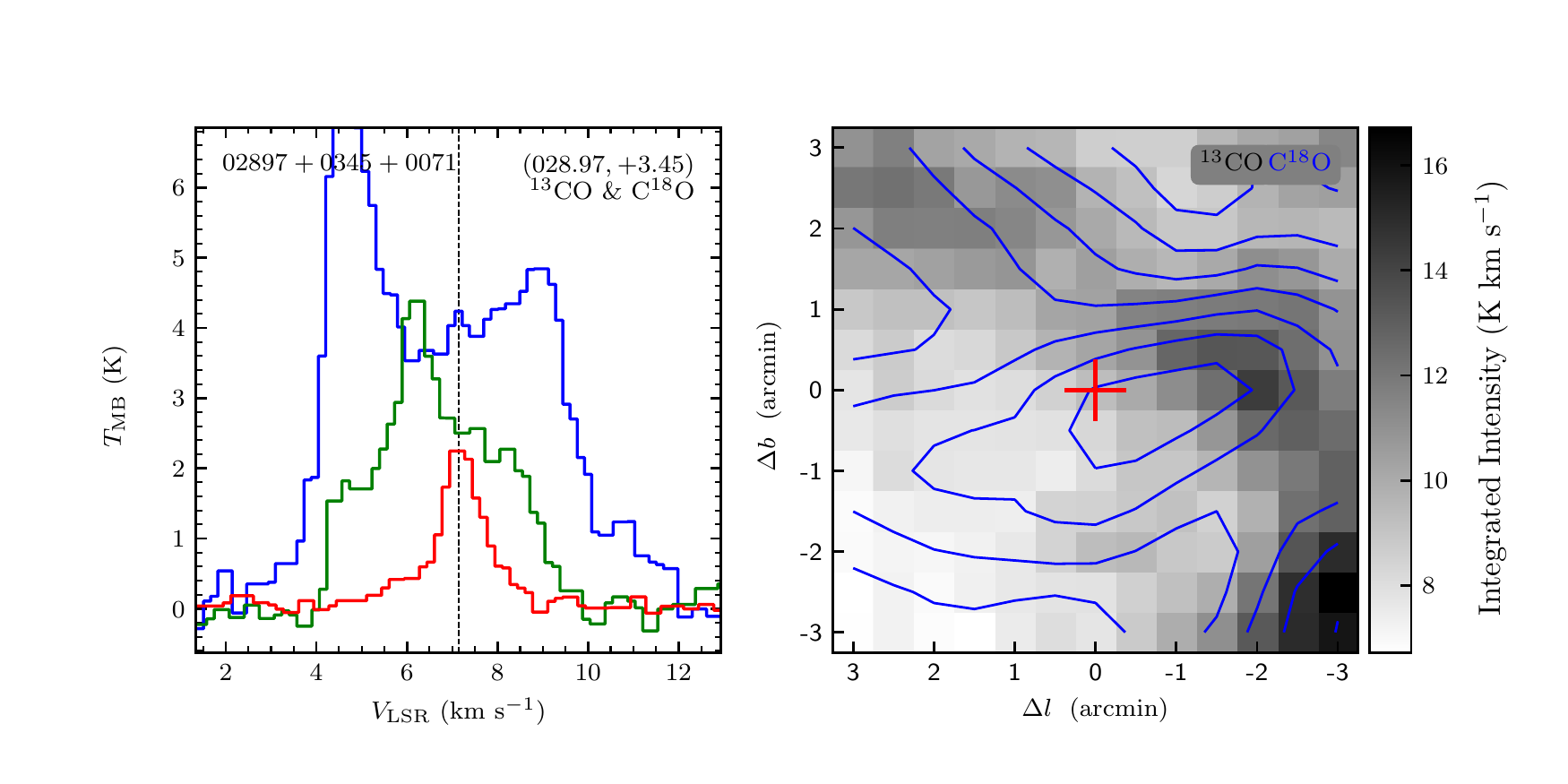}
\vspace{-0.5cm}

\includegraphics[width=9.0cm,angle=0]{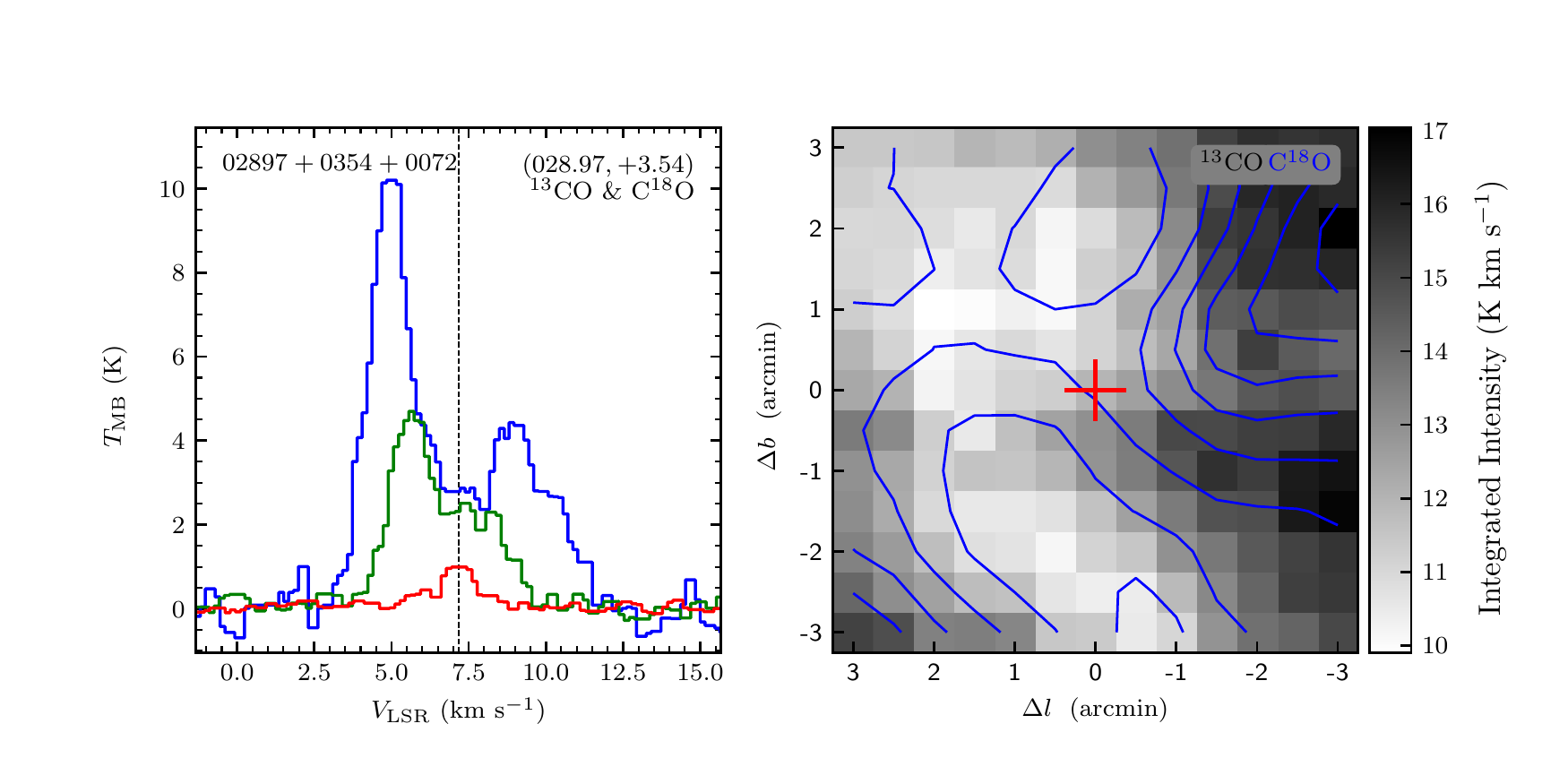}
\includegraphics[width=9.0cm,angle=0]{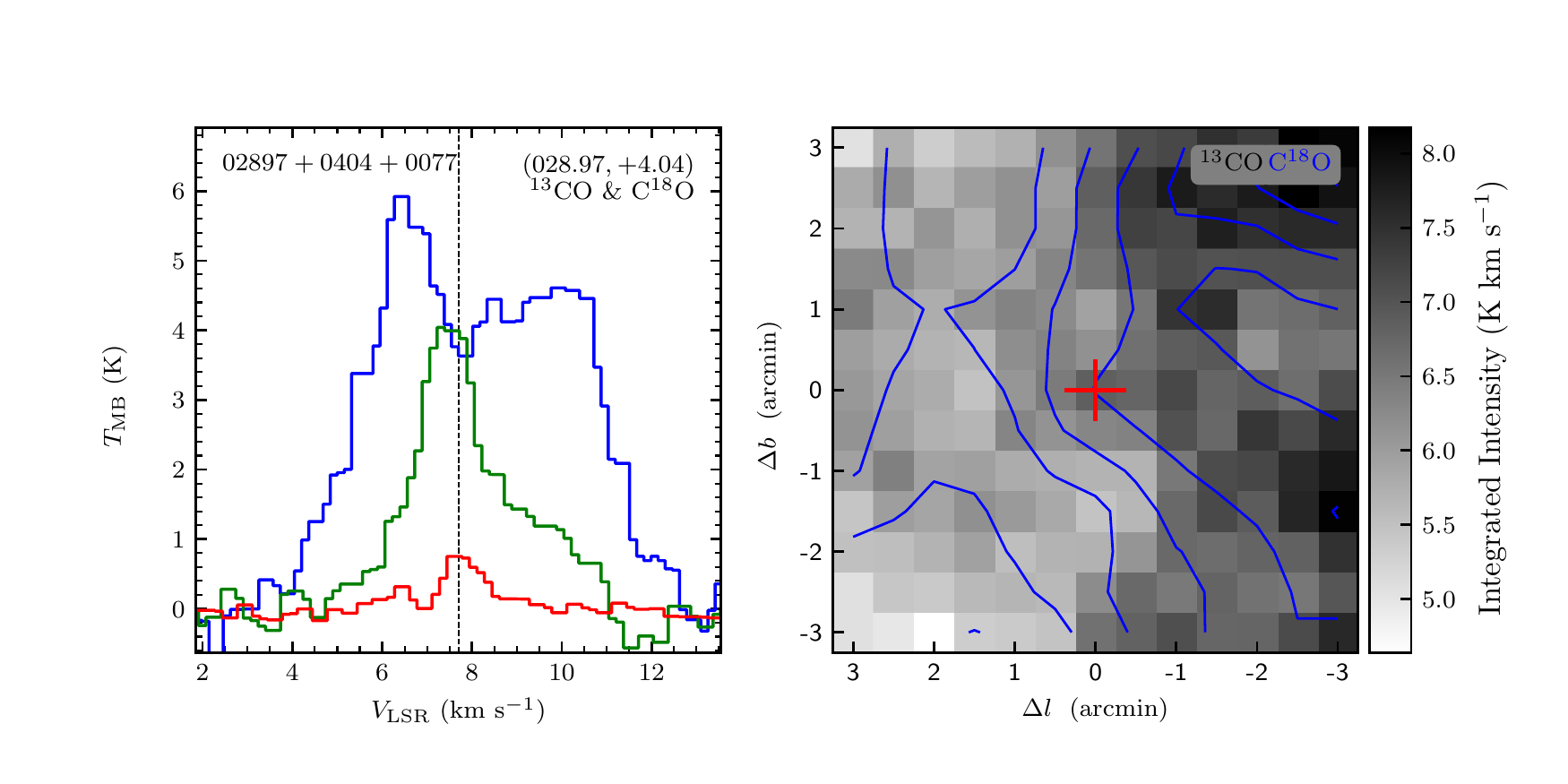}
\end{figure}
\clearpage

\begin{figure}
\includegraphics[width=9.0cm,angle=0]{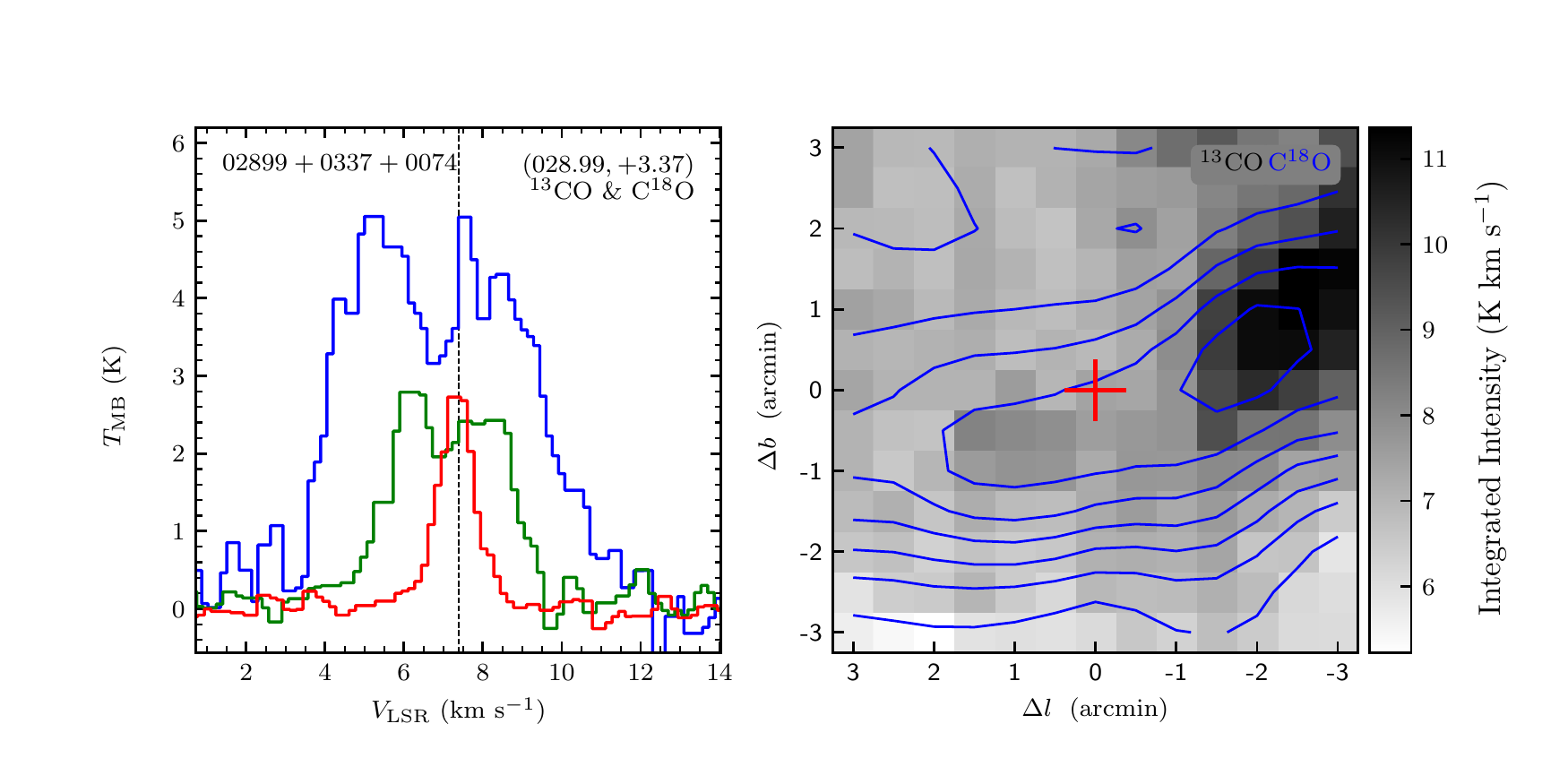}
\includegraphics[width=9.0cm,angle=0]{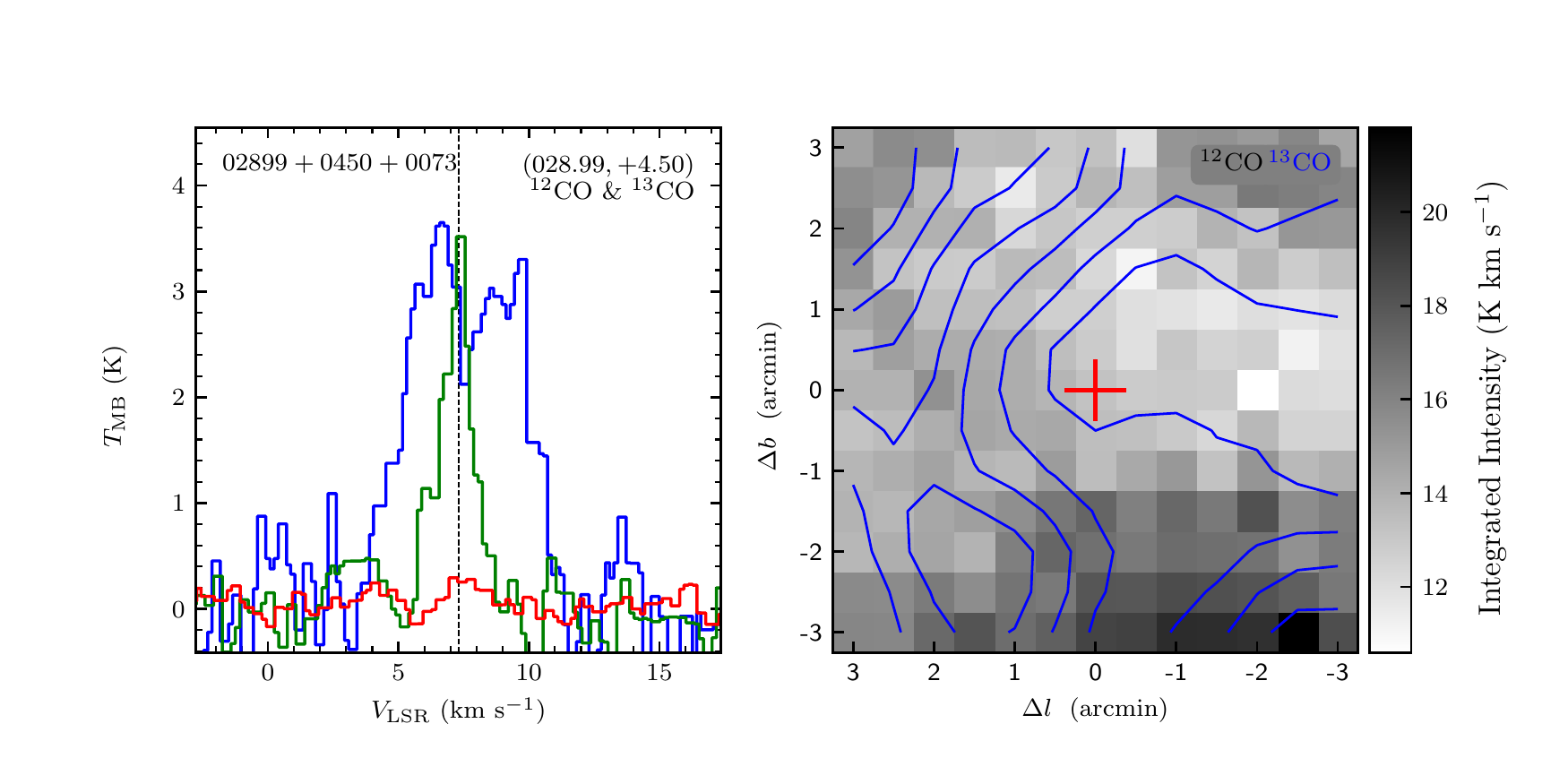}
\vspace{-0.5cm}

\includegraphics[width=9.0cm,angle=0]{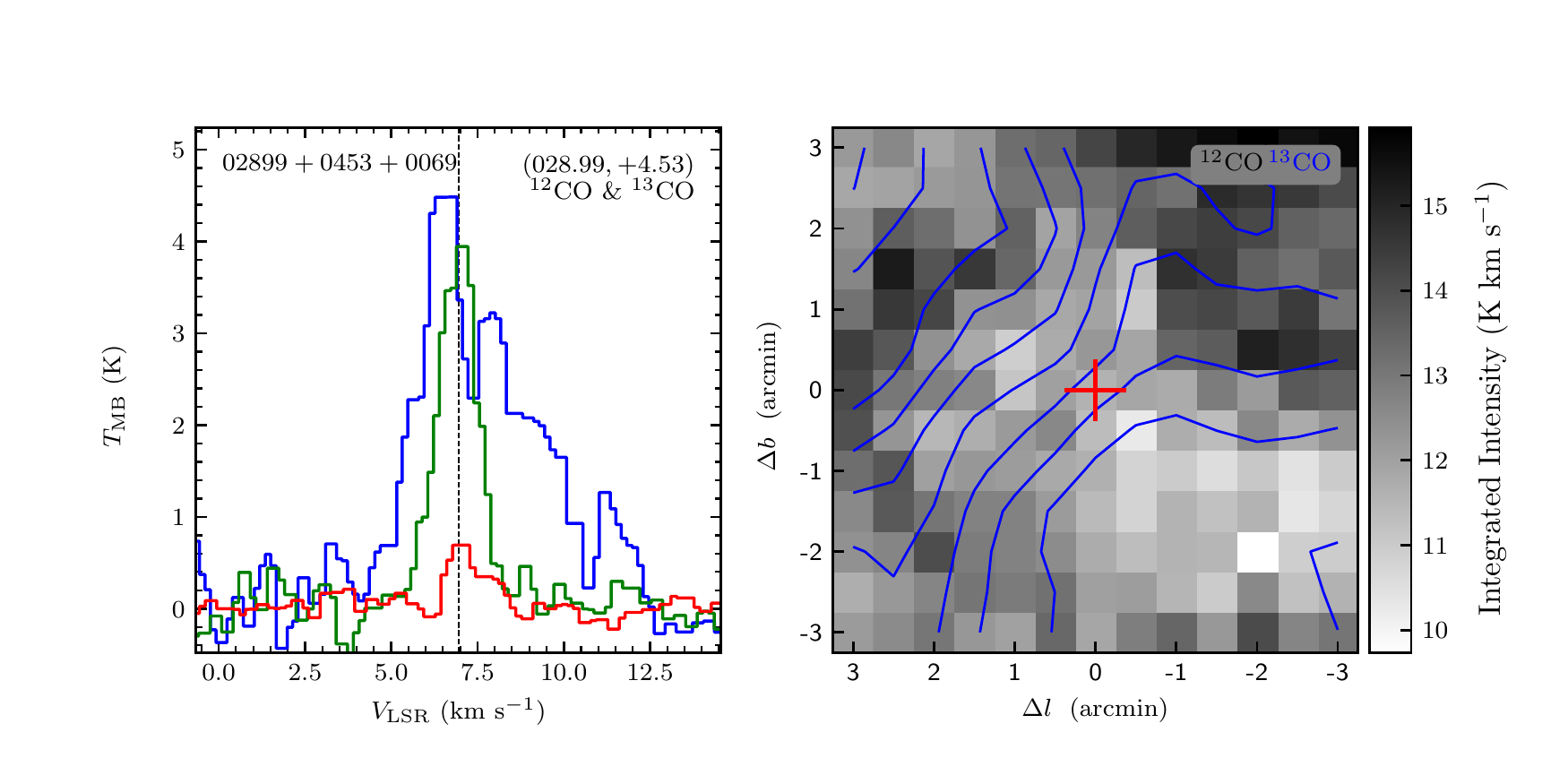}
\includegraphics[width=9.0cm,angle=0]{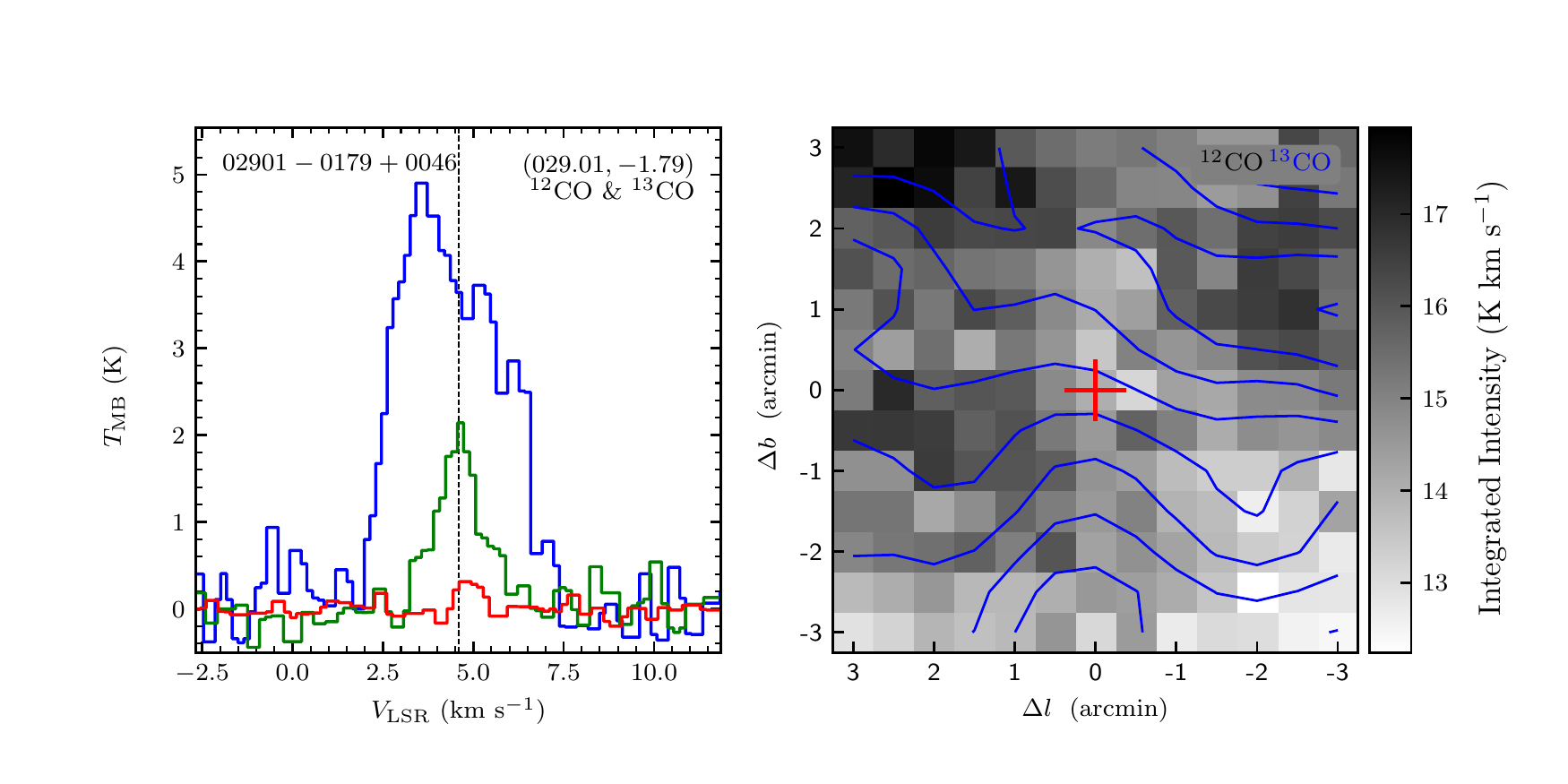}
\vspace{-0.5cm}

\includegraphics[width=9.0cm,angle=0]{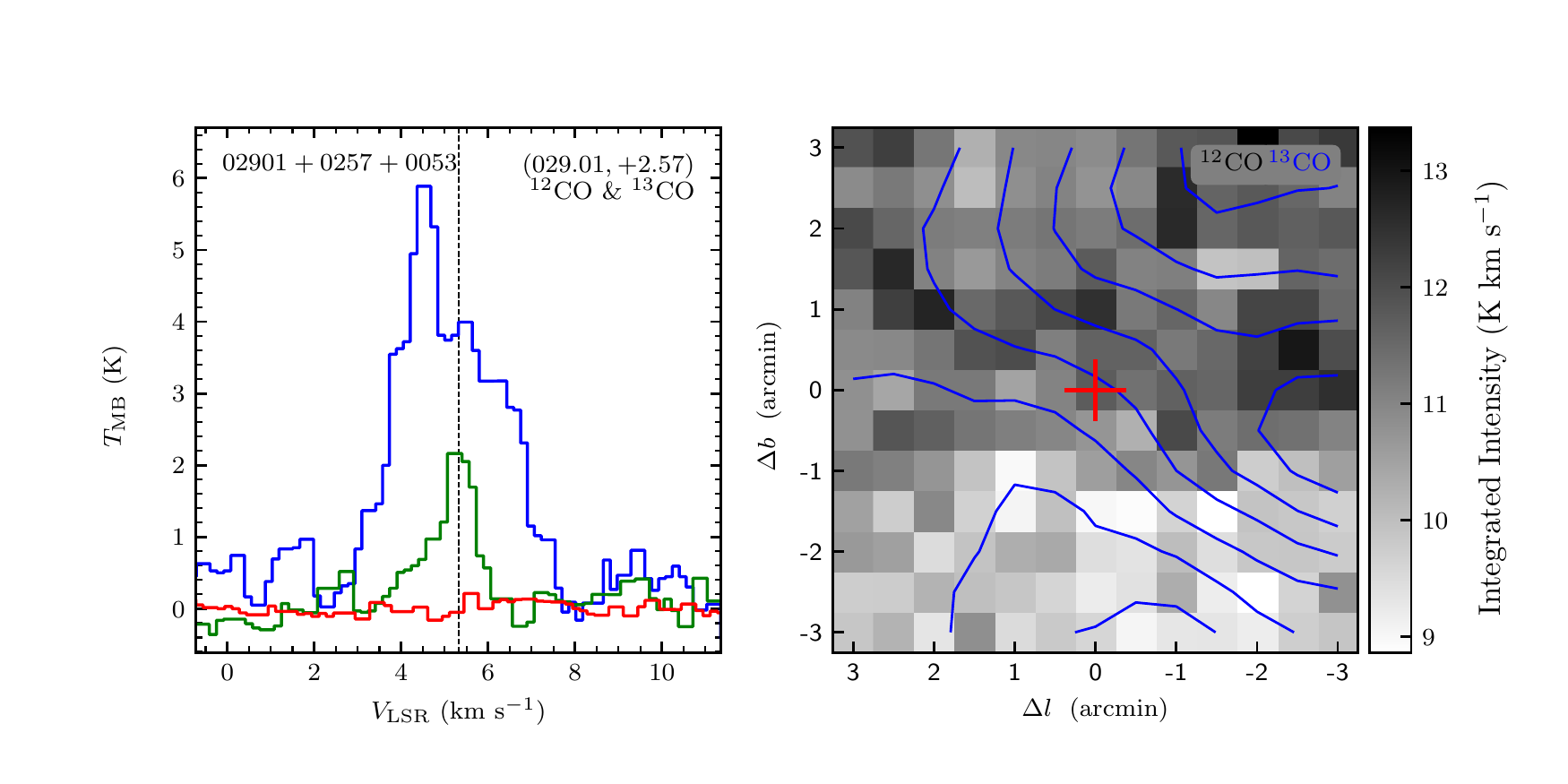}
\includegraphics[width=9.0cm,angle=0]{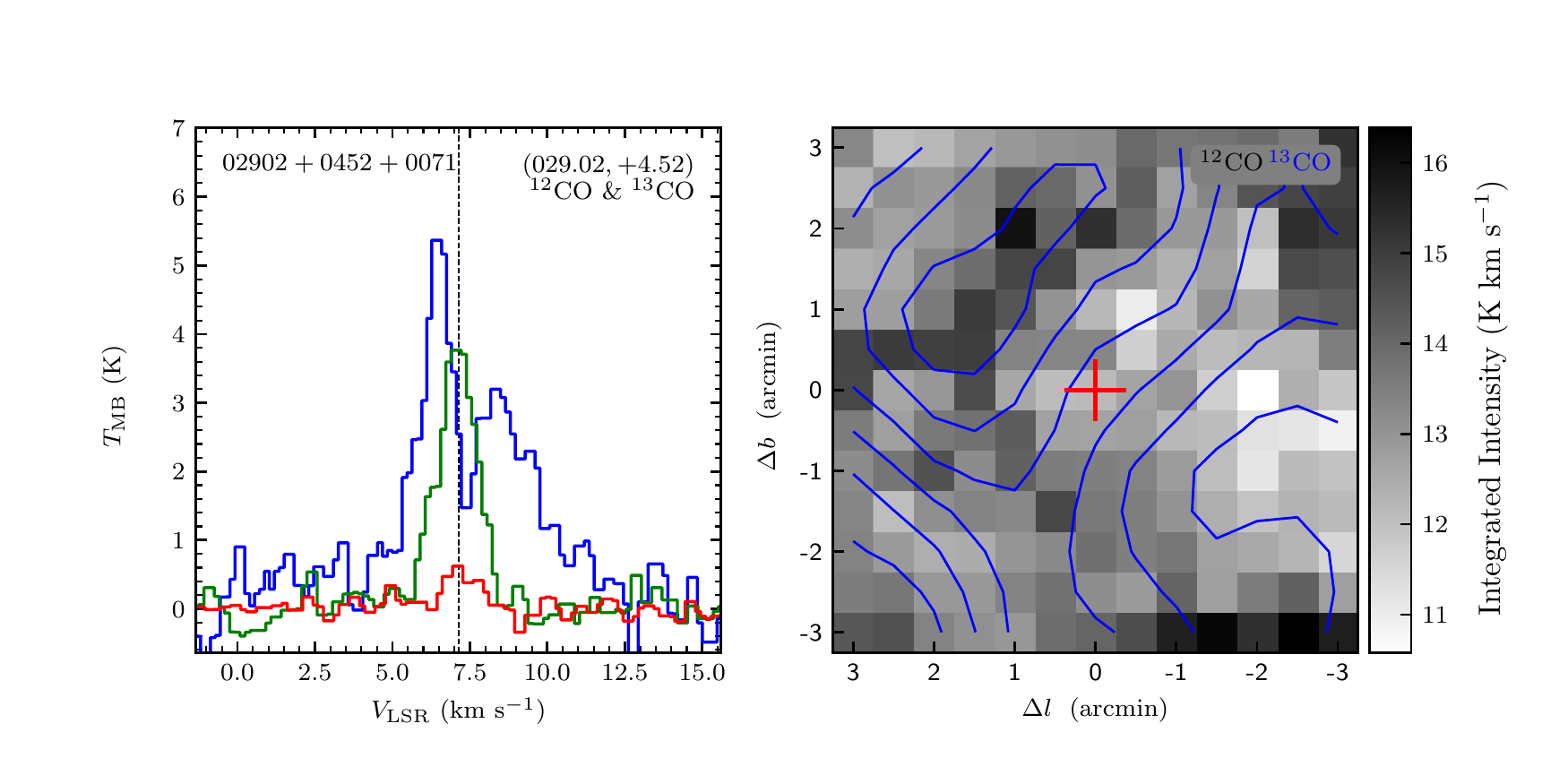}
\vspace{-0.5cm}

\includegraphics[width=9.0cm,angle=0]{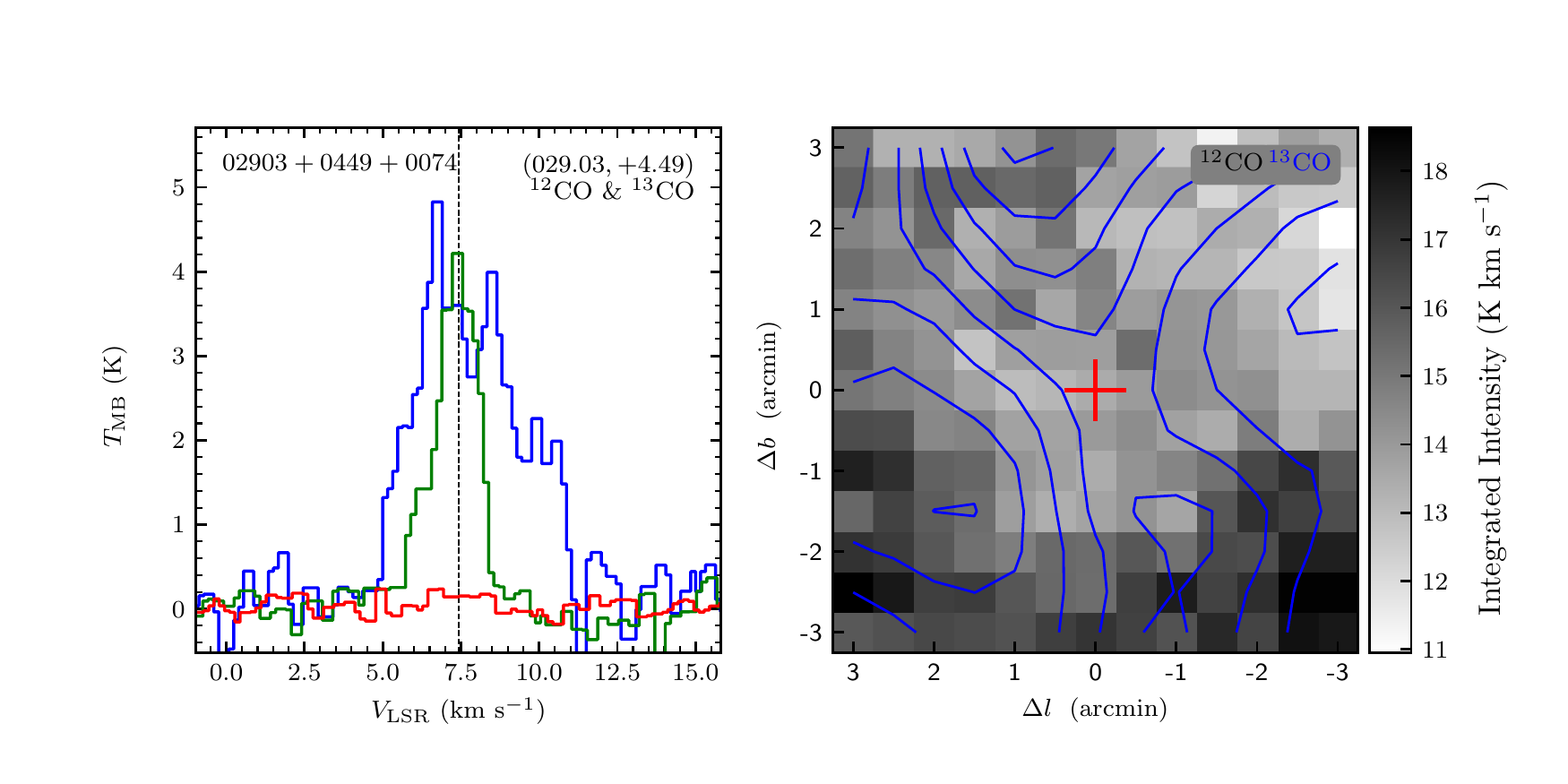}
\includegraphics[width=9.0cm,angle=0]{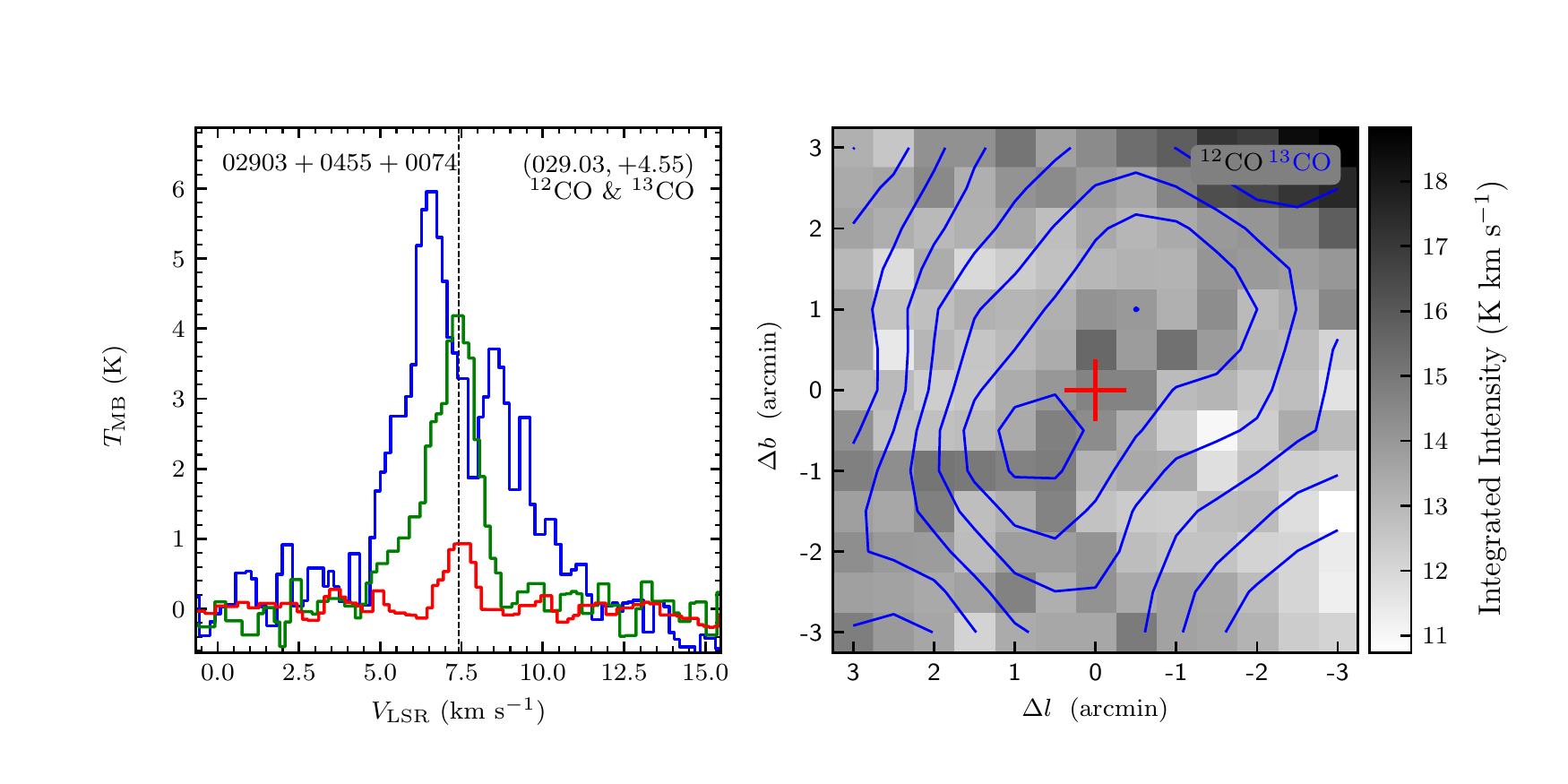}
\vspace{-0.5cm}

\includegraphics[width=9.0cm,angle=0]{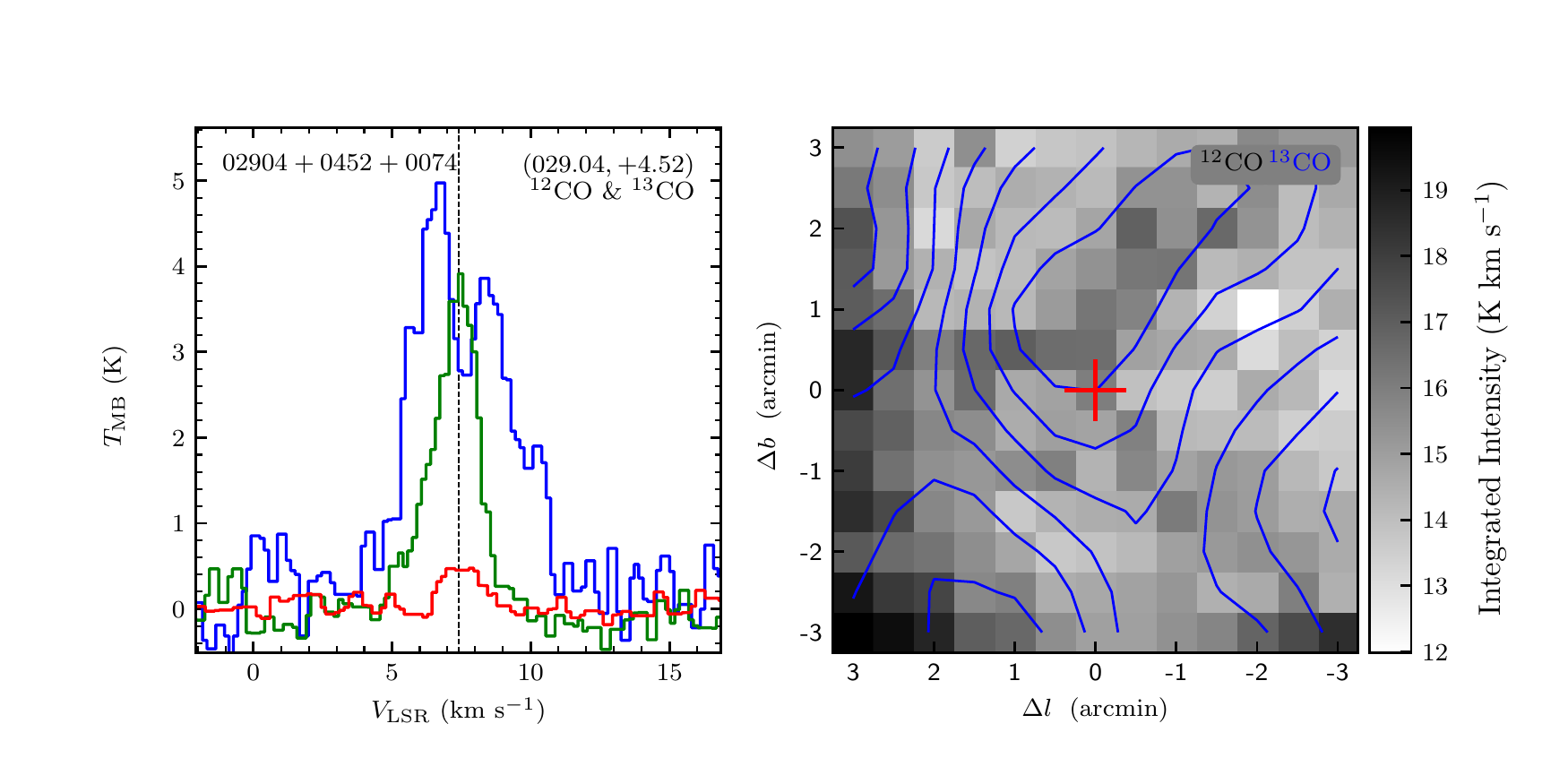}
\includegraphics[width=9.0cm,angle=0]{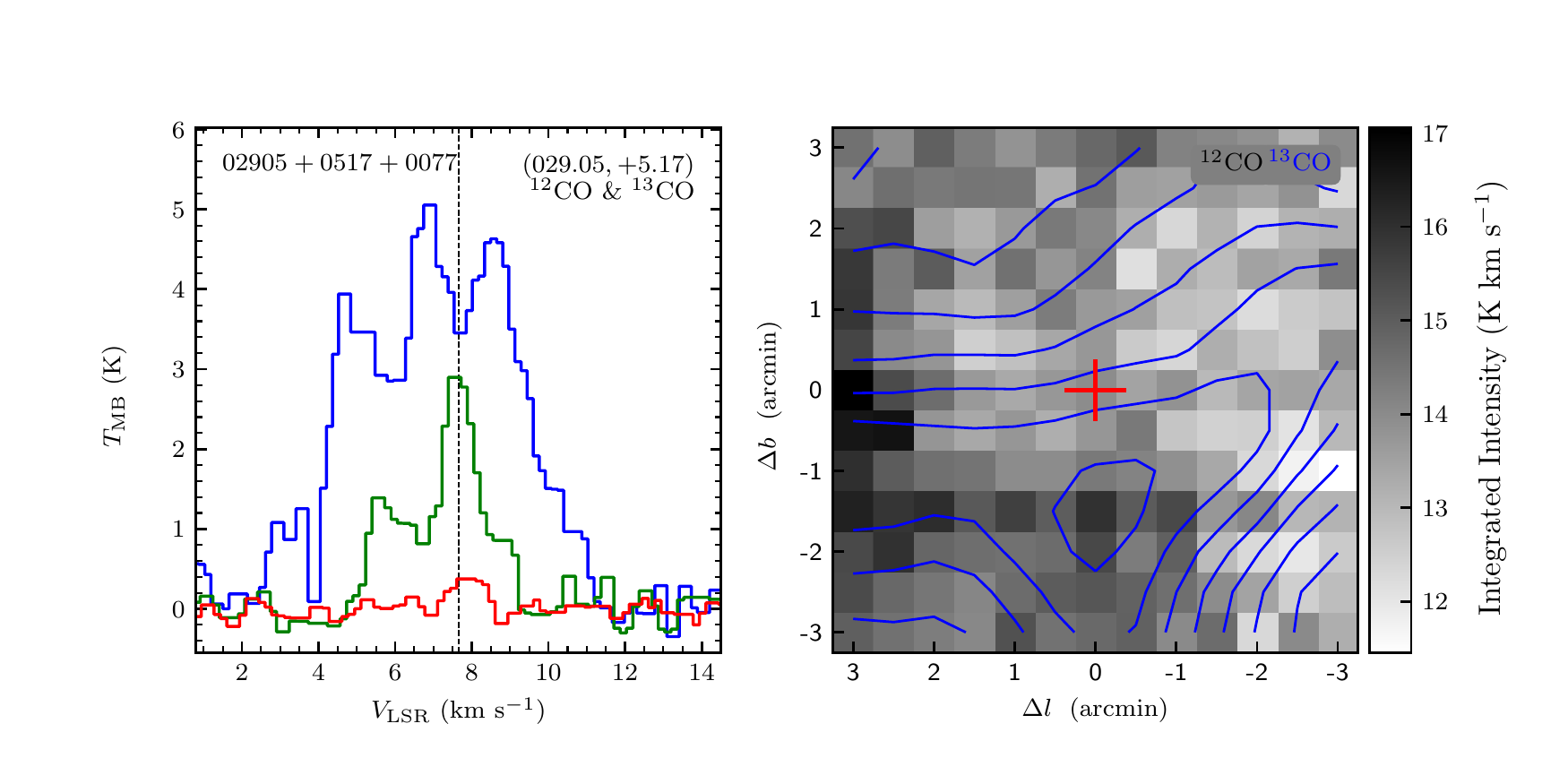}
\end{figure}
\clearpage

\begin{figure}
\includegraphics[width=9.0cm,angle=0]{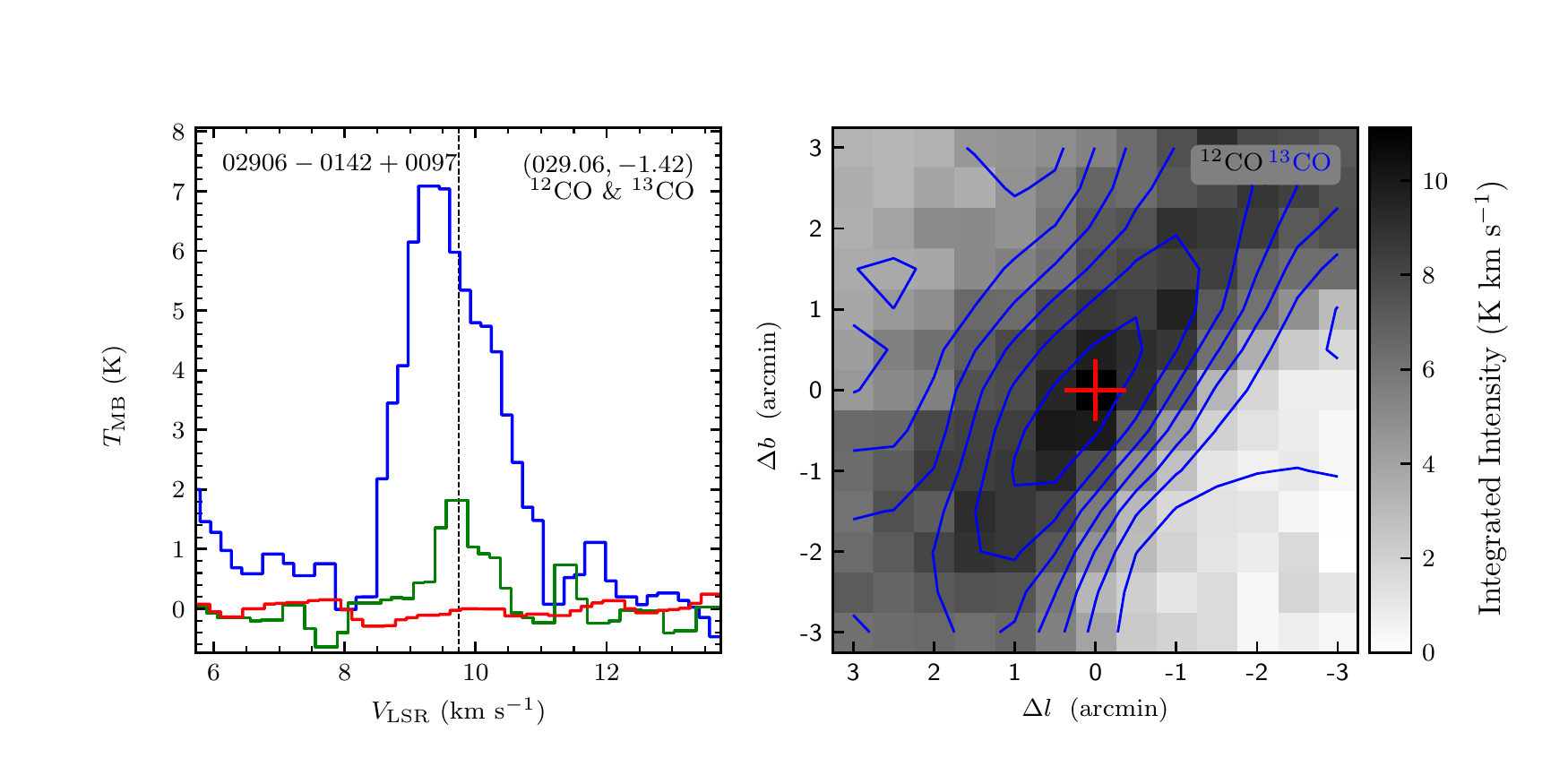}
\includegraphics[width=9.0cm,angle=0]{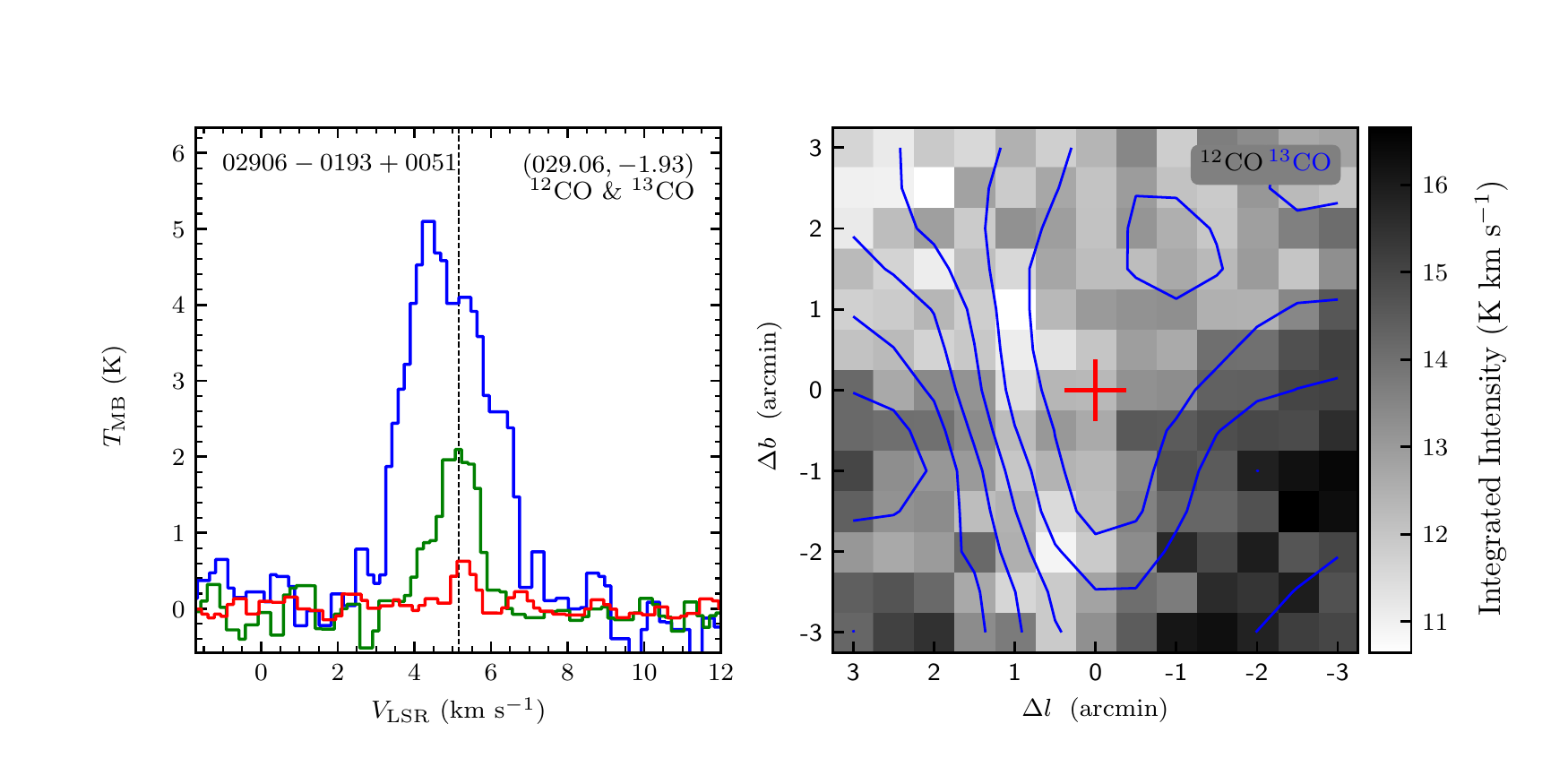}
\vspace{-0.5cm}

\includegraphics[width=9.0cm,angle=0]{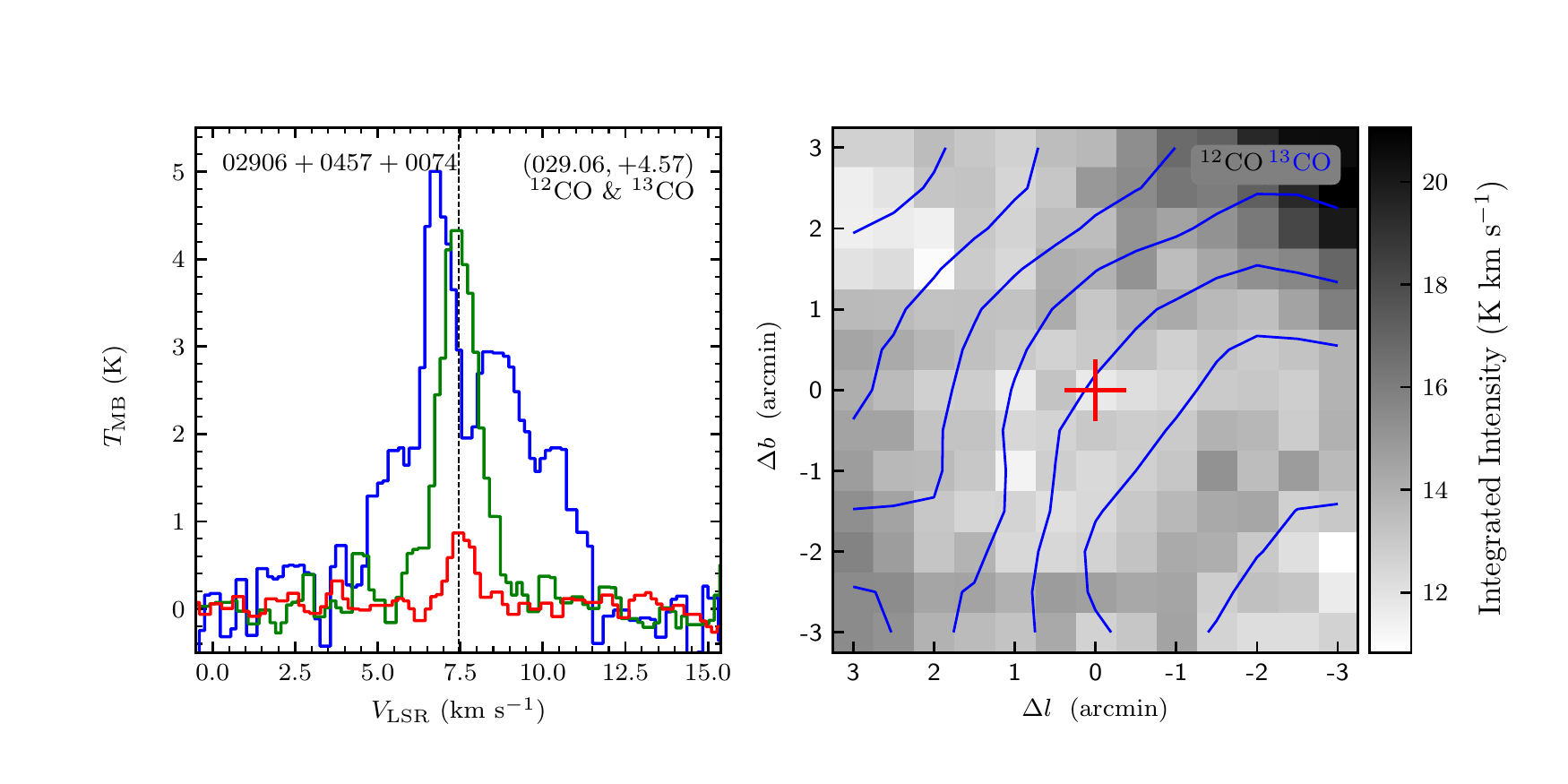}
\includegraphics[width=9.0cm,angle=0]{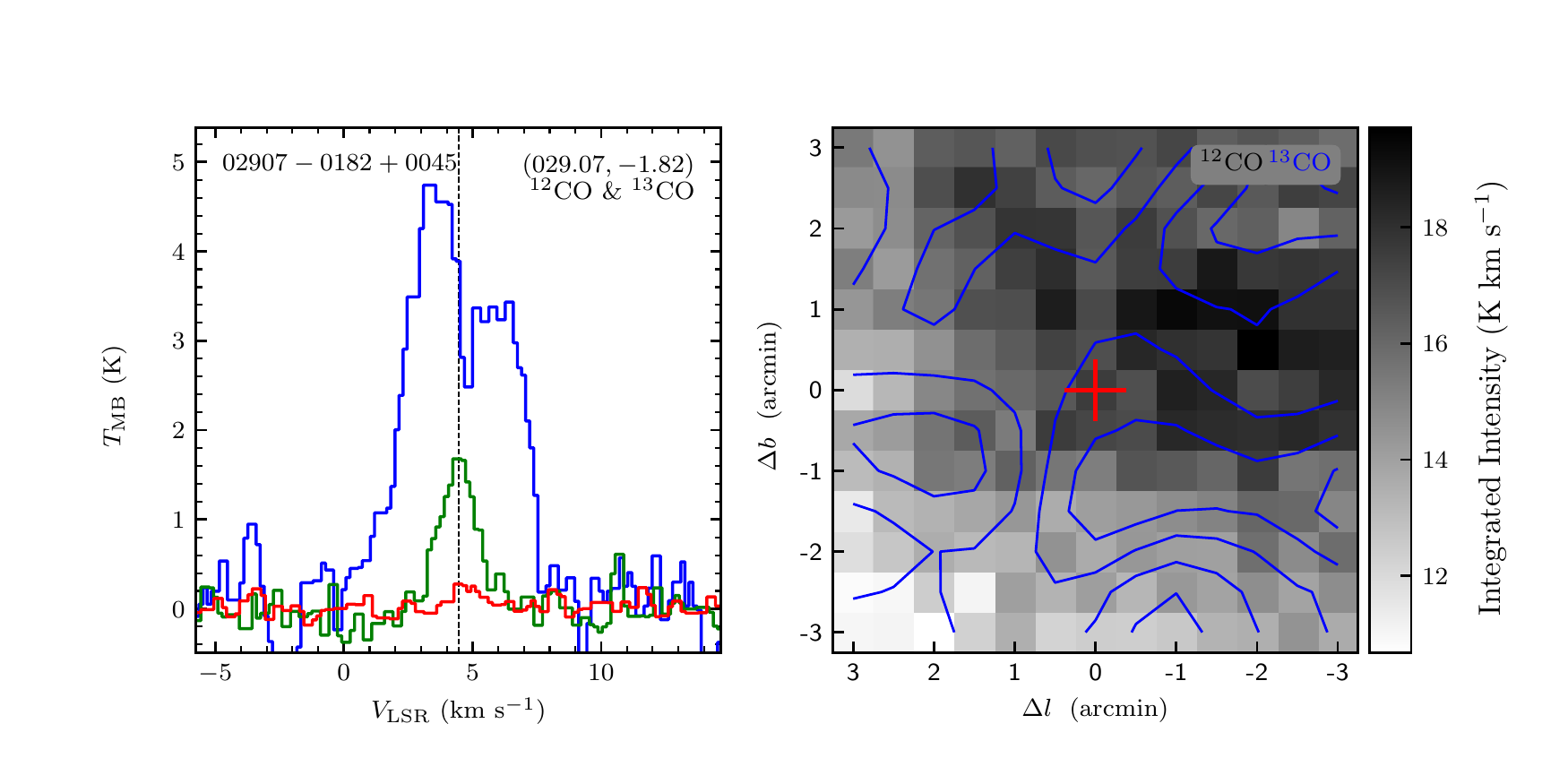}
\vspace{-0.5cm}

\includegraphics[width=9.0cm,angle=0]{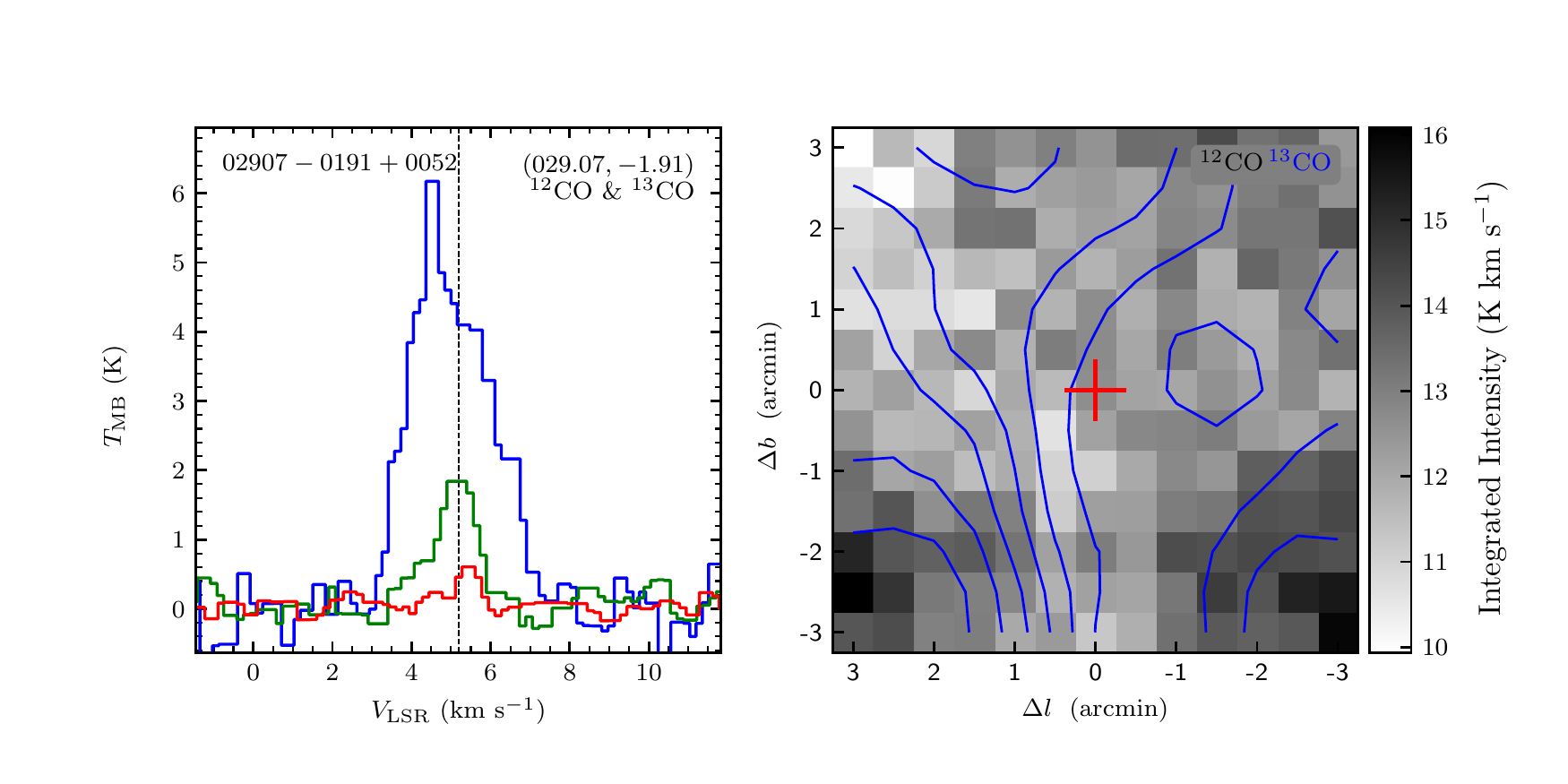}
\includegraphics[width=9.0cm,angle=0]{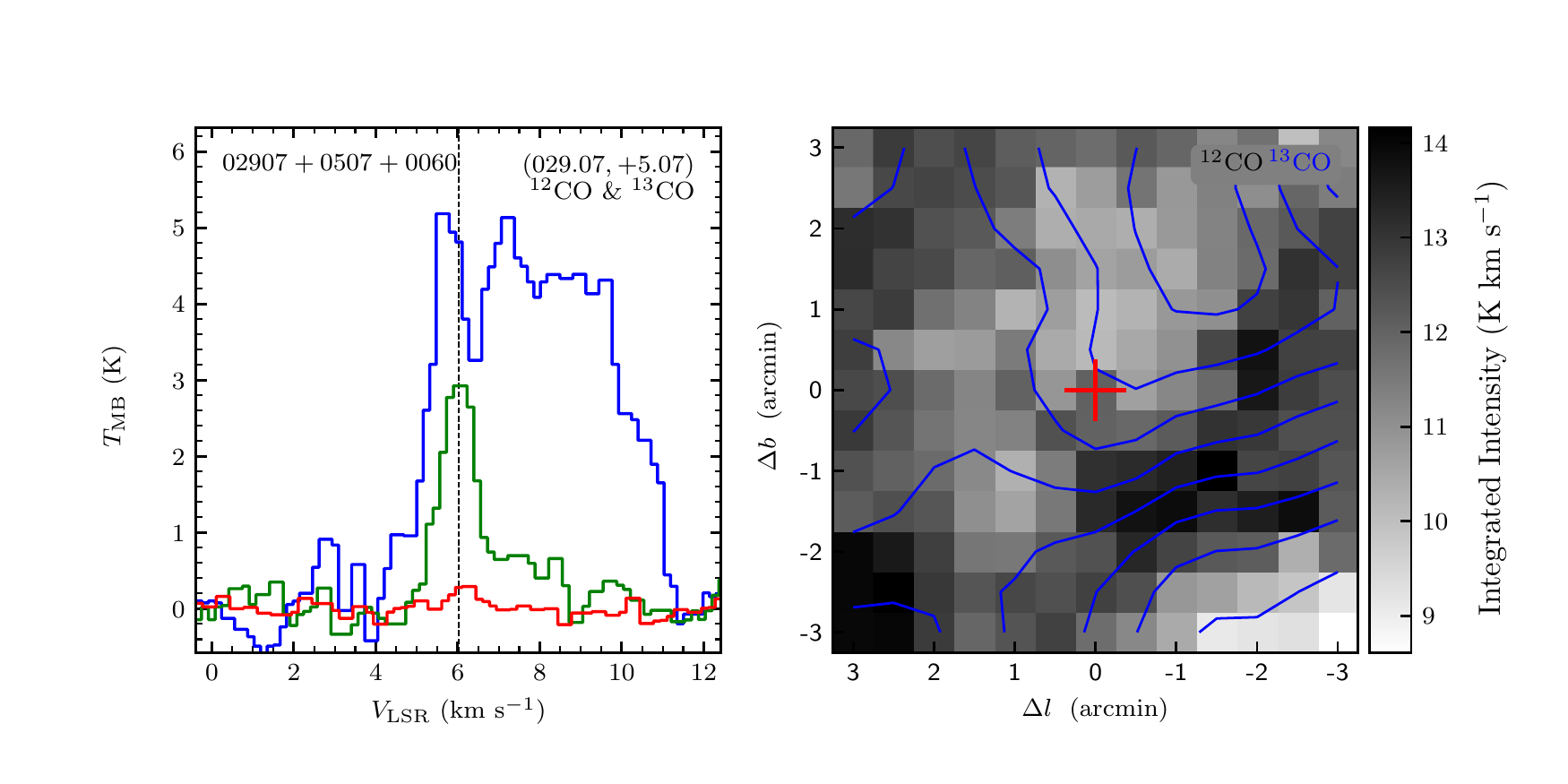}
\vspace{-0.5cm}

\includegraphics[width=9.0cm,angle=0]{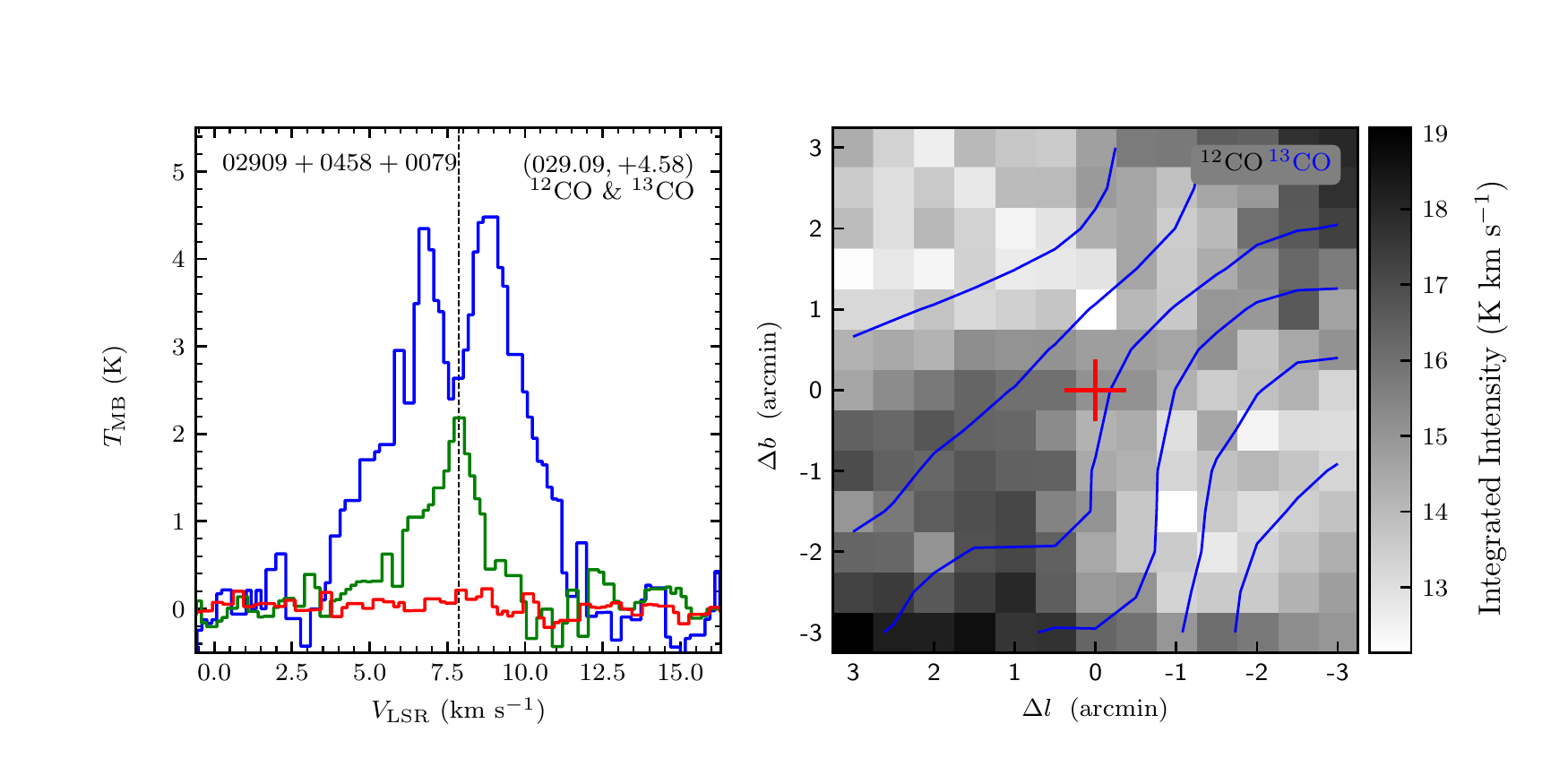}
\includegraphics[width=9.0cm,angle=0]{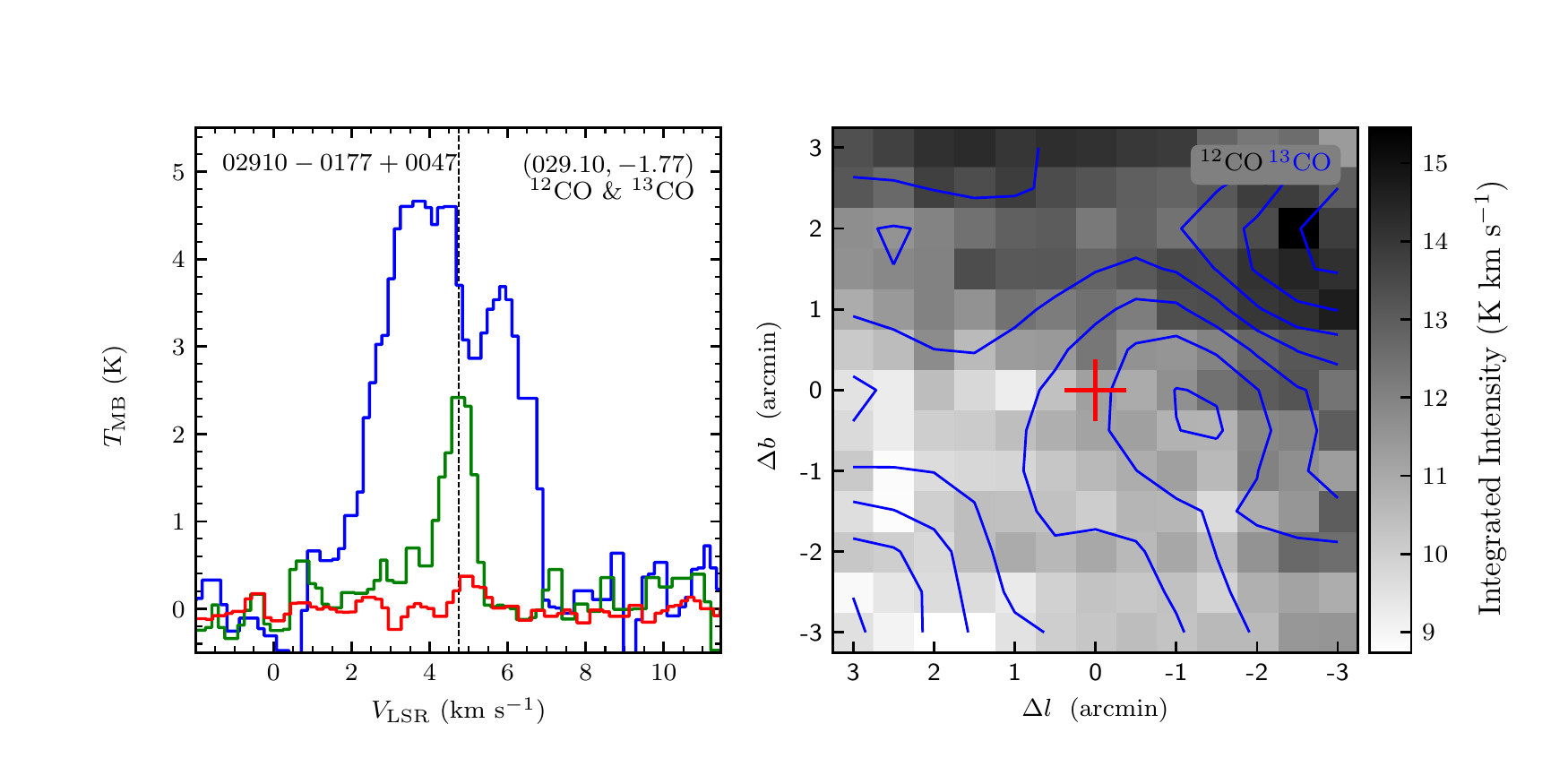}
\vspace{-0.5cm}

\includegraphics[width=9.0cm,angle=0]{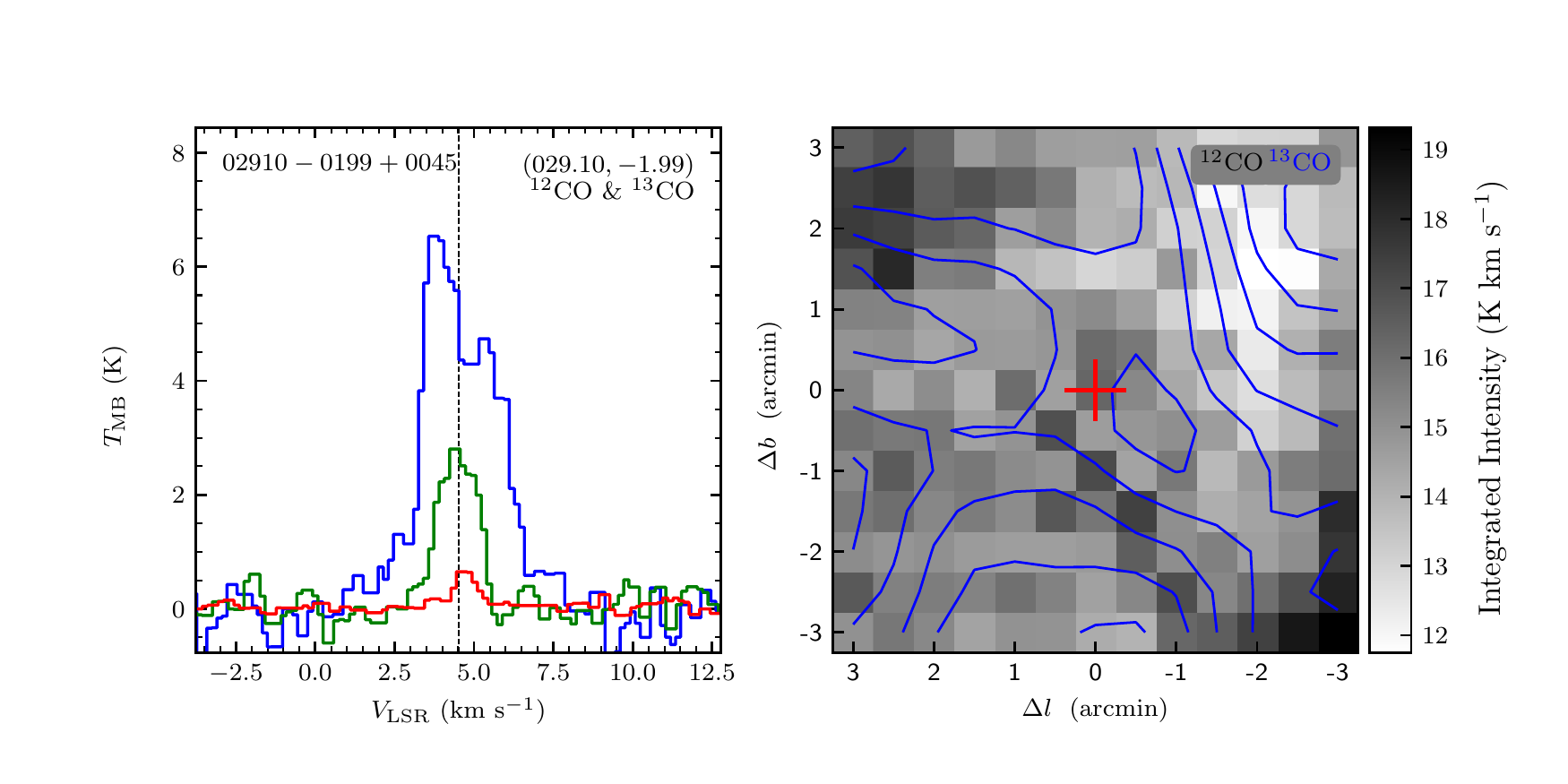}
\includegraphics[width=9.0cm,angle=0]{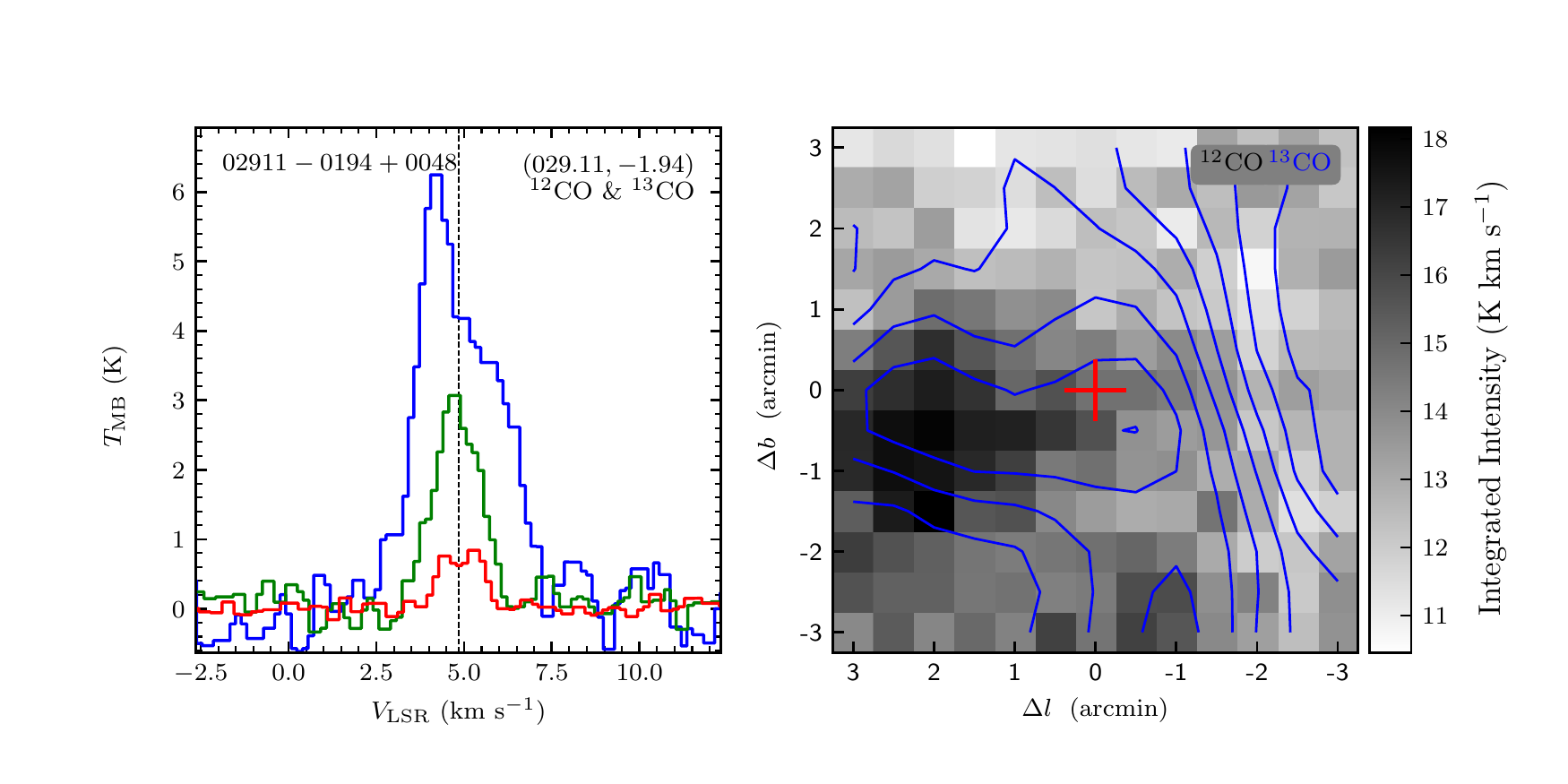}
\end{figure}
\clearpage

\begin{figure}
\includegraphics[width=9.0cm,angle=0]{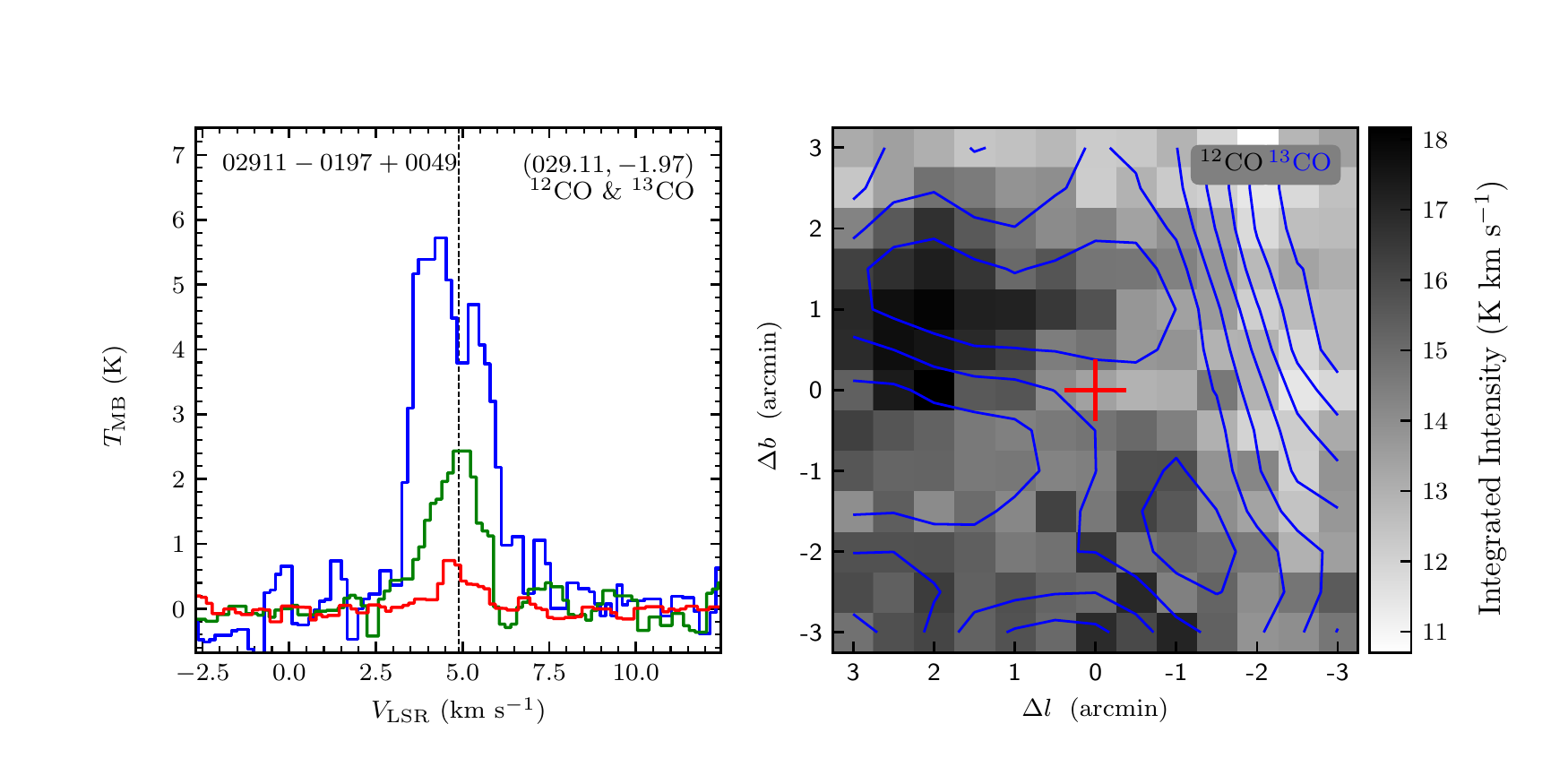}
\includegraphics[width=9.0cm,angle=0]{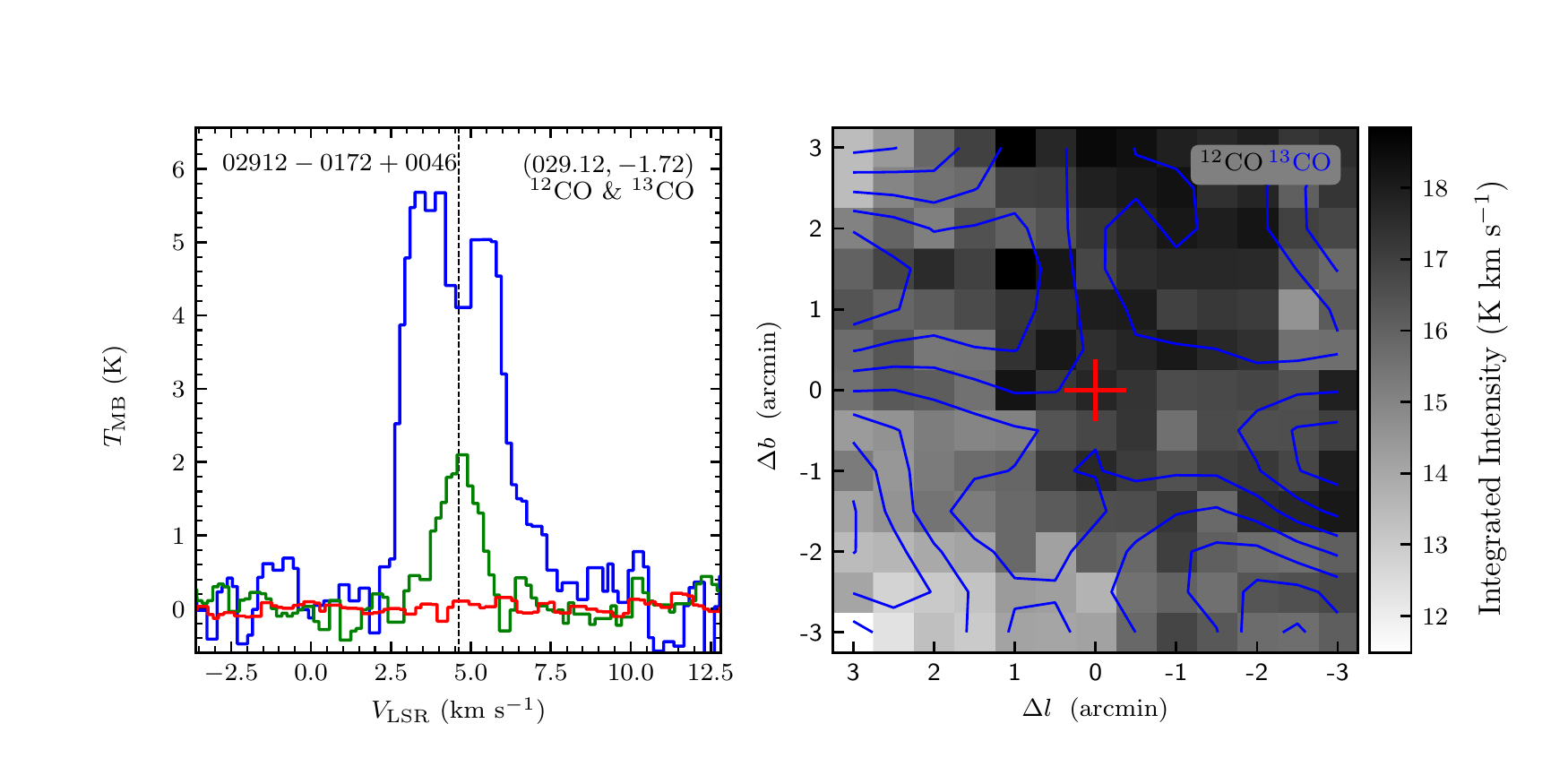}
\vspace{-0.5cm}

\includegraphics[width=9.0cm,angle=0]{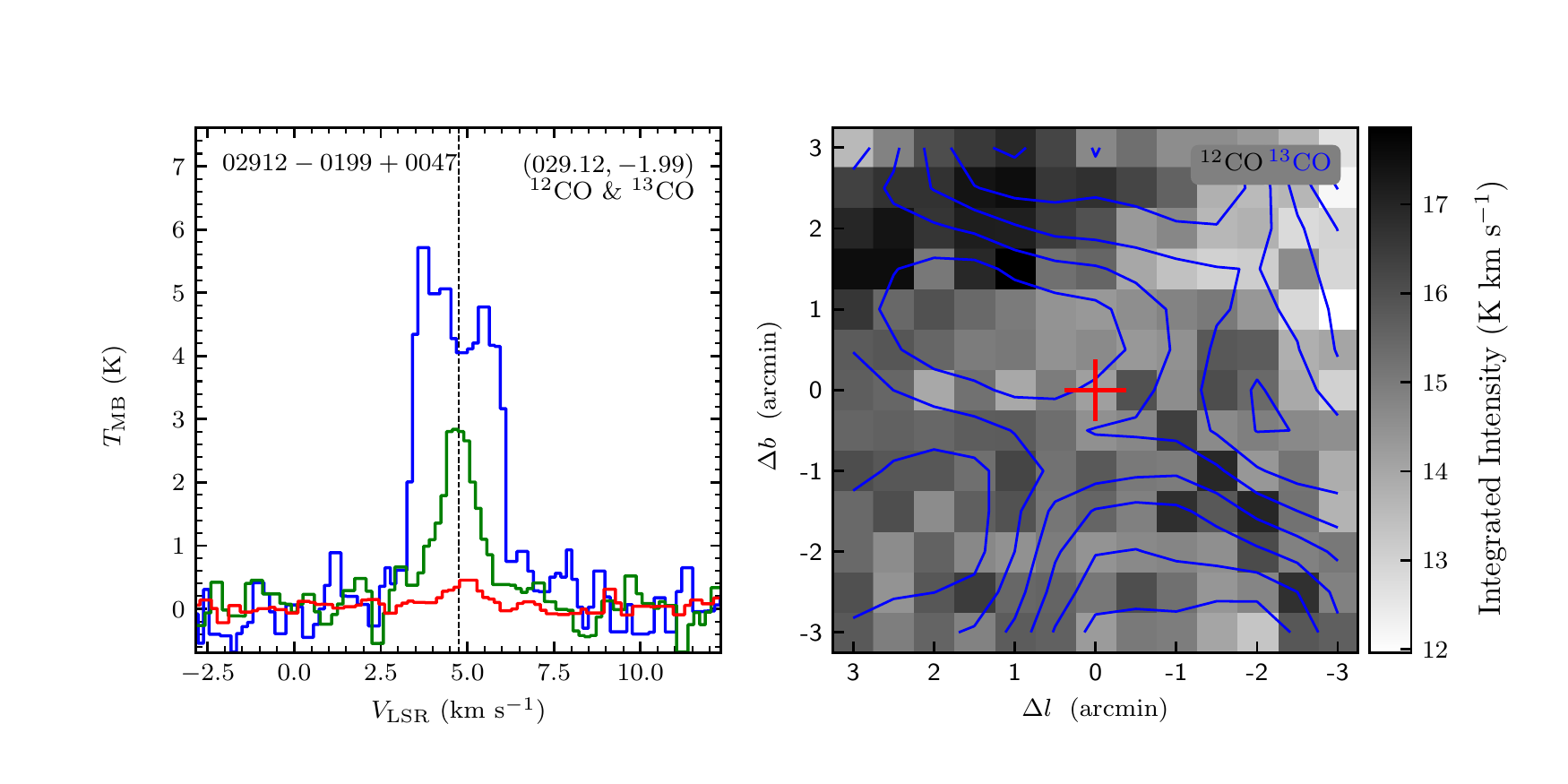}
\includegraphics[width=9.0cm,angle=0]{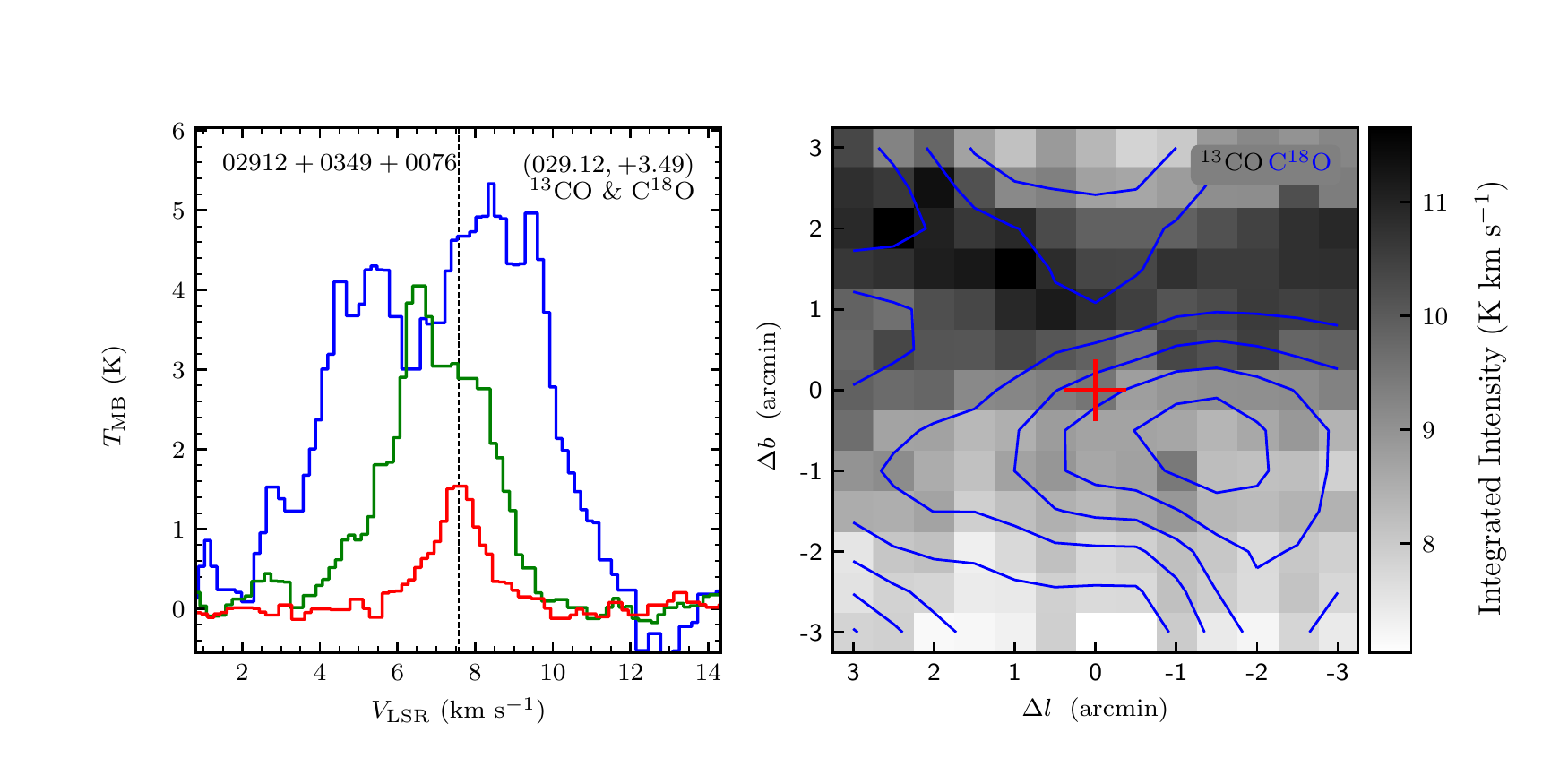}
\vspace{-0.5cm}

\includegraphics[width=9.0cm,angle=0]{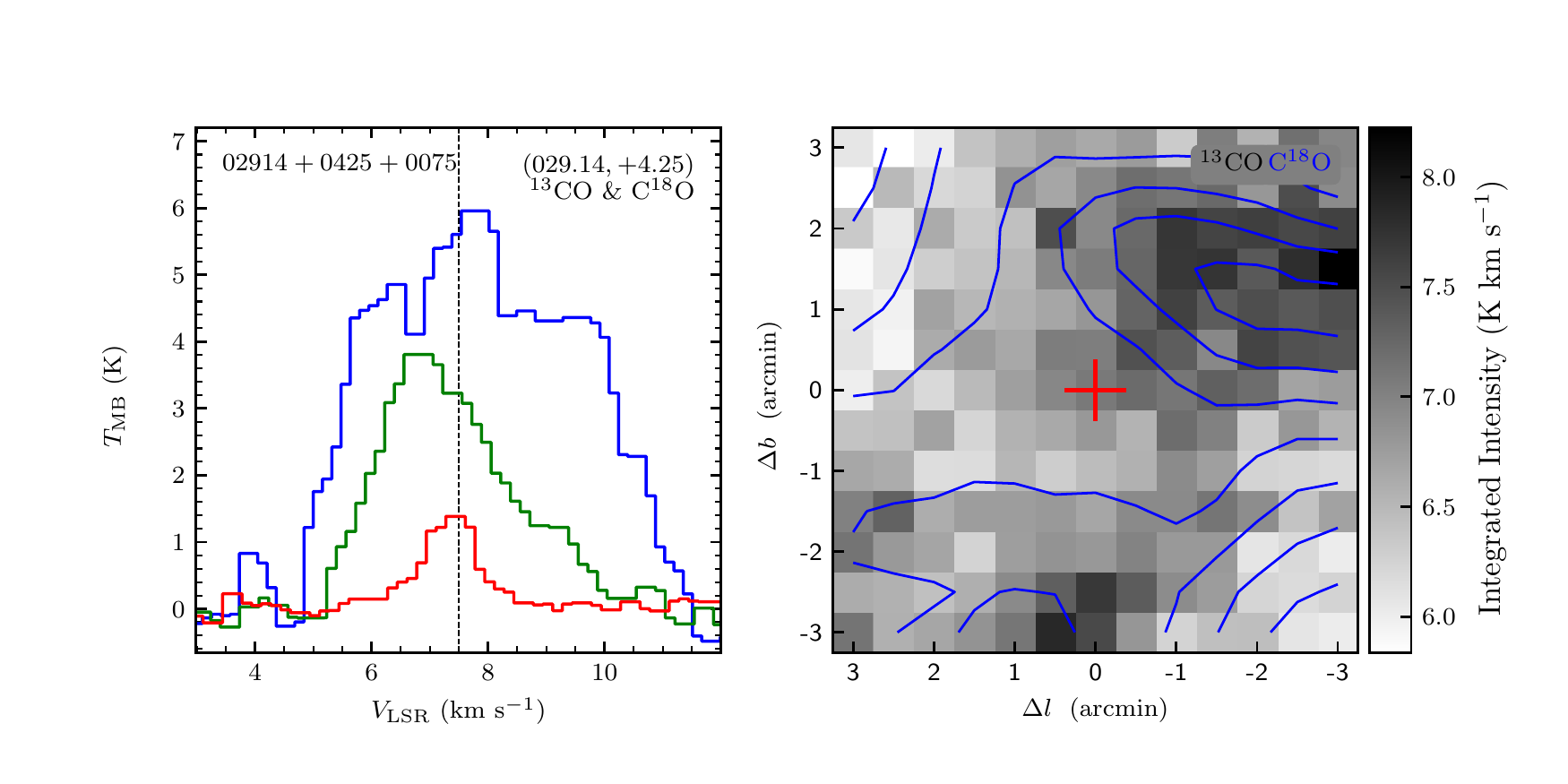}
\includegraphics[width=9.0cm,angle=0]{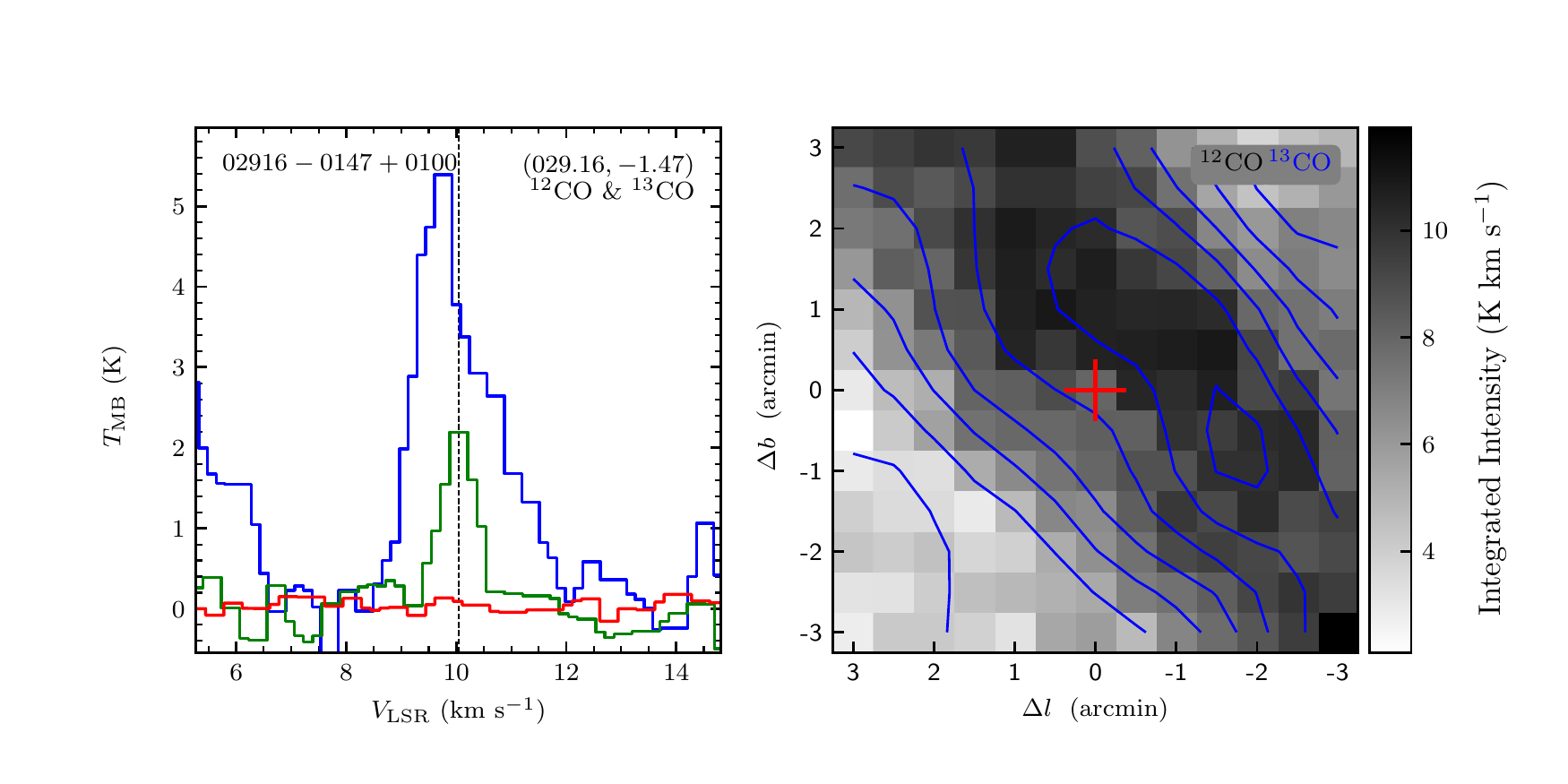}
\vspace{-0.5cm}

\includegraphics[width=9.0cm,angle=0]{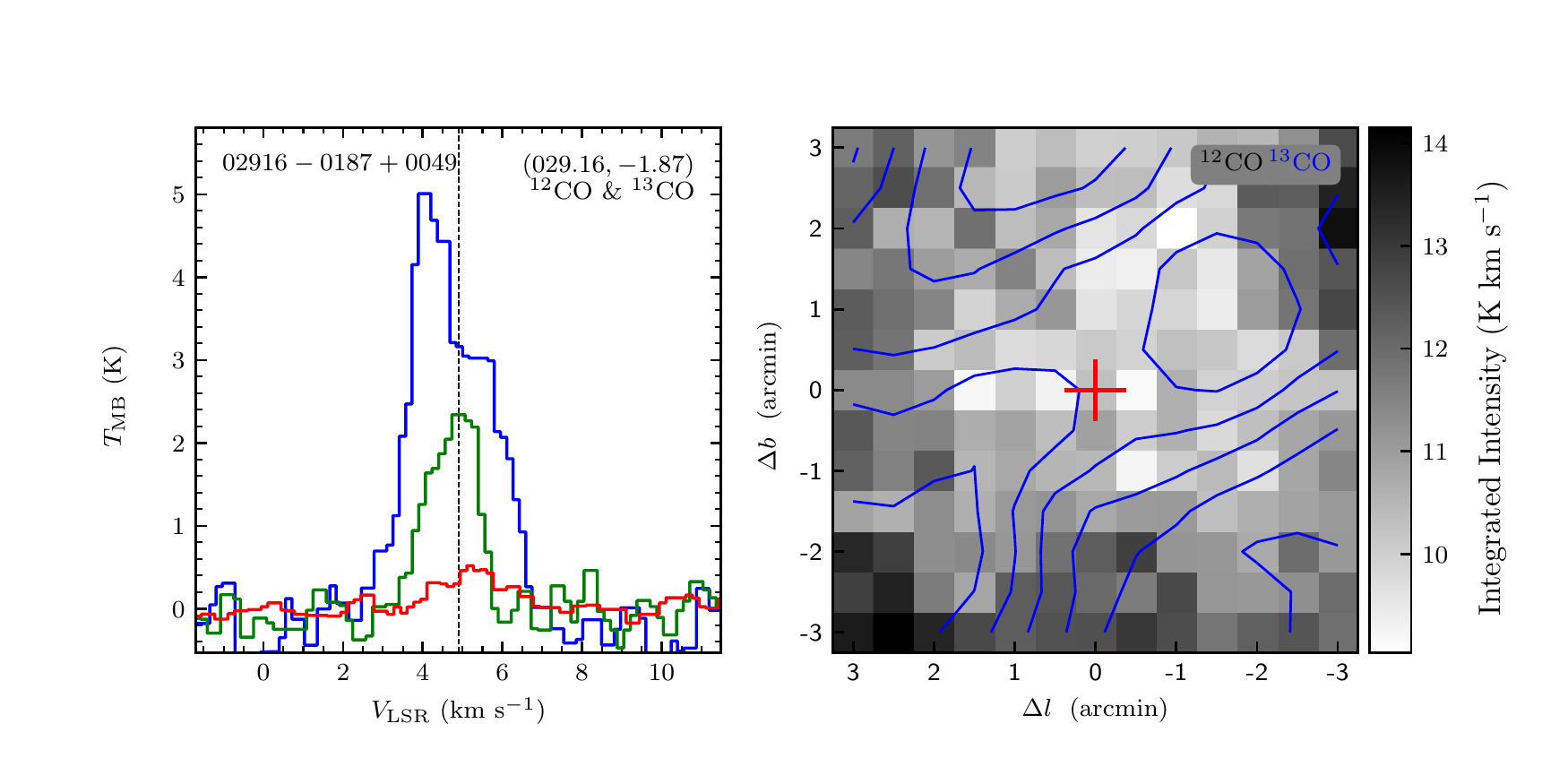}
\includegraphics[width=9.0cm,angle=0]{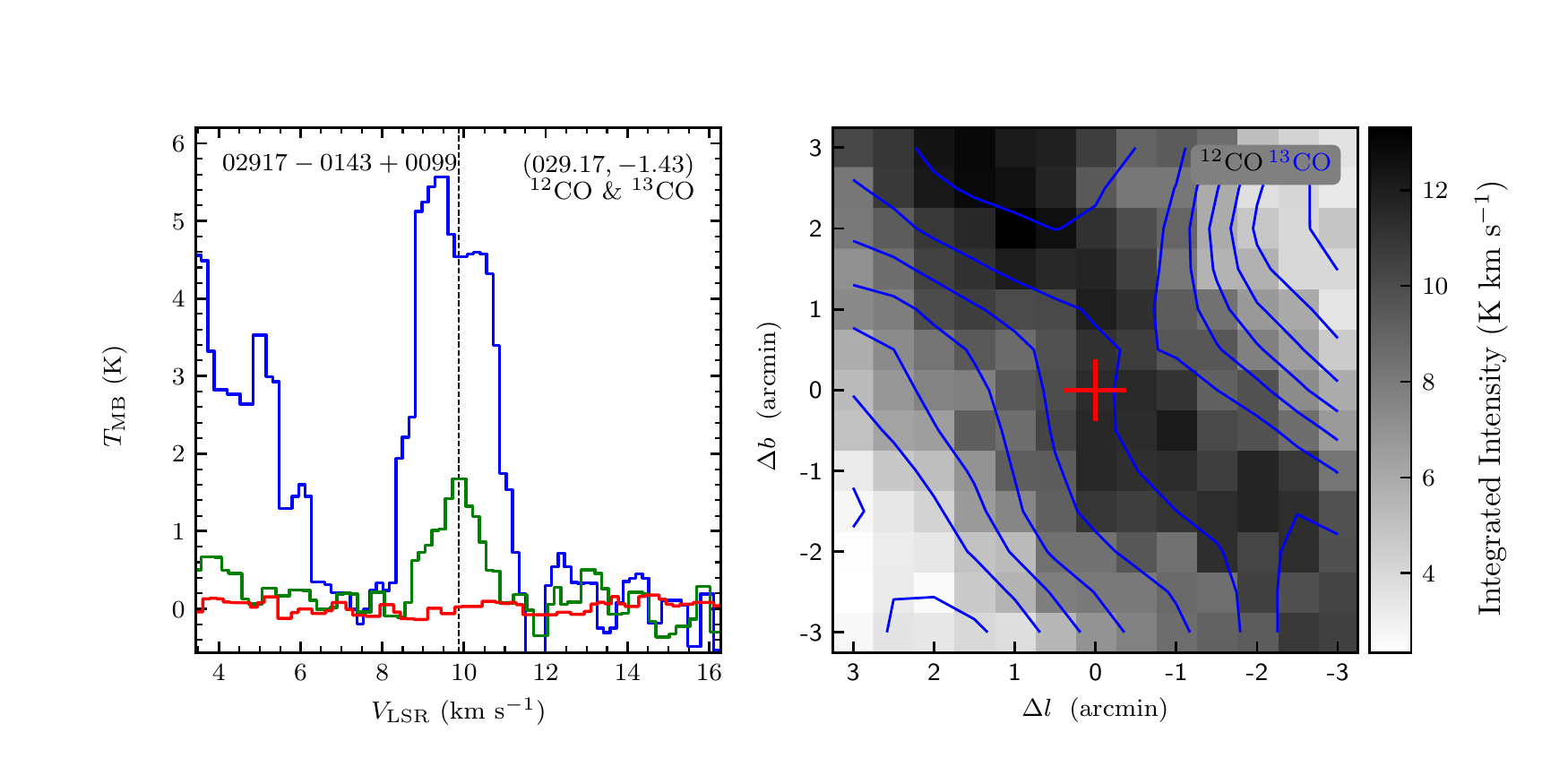}
\vspace{-0.5cm}

\includegraphics[width=9.0cm,angle=0]{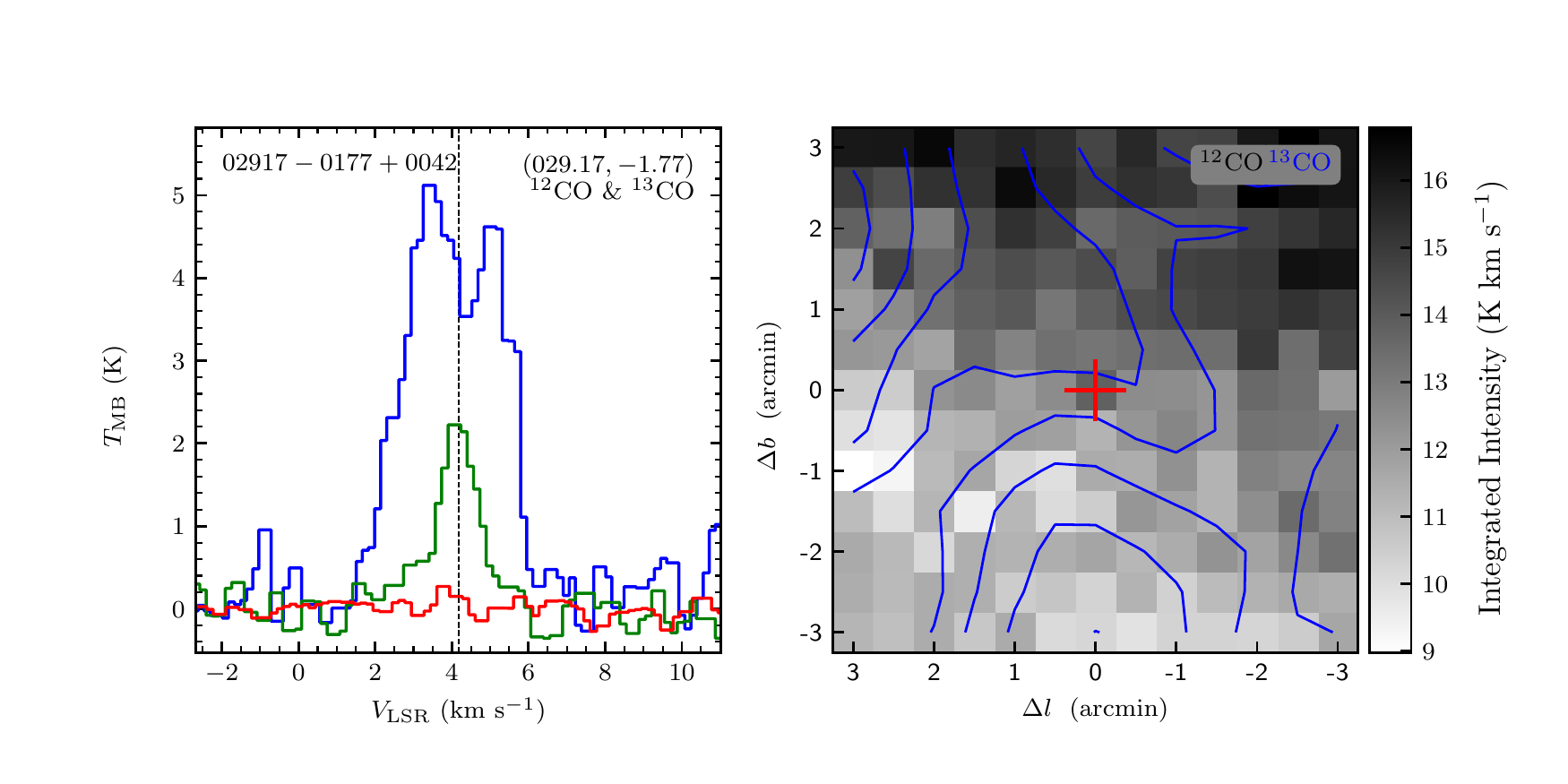}
\includegraphics[width=9.0cm,angle=0]{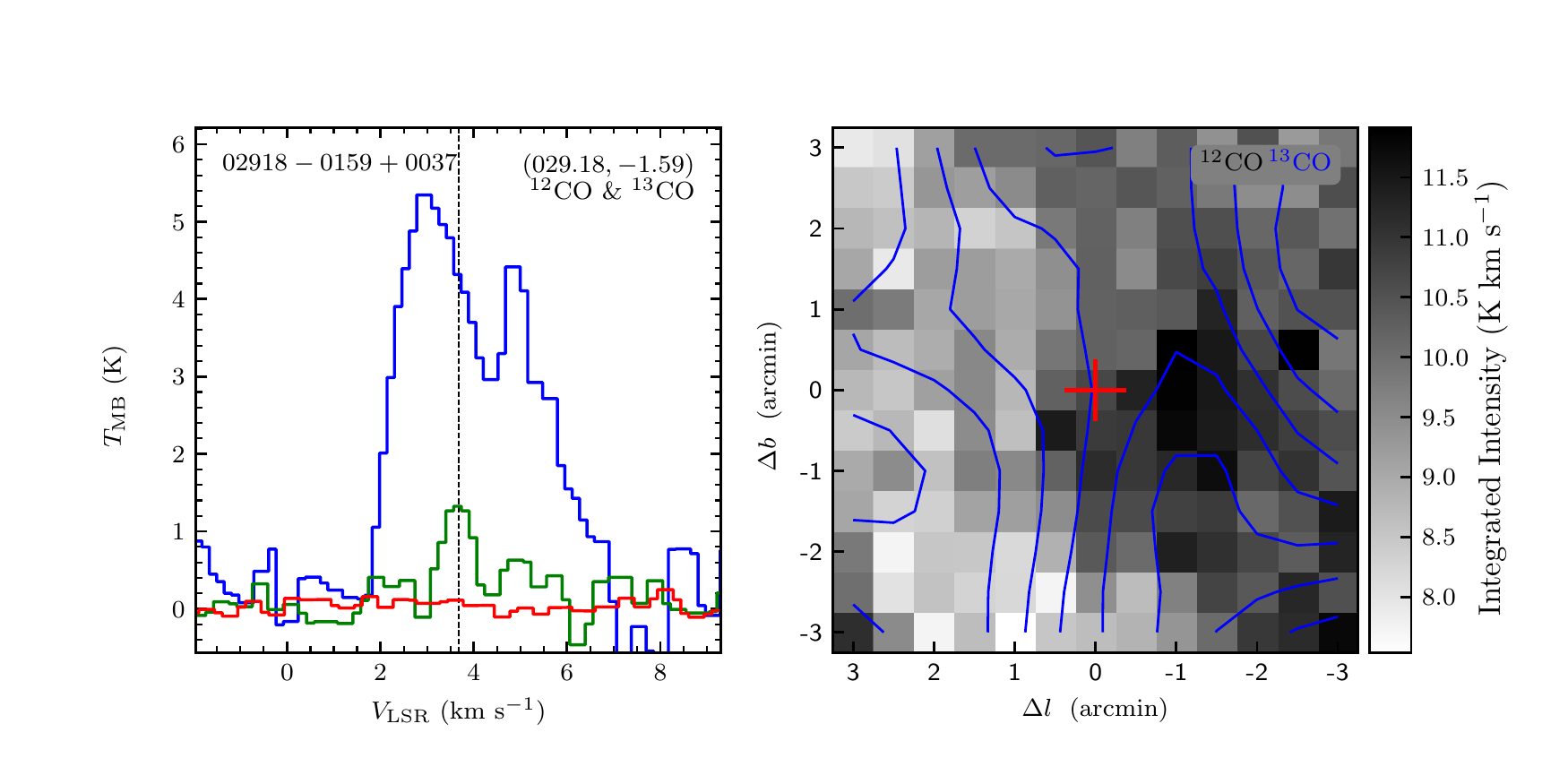}
\end{figure}
\clearpage

\begin{figure}
\includegraphics[width=9.0cm,angle=0]{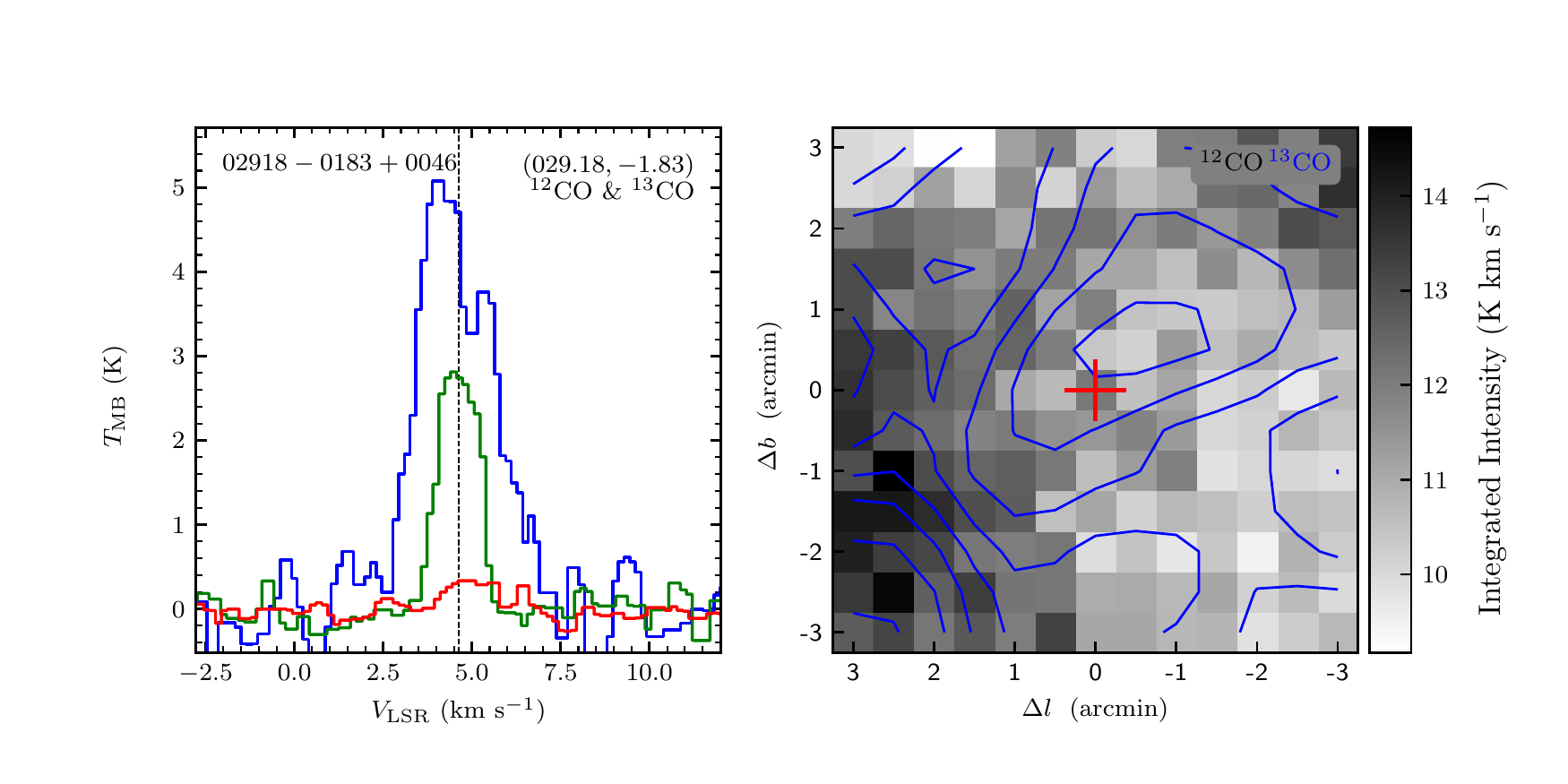}
\includegraphics[width=9.0cm,angle=0]{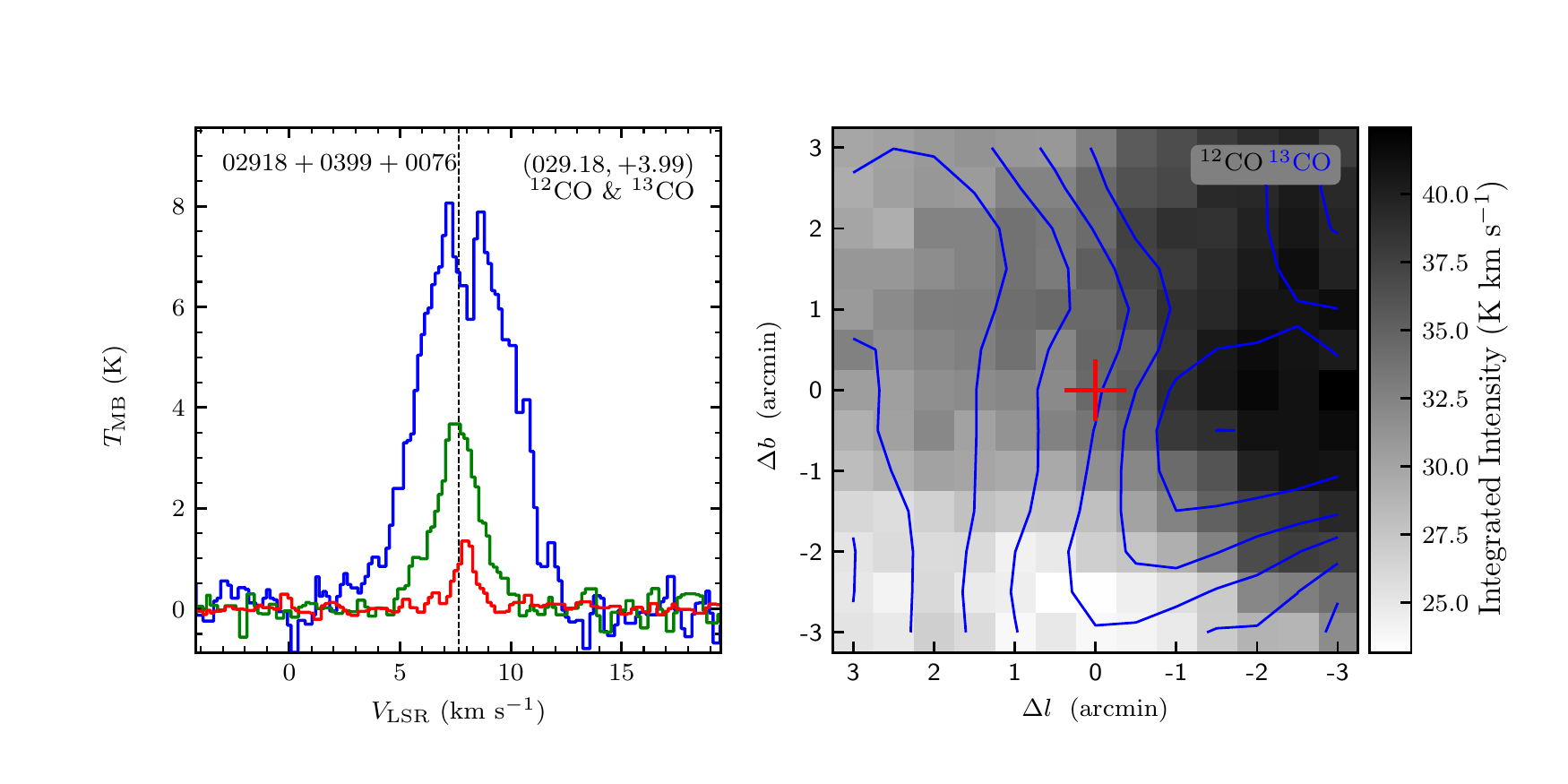}
\vspace{-0.5cm}

\includegraphics[width=9.0cm,angle=0]{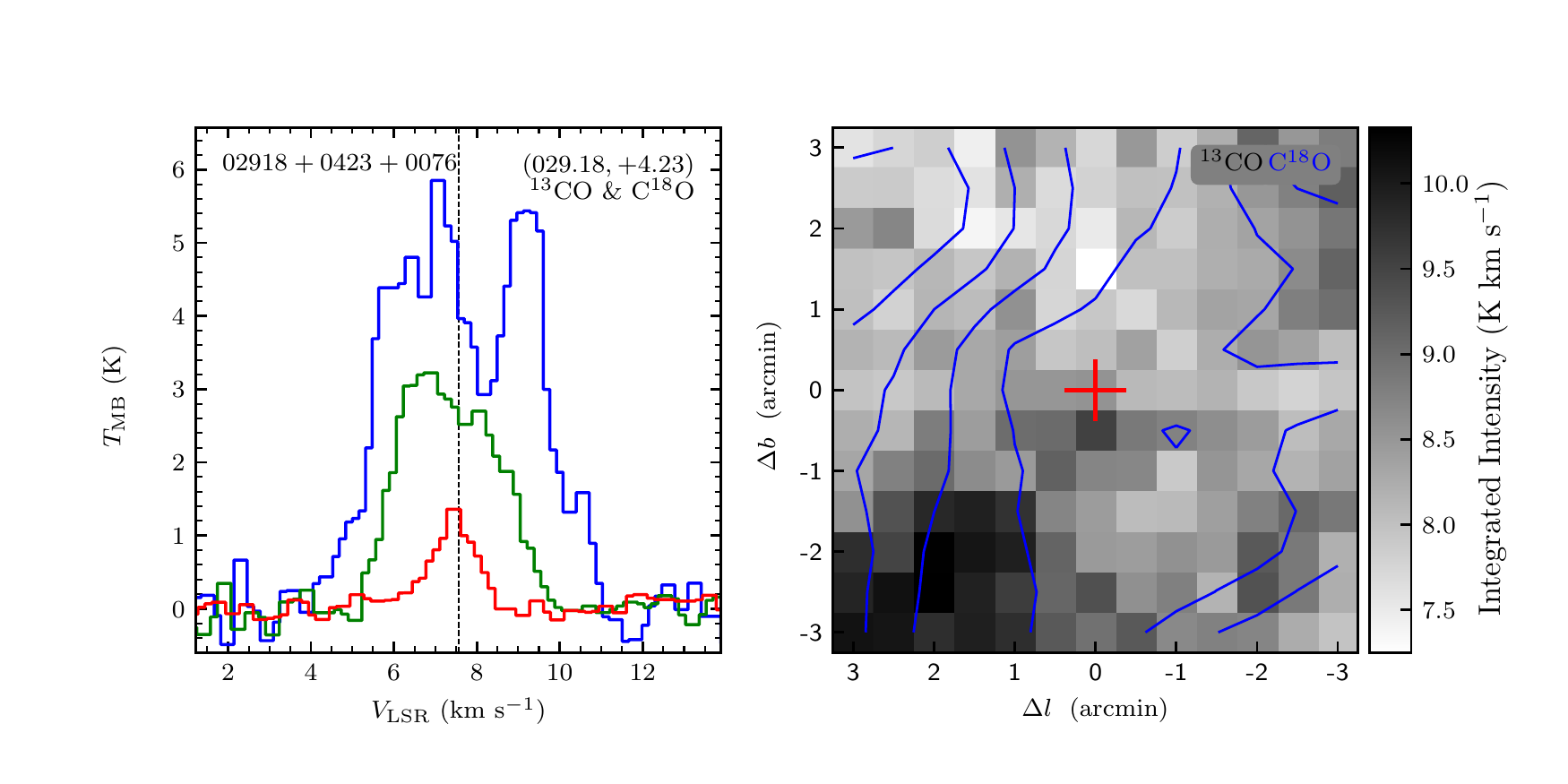}
\includegraphics[width=9.0cm,angle=0]{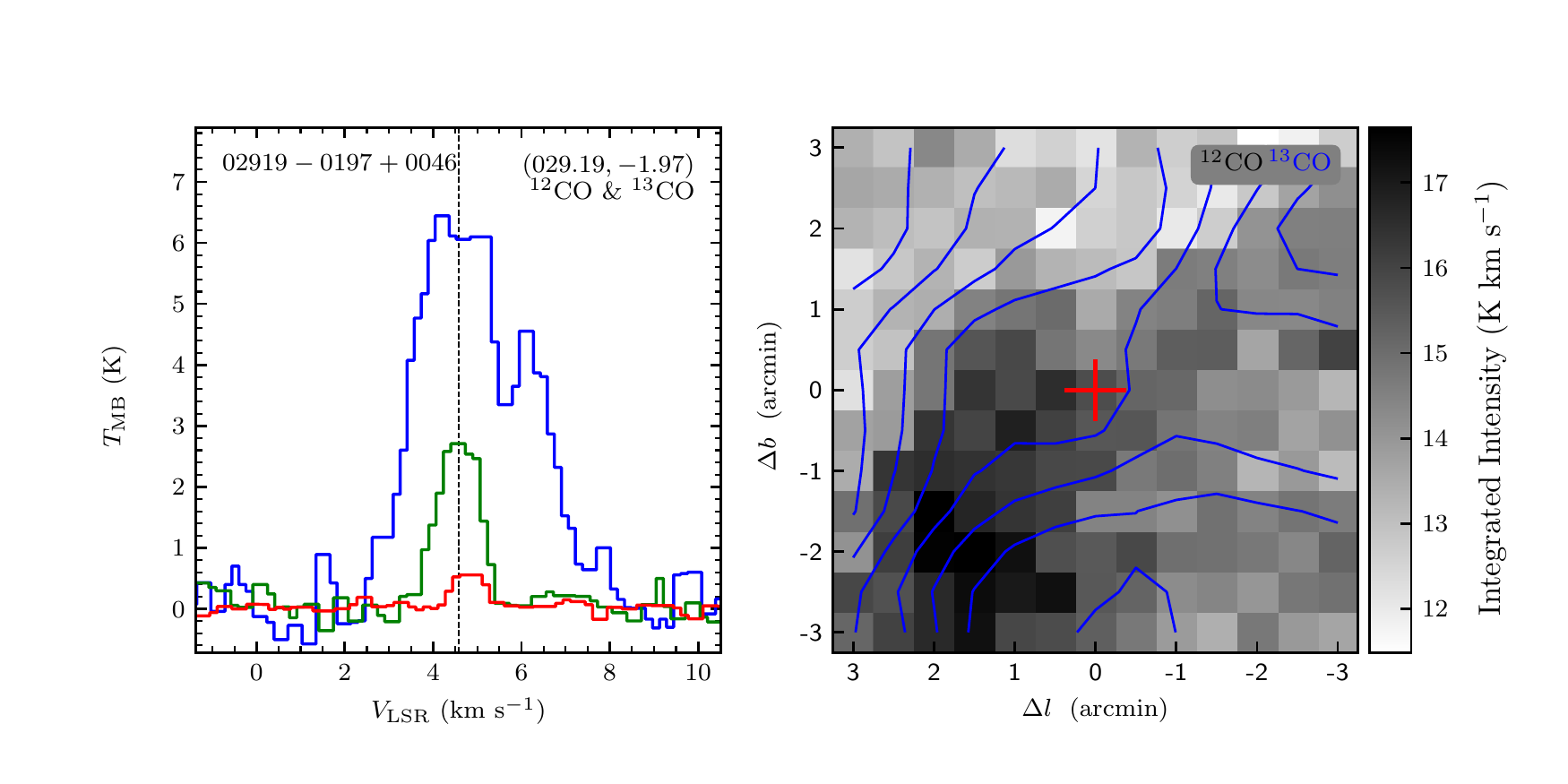}
\vspace{-0.5cm}

\includegraphics[width=9.0cm,angle=0]{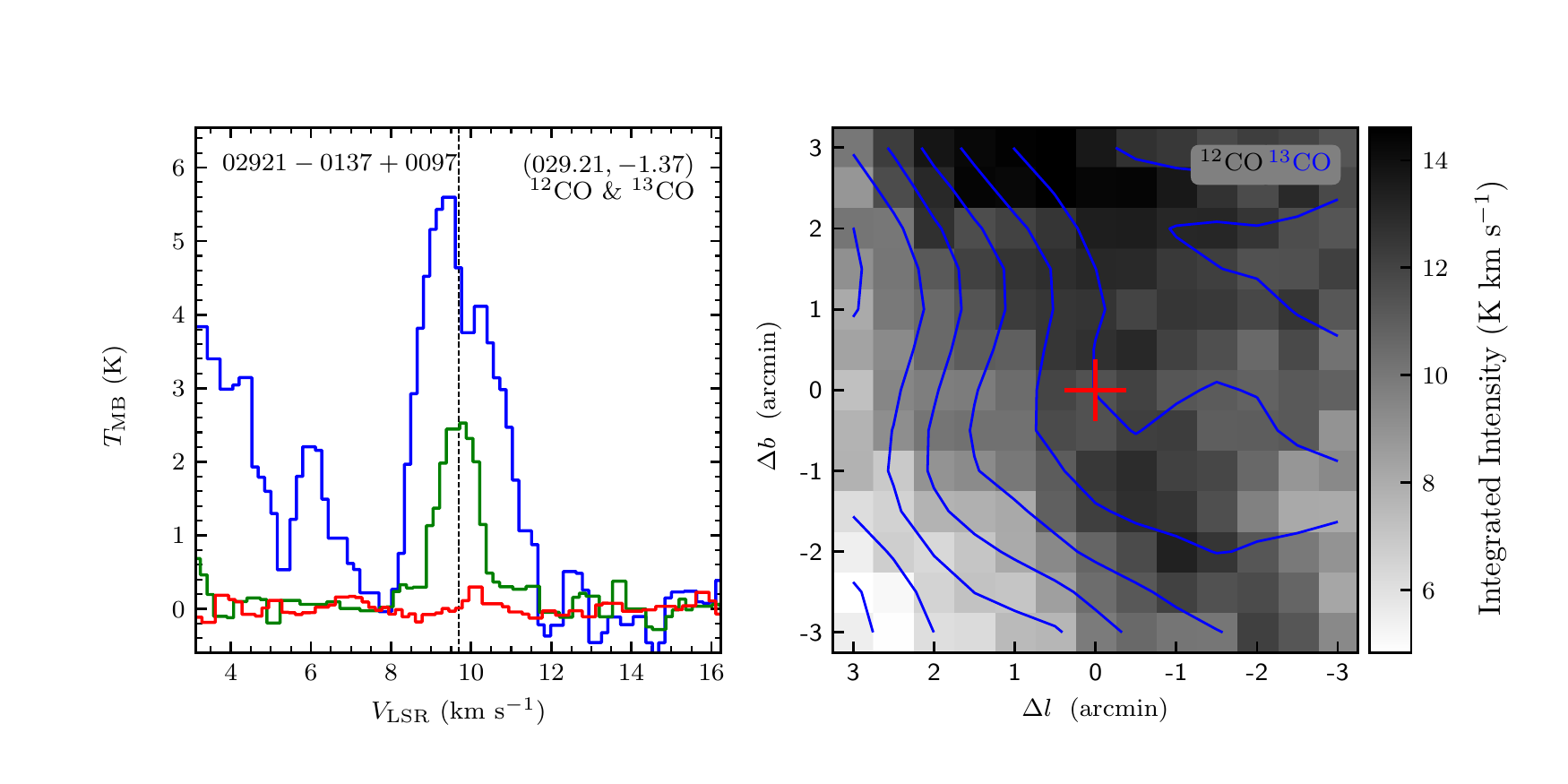}
\includegraphics[width=9.0cm,angle=0]{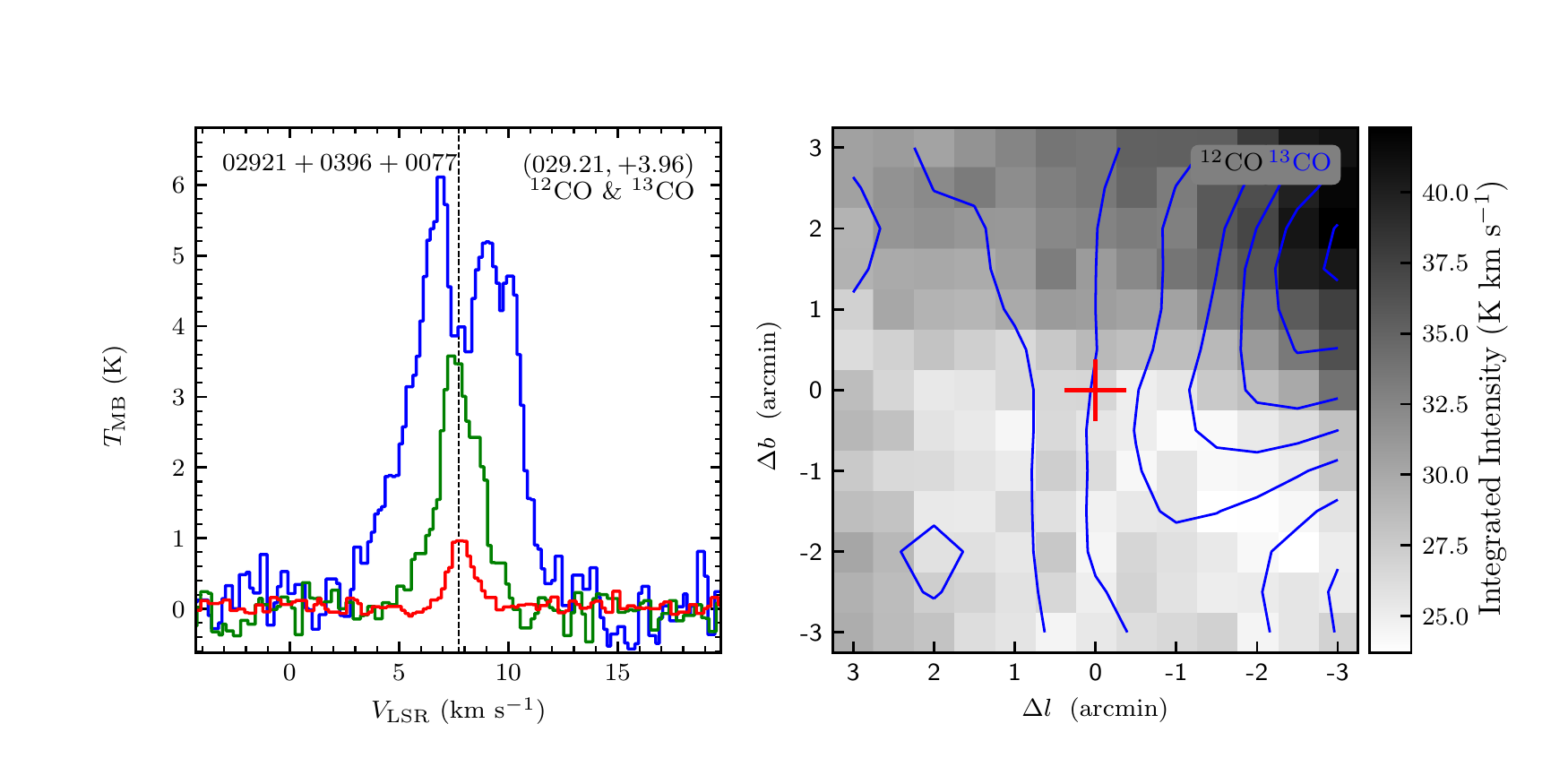}
\vspace{-0.5cm}

\includegraphics[width=9.0cm,angle=0]{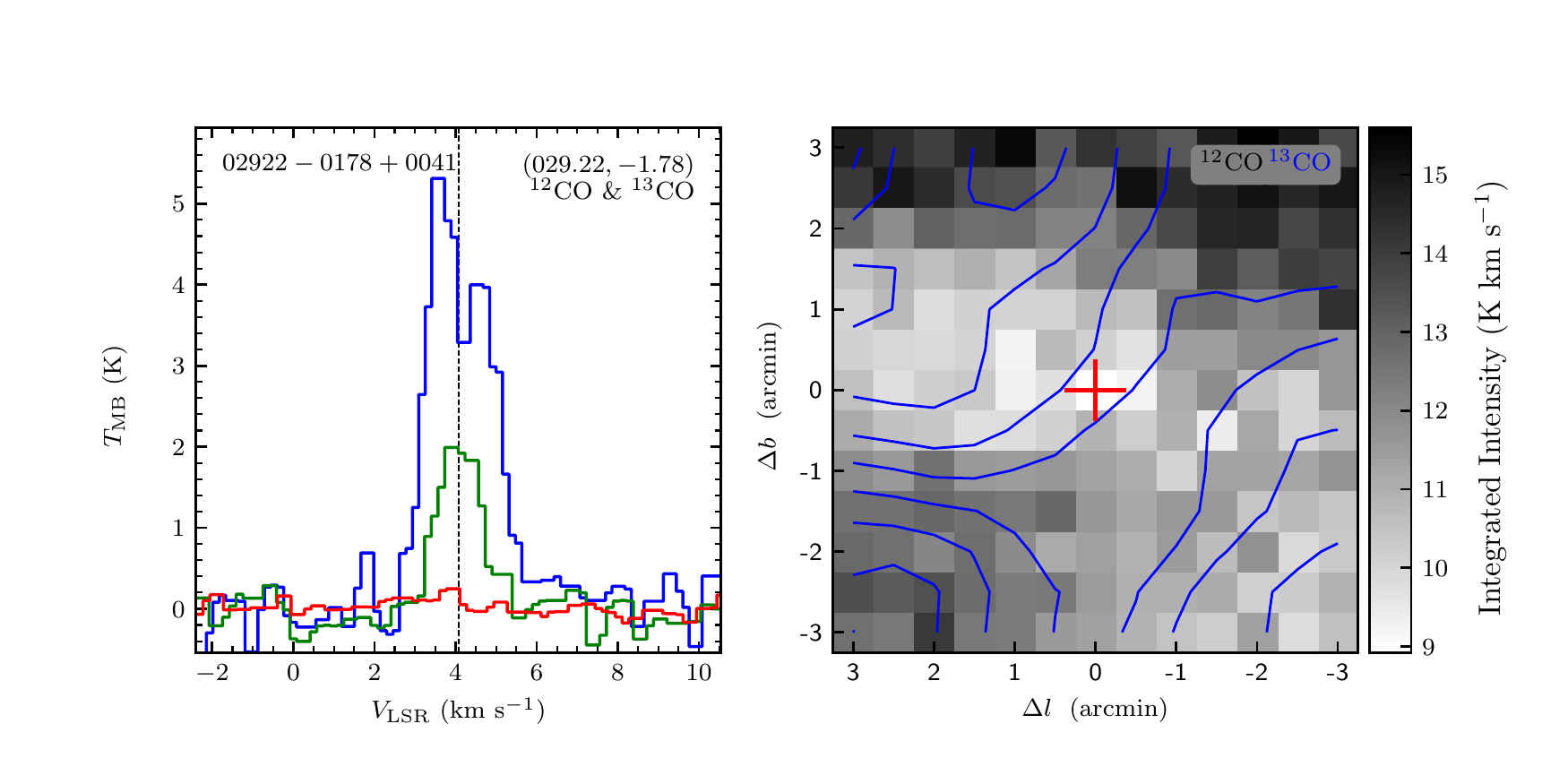}
\includegraphics[width=9.0cm,angle=0]{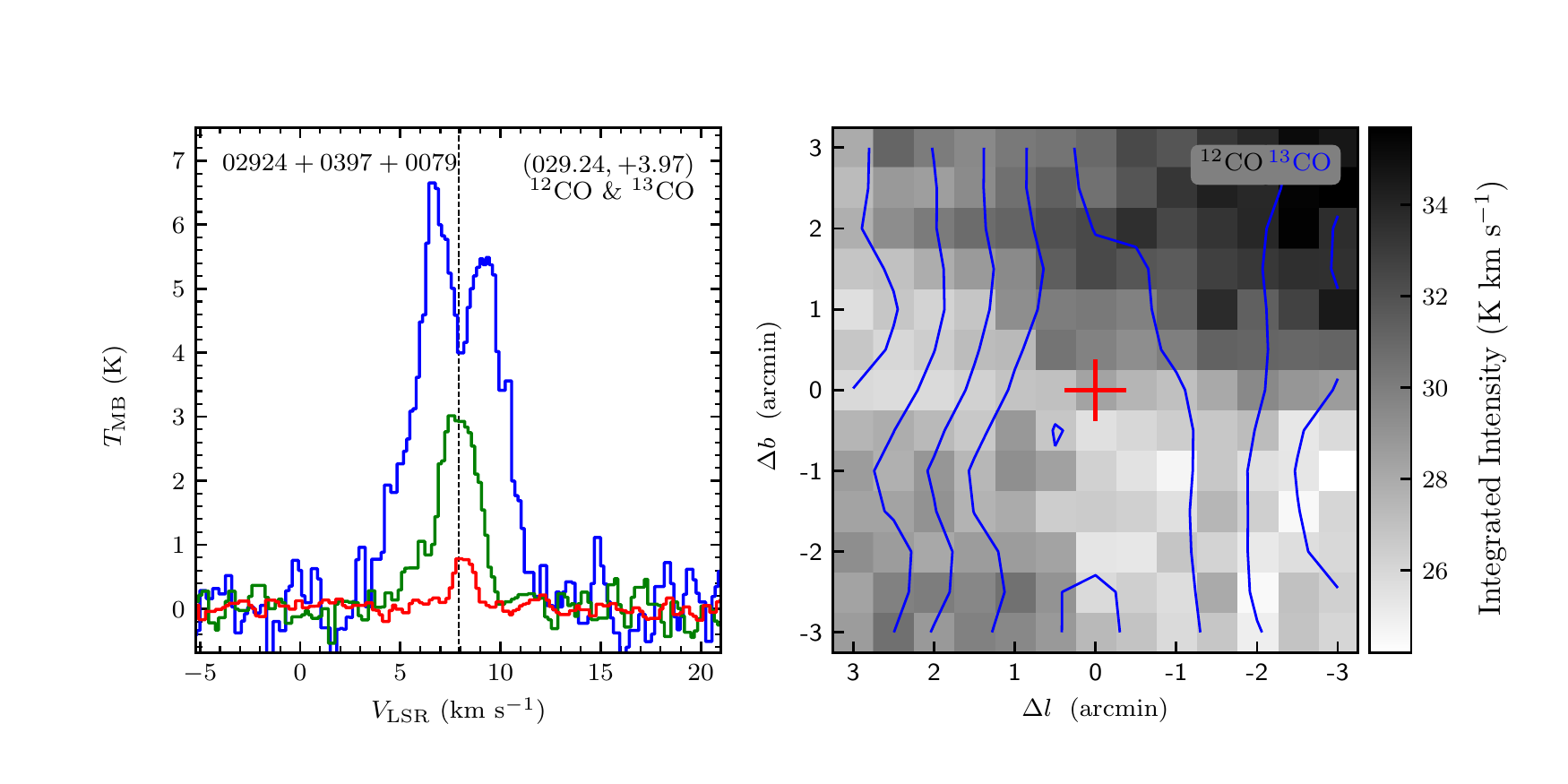}
\vspace{-0.5cm}

\includegraphics[width=9.0cm,angle=0]{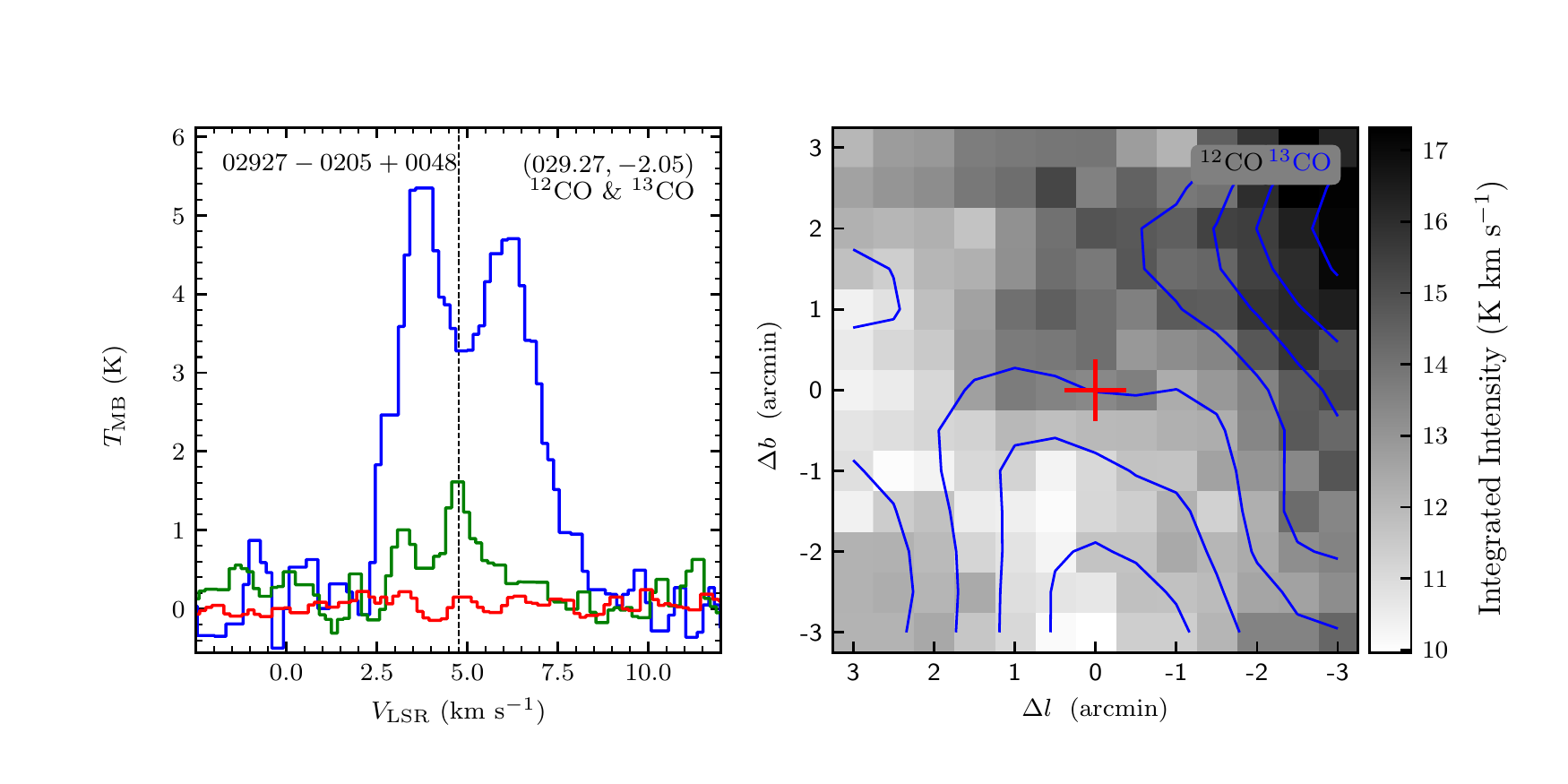}
\includegraphics[width=9.0cm,angle=0]{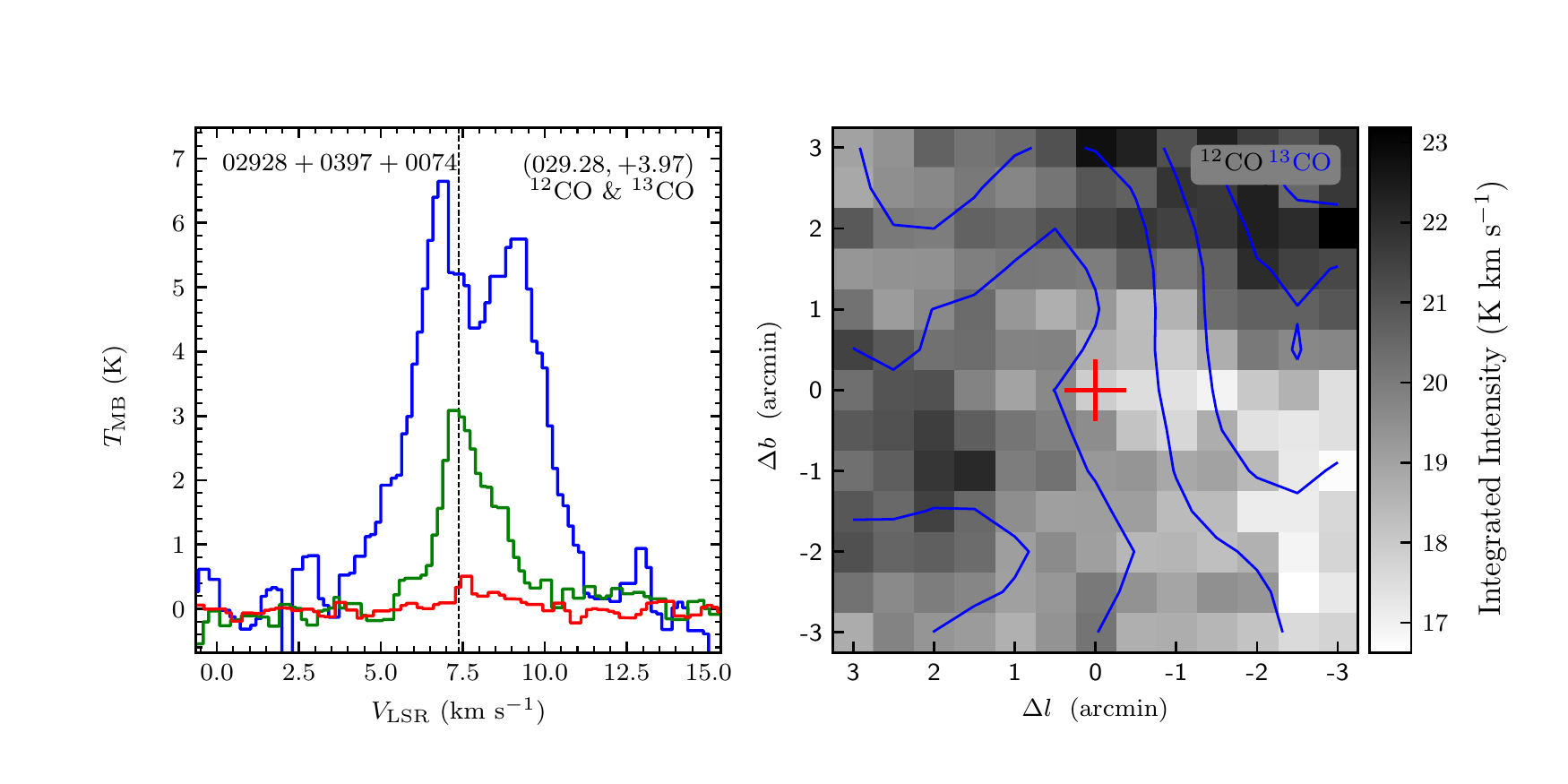}
\end{figure}
\clearpage

\begin{figure}
\includegraphics[width=9.0cm,angle=0]{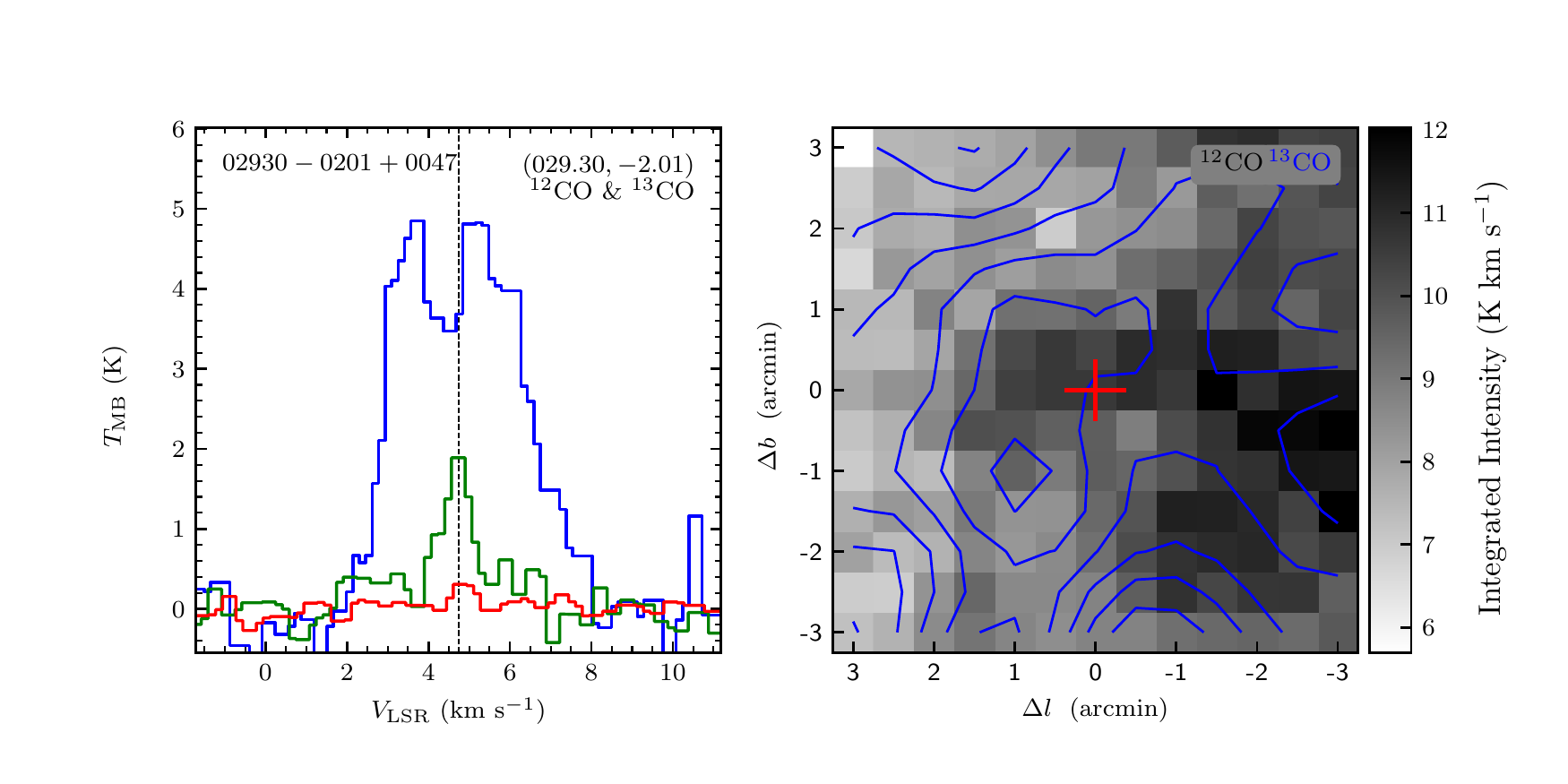}
\includegraphics[width=9.0cm,angle=0]{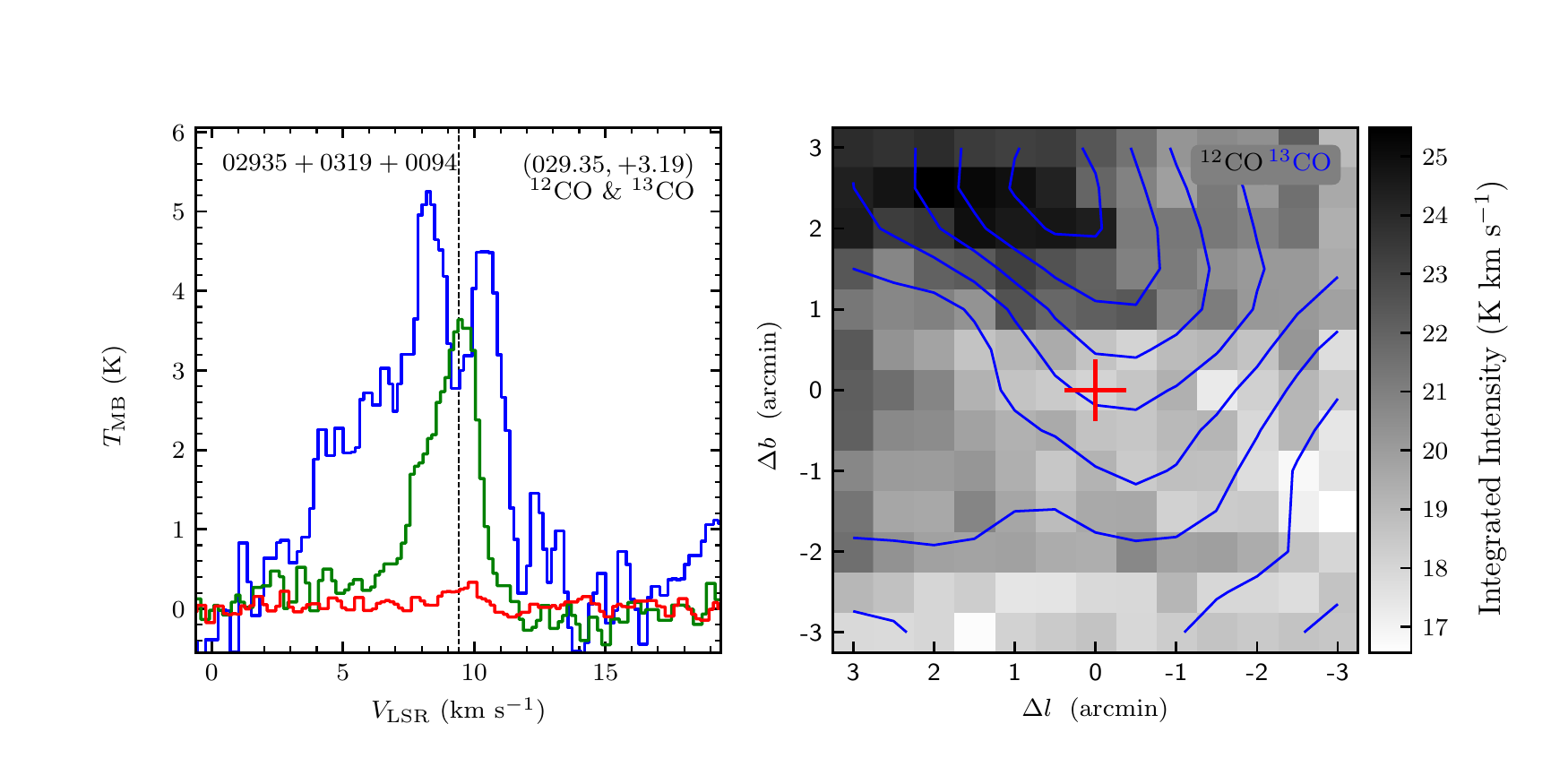}
\vspace{-0.5cm}

\includegraphics[width=9.0cm,angle=0]{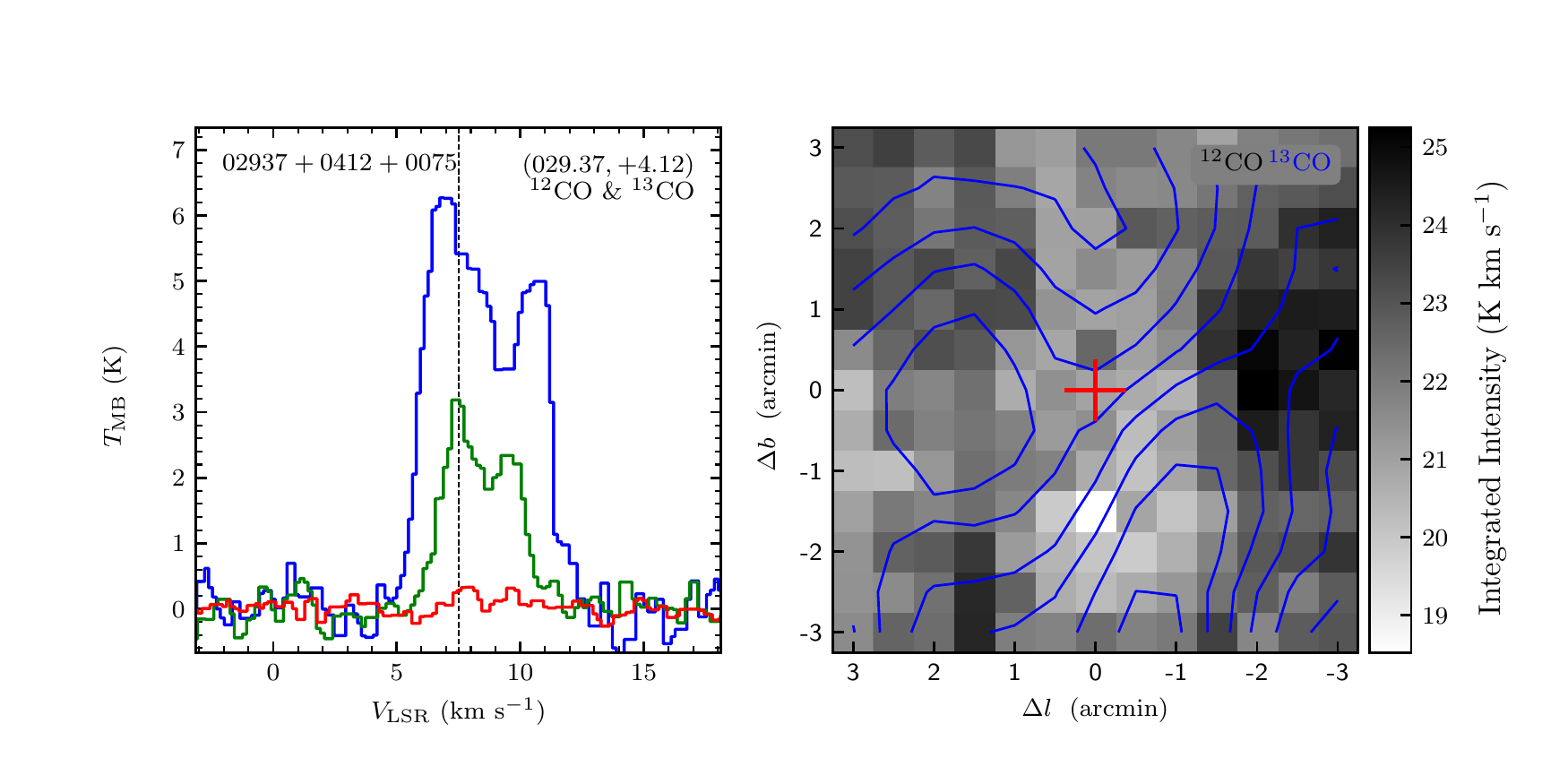}
\includegraphics[width=9.0cm,angle=0]{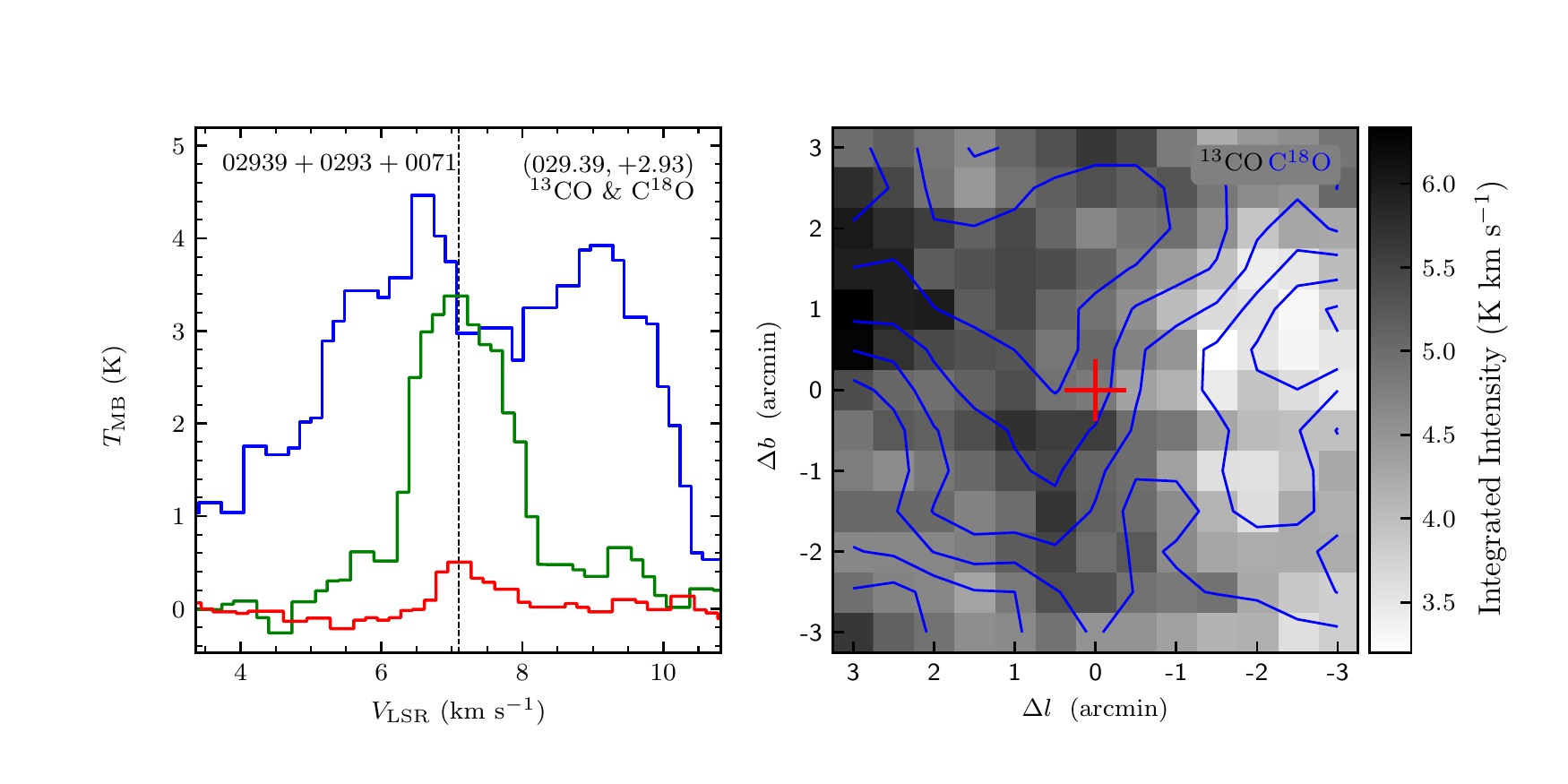}
\vspace{-0.5cm}

\includegraphics[width=9.0cm,angle=0]{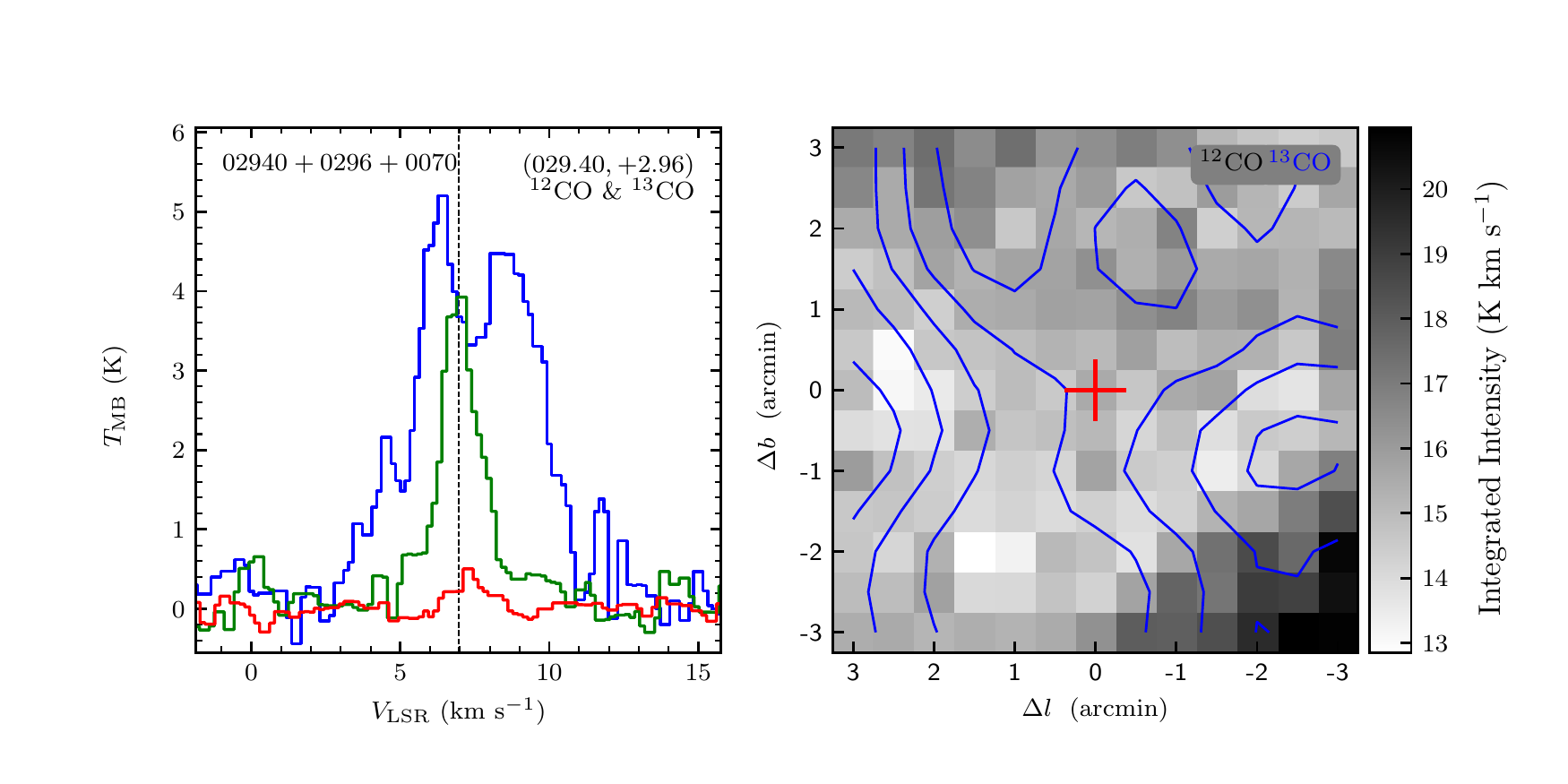}
\includegraphics[width=9.0cm,angle=0]{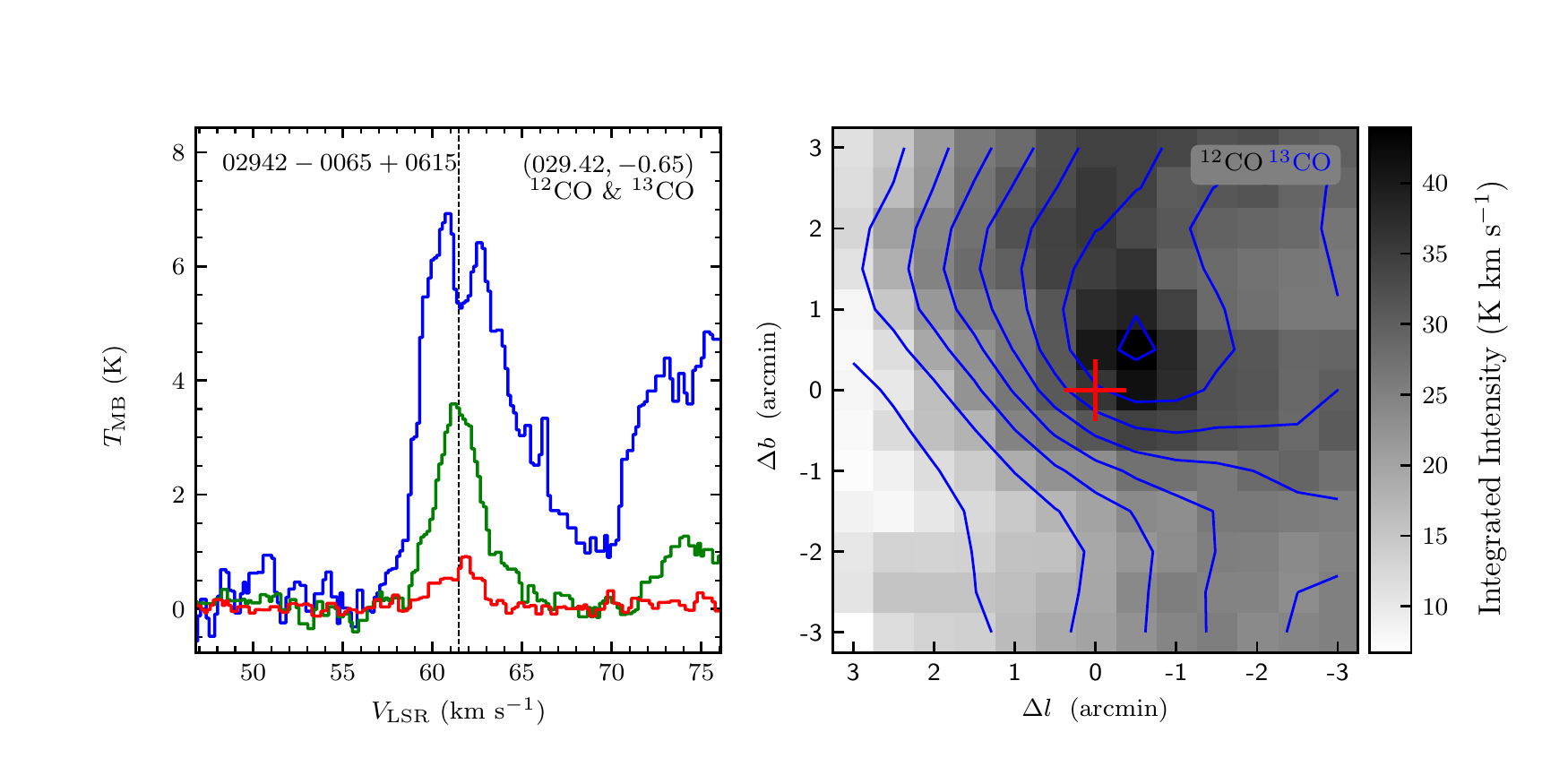}
\vspace{-0.5cm}

\includegraphics[width=9.0cm,angle=0]{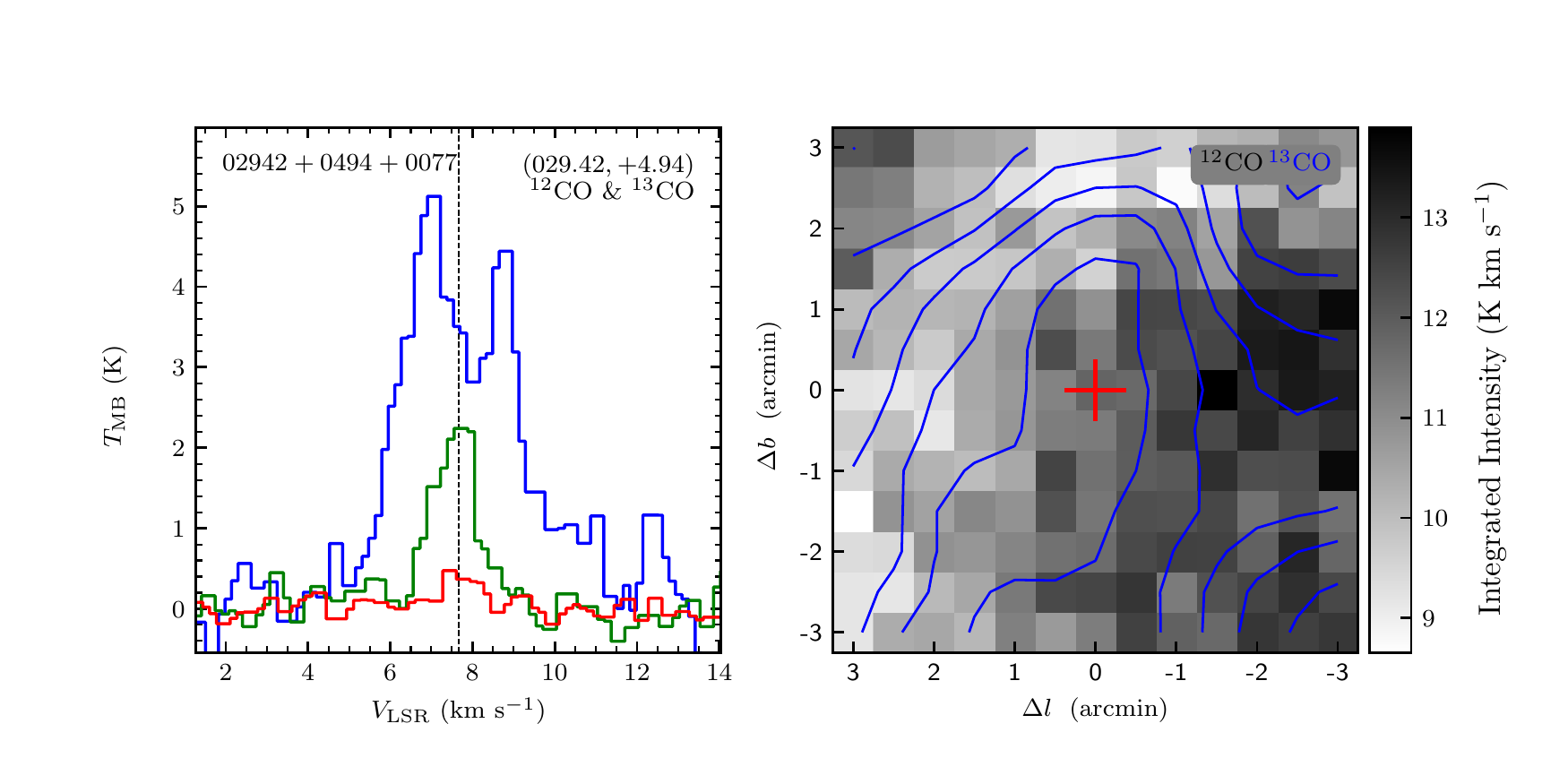}
\includegraphics[width=9.0cm,angle=0]{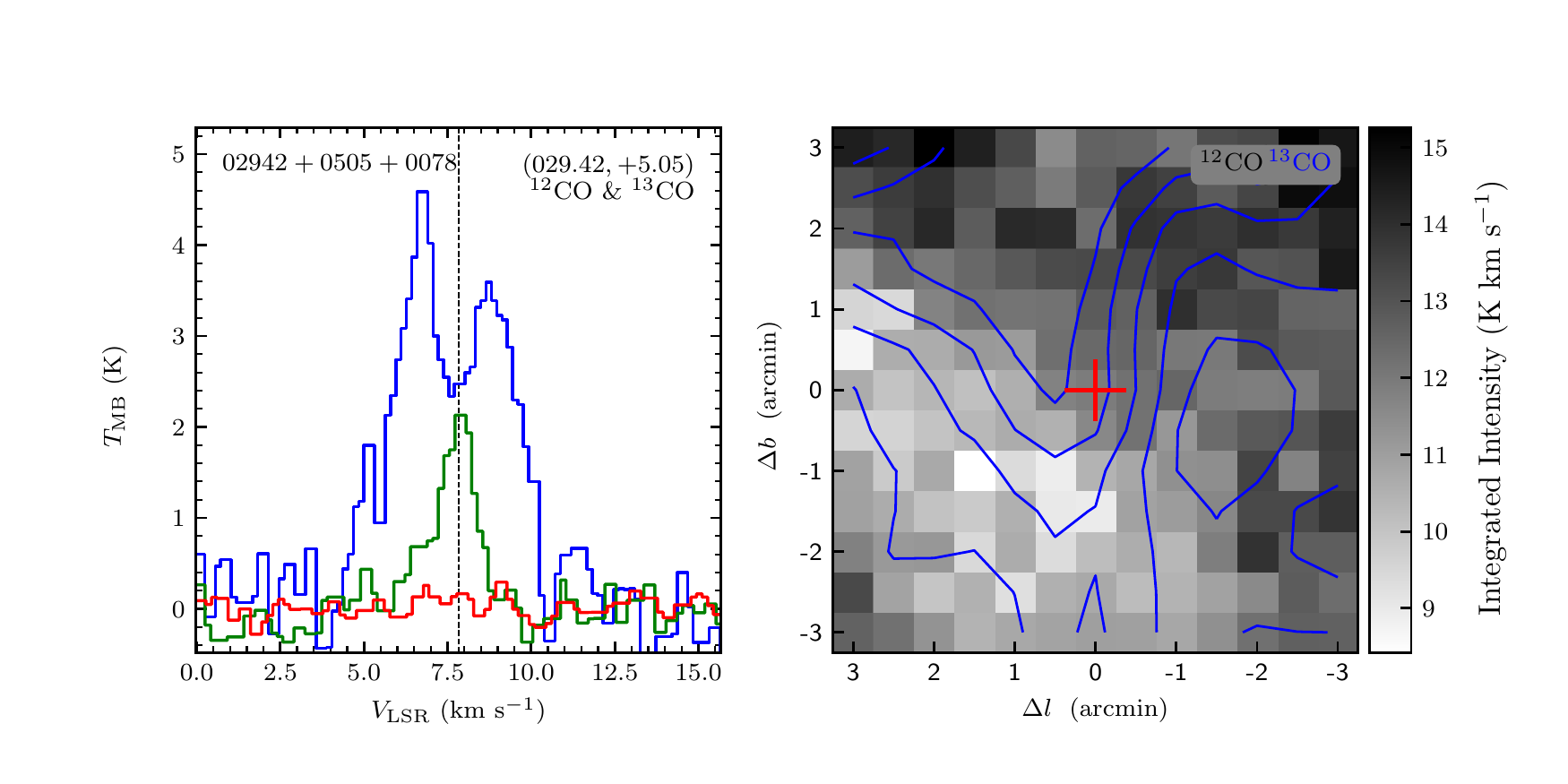}
\vspace{-0.5cm}

\includegraphics[width=9.0cm,angle=0]{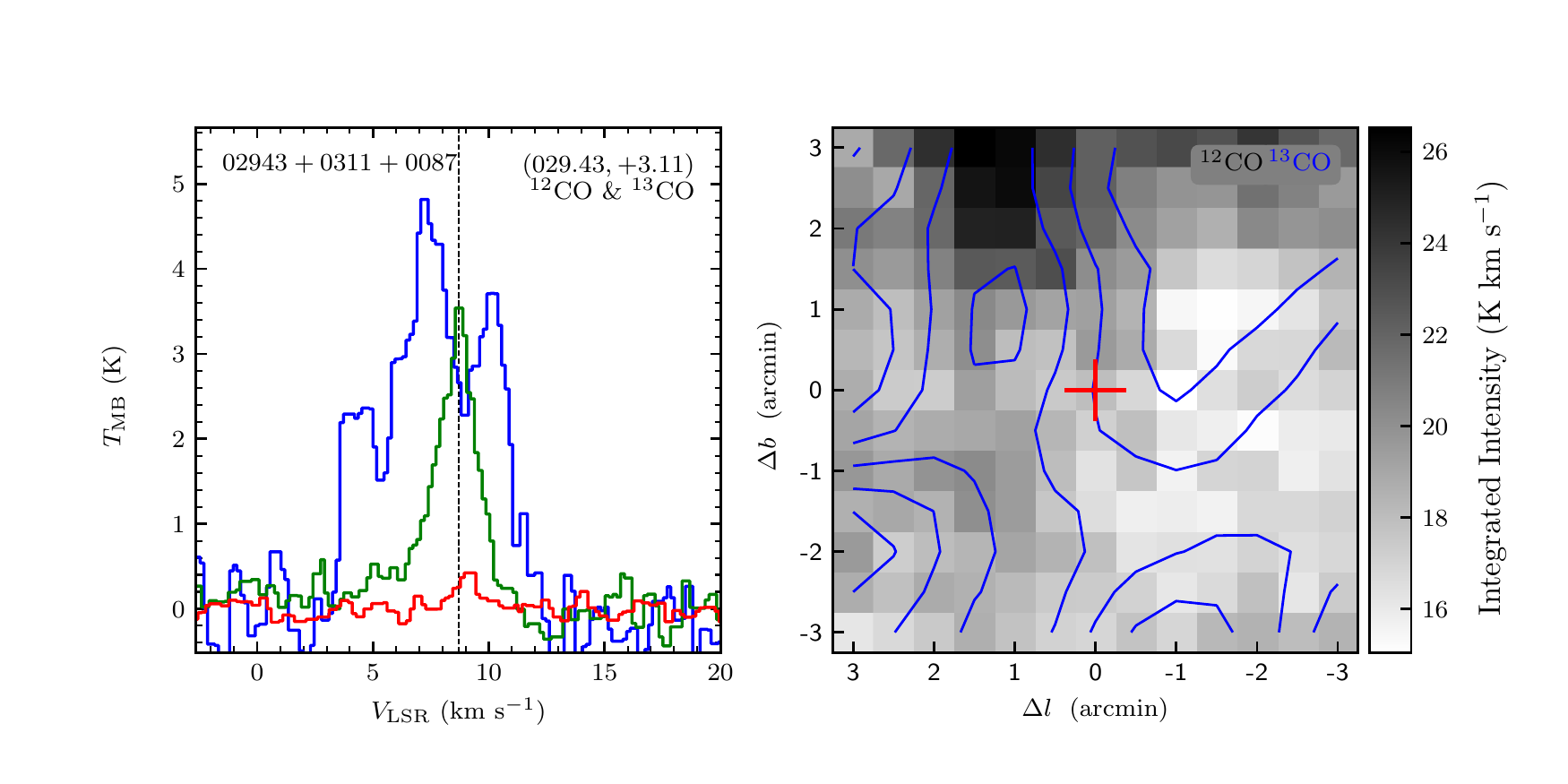}
\includegraphics[width=9.0cm,angle=0]{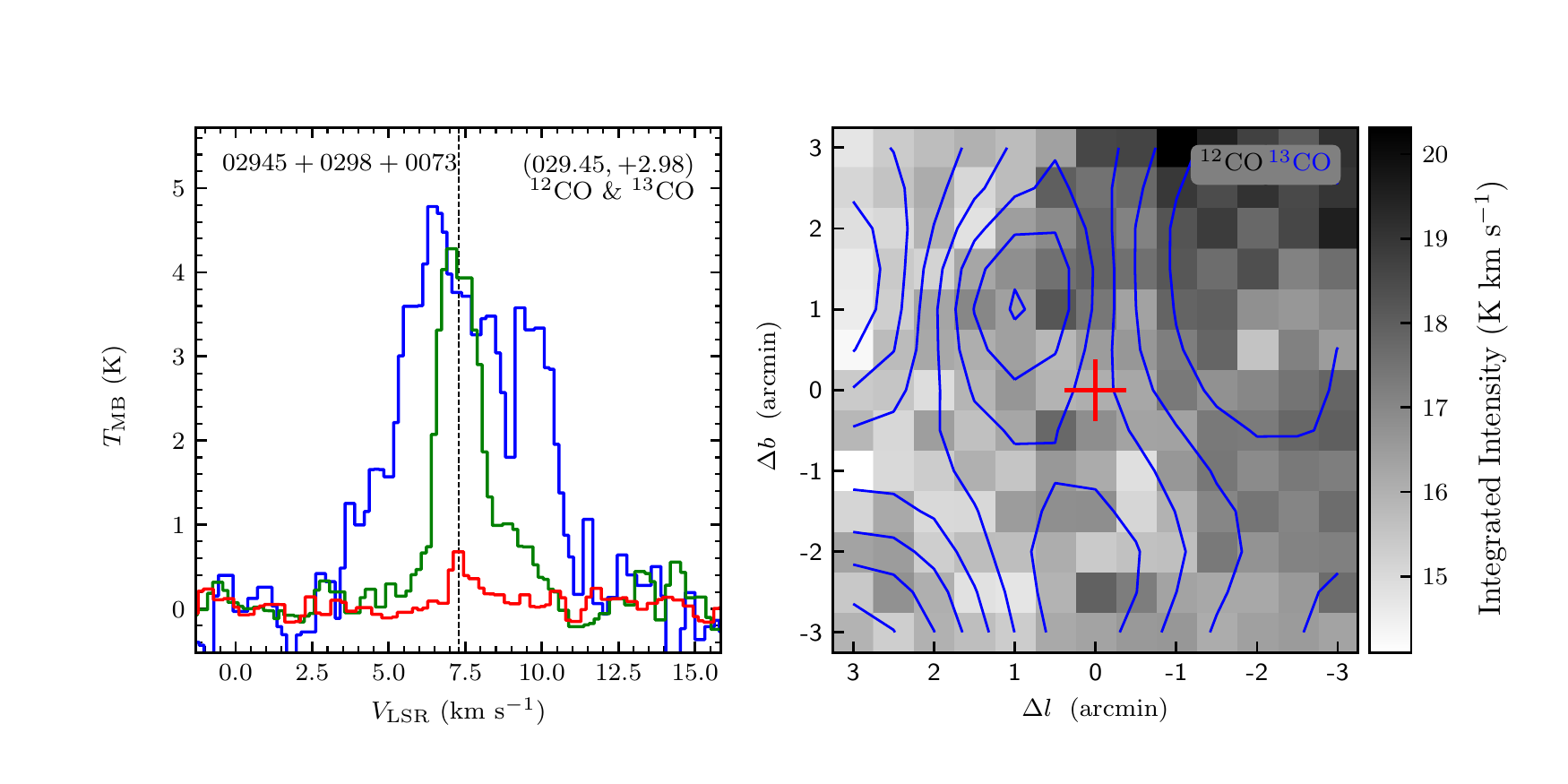}
\end{figure}
\clearpage

\begin{figure}
\includegraphics[width=9.0cm,angle=0]{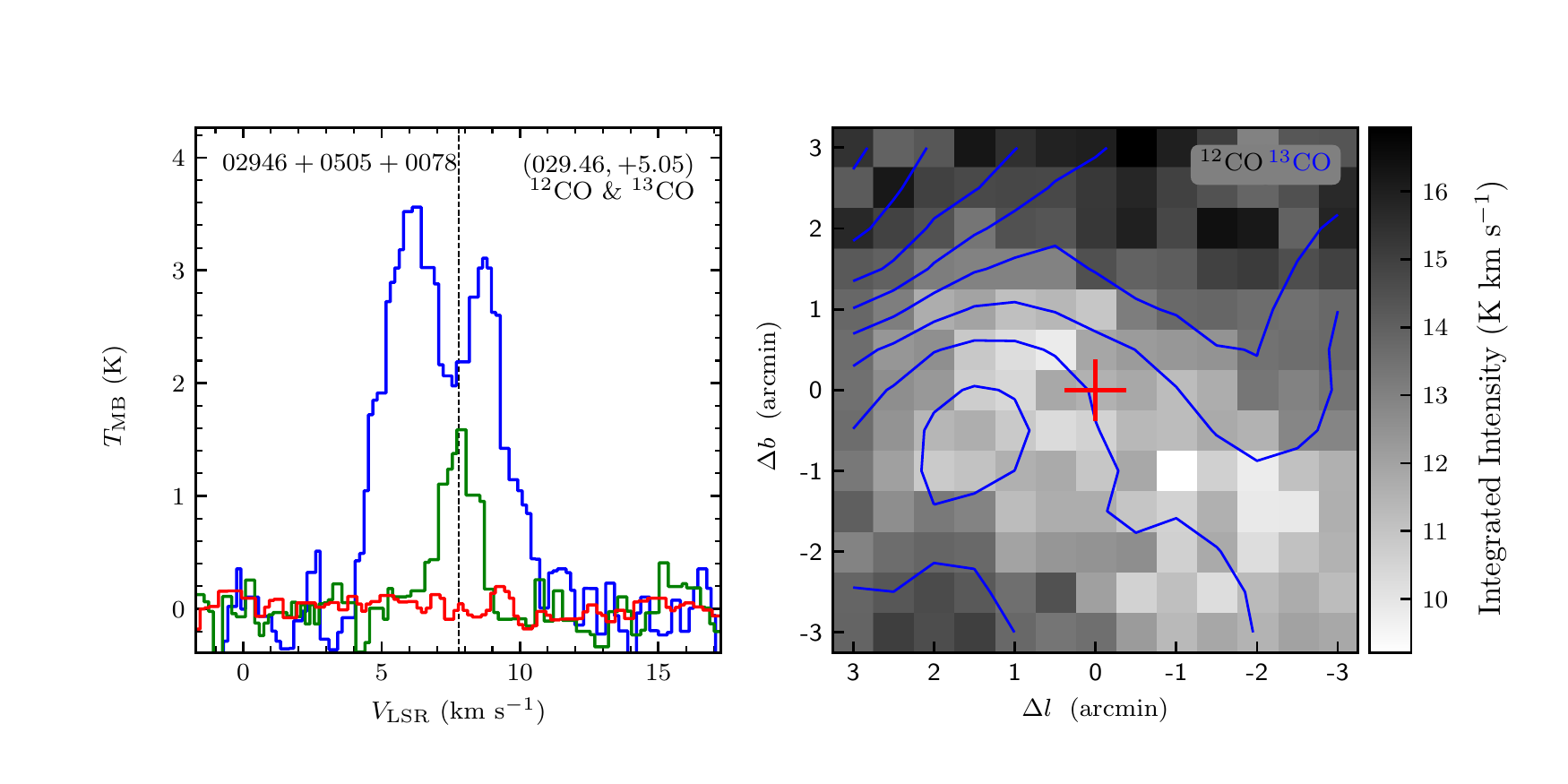}
\includegraphics[width=9.0cm,angle=0]{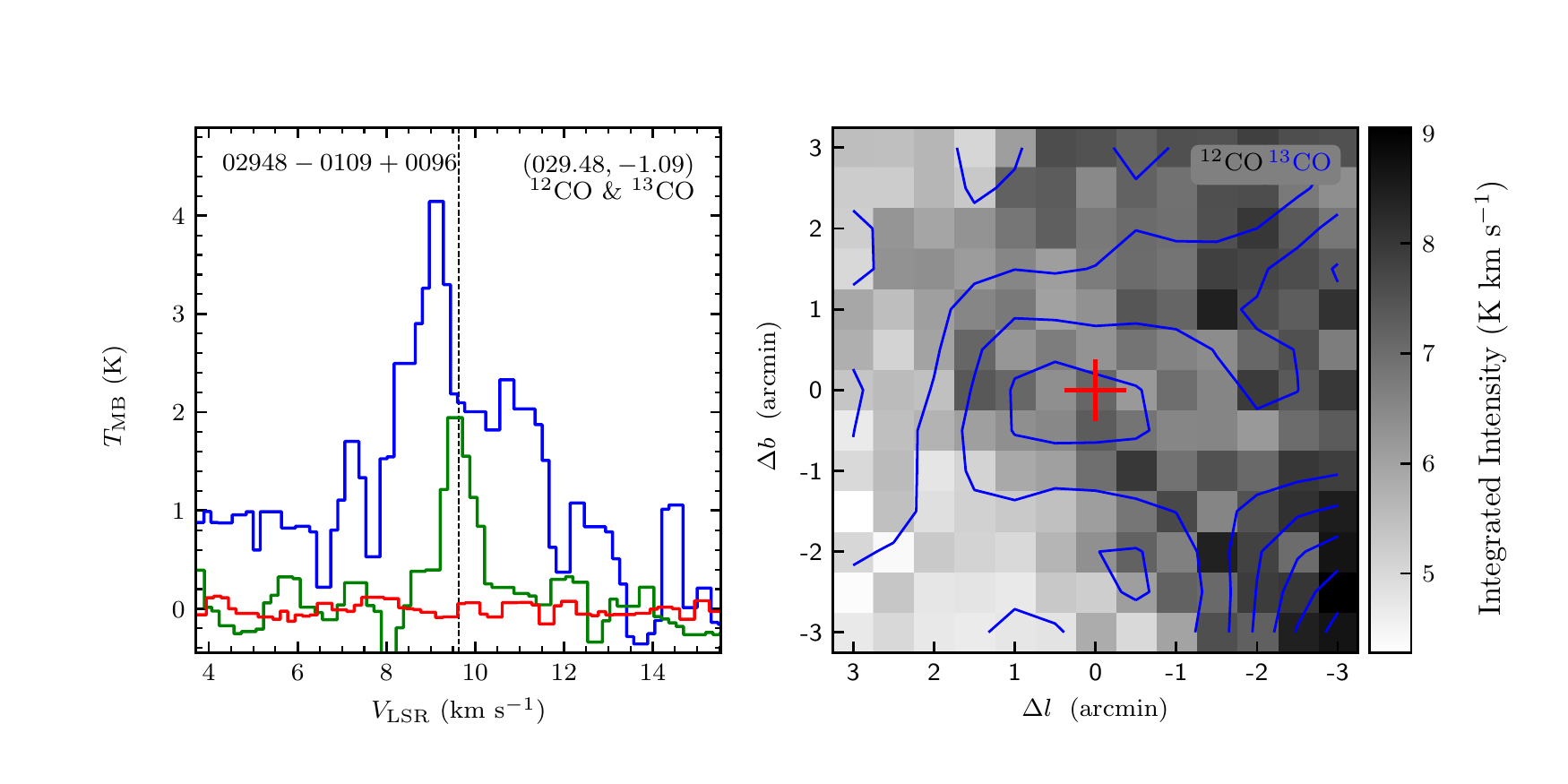}
\vspace{-0.5cm}

\includegraphics[width=9.0cm,angle=0]{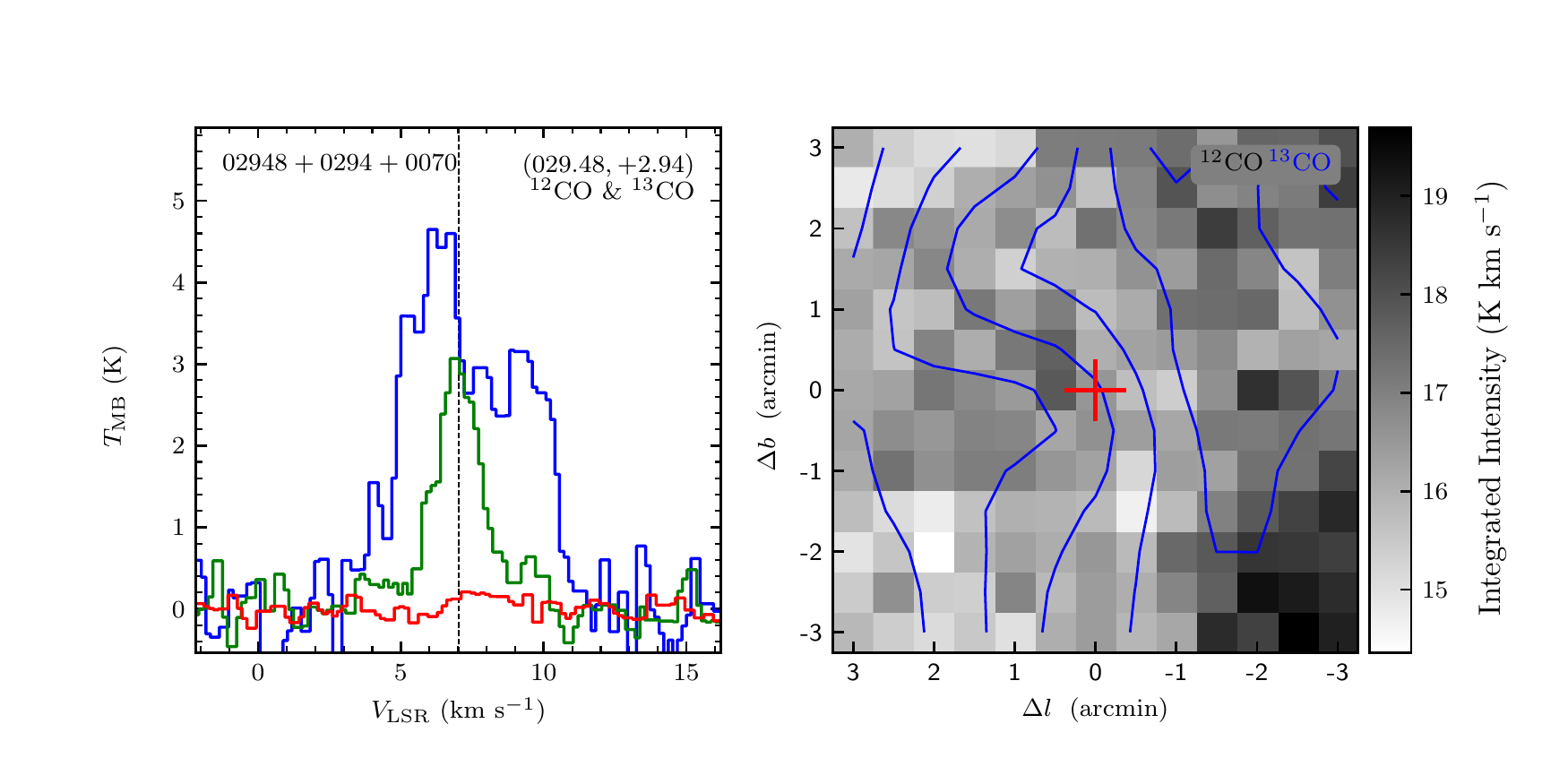}
\includegraphics[width=9.0cm,angle=0]{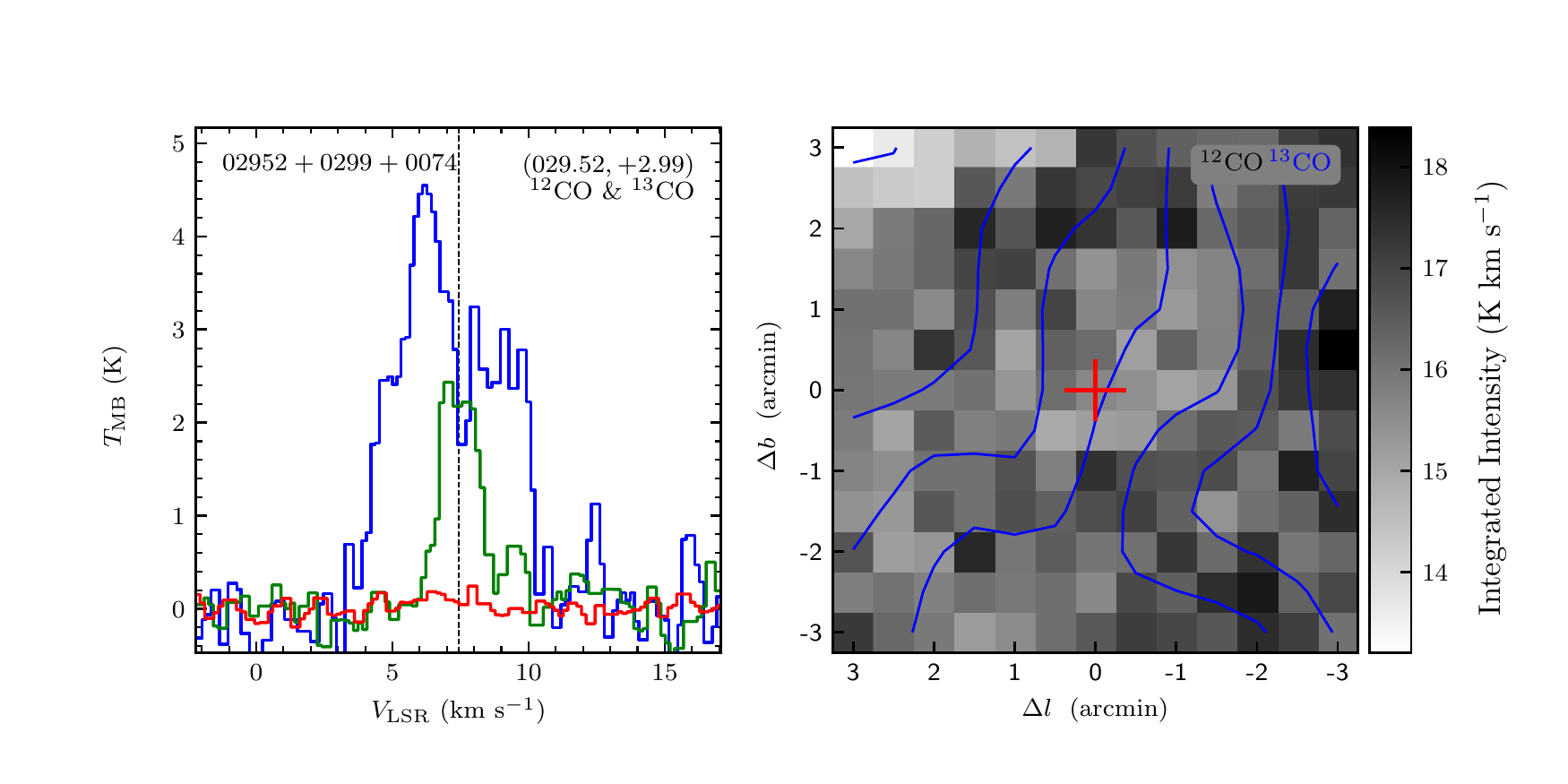}
\vspace{-0.5cm}

\includegraphics[width=9.0cm,angle=0]{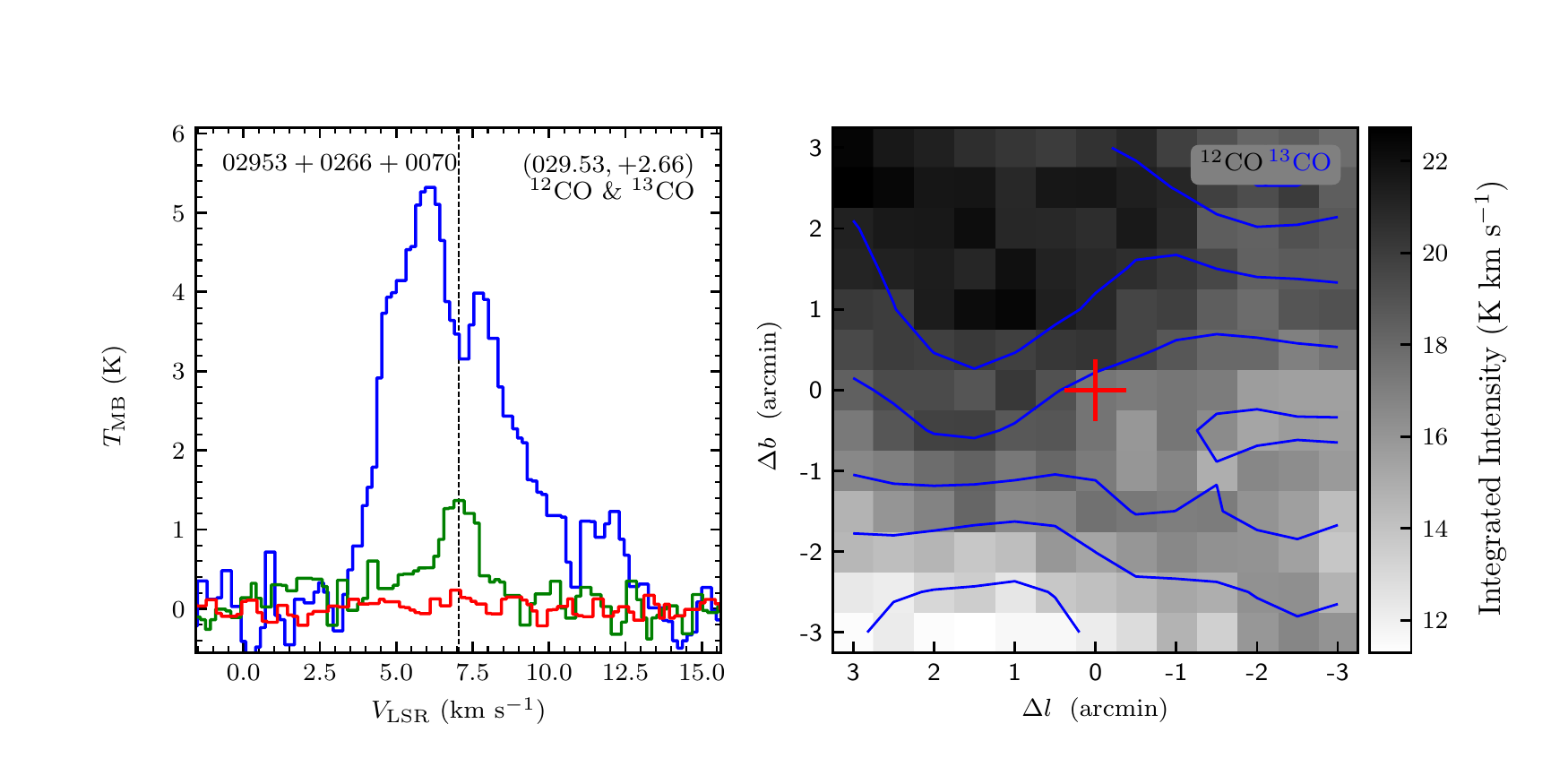}
\includegraphics[width=9.0cm,angle=0]{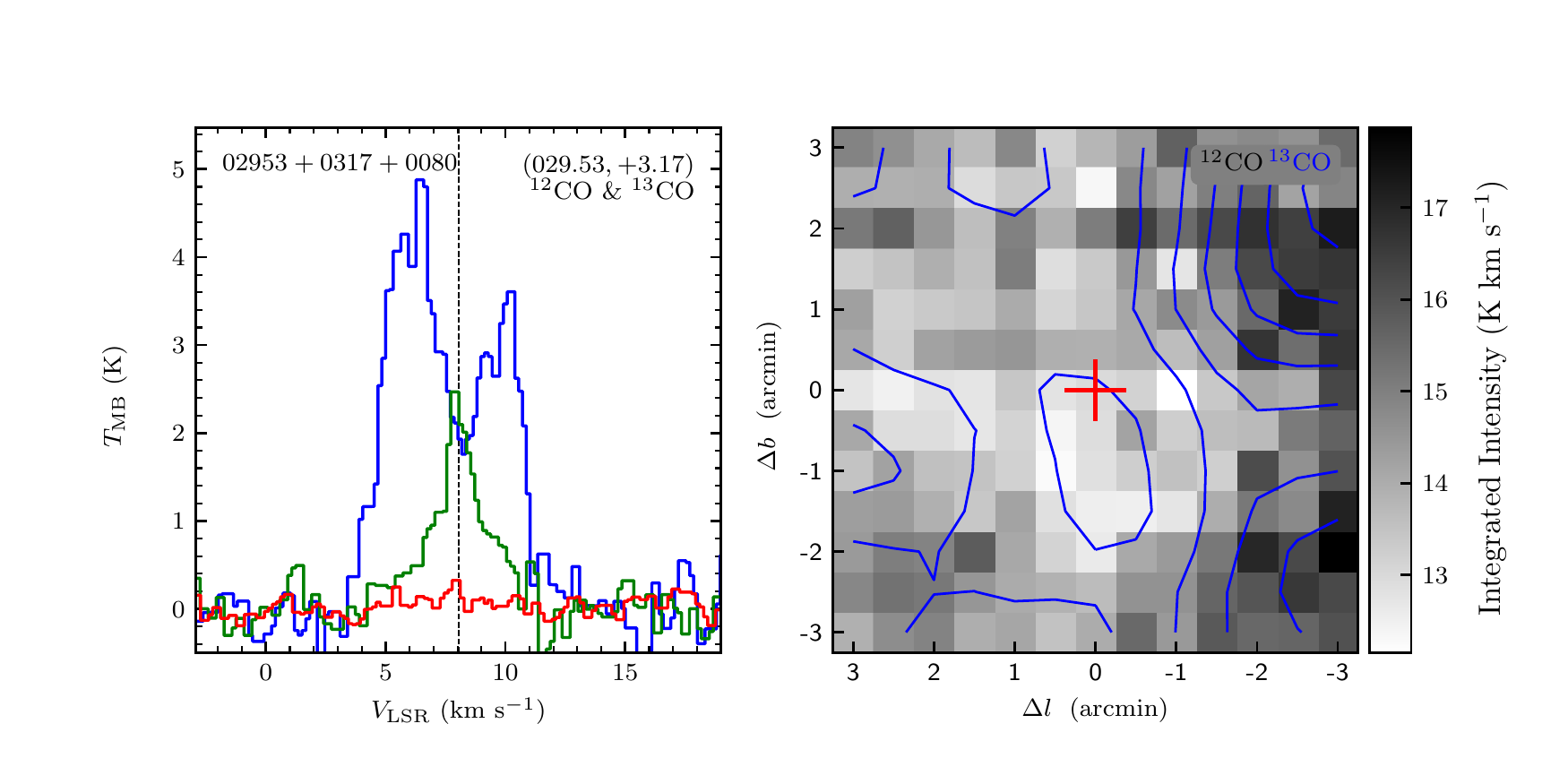}
\vspace{-0.5cm}

\includegraphics[width=9.0cm,angle=0]{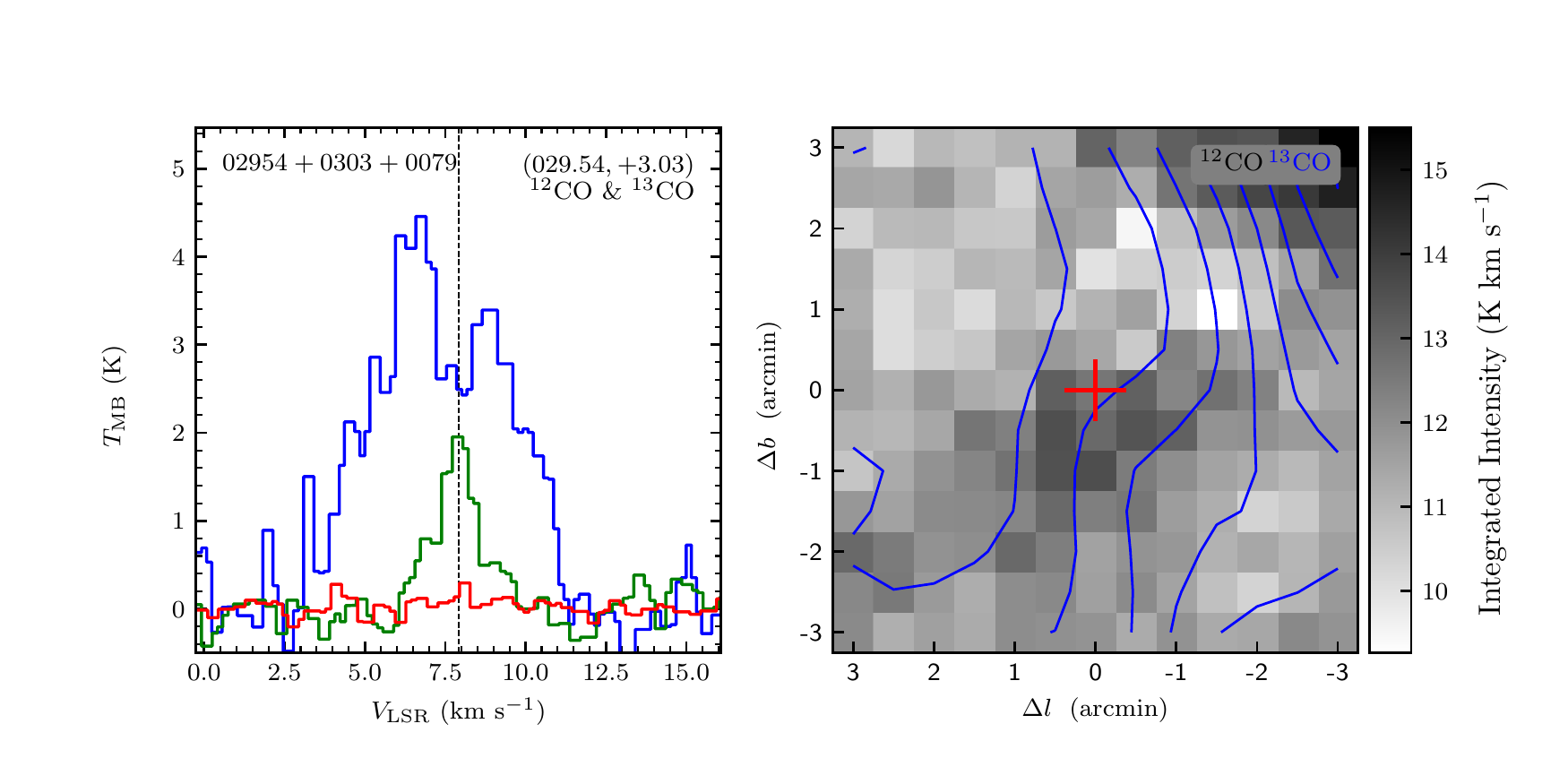}
\includegraphics[width=9.0cm,angle=0]{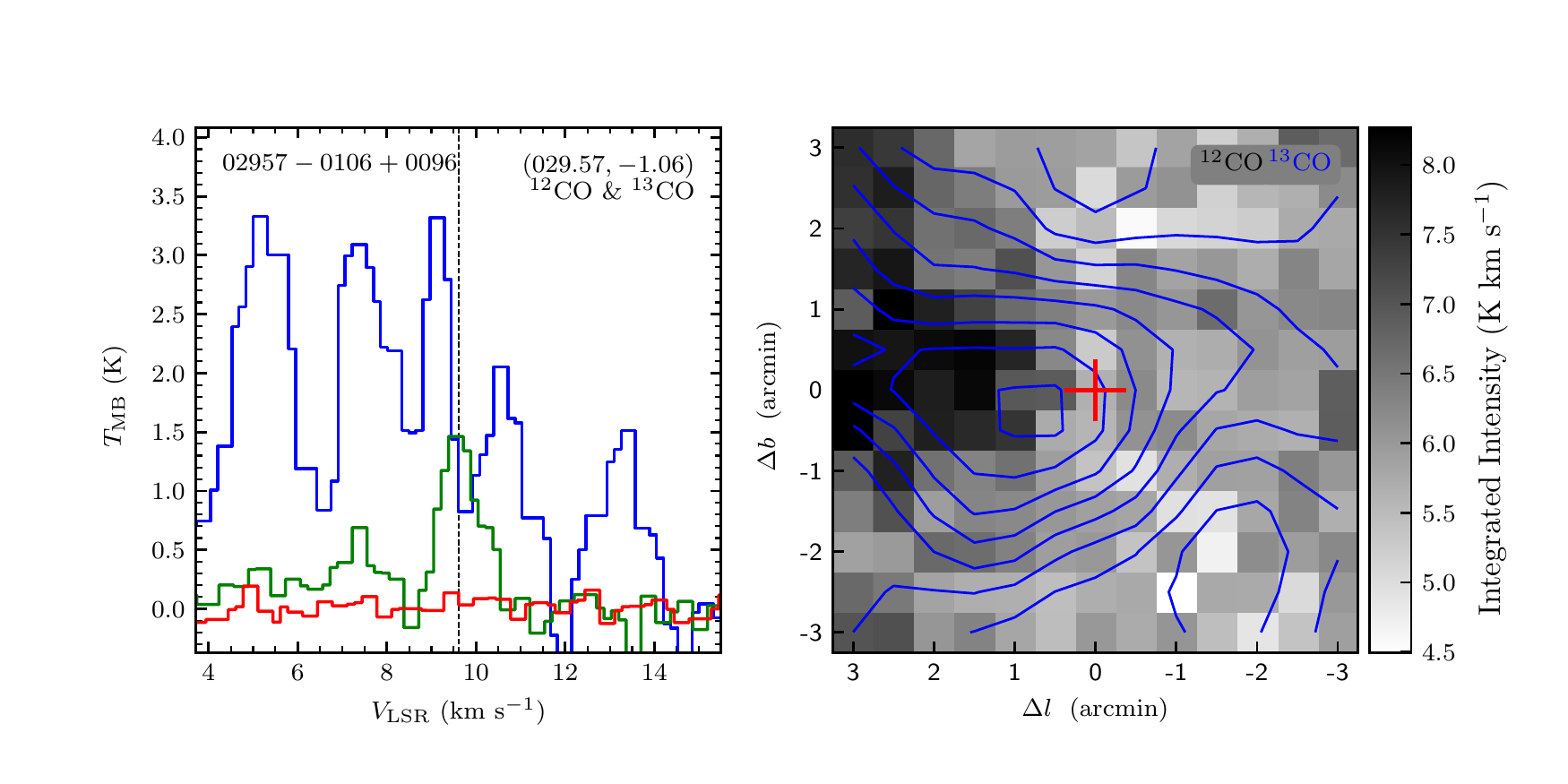}
\vspace{-0.5cm}

\includegraphics[width=9.0cm,angle=0]{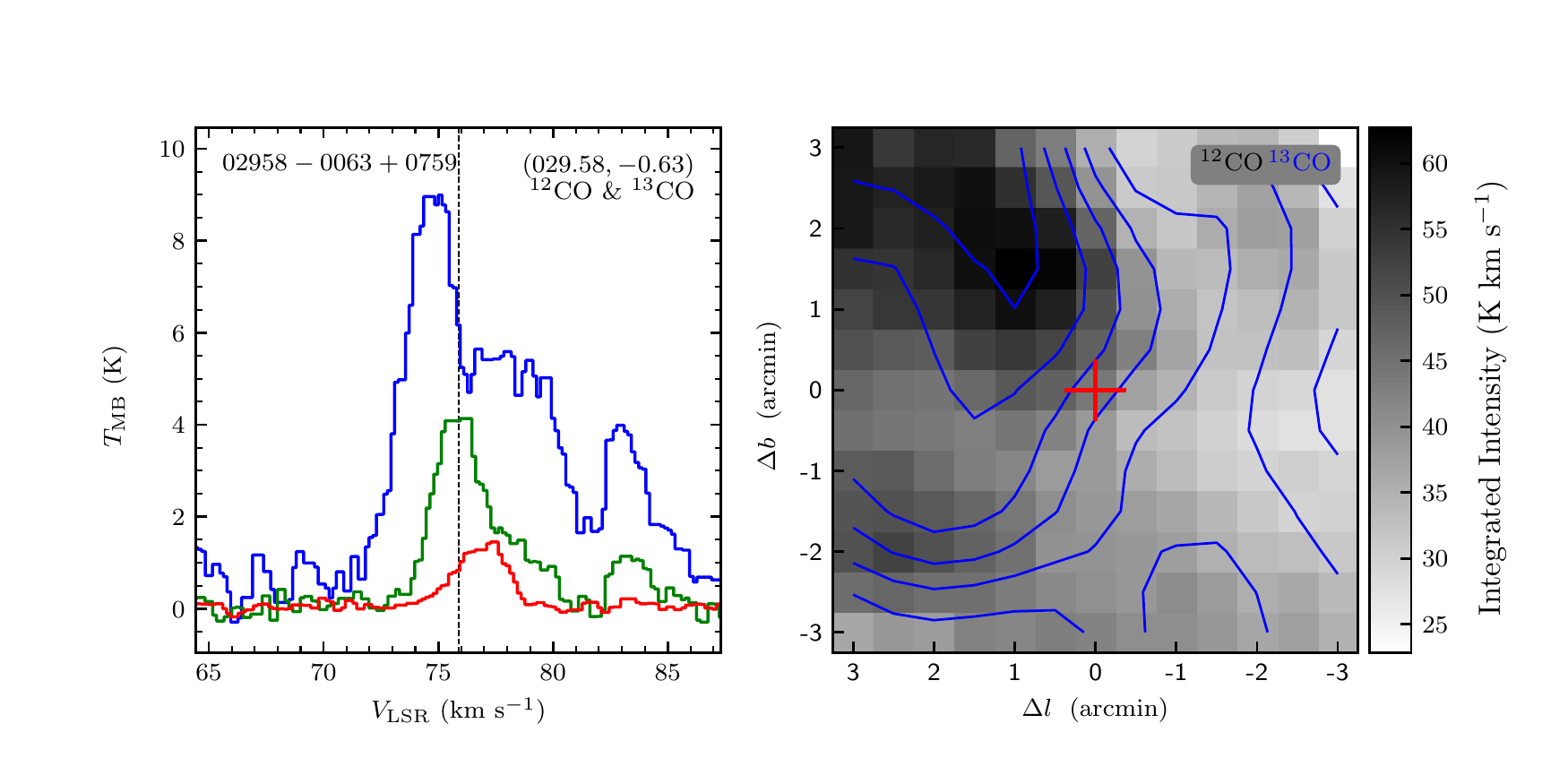}
\includegraphics[width=9.0cm,angle=0]{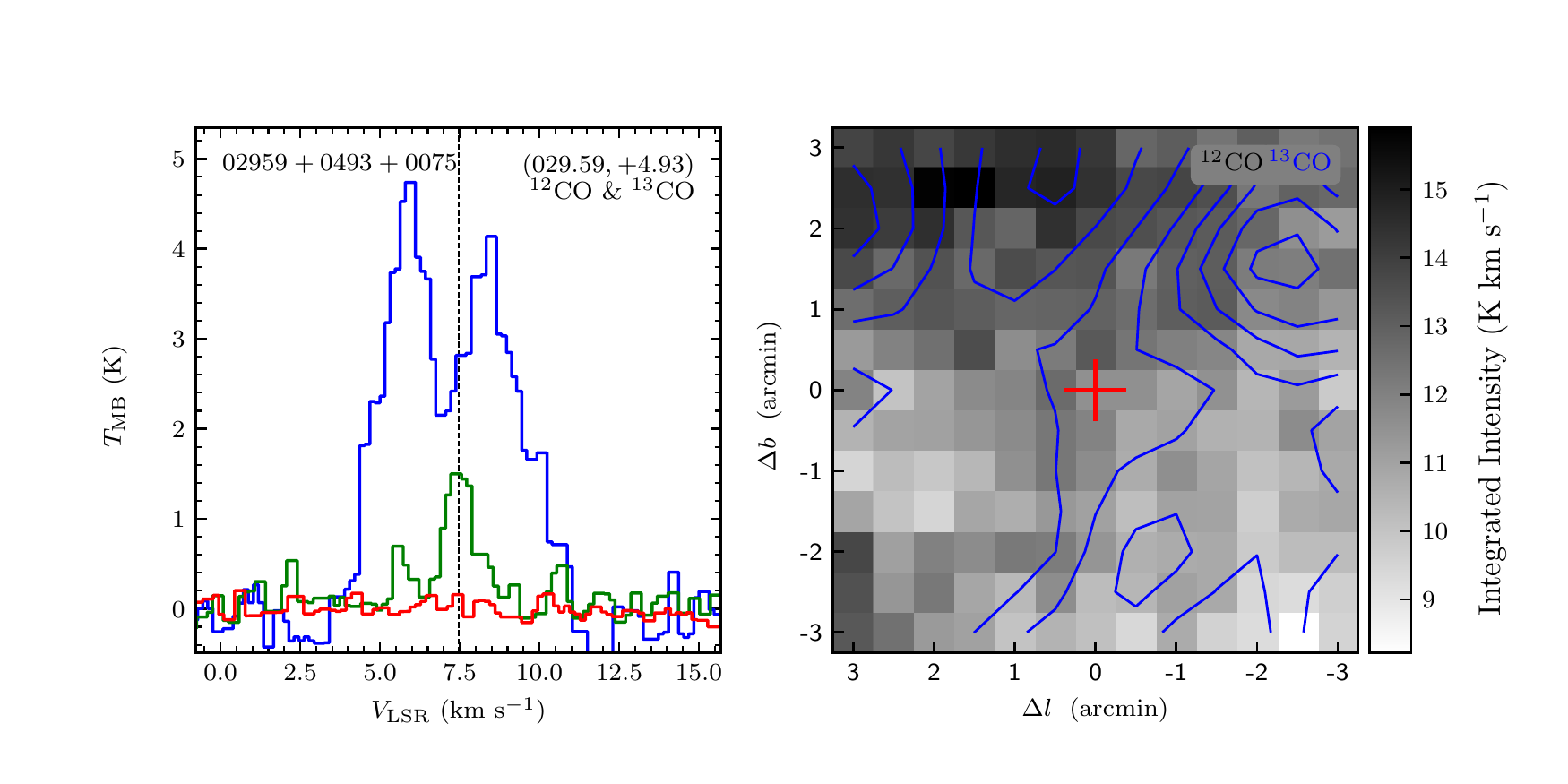}
\end{figure}
\clearpage

\begin{figure}
\includegraphics[width=9.0cm,angle=0]{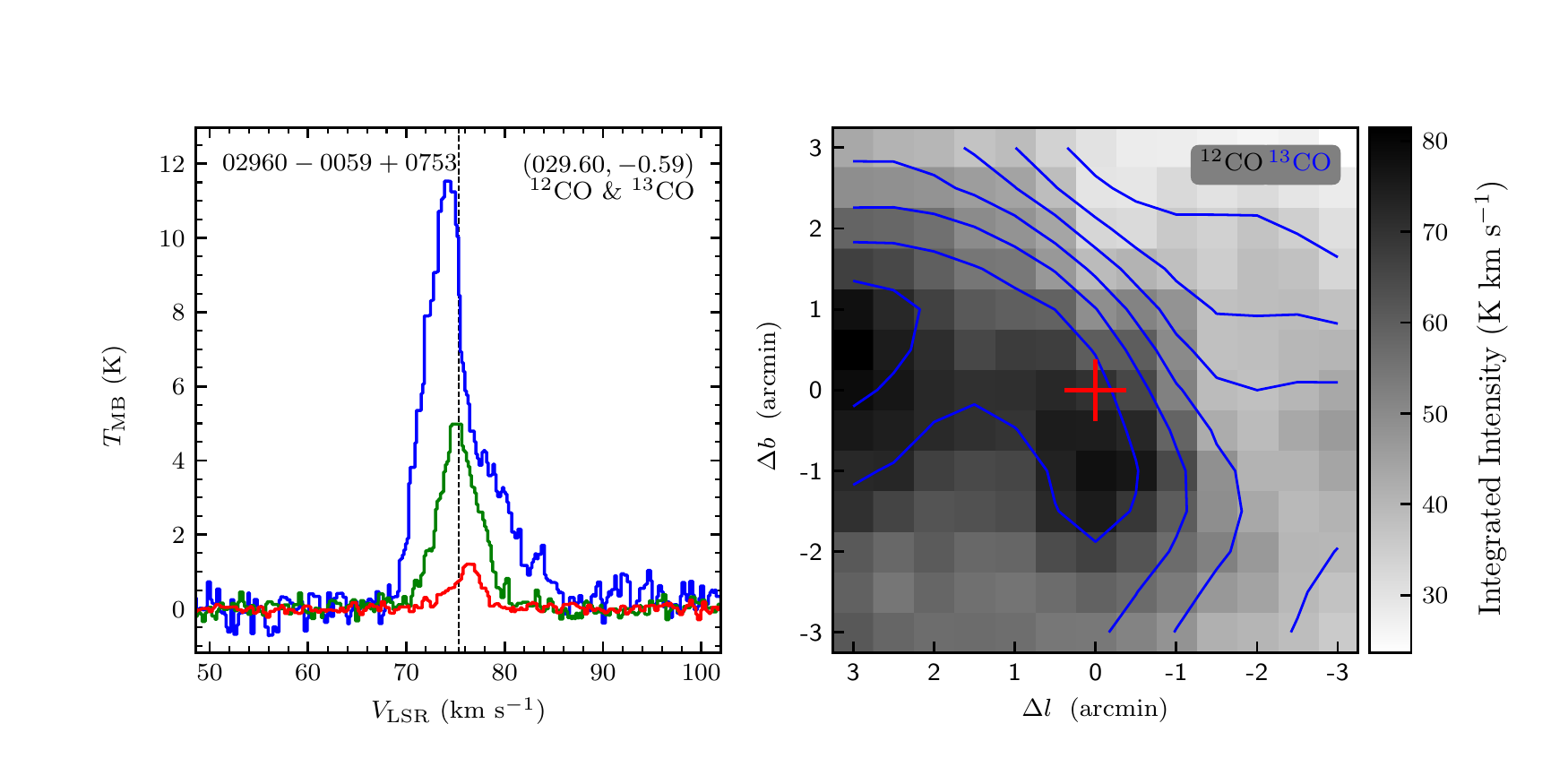}
\includegraphics[width=9.0cm,angle=0]{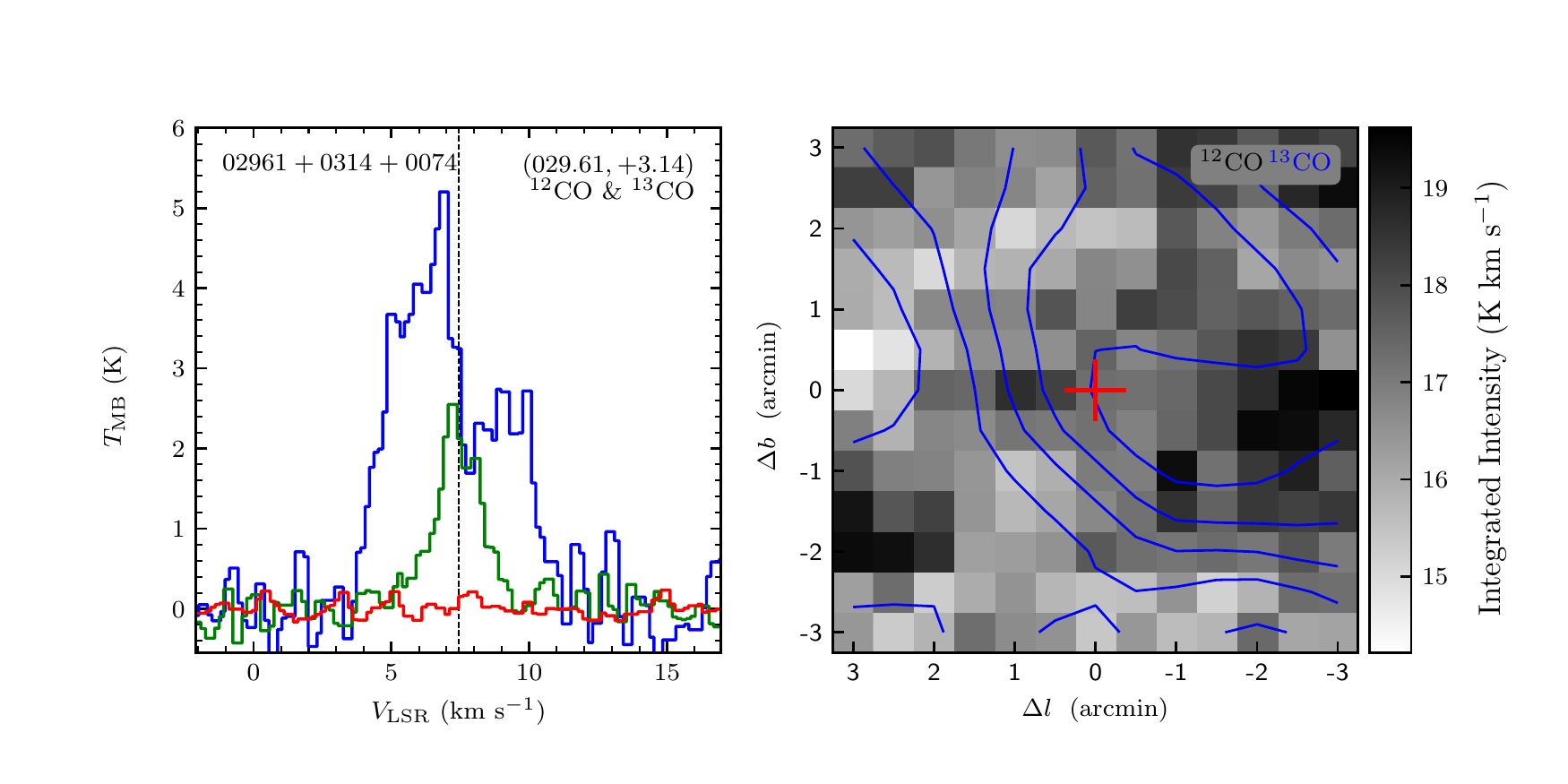}
\vspace{-0.5cm}

\includegraphics[width=9.0cm,angle=0]{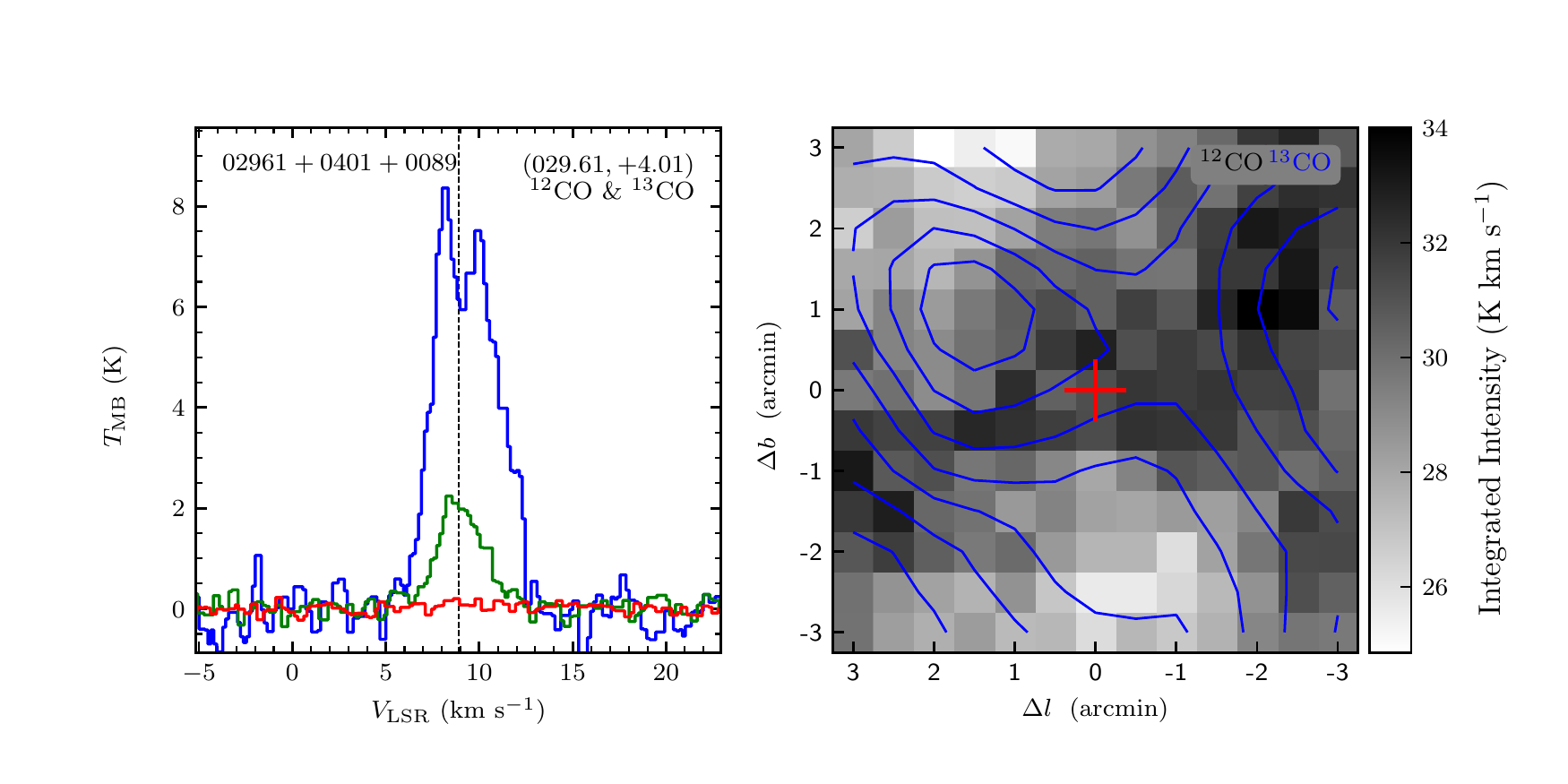}
\includegraphics[width=9.0cm,angle=0]{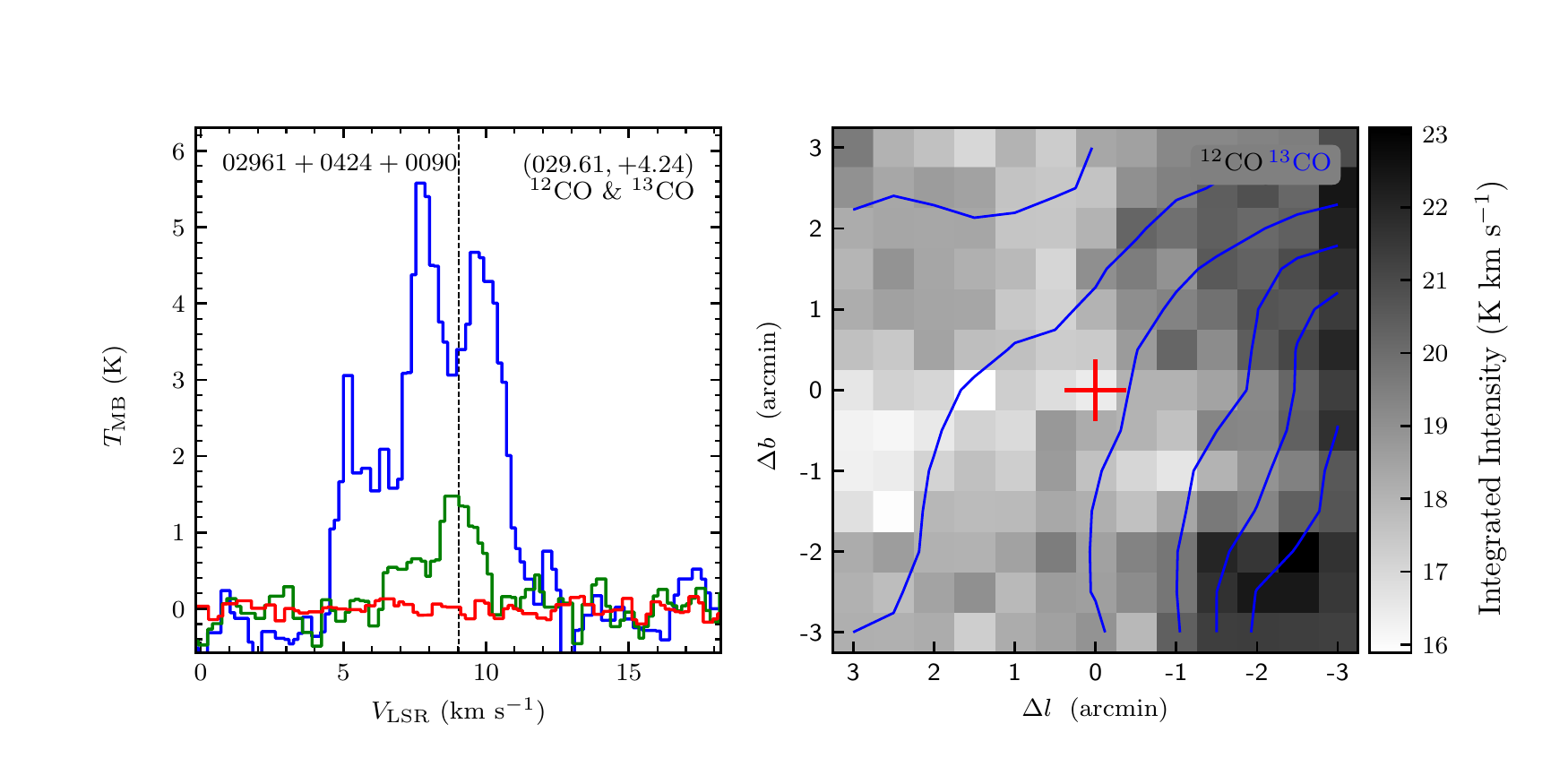}
\vspace{-0.5cm}

\includegraphics[width=9.0cm,angle=0]{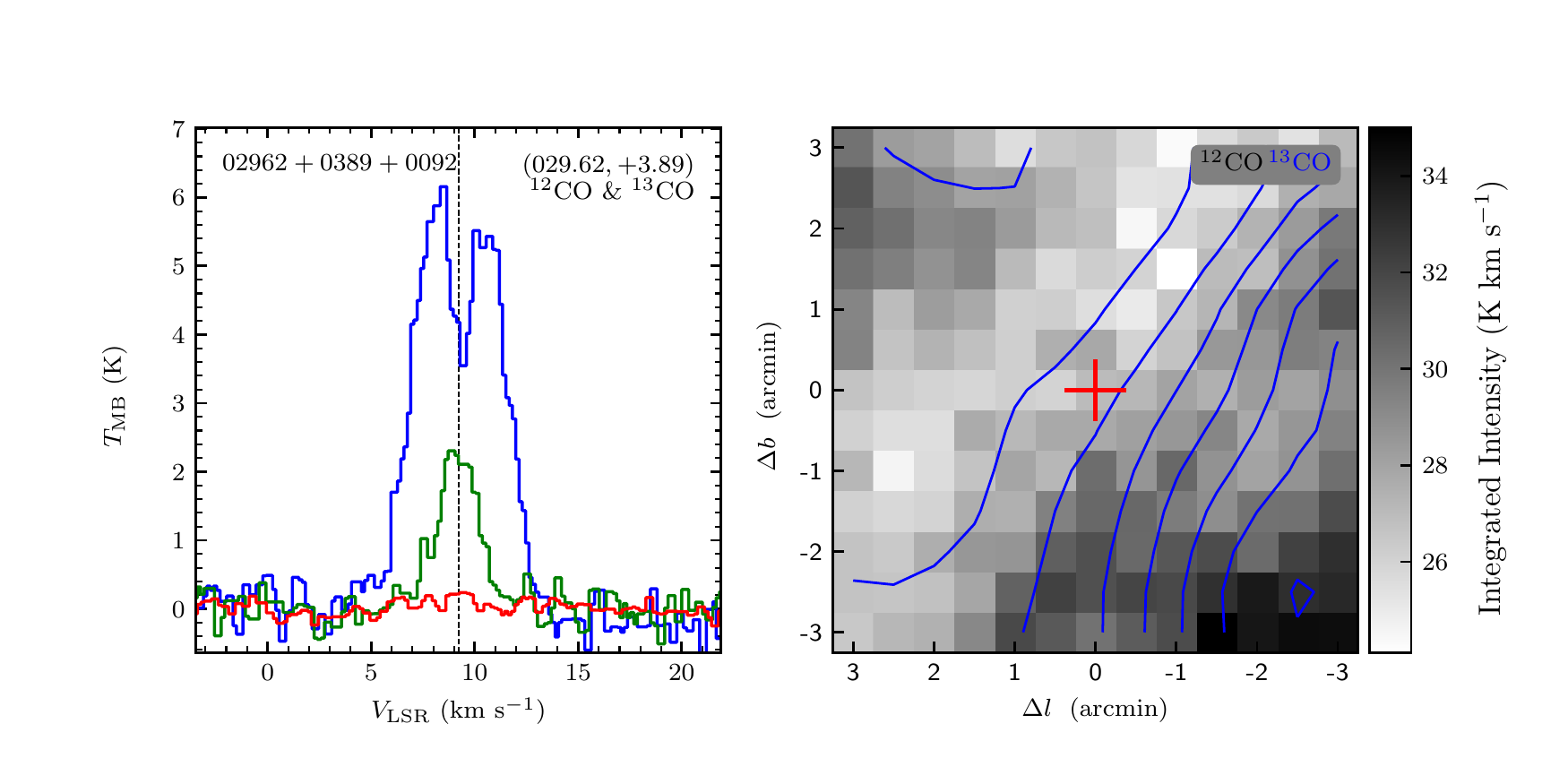}
\includegraphics[width=9.0cm,angle=0]{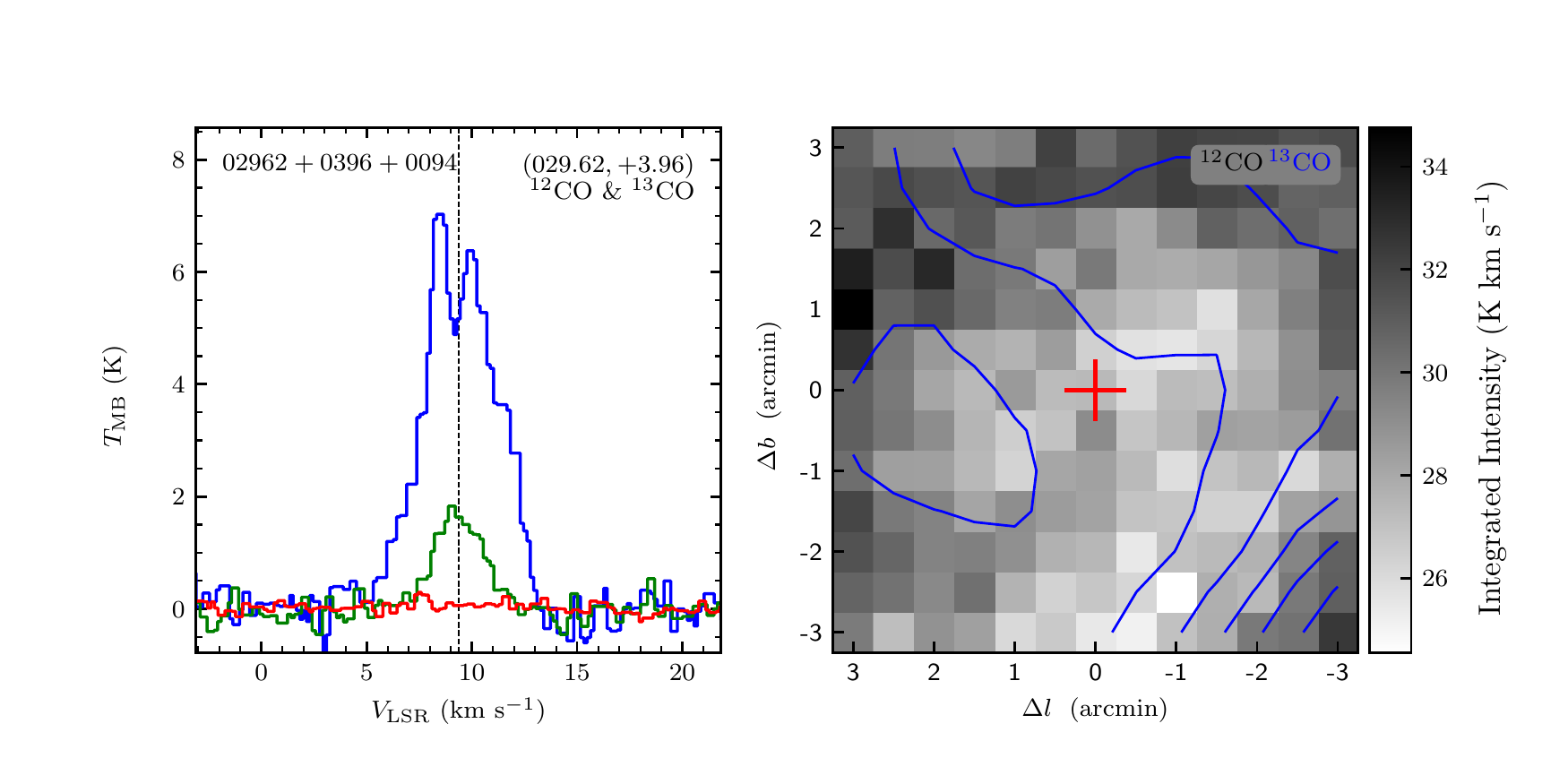}
\vspace{-0.5cm}

\includegraphics[width=9.0cm,angle=0]{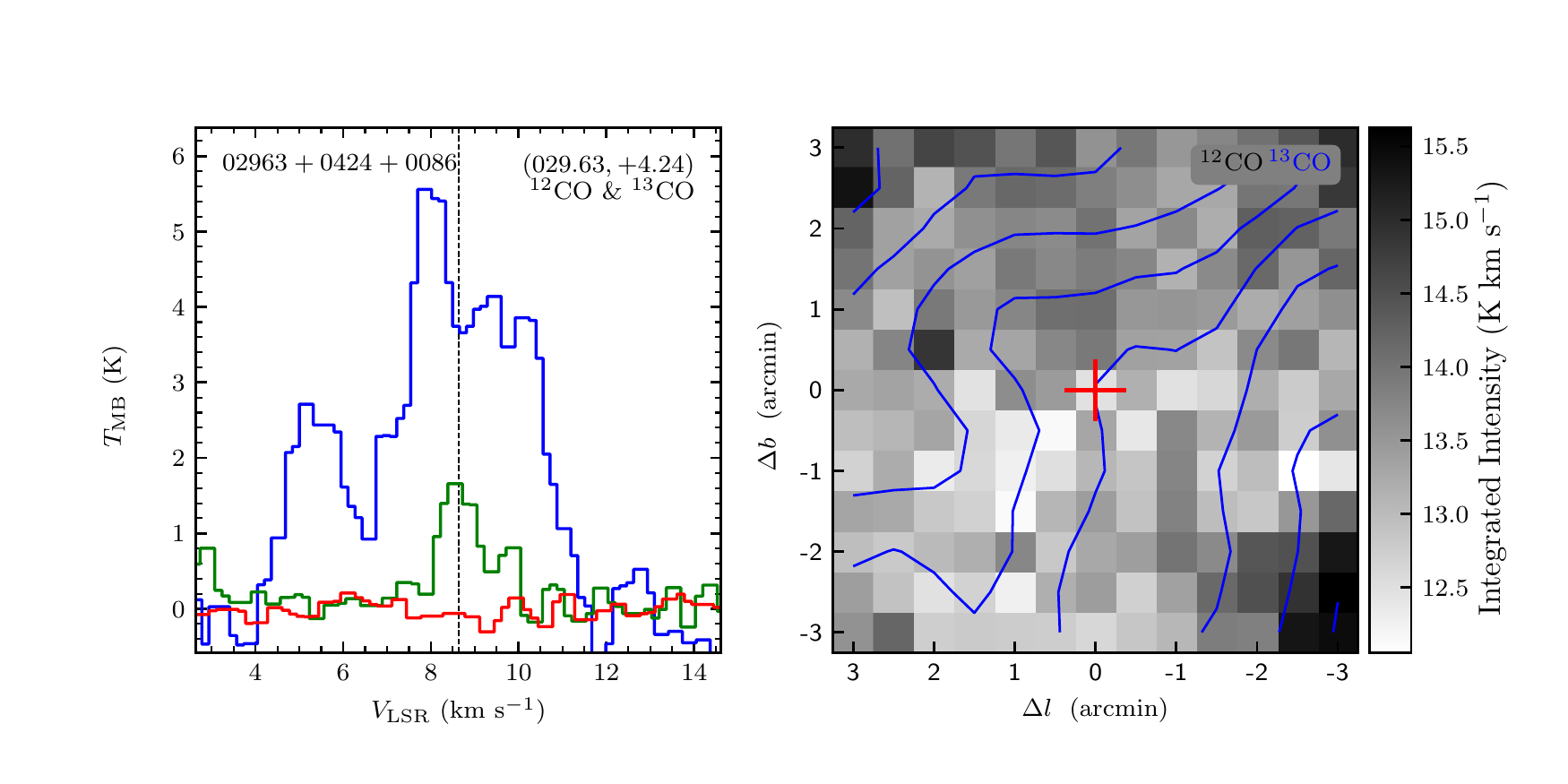}
\includegraphics[width=9.0cm,angle=0]{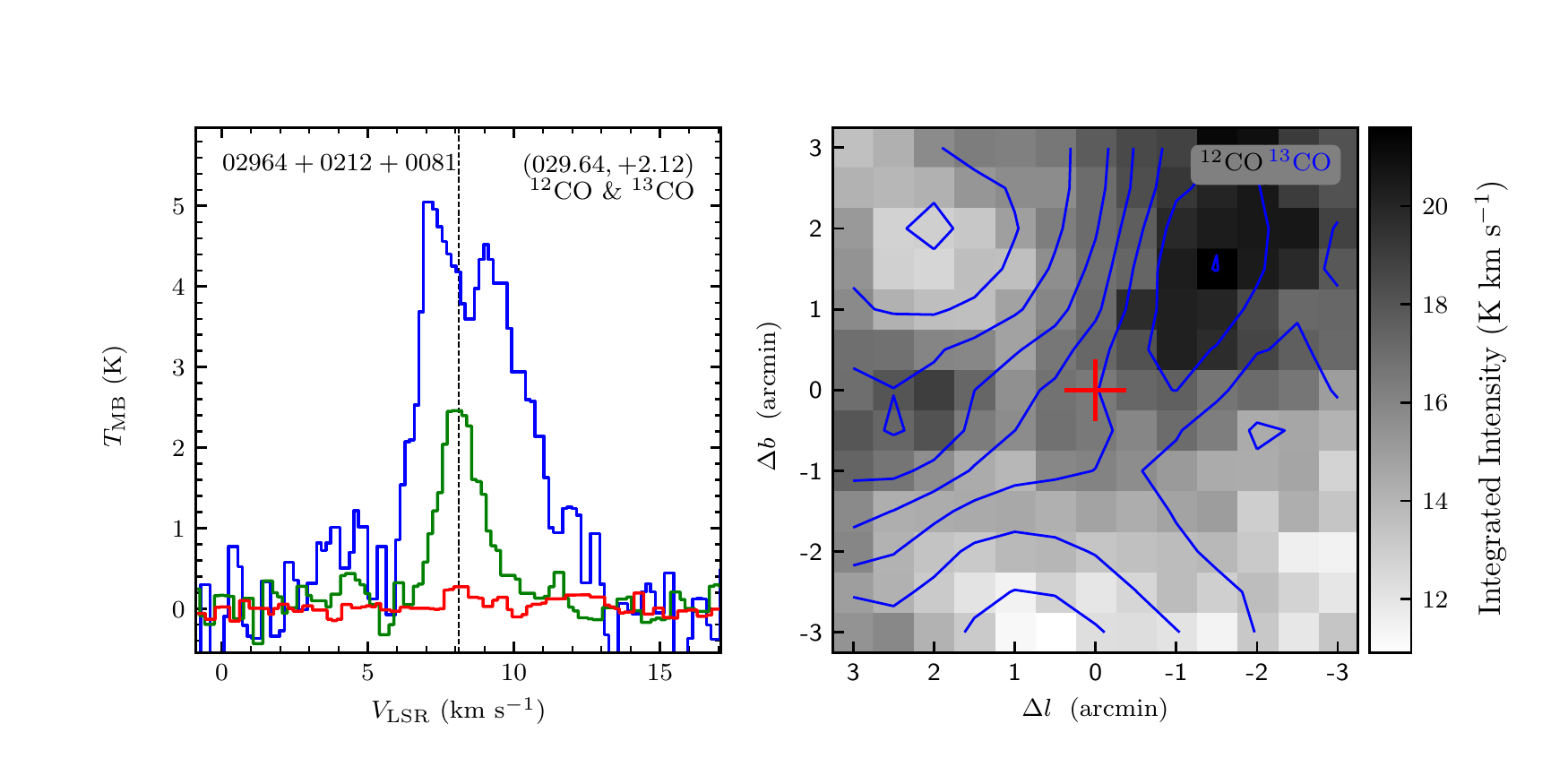}
\vspace{-0.5cm}

\includegraphics[width=9.0cm,angle=0]{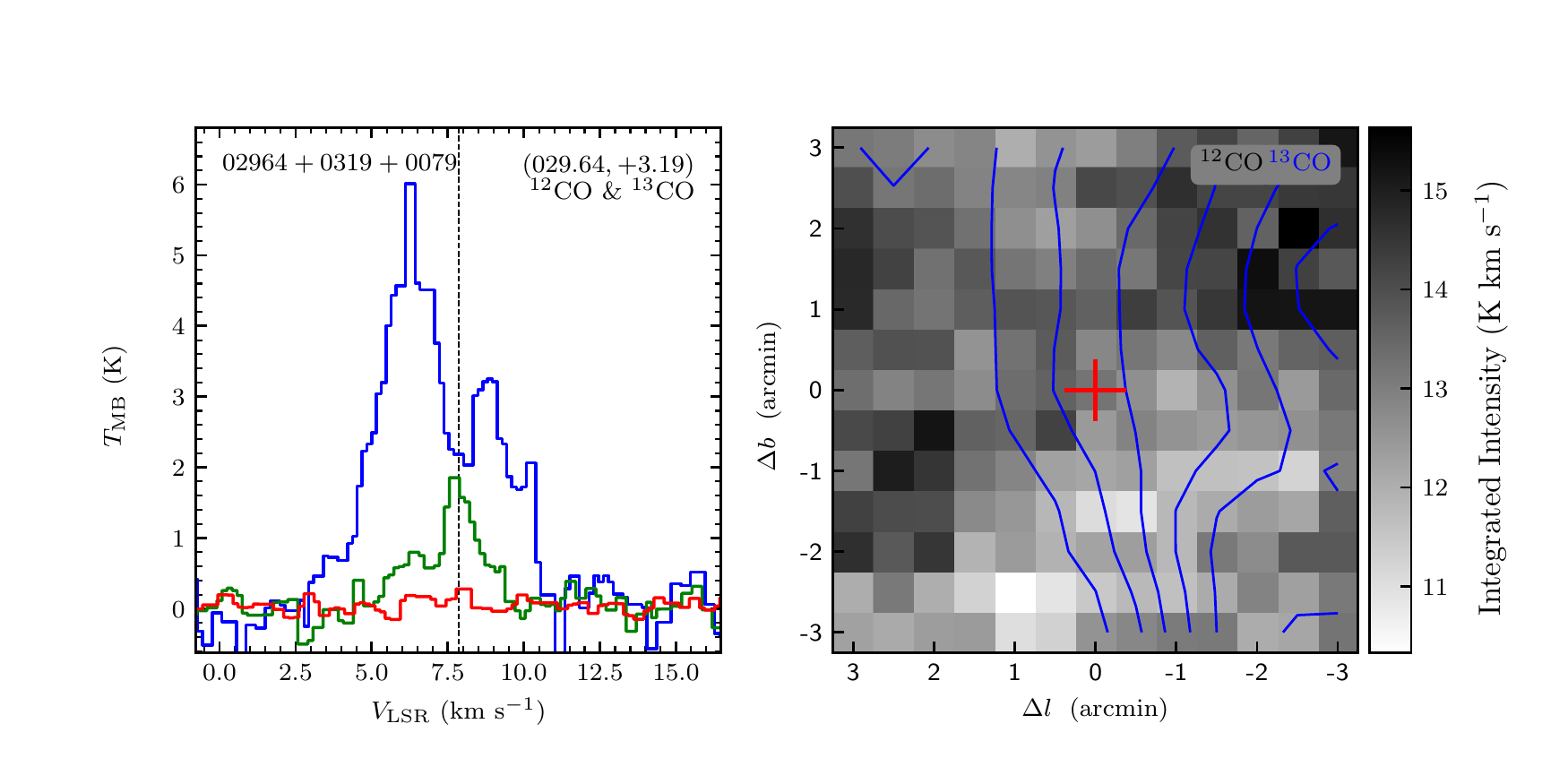}
\includegraphics[width=9.0cm,angle=0]{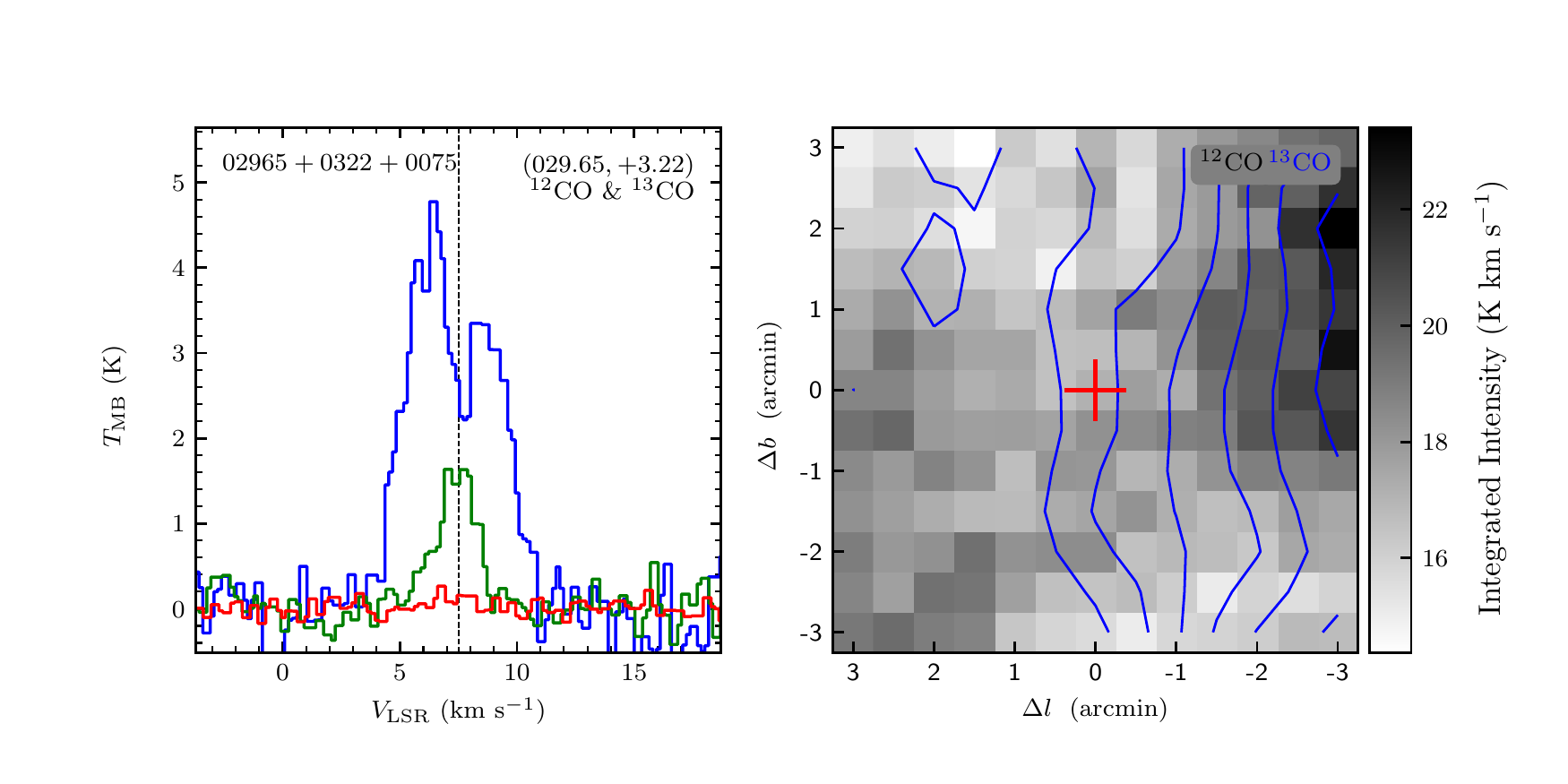}
\end{figure}
\clearpage

\begin{figure}
\includegraphics[width=9.0cm,angle=0]{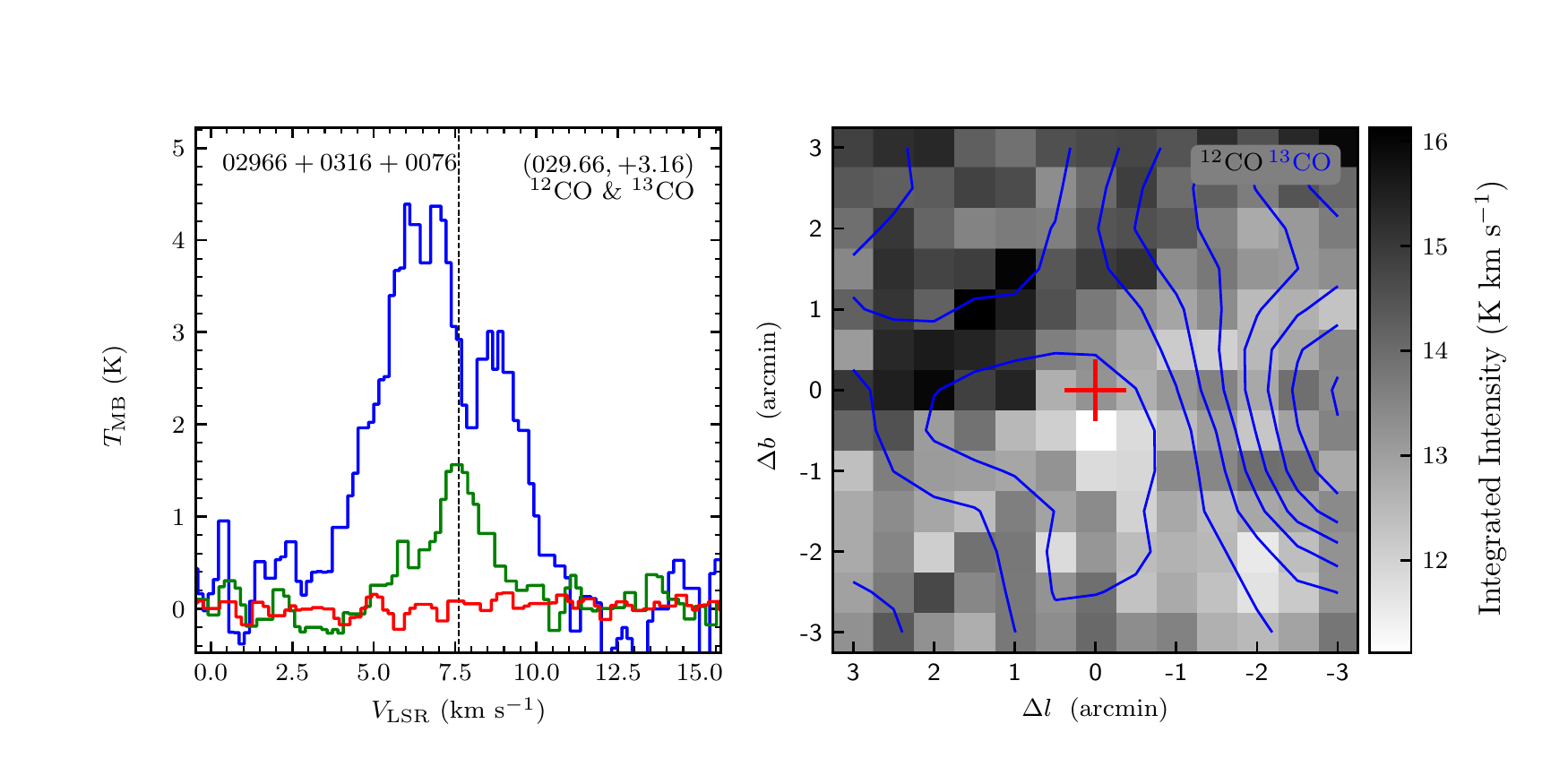}
\includegraphics[width=9.0cm,angle=0]{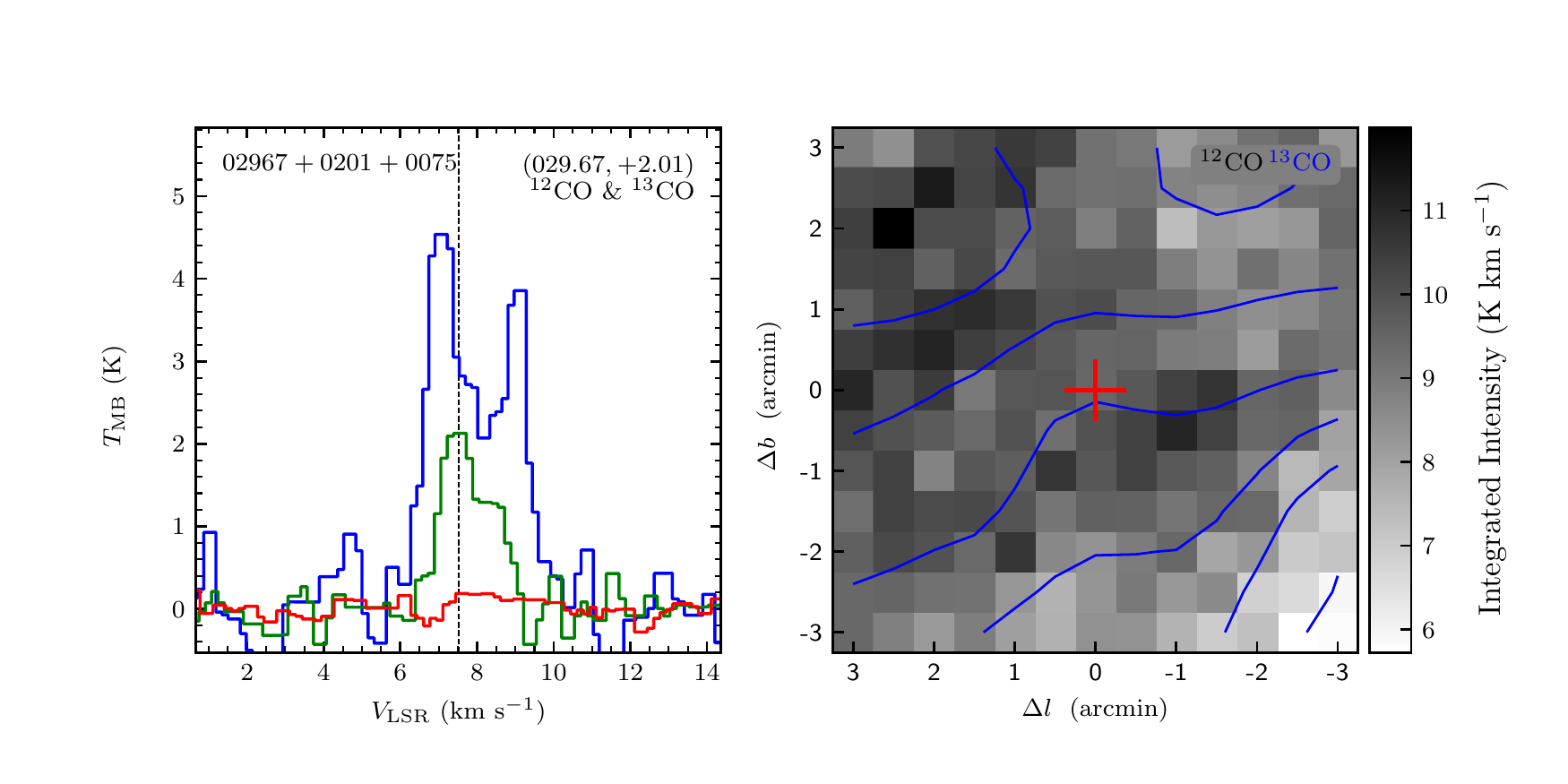}
\vspace{-0.5cm}

\includegraphics[width=9.0cm,angle=0]{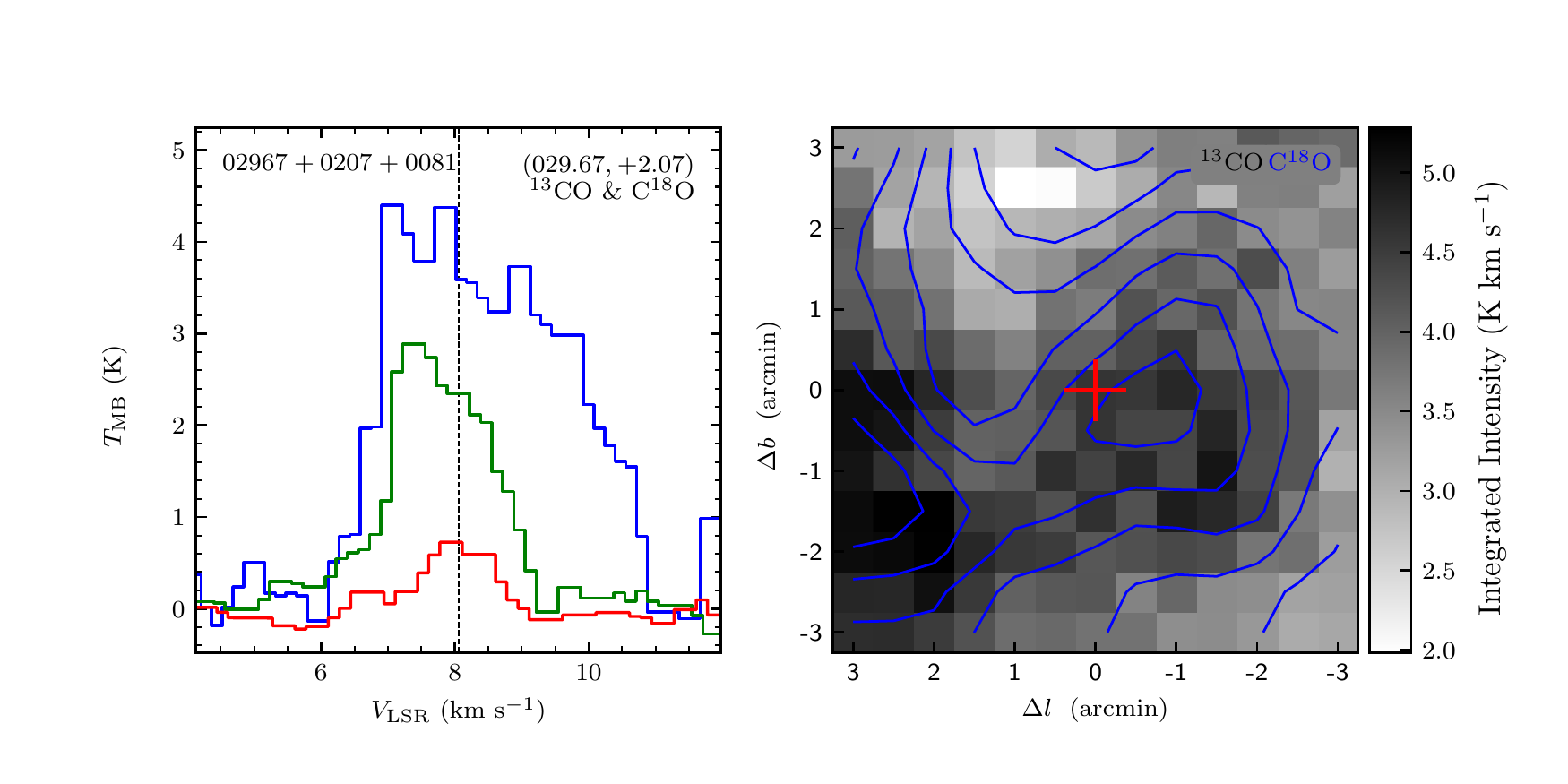}
\includegraphics[width=9.0cm,angle=0]{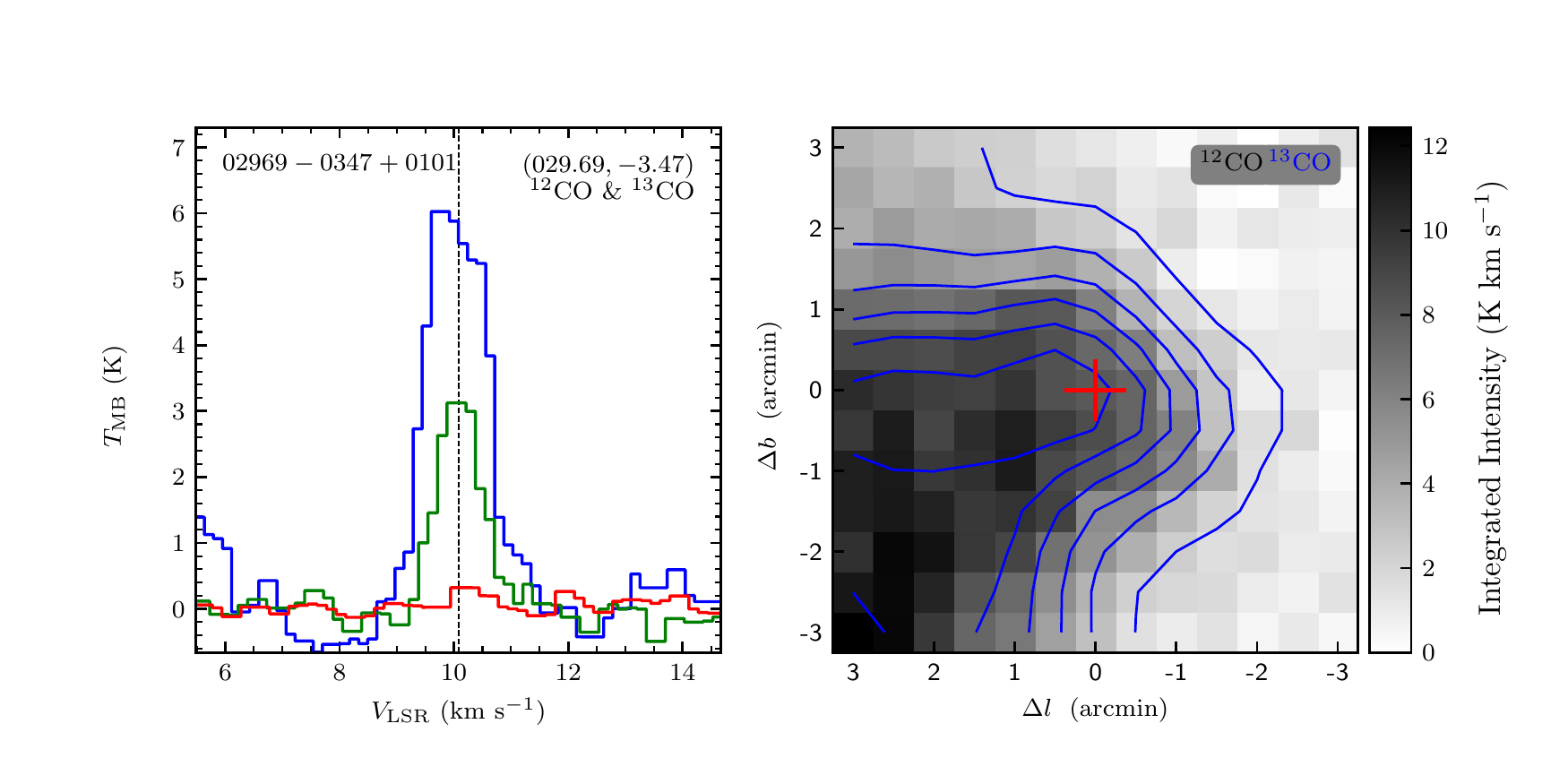}
\vspace{-0.5cm}

\includegraphics[width=9.0cm,angle=0]{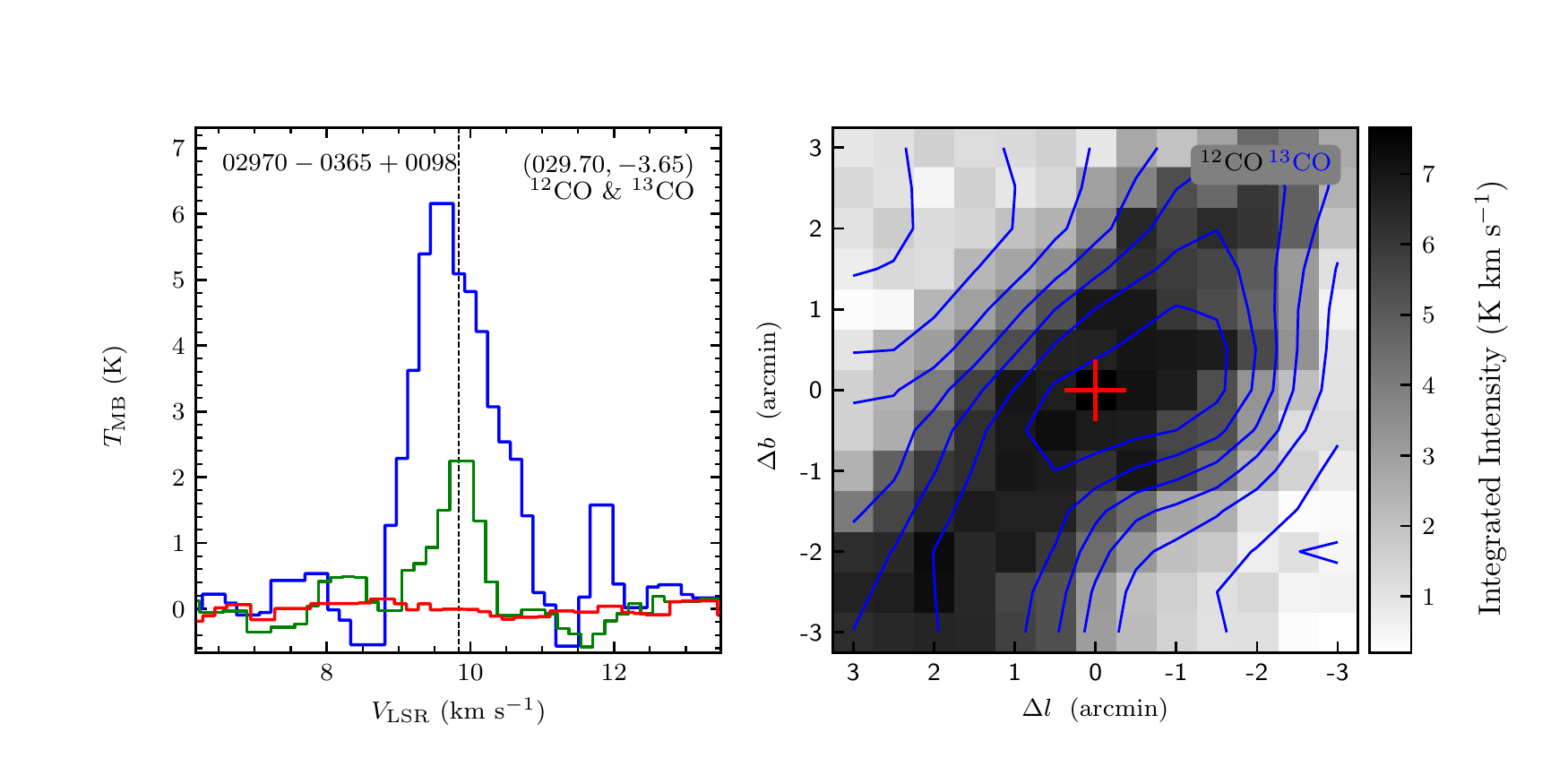}
\includegraphics[width=9.0cm,angle=0]{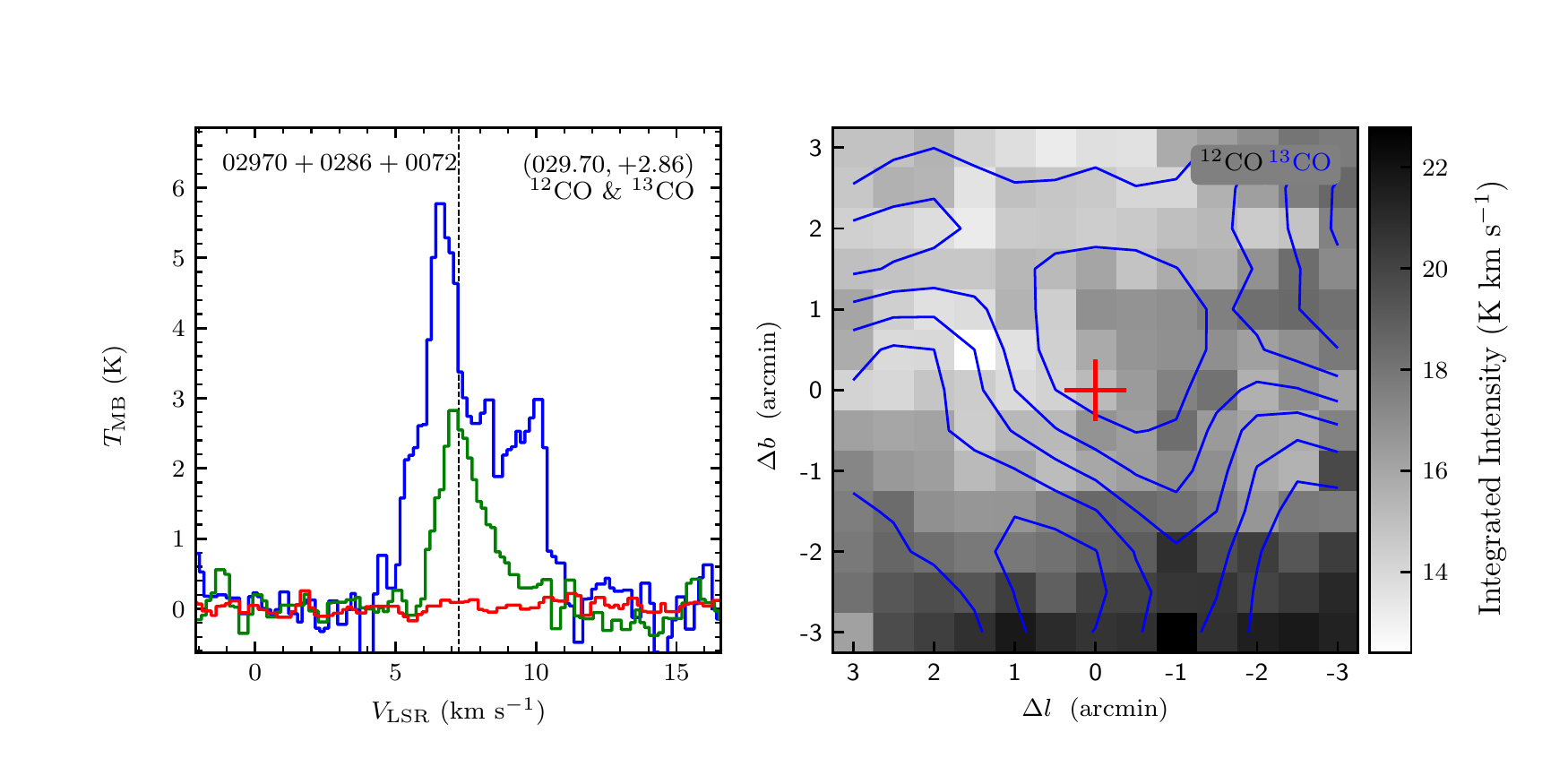}
\vspace{-0.5cm}

\includegraphics[width=9.0cm,angle=0]{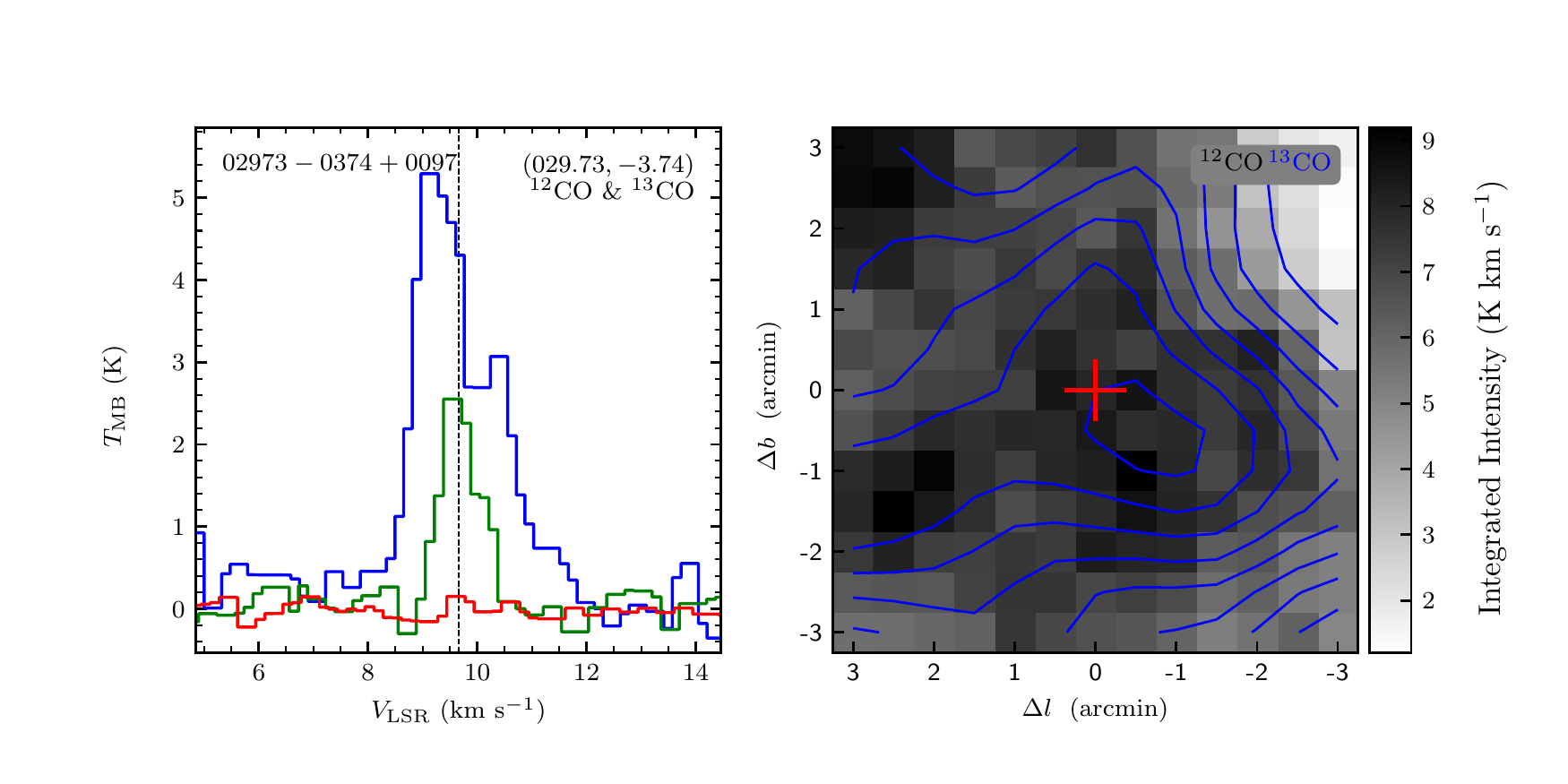}
\includegraphics[width=9.0cm,angle=0]{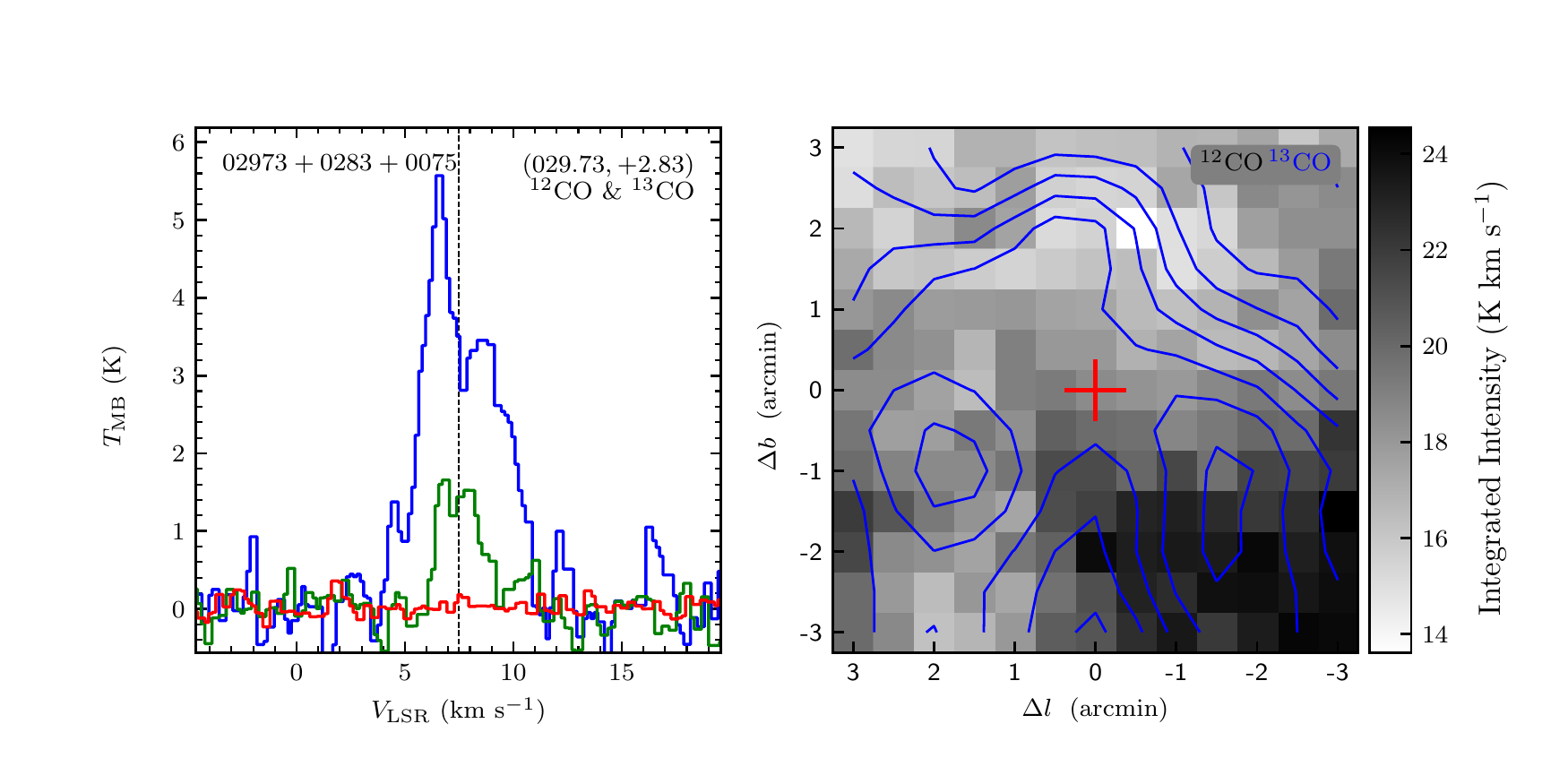}
\vspace{-0.5cm}

\includegraphics[width=9.0cm,angle=0]{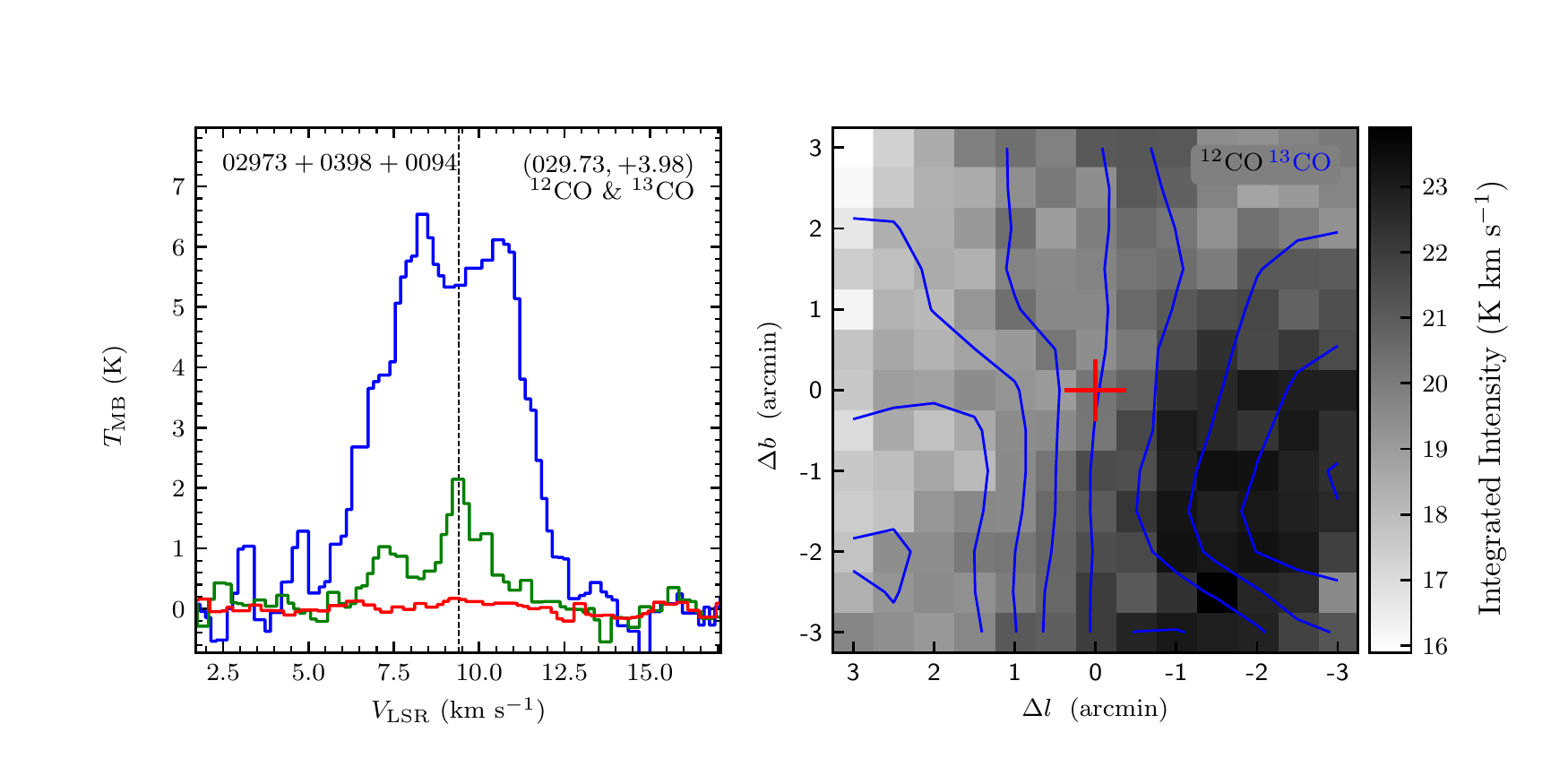}
\includegraphics[width=9.0cm,angle=0]{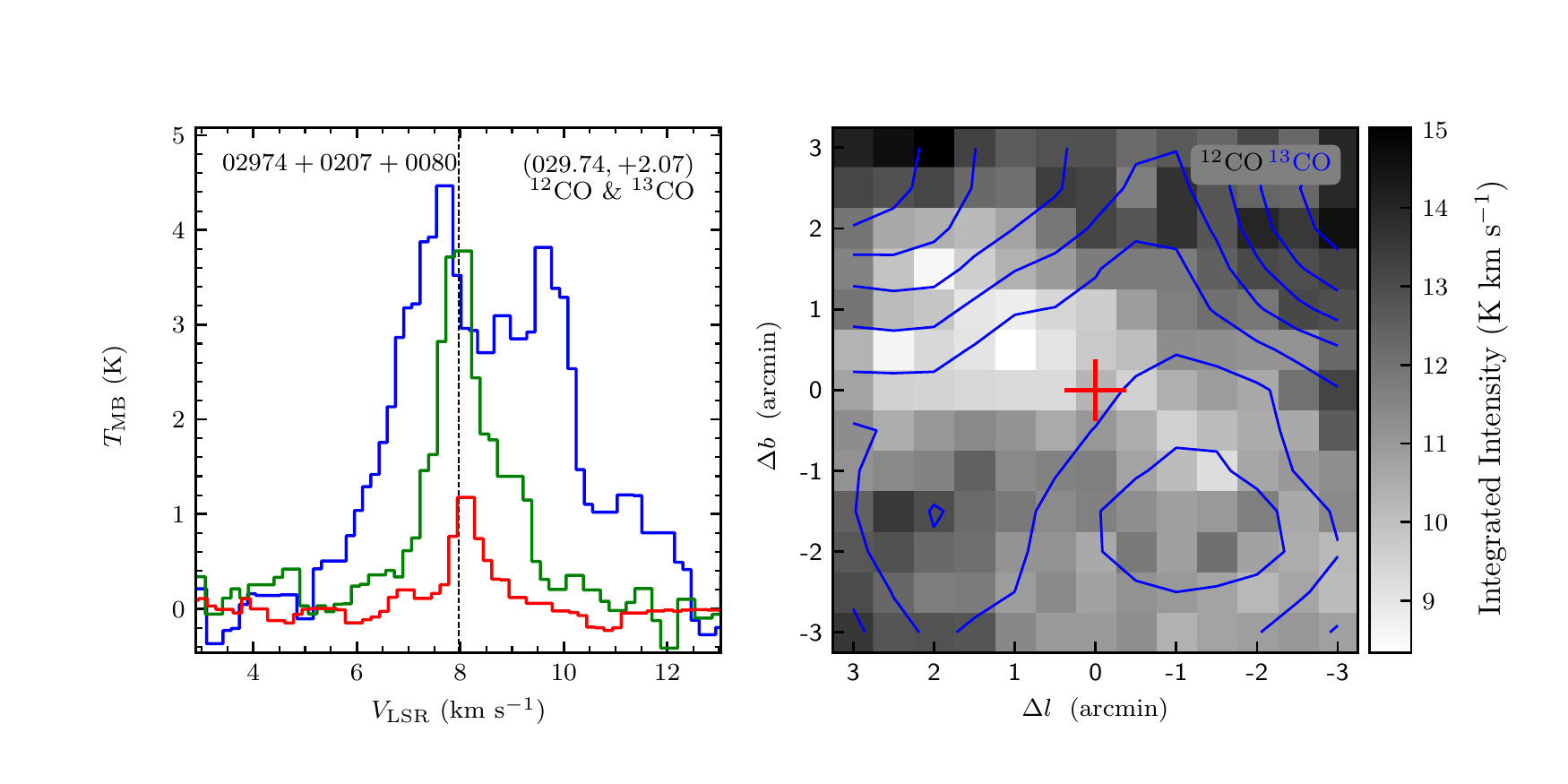}
\end{figure}
\clearpage

\begin{figure}
\includegraphics[width=9.0cm,angle=0]{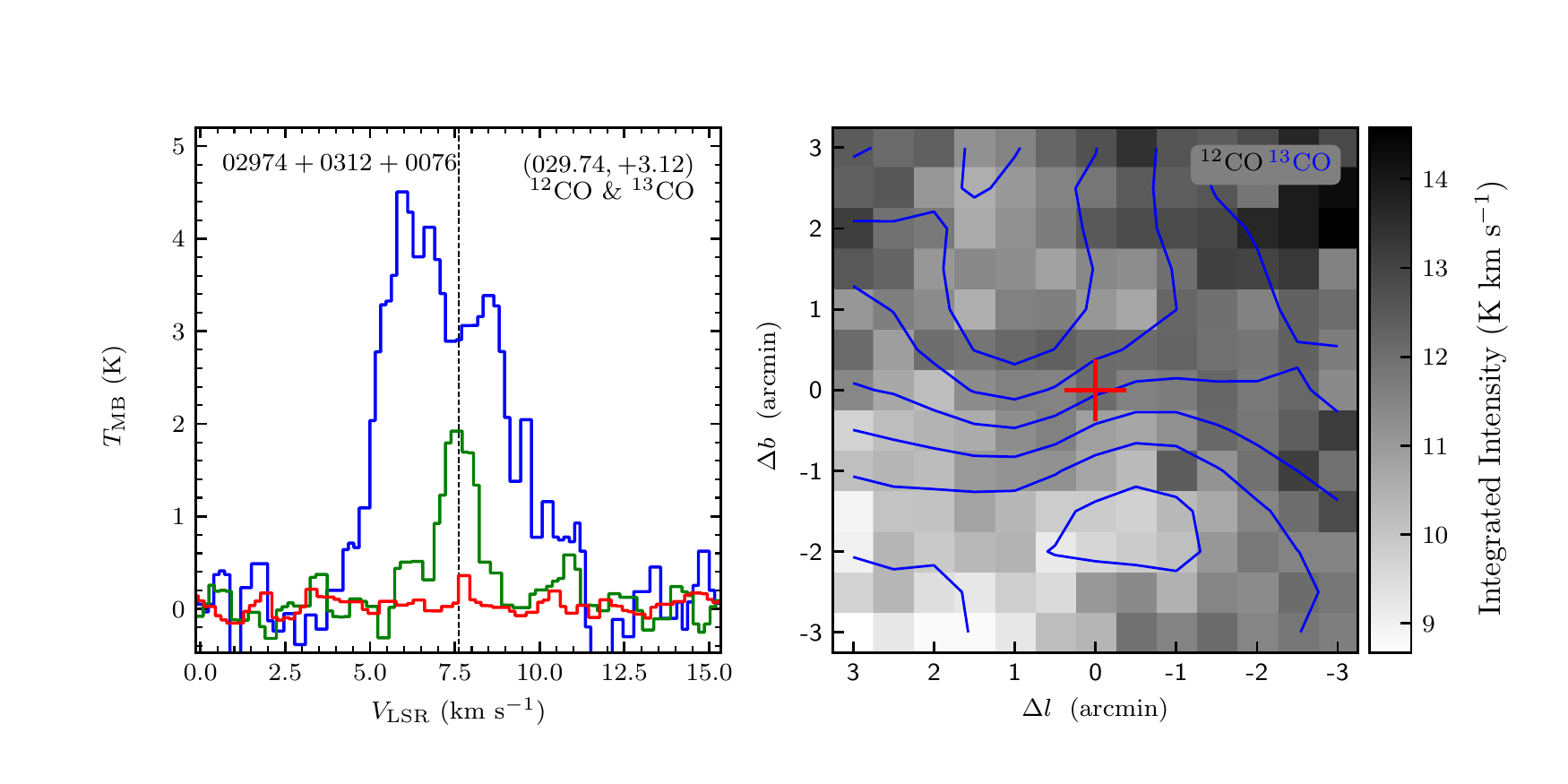}
\includegraphics[width=9.0cm,angle=0]{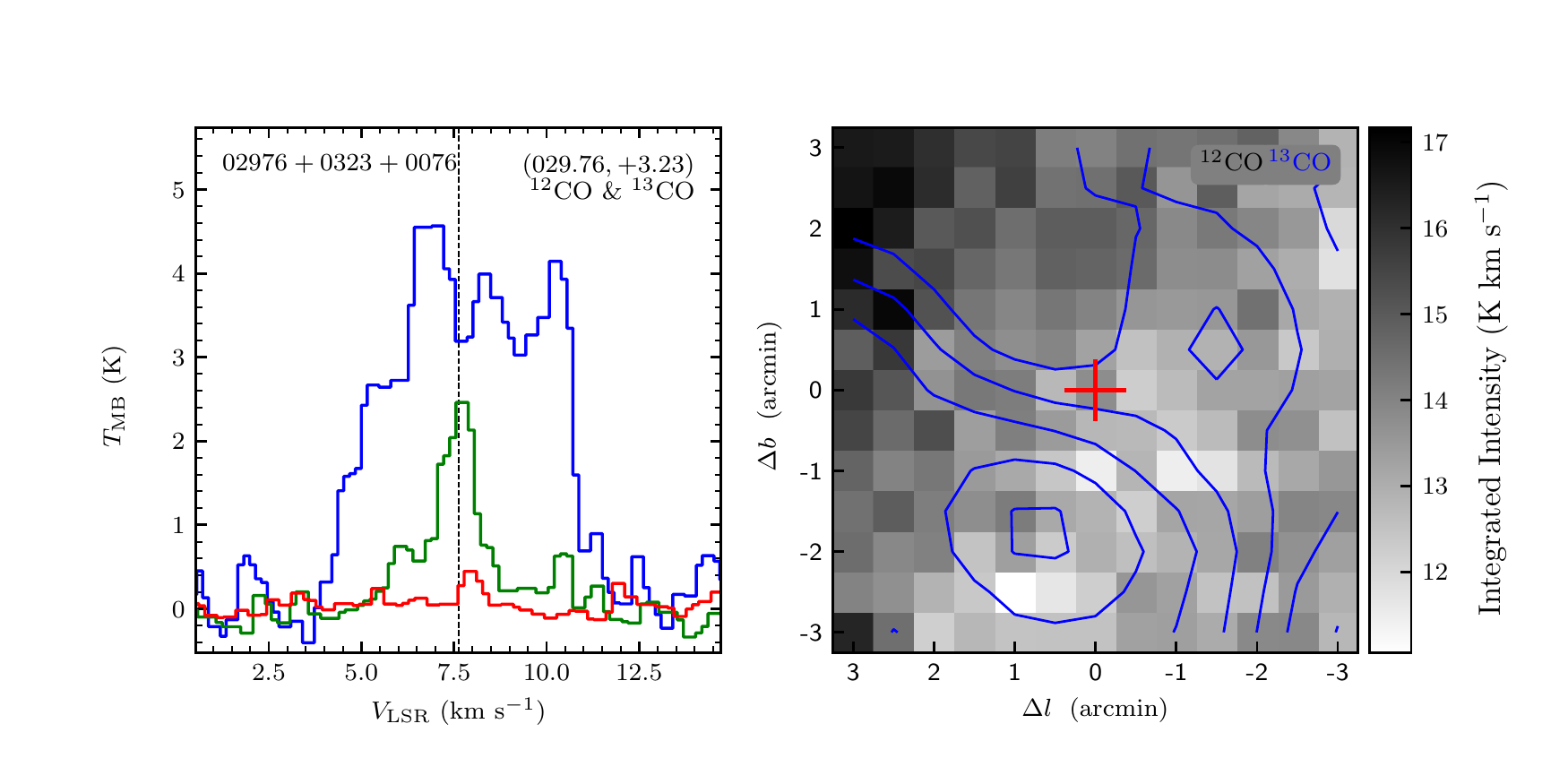}
\vspace{-0.5cm}

\includegraphics[width=9.0cm,angle=0]{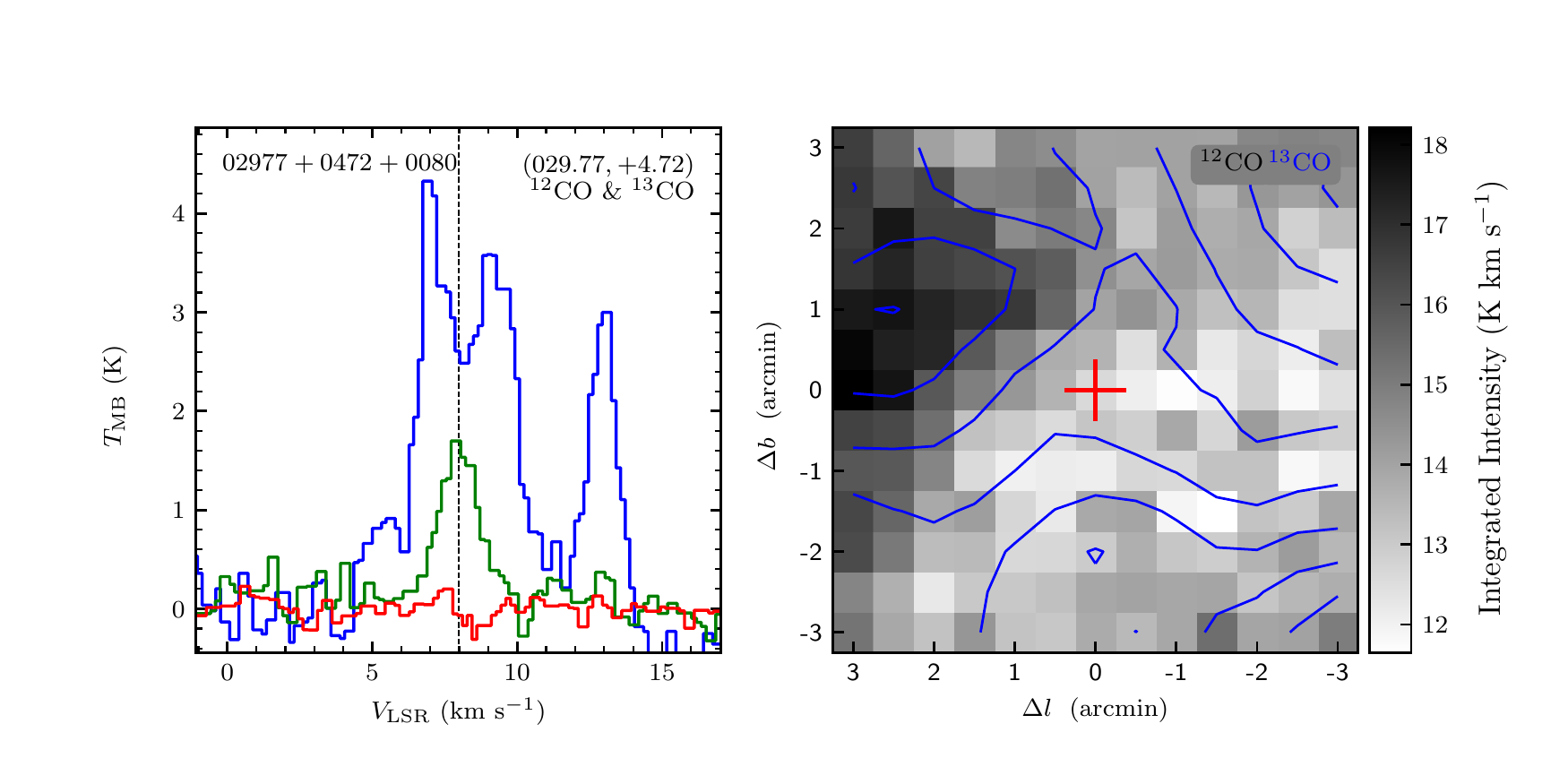}
\includegraphics[width=9.0cm,angle=0]{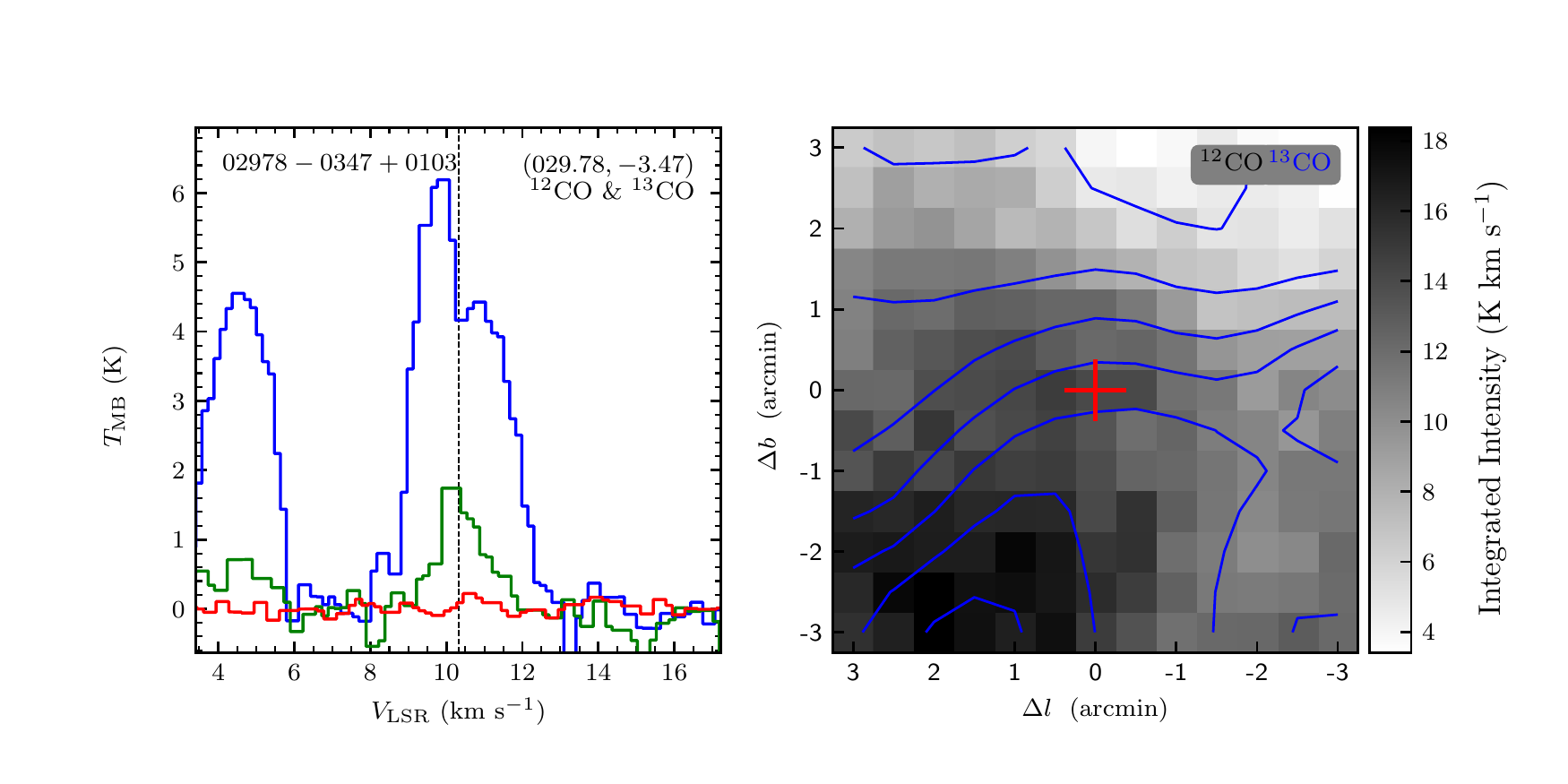}
\vspace{-0.5cm}

\includegraphics[width=9.0cm,angle=0]{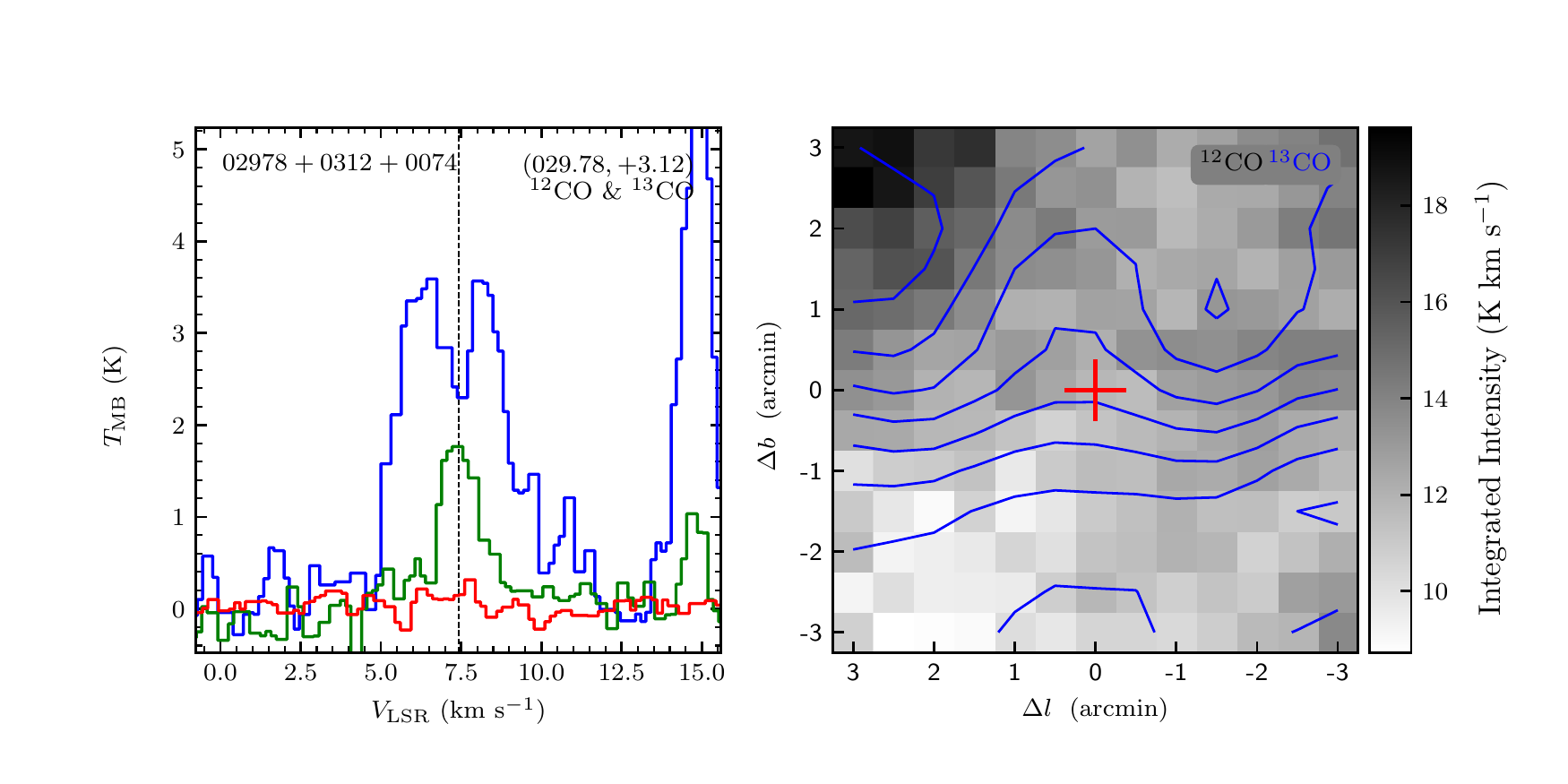}
\includegraphics[width=9.0cm,angle=0]{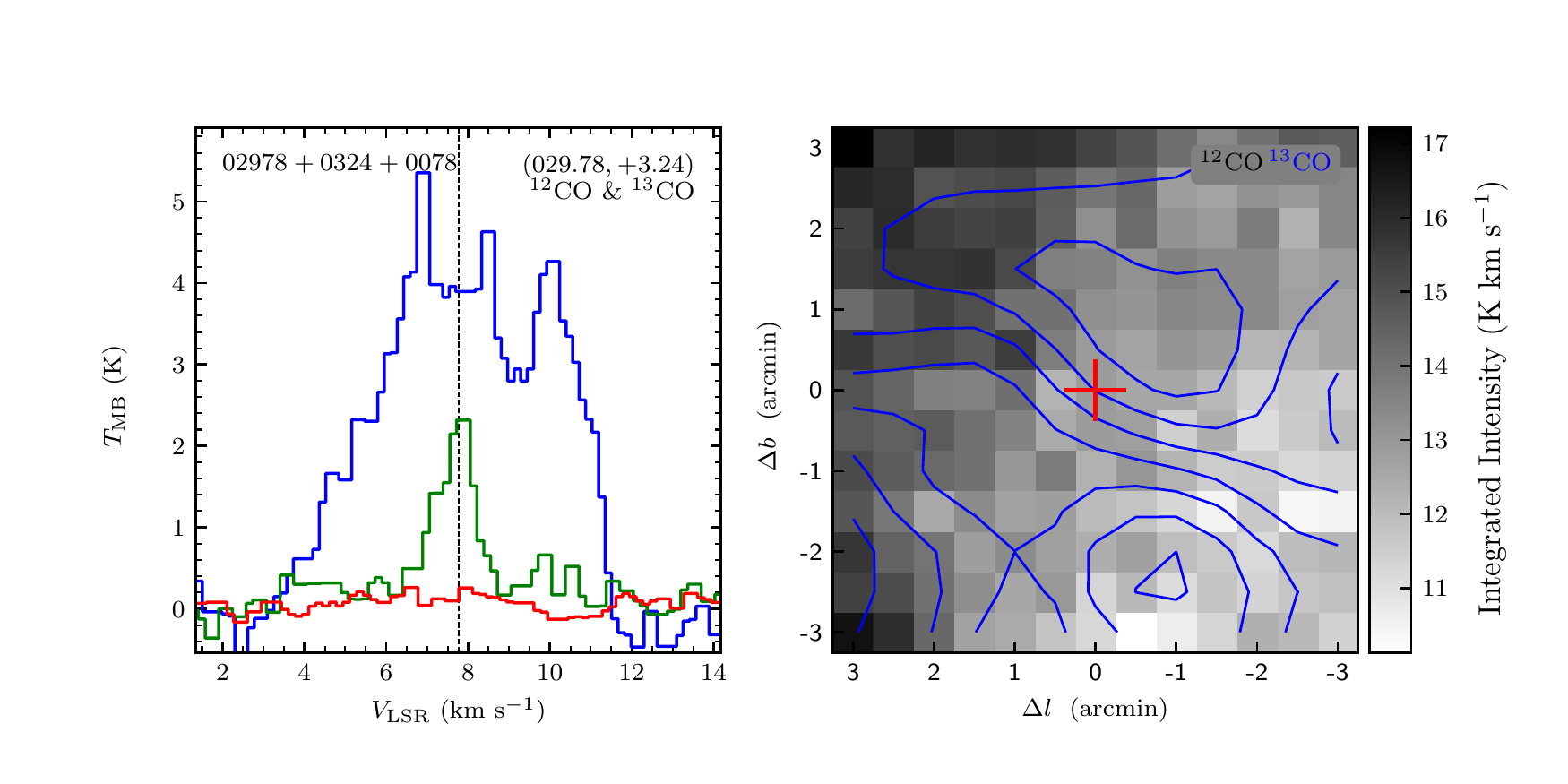}
\vspace{-0.5cm}

\includegraphics[width=9.0cm,angle=0]{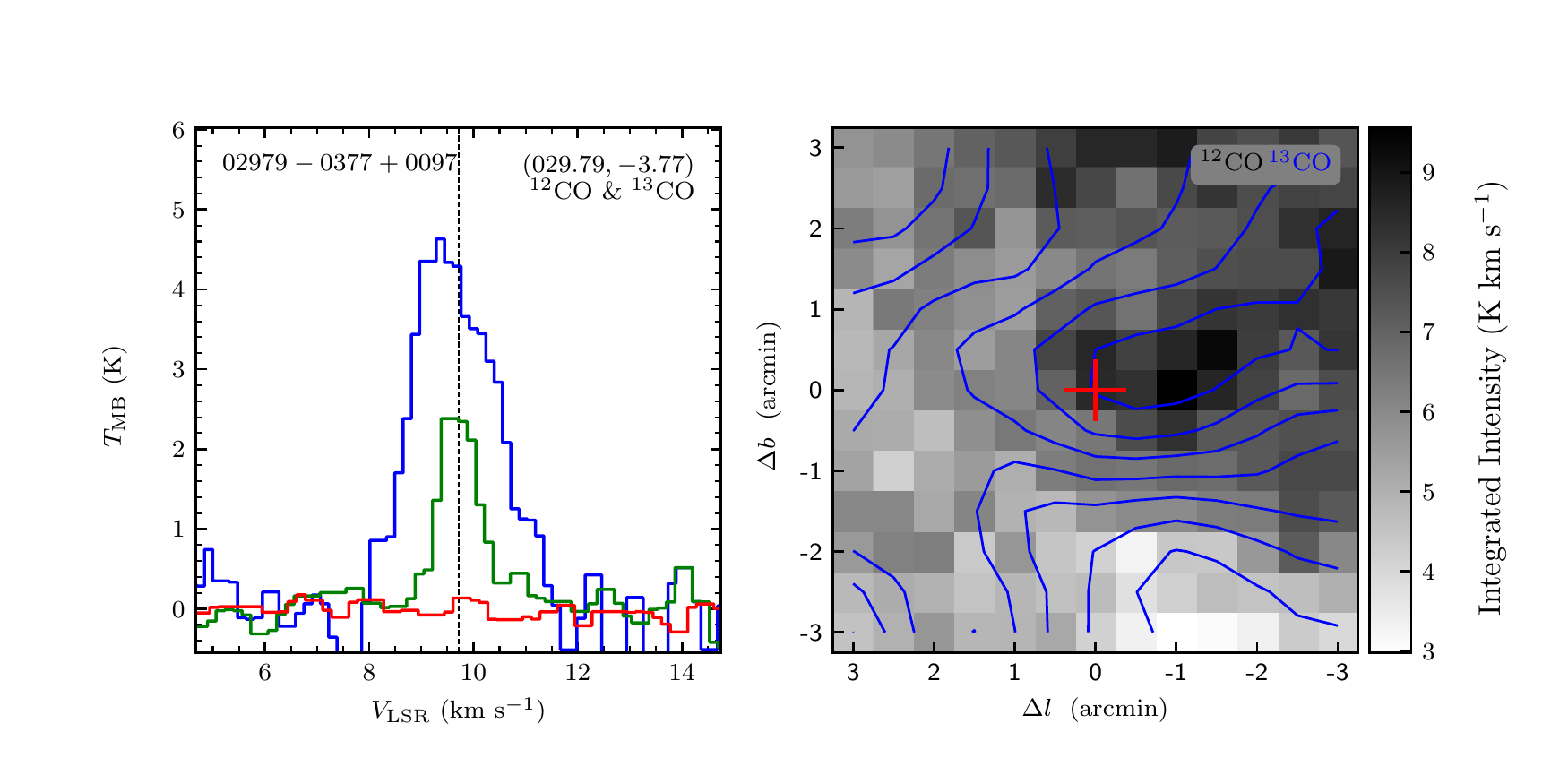}
\includegraphics[width=9.0cm,angle=0]{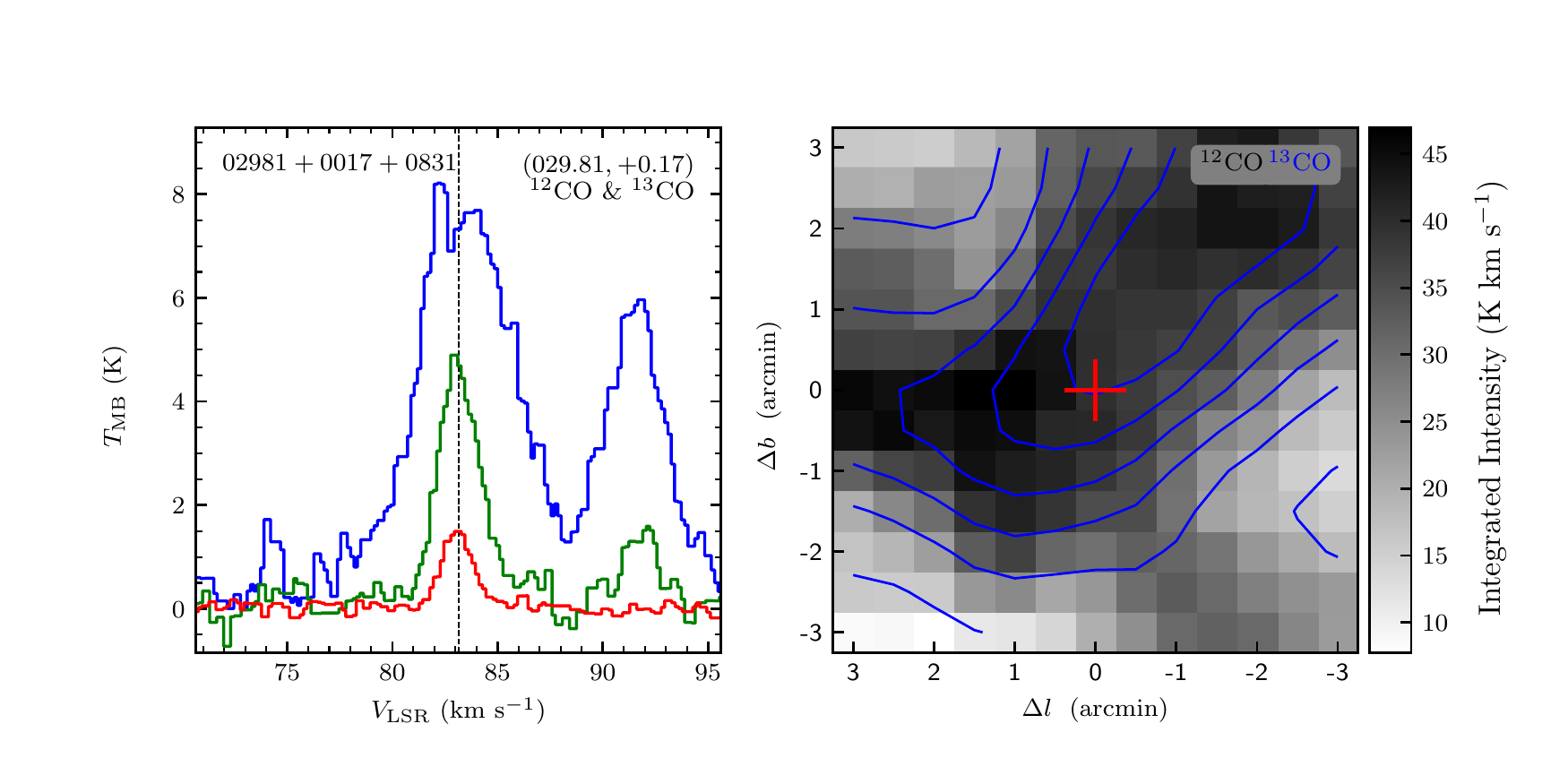}
\vspace{-0.5cm}

\includegraphics[width=9.0cm,angle=0]{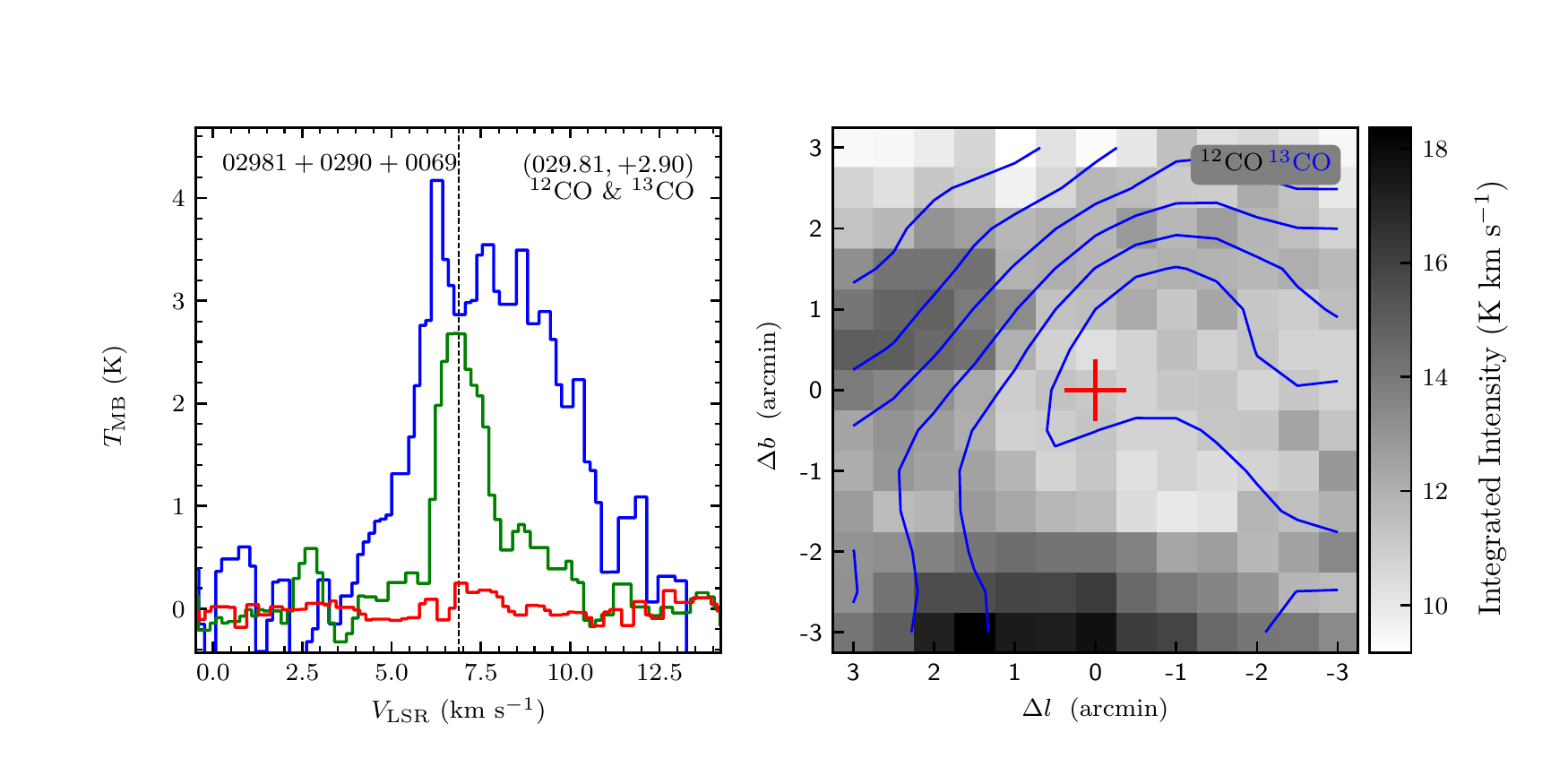}
\includegraphics[width=9.0cm,angle=0]{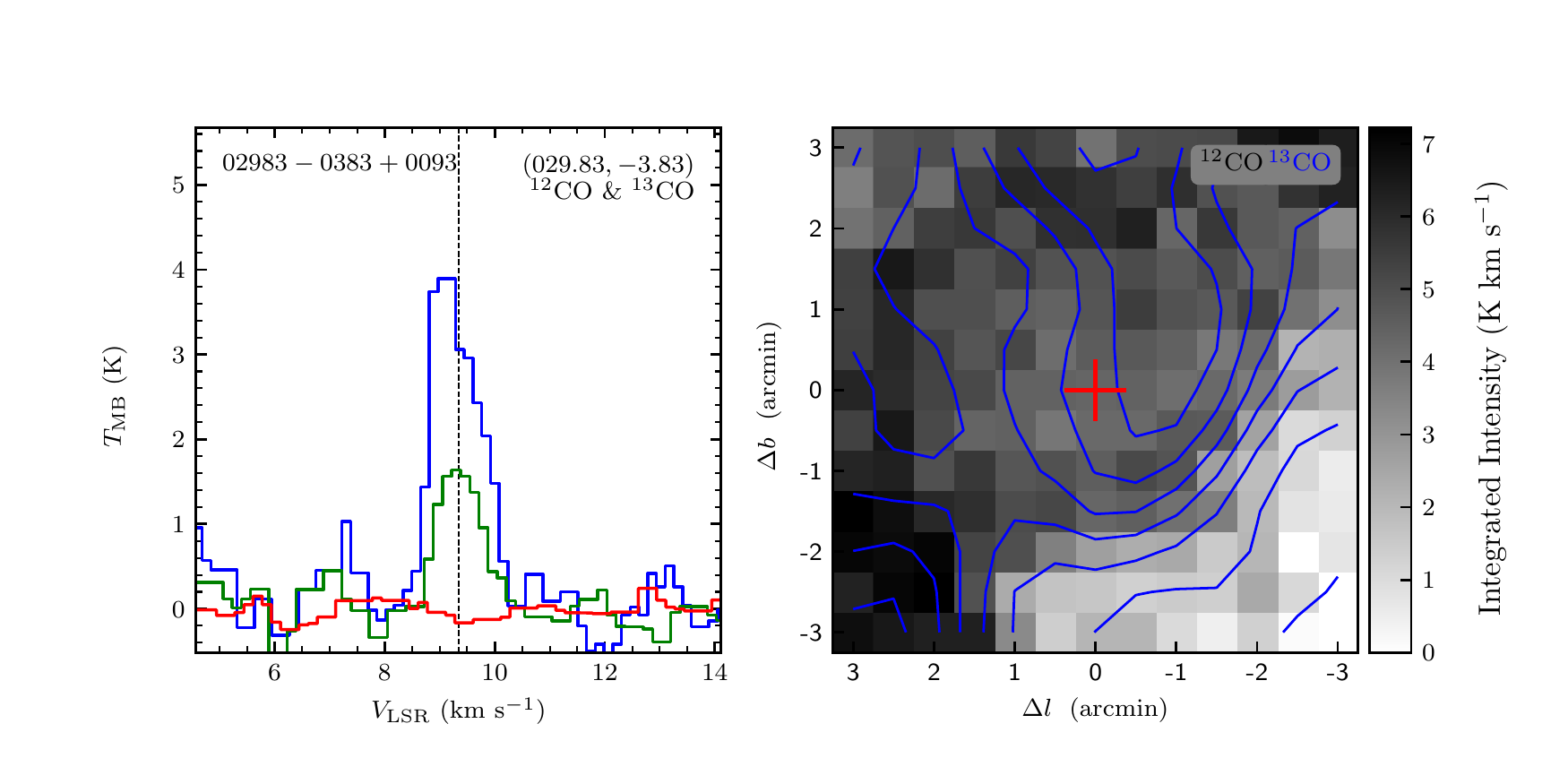}
\end{figure}
\clearpage

\begin{figure}
\includegraphics[width=9.0cm,angle=0]{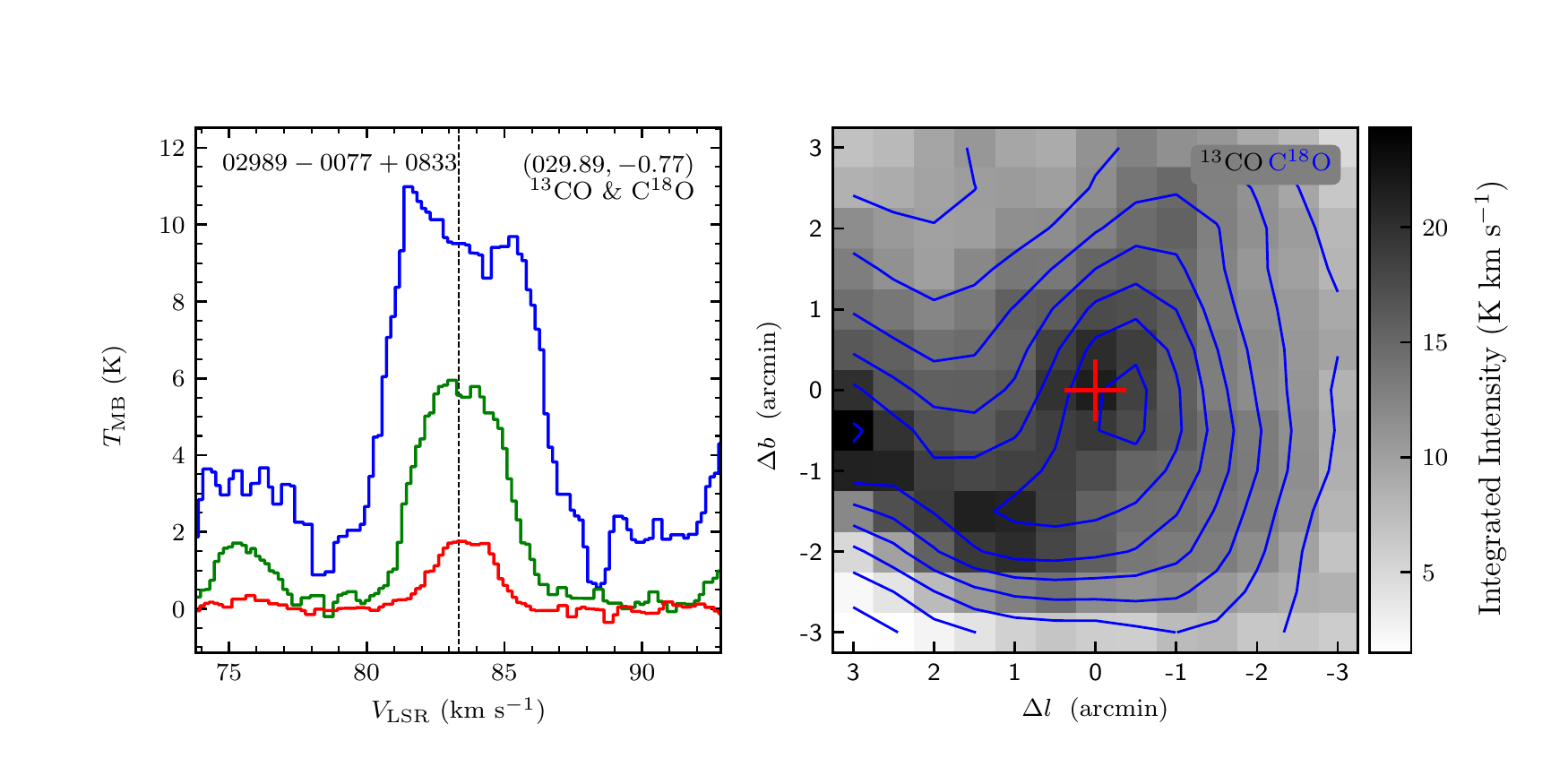}
\includegraphics[width=9.0cm,angle=0]{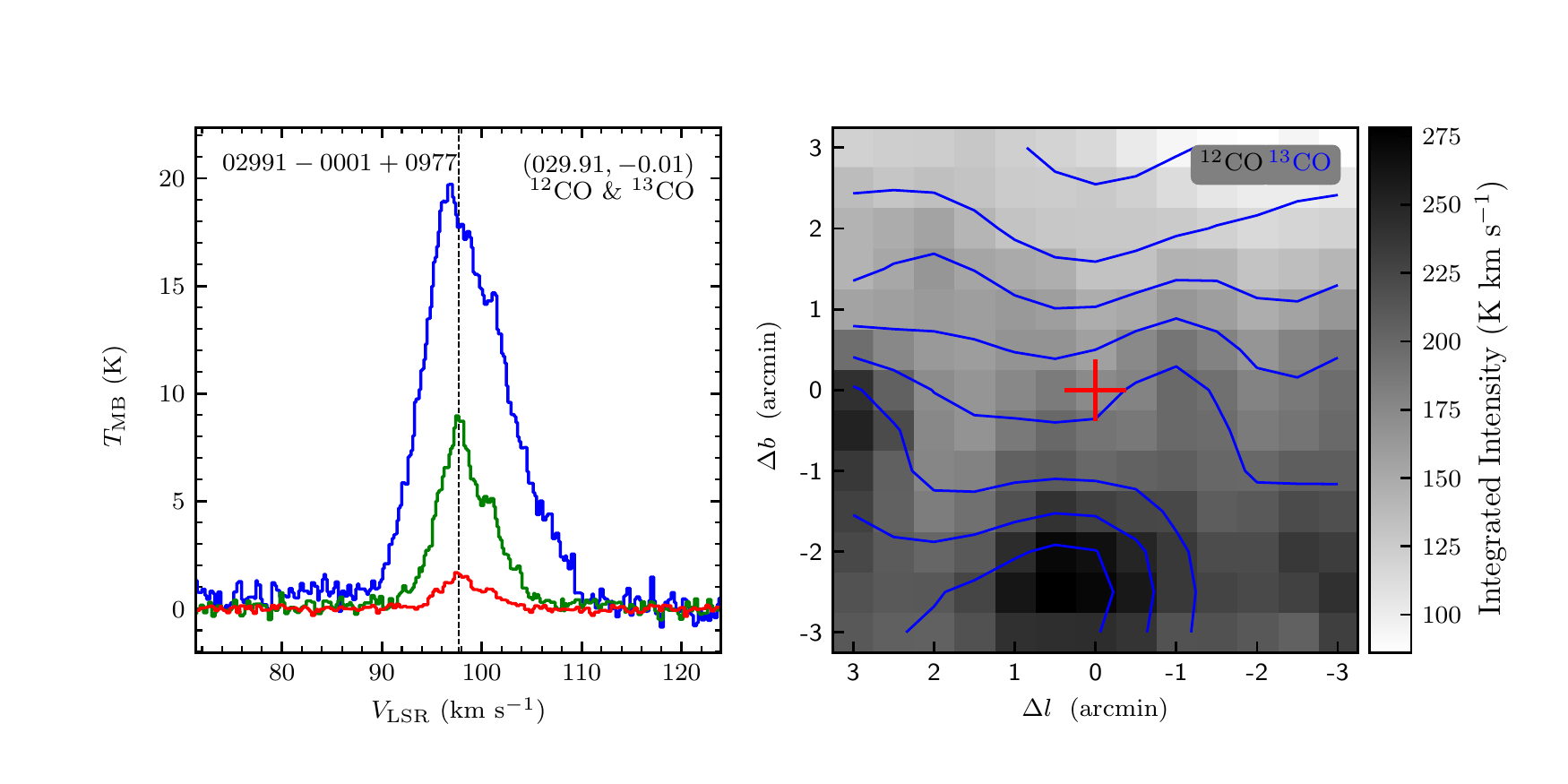}
\vspace{-0.5cm}

\includegraphics[width=9.0cm,angle=0]{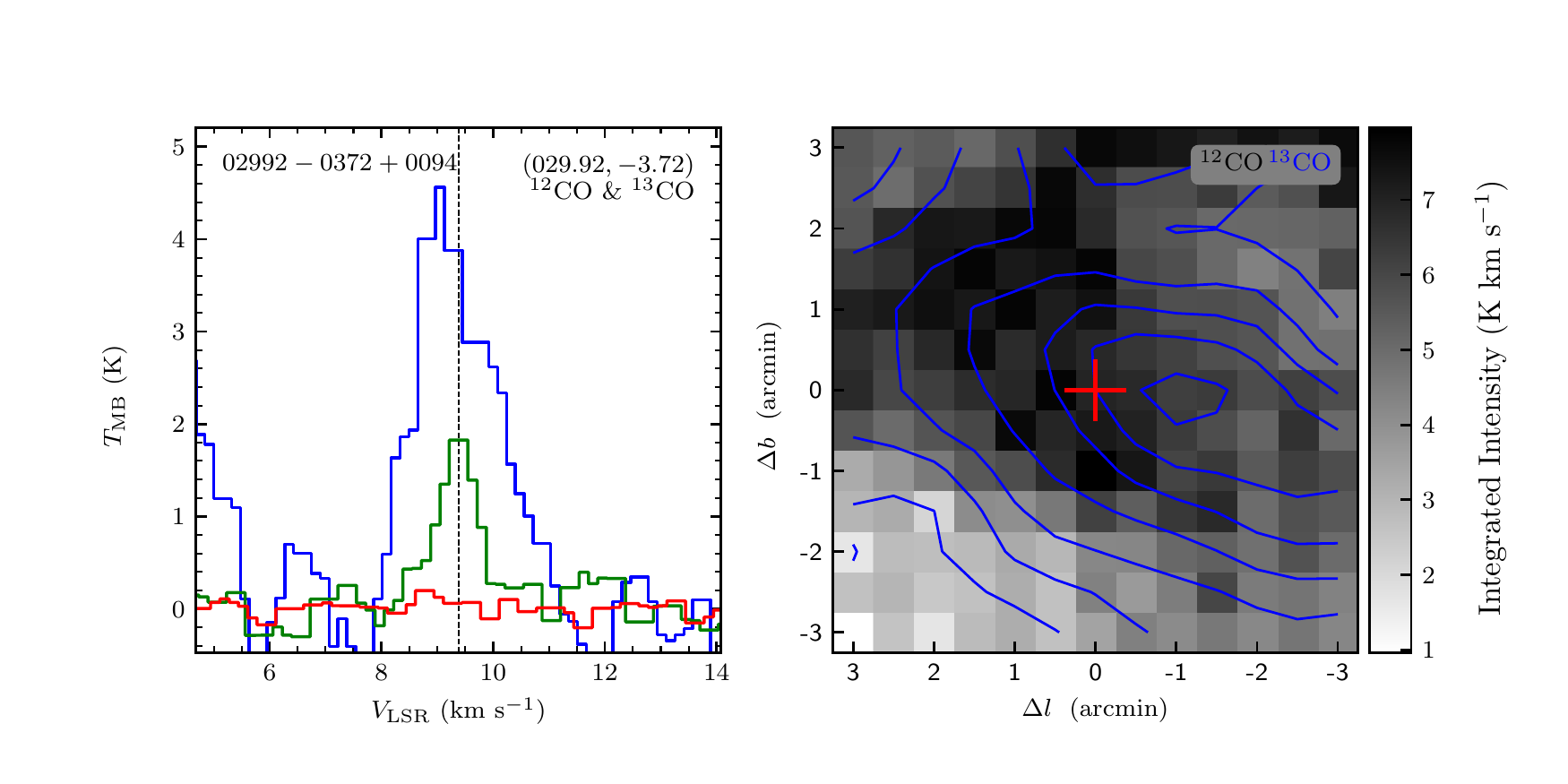}
\includegraphics[width=9.0cm,angle=0]{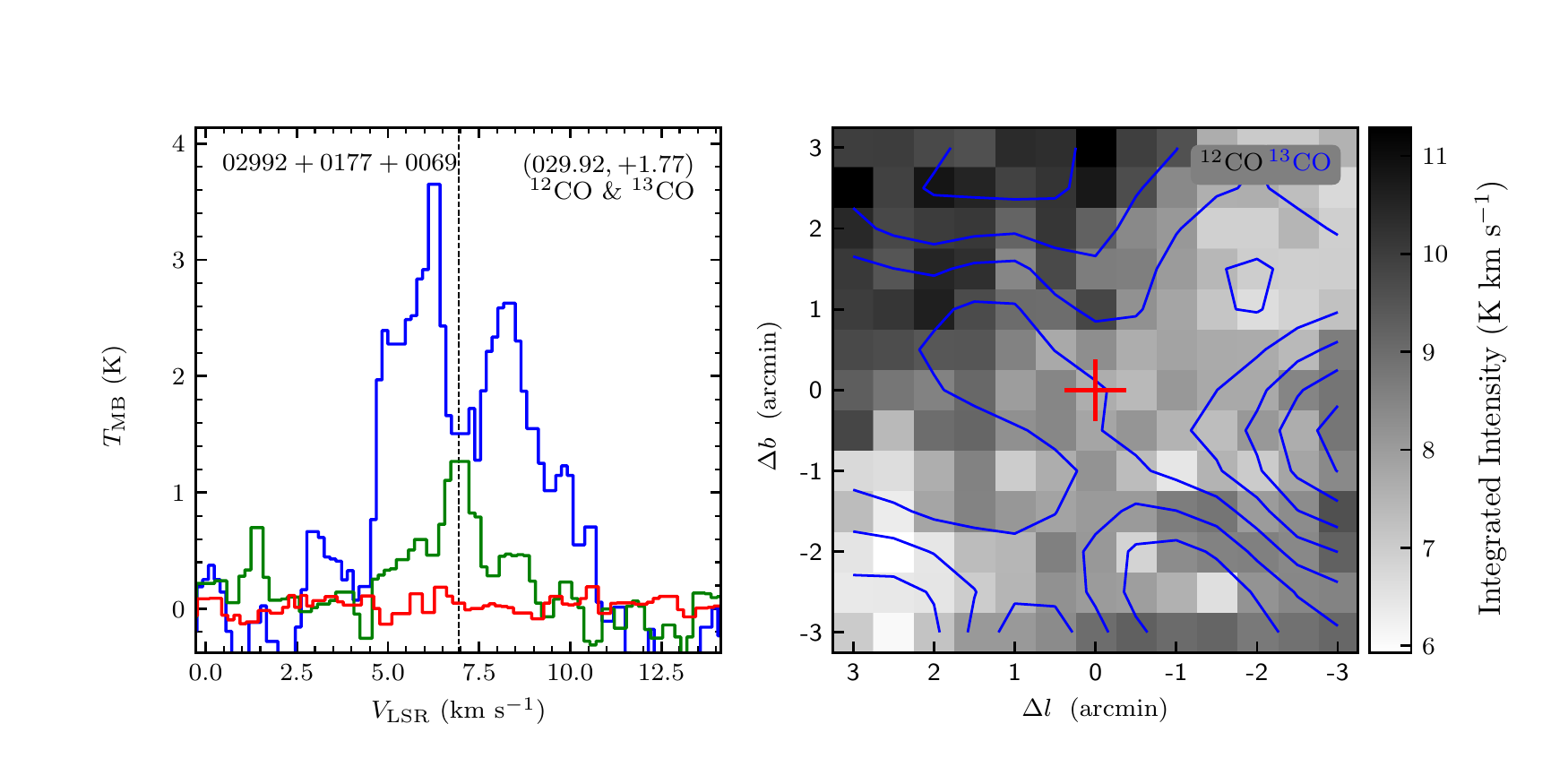}
\vspace{-0.5cm}

\includegraphics[width=9.0cm,angle=0]{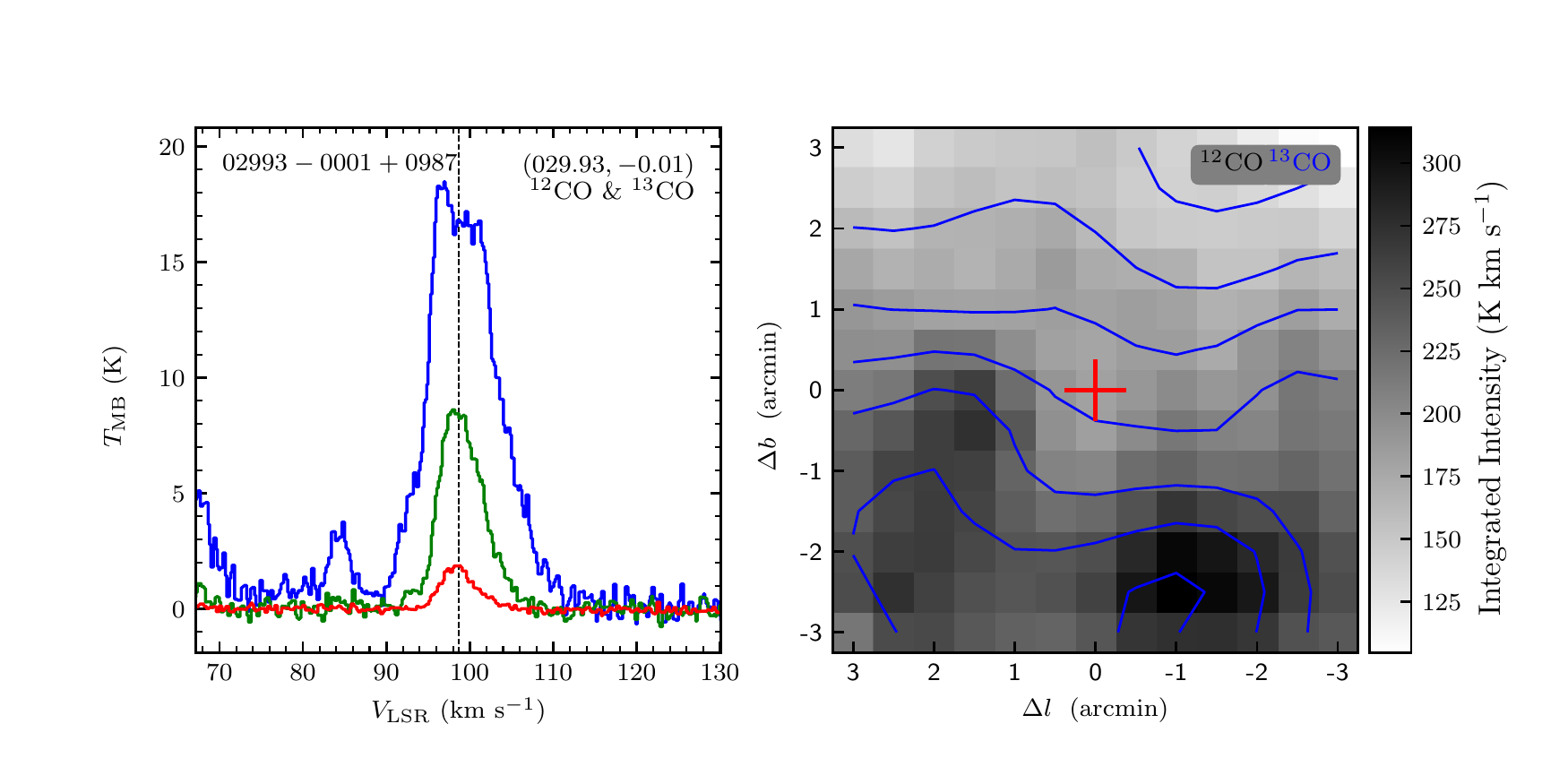}
\includegraphics[width=9.0cm,angle=0]{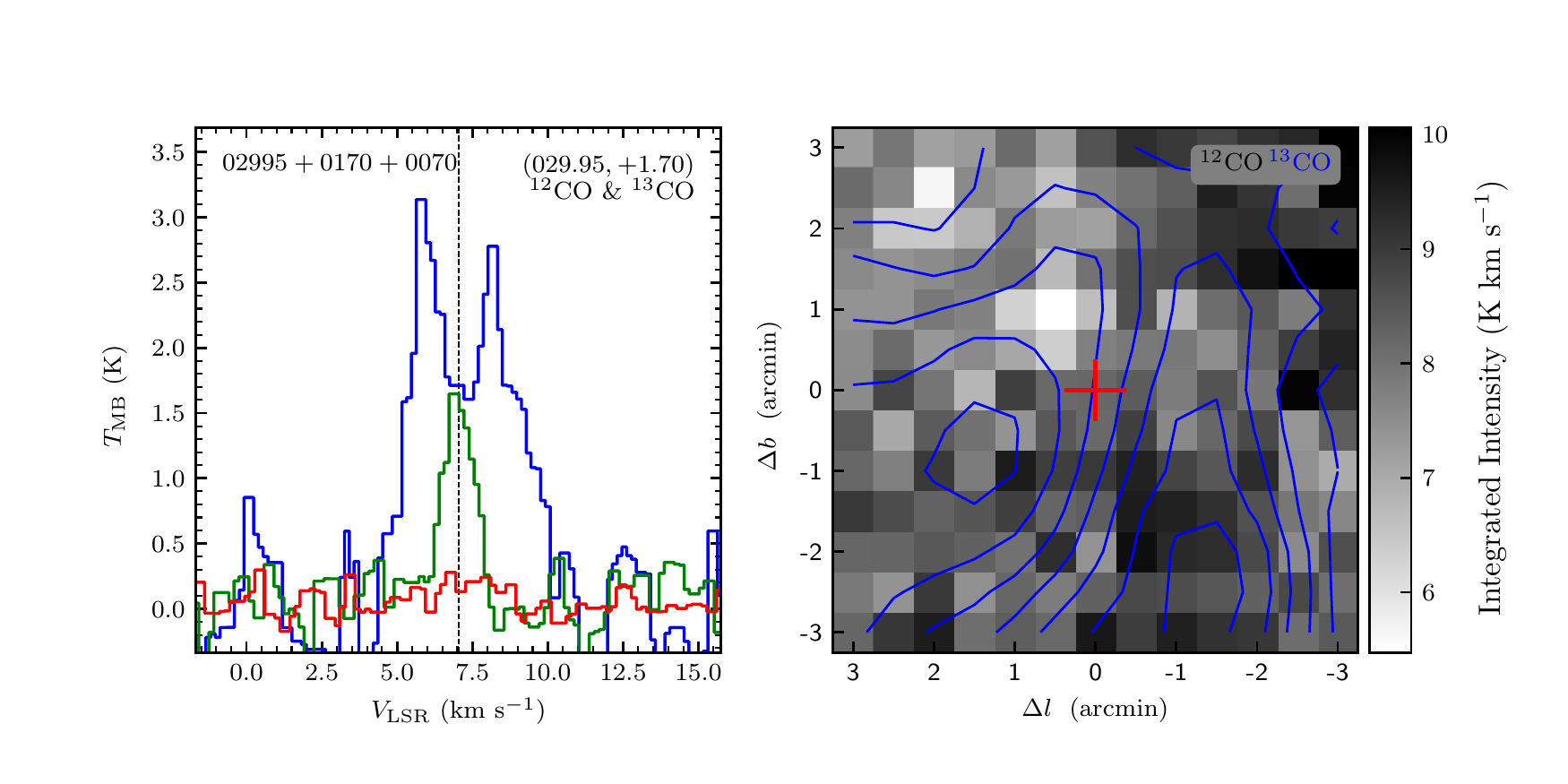}
\vspace{-0.5cm}

\includegraphics[width=9.0cm,angle=0]{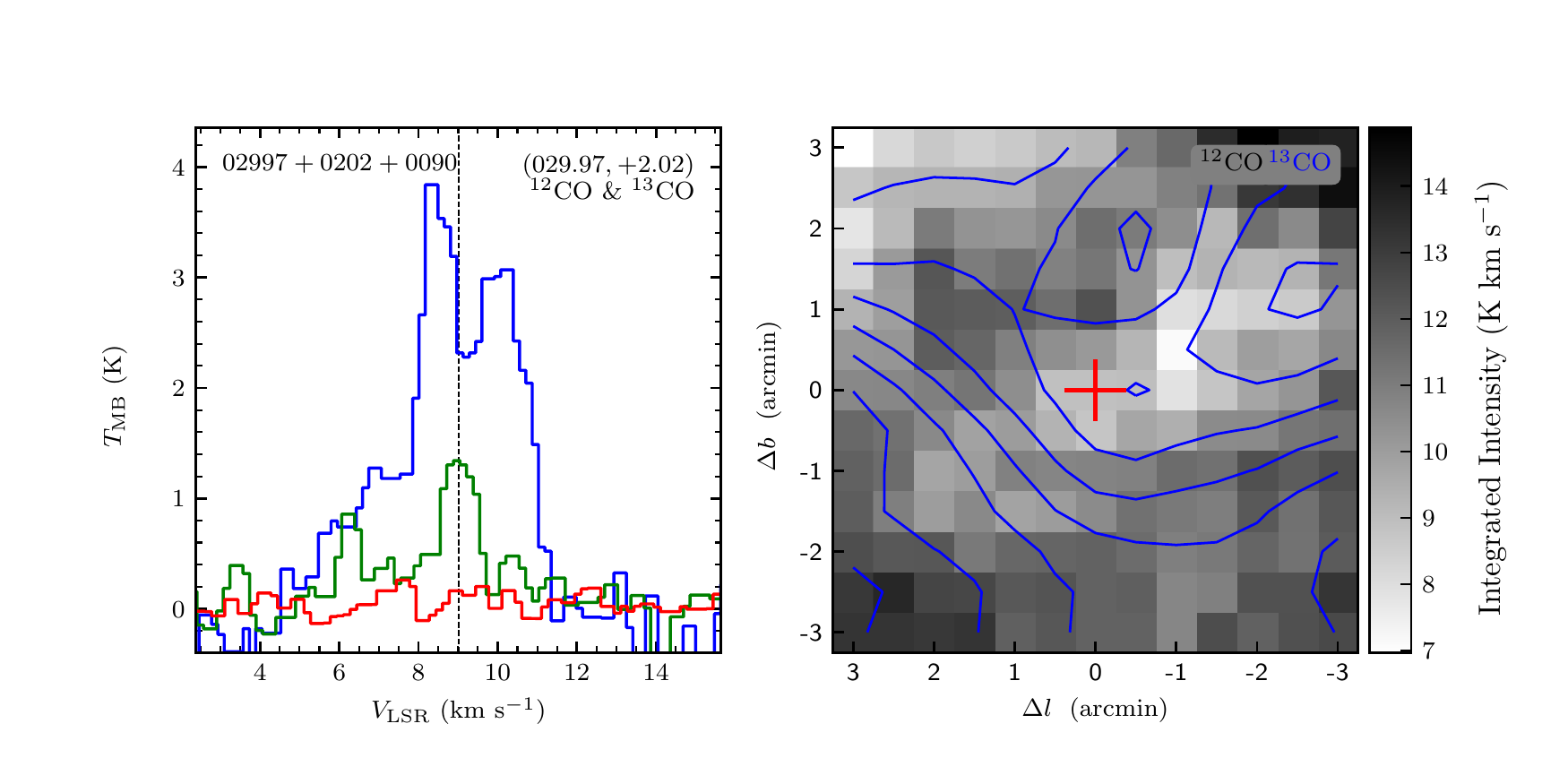}
\includegraphics[width=9.0cm,angle=0]{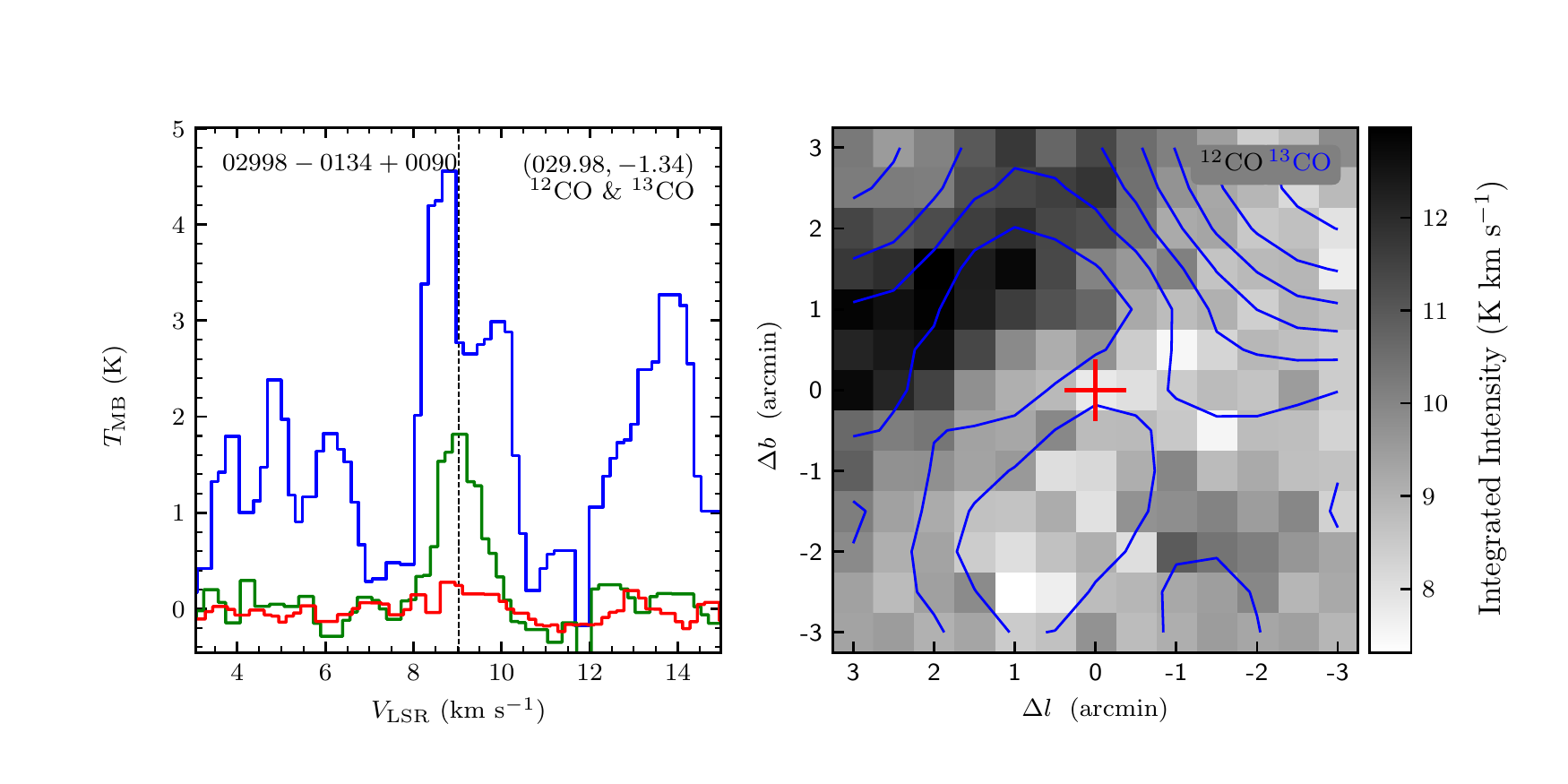}
\vspace{-0.5cm}

\includegraphics[width=9.0cm,angle=0]{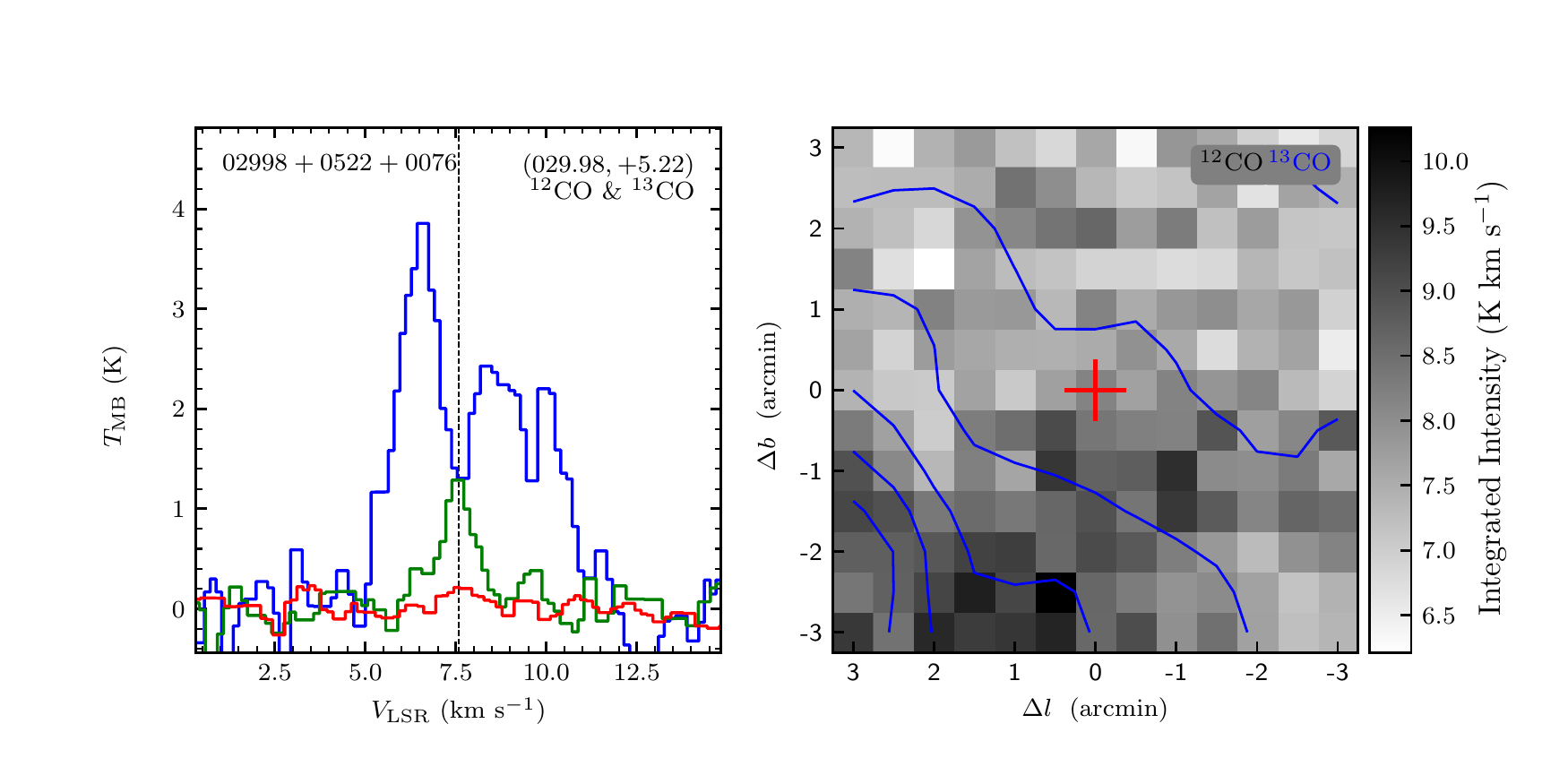}
\includegraphics[width=9.0cm,angle=0]{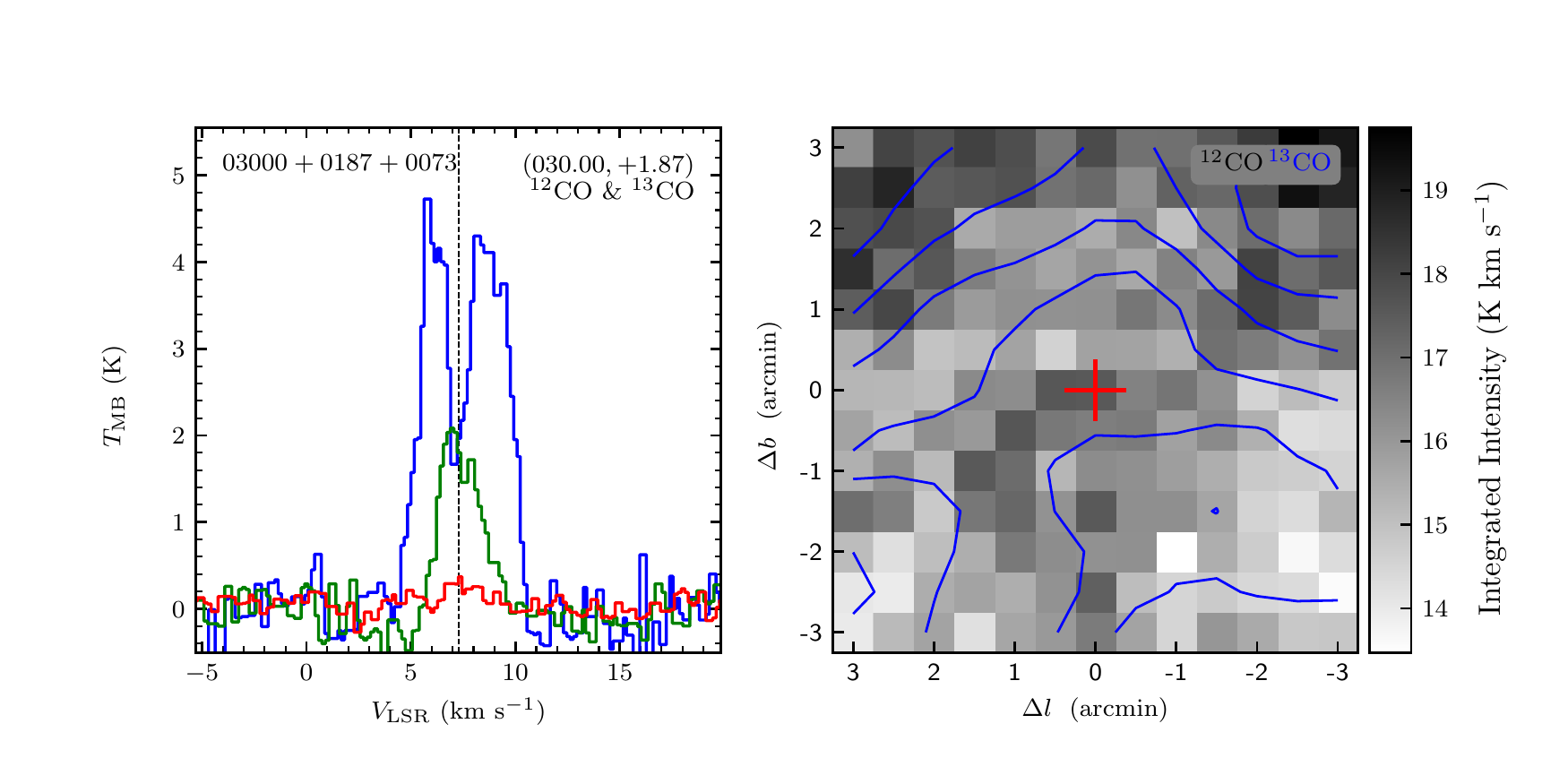}
\end{figure}
\clearpage

\begin{figure}
\includegraphics[width=9.0cm,angle=0]{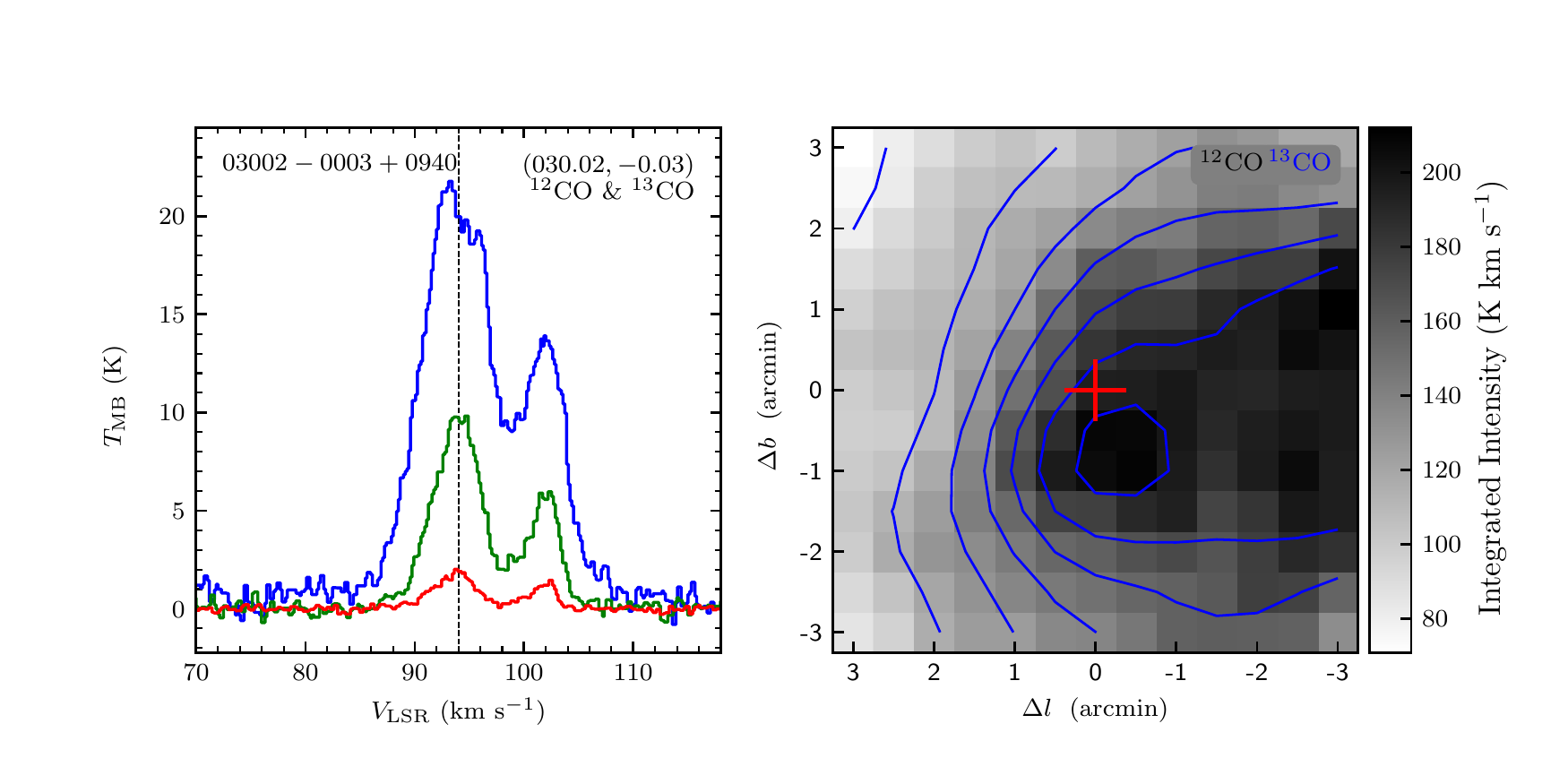}
\includegraphics[width=9.0cm,angle=0]{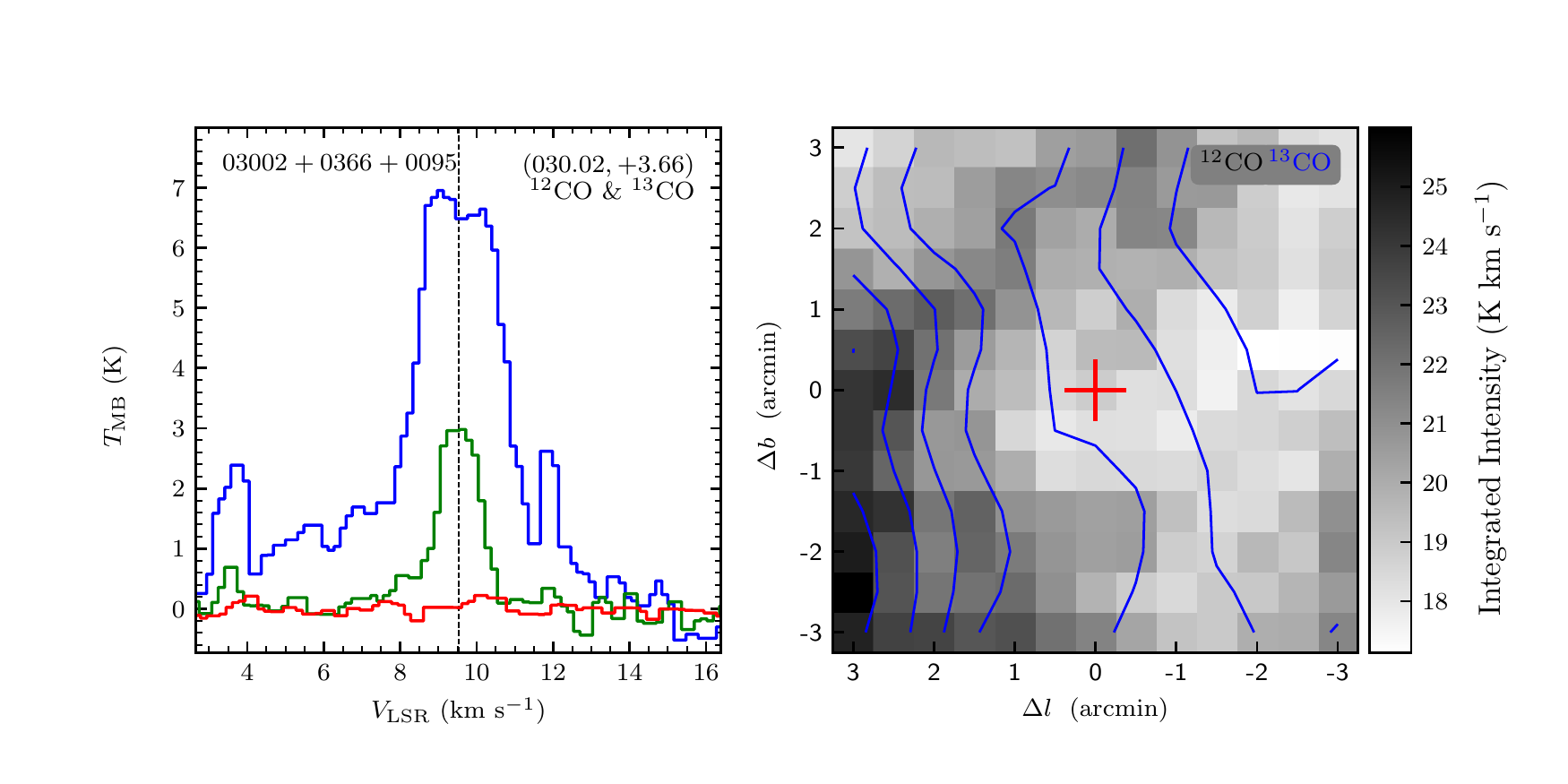}
\vspace{-0.5cm}

\includegraphics[width=9.0cm,angle=0]{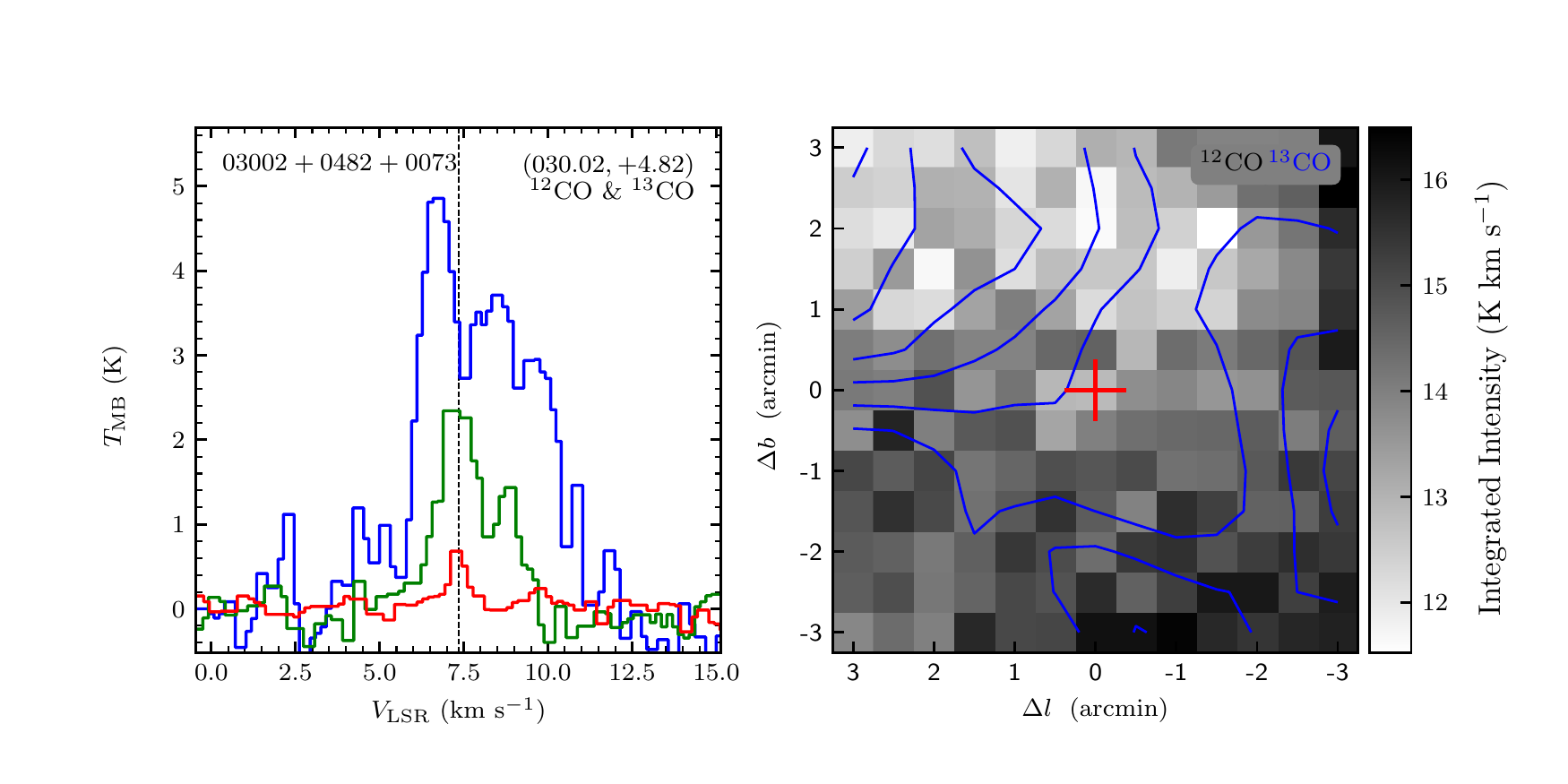}
\includegraphics[width=9.0cm,angle=0]{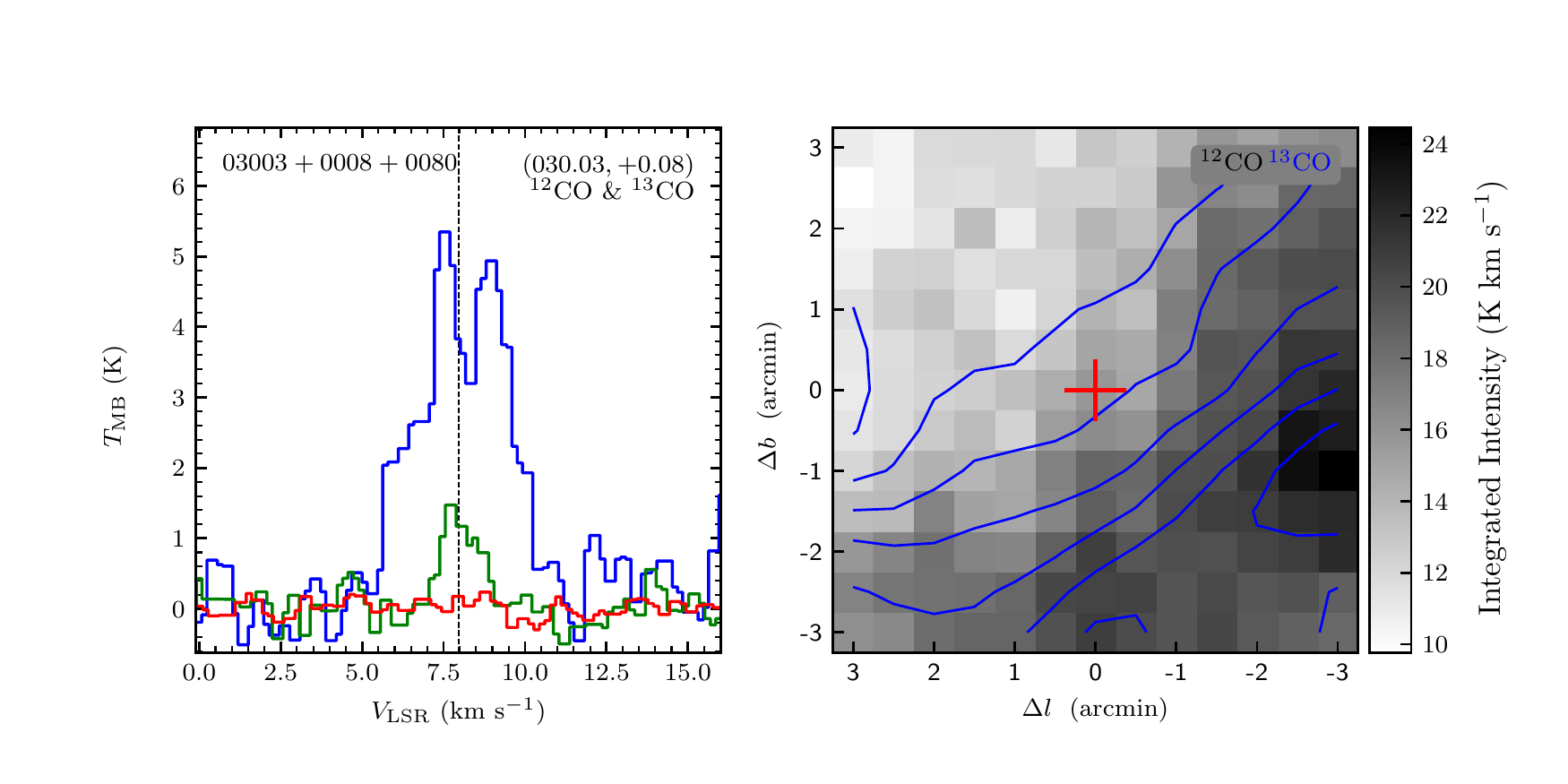}
\vspace{-0.5cm}

\includegraphics[width=9.0cm,angle=0]{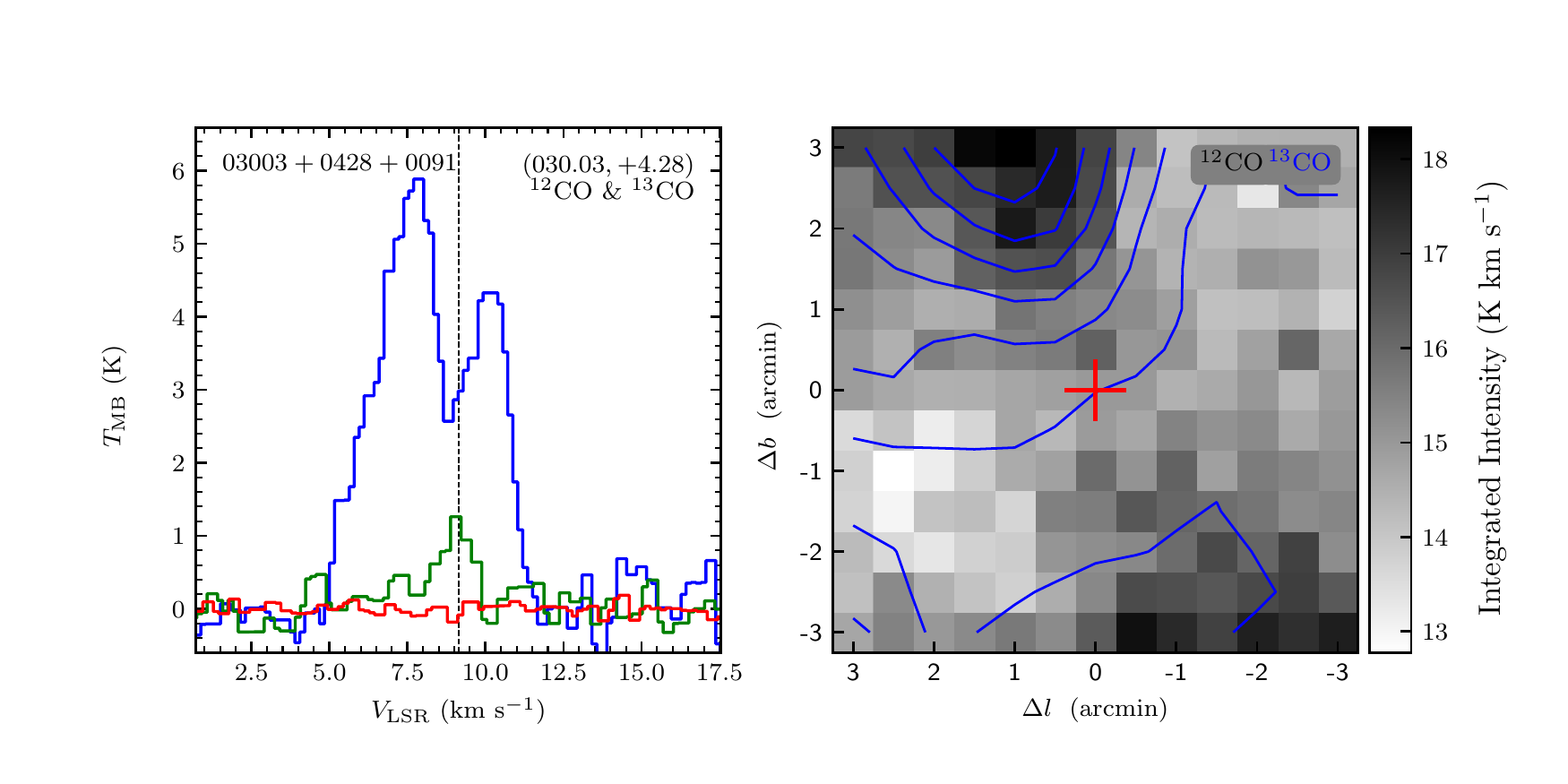}
\includegraphics[width=9.0cm,angle=0]{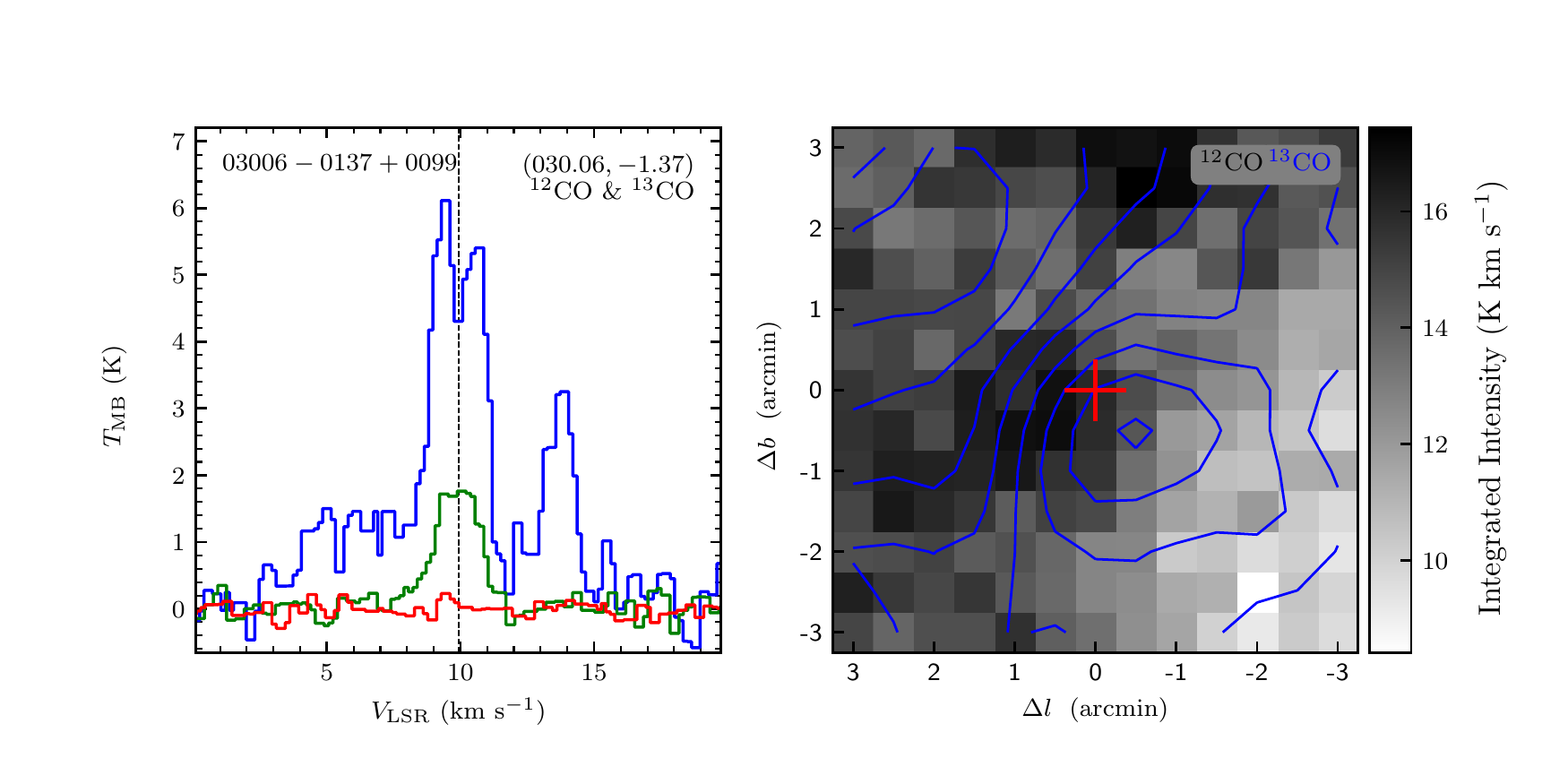}
\vspace{-0.5cm}

\includegraphics[width=9.0cm,angle=0]{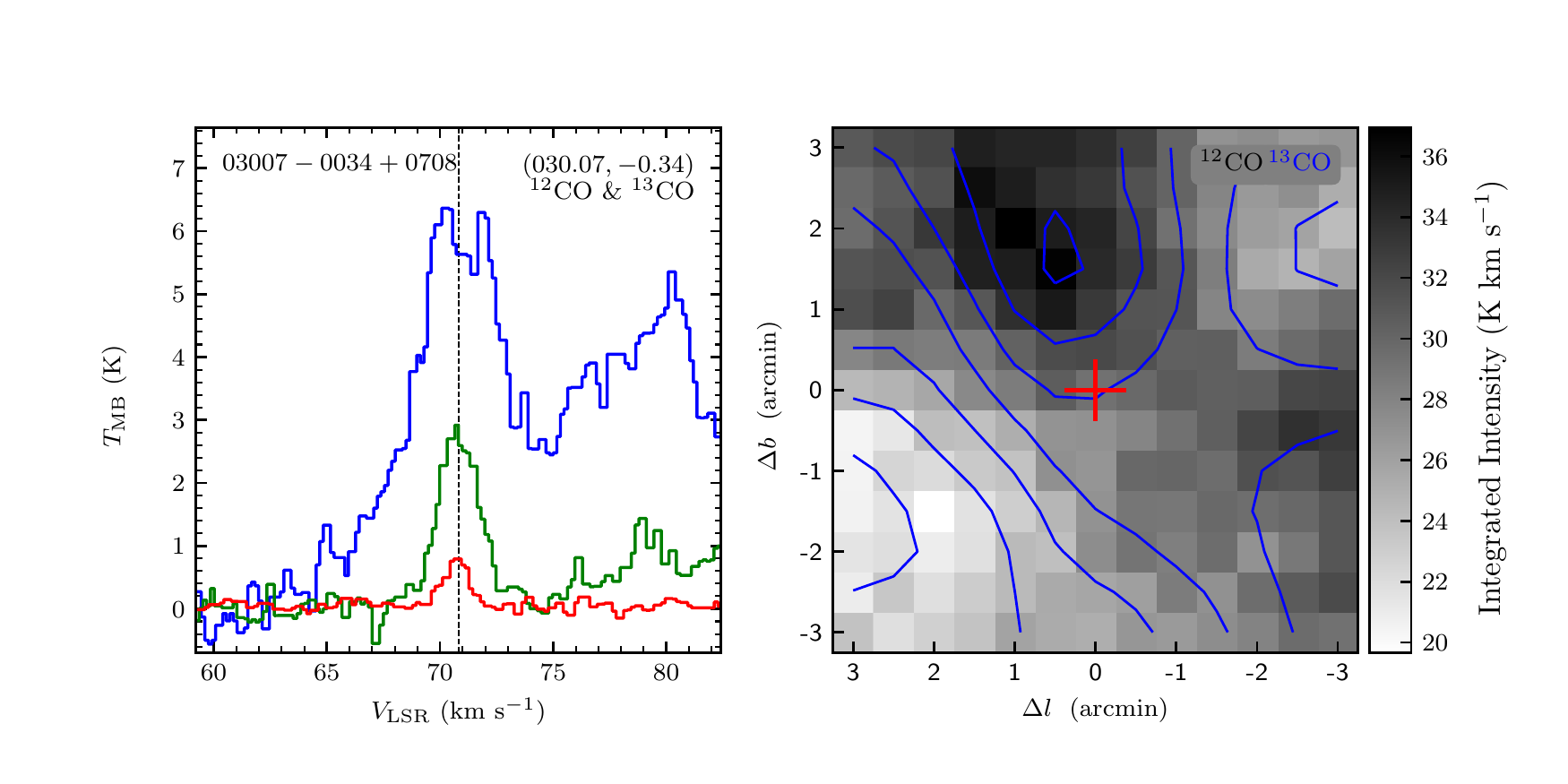}
\includegraphics[width=9.0cm,angle=0]{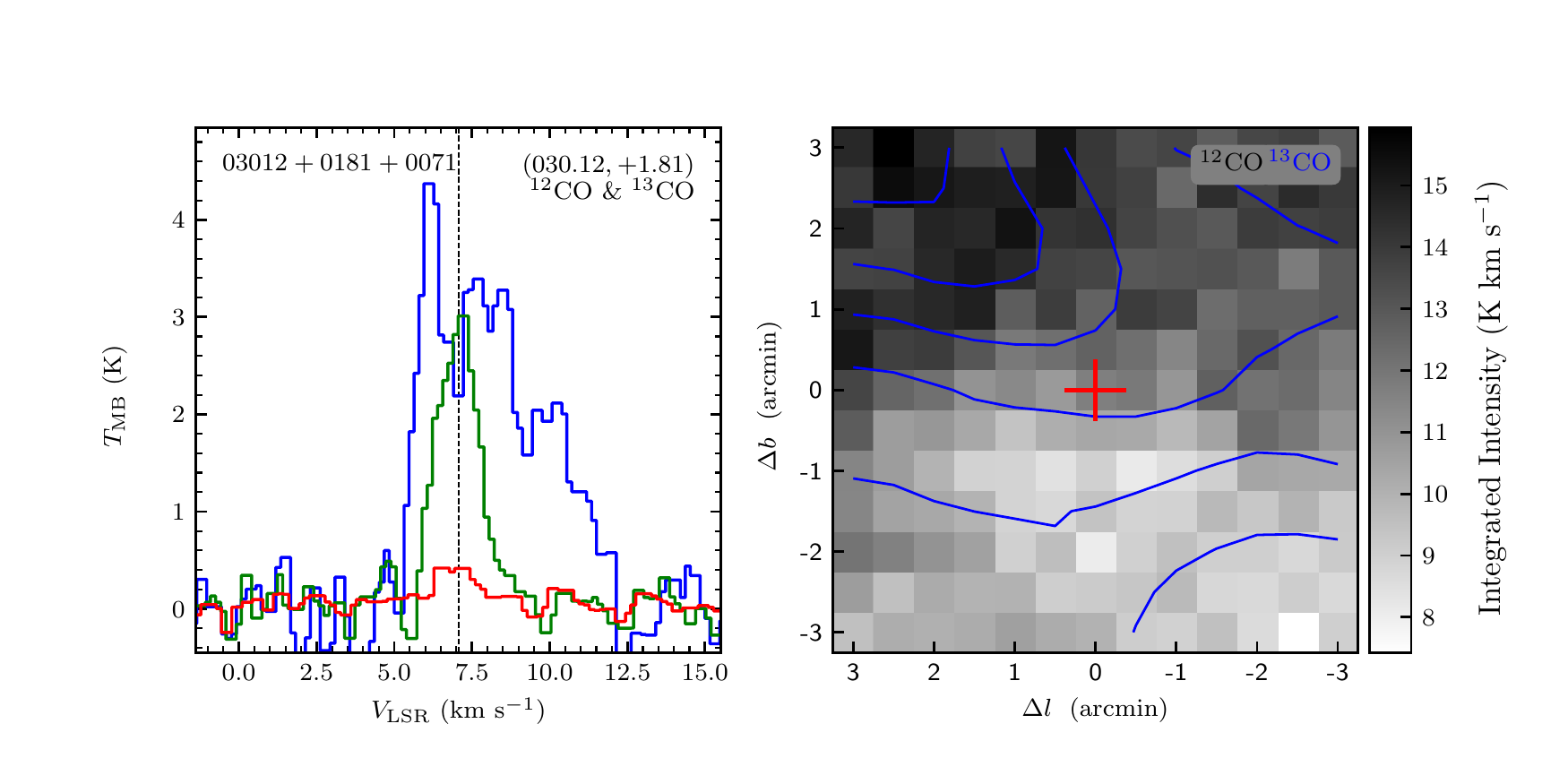}
\vspace{-0.5cm}

\includegraphics[width=9.0cm,angle=0]{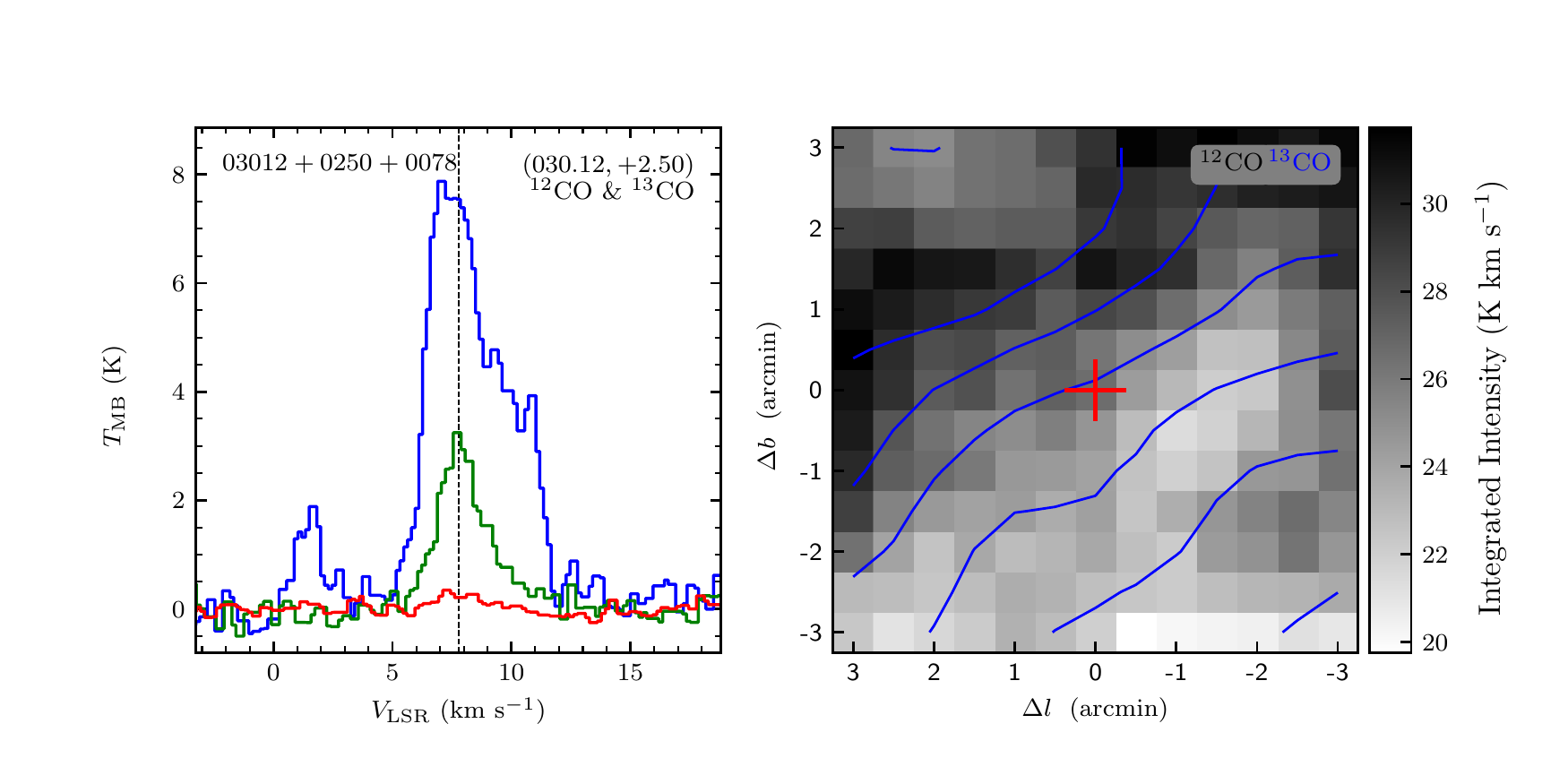}
\includegraphics[width=9.0cm,angle=0]{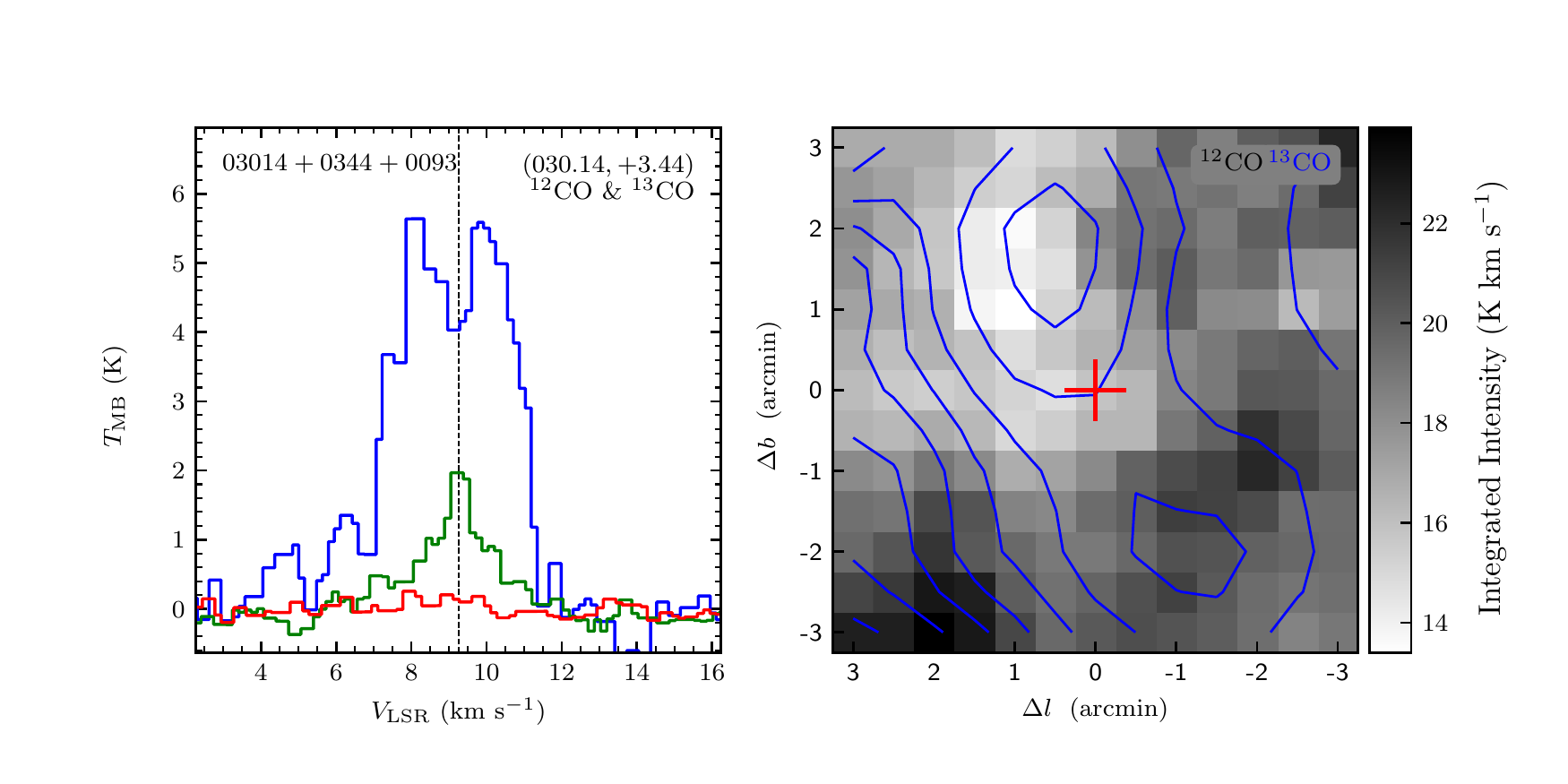}
\end{figure}
\clearpage

\begin{figure}
\includegraphics[width=9.0cm,angle=0]{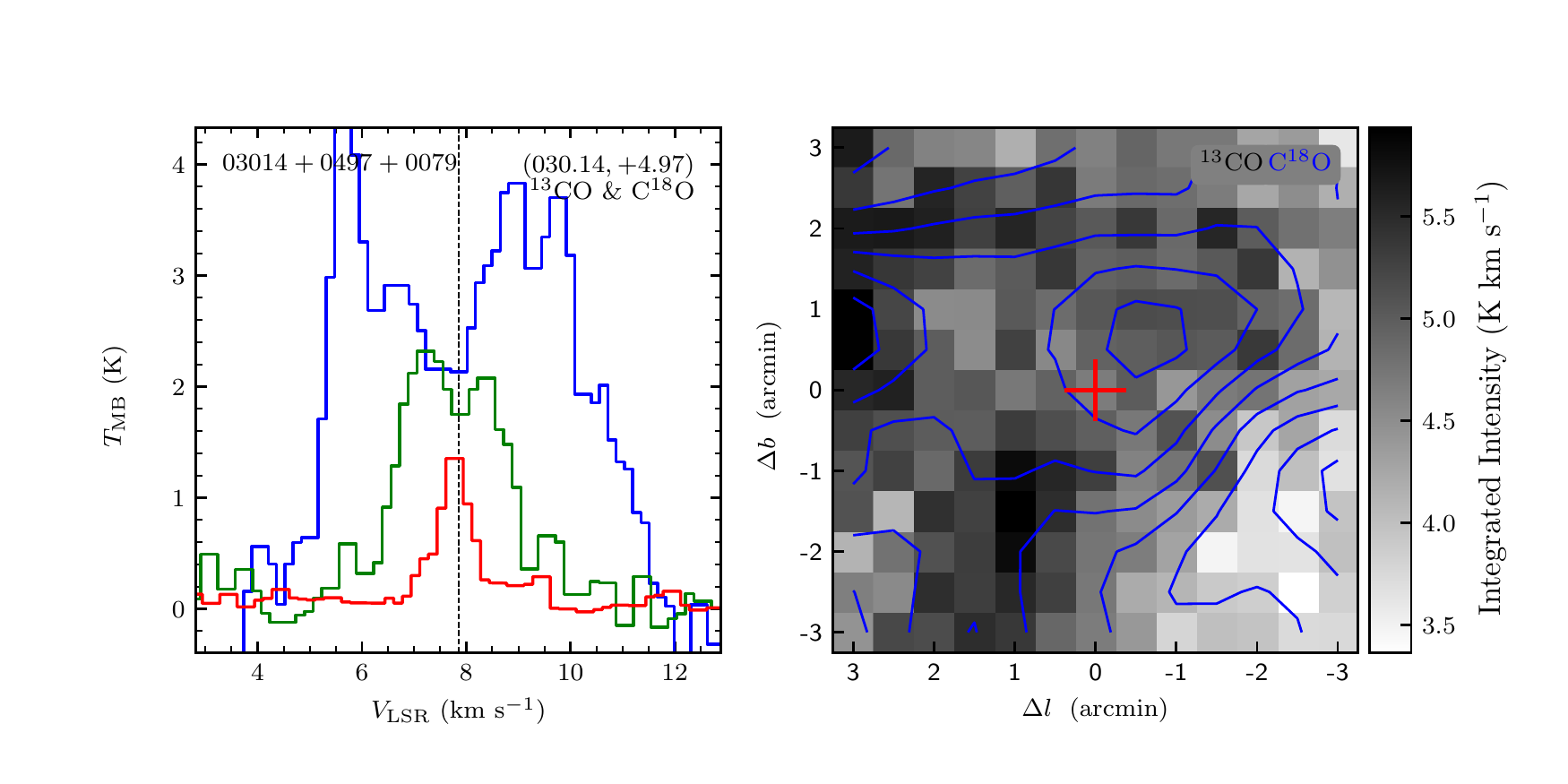}
\includegraphics[width=9.0cm,angle=0]{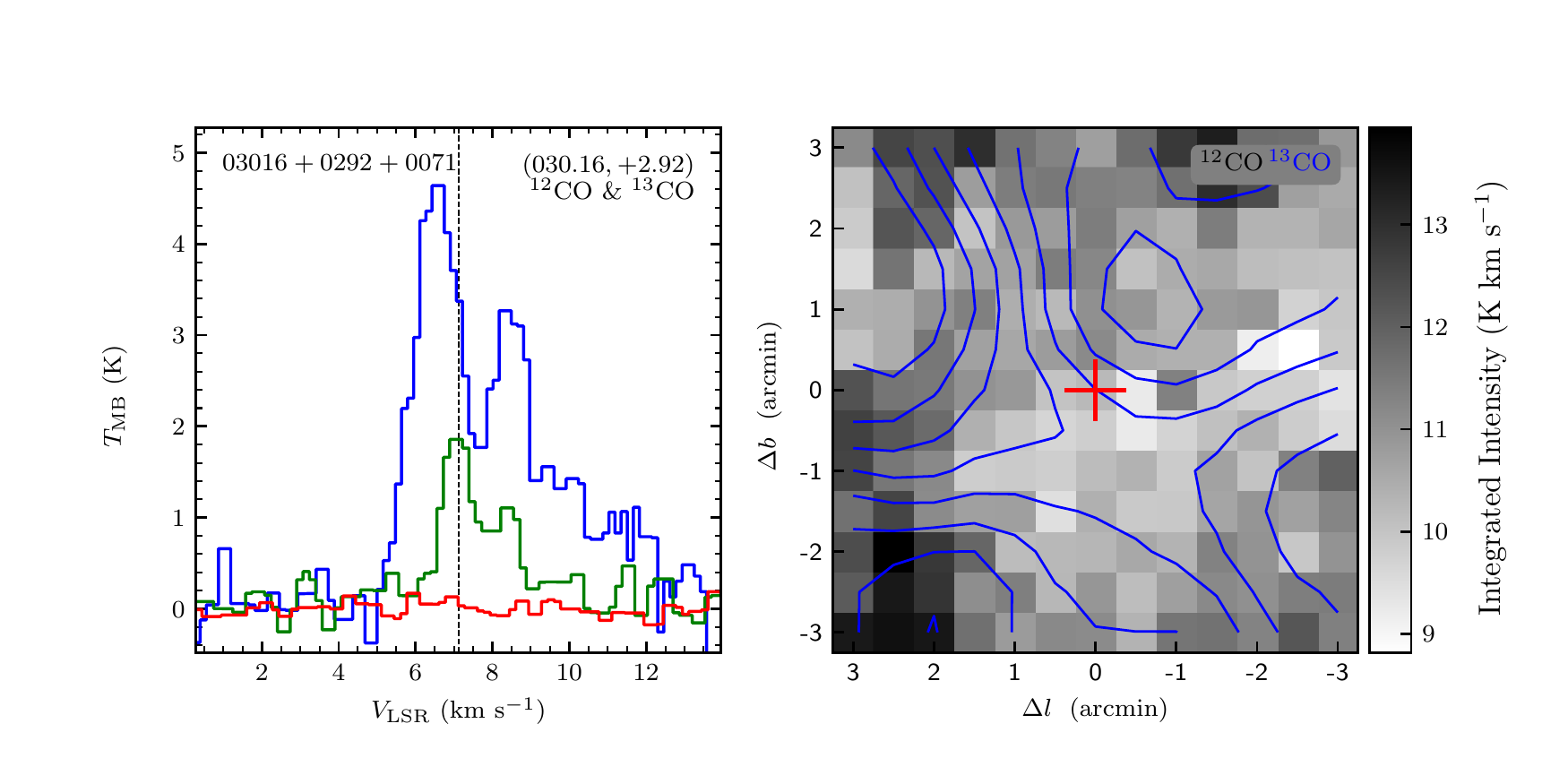}
\vspace{-0.5cm}

\includegraphics[width=9.0cm,angle=0]{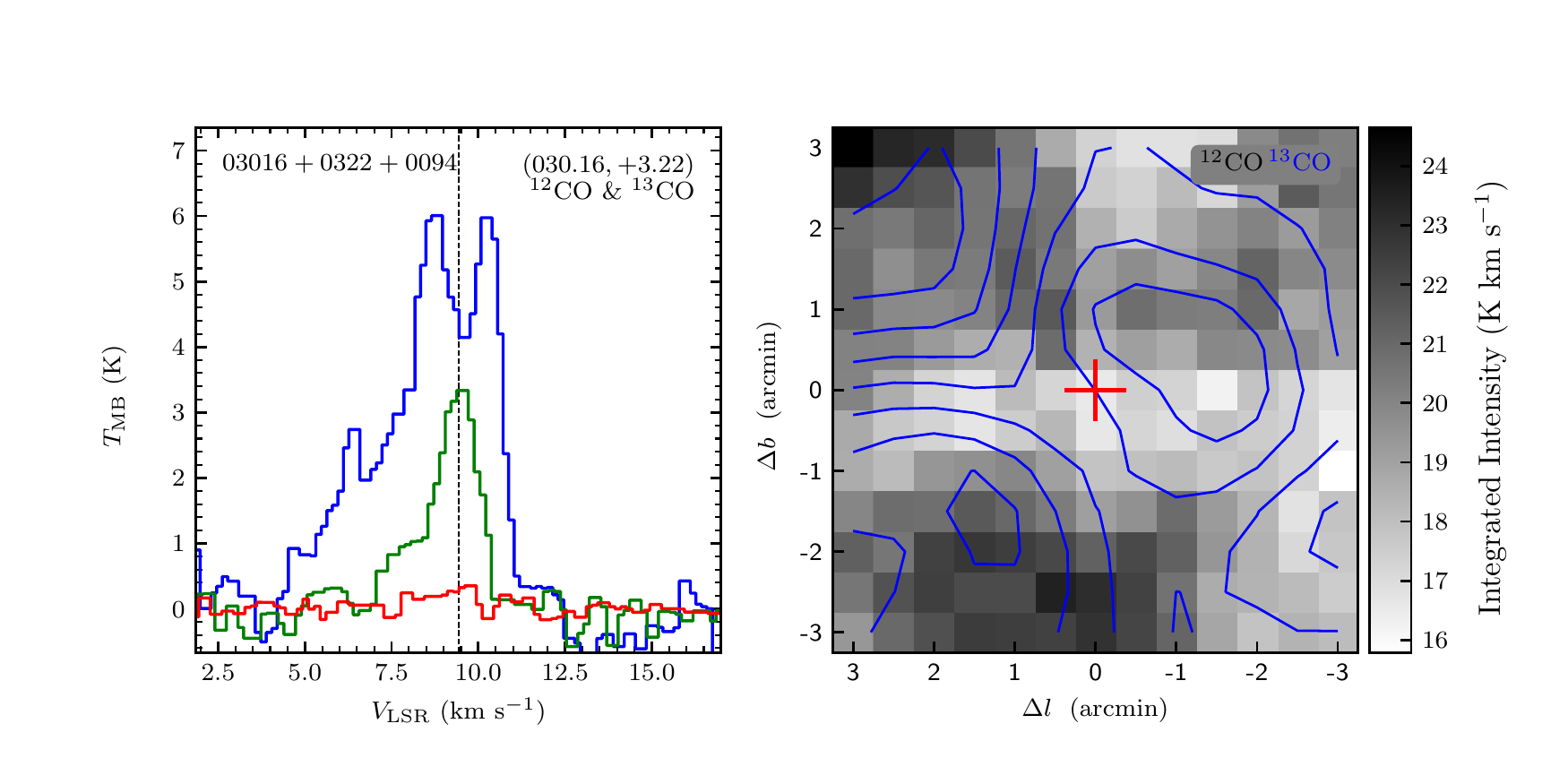}
\includegraphics[width=9.0cm,angle=0]{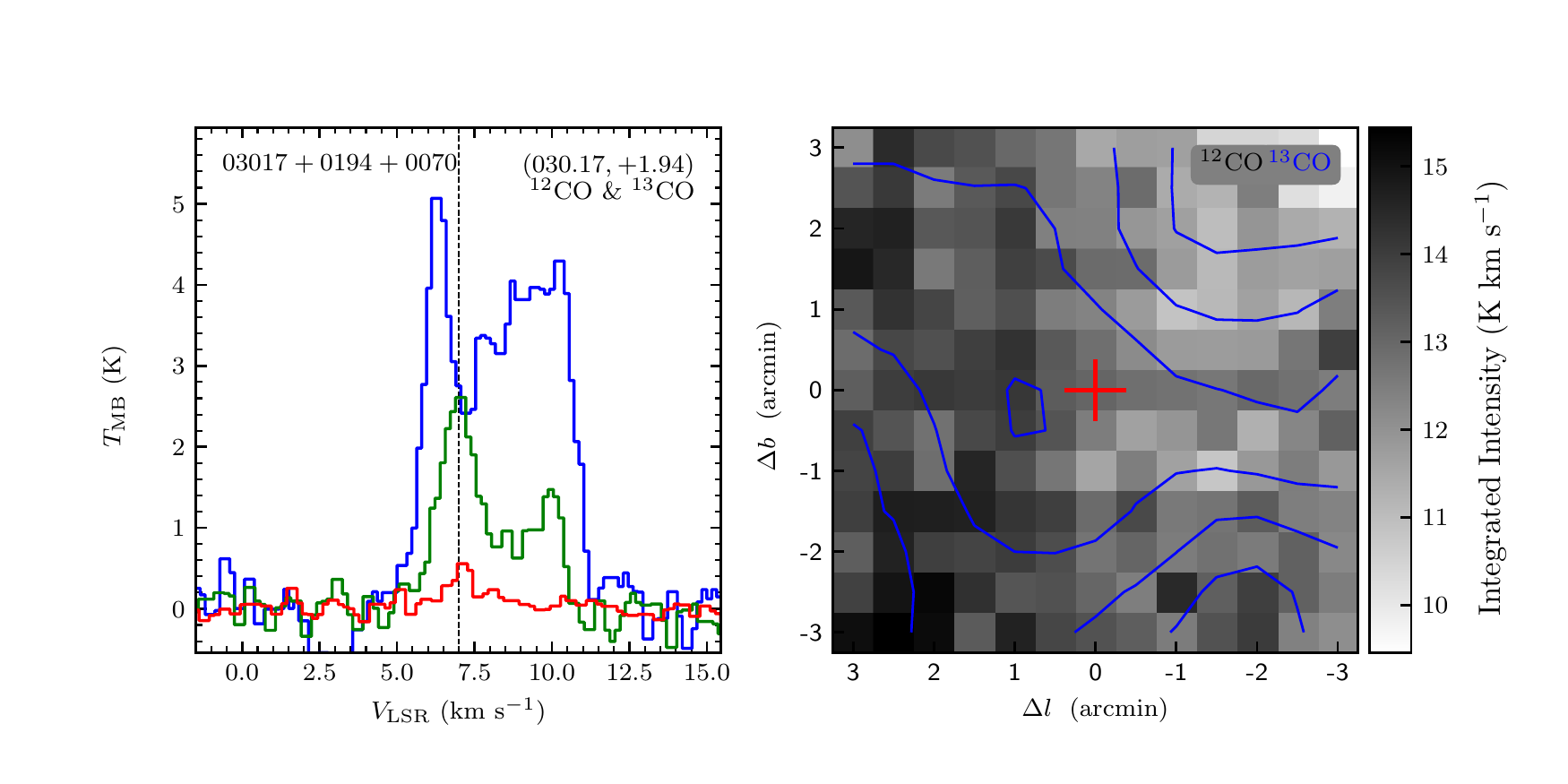}
\vspace{-0.5cm}

\includegraphics[width=9.0cm,angle=0]{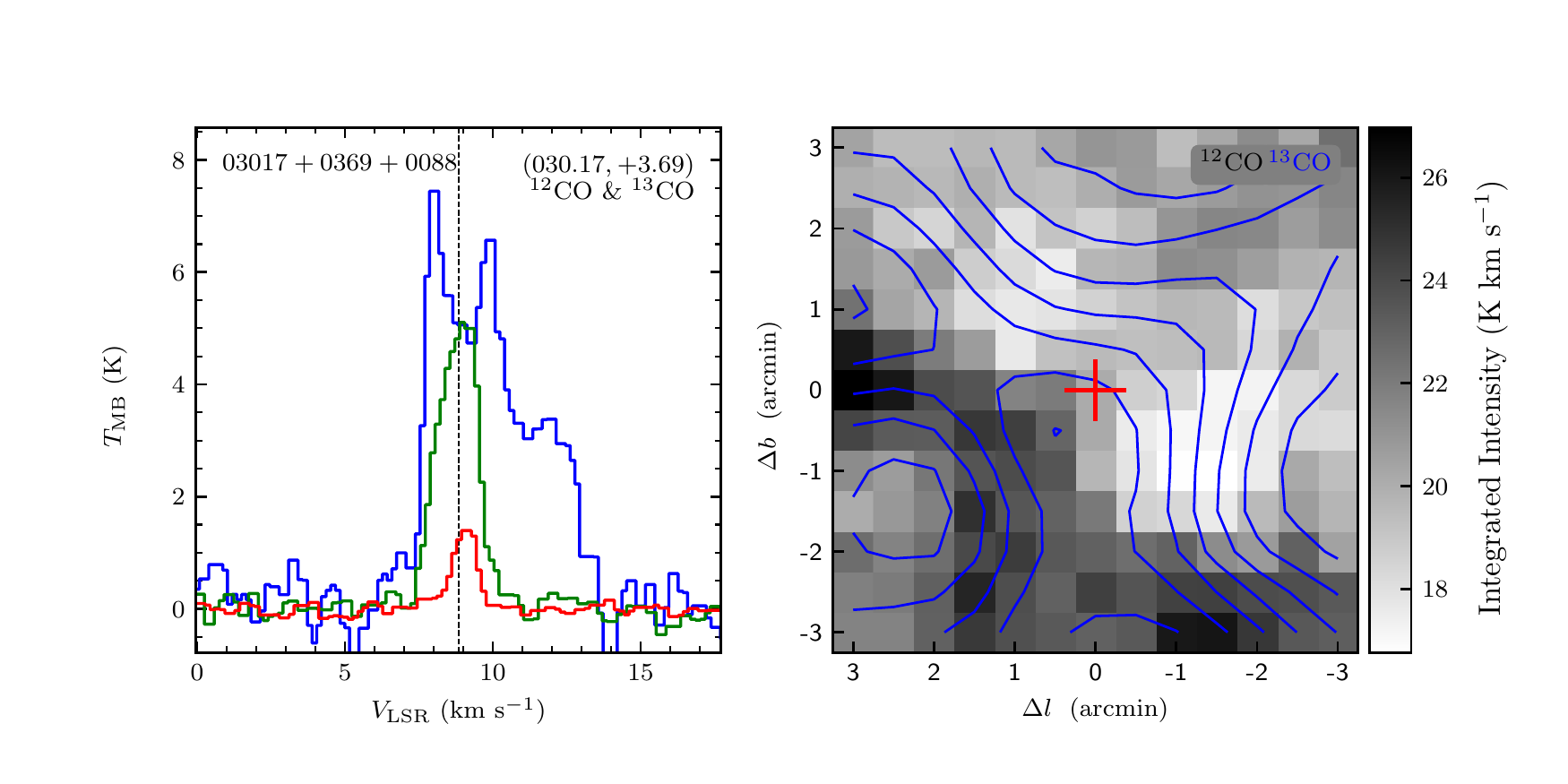}
\includegraphics[width=9.0cm,angle=0]{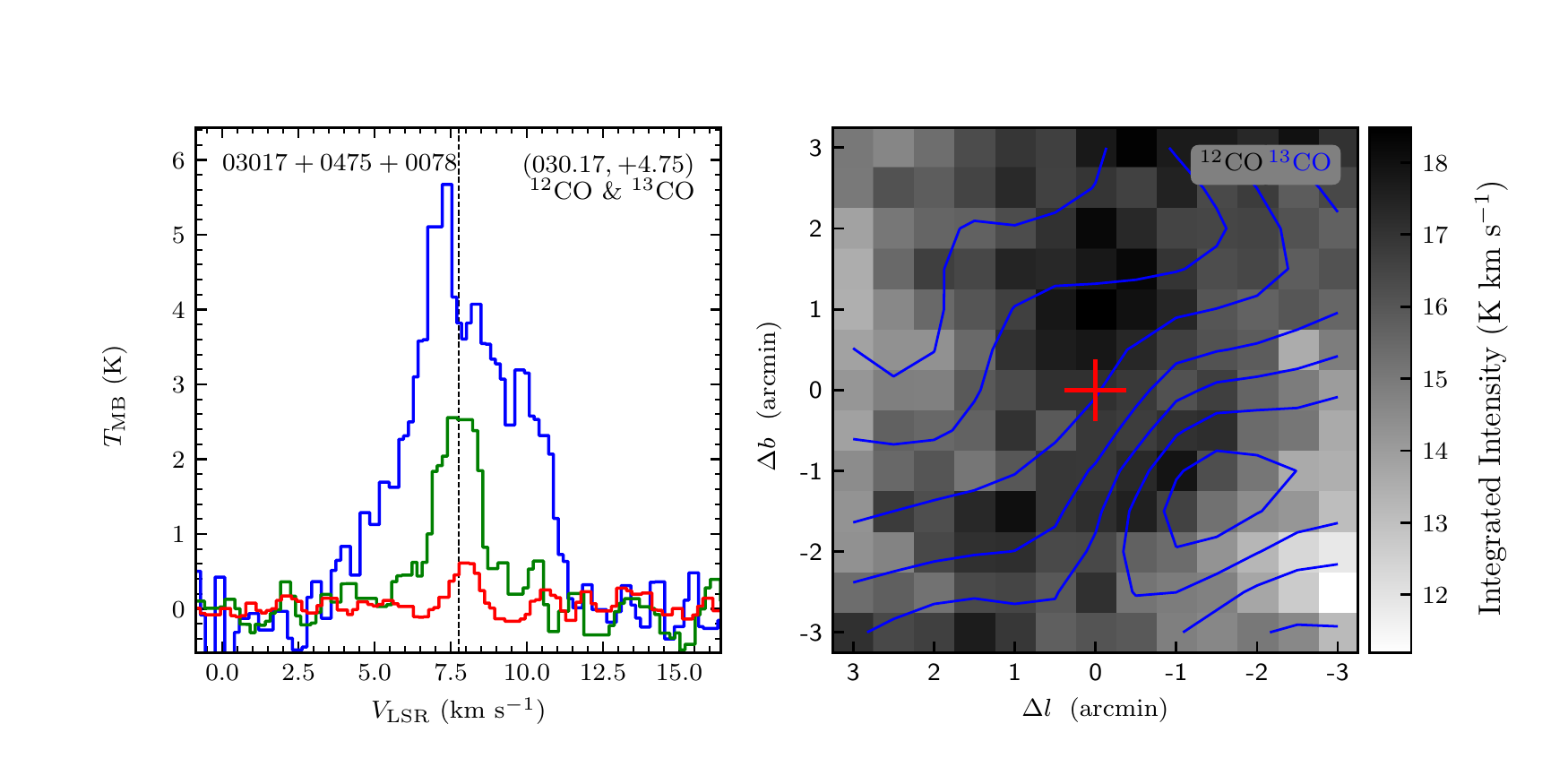}
\vspace{-0.5cm}

\includegraphics[width=9.0cm,angle=0]{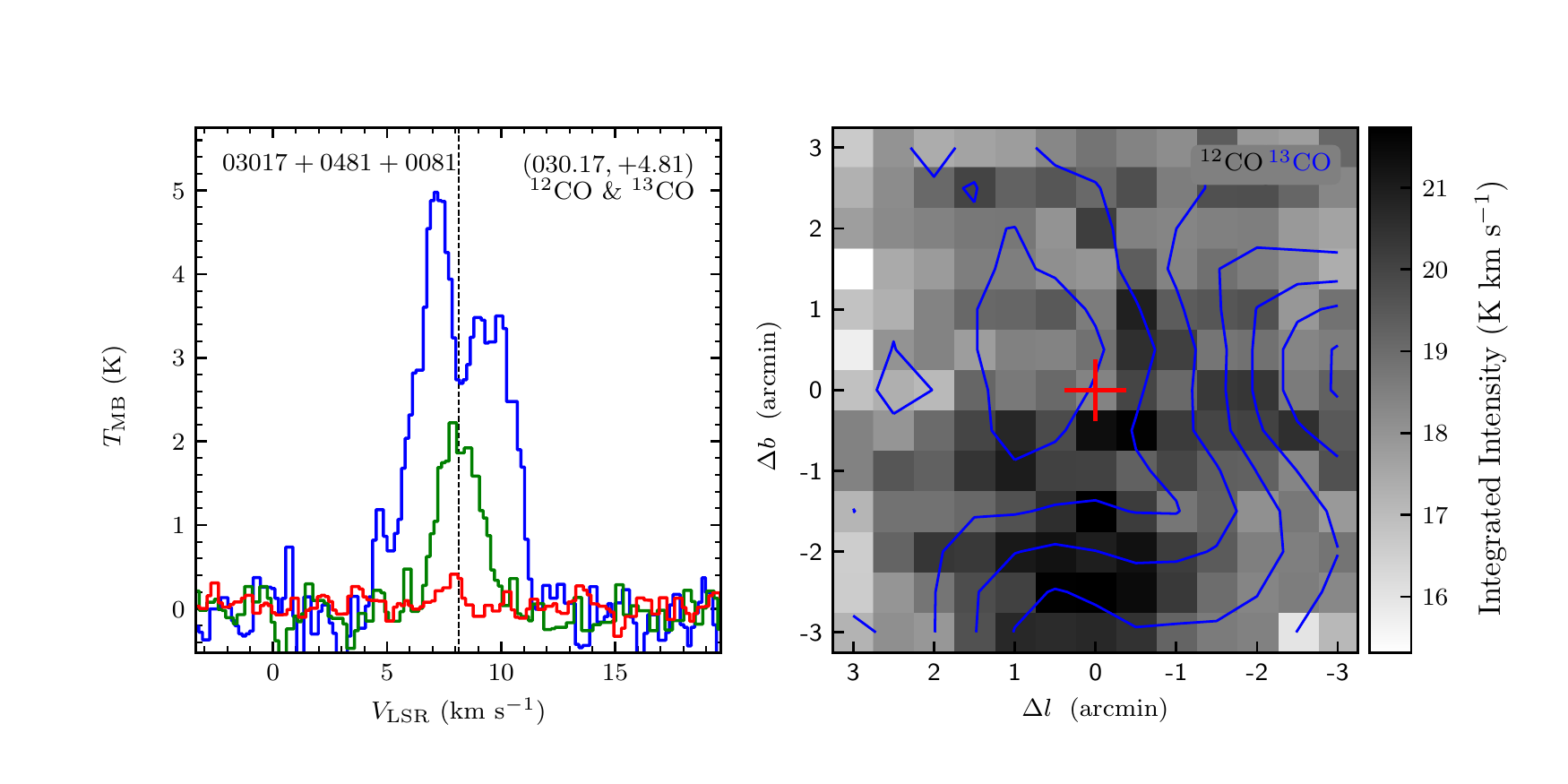}
\includegraphics[width=9.0cm,angle=0]{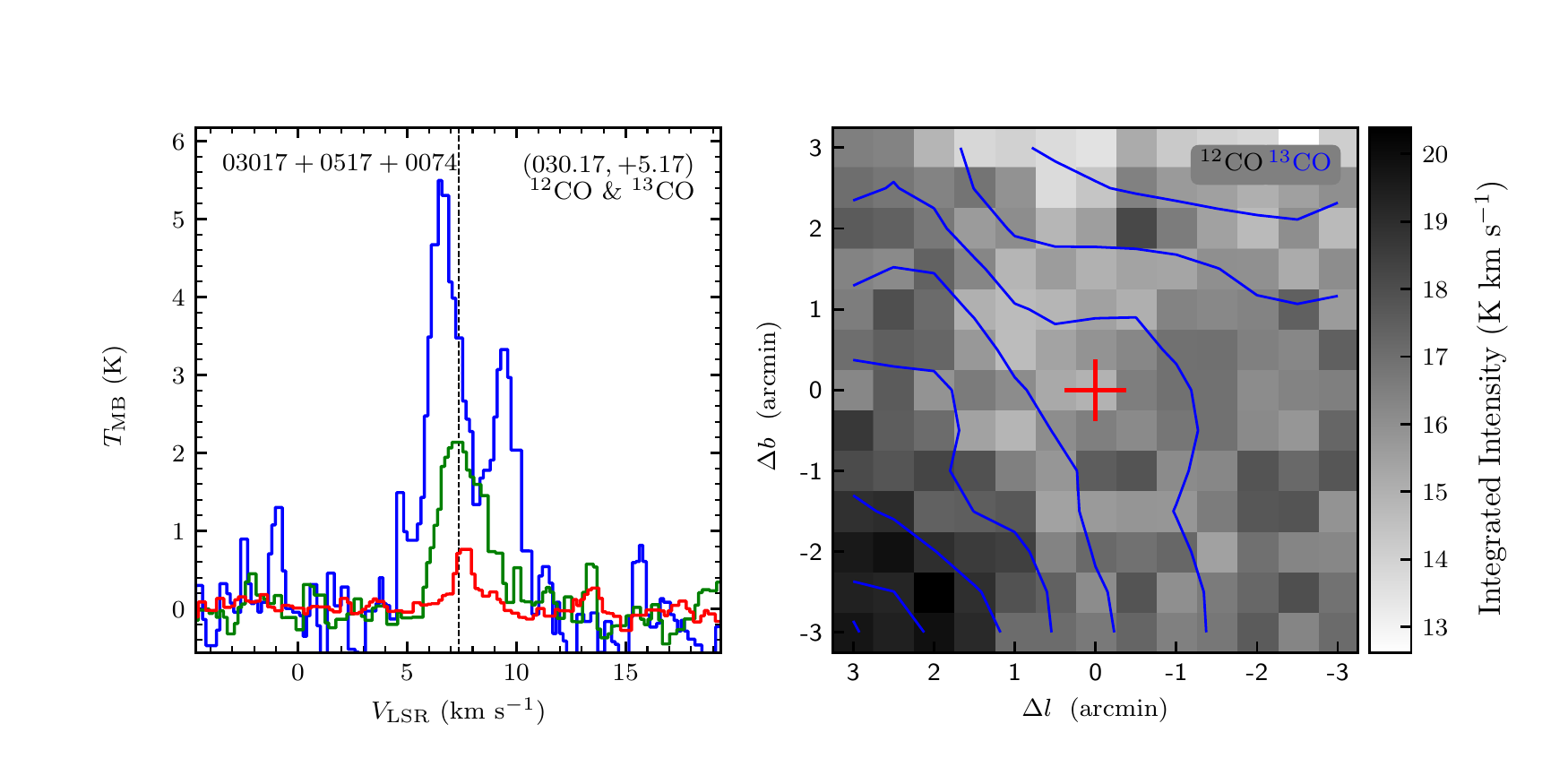}
\vspace{-0.5cm}

\includegraphics[width=9.0cm,angle=0]{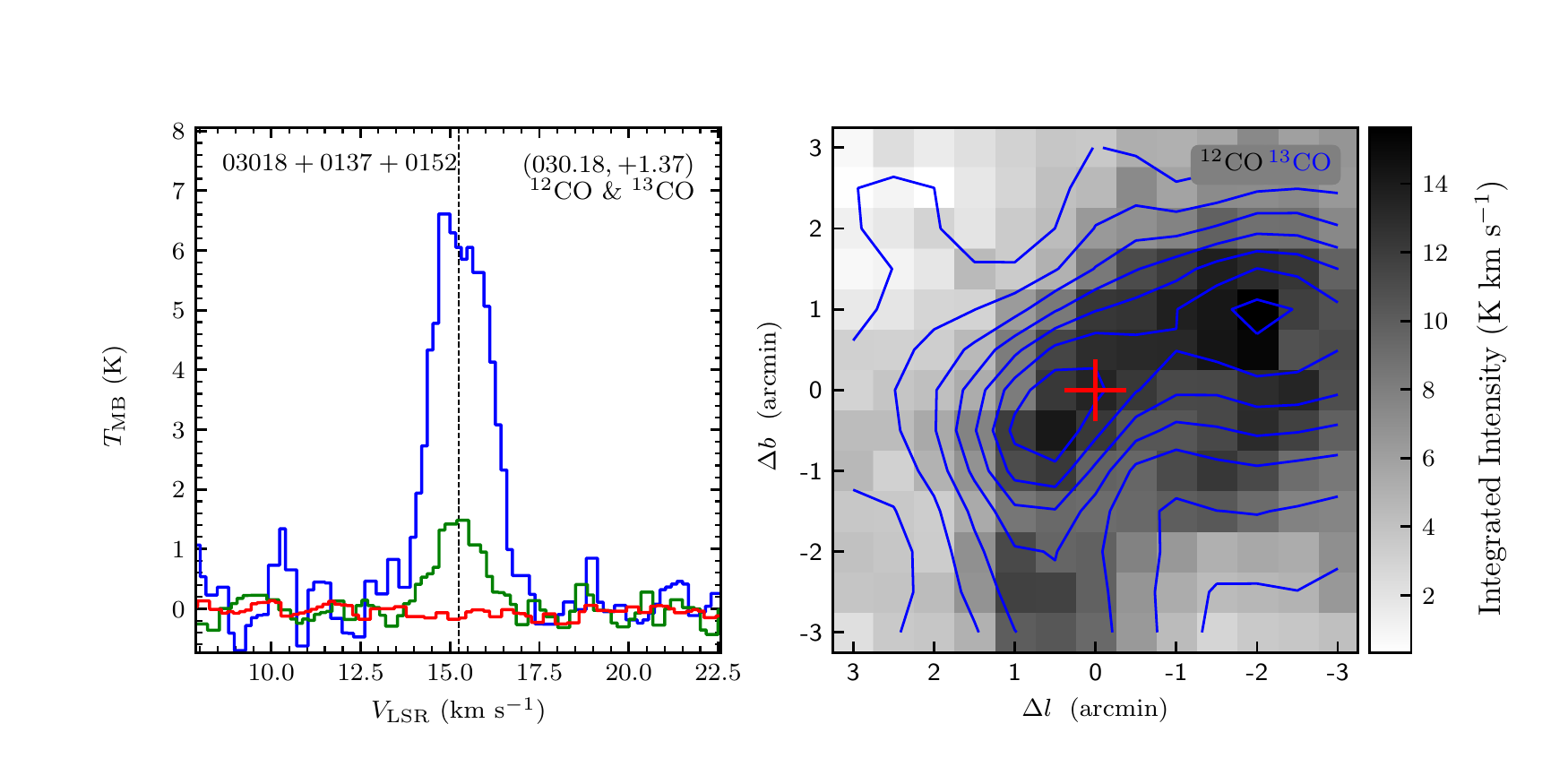}
\includegraphics[width=9.0cm,angle=0]{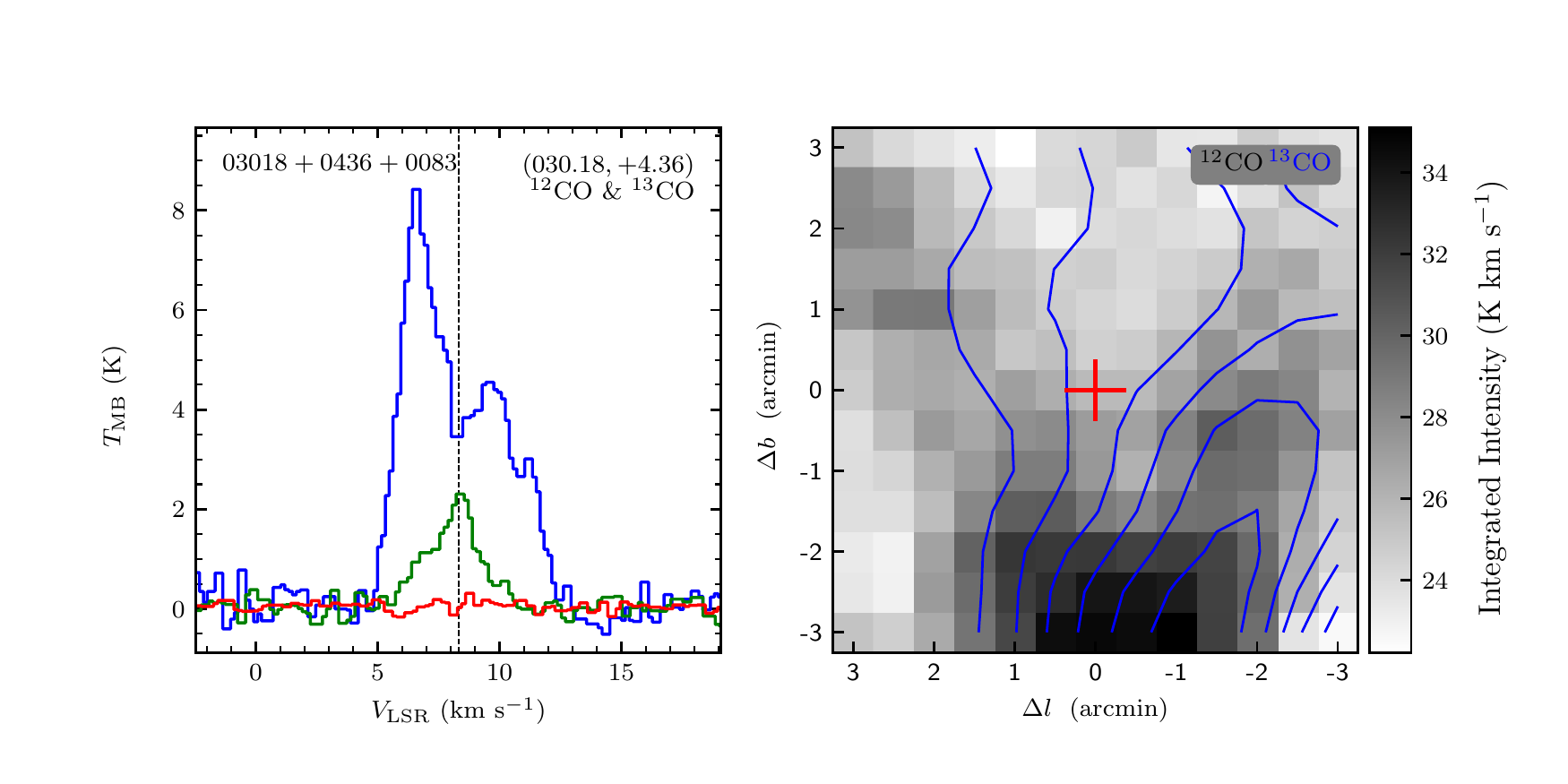}
\end{figure}
\clearpage

\begin{figure}
\includegraphics[width=9.0cm,angle=0]{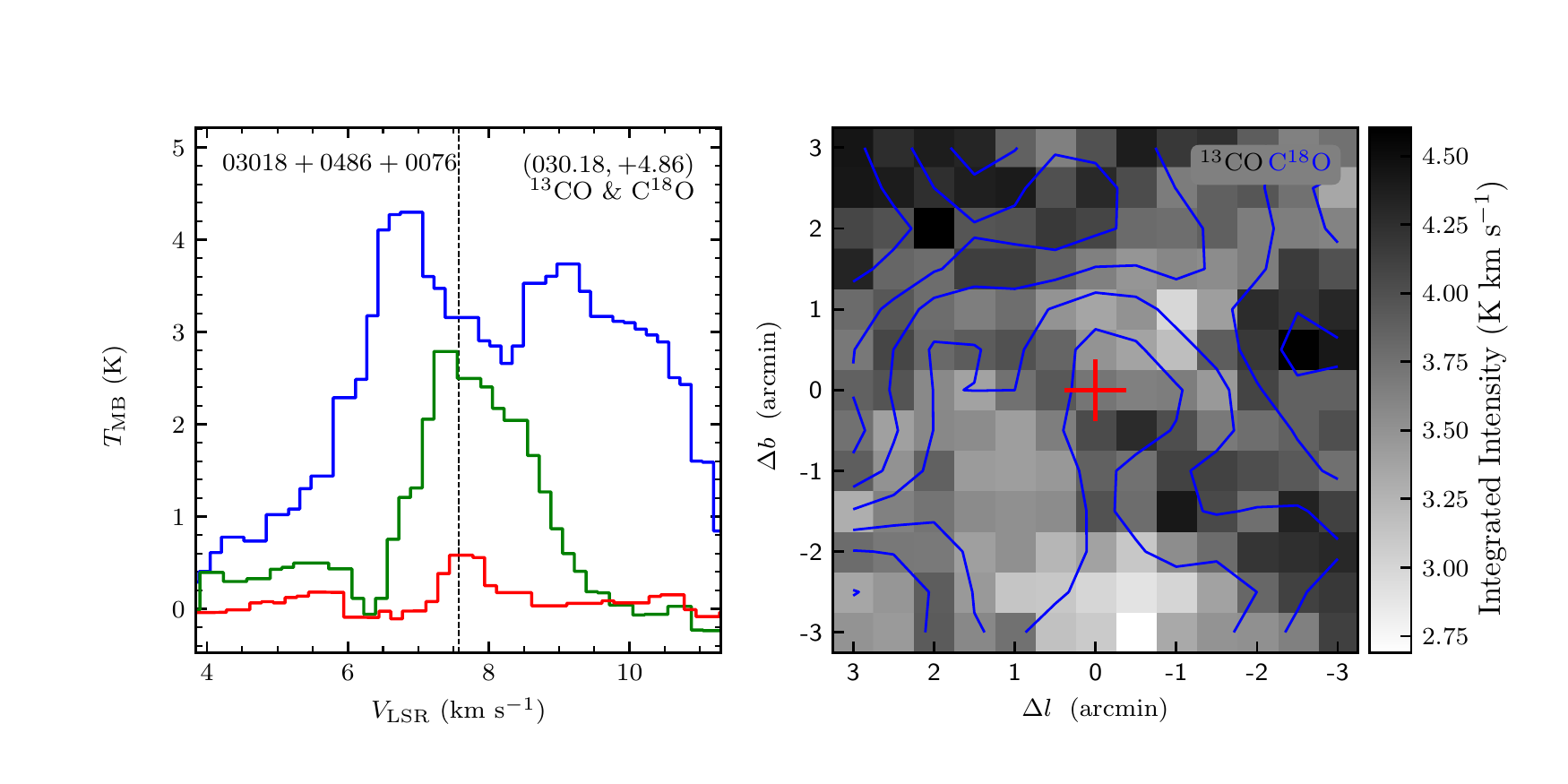}
\includegraphics[width=9.0cm,angle=0]{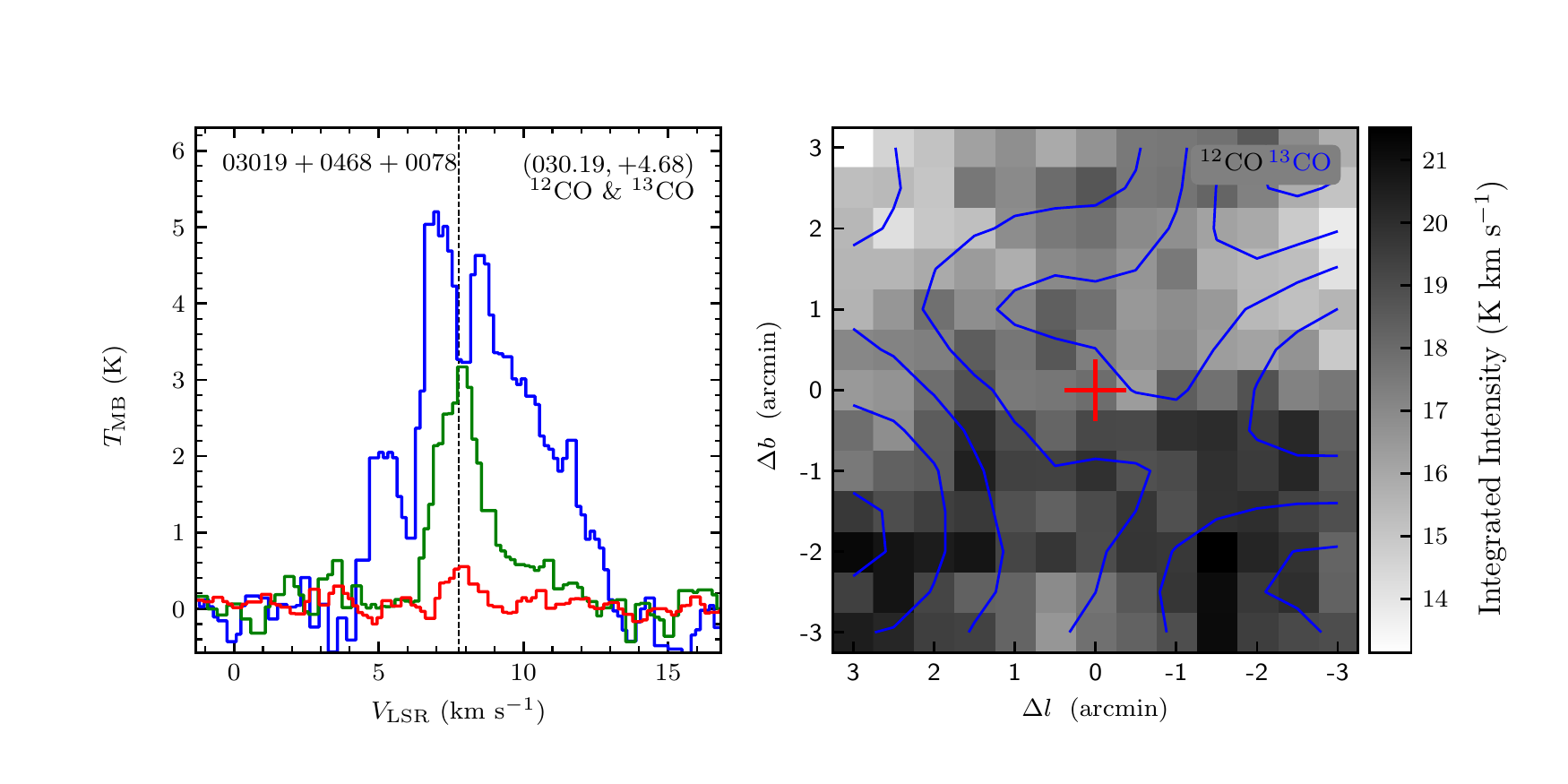}
\vspace{-0.5cm}

\includegraphics[width=9.0cm,angle=0]{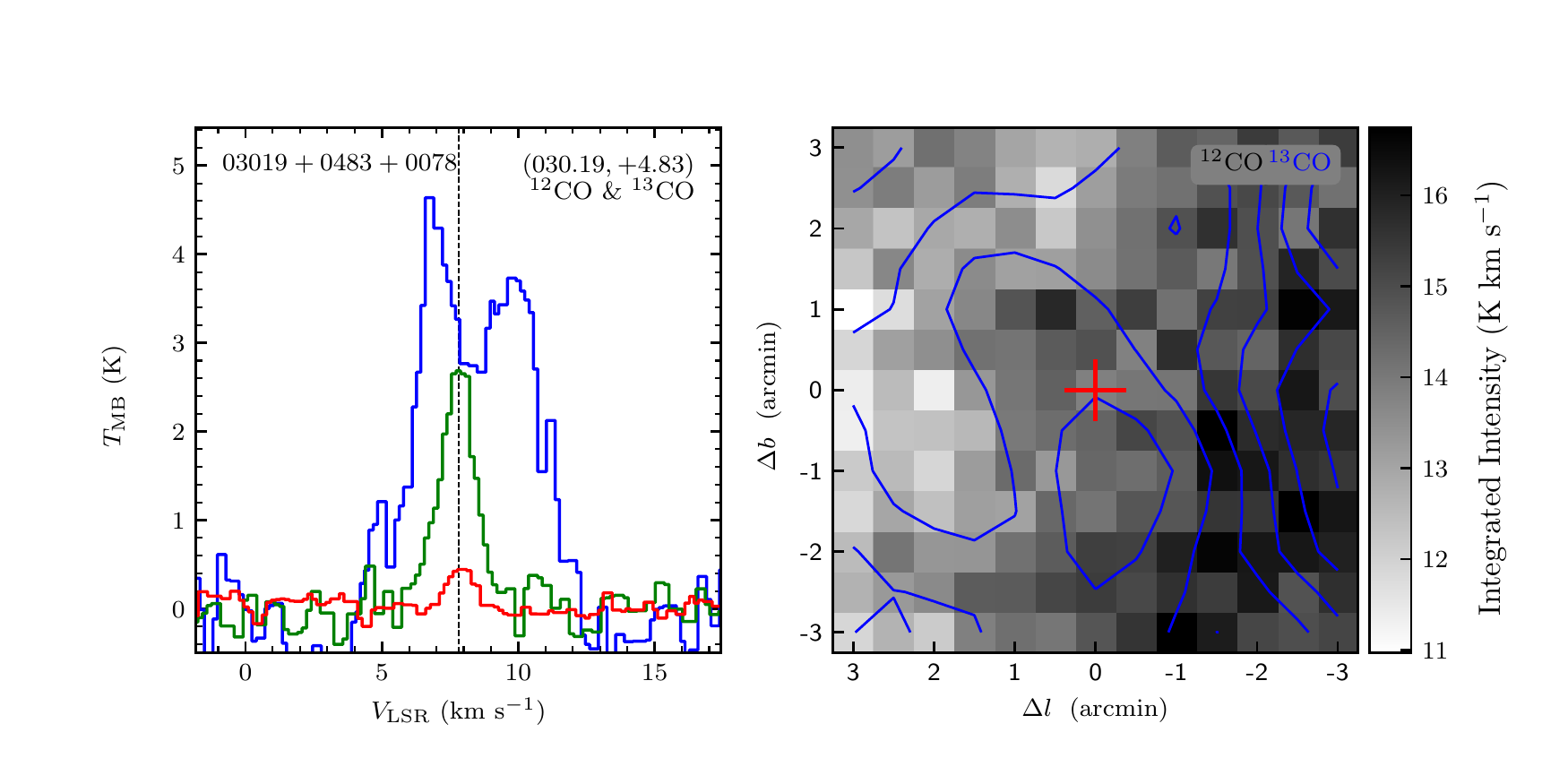}
\includegraphics[width=9.0cm,angle=0]{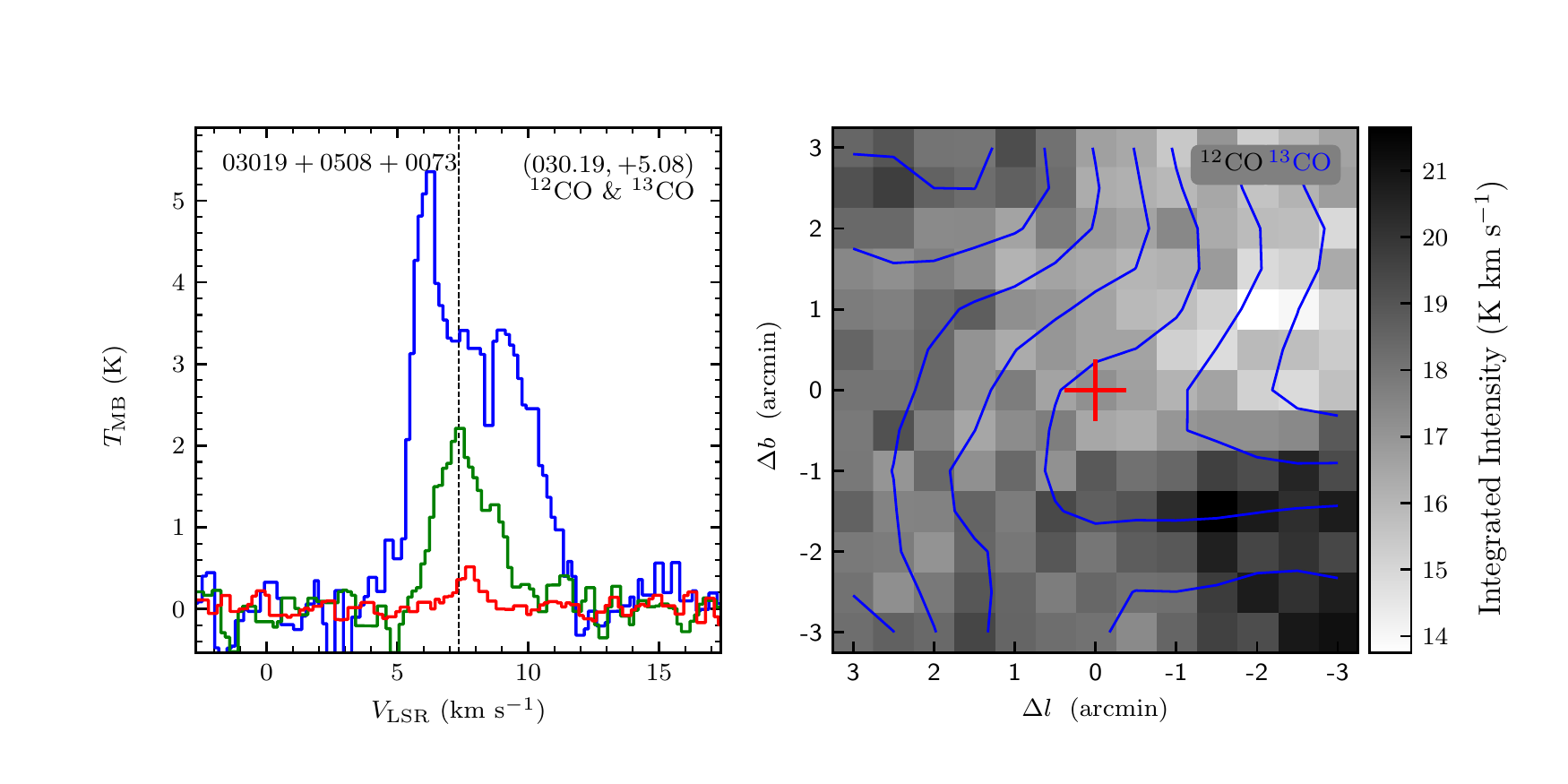}
\vspace{-0.5cm}

\includegraphics[width=9.0cm,angle=0]{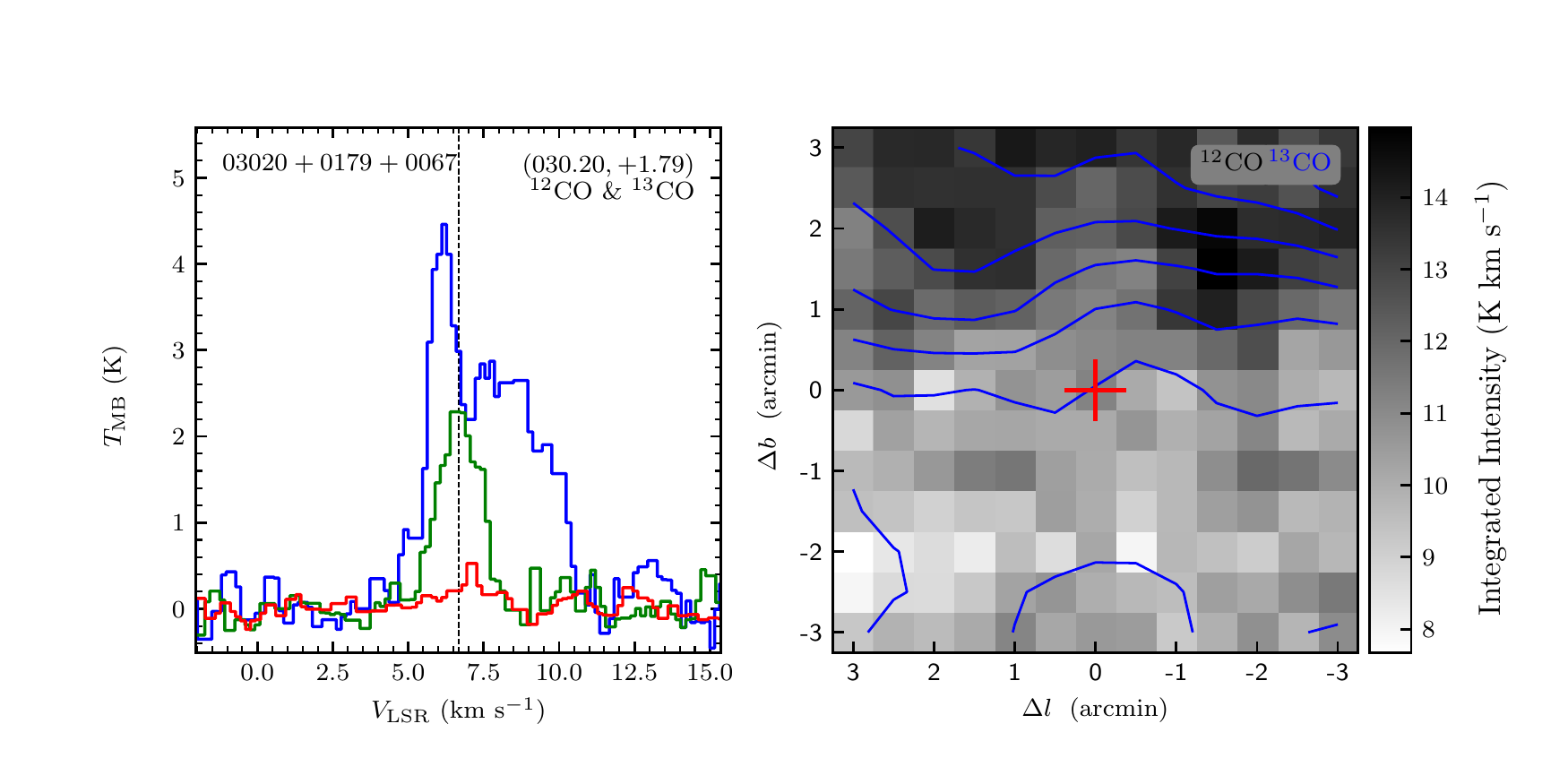}
\includegraphics[width=9.0cm,angle=0]{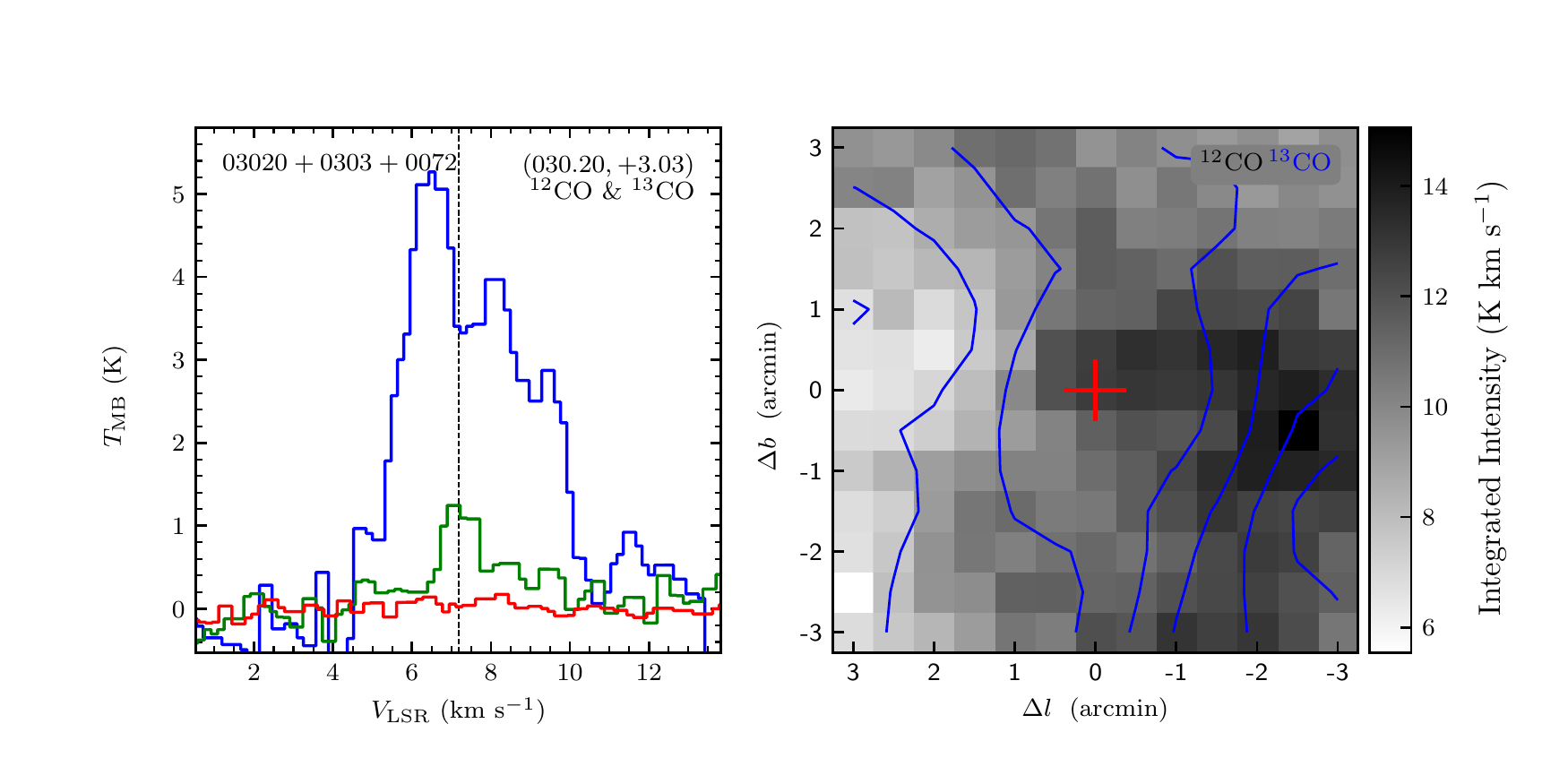}
\vspace{-0.5cm}

\includegraphics[width=9.0cm,angle=0]{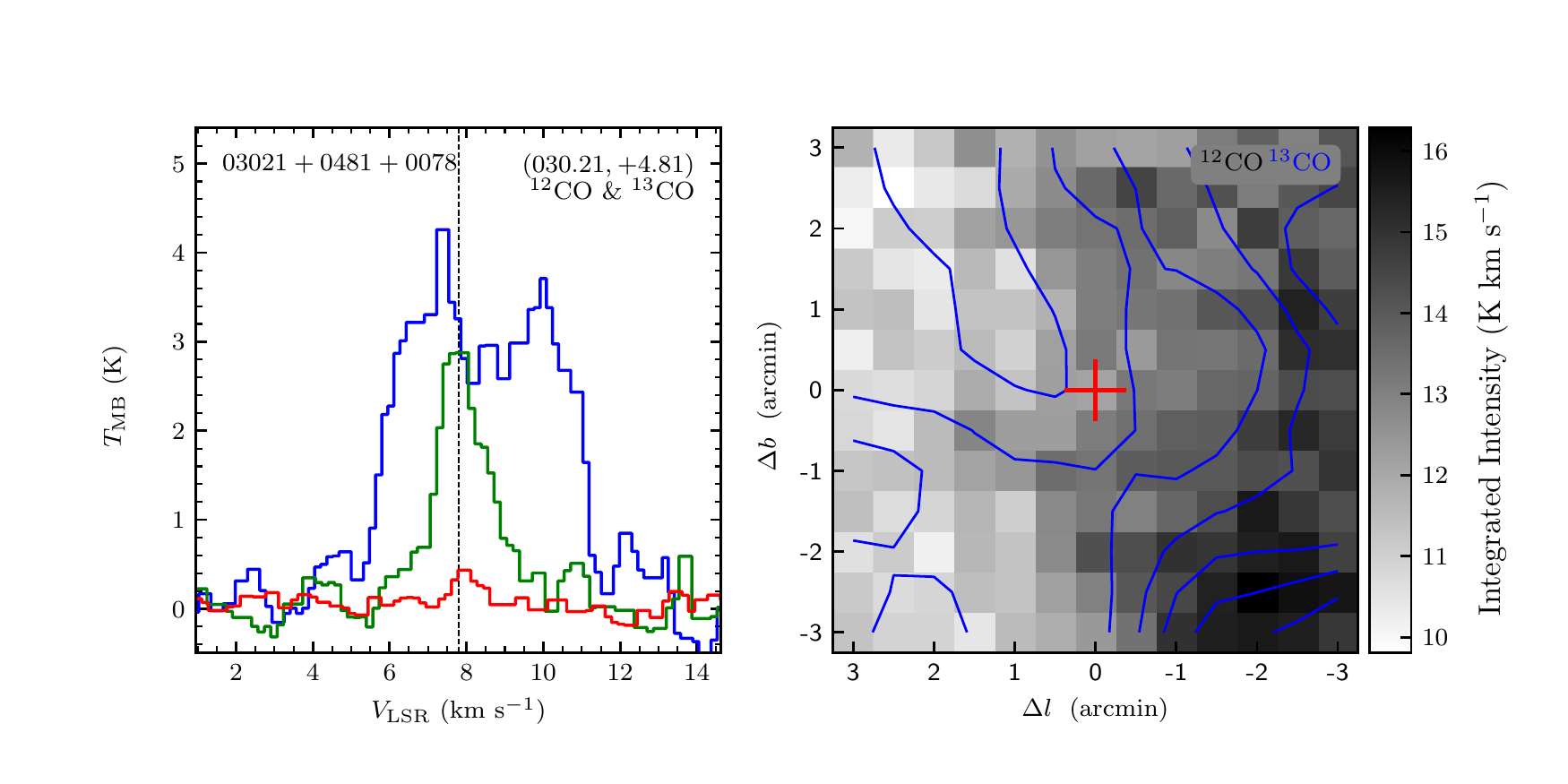}
\includegraphics[width=9.0cm,angle=0]{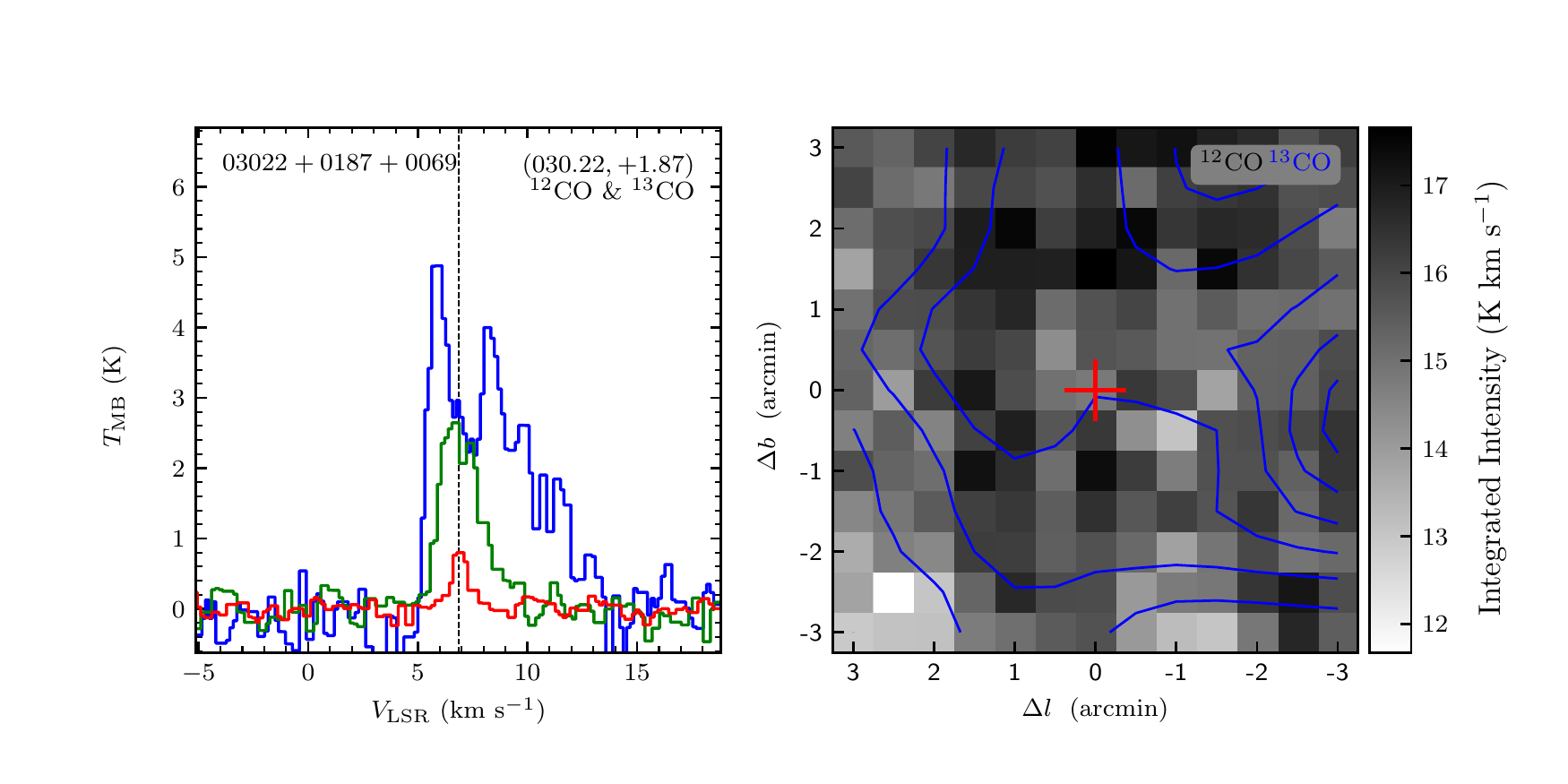}
\vspace{-0.5cm}

\includegraphics[width=9.0cm,angle=0]{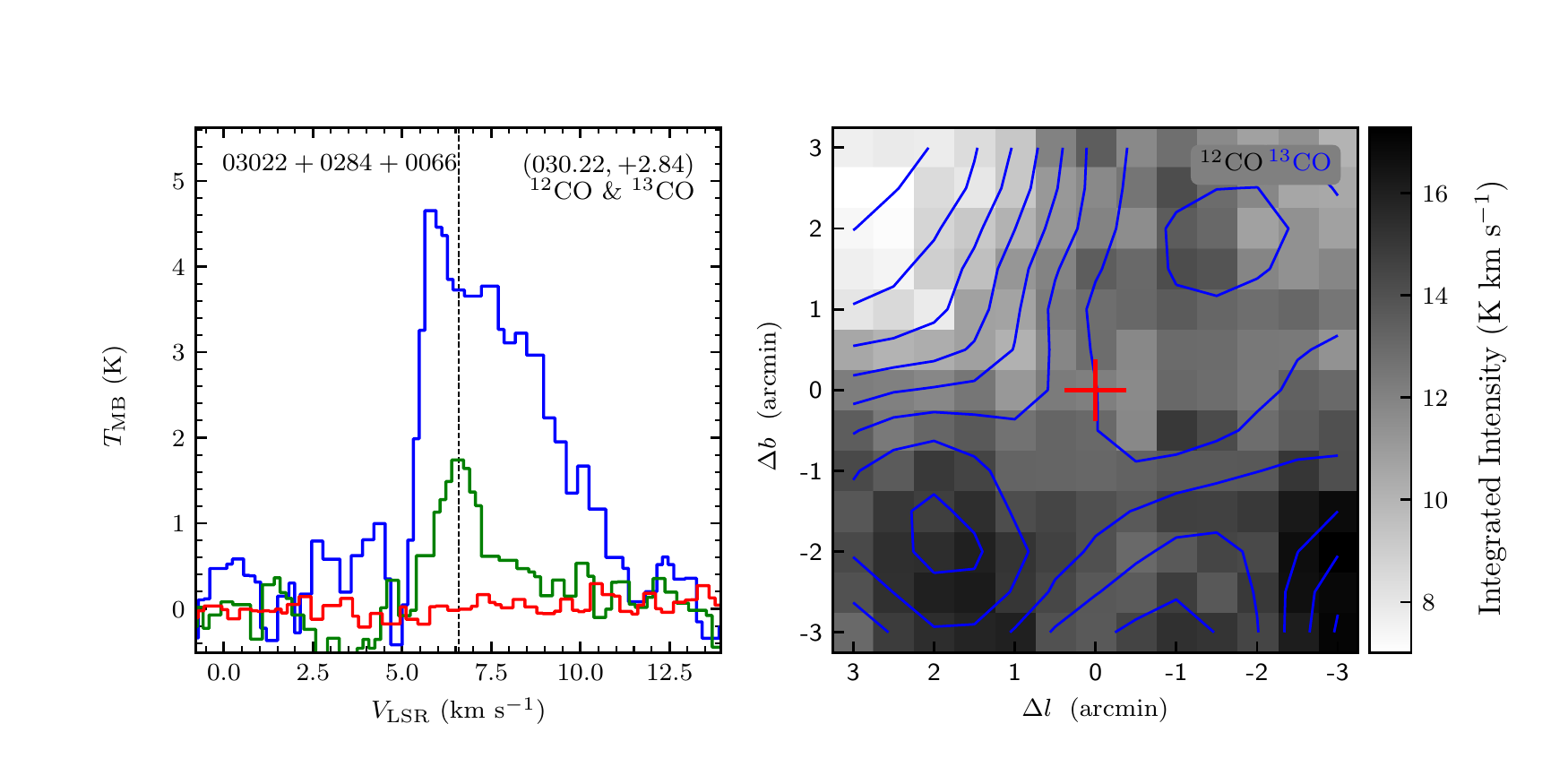}
\includegraphics[width=9.0cm,angle=0]{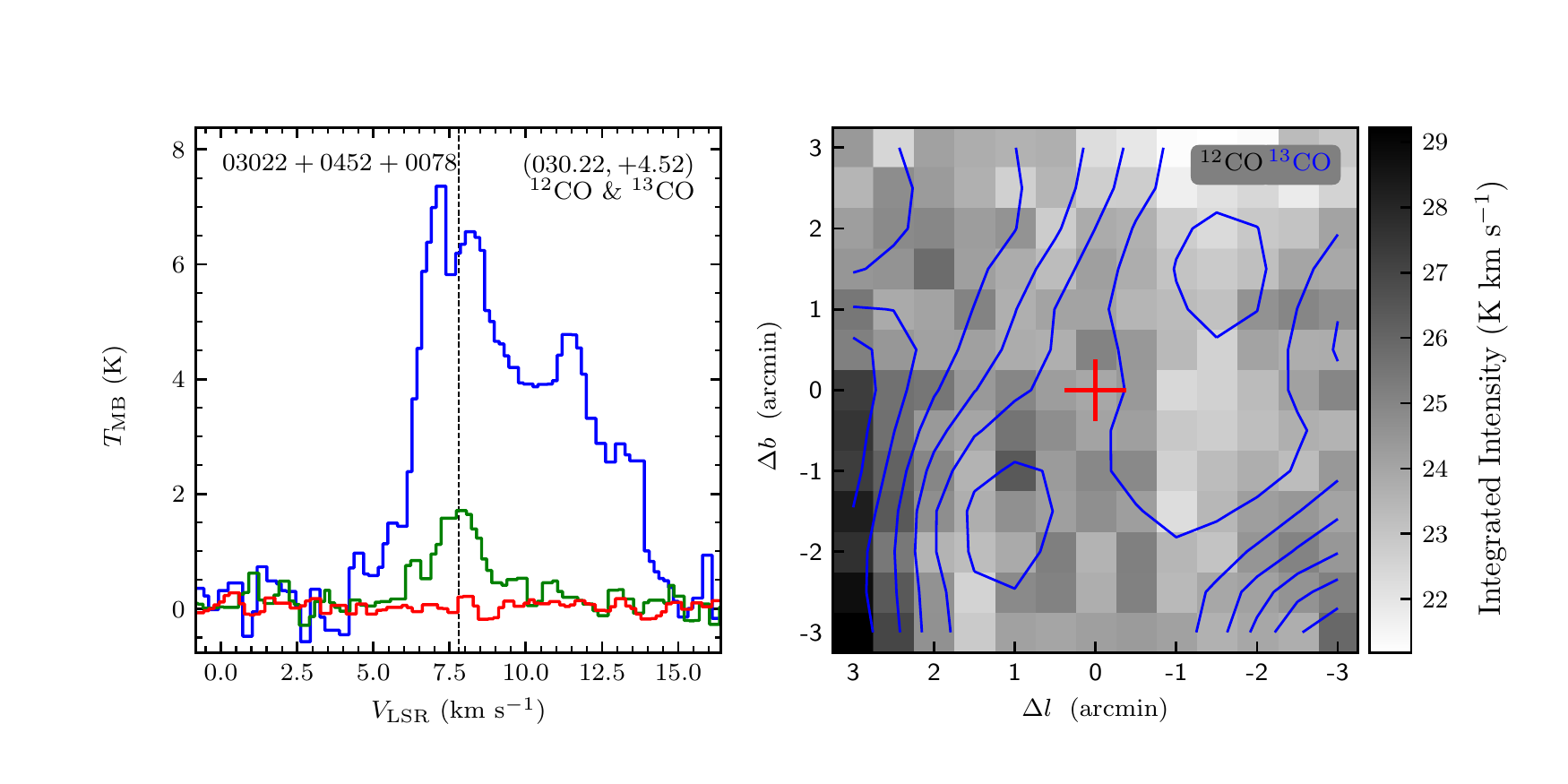}
\end{figure}
\clearpage

\begin{figure}
\includegraphics[width=9.0cm,angle=0]{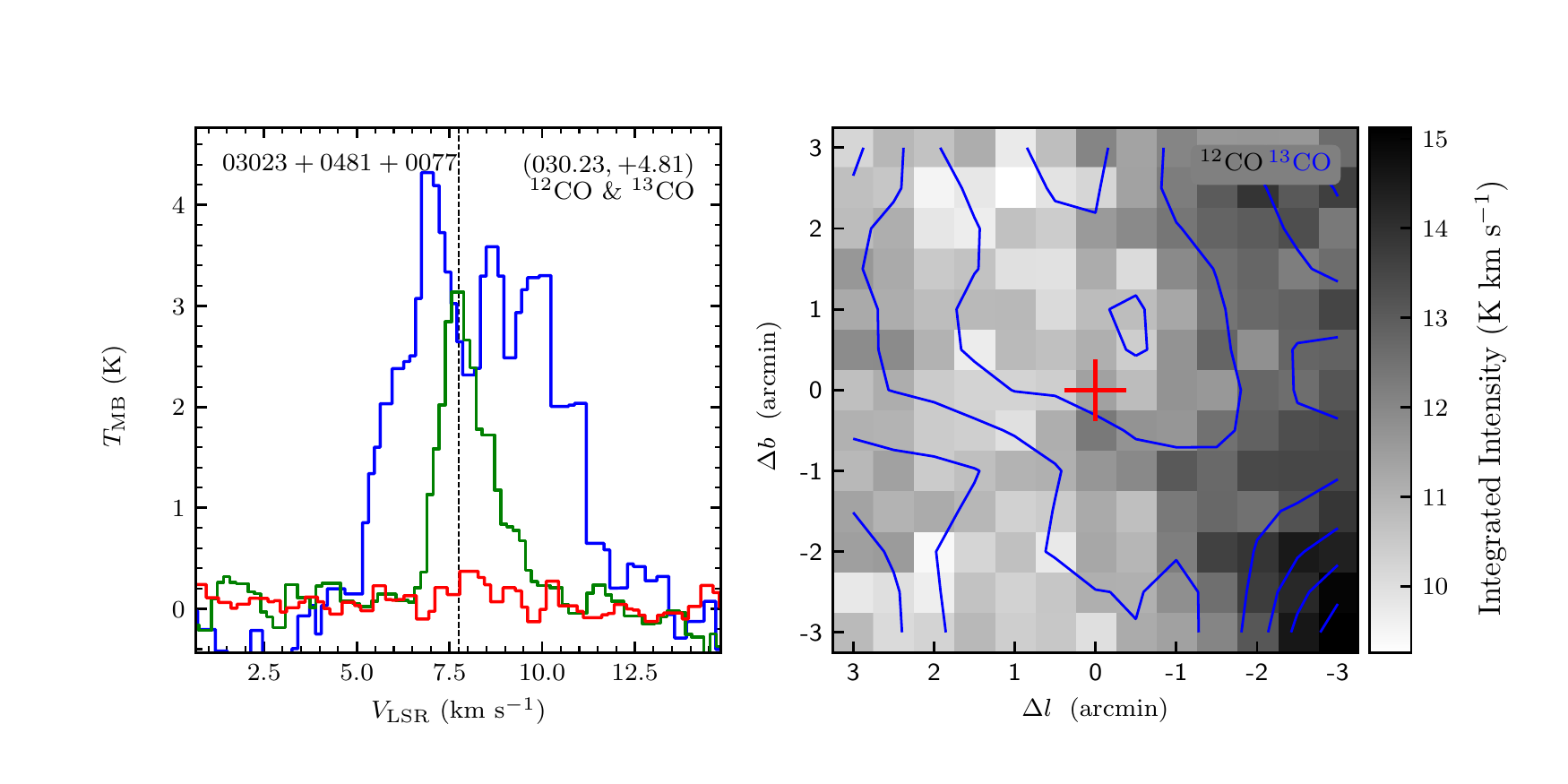}
\includegraphics[width=9.0cm,angle=0]{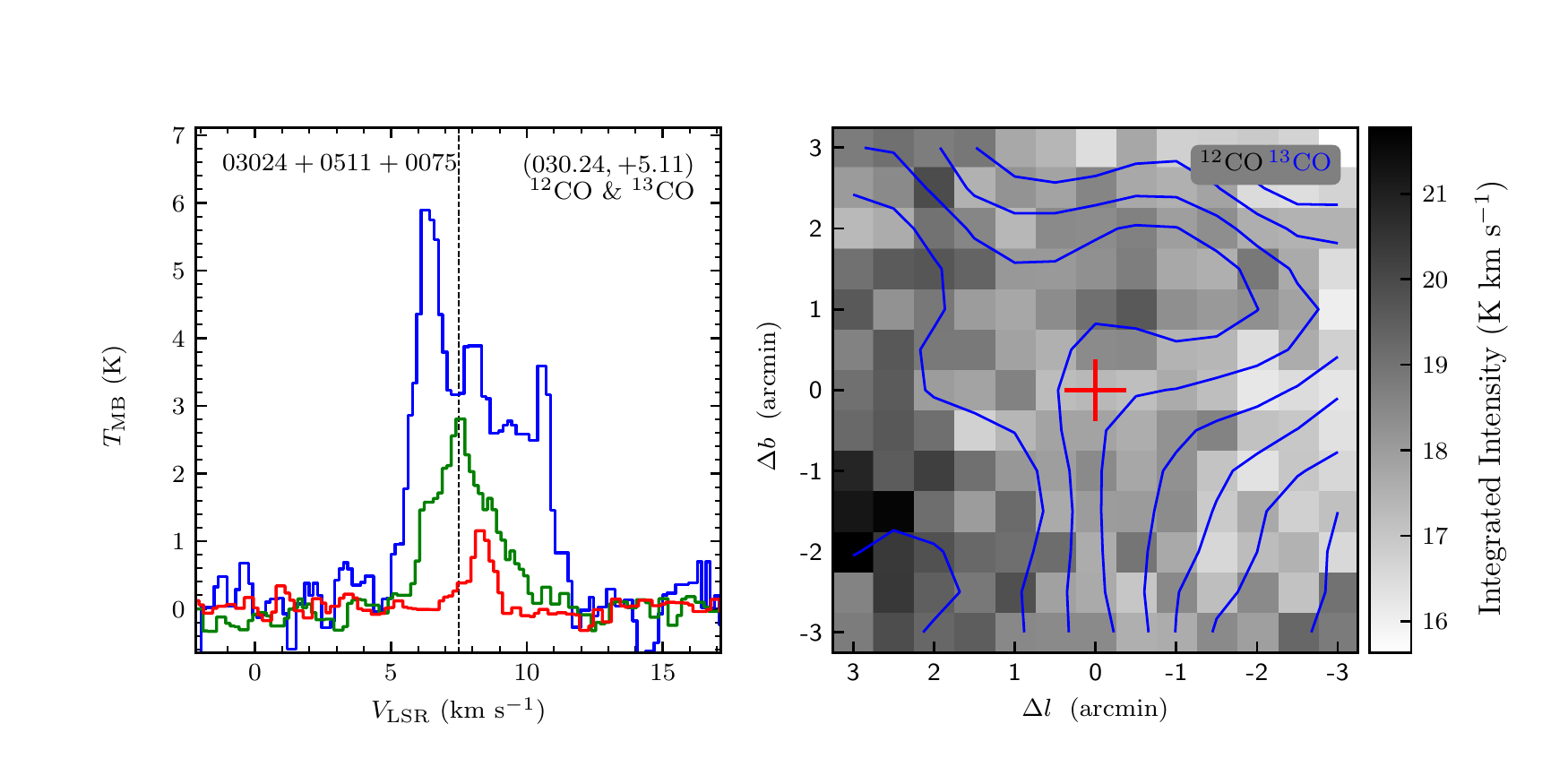}
\vspace{-0.5cm}

\includegraphics[width=9.0cm,angle=0]{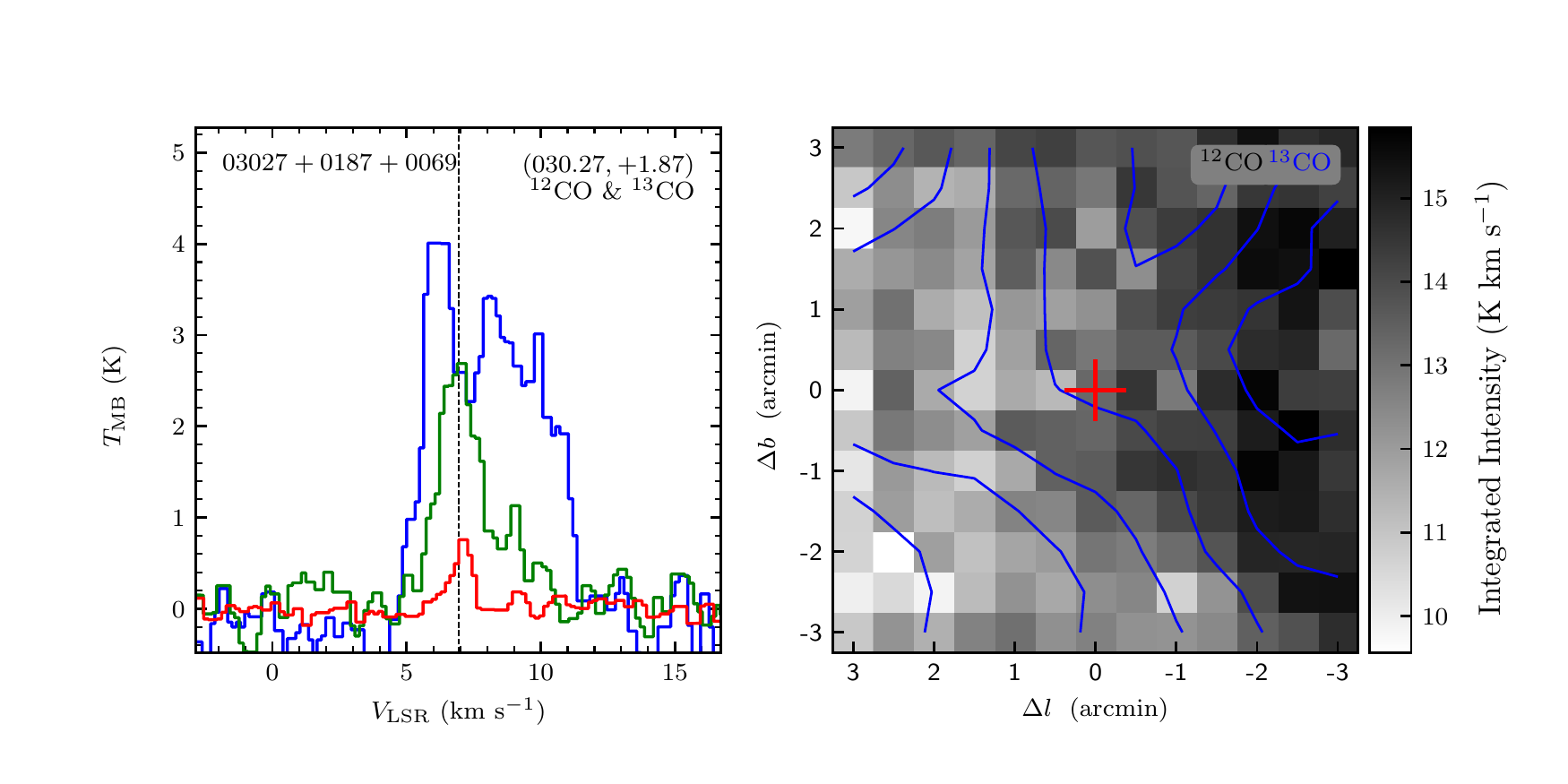}
\includegraphics[width=9.0cm,angle=0]{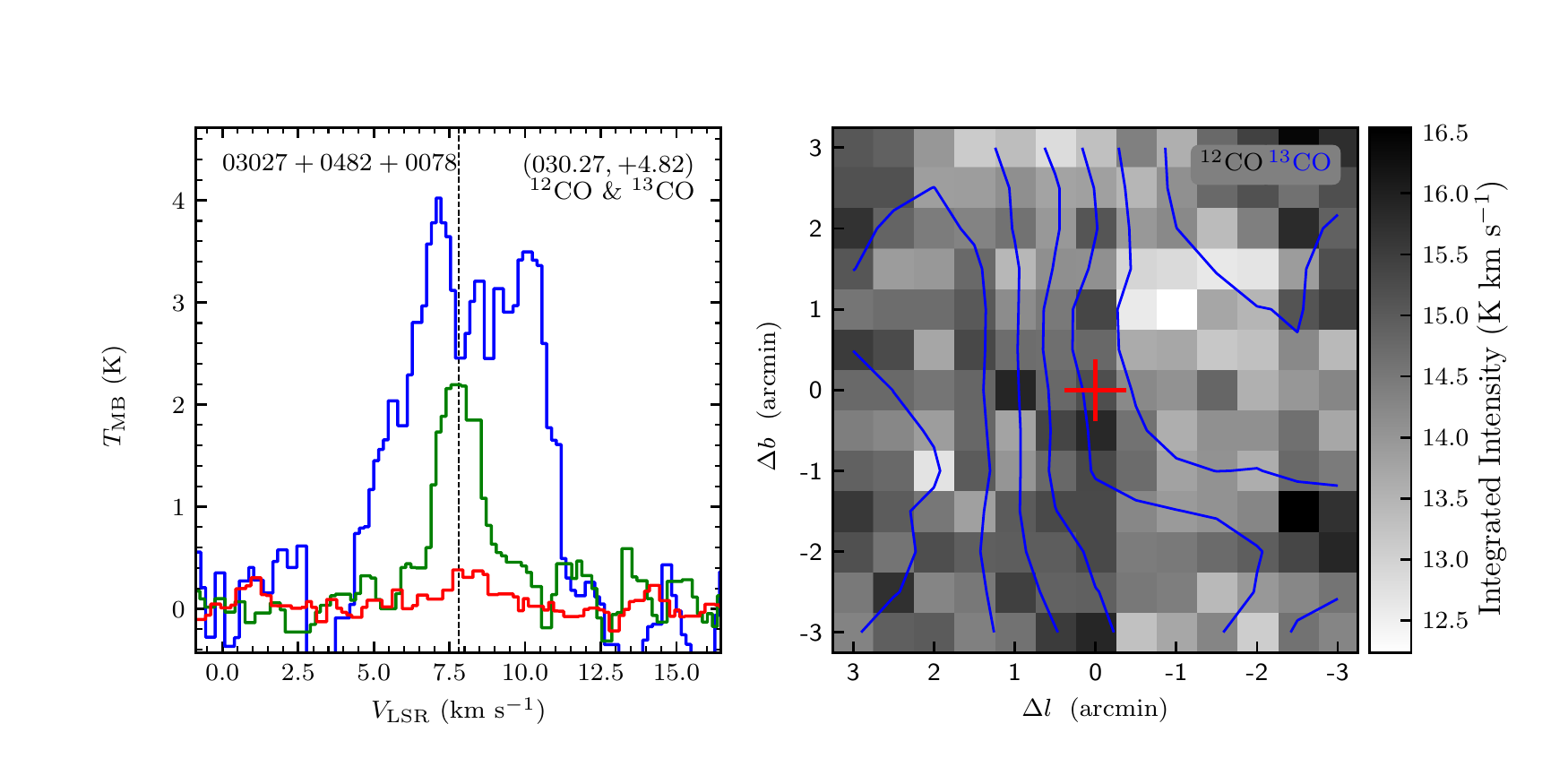}
\vspace{-0.5cm}

\includegraphics[width=9.0cm,angle=0]{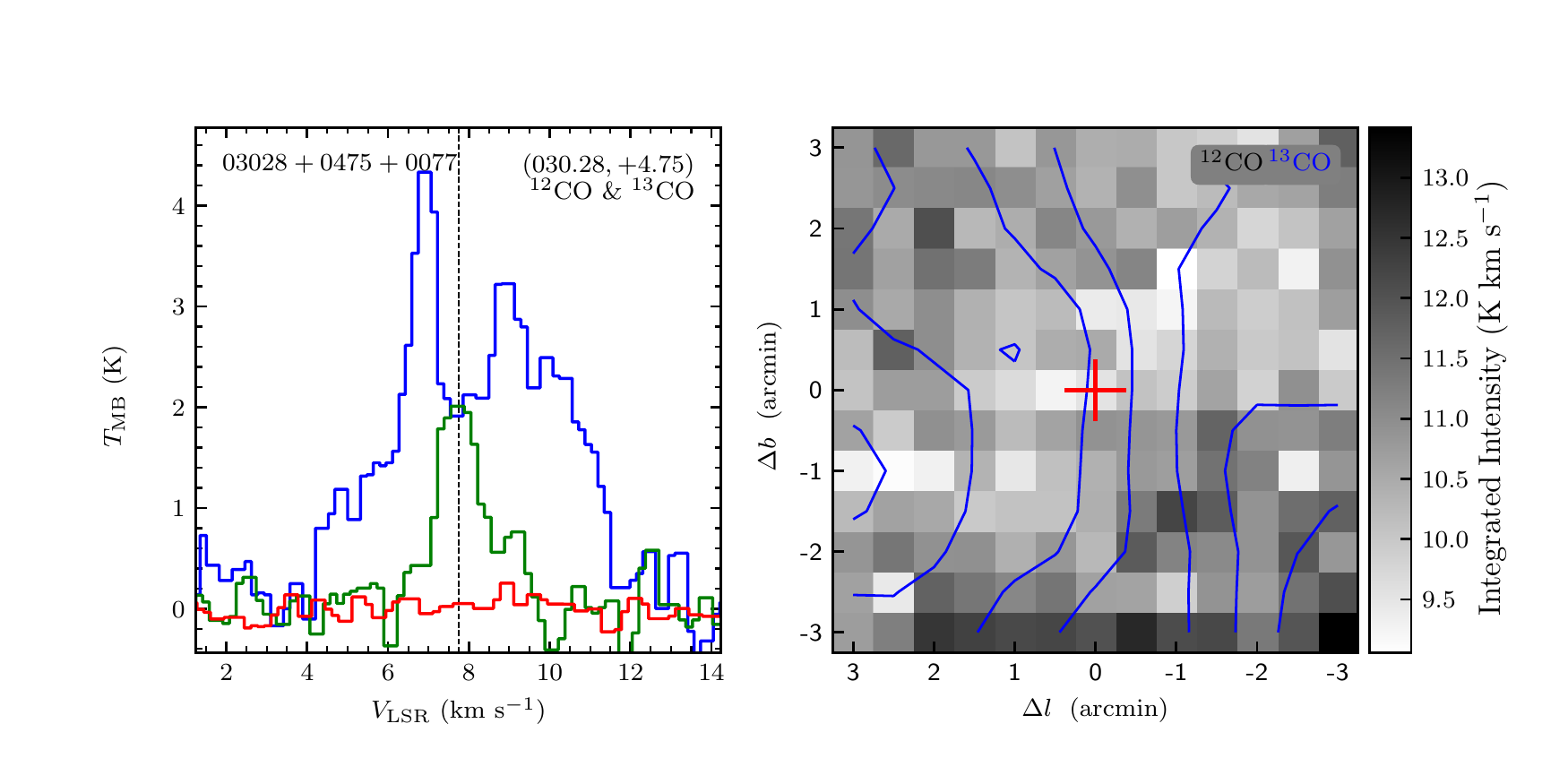}
\includegraphics[width=9.0cm,angle=0]{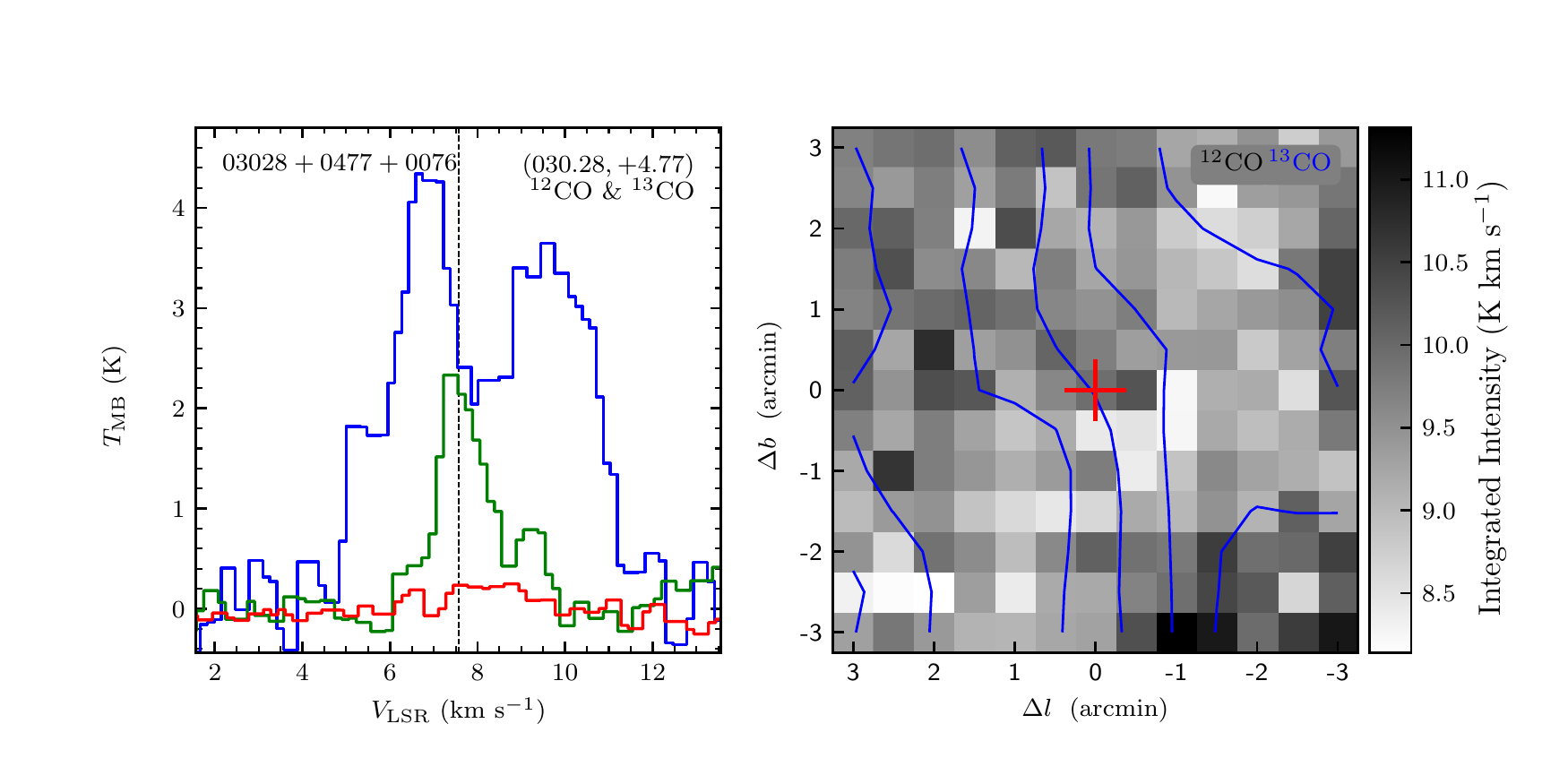}
\vspace{-0.5cm}

\includegraphics[width=9.0cm,angle=0]{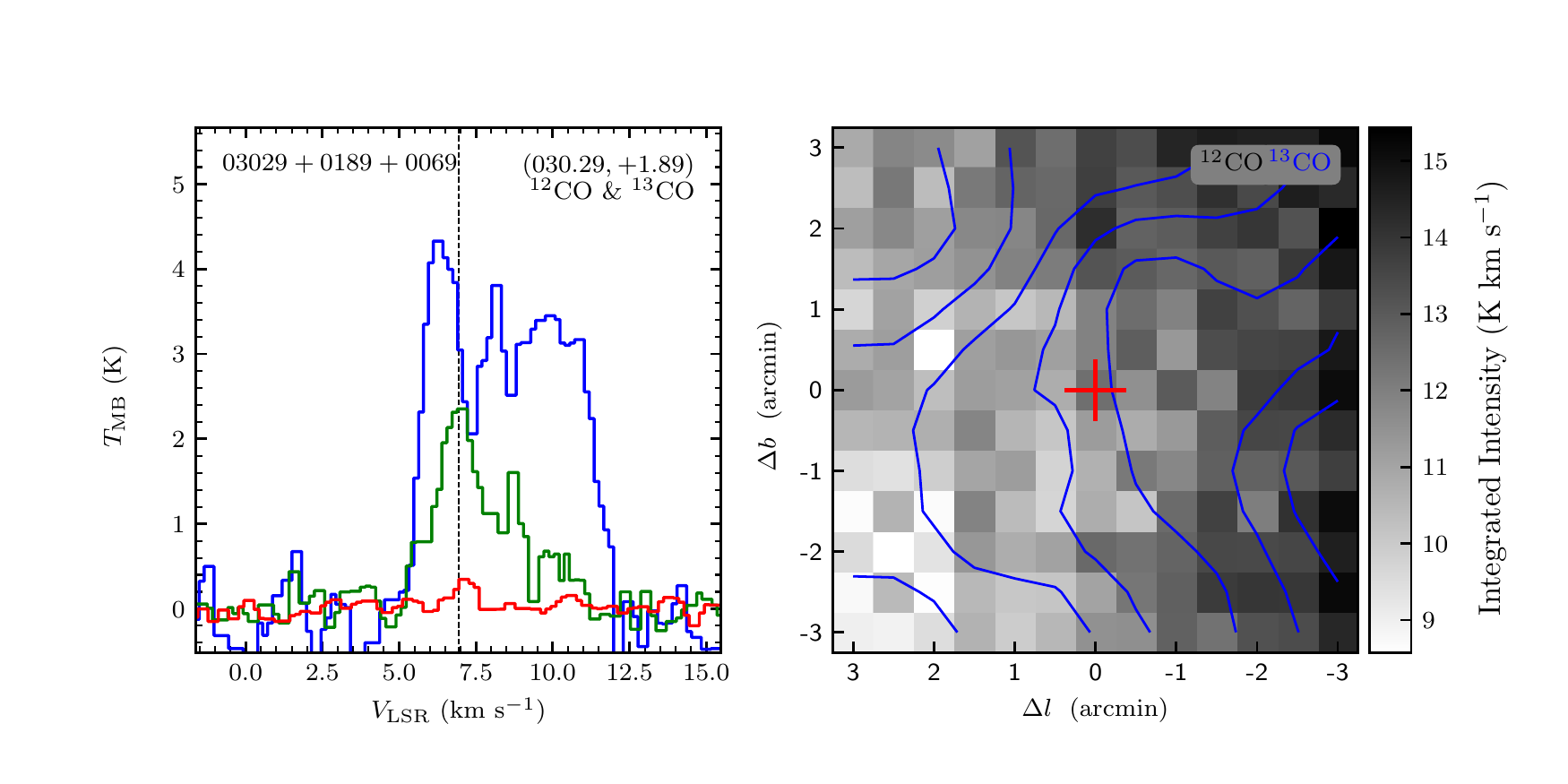}
\includegraphics[width=9.0cm,angle=0]{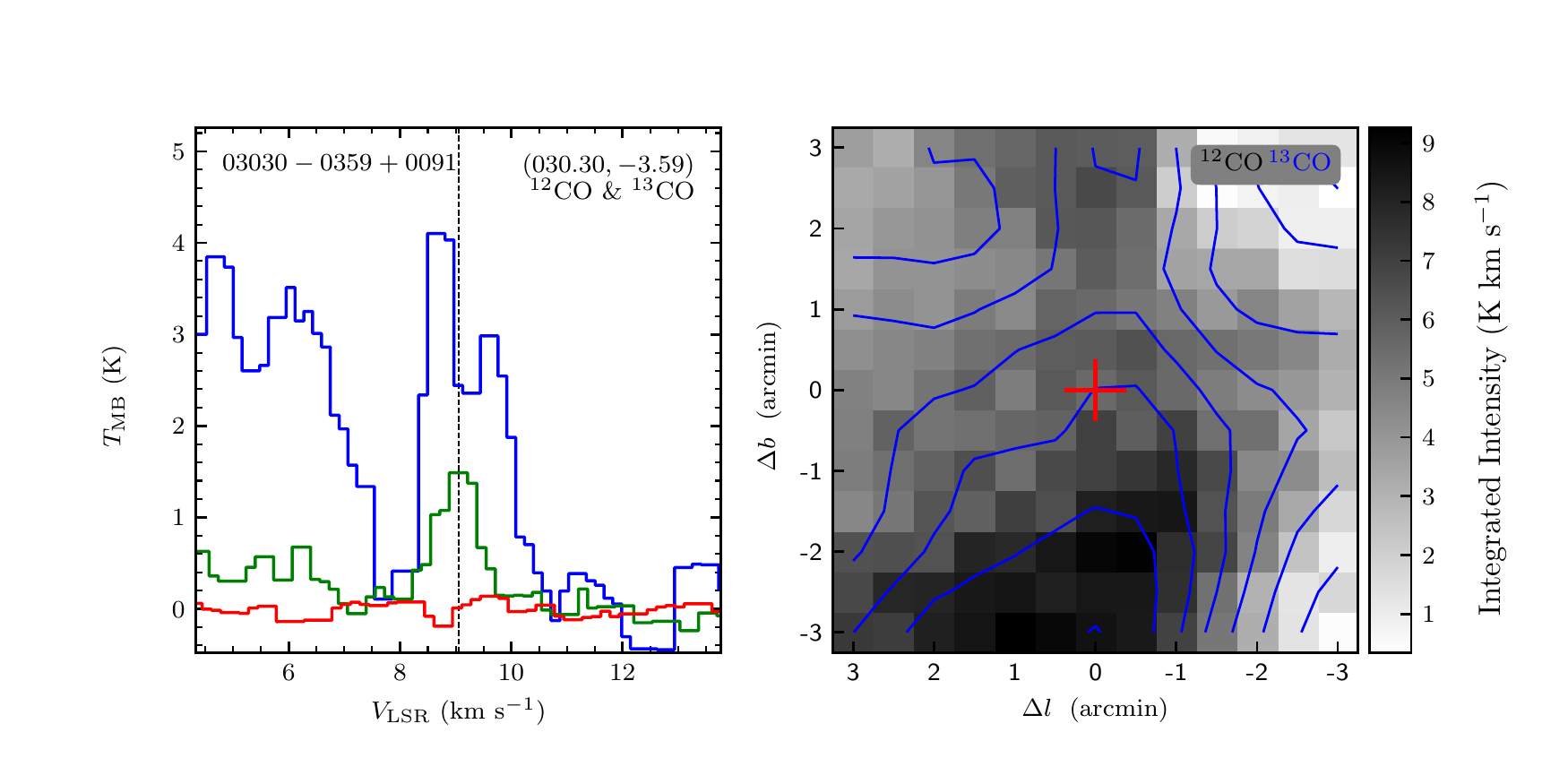}
\vspace{-0.5cm}

\includegraphics[width=9.0cm,angle=0]{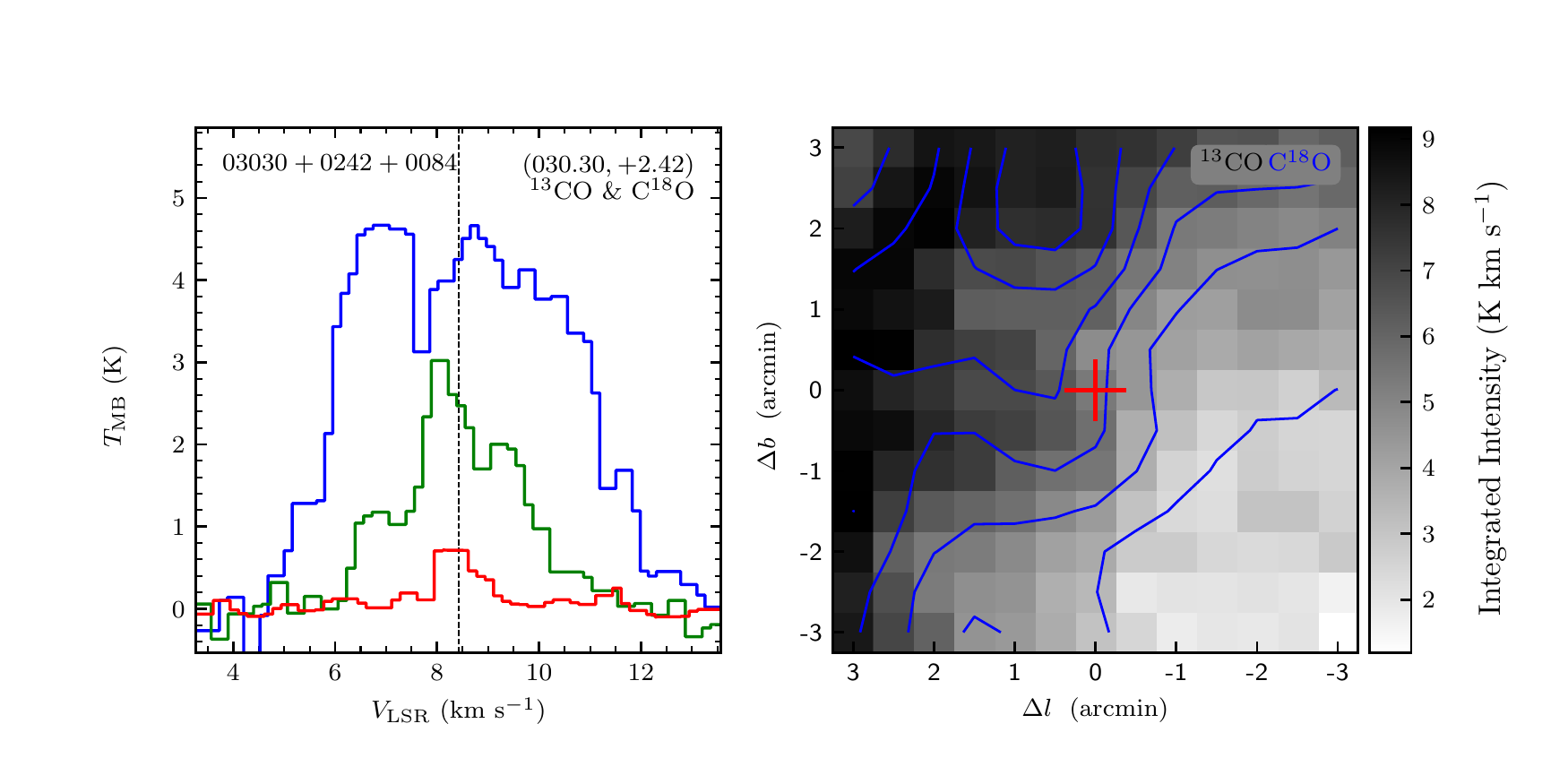}
\includegraphics[width=9.0cm,angle=0]{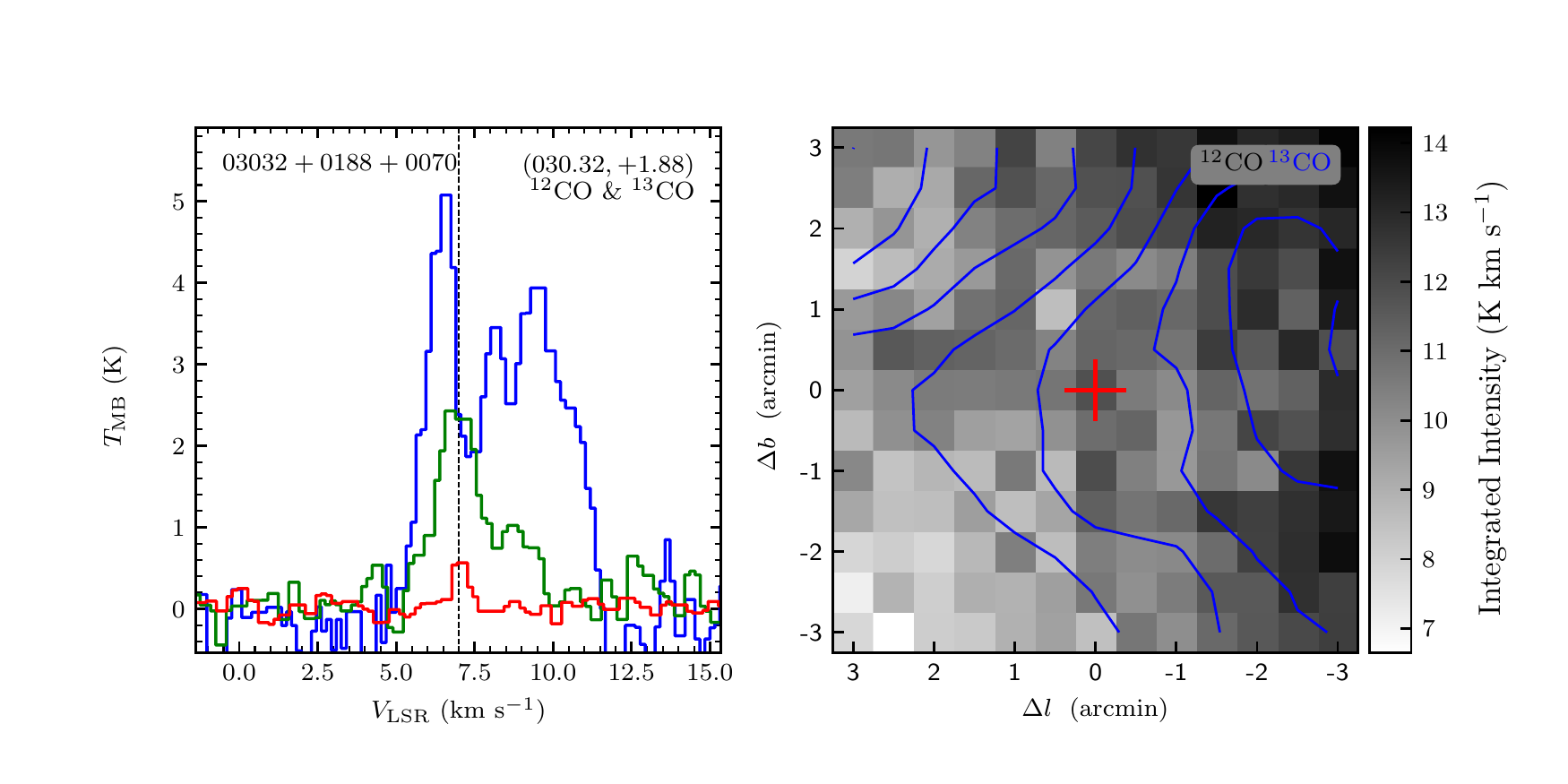}
\end{figure}
\clearpage

\begin{figure}
\includegraphics[width=9.0cm,angle=0]{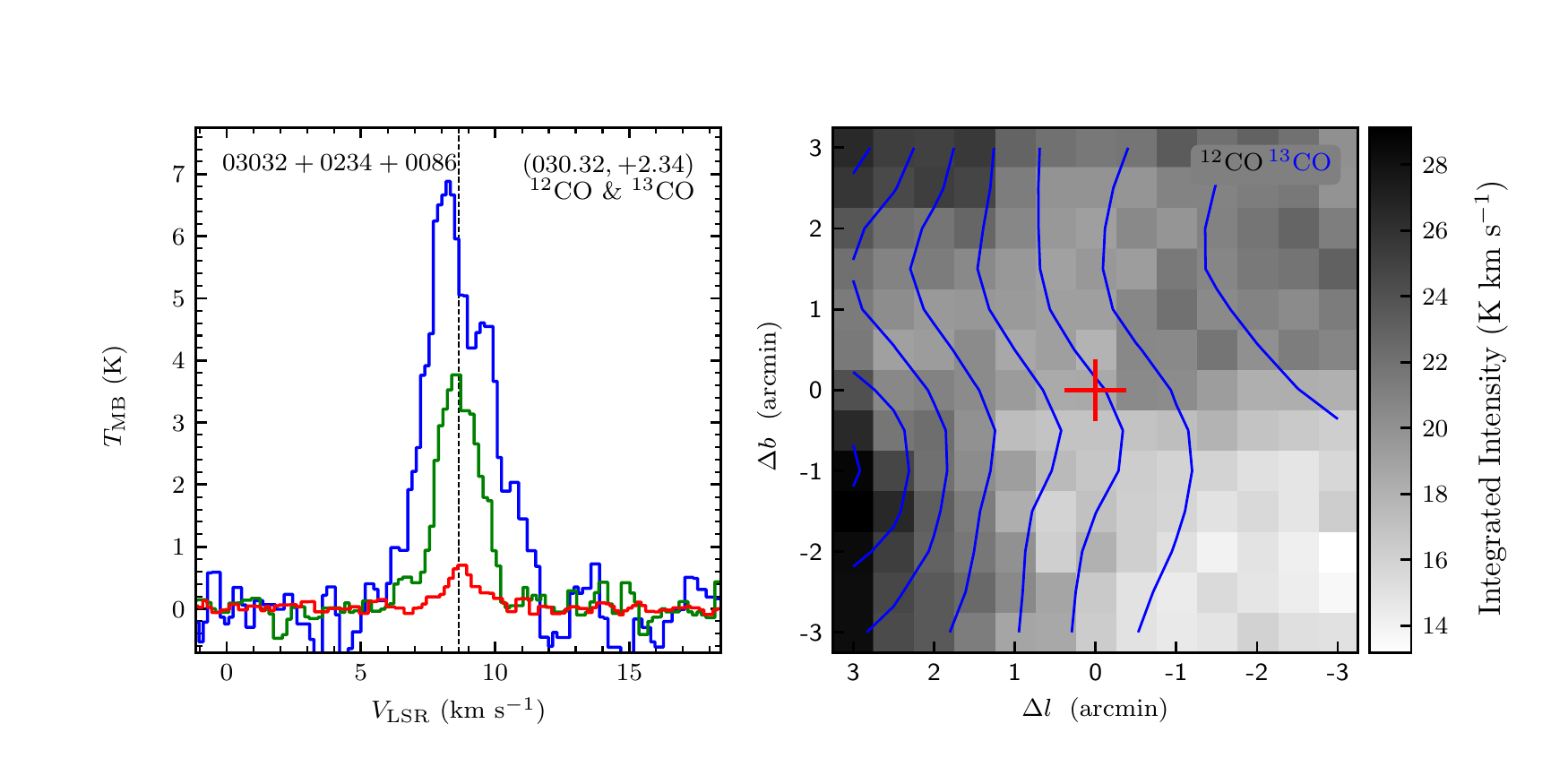}
\includegraphics[width=9.0cm,angle=0]{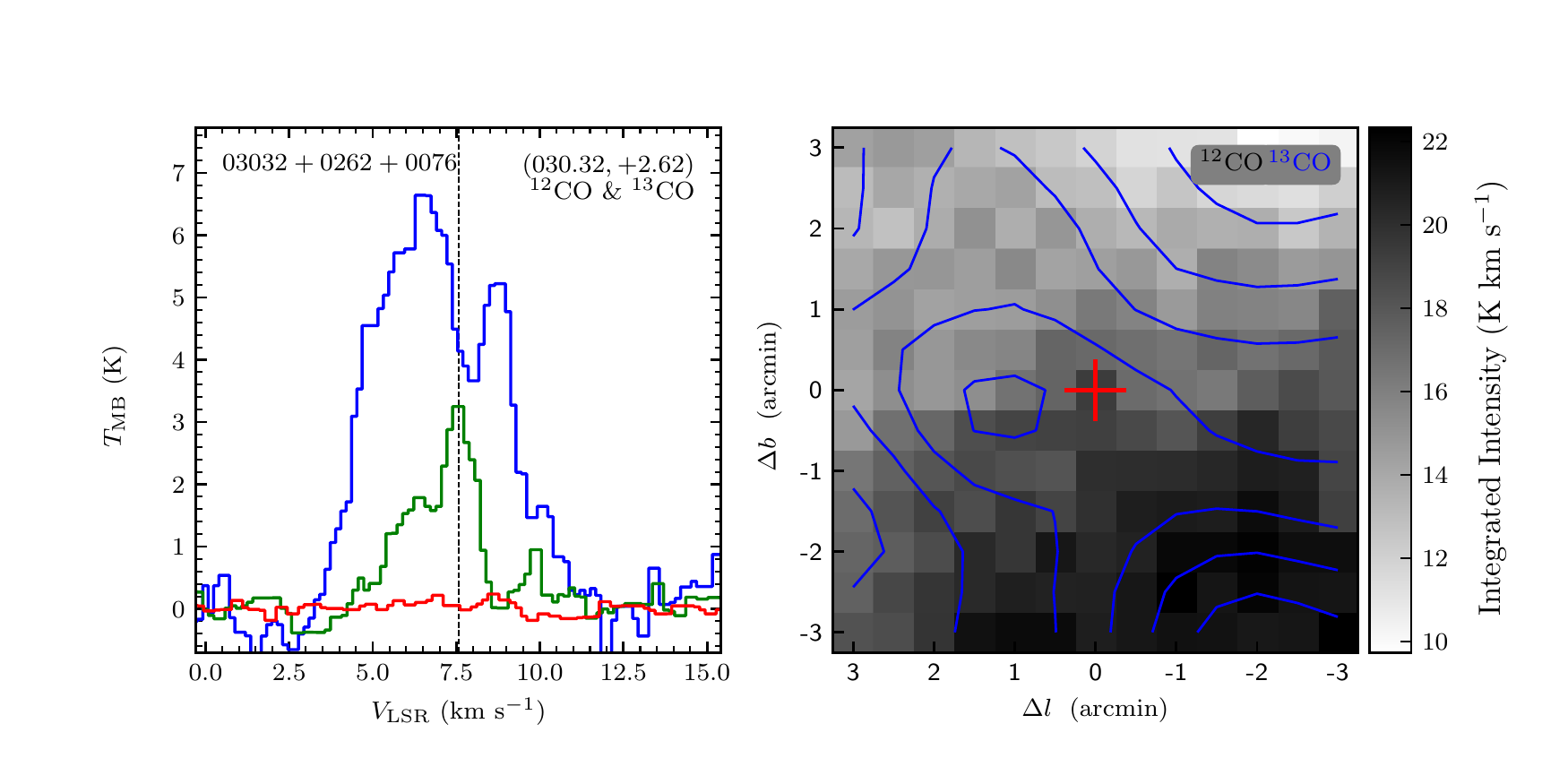}
\vspace{-0.5cm}

\includegraphics[width=9.0cm,angle=0]{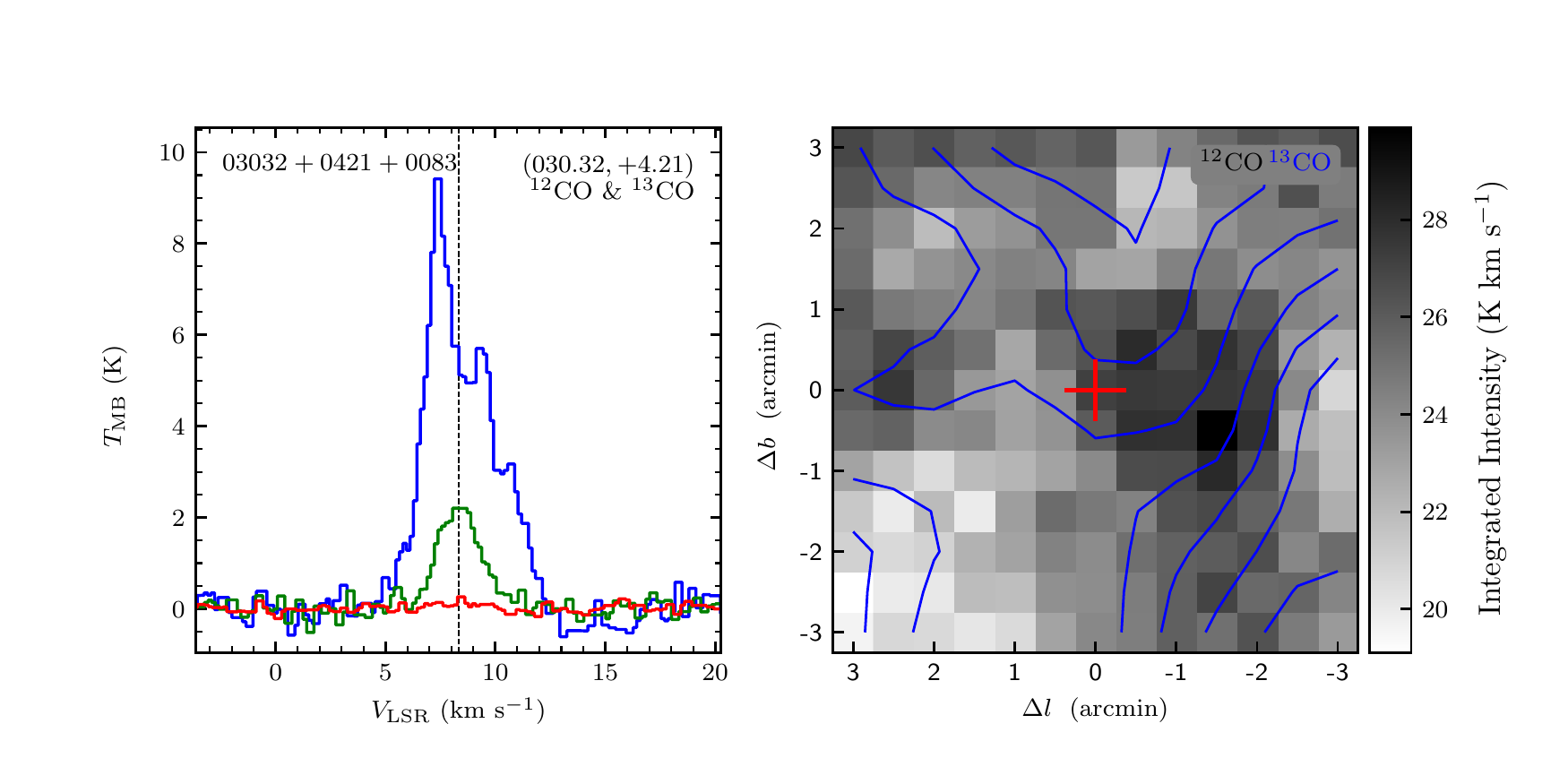}
\includegraphics[width=9.0cm,angle=0]{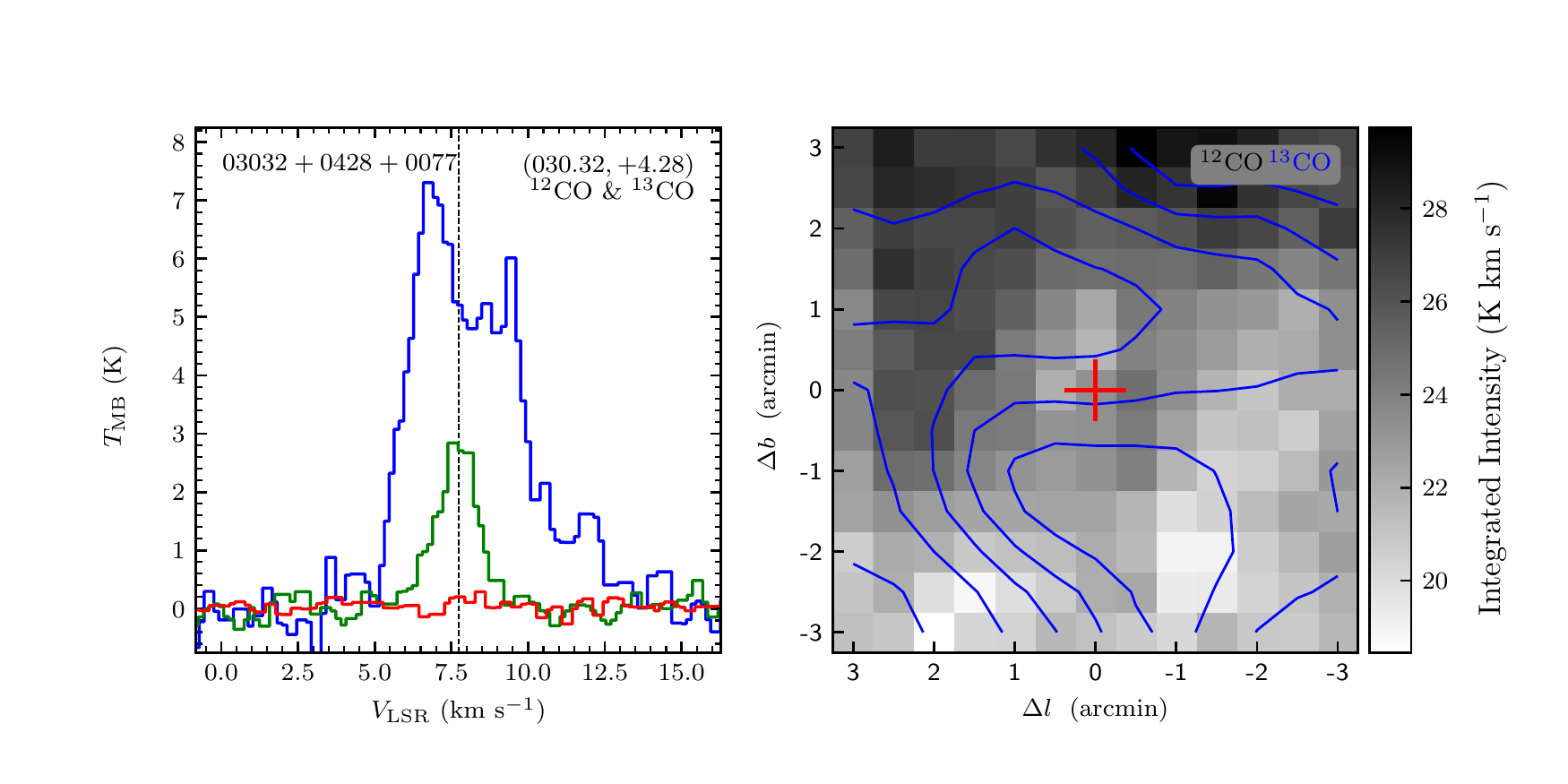}
\vspace{-0.5cm}

\includegraphics[width=9.0cm,angle=0]{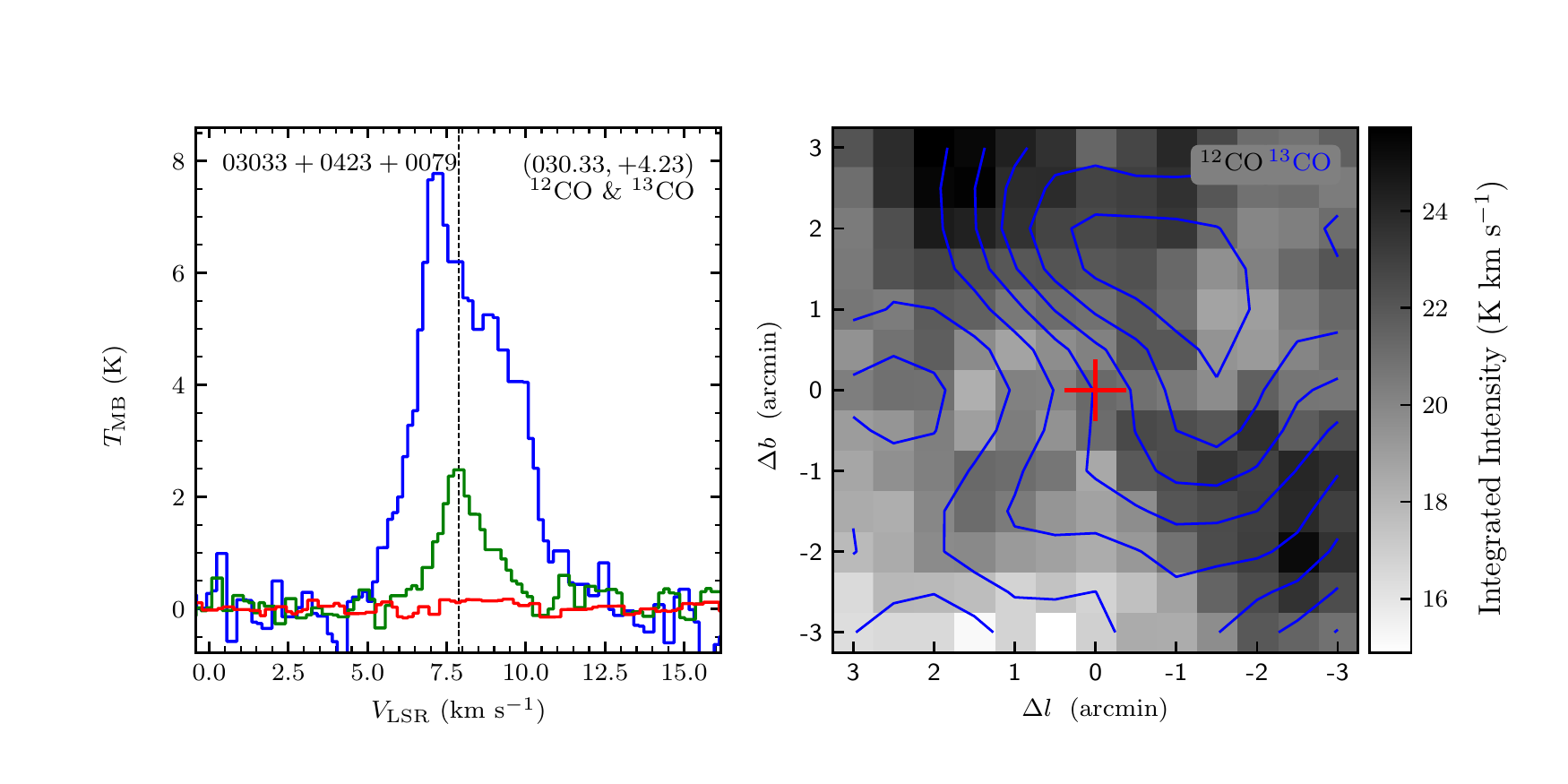}
\includegraphics[width=9.0cm,angle=0]{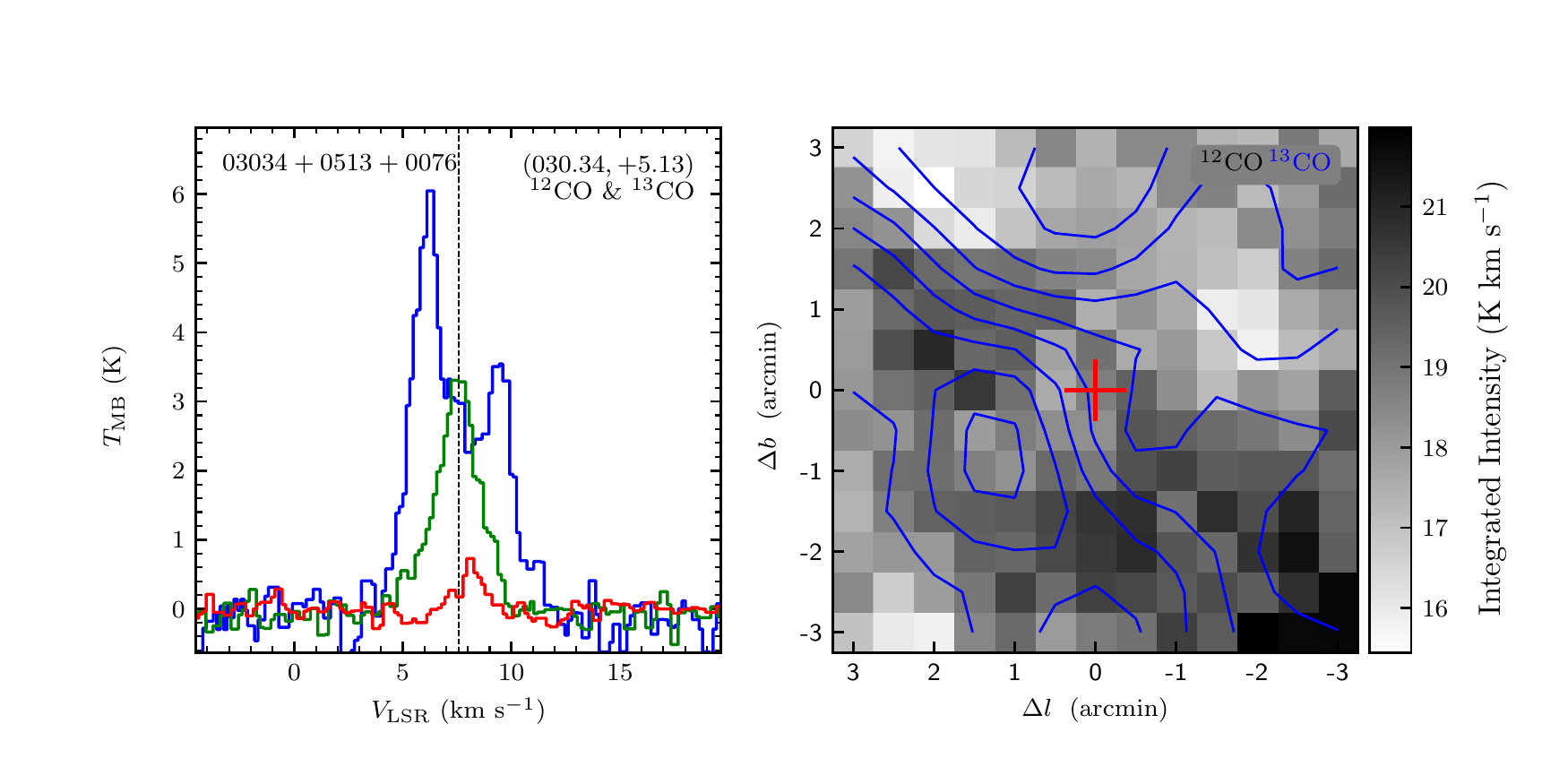}
\vspace{-0.5cm}

\includegraphics[width=9.0cm,angle=0]{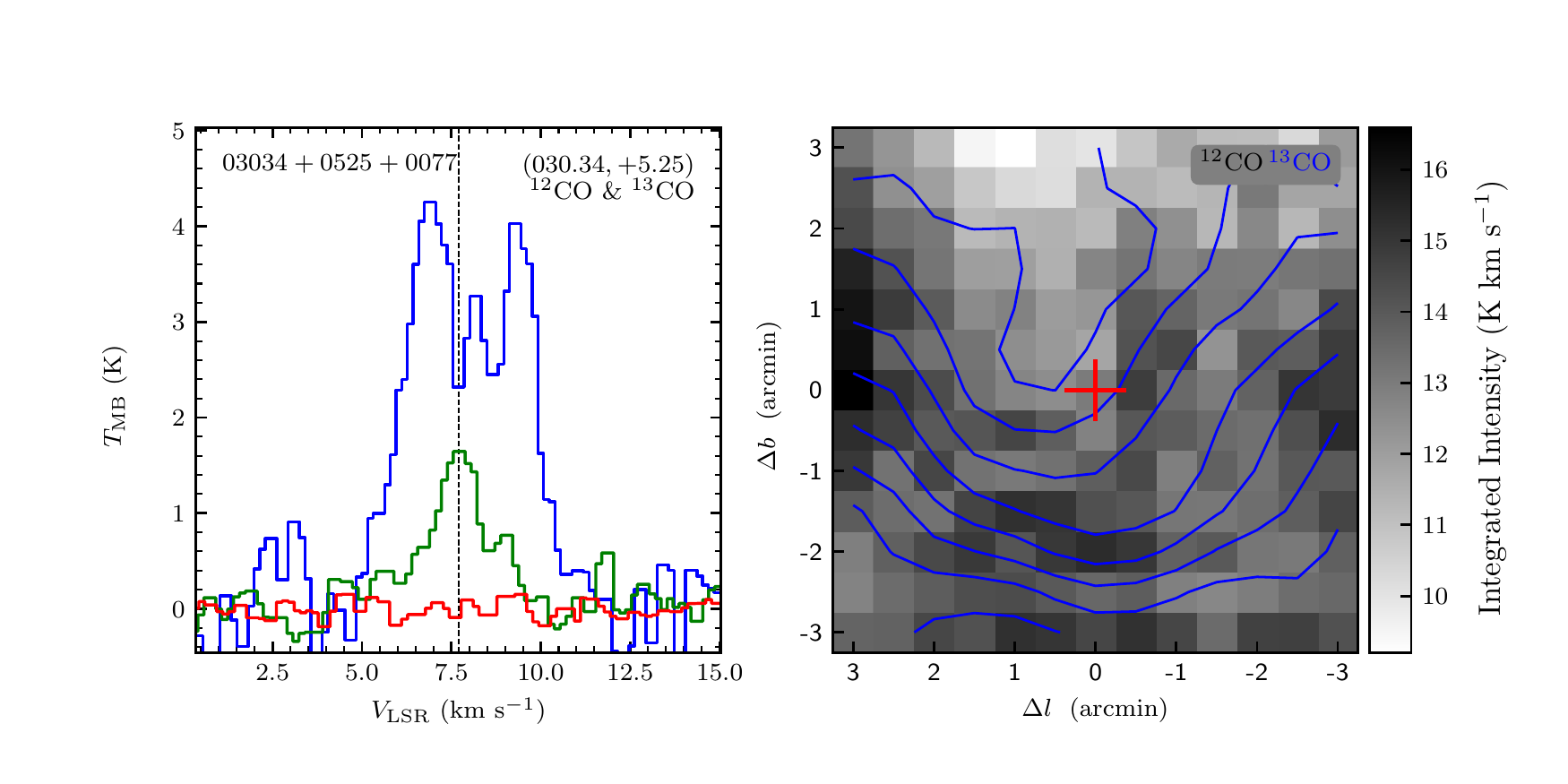}
\includegraphics[width=9.0cm,angle=0]{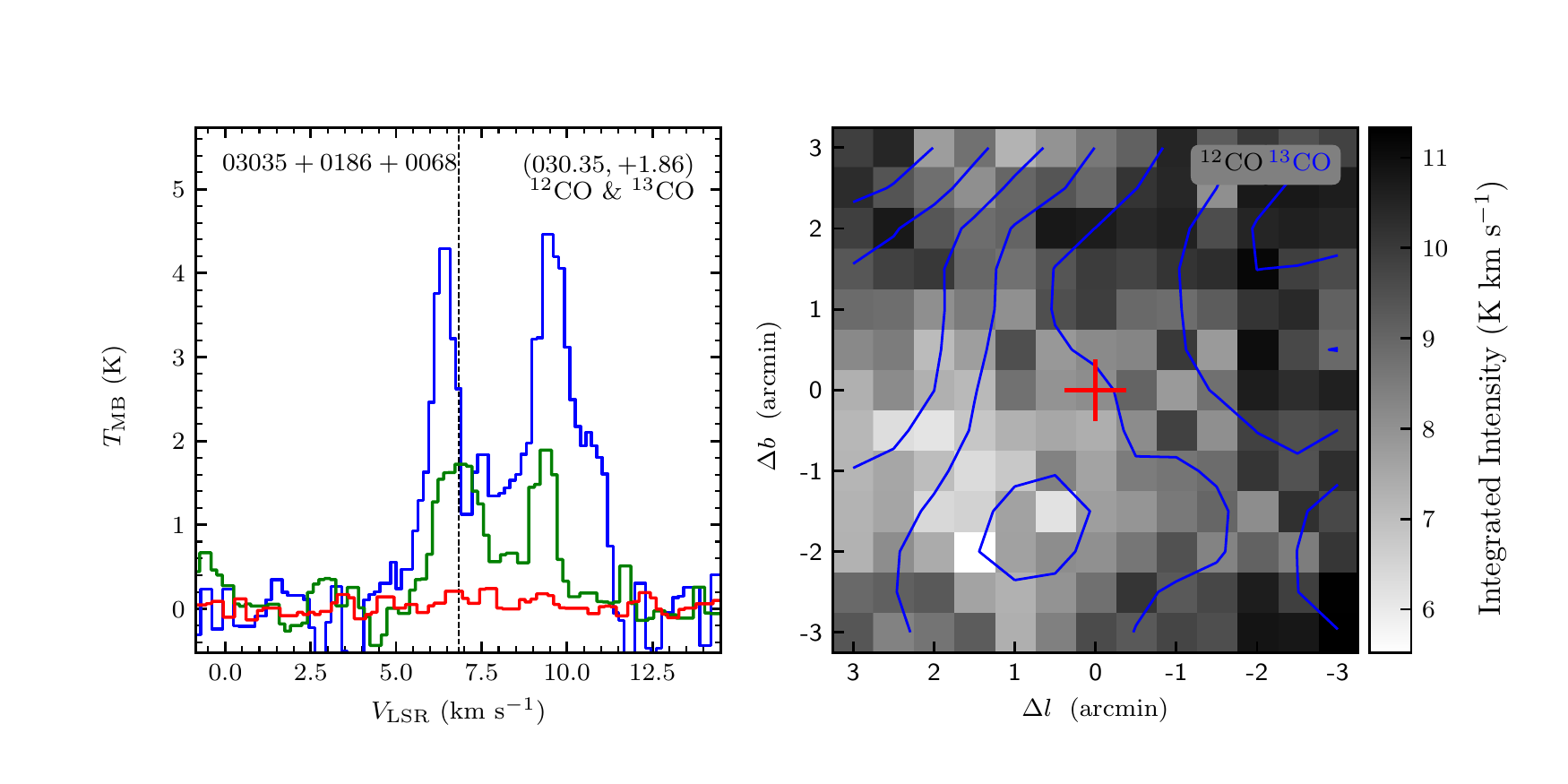}
\vspace{-0.5cm}

\includegraphics[width=9.0cm,angle=0]{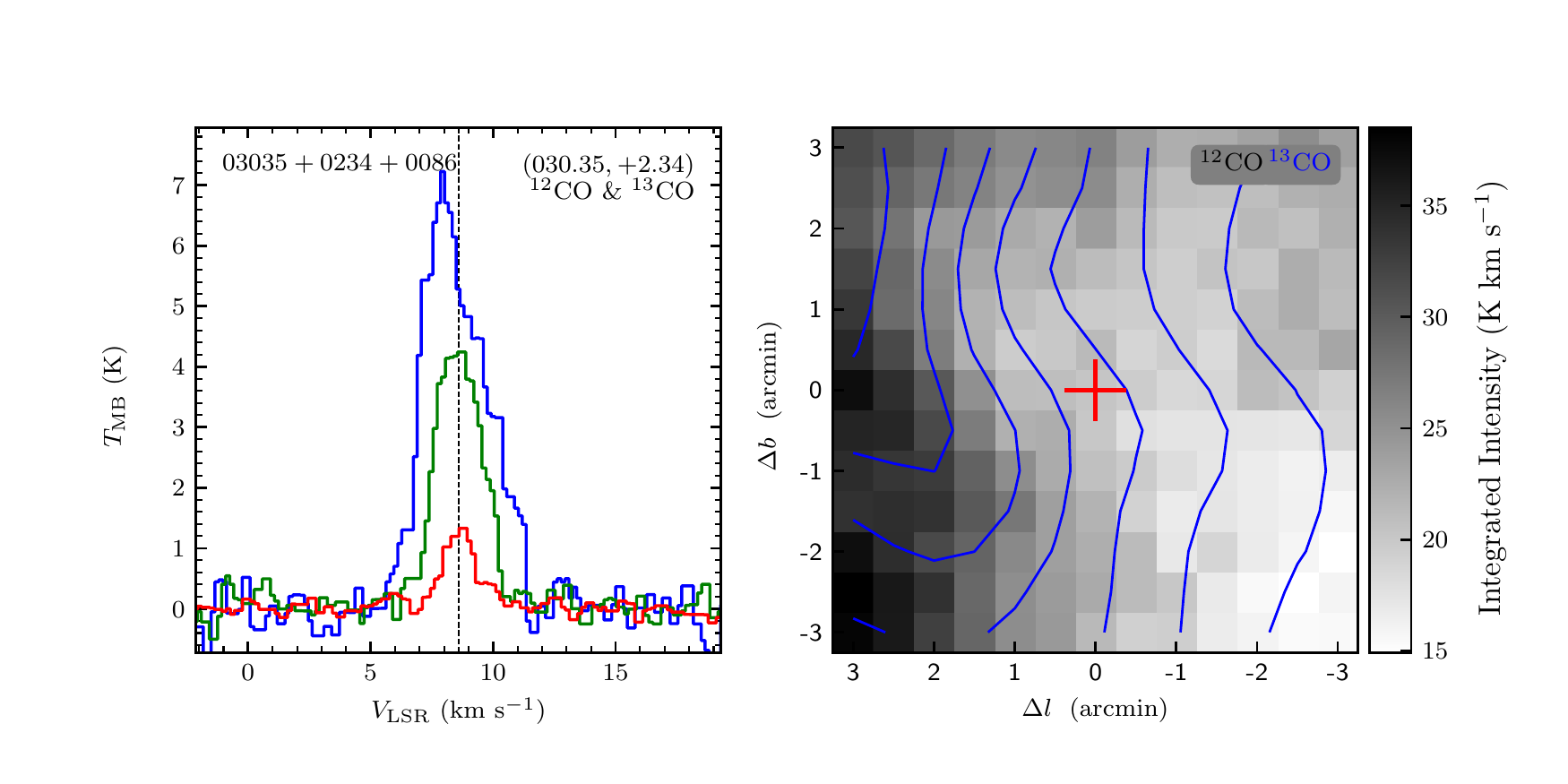}
\includegraphics[width=9.0cm,angle=0]{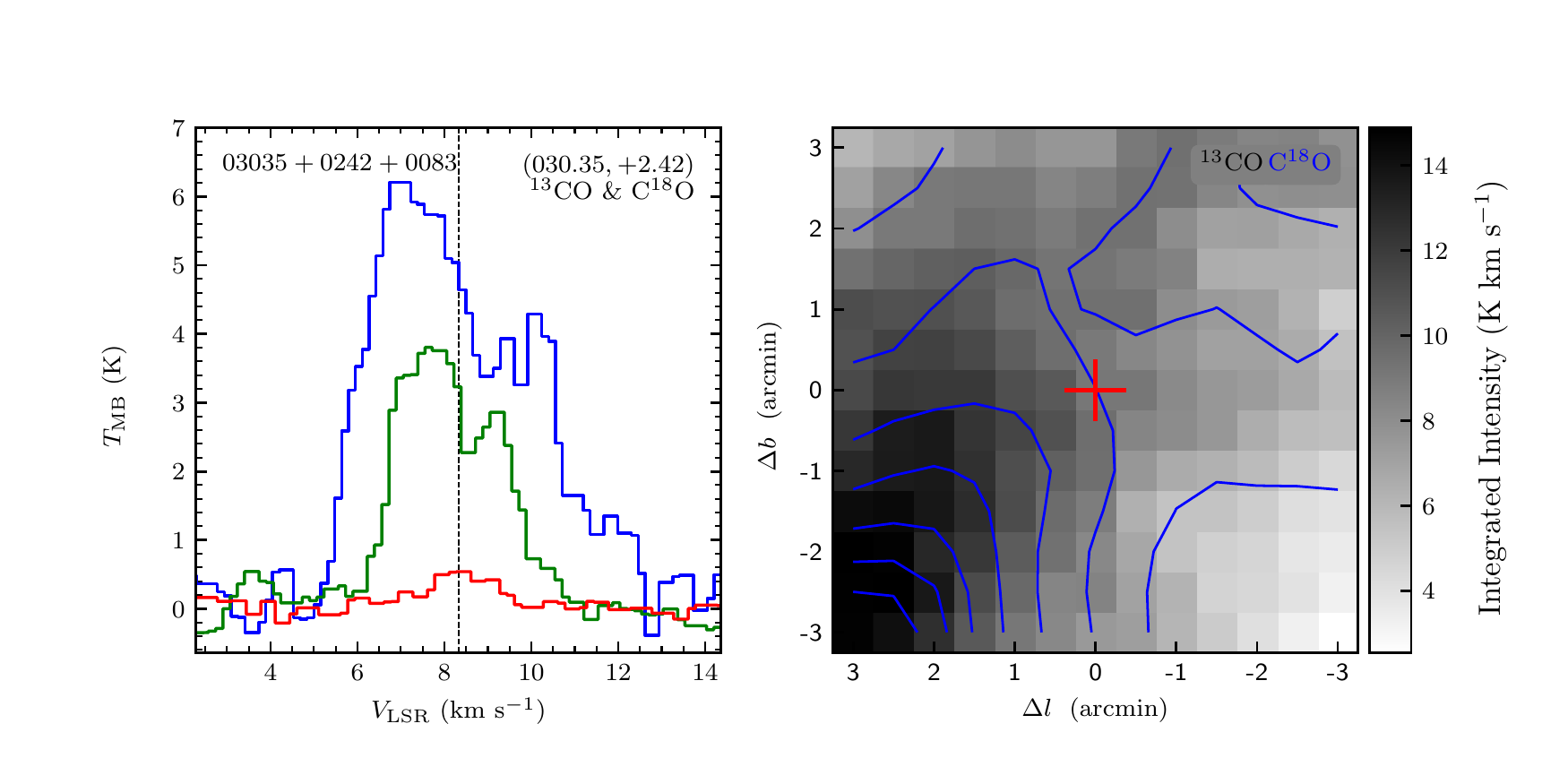}
\end{figure}
\clearpage

\begin{figure}
\includegraphics[width=9.0cm,angle=0]{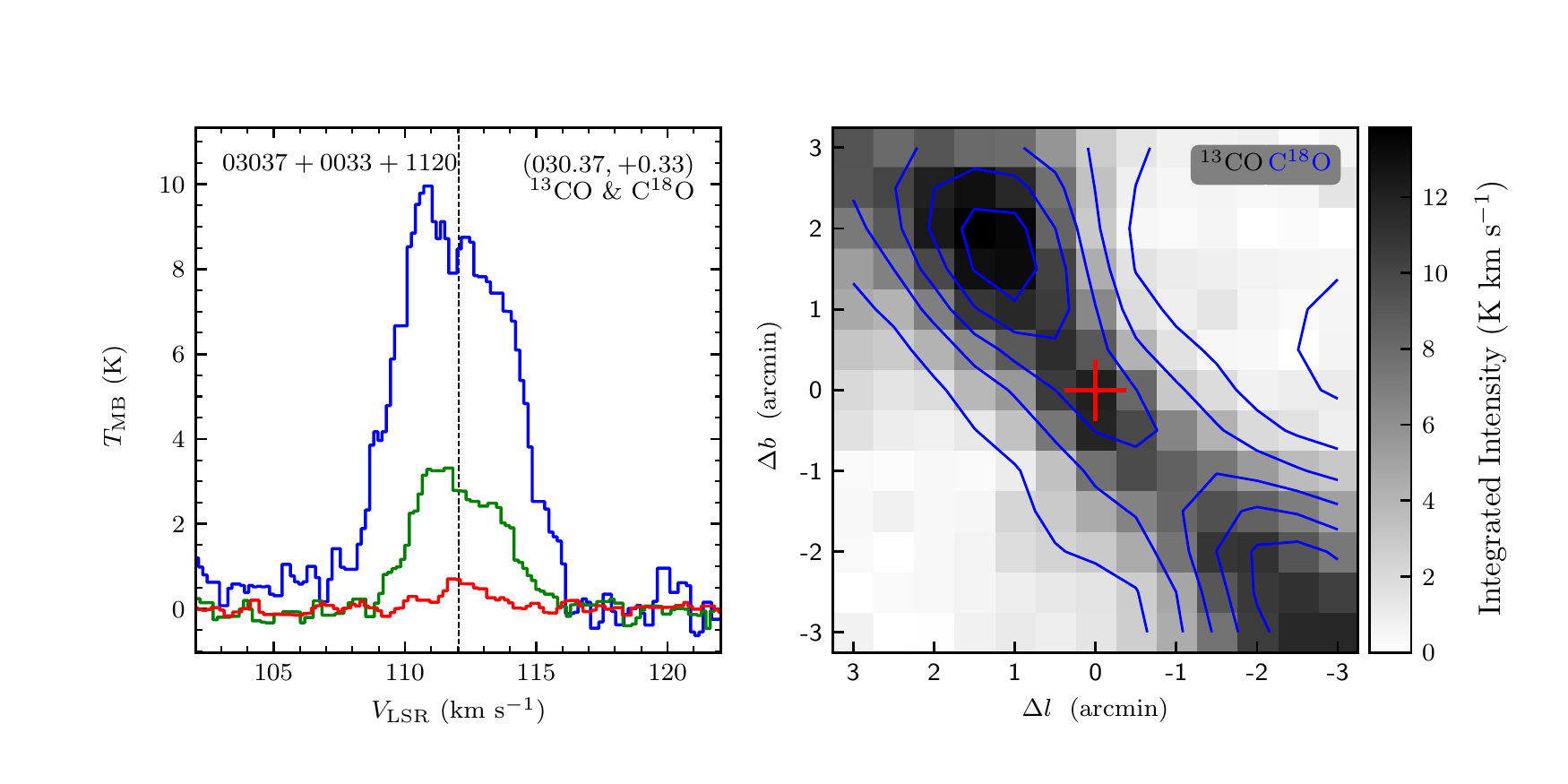}
\includegraphics[width=9.0cm,angle=0]{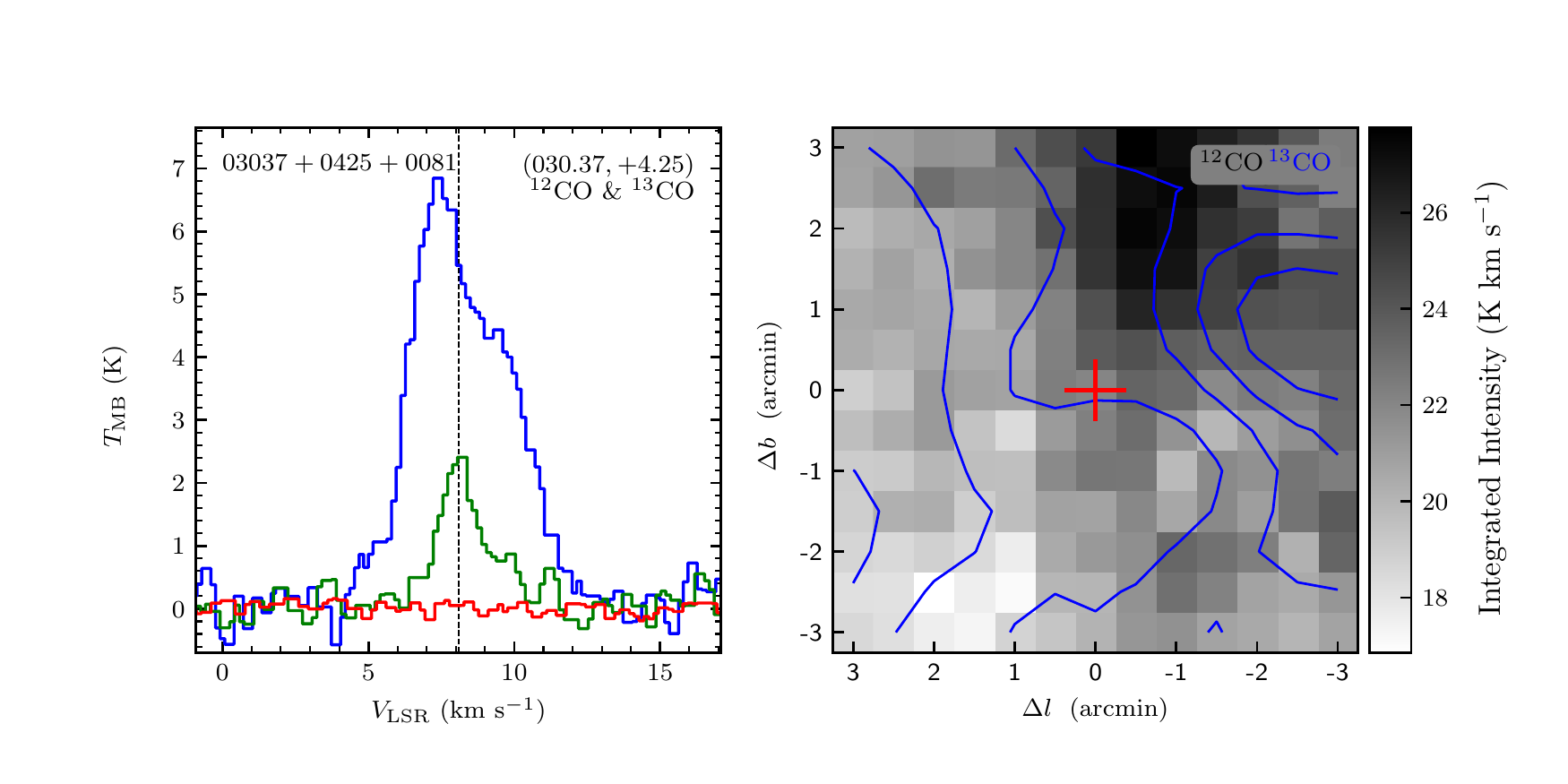}
\vspace{-0.5cm}

\includegraphics[width=9.0cm,angle=0]{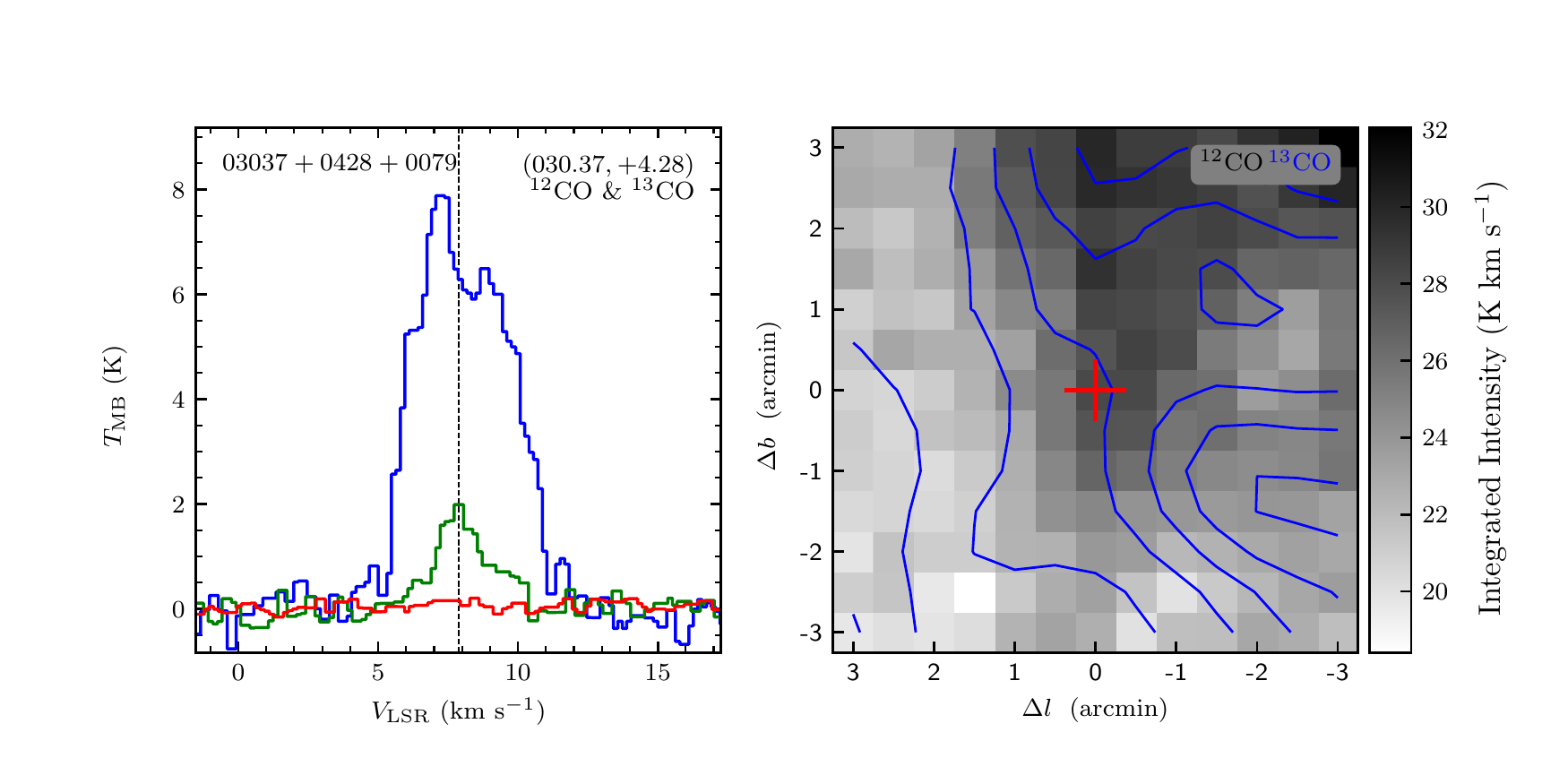}
\includegraphics[width=9.0cm,angle=0]{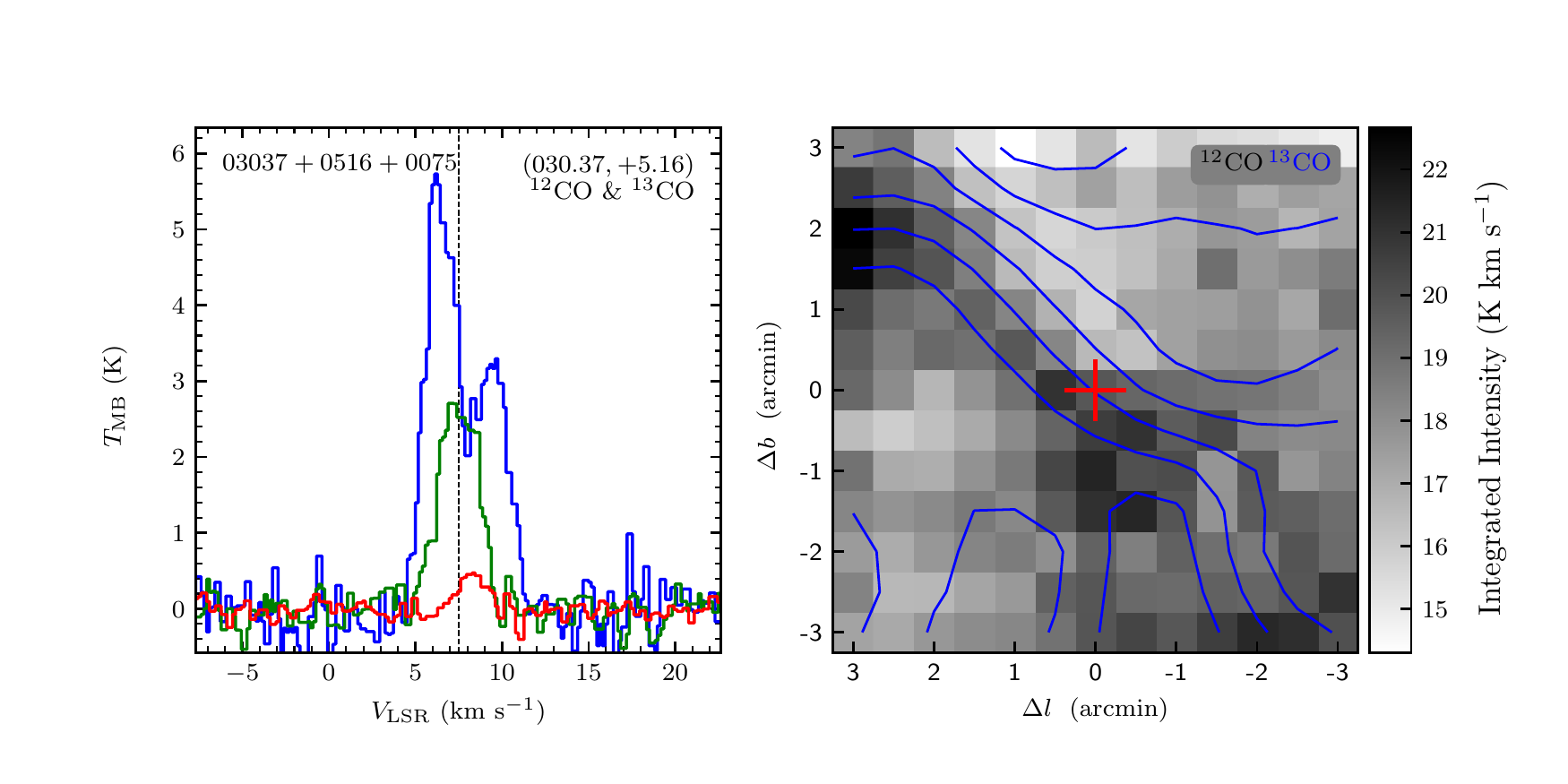}
\vspace{-0.5cm}

\includegraphics[width=9.0cm,angle=0]{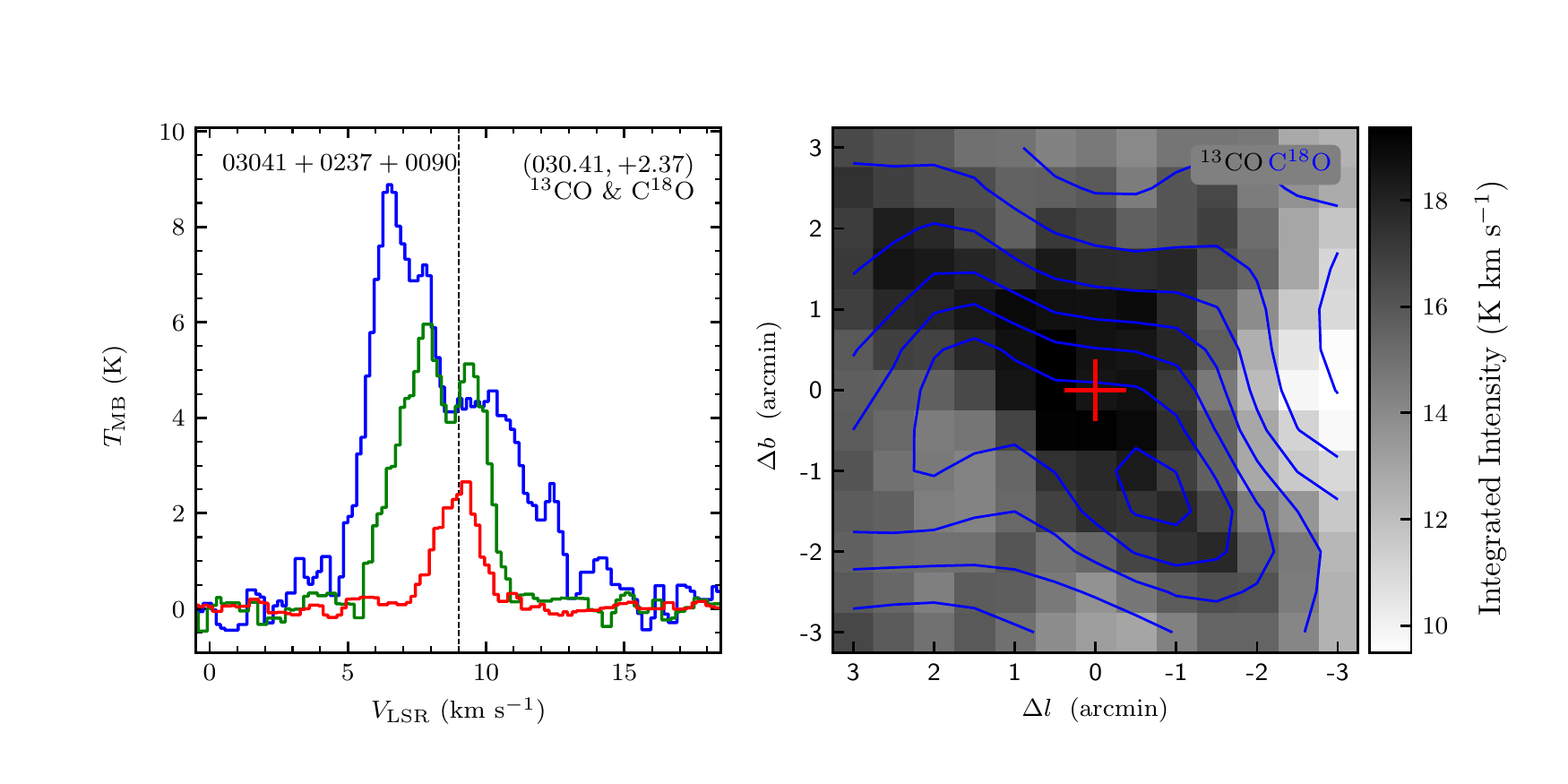}
\includegraphics[width=9.0cm,angle=0]{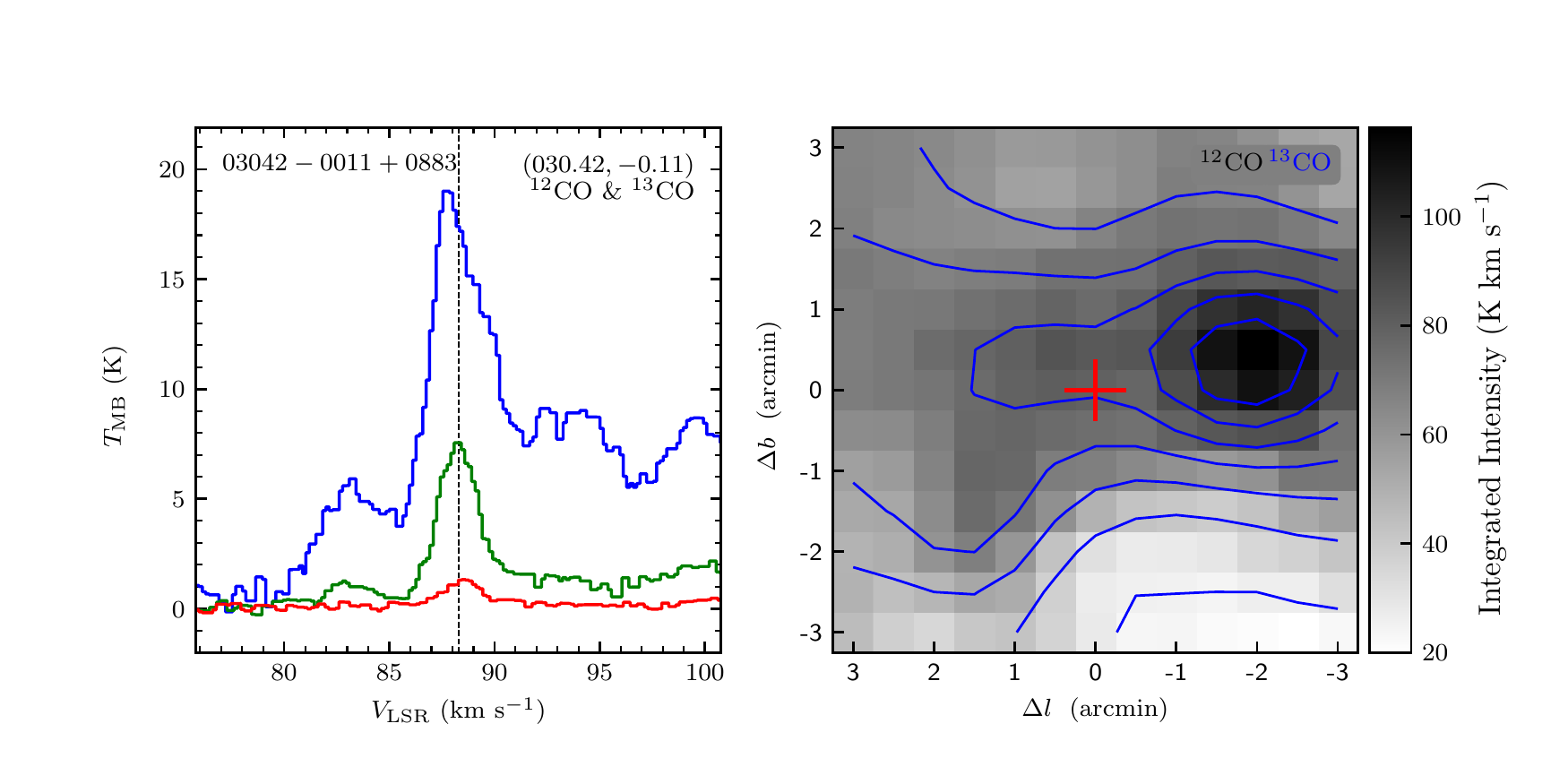}
\vspace{-0.5cm}

\includegraphics[width=9.0cm,angle=0]{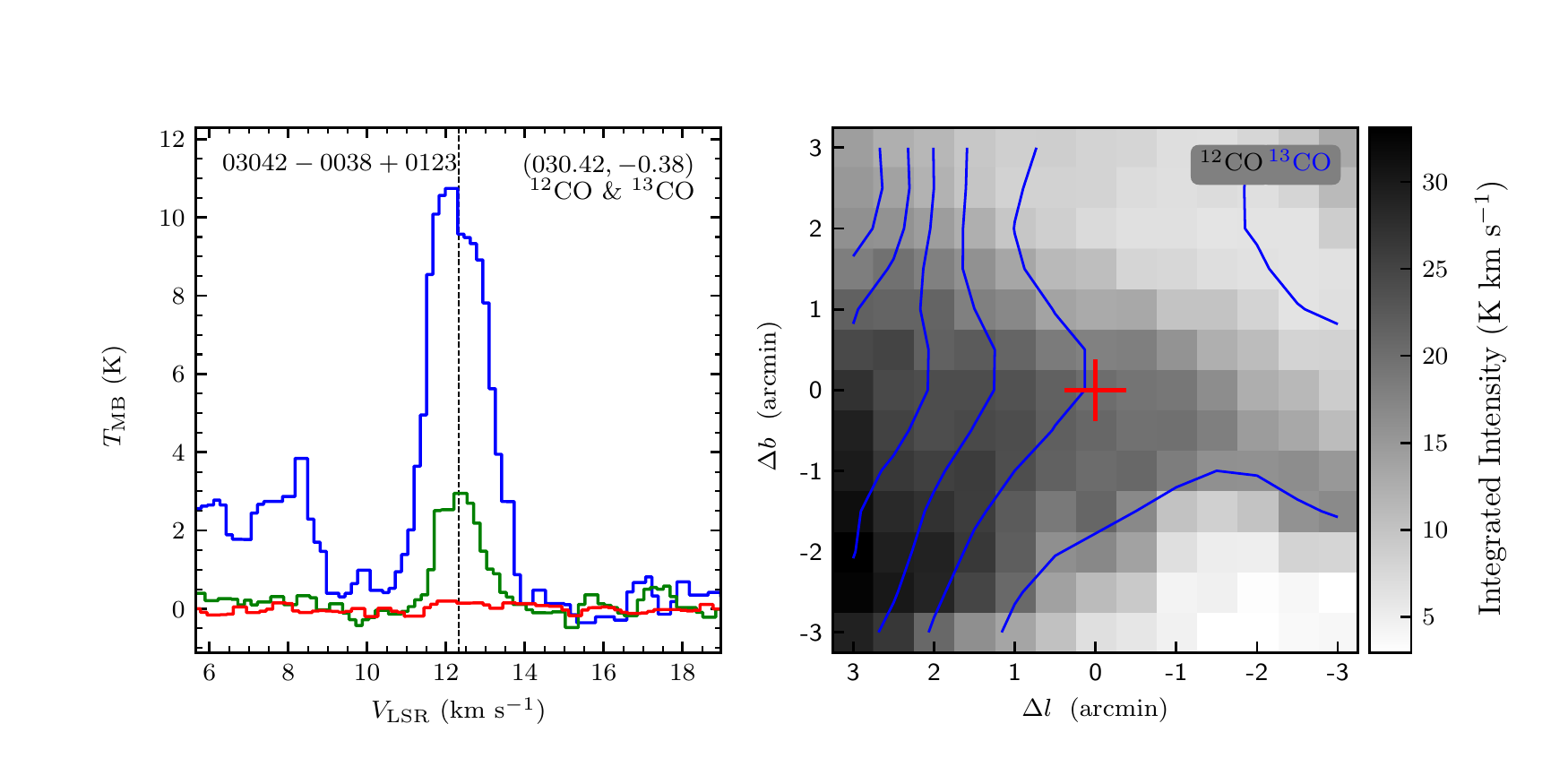}
\includegraphics[width=9.0cm,angle=0]{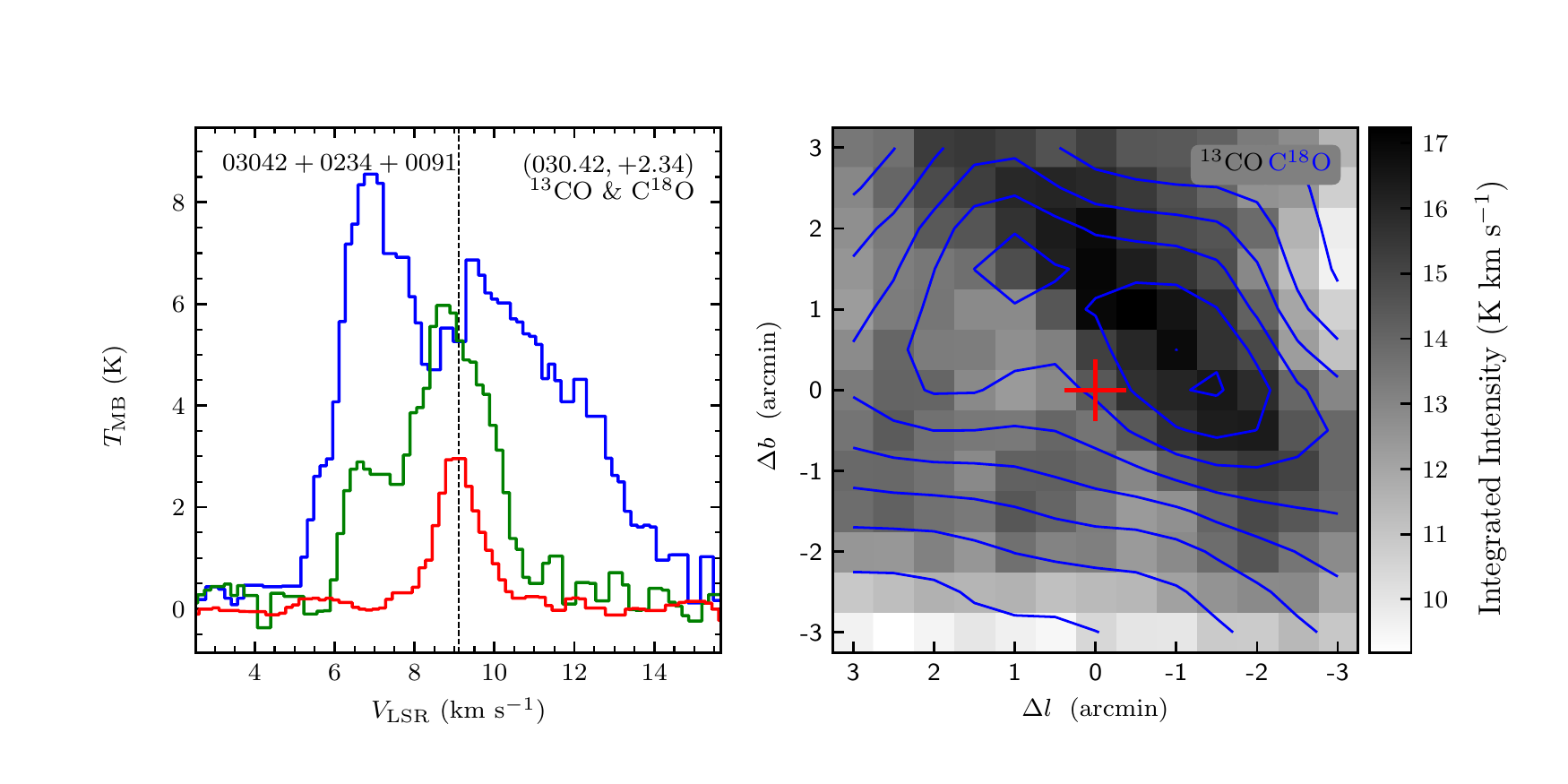}
\vspace{-0.5cm}

\includegraphics[width=9.0cm,angle=0]{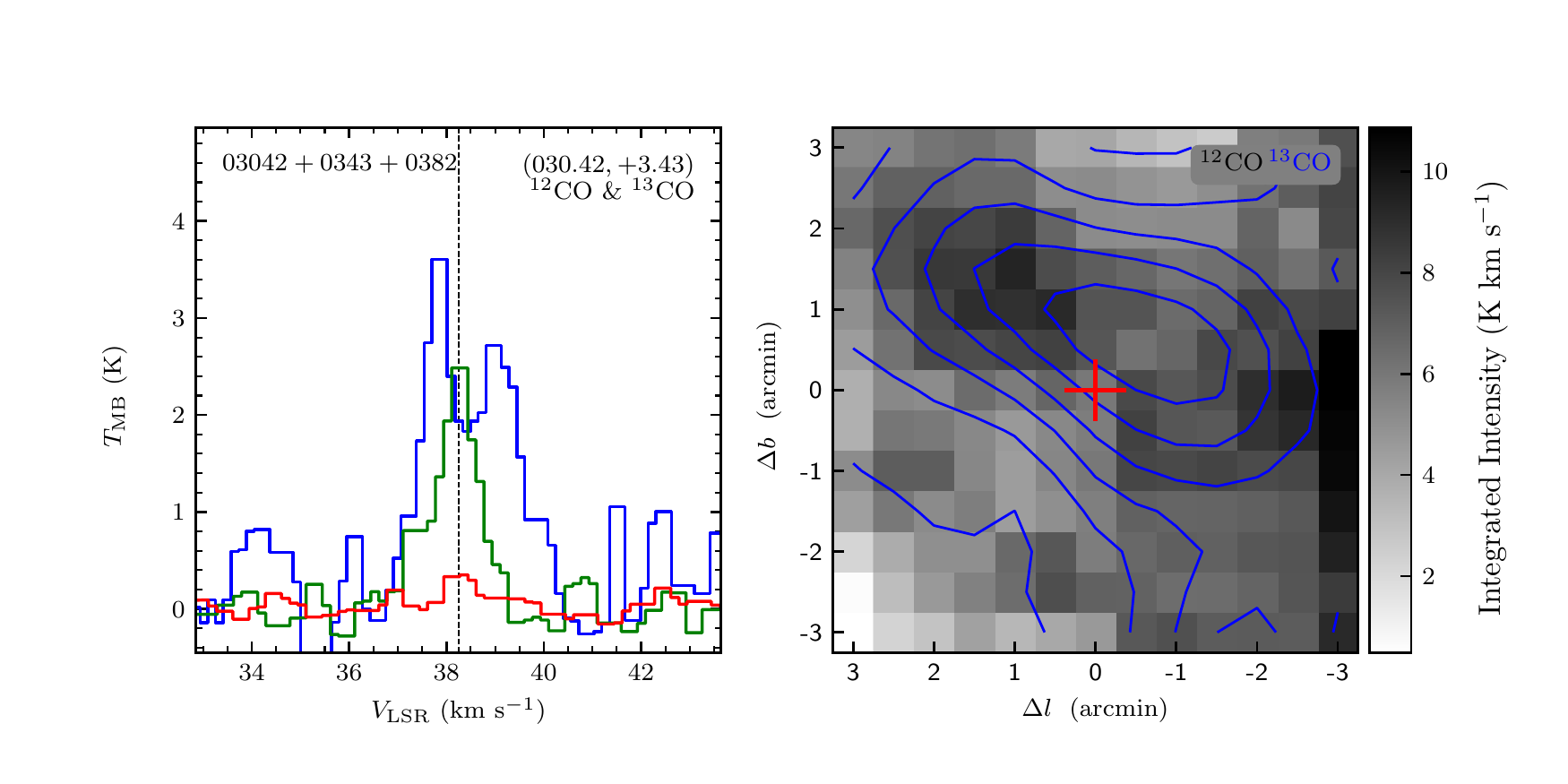}
\includegraphics[width=9.0cm,angle=0]{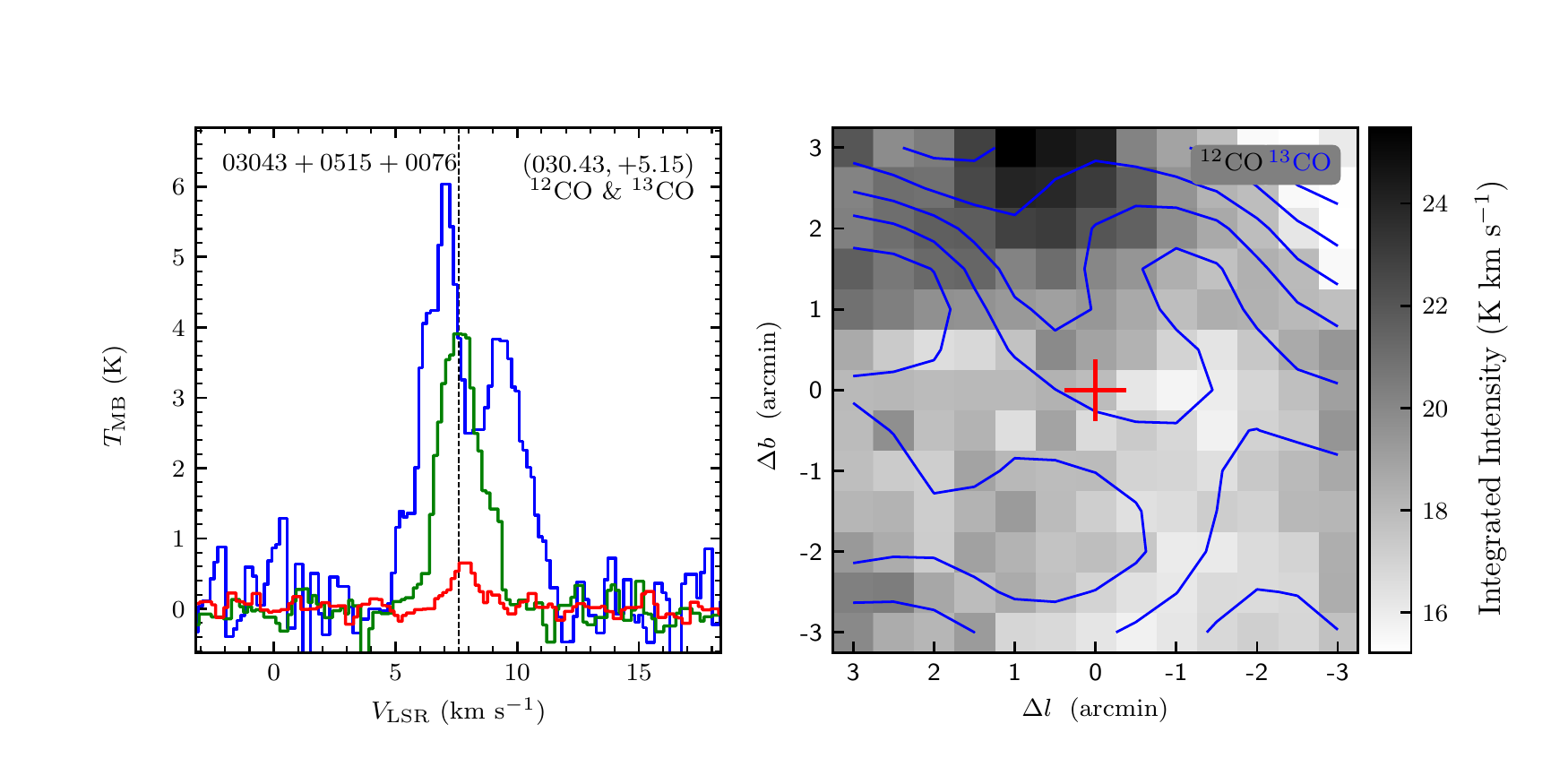}
\end{figure}
\clearpage

\begin{figure}
\includegraphics[width=9.0cm,angle=0]{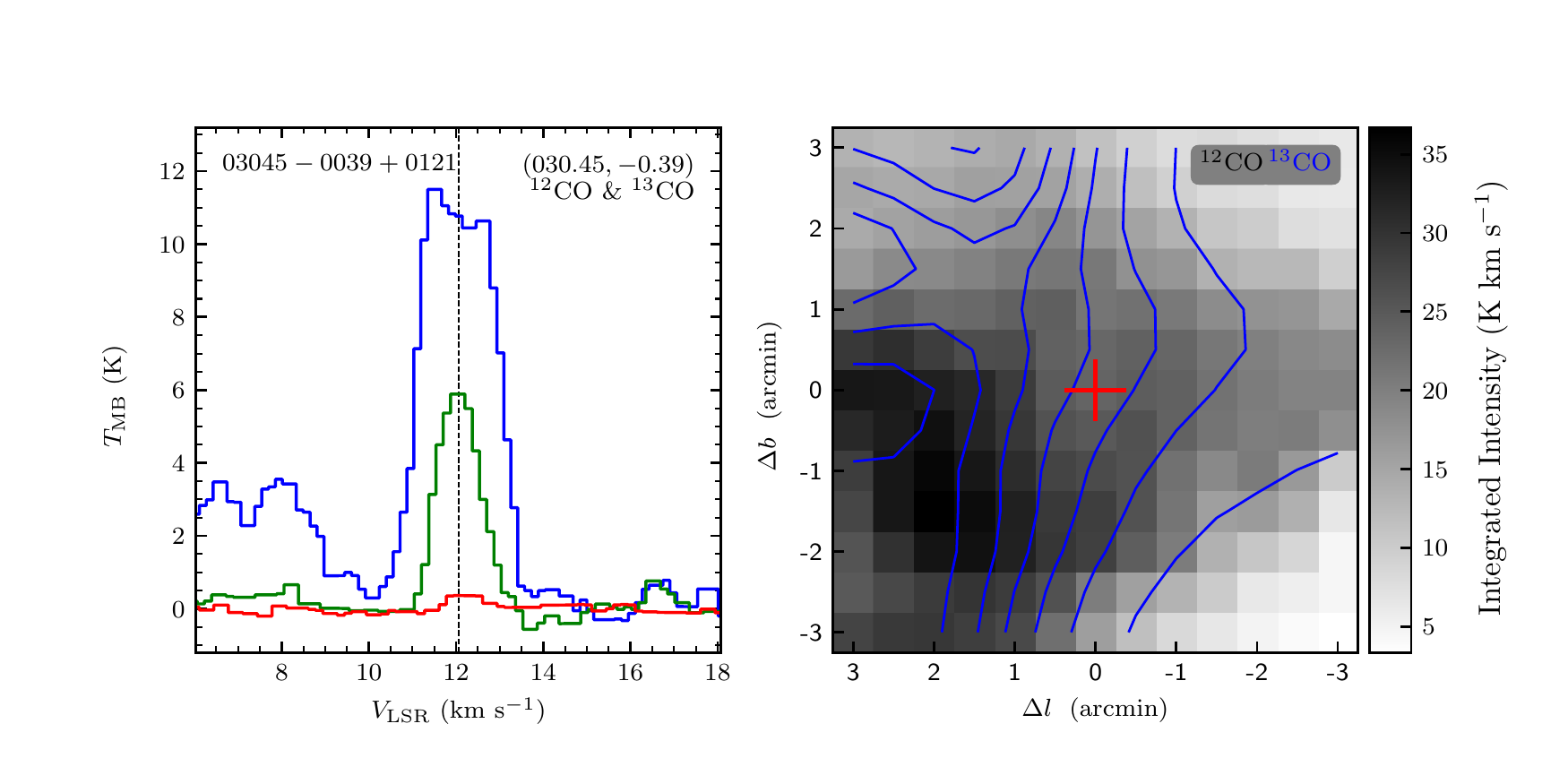}
\includegraphics[width=9.0cm,angle=0]{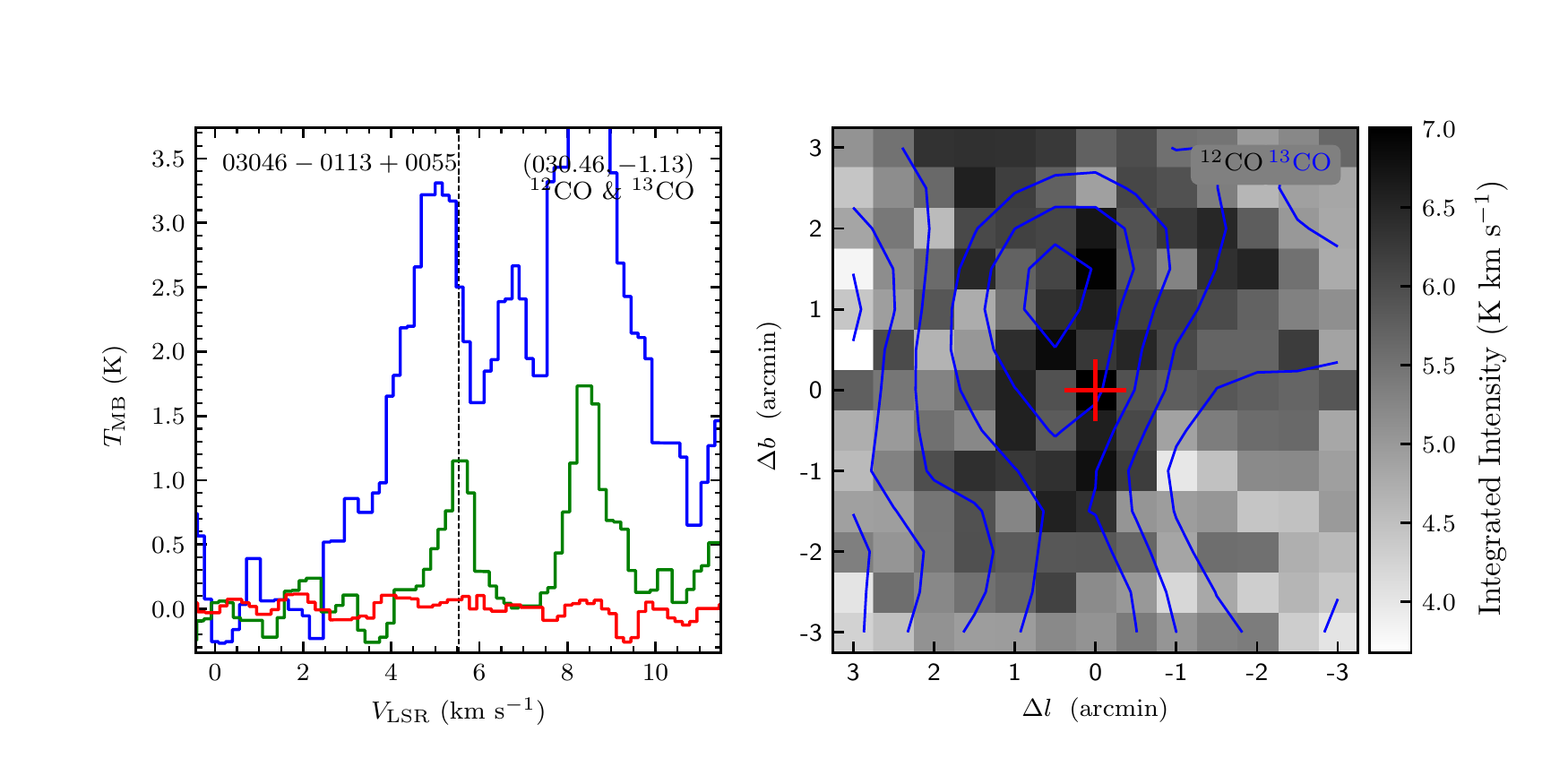}
\vspace{-0.5cm}

\includegraphics[width=9.0cm,angle=0]{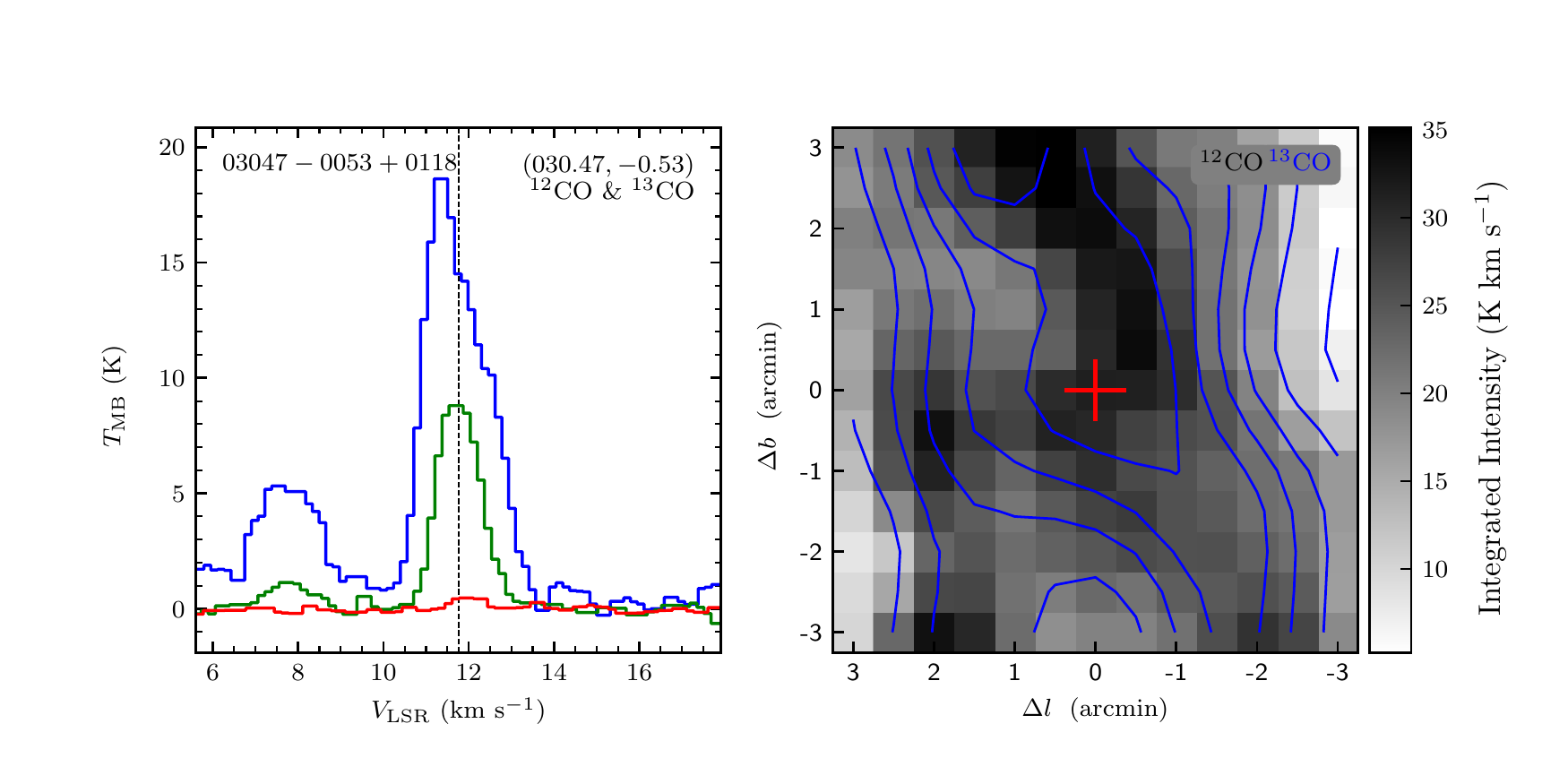}
\includegraphics[width=9.0cm,angle=0]{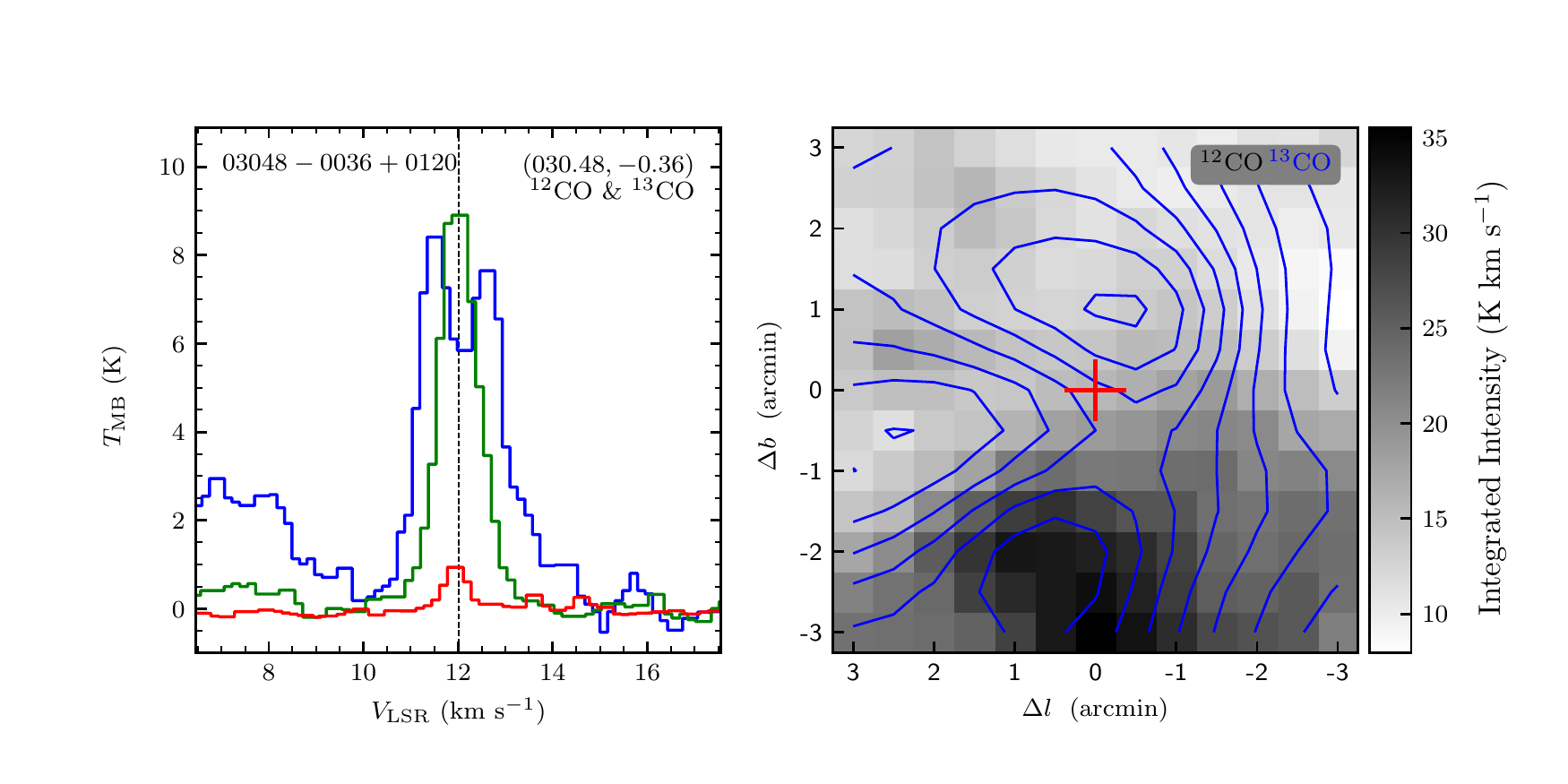}
\vspace{-0.5cm}

\includegraphics[width=9.0cm,angle=0]{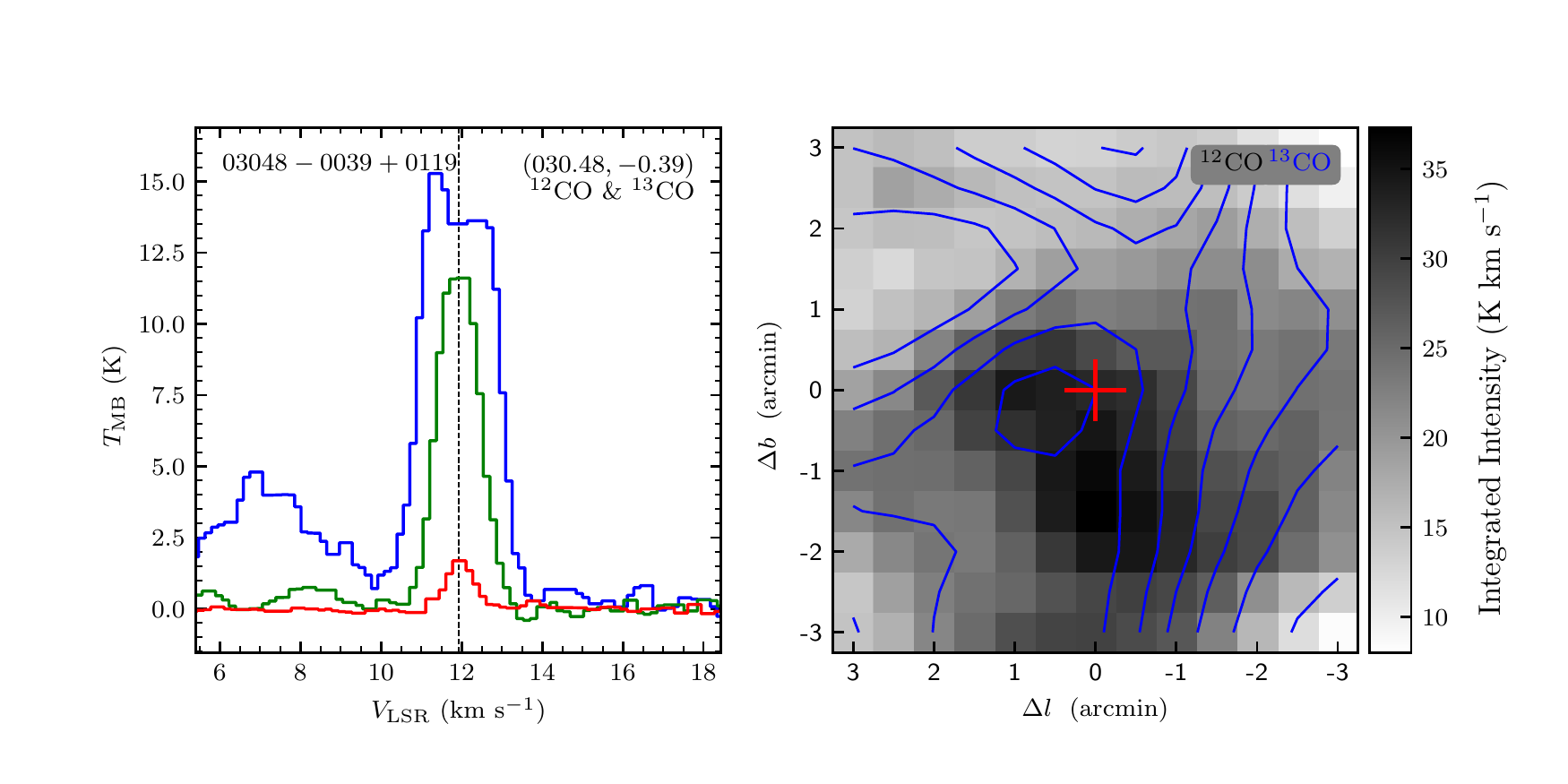}
\includegraphics[width=9.0cm,angle=0]{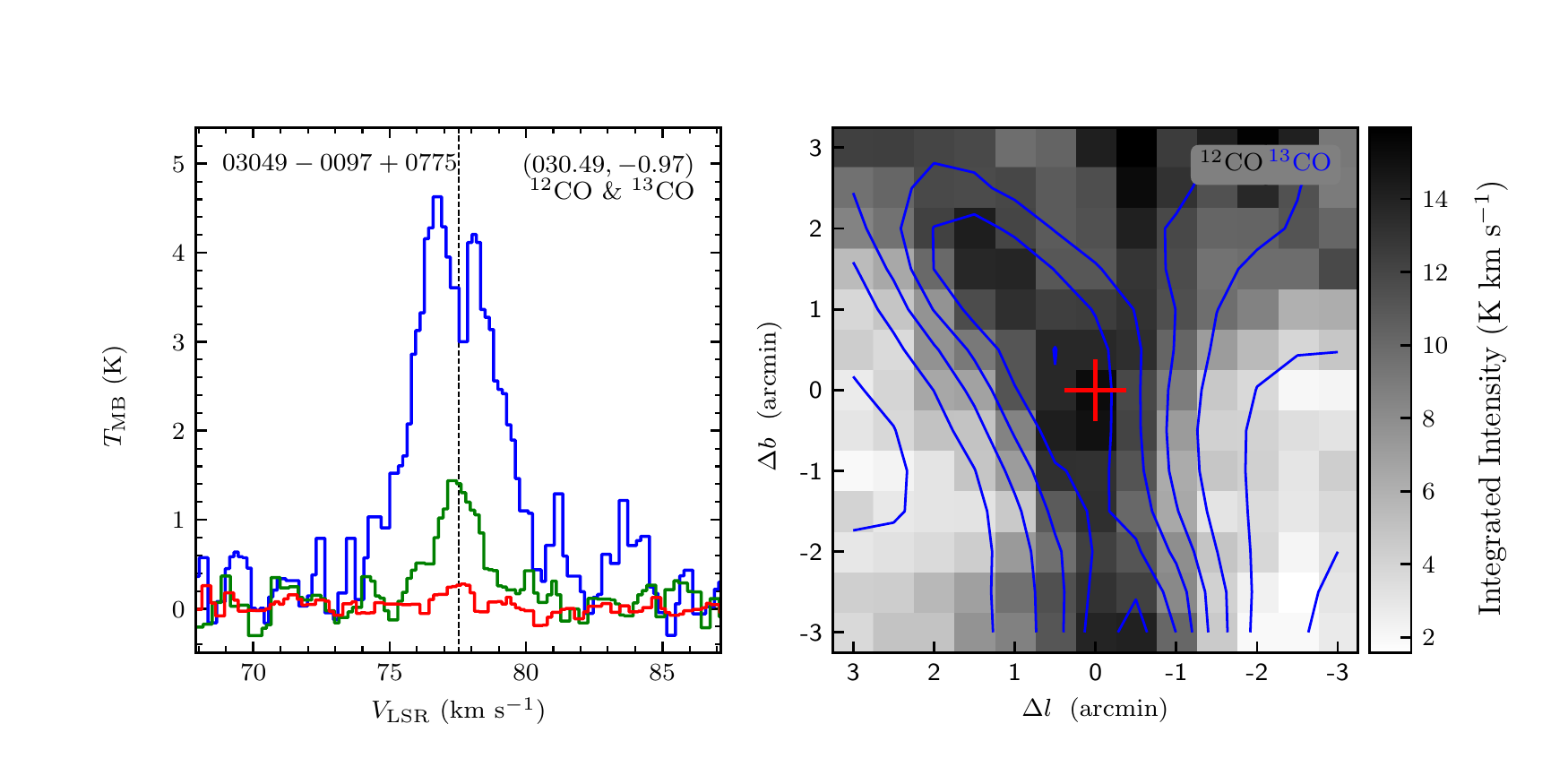}
\vspace{-0.5cm}

\includegraphics[width=9.0cm,angle=0]{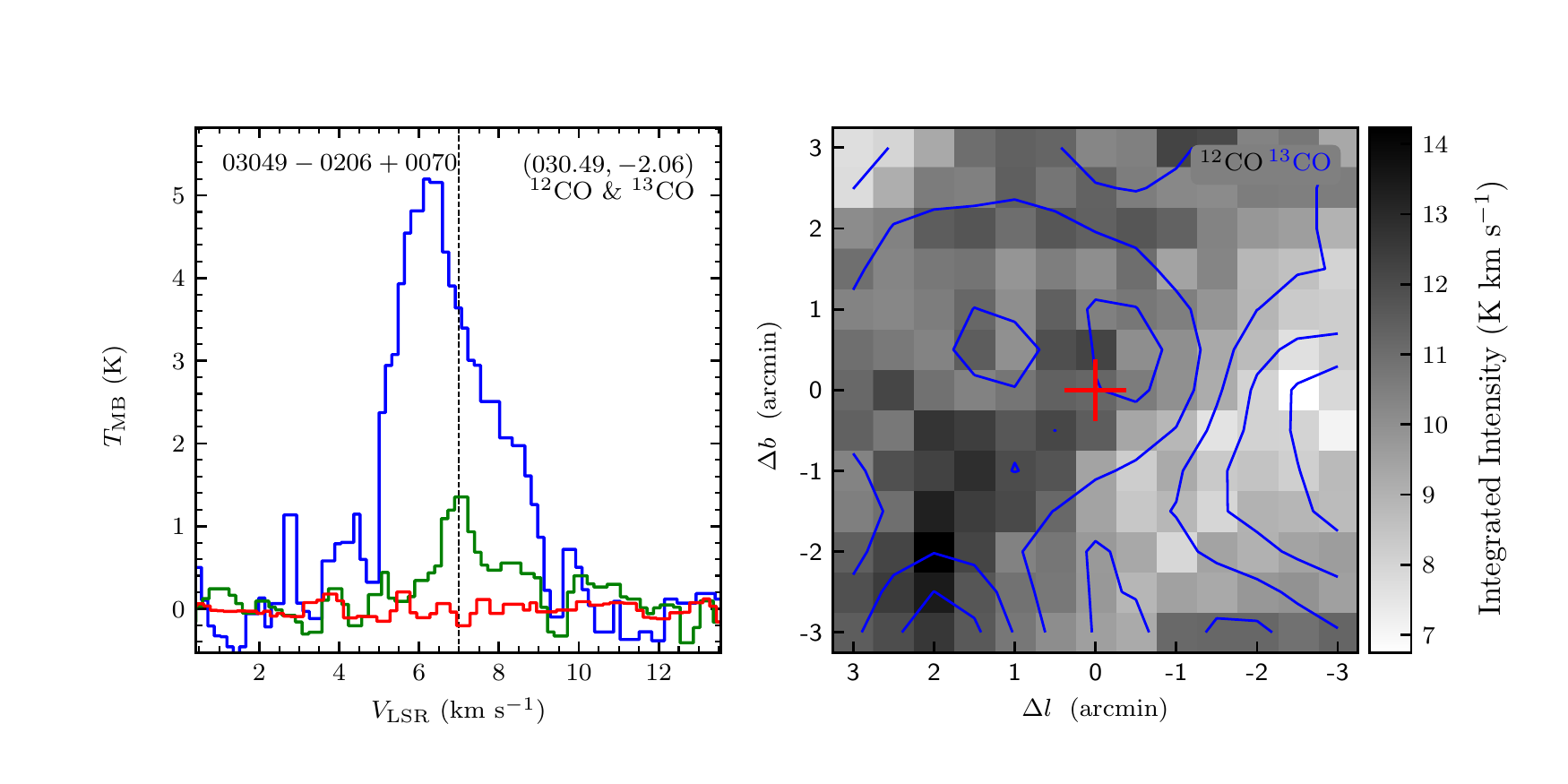}
\includegraphics[width=9.0cm,angle=0]{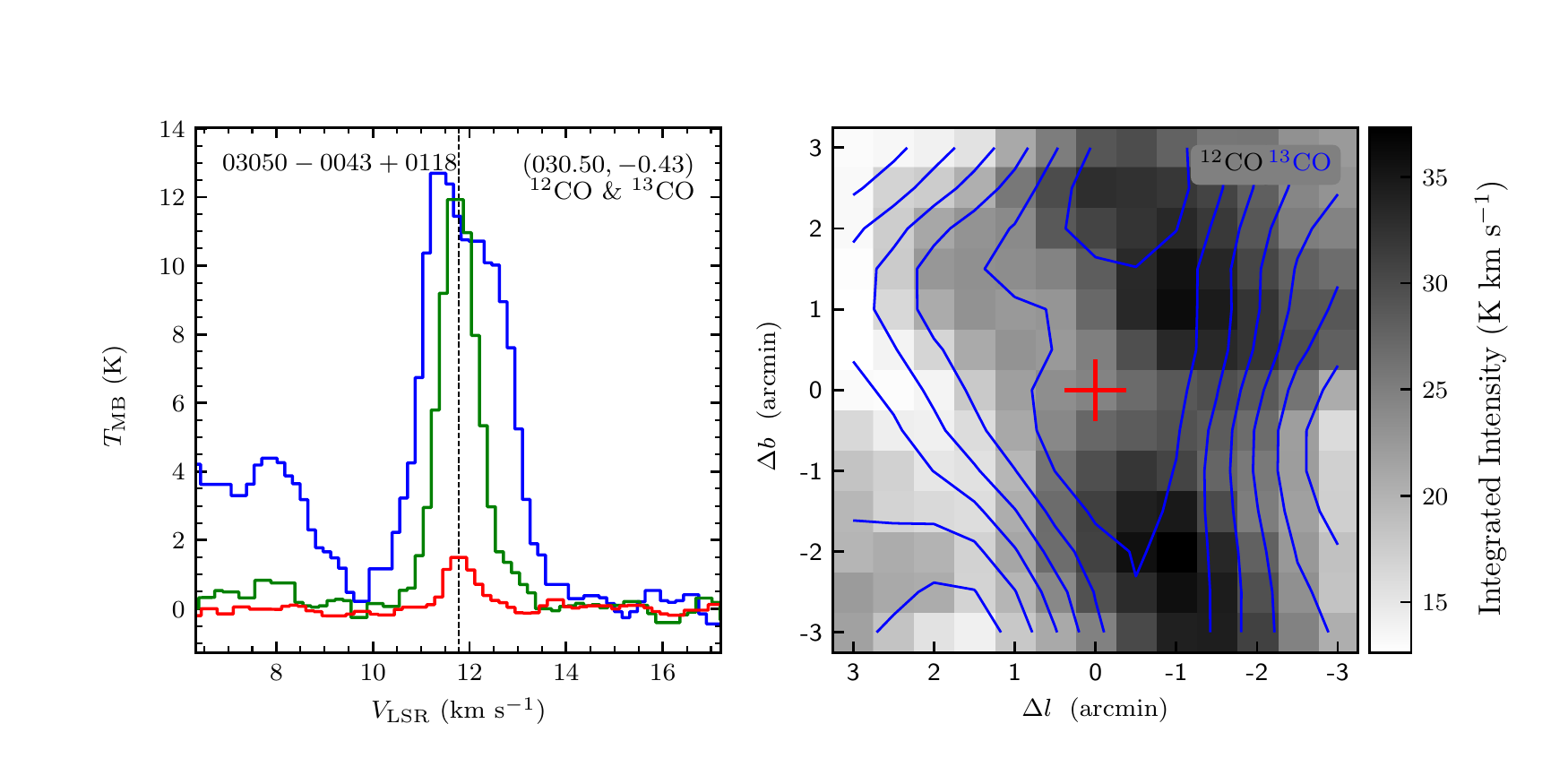}
\vspace{-0.5cm}

\includegraphics[width=9.0cm,angle=0]{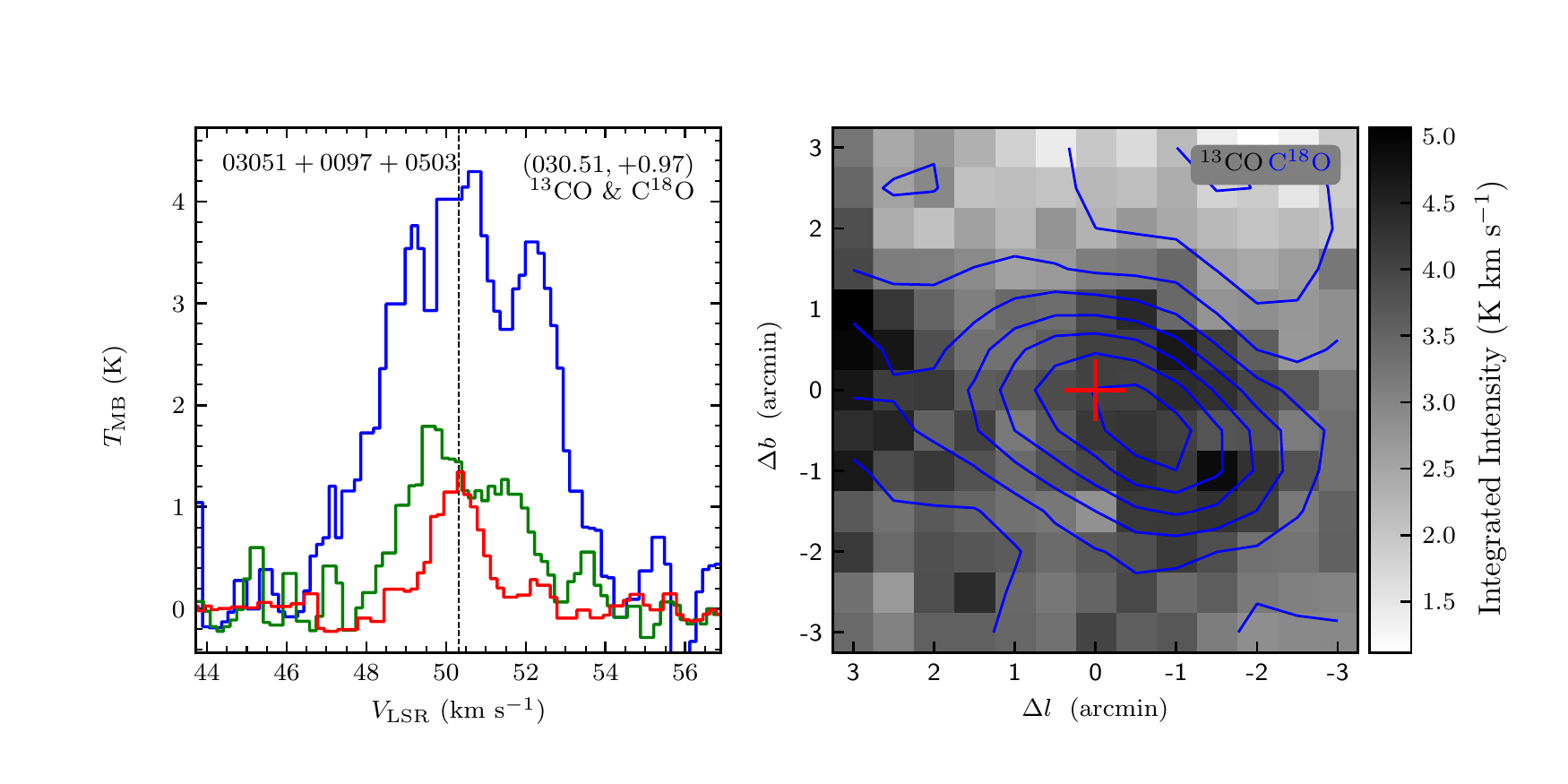}
\includegraphics[width=9.0cm,angle=0]{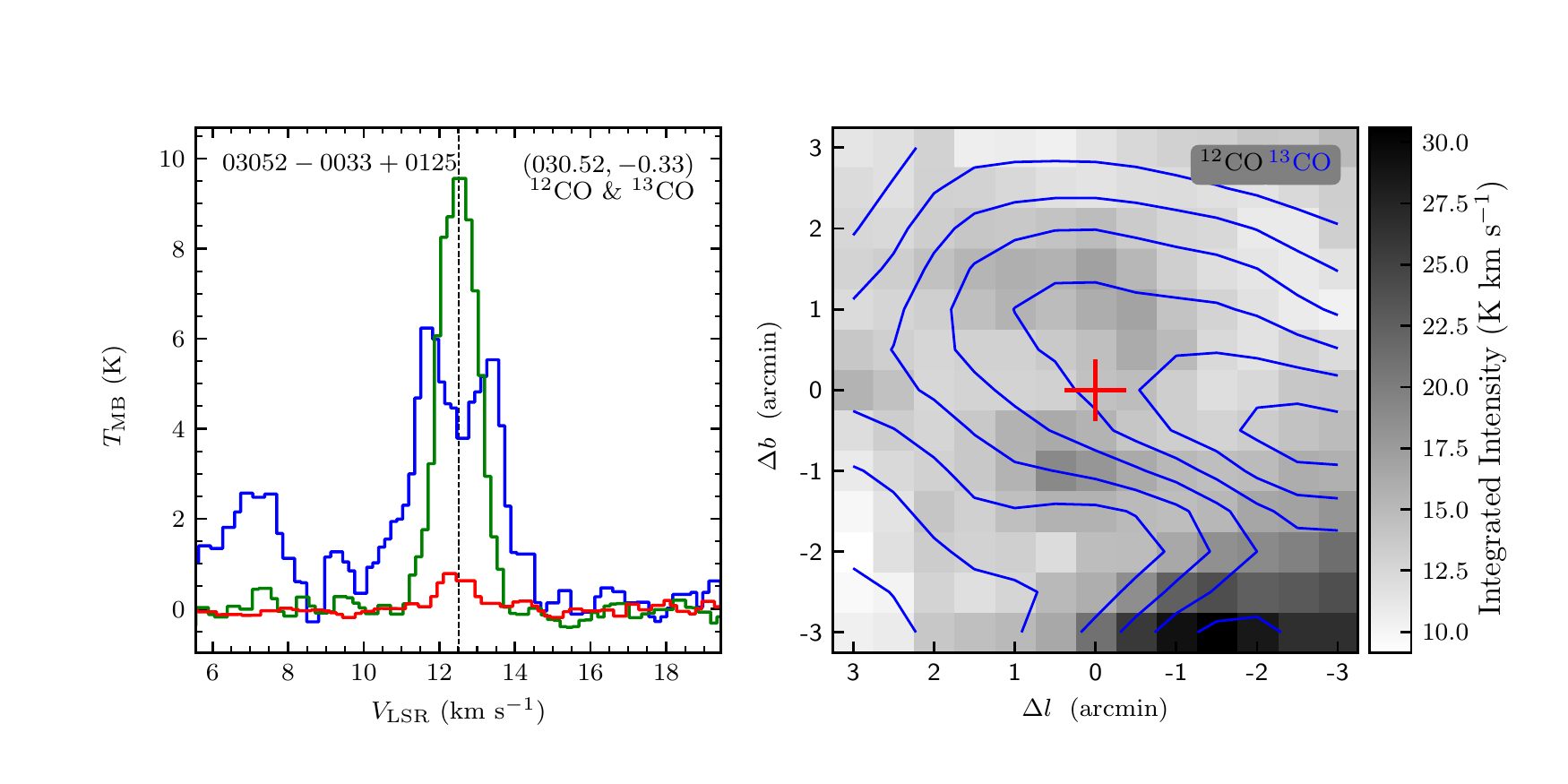}
\end{figure}
\clearpage

\begin{figure}
\includegraphics[width=9.0cm,angle=0]{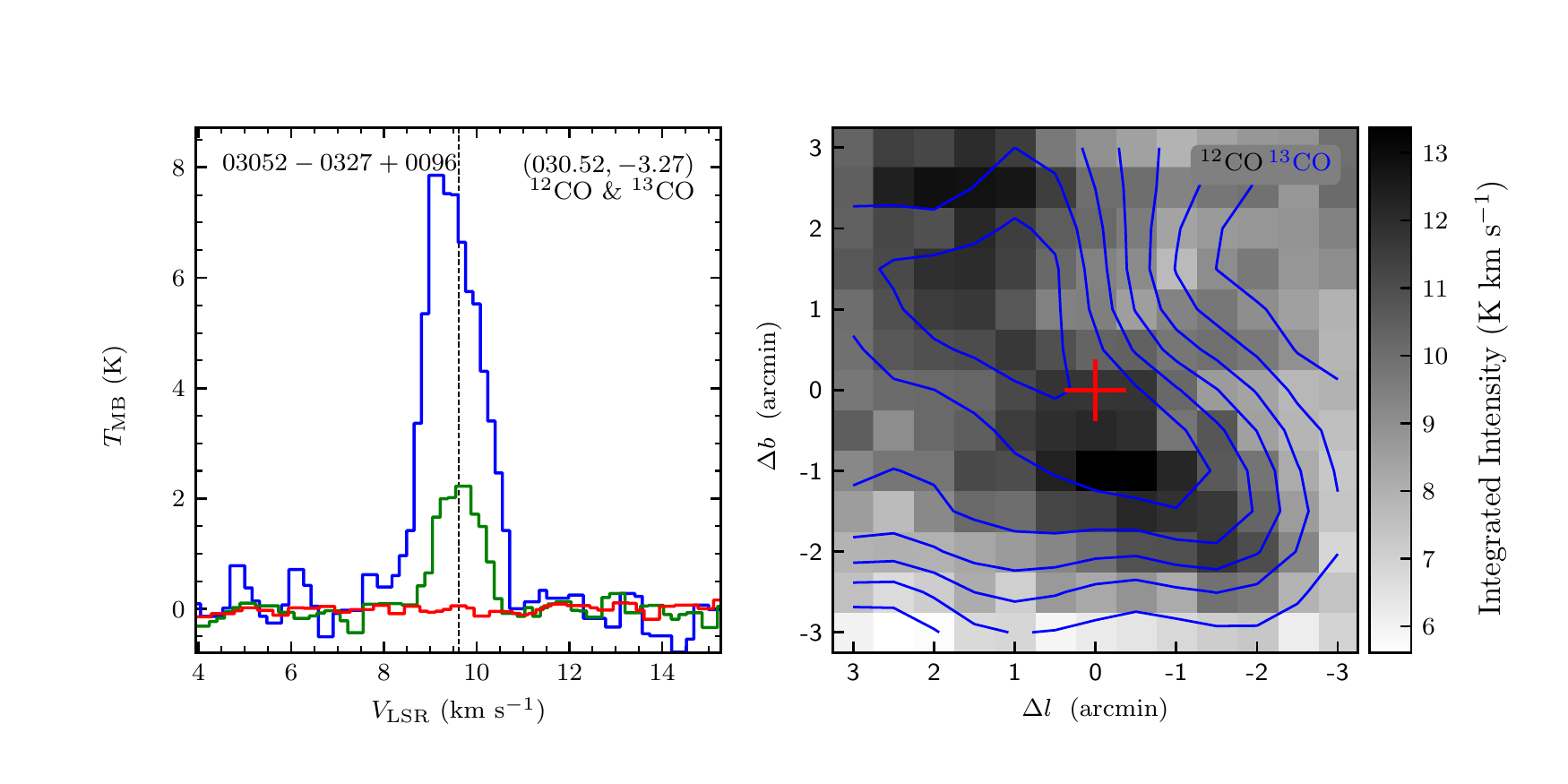}
\includegraphics[width=9.0cm,angle=0]{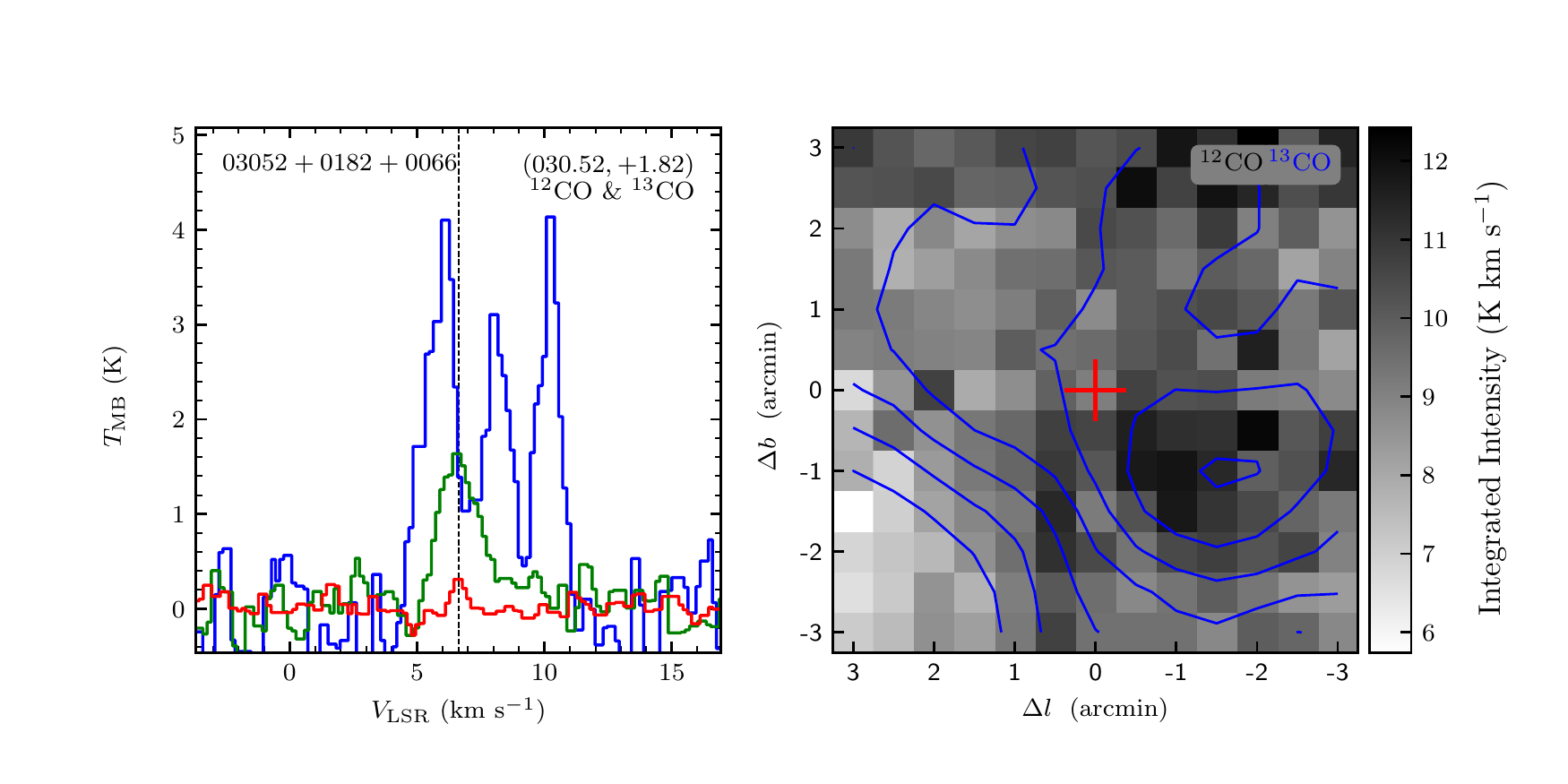}
\vspace{-0.5cm}

\includegraphics[width=9.0cm,angle=0]{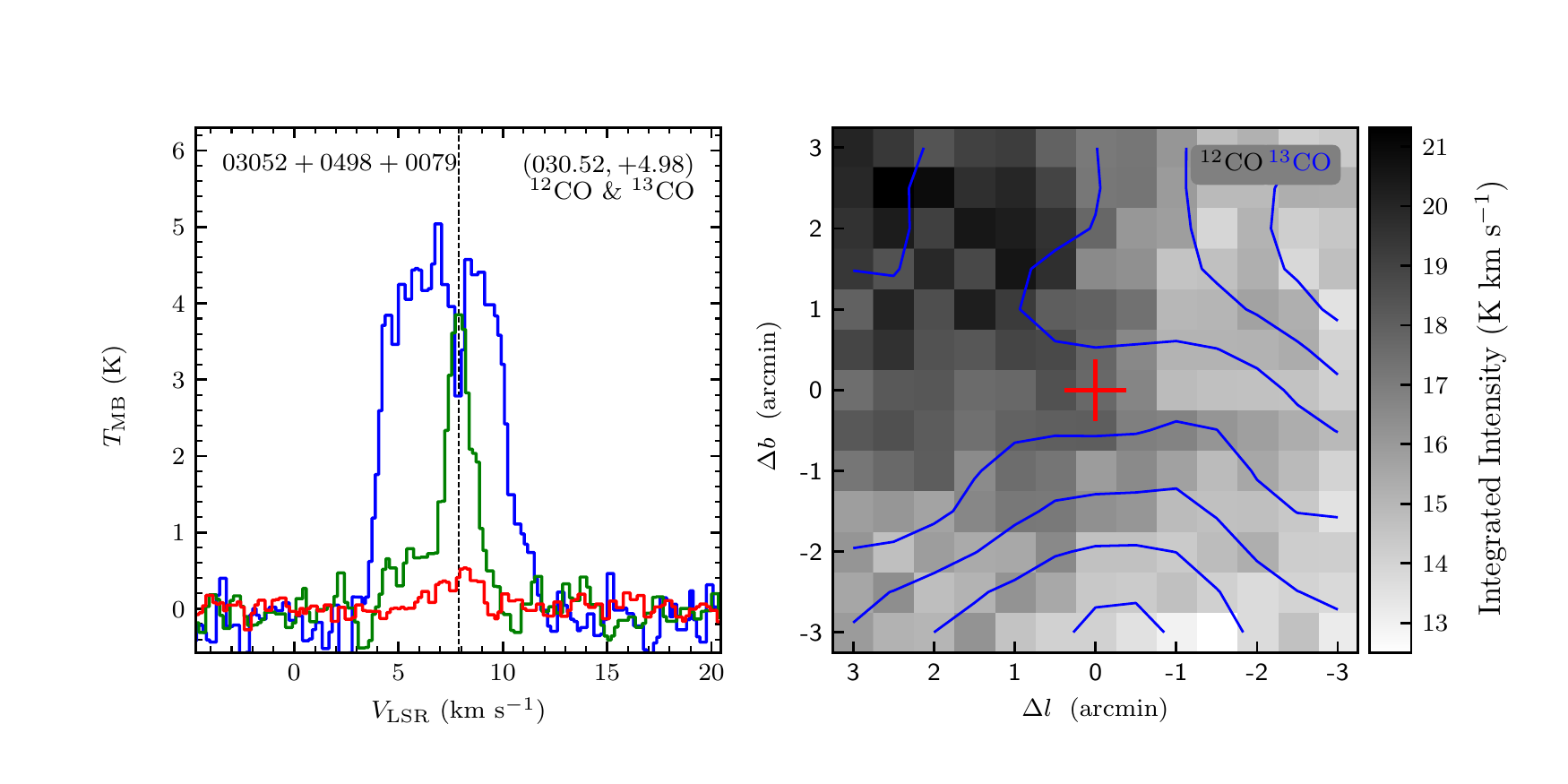}
\includegraphics[width=9.0cm,angle=0]{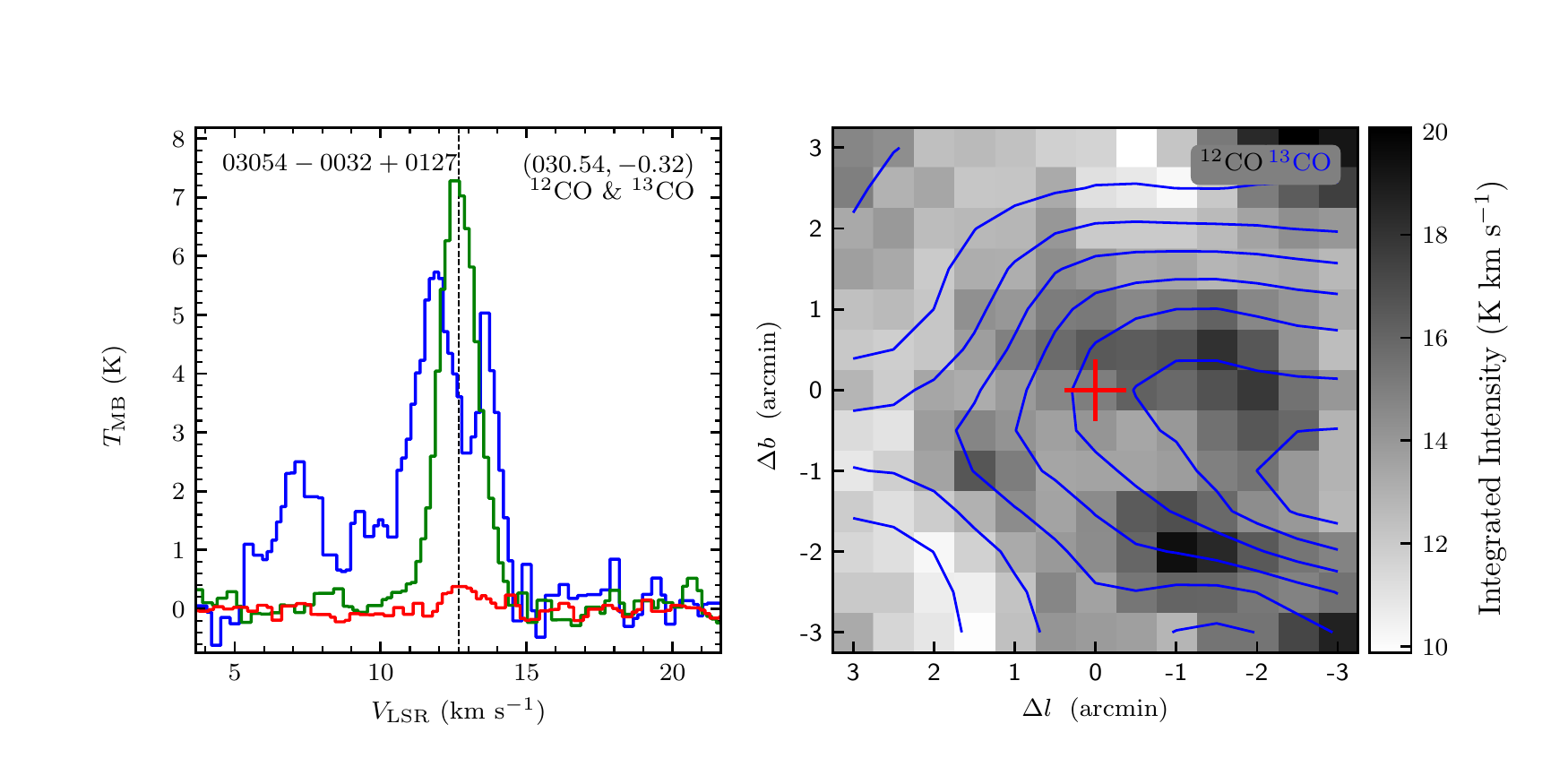}
\vspace{-0.5cm}

\includegraphics[width=9.0cm,angle=0]{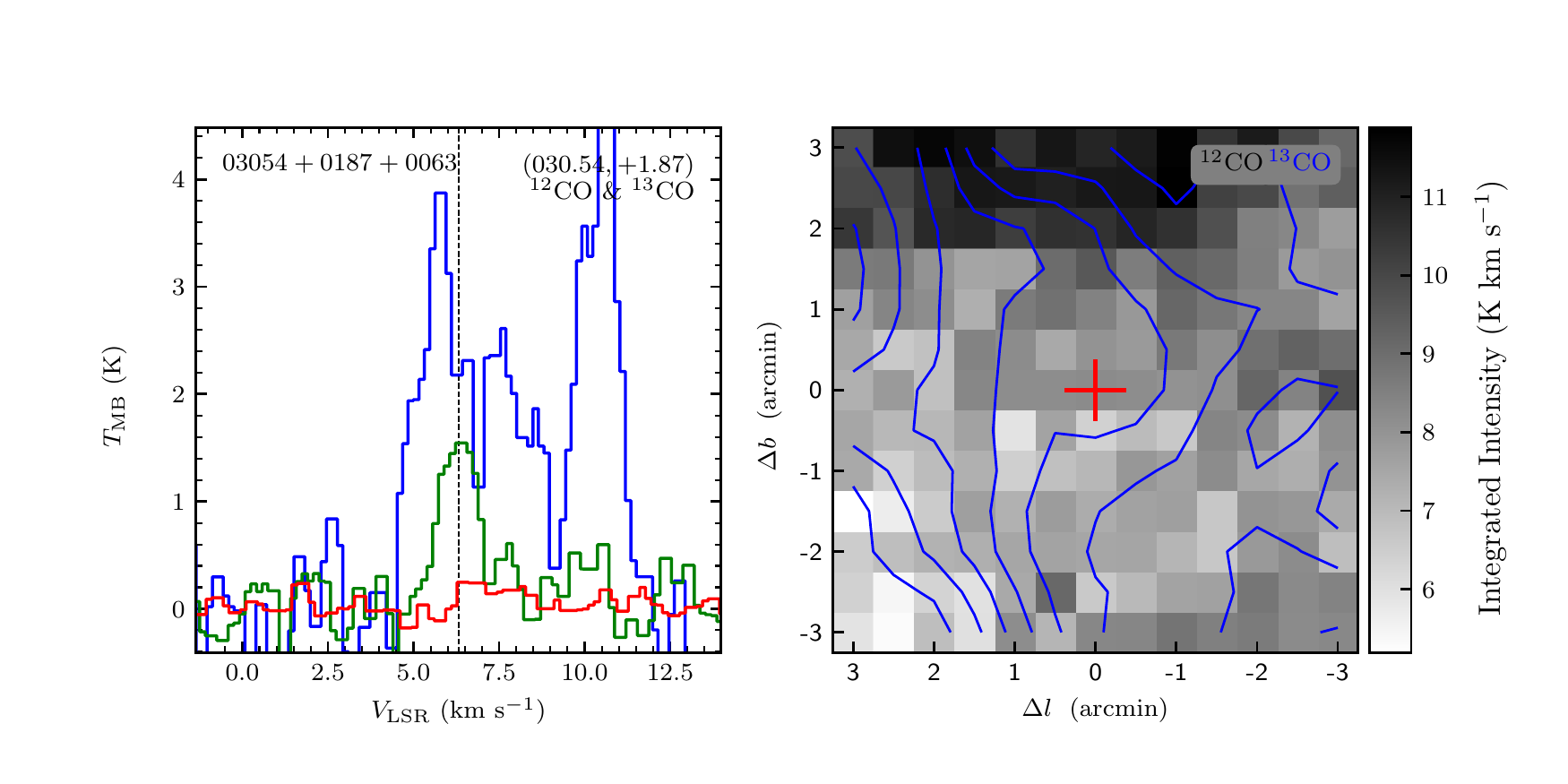}
\includegraphics[width=9.0cm,angle=0]{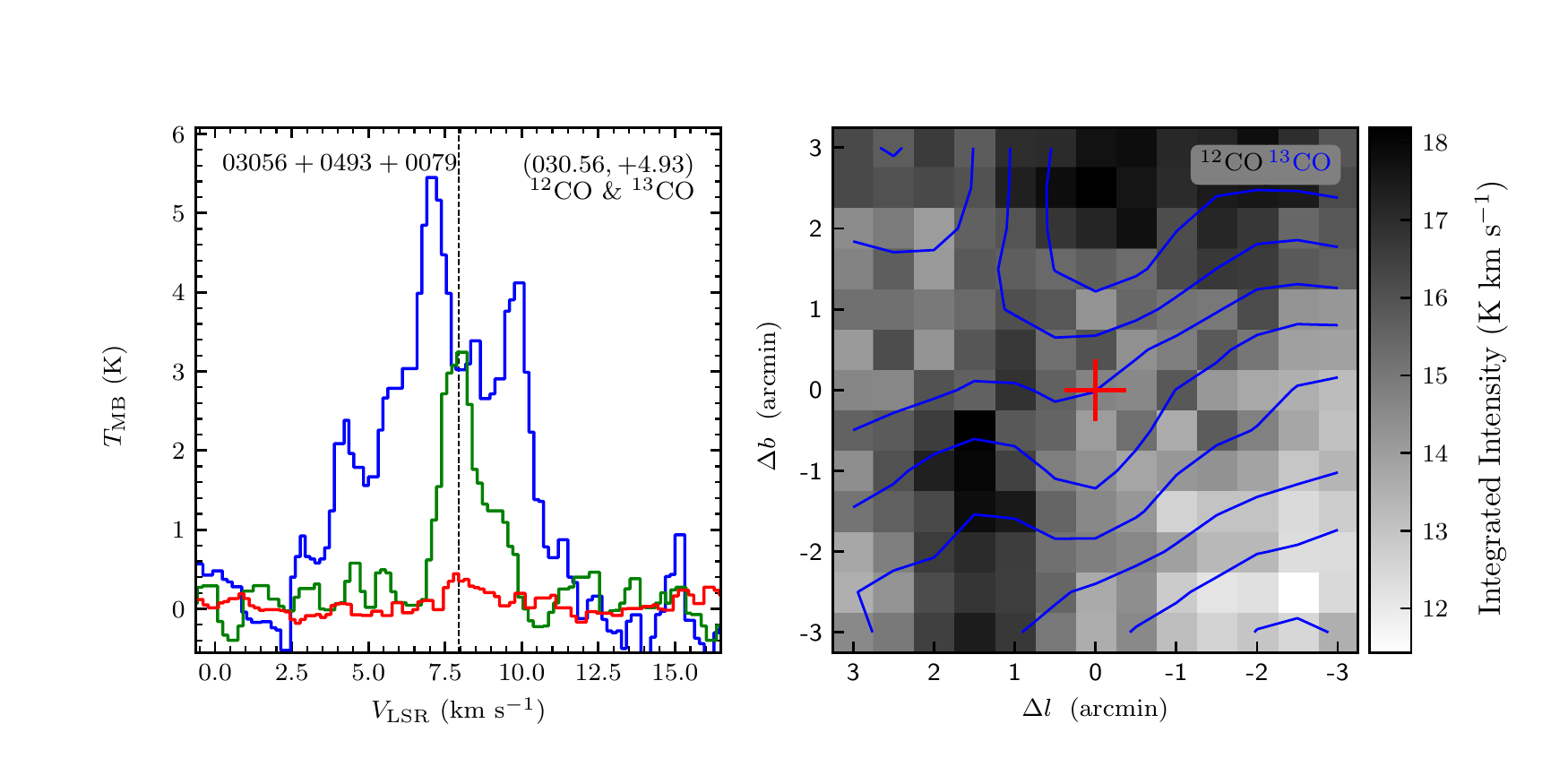}
\vspace{-0.5cm}

\includegraphics[width=9.0cm,angle=0]{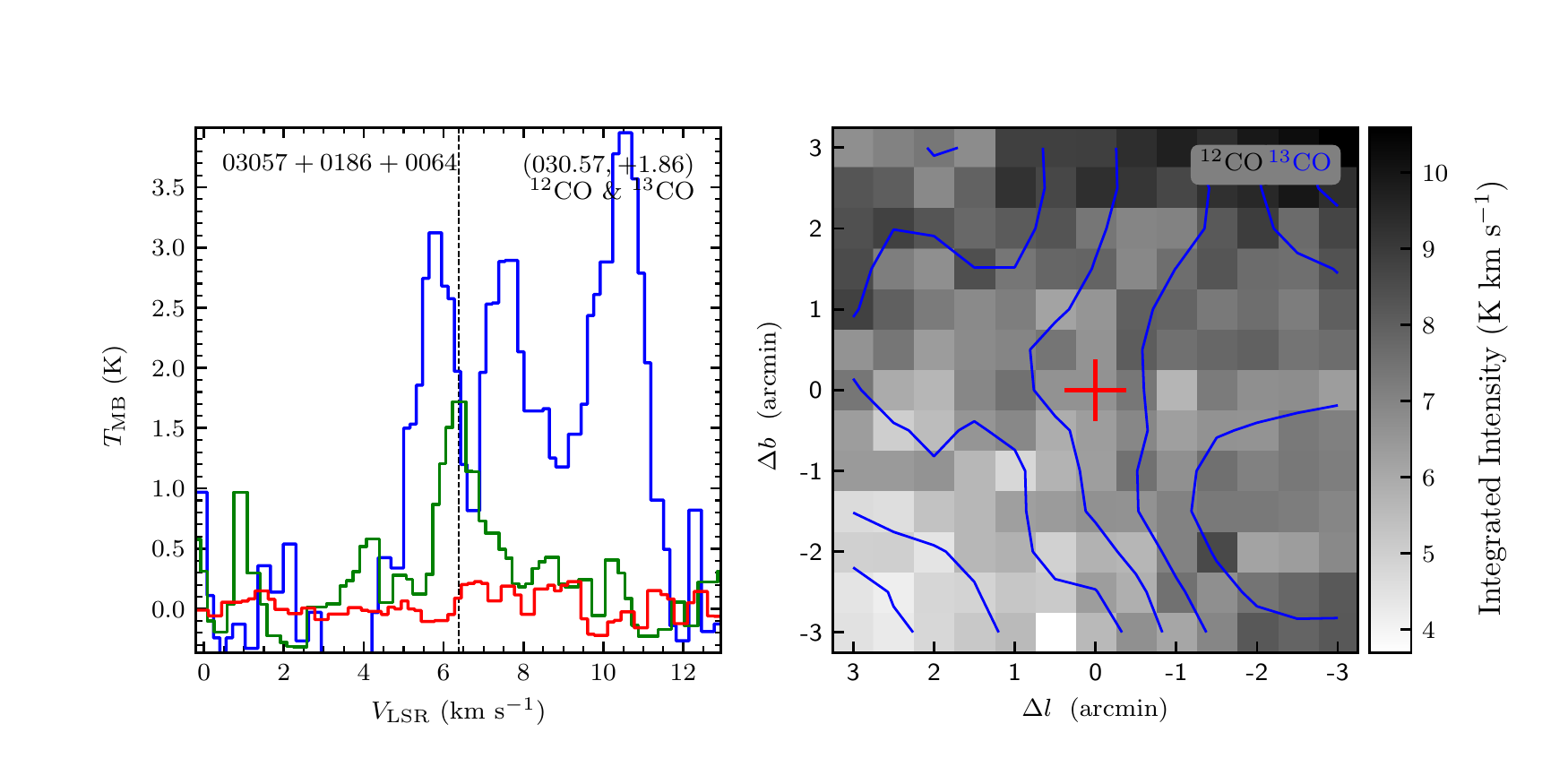}
\includegraphics[width=9.0cm,angle=0]{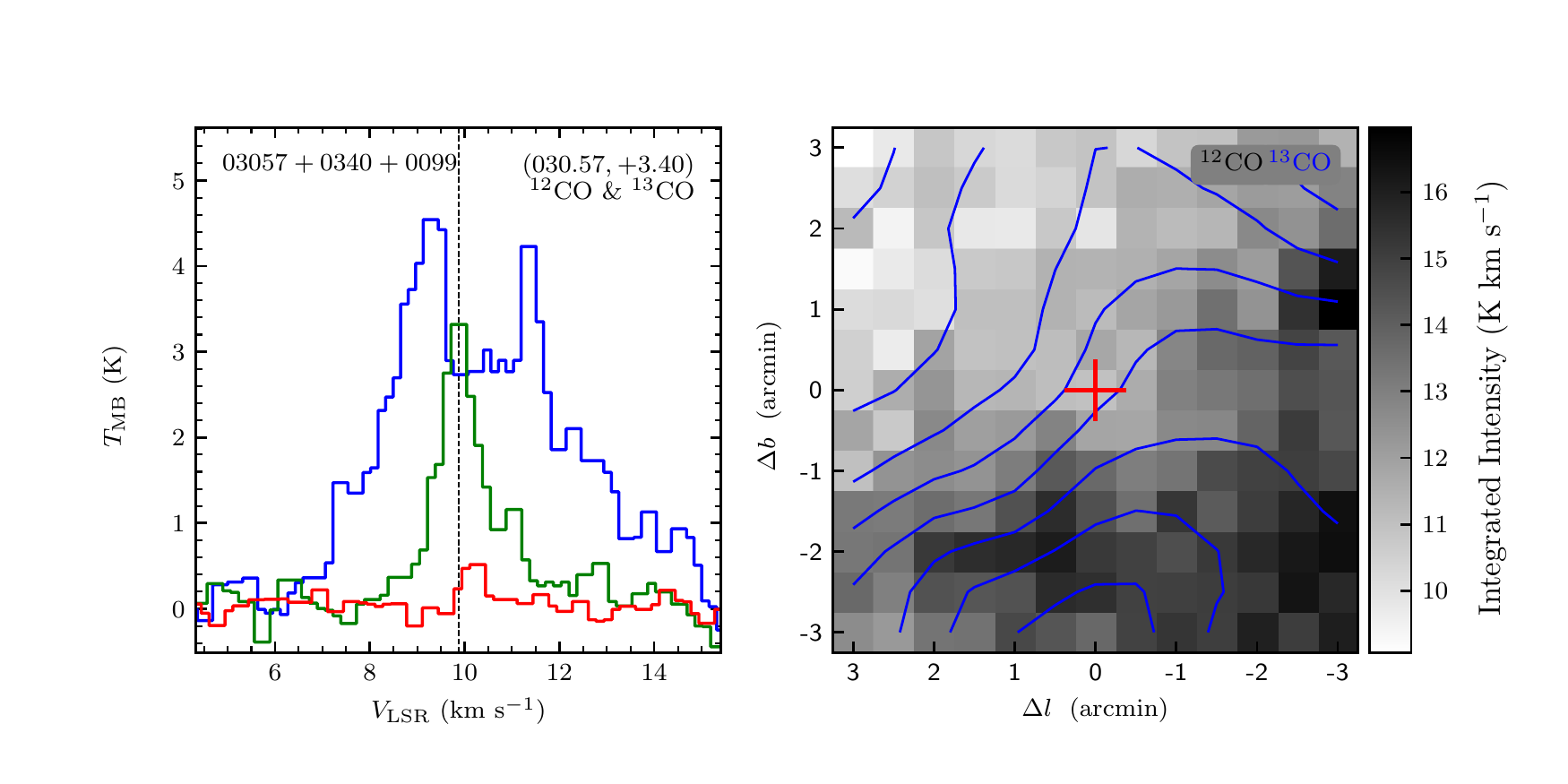}
\vspace{-0.5cm}

\includegraphics[width=9.0cm,angle=0]{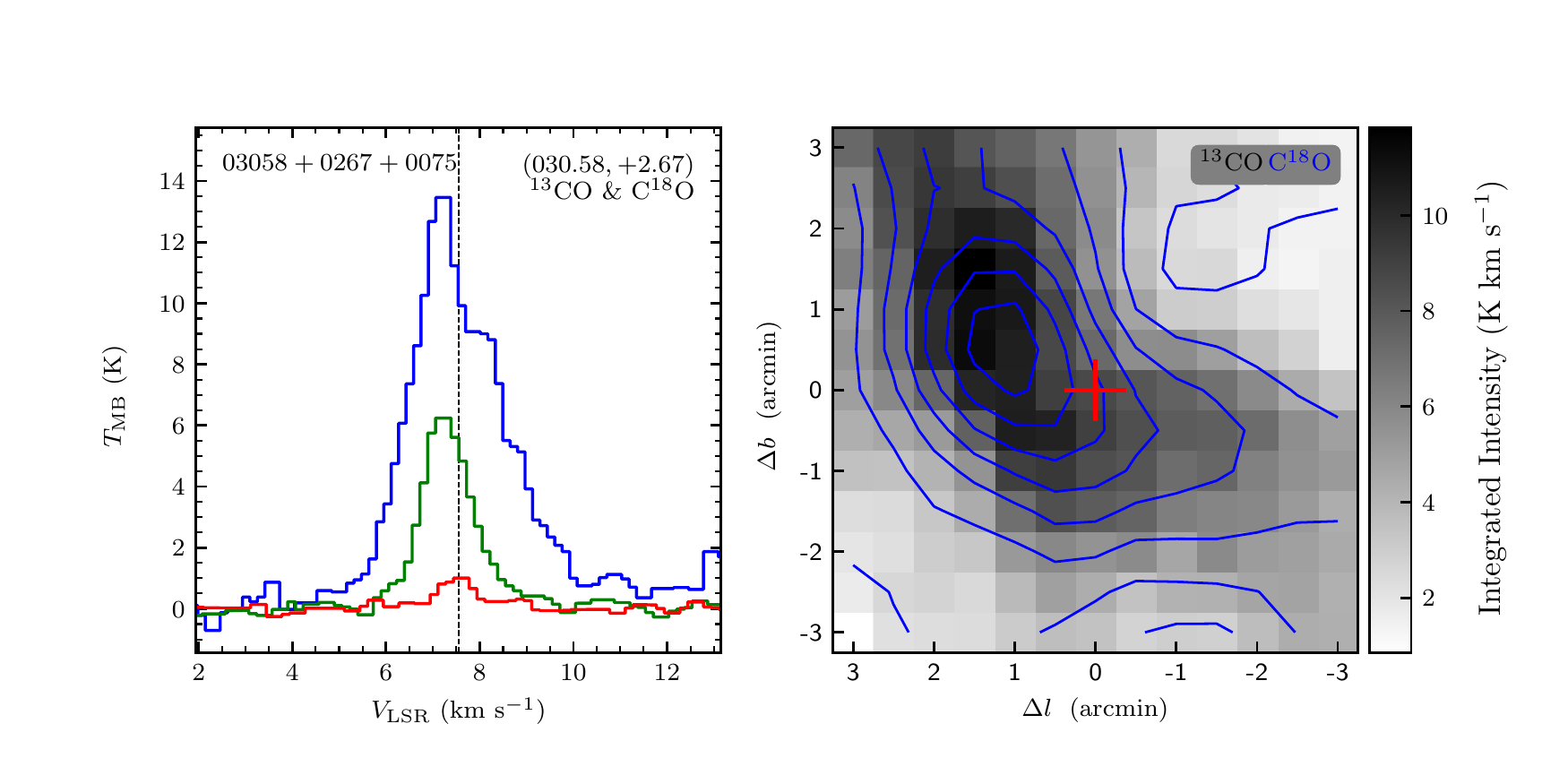}
\includegraphics[width=9.0cm,angle=0]{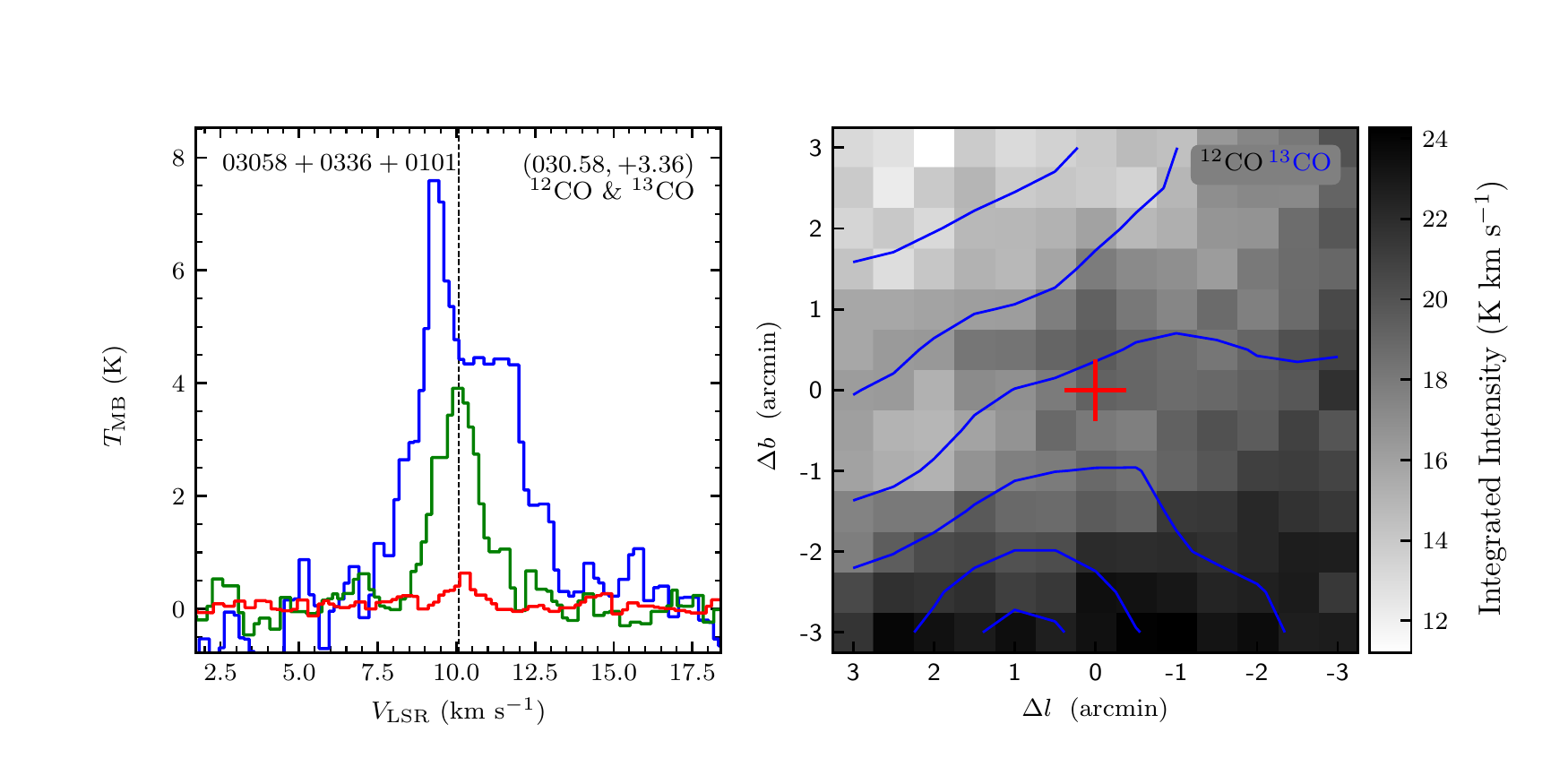}
\end{figure}
\clearpage

\begin{figure}
\includegraphics[width=9.0cm,angle=0]{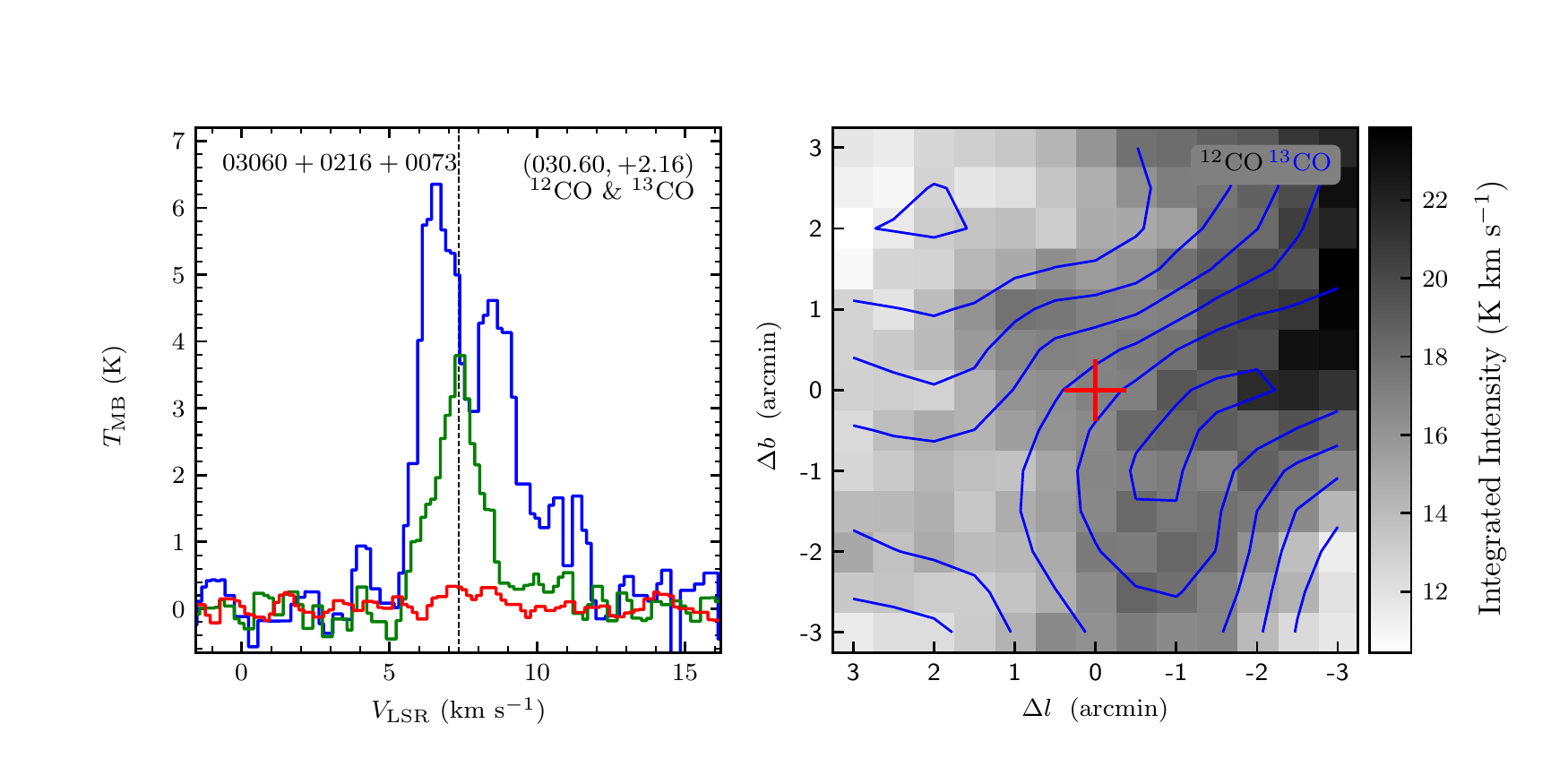}
\includegraphics[width=9.0cm,angle=0]{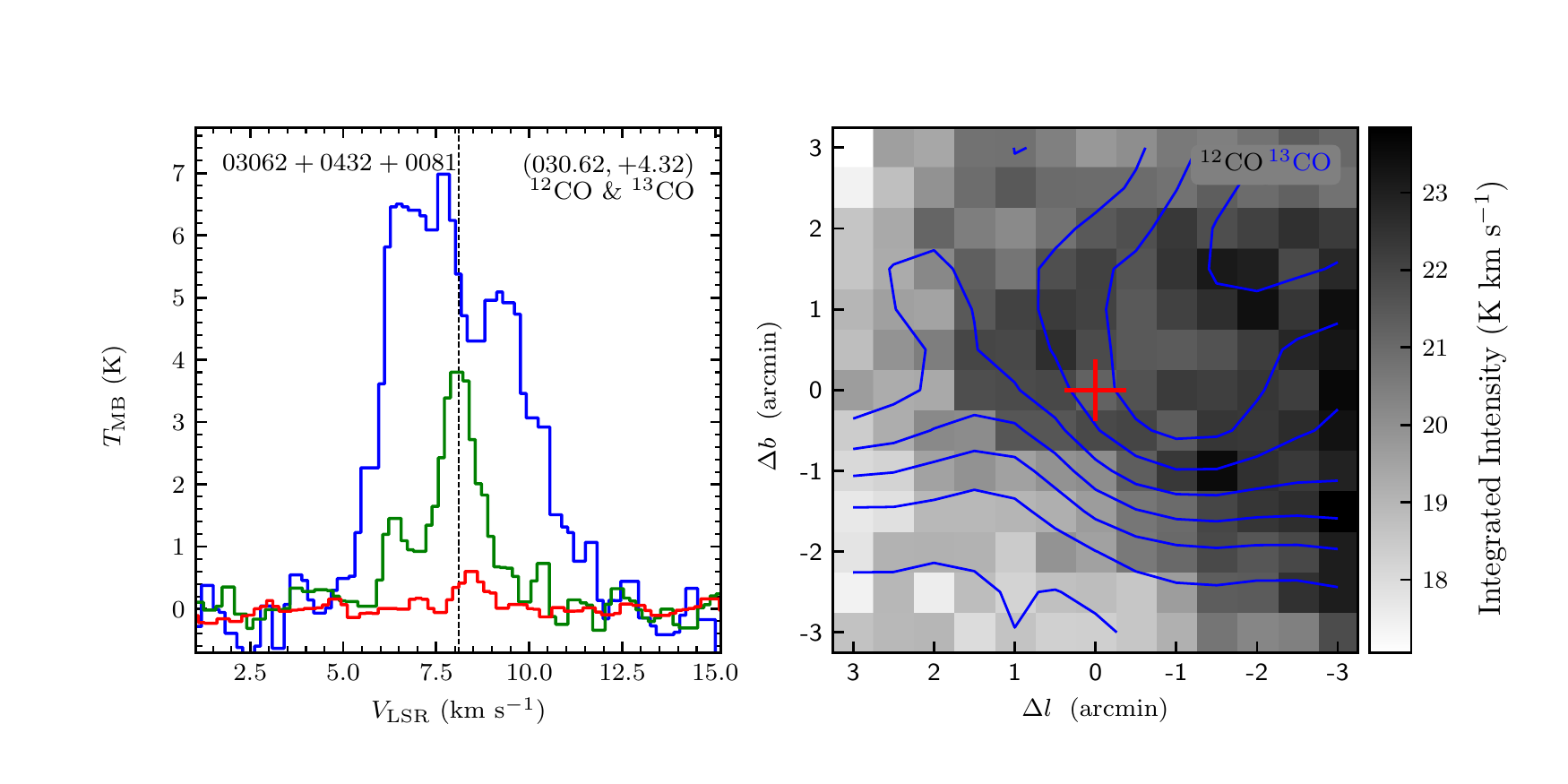}
\vspace{-0.5cm}

\includegraphics[width=9.0cm,angle=0]{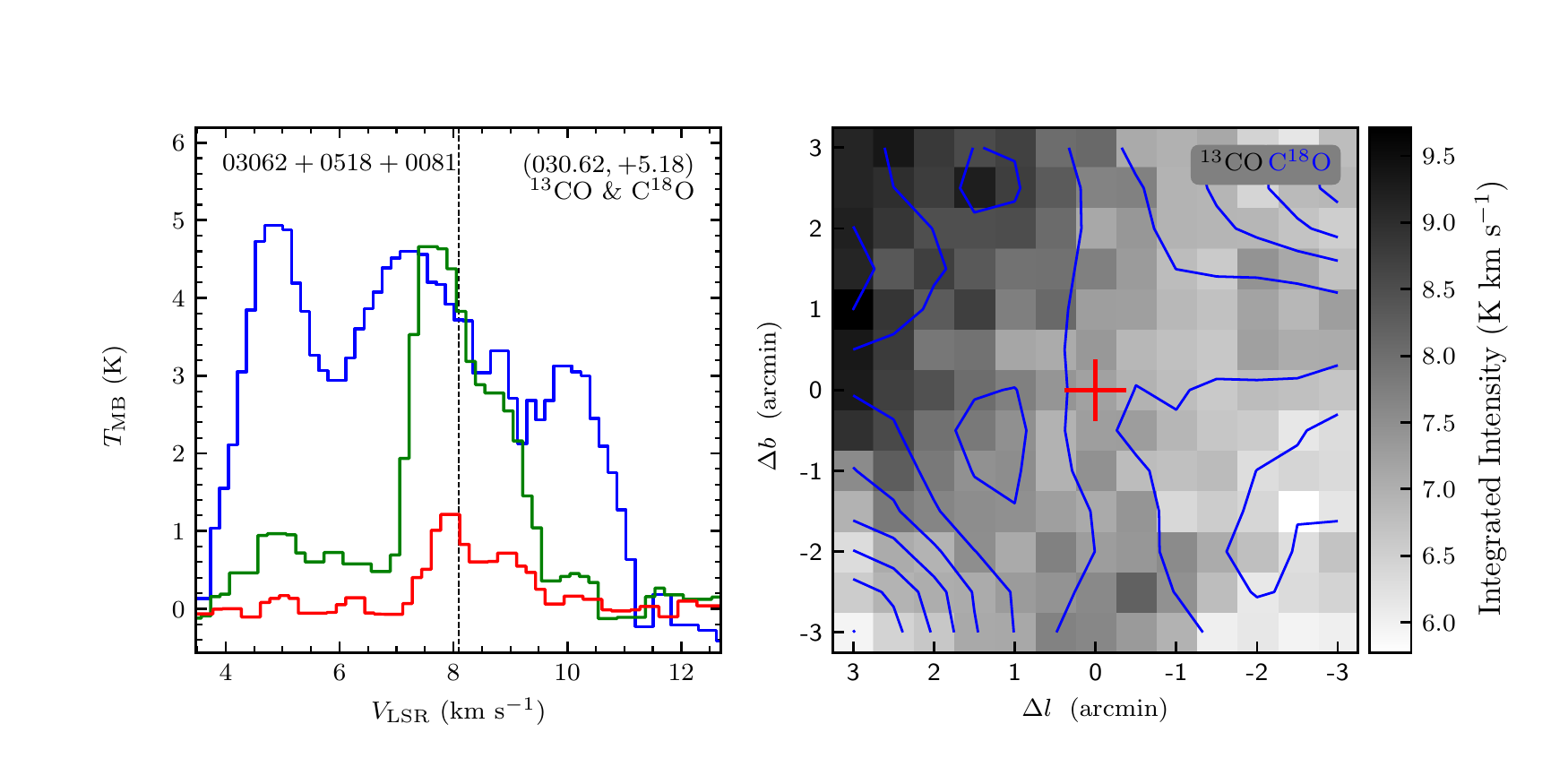}
\includegraphics[width=9.0cm,angle=0]{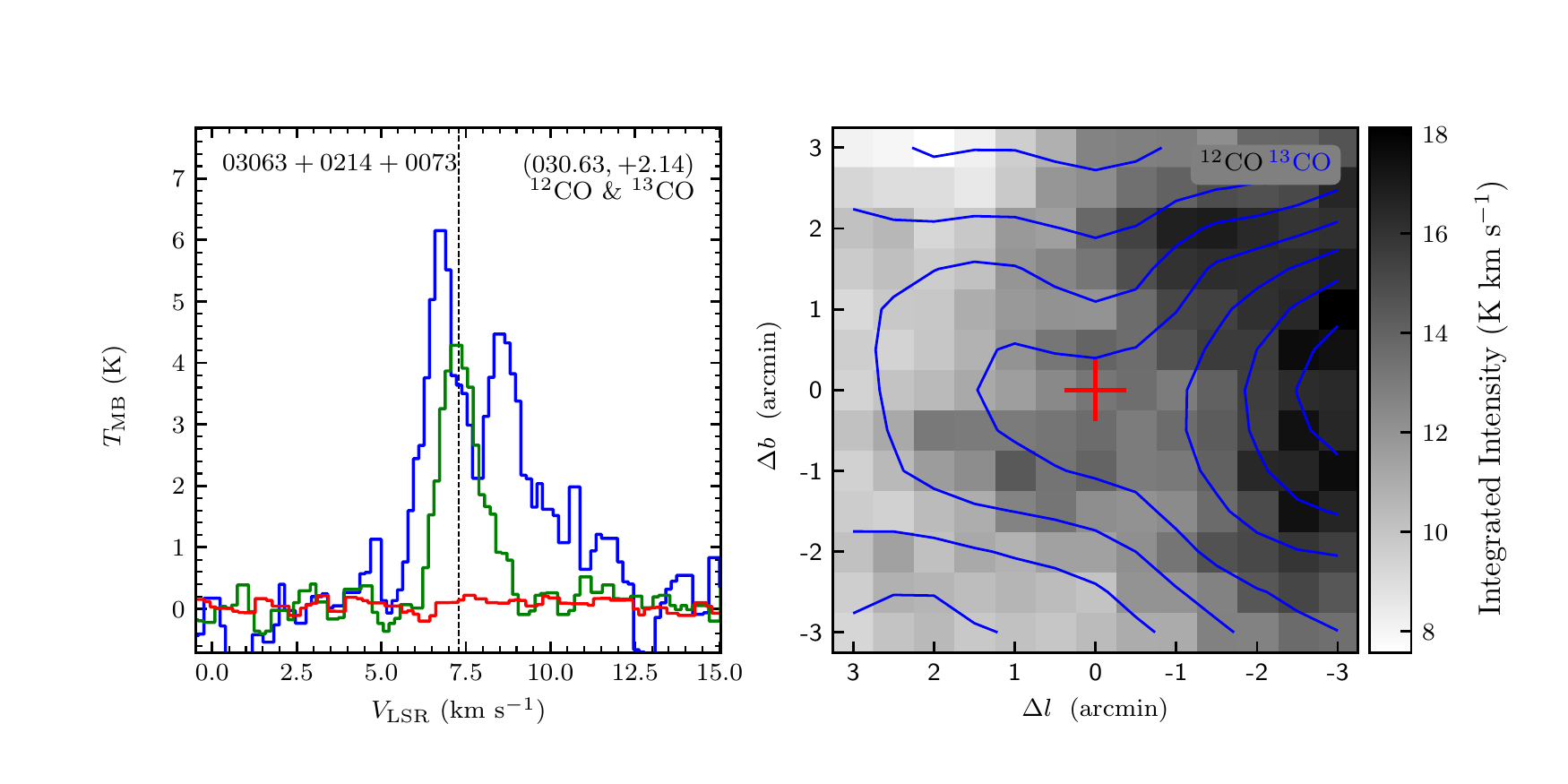}
\vspace{-0.5cm}

\includegraphics[width=9.0cm,angle=0]{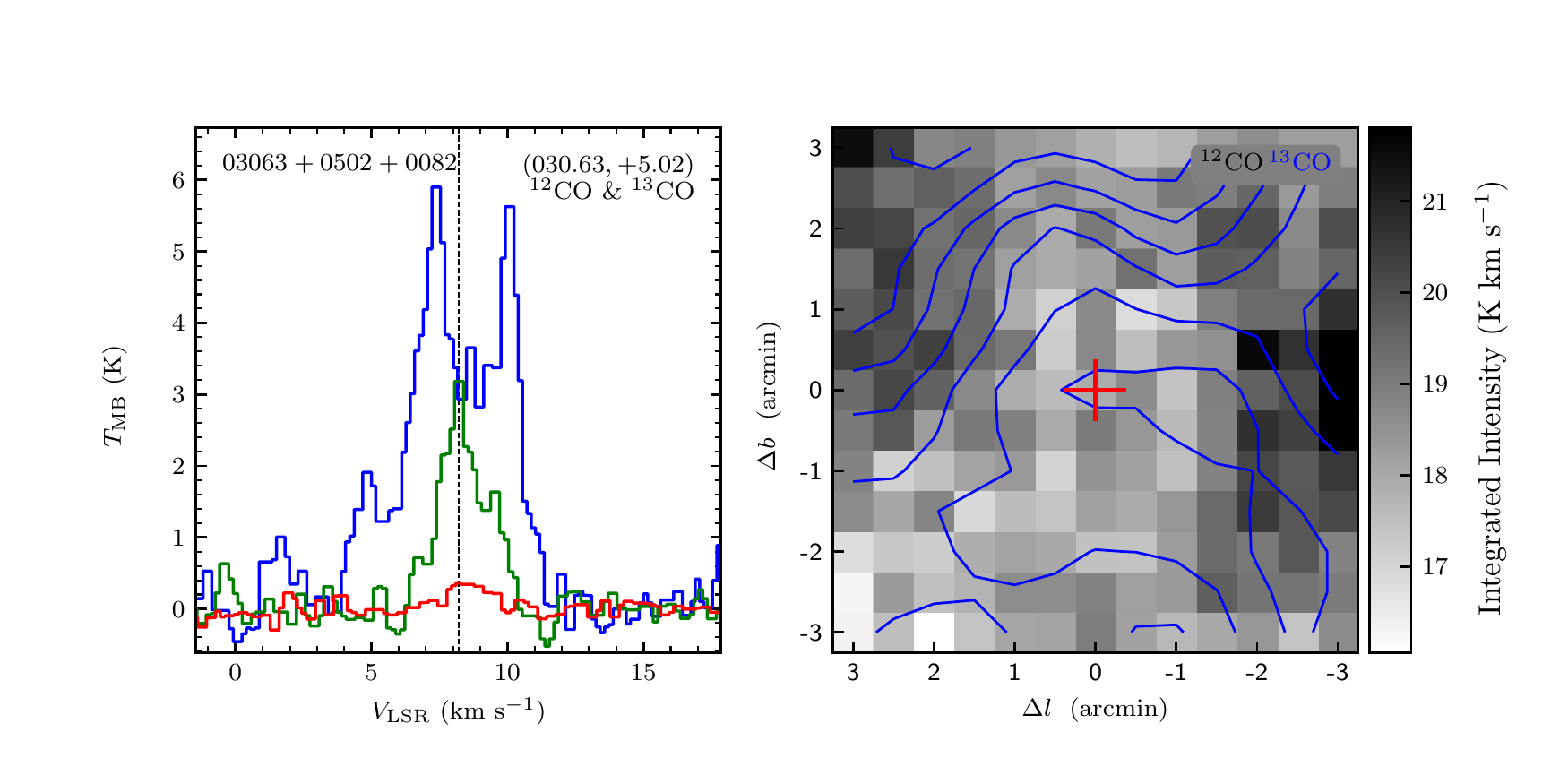}
\includegraphics[width=9.0cm,angle=0]{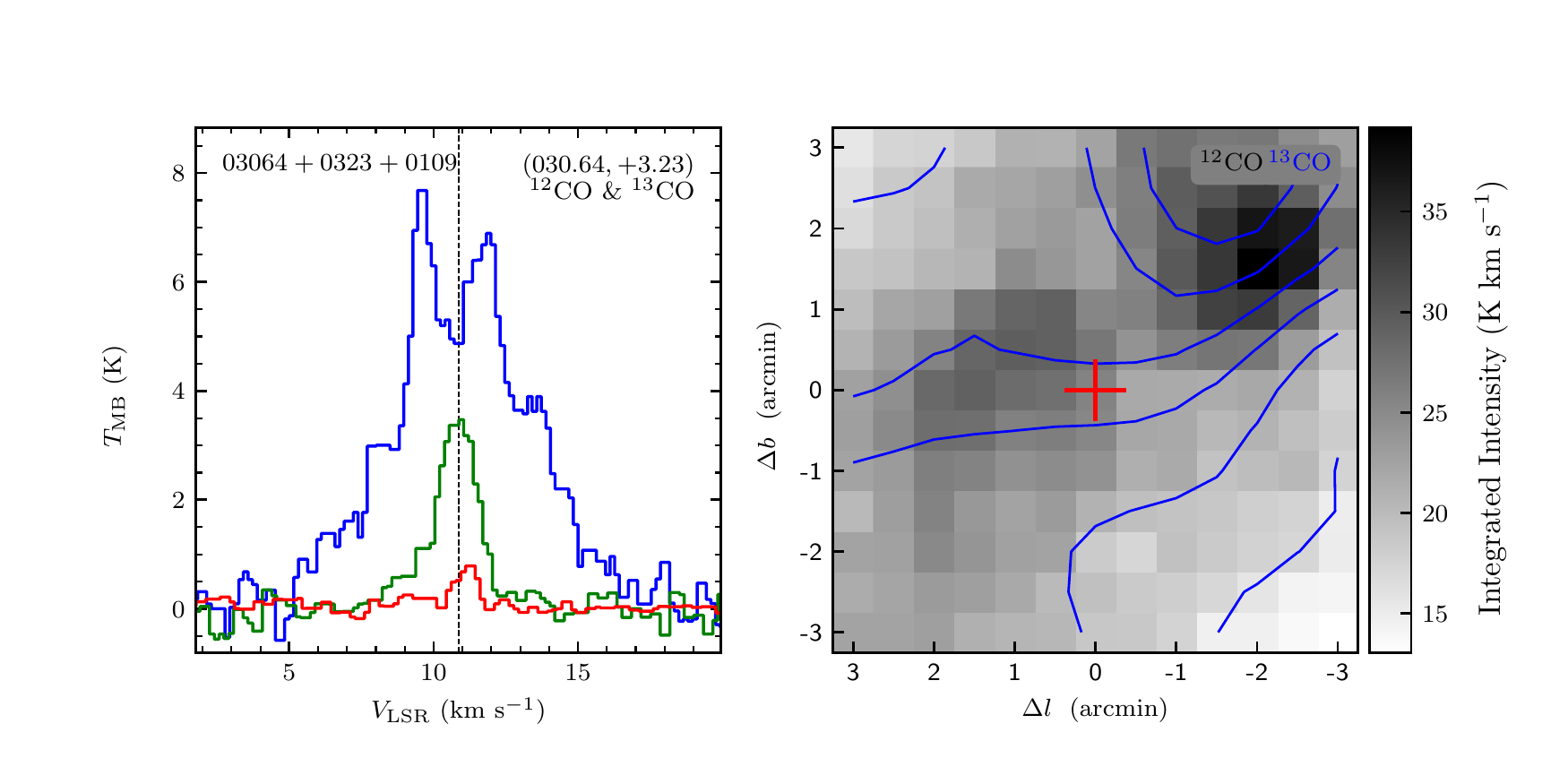}
\vspace{-0.5cm}

\includegraphics[width=9.0cm,angle=0]{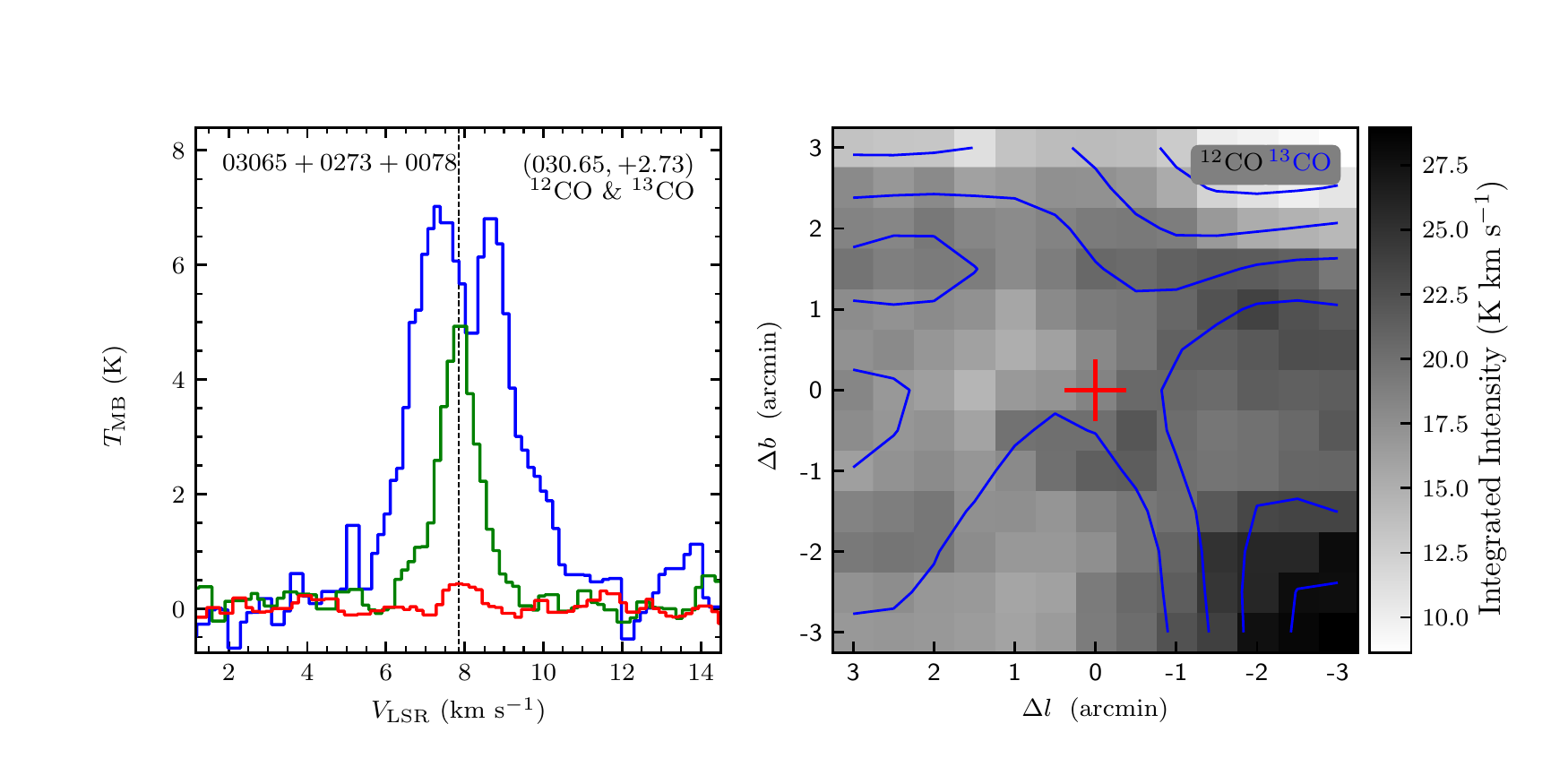}
\includegraphics[width=9.0cm,angle=0]{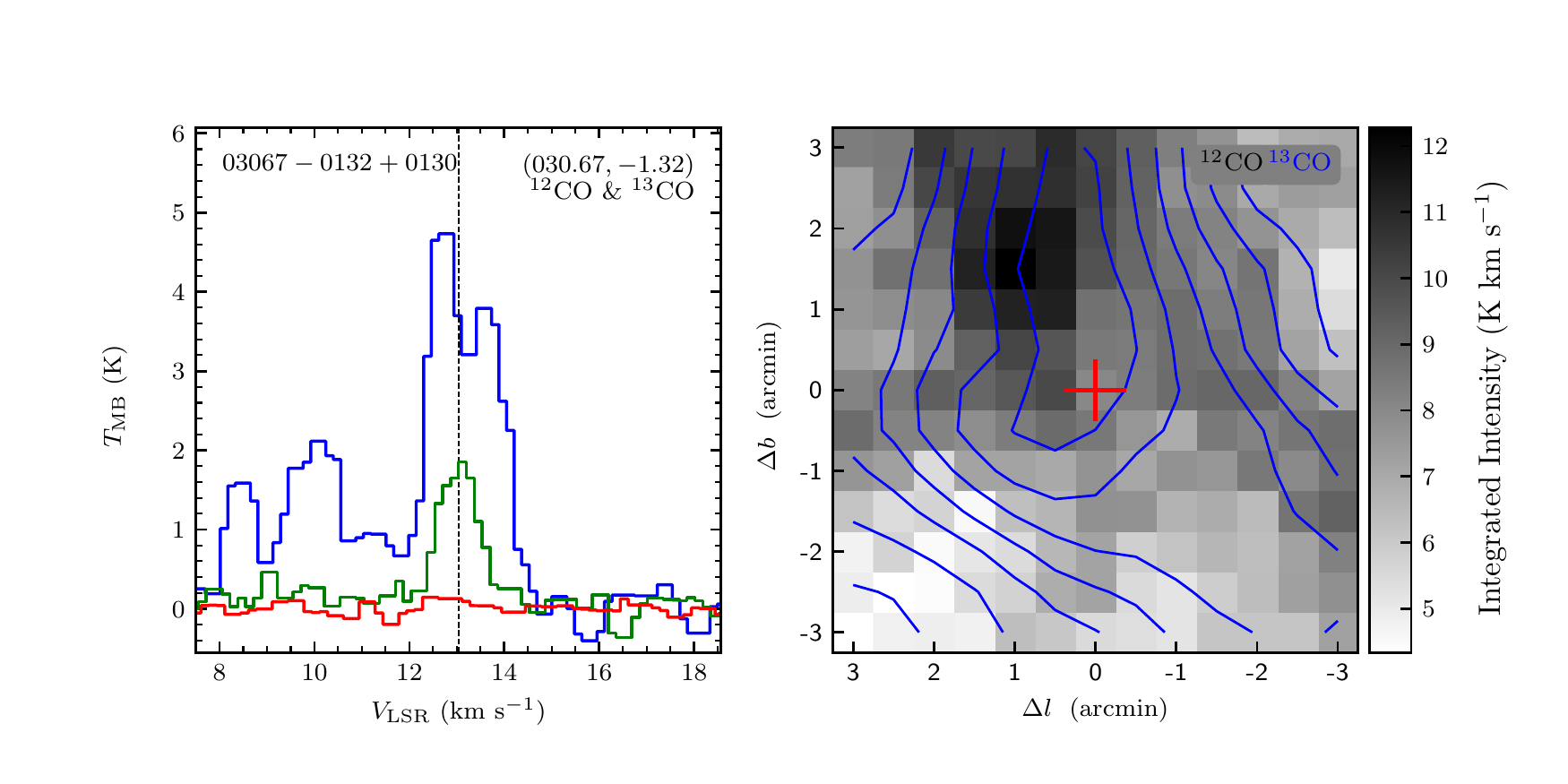}
\vspace{-0.5cm}

\includegraphics[width=9.0cm,angle=0]{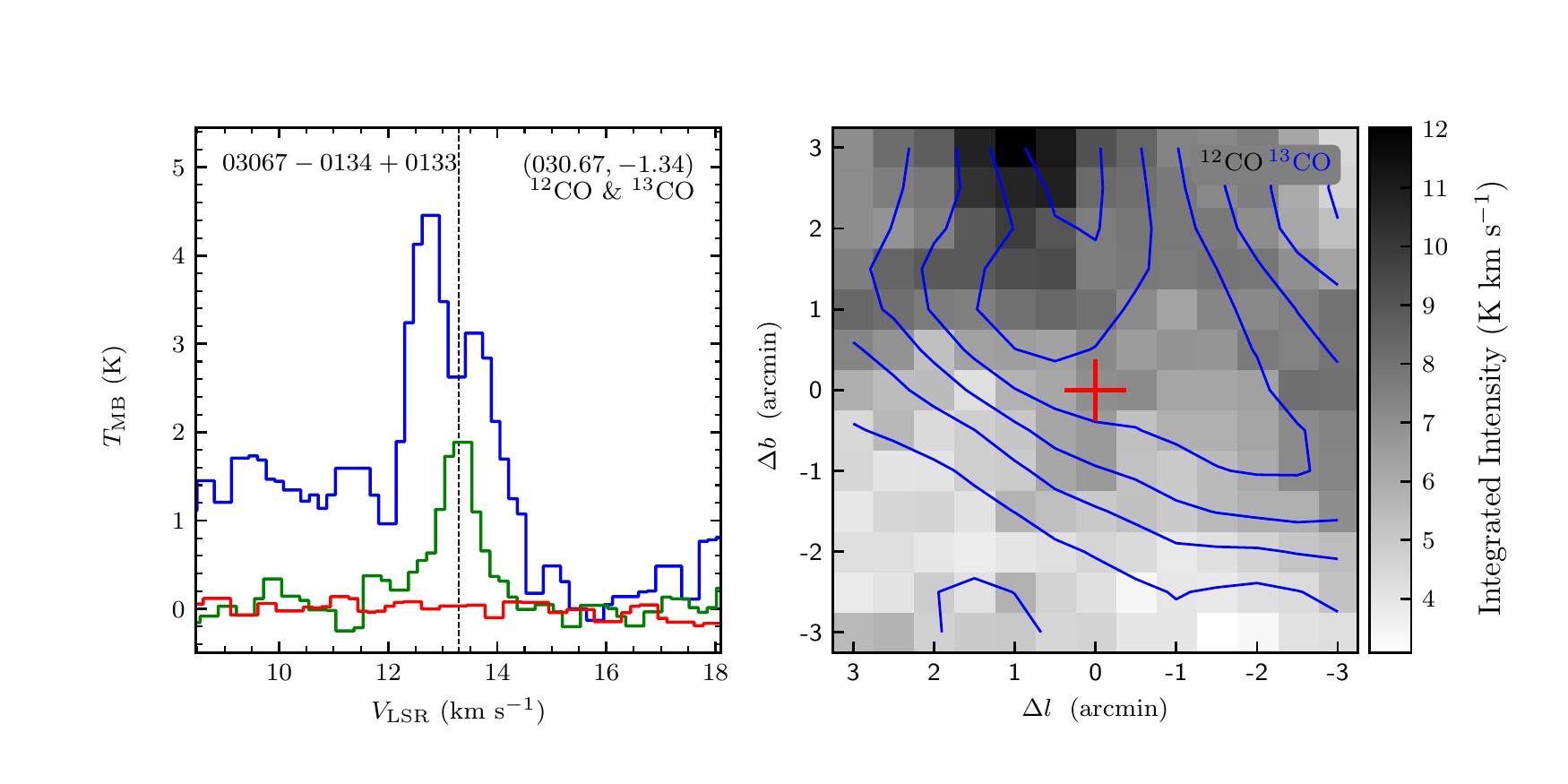}
\includegraphics[width=9.0cm,angle=0]{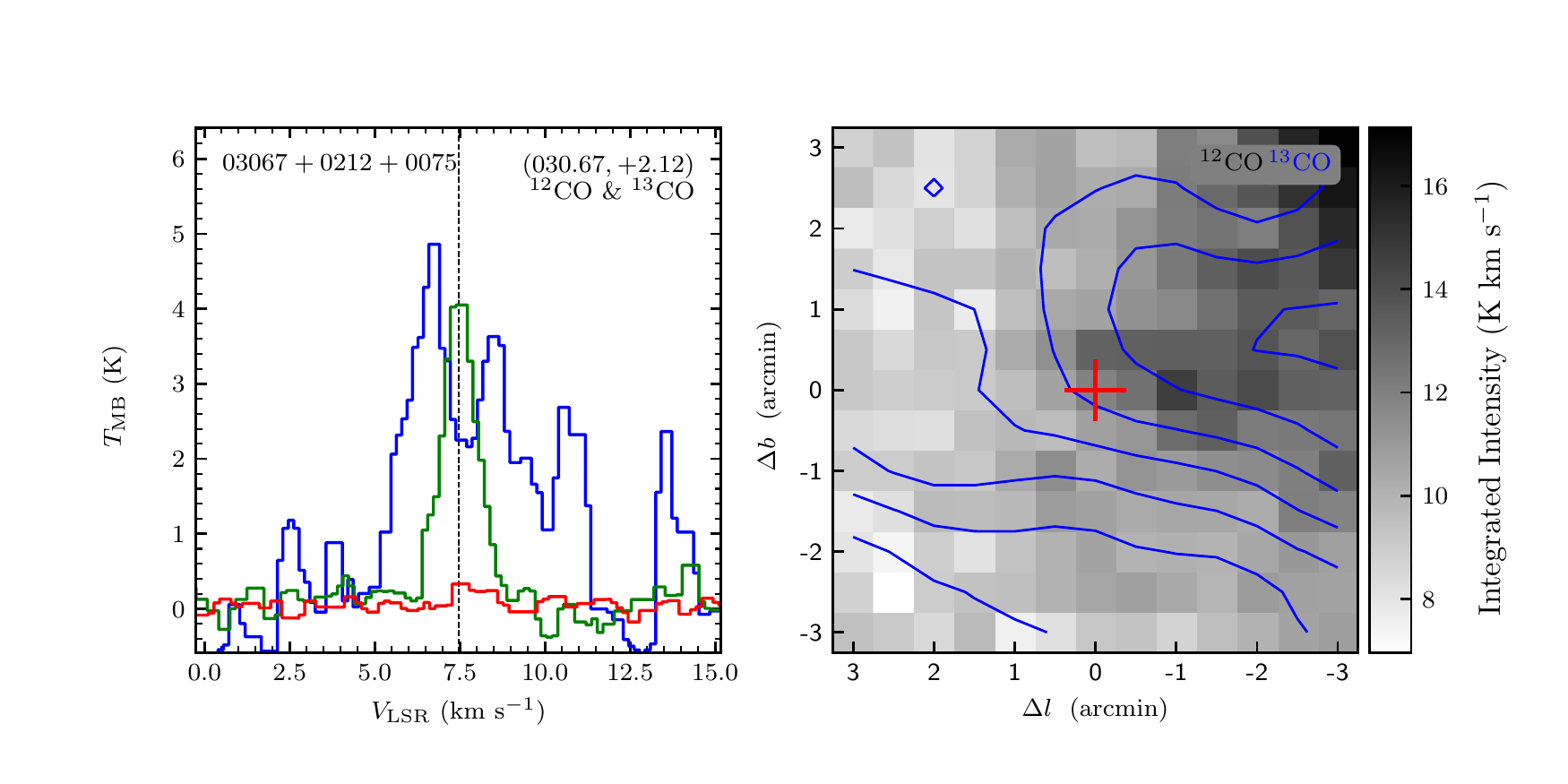}
\end{figure}
\clearpage

\begin{figure}
\includegraphics[width=9.0cm,angle=0]{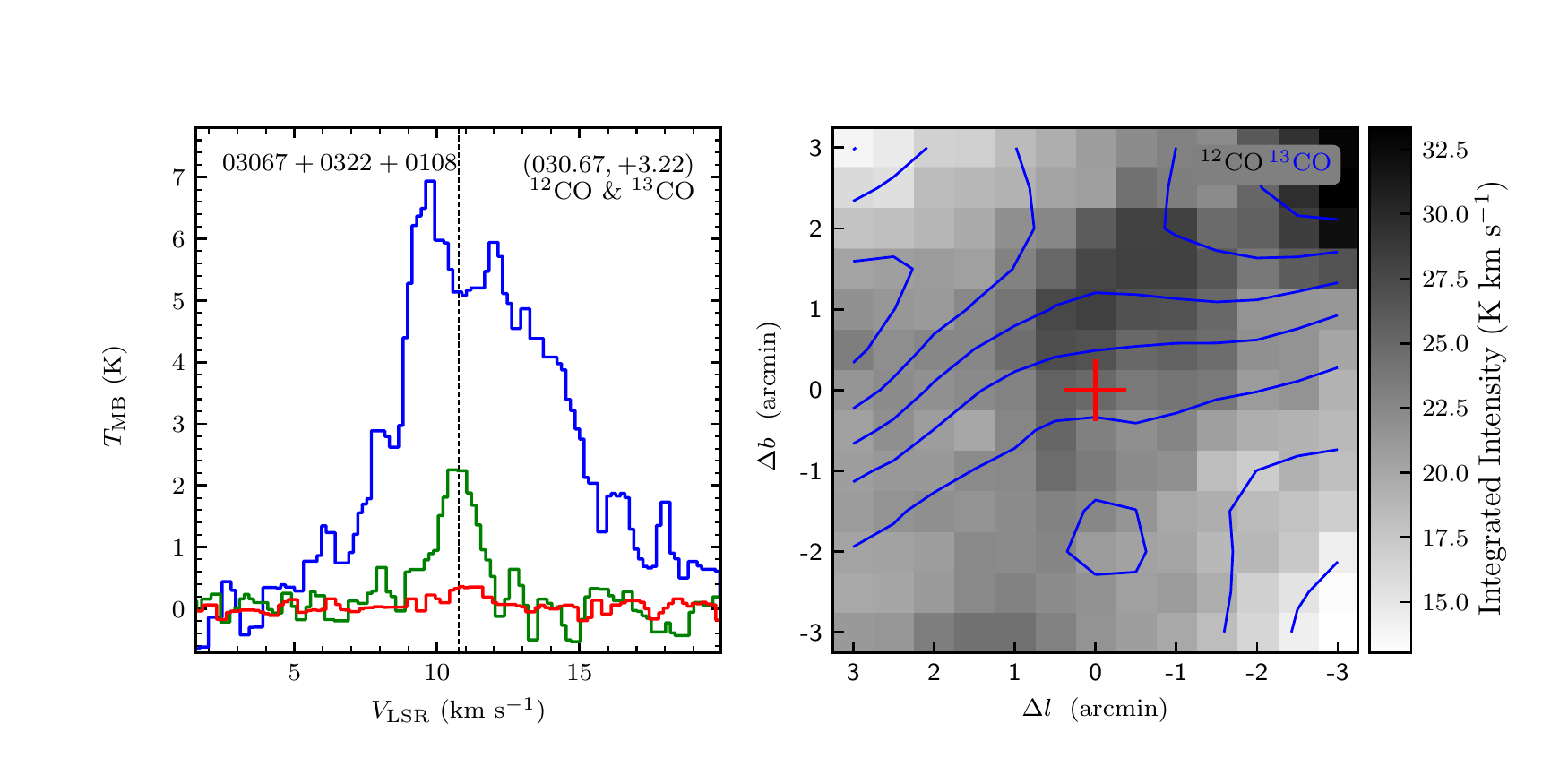}
\includegraphics[width=9.0cm,angle=0]{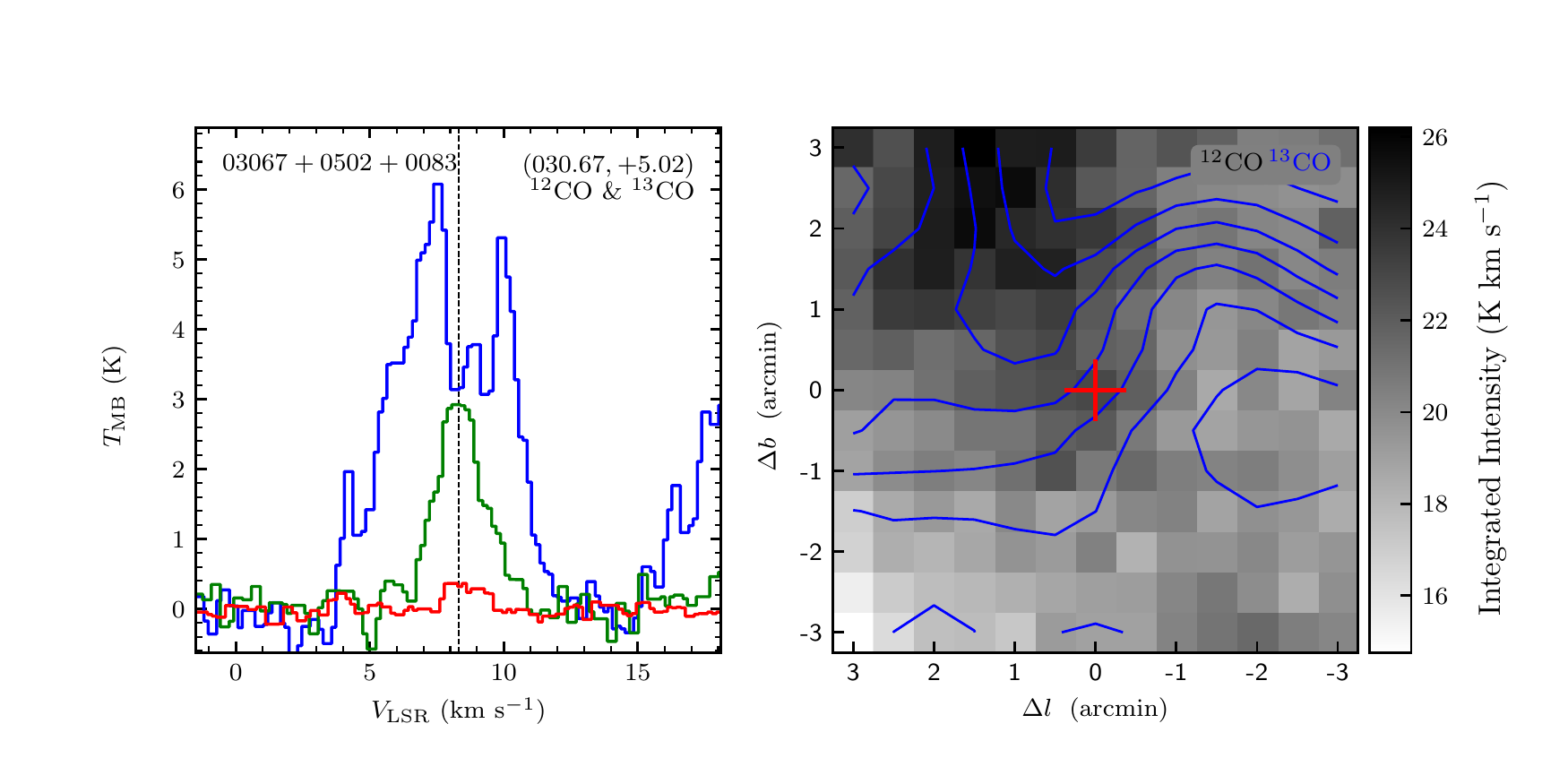}
\vspace{-0.5cm}

\includegraphics[width=9.0cm,angle=0]{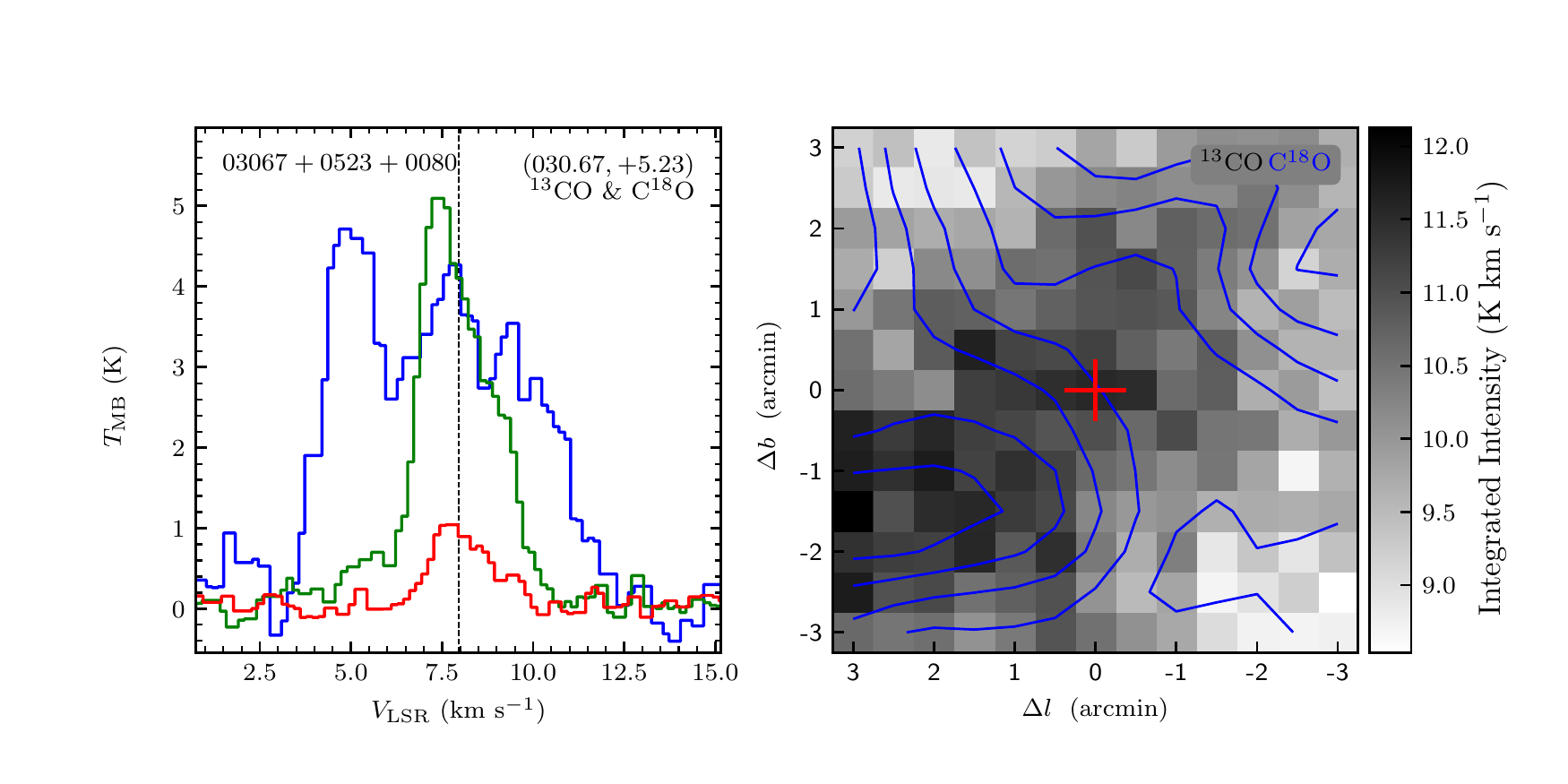}
\includegraphics[width=9.0cm,angle=0]{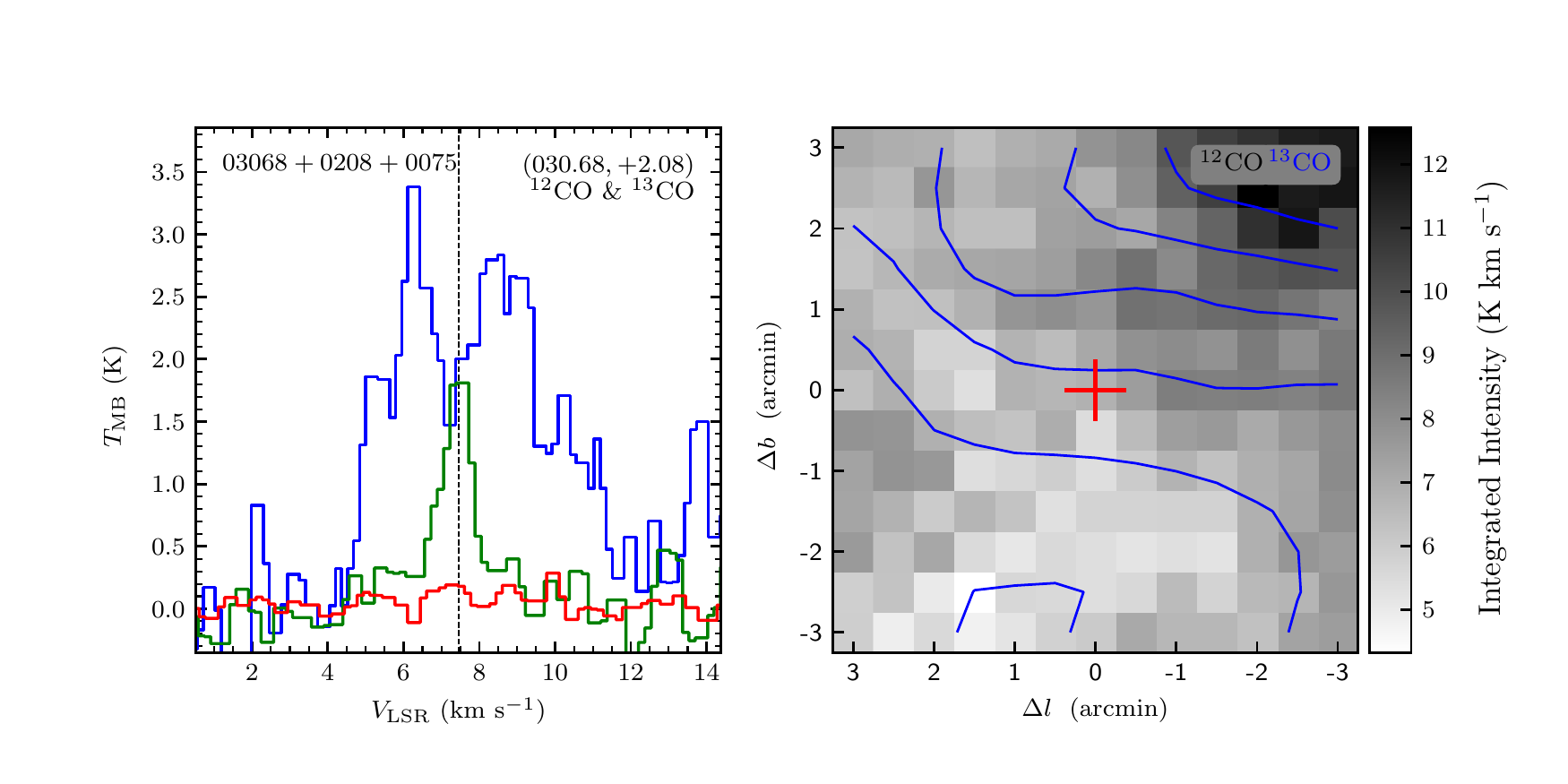}
\vspace{-0.5cm}

\includegraphics[width=9.0cm,angle=0]{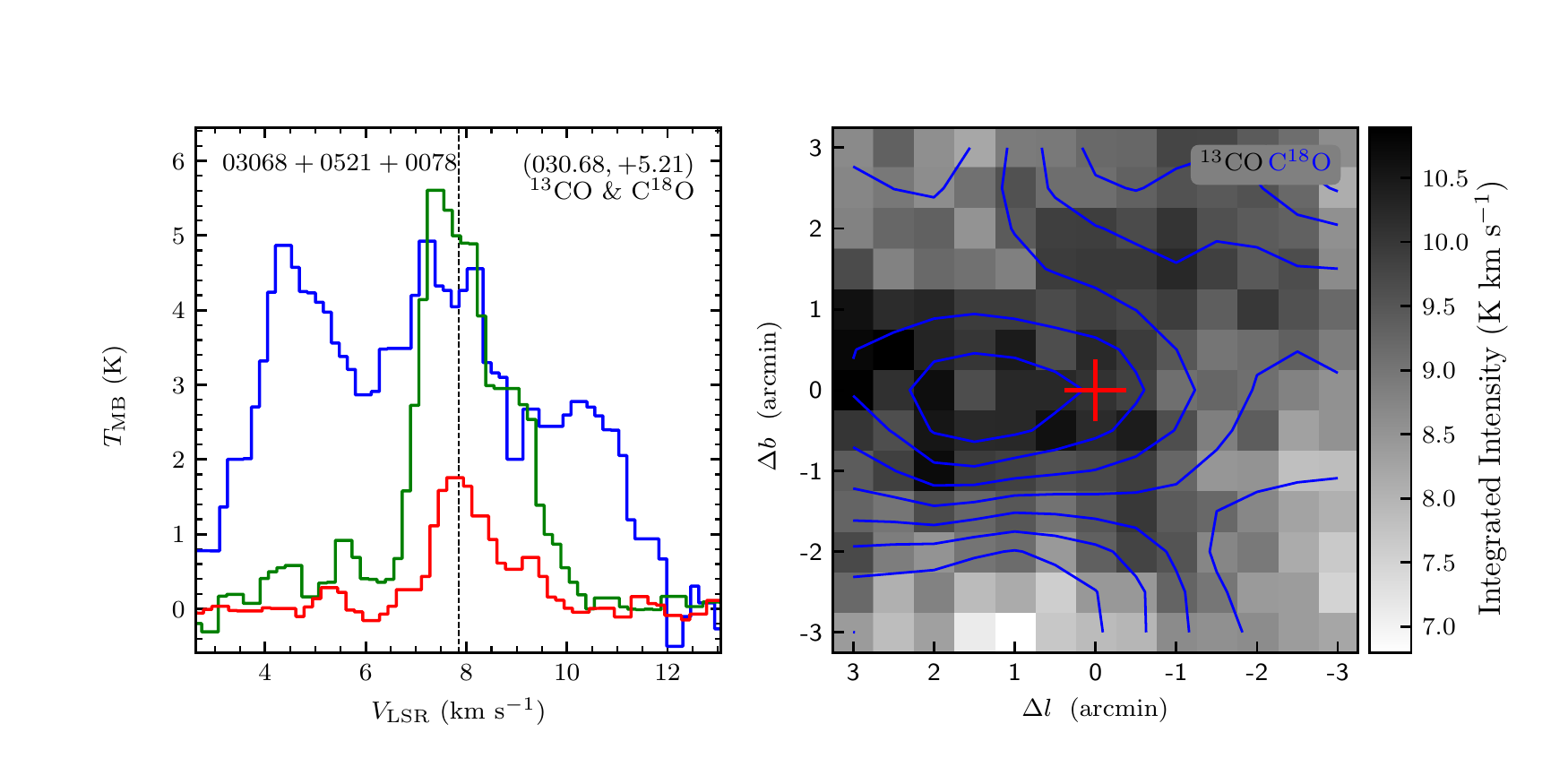}
\includegraphics[width=9.0cm,angle=0]{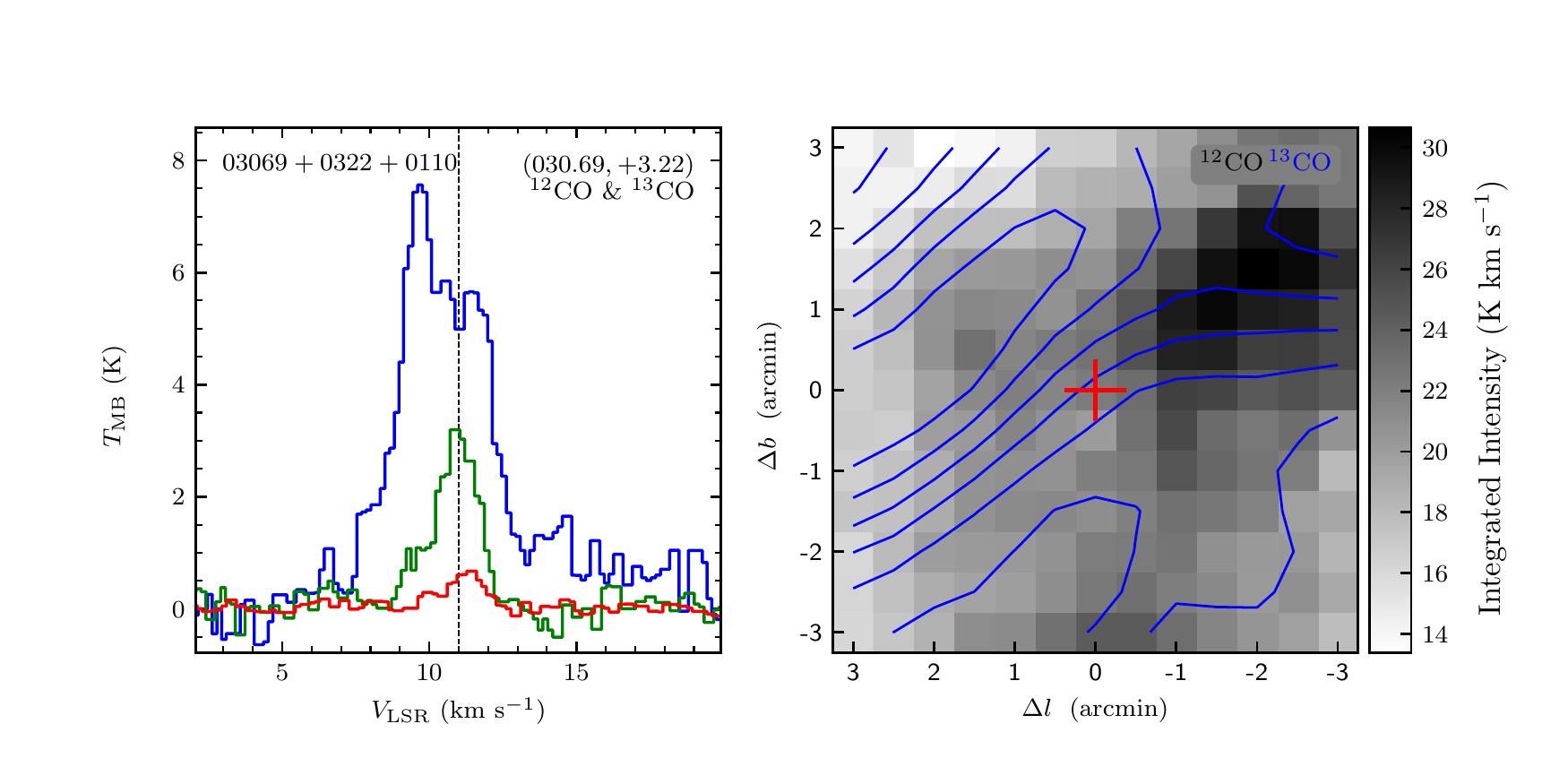}
\vspace{-0.5cm}

\includegraphics[width=9.0cm,angle=0]{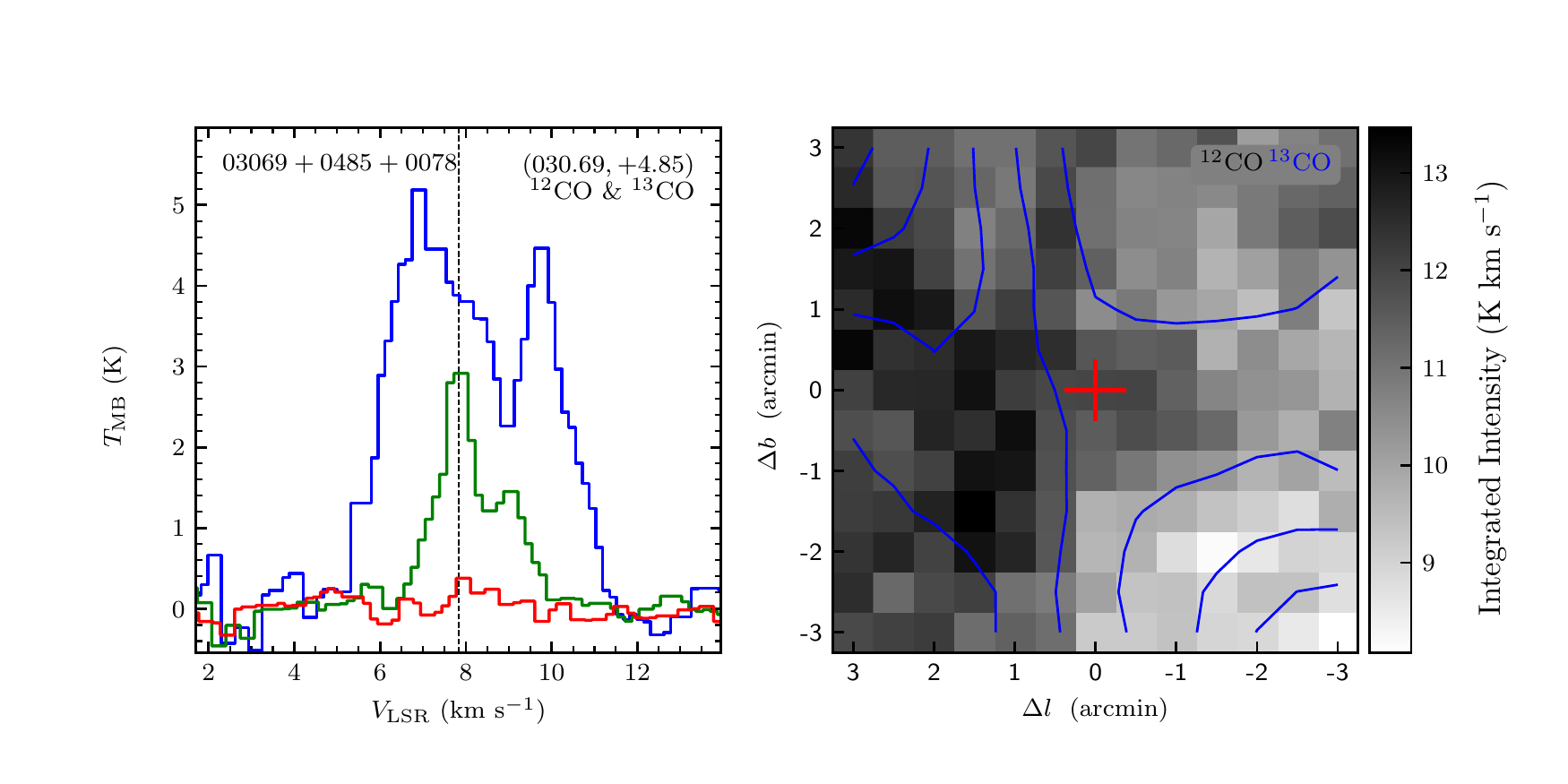}
\includegraphics[width=9.0cm,angle=0]{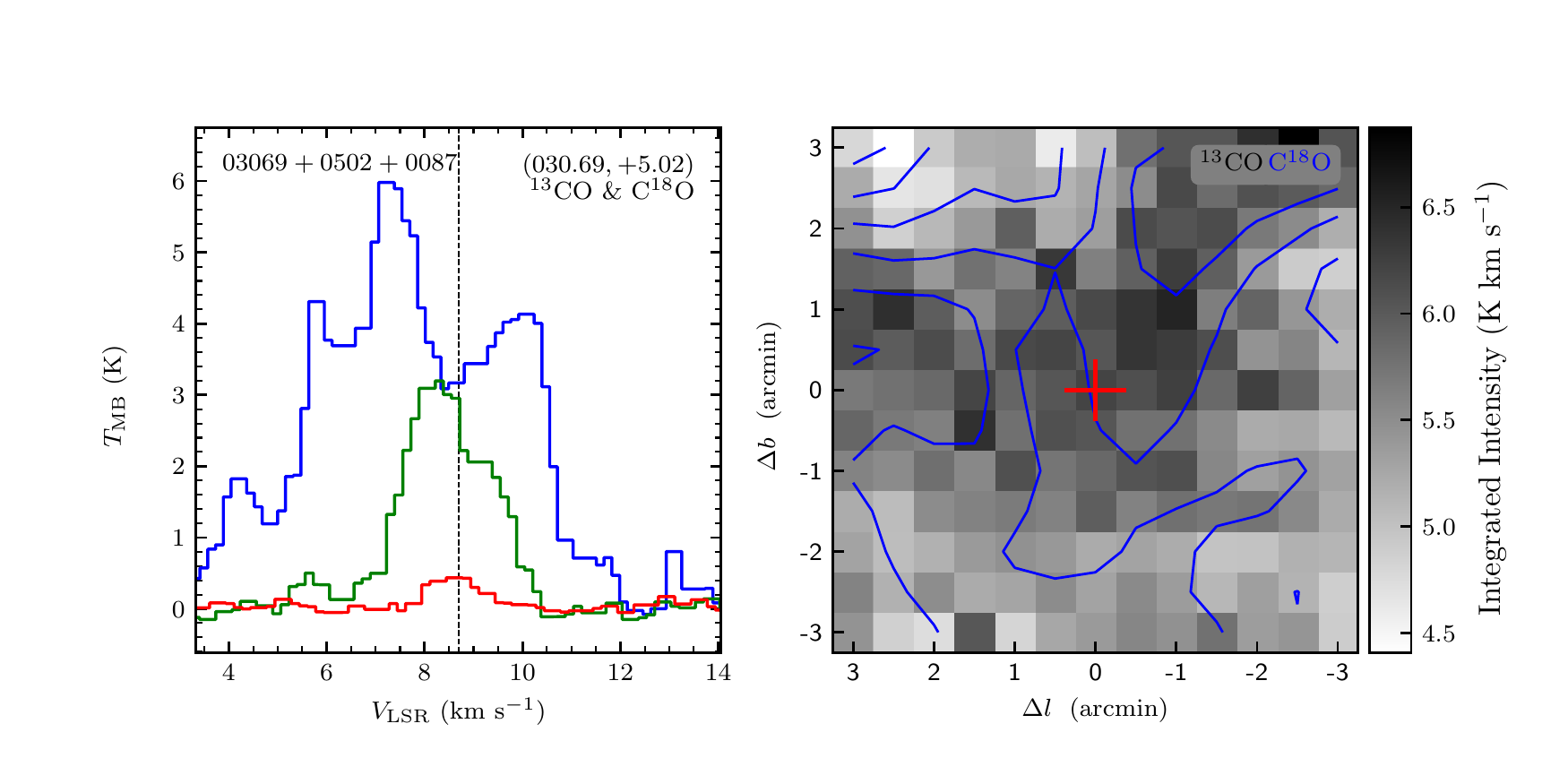}
\vspace{-0.5cm}

\includegraphics[width=9.0cm,angle=0]{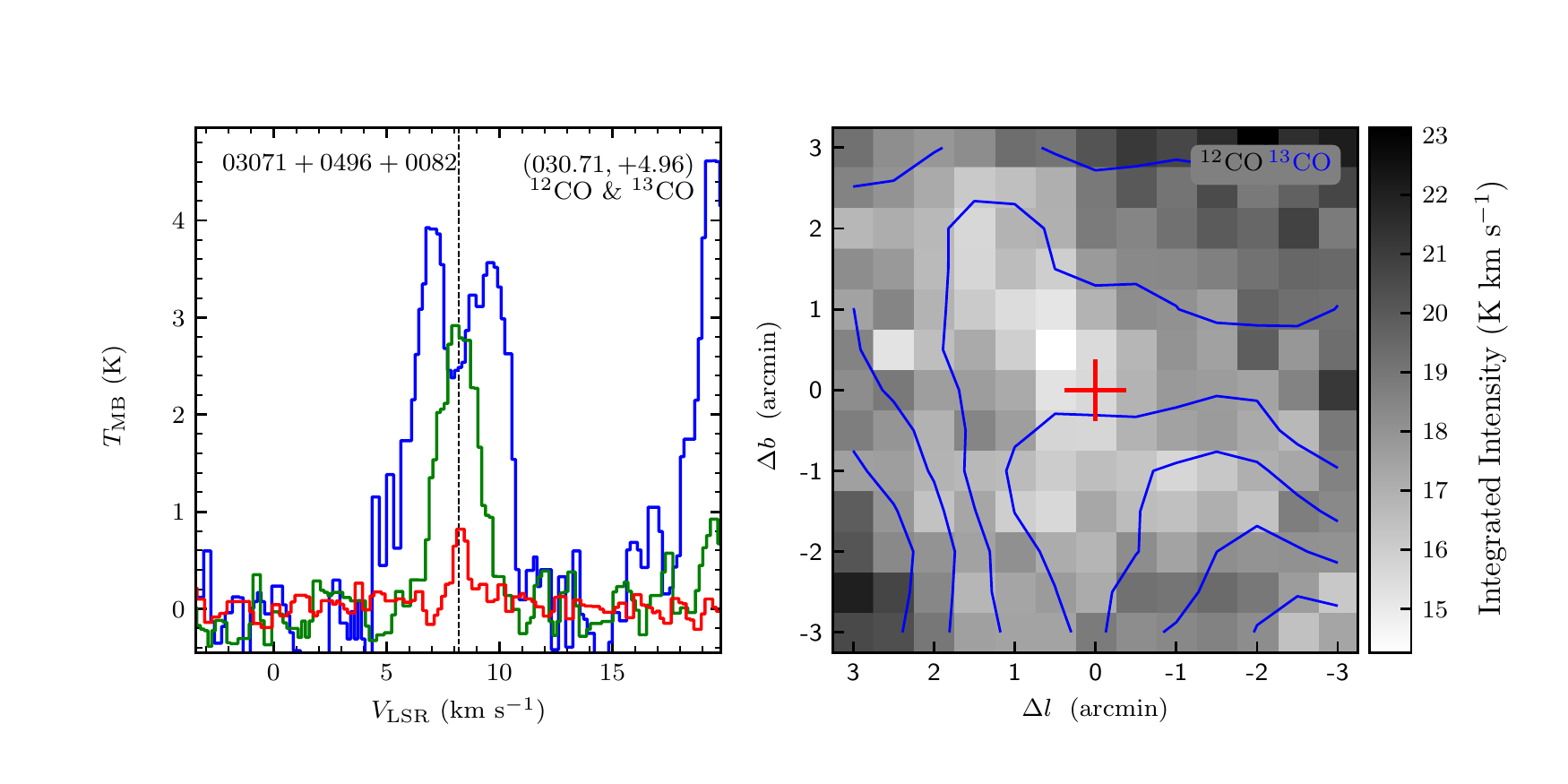}
\includegraphics[width=9.0cm,angle=0]{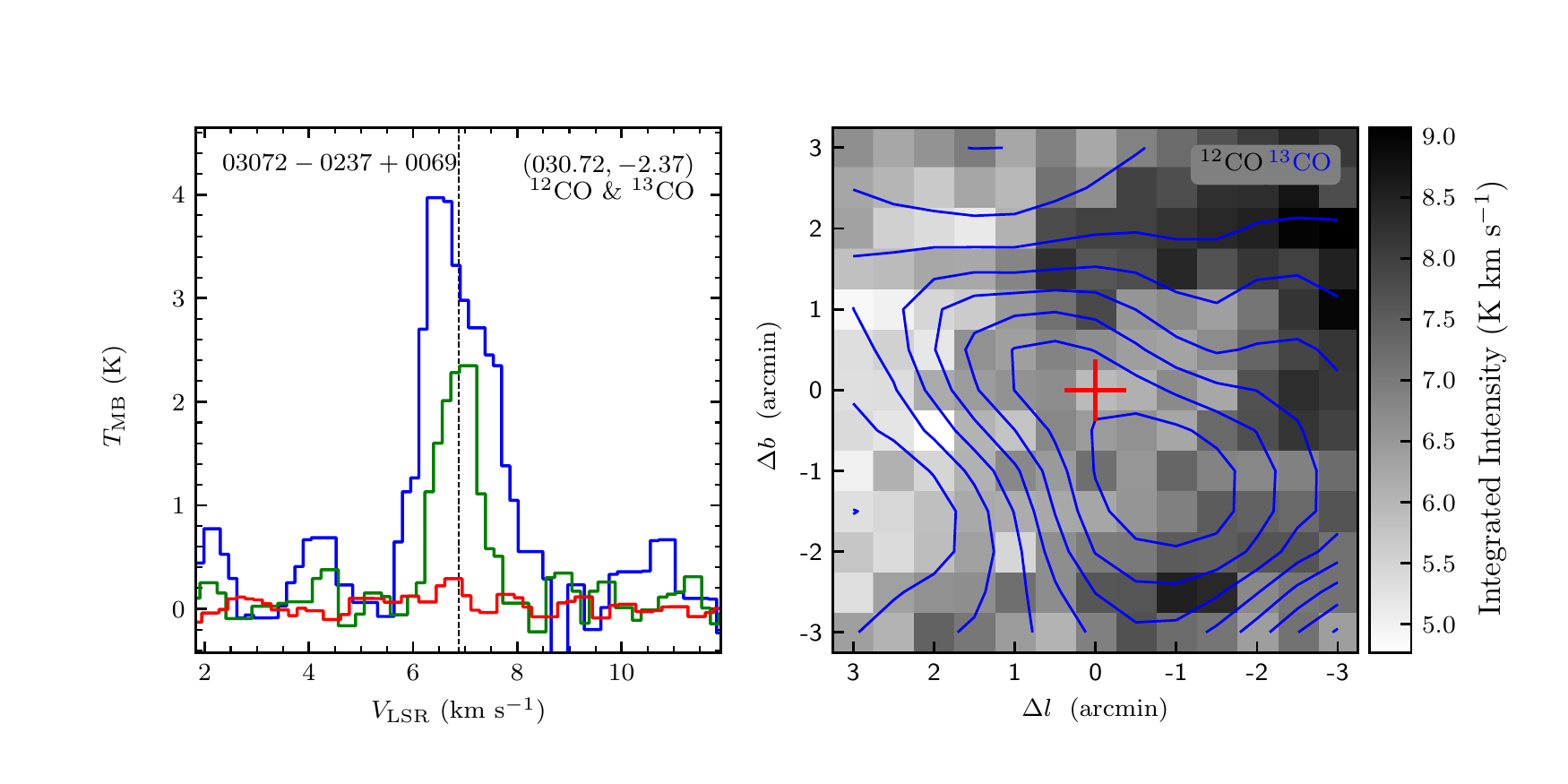}
\end{figure}
\clearpage

\begin{figure}
\includegraphics[width=9.0cm,angle=0]{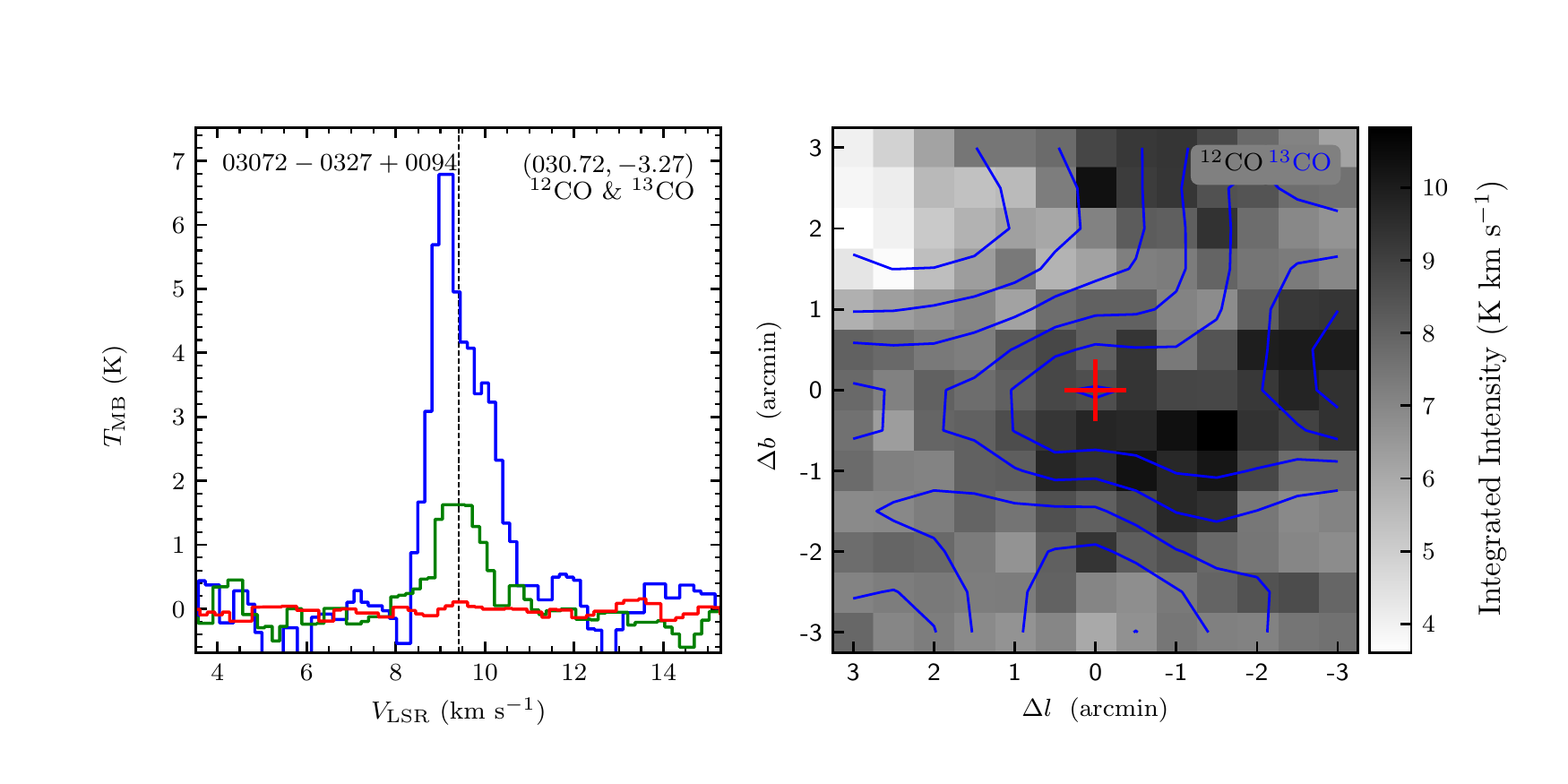}
\includegraphics[width=9.0cm,angle=0]{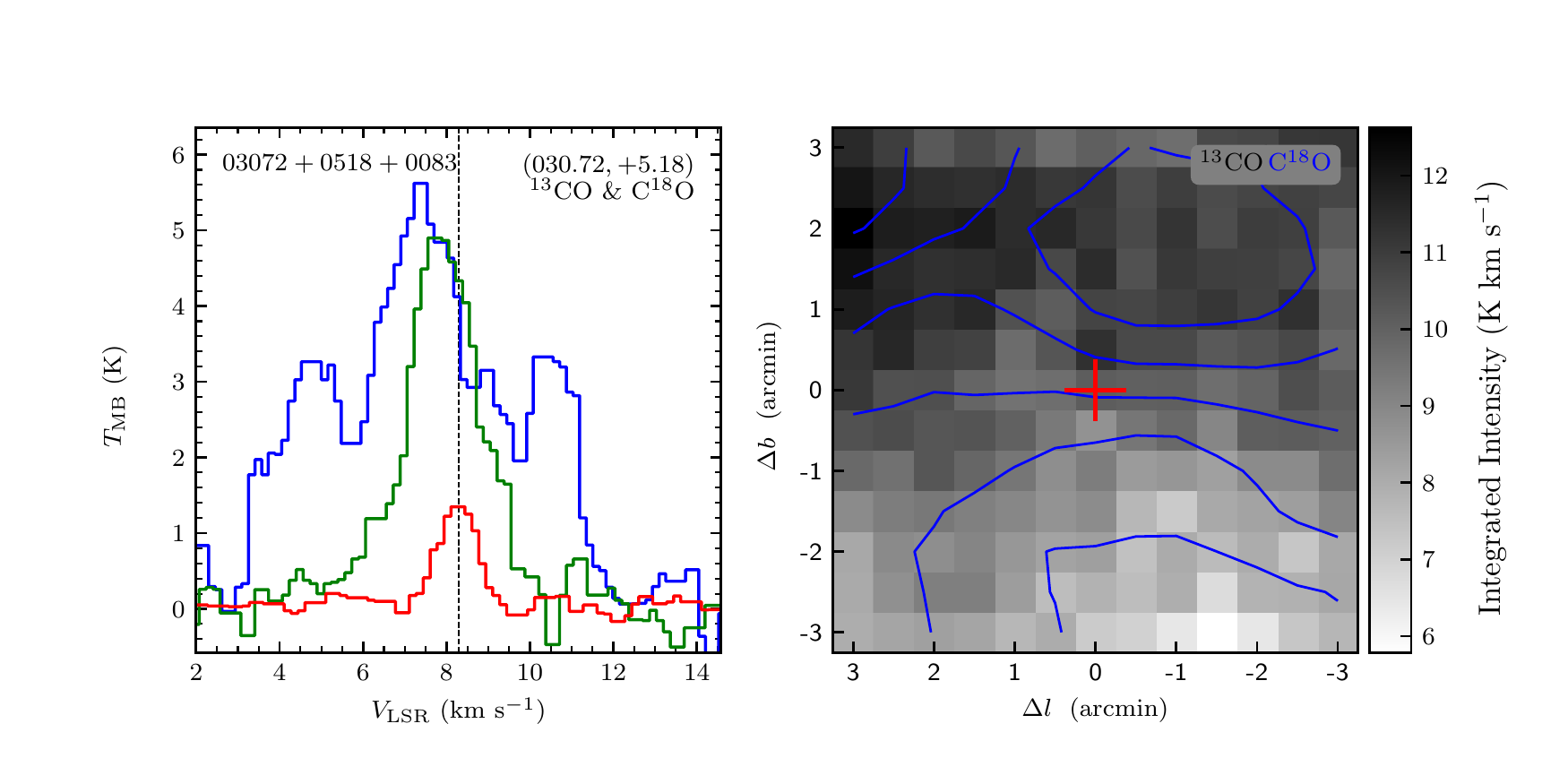}
\vspace{-0.5cm}

\includegraphics[width=9.0cm,angle=0]{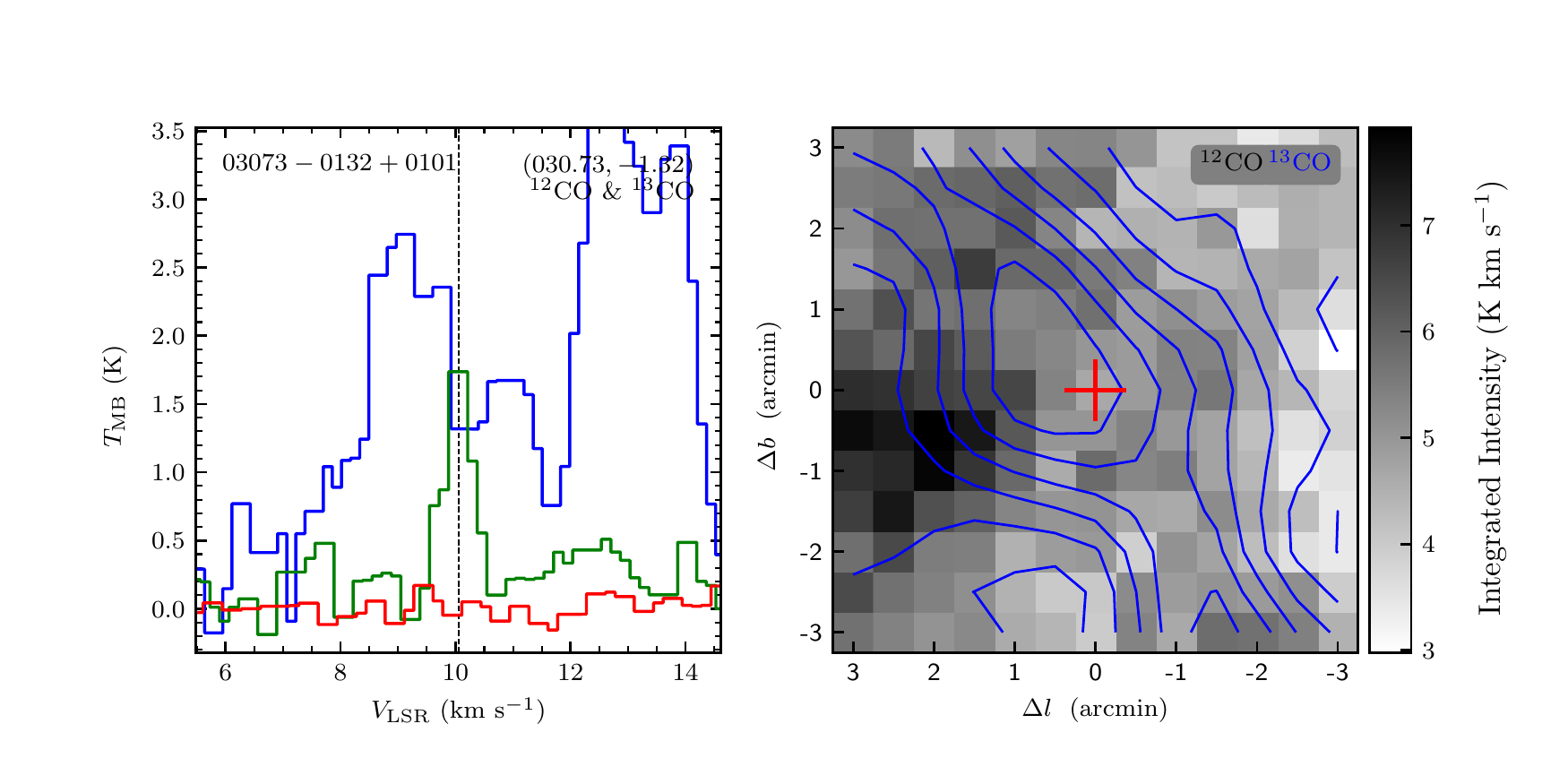}
\includegraphics[width=9.0cm,angle=0]{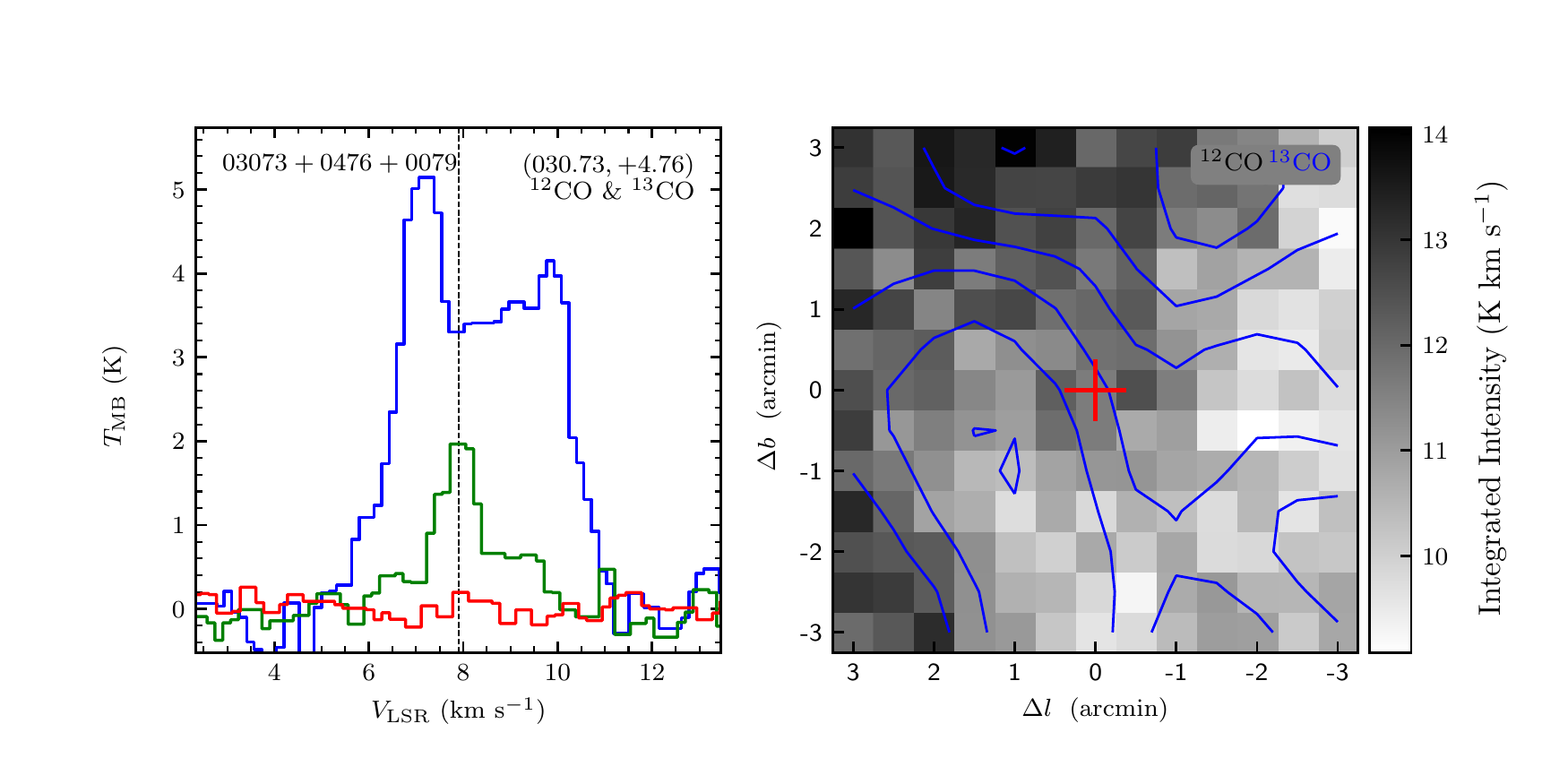}
\vspace{-0.5cm}

\includegraphics[width=9.0cm,angle=0]{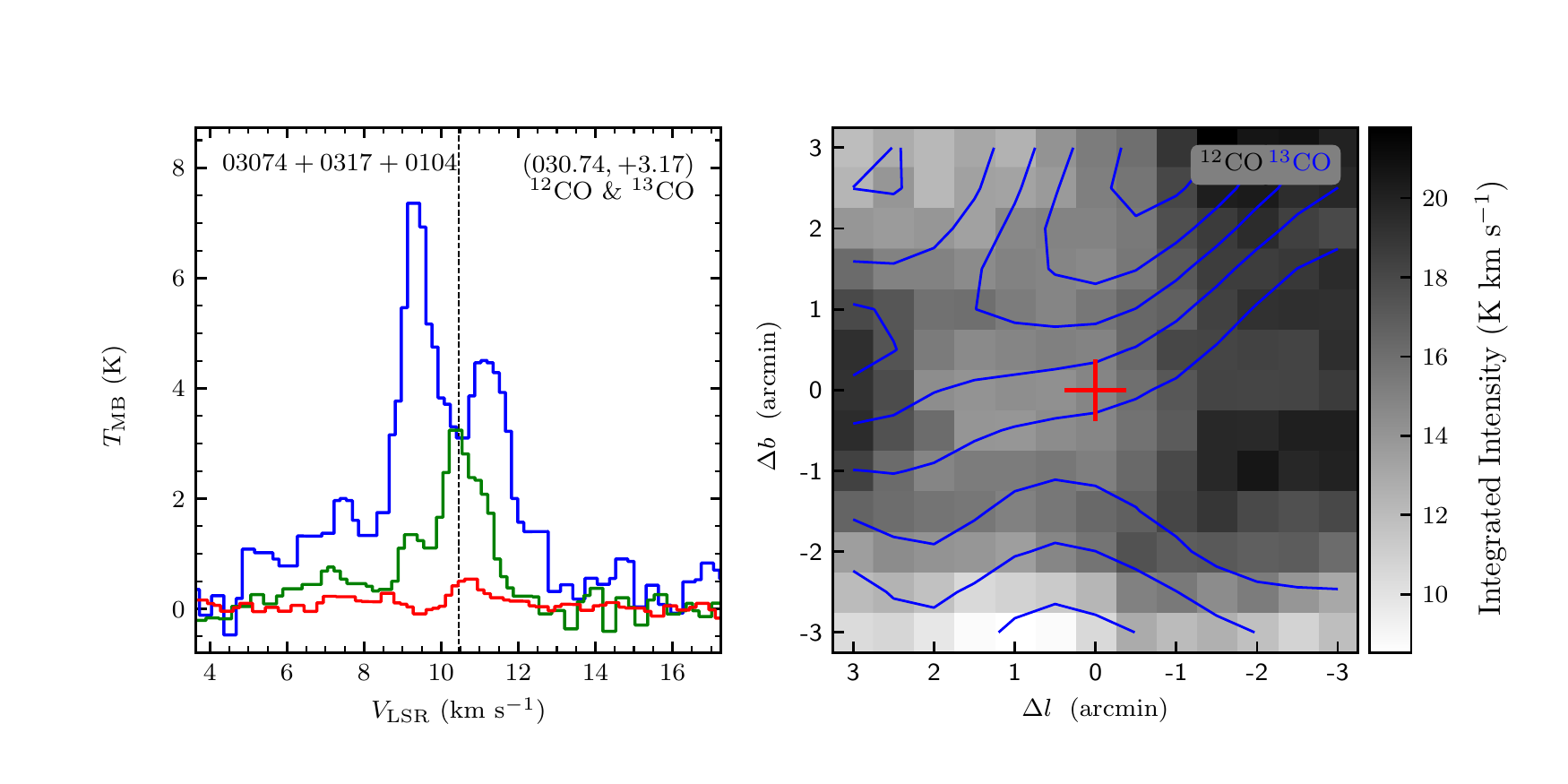}
\includegraphics[width=9.0cm,angle=0]{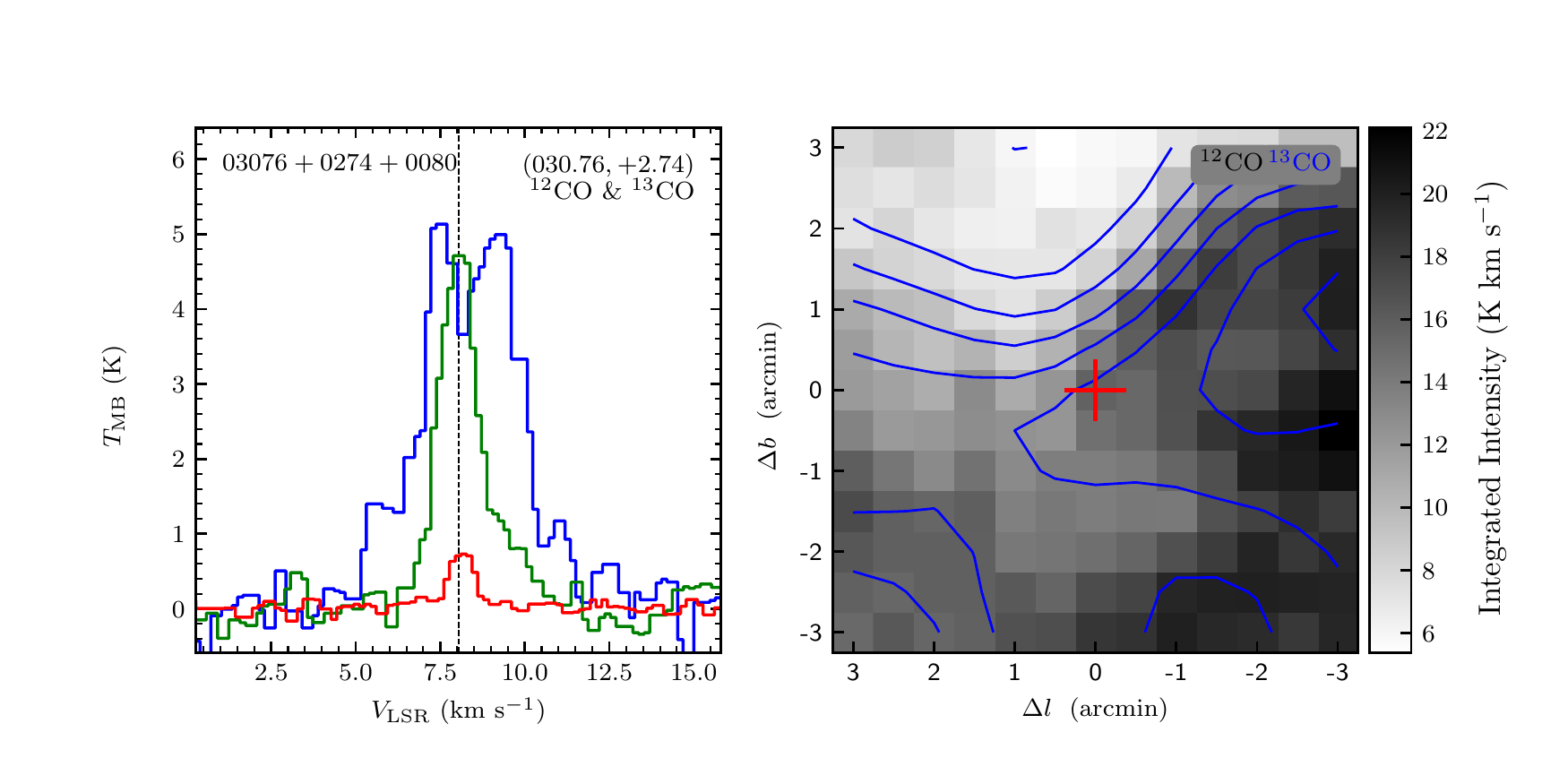}
\vspace{-0.5cm}

\includegraphics[width=9.0cm,angle=0]{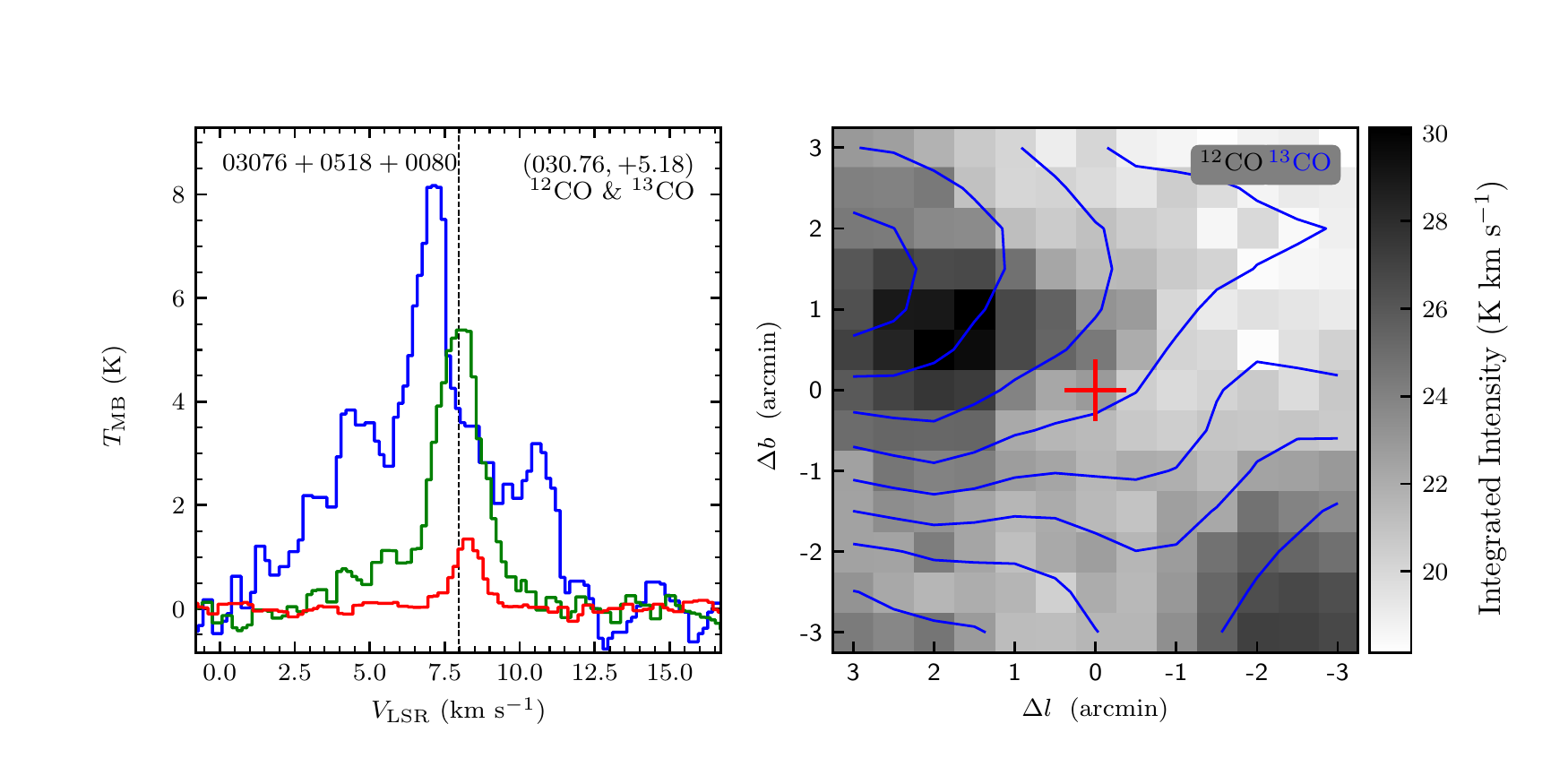}
\includegraphics[width=9.0cm,angle=0]{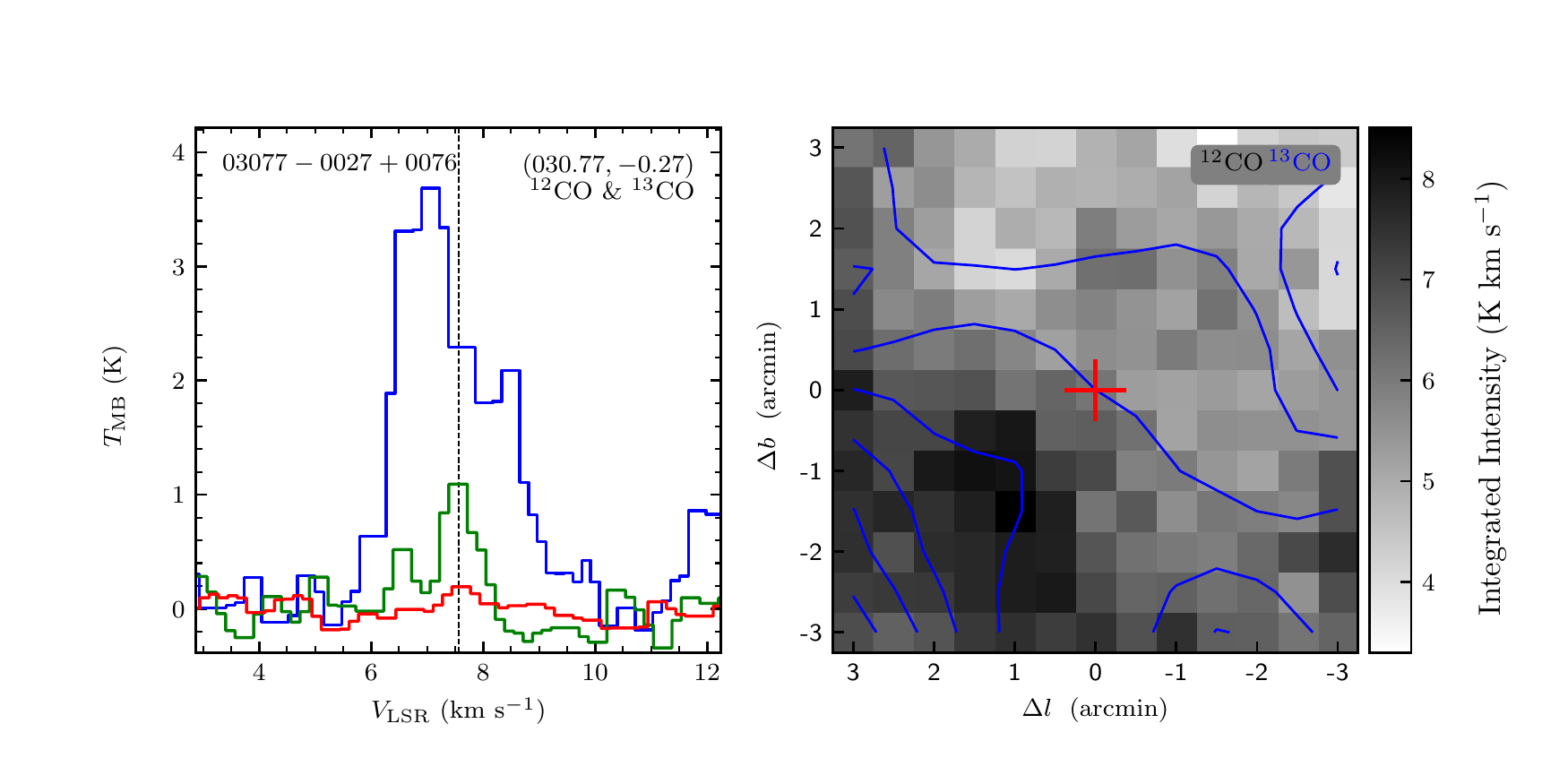}
\vspace{-0.5cm}

\includegraphics[width=9.0cm,angle=0]{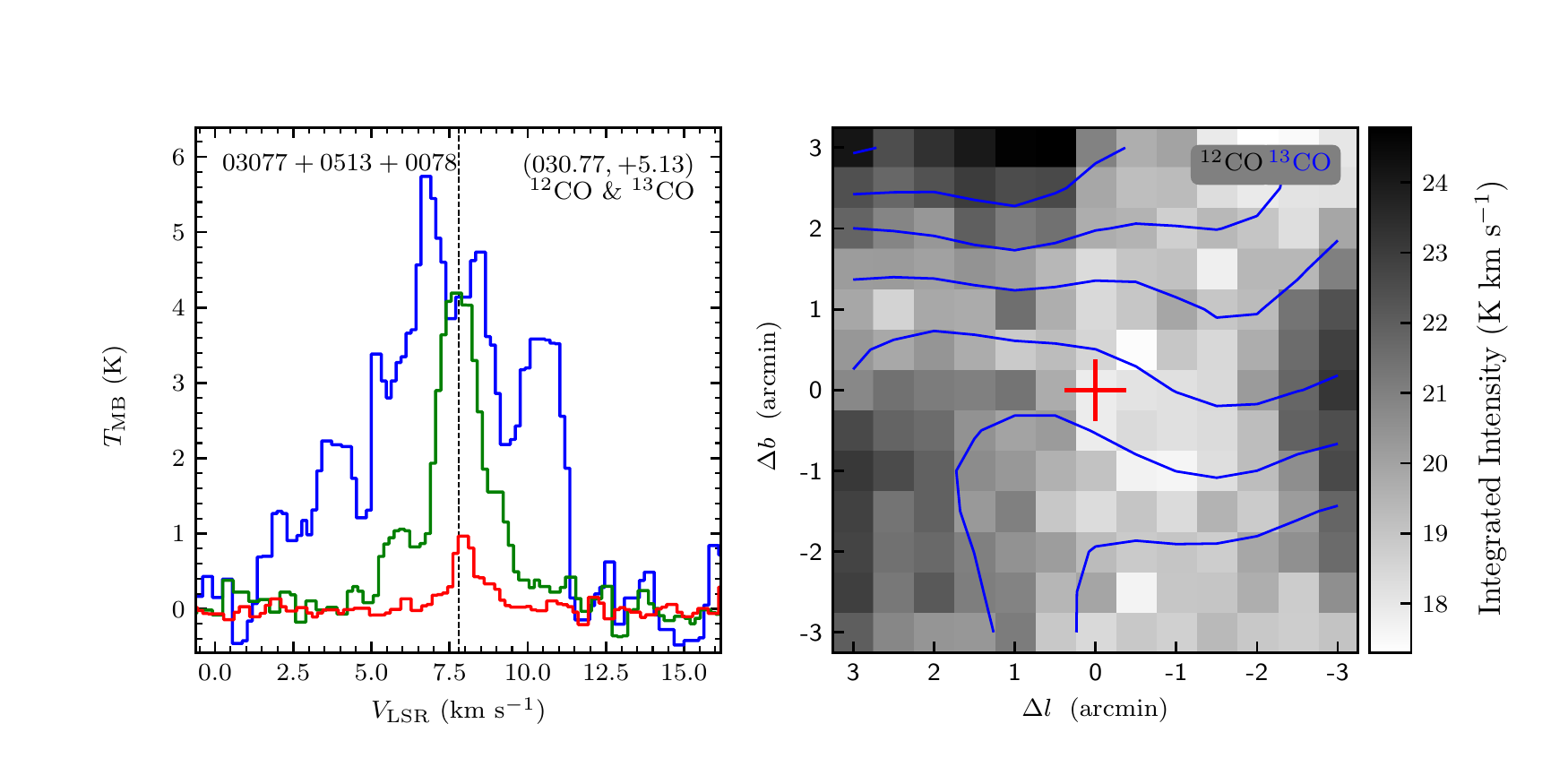}
\includegraphics[width=9.0cm,angle=0]{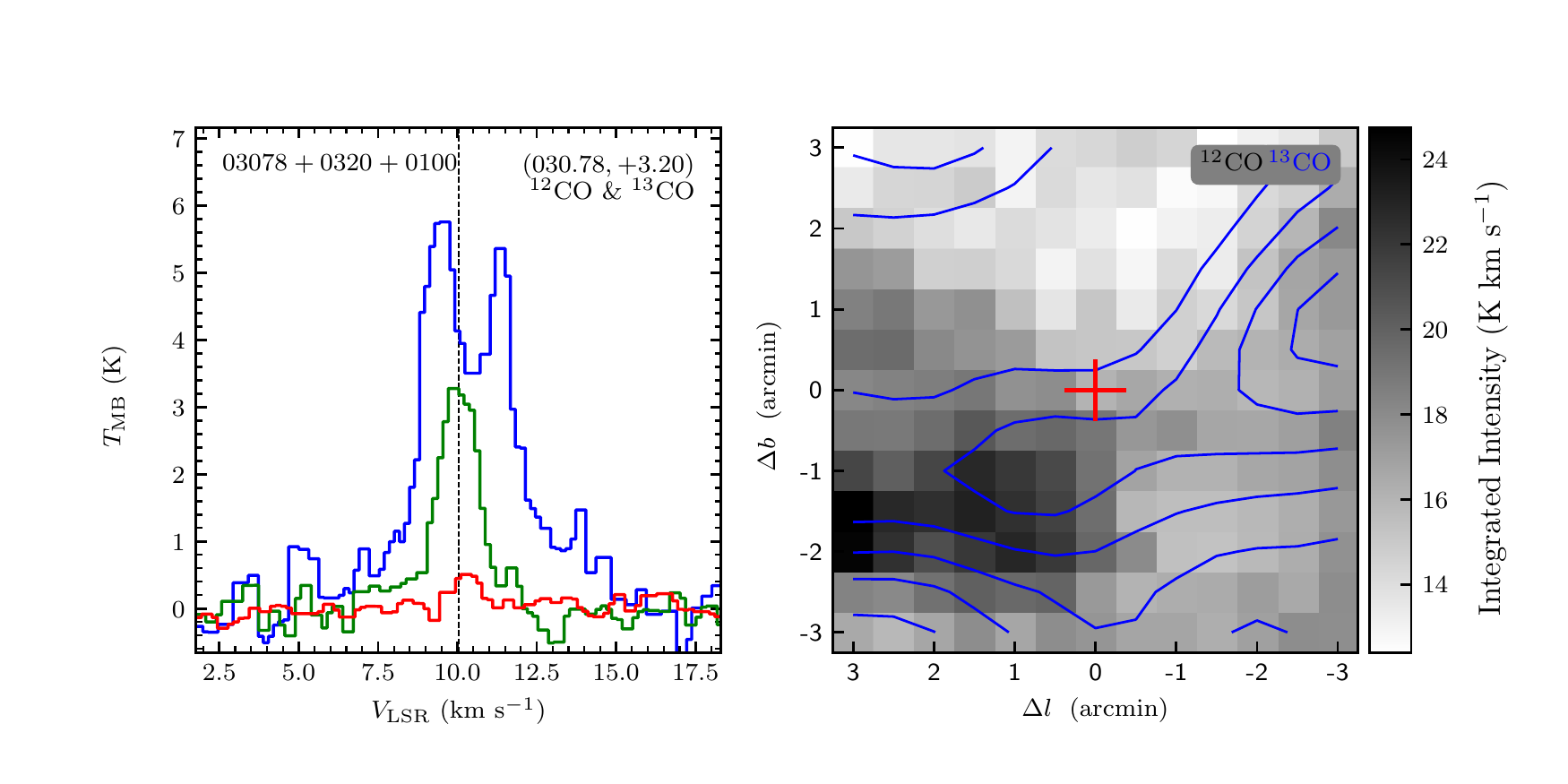}
\end{figure}
\clearpage

\begin{figure}
\includegraphics[width=9.0cm,angle=0]{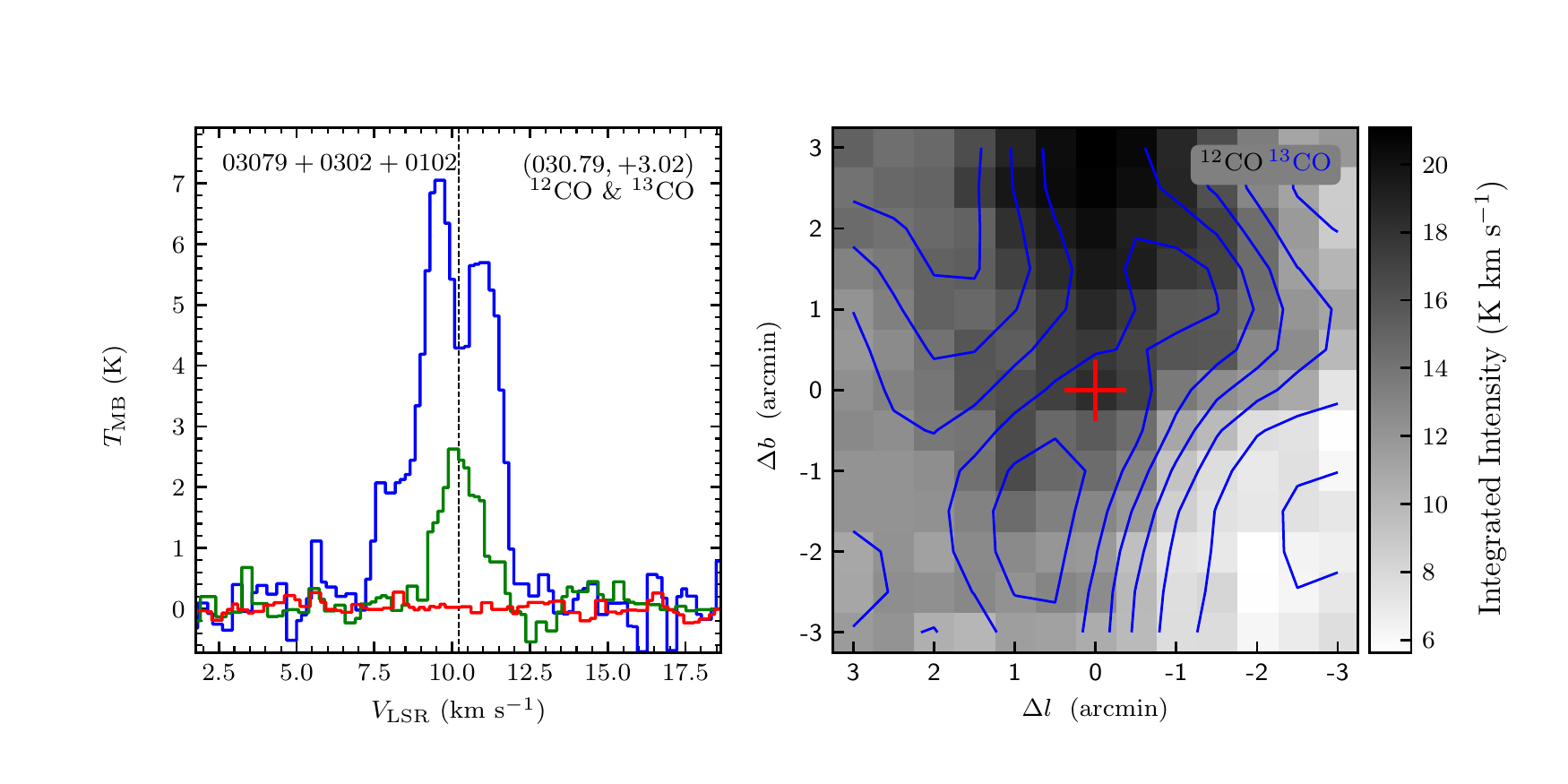}
\includegraphics[width=9.0cm,angle=0]{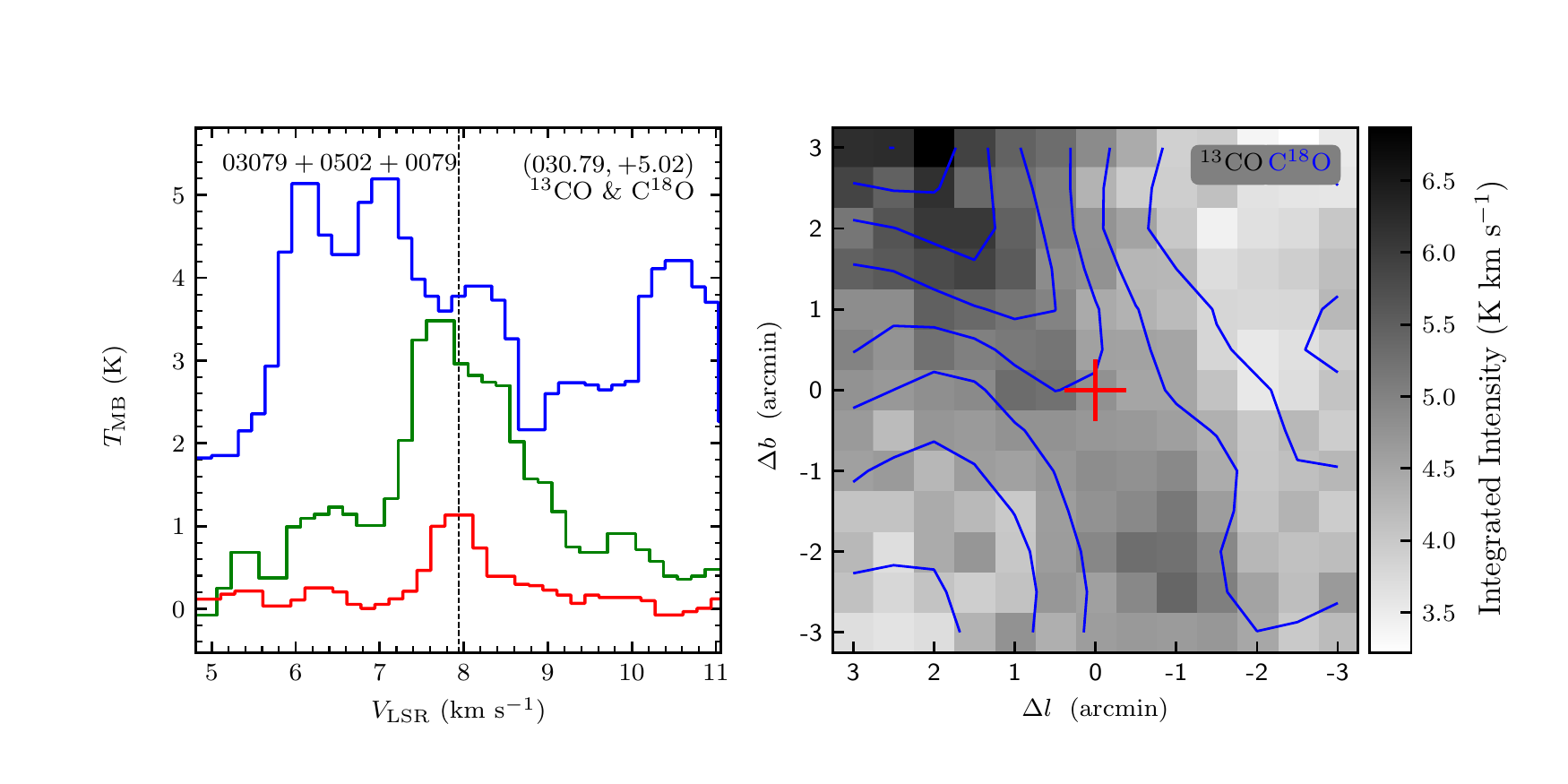}
\vspace{-0.5cm}

\includegraphics[width=9.0cm,angle=0]{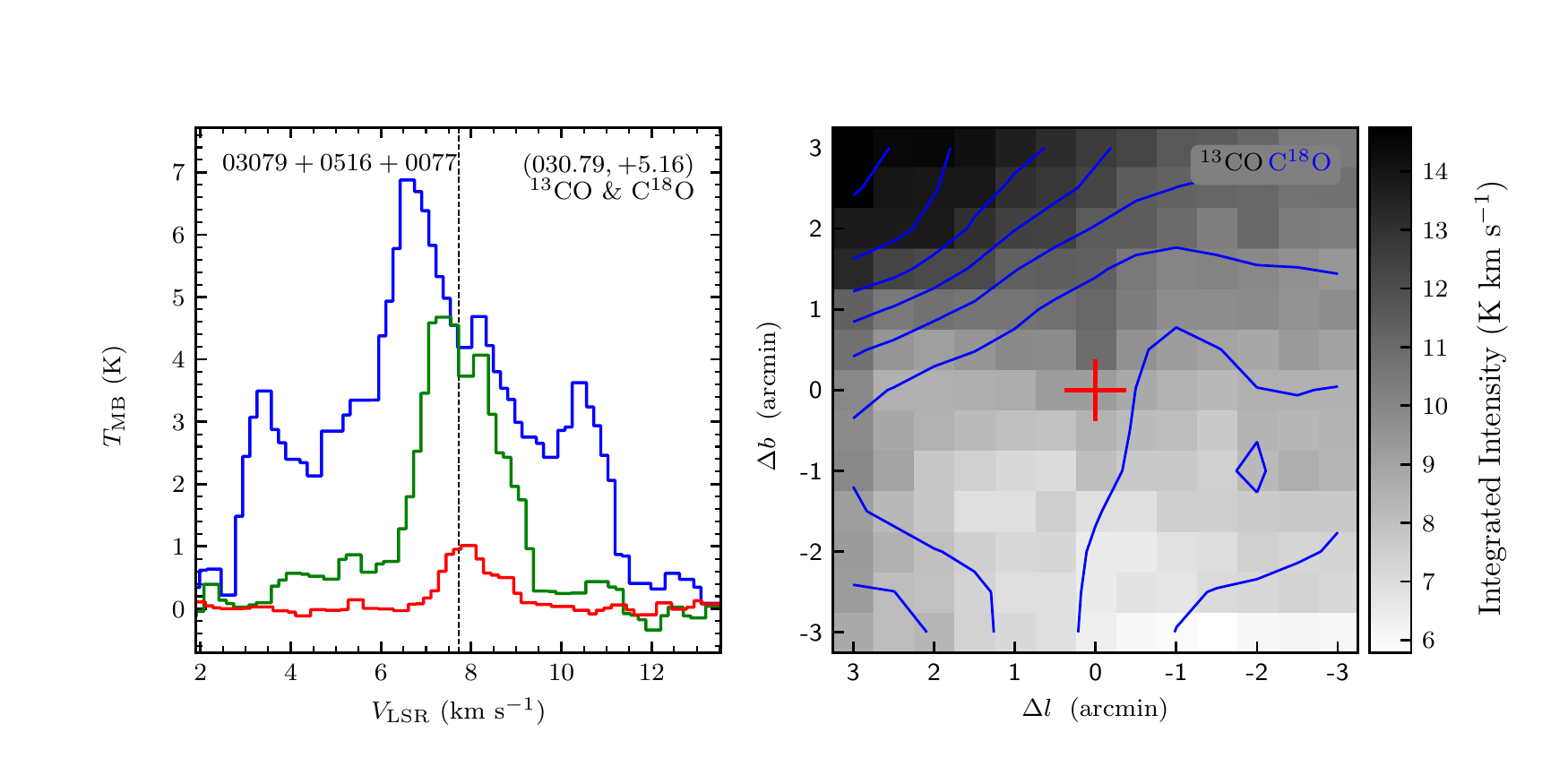}
\includegraphics[width=9.0cm,angle=0]{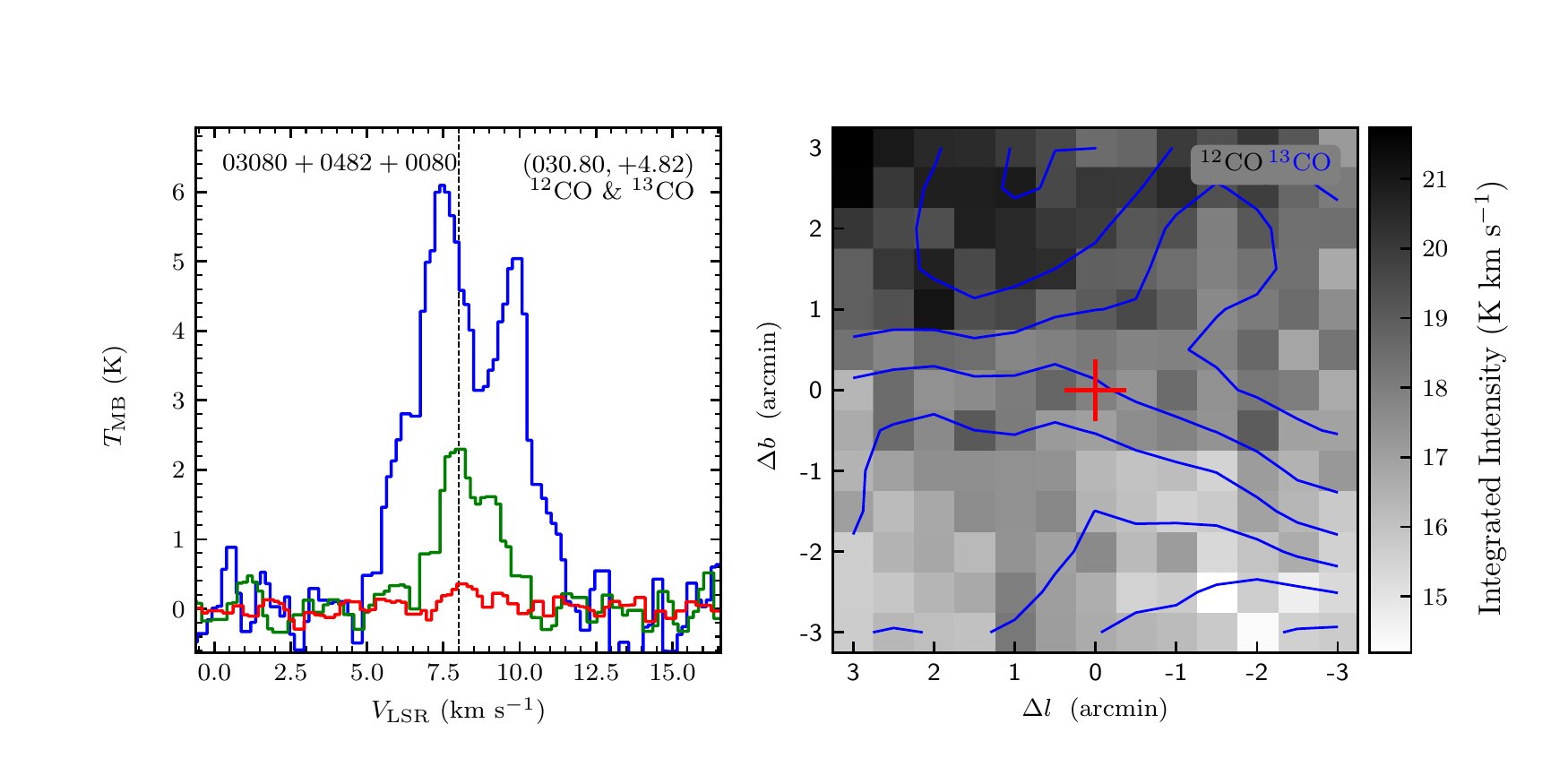}
\vspace{-0.5cm}

\includegraphics[width=9.0cm,angle=0]{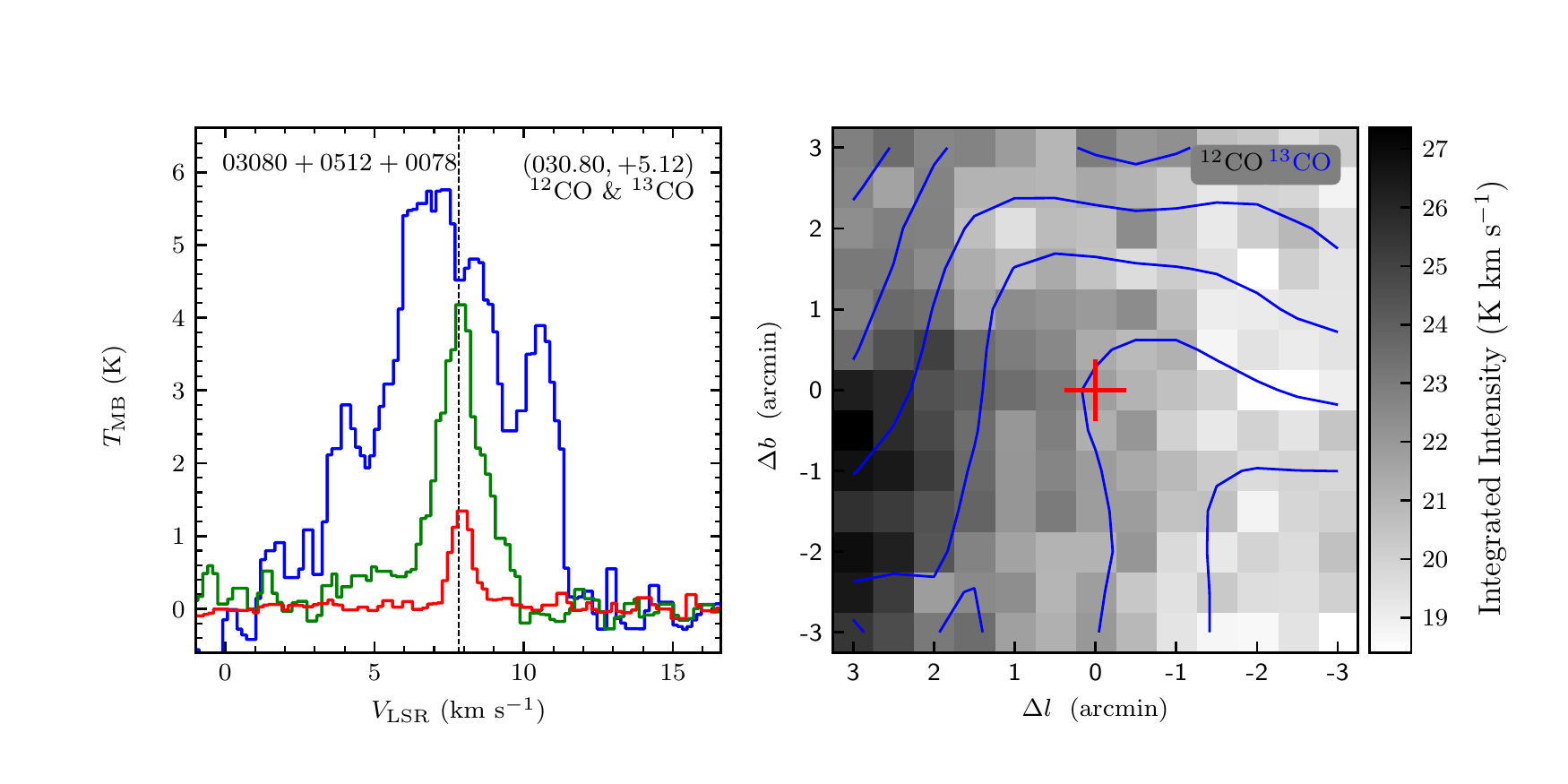}
\includegraphics[width=9.0cm,angle=0]{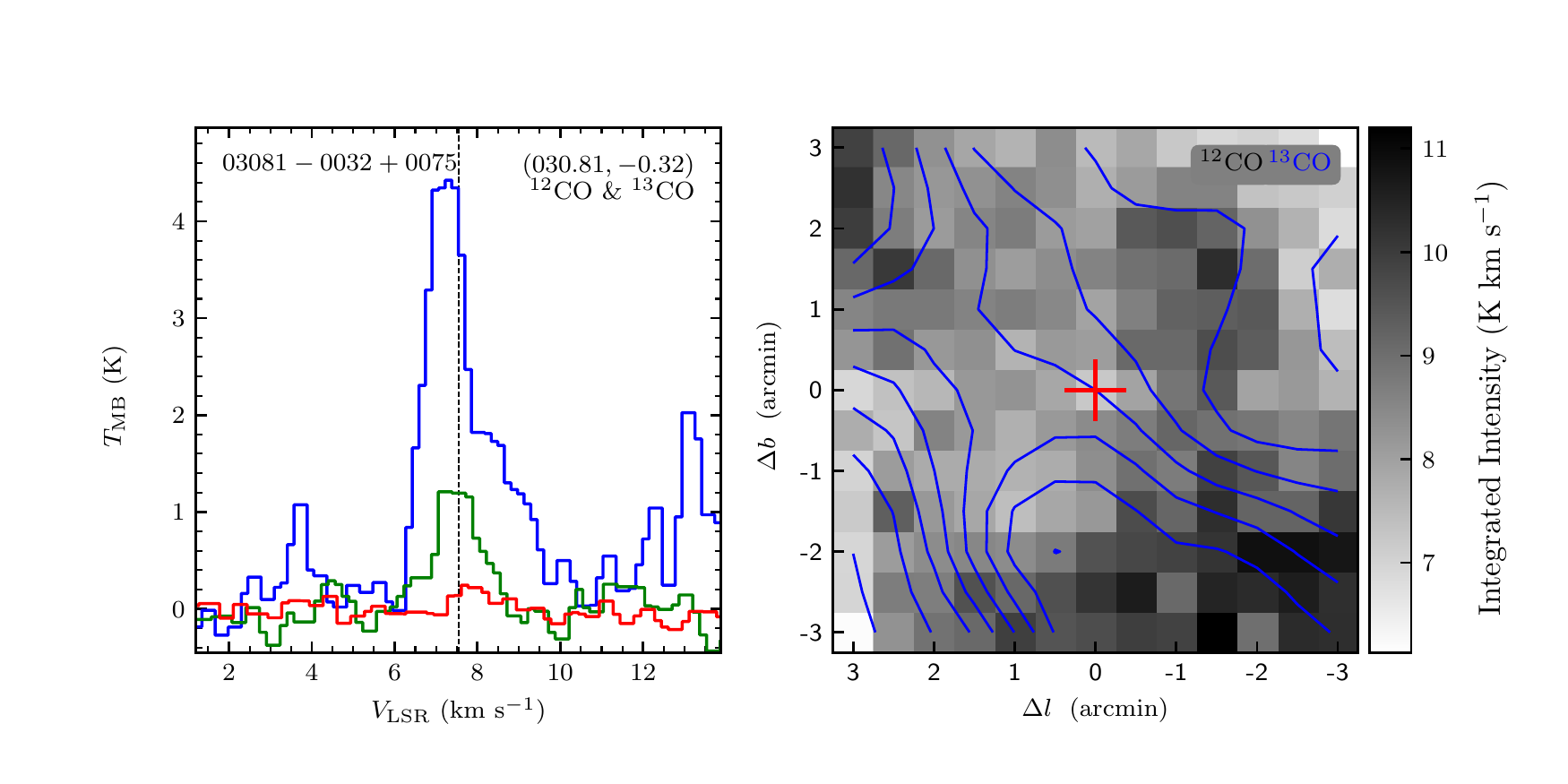}
\vspace{-0.5cm}

\includegraphics[width=9.0cm,angle=0]{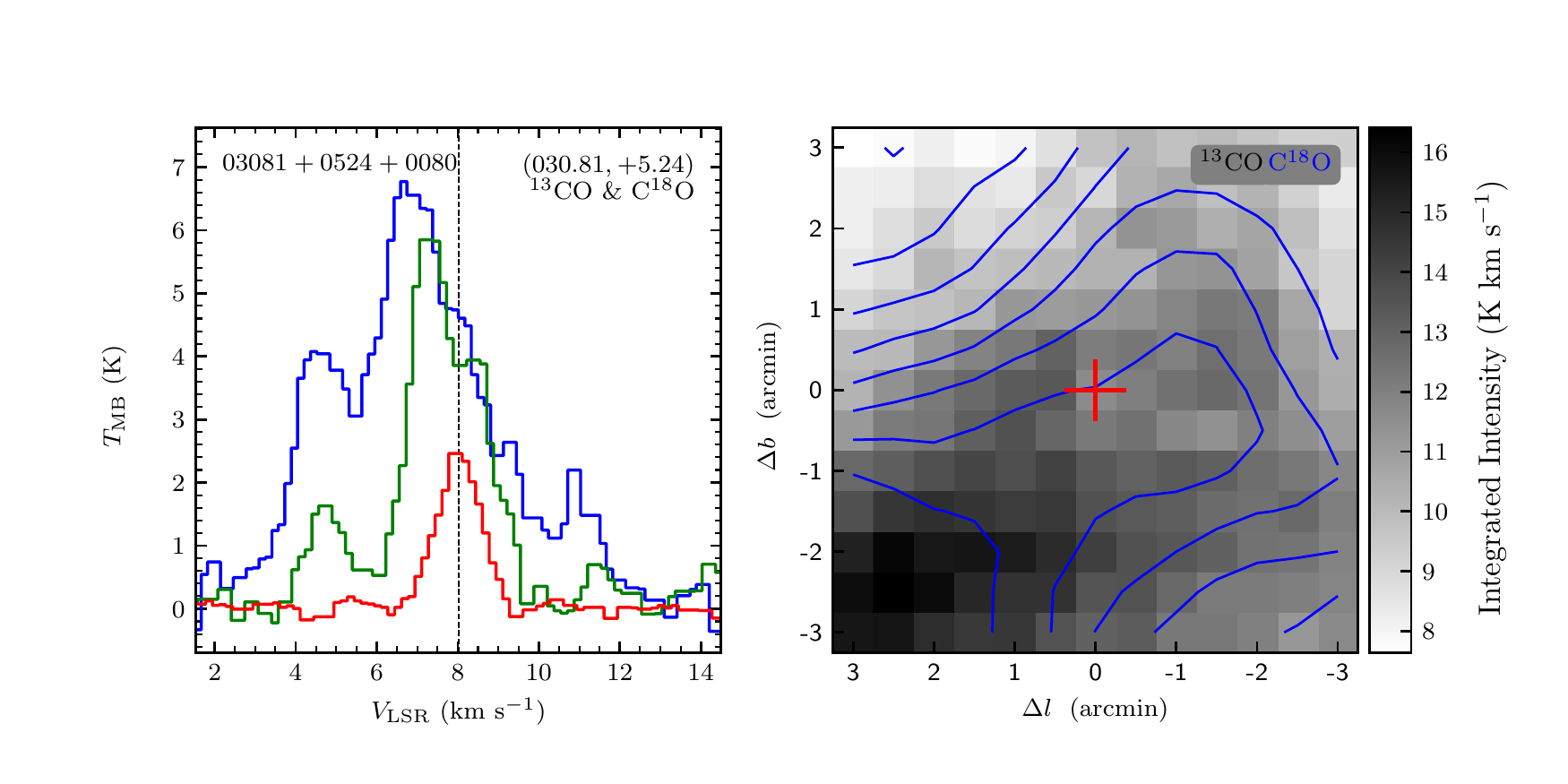}
\includegraphics[width=9.0cm,angle=0]{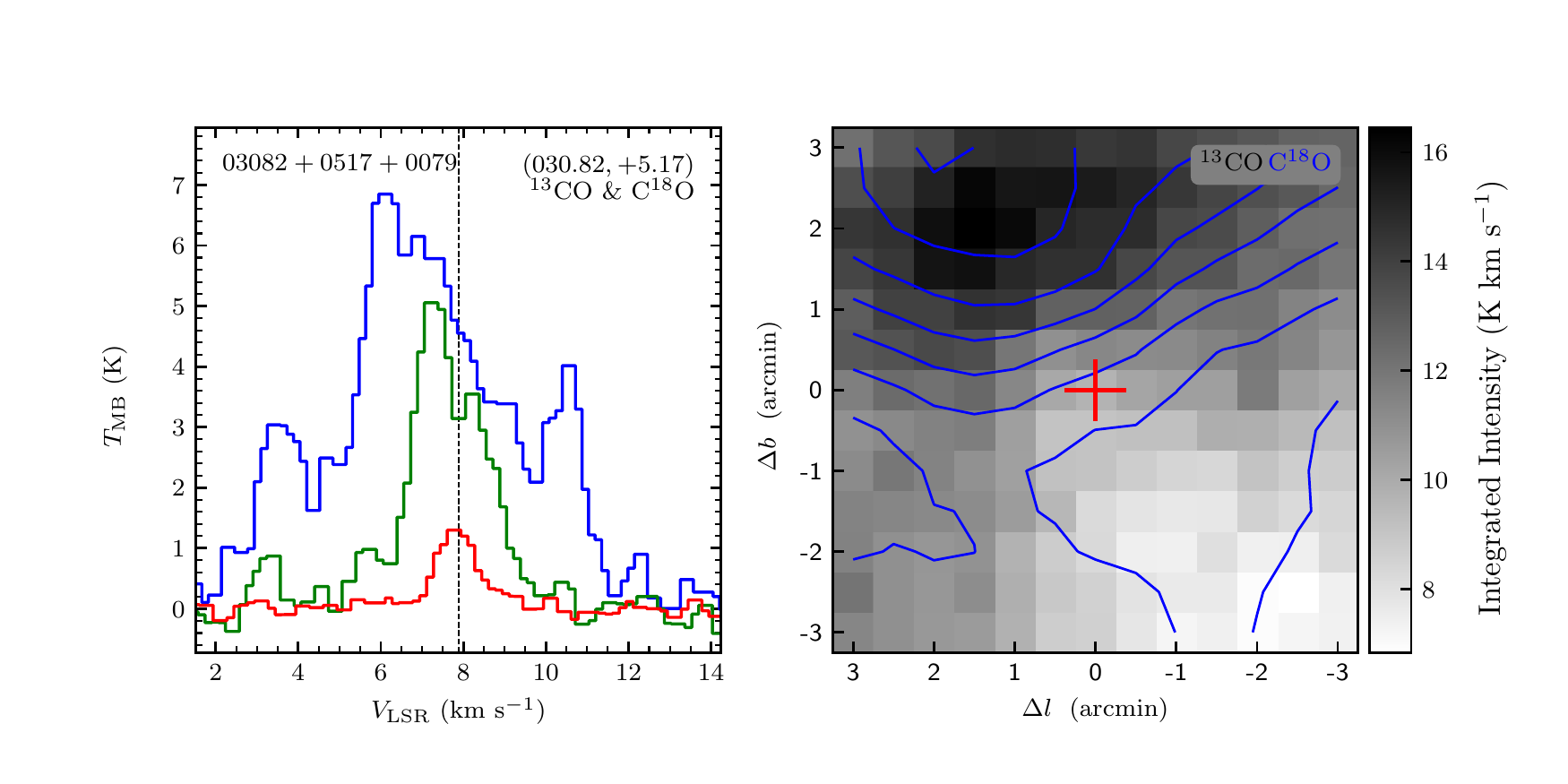}
\vspace{-0.5cm}

\includegraphics[width=9.0cm,angle=0]{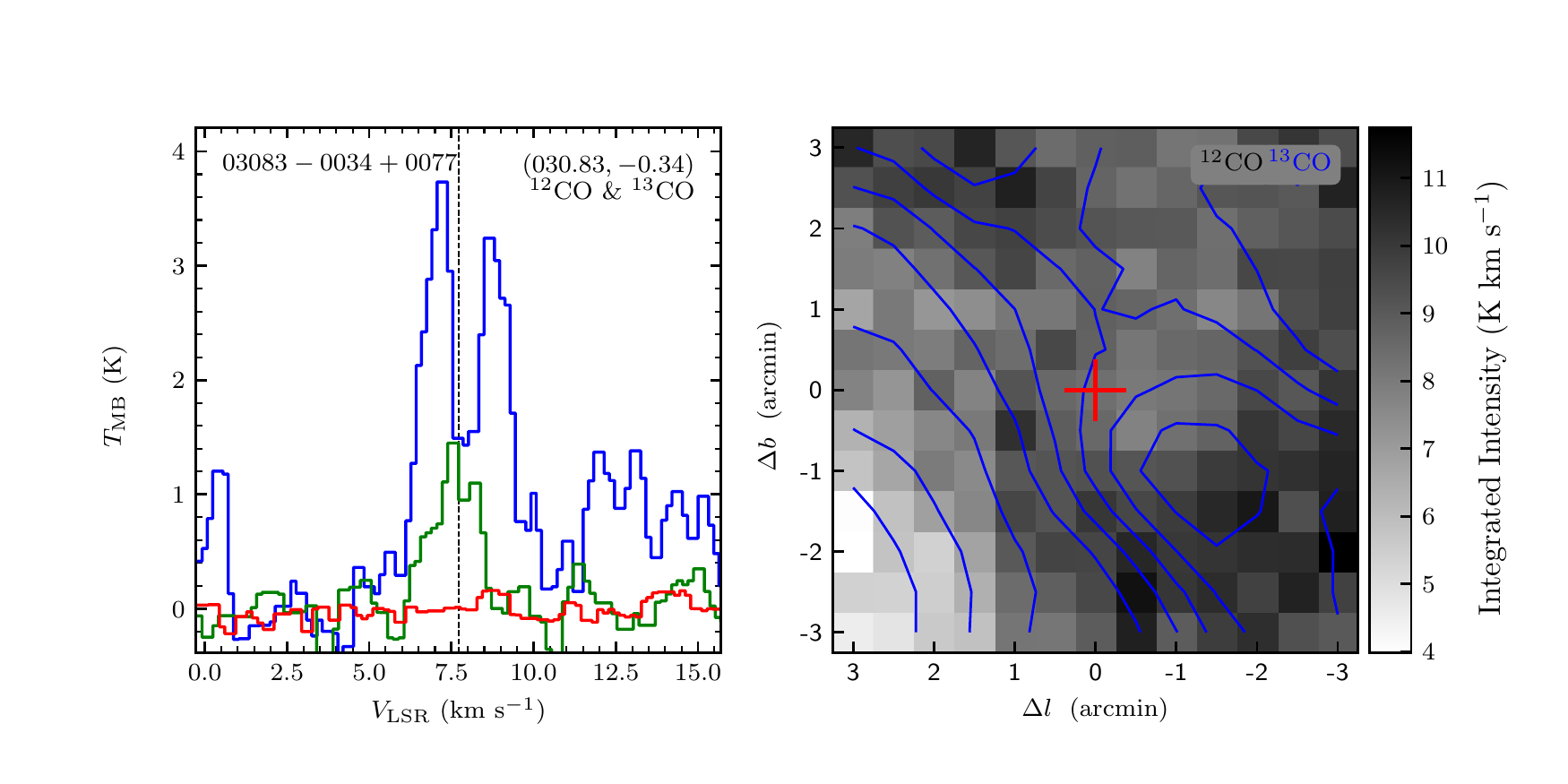}
\includegraphics[width=9.0cm,angle=0]{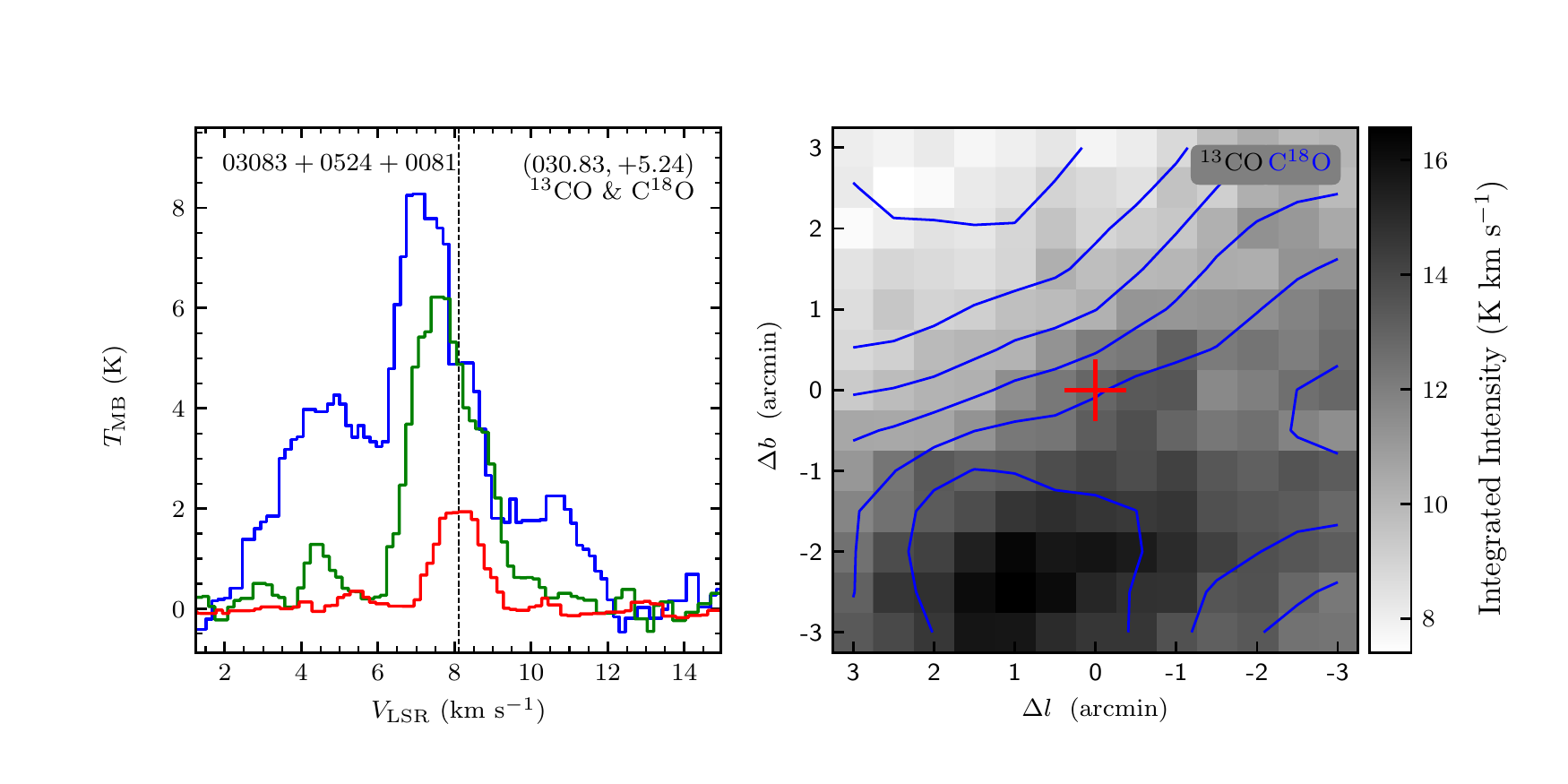}
\end{figure}
\clearpage

\begin{figure}
\includegraphics[width=9.0cm,angle=0]{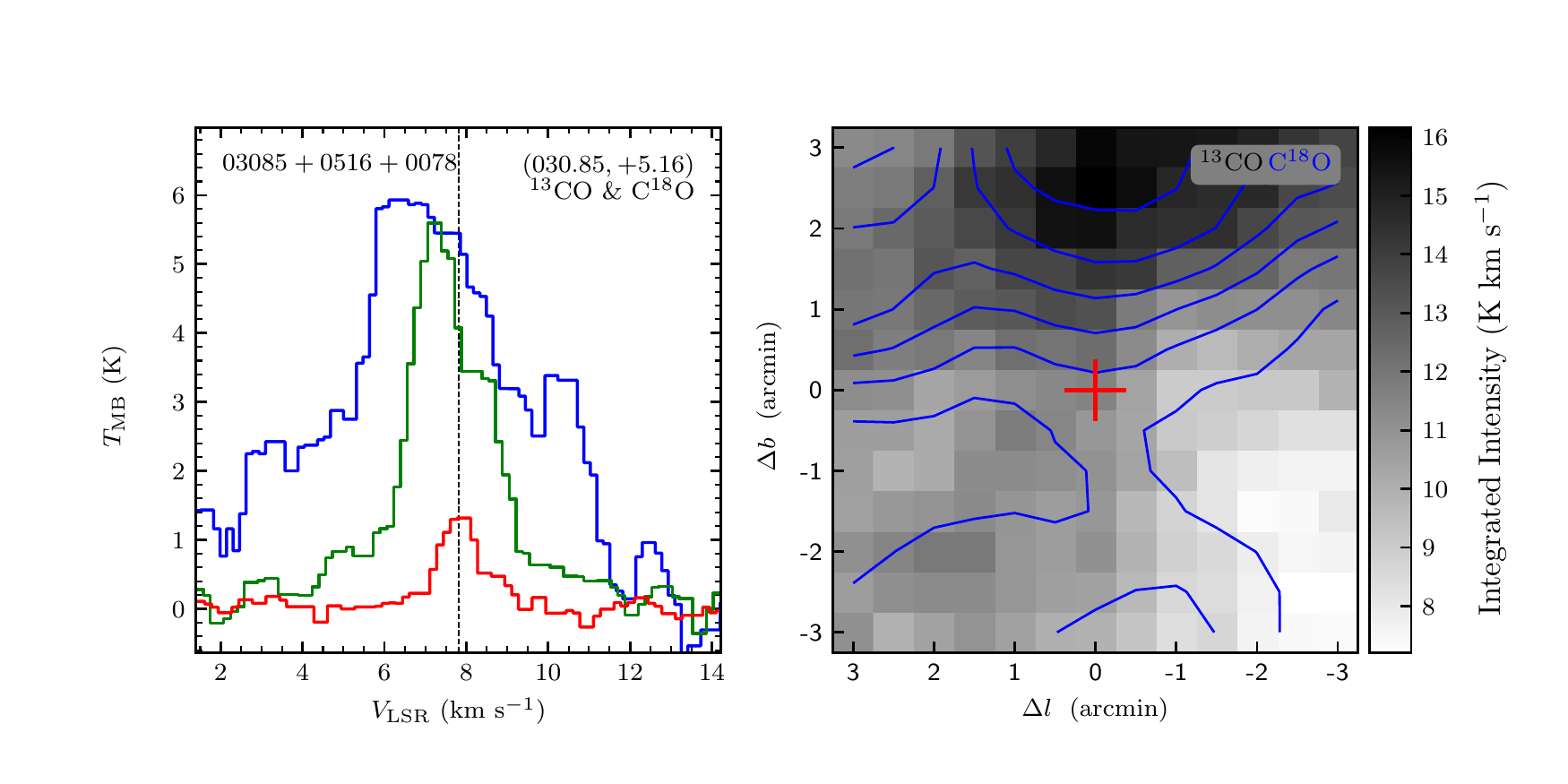}
\includegraphics[width=9.0cm,angle=0]{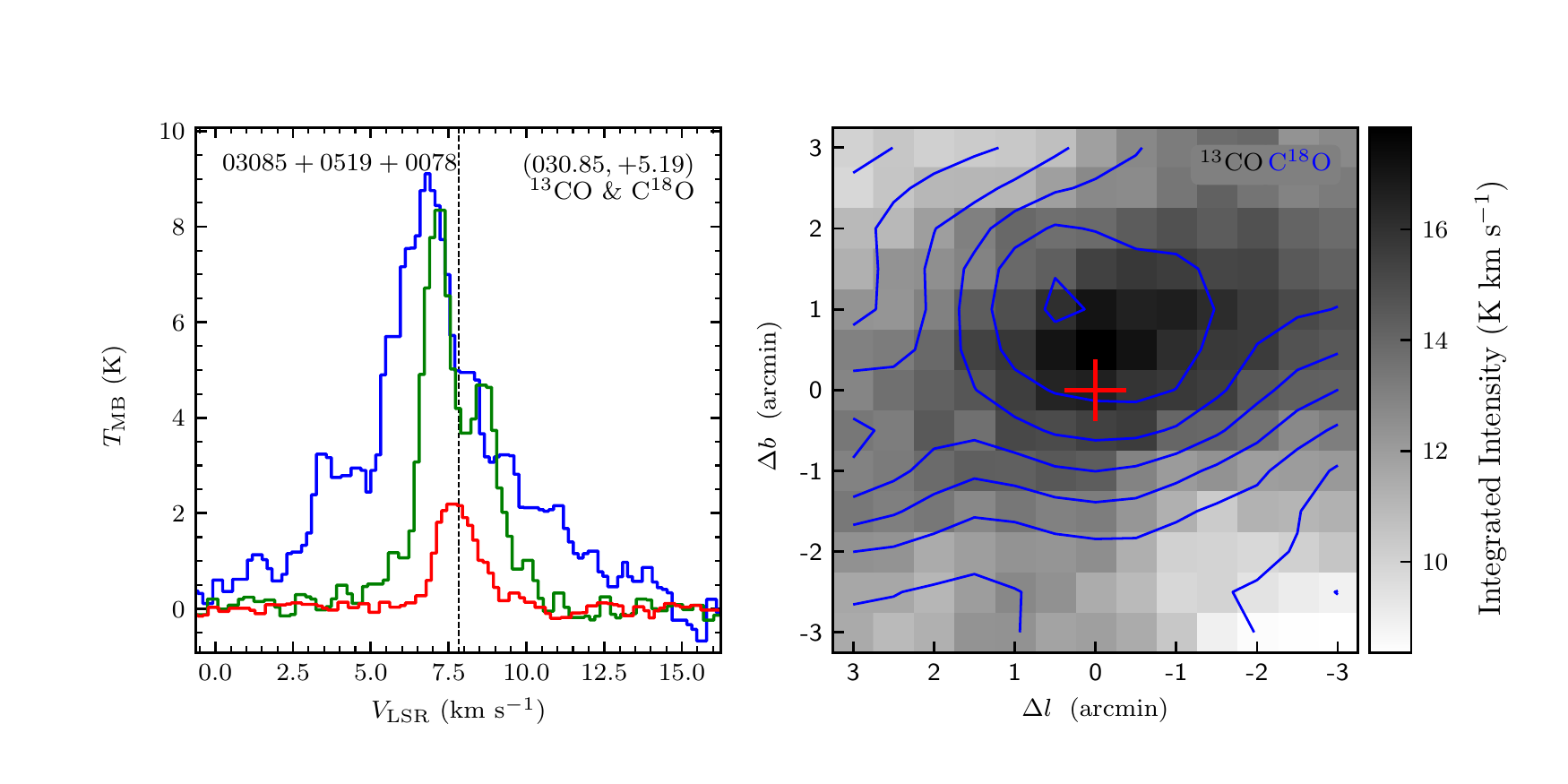}
\vspace{-0.5cm}

\includegraphics[width=9.0cm,angle=0]{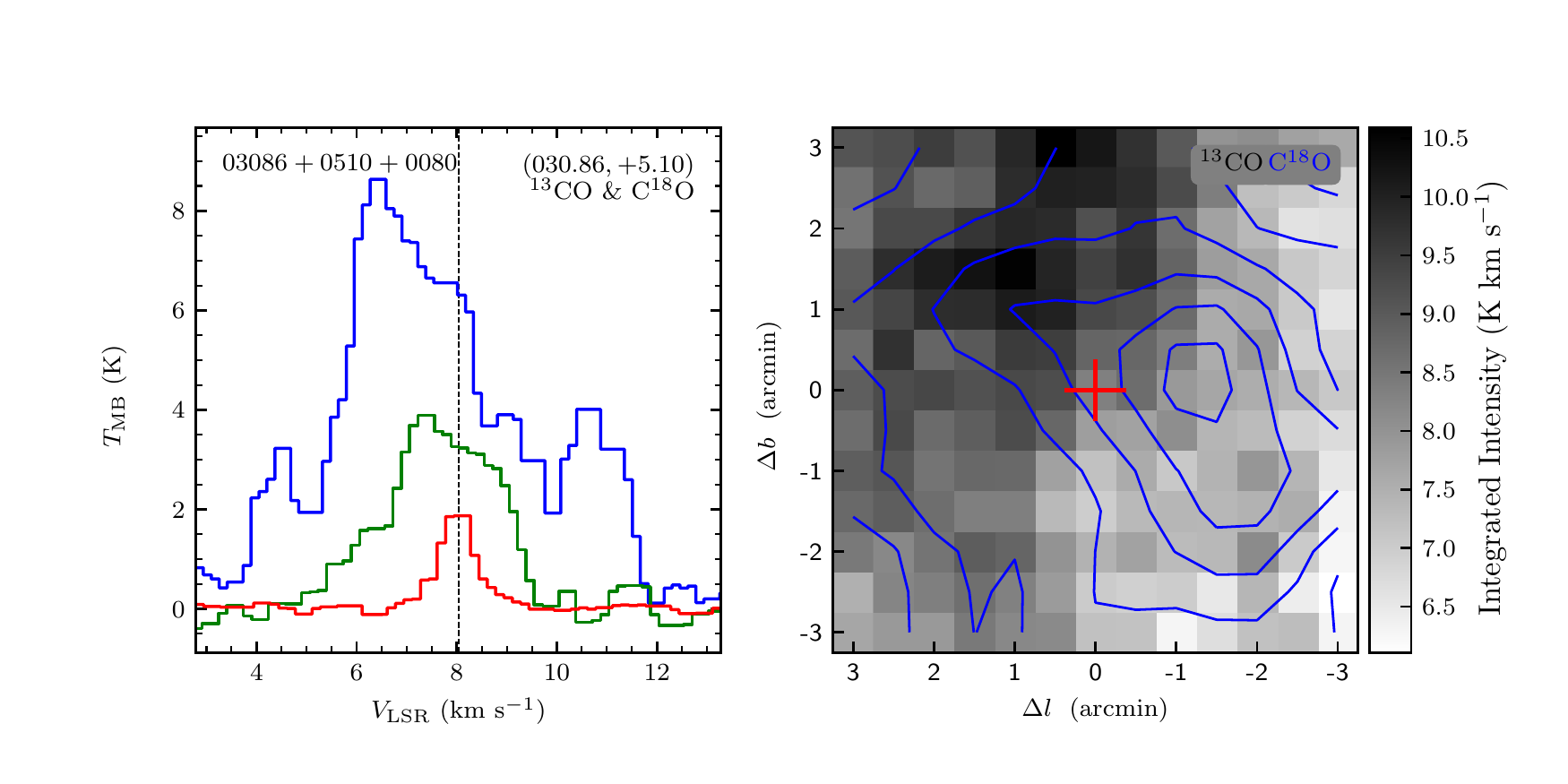}
\includegraphics[width=9.0cm,angle=0]{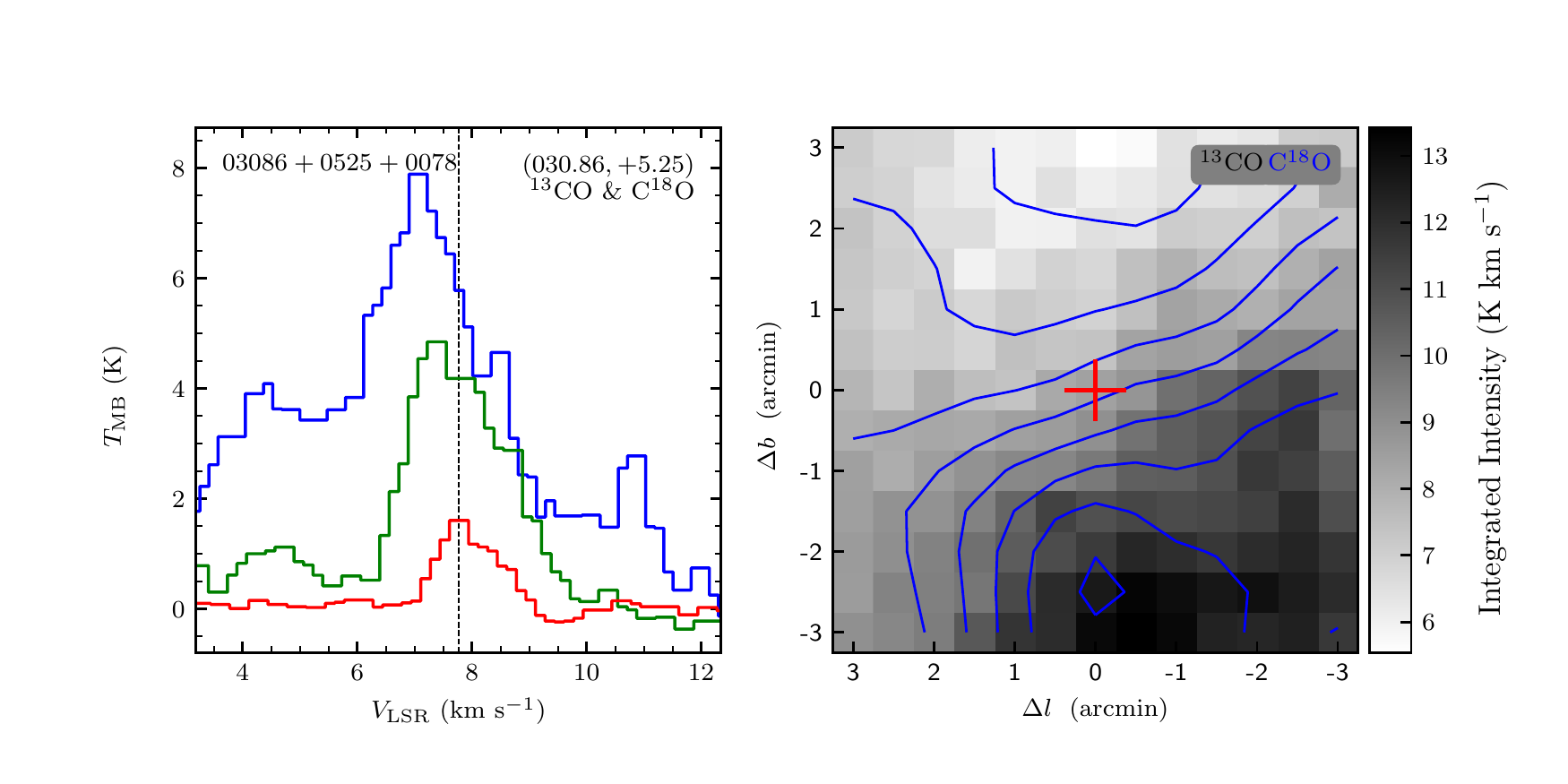}
\vspace{-0.5cm}

\includegraphics[width=9.0cm,angle=0]{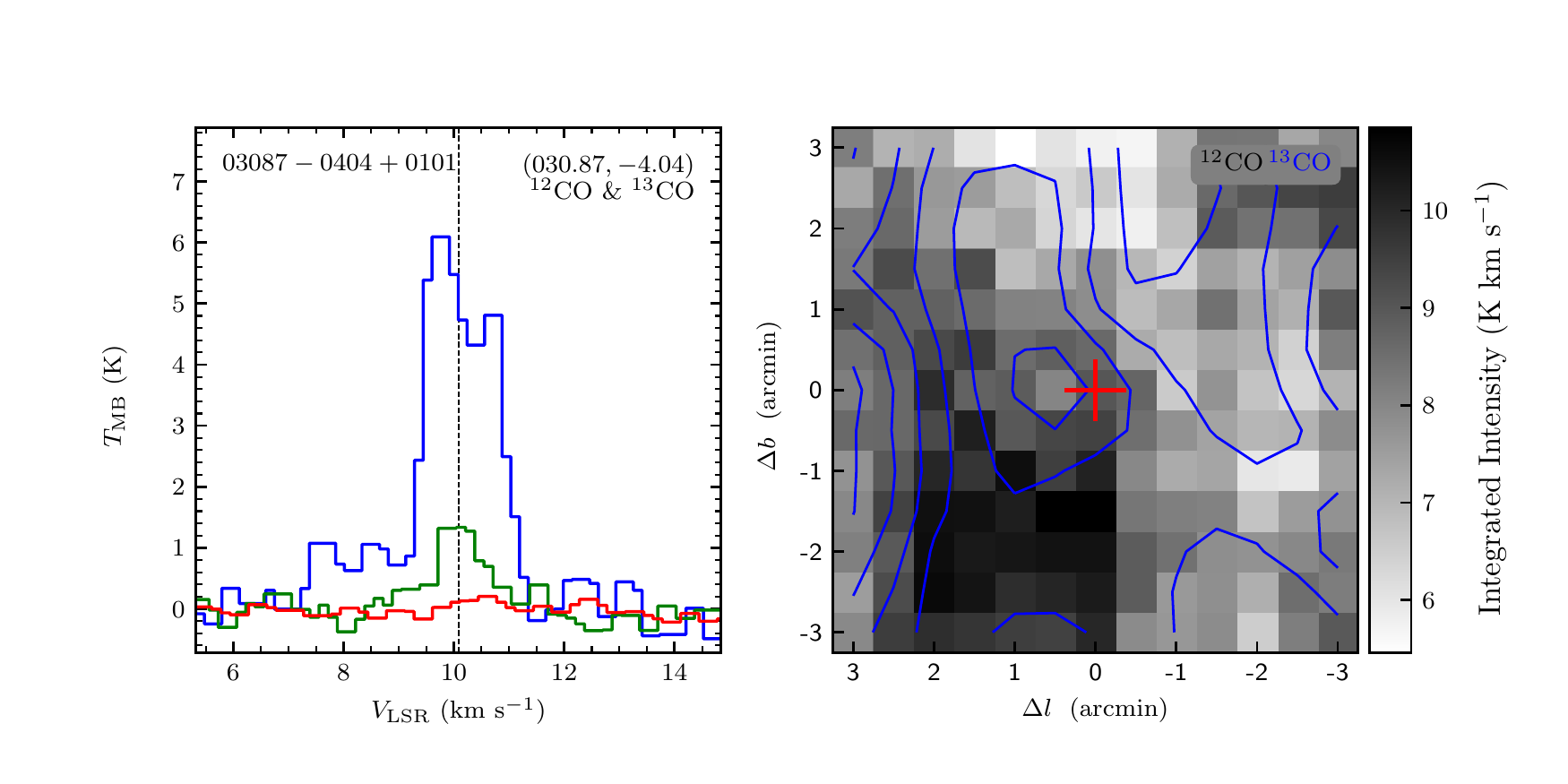}
\includegraphics[width=9.0cm,angle=0]{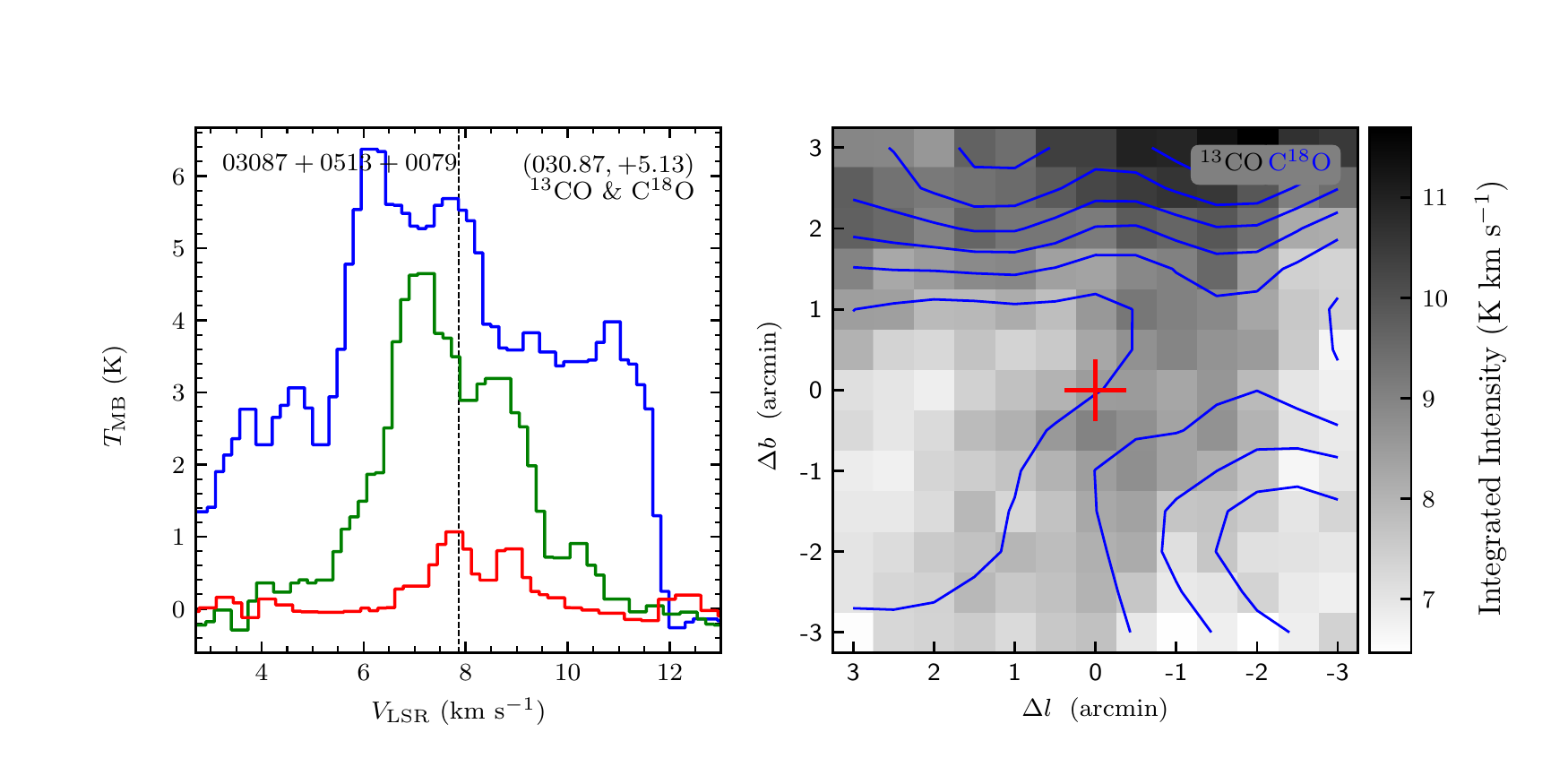}
\vspace{-0.5cm}

\includegraphics[width=9.0cm,angle=0]{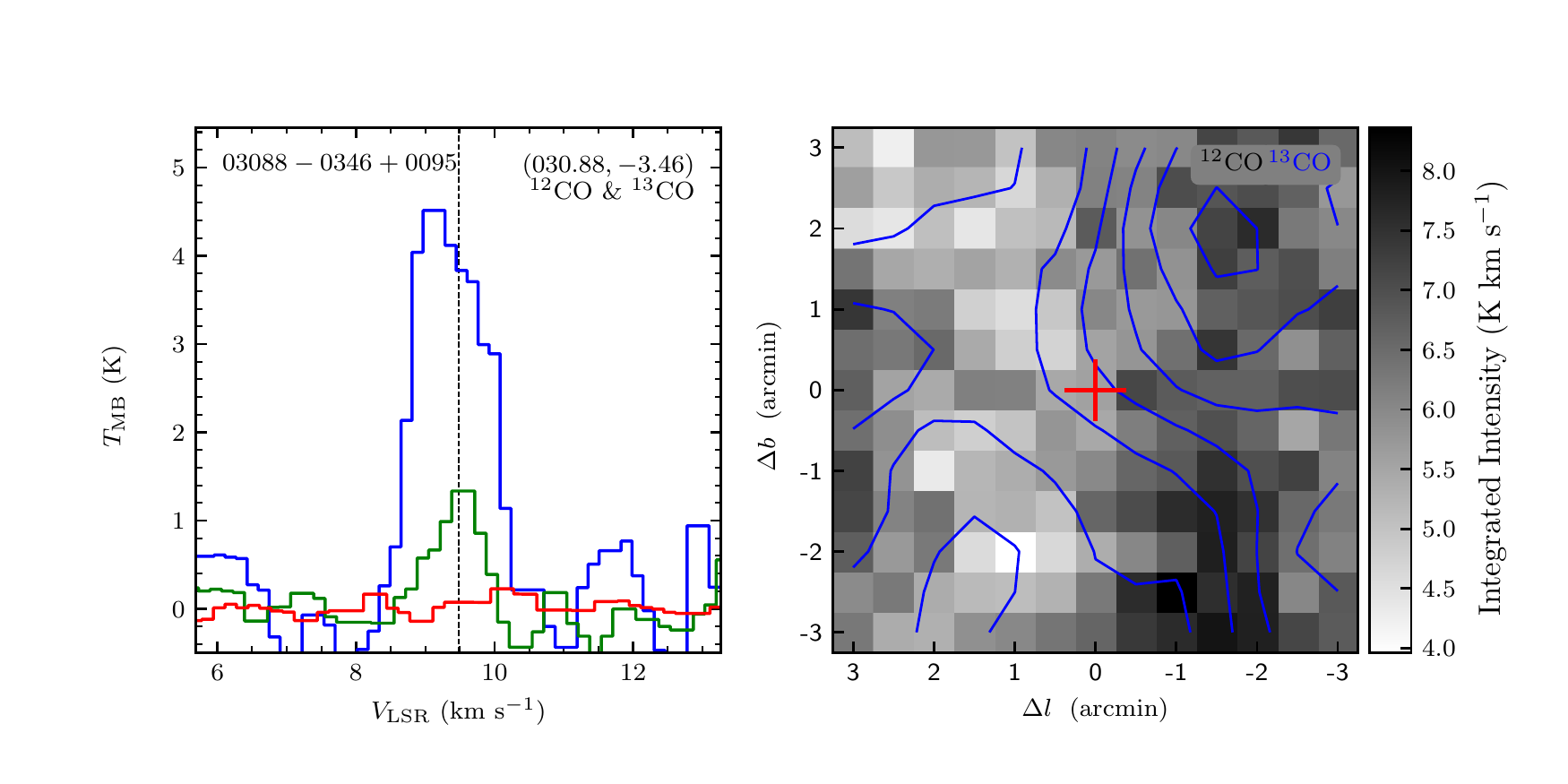}
\includegraphics[width=9.0cm,angle=0]{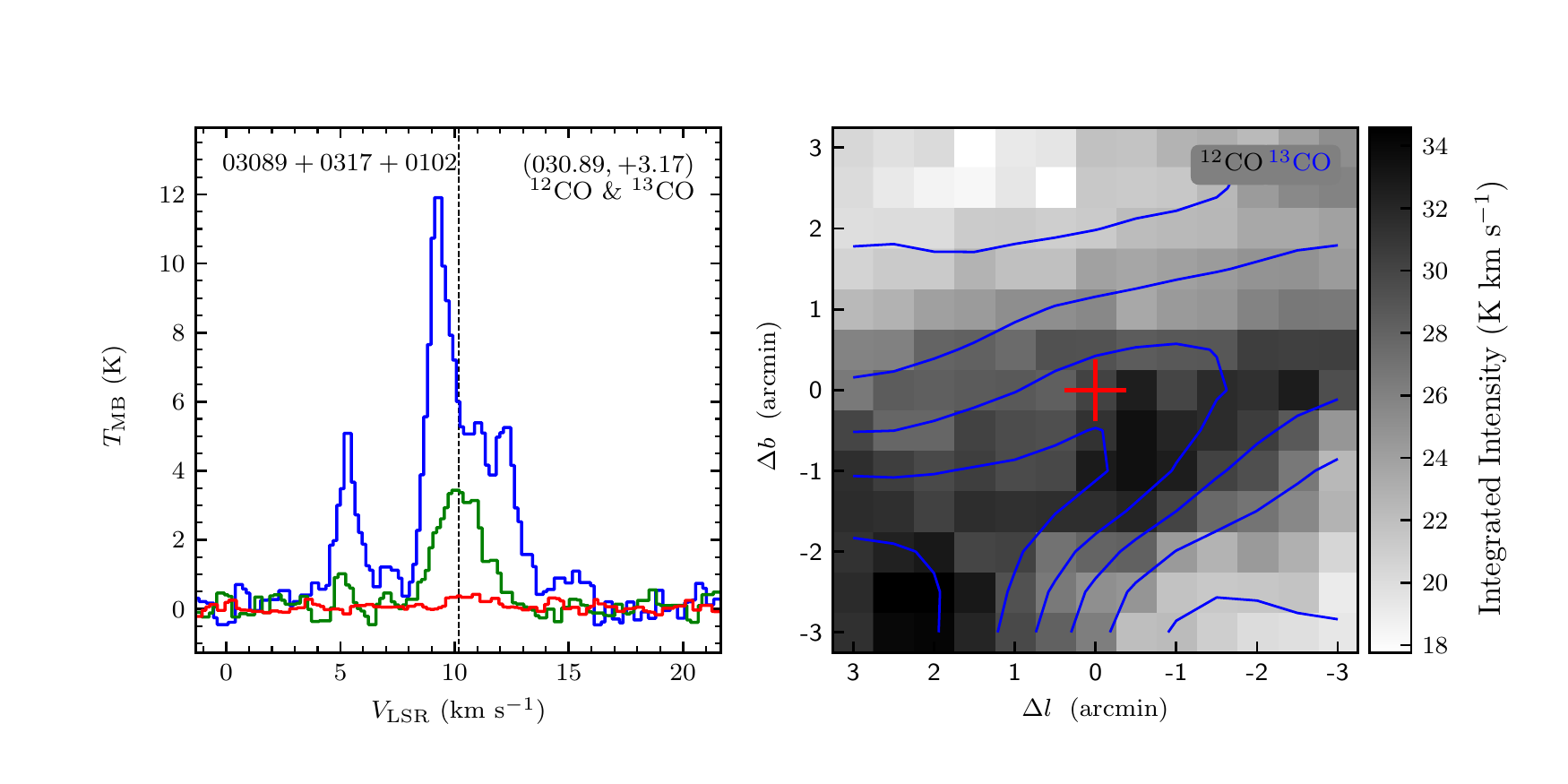}
\vspace{-0.5cm}

\includegraphics[width=9.0cm,angle=0]{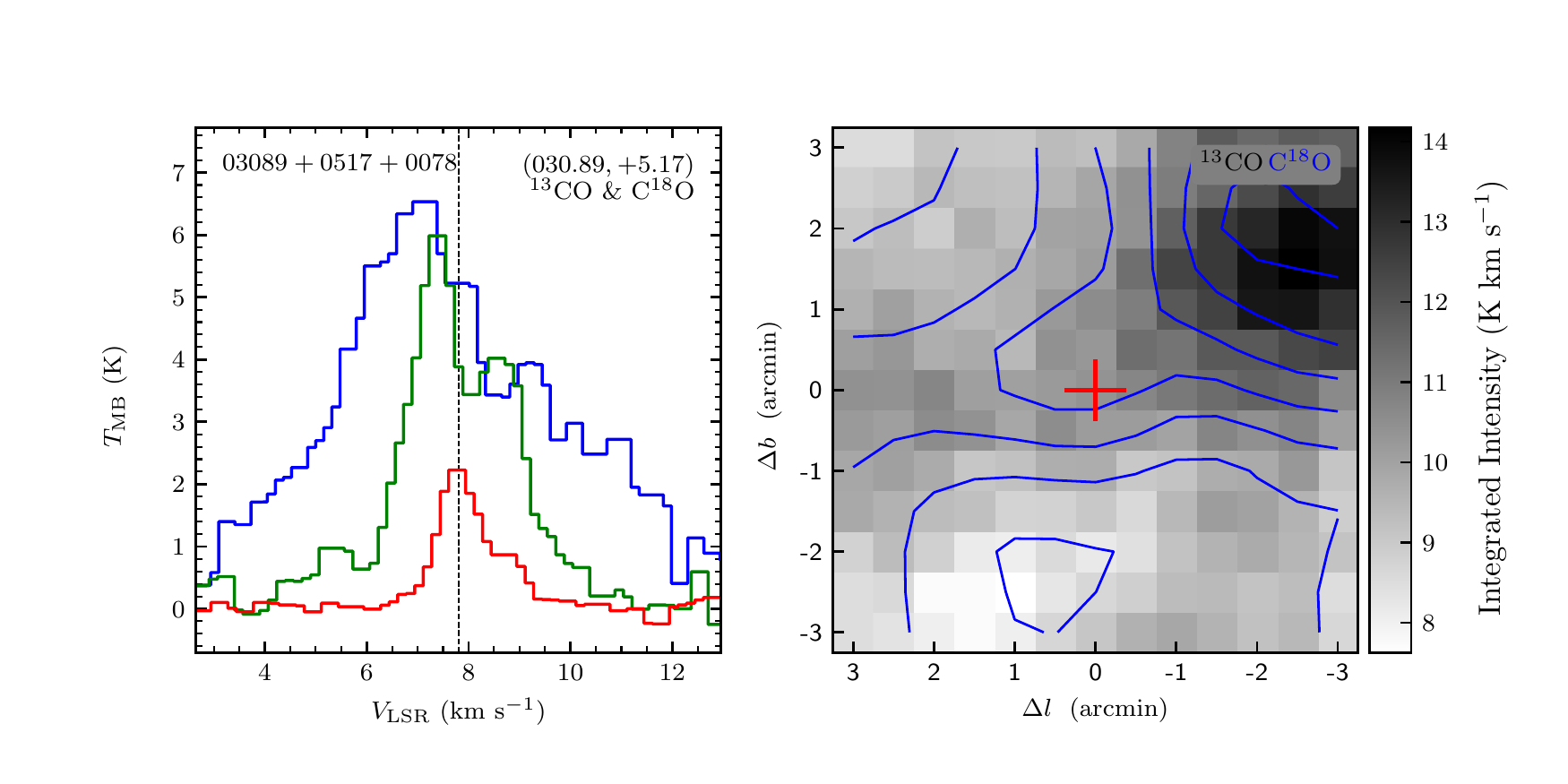}
\includegraphics[width=9.0cm,angle=0]{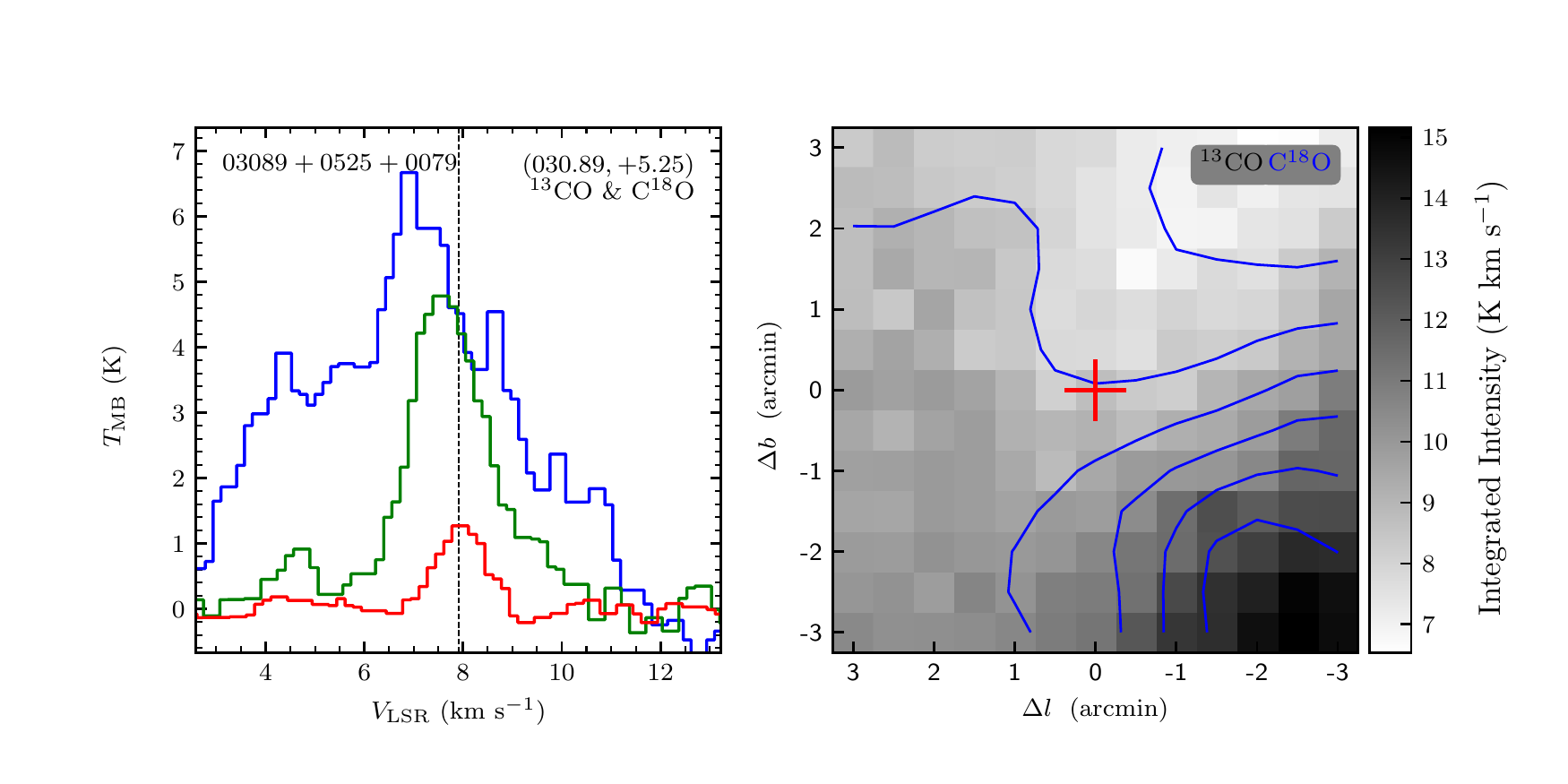}
\end{figure}
\clearpage

\begin{figure}
\includegraphics[width=9.0cm,angle=0]{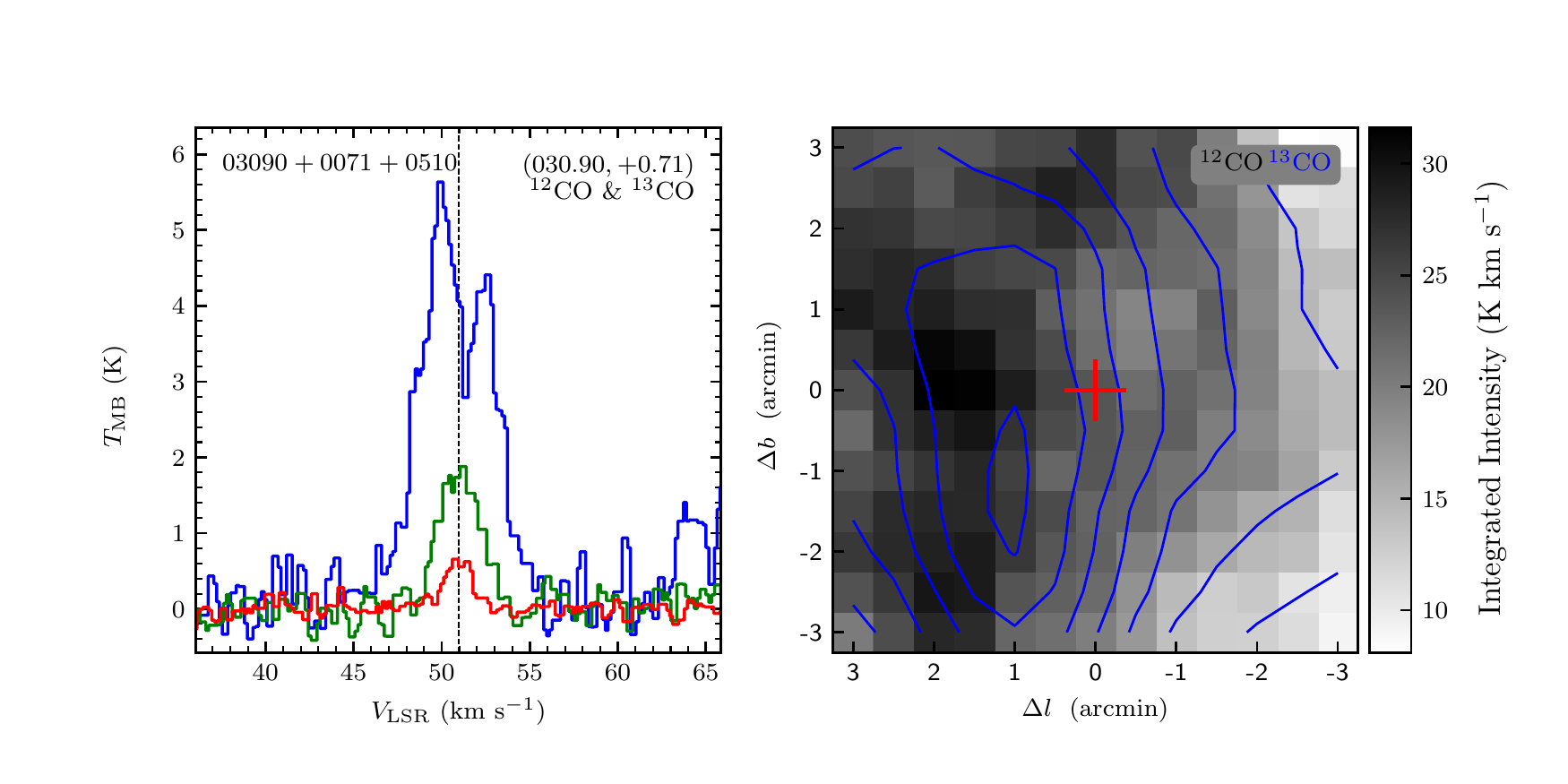}
\includegraphics[width=9.0cm,angle=0]{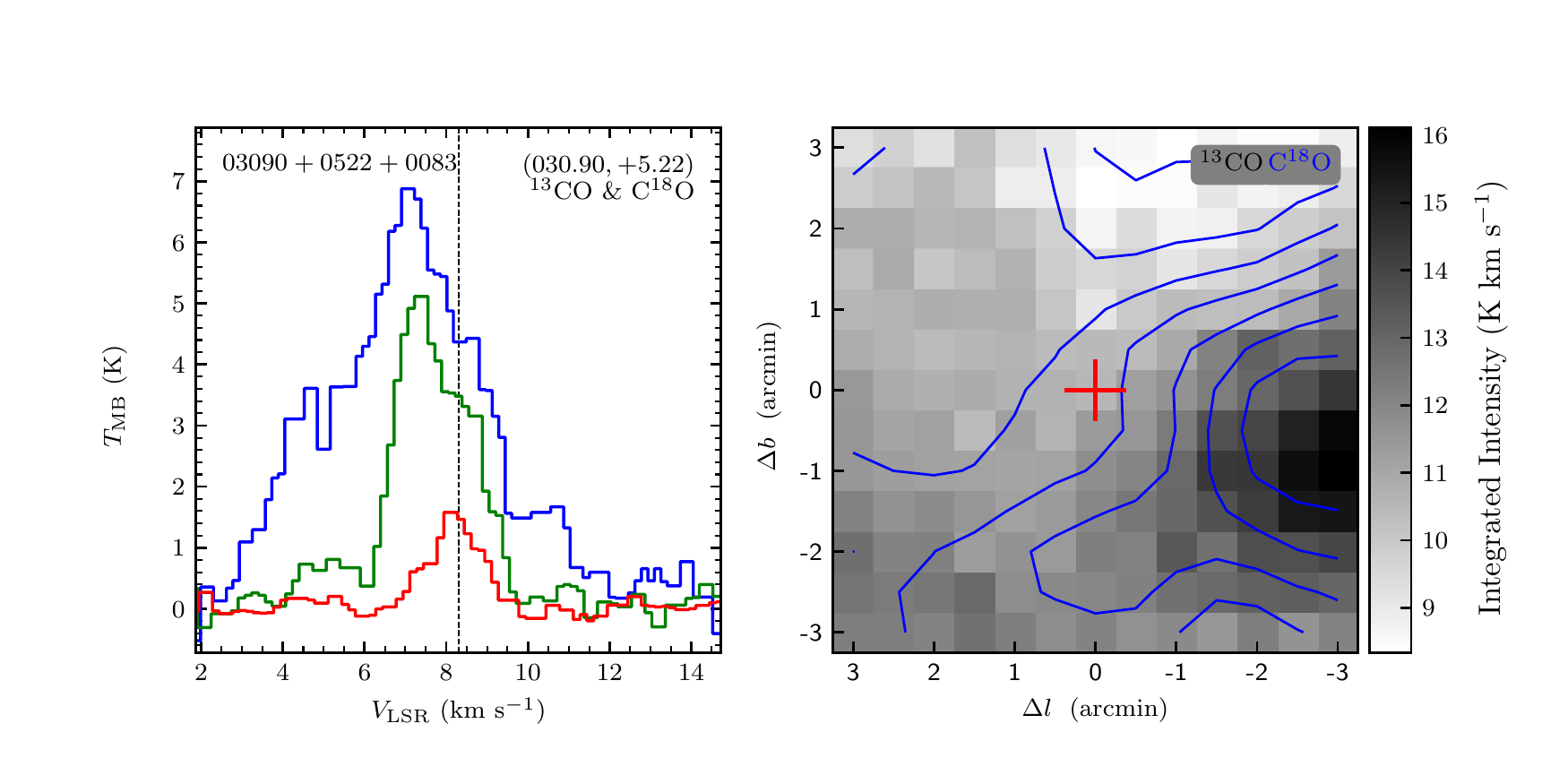}
\vspace{-0.5cm}

\includegraphics[width=9.0cm,angle=0]{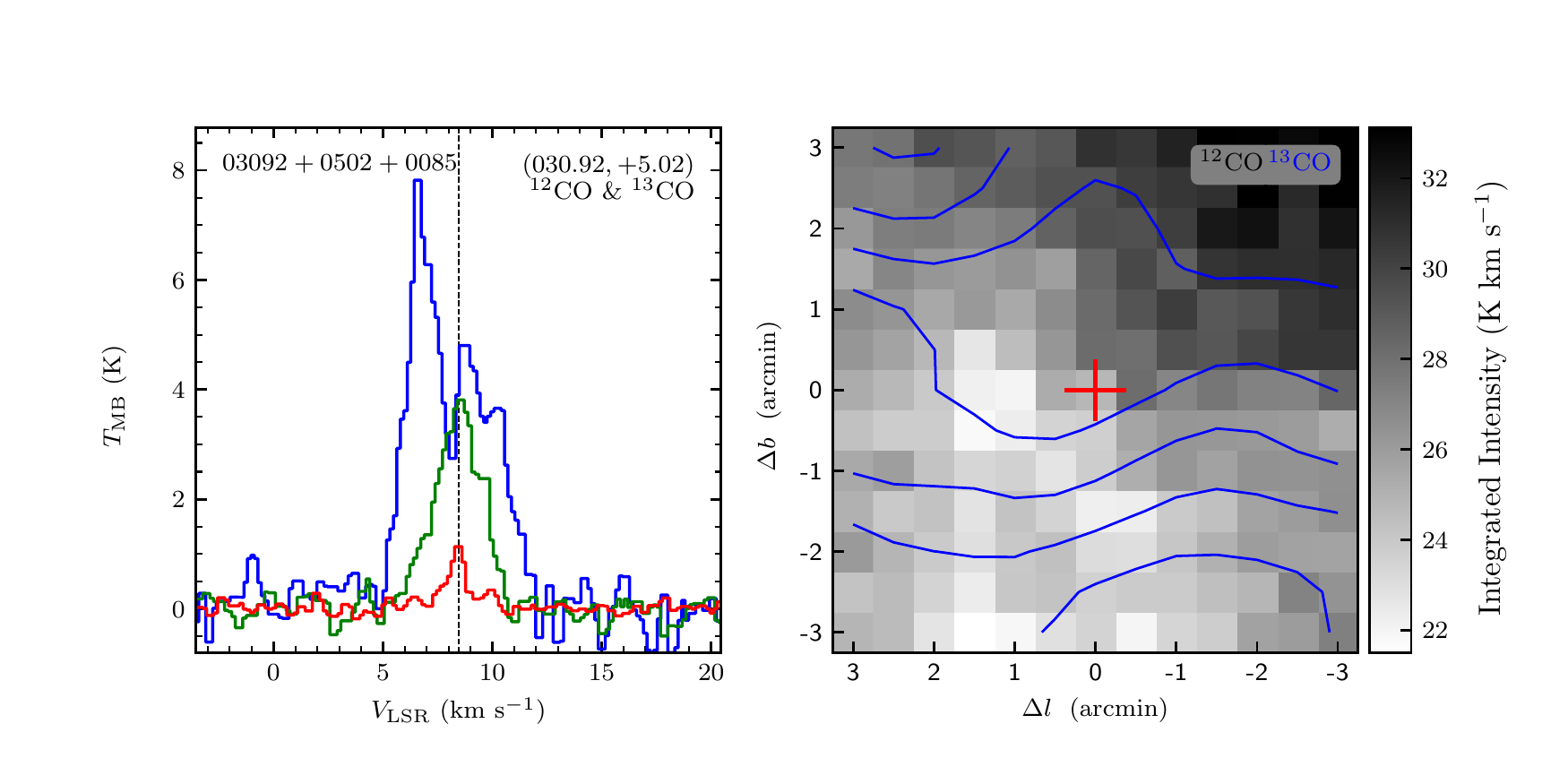}
\includegraphics[width=9.0cm,angle=0]{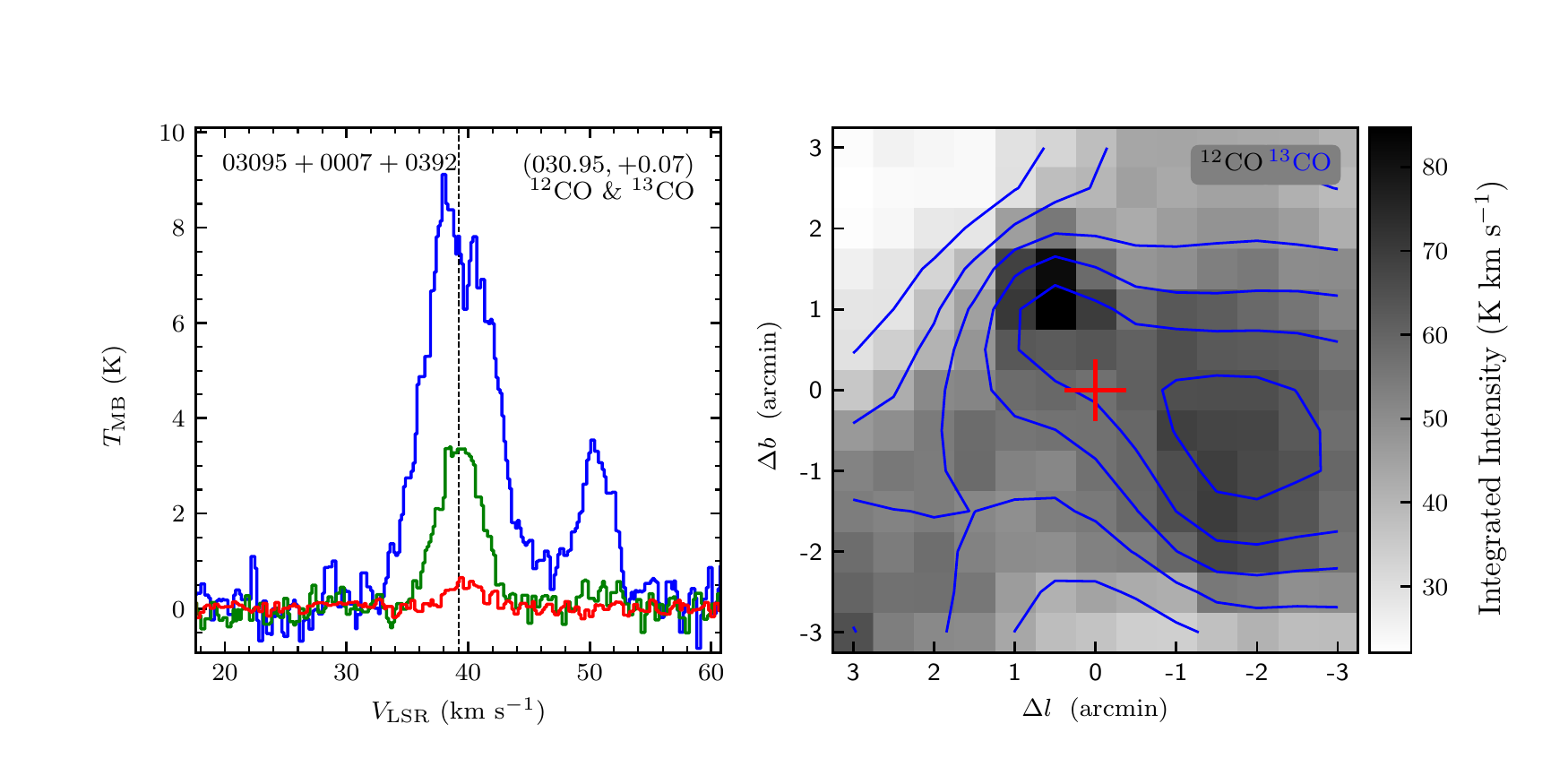}
\vspace{-0.5cm}

\includegraphics[width=9.0cm,angle=0]{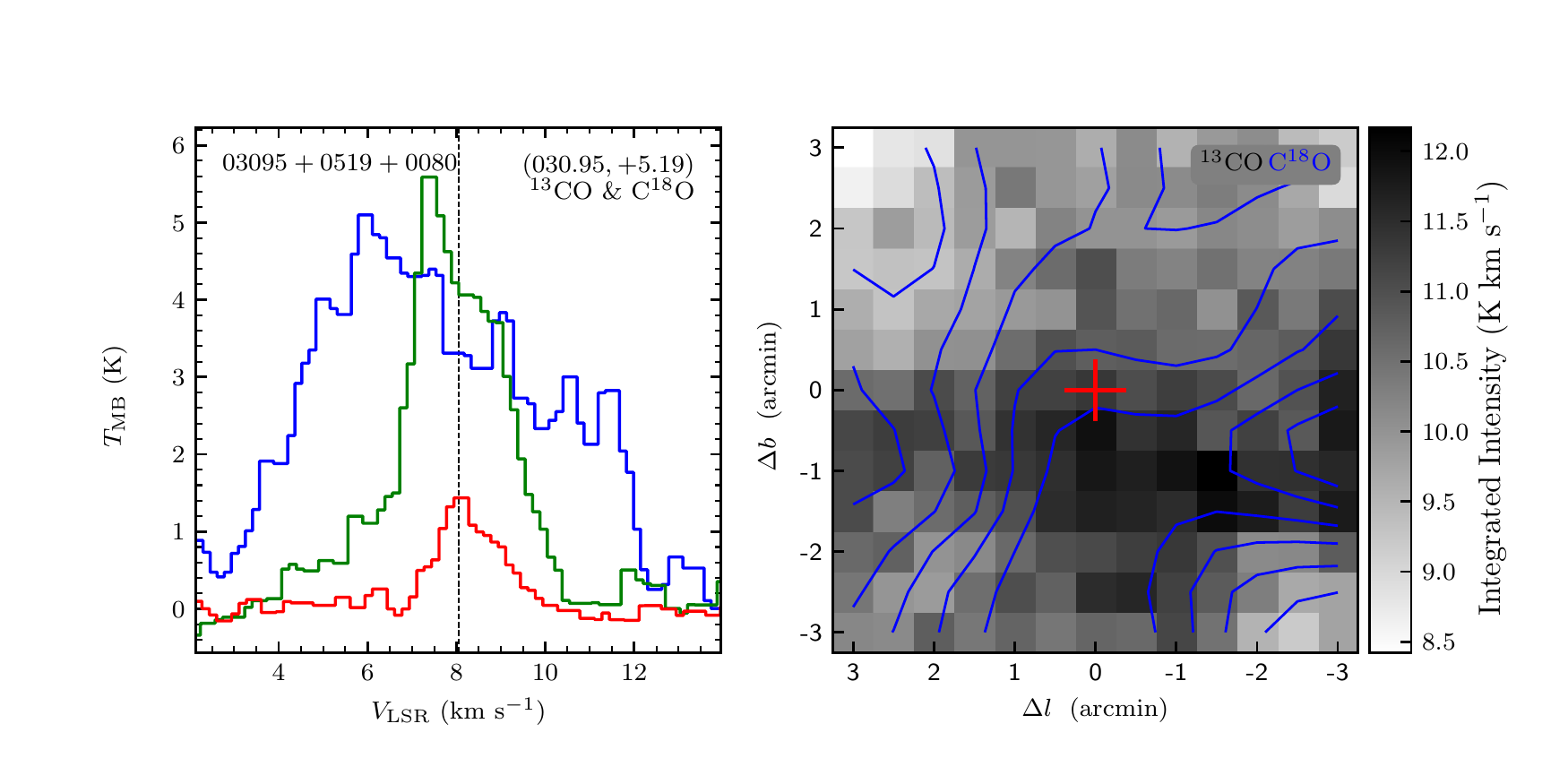}
\includegraphics[width=9.0cm,angle=0]{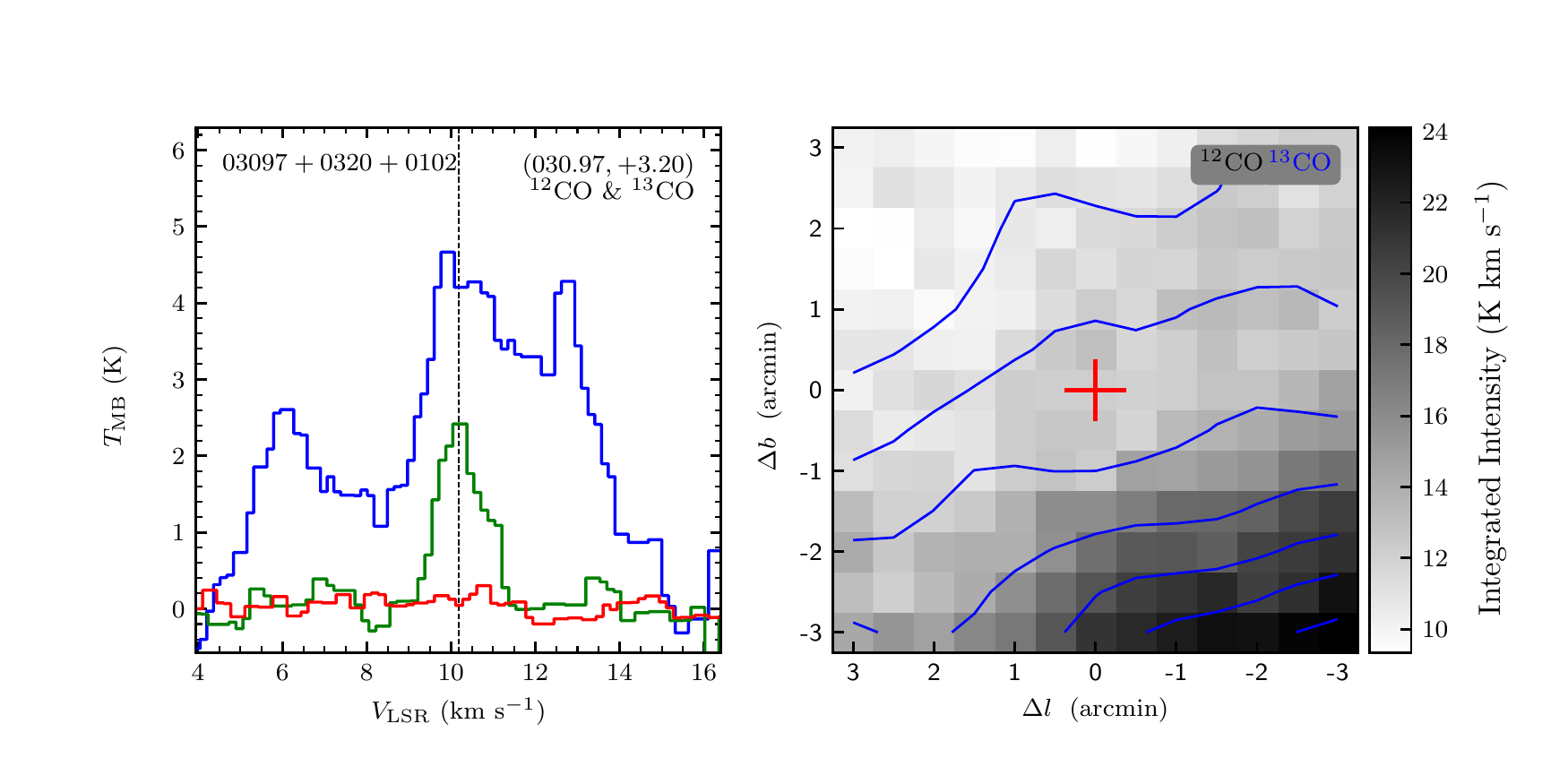}
\vspace{-0.5cm}

\includegraphics[width=9.0cm,angle=0]{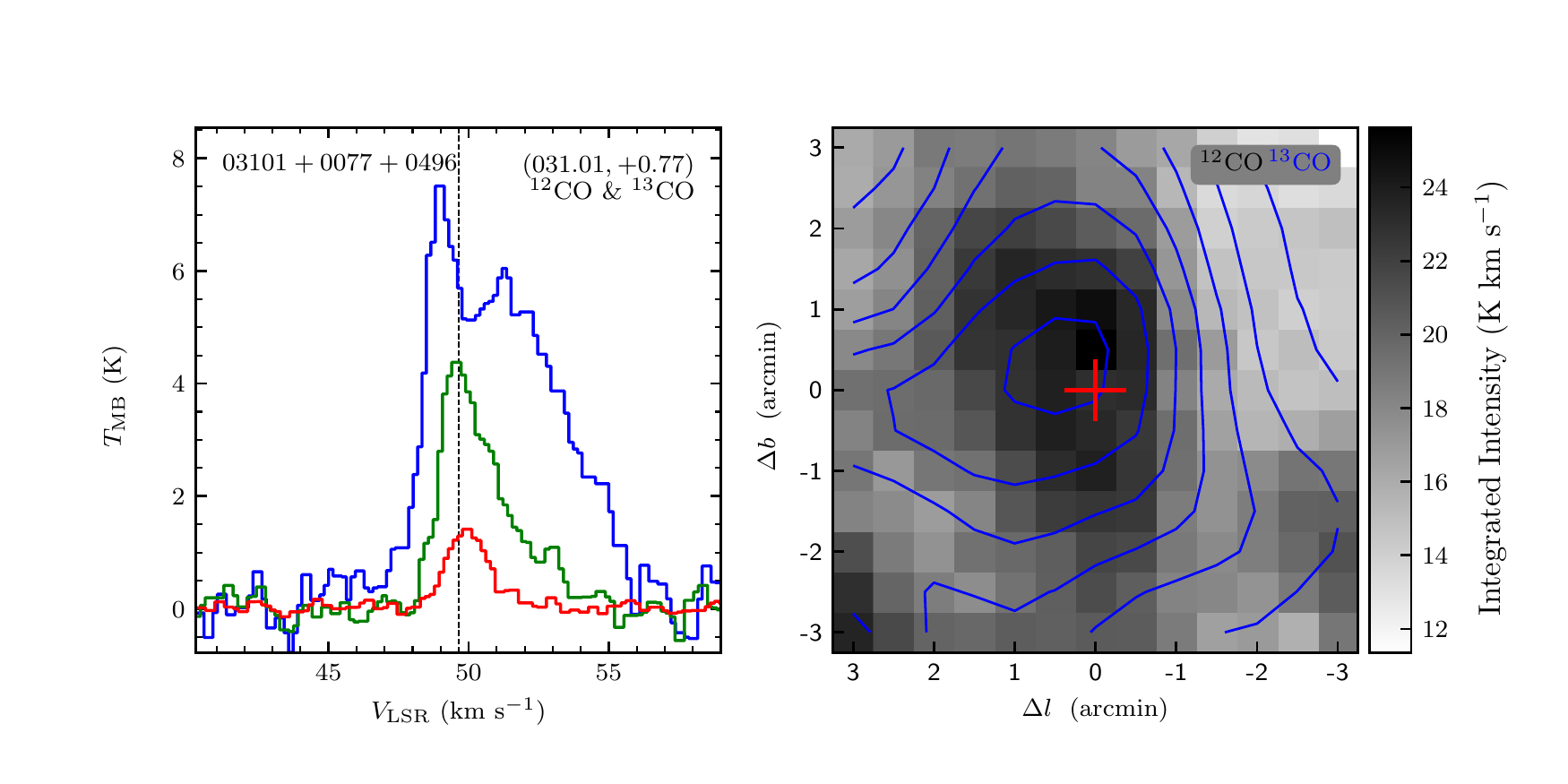}
\includegraphics[width=9.0cm,angle=0]{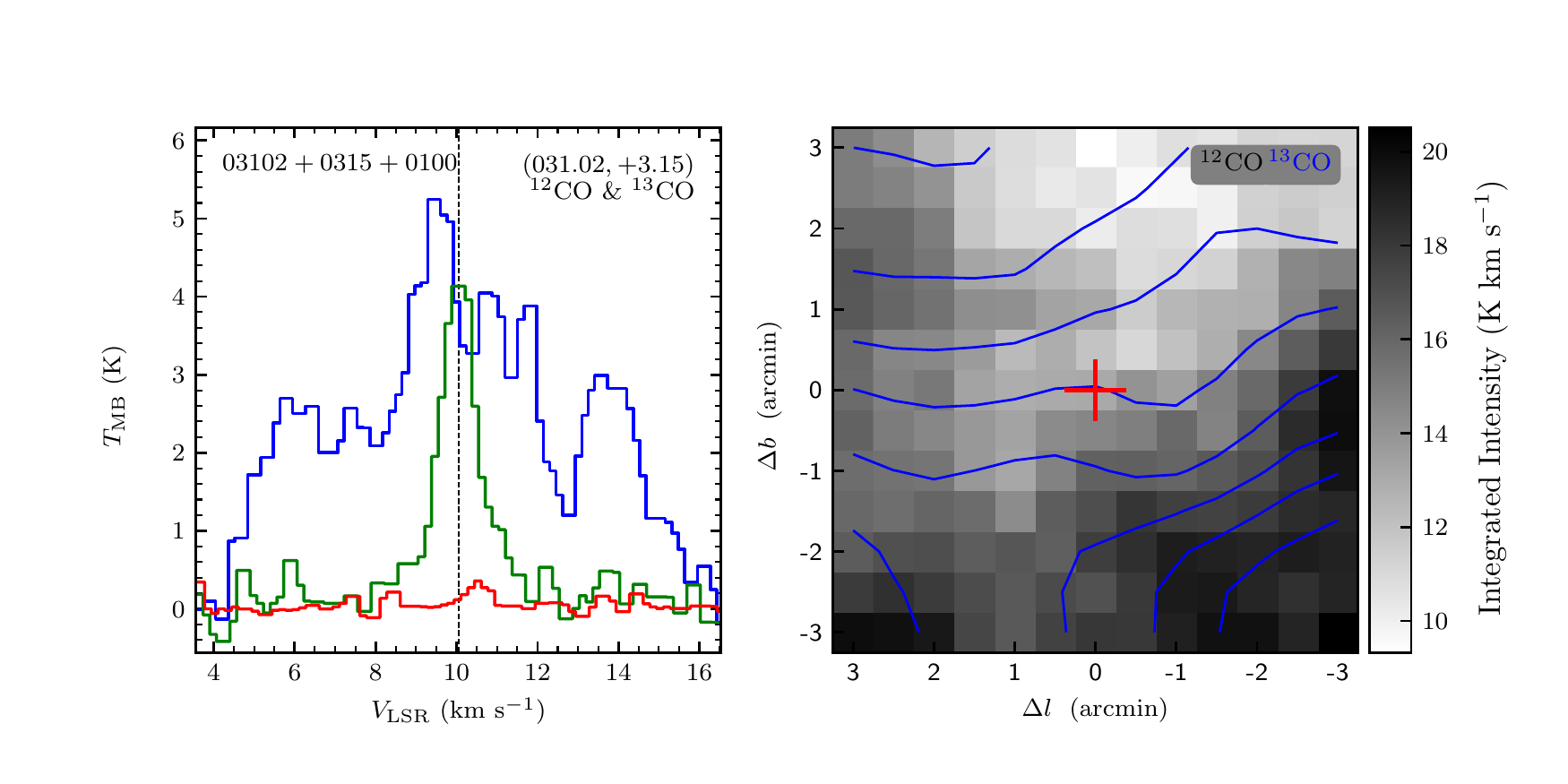}
\vspace{-0.5cm}

\includegraphics[width=9.0cm,angle=0]{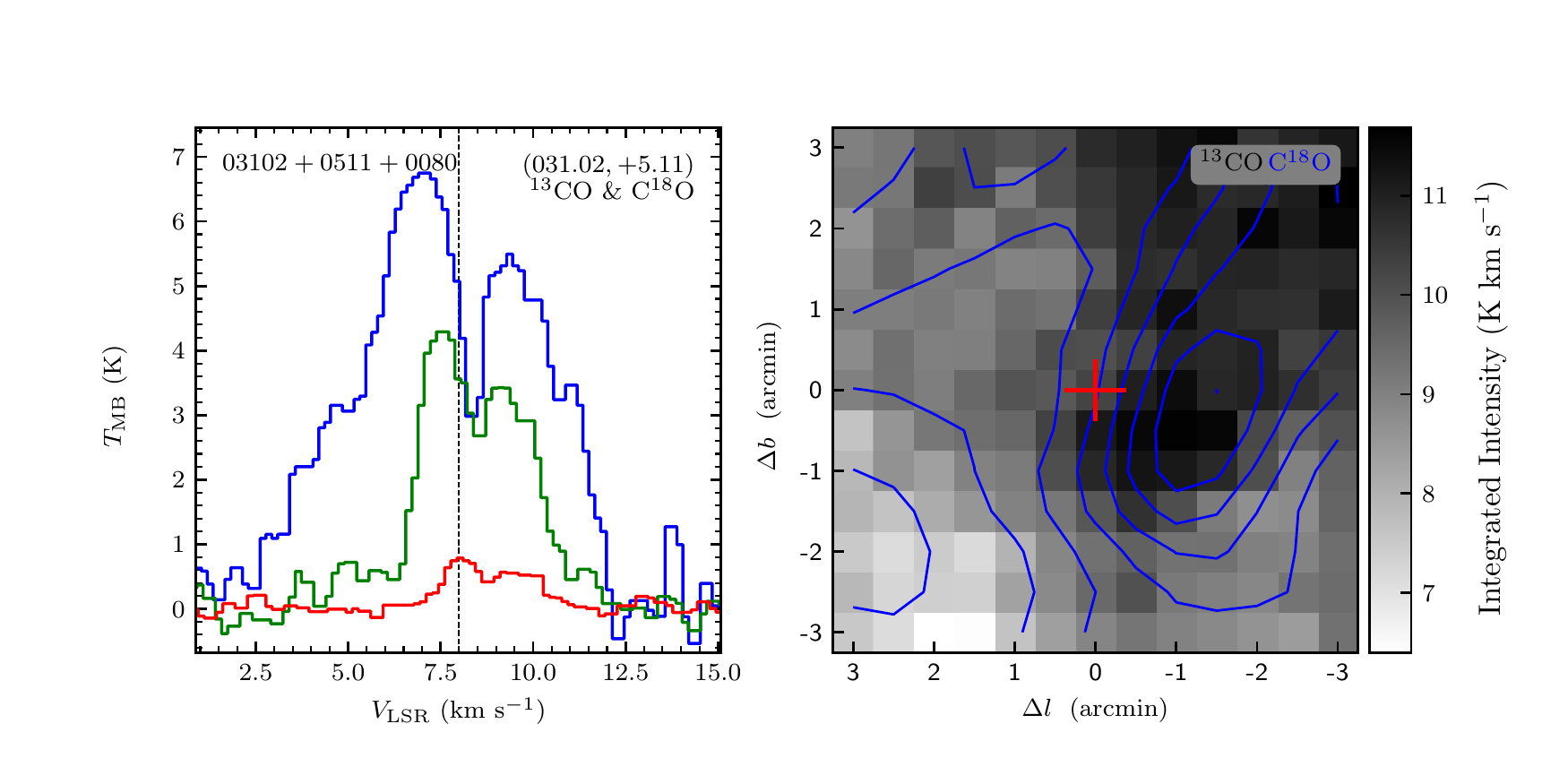}
\includegraphics[width=9.0cm,angle=0]{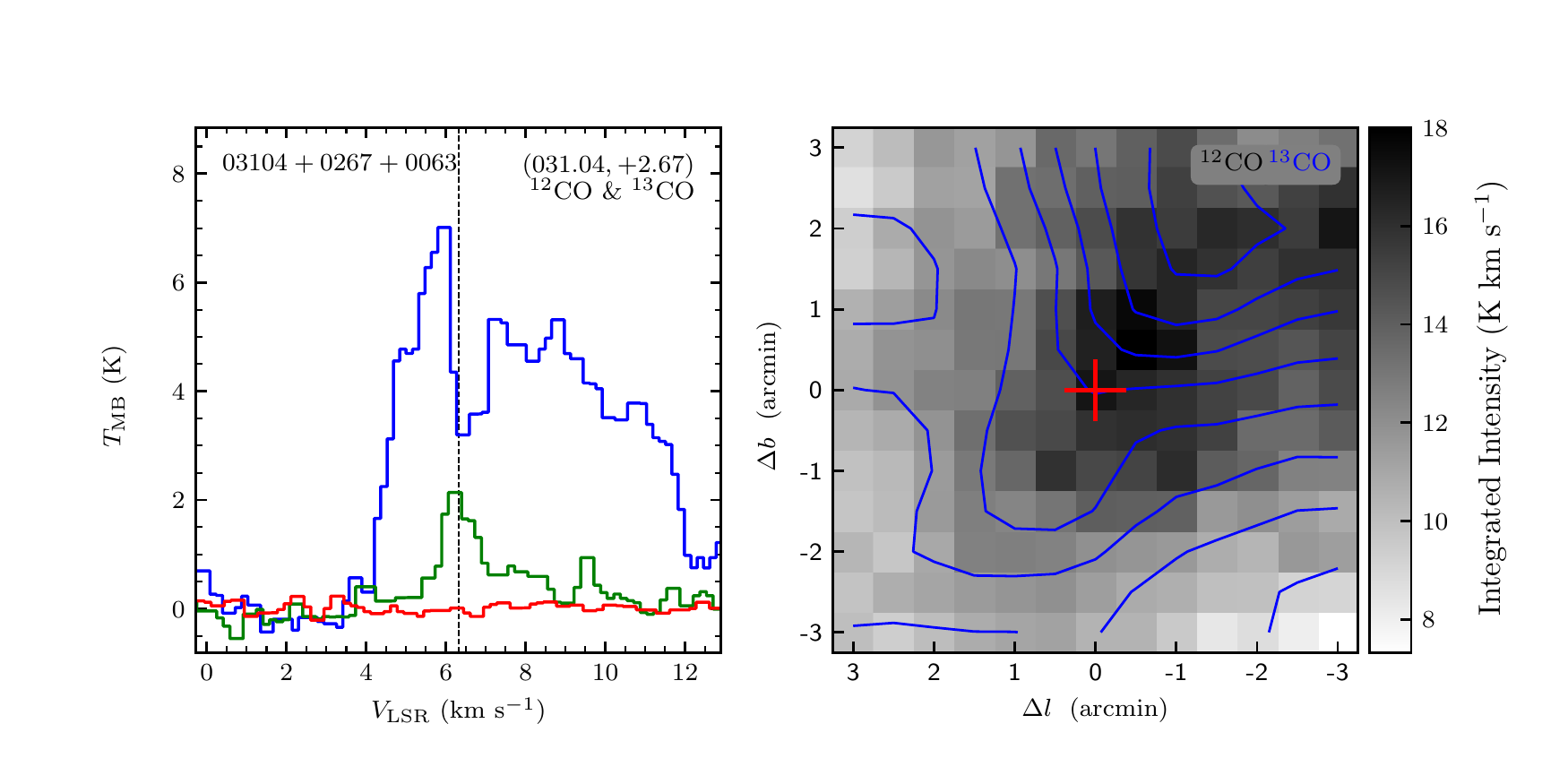}
\end{figure}
\clearpage

\begin{figure}
\includegraphics[width=9.0cm,angle=0]{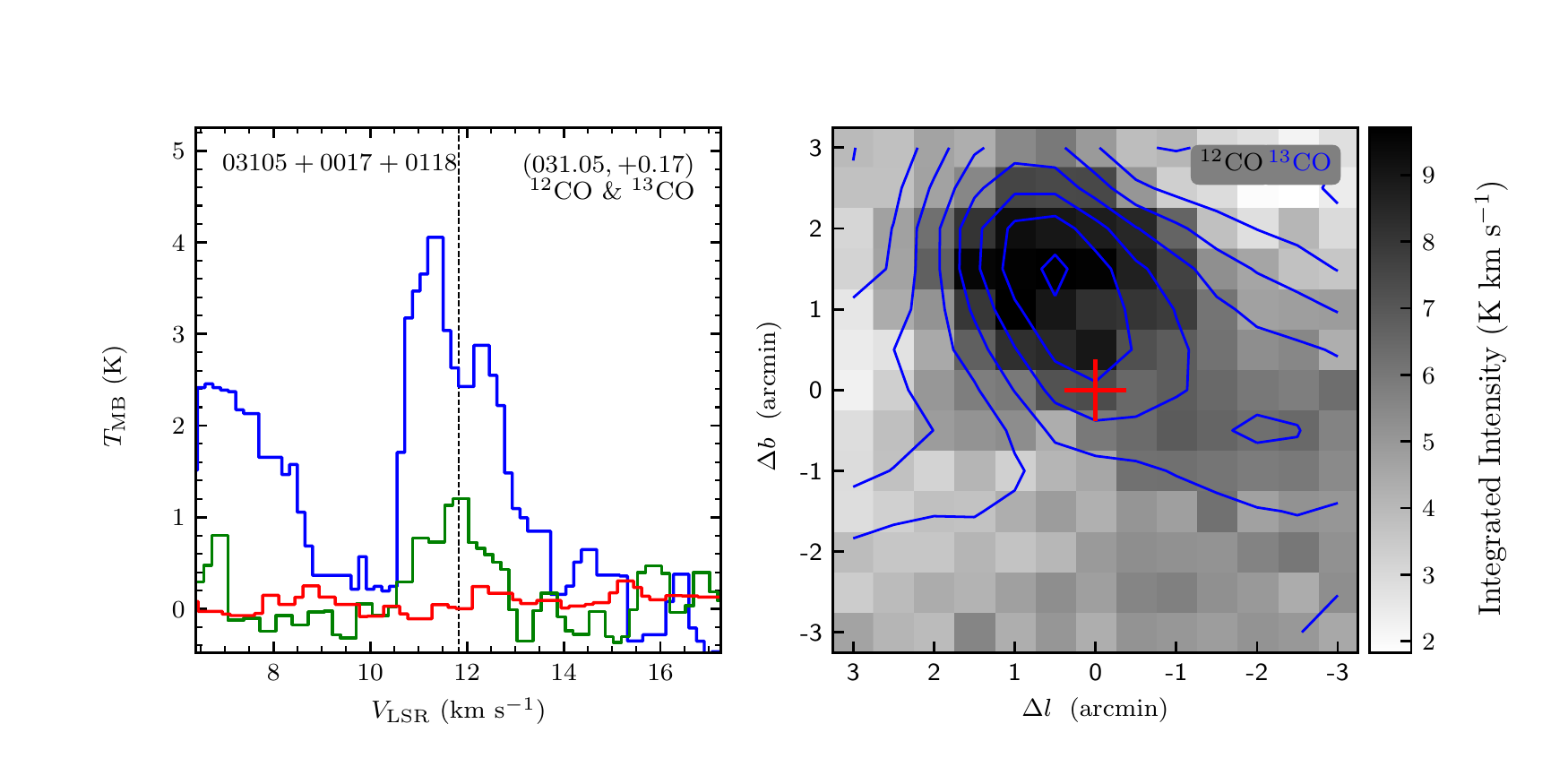}
\includegraphics[width=9.0cm,angle=0]{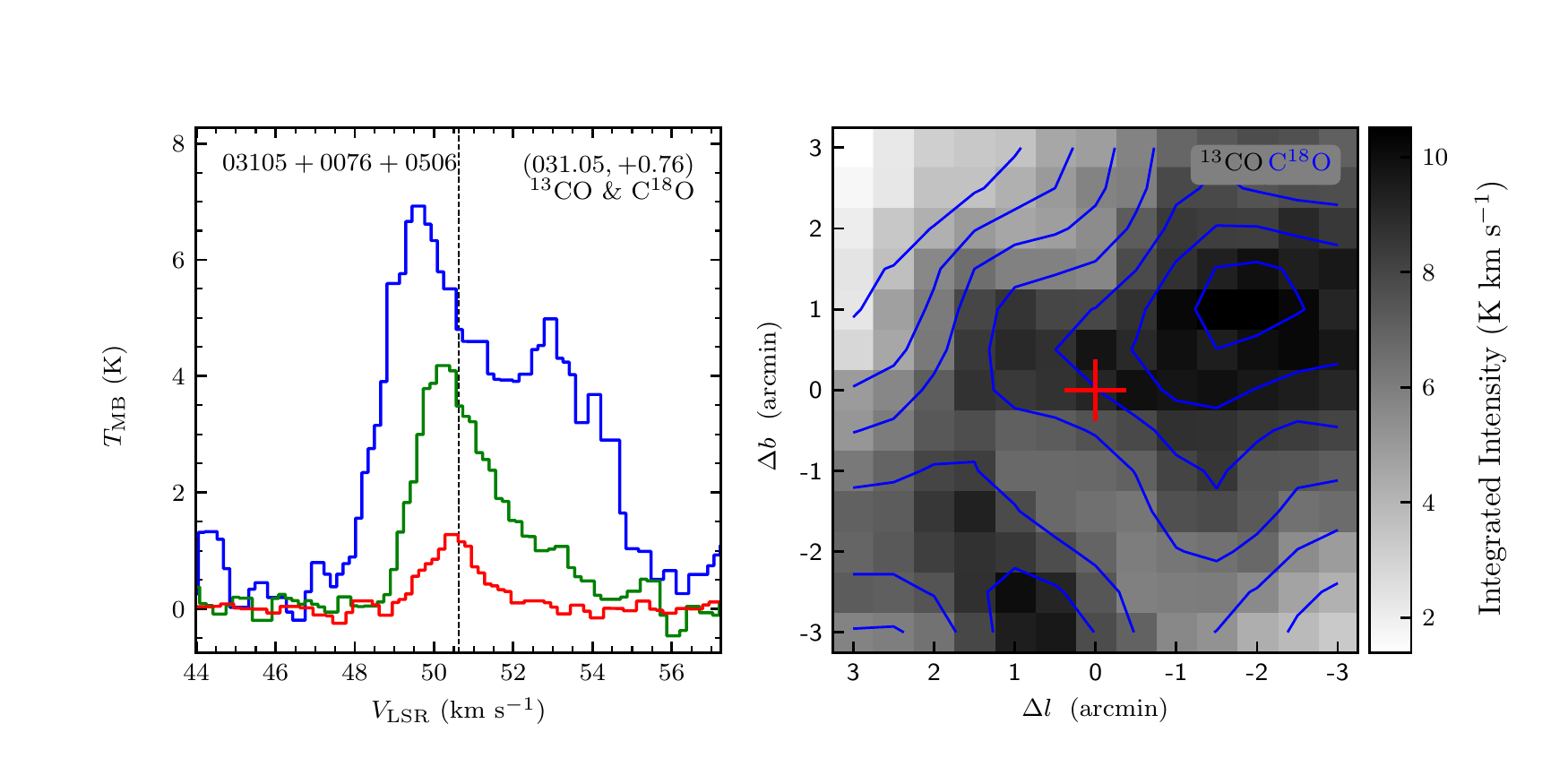}
\vspace{-0.5cm}

\includegraphics[width=9.0cm,angle=0]{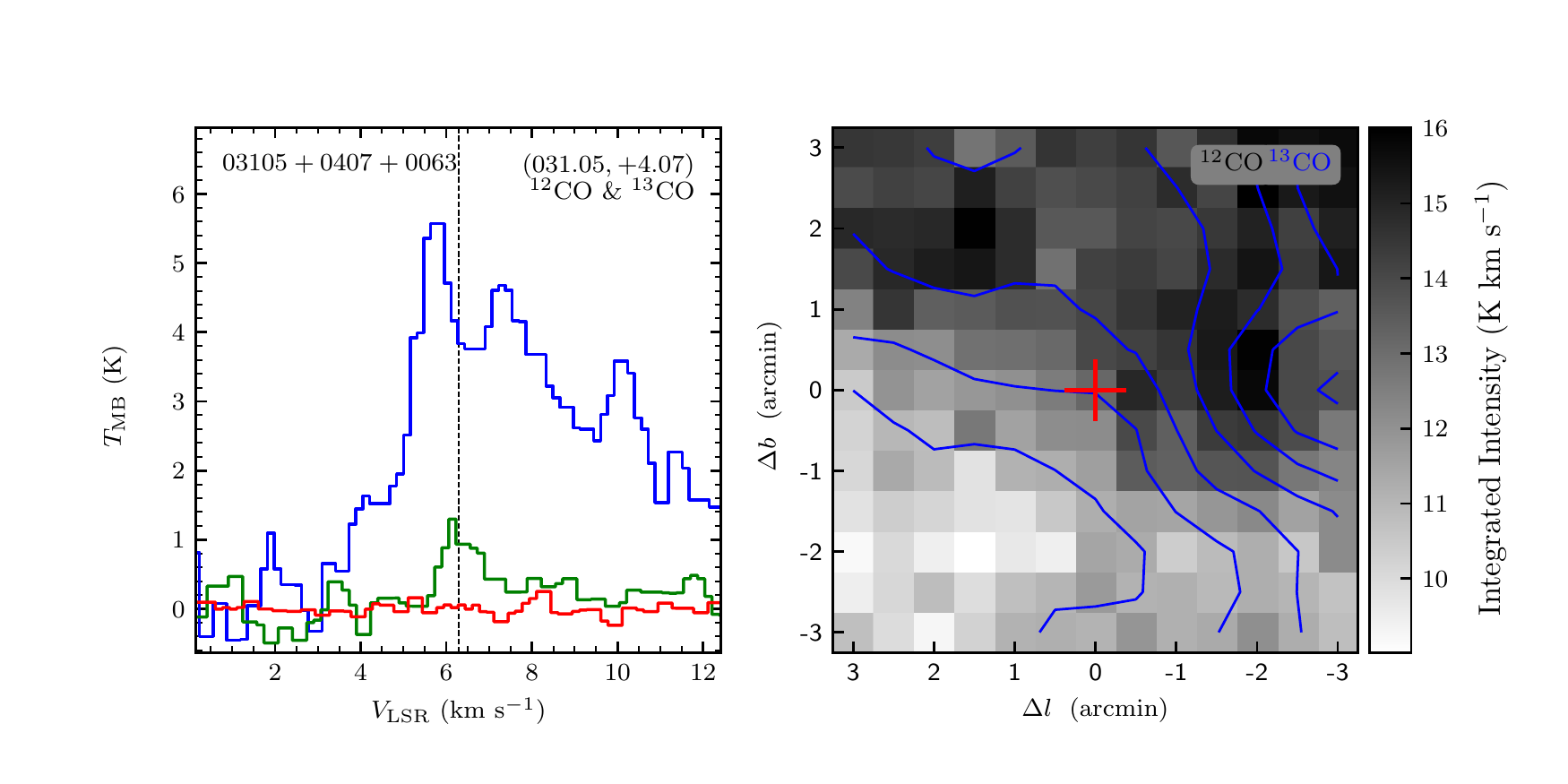}
\includegraphics[width=9.0cm,angle=0]{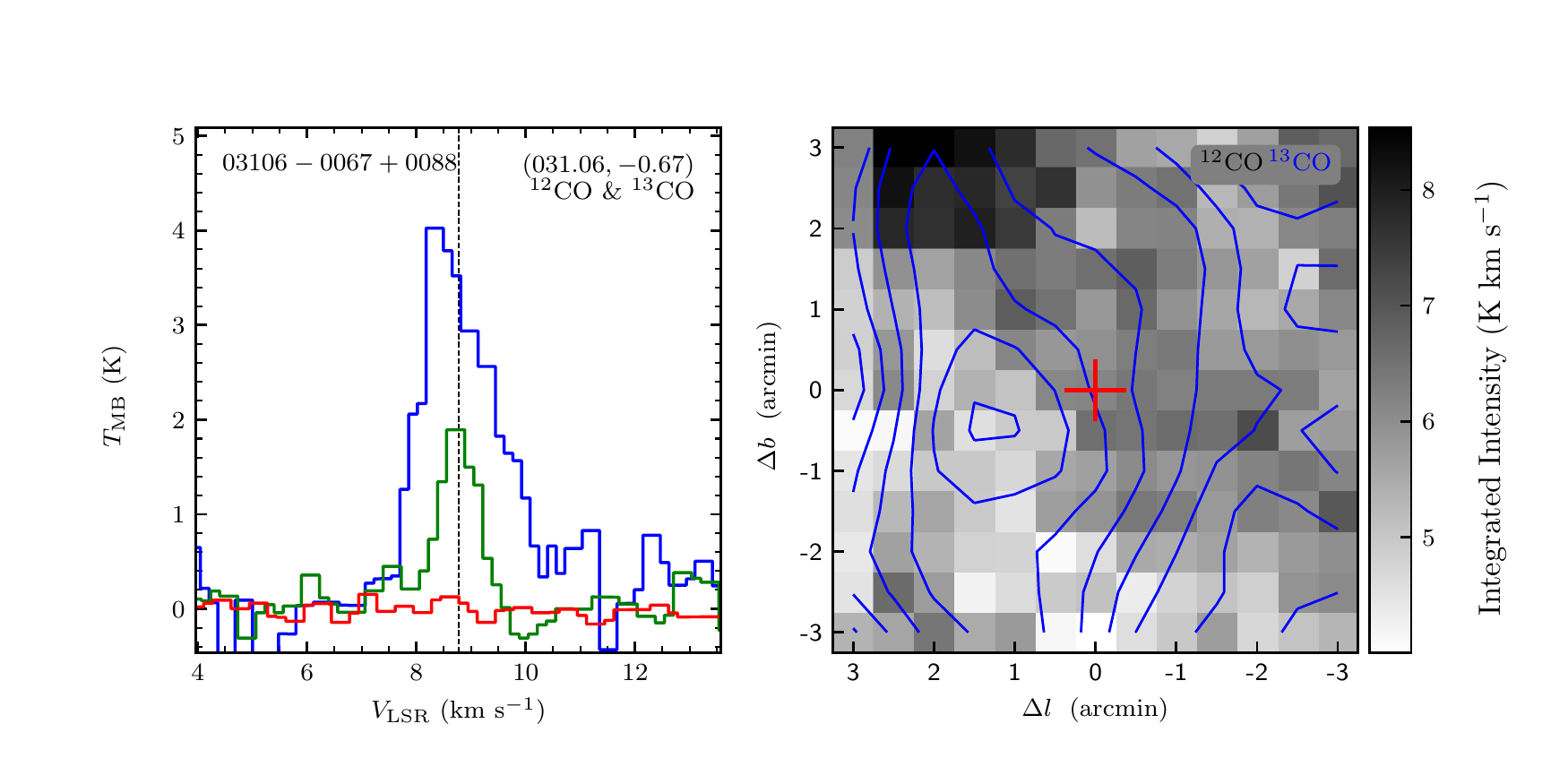}
\vspace{-0.5cm}

\includegraphics[width=9.0cm,angle=0]{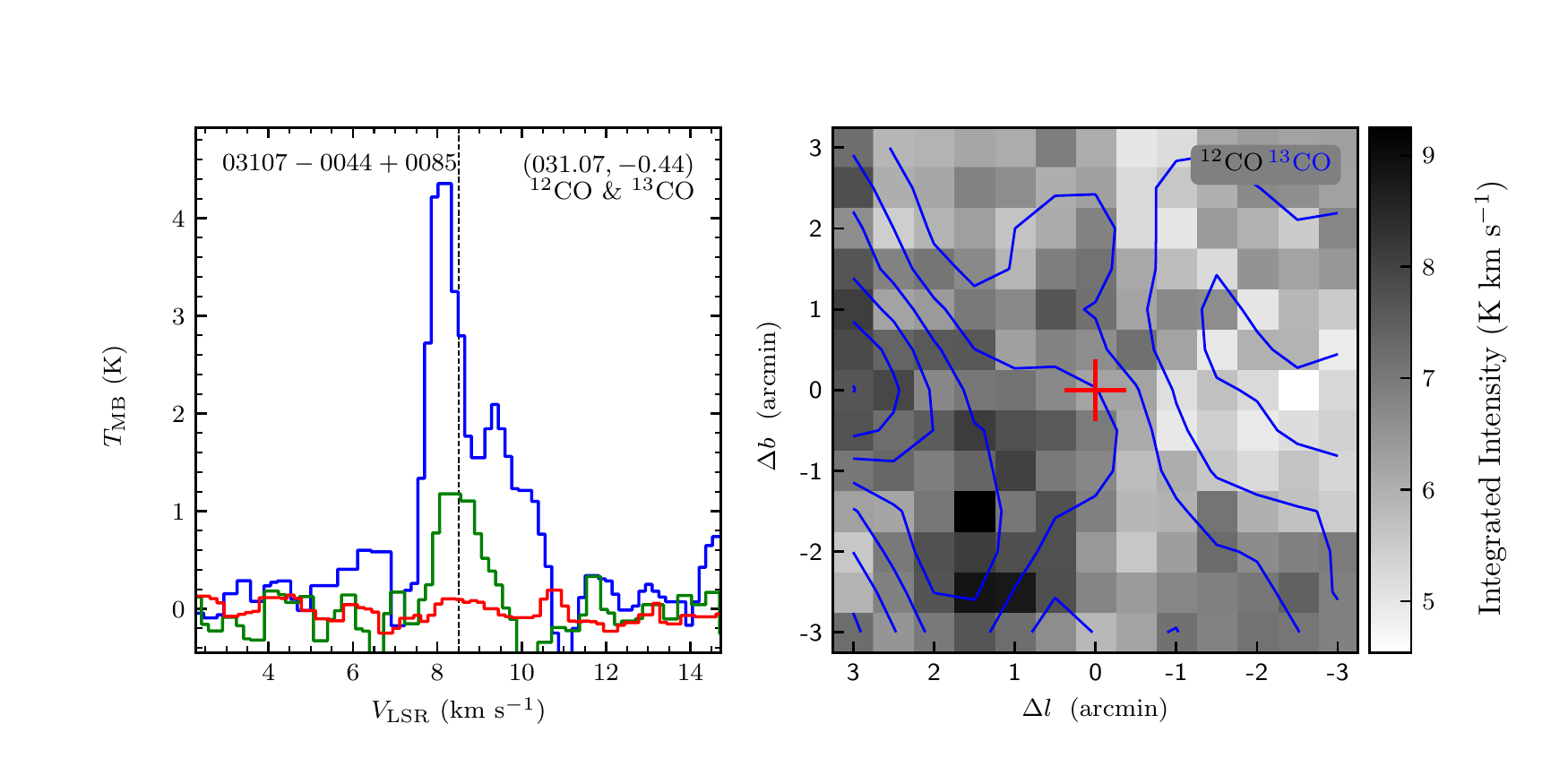}
\includegraphics[width=9.0cm,angle=0]{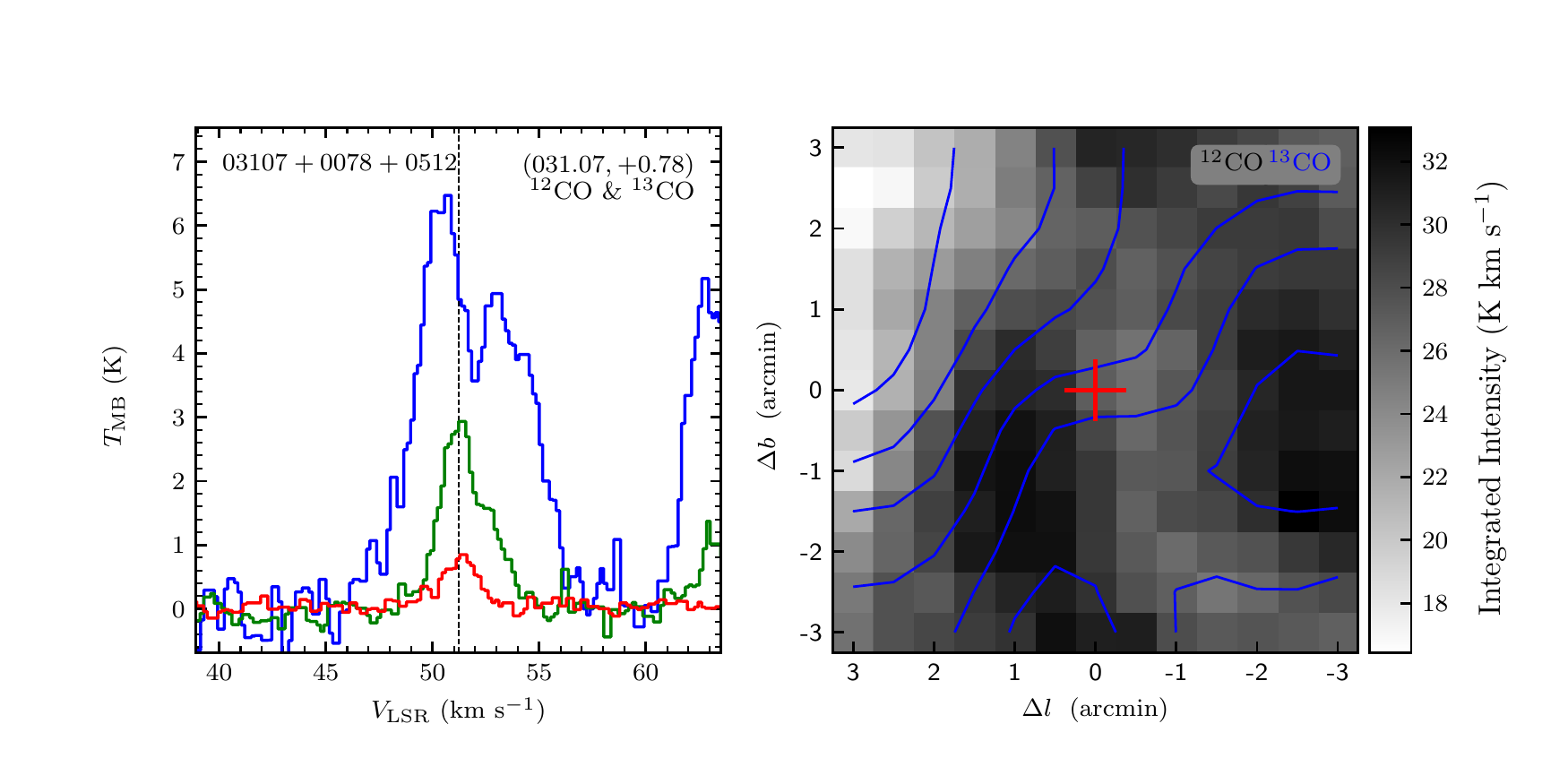}
\vspace{-0.5cm}

\includegraphics[width=9.0cm,angle=0]{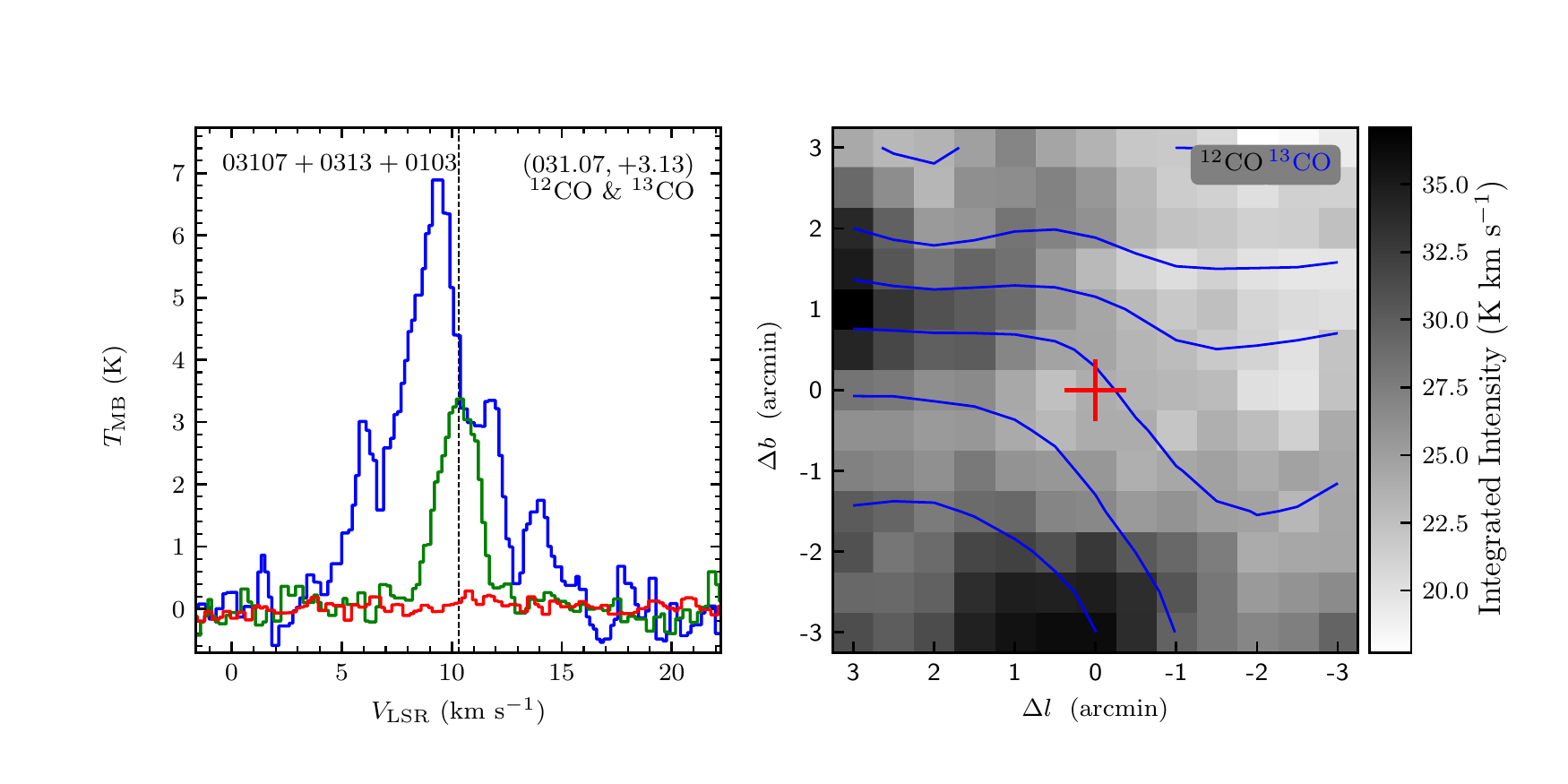}
\includegraphics[width=9.0cm,angle=0]{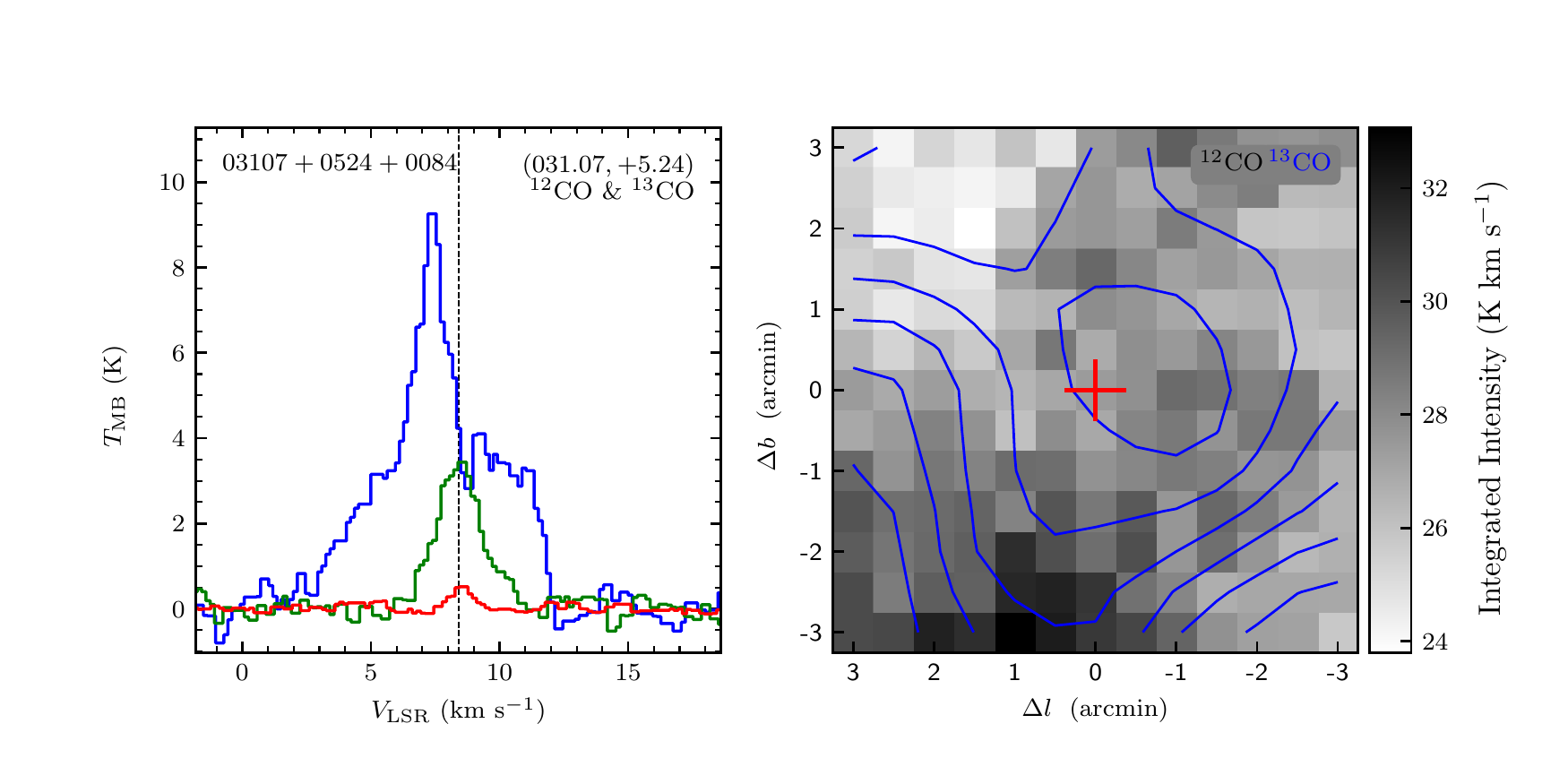}
\vspace{-0.5cm}

\includegraphics[width=9.0cm,angle=0]{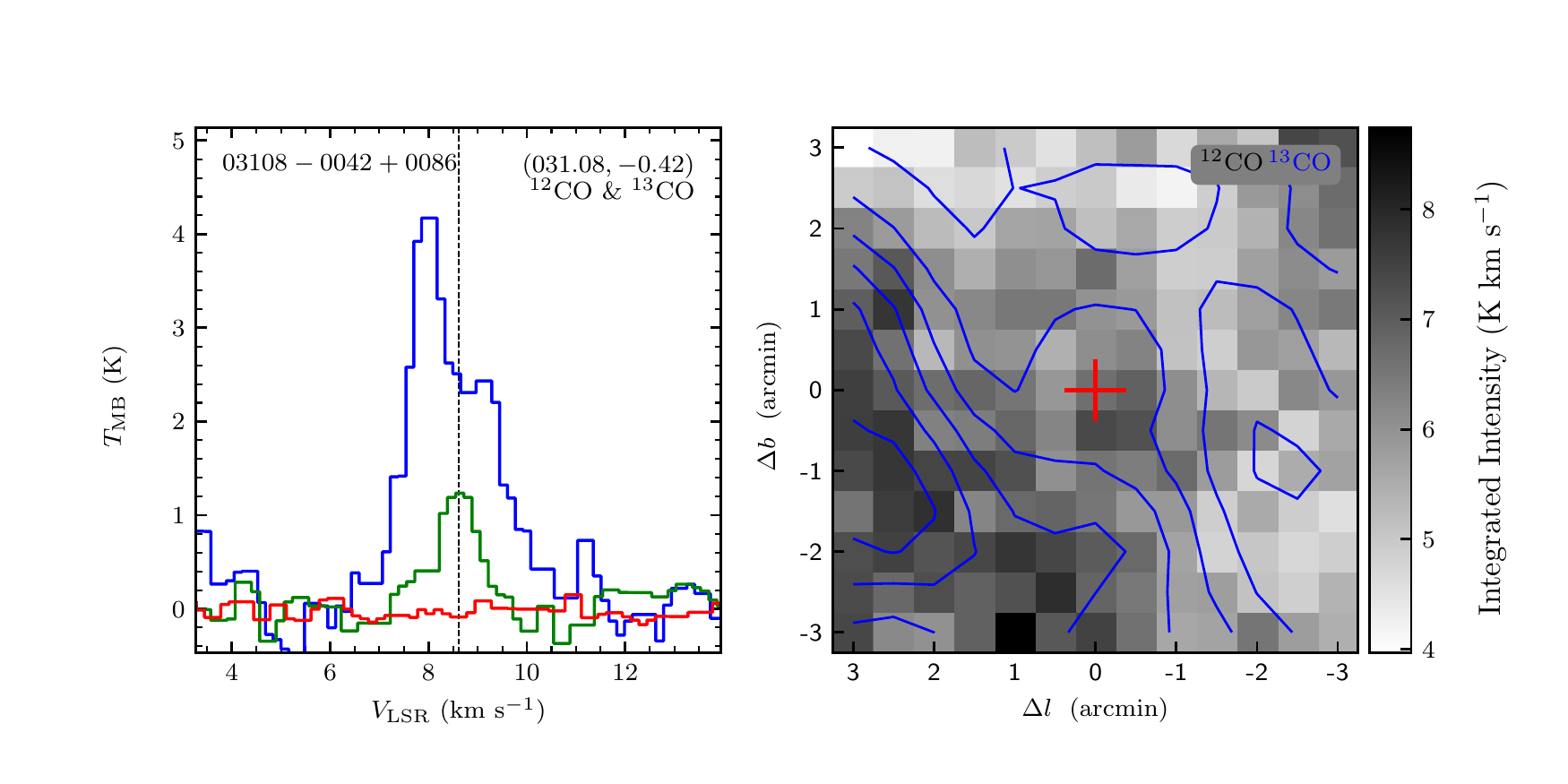}
\includegraphics[width=9.0cm,angle=0]{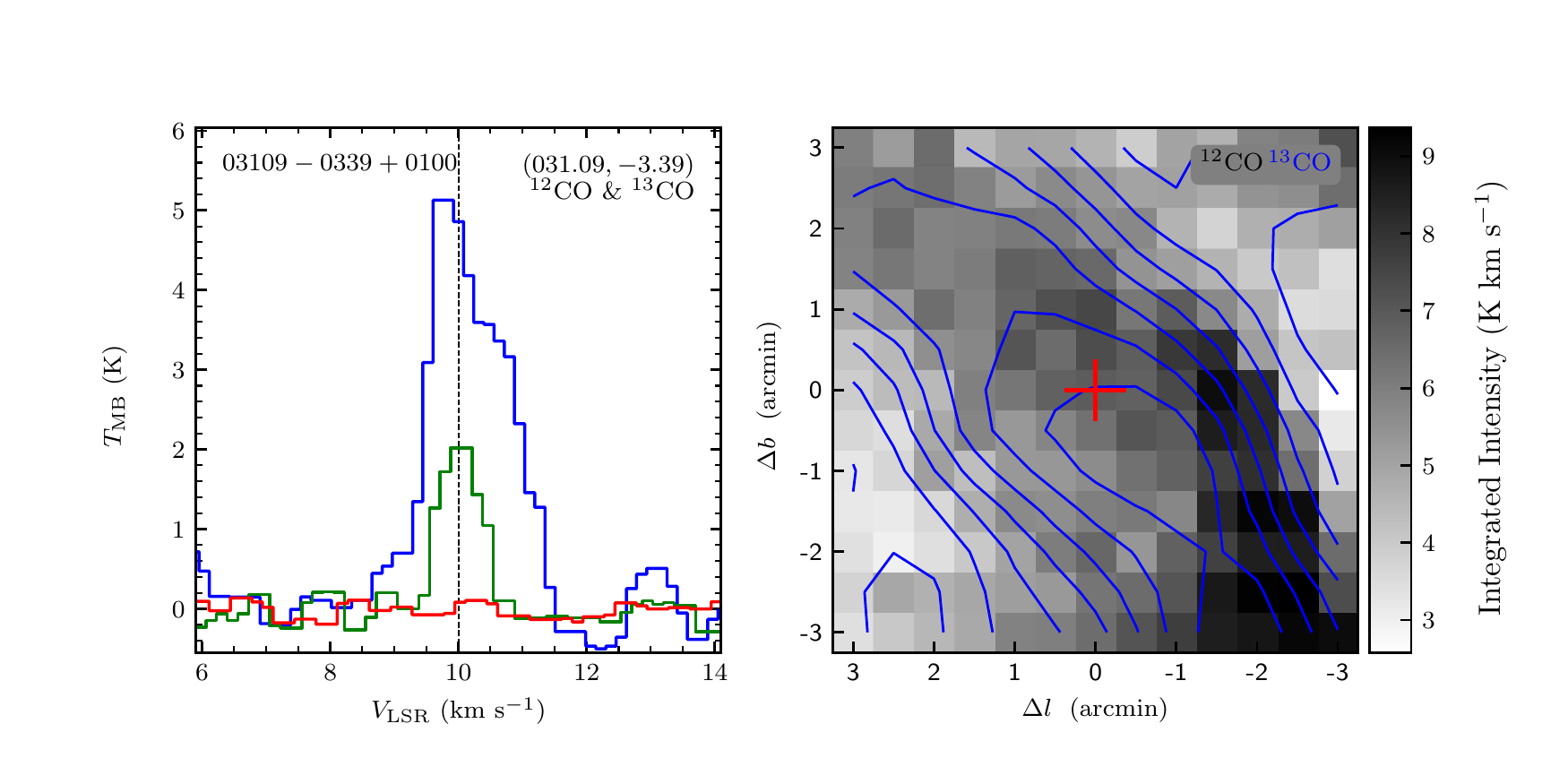}
\end{figure}
\clearpage

\begin{figure}
\includegraphics[width=9.0cm,angle=0]{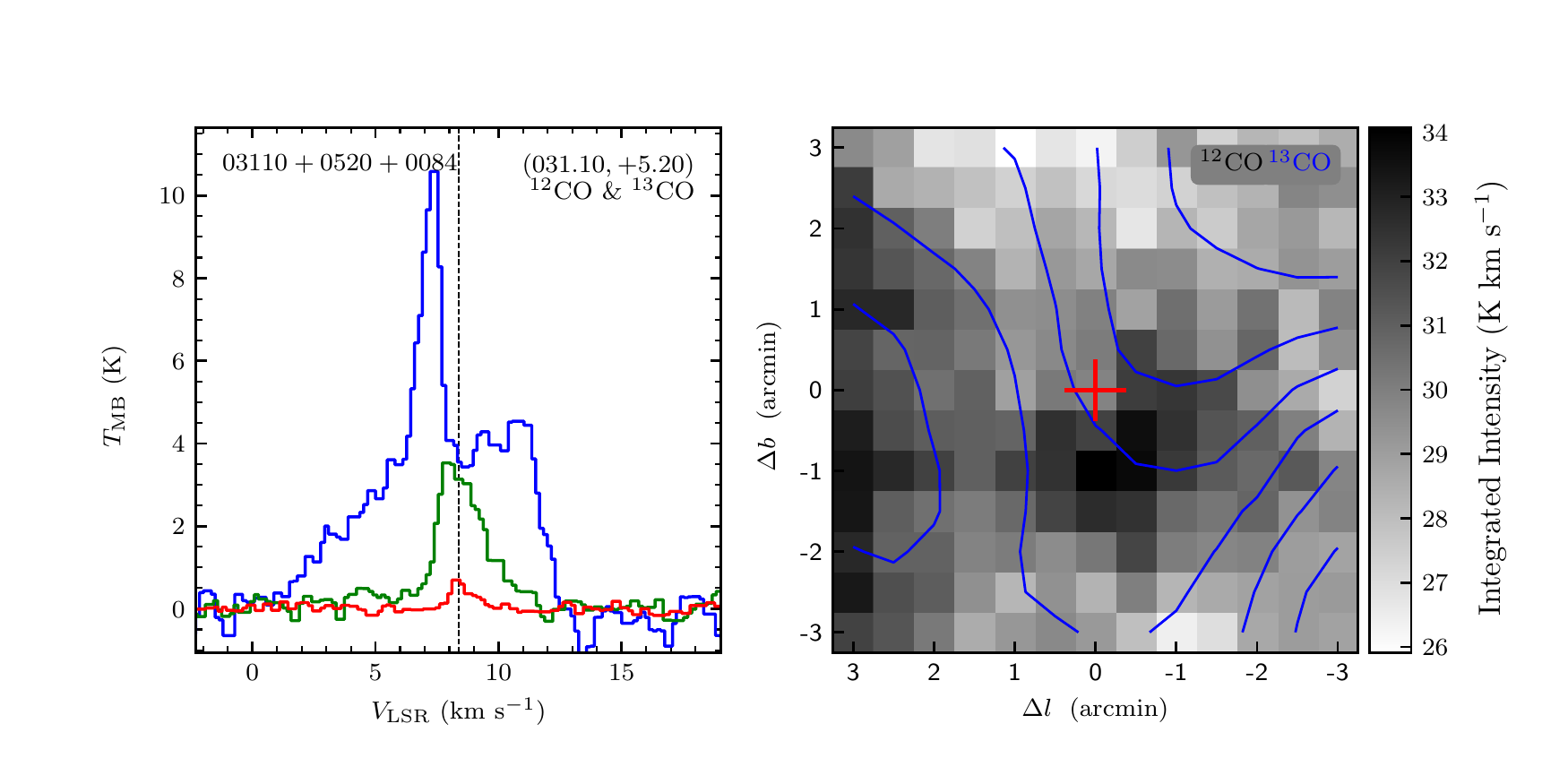}
\includegraphics[width=9.0cm,angle=0]{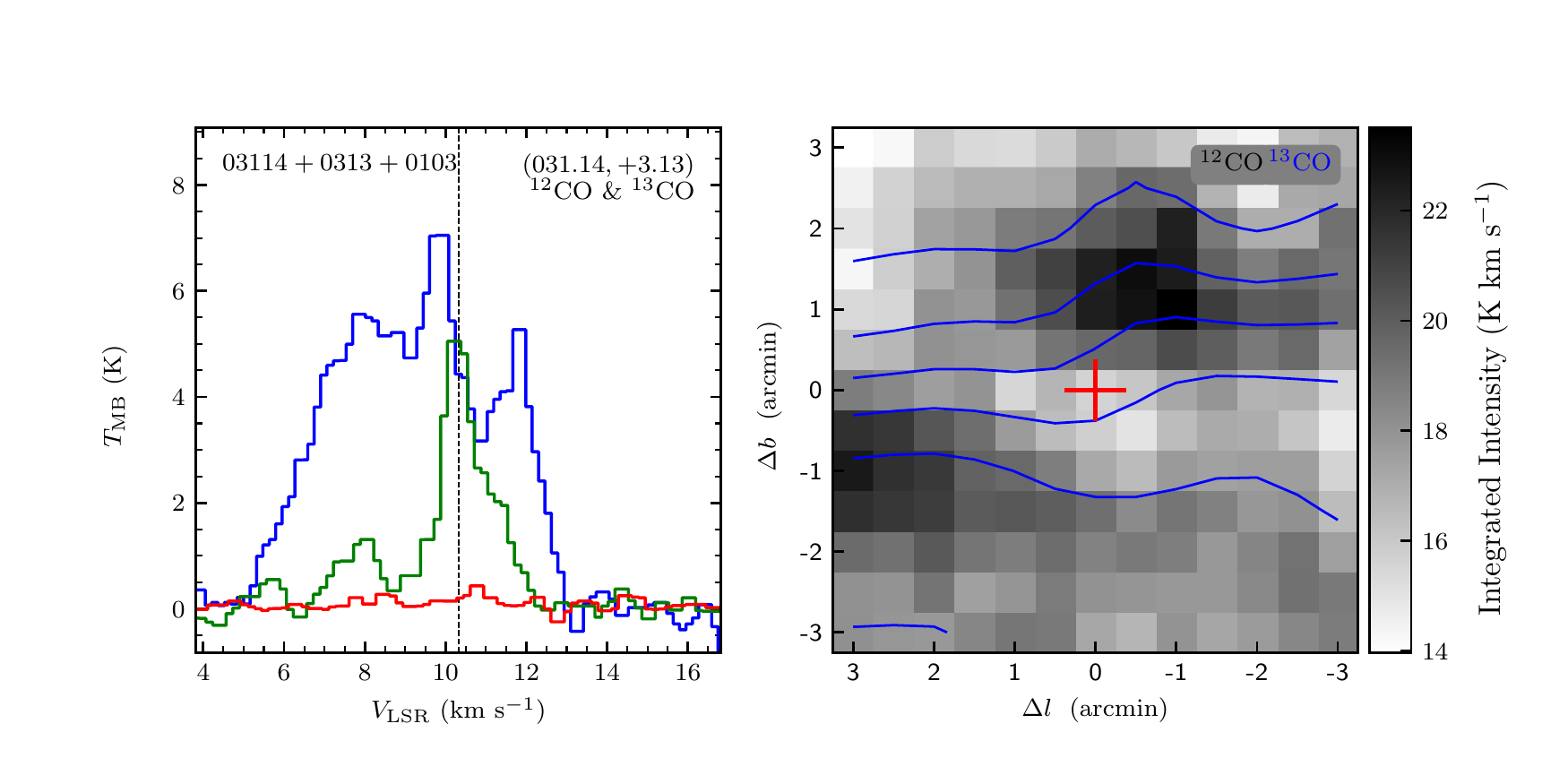}
\vspace{-0.5cm}

\includegraphics[width=9.0cm,angle=0]{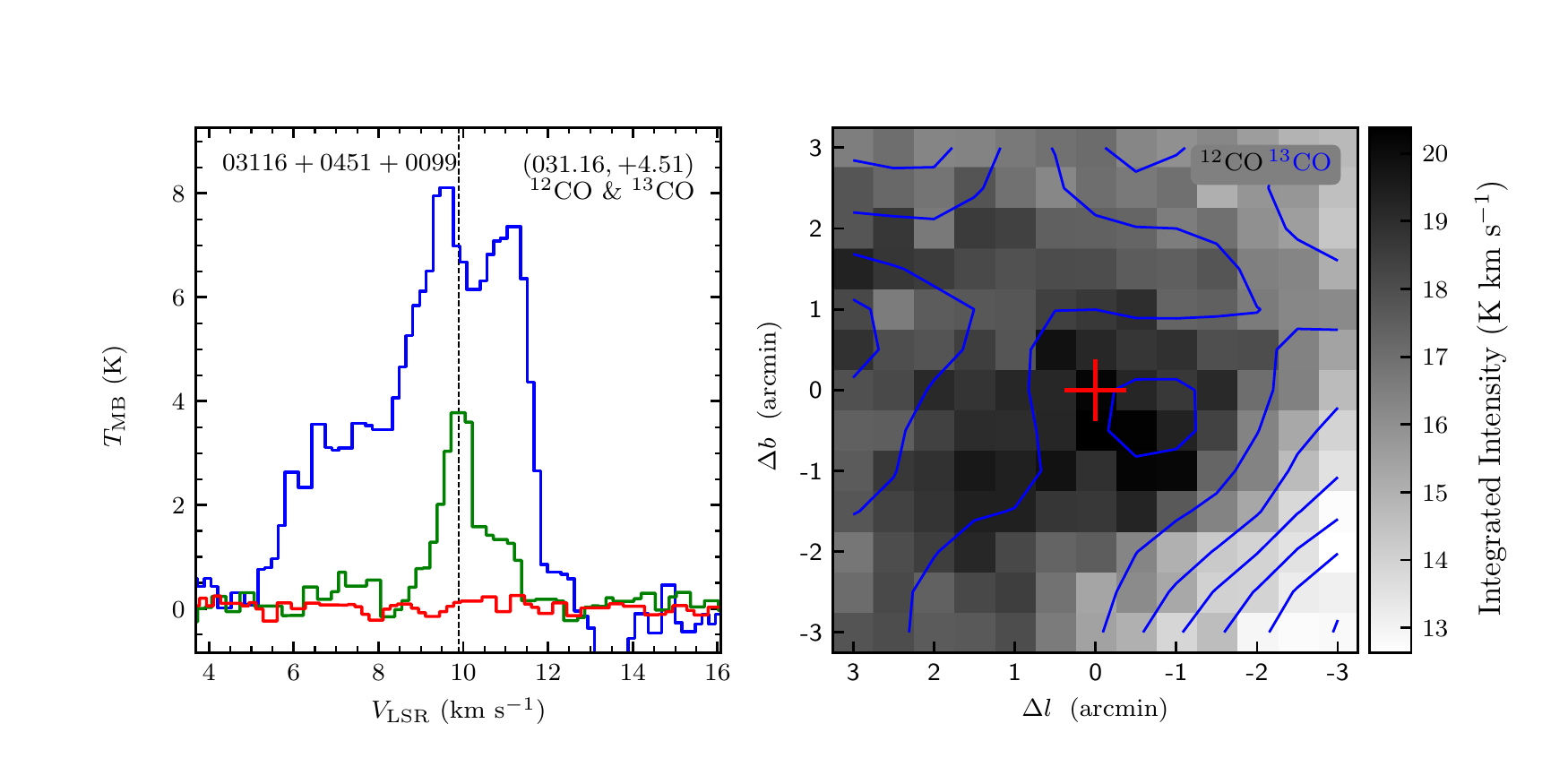}
\includegraphics[width=9.0cm,angle=0]{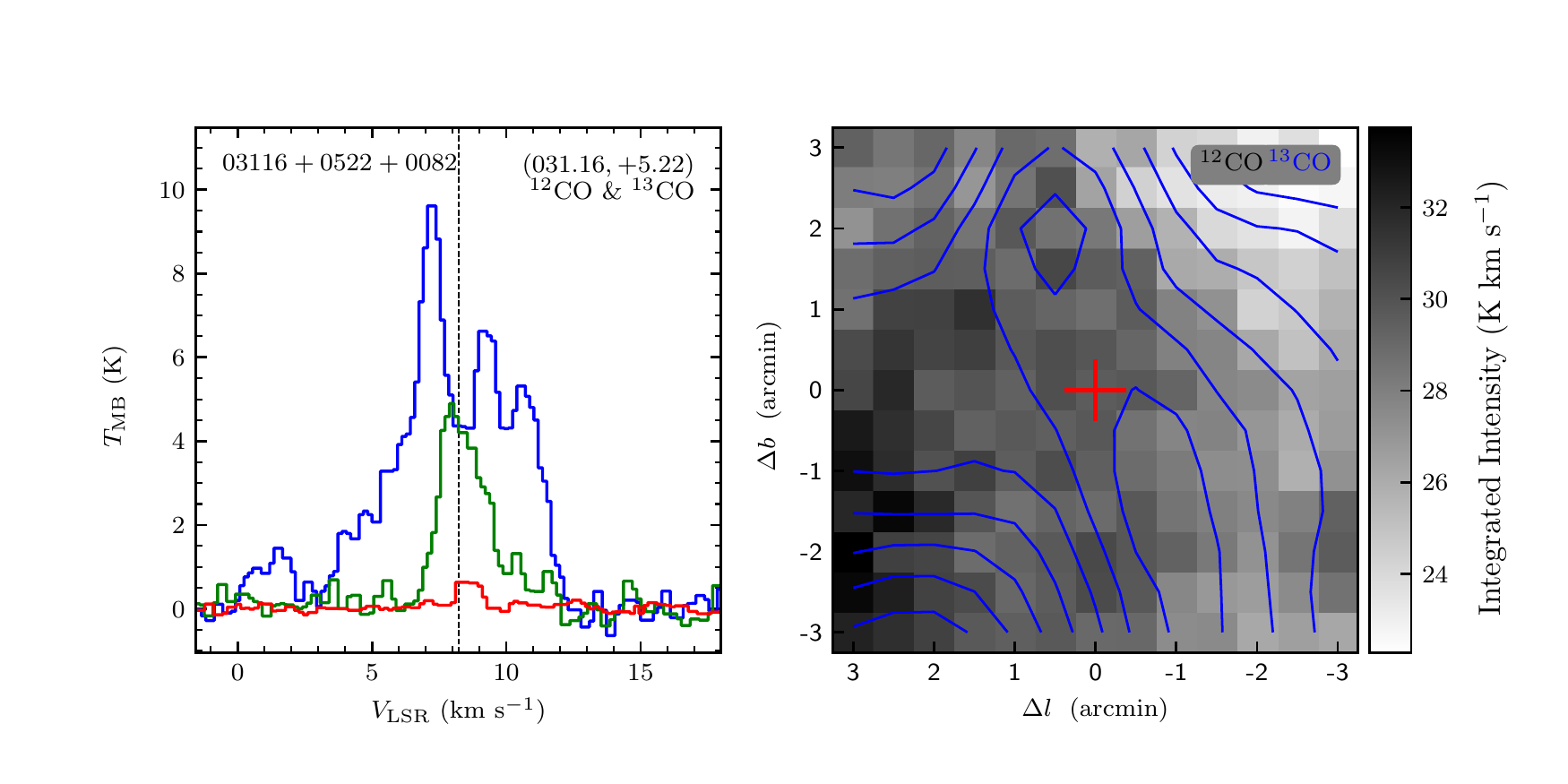}
\vspace{-0.5cm}

\includegraphics[width=9.0cm,angle=0]{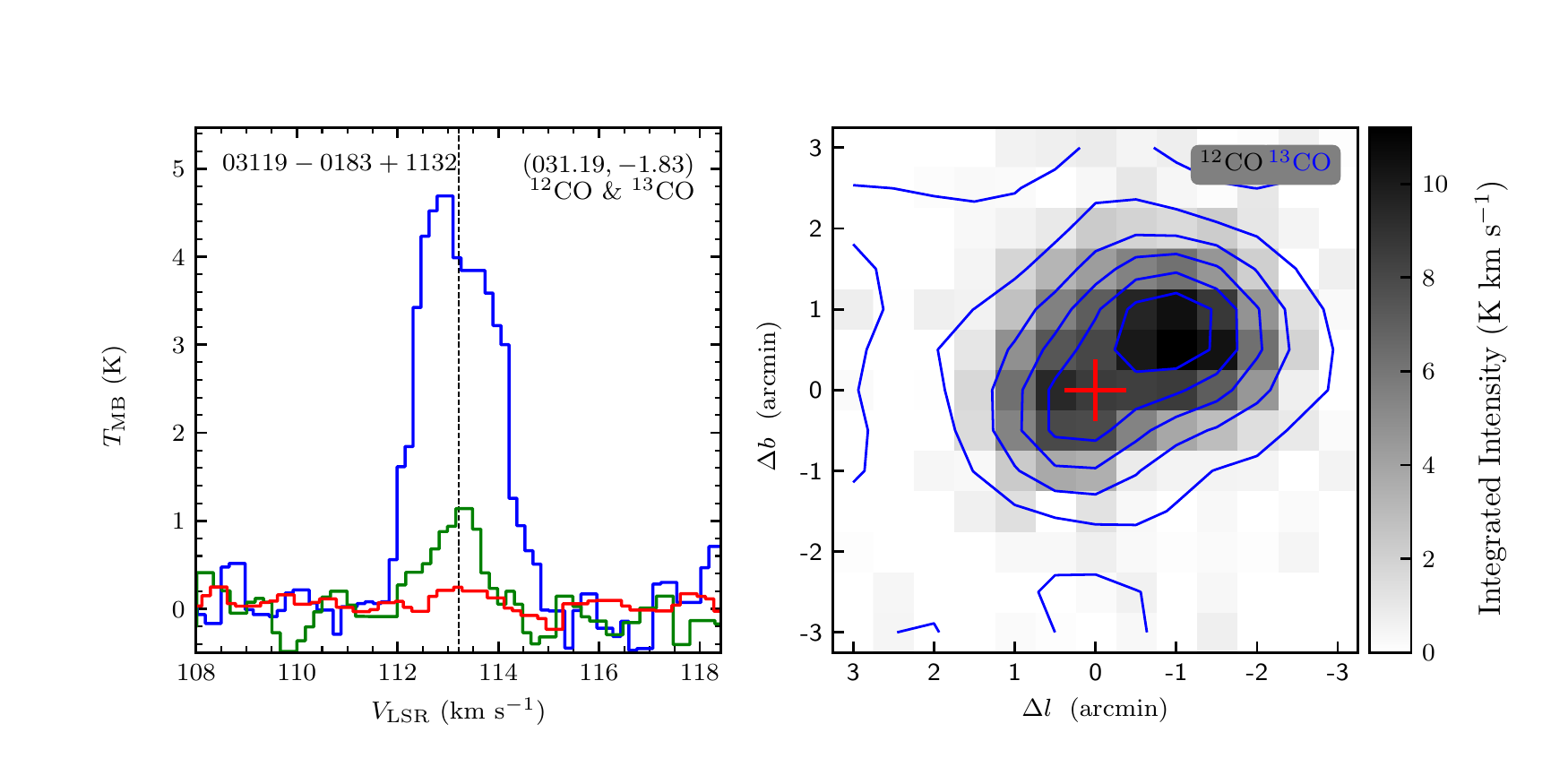}
\includegraphics[width=9.0cm,angle=0]{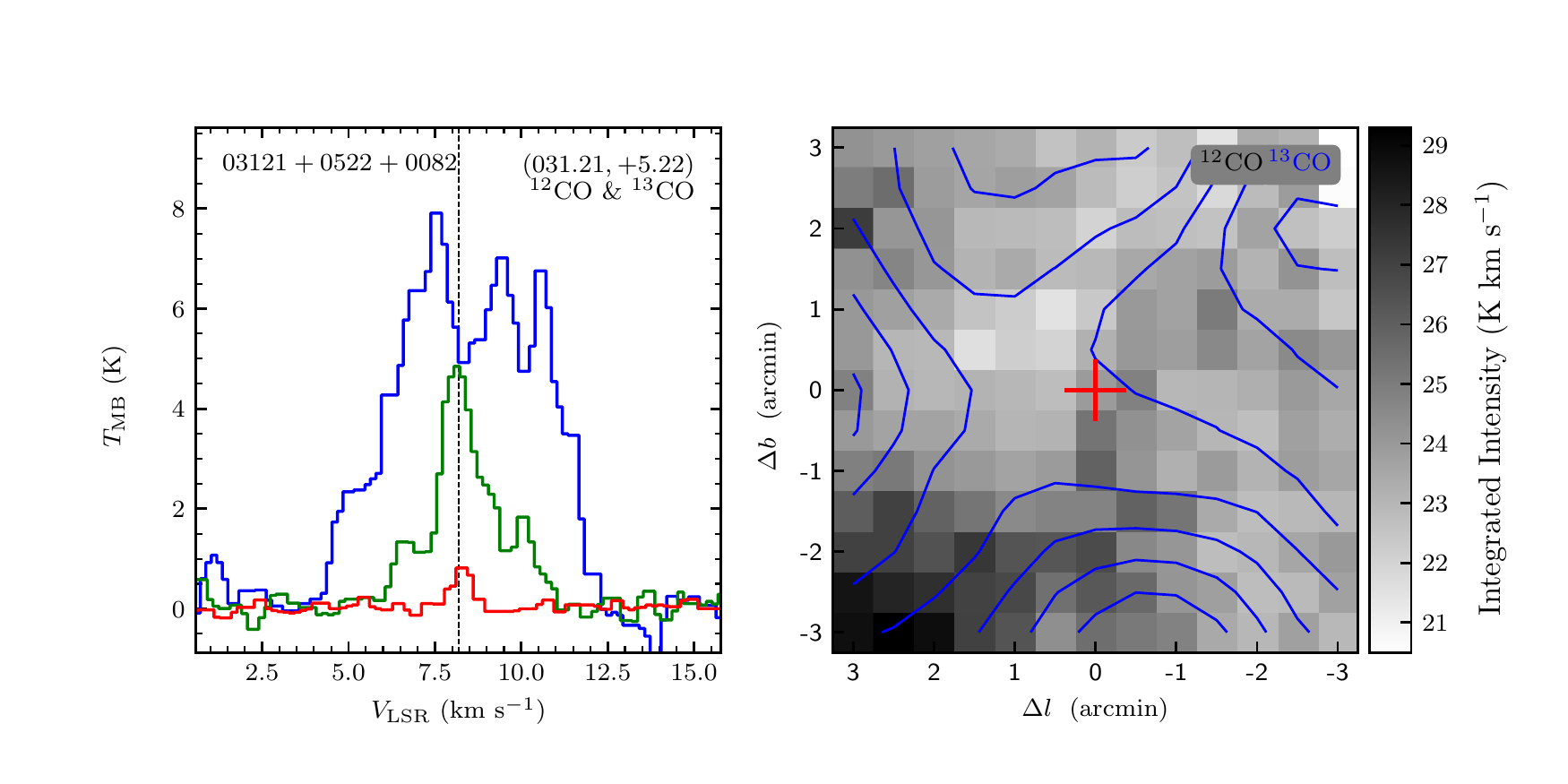}
\vspace{-0.5cm}

\includegraphics[width=9.0cm,angle=0]{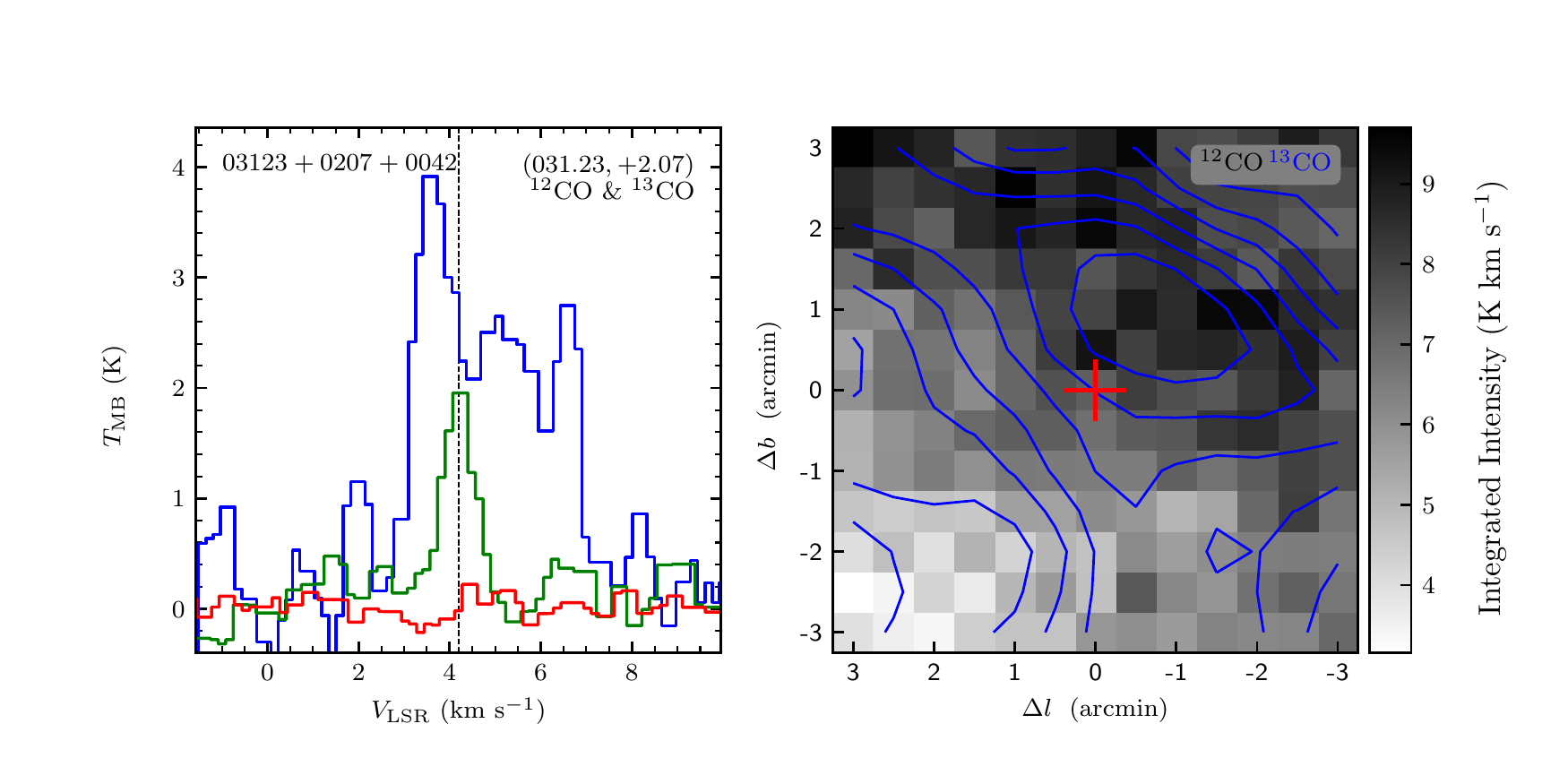}
\includegraphics[width=9.0cm,angle=0]{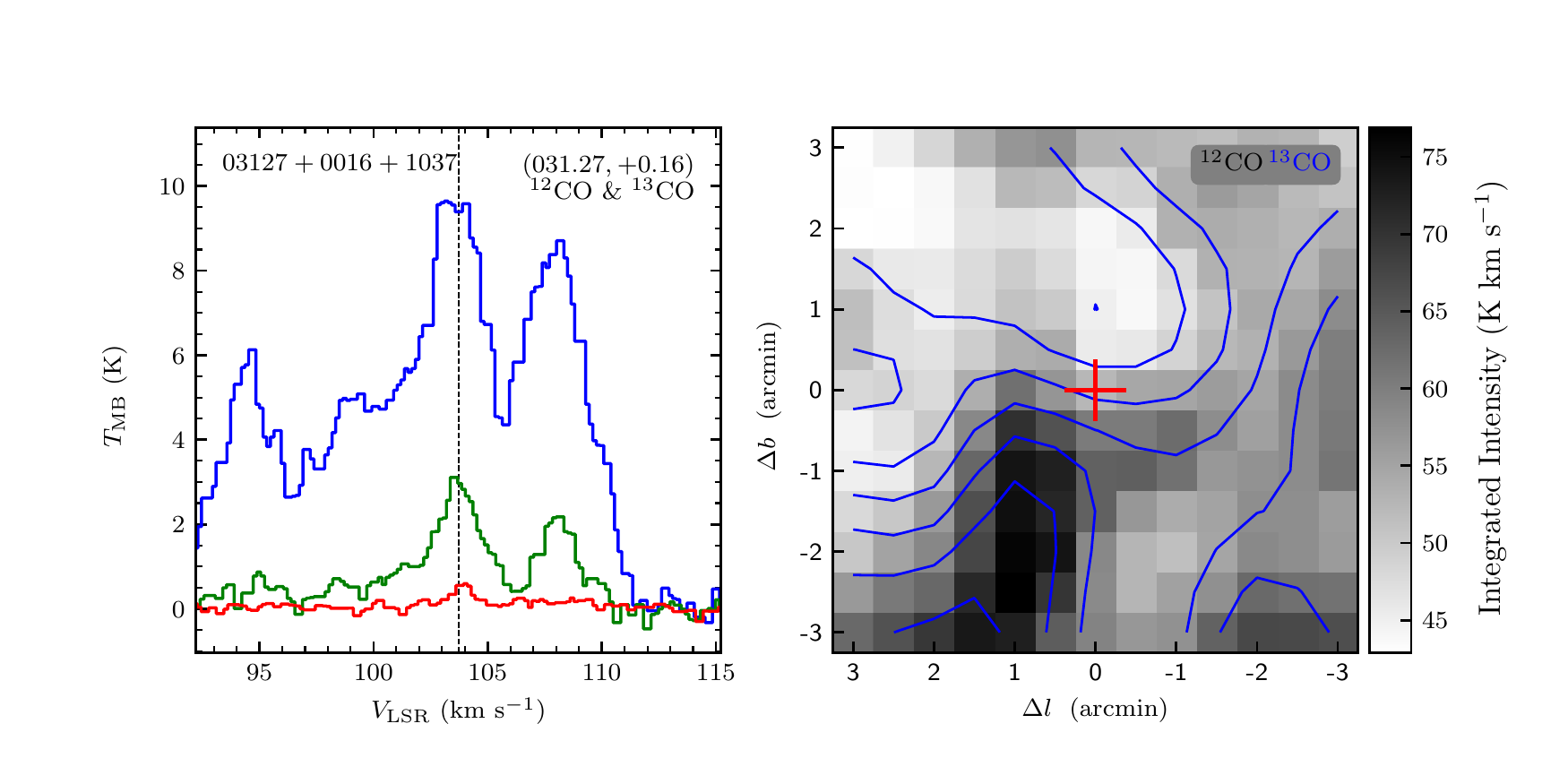}
\vspace{-0.5cm}

\includegraphics[width=9.0cm,angle=0]{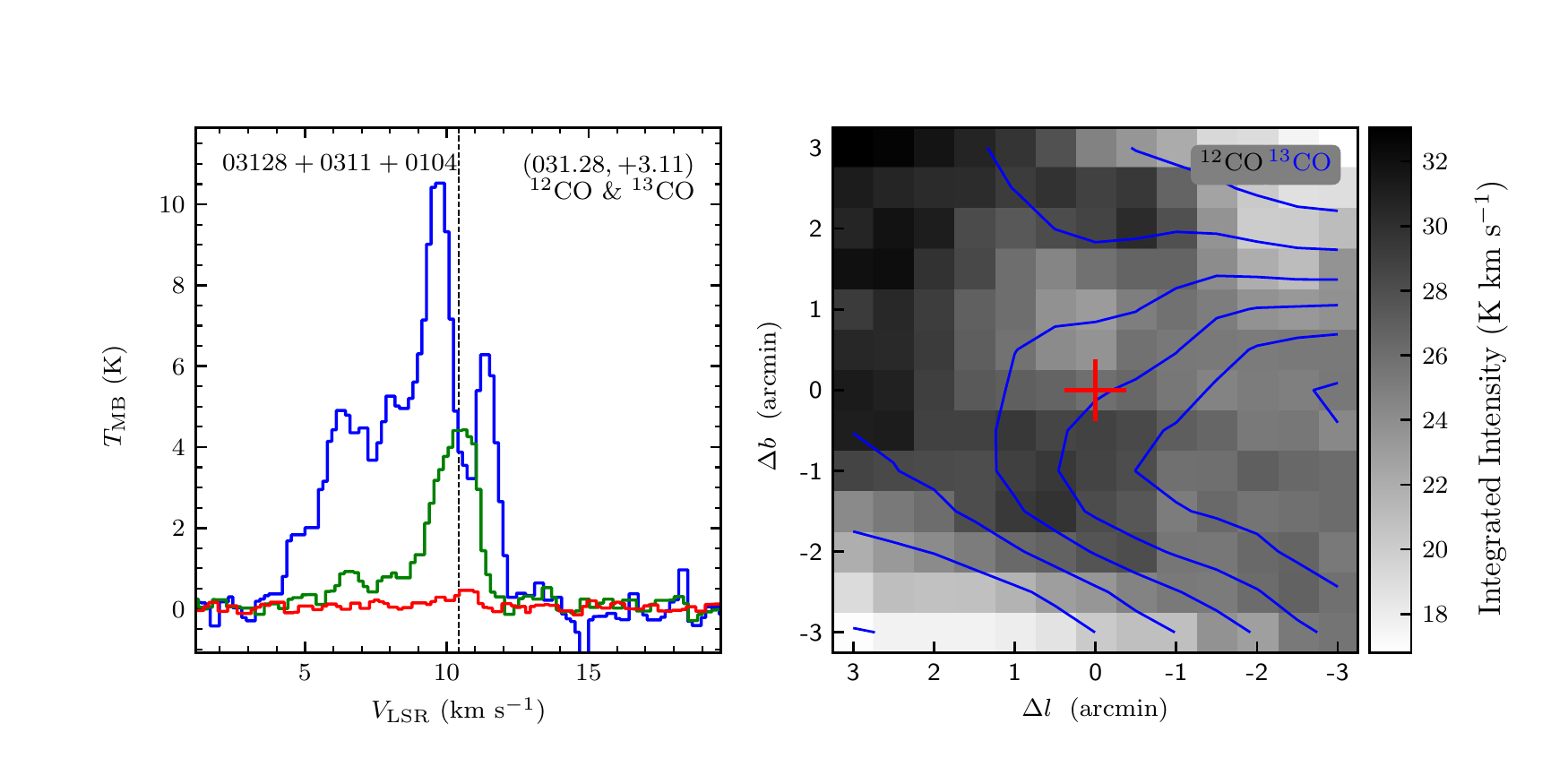}
\includegraphics[width=9.0cm,angle=0]{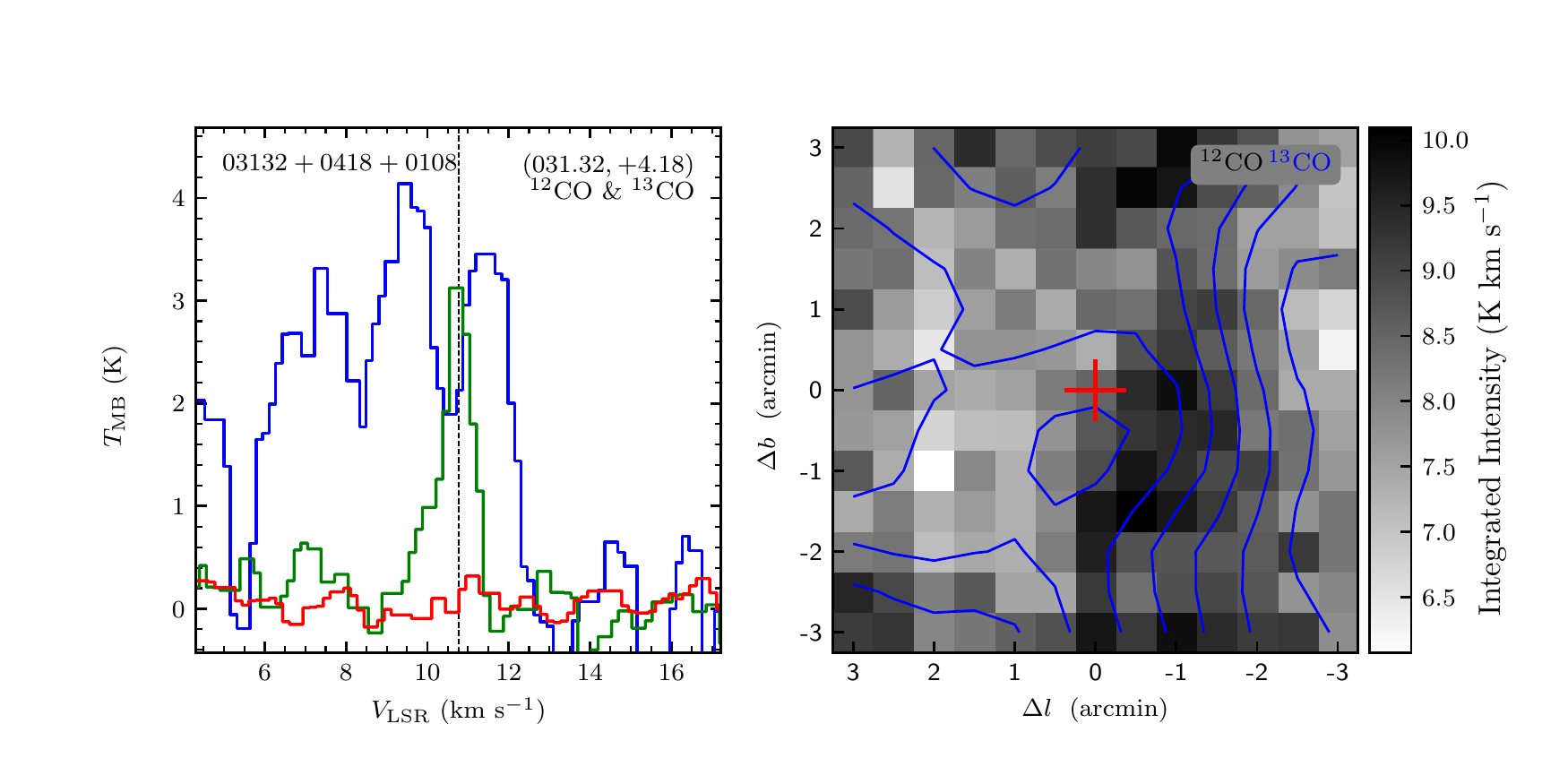}
\end{figure}
\clearpage

\begin{figure}
\includegraphics[width=9.0cm,angle=0]{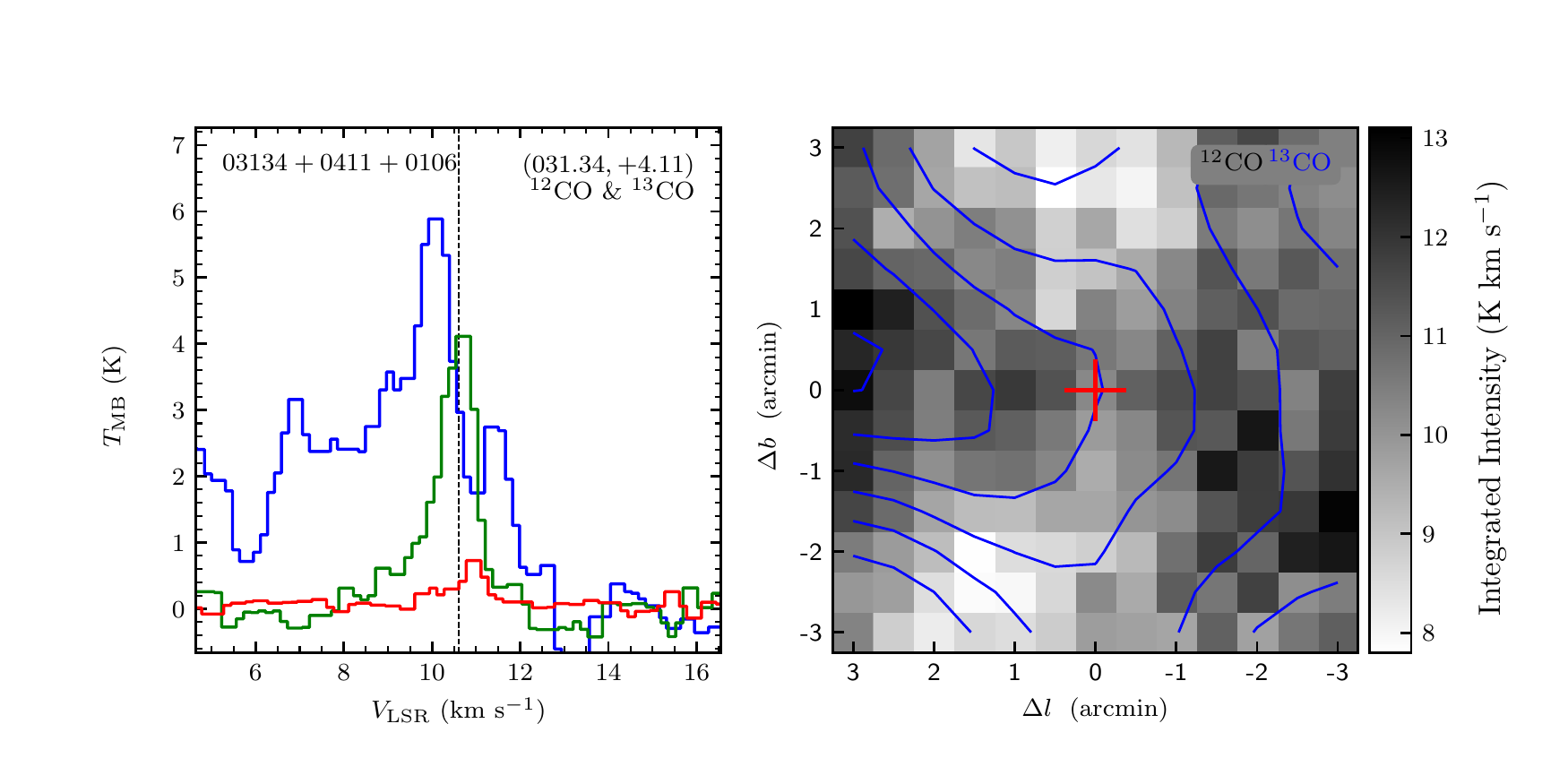}
\includegraphics[width=9.0cm,angle=0]{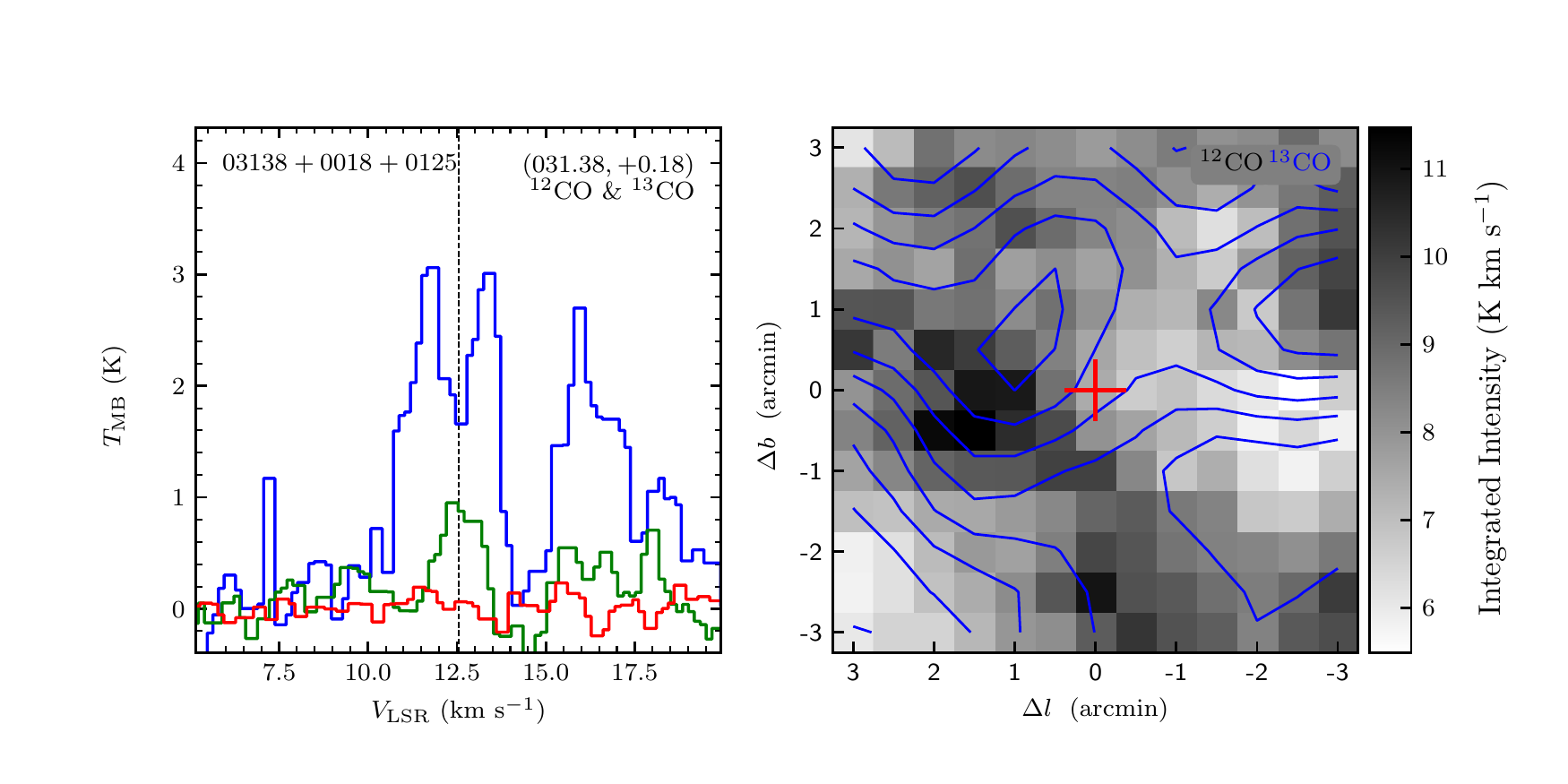}
\vspace{-0.5cm}

\includegraphics[width=9.0cm,angle=0]{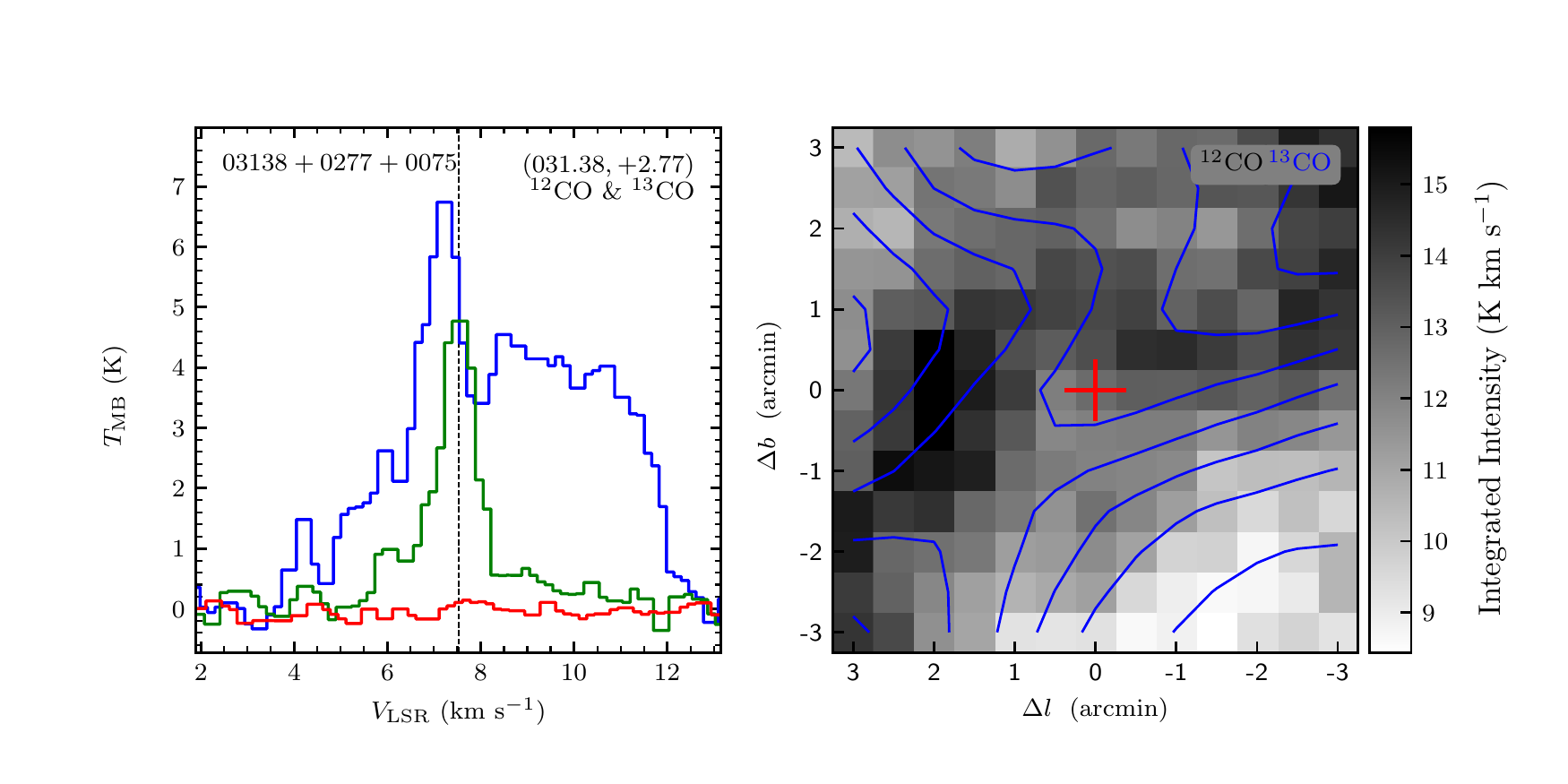}
\includegraphics[width=9.0cm,angle=0]{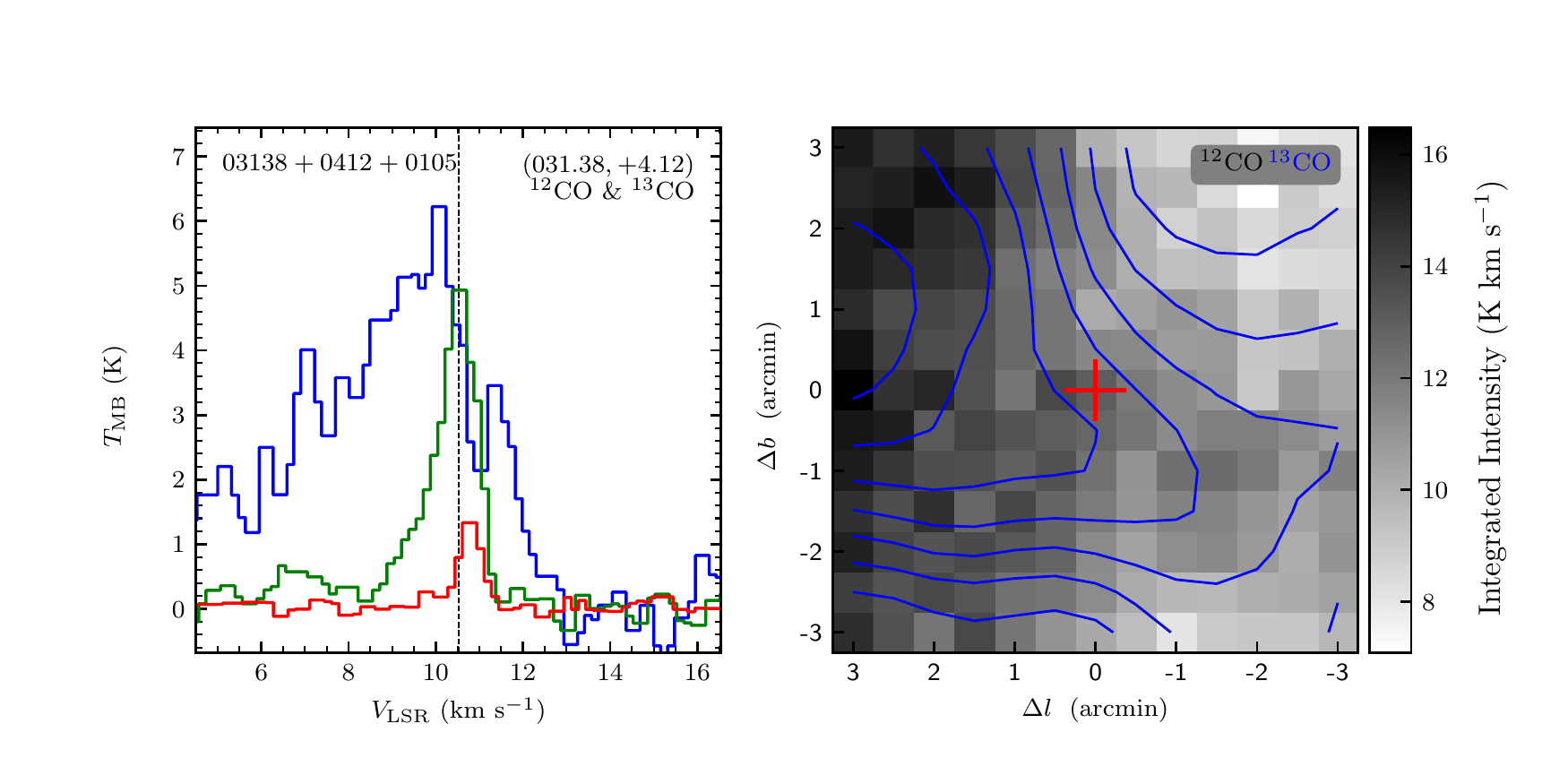}
\vspace{-0.5cm}

\includegraphics[width=9.0cm,angle=0]{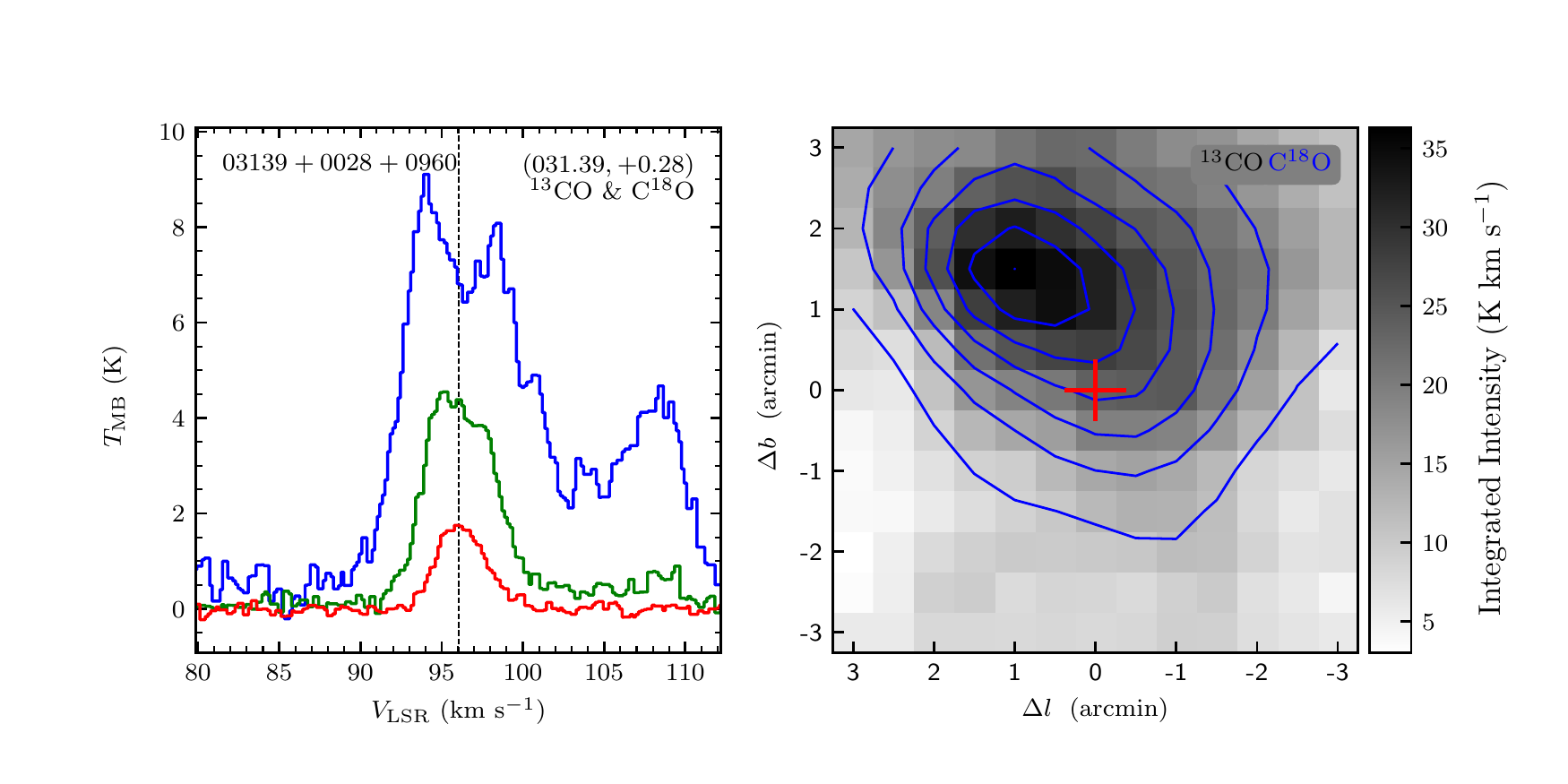}
\includegraphics[width=9.0cm,angle=0]{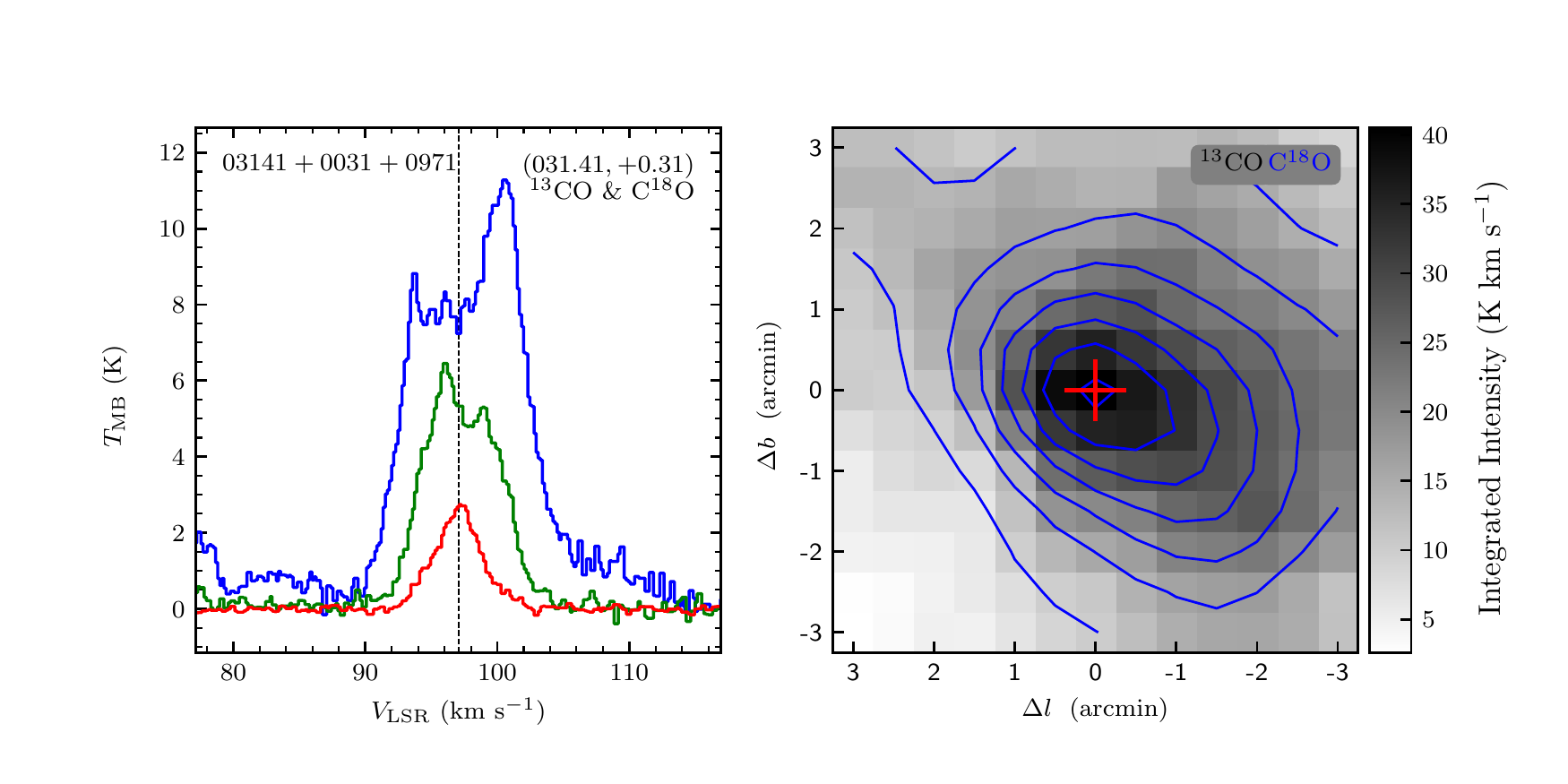}
\vspace{-0.5cm}

\includegraphics[width=9.0cm,angle=0]{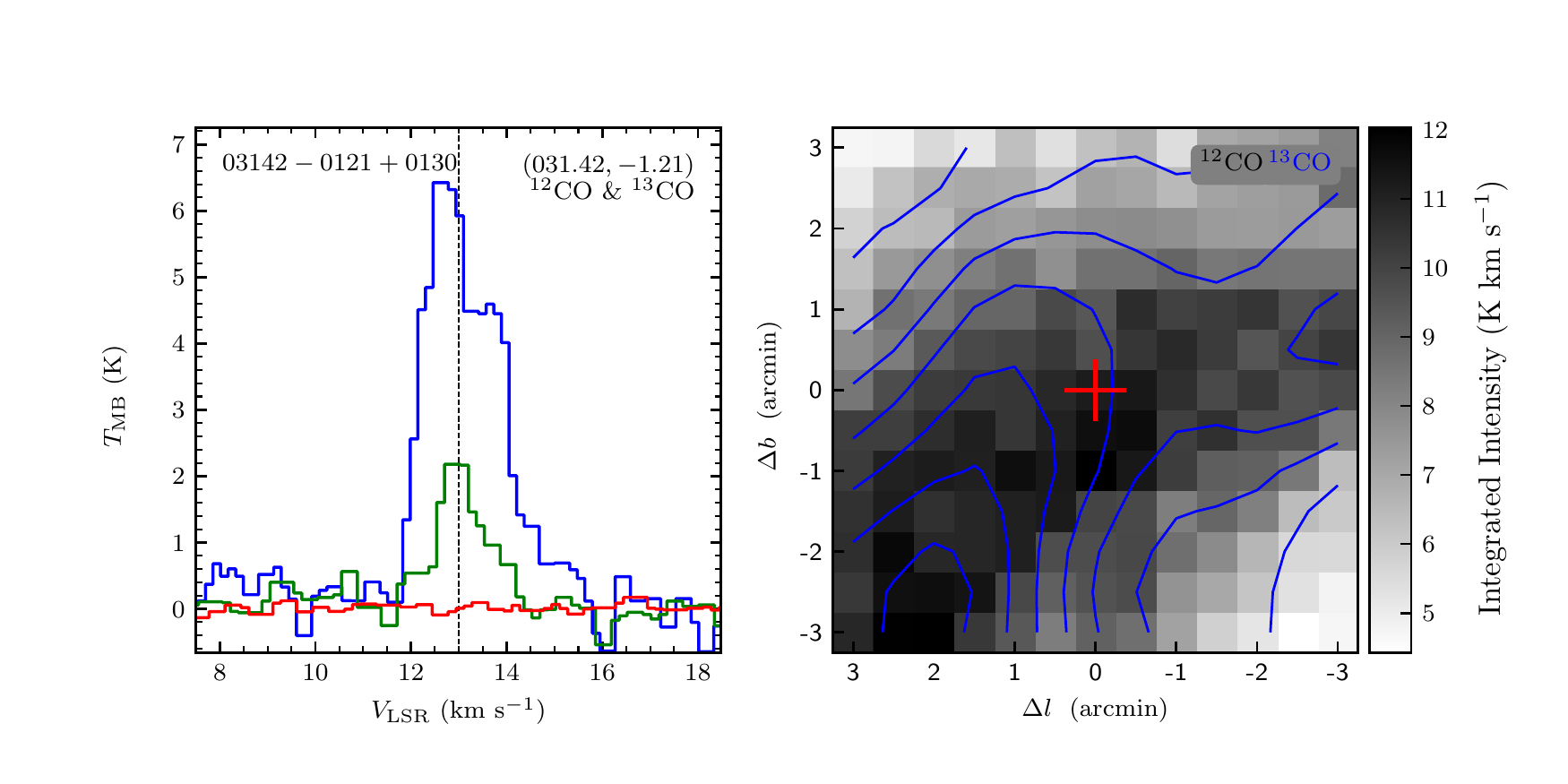}
\includegraphics[width=9.0cm,angle=0]{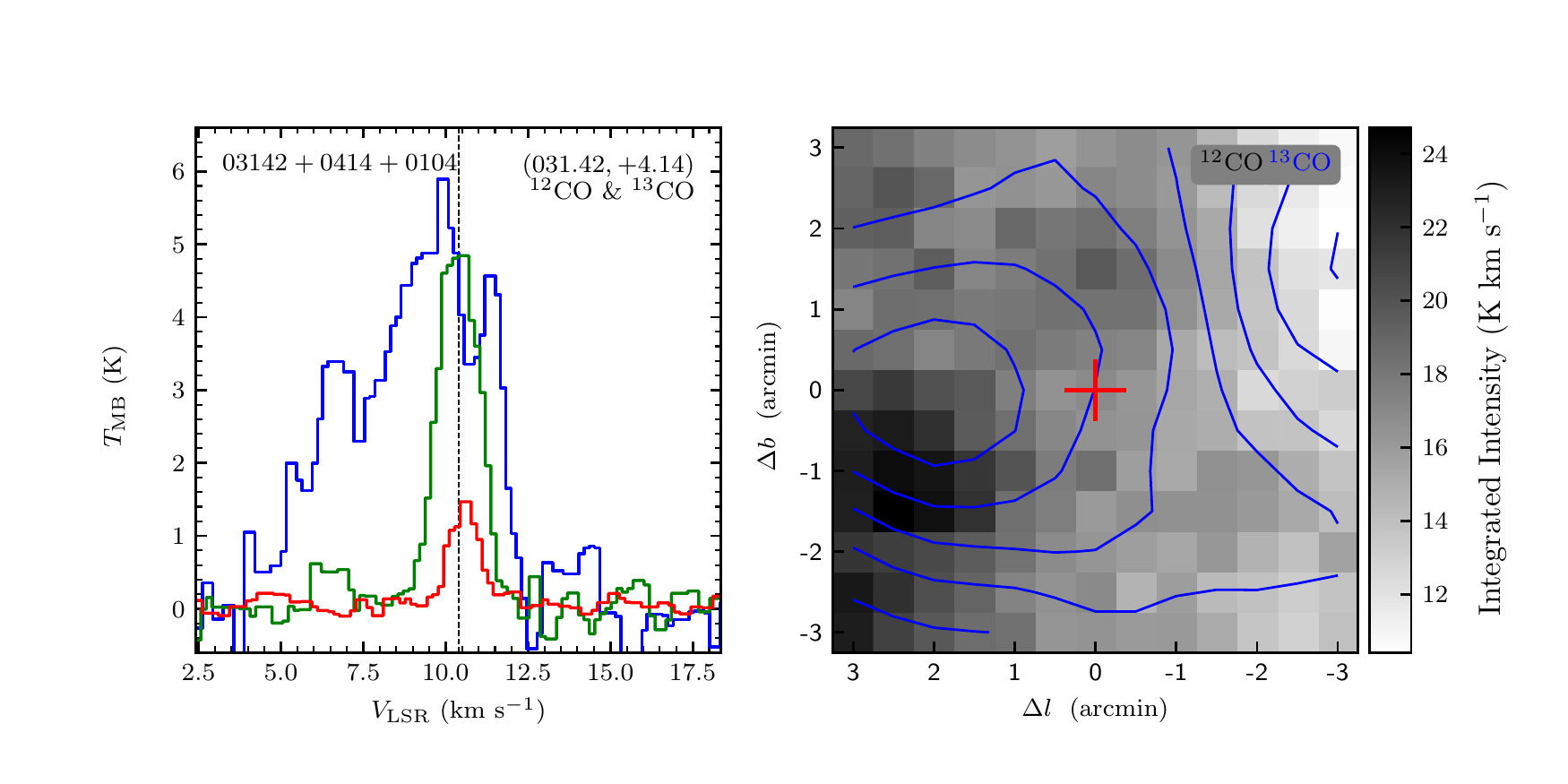}
\vspace{-0.5cm}

\includegraphics[width=9.0cm,angle=0]{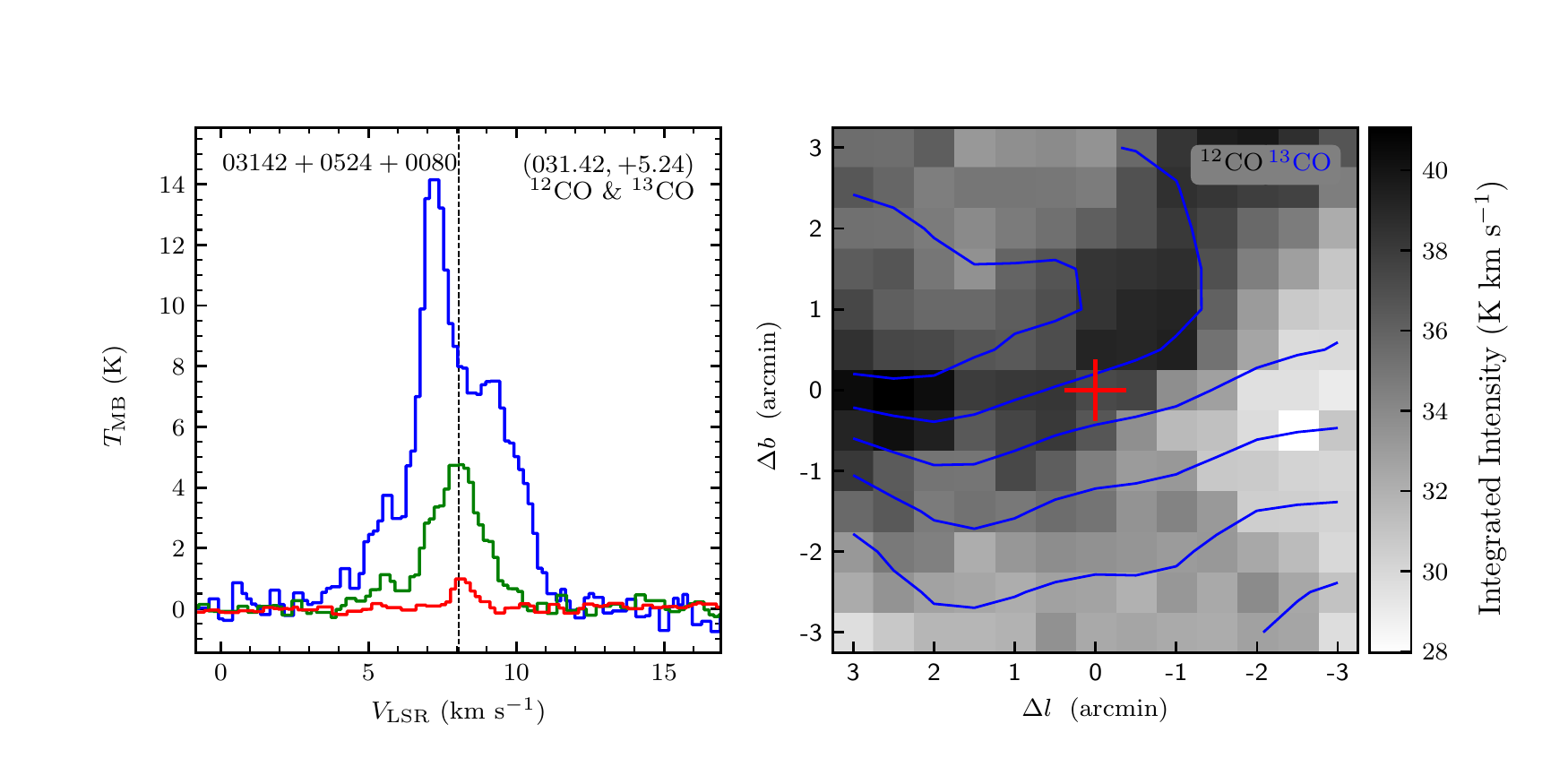}
\includegraphics[width=9.0cm,angle=0]{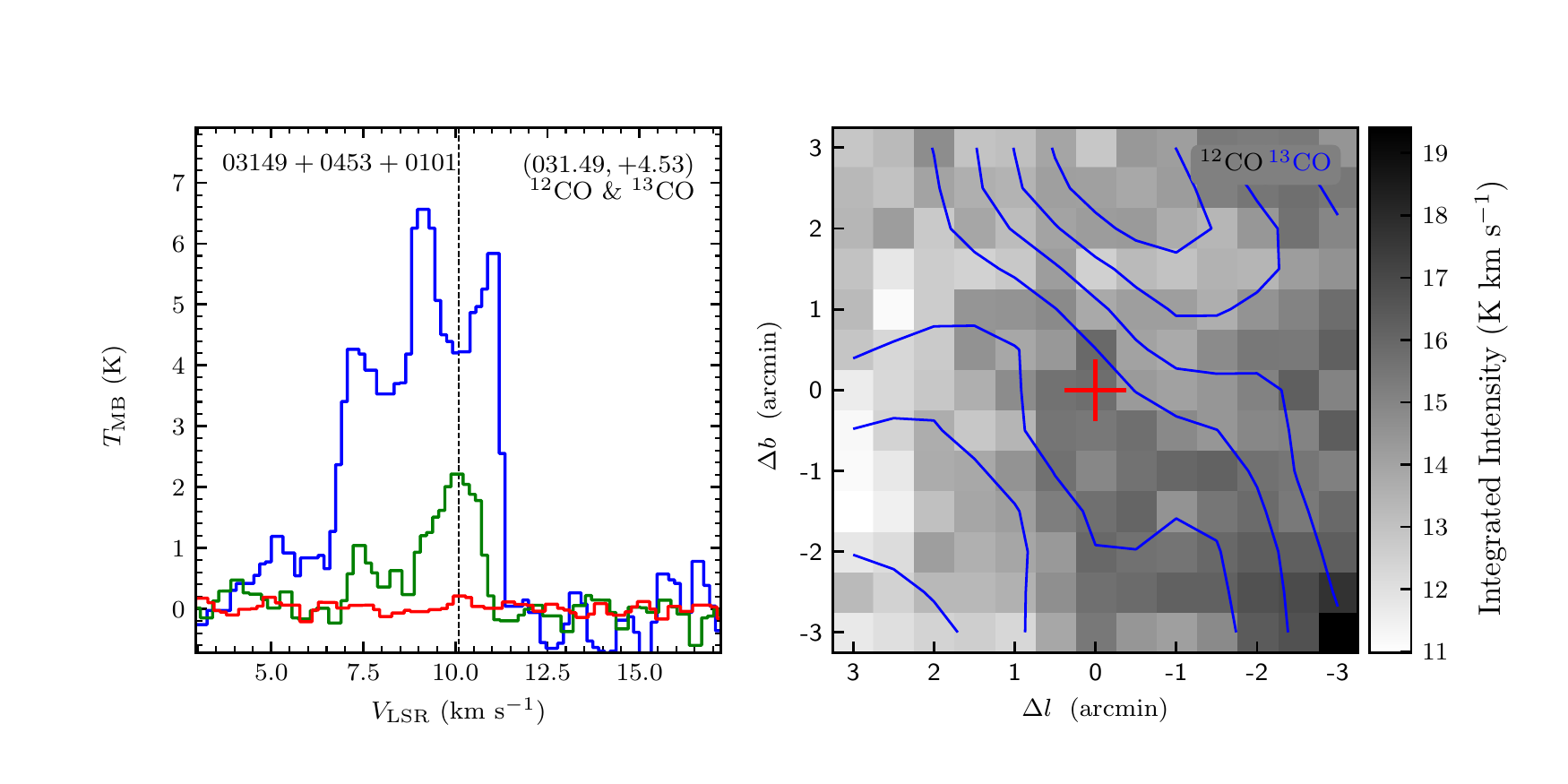}
\end{figure}
\clearpage

\begin{figure}
\includegraphics[width=9.0cm,angle=0]{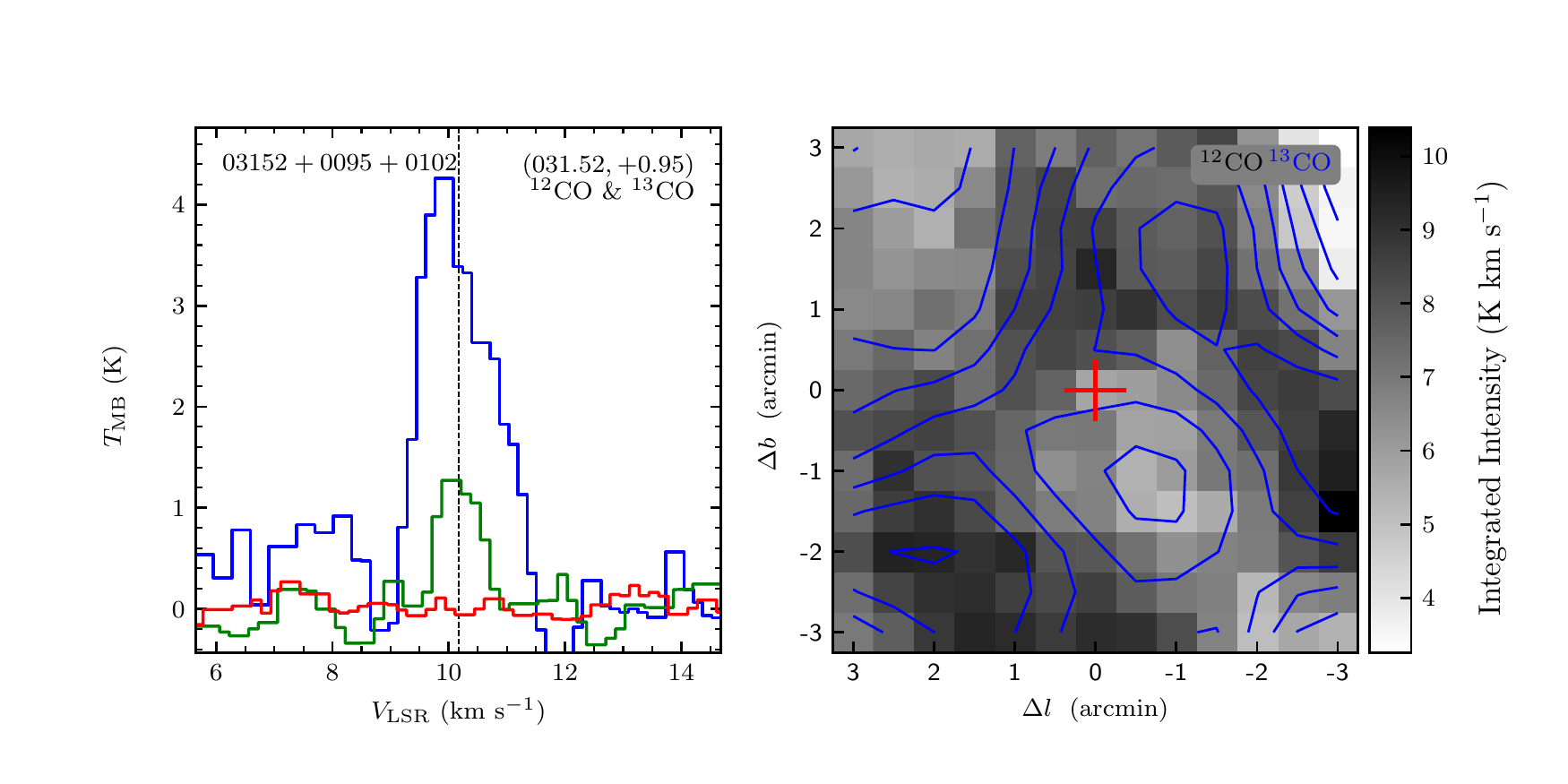}
\includegraphics[width=9.0cm,angle=0]{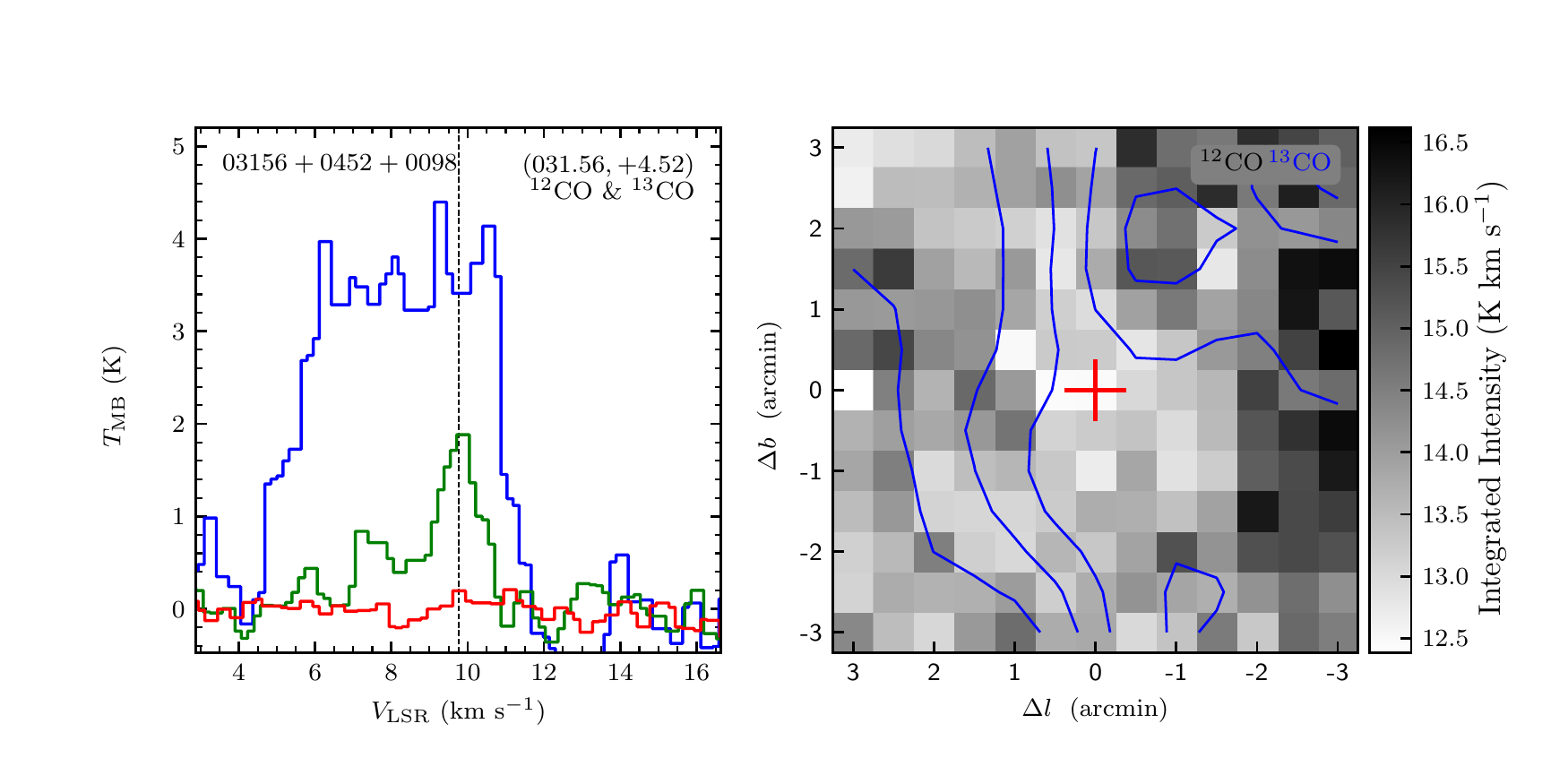}
\vspace{-0.5cm}

\includegraphics[width=9.0cm,angle=0]{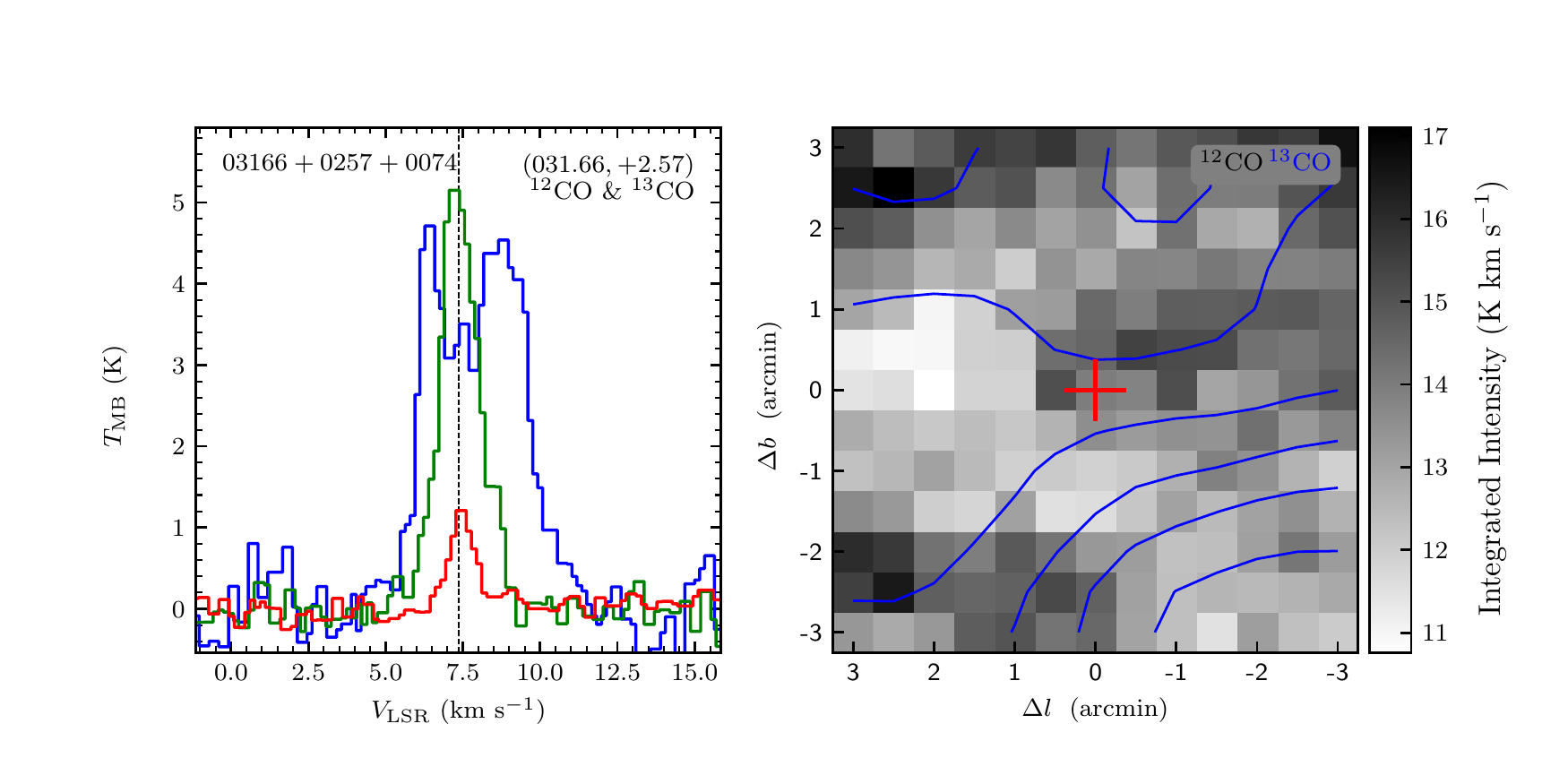}
\includegraphics[width=9.0cm,angle=0]{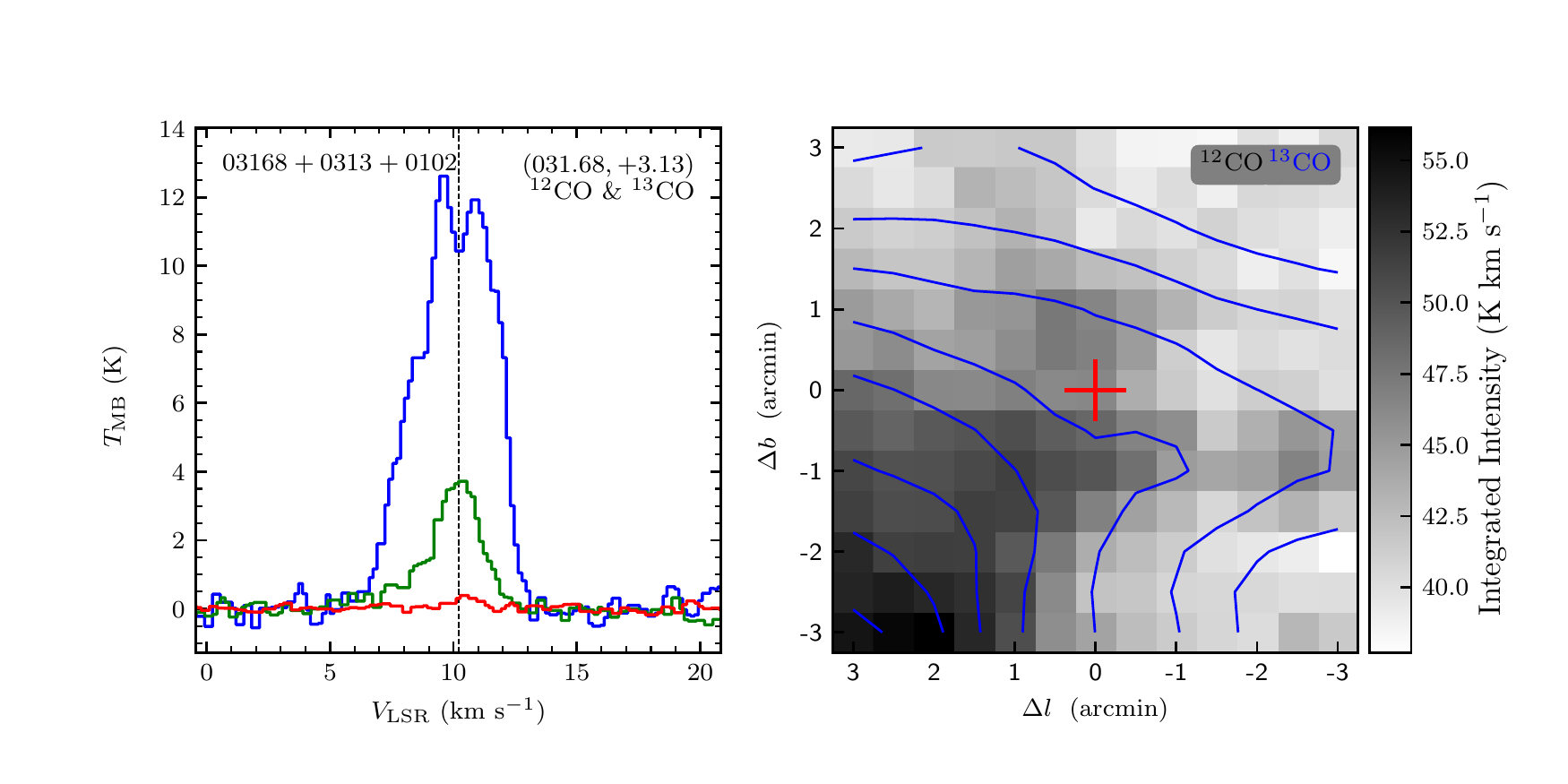}
\vspace{-0.5cm}

\includegraphics[width=9.0cm,angle=0]{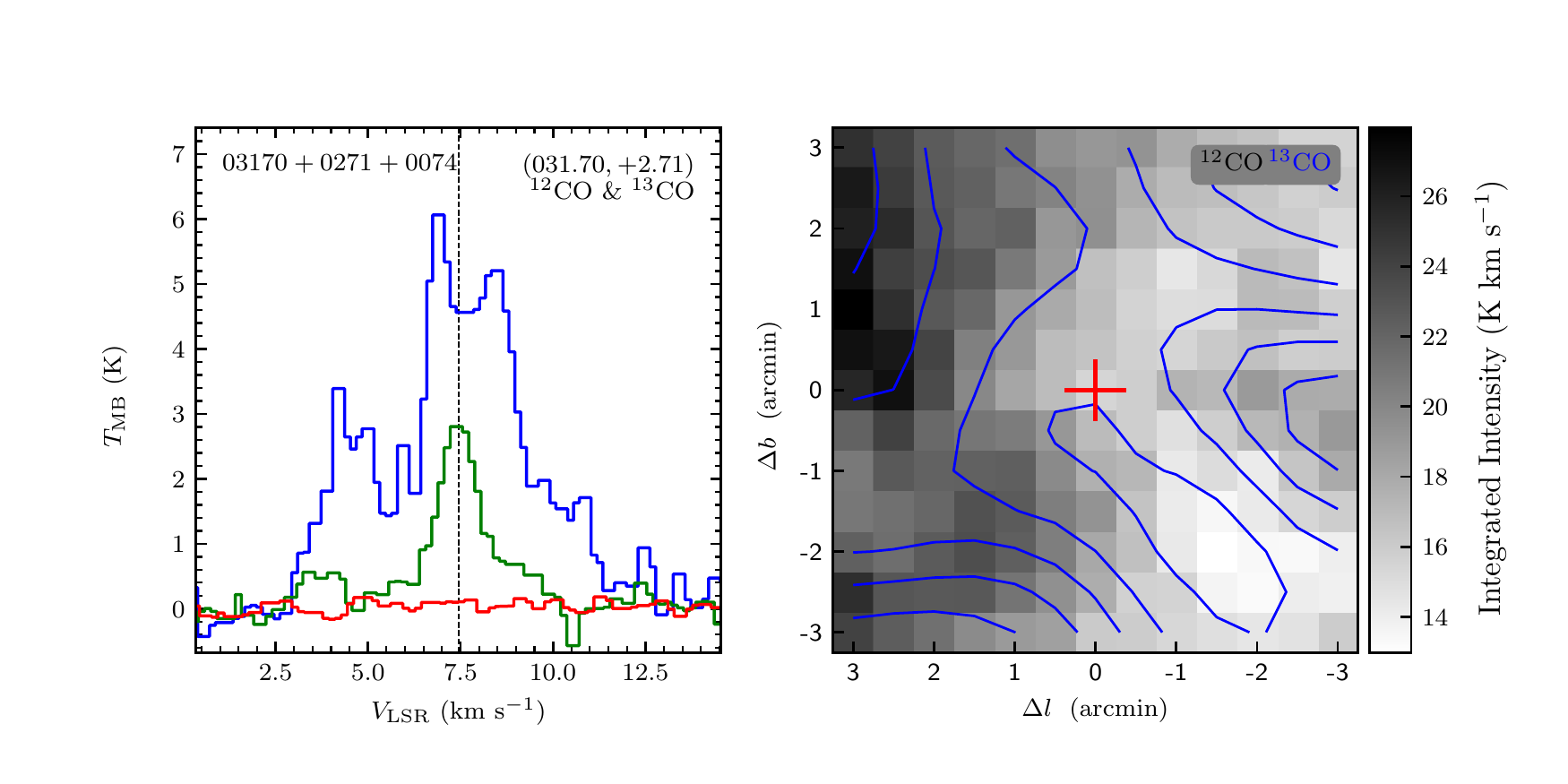}
\includegraphics[width=9.0cm,angle=0]{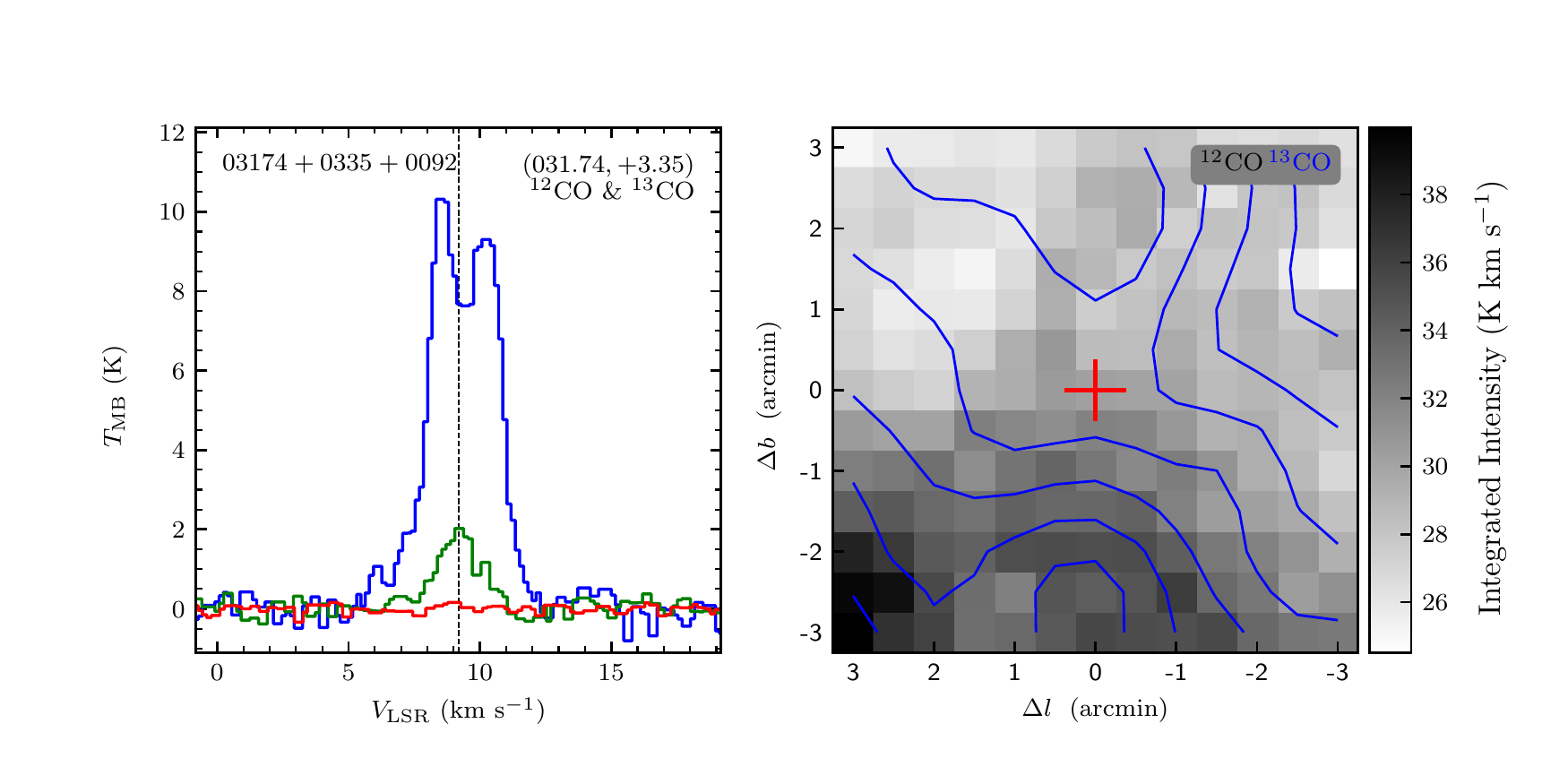}
\vspace{-0.5cm}

\includegraphics[width=9.0cm,angle=0]{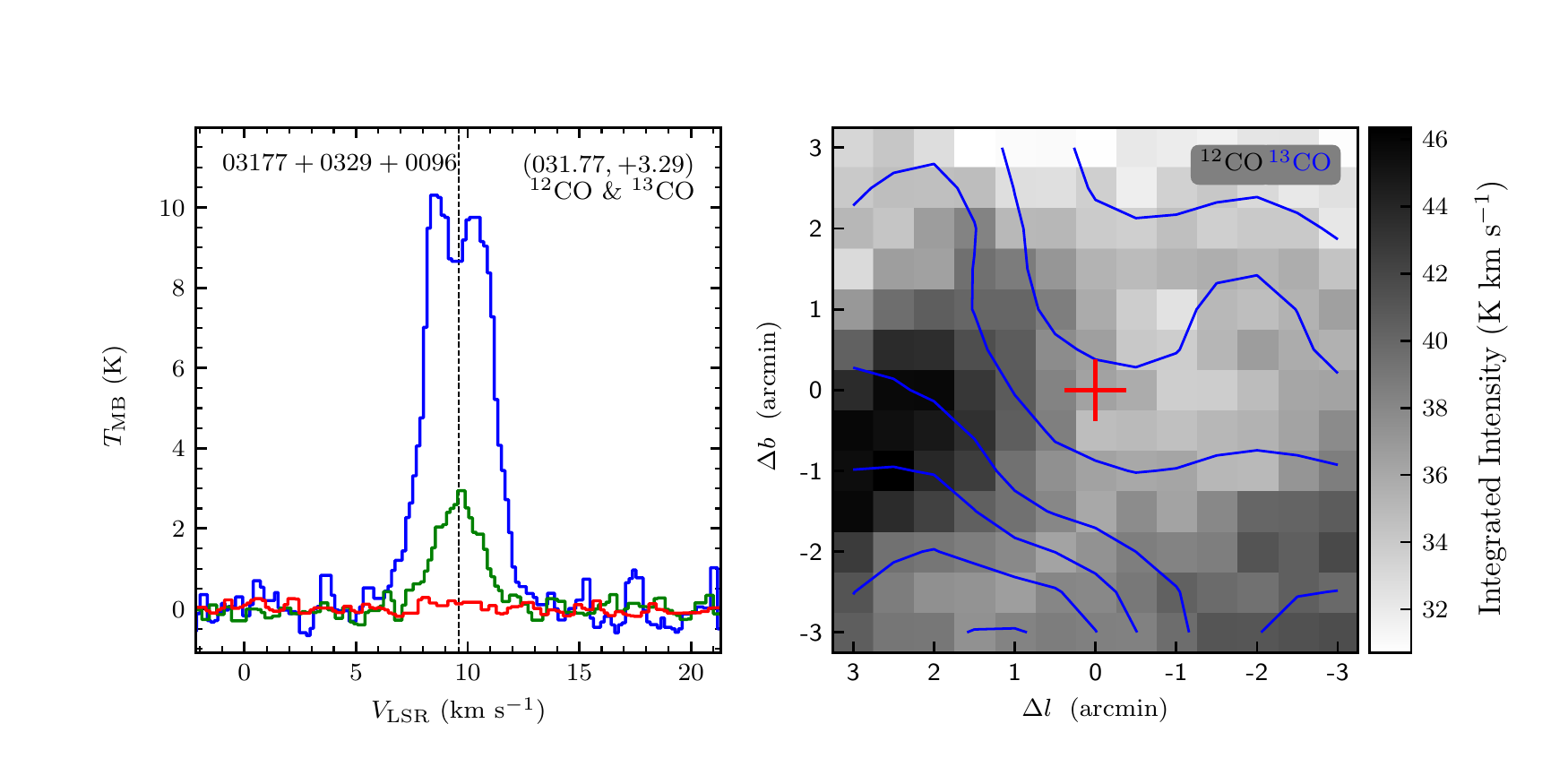}
\includegraphics[width=9.0cm,angle=0]{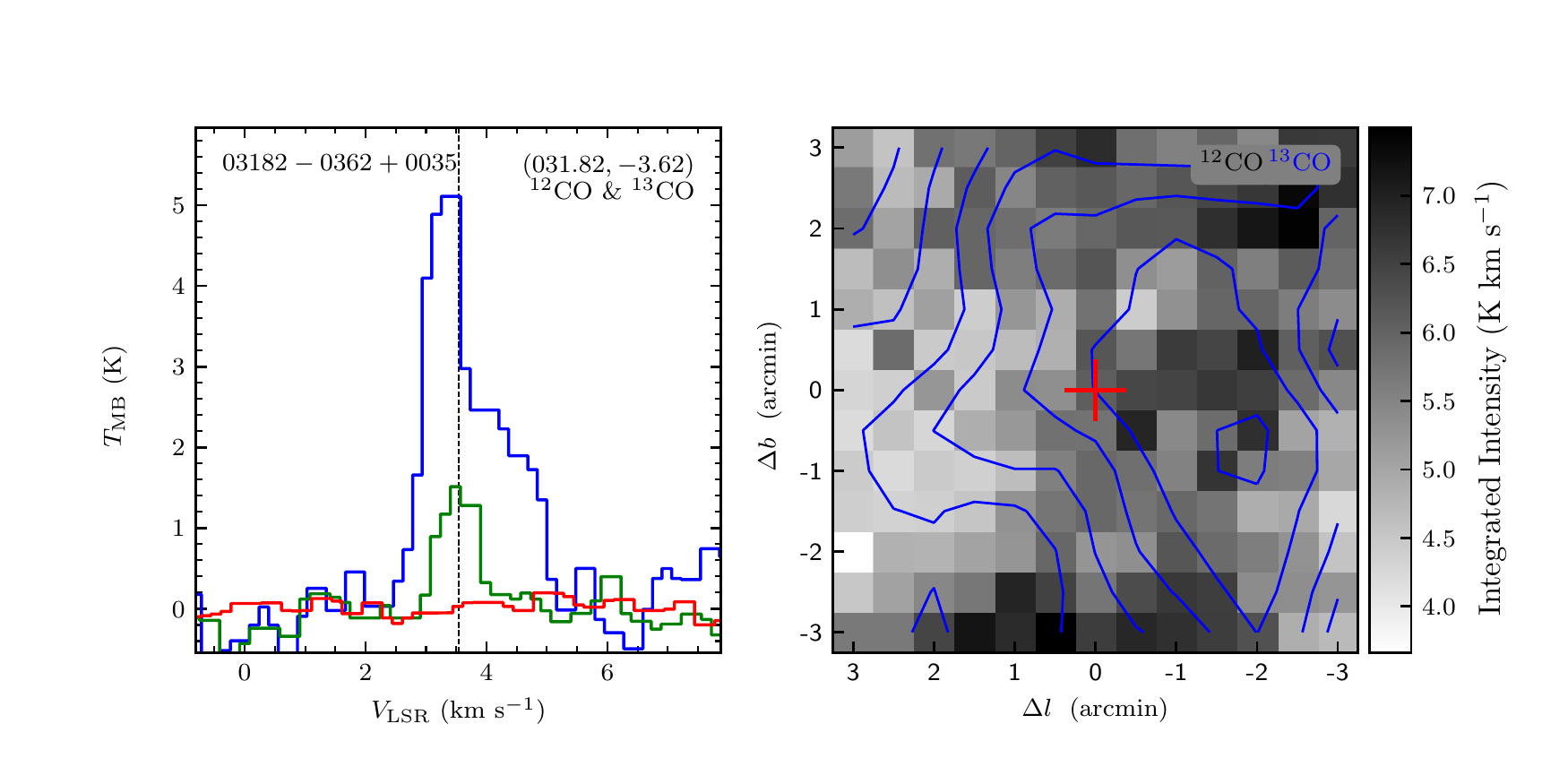}
\vspace{-0.5cm}

\includegraphics[width=9.0cm,angle=0]{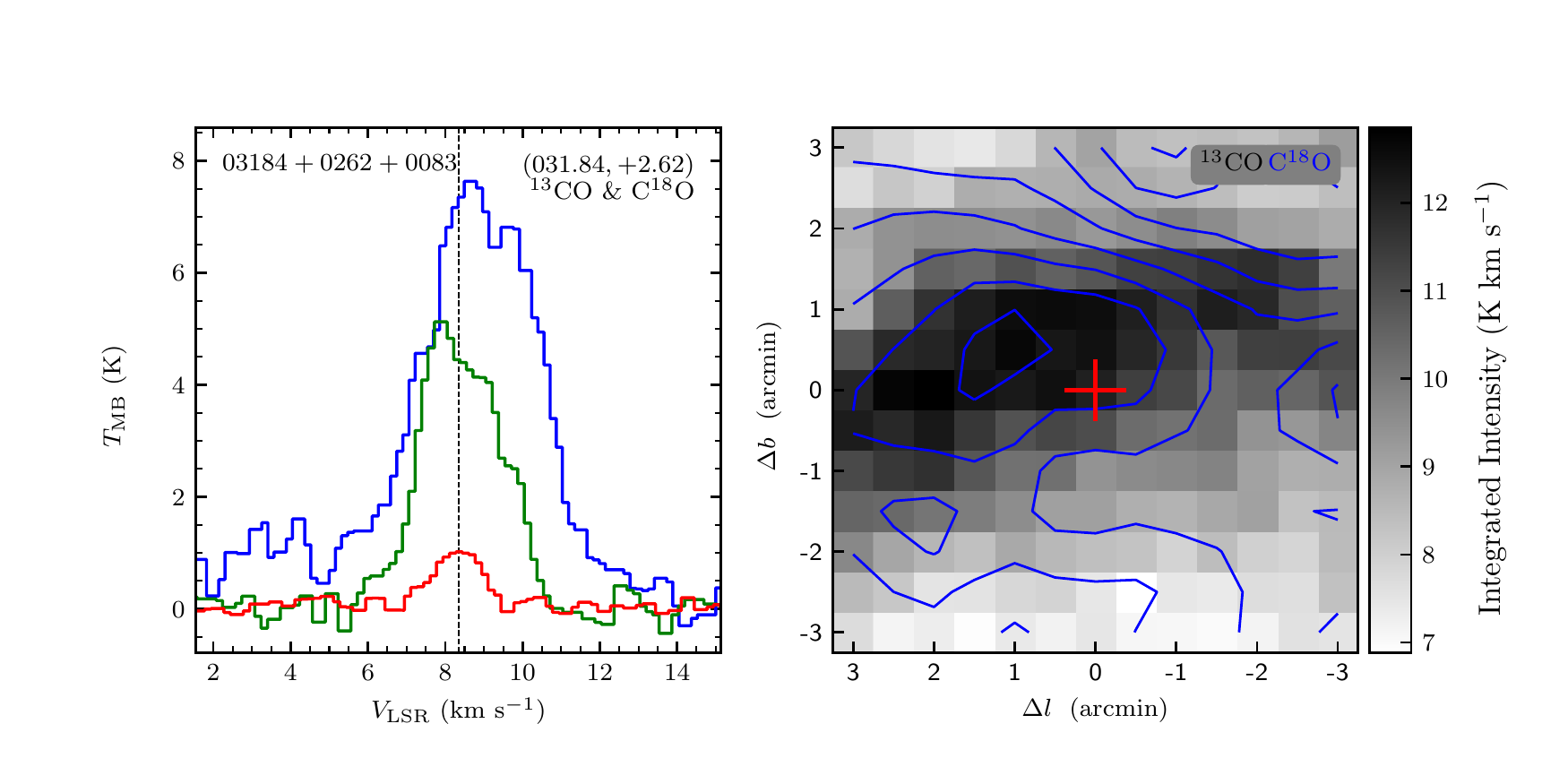}
\includegraphics[width=9.0cm,angle=0]{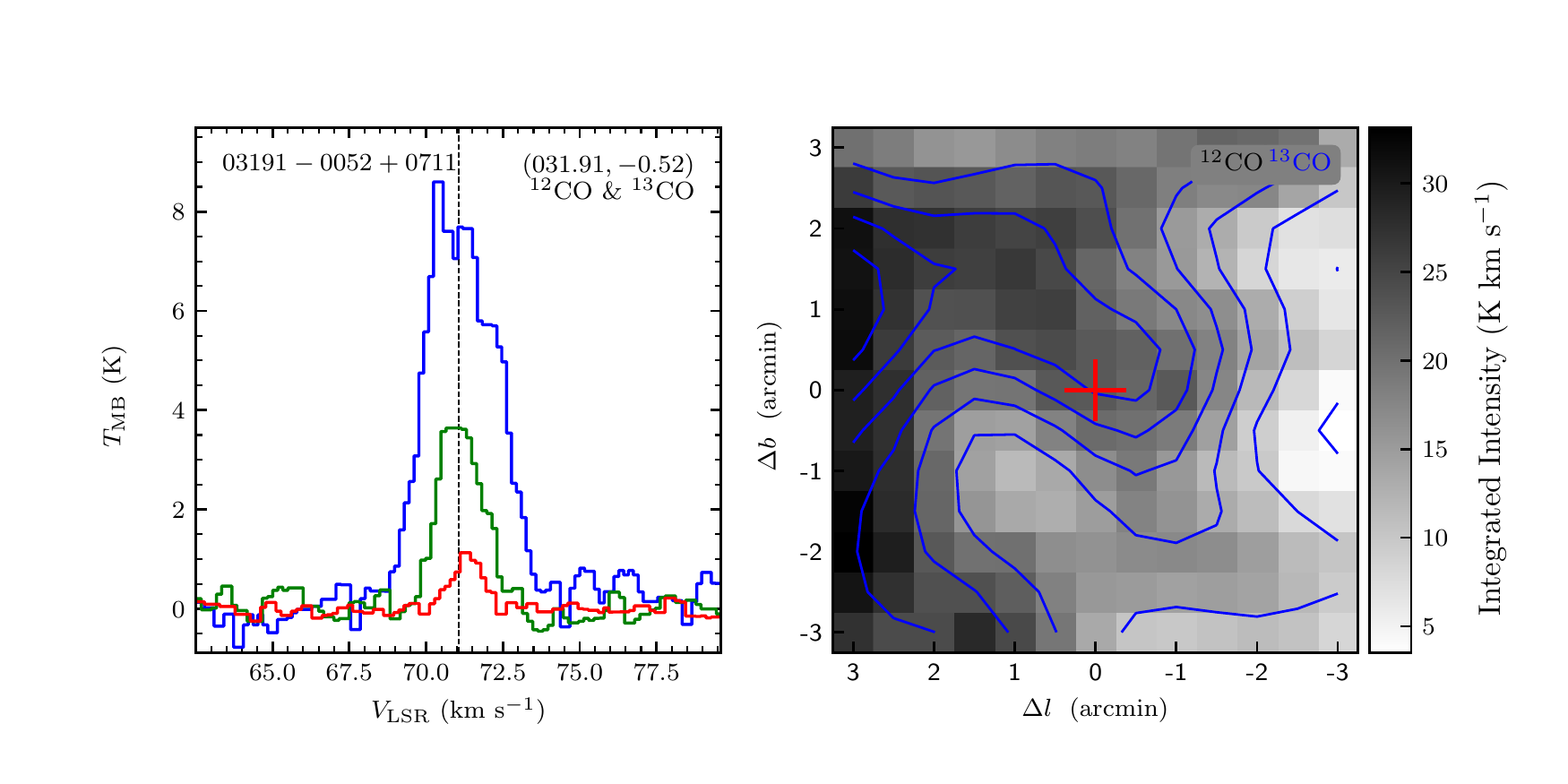}
\end{figure}
\clearpage

\begin{figure}
\includegraphics[width=9.0cm,angle=0]{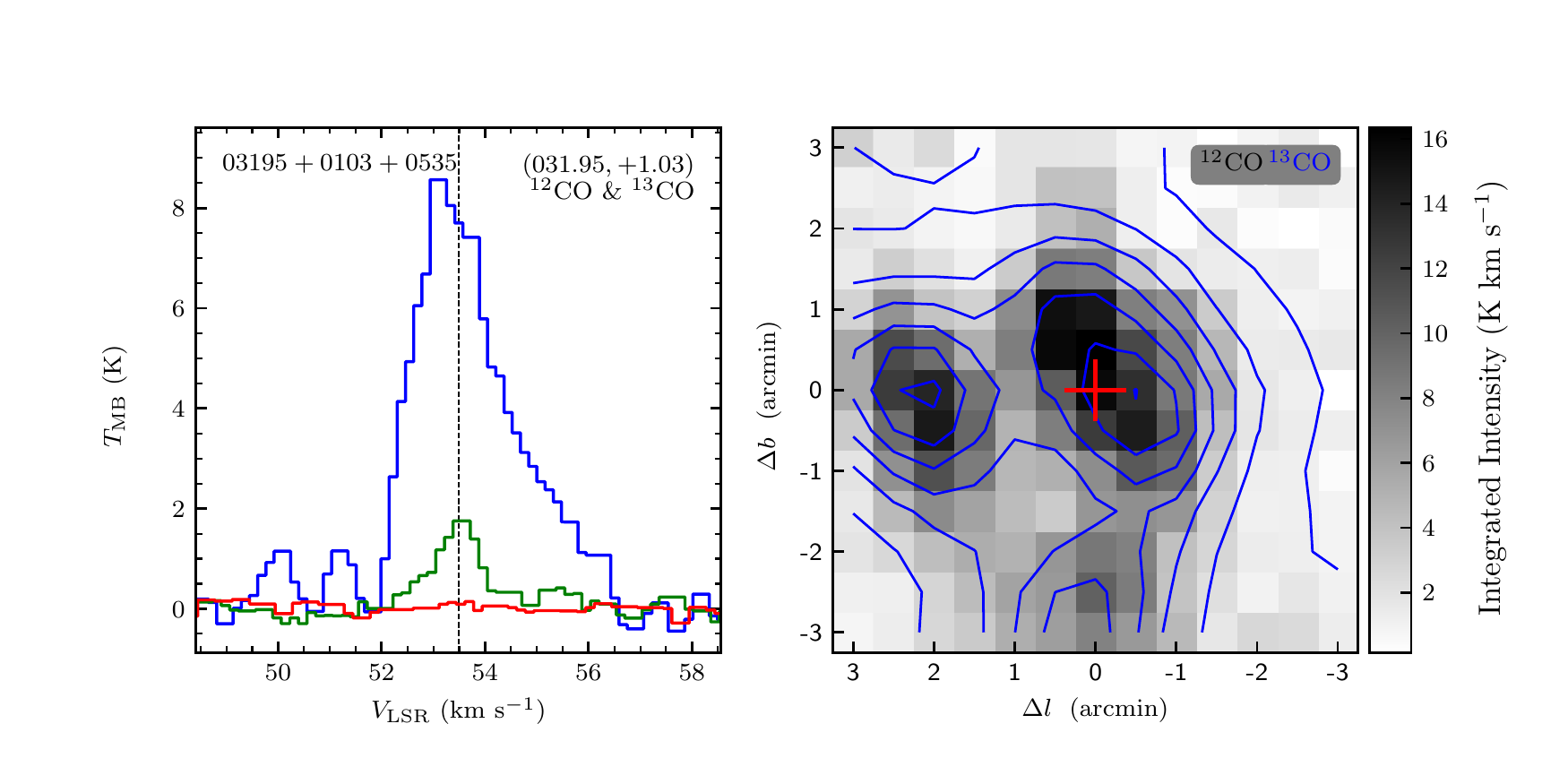}
\includegraphics[width=9.0cm,angle=0]{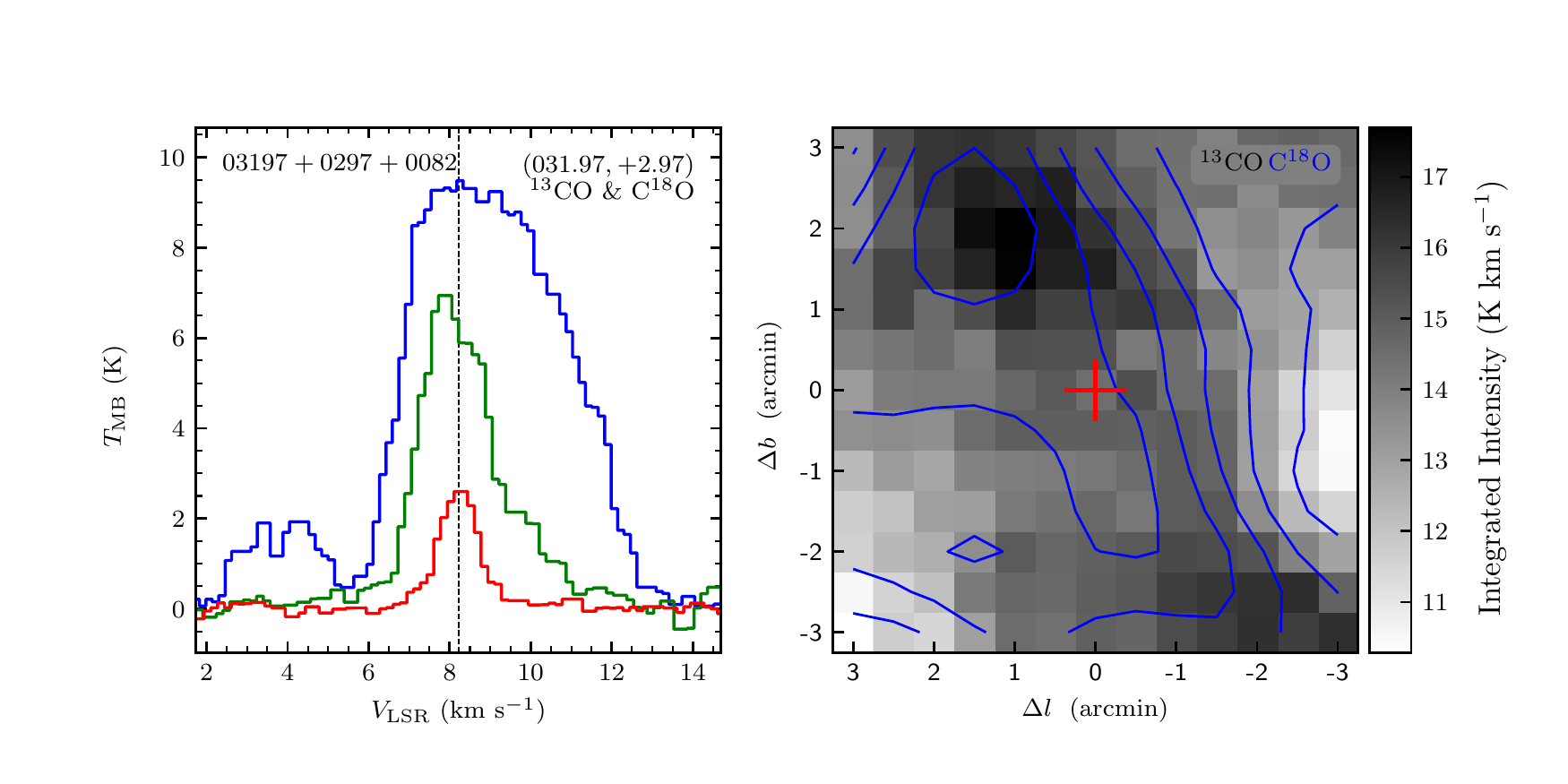}
\vspace{-0.5cm}

\includegraphics[width=9.0cm,angle=0]{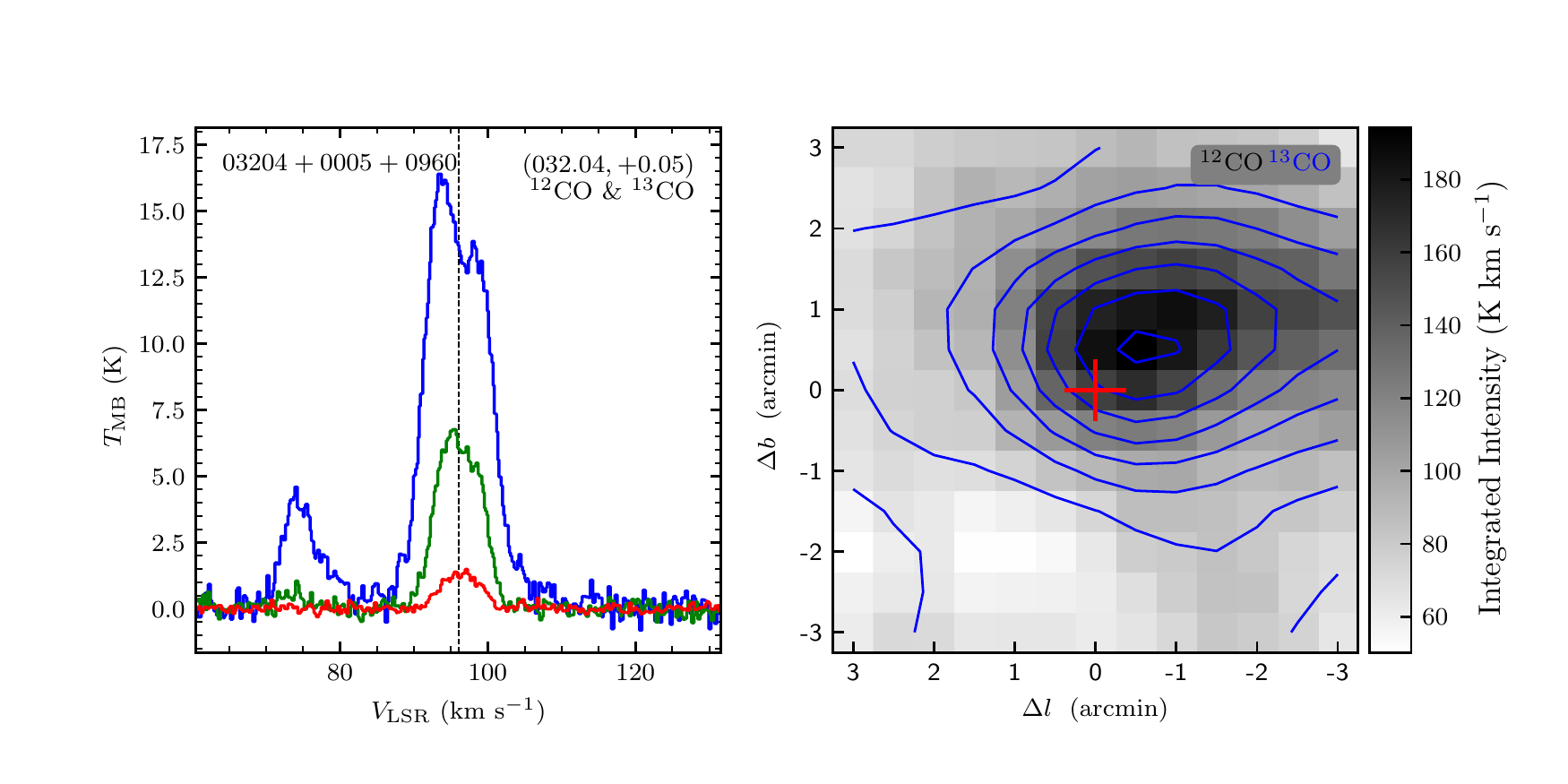}
\includegraphics[width=9.0cm,angle=0]{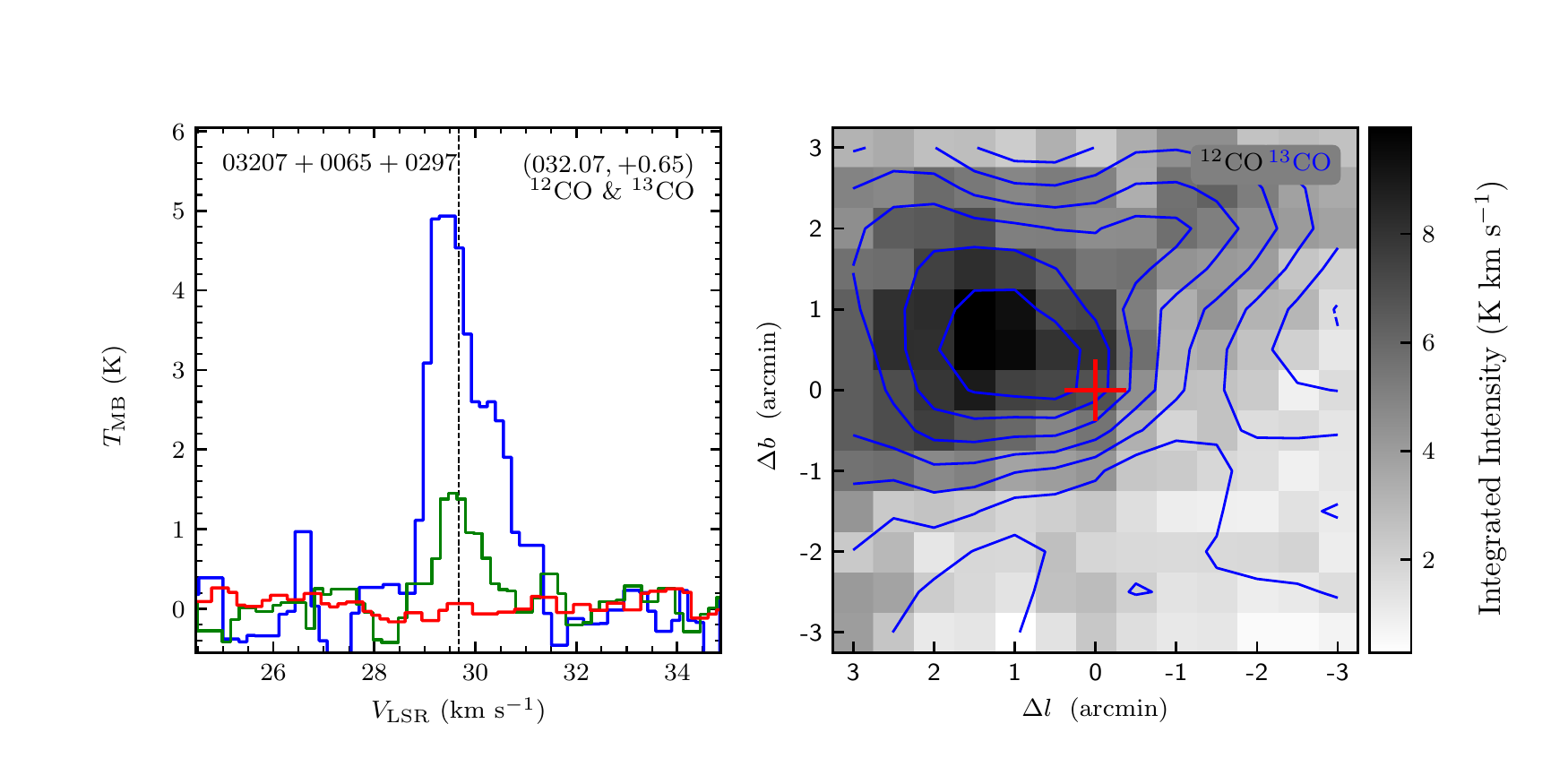}
\vspace{-0.5cm}

\includegraphics[width=9.0cm,angle=0]{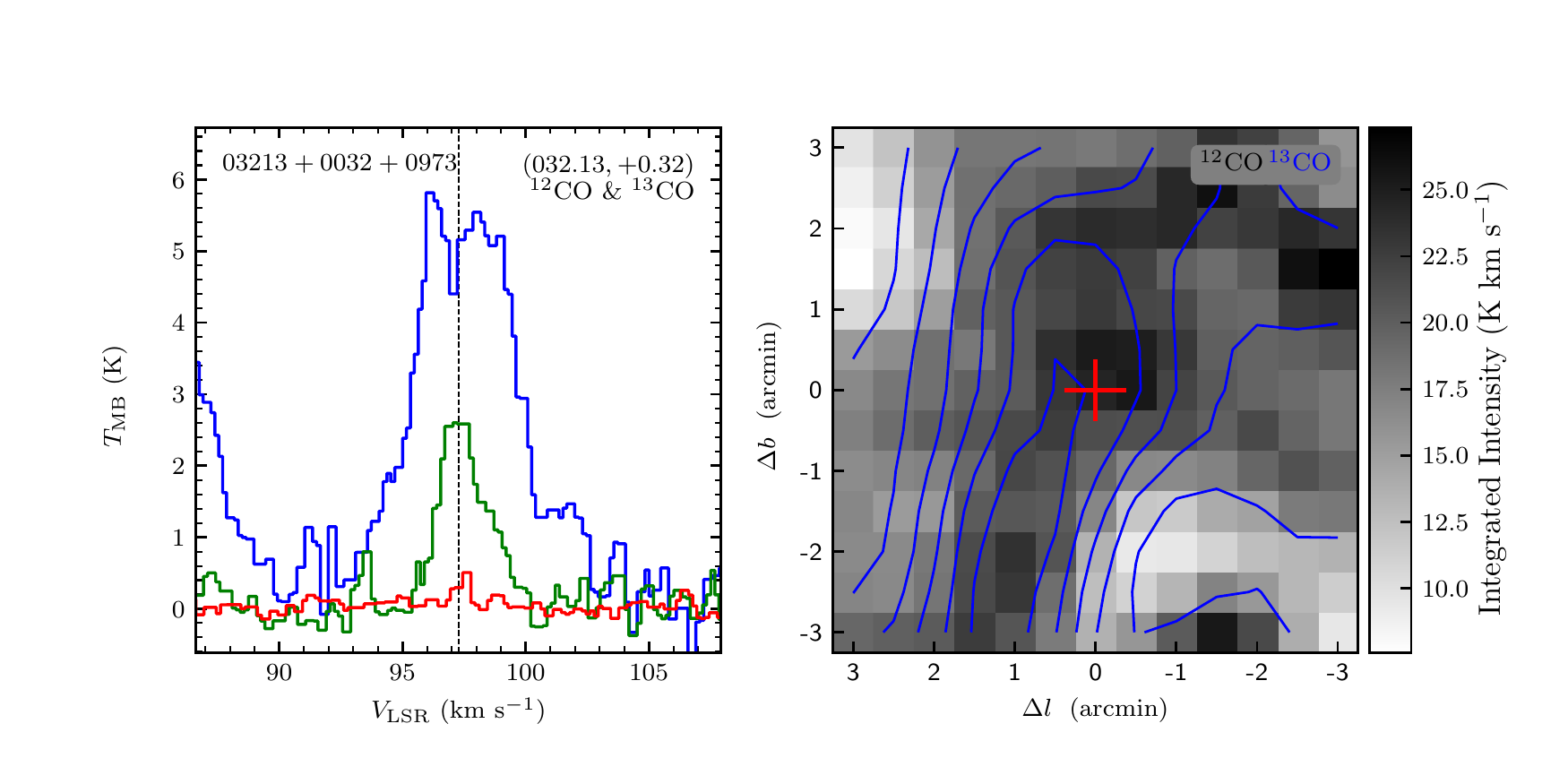}
\includegraphics[width=9.0cm,angle=0]{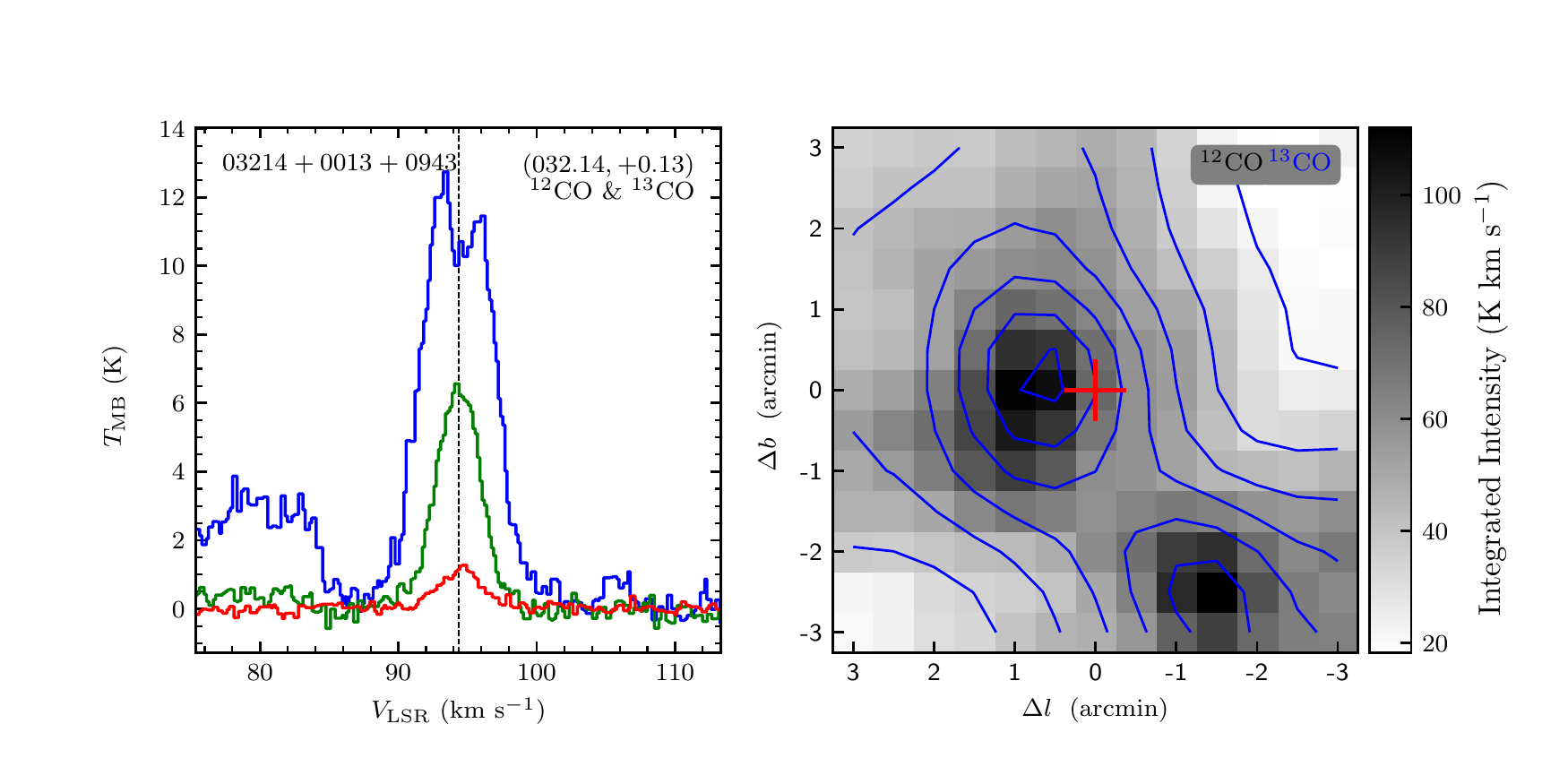}
\vspace{-0.5cm}

\includegraphics[width=9.0cm,angle=0]{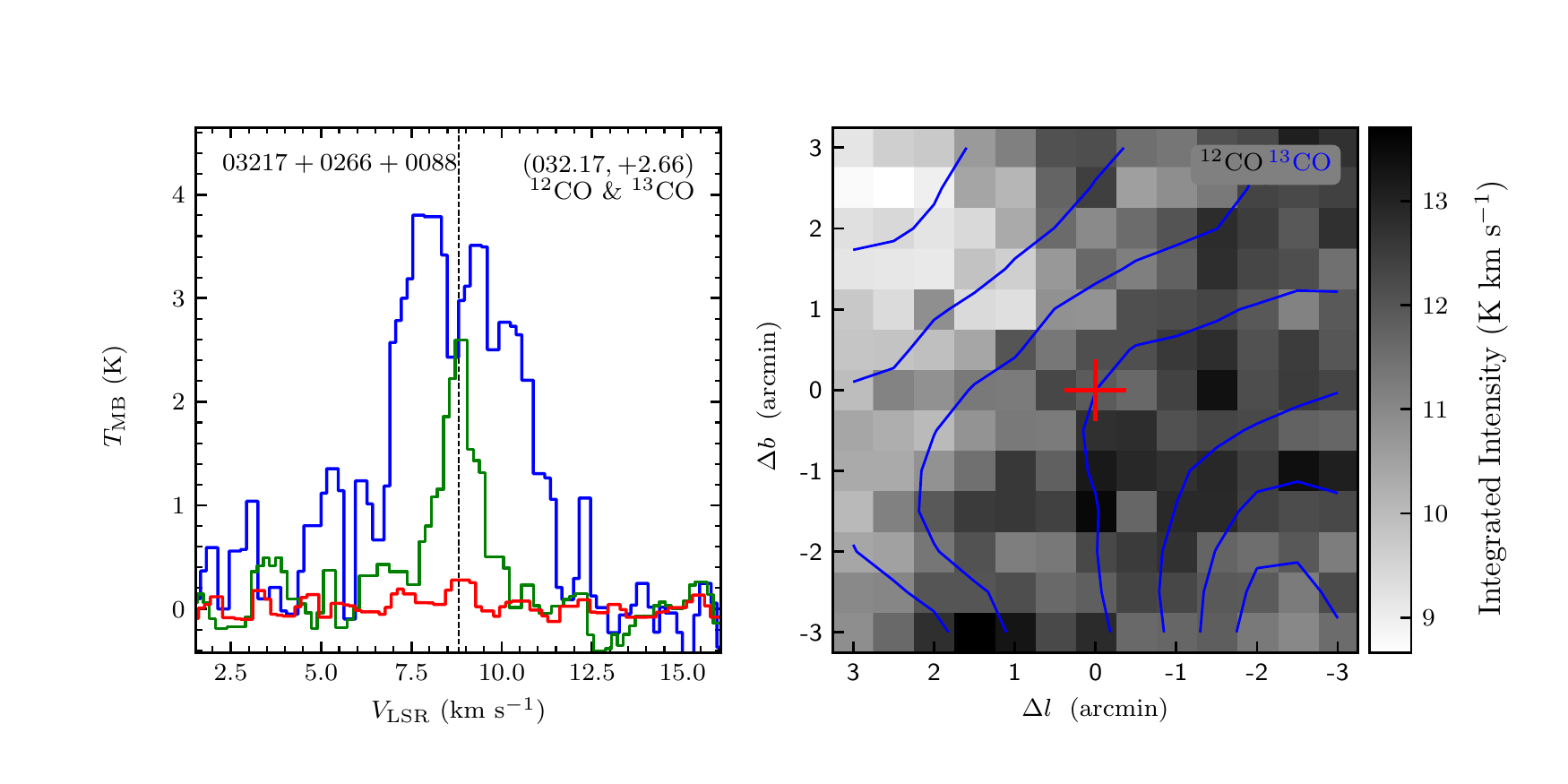}
\includegraphics[width=9.0cm,angle=0]{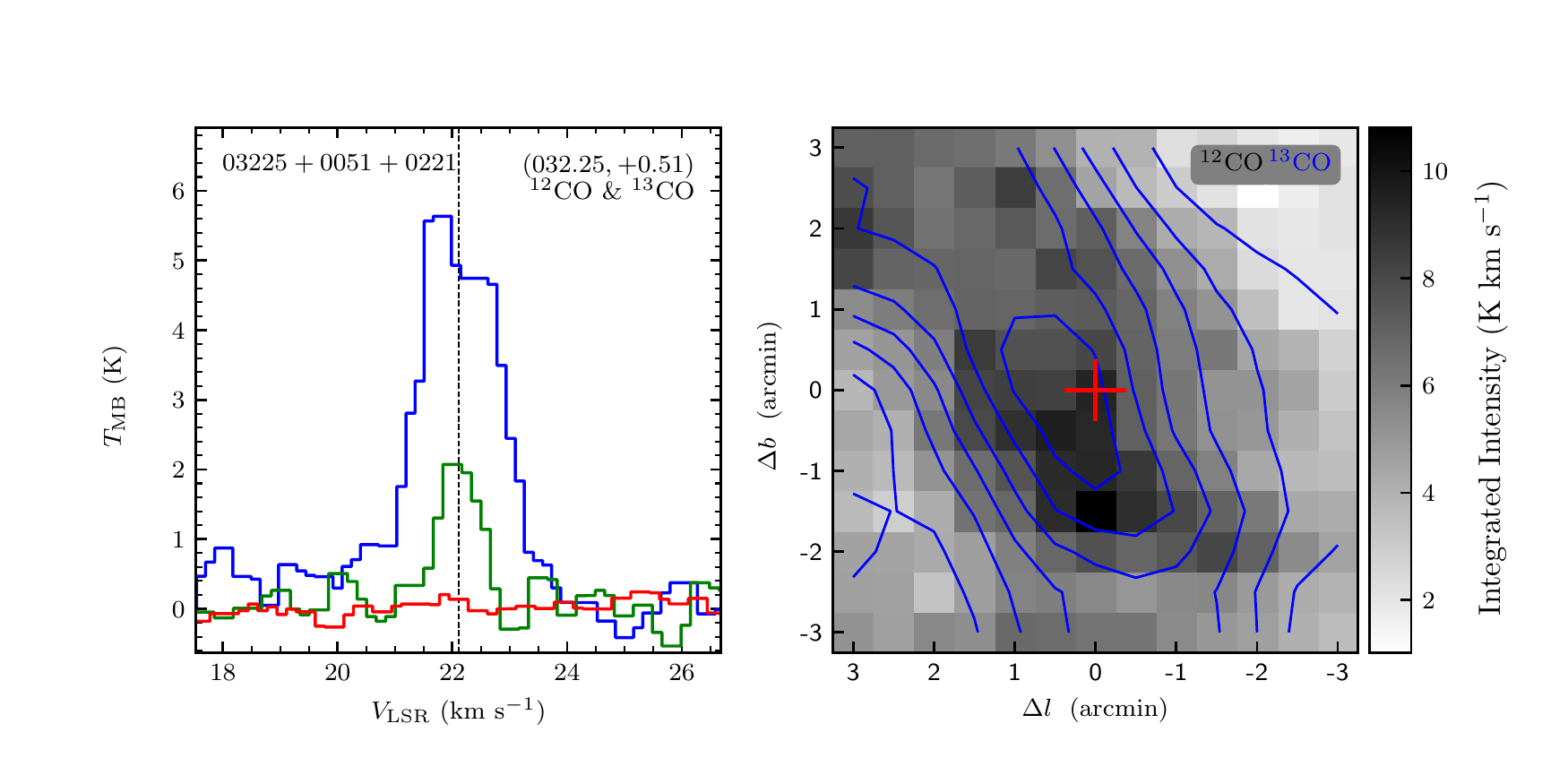}
\vspace{-0.5cm}

\includegraphics[width=9.0cm,angle=0]{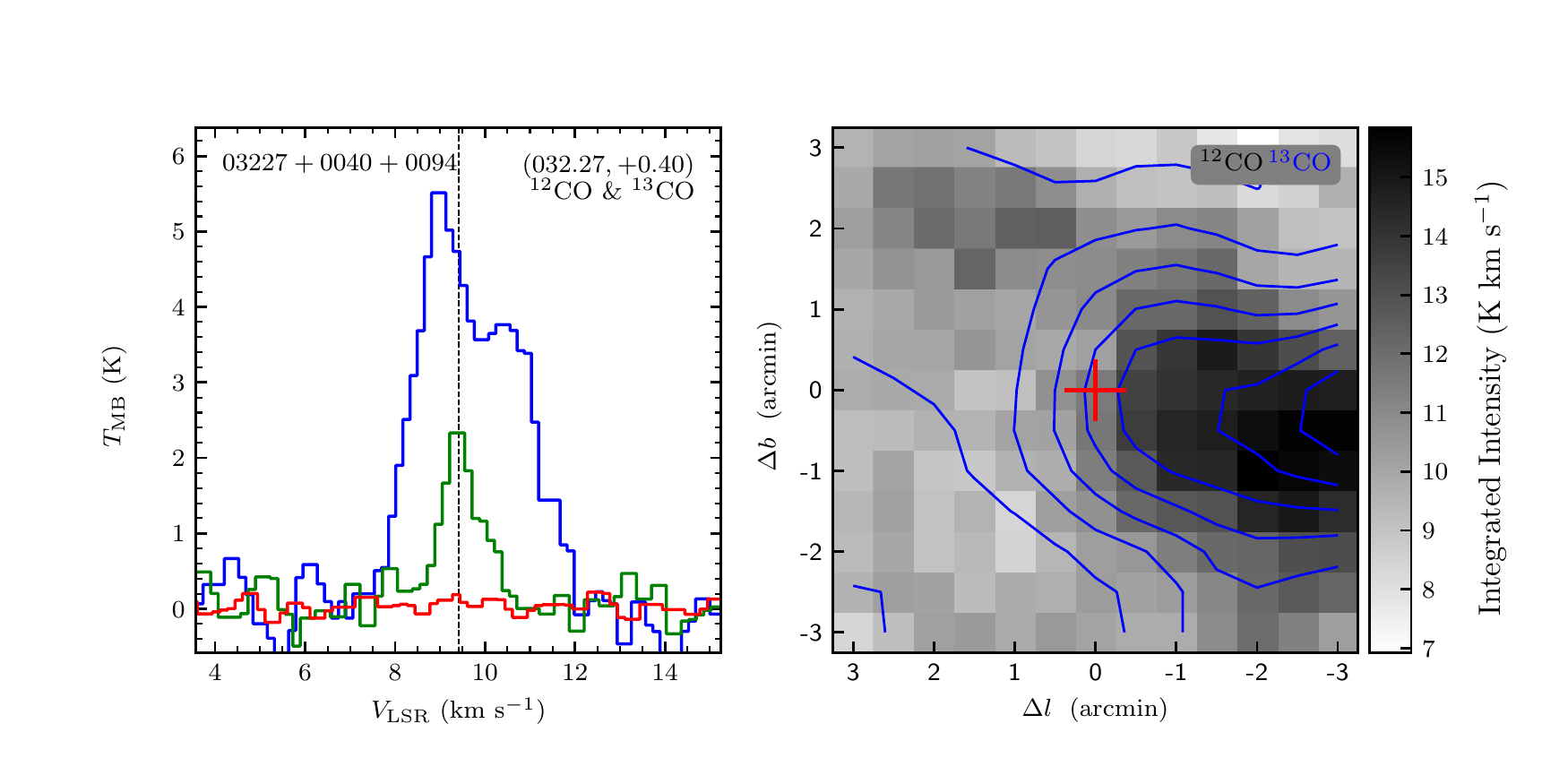}
\includegraphics[width=9.0cm,angle=0]{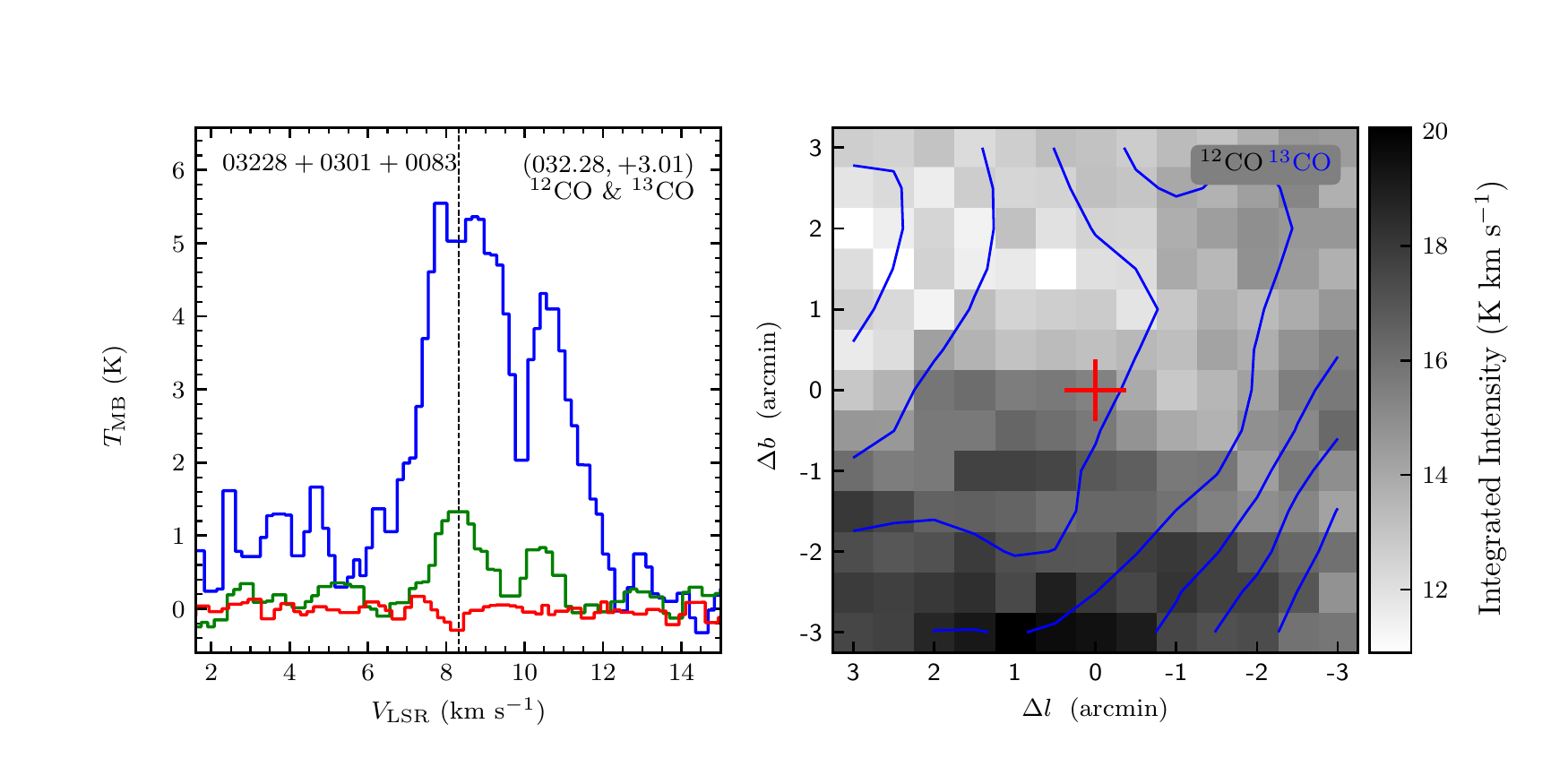}
\end{figure}
\clearpage

\begin{figure}
\includegraphics[width=9.0cm,angle=0]{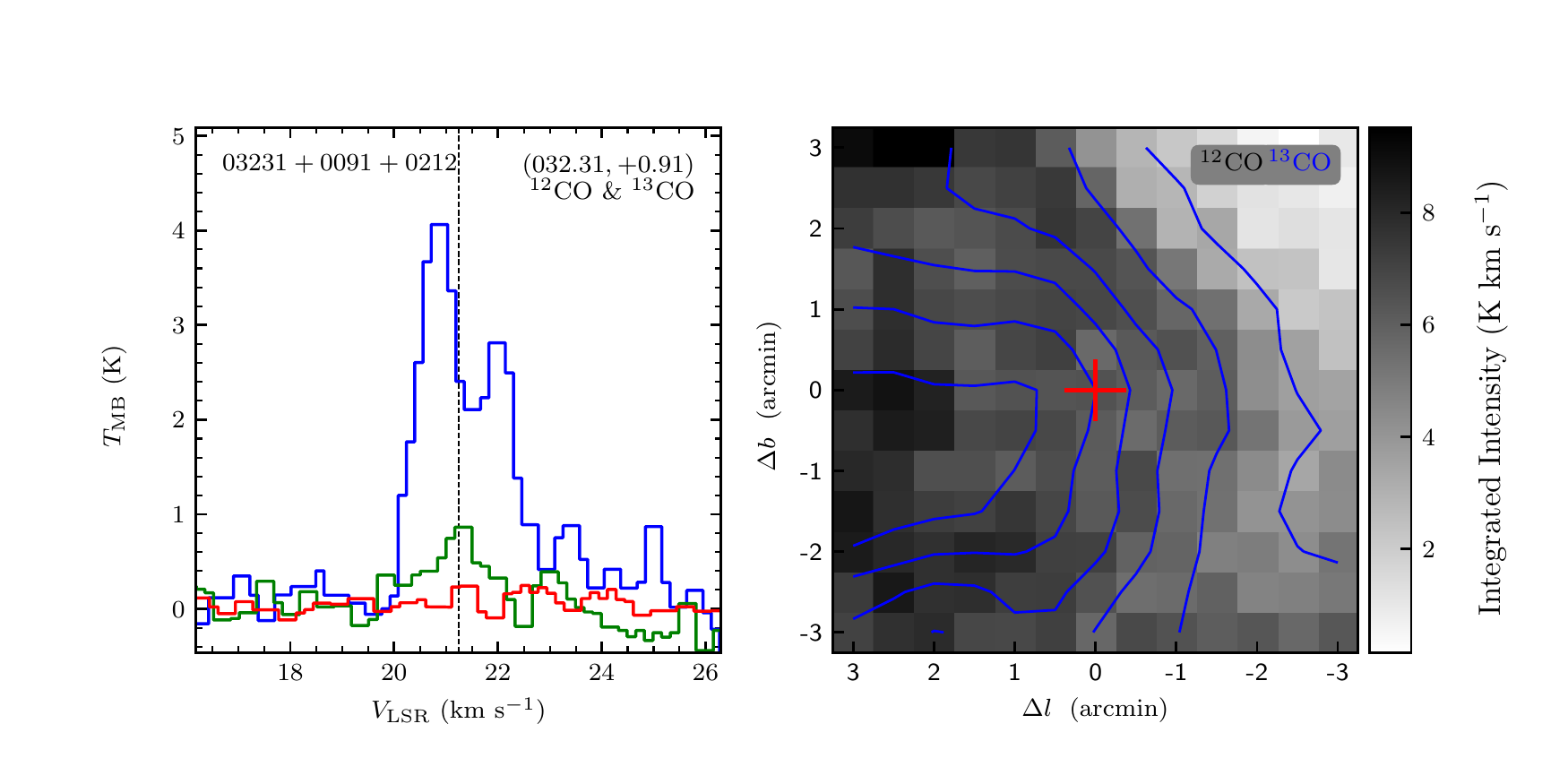}
\includegraphics[width=9.0cm,angle=0]{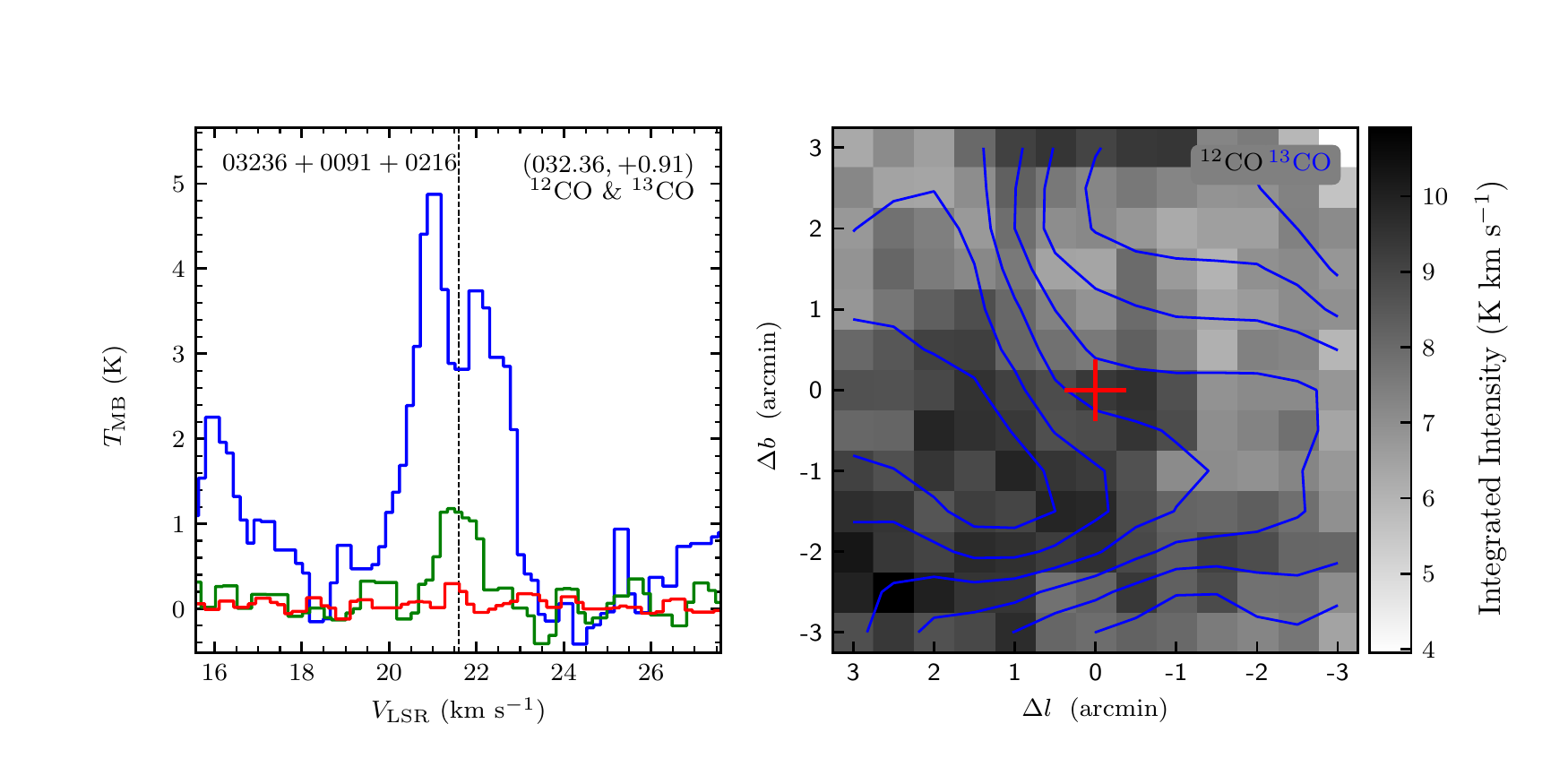}
\vspace{-0.5cm}

\includegraphics[width=9.0cm,angle=0]{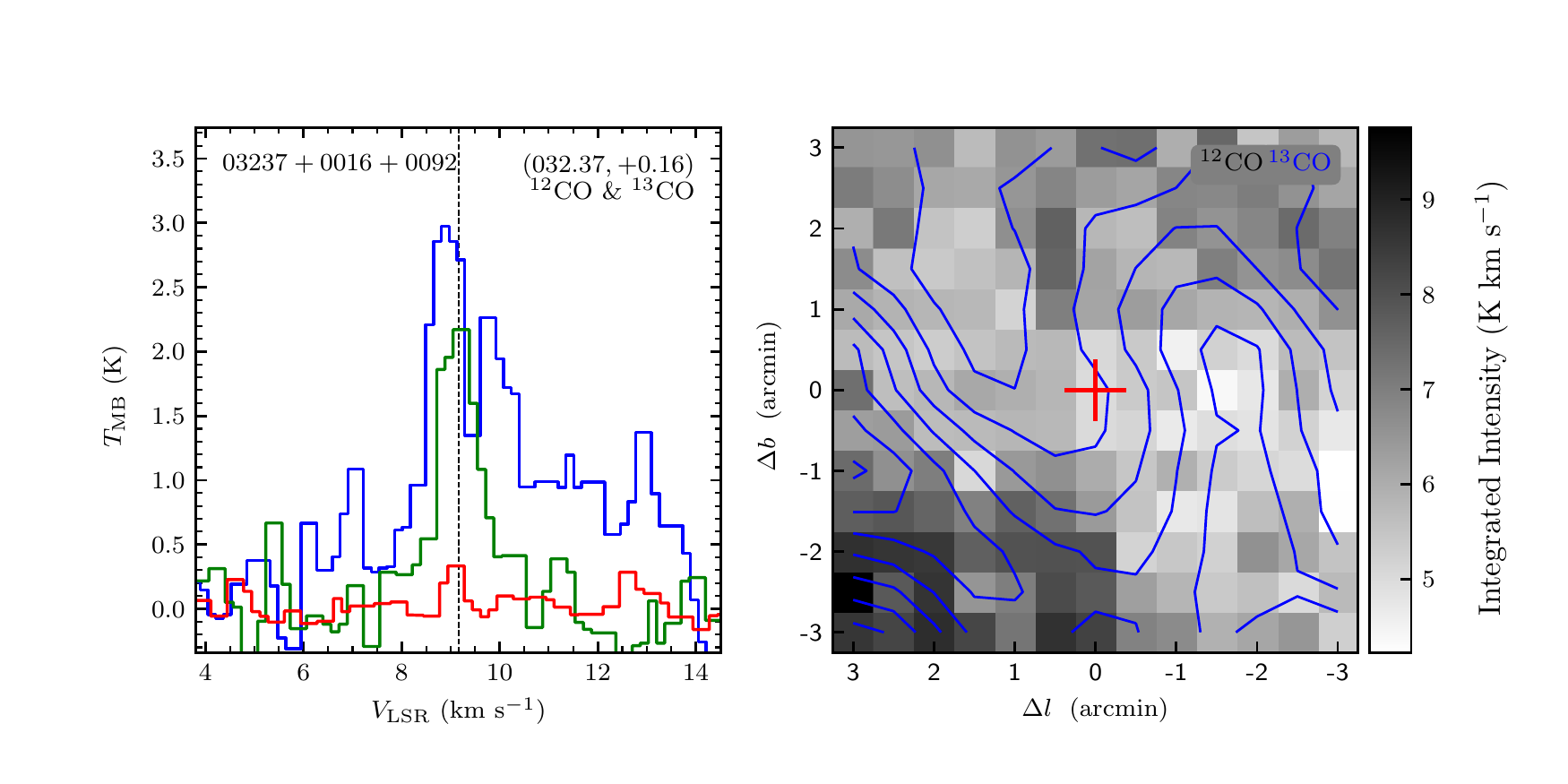}
\includegraphics[width=9.0cm,angle=0]{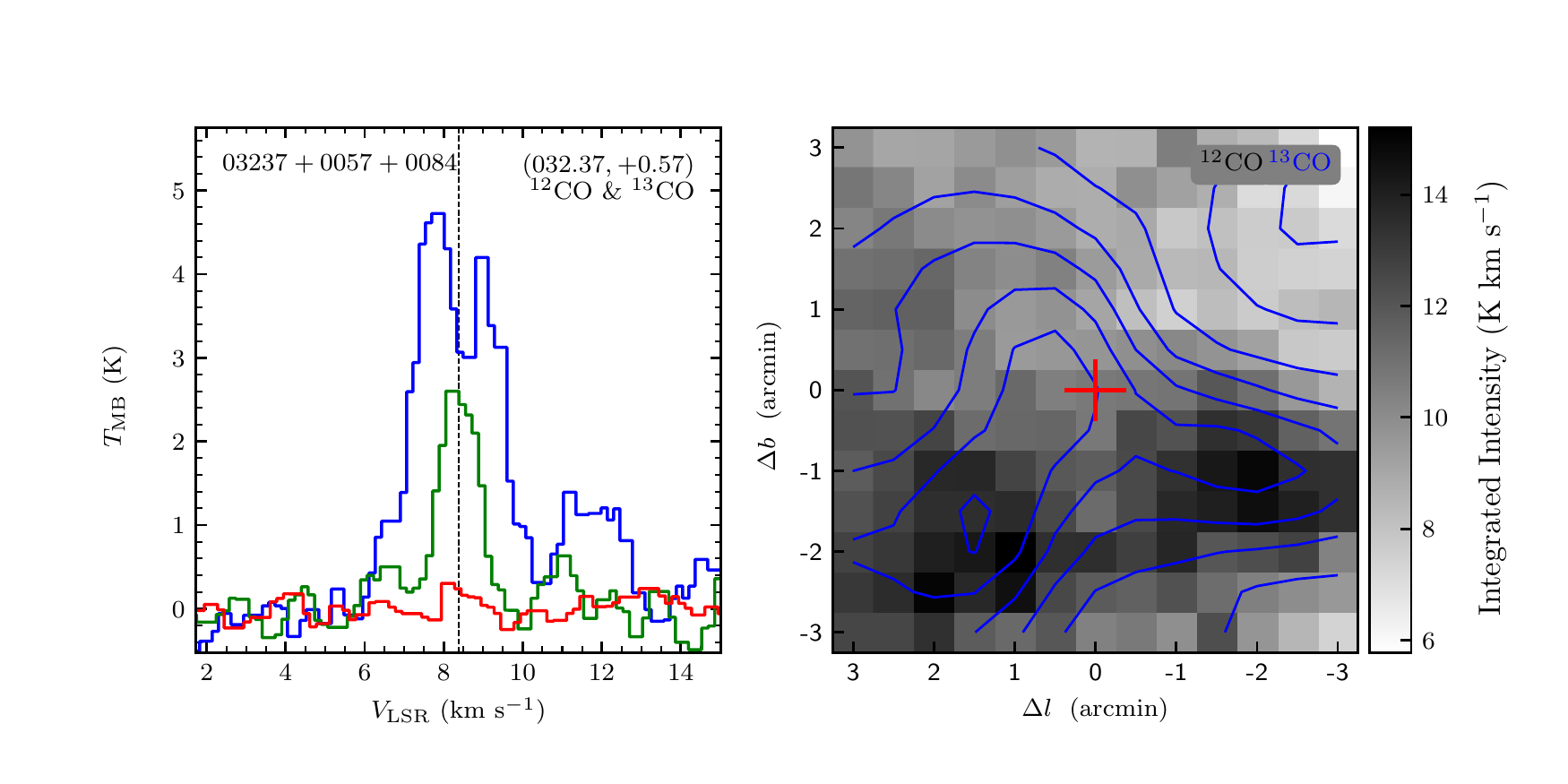}
\vspace{-0.5cm}

\includegraphics[width=9.0cm,angle=0]{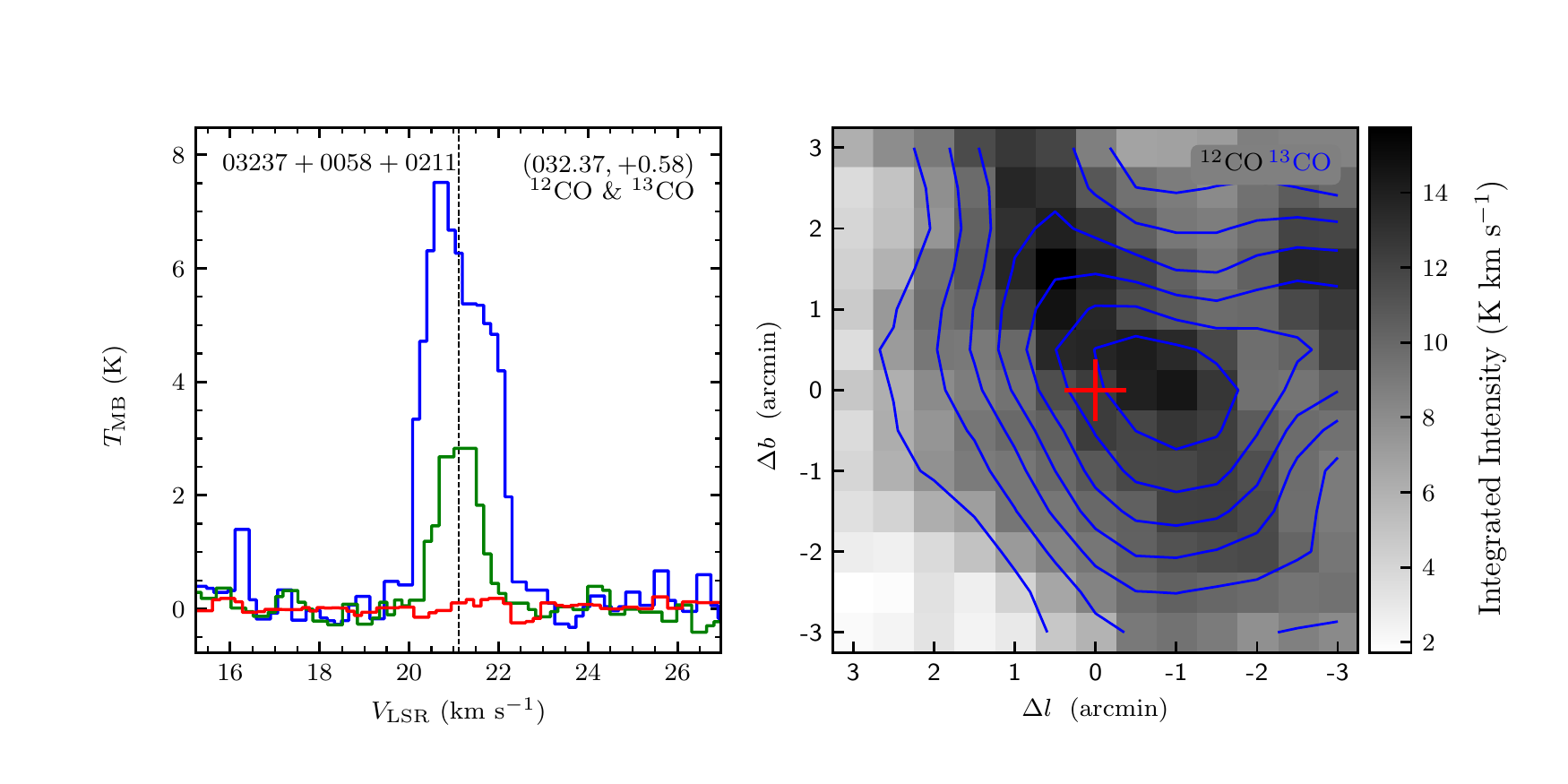}
\includegraphics[width=9.0cm,angle=0]{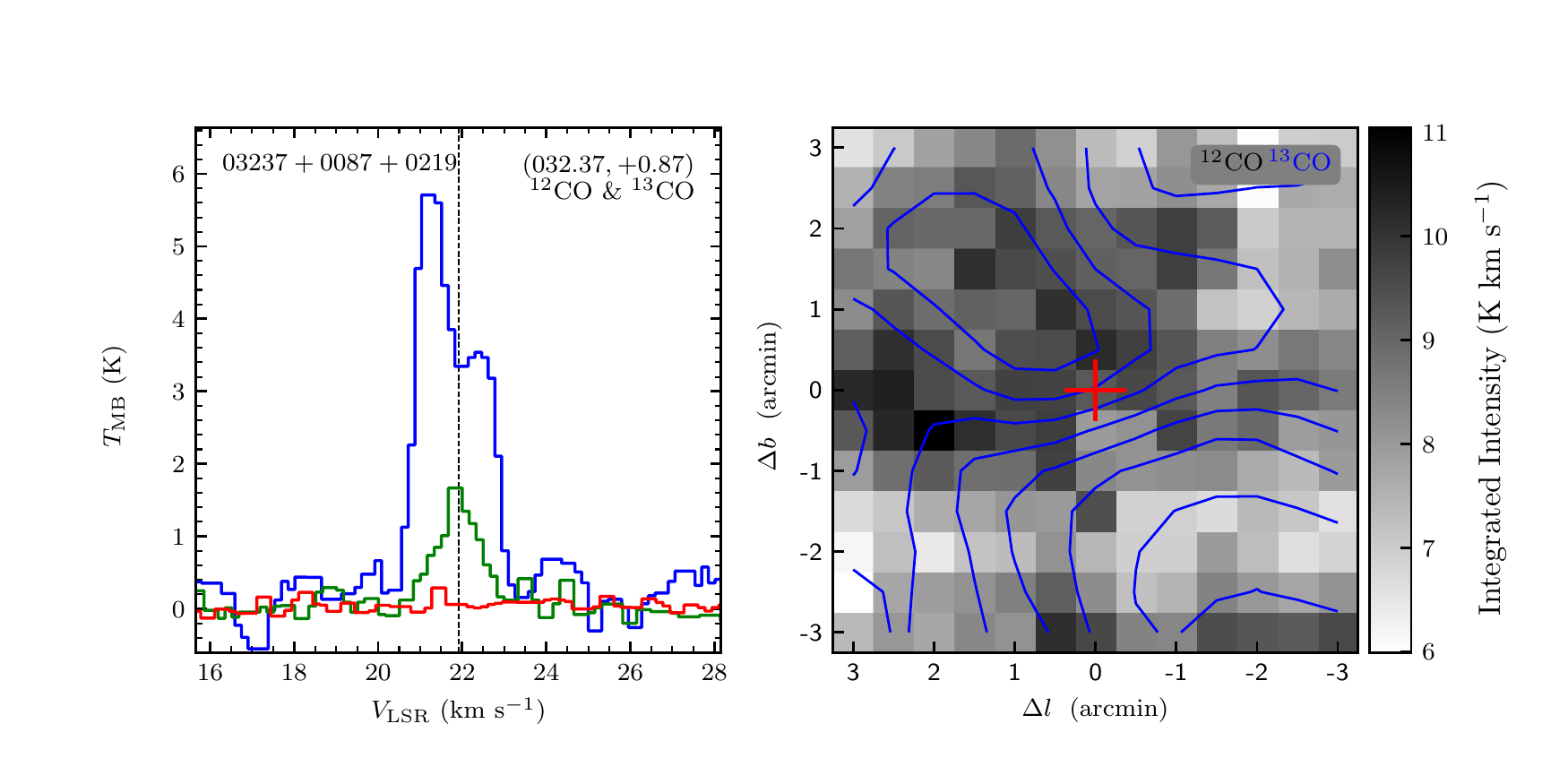}
\vspace{-0.5cm}

\includegraphics[width=9.0cm,angle=0]{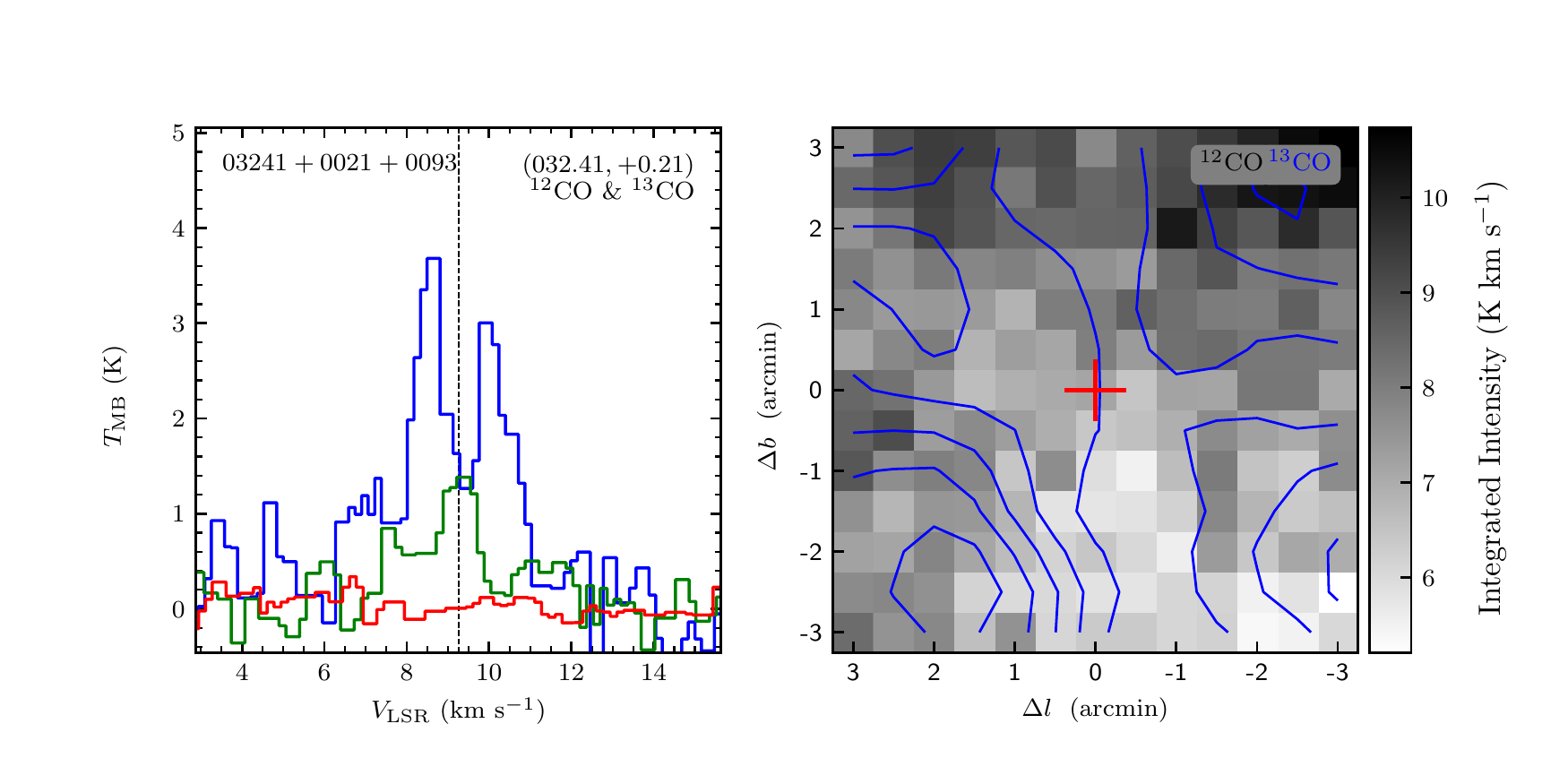}
\includegraphics[width=9.0cm,angle=0]{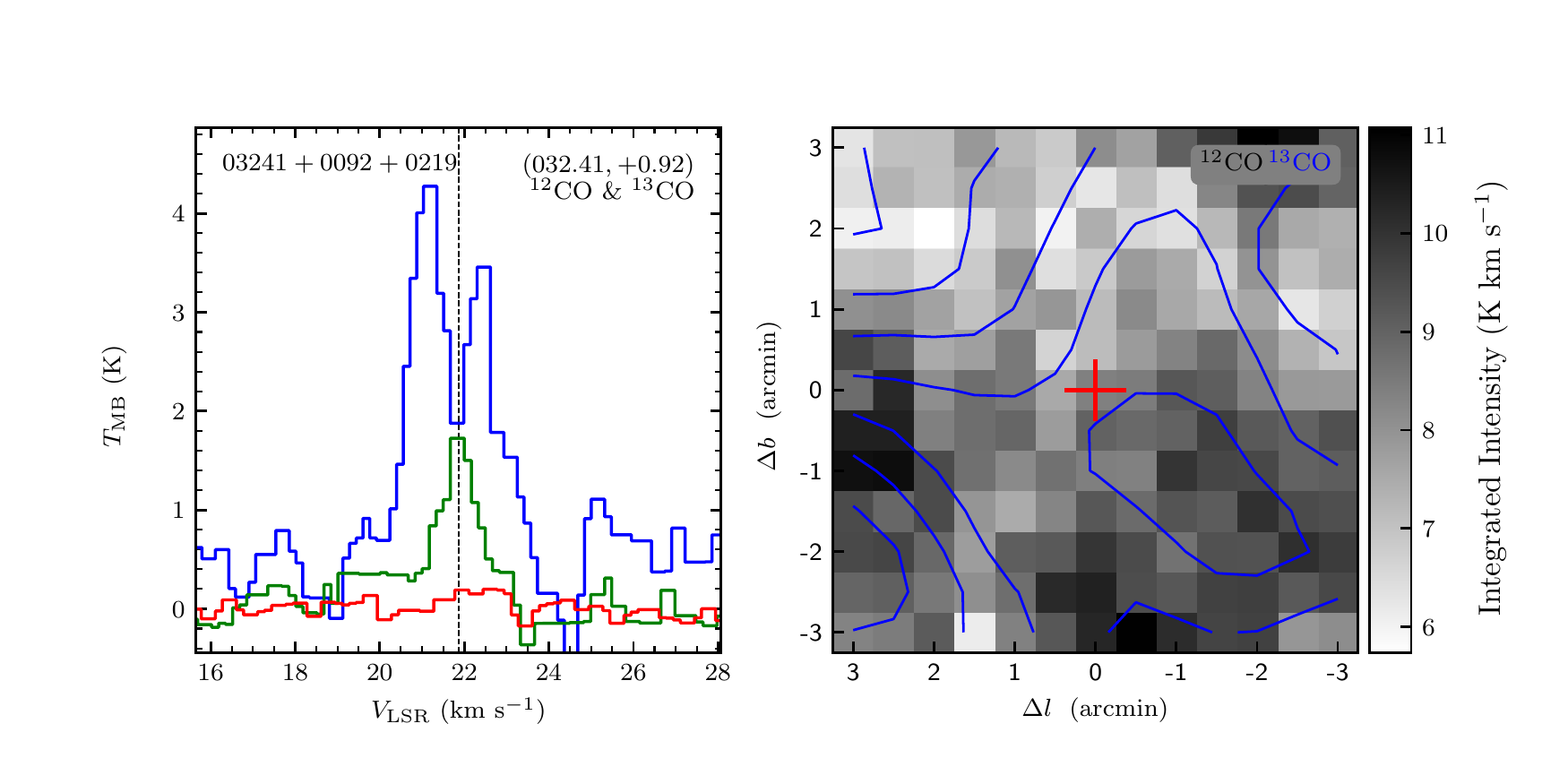}
\vspace{-0.5cm}

\includegraphics[width=9.0cm,angle=0]{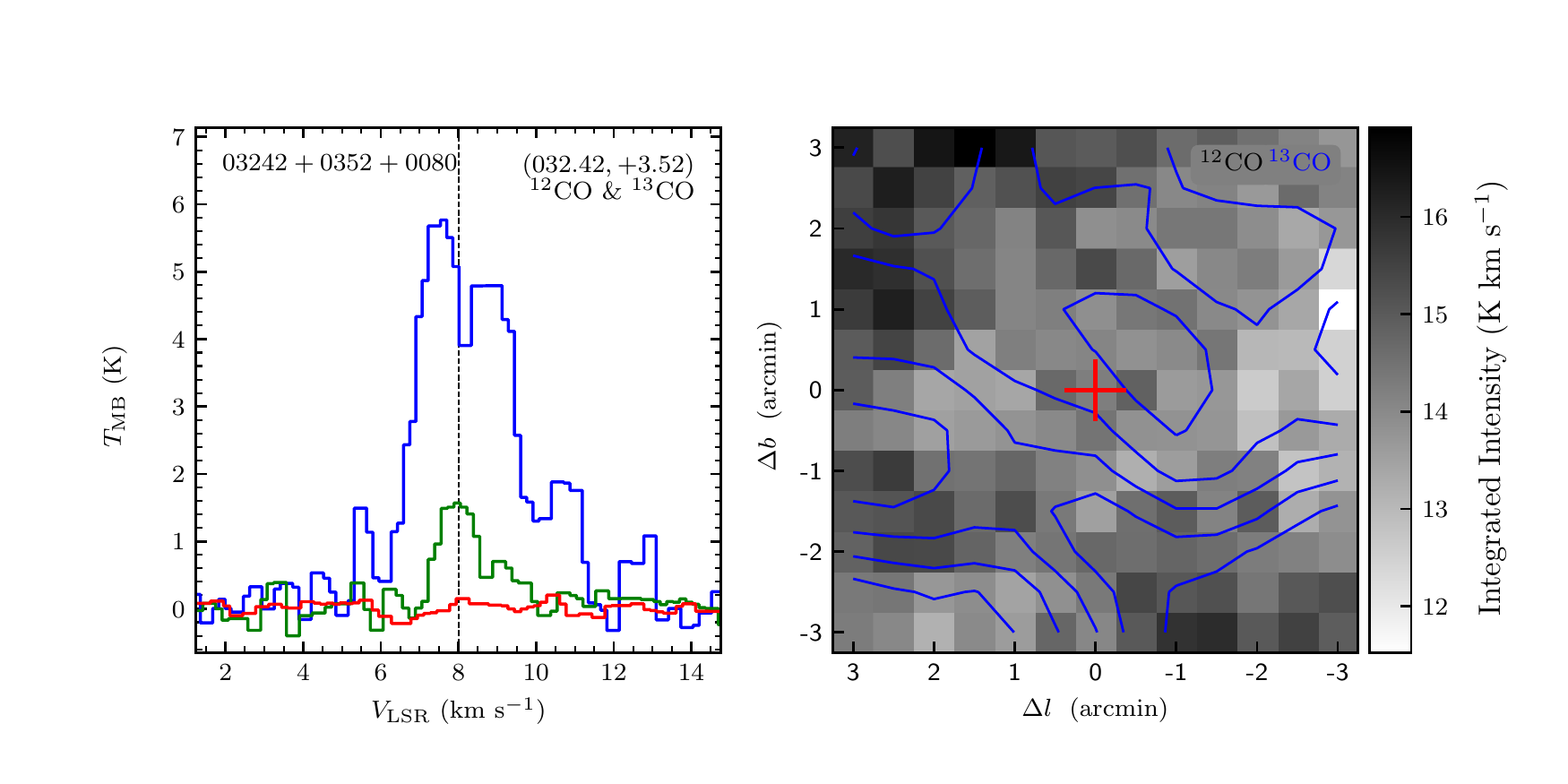}
\includegraphics[width=9.0cm,angle=0]{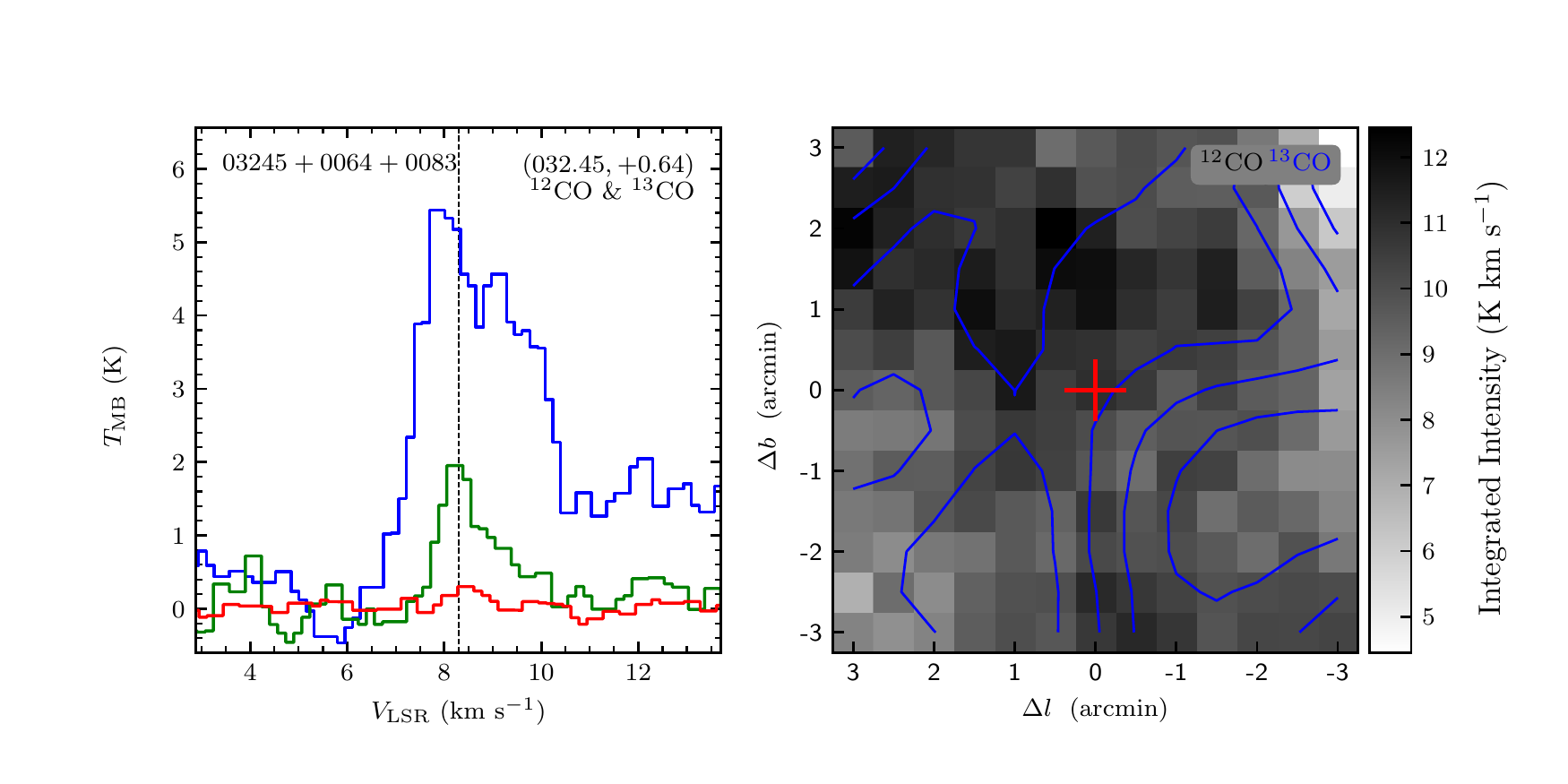}
\end{figure}
\clearpage

\begin{figure}
\includegraphics[width=9.0cm,angle=0]{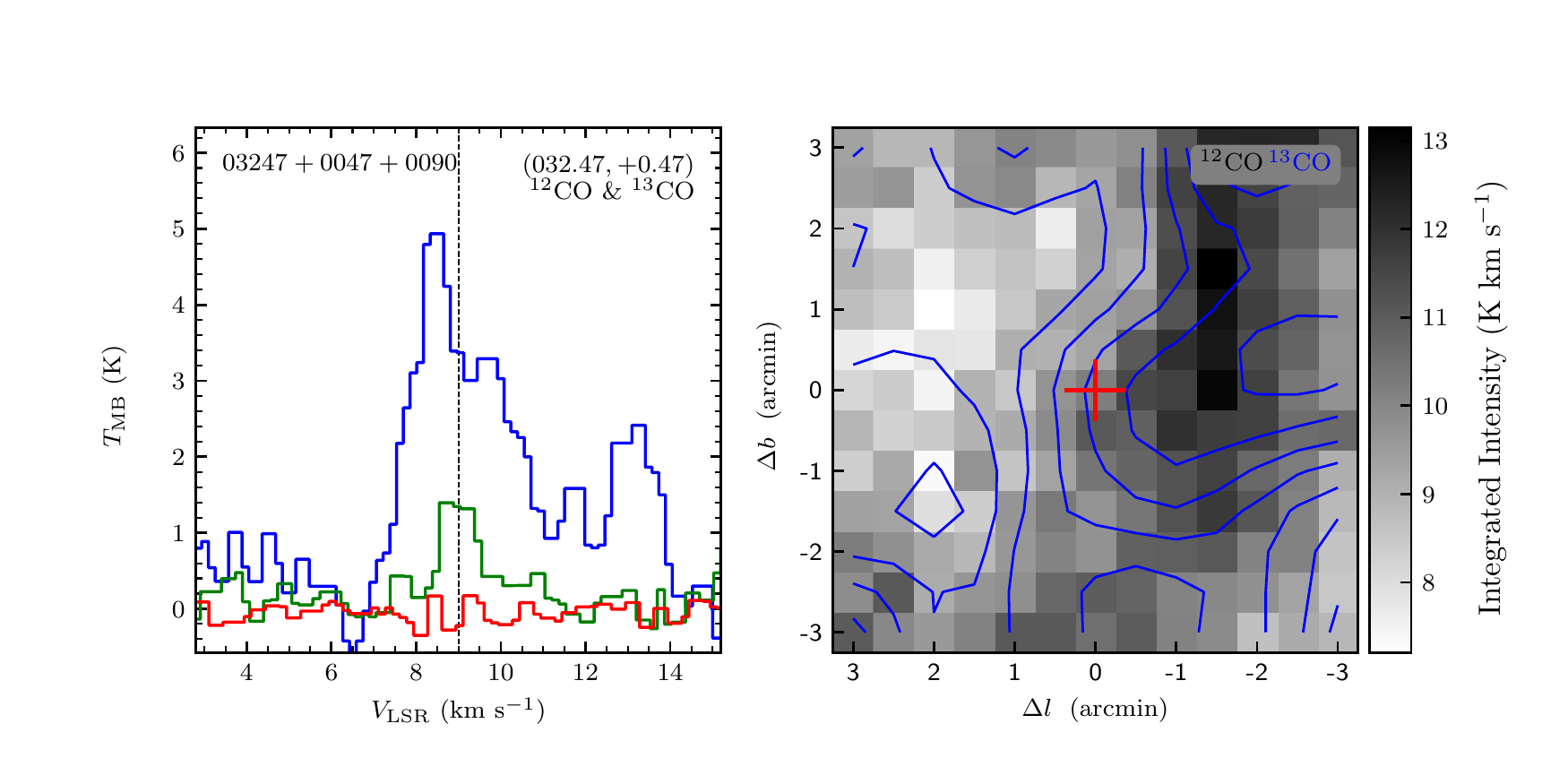}
\includegraphics[width=9.0cm,angle=0]{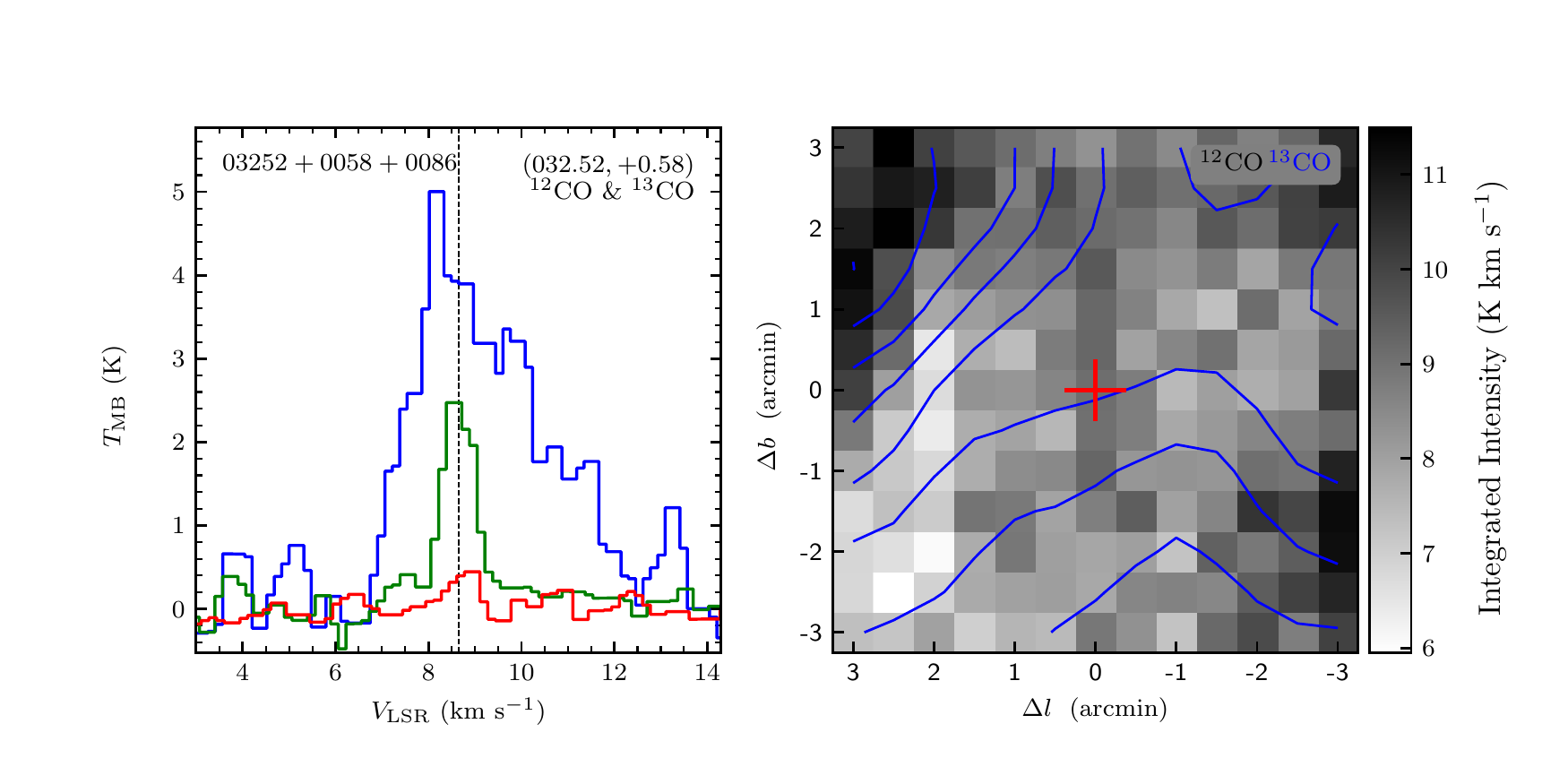}
\vspace{-0.5cm}

\includegraphics[width=9.0cm,angle=0]{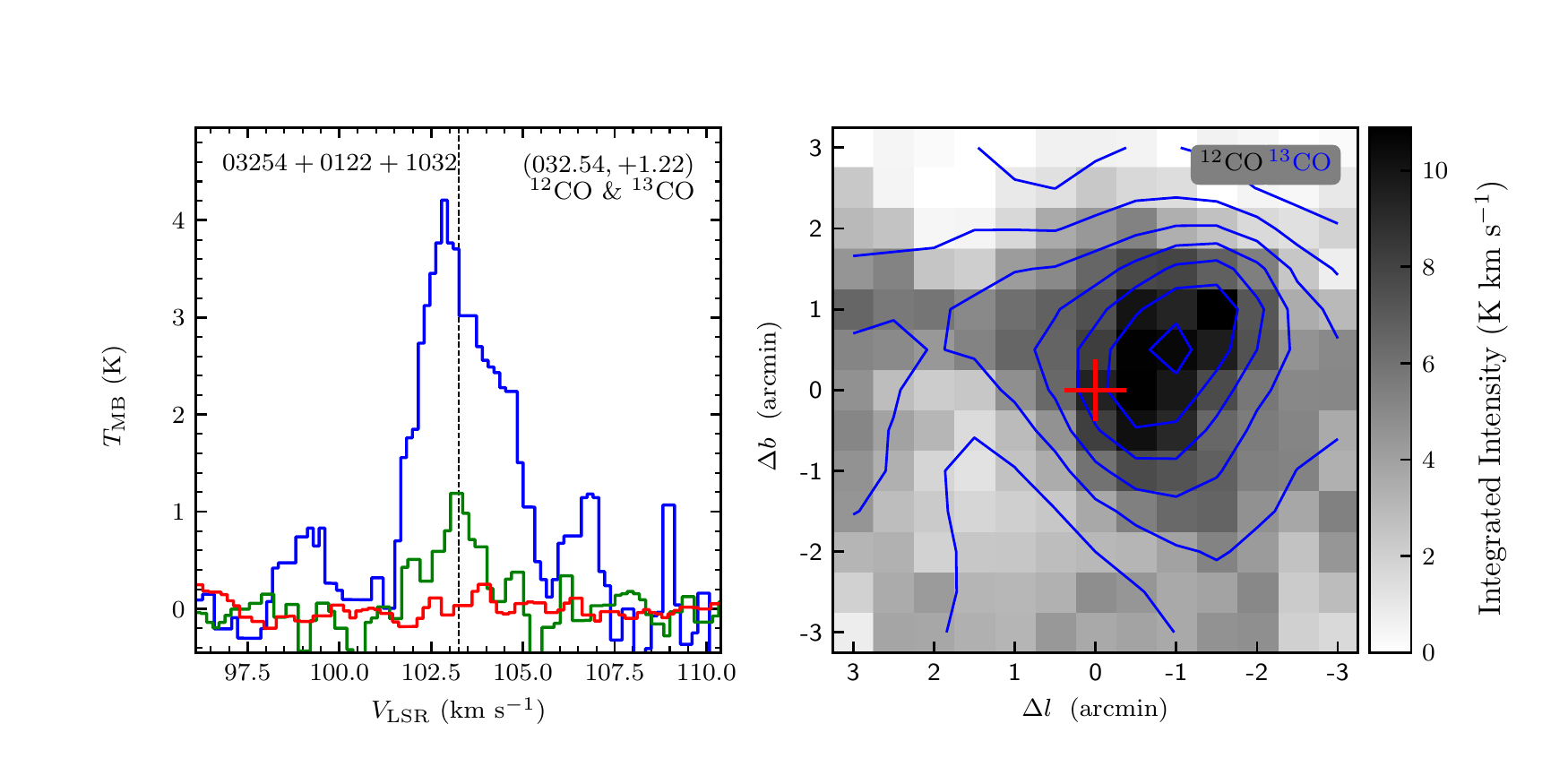}
\includegraphics[width=9.0cm,angle=0]{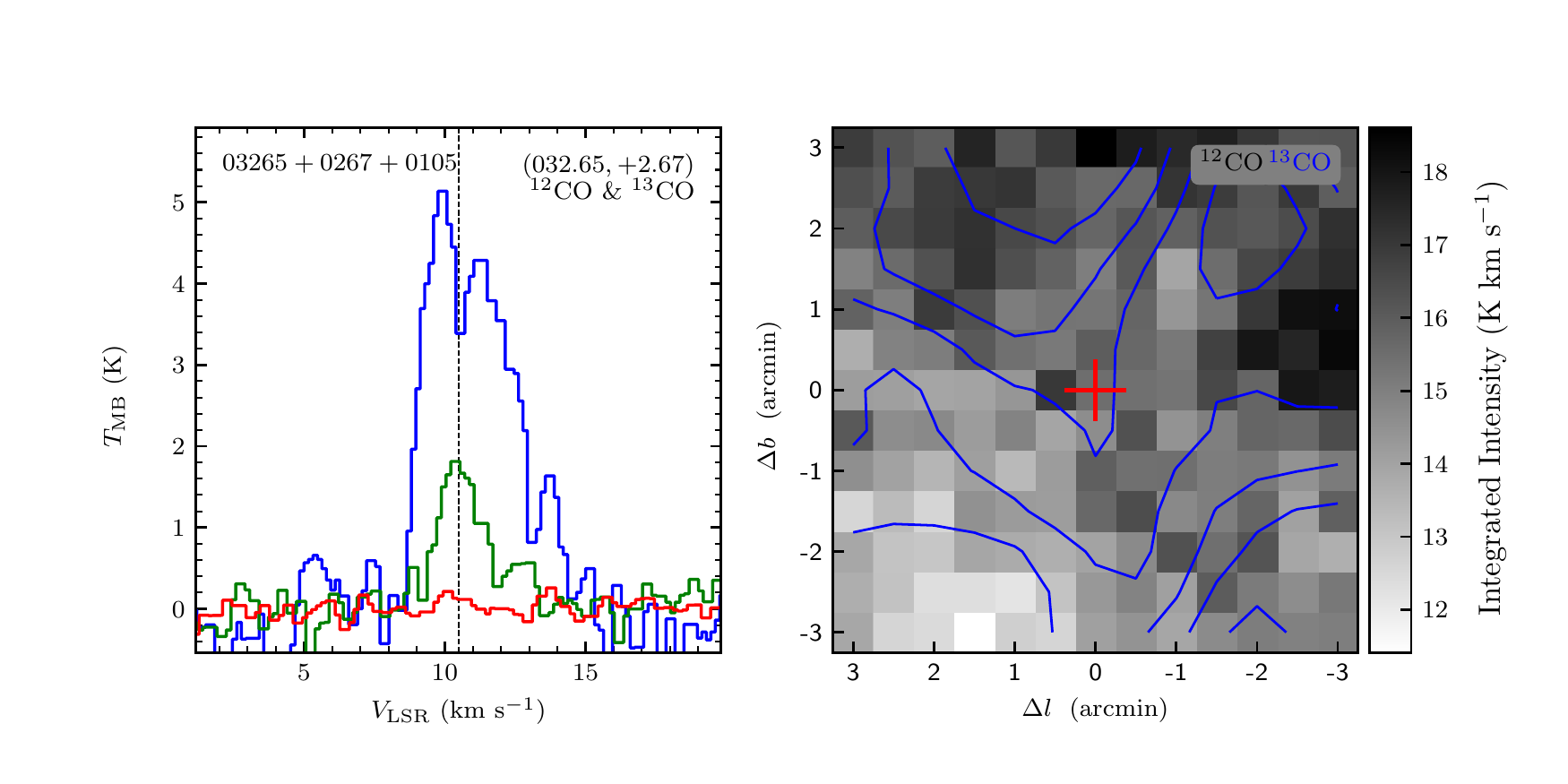}
\vspace{-0.5cm}

\includegraphics[width=9.0cm,angle=0]{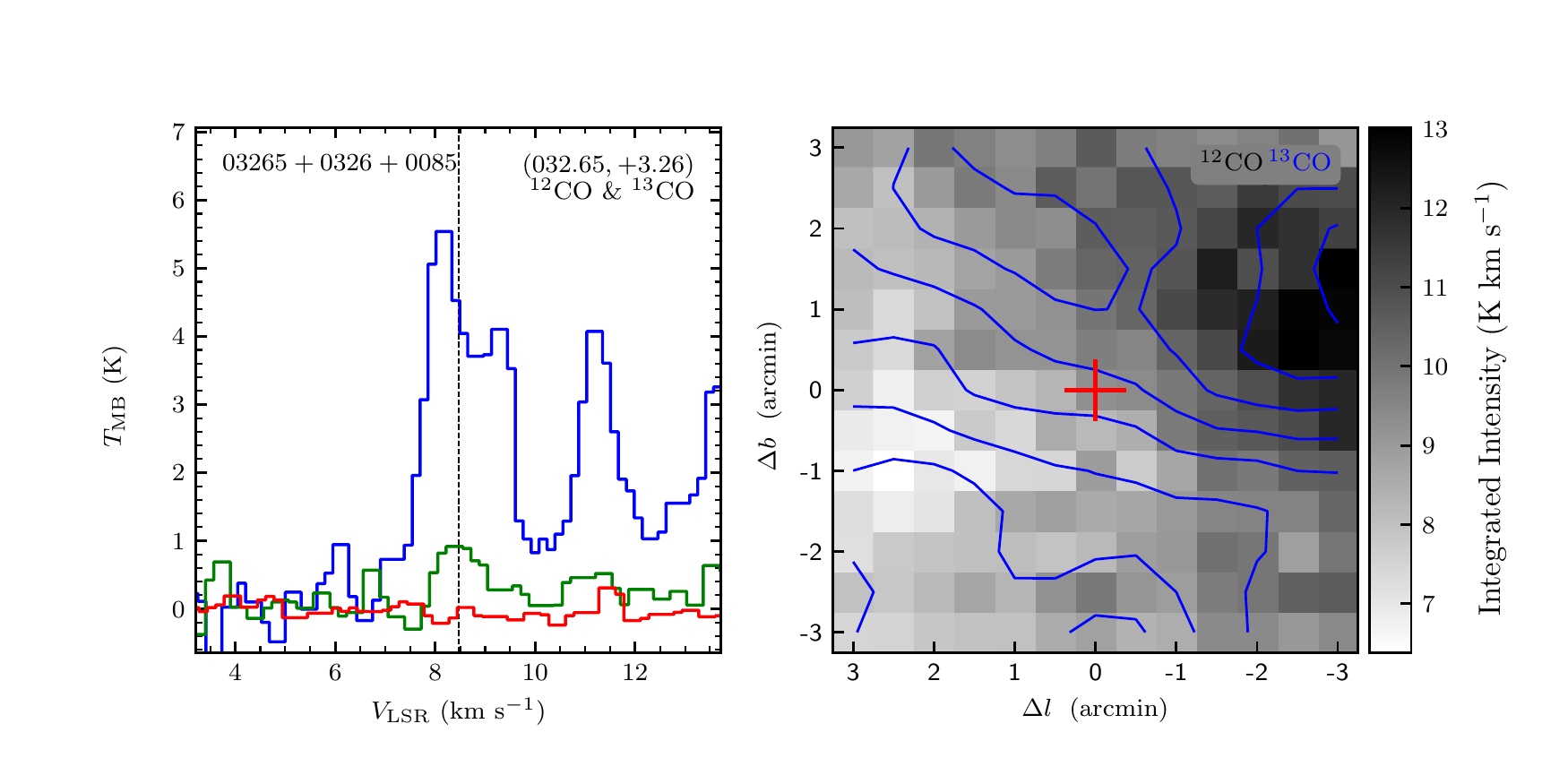}
\includegraphics[width=9.0cm,angle=0]{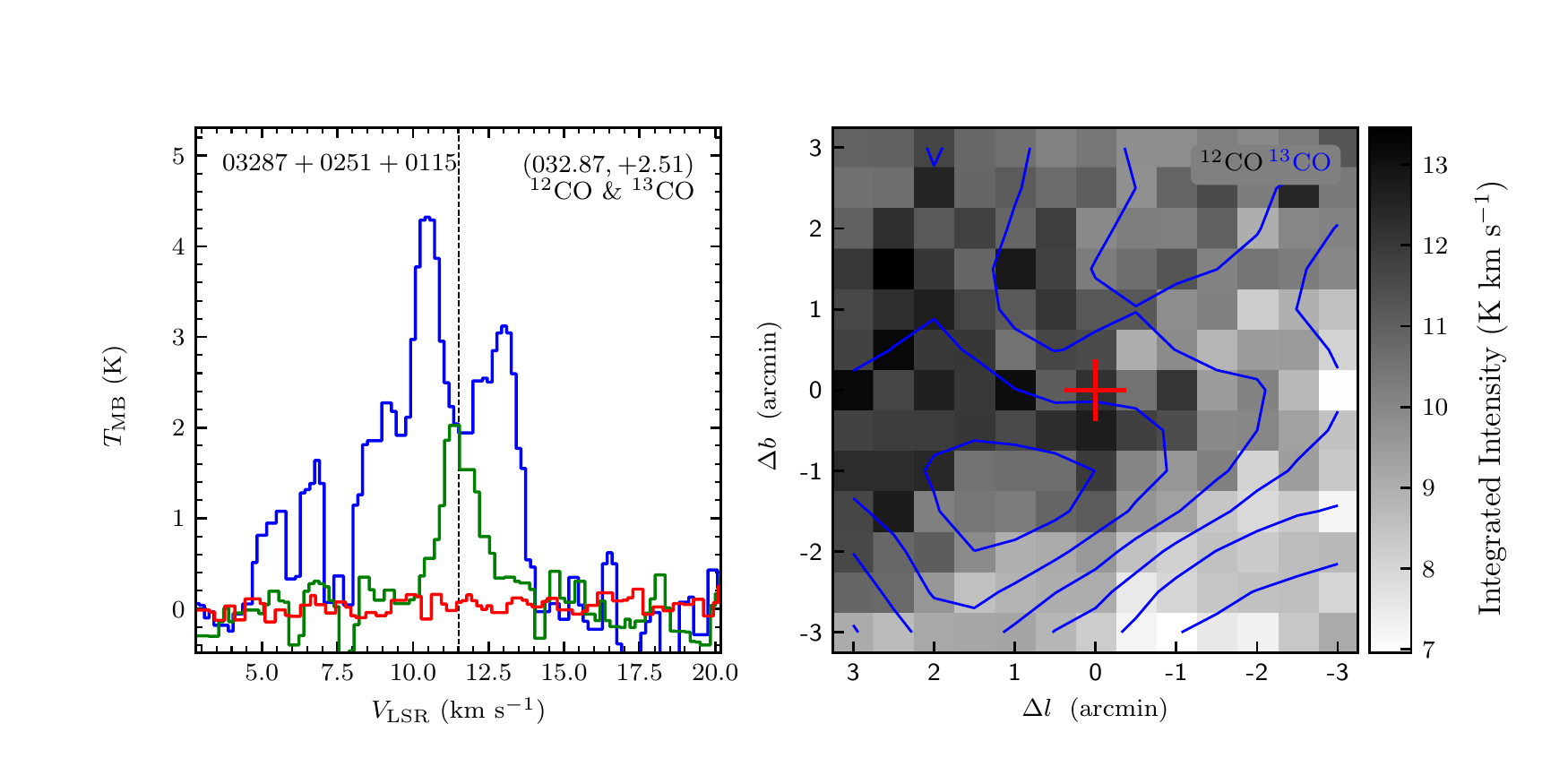}
\vspace{-0.5cm}

\includegraphics[width=9.0cm,angle=0]{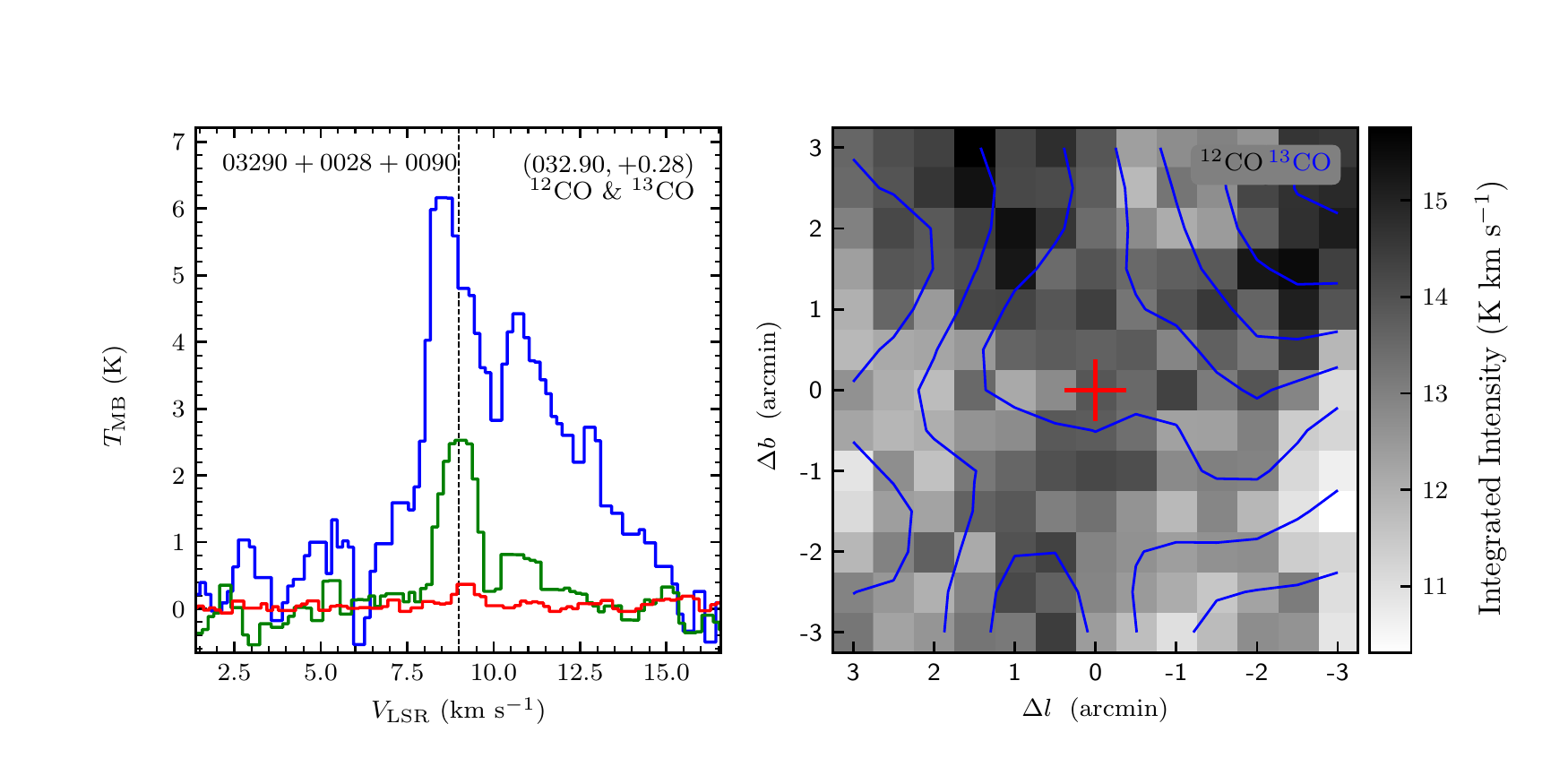}
\includegraphics[width=9.0cm,angle=0]{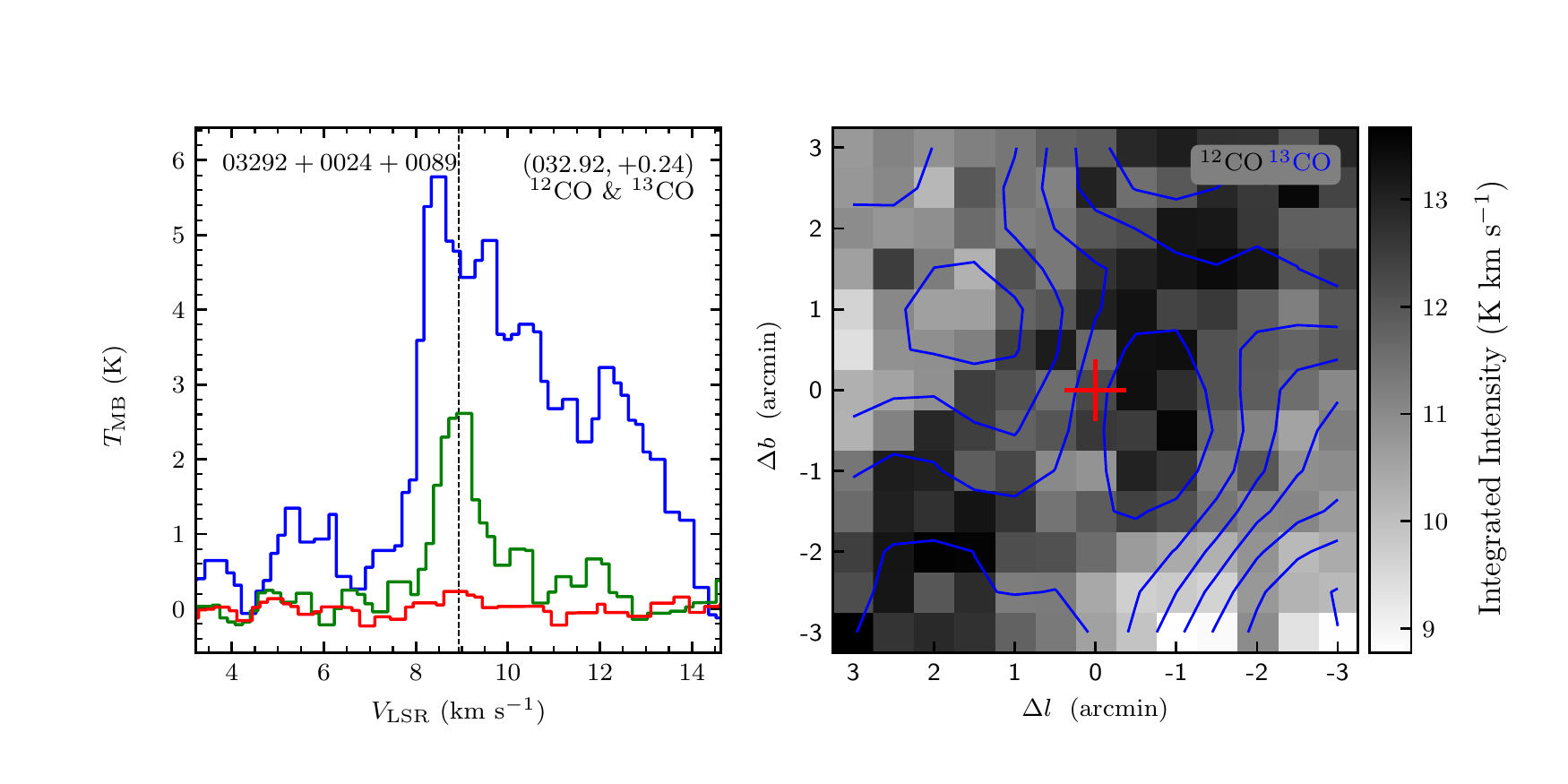}
\vspace{-0.5cm}

\includegraphics[width=9.0cm,angle=0]{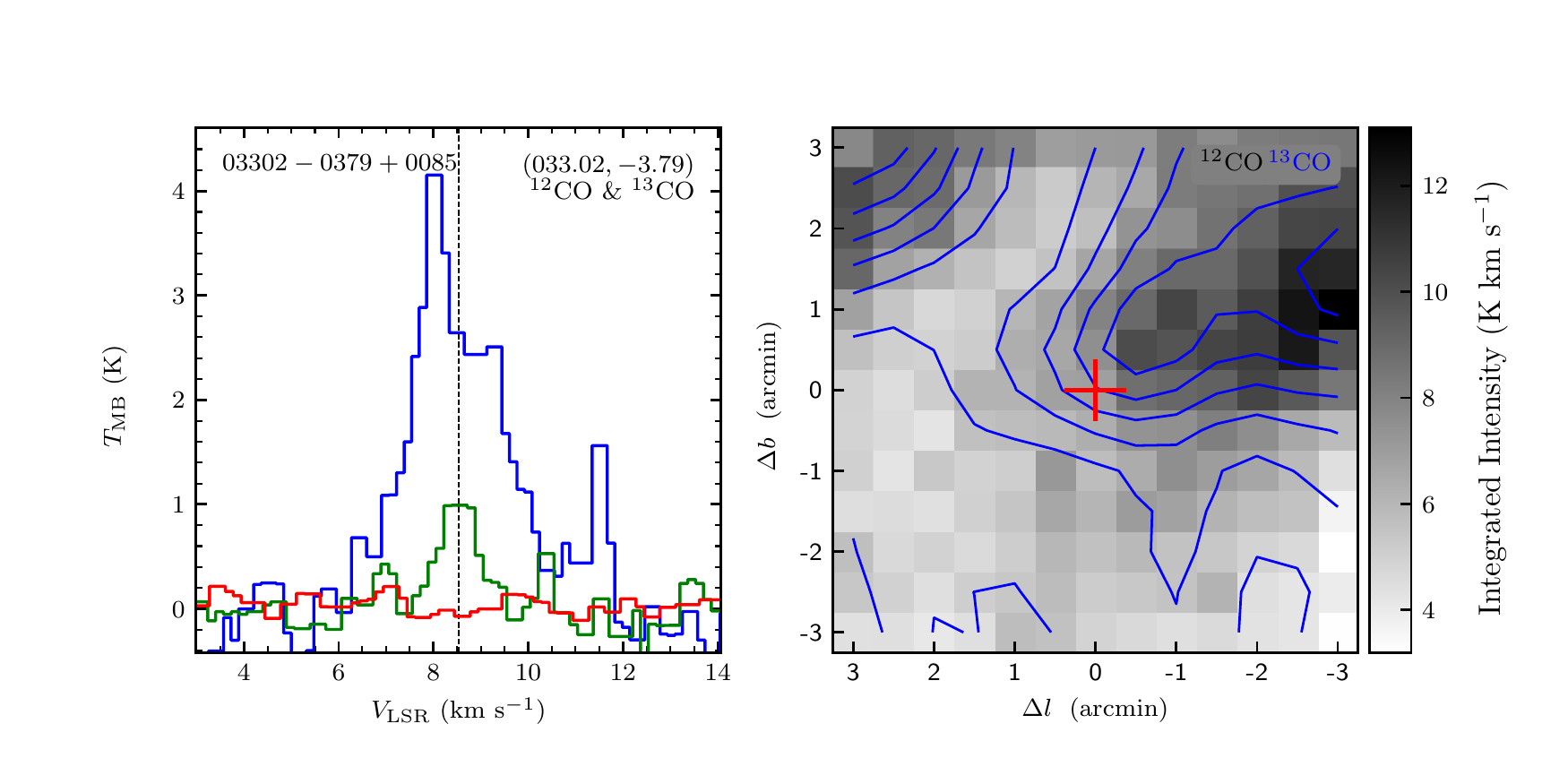}
\includegraphics[width=9.0cm,angle=0]{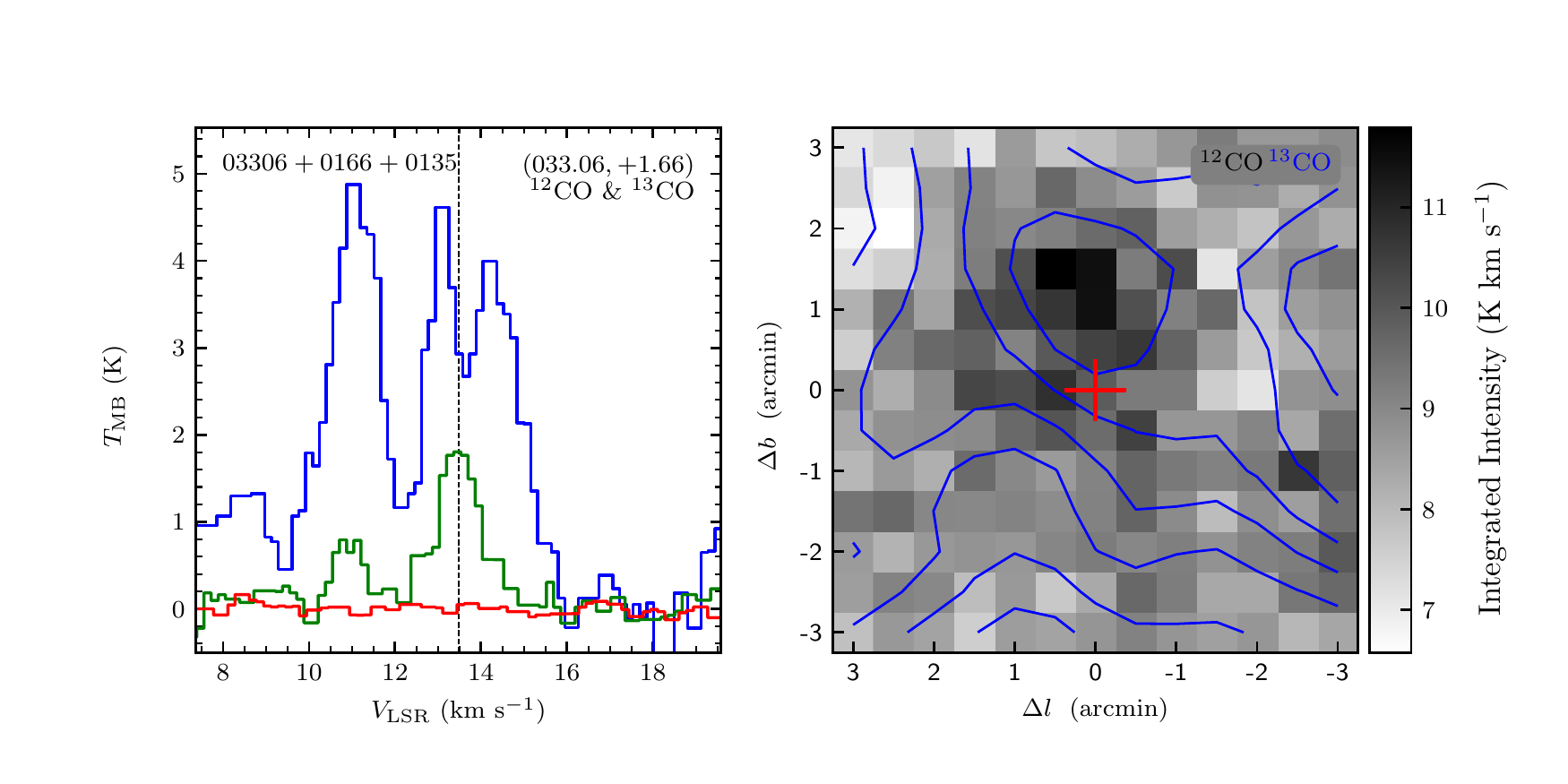}
\end{figure}
\clearpage

\begin{figure}
\includegraphics[width=9.0cm,angle=0]{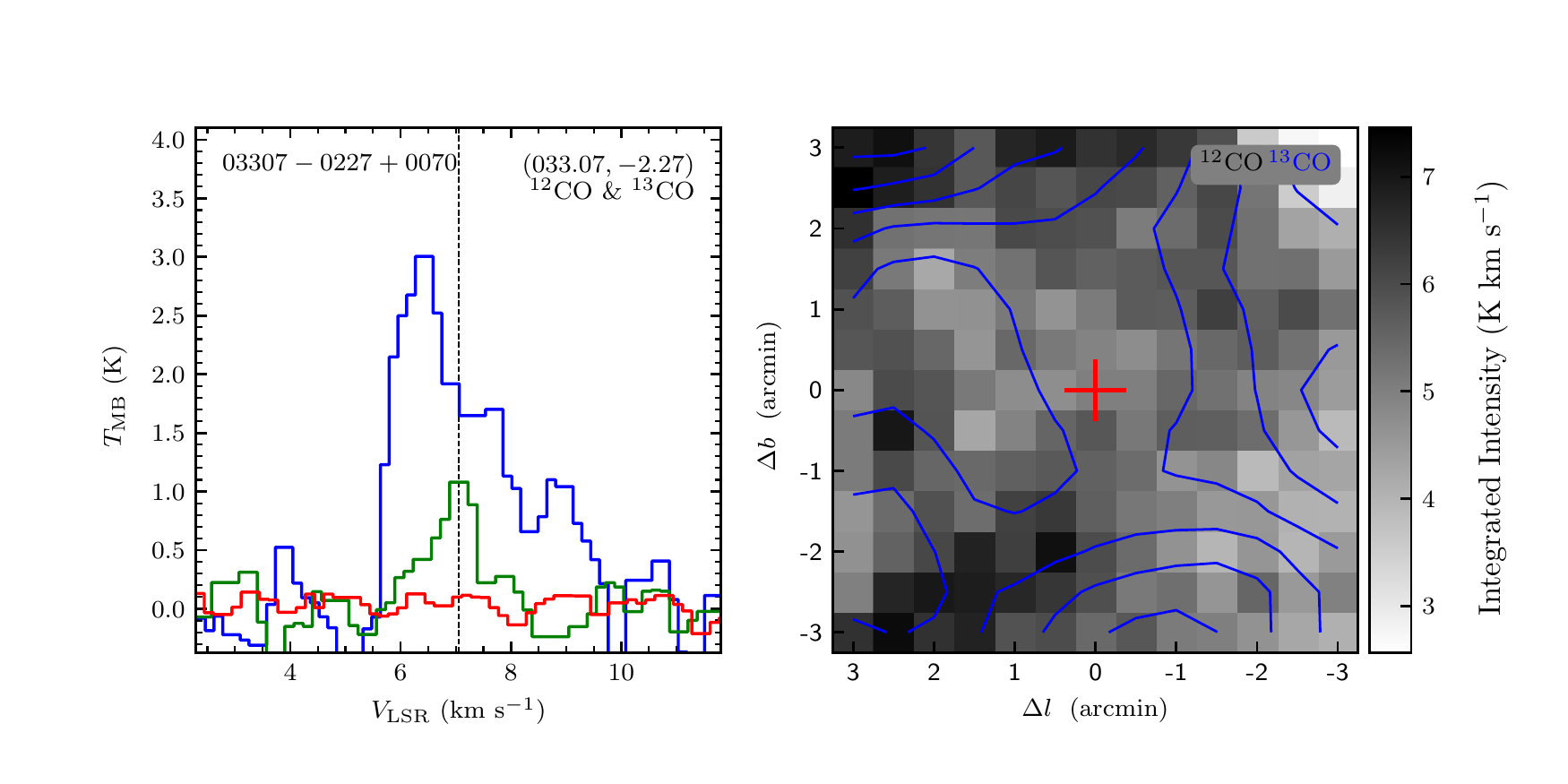}
\includegraphics[width=9.0cm,angle=0]{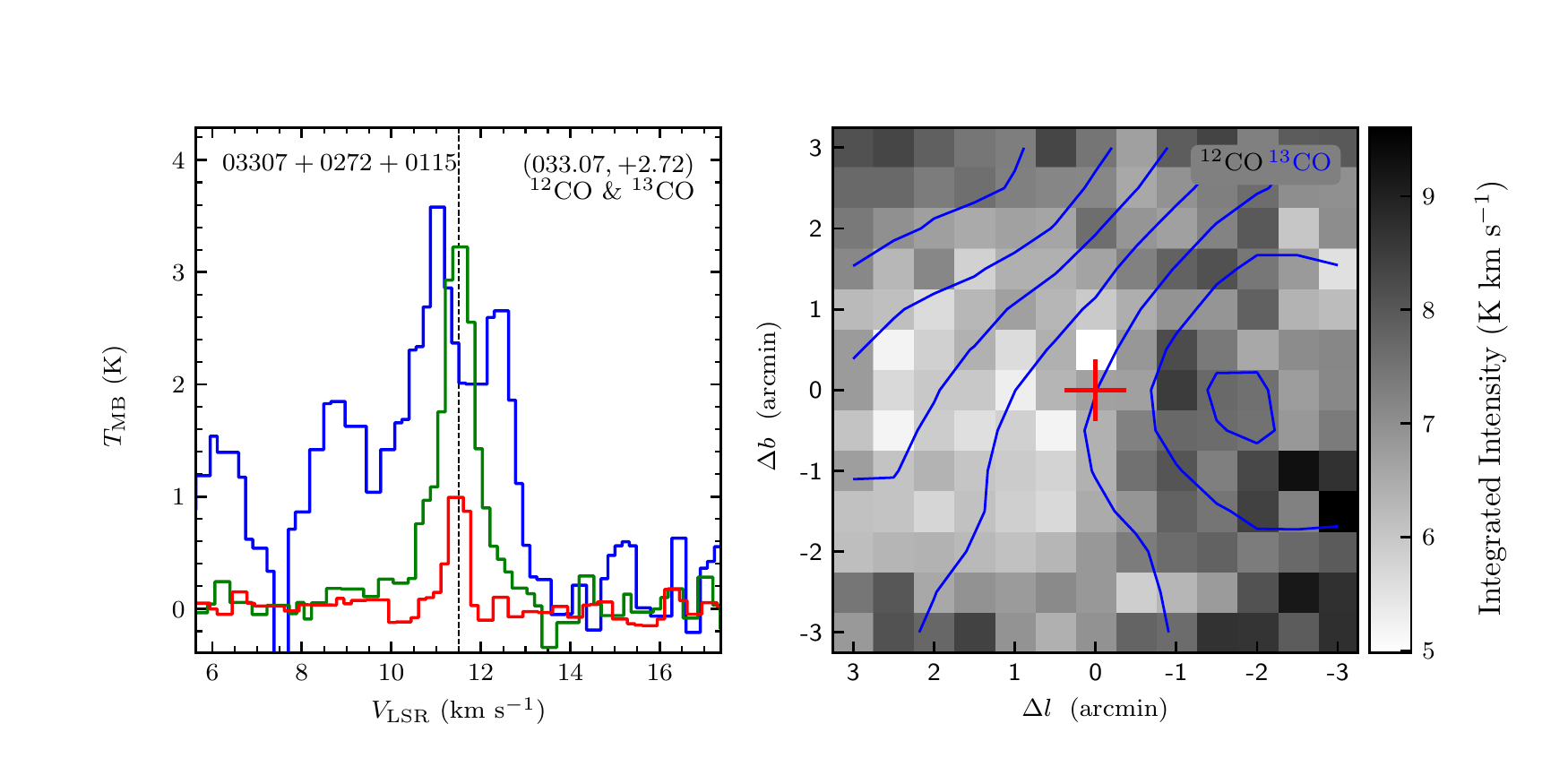}
\vspace{-0.5cm}

\includegraphics[width=9.0cm,angle=0]{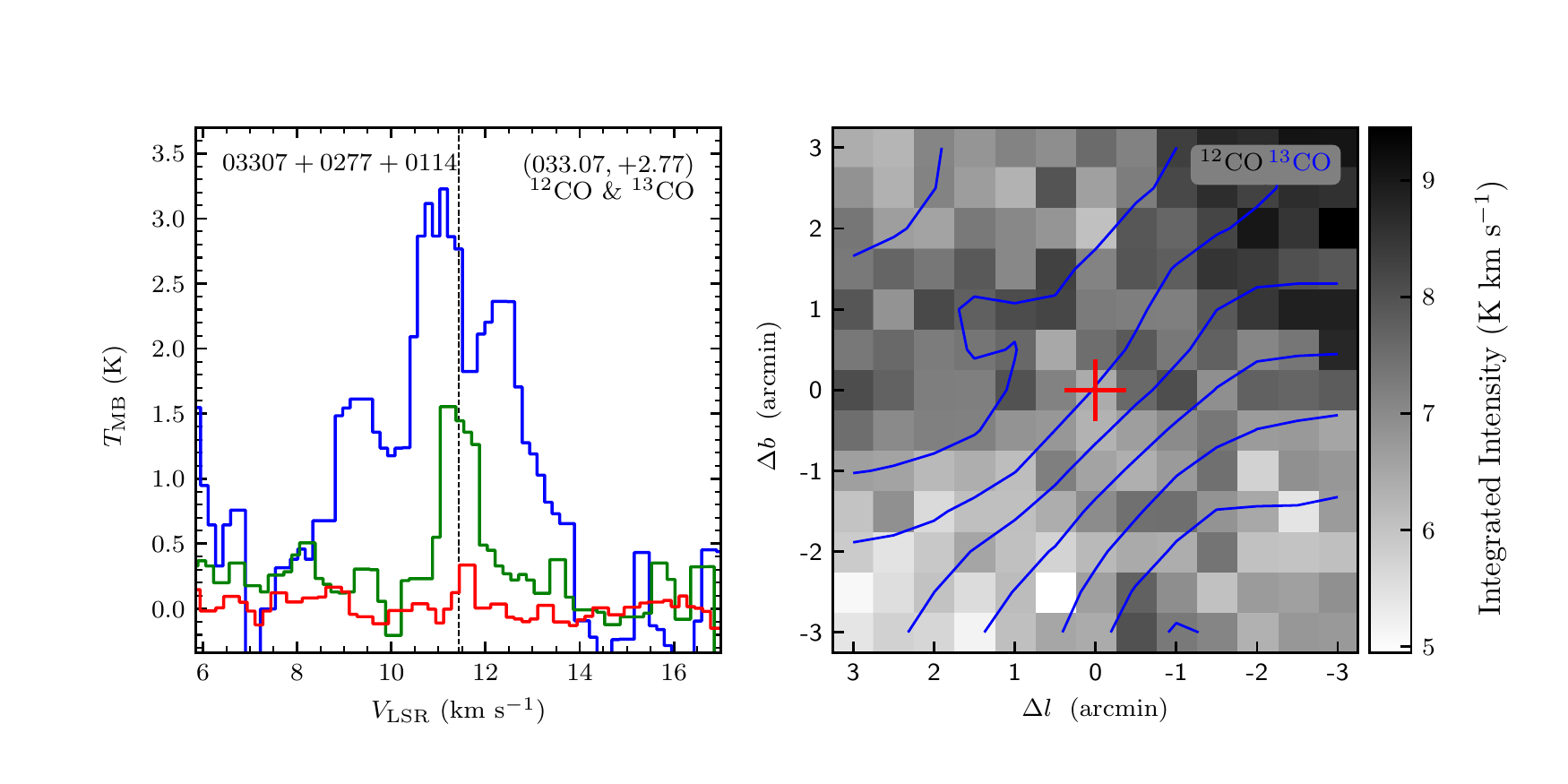}
\includegraphics[width=9.0cm,angle=0]{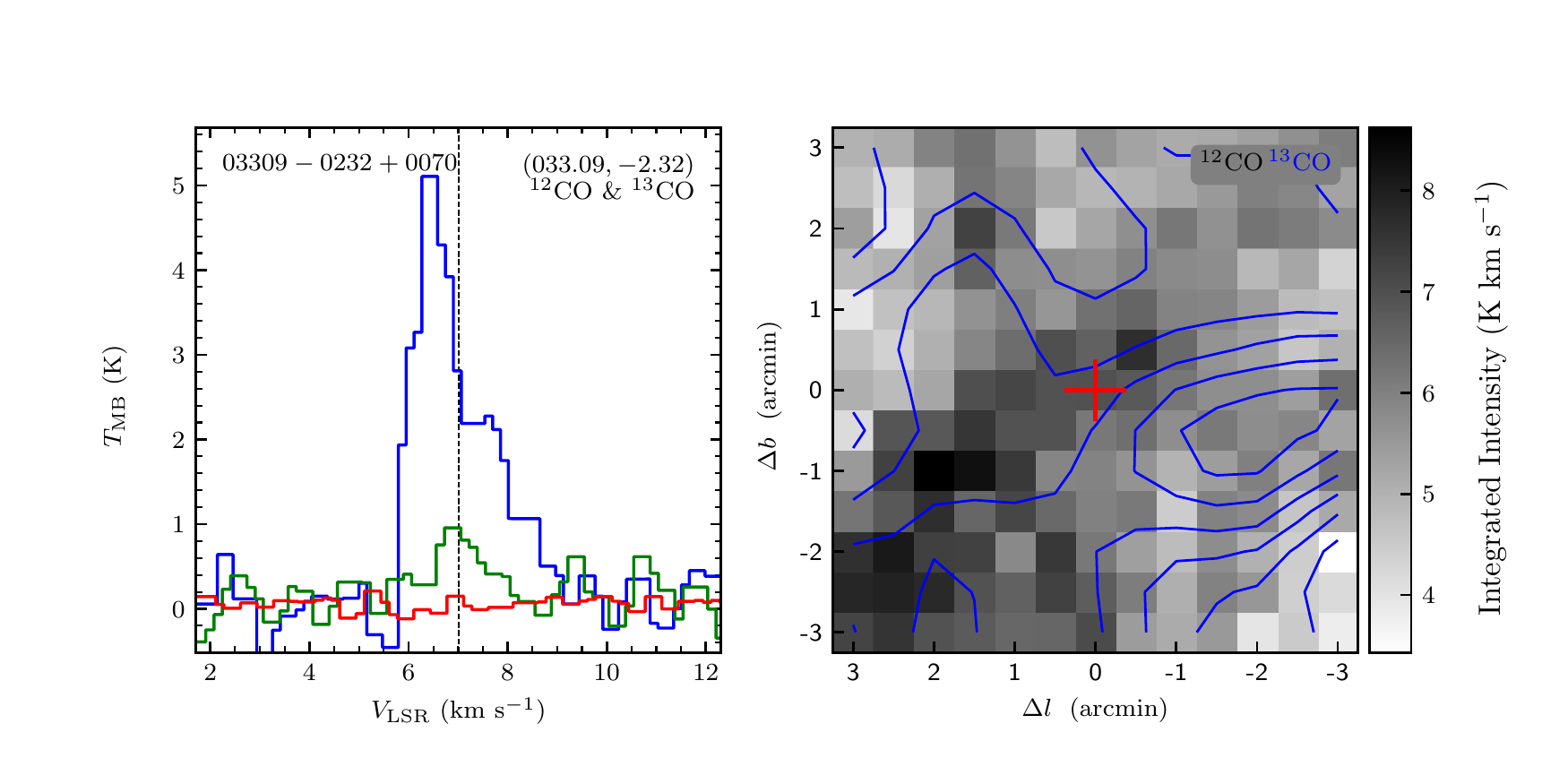}
\vspace{-0.5cm}

\includegraphics[width=9.0cm,angle=0]{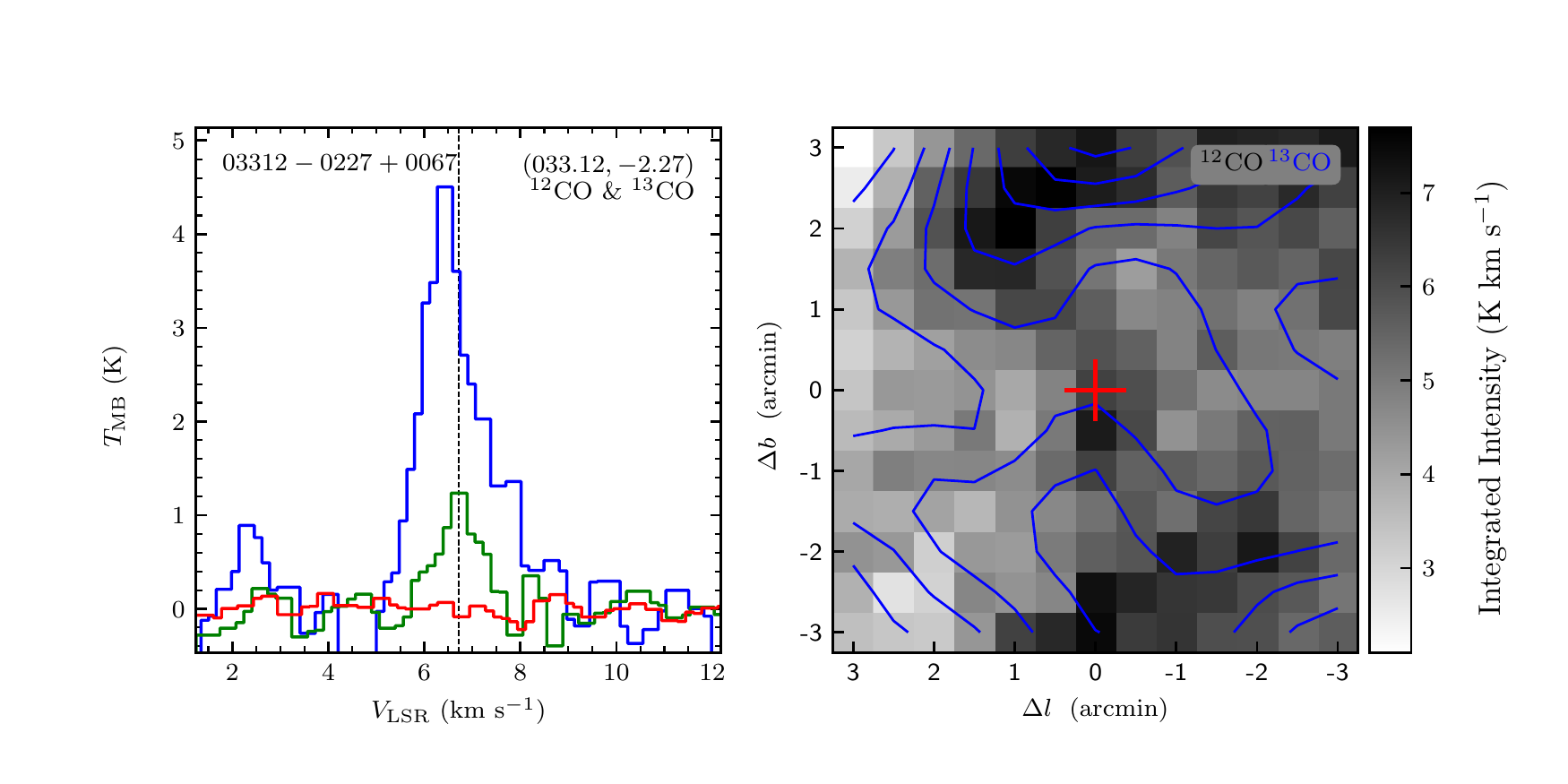}
\includegraphics[width=9.0cm,angle=0]{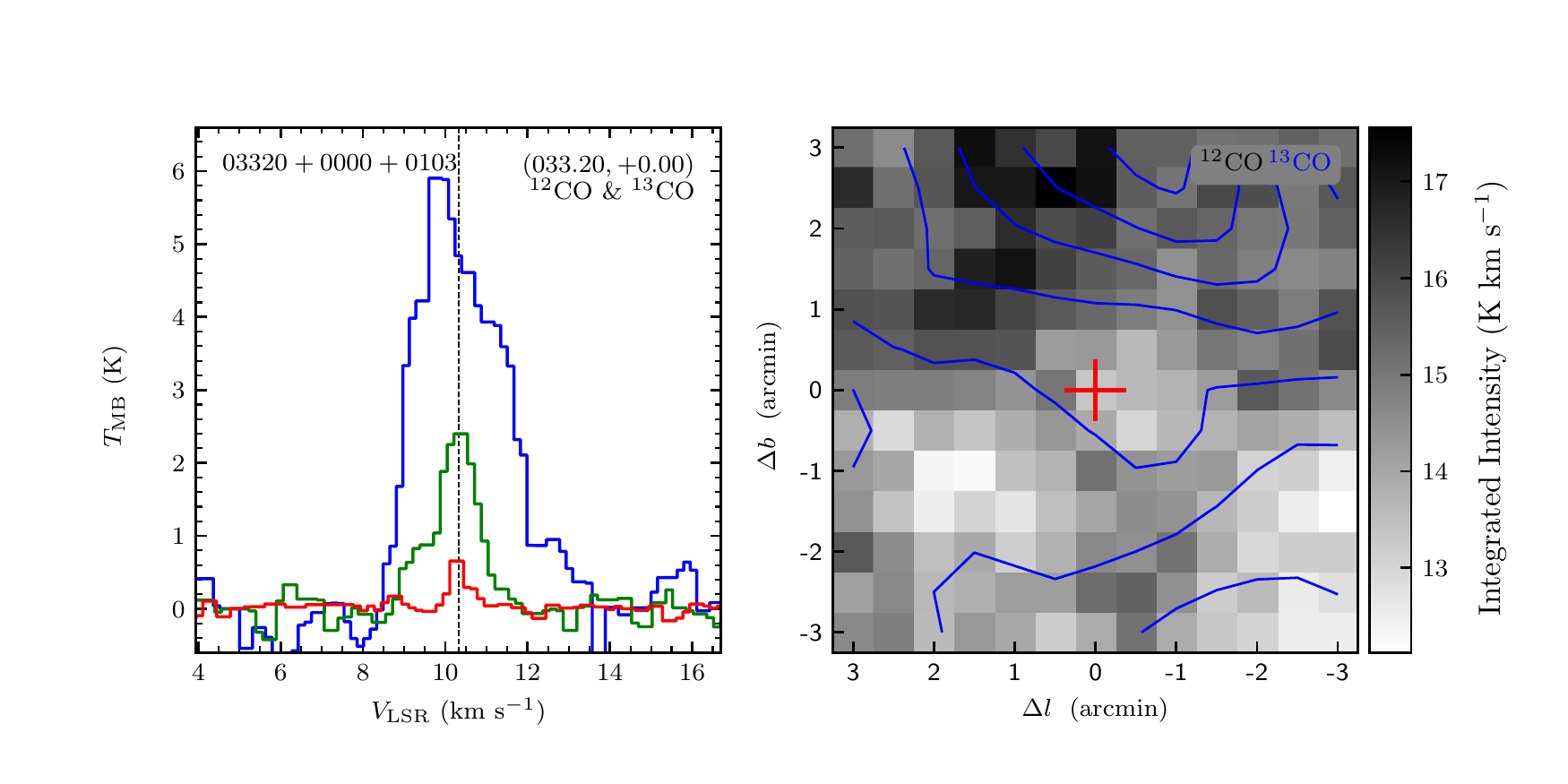}
\vspace{-0.5cm}

\includegraphics[width=9.0cm,angle=0]{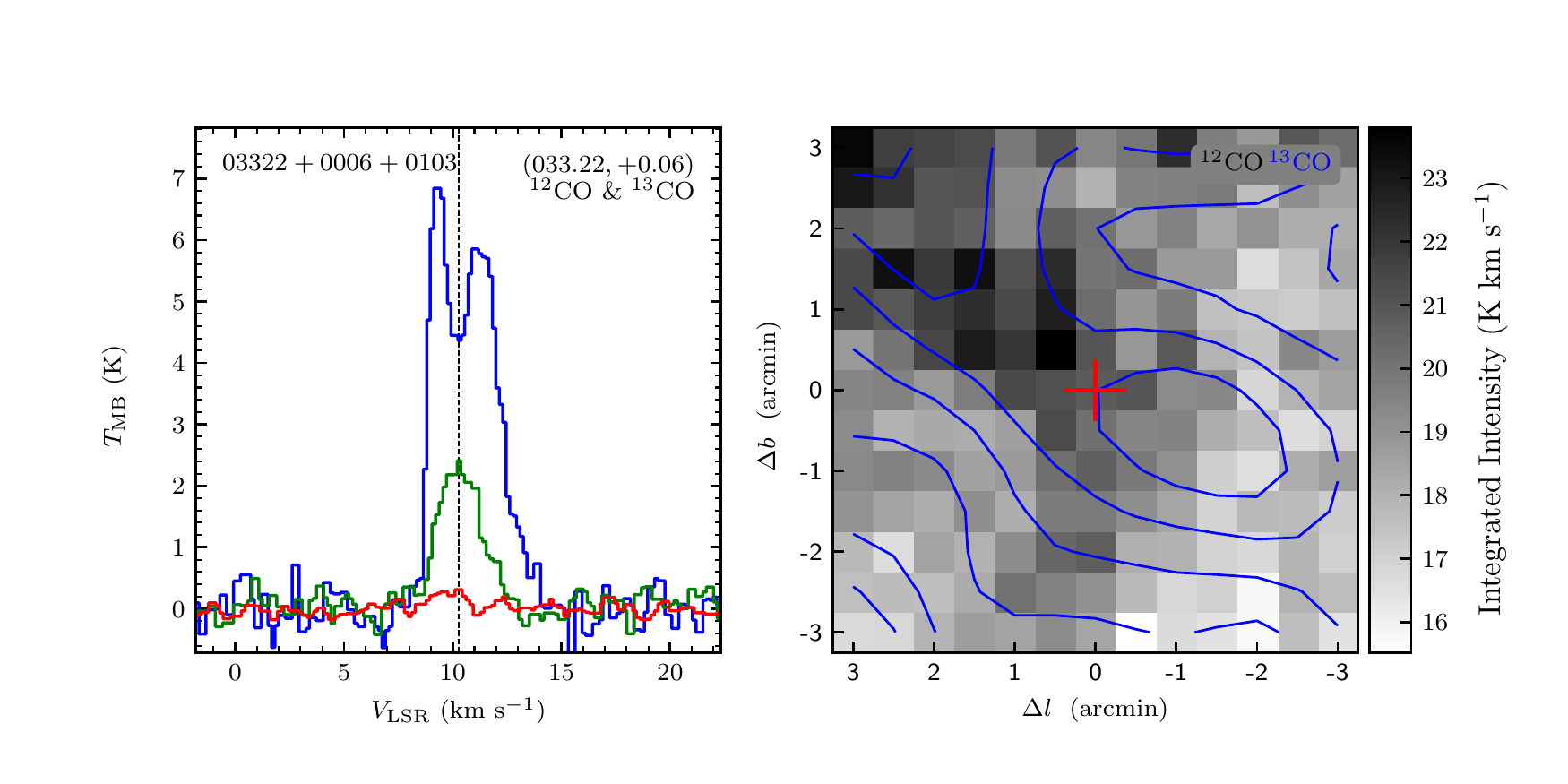}
\includegraphics[width=9.0cm,angle=0]{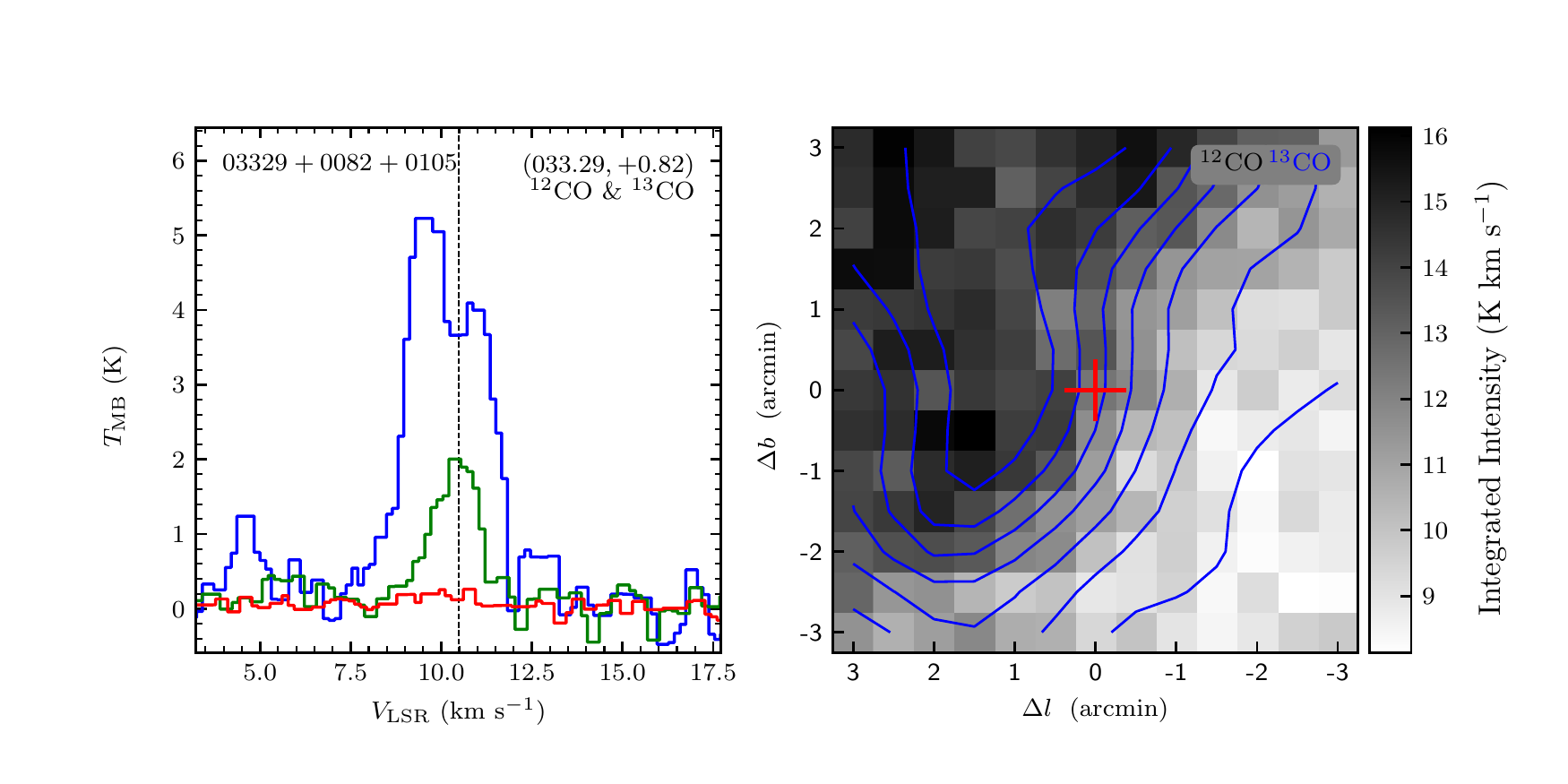}
\vspace{-0.5cm}

\includegraphics[width=9.0cm,angle=0]{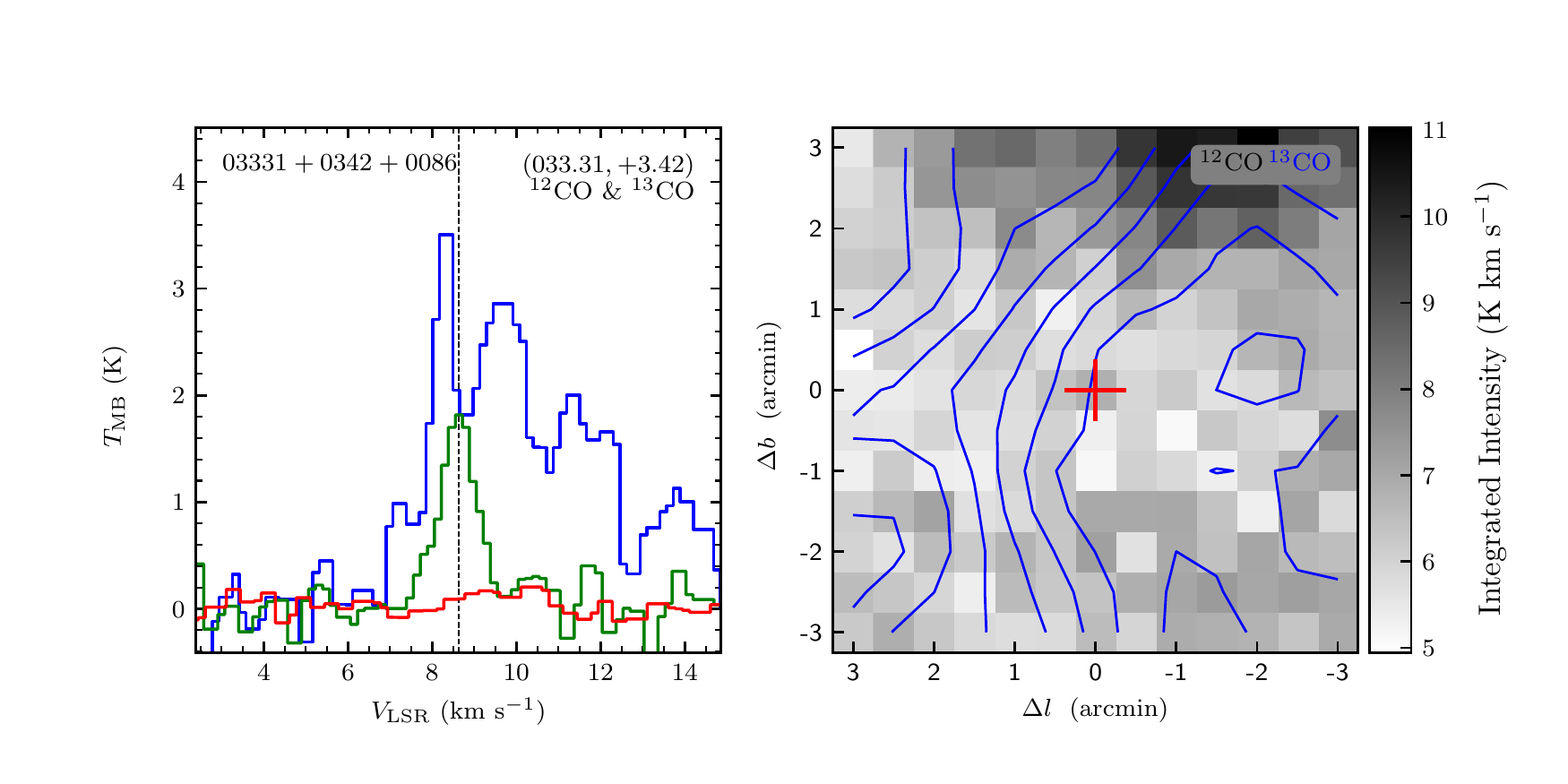}
\includegraphics[width=9.0cm,angle=0]{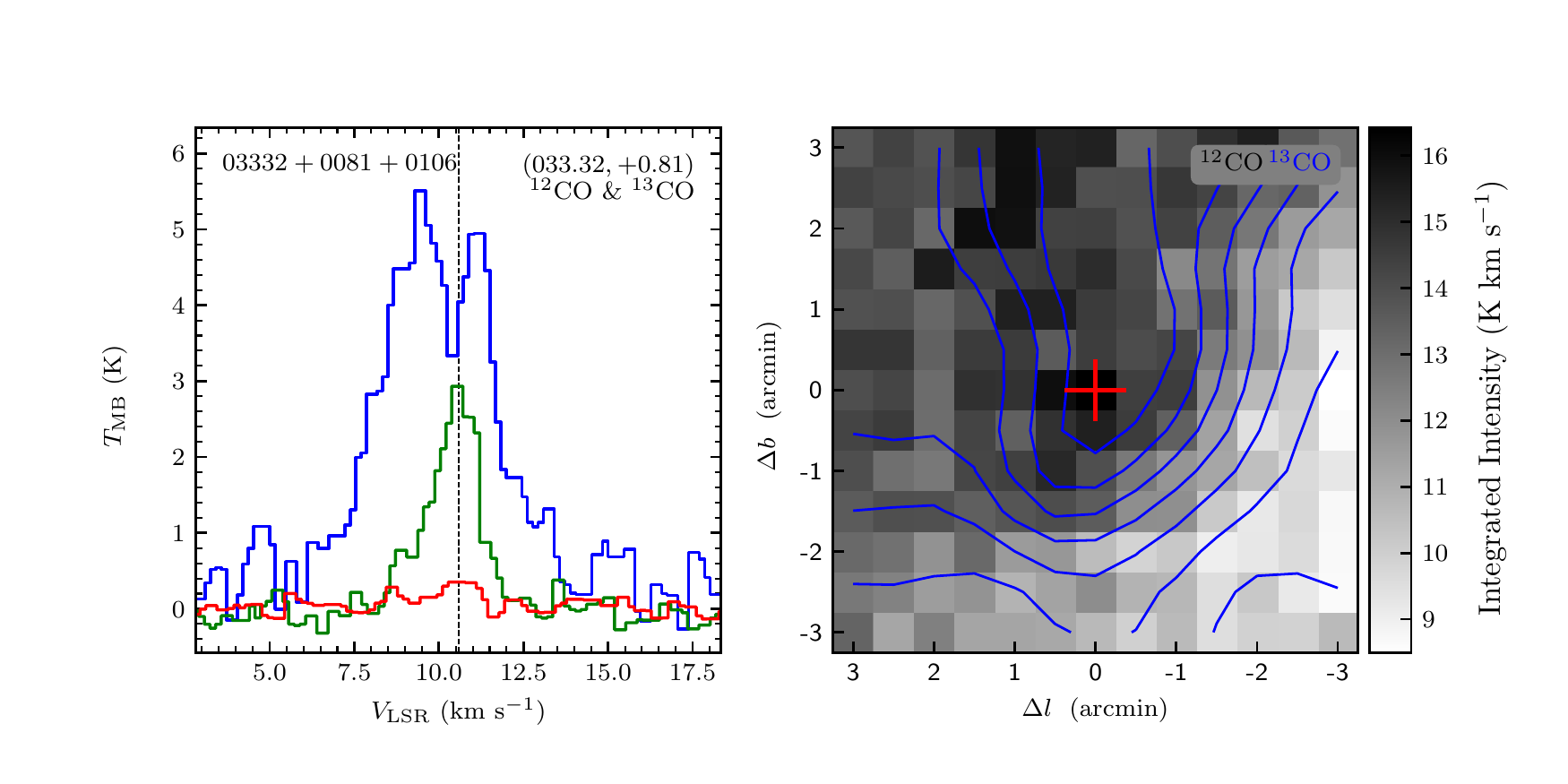}
\end{figure}
\clearpage

\begin{figure}
\includegraphics[width=9.0cm,angle=0]{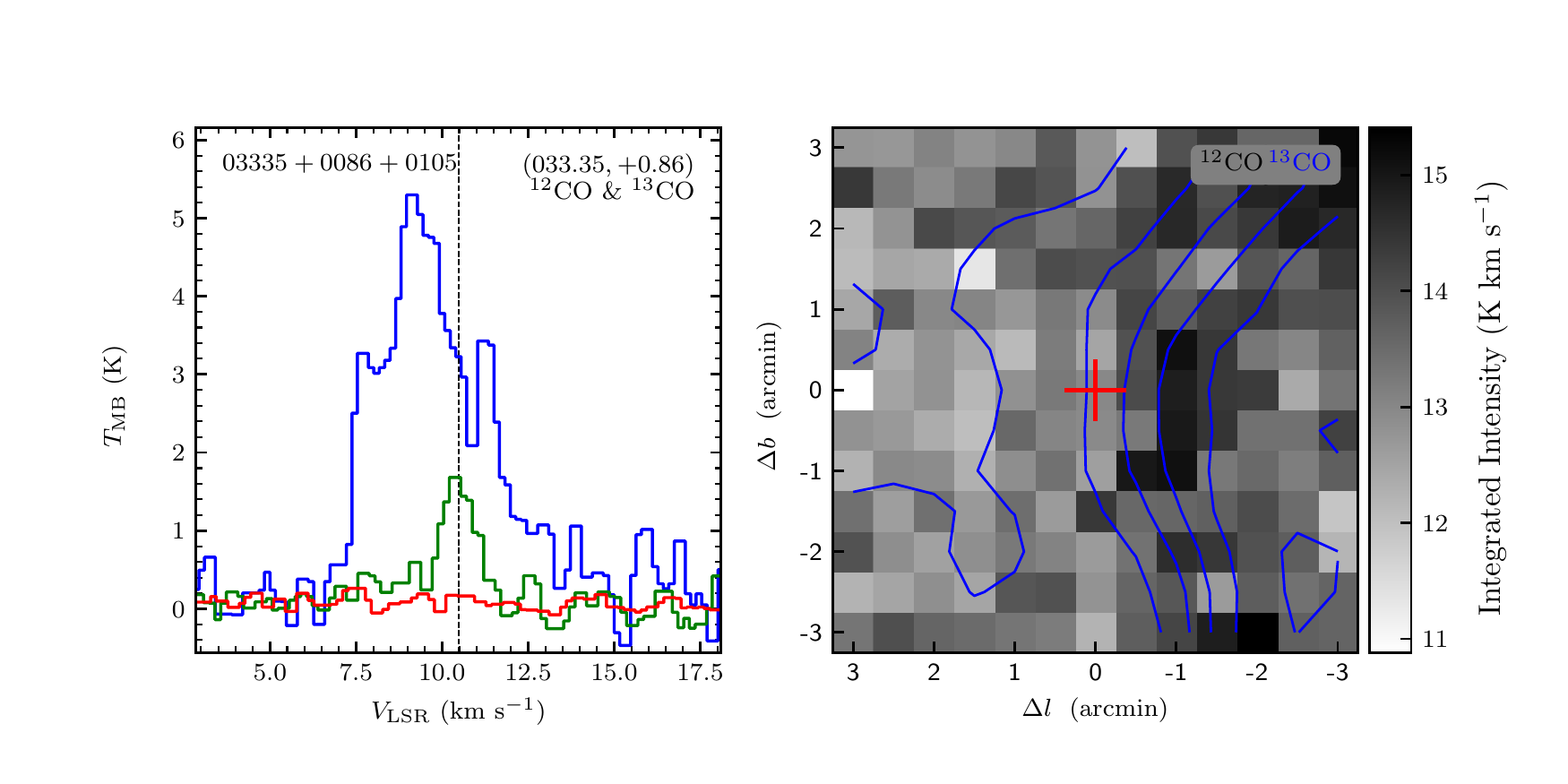}
\includegraphics[width=9.0cm,angle=0]{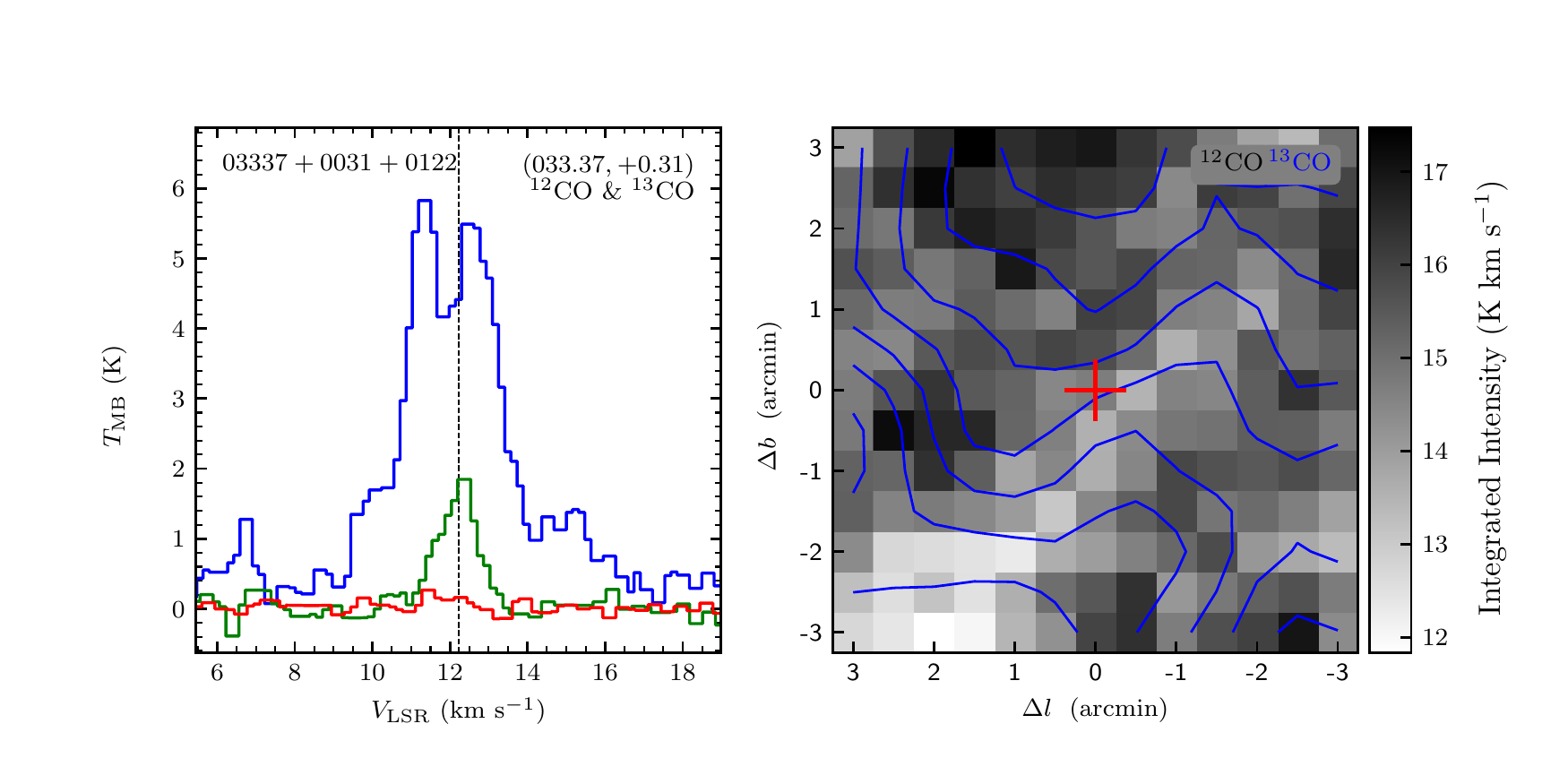}
\vspace{-0.5cm}

\includegraphics[width=9.0cm,angle=0]{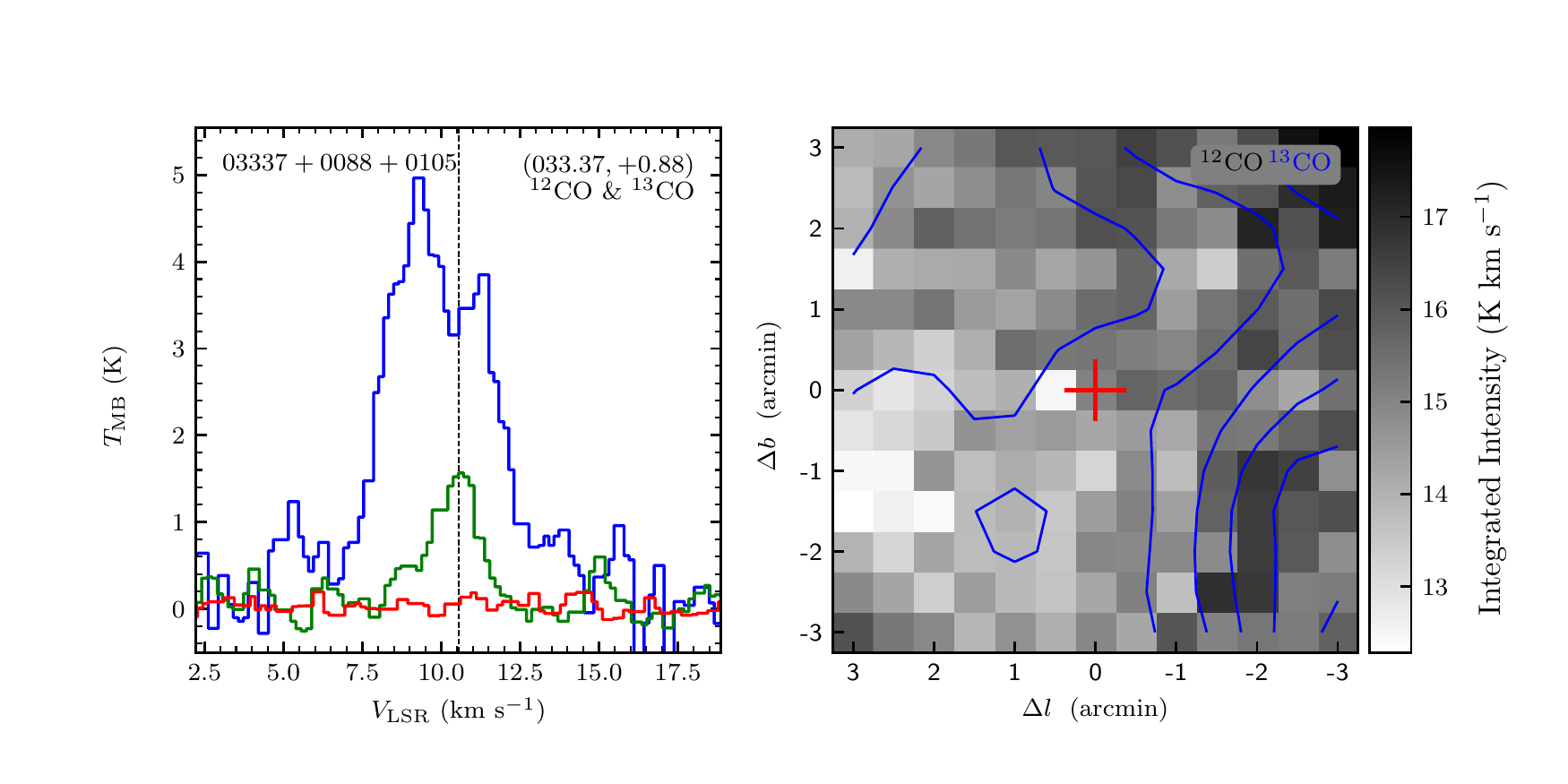}
\includegraphics[width=9.0cm,angle=0]{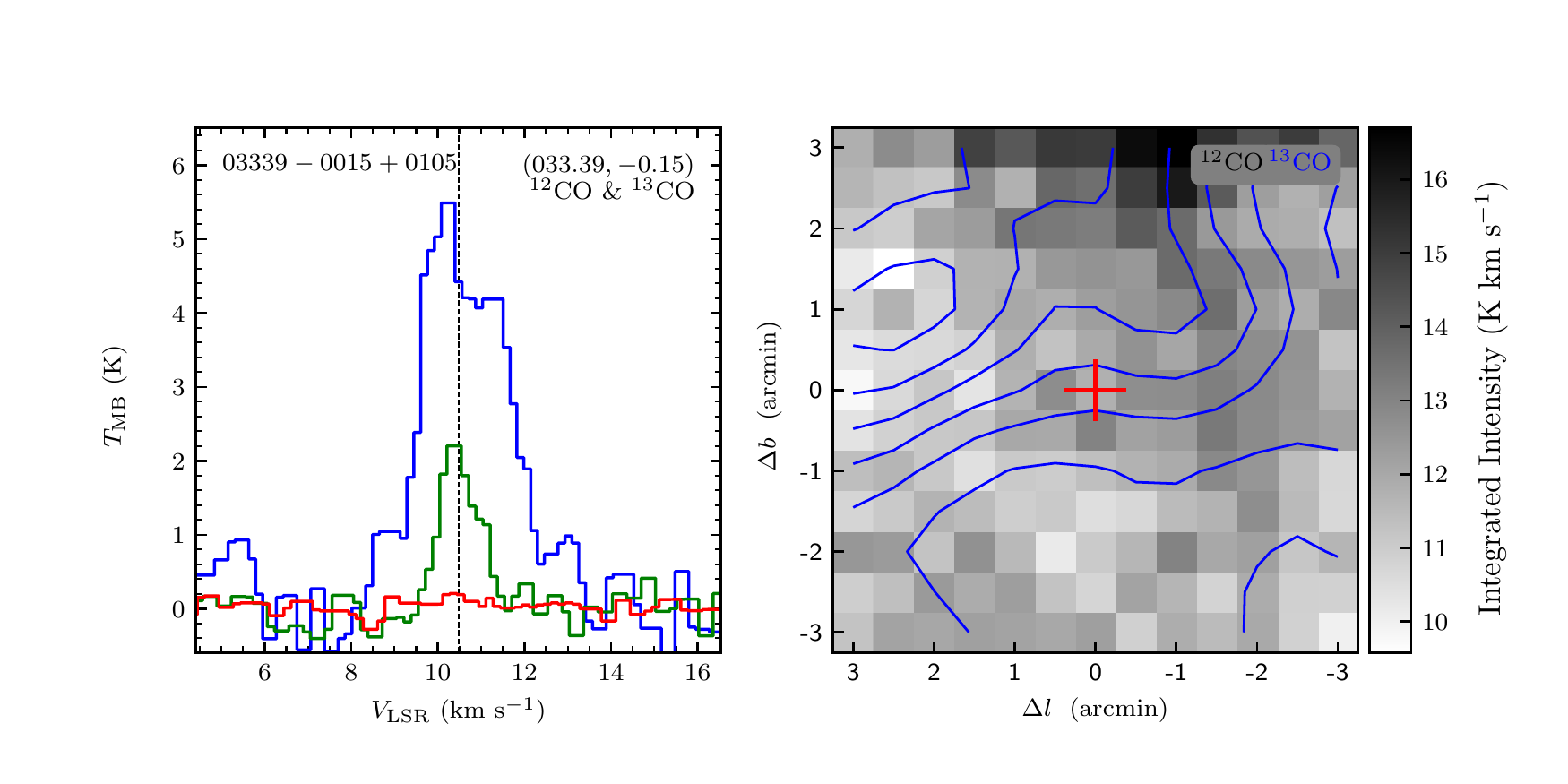}
\vspace{-0.5cm}

\includegraphics[width=9.0cm,angle=0]{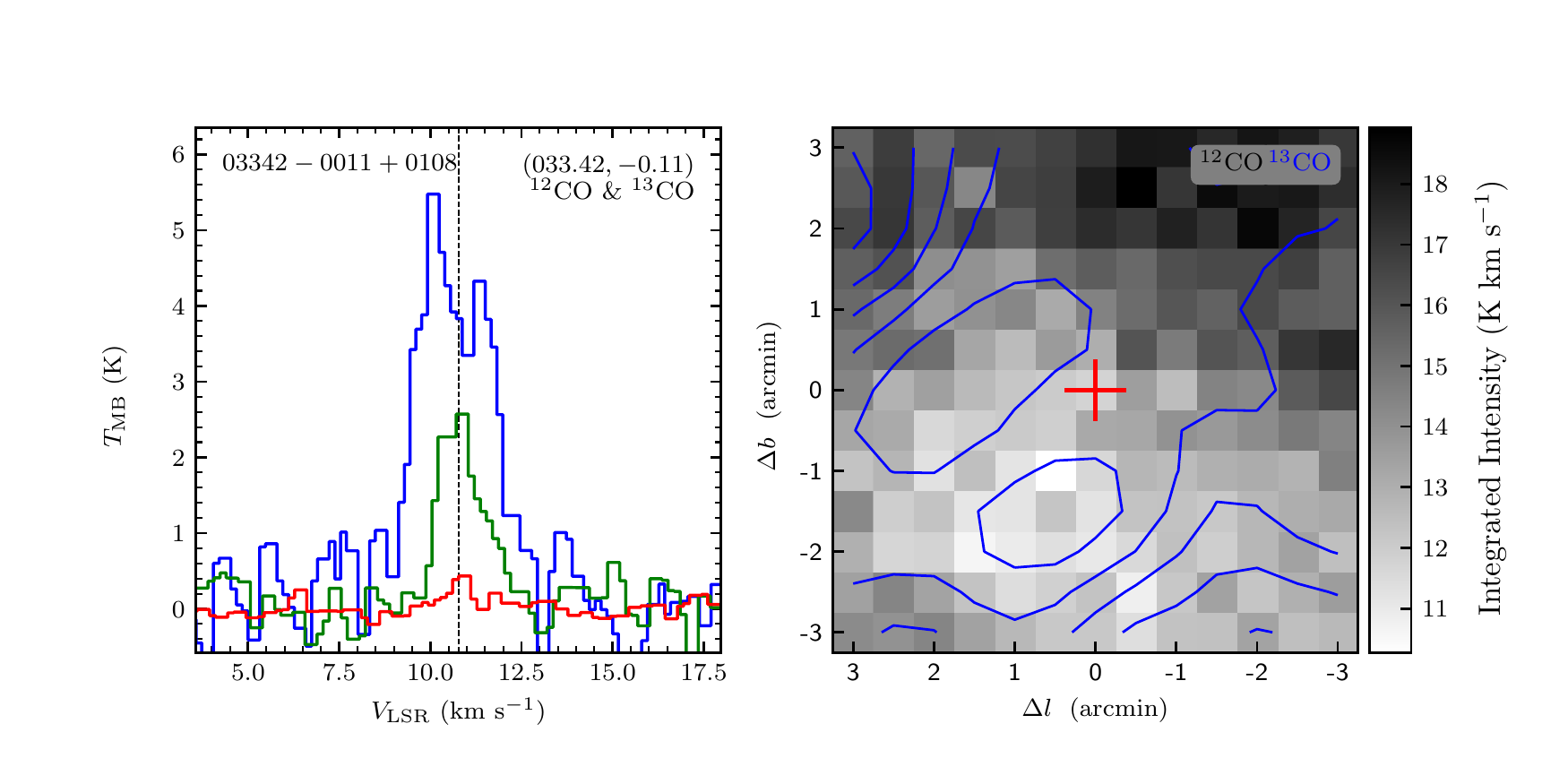}
\includegraphics[width=9.0cm,angle=0]{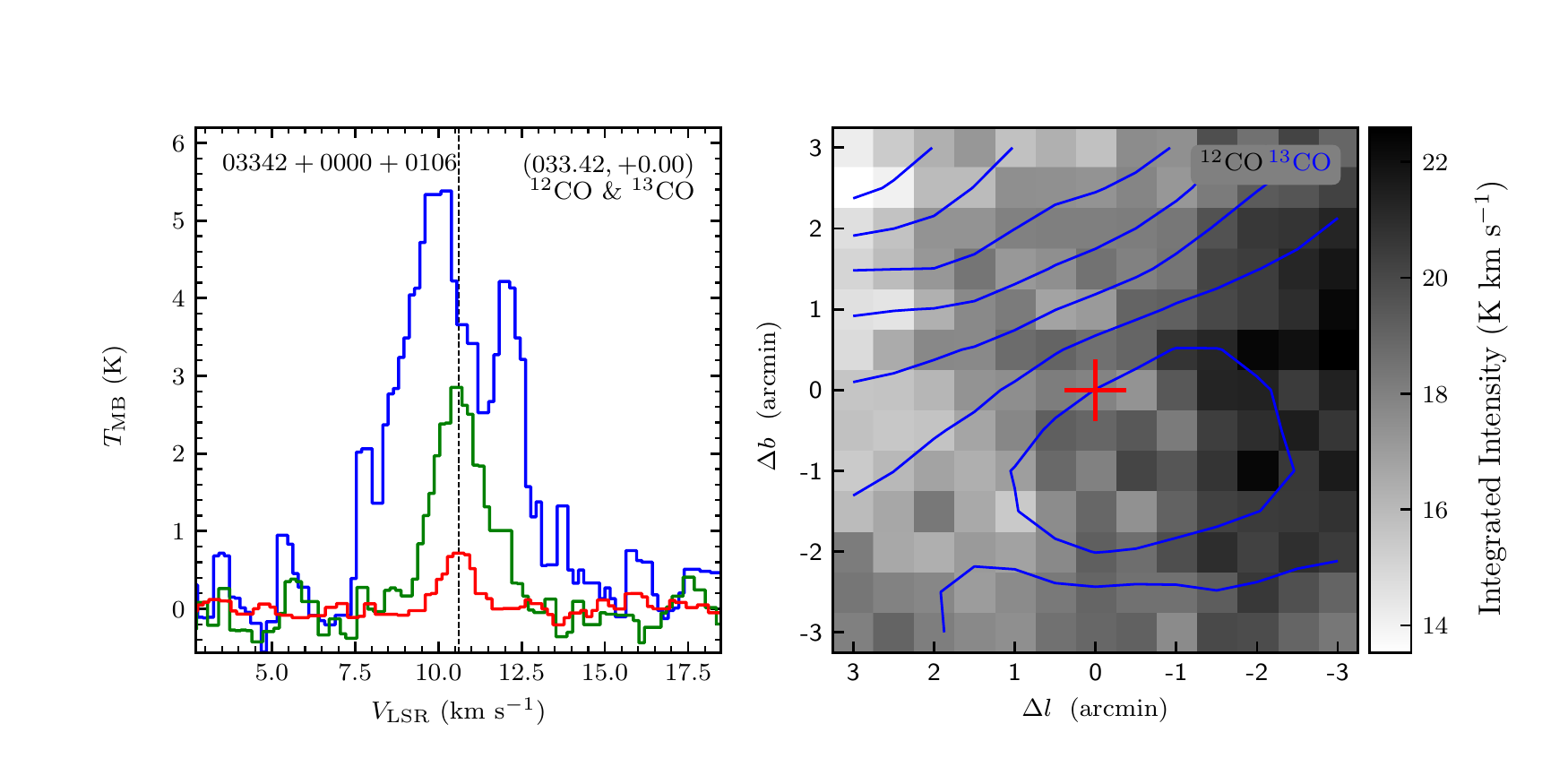}
\vspace{-0.5cm}

\includegraphics[width=9.0cm,angle=0]{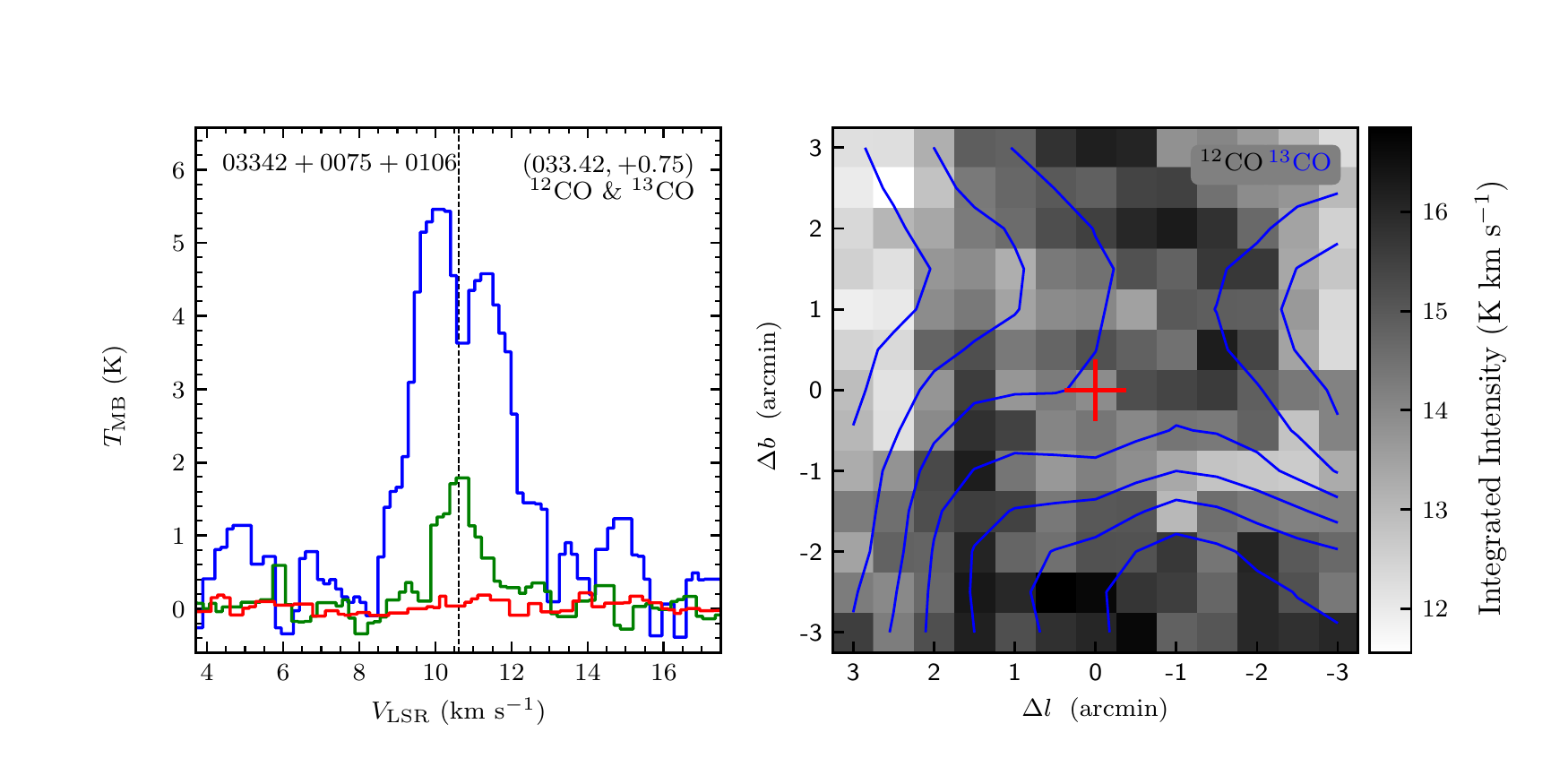}
\includegraphics[width=9.0cm,angle=0]{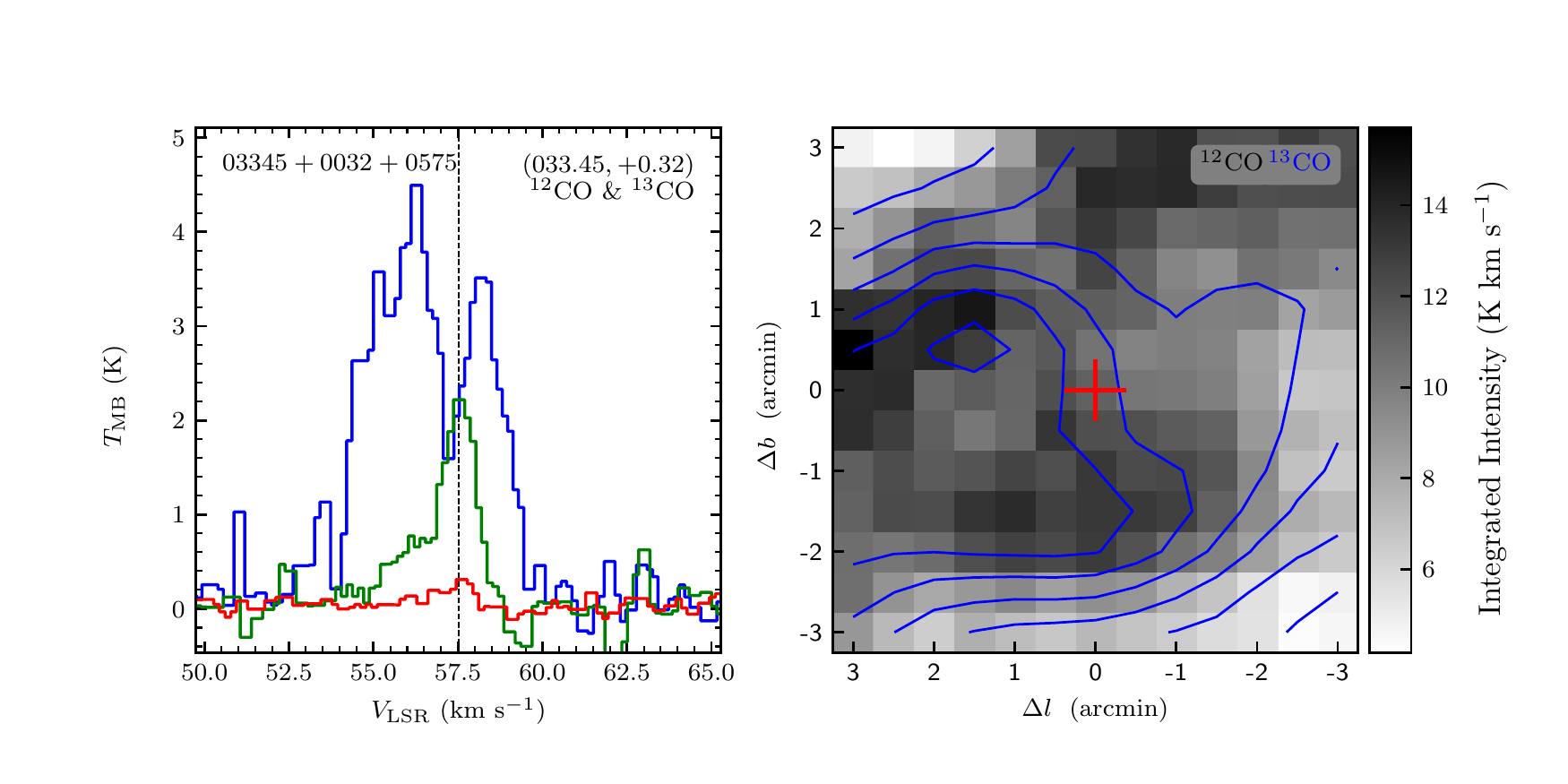}
\vspace{-0.5cm}

\includegraphics[width=9.0cm,angle=0]{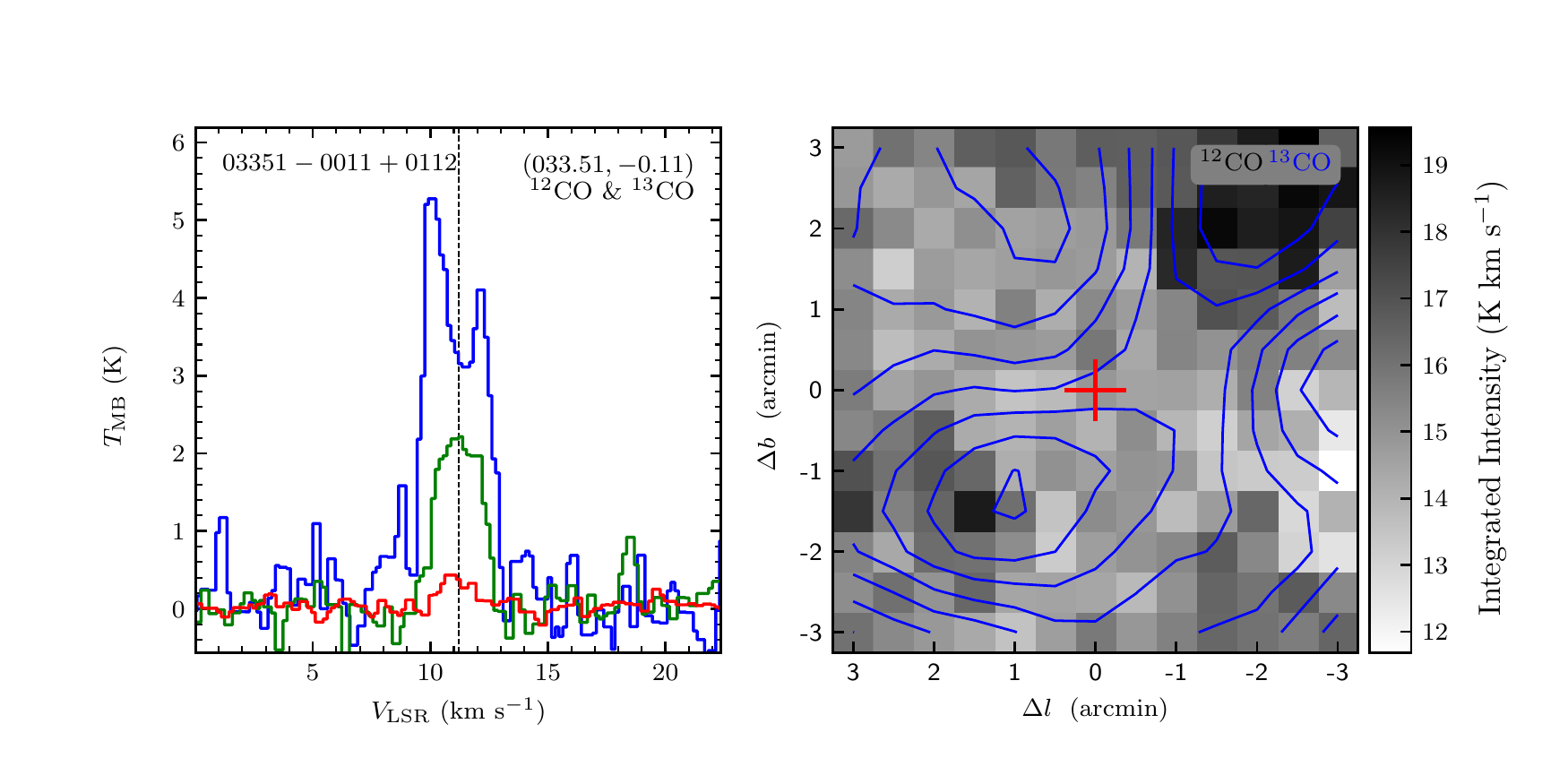}
\includegraphics[width=9.0cm,angle=0]{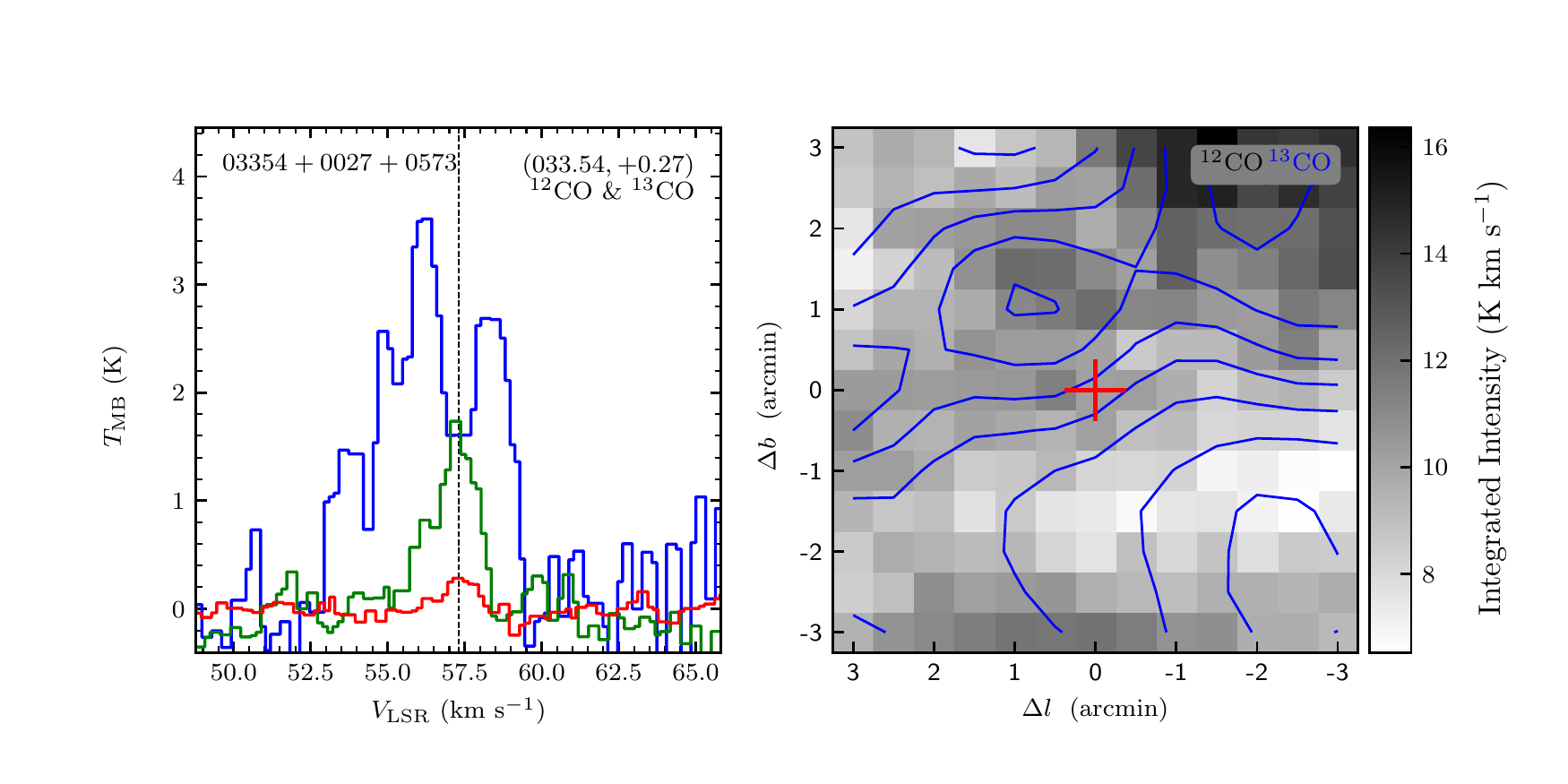}
\end{figure}
\clearpage

\begin{figure}
\includegraphics[width=9.0cm,angle=0]{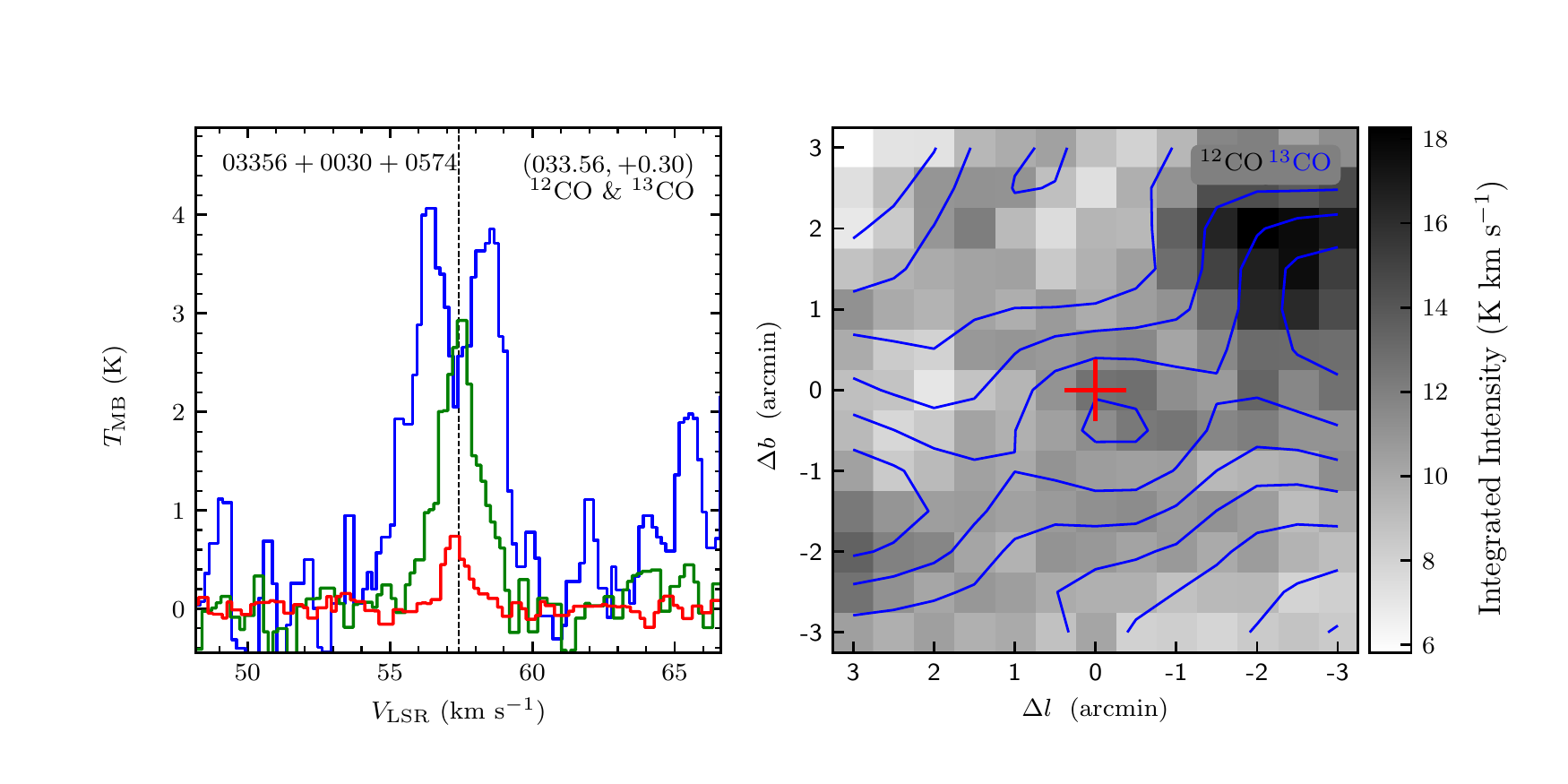}
\includegraphics[width=9.0cm,angle=0]{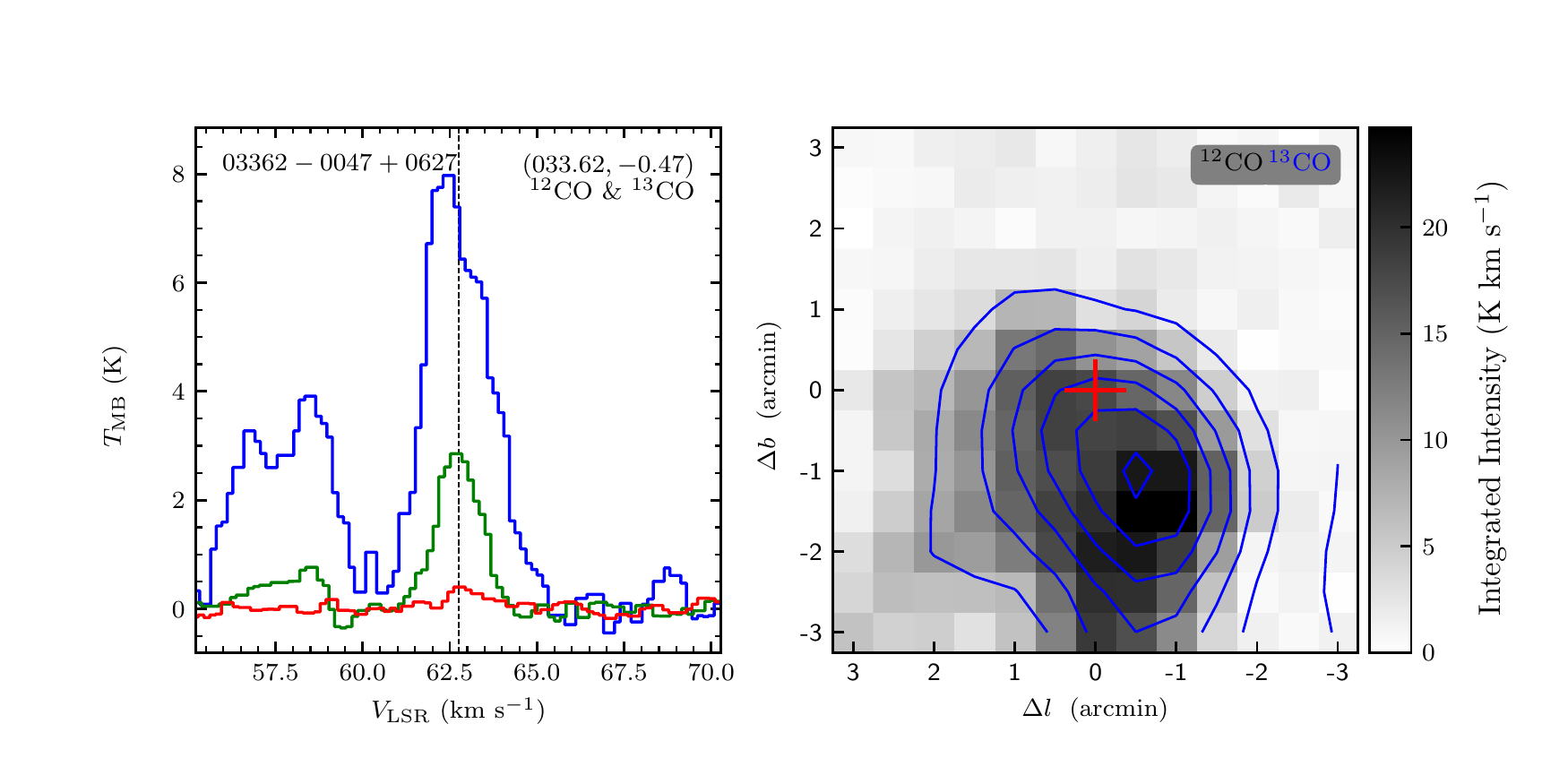}
\vspace{-0.5cm}

\includegraphics[width=9.0cm,angle=0]{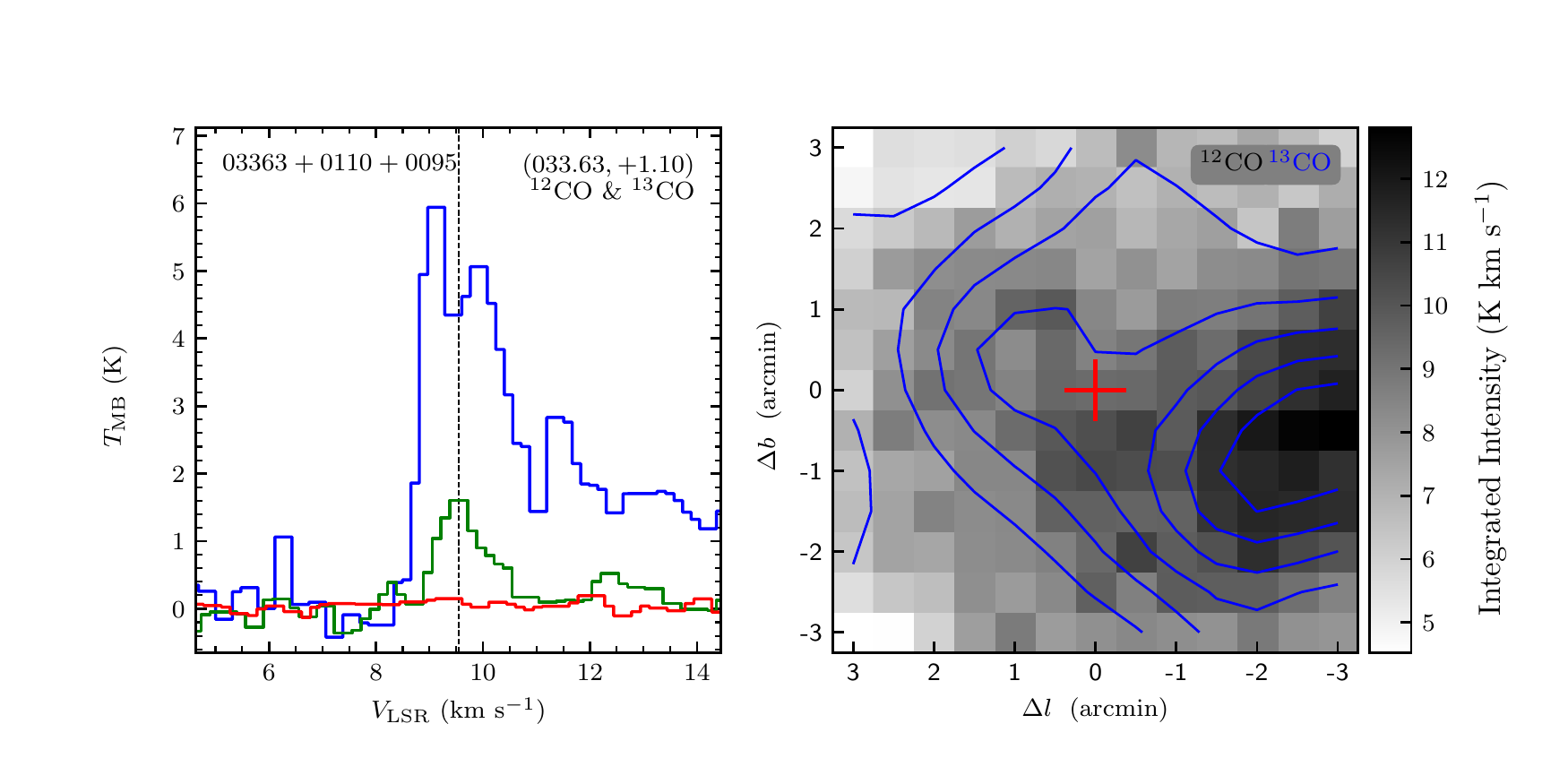}
\includegraphics[width=9.0cm,angle=0]{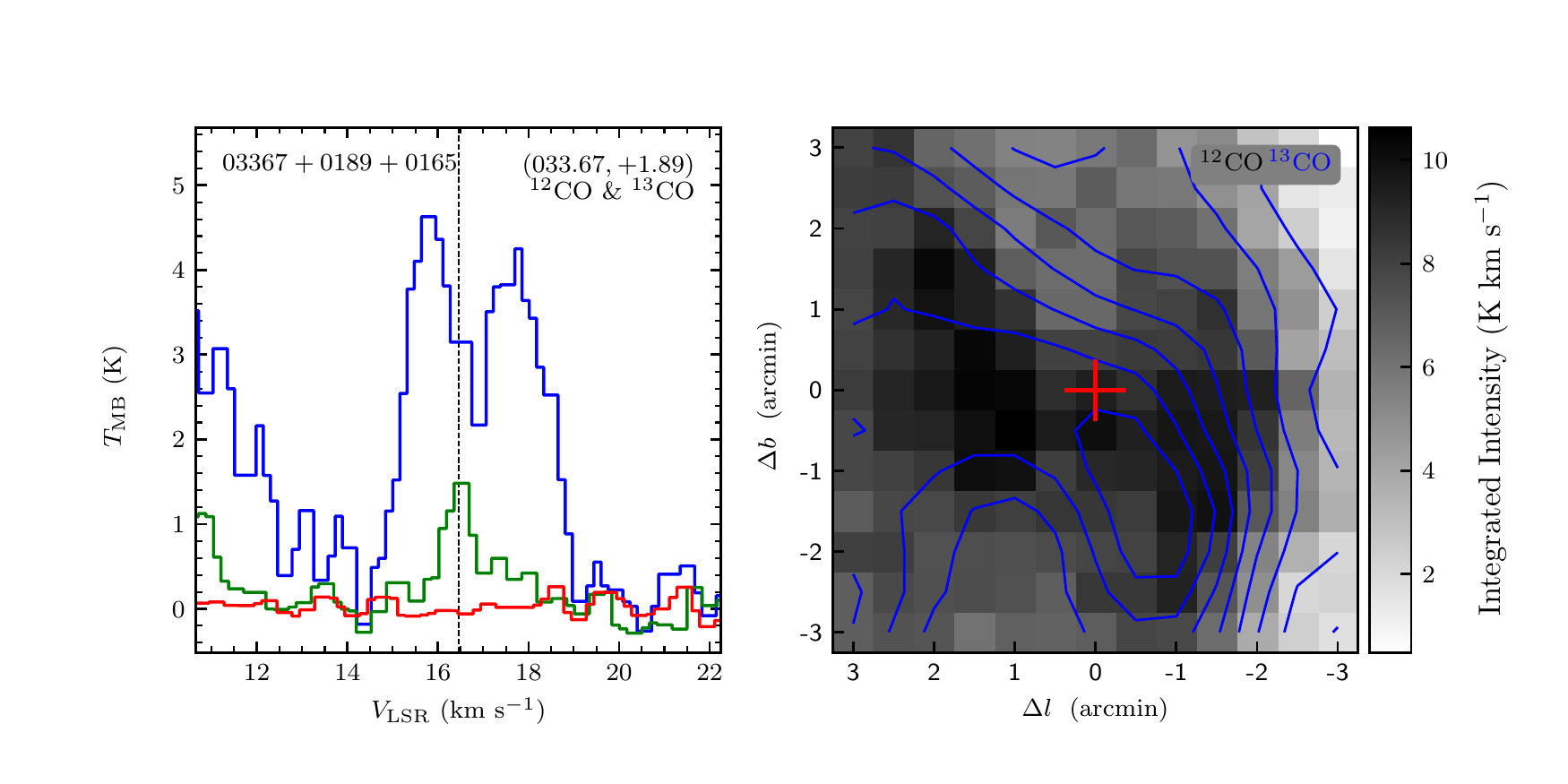}
\vspace{-0.5cm}

\includegraphics[width=9.0cm,angle=0]{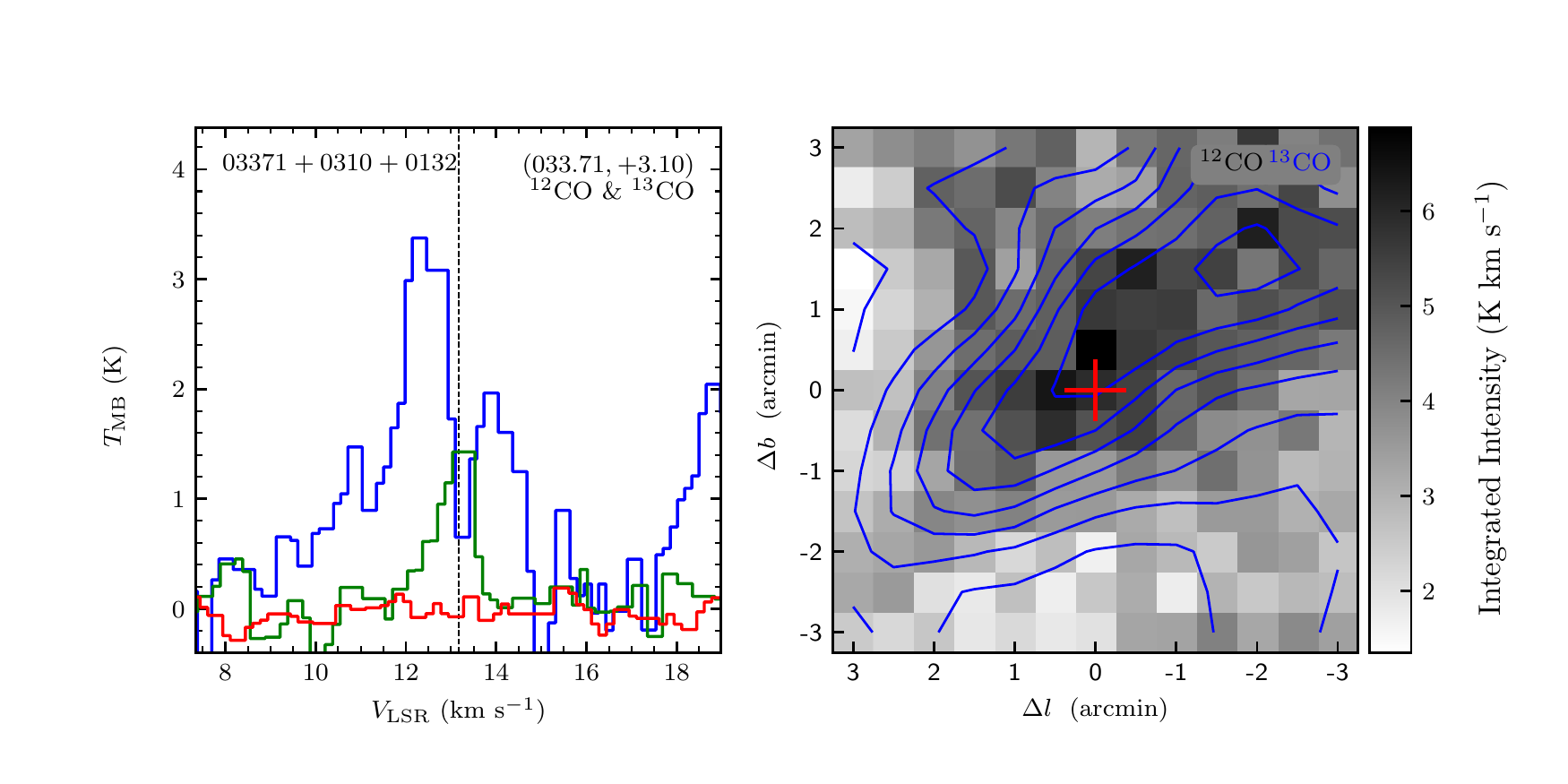}
\includegraphics[width=9.0cm,angle=0]{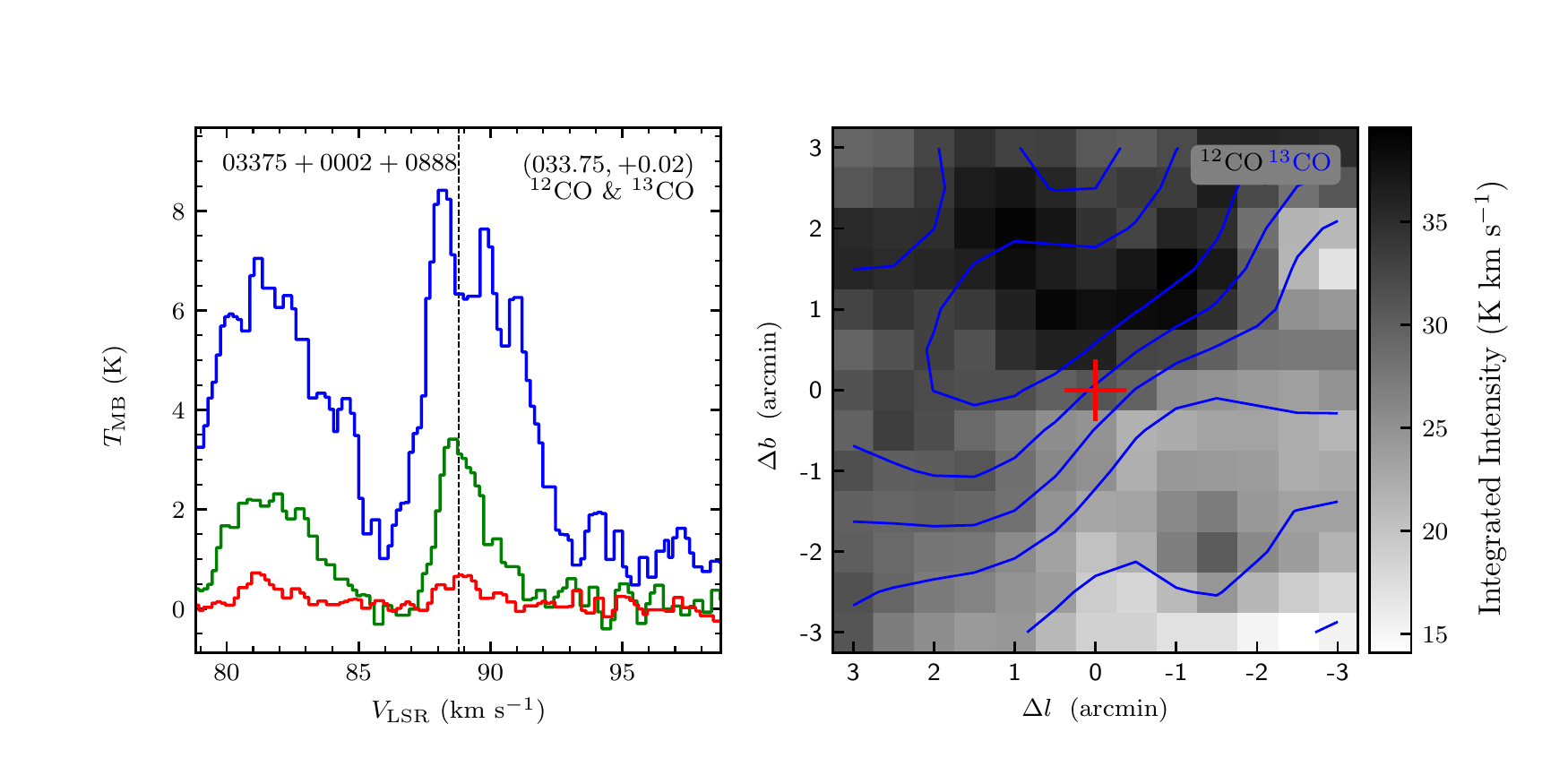}
\vspace{-0.5cm}

\includegraphics[width=9.0cm,angle=0]{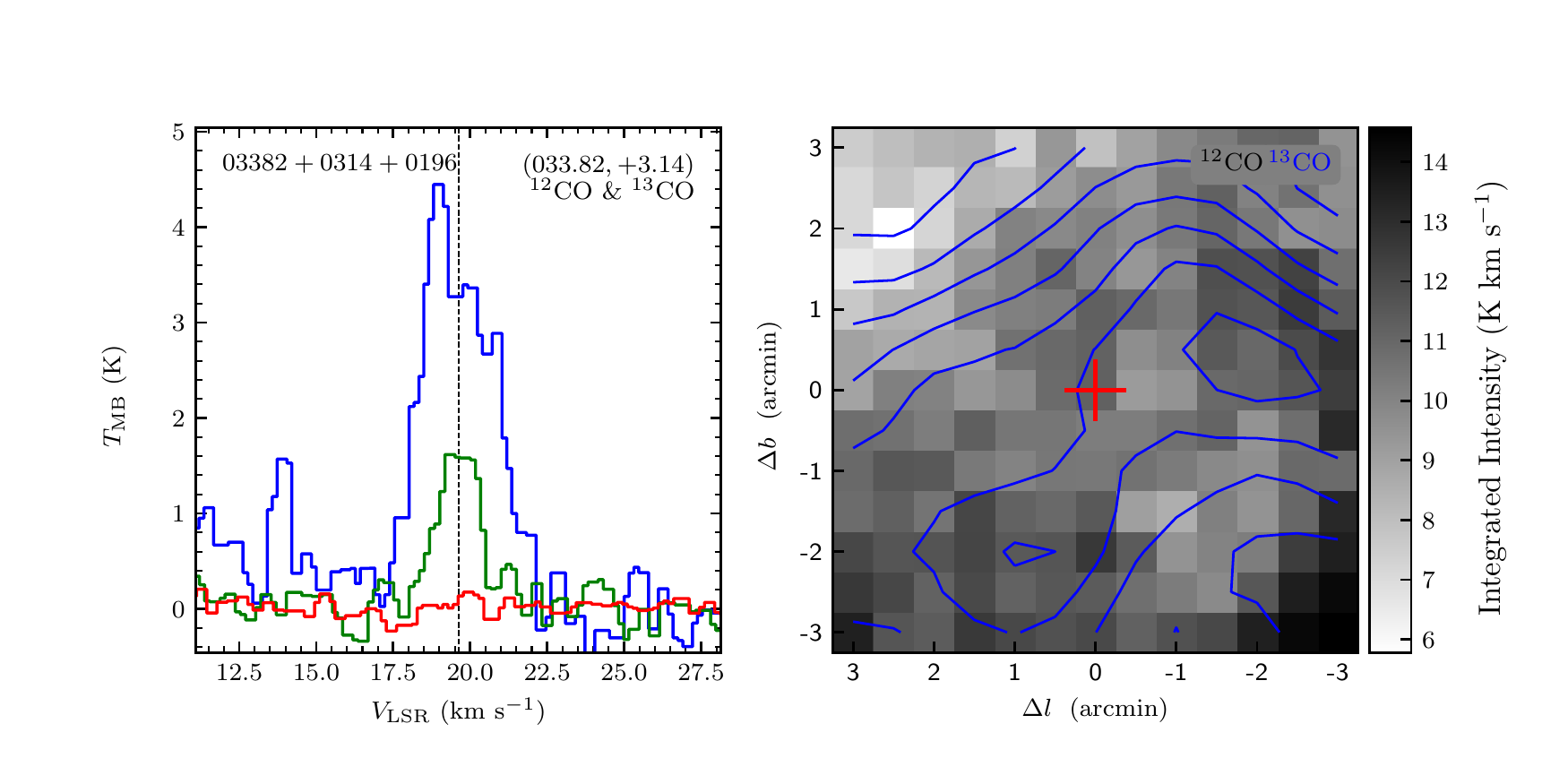}
\includegraphics[width=9.0cm,angle=0]{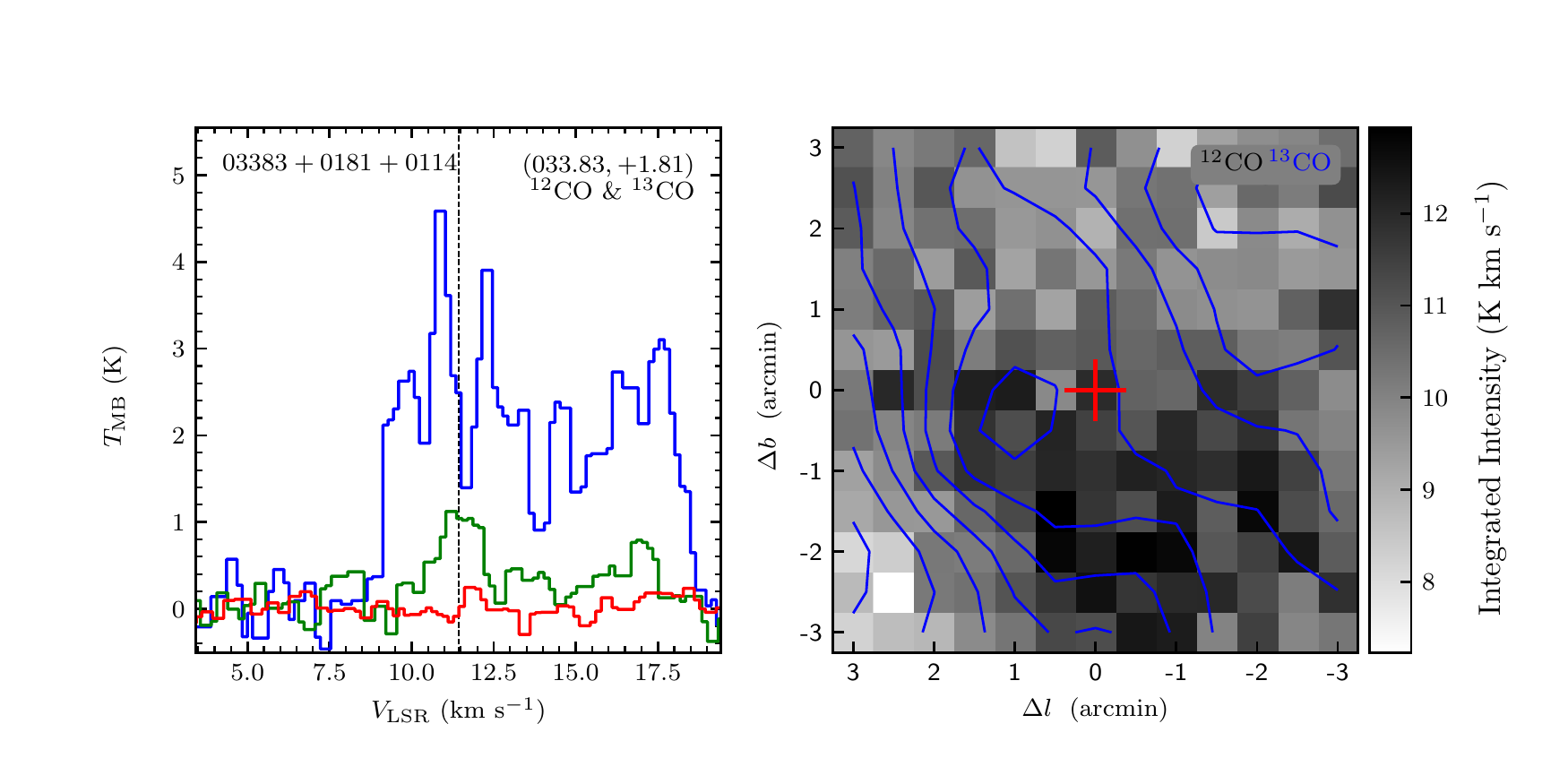}
\vspace{-0.5cm}

\includegraphics[width=9.0cm,angle=0]{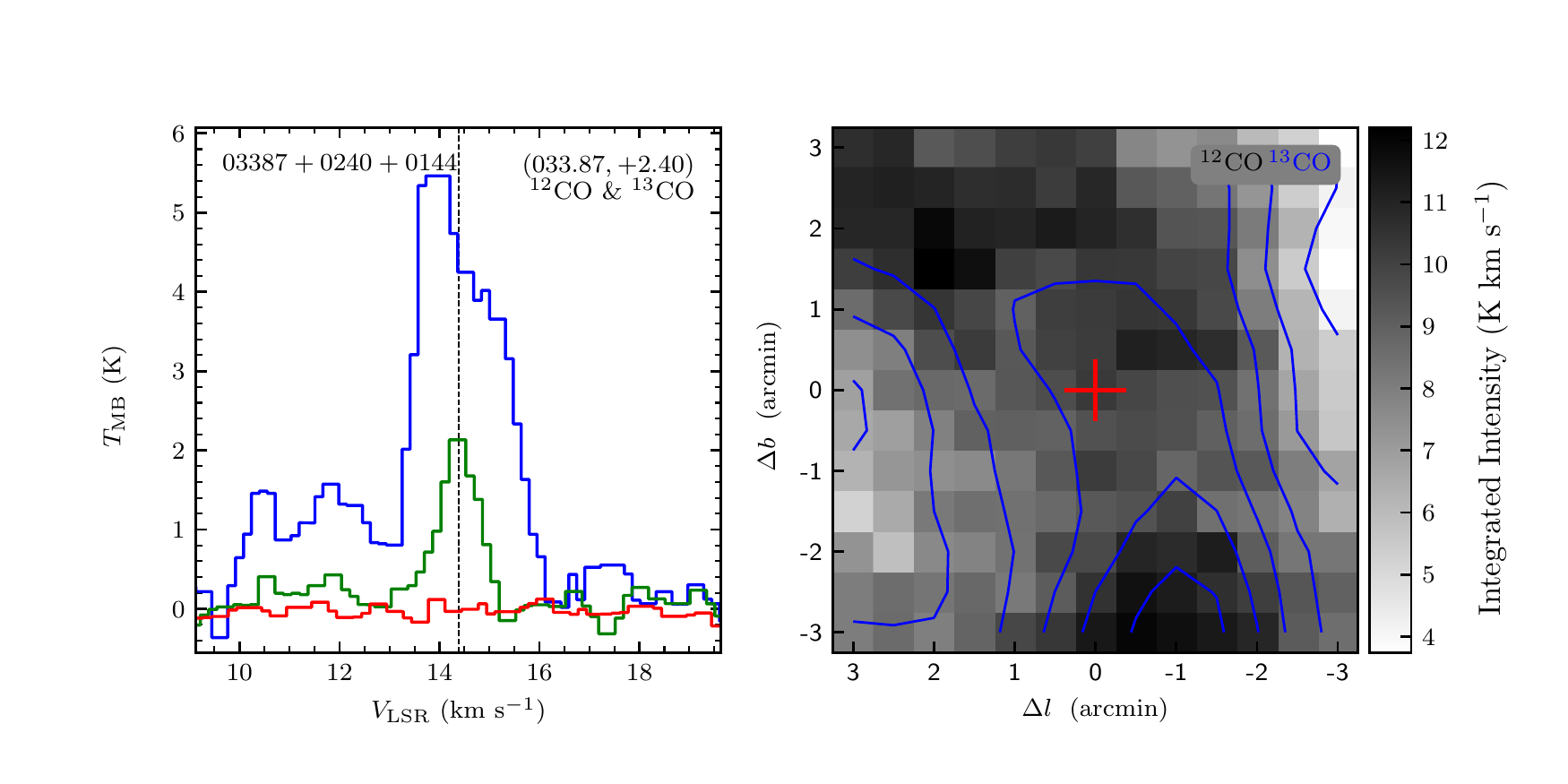}
\includegraphics[width=9.0cm,angle=0]{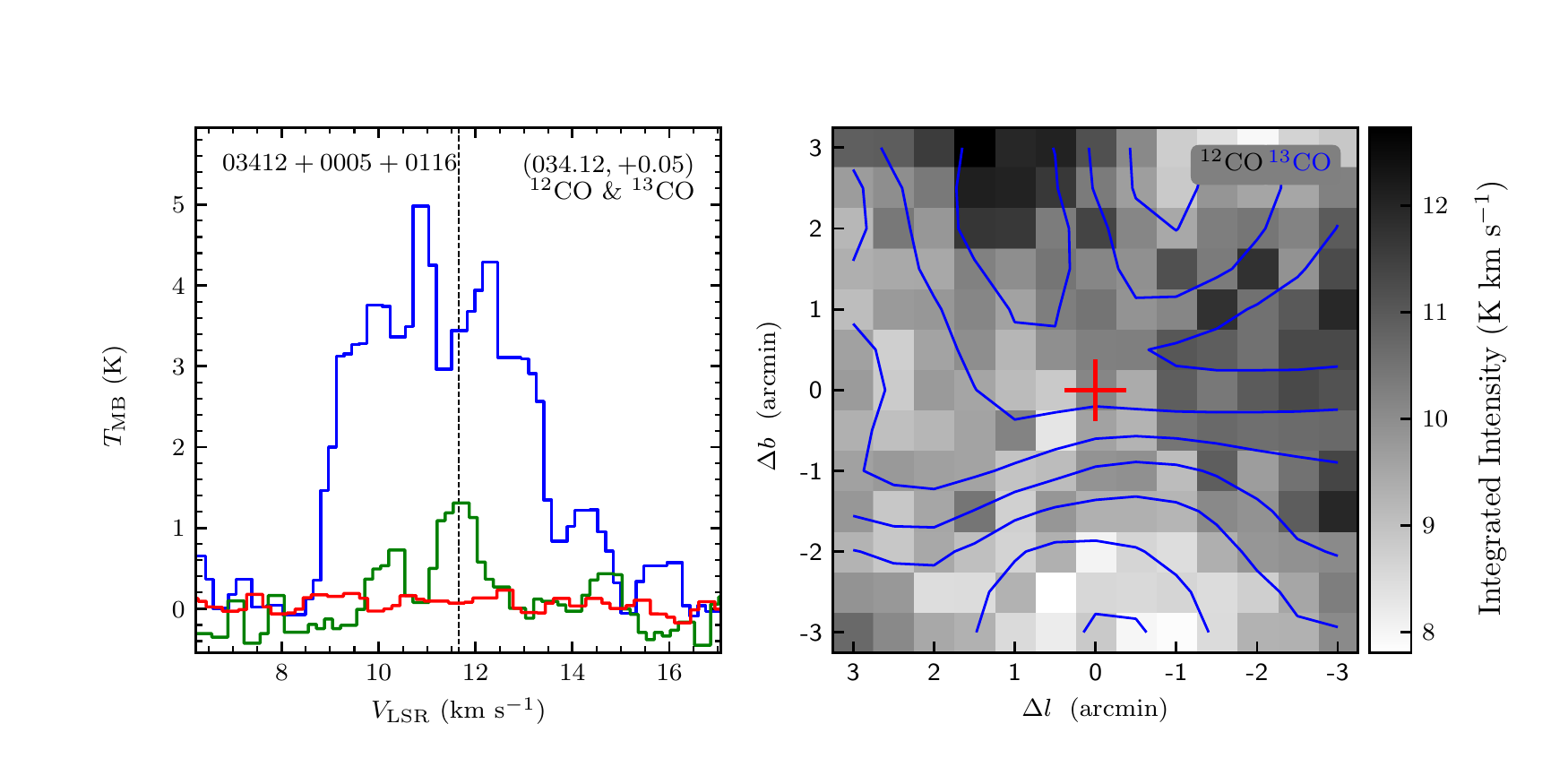}
\end{figure}
\clearpage

\begin{figure}
\includegraphics[width=9.0cm,angle=0]{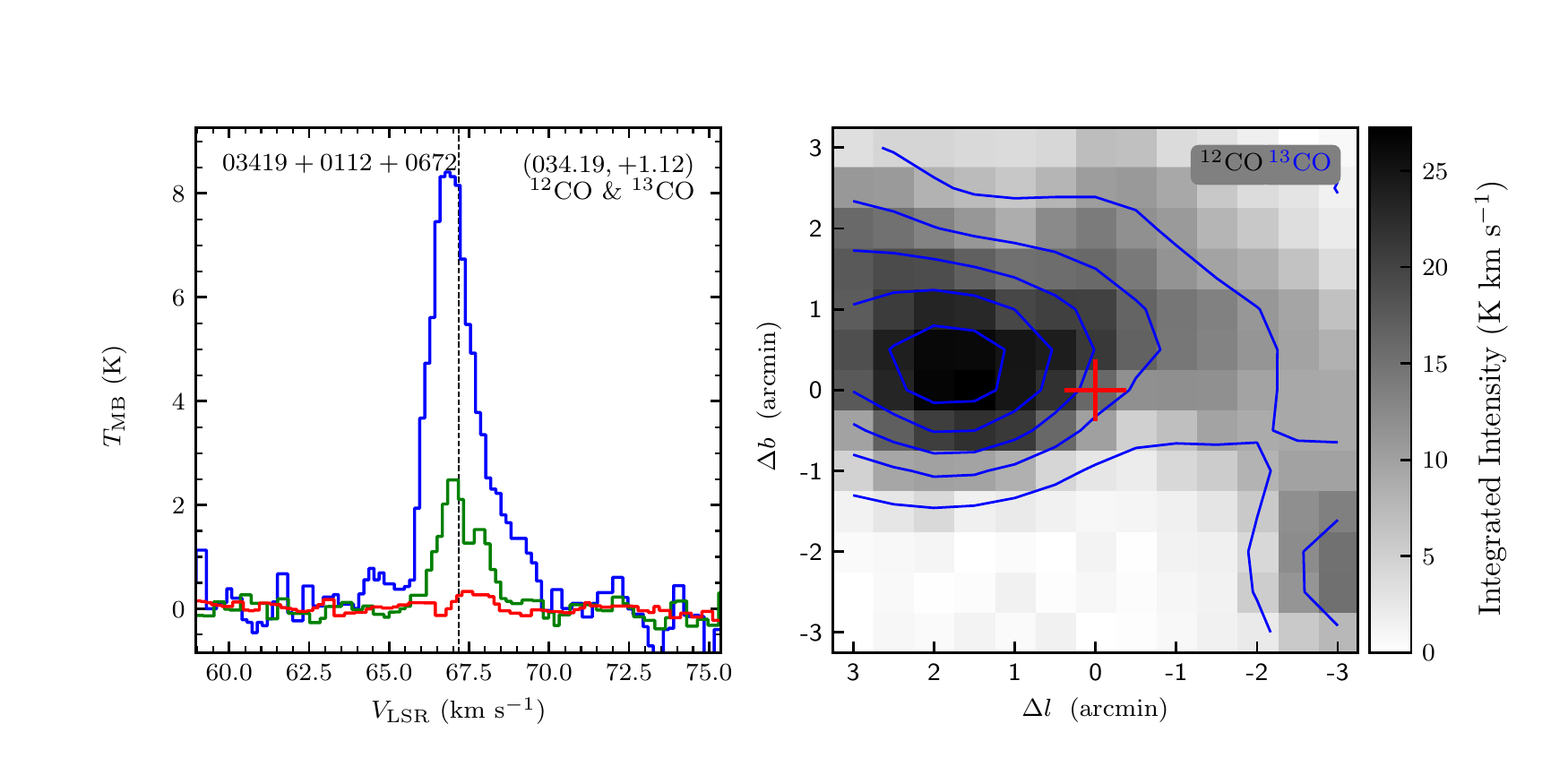}
\includegraphics[width=9.0cm,angle=0]{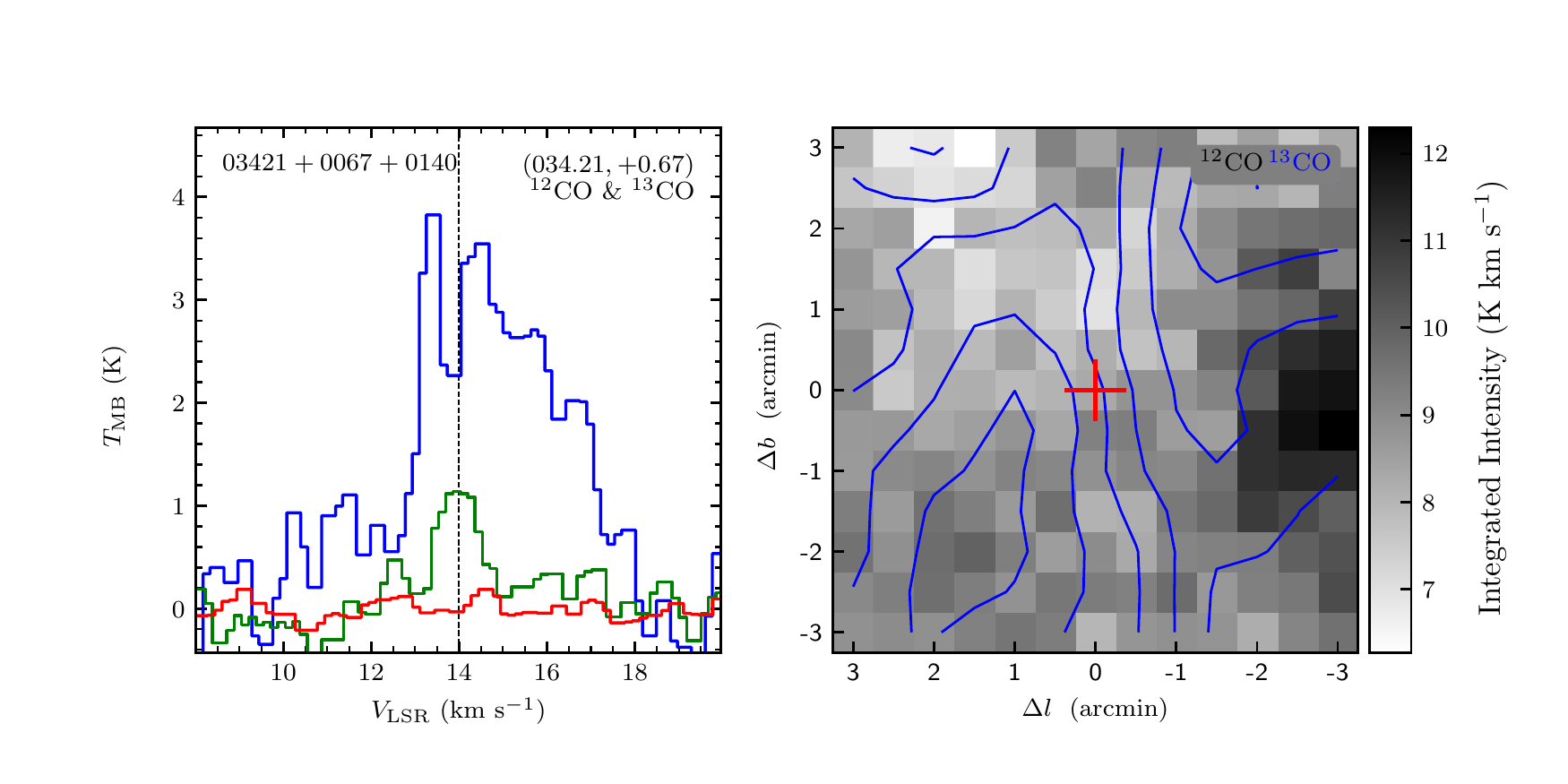}
\vspace{-0.5cm}

\includegraphics[width=9.0cm,angle=0]{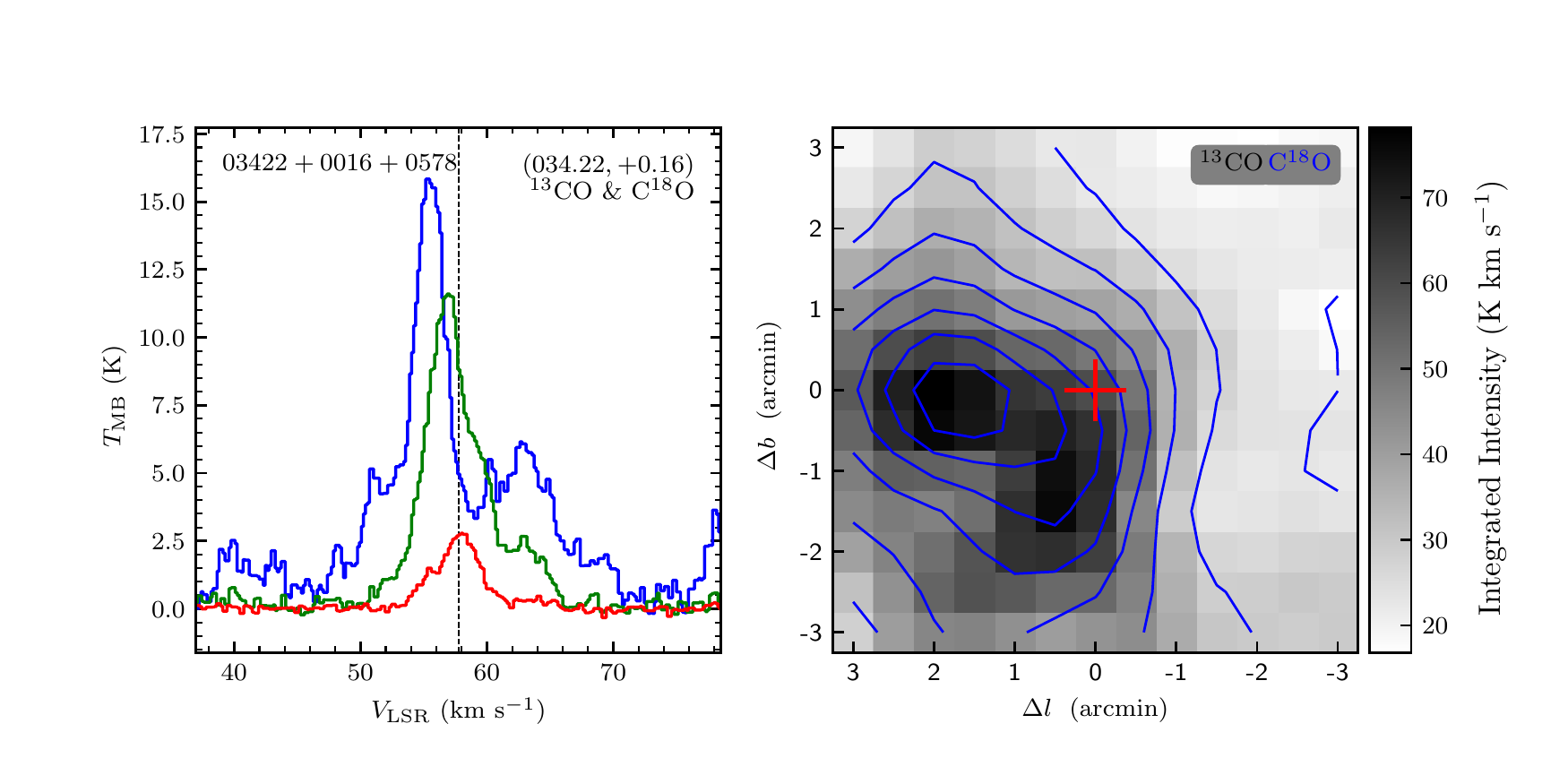}
\includegraphics[width=9.0cm,angle=0]{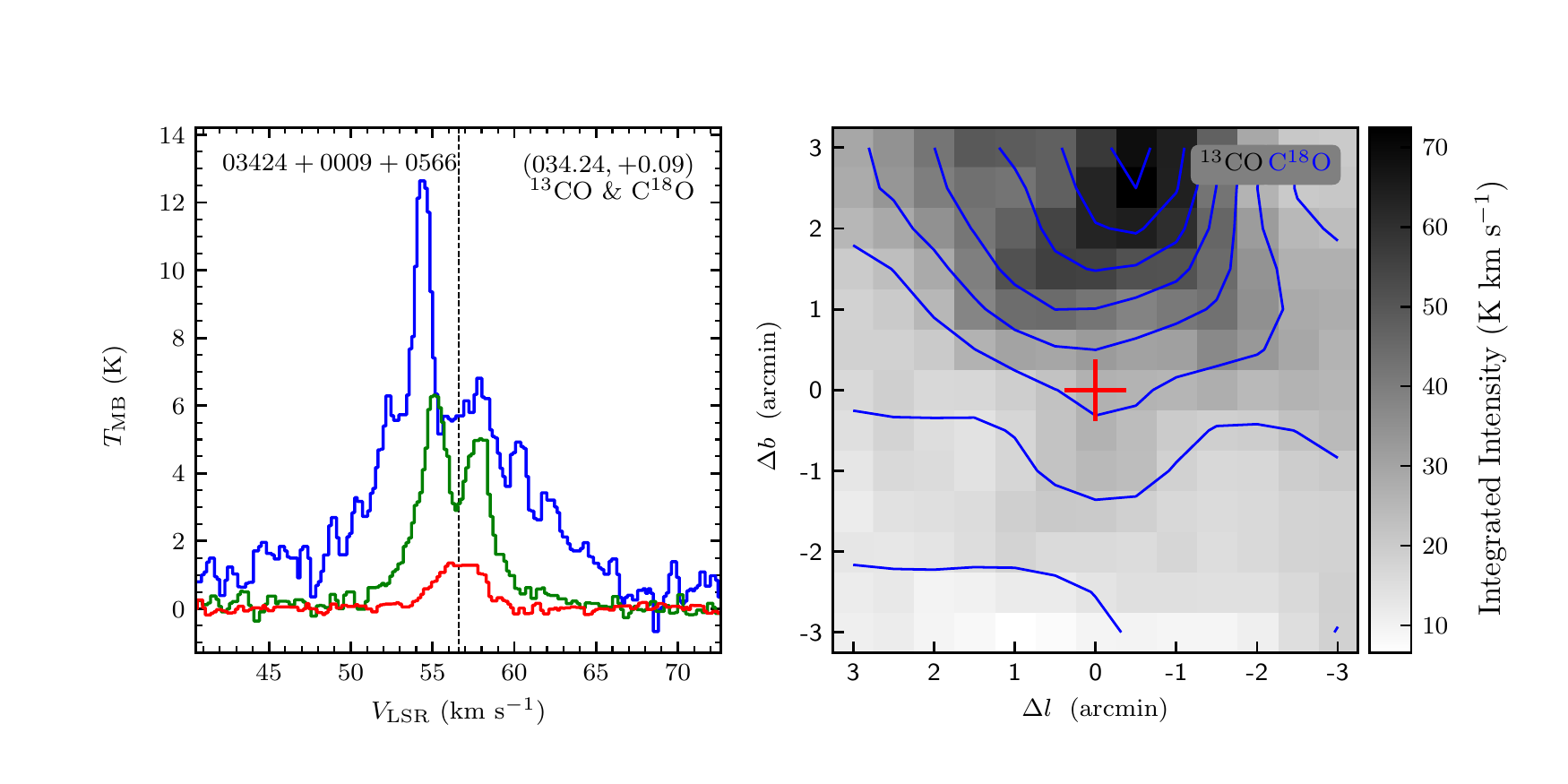}
\vspace{-0.5cm}

\includegraphics[width=9.0cm,angle=0]{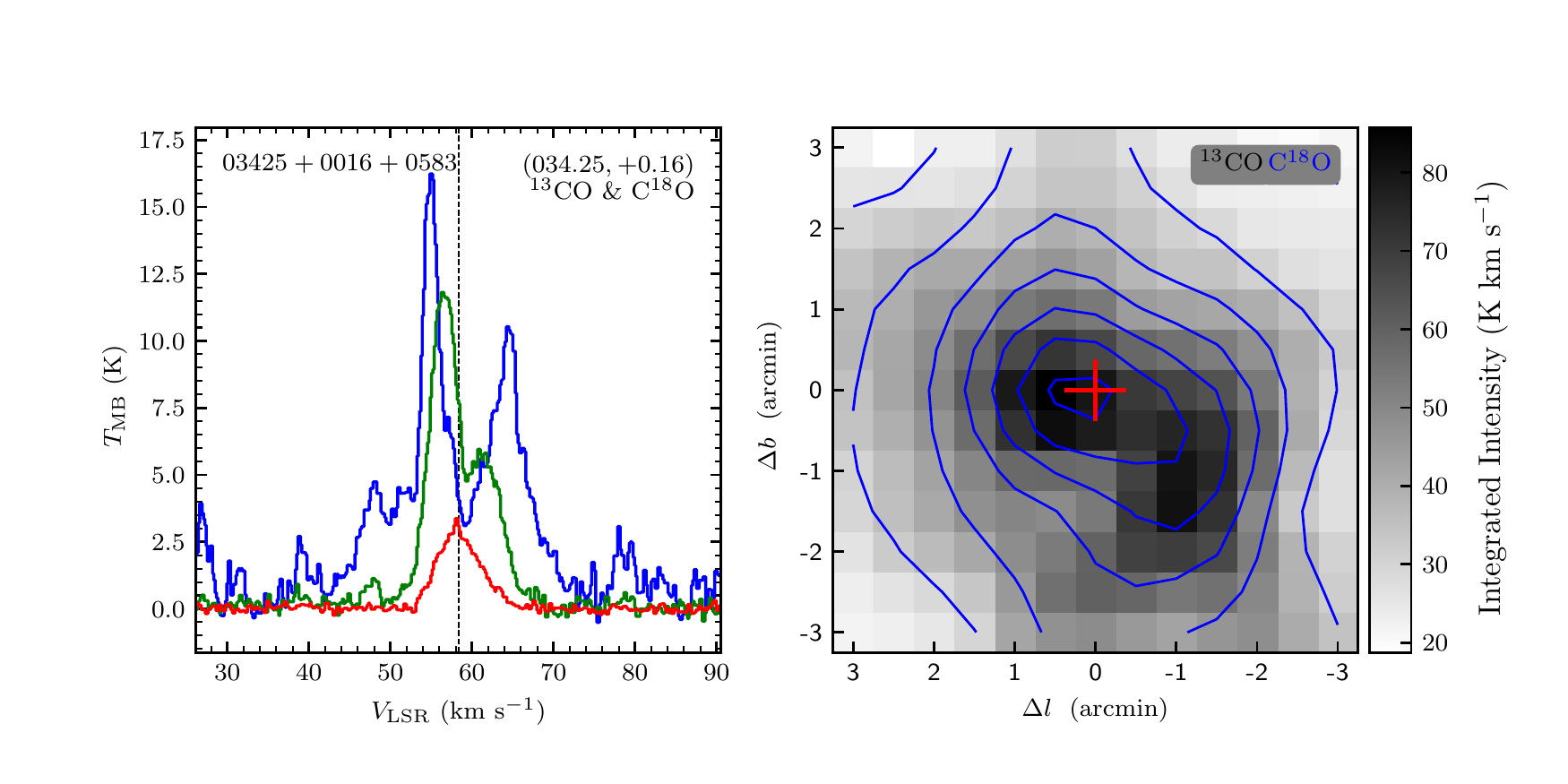}
\includegraphics[width=9.0cm,angle=0]{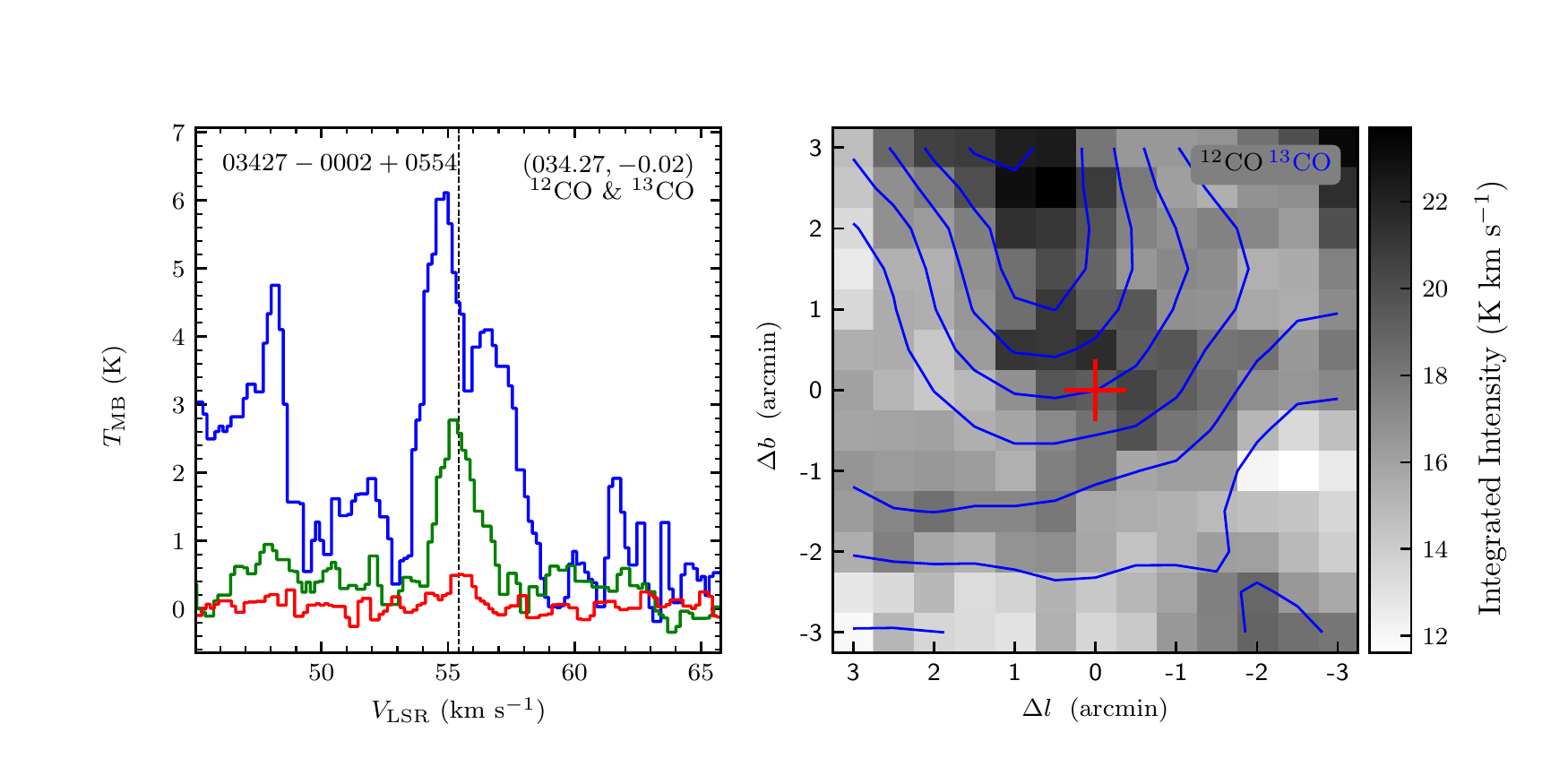}
\vspace{-0.5cm}

\includegraphics[width=9.0cm,angle=0]{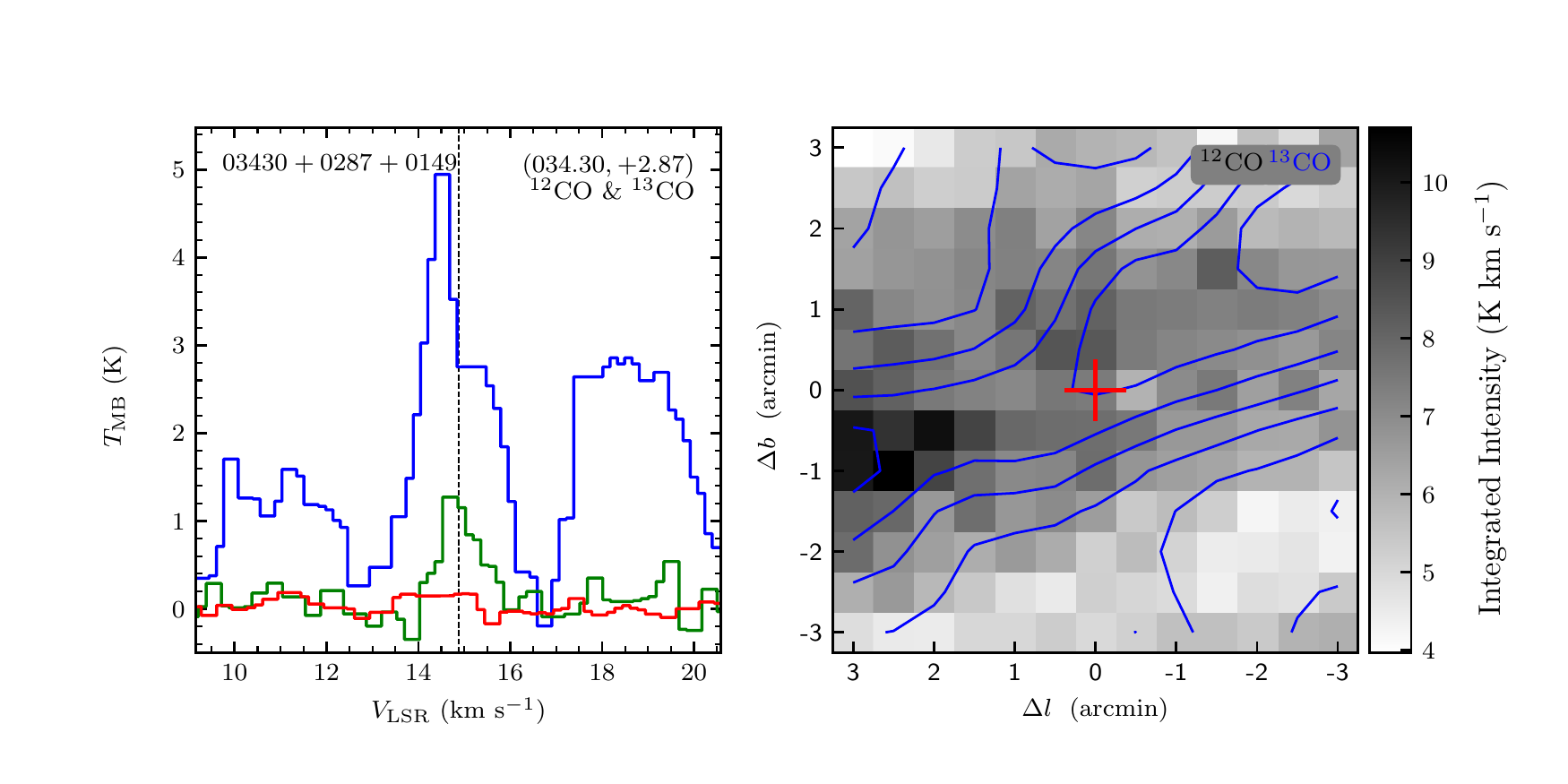}
\includegraphics[width=9.0cm,angle=0]{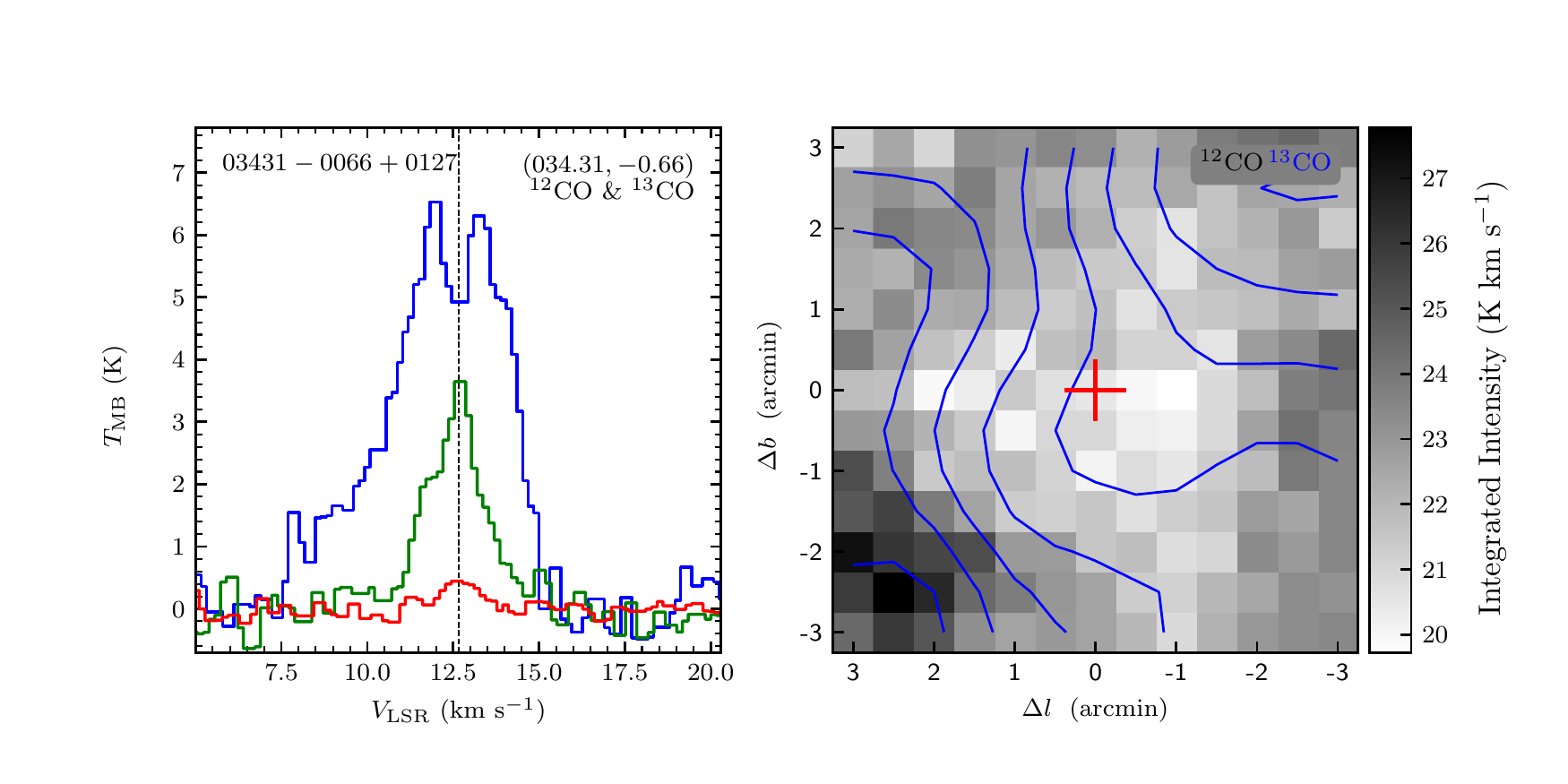}
\vspace{-0.5cm}

\includegraphics[width=9.0cm,angle=0]{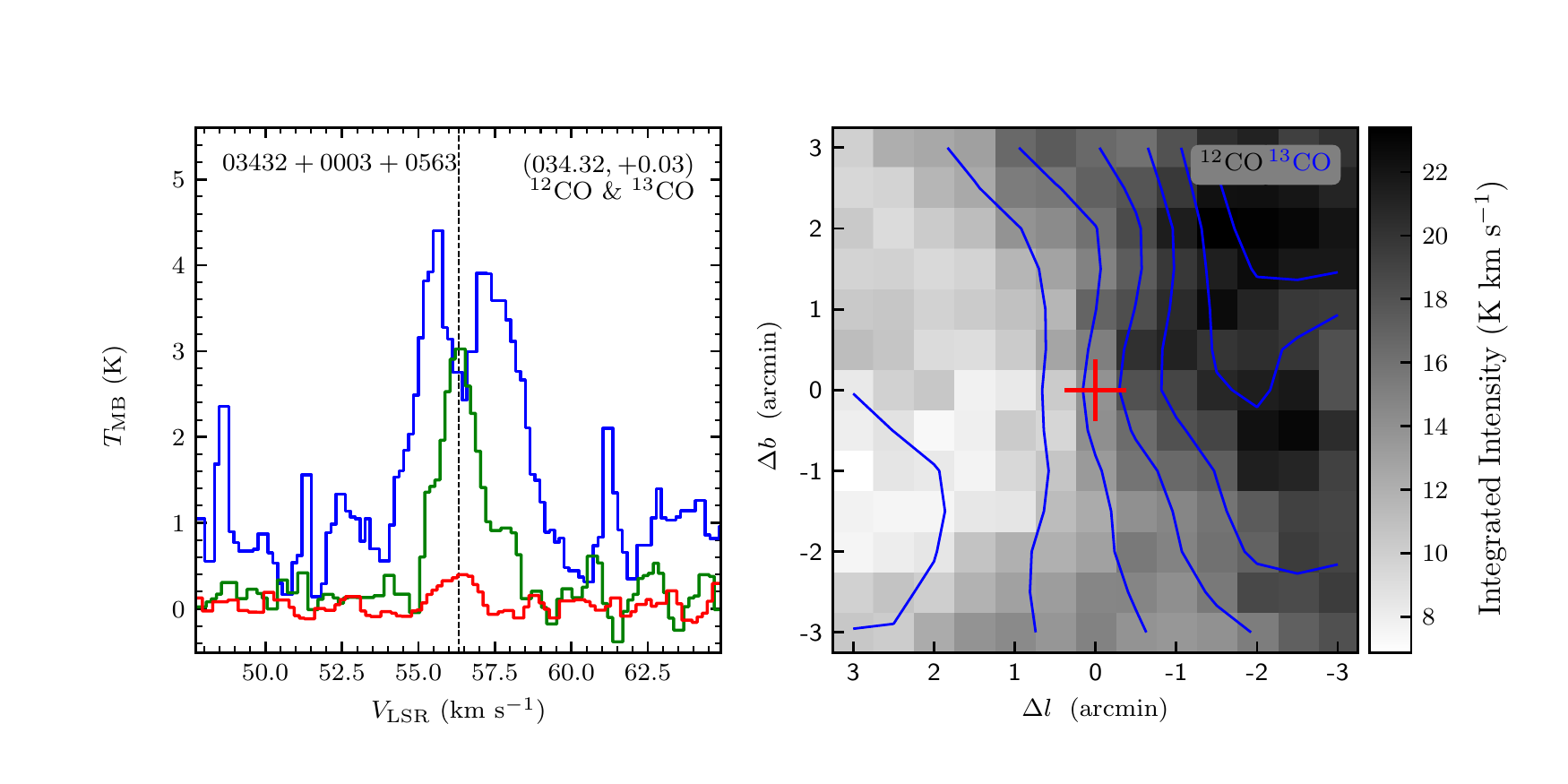}
\includegraphics[width=9.0cm,angle=0]{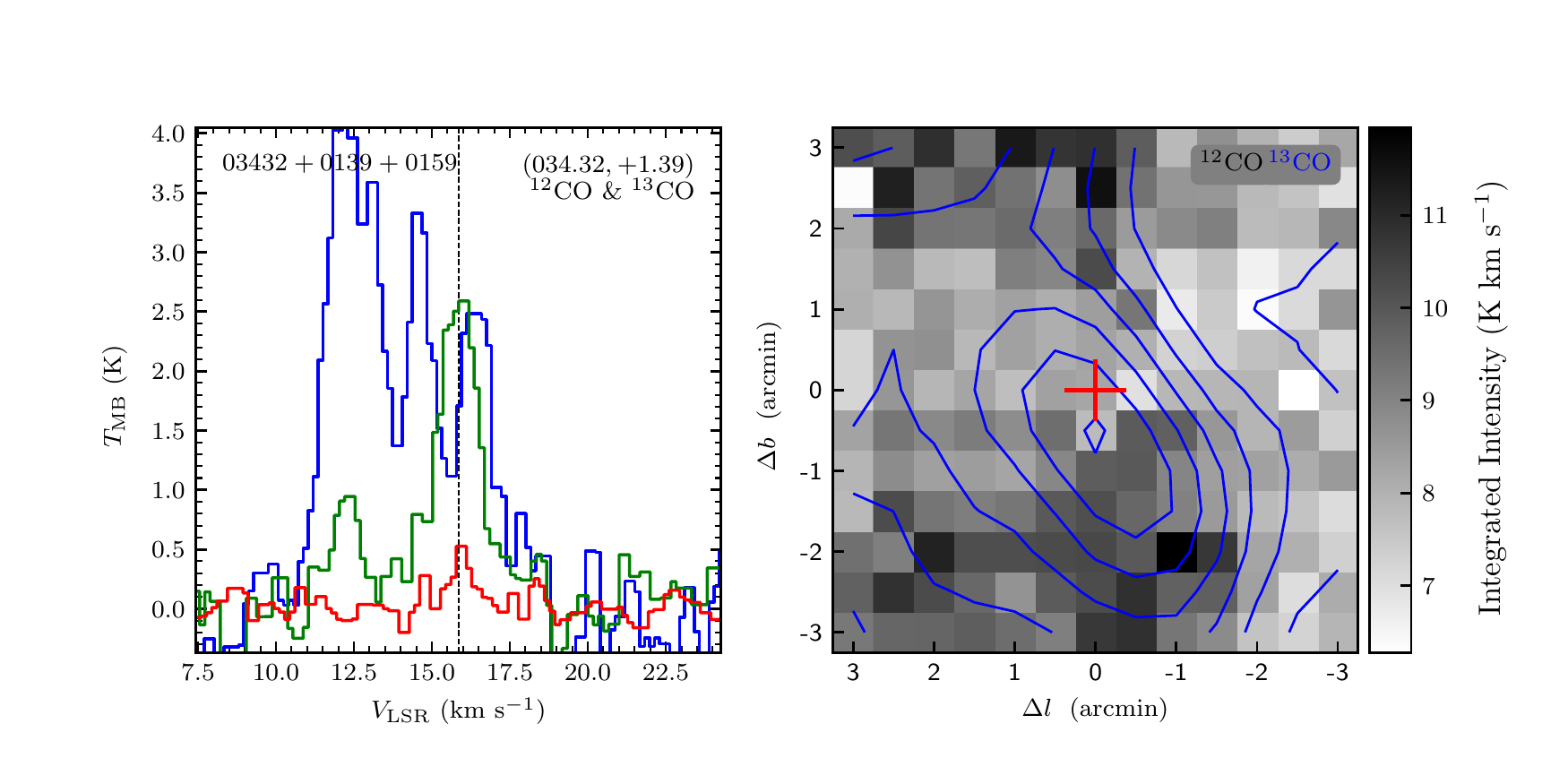}
\end{figure}
\clearpage

\begin{figure}
\includegraphics[width=9.0cm,angle=0]{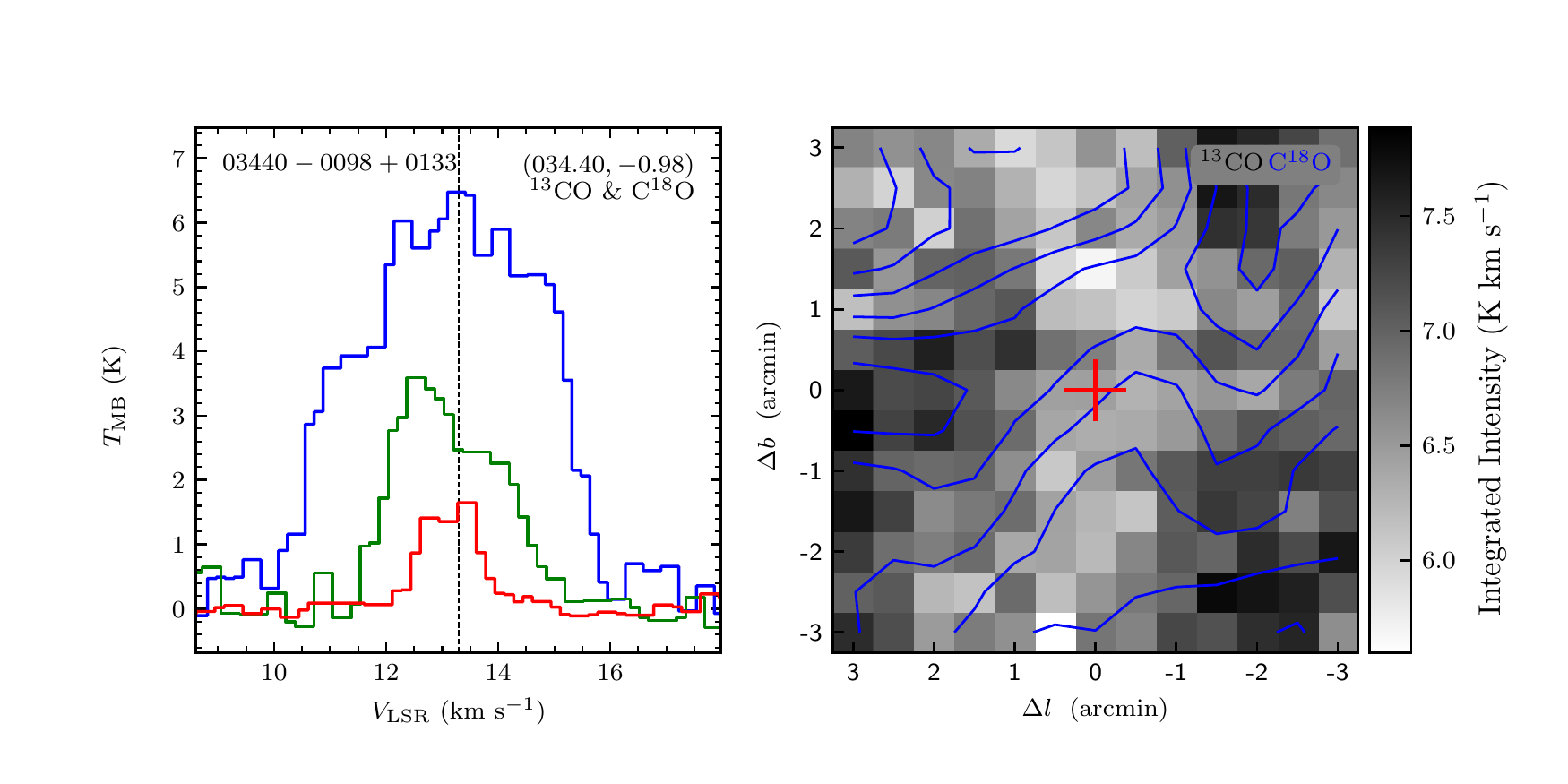}
\includegraphics[width=9.0cm,angle=0]{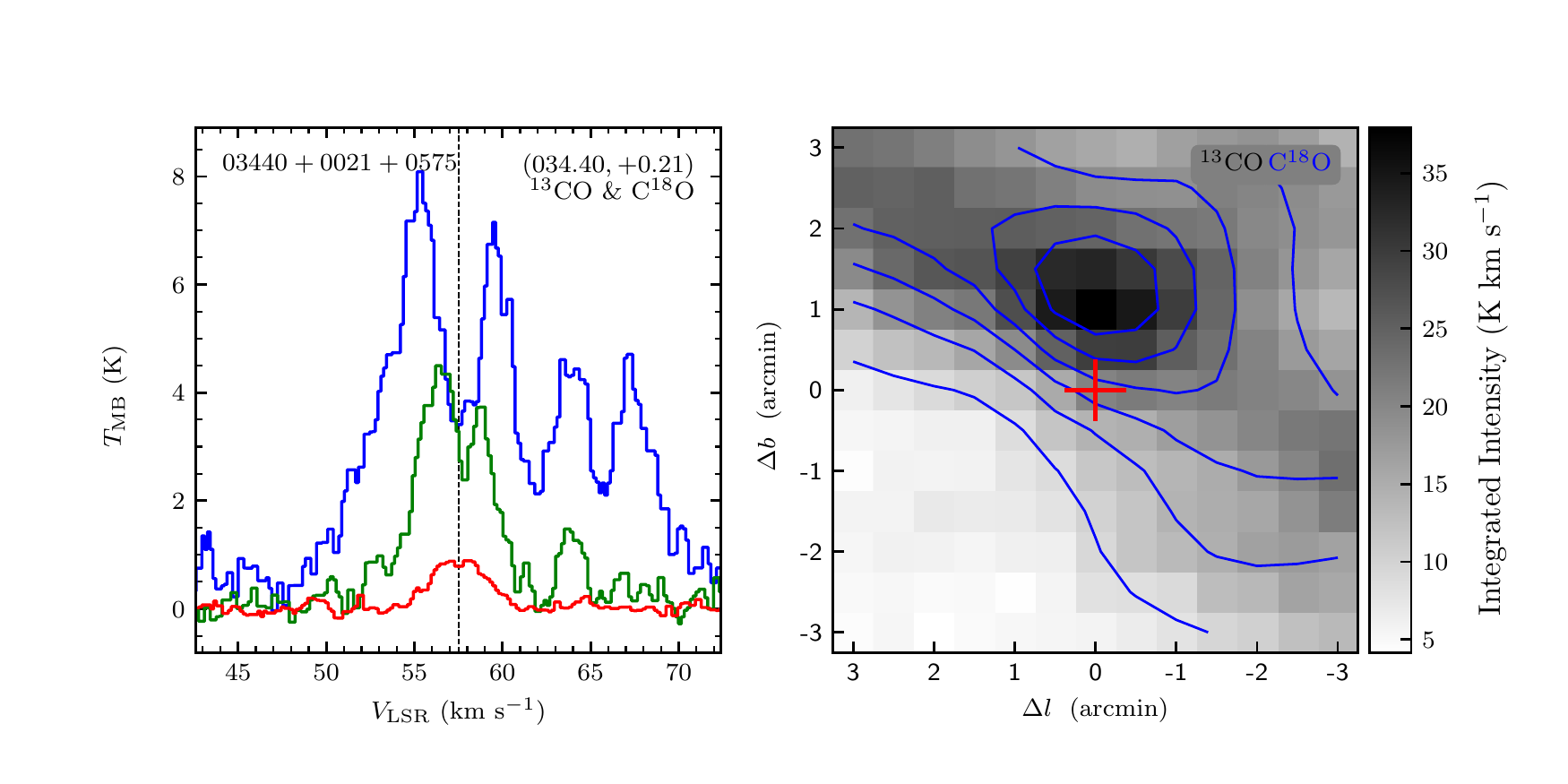}
\vspace{-0.5cm}

\includegraphics[width=9.0cm,angle=0]{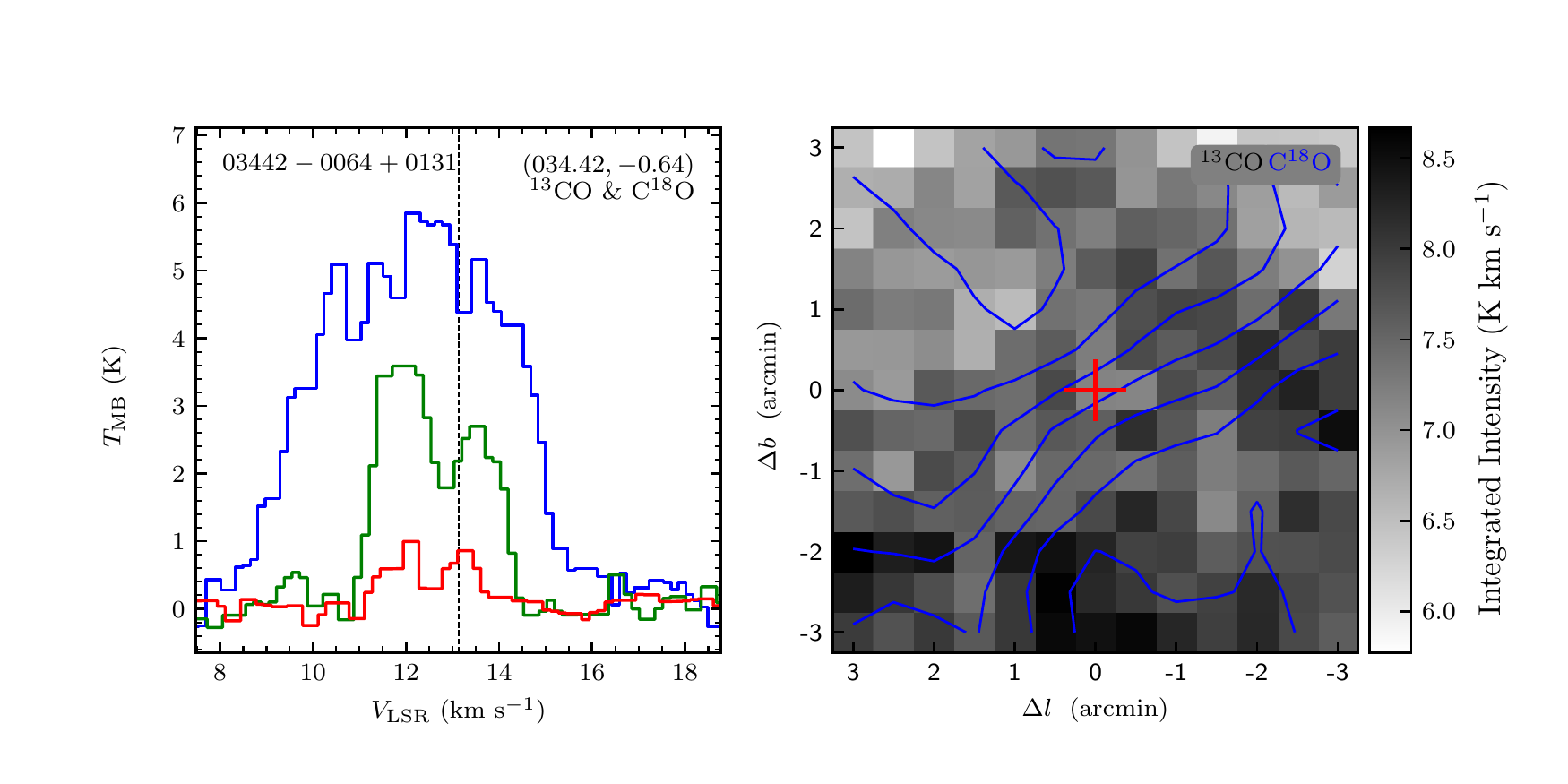}
\includegraphics[width=9.0cm,angle=0]{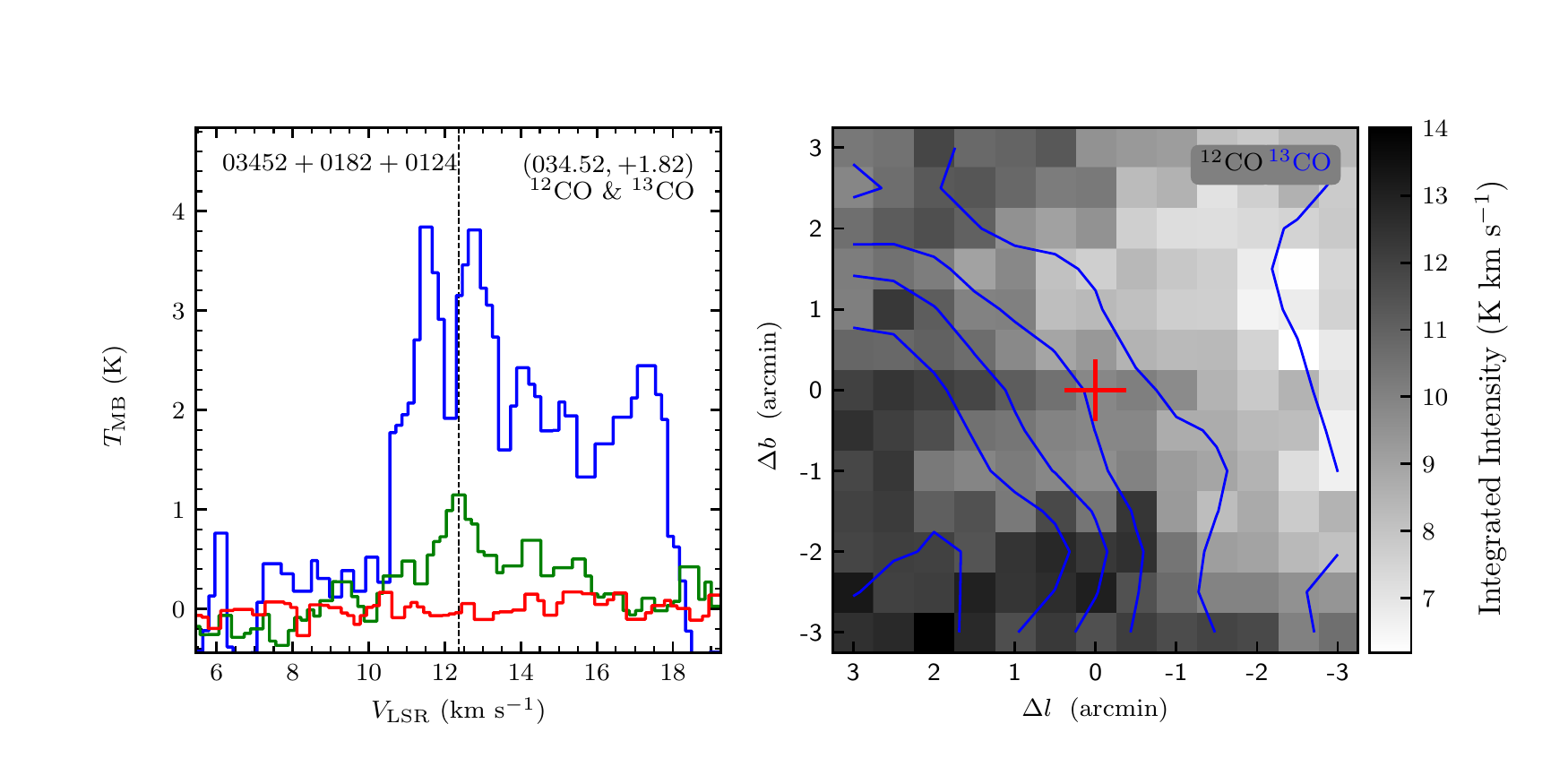}
\vspace{-0.5cm}

\includegraphics[width=9.0cm,angle=0]{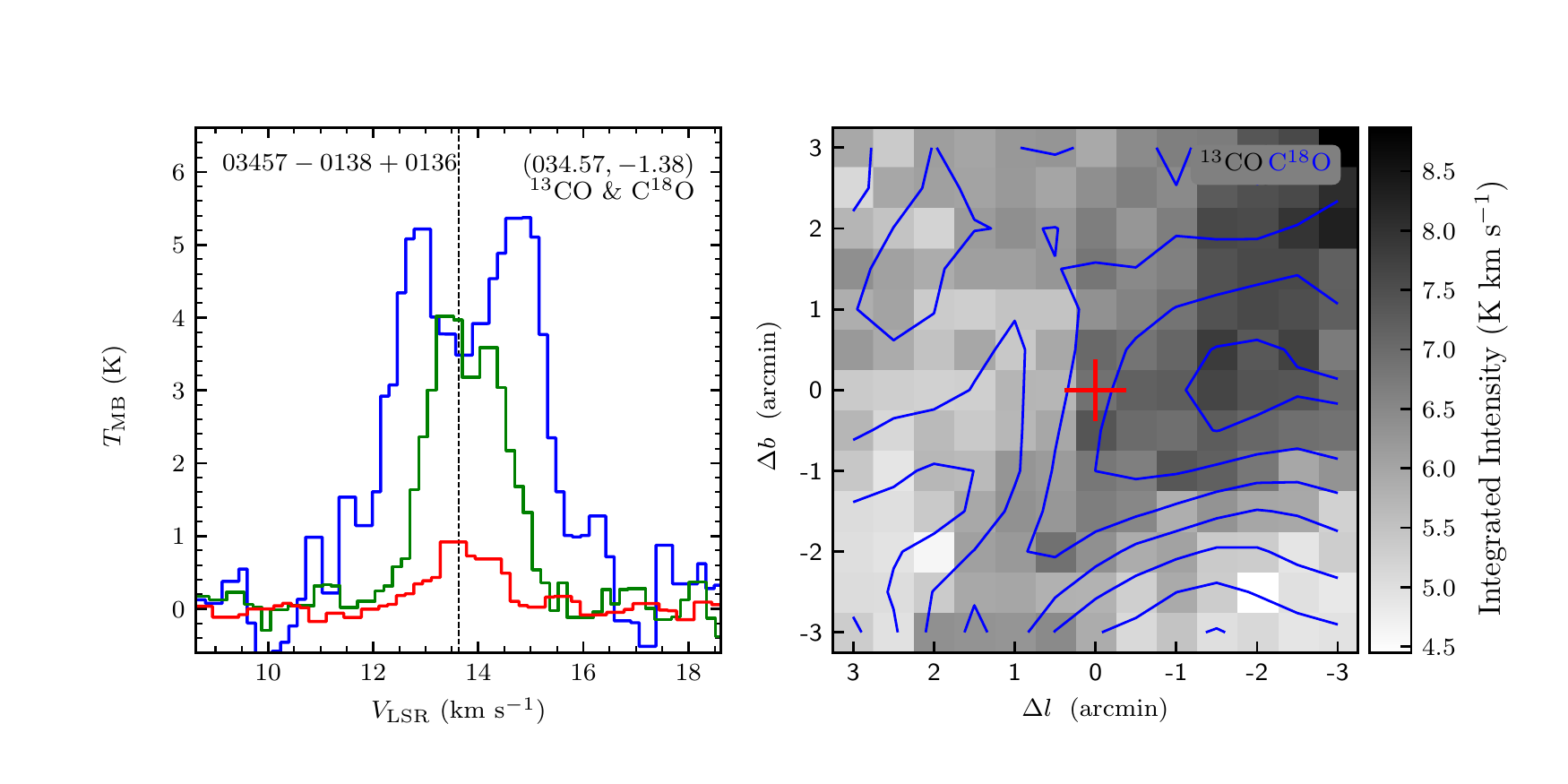}
\includegraphics[width=9.0cm,angle=0]{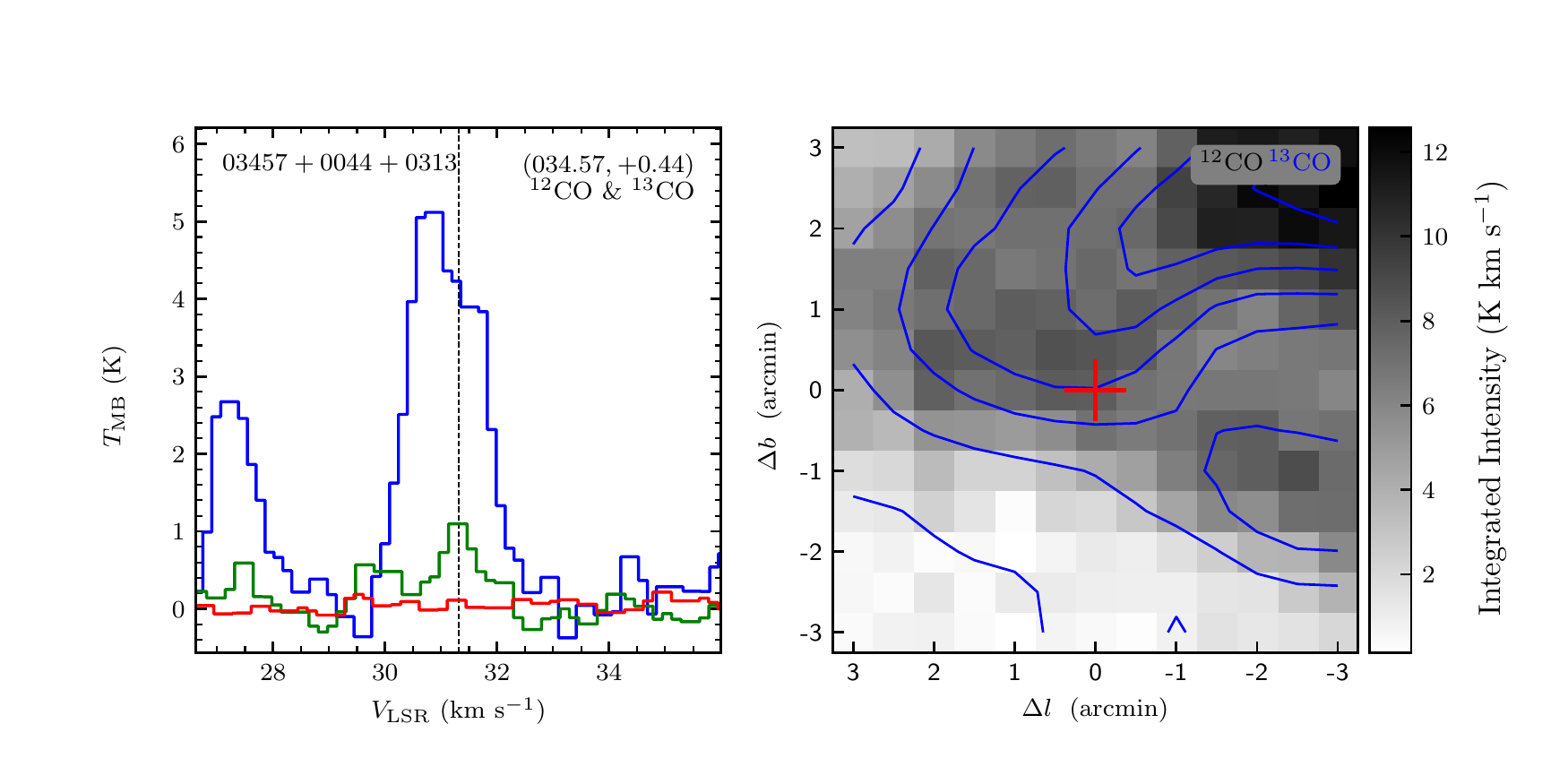}
\vspace{-0.5cm}

\includegraphics[width=9.0cm,angle=0]{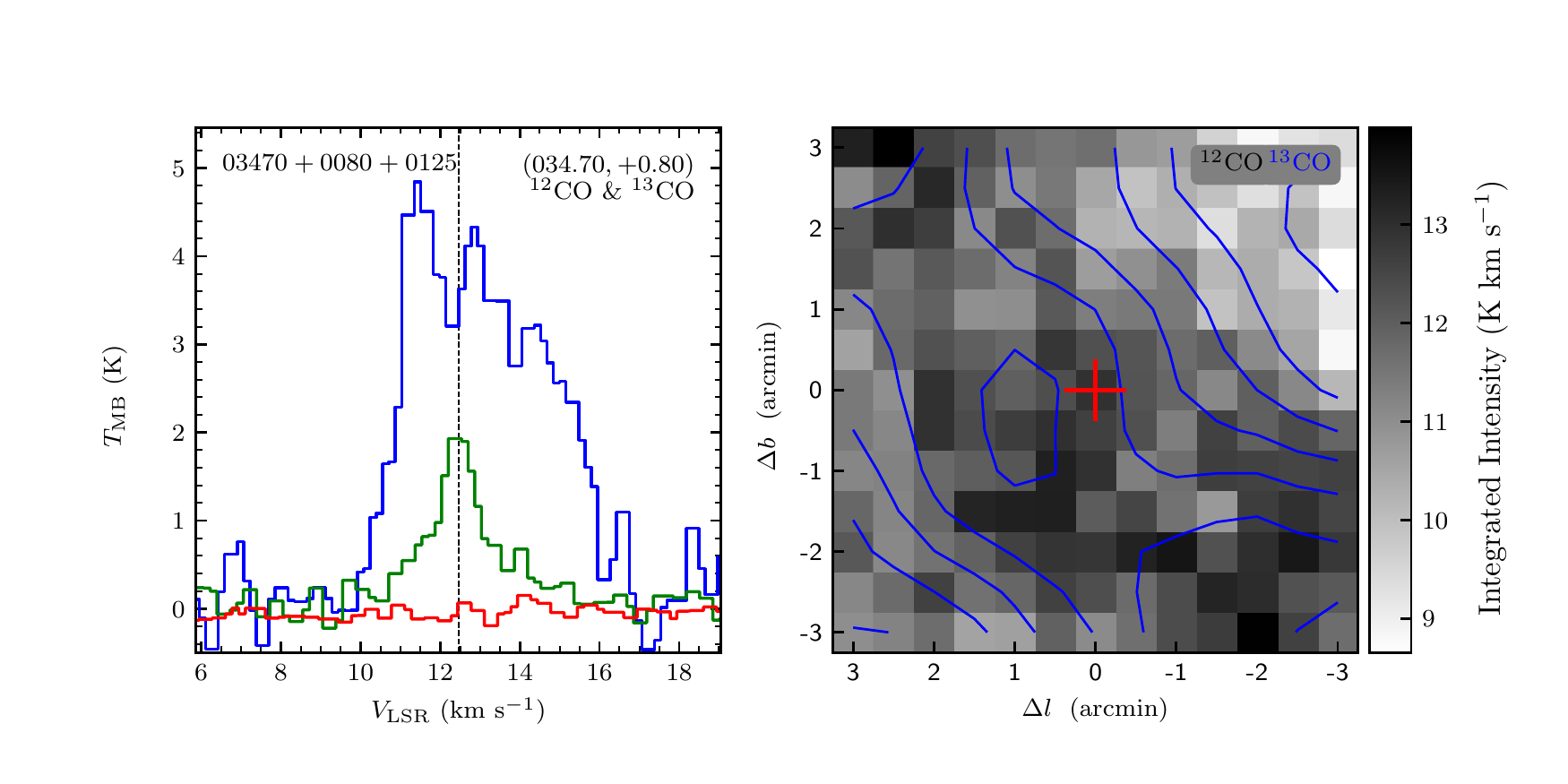}
\includegraphics[width=9.0cm,angle=0]{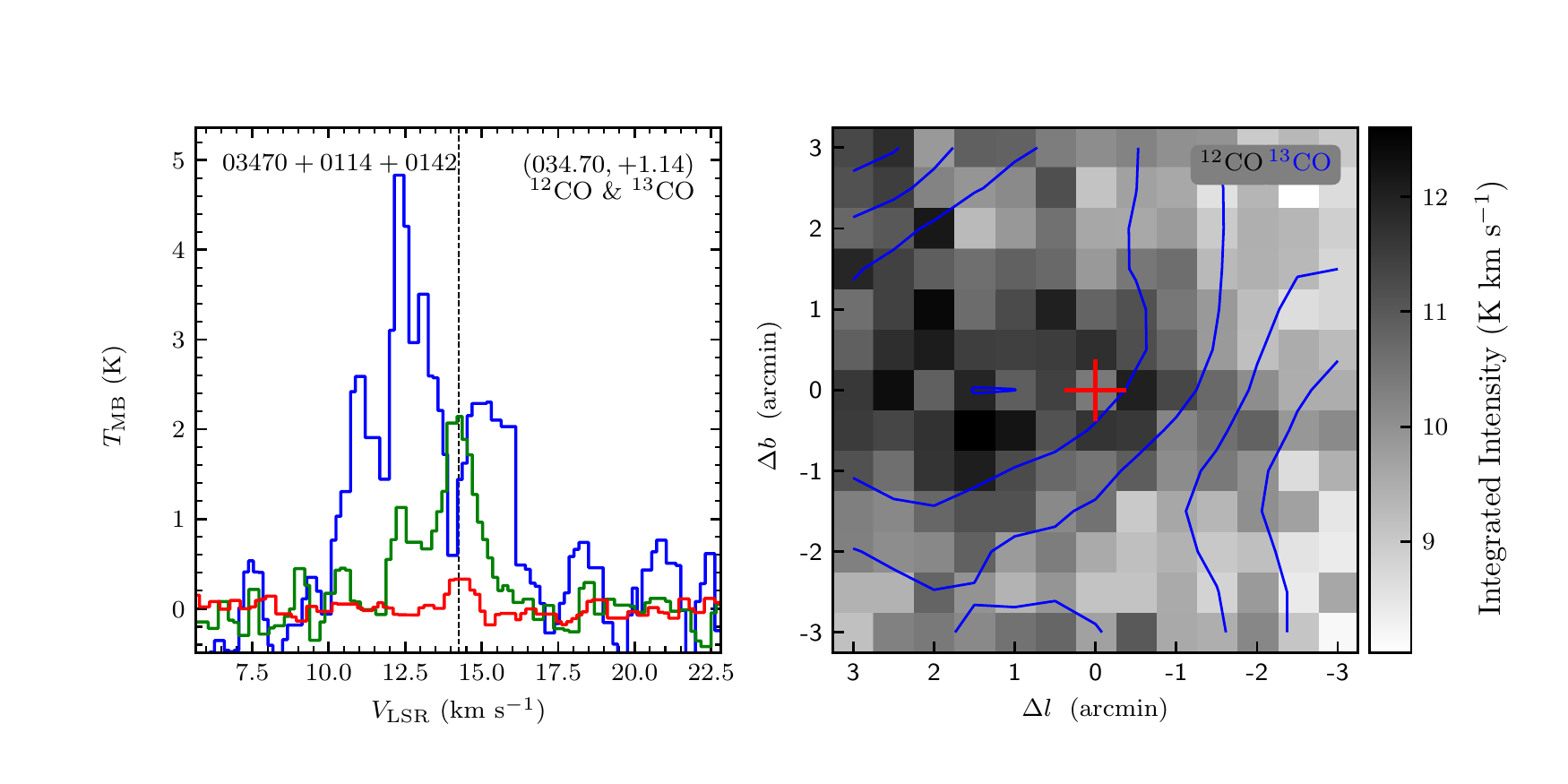}
\vspace{-0.5cm}

\includegraphics[width=9.0cm,angle=0]{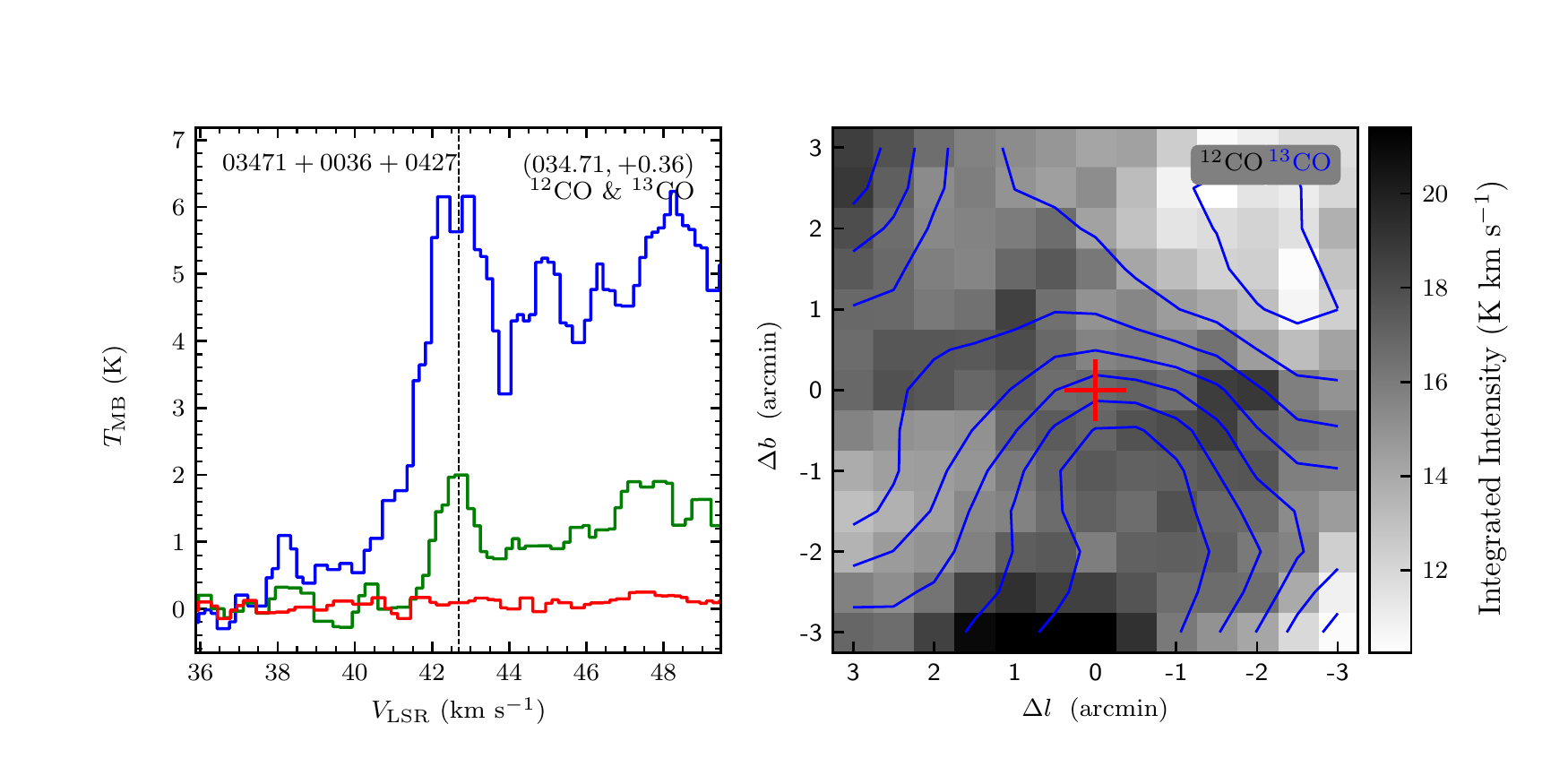}
\includegraphics[width=9.0cm,angle=0]{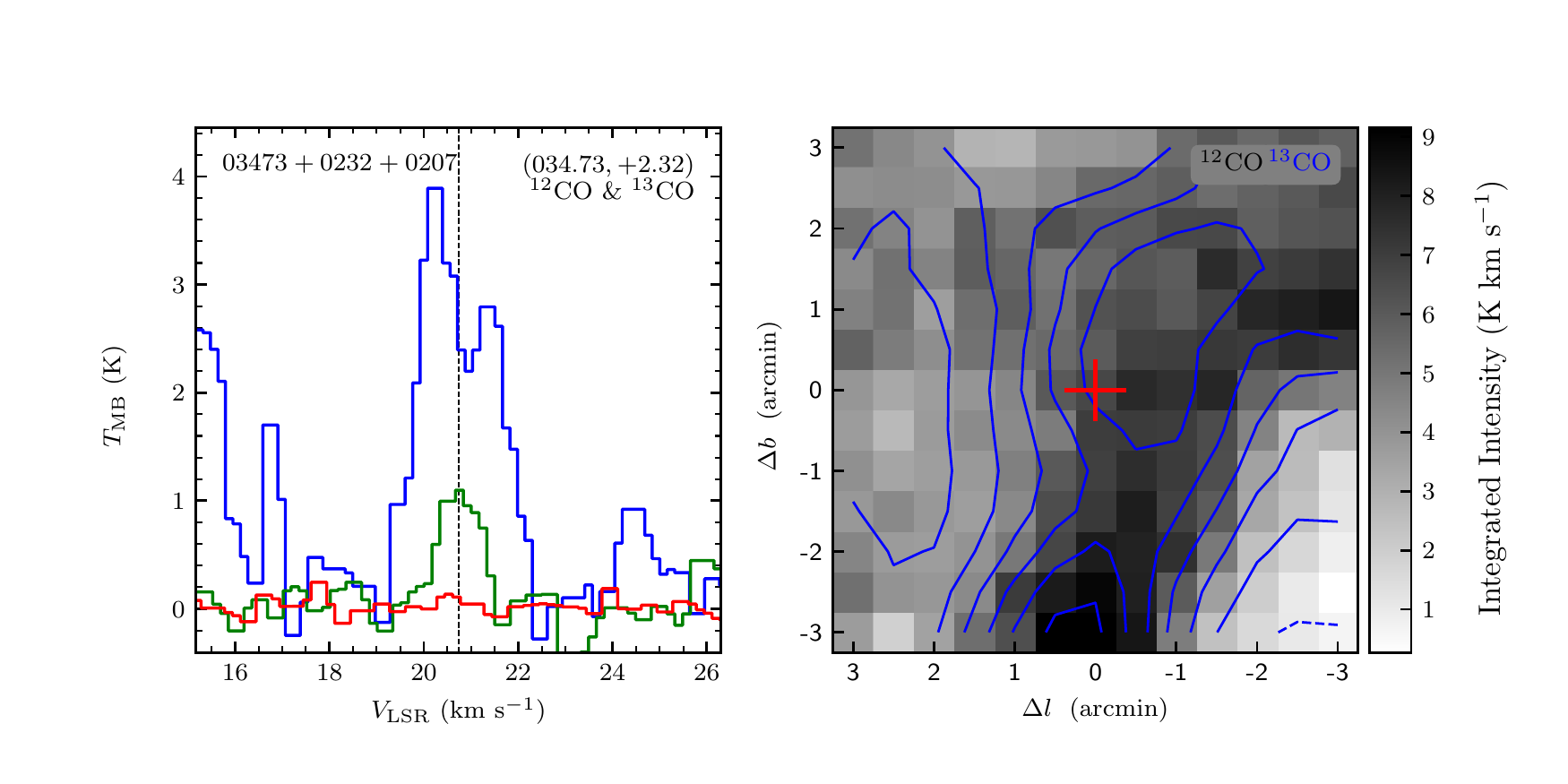}
\end{figure}
\clearpage

\begin{figure}
\includegraphics[width=9.0cm,angle=0]{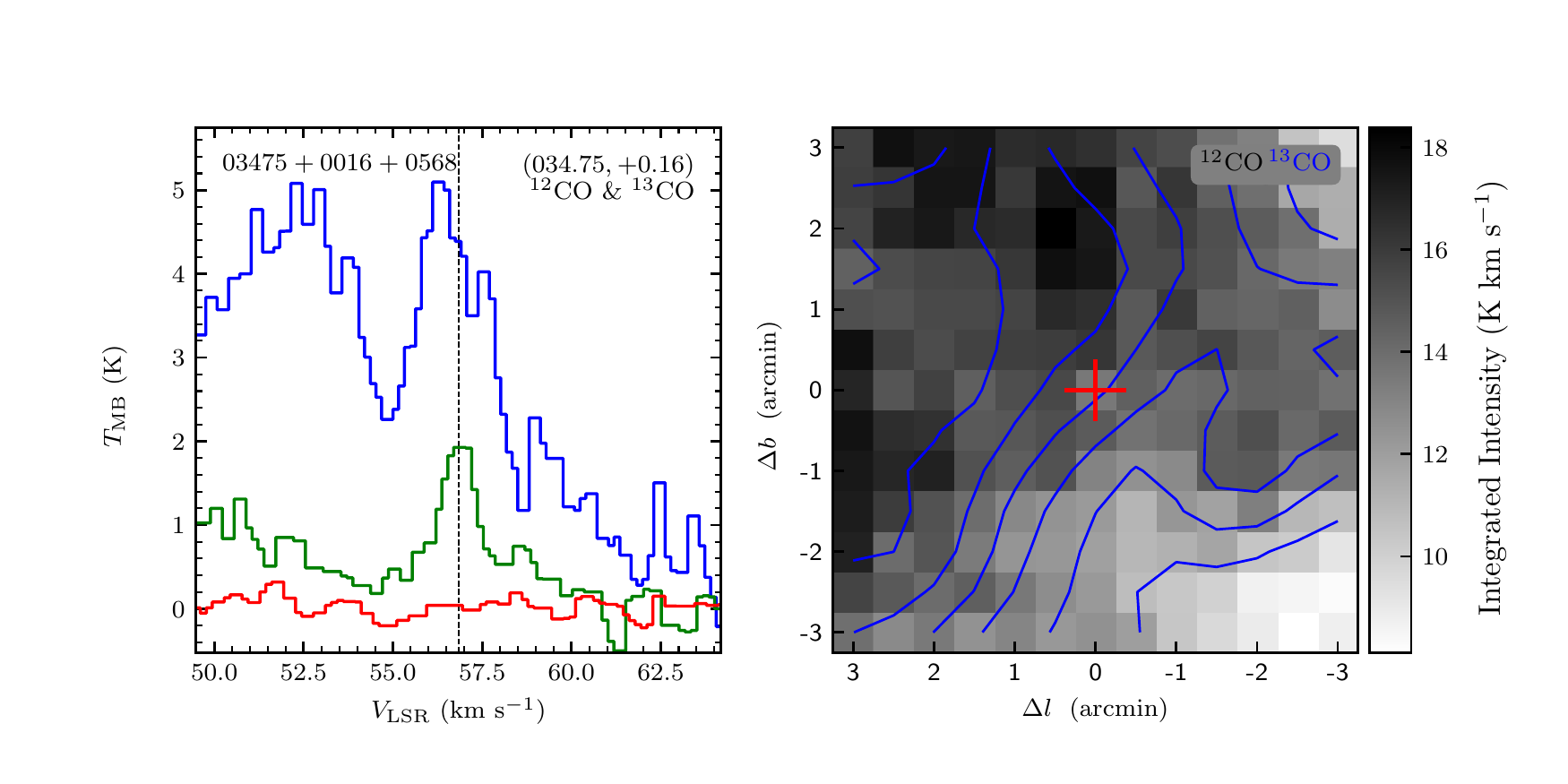}
\includegraphics[width=9.0cm,angle=0]{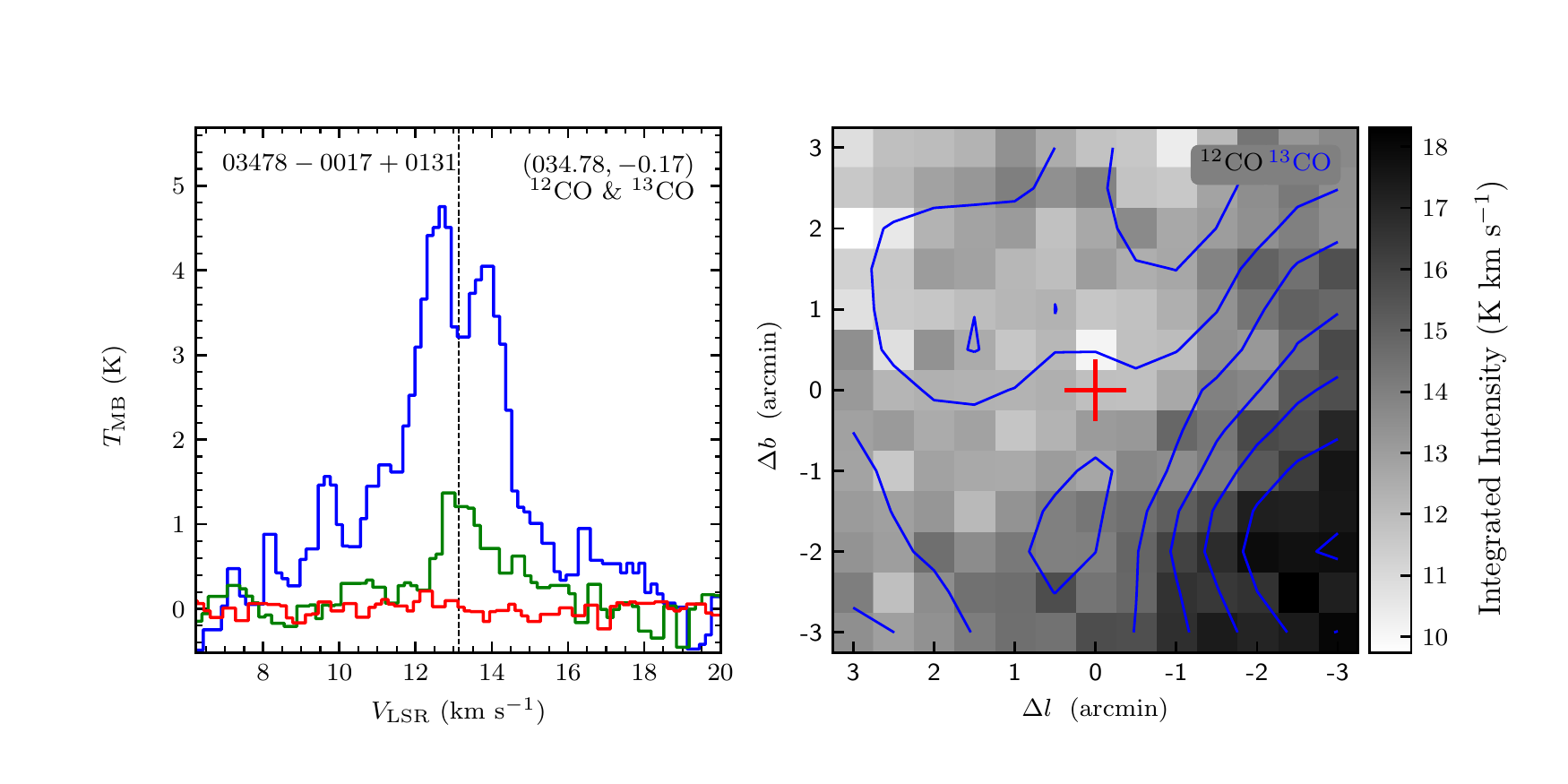}
\vspace{-0.5cm}

\includegraphics[width=9.0cm,angle=0]{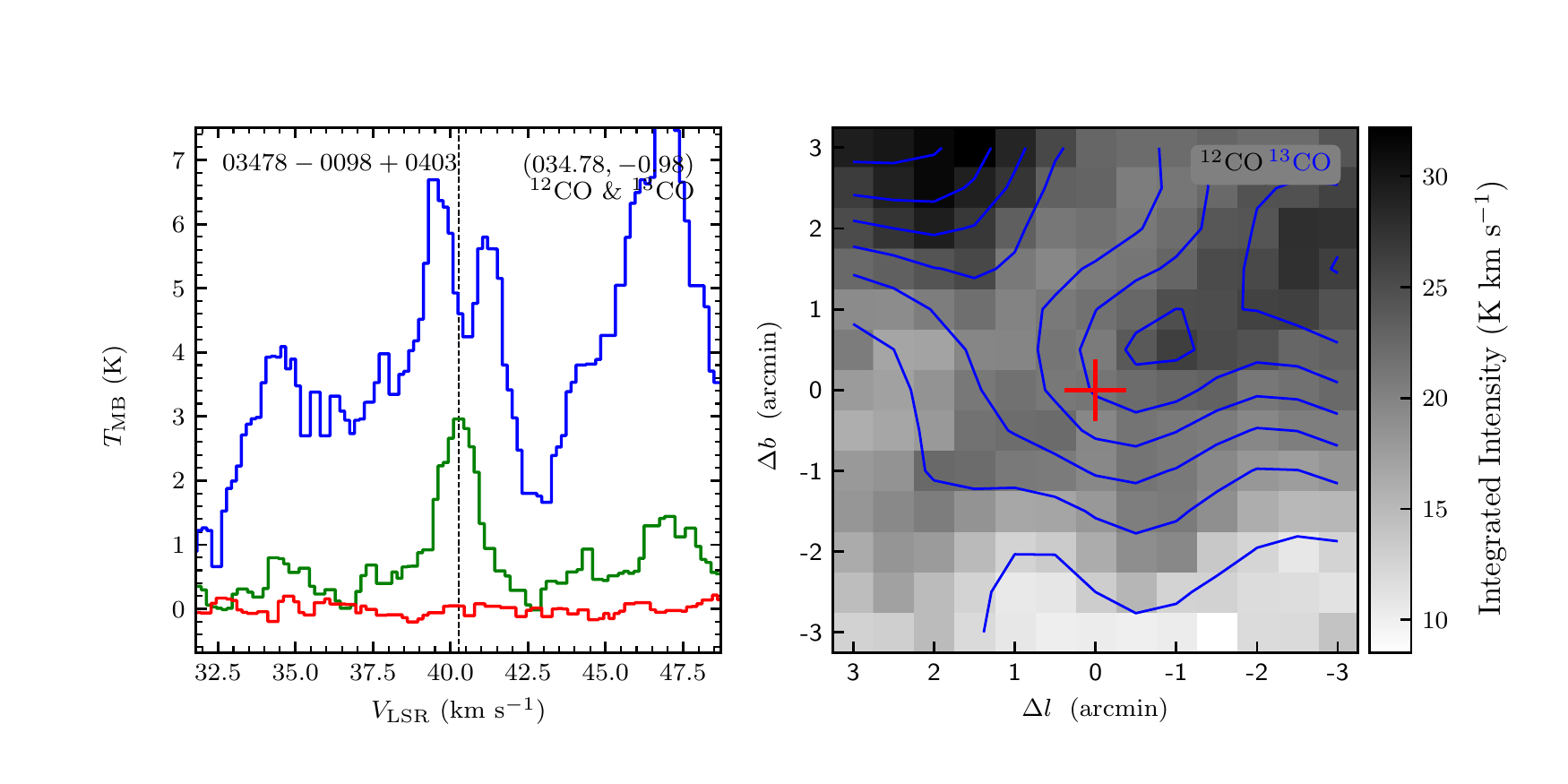}
\includegraphics[width=9.0cm,angle=0]{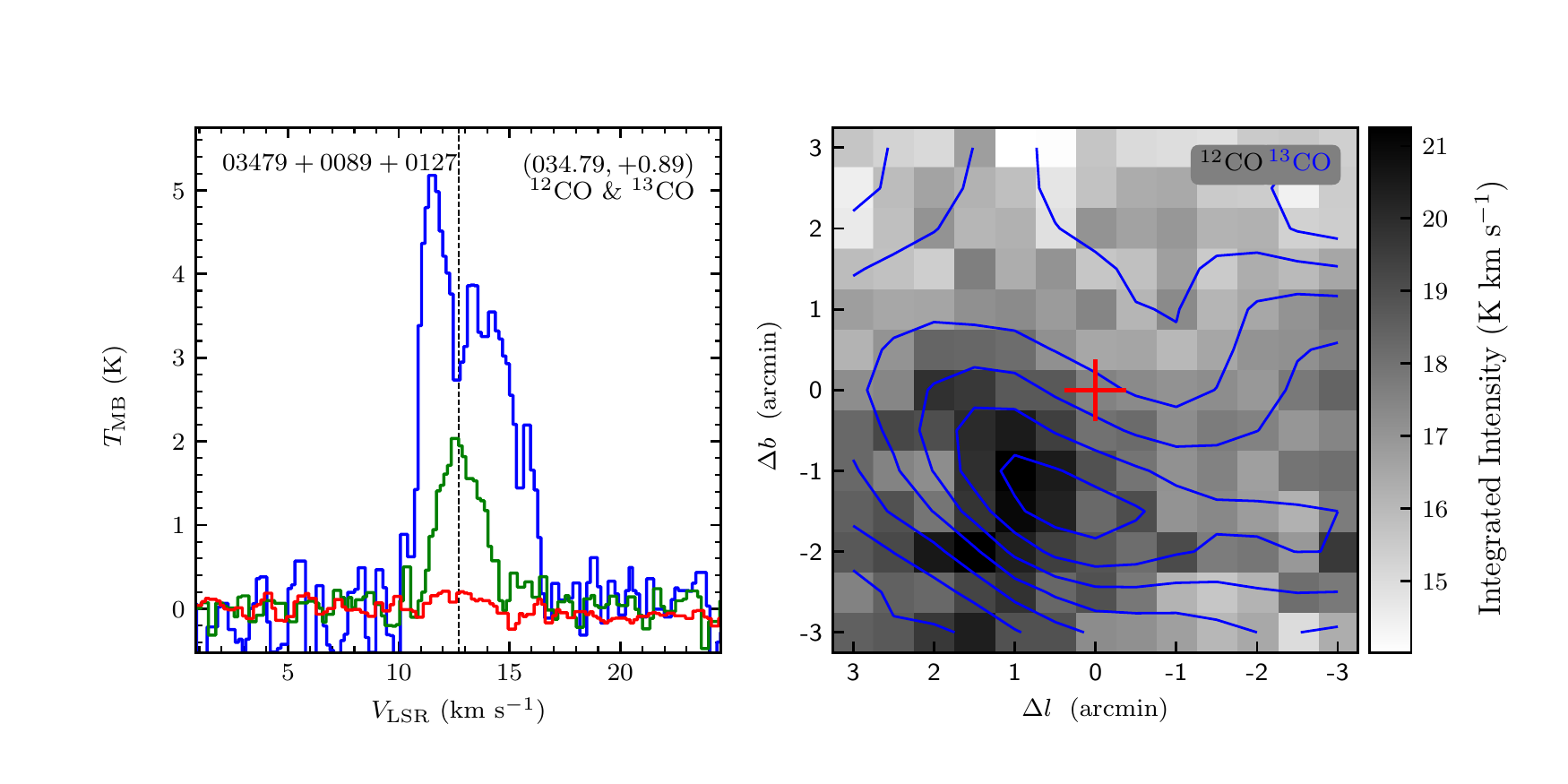}
\vspace{-0.5cm}

\includegraphics[width=9.0cm,angle=0]{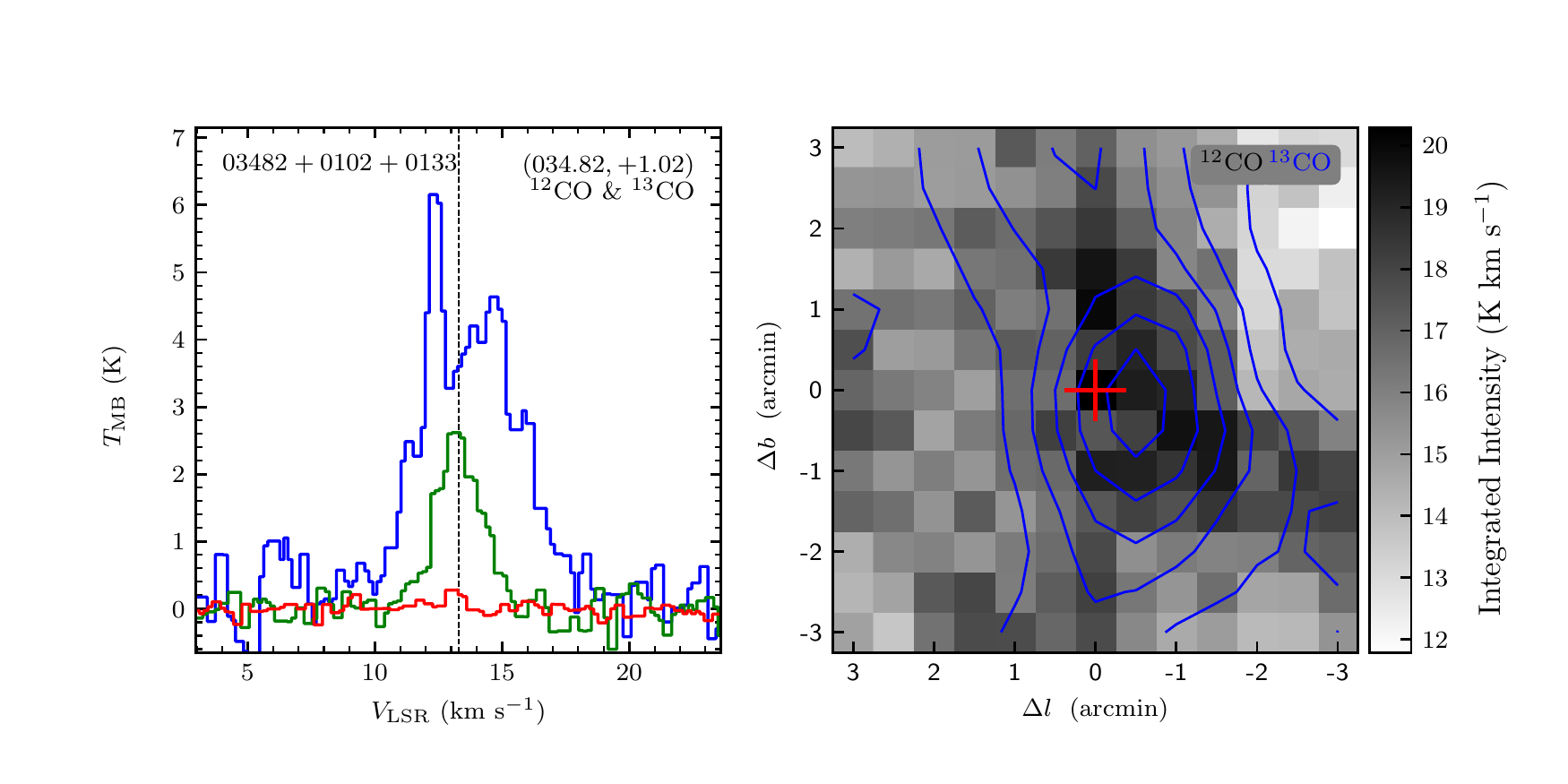}
\includegraphics[width=9.0cm,angle=0]{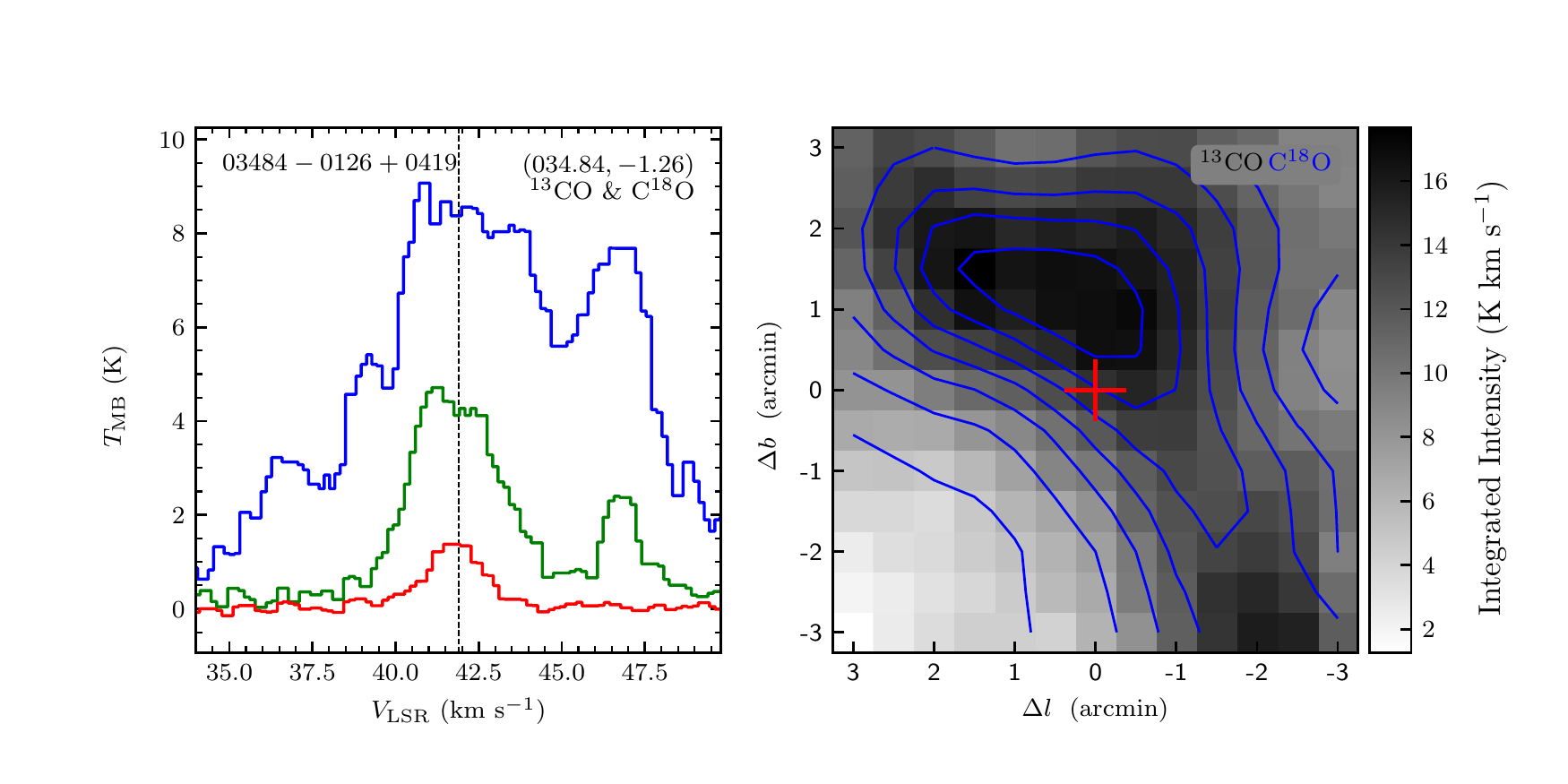}
\vspace{-0.5cm}

\includegraphics[width=9.0cm,angle=0]{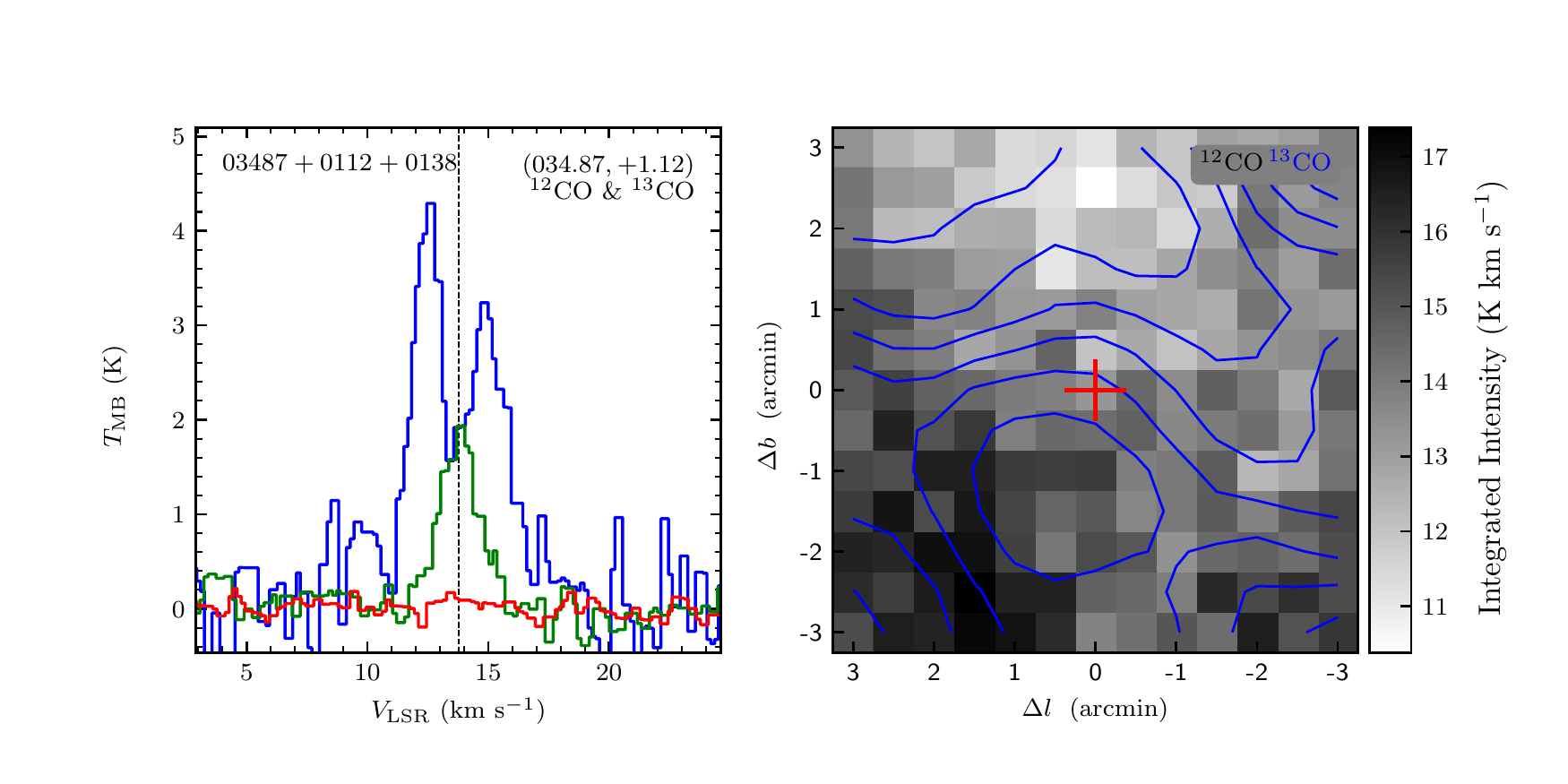}
\includegraphics[width=9.0cm,angle=0]{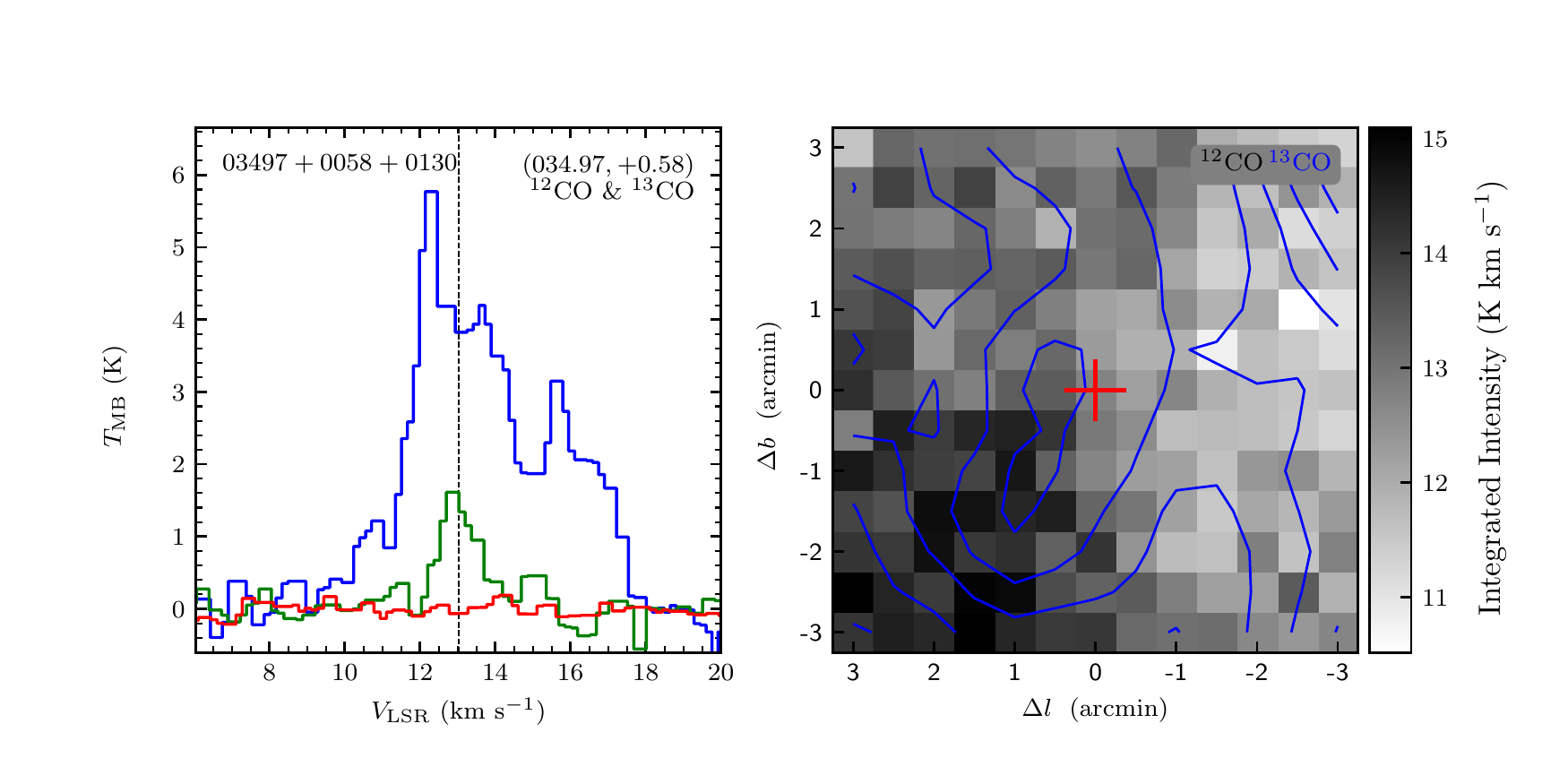}
\vspace{-0.5cm}

\includegraphics[width=9.0cm,angle=0]{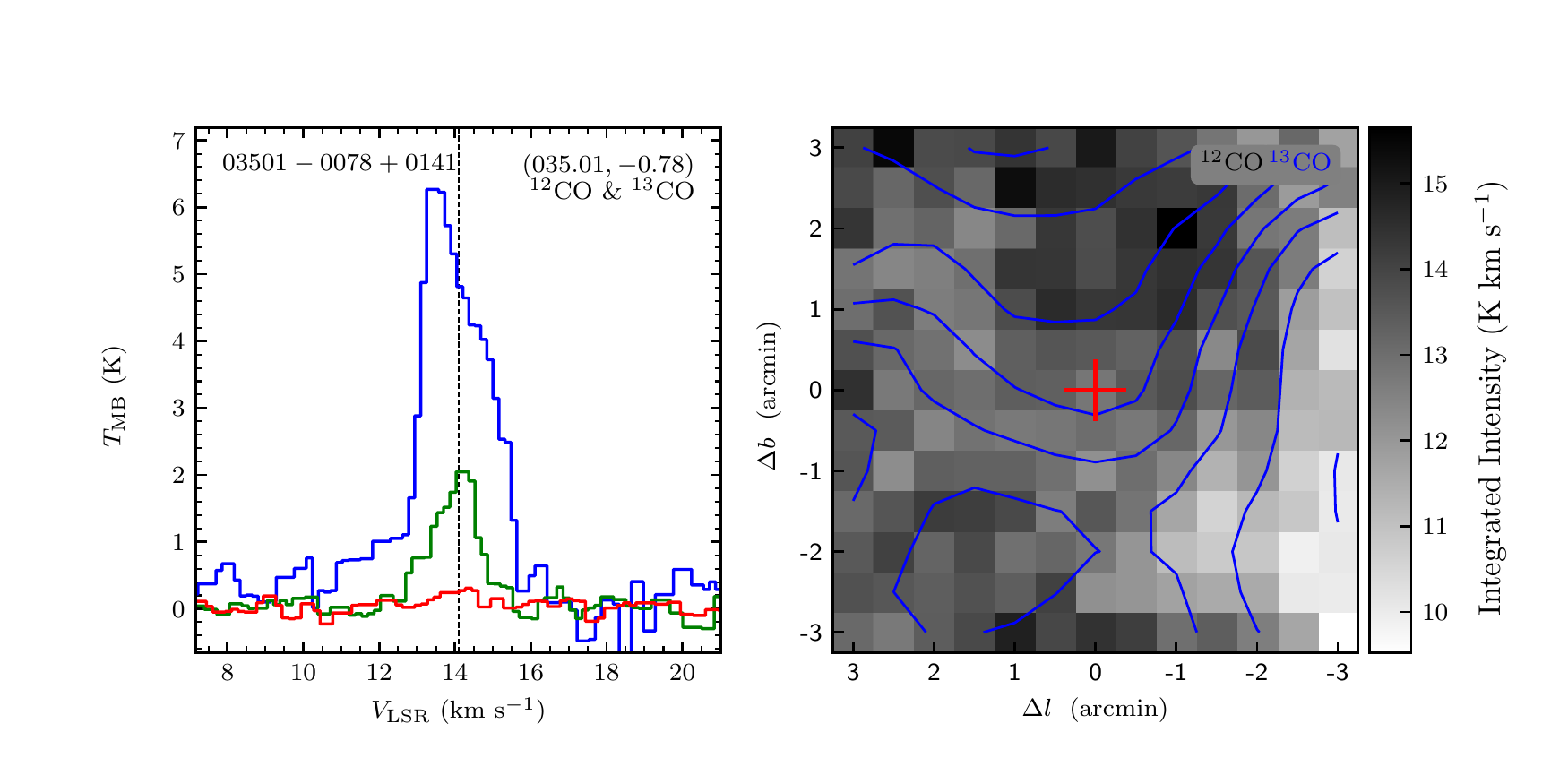}
\includegraphics[width=9.0cm,angle=0]{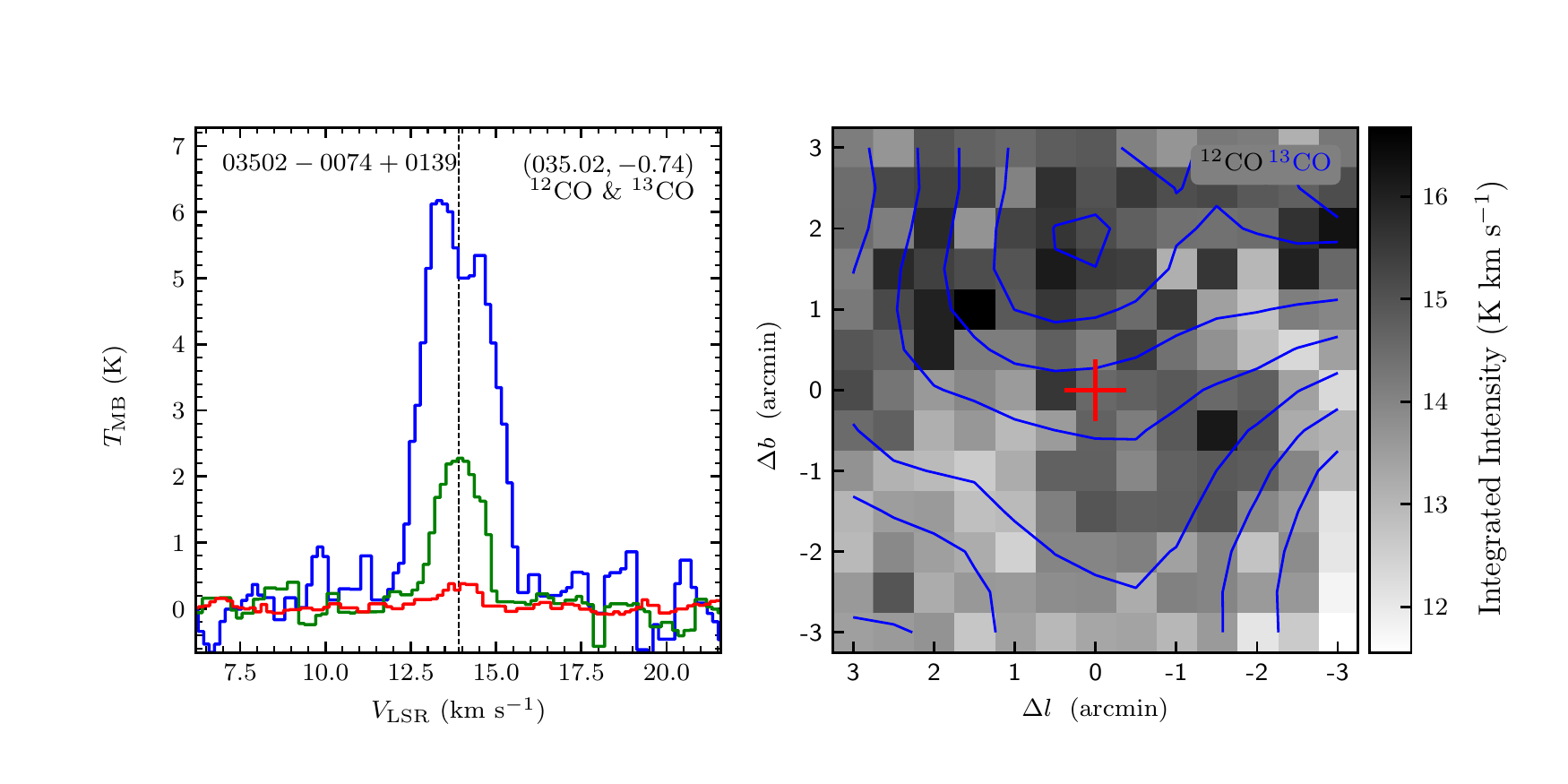}
\end{figure}
\clearpage

\begin{figure}
\includegraphics[width=9.0cm,angle=0]{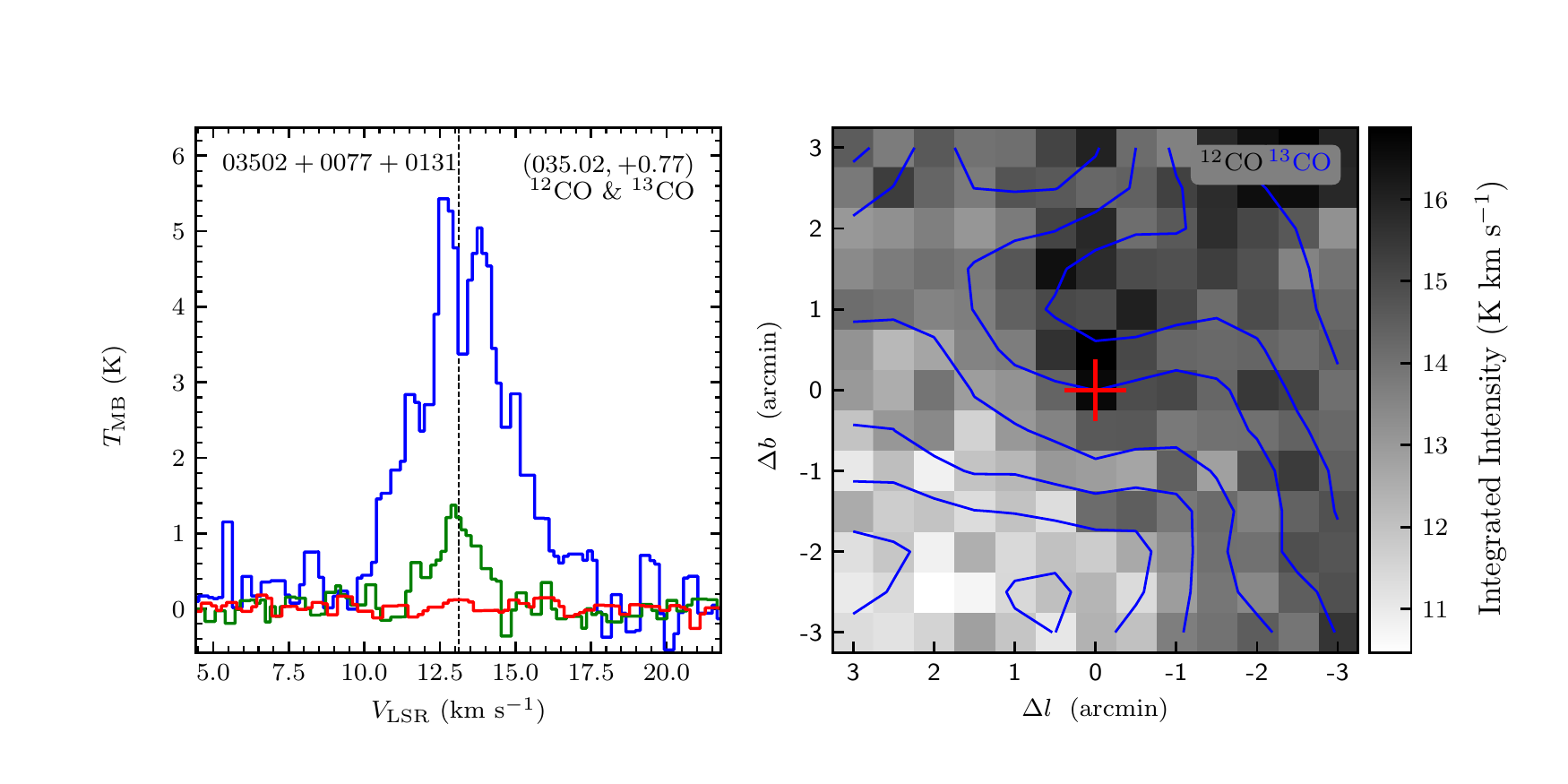}
\includegraphics[width=9.0cm,angle=0]{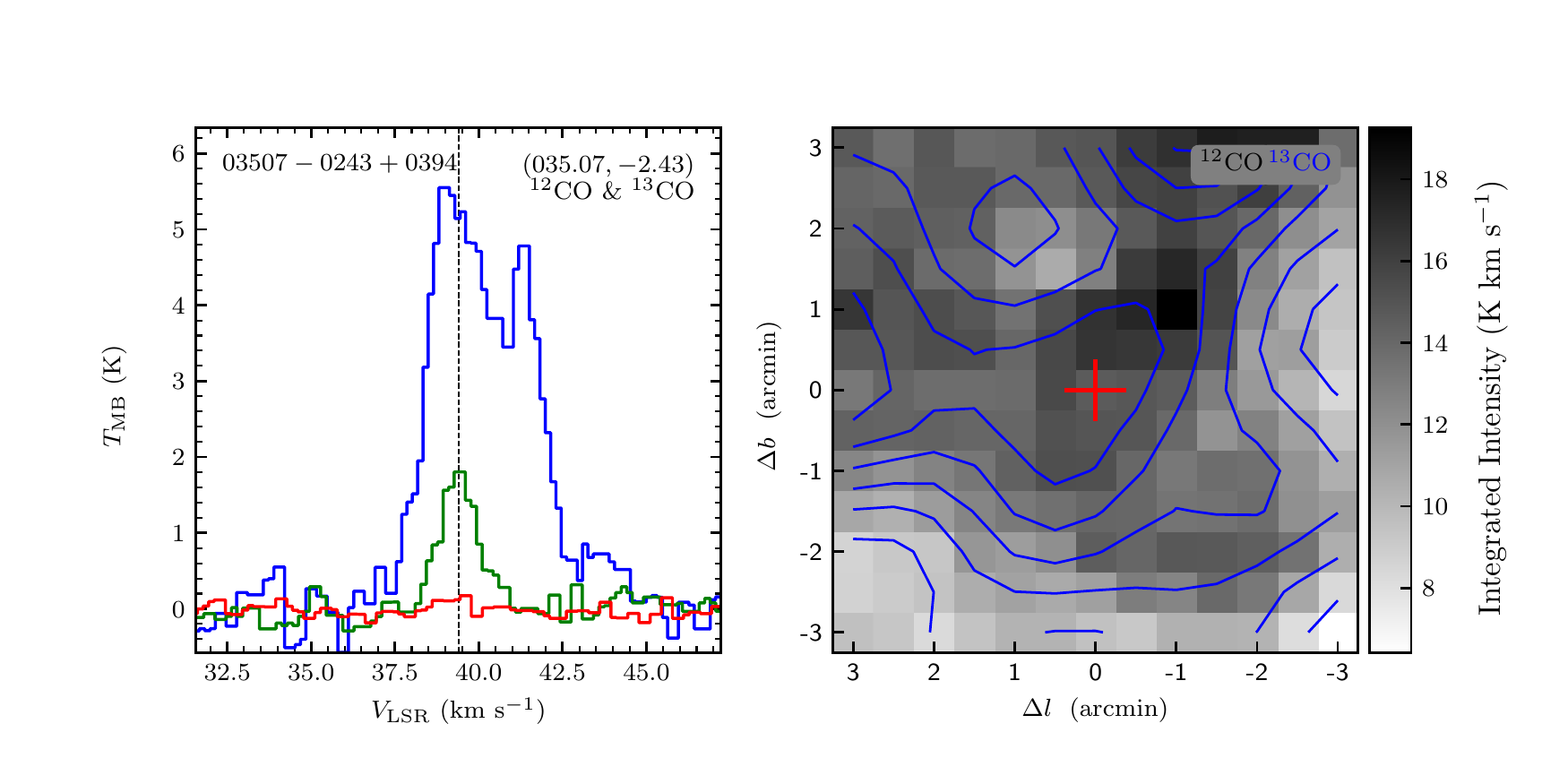}
\vspace{-0.5cm}

\includegraphics[width=9.0cm,angle=0]{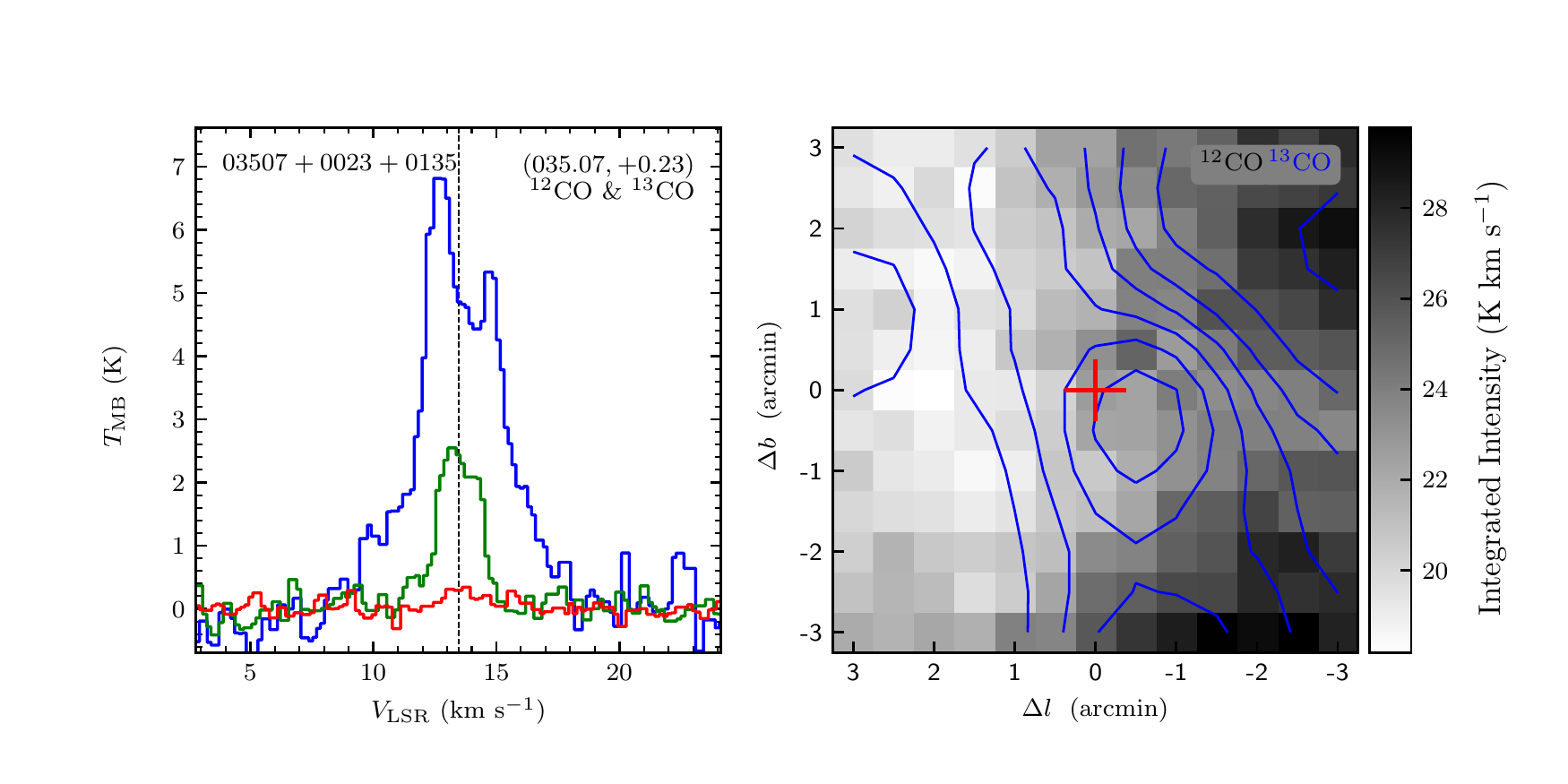}
\includegraphics[width=9.0cm,angle=0]{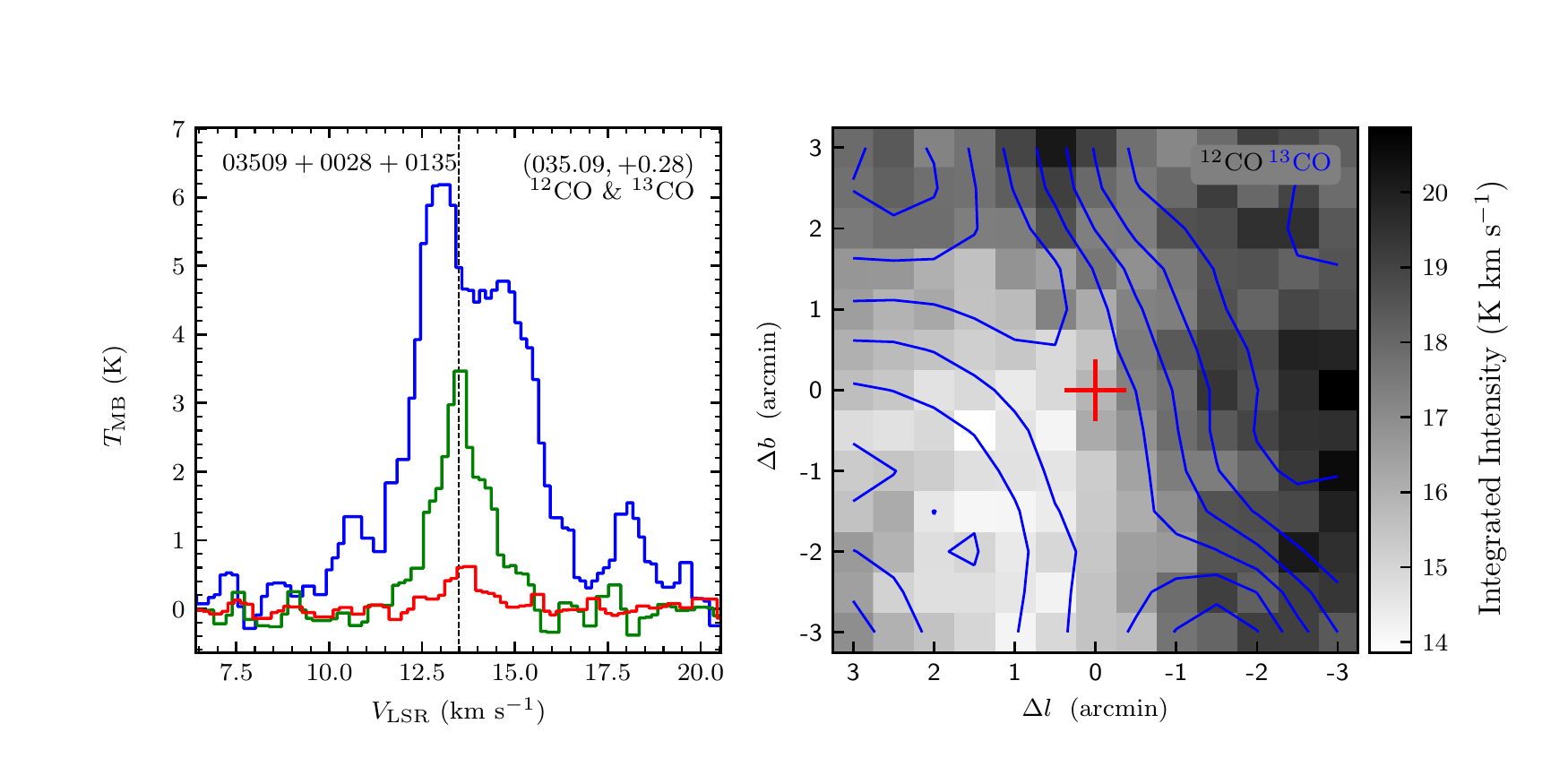}
\vspace{-0.5cm}

\includegraphics[width=9.0cm,angle=0]{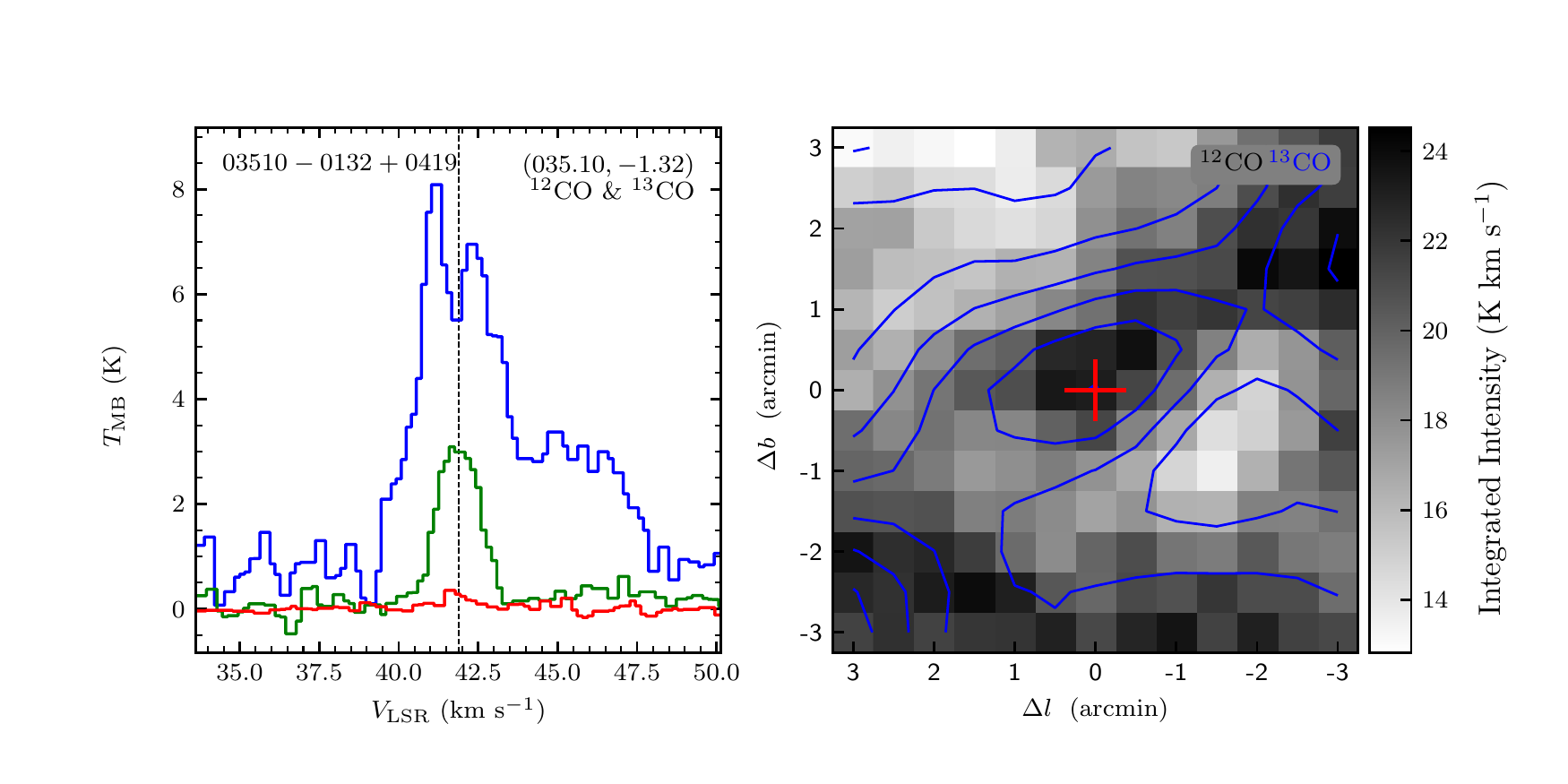}
\includegraphics[width=9.0cm,angle=0]{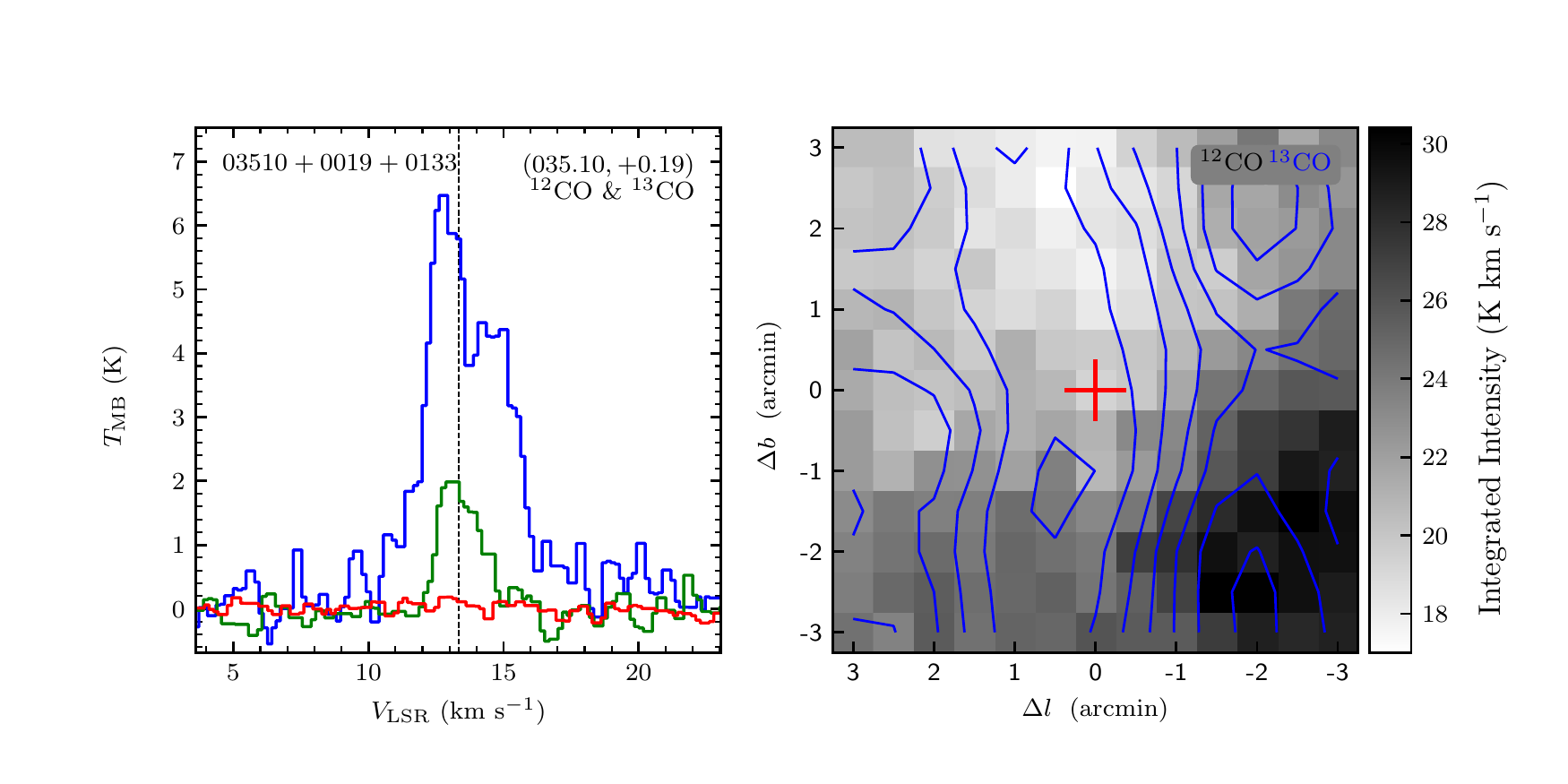}
\vspace{-0.5cm}

\includegraphics[width=9.0cm,angle=0]{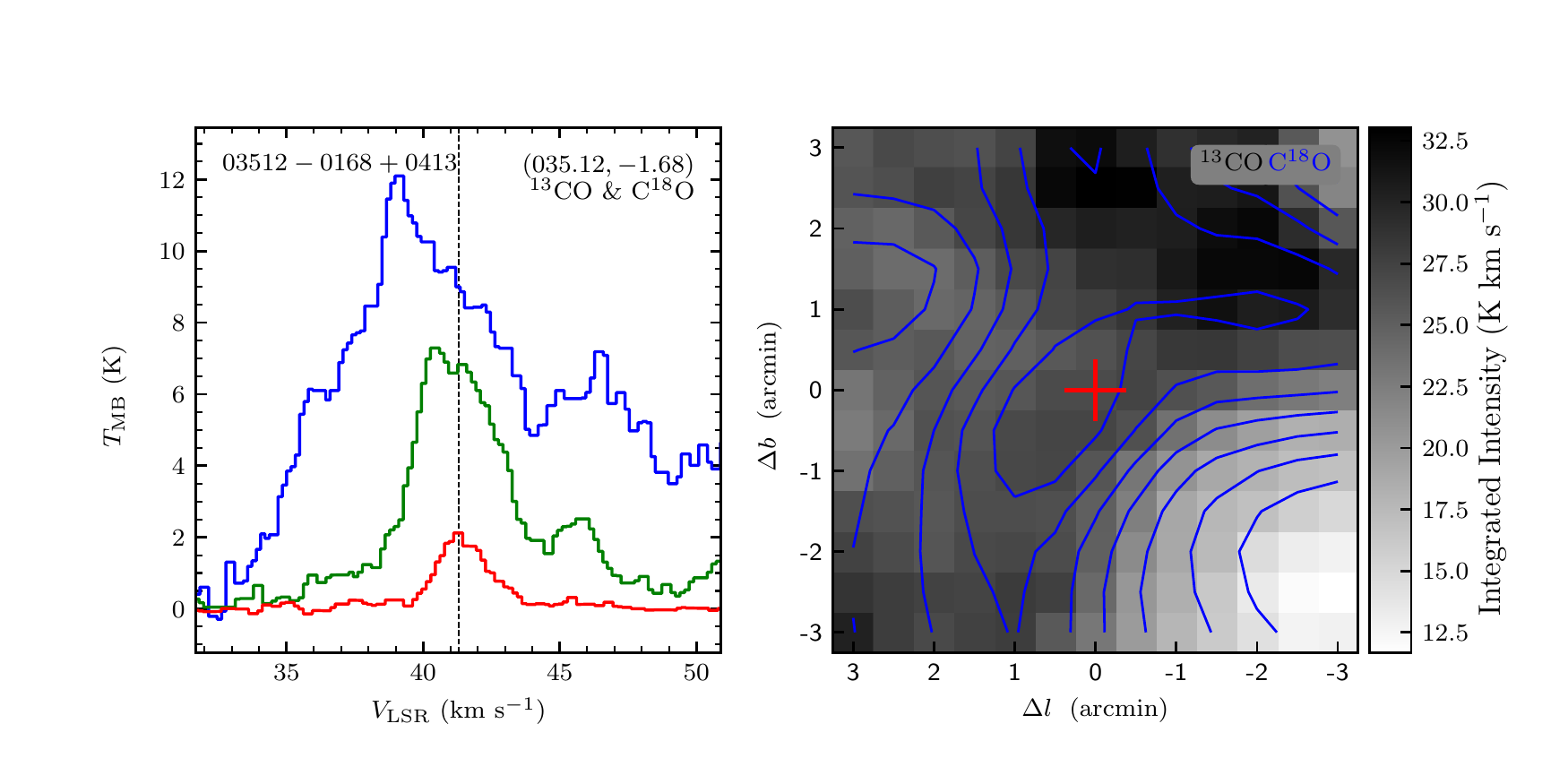}
\includegraphics[width=9.0cm,angle=0]{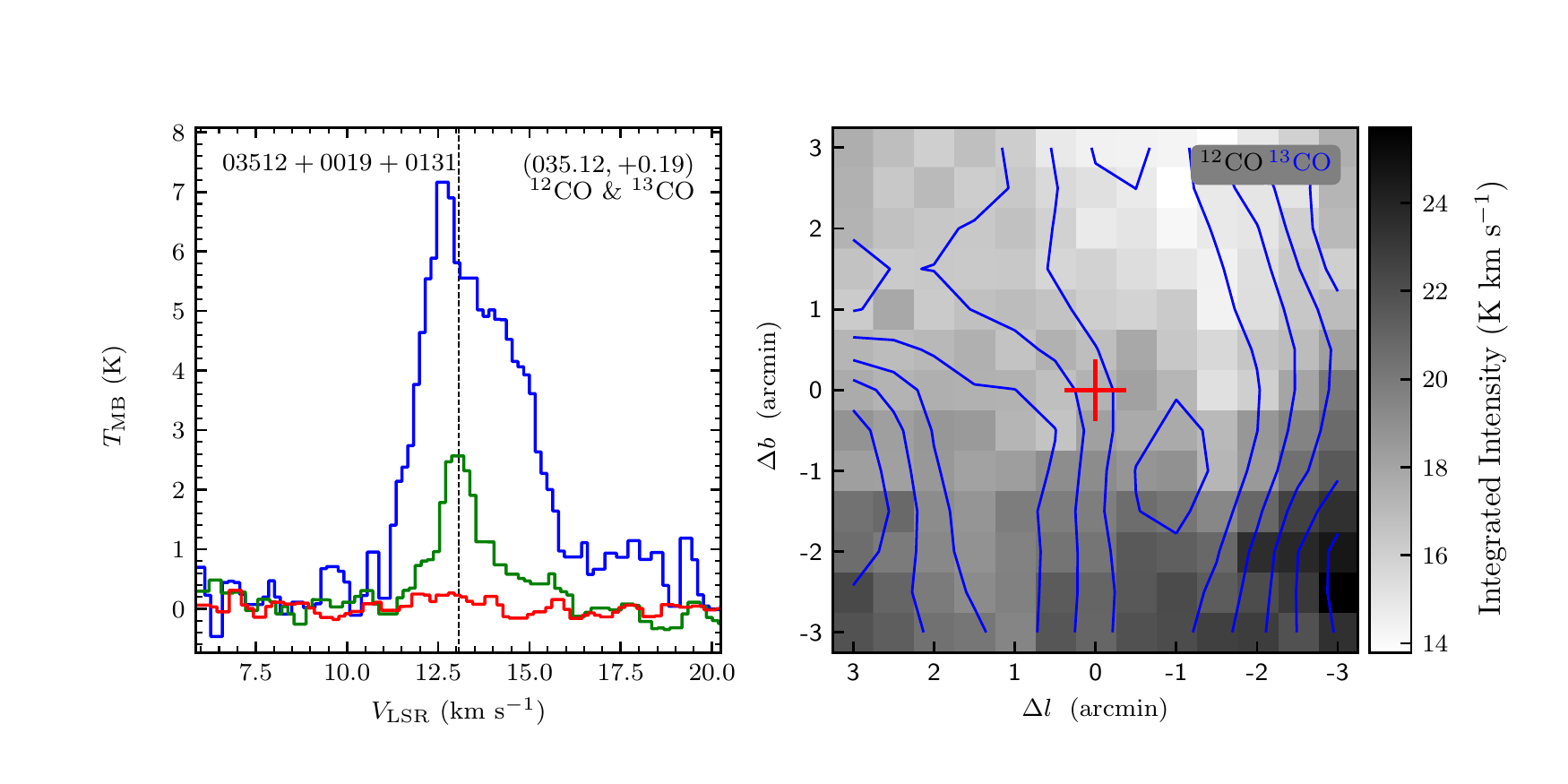}
\vspace{-0.5cm}

\includegraphics[width=9.0cm,angle=0]{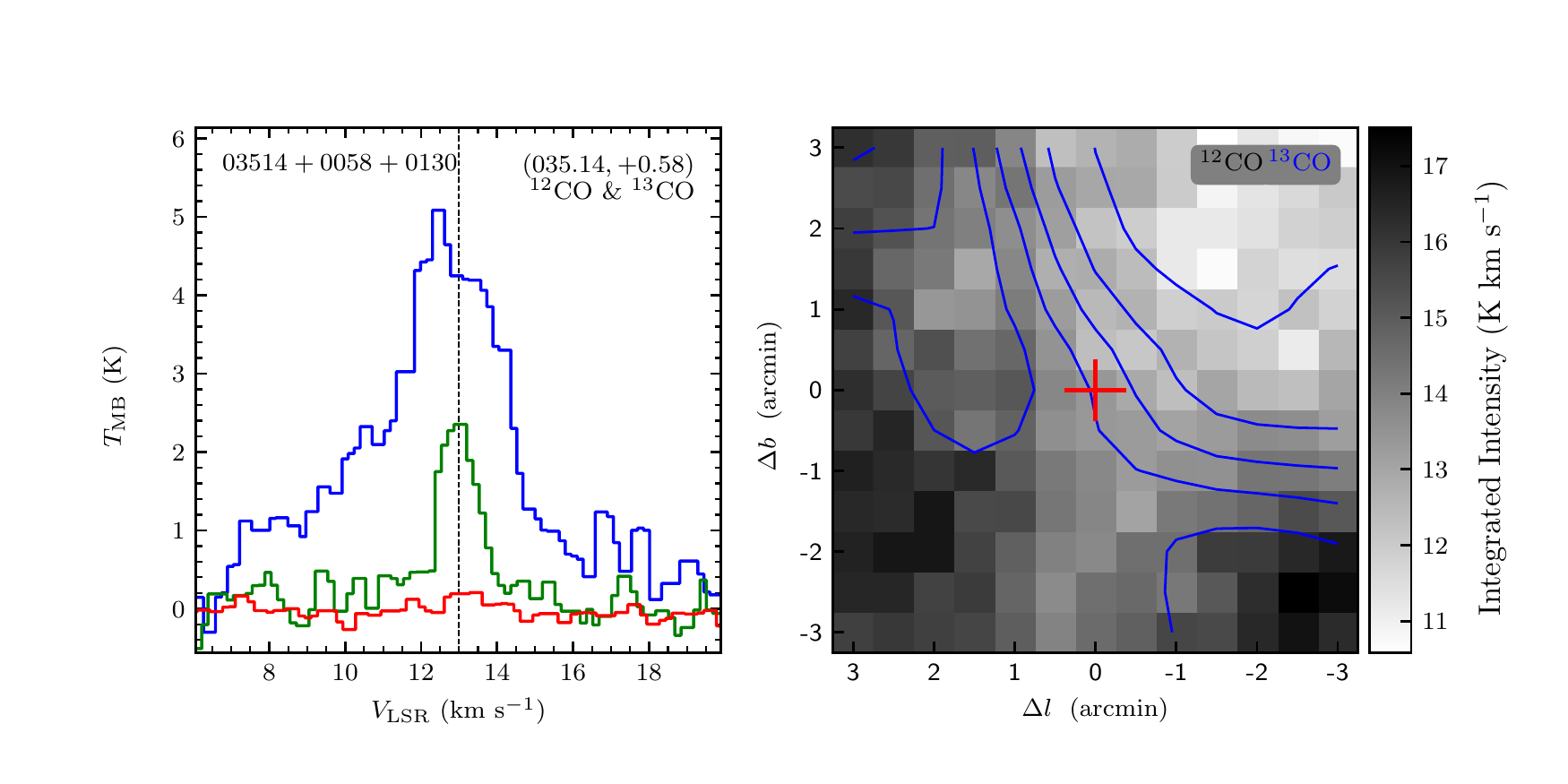}
\includegraphics[width=9.0cm,angle=0]{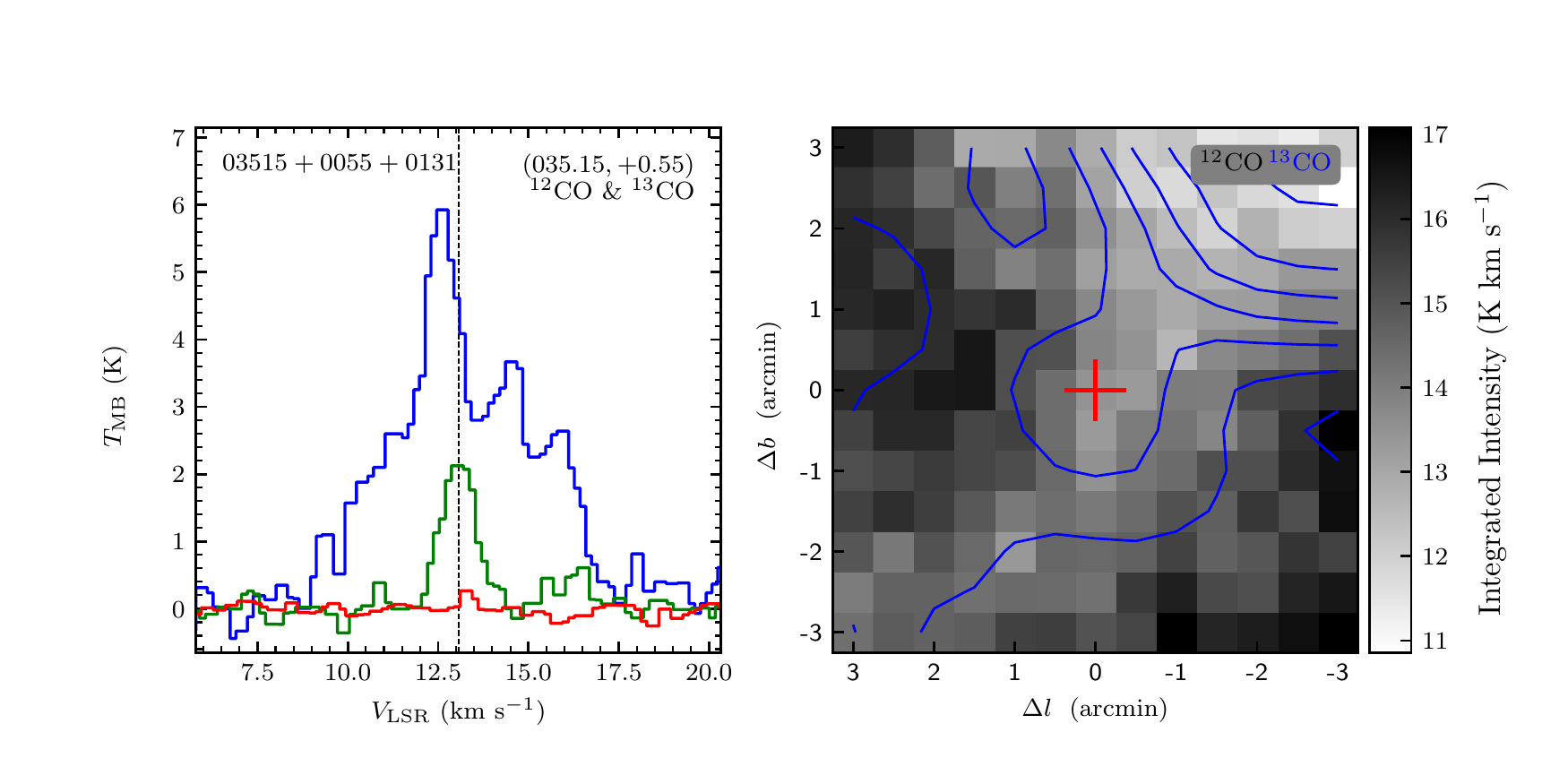}
\end{figure}
\clearpage

\begin{figure}
\includegraphics[width=9.0cm,angle=0]{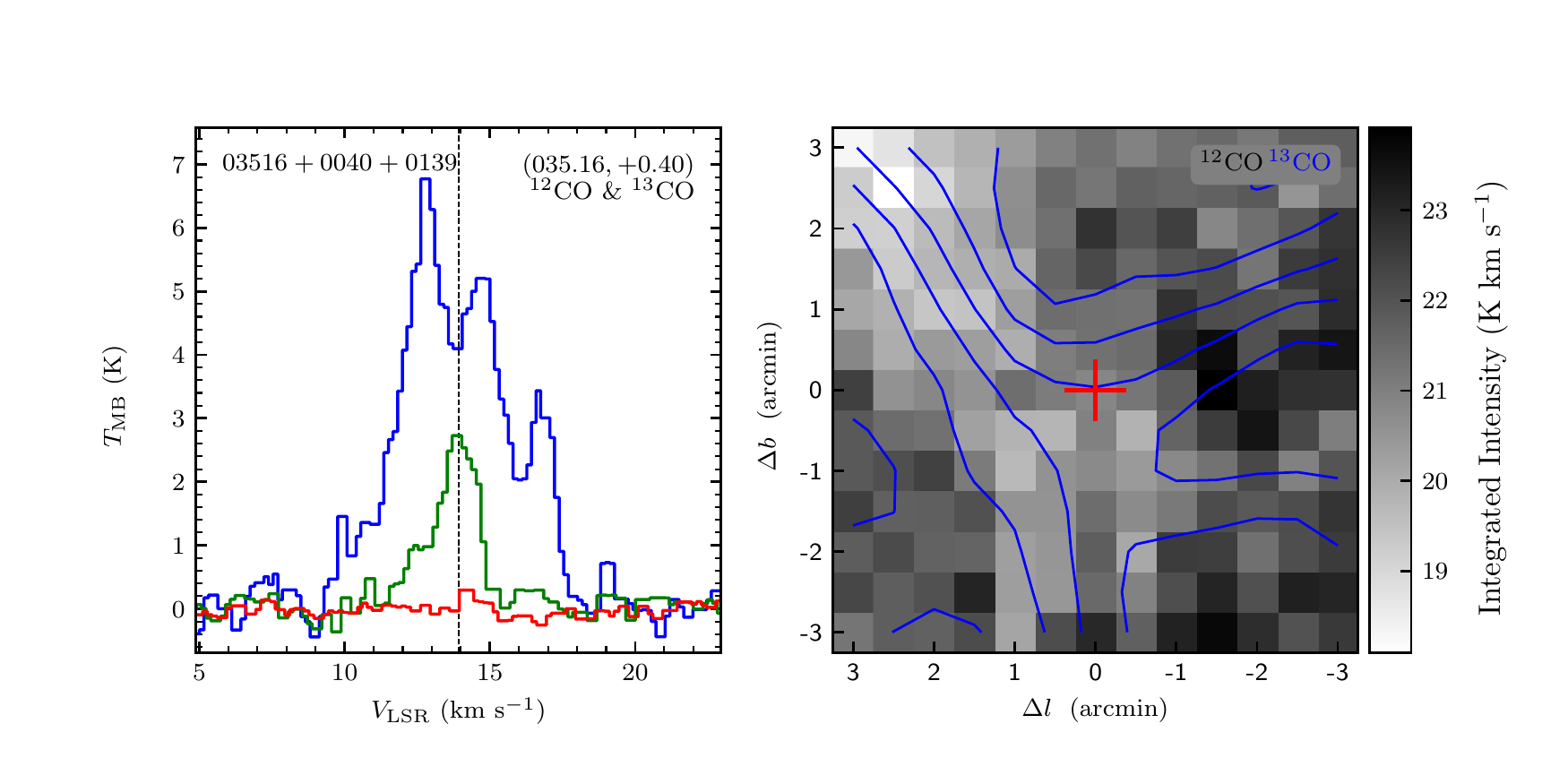}
\includegraphics[width=9.0cm,angle=0]{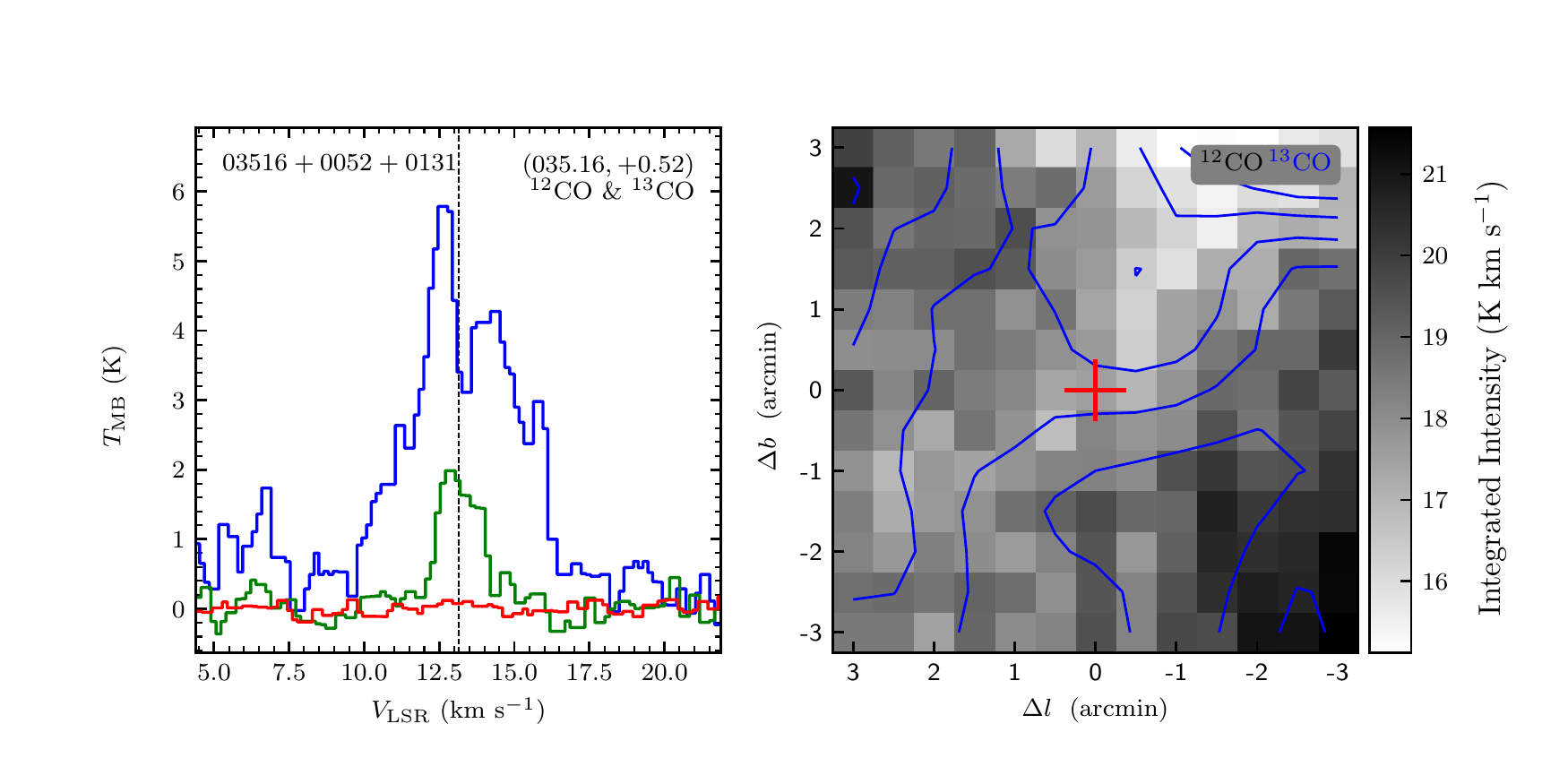}
\vspace{-0.5cm}

\includegraphics[width=9.0cm,angle=0]{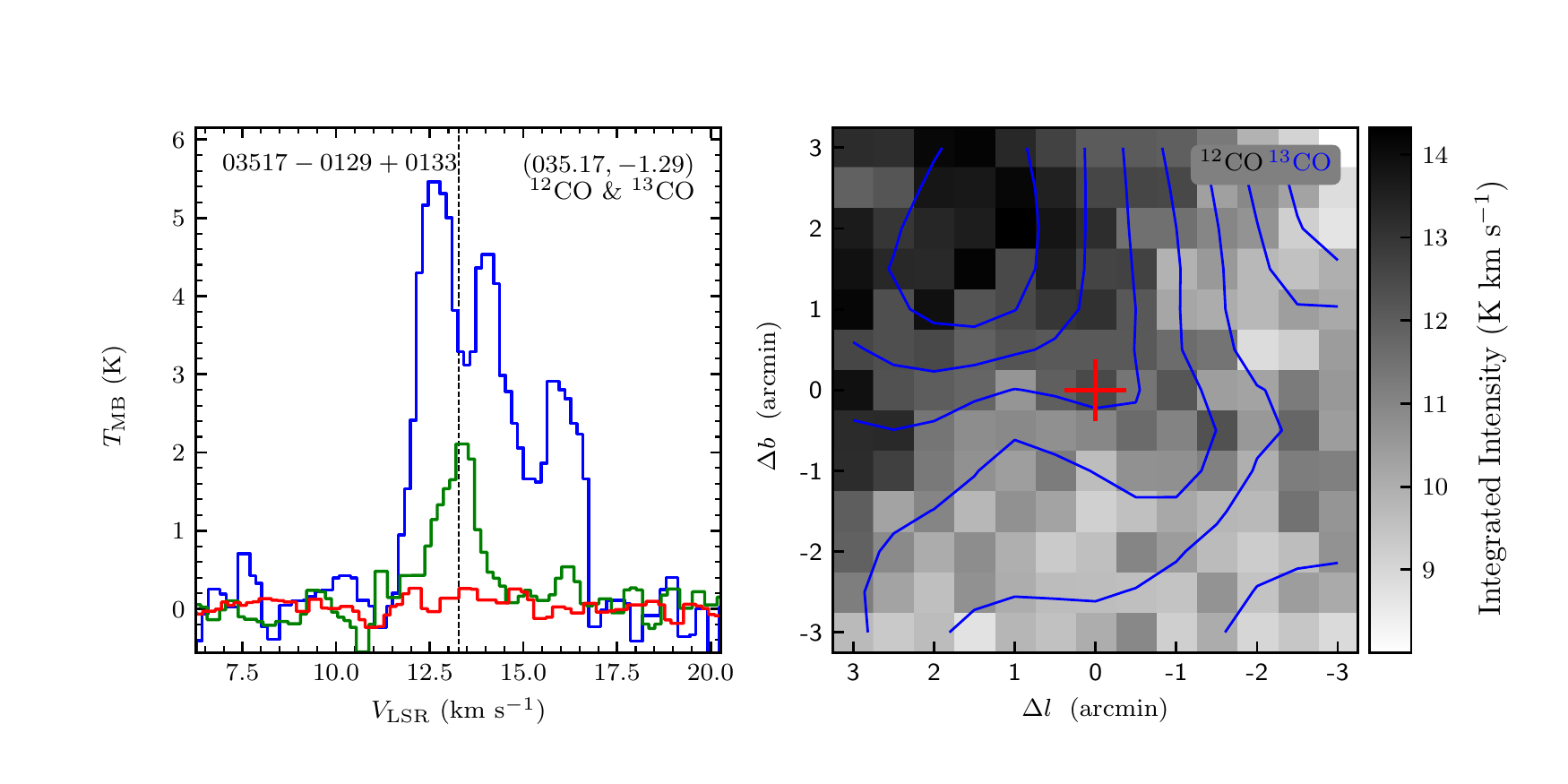}
\includegraphics[width=9.0cm,angle=0]{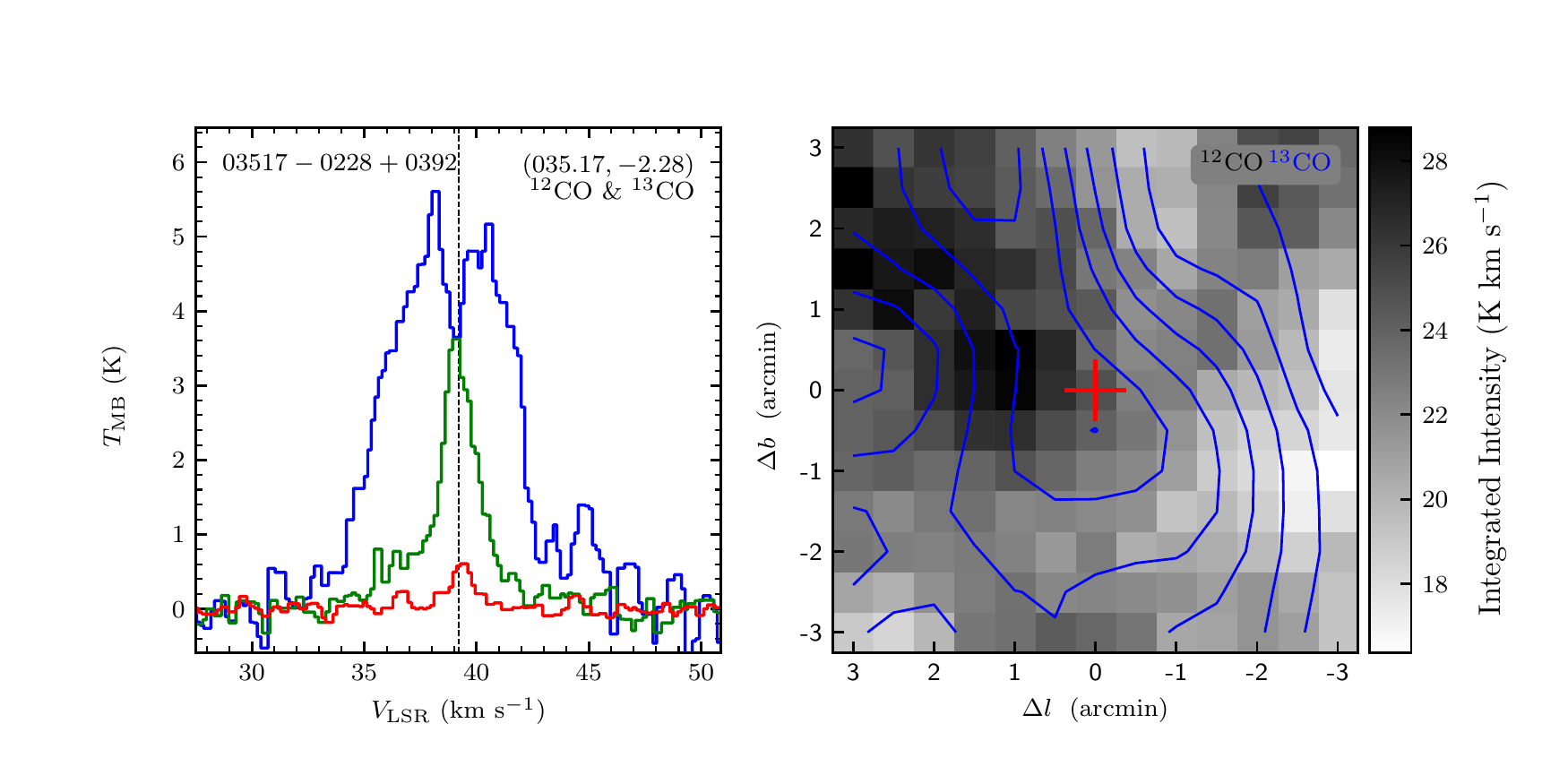}
\vspace{-0.5cm}

\includegraphics[width=9.0cm,angle=0]{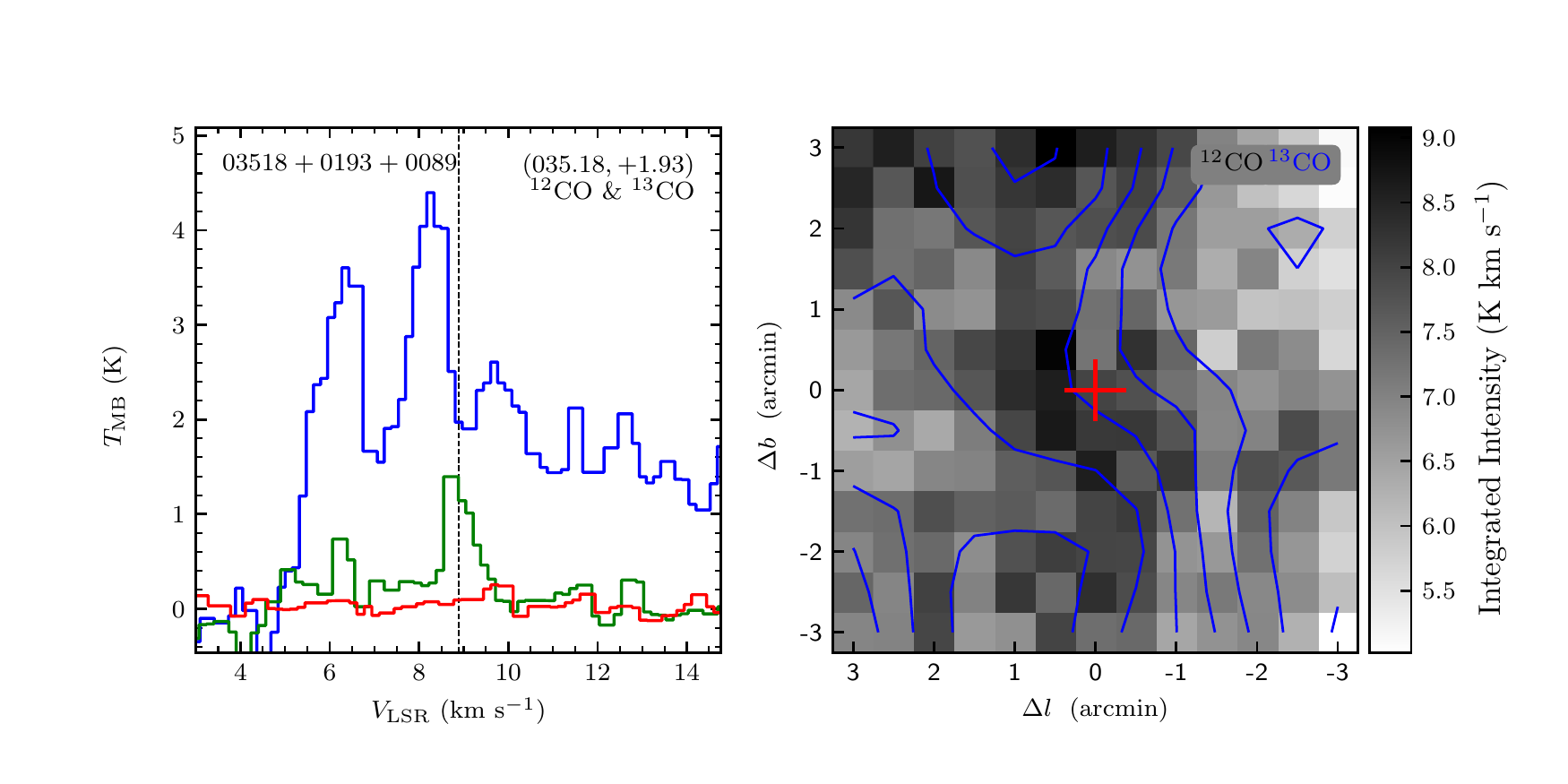}
\includegraphics[width=9.0cm,angle=0]{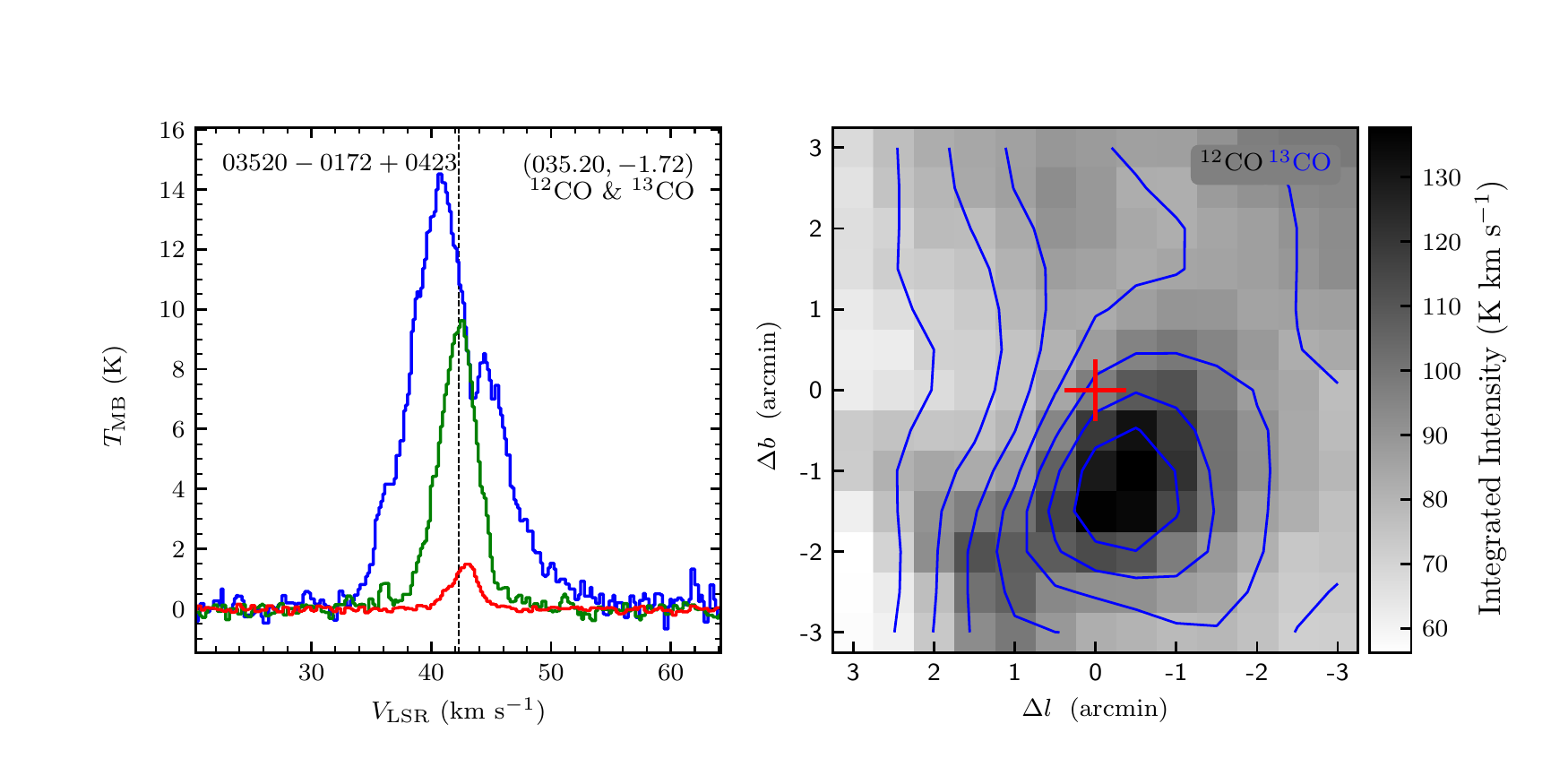}
\vspace{-0.5cm}

\includegraphics[width=9.0cm,angle=0]{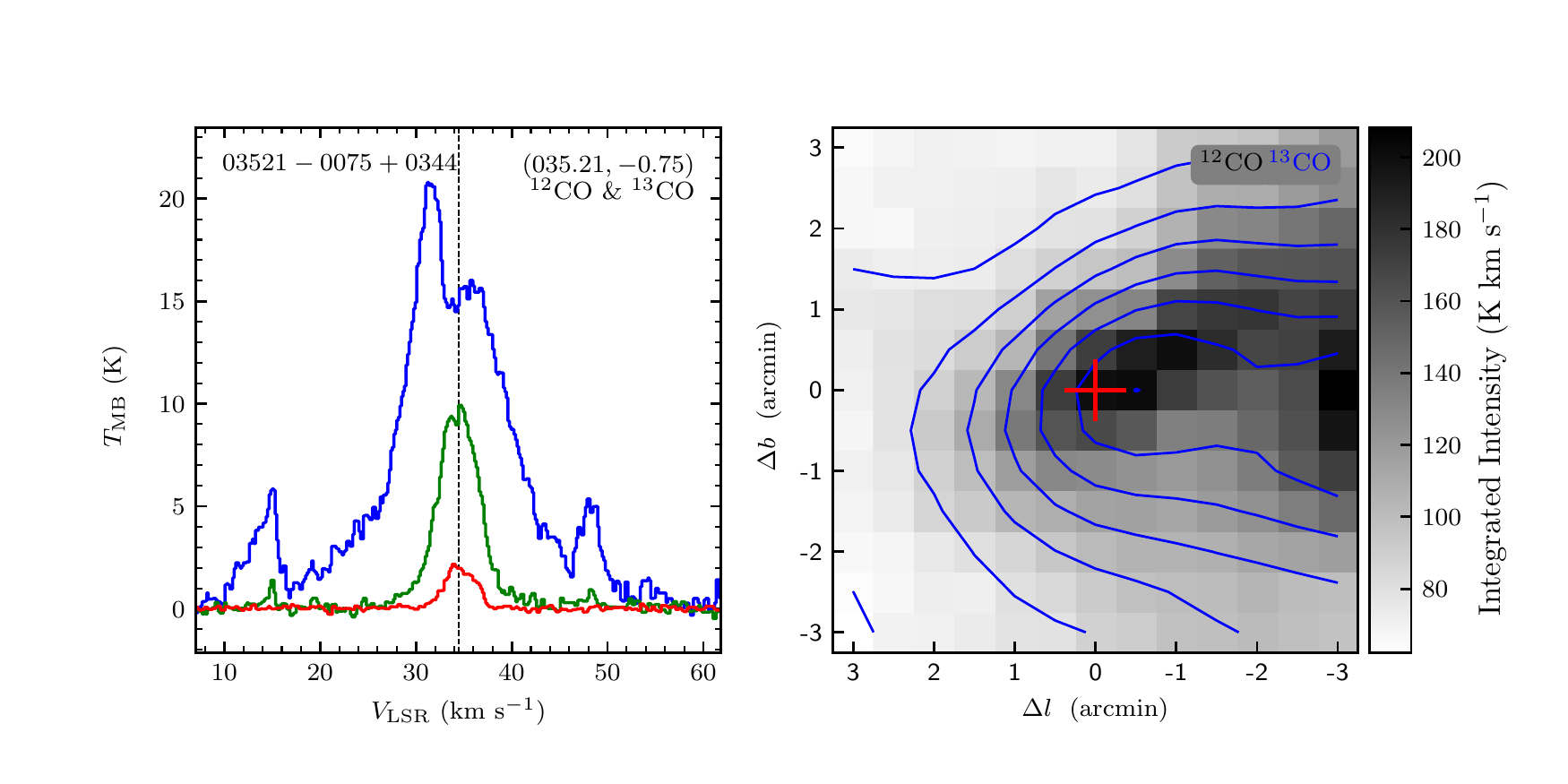}
\includegraphics[width=9.0cm,angle=0]{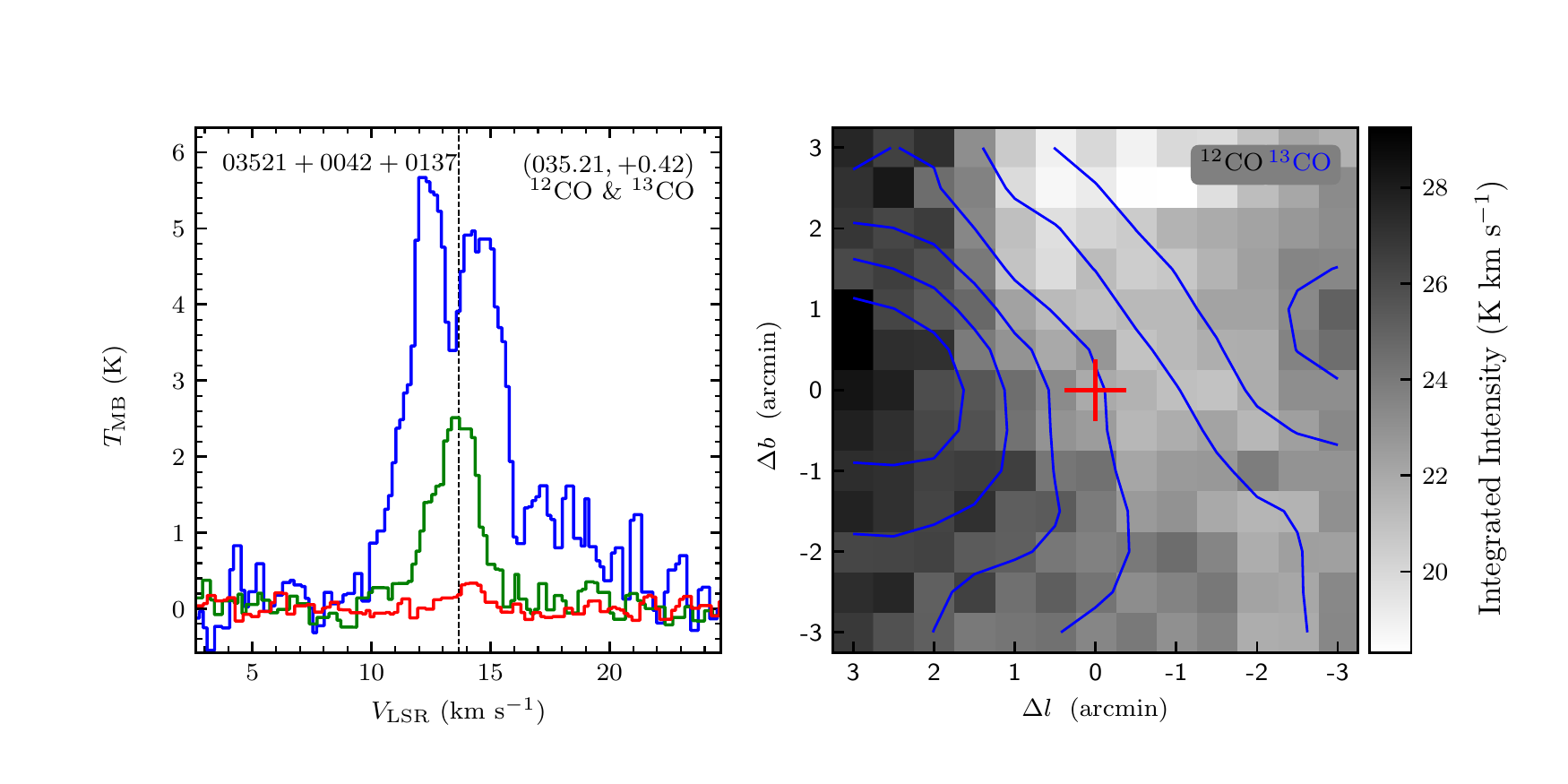}
\vspace{-0.5cm}

\includegraphics[width=9.0cm,angle=0]{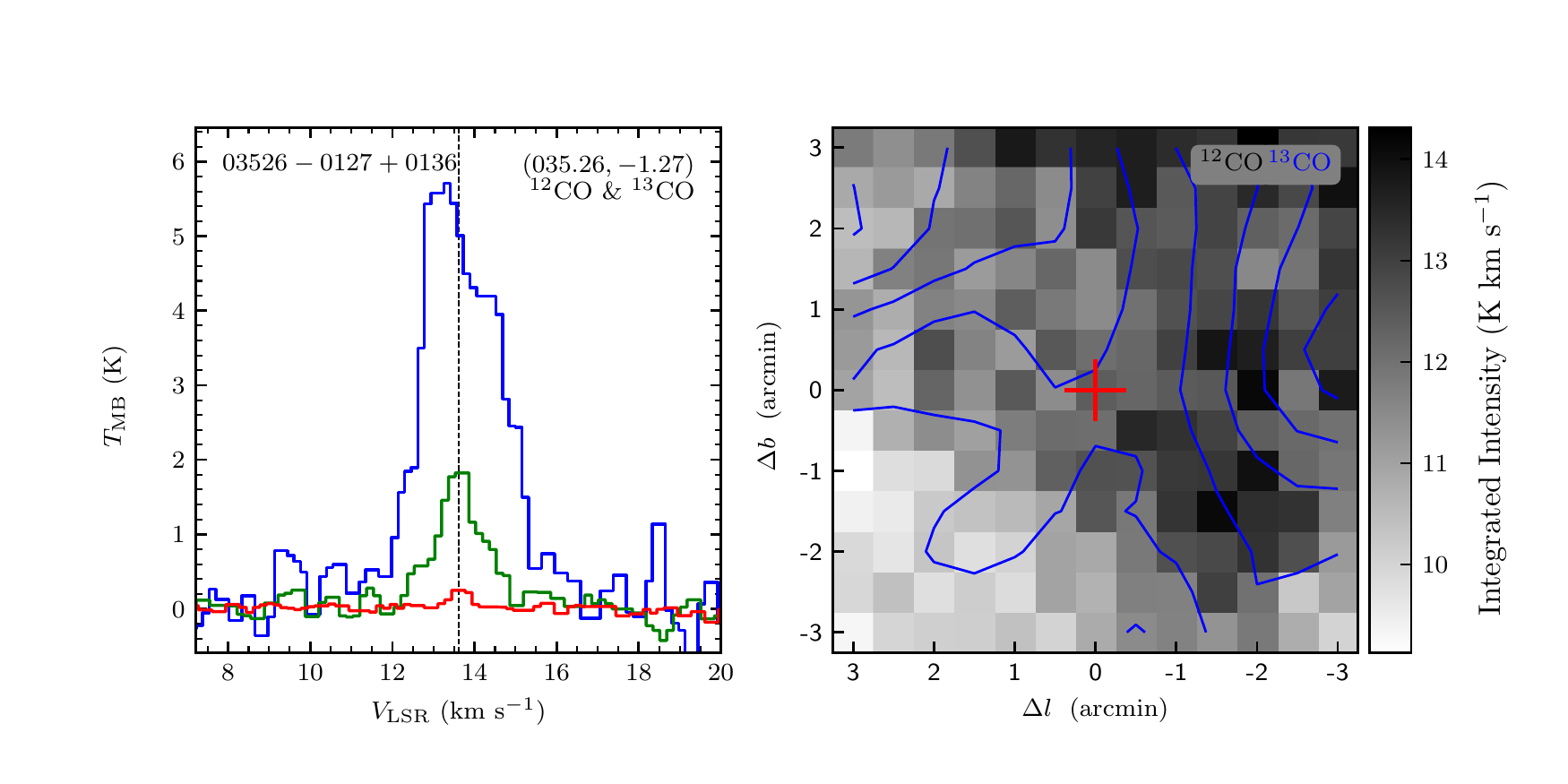}
\includegraphics[width=9.0cm,angle=0]{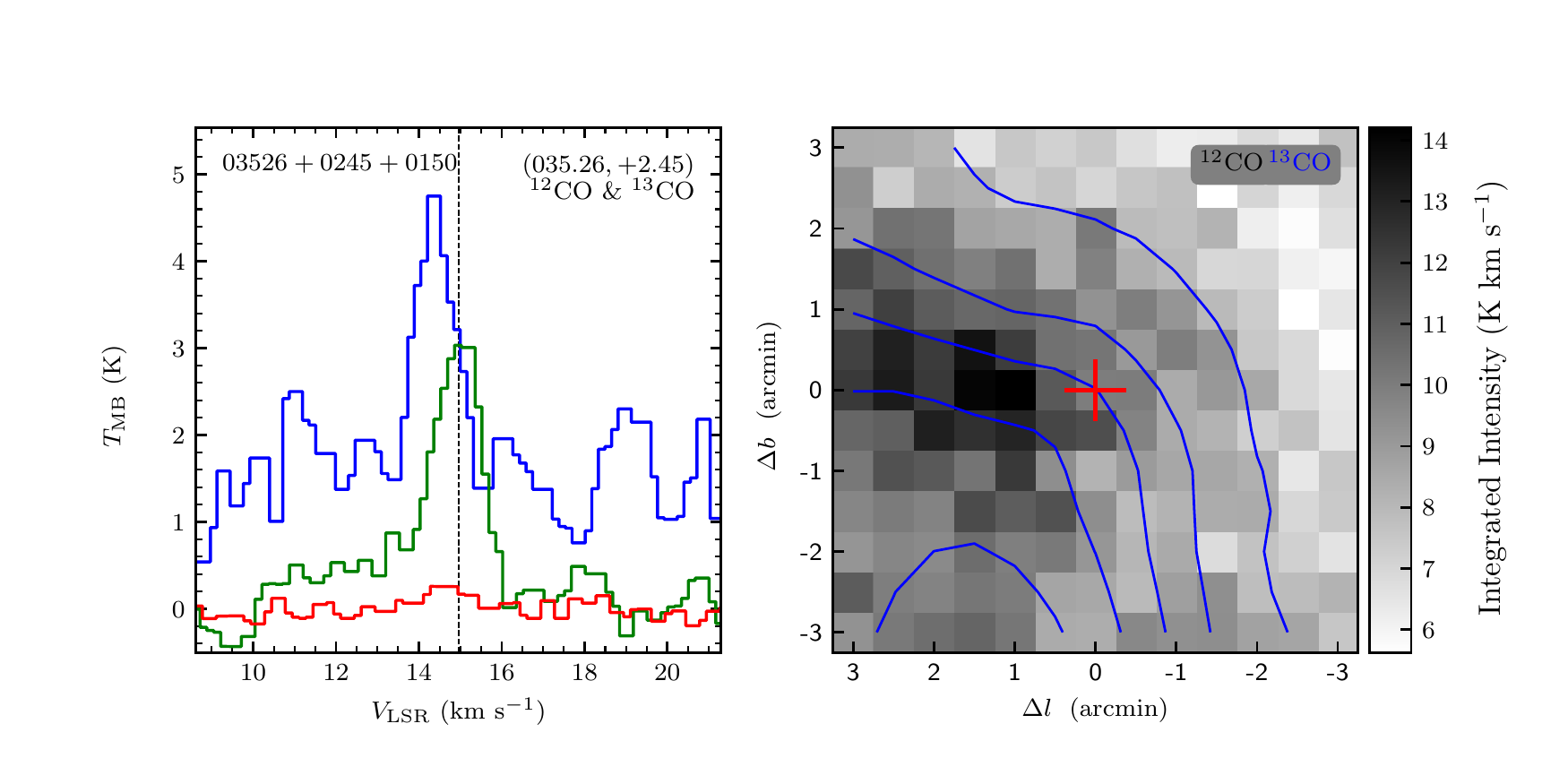}
\end{figure}
\clearpage

\begin{figure}
\includegraphics[width=9.0cm,angle=0]{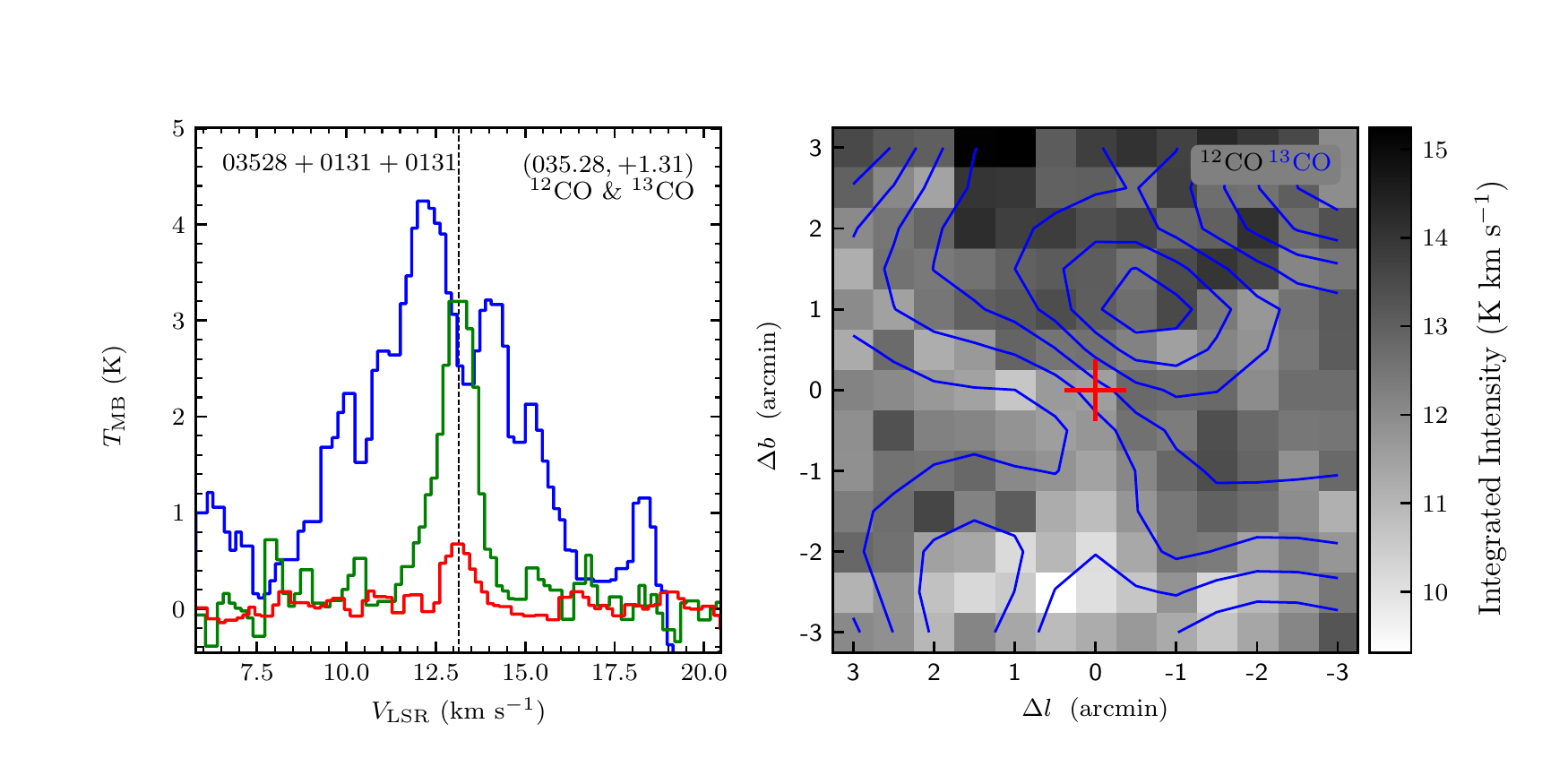}
\includegraphics[width=9.0cm,angle=0]{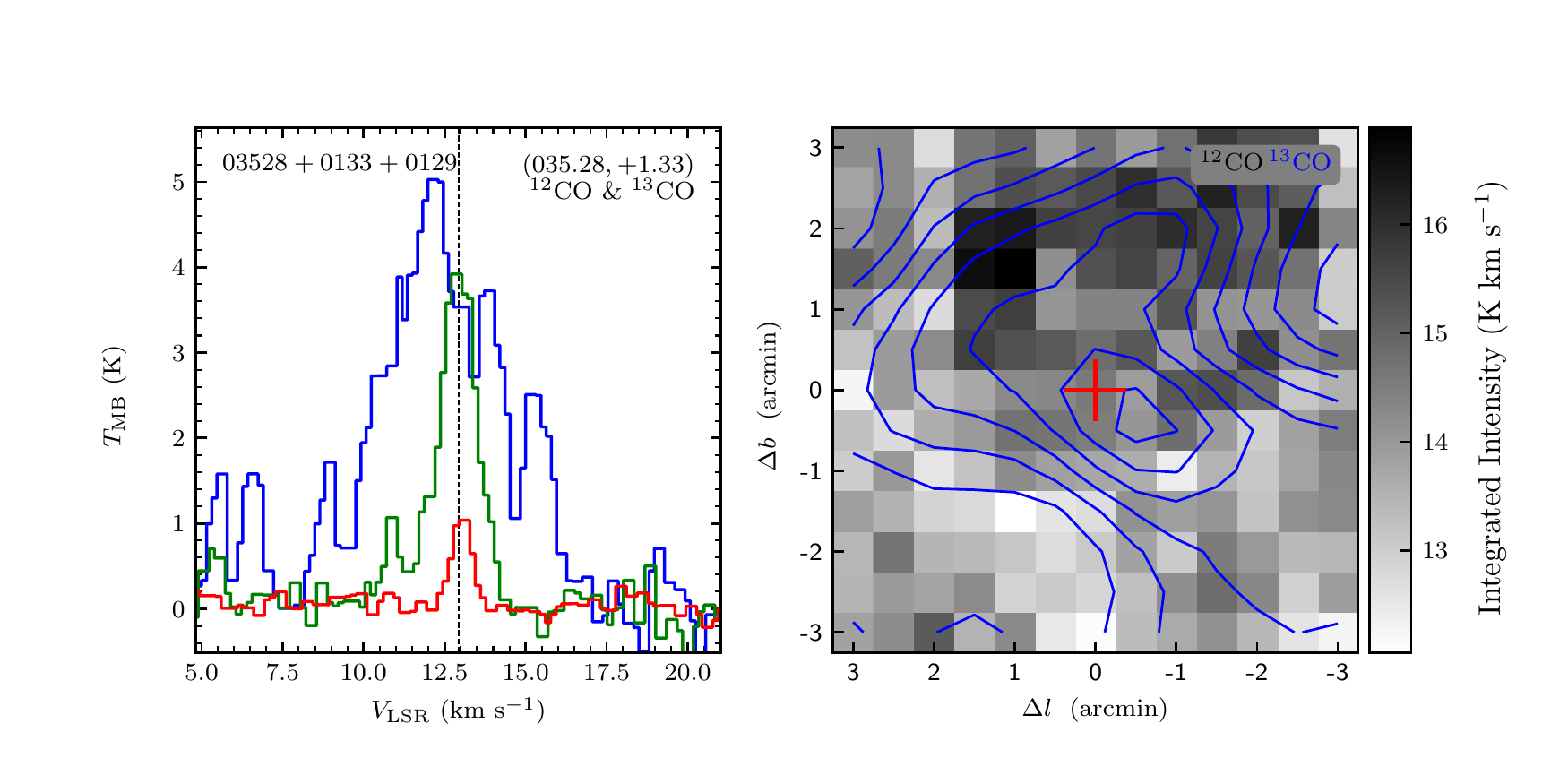}
\vspace{-0.5cm}

\includegraphics[width=9.0cm,angle=0]{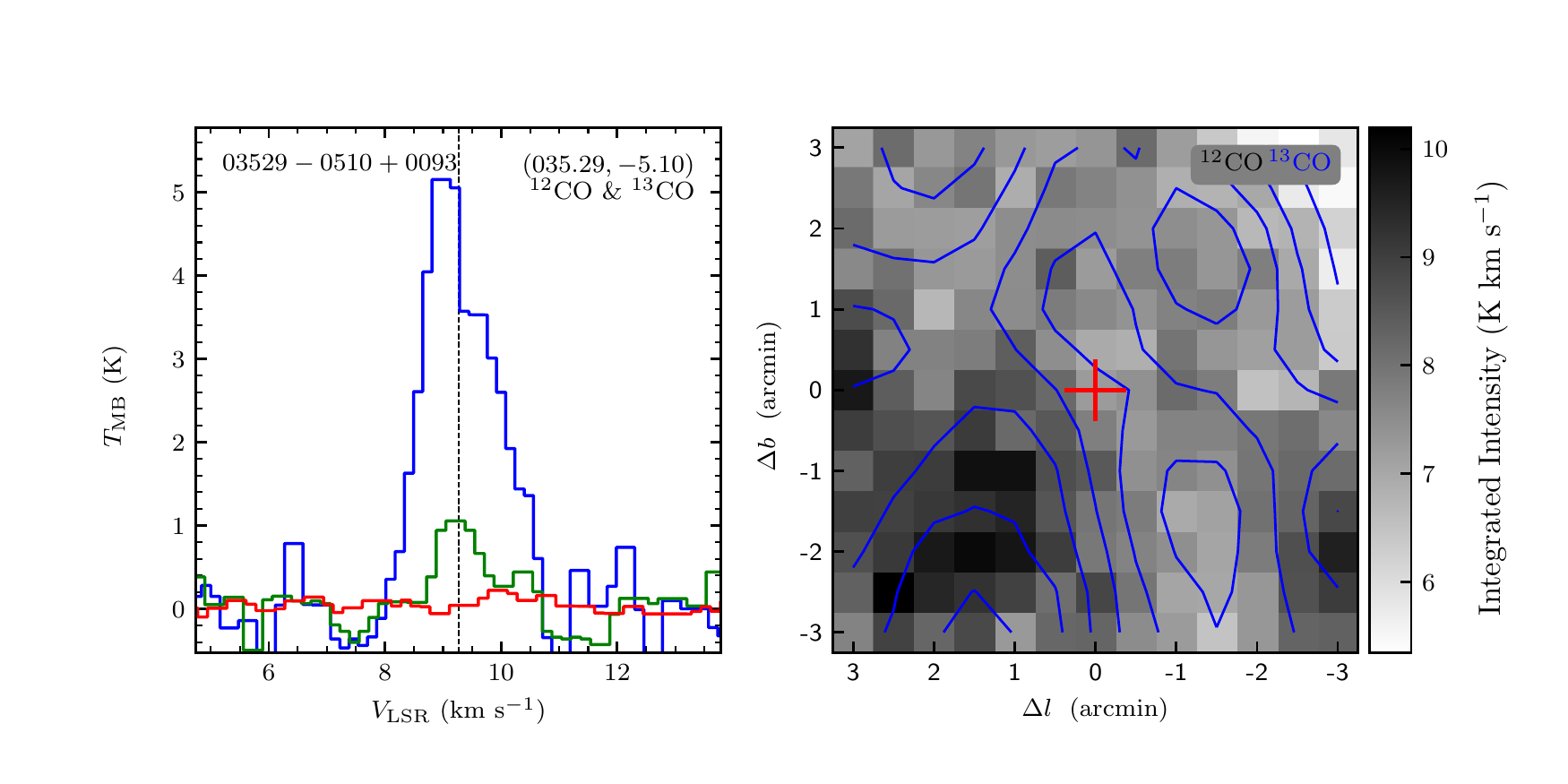}
\includegraphics[width=9.0cm,angle=0]{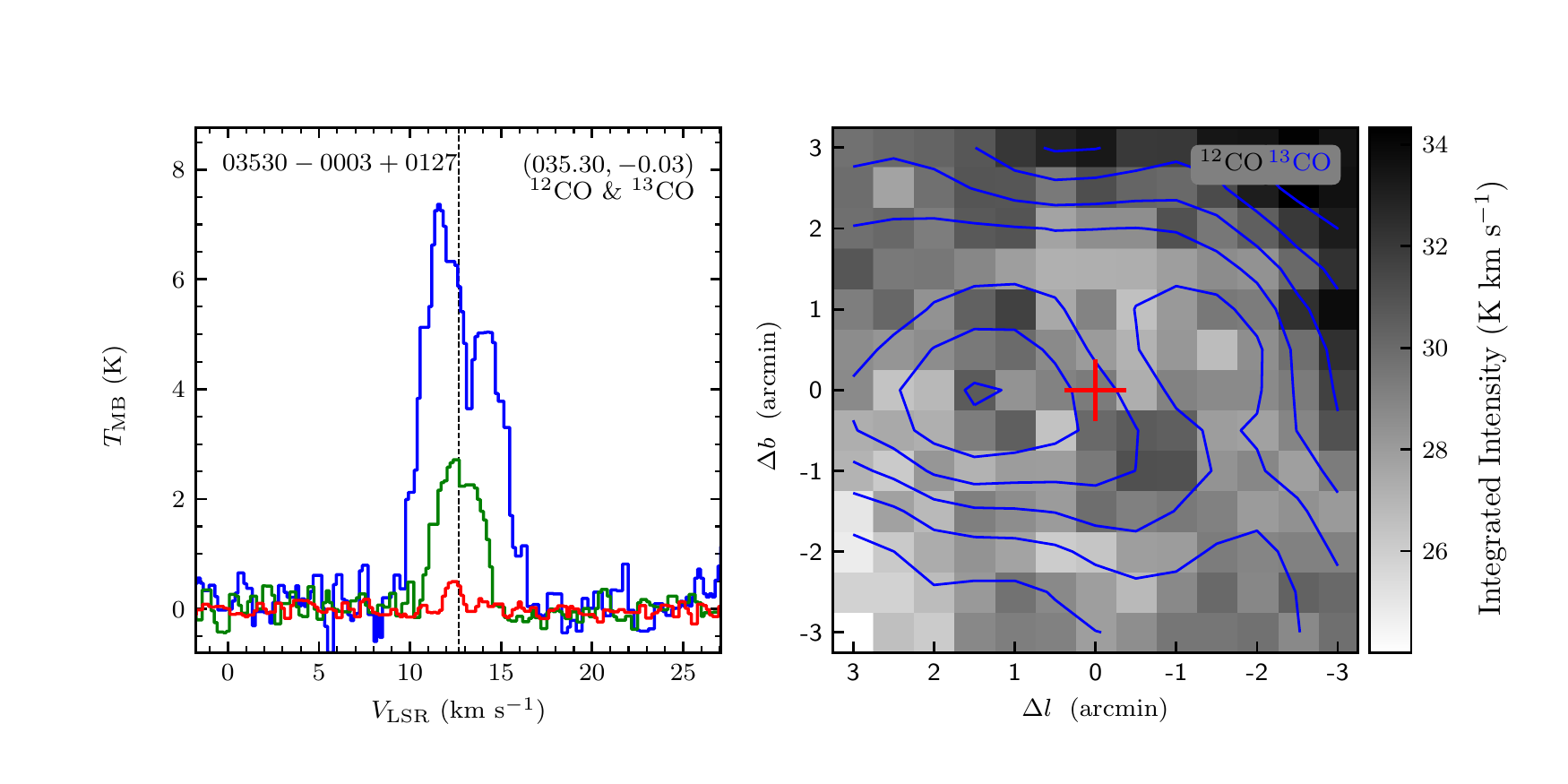}
\vspace{-0.5cm}

\includegraphics[width=9.0cm,angle=0]{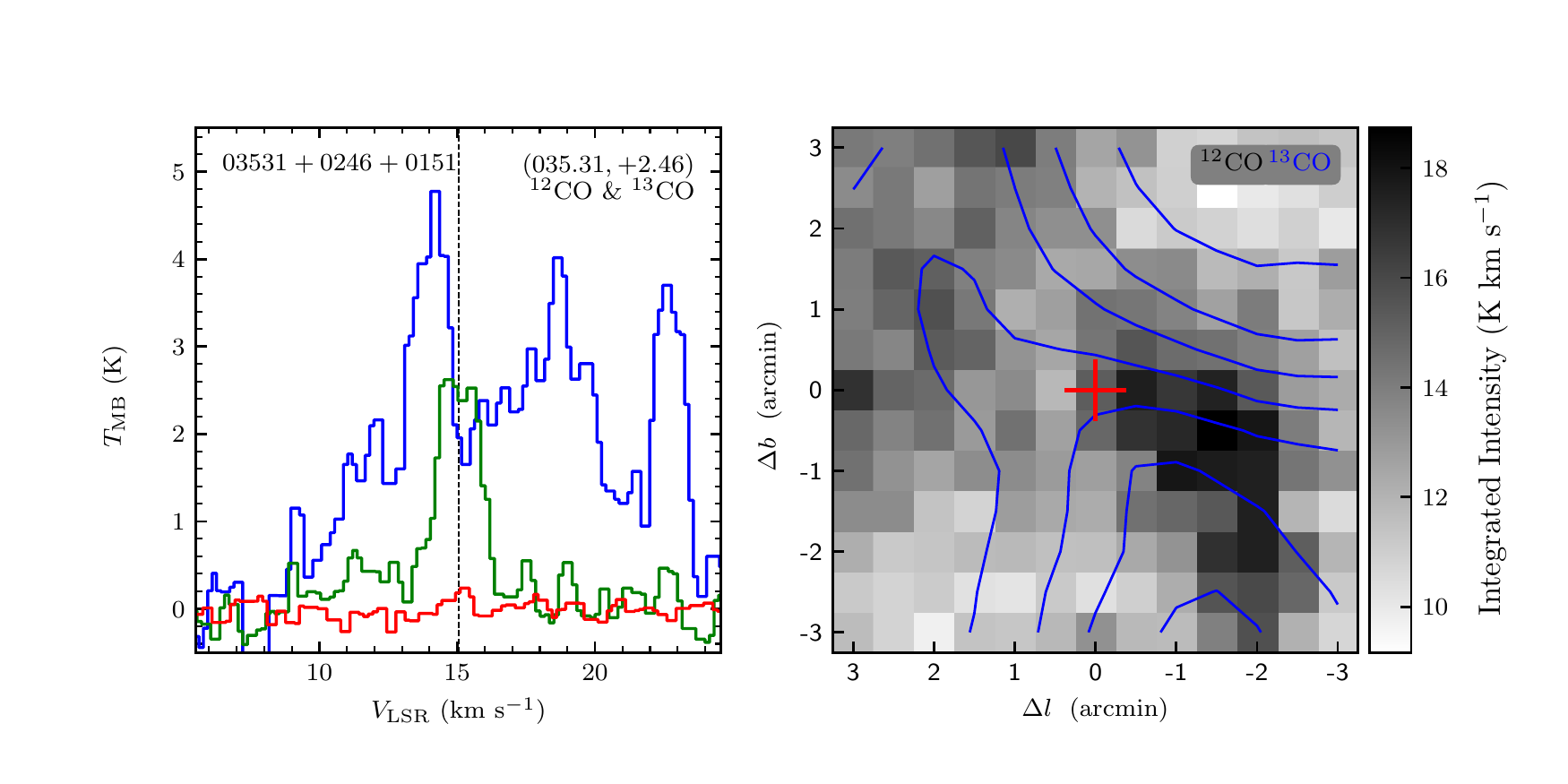}
\includegraphics[width=9.0cm,angle=0]{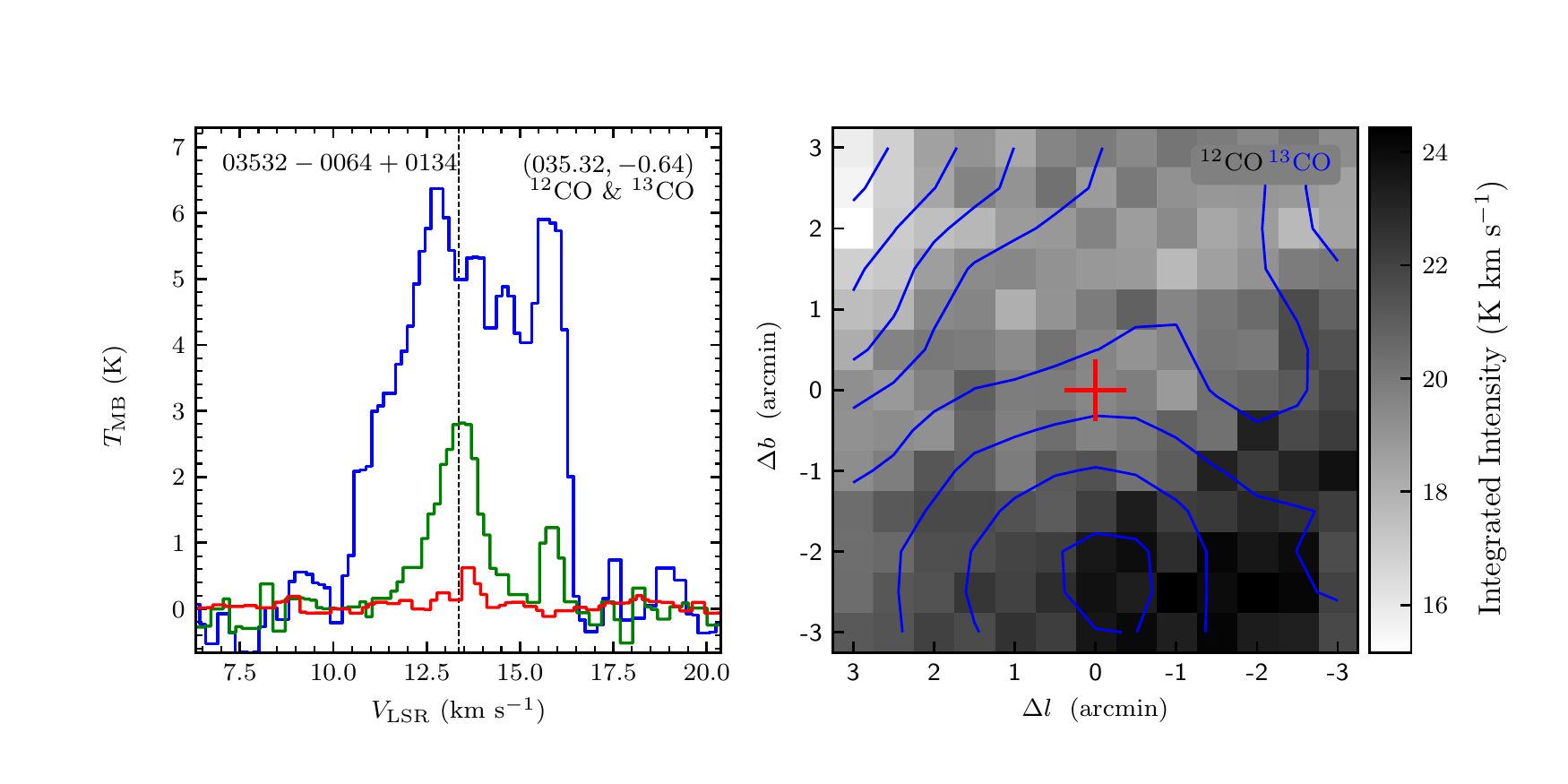}
\vspace{-0.5cm}

\includegraphics[width=9.0cm,angle=0]{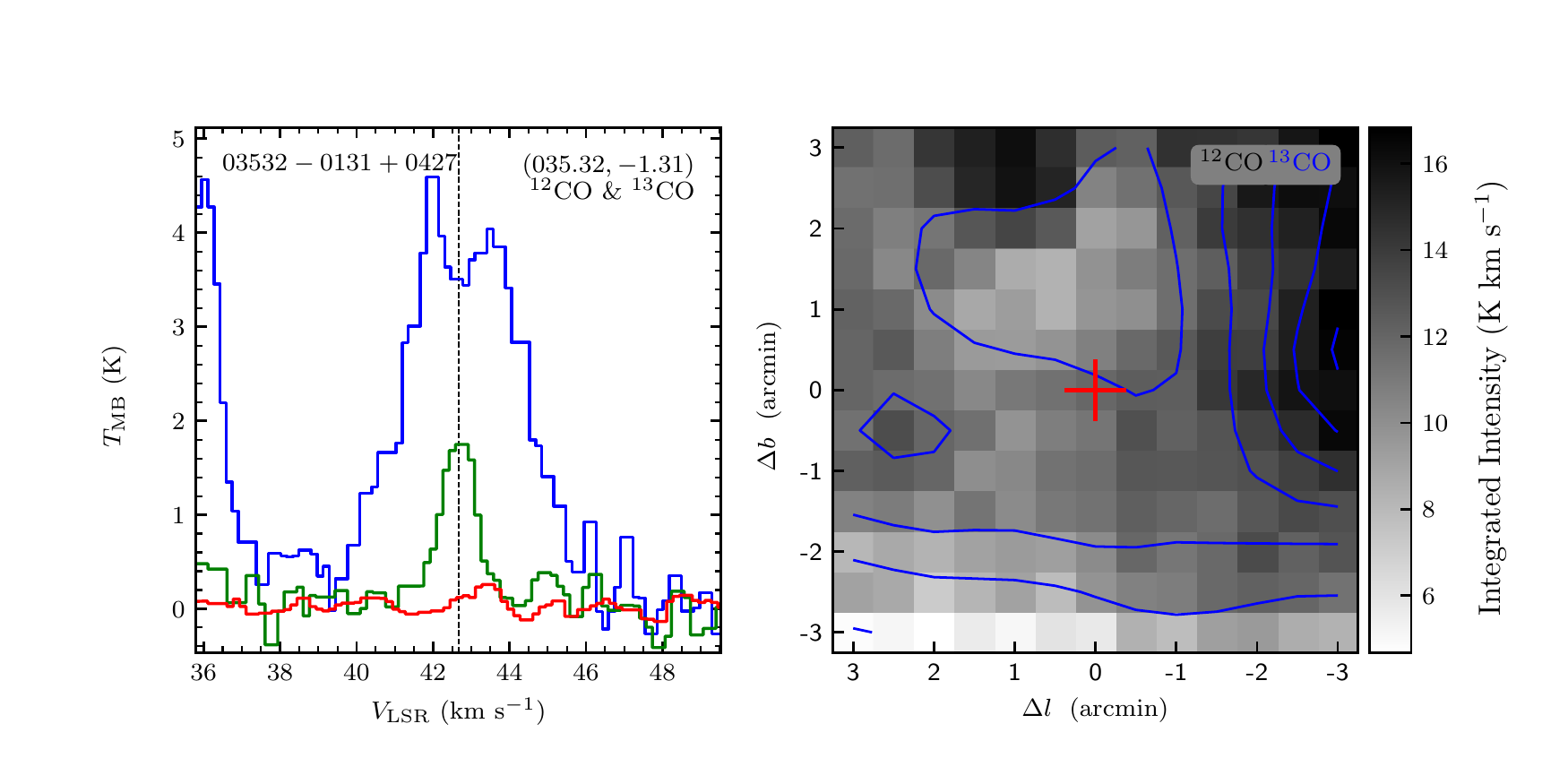}
\includegraphics[width=9.0cm,angle=0]{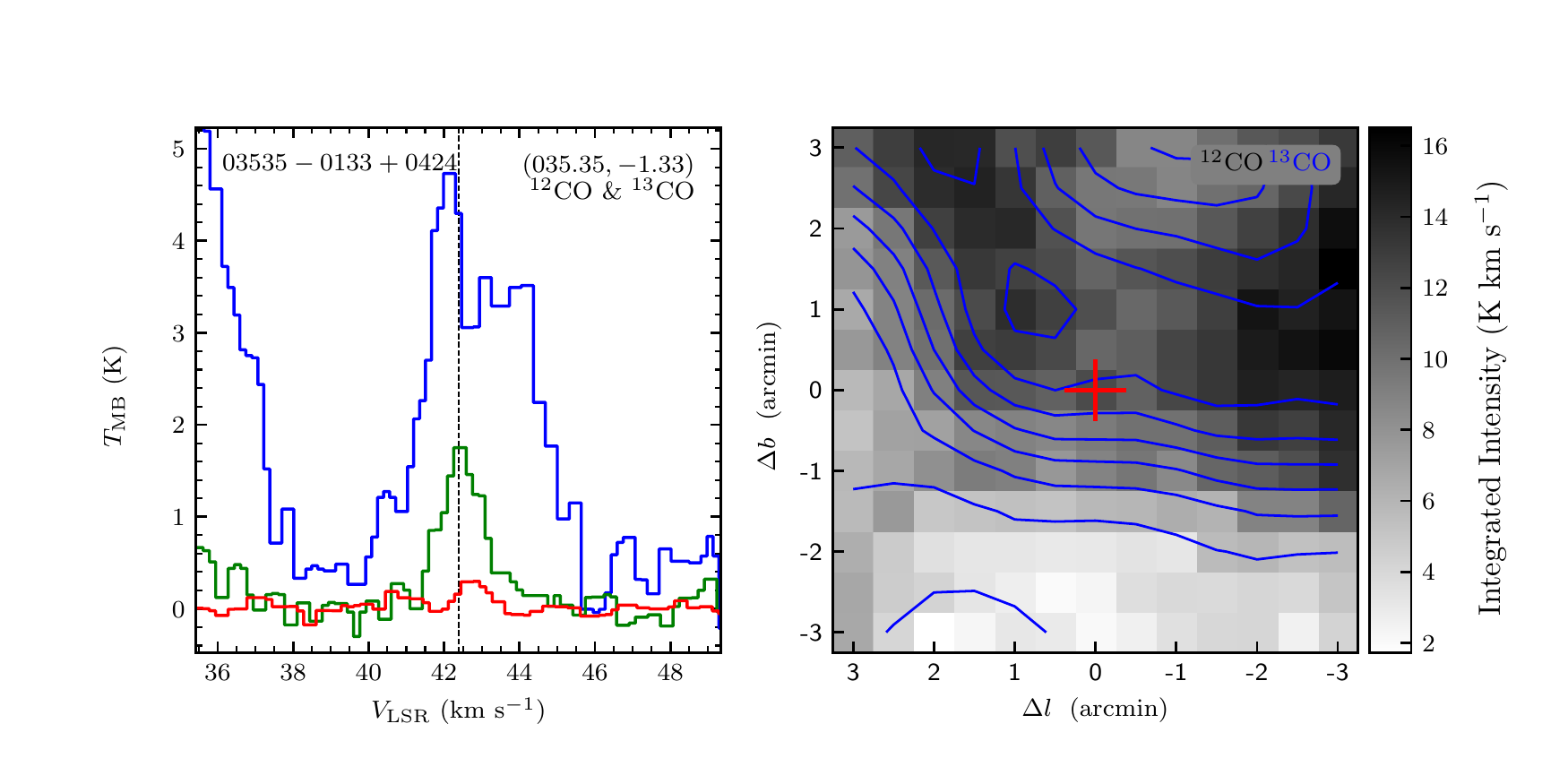}
\vspace{-0.5cm}

\includegraphics[width=9.0cm,angle=0]{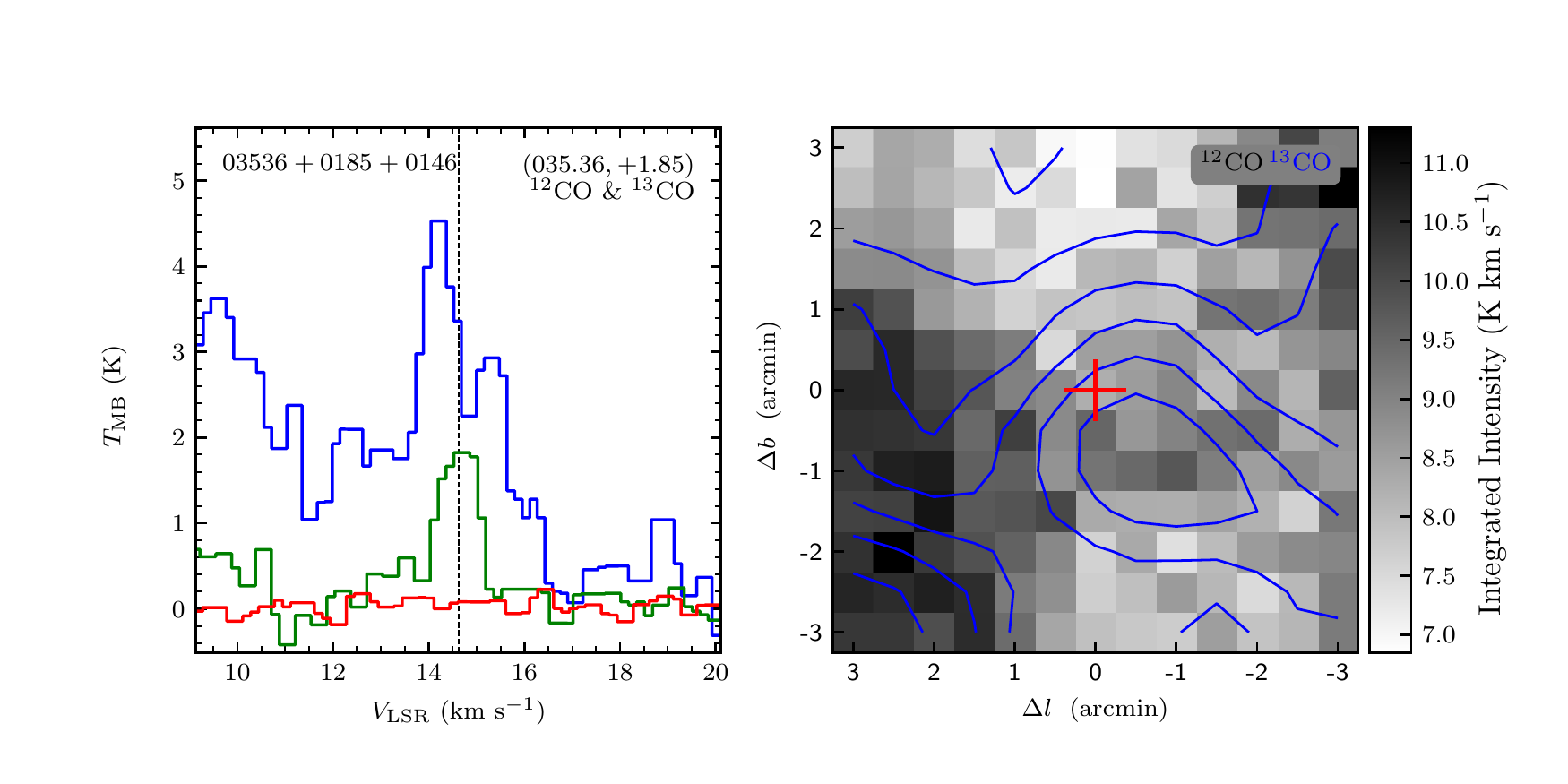}
\includegraphics[width=9.0cm,angle=0]{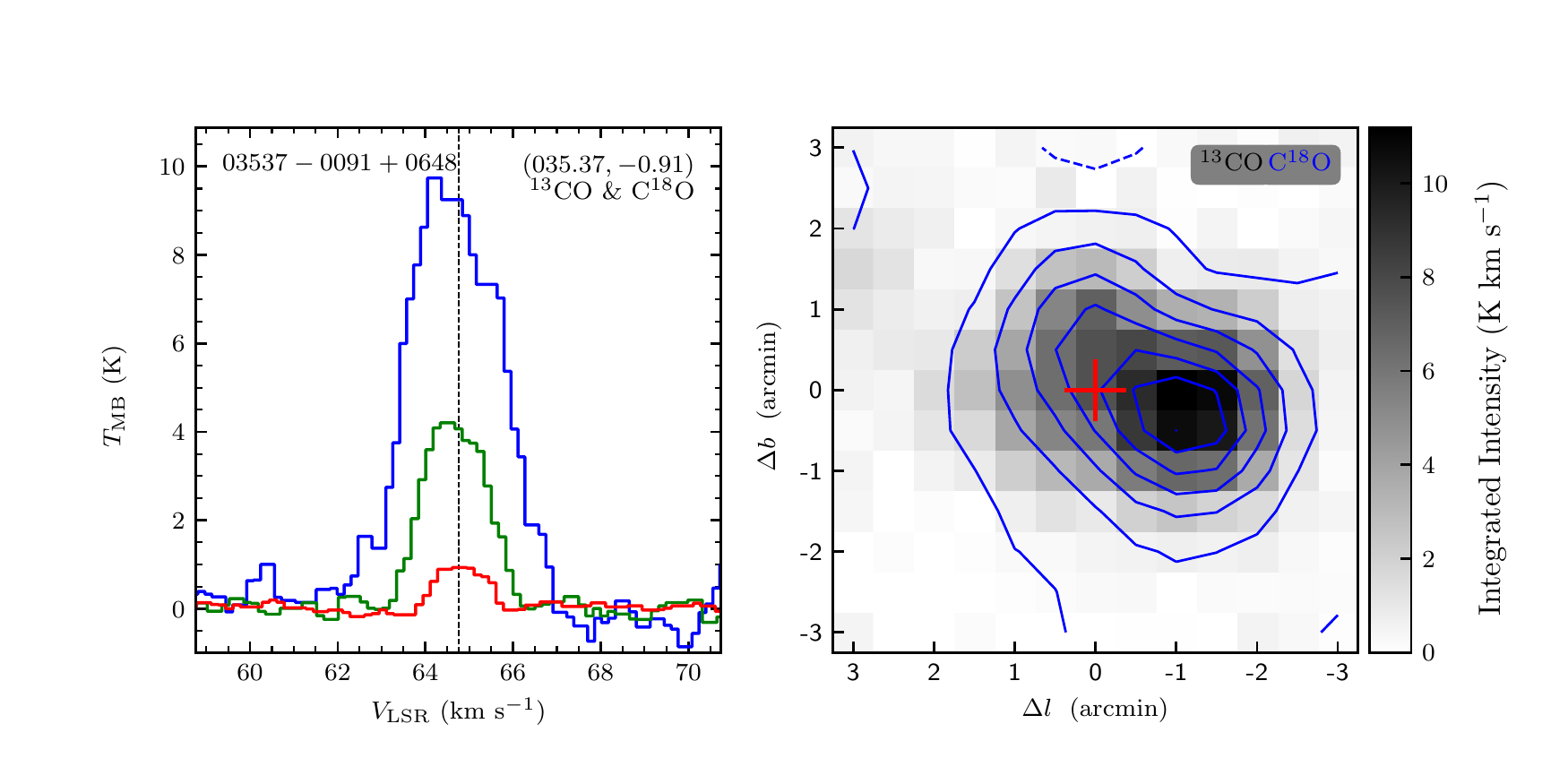}
\end{figure}
\clearpage

\begin{figure}
\includegraphics[width=9.0cm,angle=0]{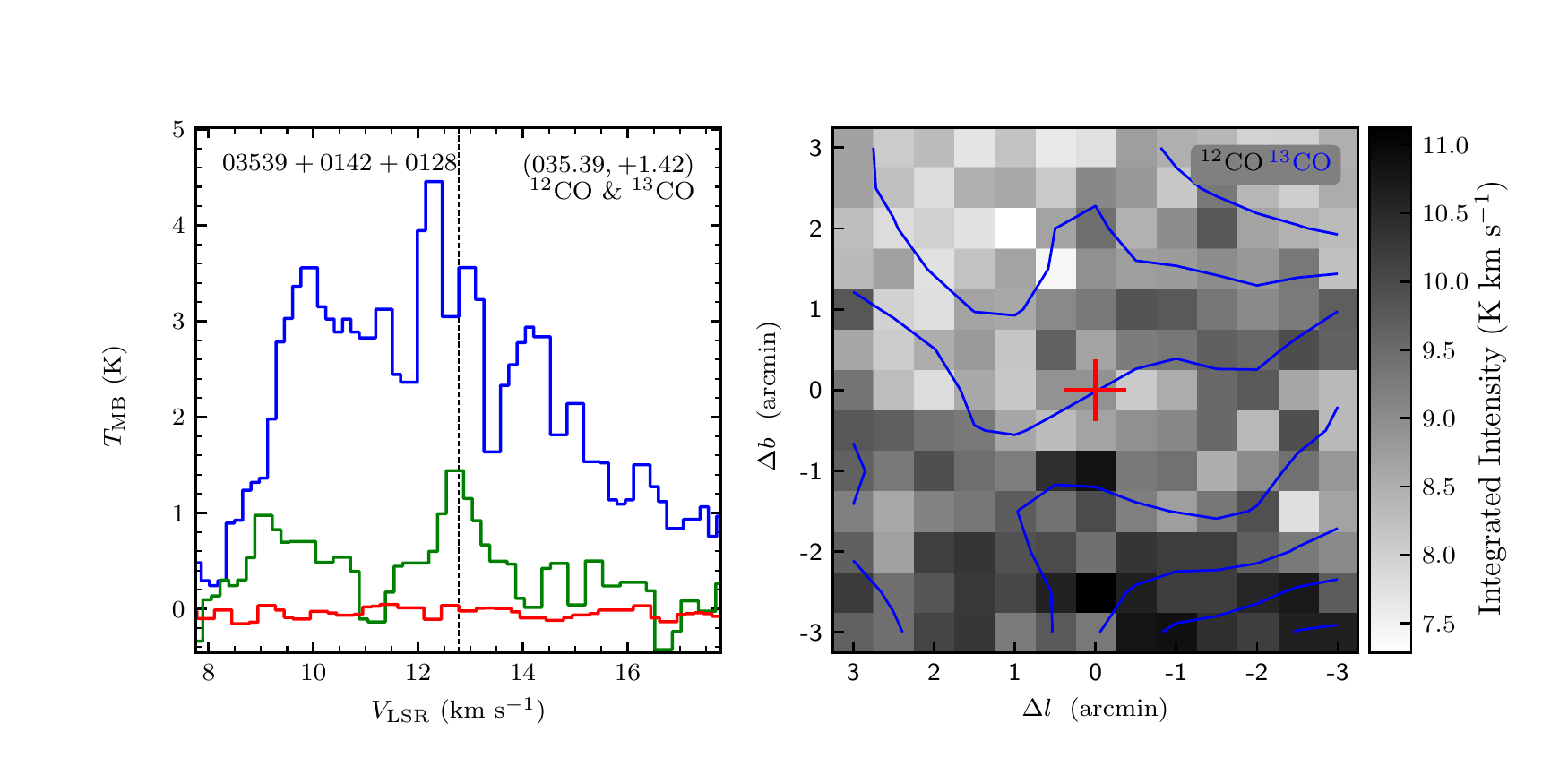}
\includegraphics[width=9.0cm,angle=0]{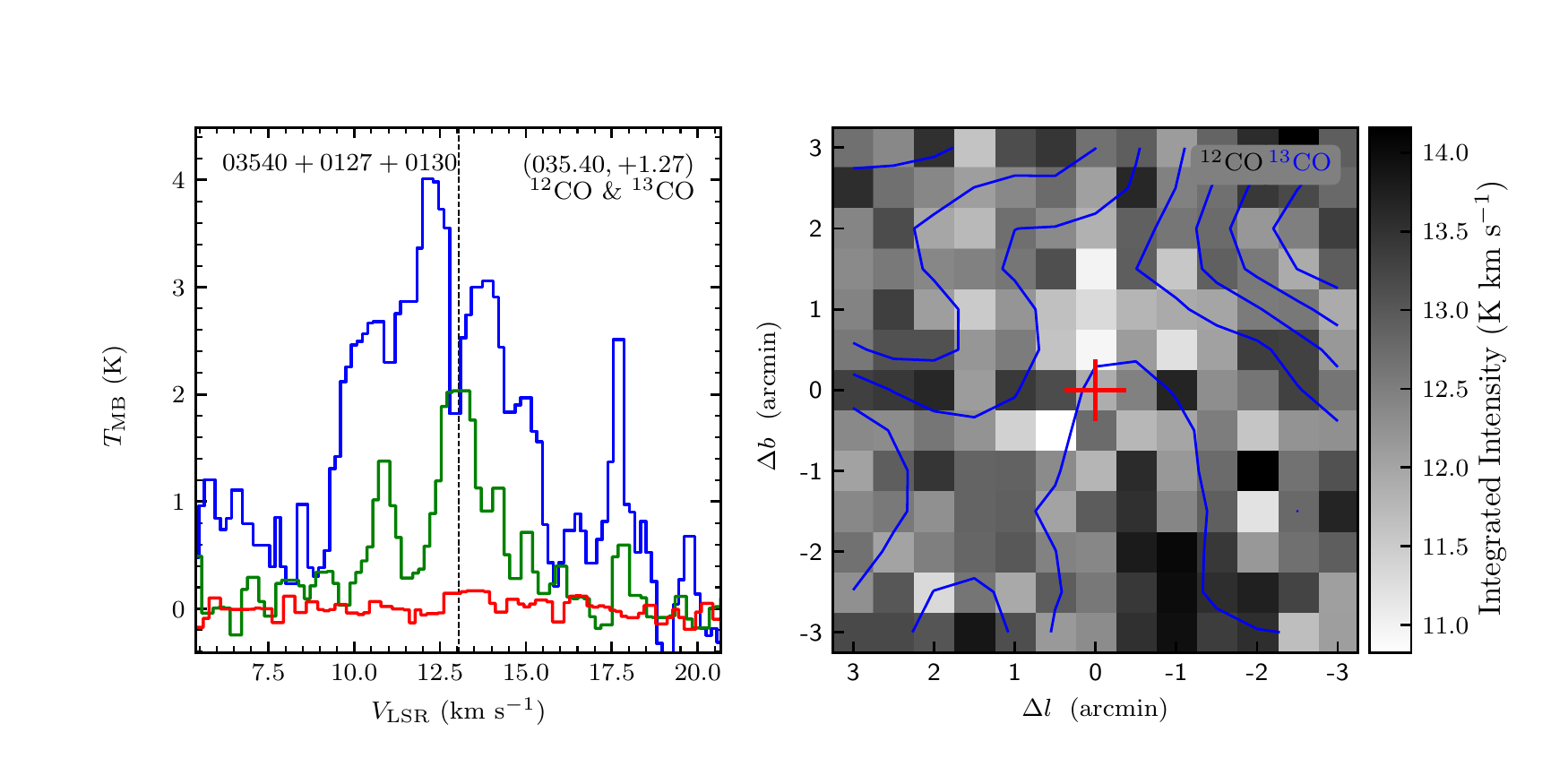}
\vspace{-0.5cm}

\includegraphics[width=9.0cm,angle=0]{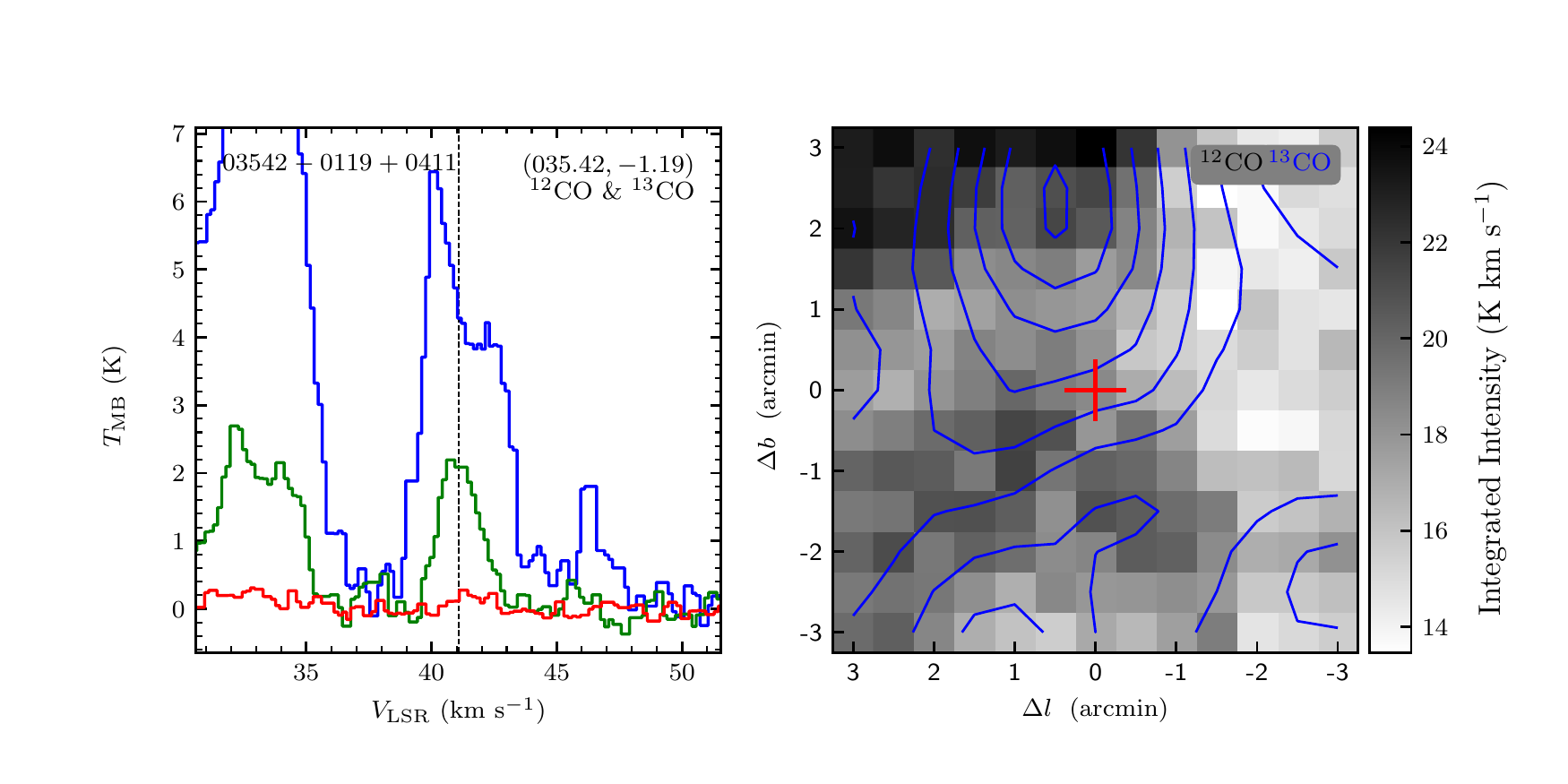}
\includegraphics[width=9.0cm,angle=0]{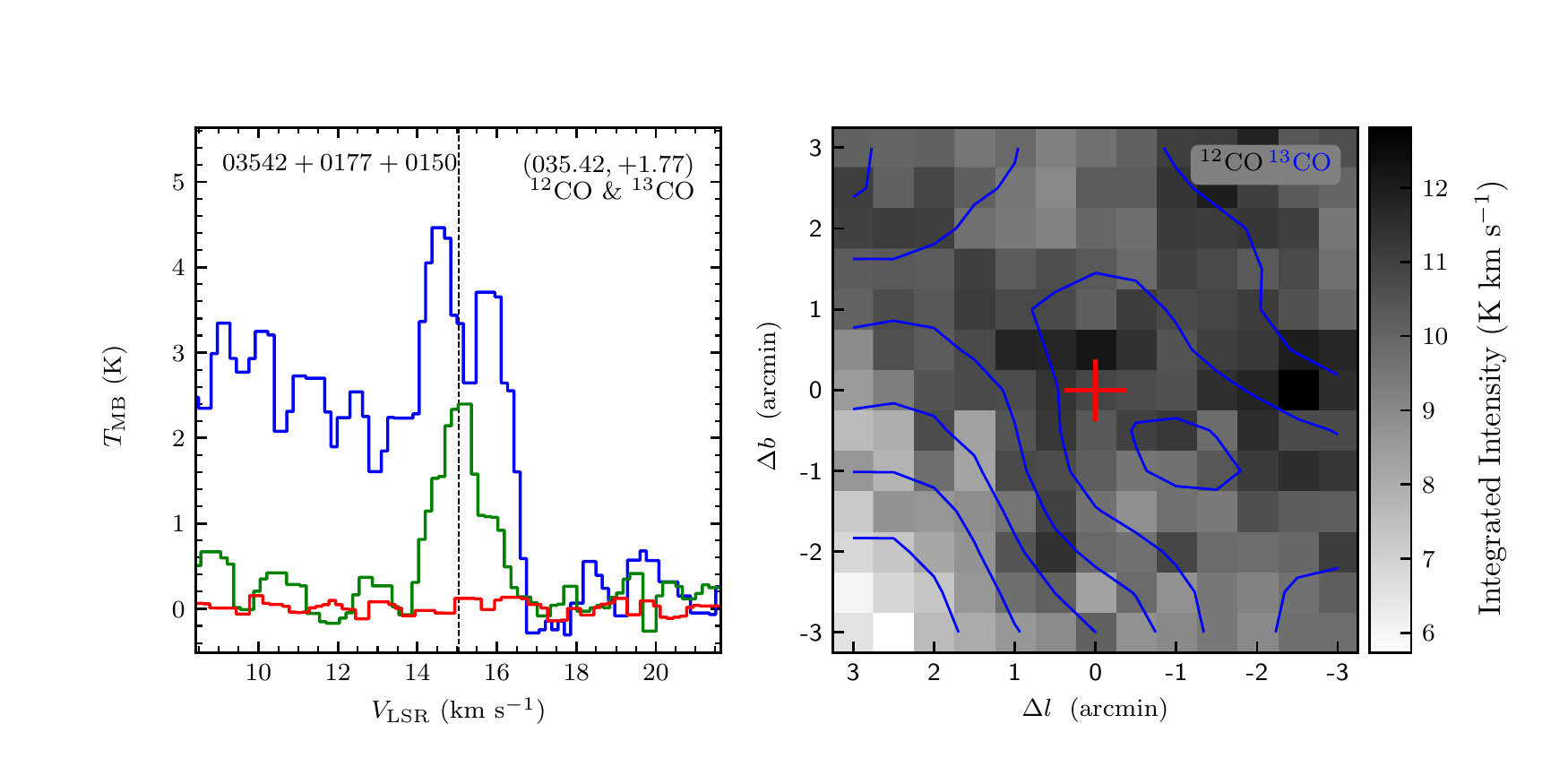}
\vspace{-0.5cm}

\includegraphics[width=9.0cm,angle=0]{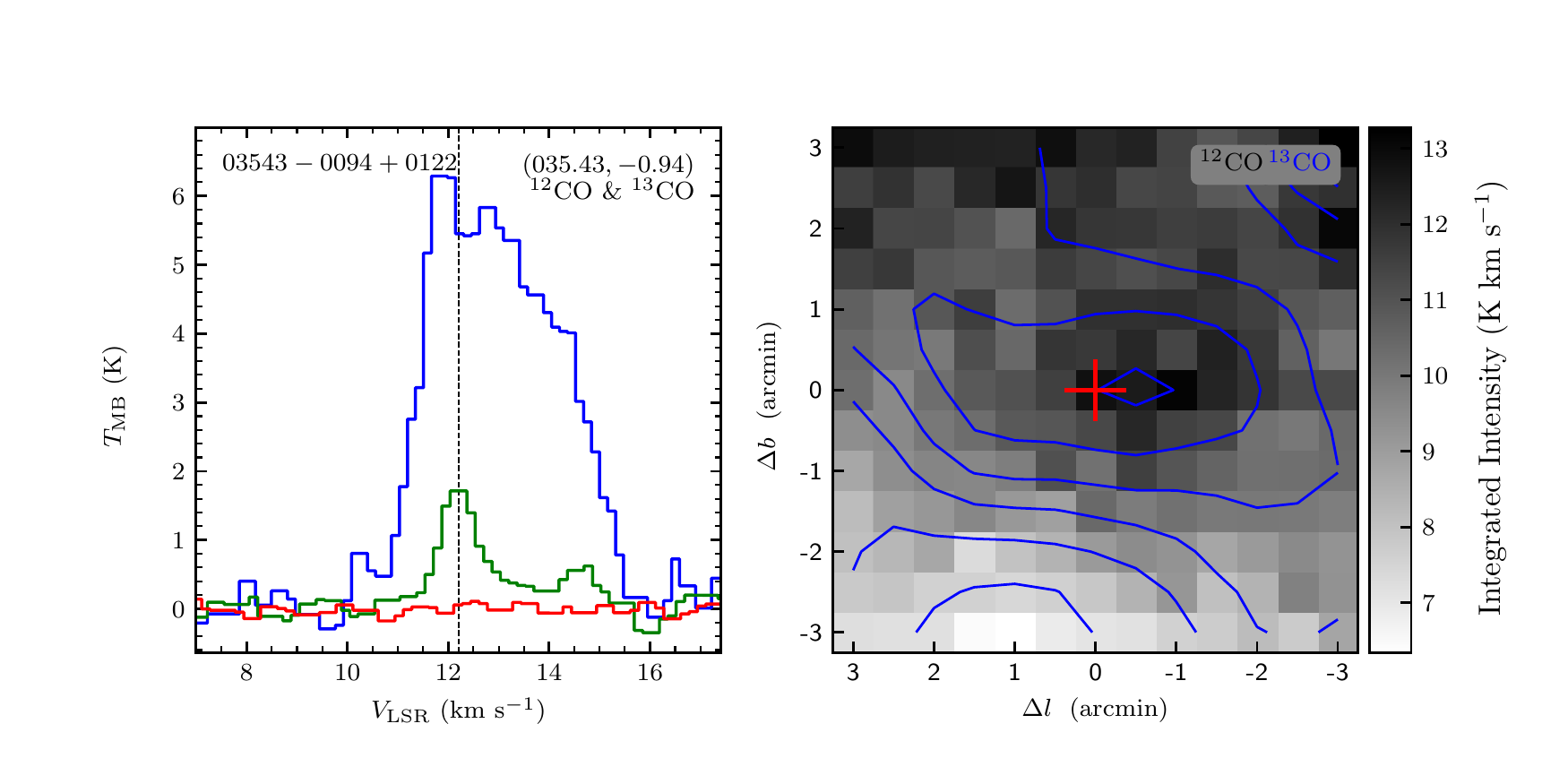}
\includegraphics[width=9.0cm,angle=0]{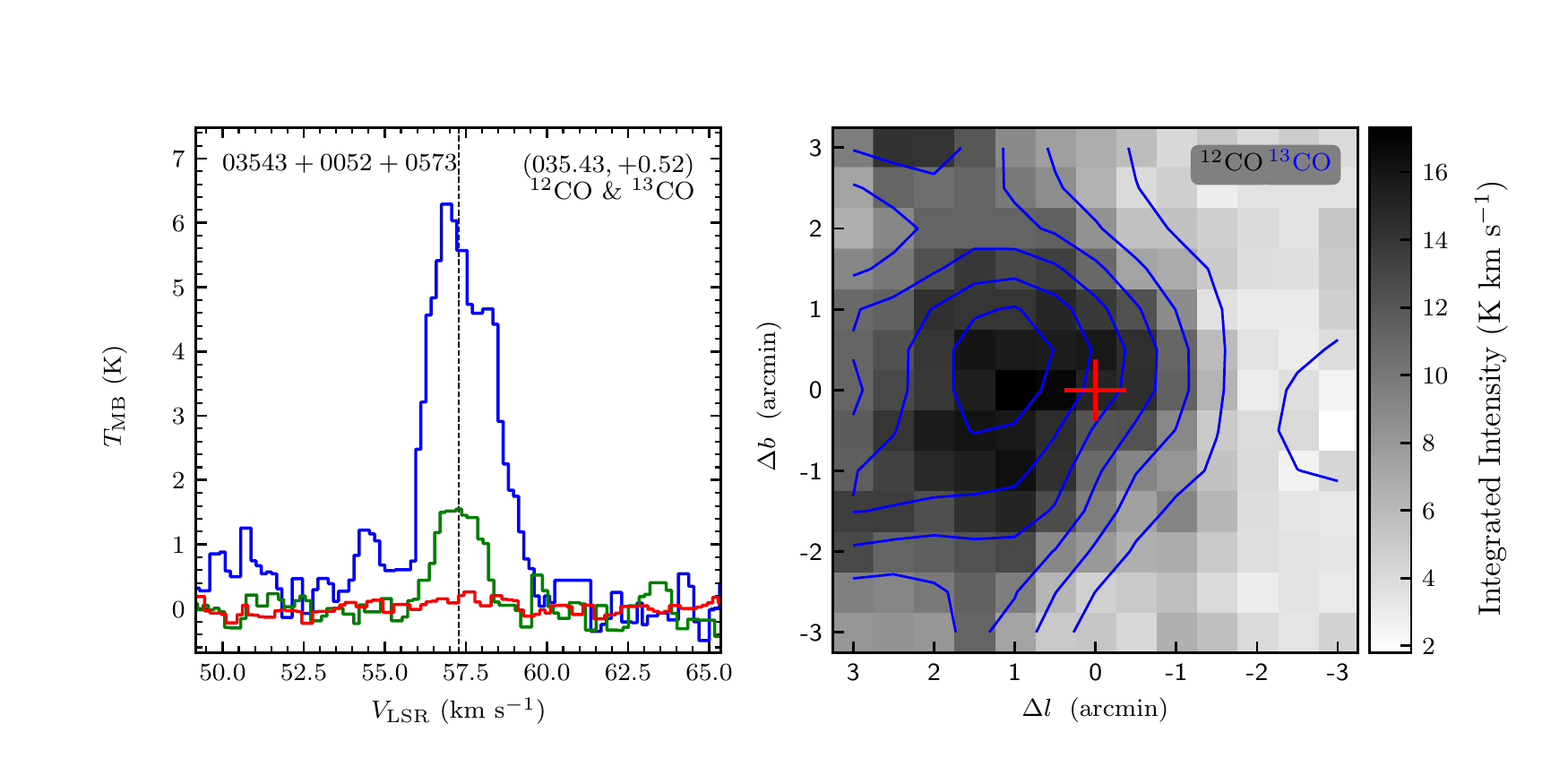}
\vspace{-0.5cm}

\includegraphics[width=9.0cm,angle=0]{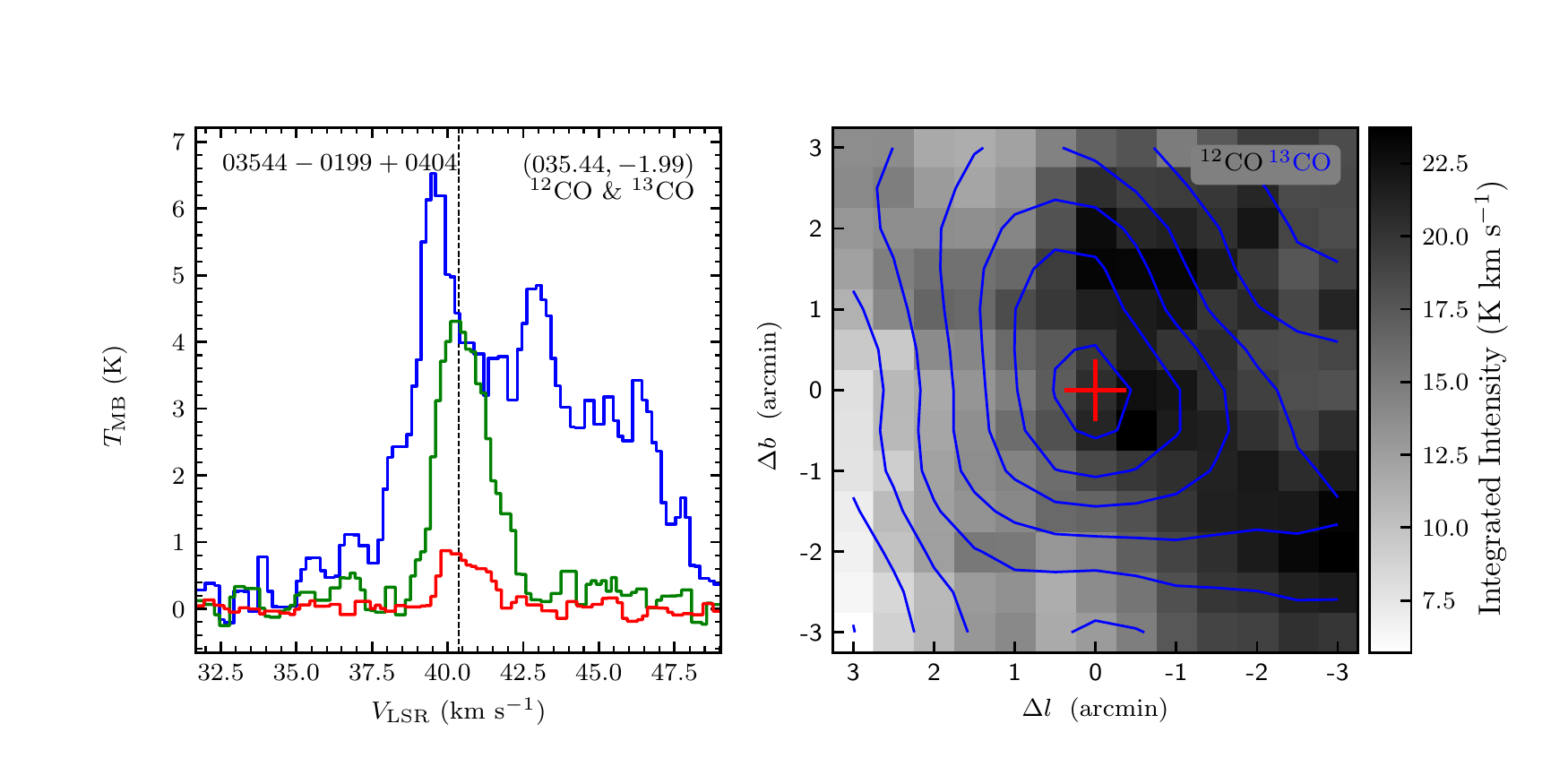}
\includegraphics[width=9.0cm,angle=0]{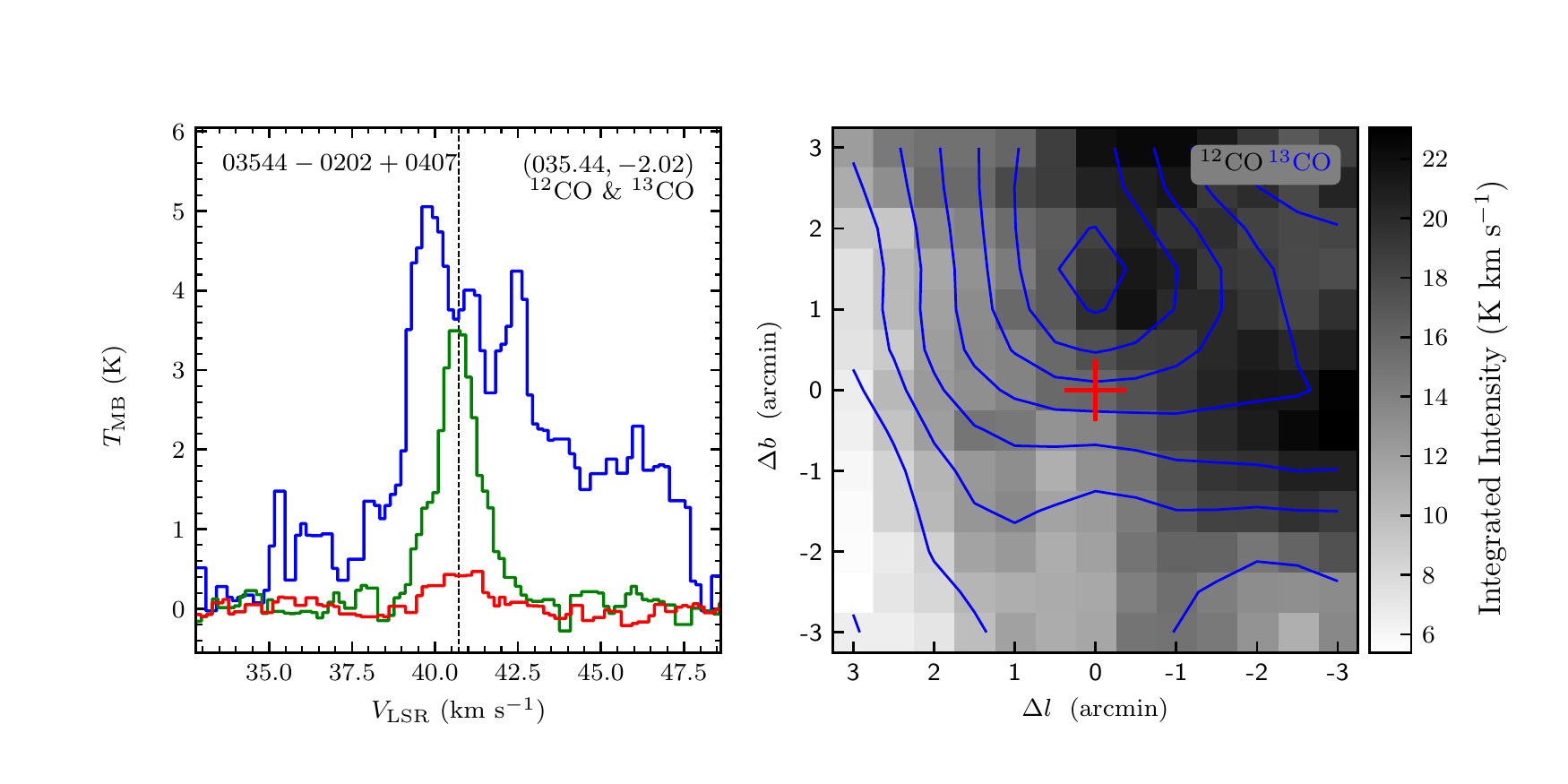}
\vspace{-0.5cm}

\includegraphics[width=9.0cm,angle=0]{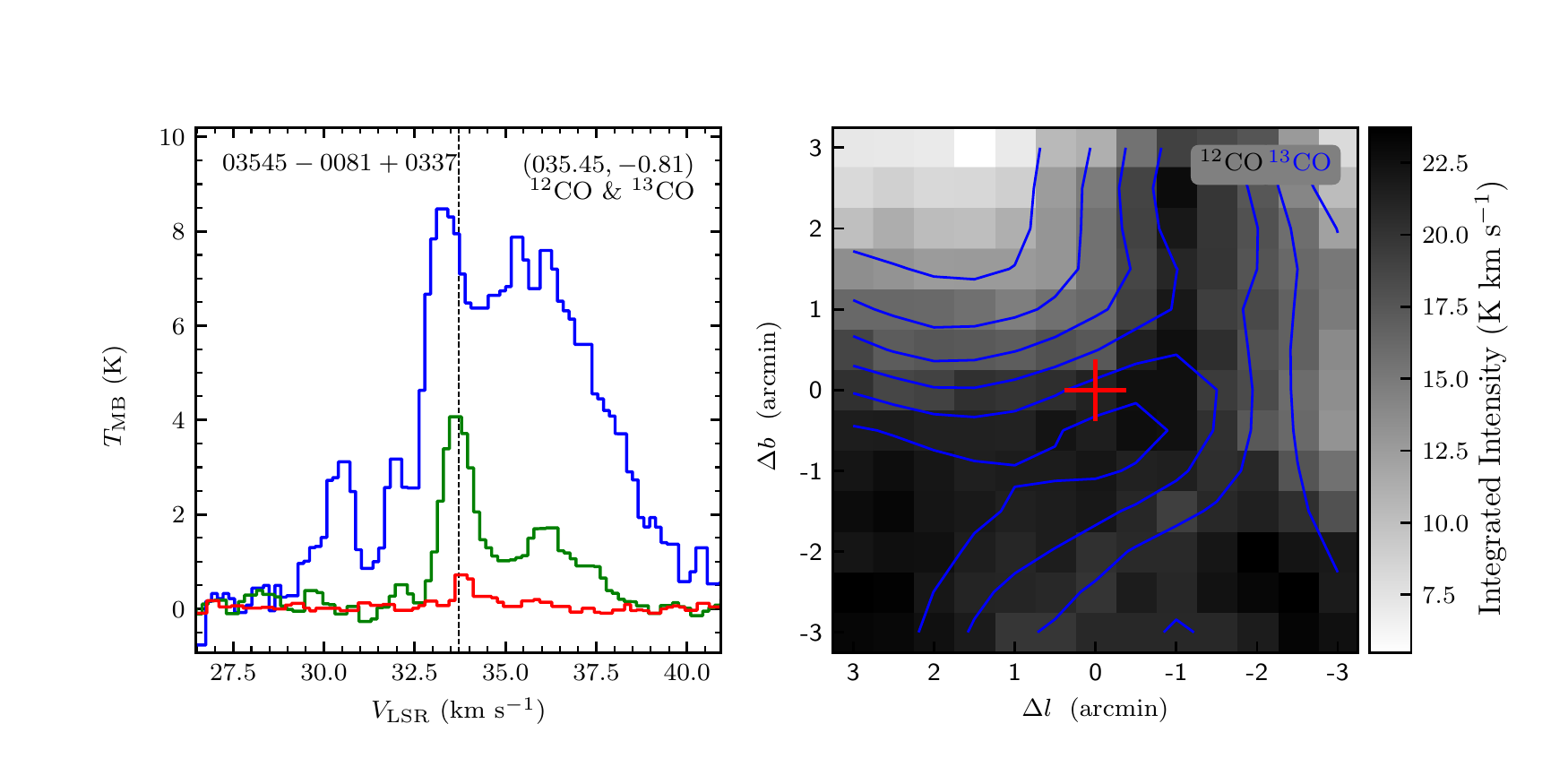}
\includegraphics[width=9.0cm,angle=0]{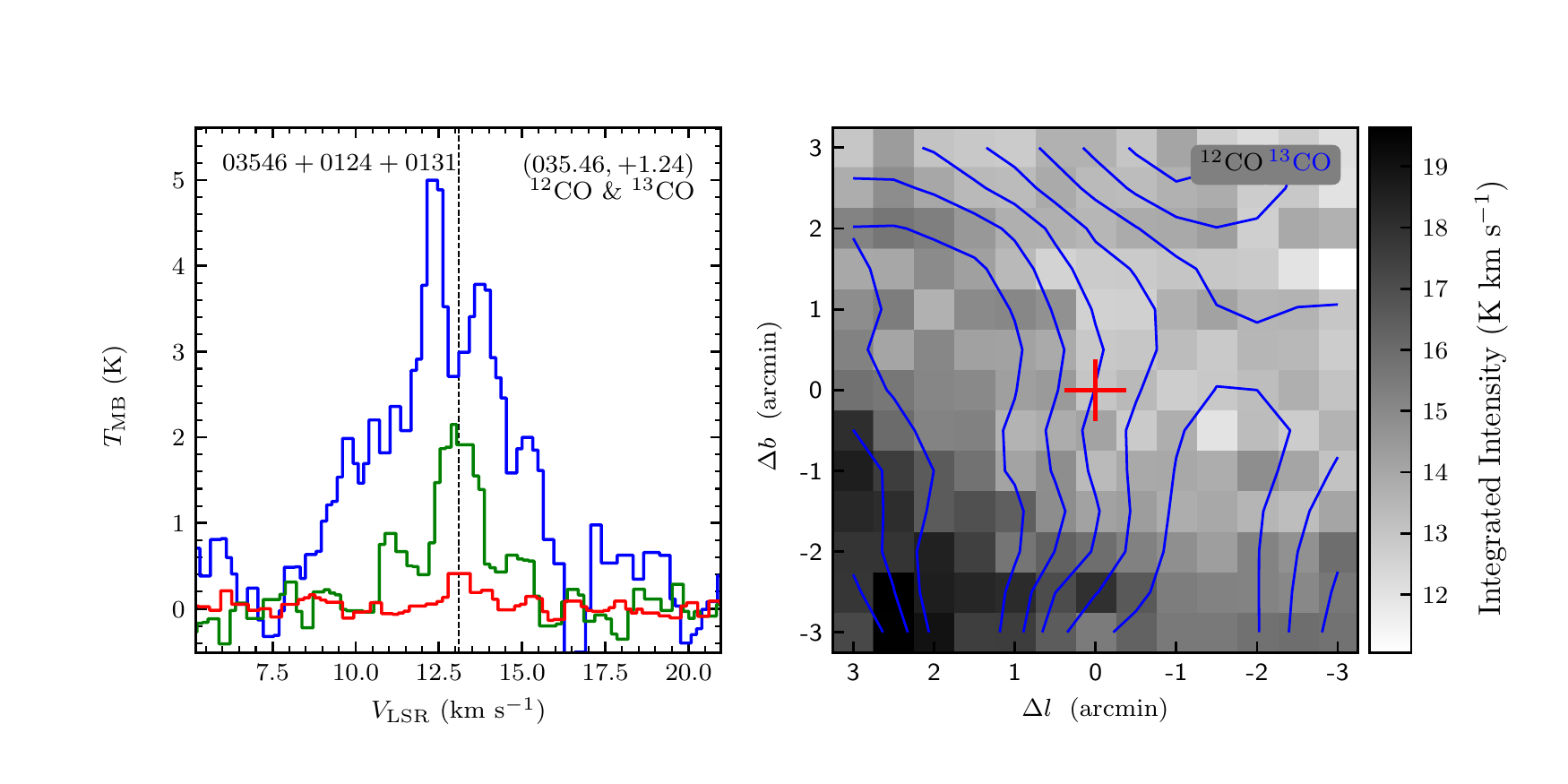}
\end{figure}
\clearpage

\begin{figure}
\includegraphics[width=9.0cm,angle=0]{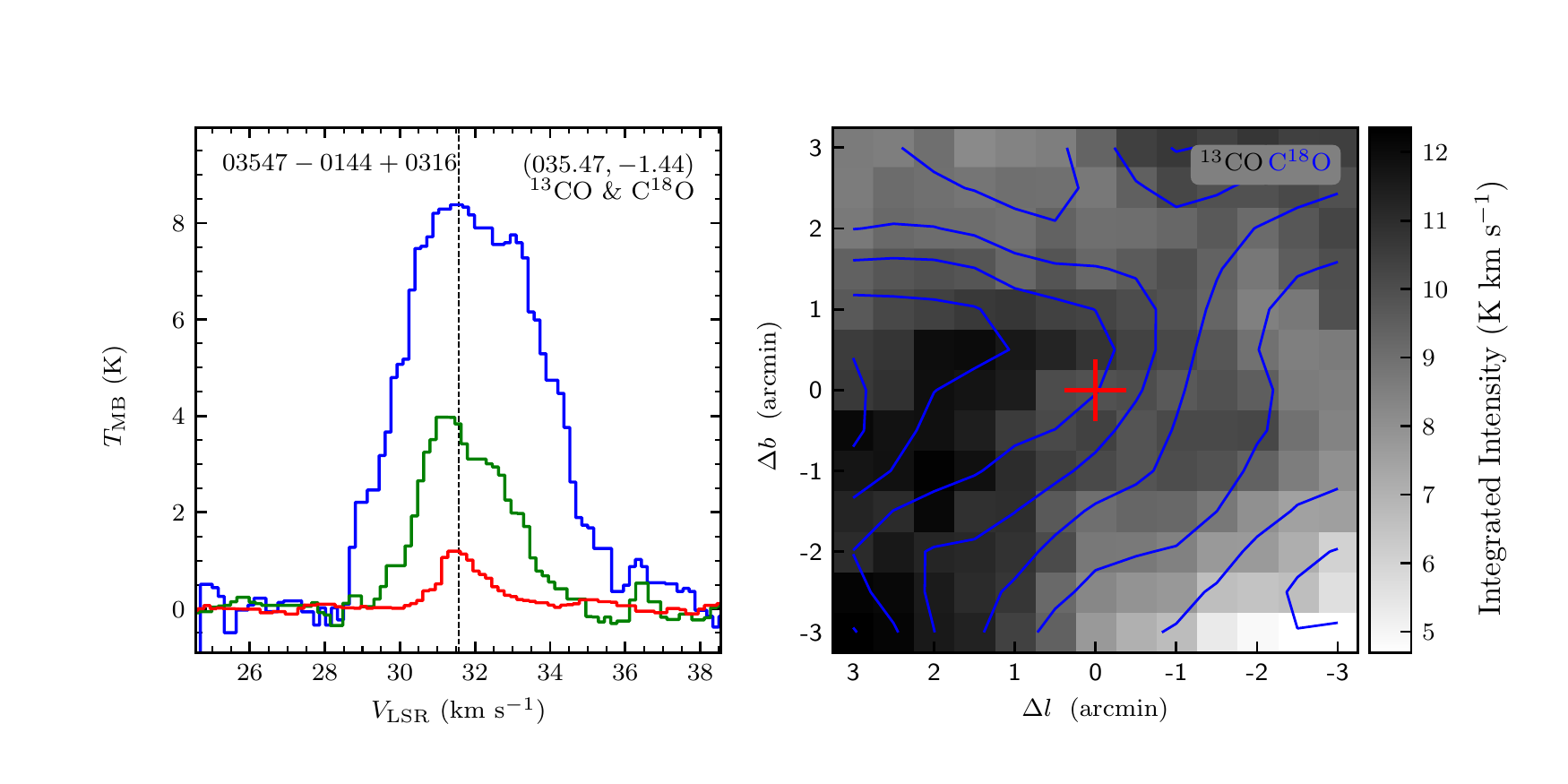}
\includegraphics[width=9.0cm,angle=0]{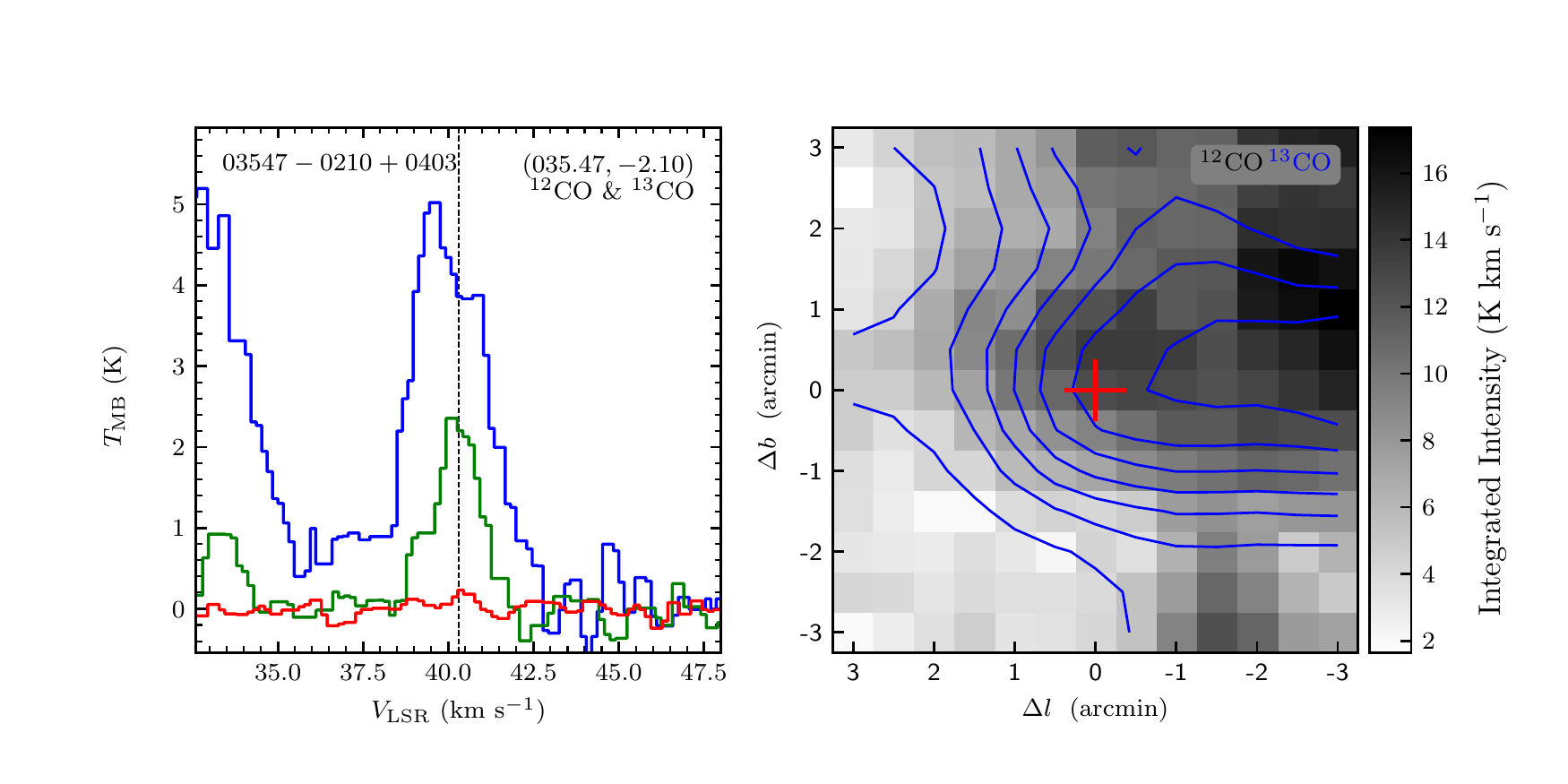}
\vspace{-0.5cm}

\includegraphics[width=9.0cm,angle=0]{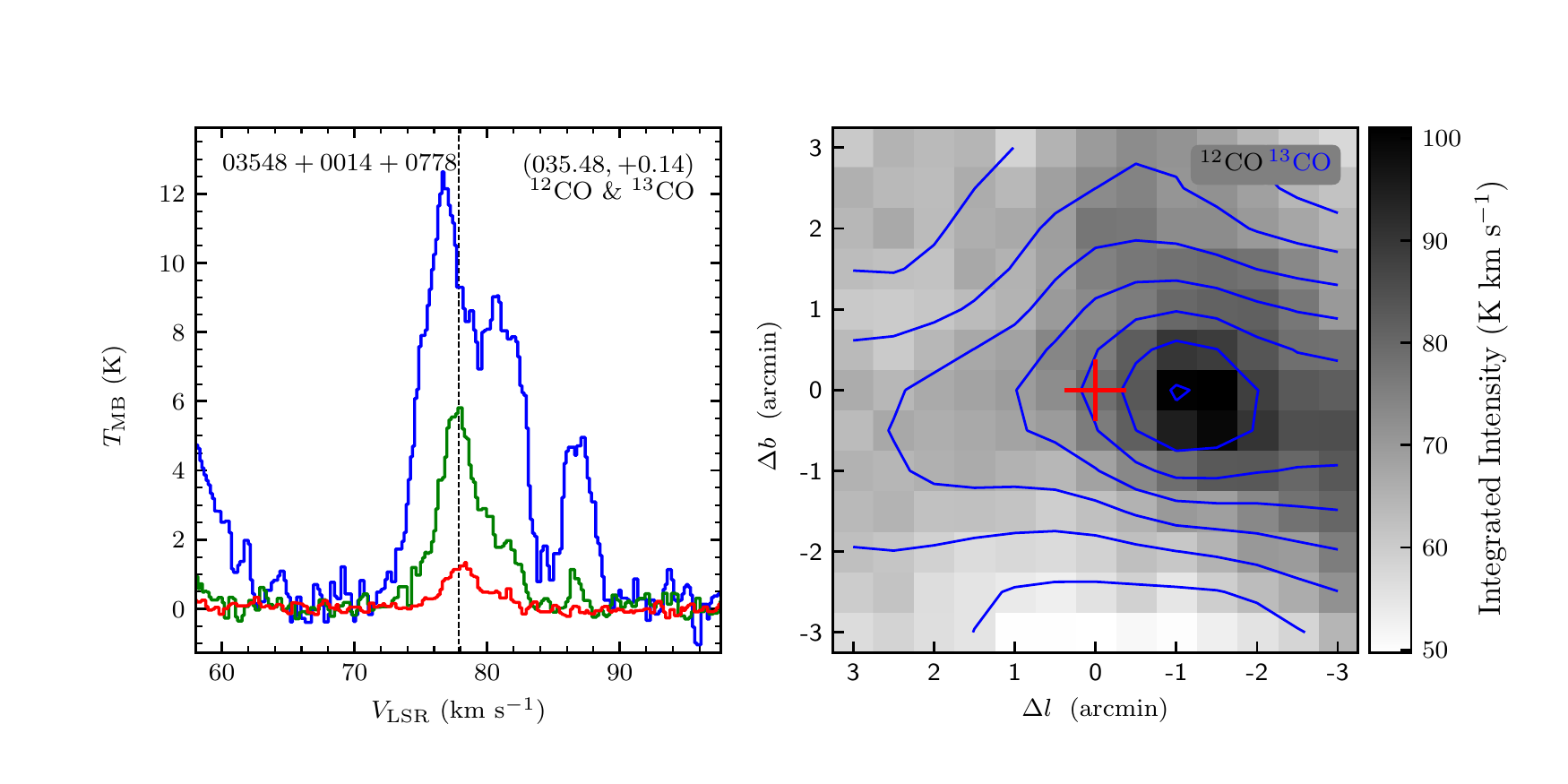}
\includegraphics[width=9.0cm,angle=0]{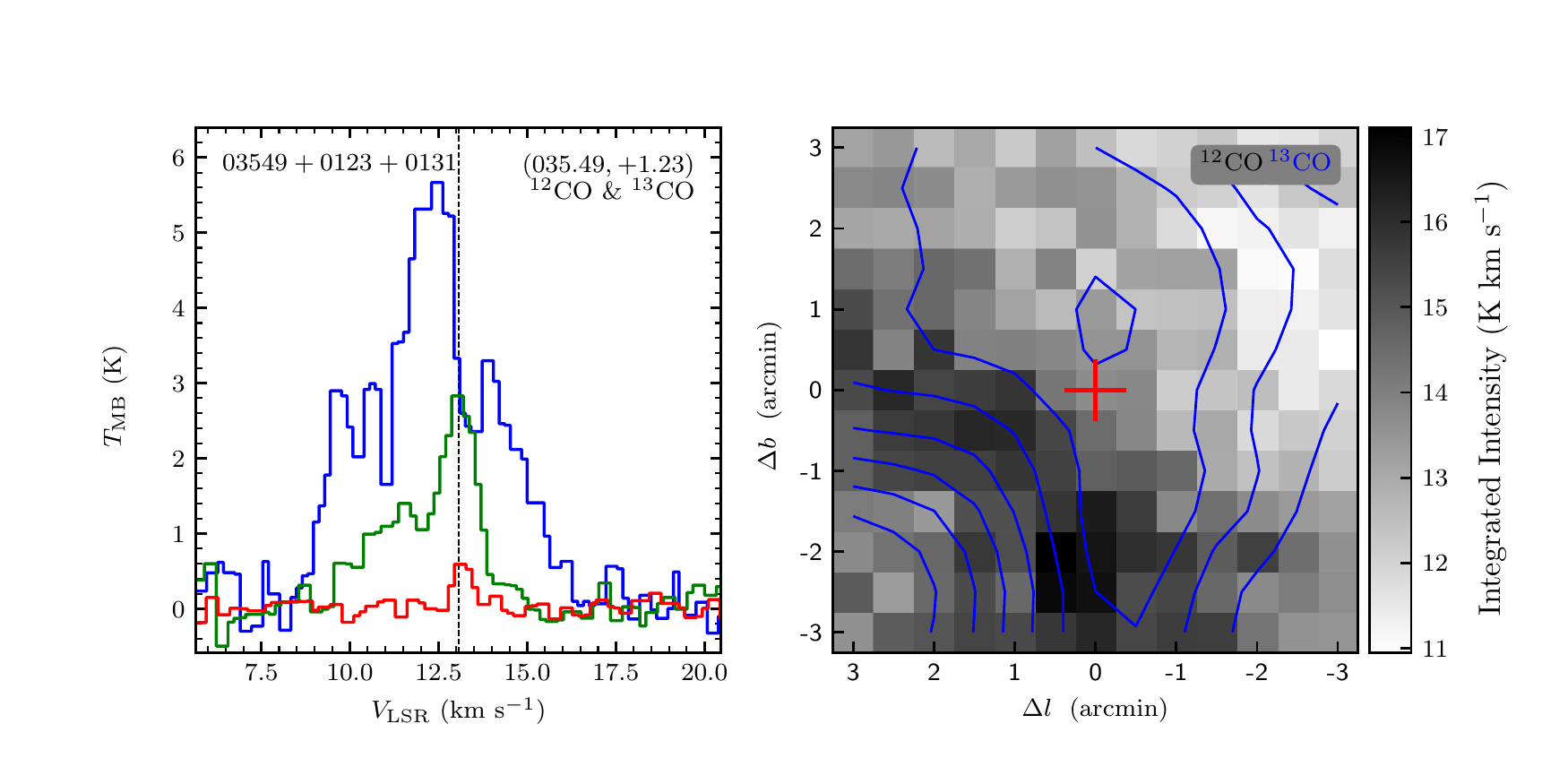}
\vspace{-0.5cm}

\includegraphics[width=9.0cm,angle=0]{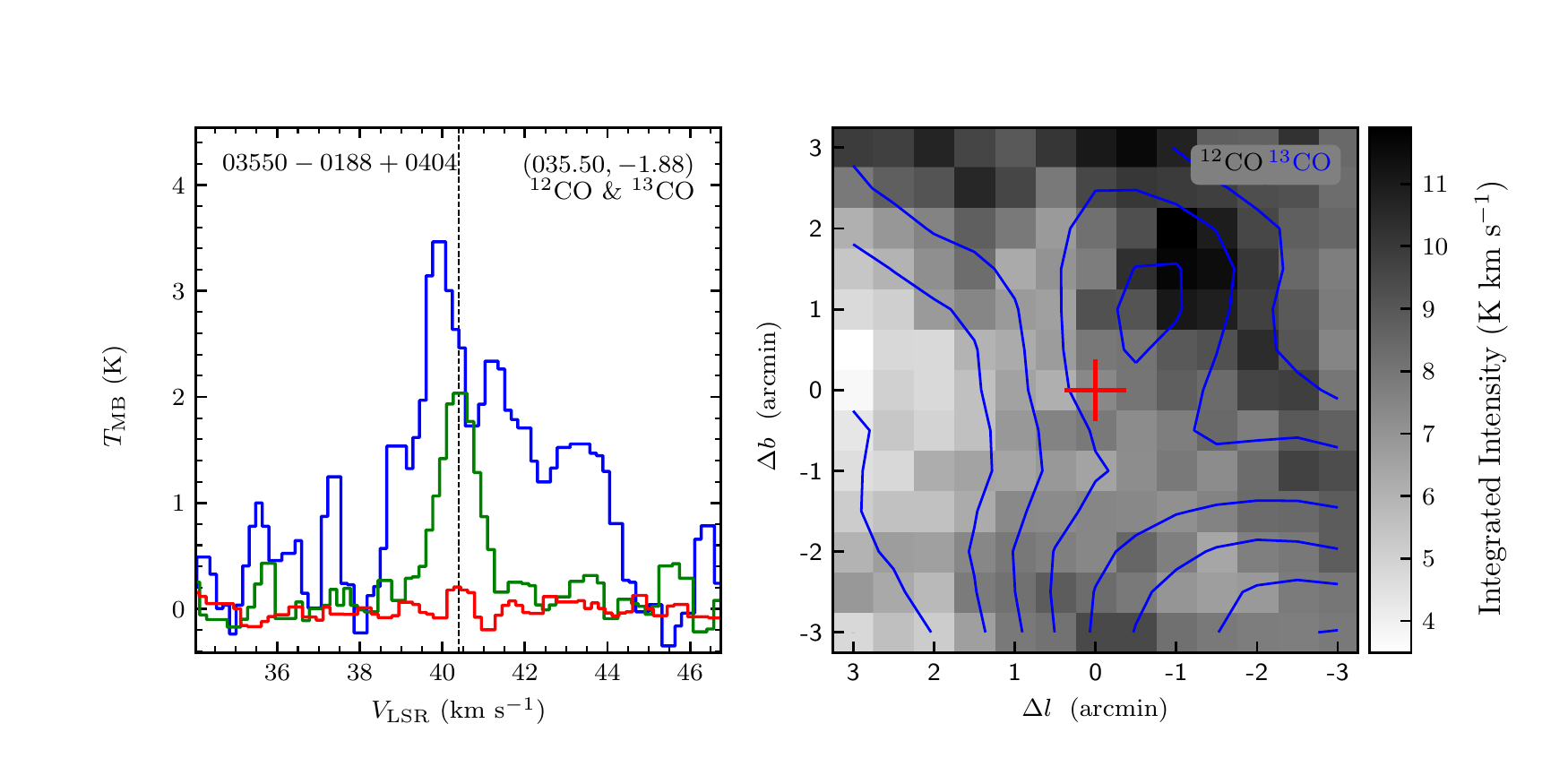}
\includegraphics[width=9.0cm,angle=0]{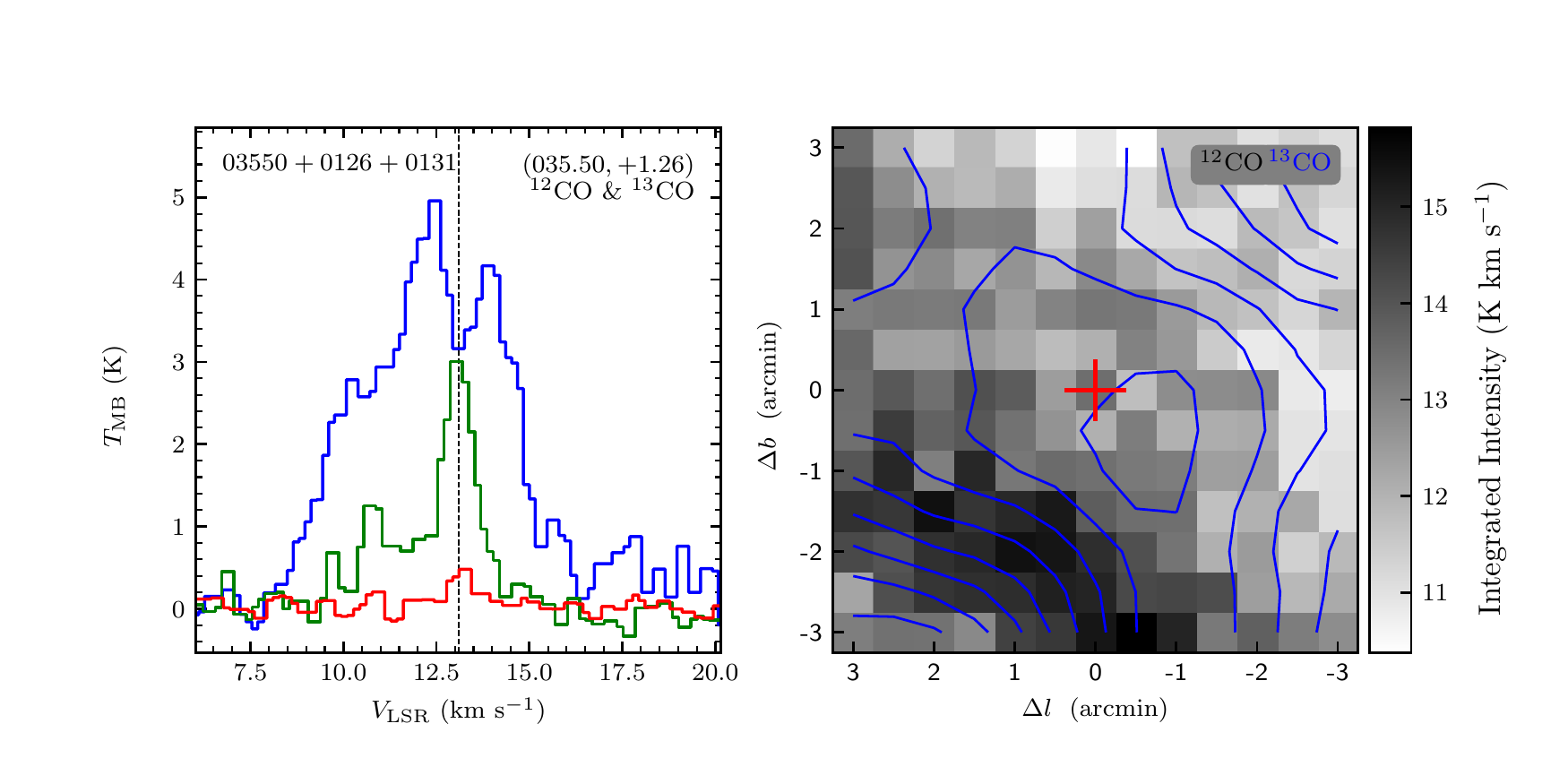}
\vspace{-0.5cm}

\includegraphics[width=9.0cm,angle=0]{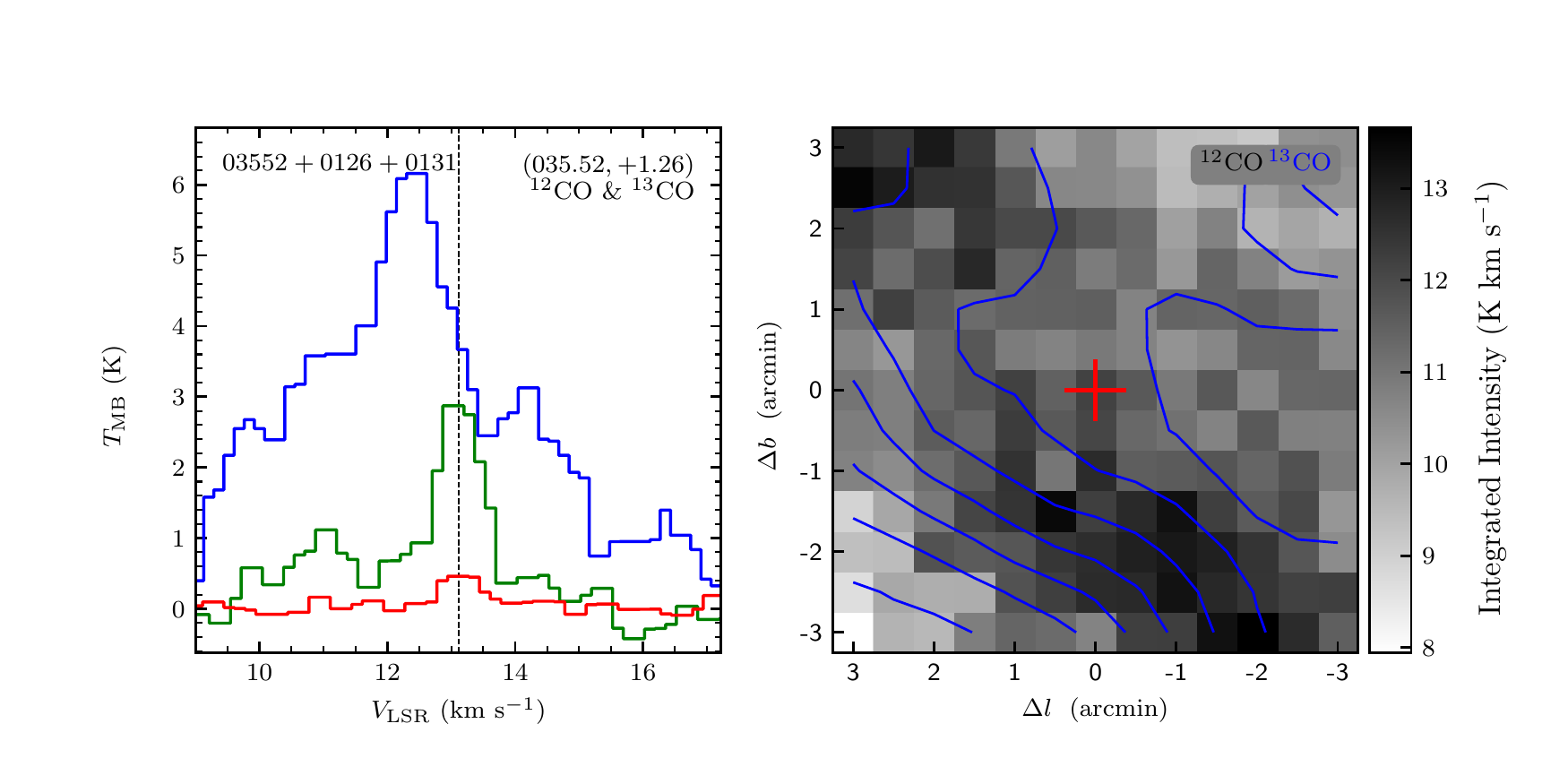}
\includegraphics[width=9.0cm,angle=0]{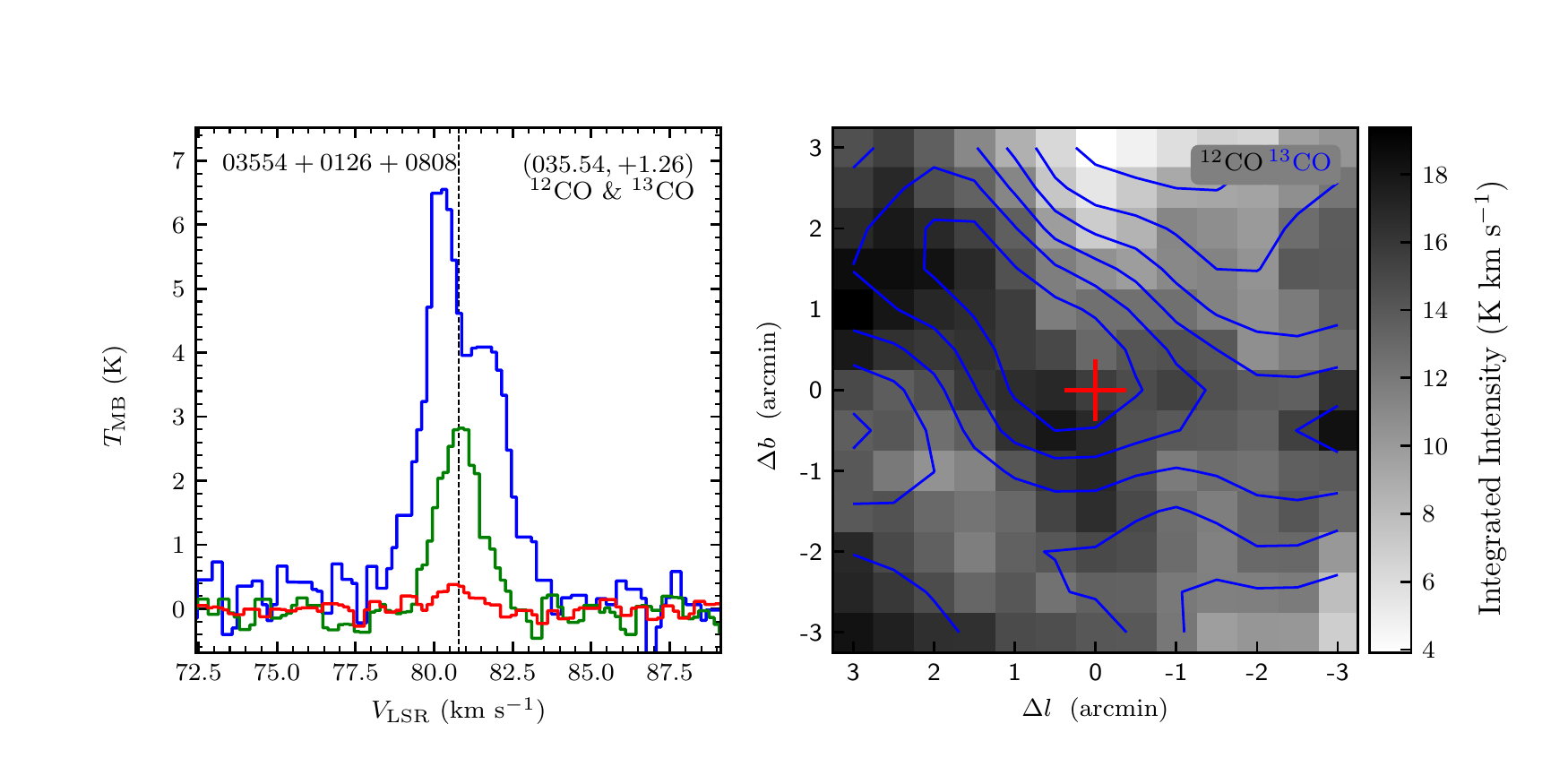}
\vspace{-0.5cm}

\includegraphics[width=9.0cm,angle=0]{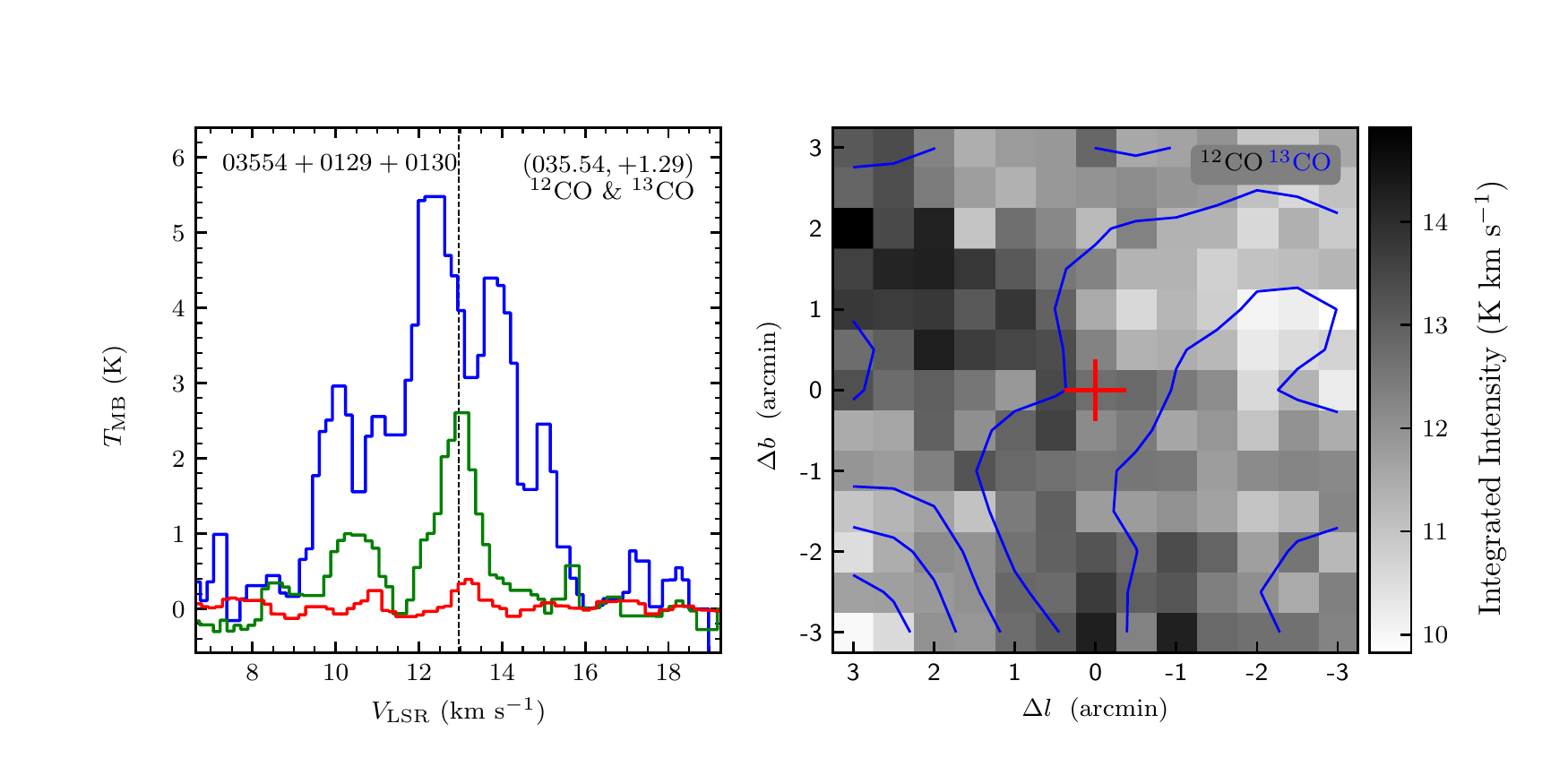}
\includegraphics[width=9.0cm,angle=0]{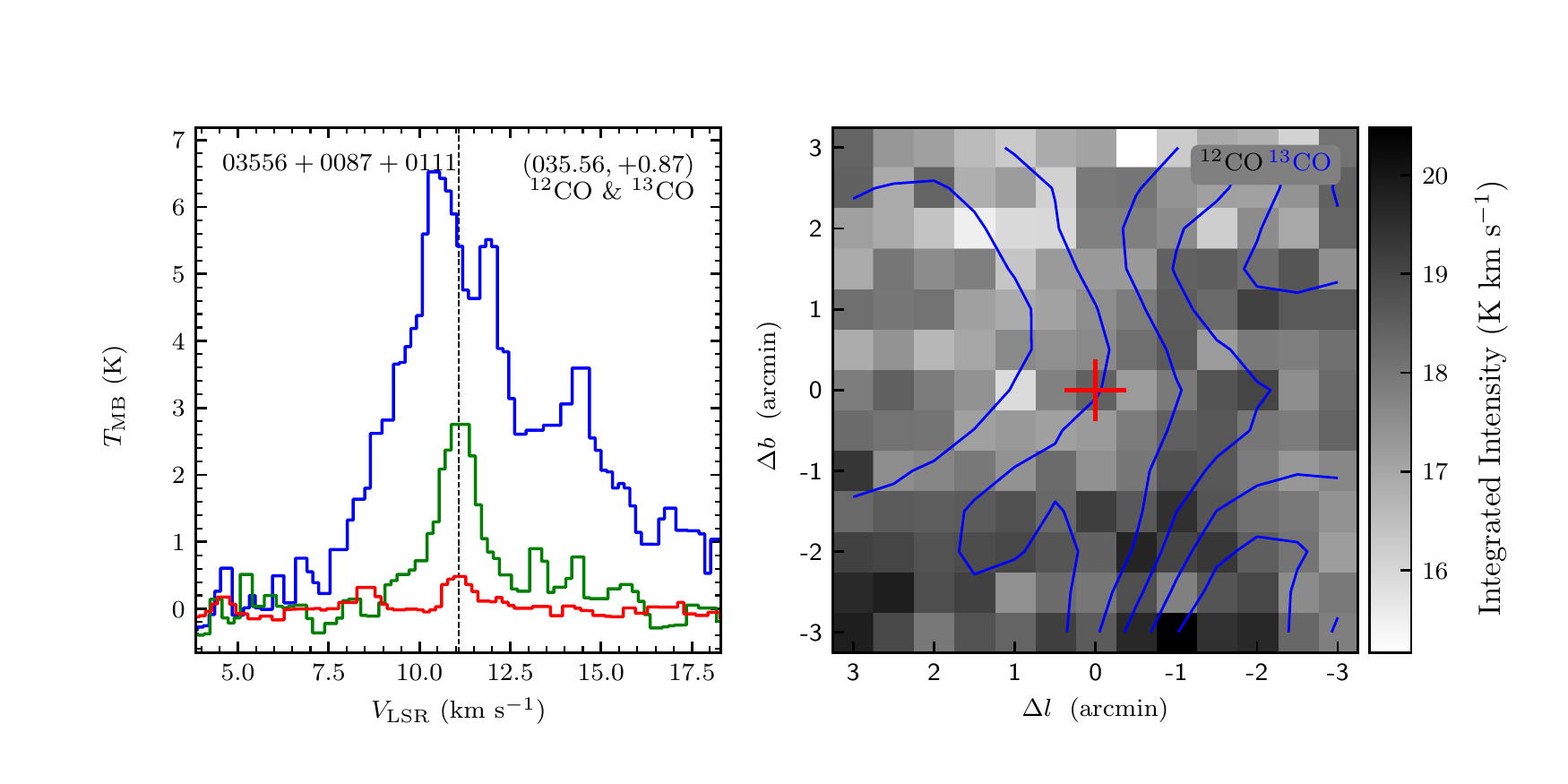}
\end{figure}
\clearpage

\begin{figure}
\includegraphics[width=9.0cm,angle=0]{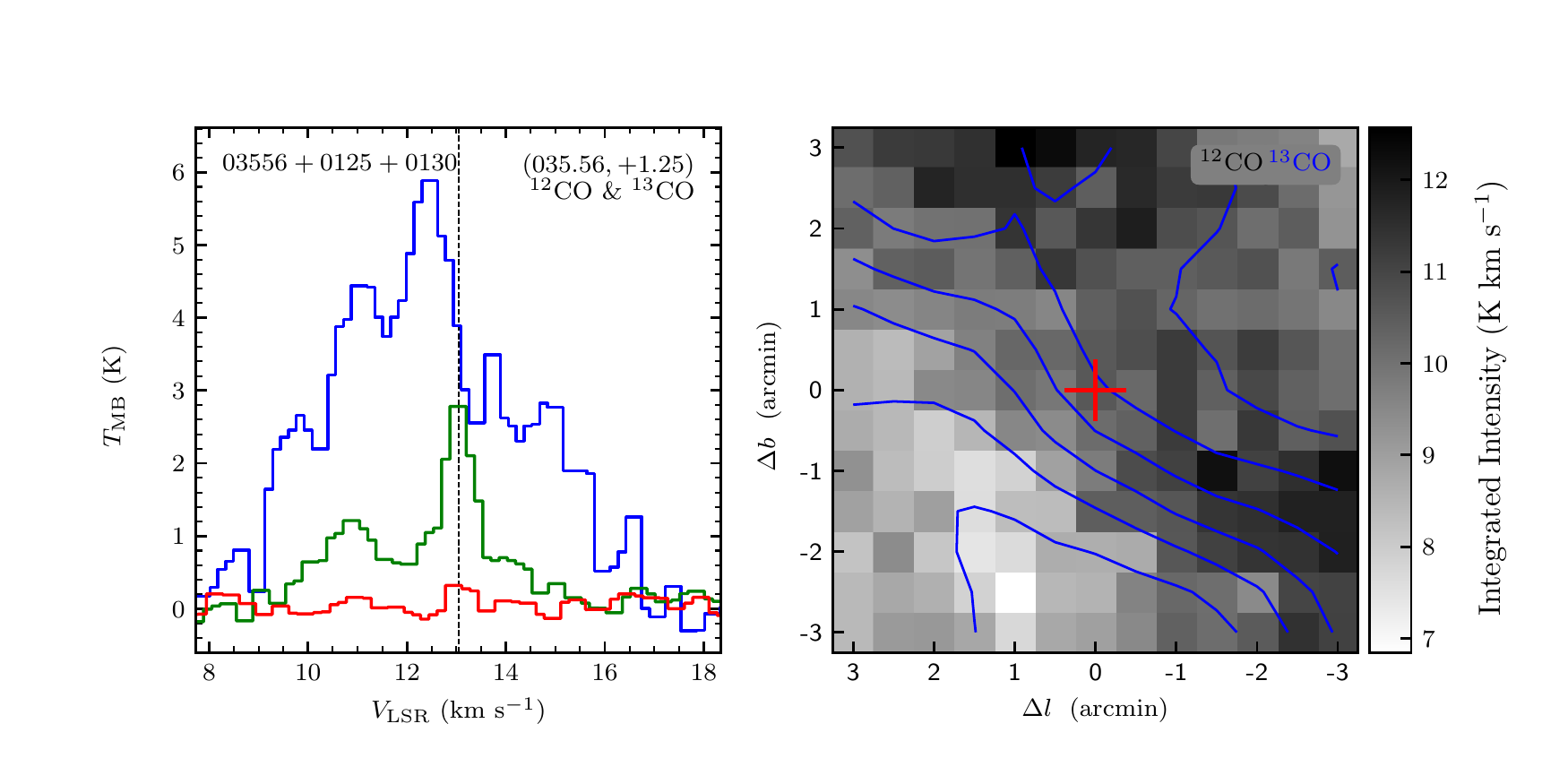}
\includegraphics[width=9.0cm,angle=0]{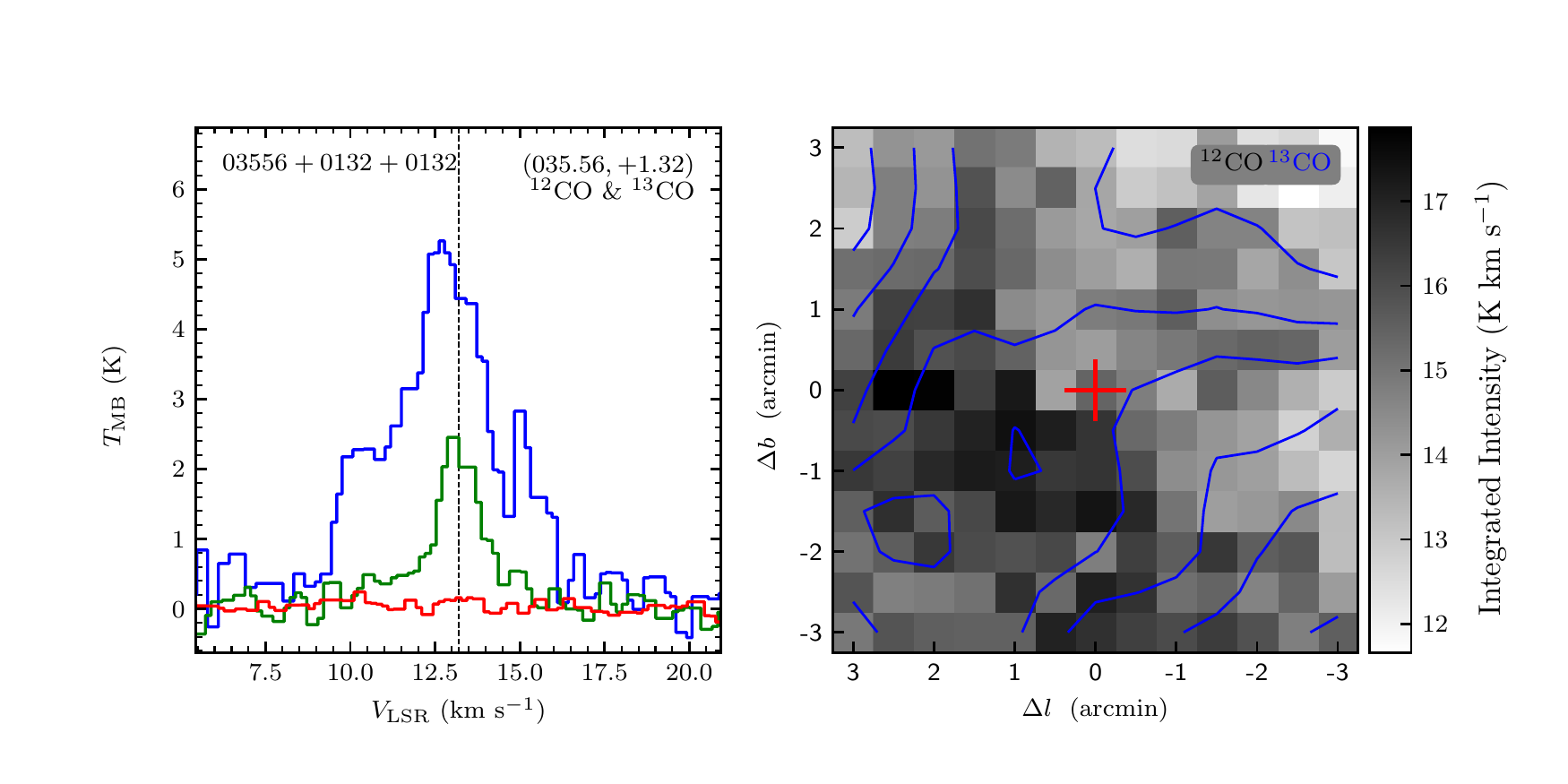}
\vspace{-0.5cm}

\includegraphics[width=9.0cm,angle=0]{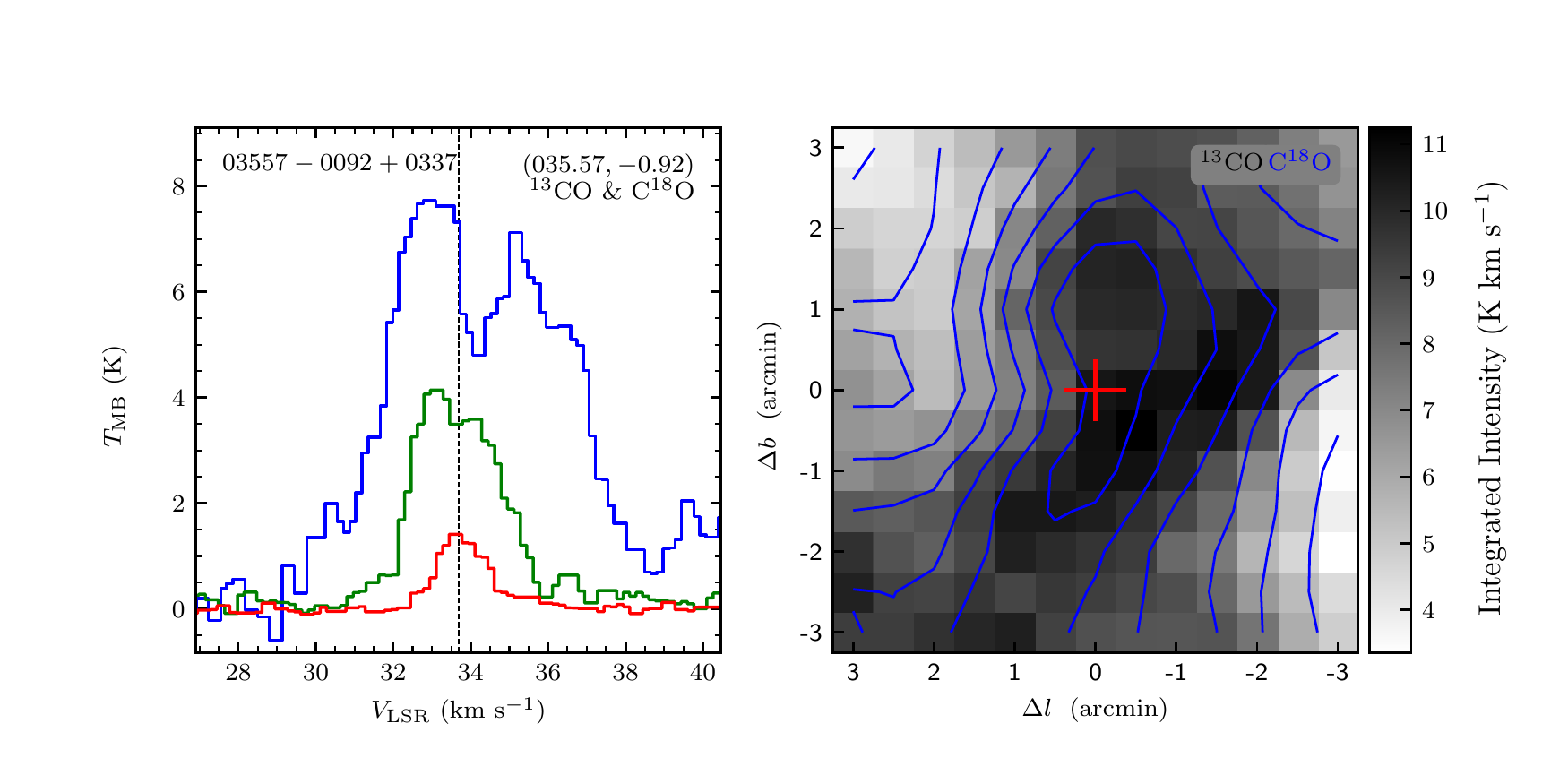}
\includegraphics[width=9.0cm,angle=0]{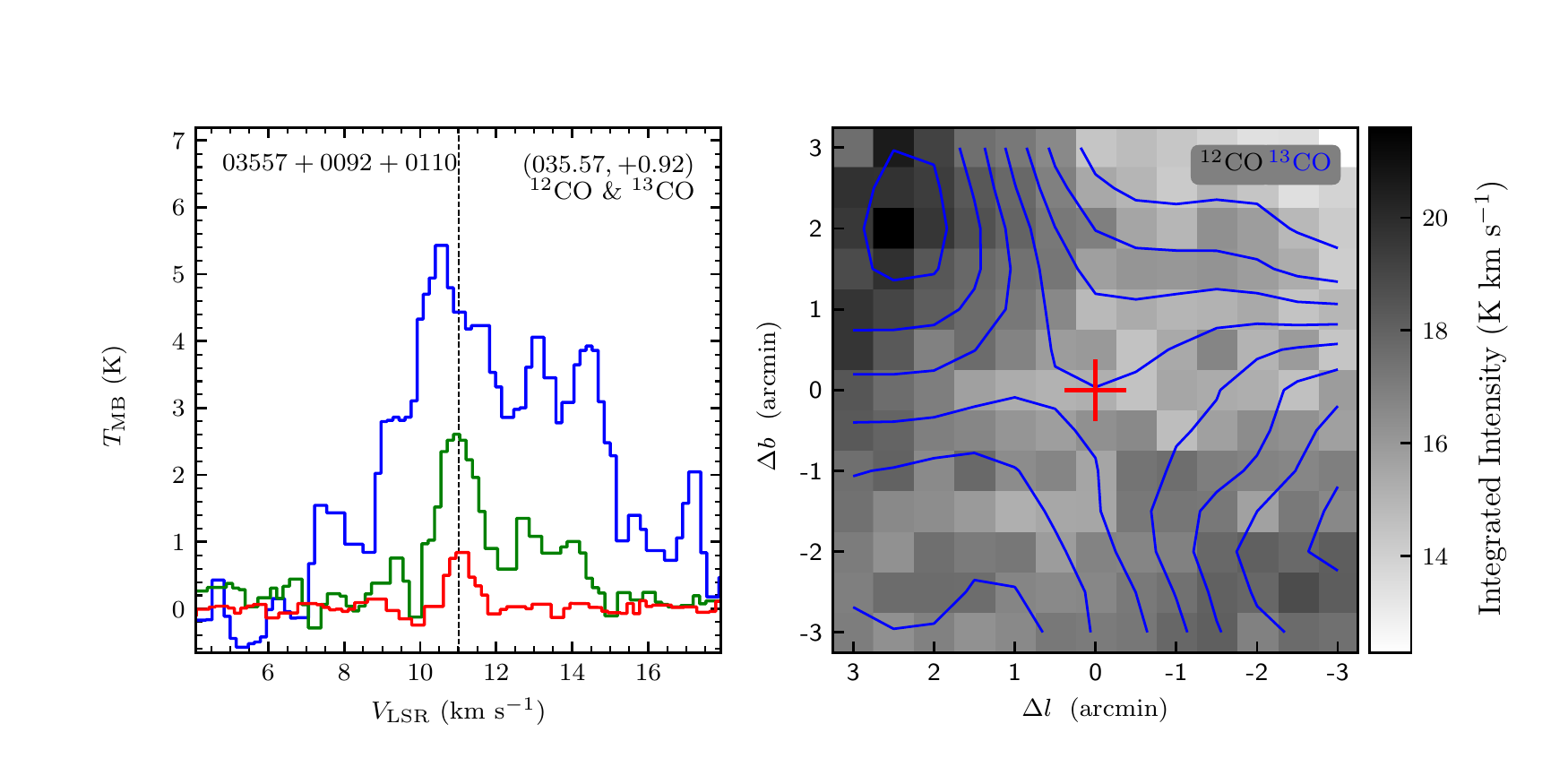}
\vspace{-0.5cm}

\includegraphics[width=9.0cm,angle=0]{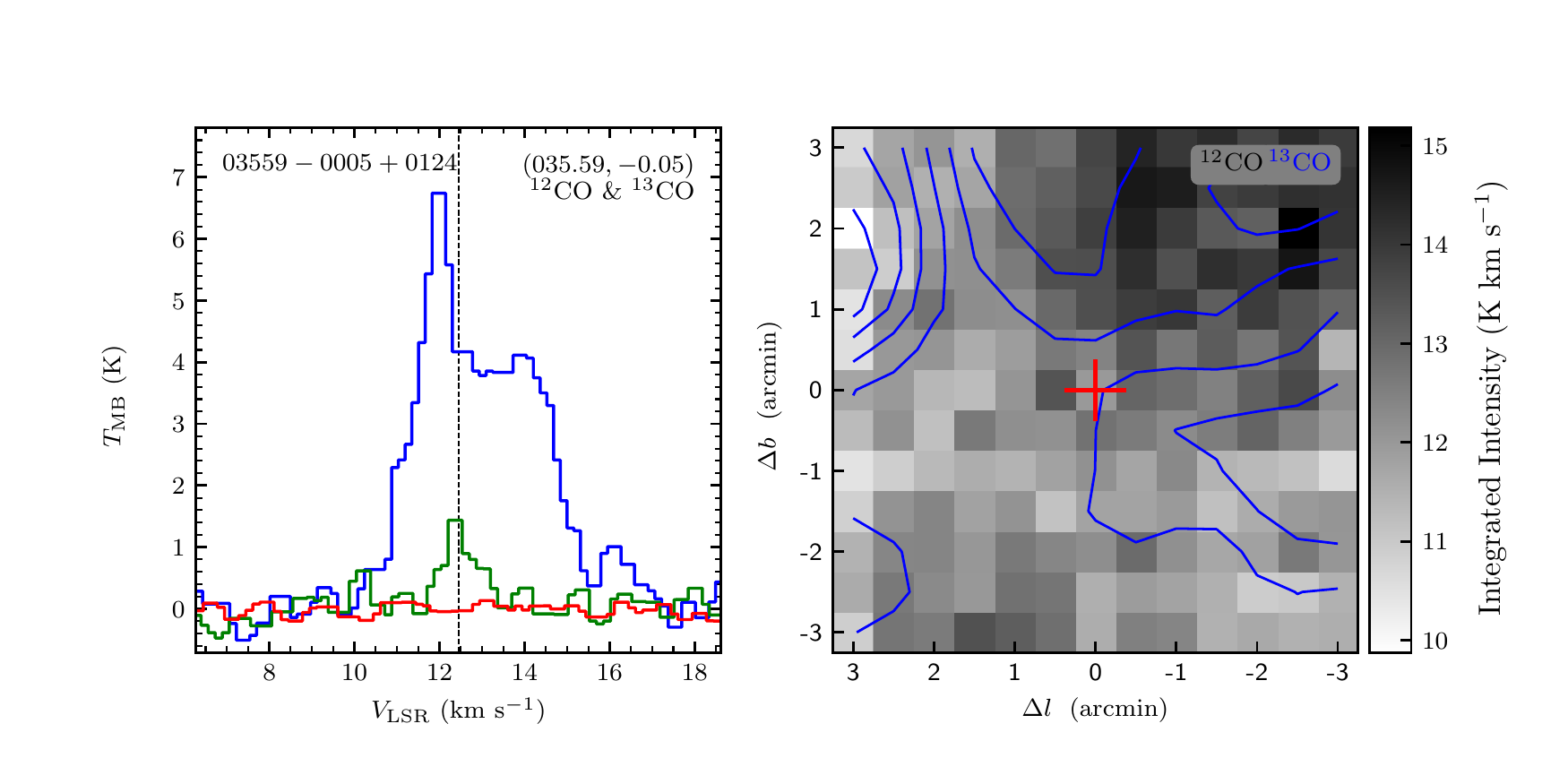}
\includegraphics[width=9.0cm,angle=0]{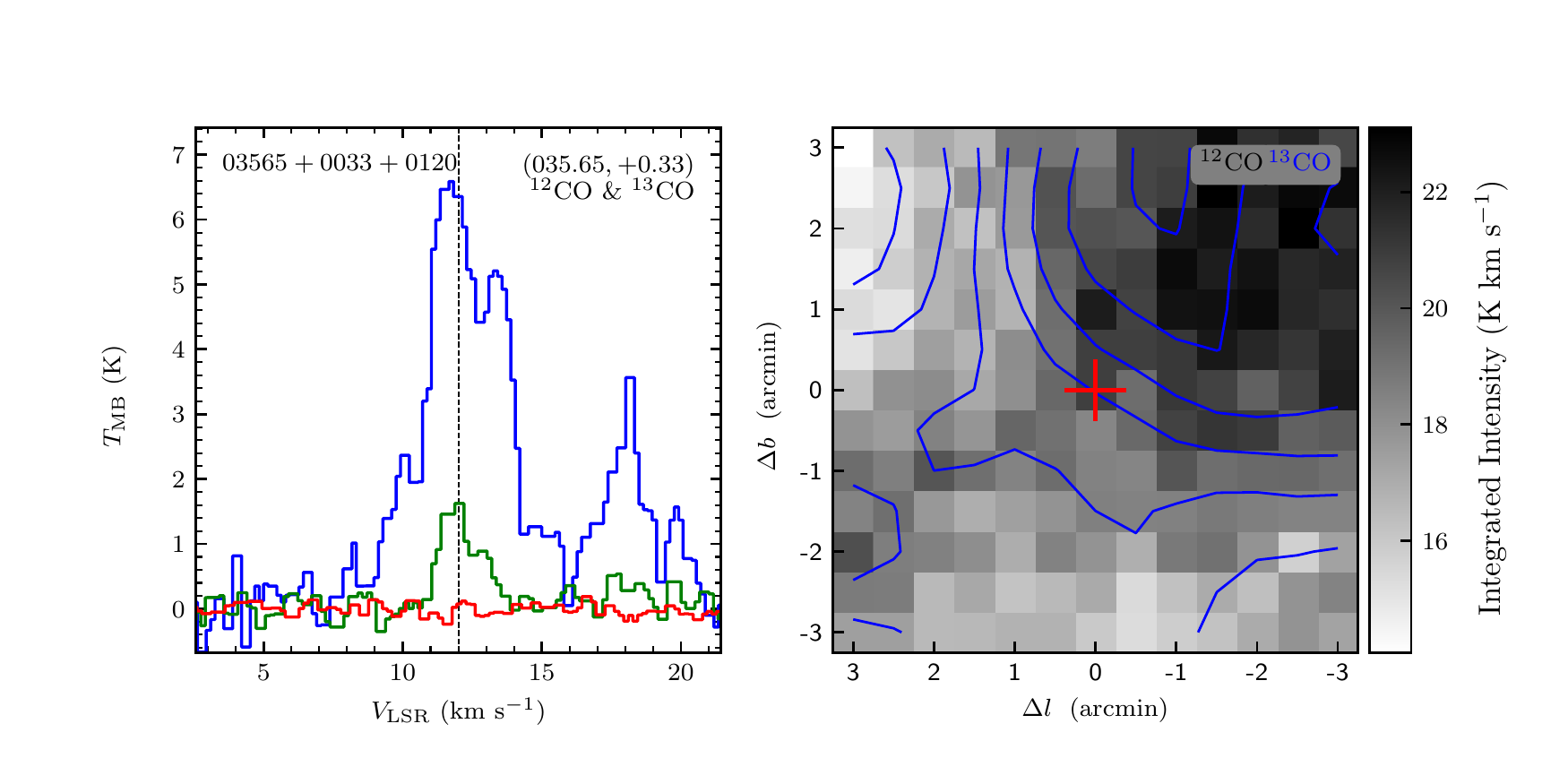}
\vspace{-0.5cm}

\includegraphics[width=9.0cm,angle=0]{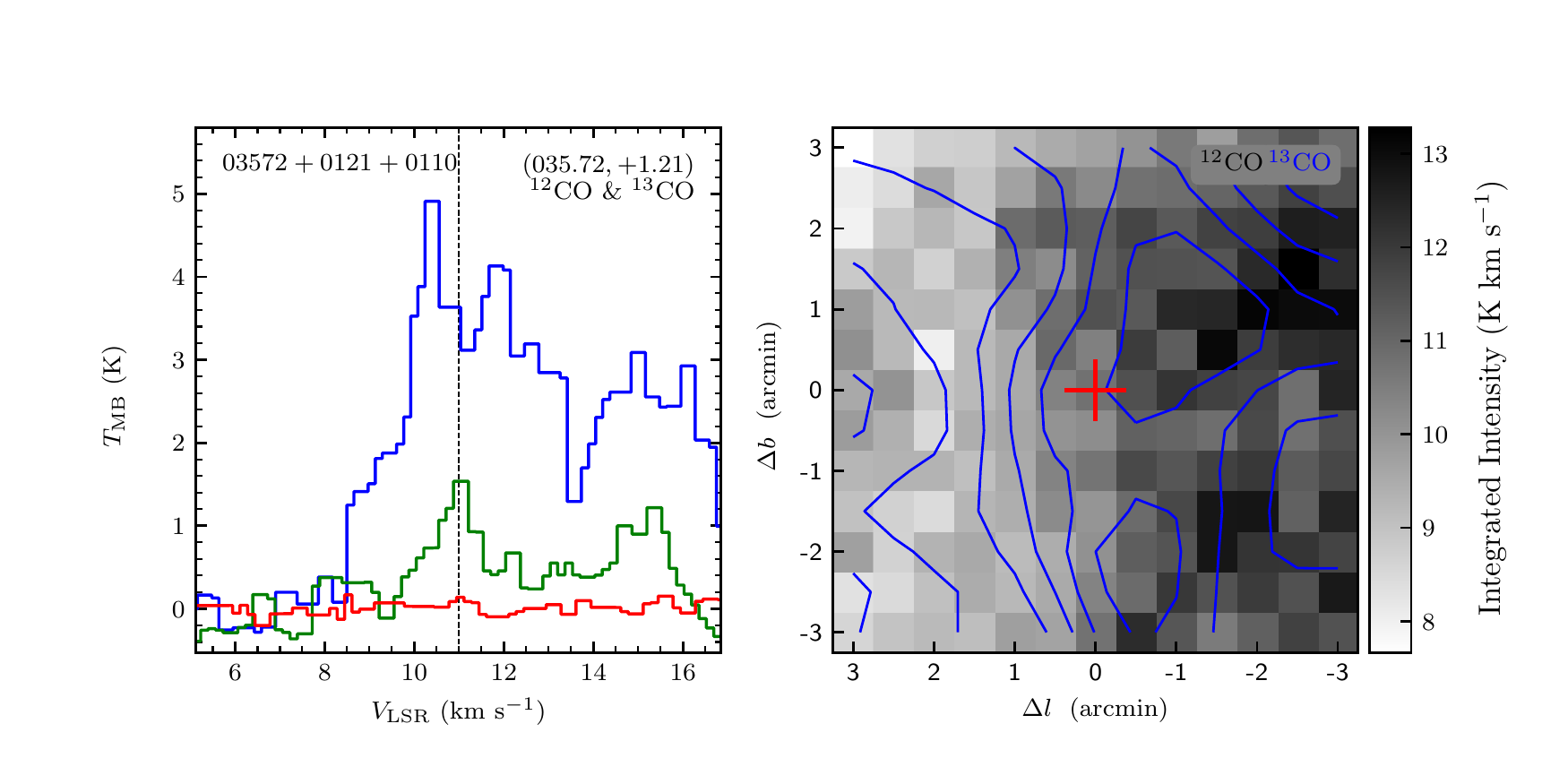}
\includegraphics[width=9.0cm,angle=0]{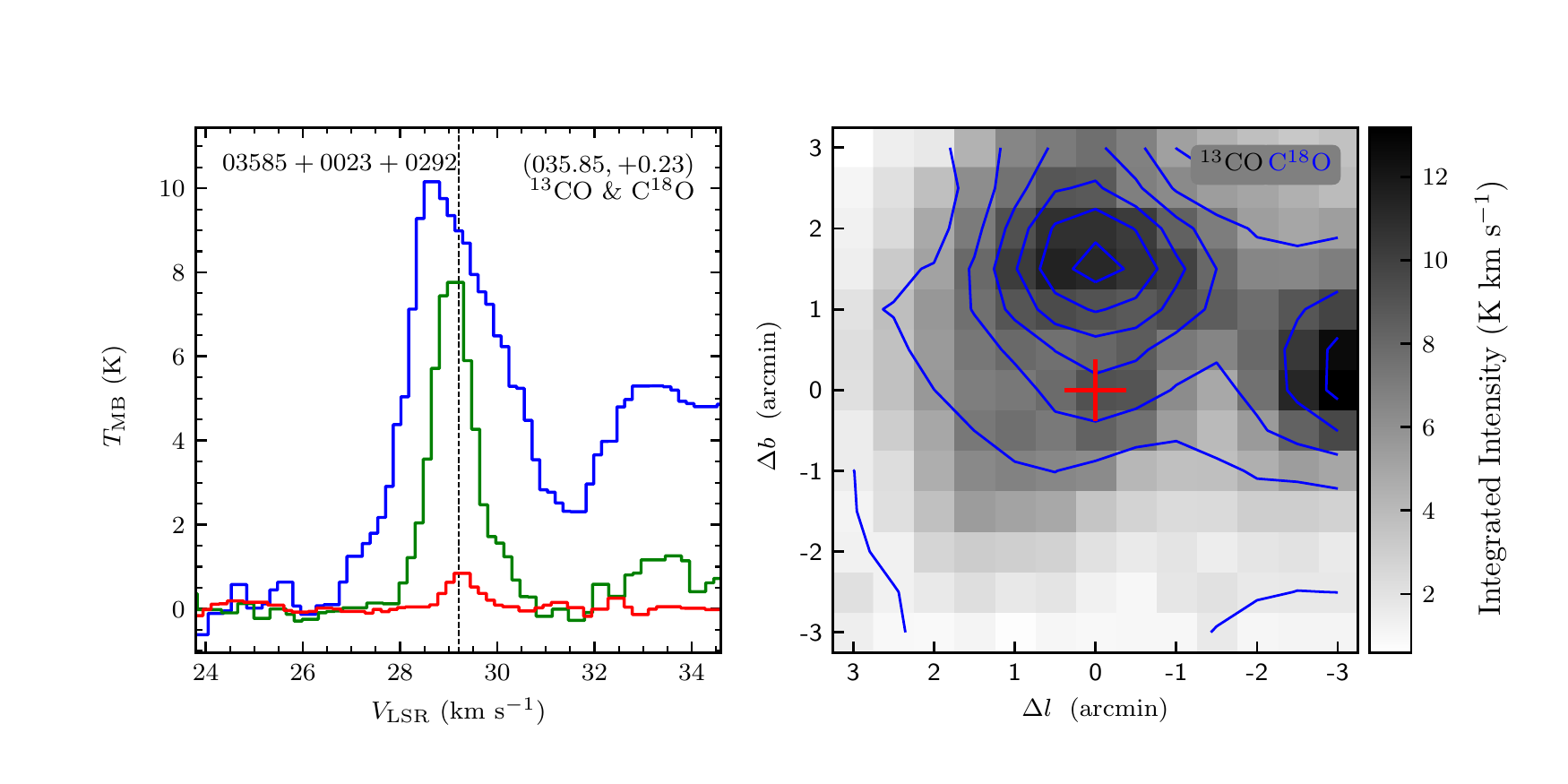}
\vspace{-0.5cm}

\includegraphics[width=9.0cm,angle=0]{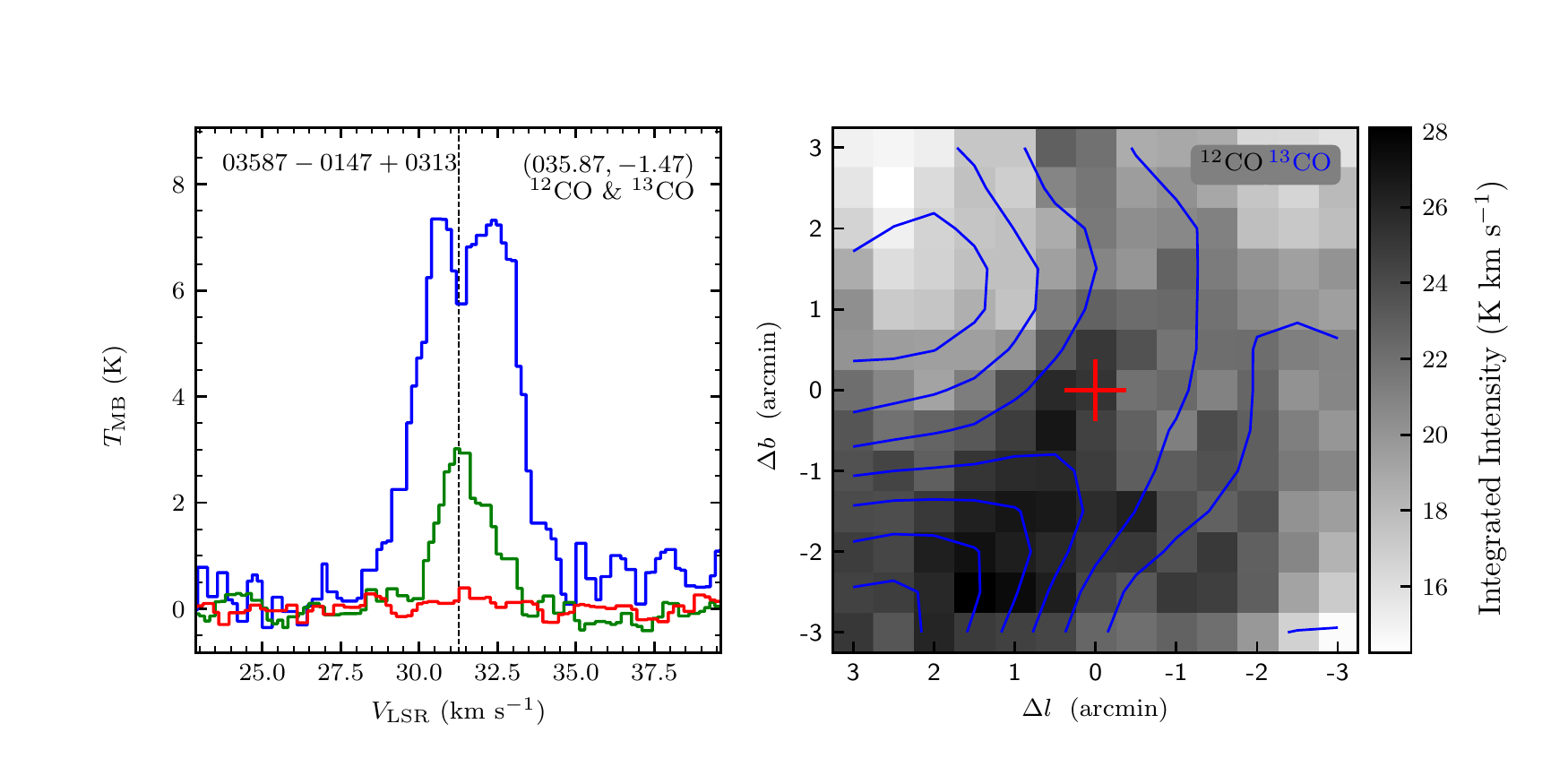}
\includegraphics[width=9.0cm,angle=0]{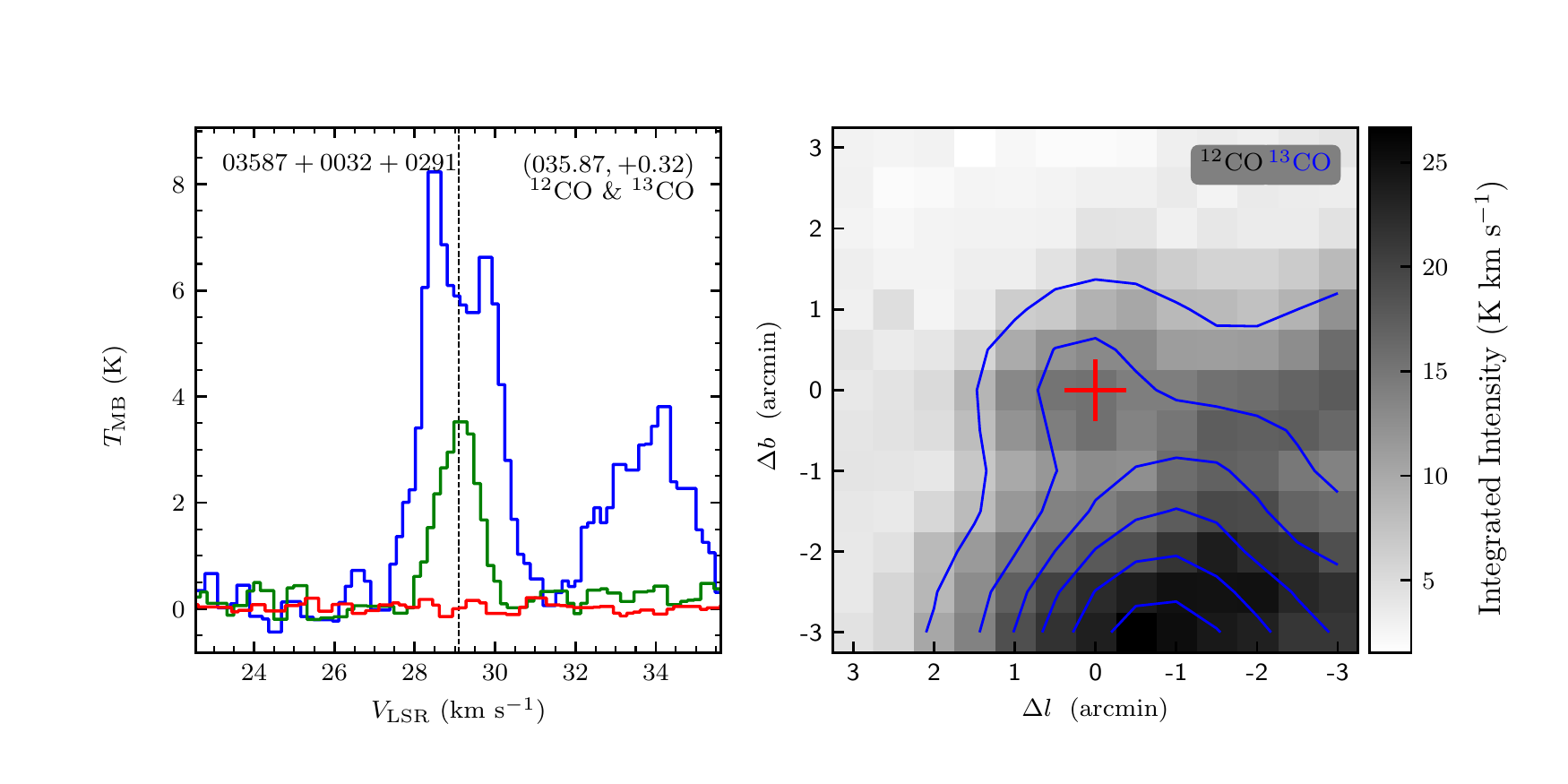}
\end{figure}
\clearpage

\begin{figure}
\includegraphics[width=9.0cm,angle=0]{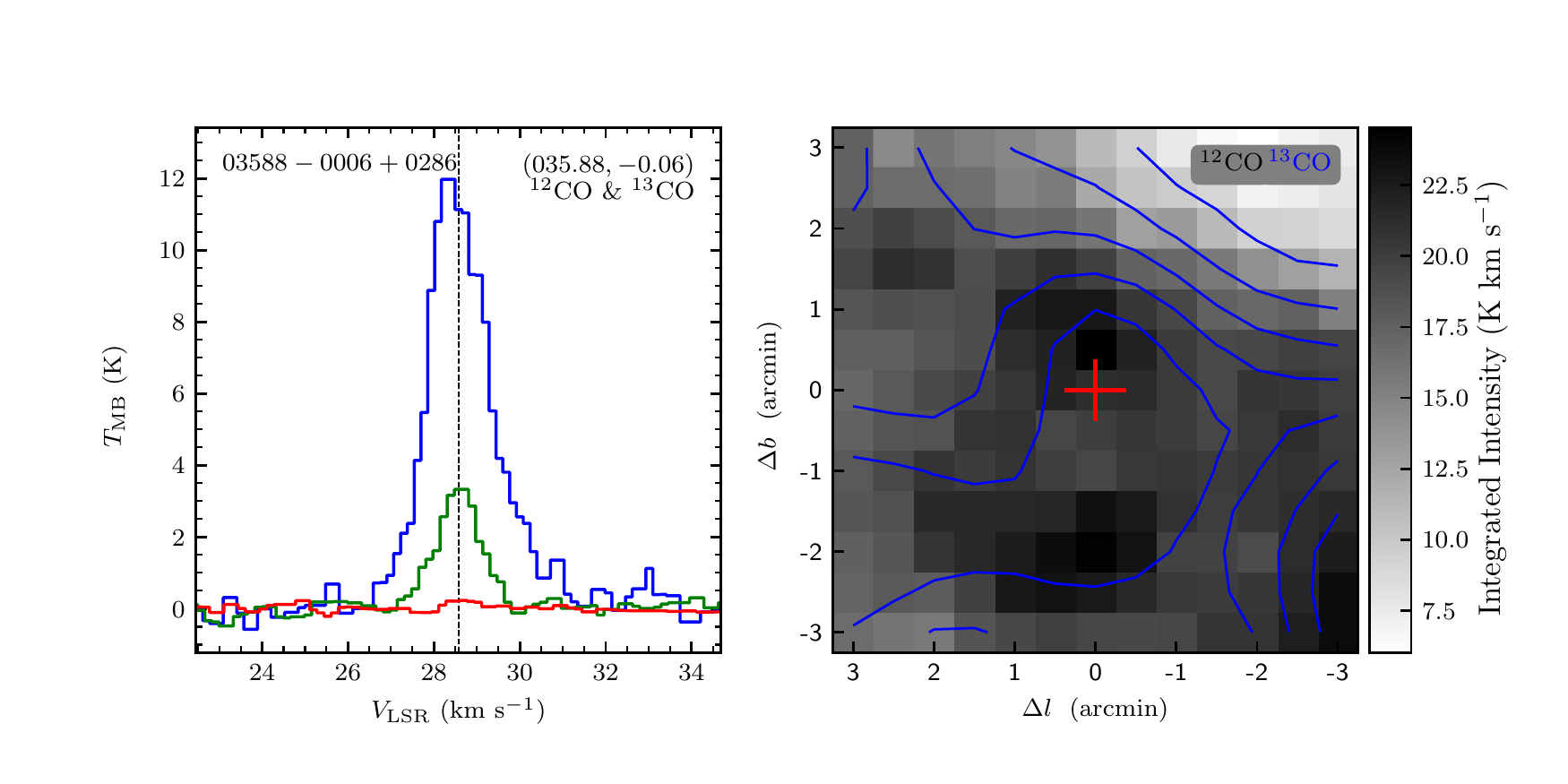}
\includegraphics[width=9.0cm,angle=0]{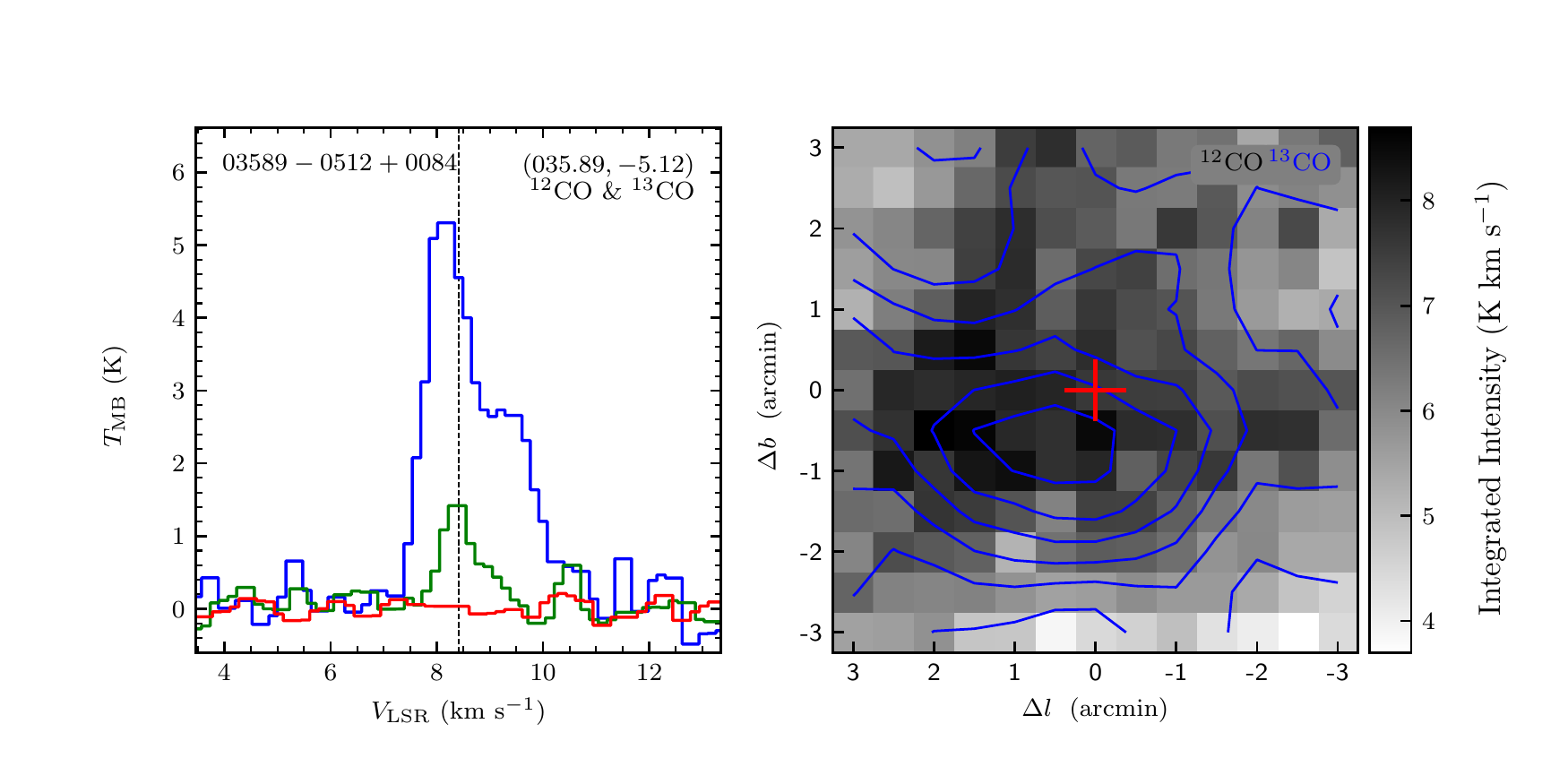}
\vspace{-0.5cm}

\includegraphics[width=9.0cm,angle=0]{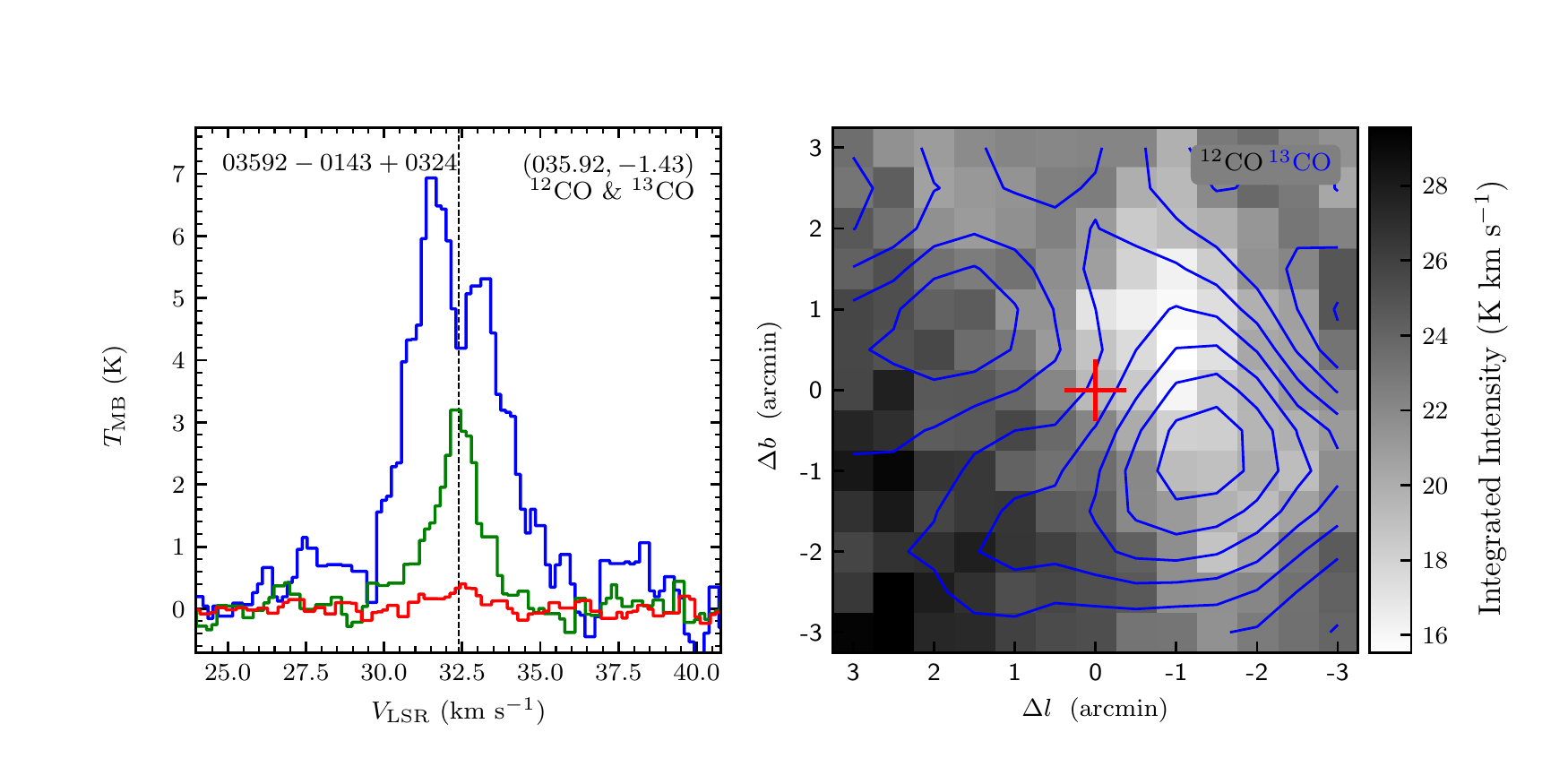}
\includegraphics[width=9.0cm,angle=0]{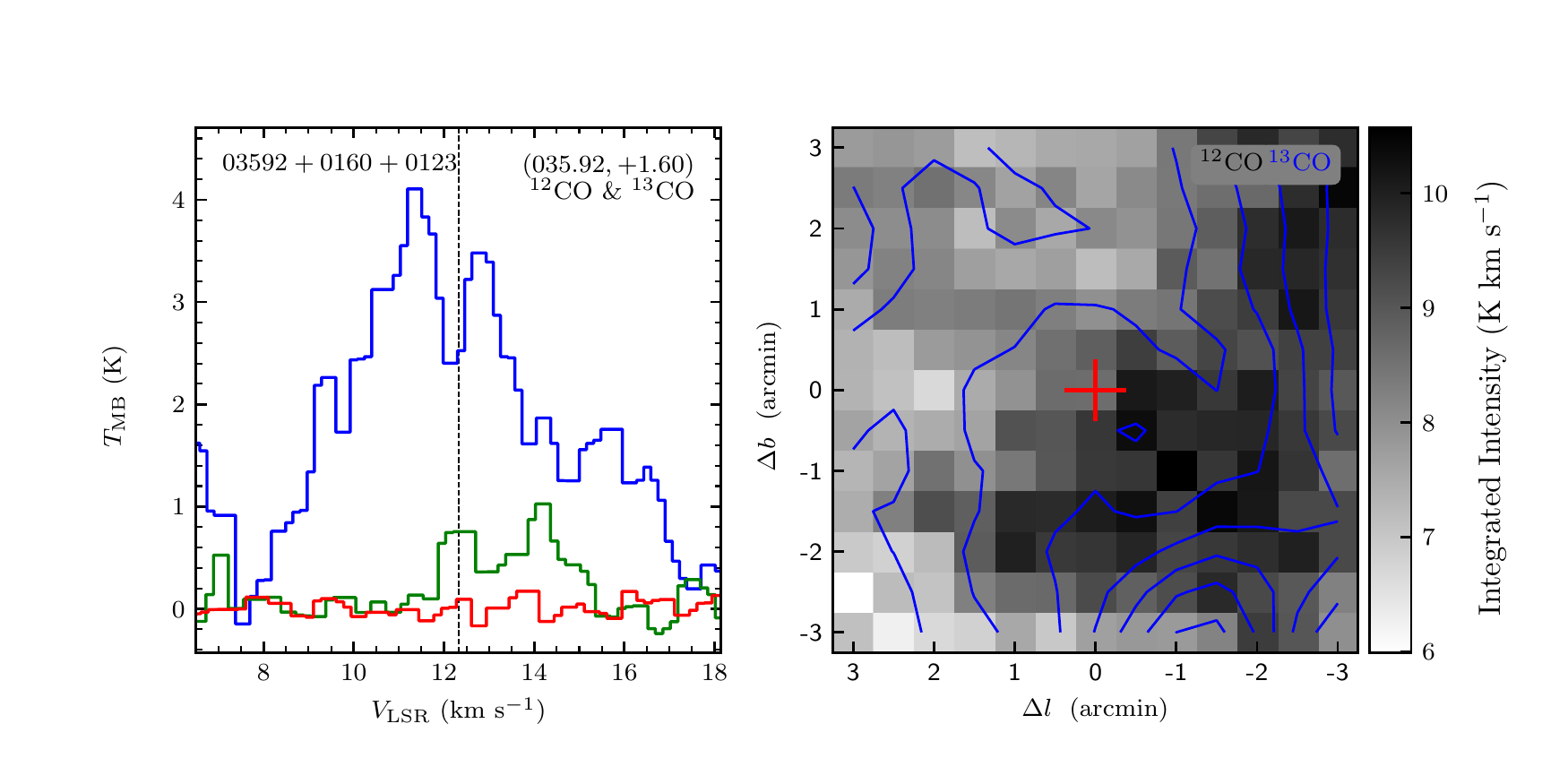}
\vspace{-0.5cm}

\includegraphics[width=9.0cm,angle=0]{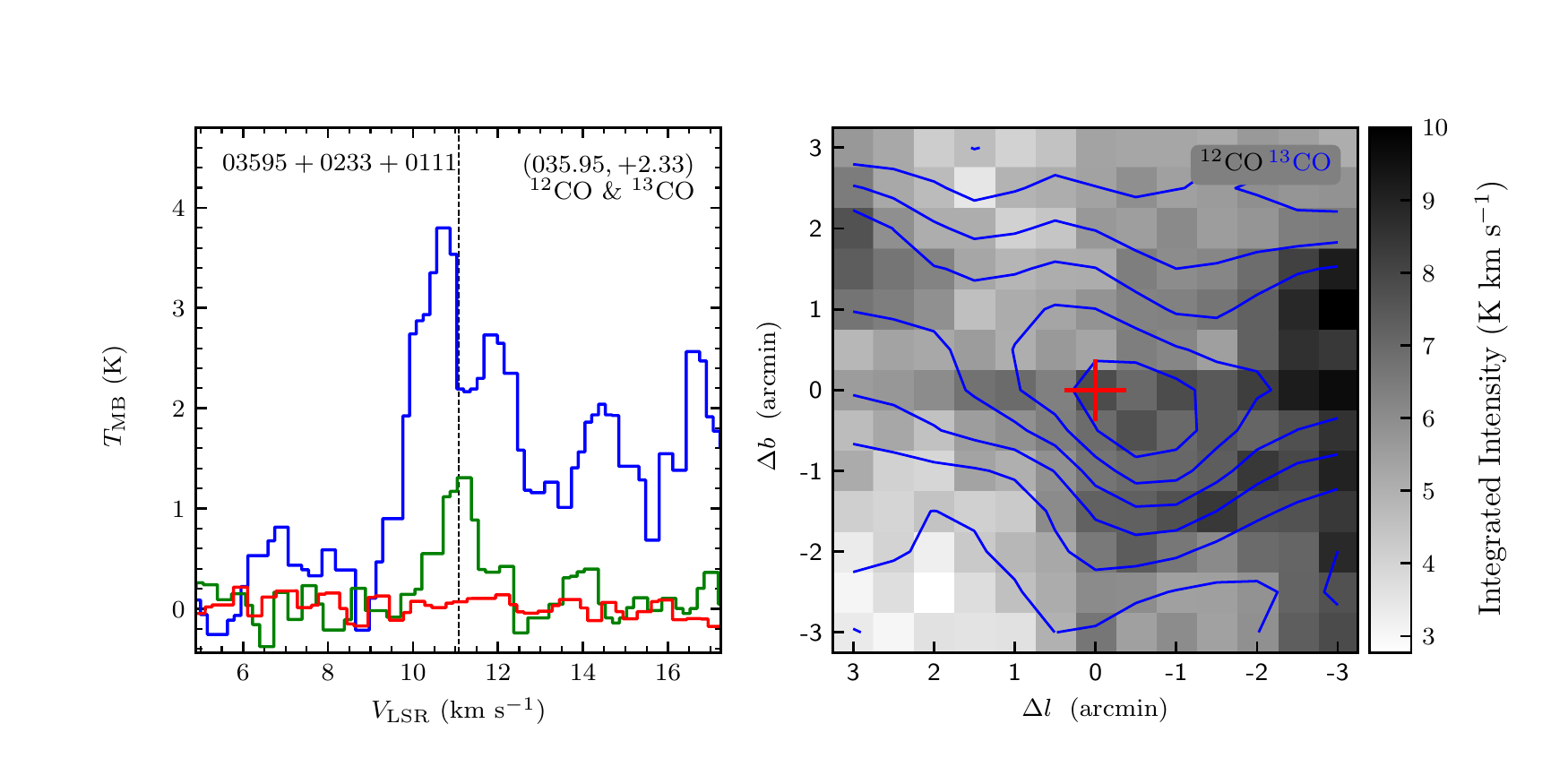}
\includegraphics[width=9.0cm,angle=0]{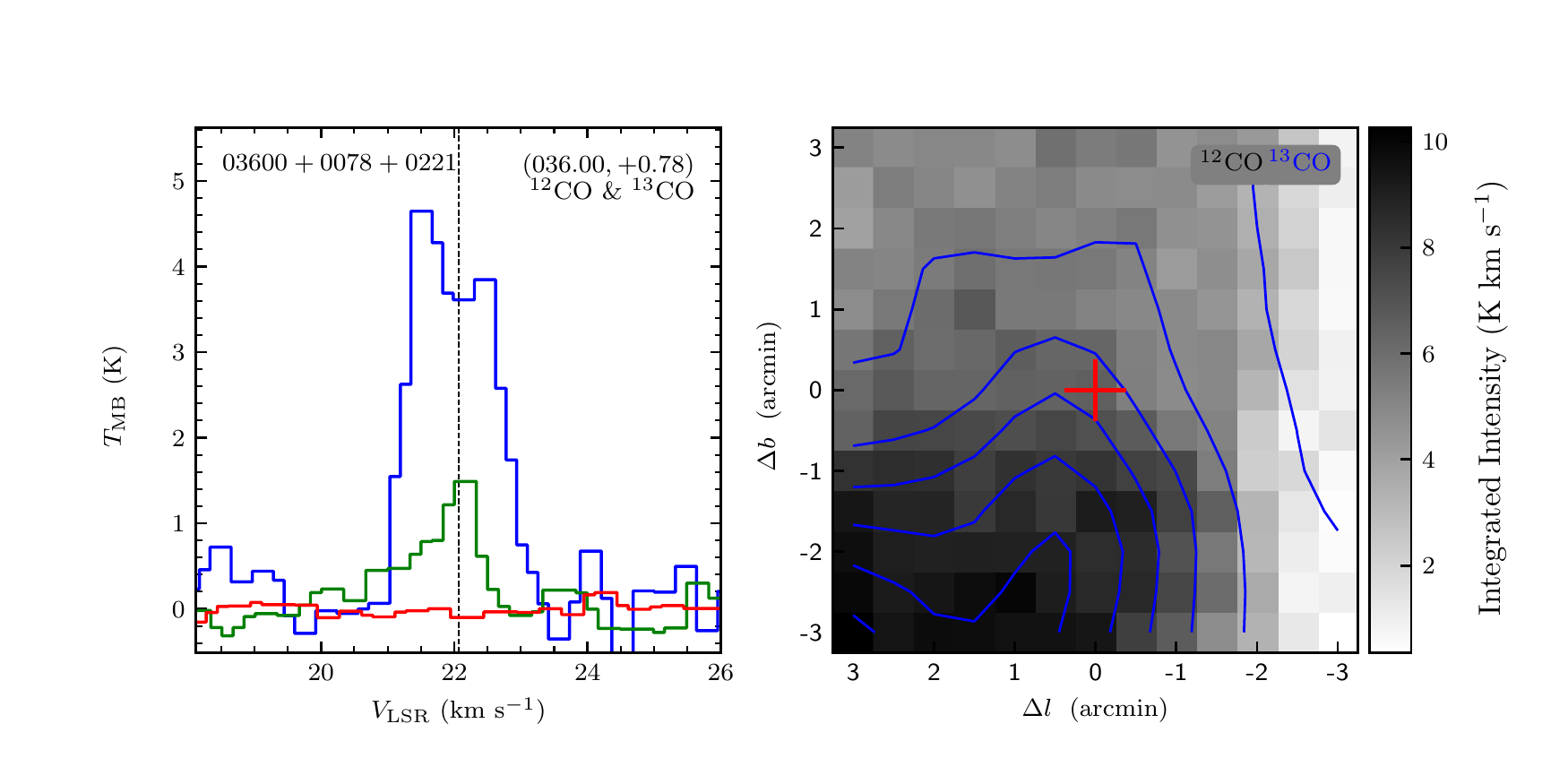}
\vspace{-0.5cm}

\includegraphics[width=9.0cm,angle=0]{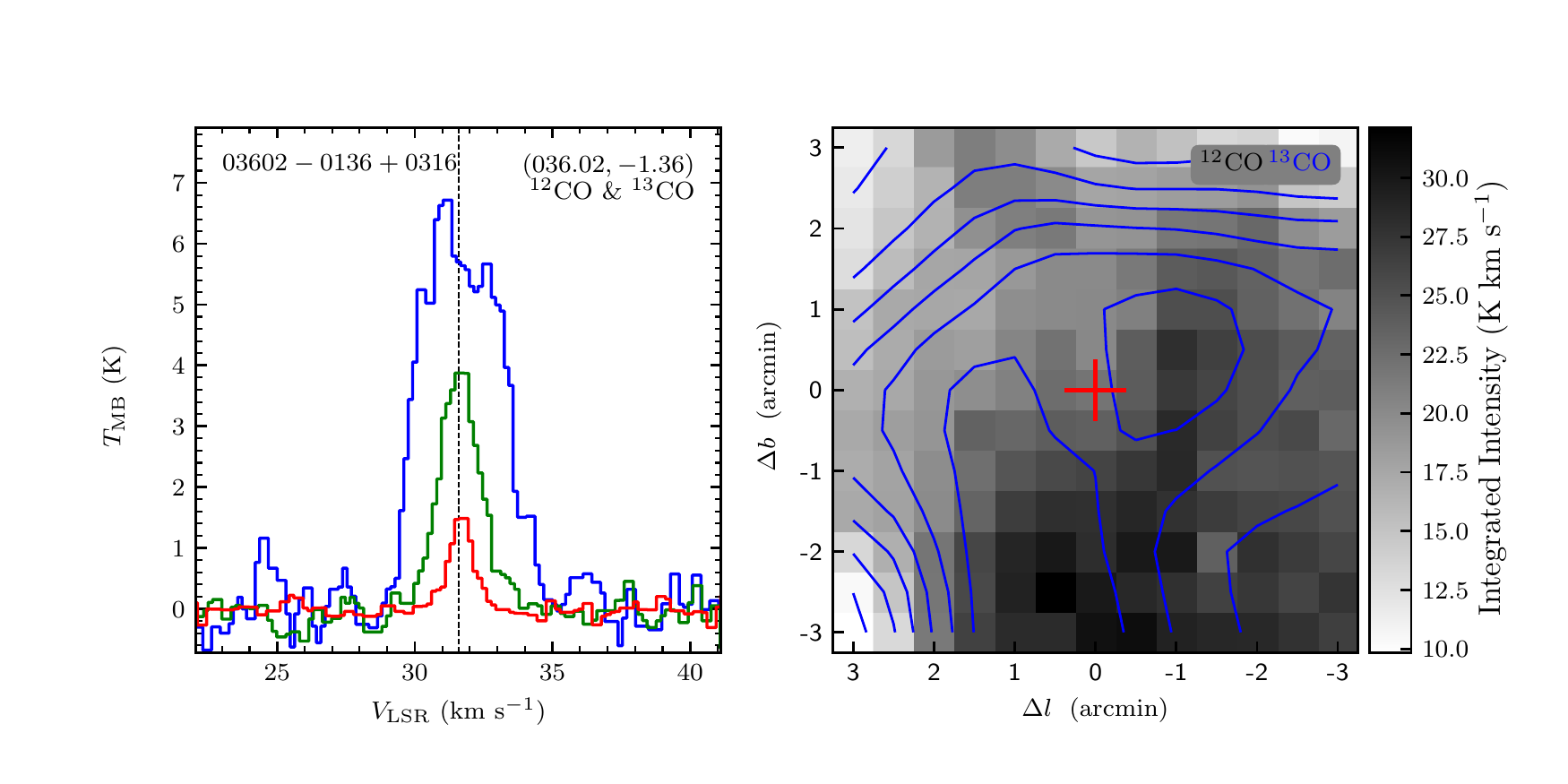}
\includegraphics[width=9.0cm,angle=0]{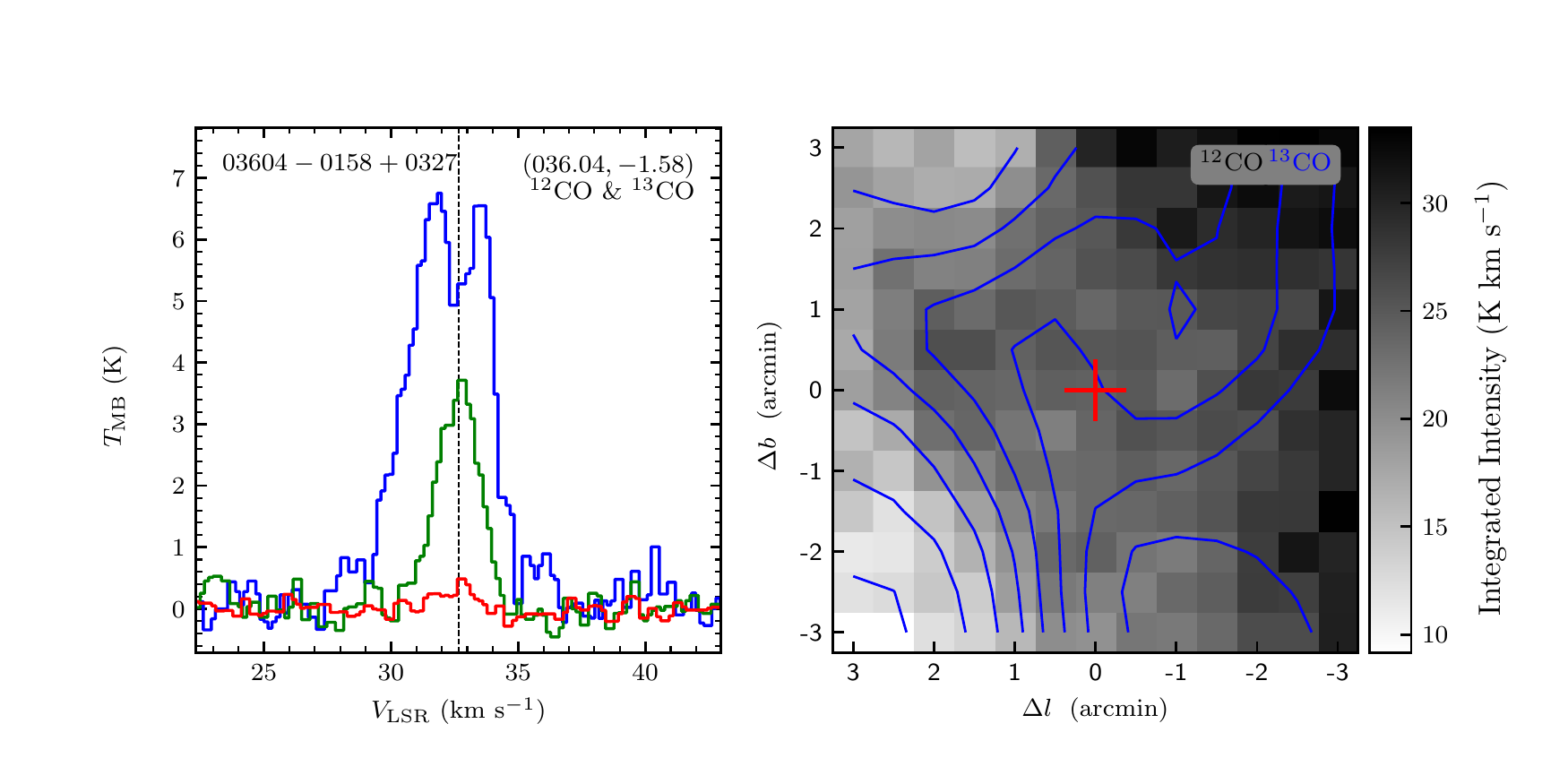}
\vspace{-0.5cm}

\includegraphics[width=9.0cm,angle=0]{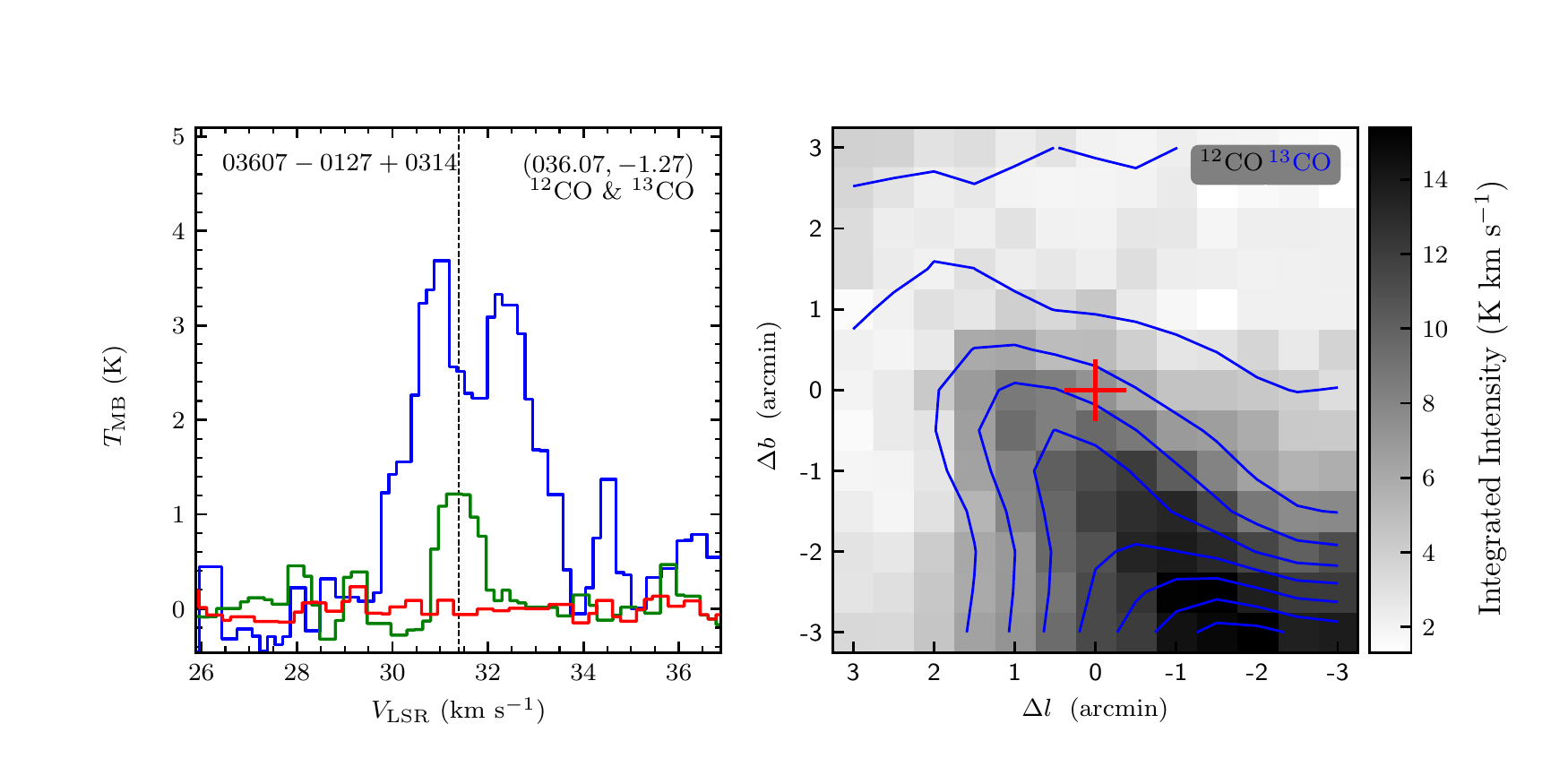}
\includegraphics[width=9.0cm,angle=0]{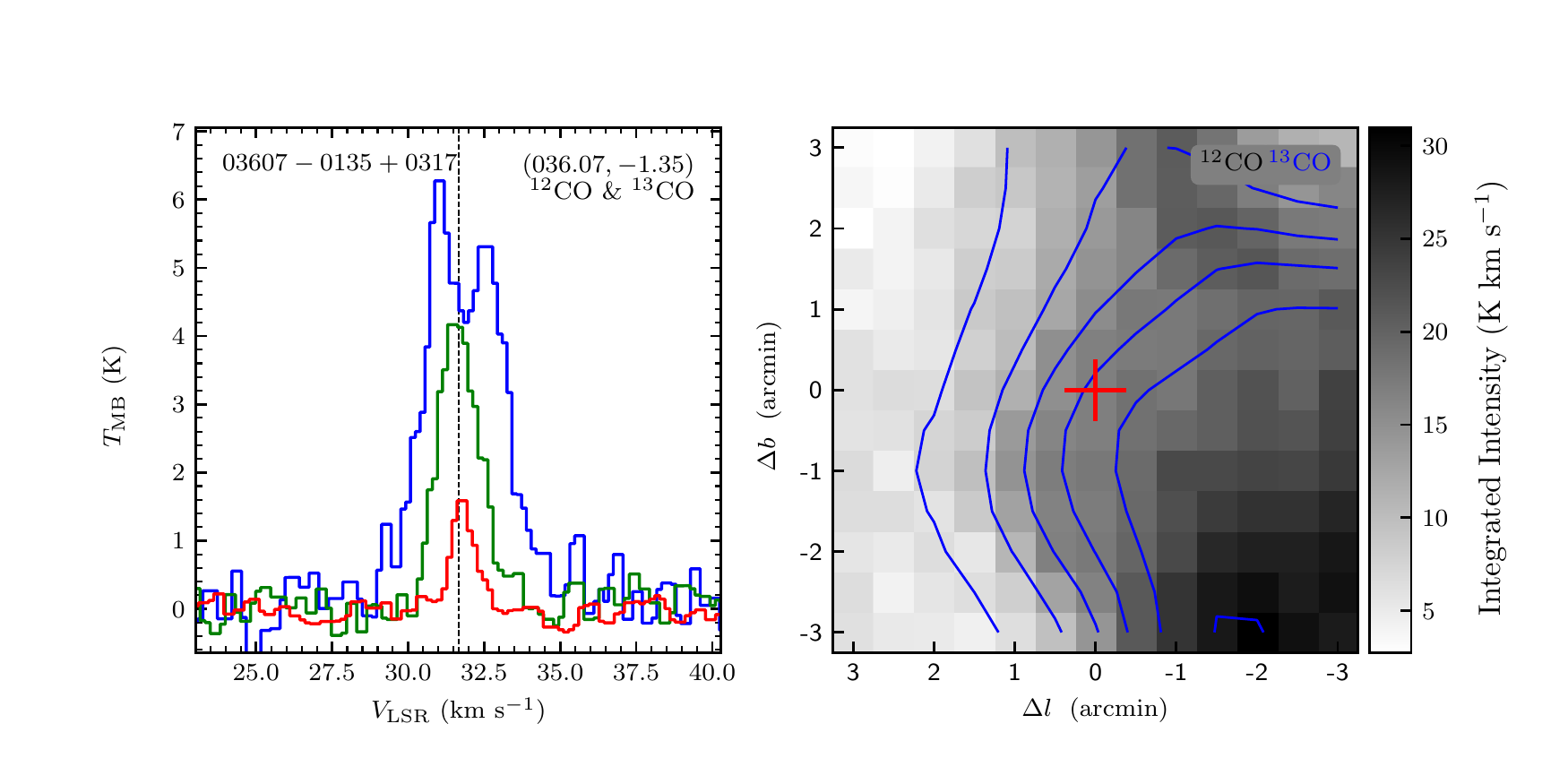}
\end{figure}
\clearpage

\begin{figure}
\includegraphics[width=9.0cm,angle=0]{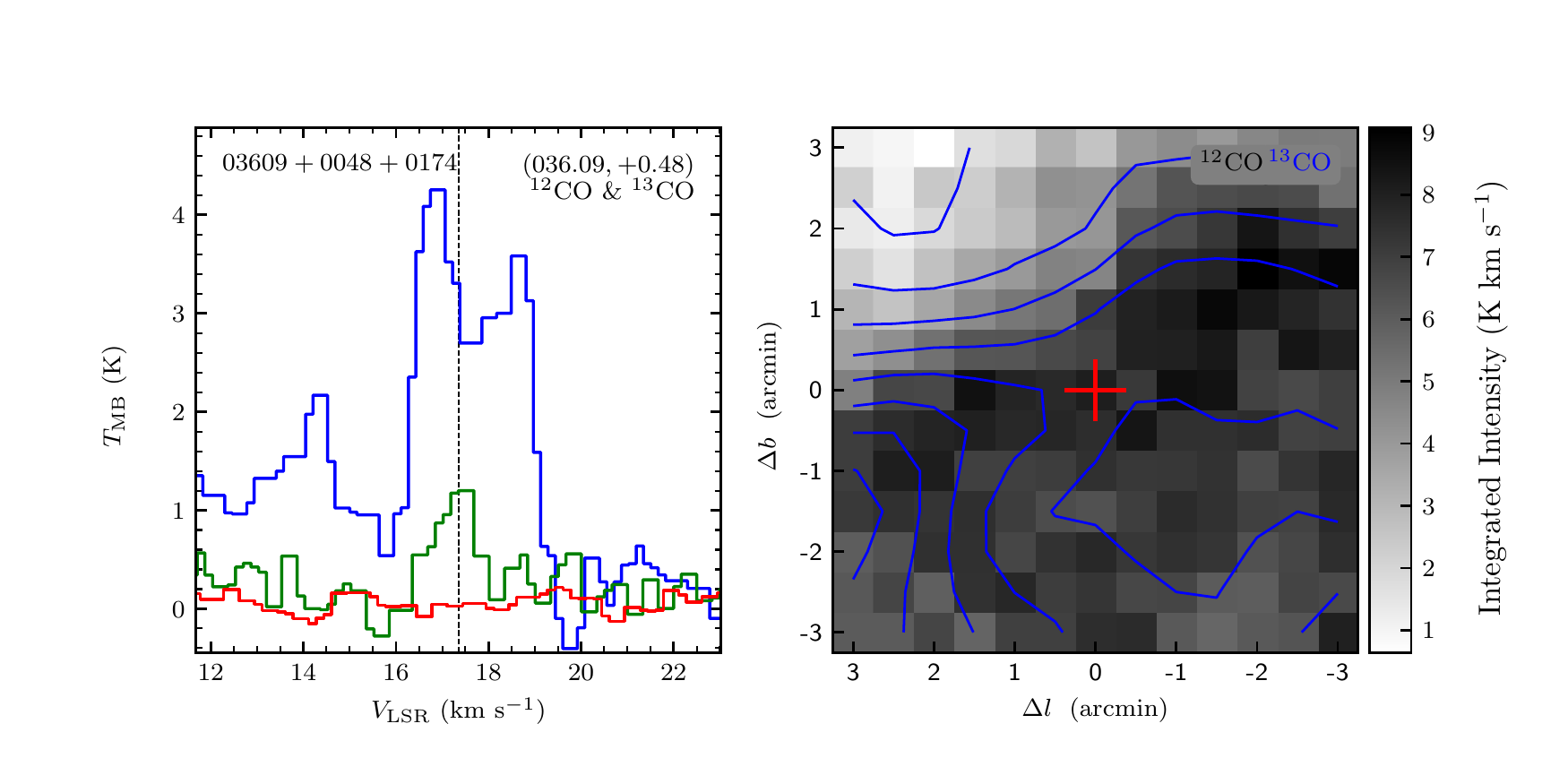}
\includegraphics[width=9.0cm,angle=0]{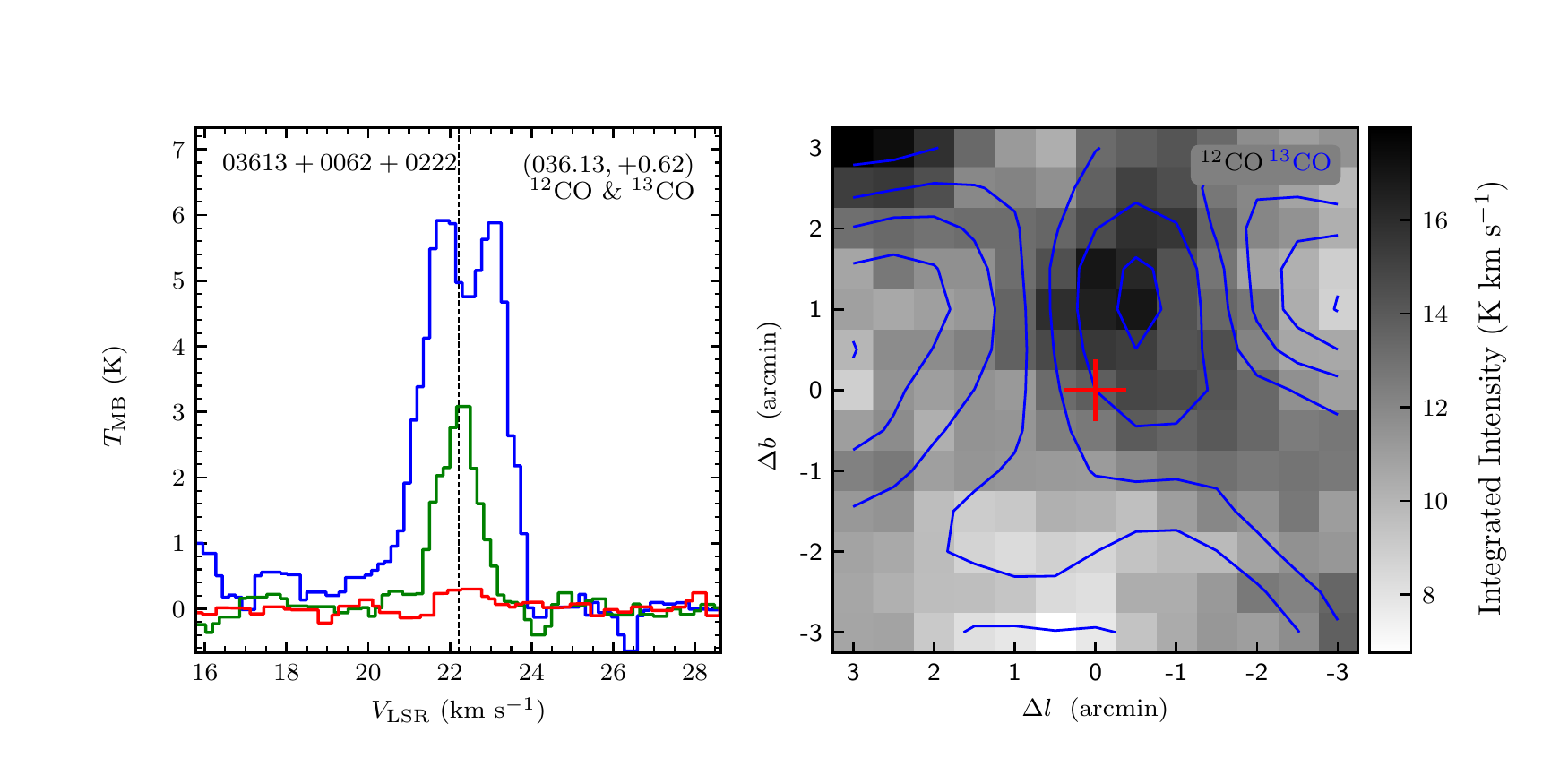}
\vspace{-0.5cm}

\includegraphics[width=9.0cm,angle=0]{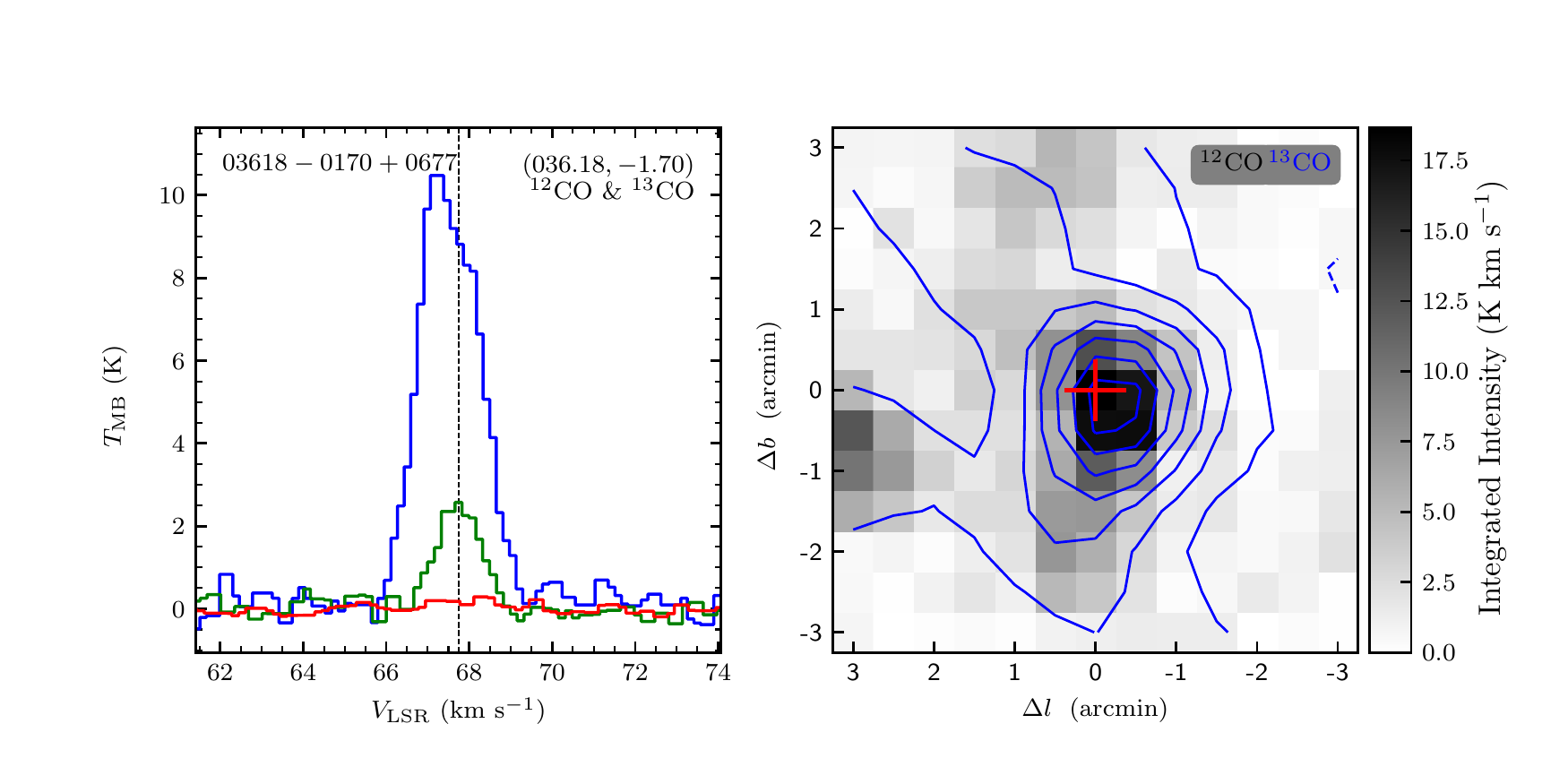}
\includegraphics[width=9.0cm,angle=0]{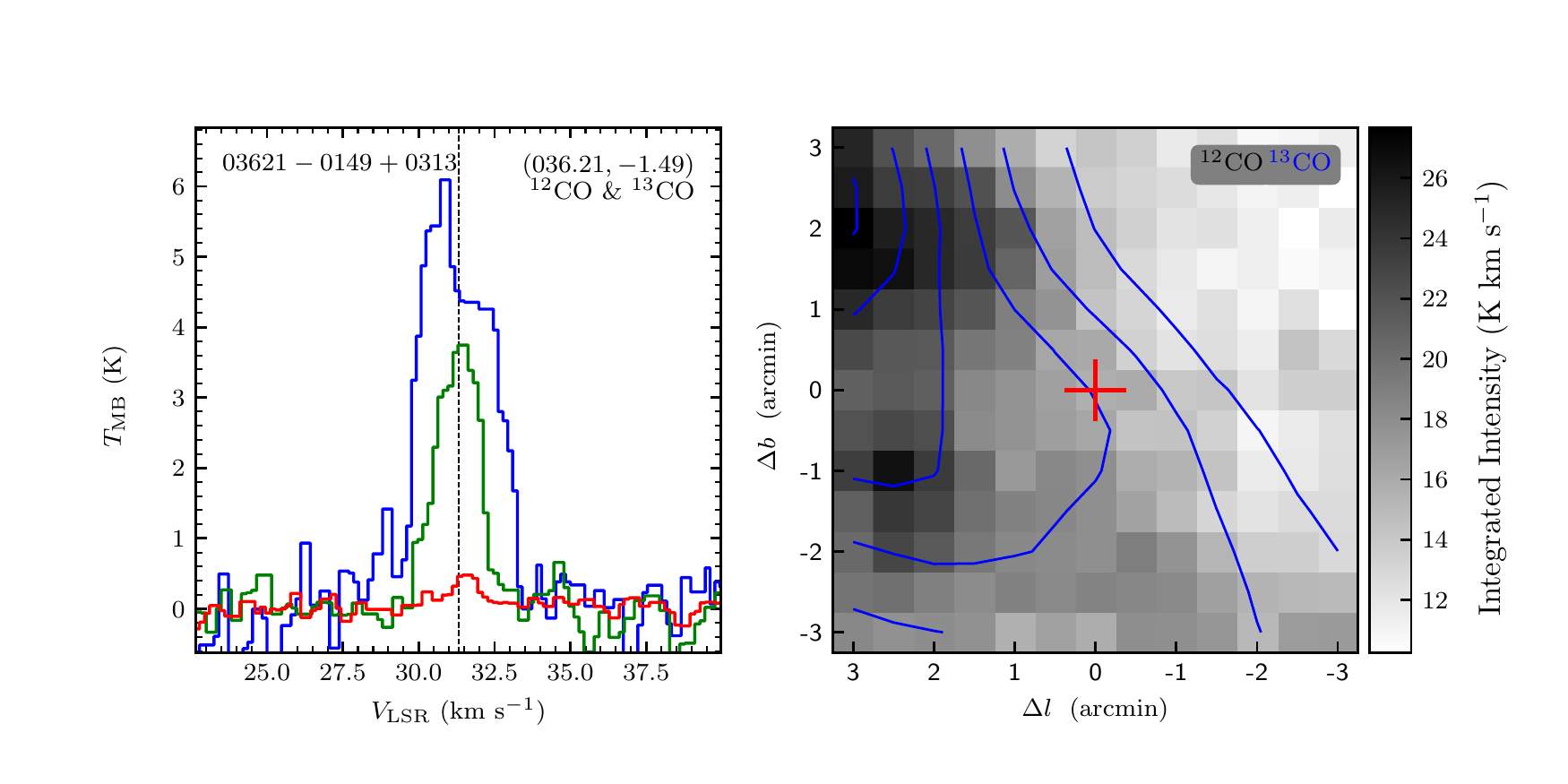}
\vspace{-0.5cm}

\includegraphics[width=9.0cm,angle=0]{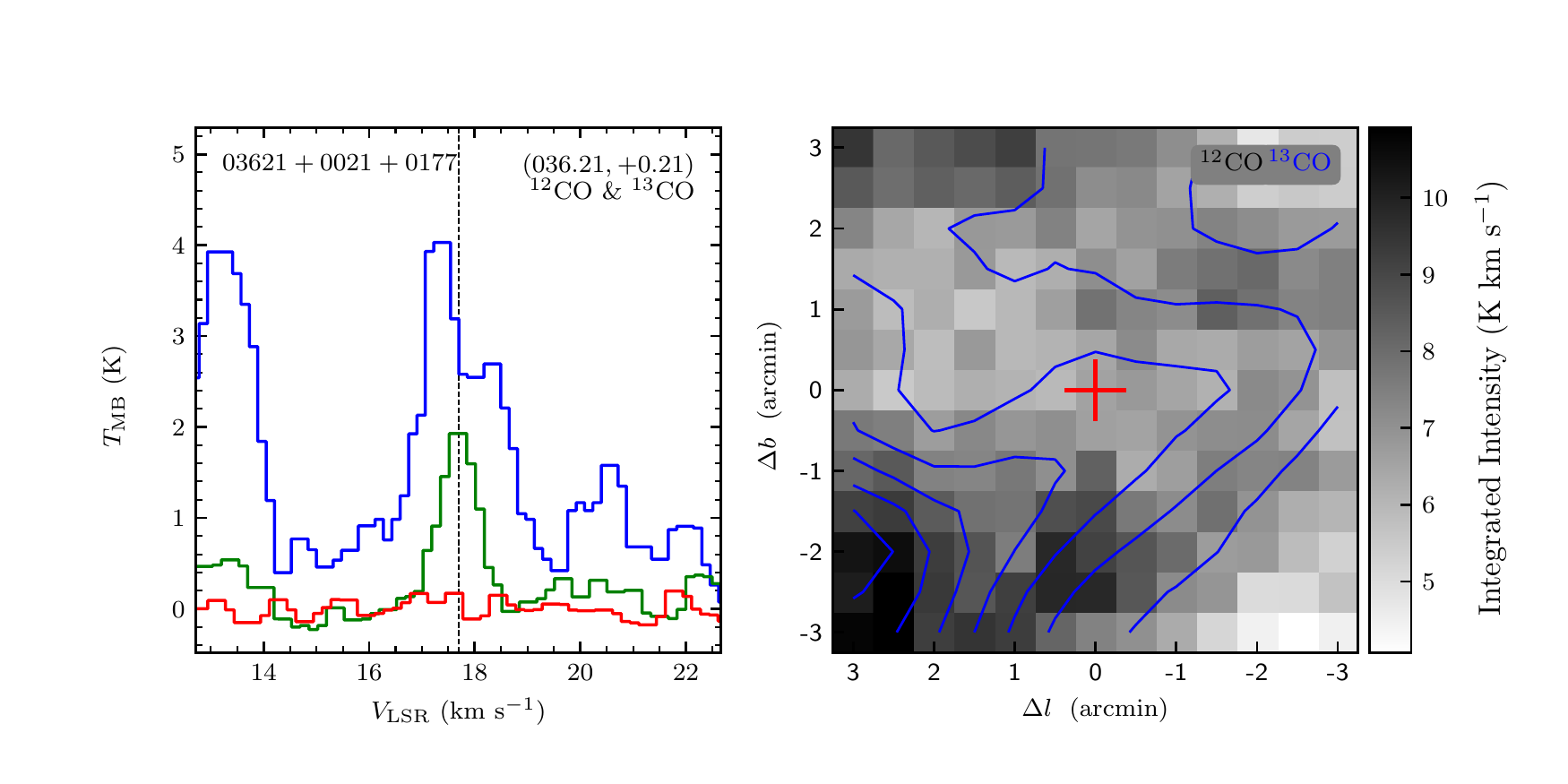}
\includegraphics[width=9.0cm,angle=0]{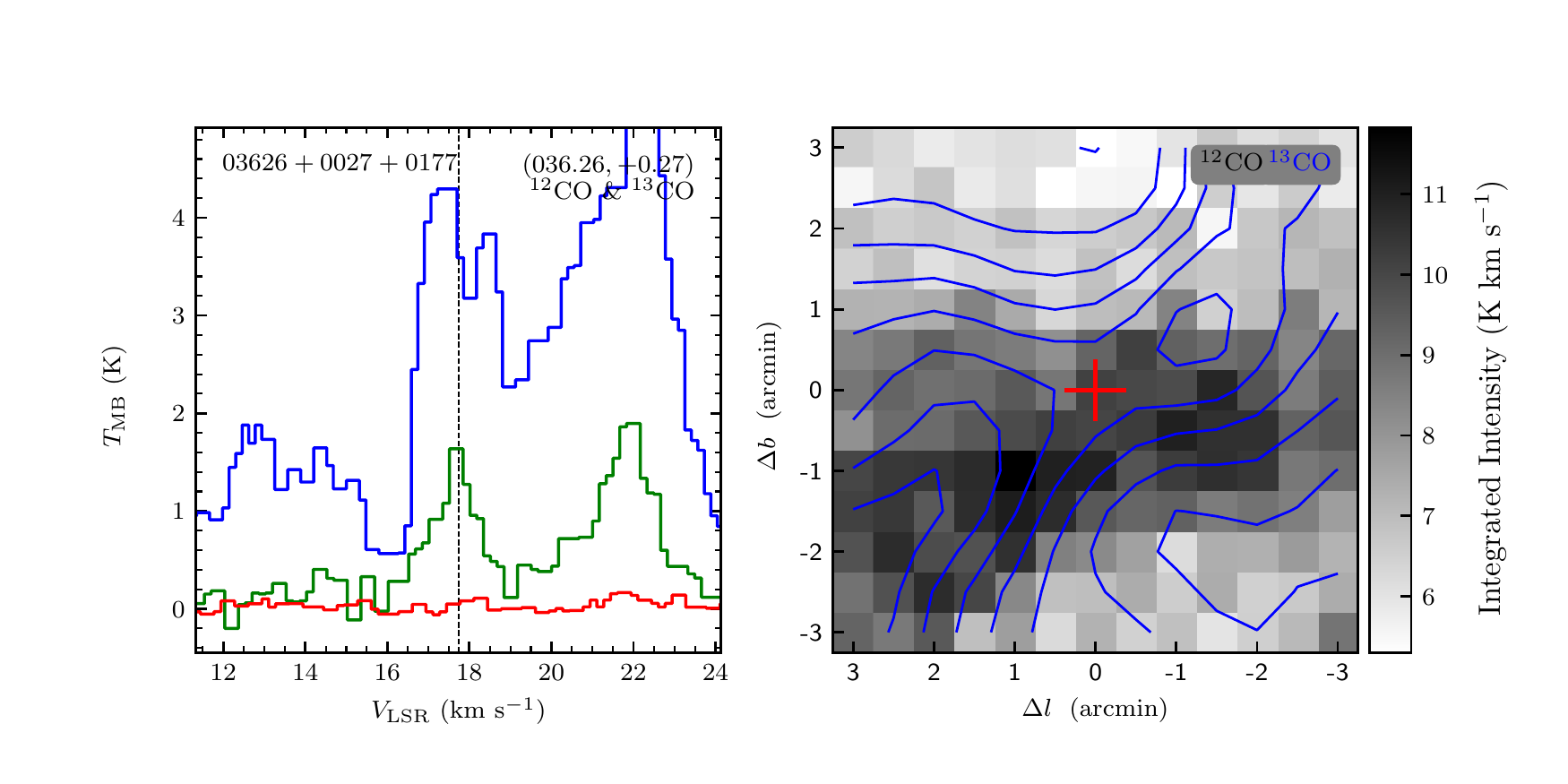}
\vspace{-0.5cm}

\includegraphics[width=9.0cm,angle=0]{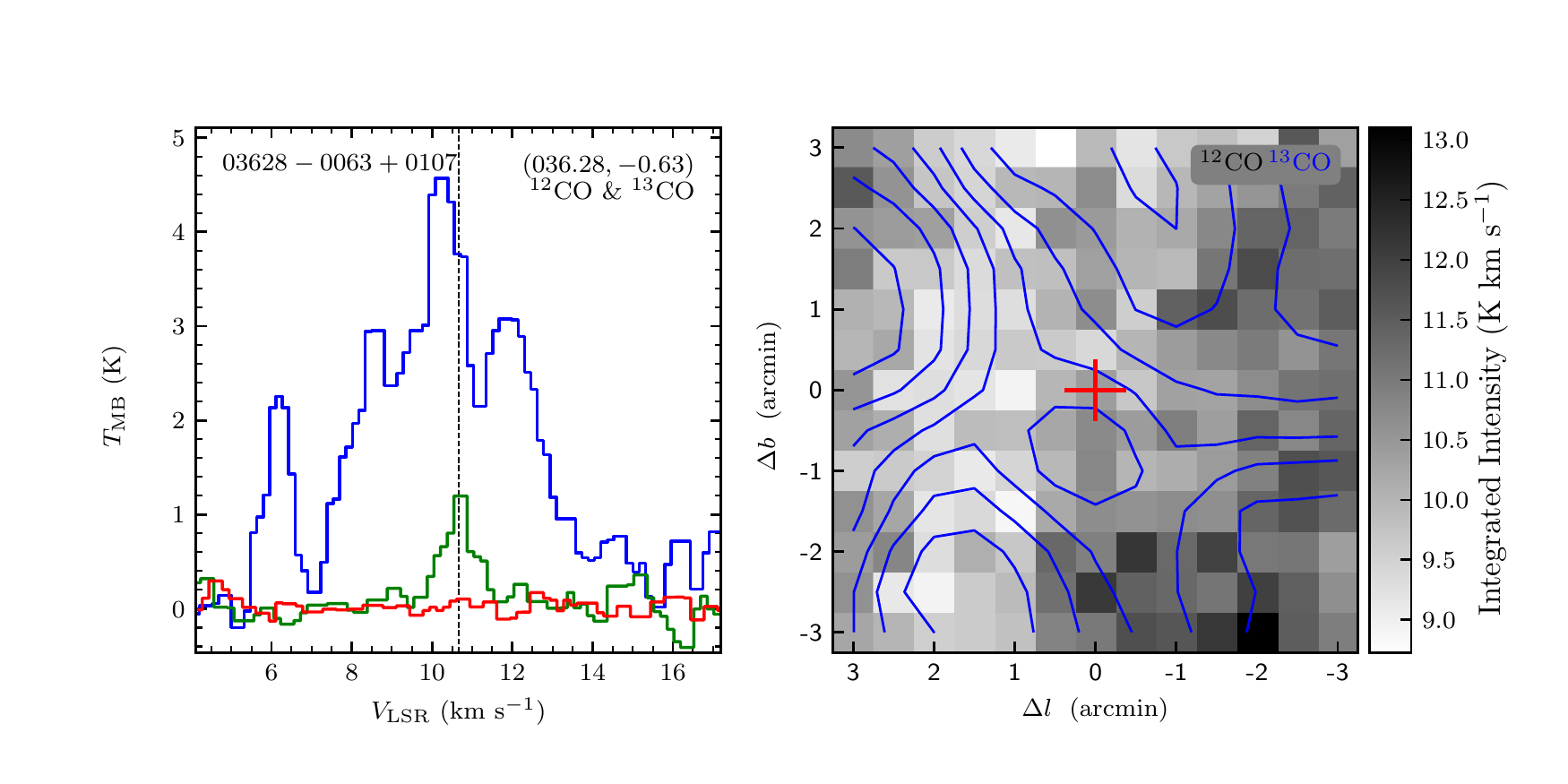}
\includegraphics[width=9.0cm,angle=0]{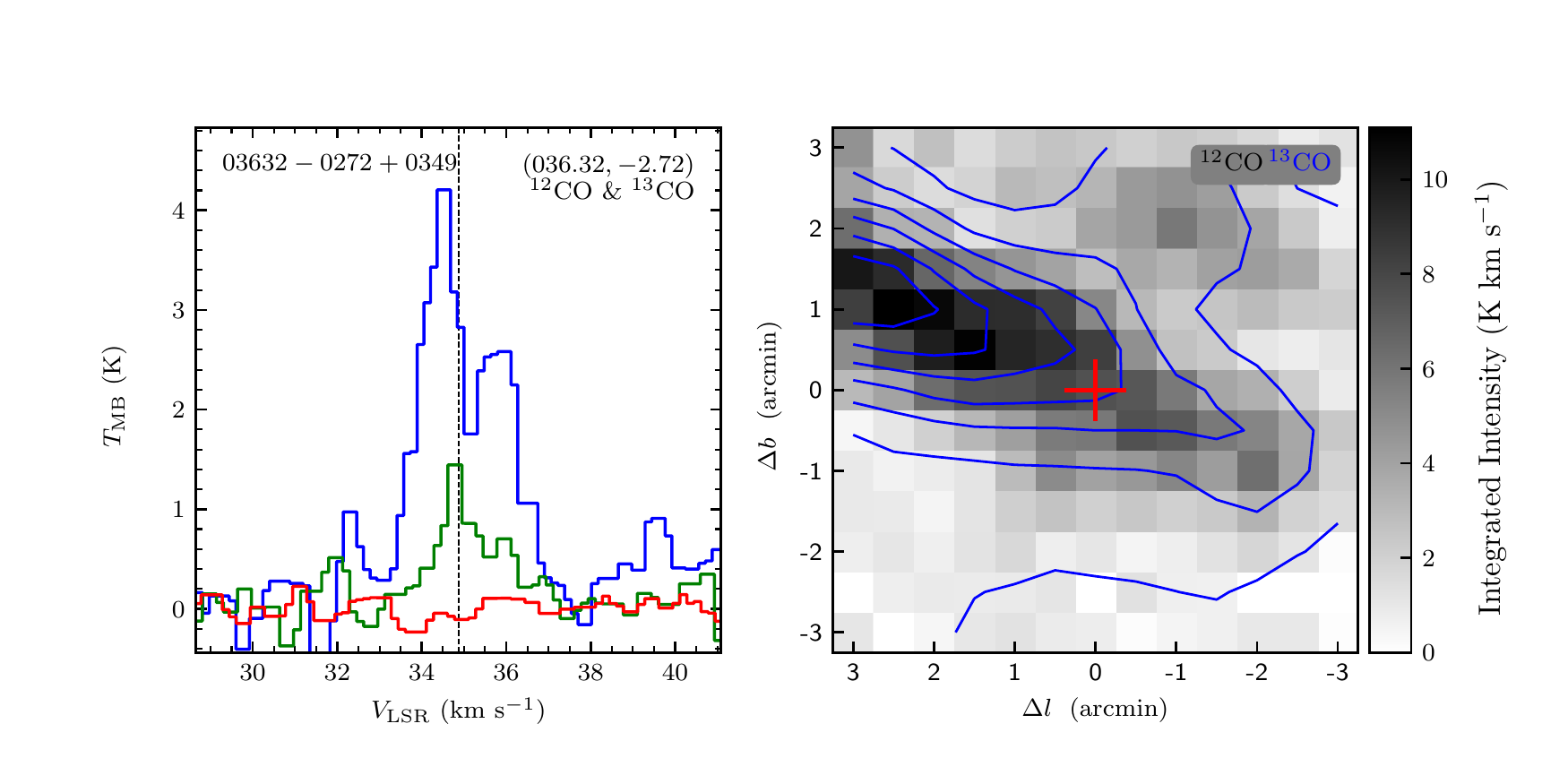}
\vspace{-0.5cm}

\includegraphics[width=9.0cm,angle=0]{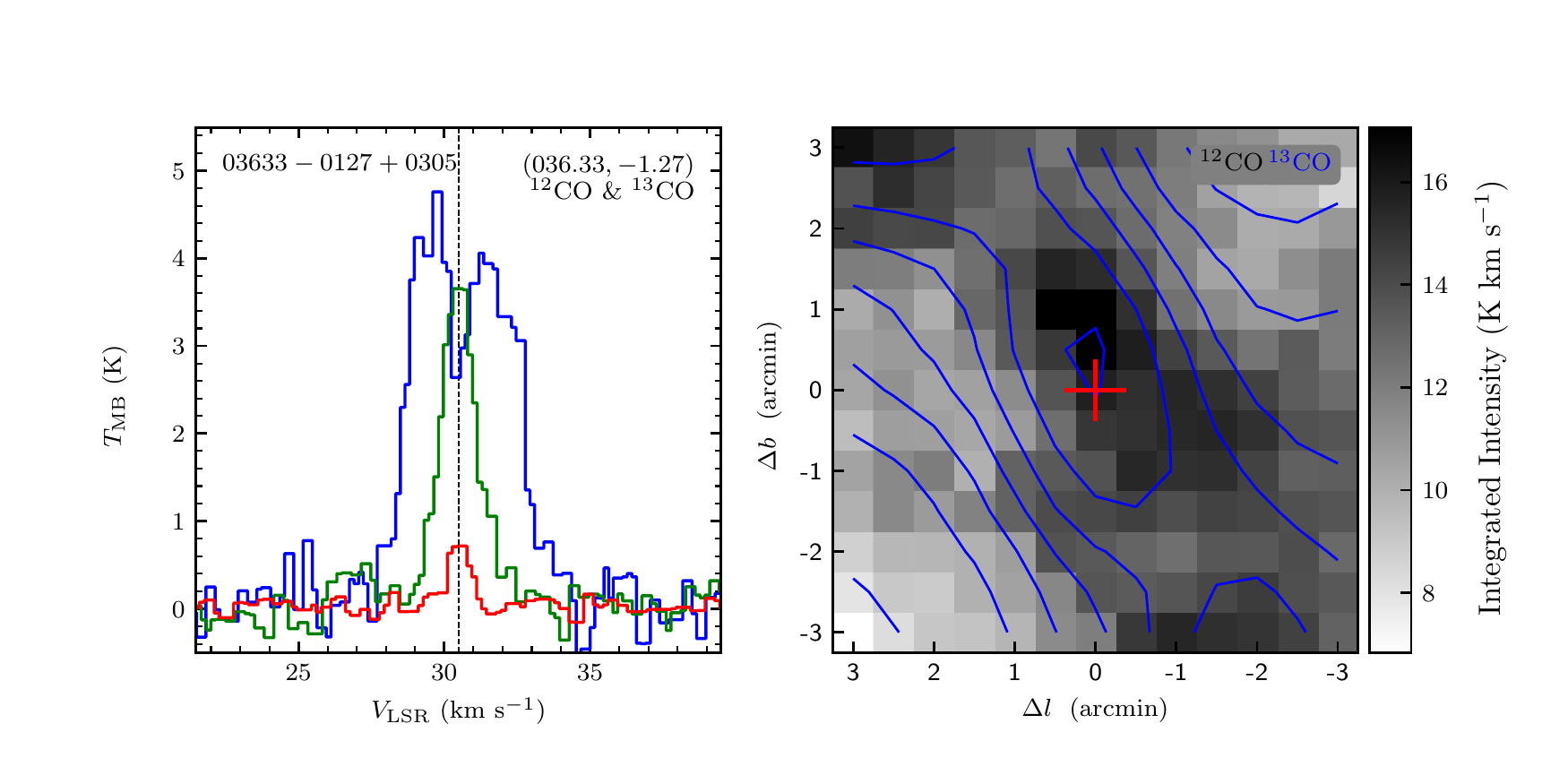}
\includegraphics[width=9.0cm,angle=0]{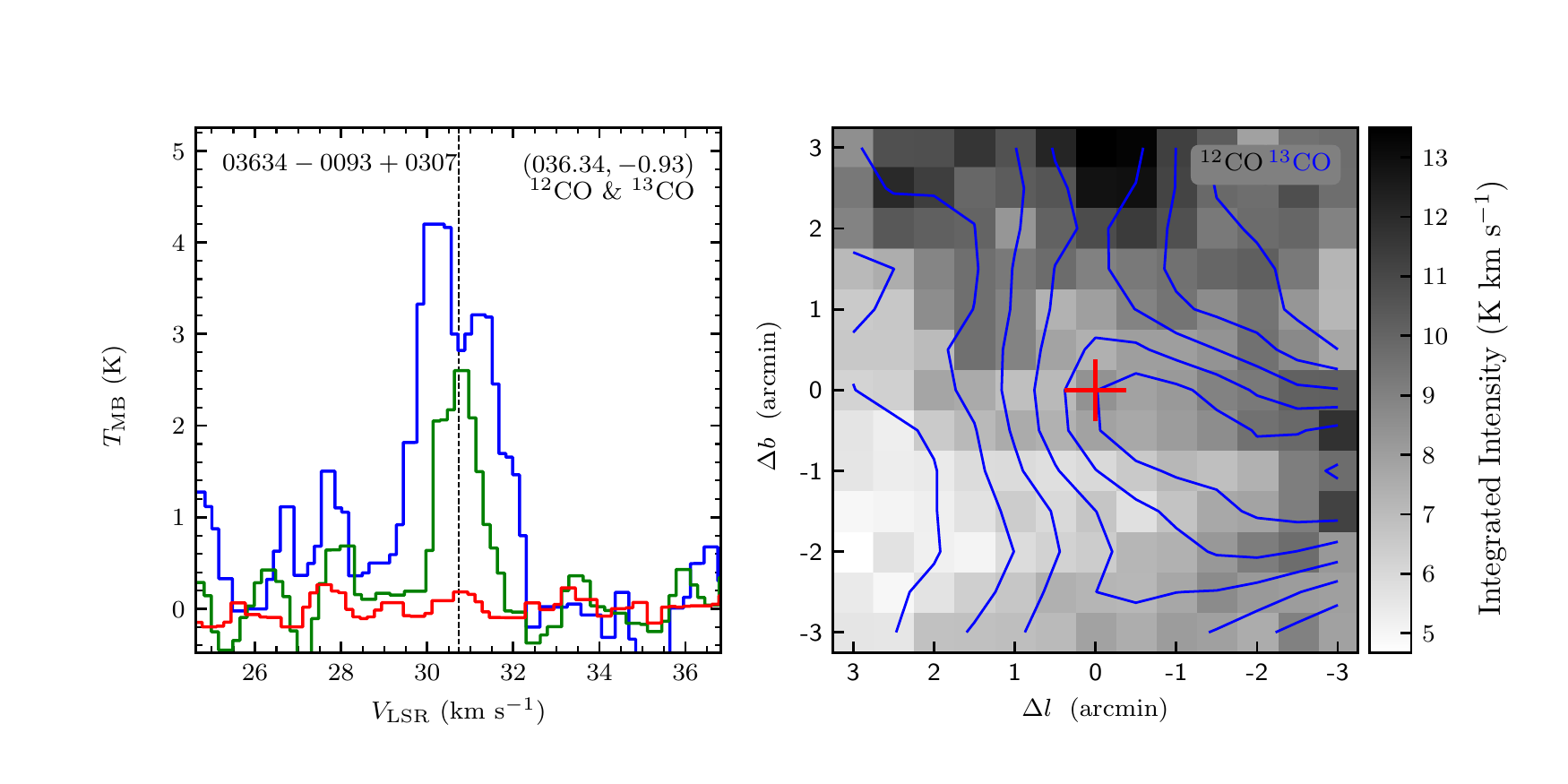}
\end{figure}
\clearpage

\begin{figure}
\includegraphics[width=9.0cm,angle=0]{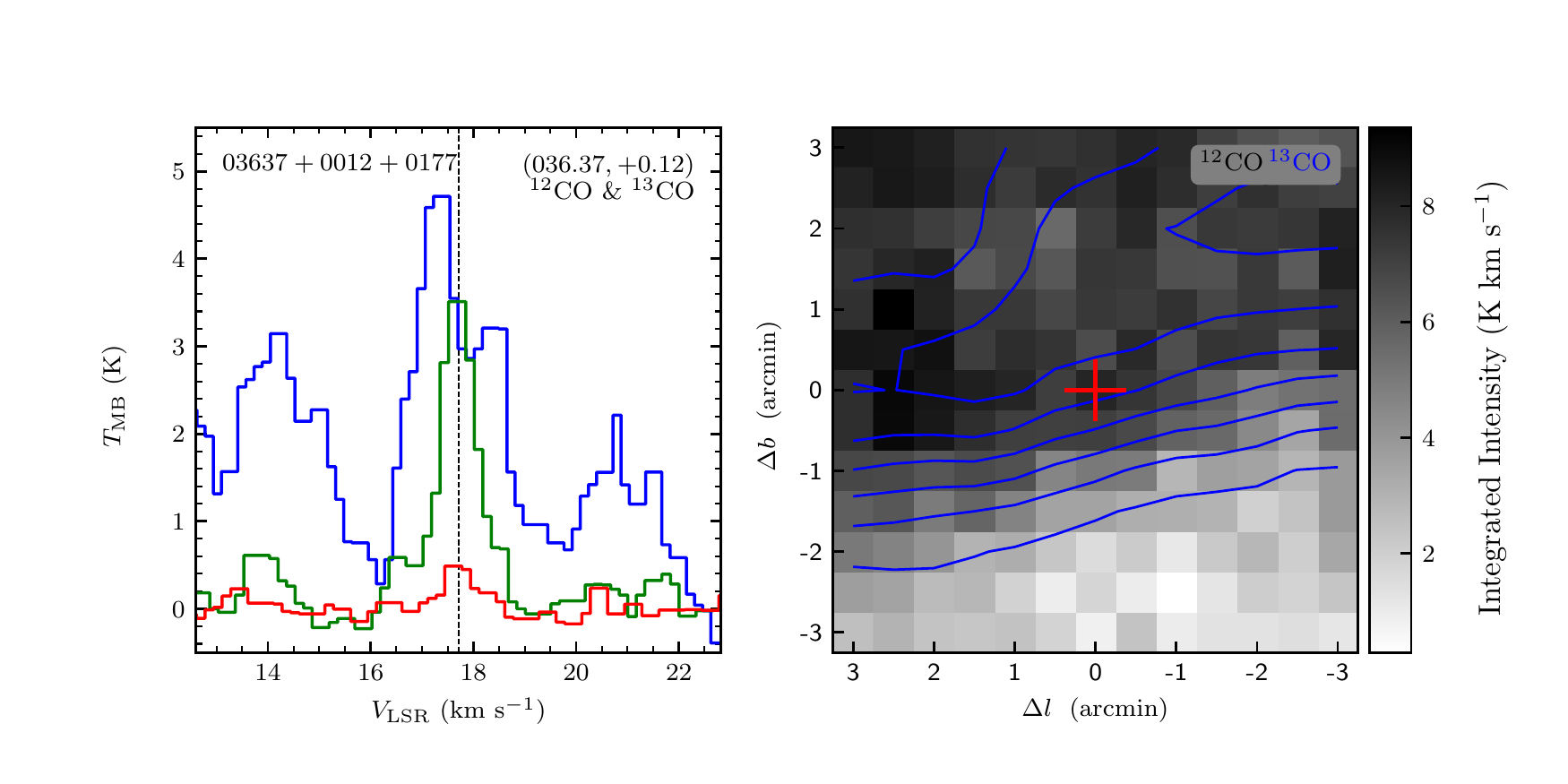}
\includegraphics[width=9.0cm,angle=0]{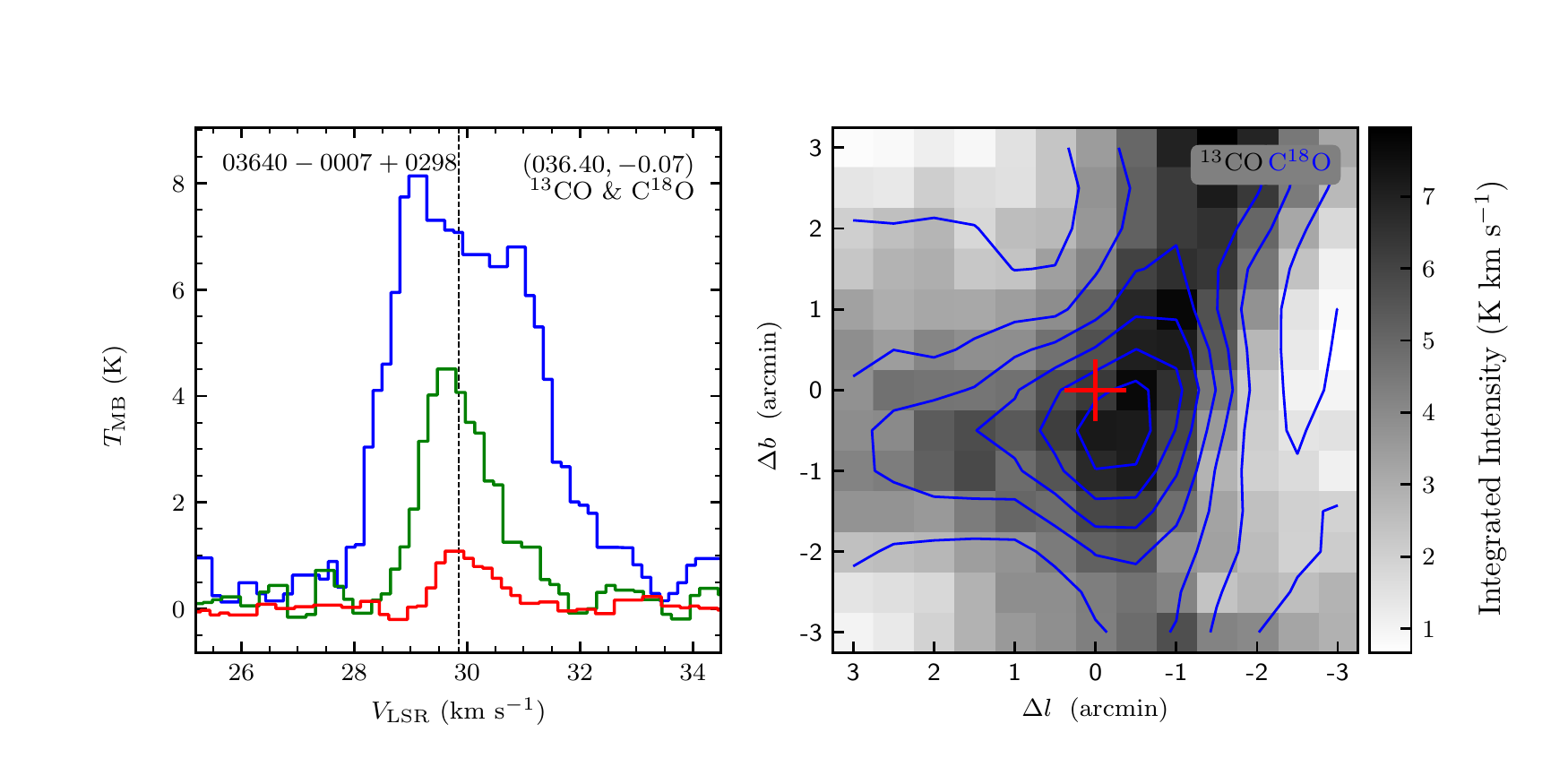}
\vspace{-0.5cm}

\includegraphics[width=9.0cm,angle=0]{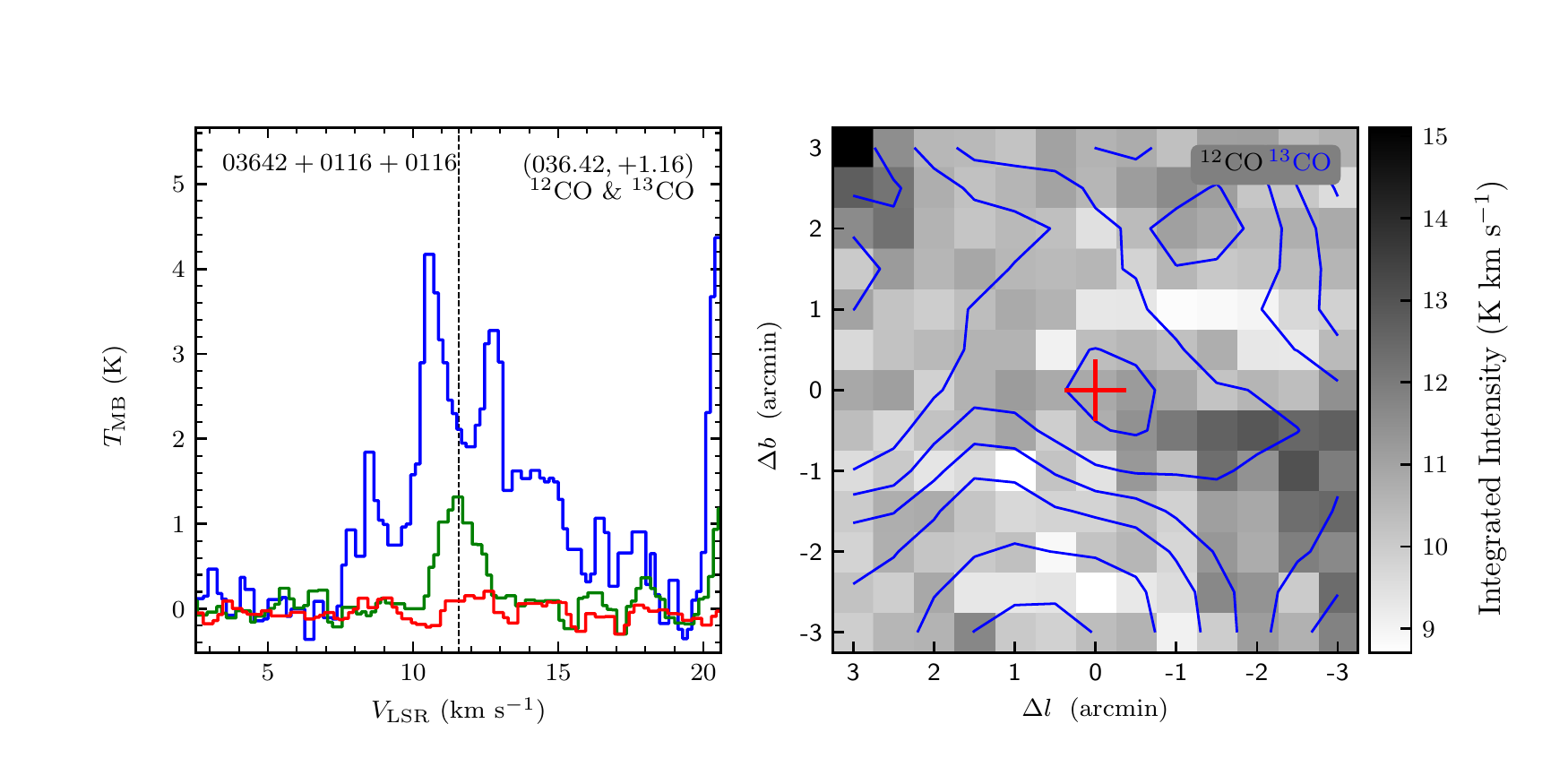}
\includegraphics[width=9.0cm,angle=0]{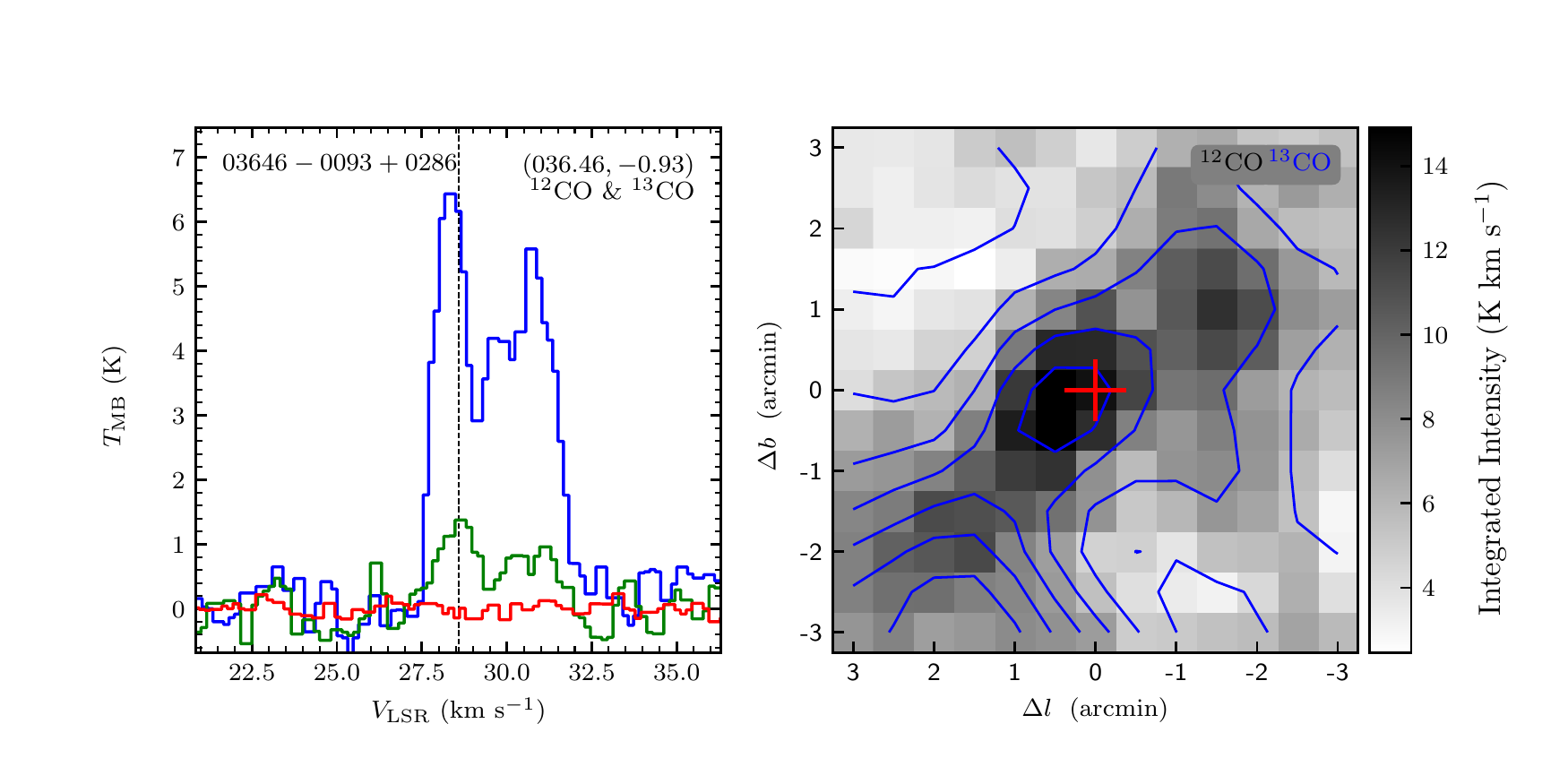}
\vspace{-0.5cm}

\includegraphics[width=9.0cm,angle=0]{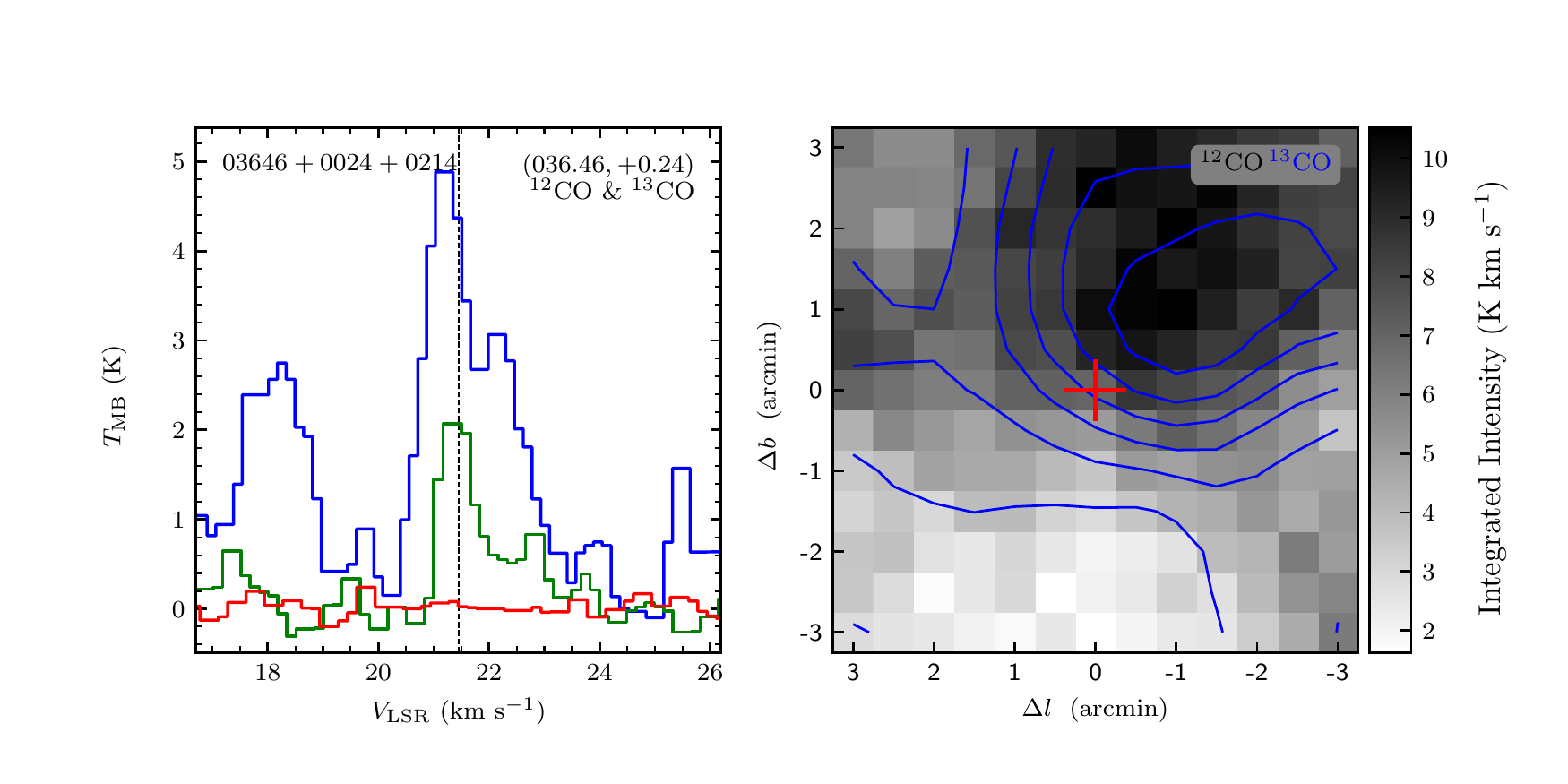}
\includegraphics[width=9.0cm,angle=0]{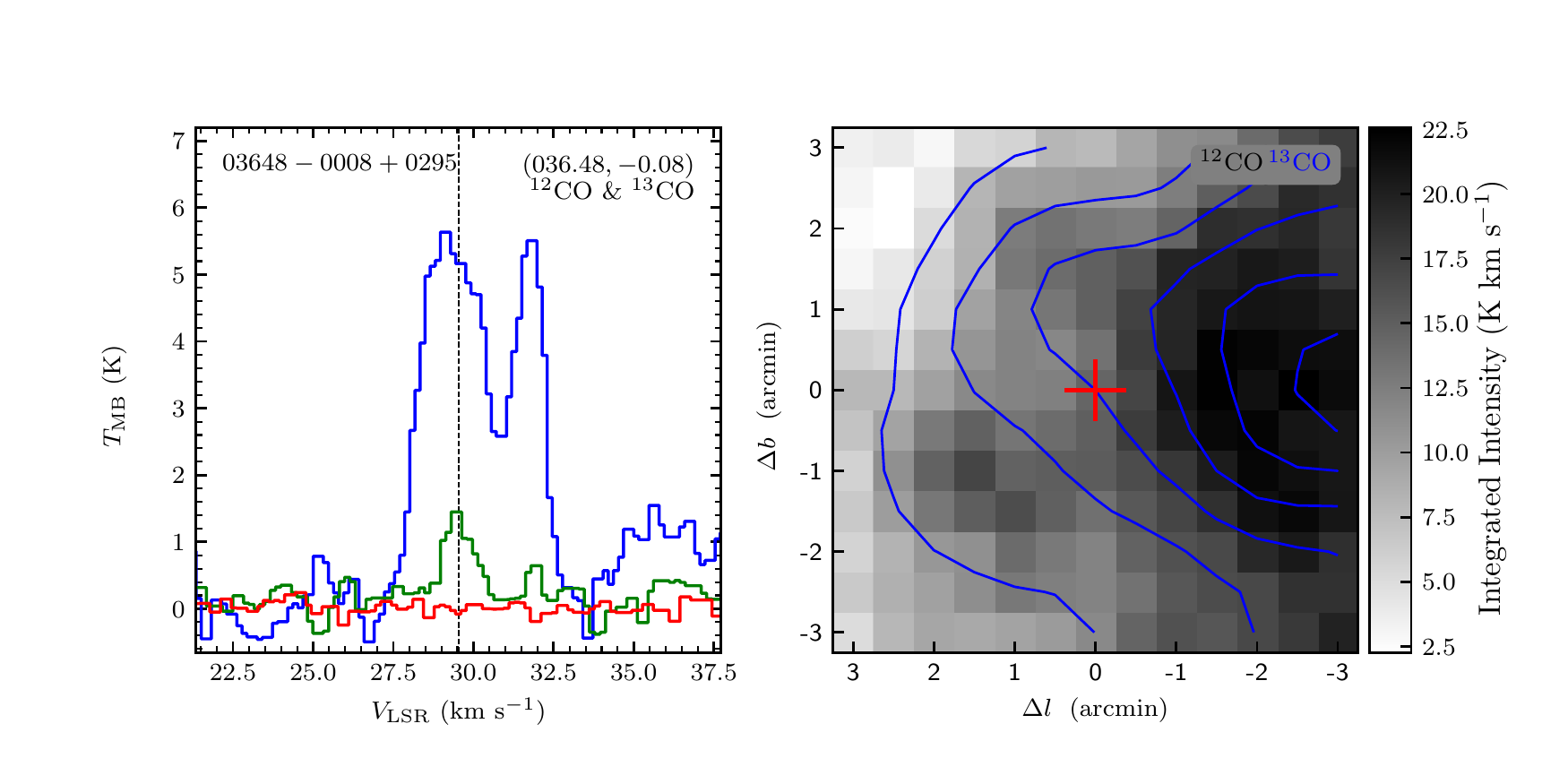}
\vspace{-0.5cm}

\includegraphics[width=9.0cm,angle=0]{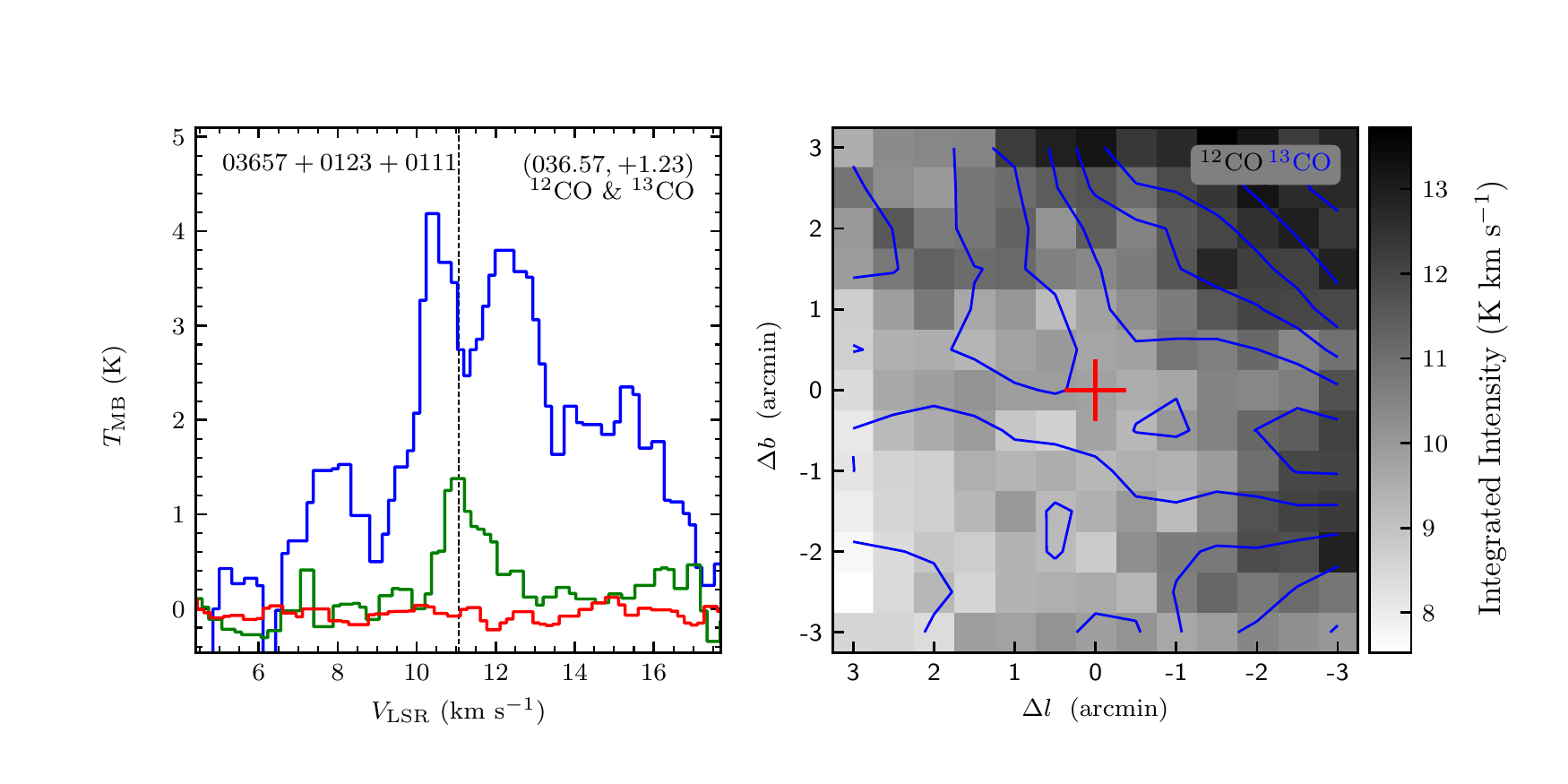}
\includegraphics[width=9.0cm,angle=0]{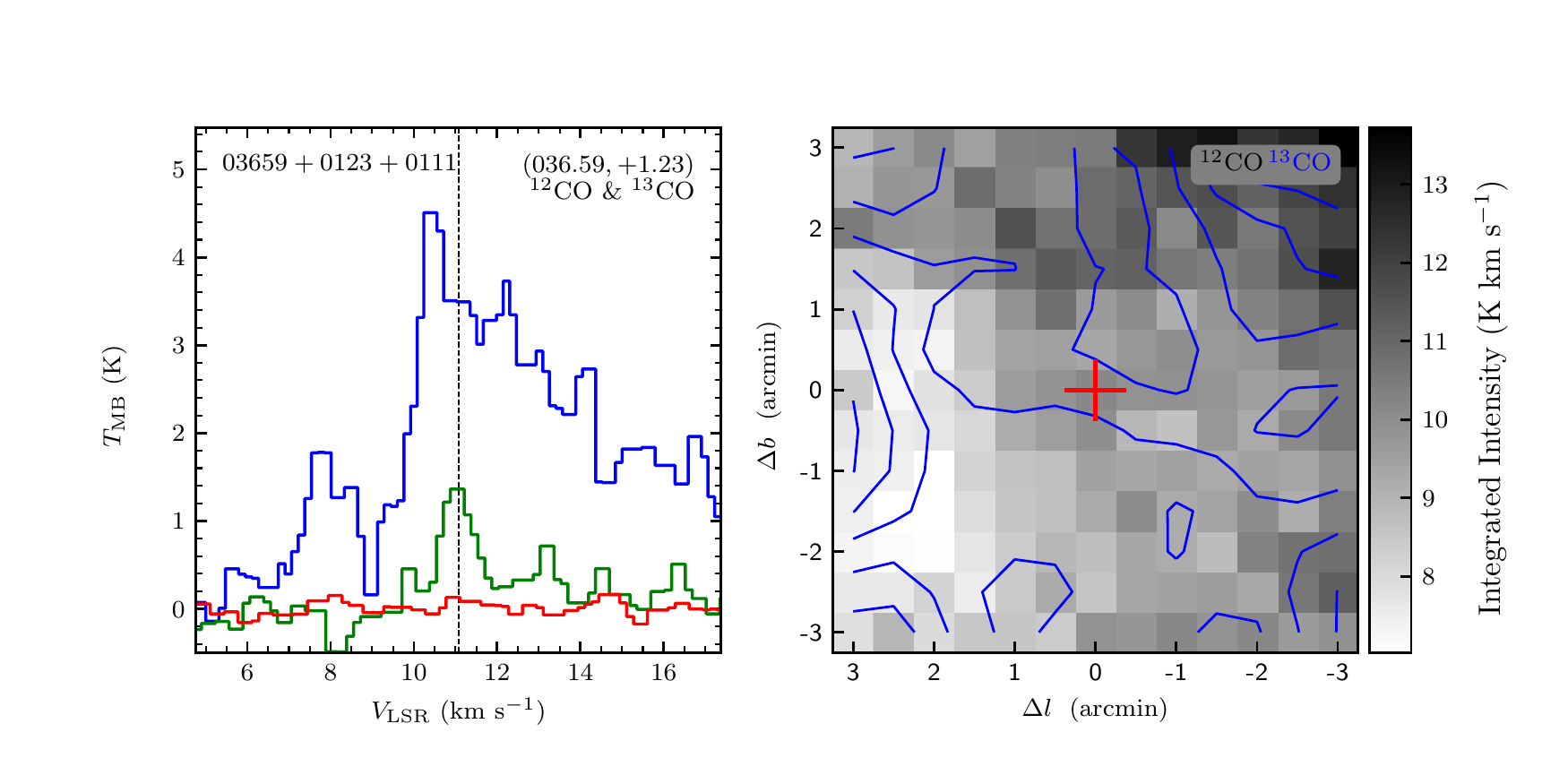}
\vspace{-0.5cm}

\includegraphics[width=9.0cm,angle=0]{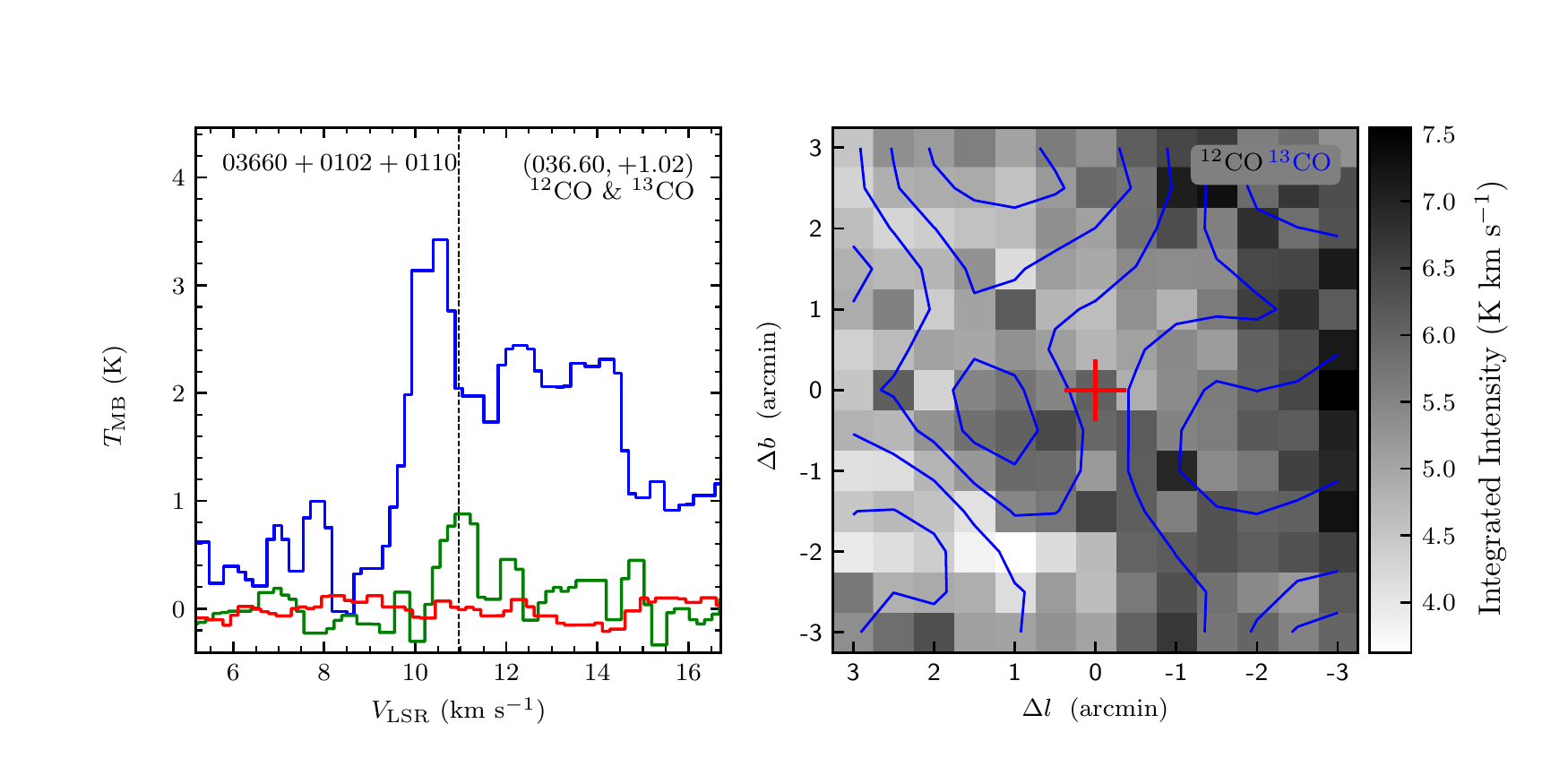}
\includegraphics[width=9.0cm,angle=0]{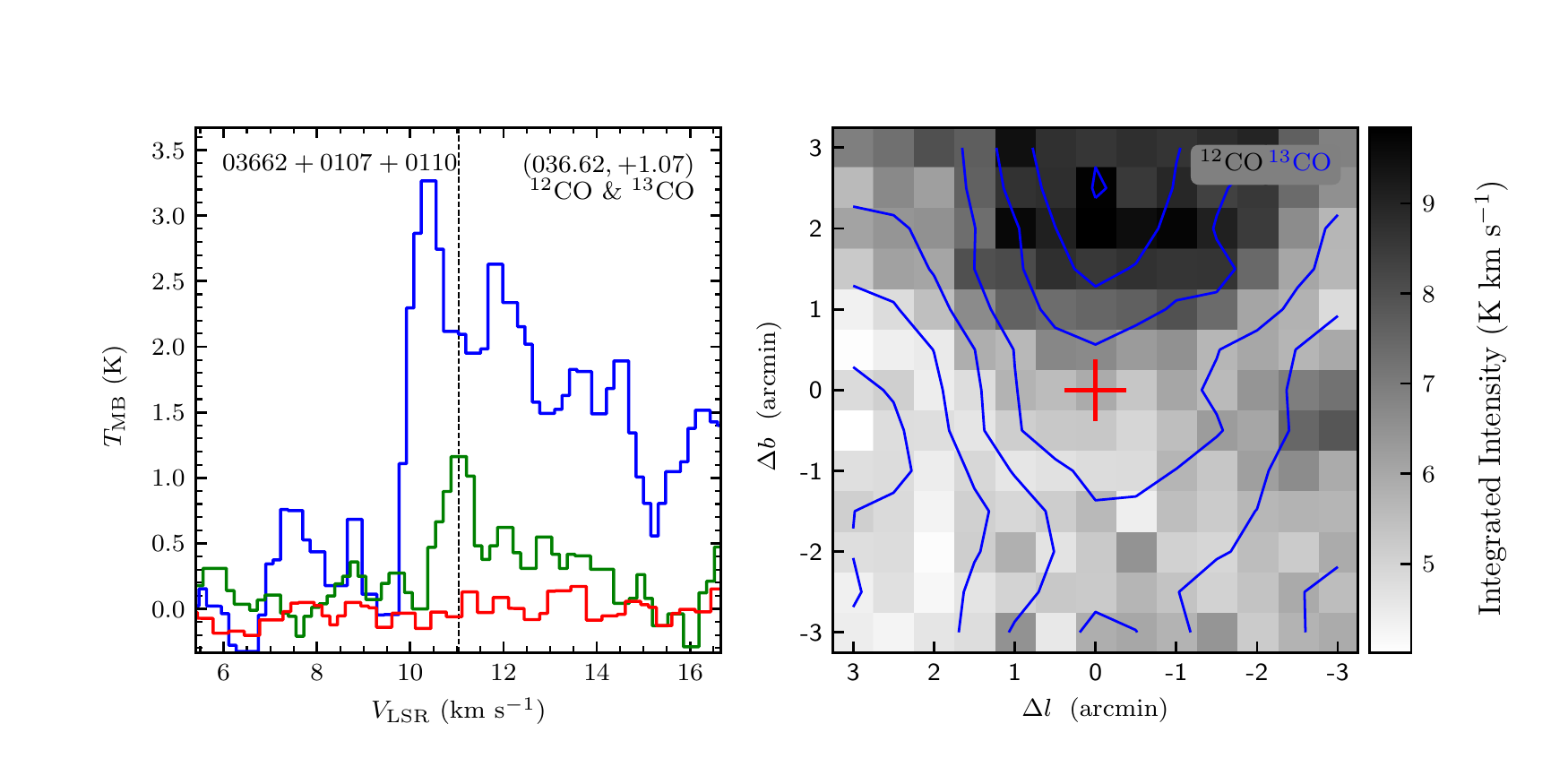}
\end{figure}
\clearpage

\begin{figure}
\includegraphics[width=9.0cm,angle=0]{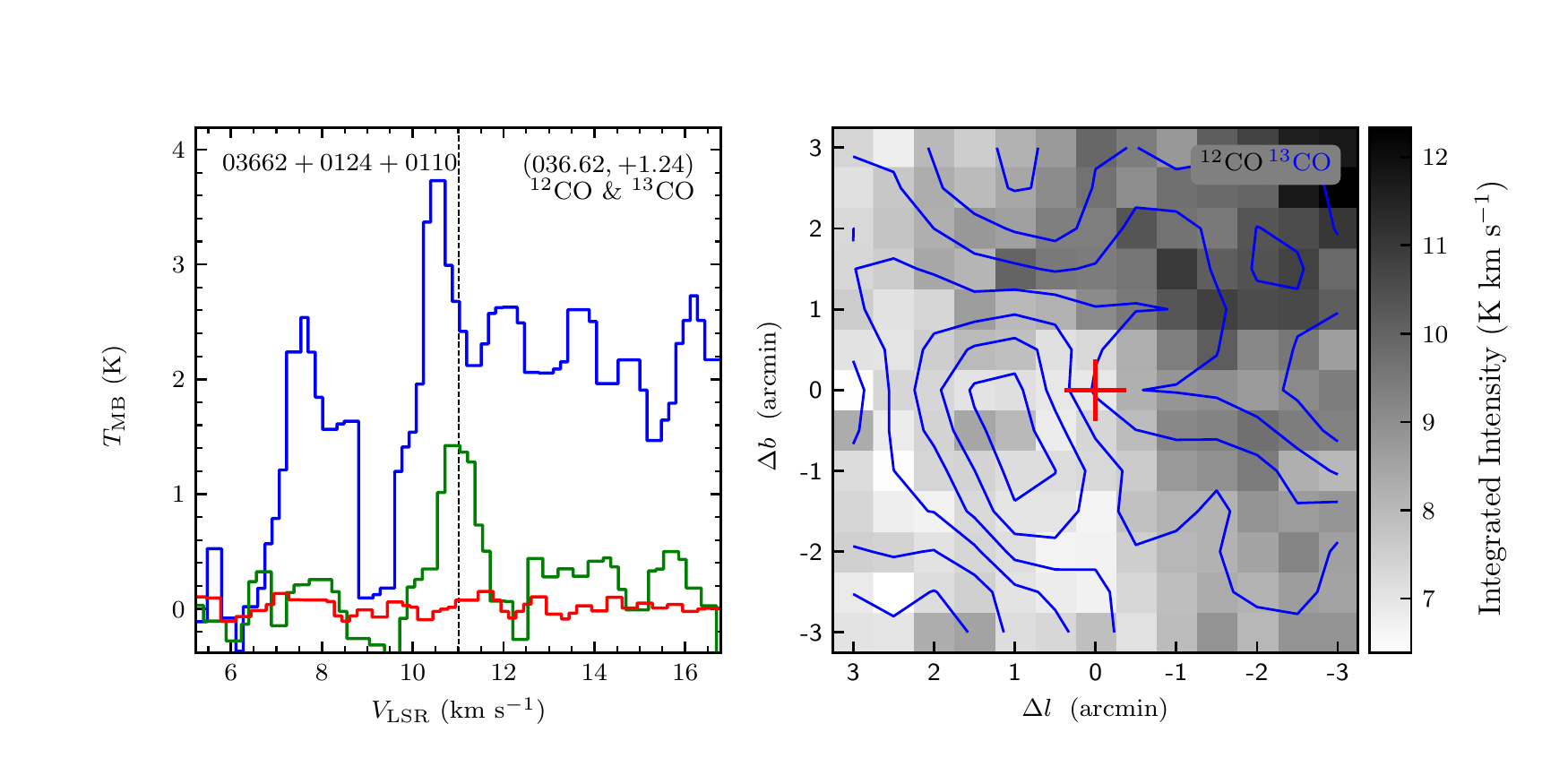}
\includegraphics[width=9.0cm,angle=0]{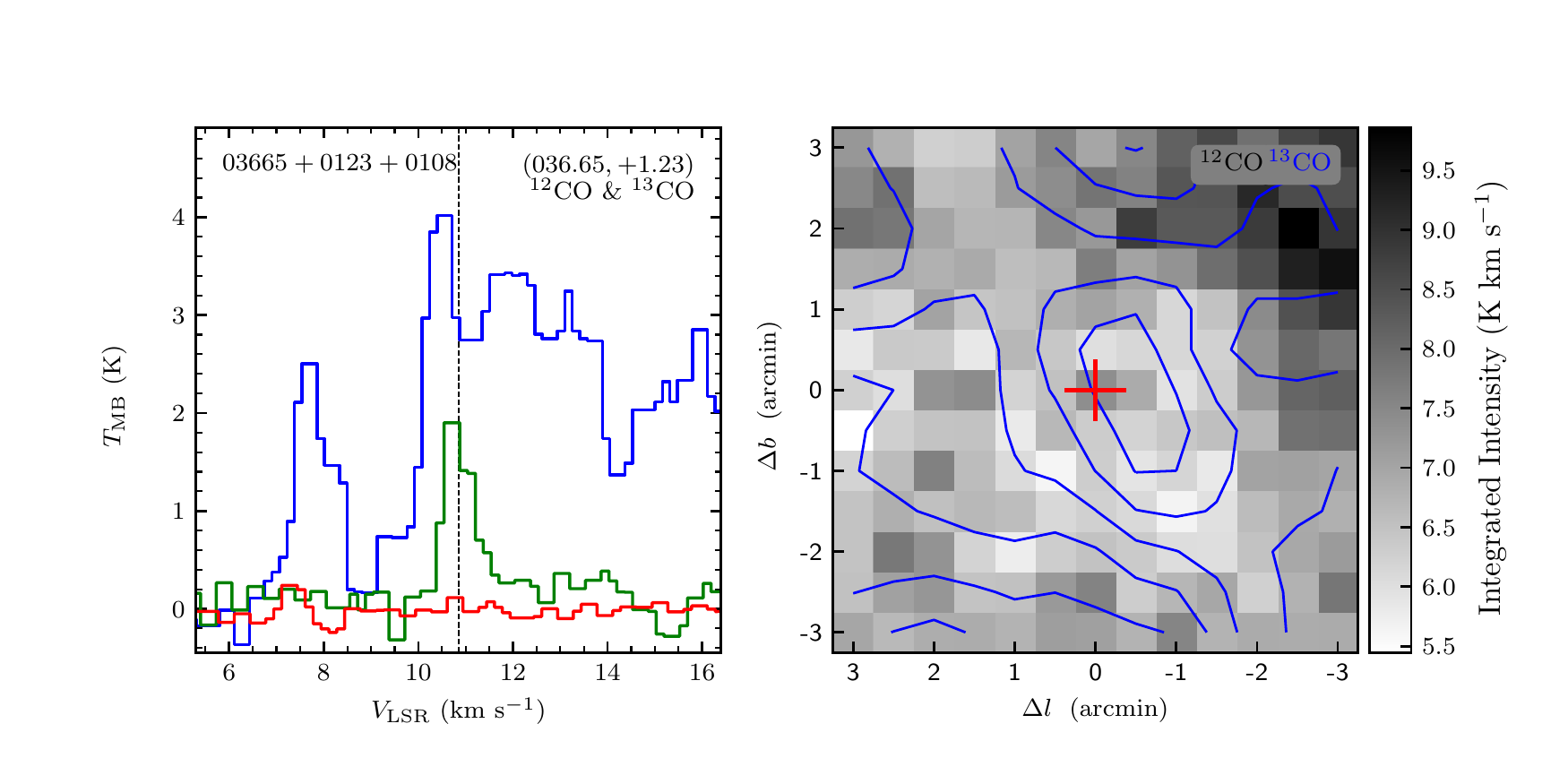}
\vspace{-0.5cm}

\includegraphics[width=9.0cm,angle=0]{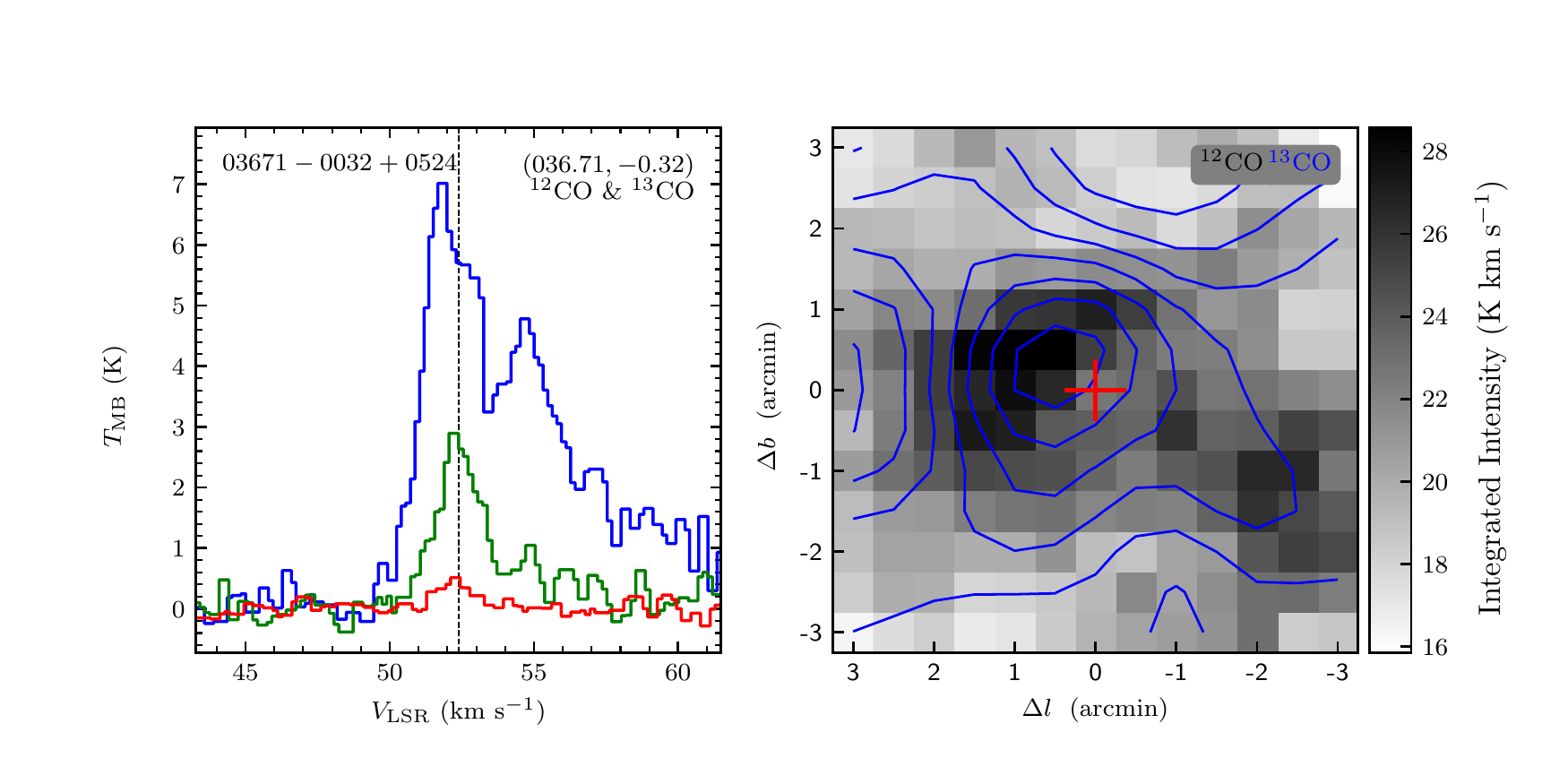}
\includegraphics[width=9.0cm,angle=0]{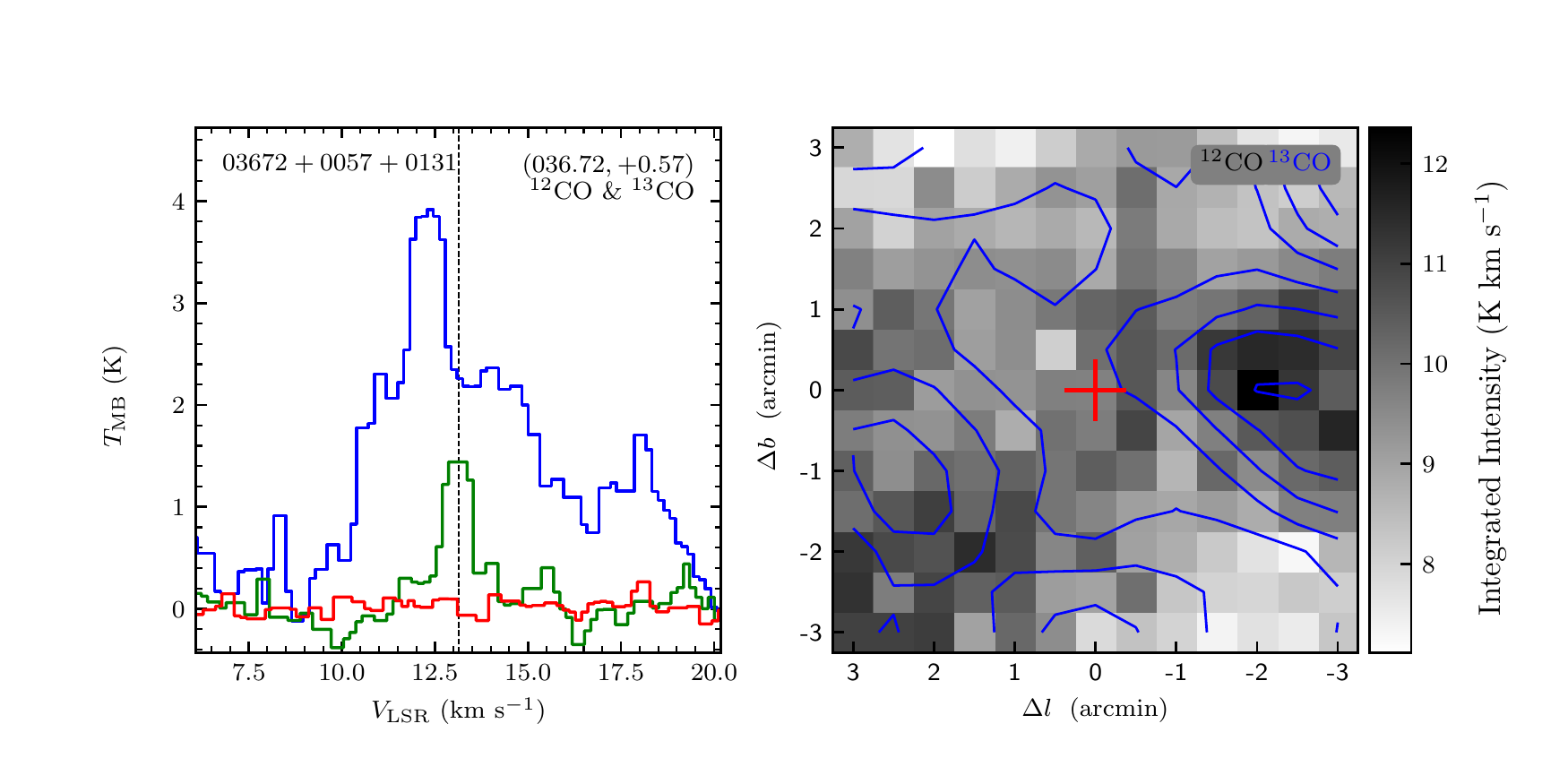}
\vspace{-0.5cm}

\includegraphics[width=9.0cm,angle=0]{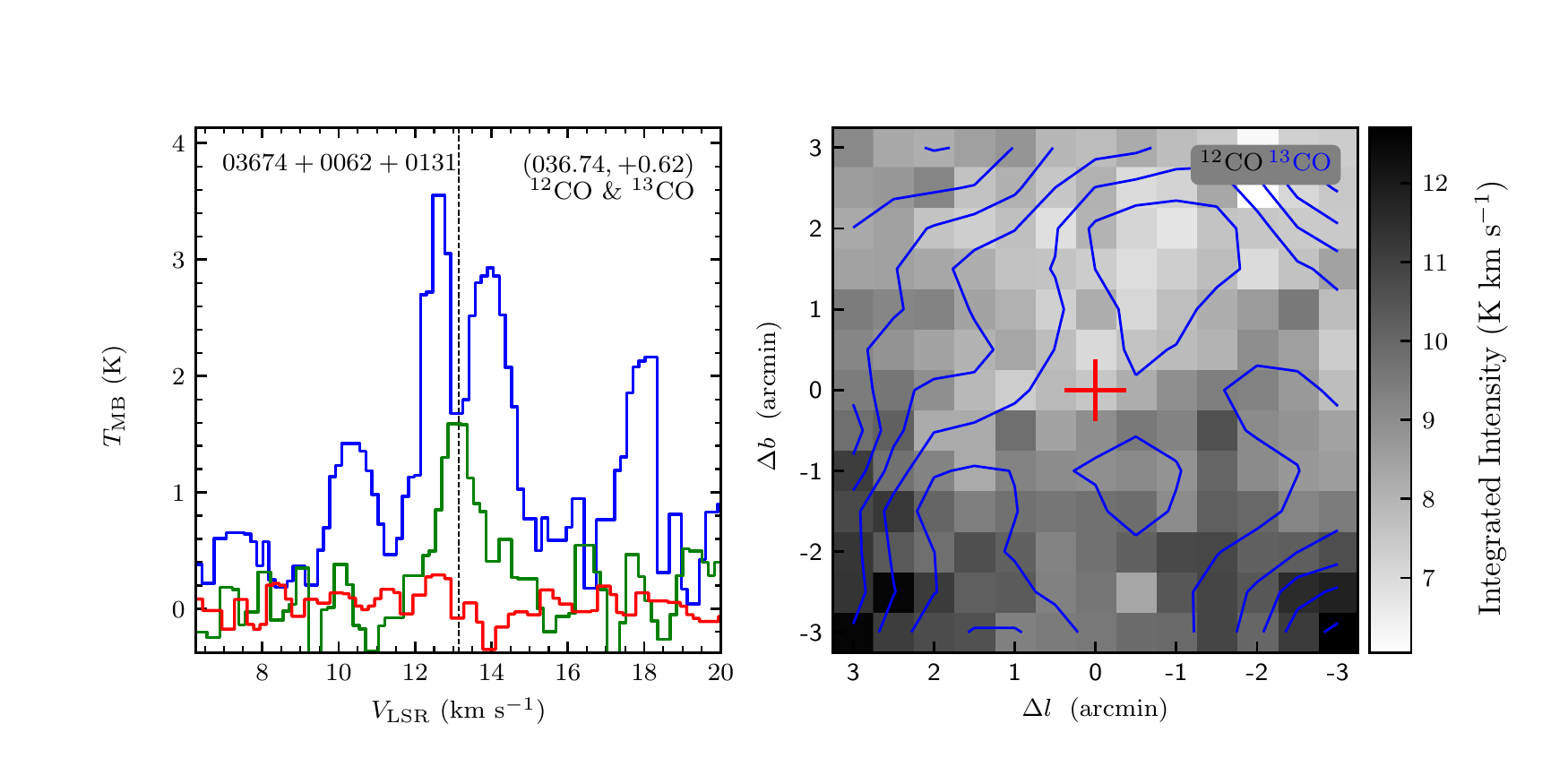}
\includegraphics[width=9.0cm,angle=0]{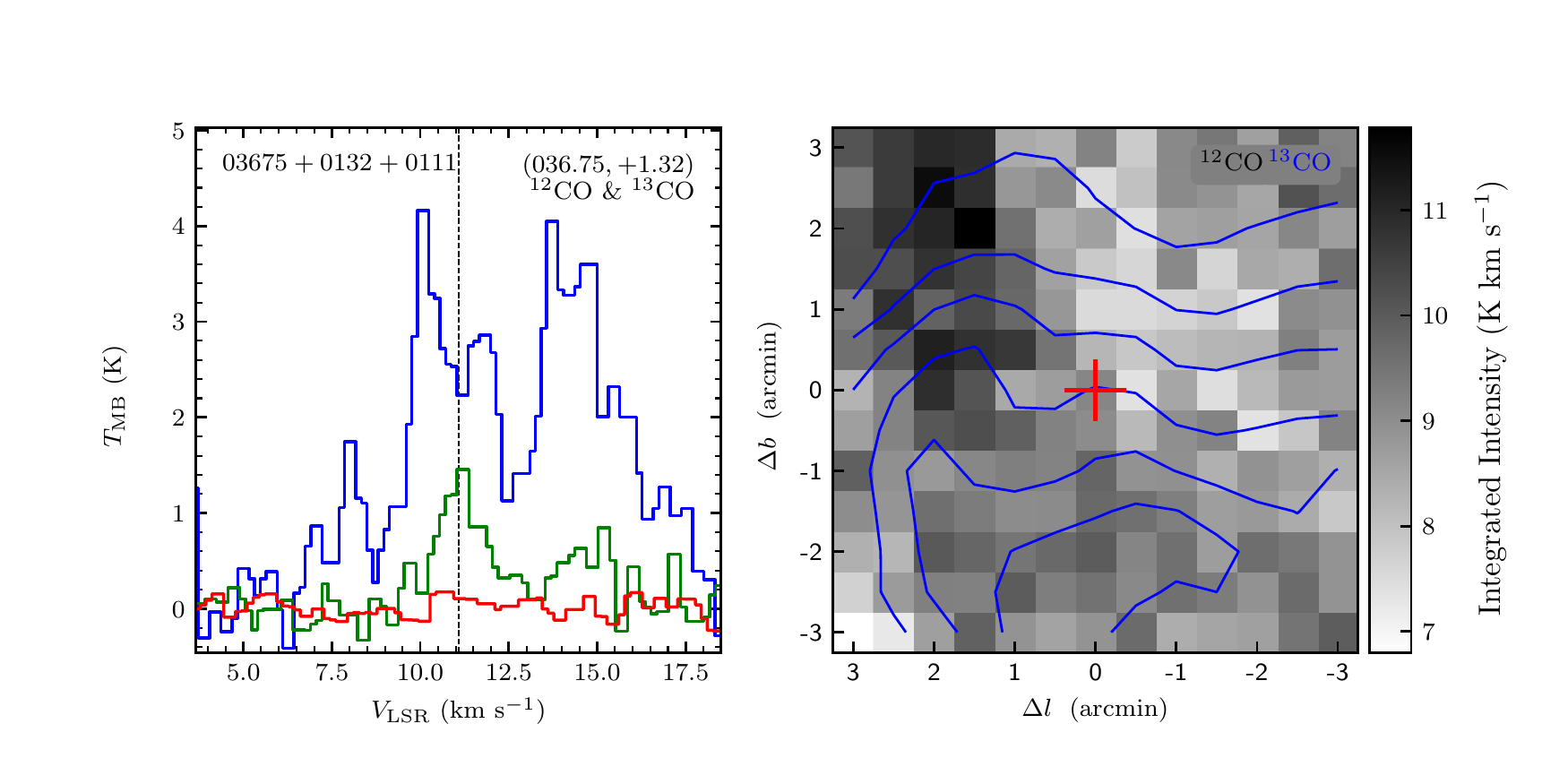}
\vspace{-0.5cm}

\includegraphics[width=9.0cm,angle=0]{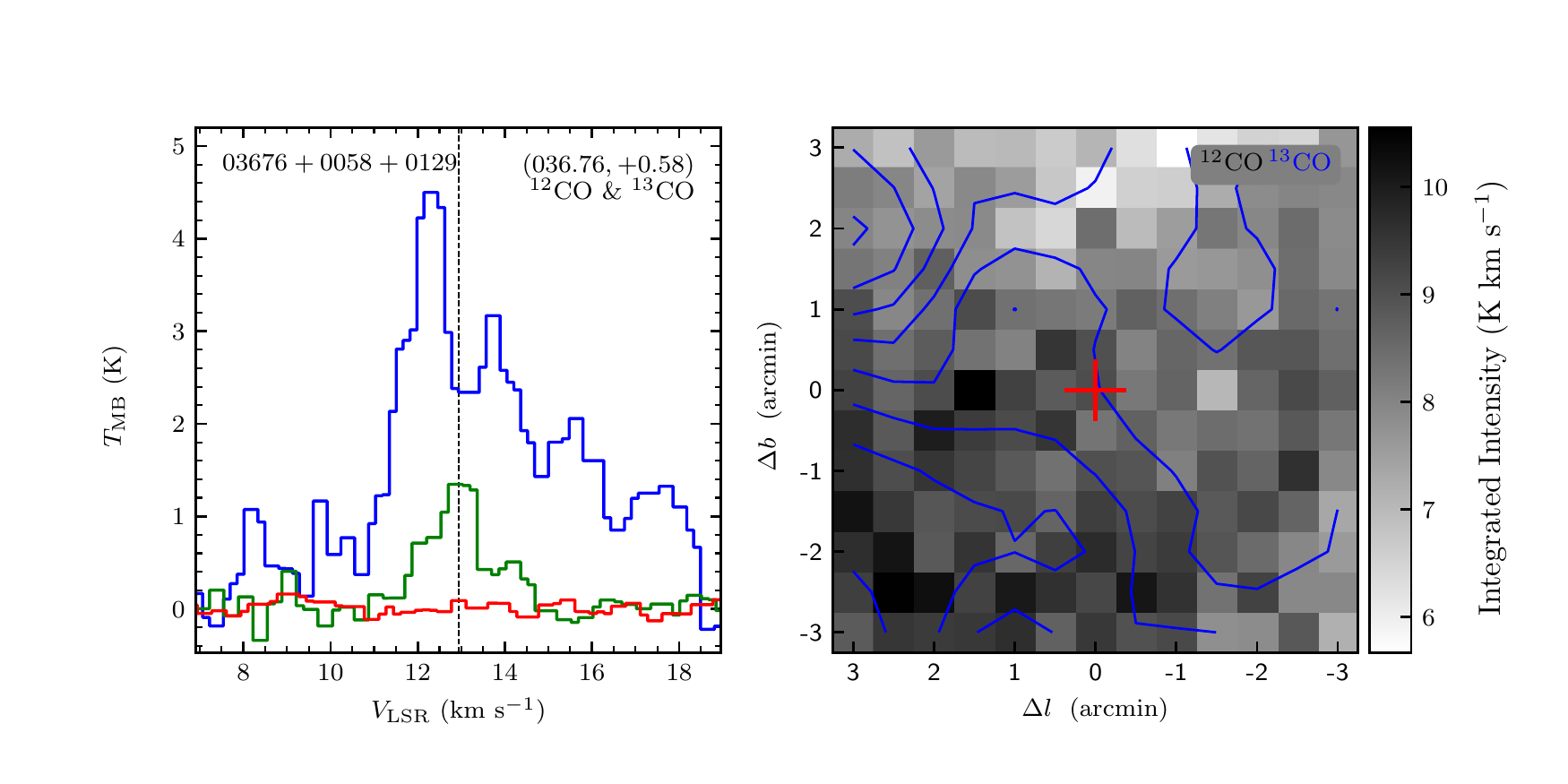}
\includegraphics[width=9.0cm,angle=0]{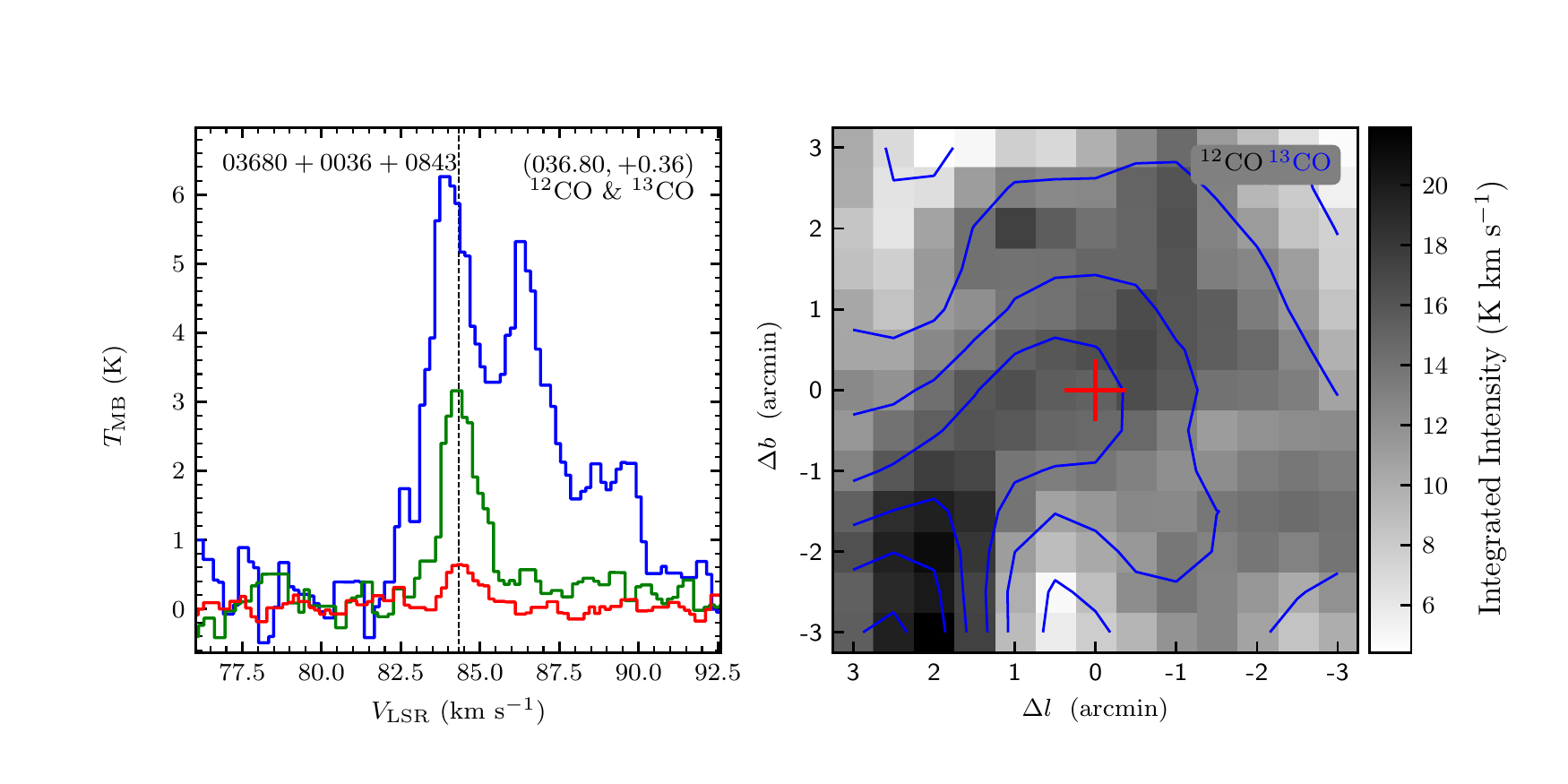}
\vspace{-0.5cm}

\includegraphics[width=9.0cm,angle=0]{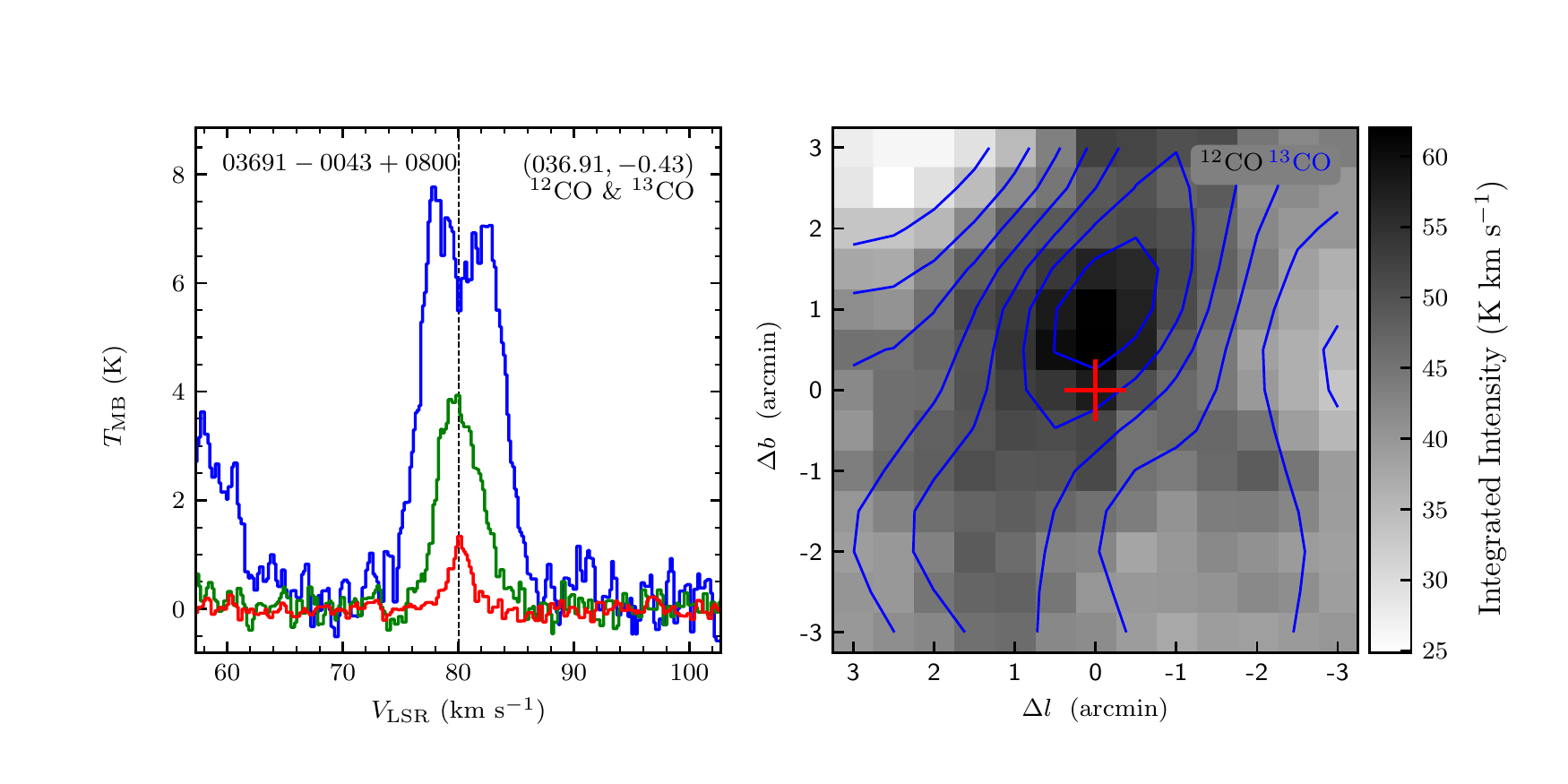}
\includegraphics[width=9.0cm,angle=0]{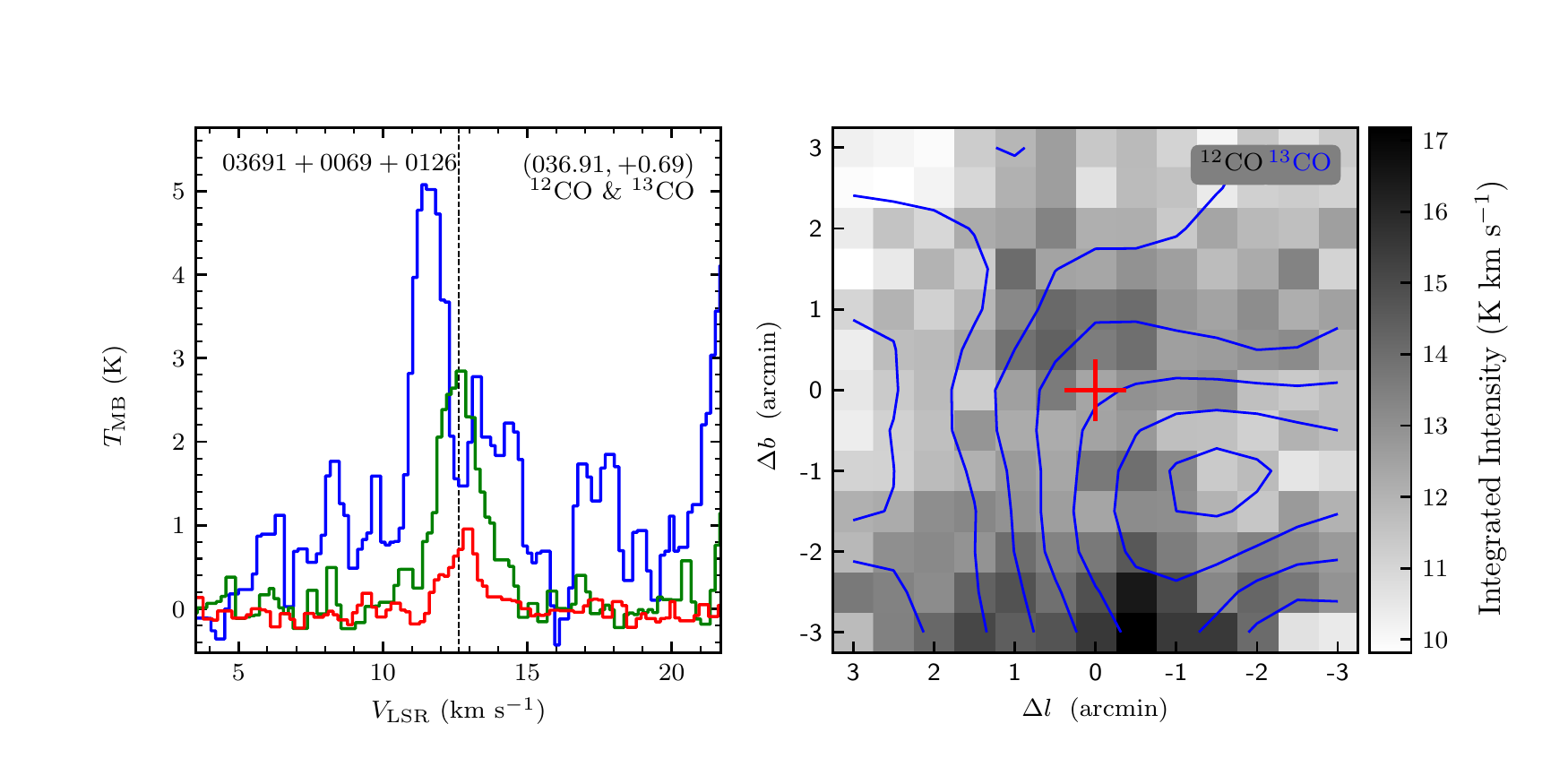}
\end{figure}
\clearpage

\begin{figure}
\includegraphics[width=9.0cm,angle=0]{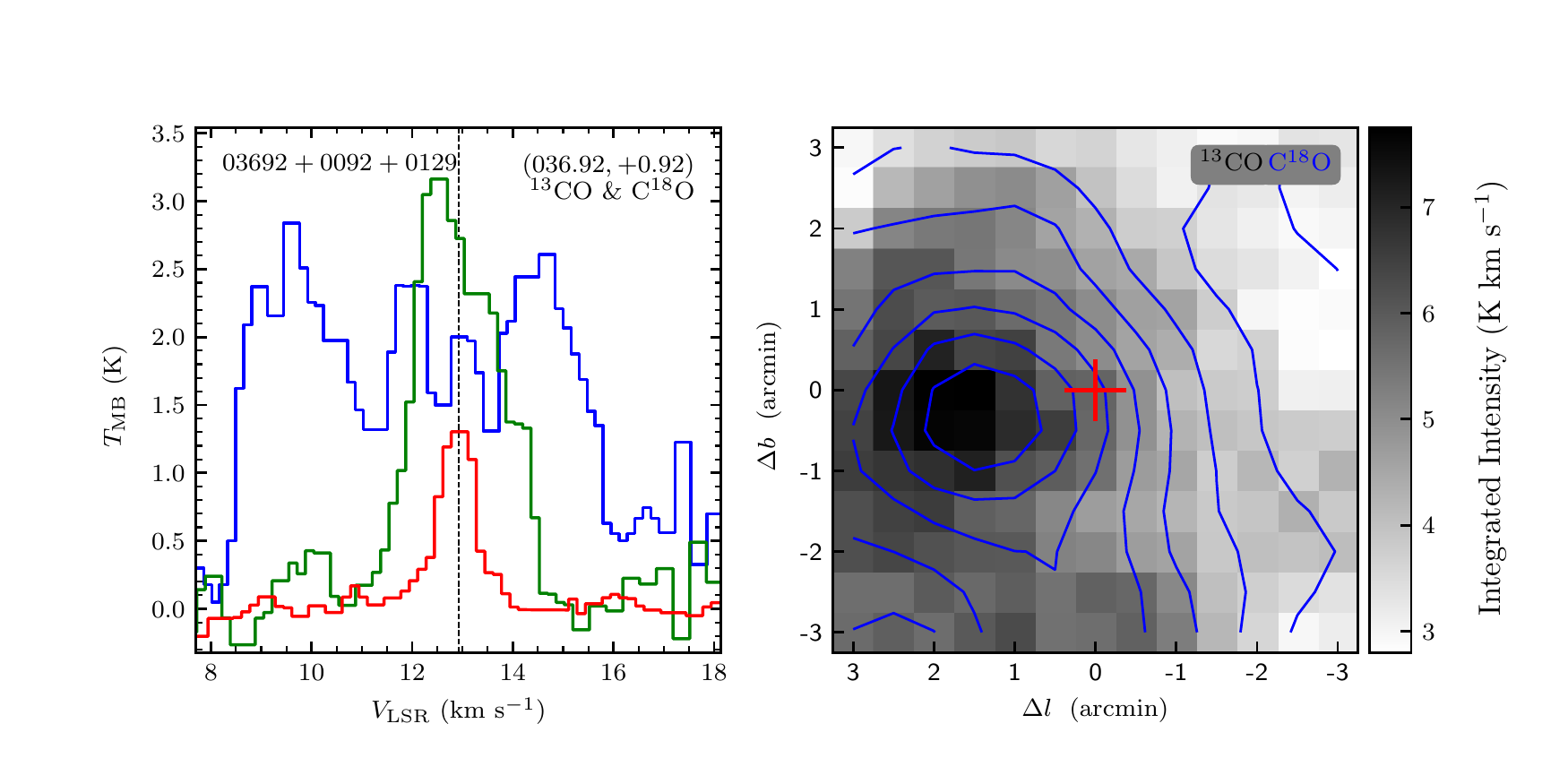}
\includegraphics[width=9.0cm,angle=0]{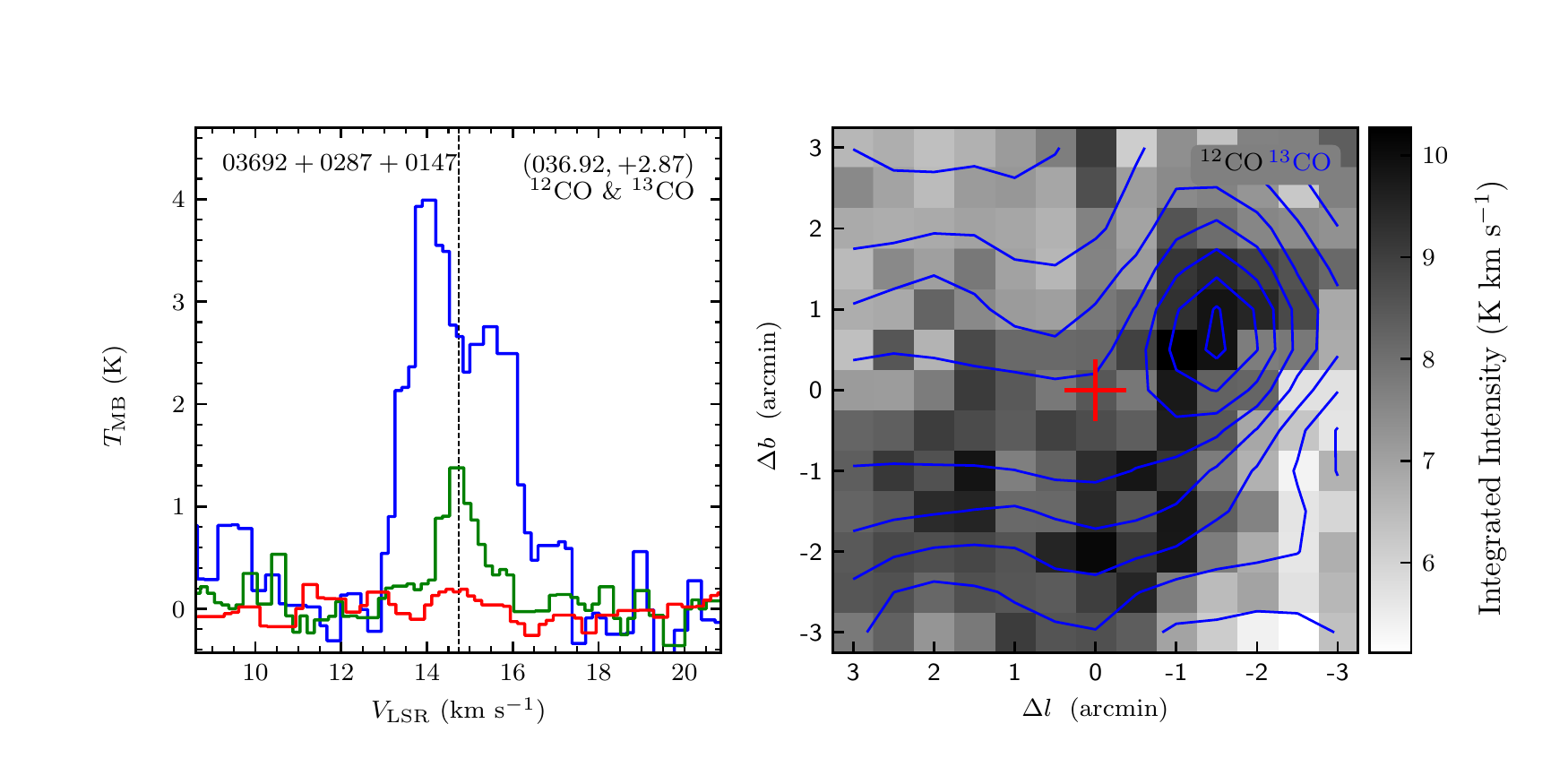}
\vspace{-0.5cm}

\includegraphics[width=9.0cm,angle=0]{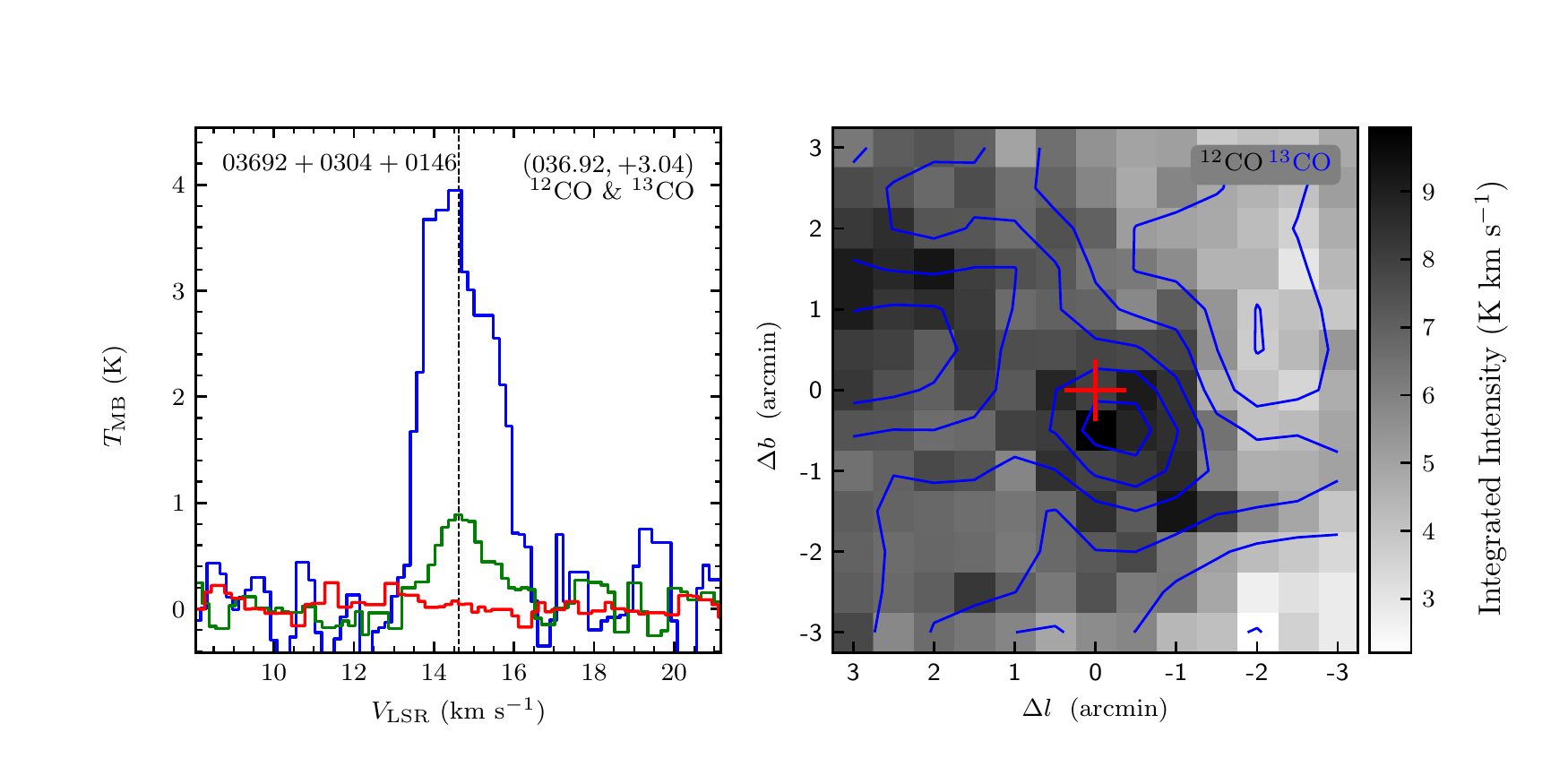}
\includegraphics[width=9.0cm,angle=0]{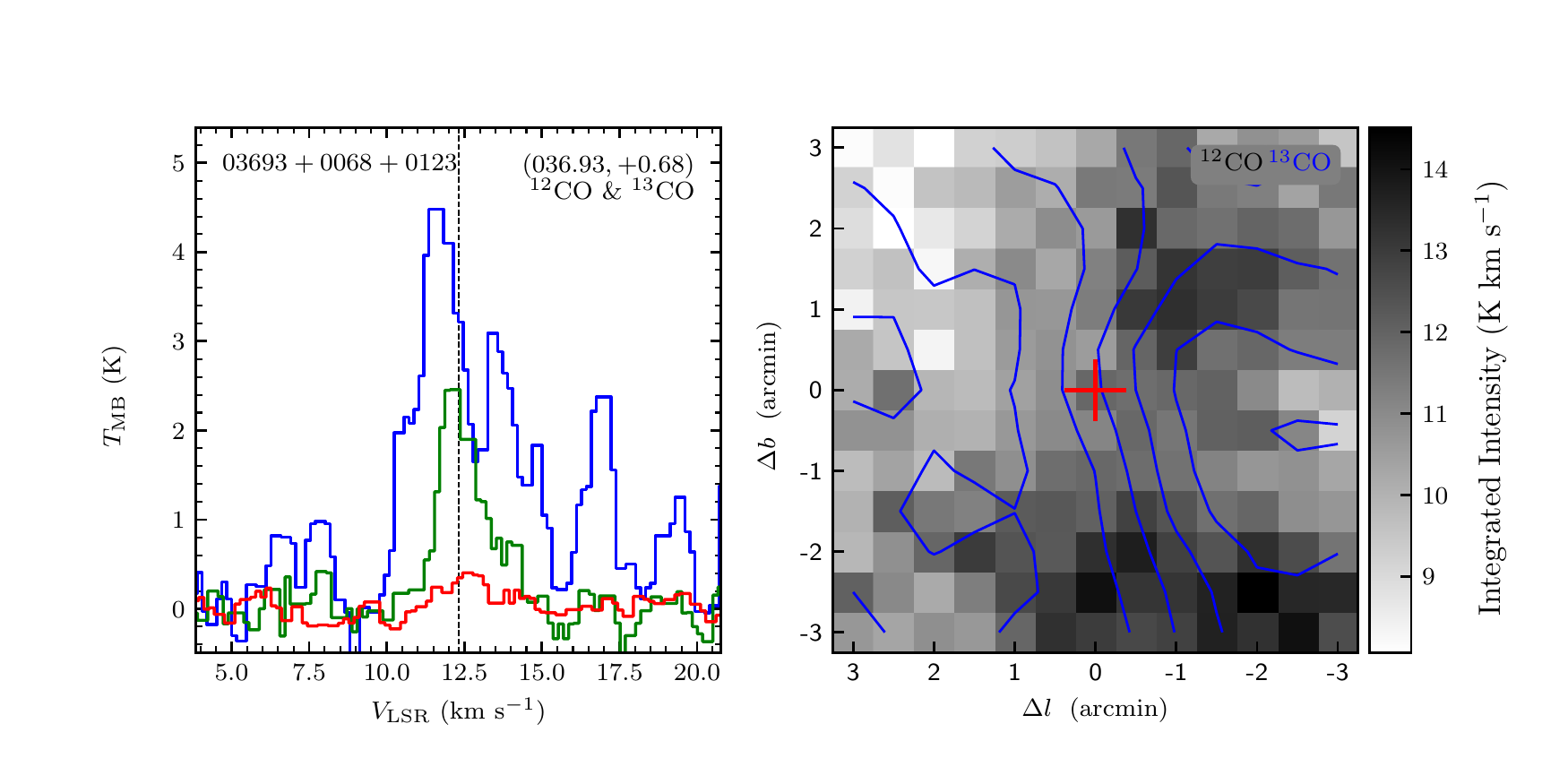}
\vspace{-0.5cm}

\includegraphics[width=9.0cm,angle=0]{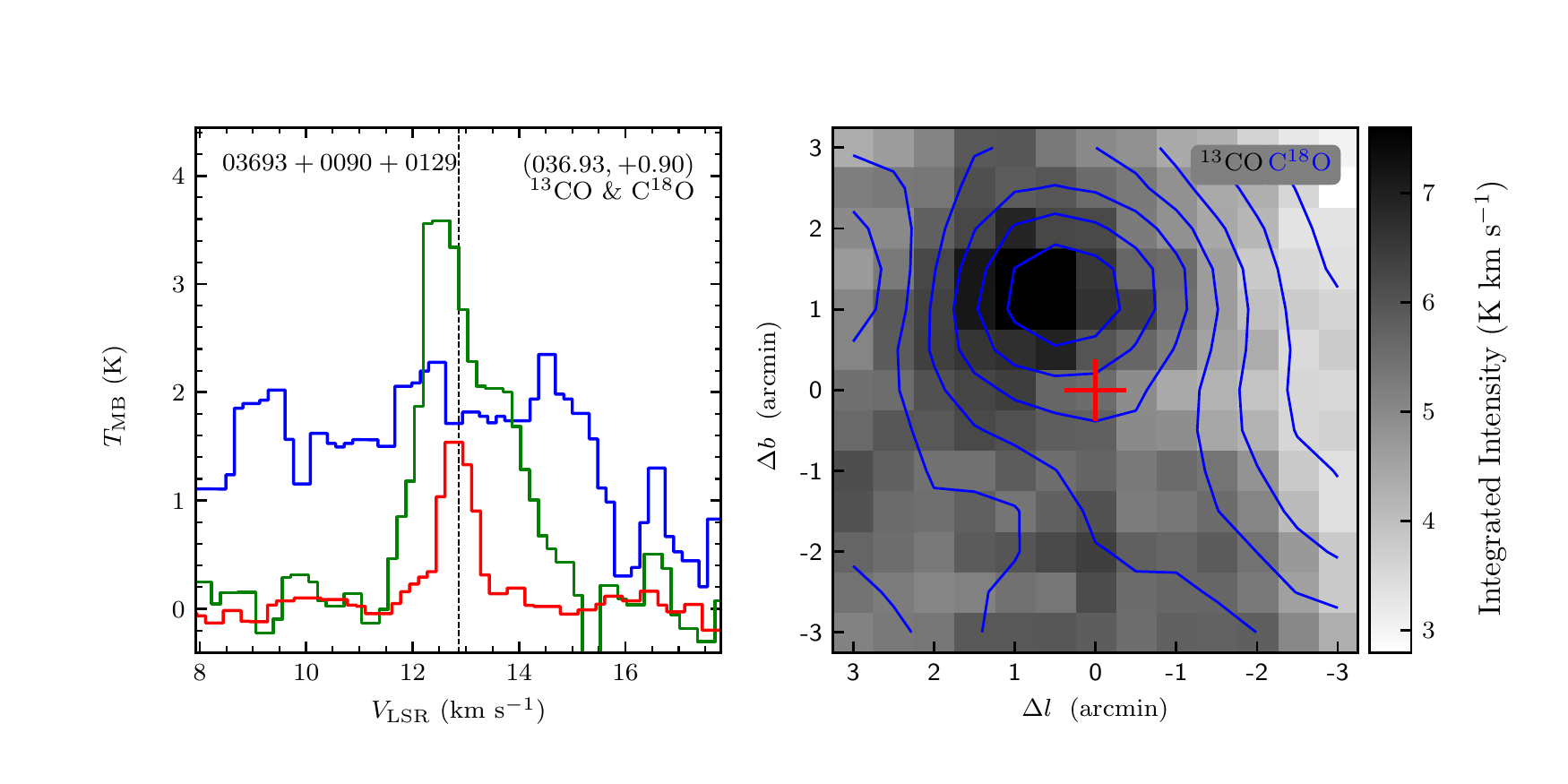}
\includegraphics[width=9.0cm,angle=0]{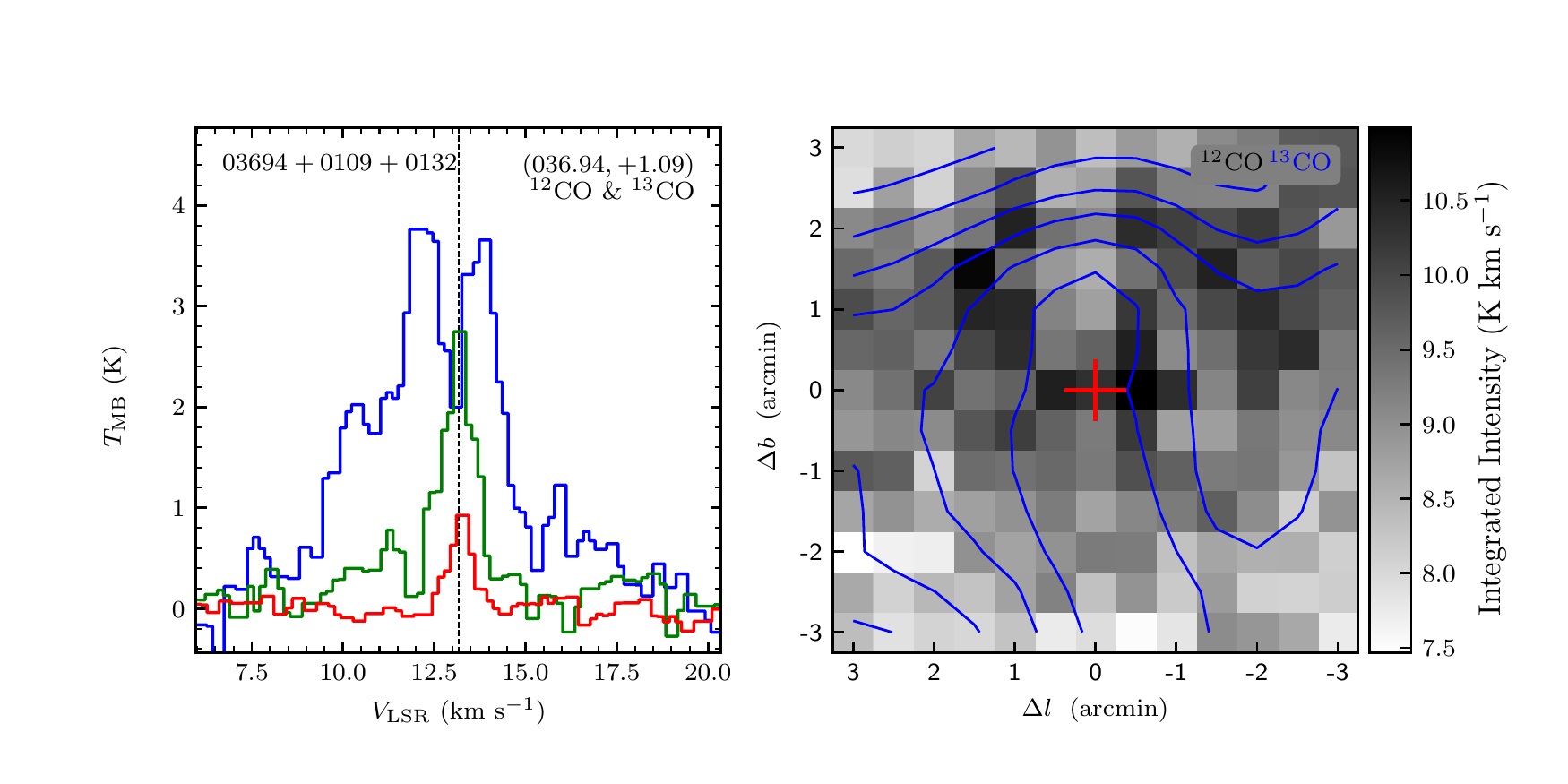}
\vspace{-0.5cm}

\includegraphics[width=9.0cm,angle=0]{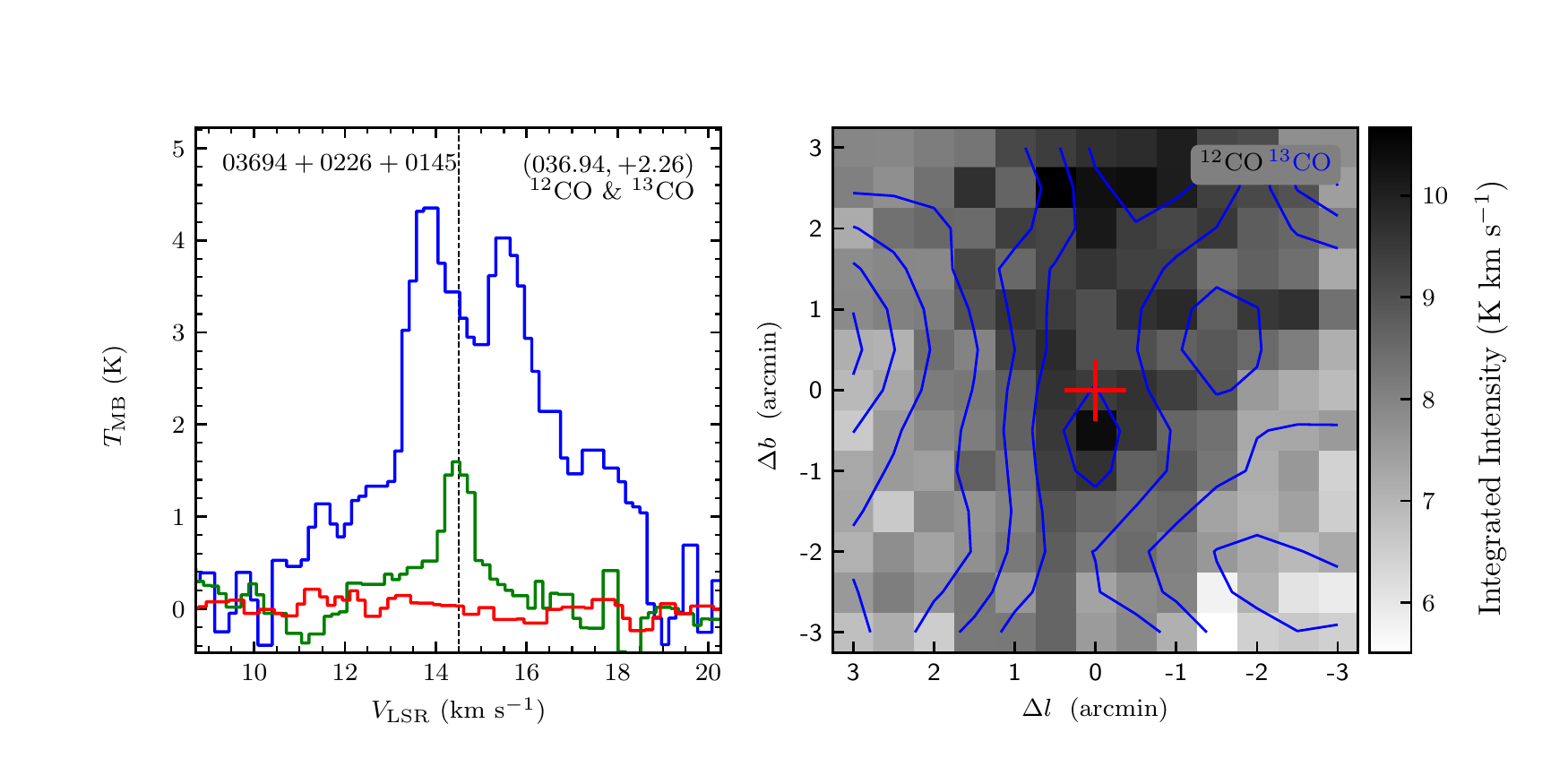}
\includegraphics[width=9.0cm,angle=0]{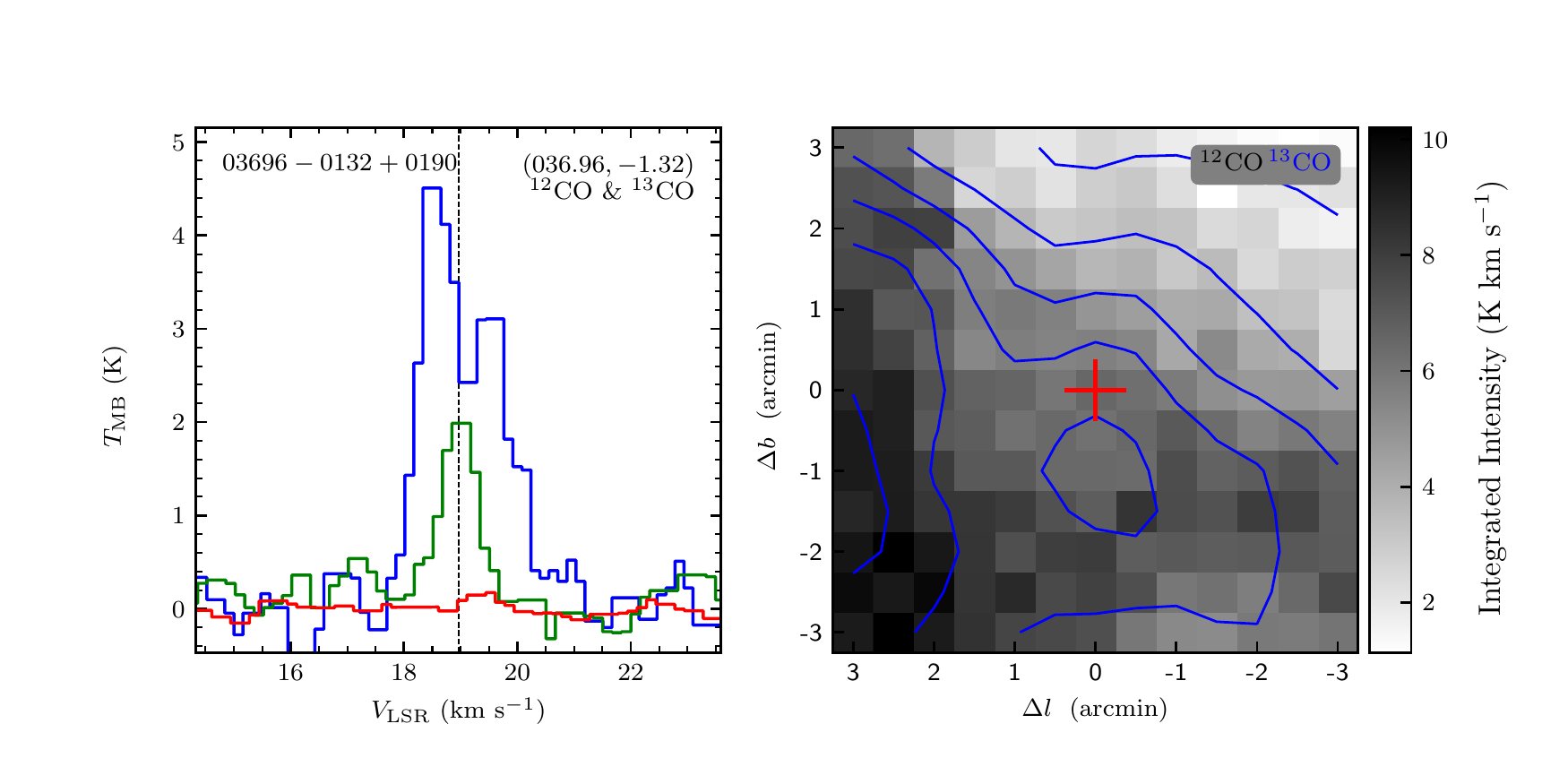}
\vspace{-0.5cm}

\includegraphics[width=9.0cm,angle=0]{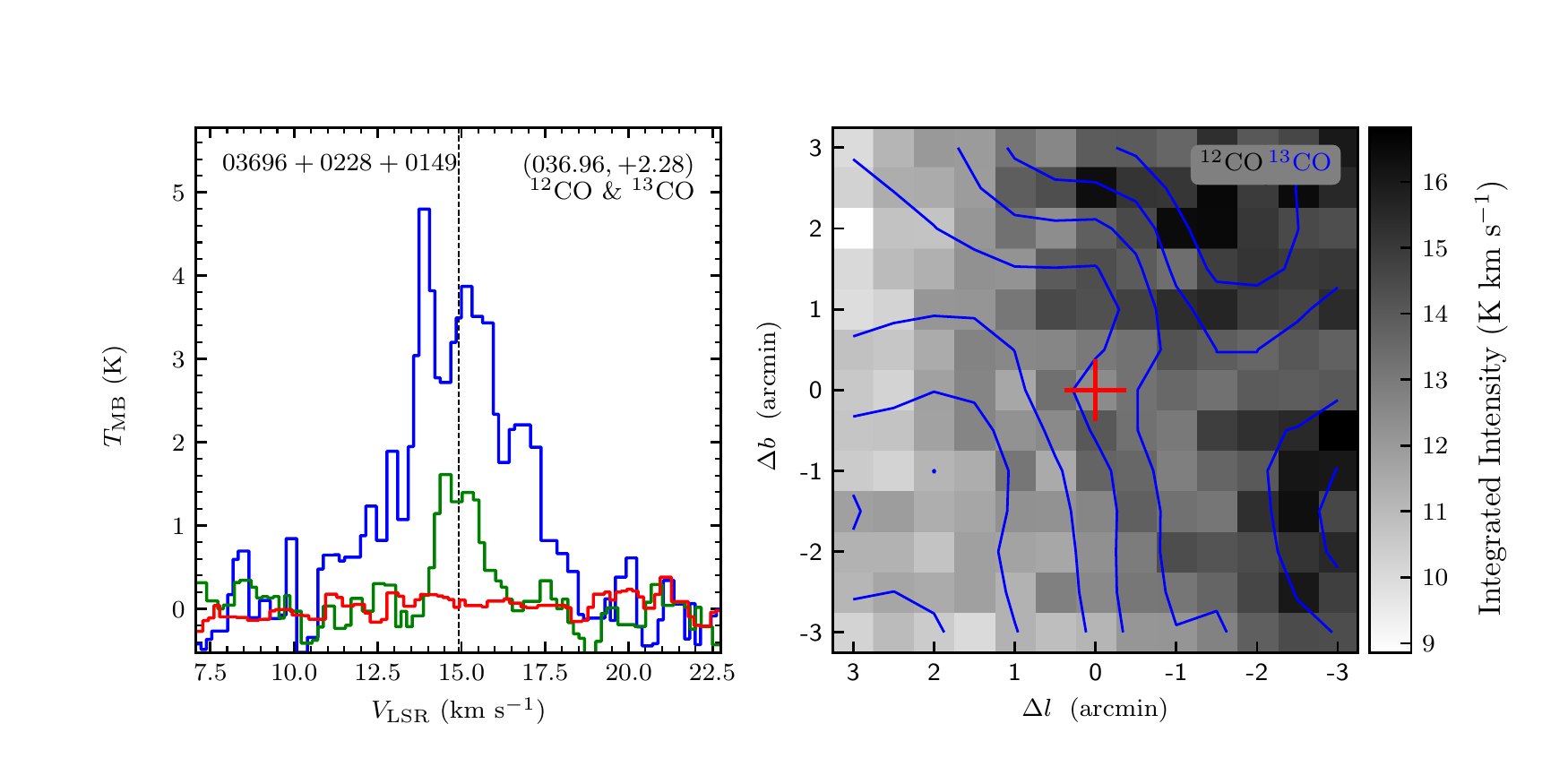}
\includegraphics[width=9.0cm,angle=0]{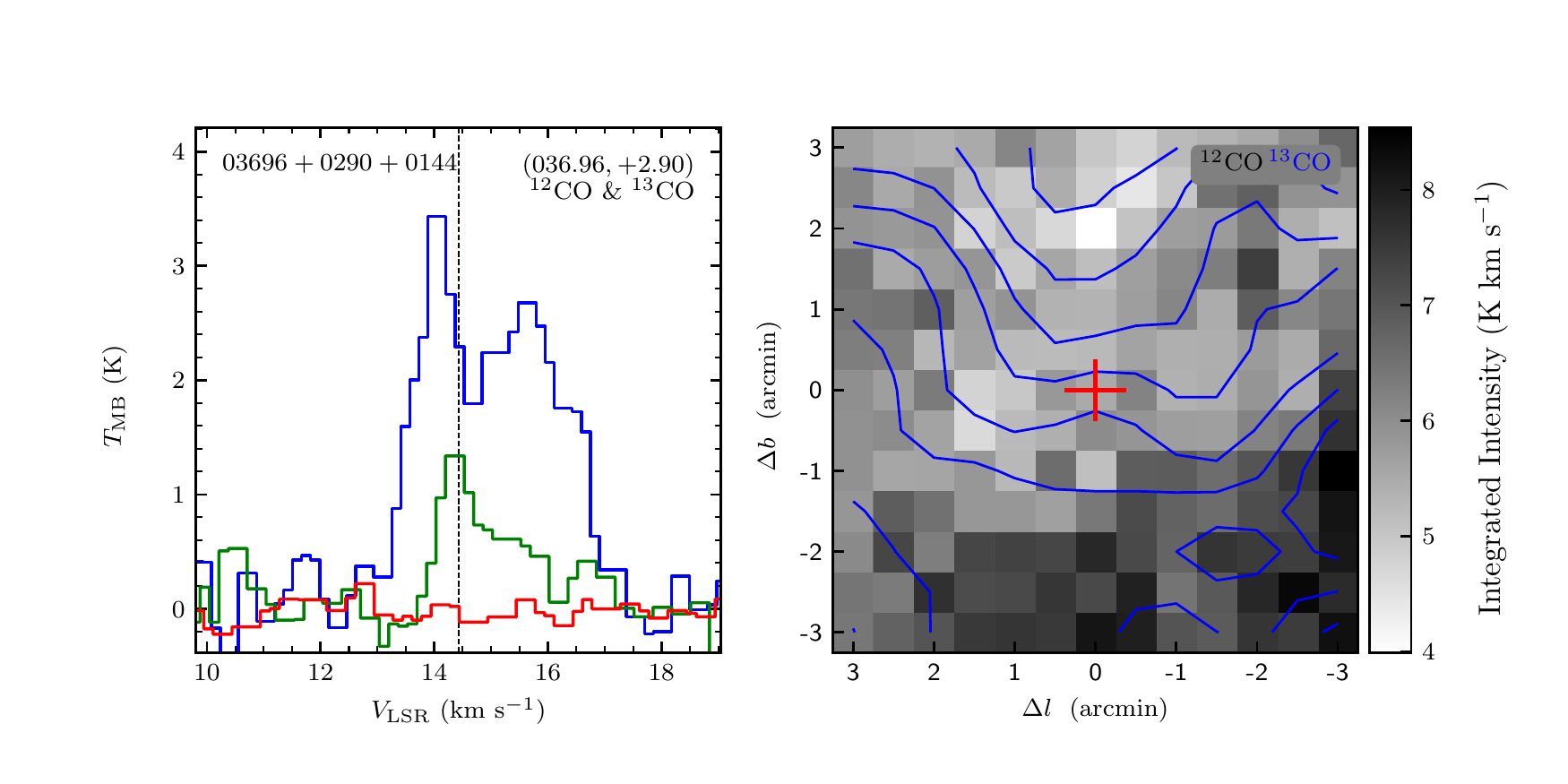}
\end{figure}
\clearpage

\begin{figure}
\includegraphics[width=9.0cm,angle=0]{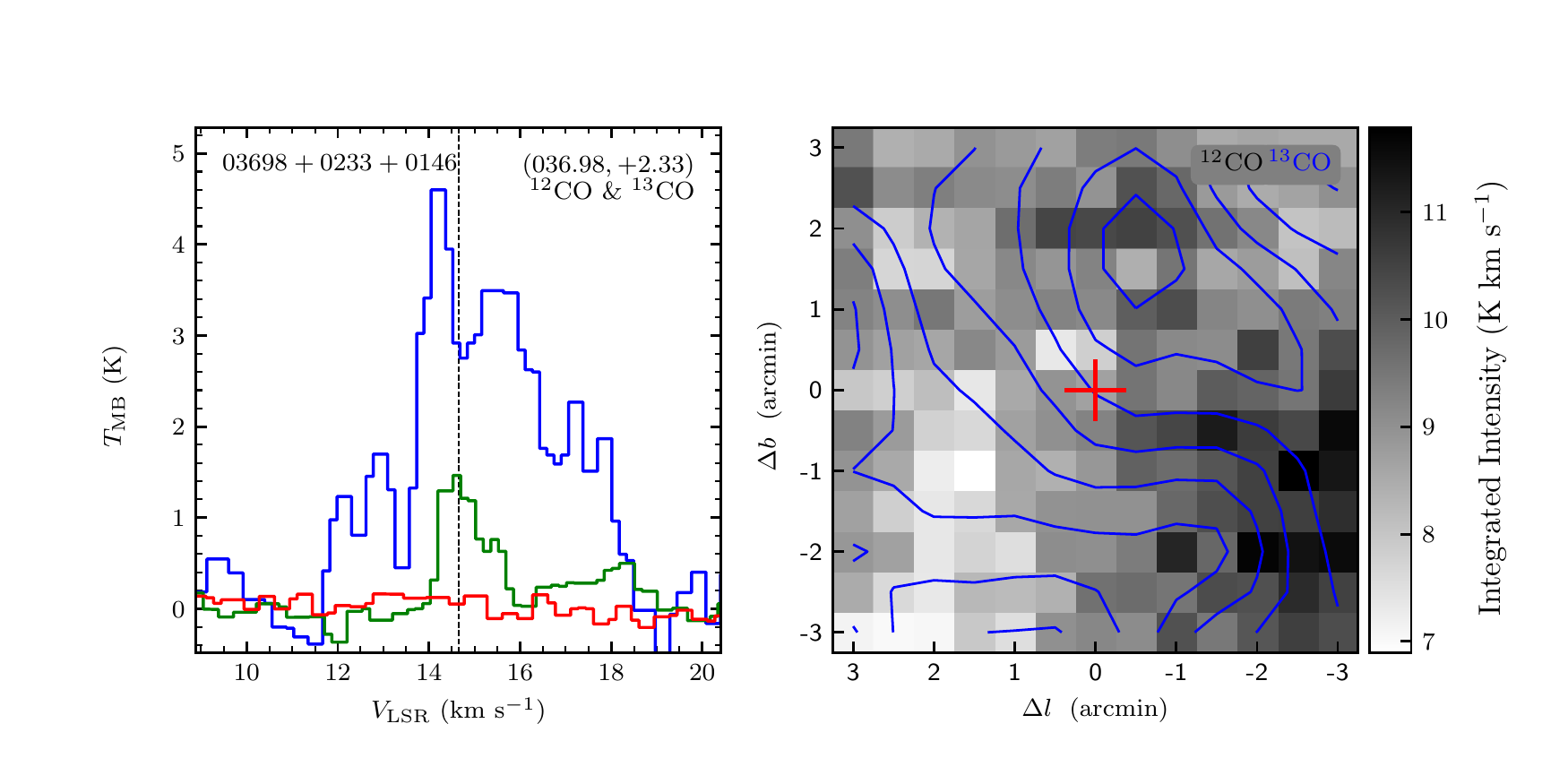}
\includegraphics[width=9.0cm,angle=0]{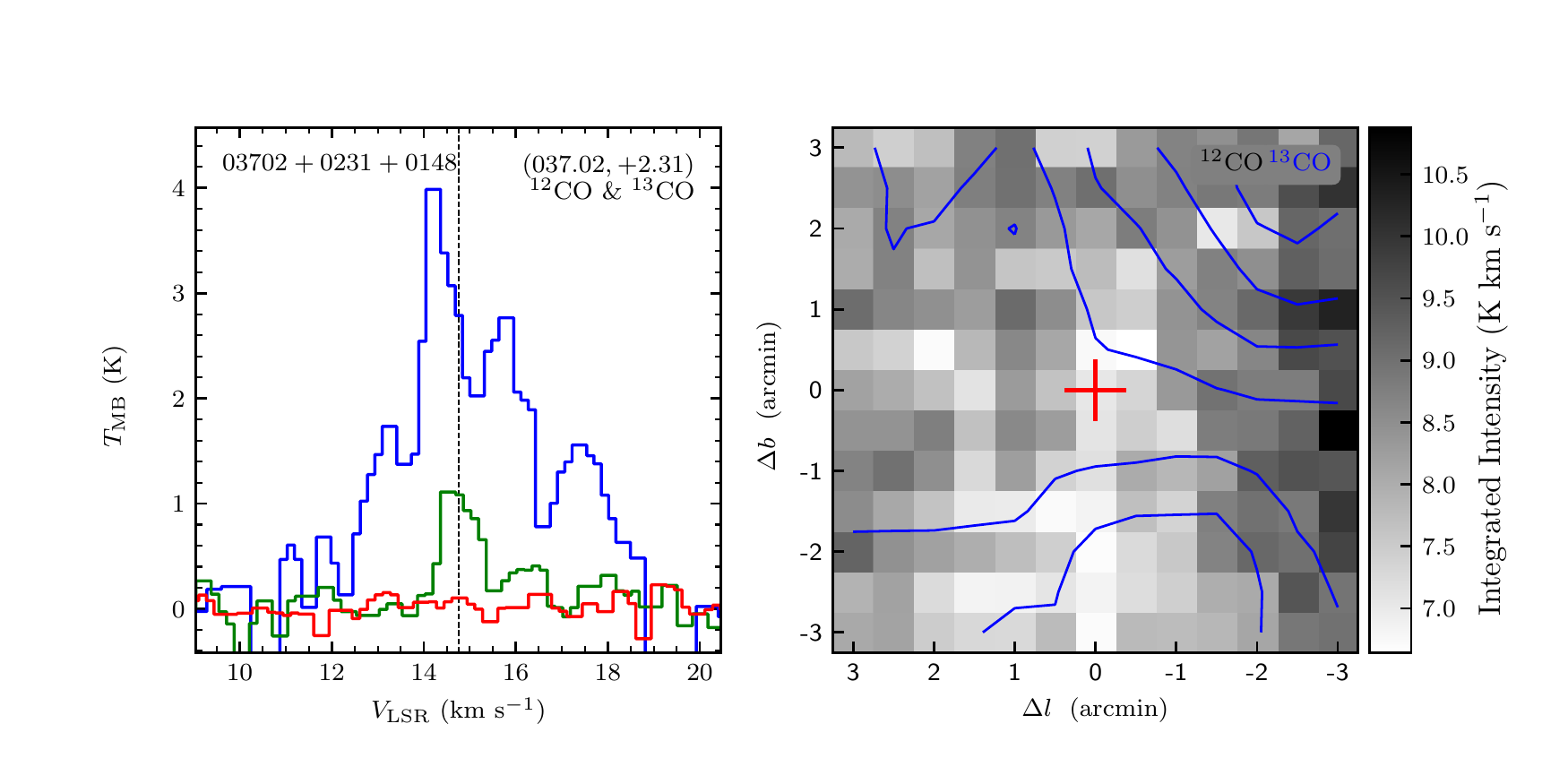}
\vspace{-0.5cm}

\includegraphics[width=9.0cm,angle=0]{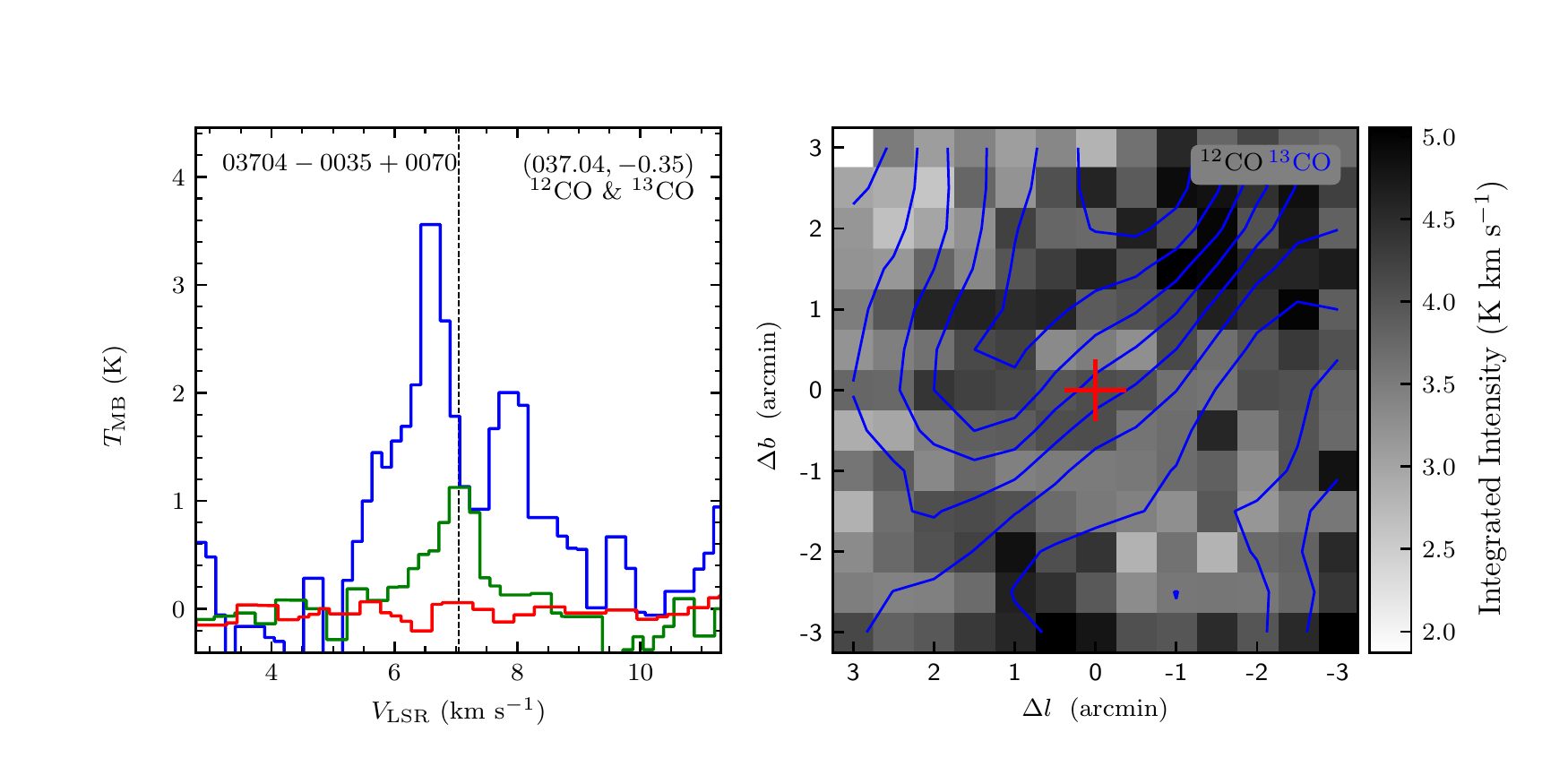}
\includegraphics[width=9.0cm,angle=0]{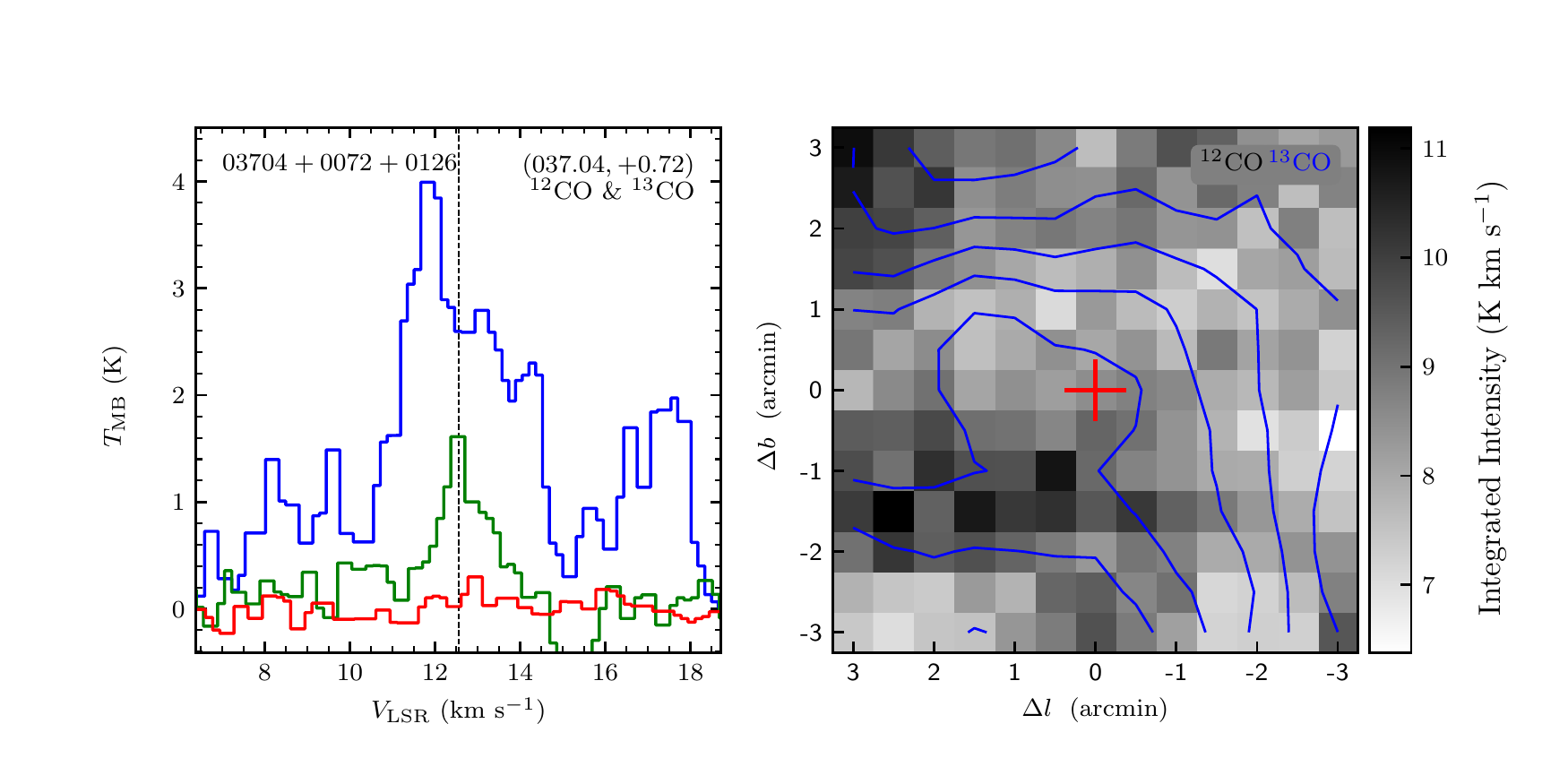}
\vspace{-0.5cm}

\includegraphics[width=9.0cm,angle=0]{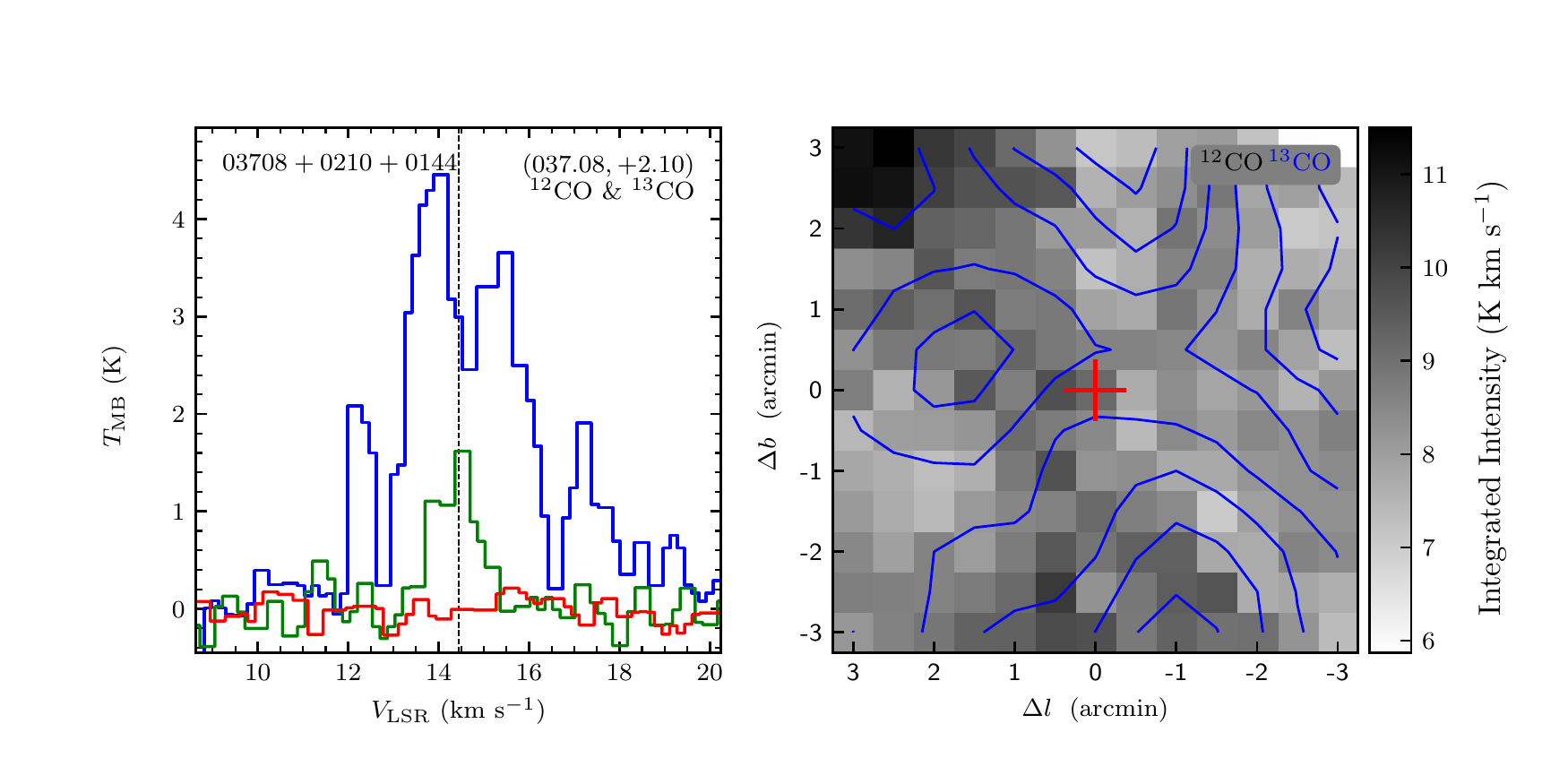}
\includegraphics[width=9.0cm,angle=0]{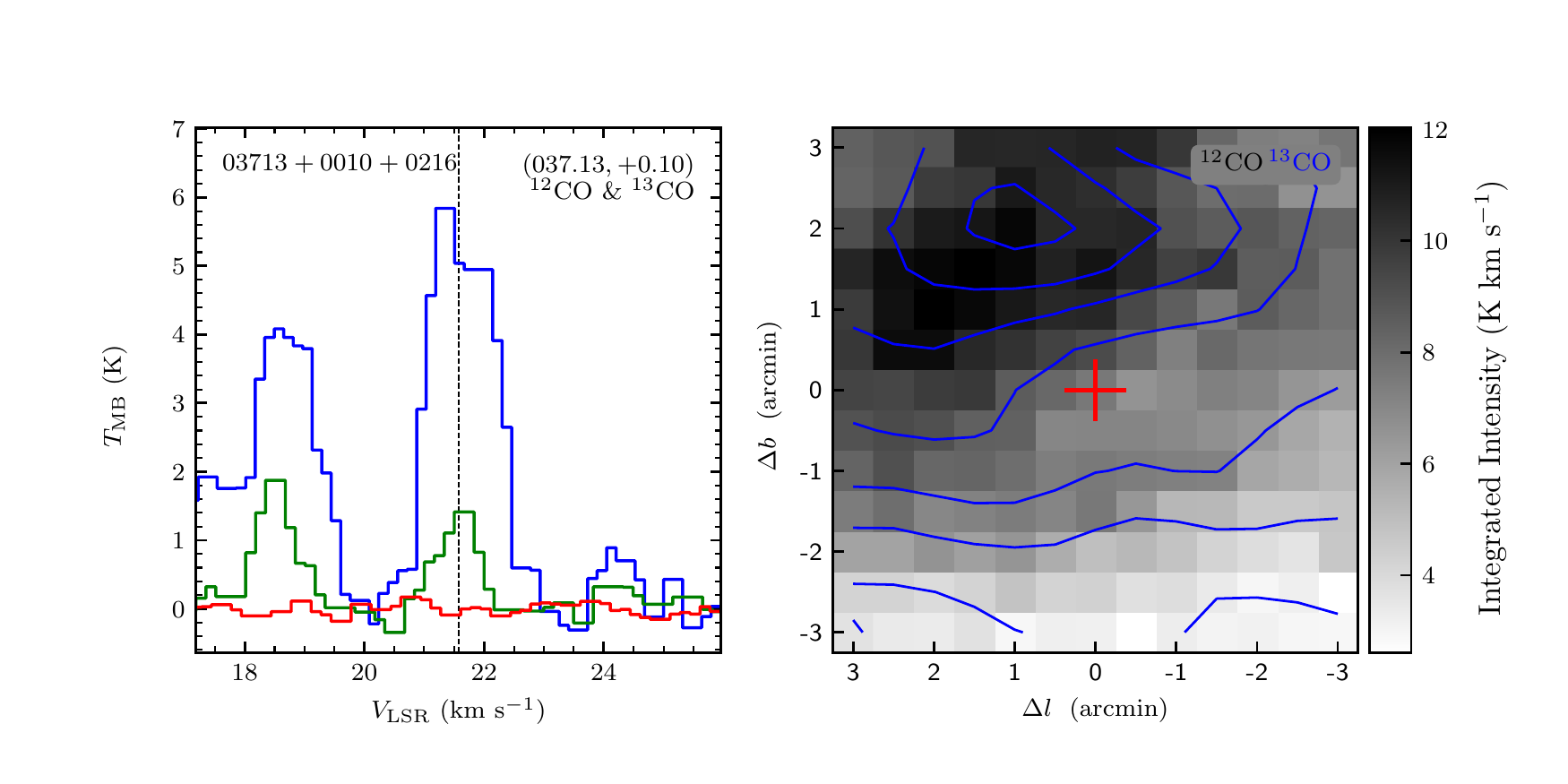}
\vspace{-0.5cm}

\includegraphics[width=9.0cm,angle=0]{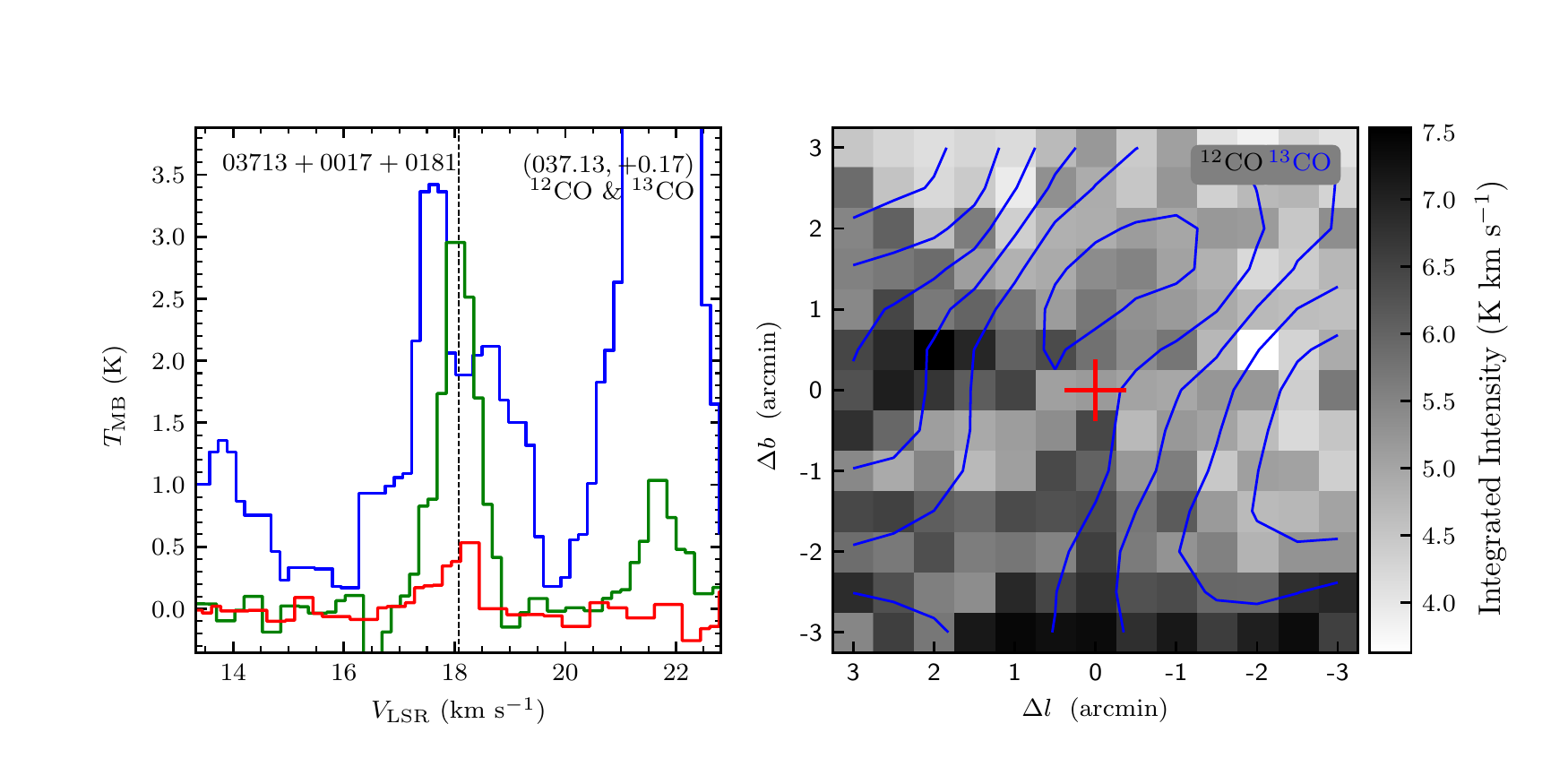}
\includegraphics[width=9.0cm,angle=0]{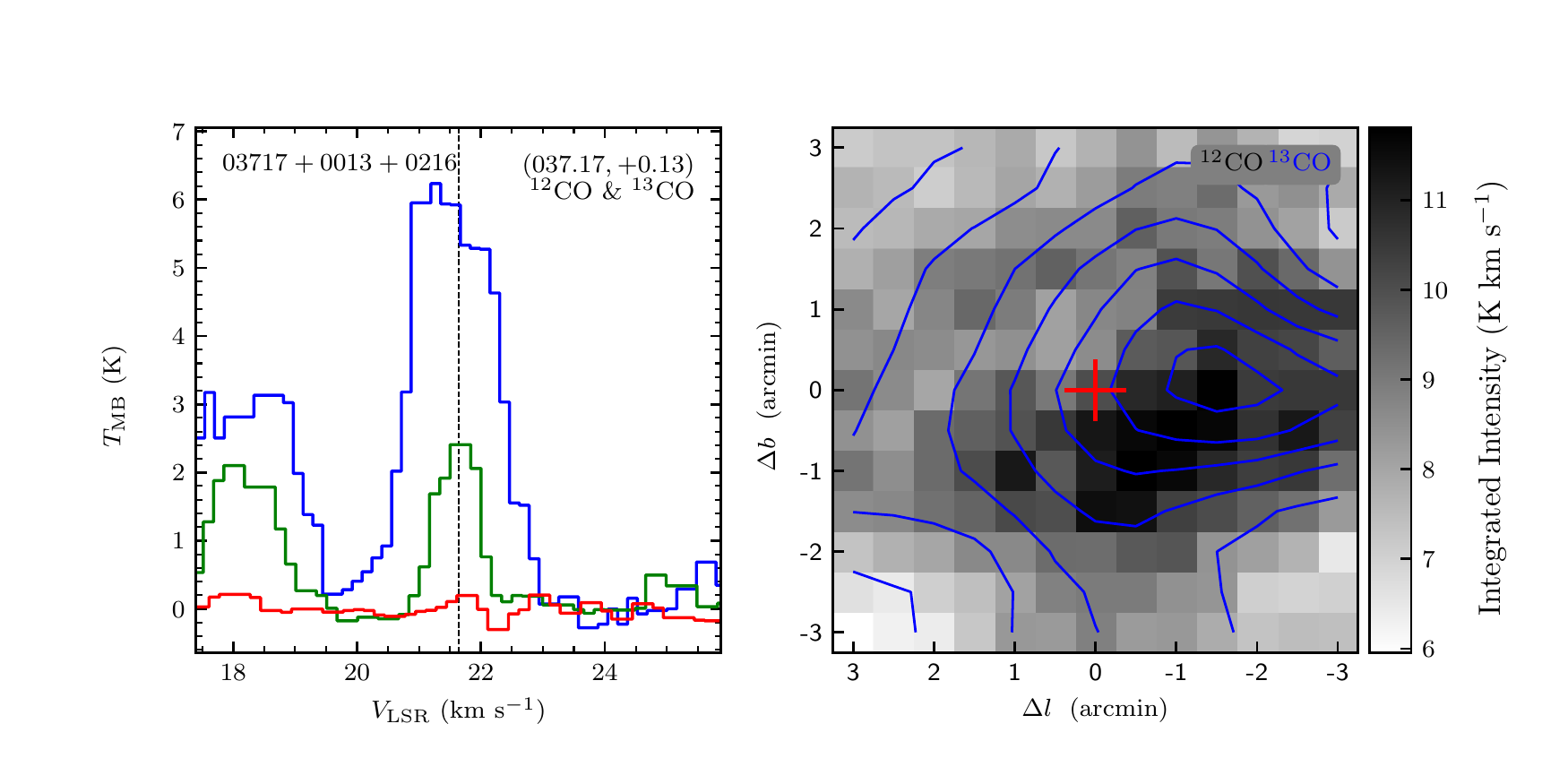}
\vspace{-0.5cm}

\includegraphics[width=9.0cm,angle=0]{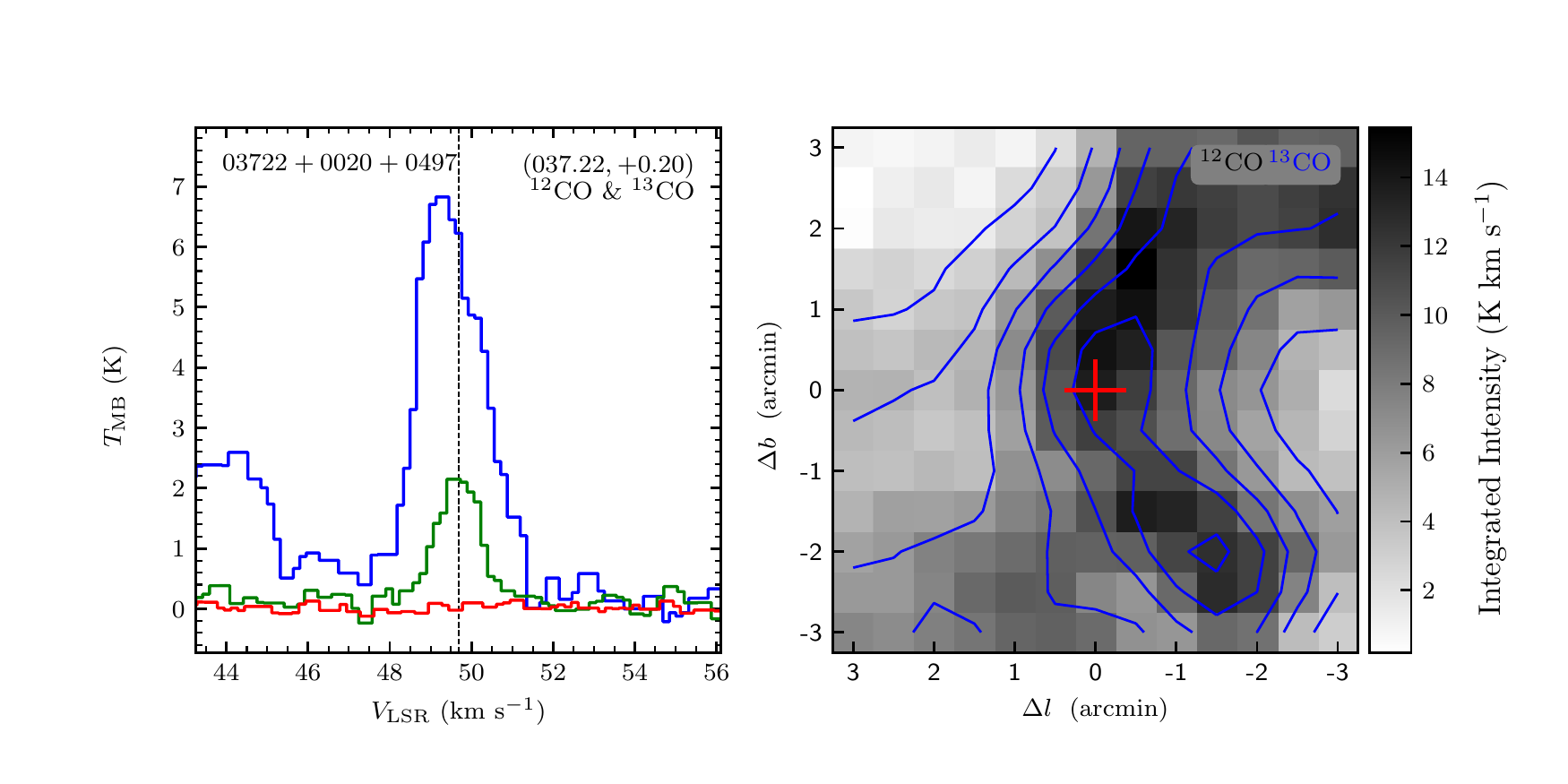}
\includegraphics[width=9.0cm,angle=0]{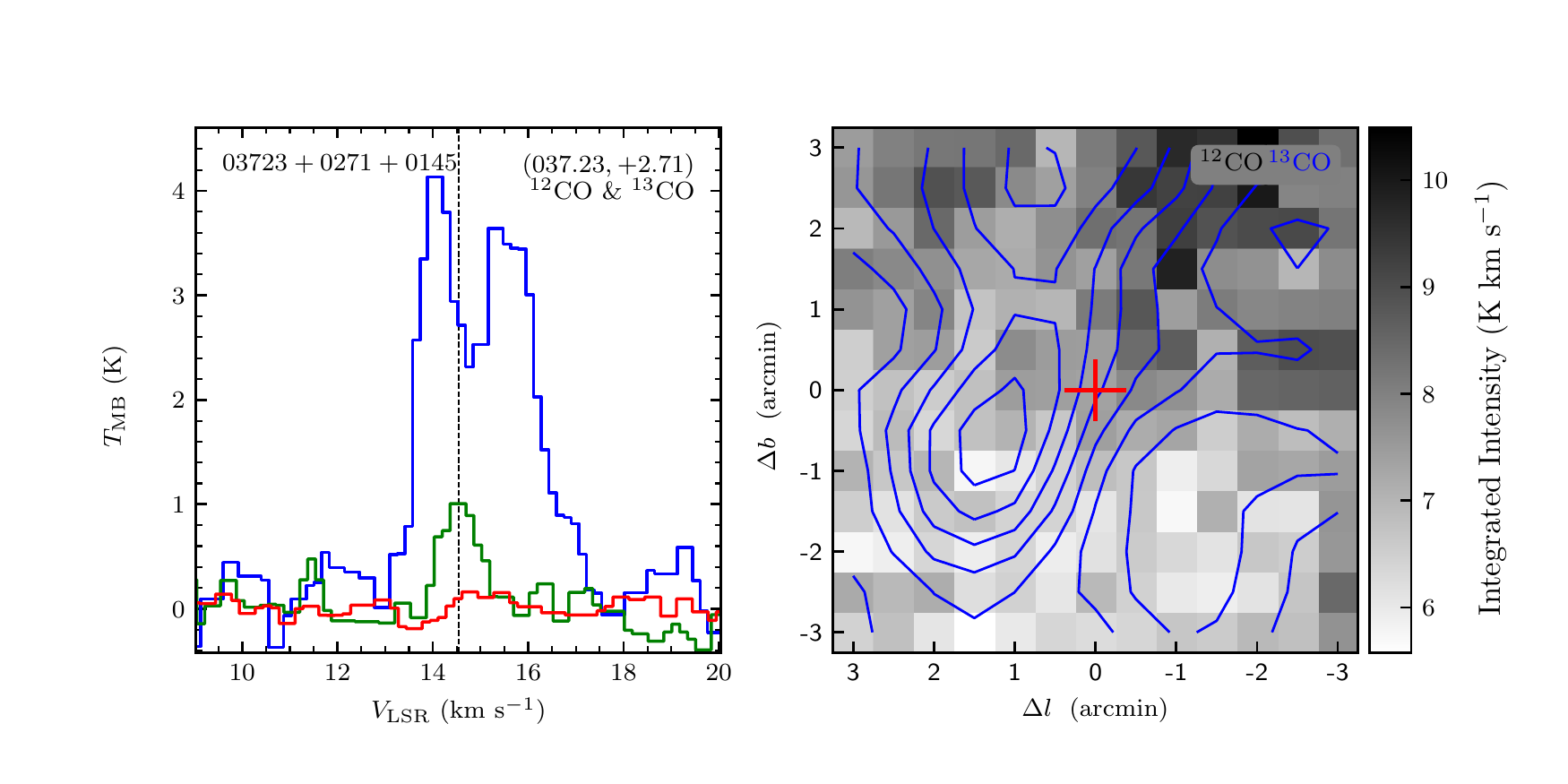}
\end{figure}
\clearpage

\begin{figure}
\includegraphics[width=9.0cm,angle=0]{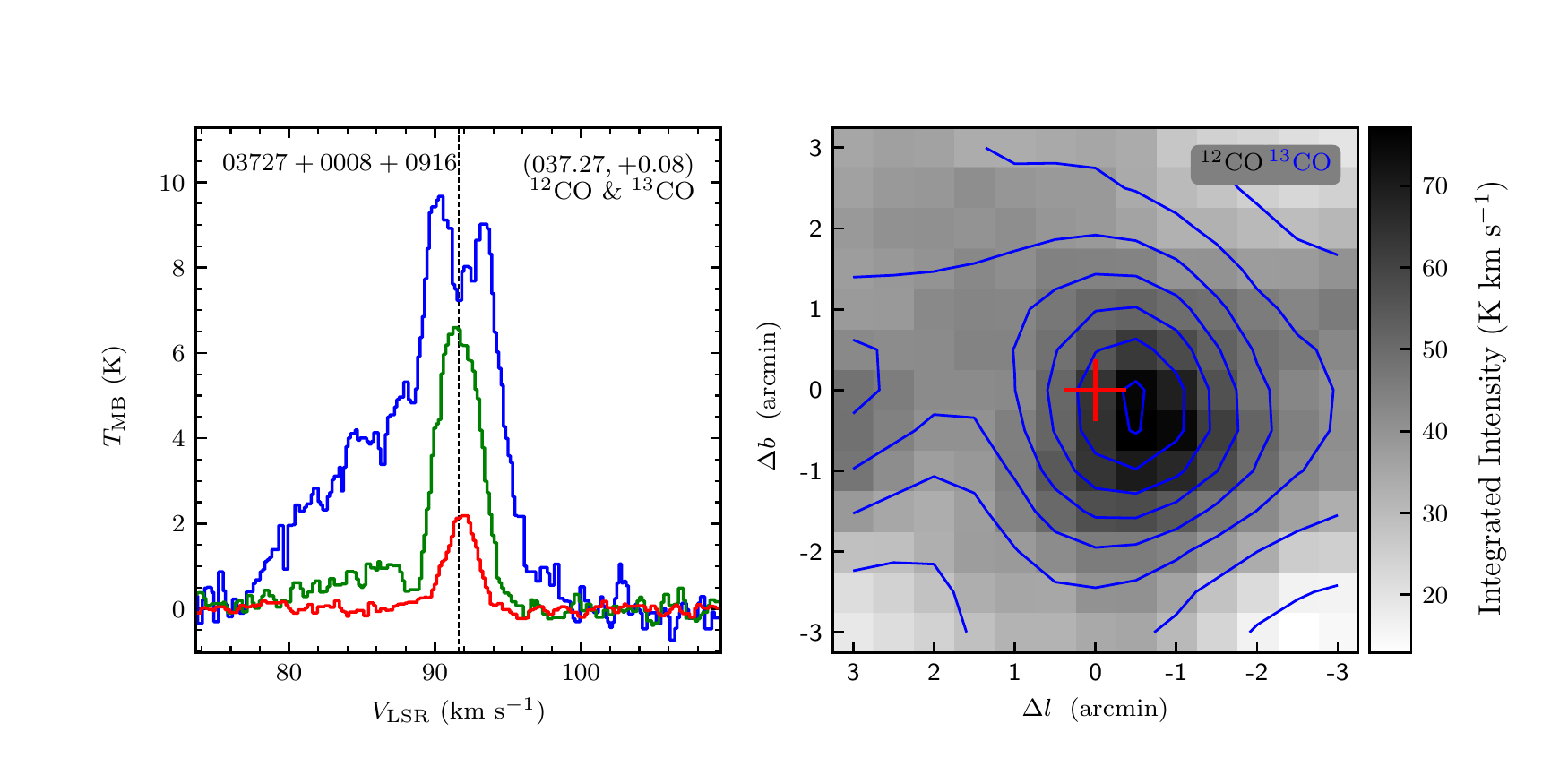}
\includegraphics[width=9.0cm,angle=0]{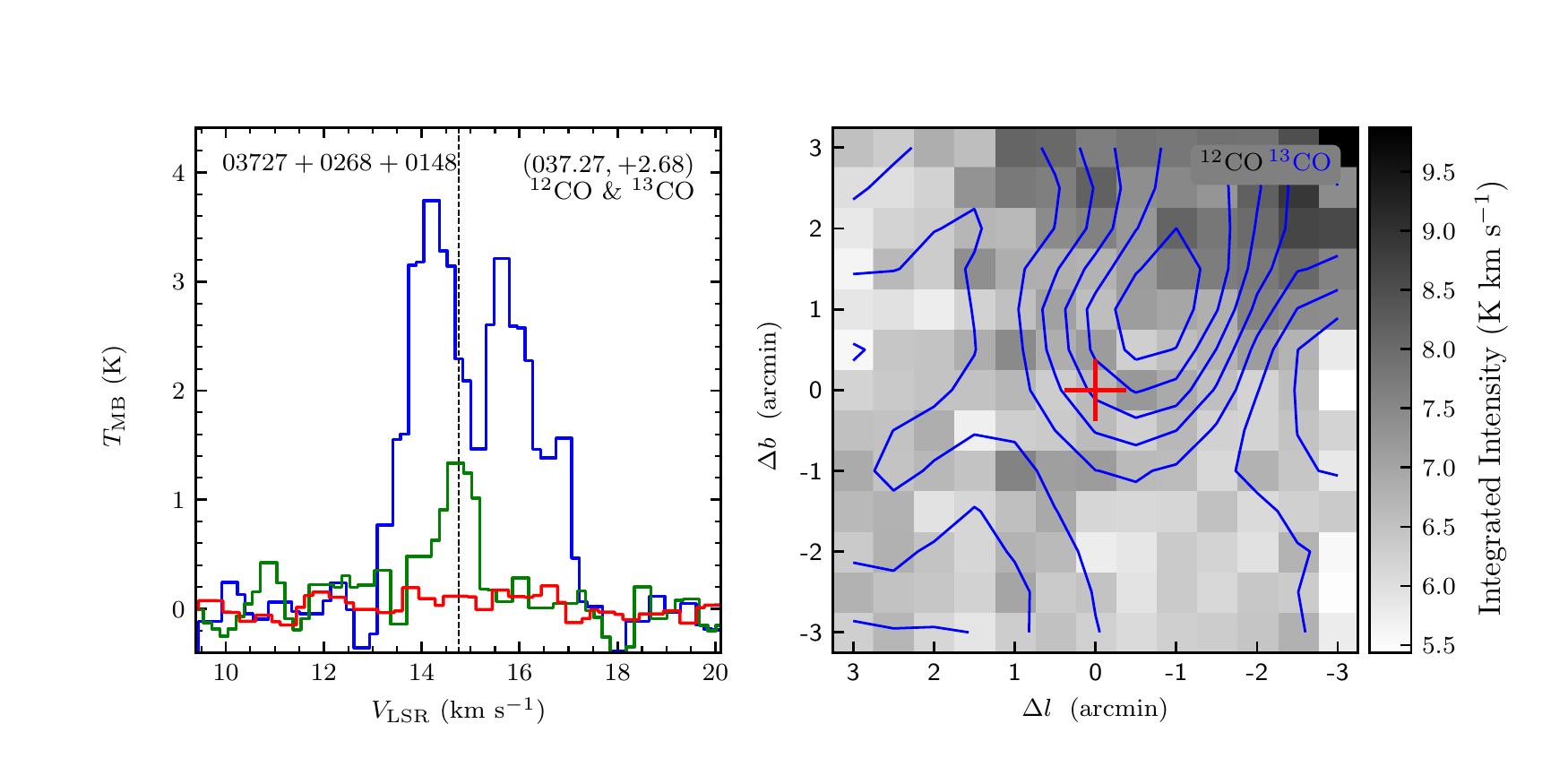}
\vspace{-0.5cm}

\includegraphics[width=9.0cm,angle=0]{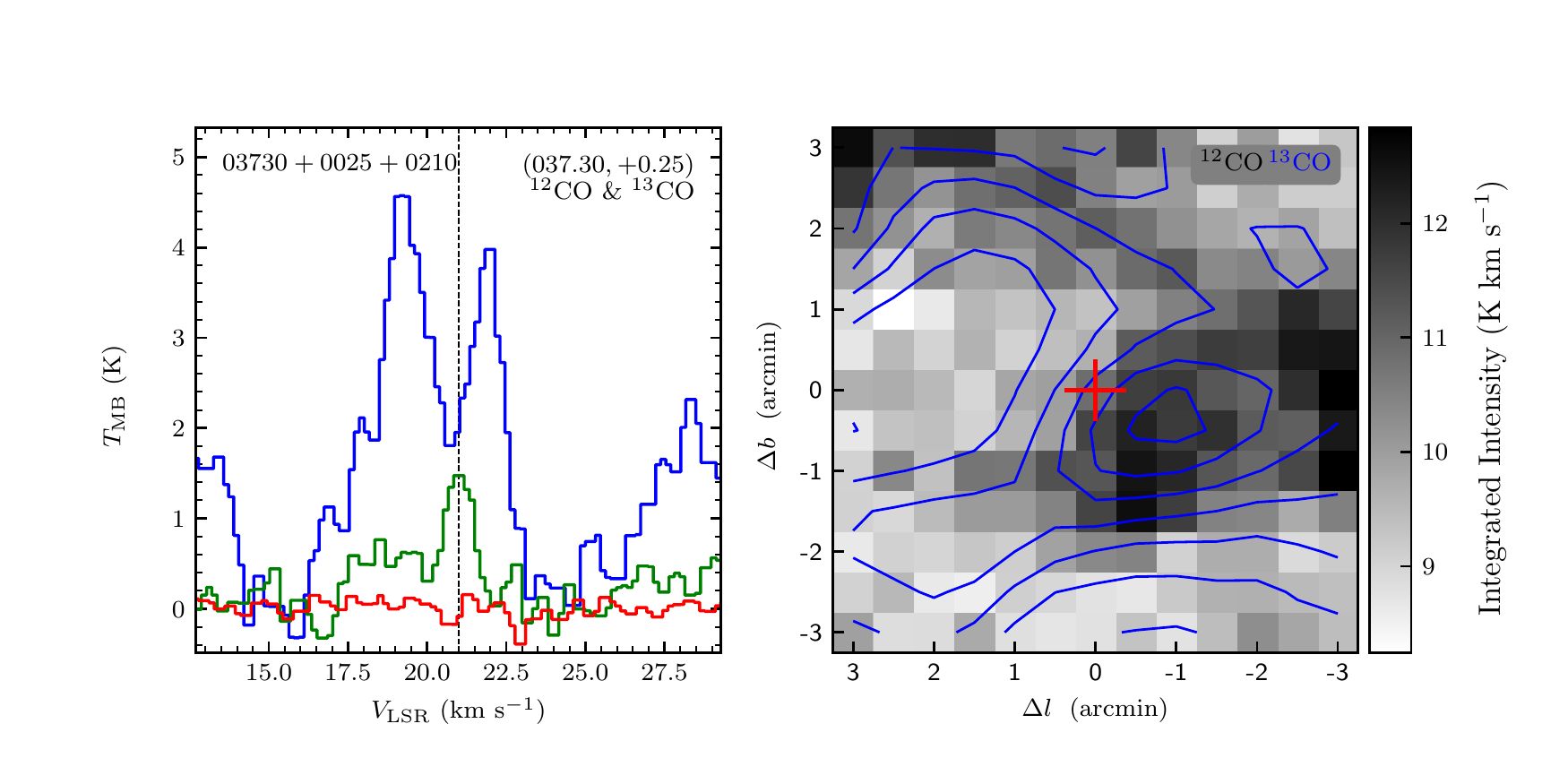}
\includegraphics[width=9.0cm,angle=0]{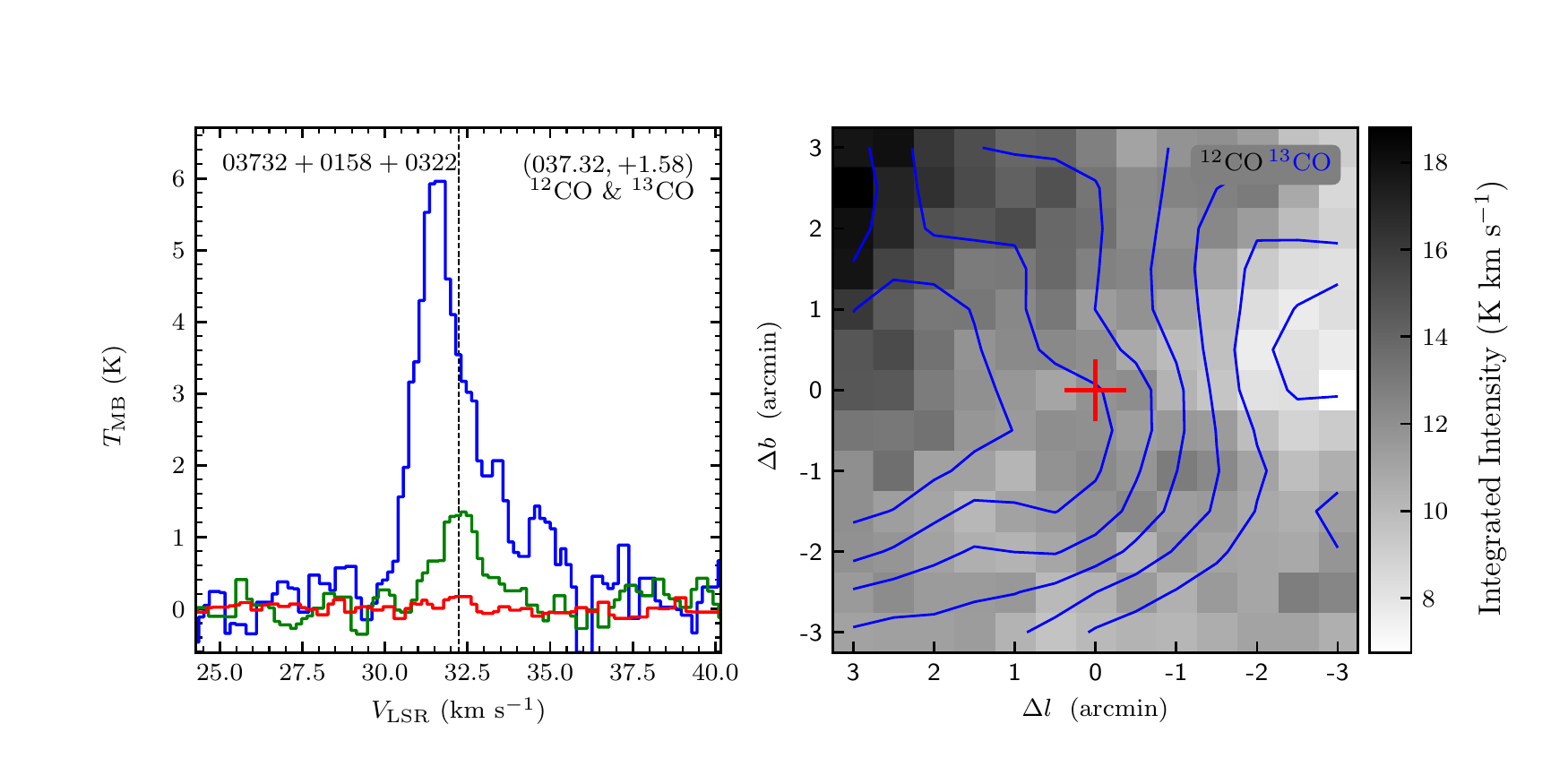}
\vspace{-0.5cm}

\includegraphics[width=9.0cm,angle=0]{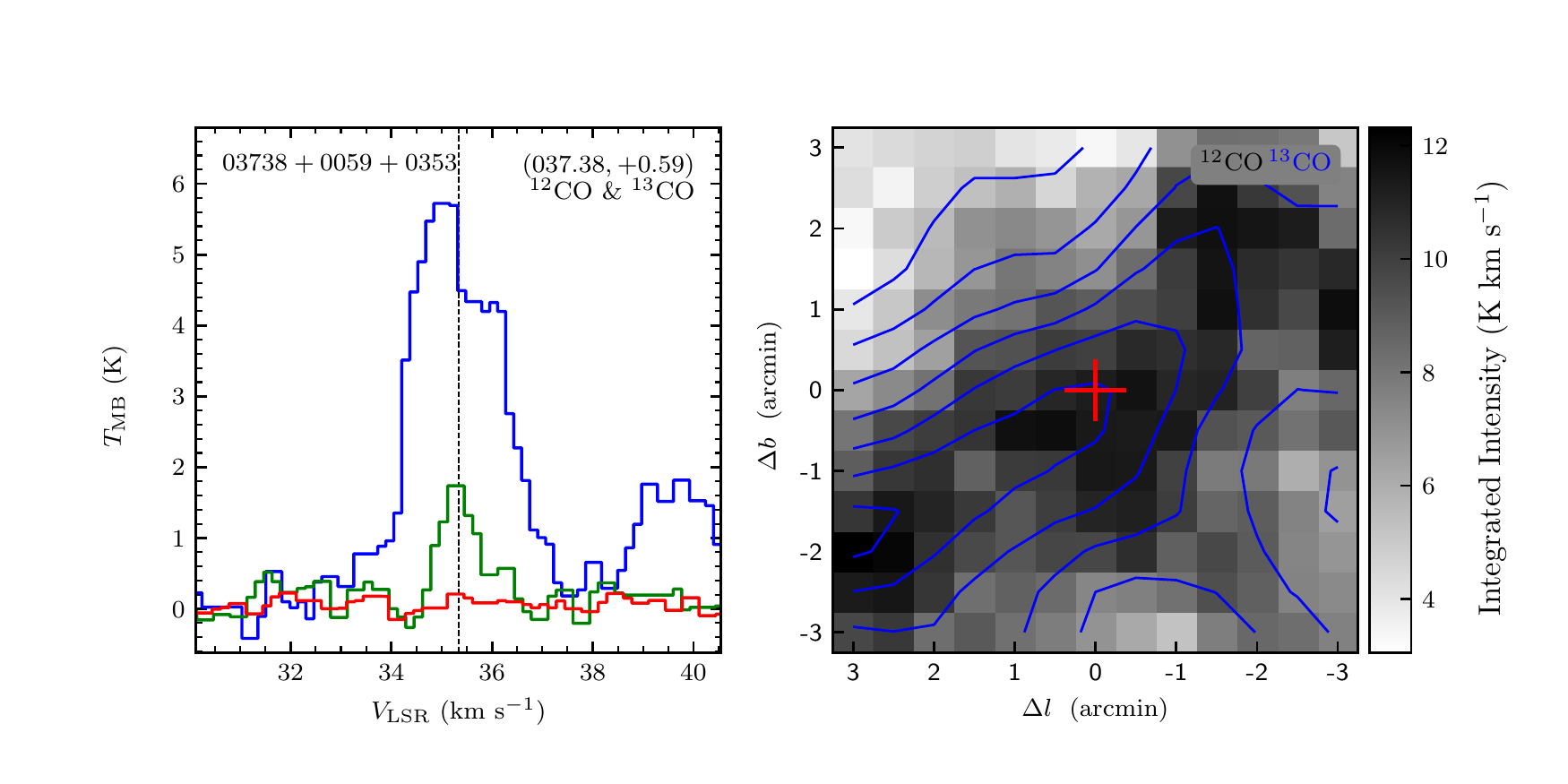}
\includegraphics[width=9.0cm,angle=0]{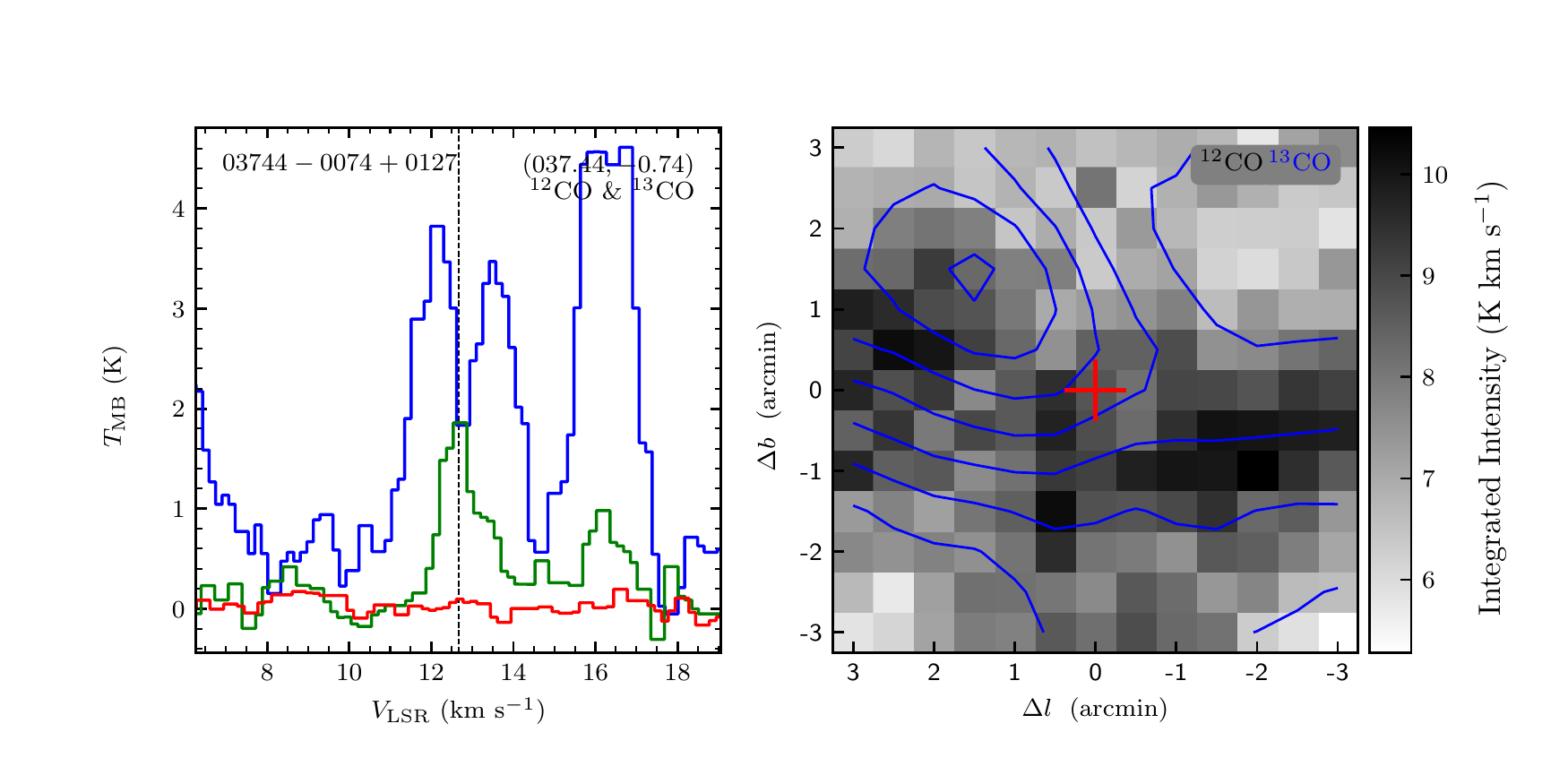}
\vspace{-0.5cm}

\includegraphics[width=9.0cm,angle=0]{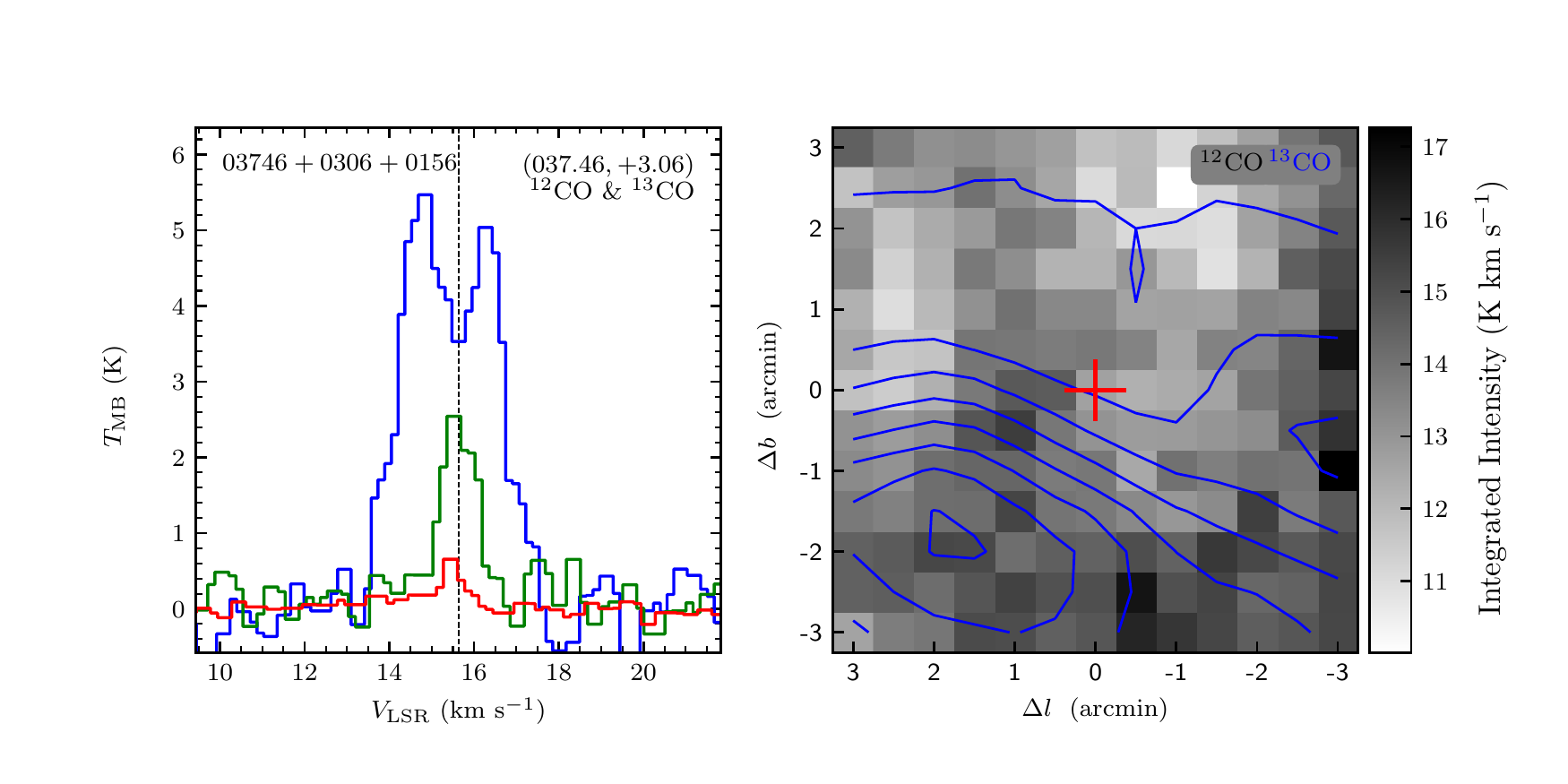}
\includegraphics[width=9.0cm,angle=0]{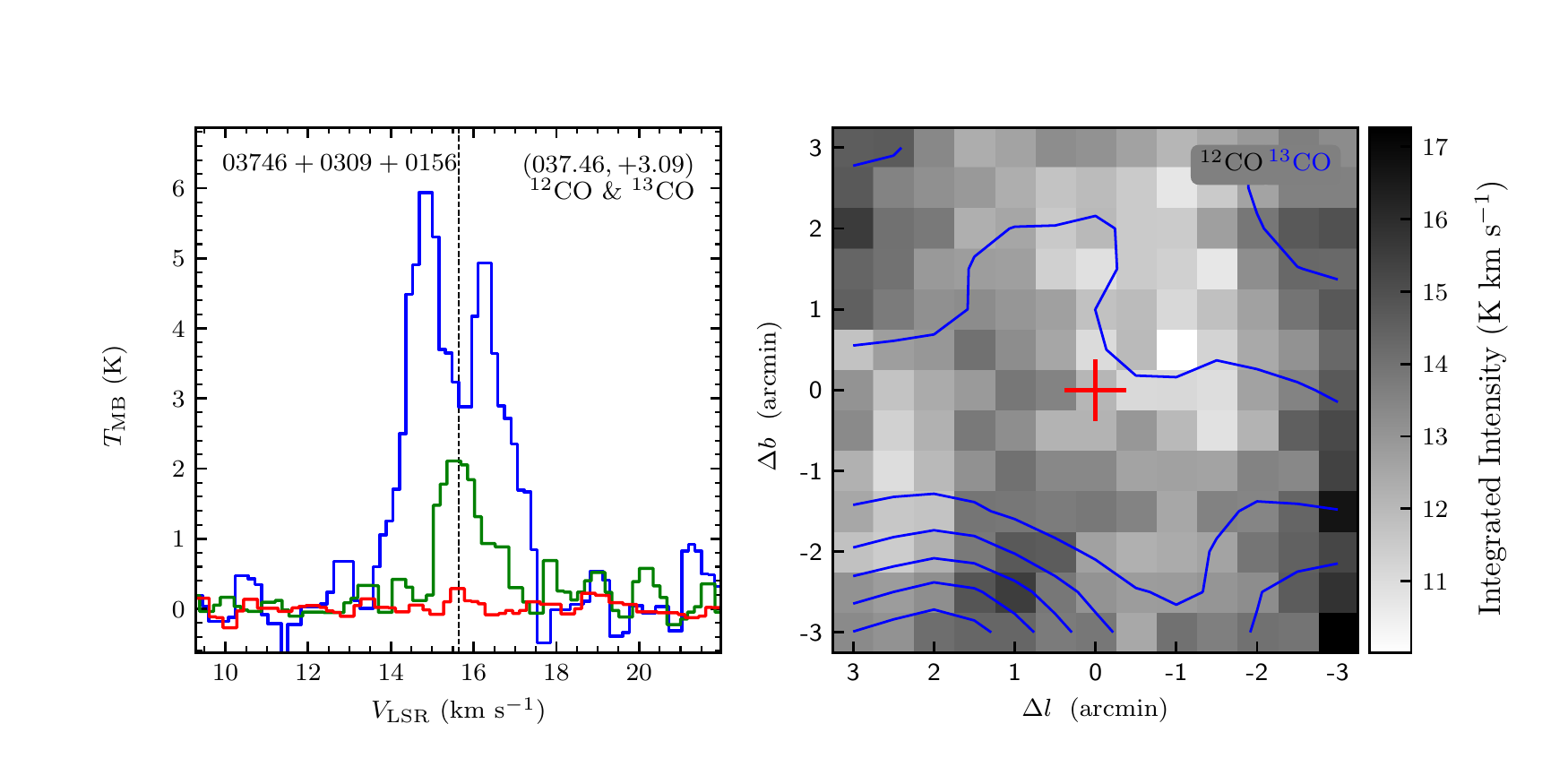}
\vspace{-0.5cm}

\includegraphics[width=9.0cm,angle=0]{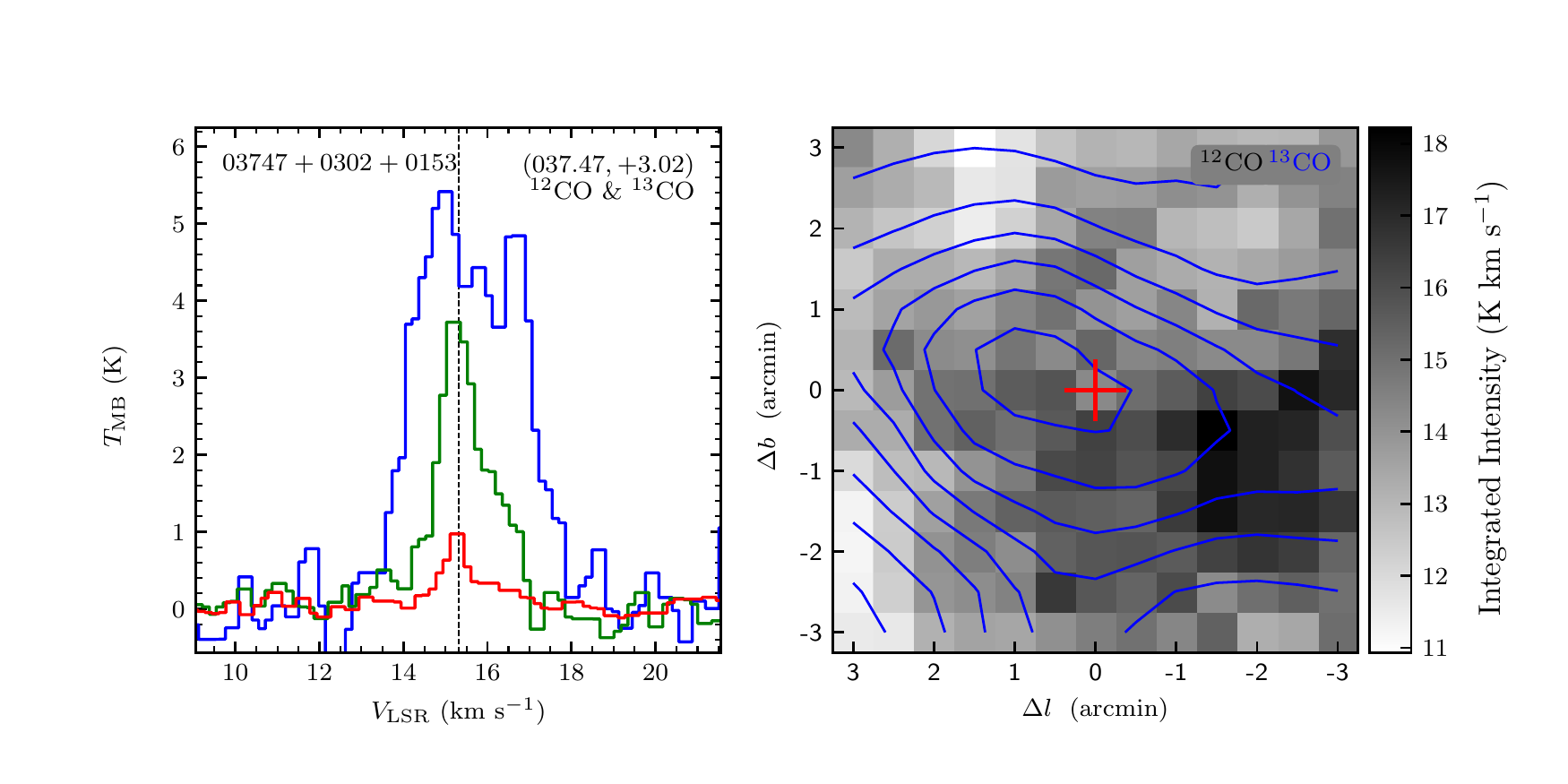}
\includegraphics[width=9.0cm,angle=0]{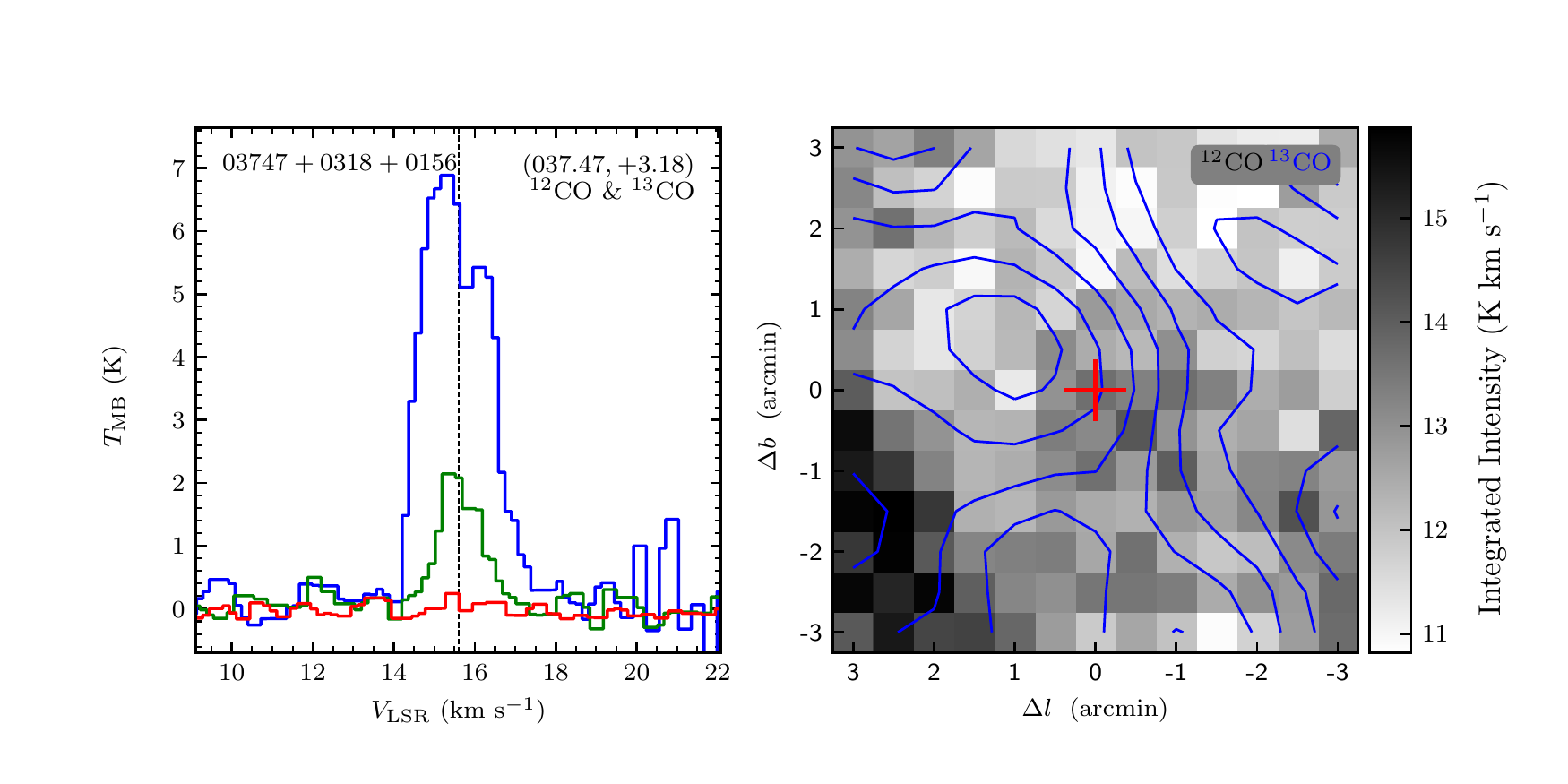}
\end{figure}
\clearpage

\begin{figure}
\includegraphics[width=9.0cm,angle=0]{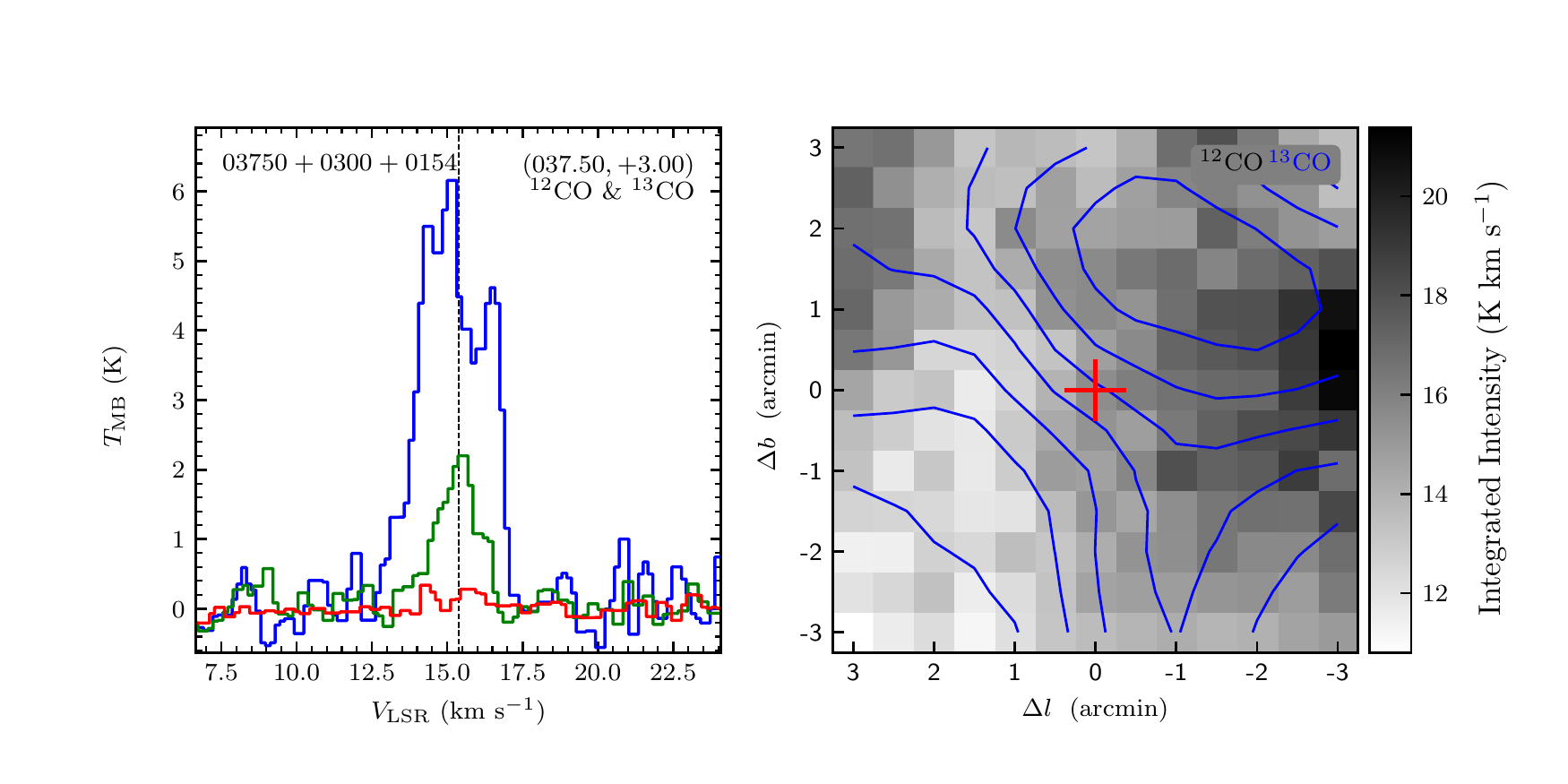}
\includegraphics[width=9.0cm,angle=0]{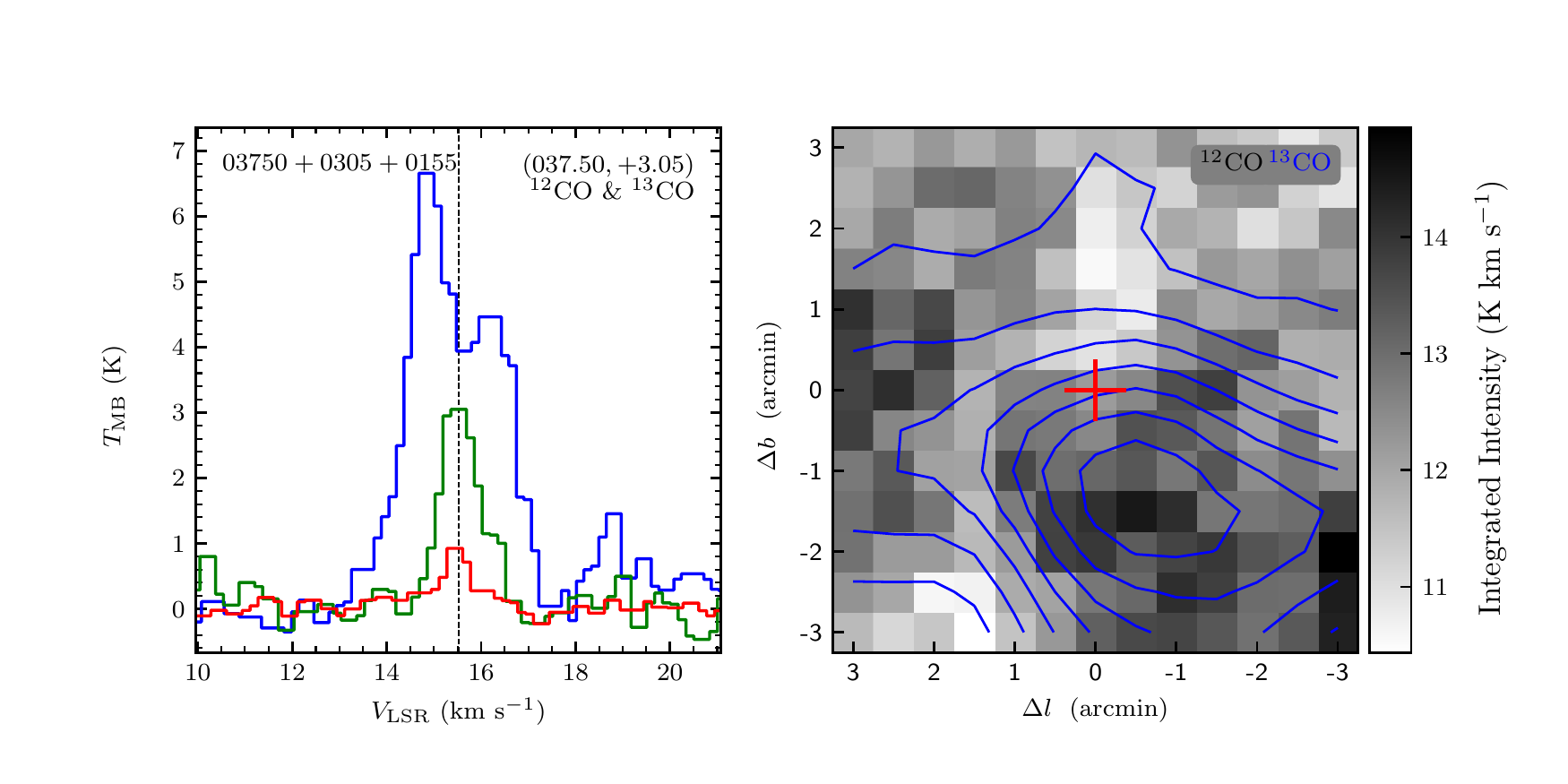}
\vspace{-0.5cm}

\includegraphics[width=9.0cm,angle=0]{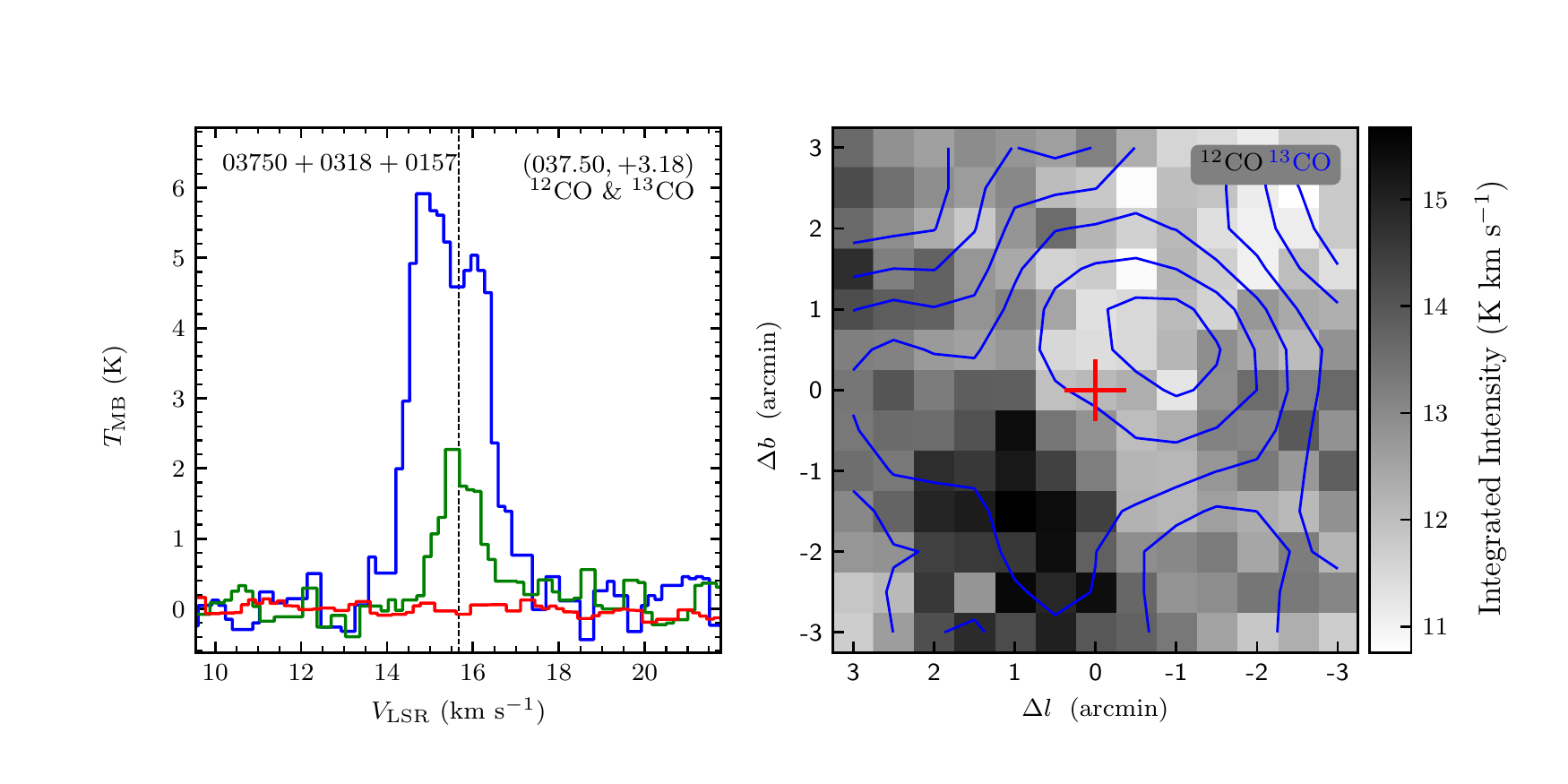}
\includegraphics[width=9.0cm,angle=0]{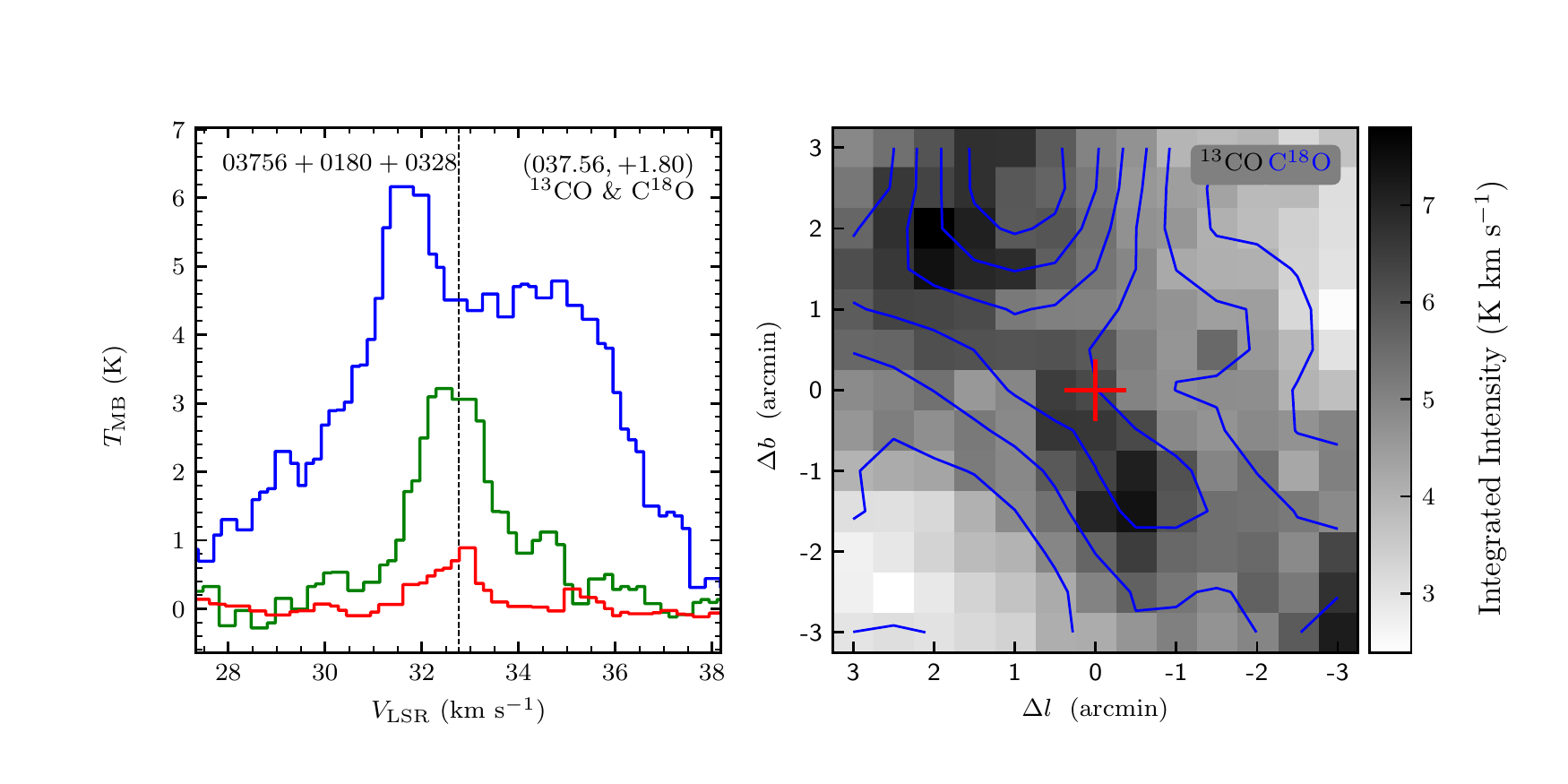}
\vspace{-0.5cm}

\includegraphics[width=9.0cm,angle=0]{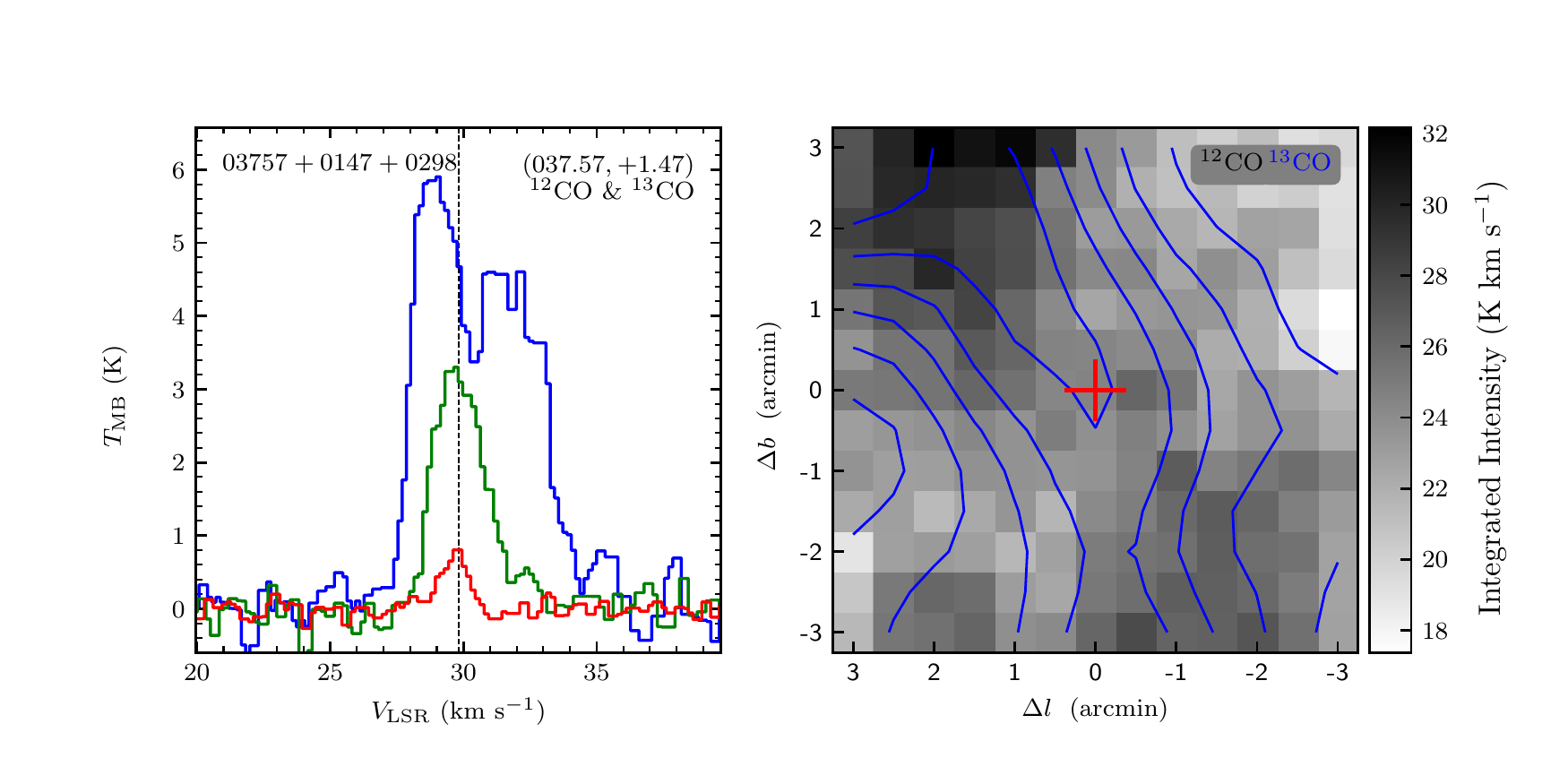}
\includegraphics[width=9.0cm,angle=0]{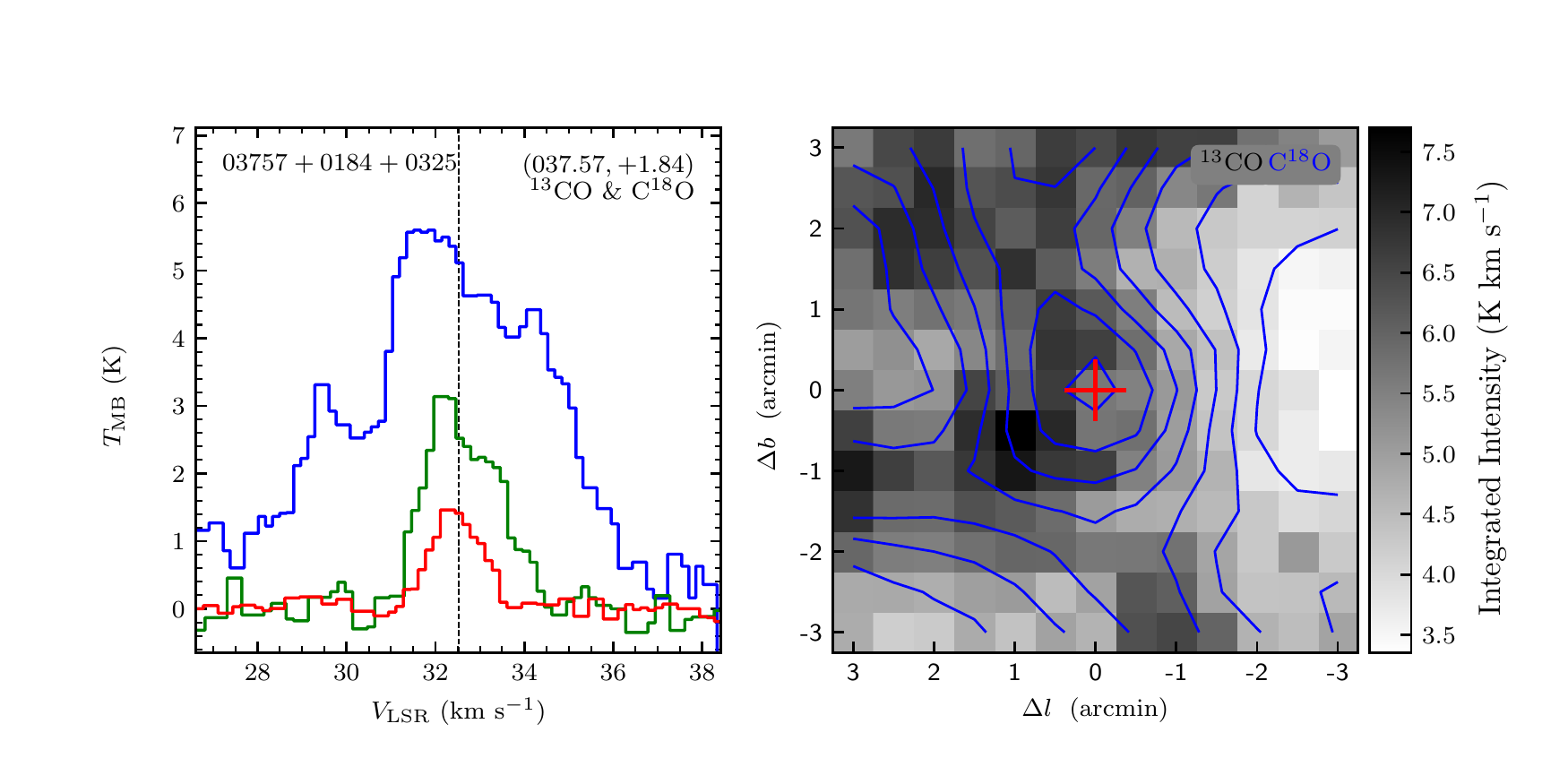}
\vspace{-0.5cm}

\includegraphics[width=9.0cm,angle=0]{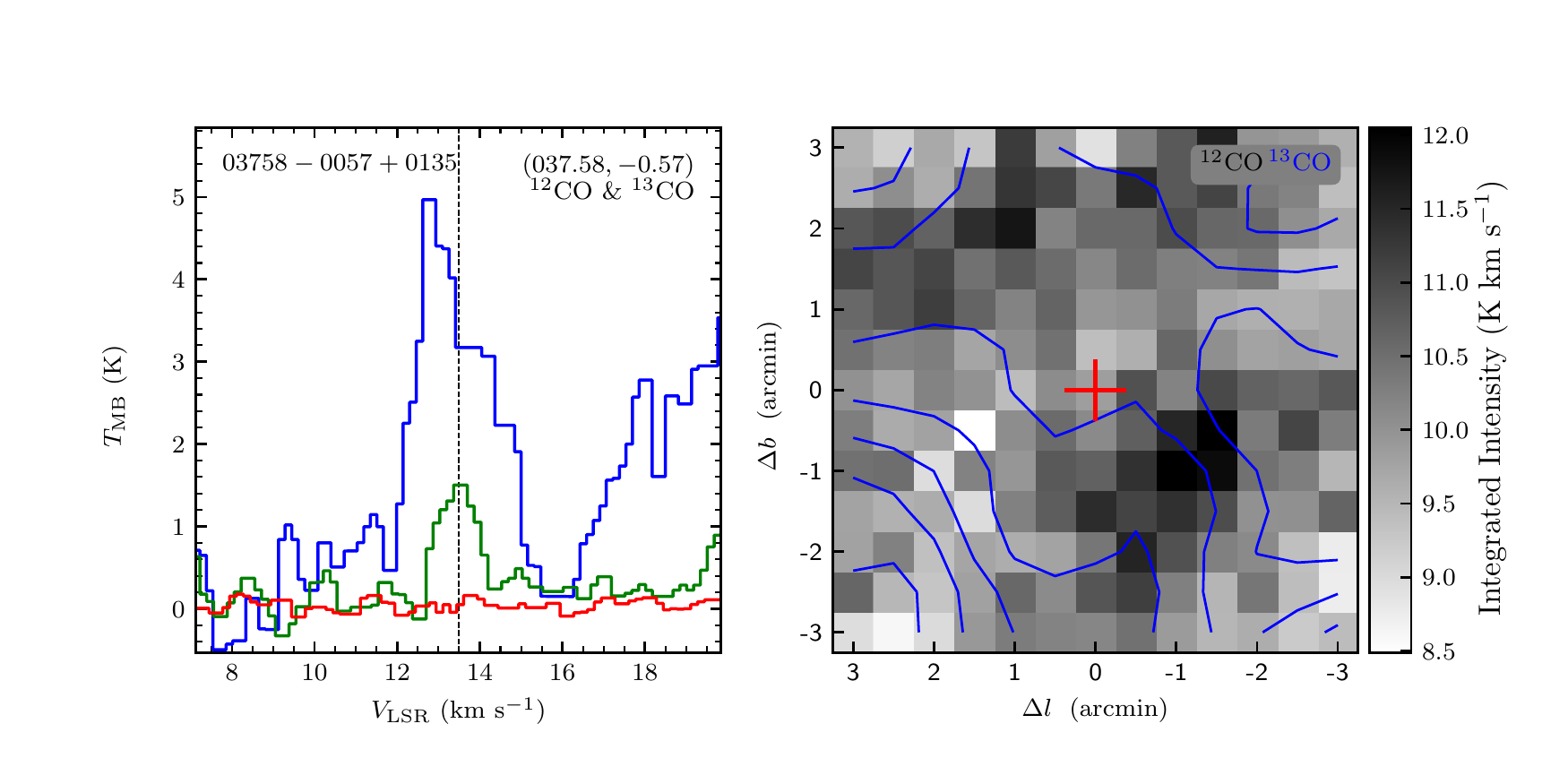}
\includegraphics[width=9.0cm,angle=0]{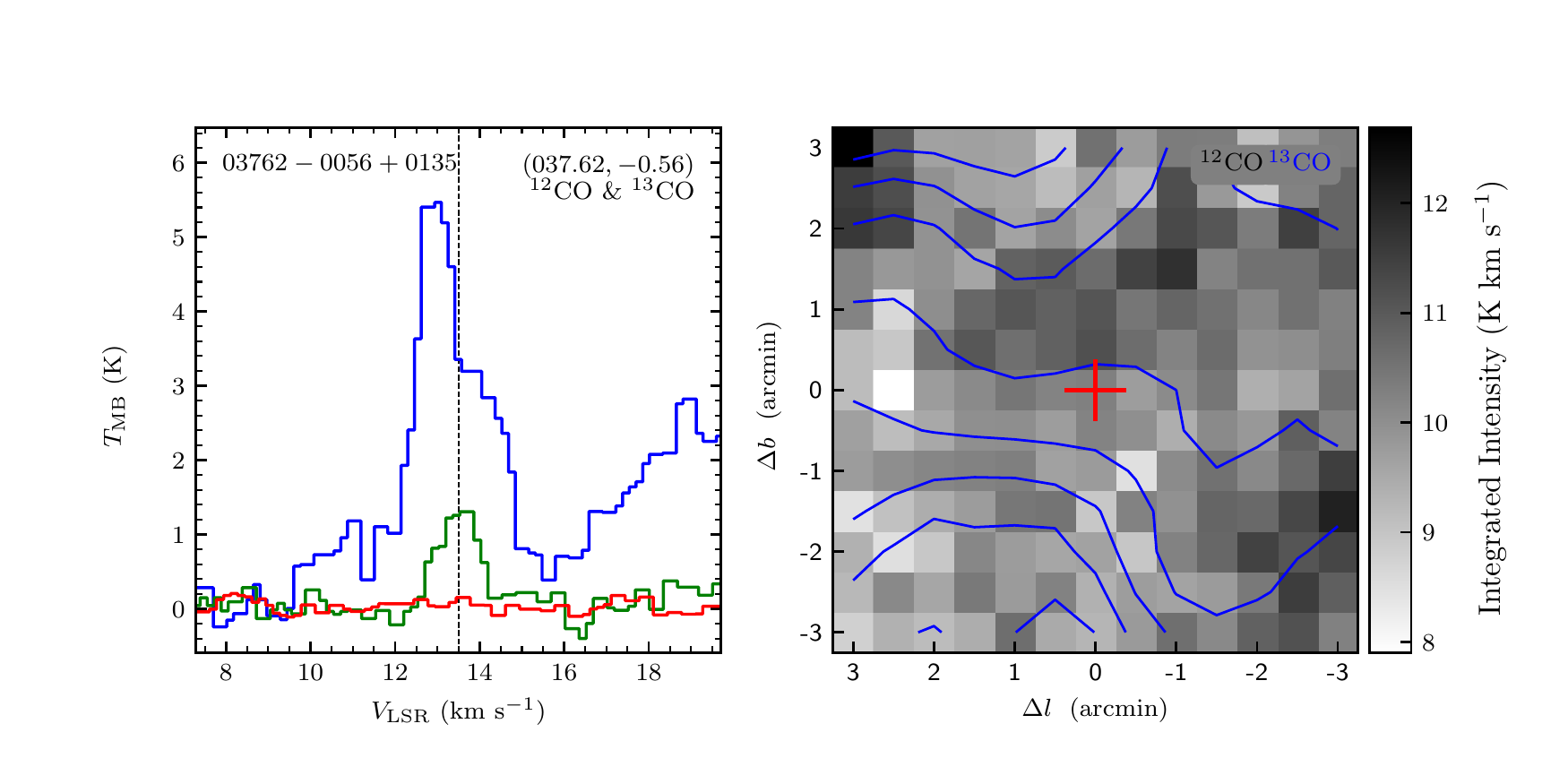}
\vspace{-0.5cm}

\includegraphics[width=9.0cm,angle=0]{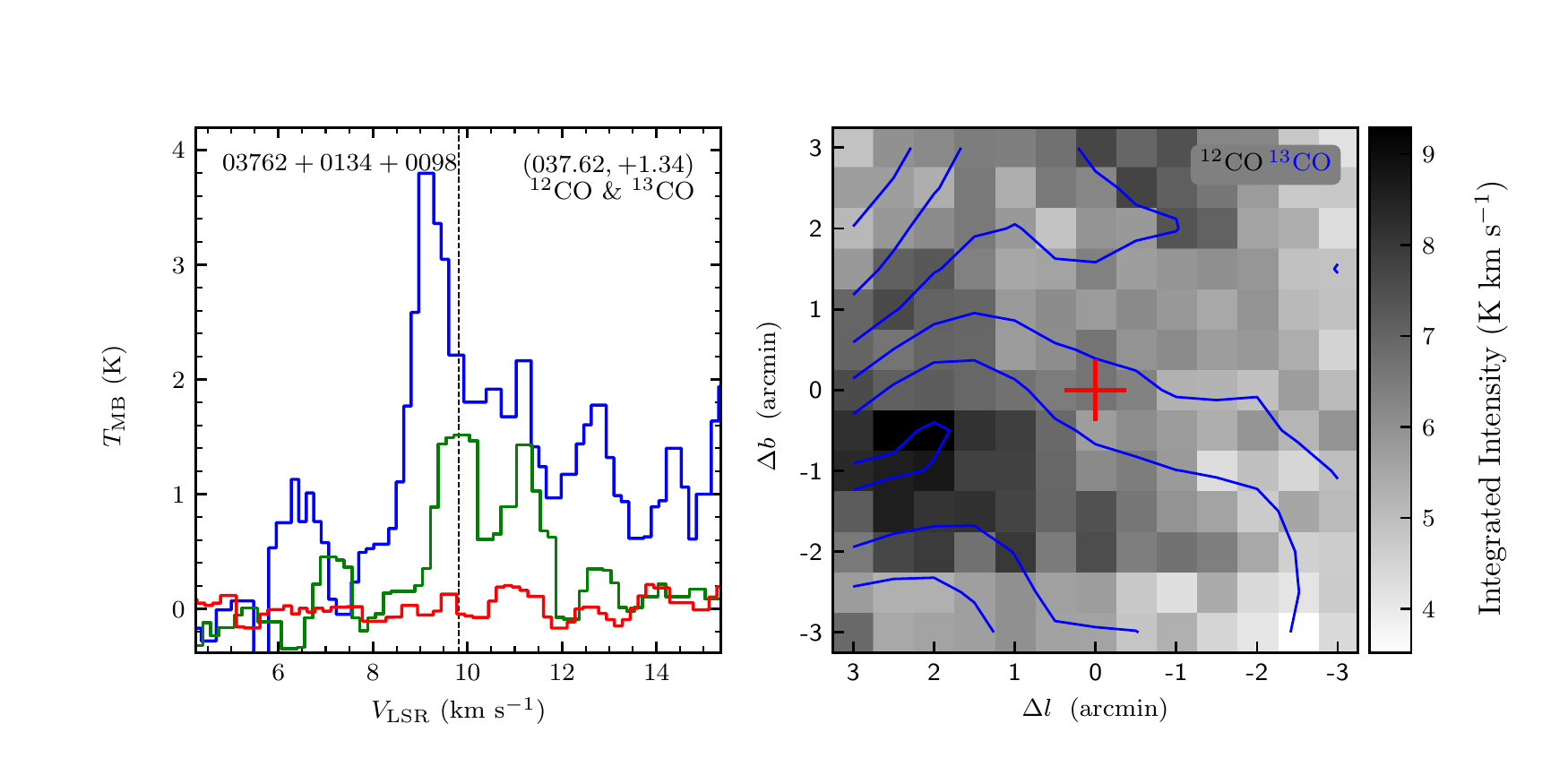}
\includegraphics[width=9.0cm,angle=0]{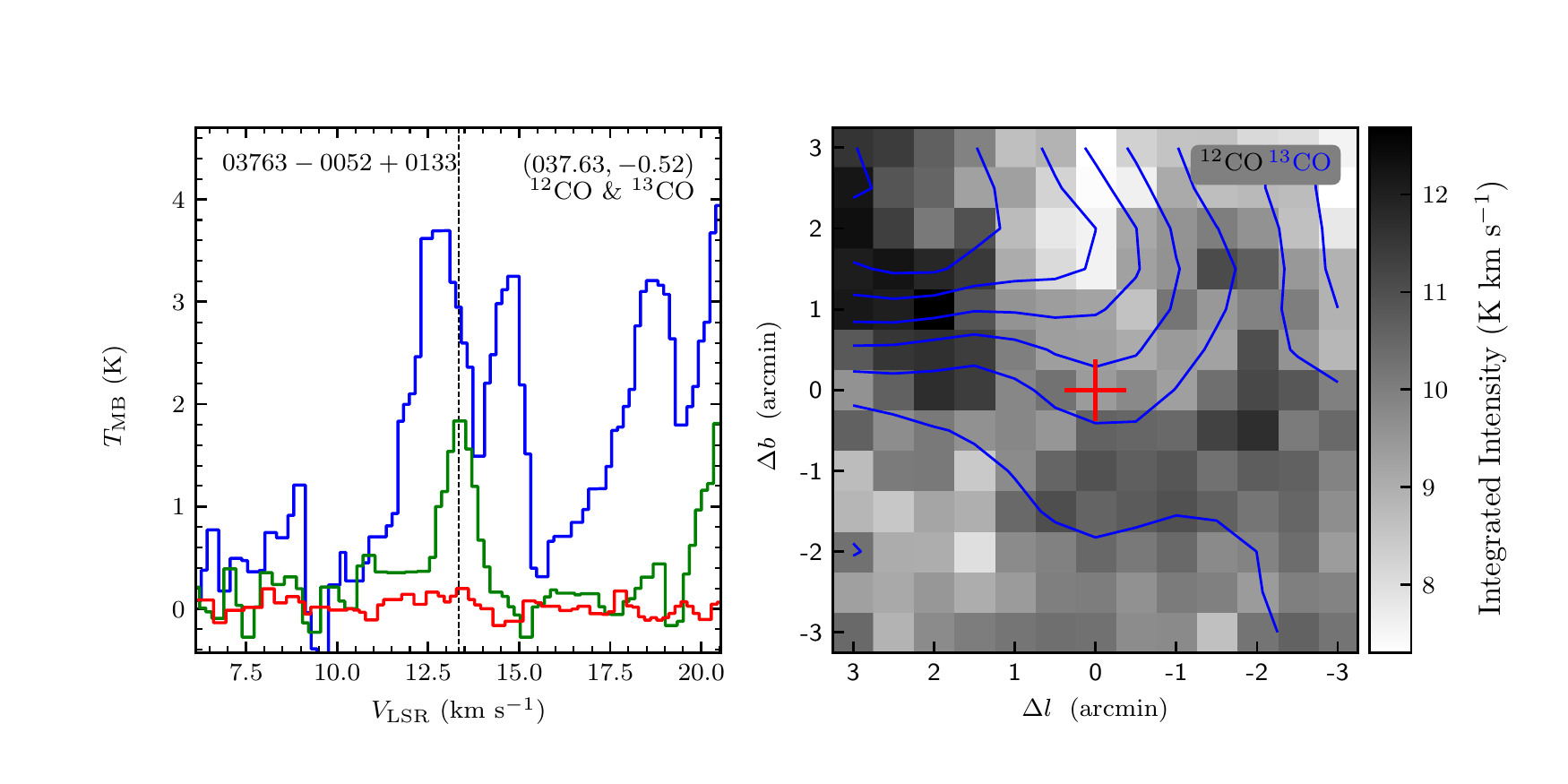}
\end{figure}
\clearpage

\begin{figure}
\includegraphics[width=9.0cm,angle=0]{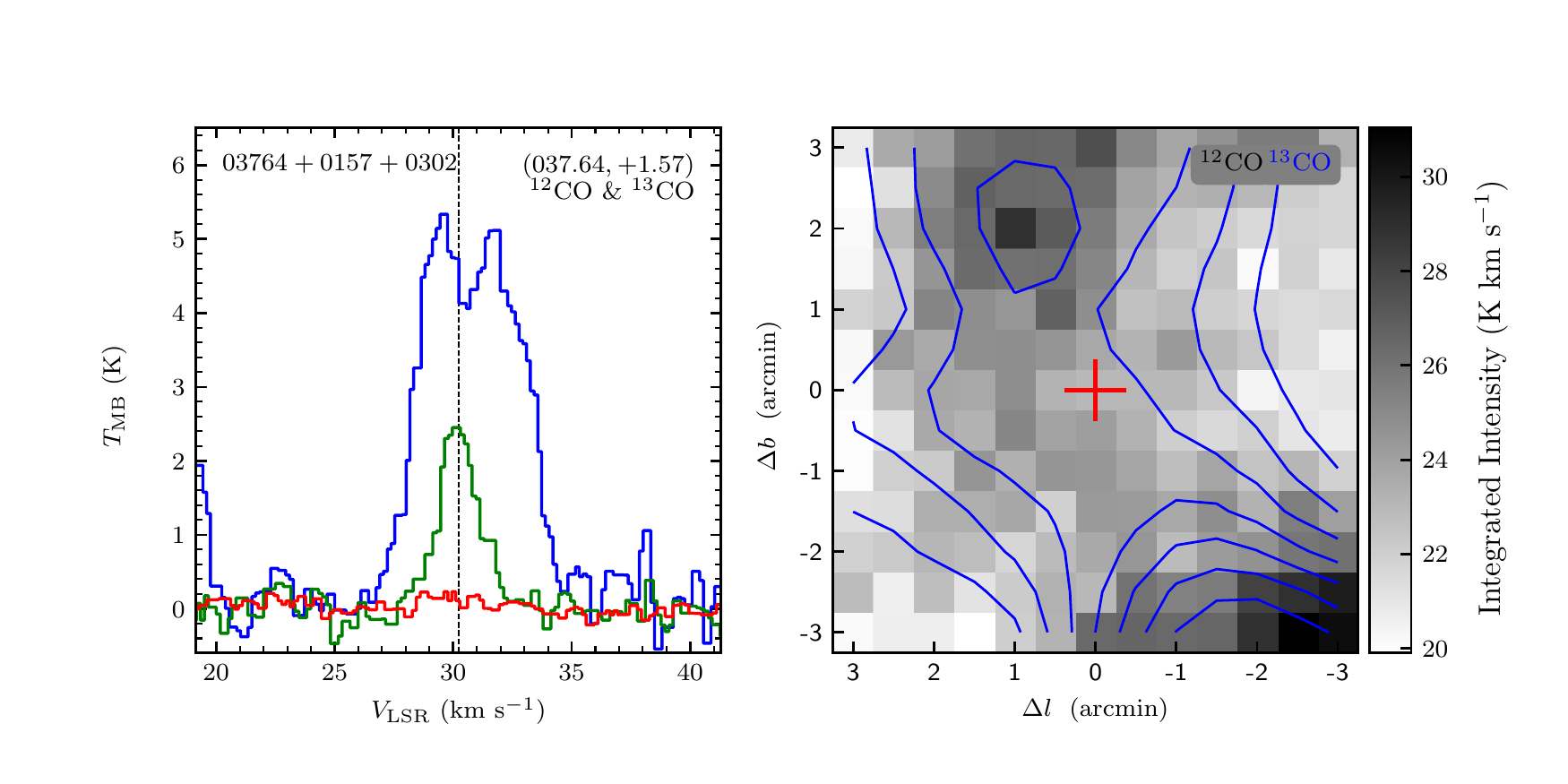}
\includegraphics[width=9.0cm,angle=0]{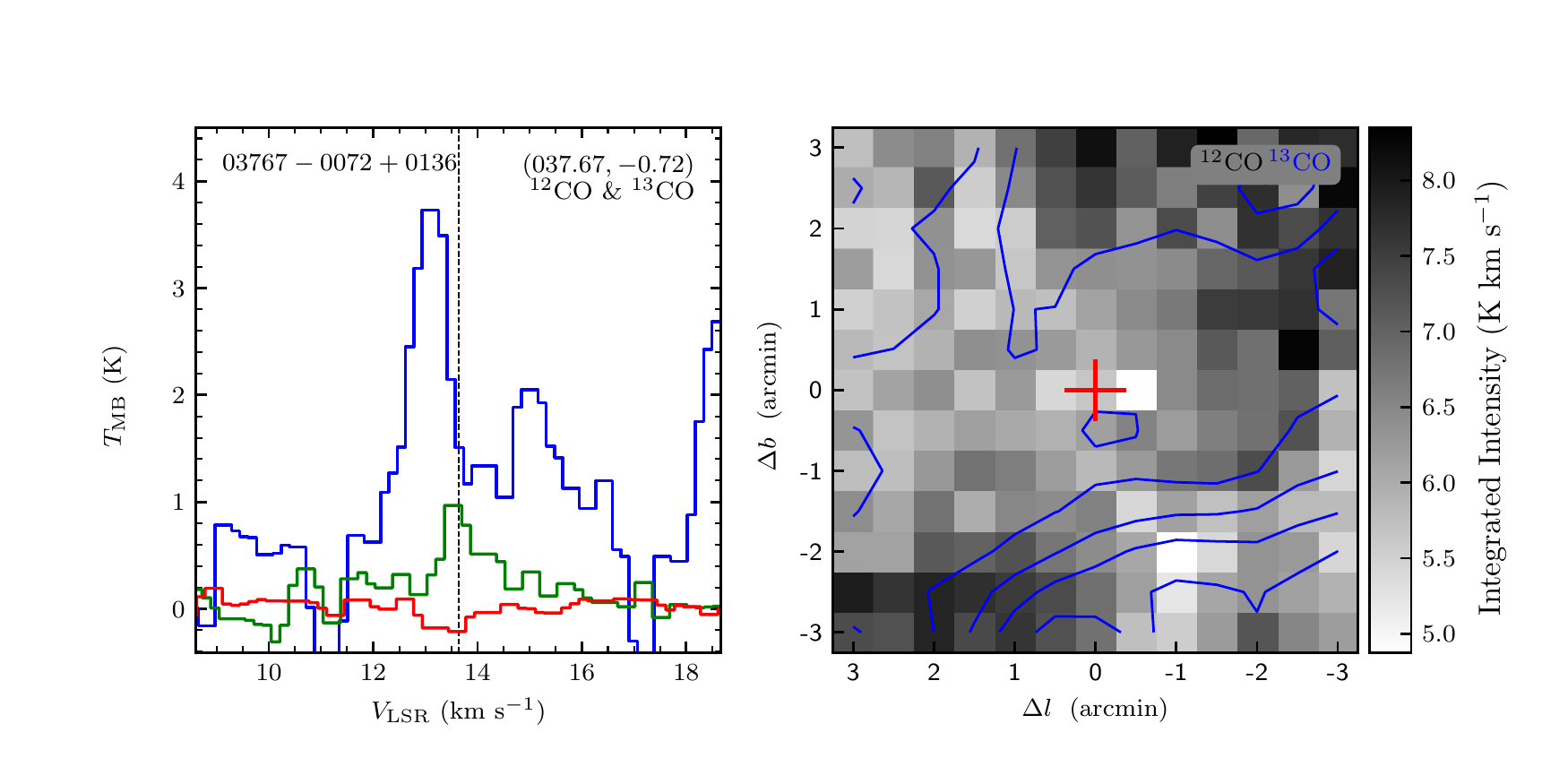}
\vspace{-0.5cm}

\includegraphics[width=9.0cm,angle=0]{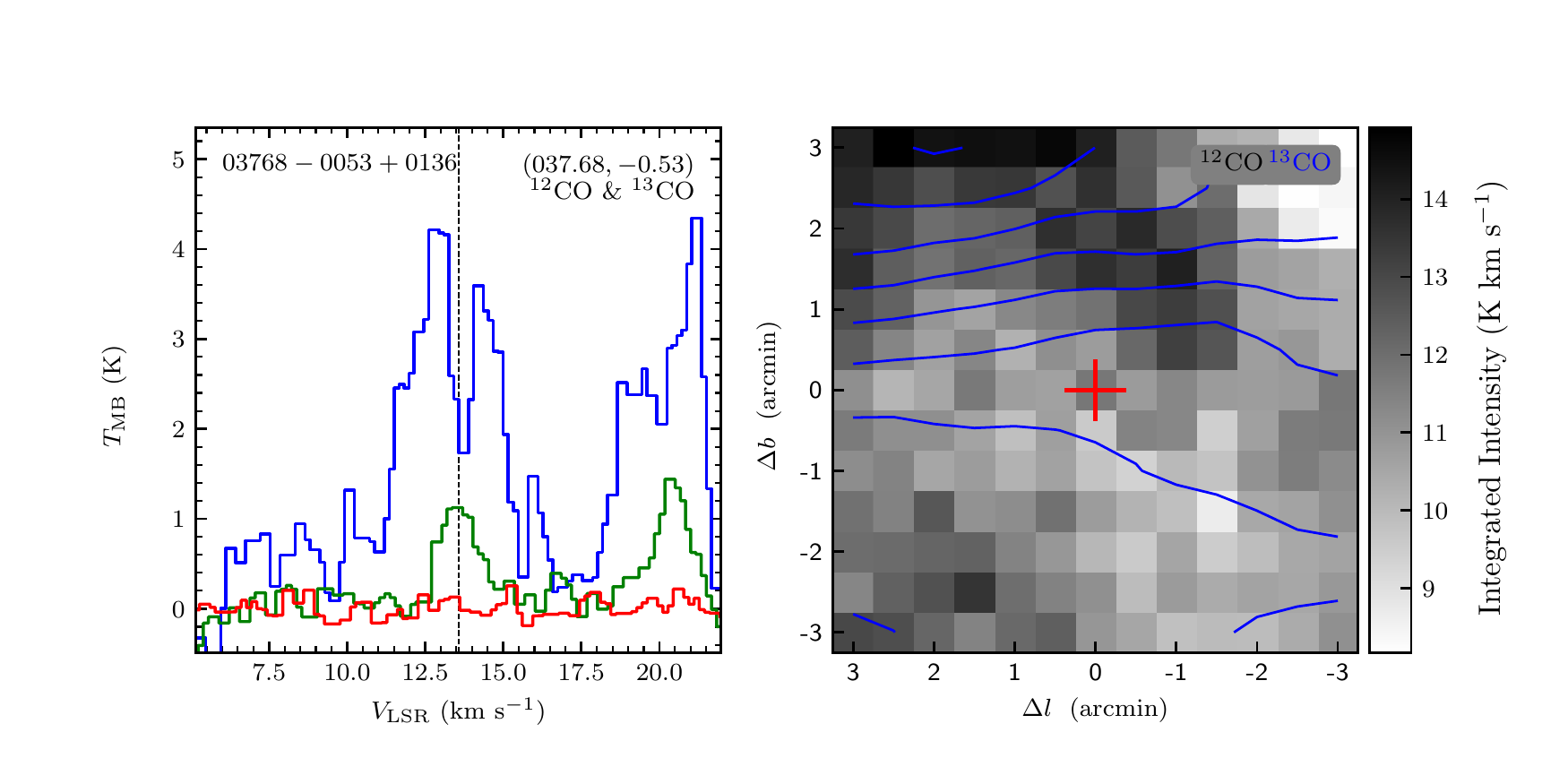}
\includegraphics[width=9.0cm,angle=0]{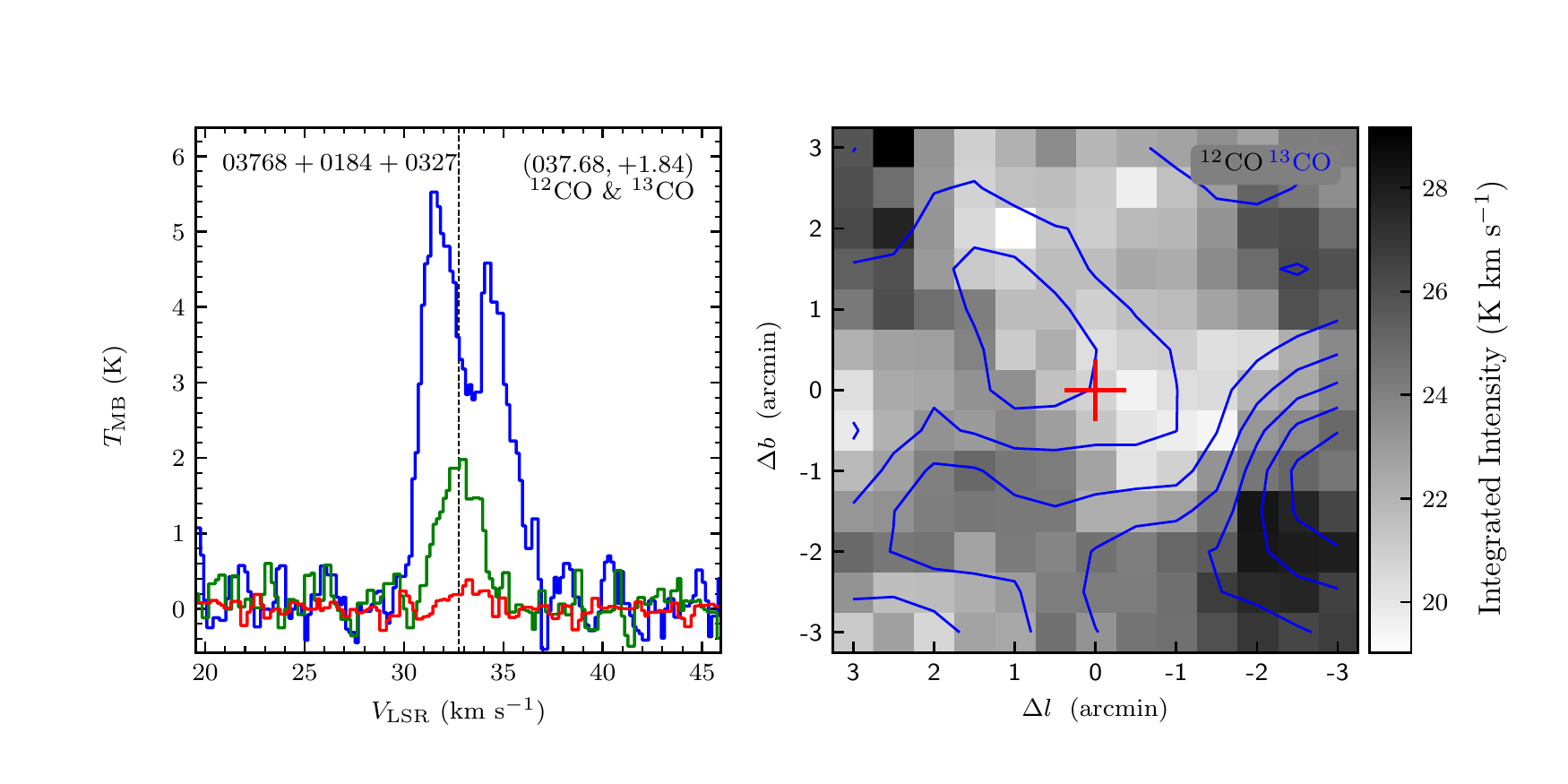}
\vspace{-0.5cm}

\includegraphics[width=9.0cm,angle=0]{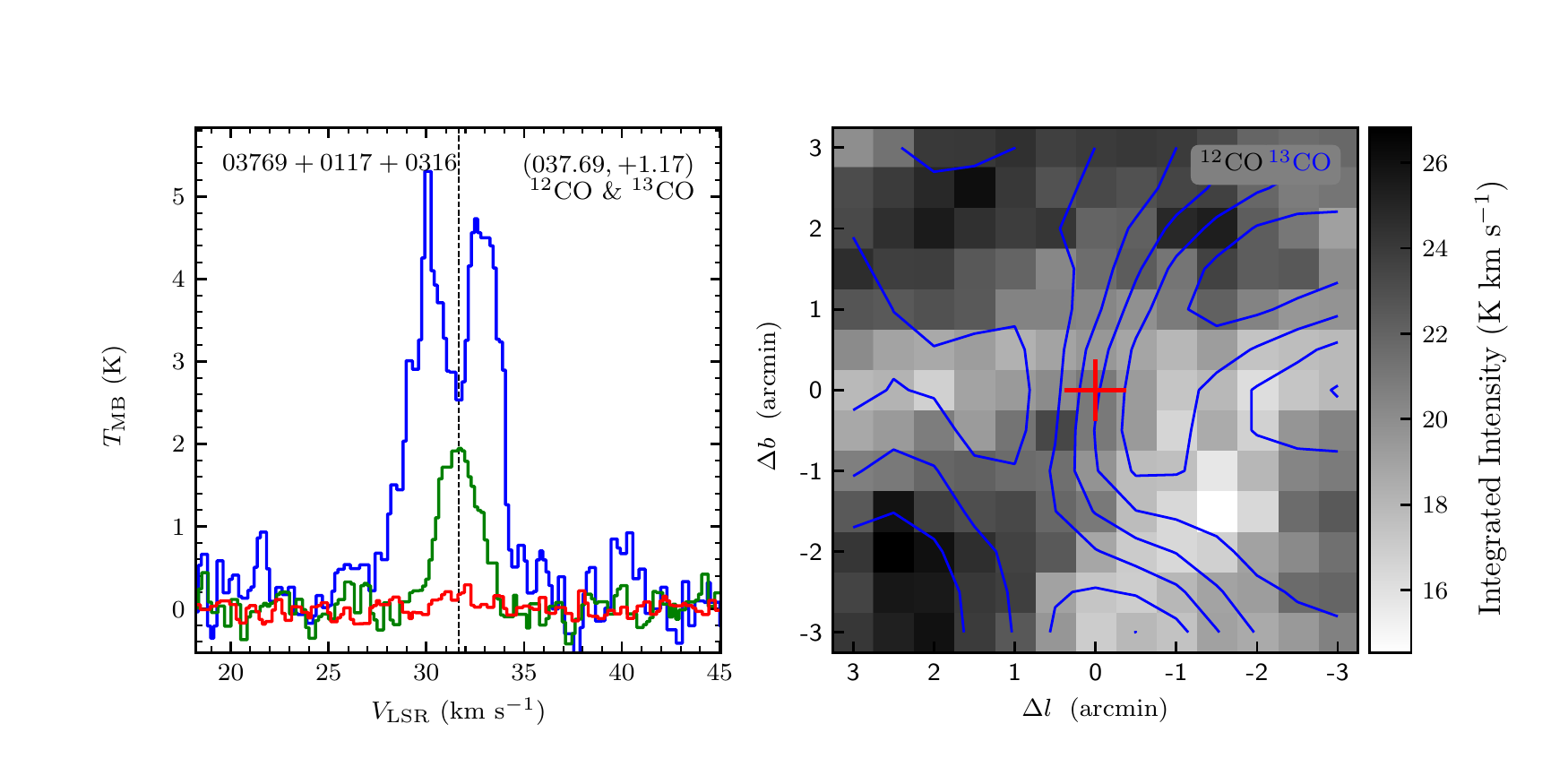}
\includegraphics[width=9.0cm,angle=0]{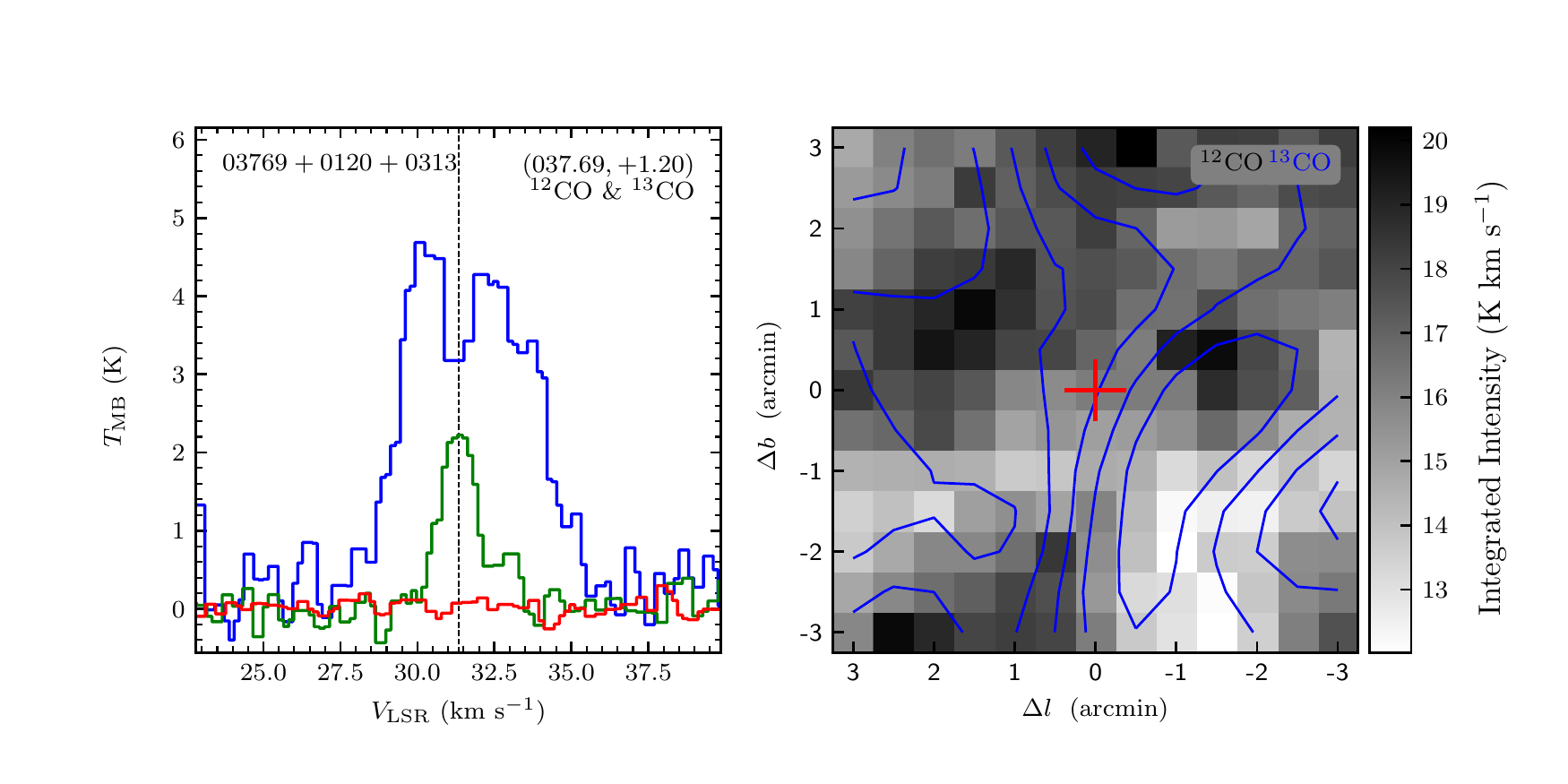}
\vspace{-0.5cm}

\includegraphics[width=9.0cm,angle=0]{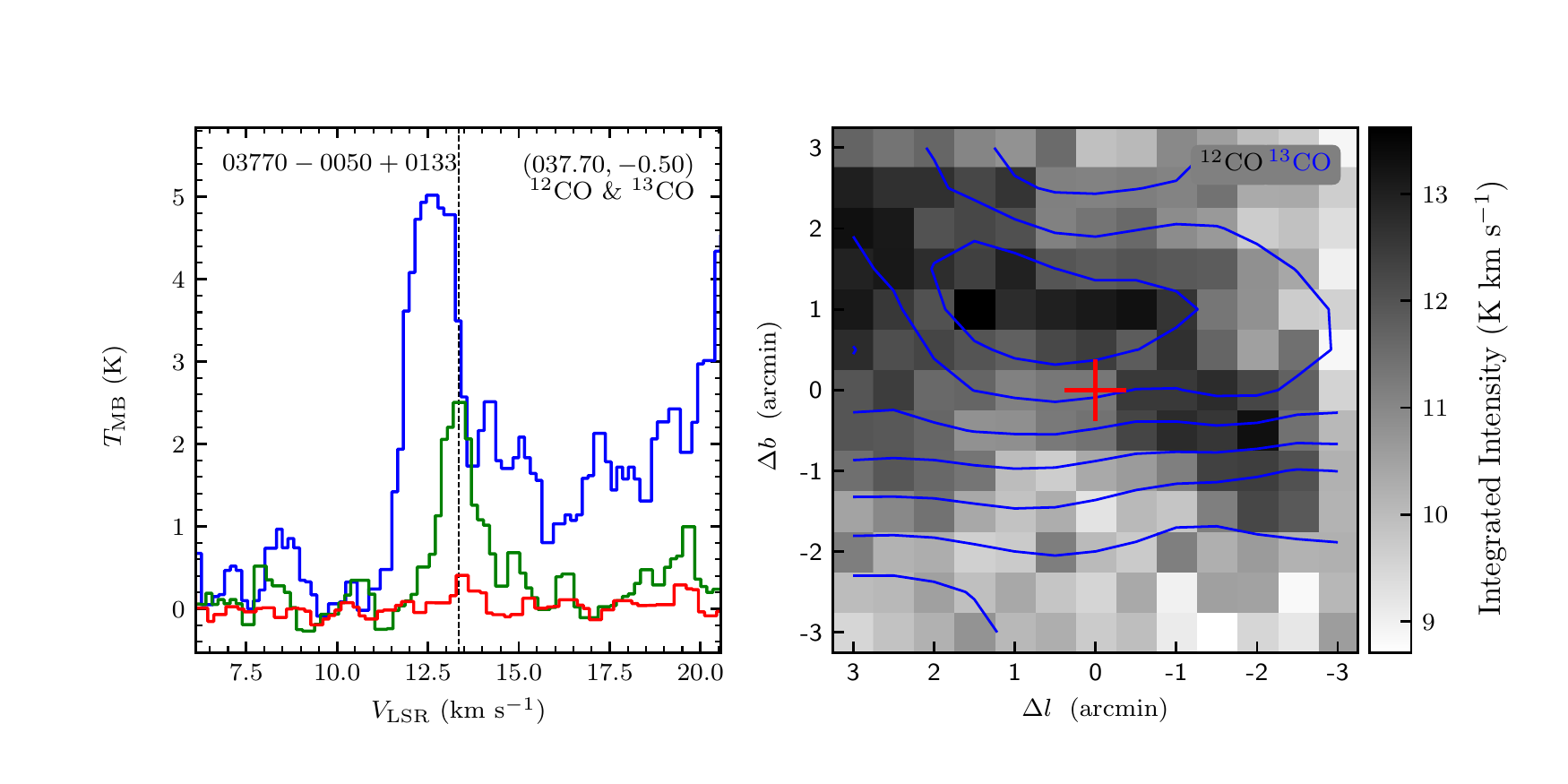}
\includegraphics[width=9.0cm,angle=0]{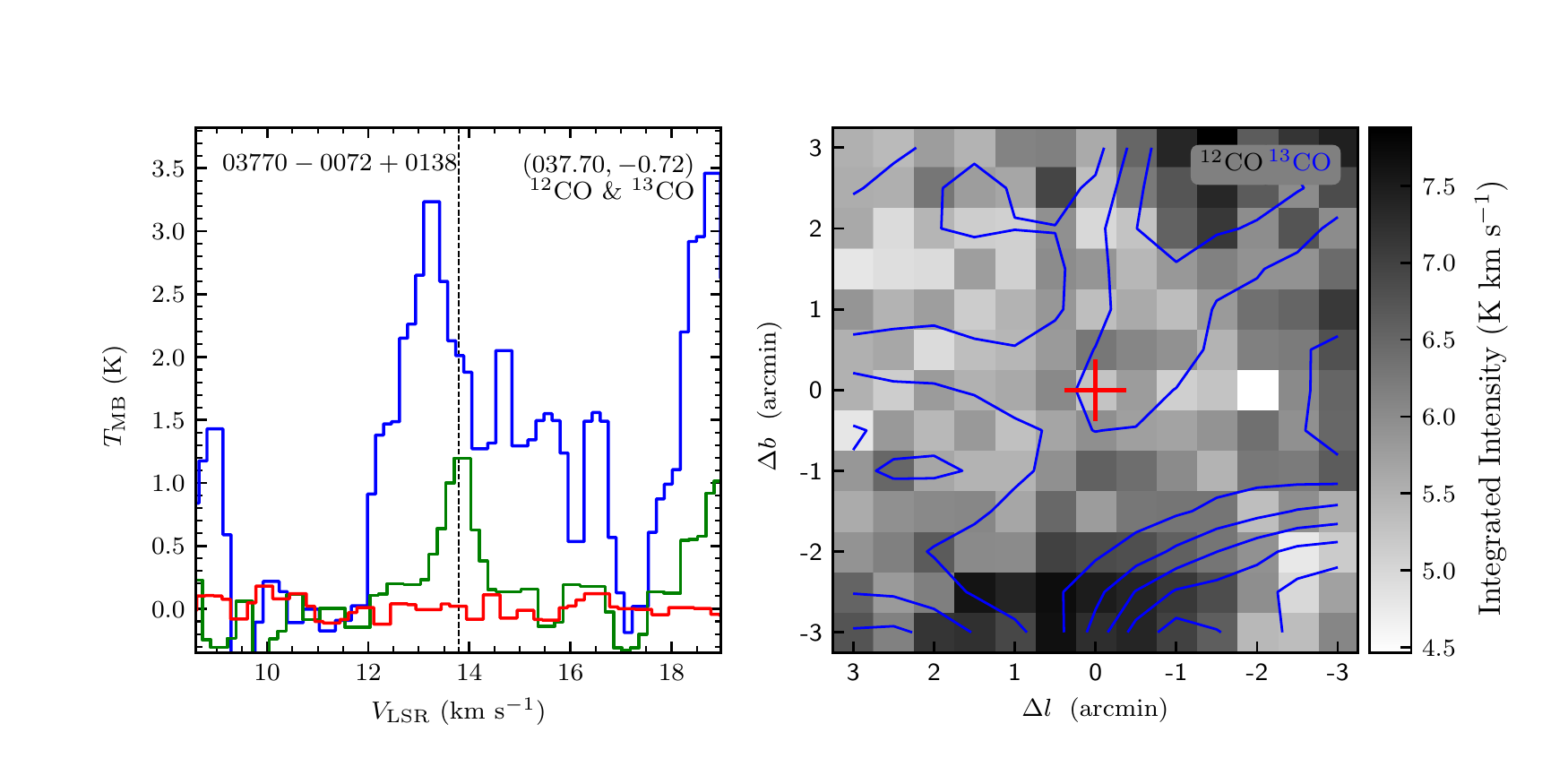}
\vspace{-0.5cm}

\includegraphics[width=9.0cm,angle=0]{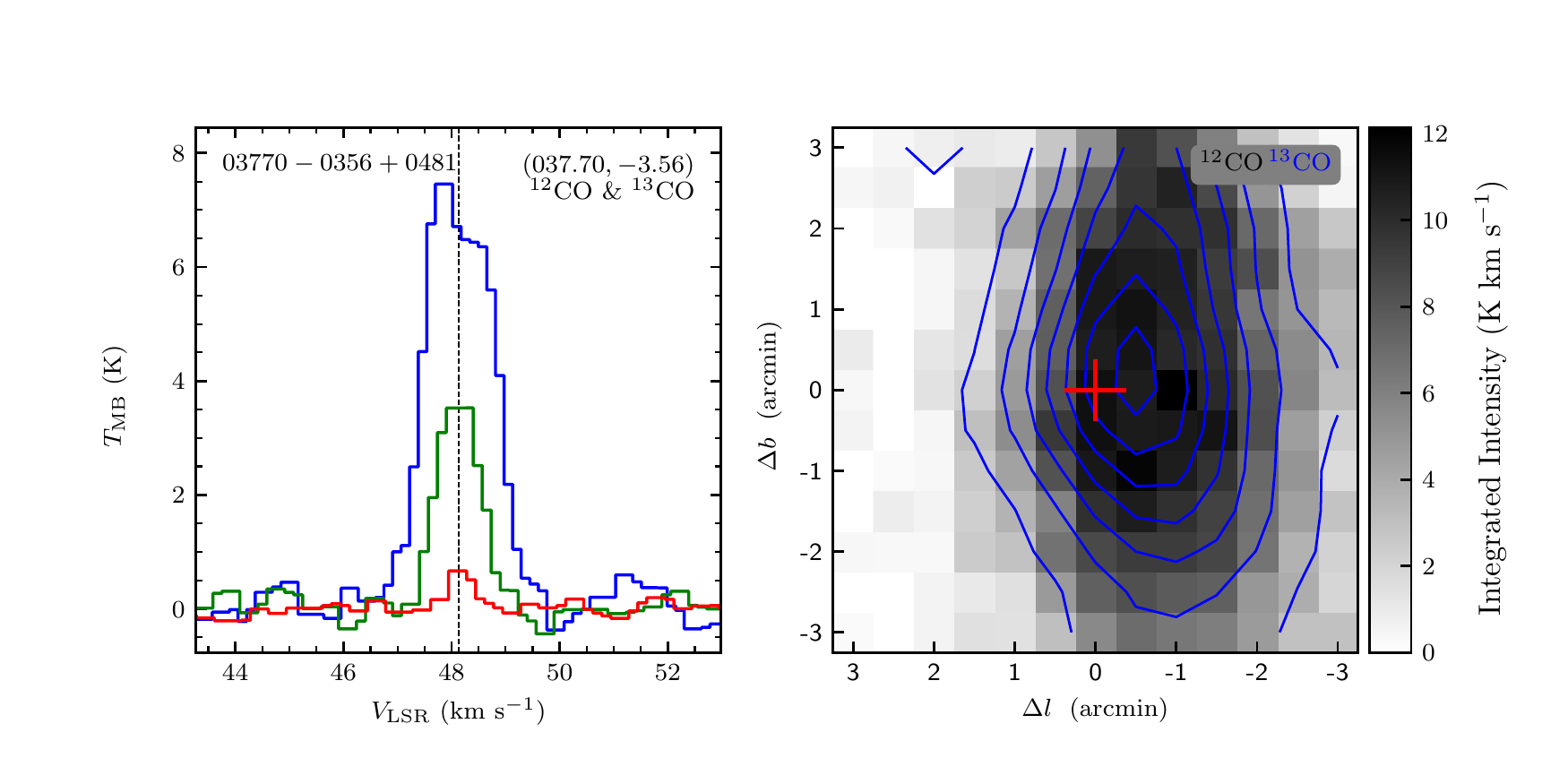}
\includegraphics[width=9.0cm,angle=0]{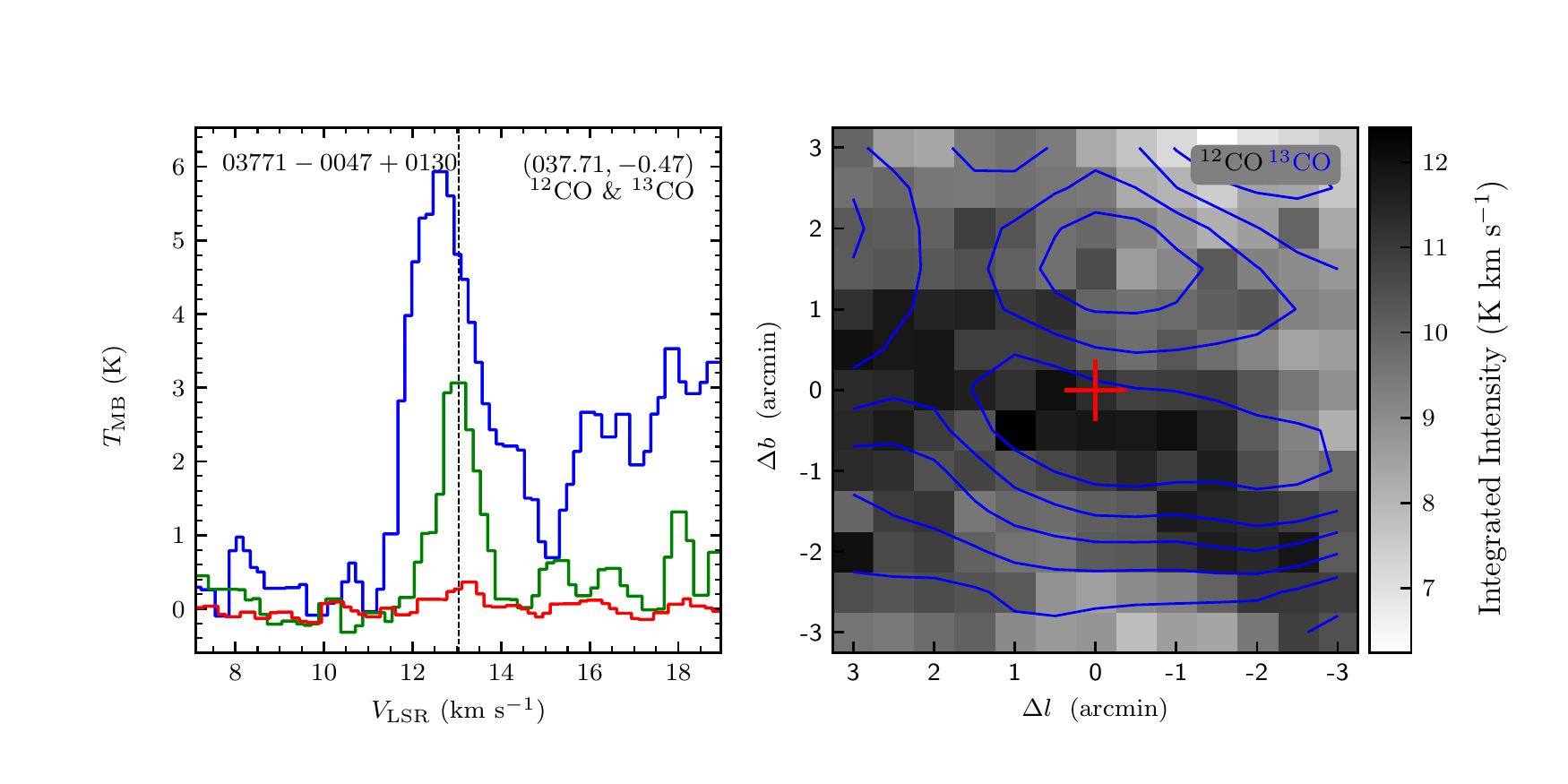}
\end{figure}
\clearpage

\begin{figure}
\includegraphics[width=9.0cm,angle=0]{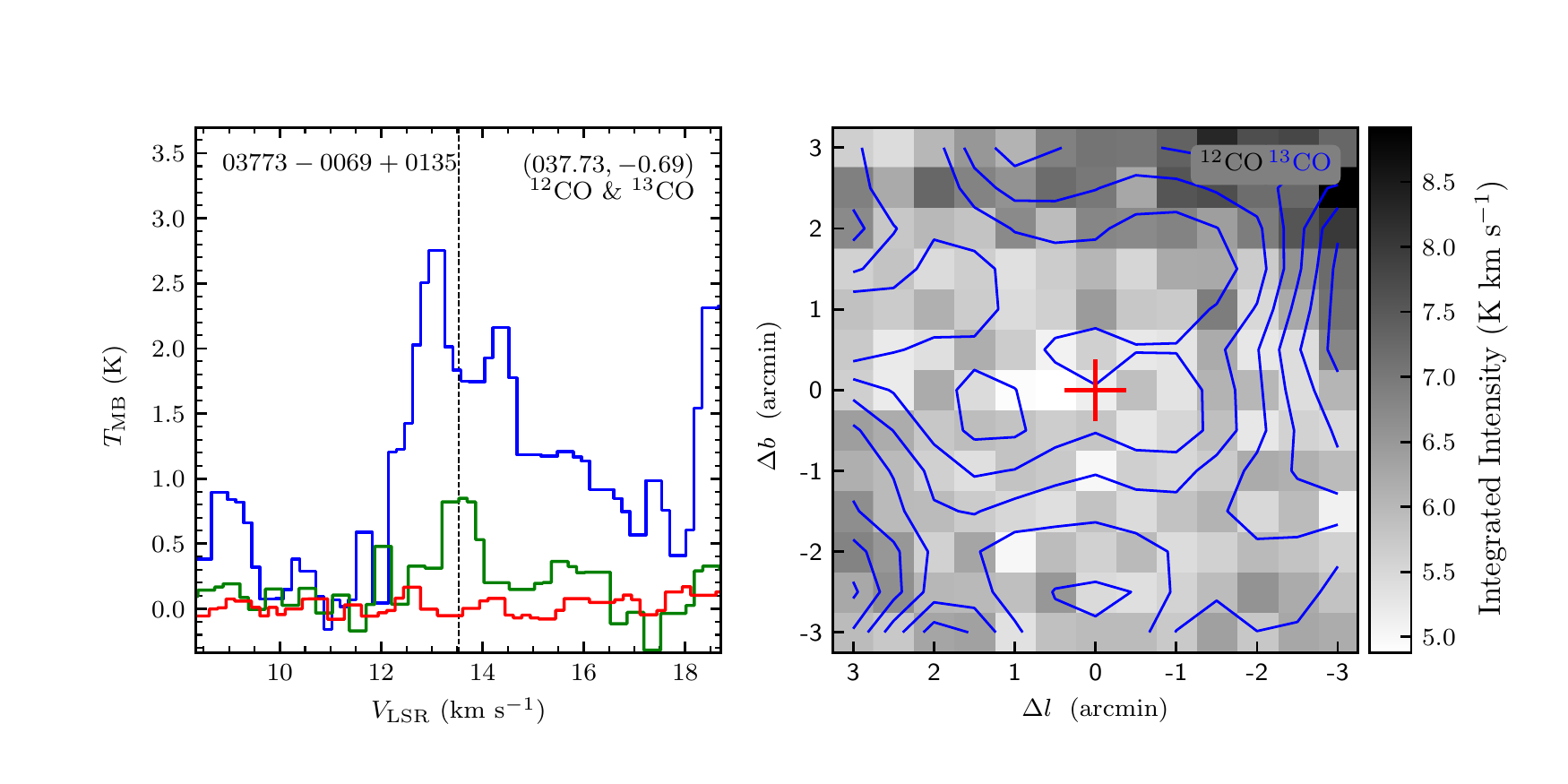}
\includegraphics[width=9.0cm,angle=0]{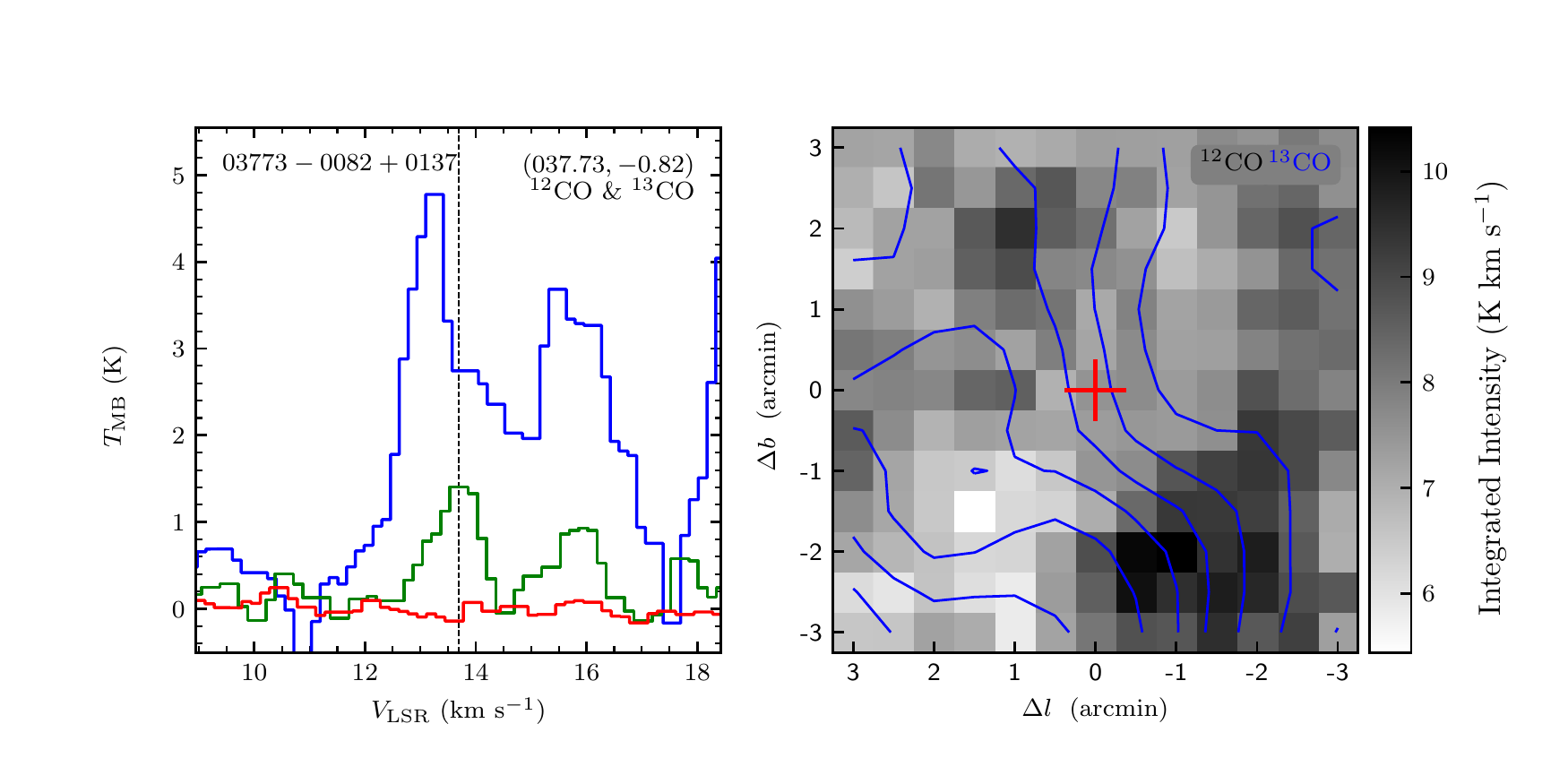}
\vspace{-0.5cm}

\includegraphics[width=9.0cm,angle=0]{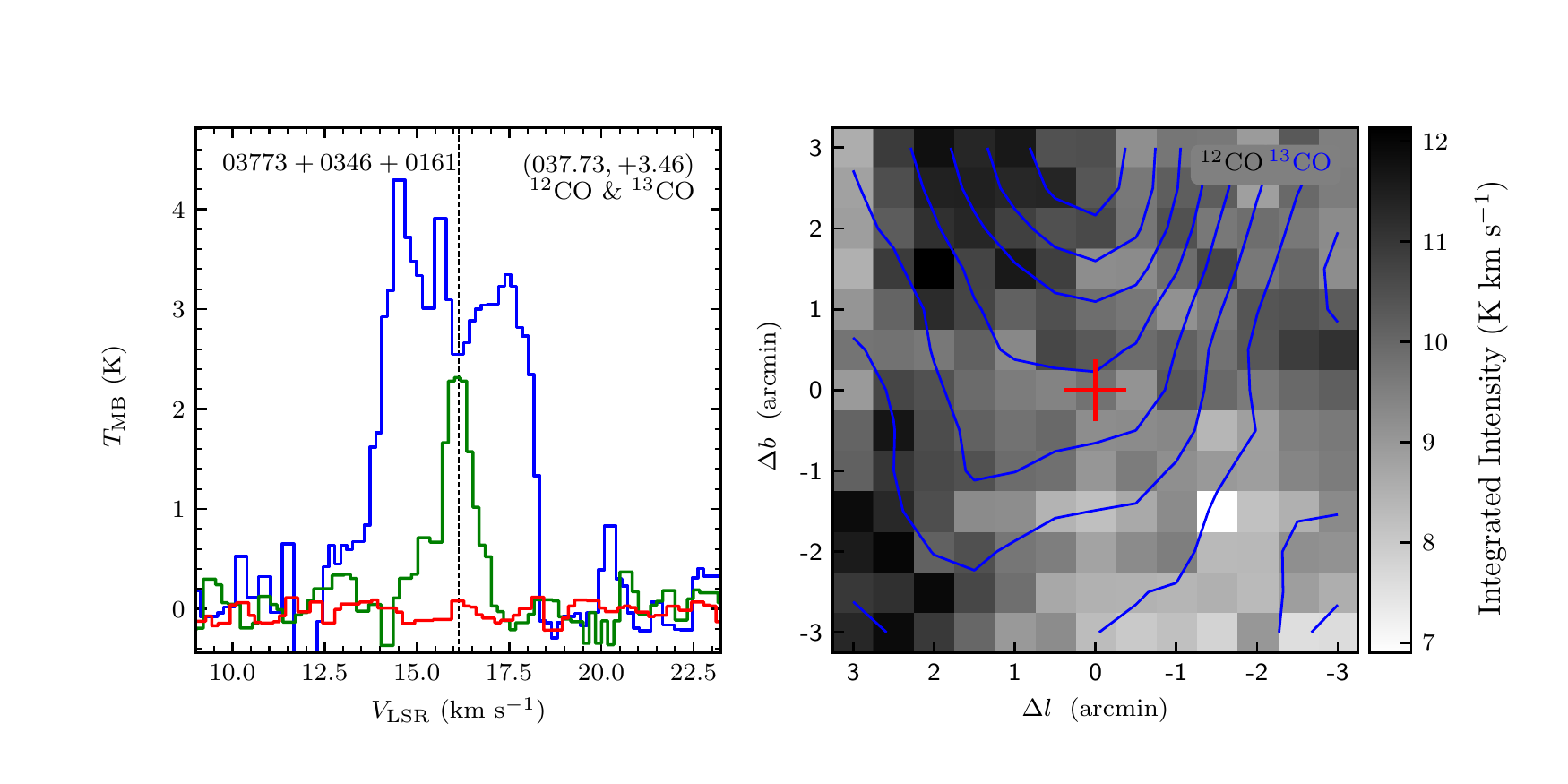}
\includegraphics[width=9.0cm,angle=0]{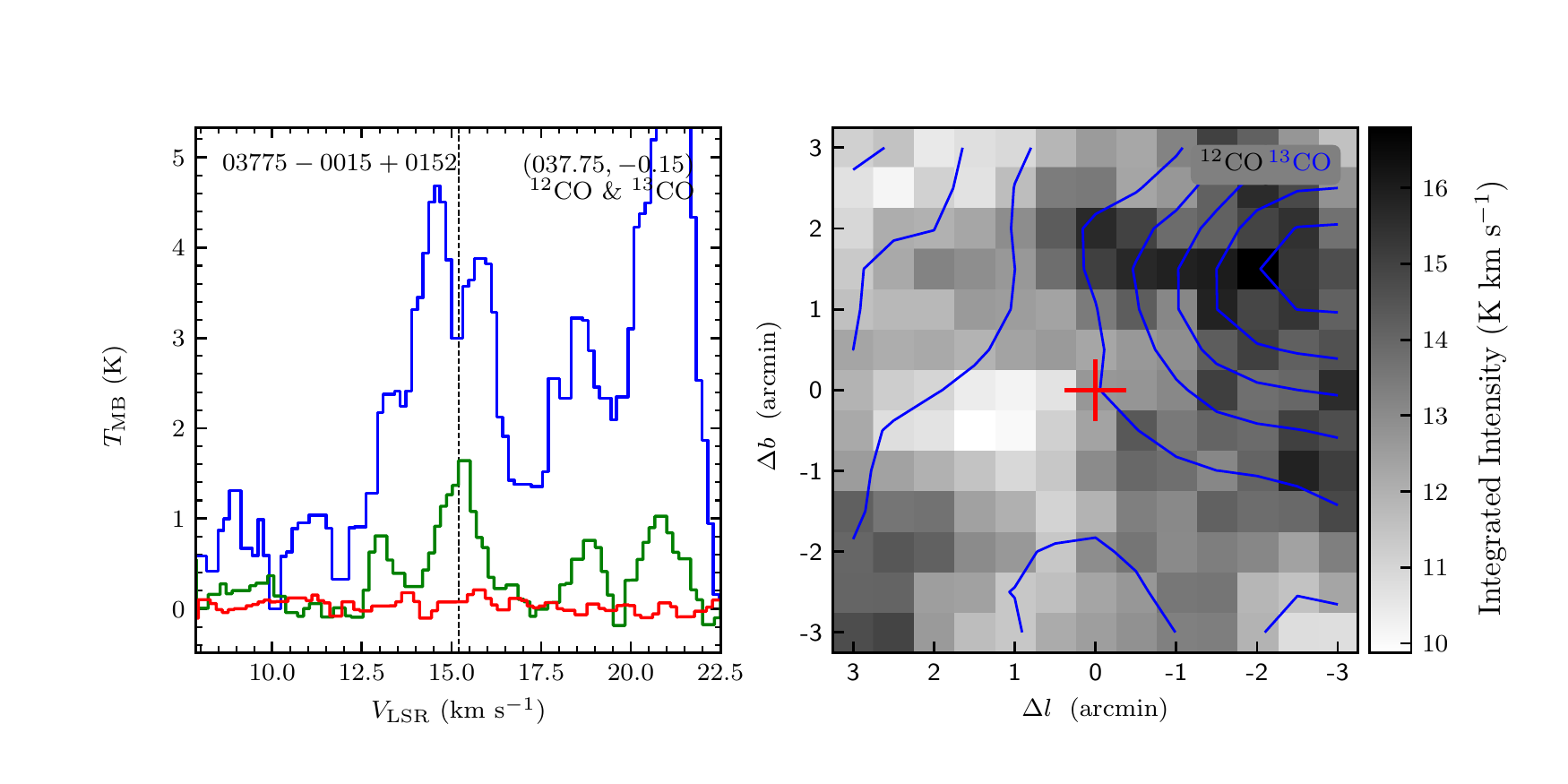}
\vspace{-0.5cm}

\includegraphics[width=9.0cm,angle=0]{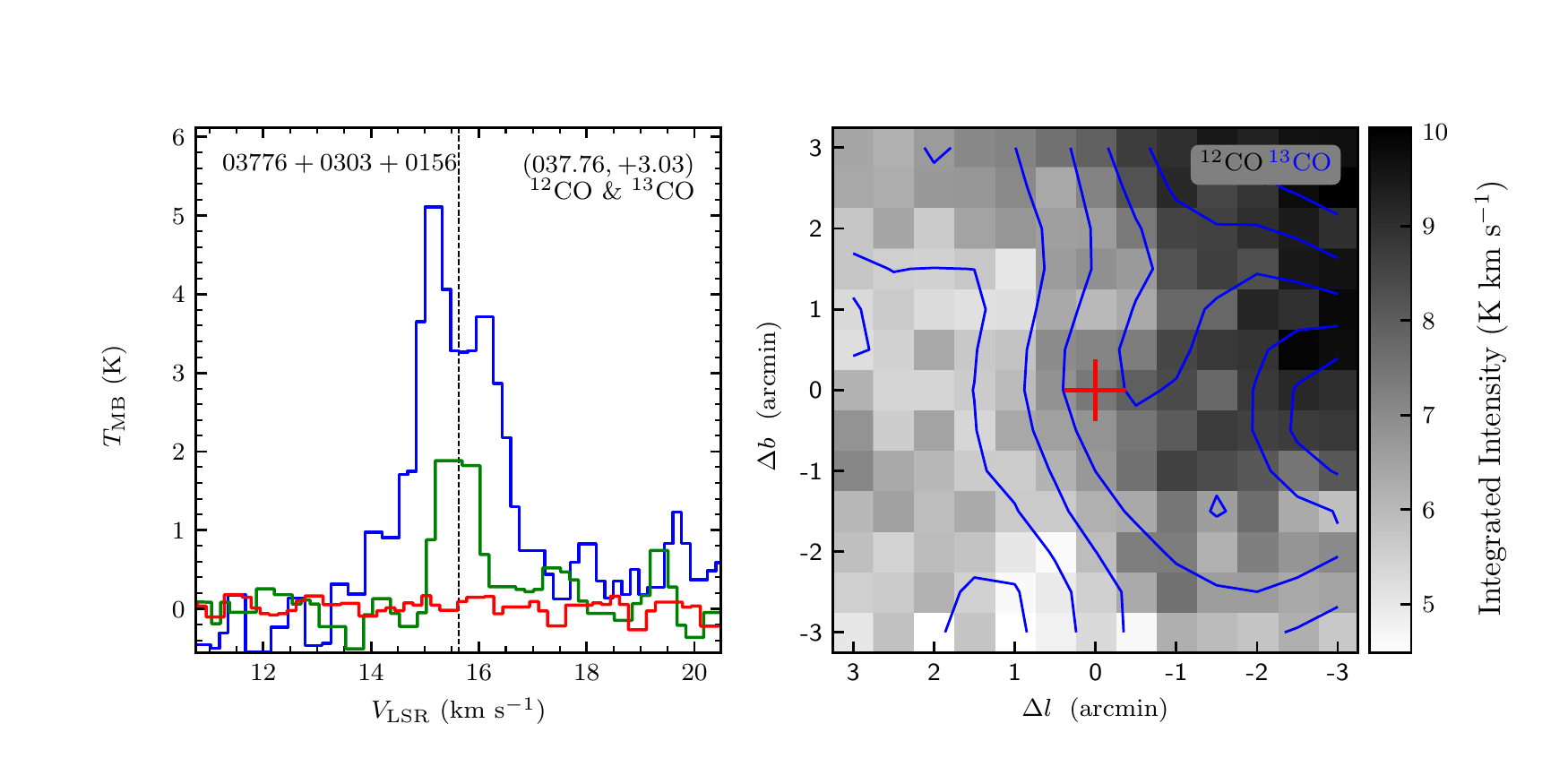}
\includegraphics[width=9.0cm,angle=0]{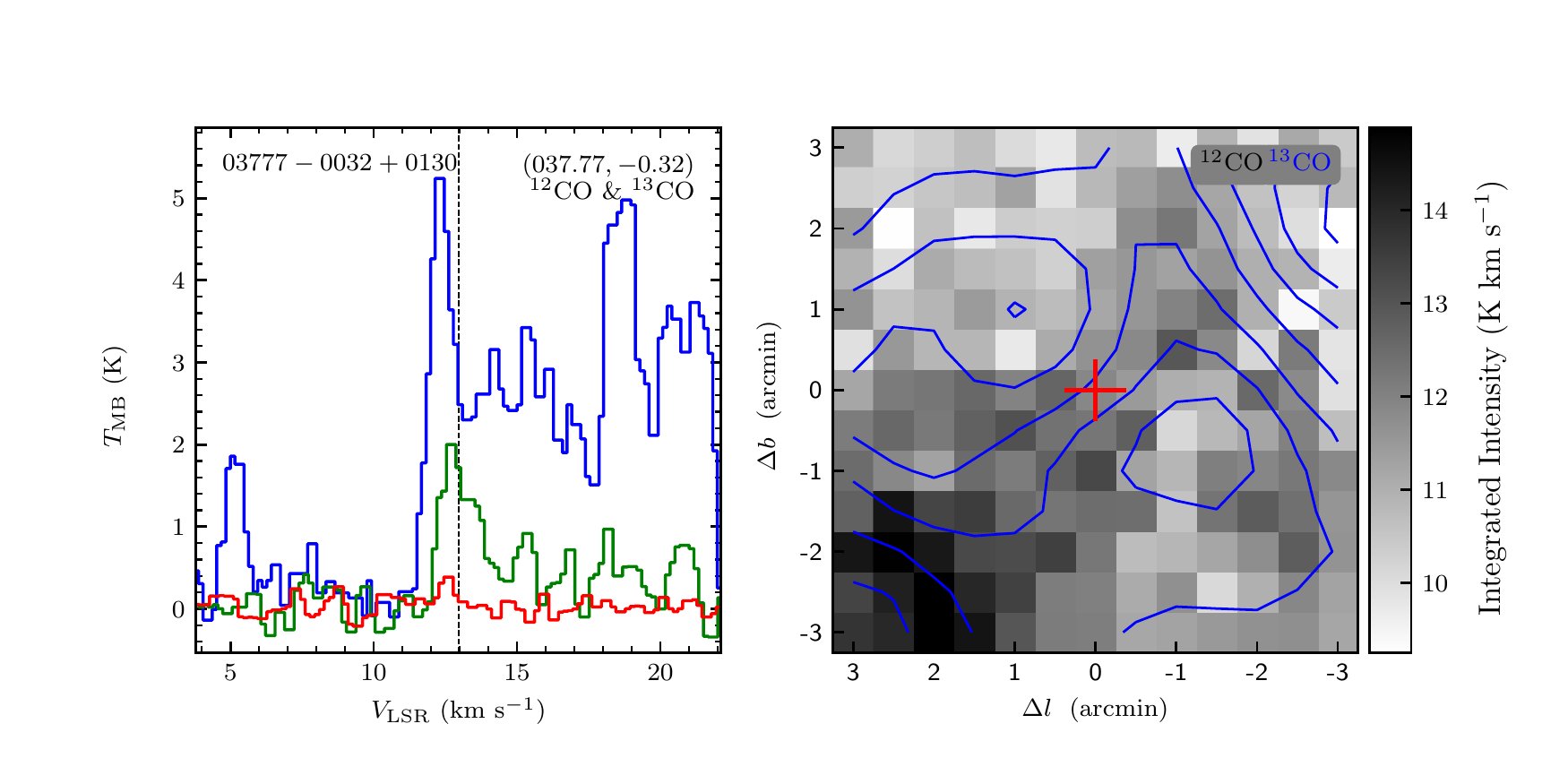}
\vspace{-0.5cm}

\includegraphics[width=9.0cm,angle=0]{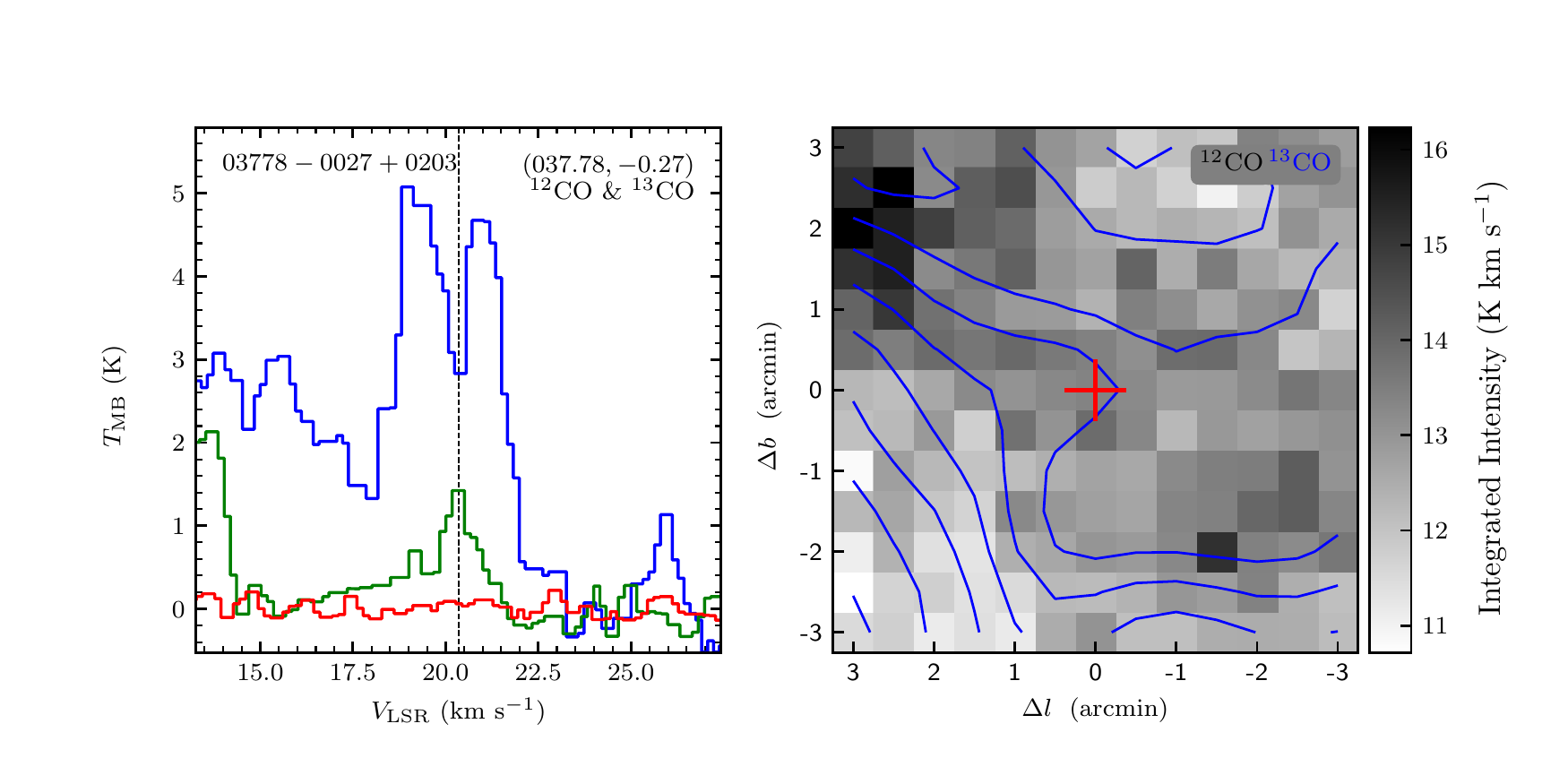}
\includegraphics[width=9.0cm,angle=0]{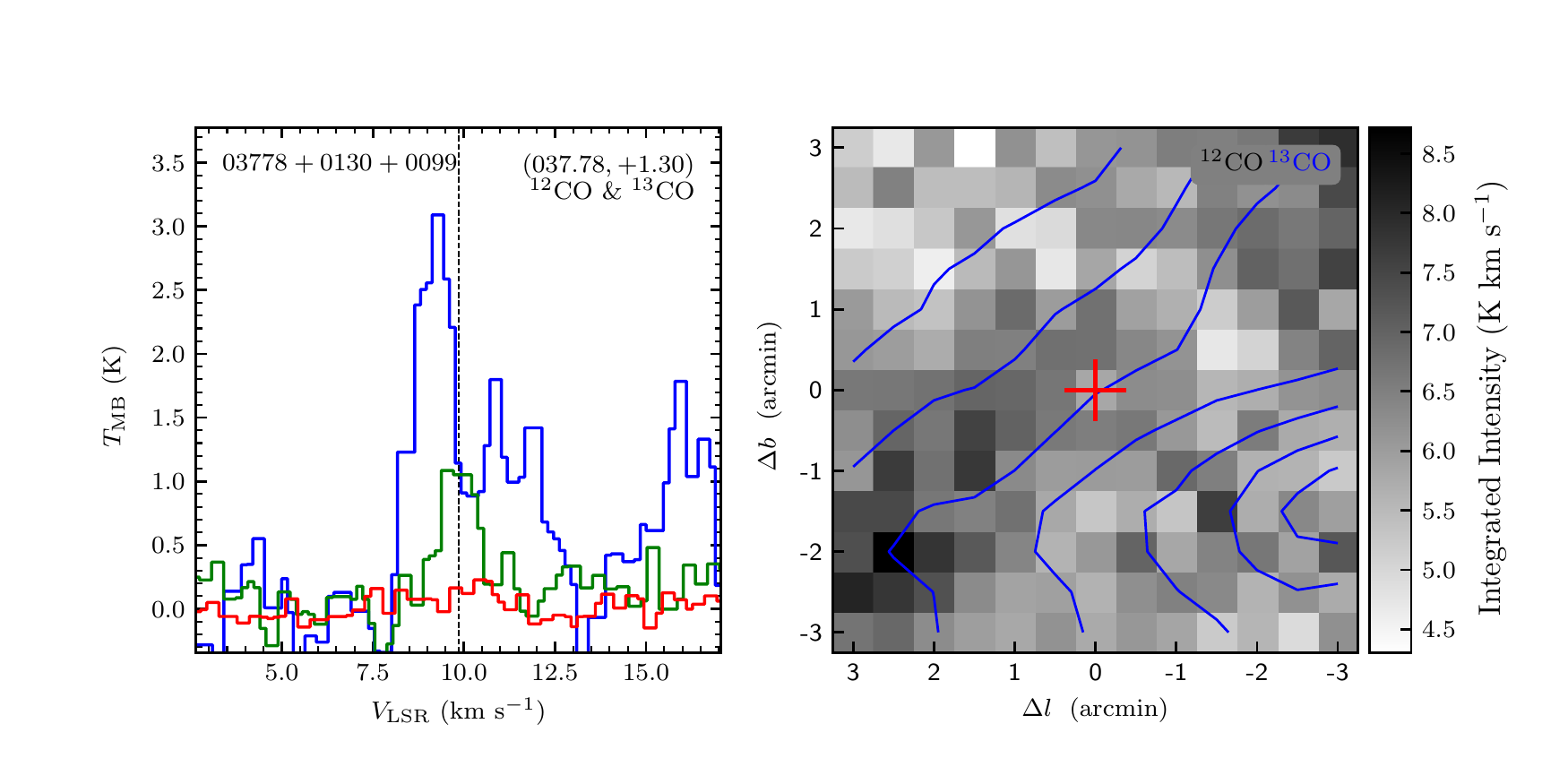}
\vspace{-0.5cm}

\includegraphics[width=9.0cm,angle=0]{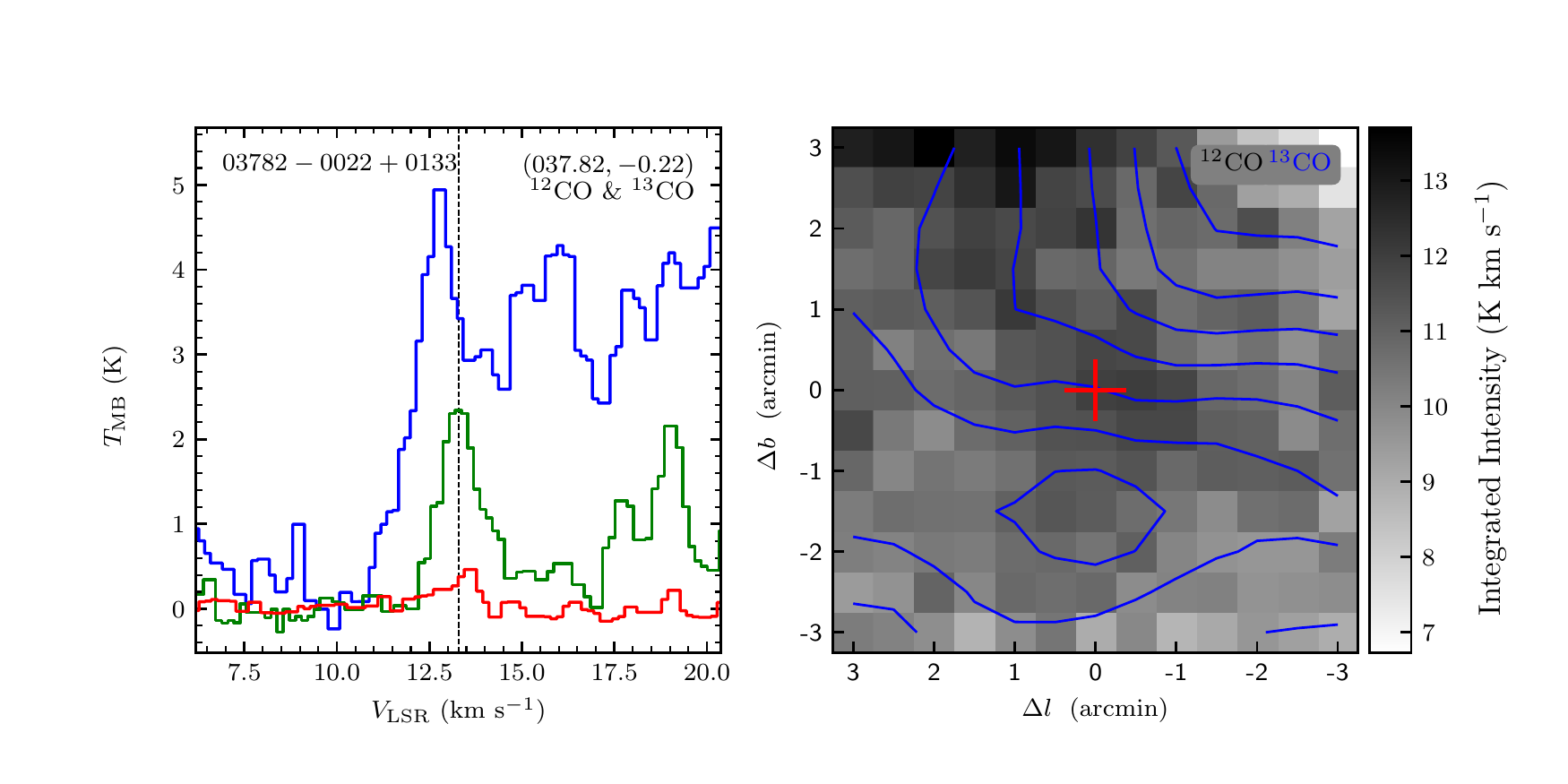}
\includegraphics[width=9.0cm,angle=0]{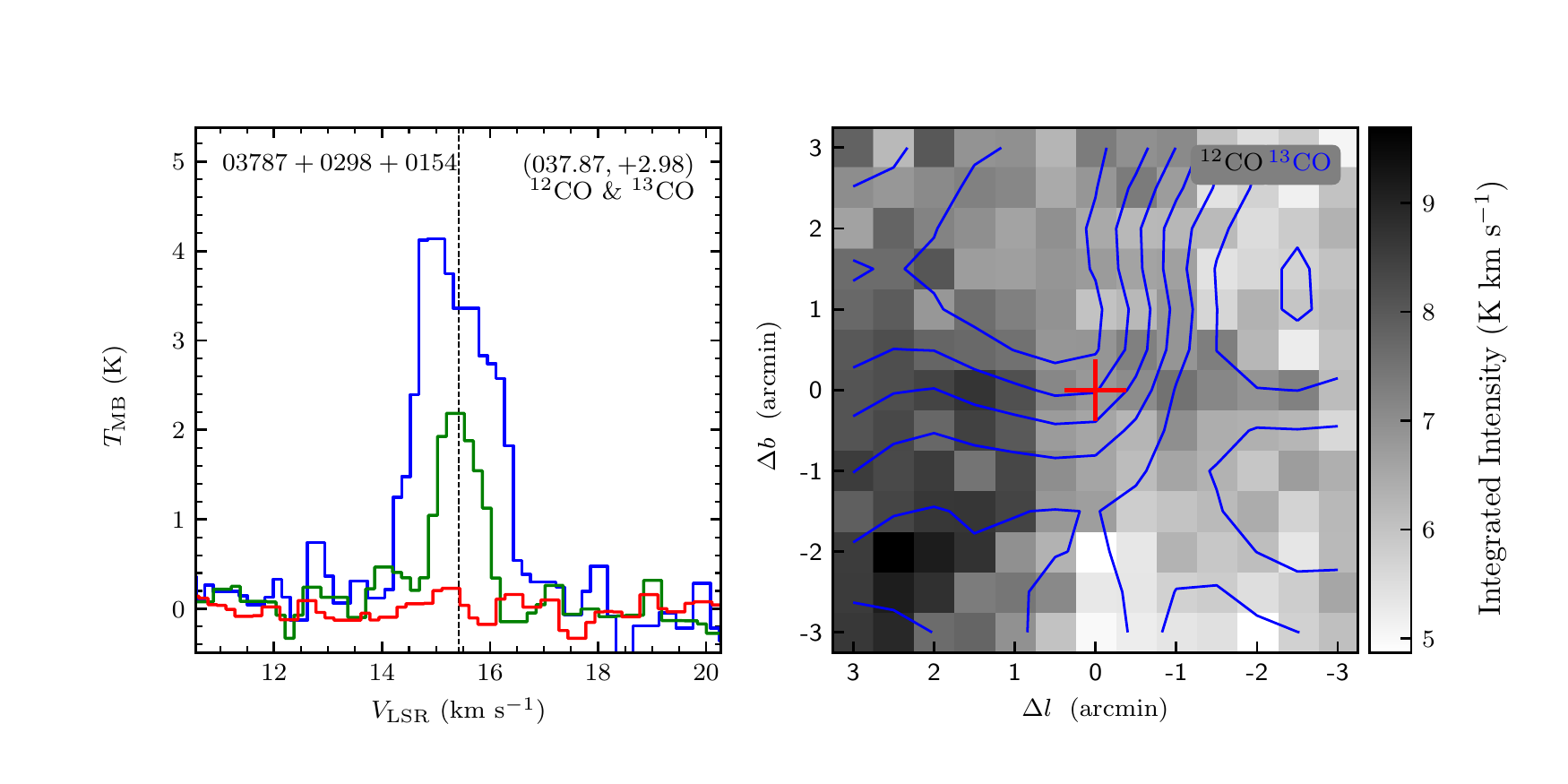}
\end{figure}
\clearpage

\begin{figure}
\includegraphics[width=9.0cm,angle=0]{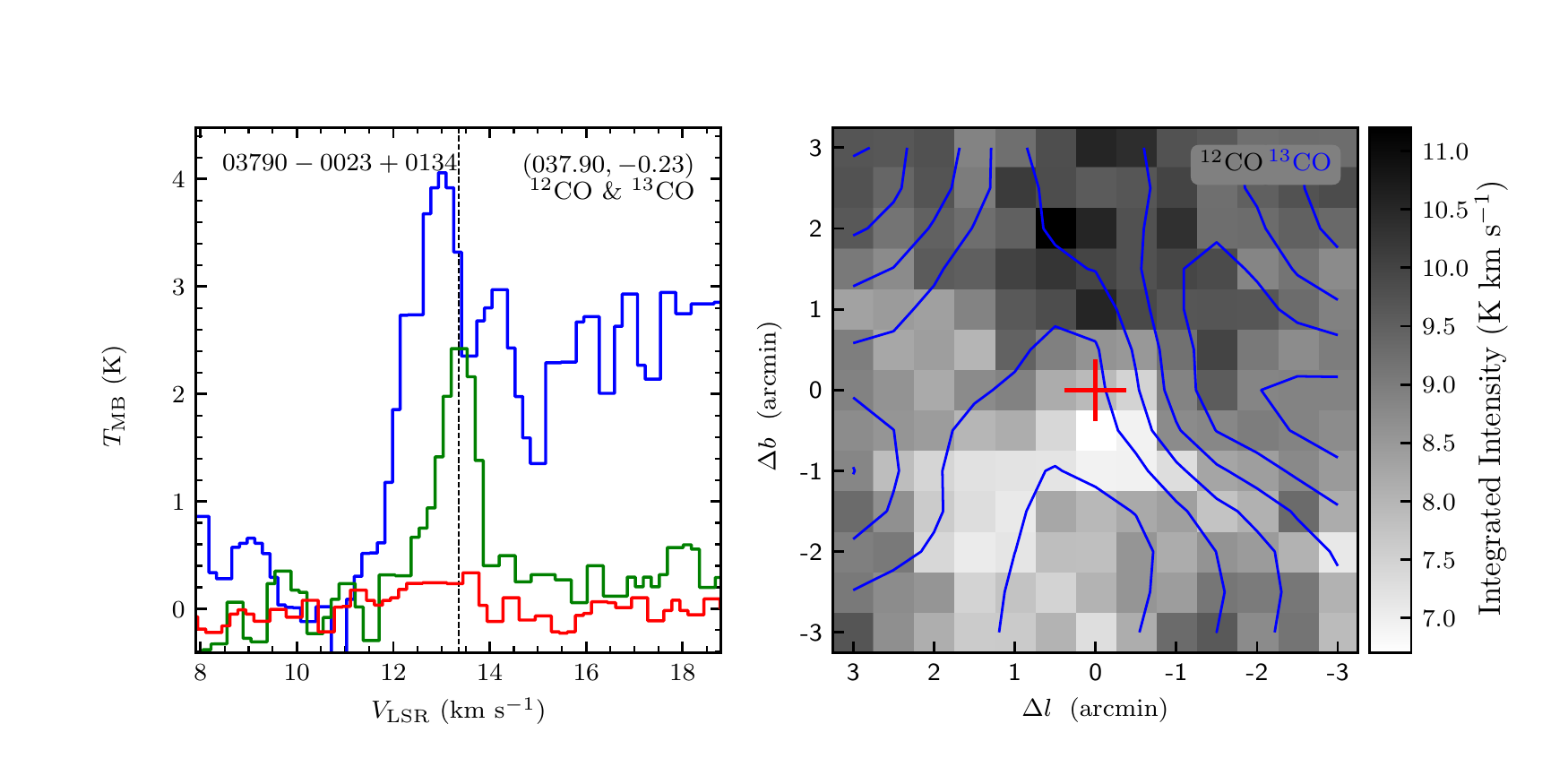}
\includegraphics[width=9.0cm,angle=0]{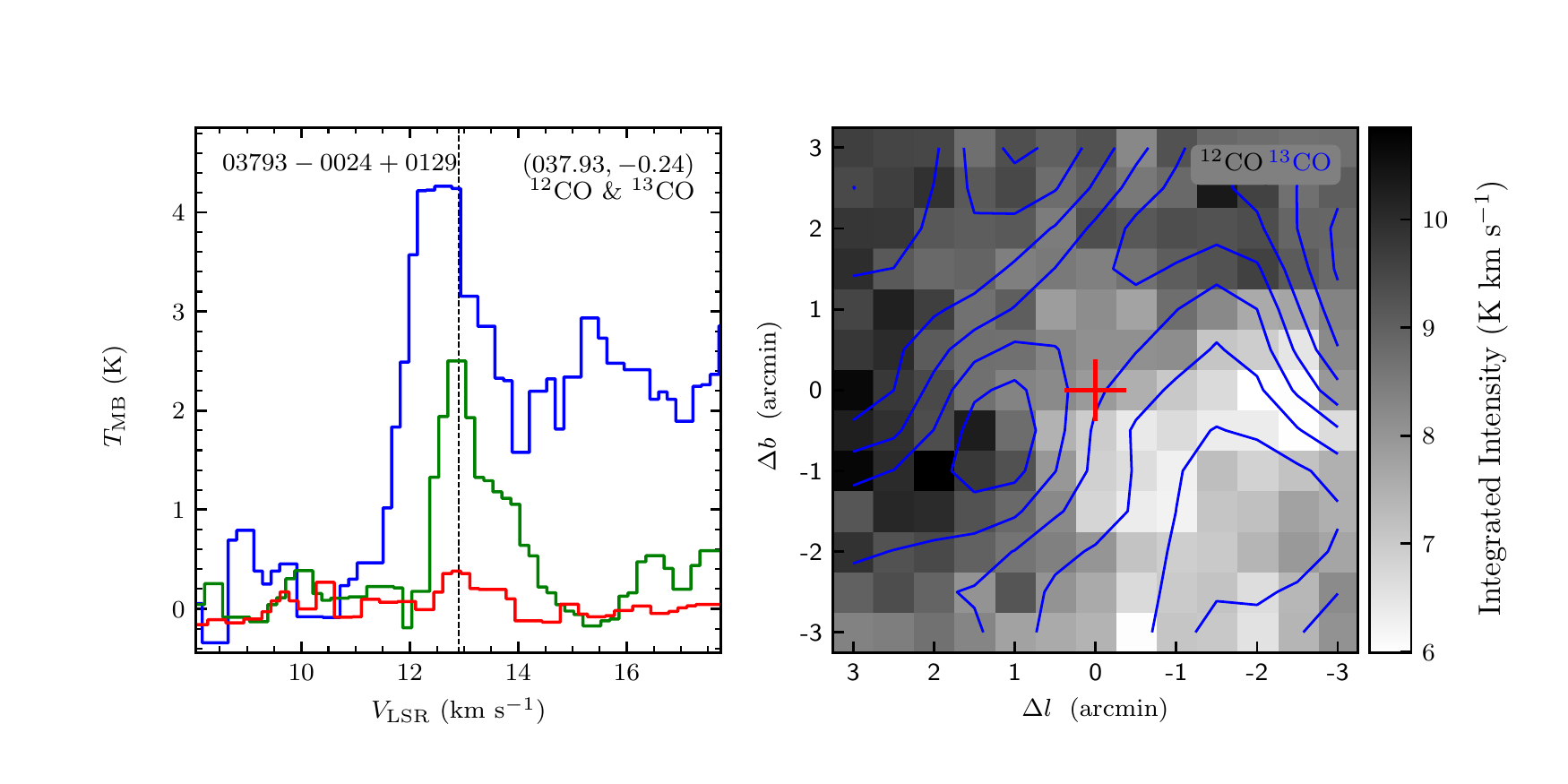}
\vspace{-0.5cm}

\includegraphics[width=9.0cm,angle=0]{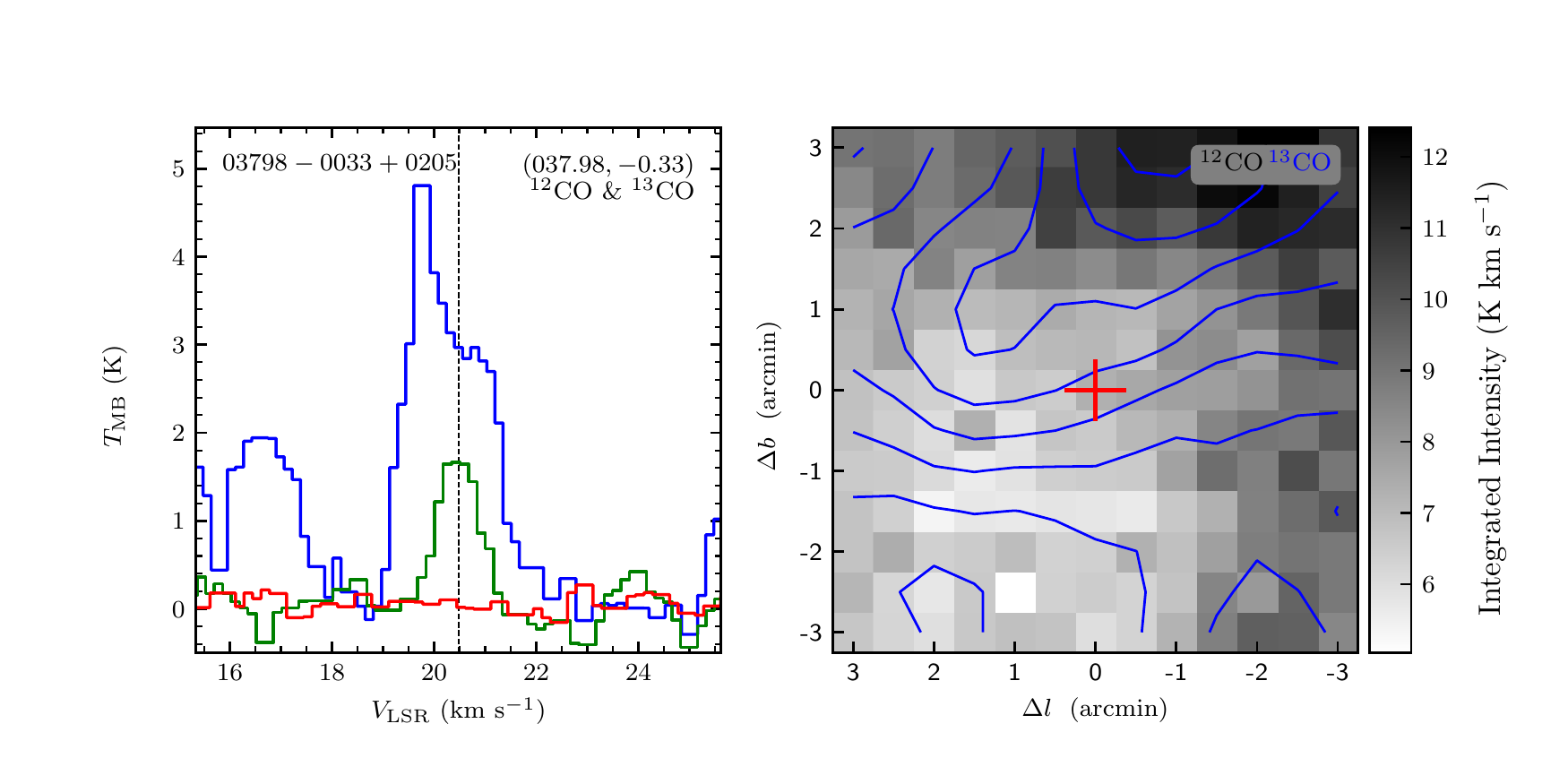}
\includegraphics[width=9.0cm,angle=0]{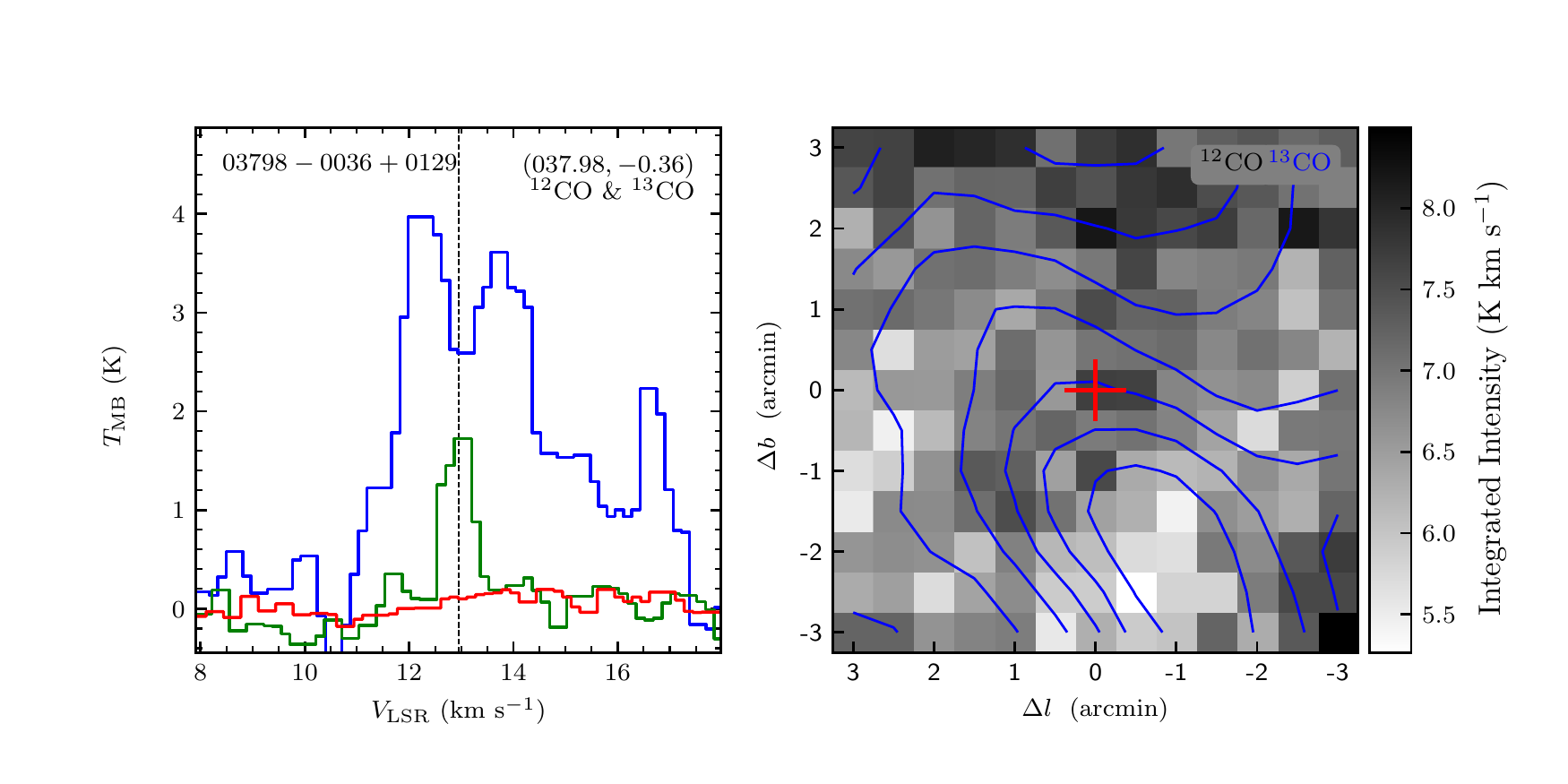}
\vspace{-0.5cm}

\includegraphics[width=9.0cm,angle=0]{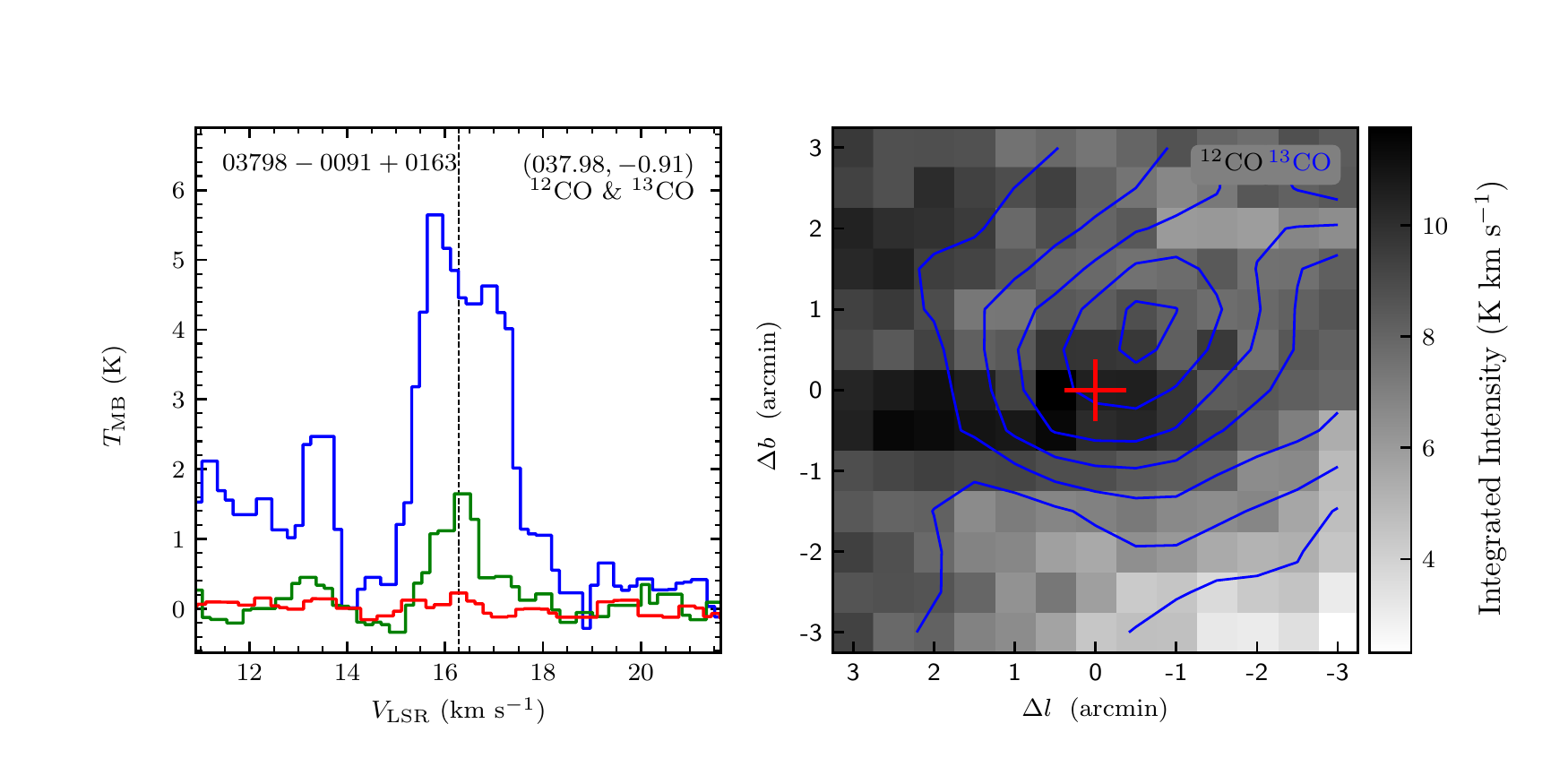}
\includegraphics[width=9.0cm,angle=0]{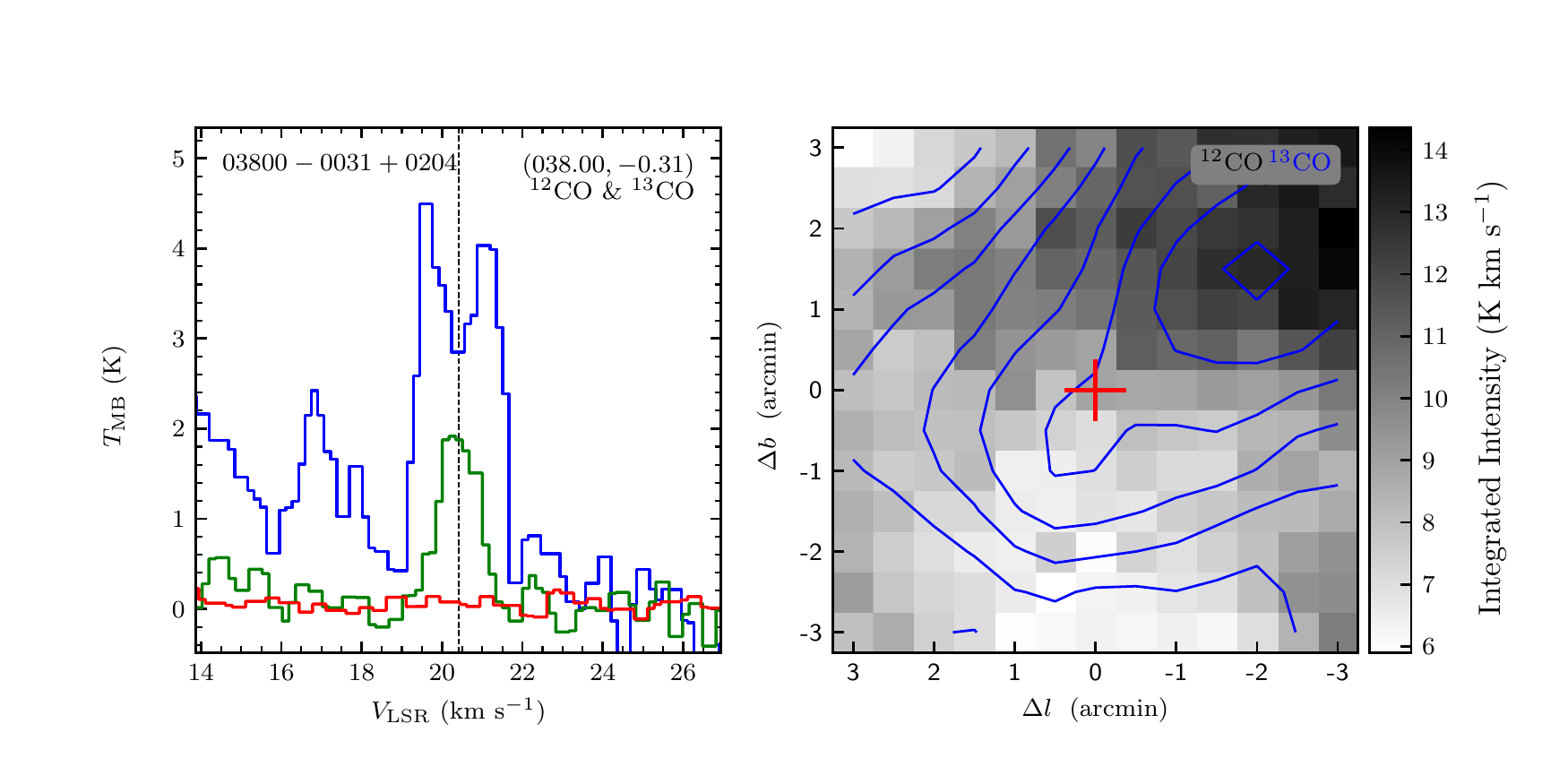}
\vspace{-0.5cm}

\includegraphics[width=9.0cm,angle=0]{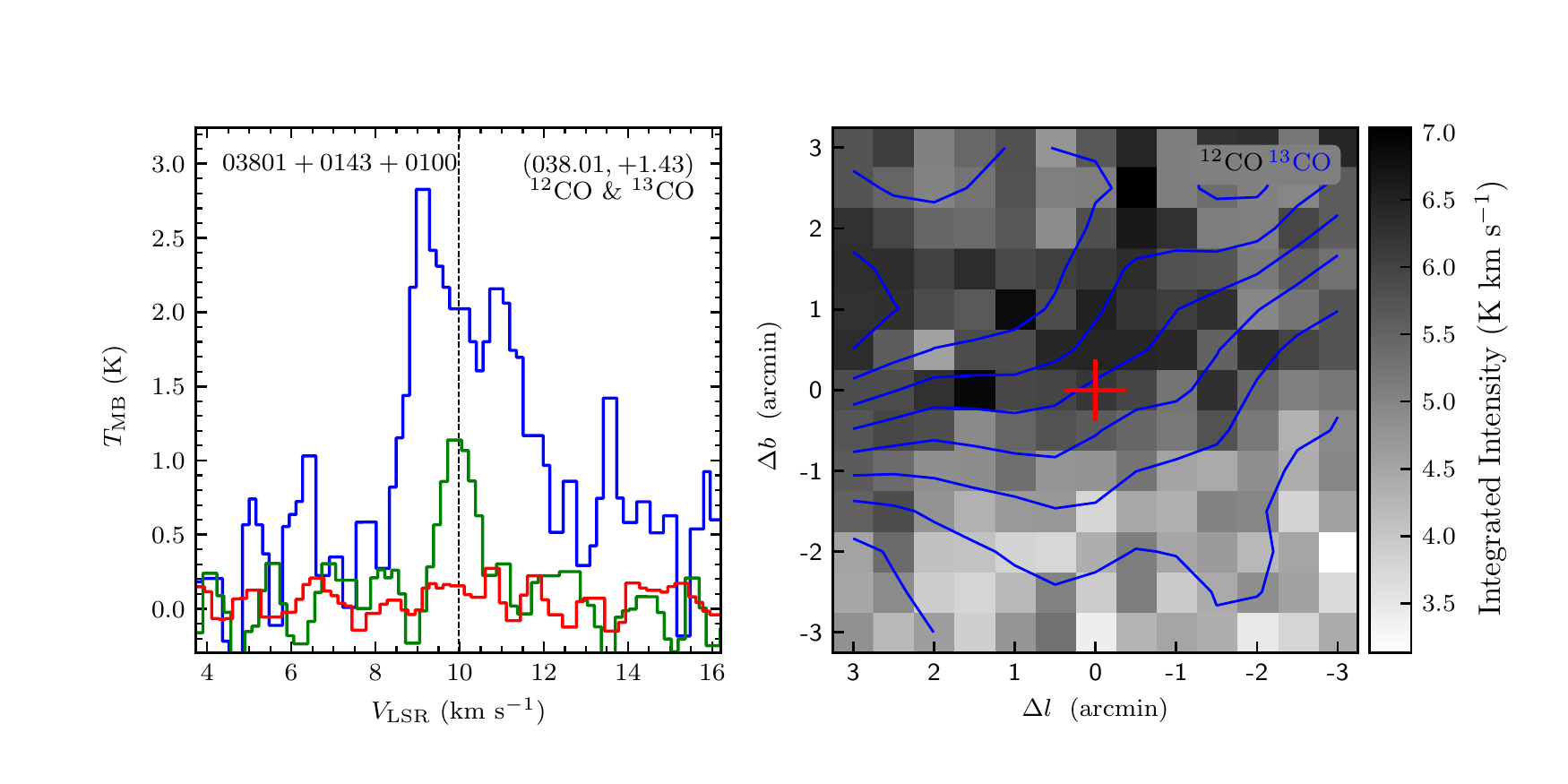}
\includegraphics[width=9.0cm,angle=0]{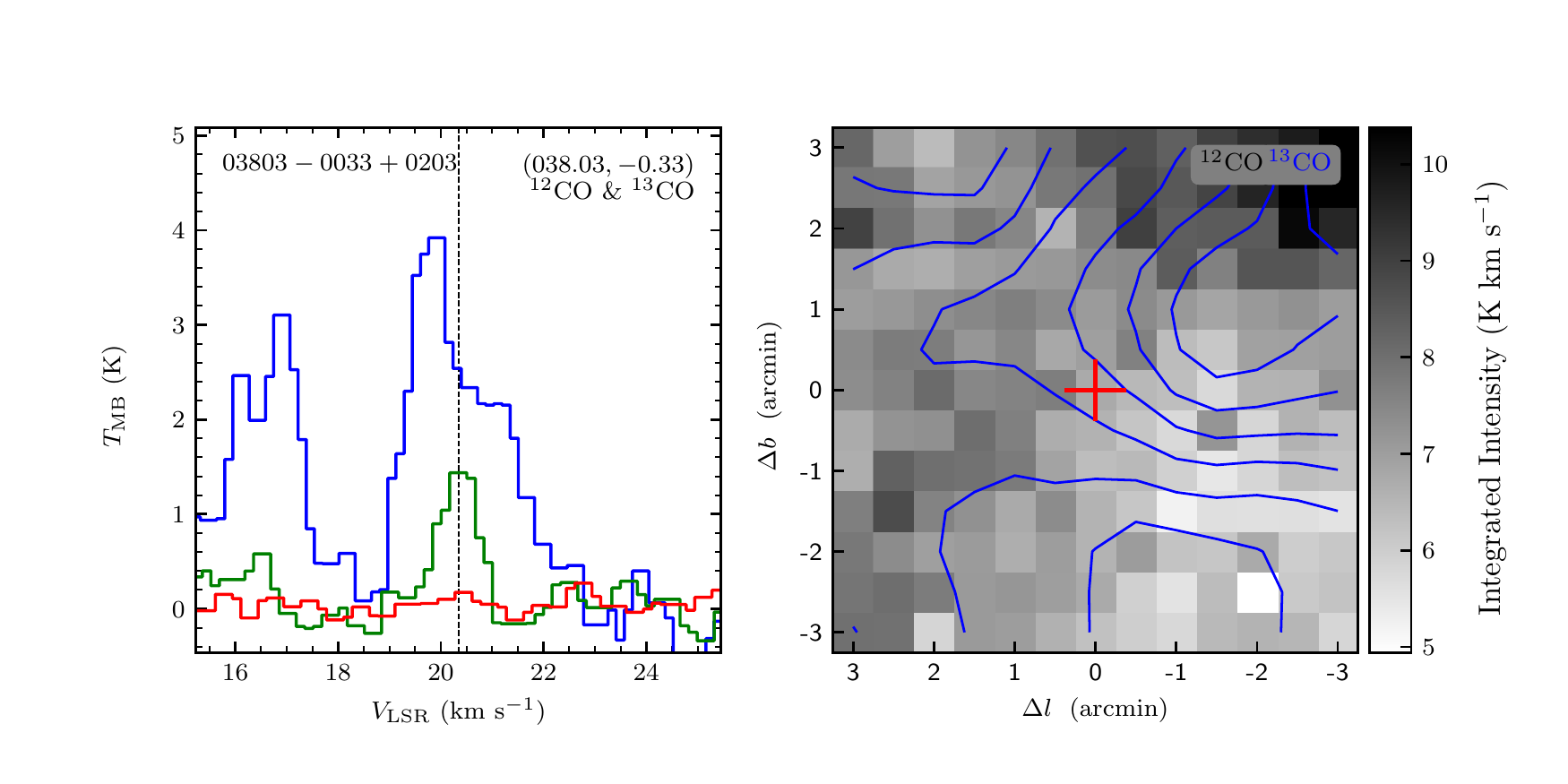}
\vspace{-0.5cm}

\includegraphics[width=9.0cm,angle=0]{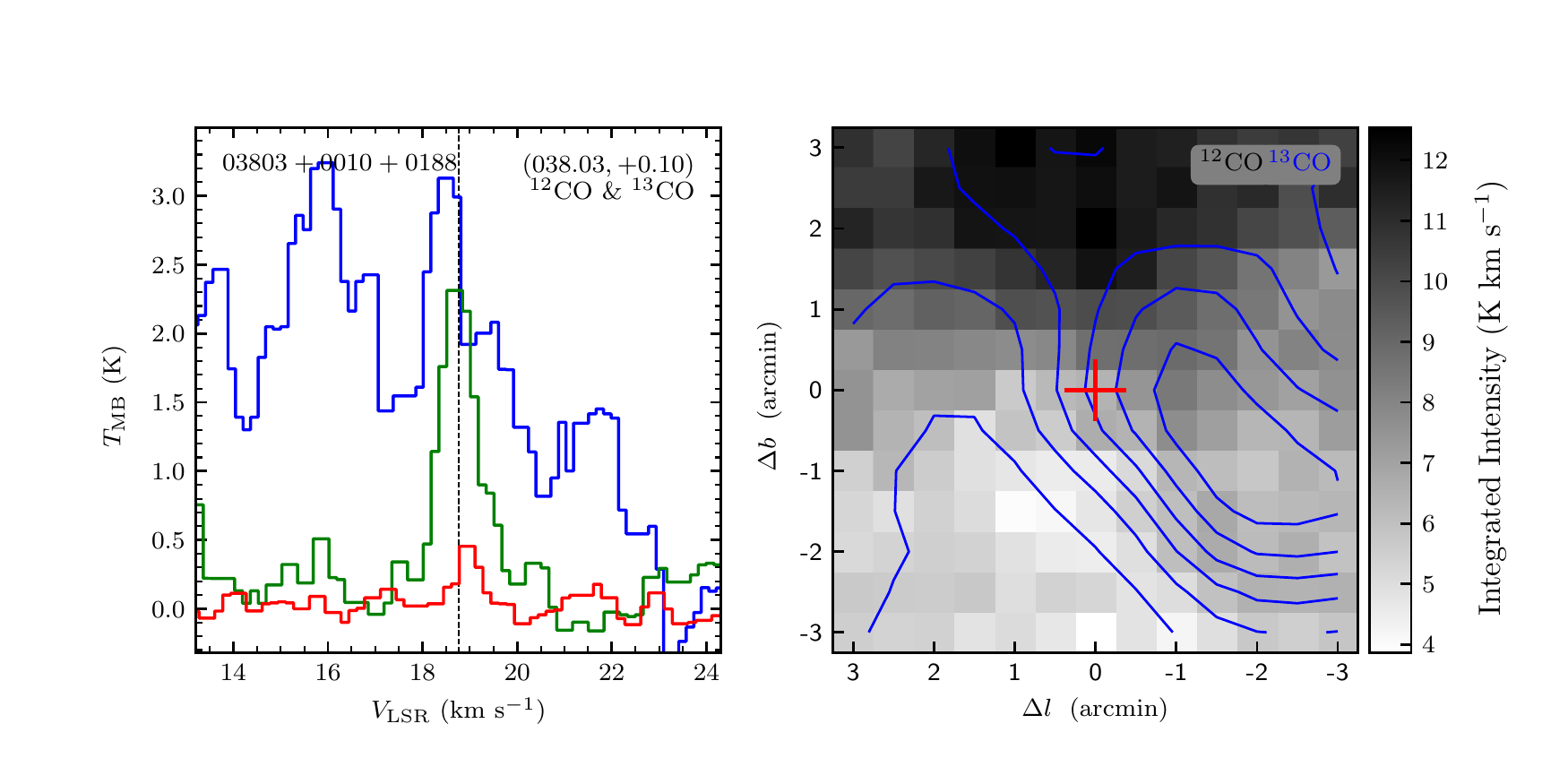}
\includegraphics[width=9.0cm,angle=0]{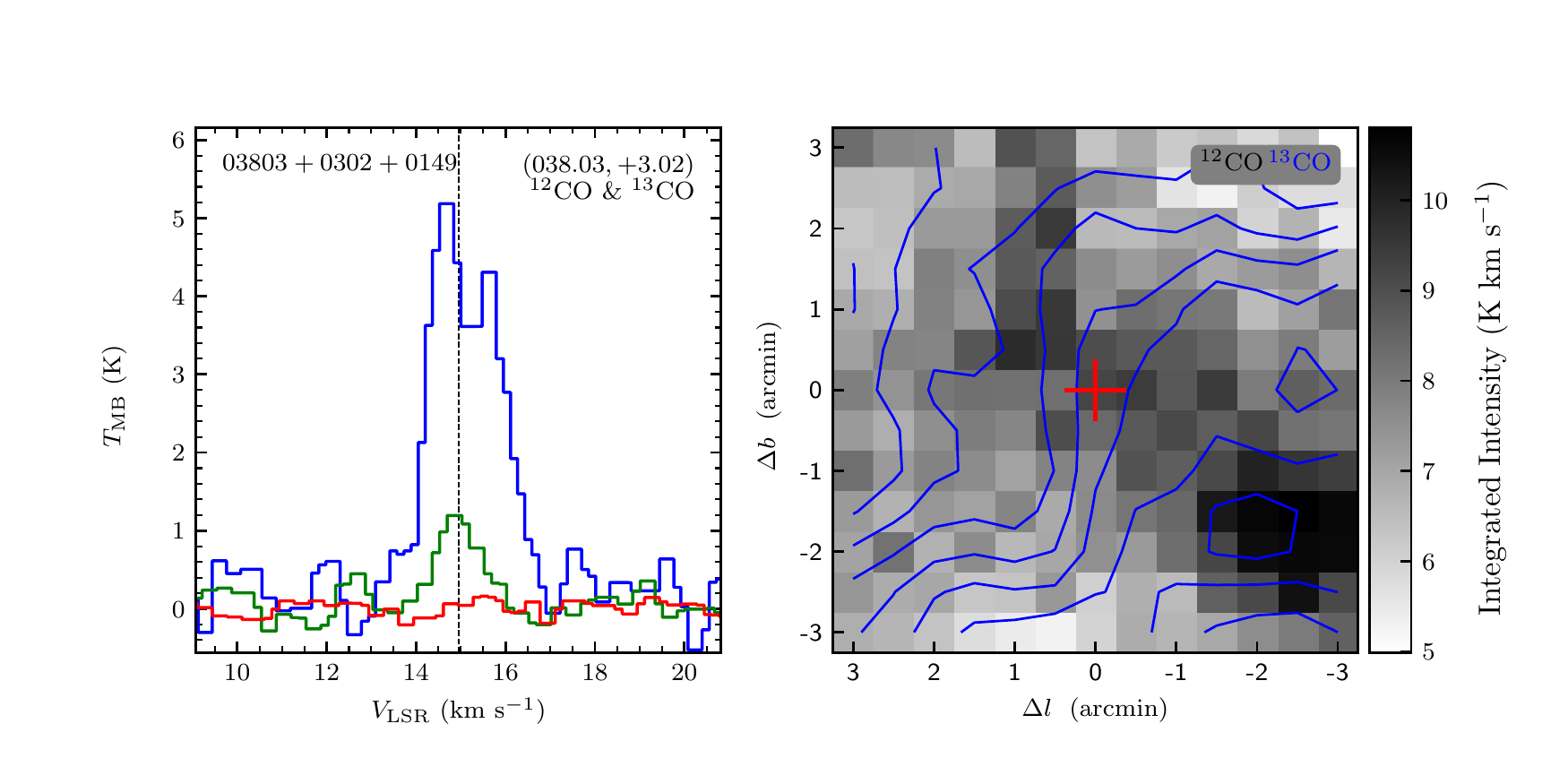}
\end{figure}
\clearpage

\begin{figure}
\includegraphics[width=9.0cm,angle=0]{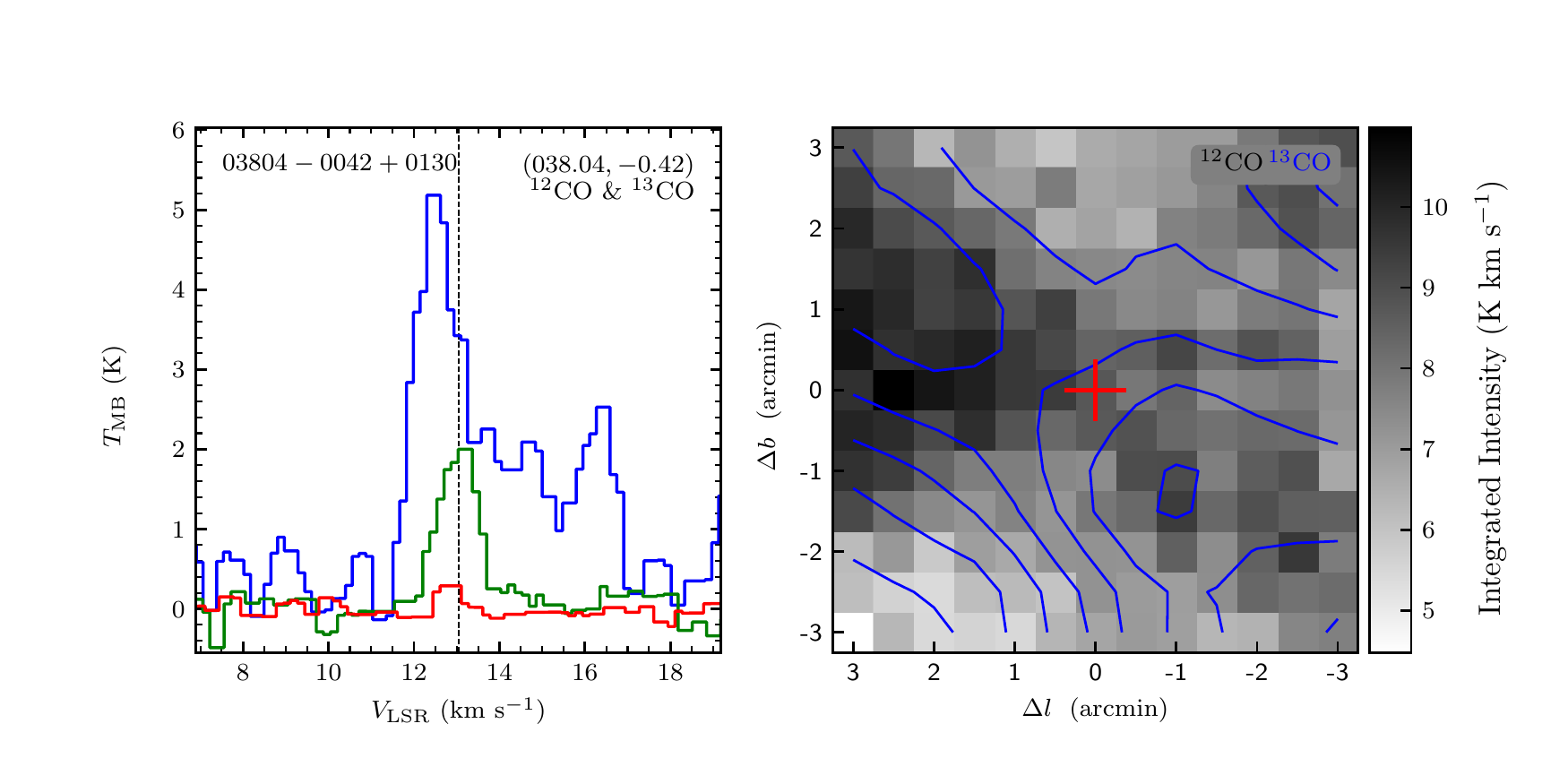}
\includegraphics[width=9.0cm,angle=0]{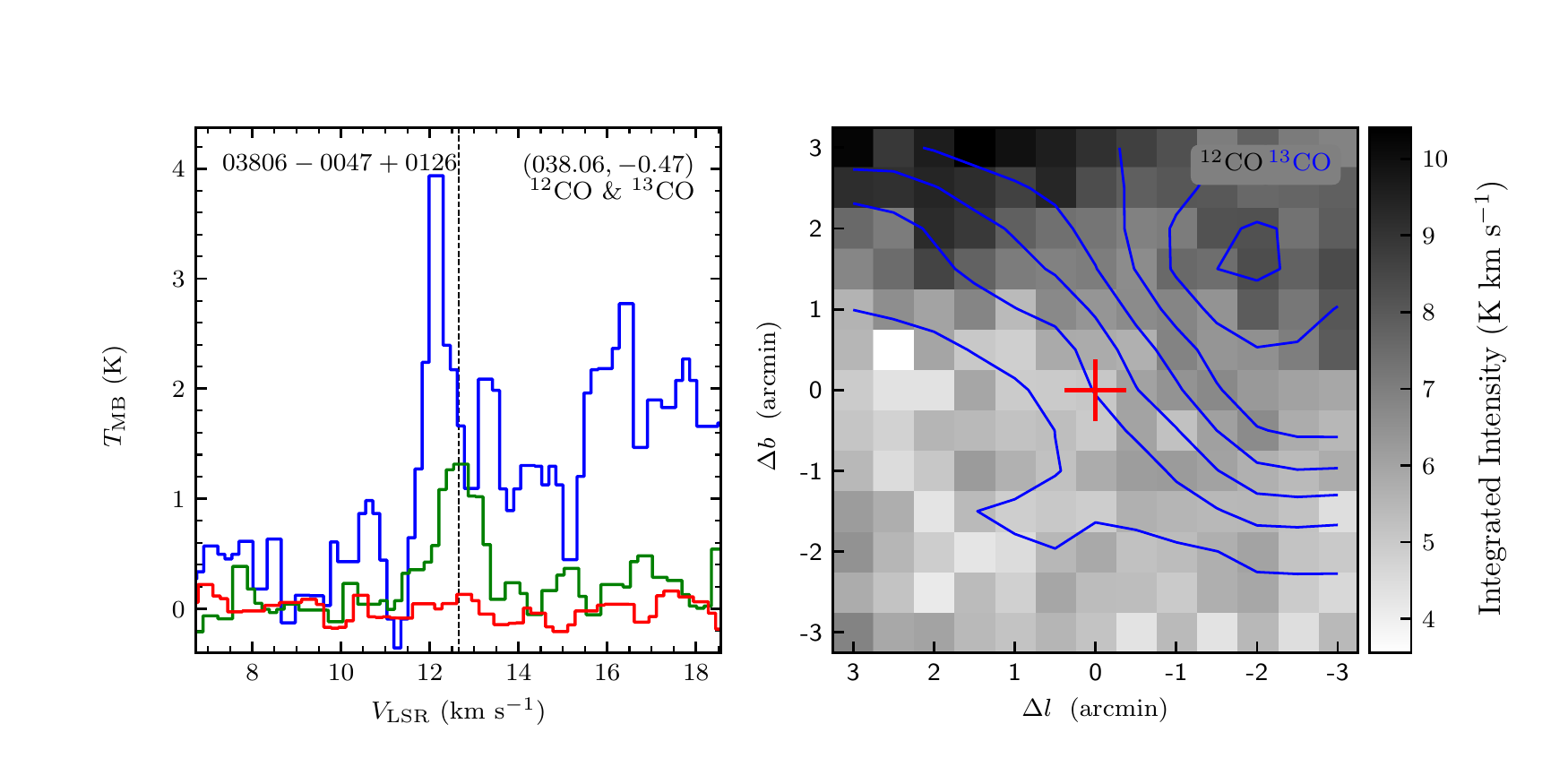}
\vspace{-0.5cm}

\includegraphics[width=9.0cm,angle=0]{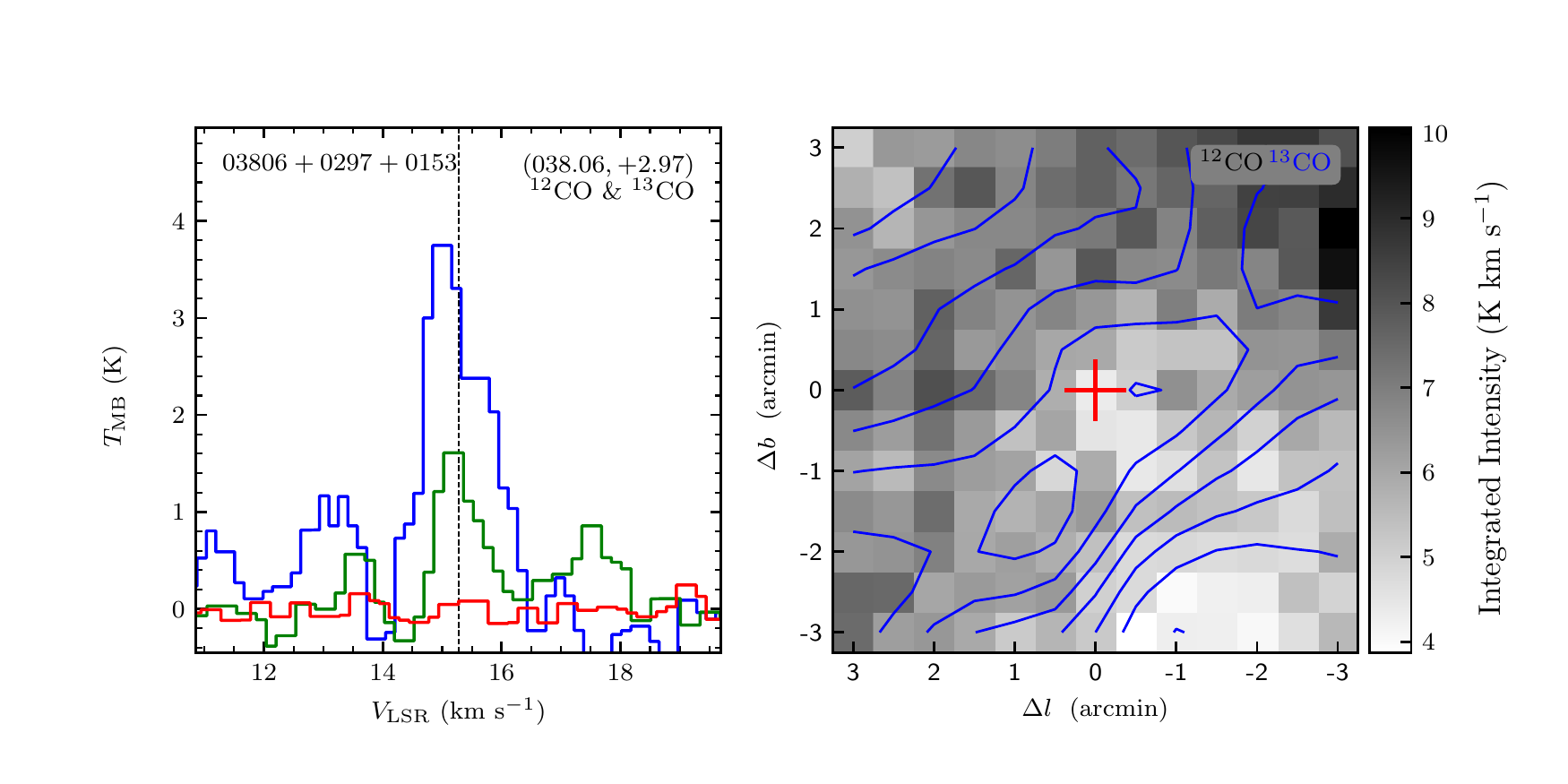}
\includegraphics[width=9.0cm,angle=0]{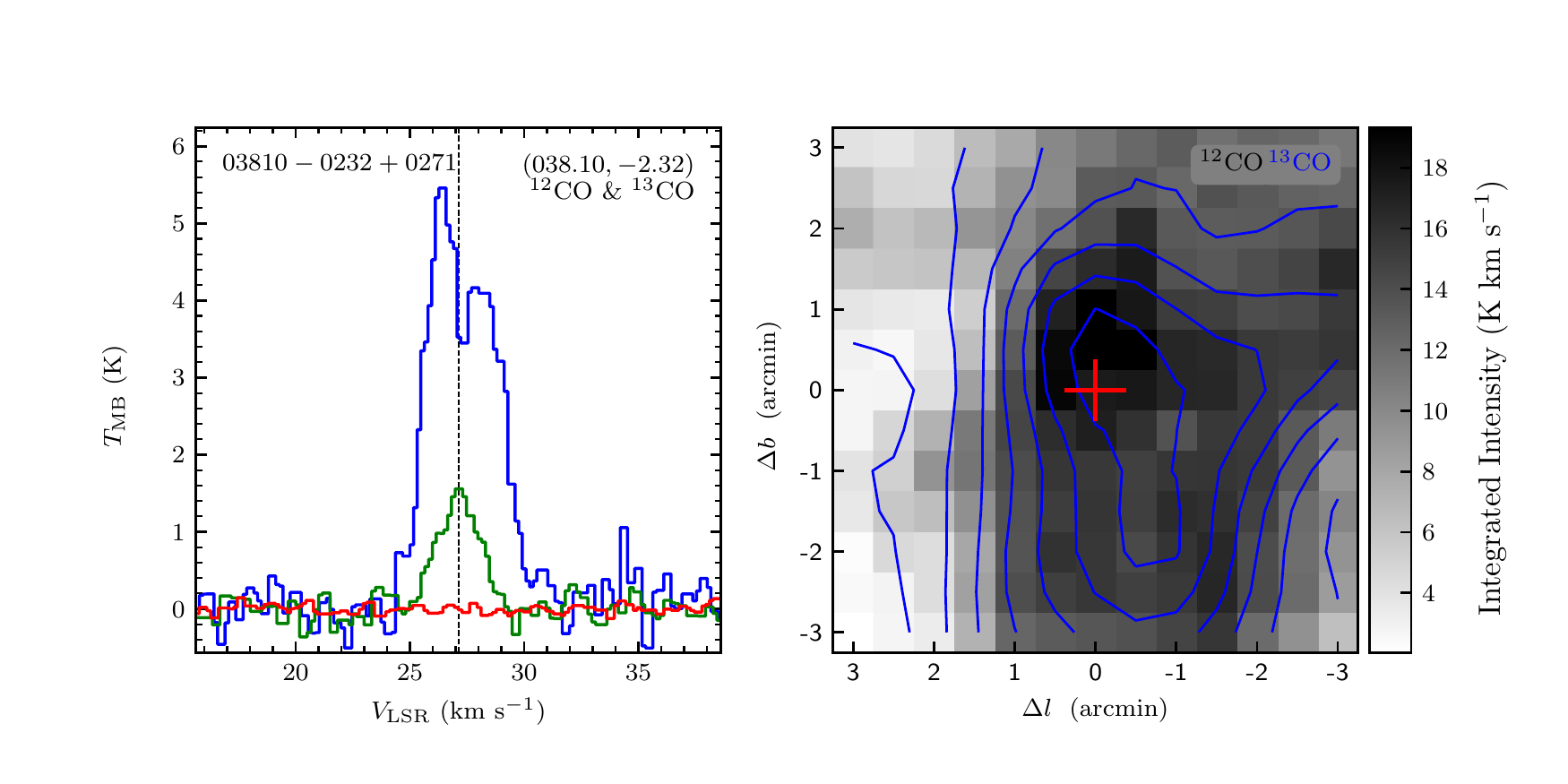}
\vspace{-0.5cm}

\includegraphics[width=9.0cm,angle=0]{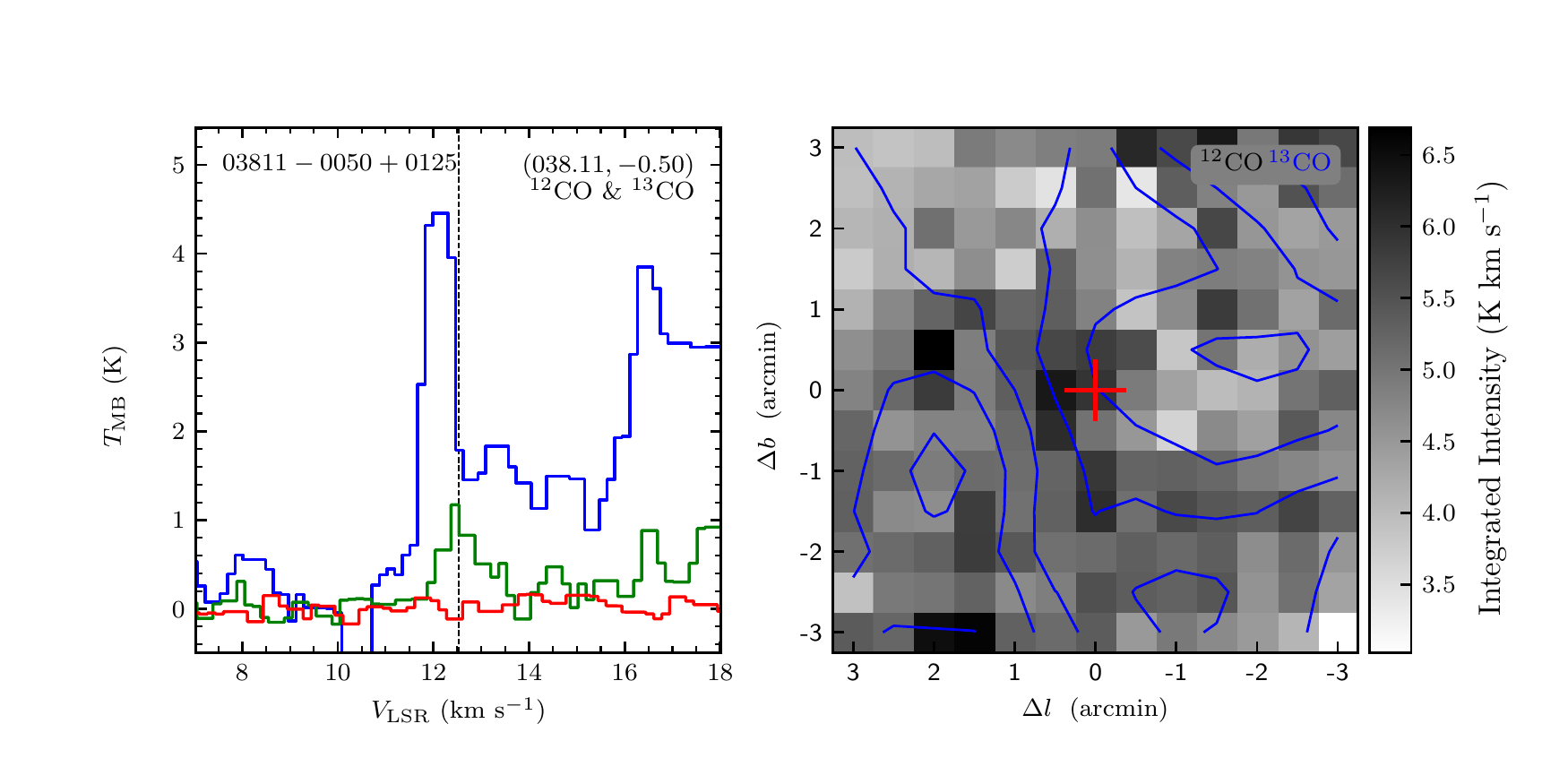}
\includegraphics[width=9.0cm,angle=0]{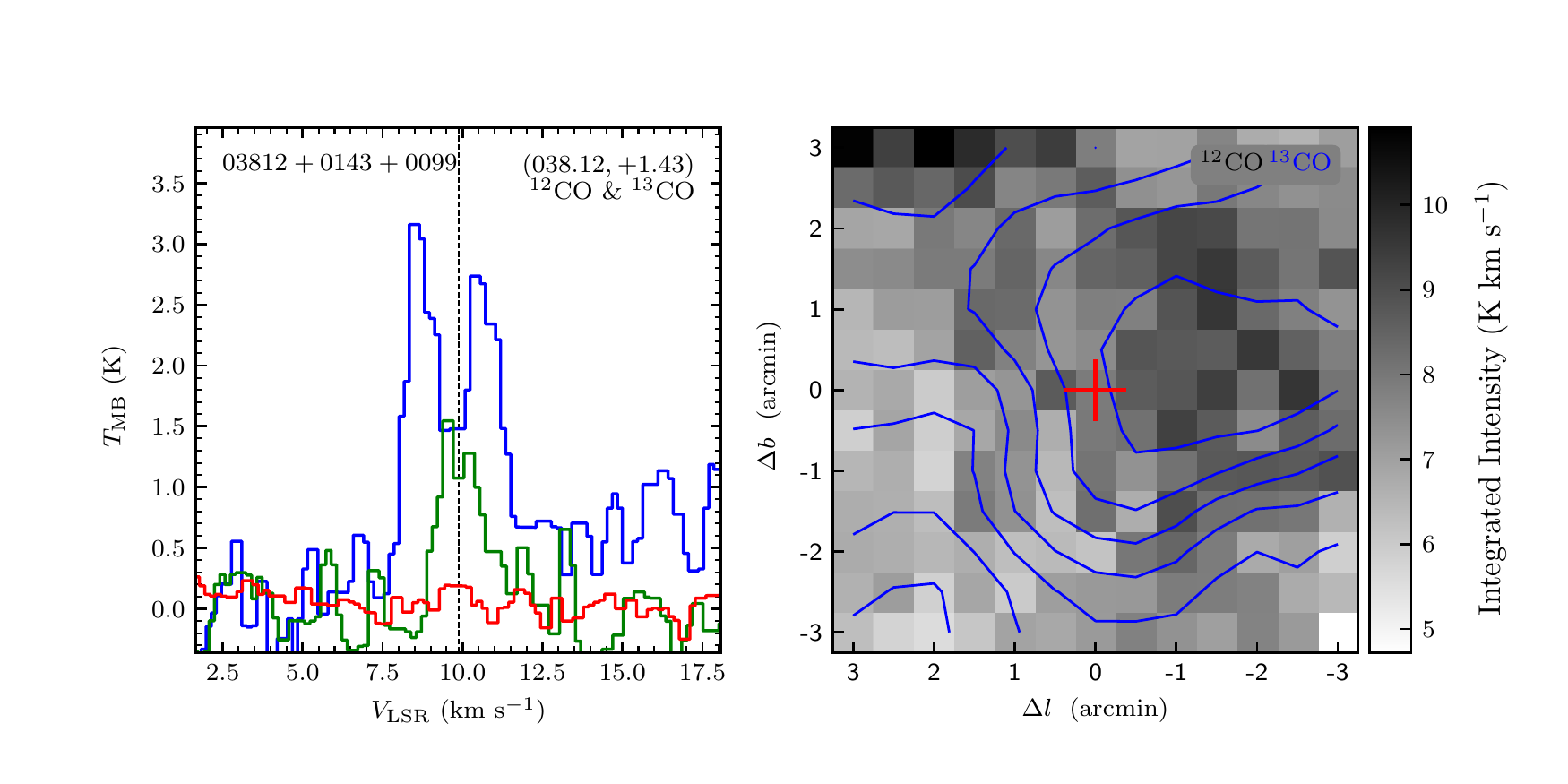}
\vspace{-0.5cm}

\includegraphics[width=9.0cm,angle=0]{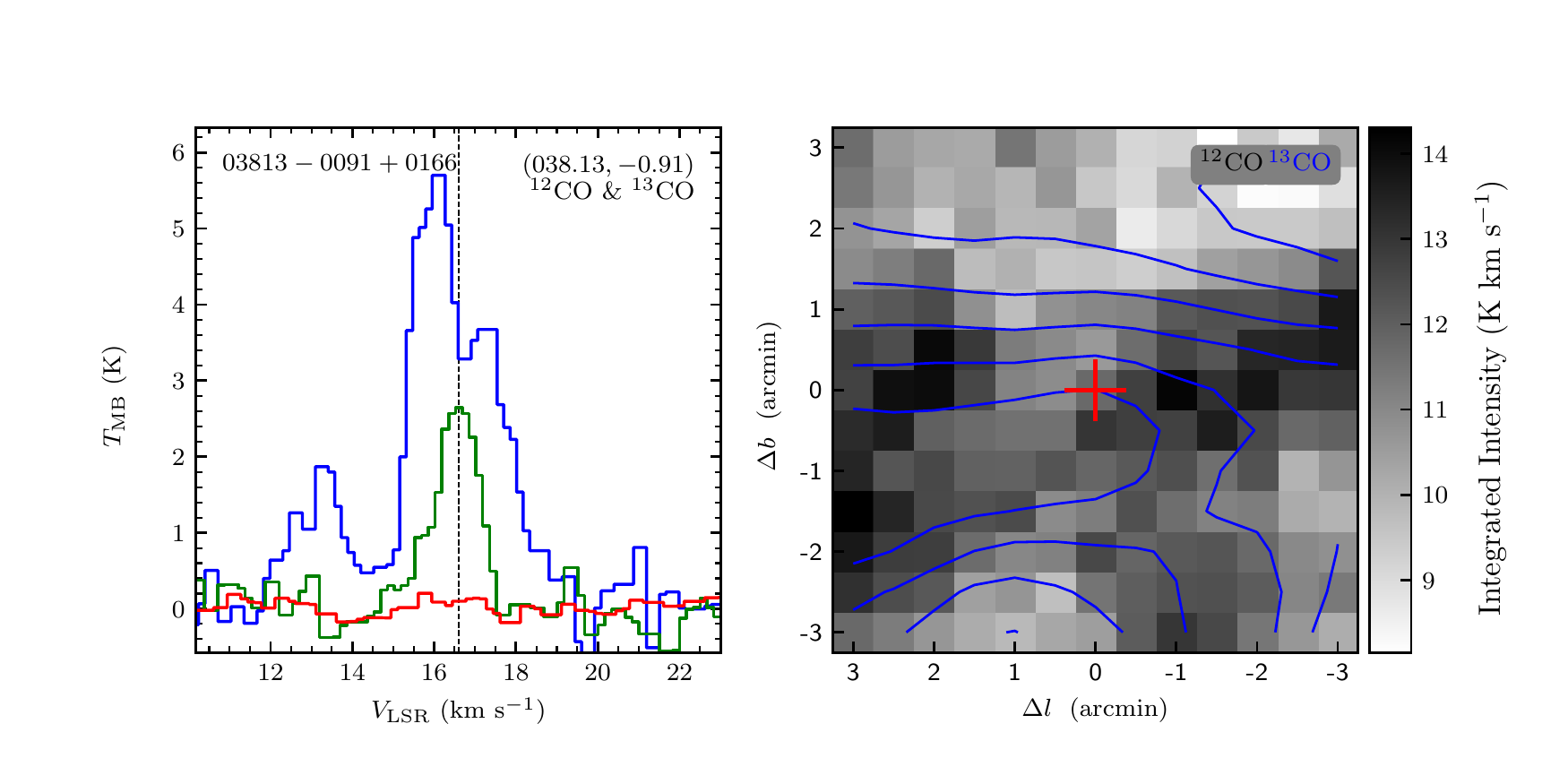}
\includegraphics[width=9.0cm,angle=0]{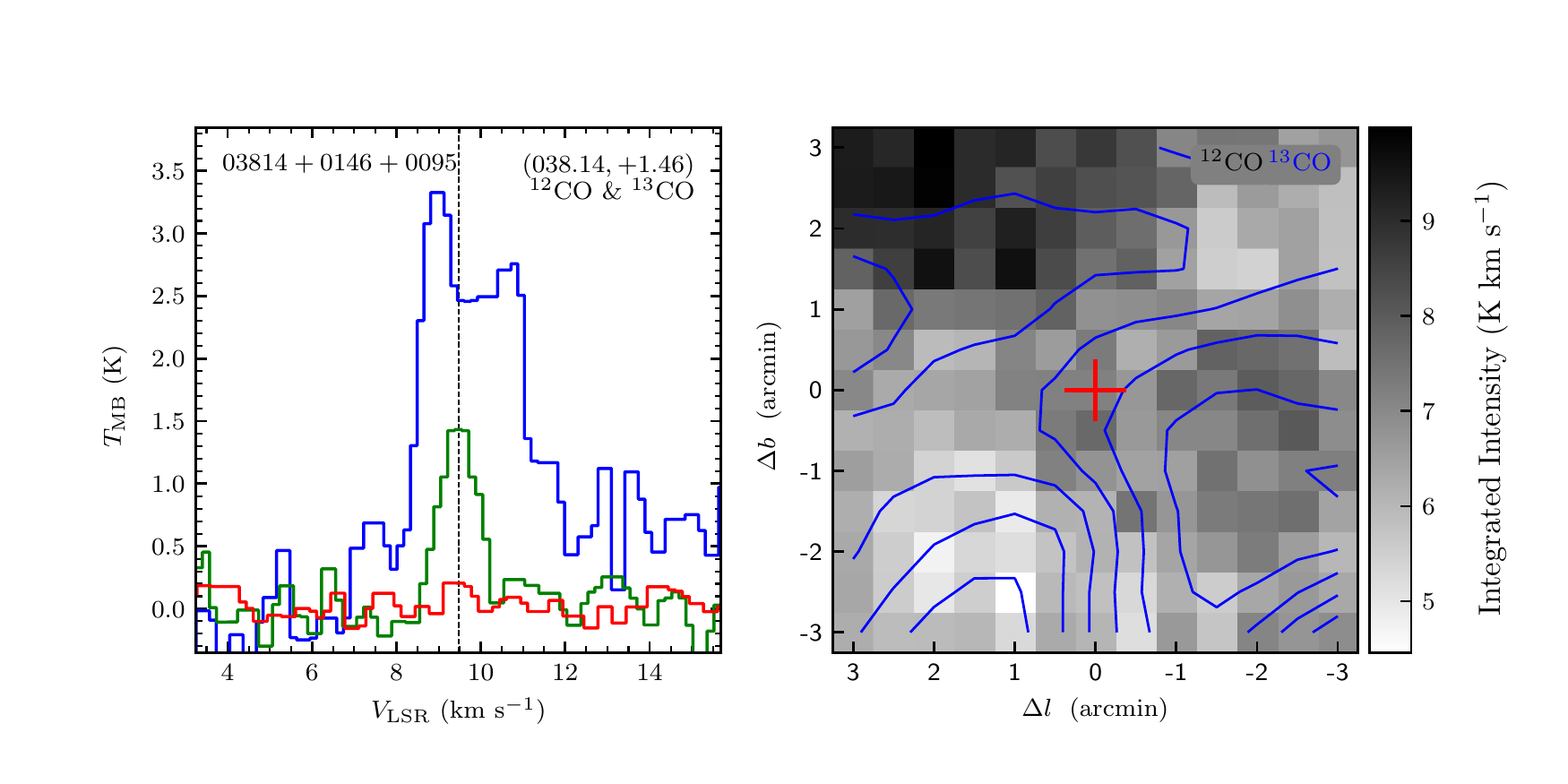}
\vspace{-0.5cm}

\includegraphics[width=9.0cm,angle=0]{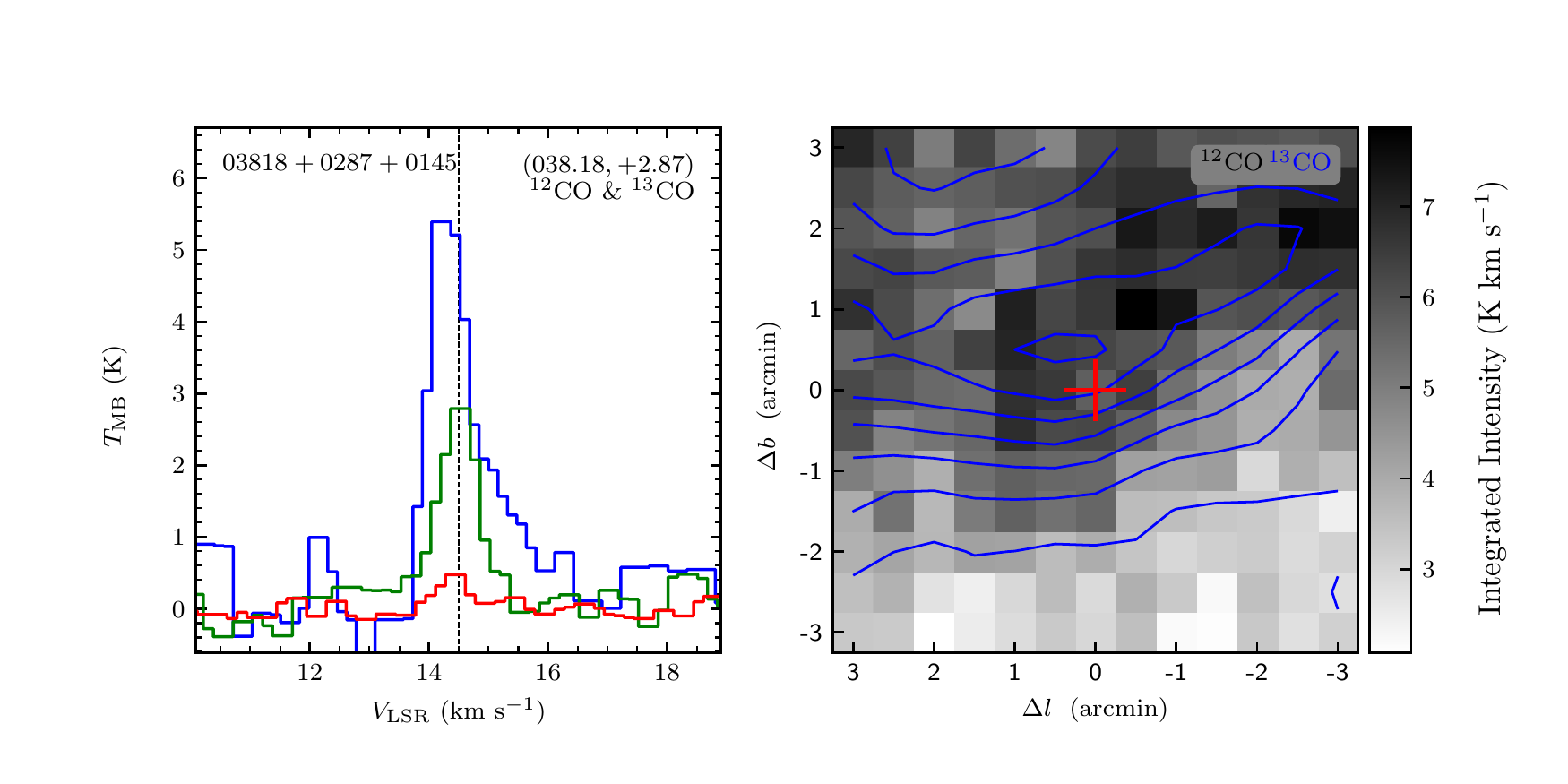}
\includegraphics[width=9.0cm,angle=0]{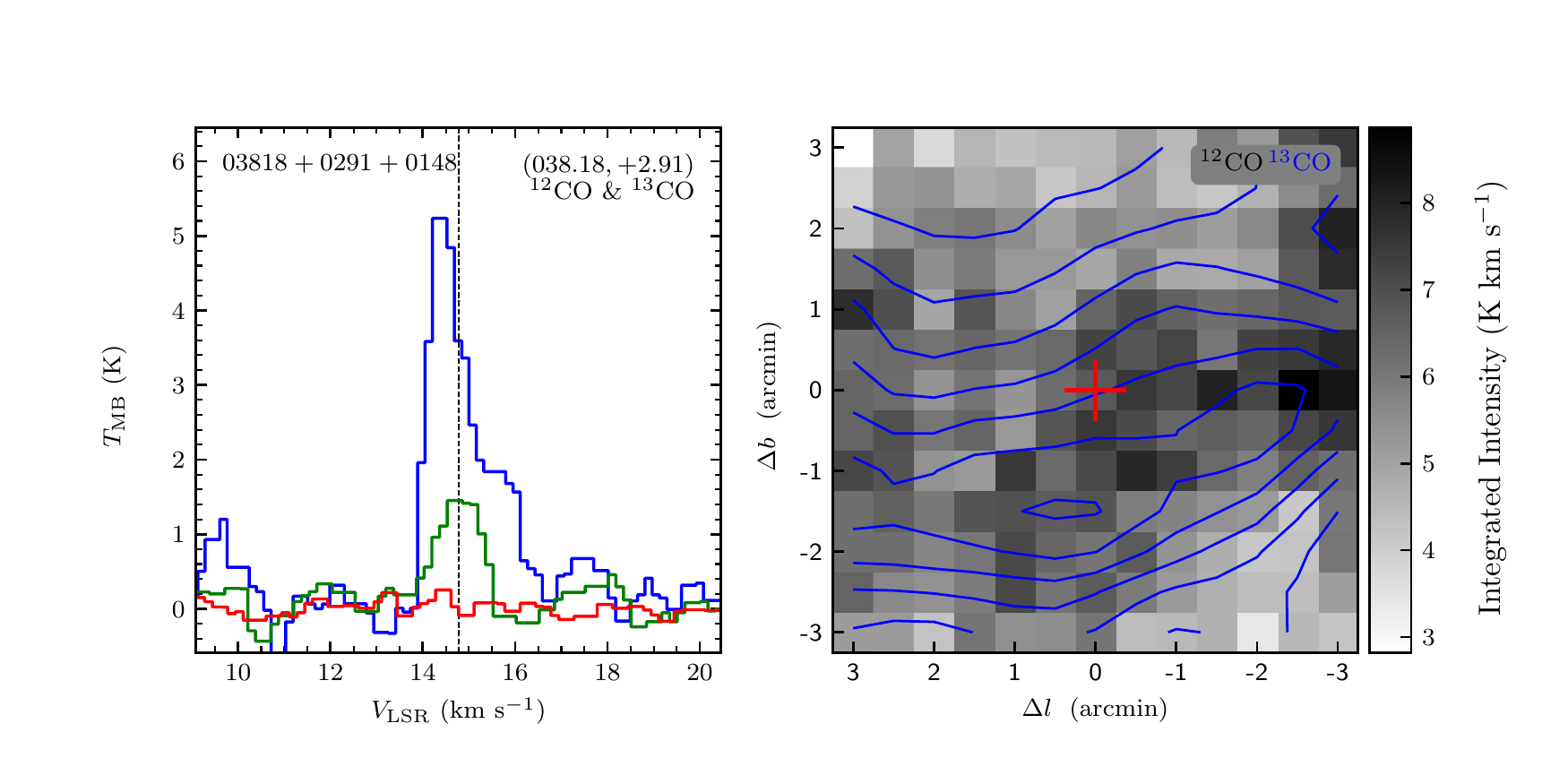}
\end{figure}
\clearpage

\begin{figure}
\includegraphics[width=9.0cm,angle=0]{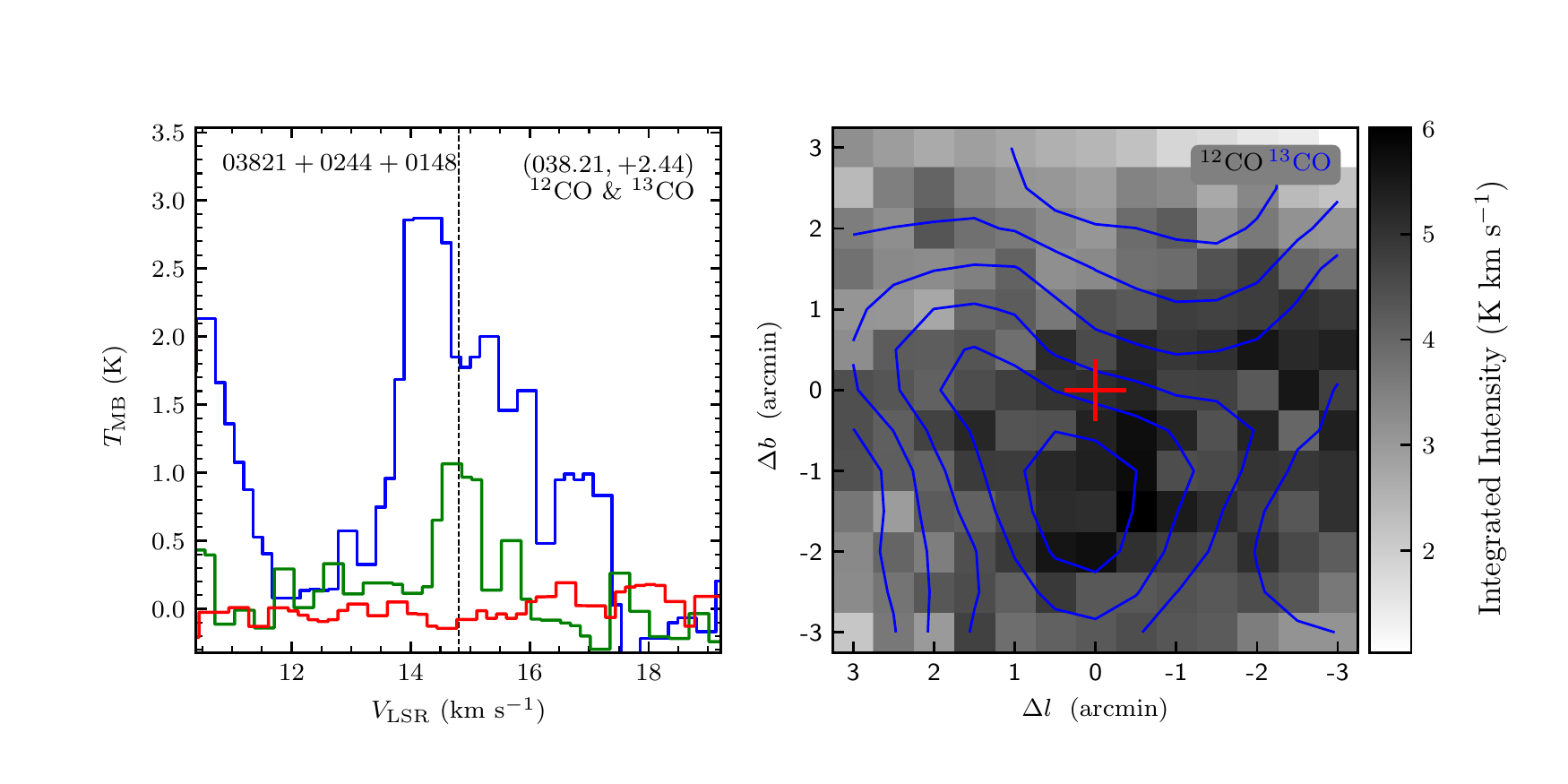}
\includegraphics[width=9.0cm,angle=0]{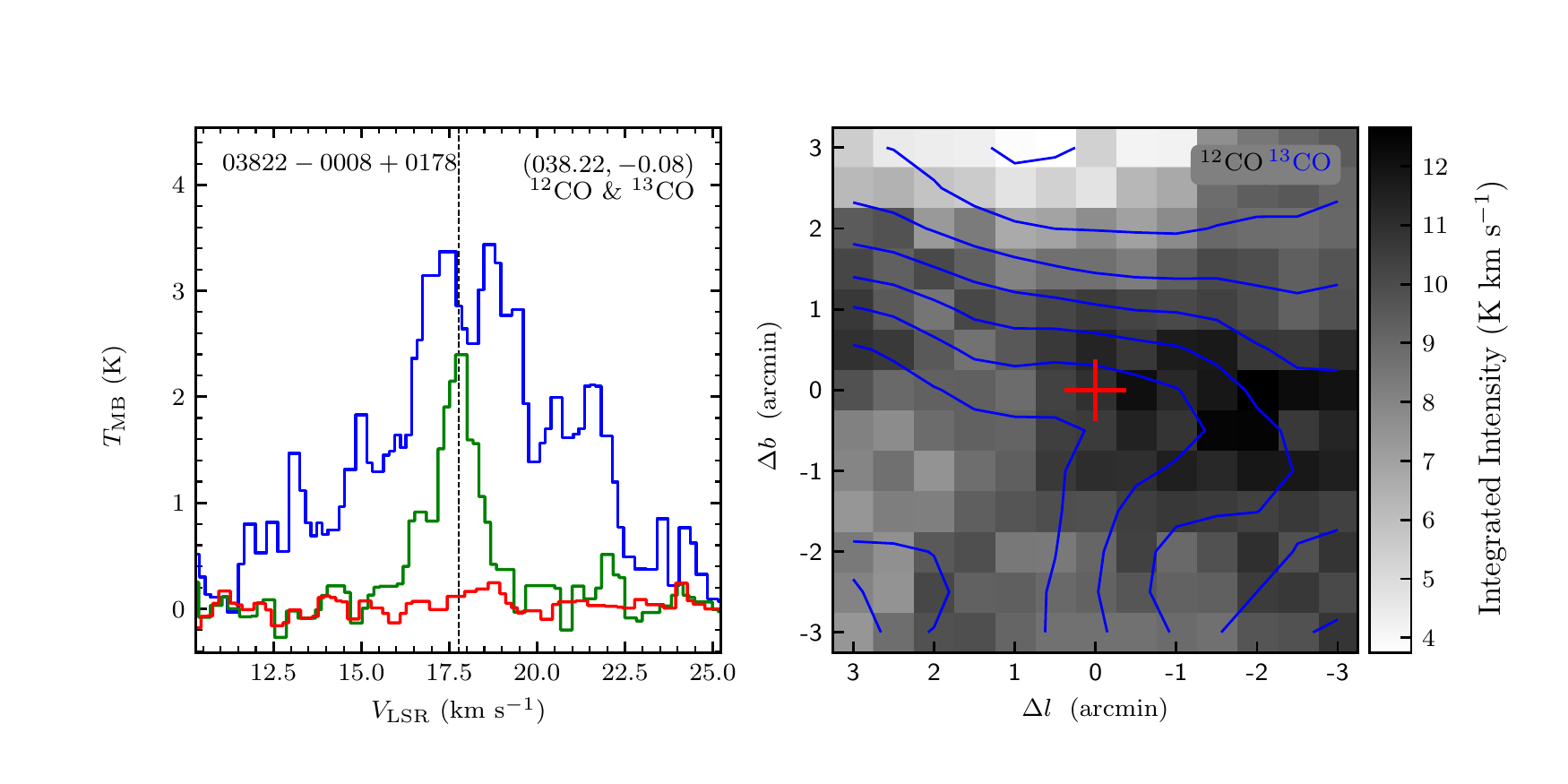}
\vspace{-0.5cm}

\includegraphics[width=9.0cm,angle=0]{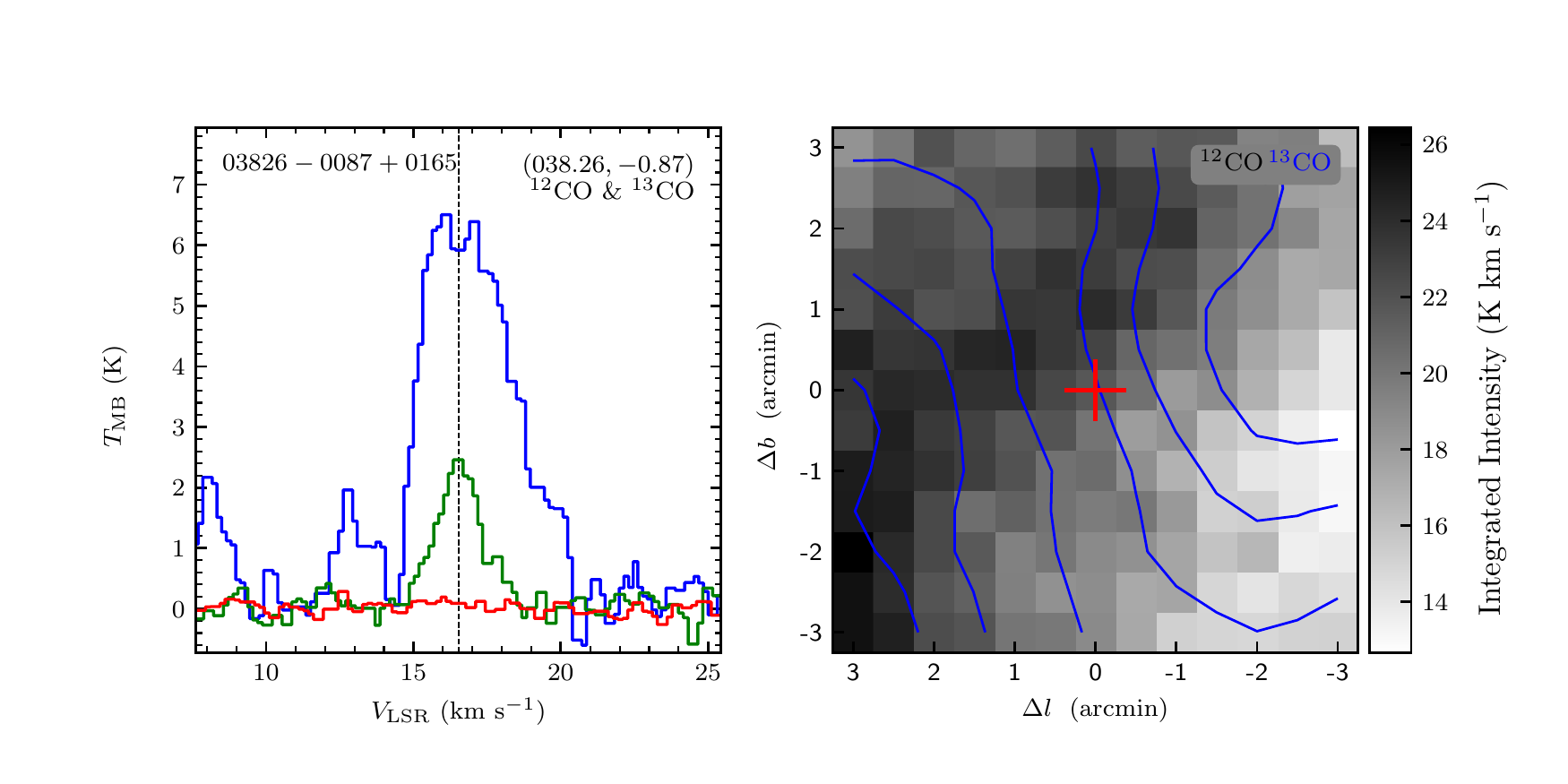}
\includegraphics[width=9.0cm,angle=0]{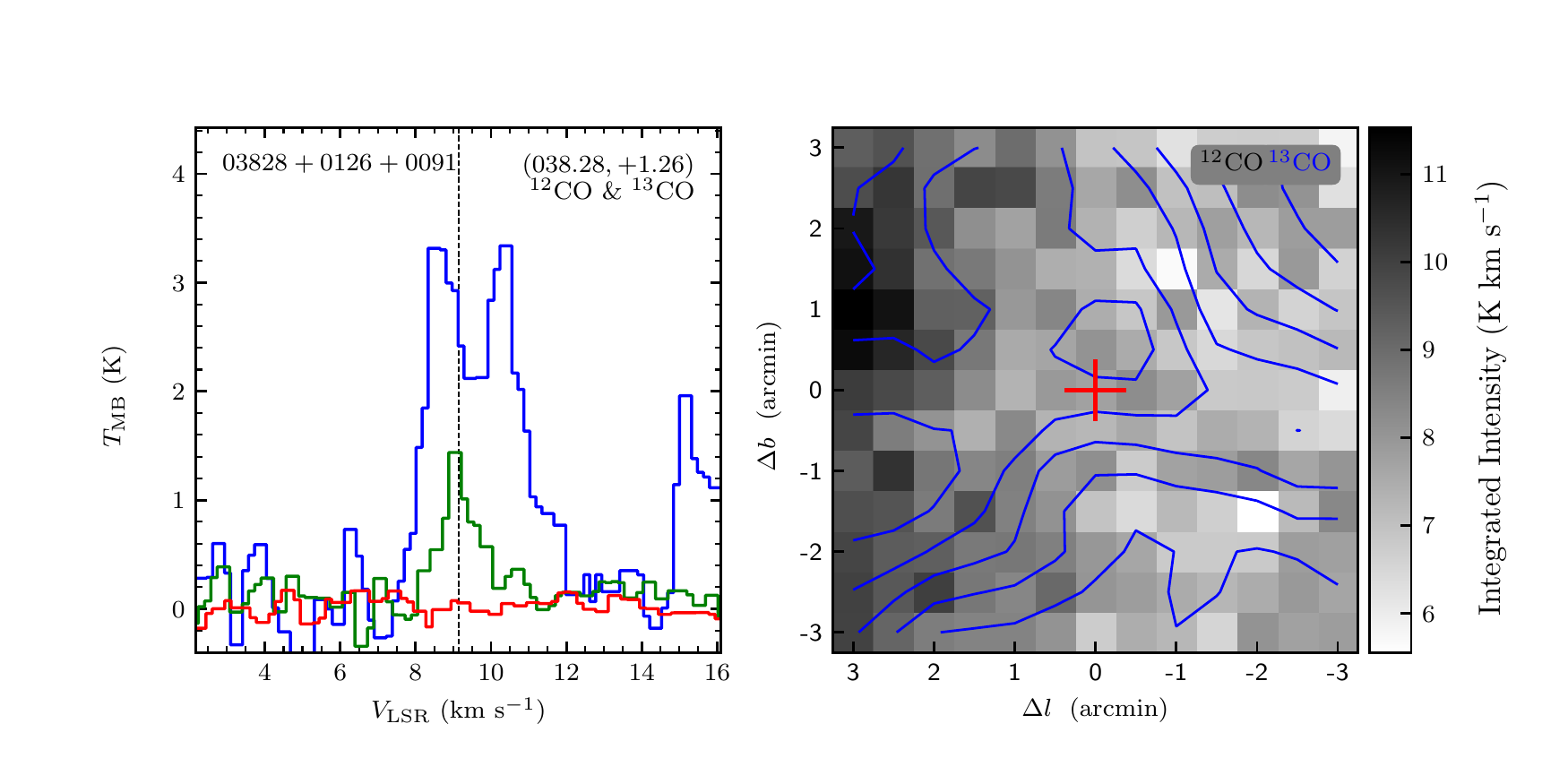}
\vspace{-0.5cm}

\includegraphics[width=9.0cm,angle=0]{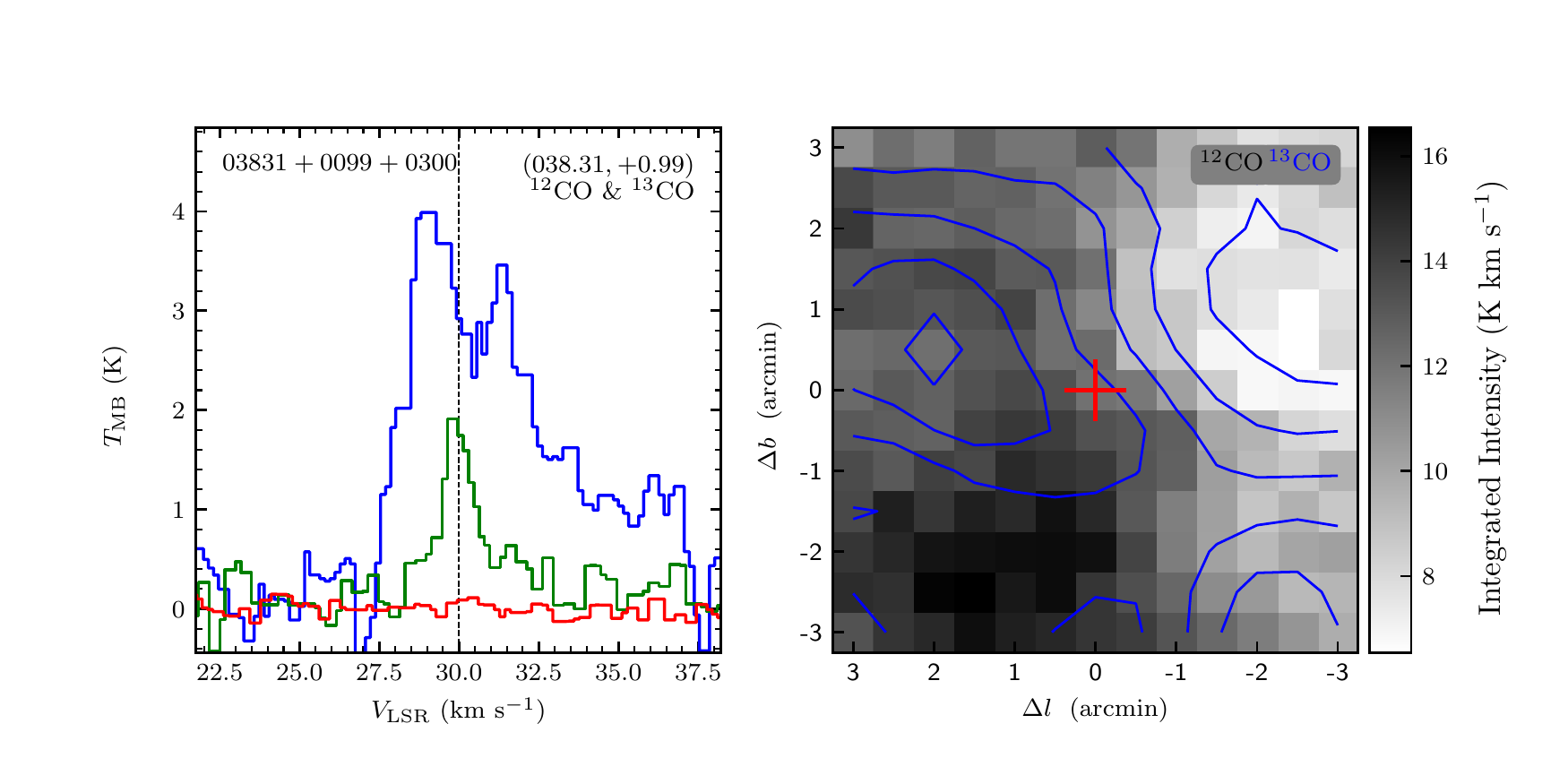}
\includegraphics[width=9.0cm,angle=0]{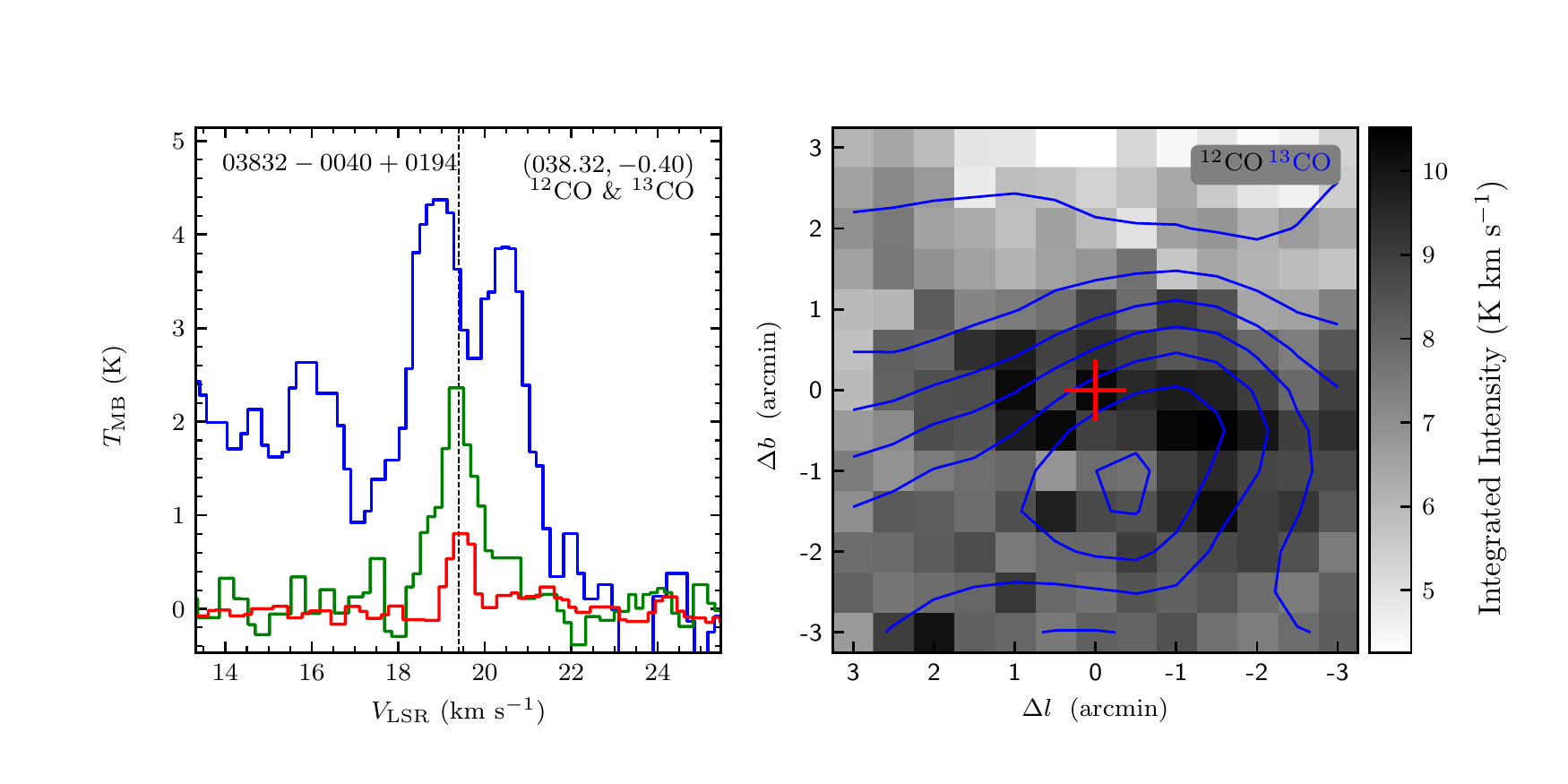}
\vspace{-0.5cm}

\includegraphics[width=9.0cm,angle=0]{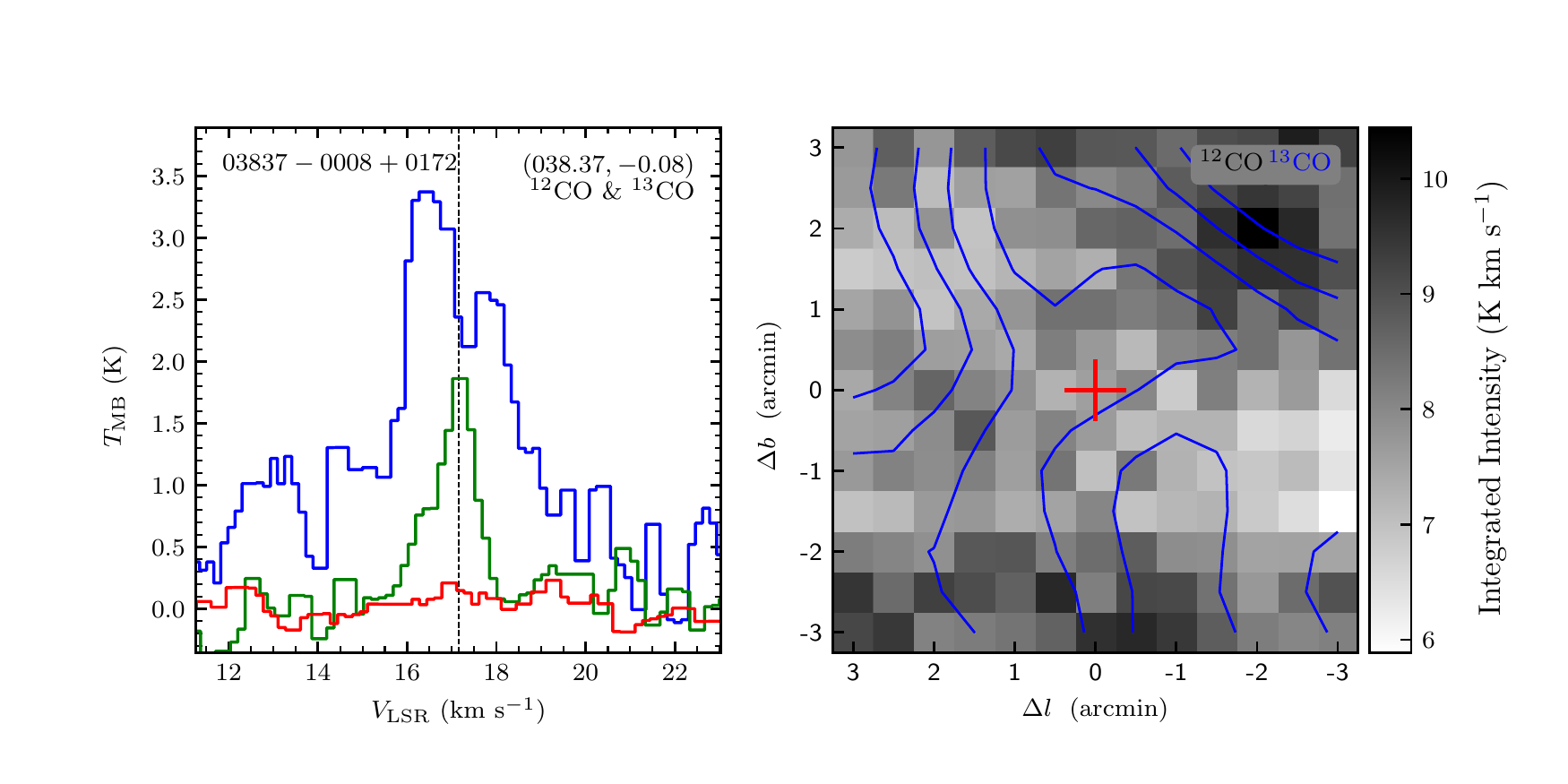}
\includegraphics[width=9.0cm,angle=0]{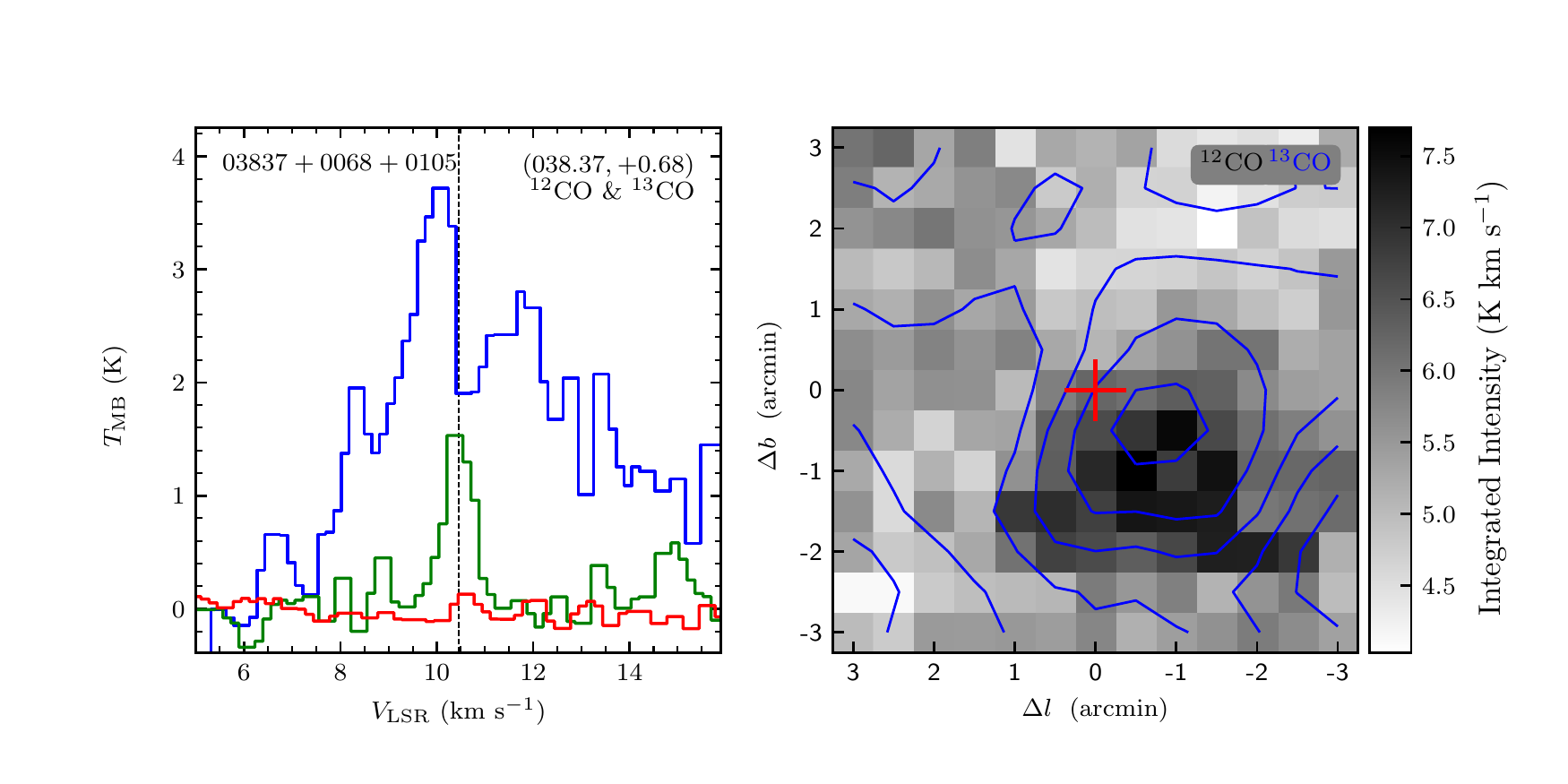}
\vspace{-0.5cm}

\includegraphics[width=9.0cm,angle=0]{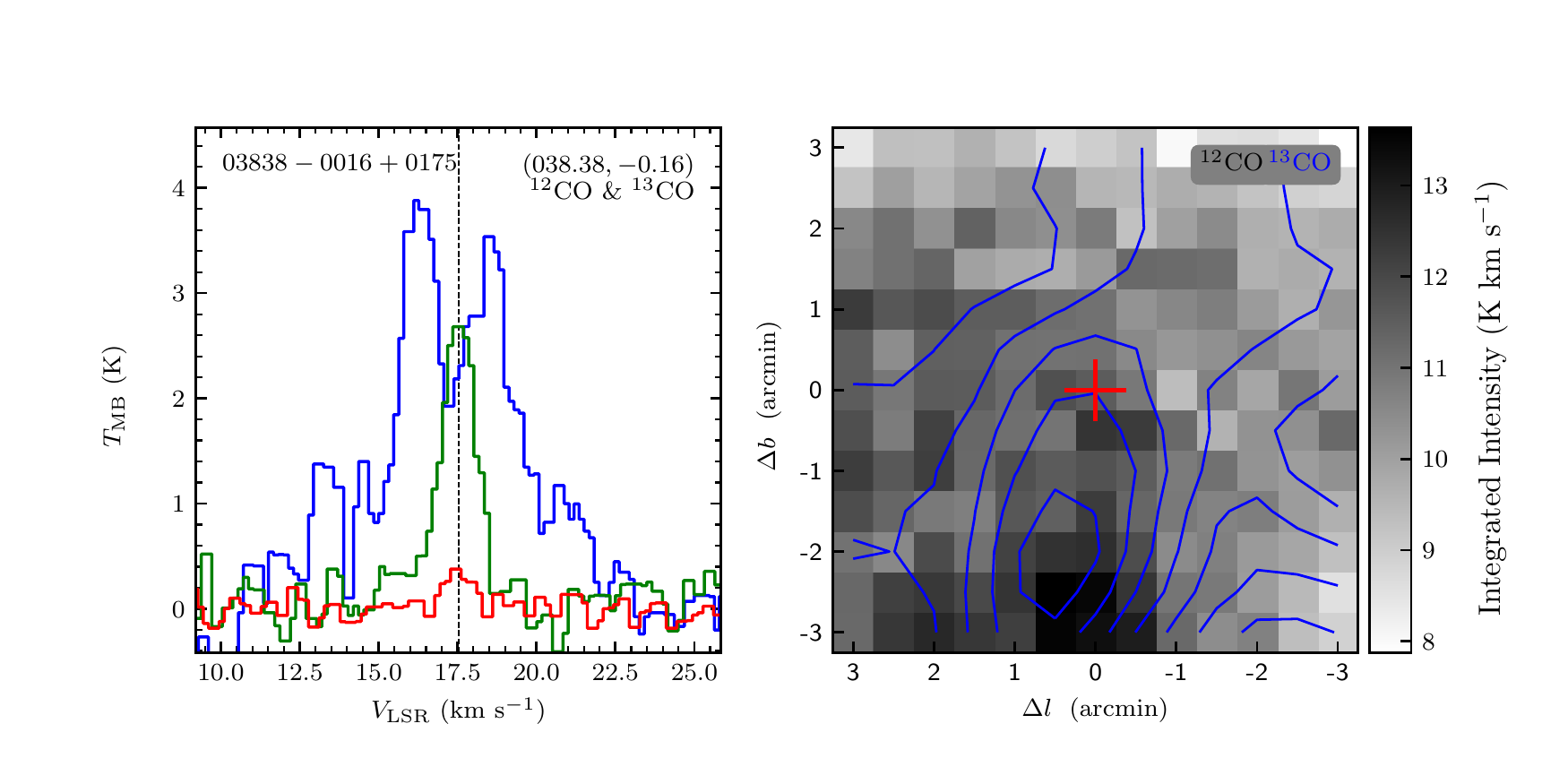}
\includegraphics[width=9.0cm,angle=0]{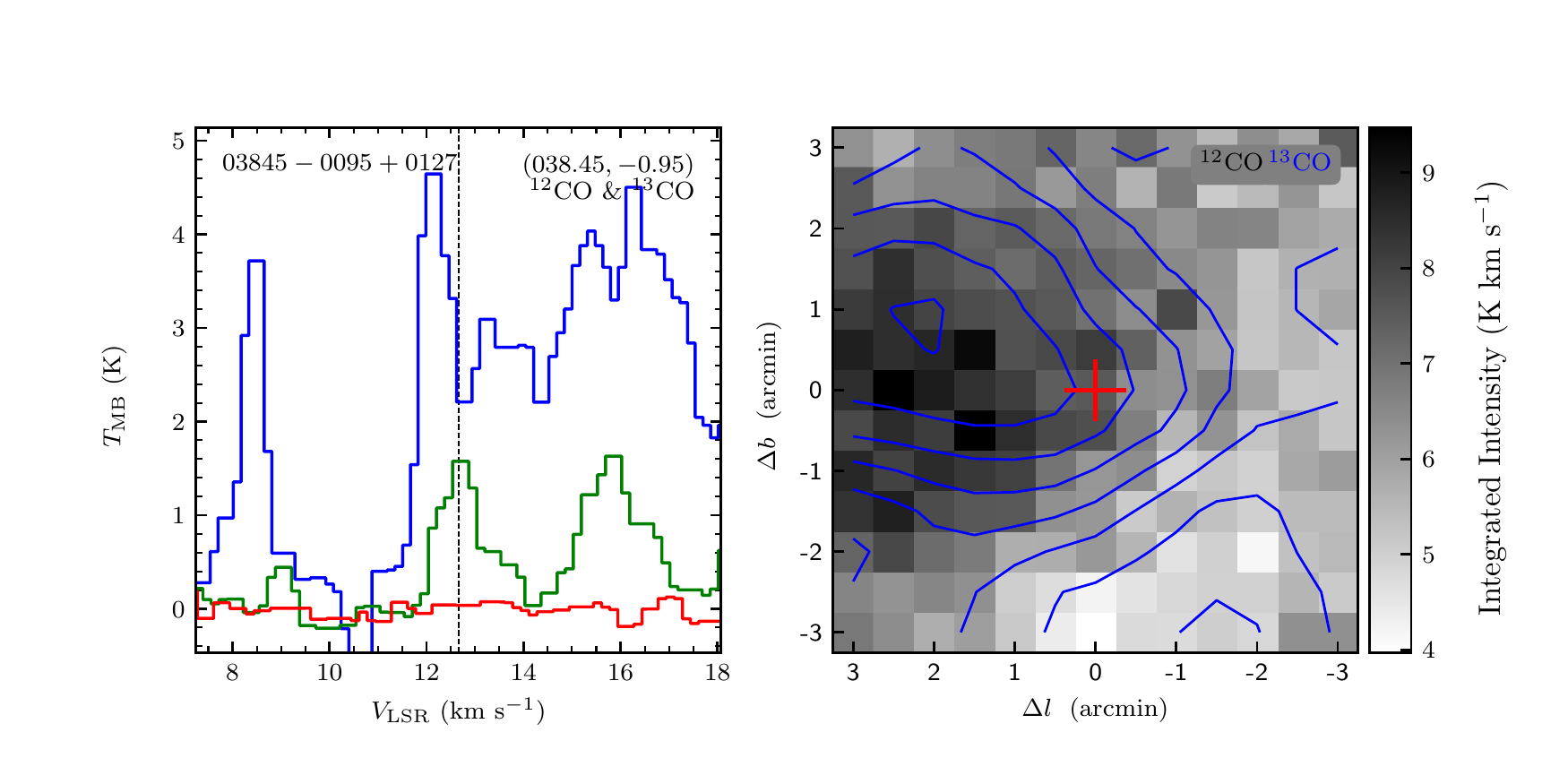}
\end{figure}
\clearpage

\begin{figure}
\includegraphics[width=9.0cm,angle=0]{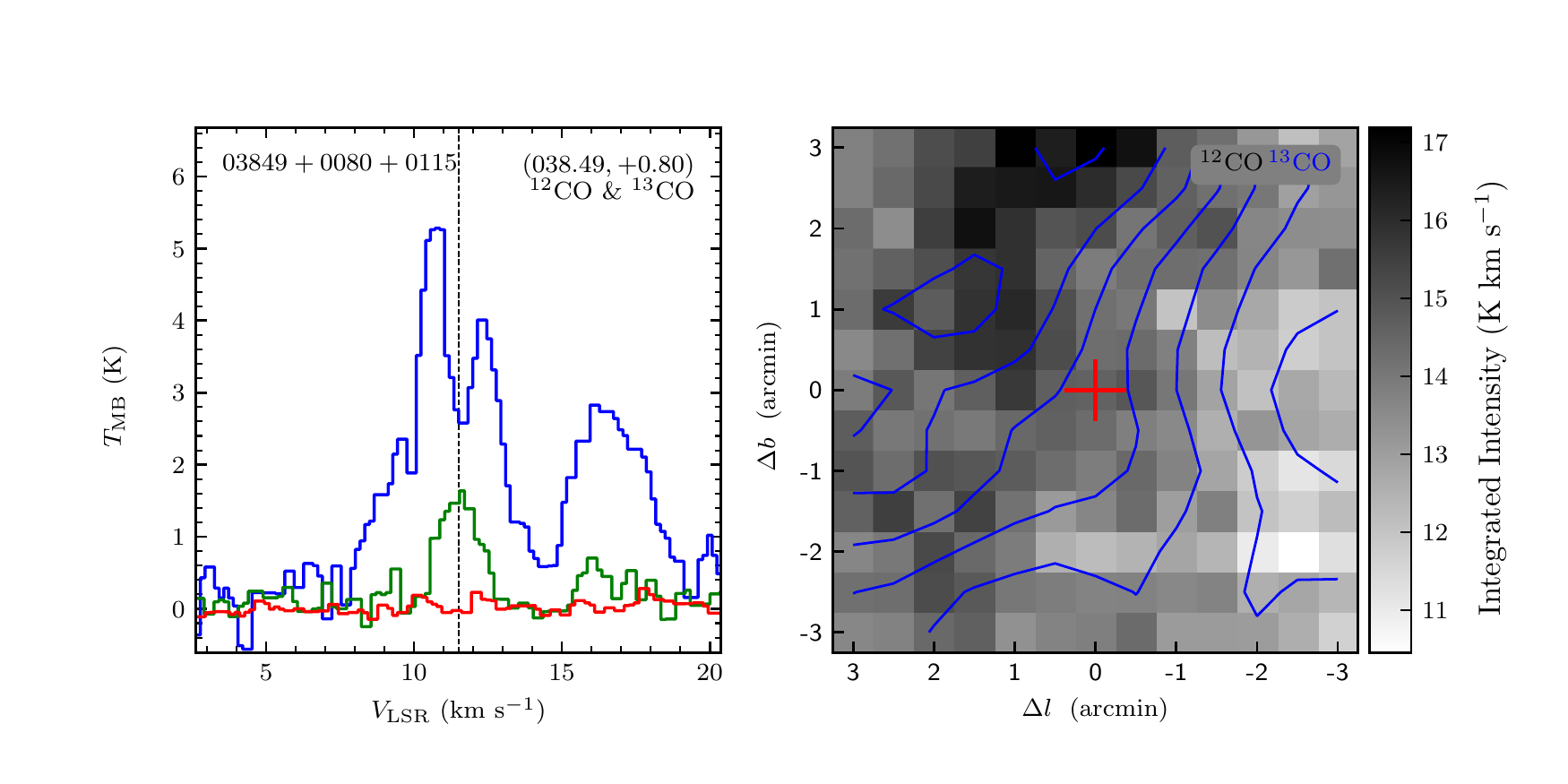}
\includegraphics[width=9.0cm,angle=0]{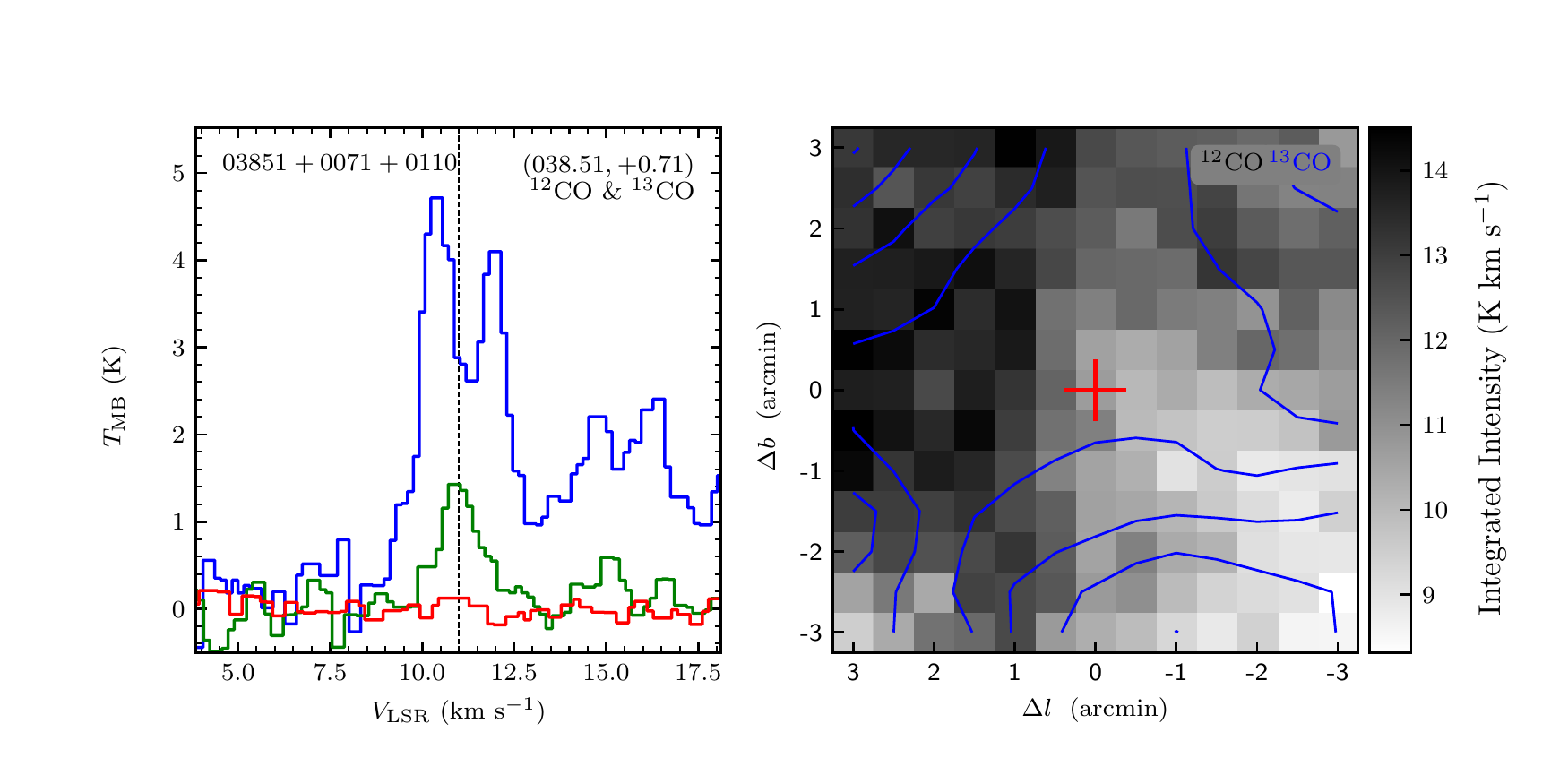}
\vspace{-0.5cm}

\includegraphics[width=9.0cm,angle=0]{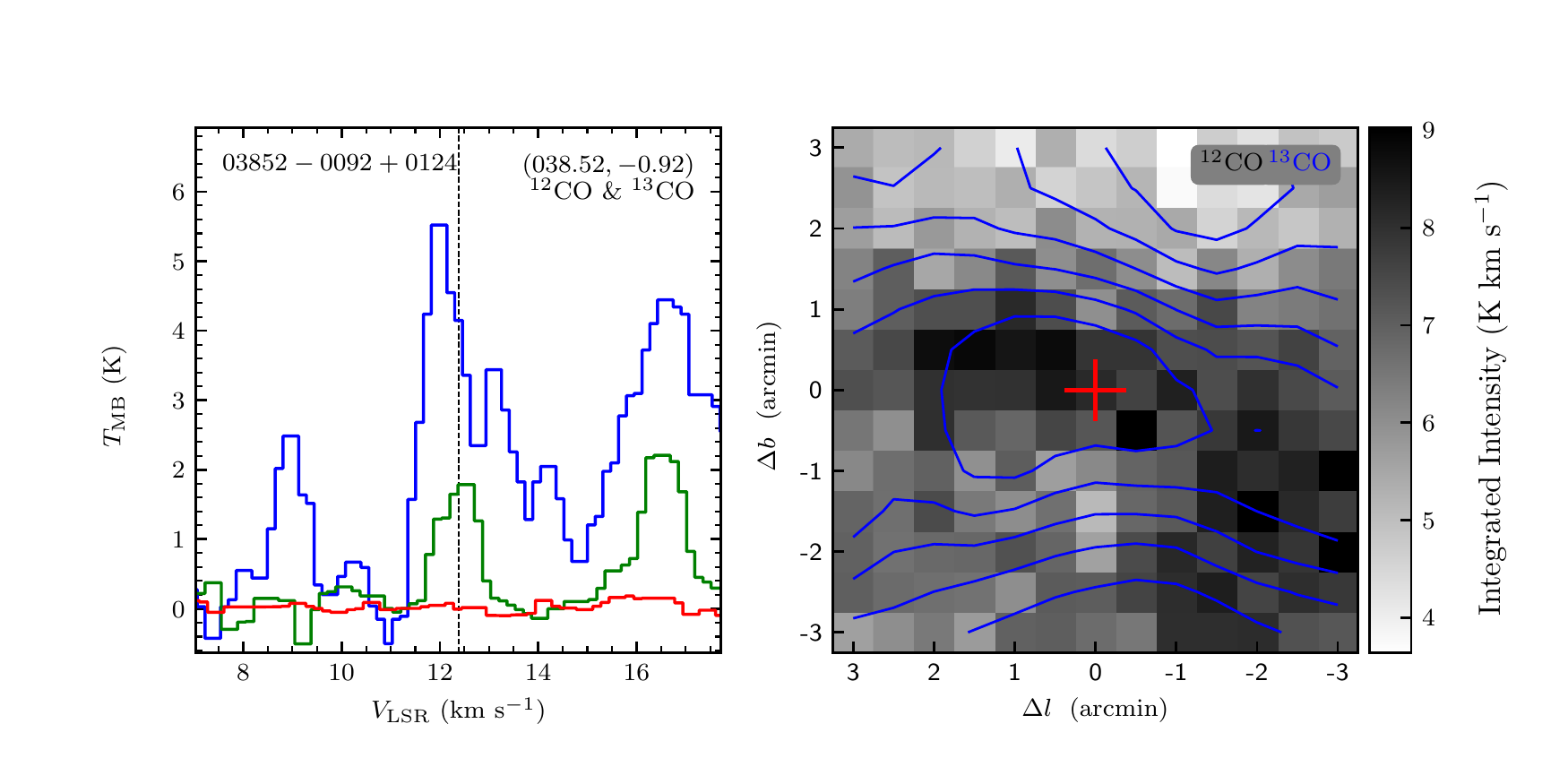}
\includegraphics[width=9.0cm,angle=0]{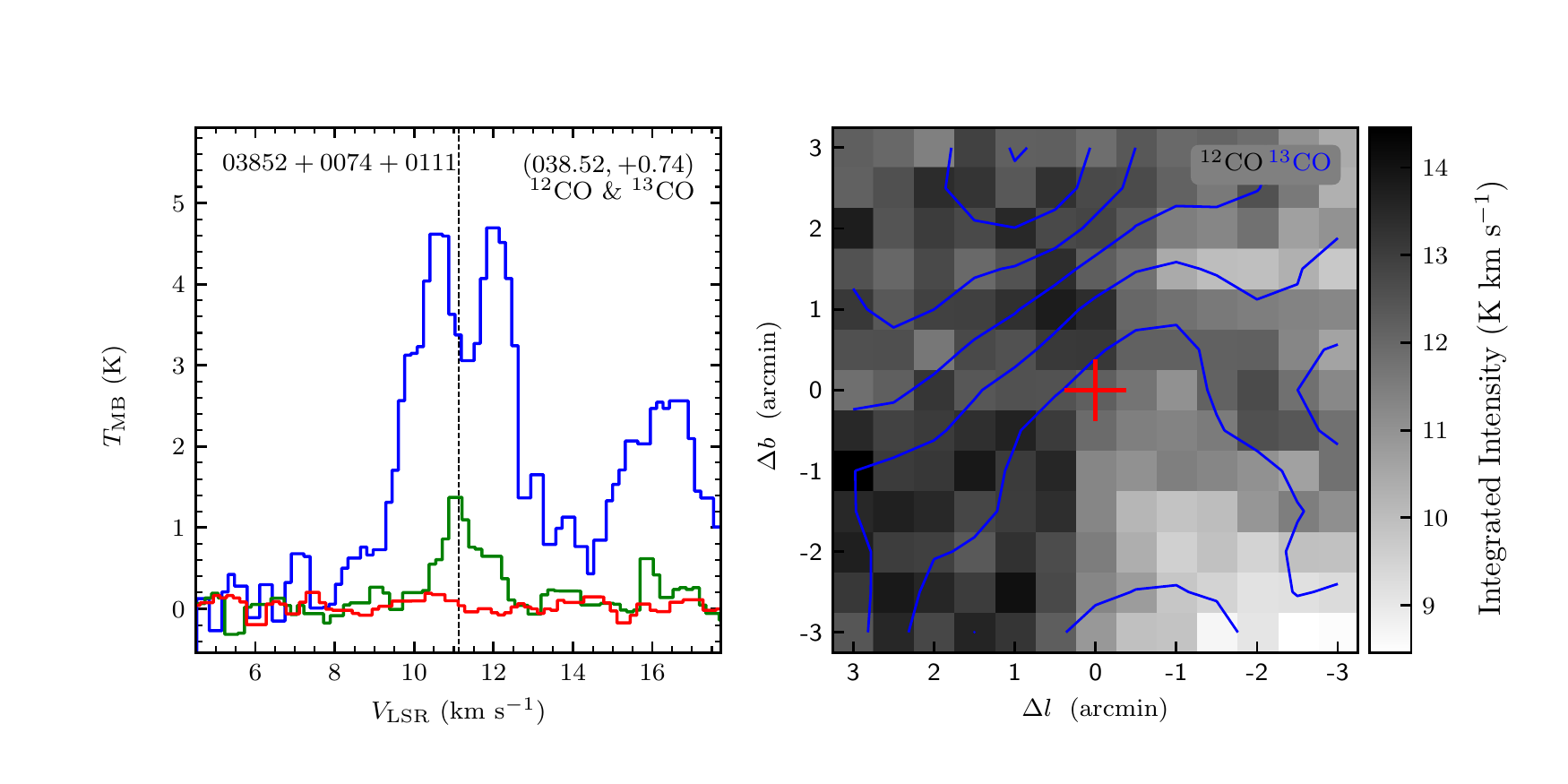}
\vspace{-0.5cm}

\includegraphics[width=9.0cm,angle=0]{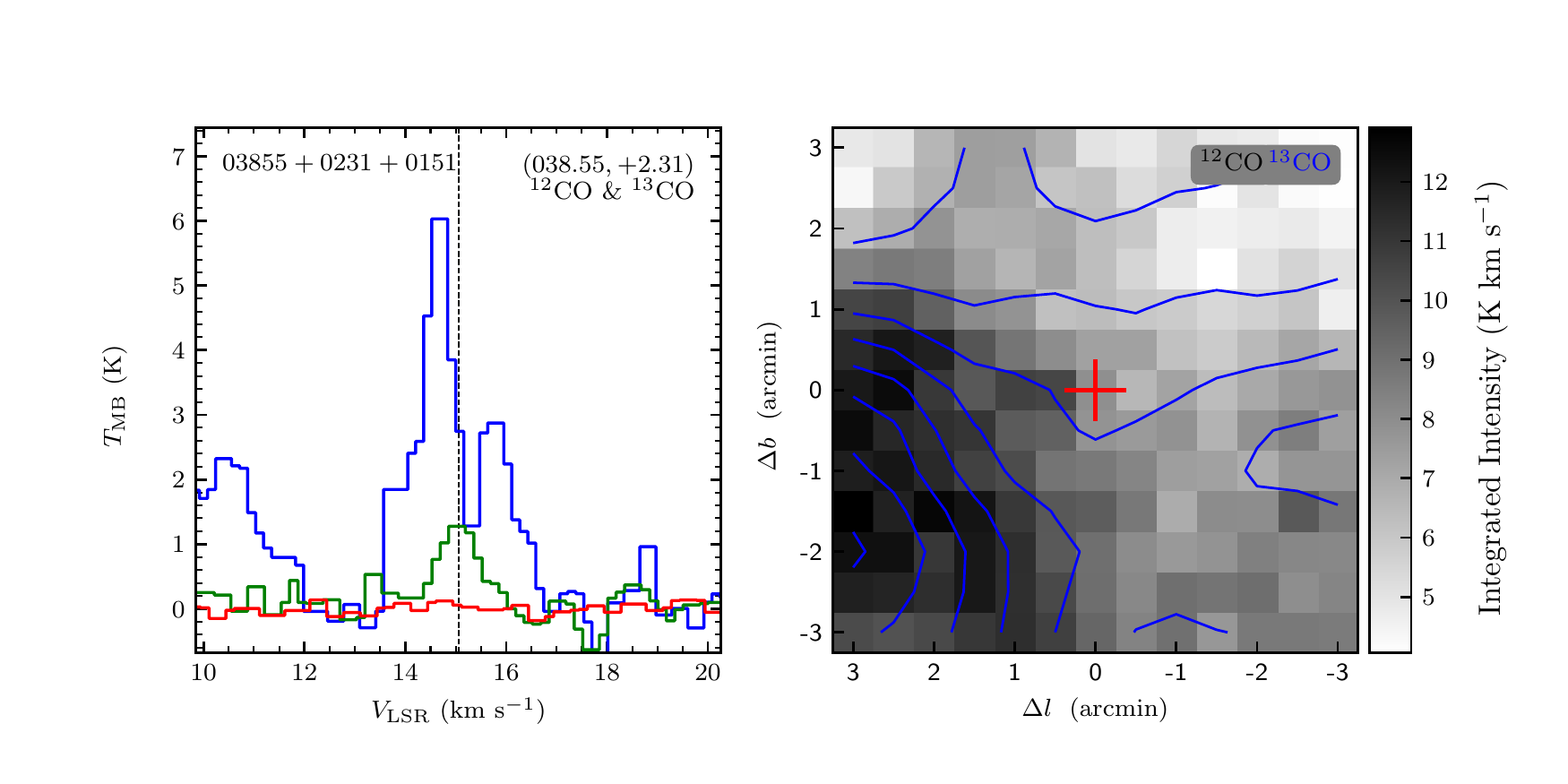}
\includegraphics[width=9.0cm,angle=0]{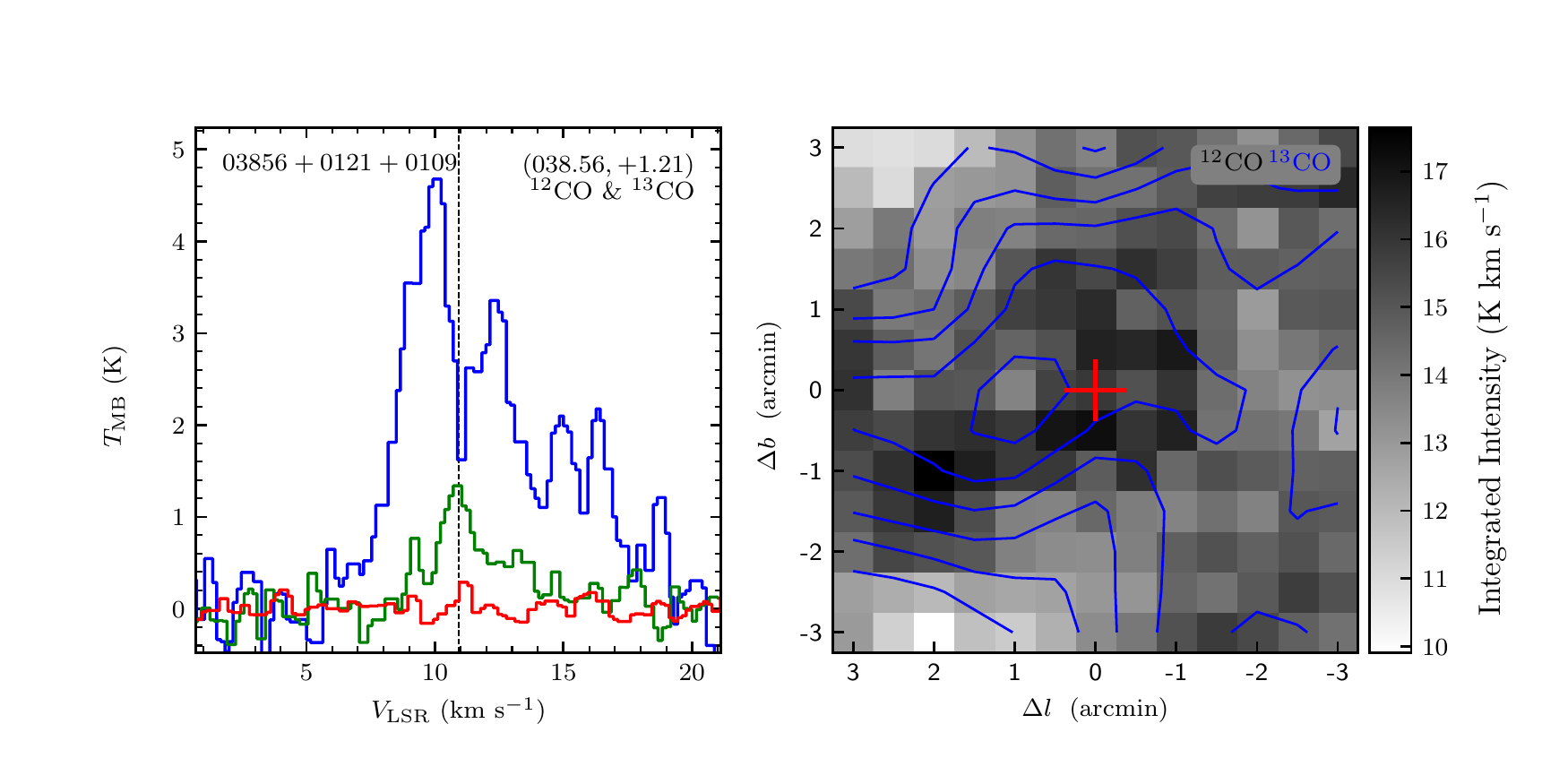}
\vspace{-0.5cm}

\includegraphics[width=9.0cm,angle=0]{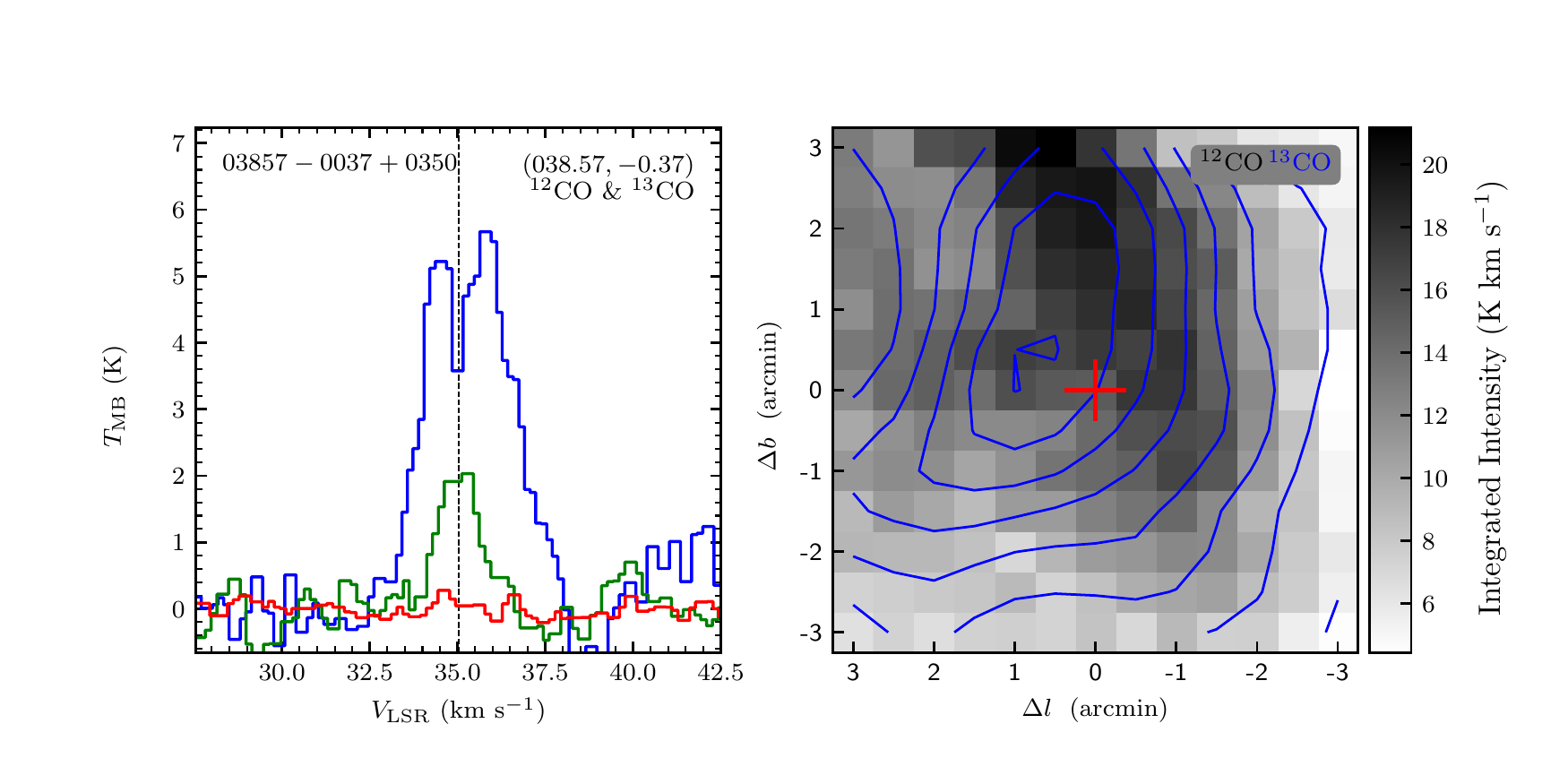}
\includegraphics[width=9.0cm,angle=0]{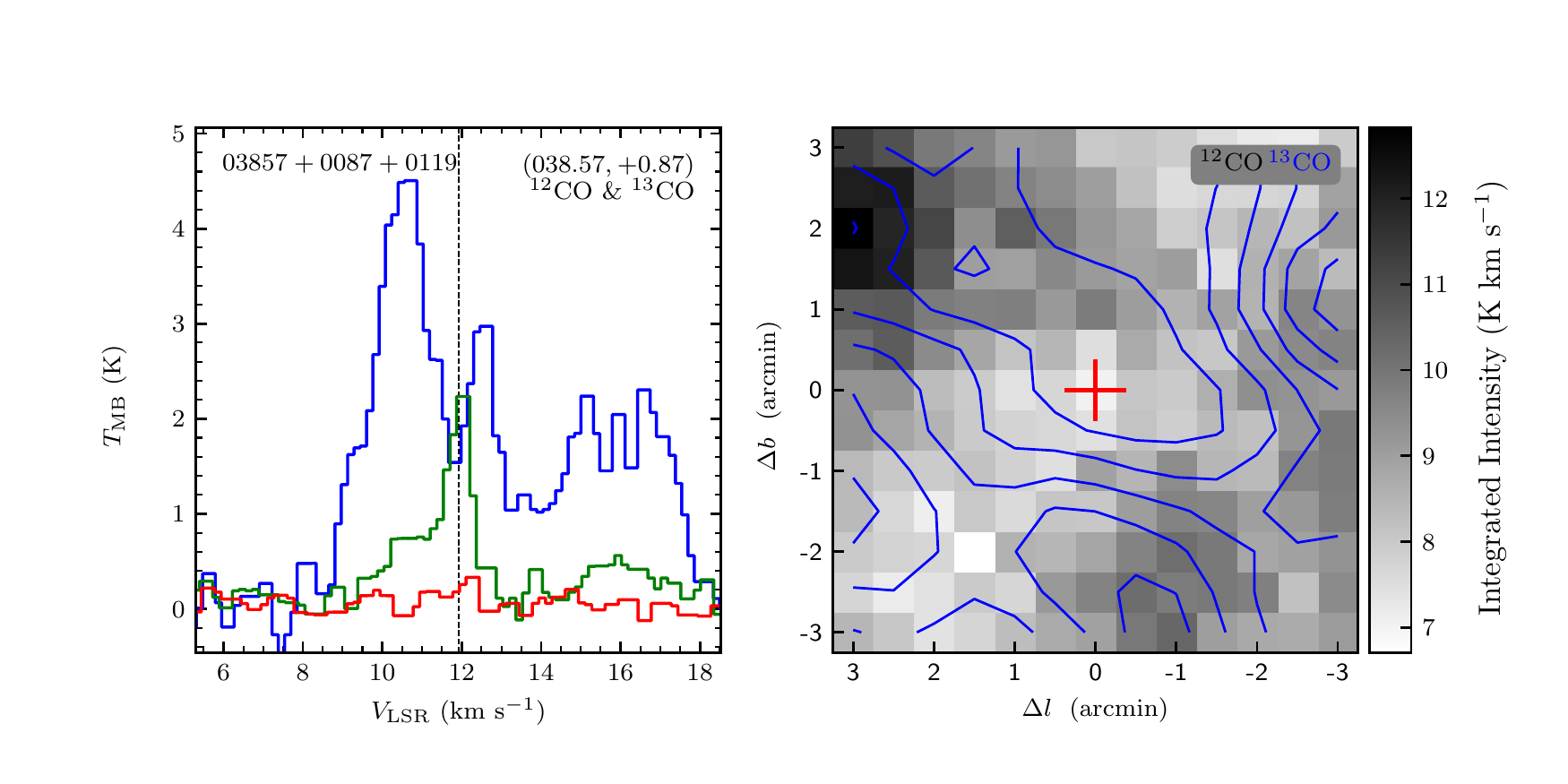}
\vspace{-0.5cm}

\includegraphics[width=9.0cm,angle=0]{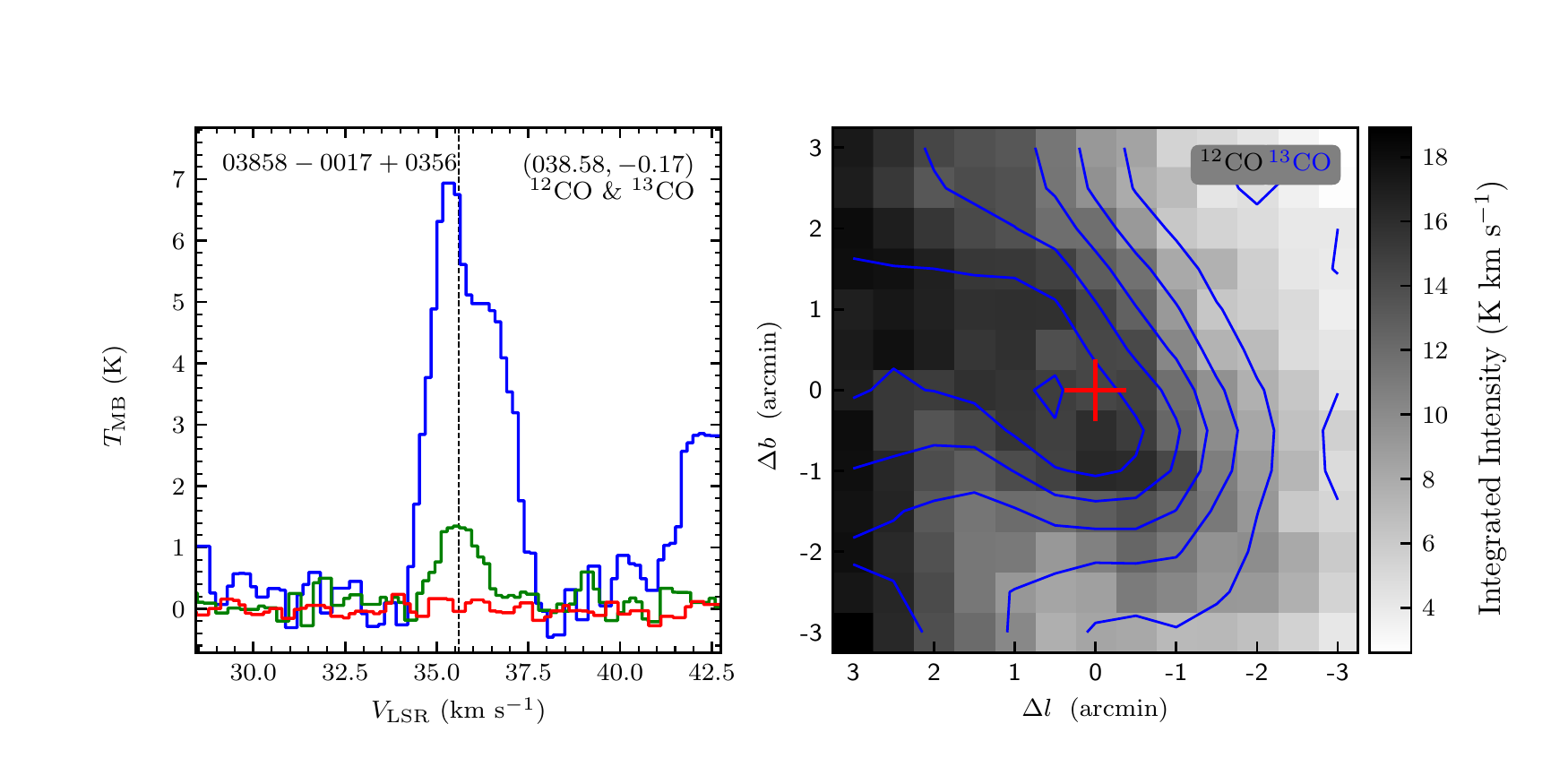}
\includegraphics[width=9.0cm,angle=0]{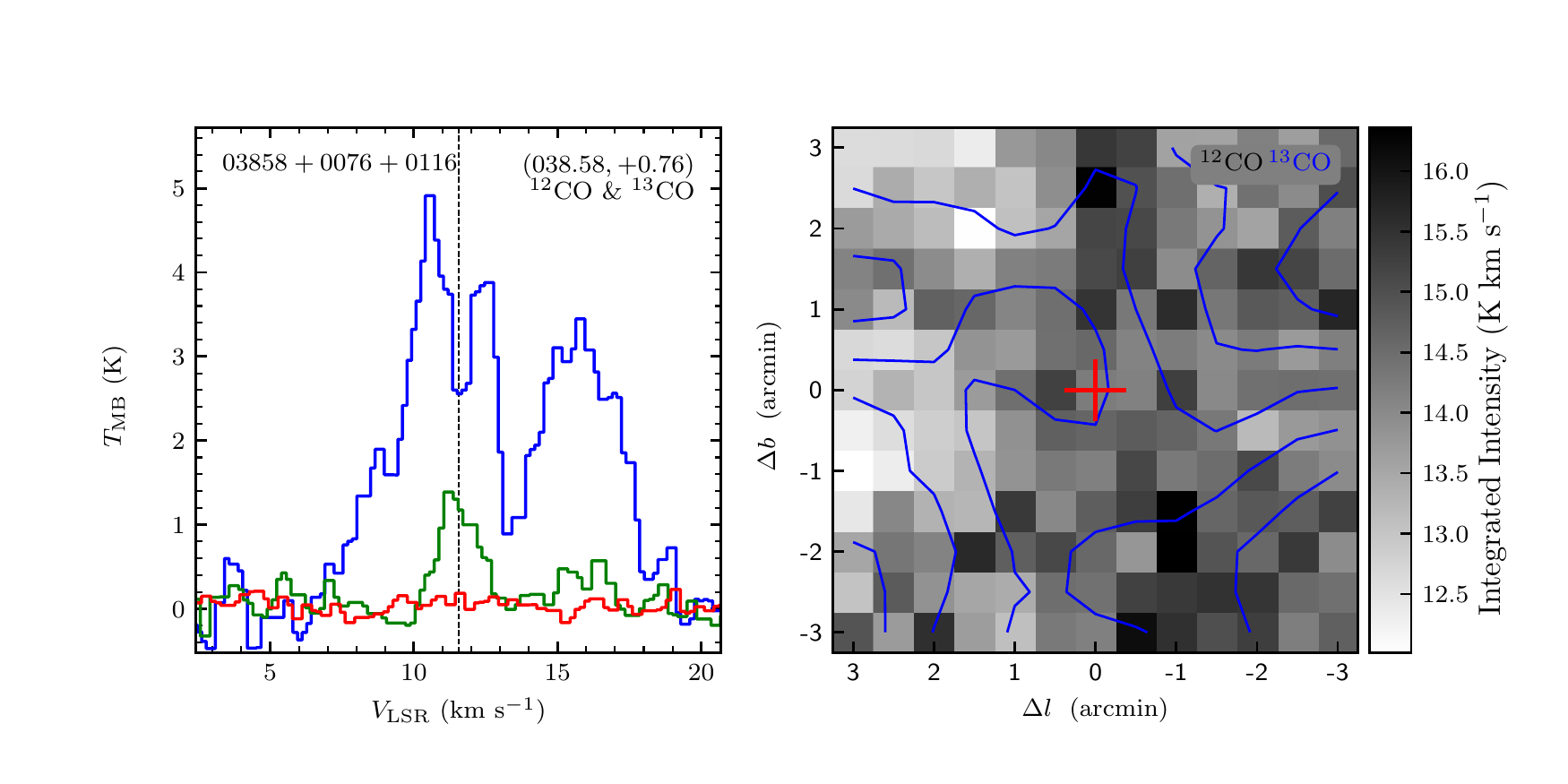}
\end{figure}
\clearpage

\begin{figure}
\includegraphics[width=9.0cm,angle=0]{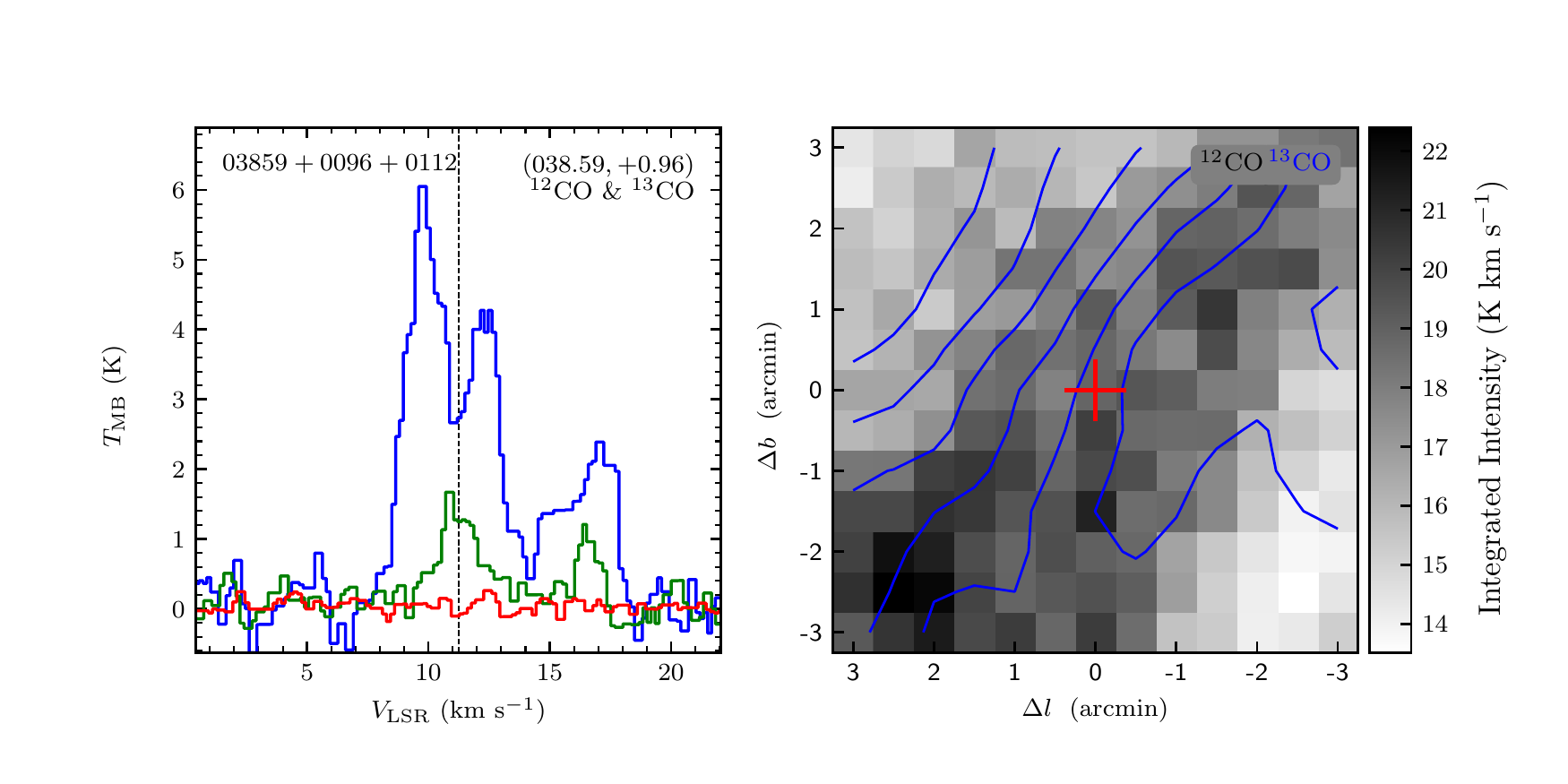}
\includegraphics[width=9.0cm,angle=0]{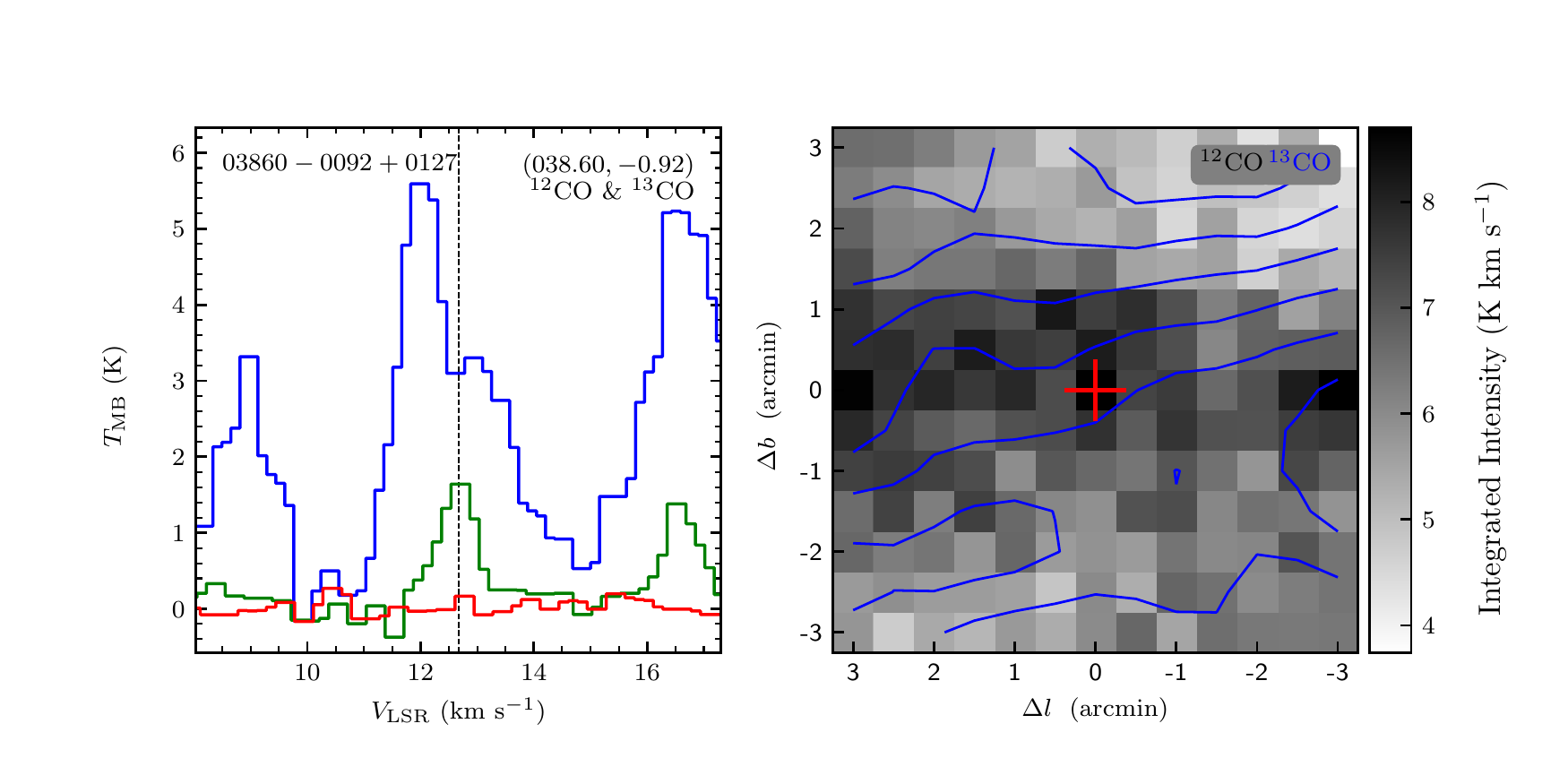}
\vspace{-0.5cm}

\includegraphics[width=9.0cm,angle=0]{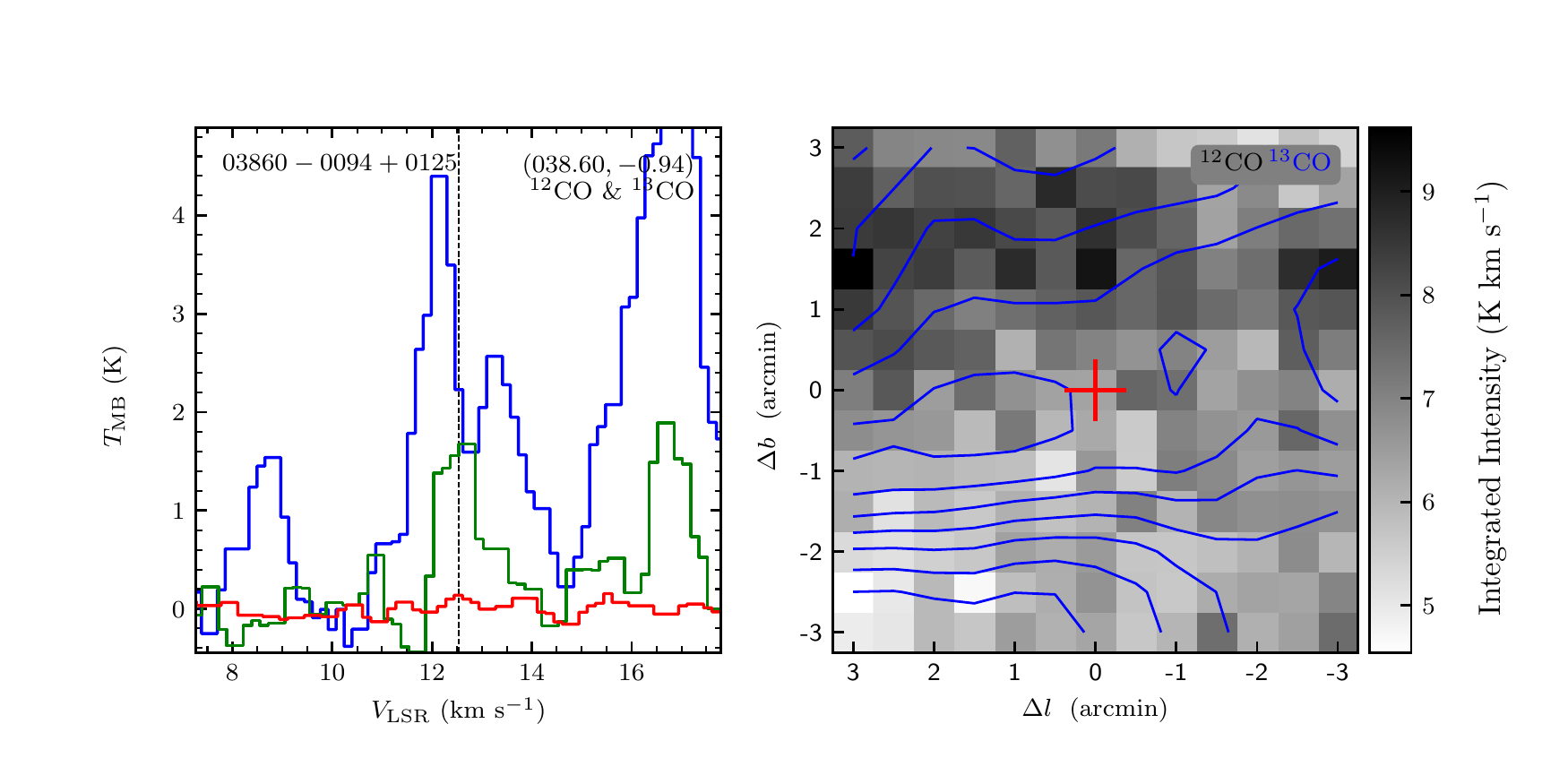}
\includegraphics[width=9.0cm,angle=0]{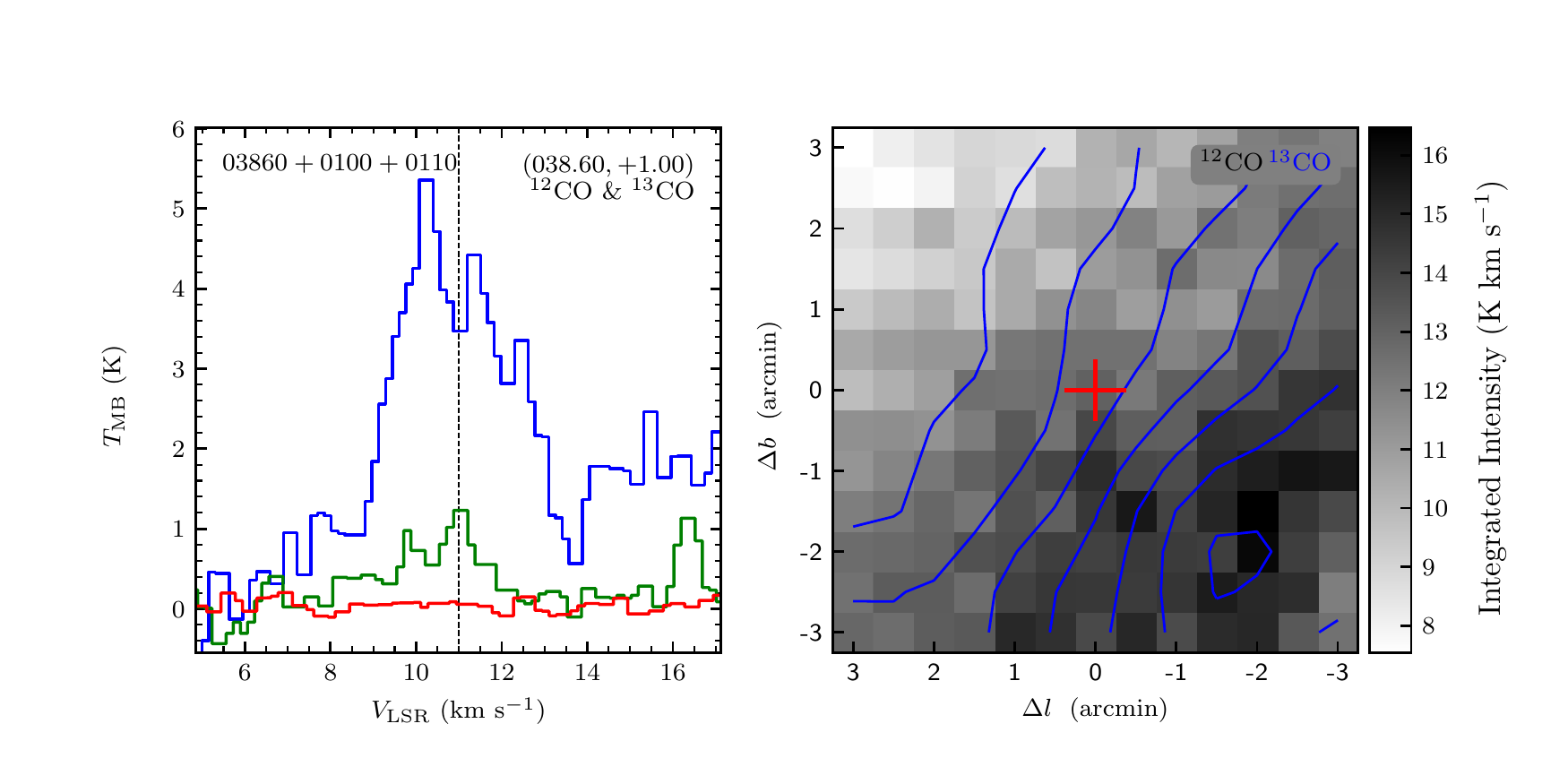}
\vspace{-0.5cm}

\includegraphics[width=9.0cm,angle=0]{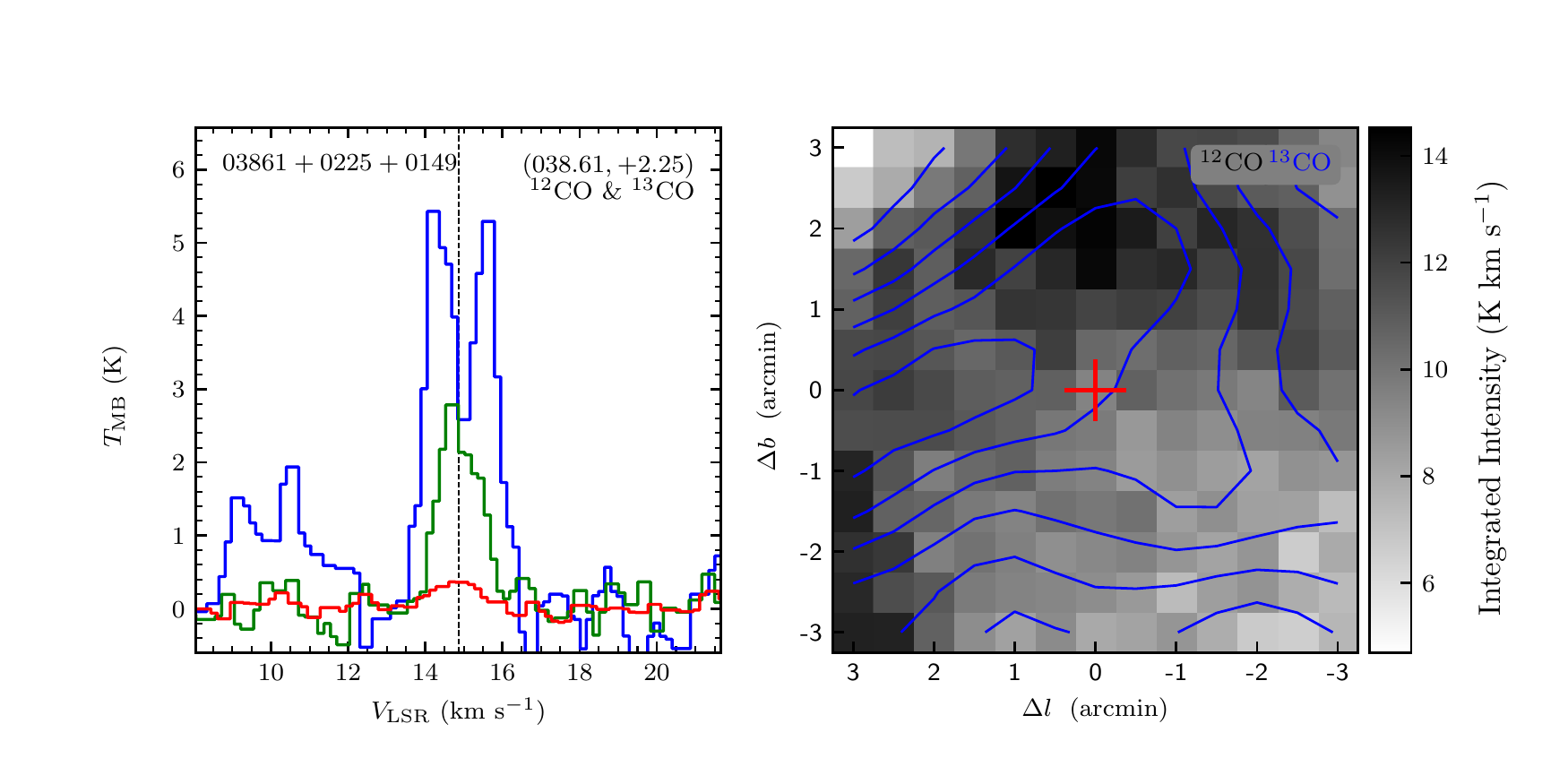}
\includegraphics[width=9.0cm,angle=0]{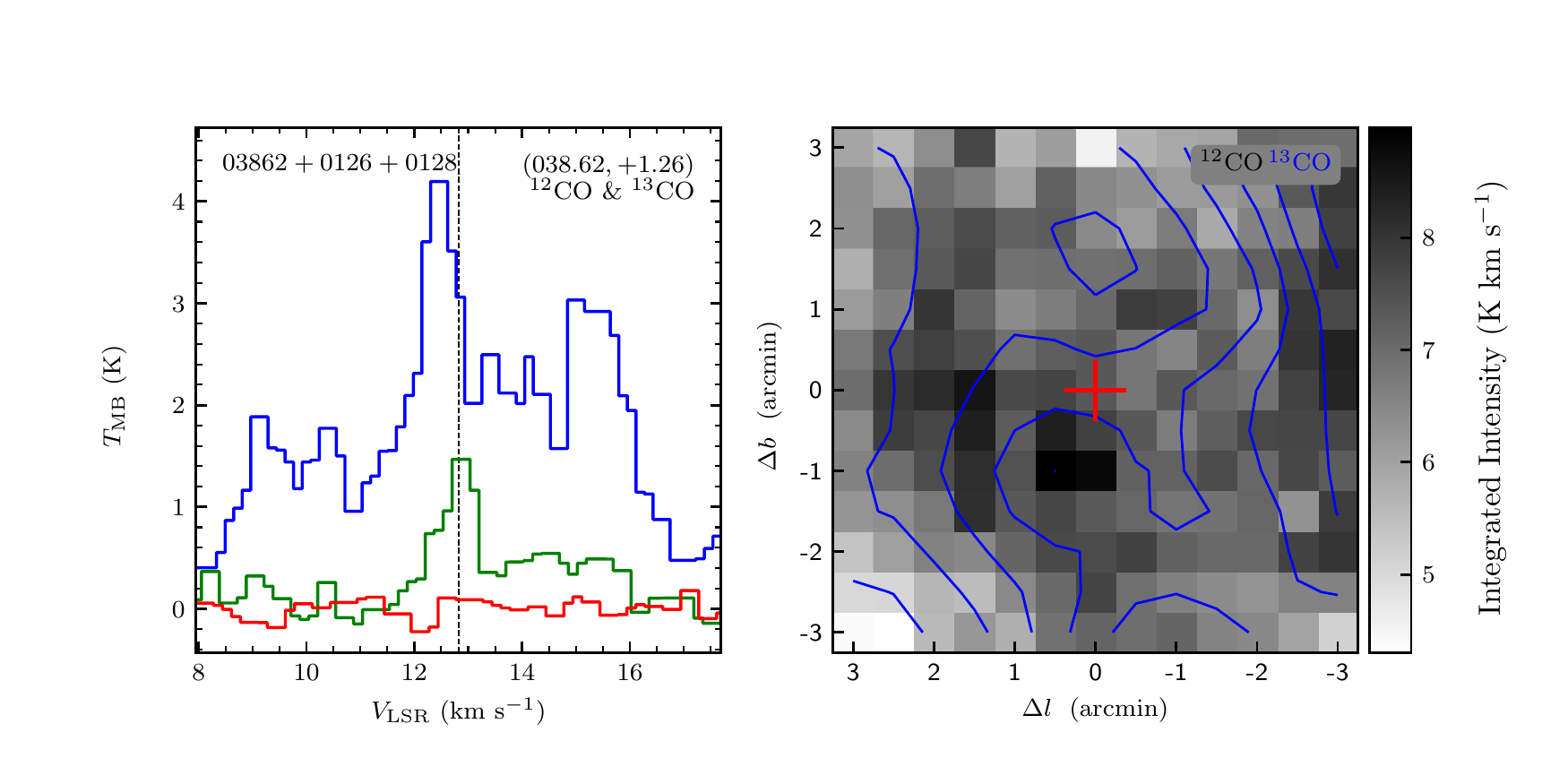}
\vspace{-0.5cm}

\includegraphics[width=9.0cm,angle=0]{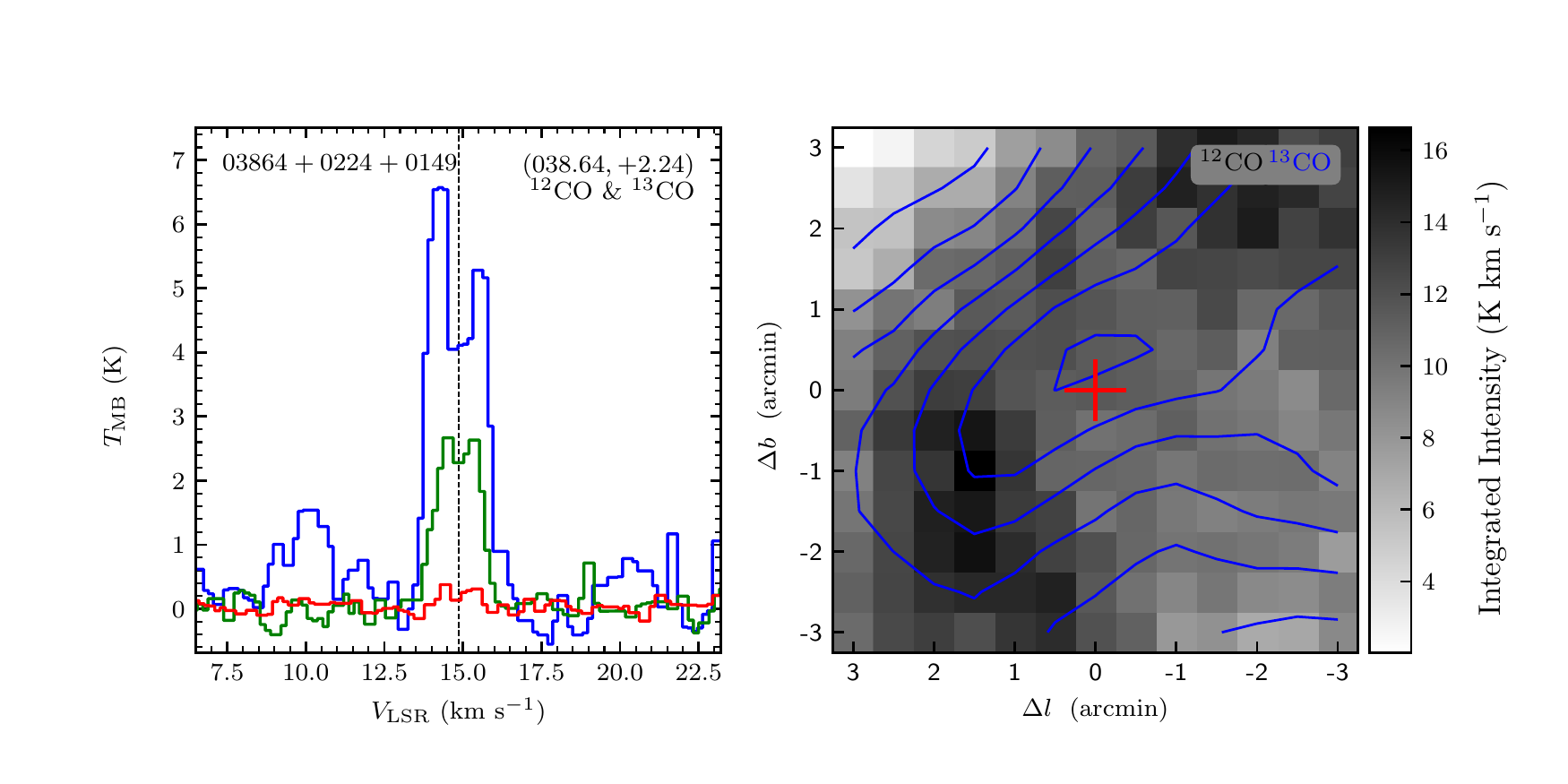}
\includegraphics[width=9.0cm,angle=0]{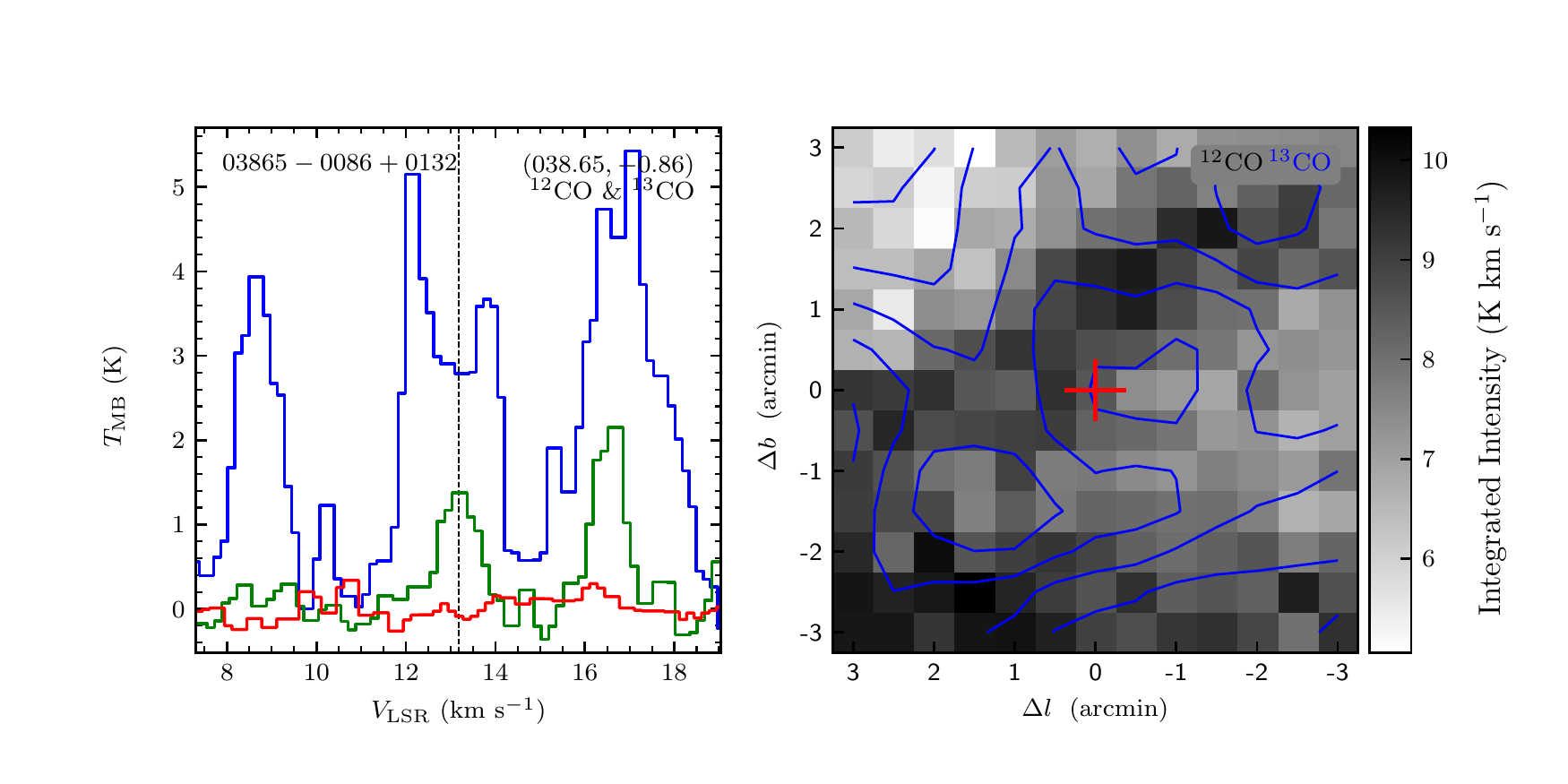}
\vspace{-0.5cm}

\includegraphics[width=9.0cm,angle=0]{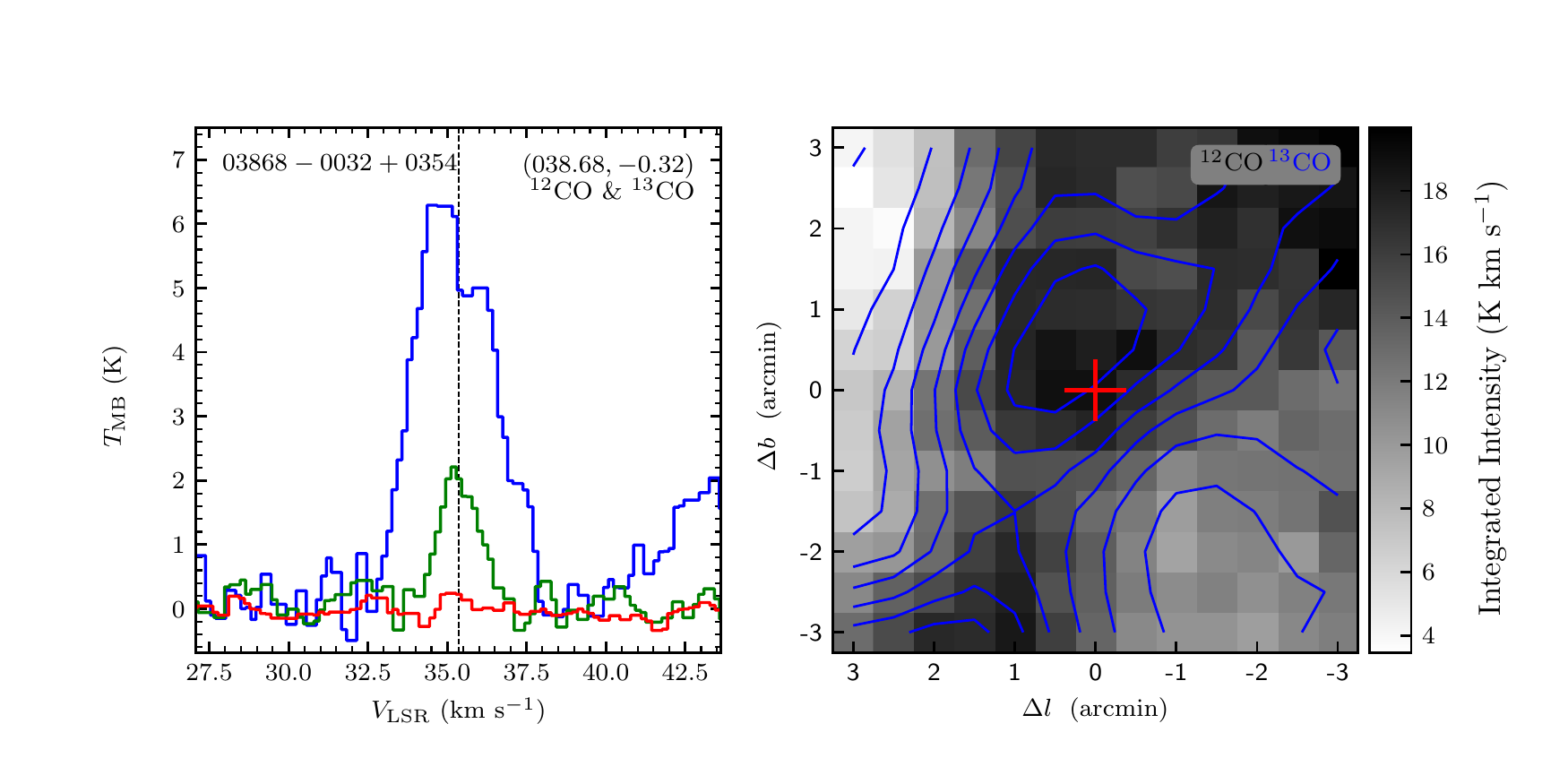}
\includegraphics[width=9.0cm,angle=0]{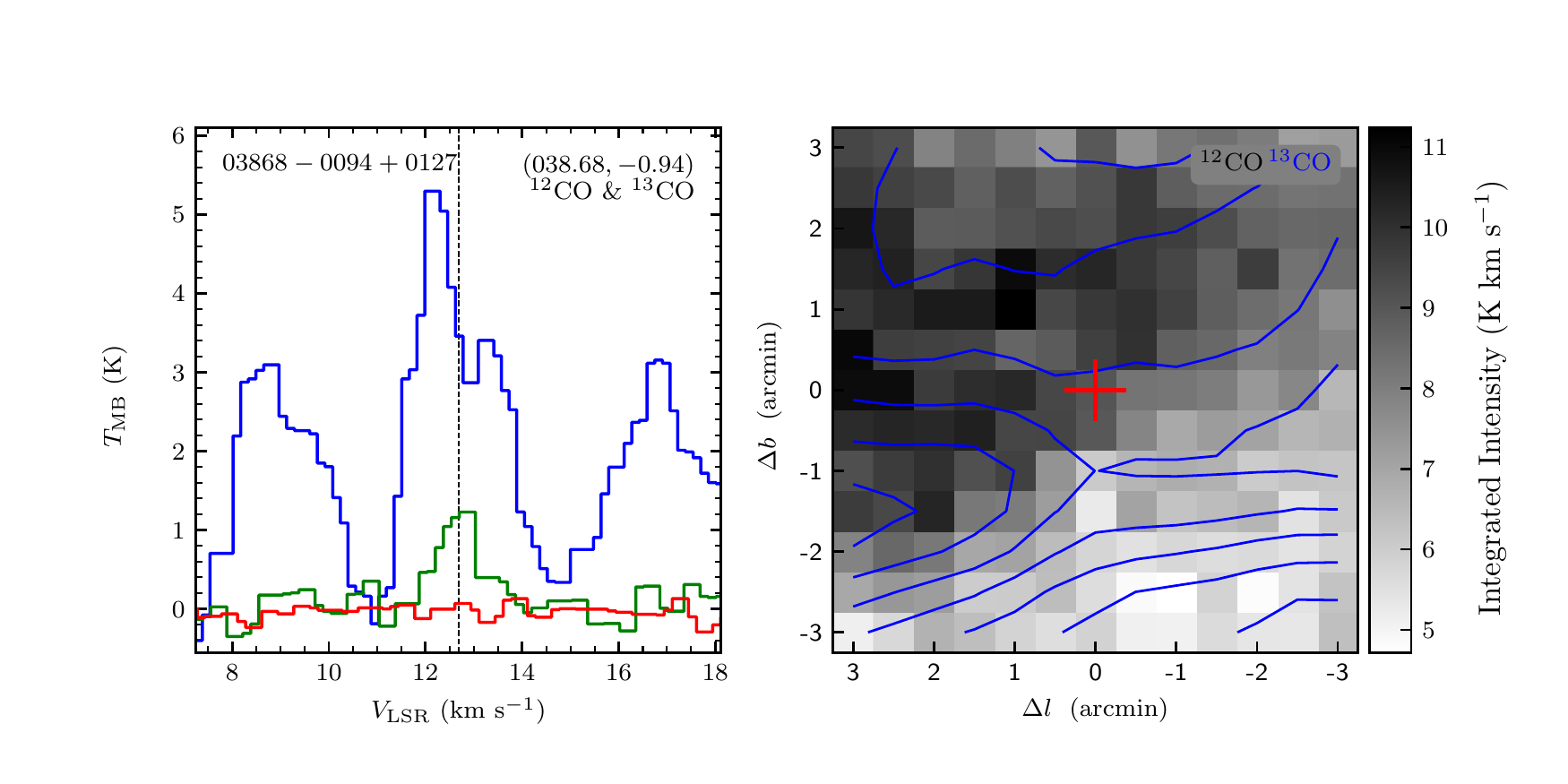}
\end{figure}
\clearpage

\begin{figure}
\includegraphics[width=9.0cm,angle=0]{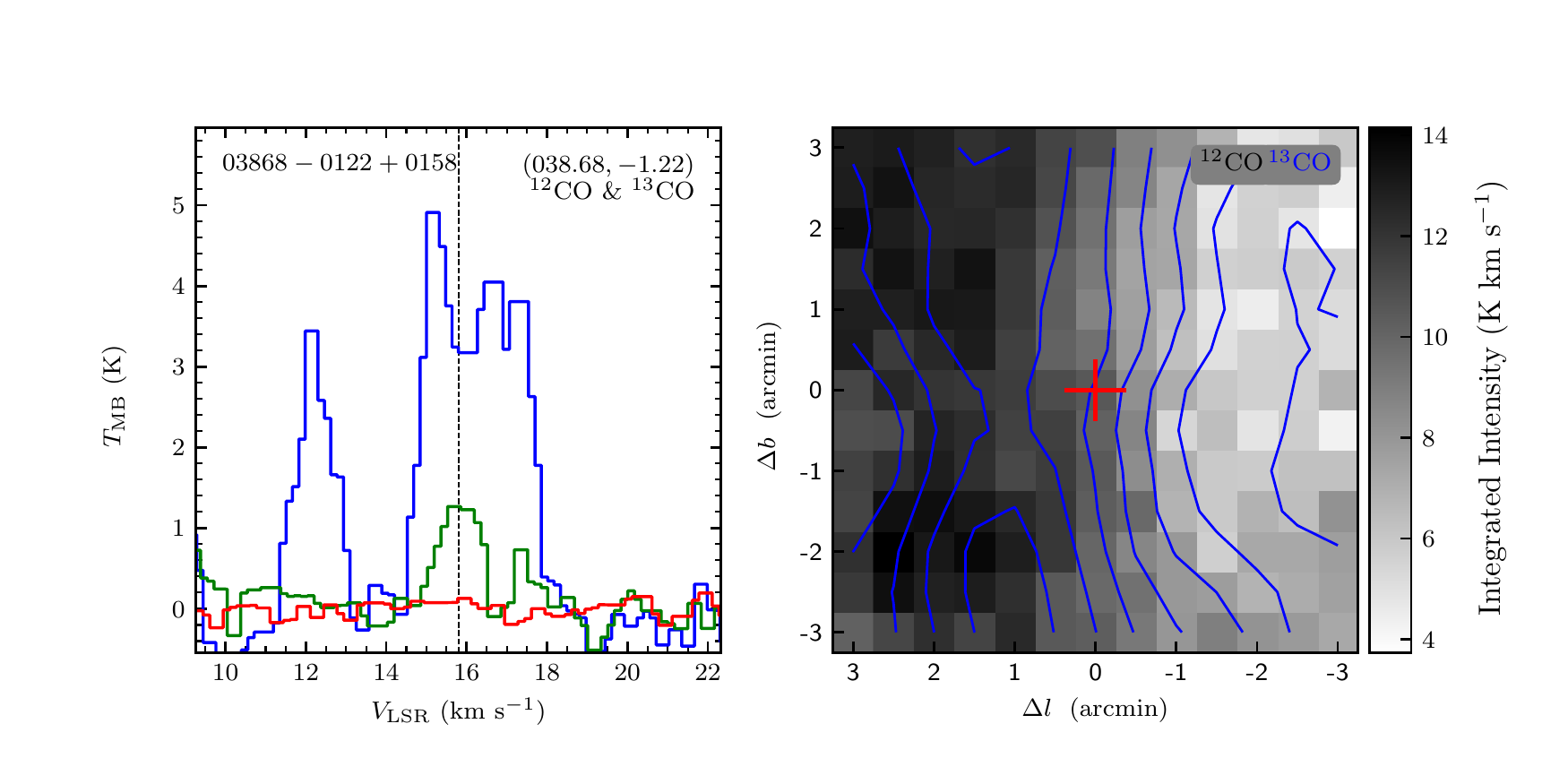}
\includegraphics[width=9.0cm,angle=0]{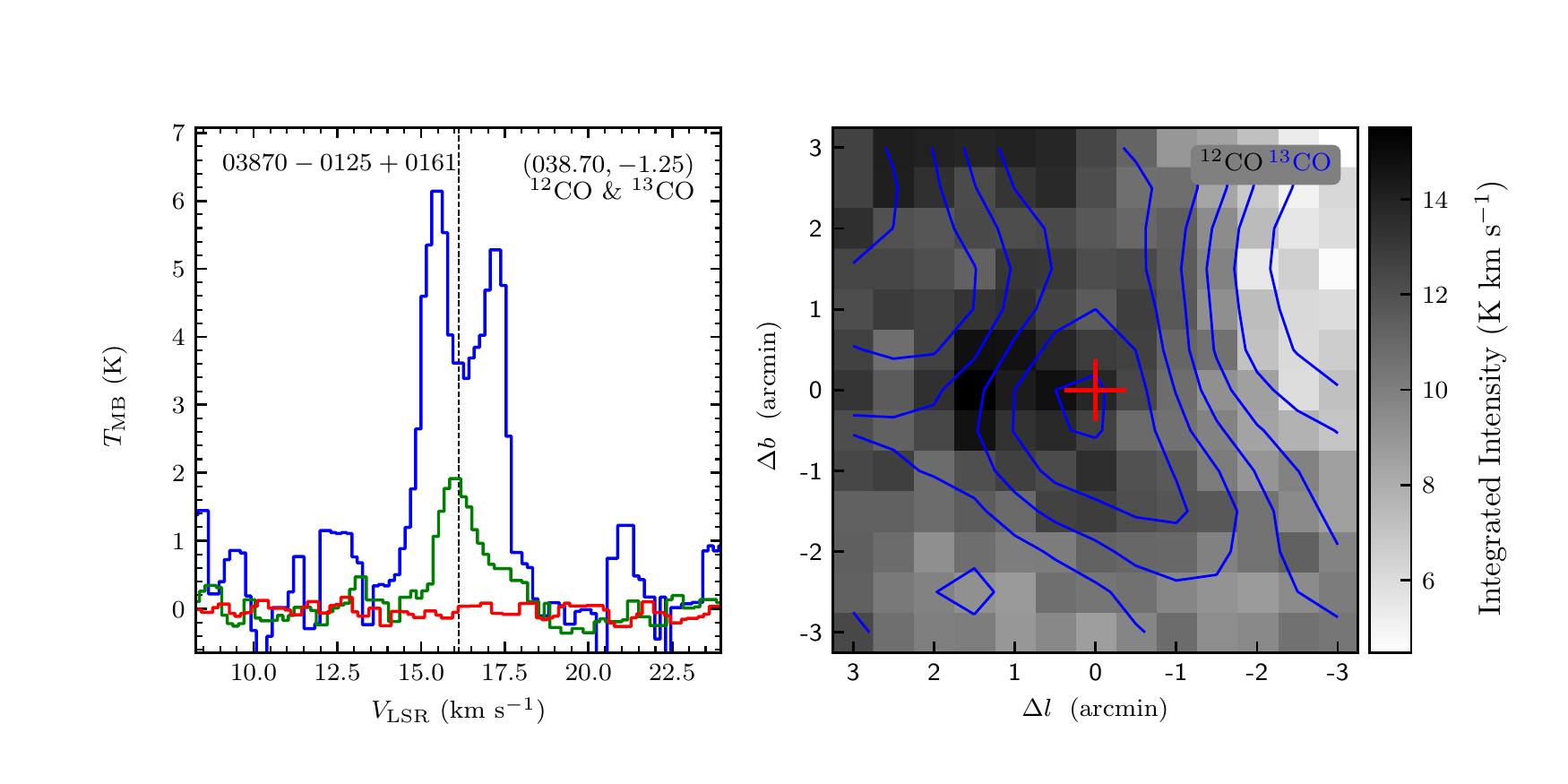}
\vspace{-0.5cm}

\includegraphics[width=9.0cm,angle=0]{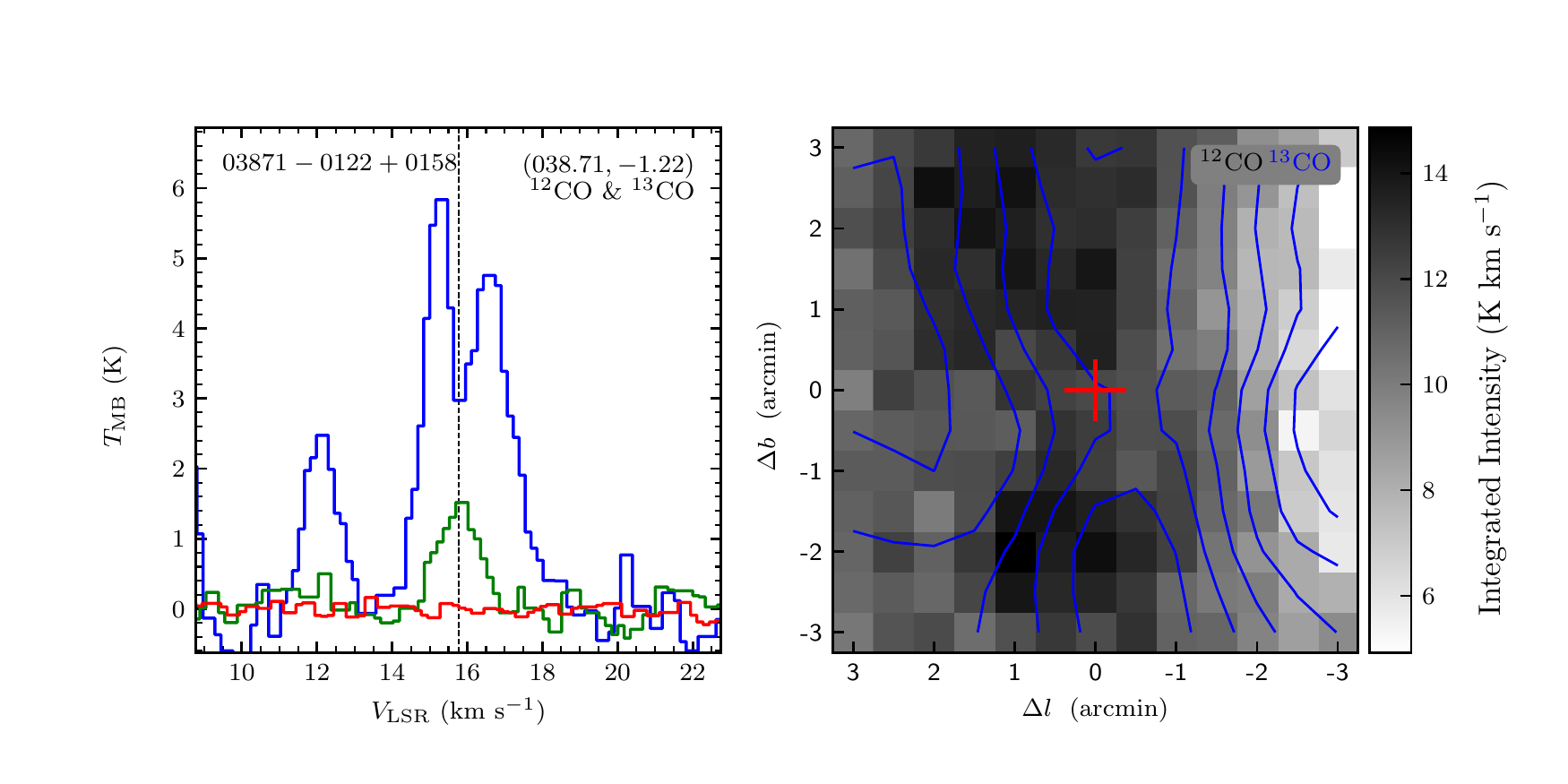}
\includegraphics[width=9.0cm,angle=0]{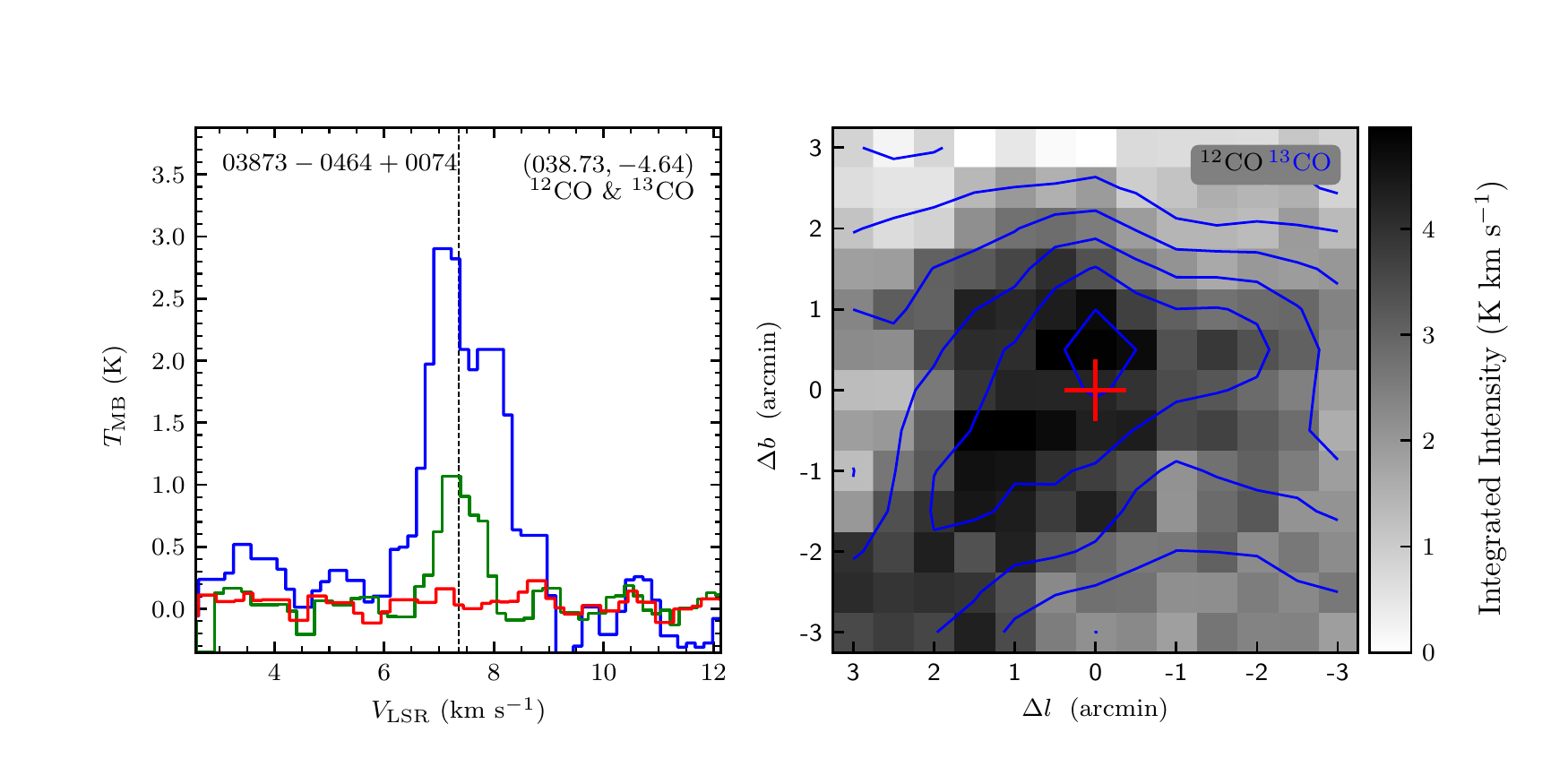}
\vspace{-0.5cm}

\includegraphics[width=9.0cm,angle=0]{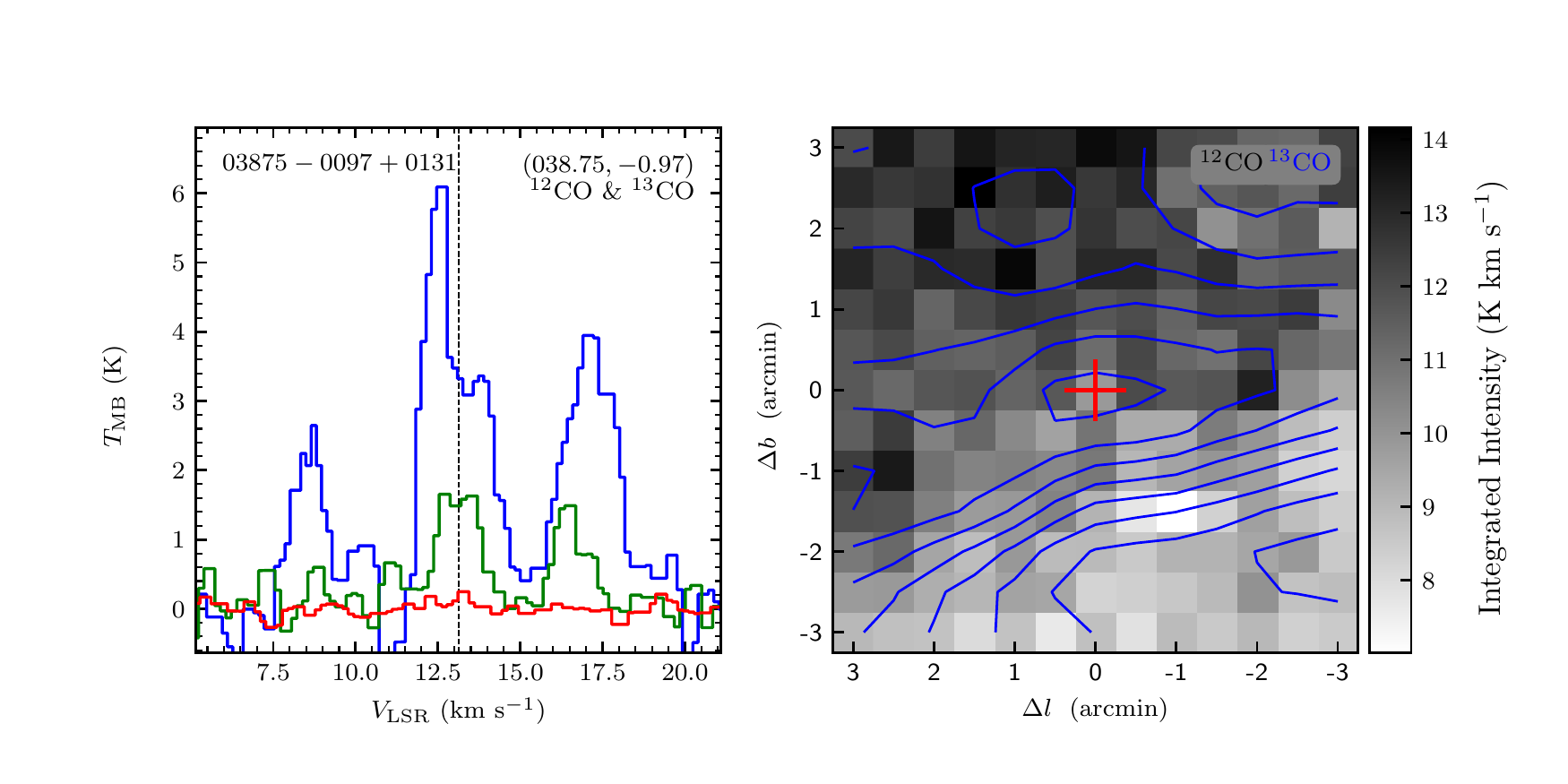}
\includegraphics[width=9.0cm,angle=0]{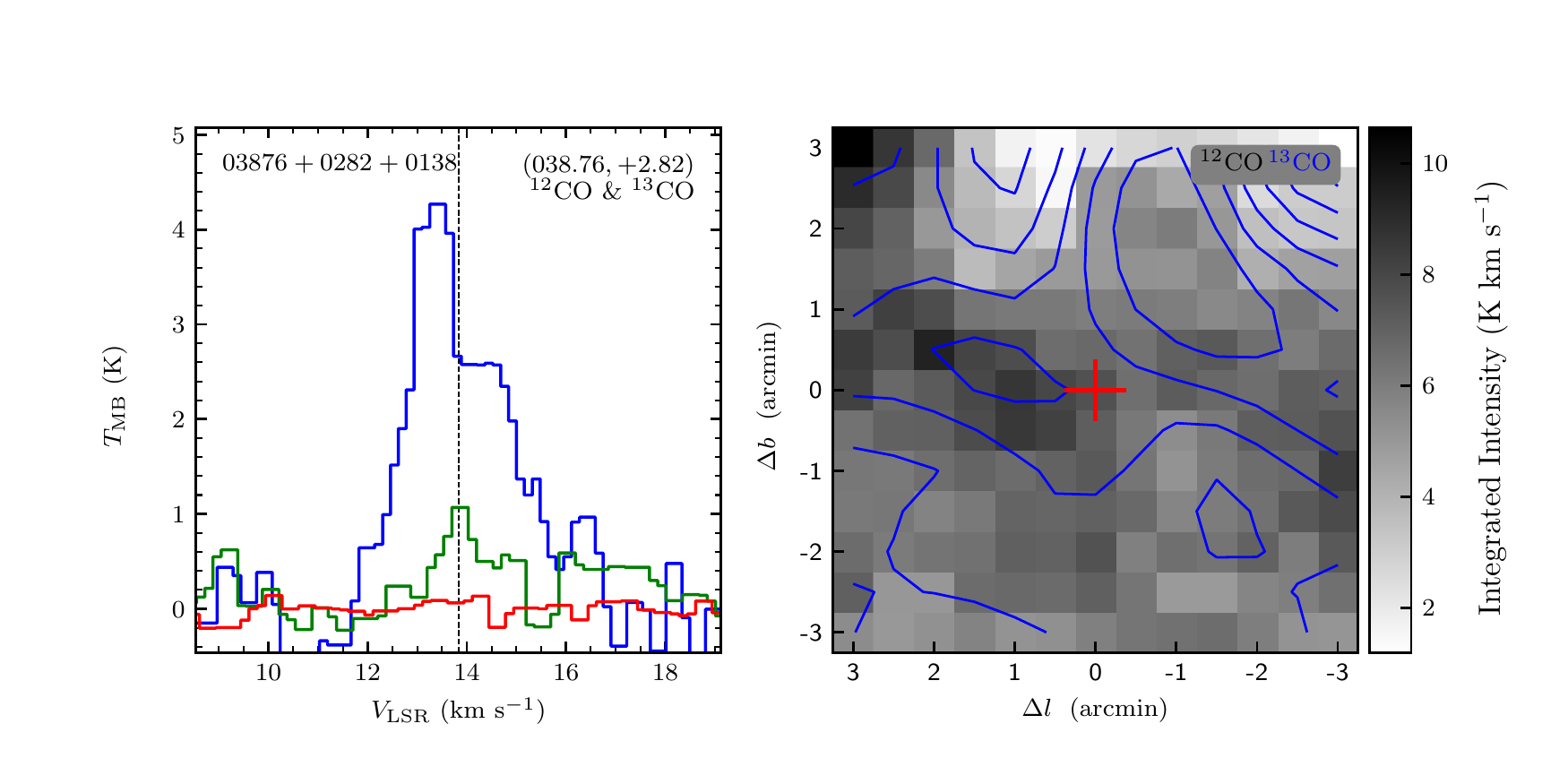}
\vspace{-0.5cm}

\includegraphics[width=9.0cm,angle=0]{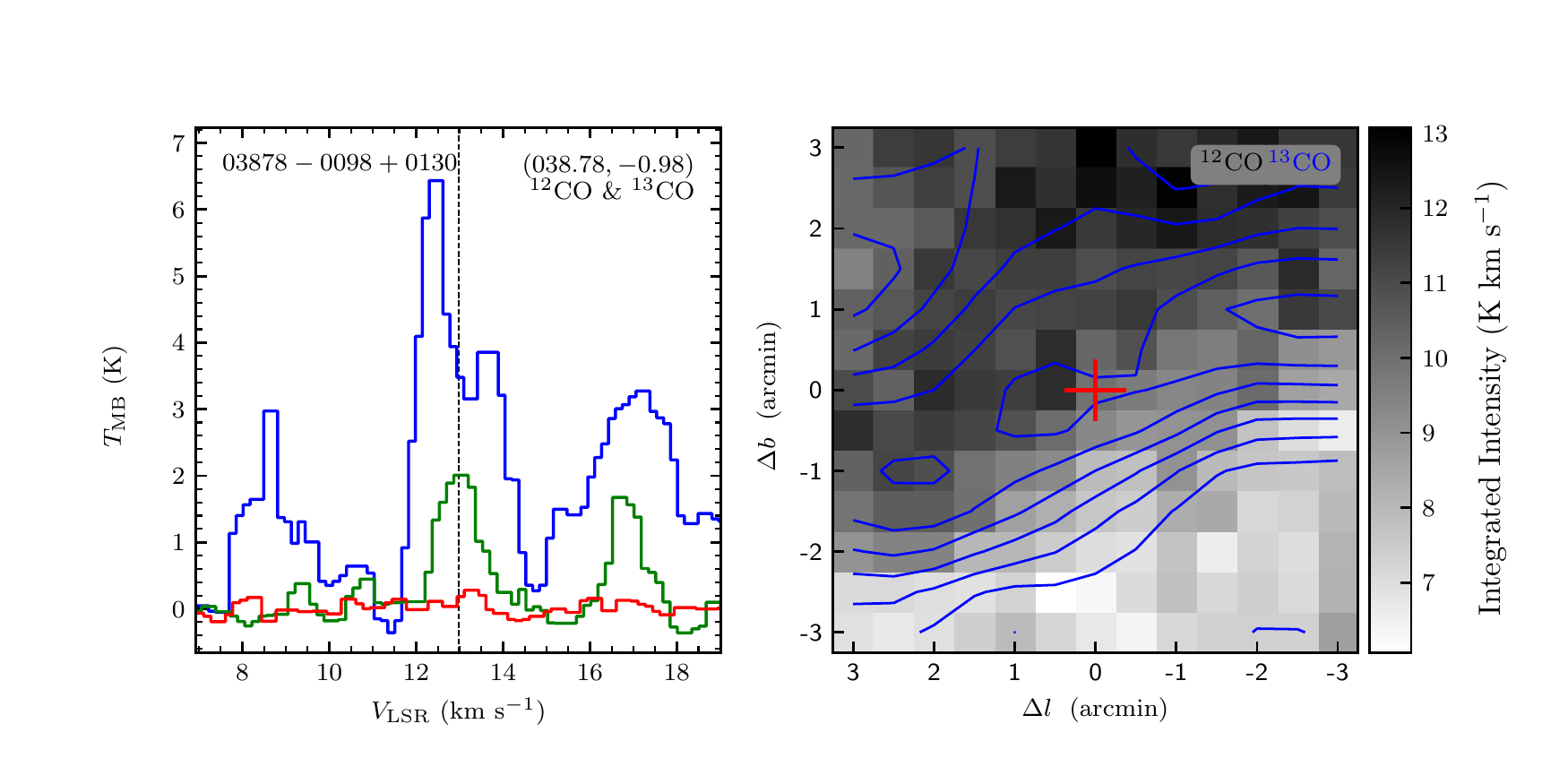}
\includegraphics[width=9.0cm,angle=0]{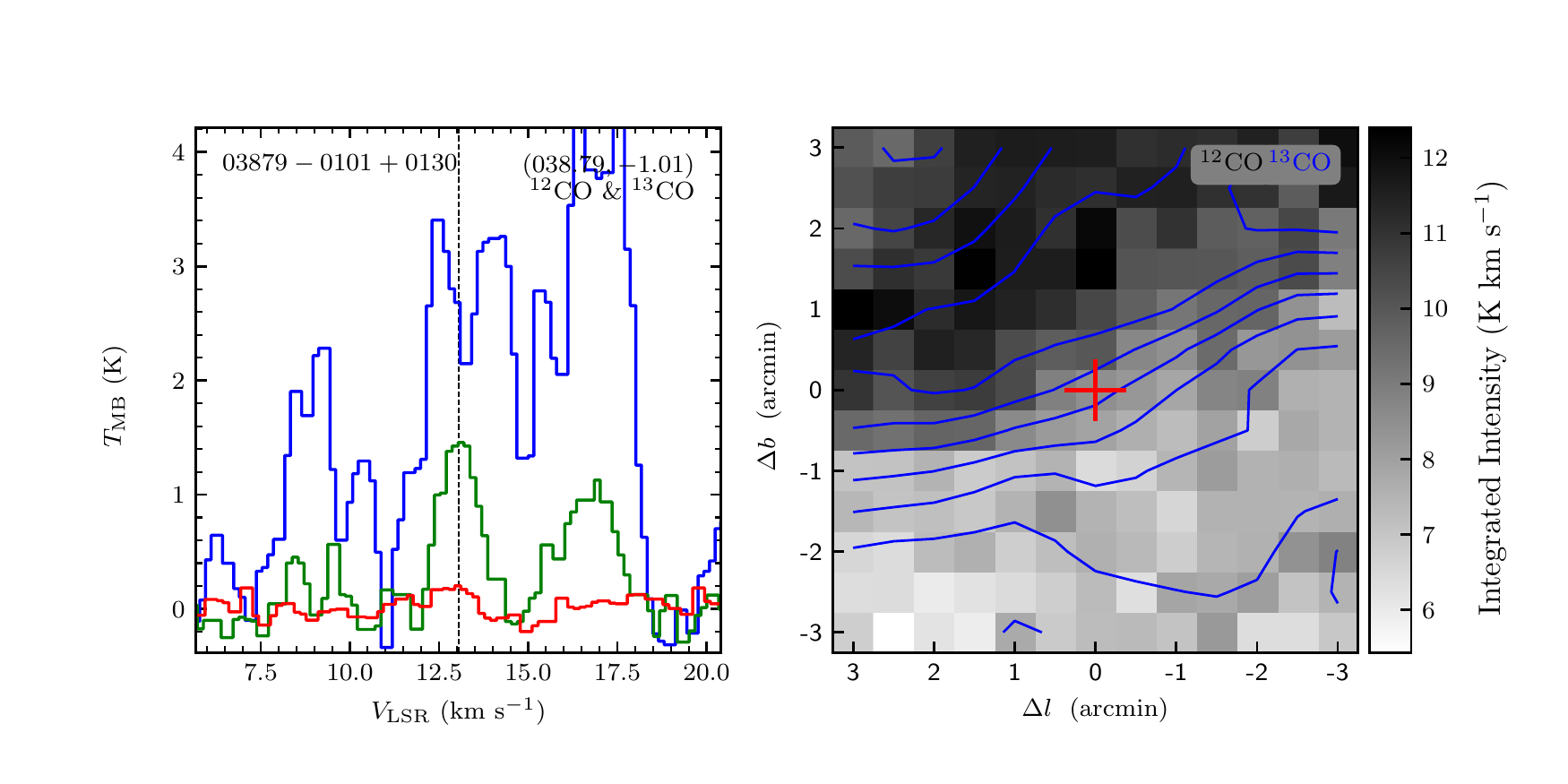}
\vspace{-0.5cm}

\includegraphics[width=9.0cm,angle=0]{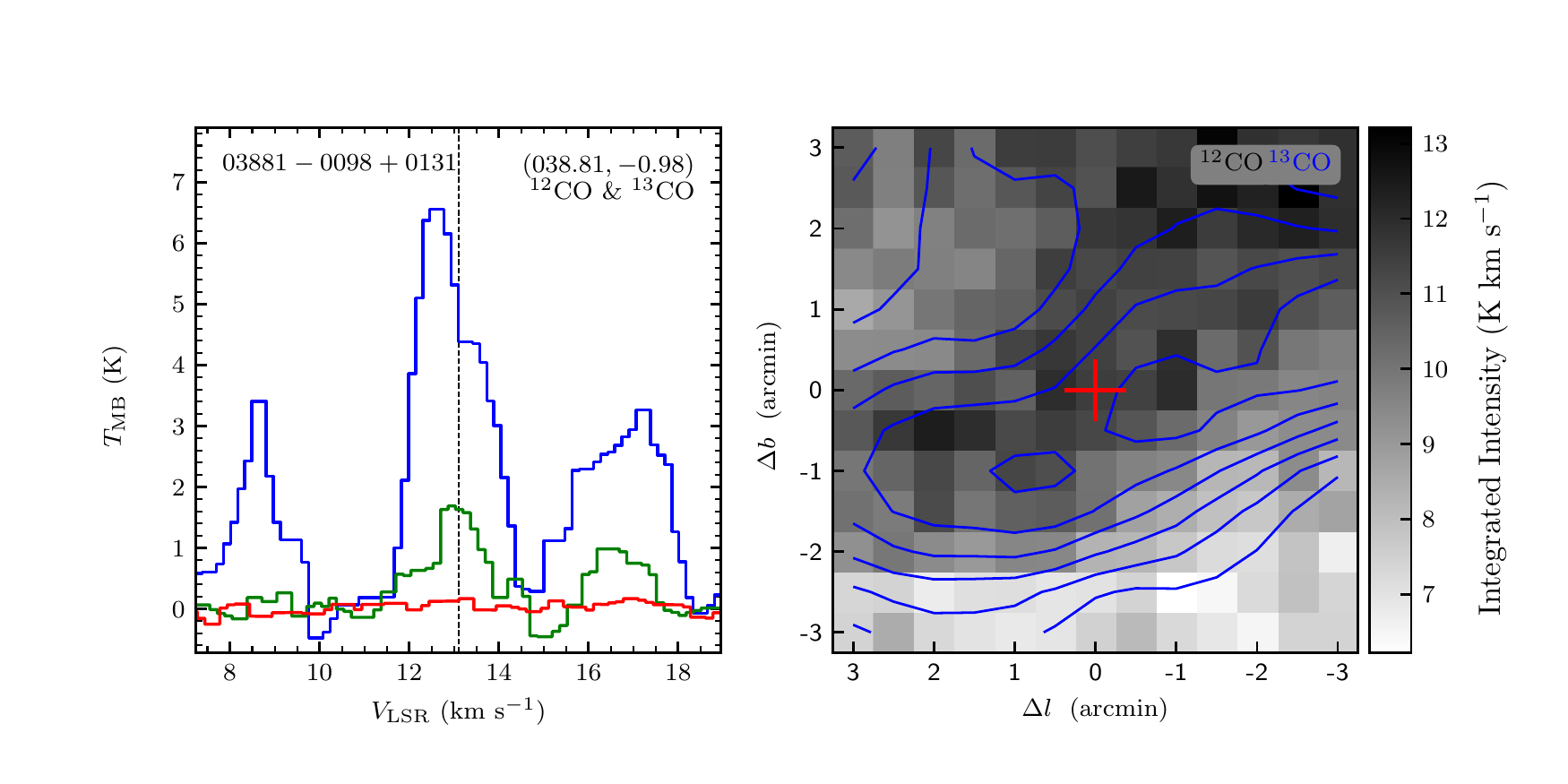}
\includegraphics[width=9.0cm,angle=0]{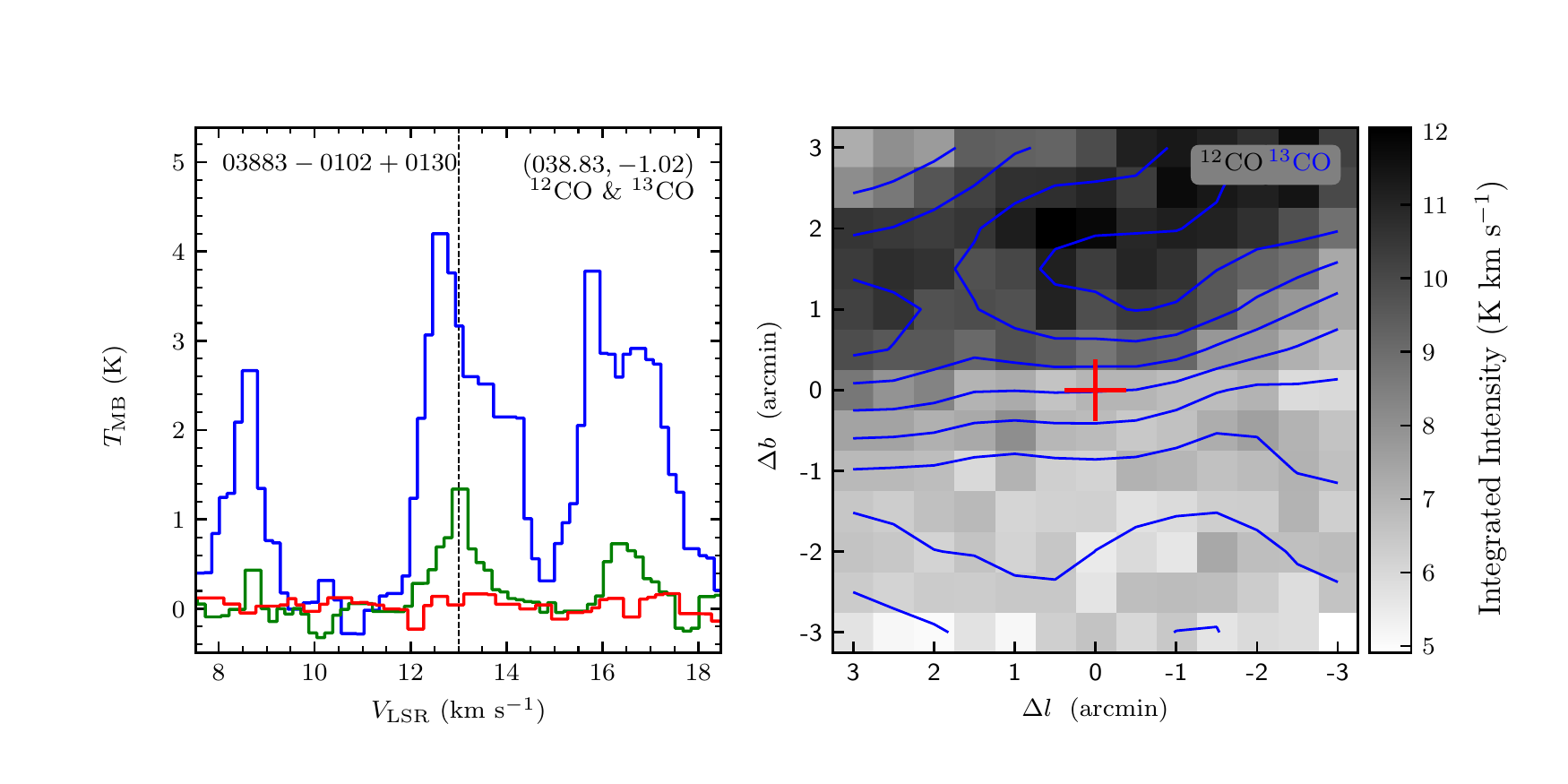}
\end{figure}
\clearpage

\begin{figure}
\includegraphics[width=9.0cm,angle=0]{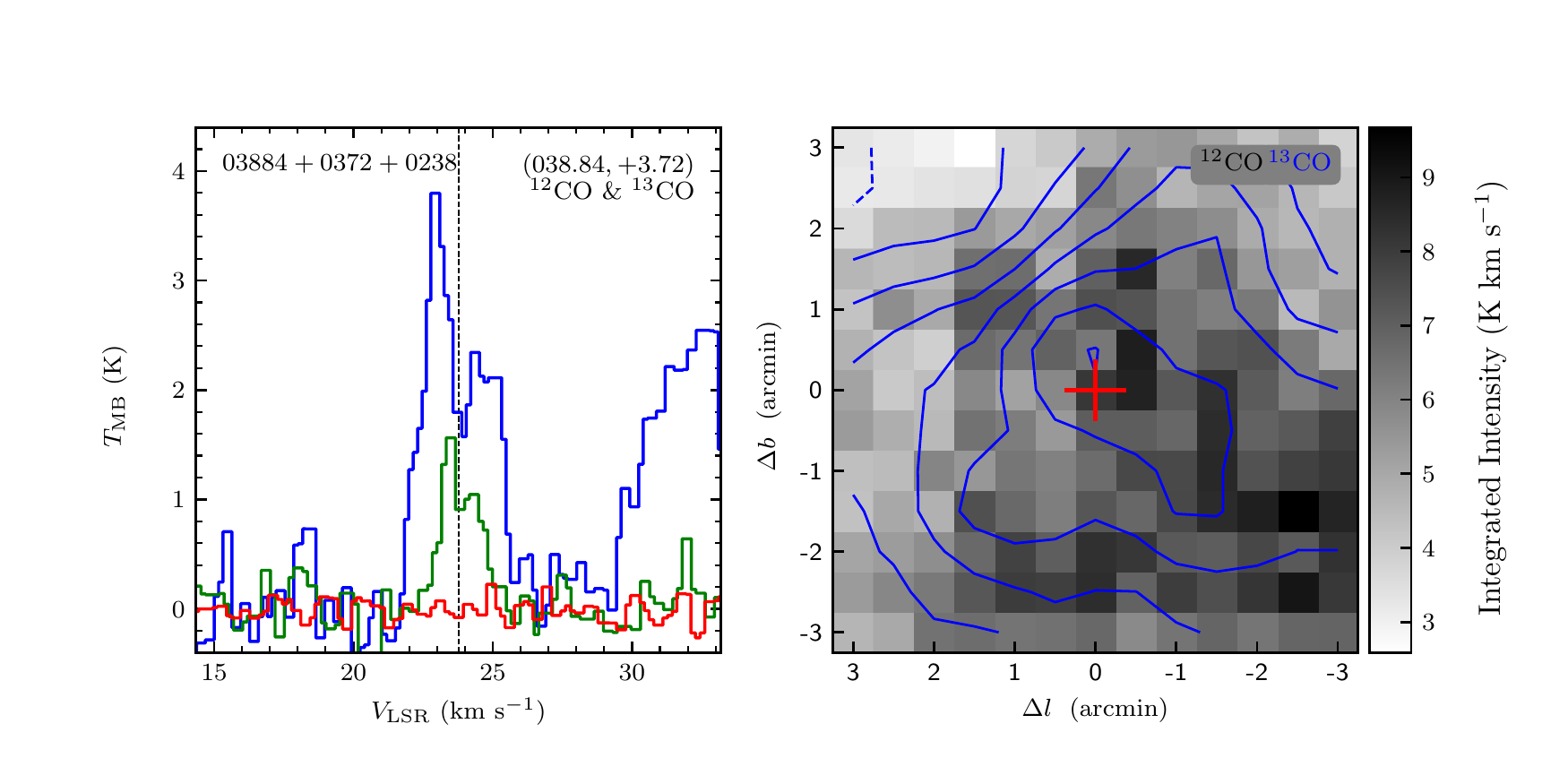}
\includegraphics[width=9.0cm,angle=0]{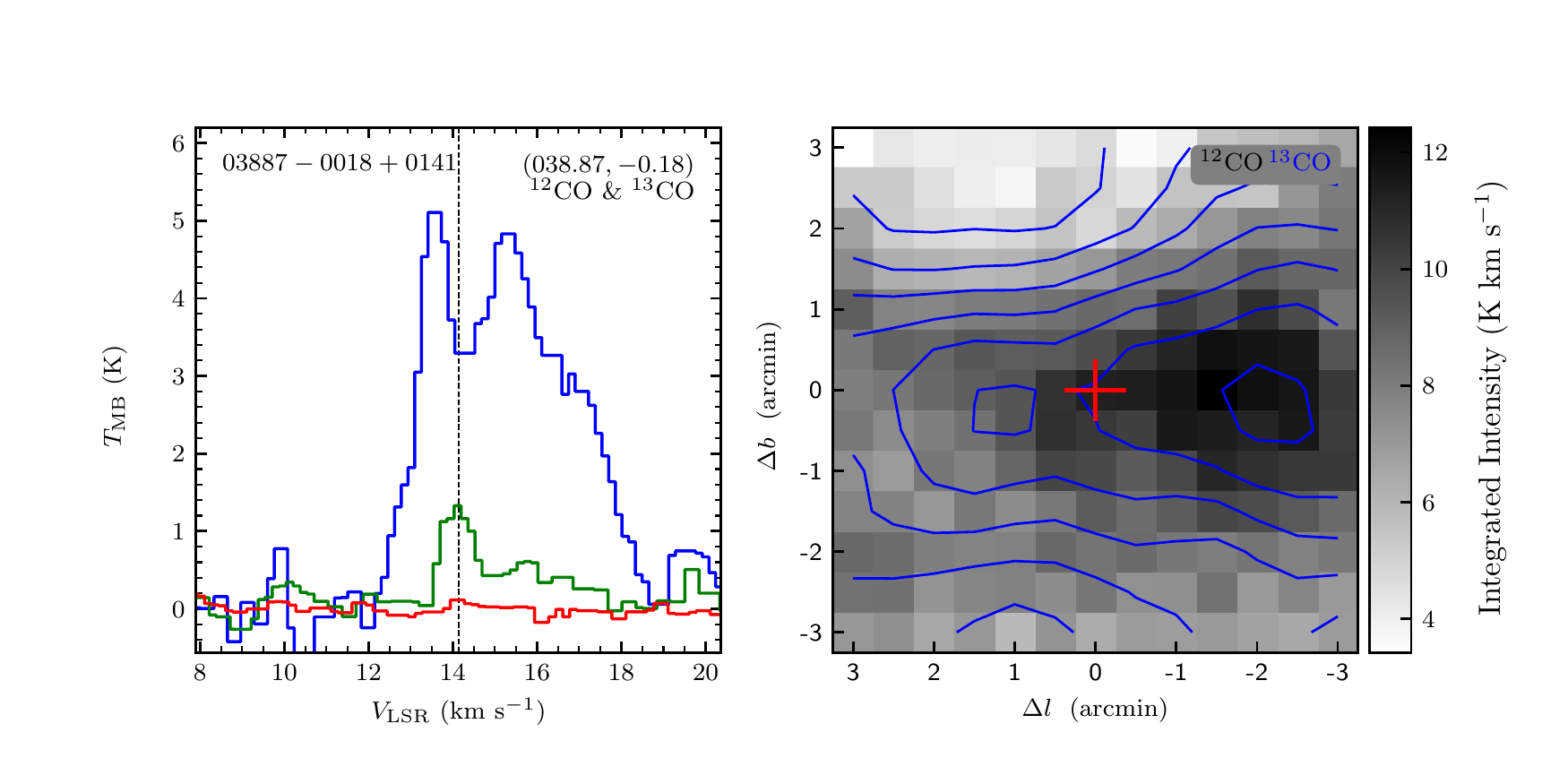}
\vspace{-0.5cm}

\includegraphics[width=9.0cm,angle=0]{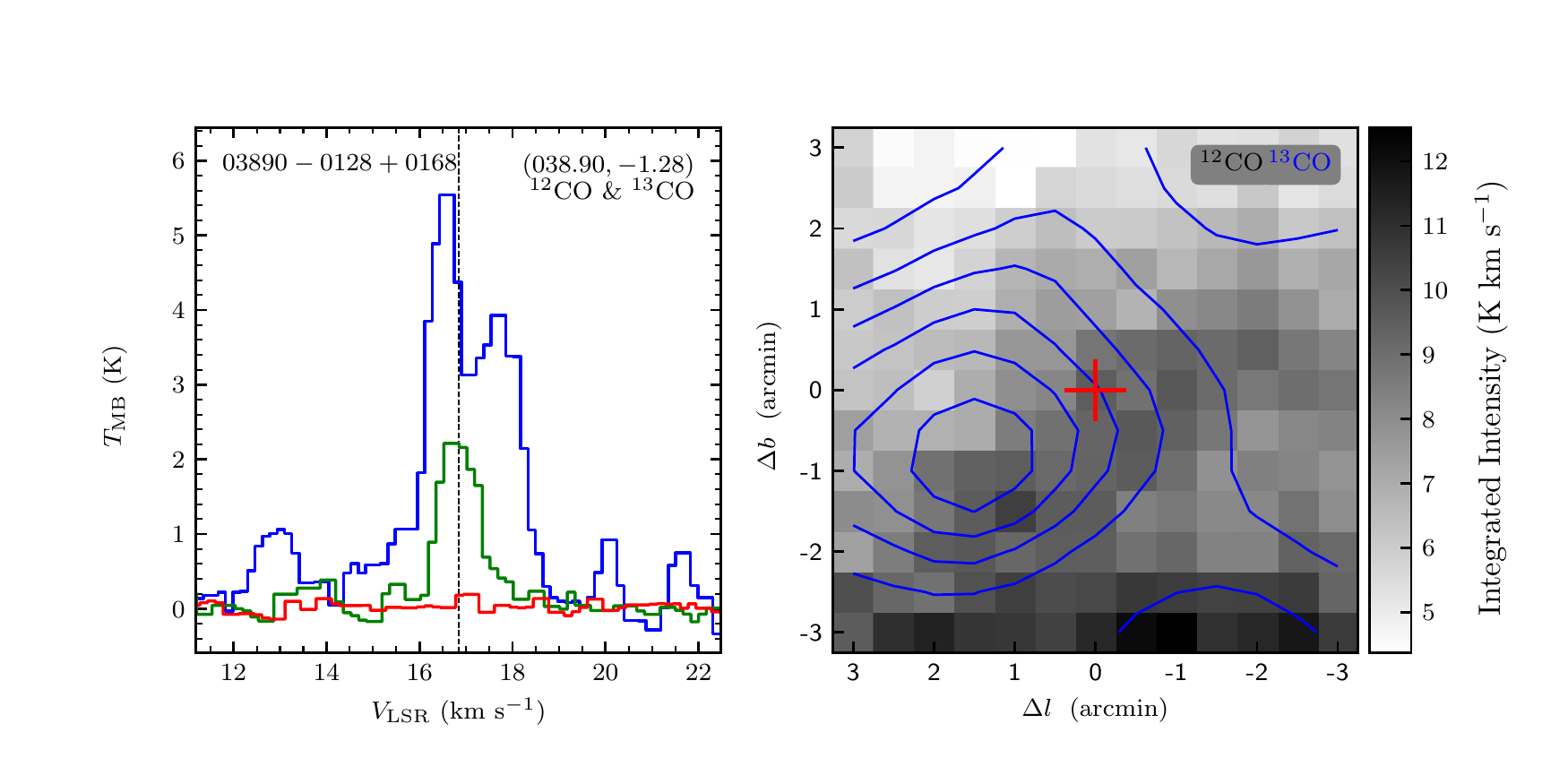}
\includegraphics[width=9.0cm,angle=0]{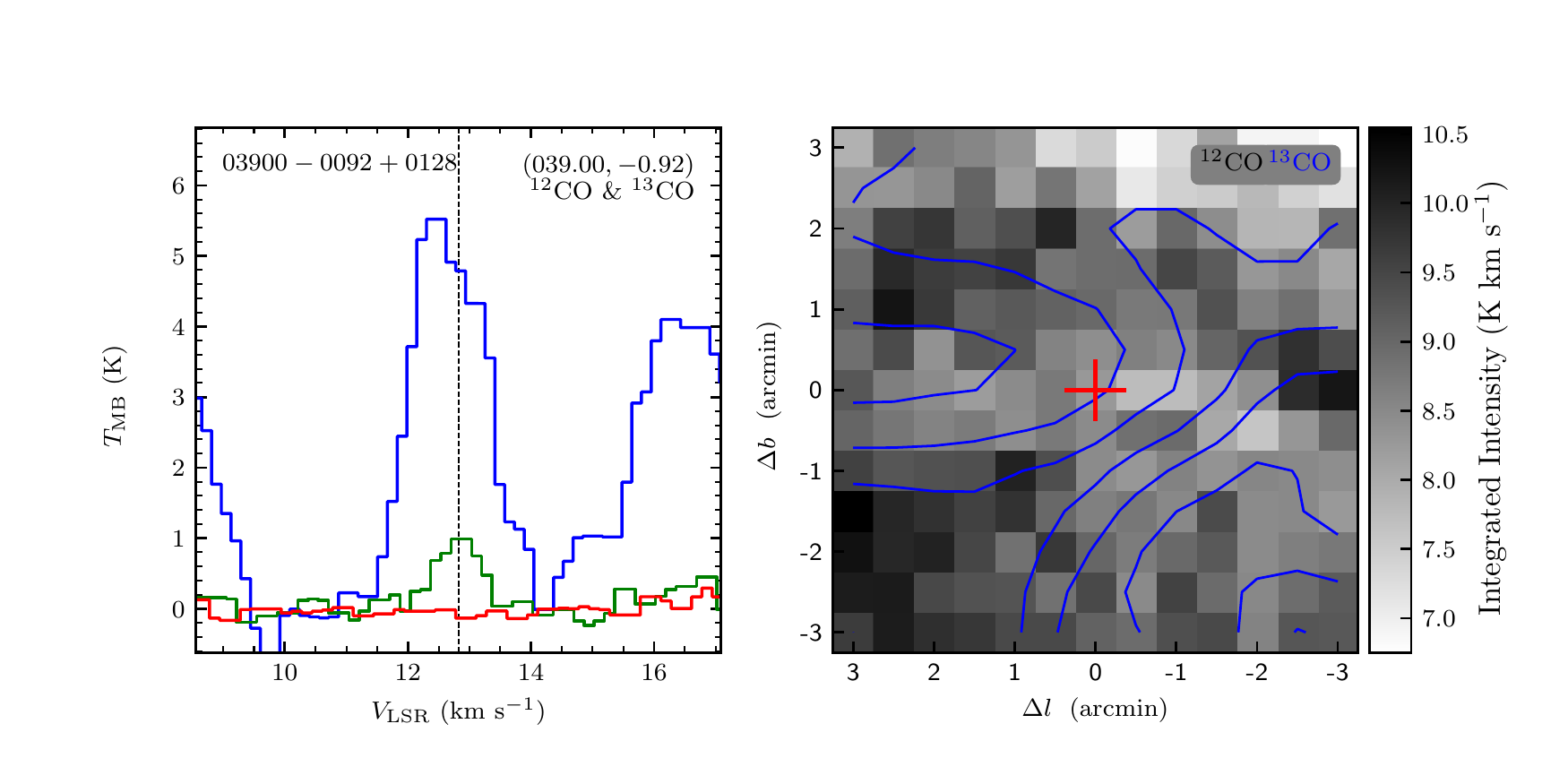}
\vspace{-0.5cm}

\includegraphics[width=9.0cm,angle=0]{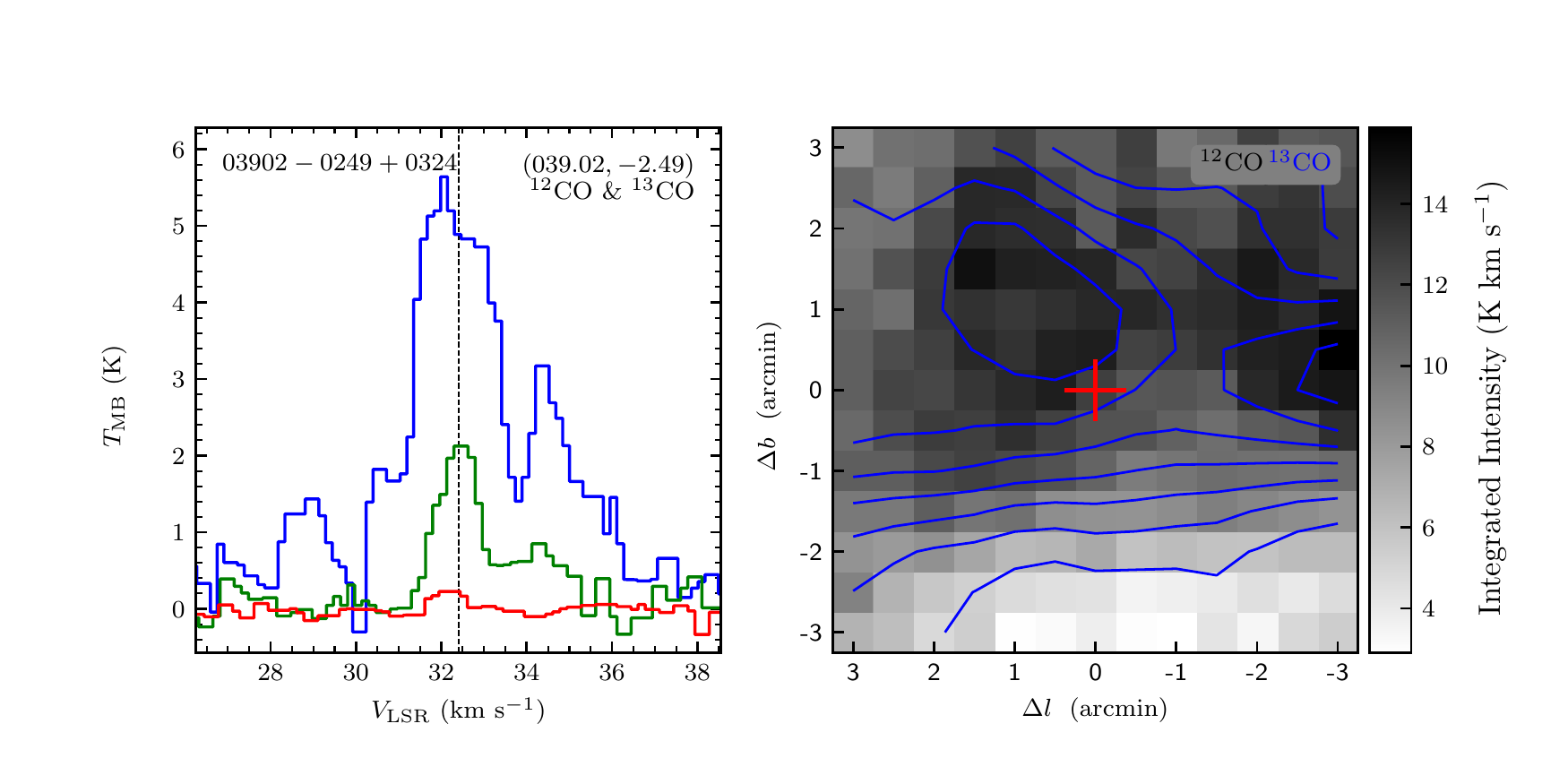}
\includegraphics[width=9.0cm,angle=0]{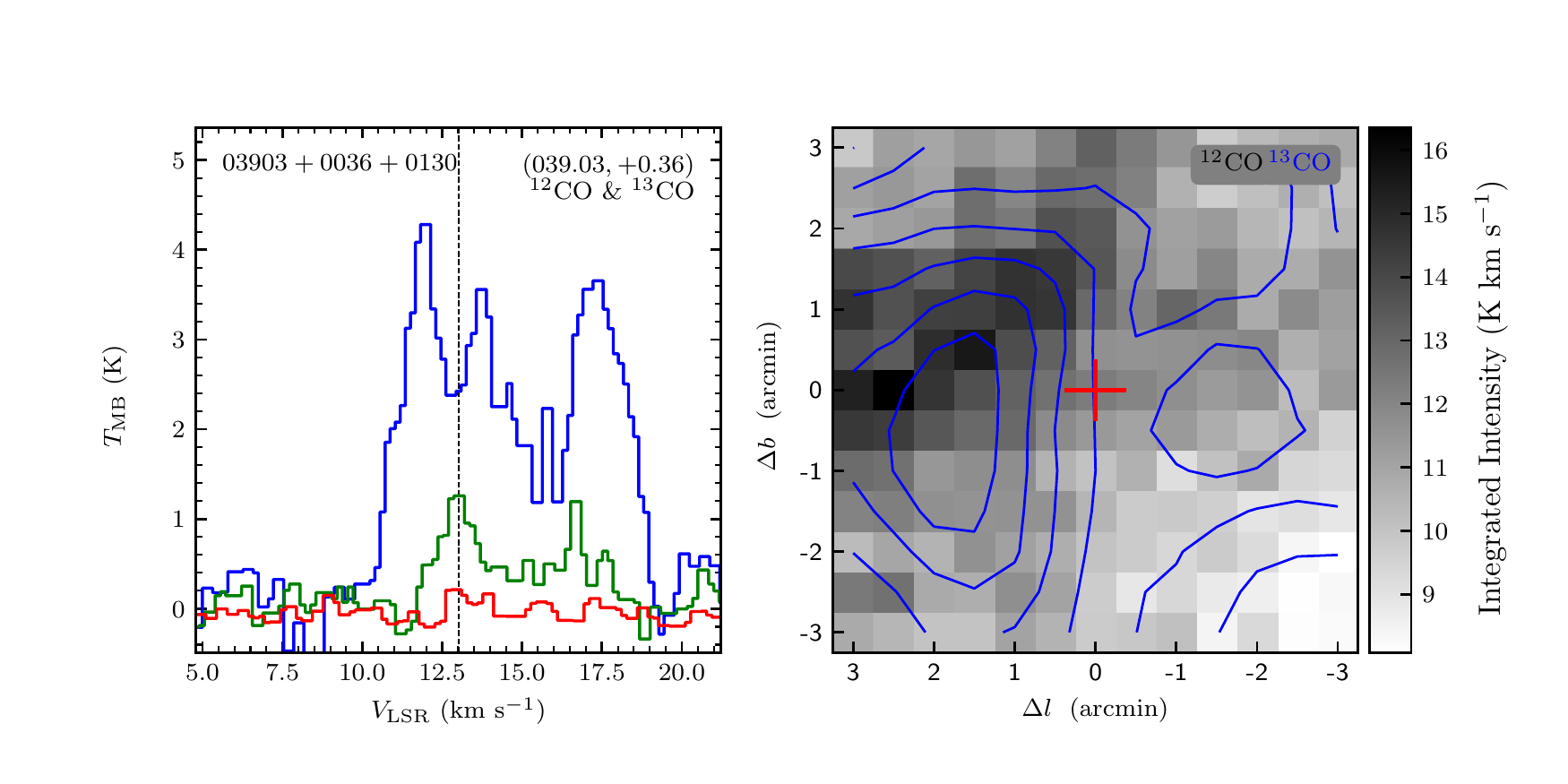}
\vspace{-0.5cm}

\includegraphics[width=9.0cm,angle=0]{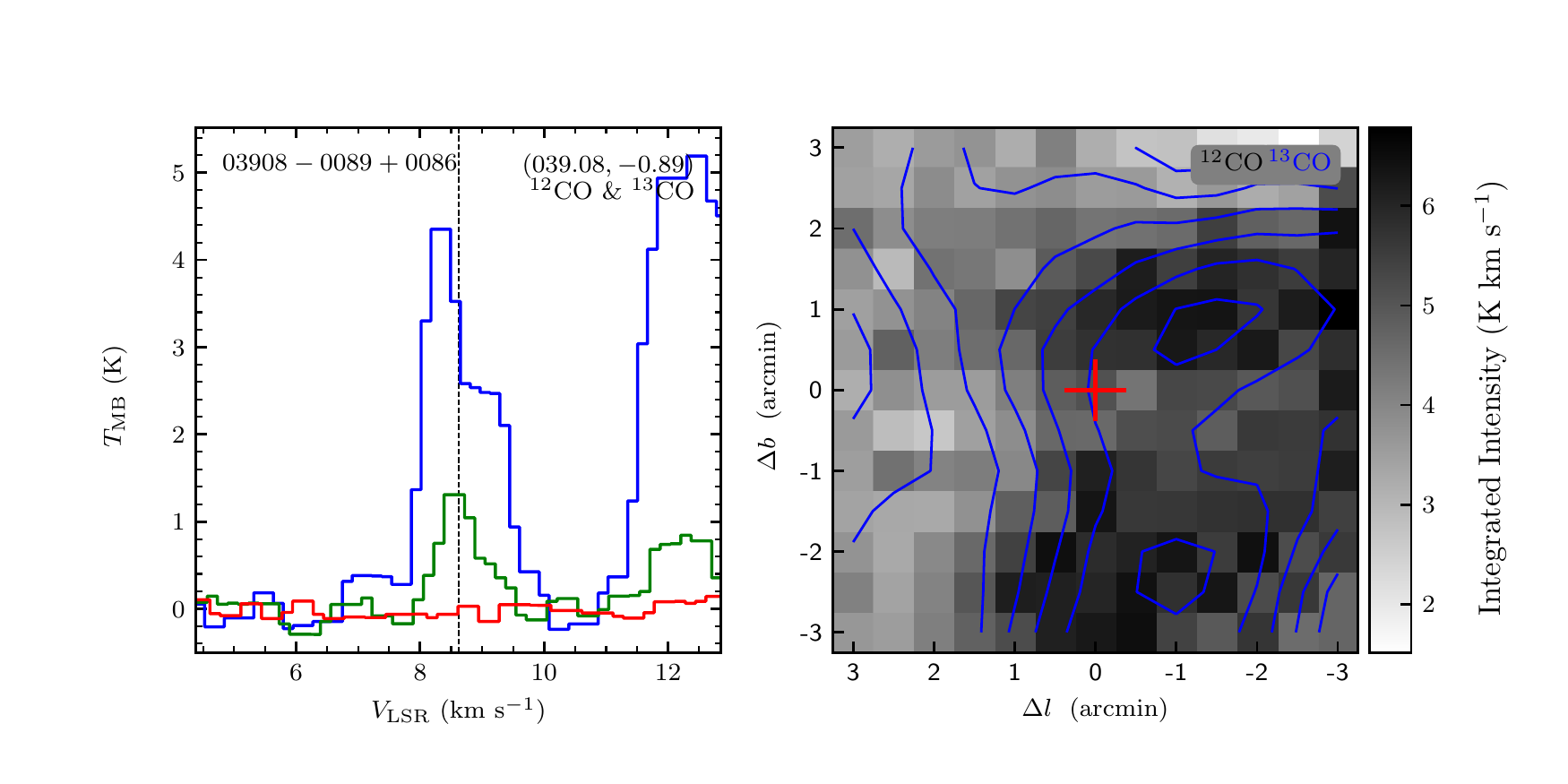}
\includegraphics[width=9.0cm,angle=0]{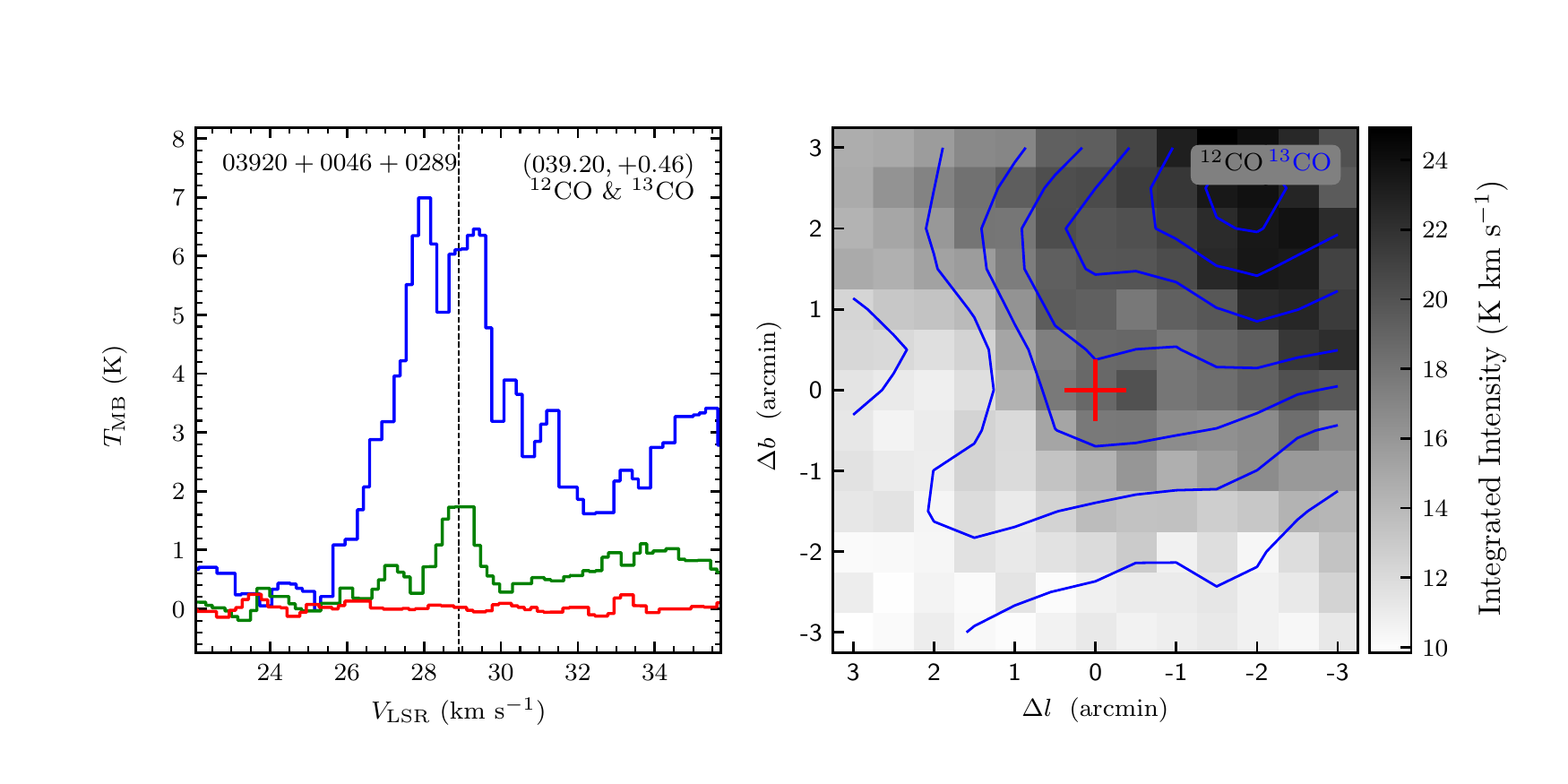}
\vspace{-0.5cm}

\includegraphics[width=9.0cm,angle=0]{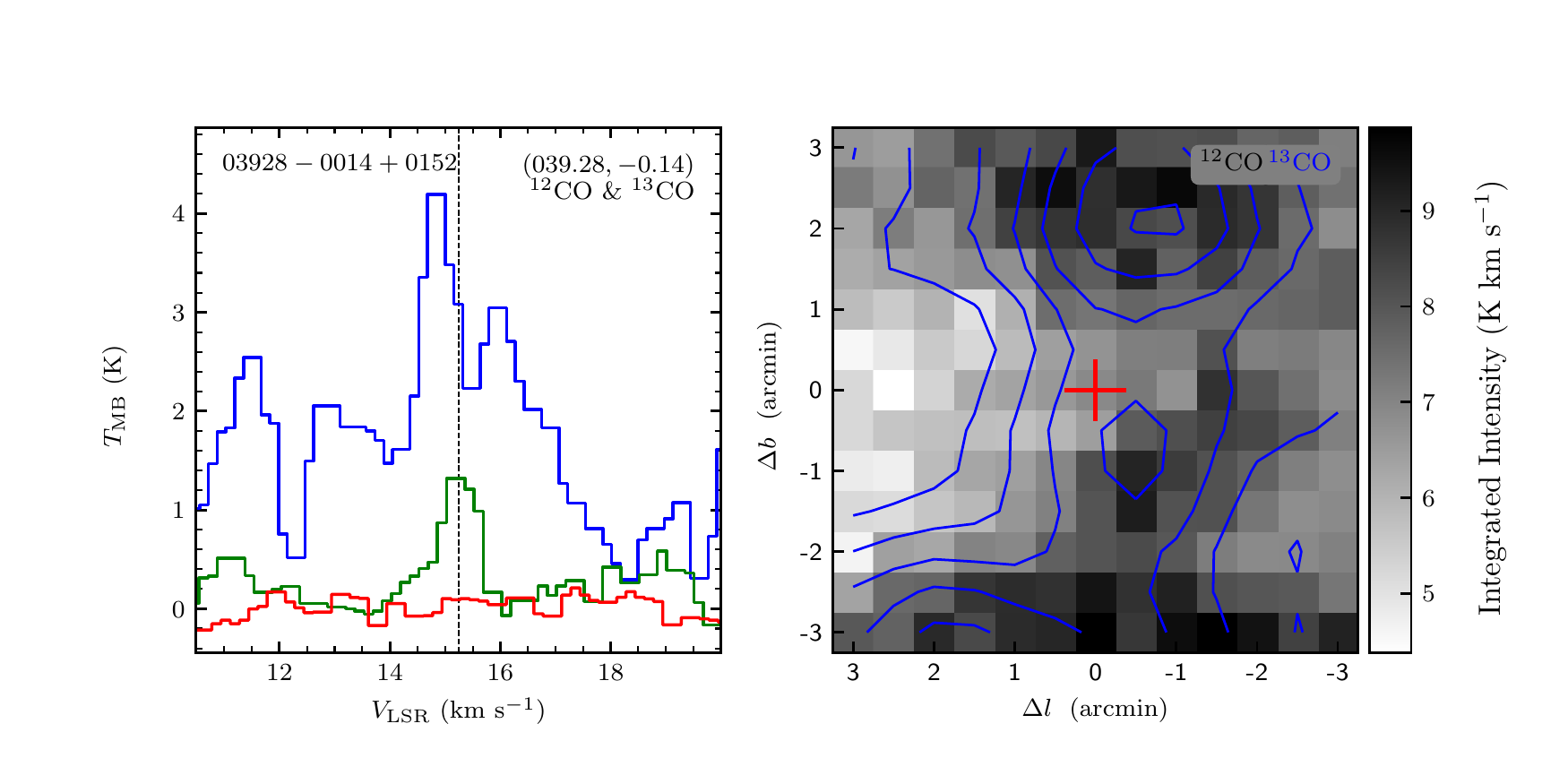}
\includegraphics[width=9.0cm,angle=0]{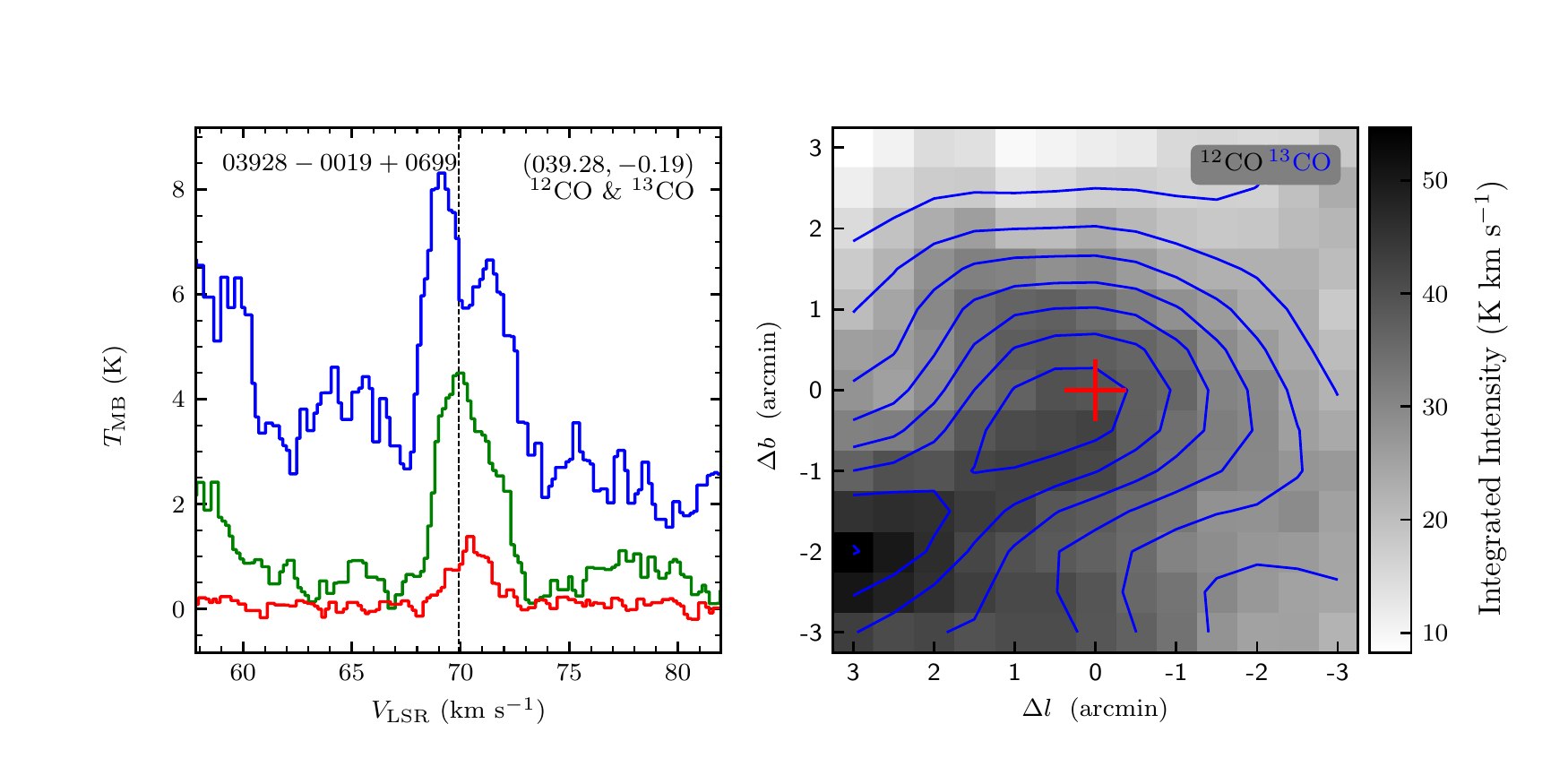}
\end{figure}
\clearpage

\begin{figure}
\includegraphics[width=9.0cm,angle=0]{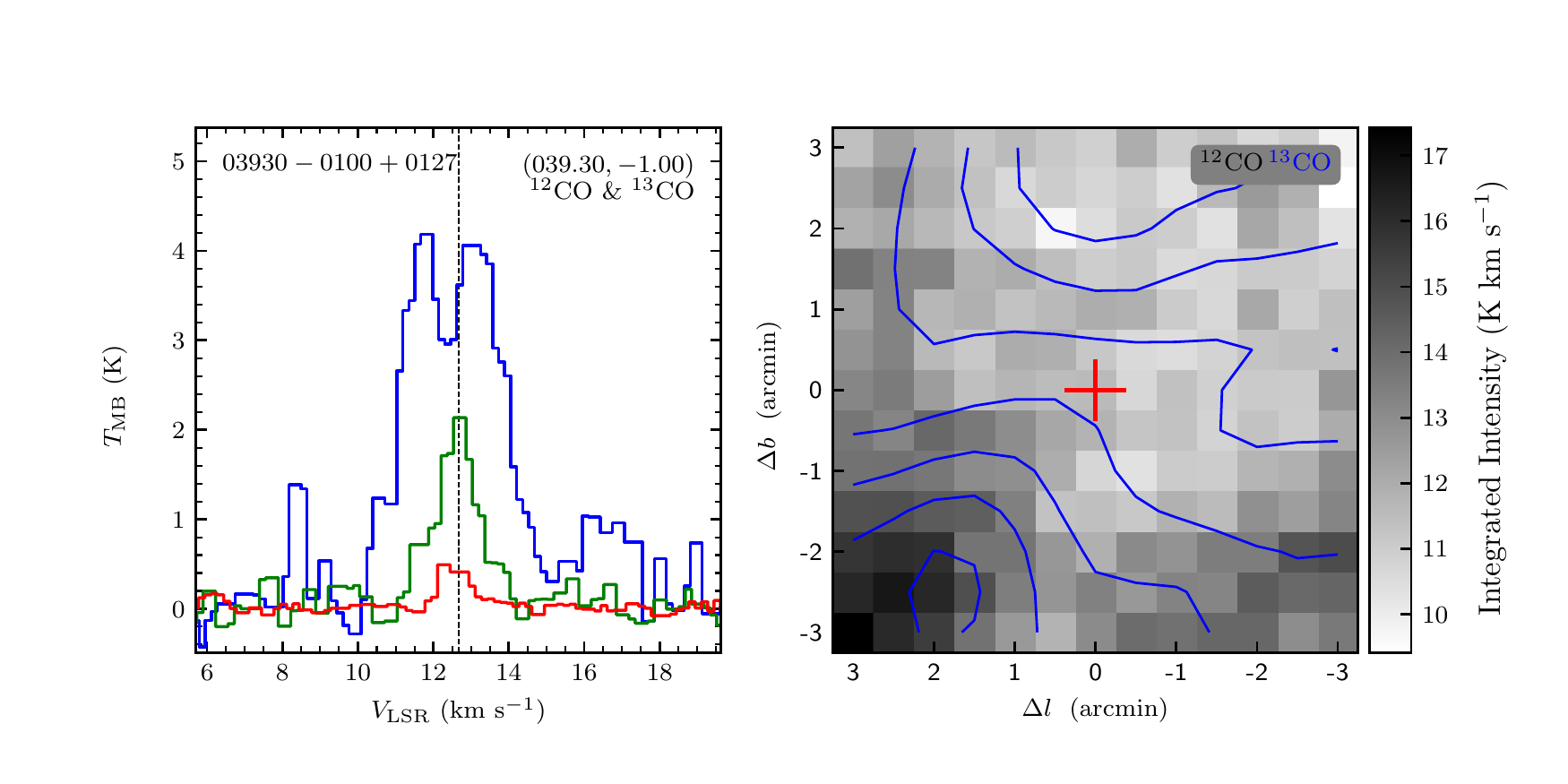}
\includegraphics[width=9.0cm,angle=0]{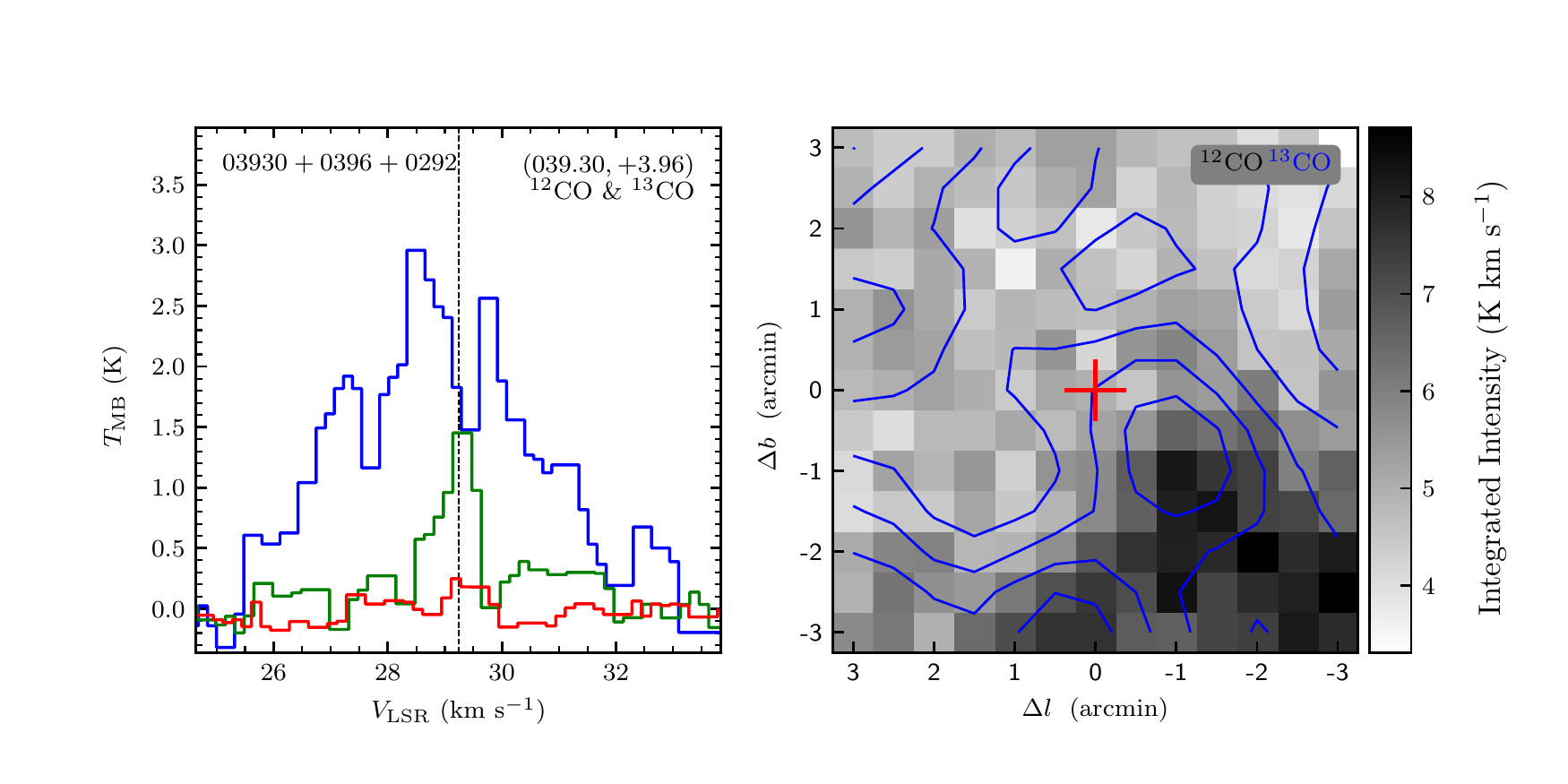}
\vspace{-0.5cm}

\includegraphics[width=9.0cm,angle=0]{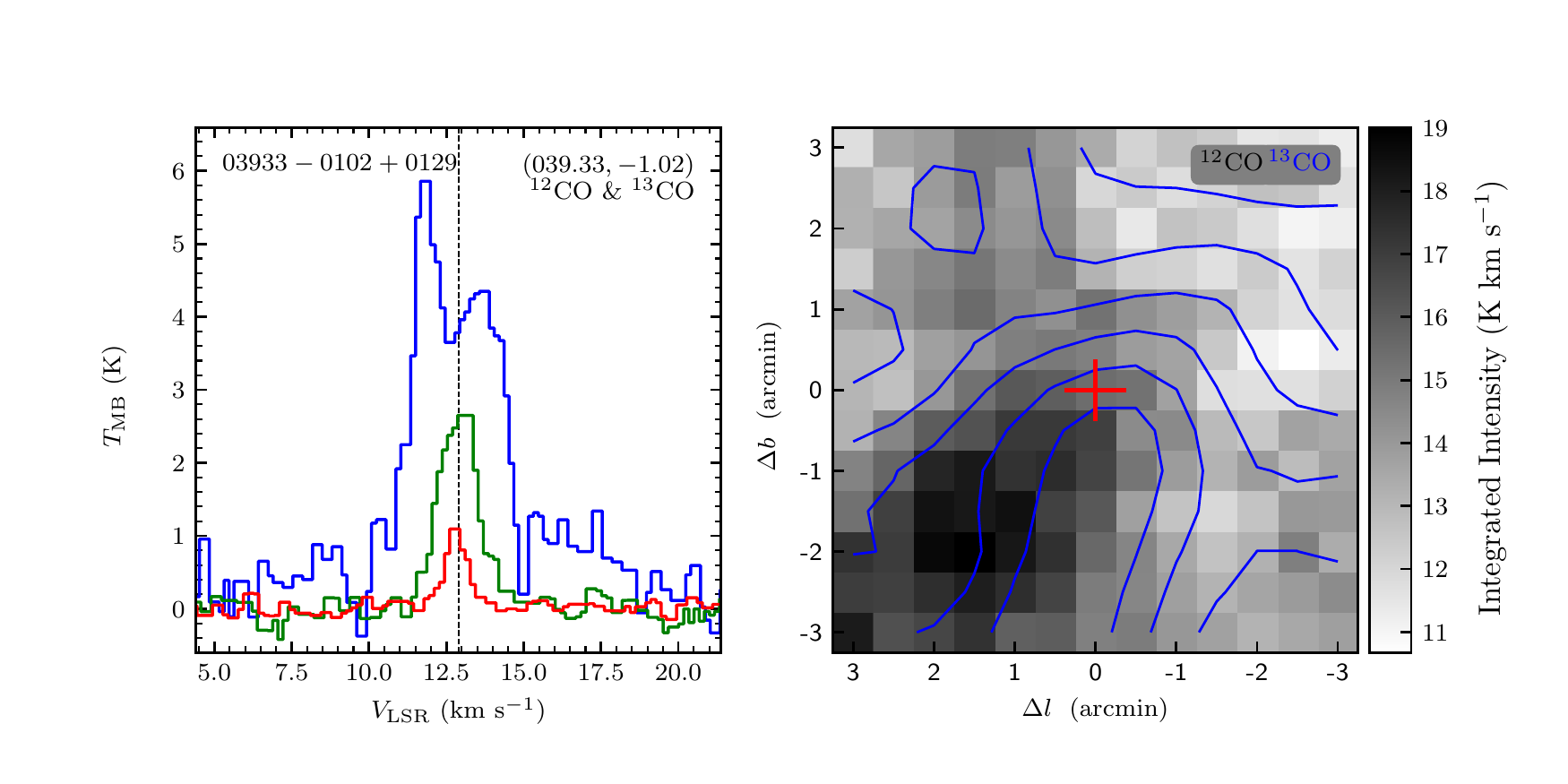}
\includegraphics[width=9.0cm,angle=0]{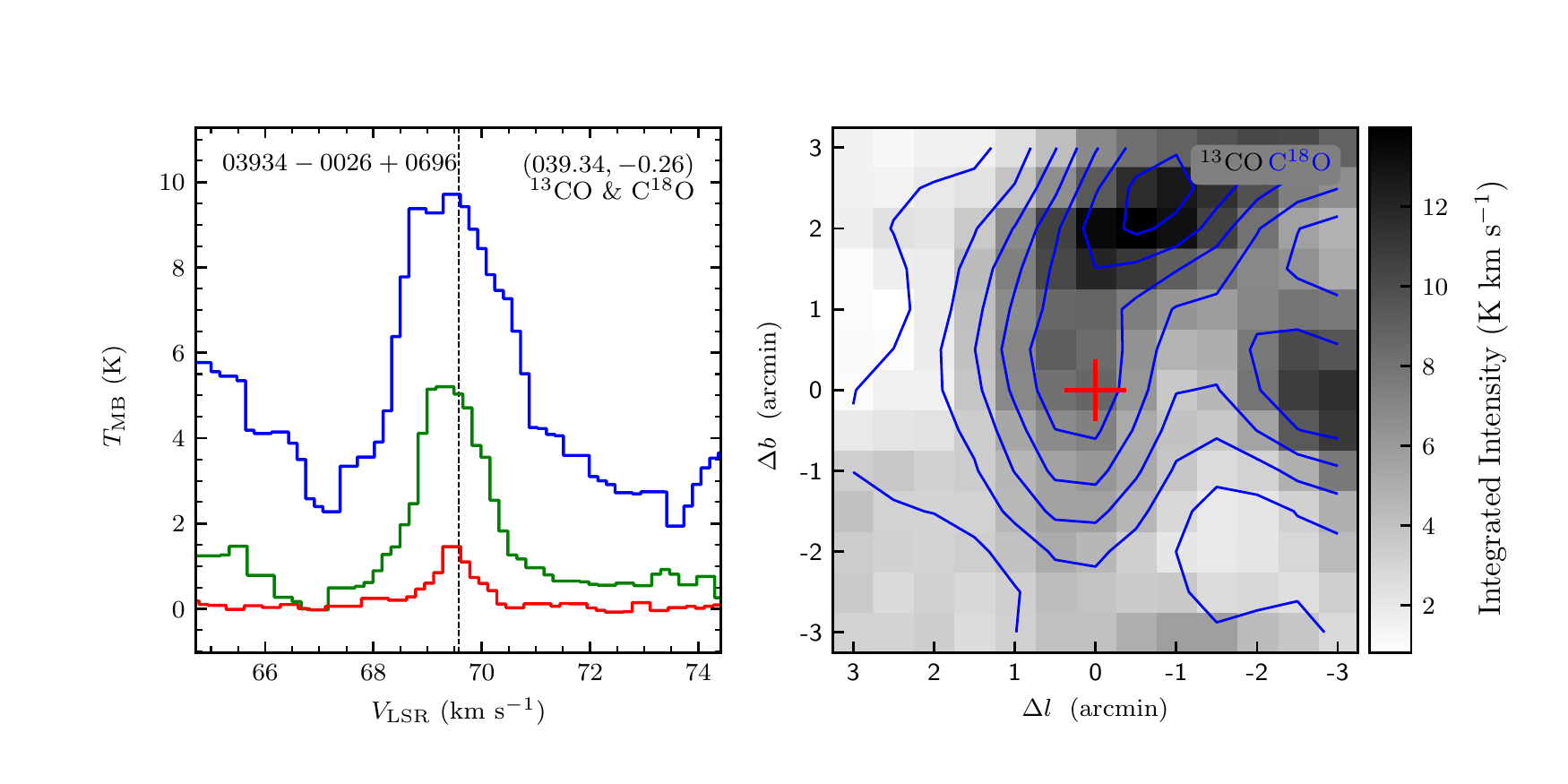}
\vspace{-0.5cm}

\includegraphics[width=9.0cm,angle=0]{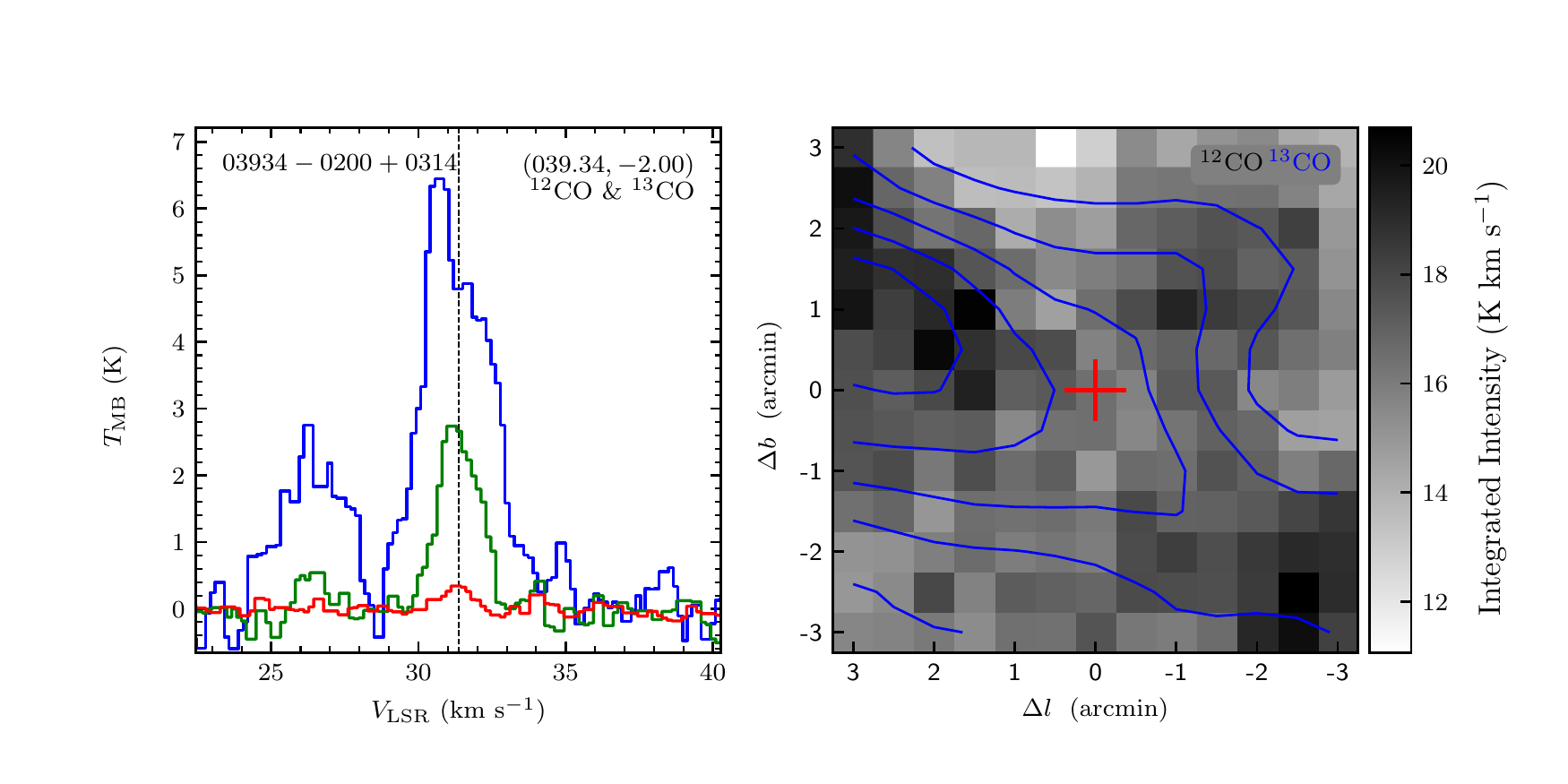}
\includegraphics[width=9.0cm,angle=0]{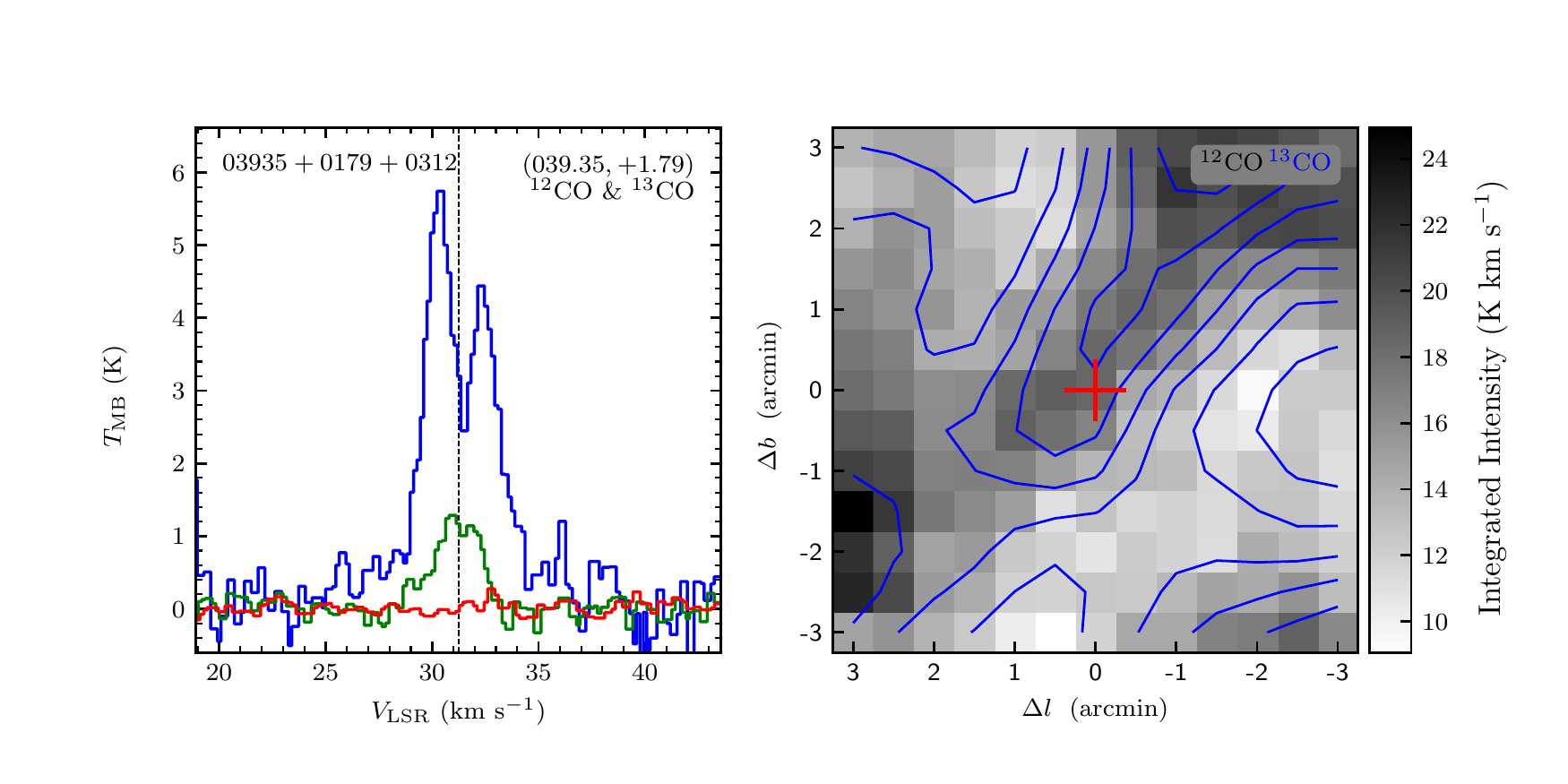}
\vspace{-0.5cm}

\includegraphics[width=9.0cm,angle=0]{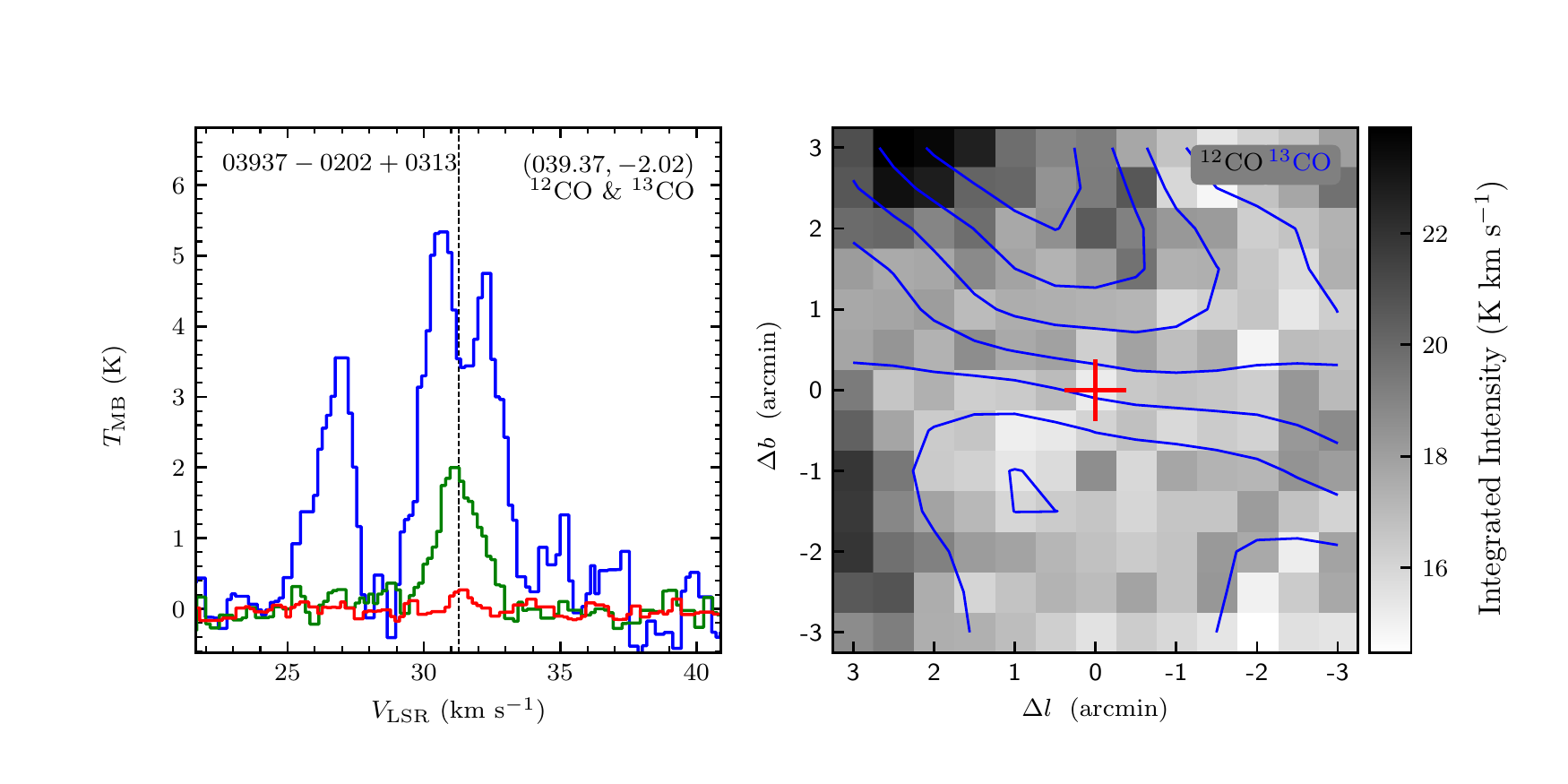}
\includegraphics[width=9.0cm,angle=0]{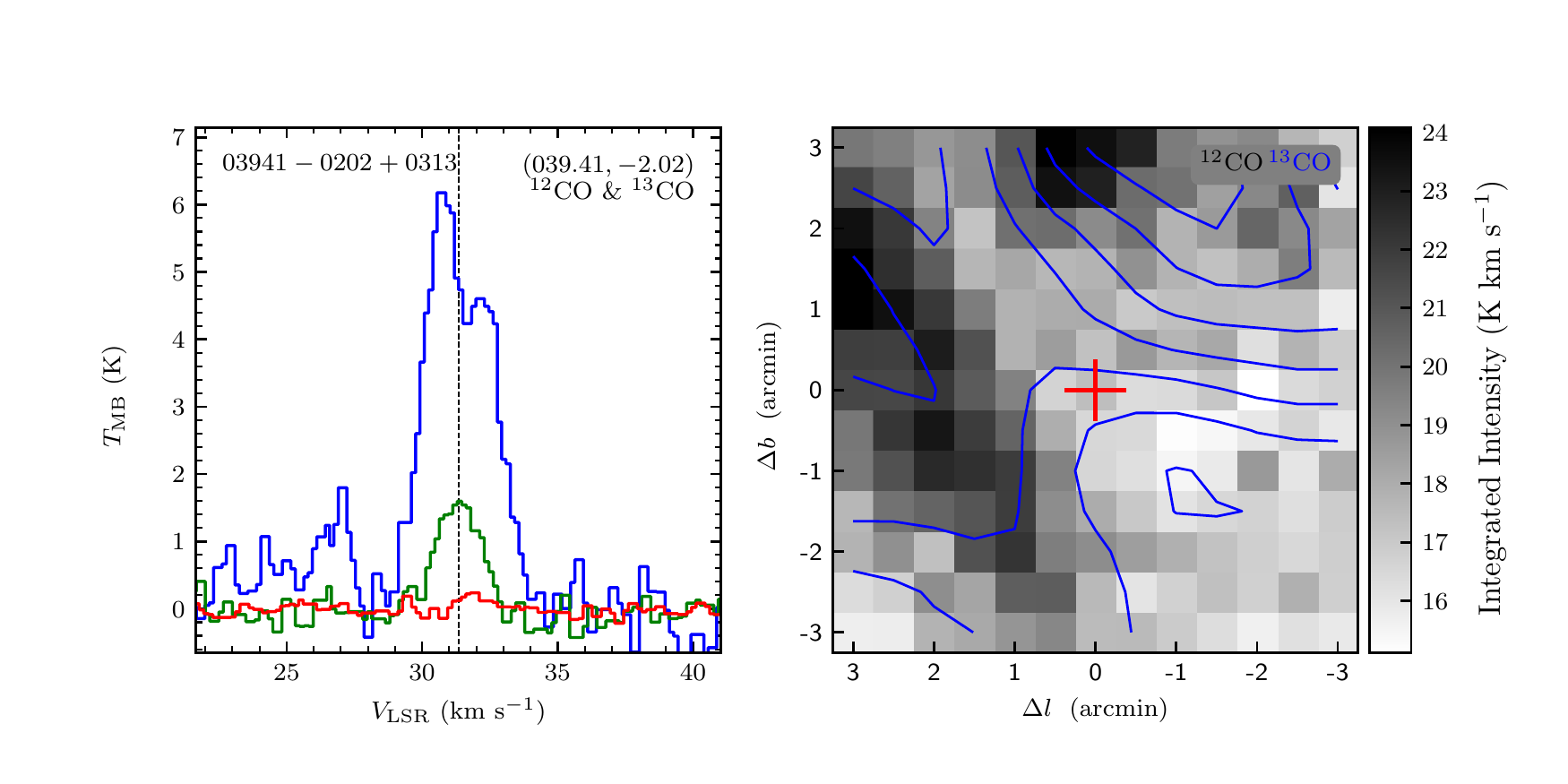}
\vspace{-0.5cm}

\includegraphics[width=9.0cm,angle=0]{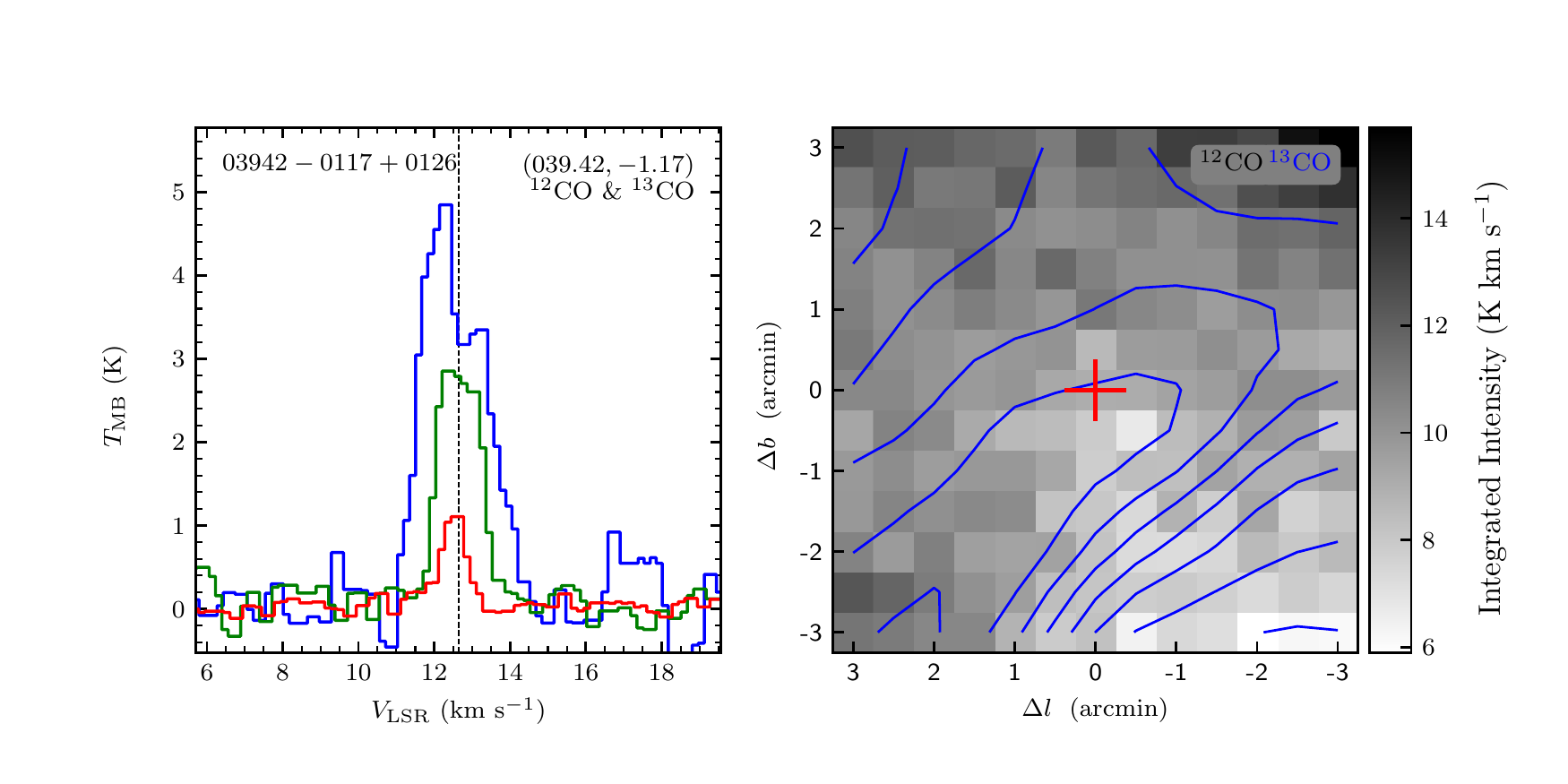}
\includegraphics[width=9.0cm,angle=0]{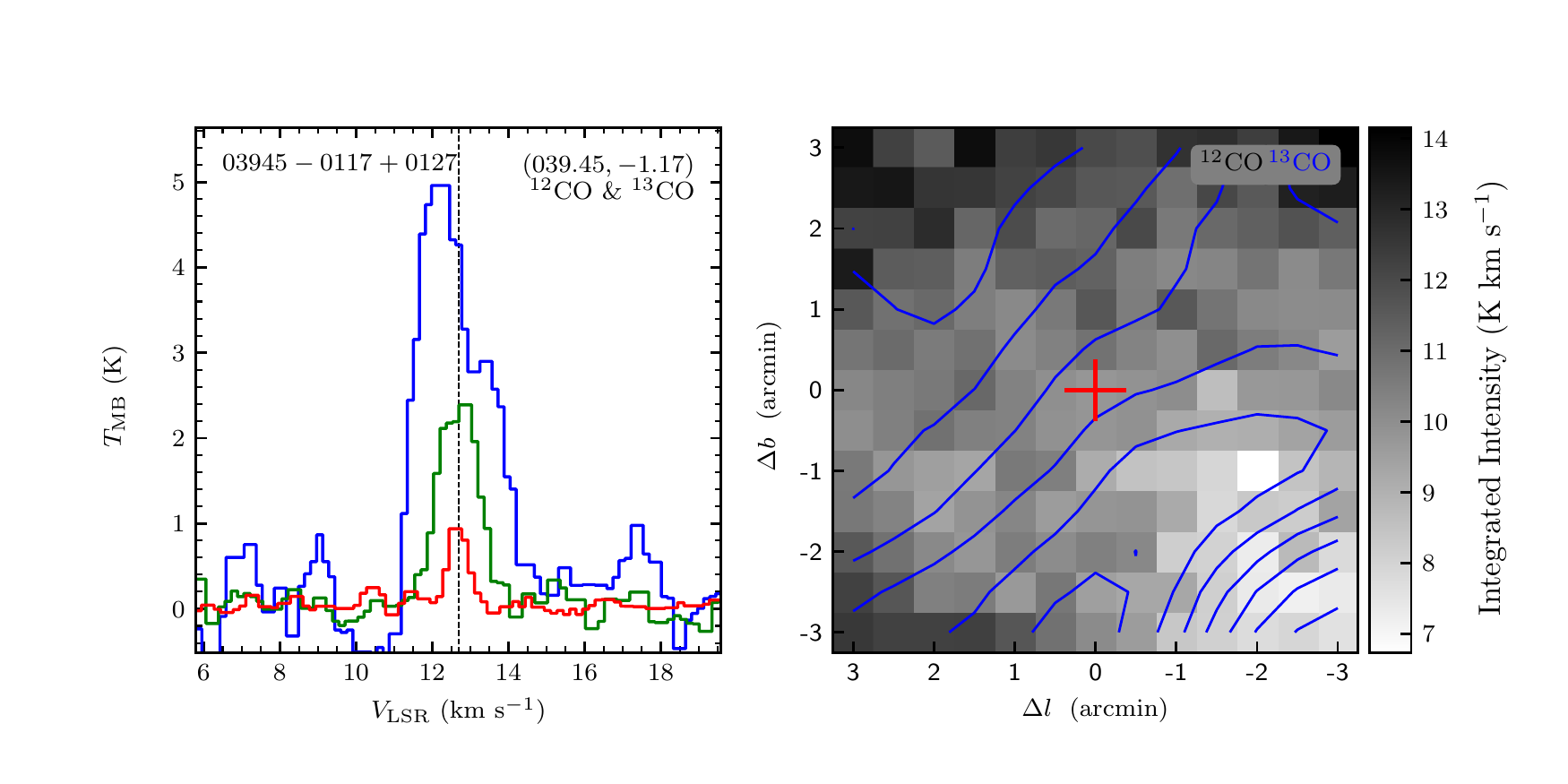}
\end{figure}
\clearpage

\begin{figure}
\includegraphics[width=9.0cm,angle=0]{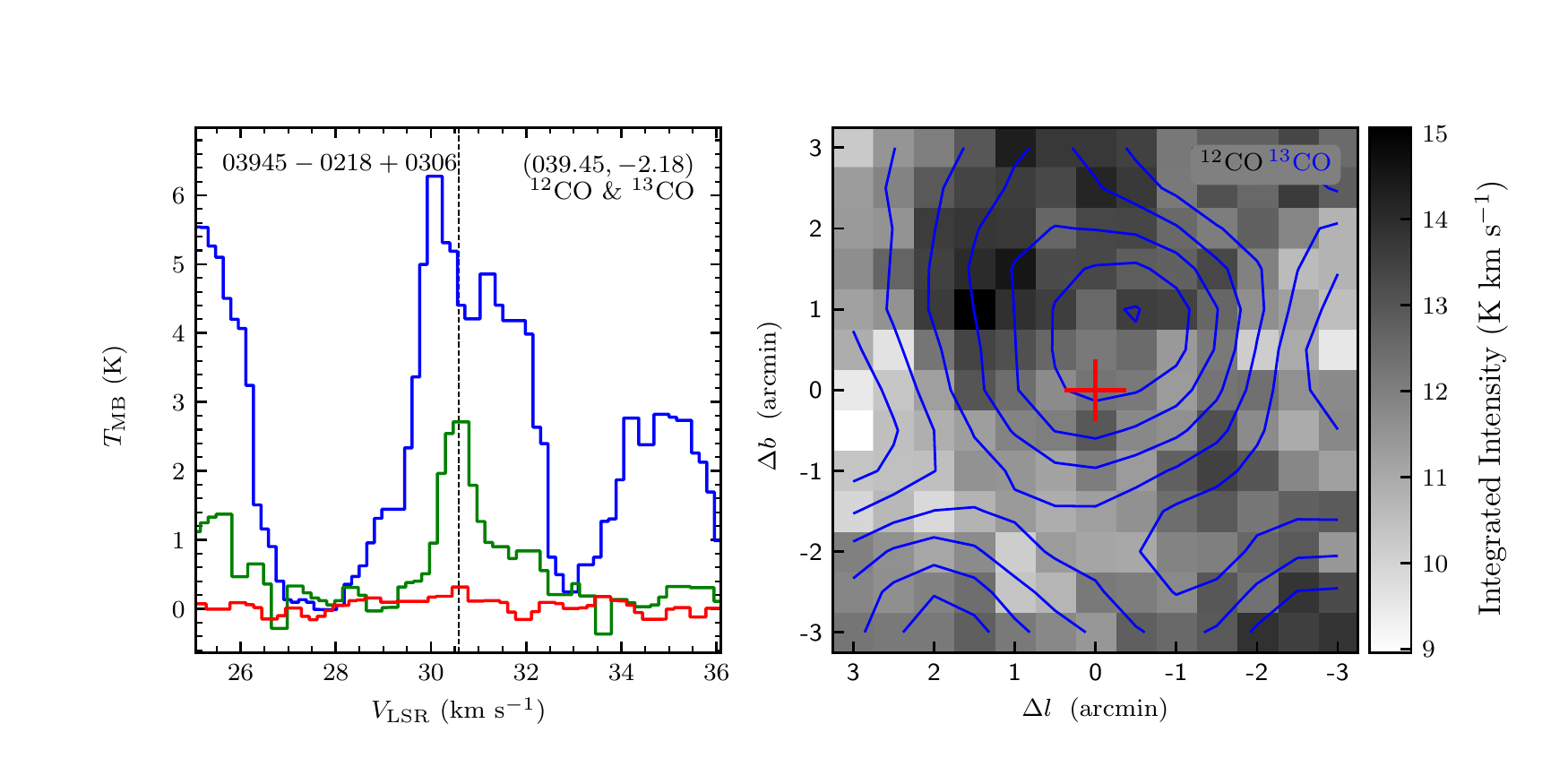}
\includegraphics[width=9.0cm,angle=0]{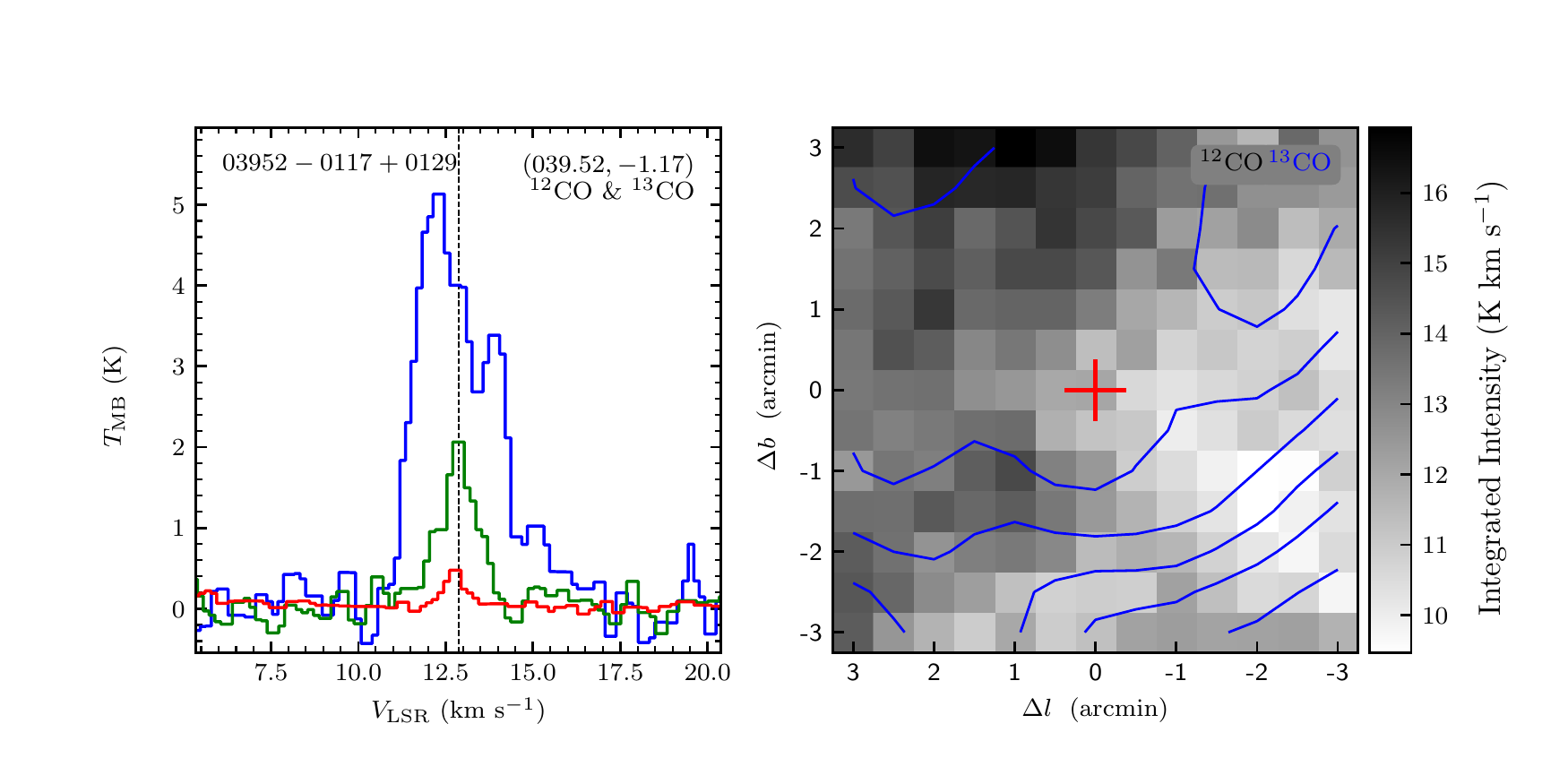}
\vspace{-0.5cm}

\includegraphics[width=9.0cm,angle=0]{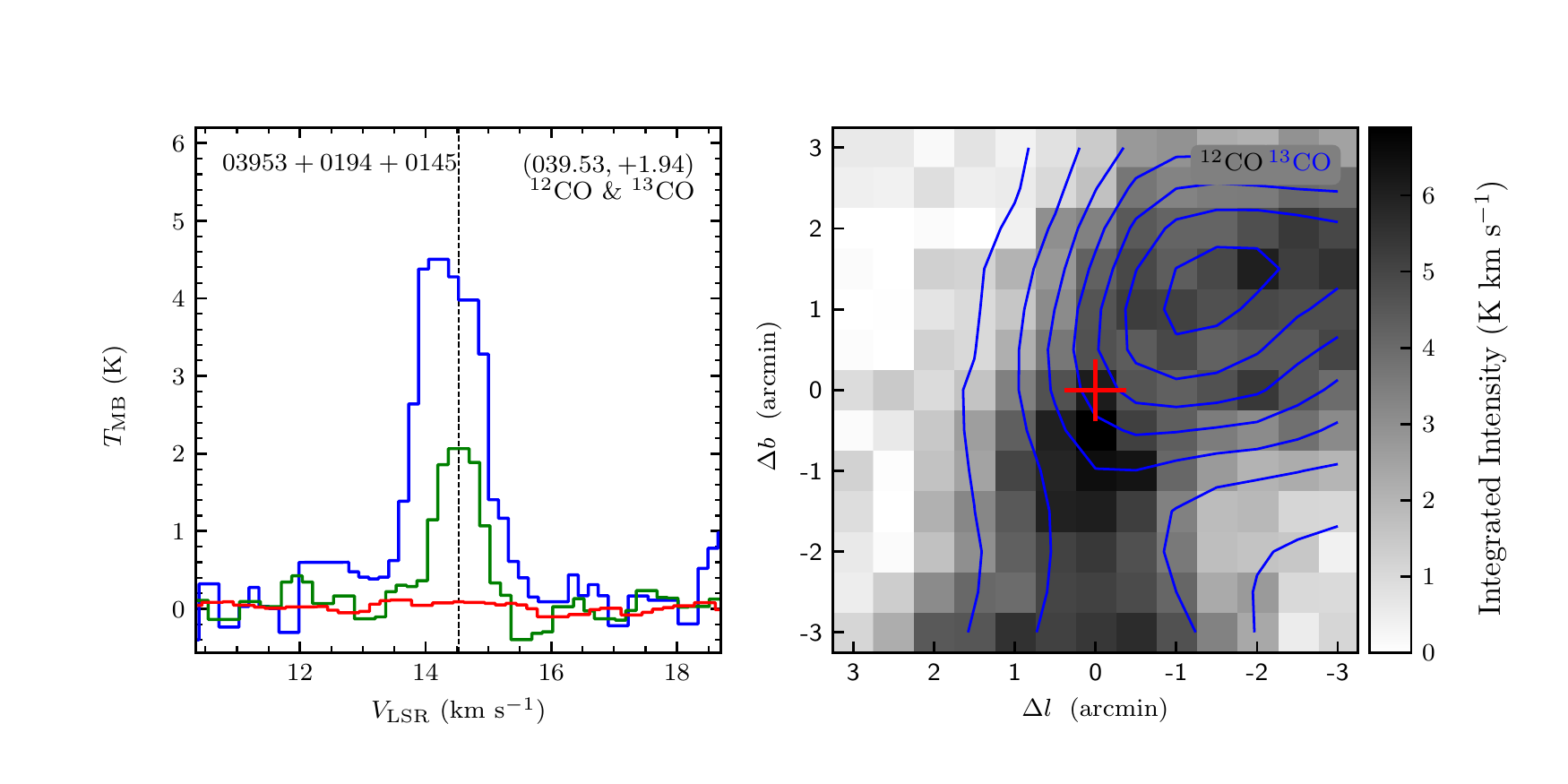}
\includegraphics[width=9.0cm,angle=0]{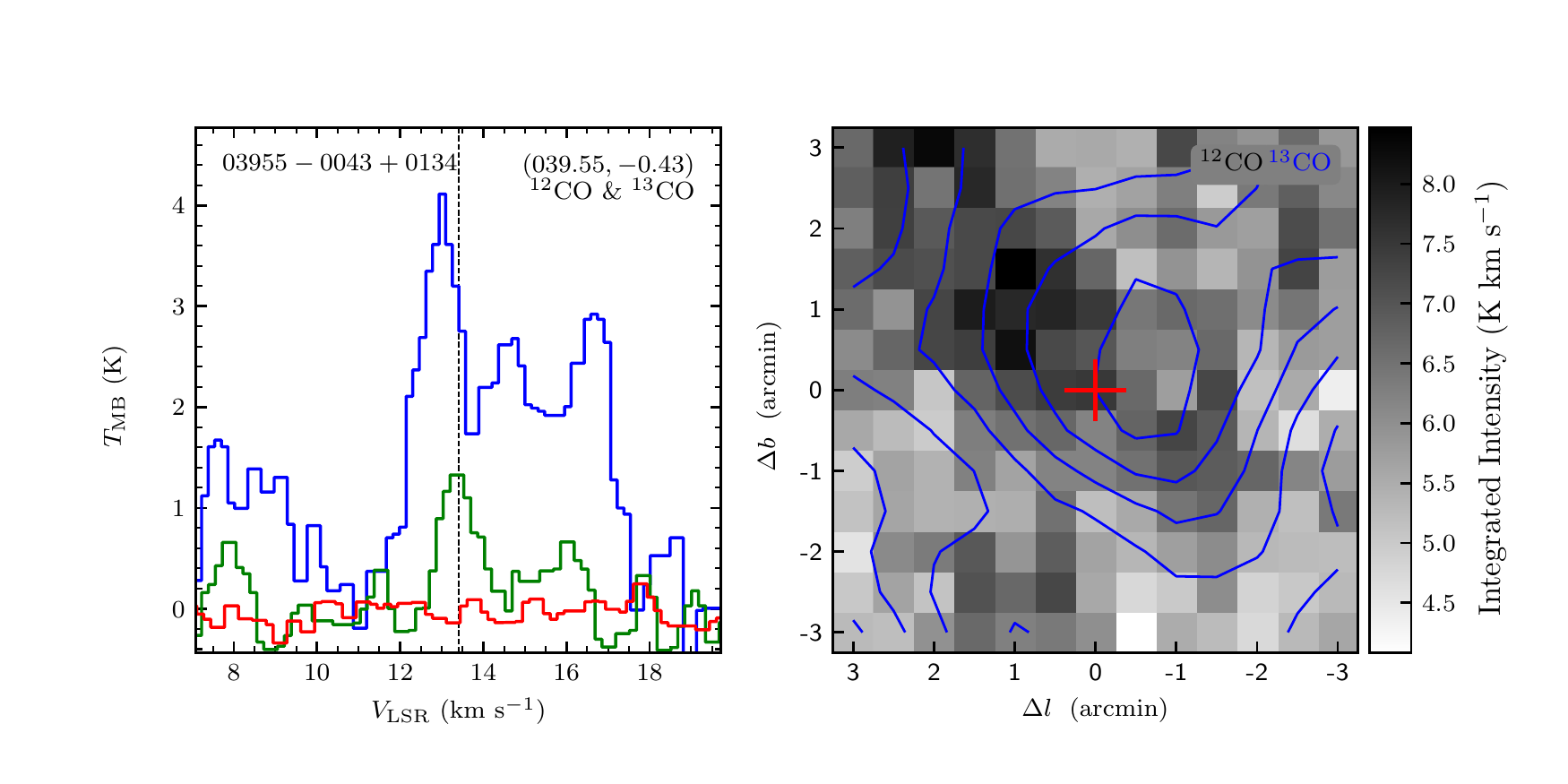}
\vspace{-0.5cm}

\includegraphics[width=9.0cm,angle=0]{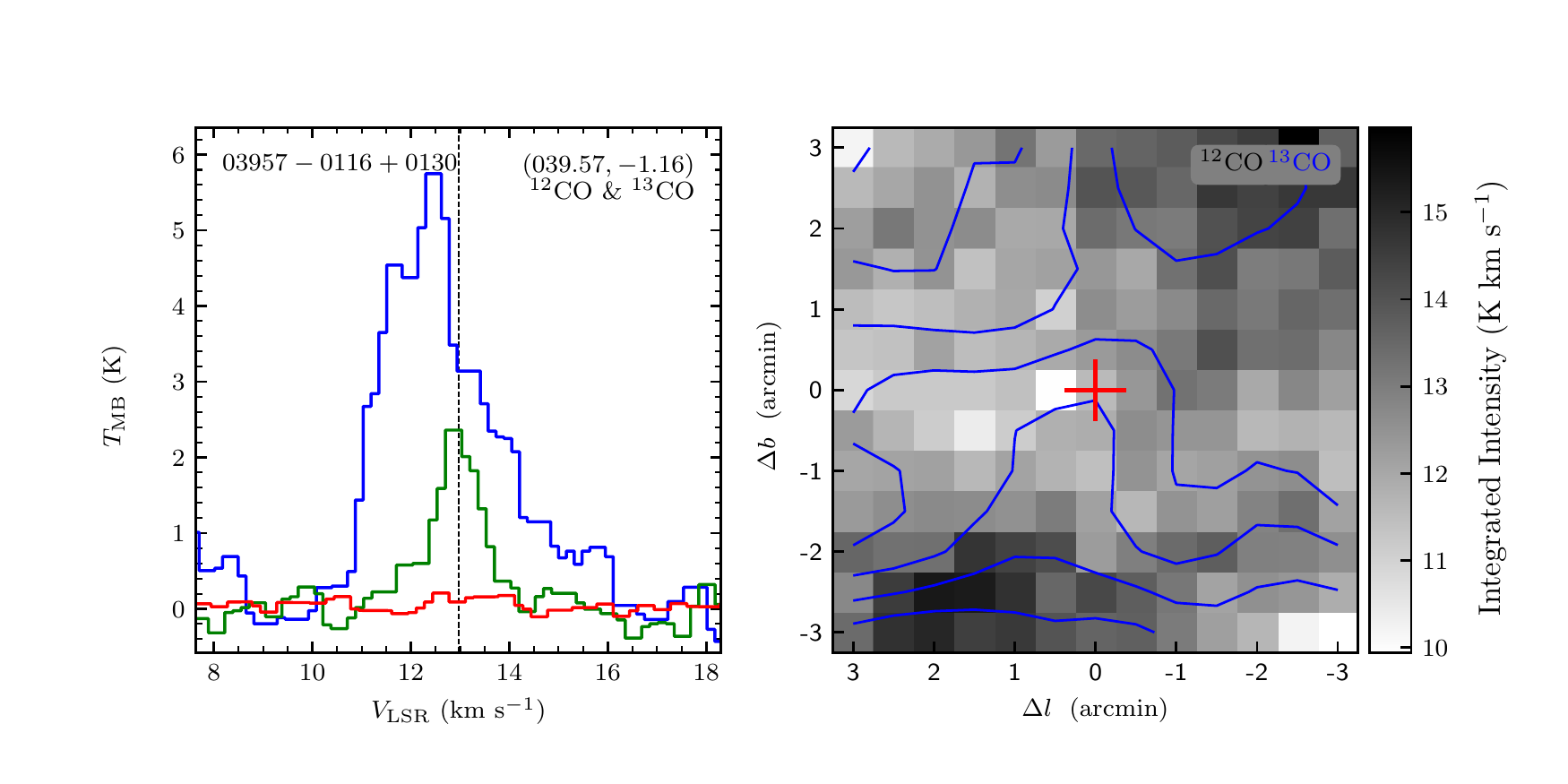}
\includegraphics[width=9.0cm,angle=0]{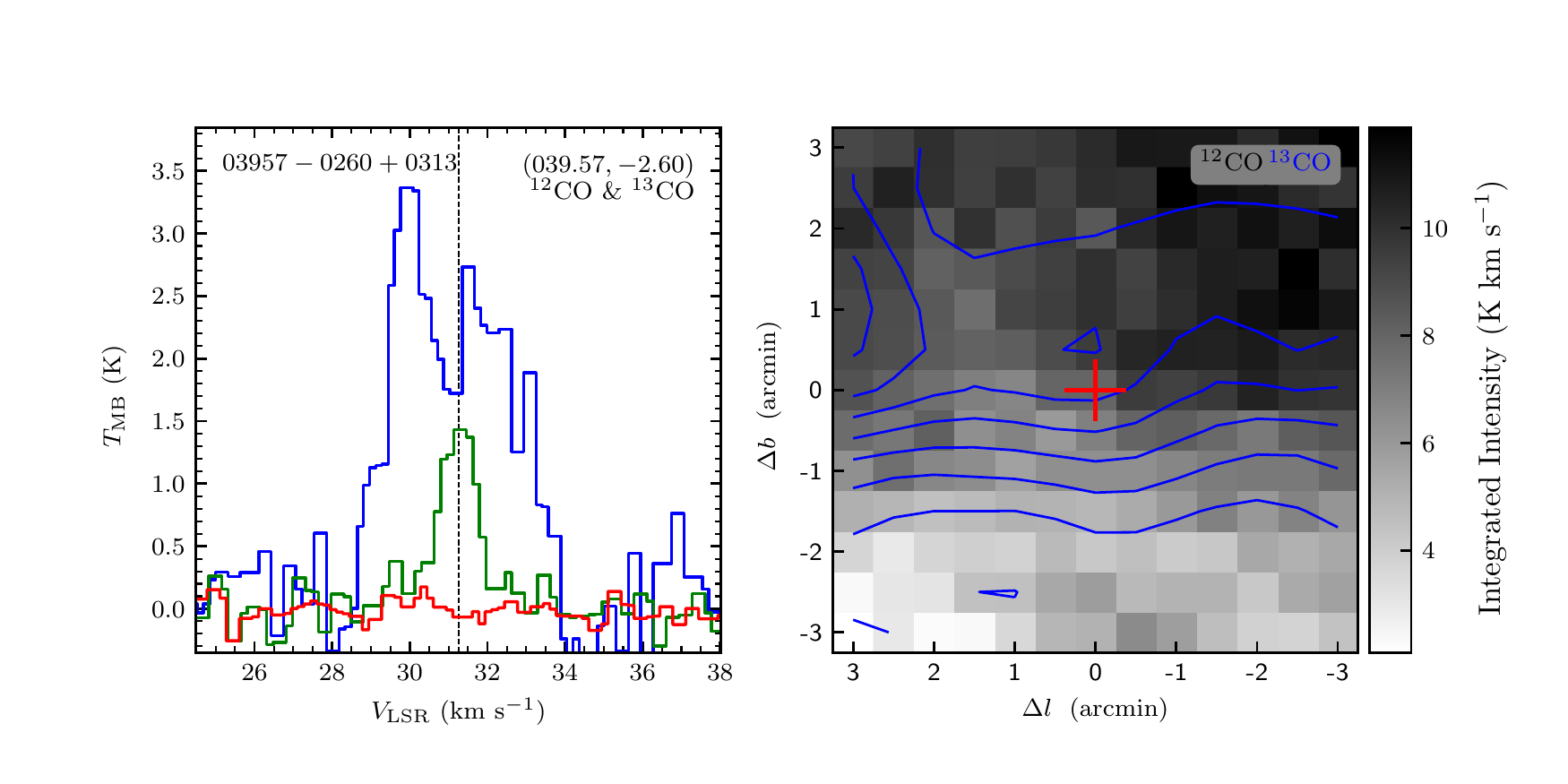}
\vspace{-0.5cm}

\includegraphics[width=9.0cm,angle=0]{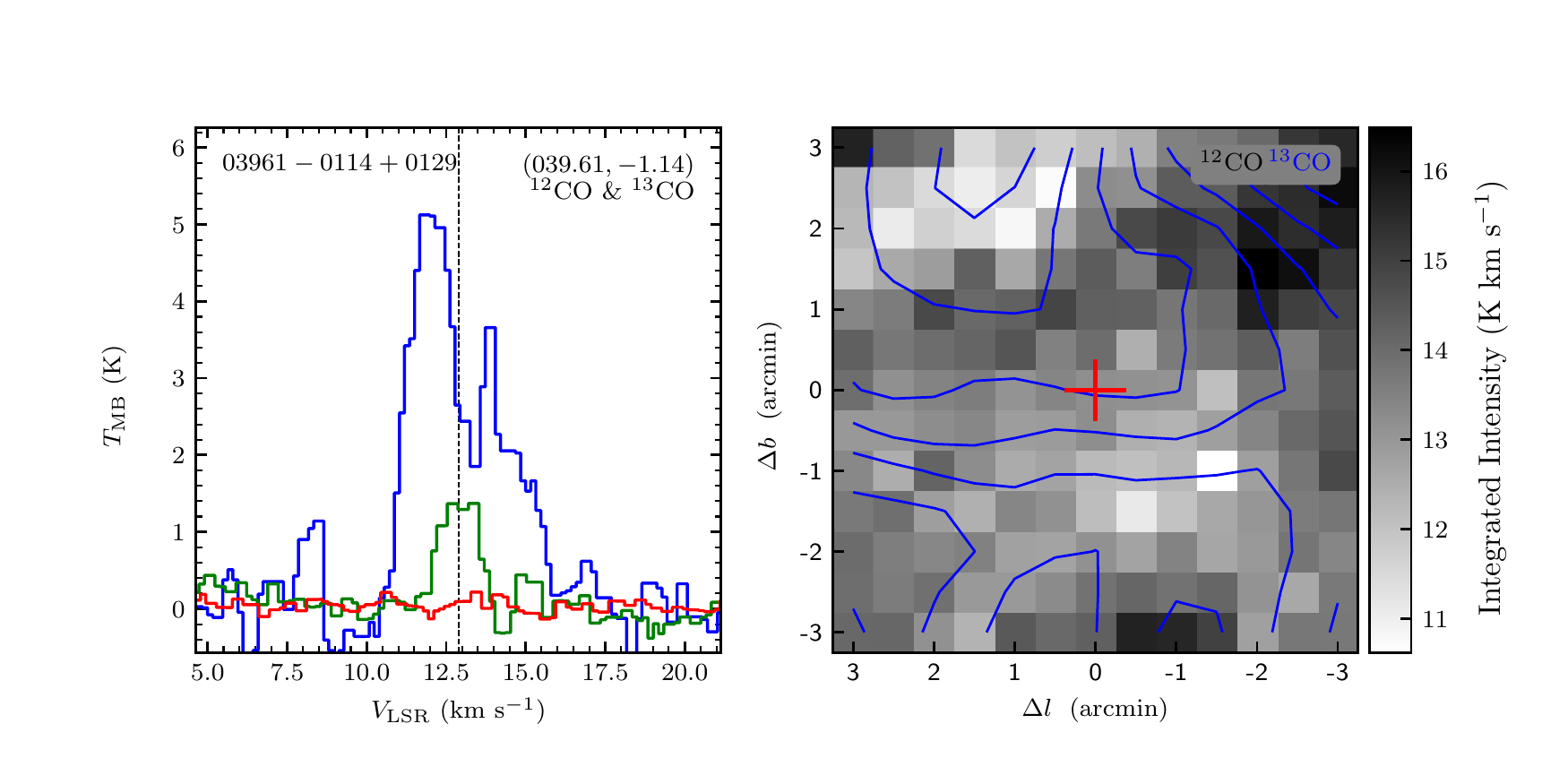}
\includegraphics[width=9.0cm,angle=0]{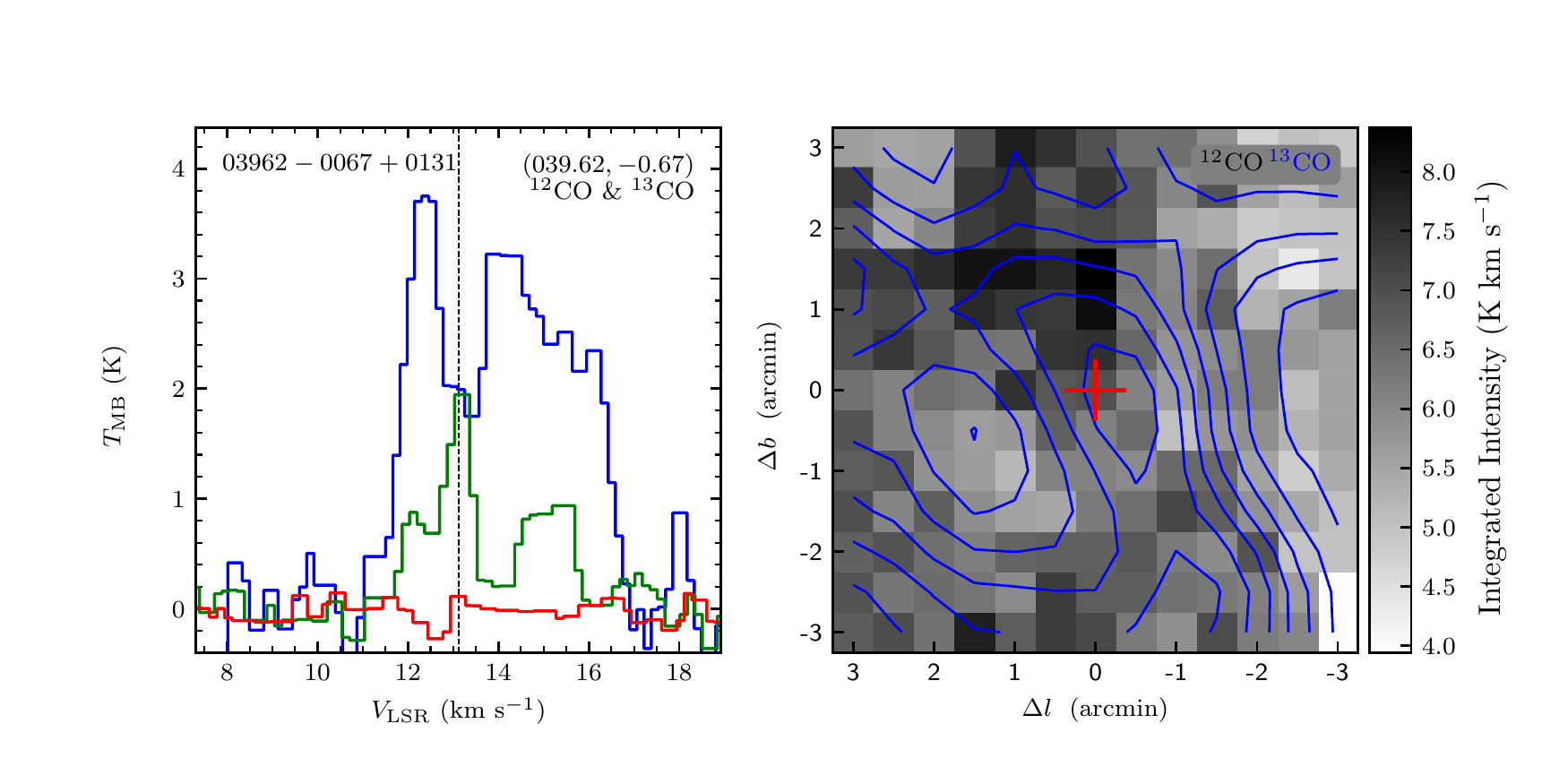}
\vspace{-0.5cm}

\includegraphics[width=9.0cm,angle=0]{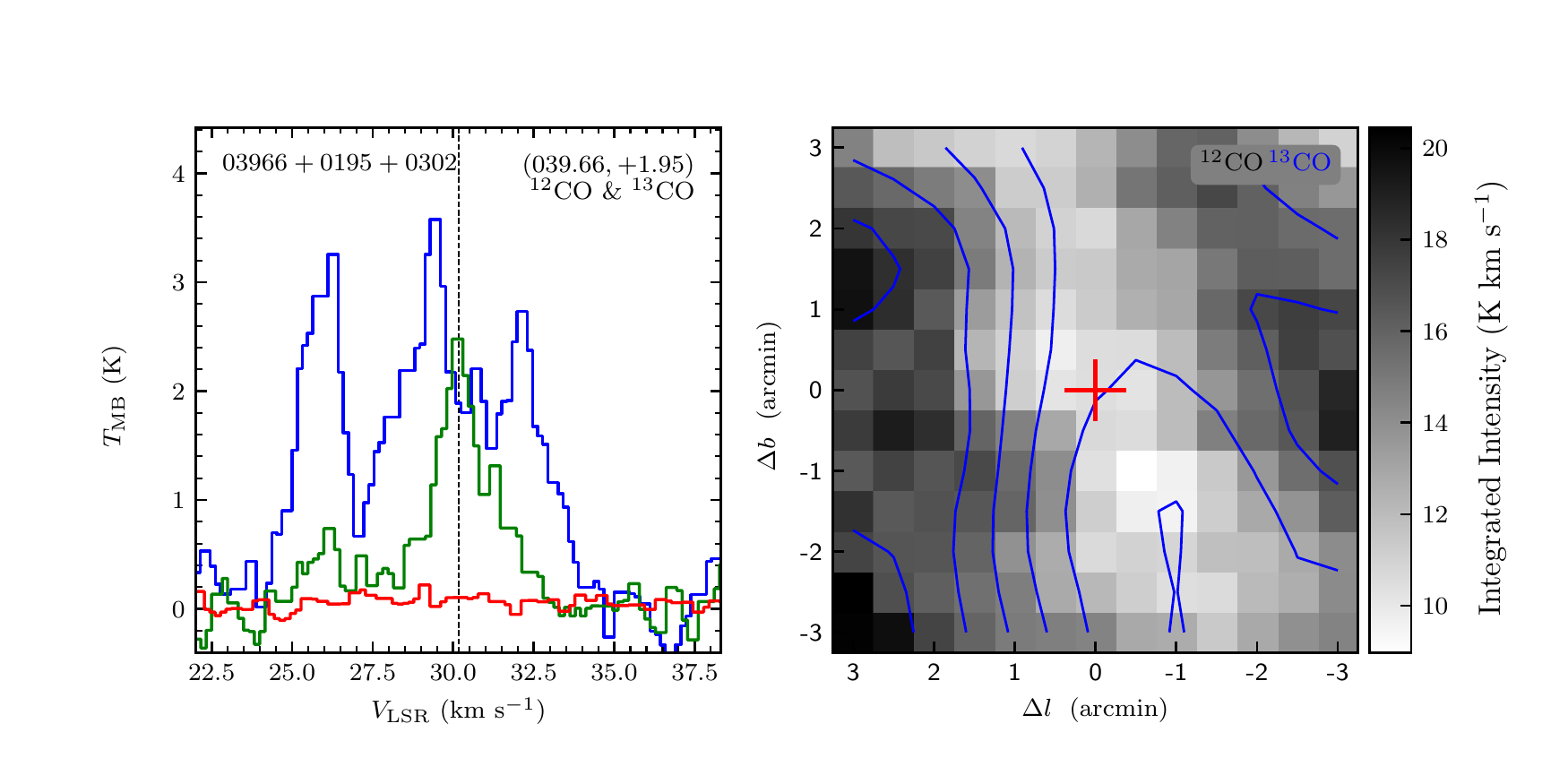}
\includegraphics[width=9.0cm,angle=0]{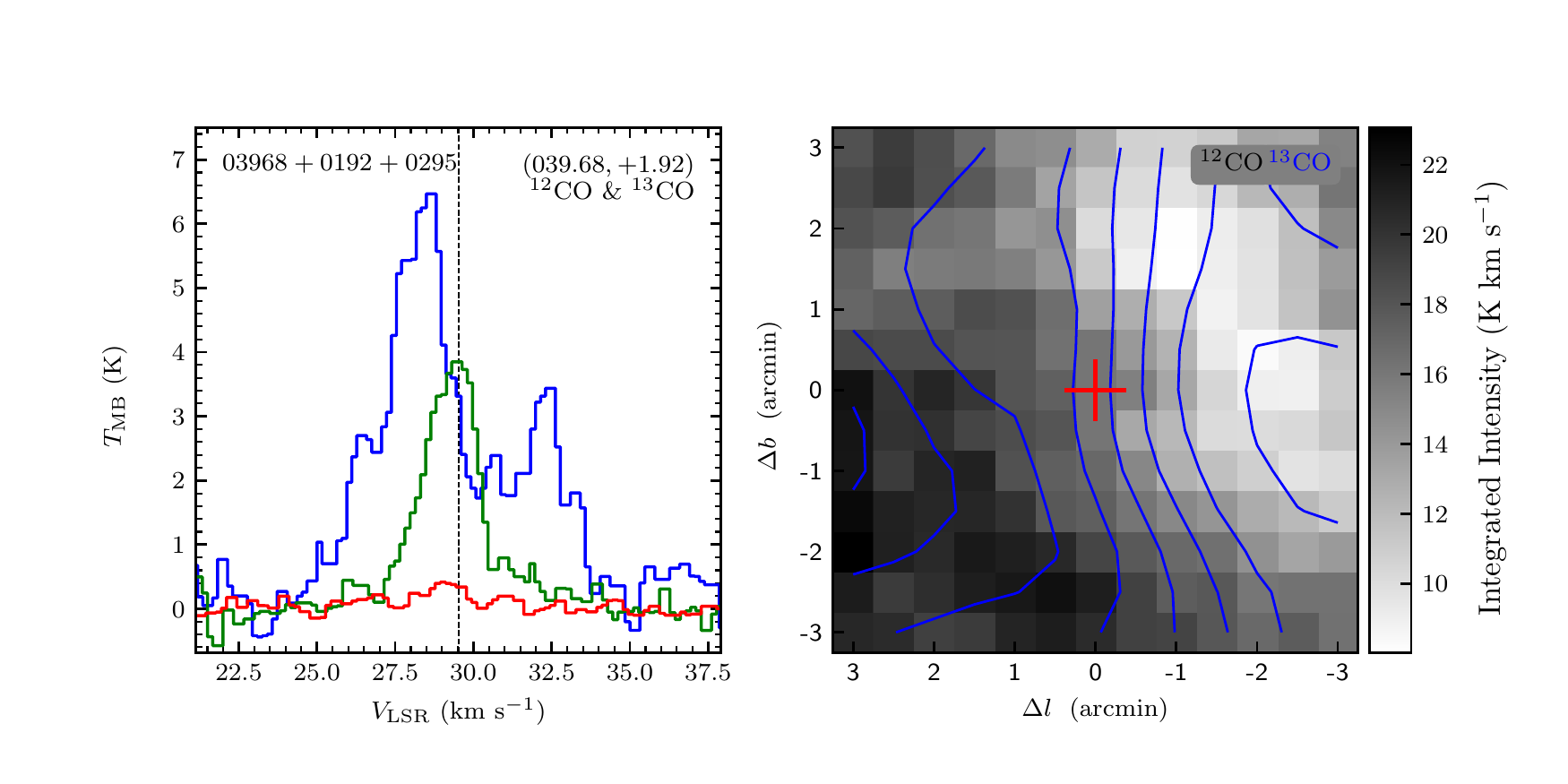}
\end{figure}
\clearpage

\begin{figure}
\includegraphics[width=9.0cm,angle=0]{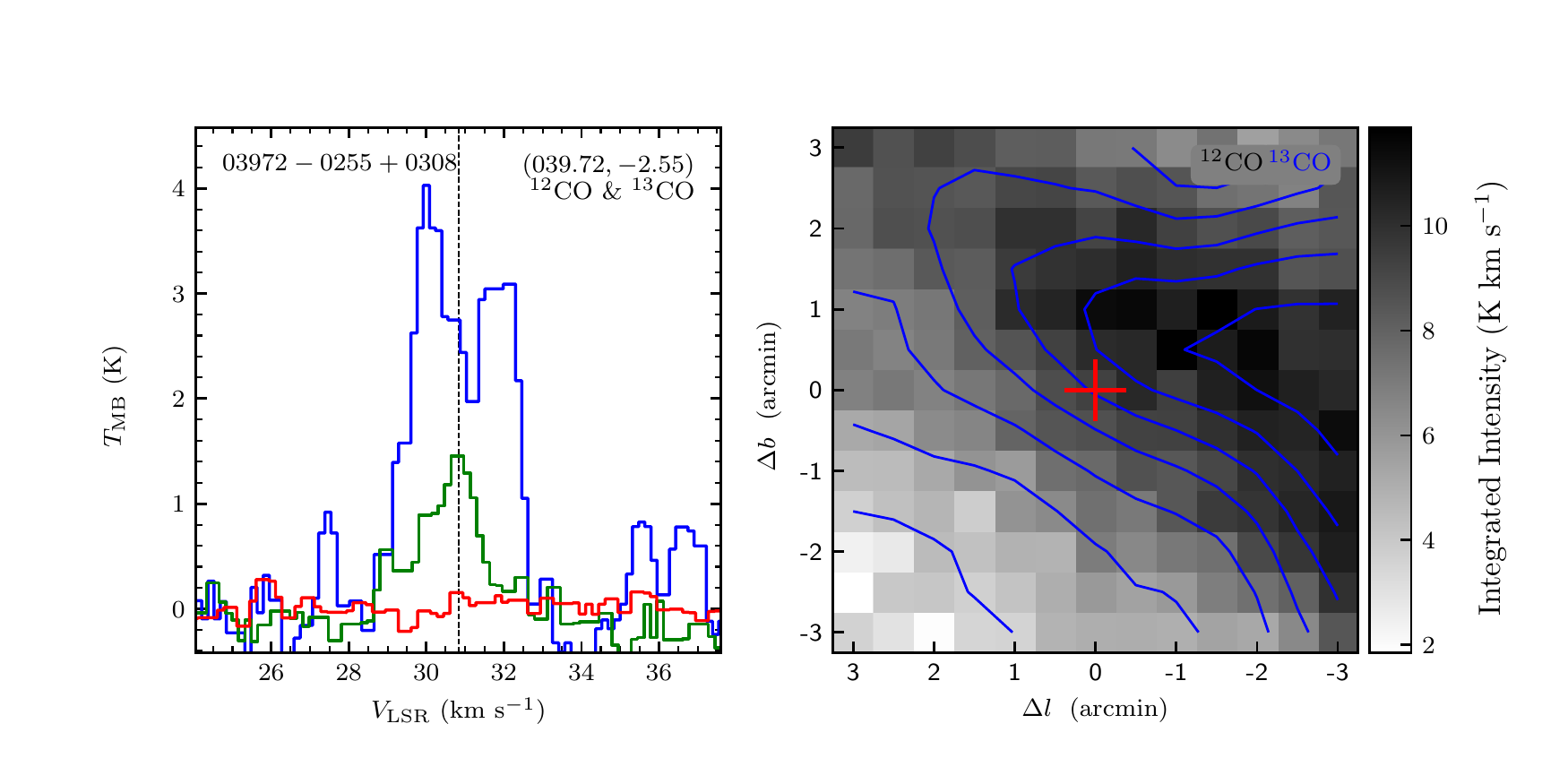}
\includegraphics[width=9.0cm,angle=0]{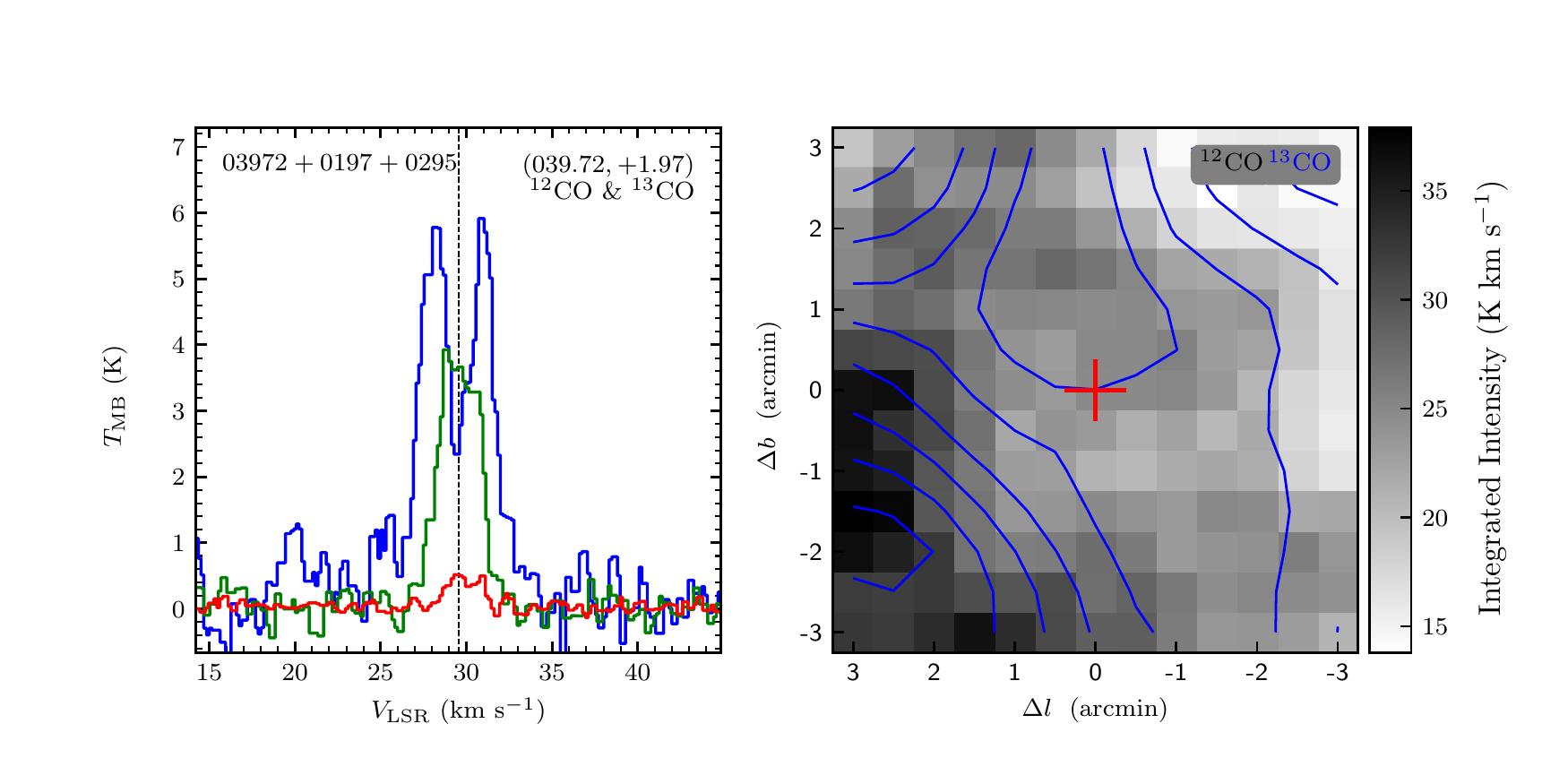}
\vspace{-0.5cm}

\includegraphics[width=9.0cm,angle=0]{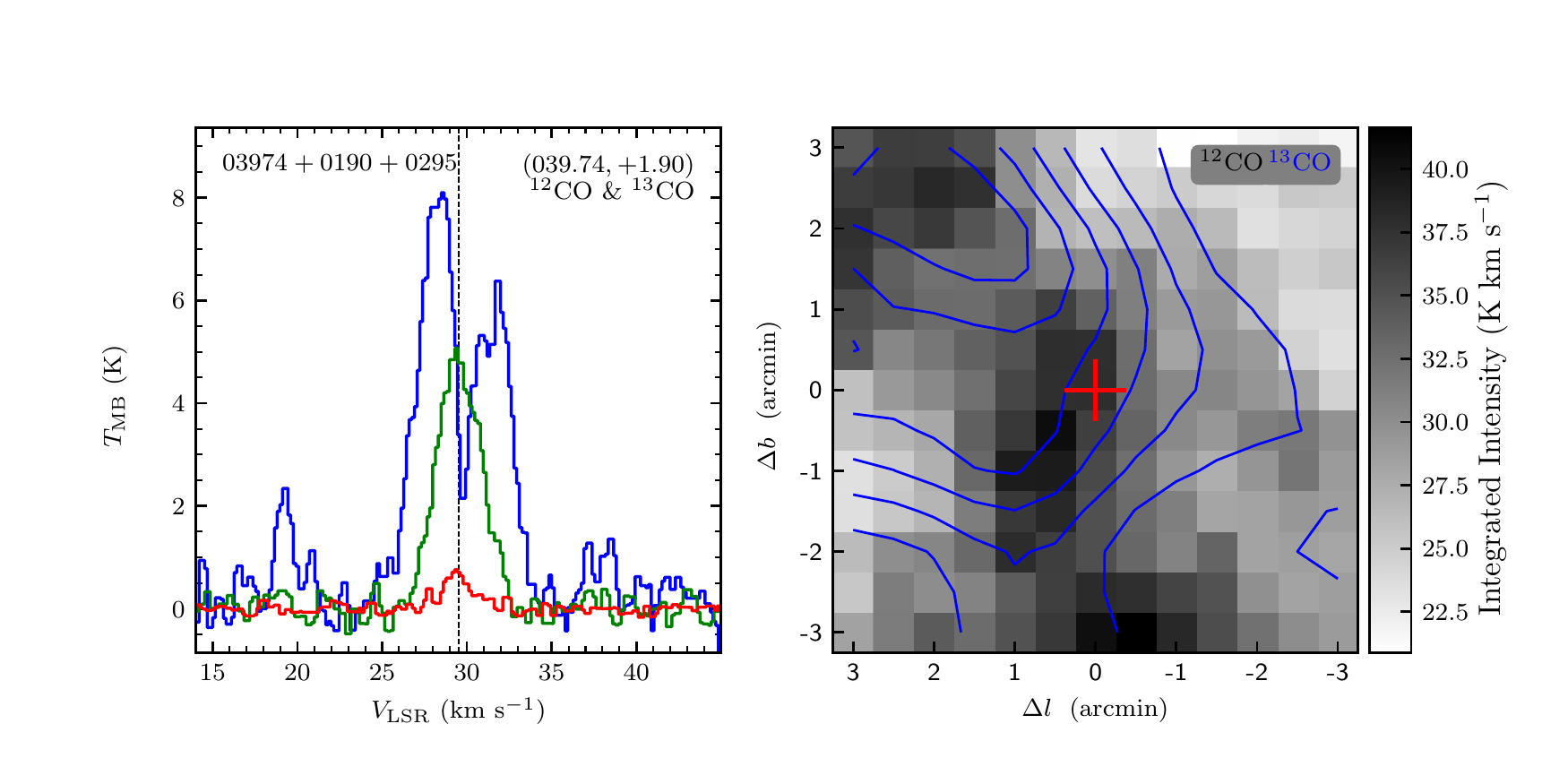}
\includegraphics[width=9.0cm,angle=0]{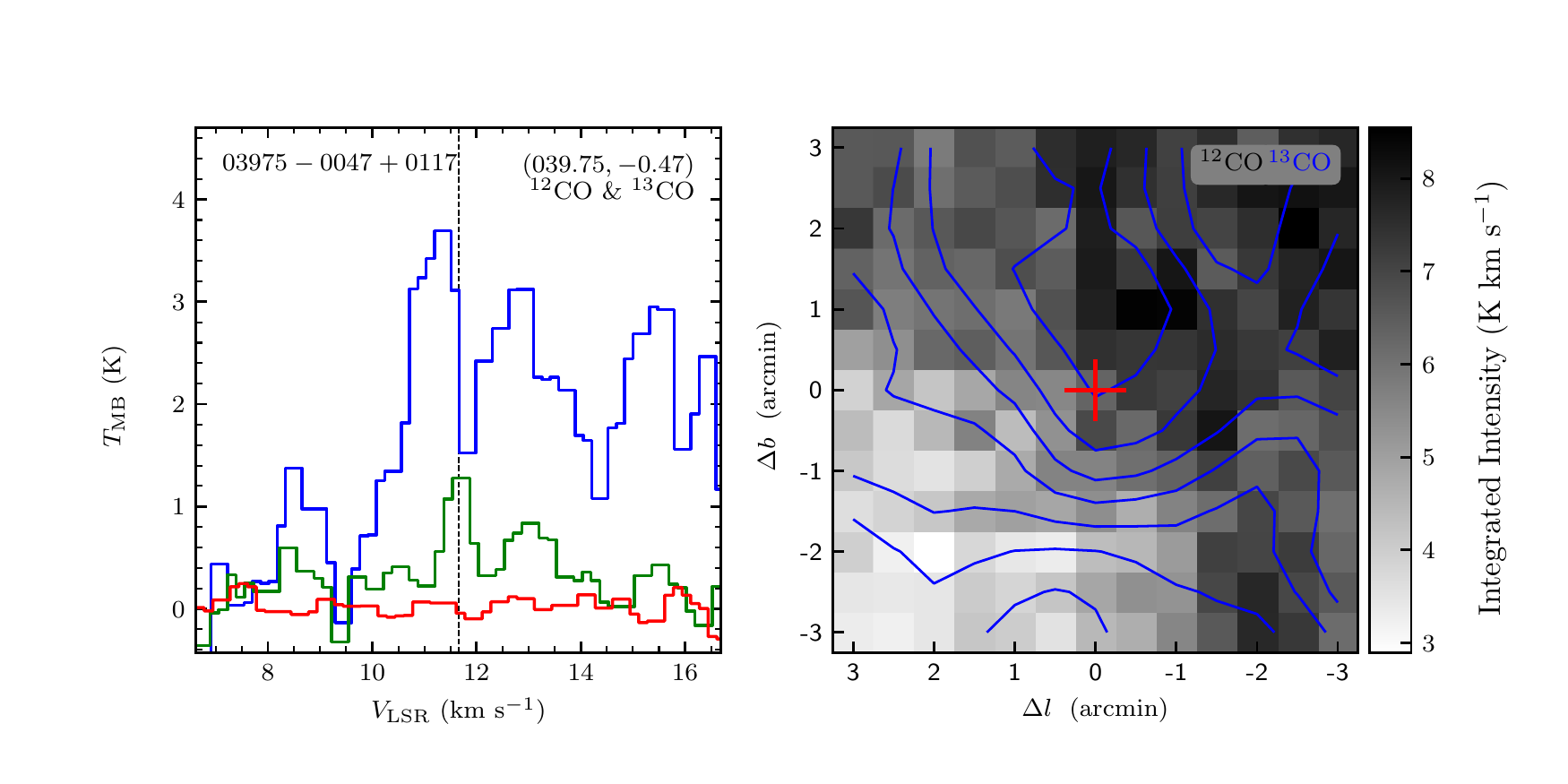}
\vspace{-0.5cm}

\includegraphics[width=9.0cm,angle=0]{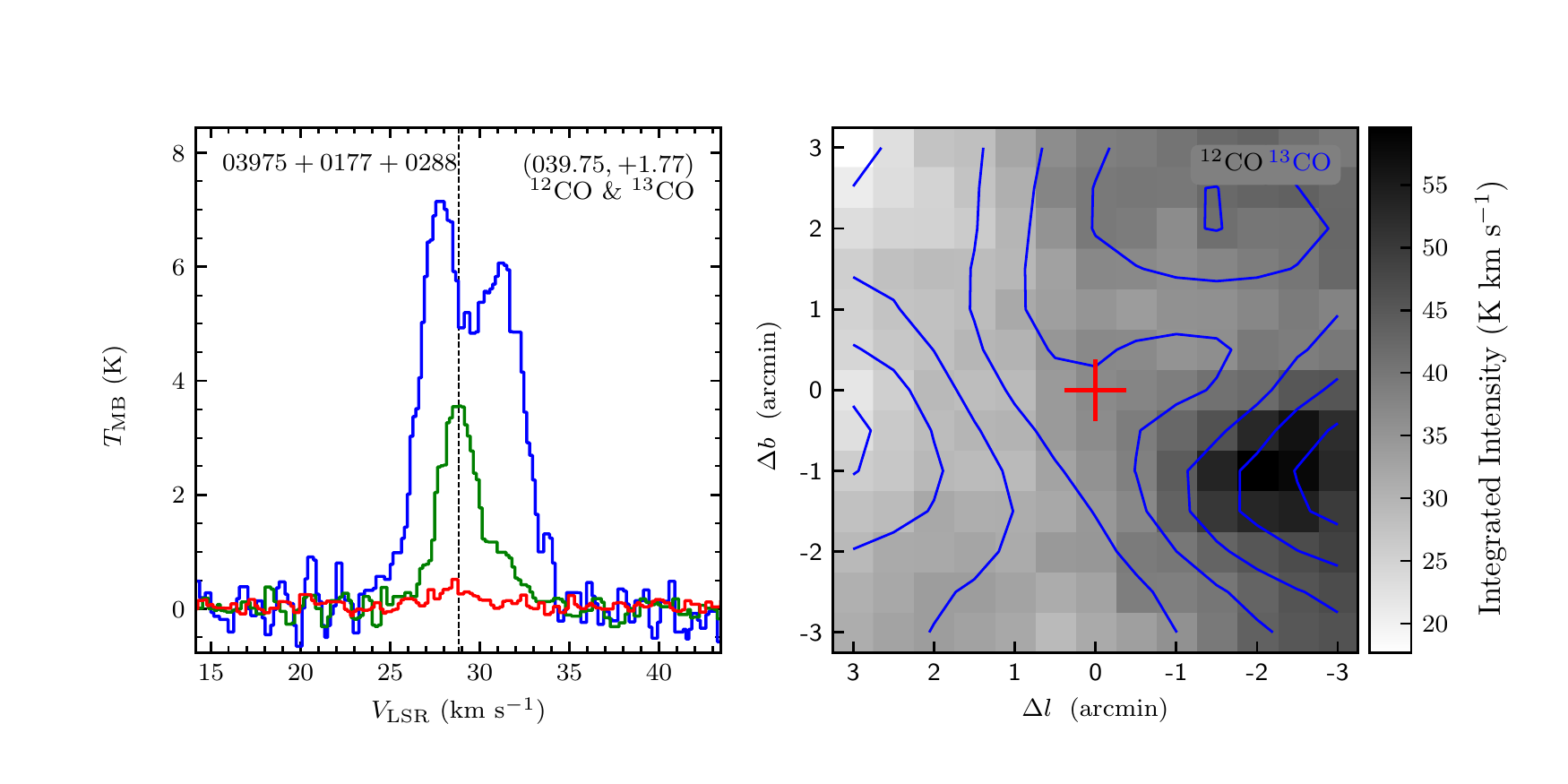}
\includegraphics[width=9.0cm,angle=0]{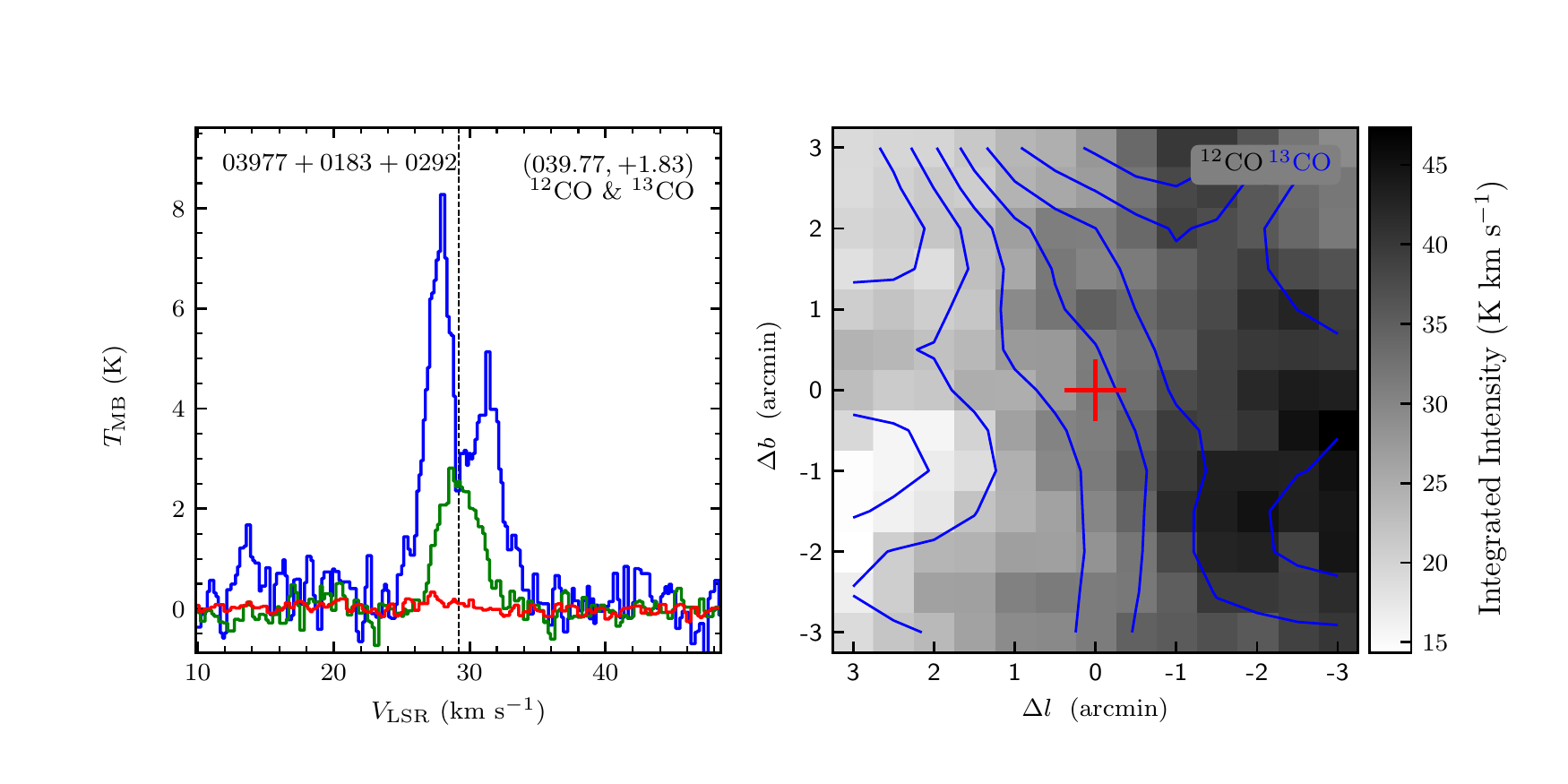}
\vspace{-0.5cm}

\includegraphics[width=9.0cm,angle=0]{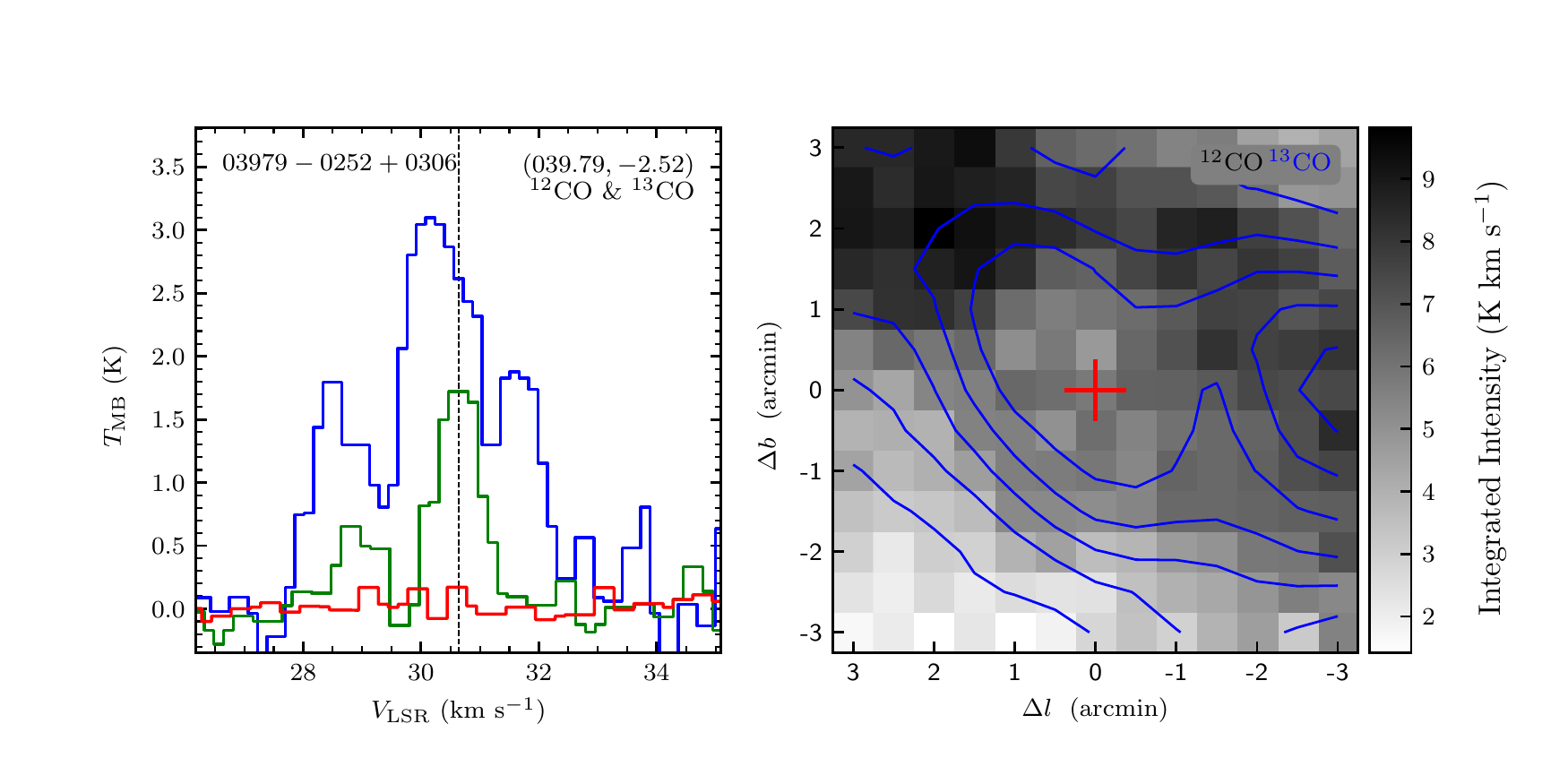}
\includegraphics[width=9.0cm,angle=0]{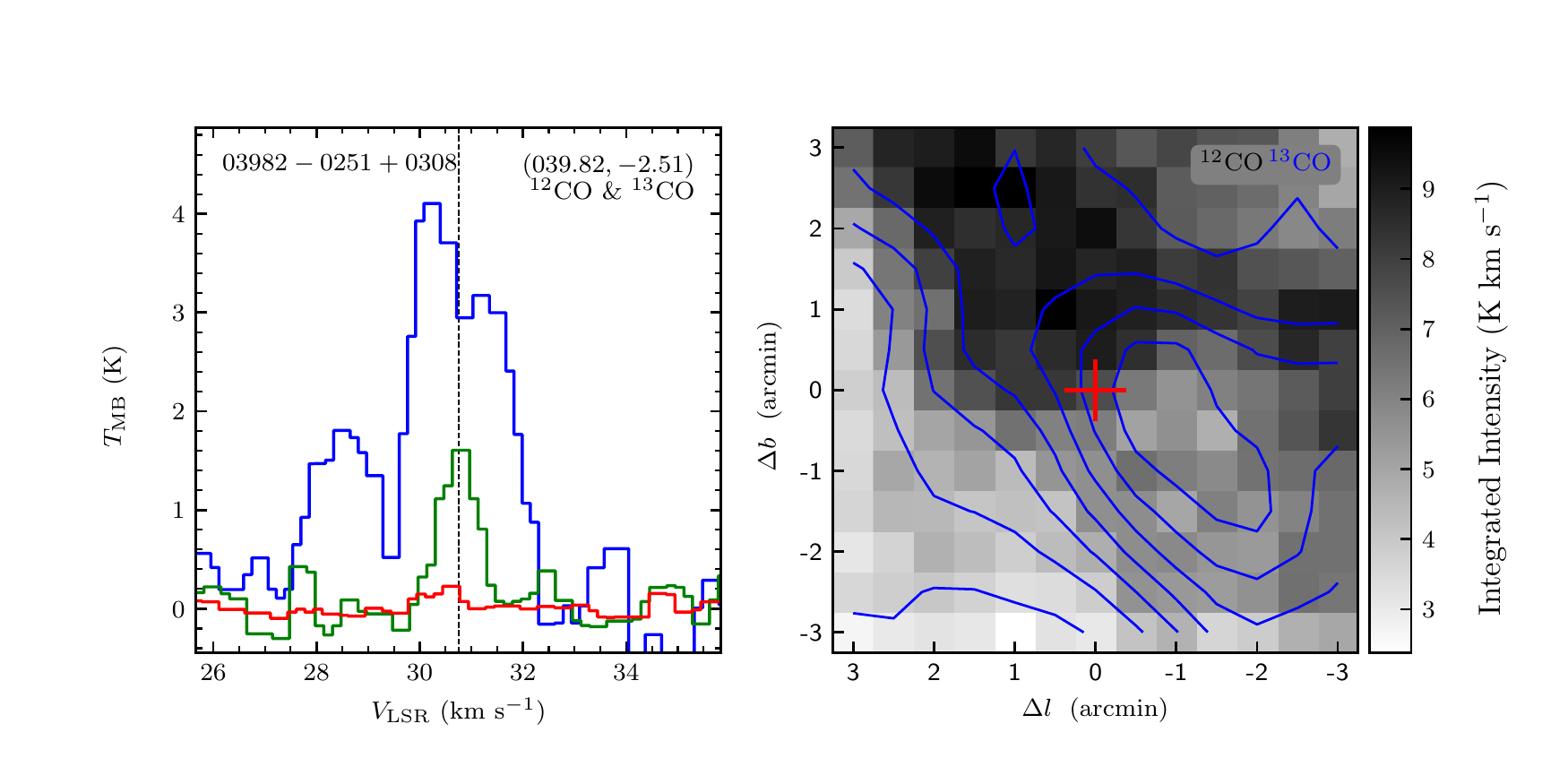}
\vspace{-0.5cm}

\includegraphics[width=9.0cm,angle=0]{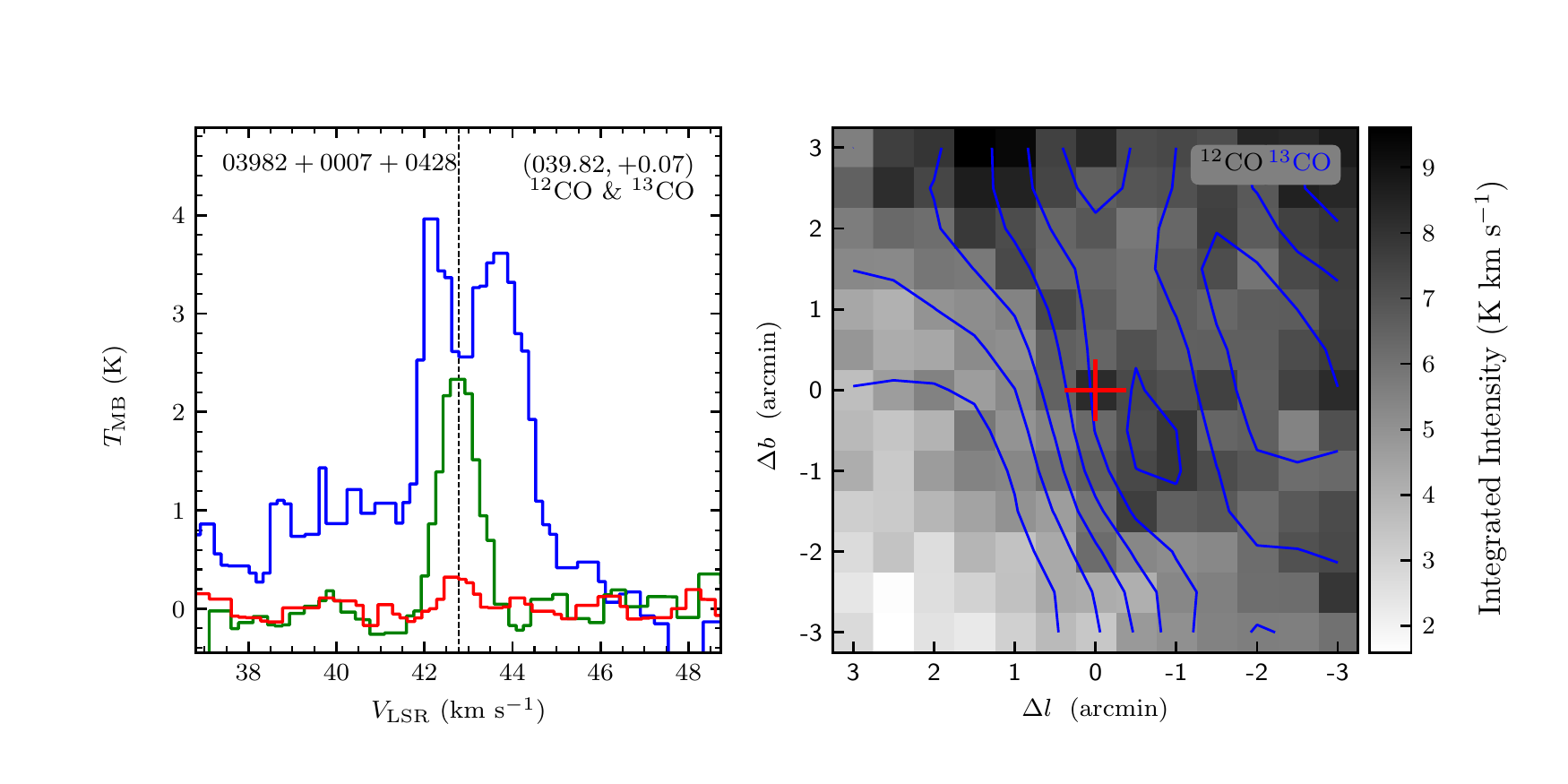}
\includegraphics[width=9.0cm,angle=0]{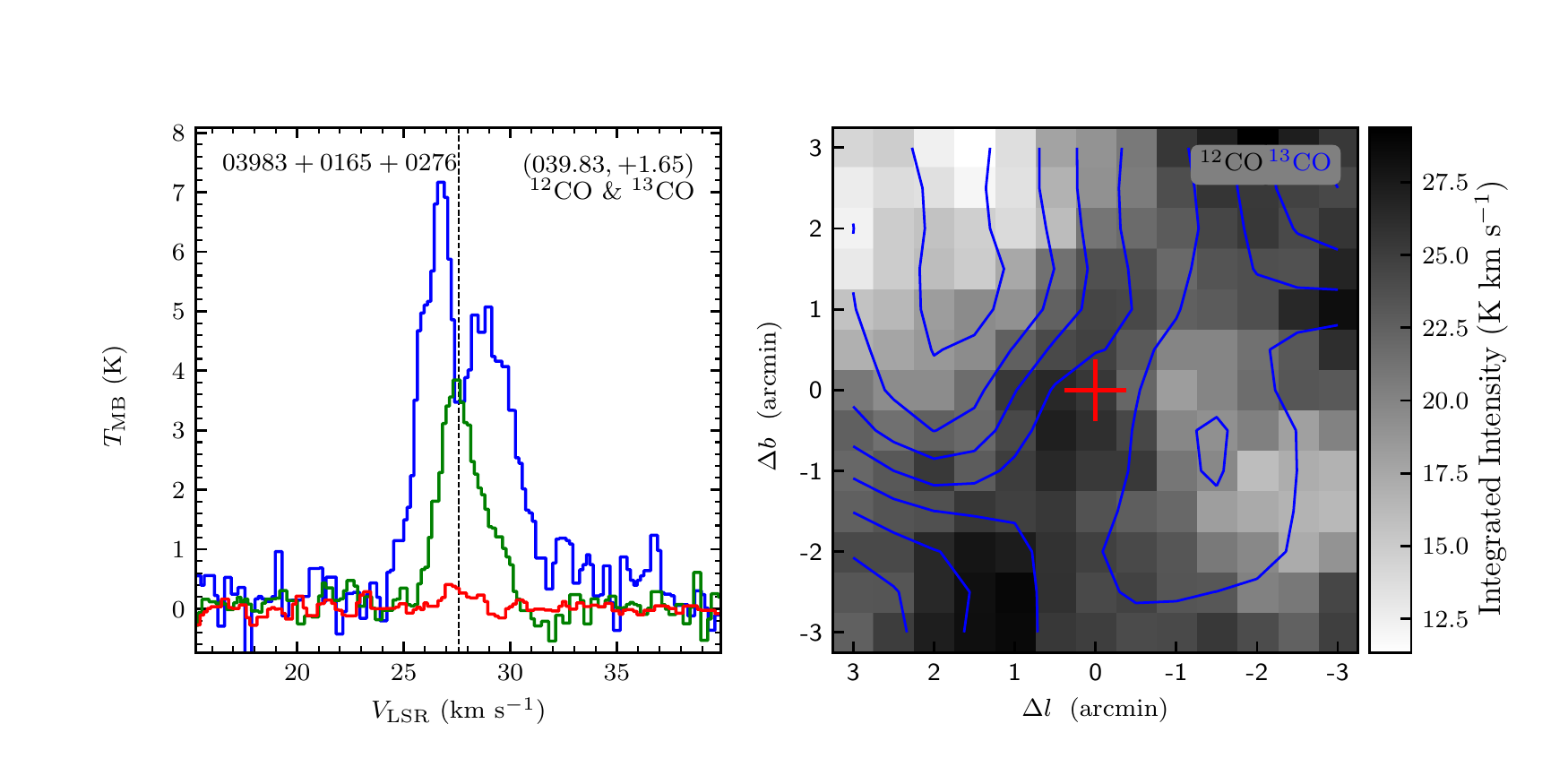}
\end{figure}
\clearpage

\begin{figure}
\includegraphics[width=9.0cm,angle=0]{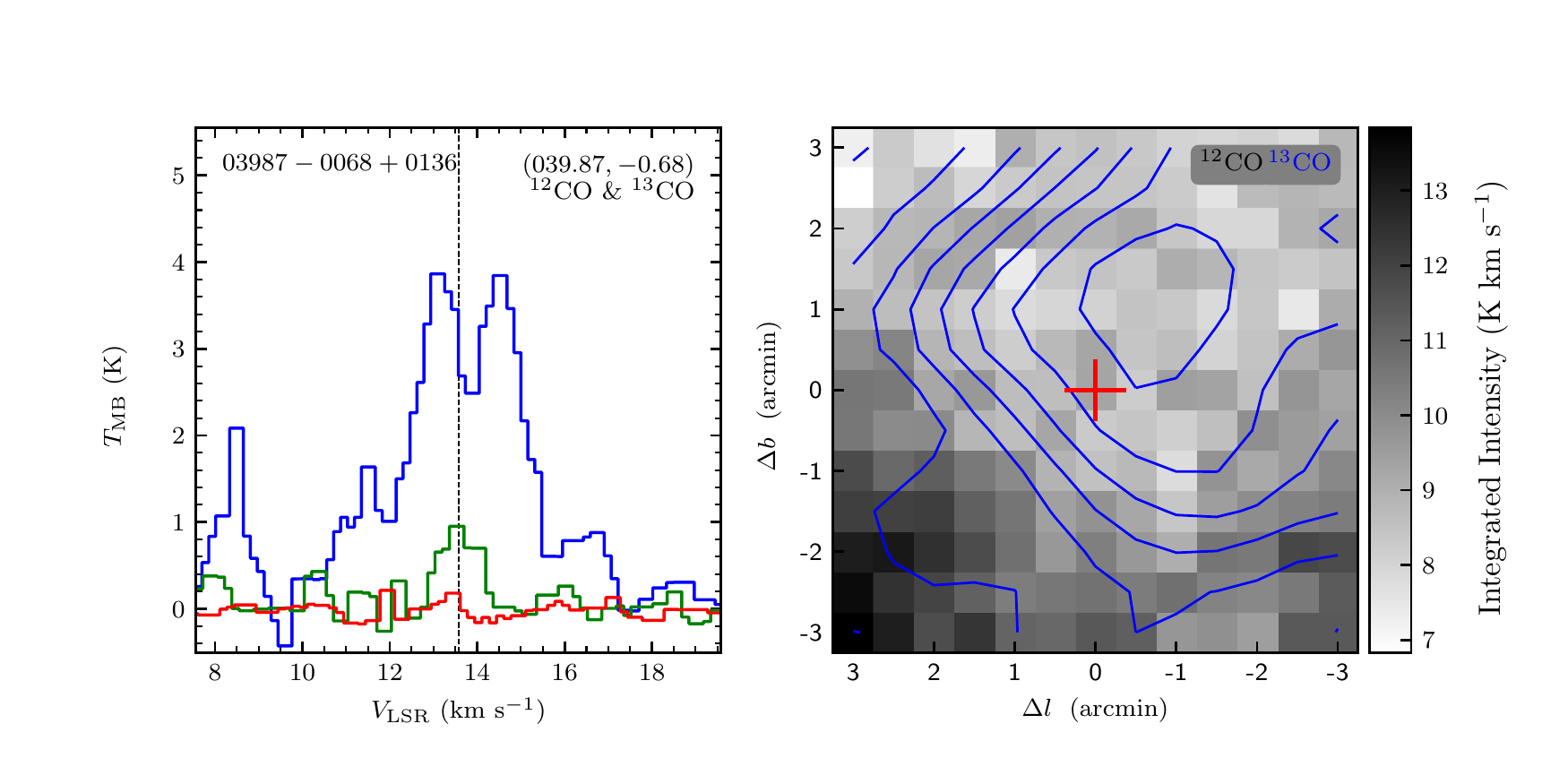}
\includegraphics[width=9.0cm,angle=0]{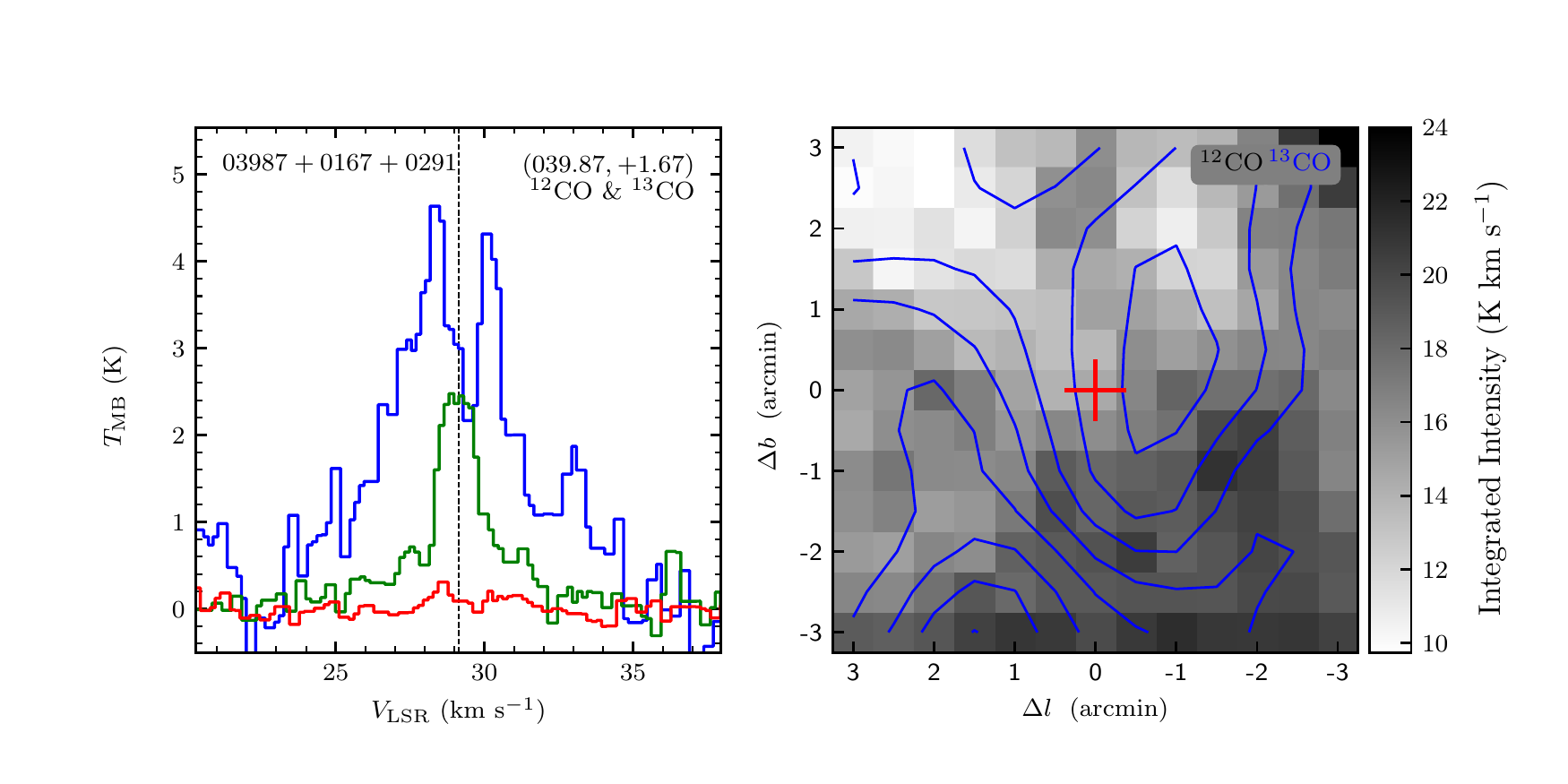}
\vspace{-0.5cm}

\includegraphics[width=9.0cm,angle=0]{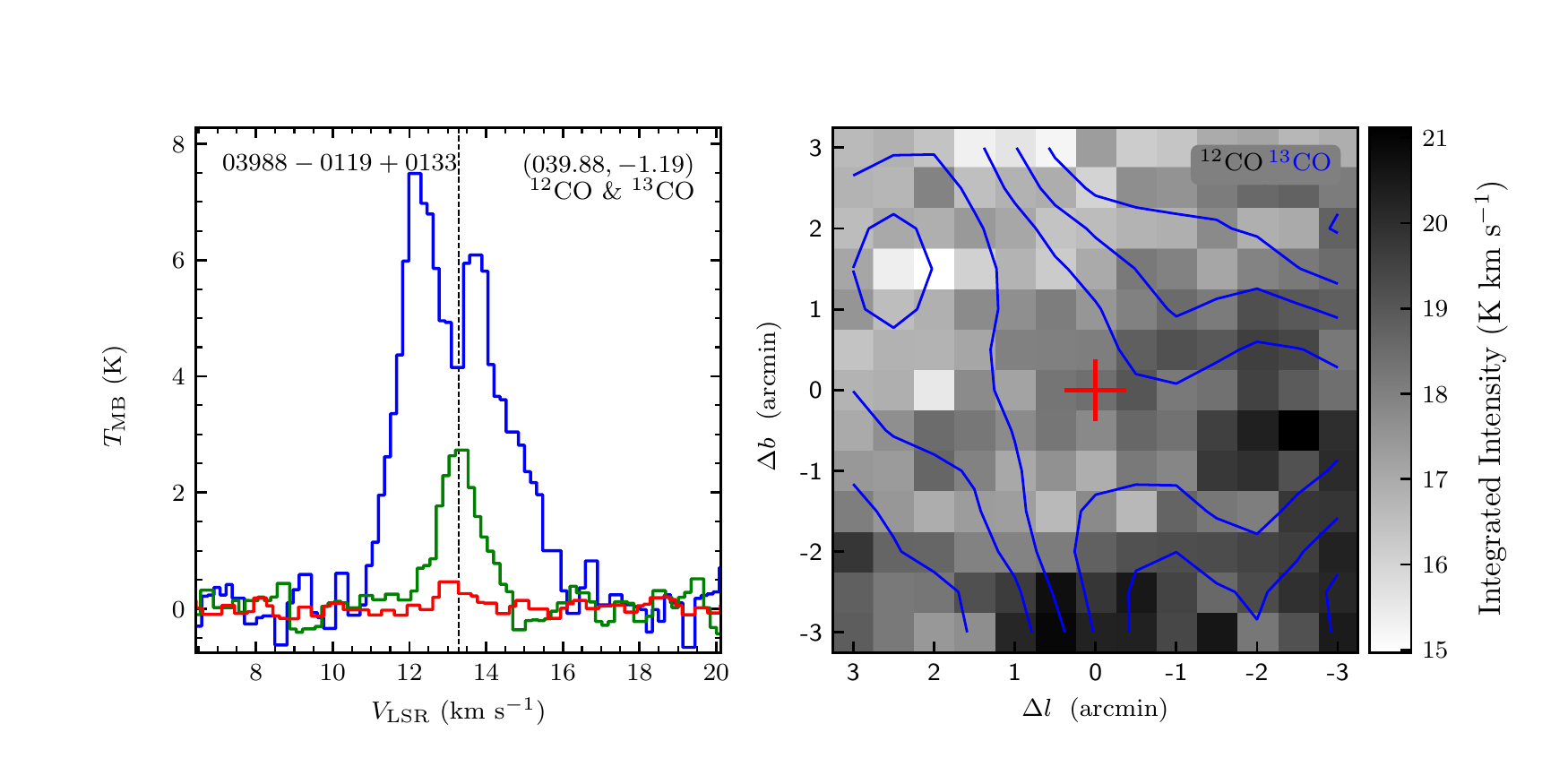}
\includegraphics[width=9.0cm,angle=0]{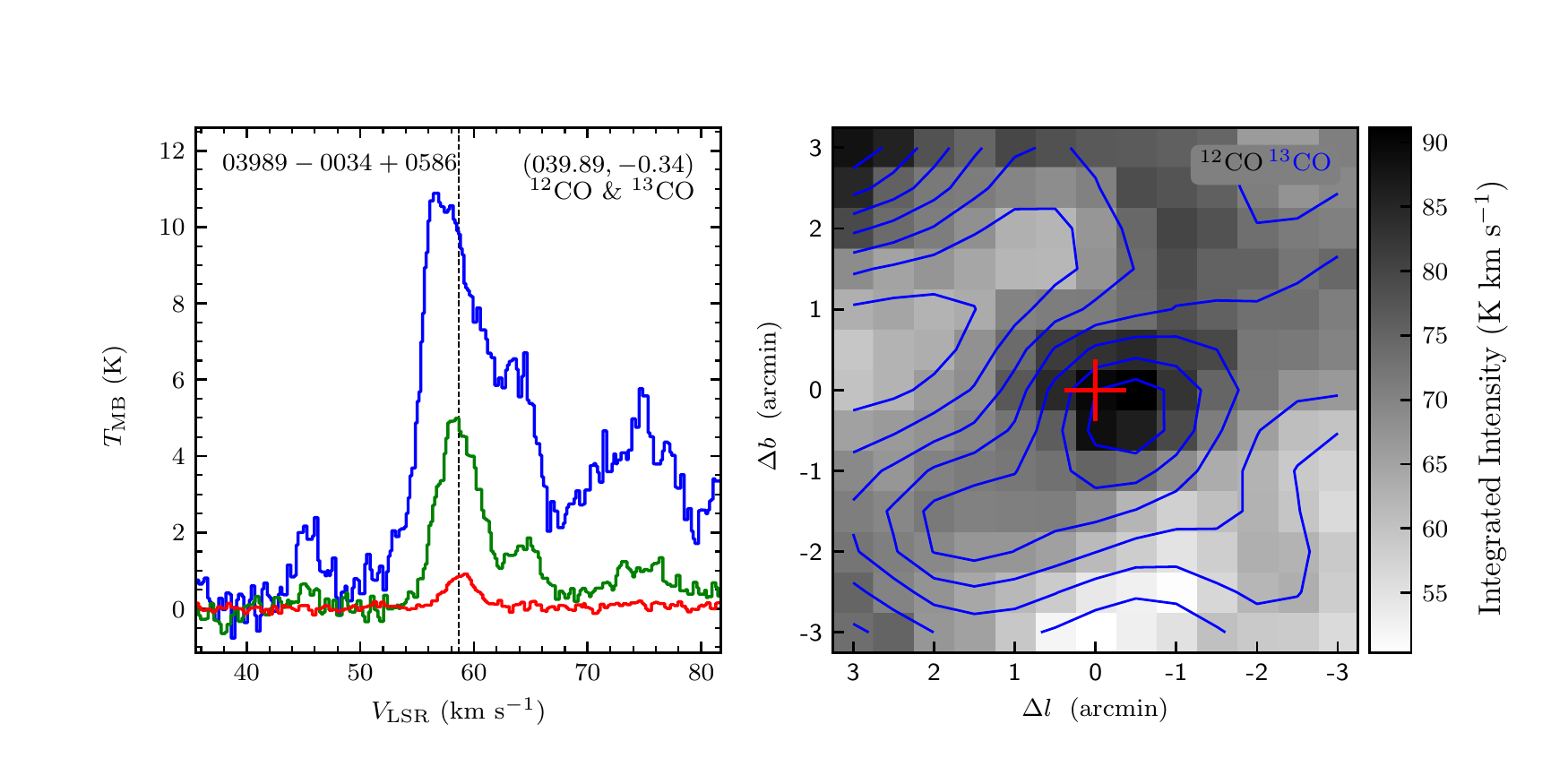}
\vspace{-0.5cm}

\includegraphics[width=9.0cm,angle=0]{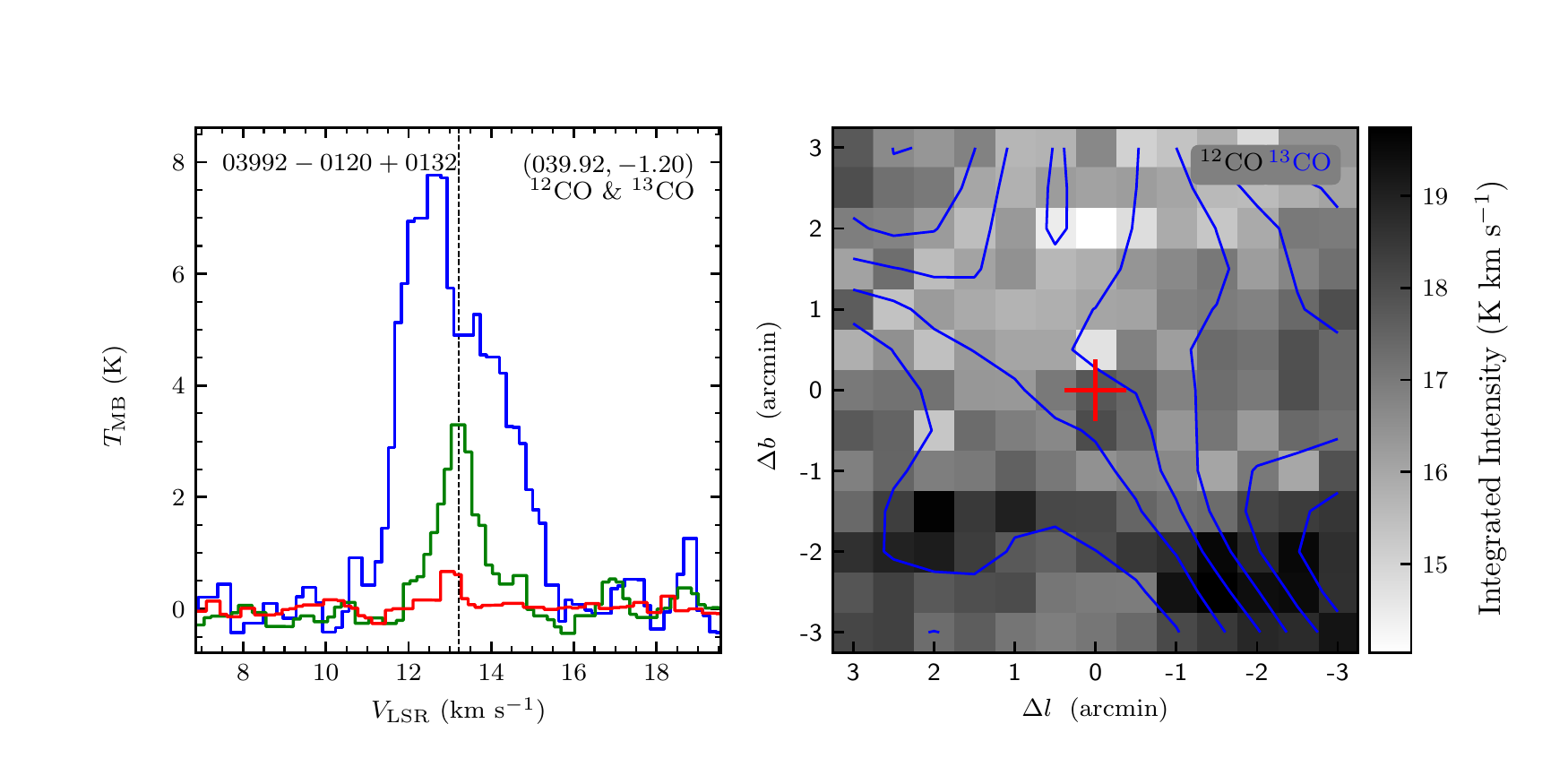}
\includegraphics[width=9.0cm,angle=0]{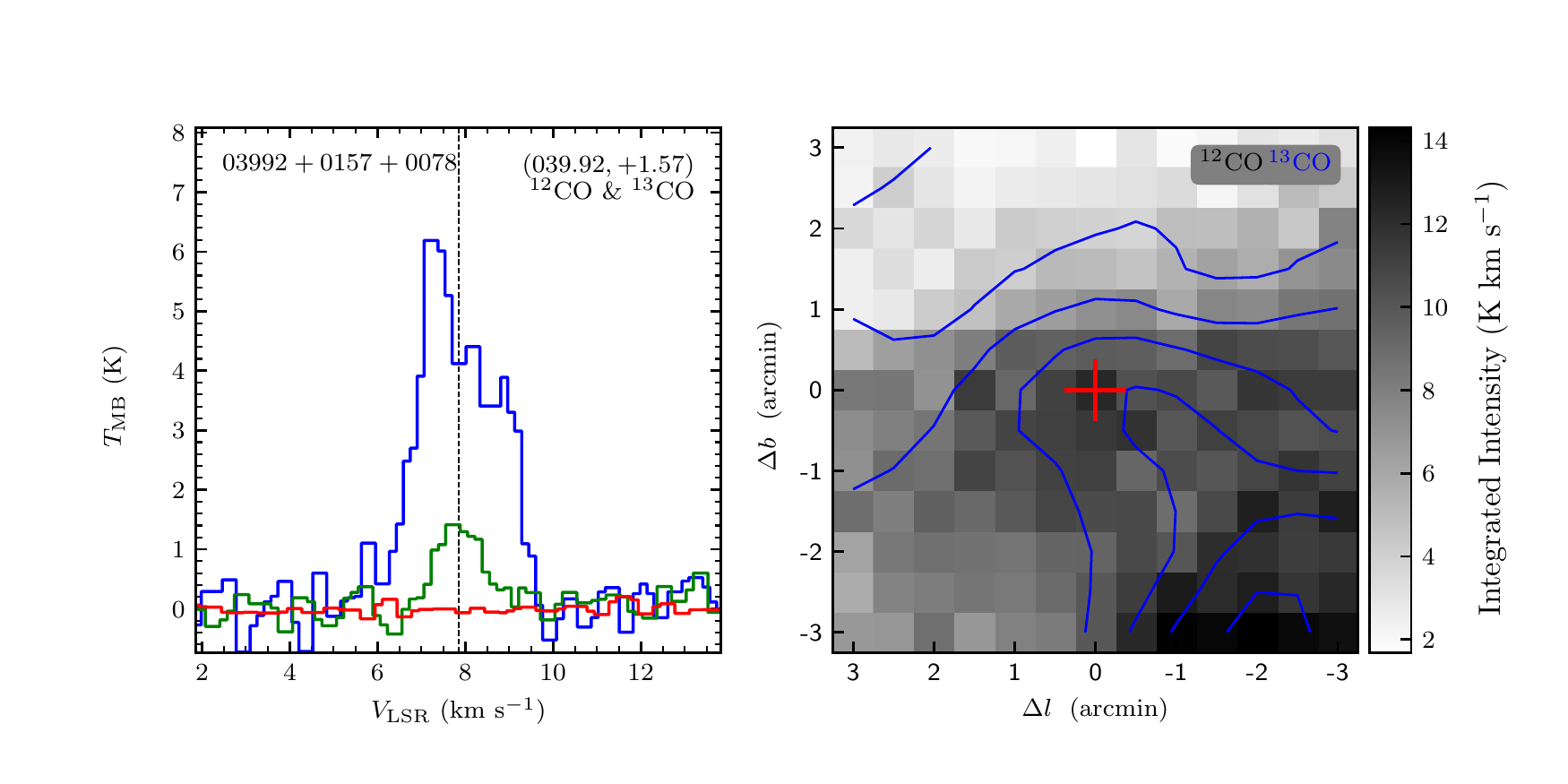}
\vspace{-0.5cm}

\includegraphics[width=9.0cm,angle=0]{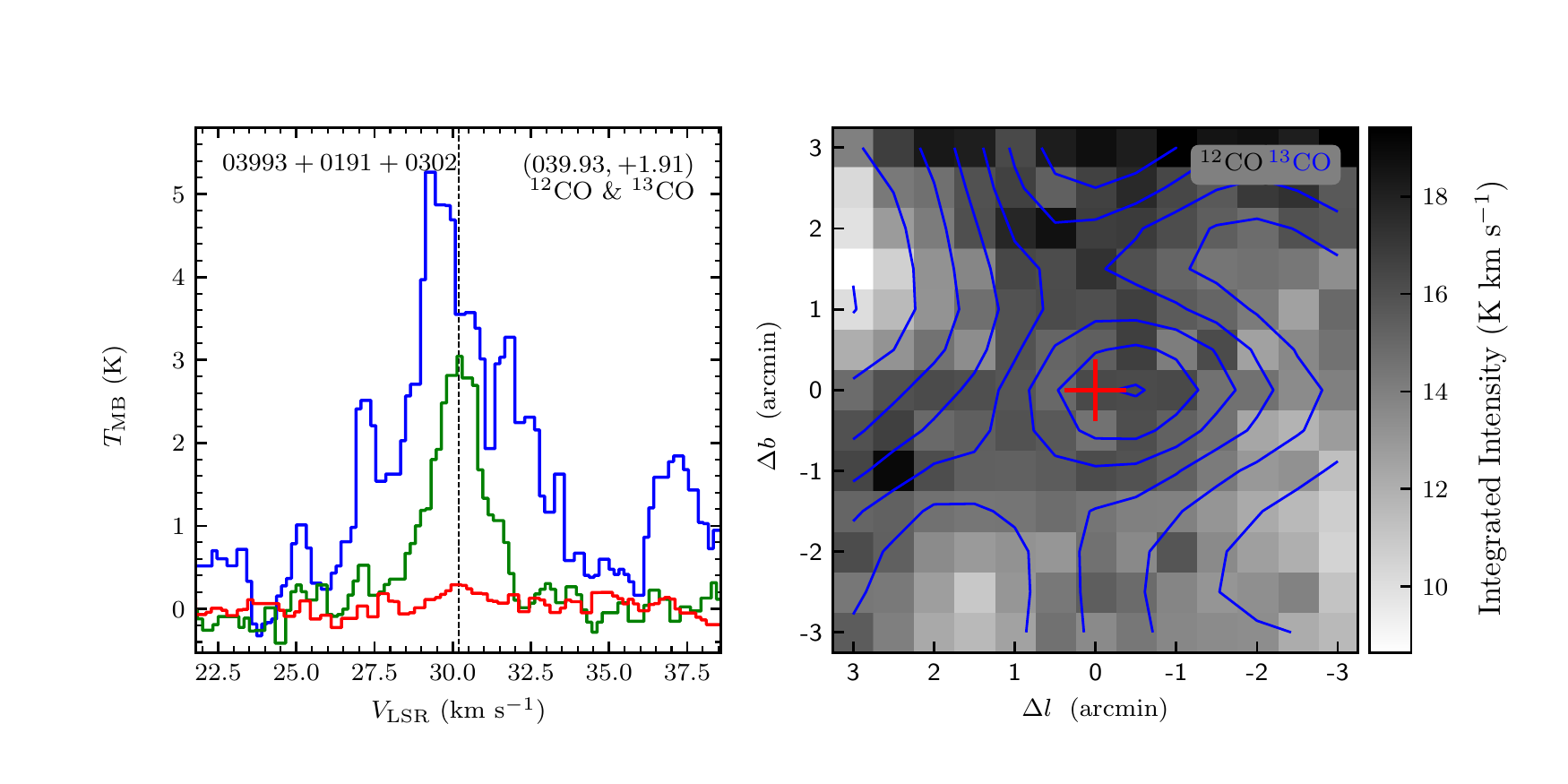}
\includegraphics[width=9.0cm,angle=0]{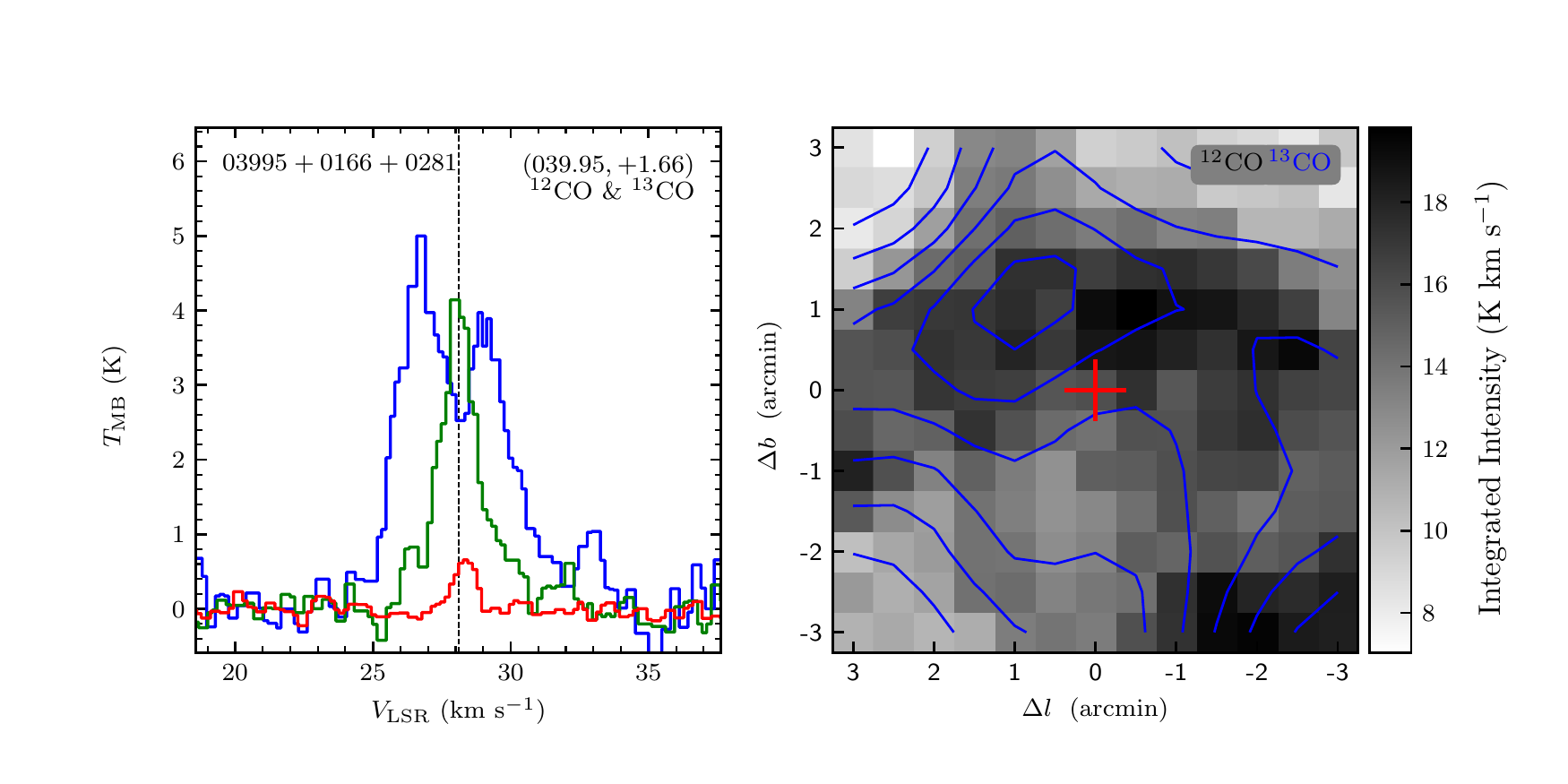}
\vspace{-0.5cm}

\includegraphics[width=9.0cm,angle=0]{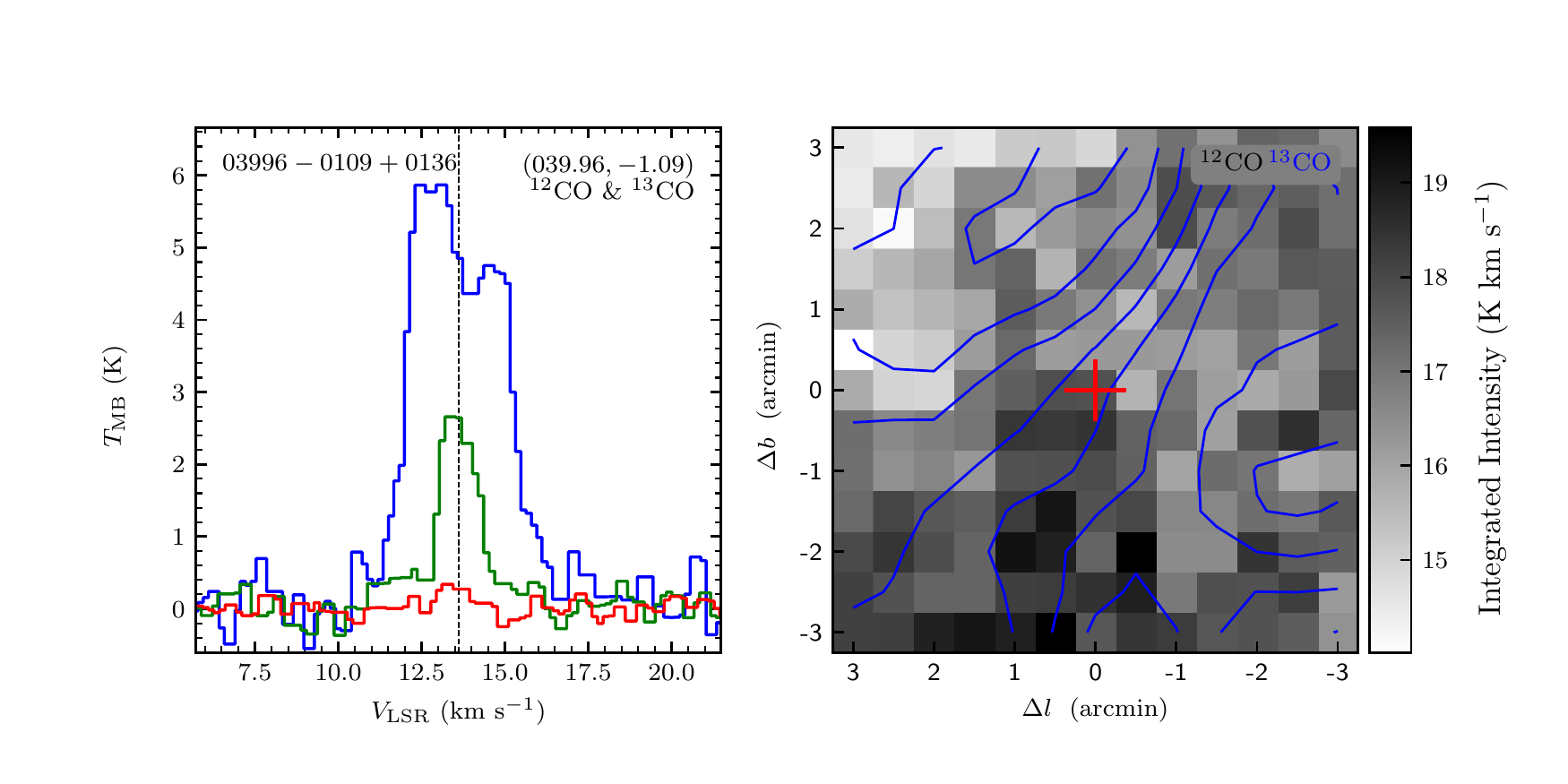}
\includegraphics[width=9.0cm,angle=0]{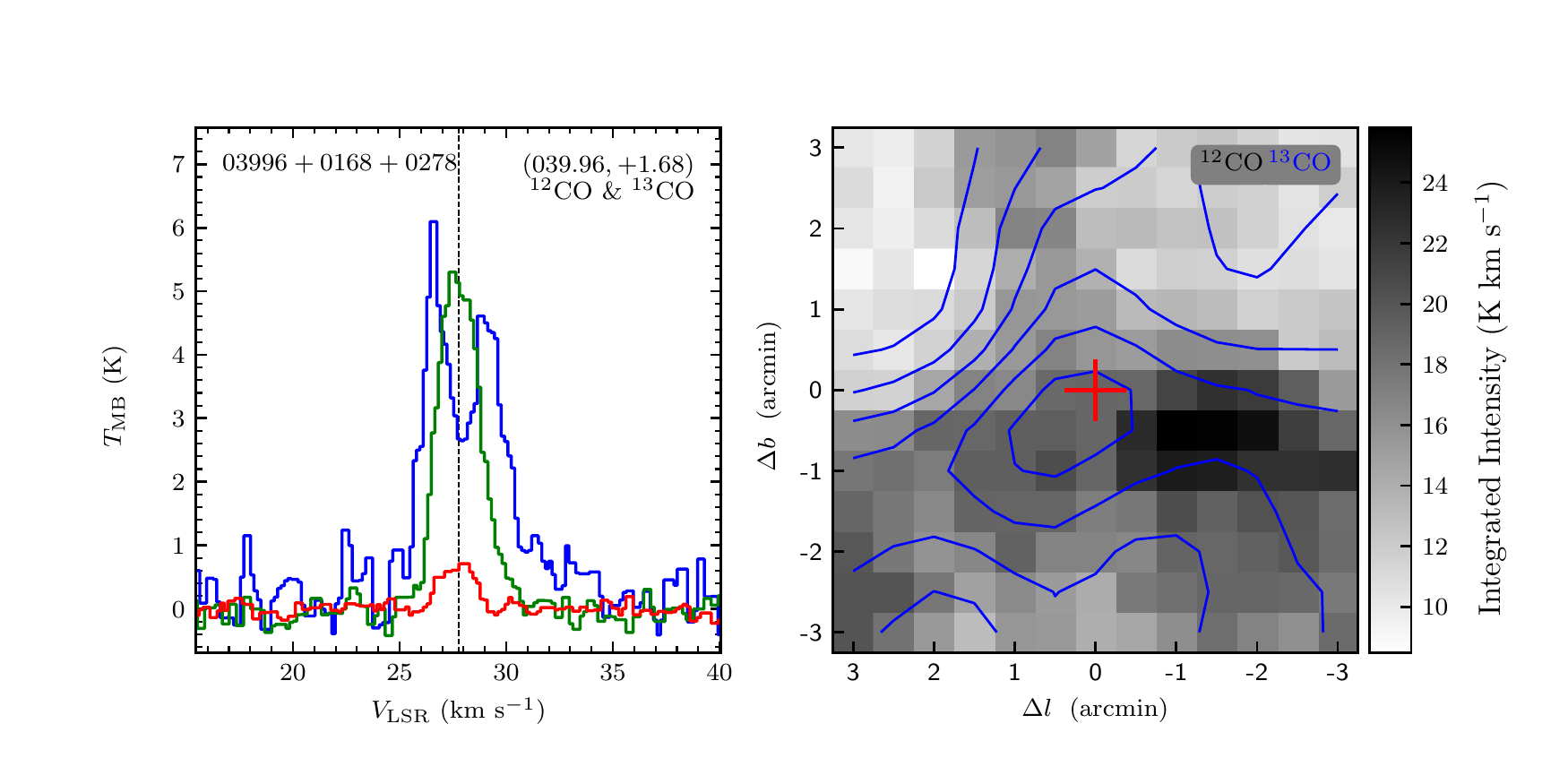}
\end{figure}
\clearpage

\begin{figure}
\includegraphics[width=9.0cm,angle=0]{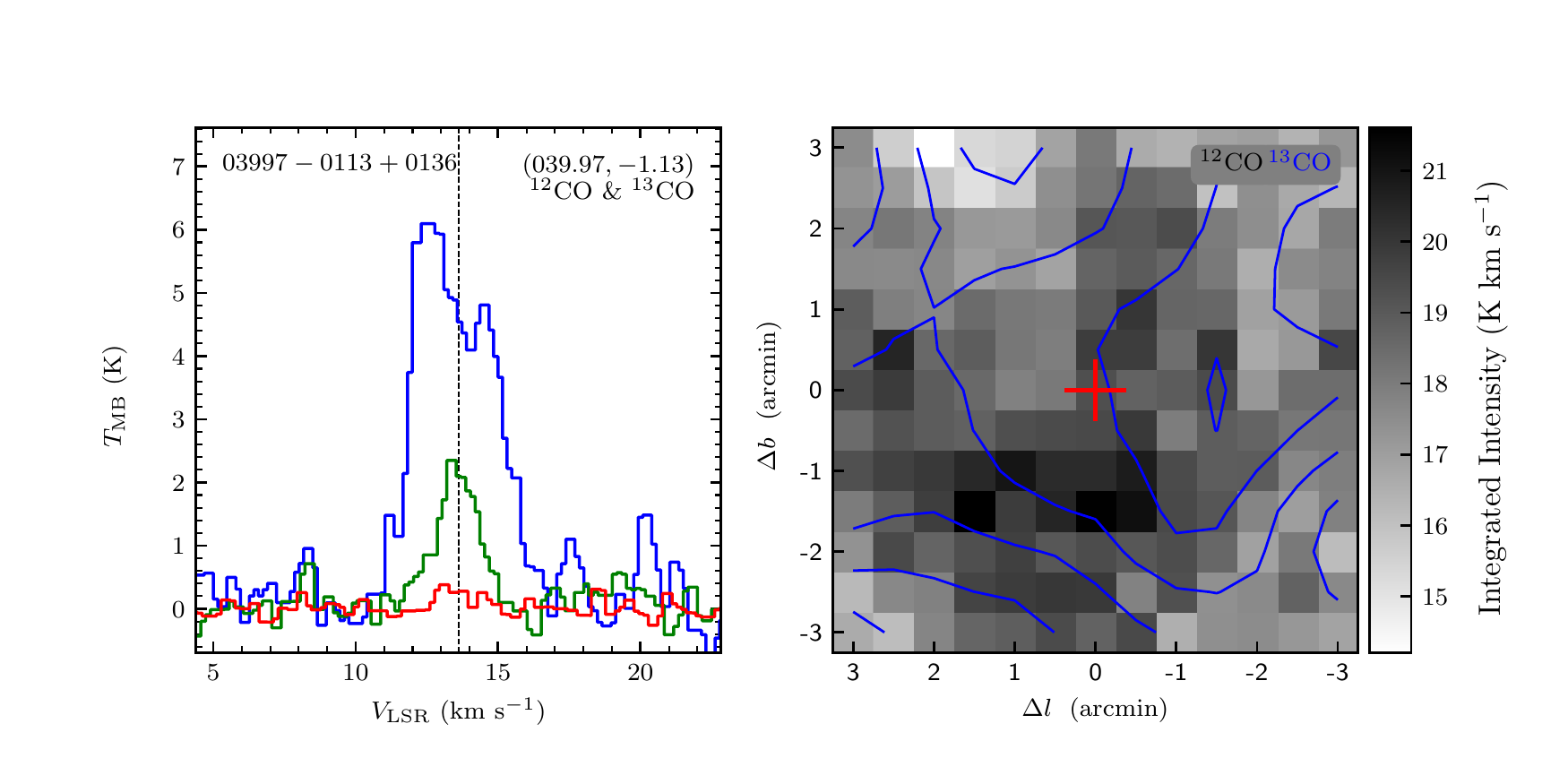}
\includegraphics[width=9.0cm,angle=0]{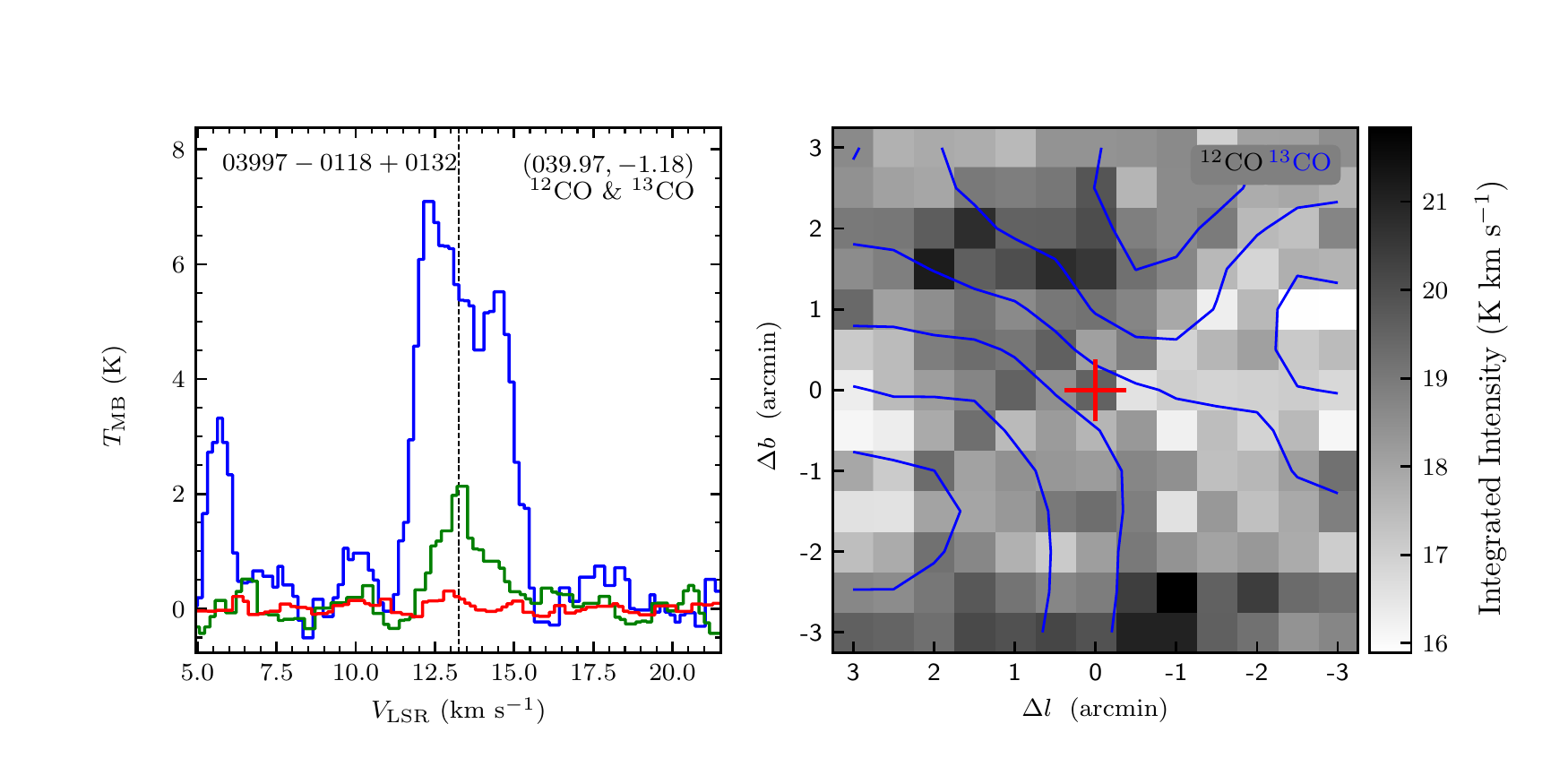}
\vspace{-0.5cm}

\includegraphics[width=9.0cm,angle=0]{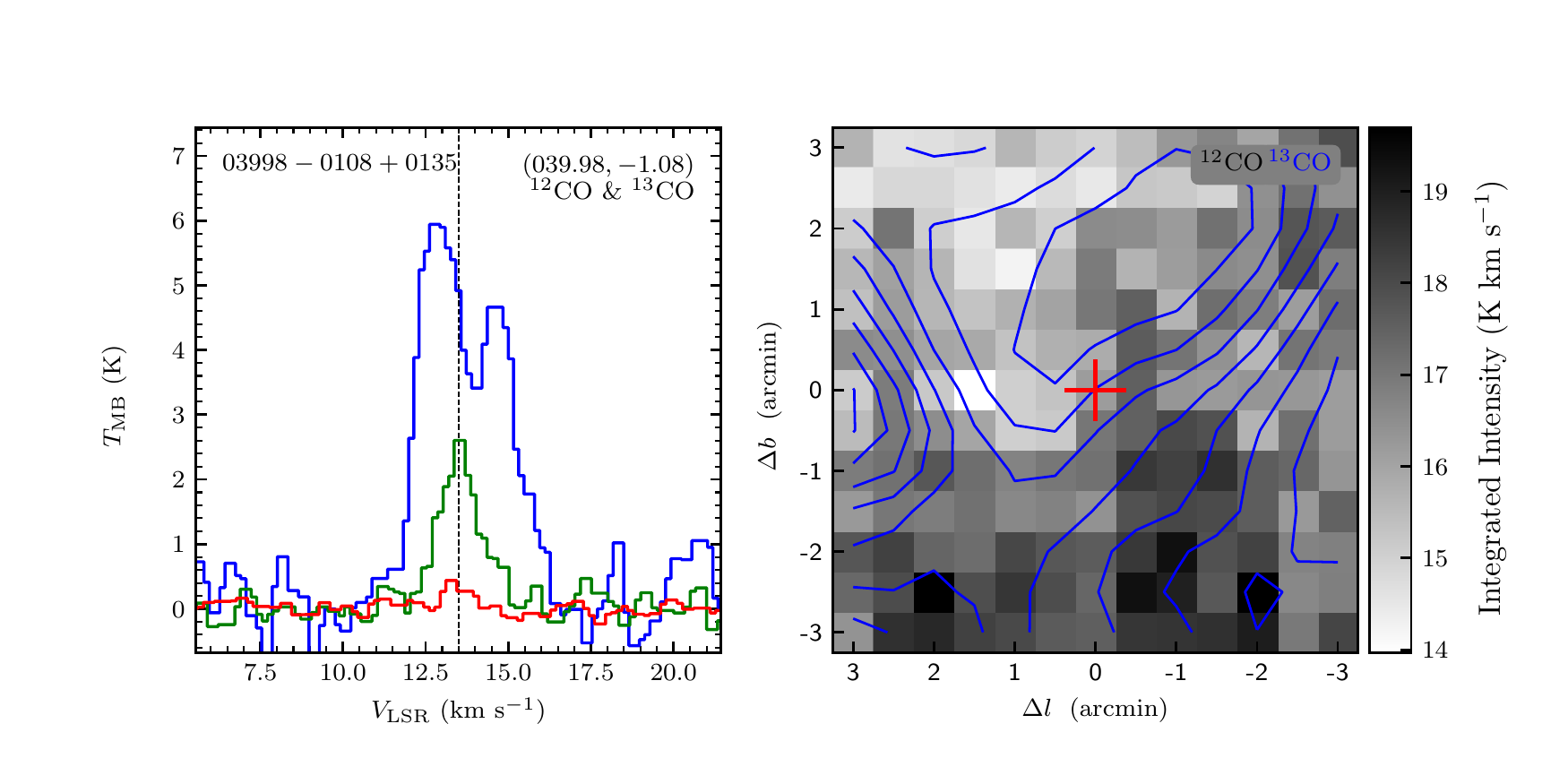}
\includegraphics[width=9.0cm,angle=0]{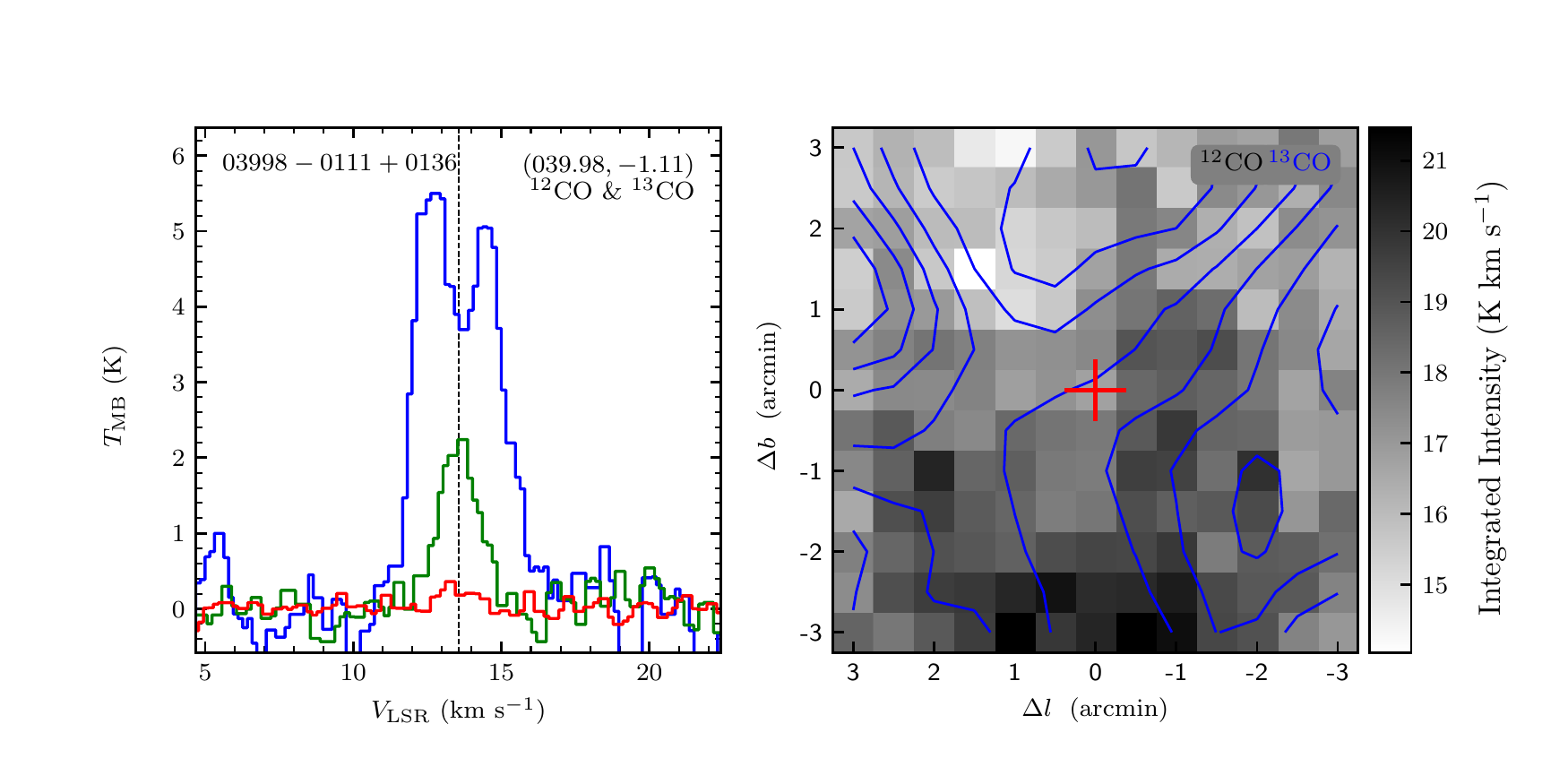}
\vspace{-0.5cm}

\includegraphics[width=9.0cm,angle=0]{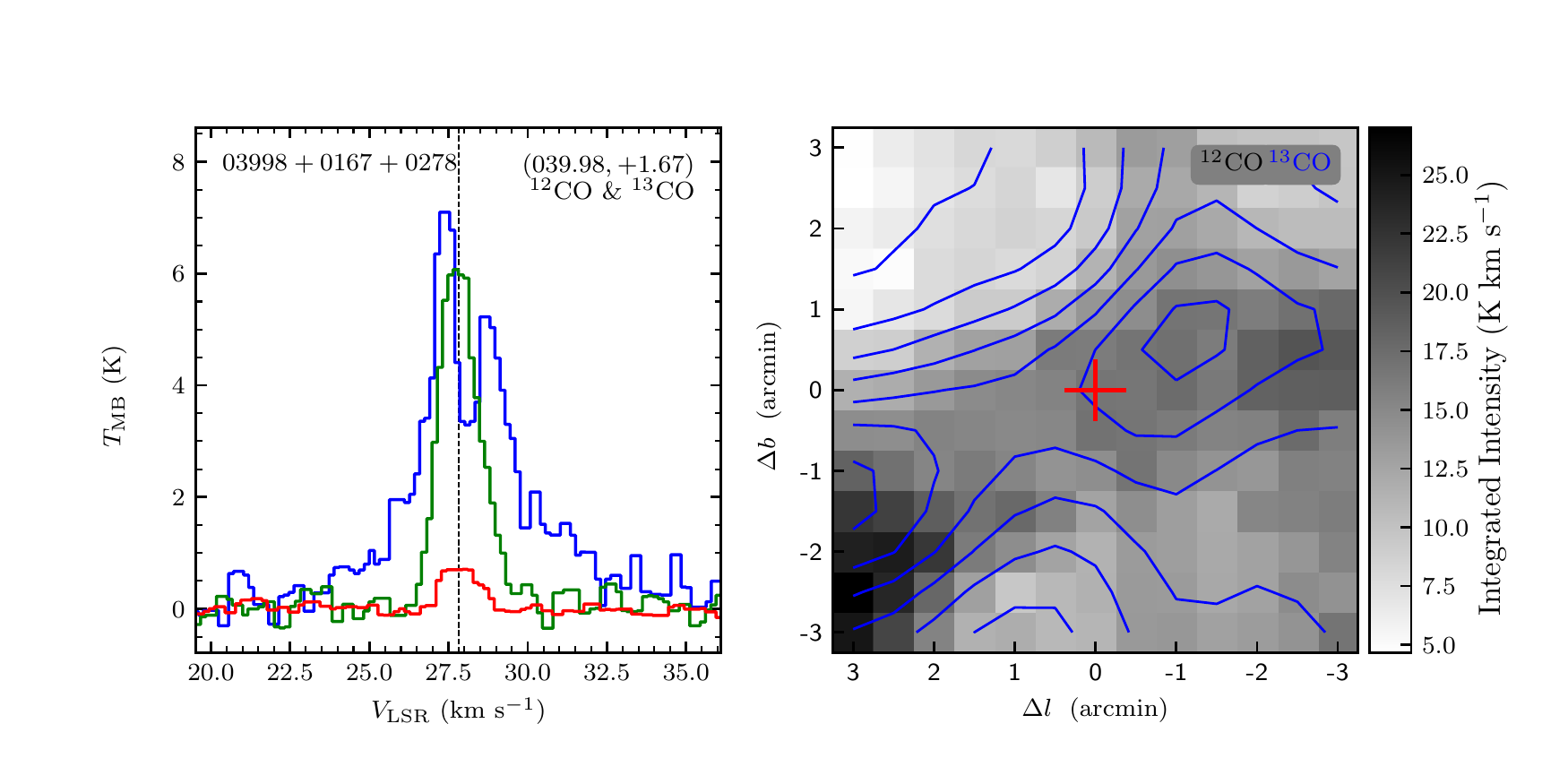}
\includegraphics[width=9.0cm,angle=0]{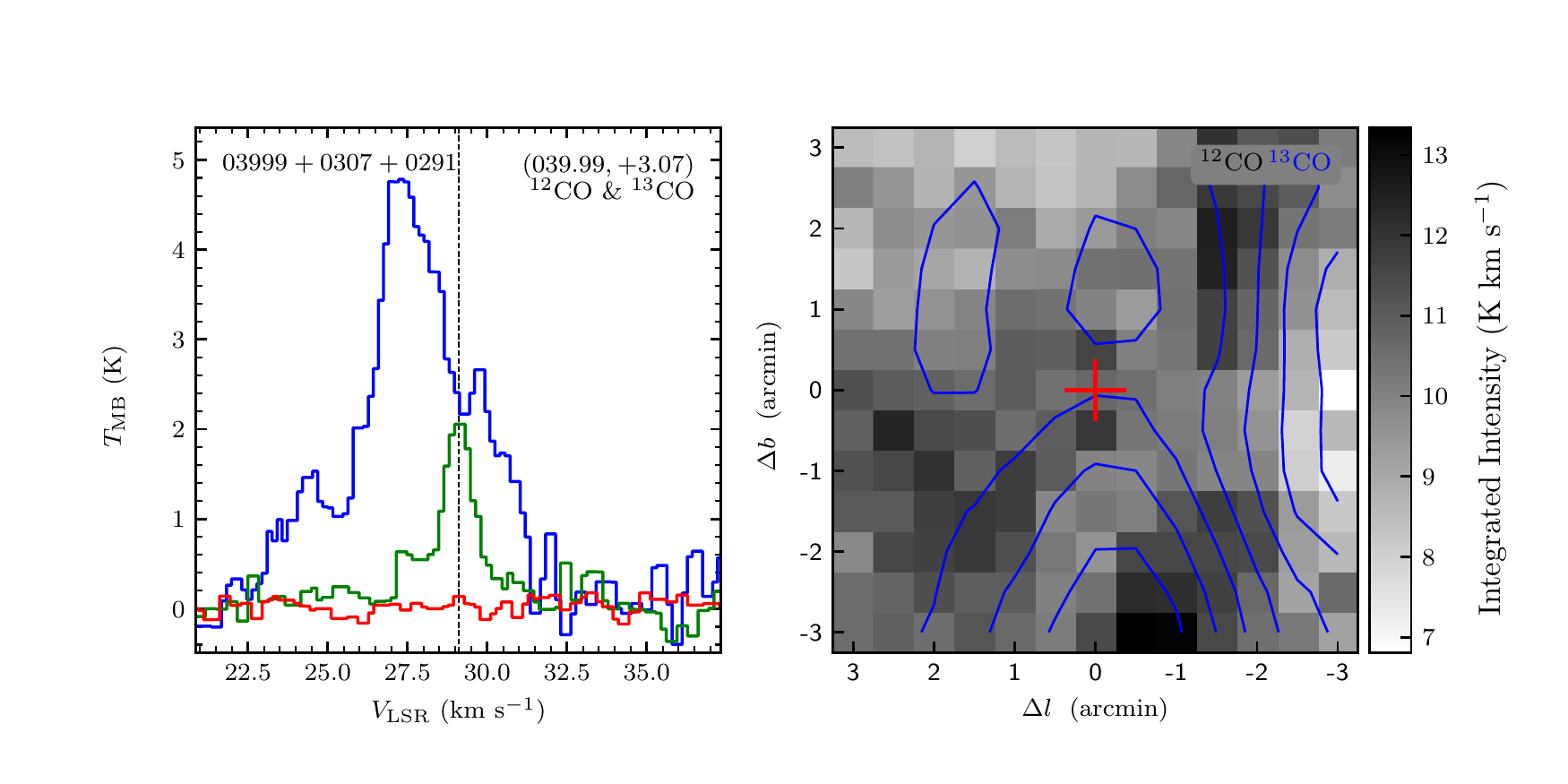}
\vspace{-0.5cm}

\includegraphics[width=9.0cm,angle=0]{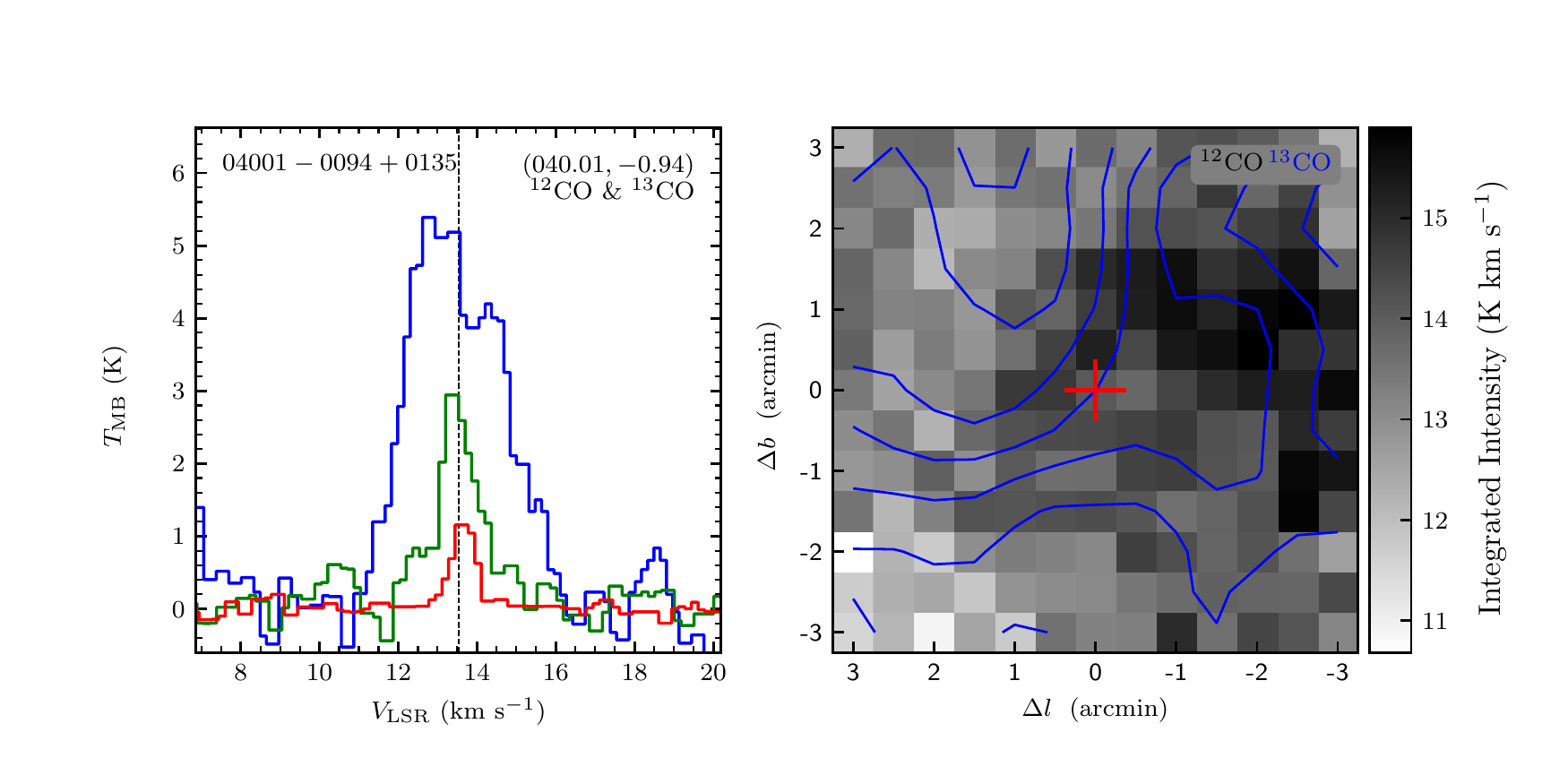}
\includegraphics[width=9.0cm,angle=0]{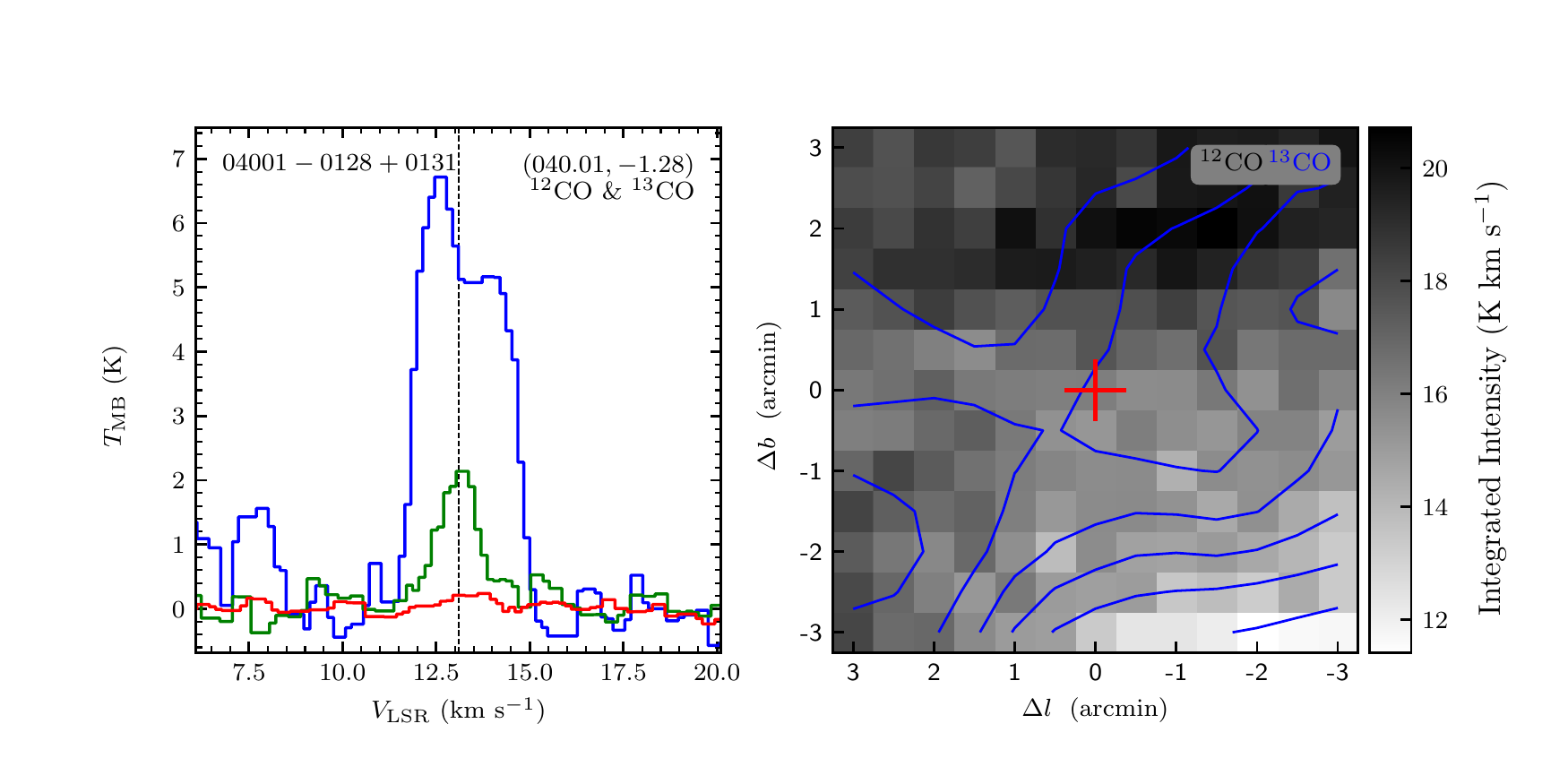}
\vspace{-0.5cm}

\includegraphics[width=9.0cm,angle=0]{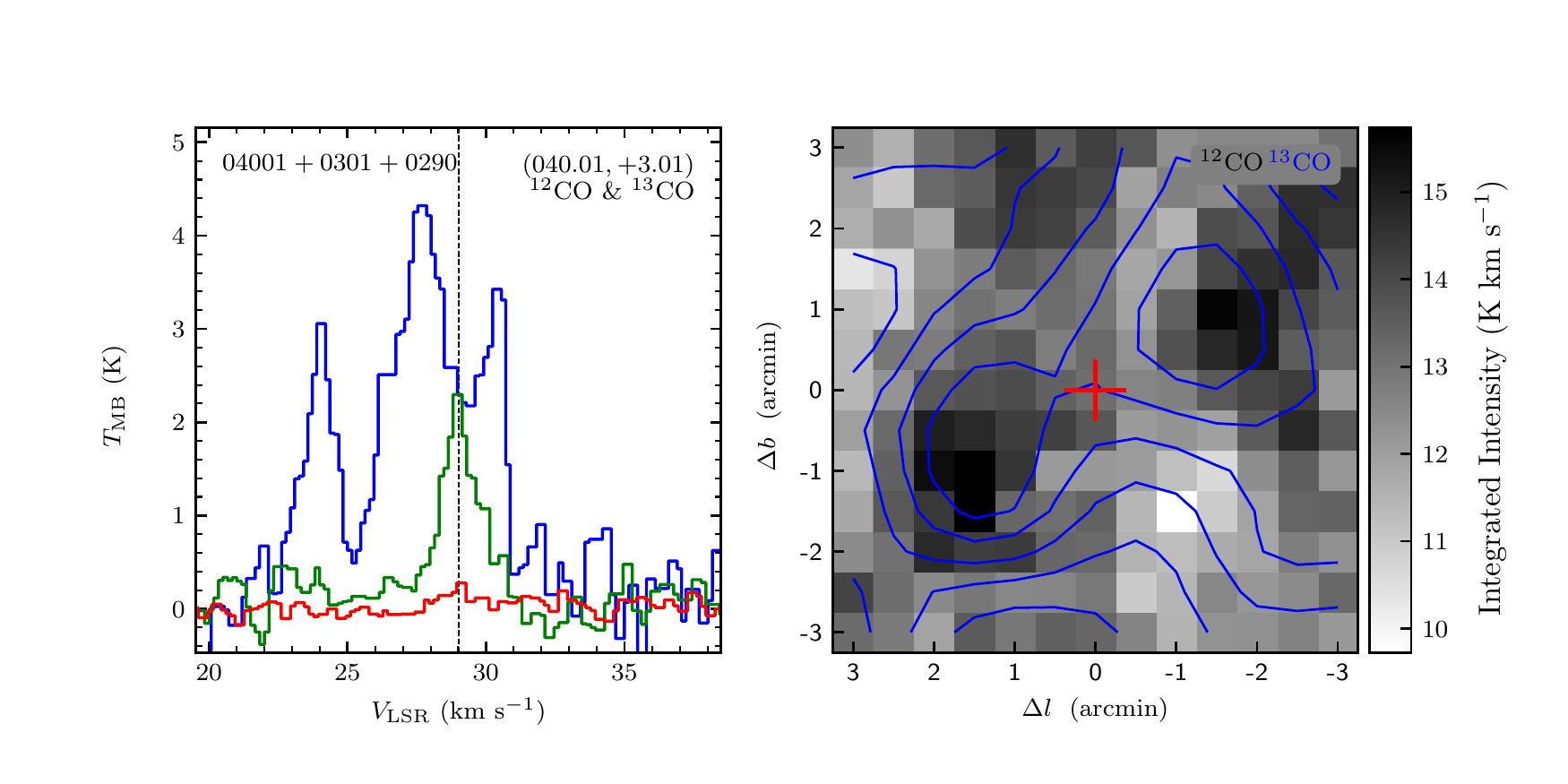}
\includegraphics[width=9.0cm,angle=0]{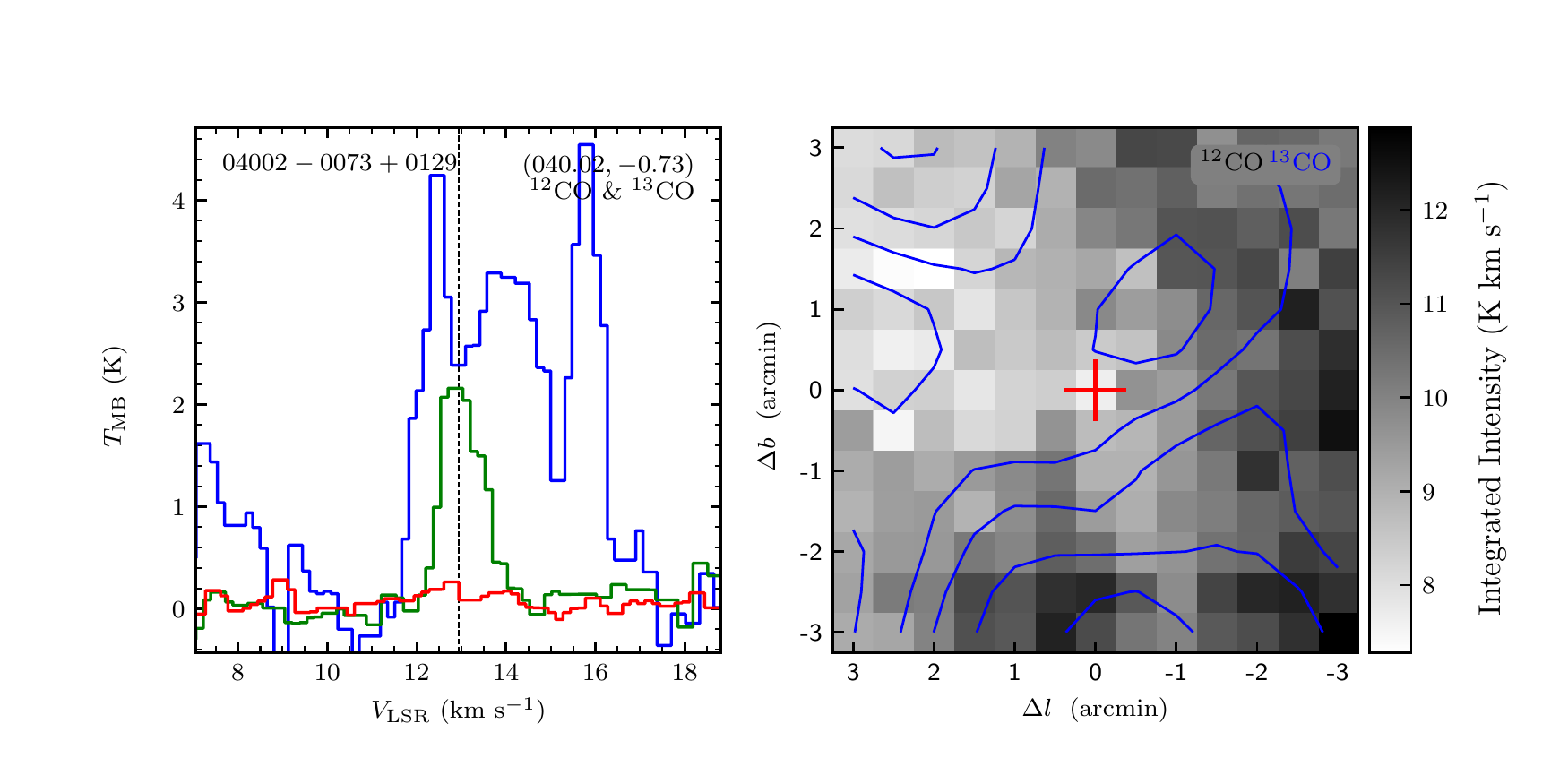}
\end{figure}
\clearpage

\begin{figure}
\includegraphics[width=9.0cm,angle=0]{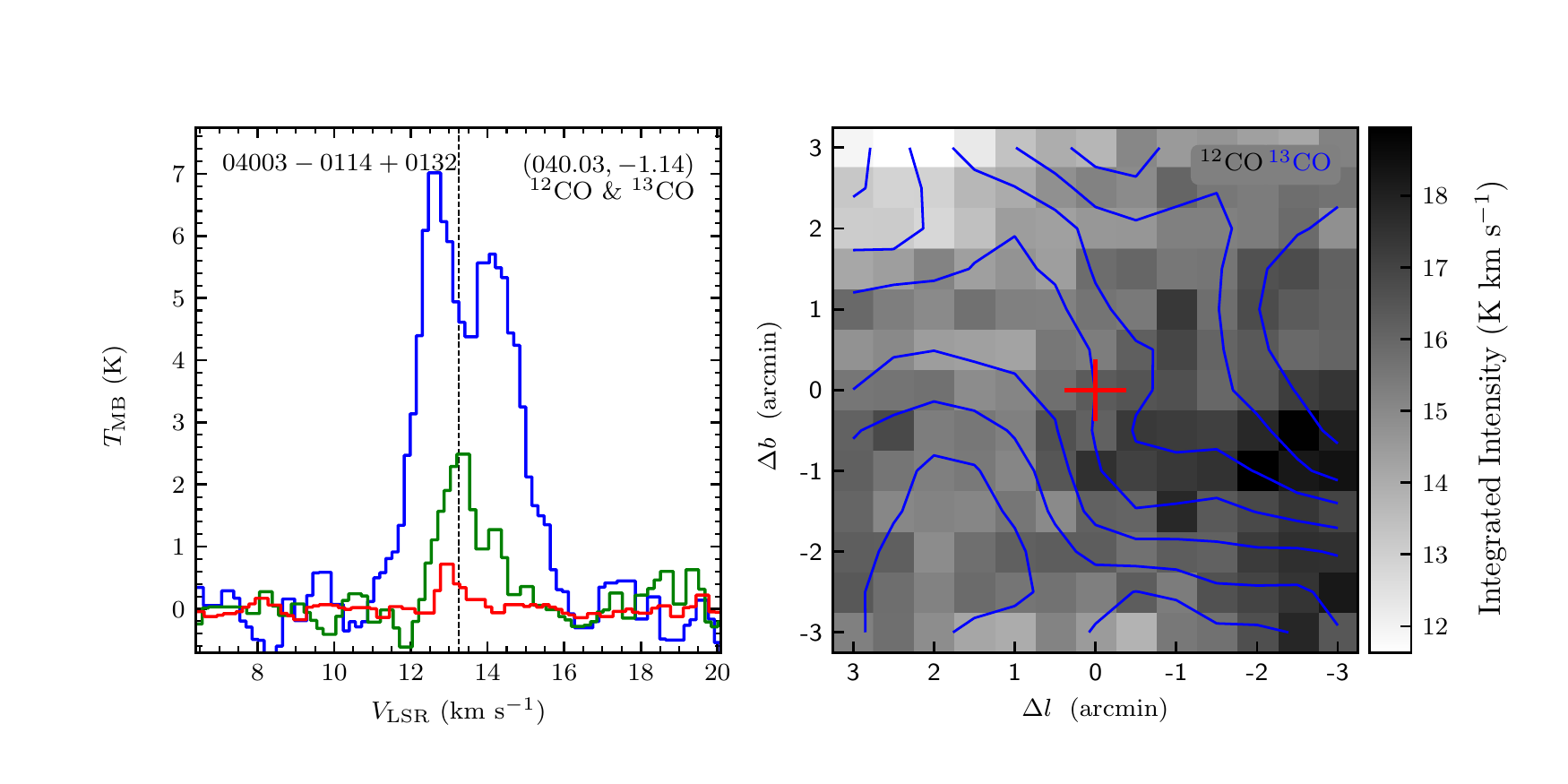}
\includegraphics[width=9.0cm,angle=0]{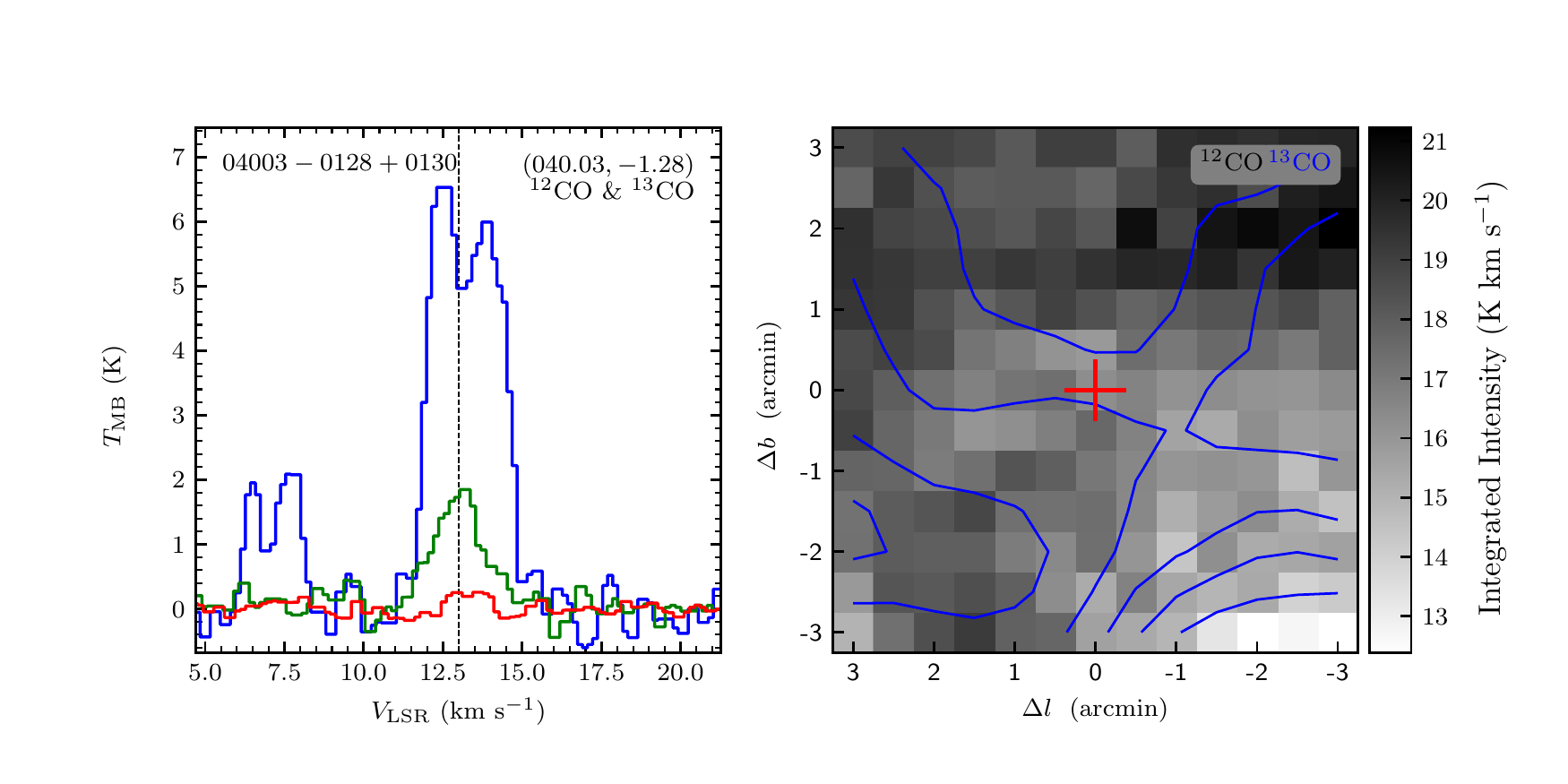}
\vspace{-0.5cm}

\includegraphics[width=9.0cm,angle=0]{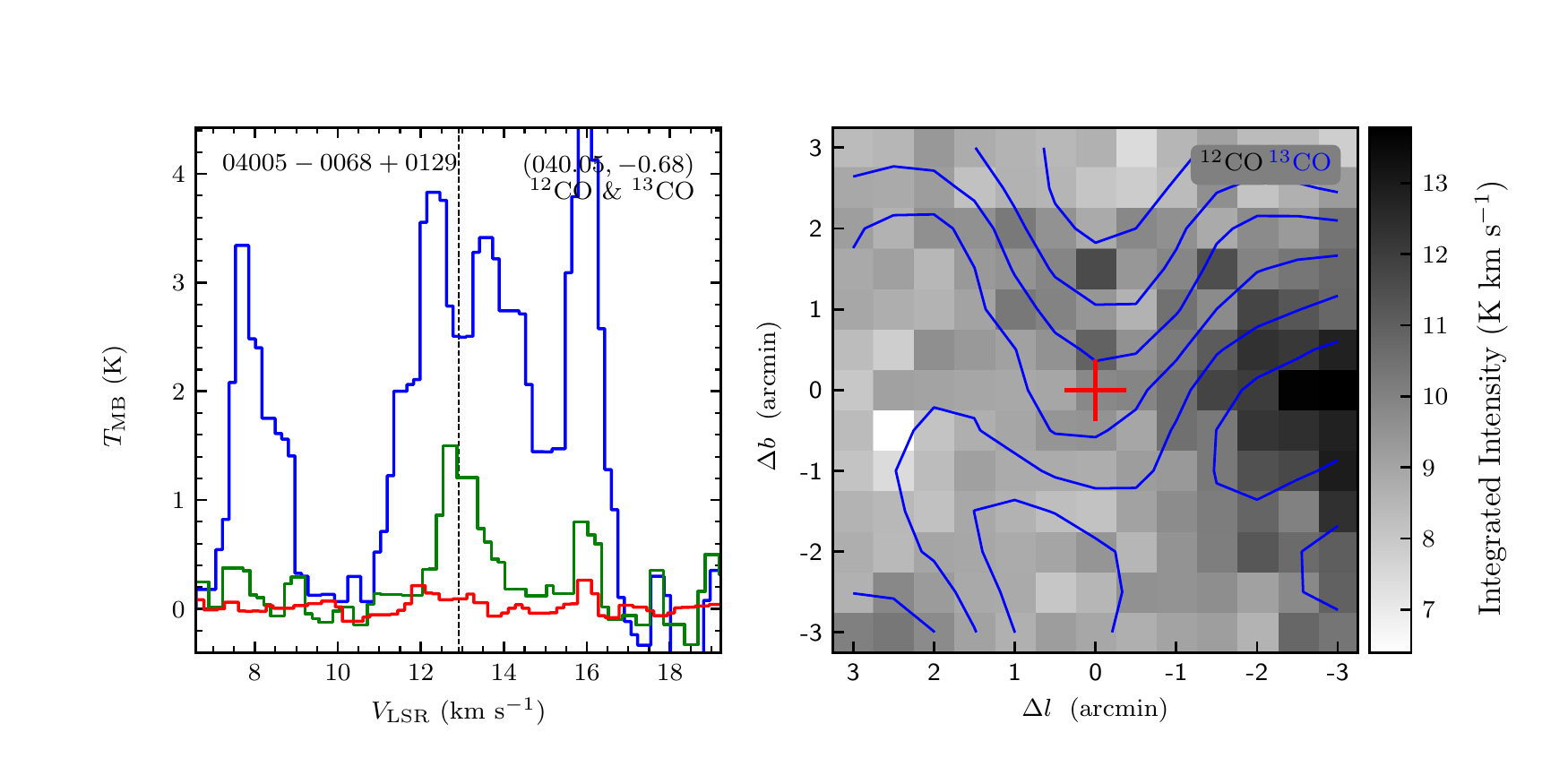}
\includegraphics[width=9.0cm,angle=0]{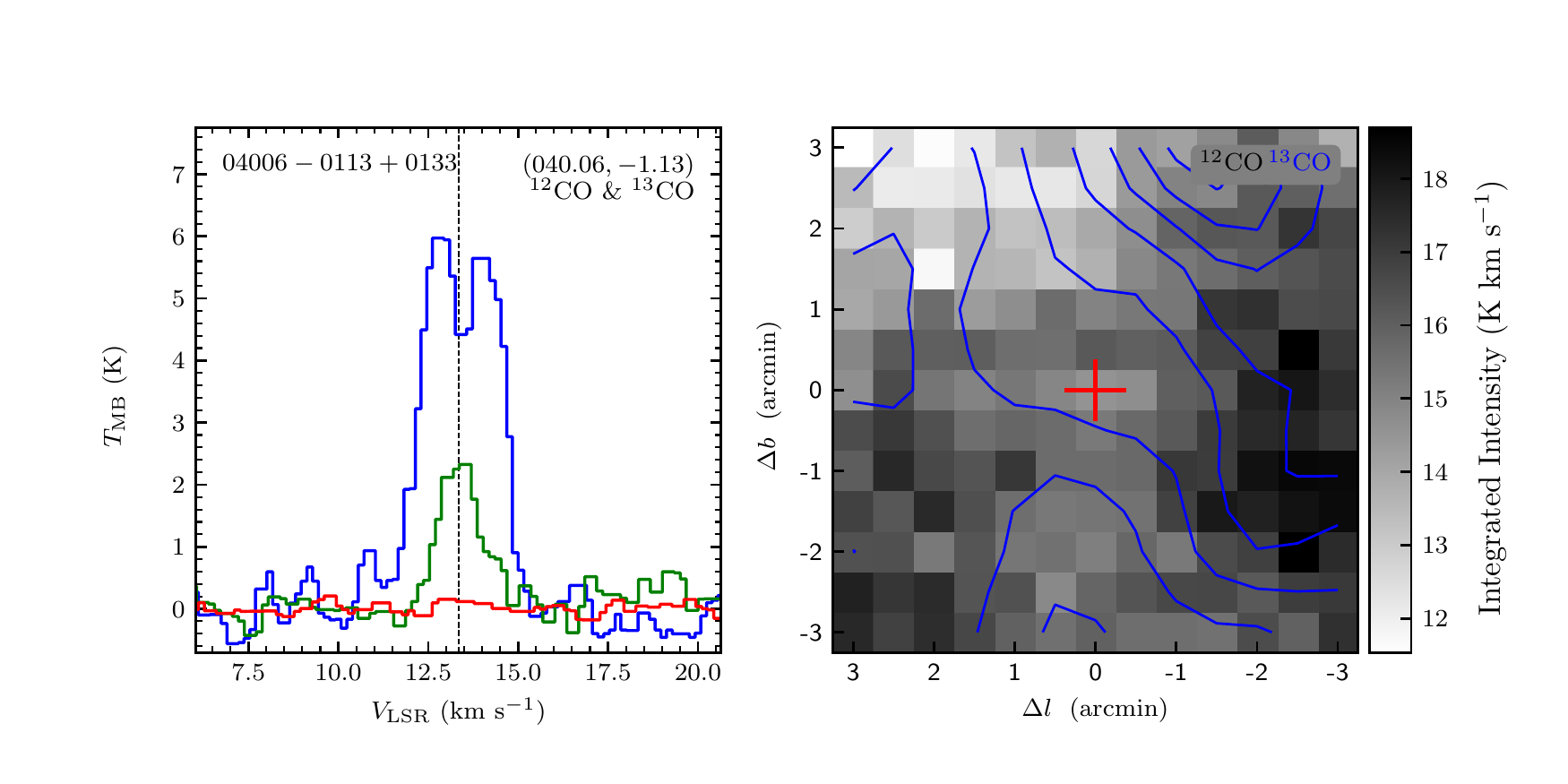}
\vspace{-0.5cm}

\includegraphics[width=9.0cm,angle=0]{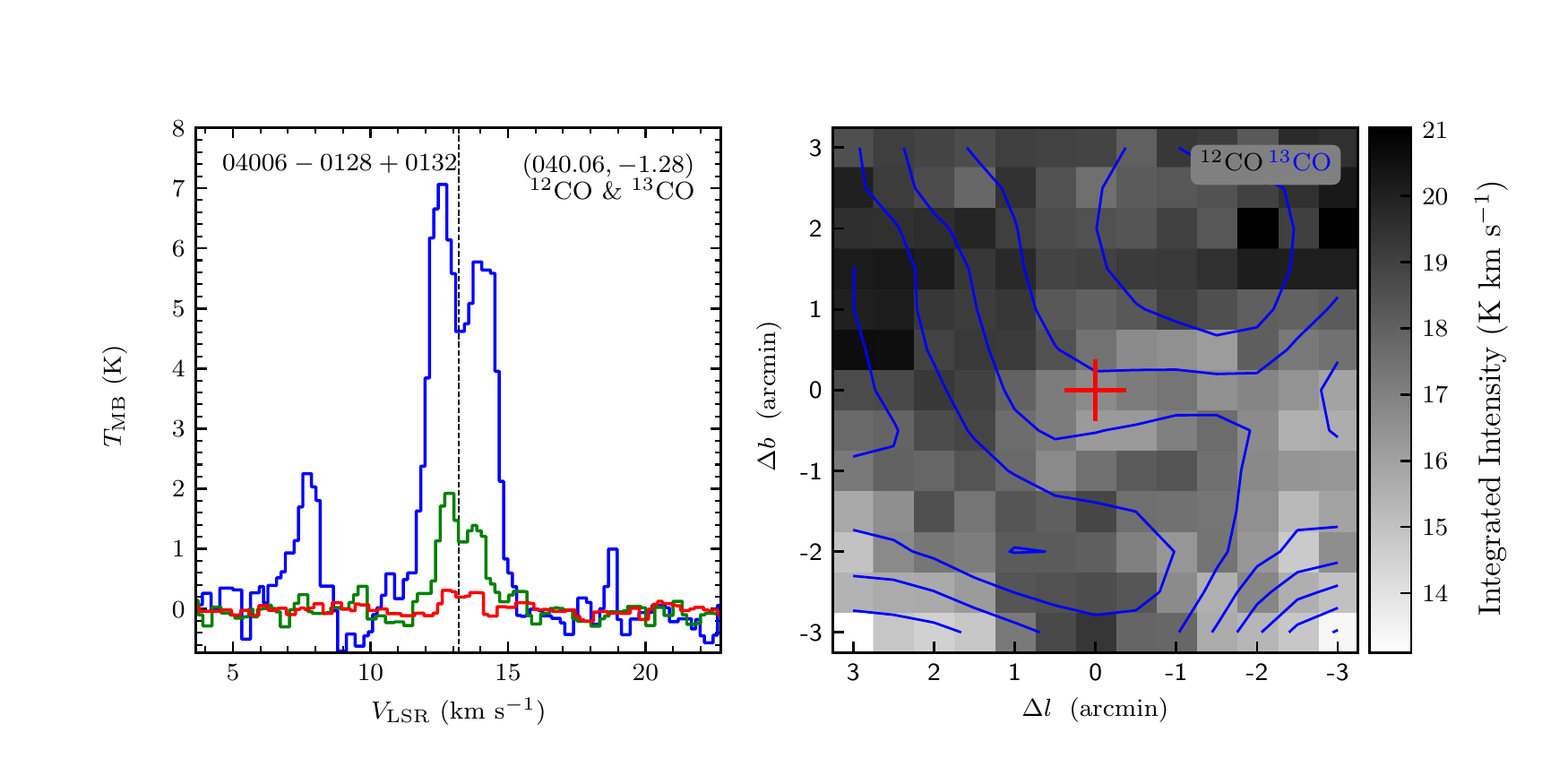}
\includegraphics[width=9.0cm,angle=0]{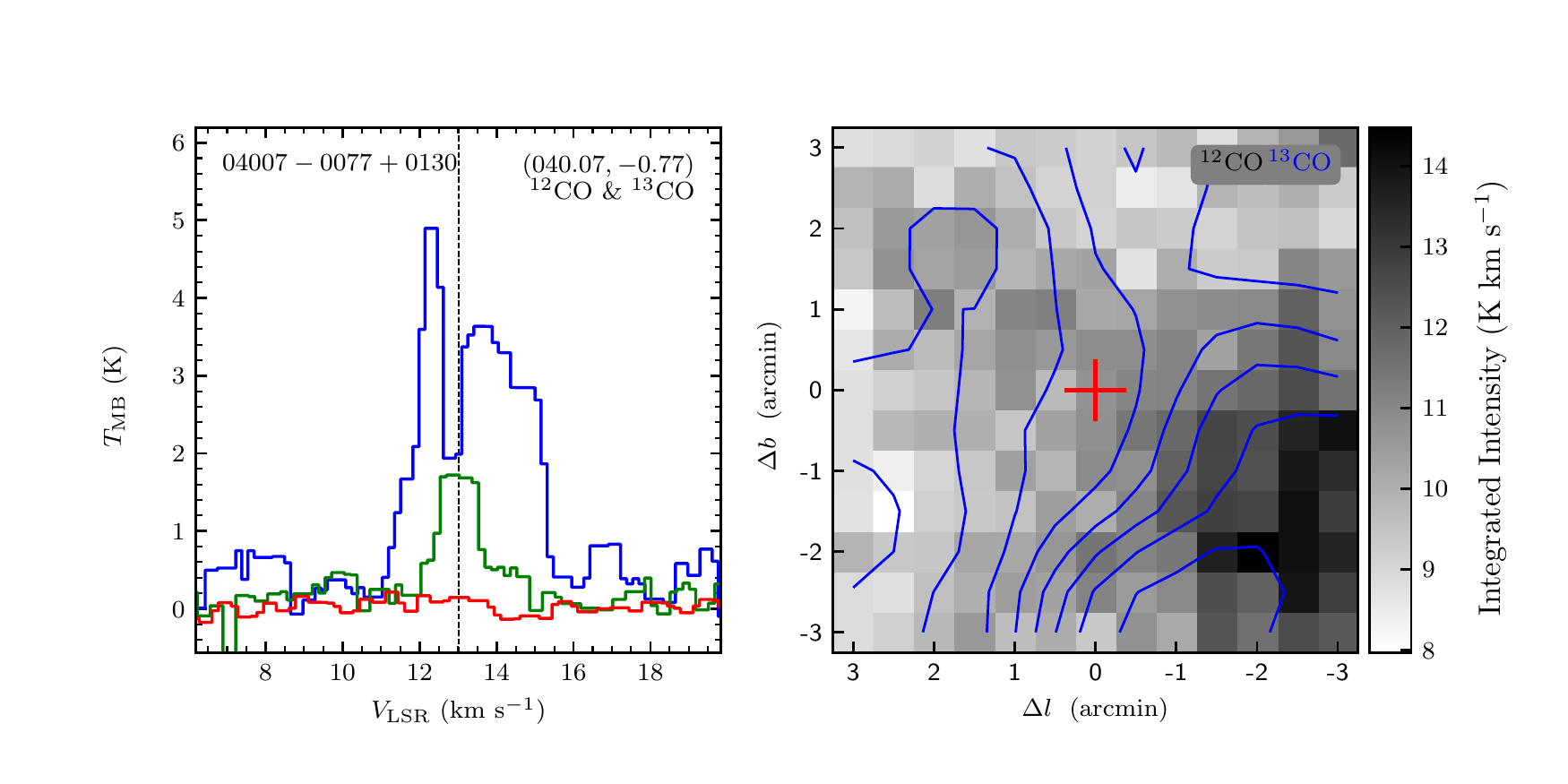}
\vspace{-0.5cm}

\includegraphics[width=9.0cm,angle=0]{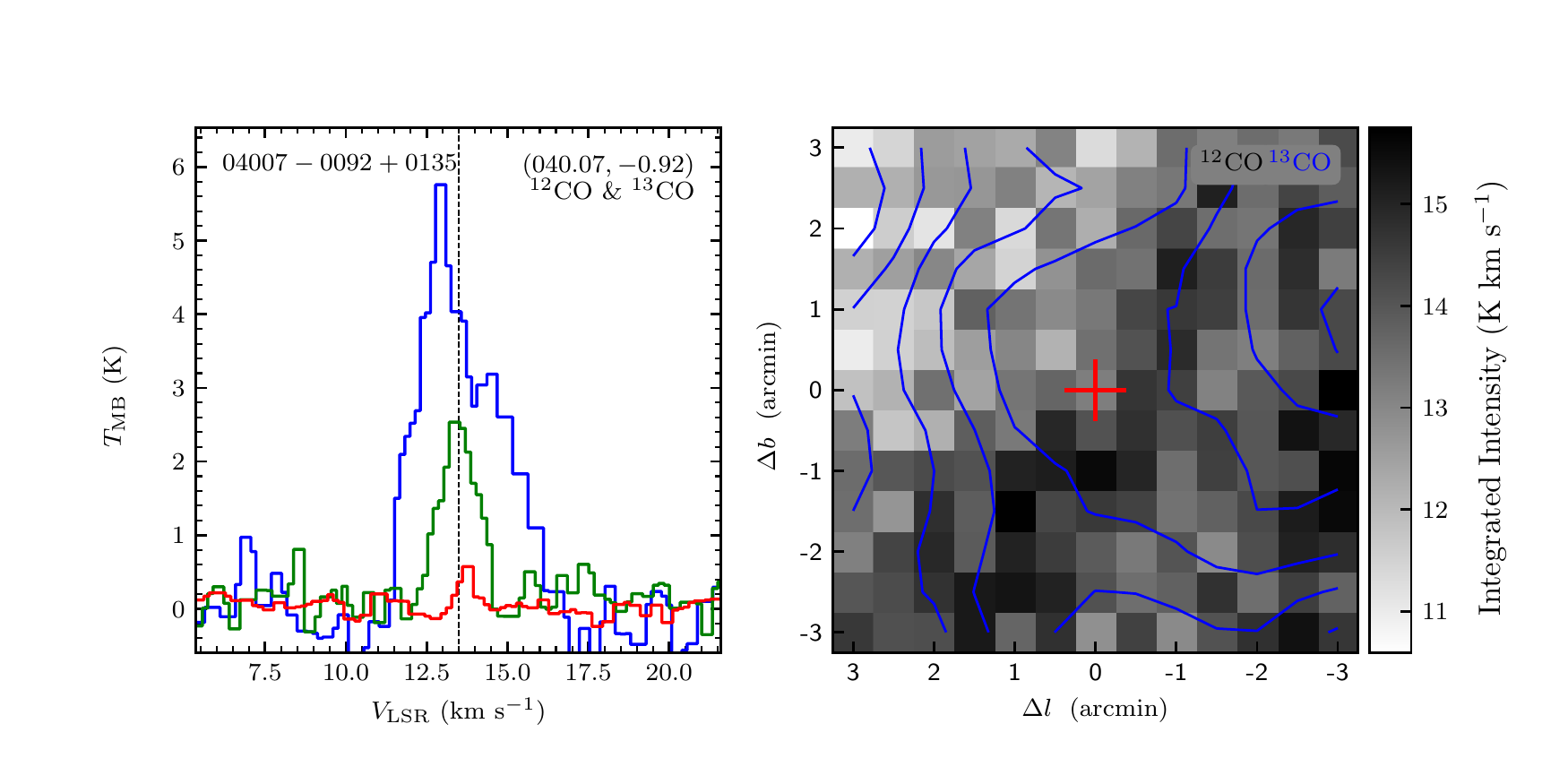}
\includegraphics[width=9.0cm,angle=0]{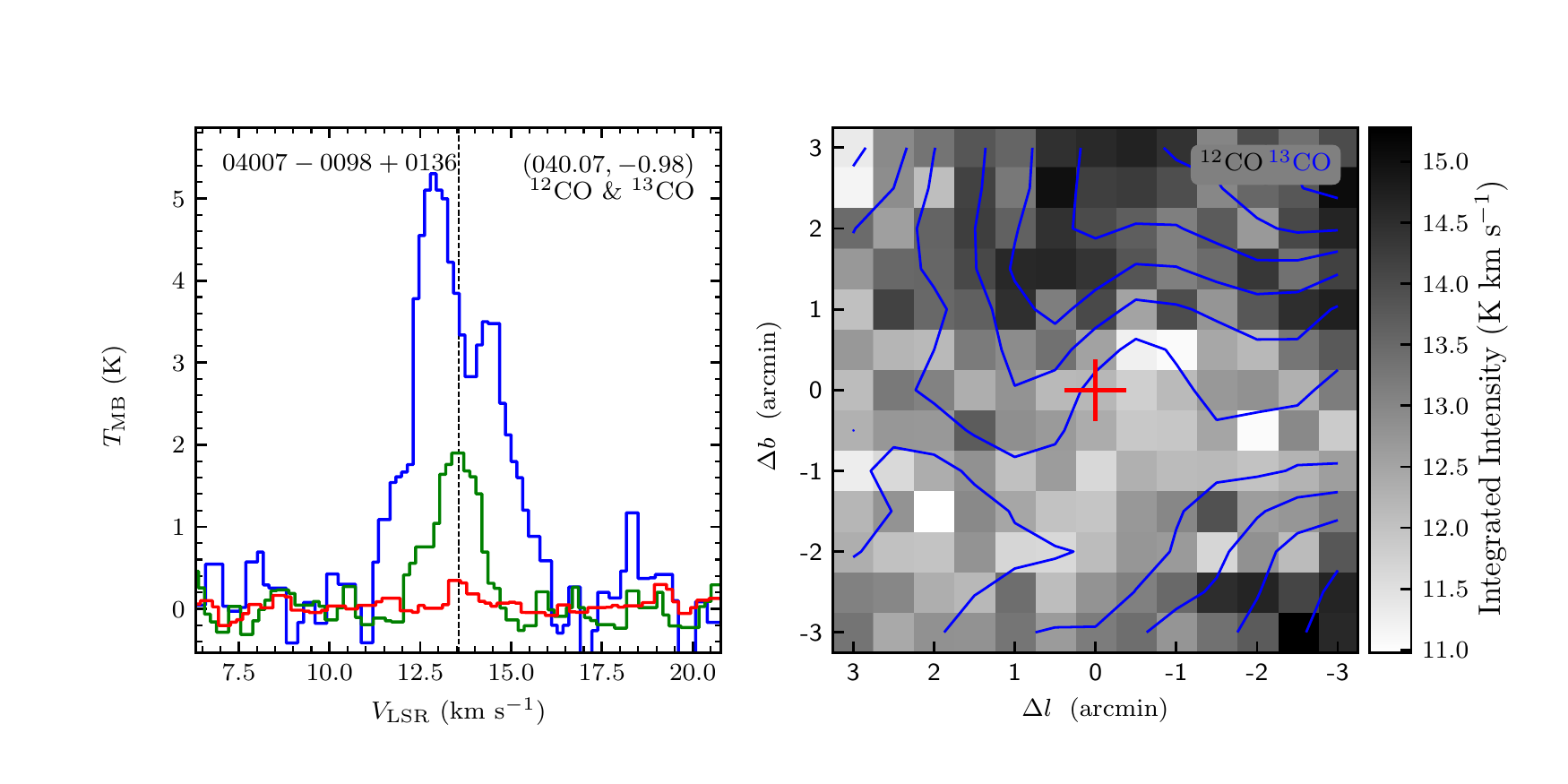}
\vspace{-0.5cm}

\includegraphics[width=9.0cm,angle=0]{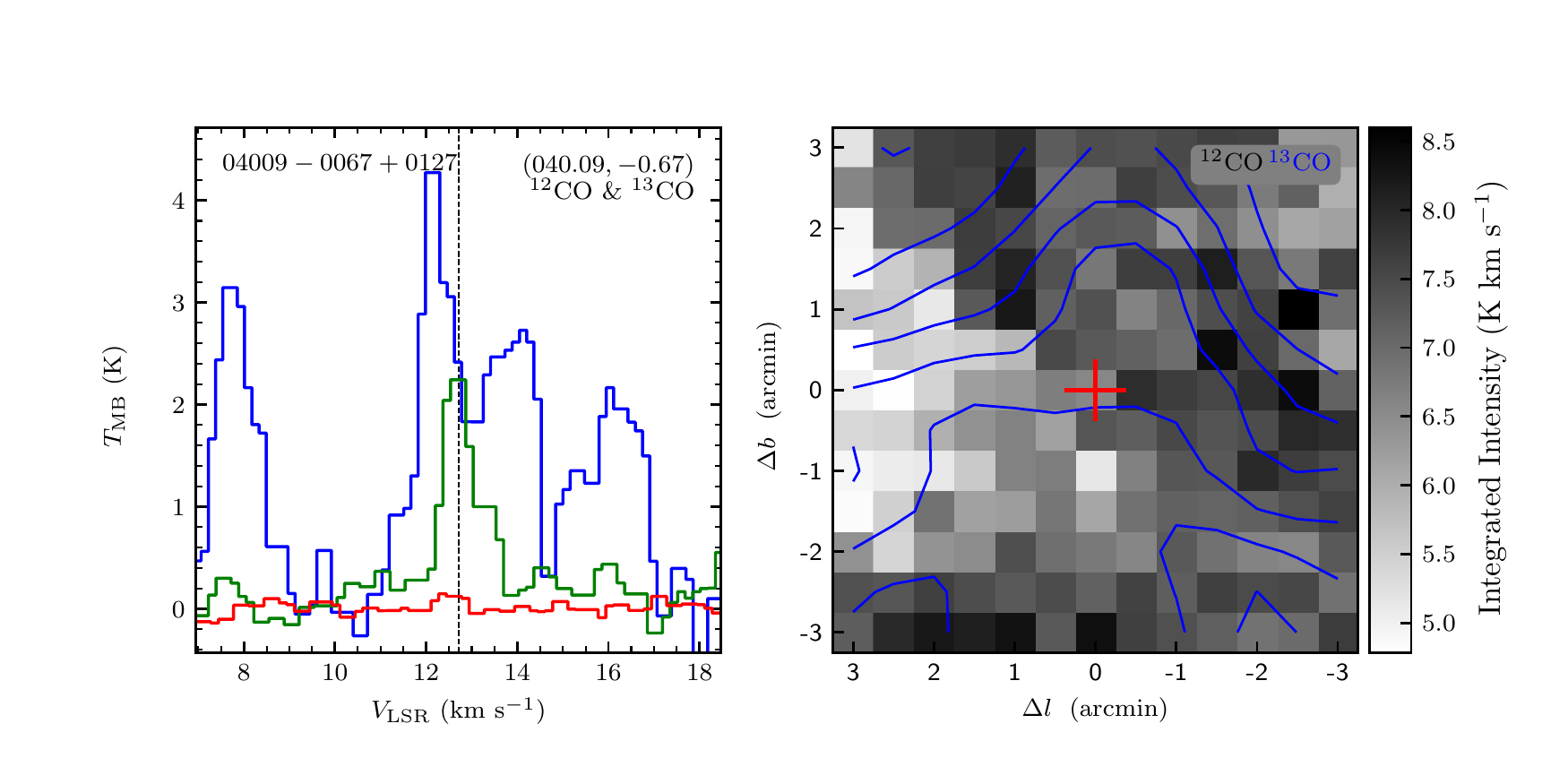}
\includegraphics[width=9.0cm,angle=0]{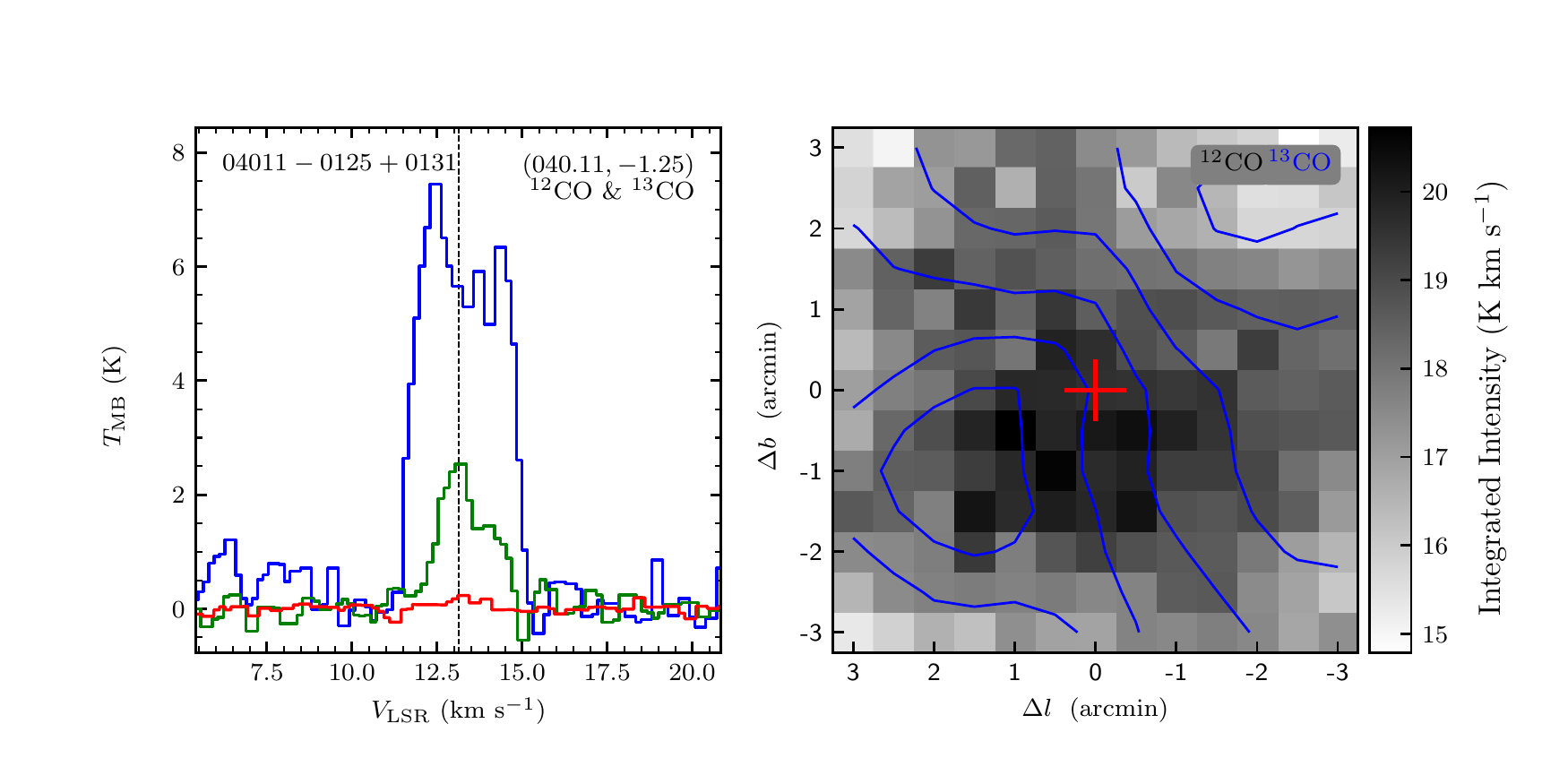}
\end{figure}
\clearpage

\begin{figure}
\includegraphics[width=9.0cm,angle=0]{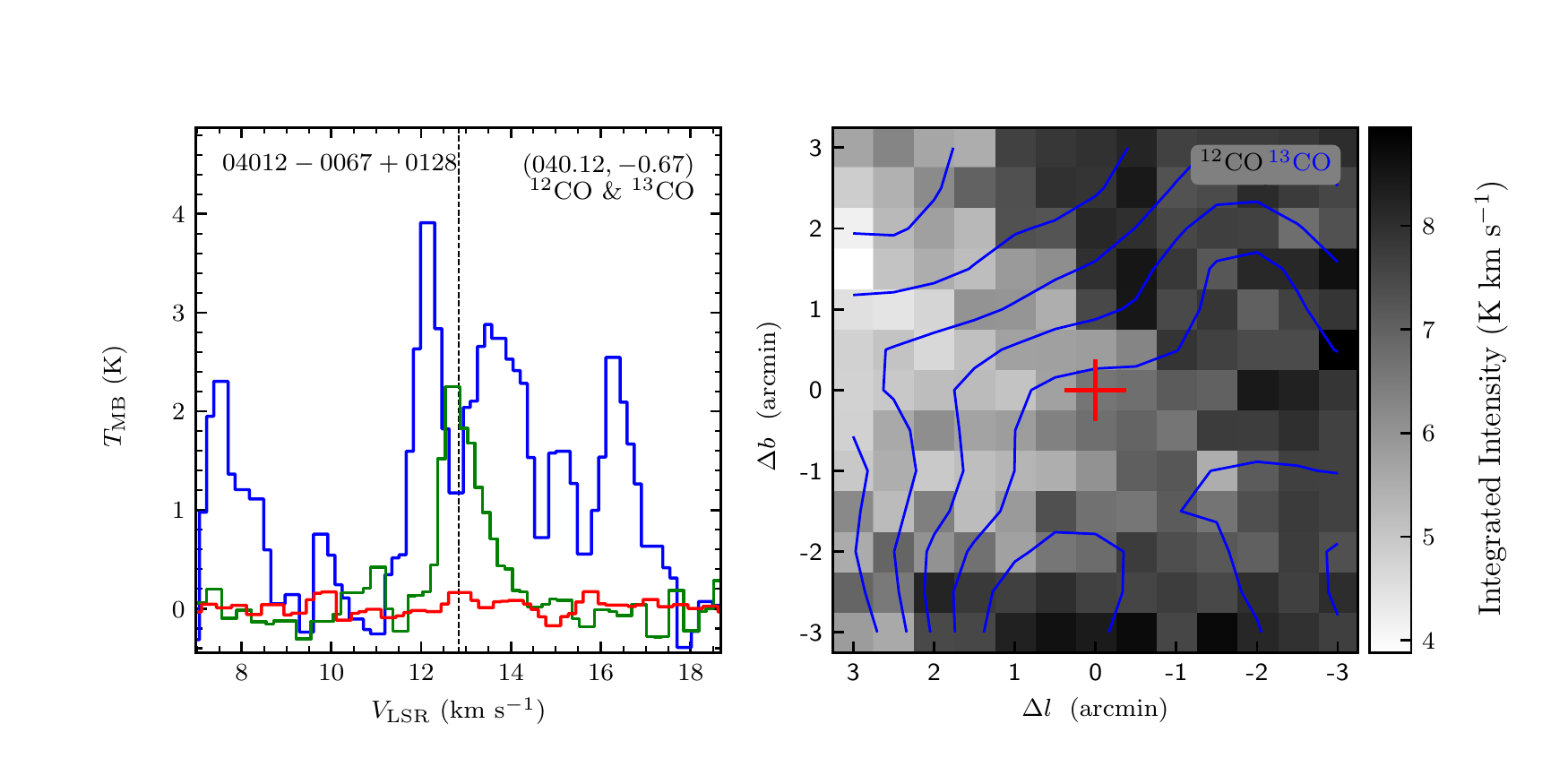}
\includegraphics[width=9.0cm,angle=0]{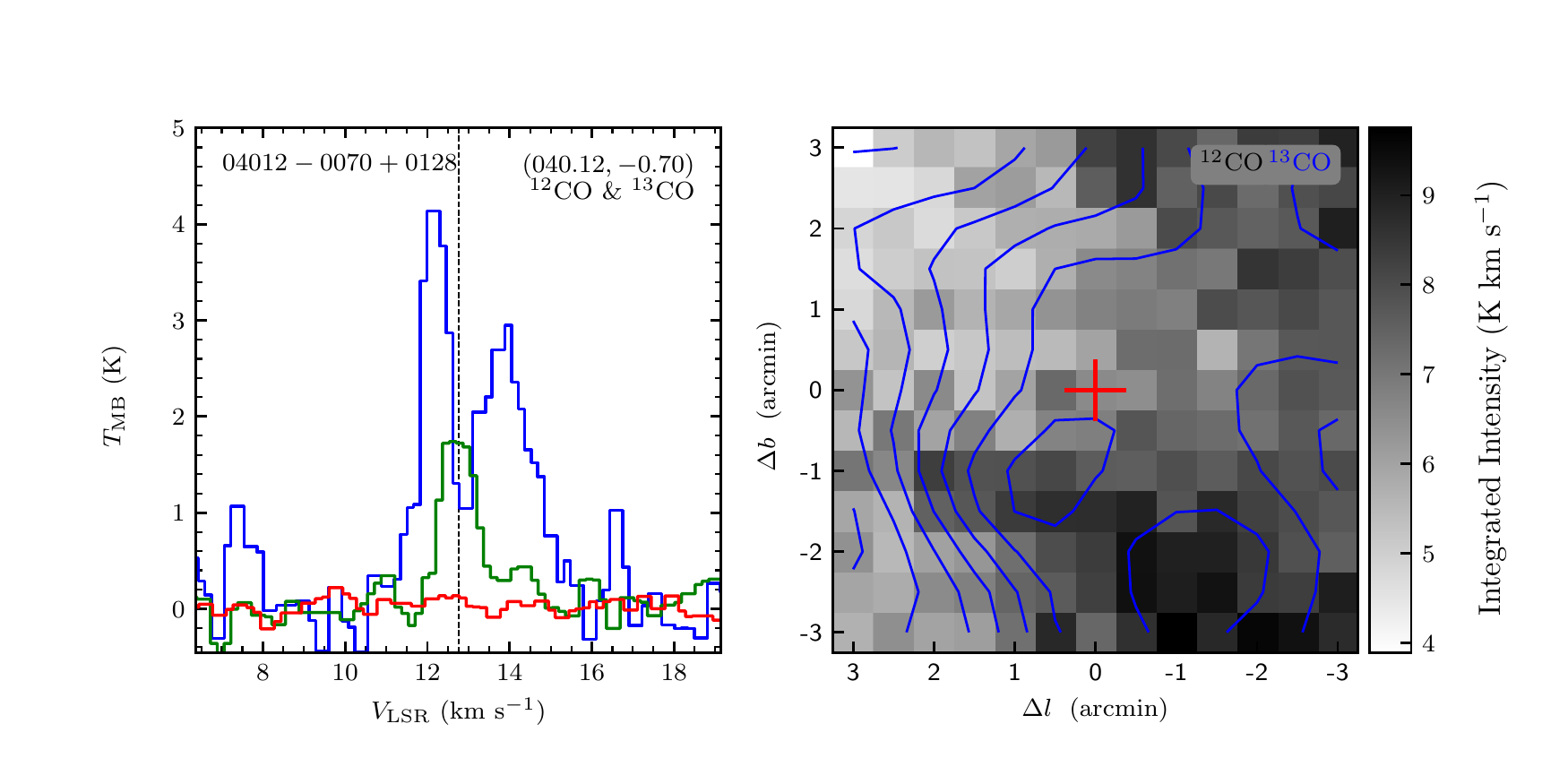}
\vspace{-0.5cm}

\includegraphics[width=9.0cm,angle=0]{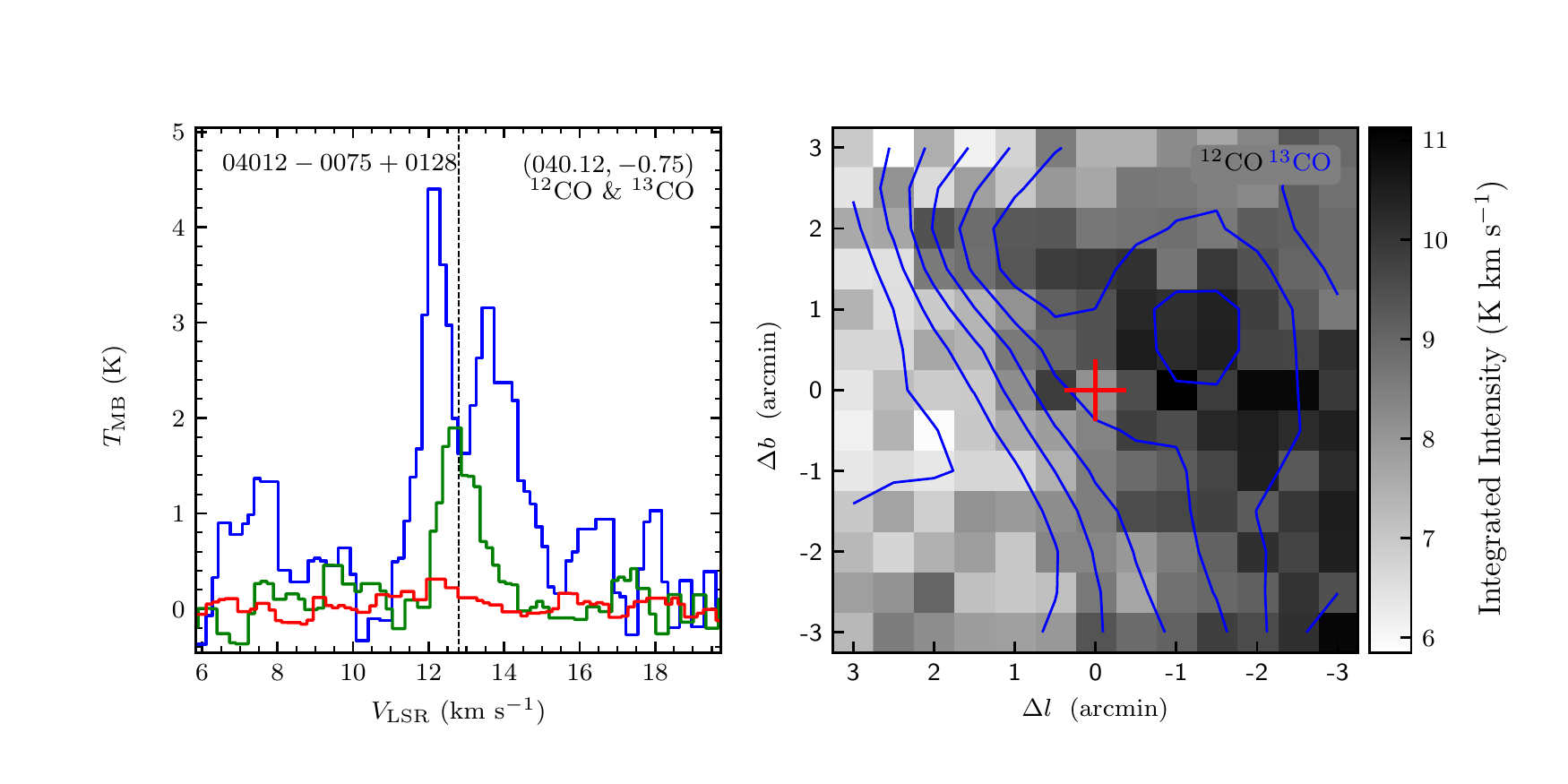}
\includegraphics[width=9.0cm,angle=0]{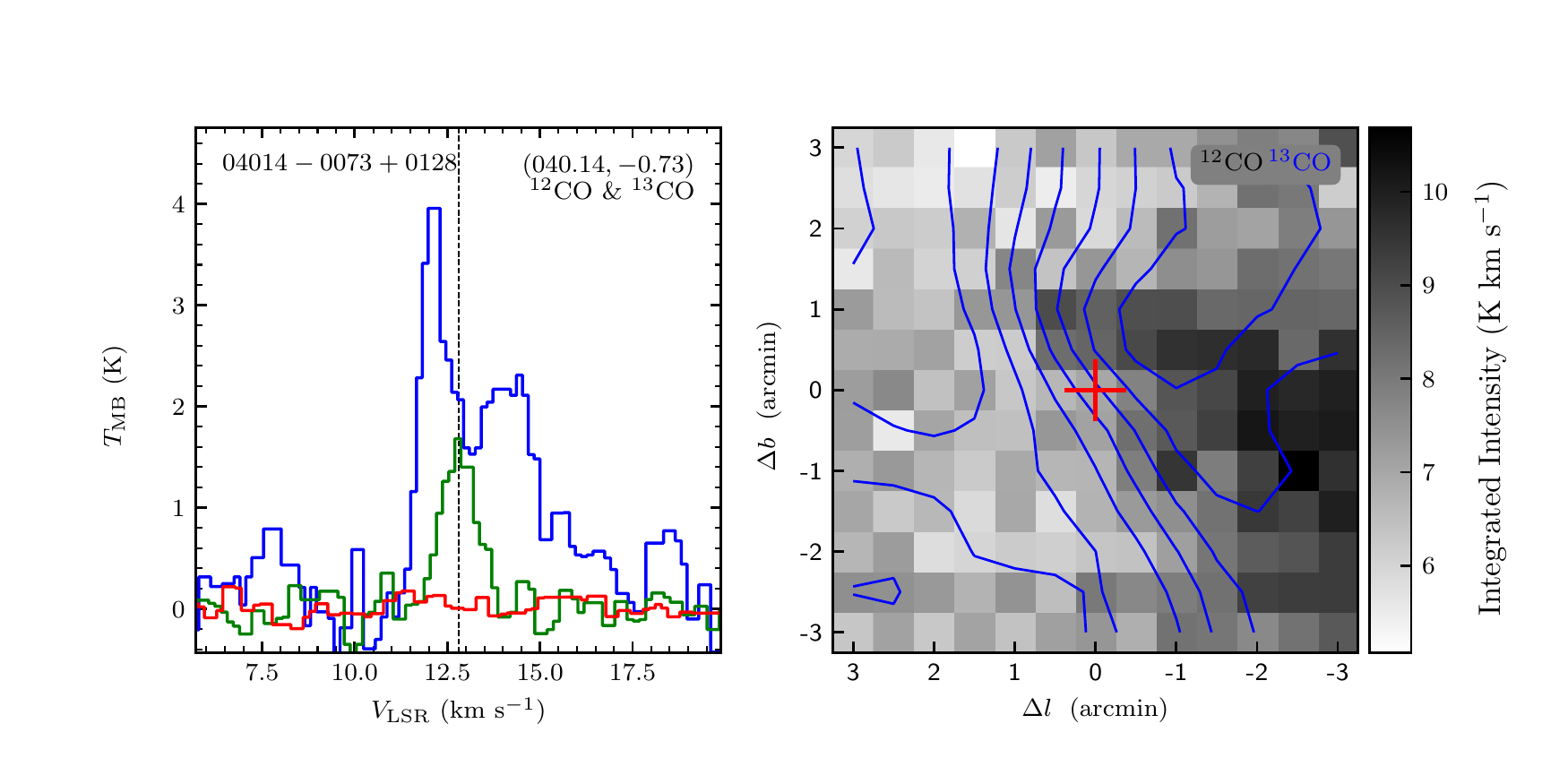}
\vspace{-0.5cm}

\includegraphics[width=9.0cm,angle=0]{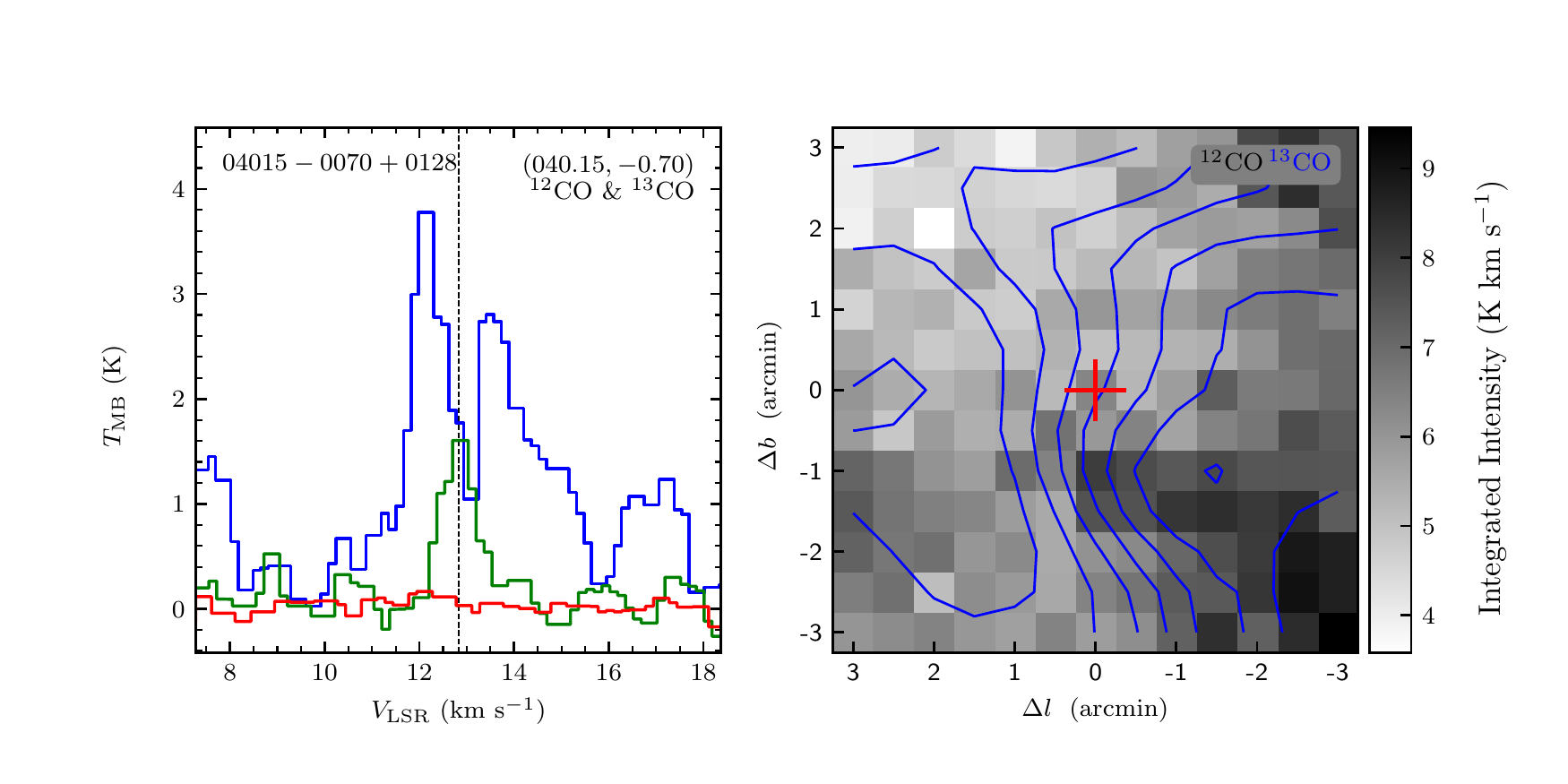}
\includegraphics[width=9.0cm,angle=0]{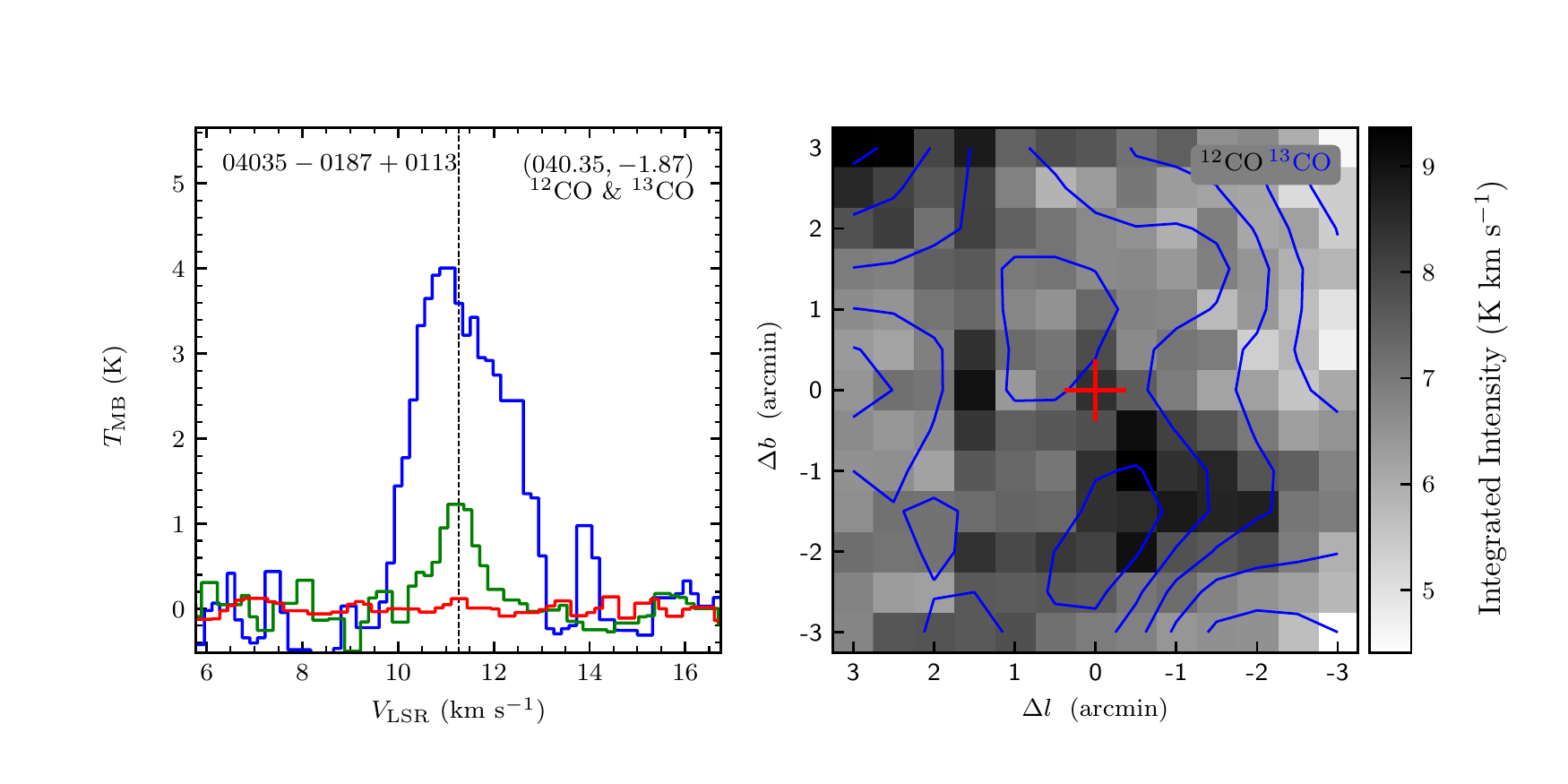}
\vspace{-0.5cm}

\includegraphics[width=9.0cm,angle=0]{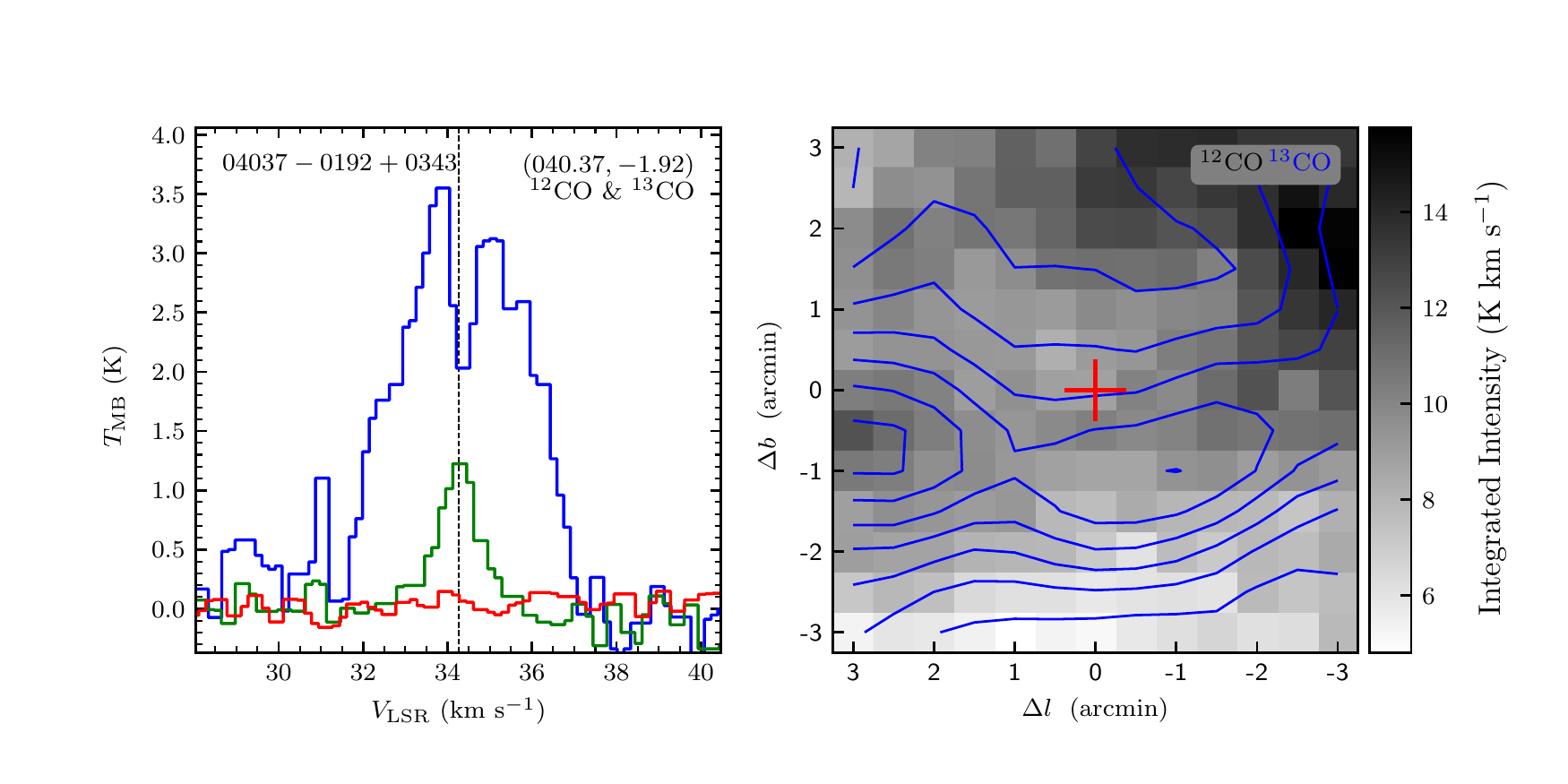}
\includegraphics[width=9.0cm,angle=0]{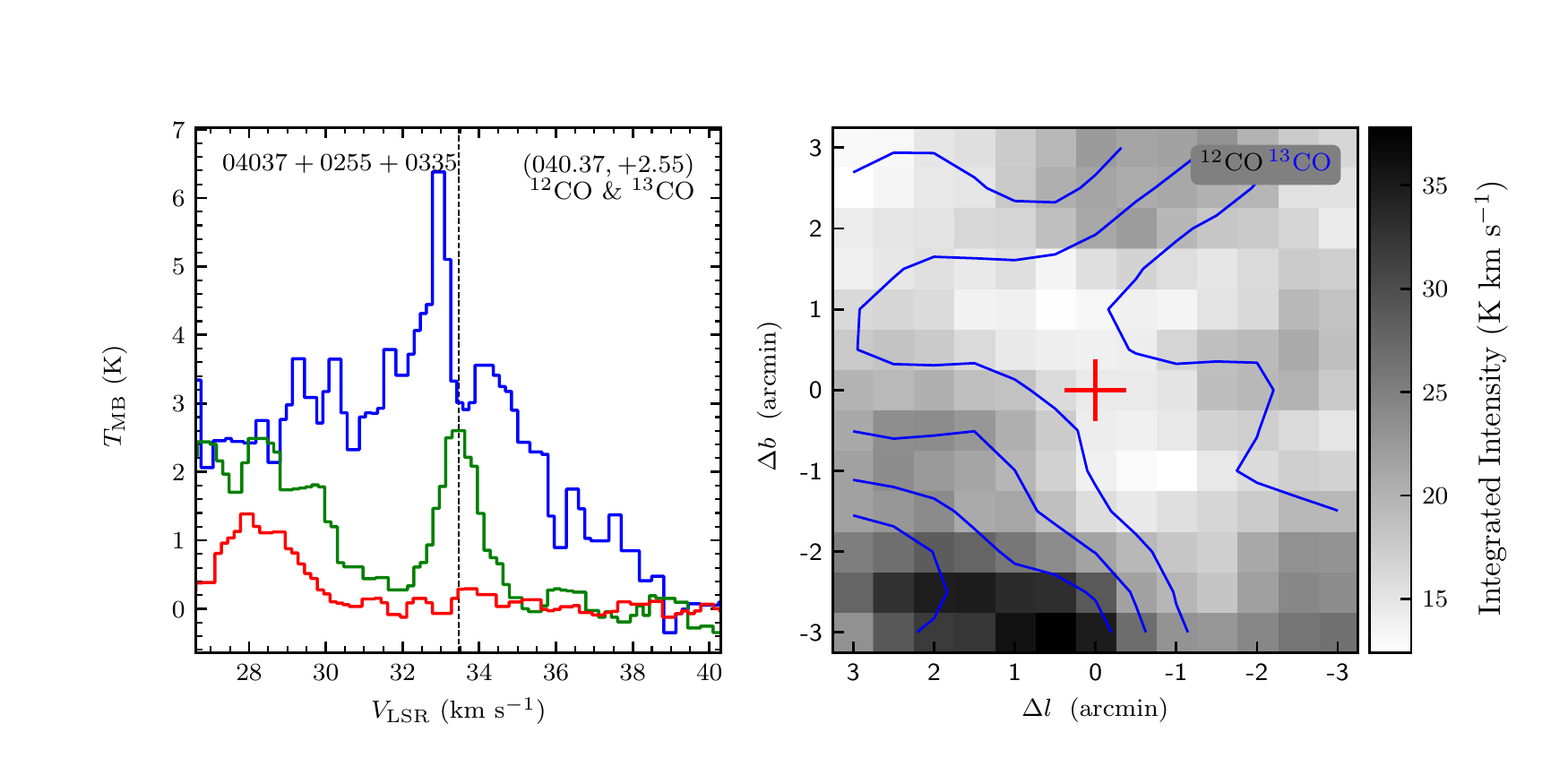}
\vspace{-0.5cm}

\includegraphics[width=9.0cm,angle=0]{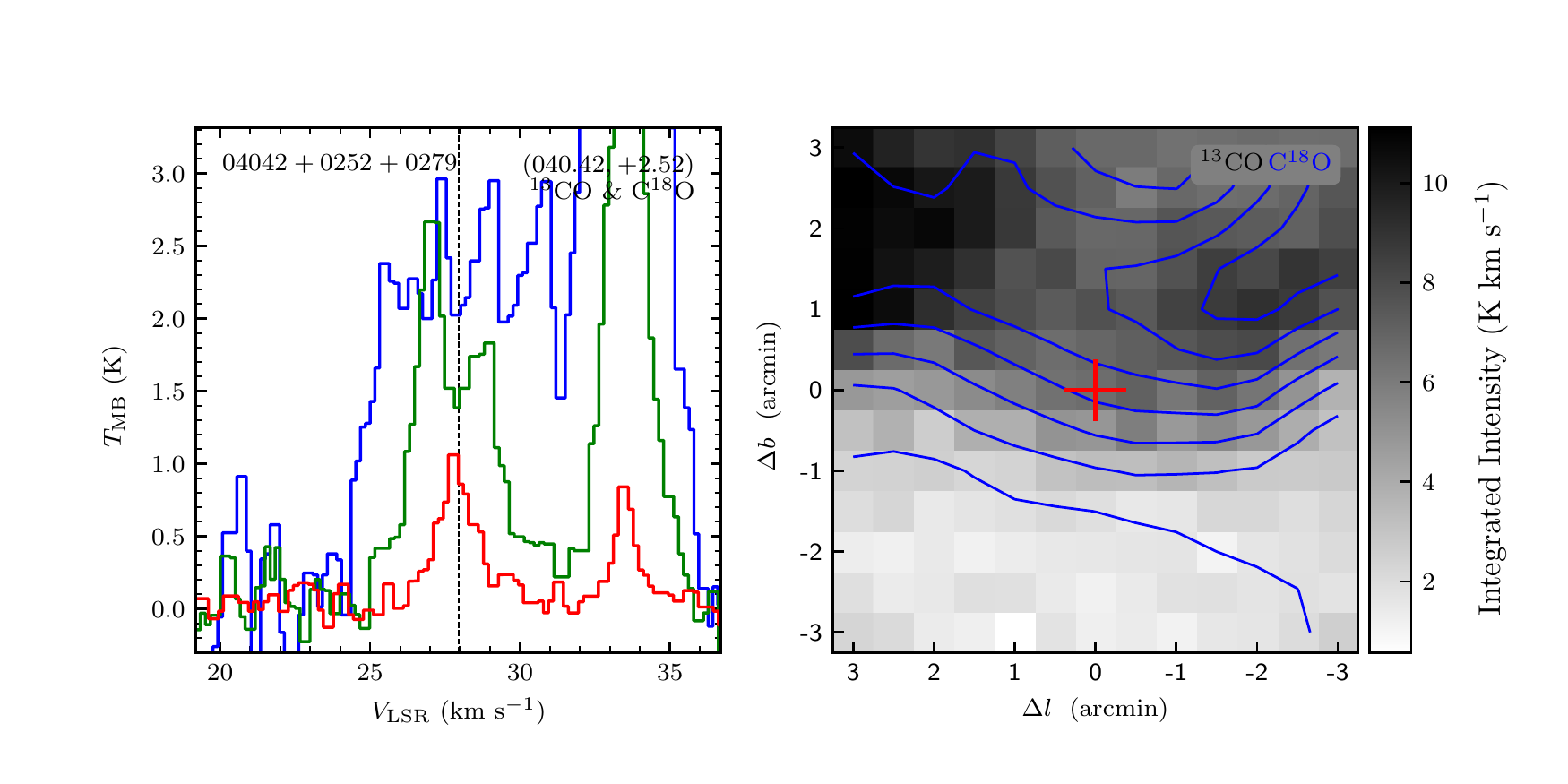}
\includegraphics[width=9.0cm,angle=0]{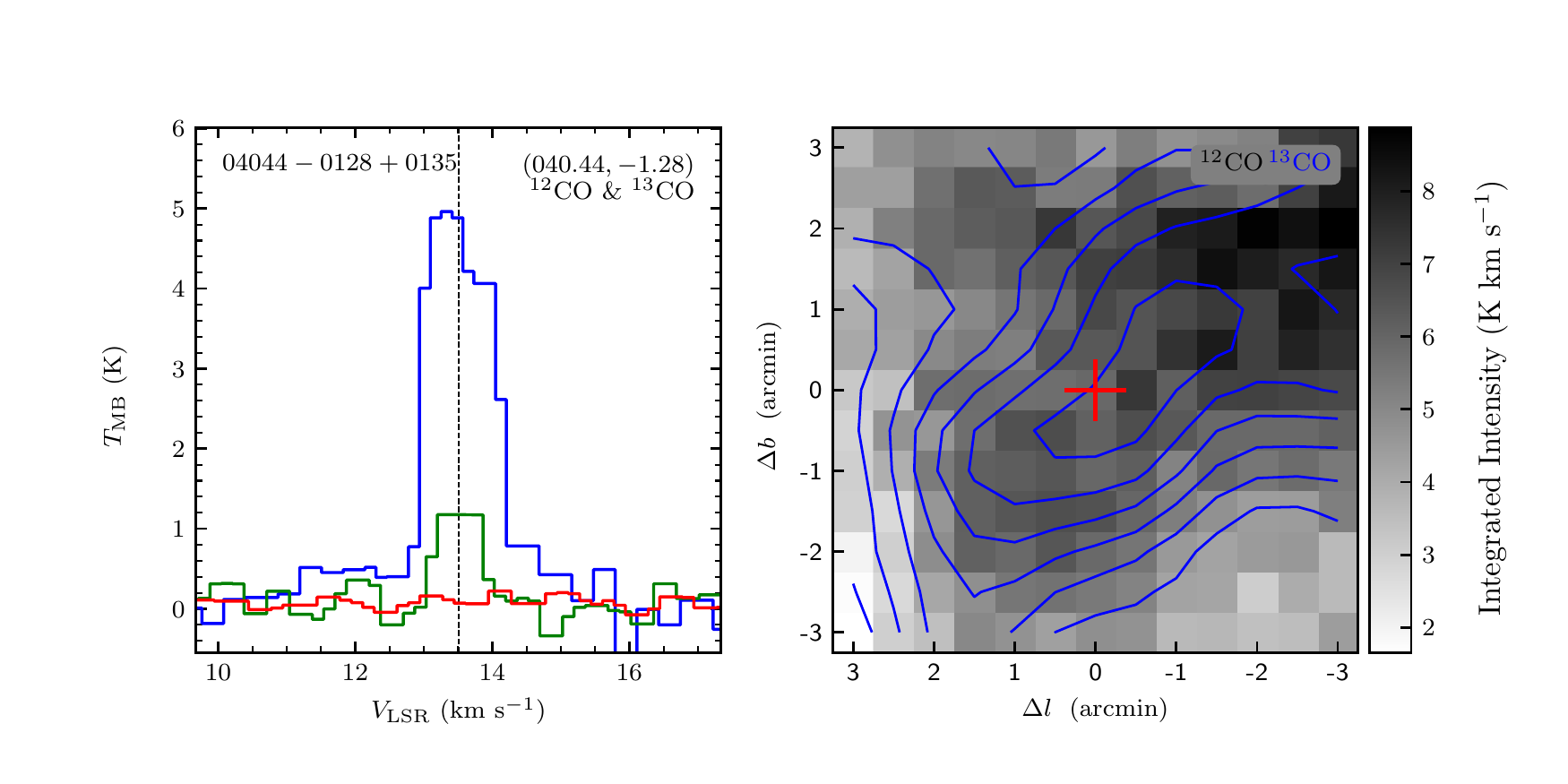}
\end{figure}
\clearpage

\begin{figure}
\includegraphics[width=9.0cm,angle=0]{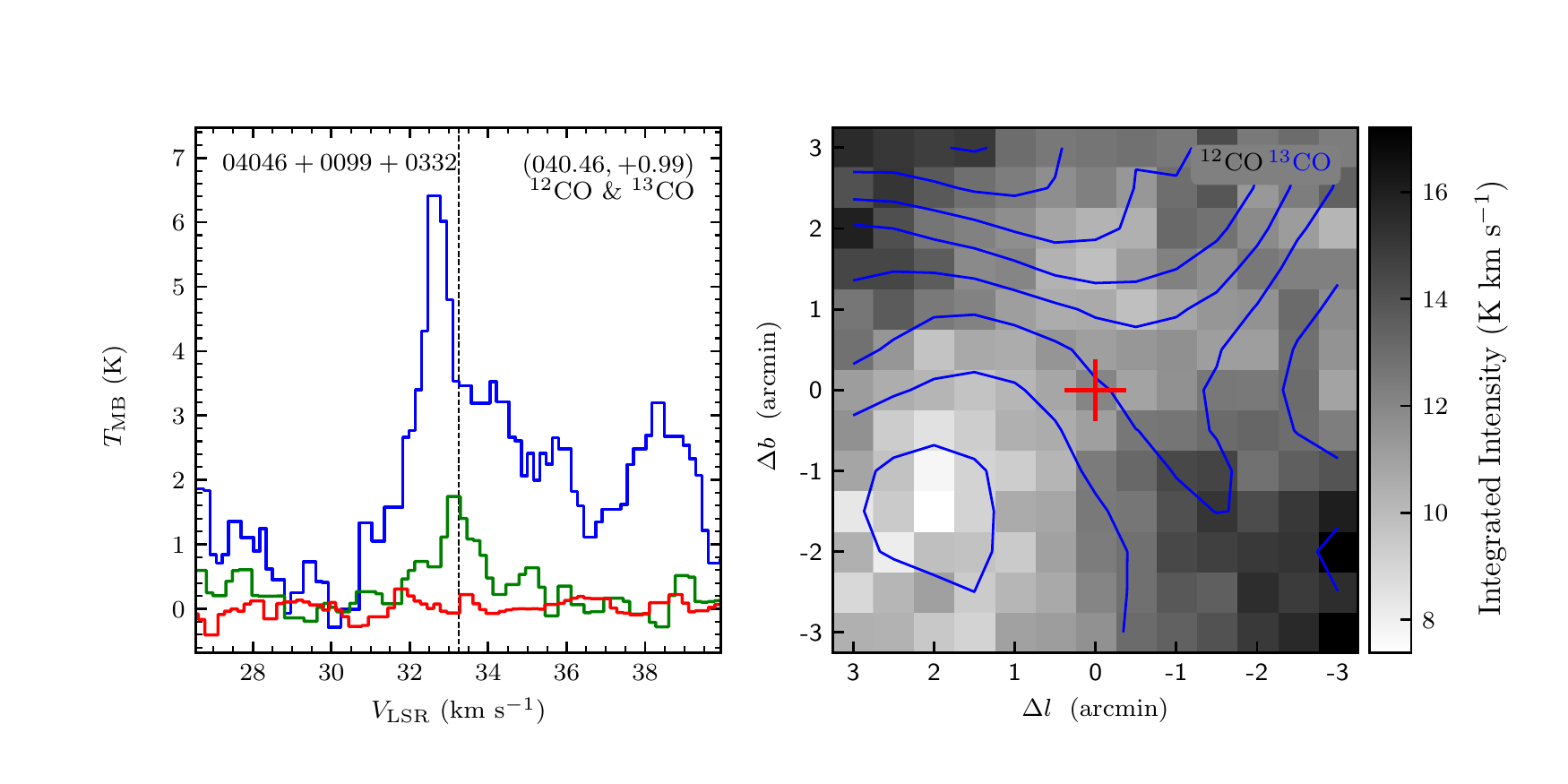}
\includegraphics[width=9.0cm,angle=0]{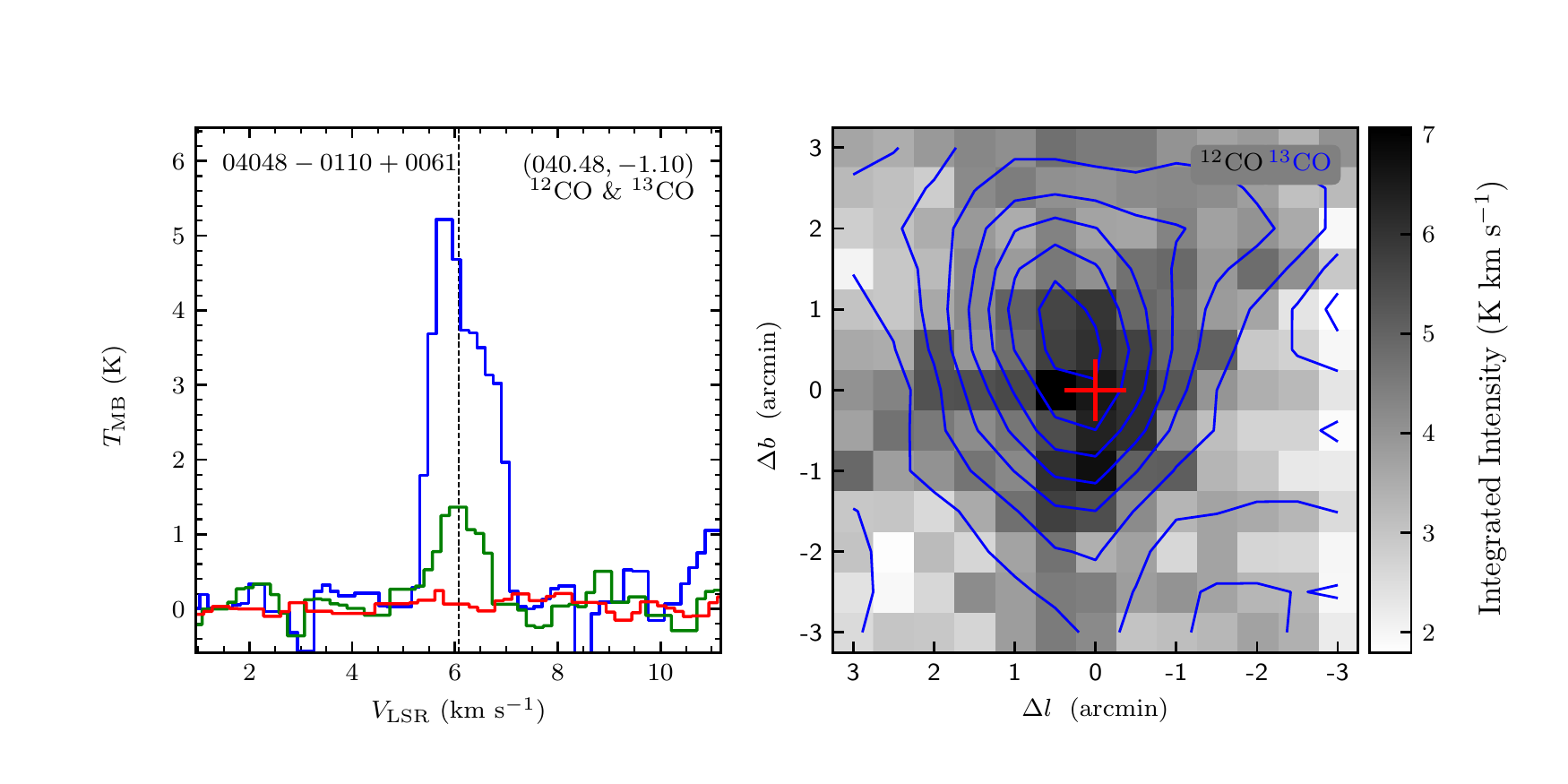}
\vspace{-0.5cm}

\includegraphics[width=9.0cm,angle=0]{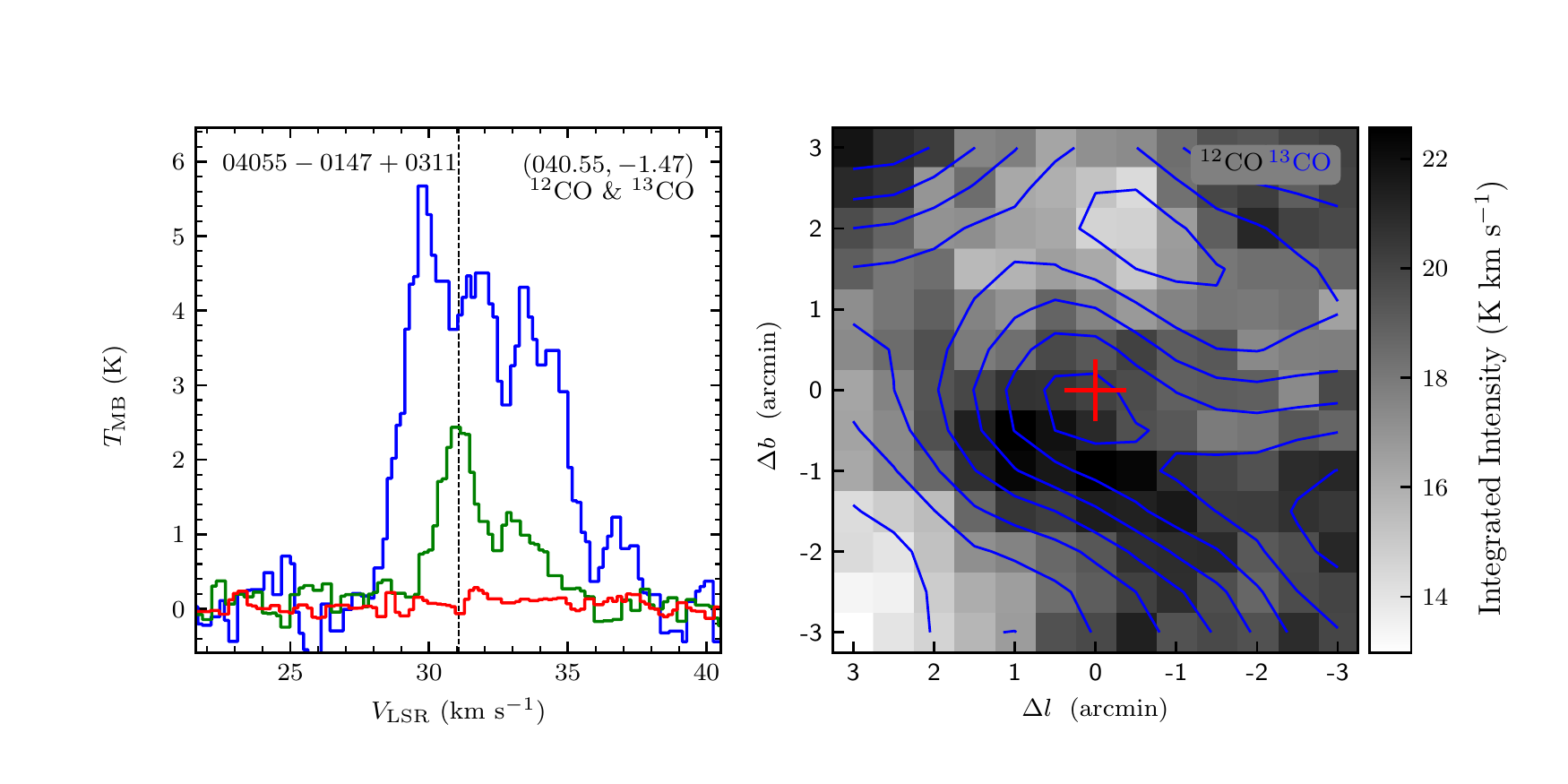}
\includegraphics[width=9.0cm,angle=0]{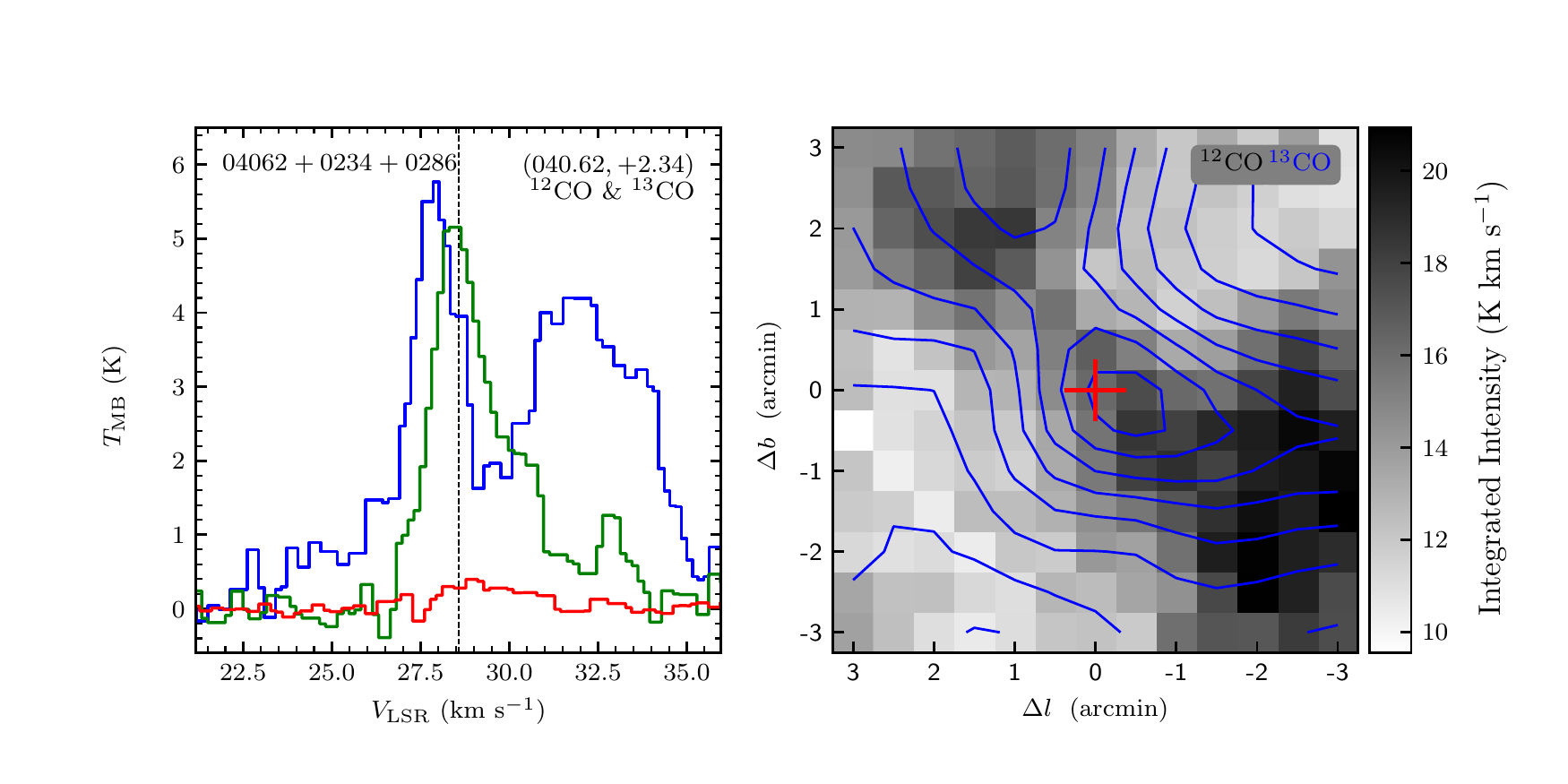}
\vspace{-0.5cm}

\includegraphics[width=9.0cm,angle=0]{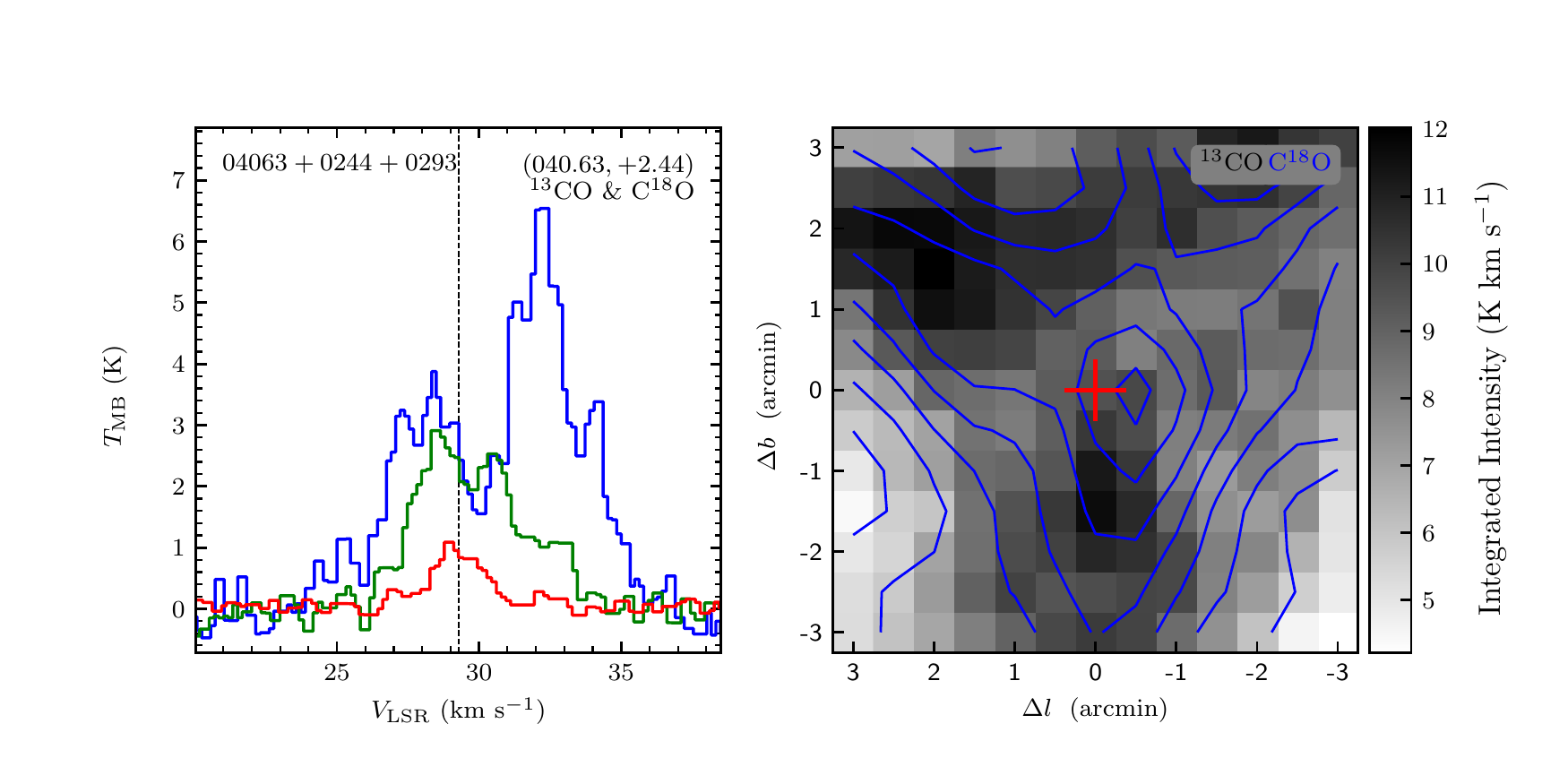}
\includegraphics[width=9.0cm,angle=0]{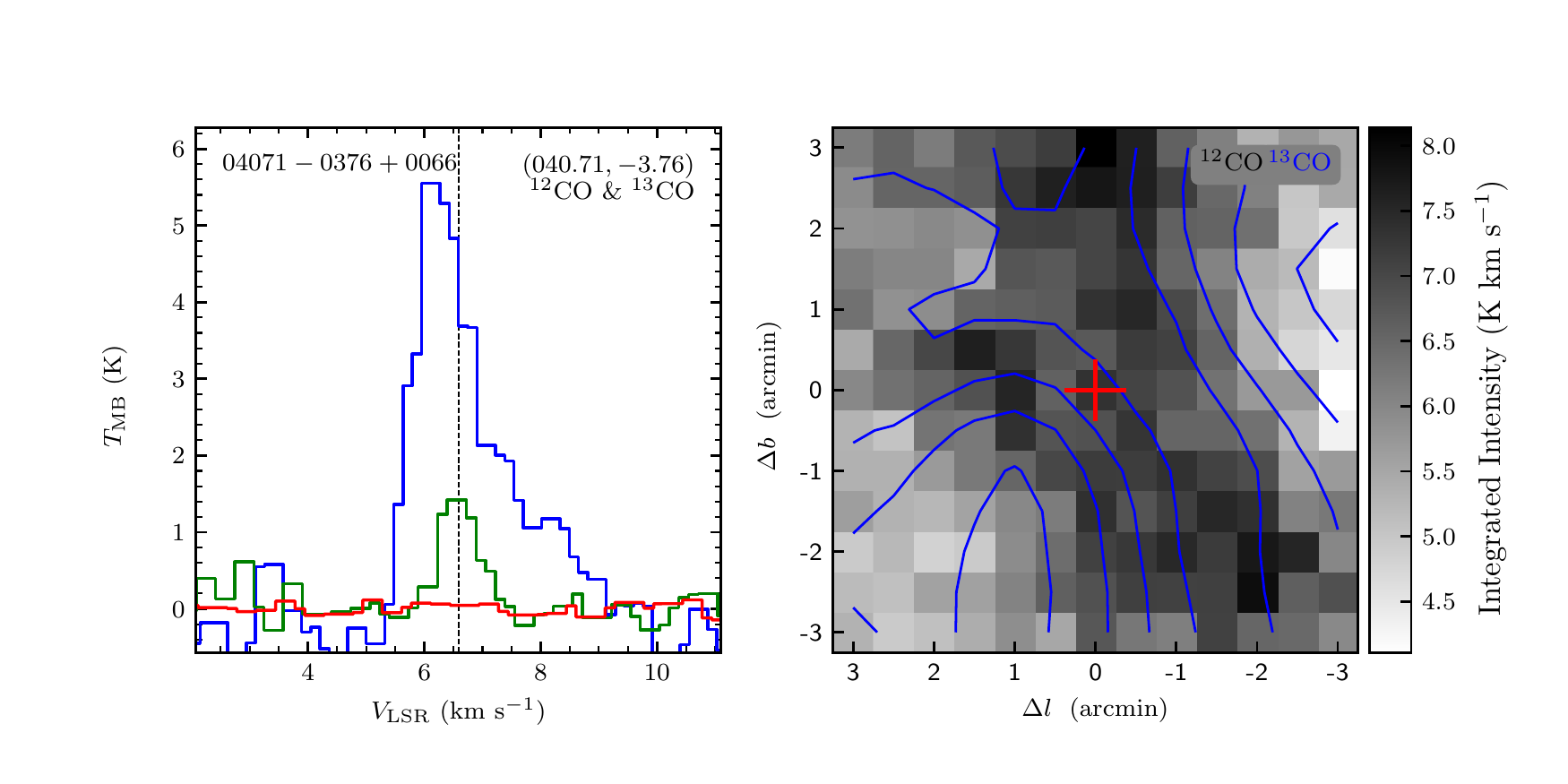}
\vspace{-0.5cm}

\includegraphics[width=9.0cm,angle=0]{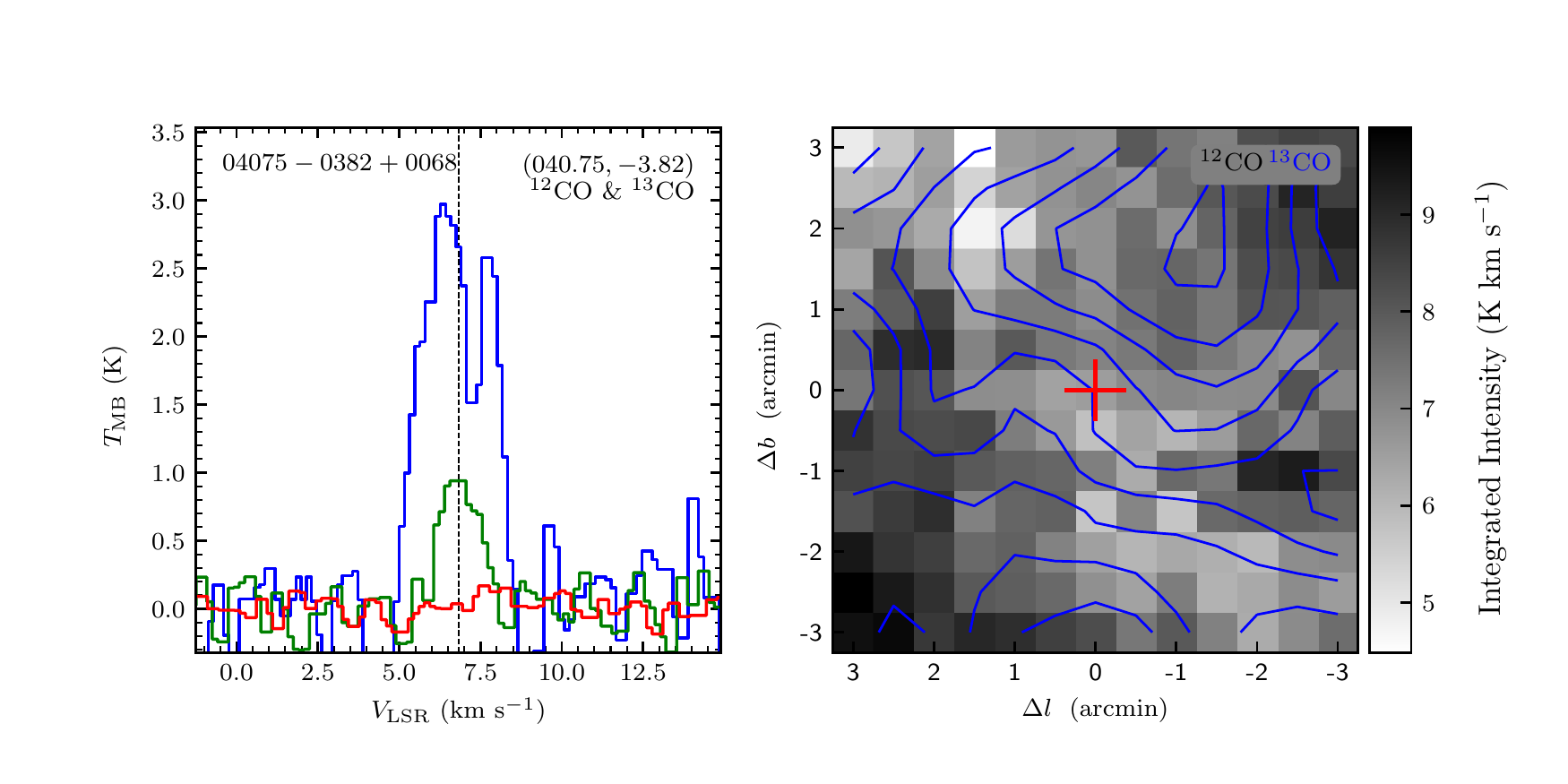}
\includegraphics[width=9.0cm,angle=0]{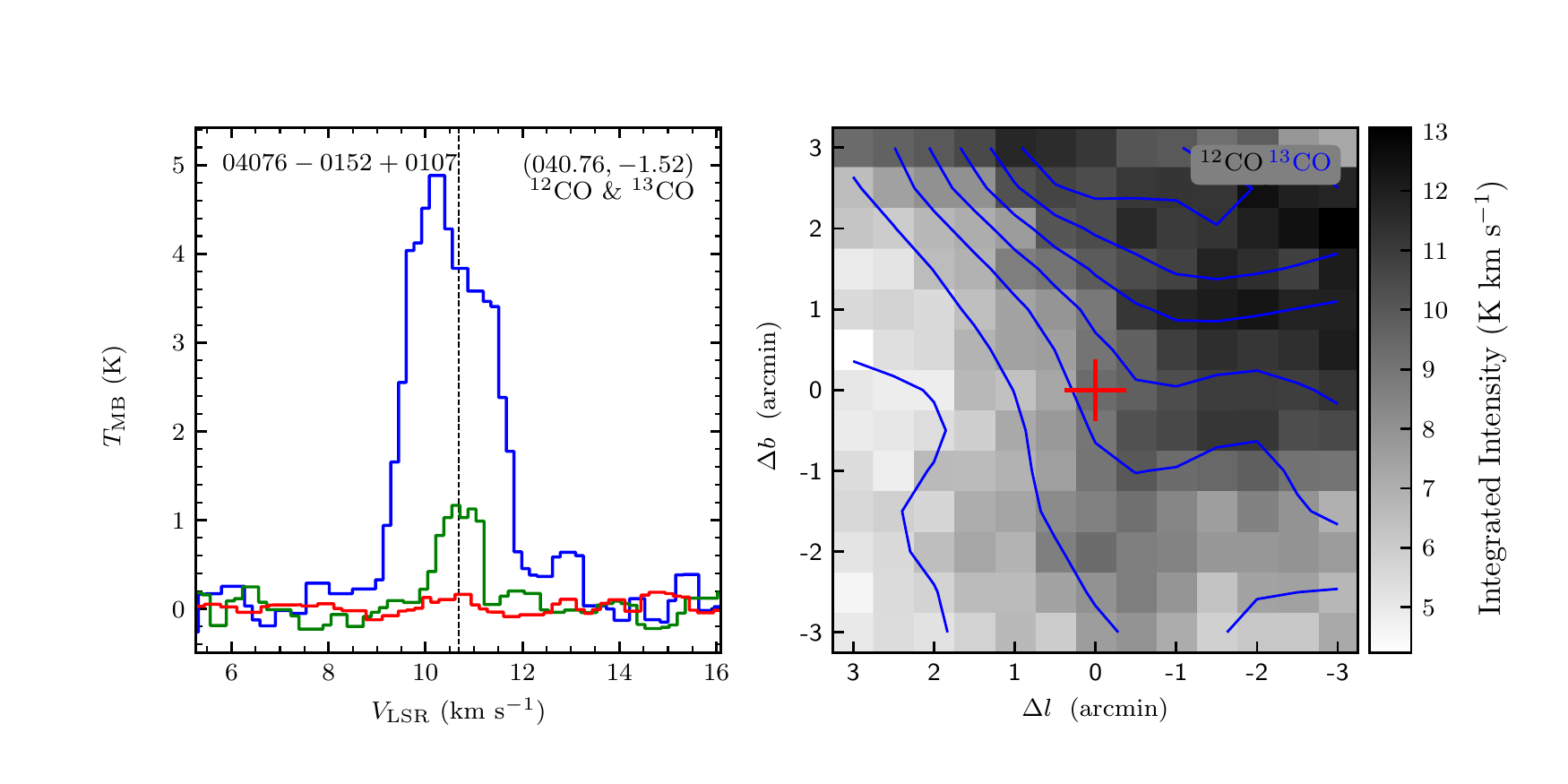}
\vspace{-0.5cm}

\includegraphics[width=9.0cm,angle=0]{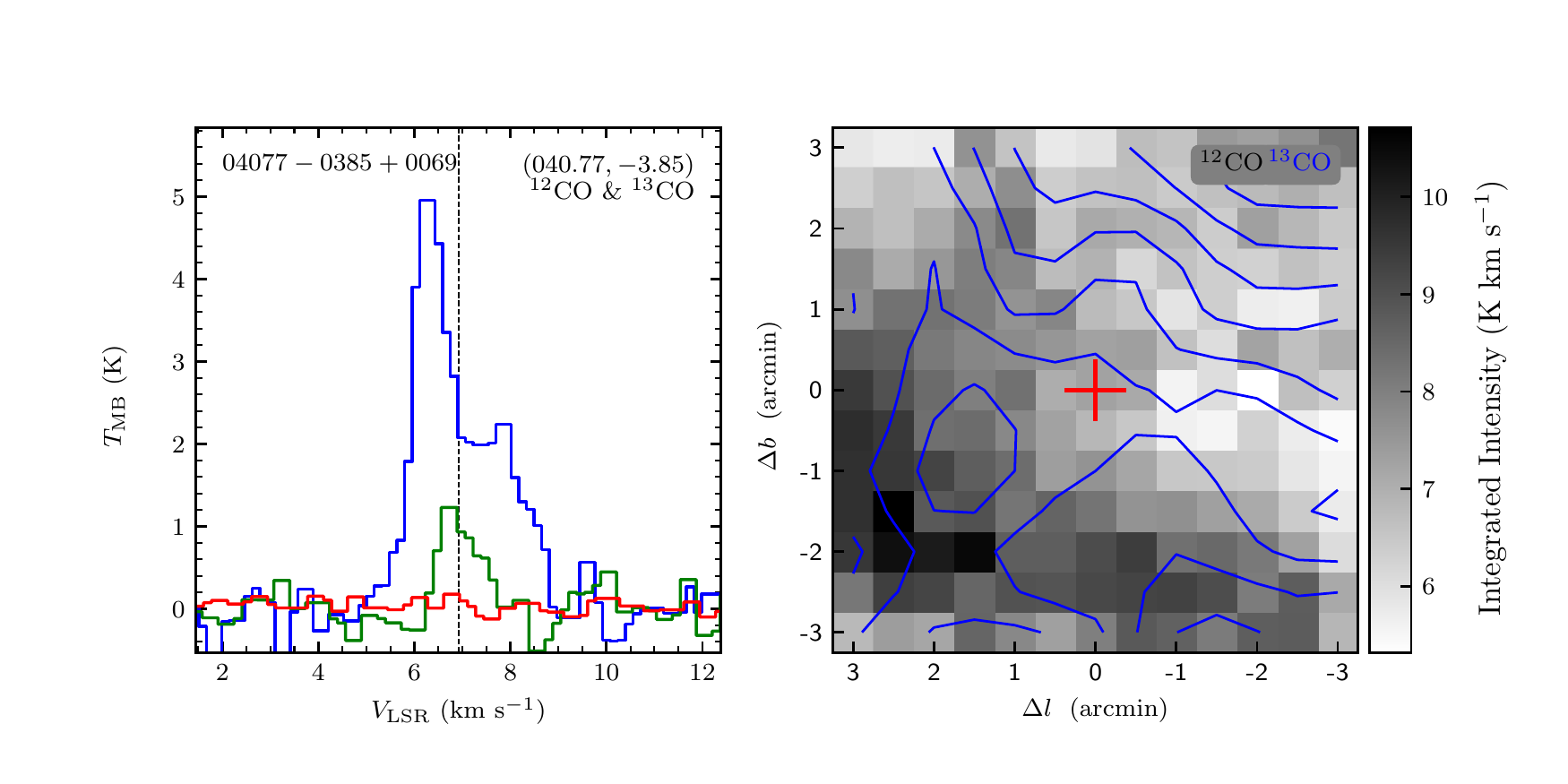}
\includegraphics[width=9.0cm,angle=0]{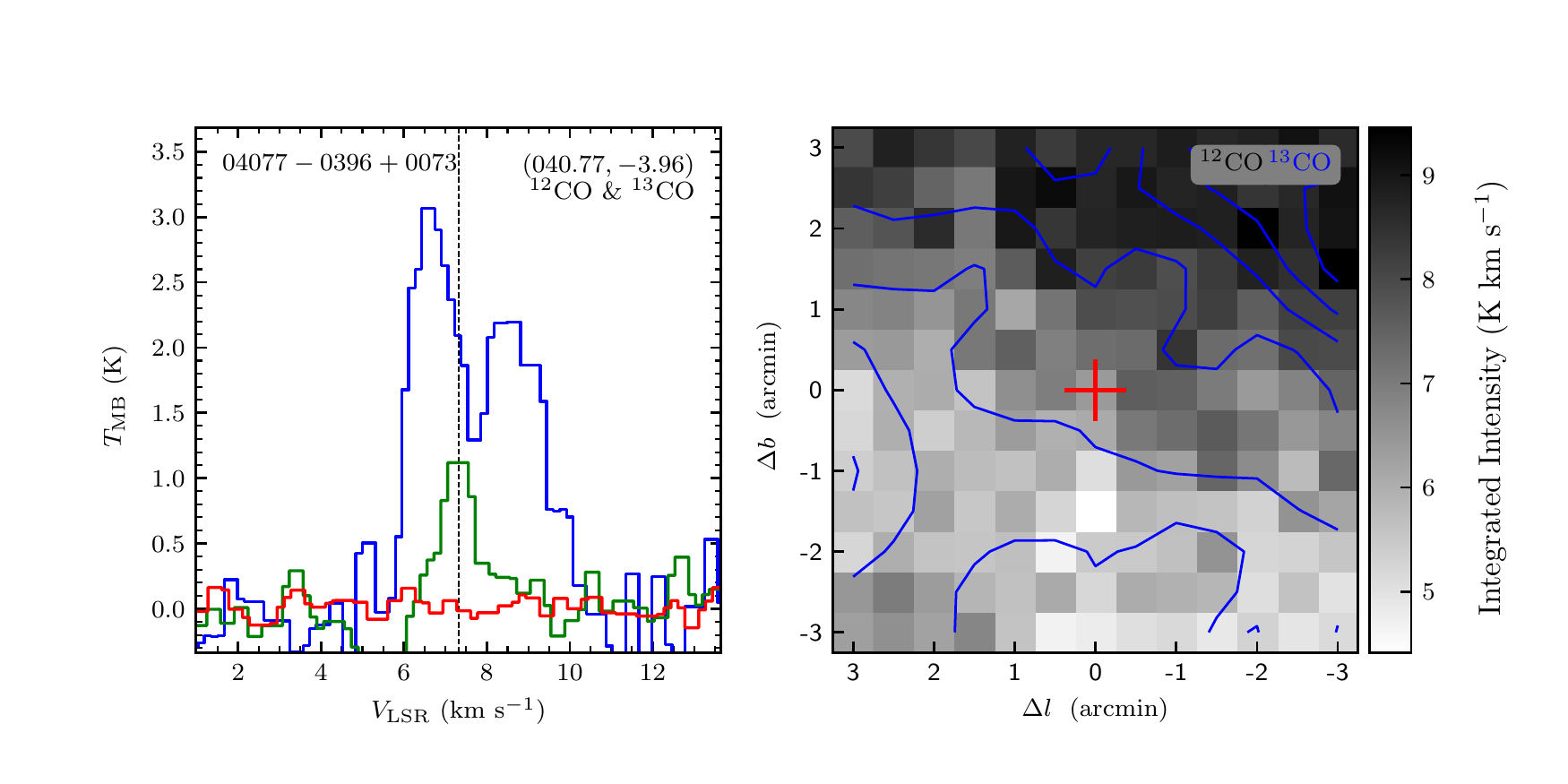}
\end{figure}
\clearpage

\begin{figure}
\includegraphics[width=9.0cm,angle=0]{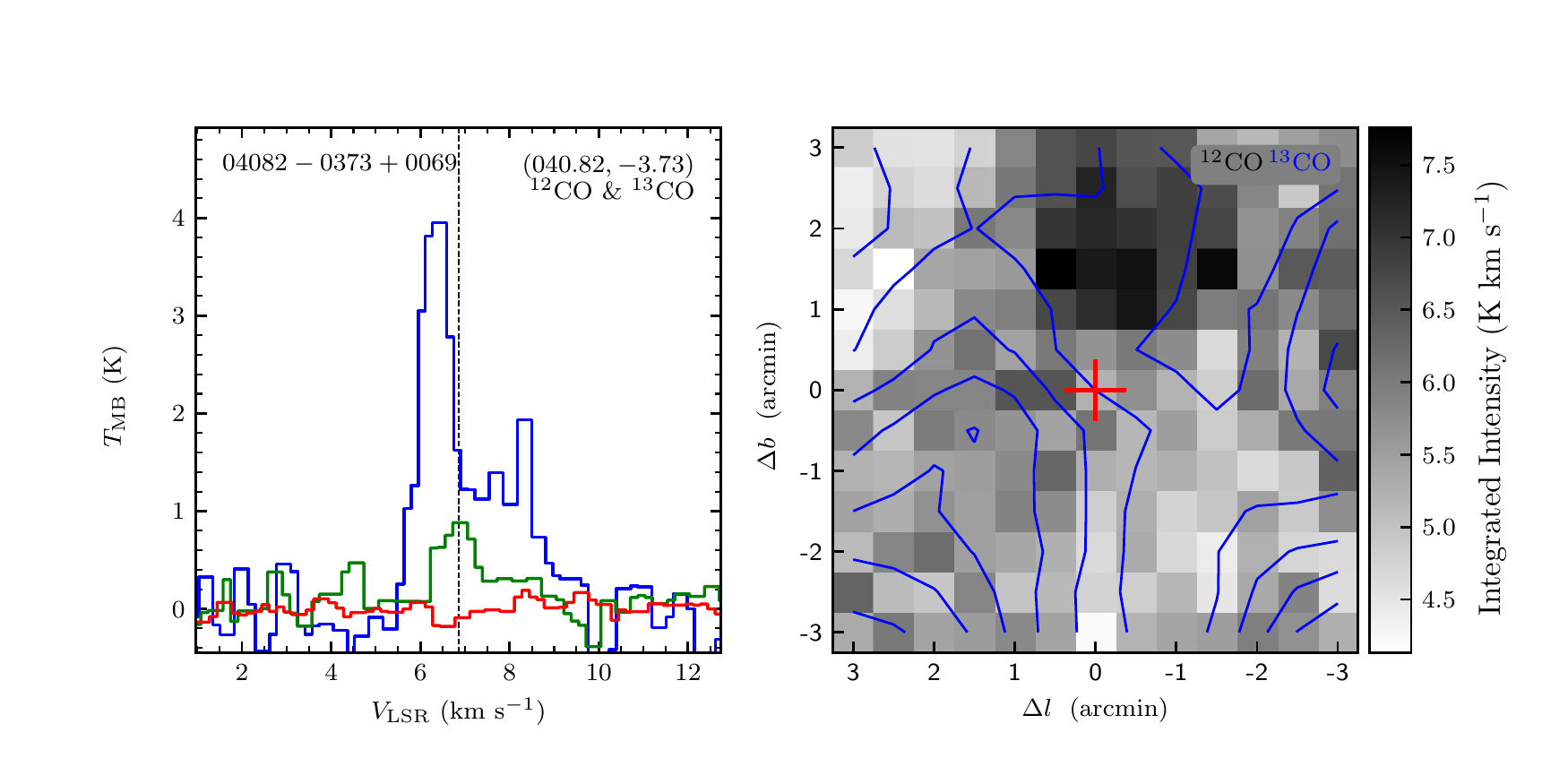}
\includegraphics[width=9.0cm,angle=0]{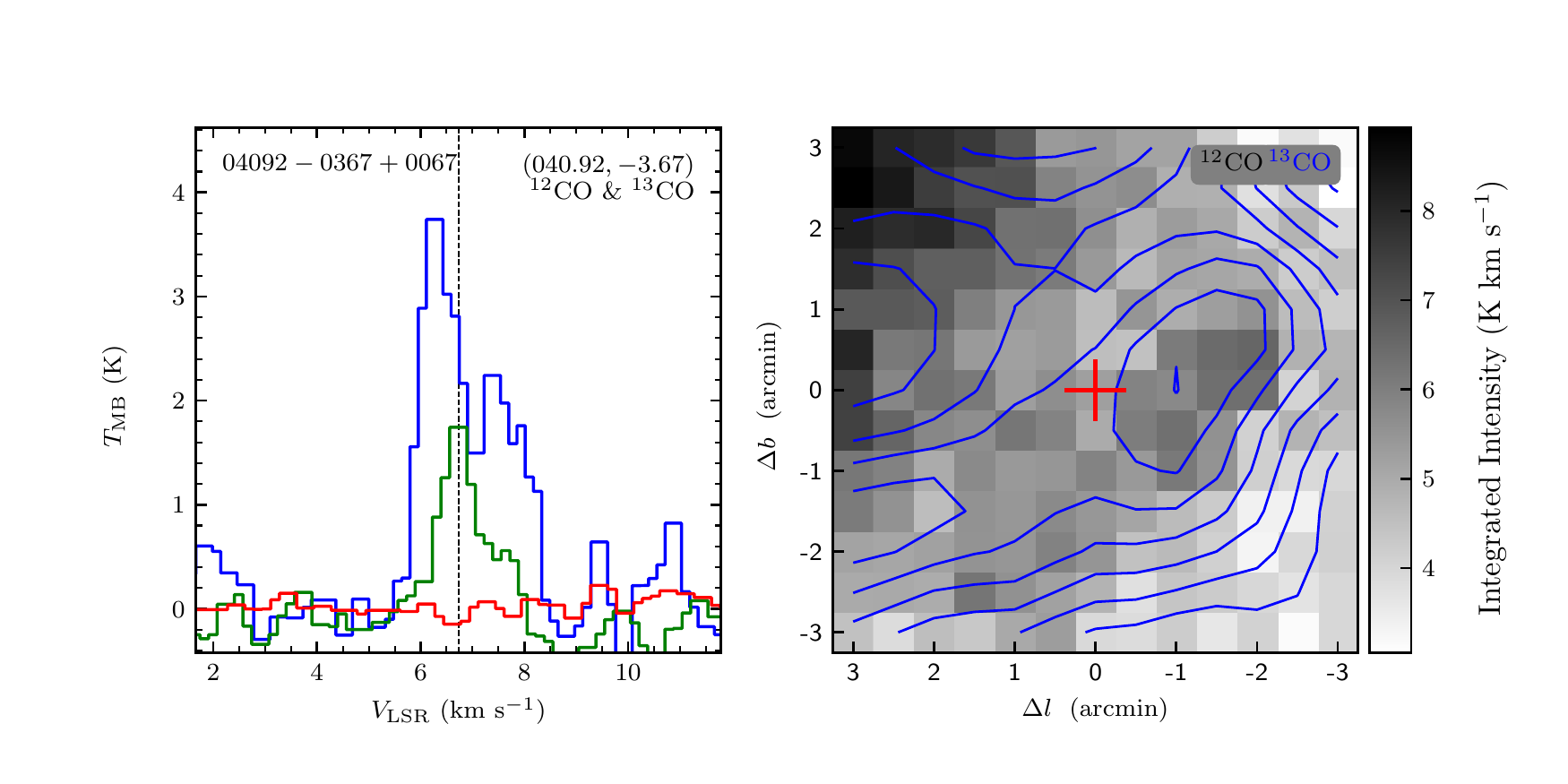}
\vspace{-0.5cm}

\includegraphics[width=9.0cm,angle=0]{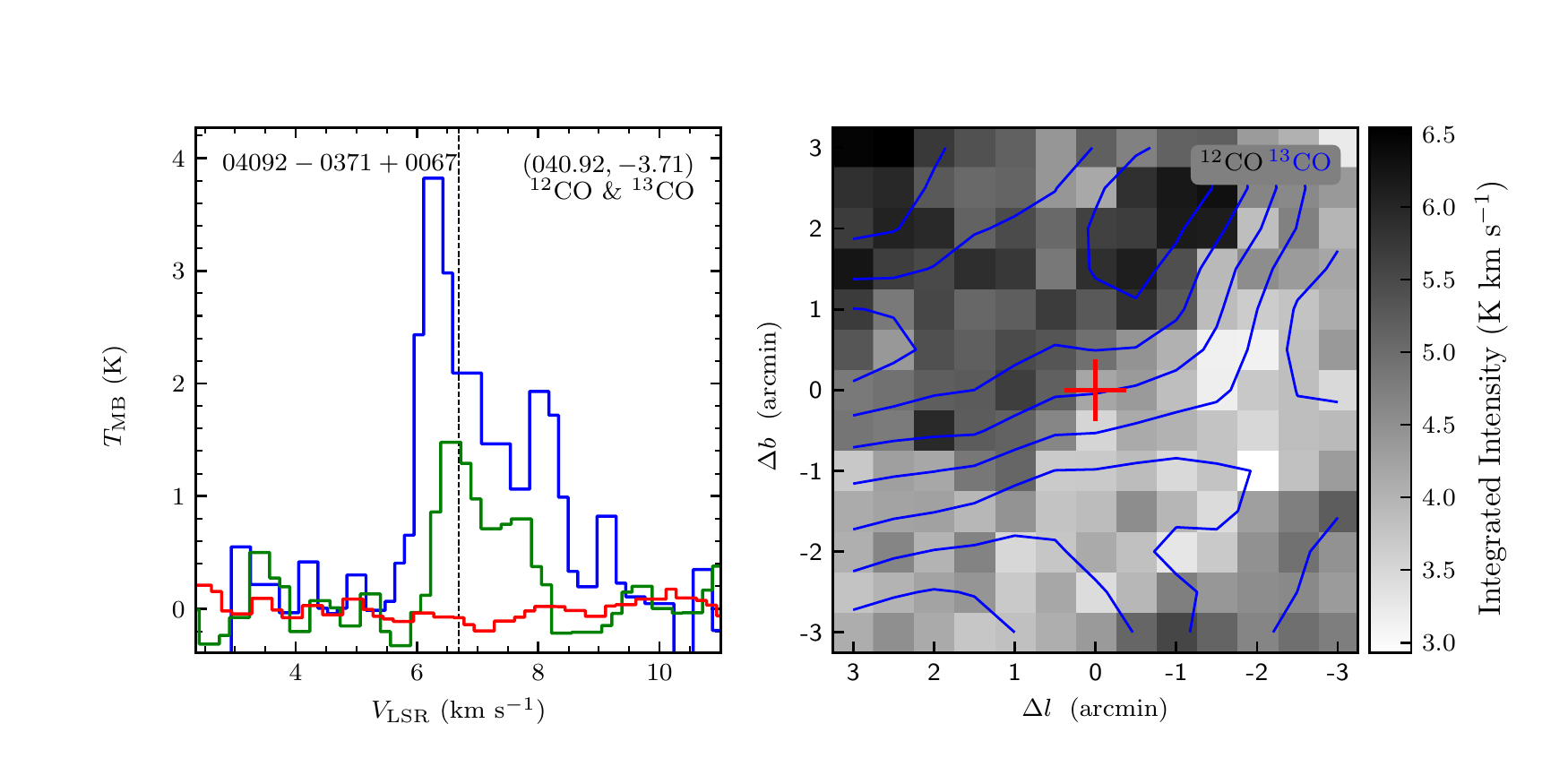}
\includegraphics[width=9.0cm,angle=0]{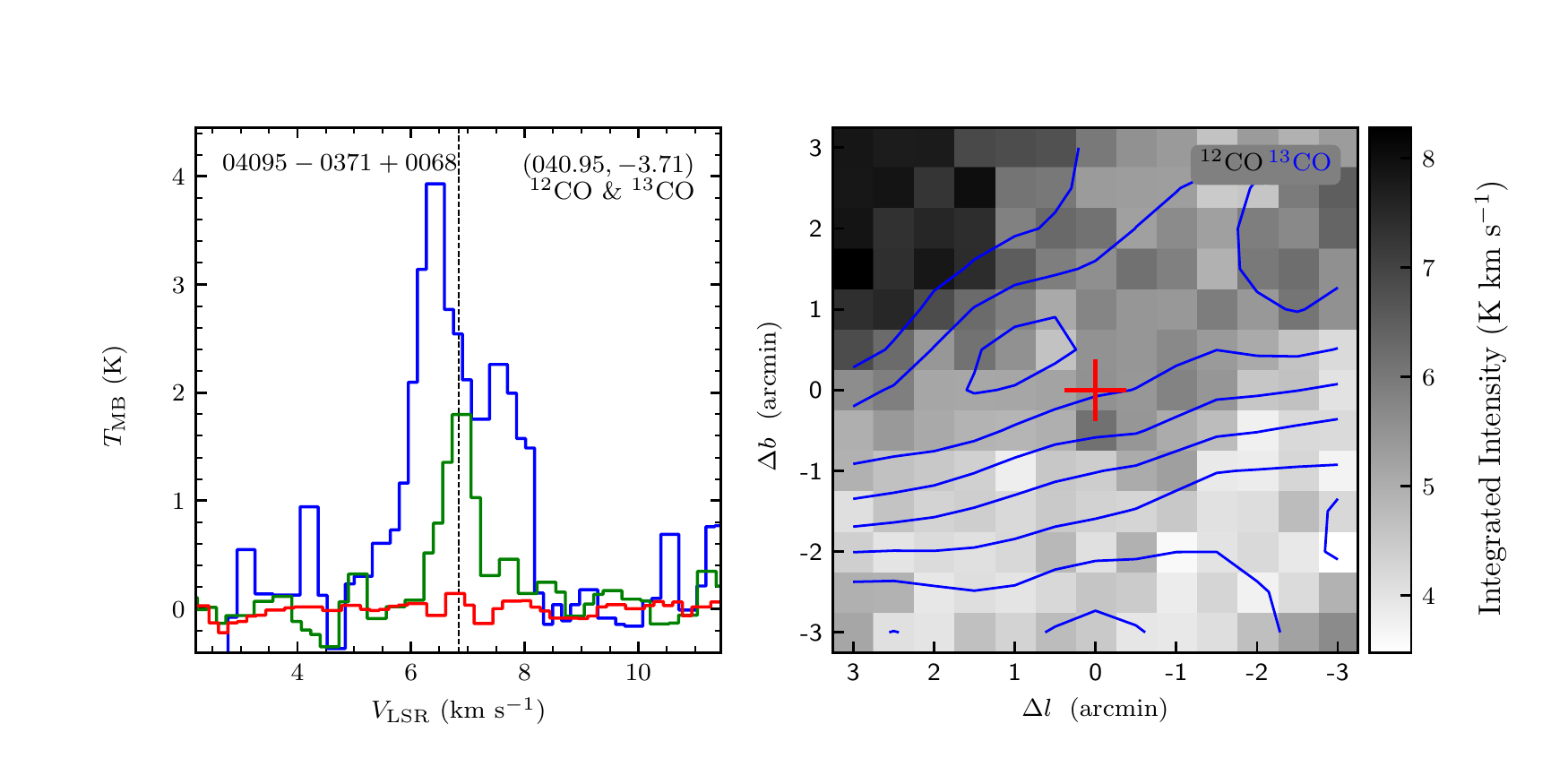}
\vspace{-0.5cm}

\includegraphics[width=9.0cm,angle=0]{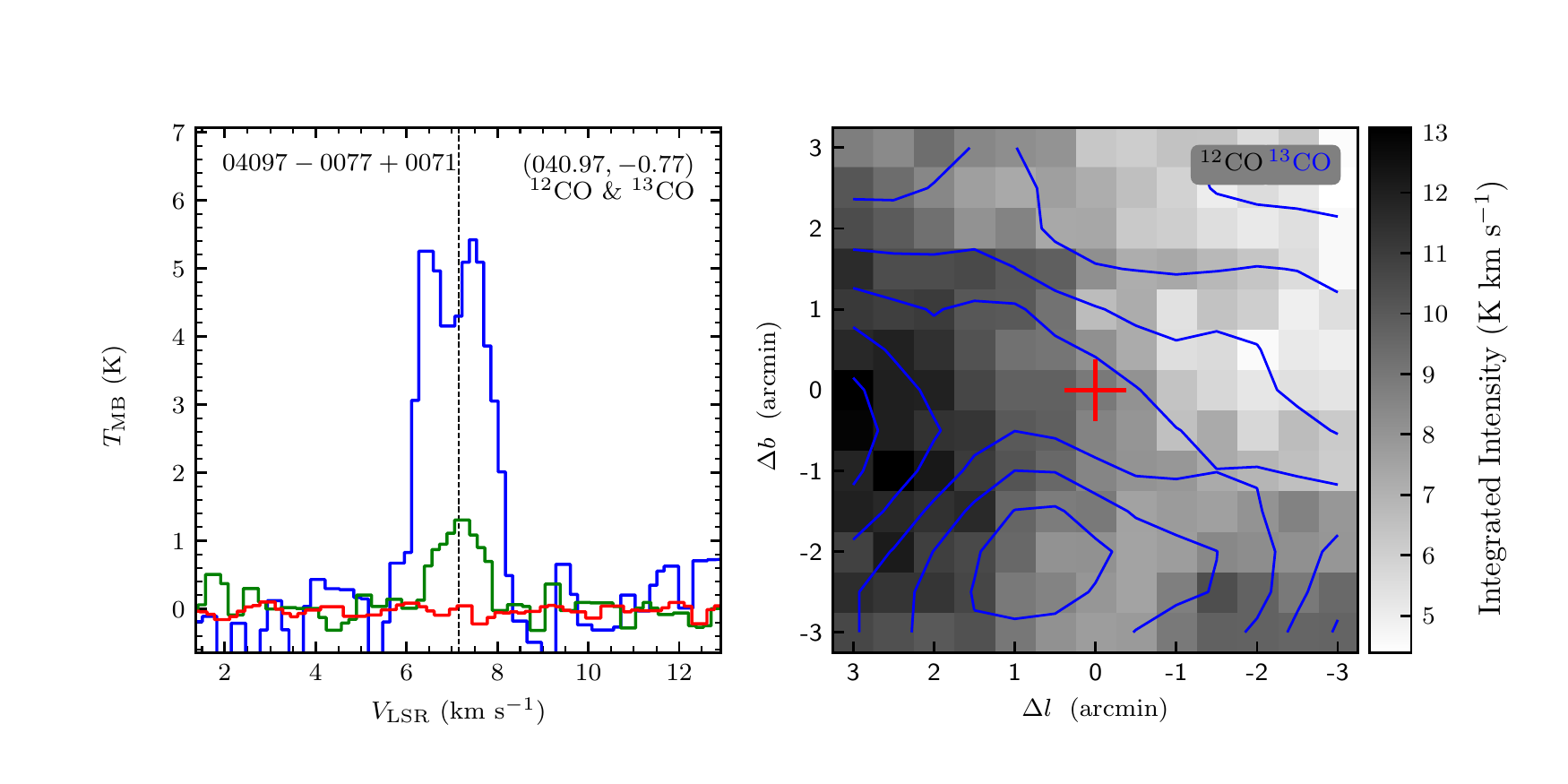}
\includegraphics[width=9.0cm,angle=0]{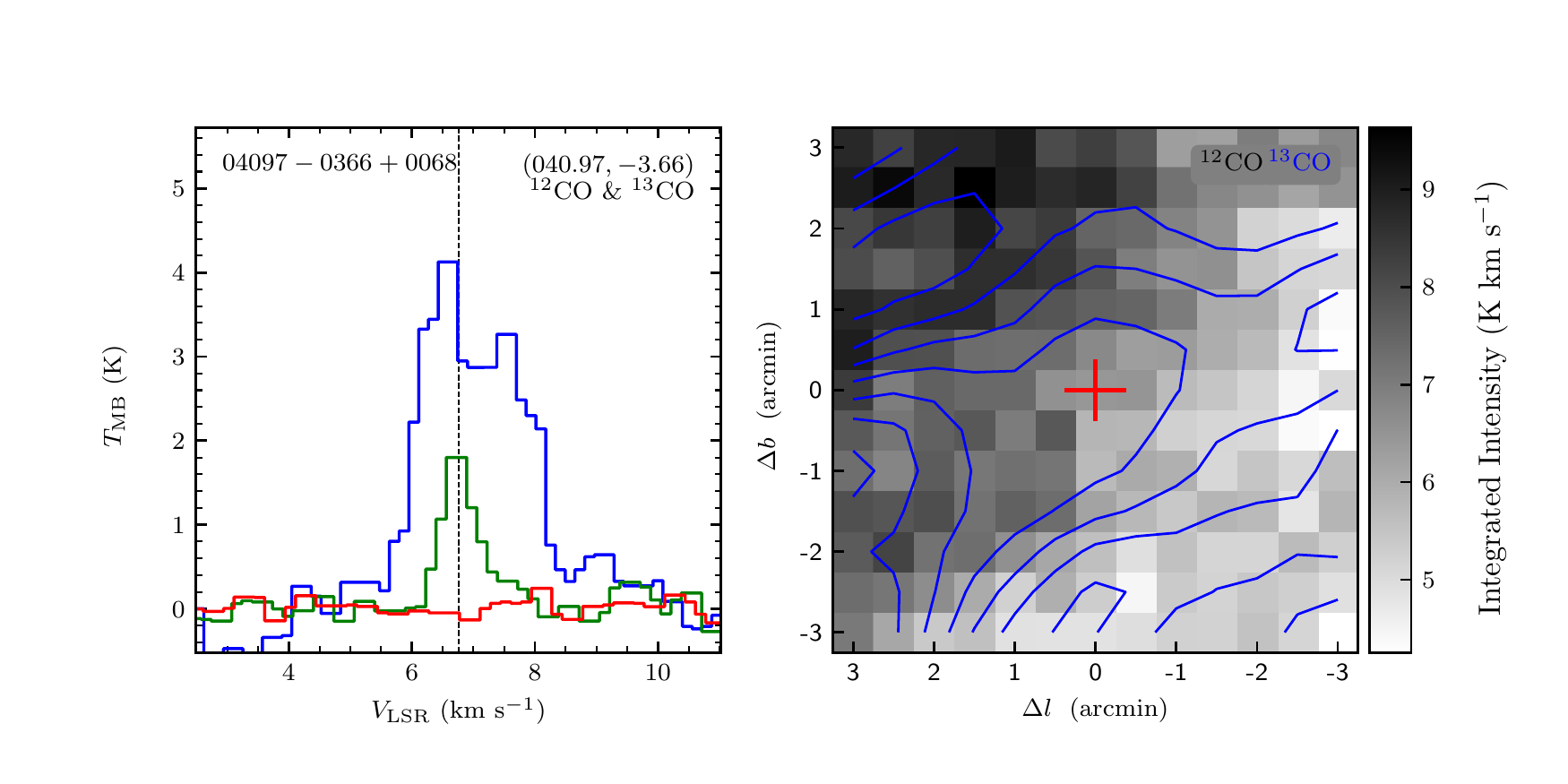}
\vspace{-0.5cm}

\includegraphics[width=9.0cm,angle=0]{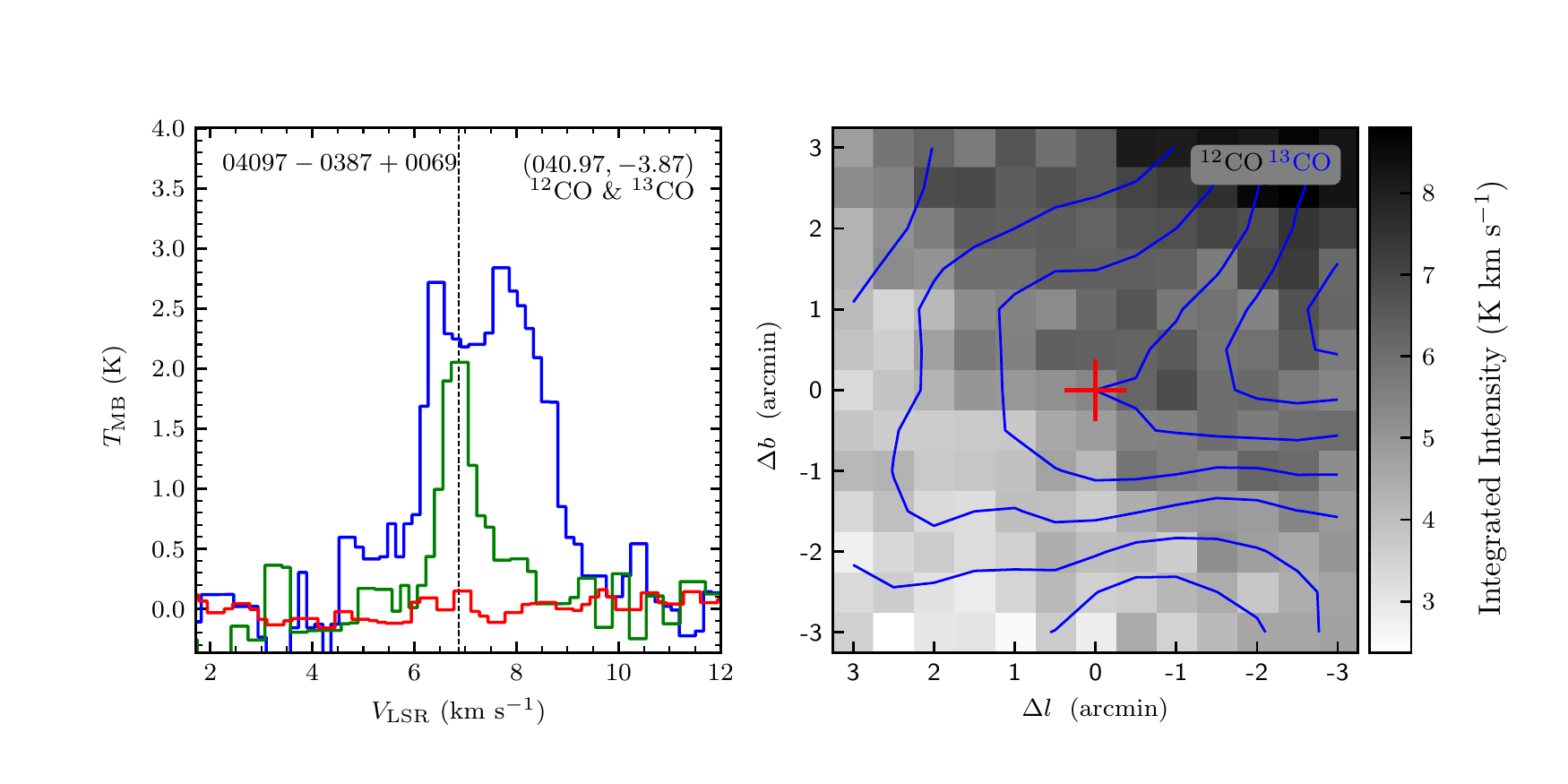}
\includegraphics[width=9.0cm,angle=0]{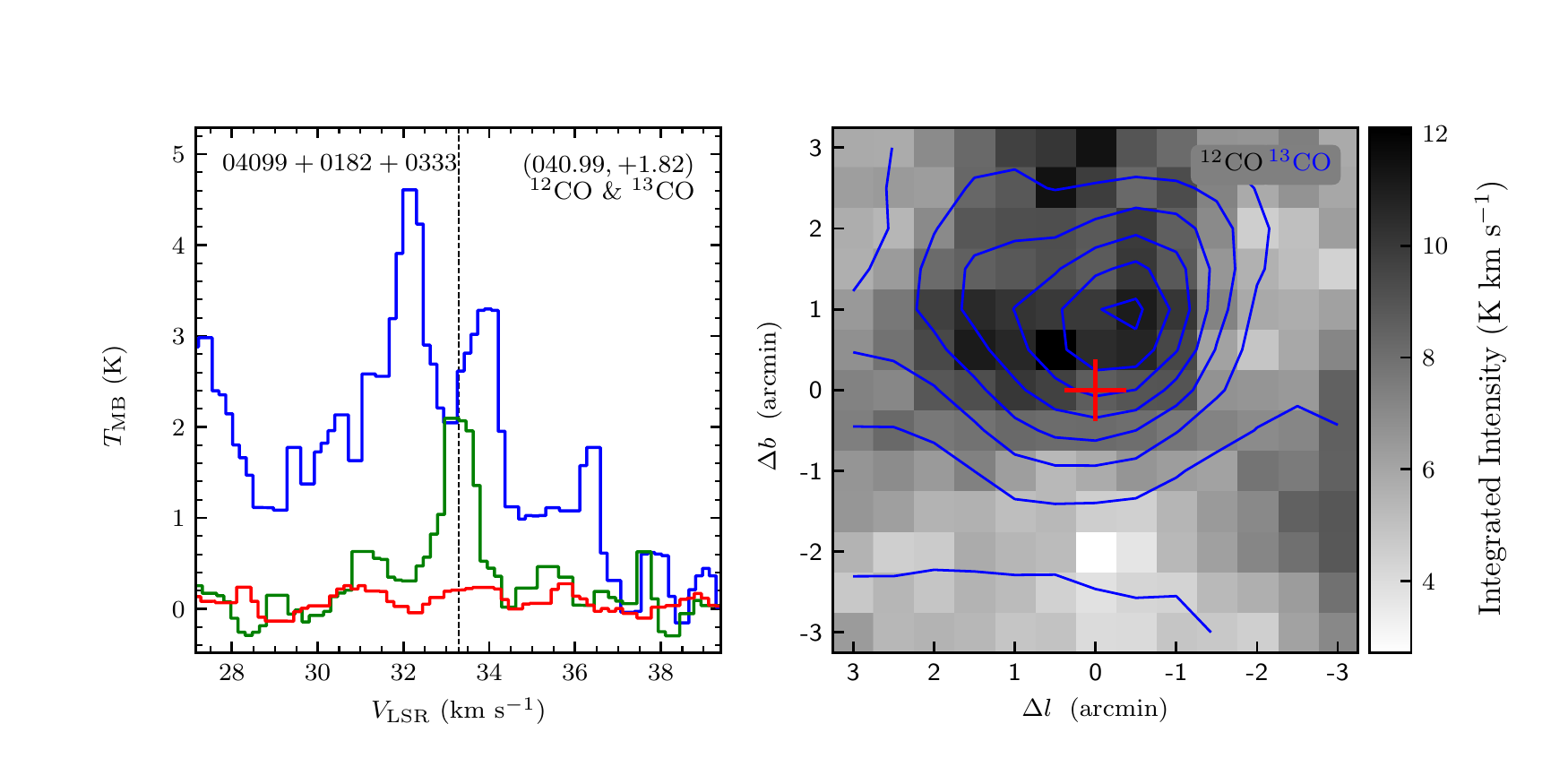}
\vspace{-0.5cm}

\includegraphics[width=9.0cm,angle=0]{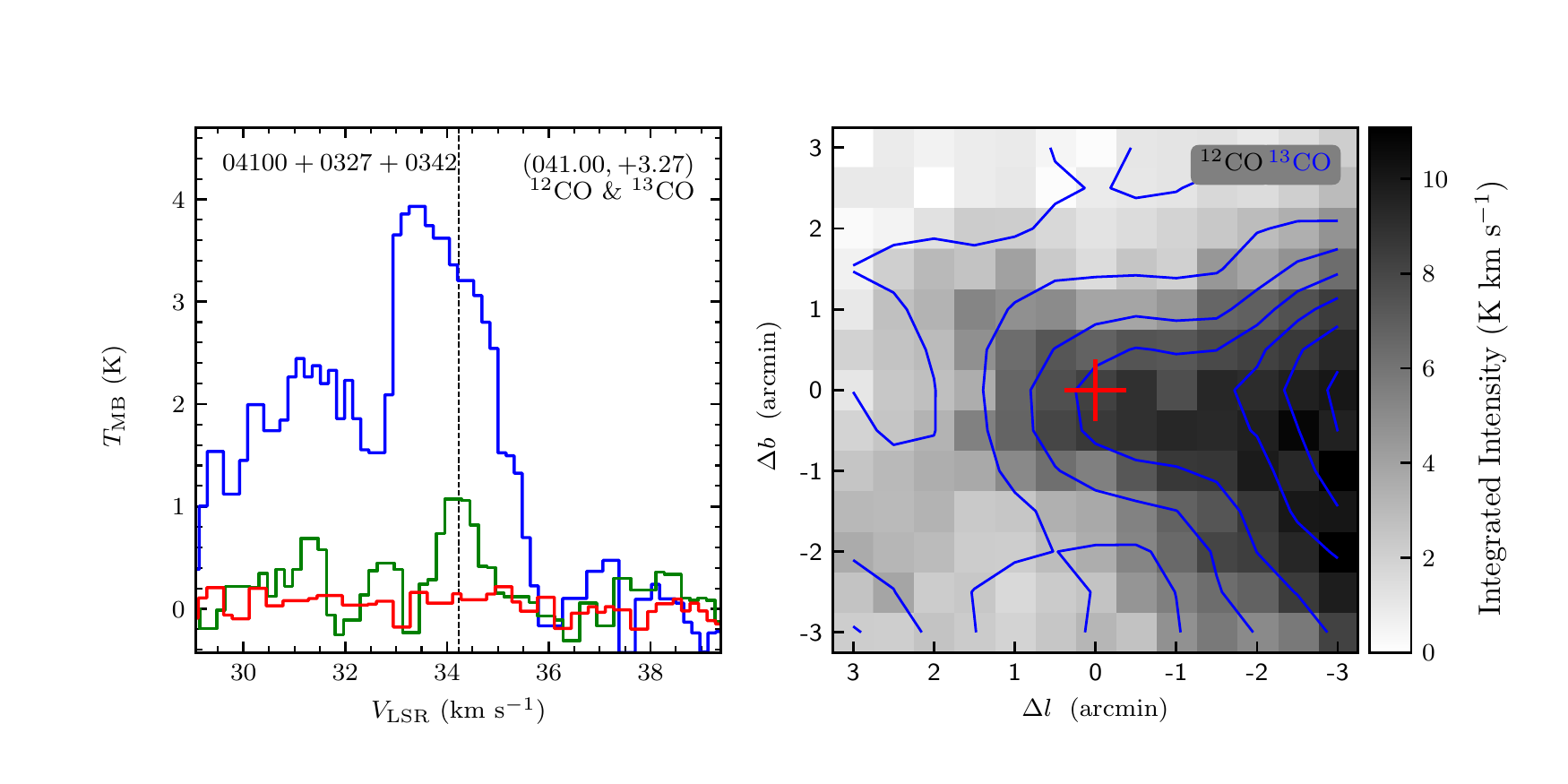}
\includegraphics[width=9.0cm,angle=0]{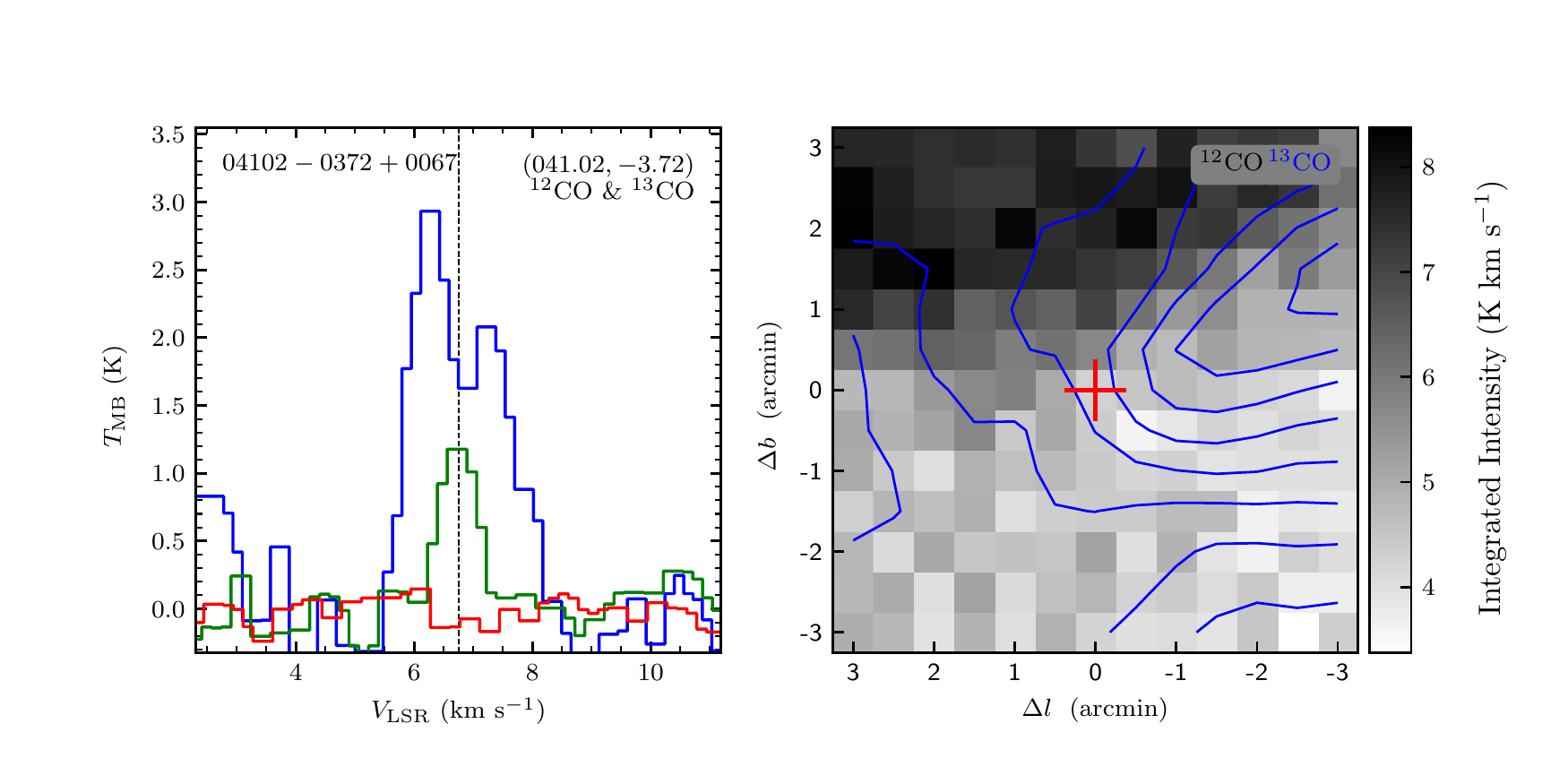}
\end{figure}
\clearpage

\begin{figure}
\includegraphics[width=9.0cm,angle=0]{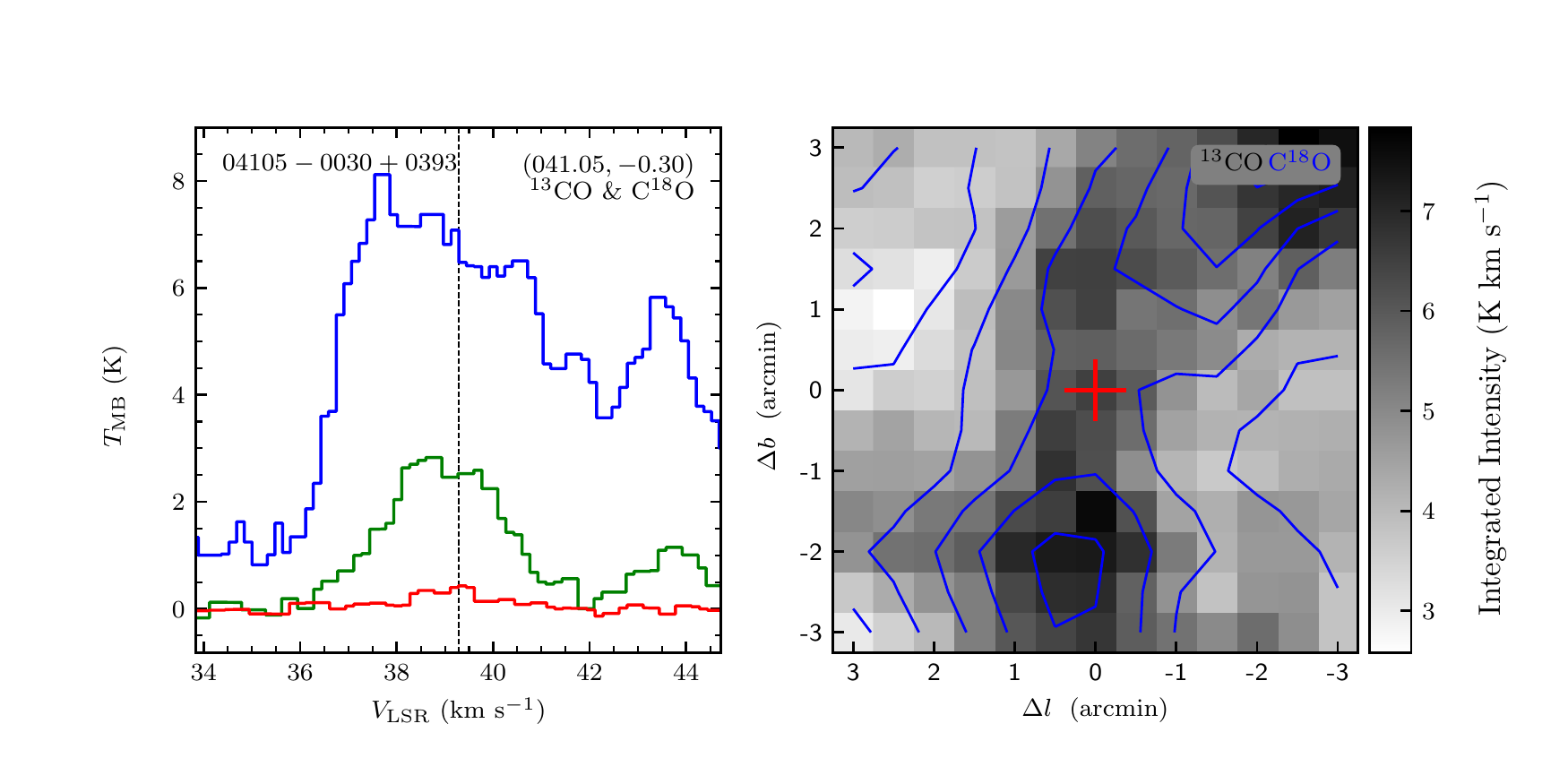}
\includegraphics[width=9.0cm,angle=0]{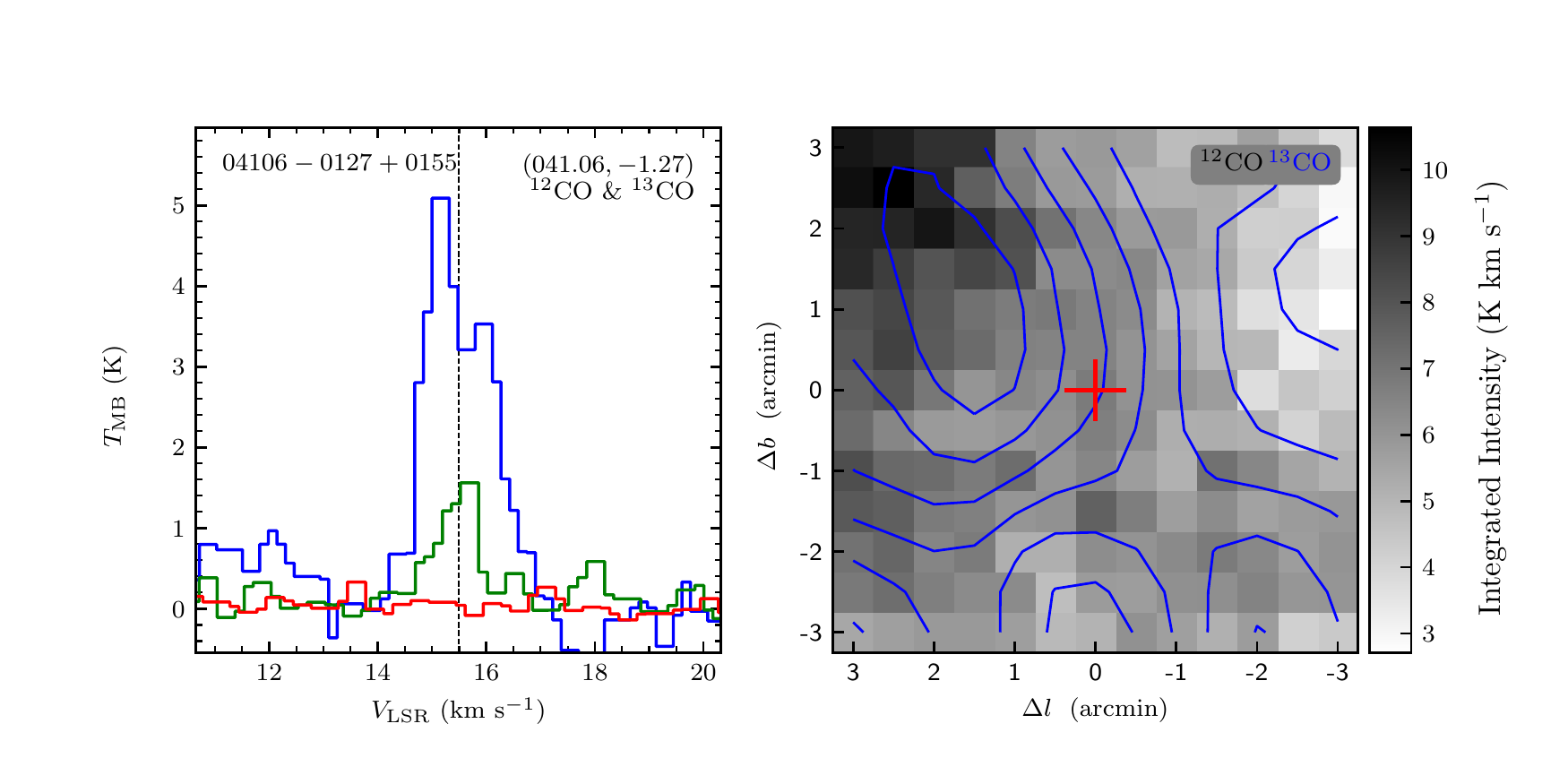}
\vspace{-0.5cm}

\includegraphics[width=9.0cm,angle=0]{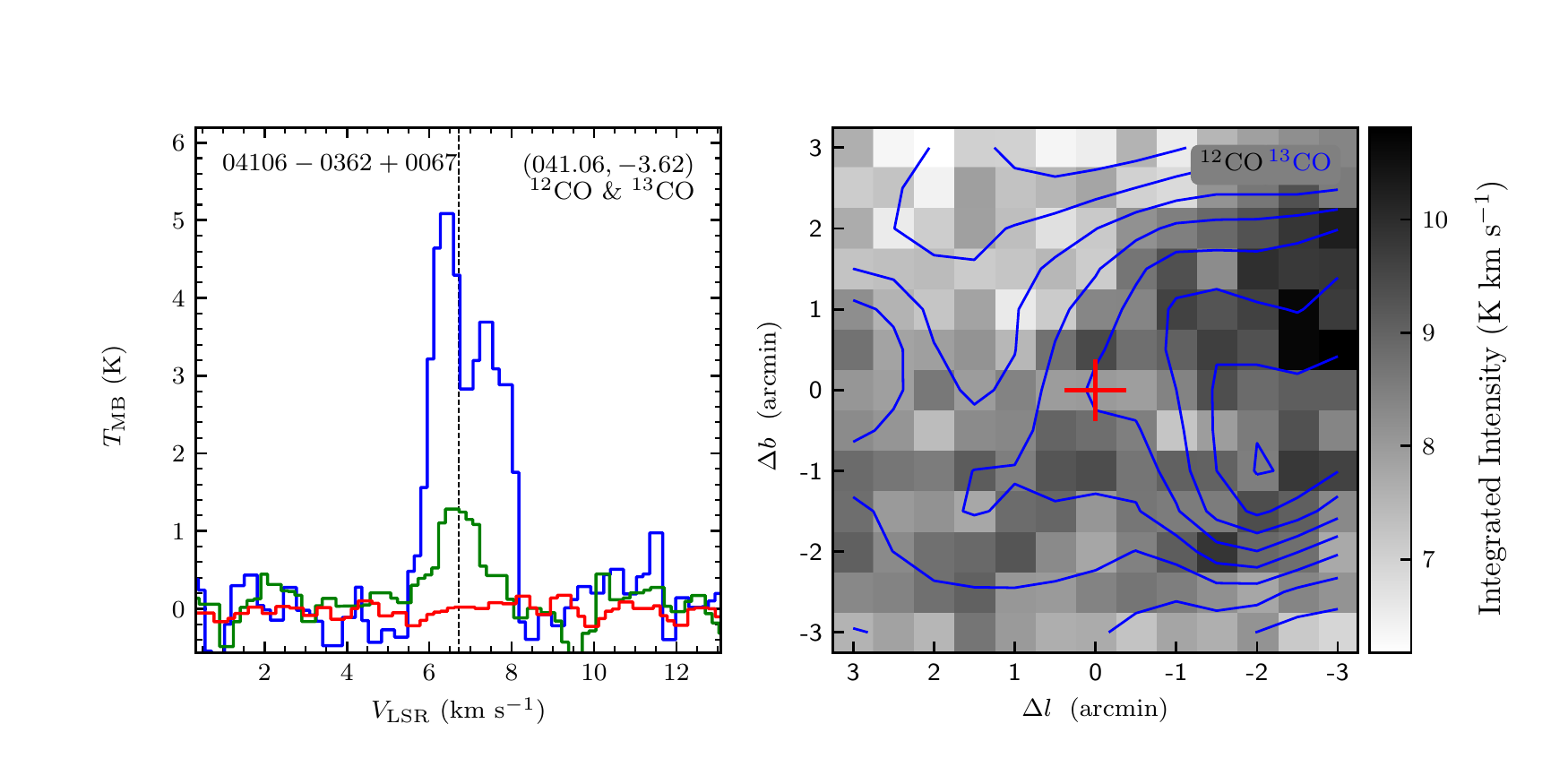}
\includegraphics[width=9.0cm,angle=0]{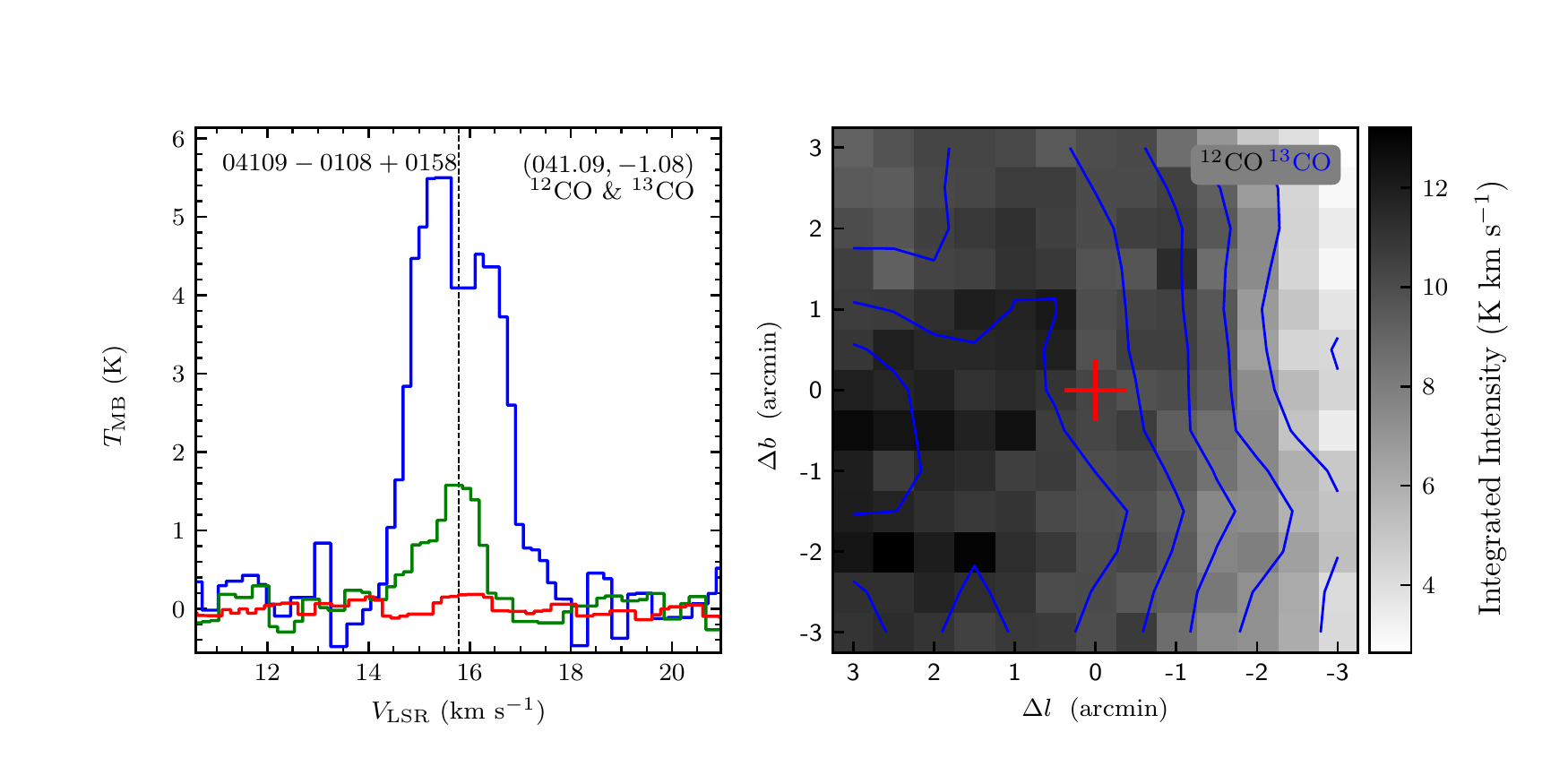}
\vspace{-0.5cm}

\includegraphics[width=9.0cm,angle=0]{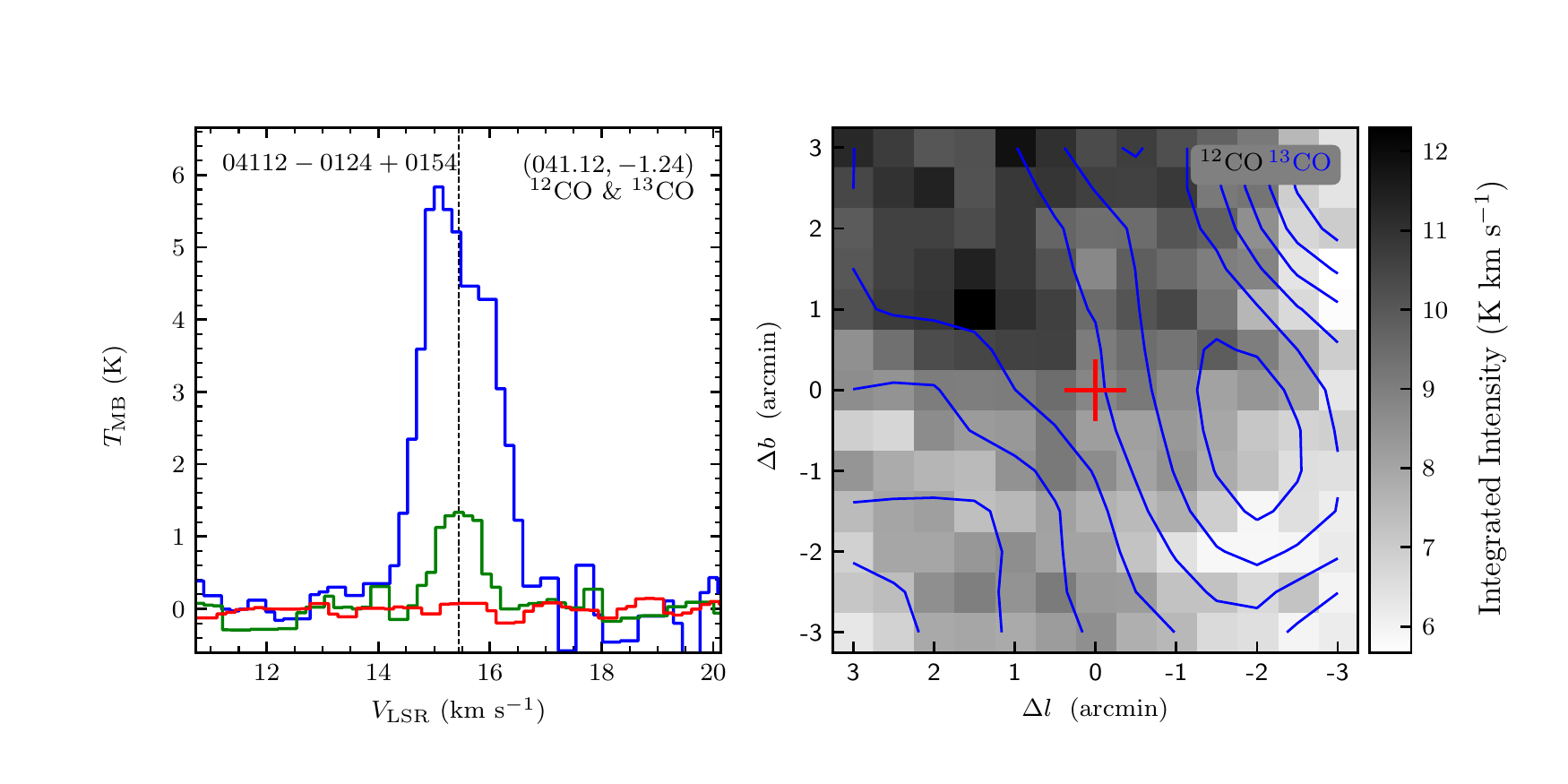}
\includegraphics[width=9.0cm,angle=0]{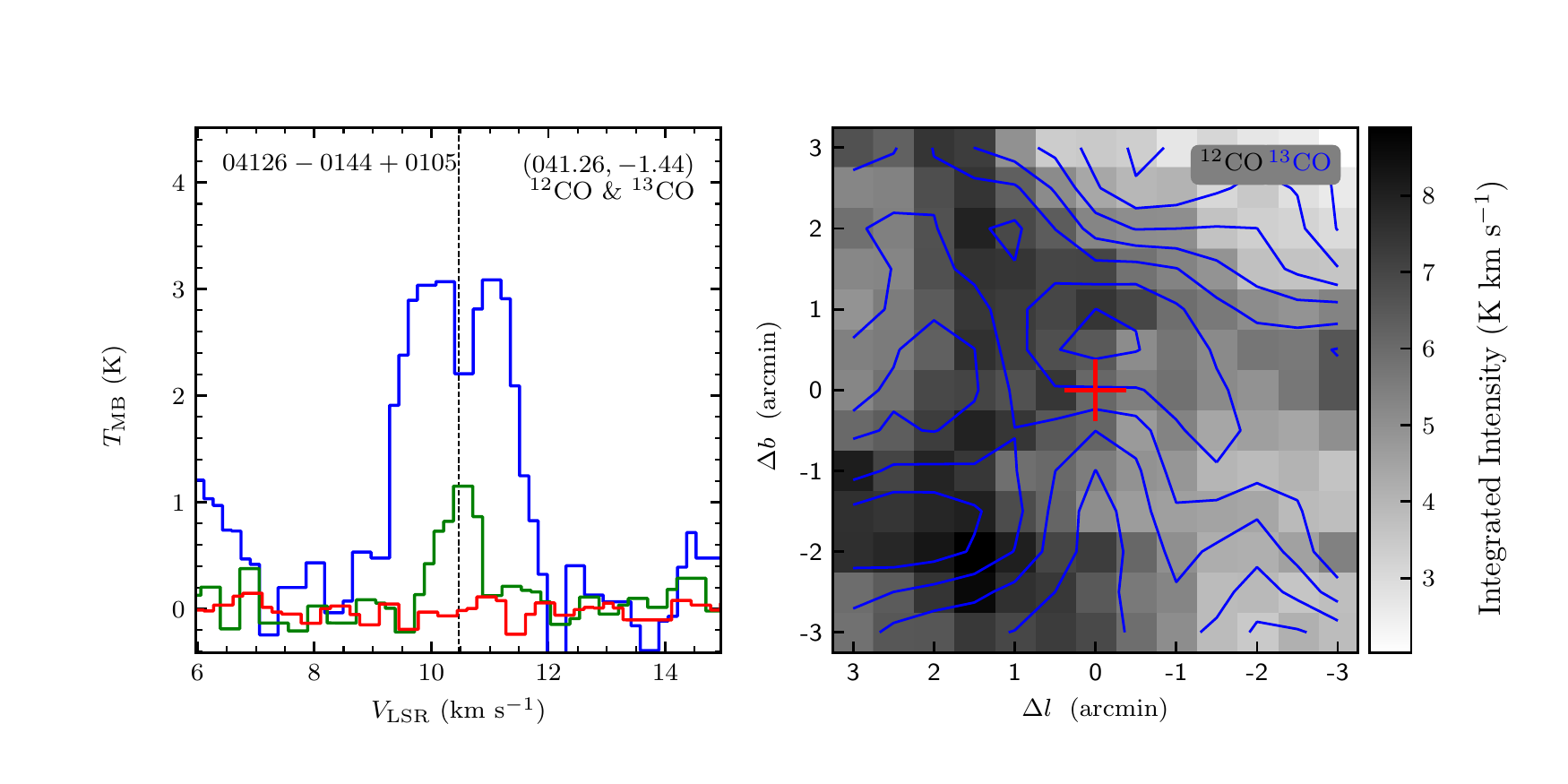}
\vspace{-0.5cm}

\includegraphics[width=9.0cm,angle=0]{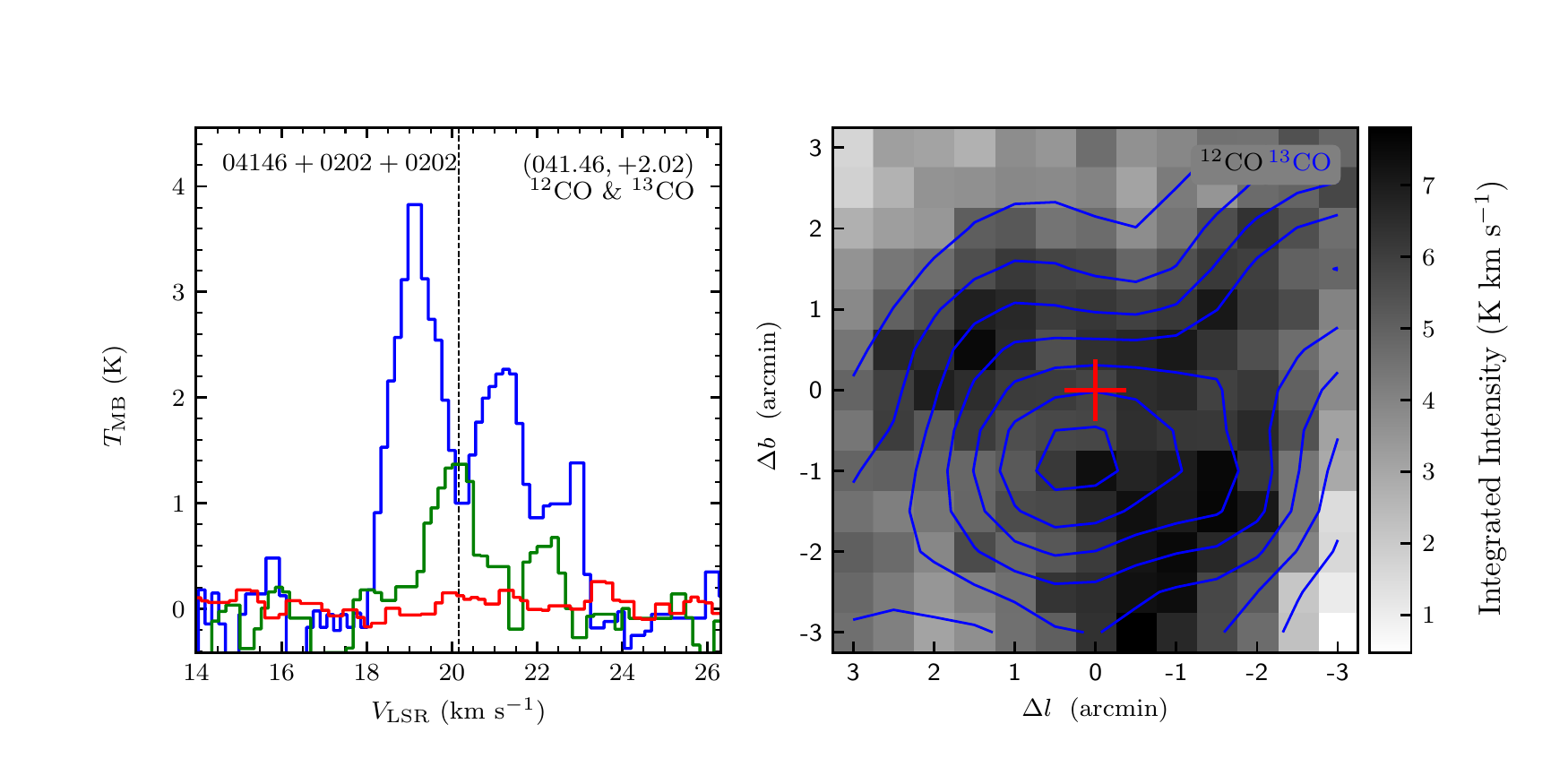}
\includegraphics[width=9.0cm,angle=0]{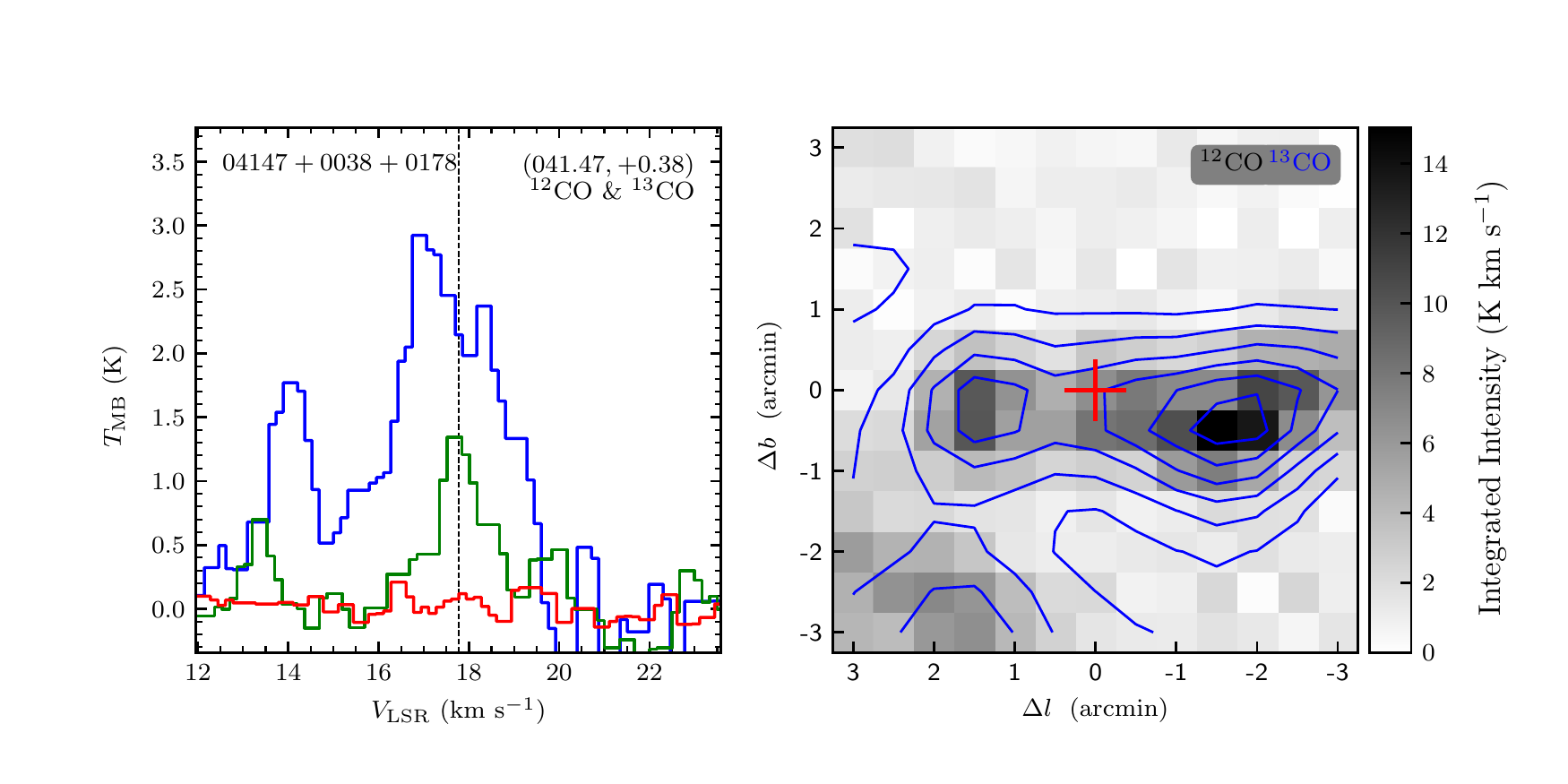}
\vspace{-0.5cm}

\includegraphics[width=9.0cm,angle=0]{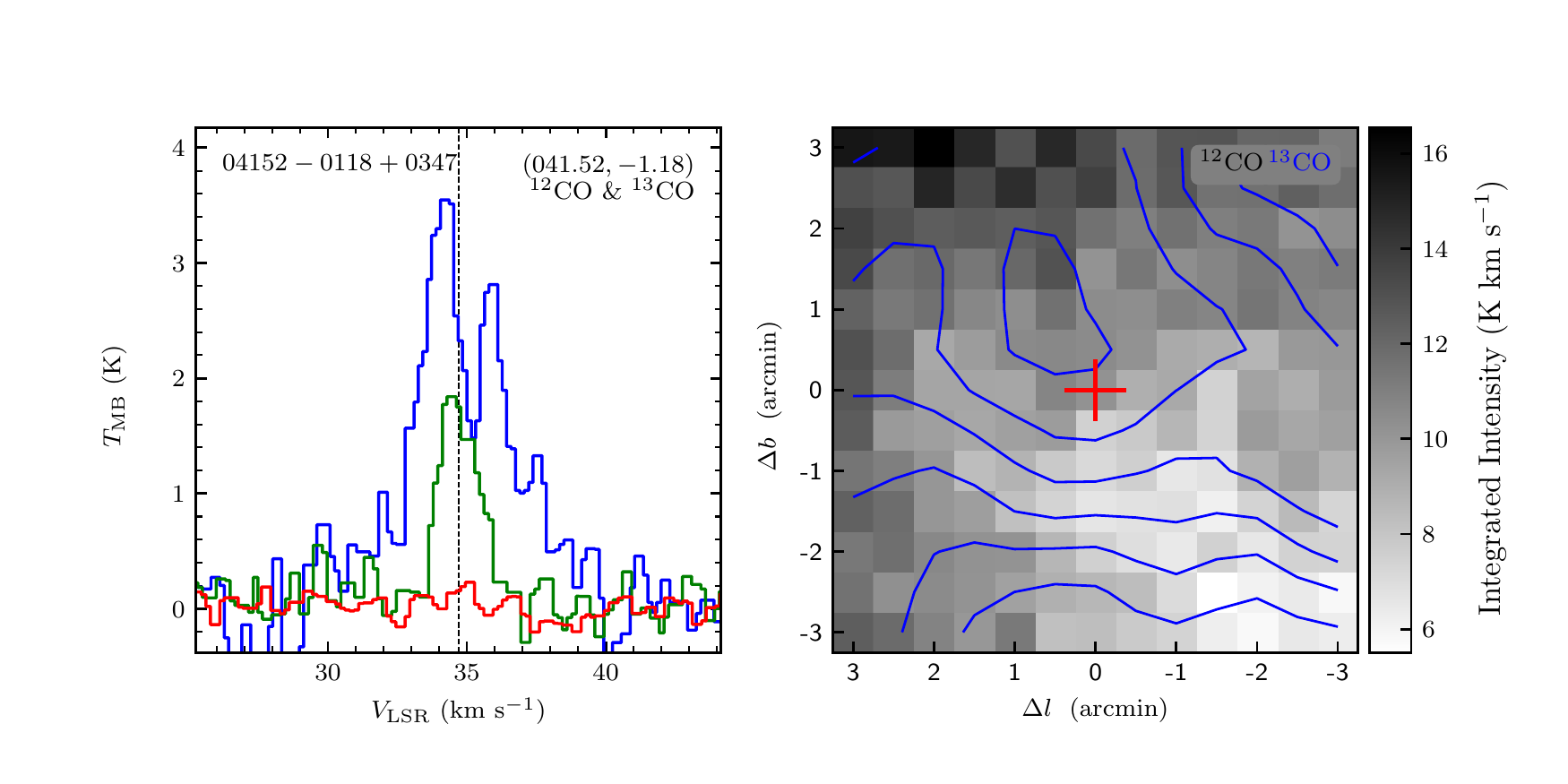}
\includegraphics[width=9.0cm,angle=0]{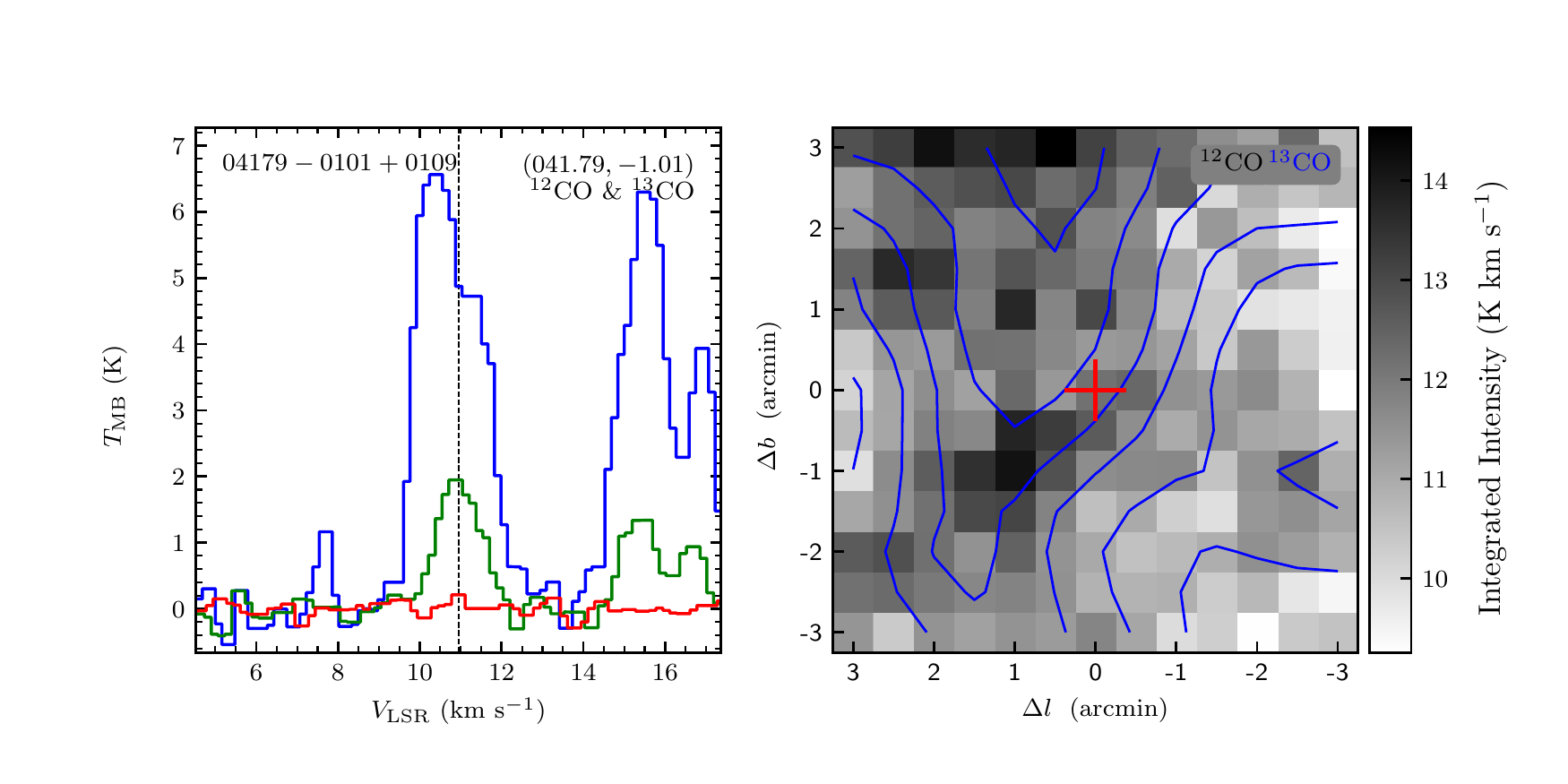}
\end{figure}
\clearpage

\begin{figure}
\includegraphics[width=9.0cm,angle=0]{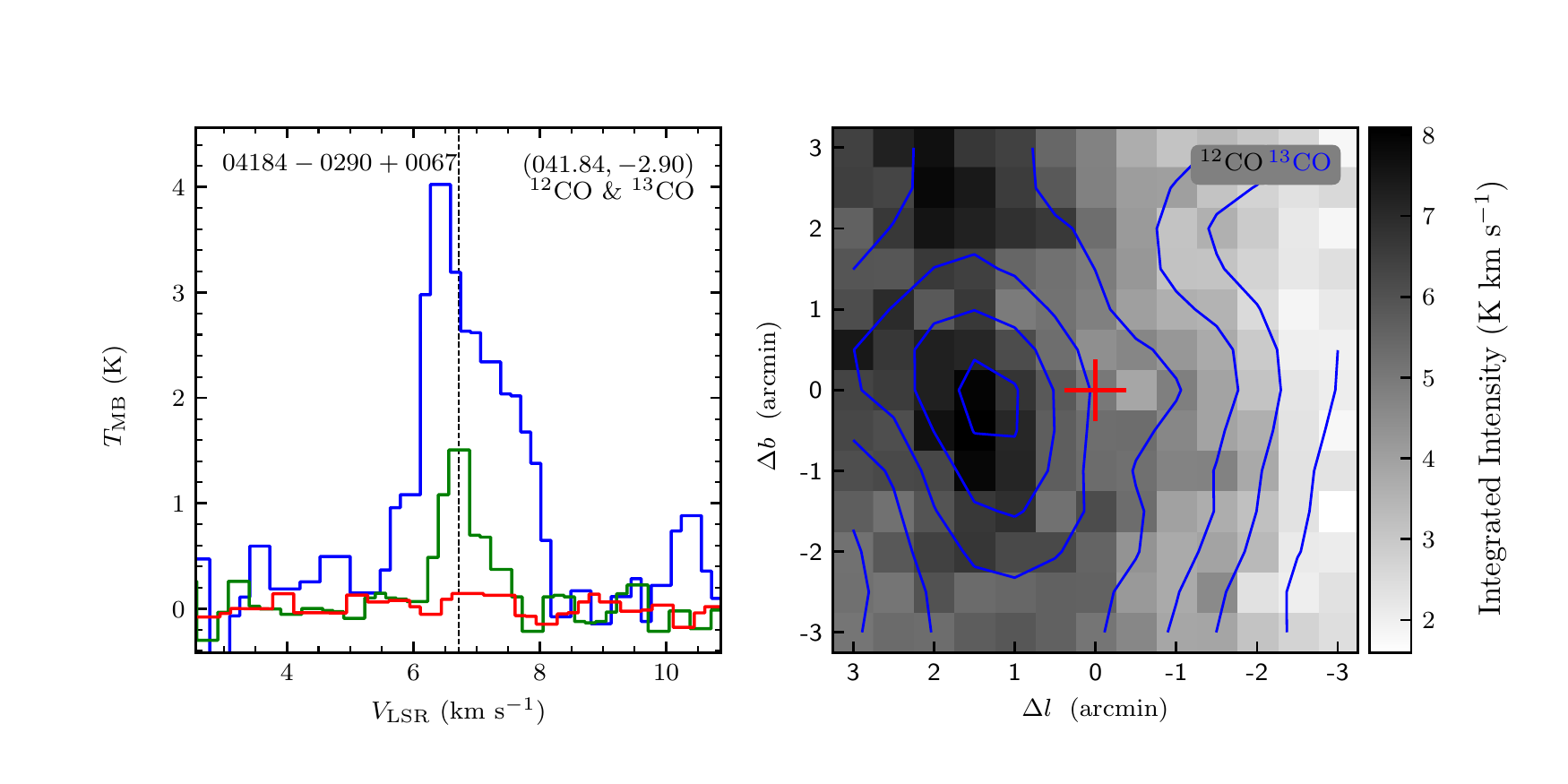}
\includegraphics[width=9.0cm,angle=0]{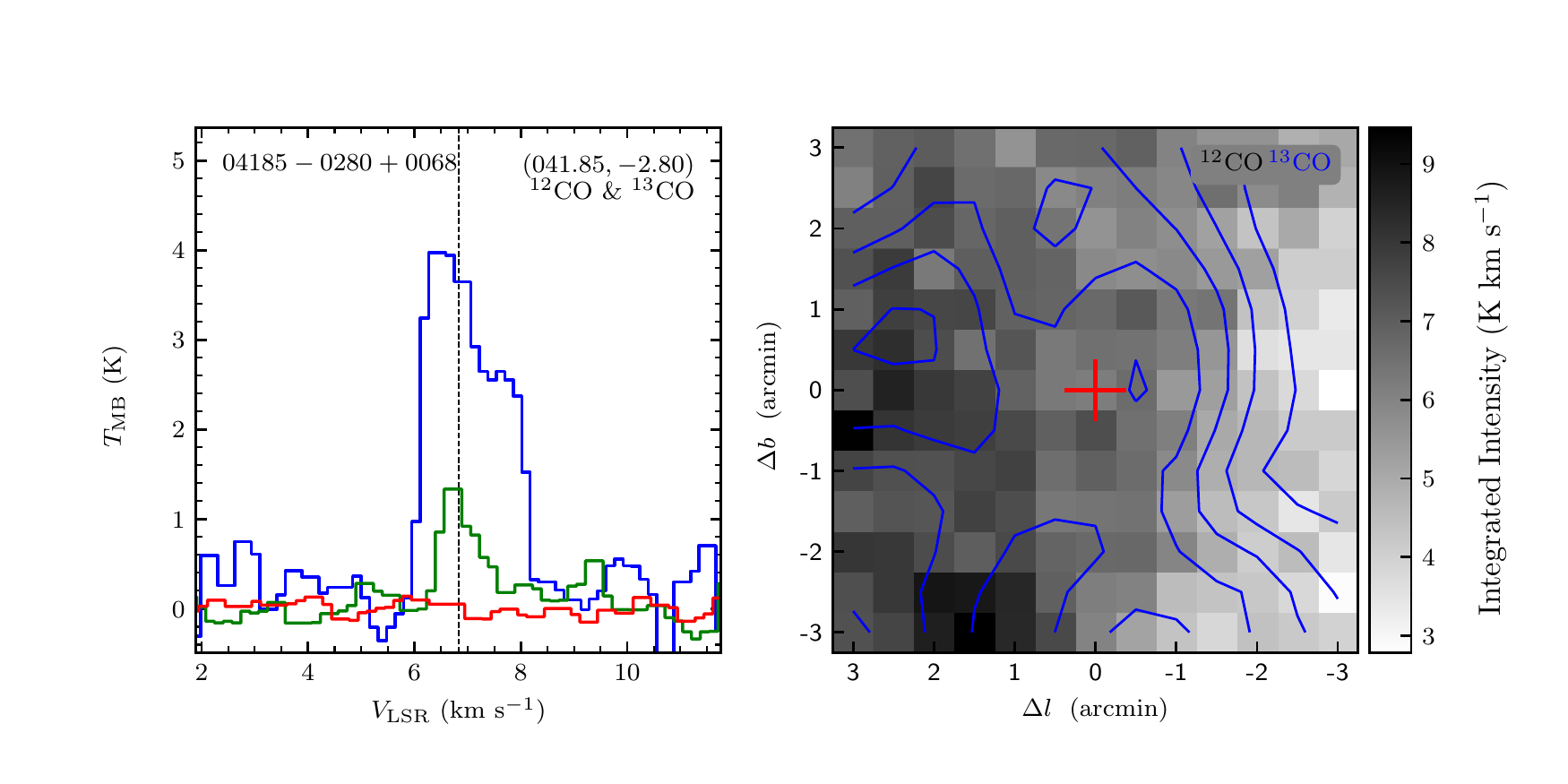}
\vspace{-0.5cm}

\includegraphics[width=9.0cm,angle=0]{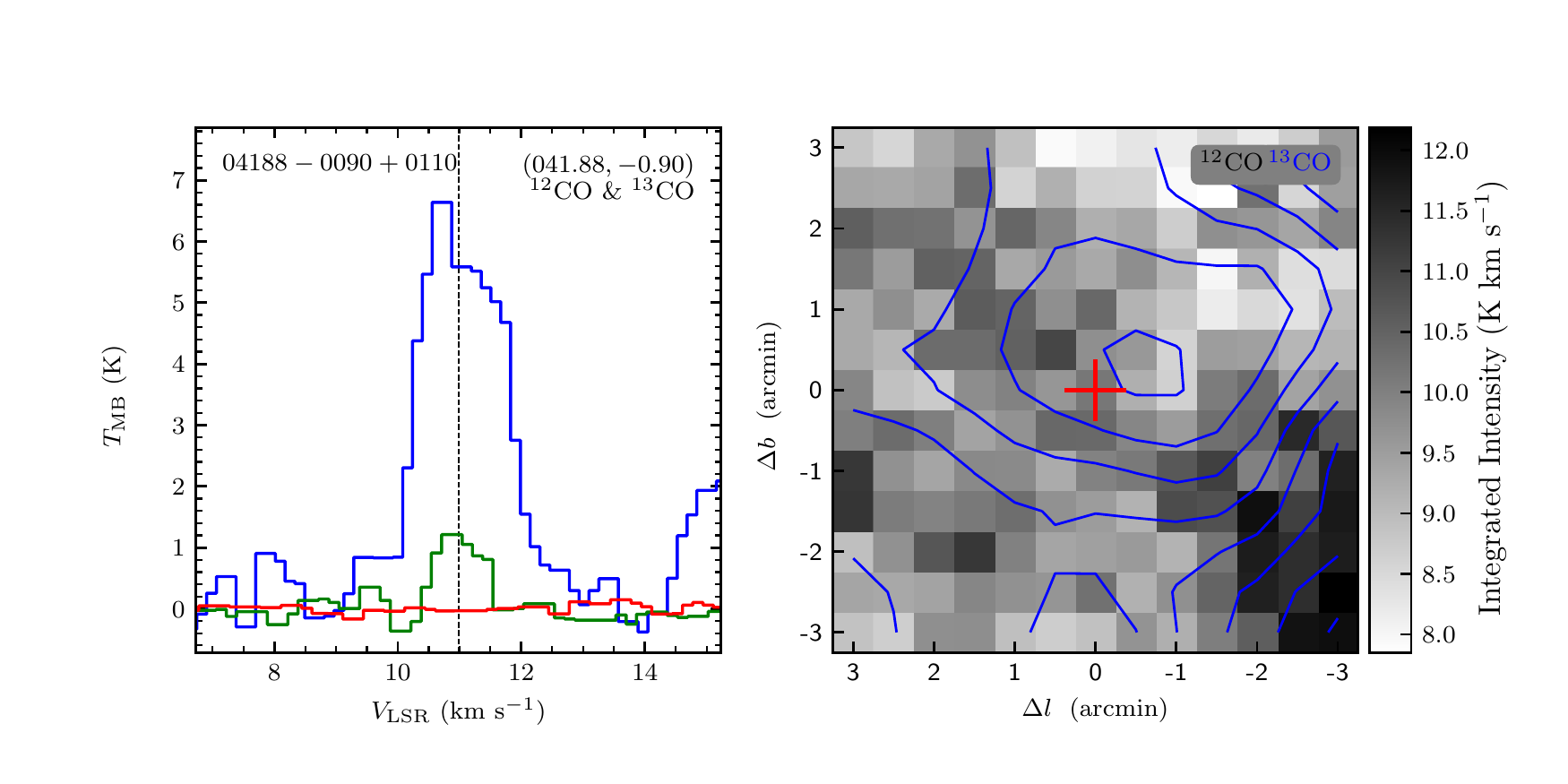}
\includegraphics[width=9.0cm,angle=0]{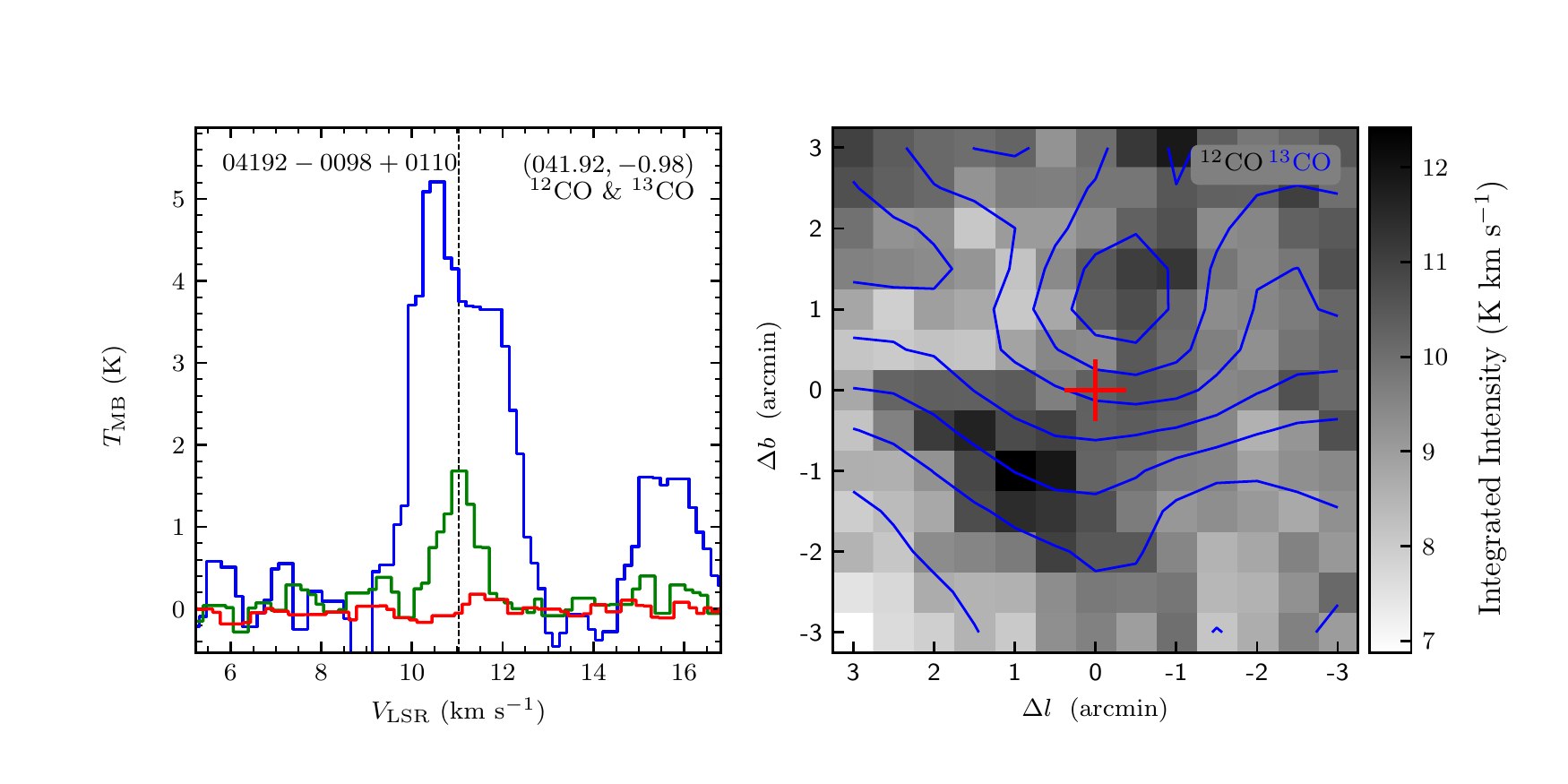}
\vspace{-0.5cm}

\includegraphics[width=9.0cm,angle=0]{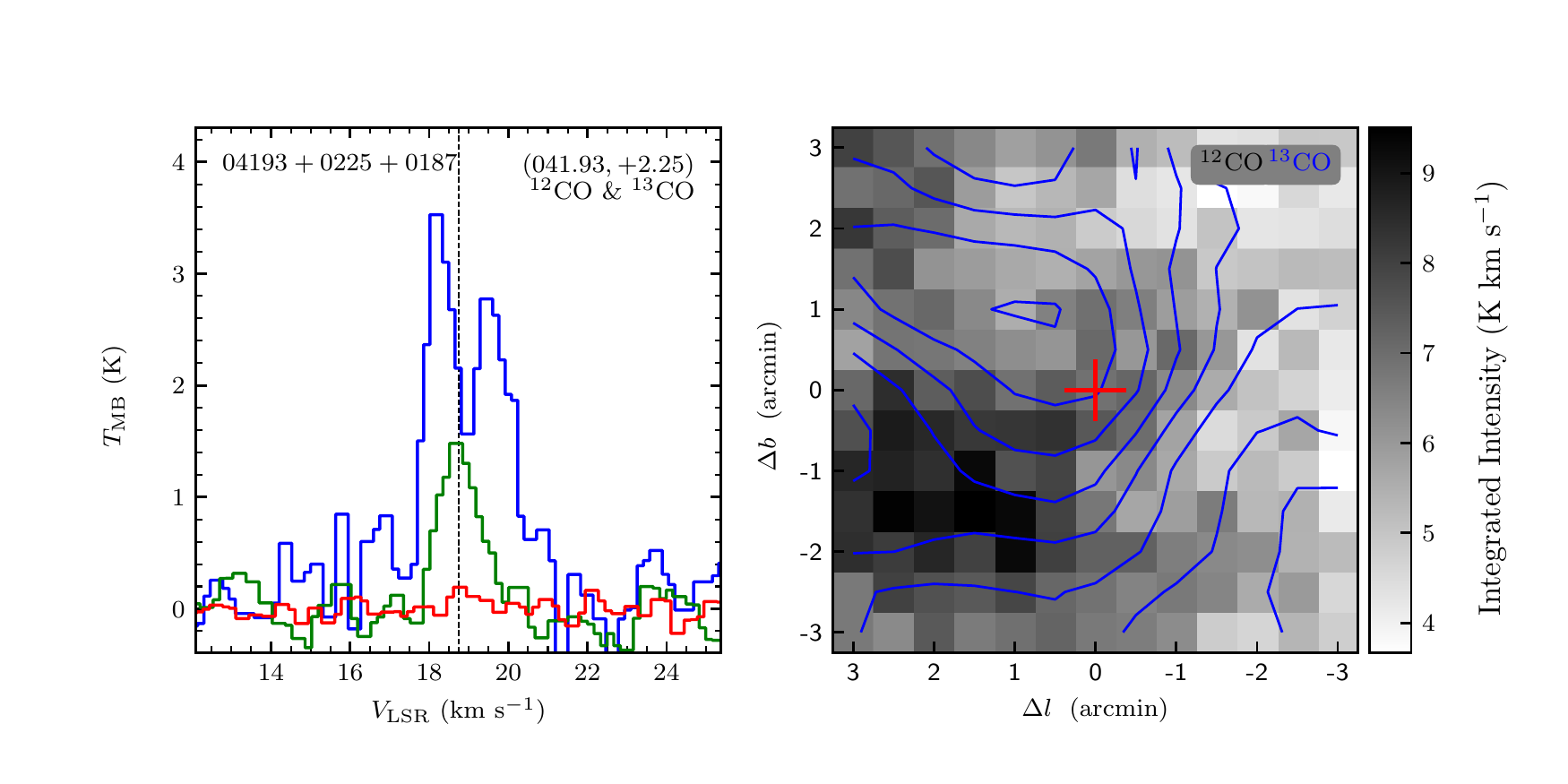}
\includegraphics[width=9.0cm,angle=0]{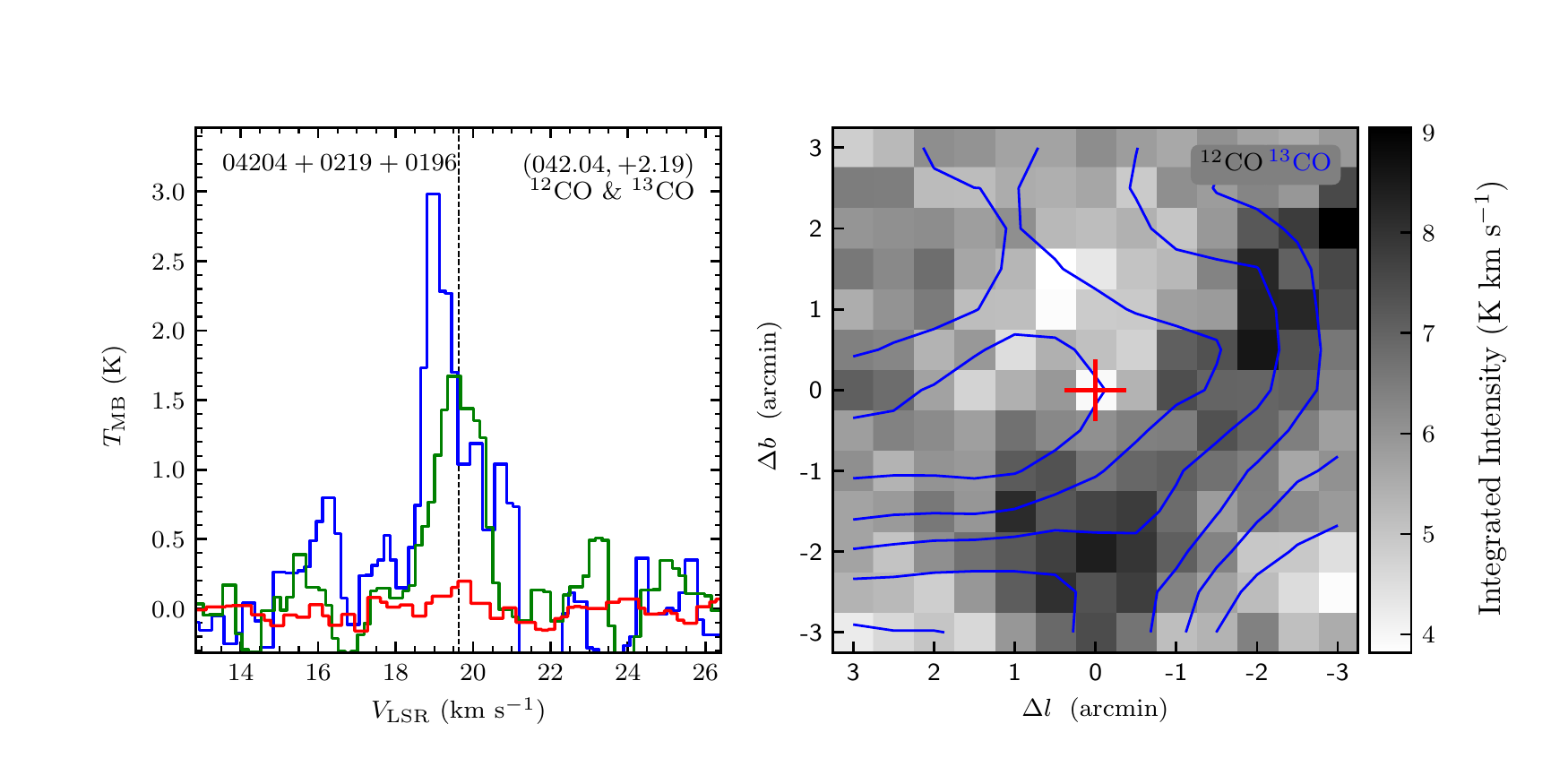}
\vspace{-0.5cm}

\includegraphics[width=9.0cm,angle=0]{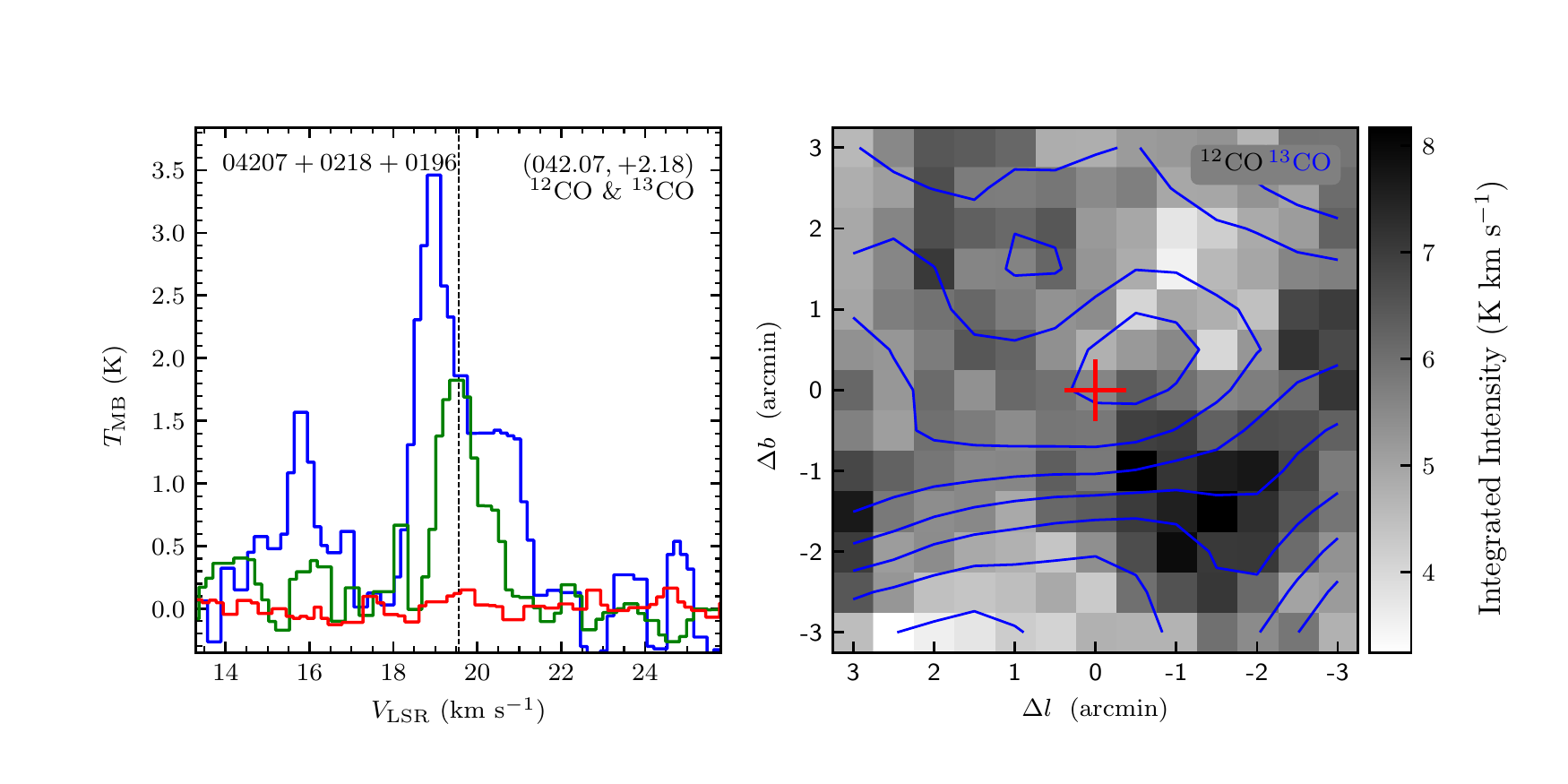}
\includegraphics[width=9.0cm,angle=0]{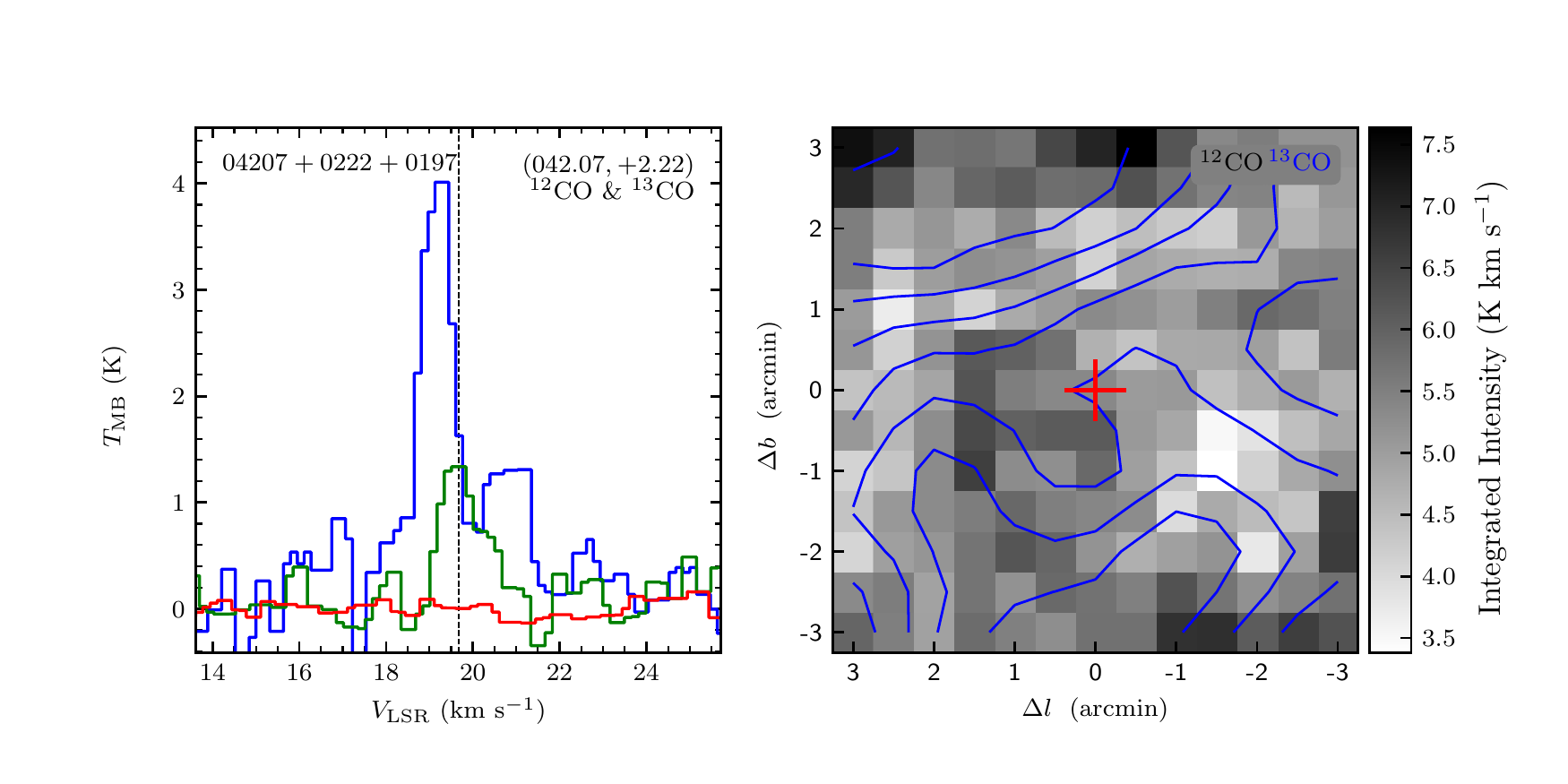}
\vspace{-0.5cm}

\includegraphics[width=9.0cm,angle=0]{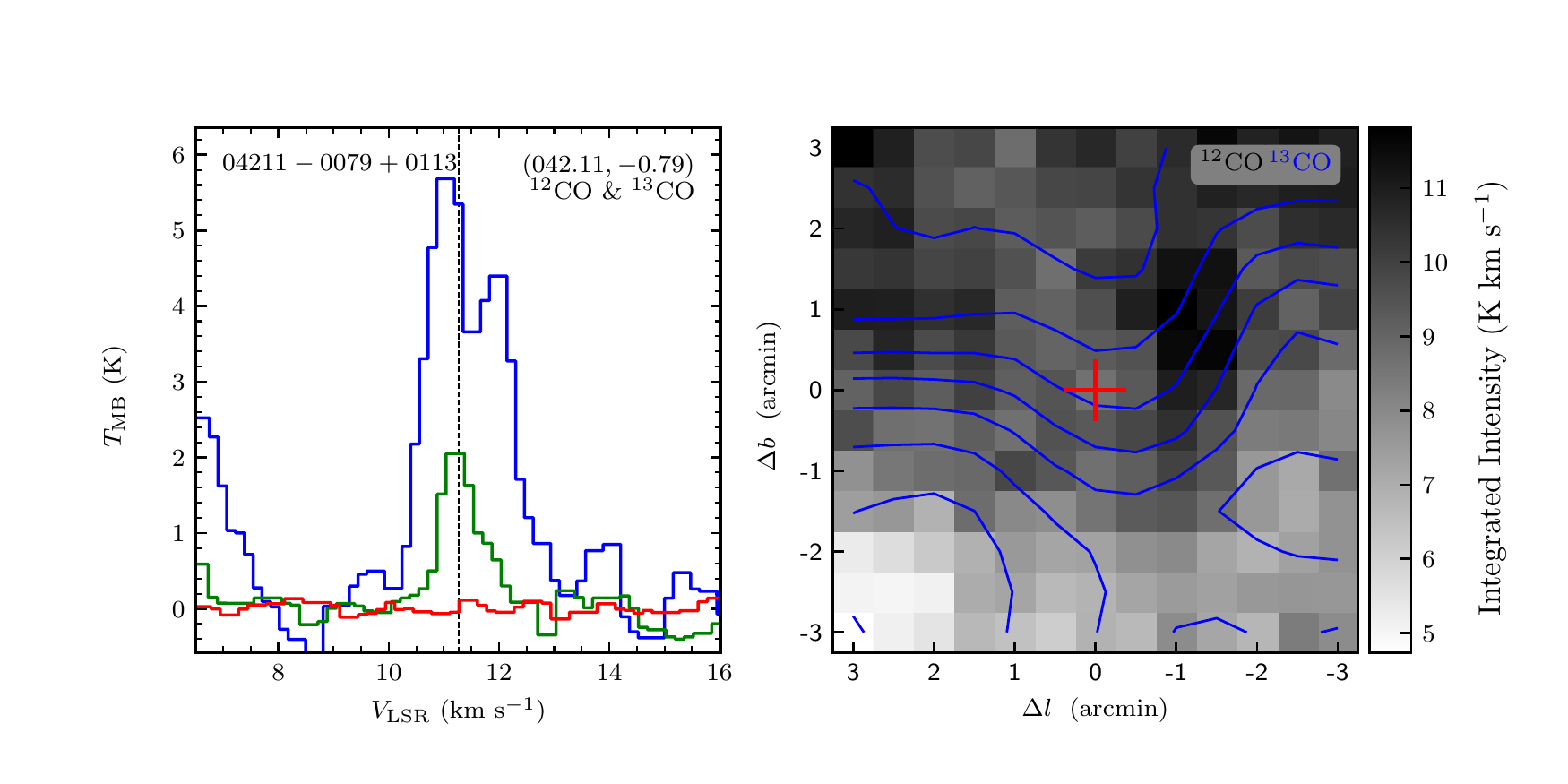}
\includegraphics[width=9.0cm,angle=0]{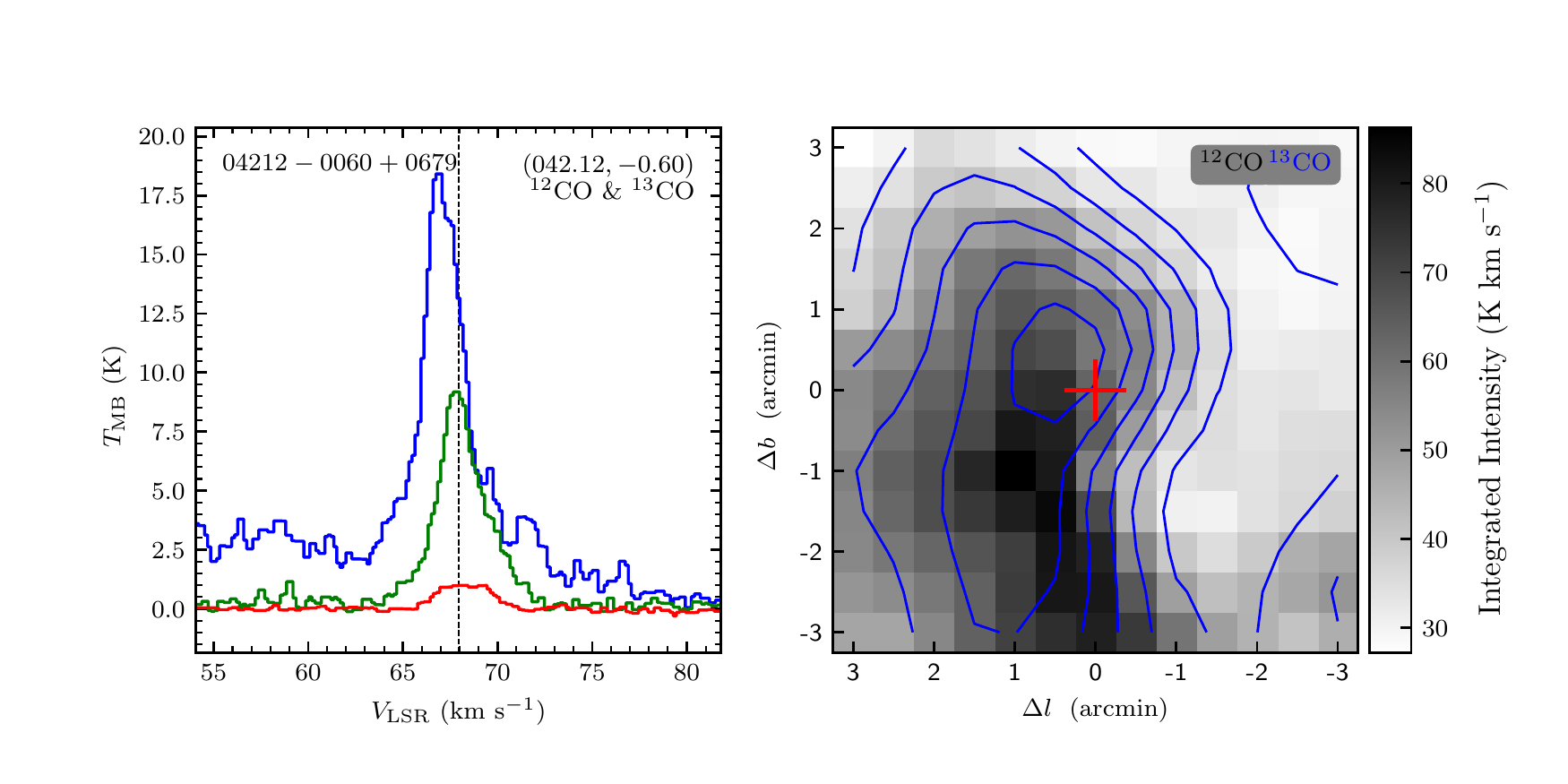}
\end{figure}
\clearpage

\begin{figure}
\includegraphics[width=9.0cm,angle=0]{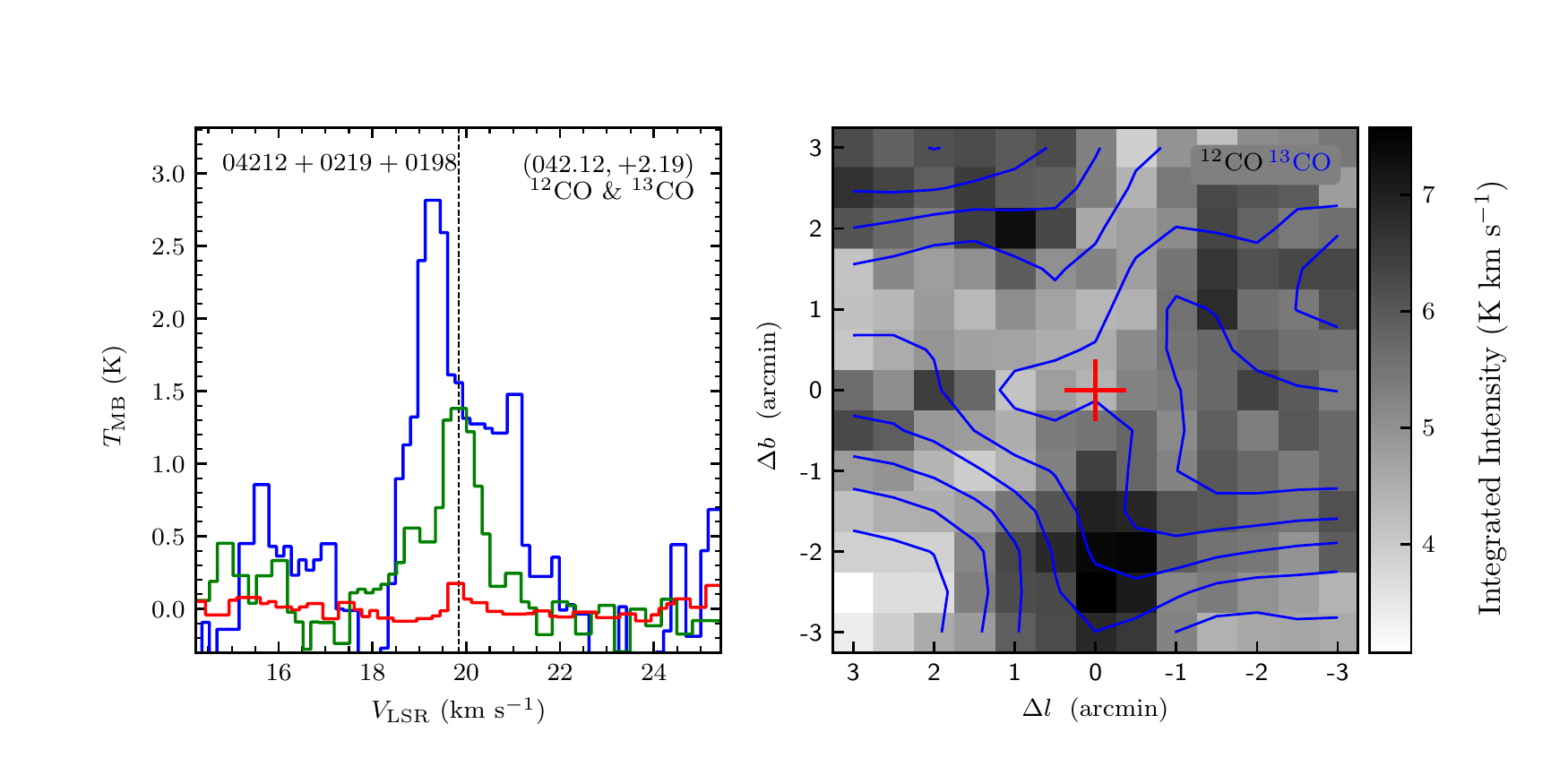}
\includegraphics[width=9.0cm,angle=0]{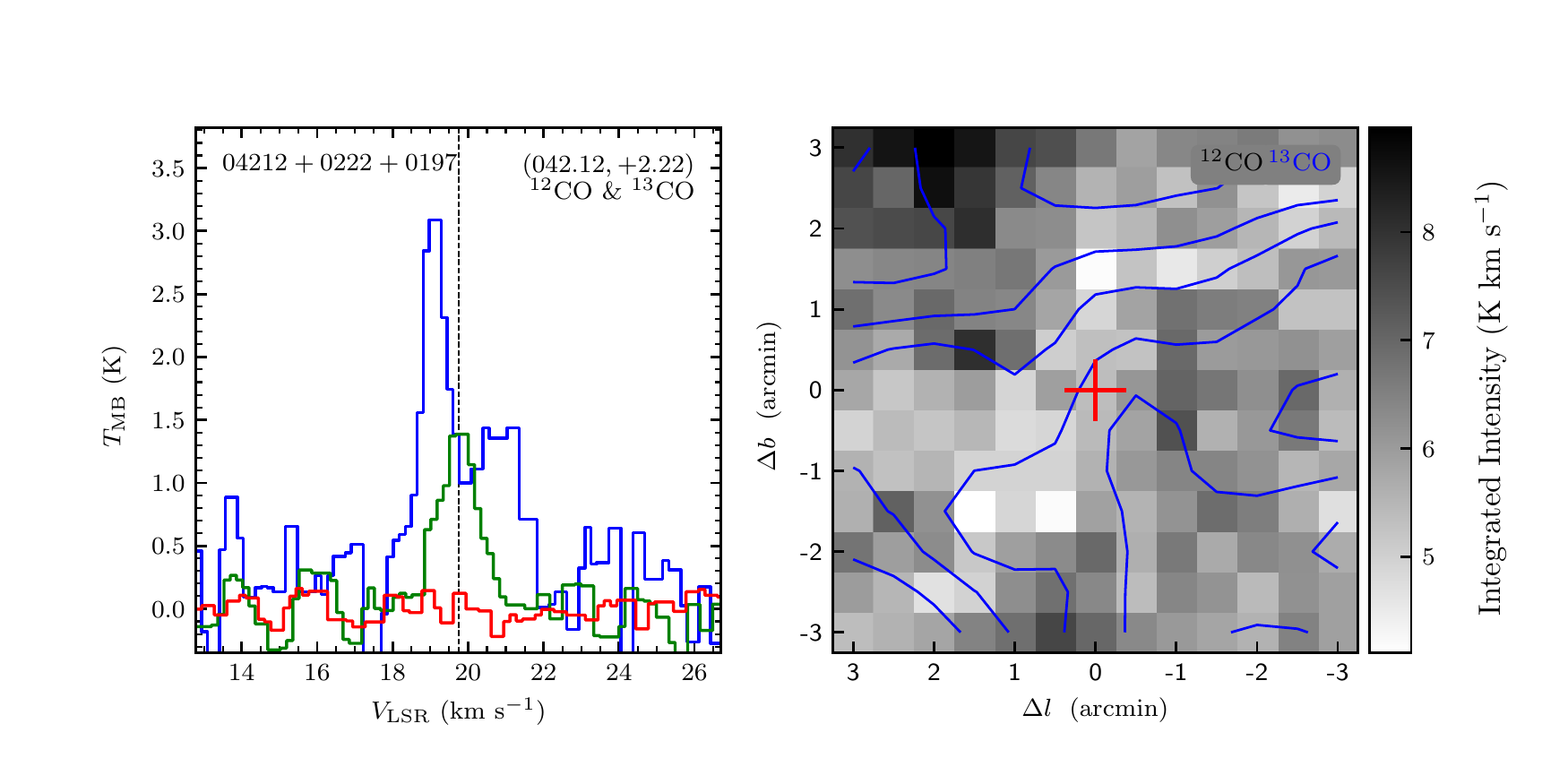}
\vspace{-0.5cm}

\includegraphics[width=9.0cm,angle=0]{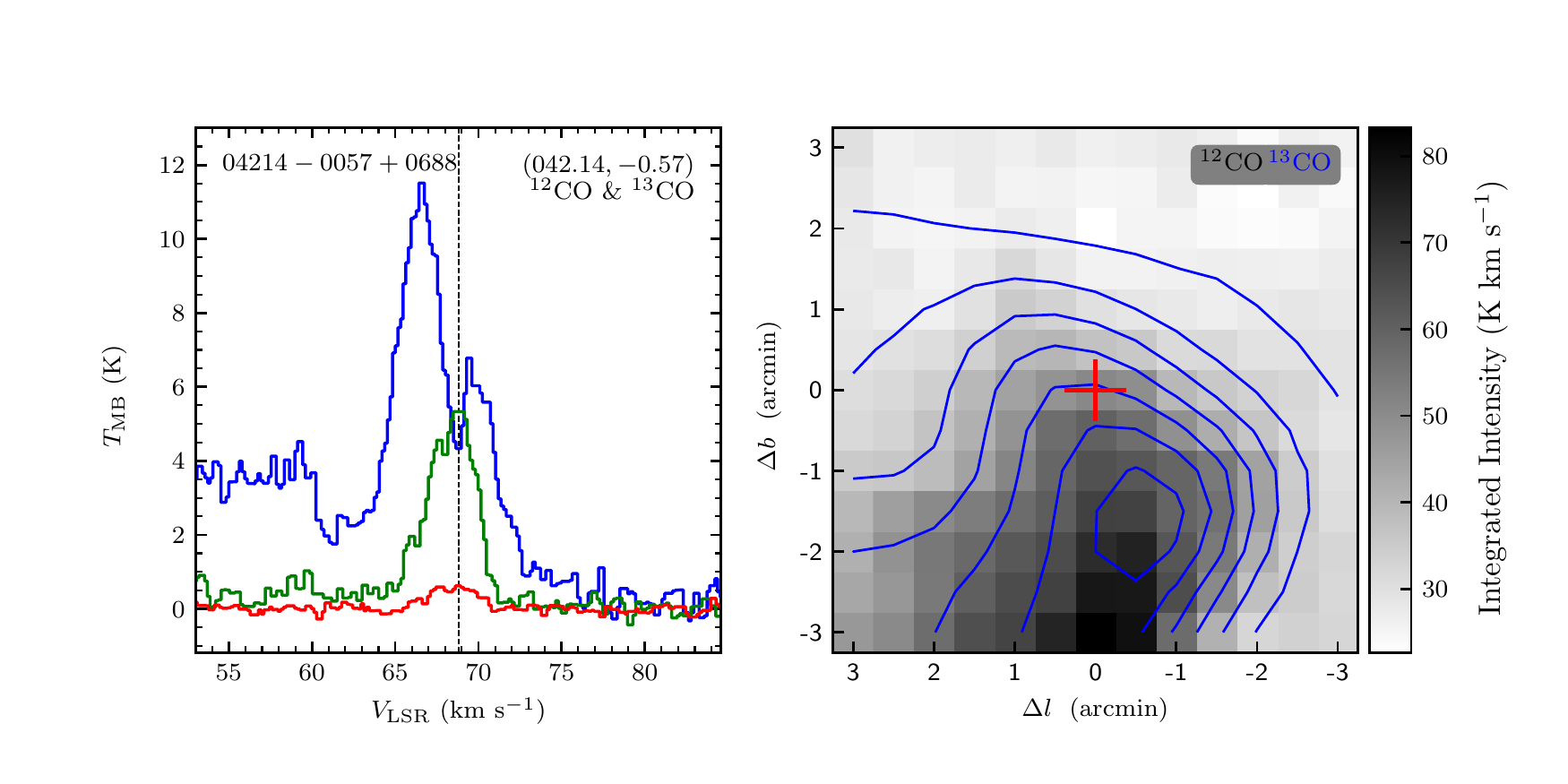}
\includegraphics[width=9.0cm,angle=0]{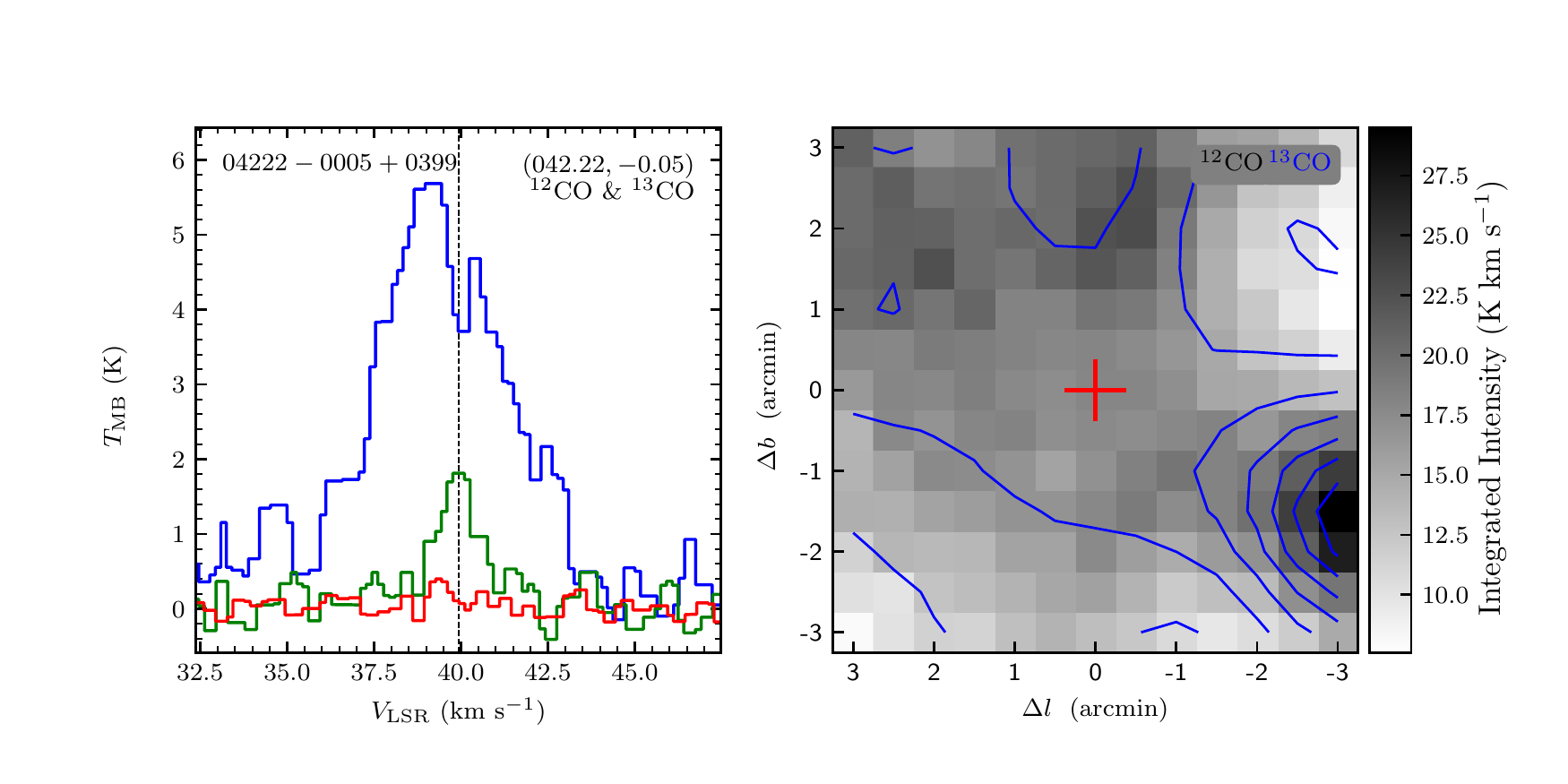}
\vspace{-0.5cm}

\includegraphics[width=9.0cm,angle=0]{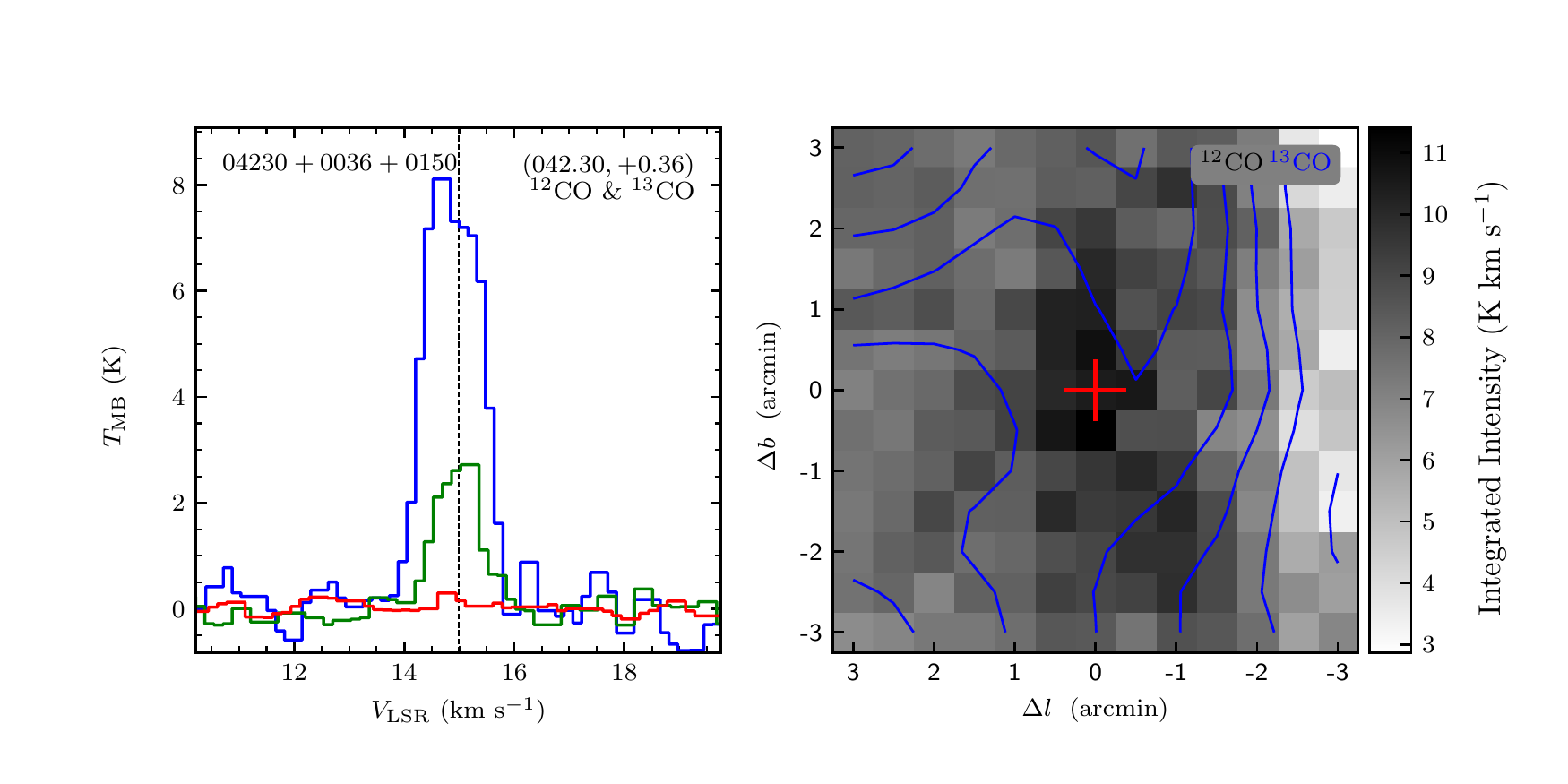}
\includegraphics[width=9.0cm,angle=0]{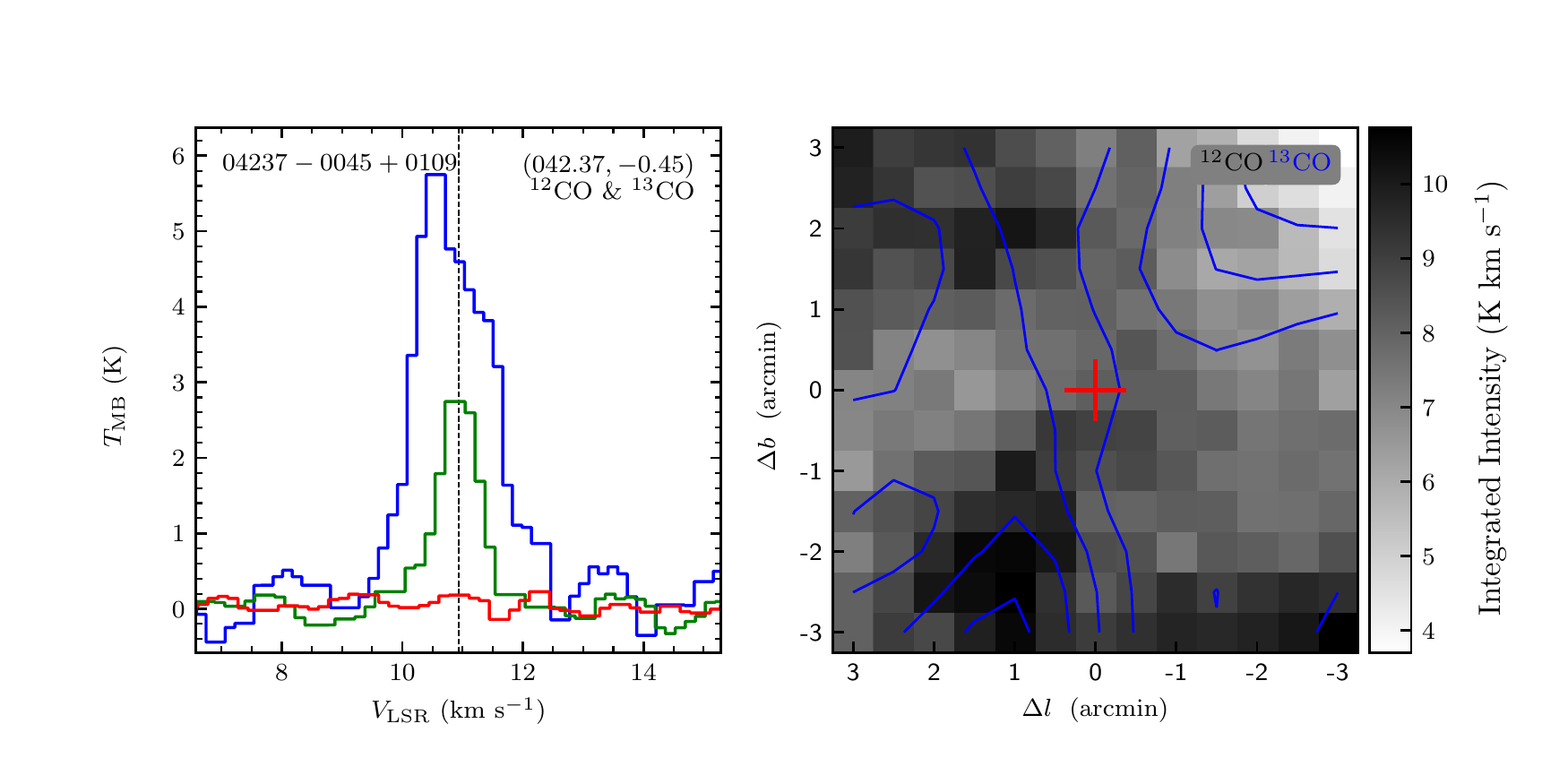}
\vspace{-0.5cm}

\includegraphics[width=9.0cm,angle=0]{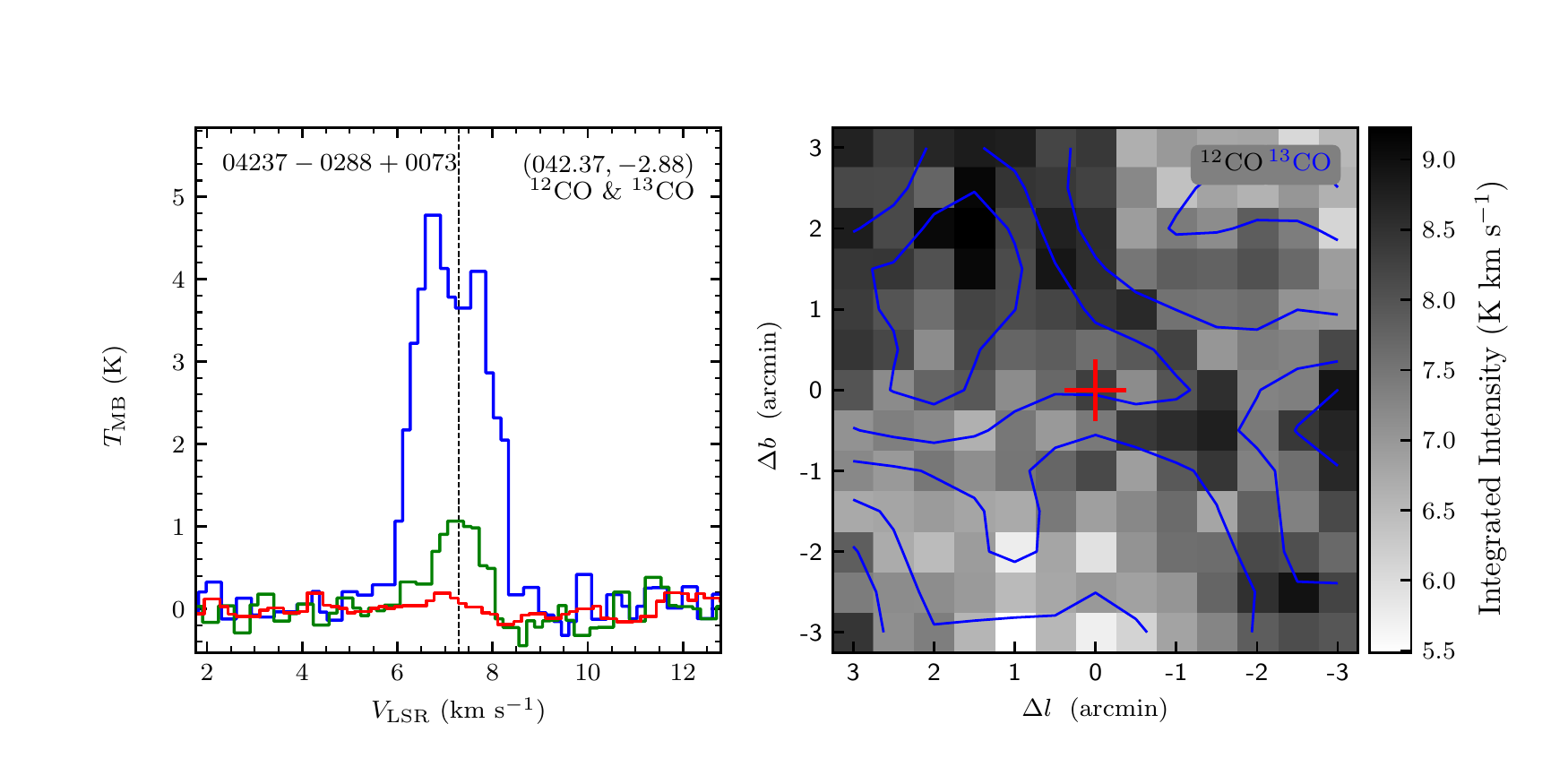}
\includegraphics[width=9.0cm,angle=0]{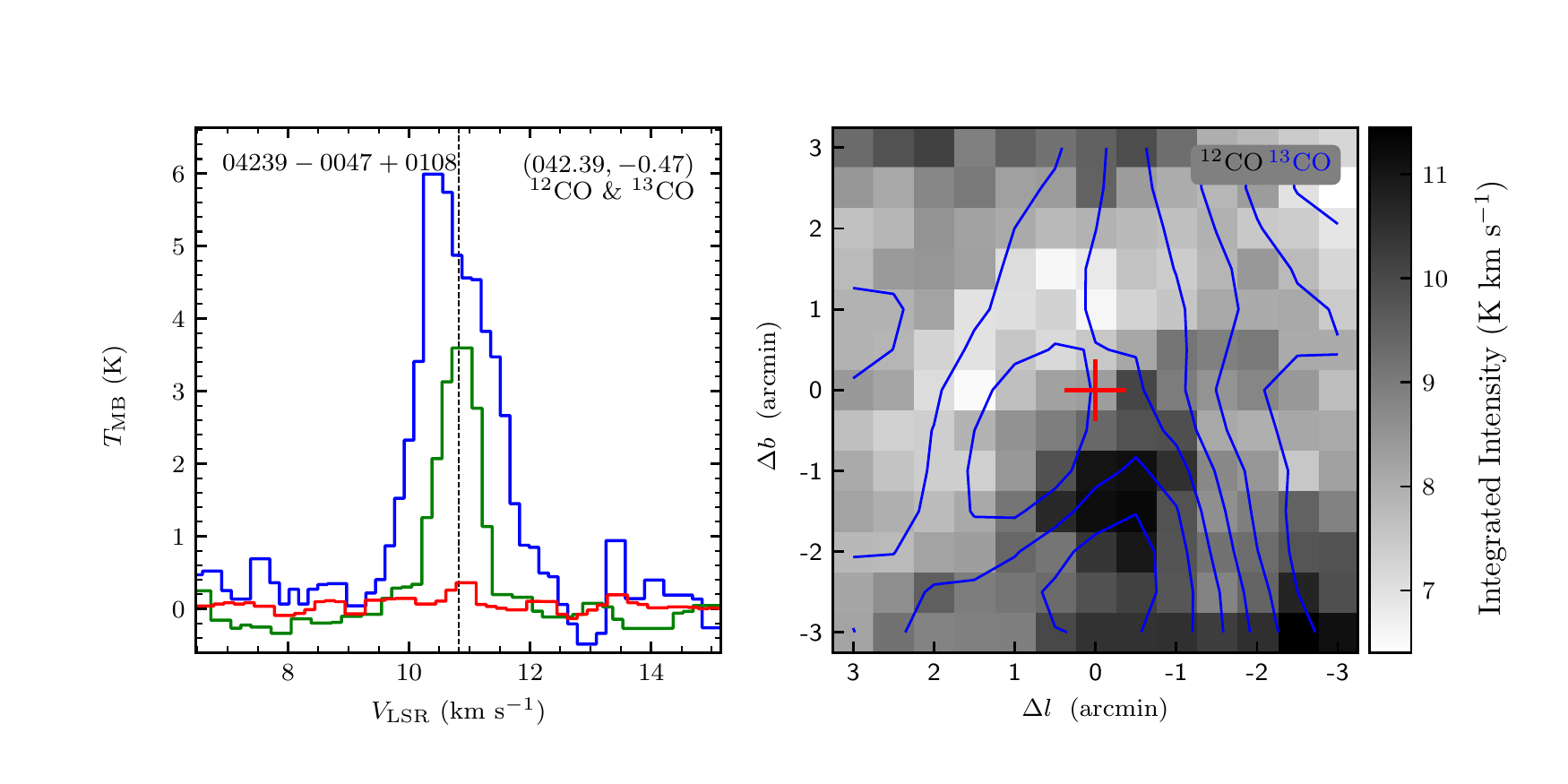}
\vspace{-0.5cm}

\includegraphics[width=9.0cm,angle=0]{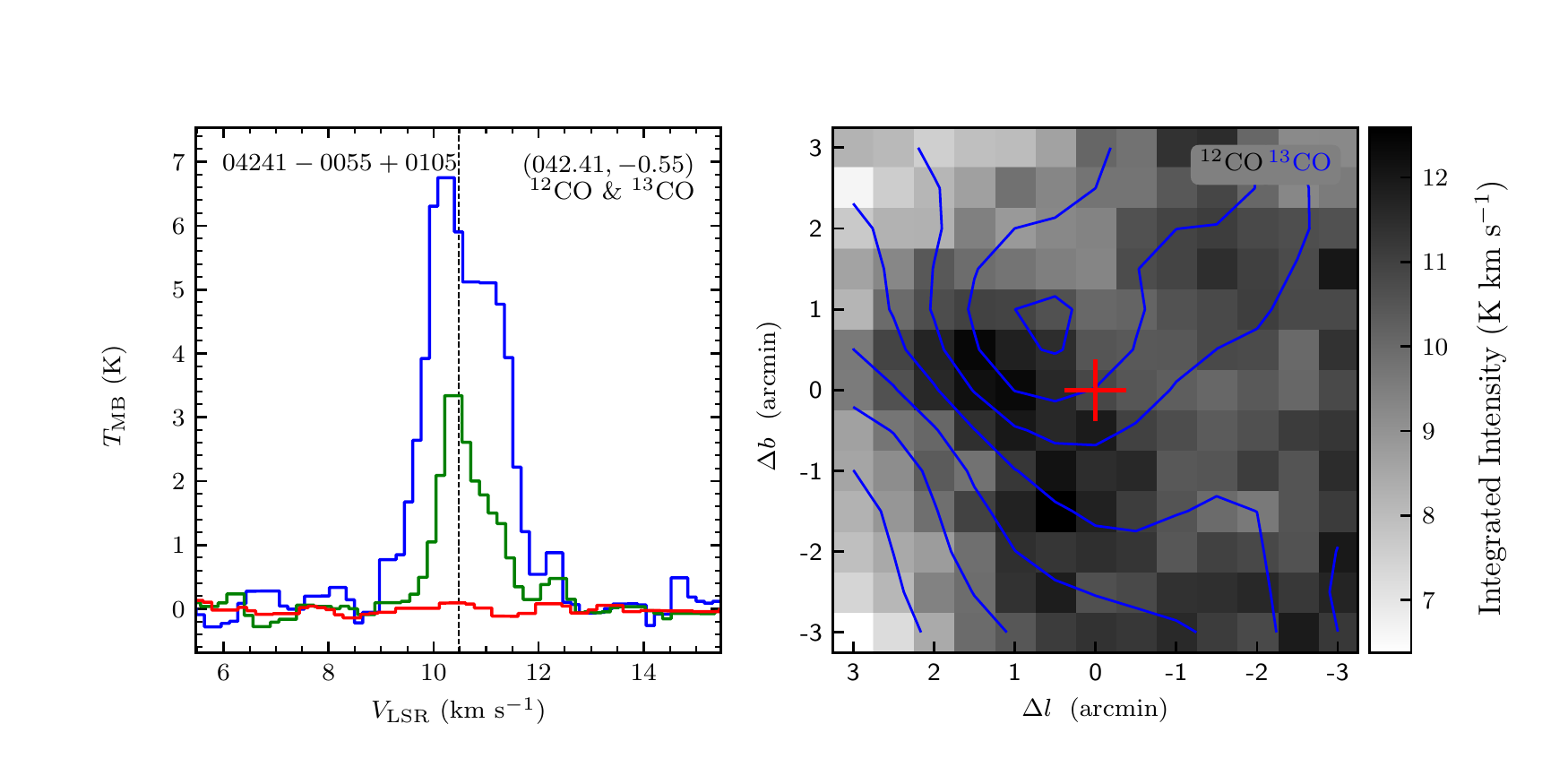}
\includegraphics[width=9.0cm,angle=0]{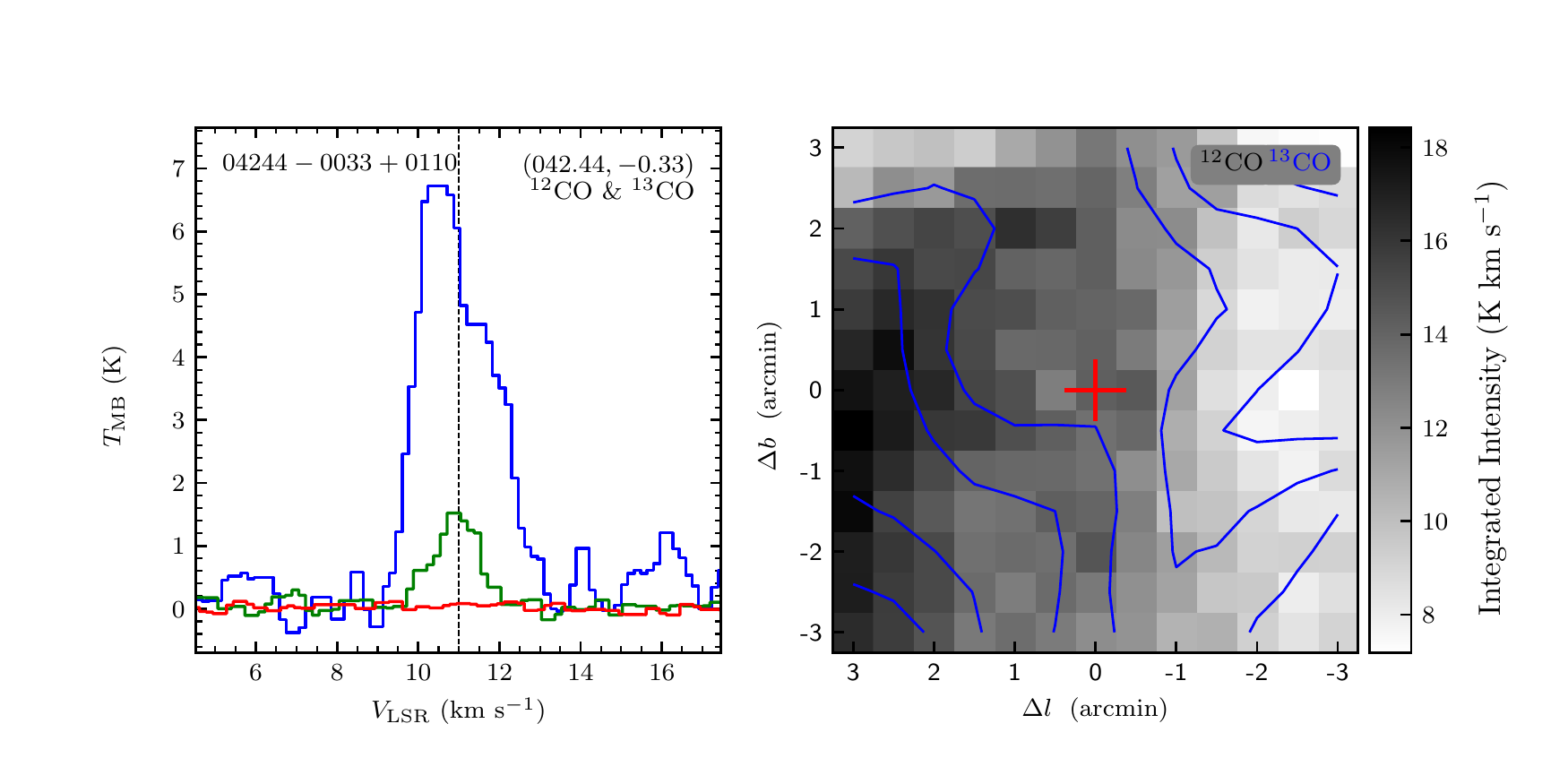}
\end{figure}
\clearpage

\begin{figure}
\includegraphics[width=9.0cm,angle=0]{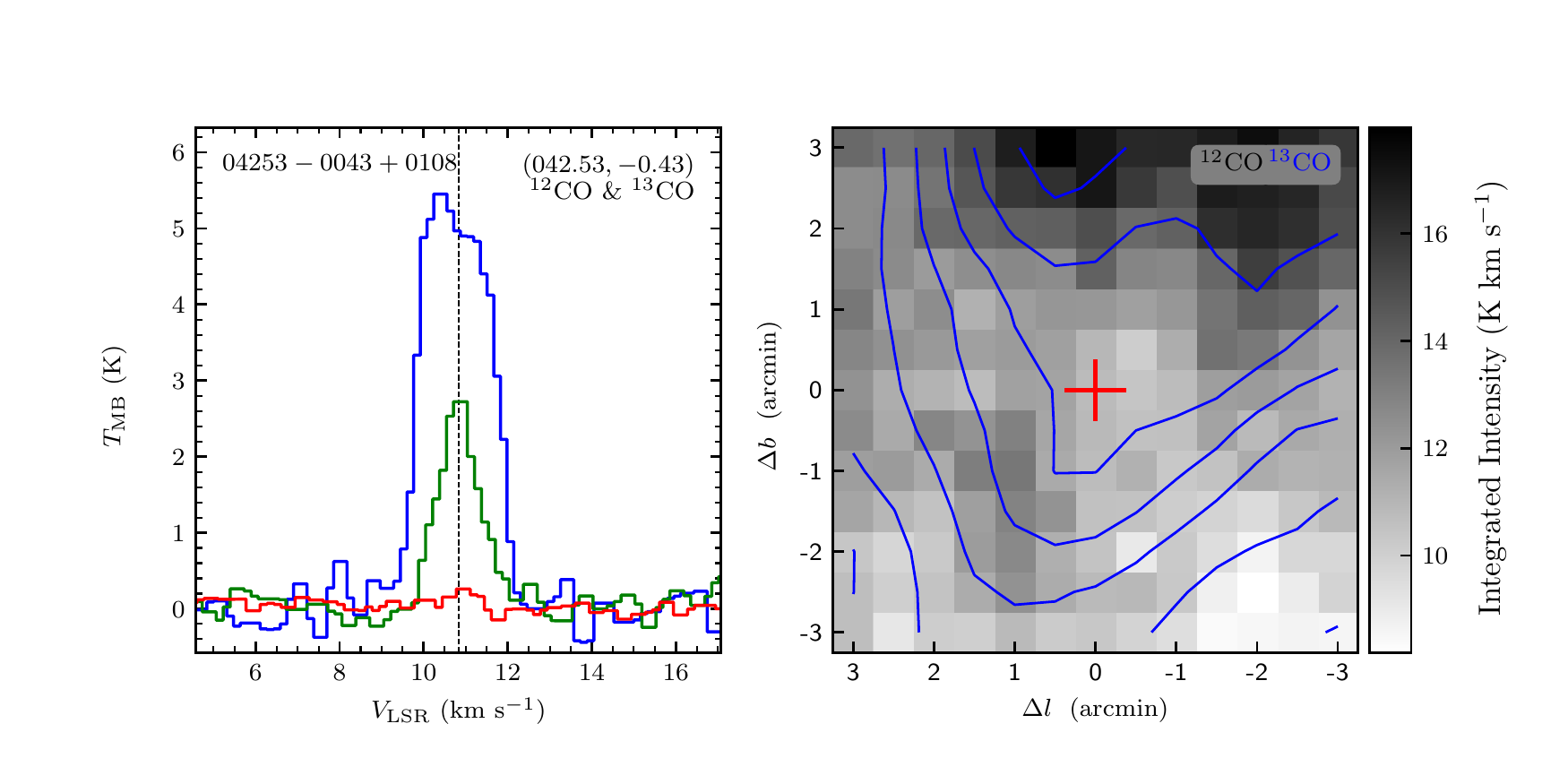}
\includegraphics[width=9.0cm,angle=0]{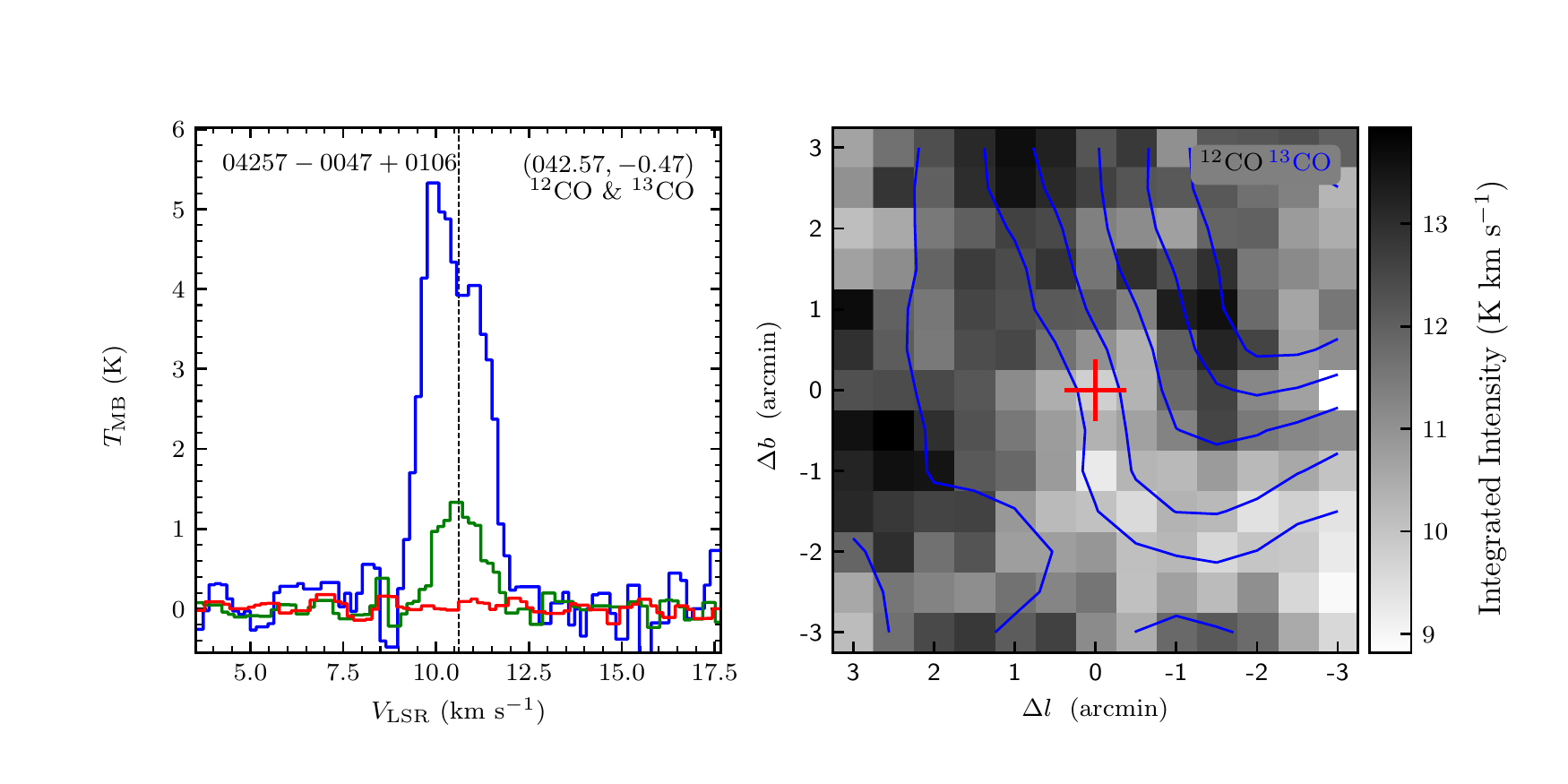}
\vspace{-0.5cm}

\includegraphics[width=9.0cm,angle=0]{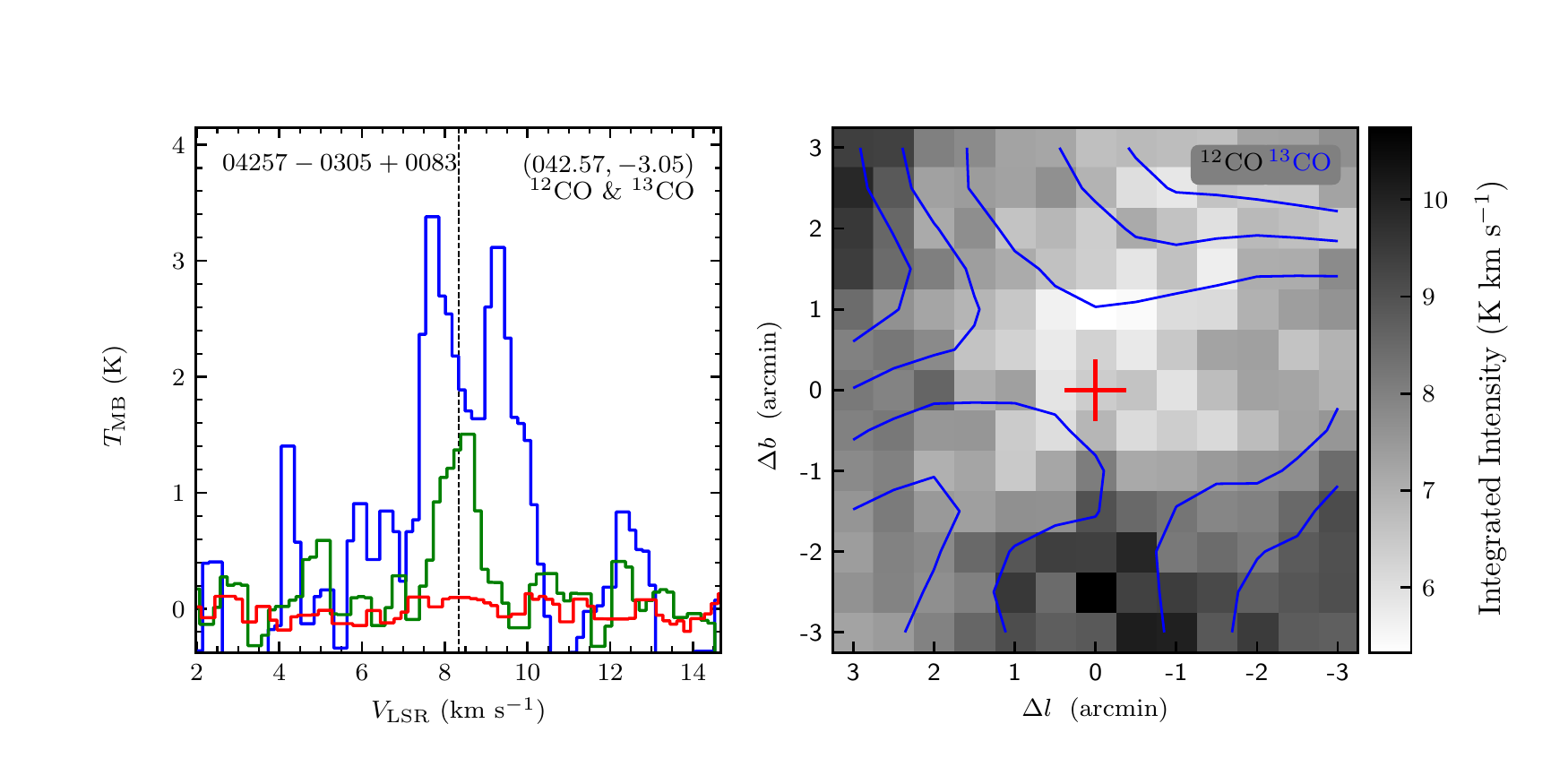}
\includegraphics[width=9.0cm,angle=0]{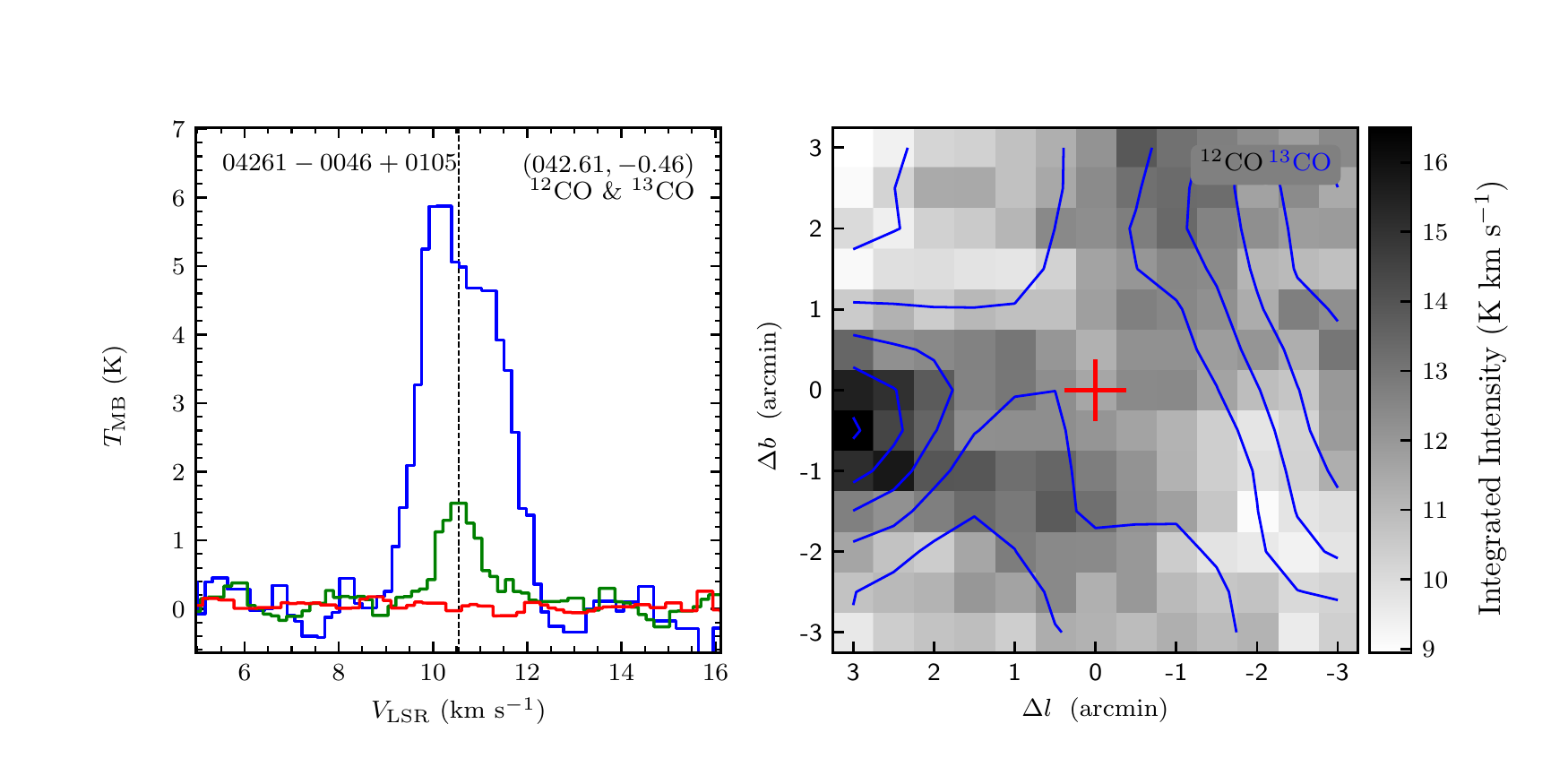}
\vspace{-0.5cm}

\includegraphics[width=9.0cm,angle=0]{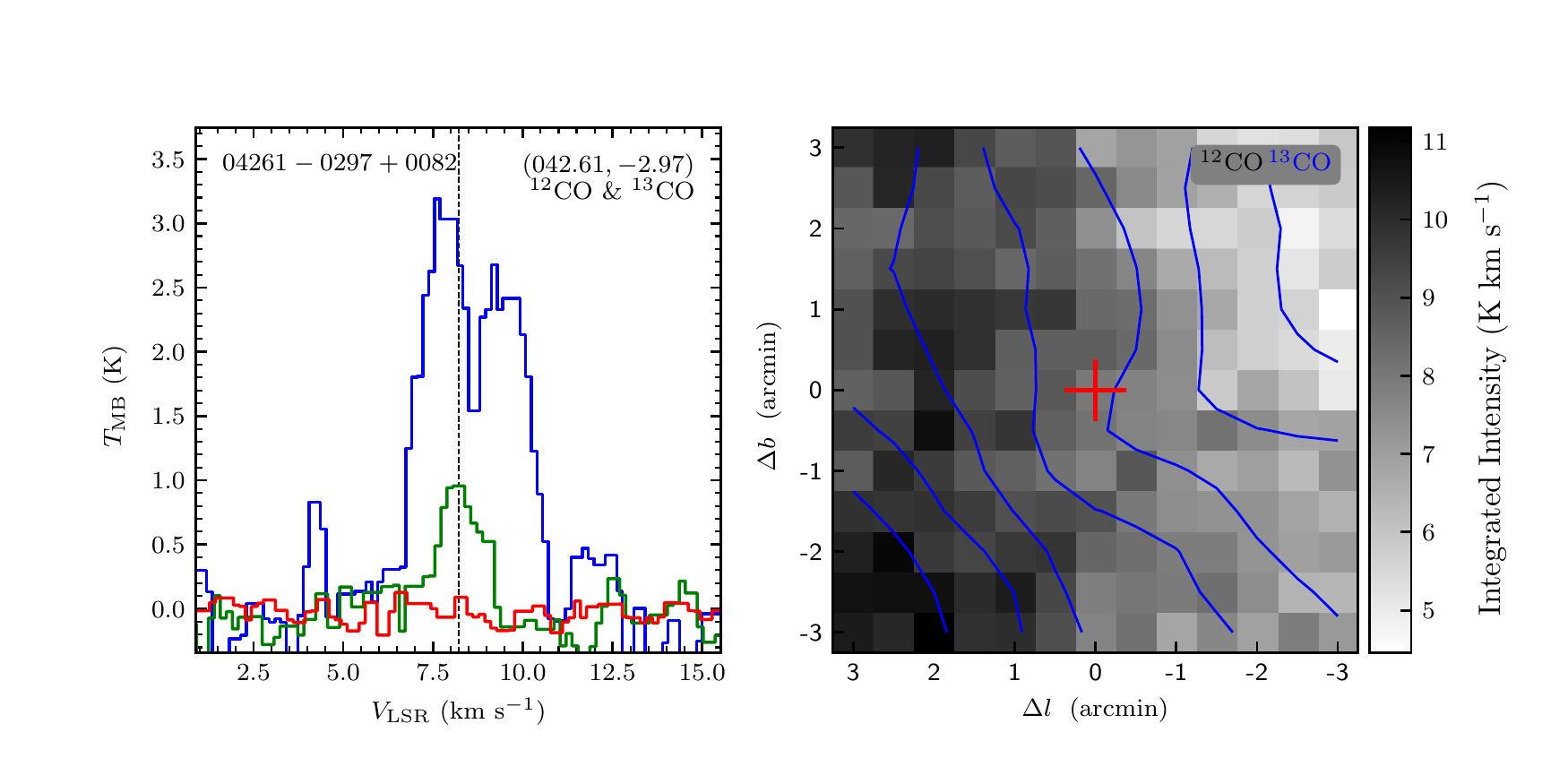}
\includegraphics[width=9.0cm,angle=0]{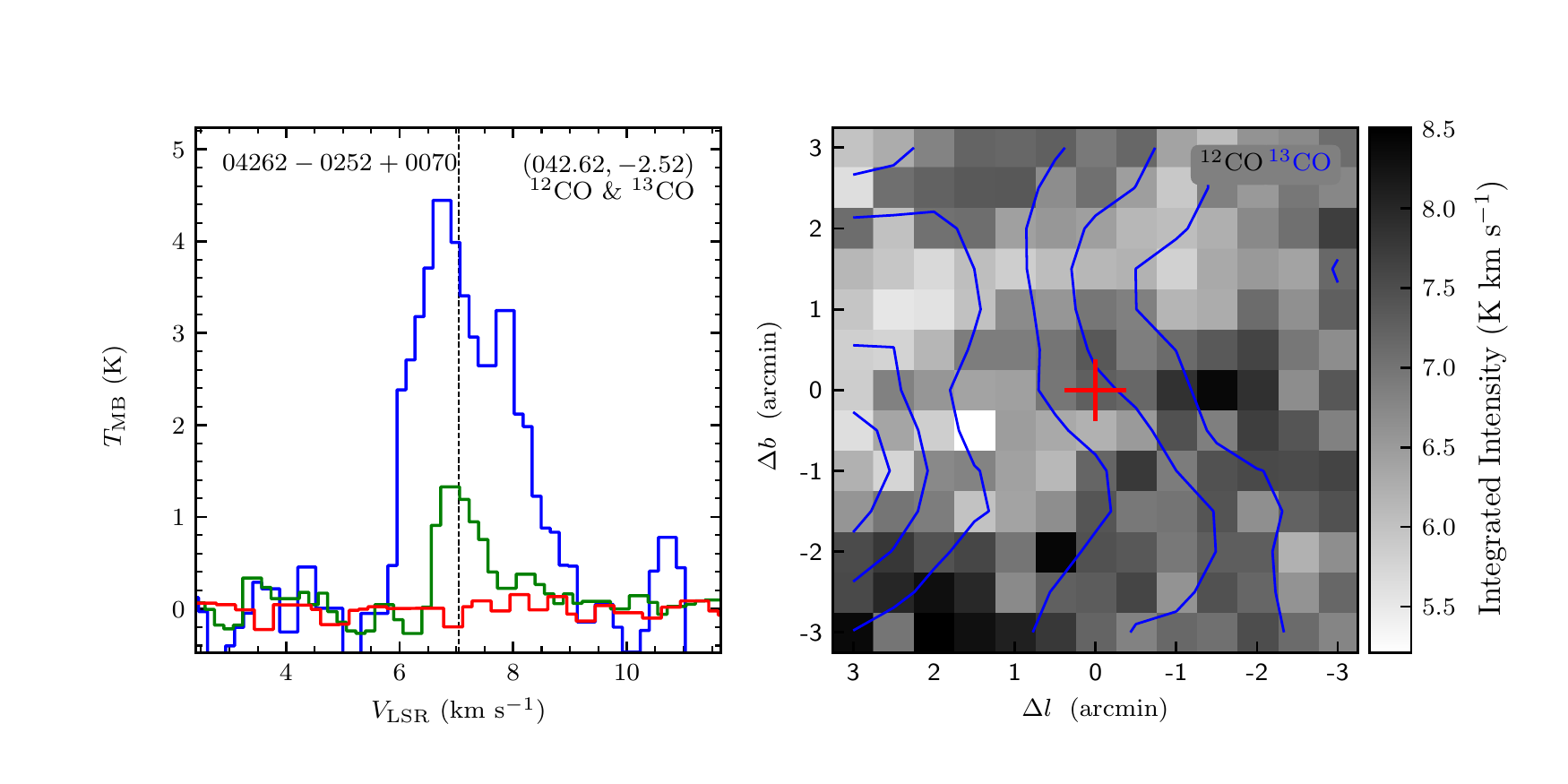}
\vspace{-0.5cm}

\includegraphics[width=9.0cm,angle=0]{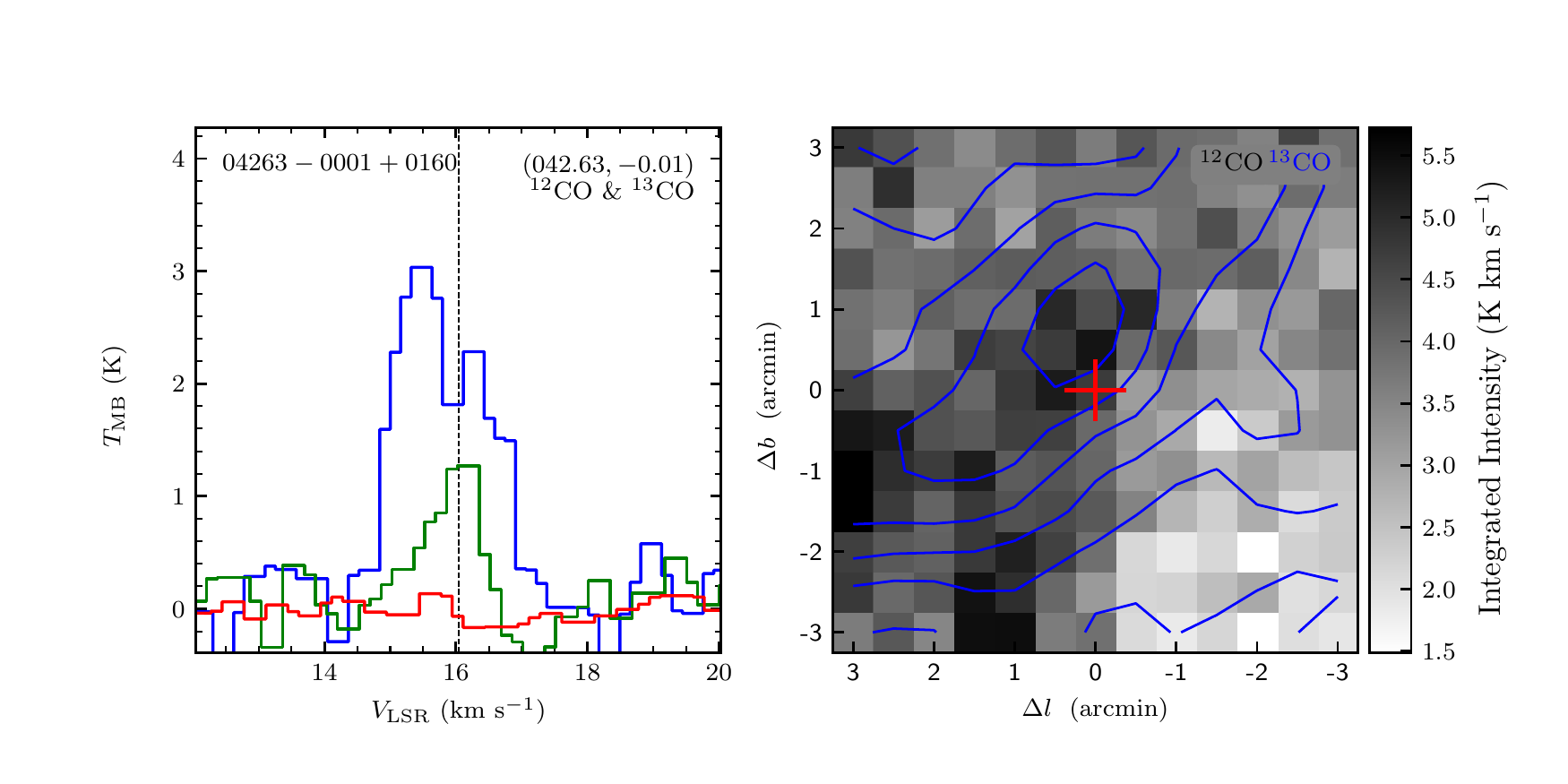}
\includegraphics[width=9.0cm,angle=0]{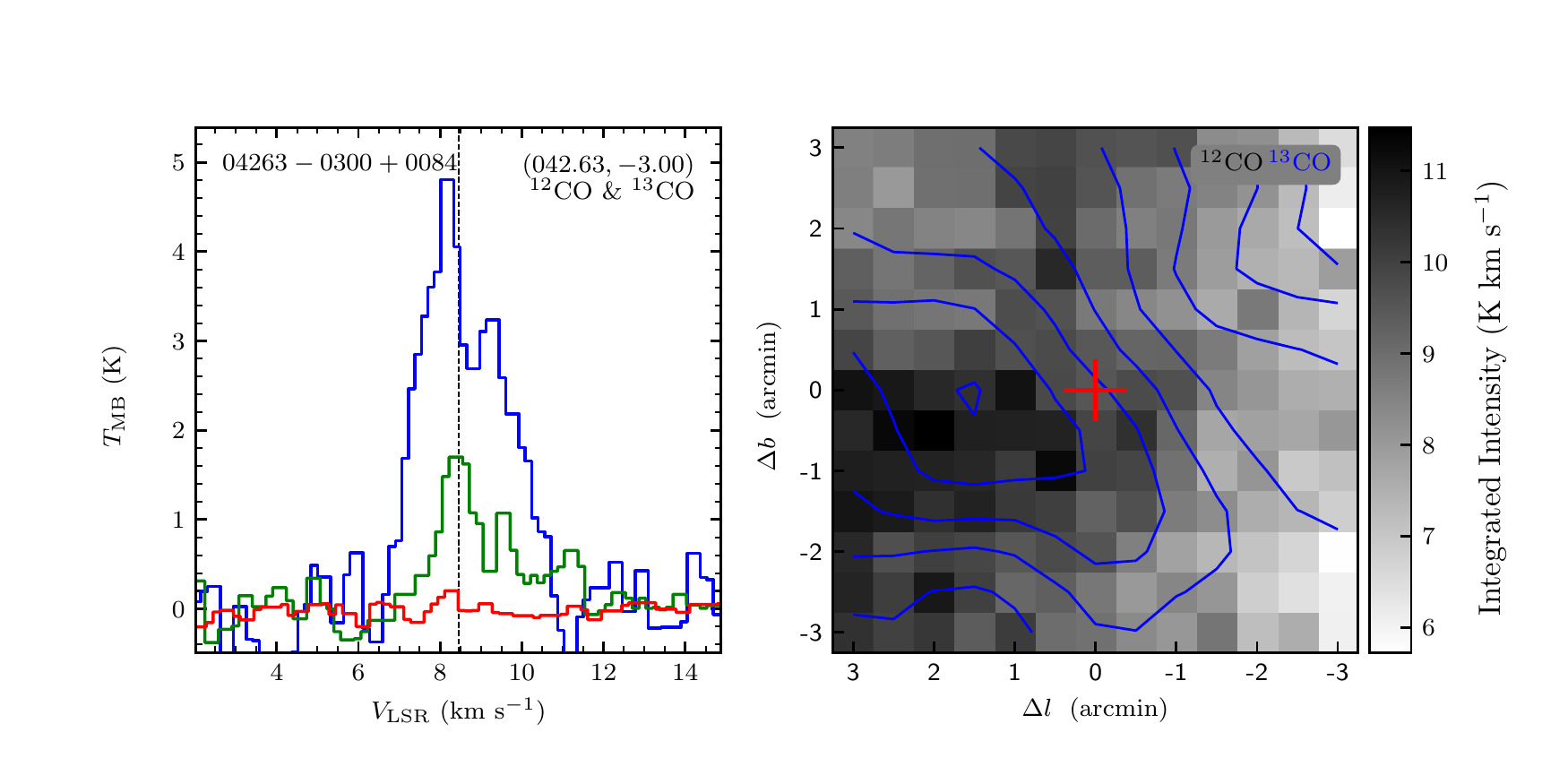}
\vspace{-0.5cm}

\includegraphics[width=9.0cm,angle=0]{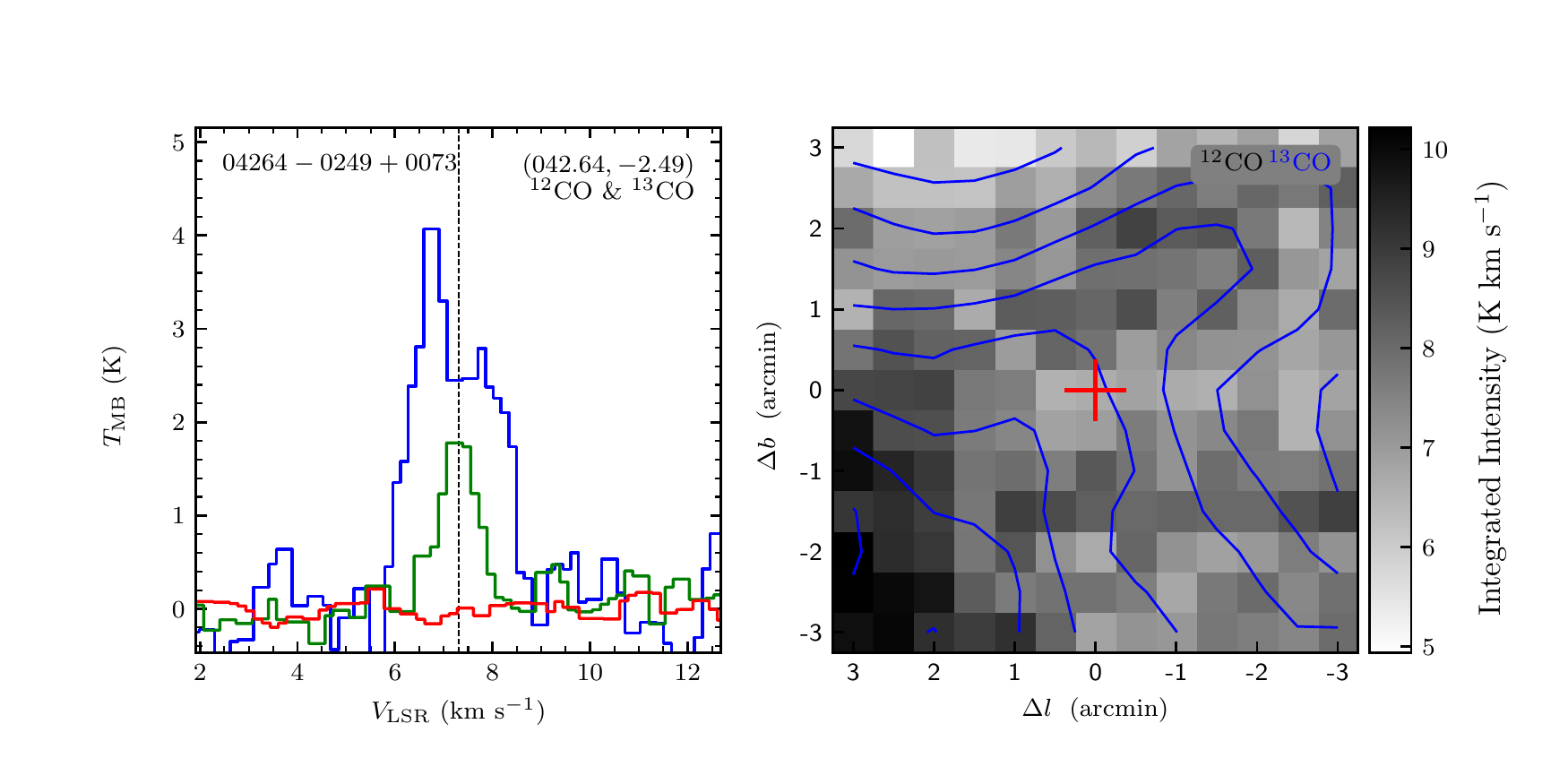}
\includegraphics[width=9.0cm,angle=0]{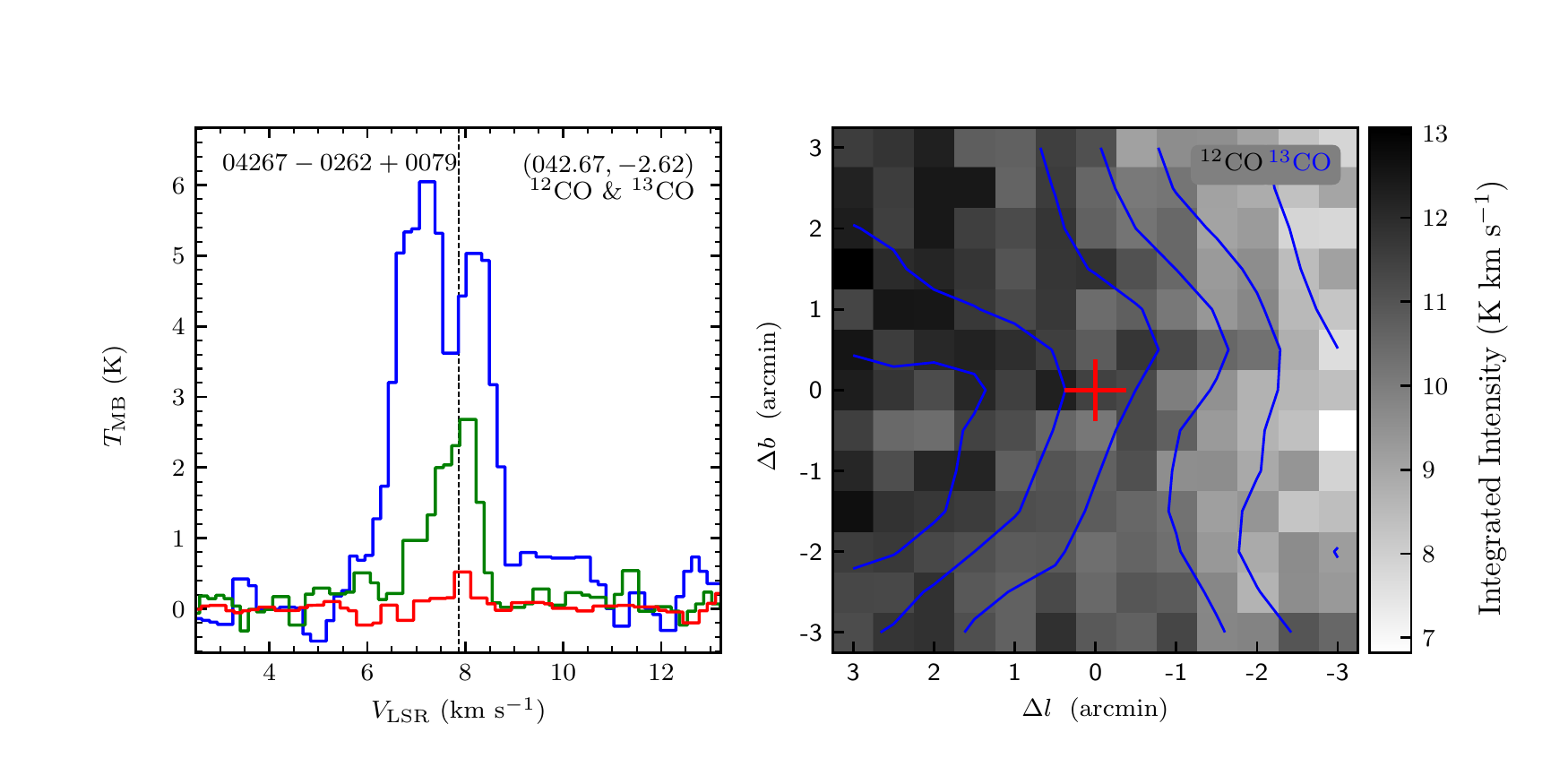}
\end{figure}
\clearpage

\begin{figure}
\includegraphics[width=9.0cm,angle=0]{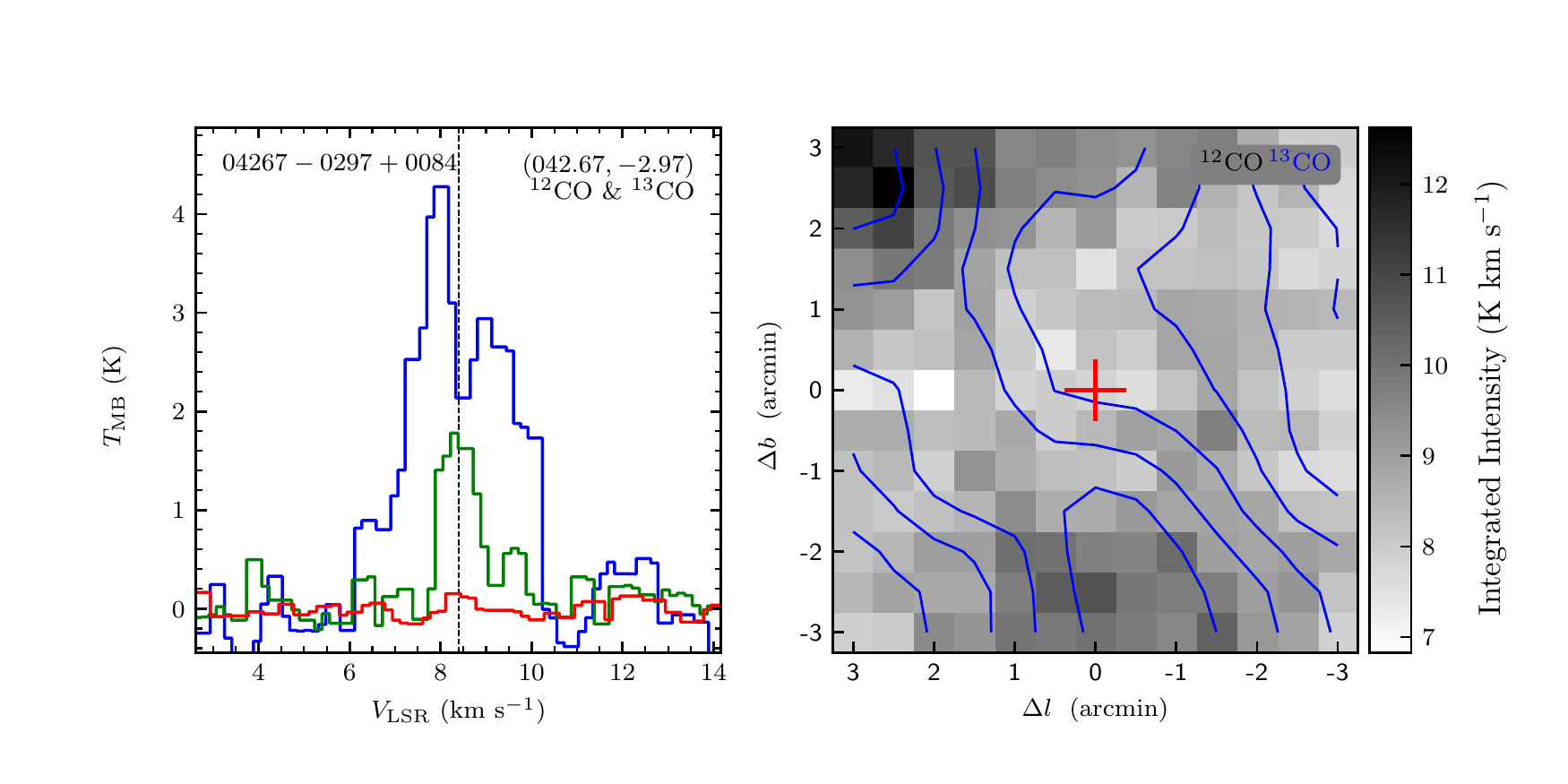}
\includegraphics[width=9.0cm,angle=0]{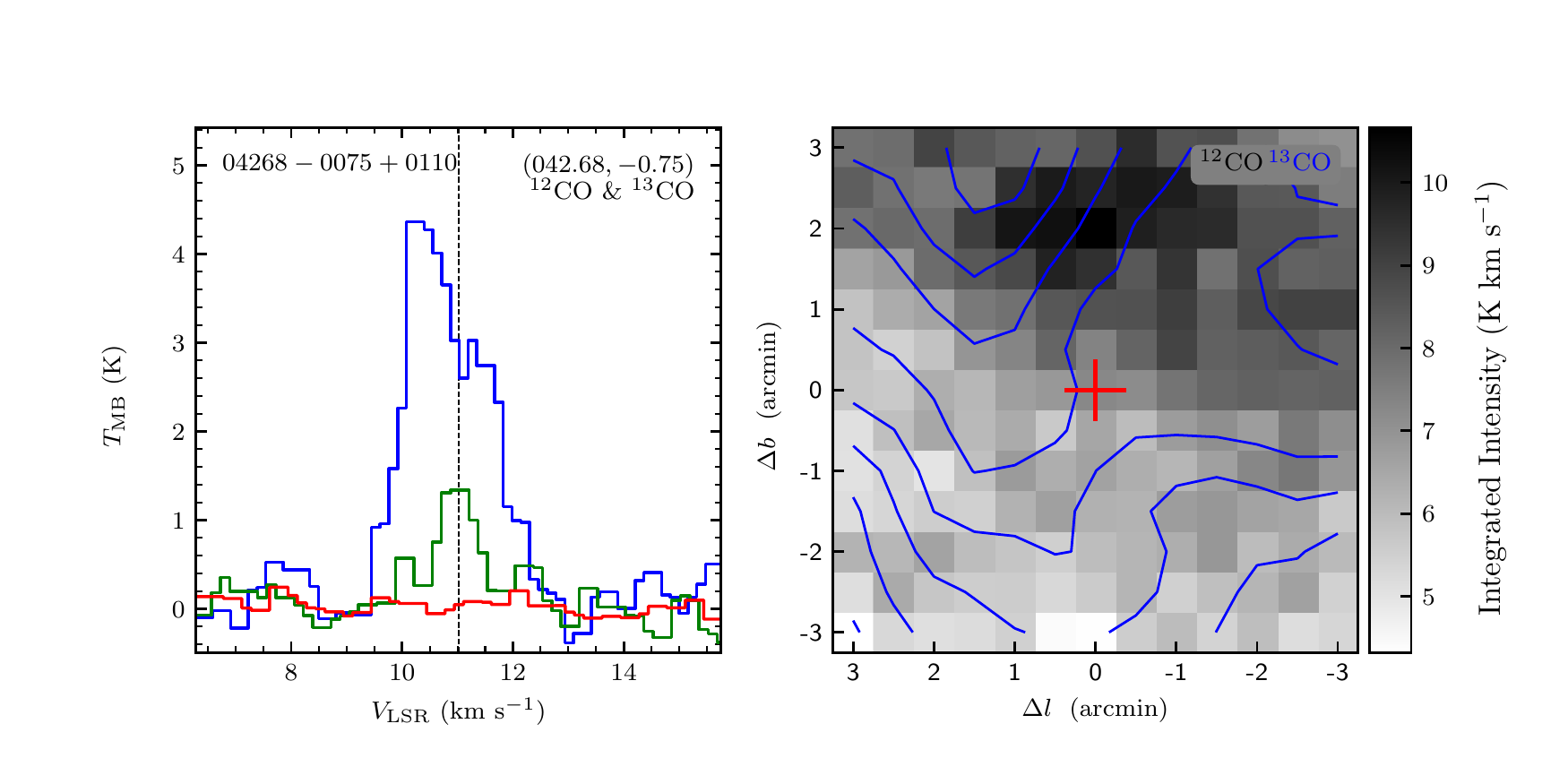}
\vspace{-0.5cm}

\includegraphics[width=9.0cm,angle=0]{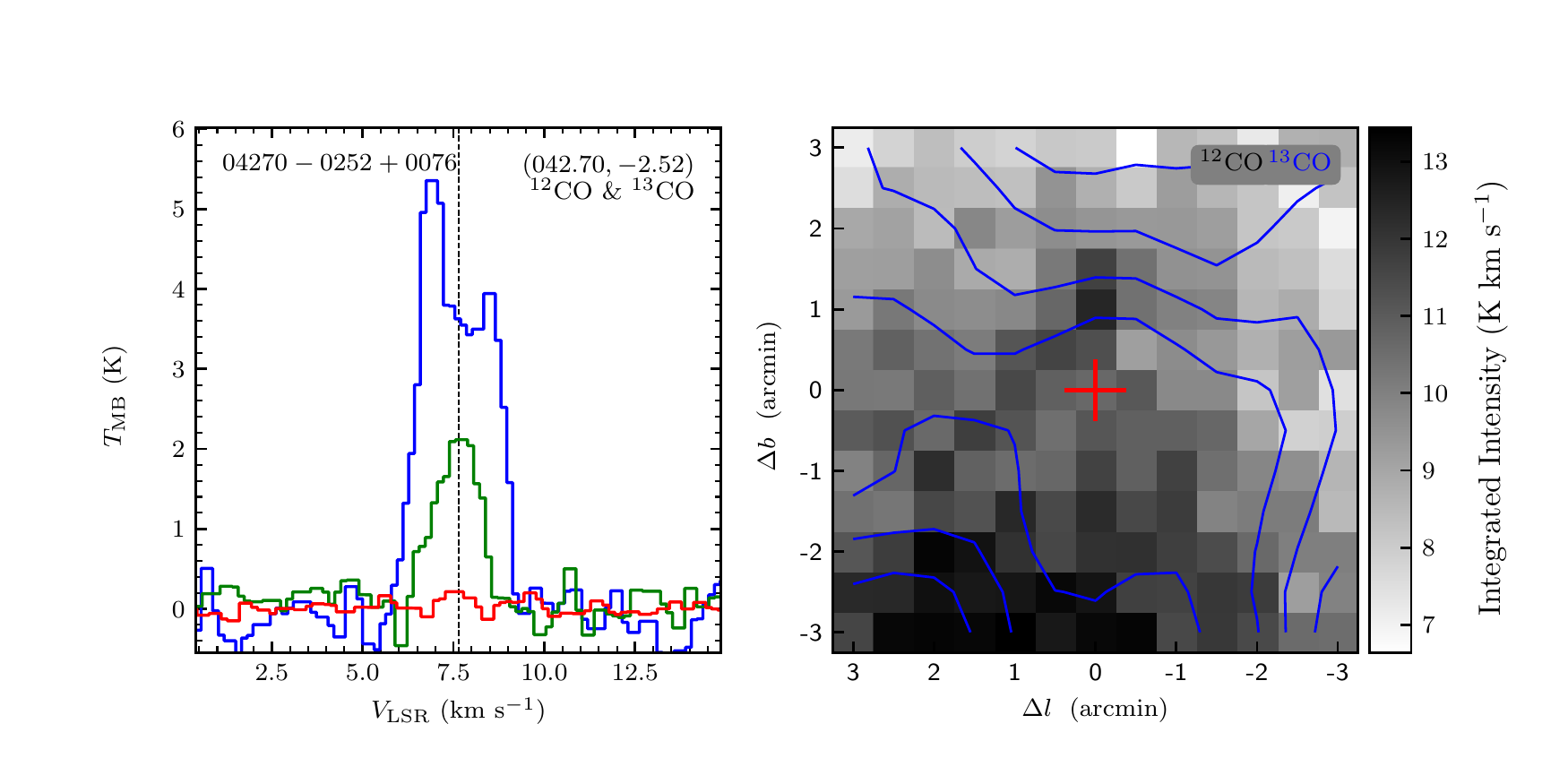}
\includegraphics[width=9.0cm,angle=0]{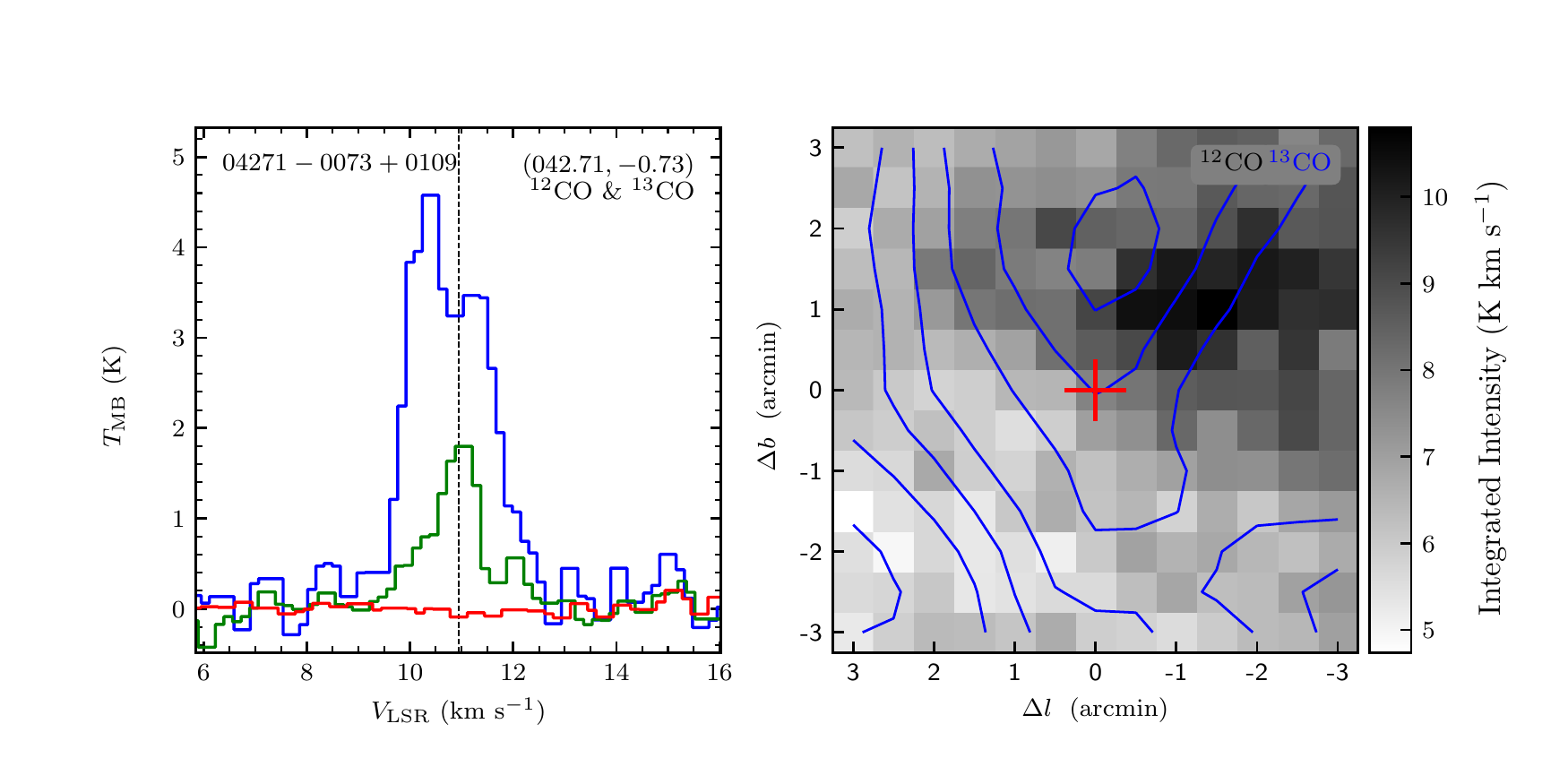}
\vspace{-0.5cm}

\includegraphics[width=9.0cm,angle=0]{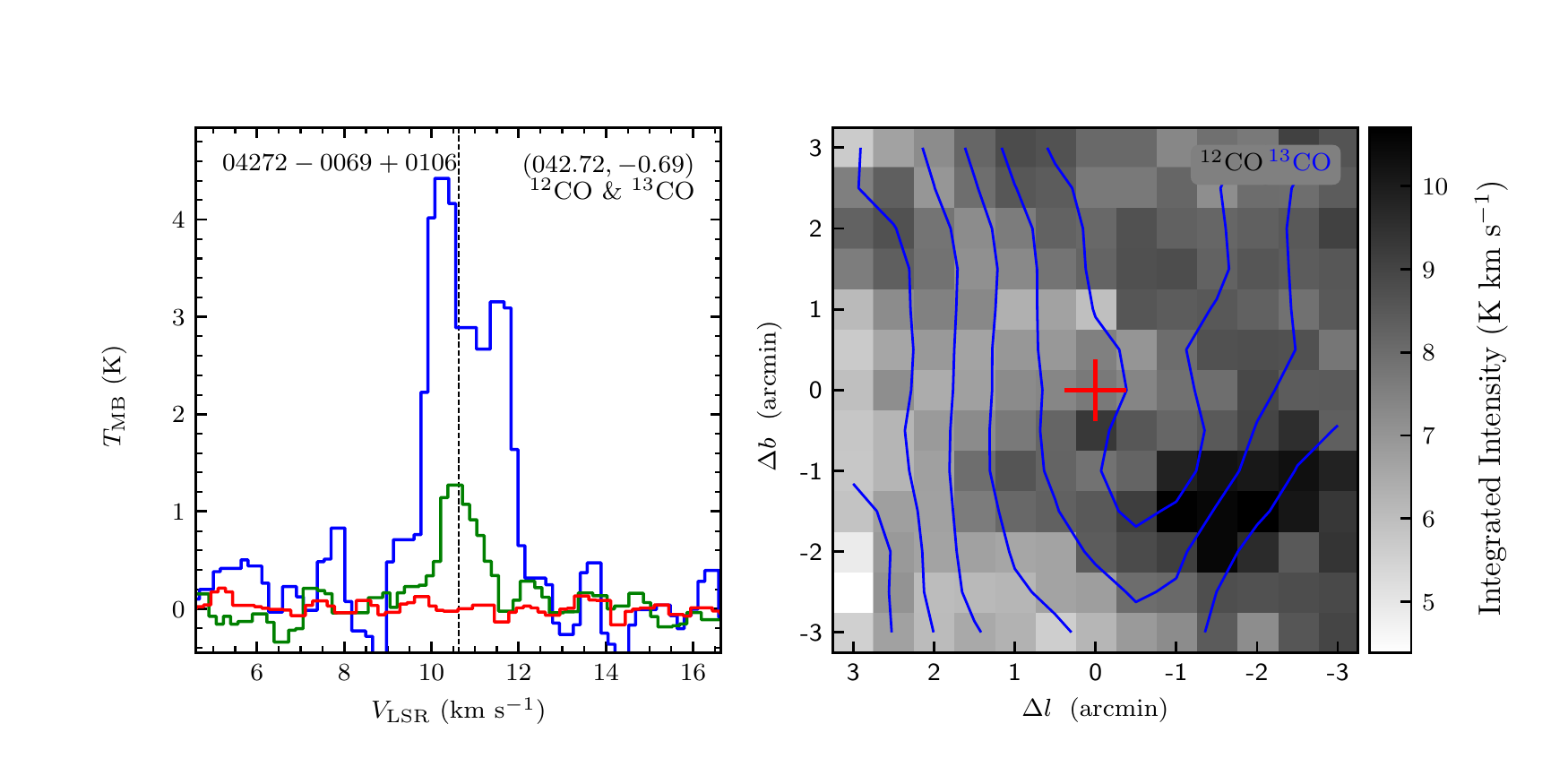}
\includegraphics[width=9.0cm,angle=0]{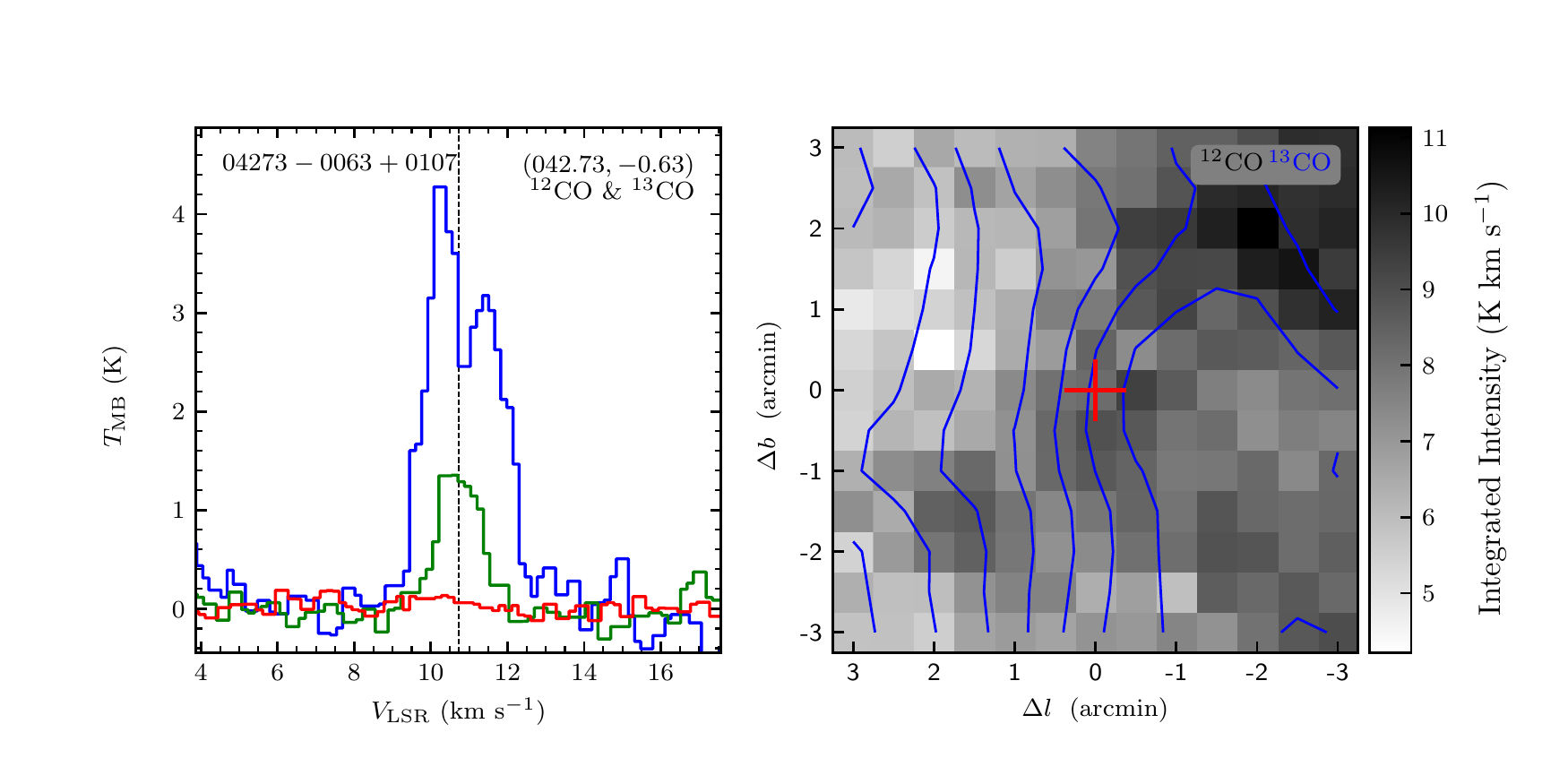}
\vspace{-0.5cm}

\includegraphics[width=9.0cm,angle=0]{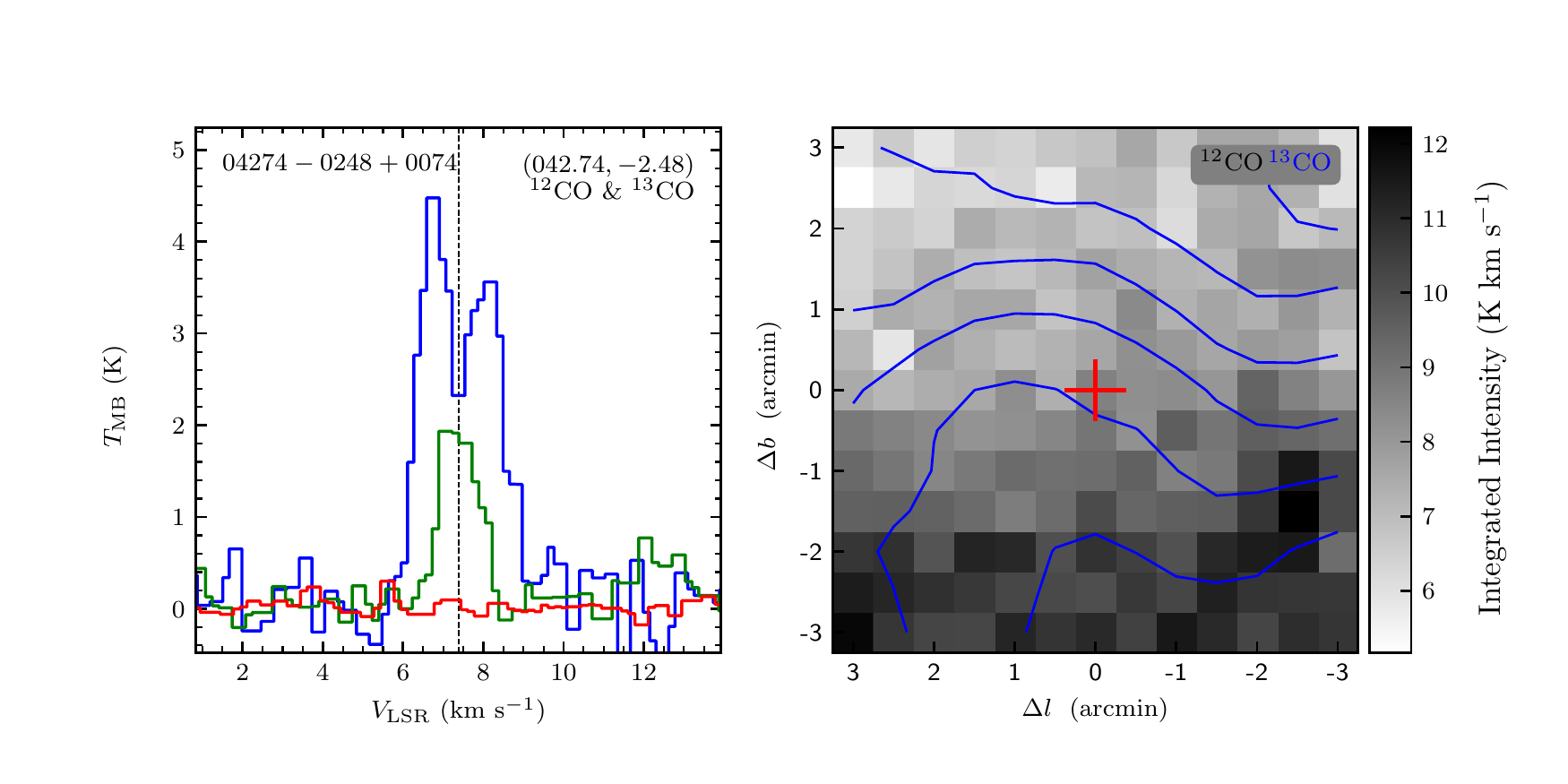}
\includegraphics[width=9.0cm,angle=0]{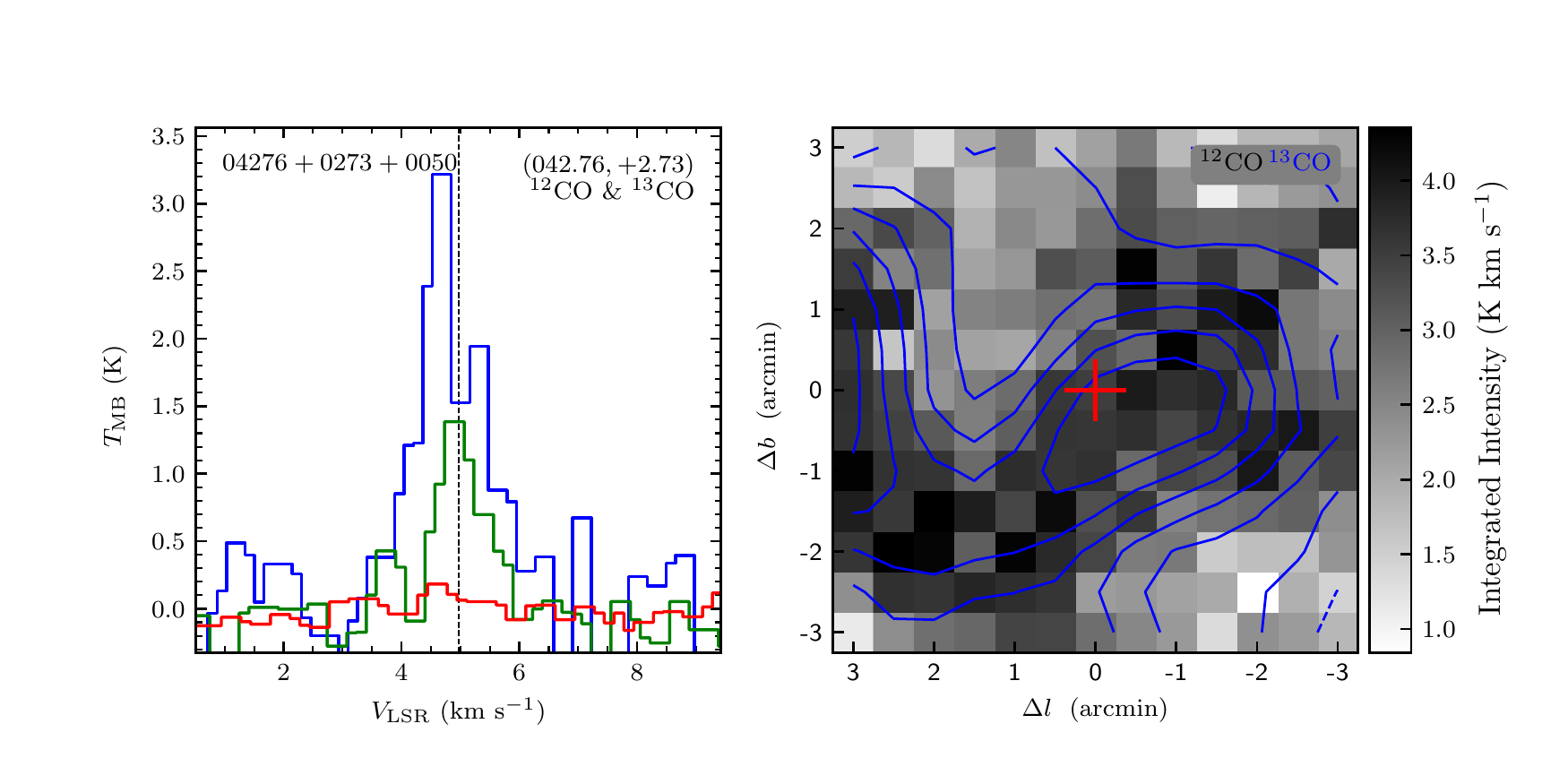}
\vspace{-0.5cm}

\includegraphics[width=9.0cm,angle=0]{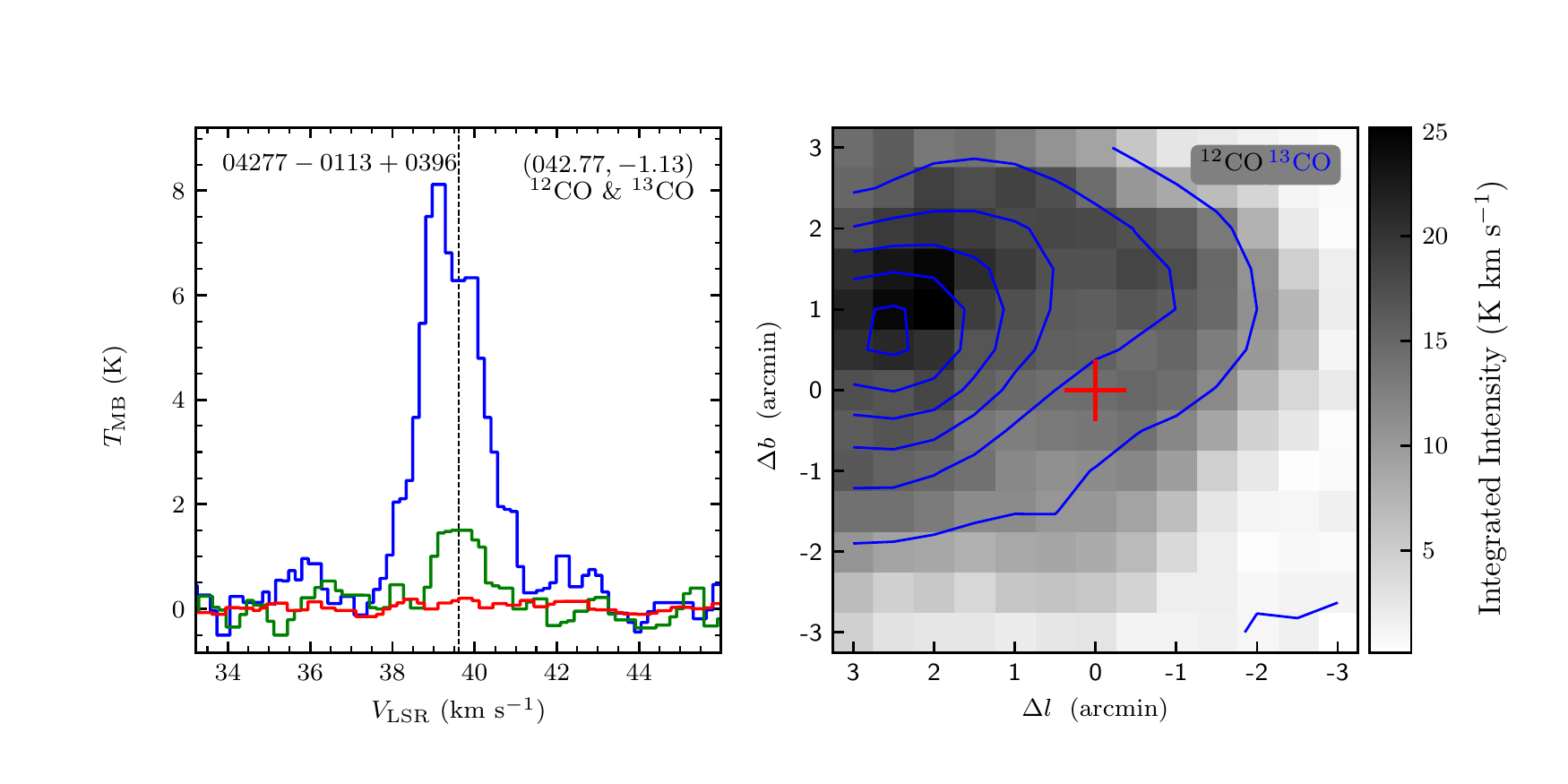}
\includegraphics[width=9.0cm,angle=0]{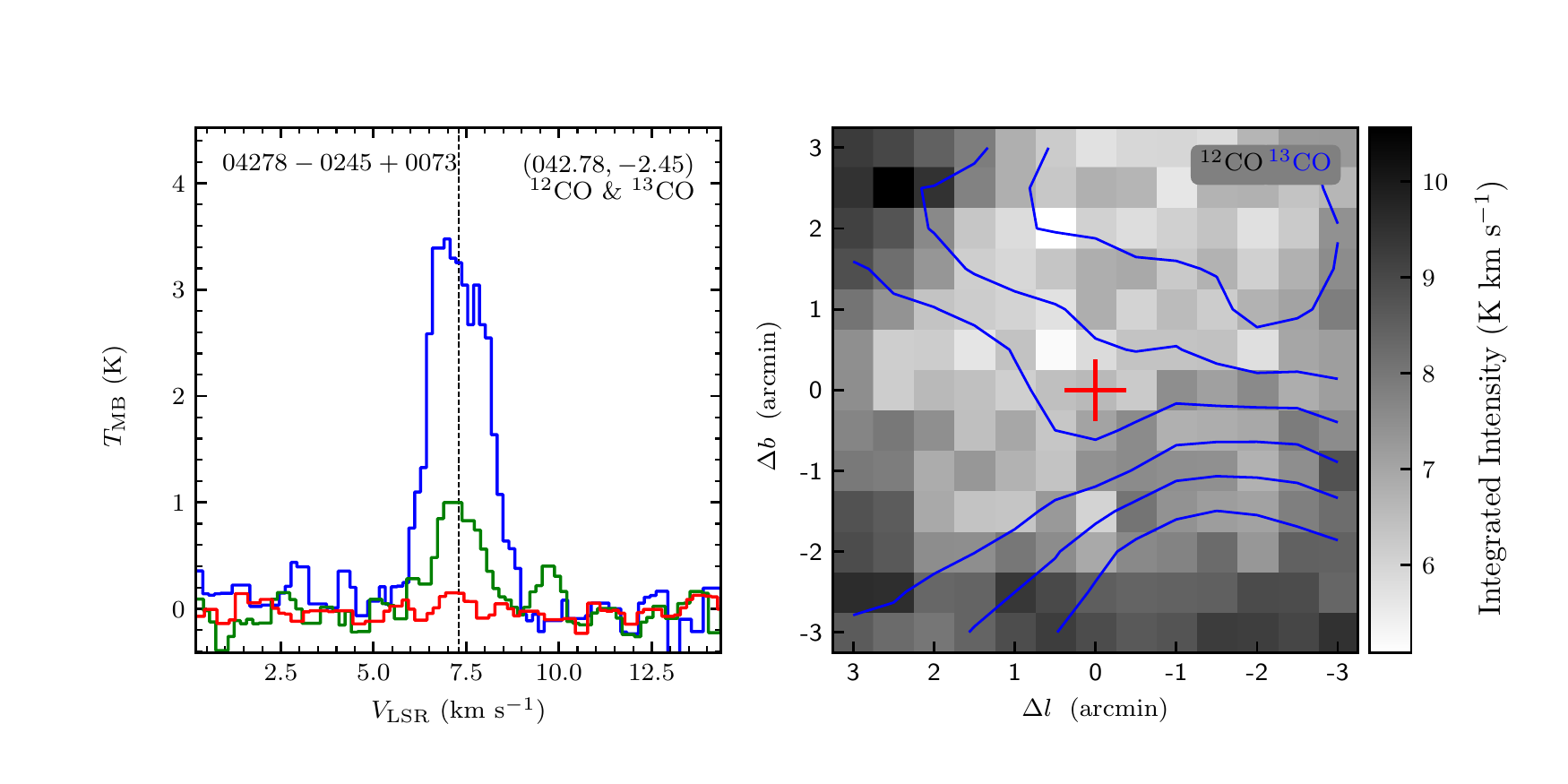}
\end{figure}
\clearpage

\begin{figure}
\includegraphics[width=9.0cm,angle=0]{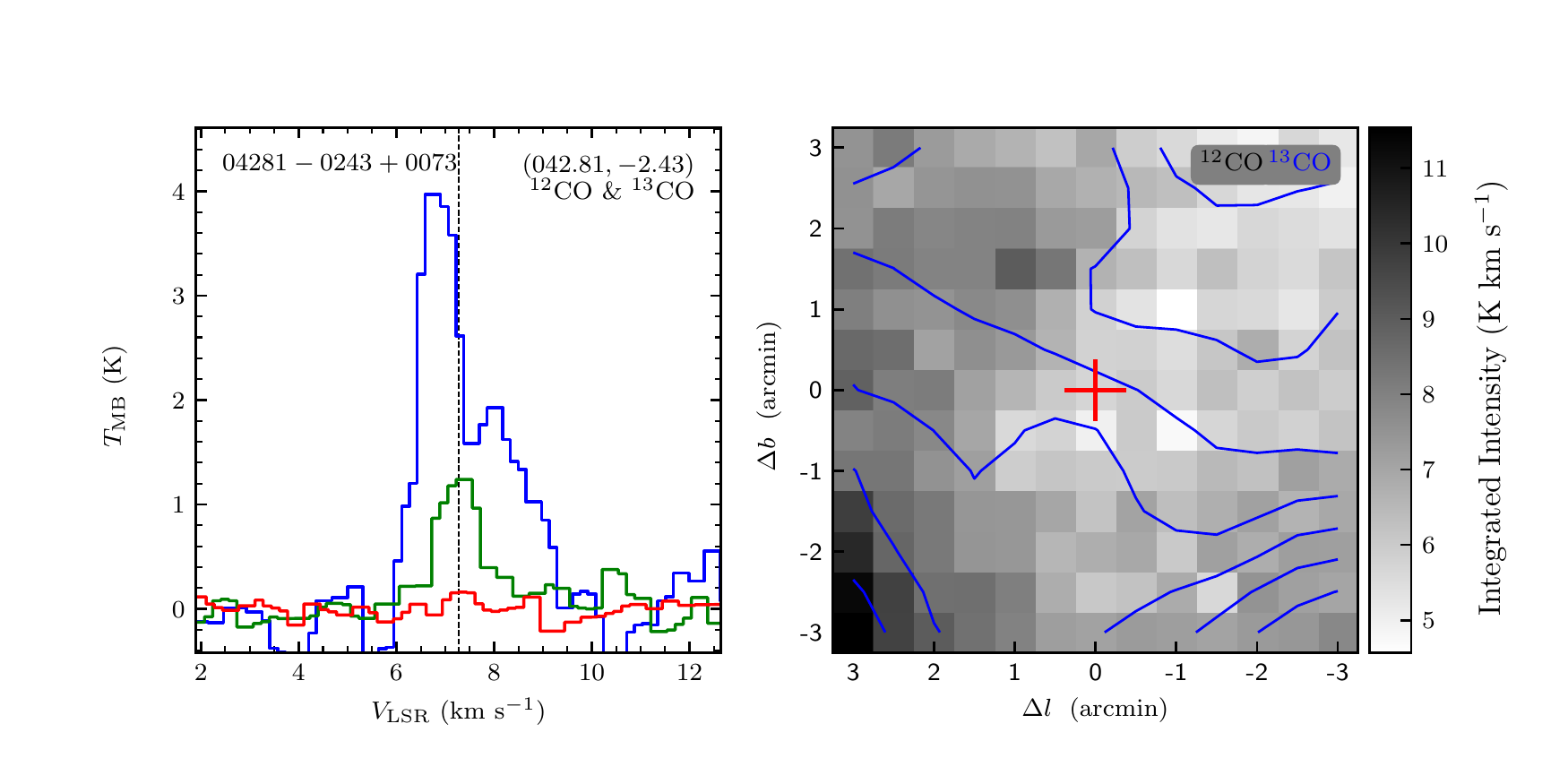}
\includegraphics[width=9.0cm,angle=0]{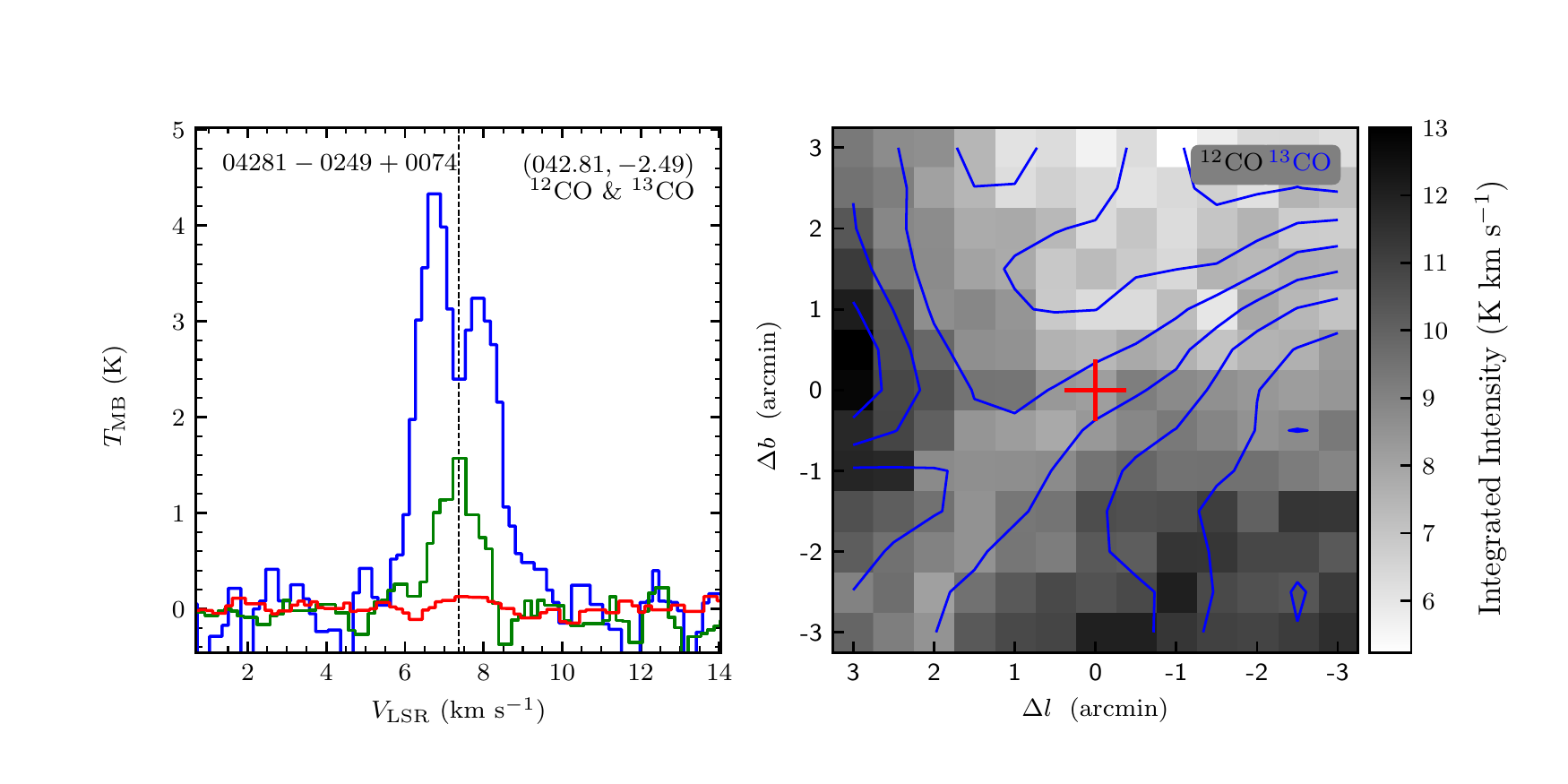}
\vspace{-0.5cm}

\includegraphics[width=9.0cm,angle=0]{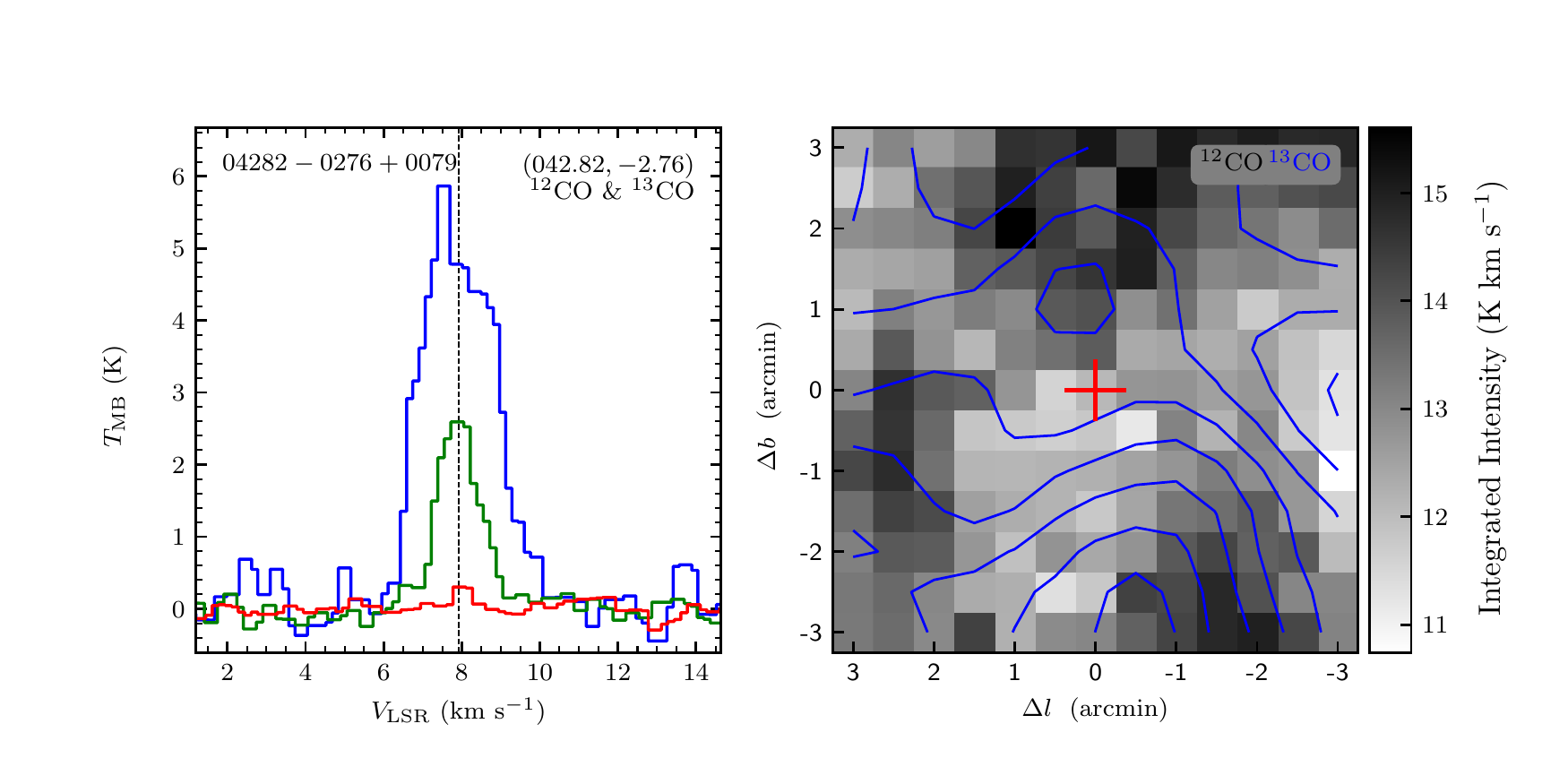}
\includegraphics[width=9.0cm,angle=0]{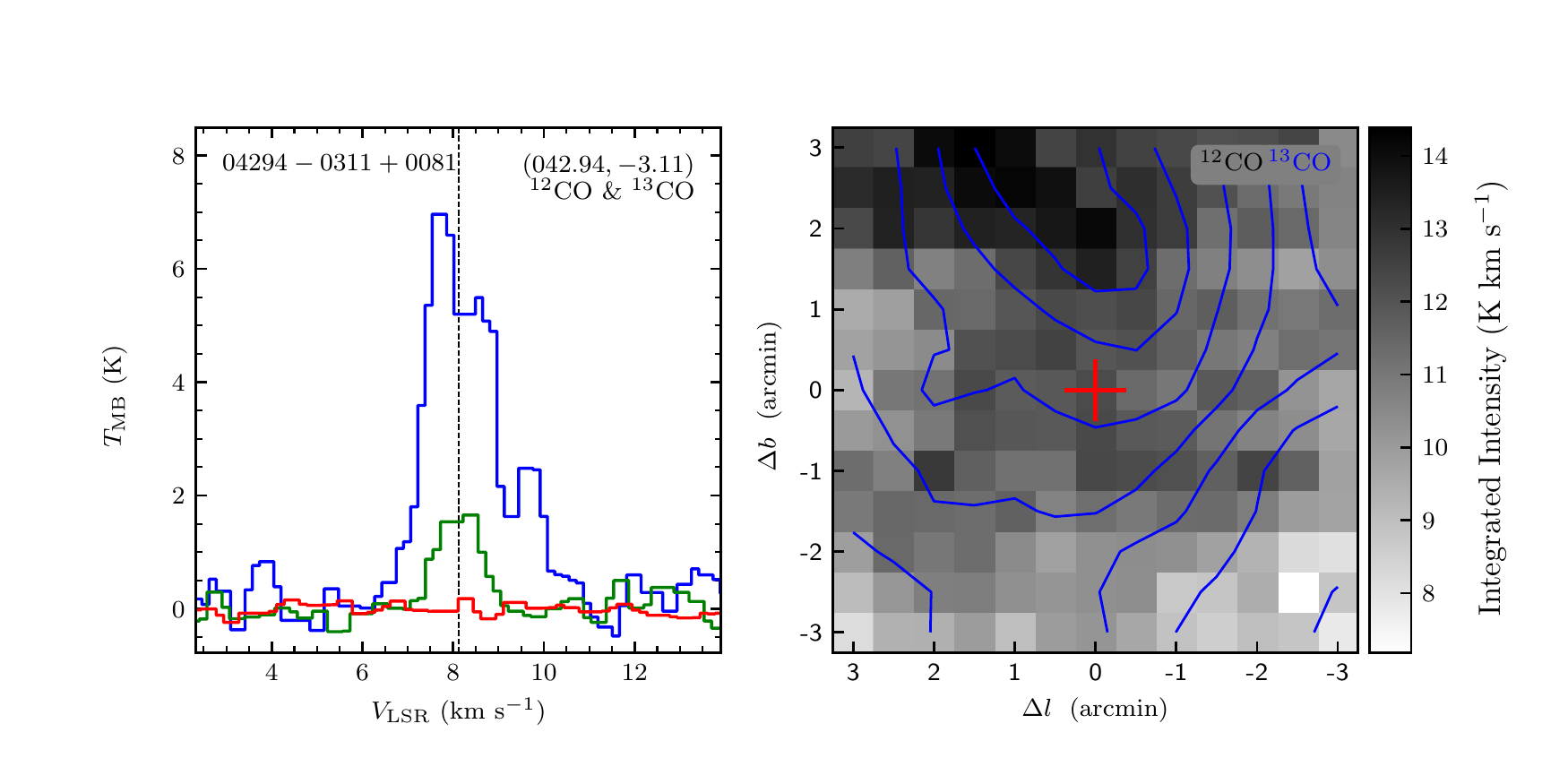}
\vspace{-0.5cm}

\includegraphics[width=9.0cm,angle=0]{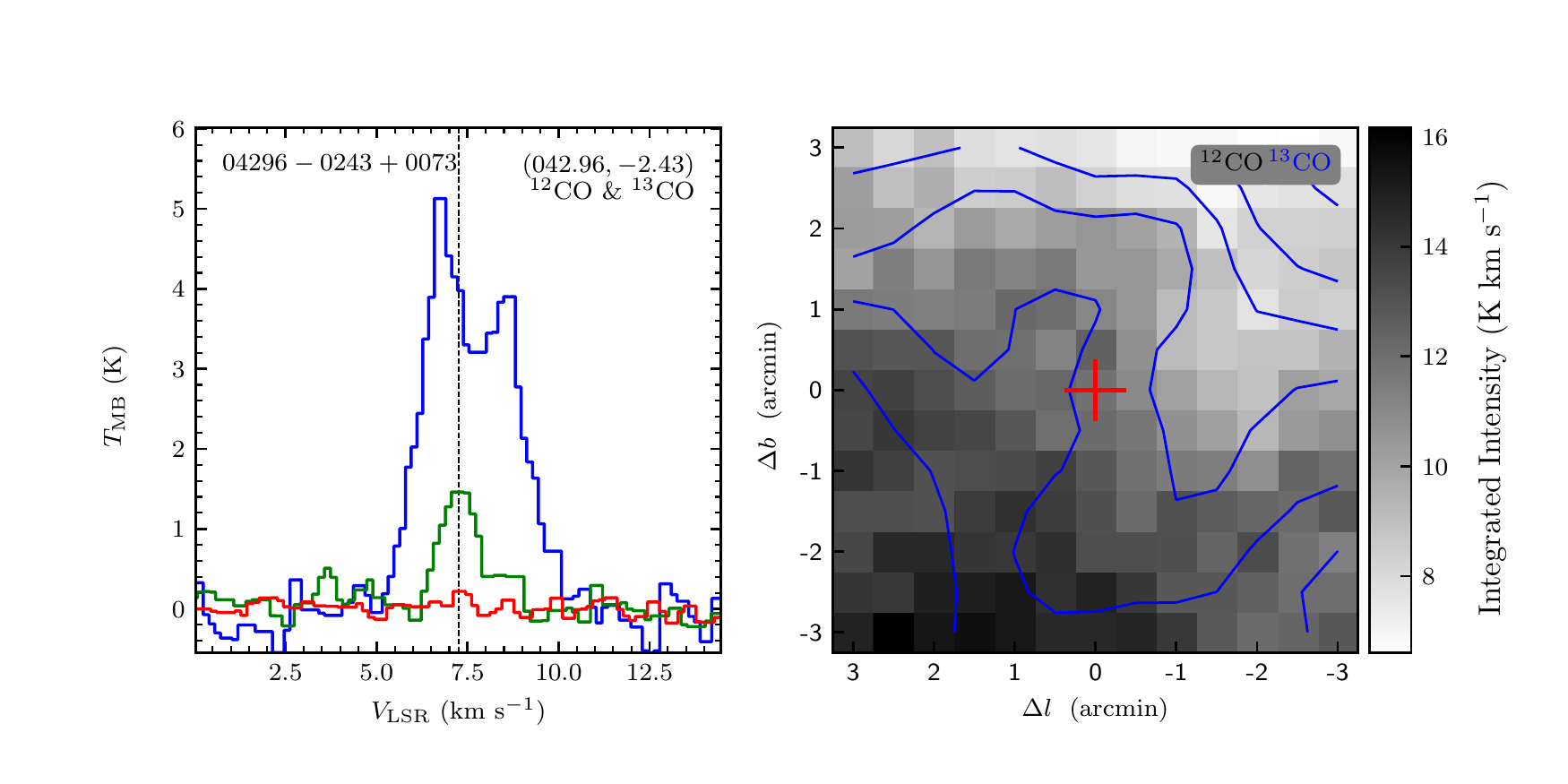}
\includegraphics[width=9.0cm,angle=0]{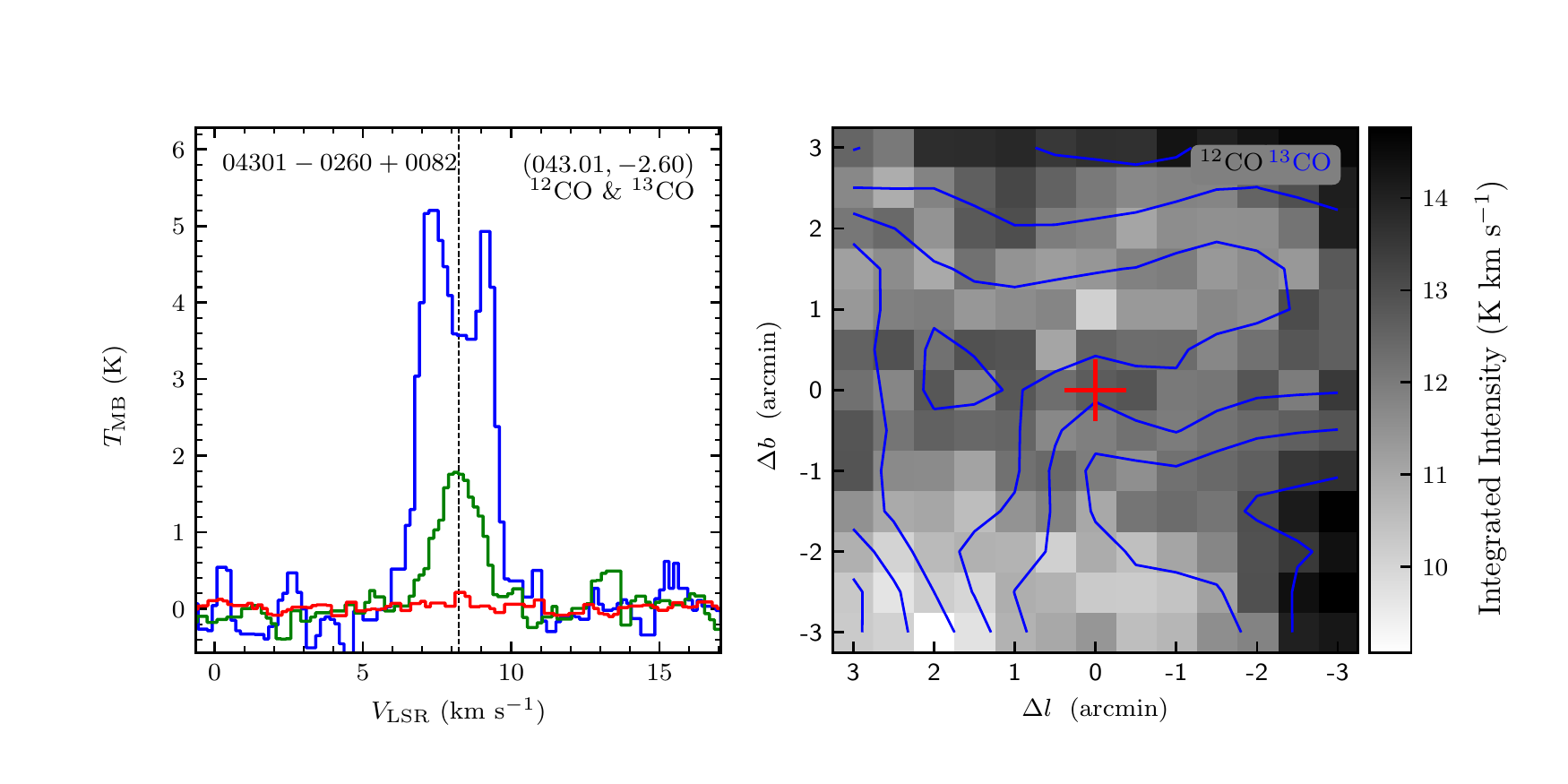}
\vspace{-0.5cm}

\includegraphics[width=9.0cm,angle=0]{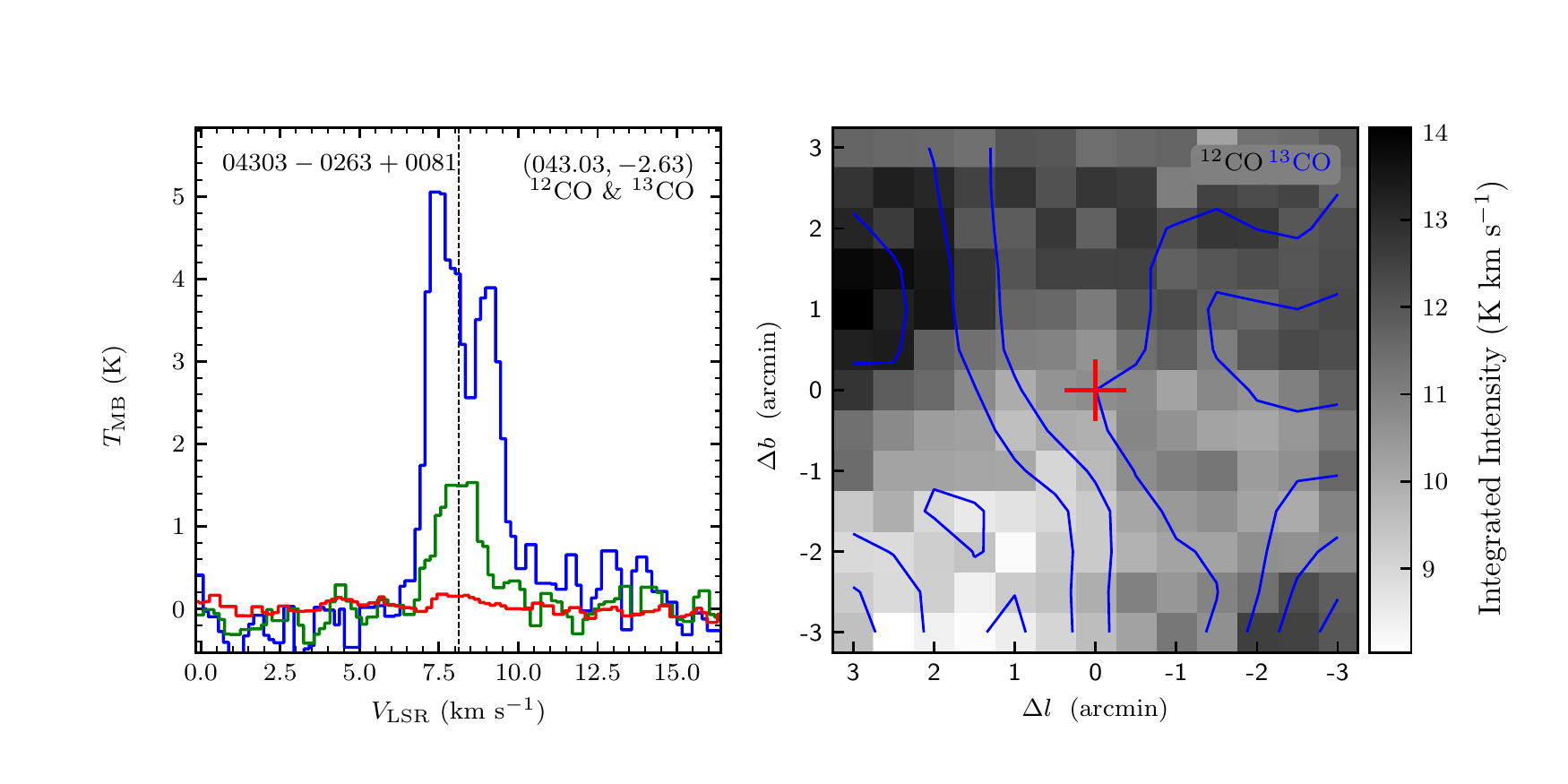}
\includegraphics[width=9.0cm,angle=0]{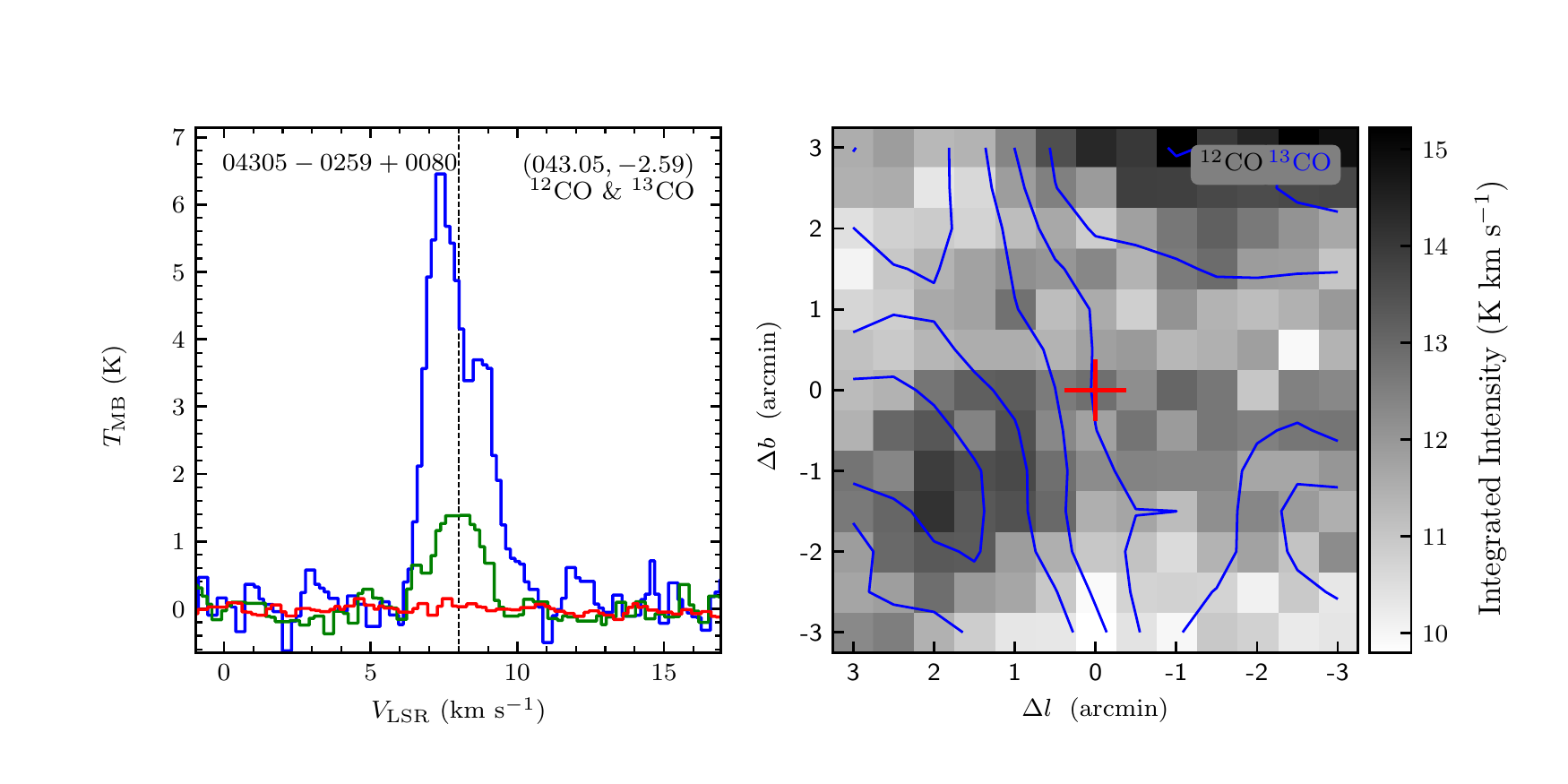}
\vspace{-0.5cm}

\includegraphics[width=9.0cm,angle=0]{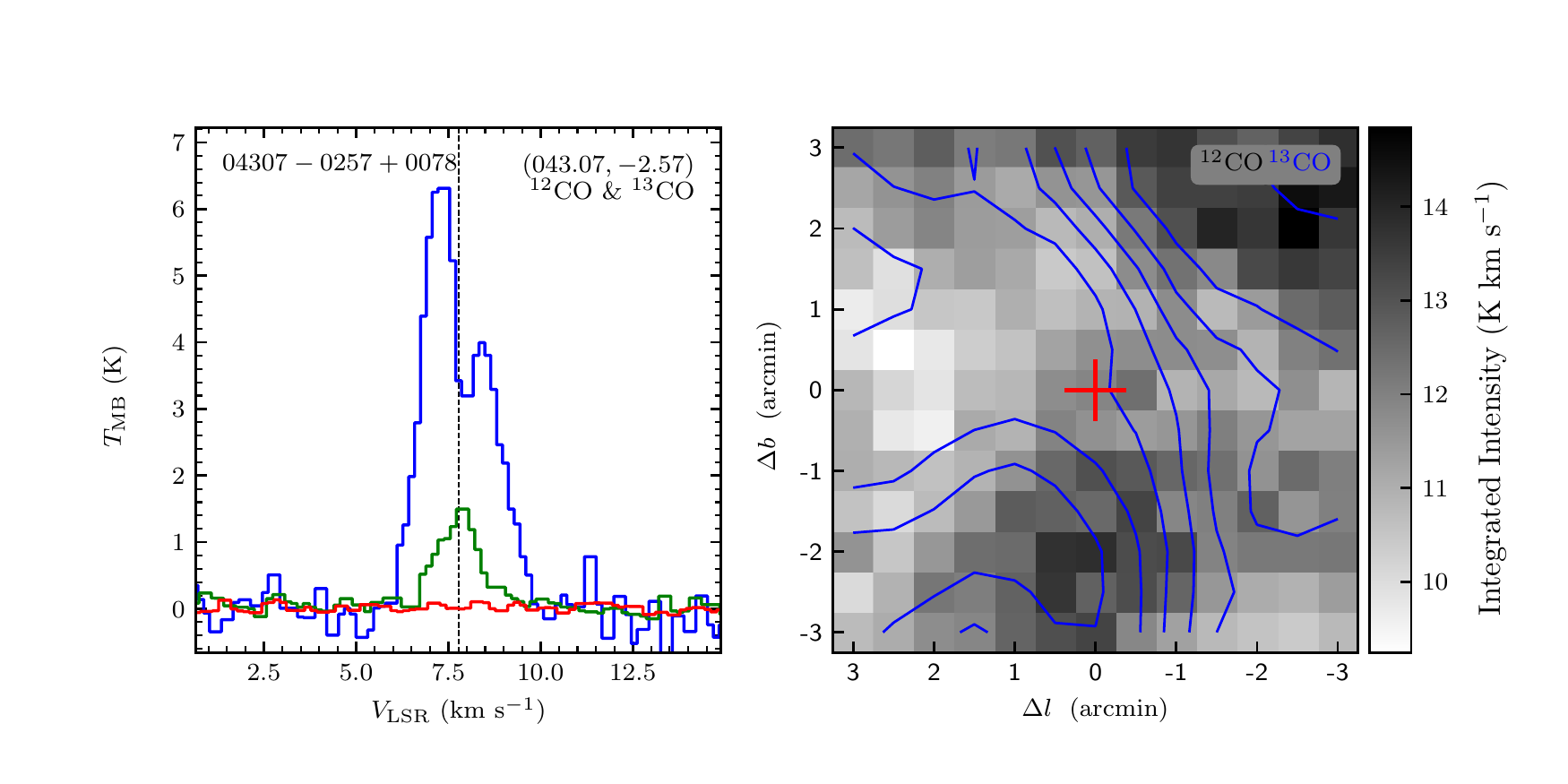}
\includegraphics[width=9.0cm,angle=0]{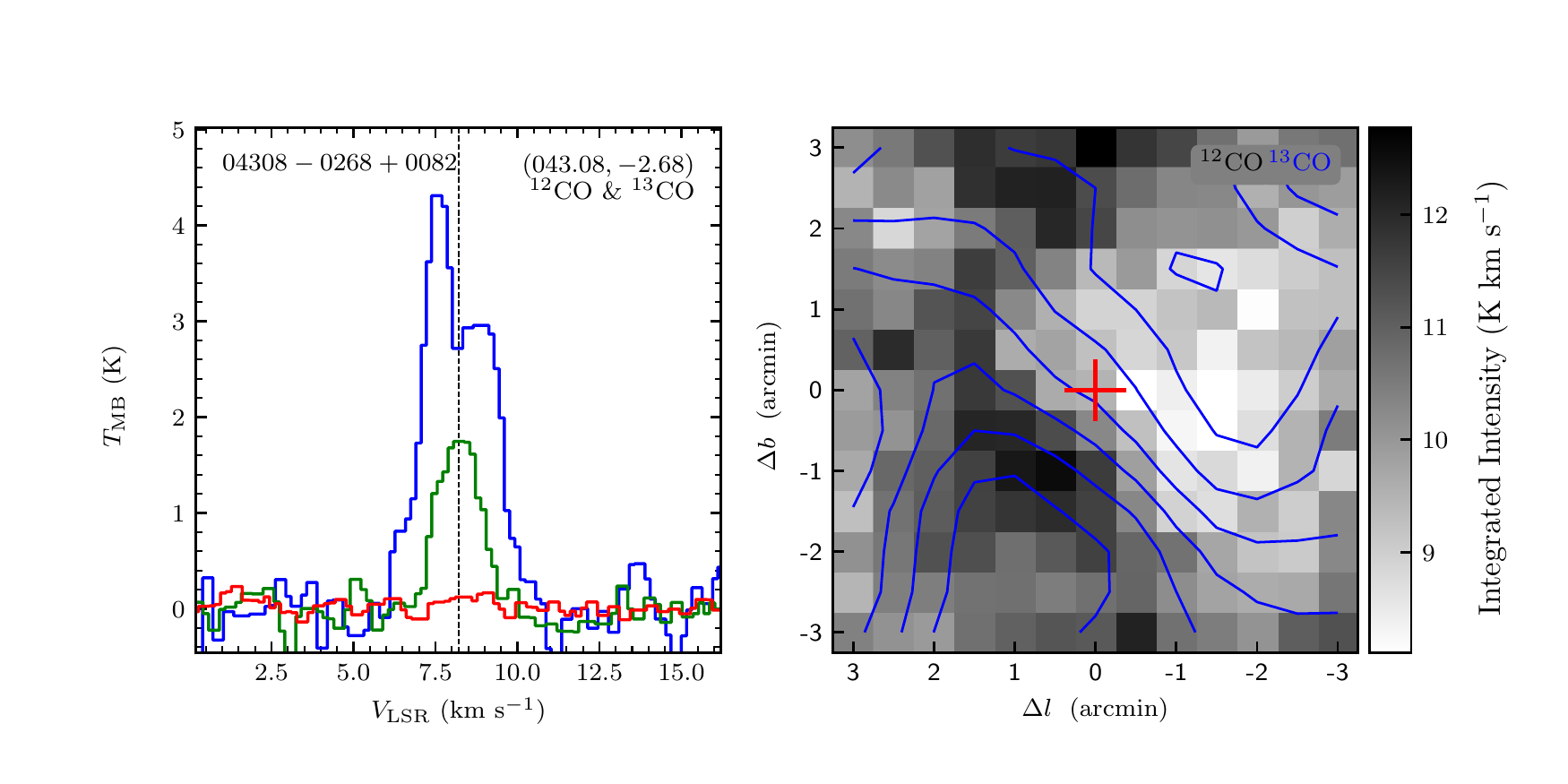}
\end{figure}
\clearpage

\begin{figure}
\includegraphics[width=9.0cm,angle=0]{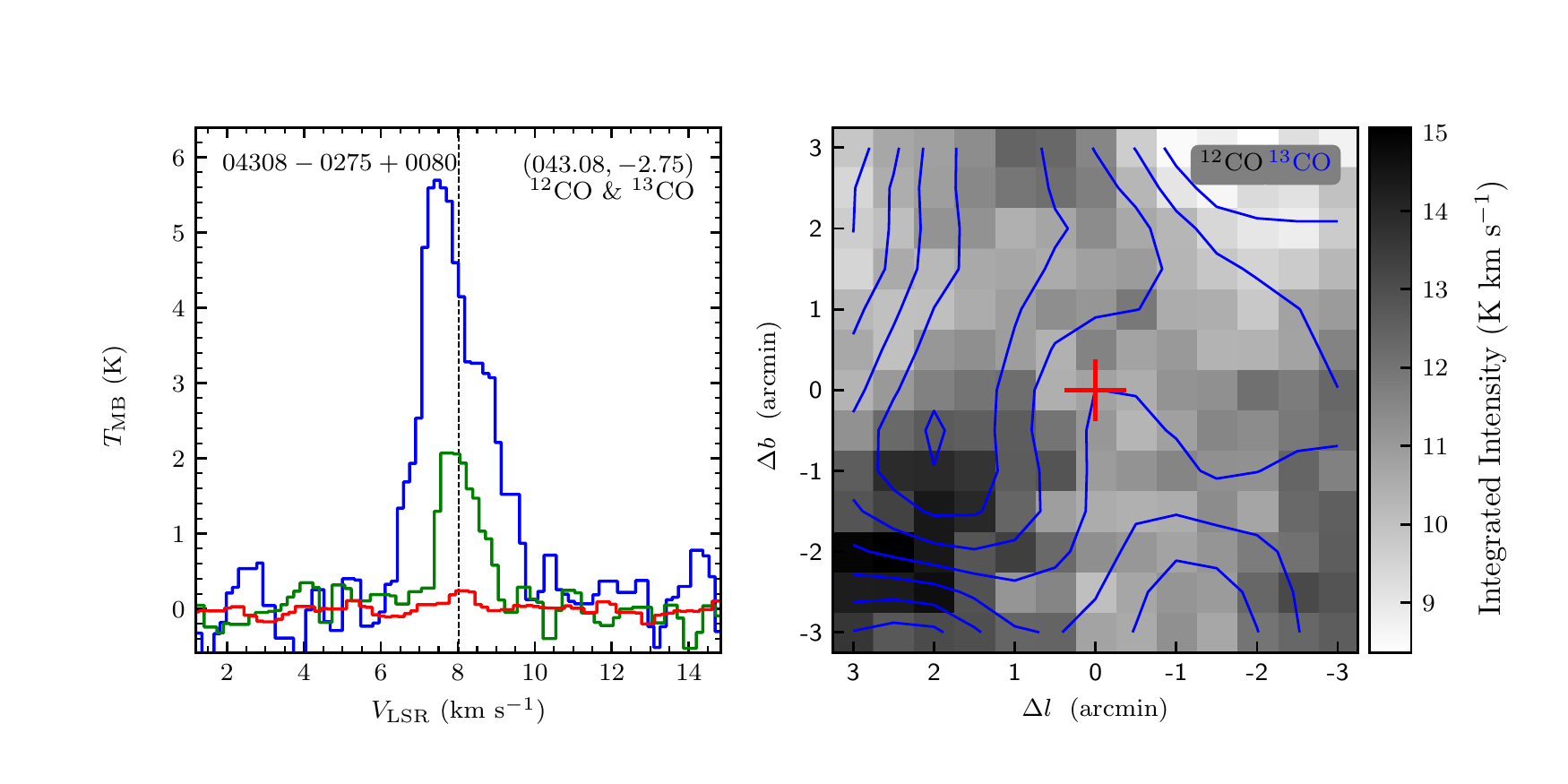}
\includegraphics[width=9.0cm,angle=0]{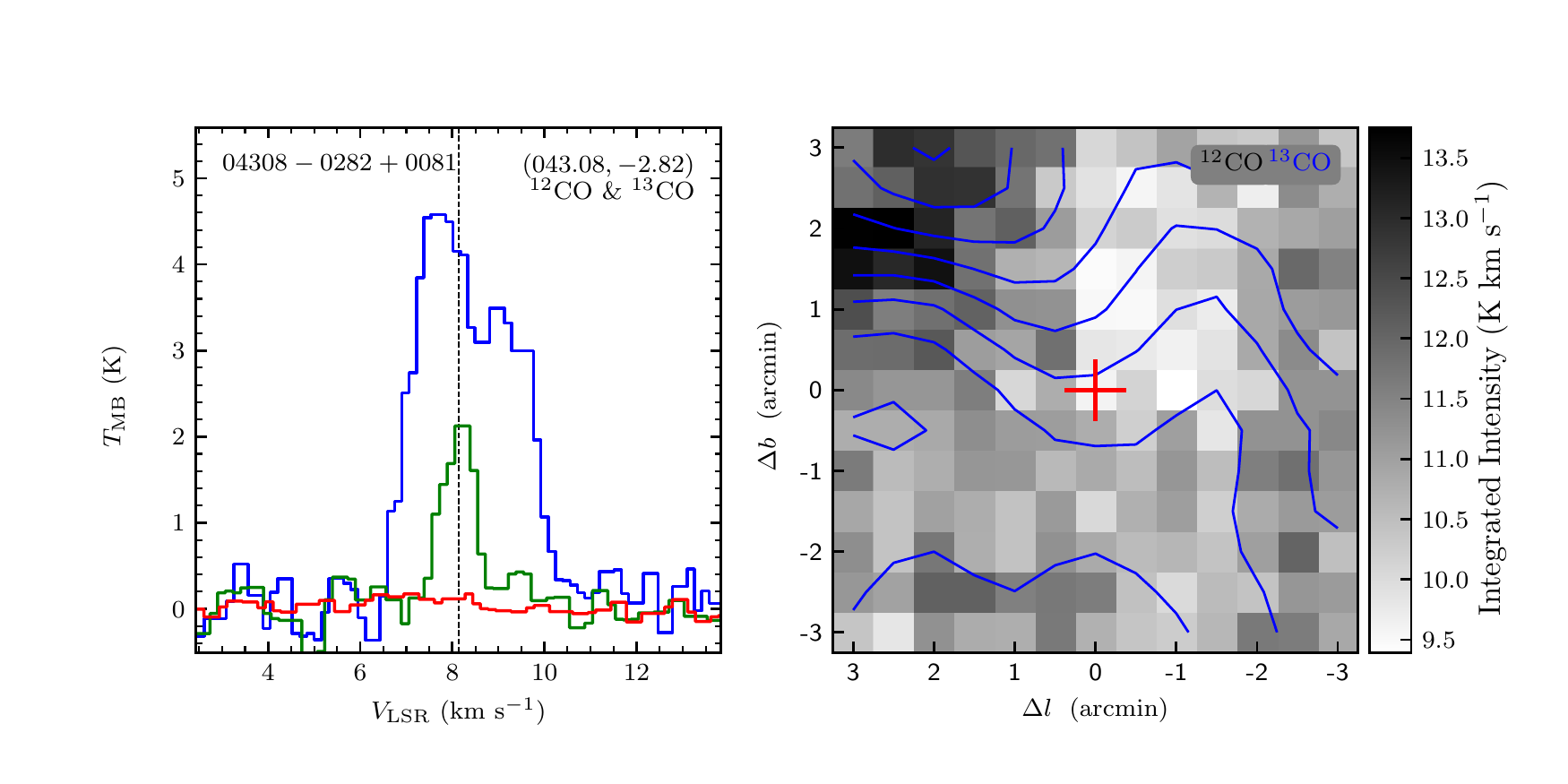}
\vspace{-0.5cm}

\includegraphics[width=9.0cm,angle=0]{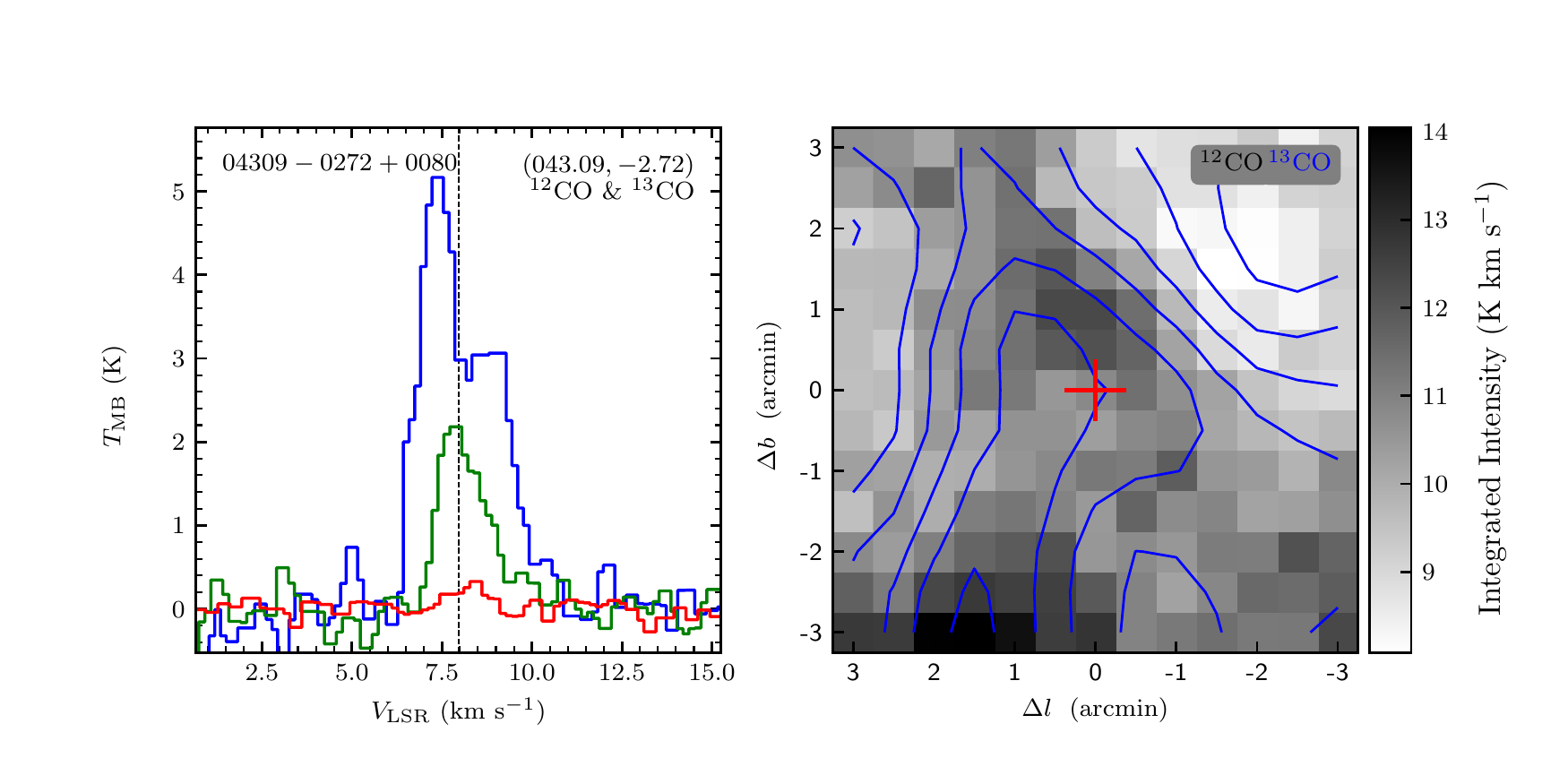}
\includegraphics[width=9.0cm,angle=0]{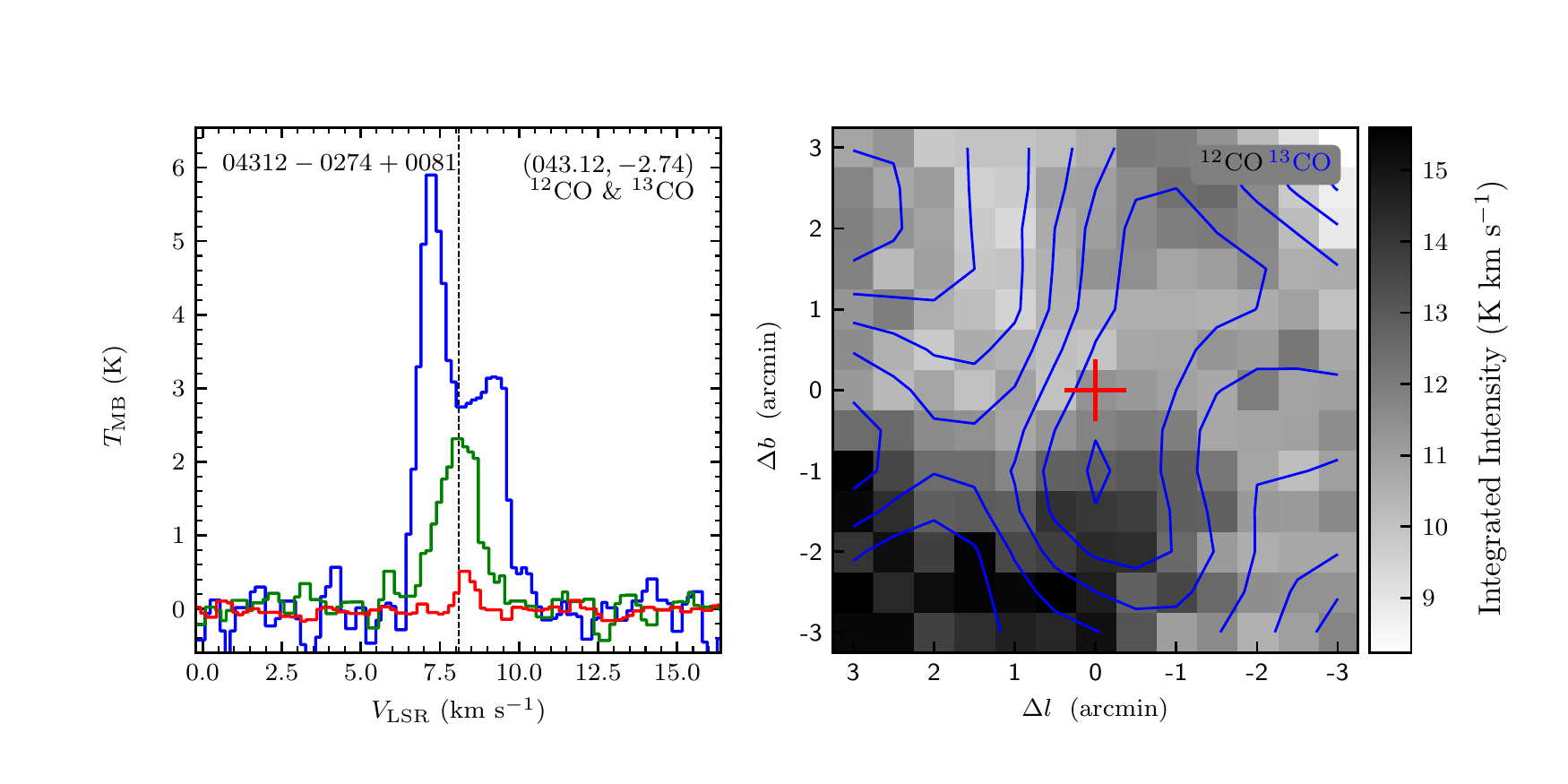}
\vspace{-0.5cm}

\includegraphics[width=9.0cm,angle=0]{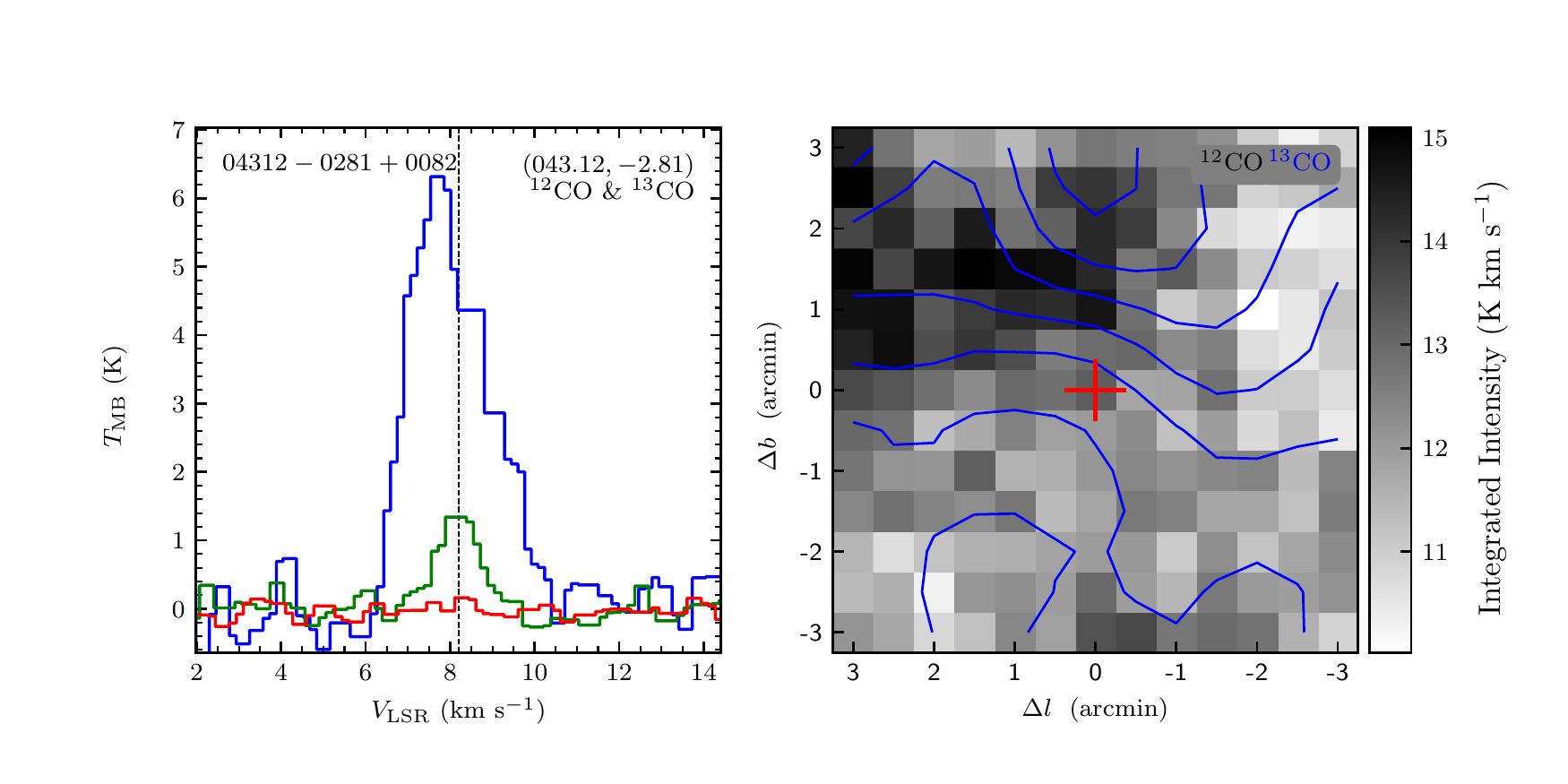}
\includegraphics[width=9.0cm,angle=0]{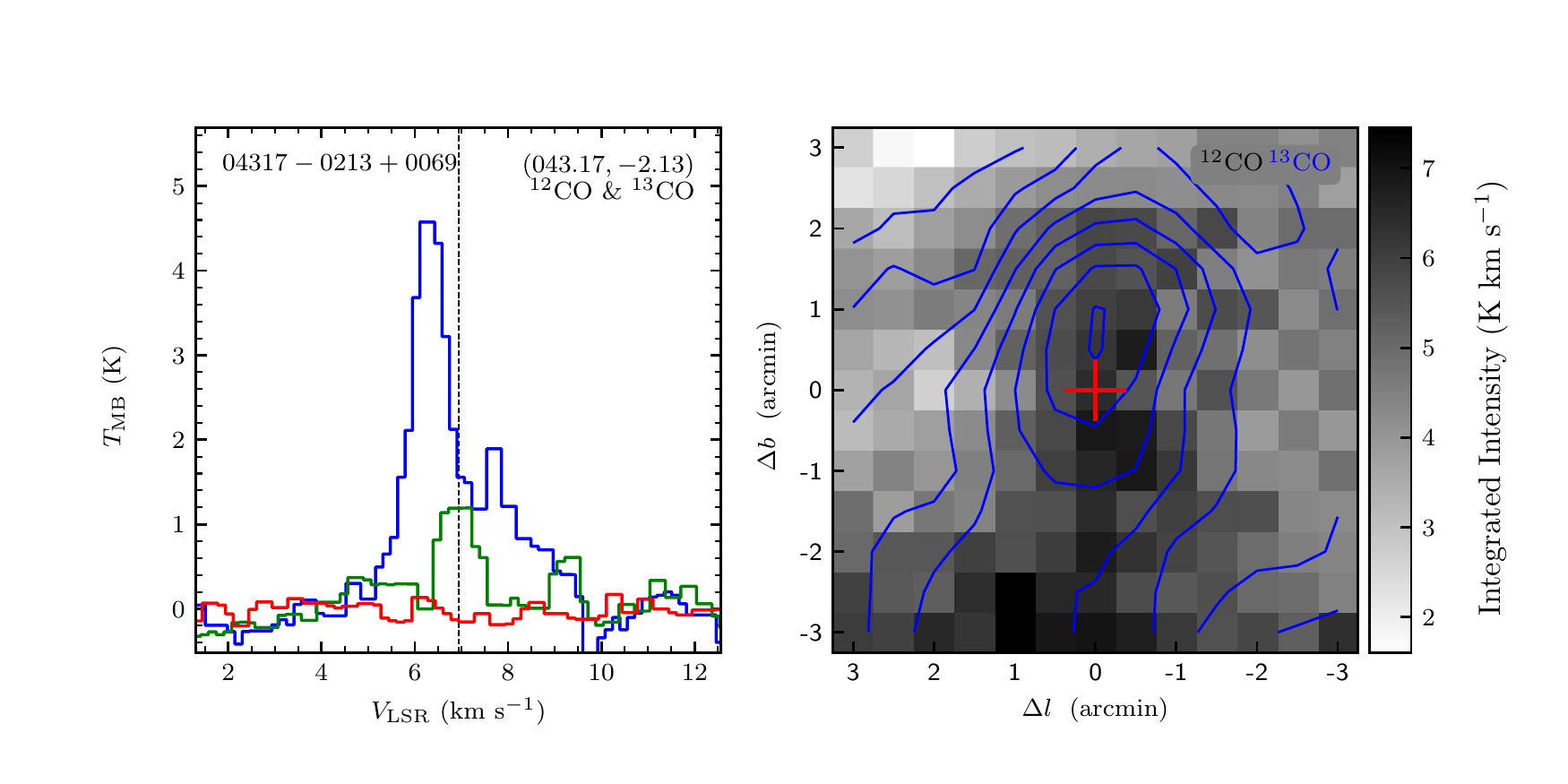}
\vspace{-0.5cm}

\includegraphics[width=9.0cm,angle=0]{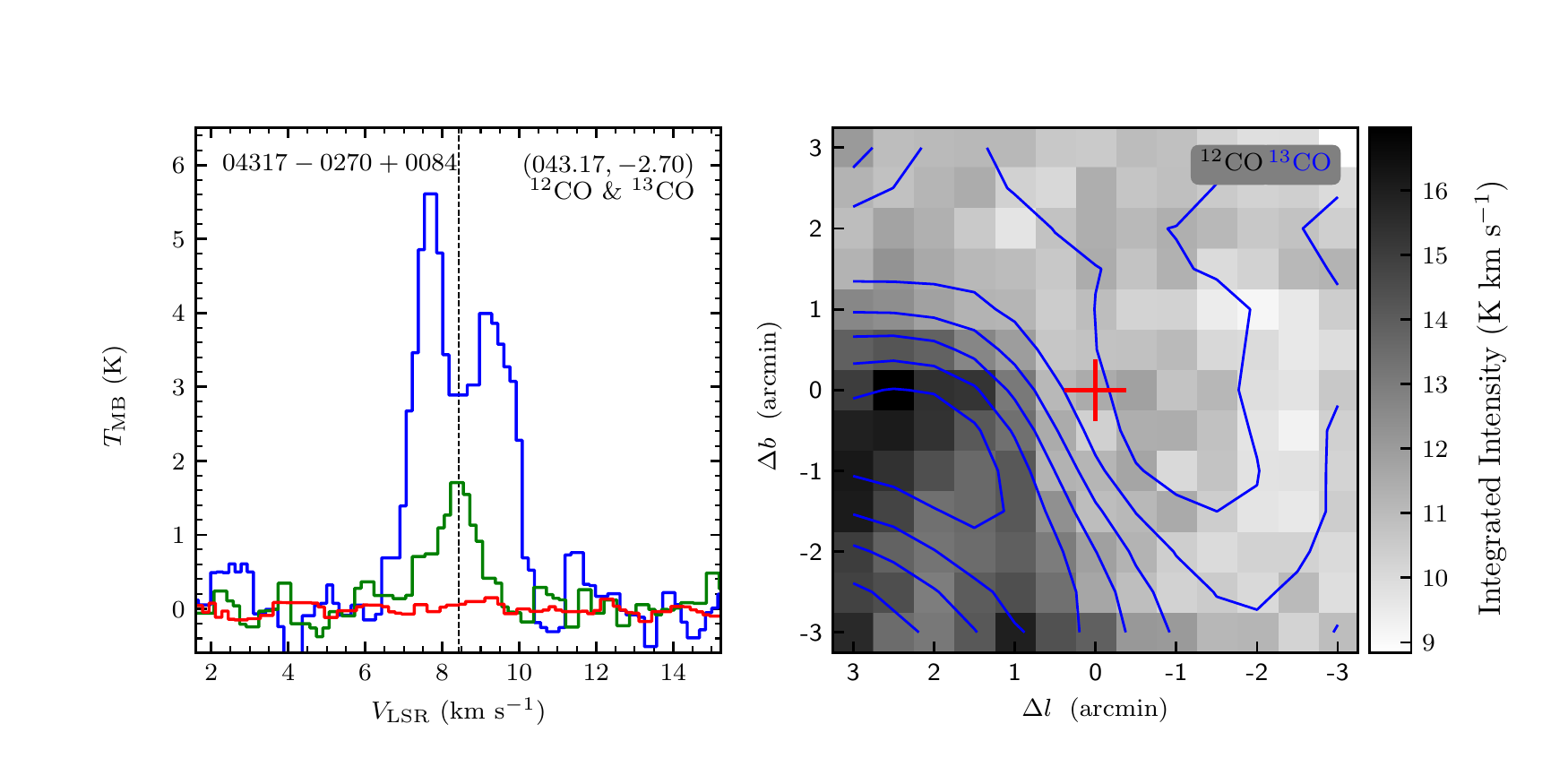}
\includegraphics[width=9.0cm,angle=0]{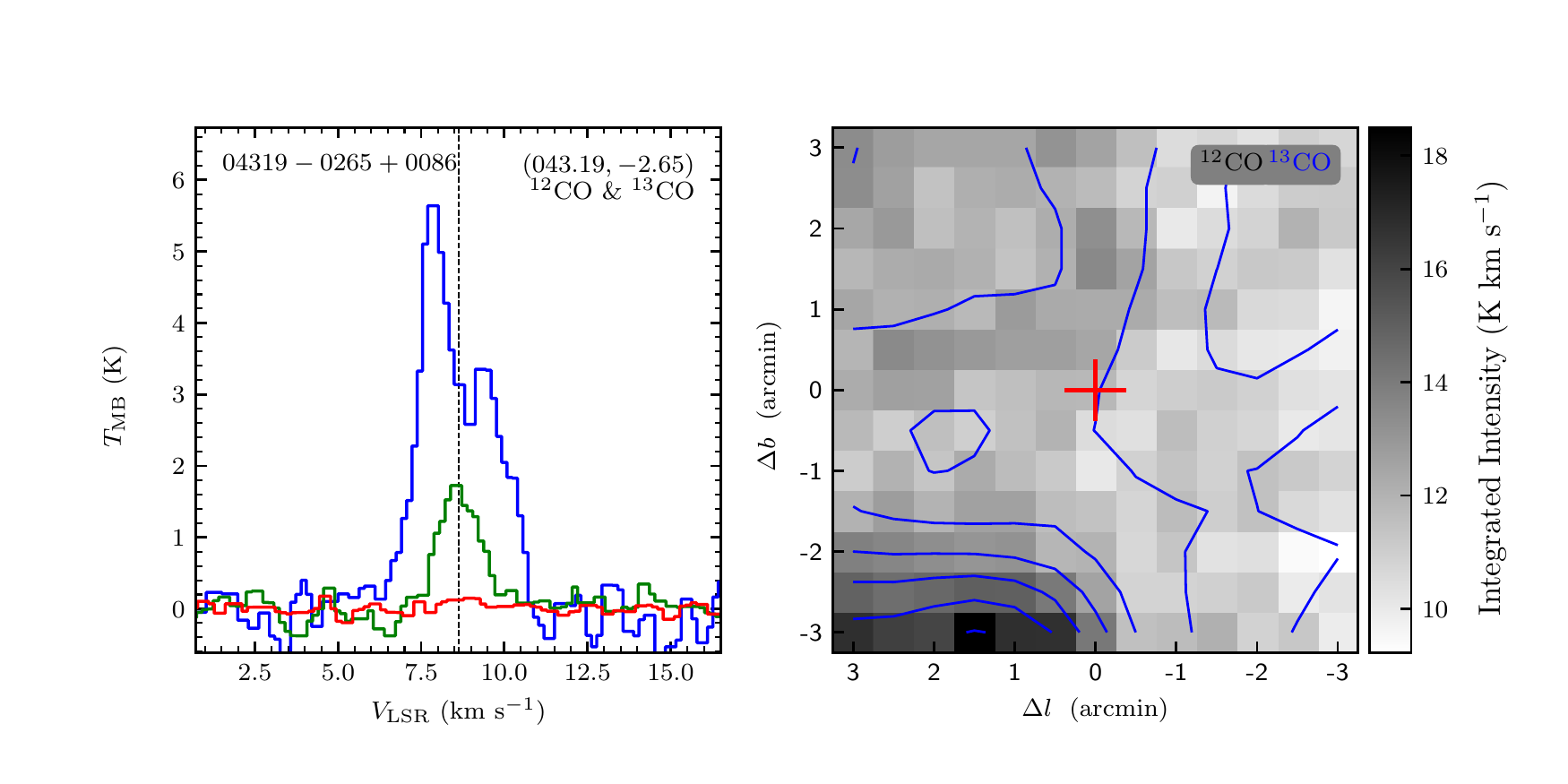}
\vspace{-0.5cm}

\includegraphics[width=9.0cm,angle=0]{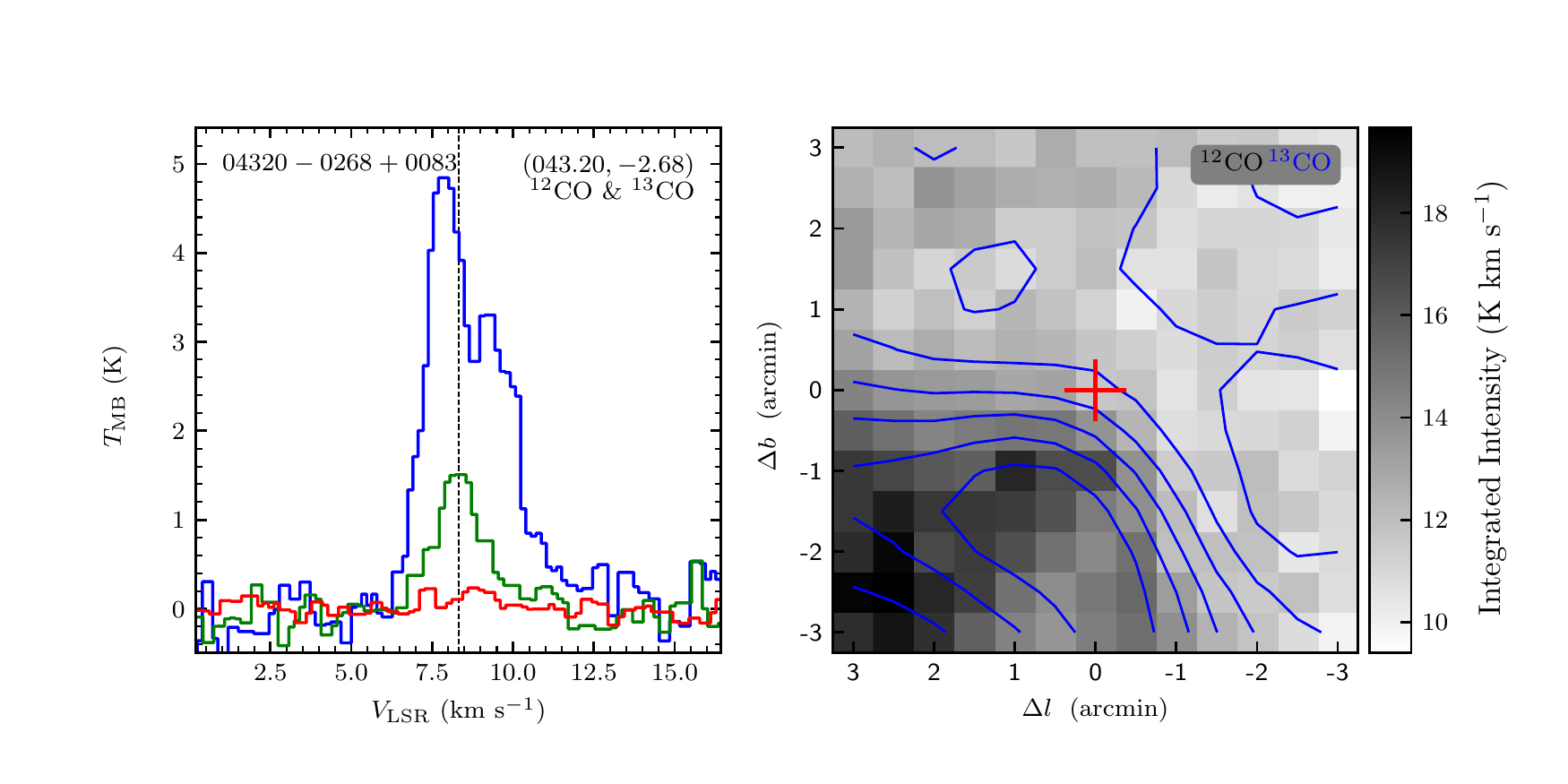}
\includegraphics[width=9.0cm,angle=0]{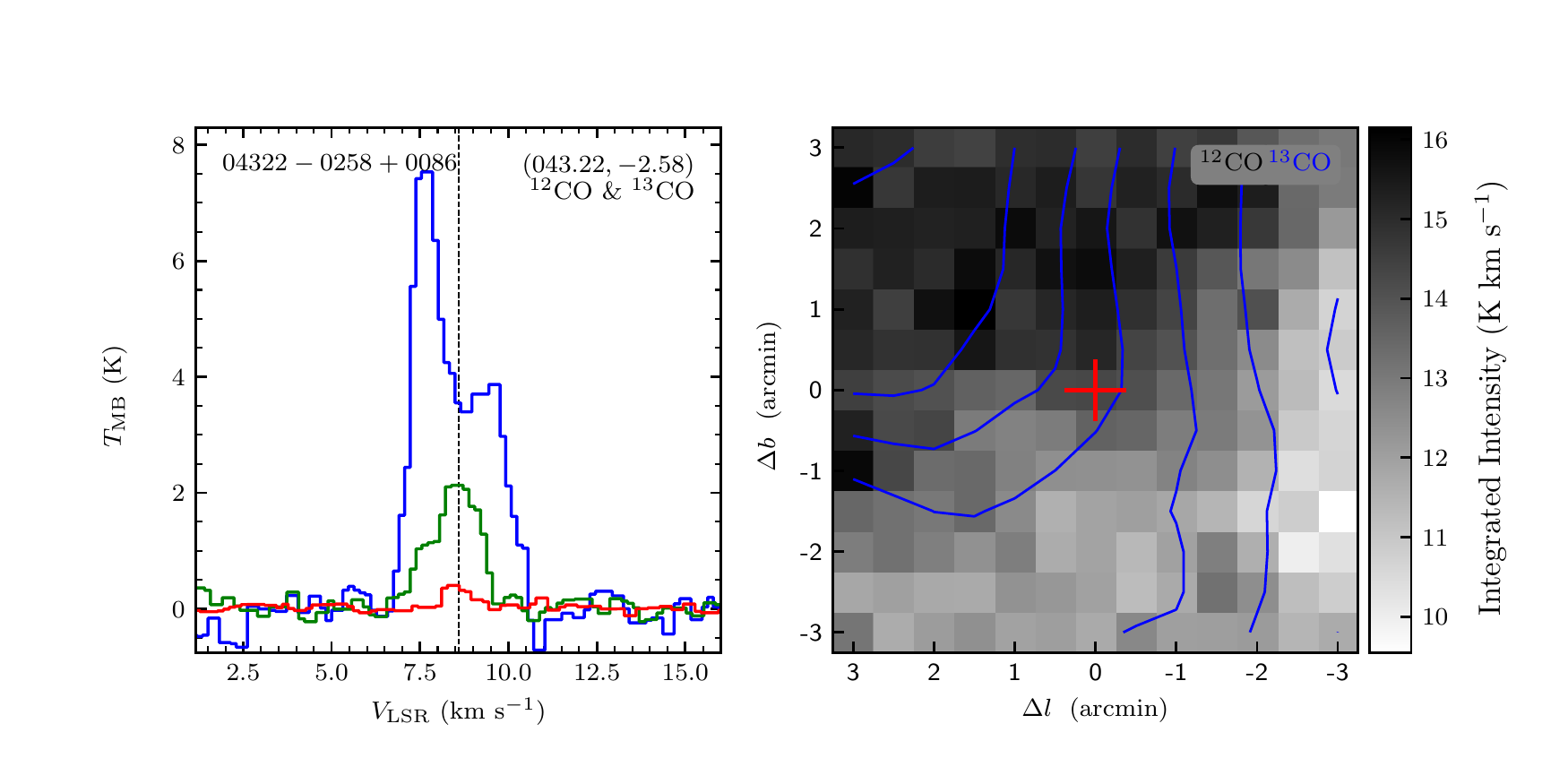}
\end{figure}
\clearpage

\begin{figure}
\includegraphics[width=9.0cm,angle=0]{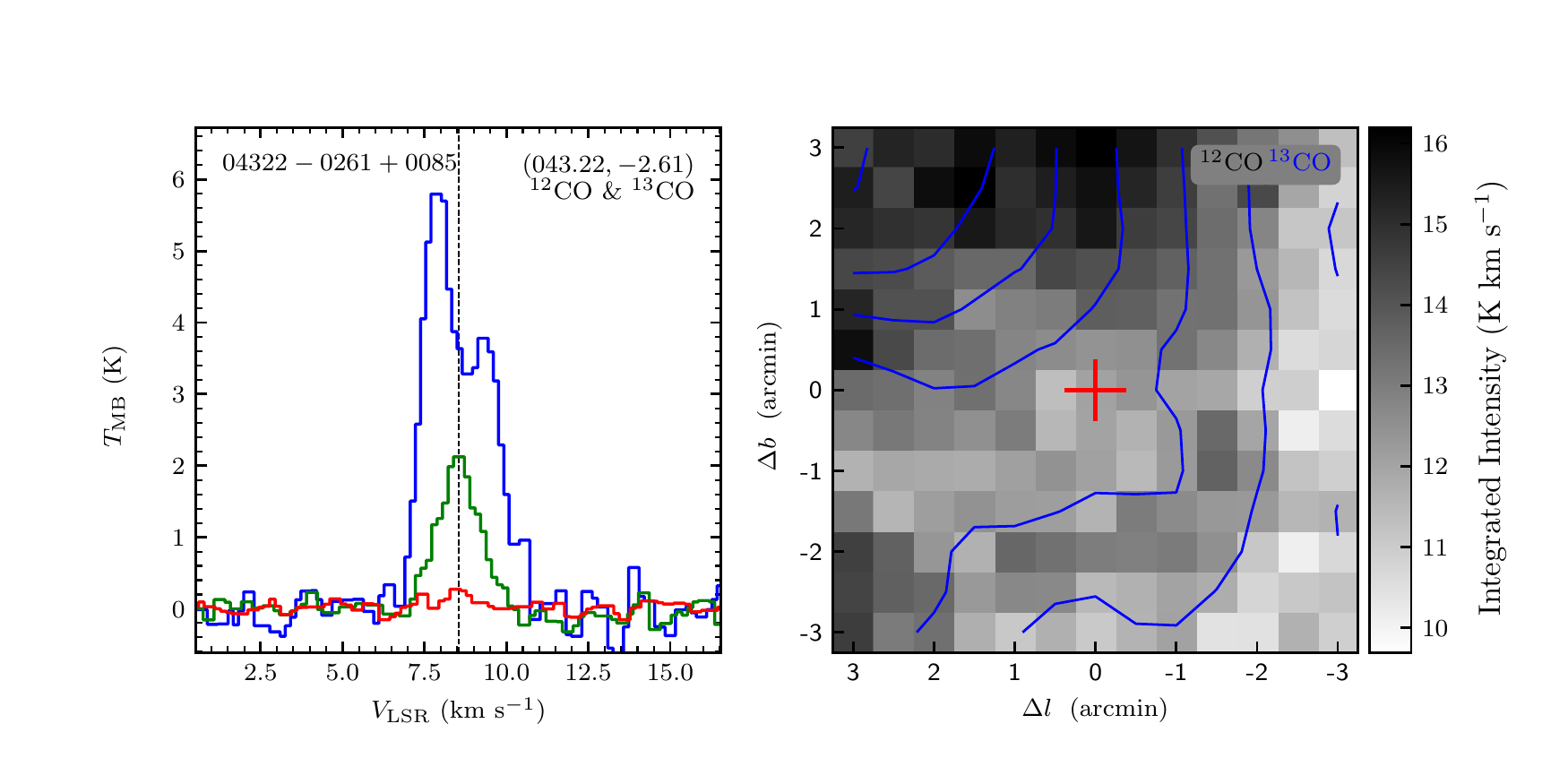}
\includegraphics[width=9.0cm,angle=0]{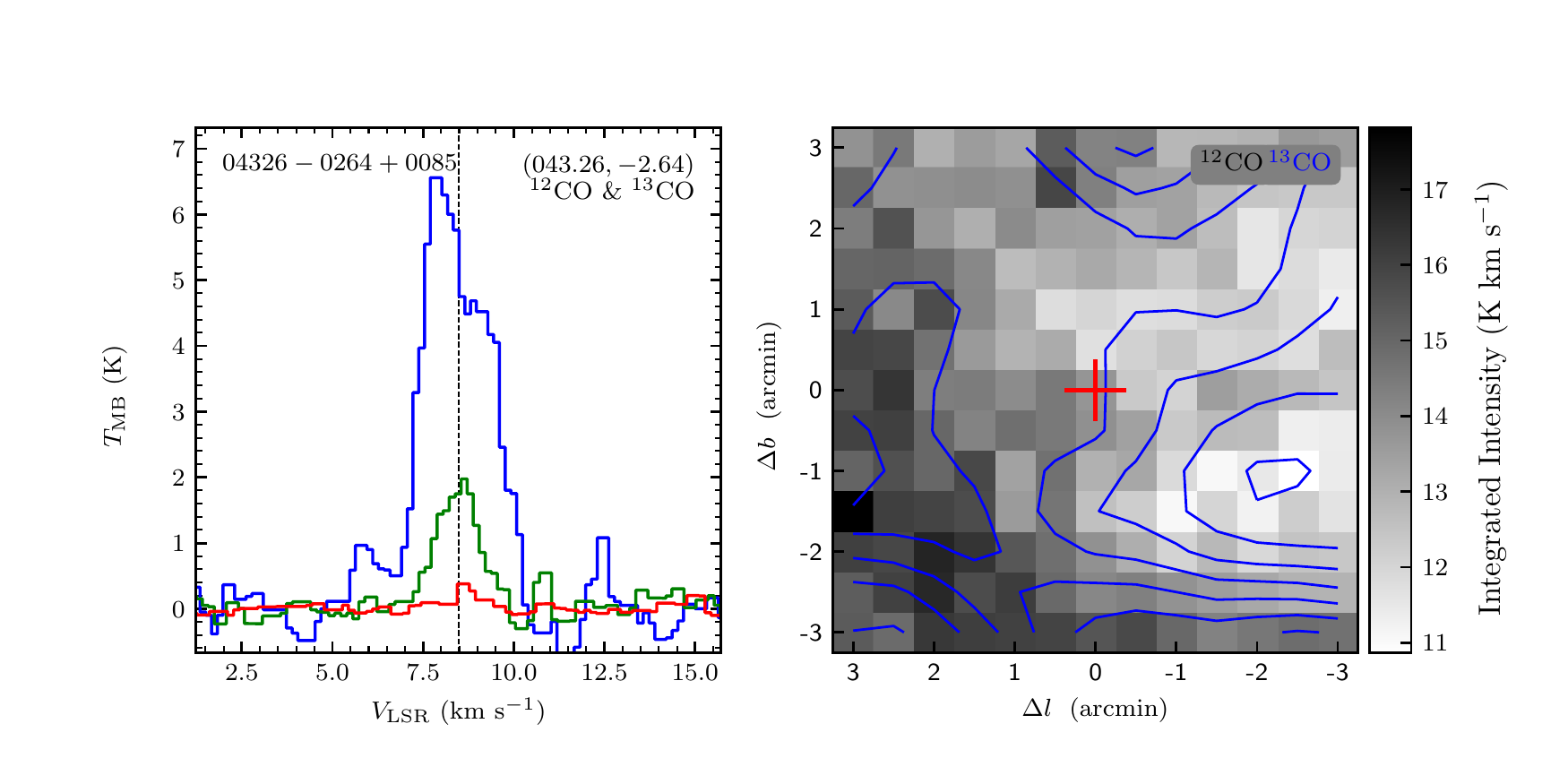}
\vspace{-0.5cm}

\includegraphics[width=9.0cm,angle=0]{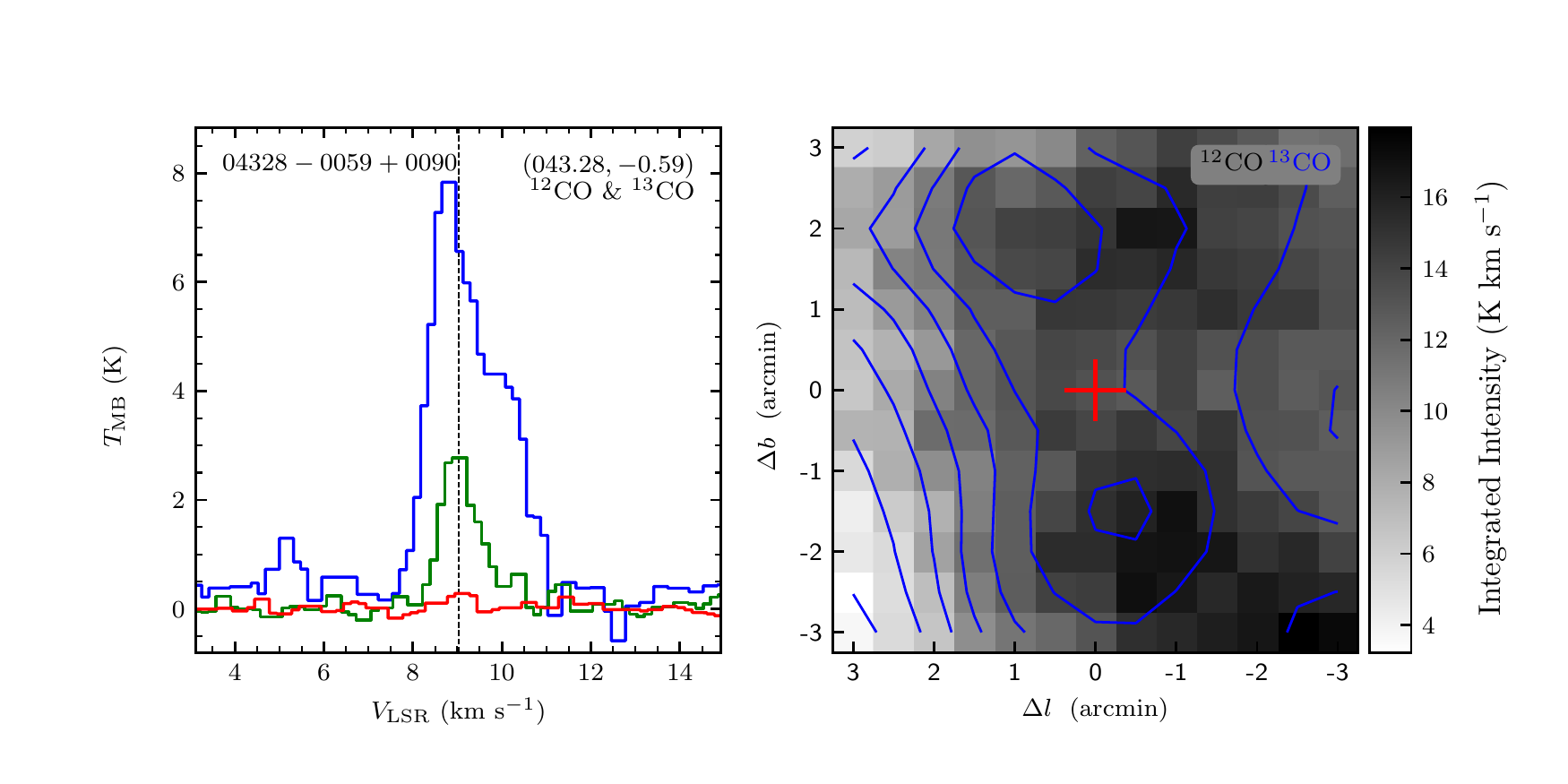}
\includegraphics[width=9.0cm,angle=0]{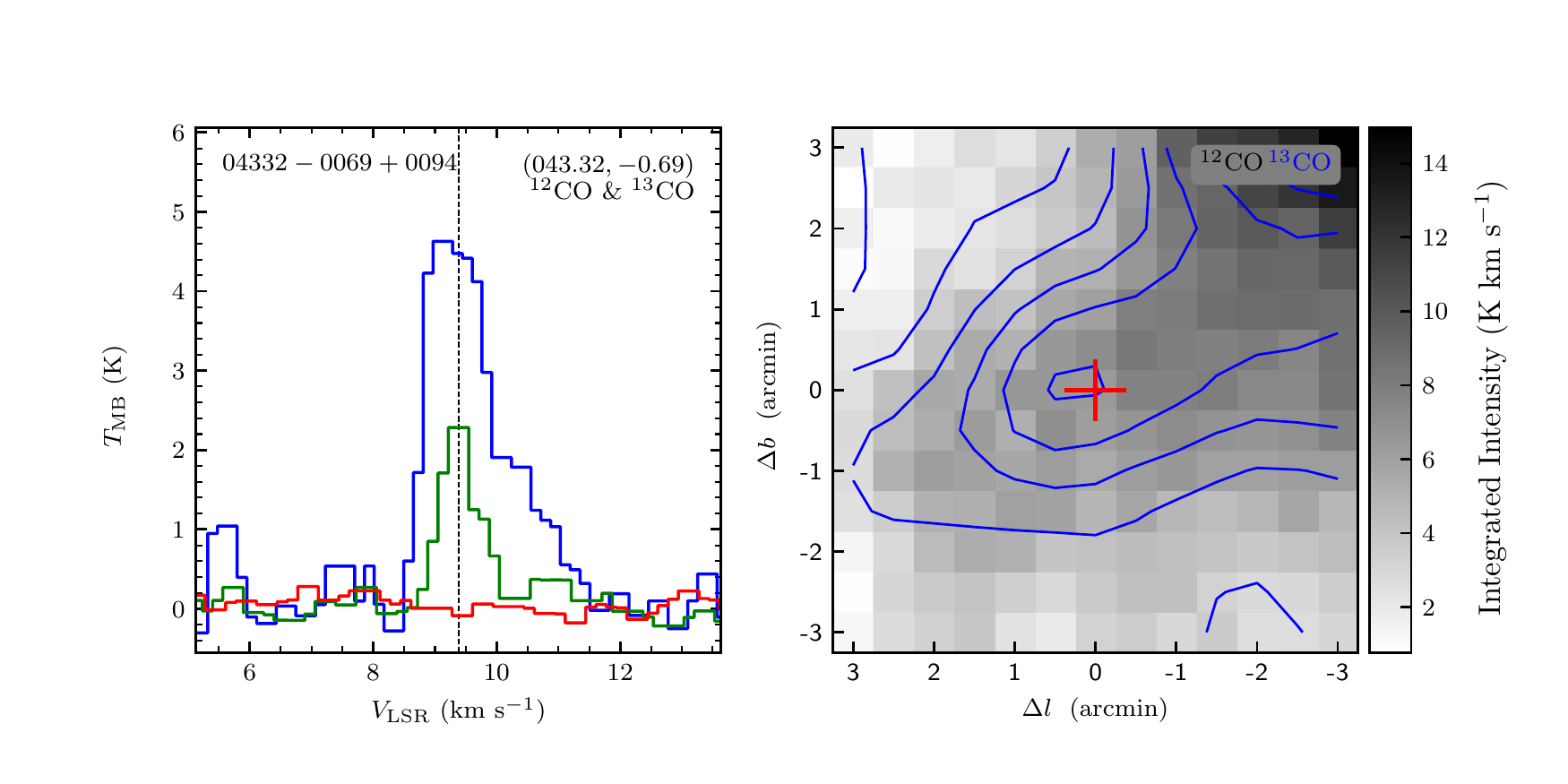}
\vspace{-0.5cm}

\includegraphics[width=9.0cm,angle=0]{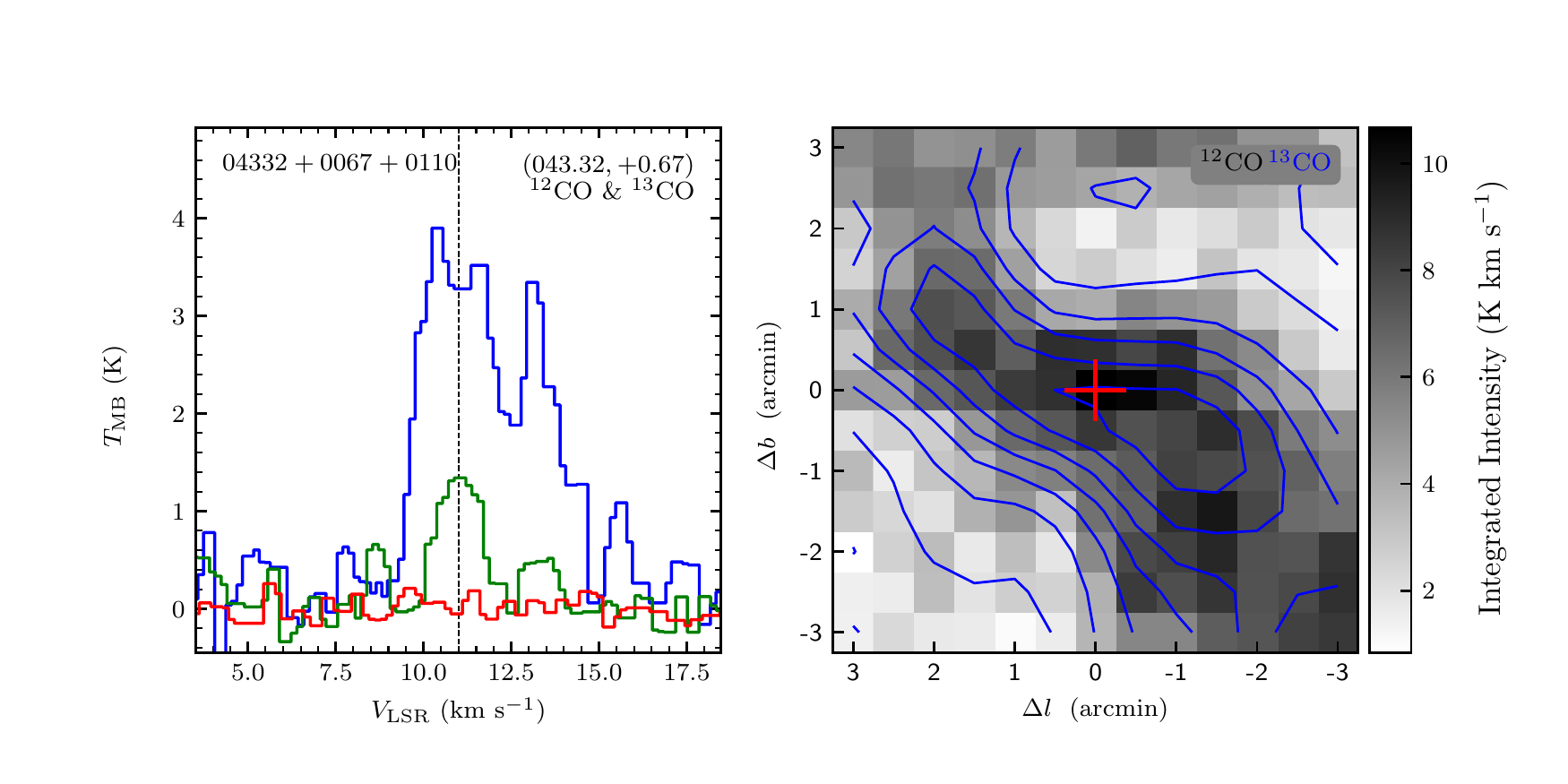}
\includegraphics[width=9.0cm,angle=0]{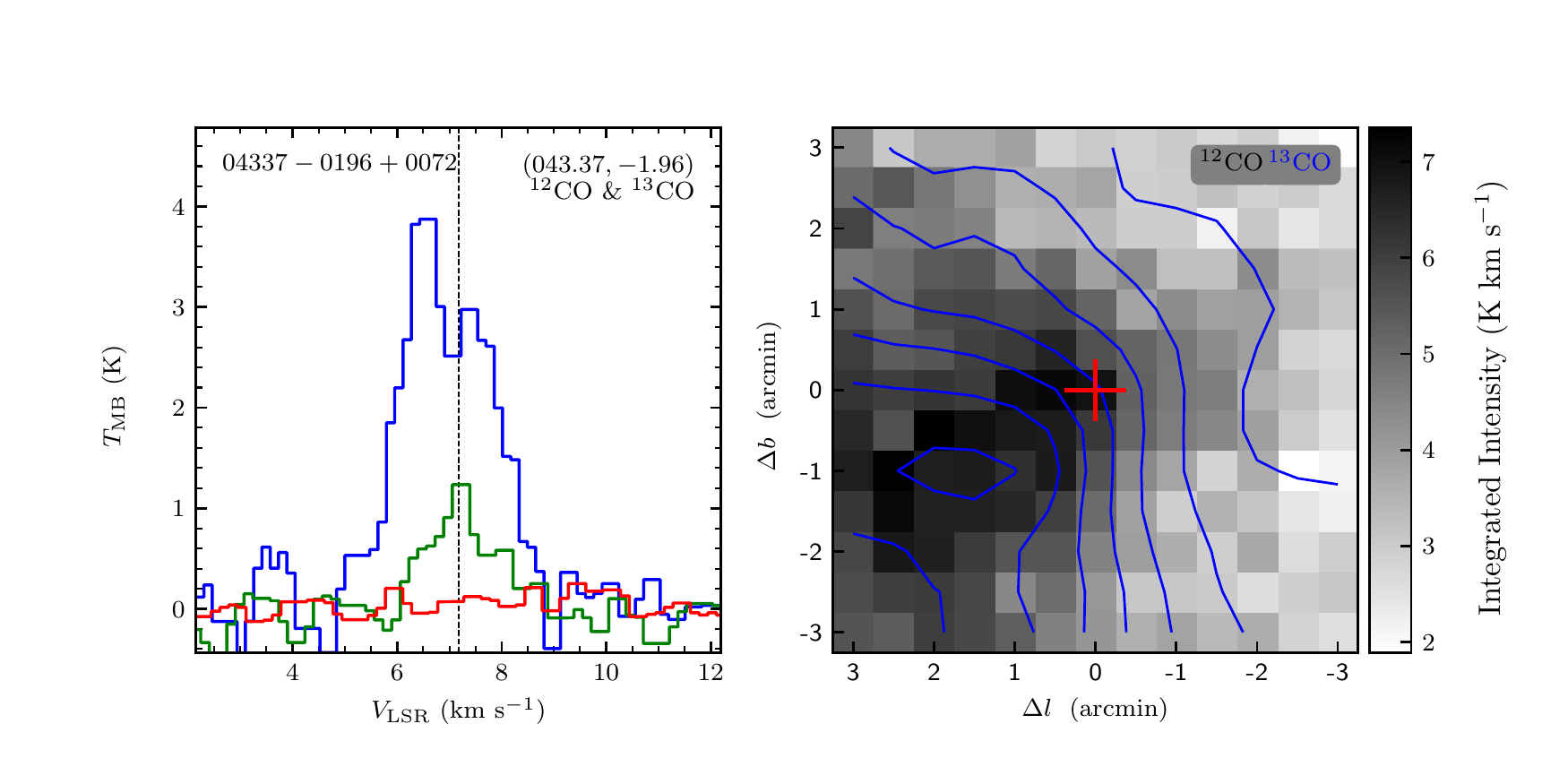}
\vspace{-0.5cm}

\includegraphics[width=9.0cm,angle=0]{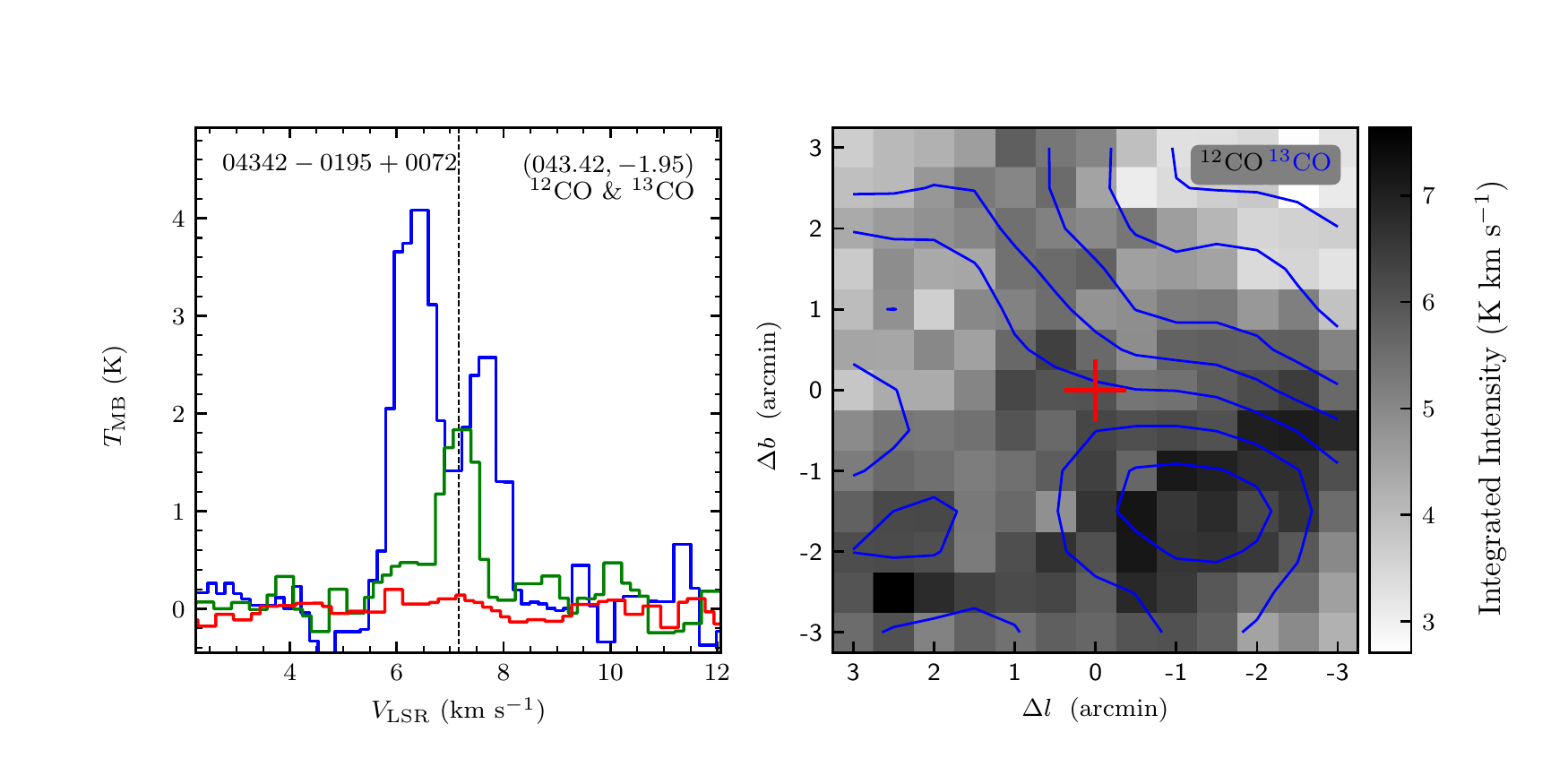}
\includegraphics[width=9.0cm,angle=0]{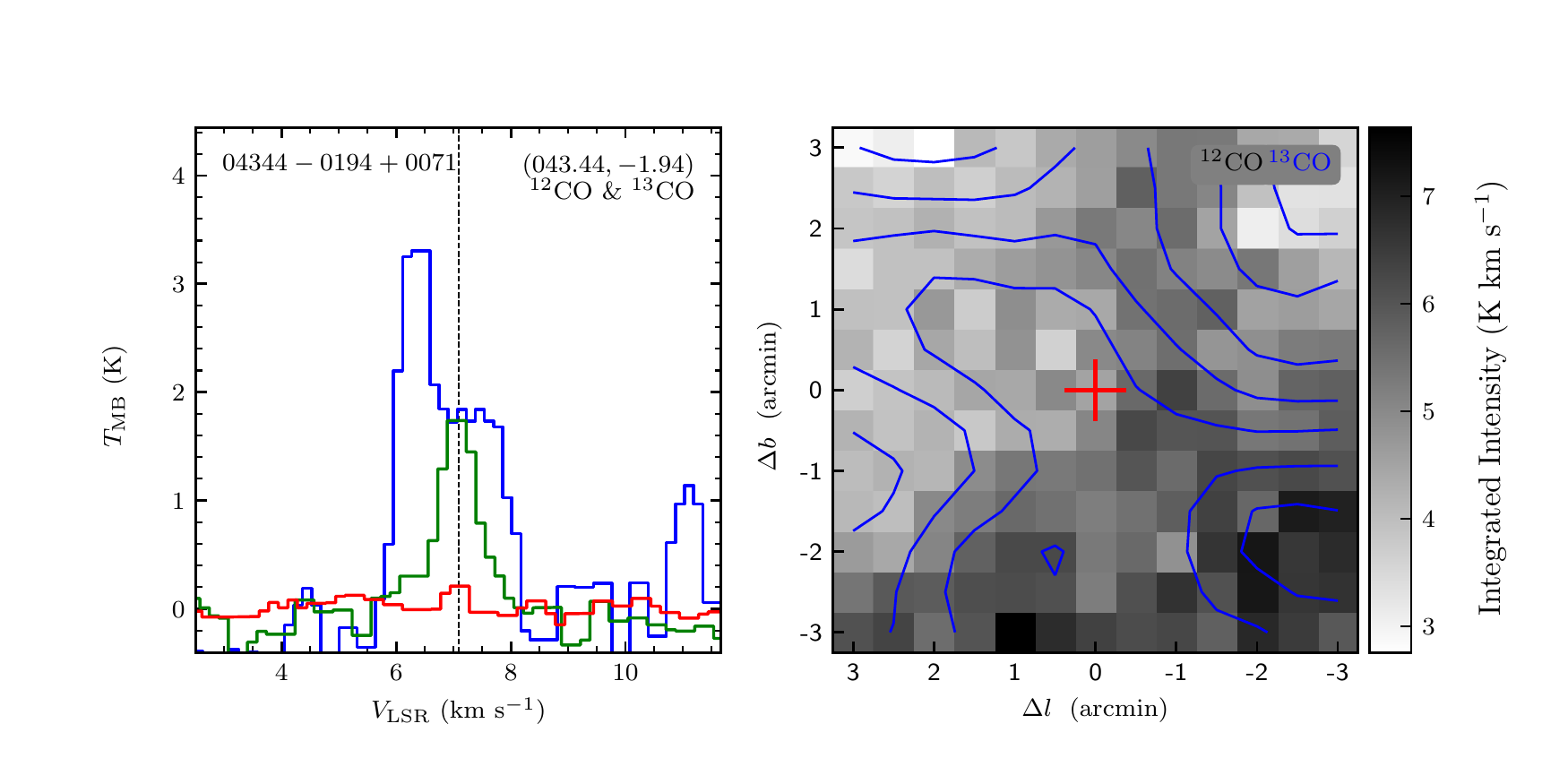}
\vspace{-0.5cm}

\includegraphics[width=9.0cm,angle=0]{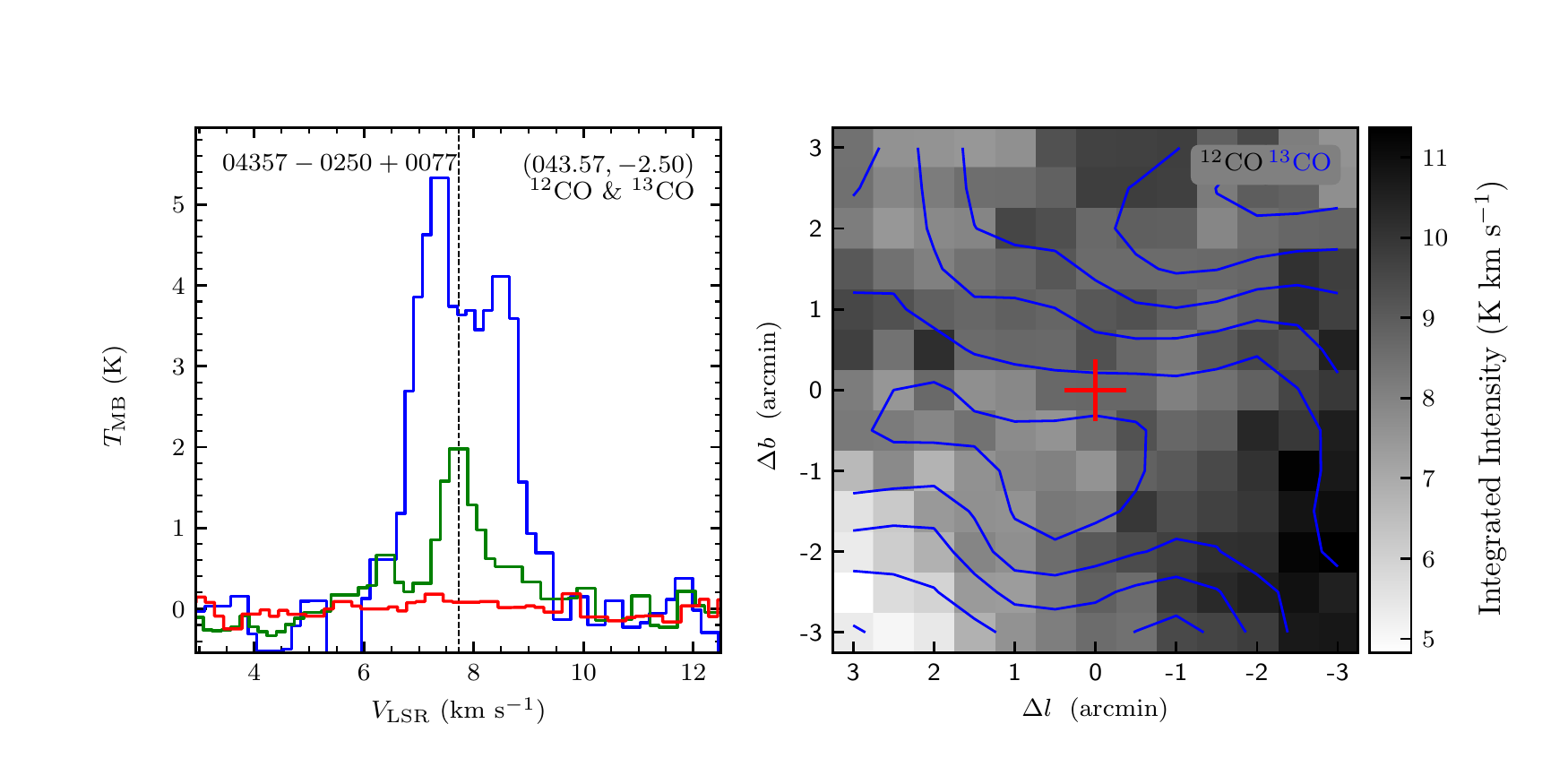}
\includegraphics[width=9.0cm,angle=0]{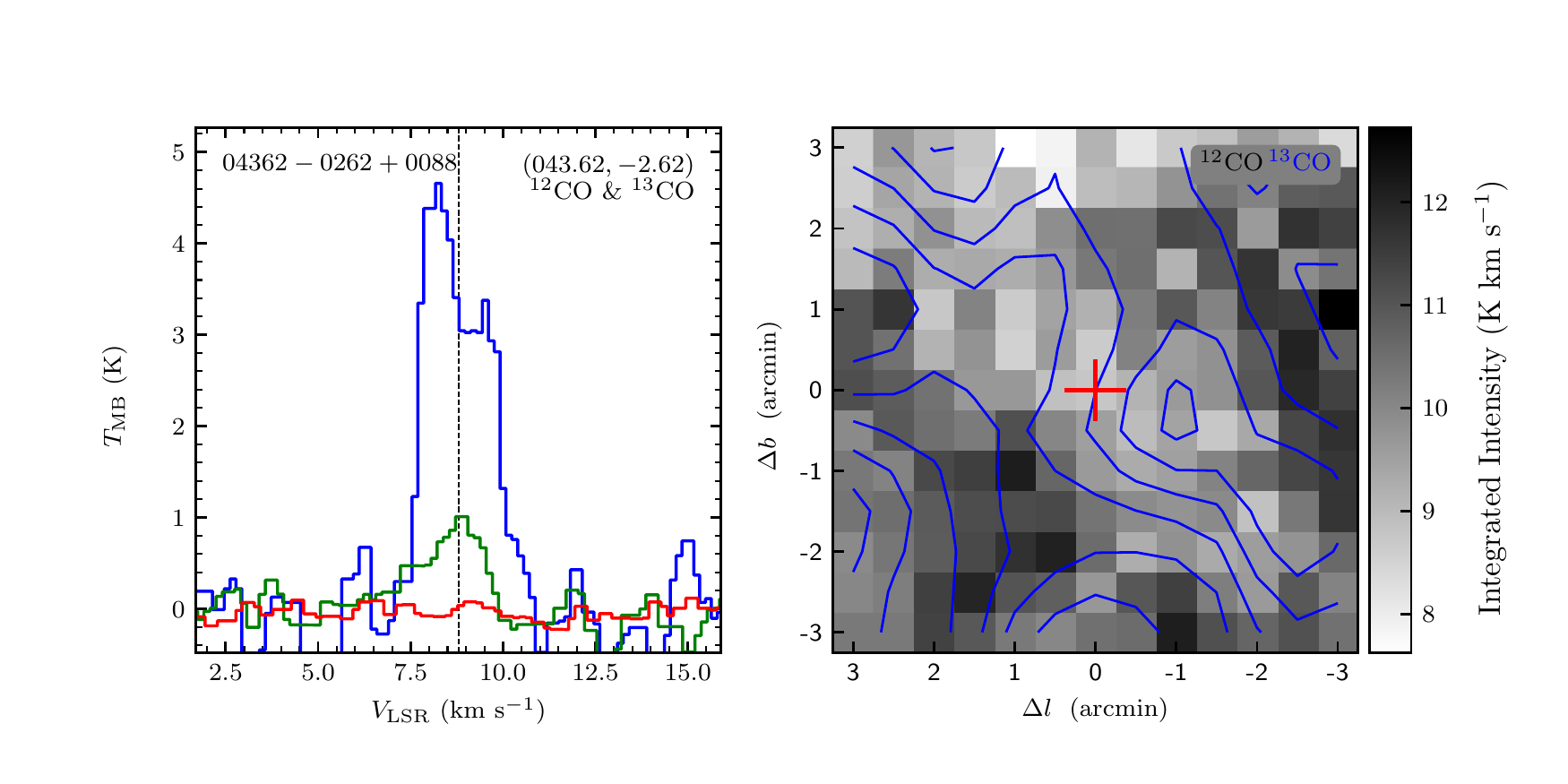}
\end{figure}
\clearpage

\begin{figure}
\includegraphics[width=9.0cm,angle=0]{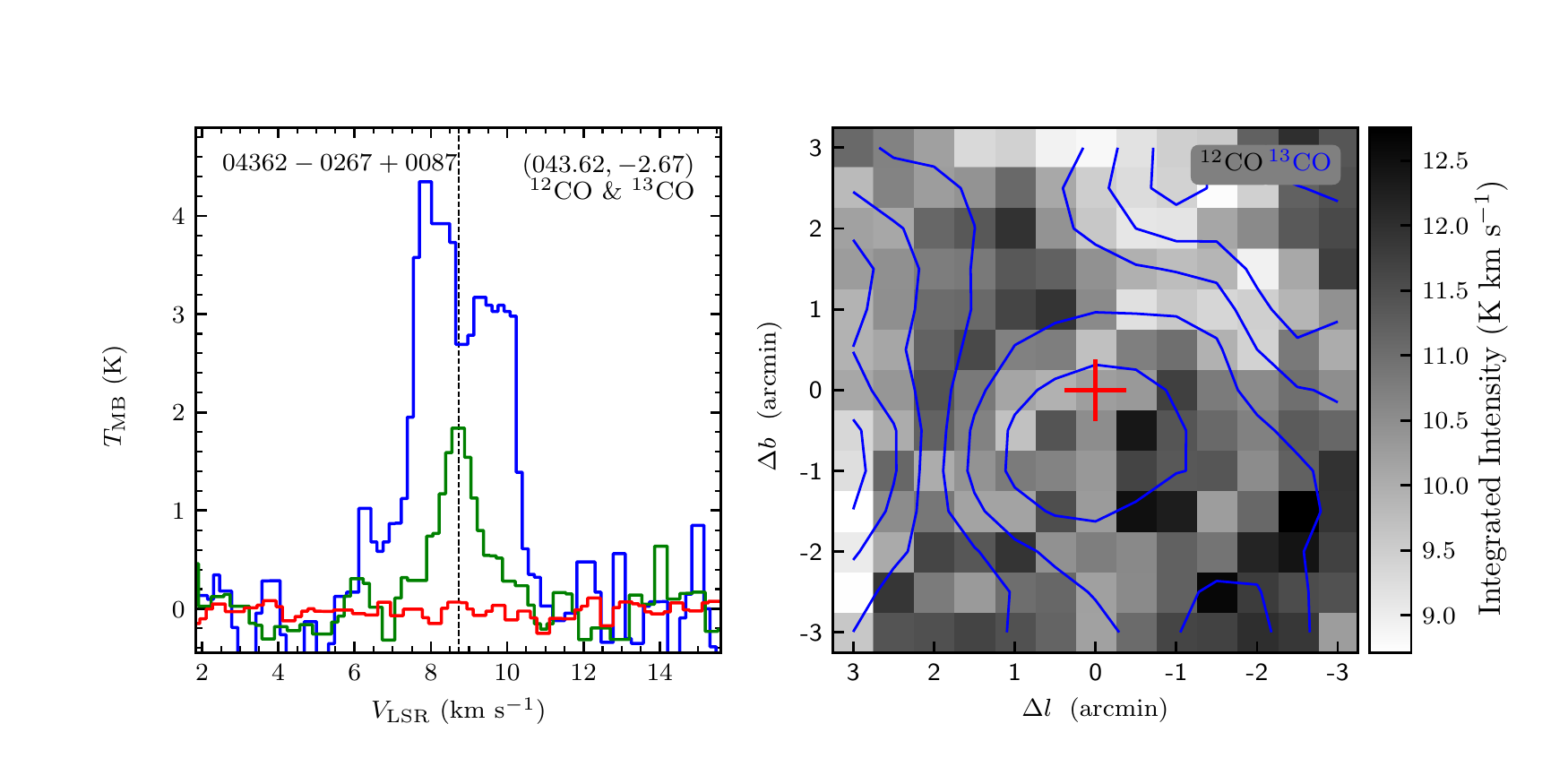}
\includegraphics[width=9.0cm,angle=0]{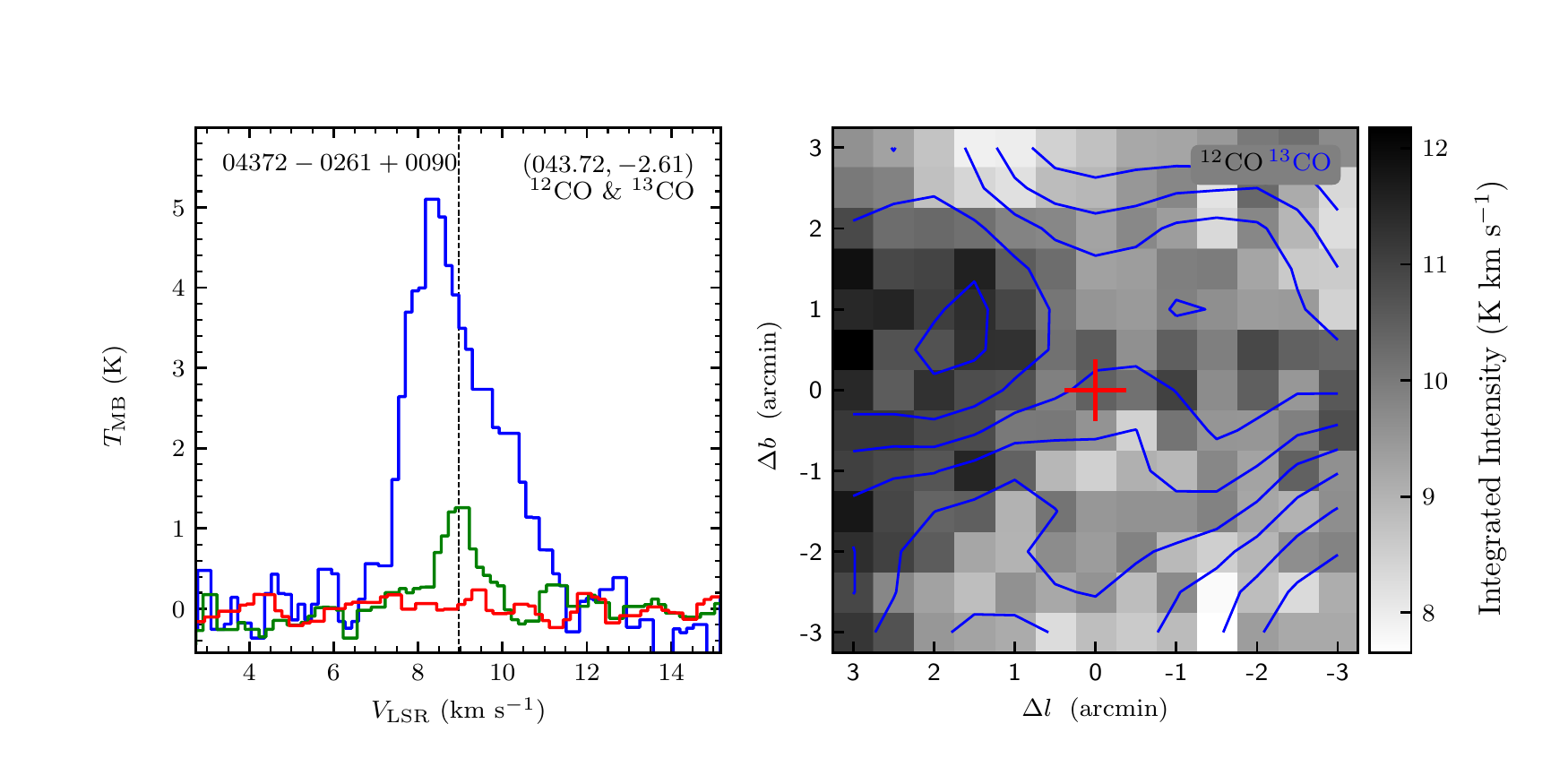}
\vspace{-0.5cm}

\includegraphics[width=9.0cm,angle=0]{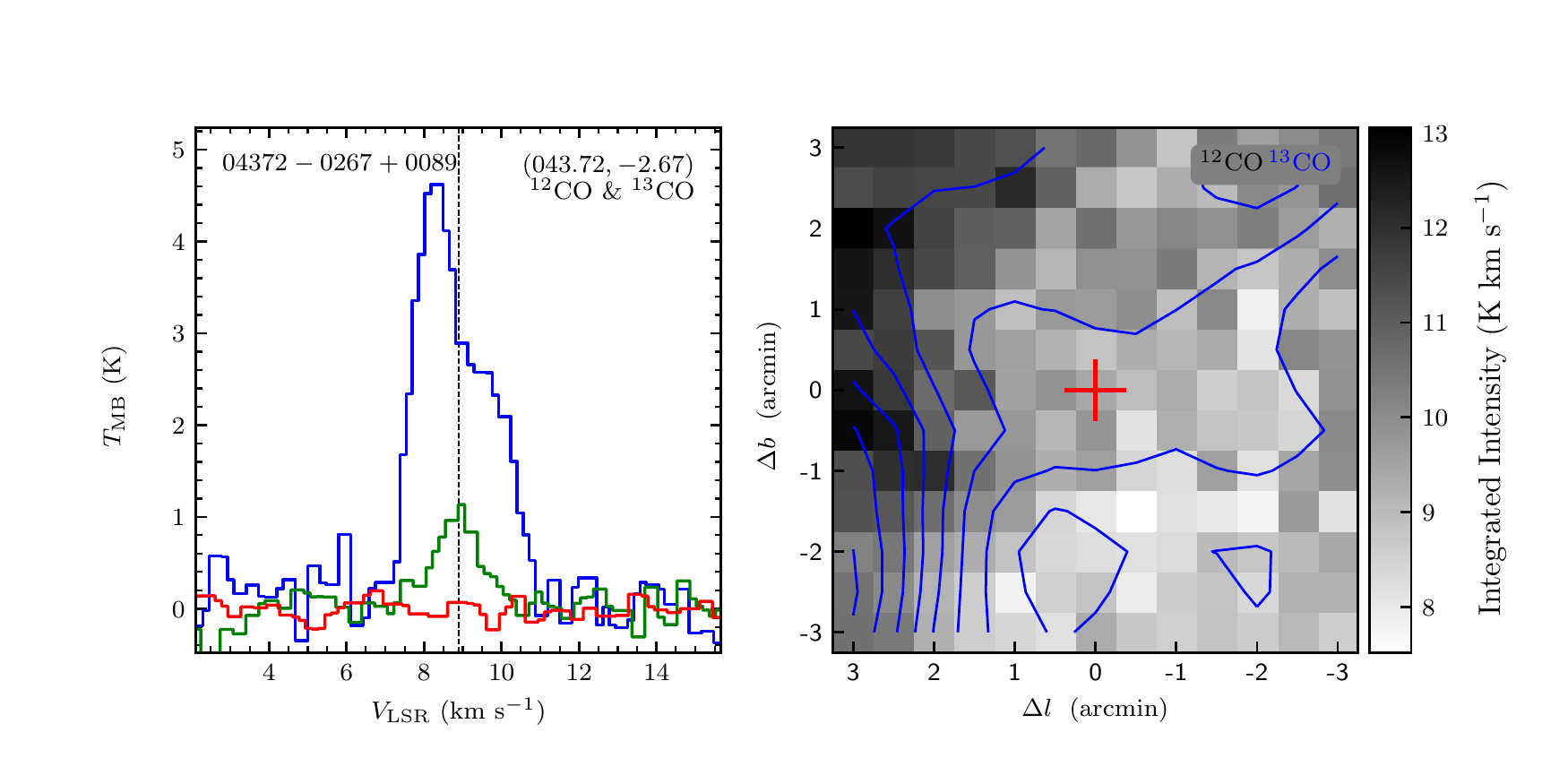}
\includegraphics[width=9.0cm,angle=0]{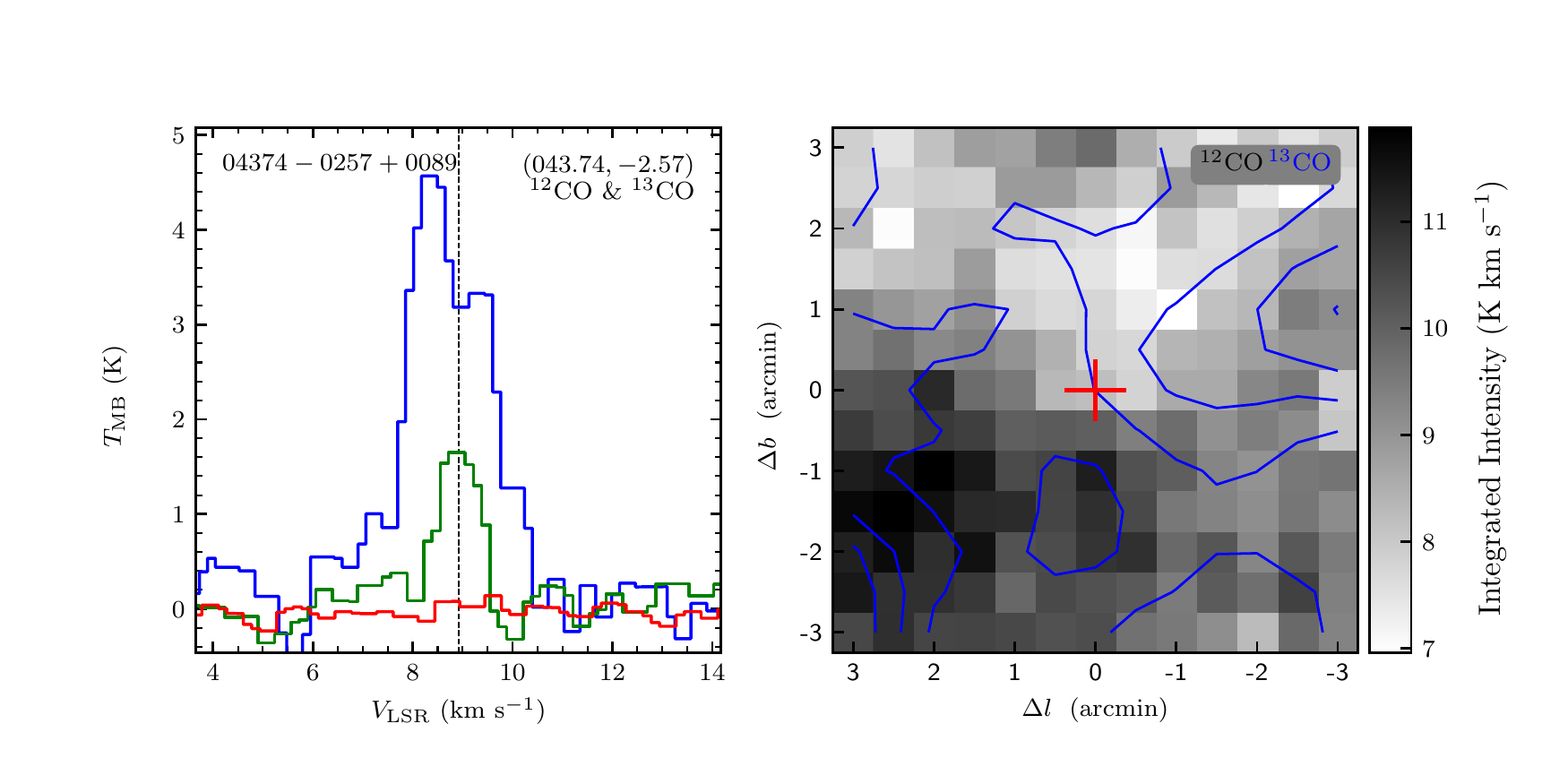}
\vspace{-0.5cm}

\includegraphics[width=9.0cm,angle=0]{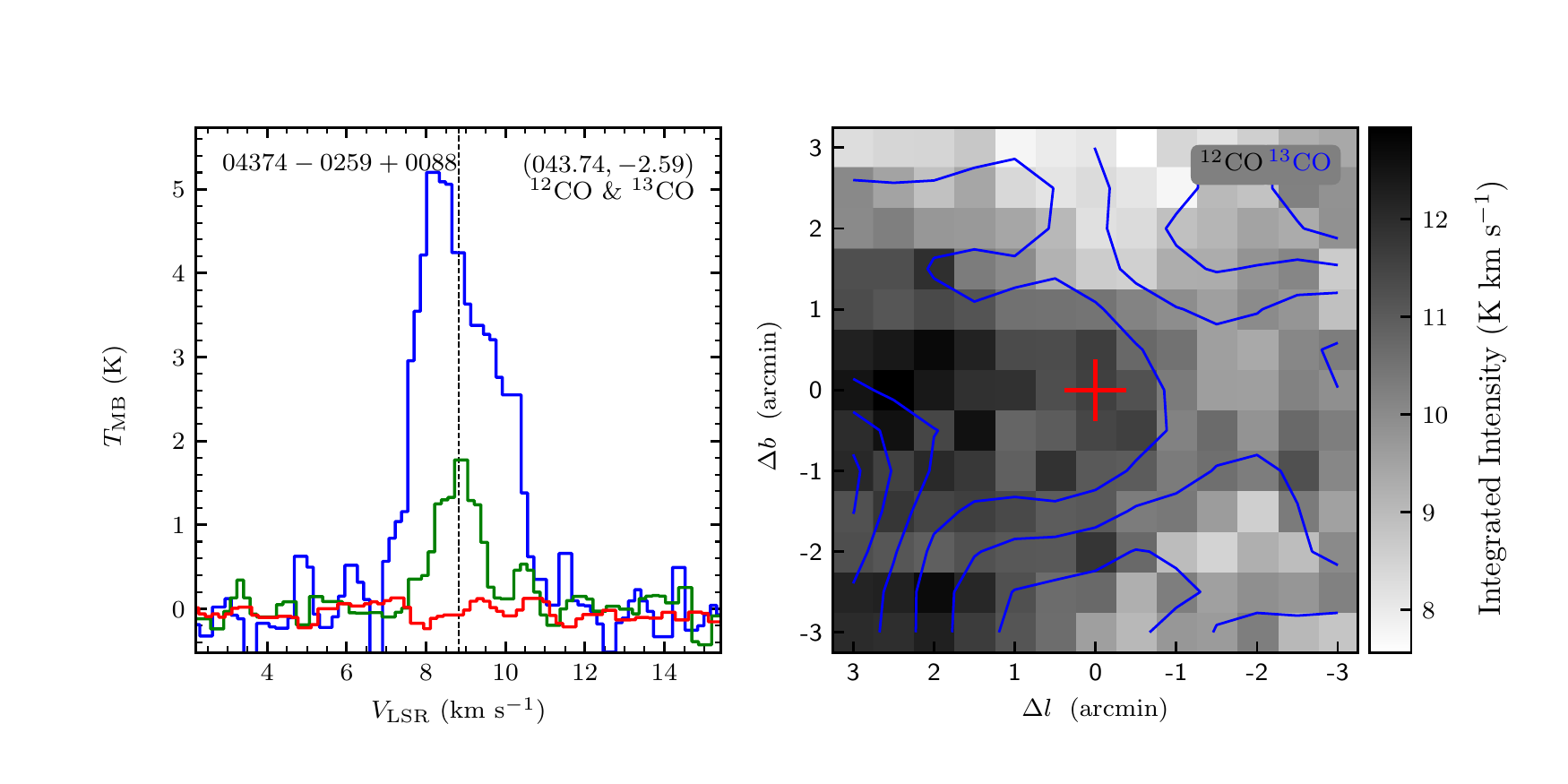}
\includegraphics[width=9.0cm,angle=0]{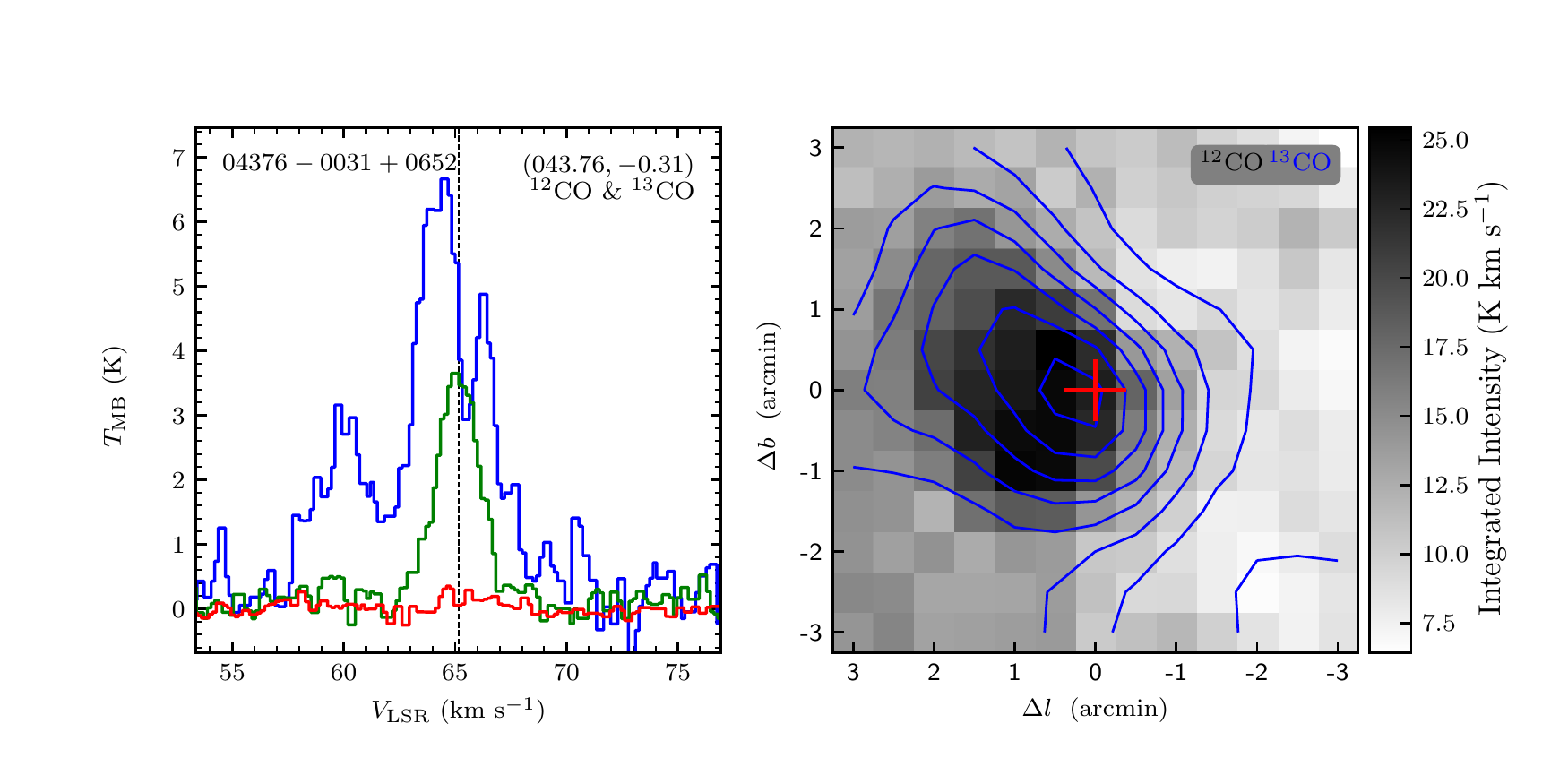}
\vspace{-0.5cm}

\includegraphics[width=9.0cm,angle=0]{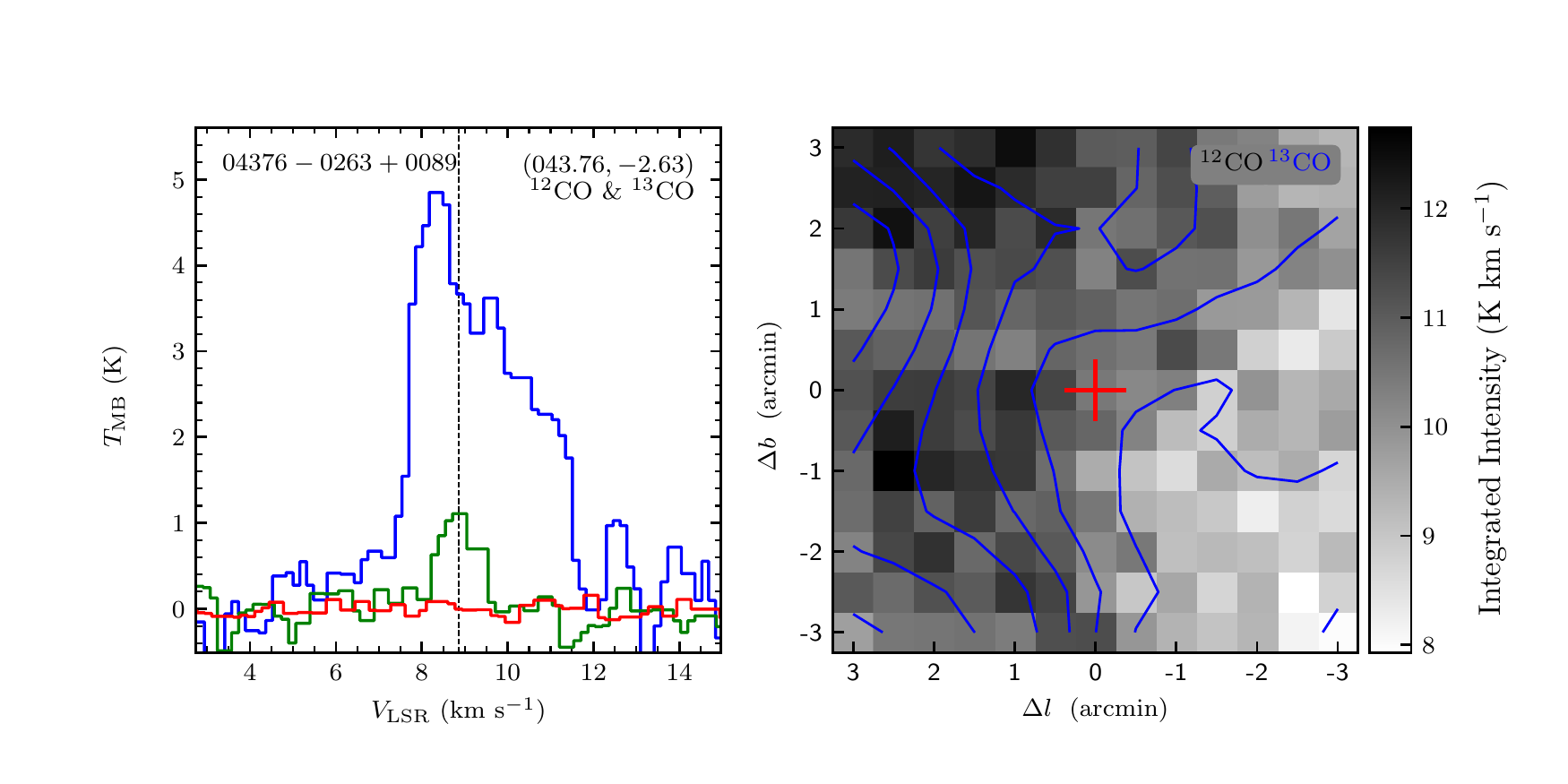}
\includegraphics[width=9.0cm,angle=0]{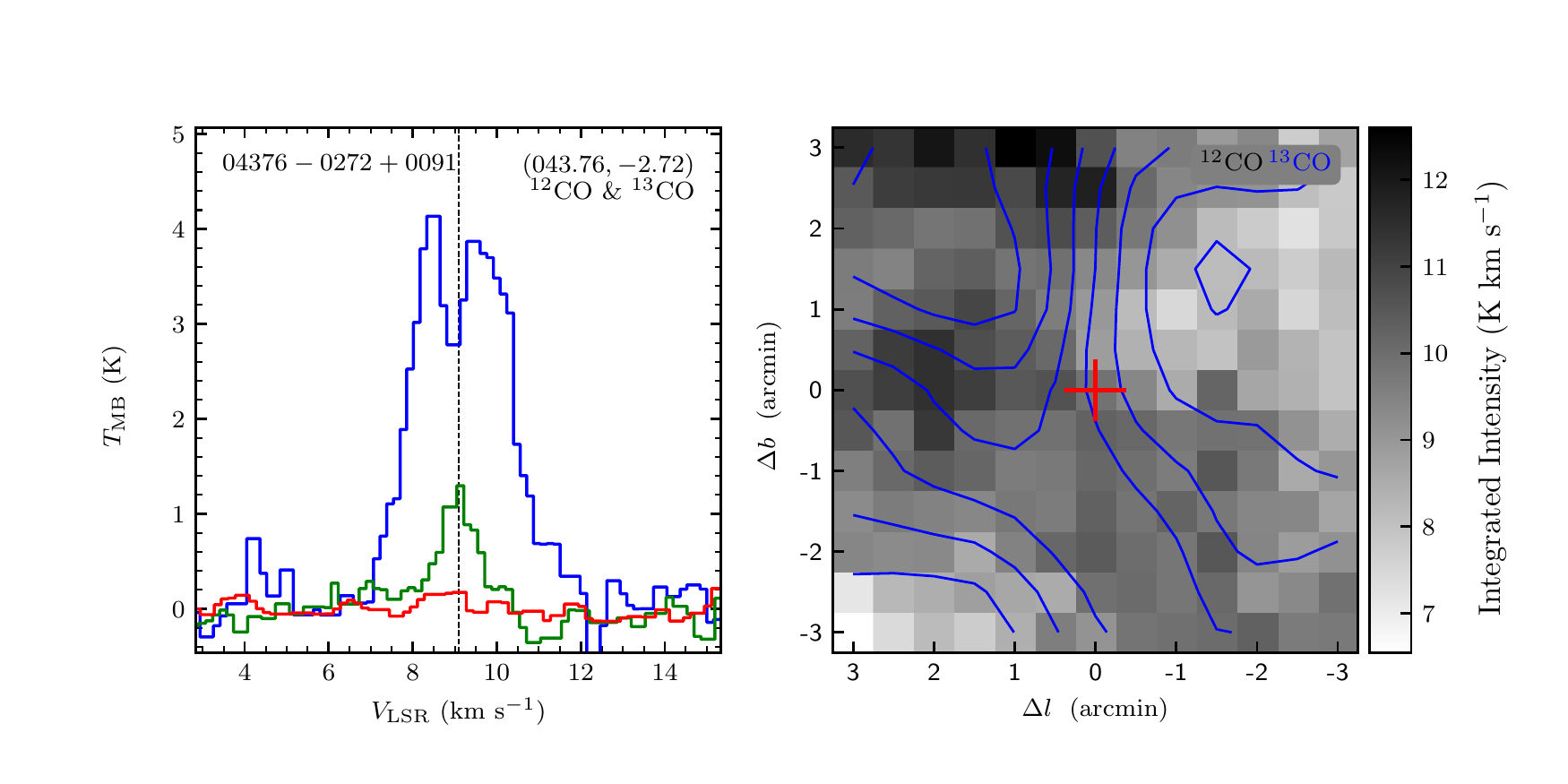}
\vspace{-0.5cm}

\includegraphics[width=9.0cm,angle=0]{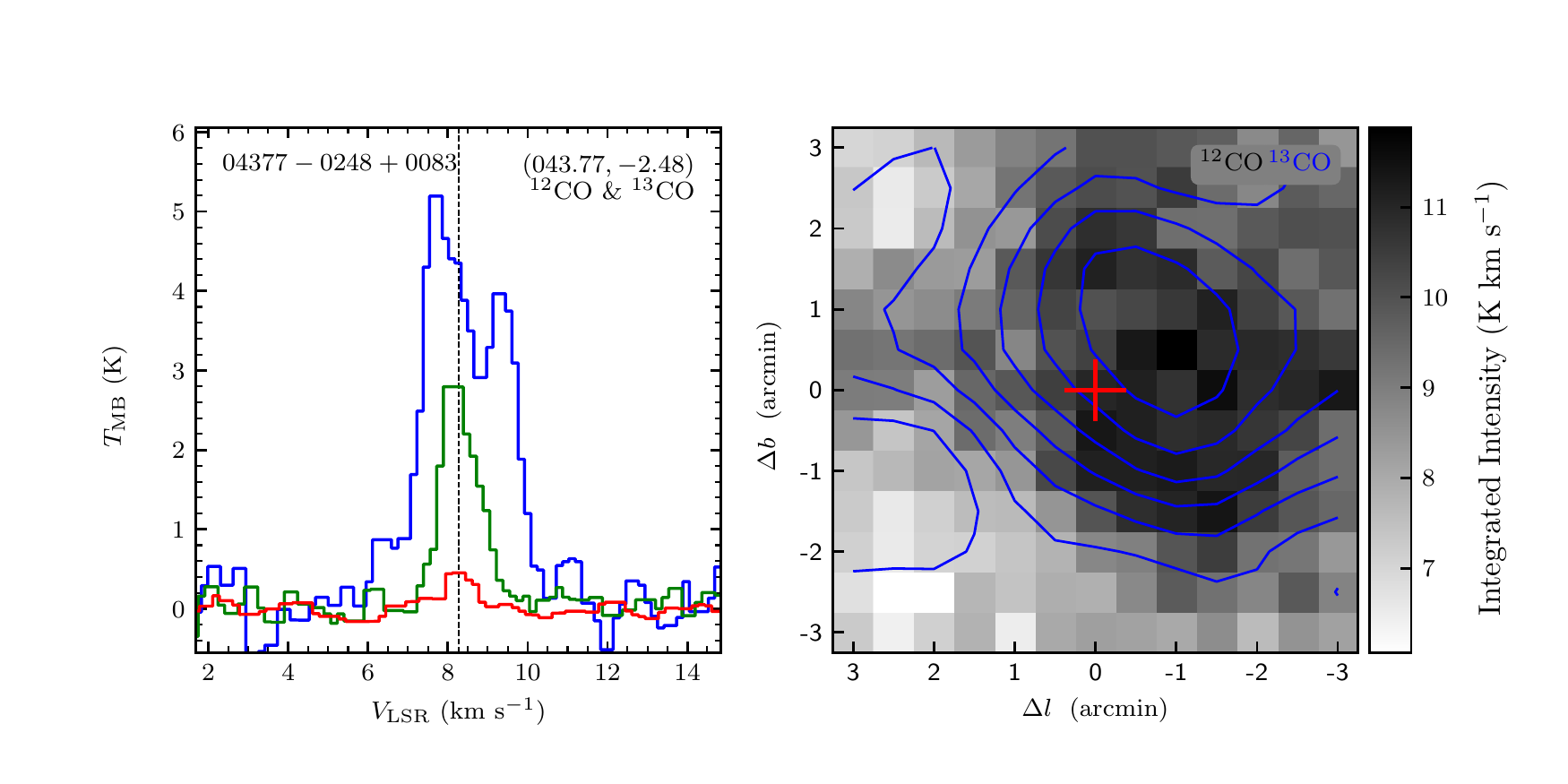}
\includegraphics[width=9.0cm,angle=0]{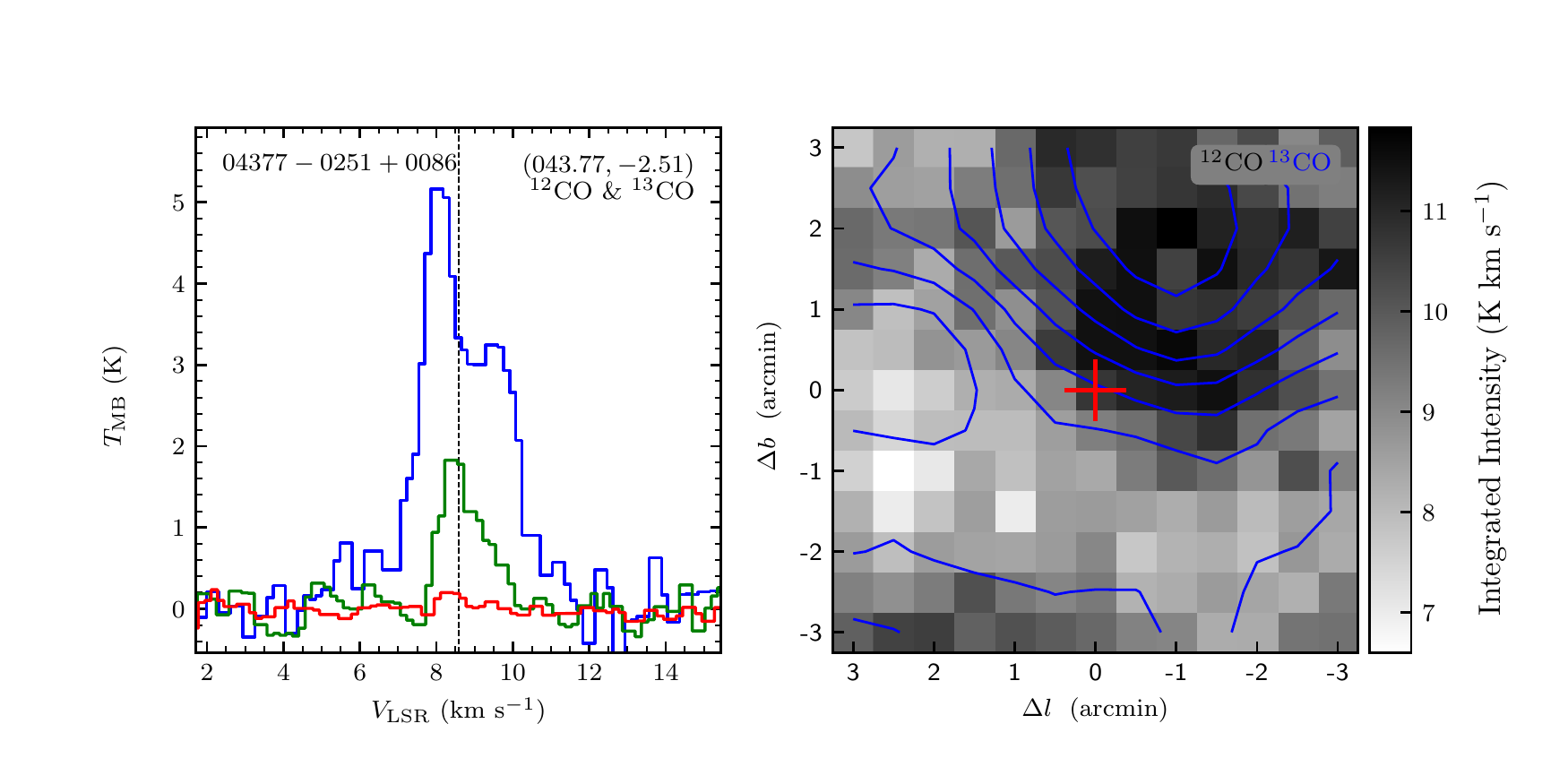}
\end{figure}
\clearpage

\begin{figure}
\includegraphics[width=9.0cm,angle=0]{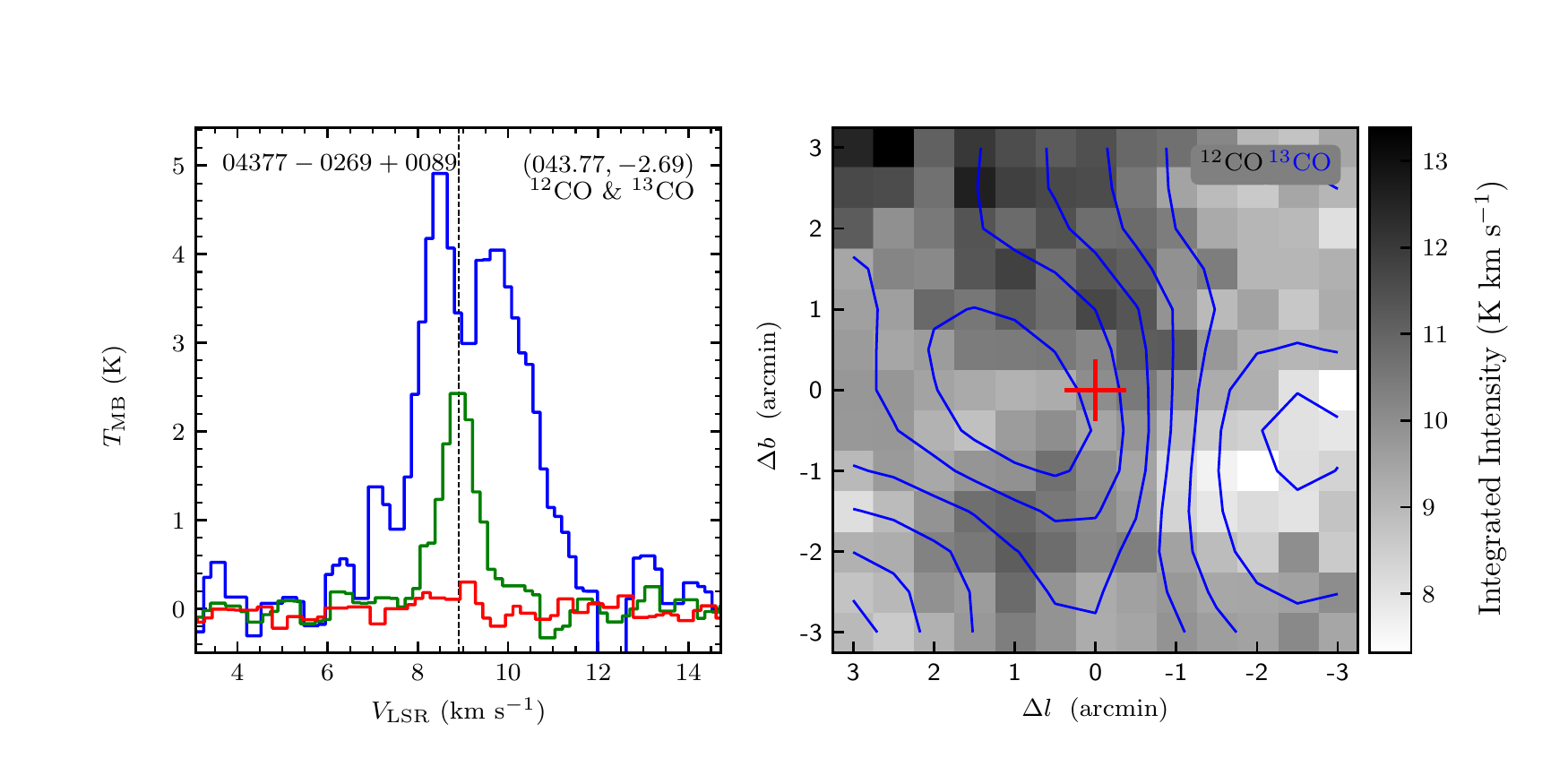}
\includegraphics[width=9.0cm,angle=0]{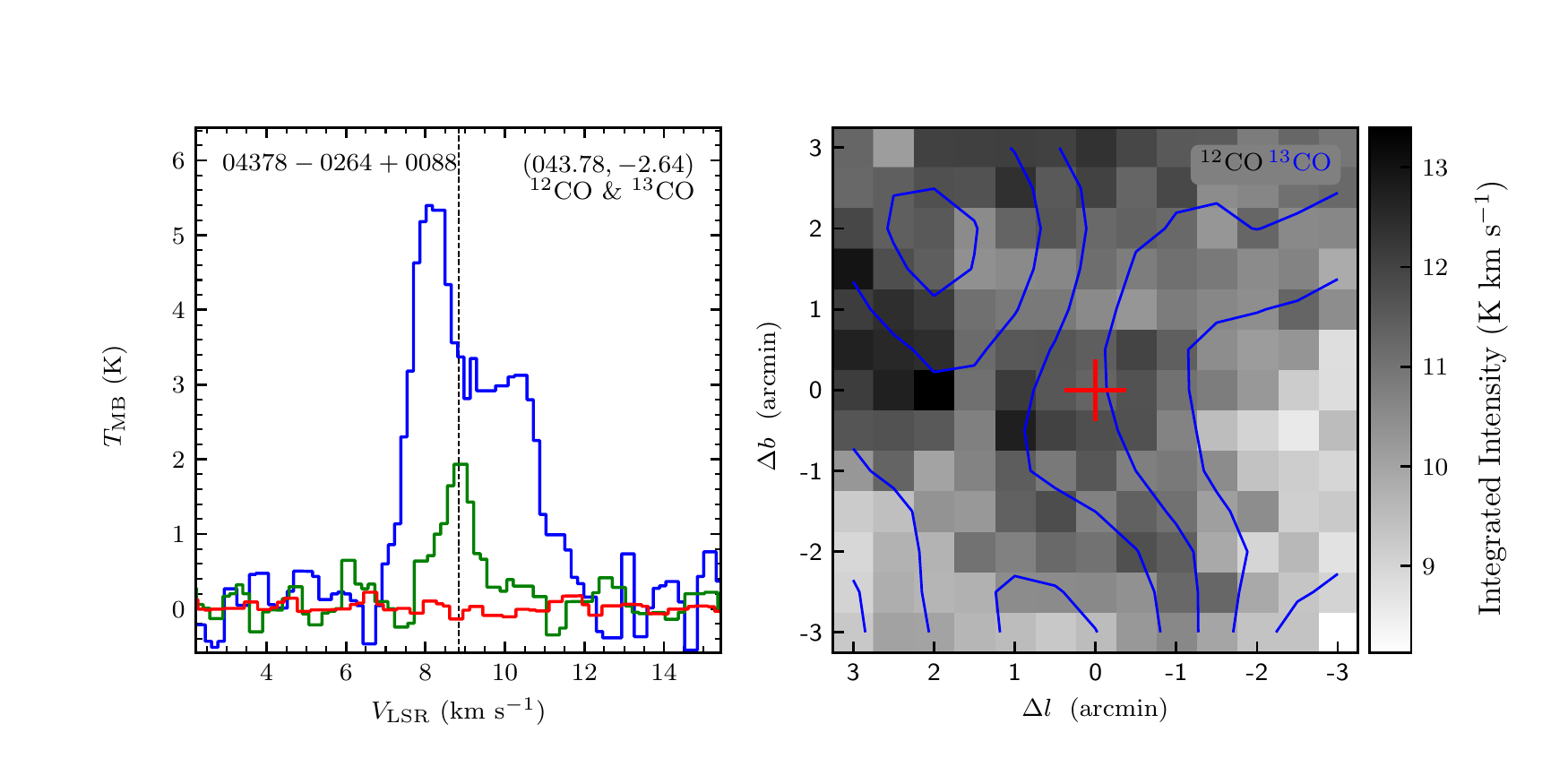}
\vspace{-0.5cm}

\includegraphics[width=9.0cm,angle=0]{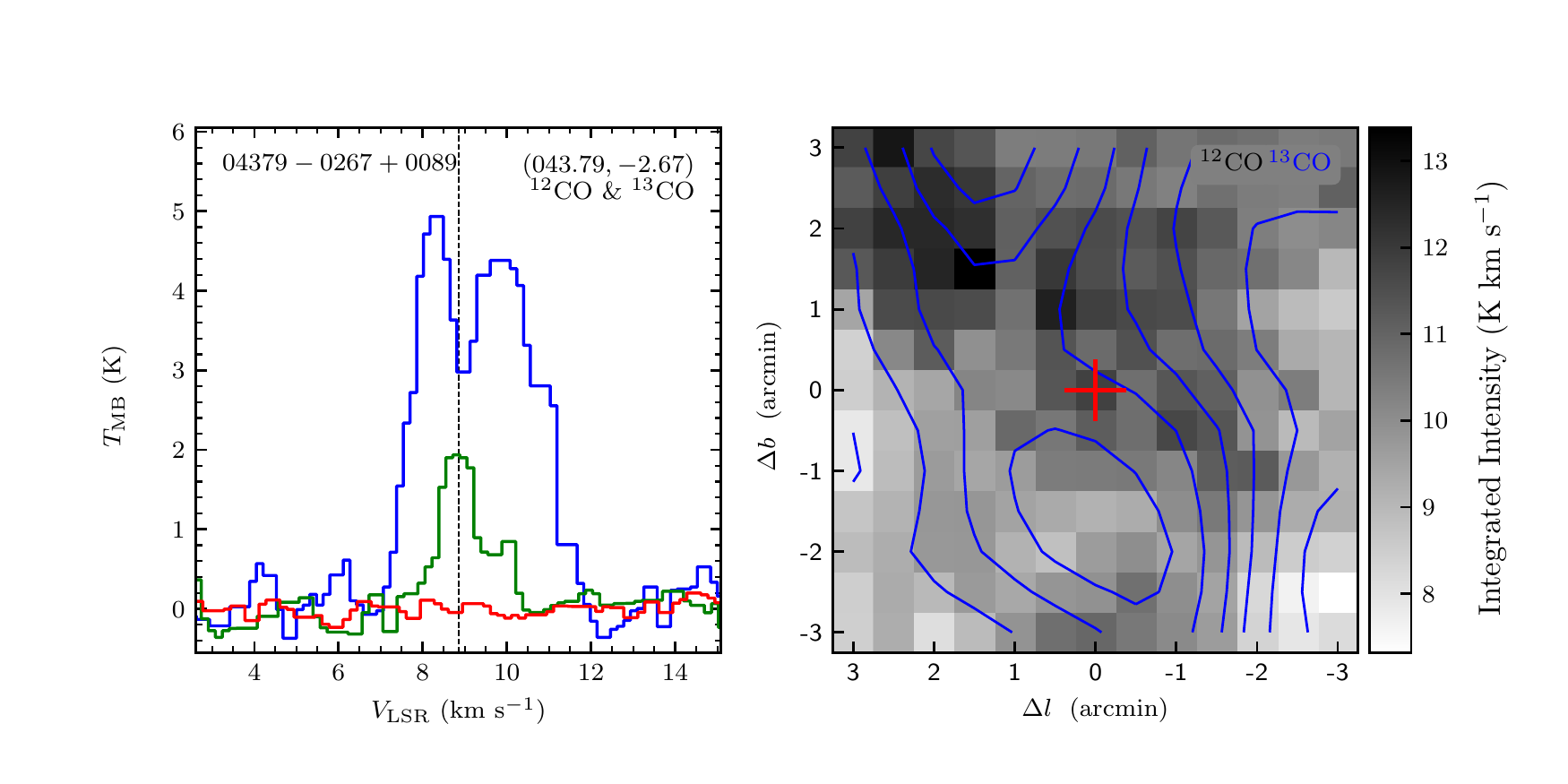}
\includegraphics[width=9.0cm,angle=0]{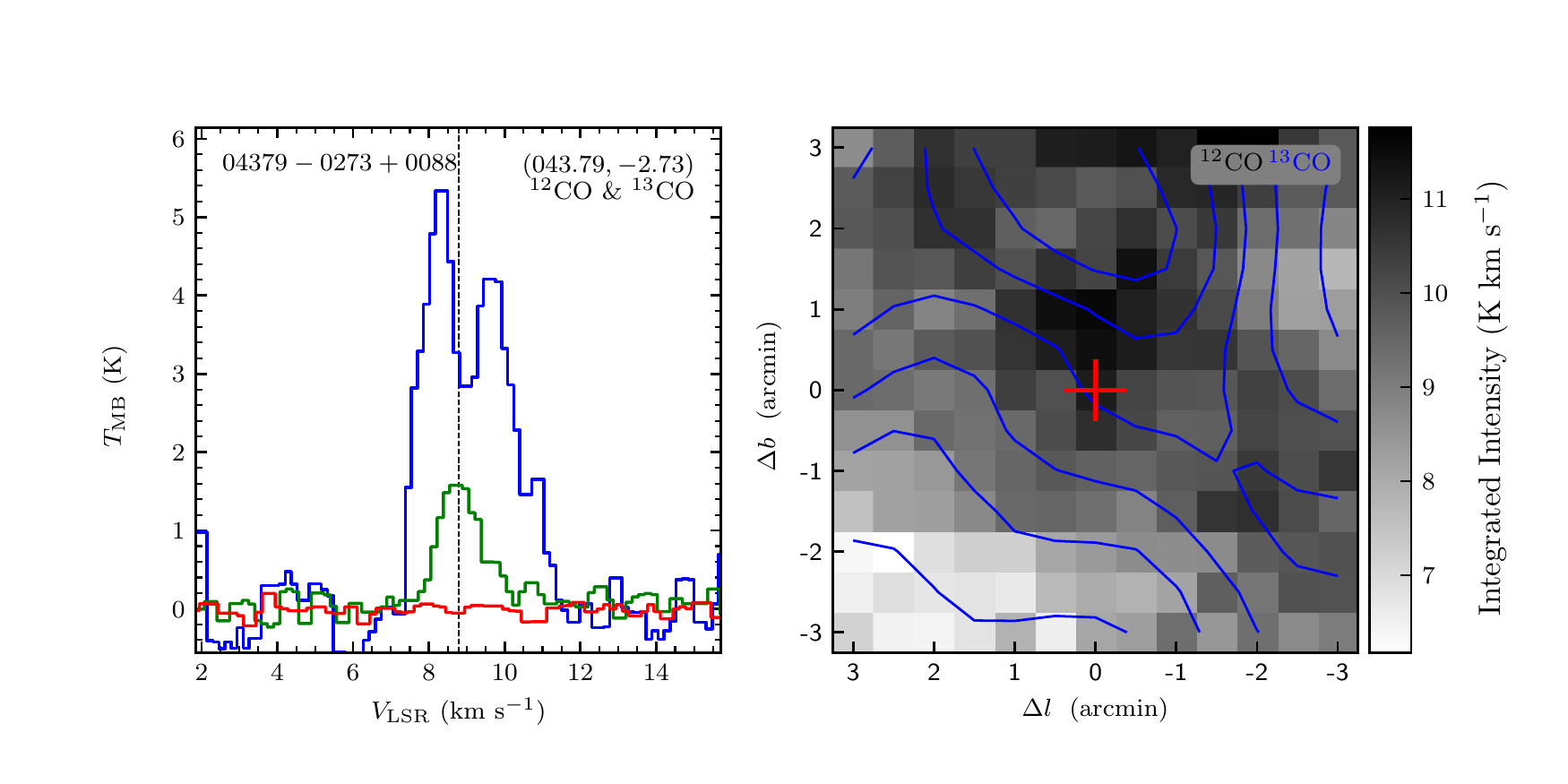}
\vspace{-0.5cm}

\includegraphics[width=9.0cm,angle=0]{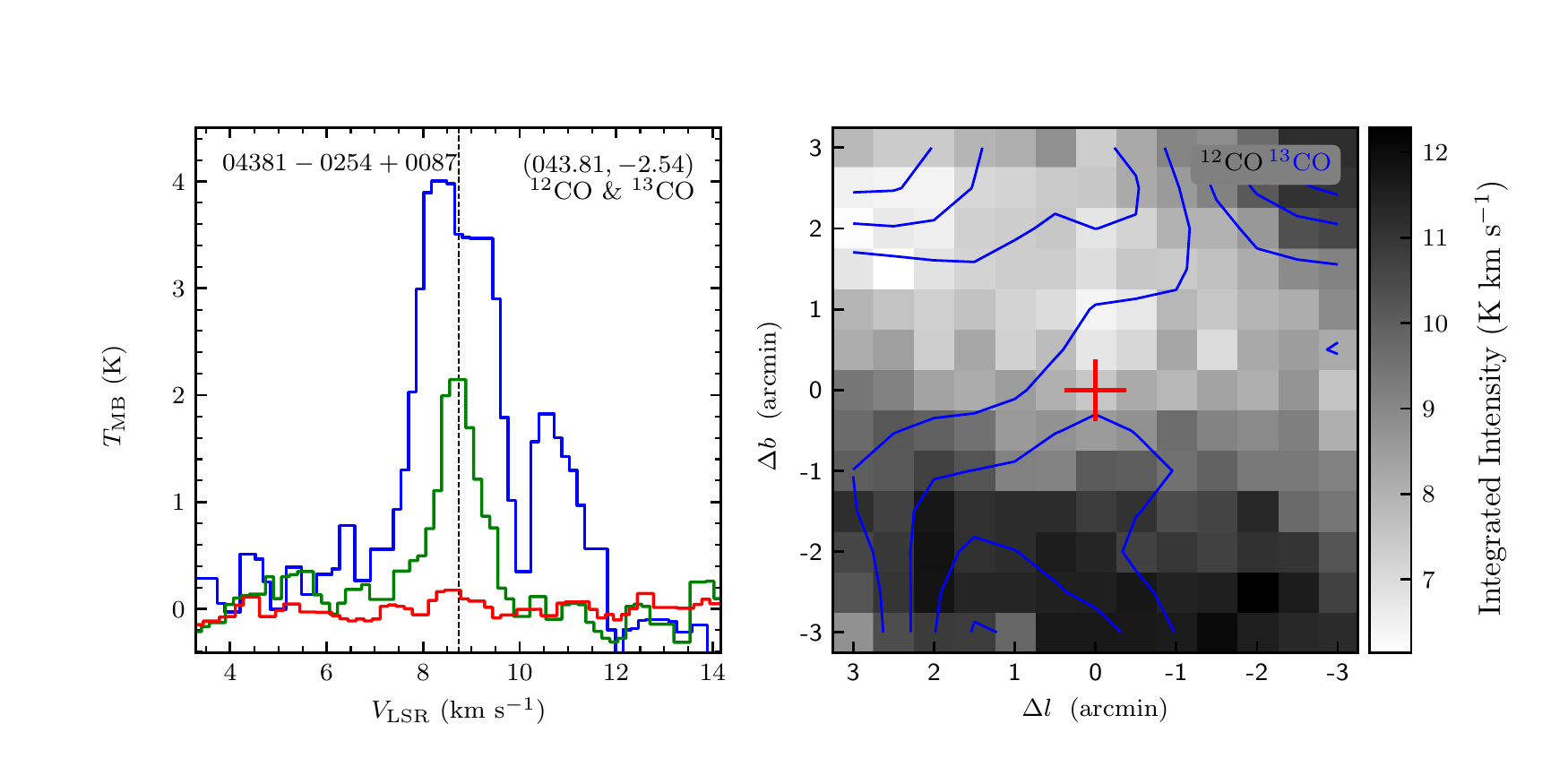}
\includegraphics[width=9.0cm,angle=0]{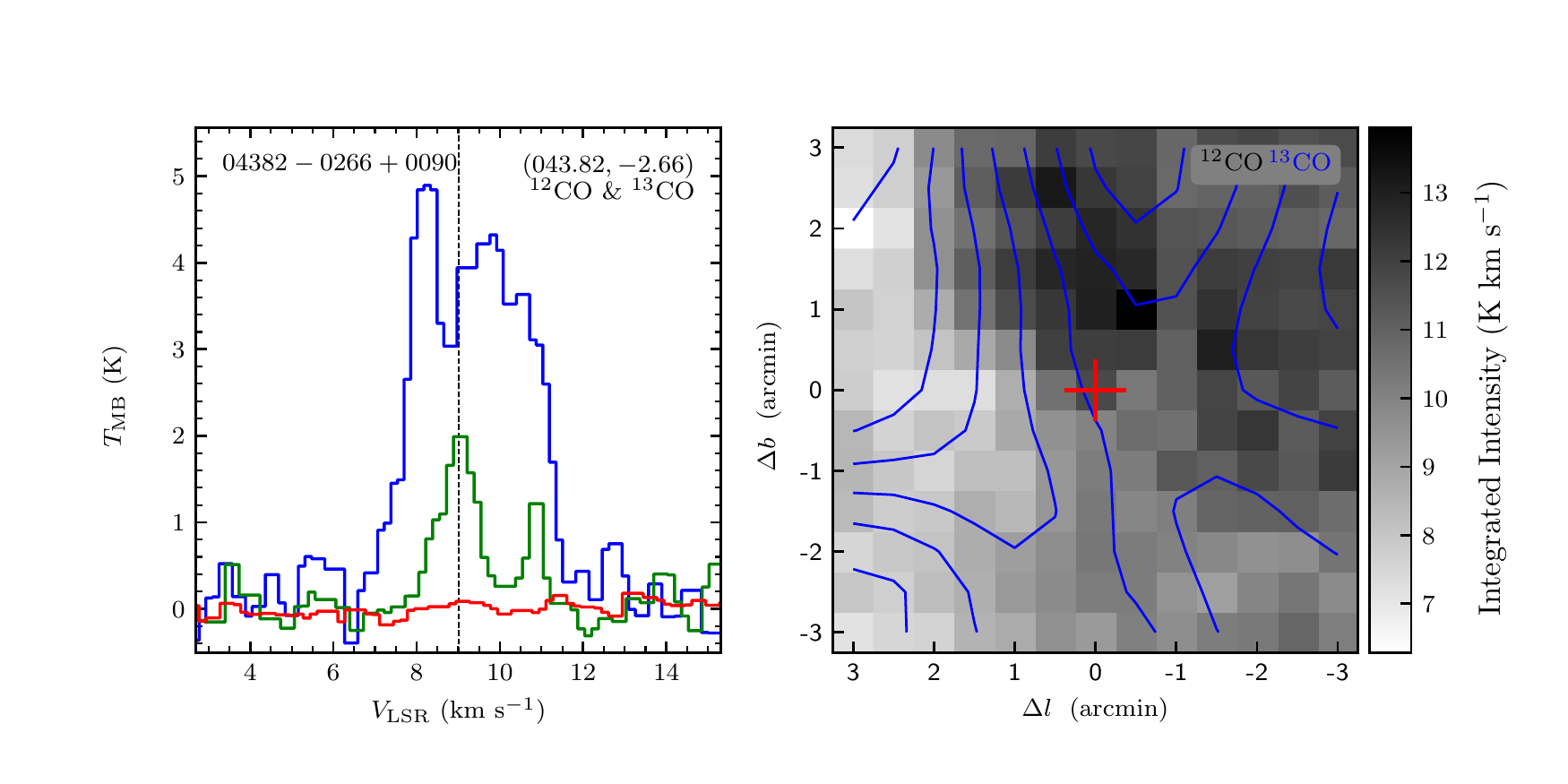}
\vspace{-0.5cm}

\includegraphics[width=9.0cm,angle=0]{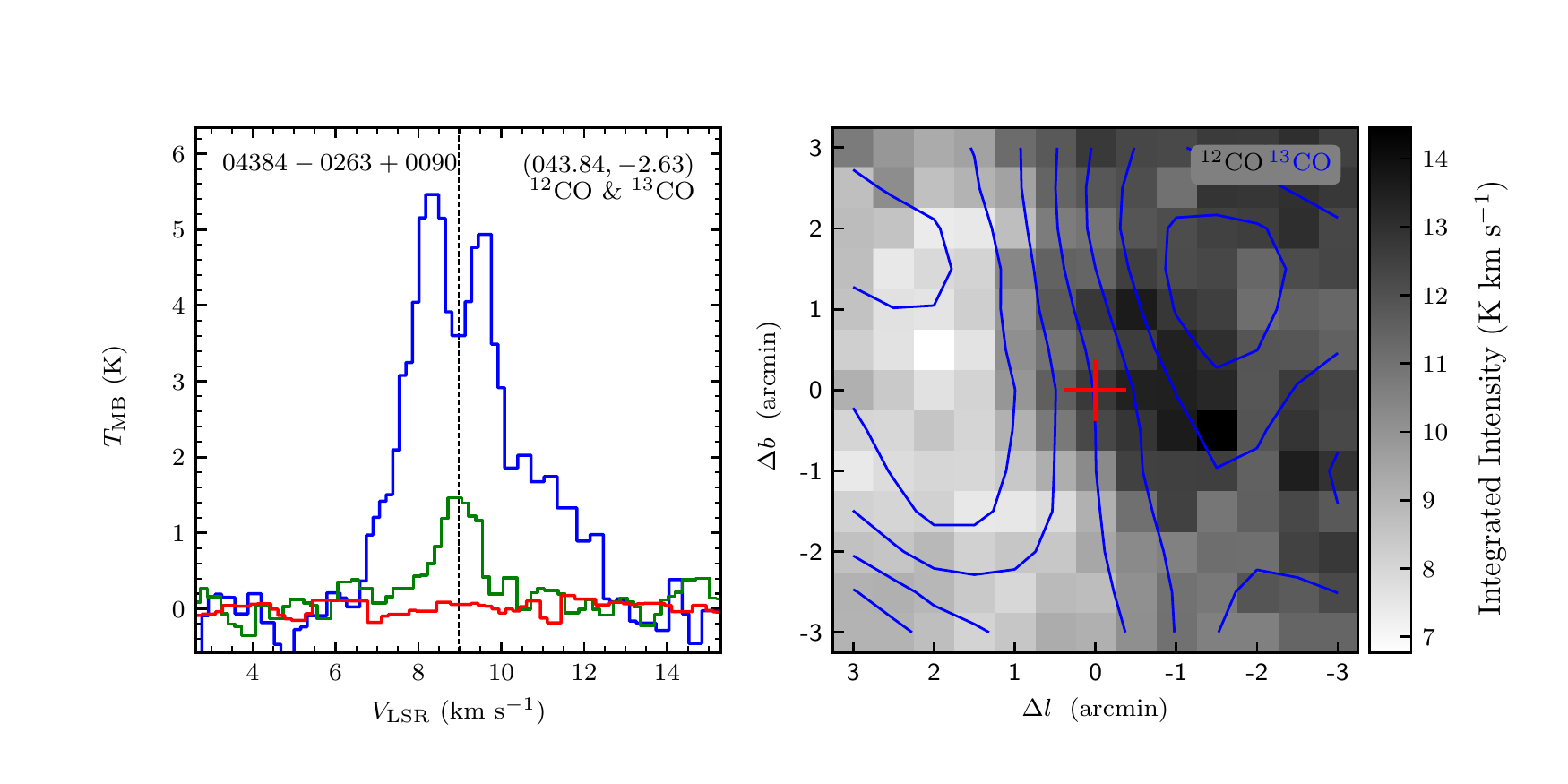}
\includegraphics[width=9.0cm,angle=0]{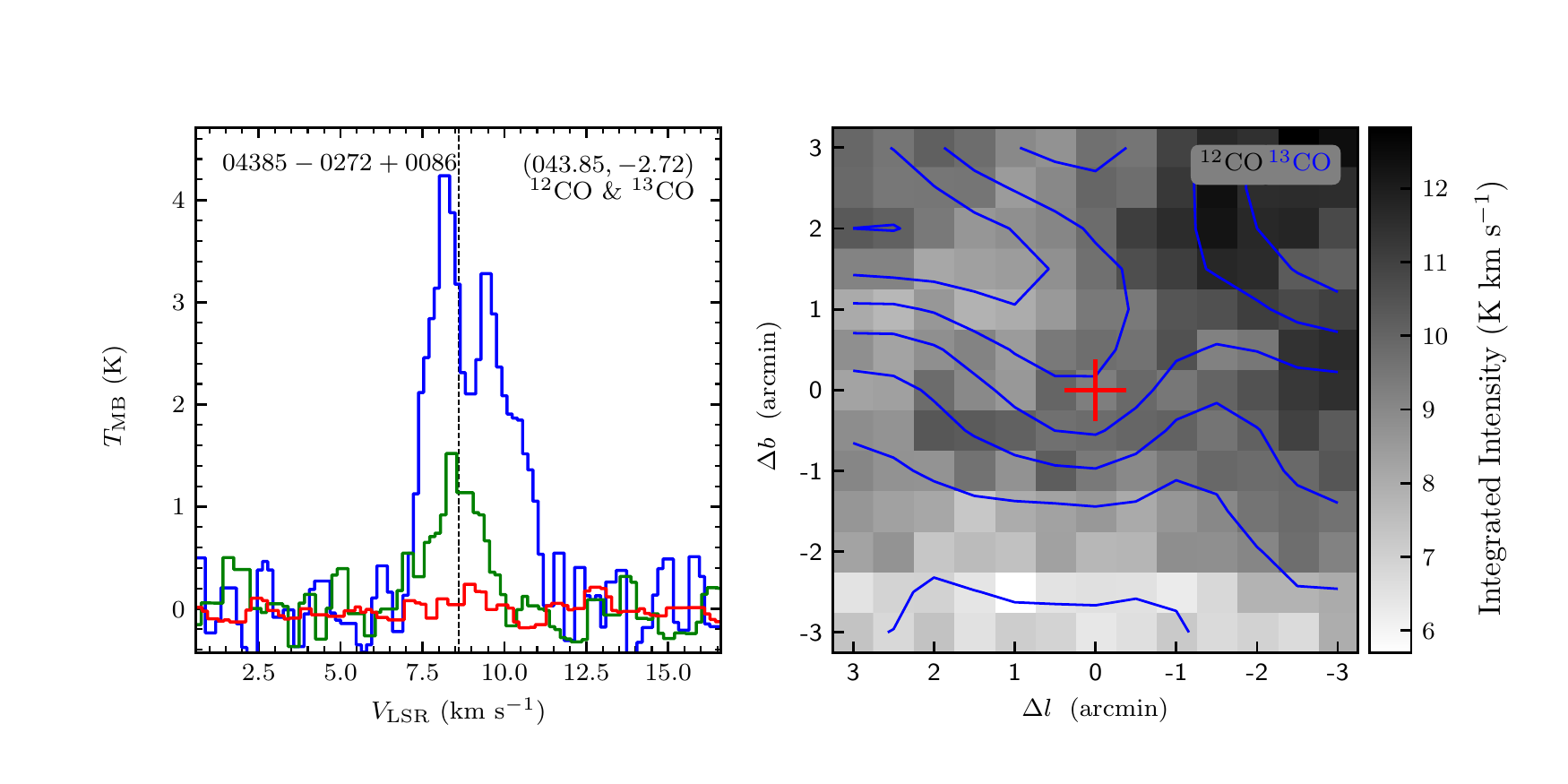}
\vspace{-0.5cm}

\includegraphics[width=9.0cm,angle=0]{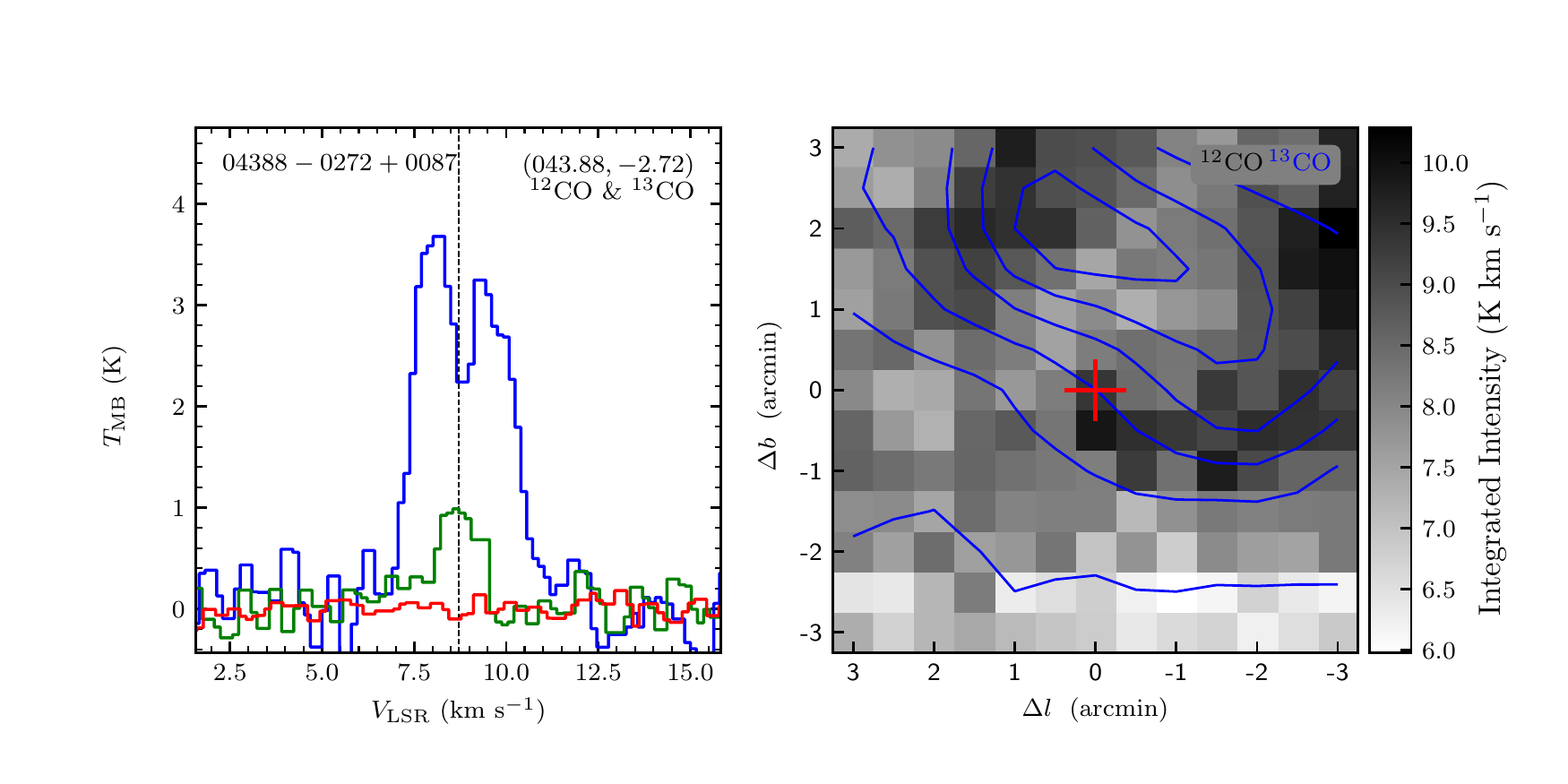}
\includegraphics[width=9.0cm,angle=0]{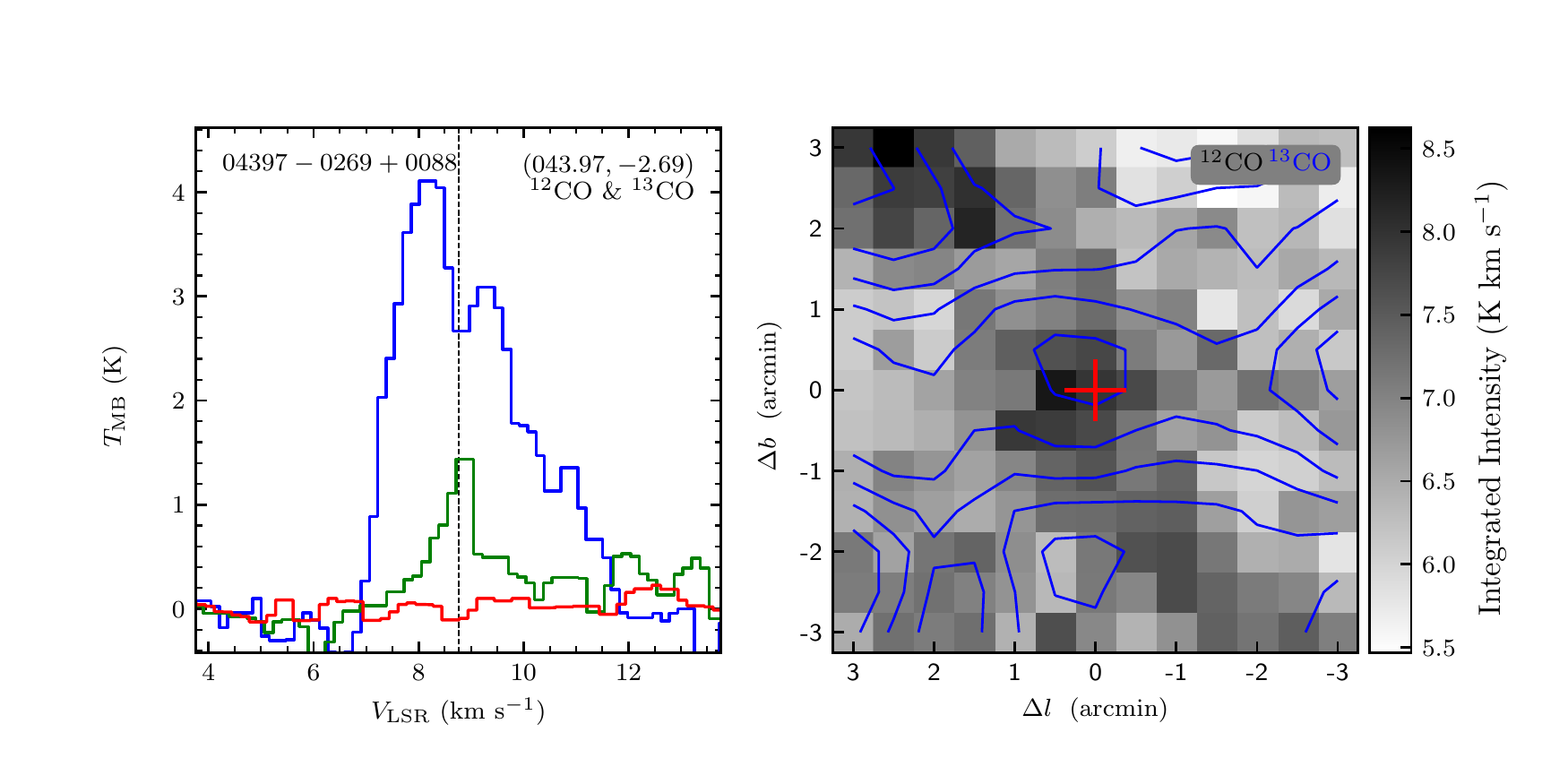}
\end{figure}
\clearpage

\begin{figure}
\includegraphics[width=9.0cm,angle=0]{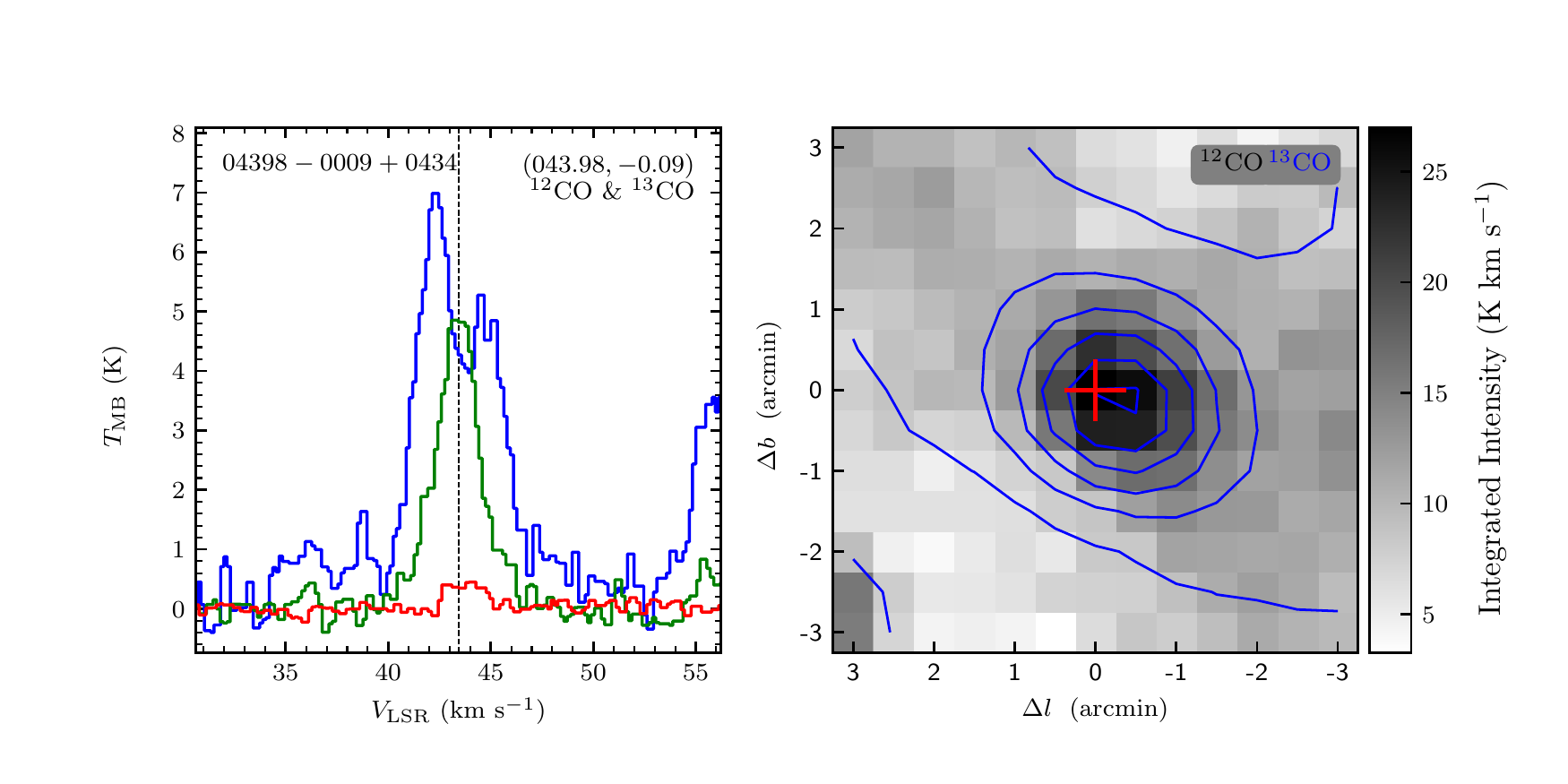}
\includegraphics[width=9.0cm,angle=0]{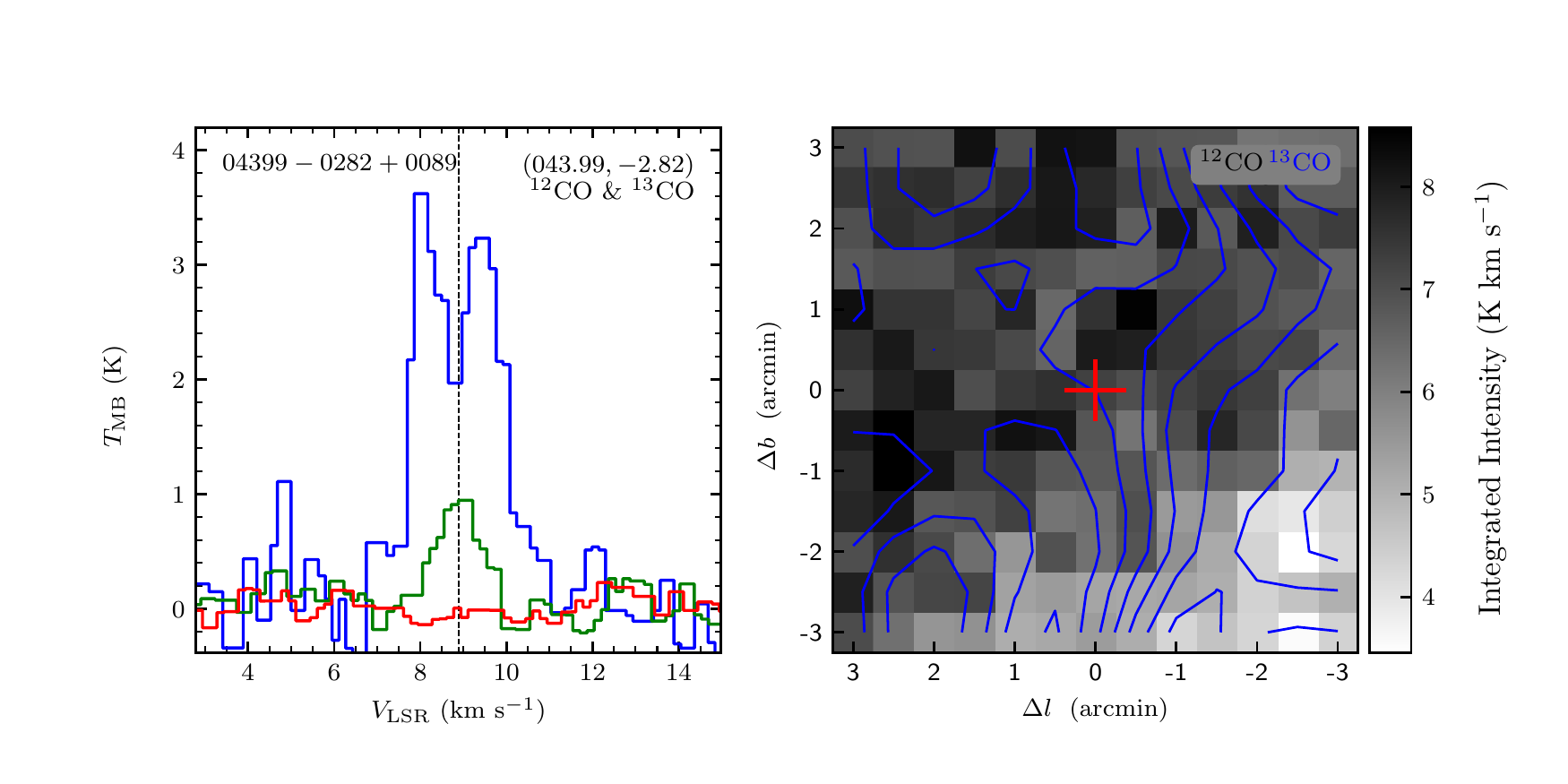}
\vspace{-0.5cm}

\includegraphics[width=9.0cm,angle=0]{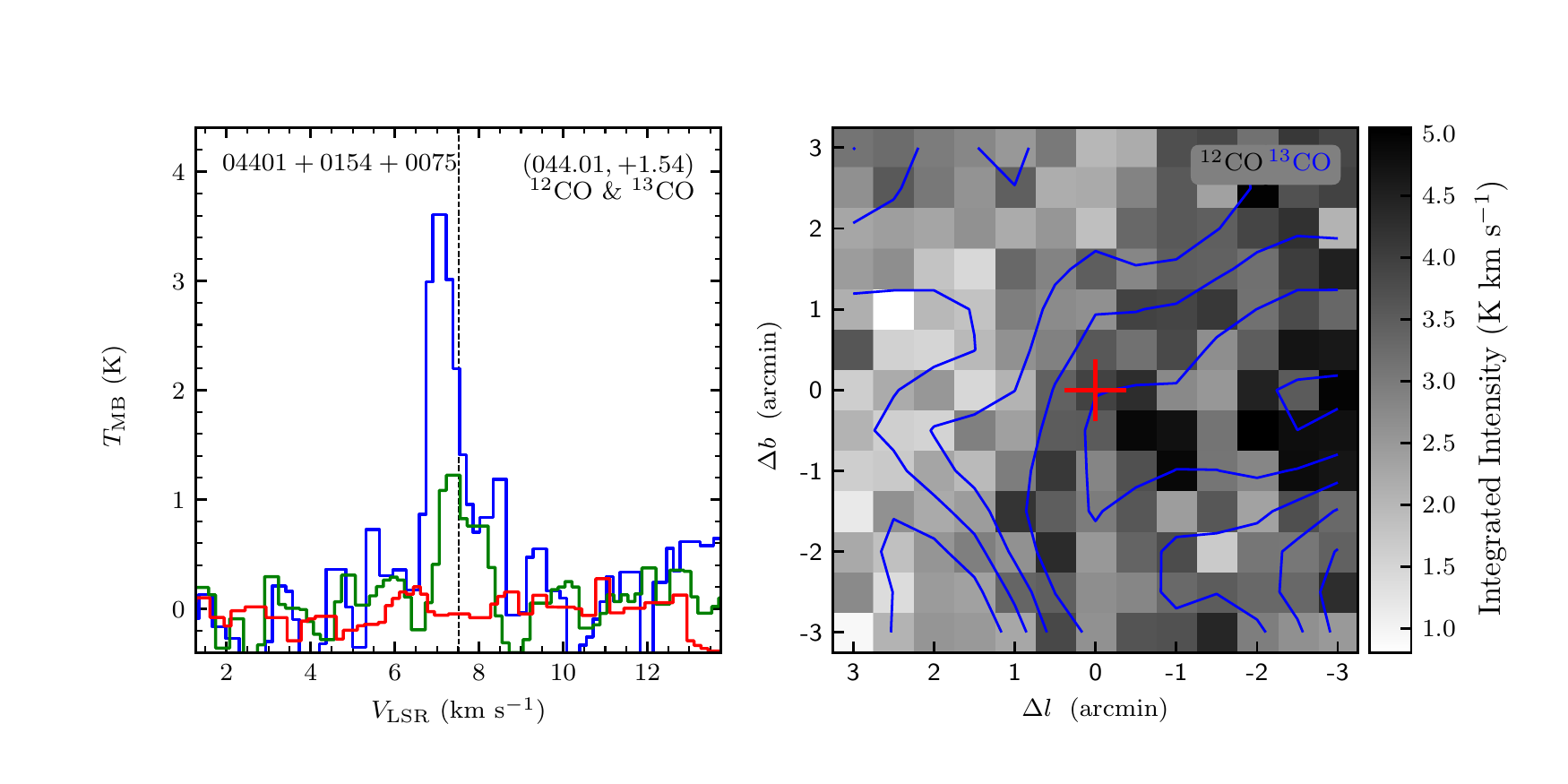}
\includegraphics[width=9.0cm,angle=0]{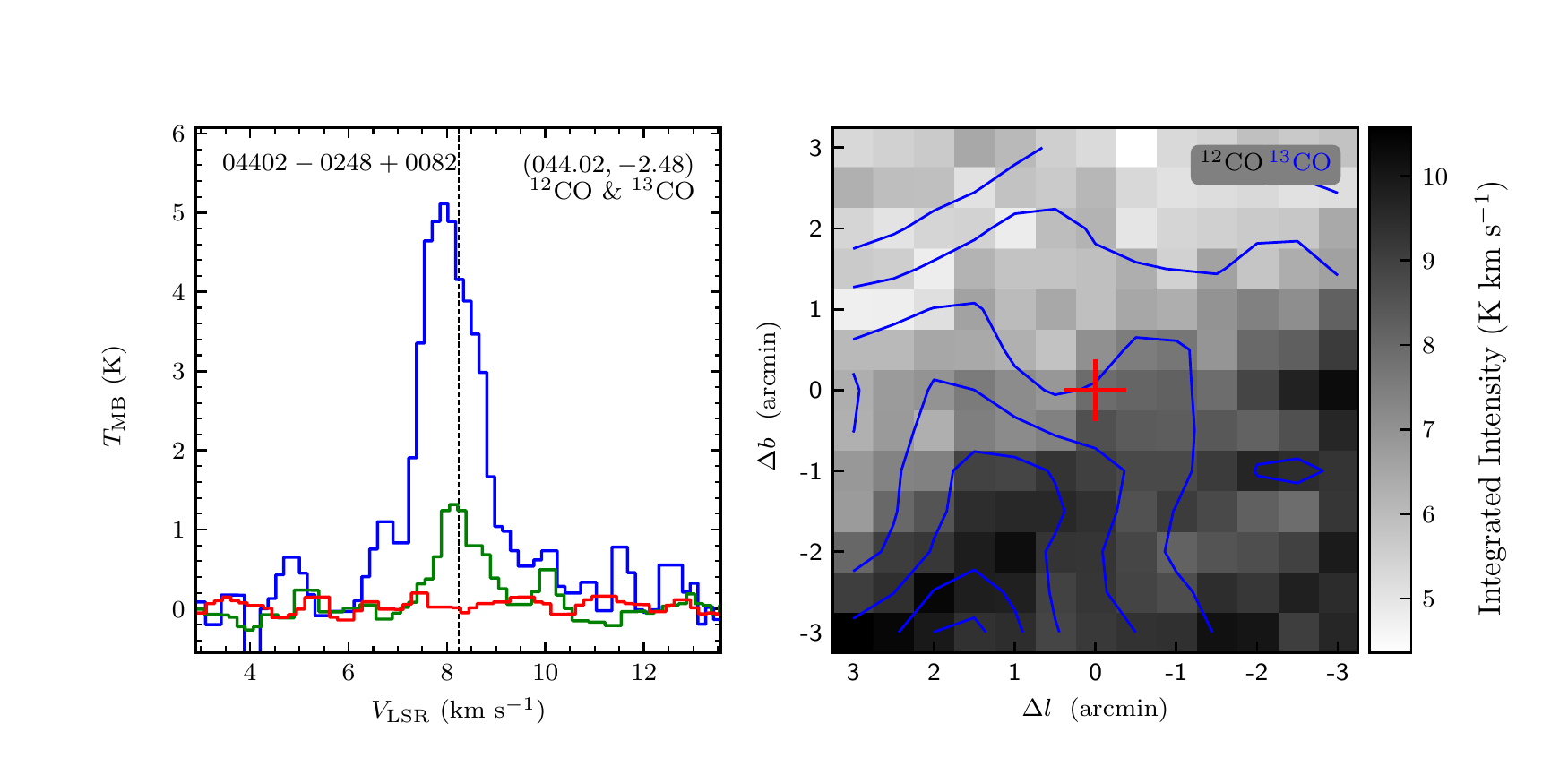}
\vspace{-0.5cm}

\includegraphics[width=9.0cm,angle=0]{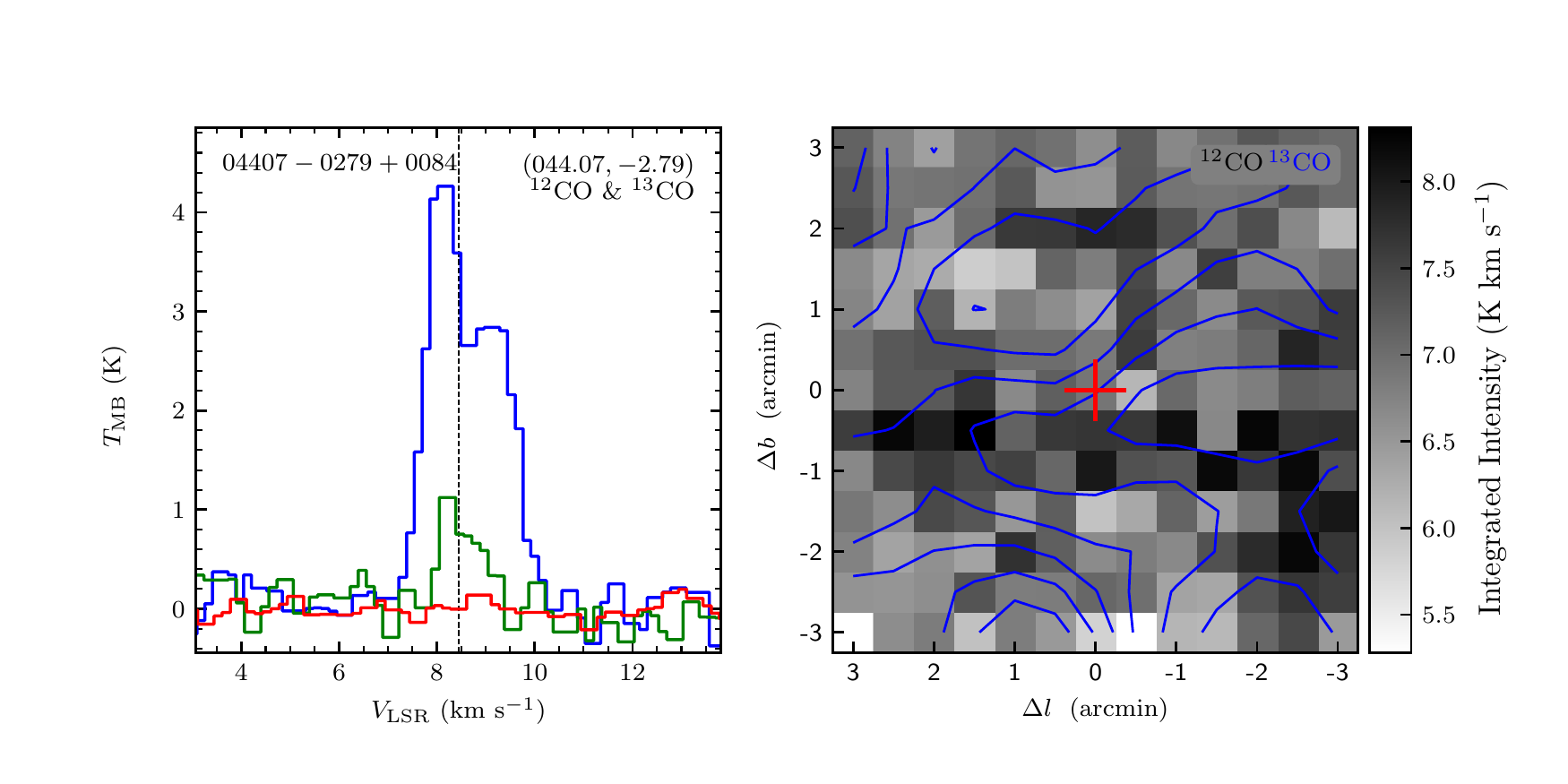}
\includegraphics[width=9.0cm,angle=0]{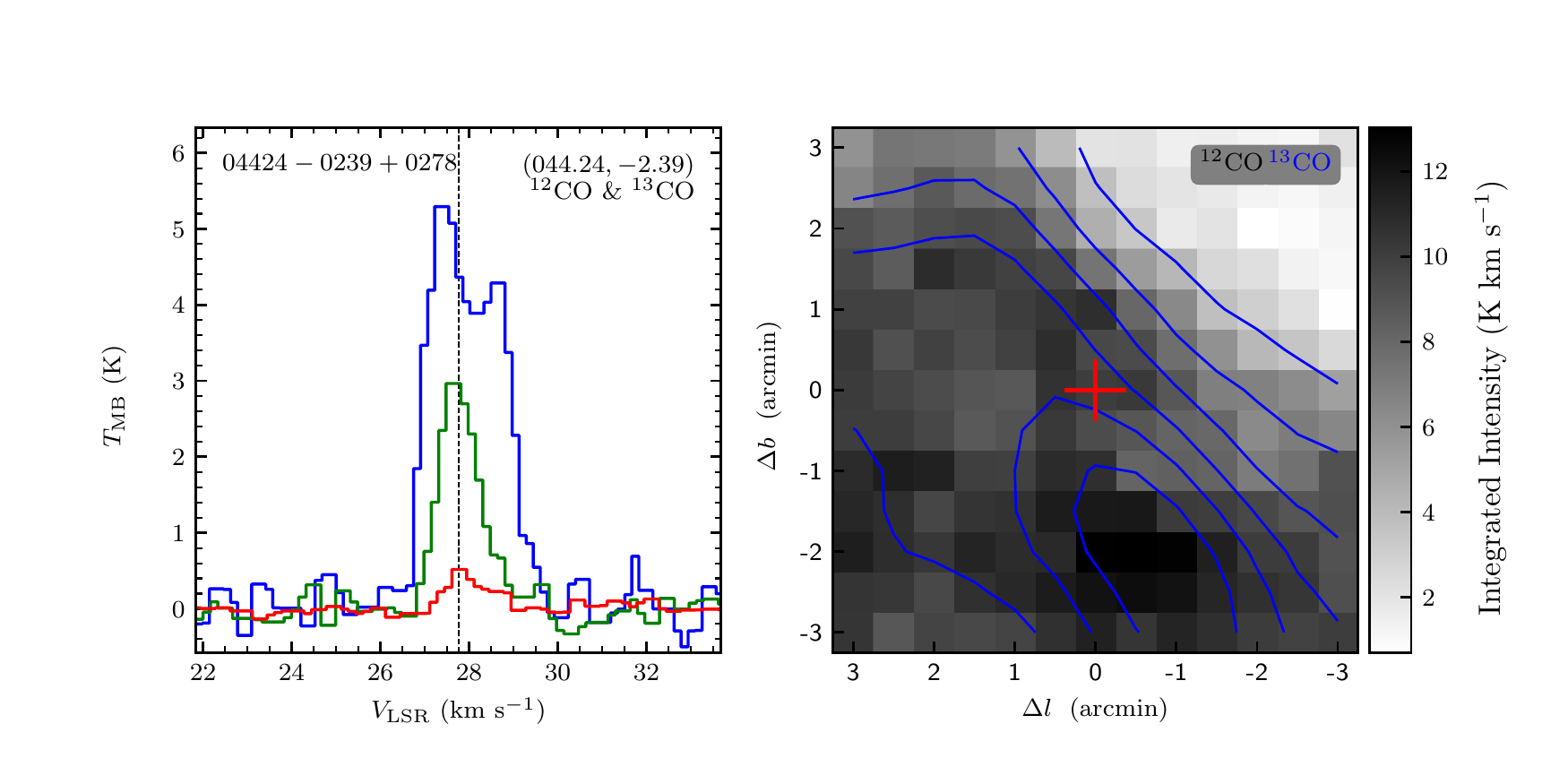}
\vspace{-0.5cm}

\includegraphics[width=9.0cm,angle=0]{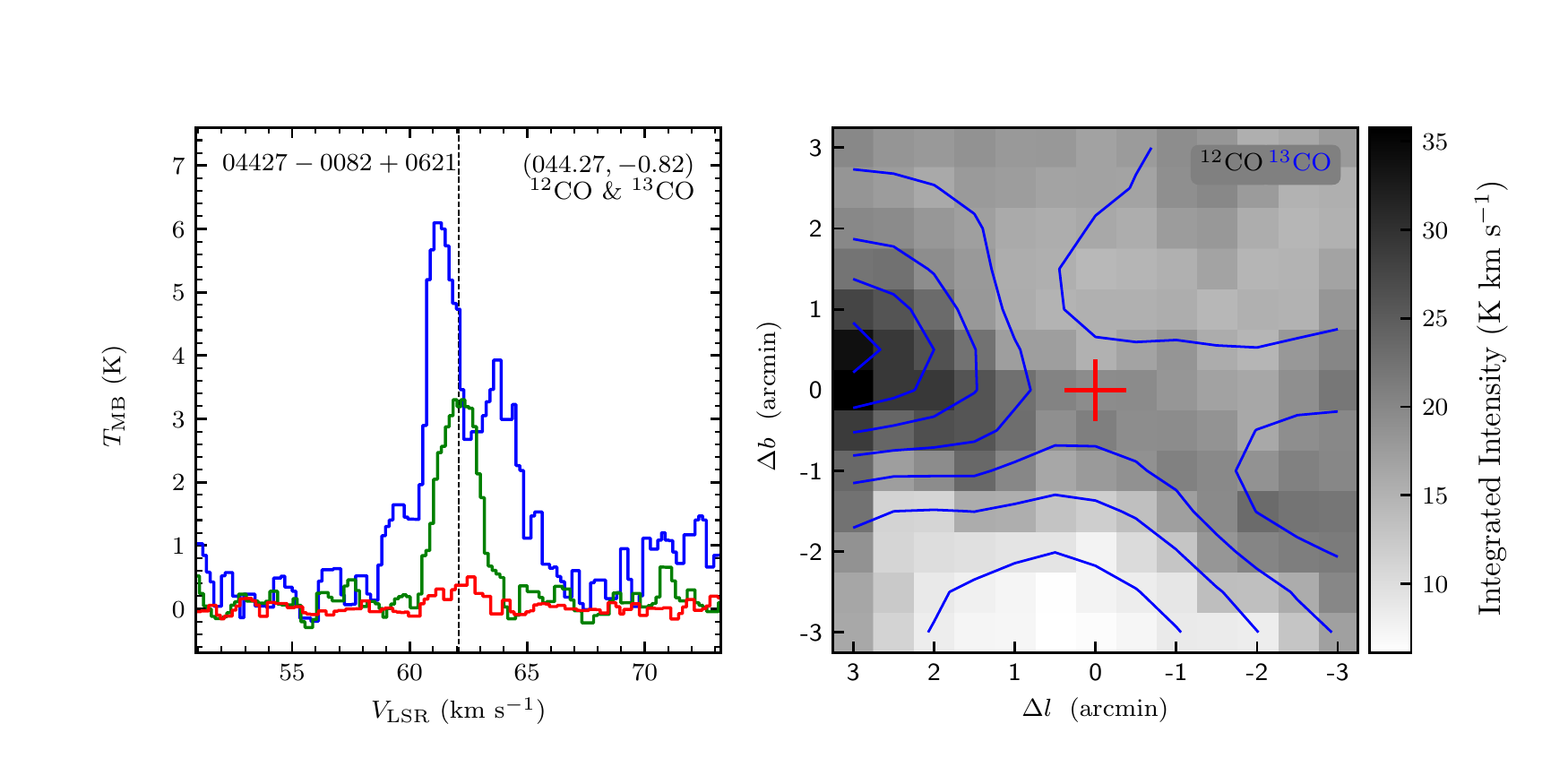}
\includegraphics[width=9.0cm,angle=0]{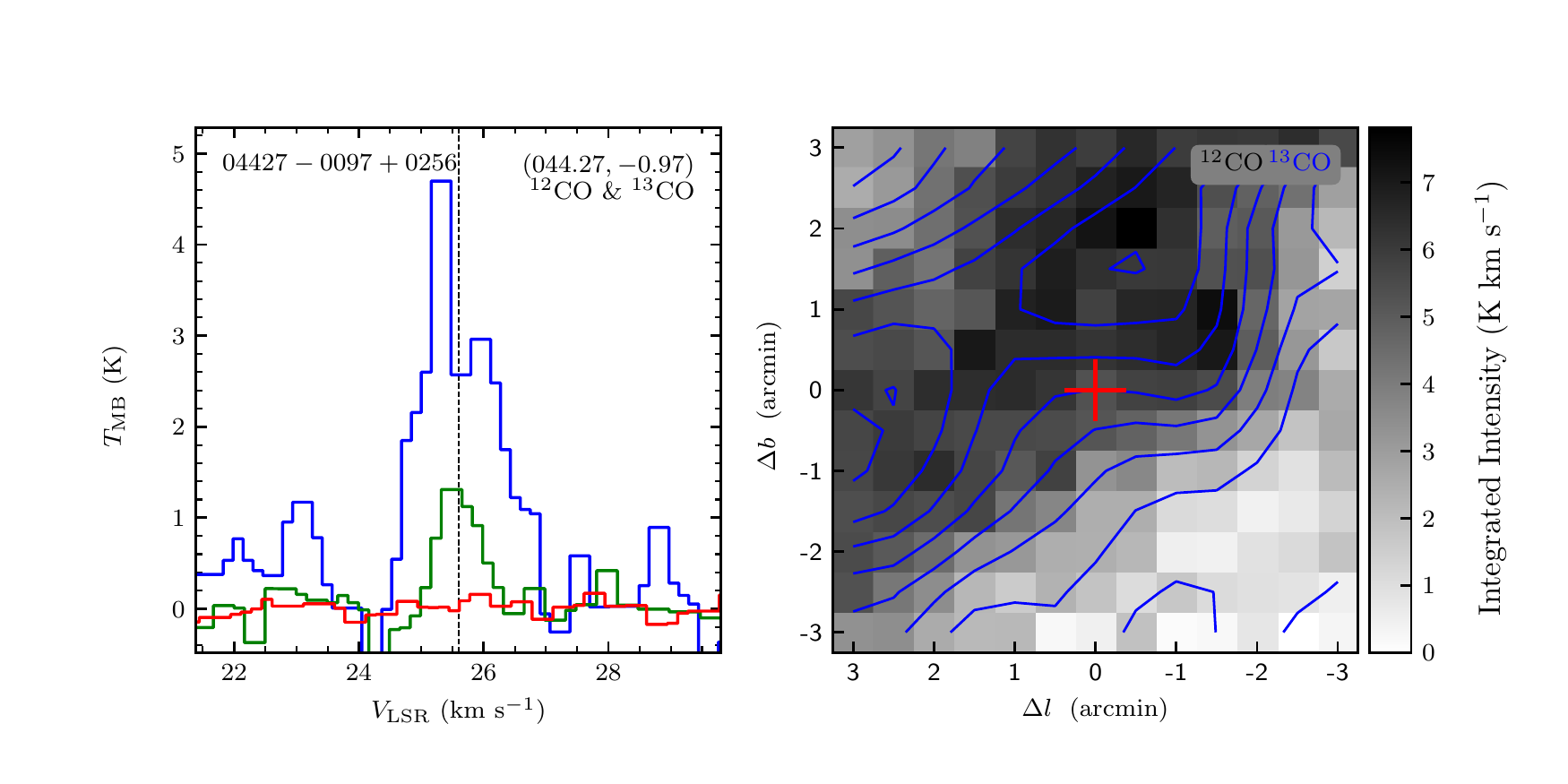}
\vspace{-0.5cm}

\includegraphics[width=9.0cm,angle=0]{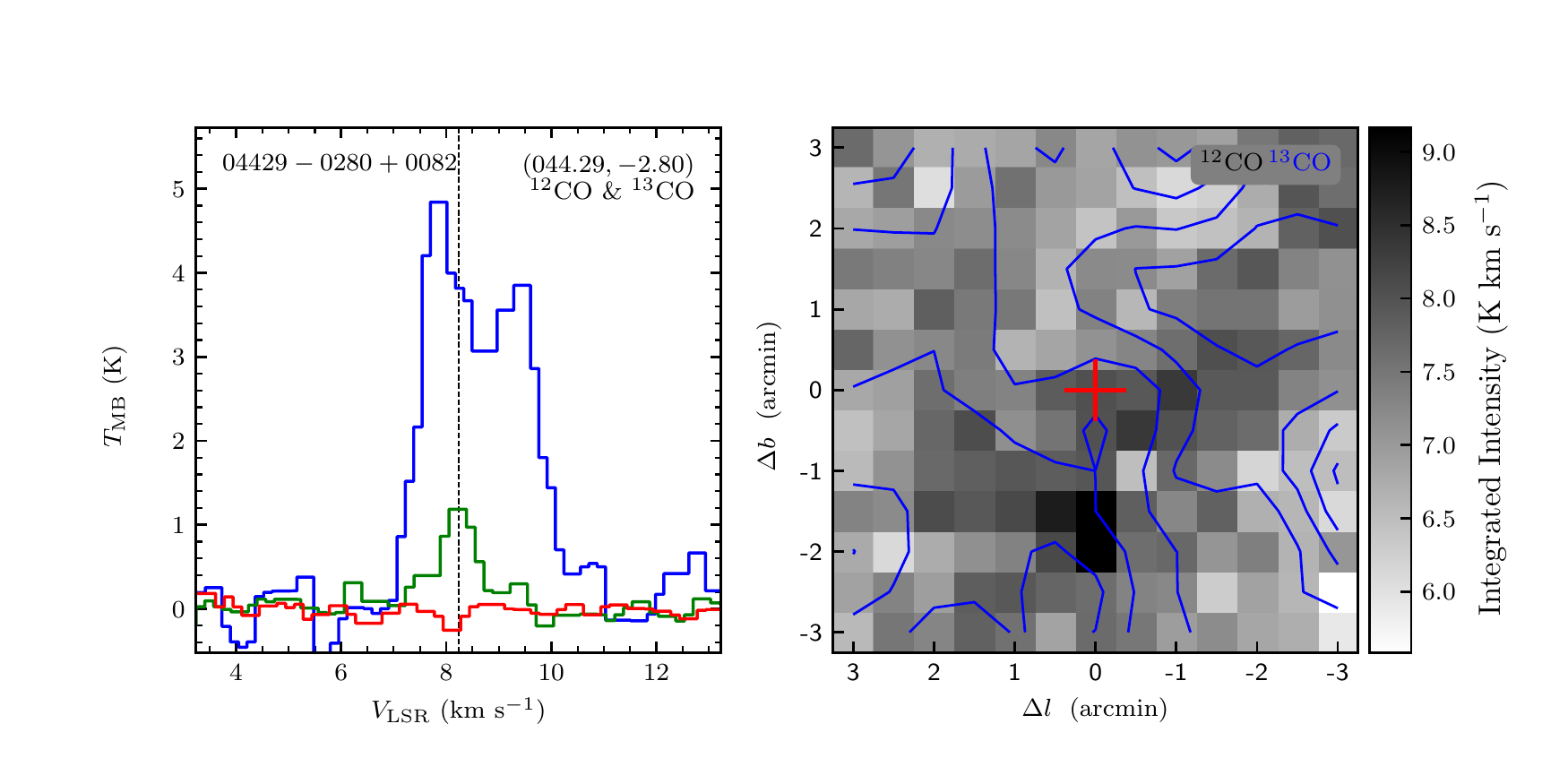}
\includegraphics[width=9.0cm,angle=0]{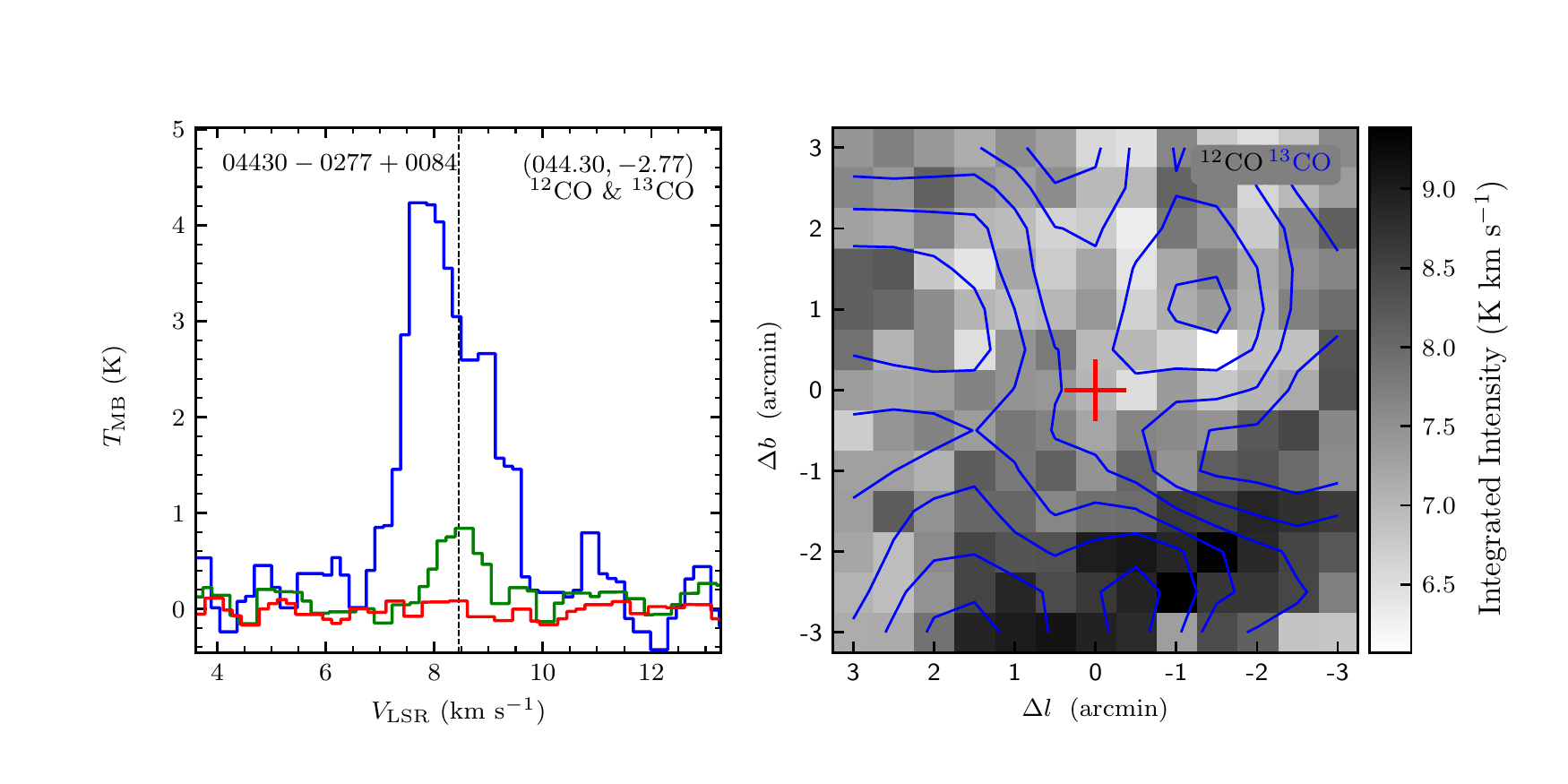}
\end{figure}
\clearpage

\begin{figure}
\includegraphics[width=9.0cm,angle=0]{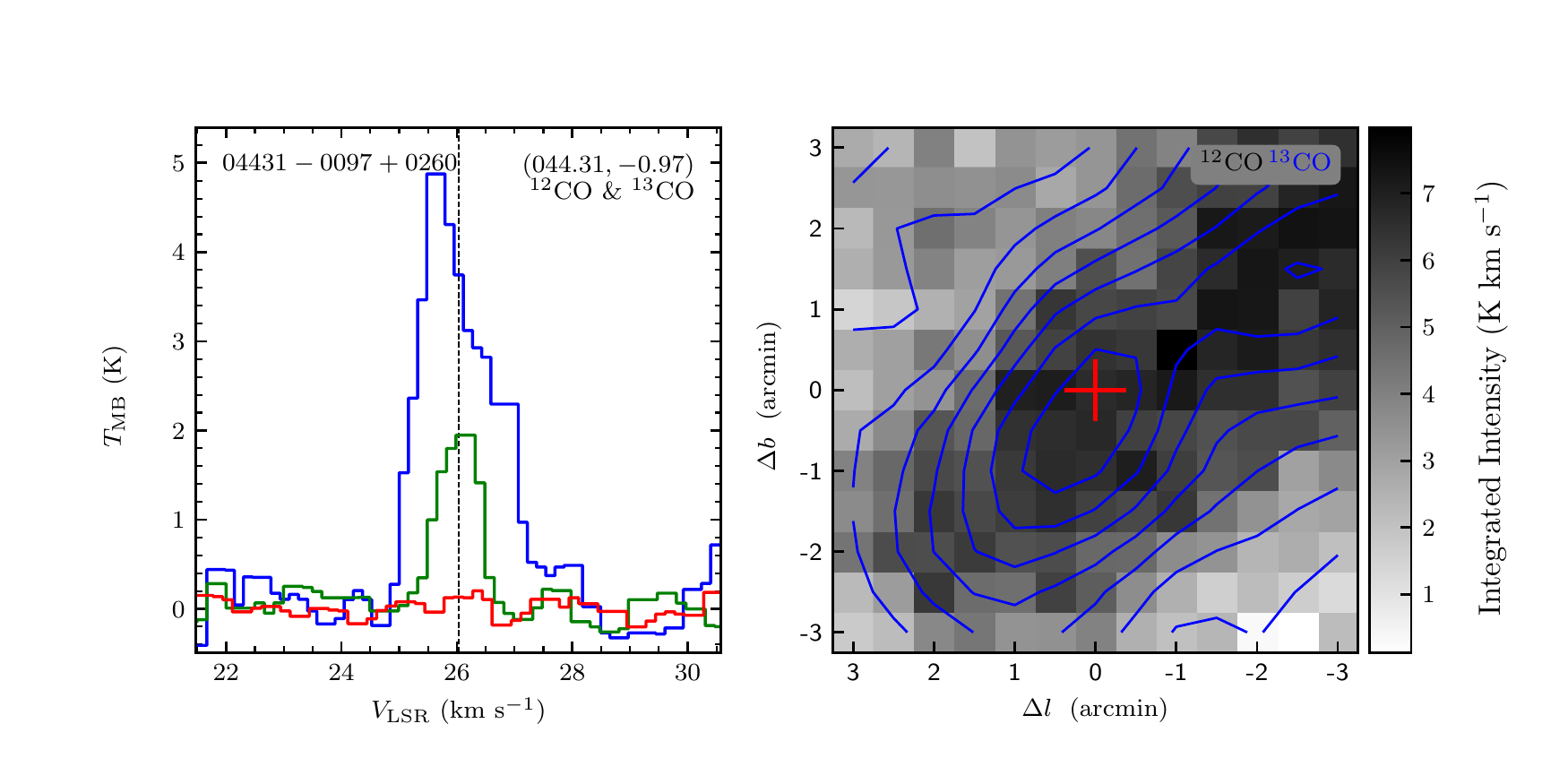}
\includegraphics[width=9.0cm,angle=0]{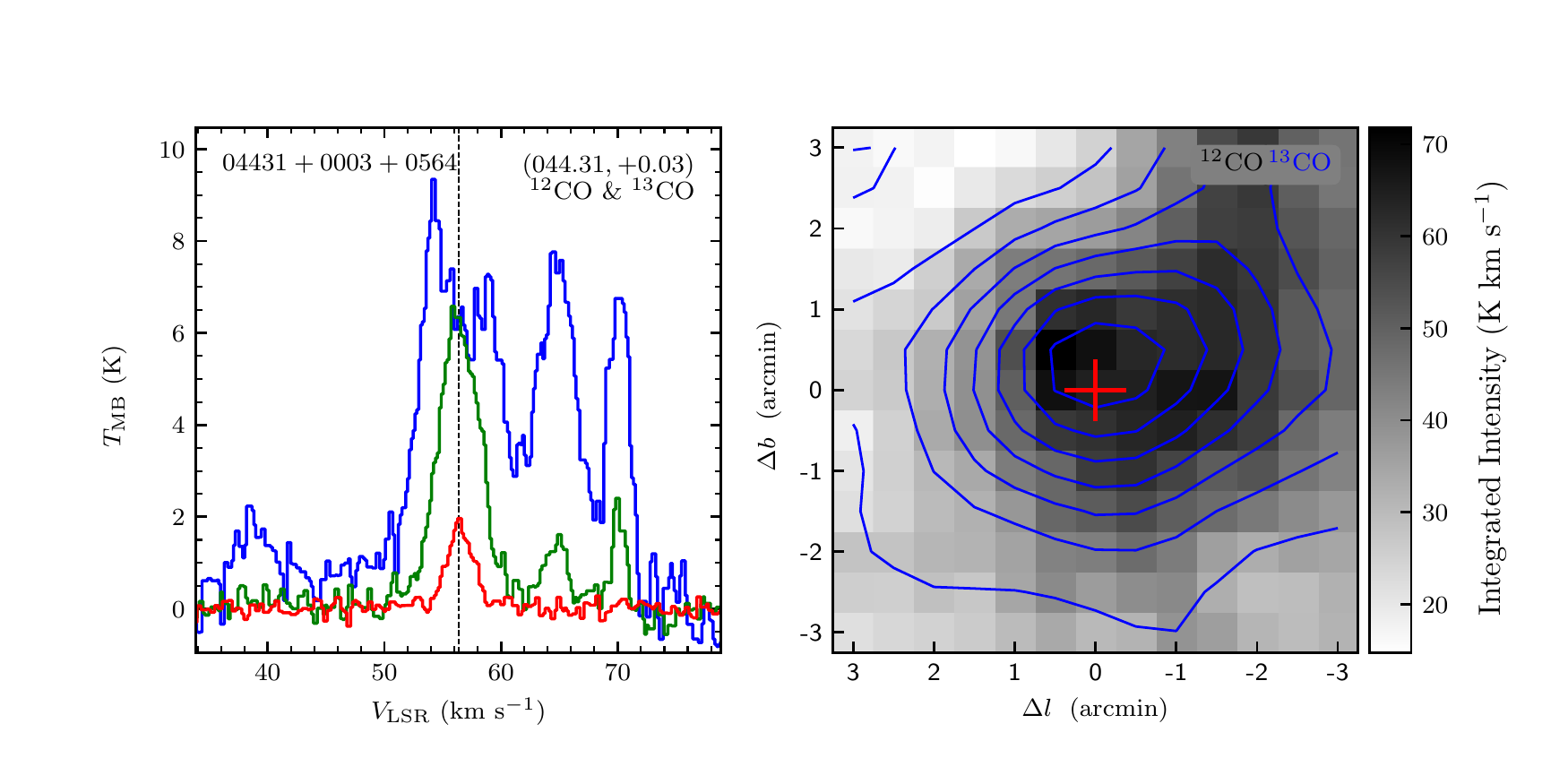}
\vspace{-0.5cm}

\includegraphics[width=9.0cm,angle=0]{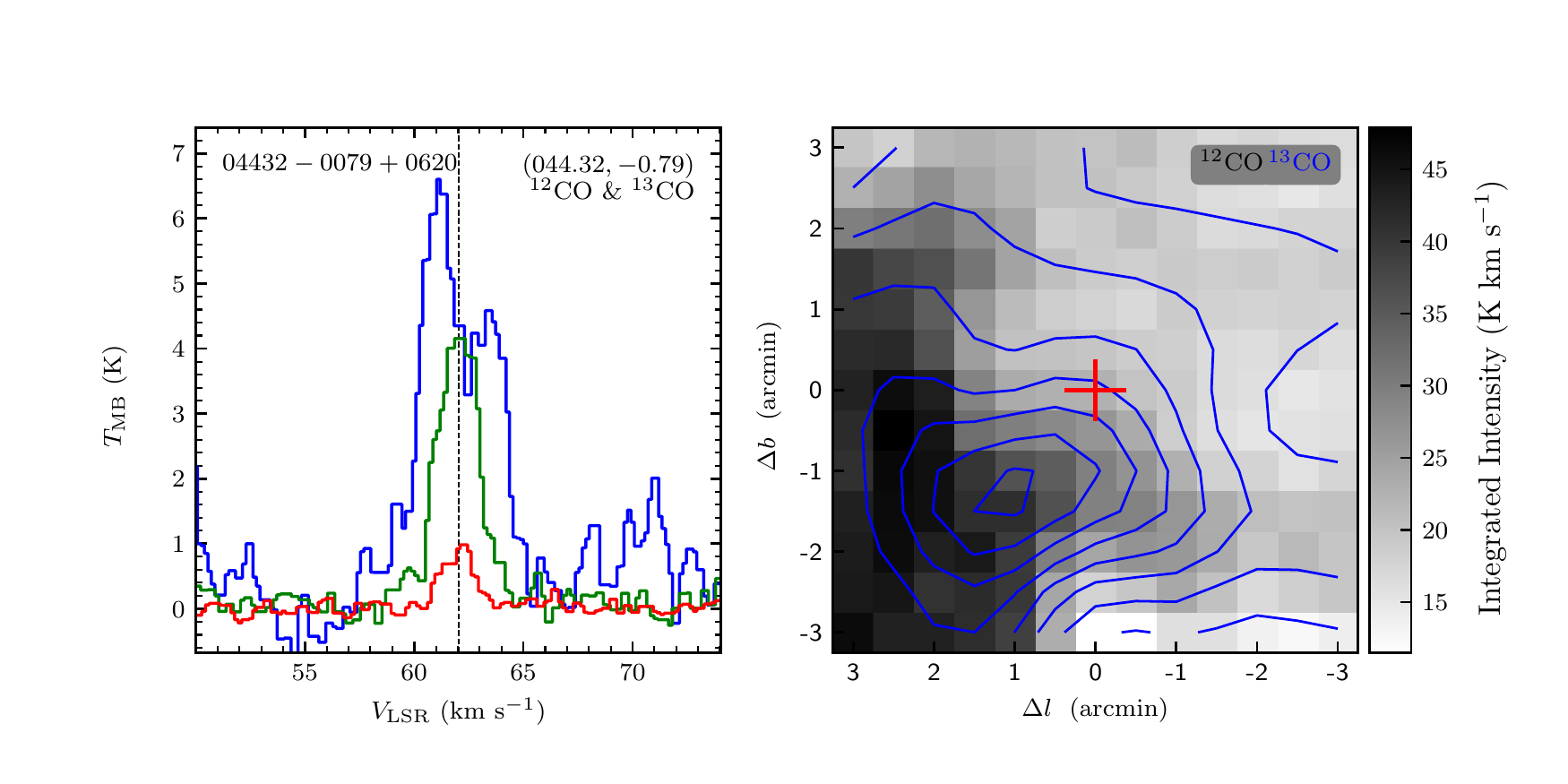}
\includegraphics[width=9.0cm,angle=0]{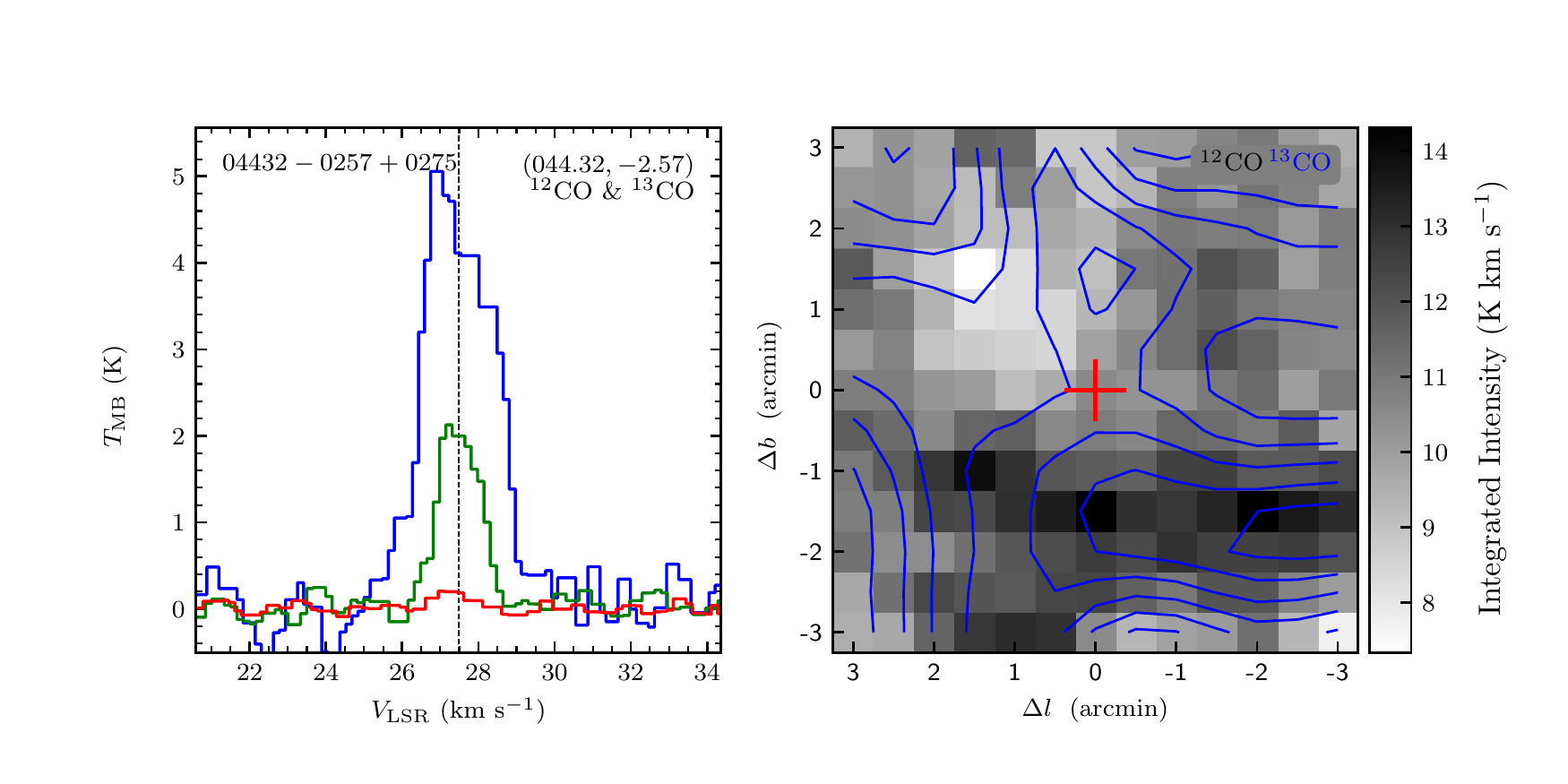}
\vspace{-0.5cm}

\includegraphics[width=9.0cm,angle=0]{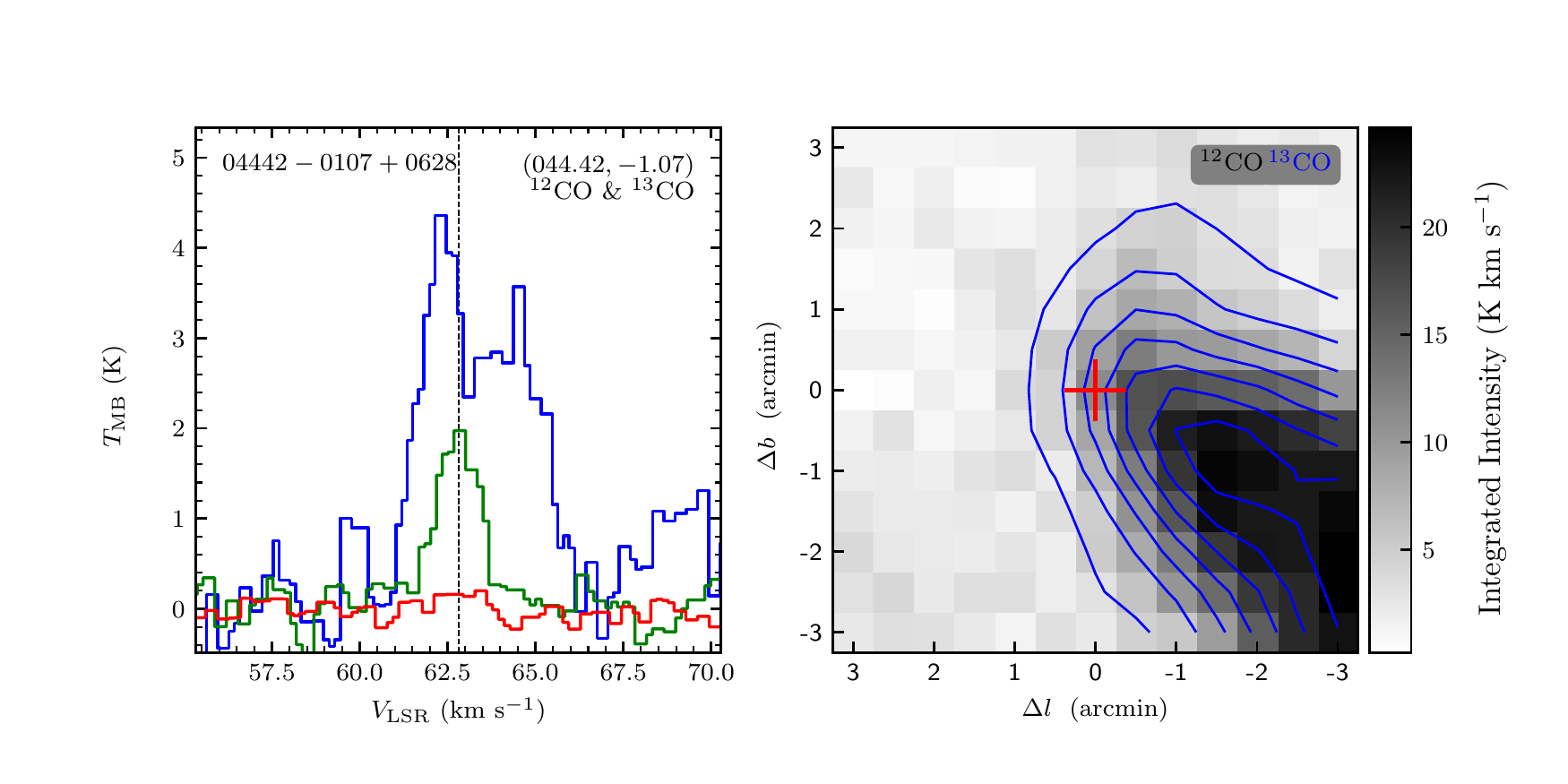}
\includegraphics[width=9.0cm,angle=0]{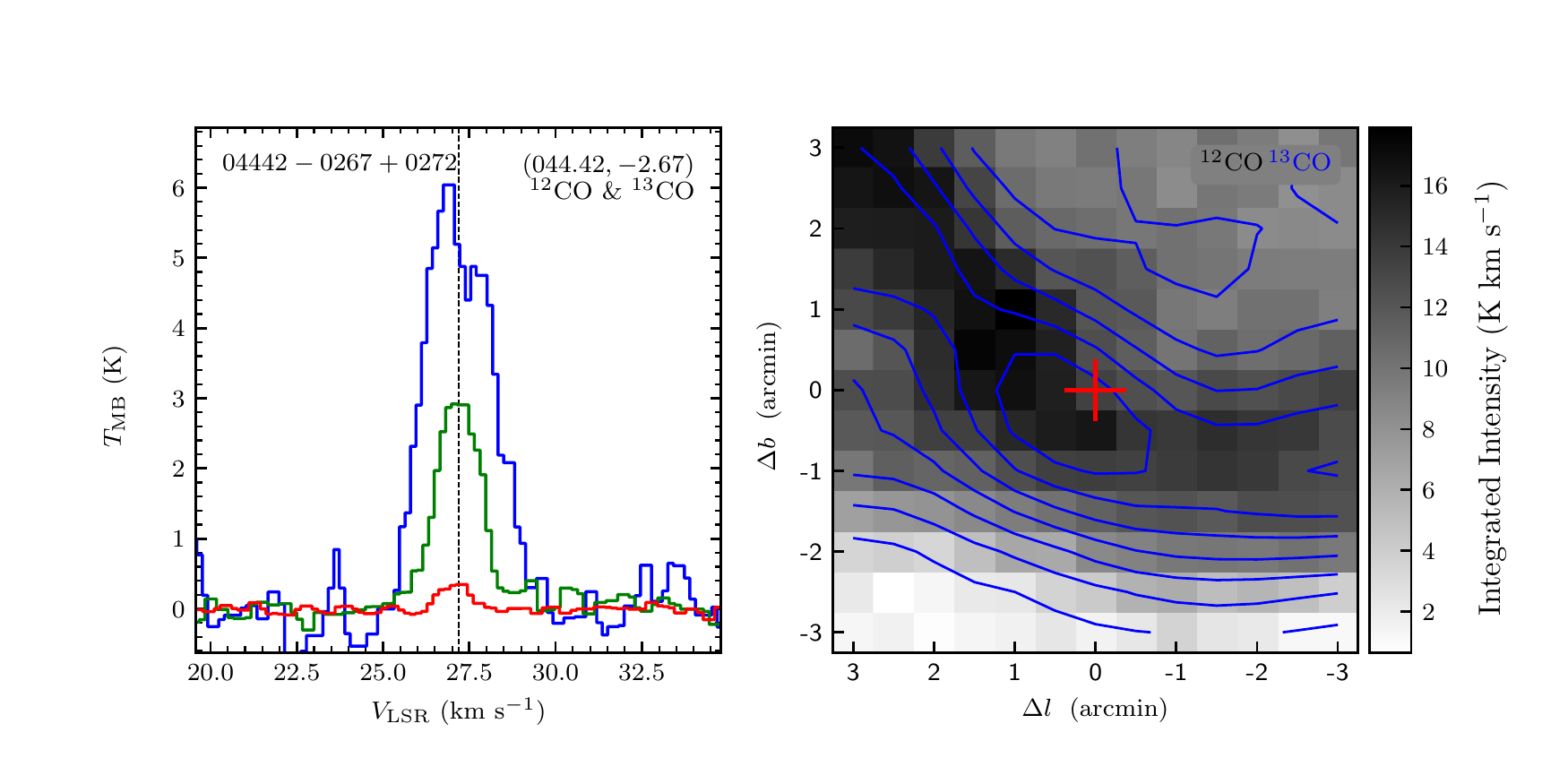}
\vspace{-0.5cm}

\includegraphics[width=9.0cm,angle=0]{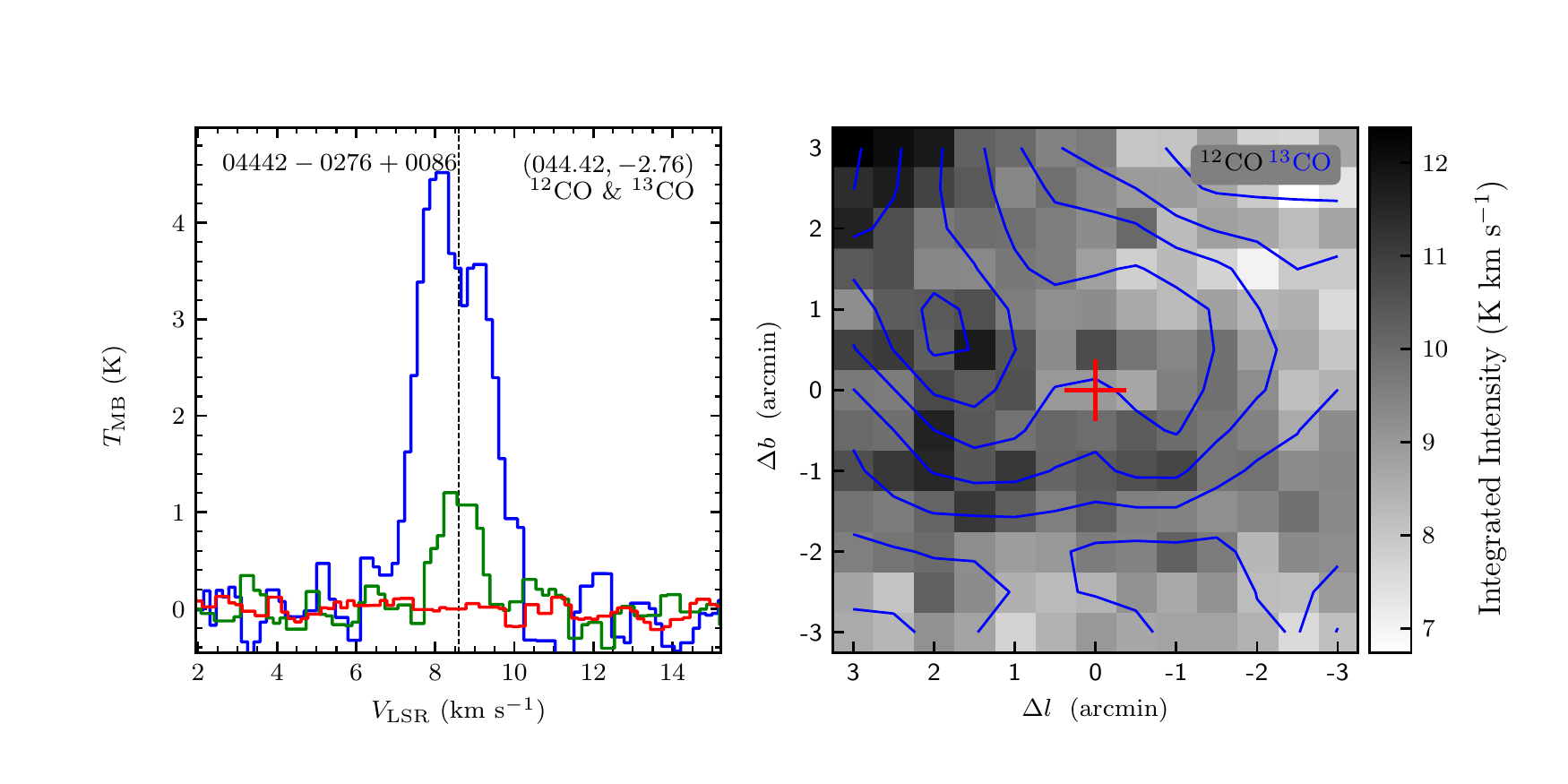}
\includegraphics[width=9.0cm,angle=0]{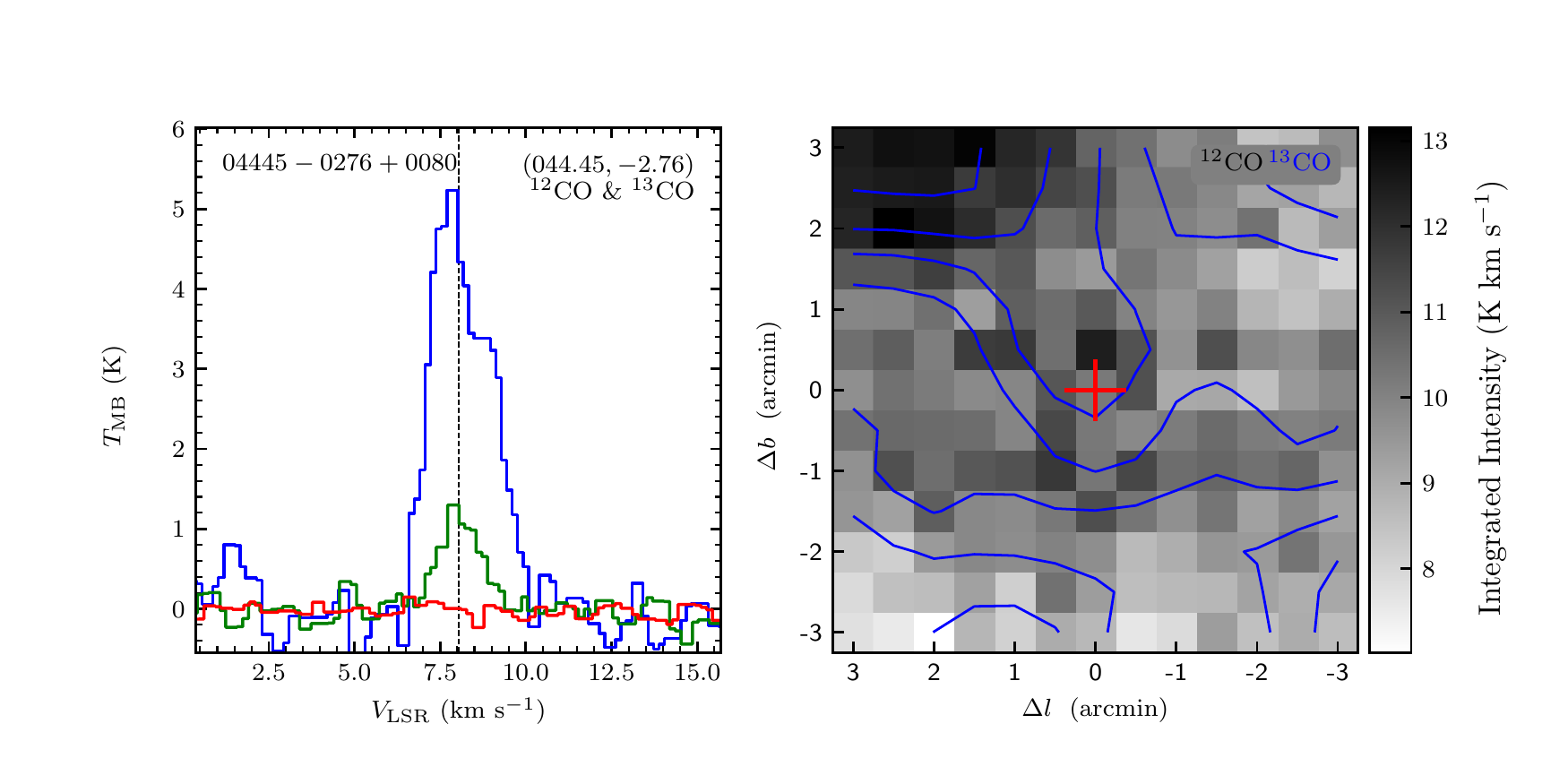}
\vspace{-0.5cm}

\includegraphics[width=9.0cm,angle=0]{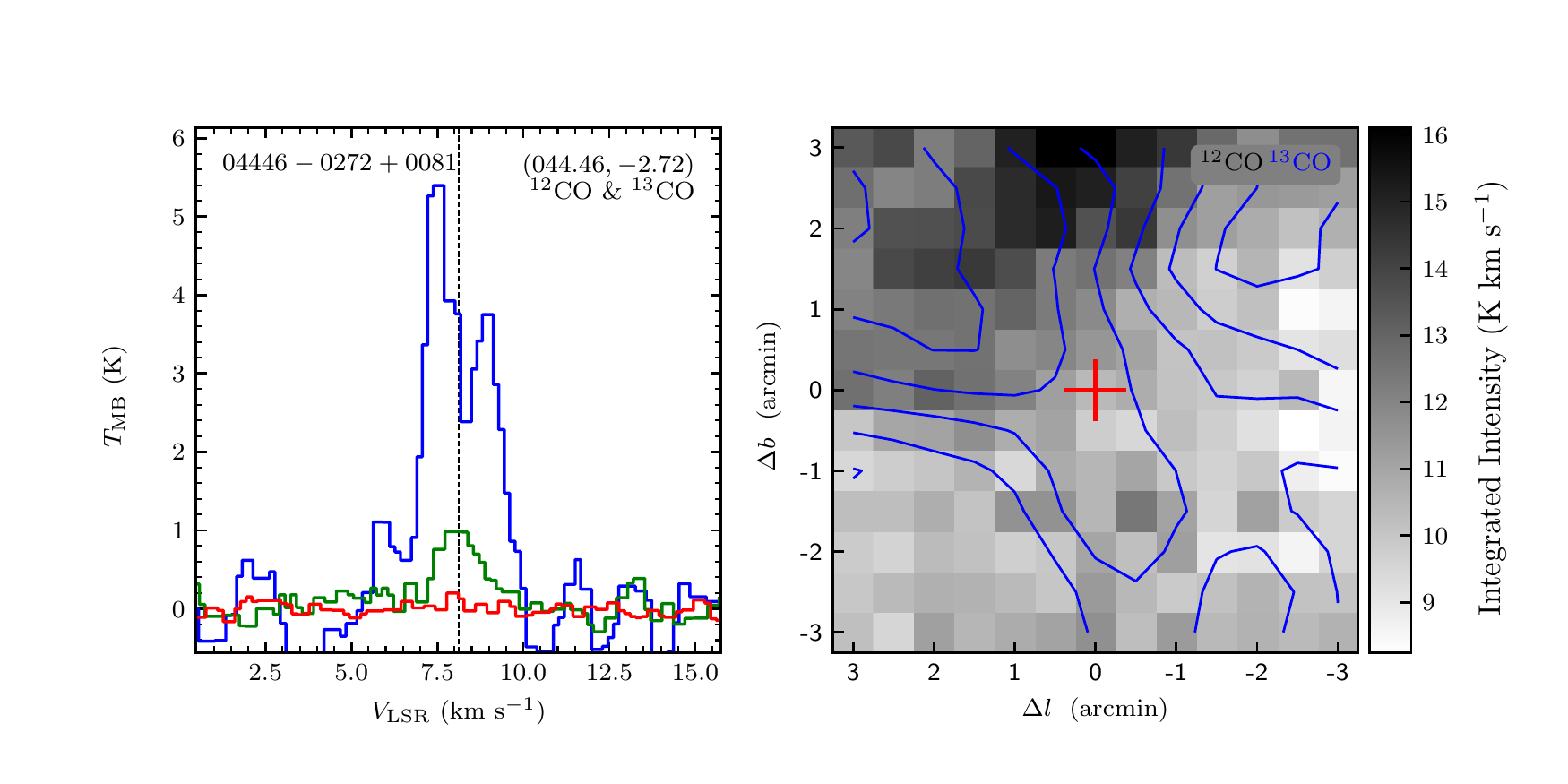}
\includegraphics[width=9.0cm,angle=0]{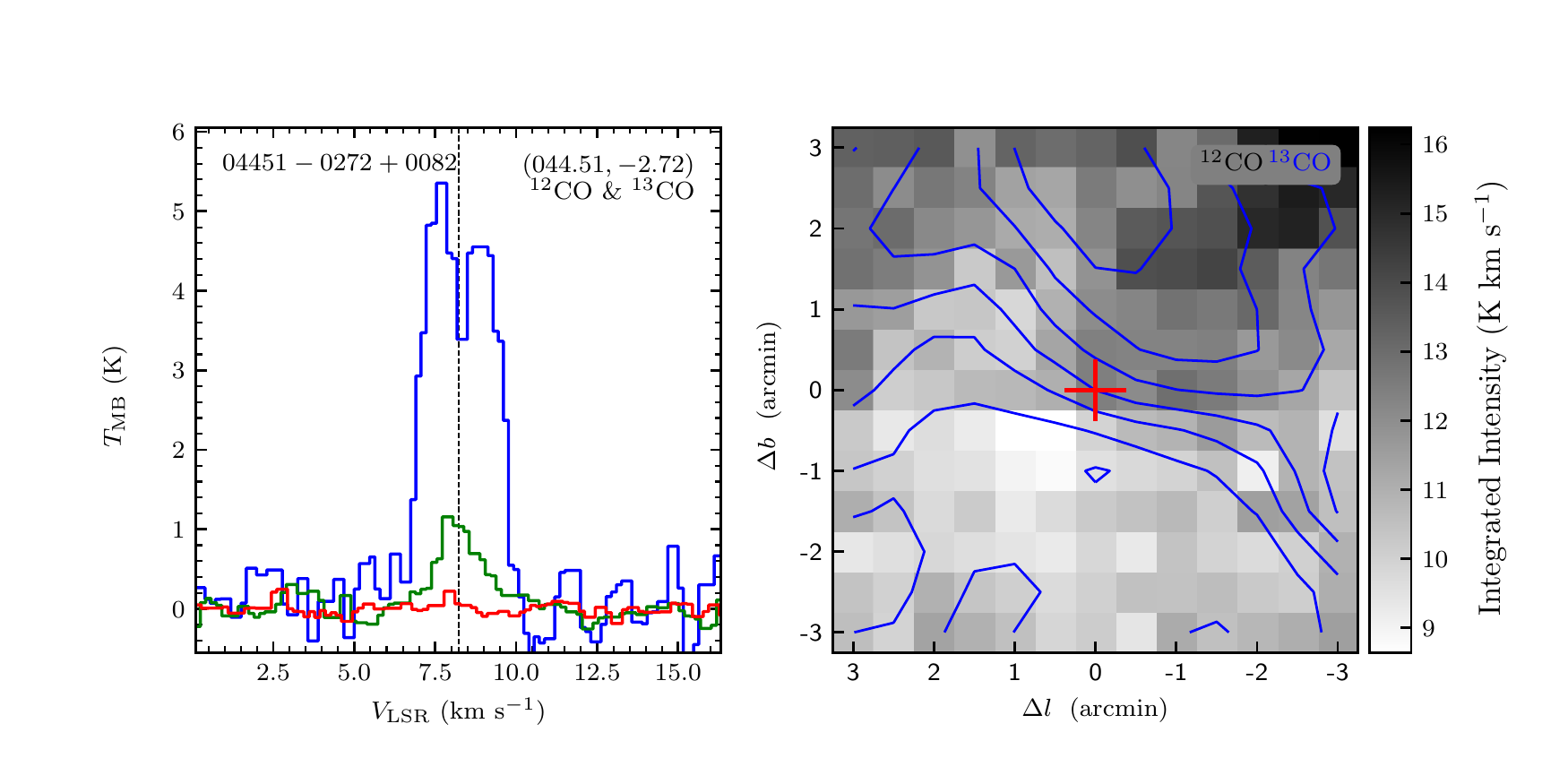}
\end{figure}
\clearpage

\begin{figure}
\includegraphics[width=9.0cm,angle=0]{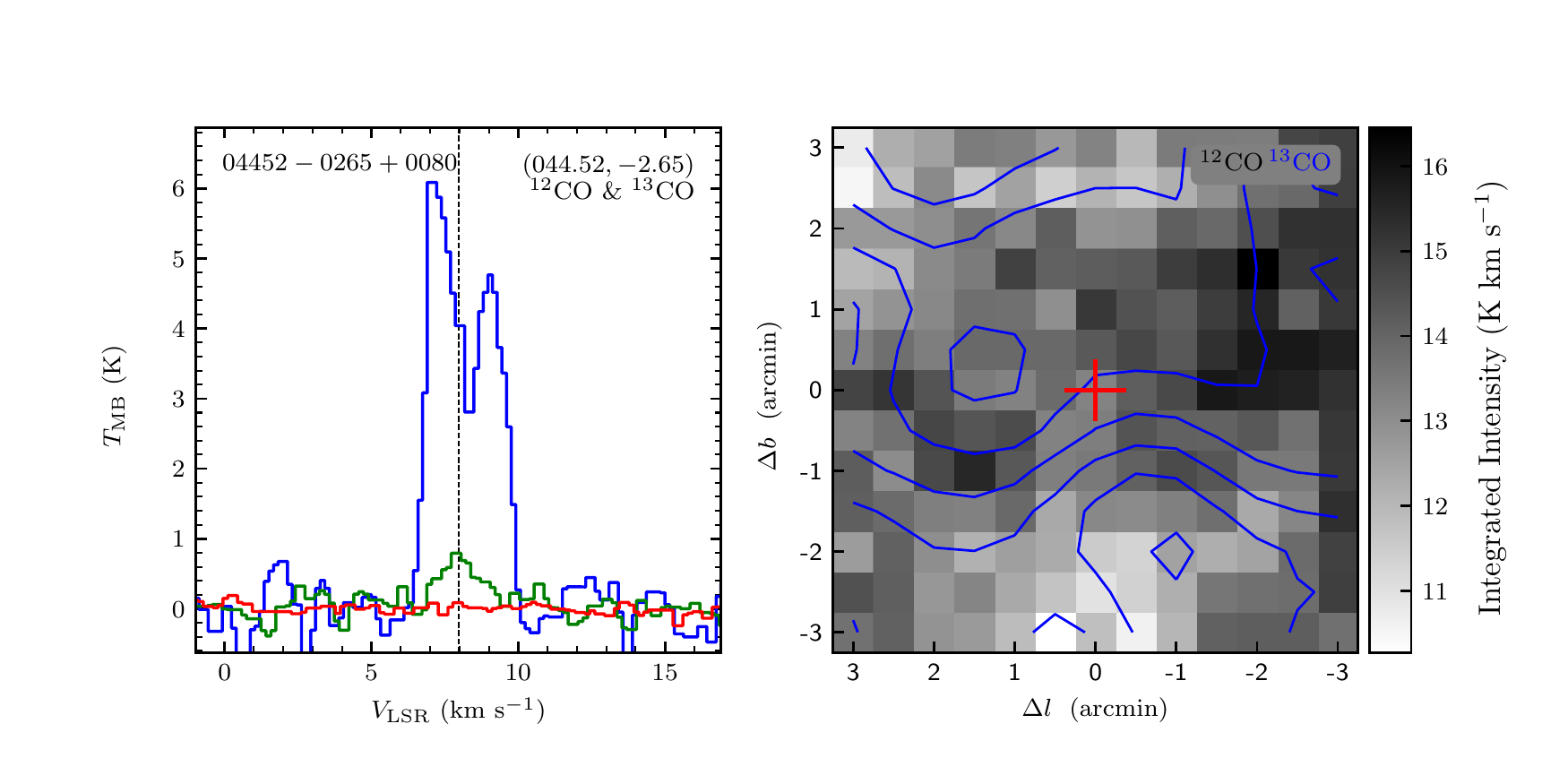}
\includegraphics[width=9.0cm,angle=0]{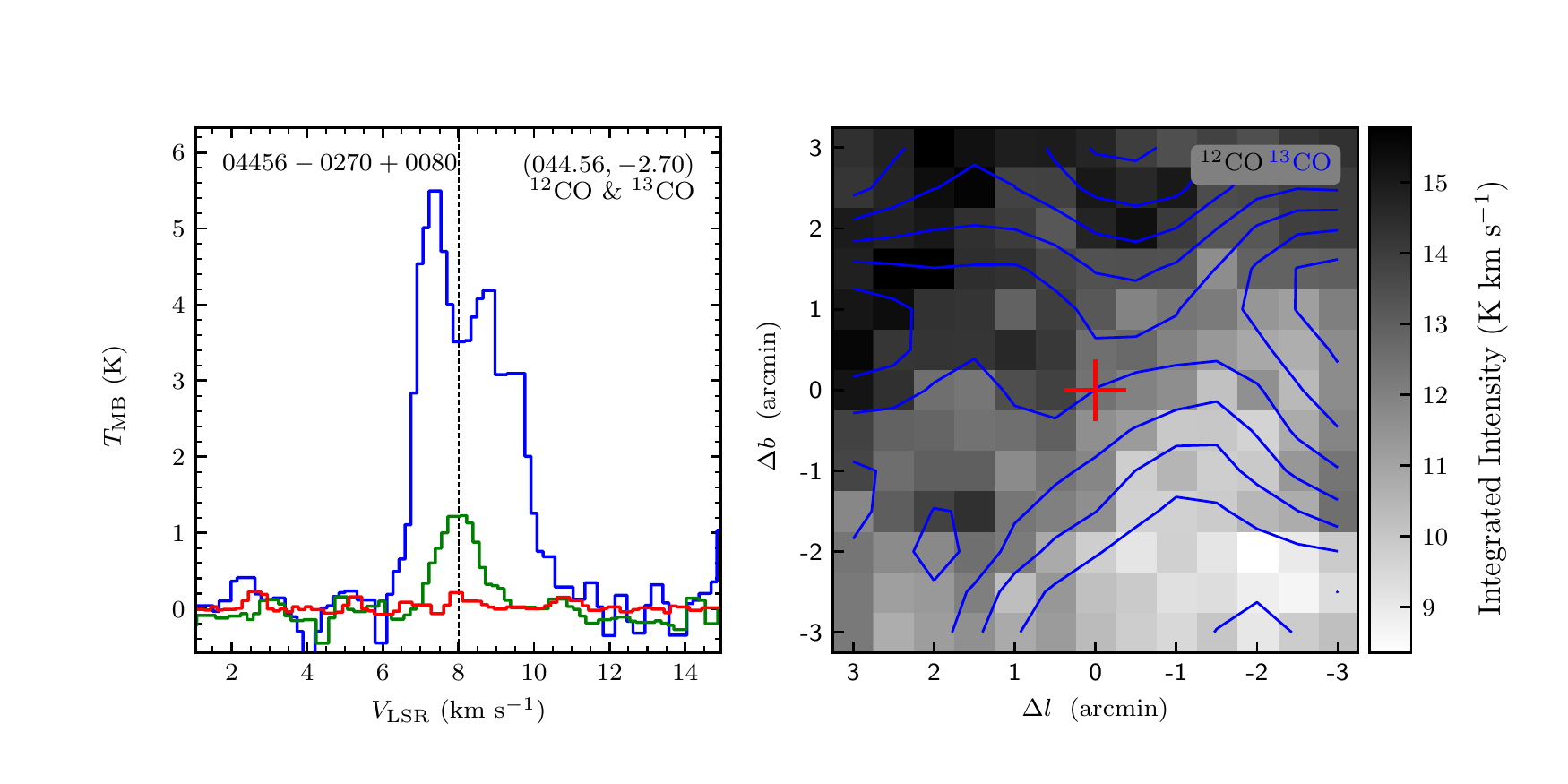}
\vspace{-0.5cm}

\includegraphics[width=9.0cm,angle=0]{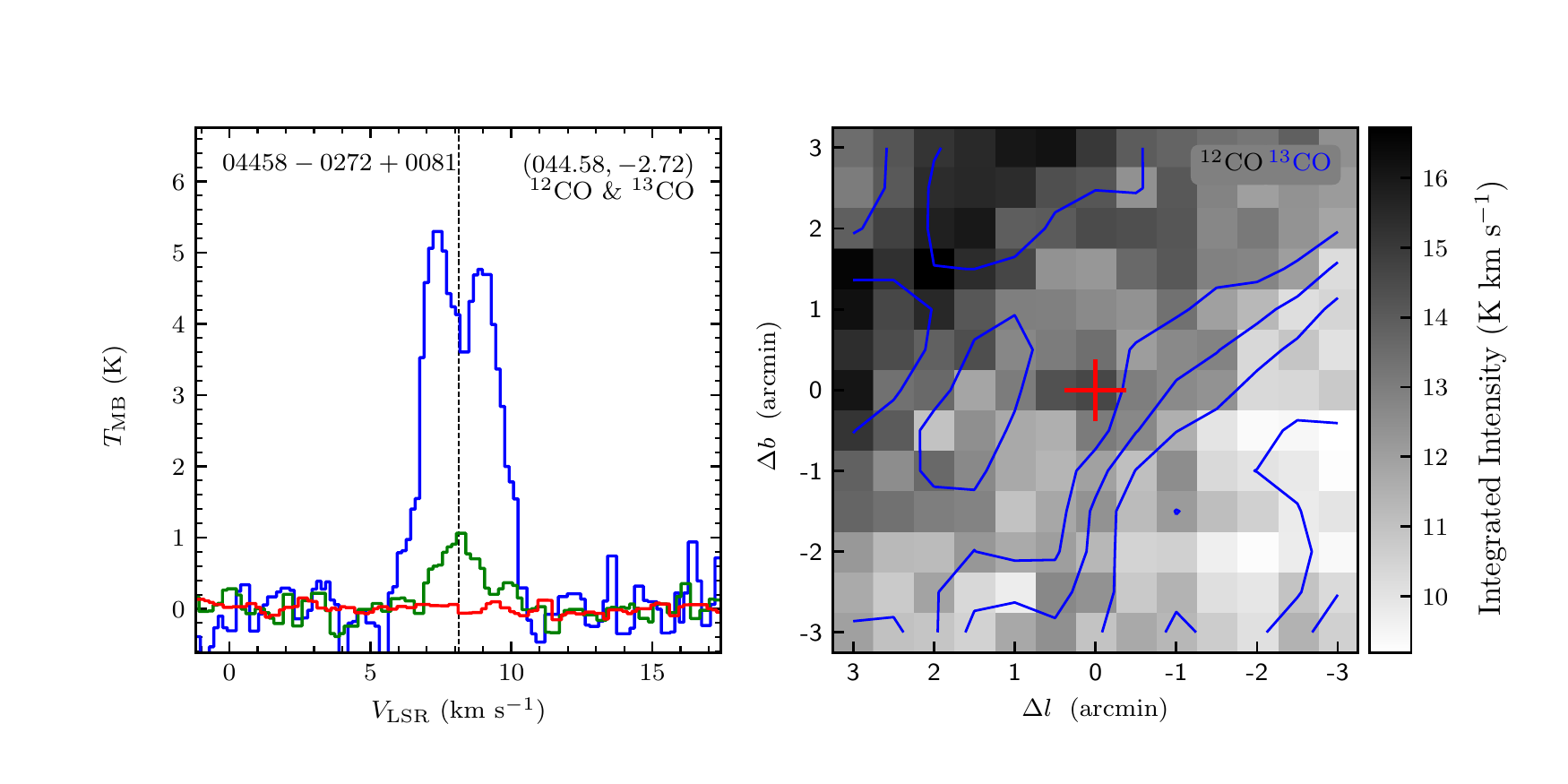}
\includegraphics[width=9.0cm,angle=0]{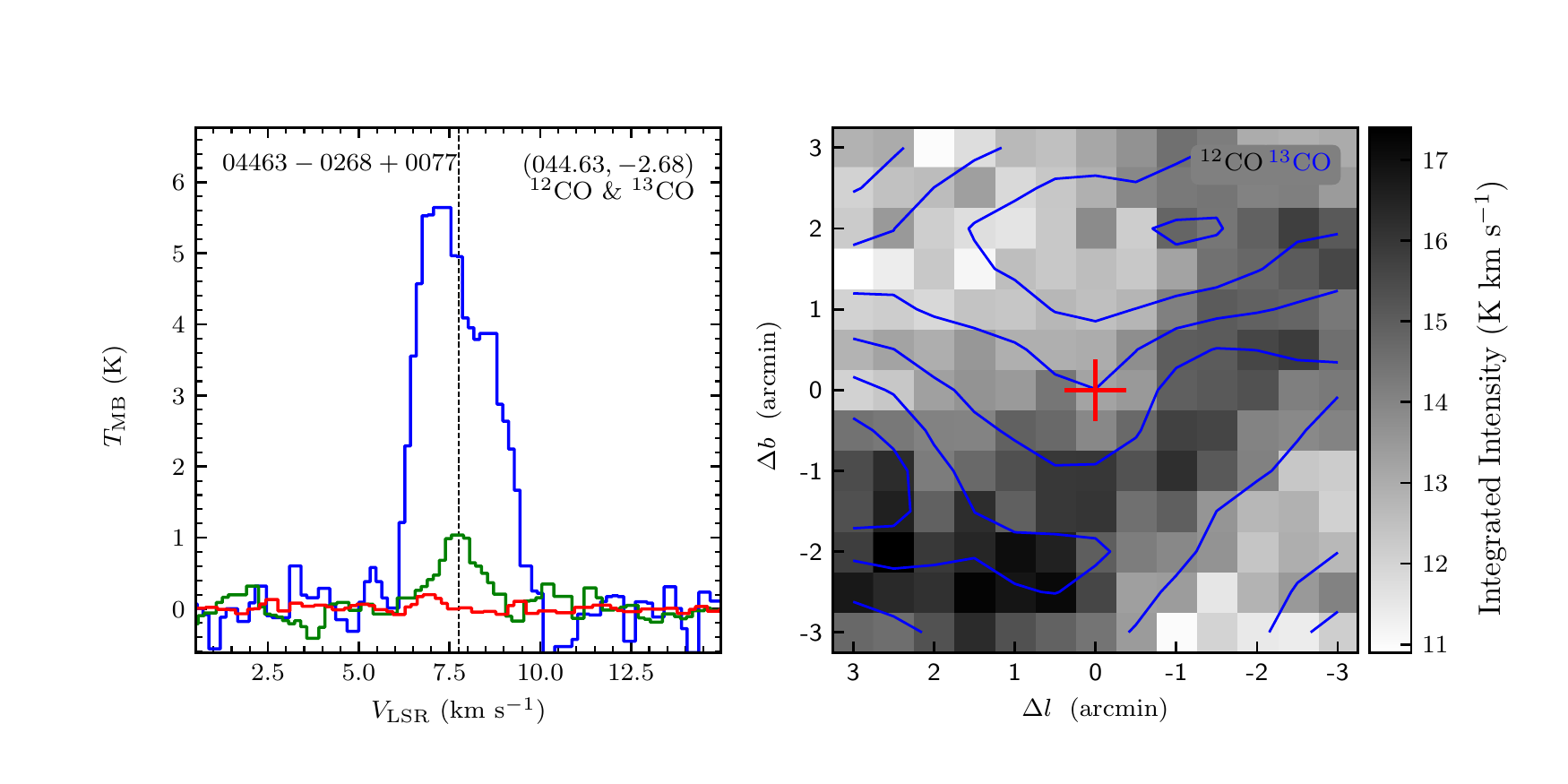}
\vspace{-0.5cm}

\includegraphics[width=9.0cm,angle=0]{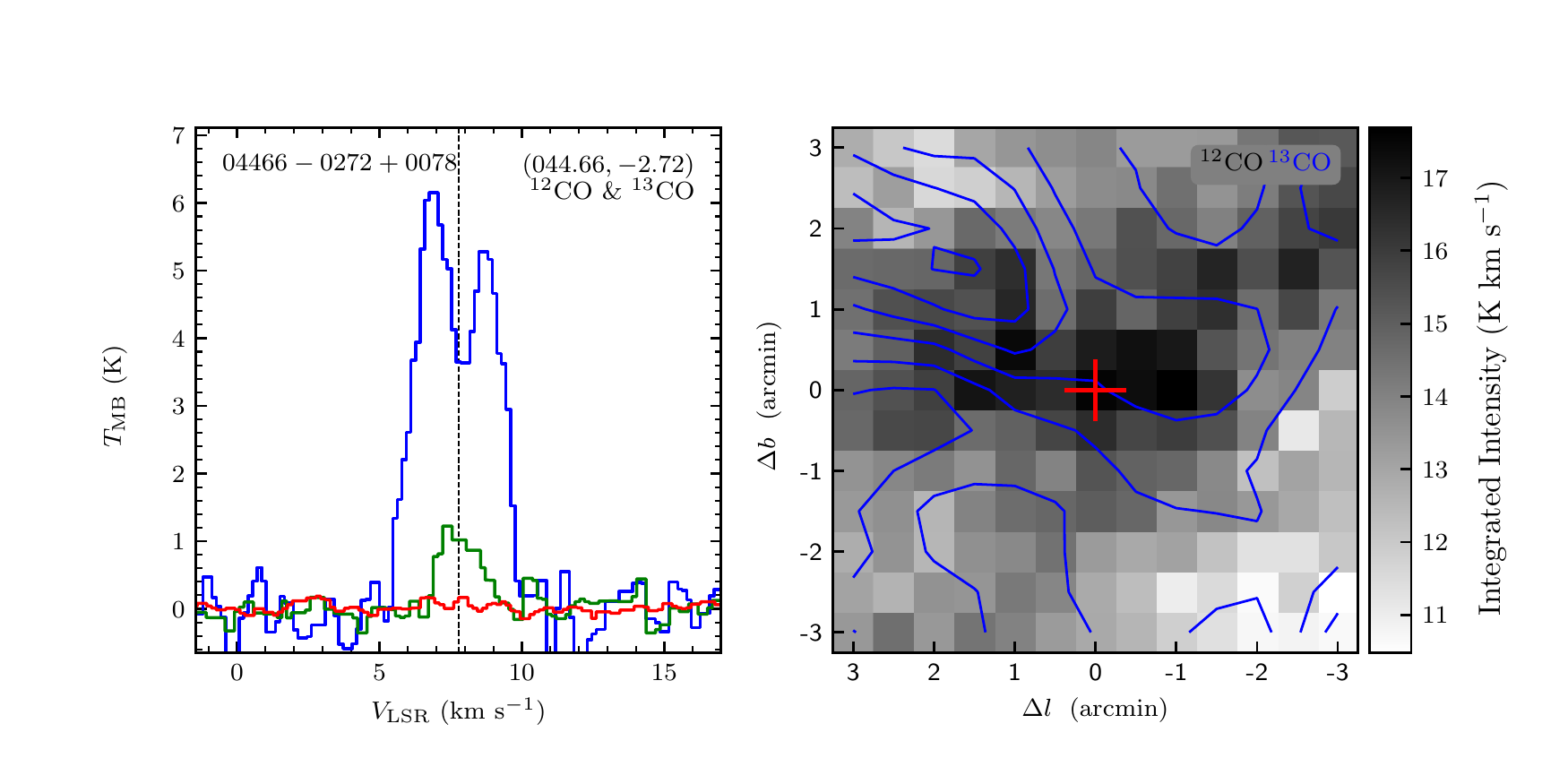}
\includegraphics[width=9.0cm,angle=0]{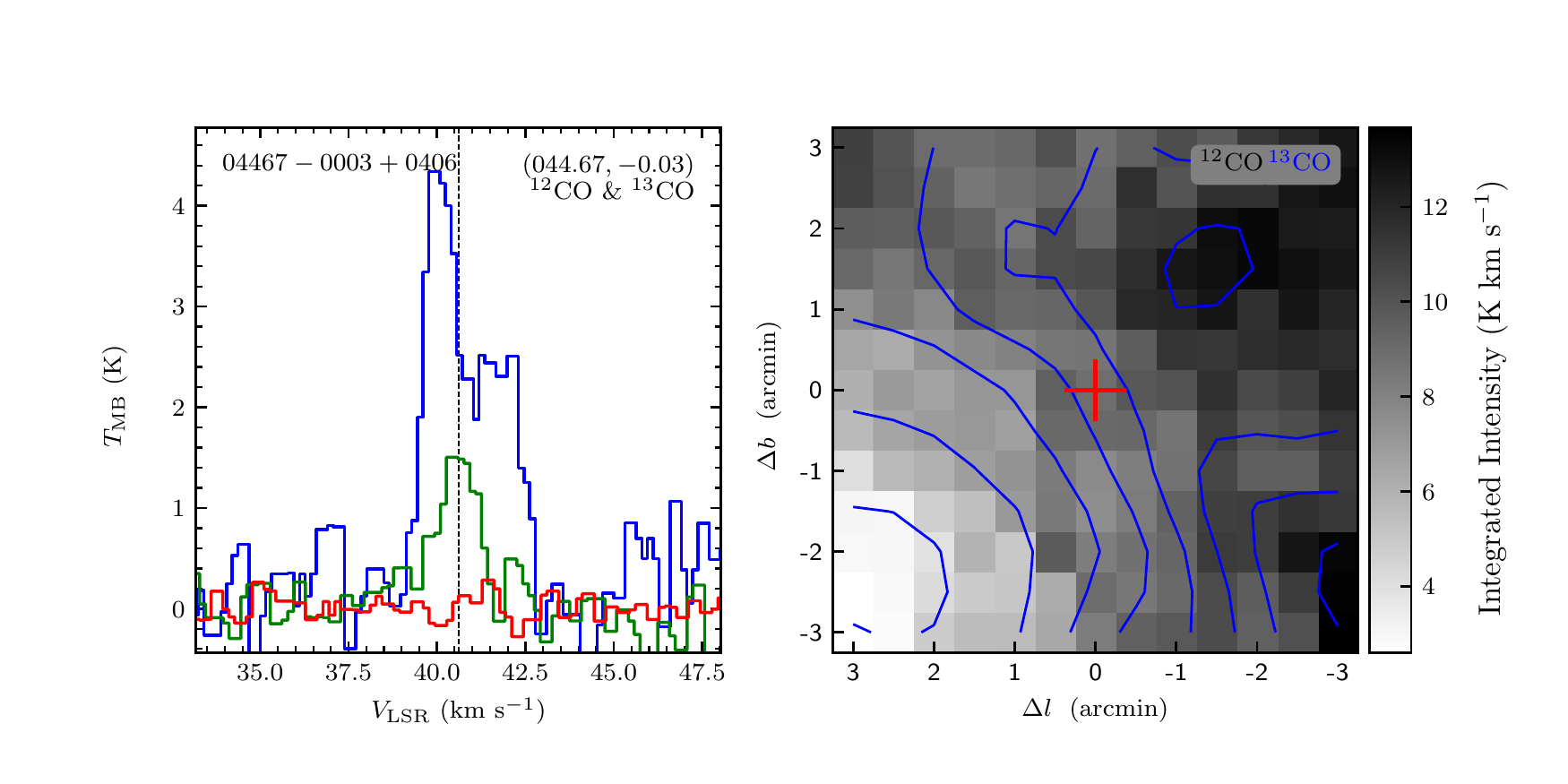}
\vspace{-0.5cm}

\includegraphics[width=9.0cm,angle=0]{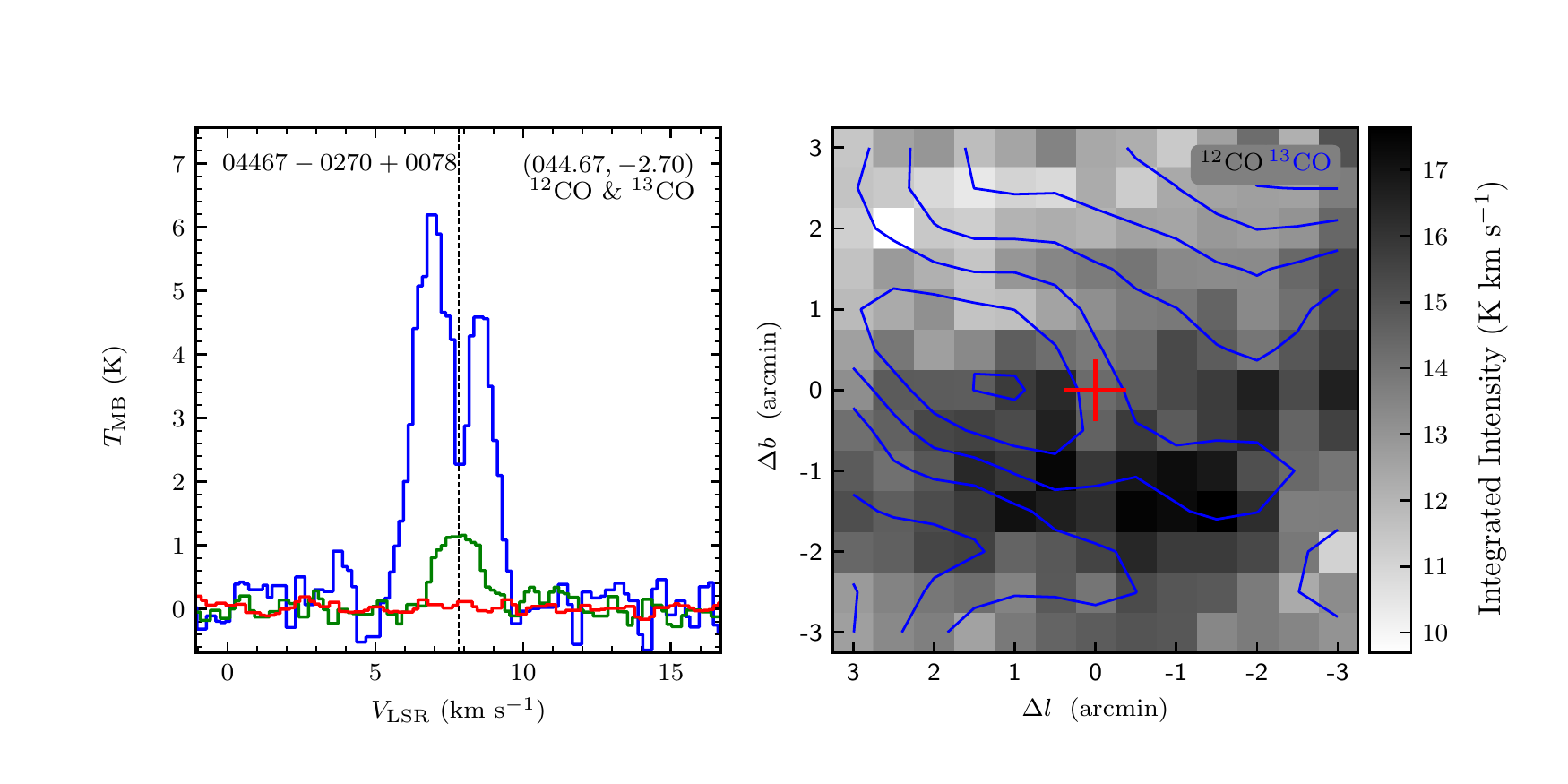}
\includegraphics[width=9.0cm,angle=0]{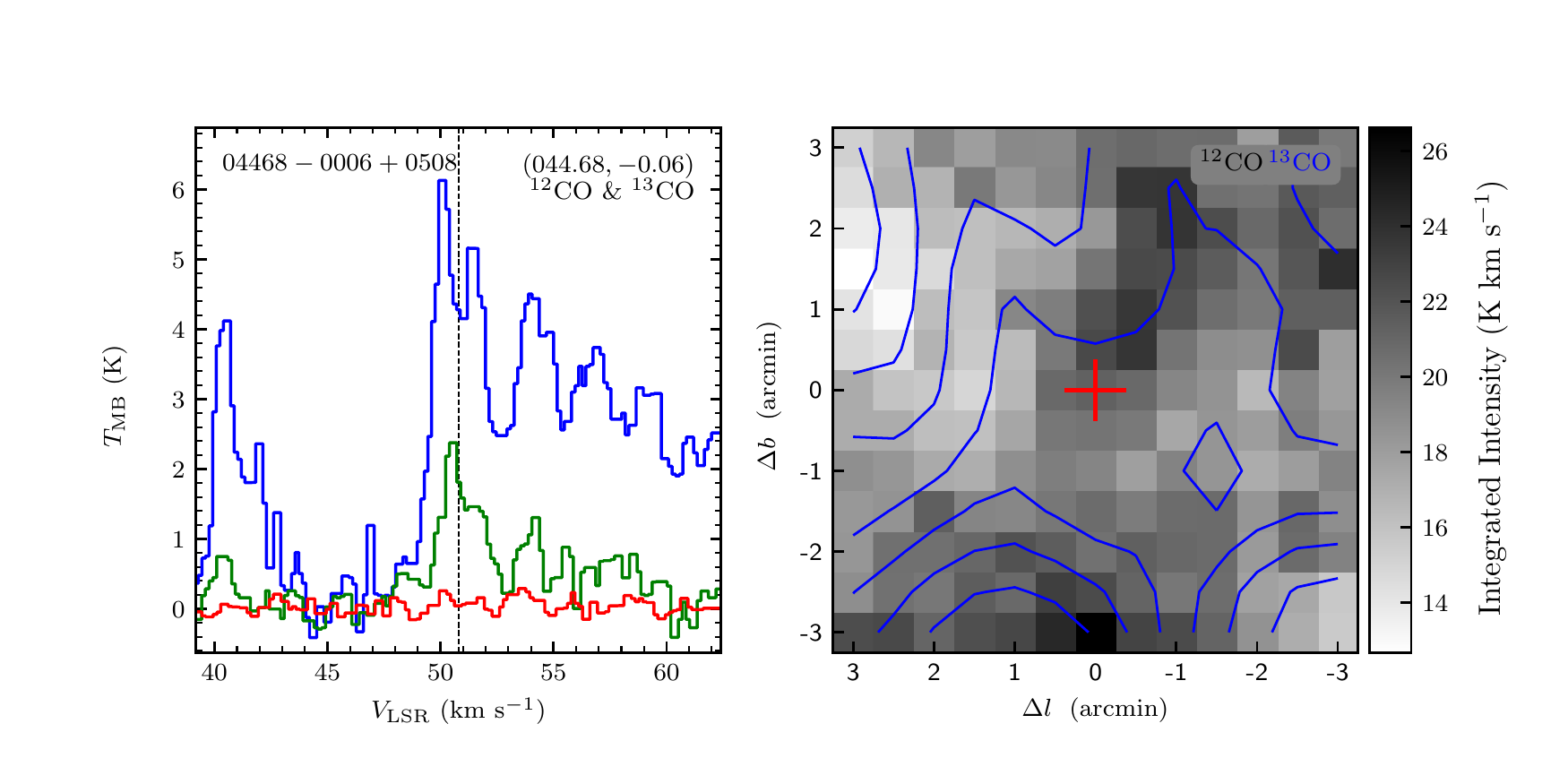}
\vspace{-0.5cm}

\includegraphics[width=9.0cm,angle=0]{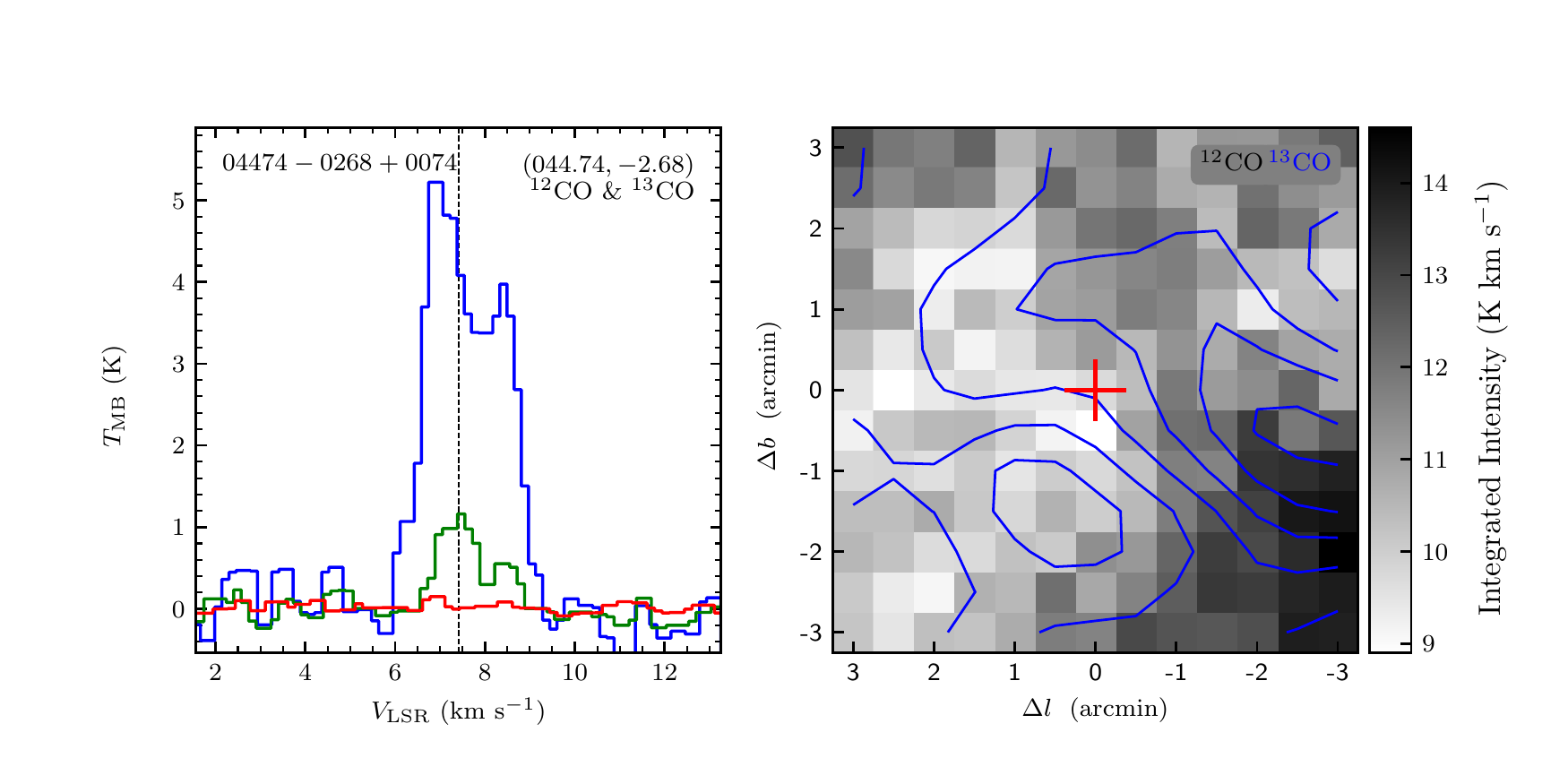}
\includegraphics[width=9.0cm,angle=0]{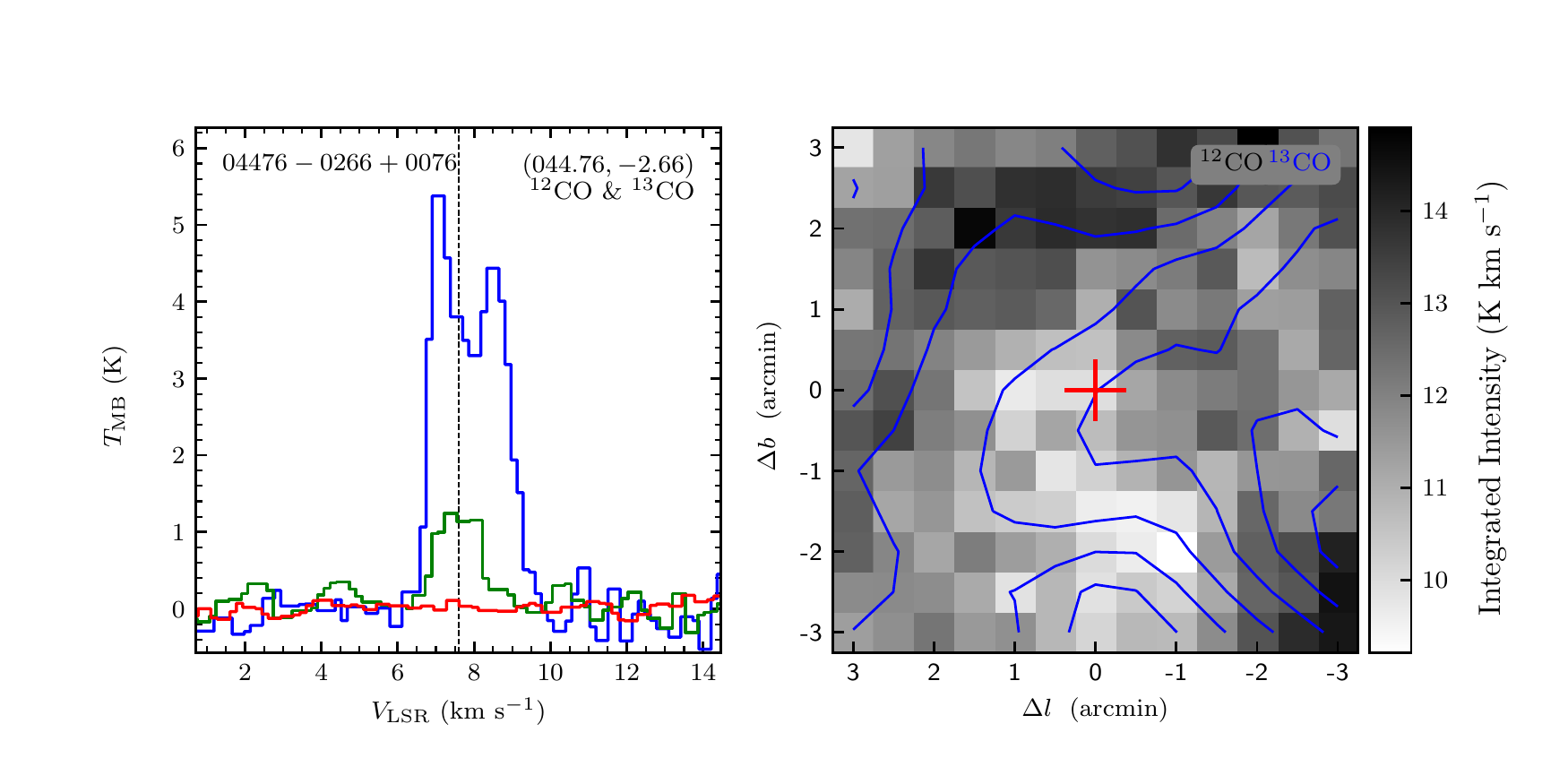}
\end{figure}
\clearpage

\begin{figure}
\includegraphics[width=9.0cm,angle=0]{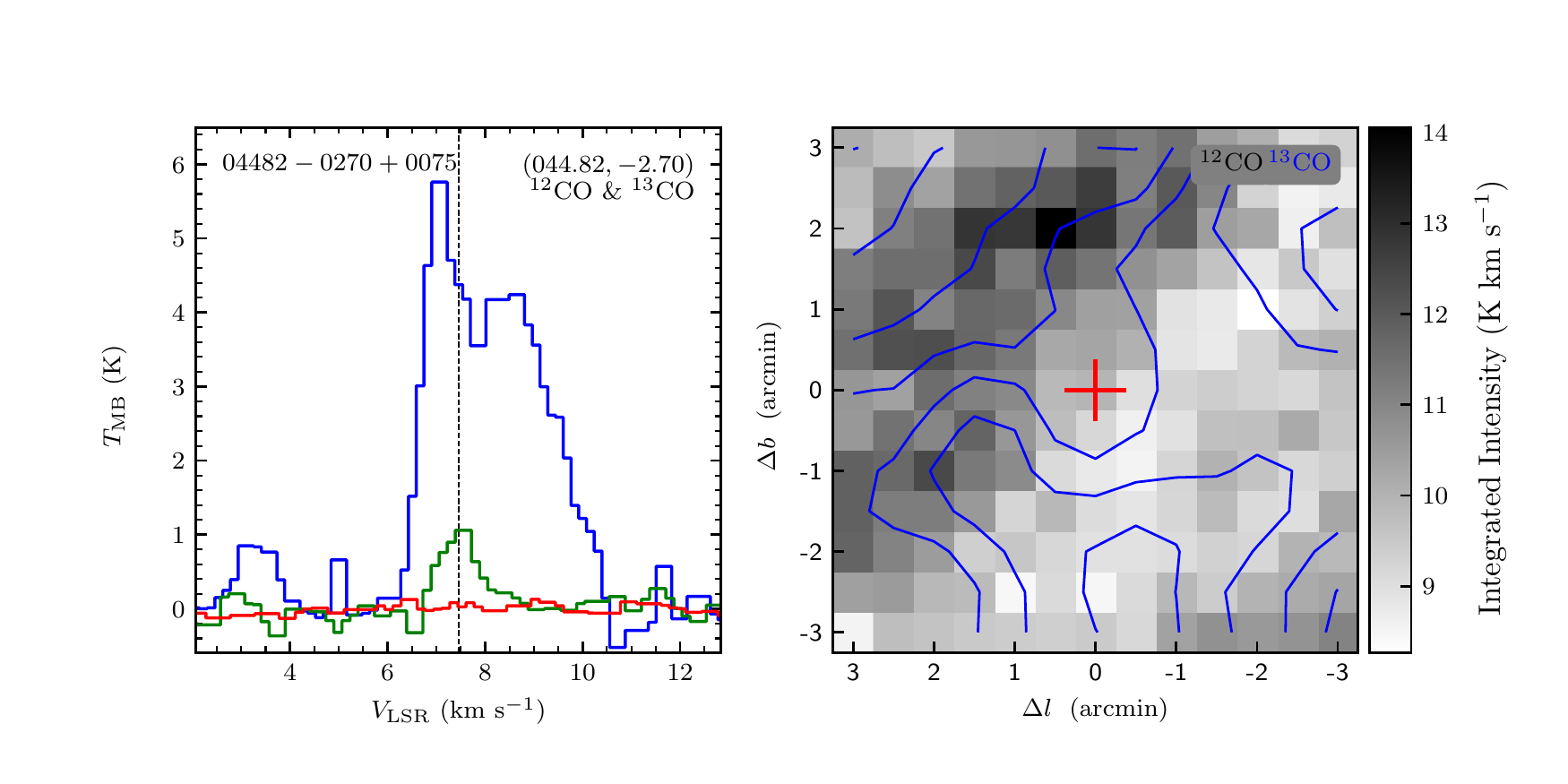}
\includegraphics[width=9.0cm,angle=0]{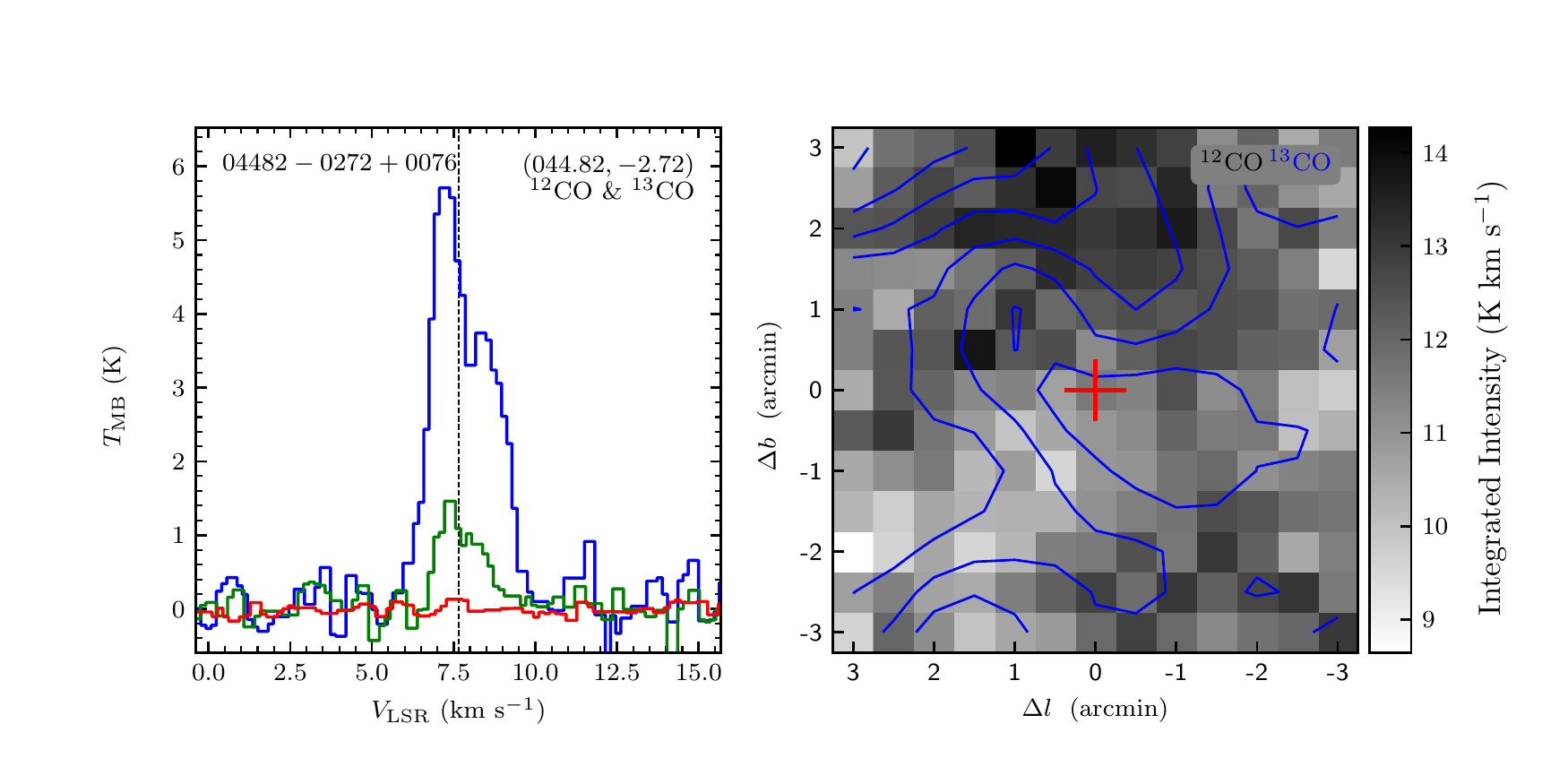}
\vspace{-0.5cm}

\includegraphics[width=9.0cm,angle=0]{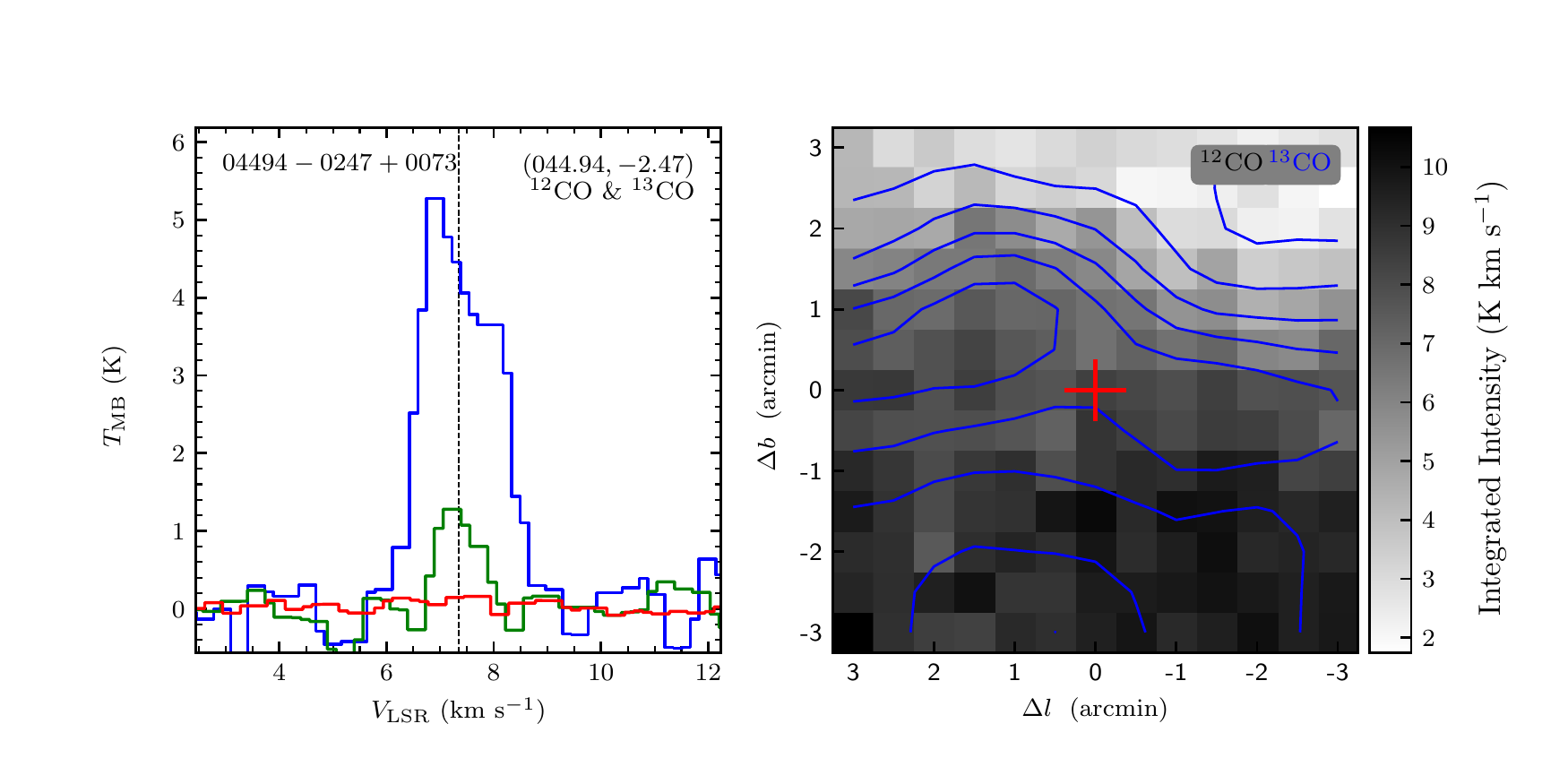}
\includegraphics[width=9.0cm,angle=0]{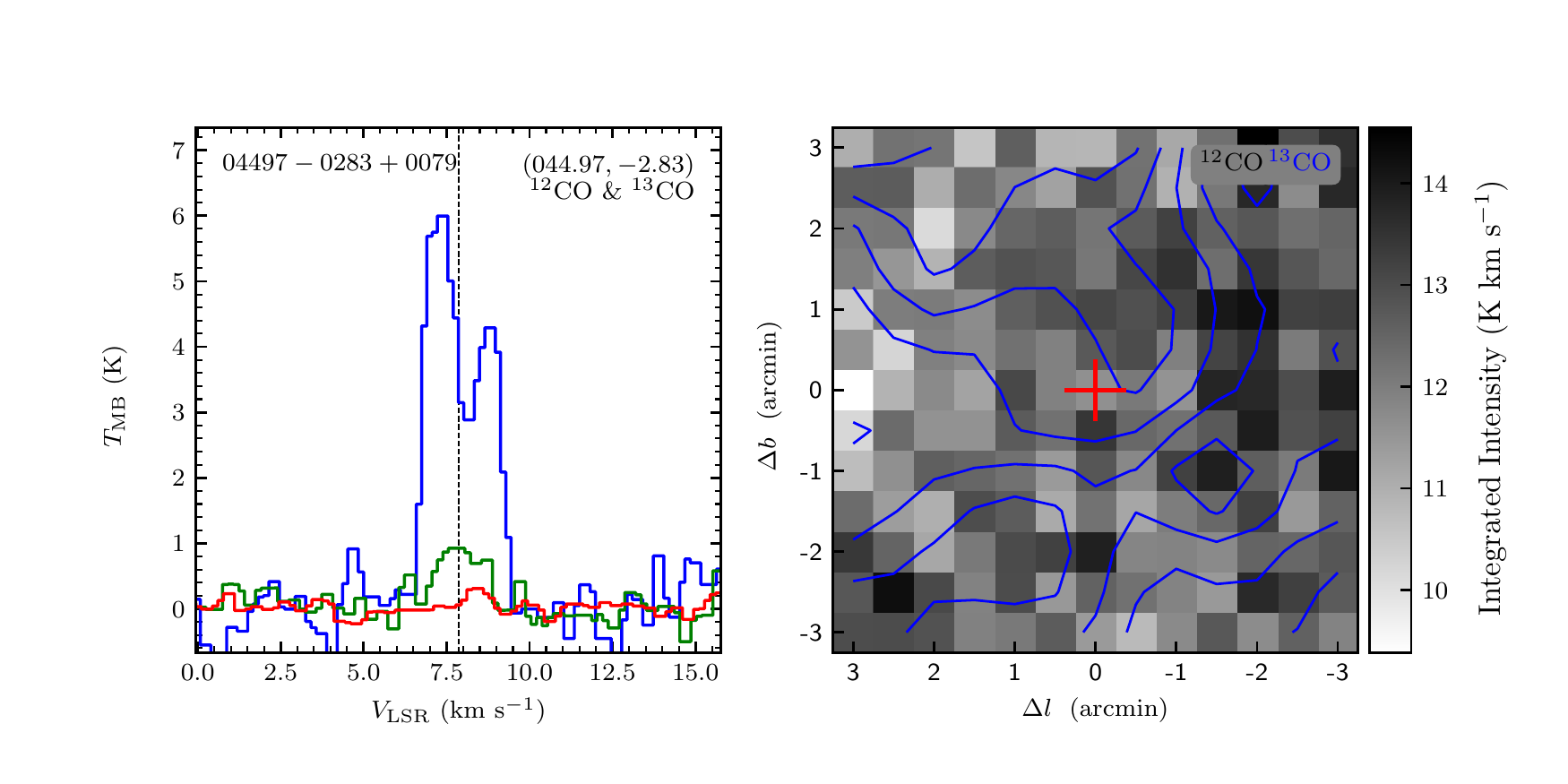}
\vspace{-0.5cm}

\includegraphics[width=9.0cm,angle=0]{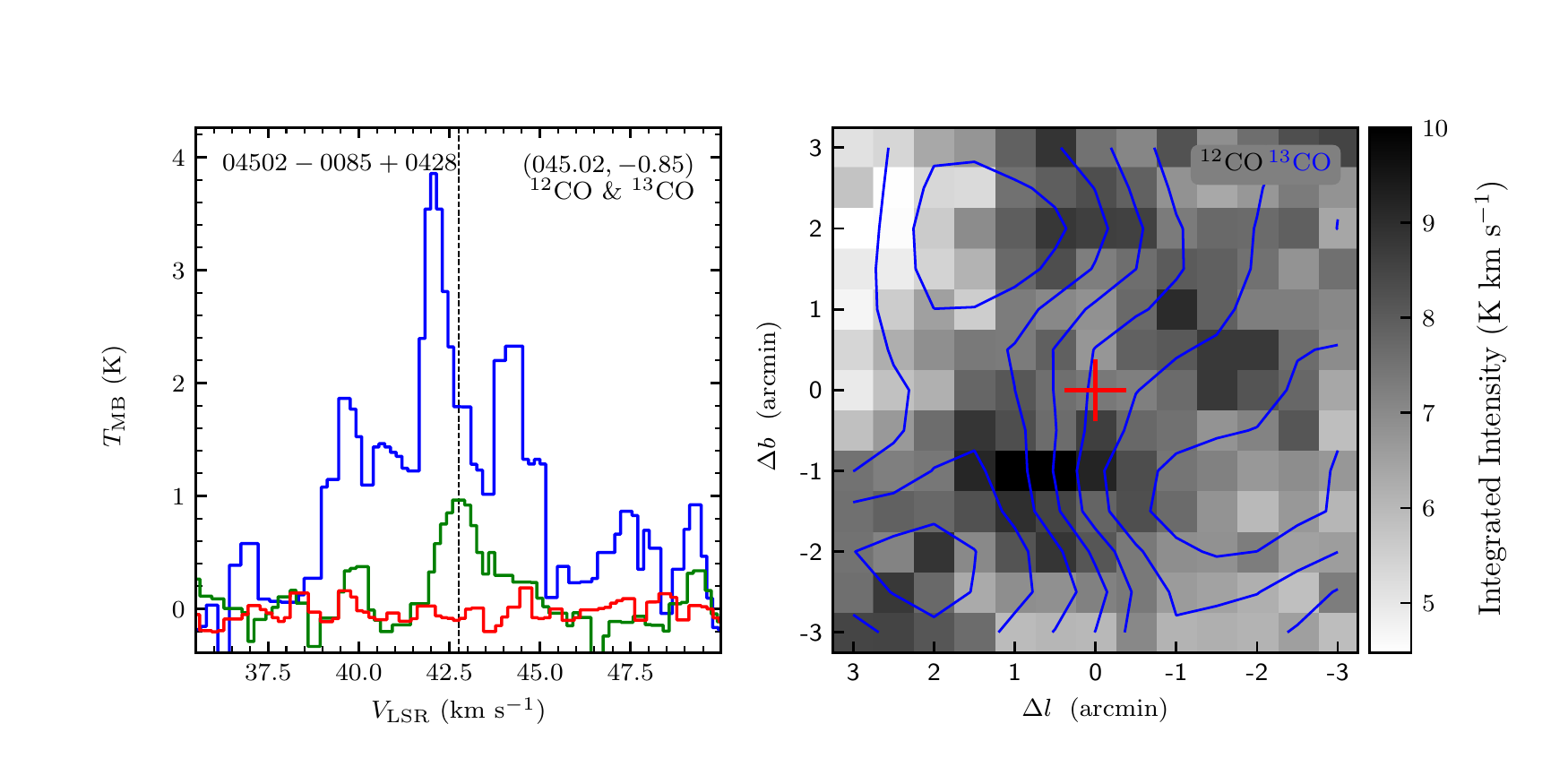}
\includegraphics[width=9.0cm,angle=0]{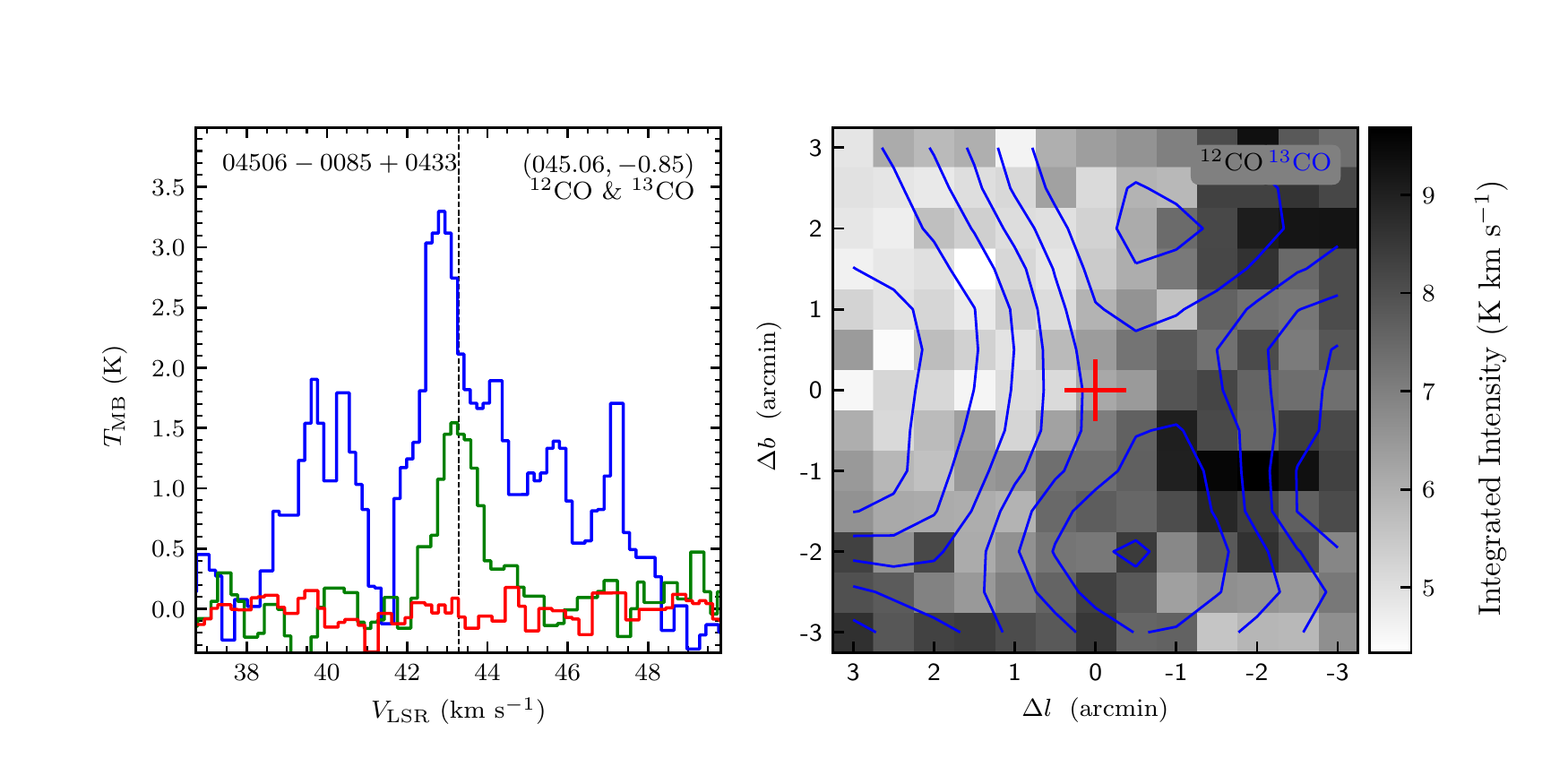}
\vspace{-0.5cm}

\includegraphics[width=9.0cm,angle=0]{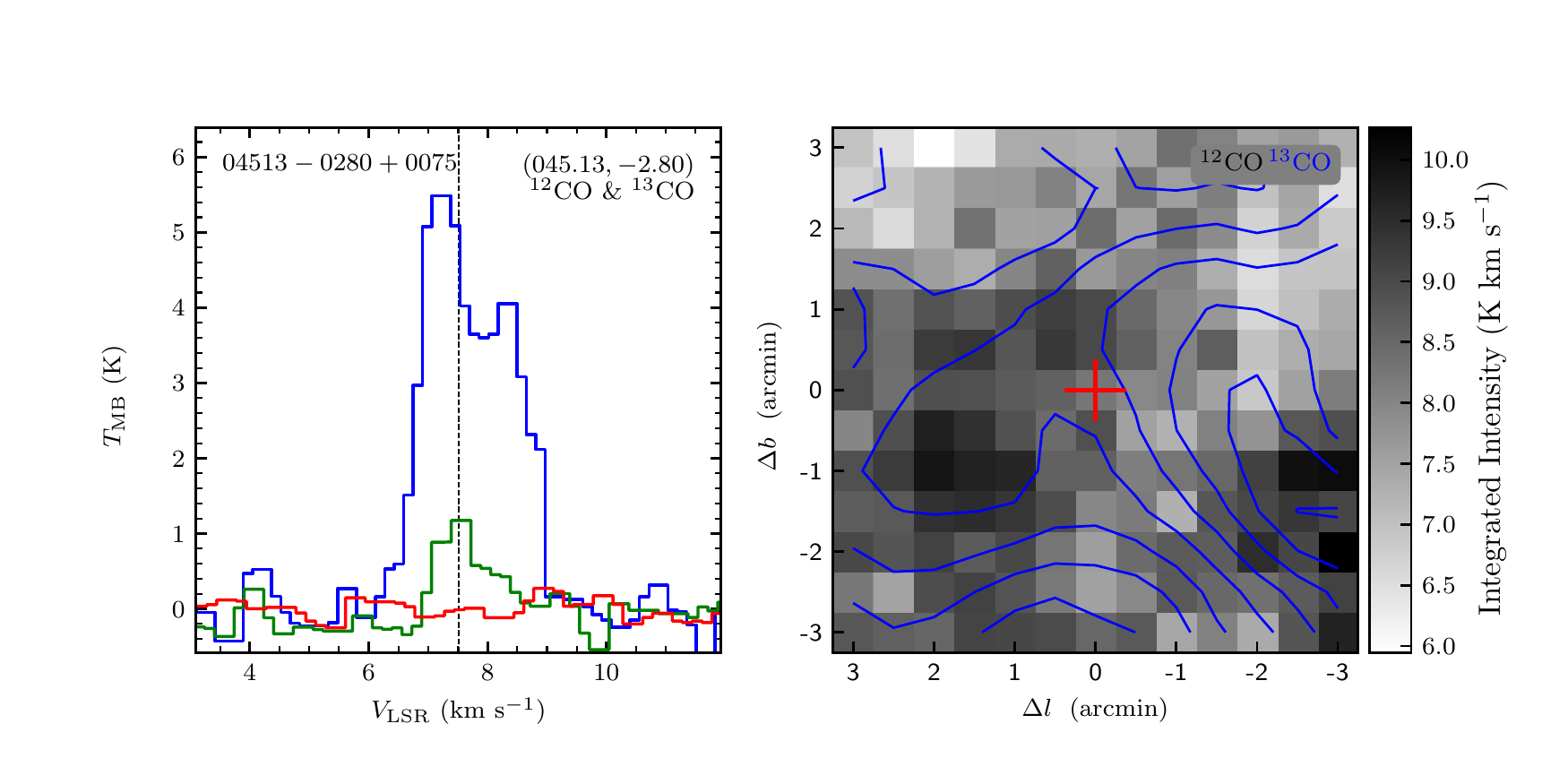}
\includegraphics[width=9.0cm,angle=0]{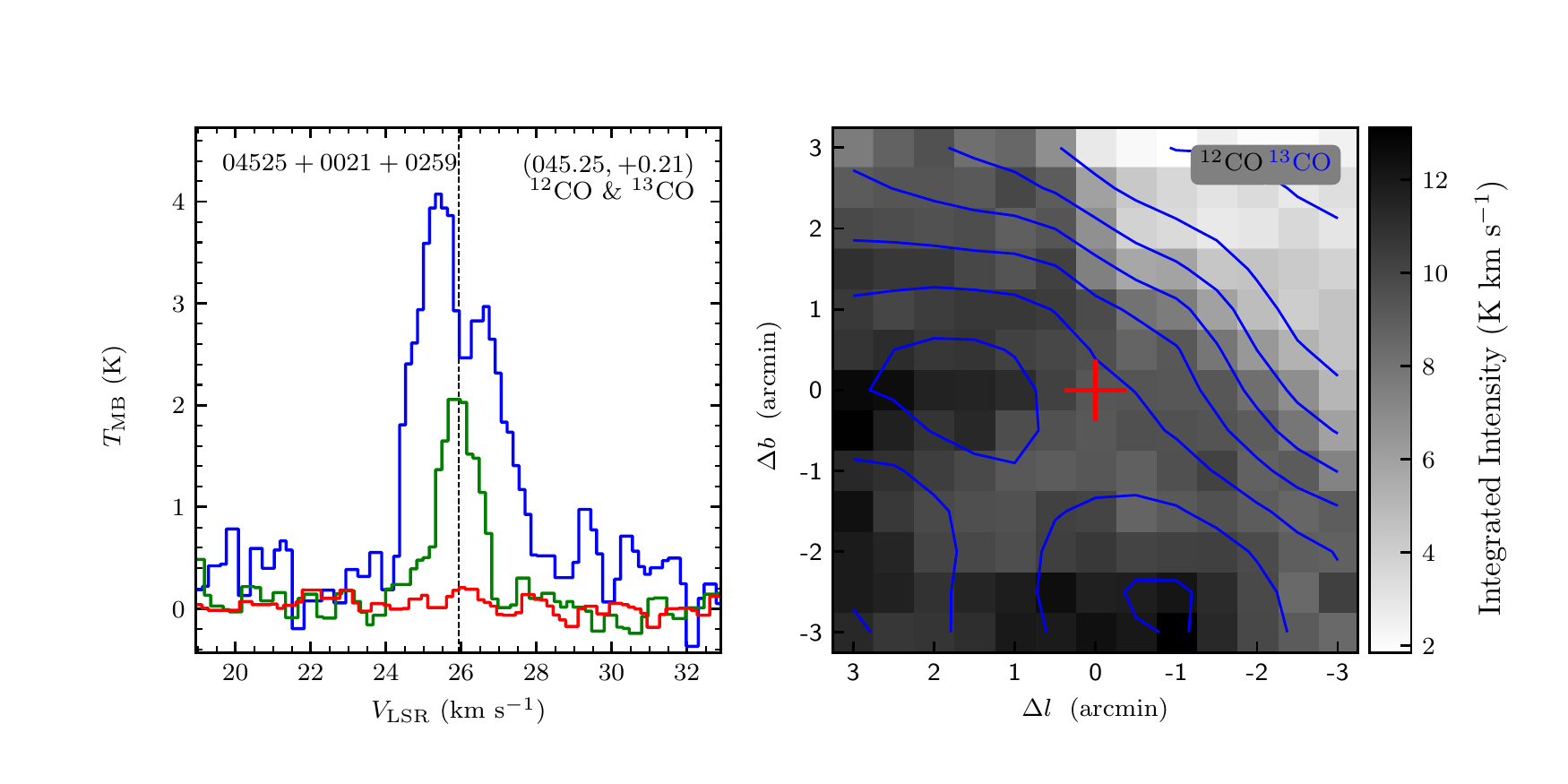}
\vspace{-0.5cm}

\includegraphics[width=9.0cm,angle=0]{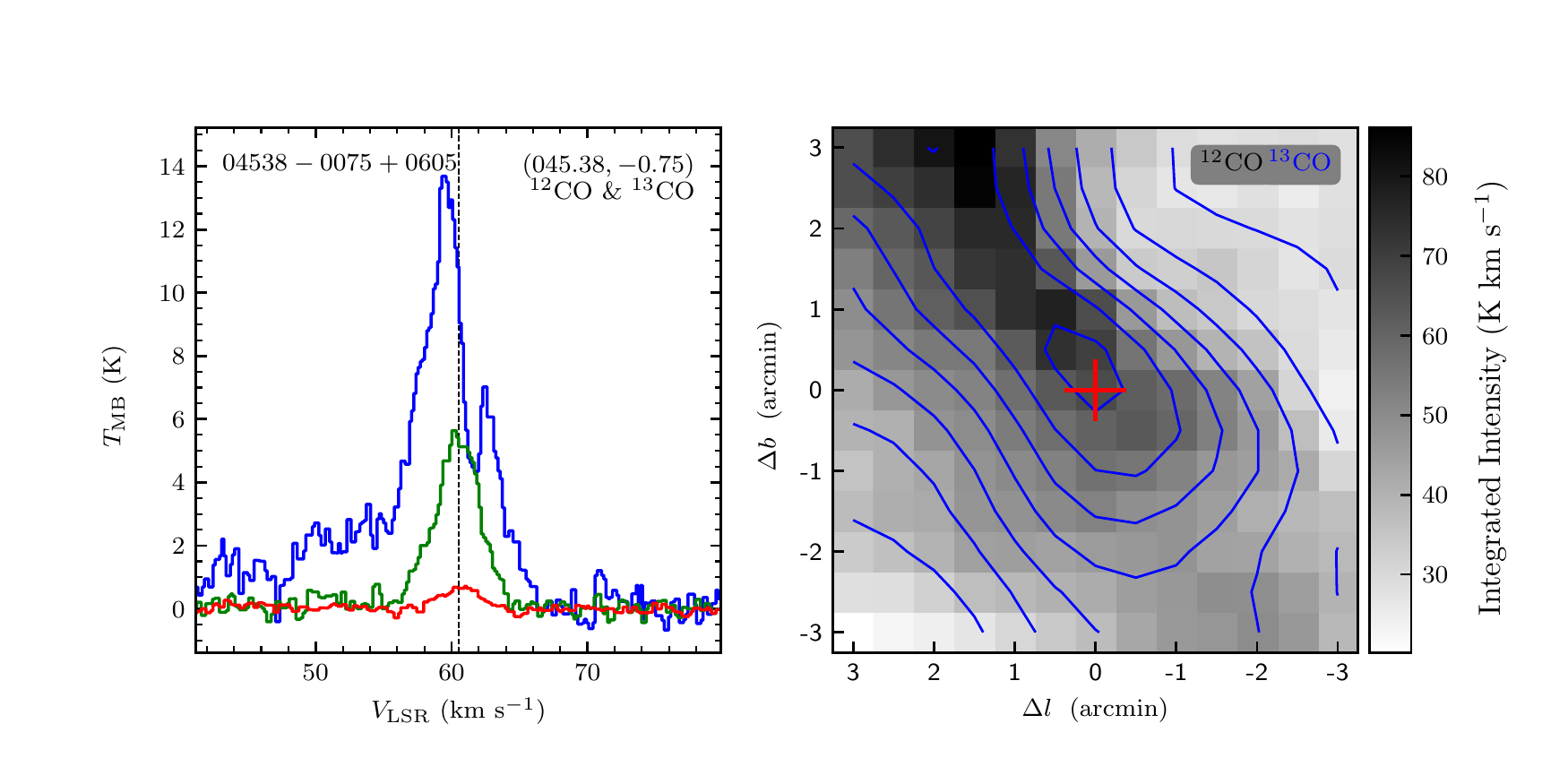}
\includegraphics[width=9.0cm,angle=0]{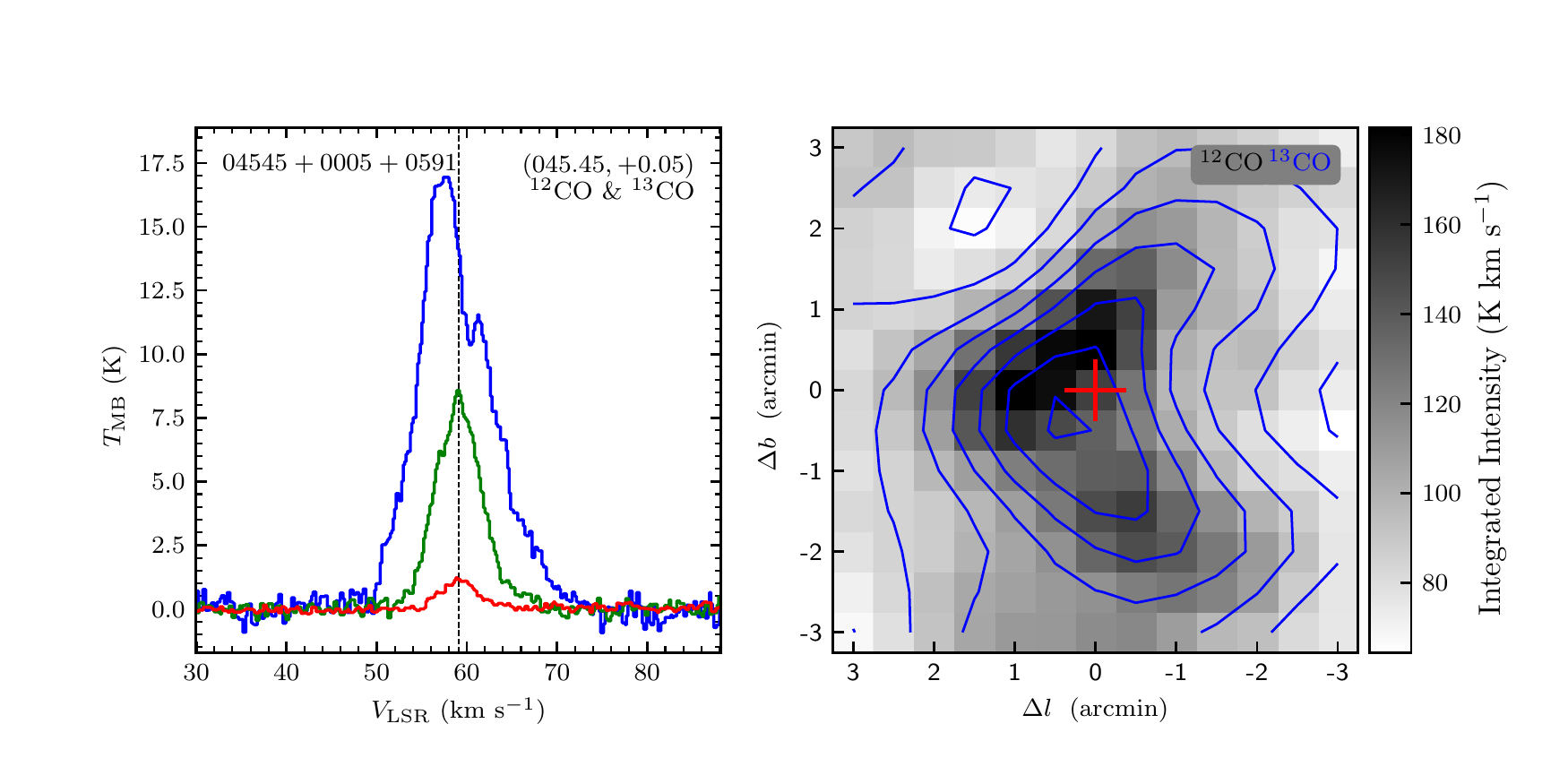}
\end{figure}
\clearpage

\begin{figure}
\includegraphics[width=9.0cm,angle=0]{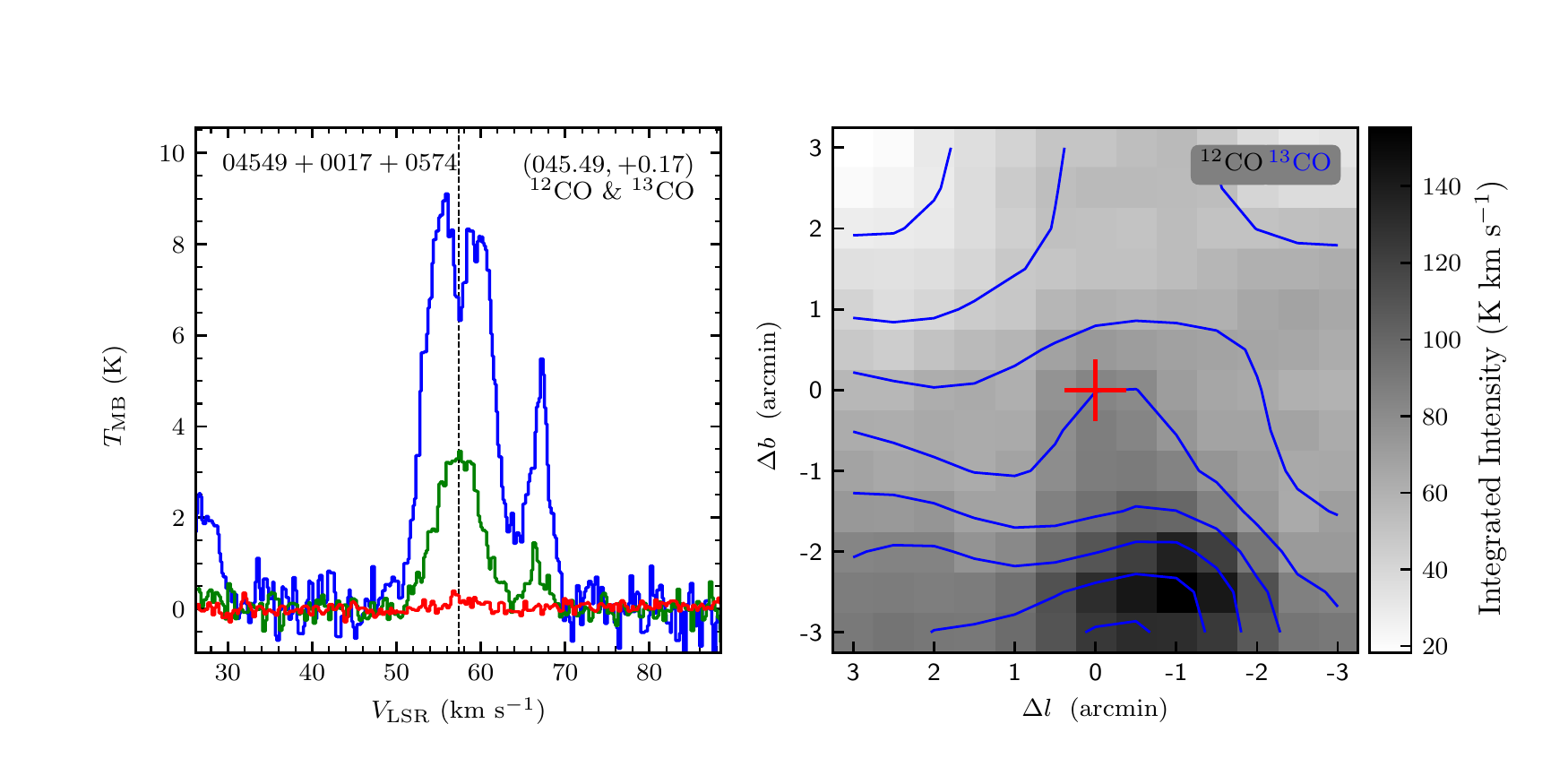}
\includegraphics[width=9.0cm,angle=0]{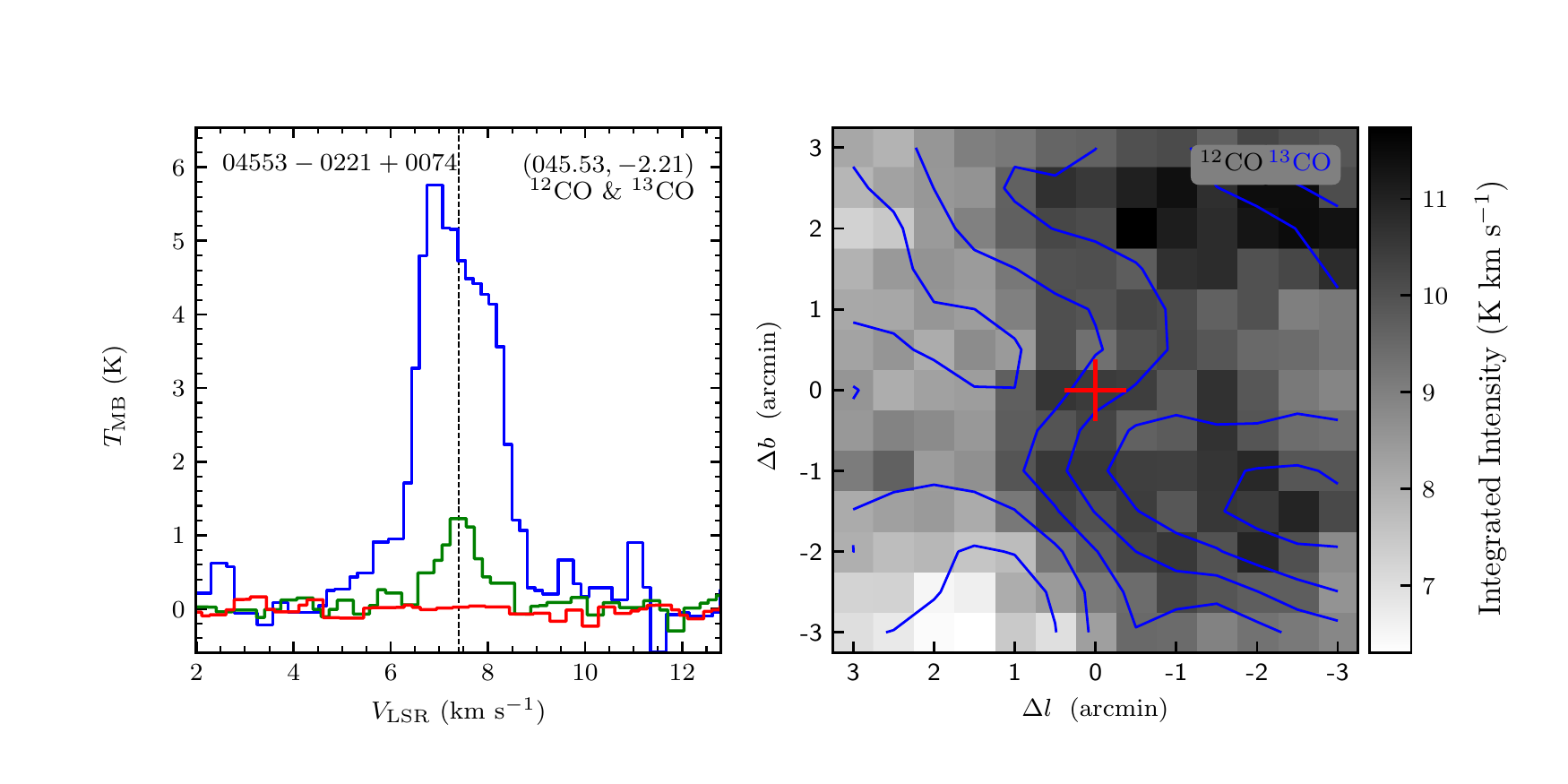}
\vspace{-0.5cm}

\includegraphics[width=9.0cm,angle=0]{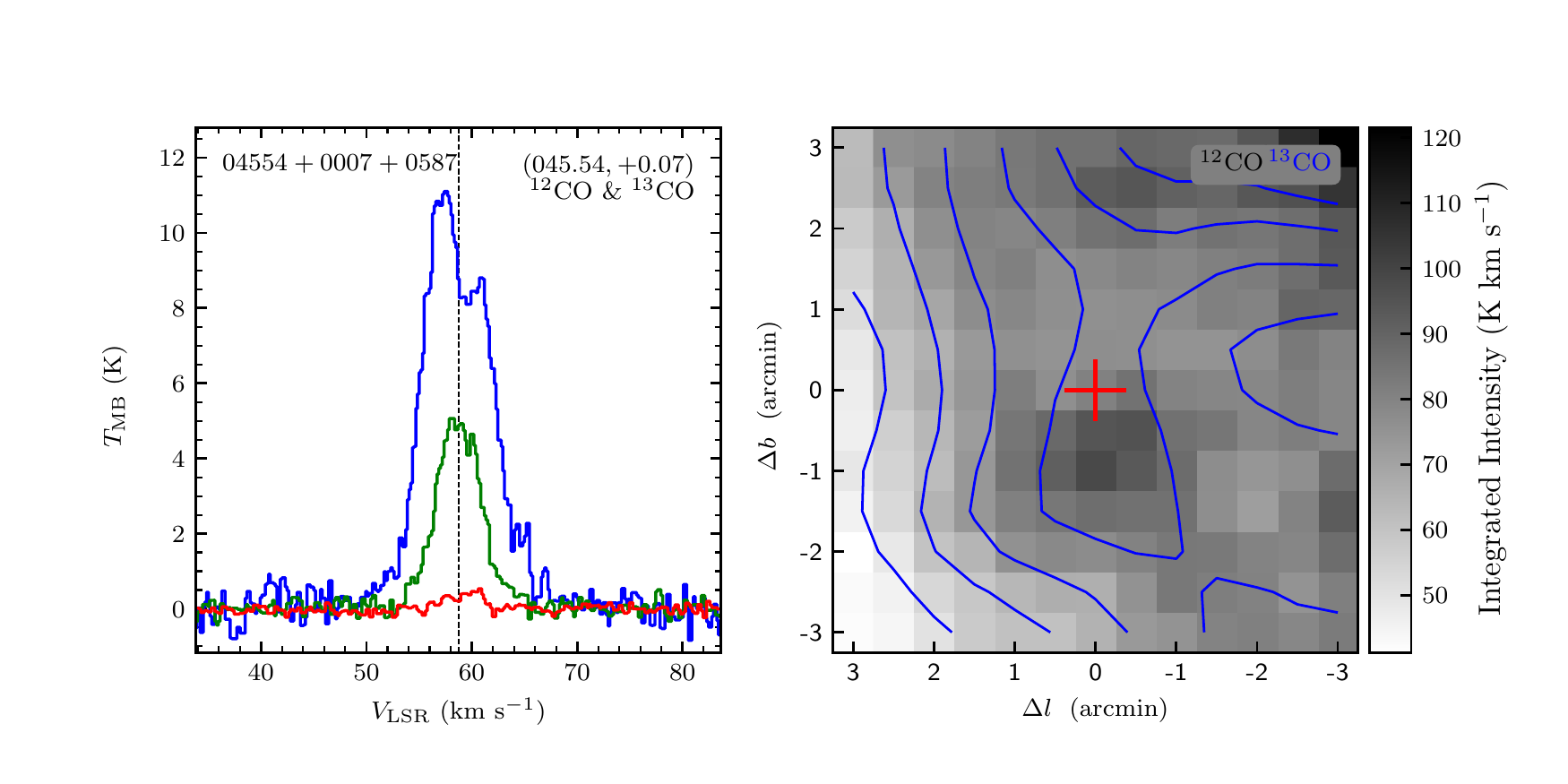}
\includegraphics[width=9.0cm,angle=0]{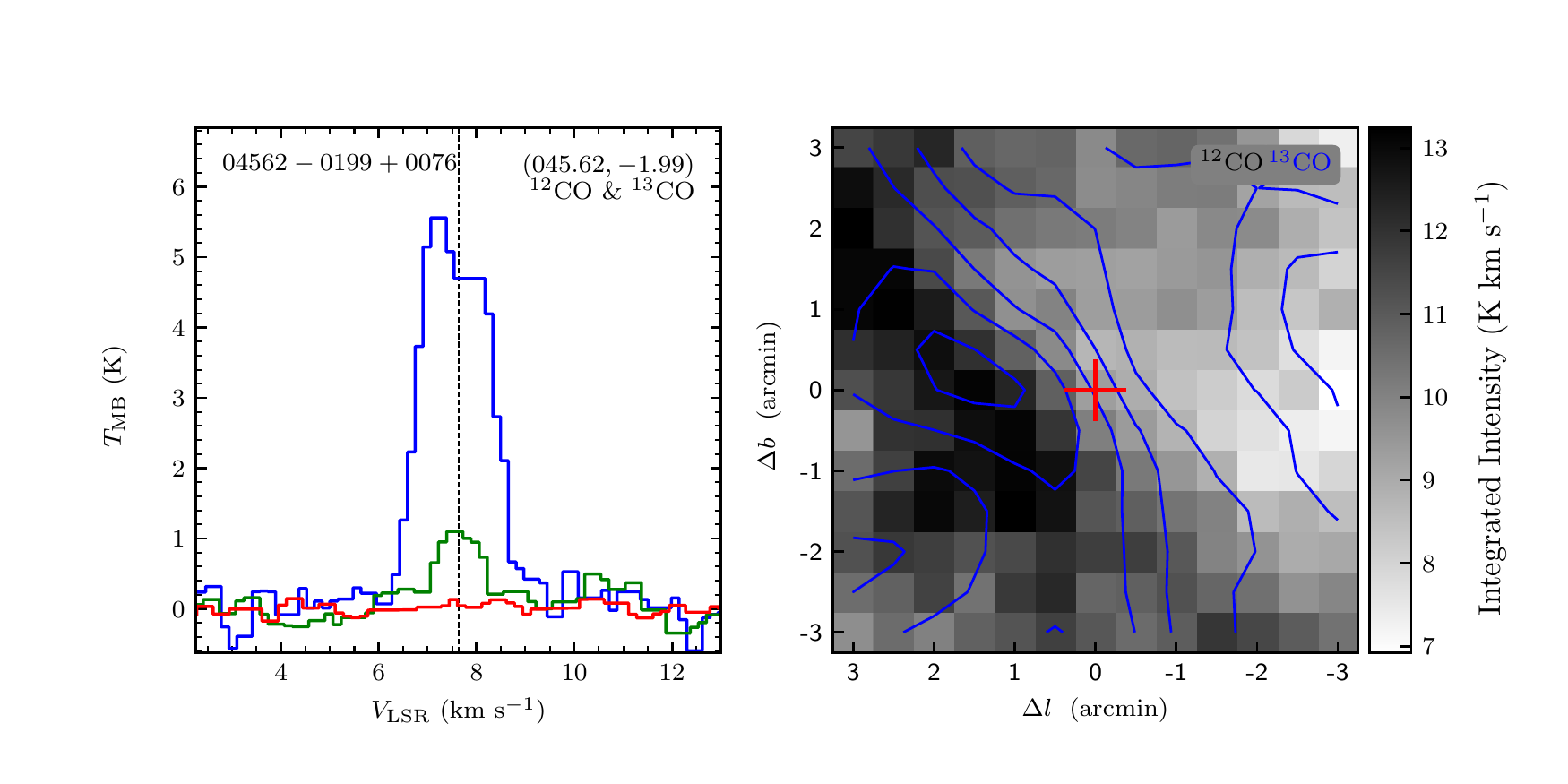}
\vspace{-0.5cm}

\includegraphics[width=9.0cm,angle=0]{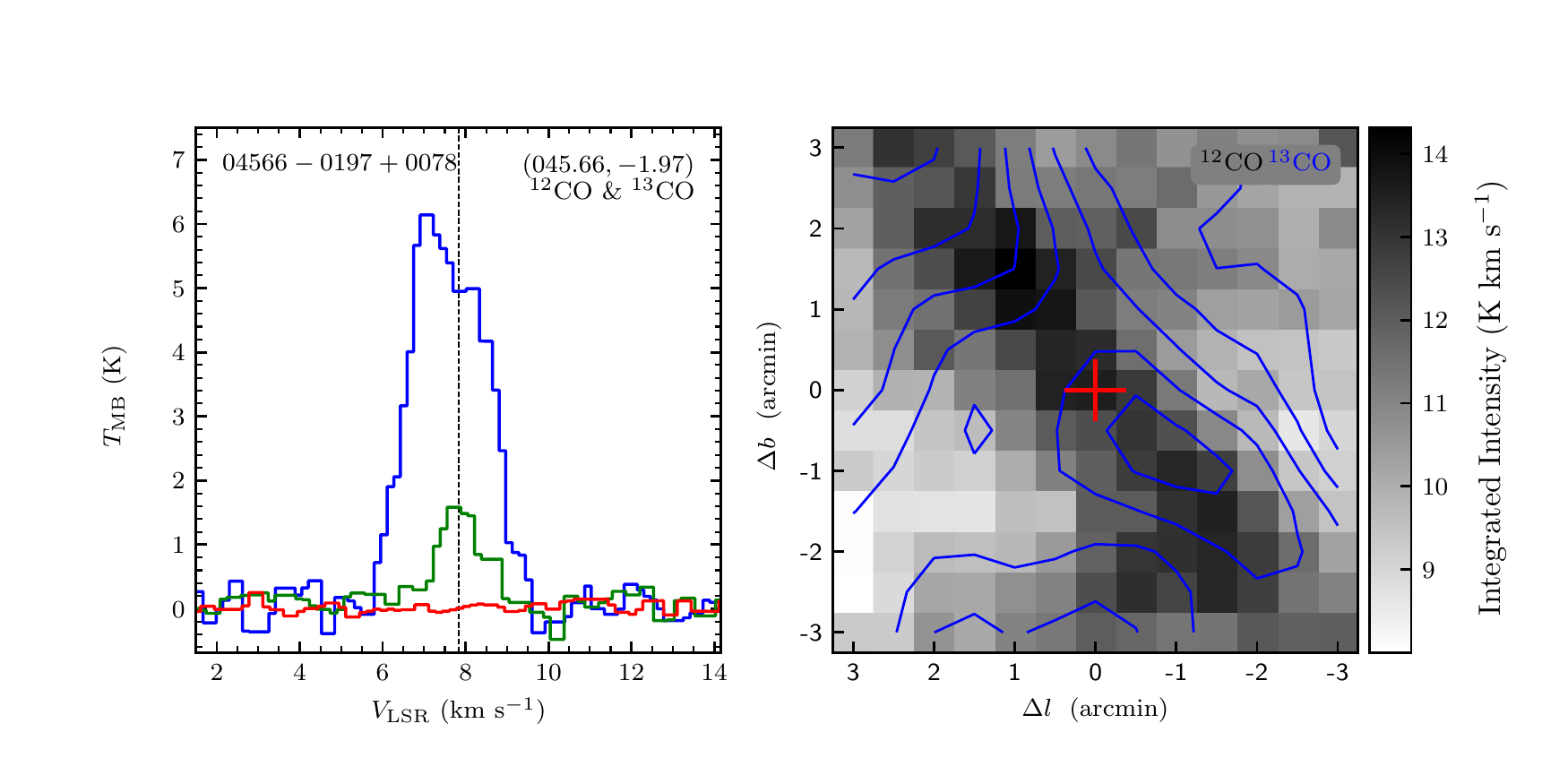}
\includegraphics[width=9.0cm,angle=0]{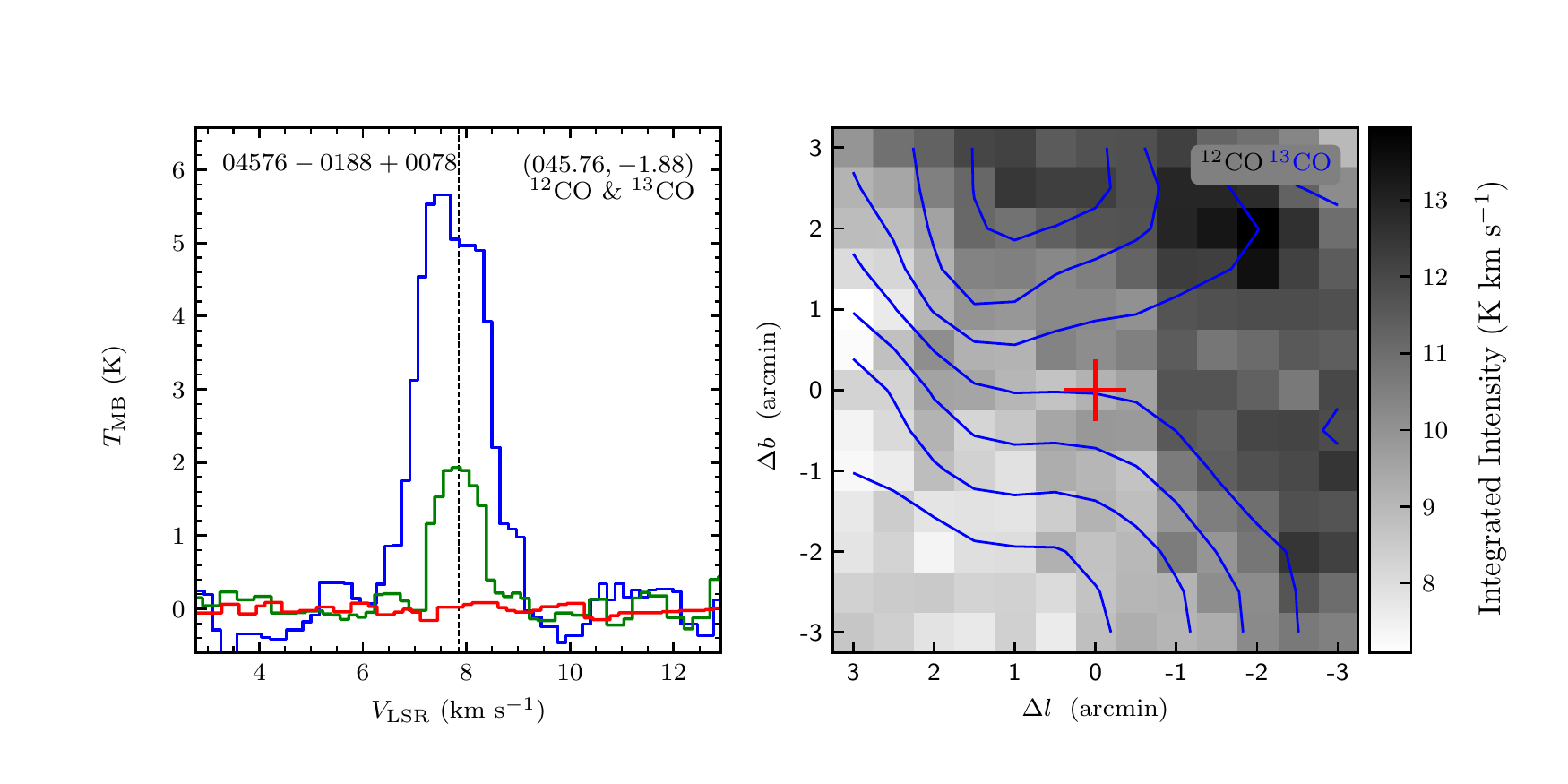}
\vspace{-0.5cm}

\includegraphics[width=9.0cm,angle=0]{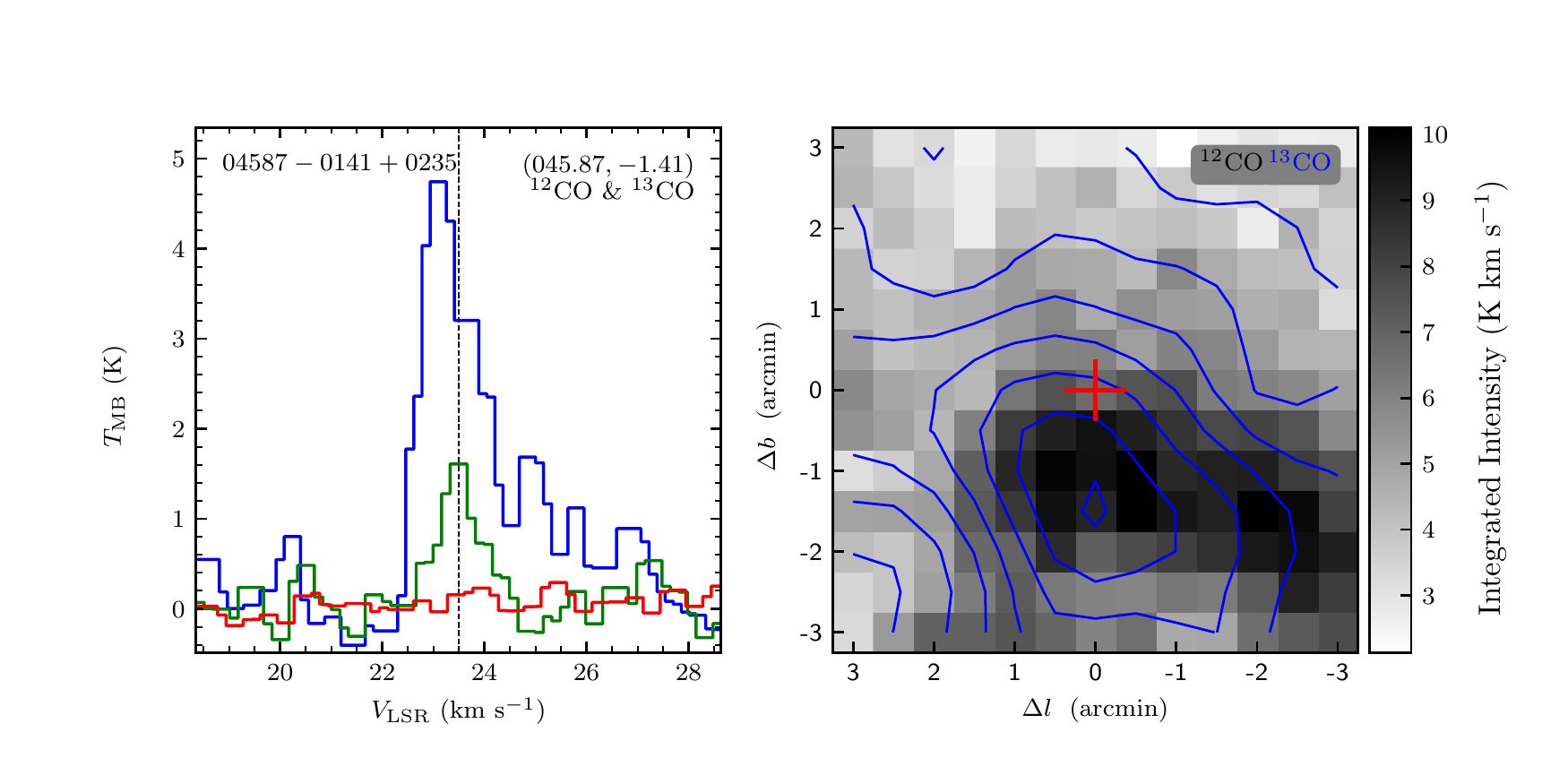}
\includegraphics[width=9.0cm,angle=0]{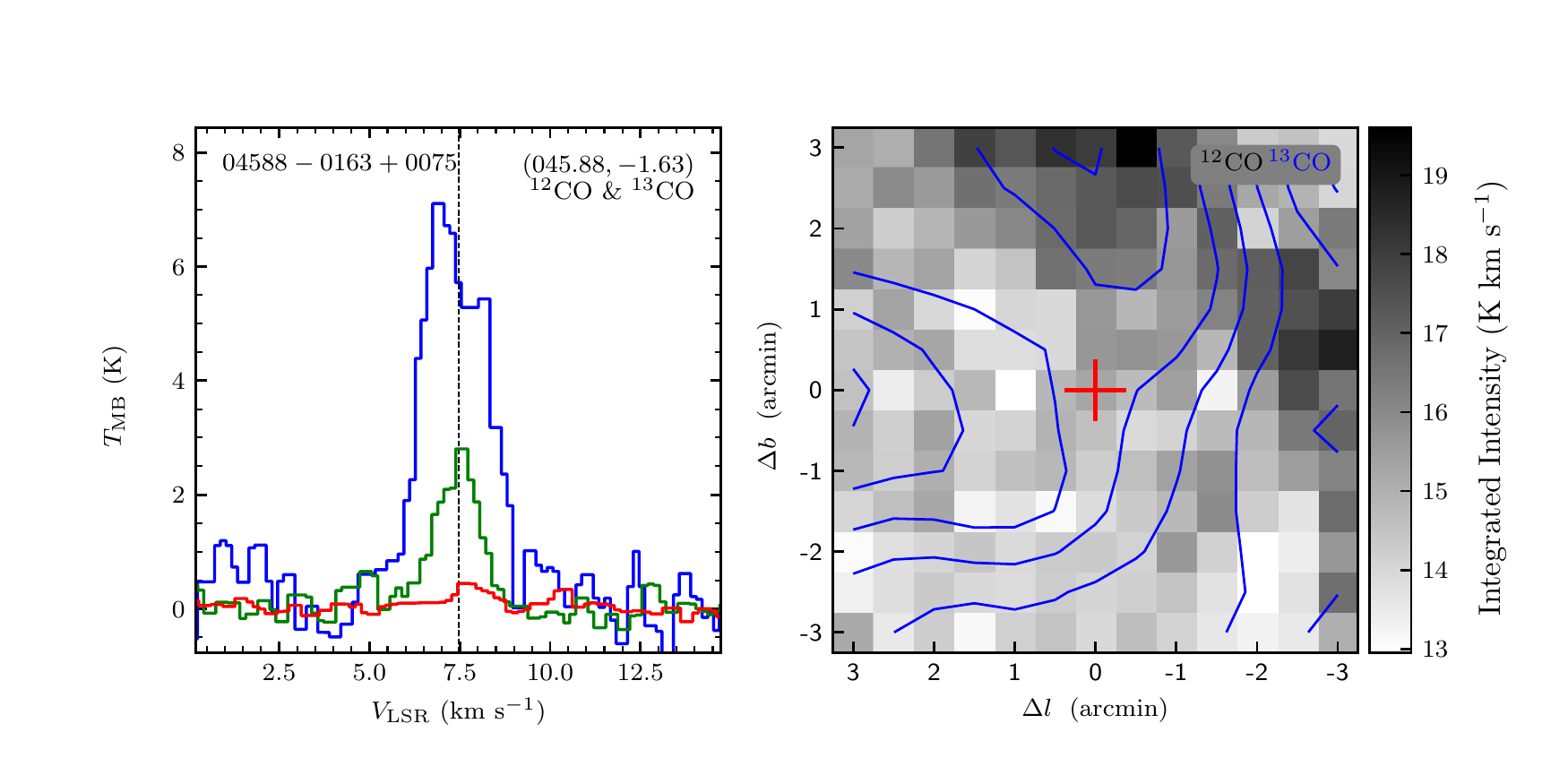}
\vspace{-0.5cm}

\includegraphics[width=9.0cm,angle=0]{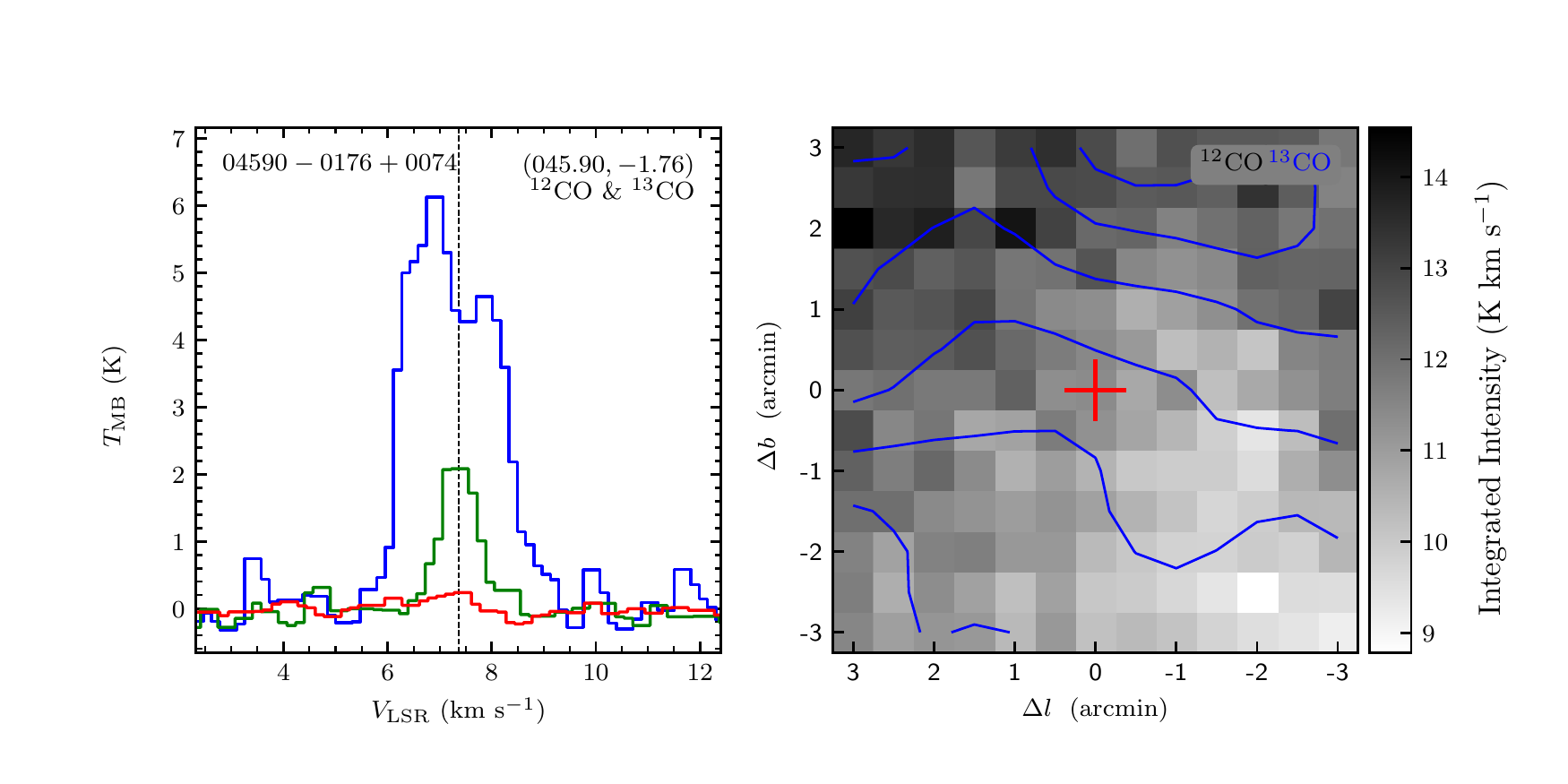}
\includegraphics[width=9.0cm,angle=0]{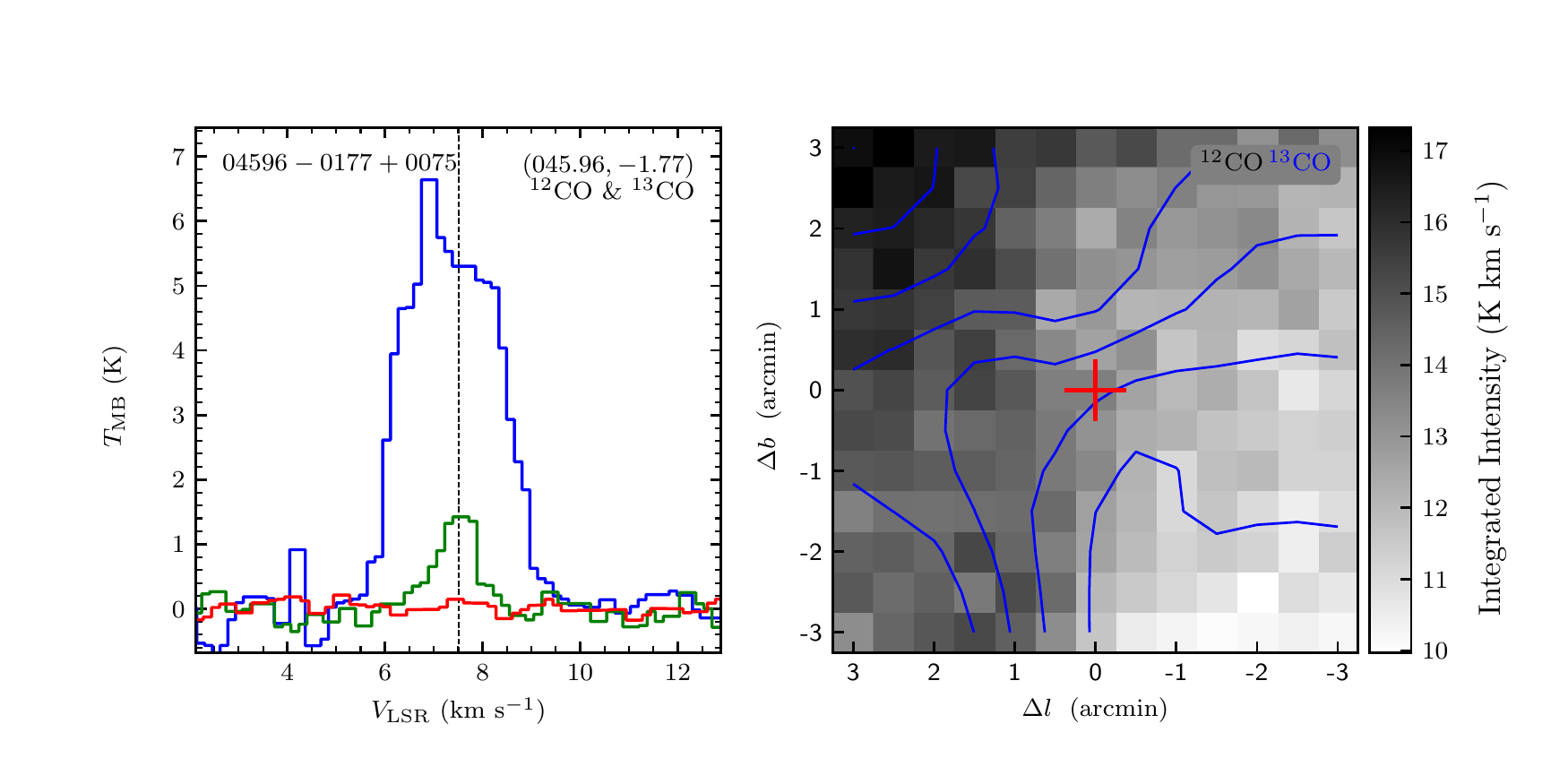}
\end{figure}
\clearpage

\begin{figure}
\includegraphics[width=9.0cm,angle=0]{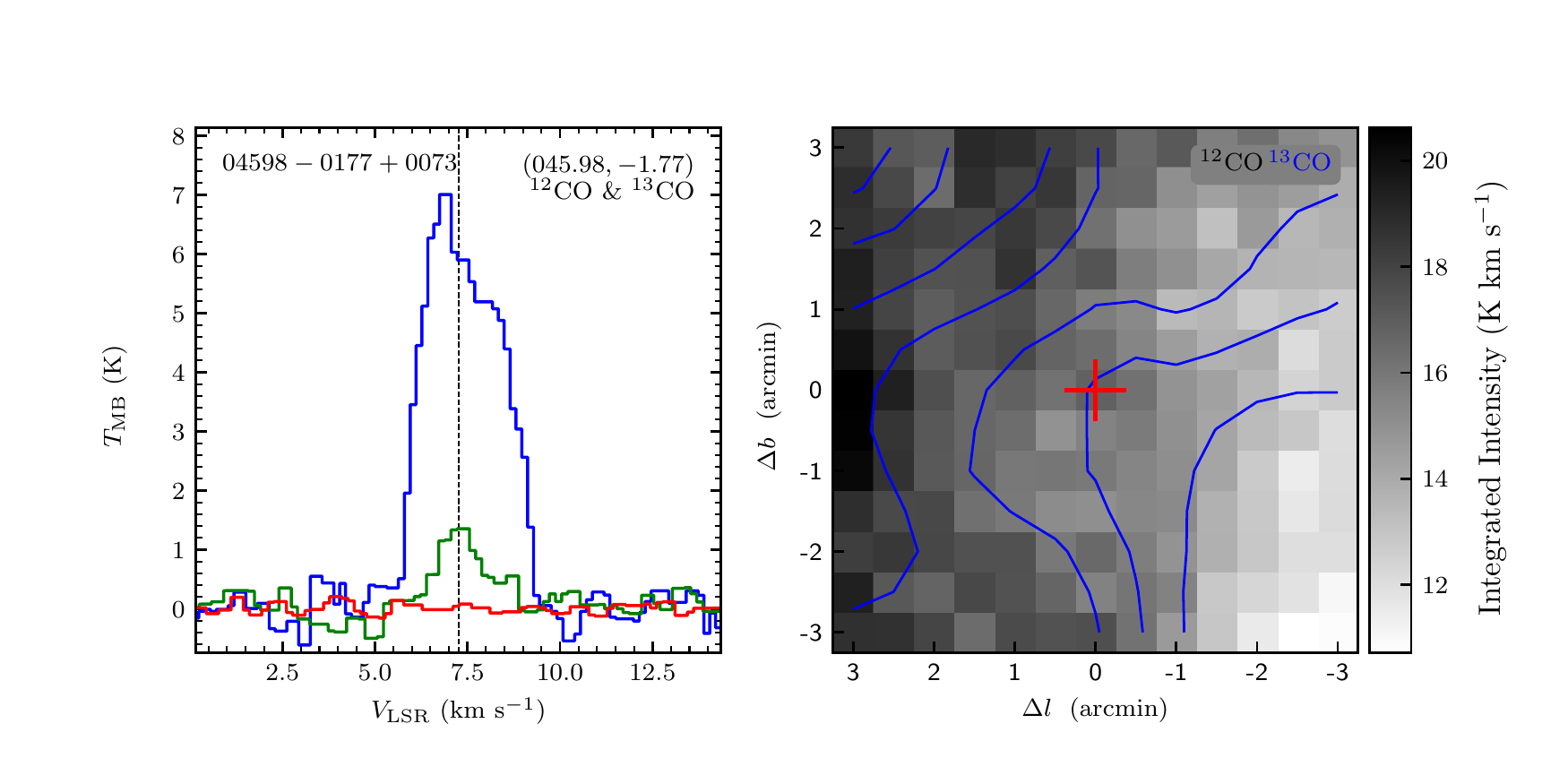}
\includegraphics[width=9.0cm,angle=0]{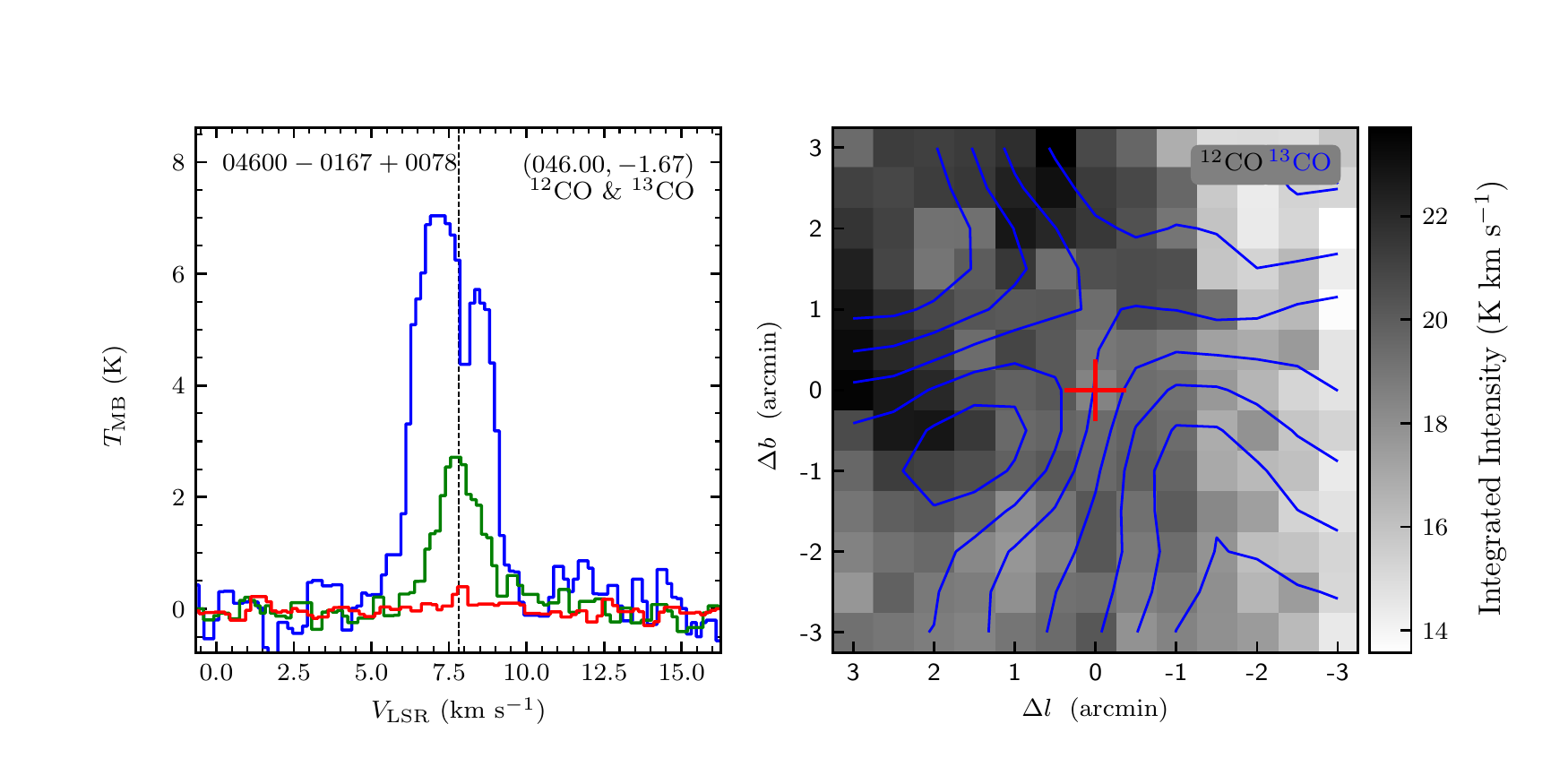}
\vspace{-0.5cm}

\includegraphics[width=9.0cm,angle=0]{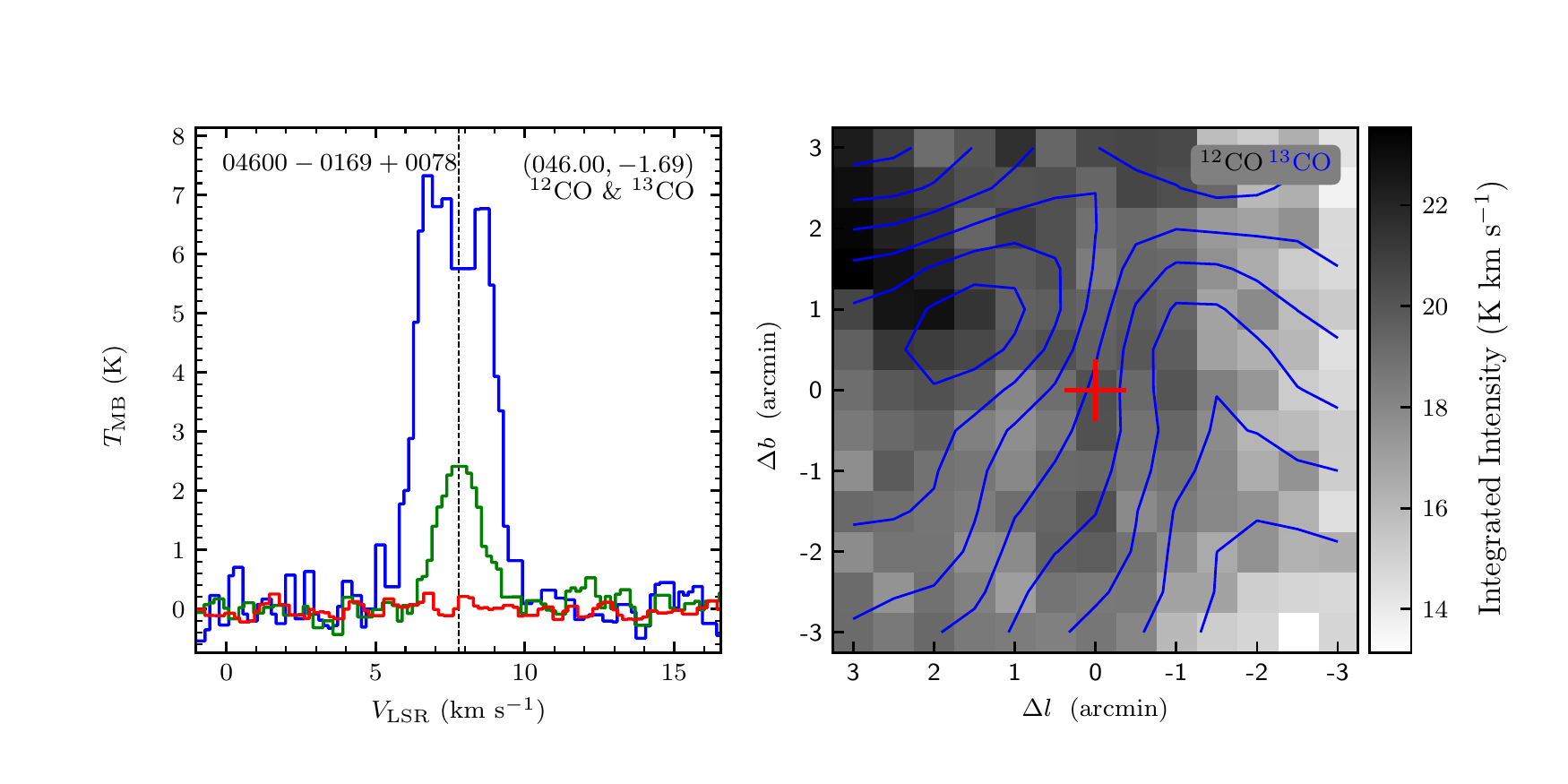}
\includegraphics[width=9.0cm,angle=0]{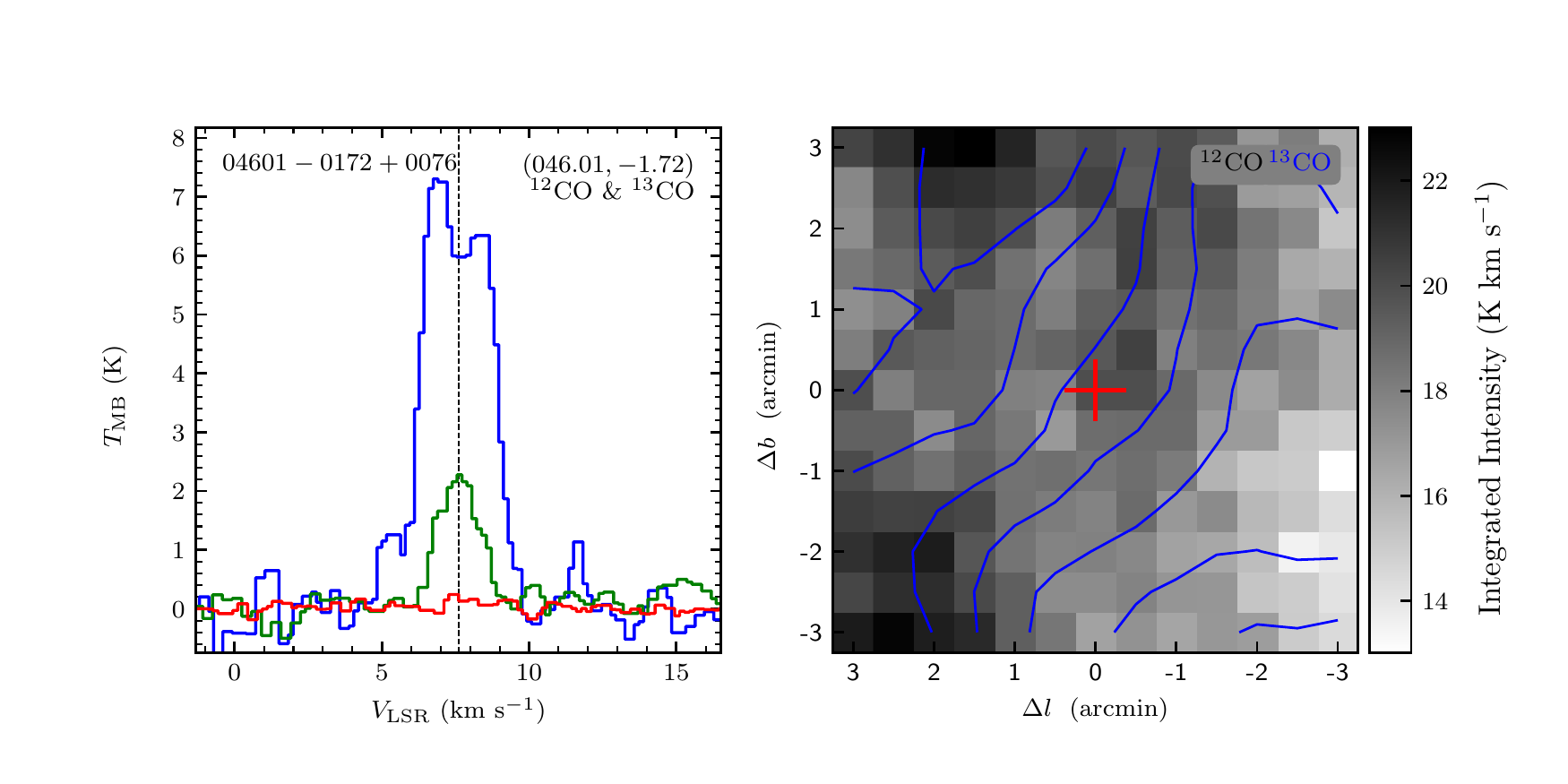}
\vspace{-0.5cm}

\includegraphics[width=9.0cm,angle=0]{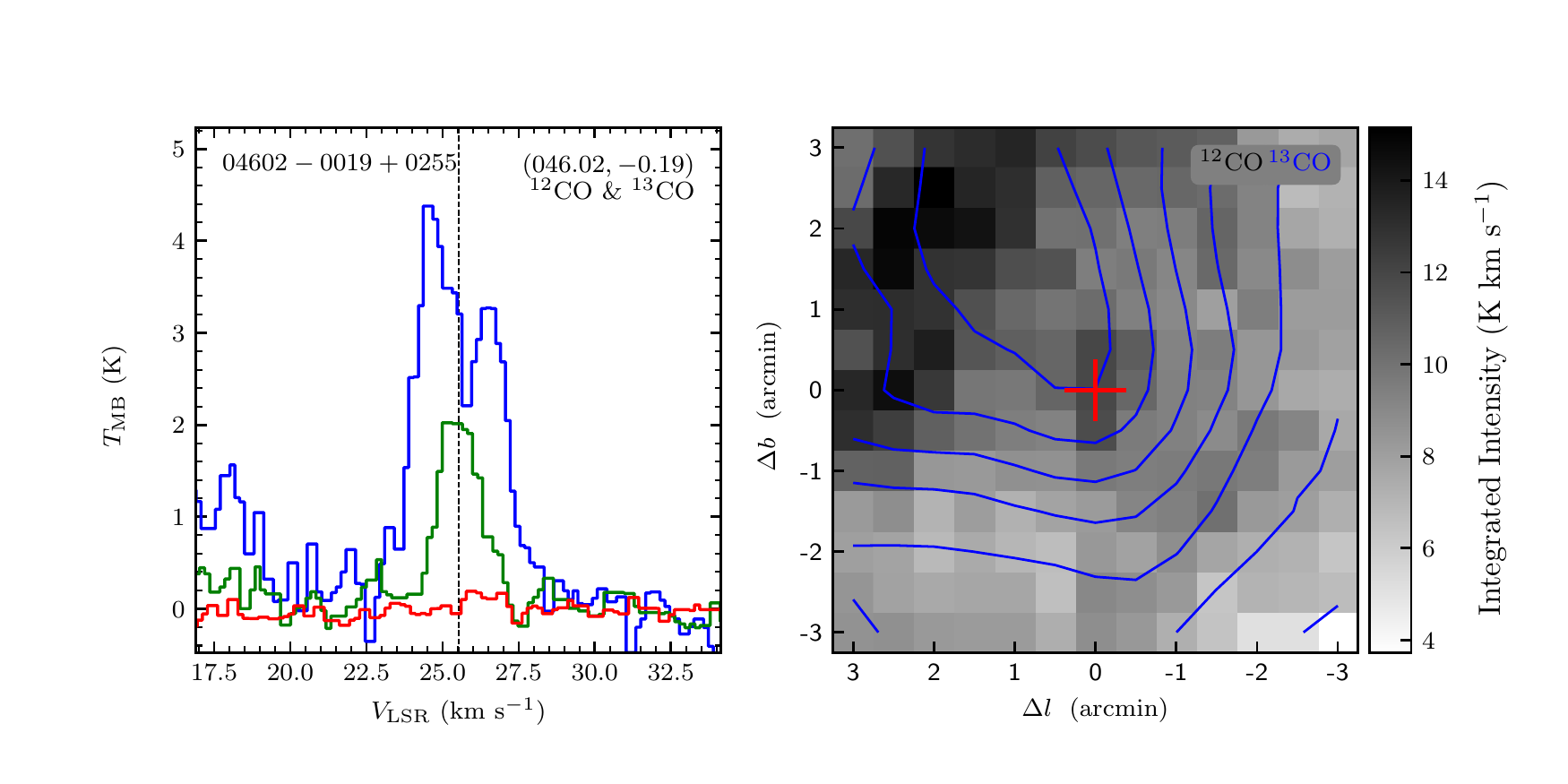}
\includegraphics[width=9.0cm,angle=0]{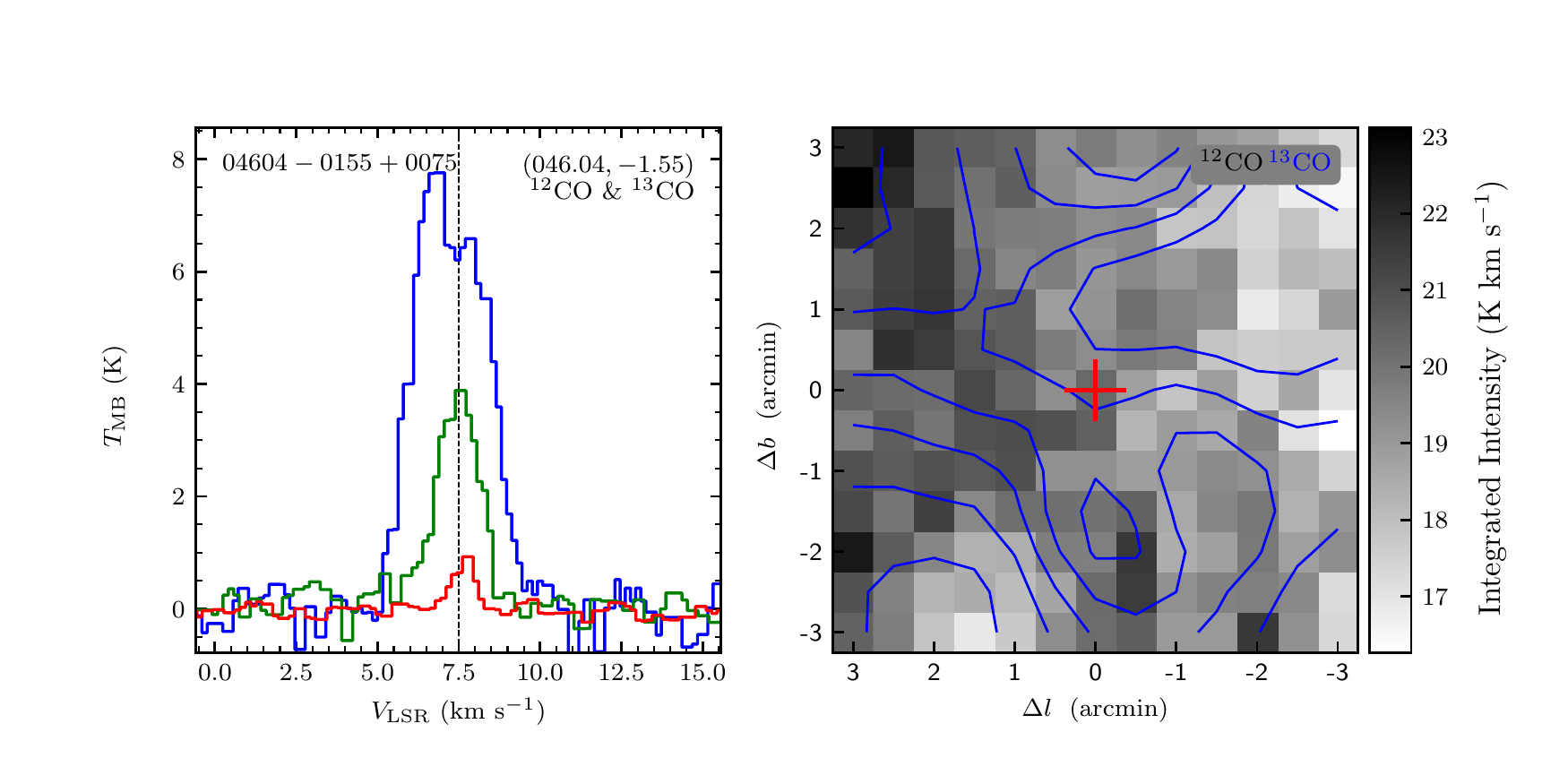}
\vspace{-0.5cm}

\includegraphics[width=9.0cm,angle=0]{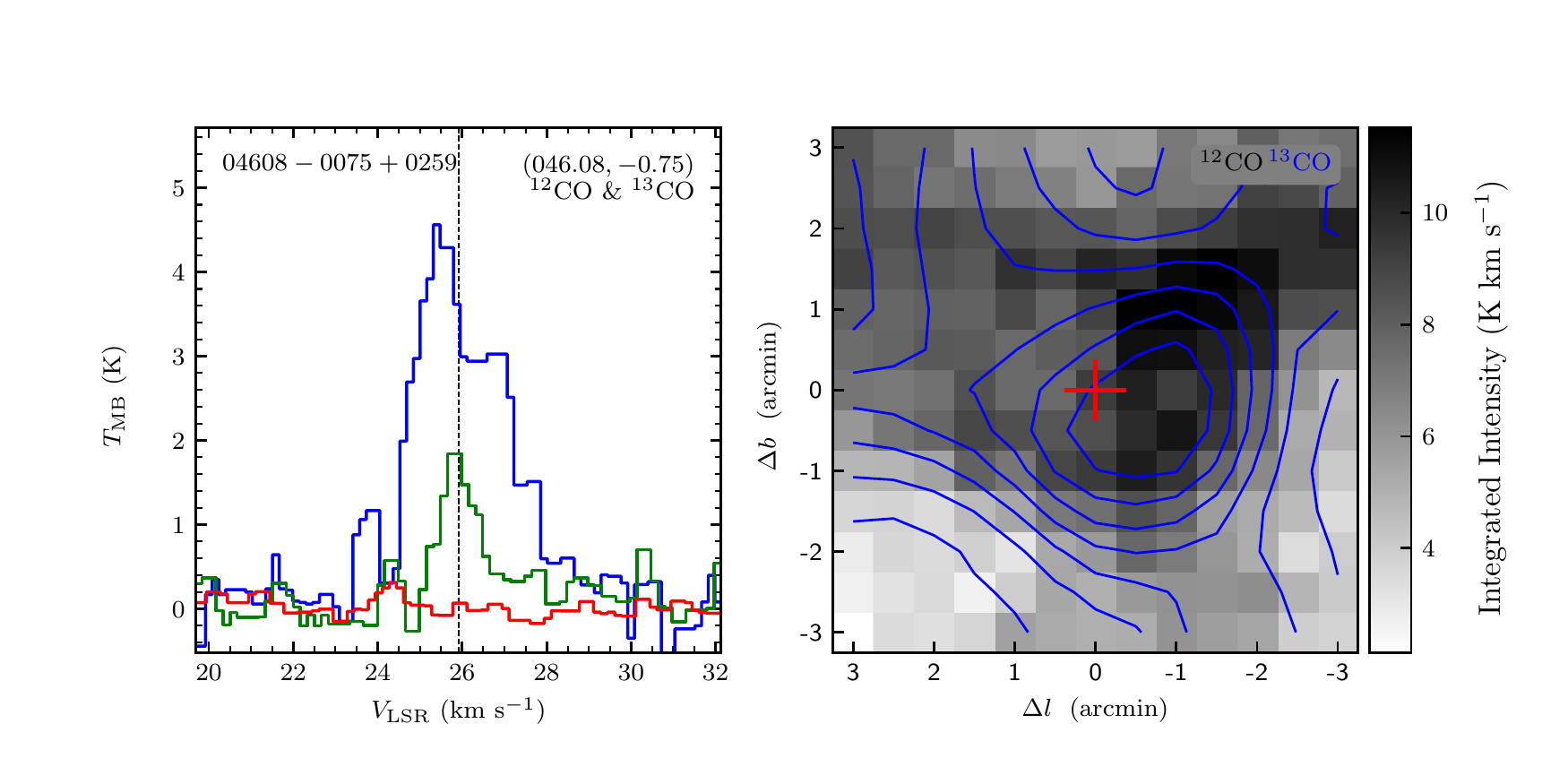}
\includegraphics[width=9.0cm,angle=0]{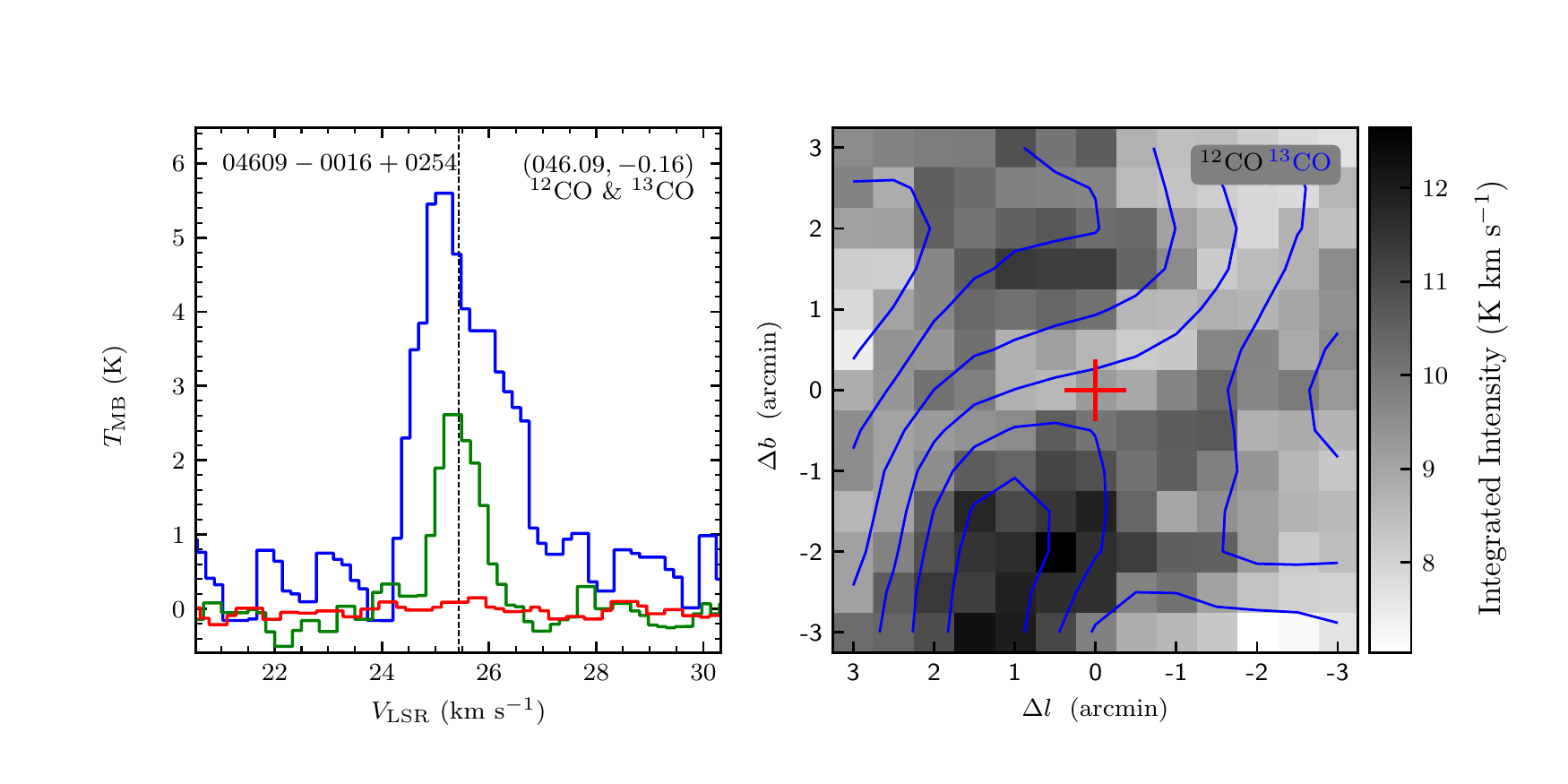}
\vspace{-0.5cm}

\includegraphics[width=9.0cm,angle=0]{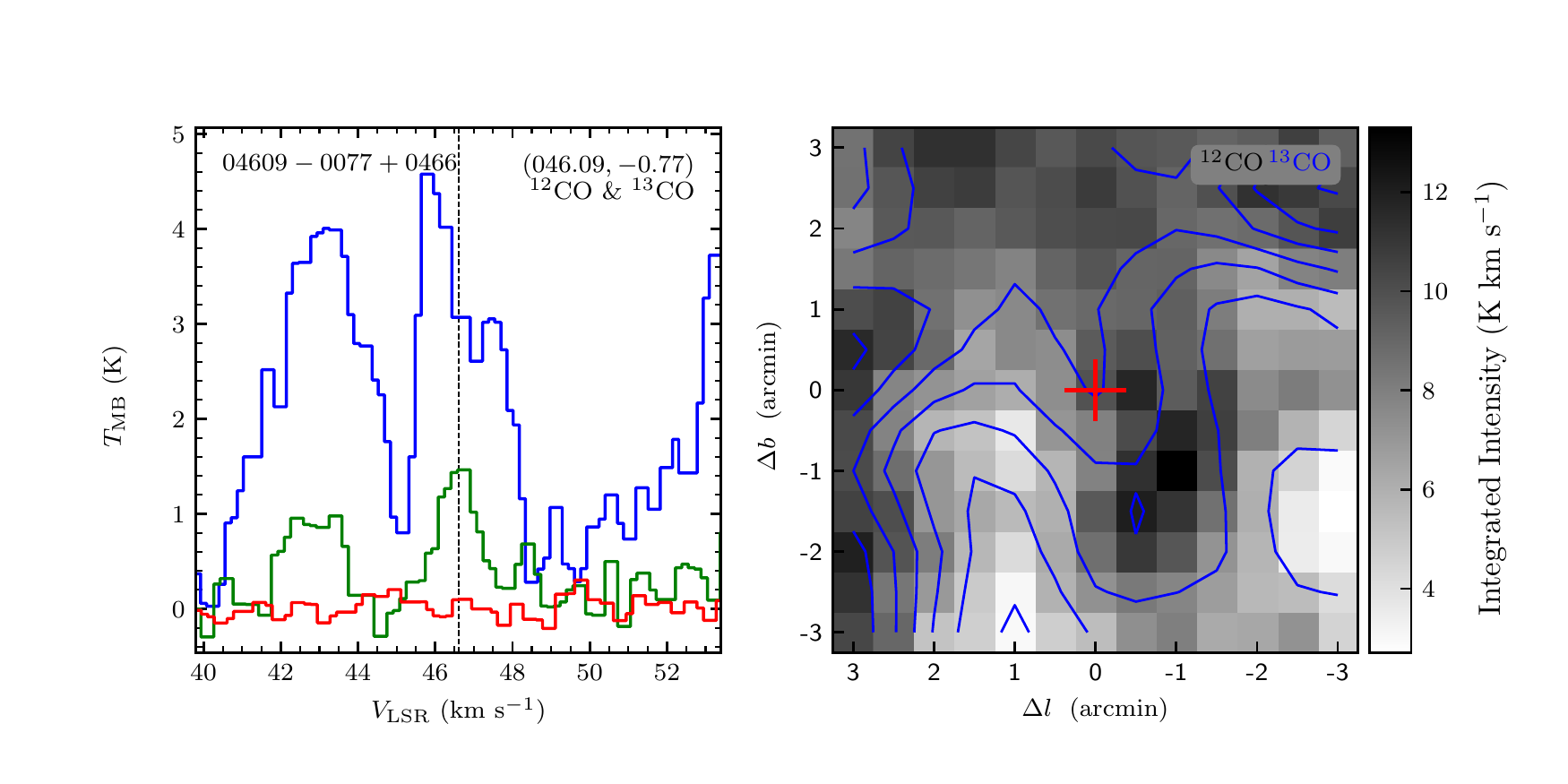}
\includegraphics[width=9.0cm,angle=0]{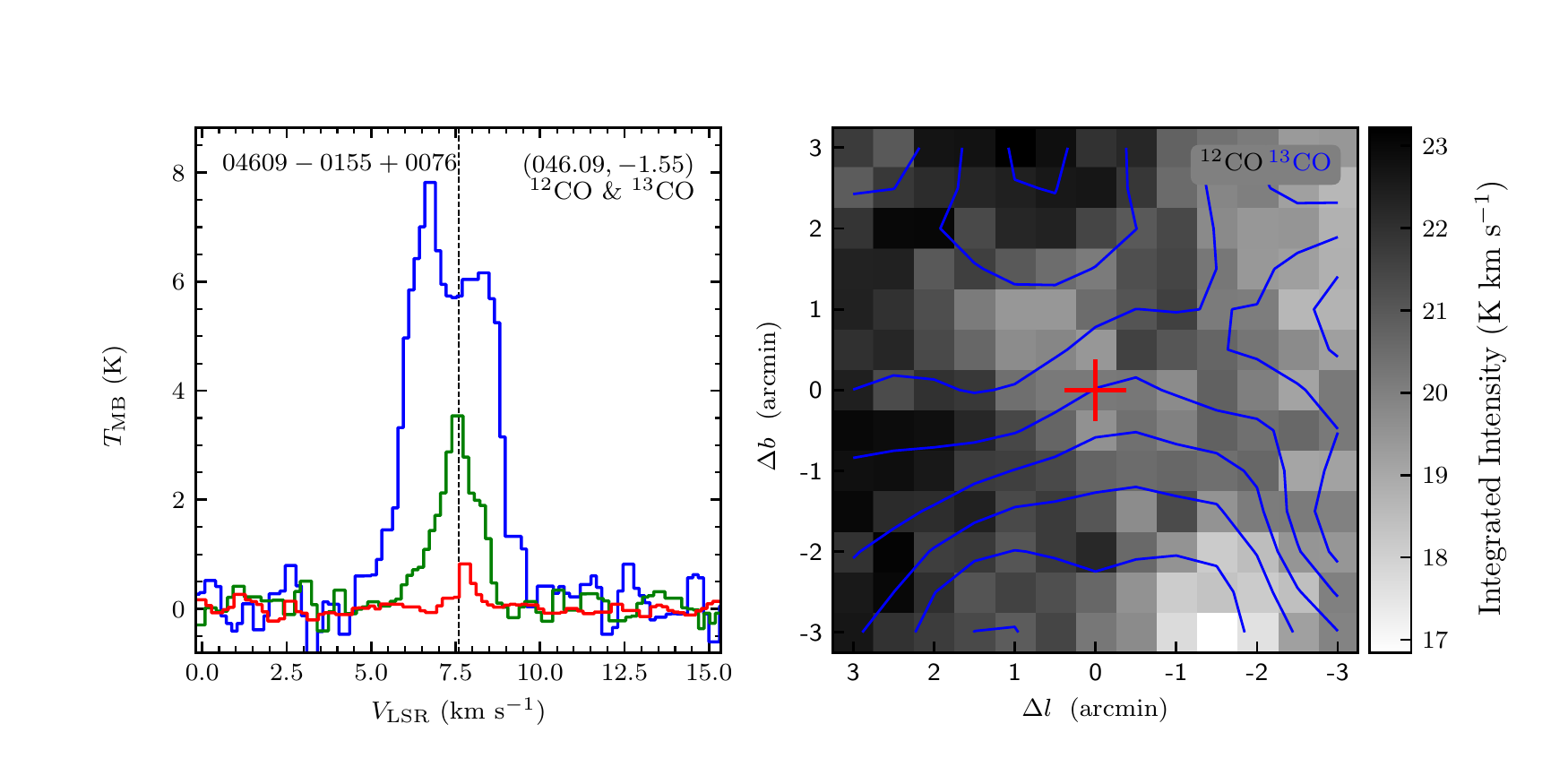}
\end{figure}
\clearpage

\begin{figure}
\includegraphics[width=9.0cm,angle=0]{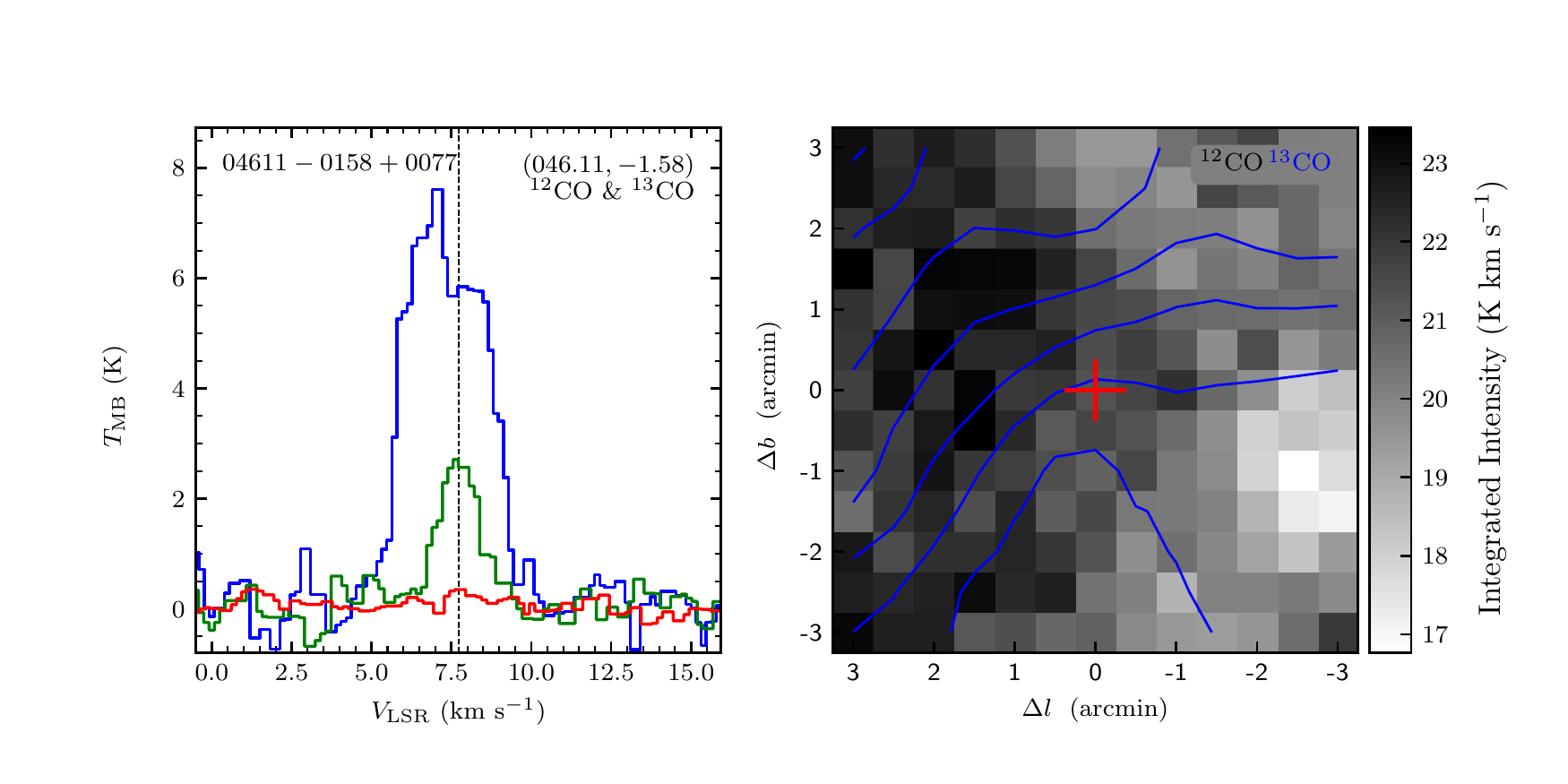}
\includegraphics[width=9.0cm,angle=0]{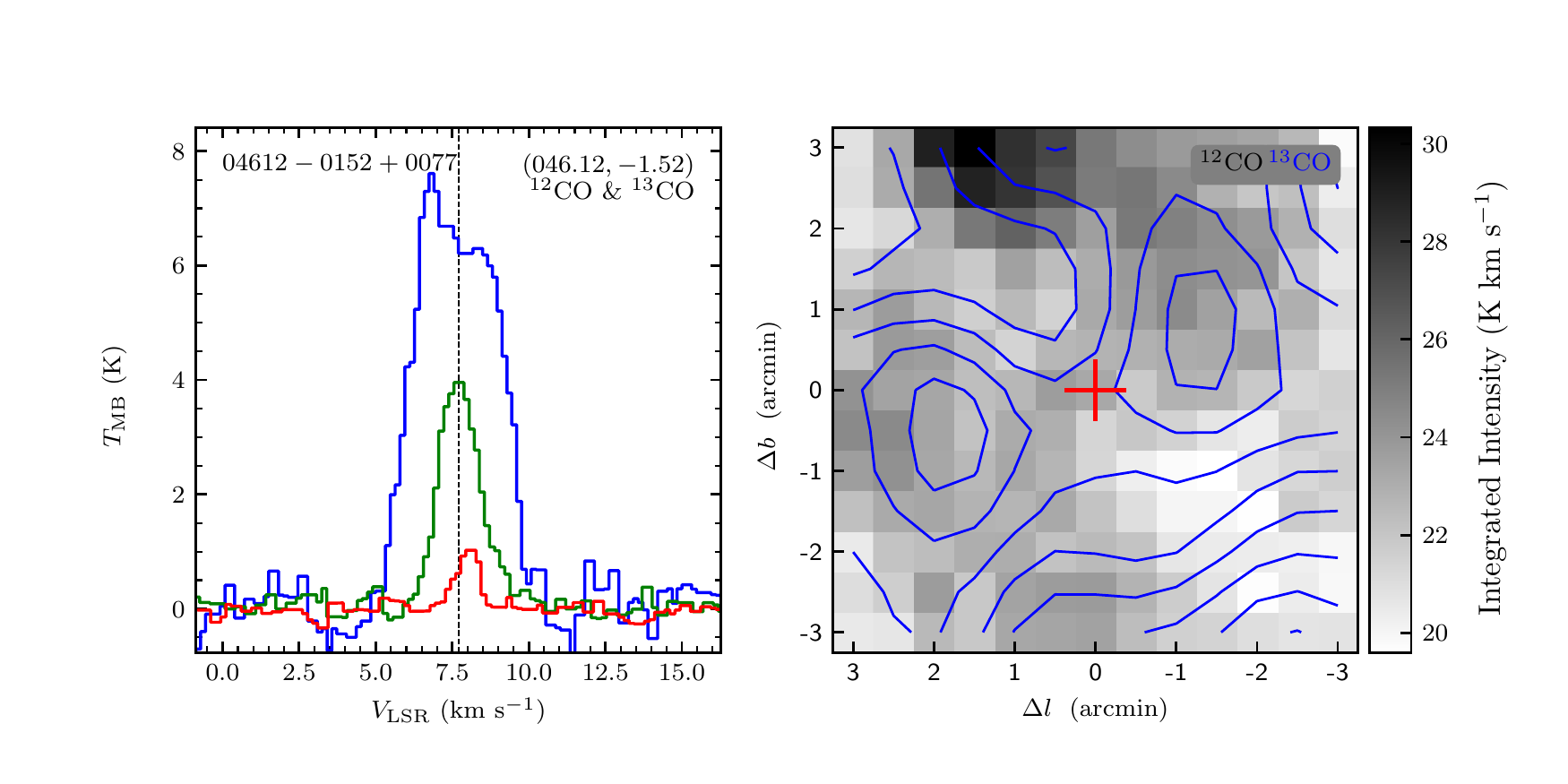}
\vspace{-0.5cm}

\includegraphics[width=9.0cm,angle=0]{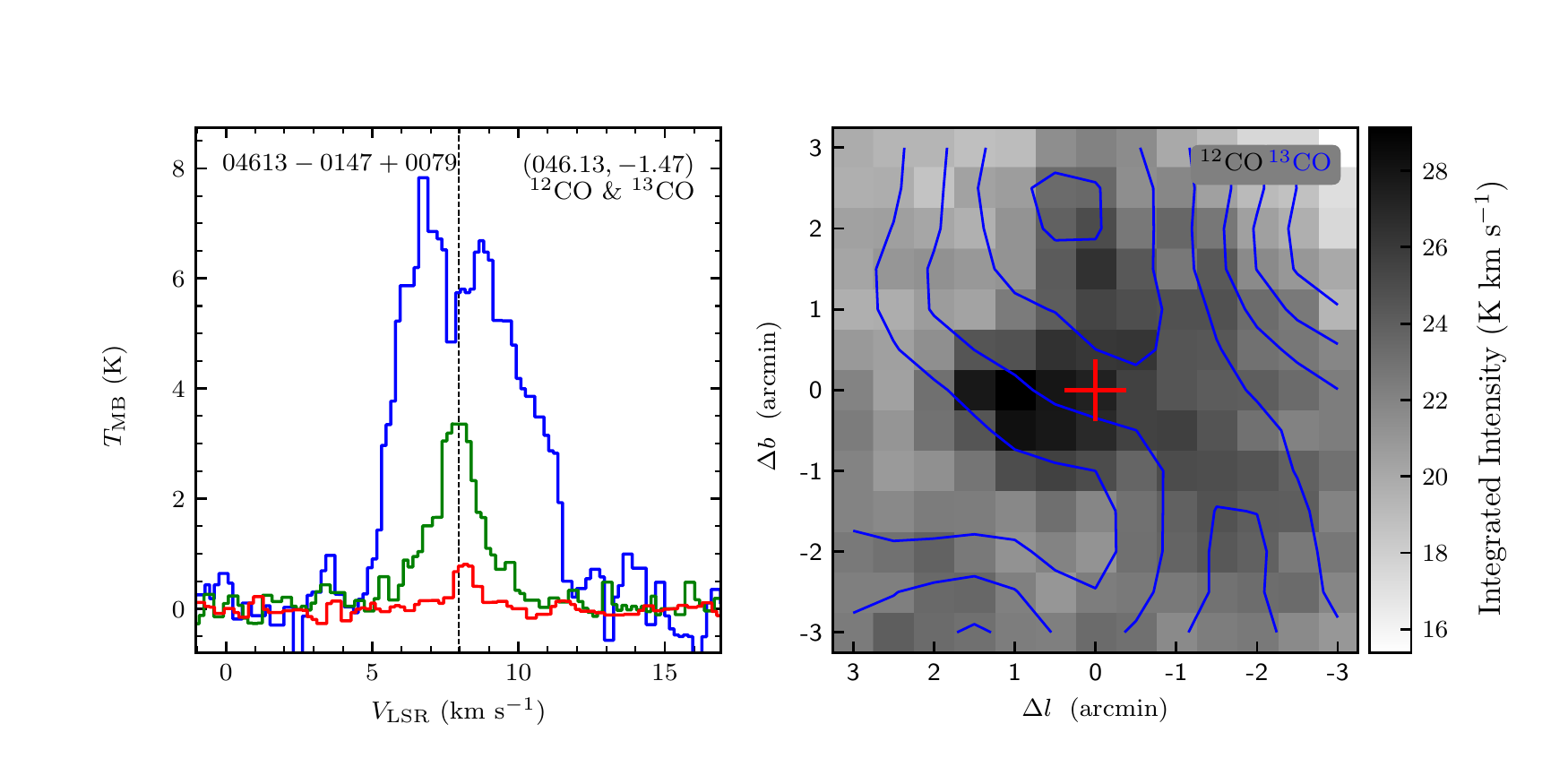}
\includegraphics[width=9.0cm,angle=0]{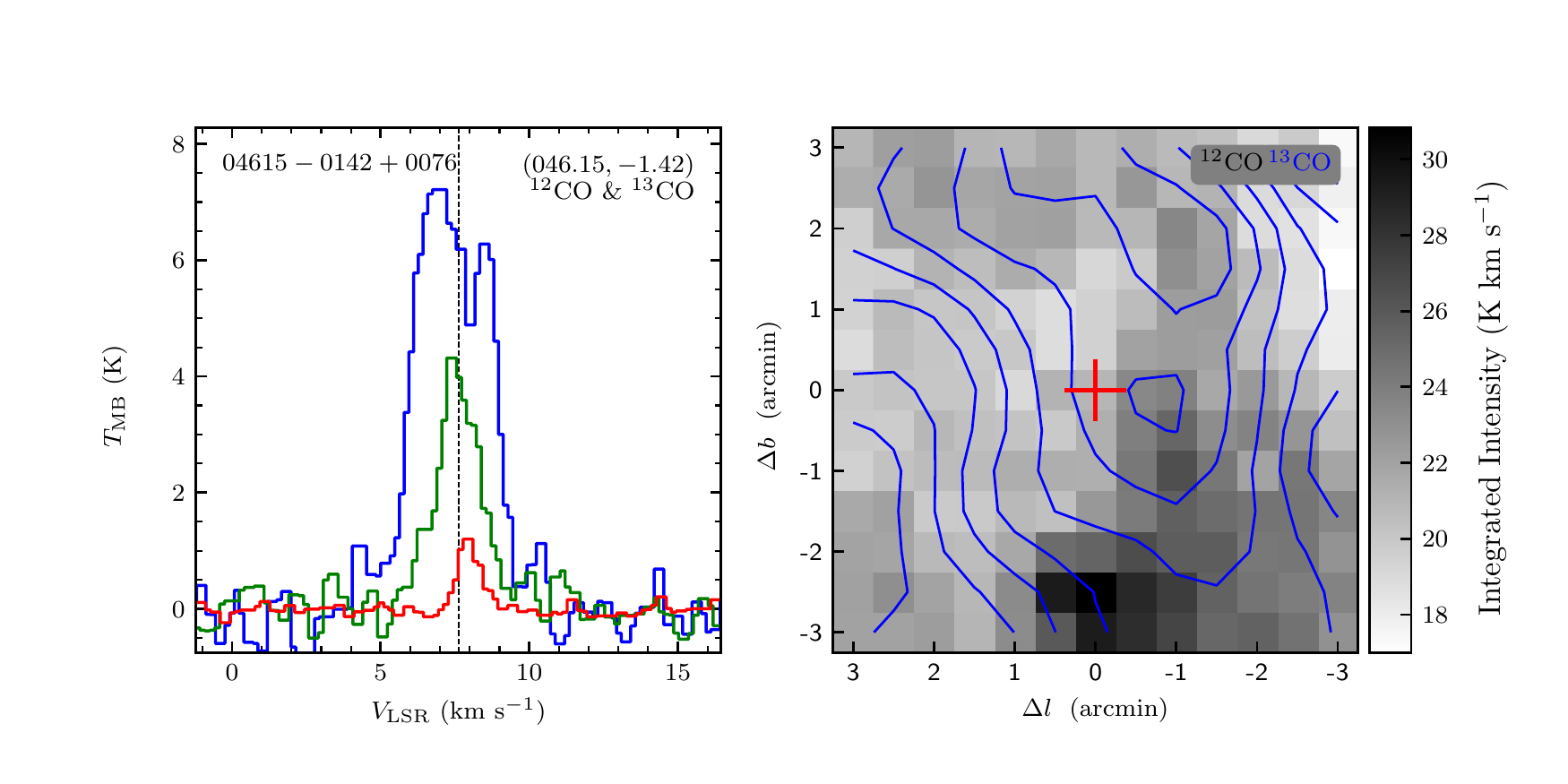}
\vspace{-0.5cm}

\includegraphics[width=9.0cm,angle=0]{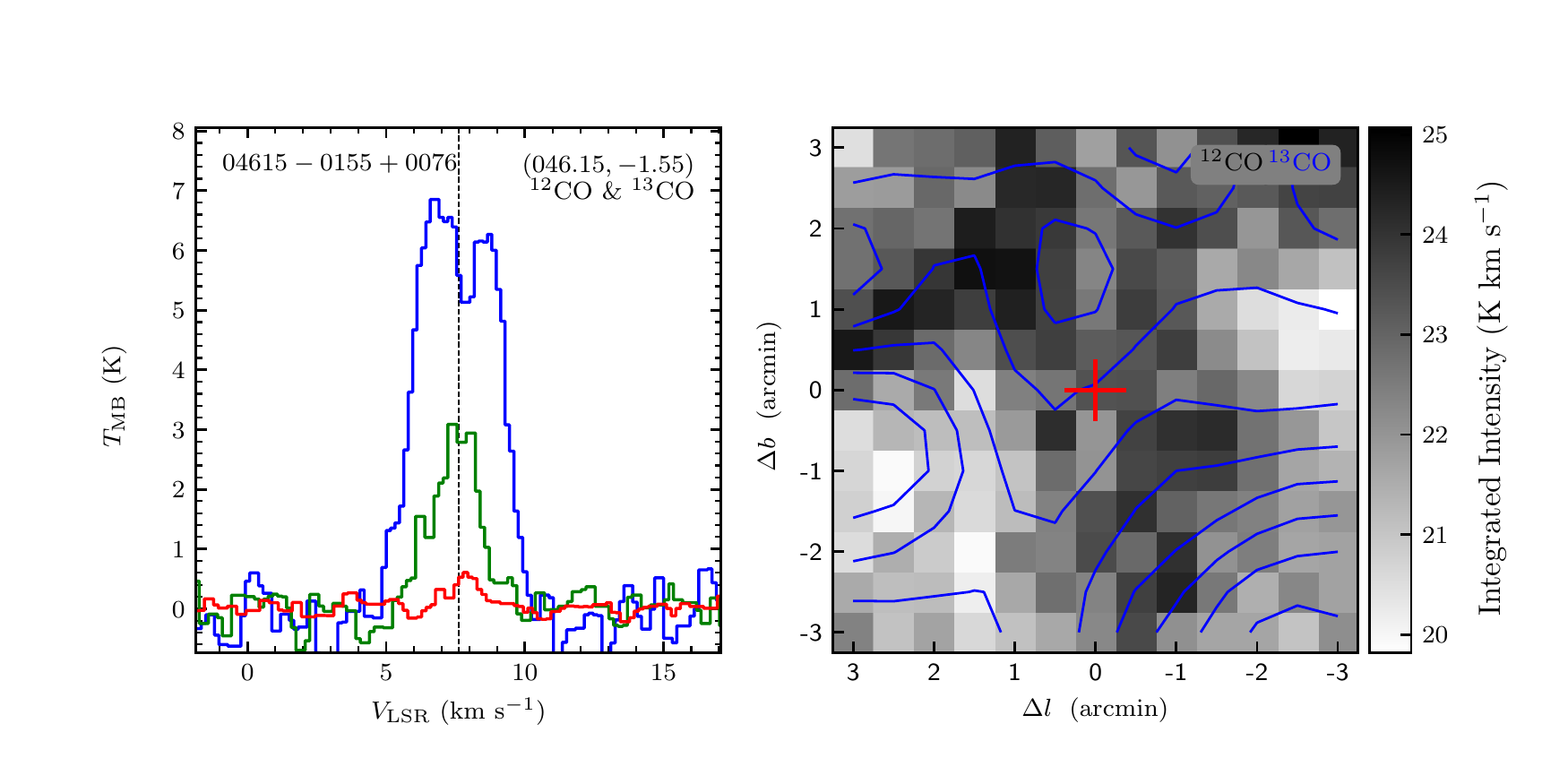}
\includegraphics[width=9.0cm,angle=0]{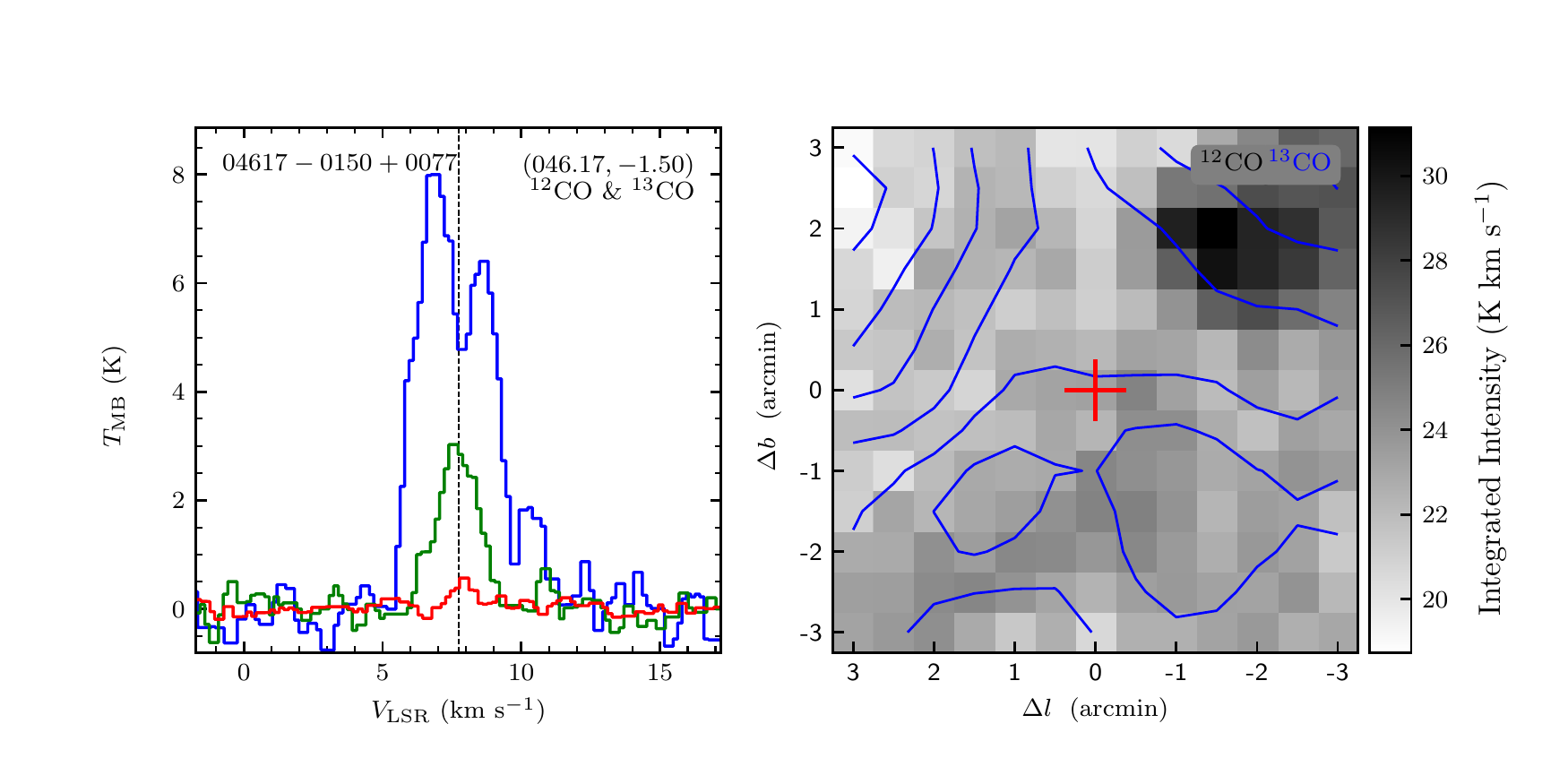}
\vspace{-0.5cm}

\includegraphics[width=9.0cm,angle=0]{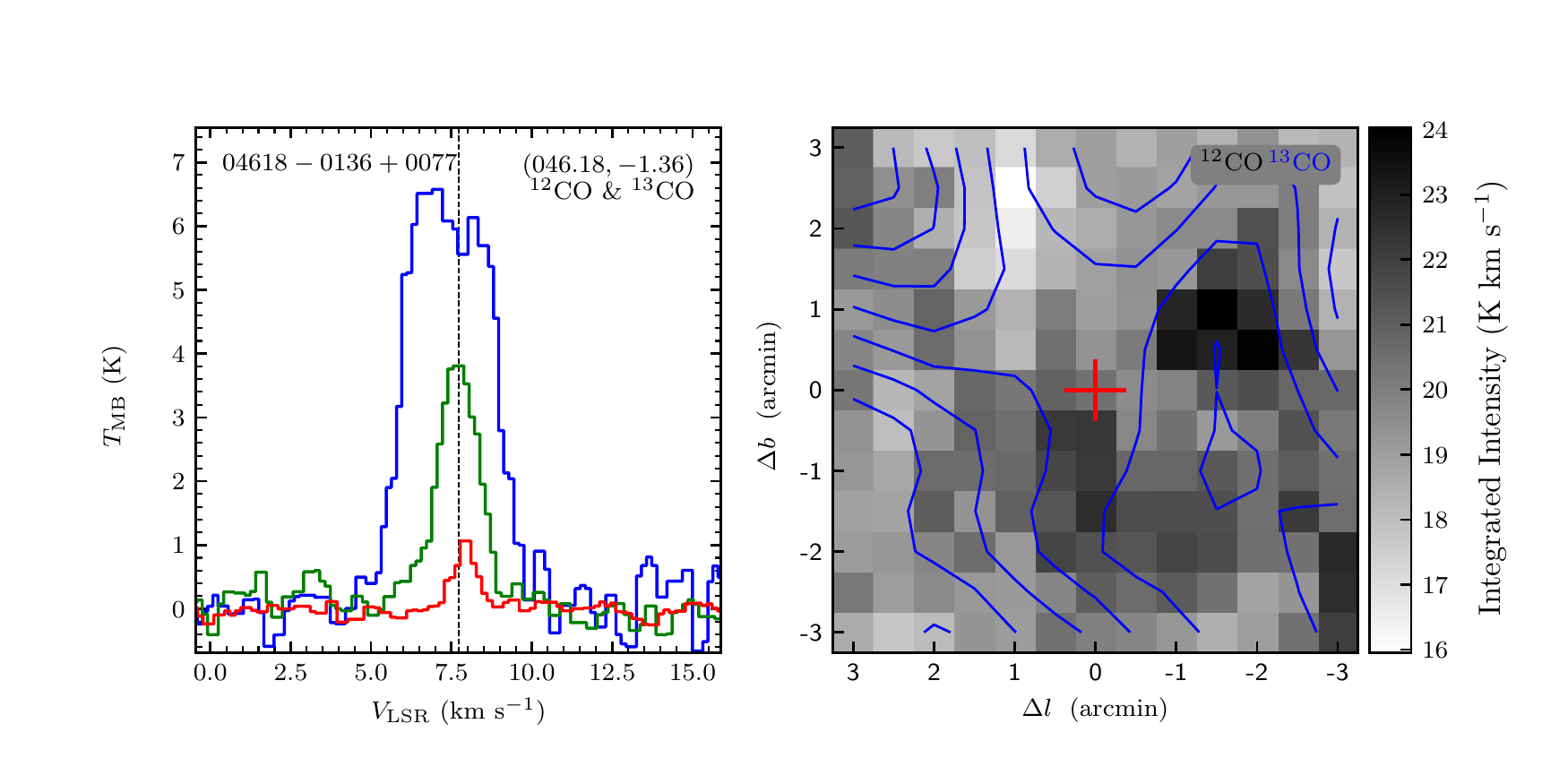}
\includegraphics[width=9.0cm,angle=0]{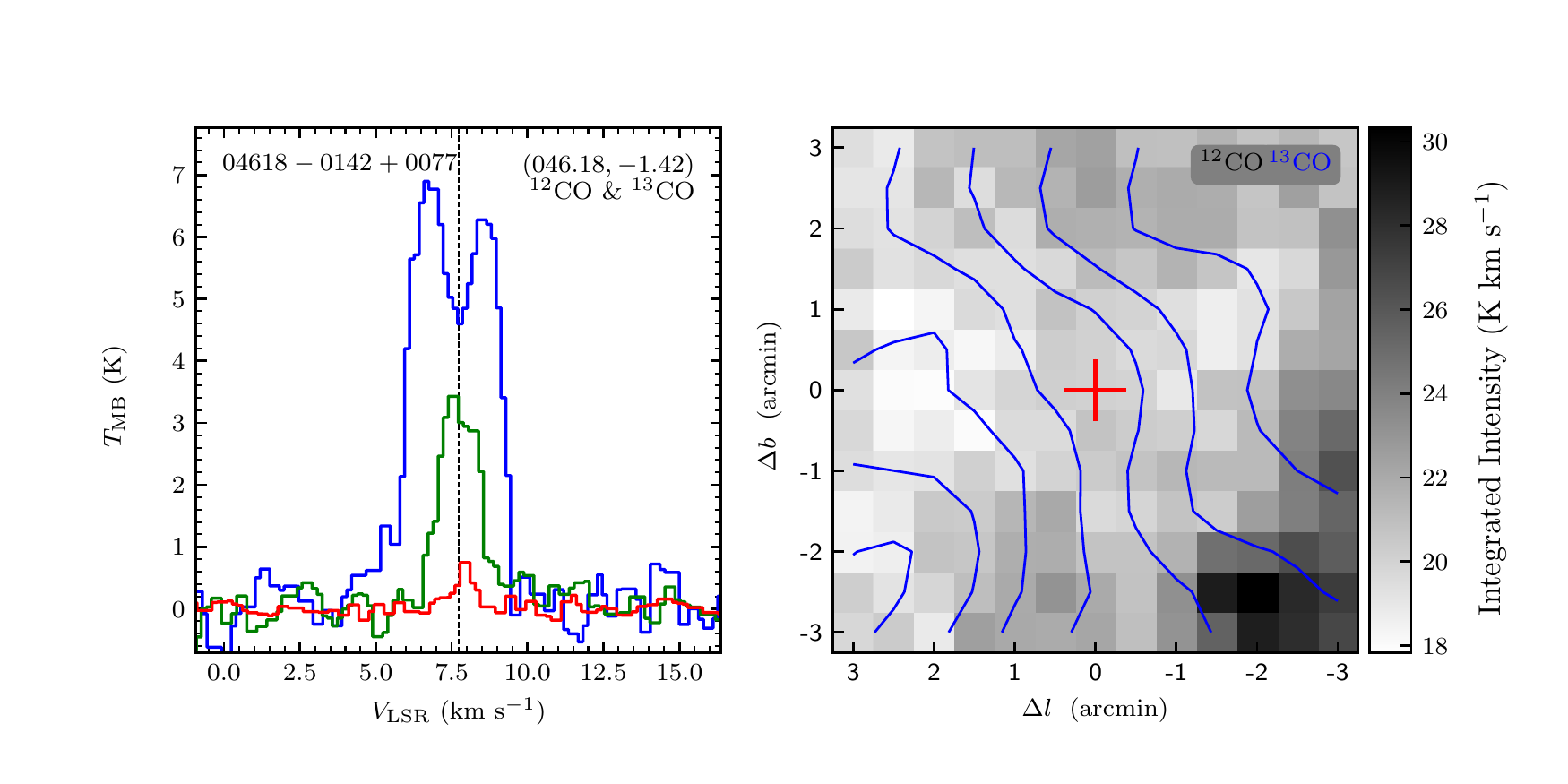}
\vspace{-0.5cm}

\includegraphics[width=9.0cm,angle=0]{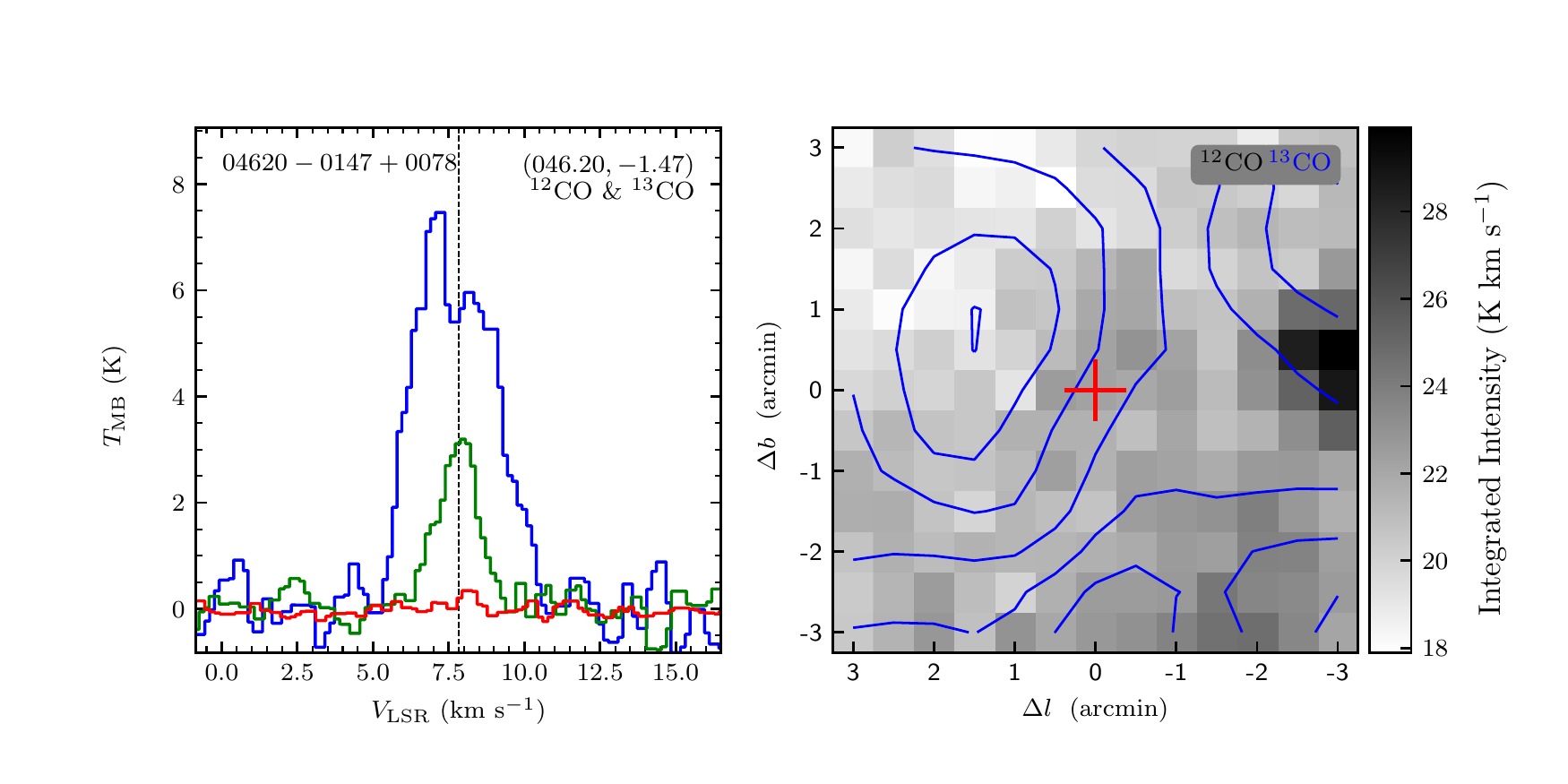}
\includegraphics[width=9.0cm,angle=0]{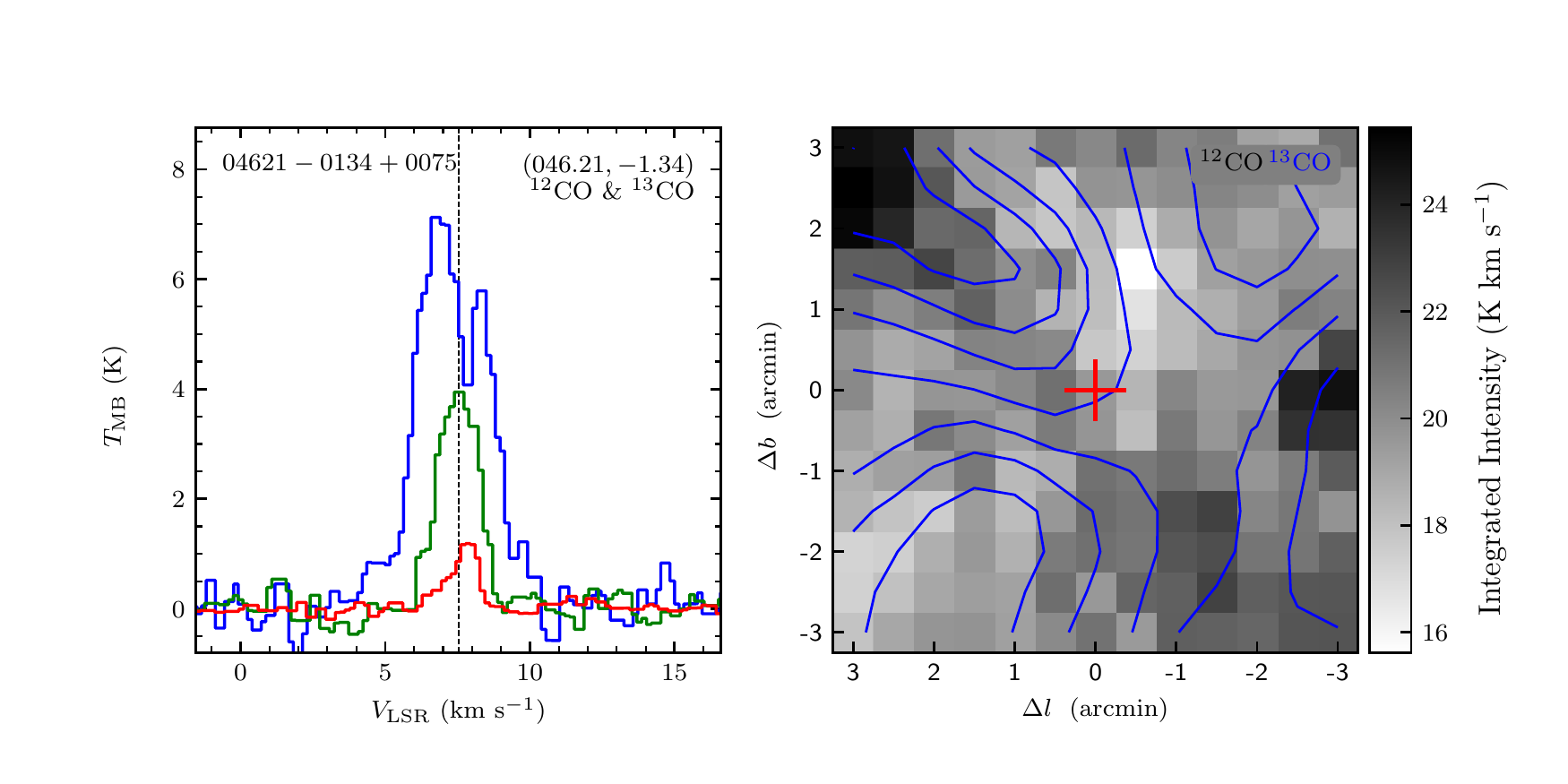}
\end{figure}
\clearpage

\begin{figure}
\includegraphics[width=9.0cm,angle=0]{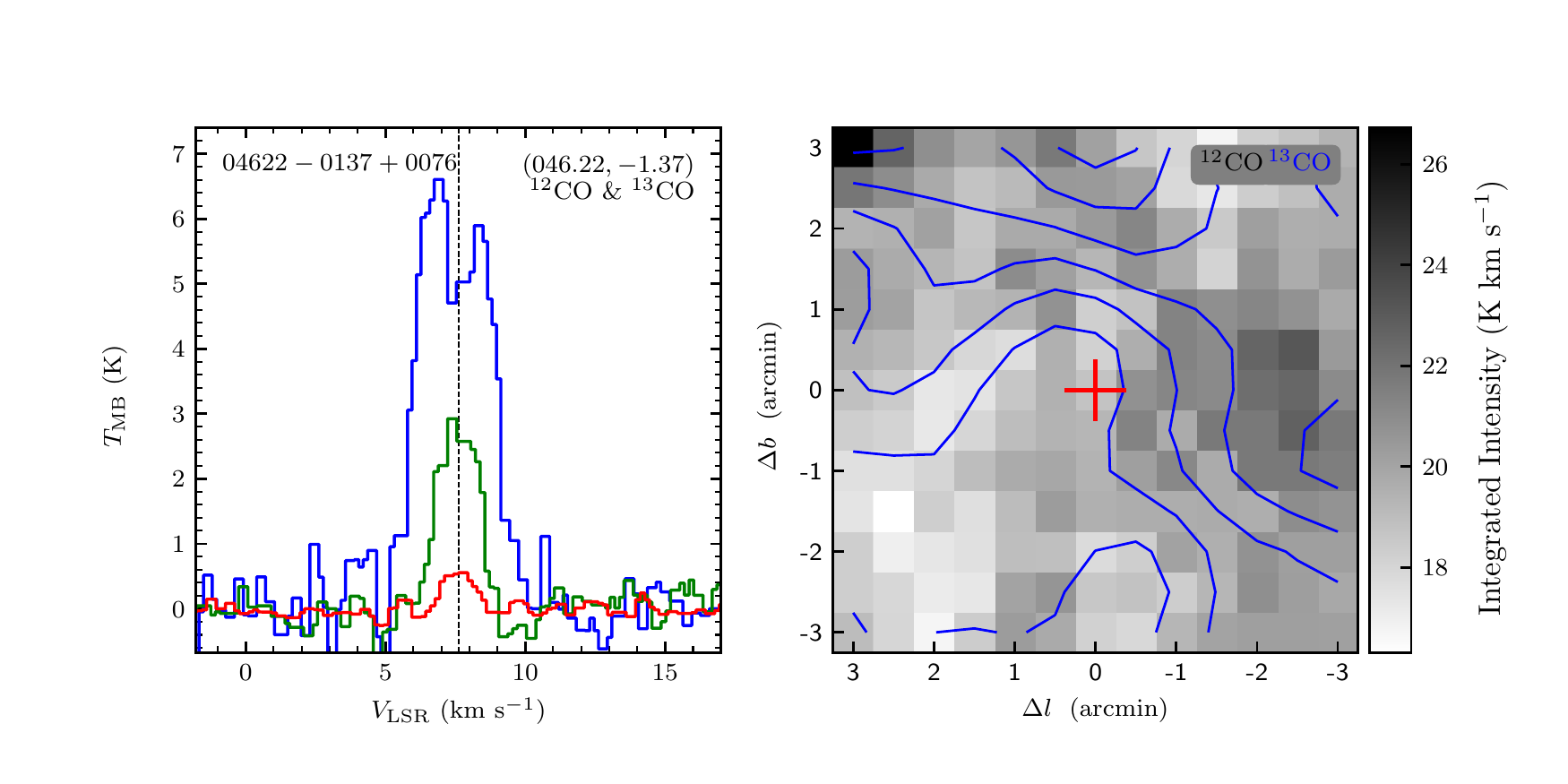}
\includegraphics[width=9.0cm,angle=0]{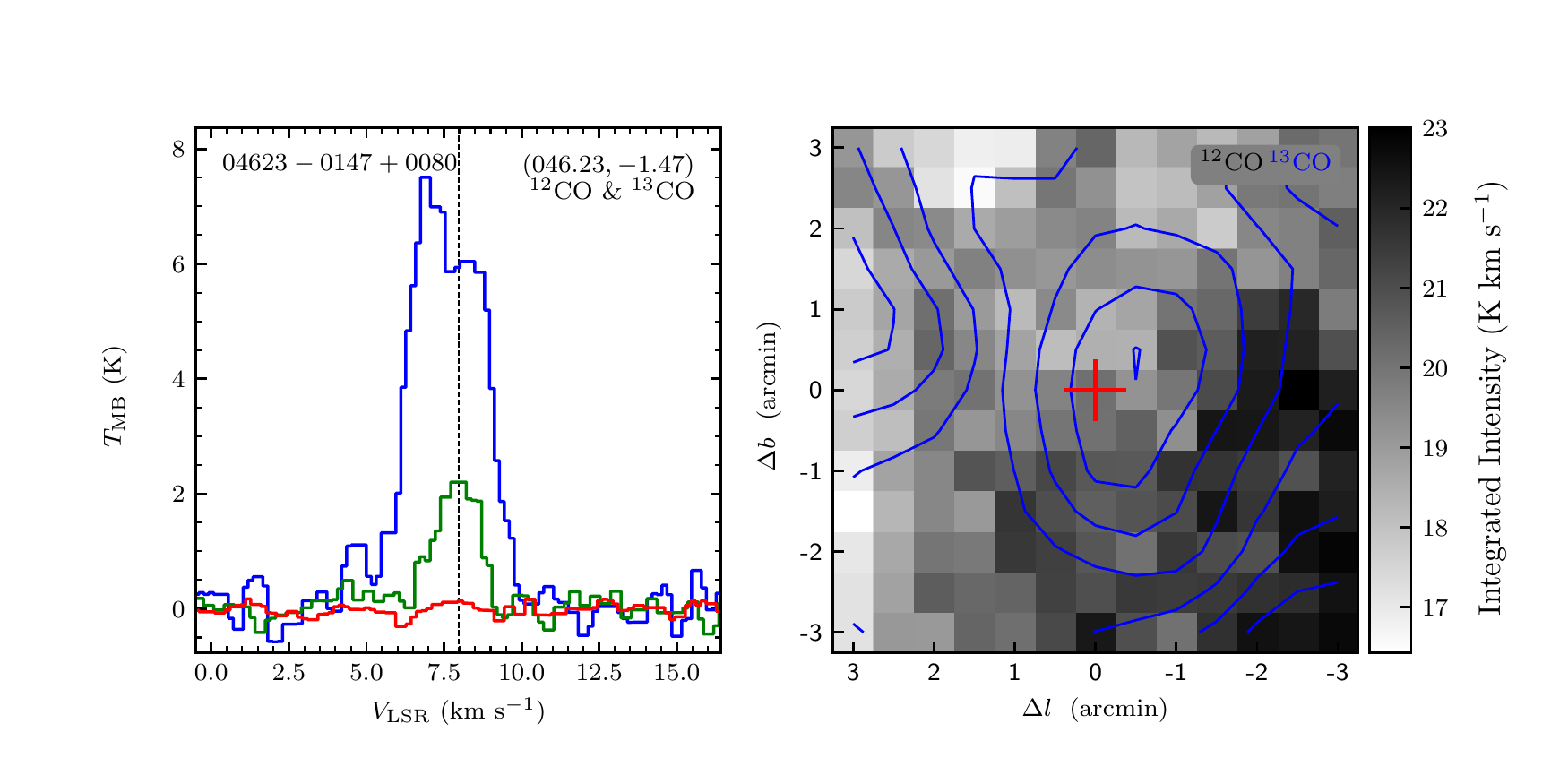}
\vspace{-0.5cm}

\includegraphics[width=9.0cm,angle=0]{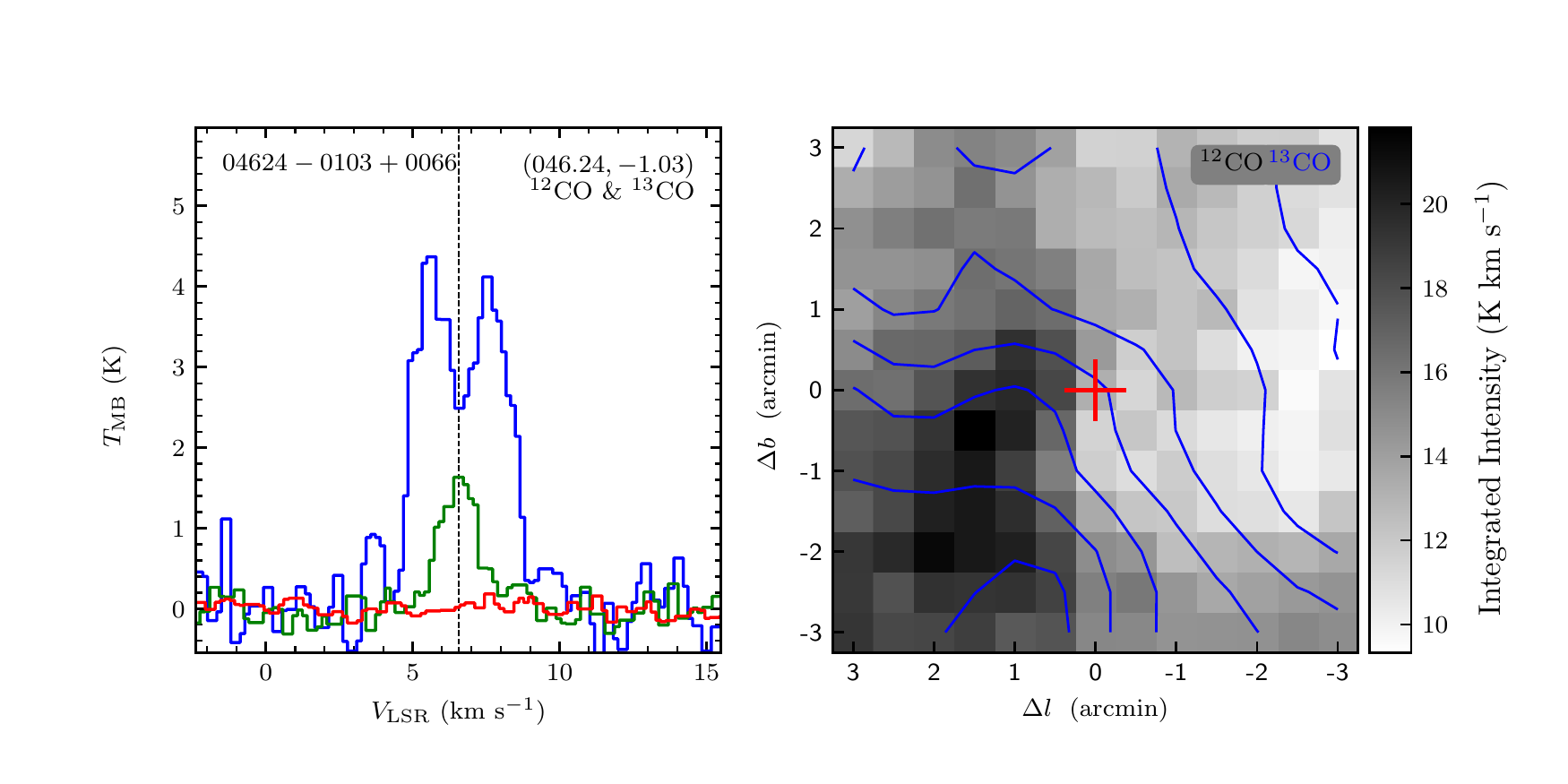}
\includegraphics[width=9.0cm,angle=0]{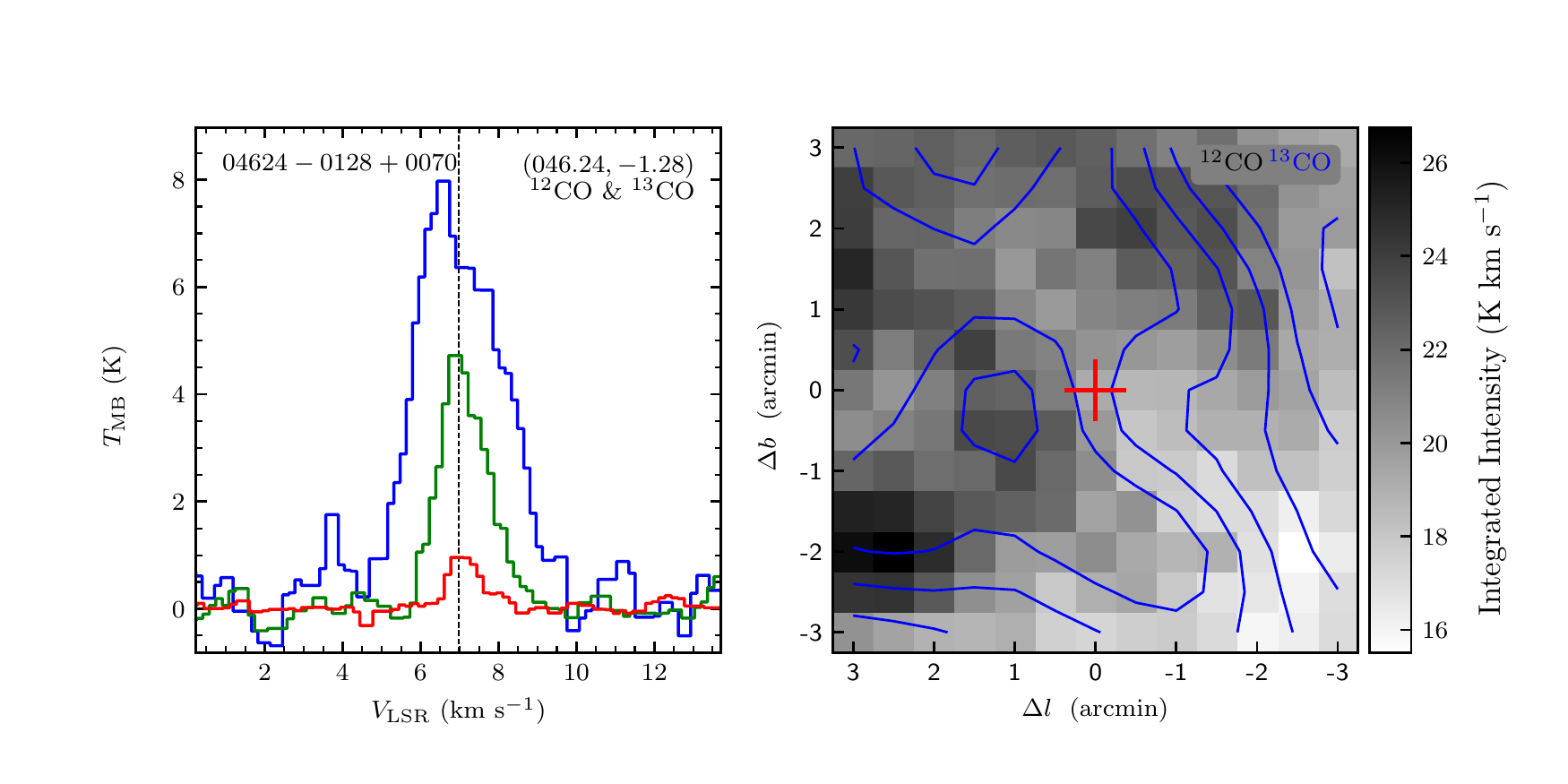}
\vspace{-0.5cm}

\includegraphics[width=9.0cm,angle=0]{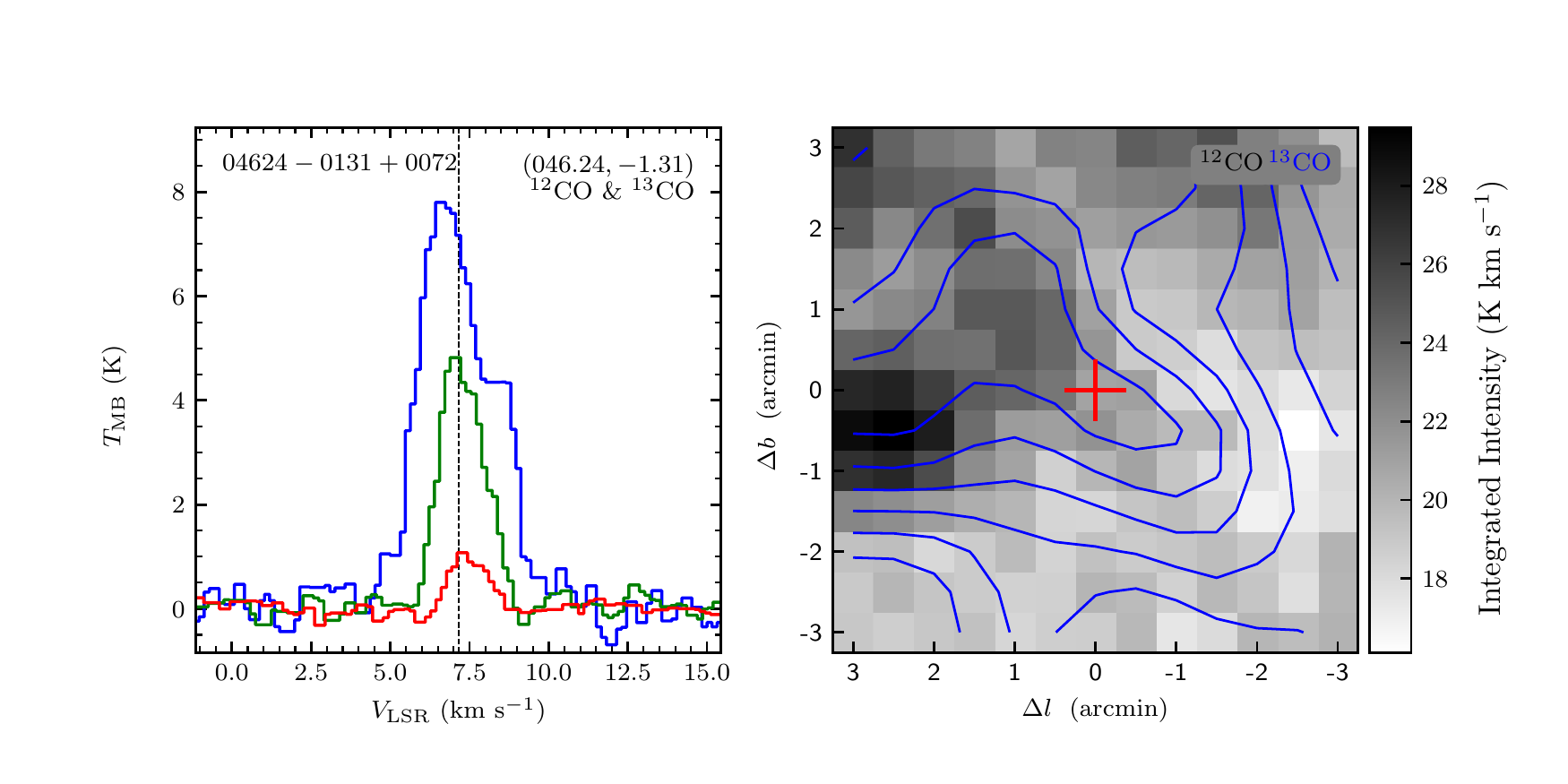}
\includegraphics[width=9.0cm,angle=0]{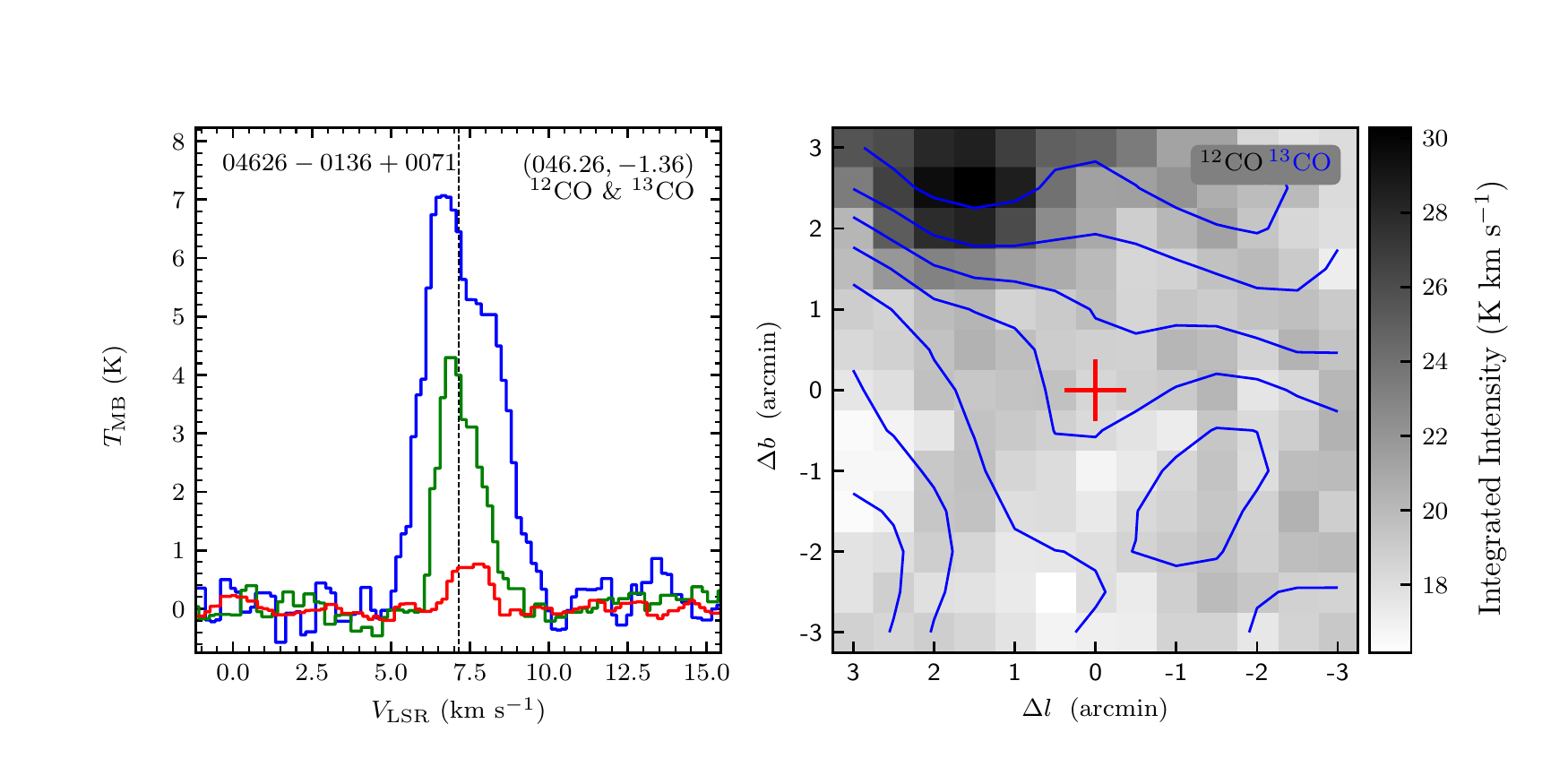}
\vspace{-0.5cm}

\includegraphics[width=9.0cm,angle=0]{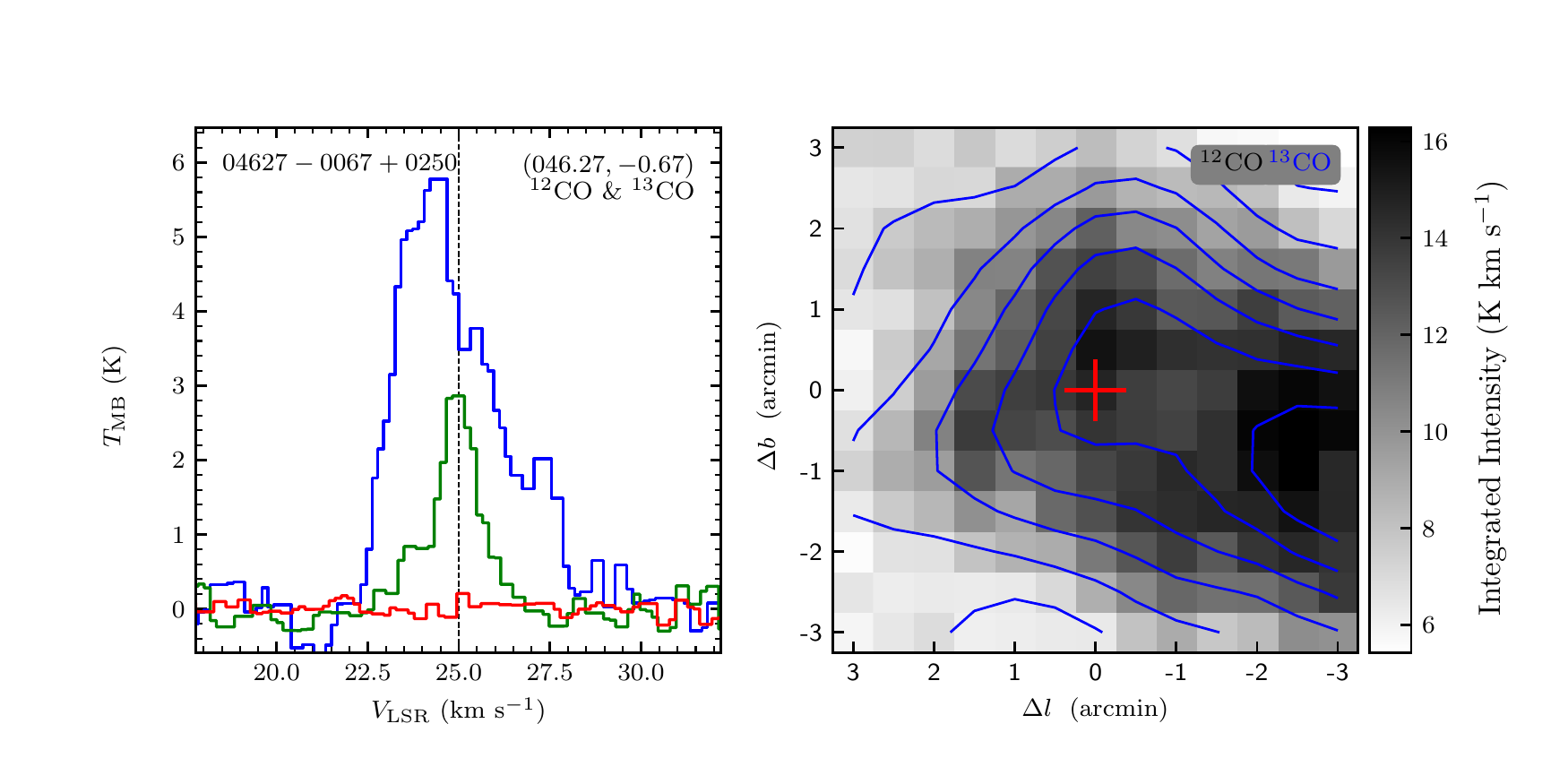}
\includegraphics[width=9.0cm,angle=0]{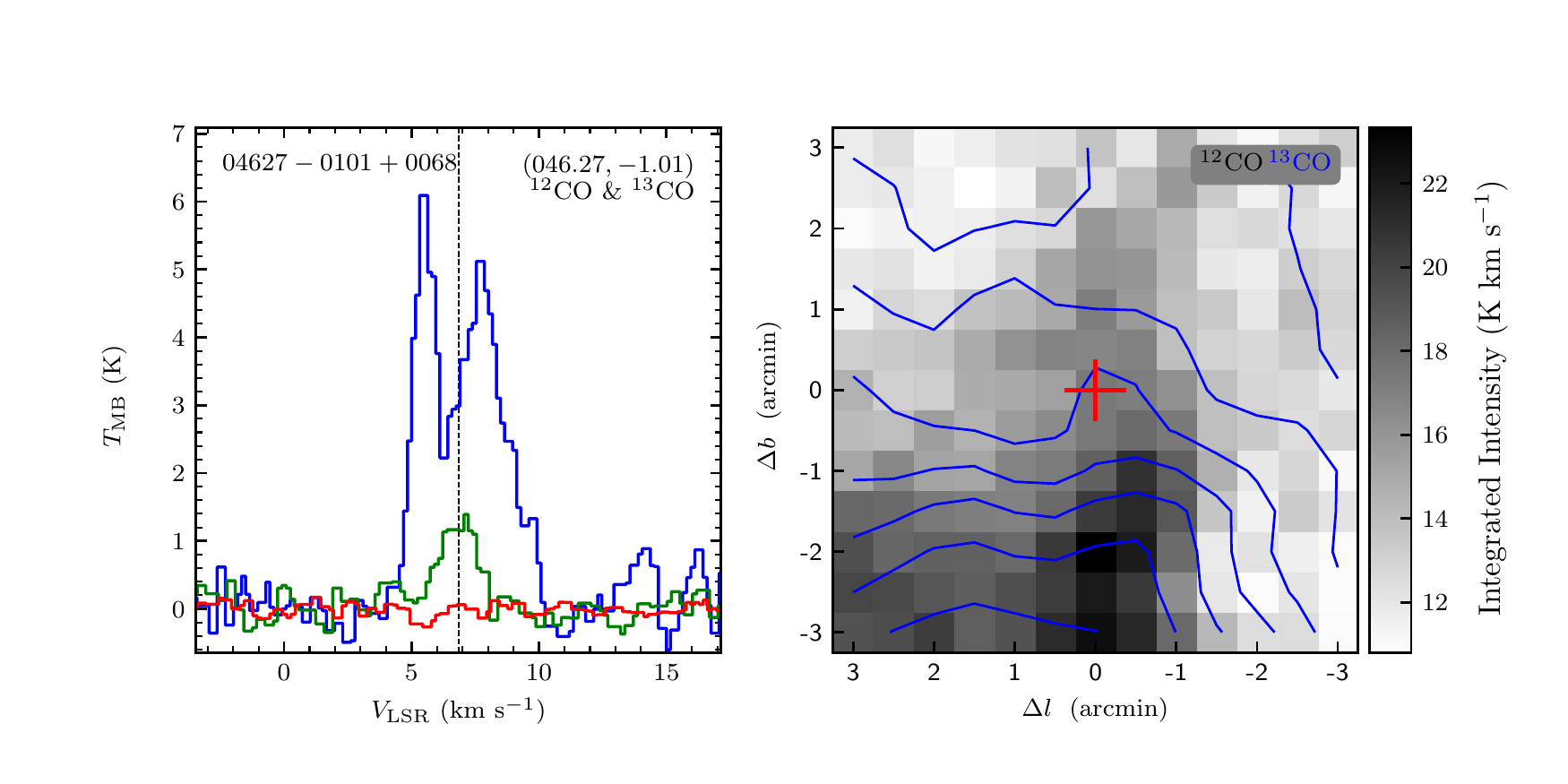}
\vspace{-0.5cm}

\includegraphics[width=9.0cm,angle=0]{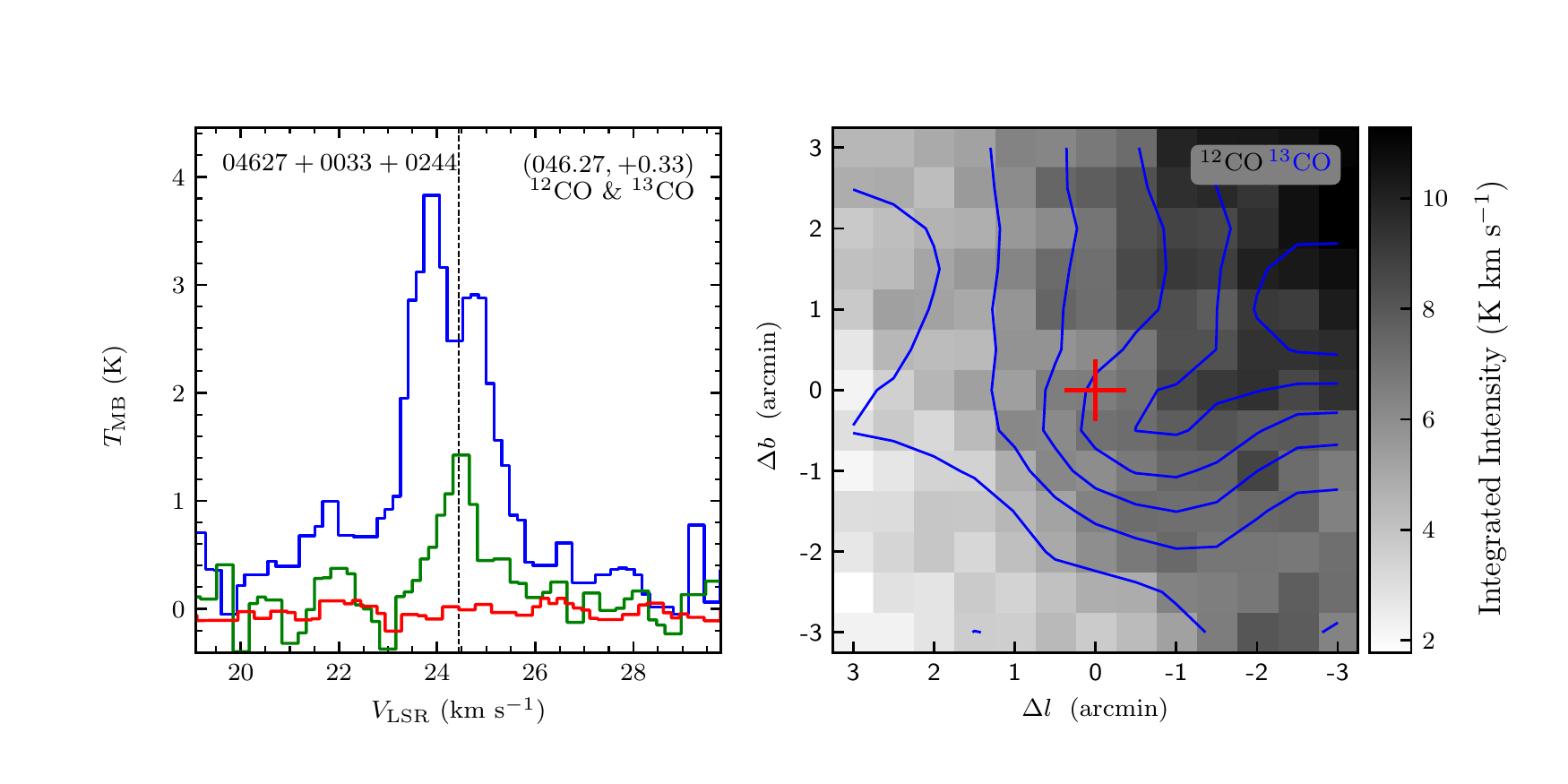}
\includegraphics[width=9.0cm,angle=0]{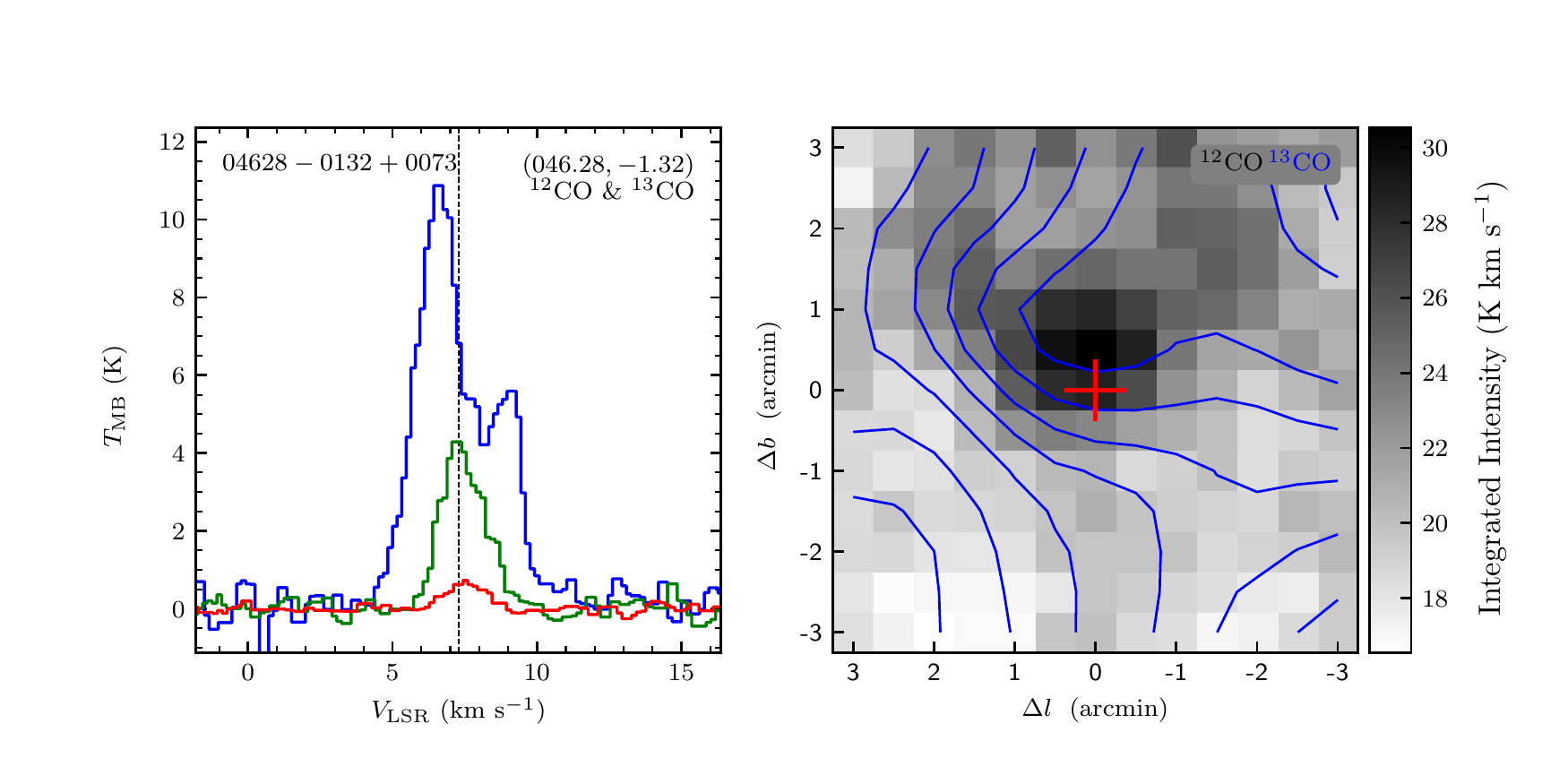}
\end{figure}
\clearpage

\begin{figure}
\includegraphics[width=9.0cm,angle=0]{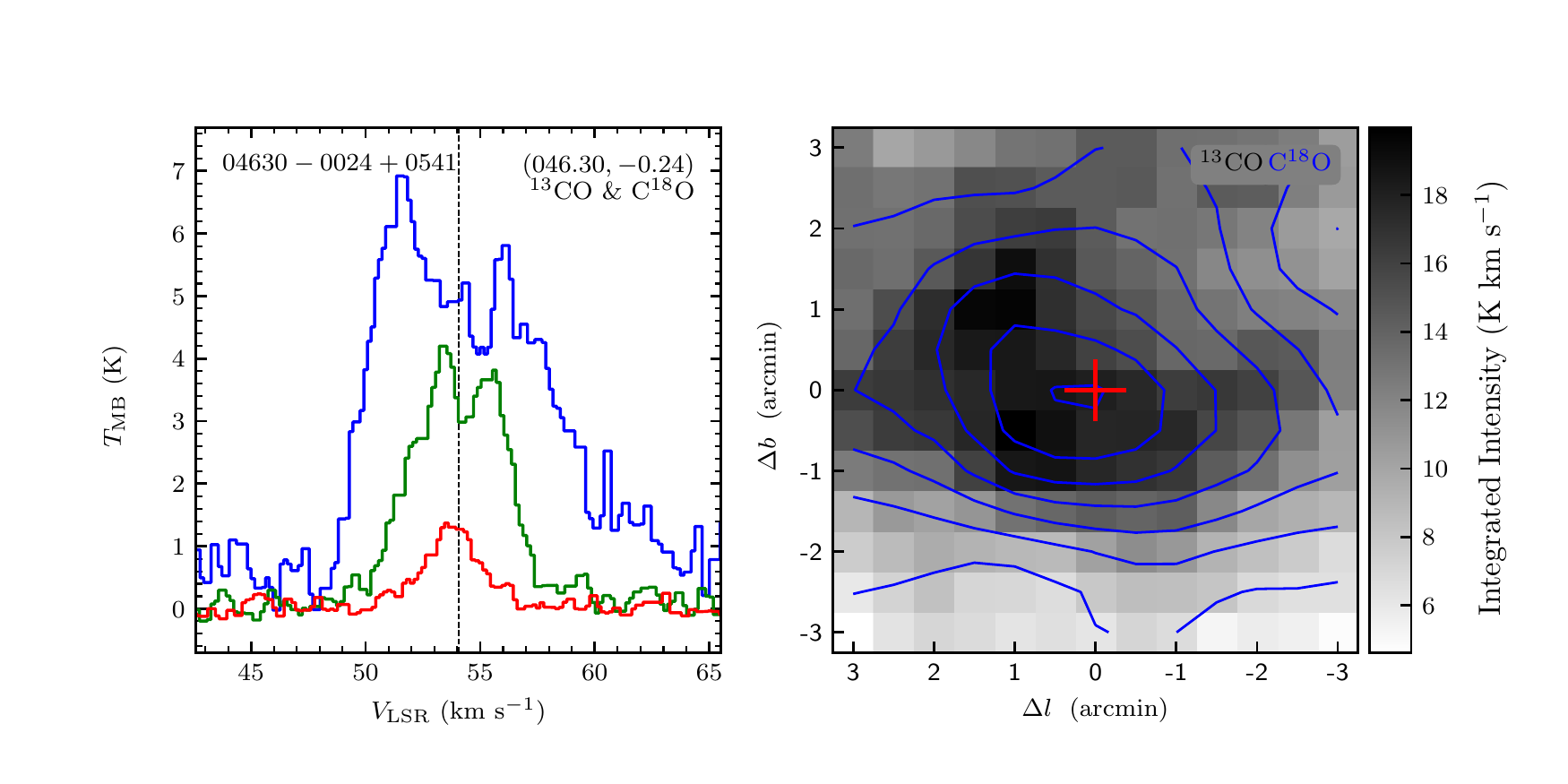}
\includegraphics[width=9.0cm,angle=0]{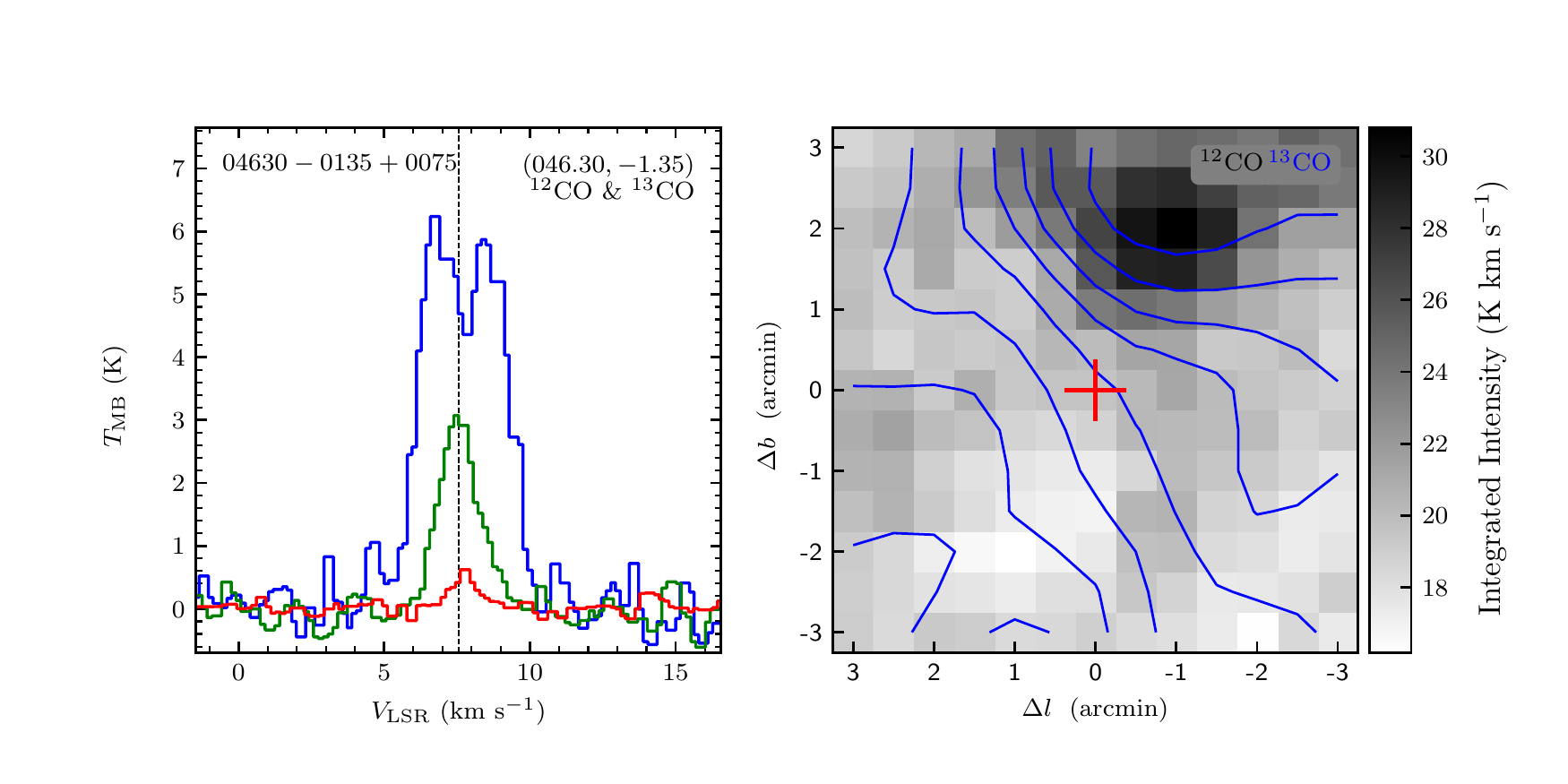}
\vspace{-0.5cm}

\includegraphics[width=9.0cm,angle=0]{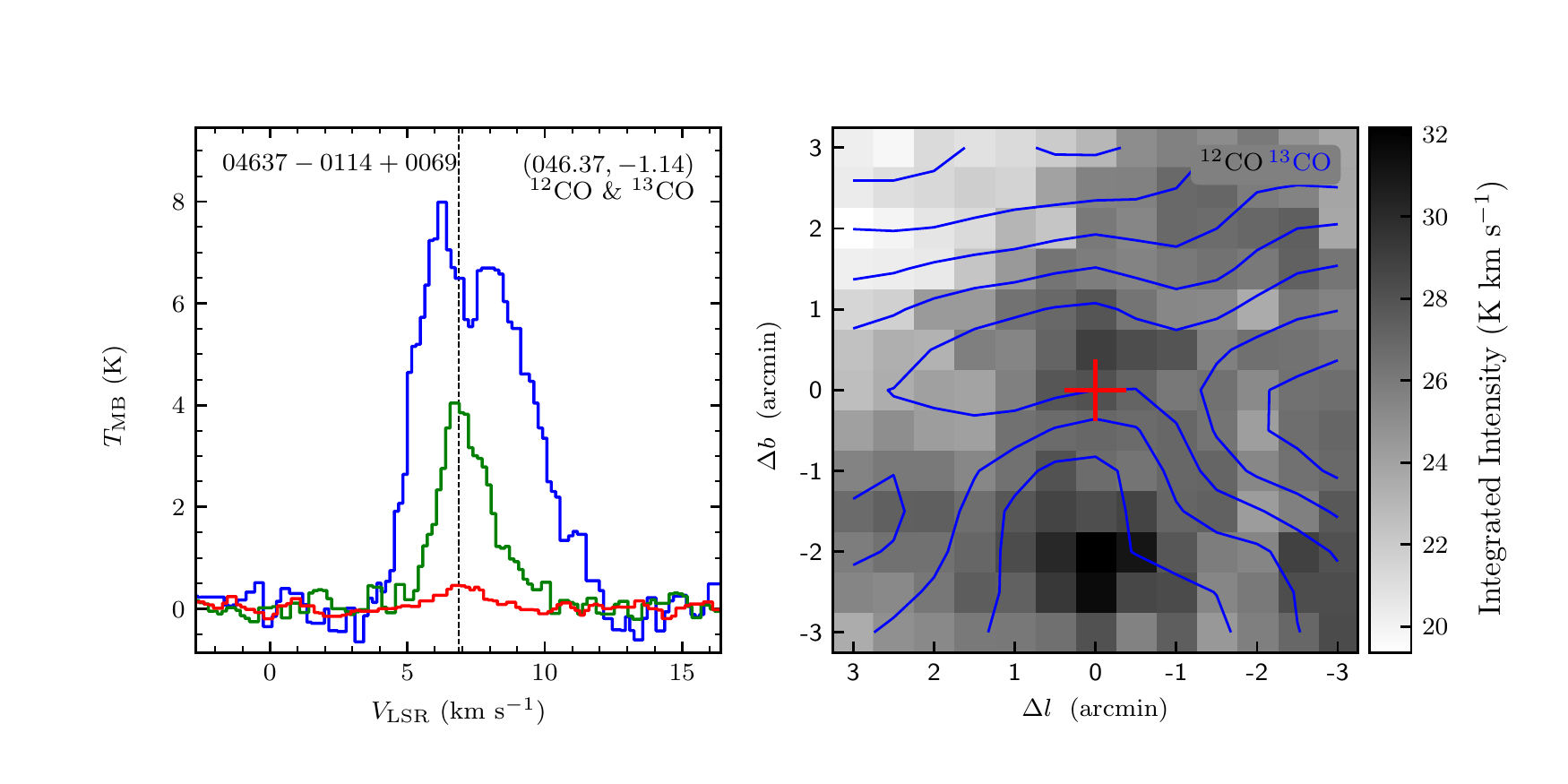}
\includegraphics[width=9.0cm,angle=0]{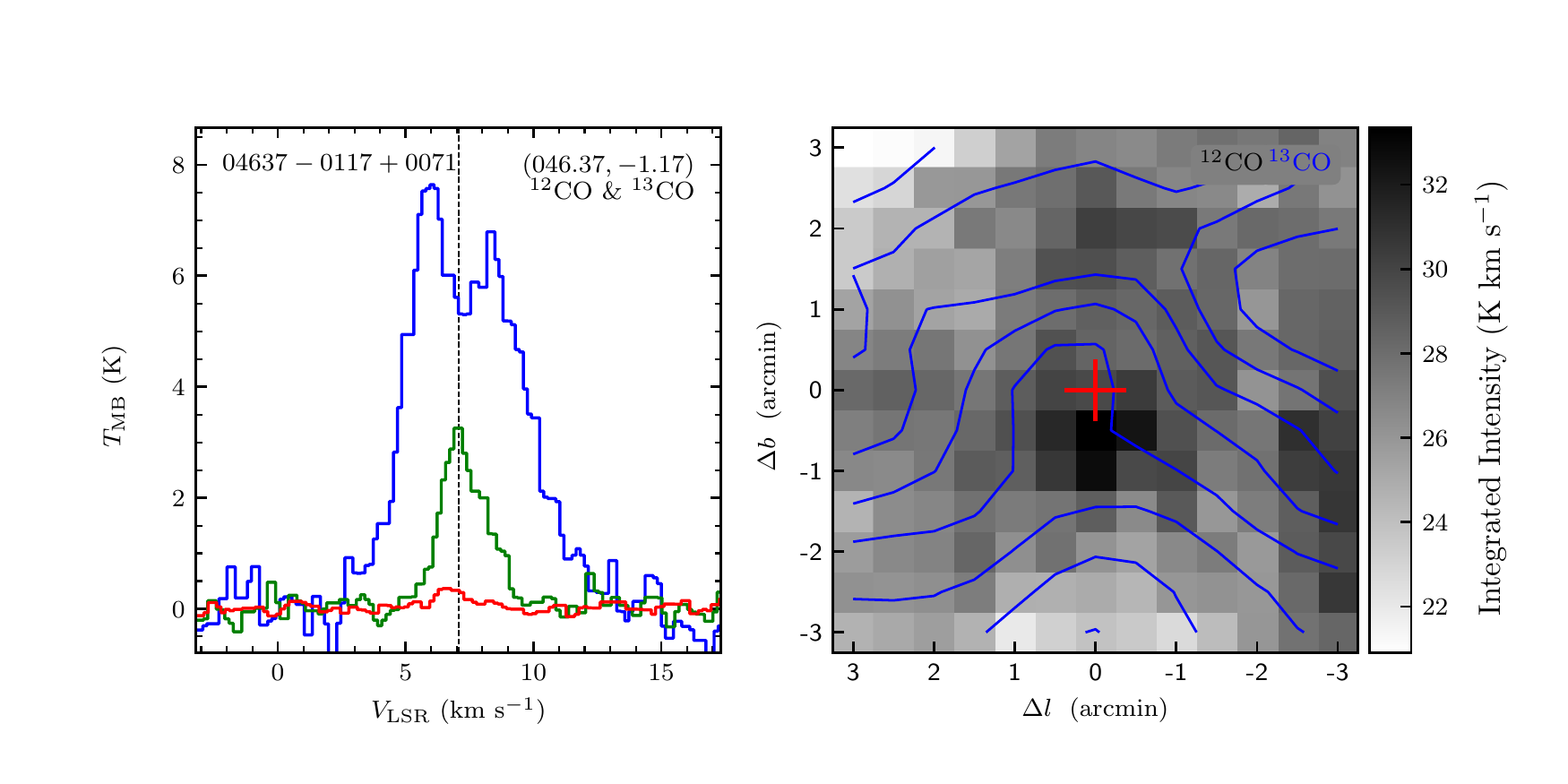}
\vspace{-0.5cm}

\includegraphics[width=9.0cm,angle=0]{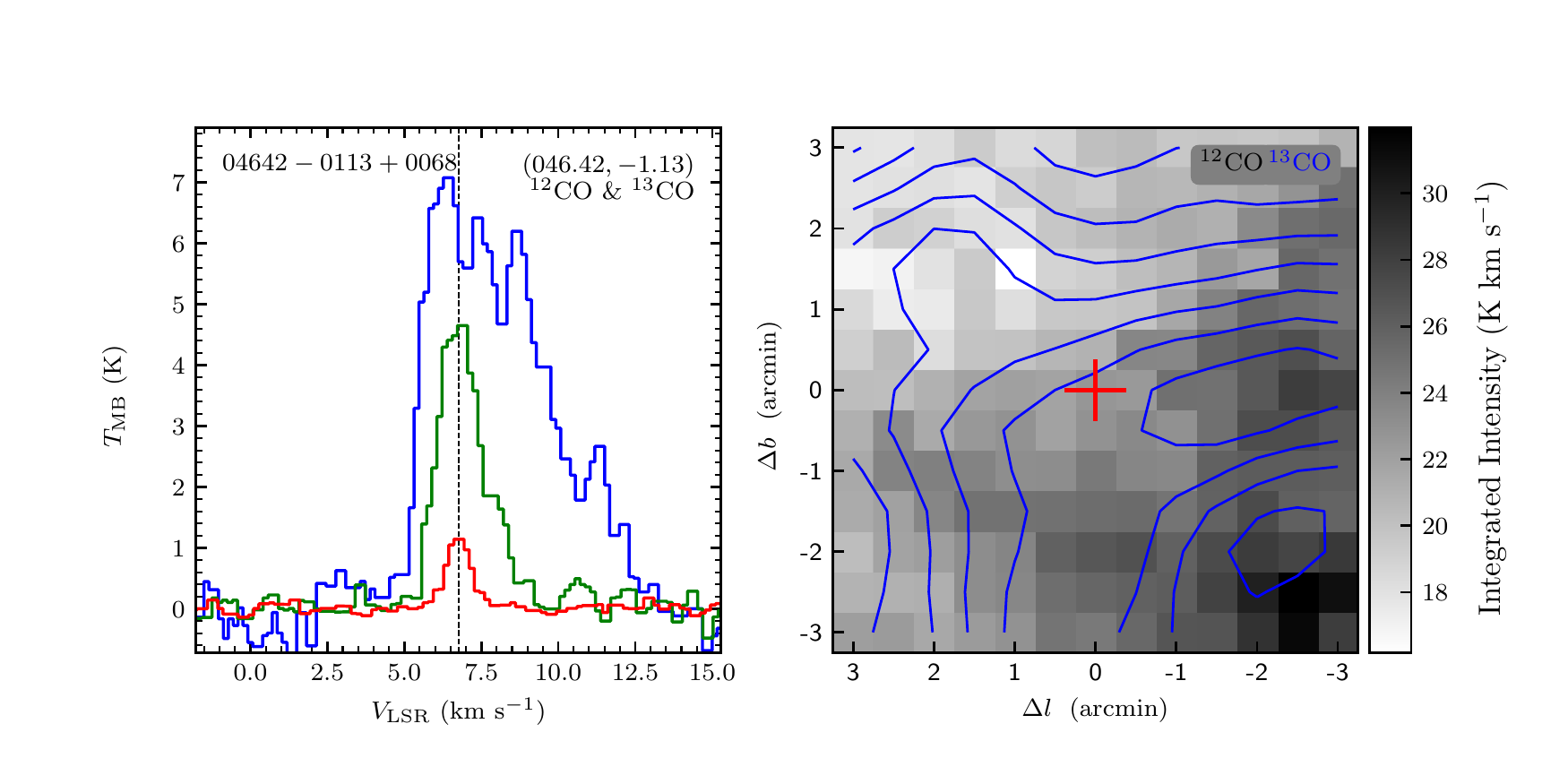}
\includegraphics[width=9.0cm,angle=0]{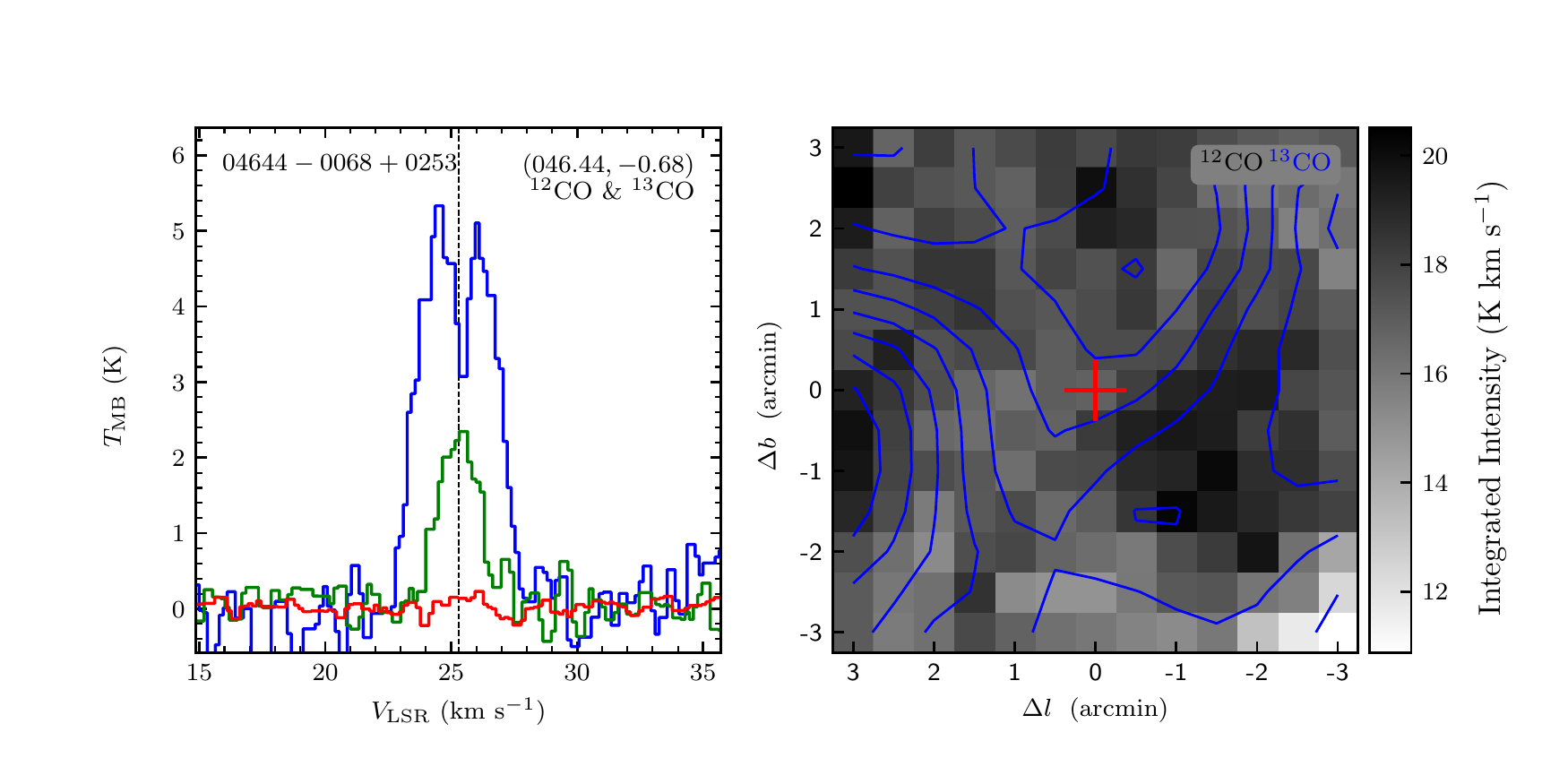}
\vspace{-0.5cm}

\includegraphics[width=9.0cm,angle=0]{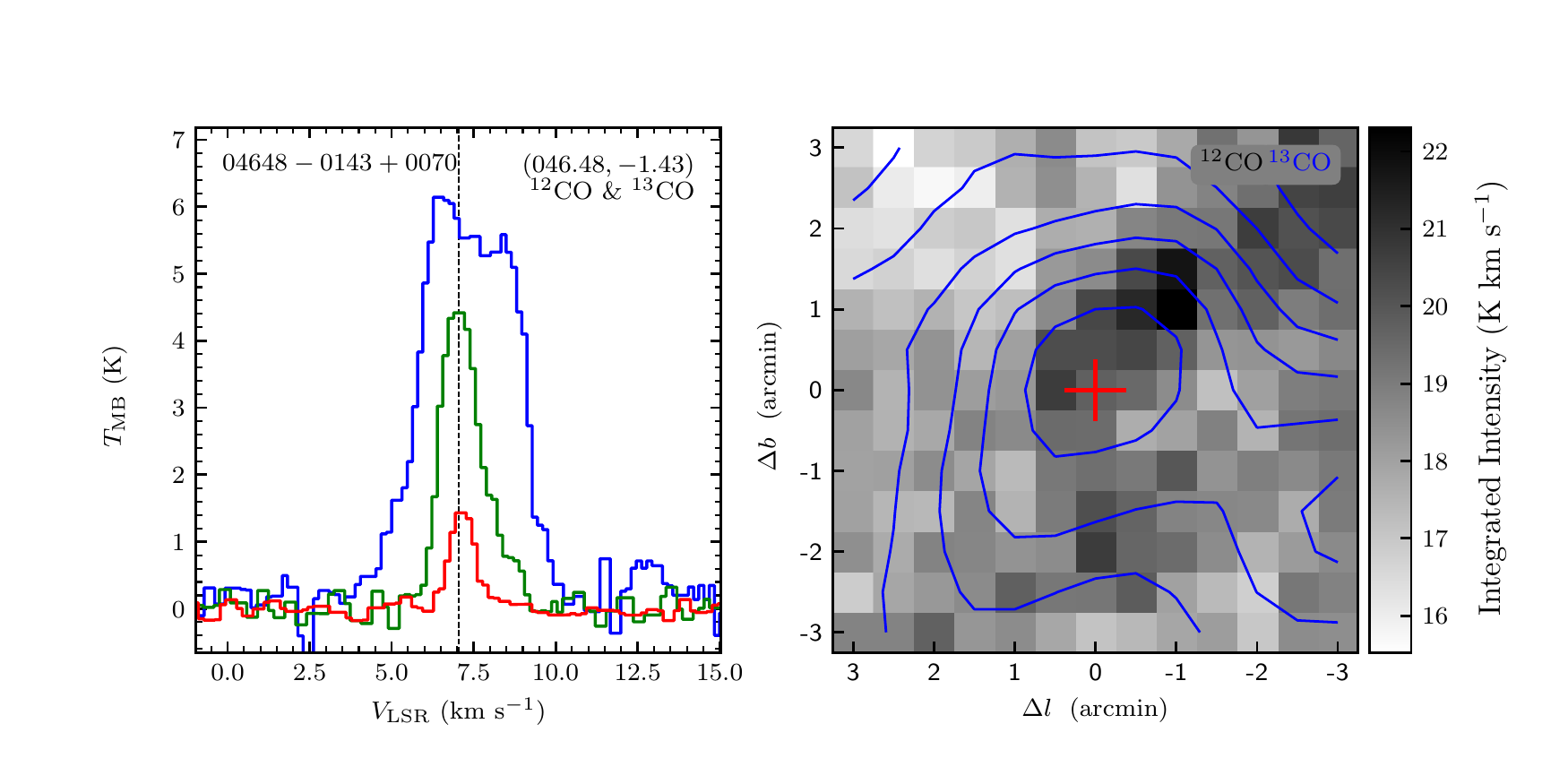}
\includegraphics[width=9.0cm,angle=0]{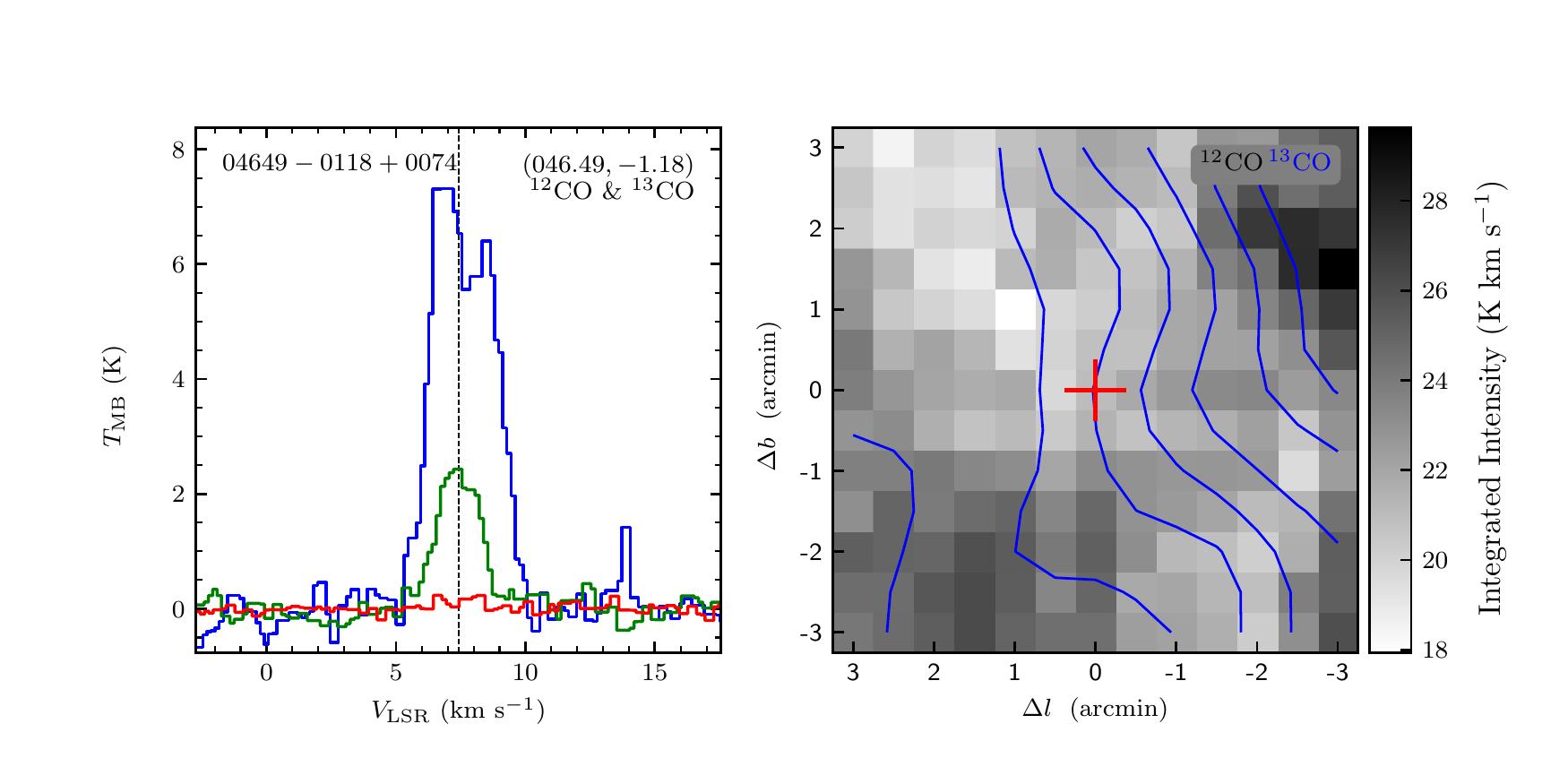}
\vspace{-0.5cm}

\includegraphics[width=9.0cm,angle=0]{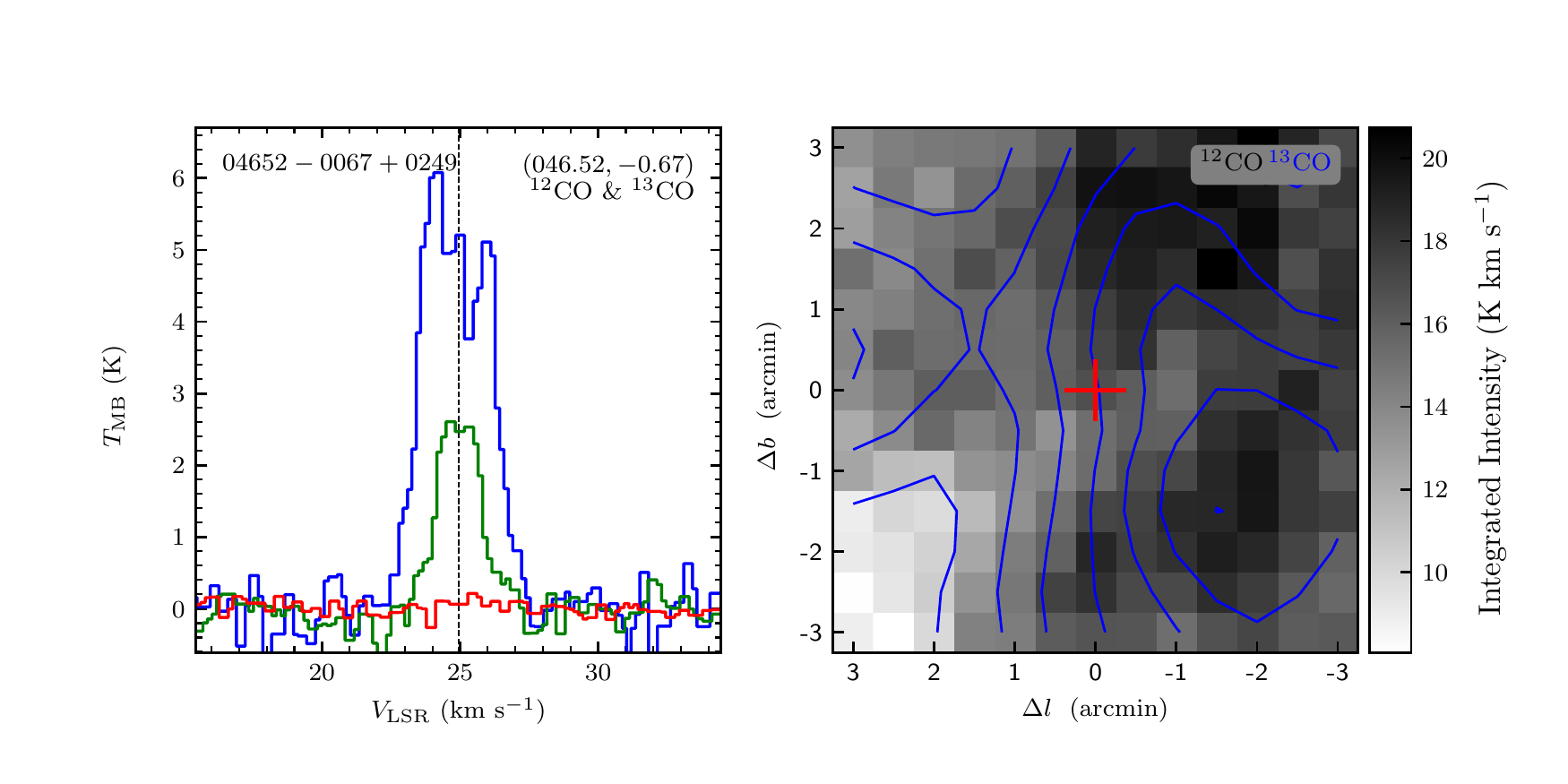}
\includegraphics[width=9.0cm,angle=0]{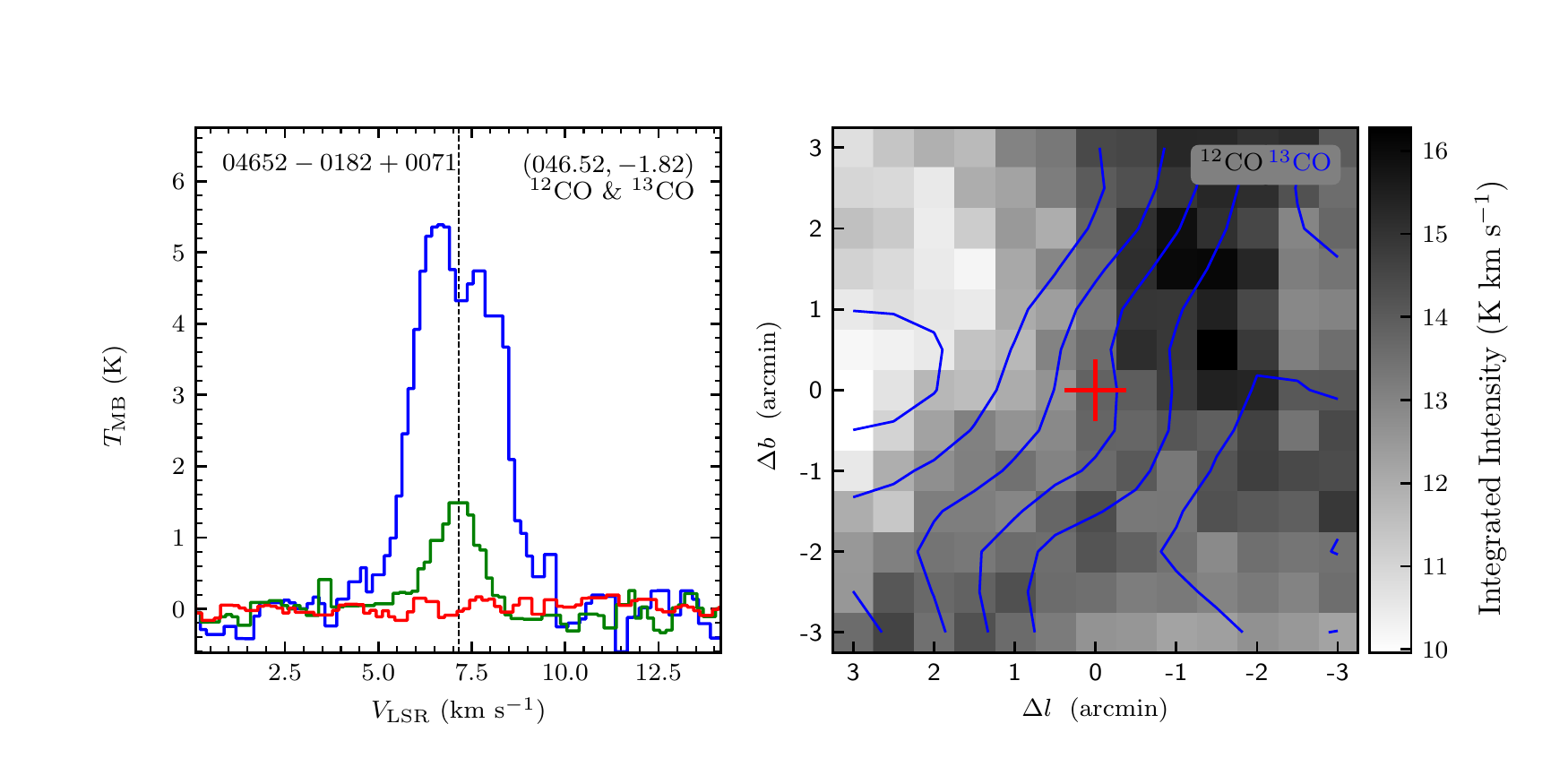}
\end{figure}
\clearpage

\begin{figure}
\includegraphics[width=9.0cm,angle=0]{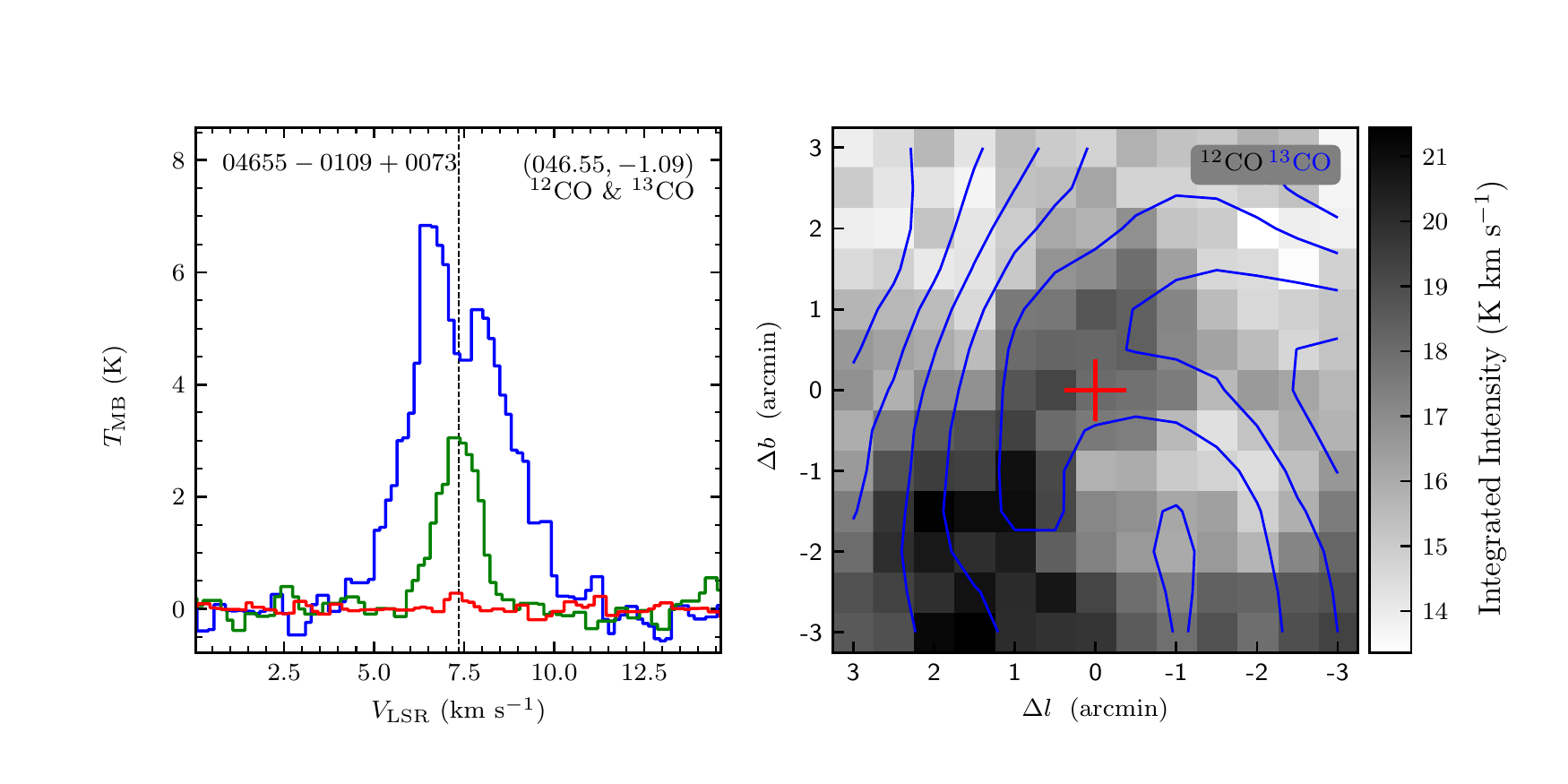}
\includegraphics[width=9.0cm,angle=0]{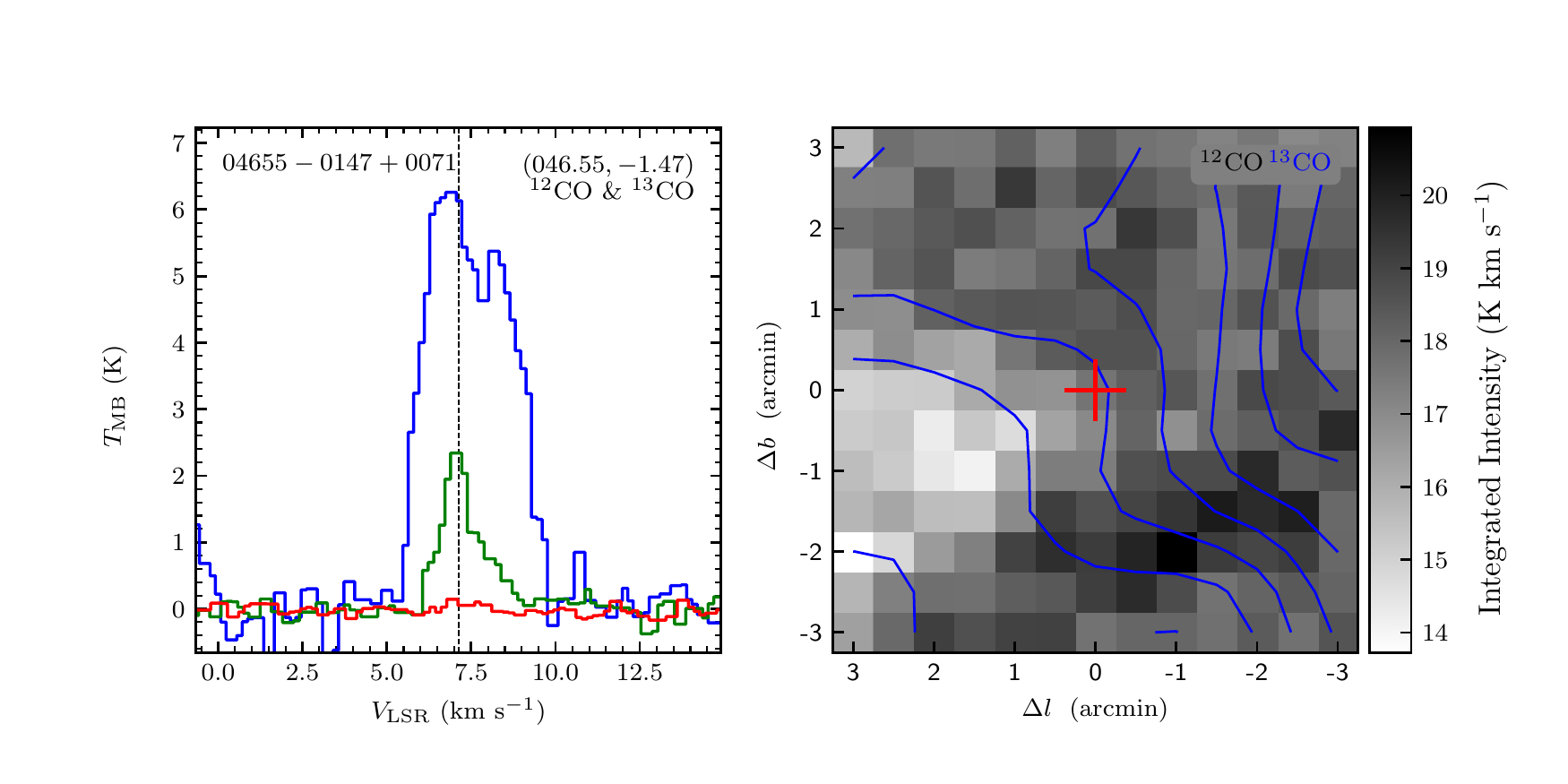}
\vspace{-0.5cm}

\includegraphics[width=9.0cm,angle=0]{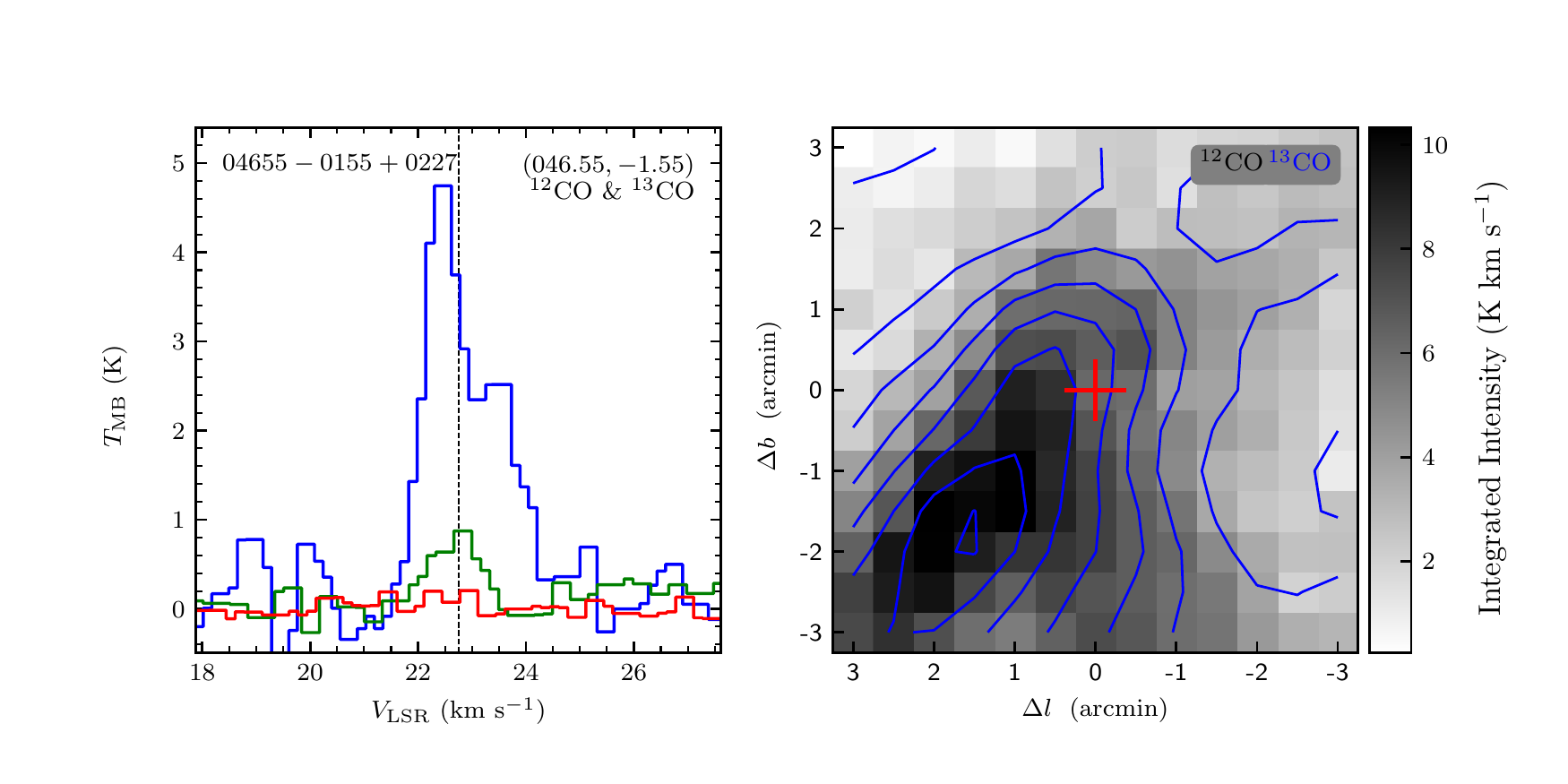}
\includegraphics[width=9.0cm,angle=0]{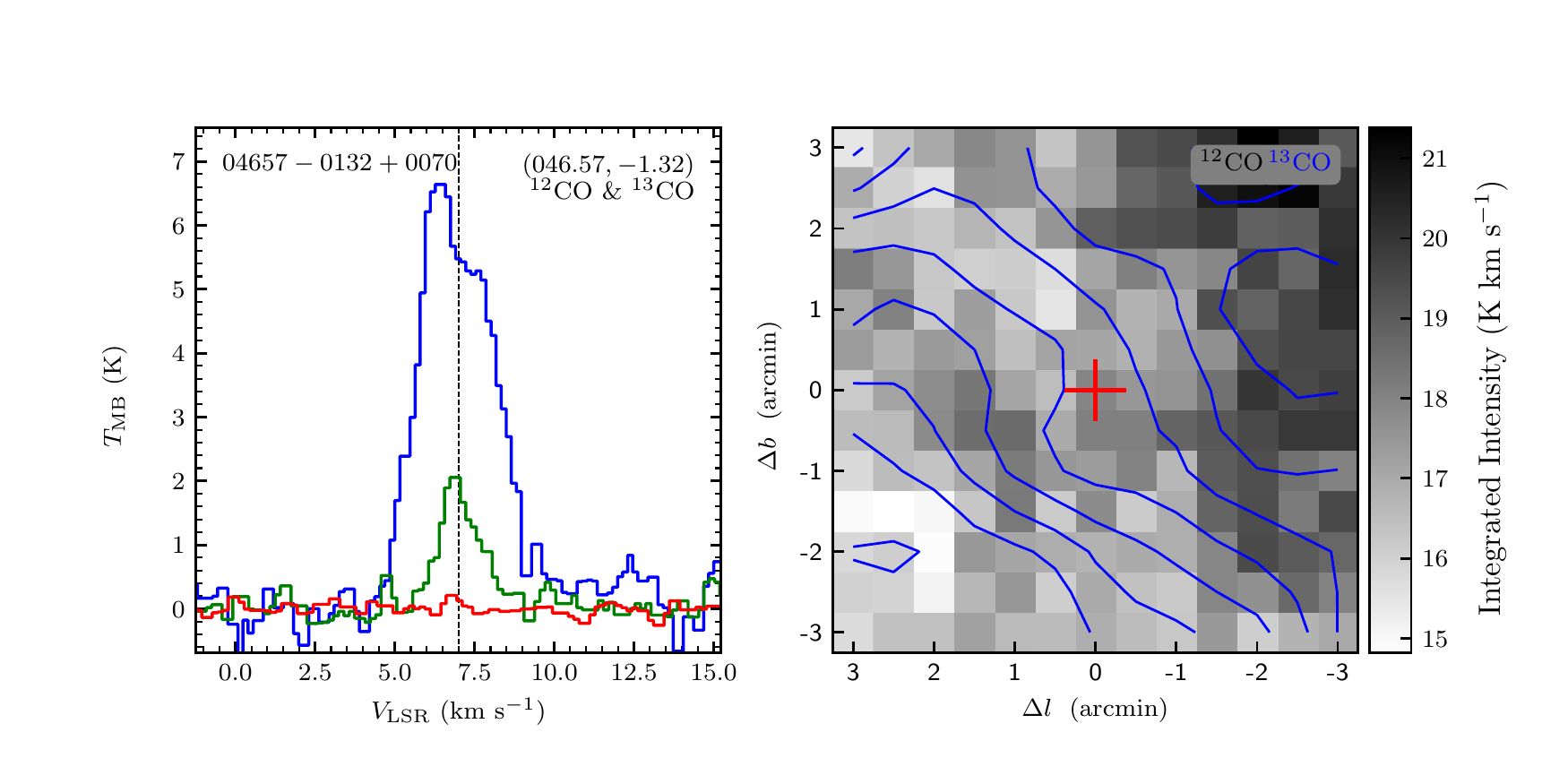}
\vspace{-0.5cm}

\includegraphics[width=9.0cm,angle=0]{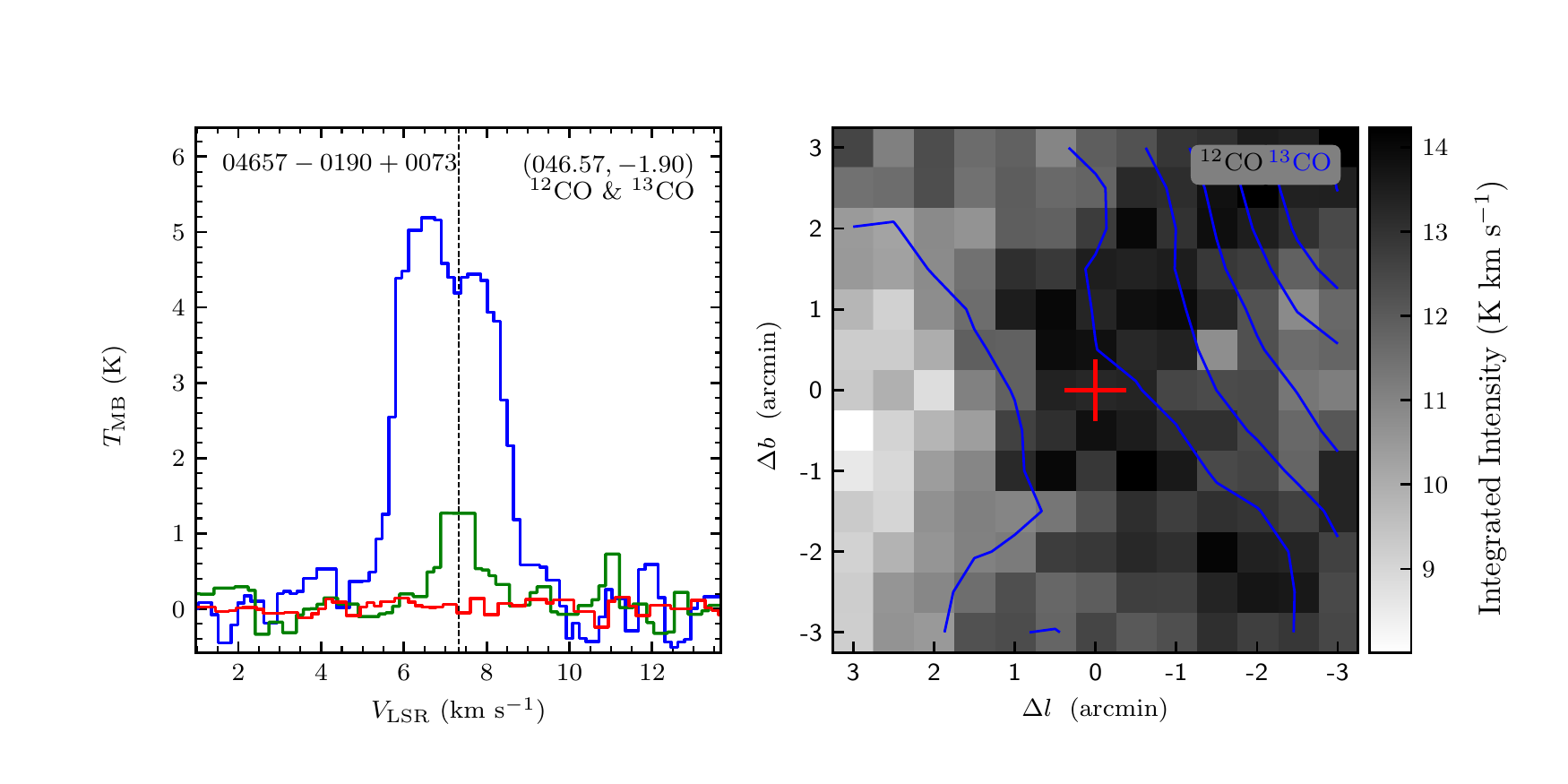}
\includegraphics[width=9.0cm,angle=0]{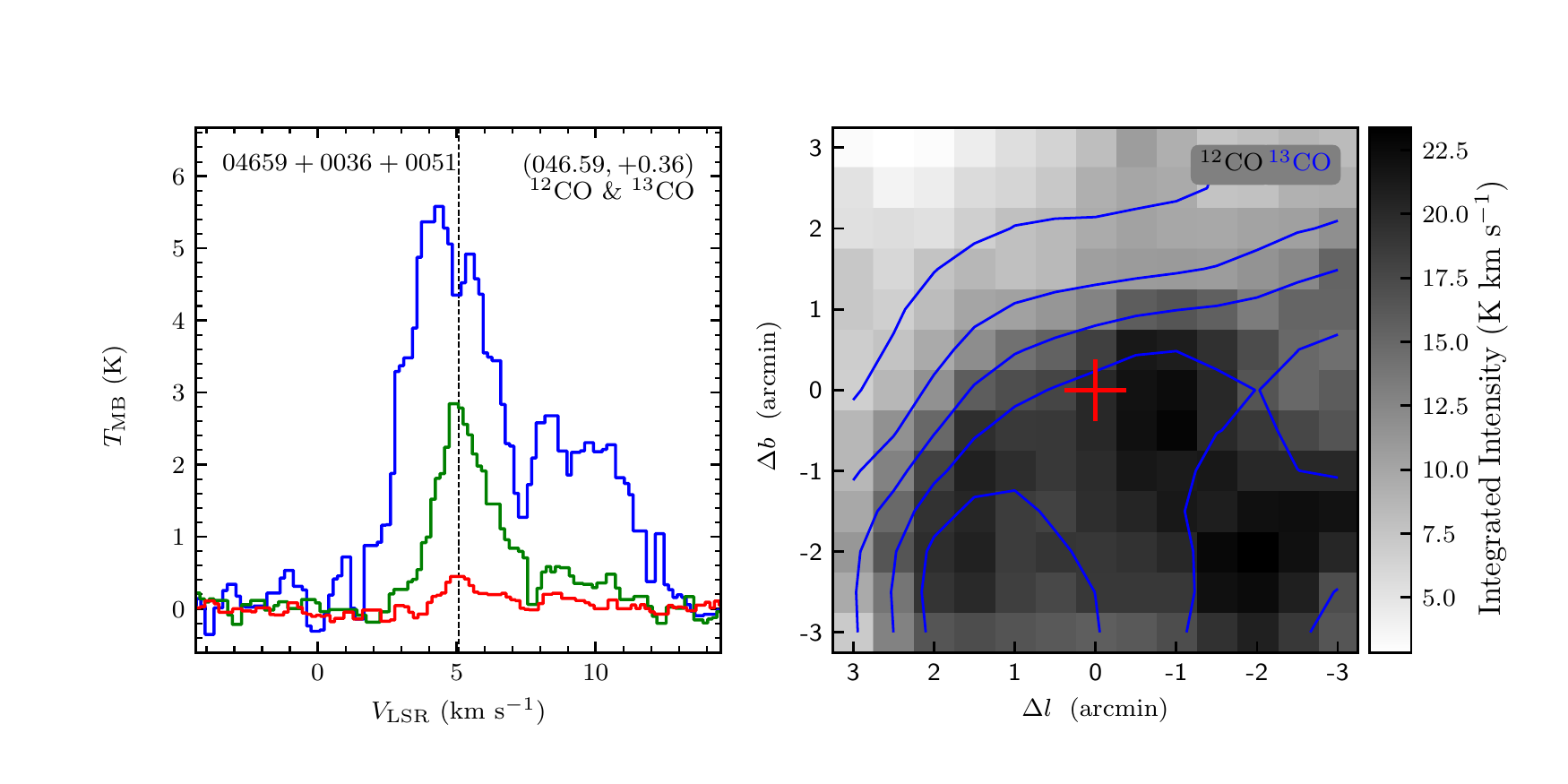}
\vspace{-0.5cm}

\includegraphics[width=9.0cm,angle=0]{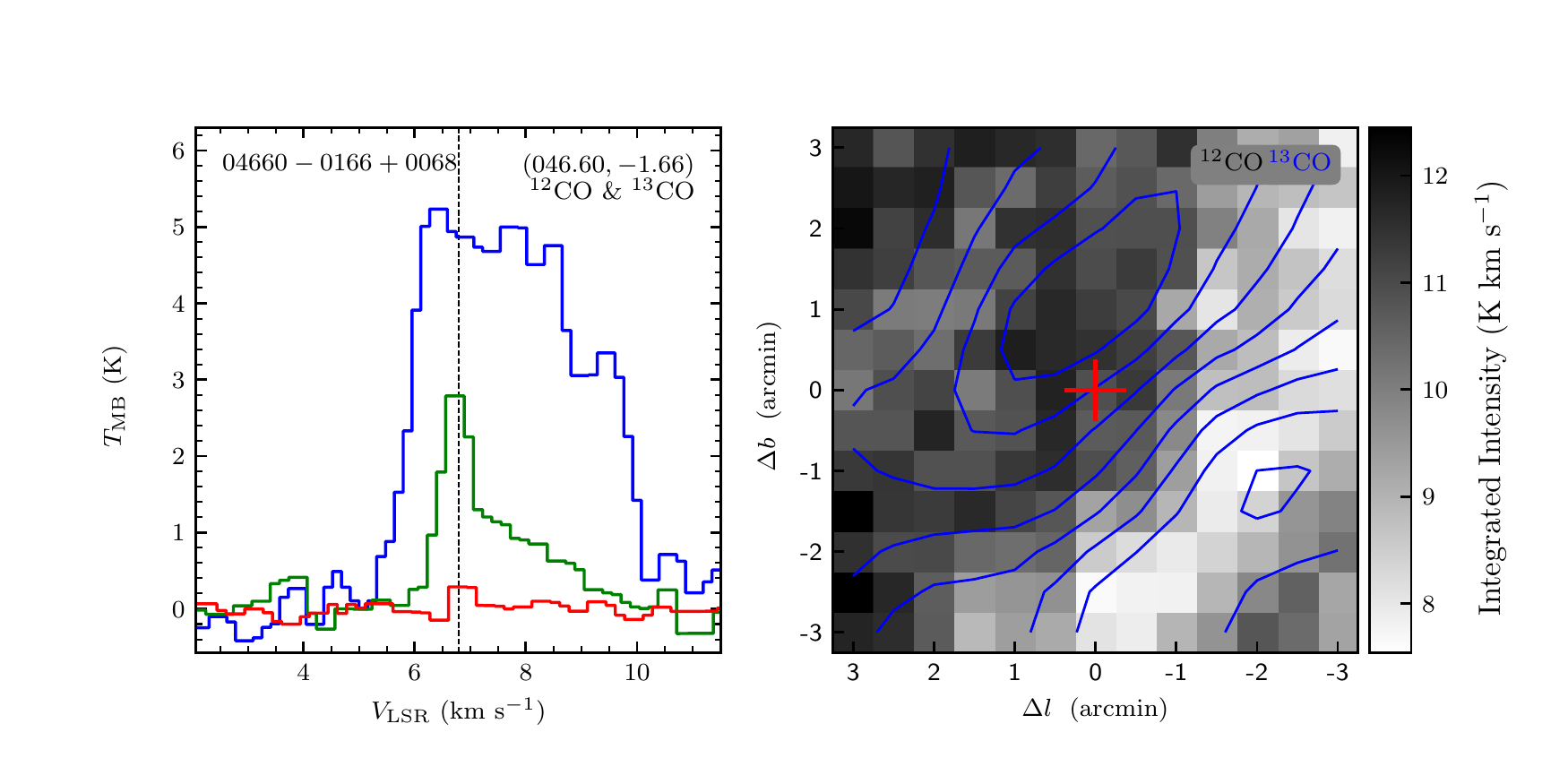}
\includegraphics[width=9.0cm,angle=0]{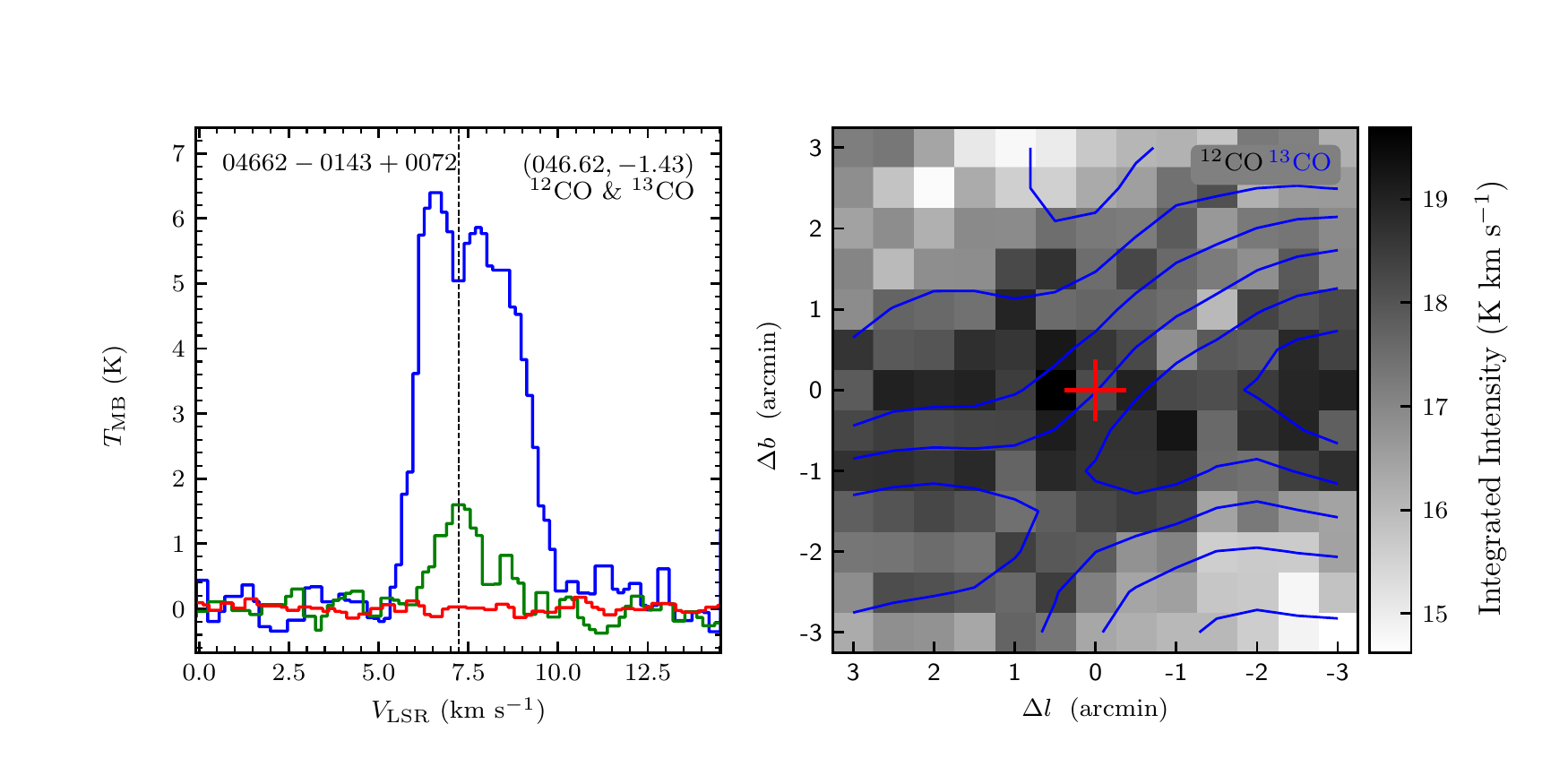}
\vspace{-0.5cm}

\includegraphics[width=9.0cm,angle=0]{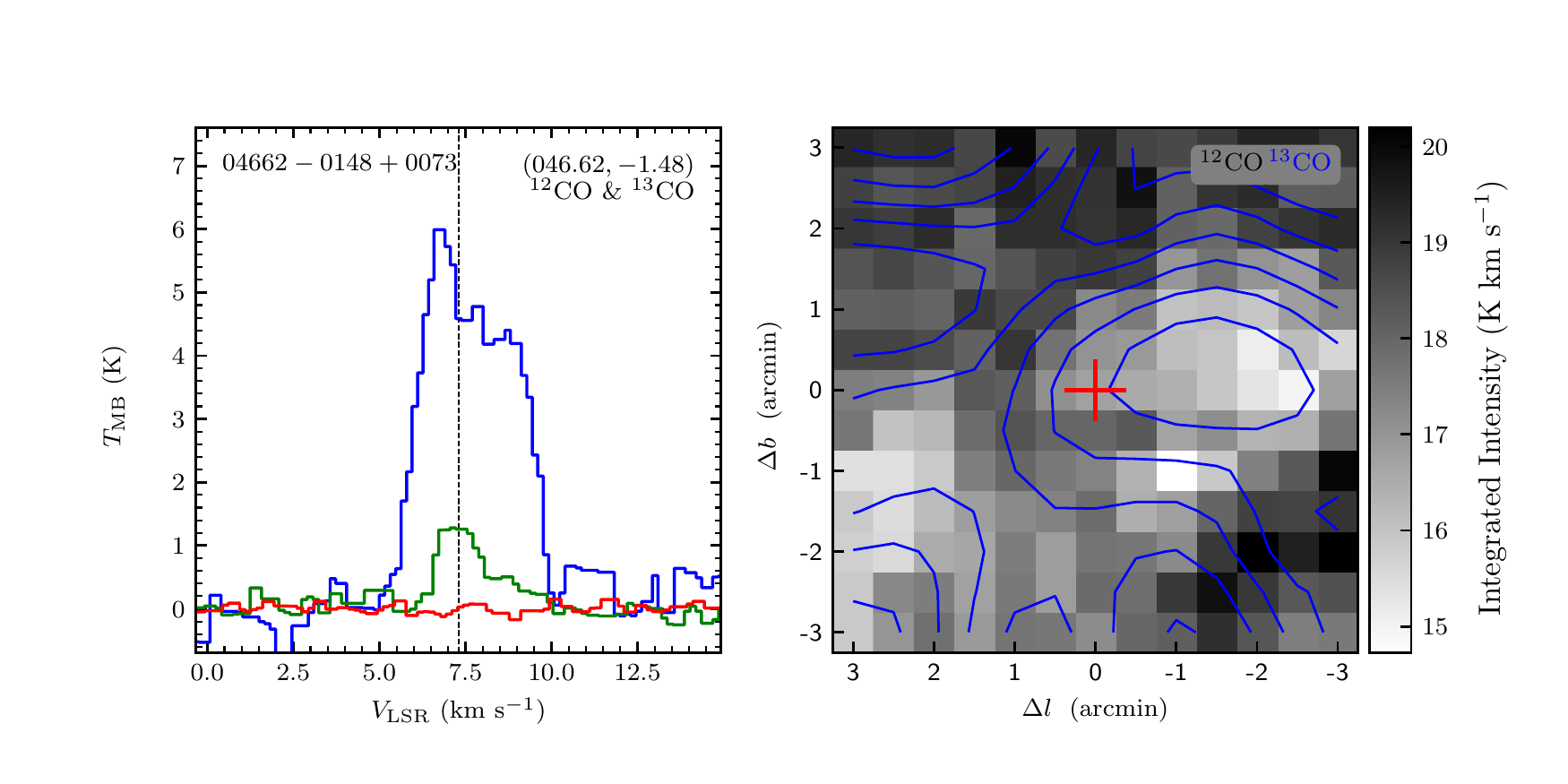}
\includegraphics[width=9.0cm,angle=0]{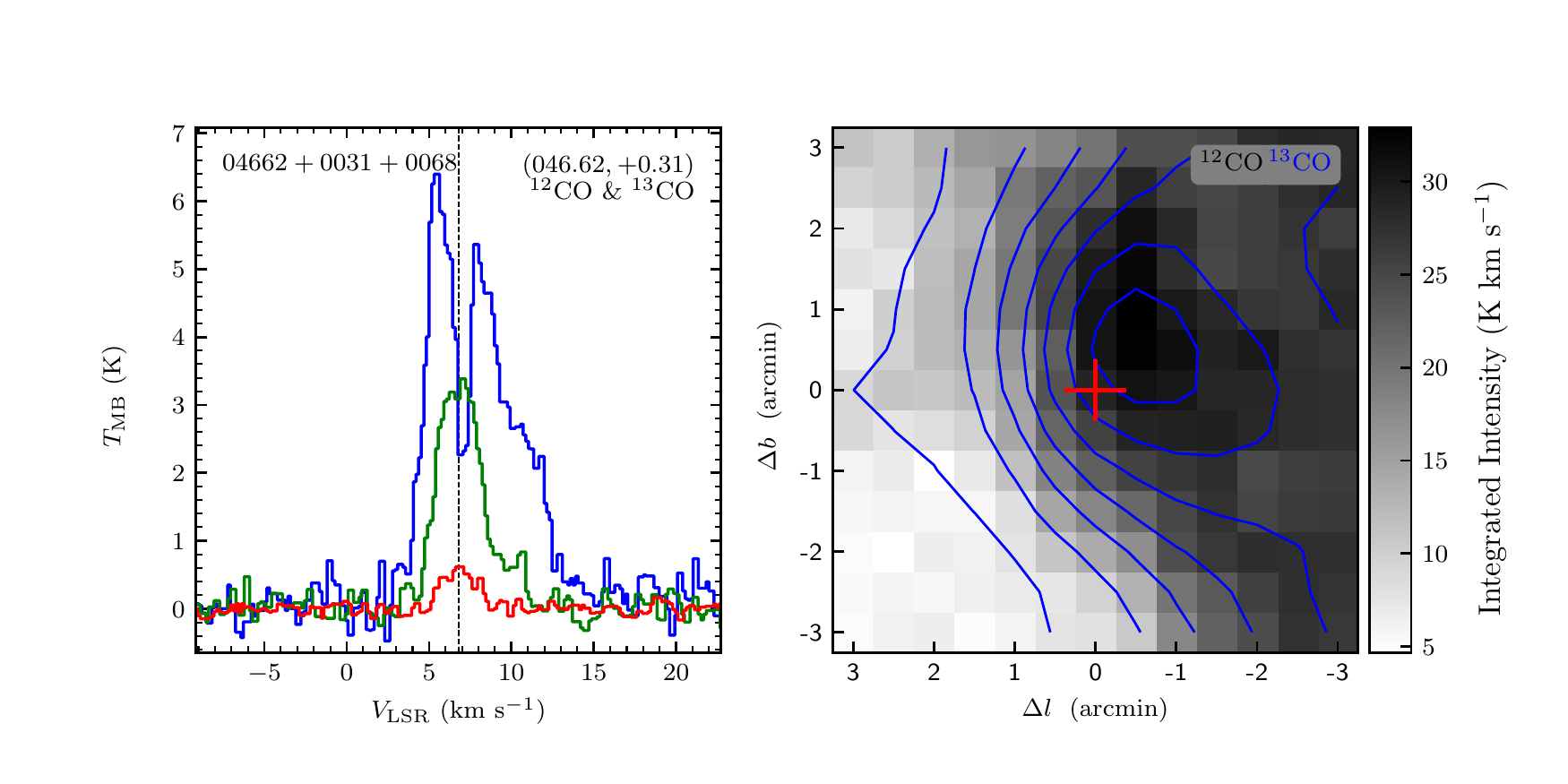}
\end{figure}
\clearpage

\begin{figure}
\includegraphics[width=9.0cm,angle=0]{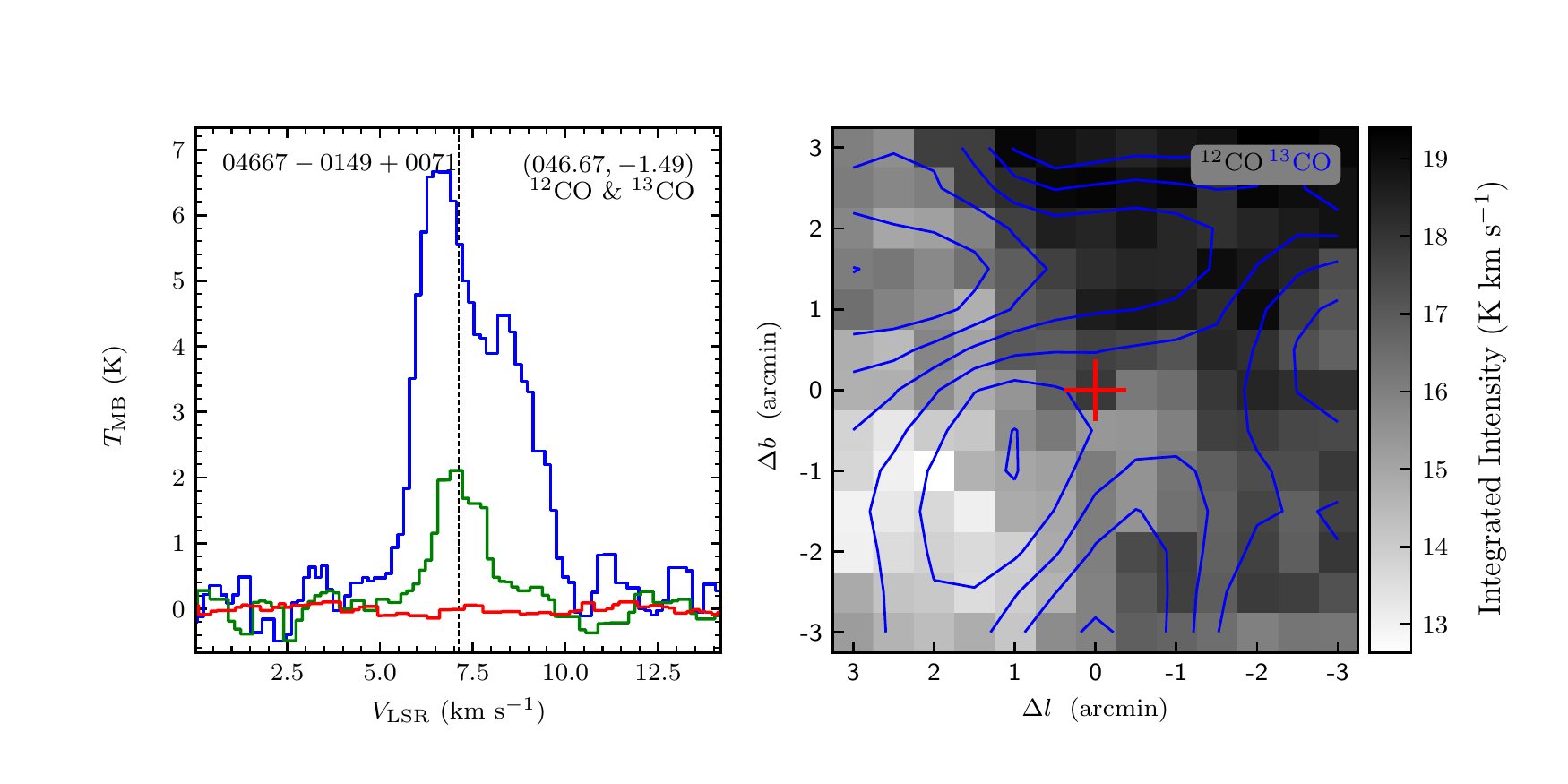}
\includegraphics[width=9.0cm,angle=0]{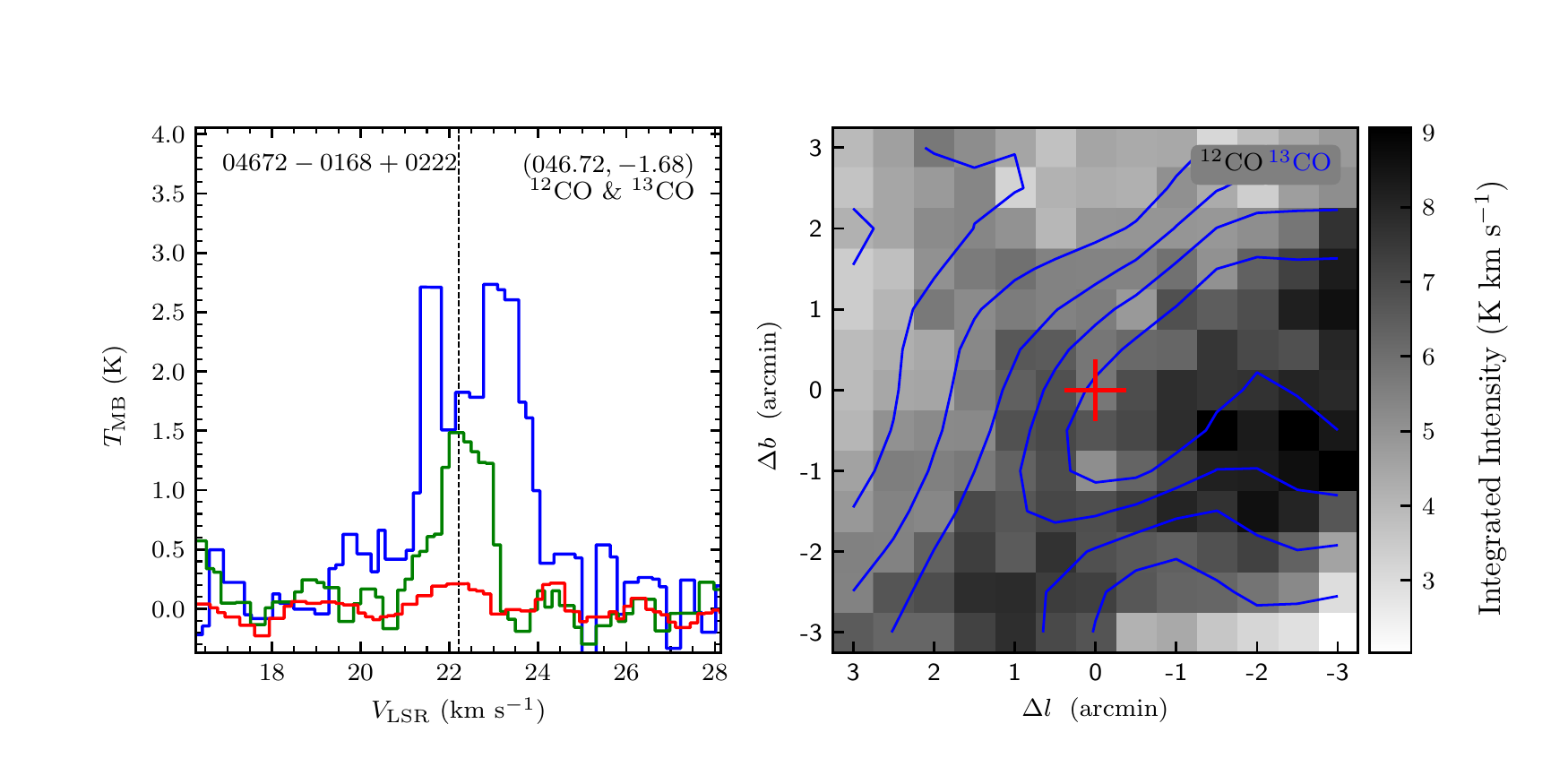}
\vspace{-0.5cm}

\includegraphics[width=9.0cm,angle=0]{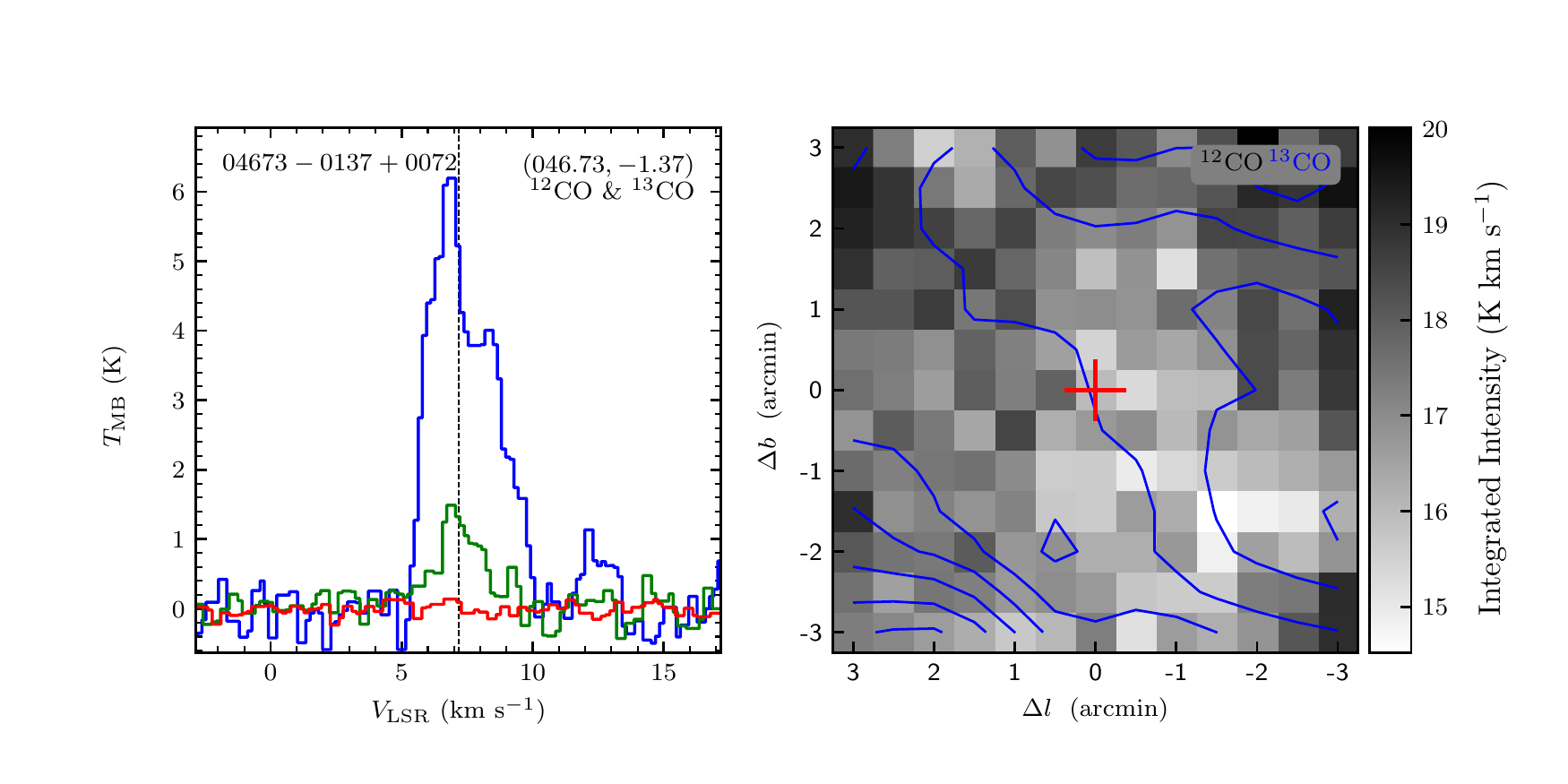}
\includegraphics[width=9.0cm,angle=0]{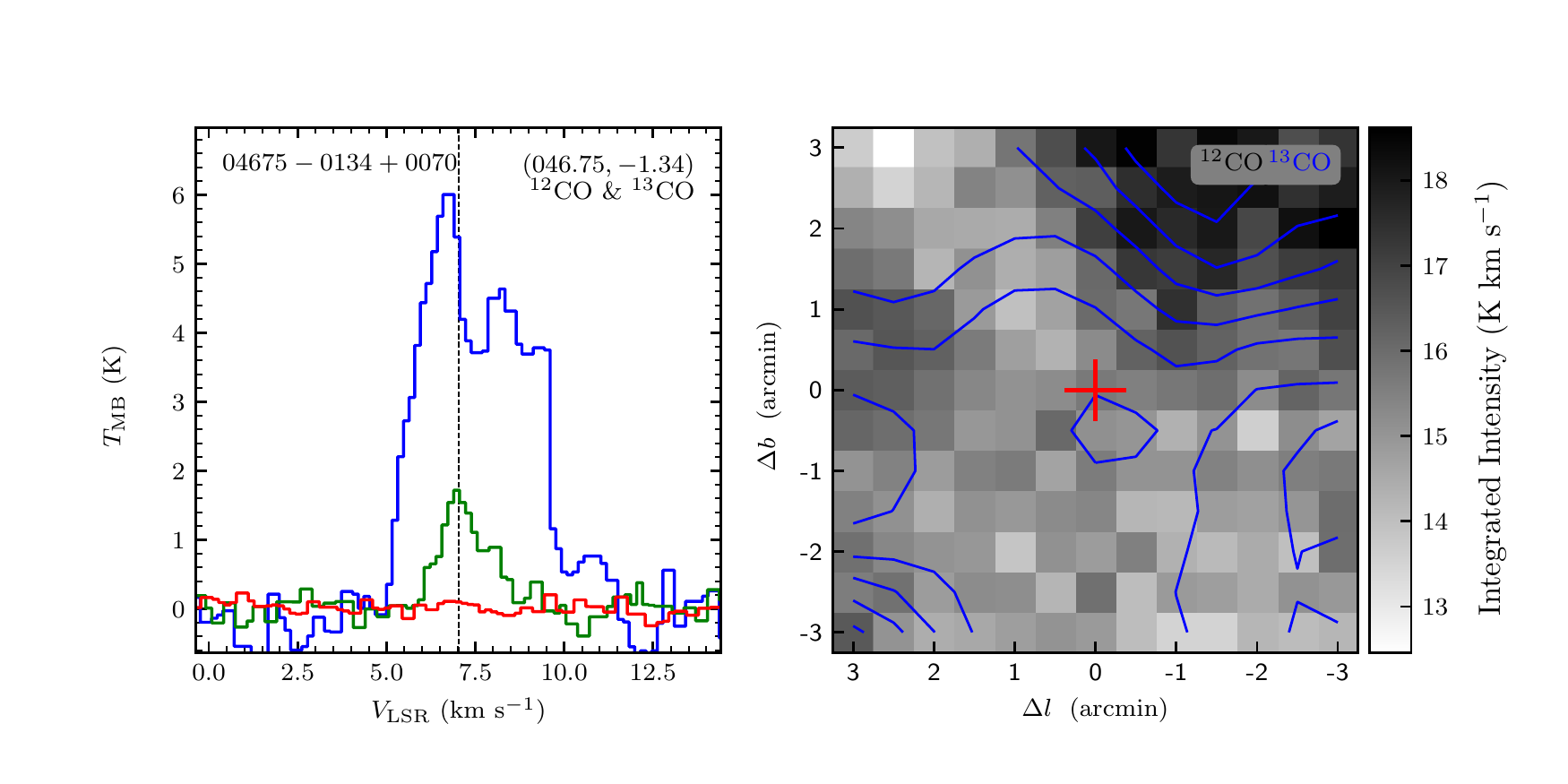}
\vspace{-0.5cm}

\includegraphics[width=9.0cm,angle=0]{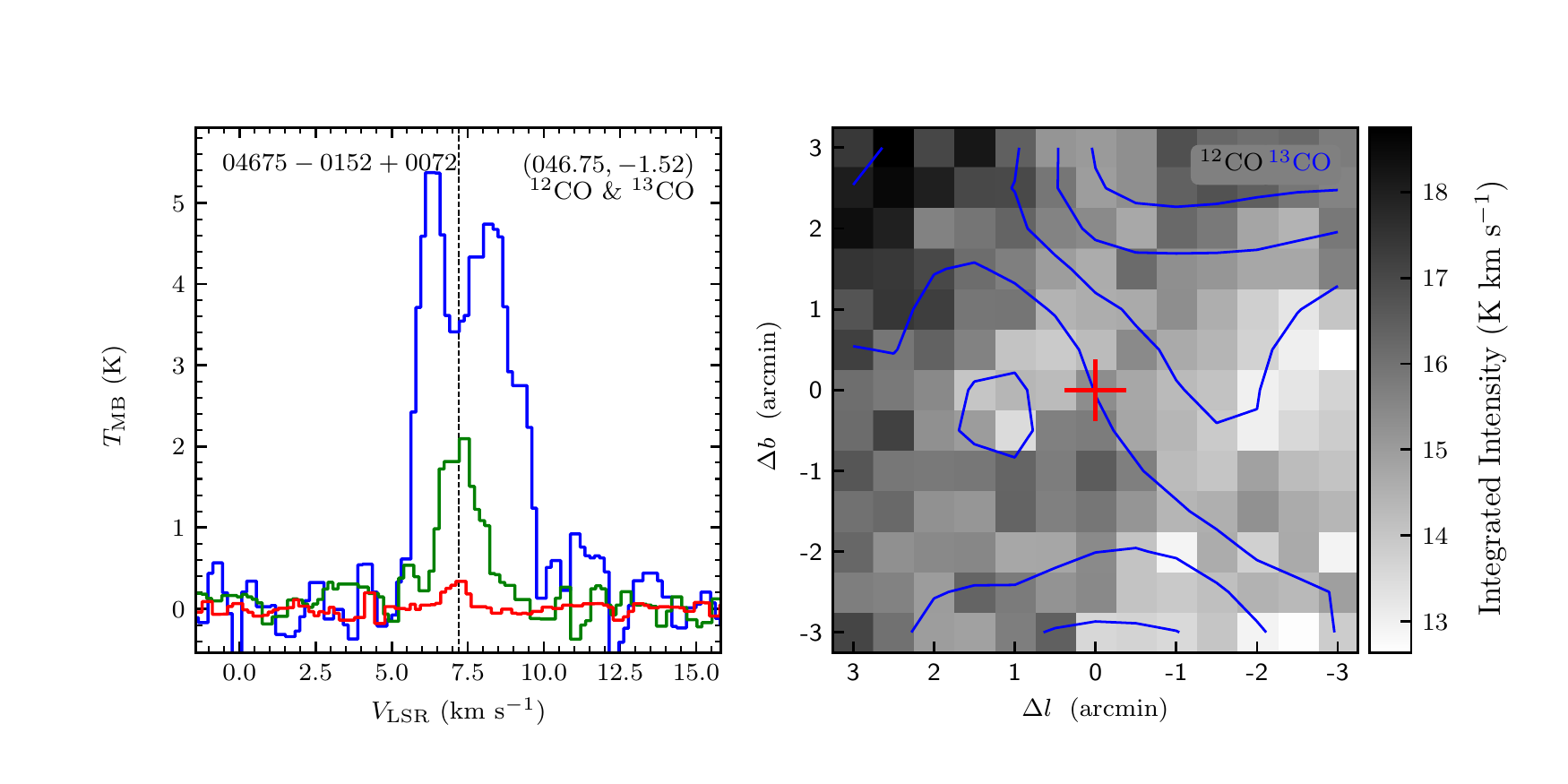}
\includegraphics[width=9.0cm,angle=0]{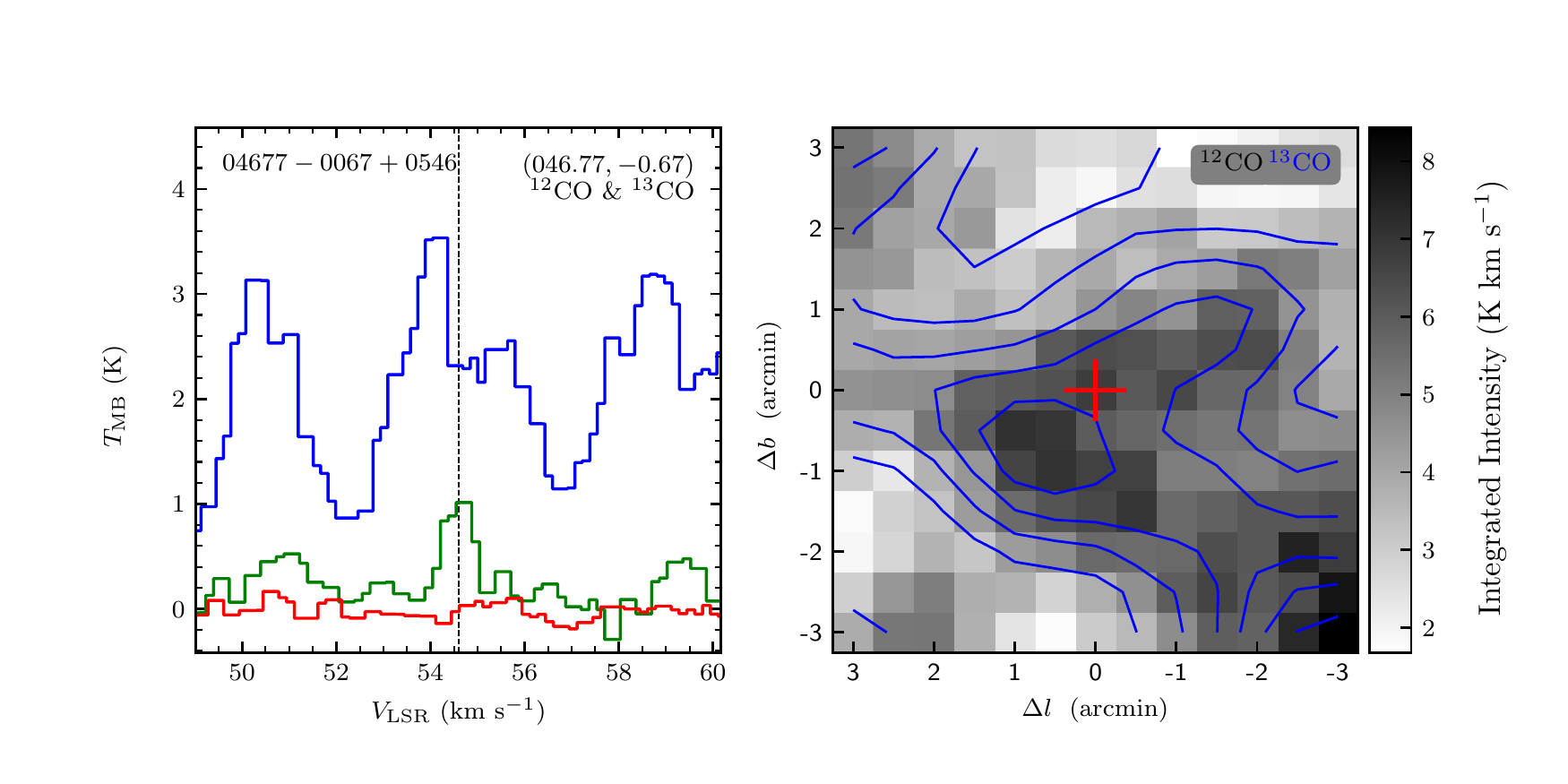}
\vspace{-0.5cm}

\includegraphics[width=9.0cm,angle=0]{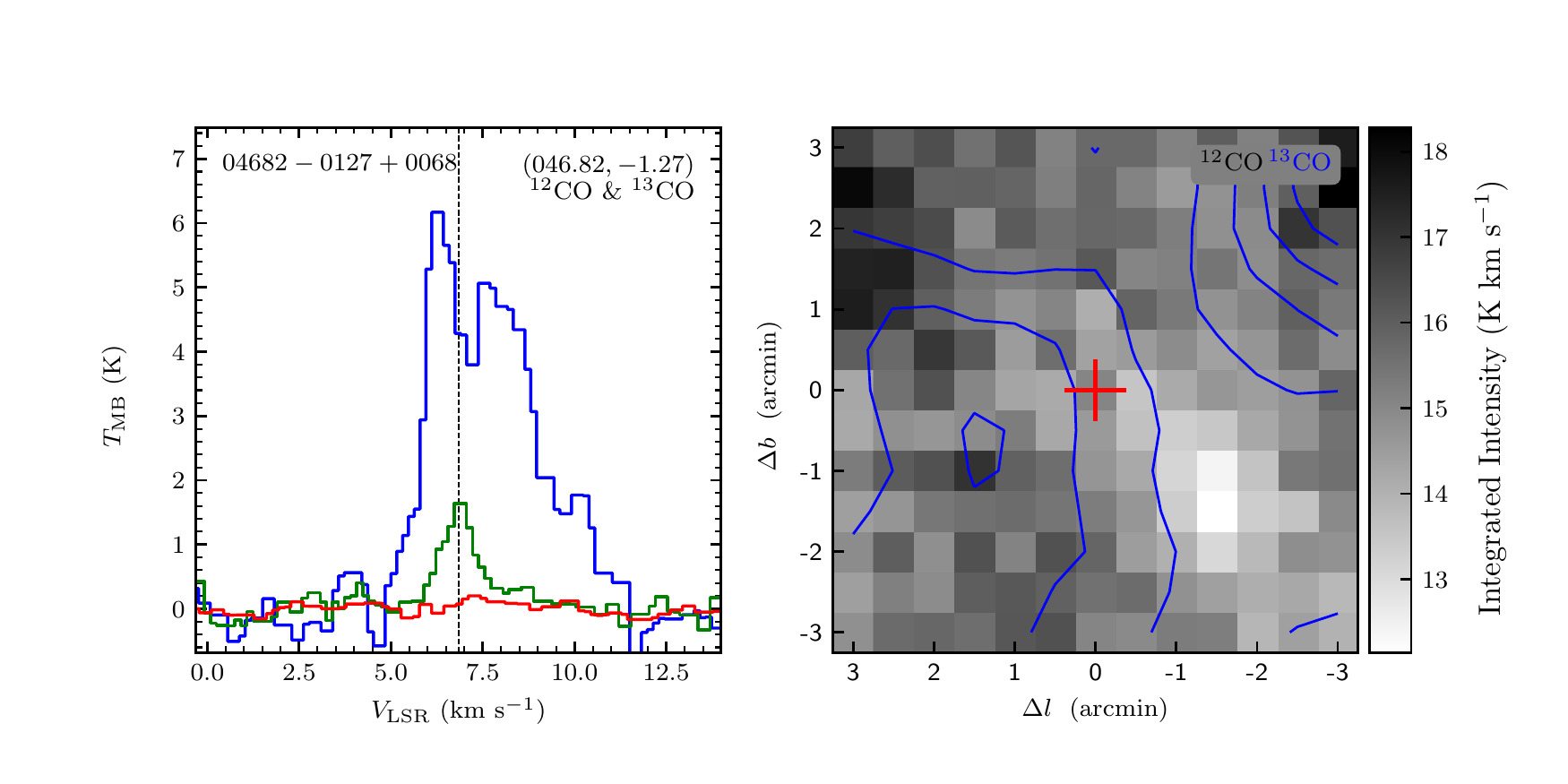}
\includegraphics[width=9.0cm,angle=0]{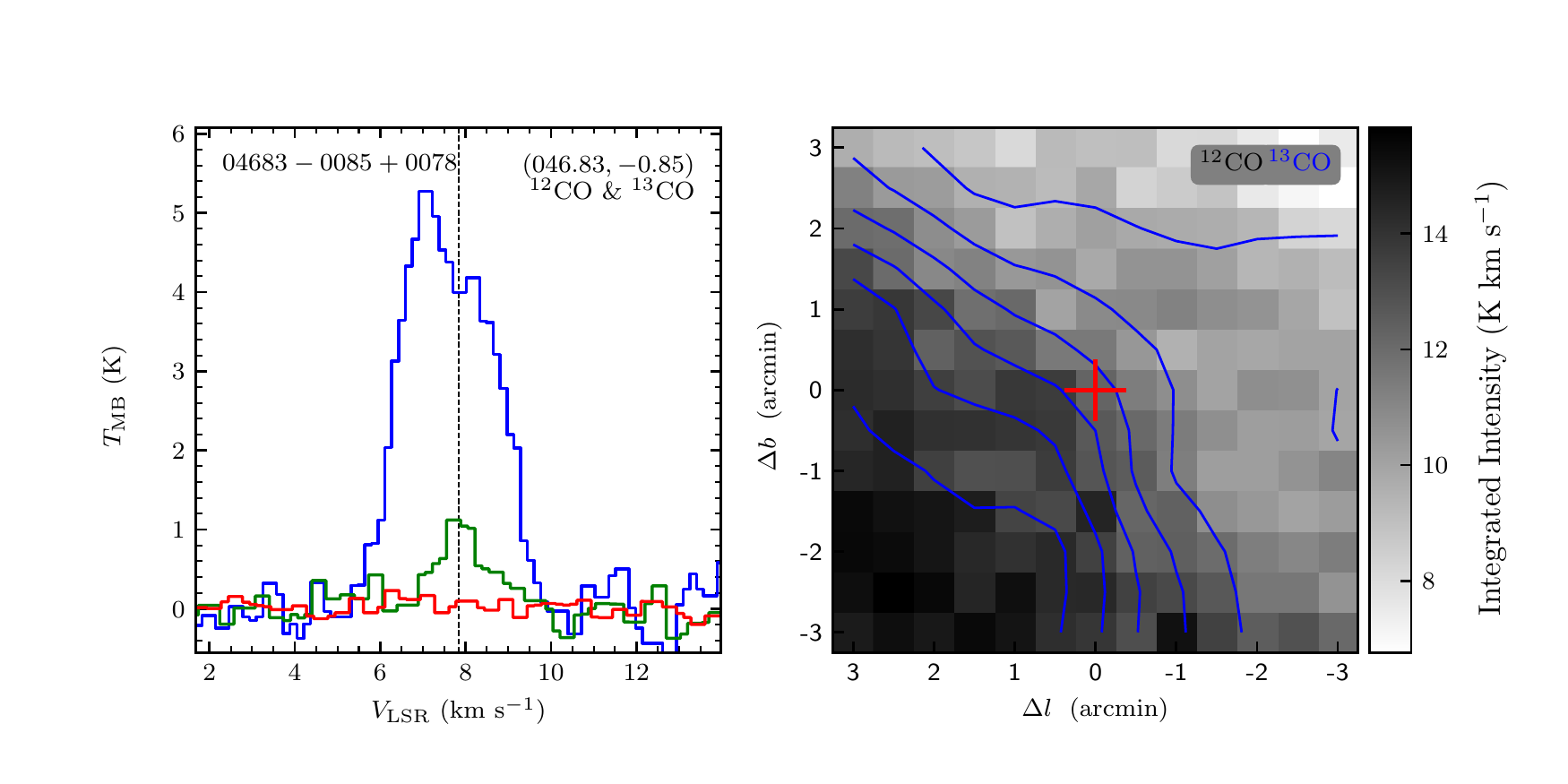}
\vspace{-0.5cm}

\includegraphics[width=9.0cm,angle=0]{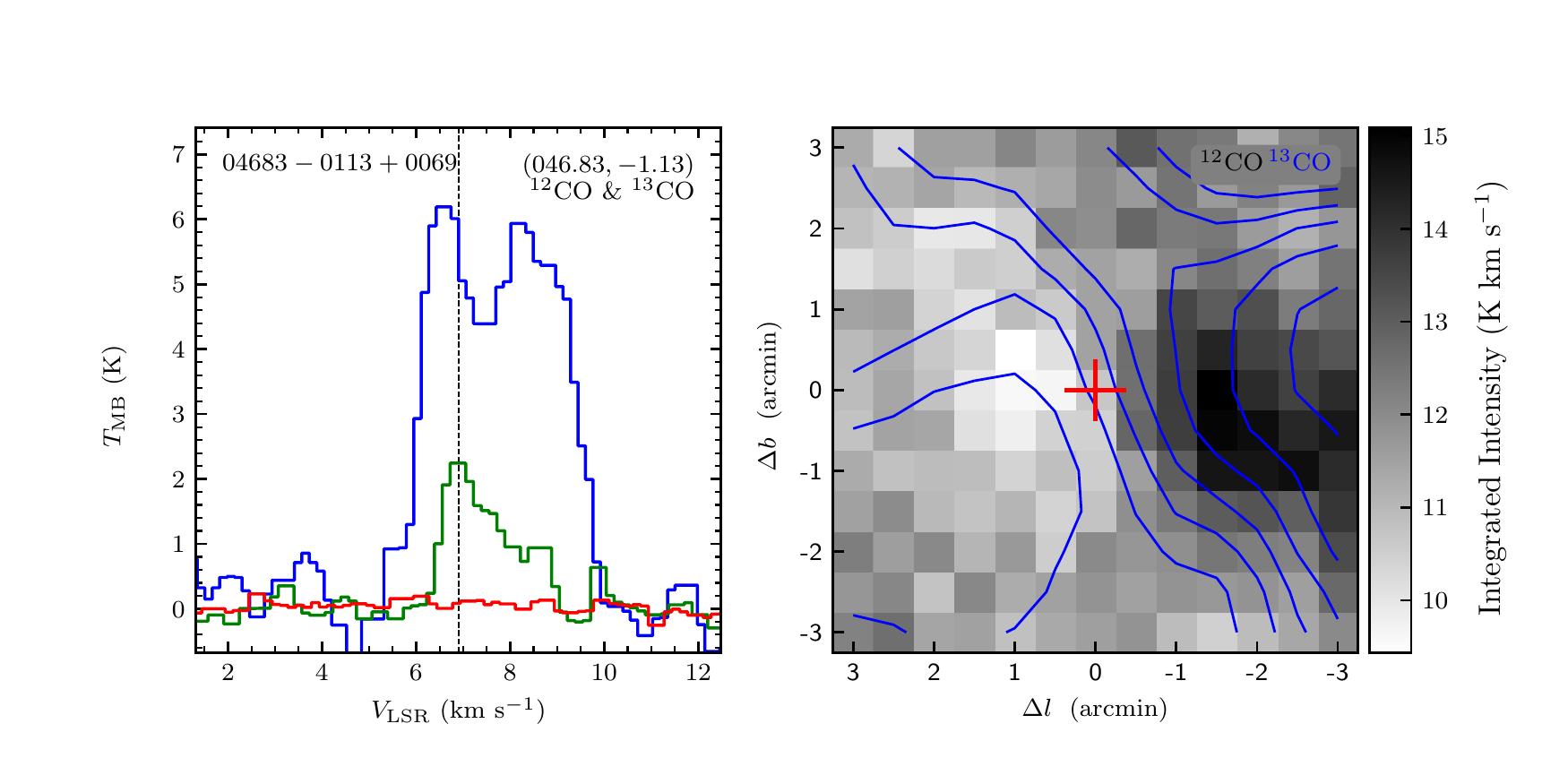}
\includegraphics[width=9.0cm,angle=0]{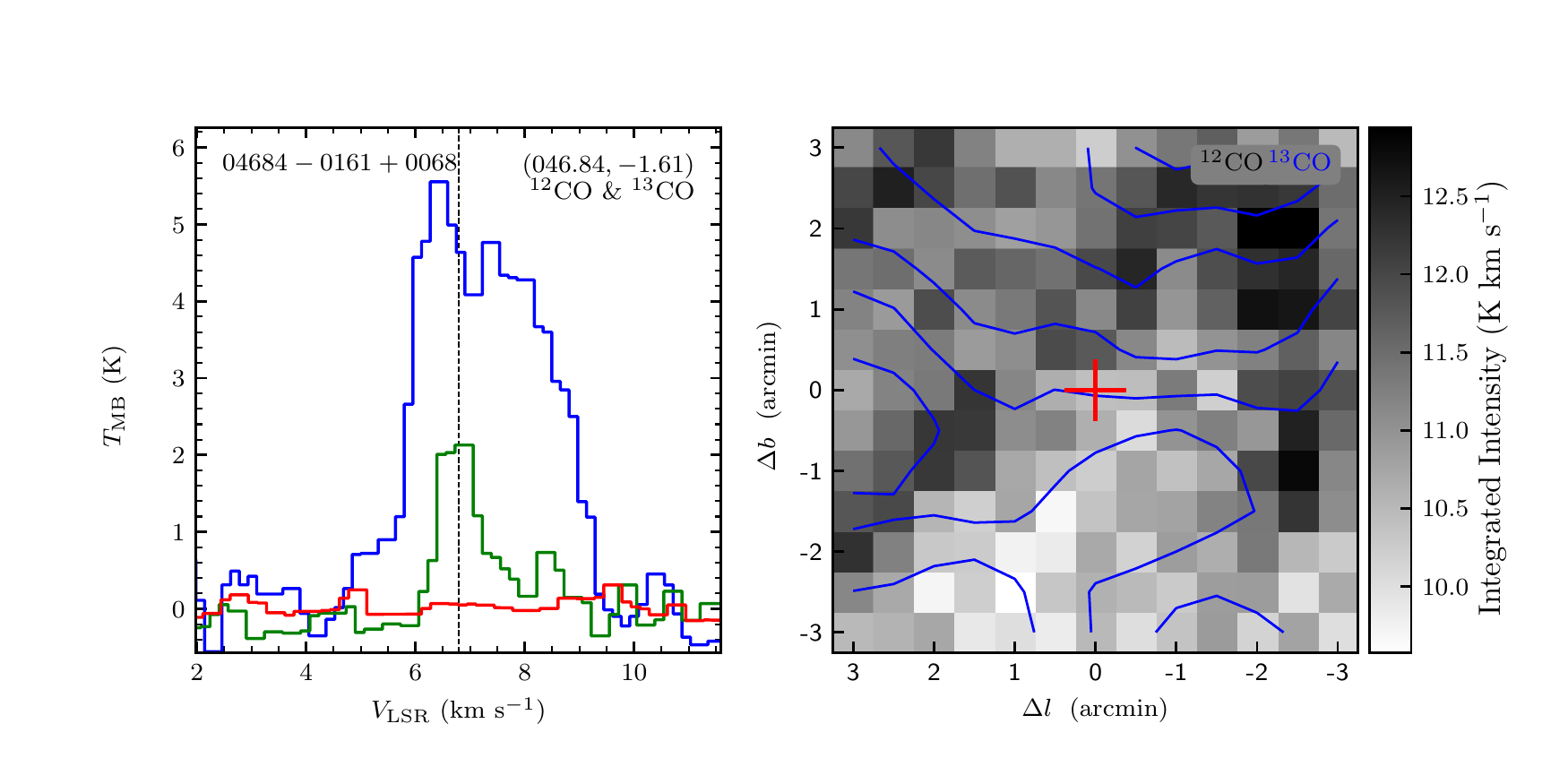}
\end{figure}
\clearpage

\begin{figure}
\includegraphics[width=9.0cm,angle=0]{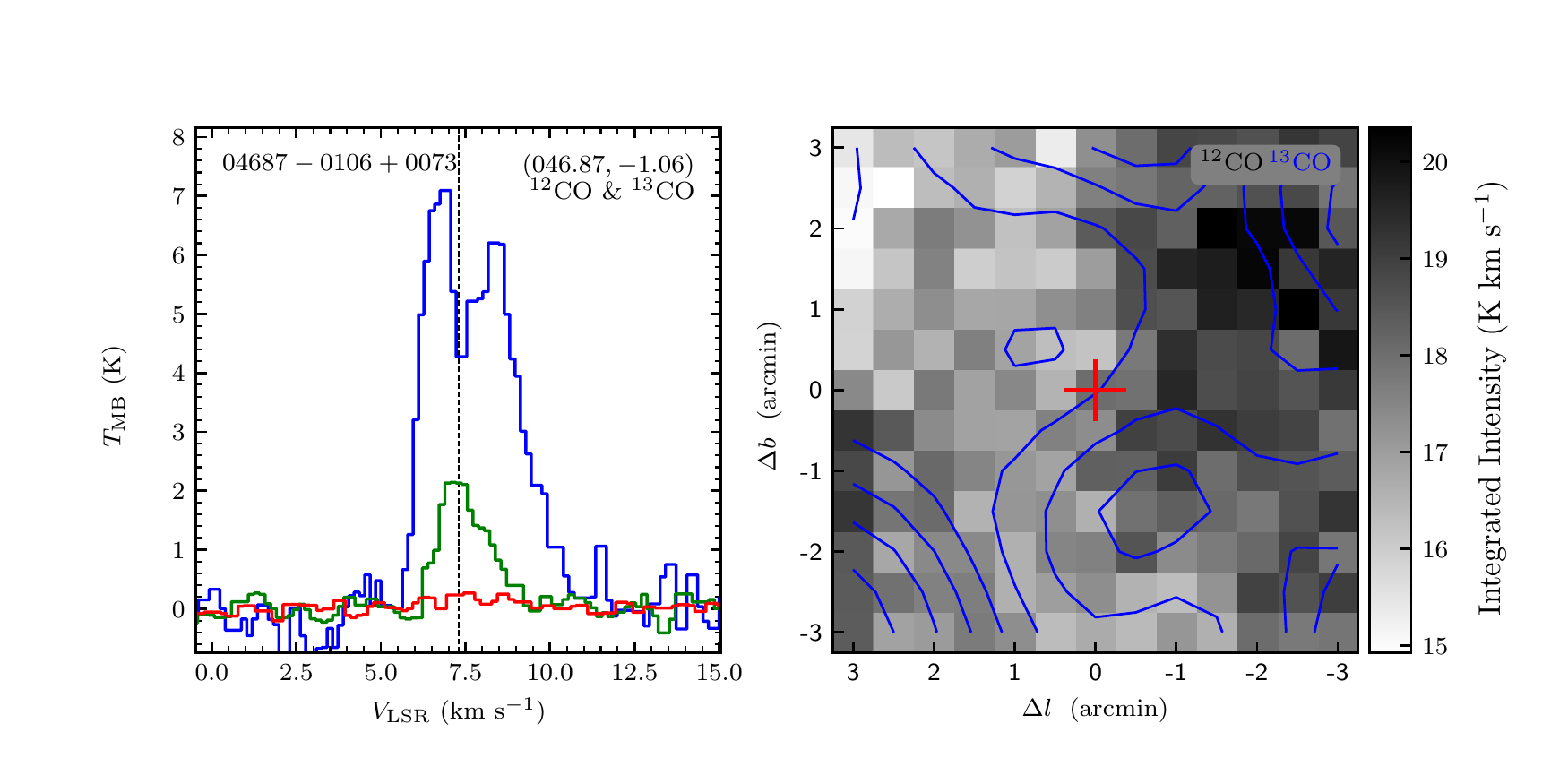}
\includegraphics[width=9.0cm,angle=0]{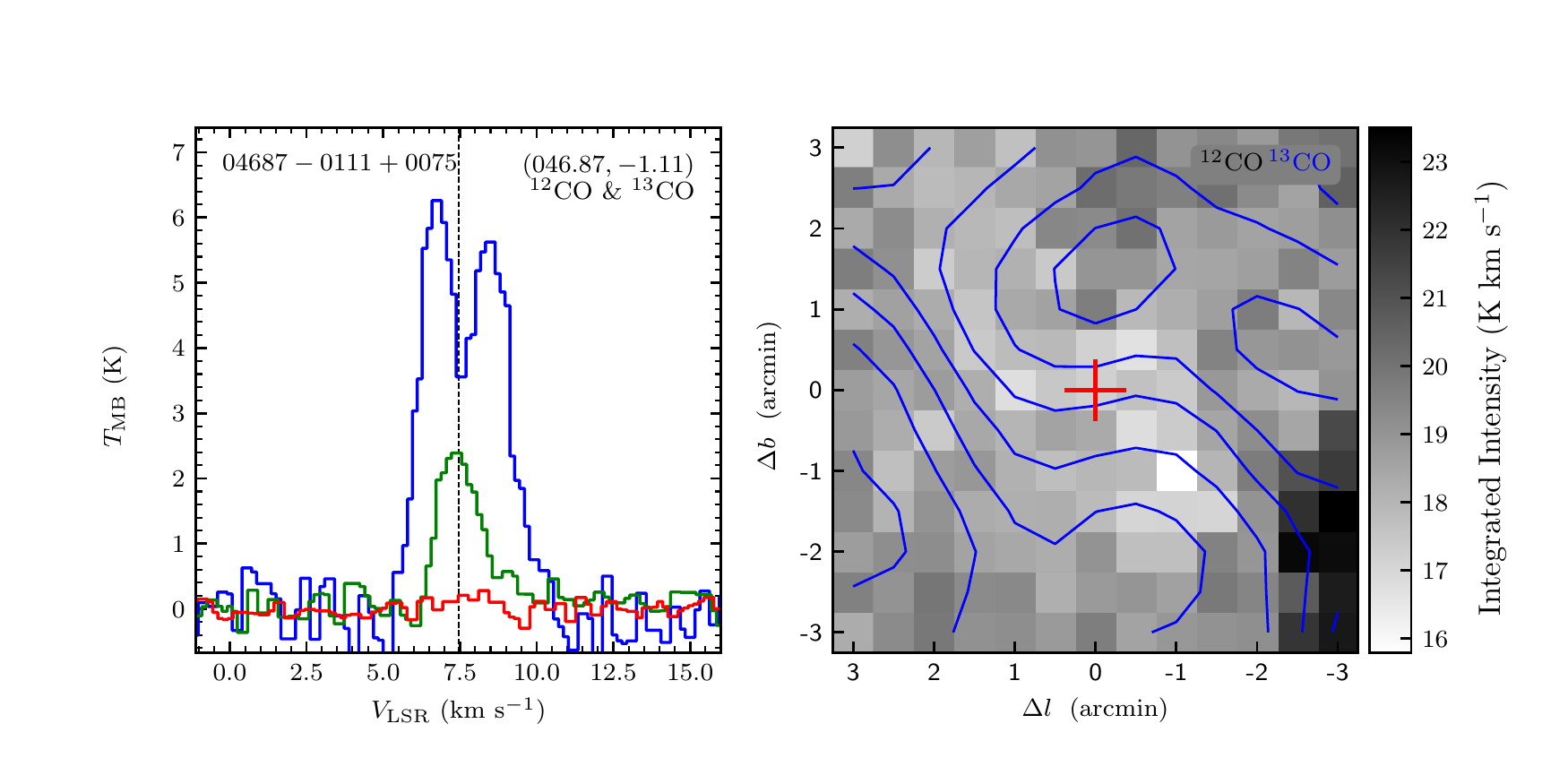}
\vspace{-0.5cm}

\includegraphics[width=9.0cm,angle=0]{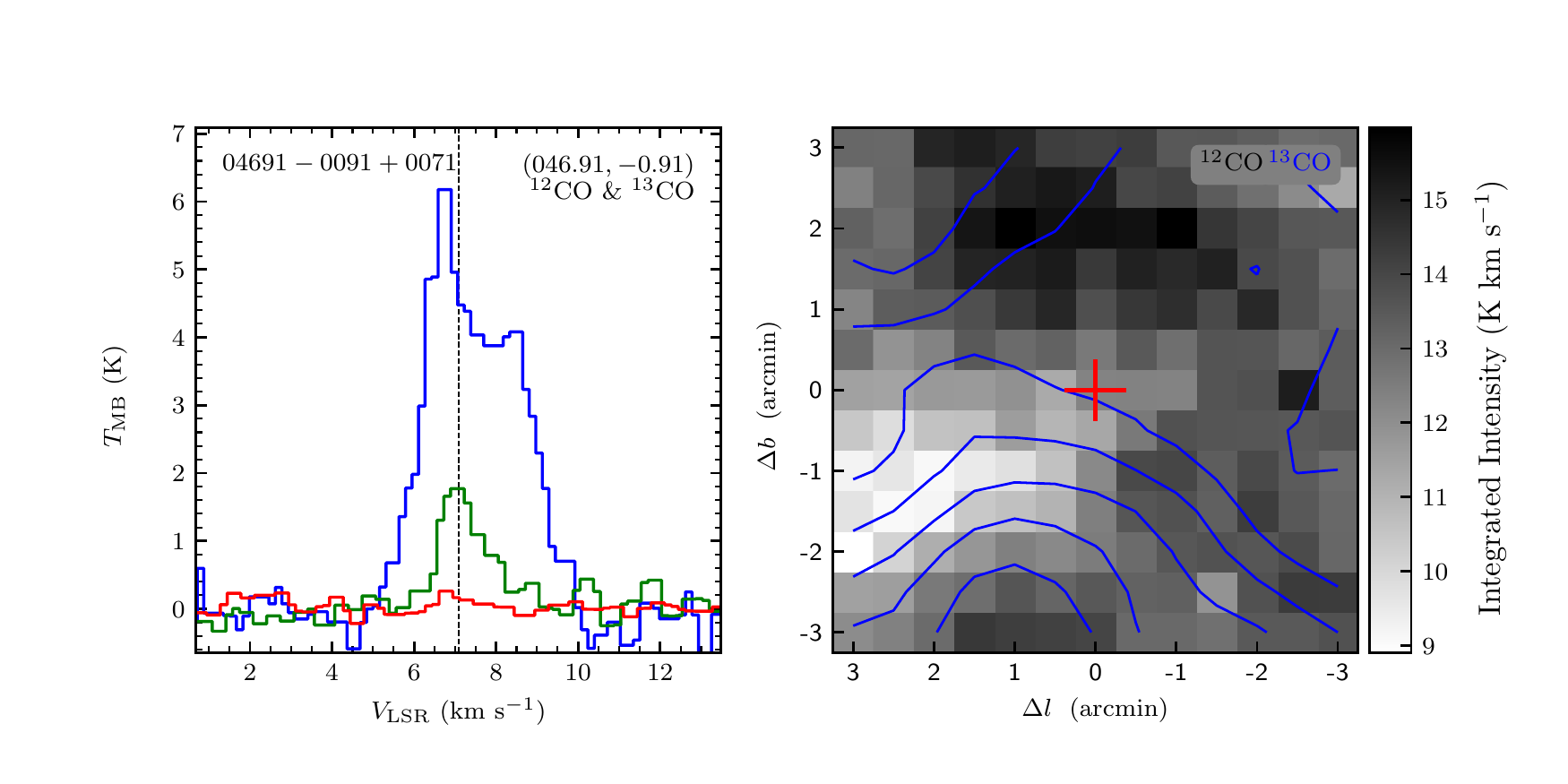}
\includegraphics[width=9.0cm,angle=0]{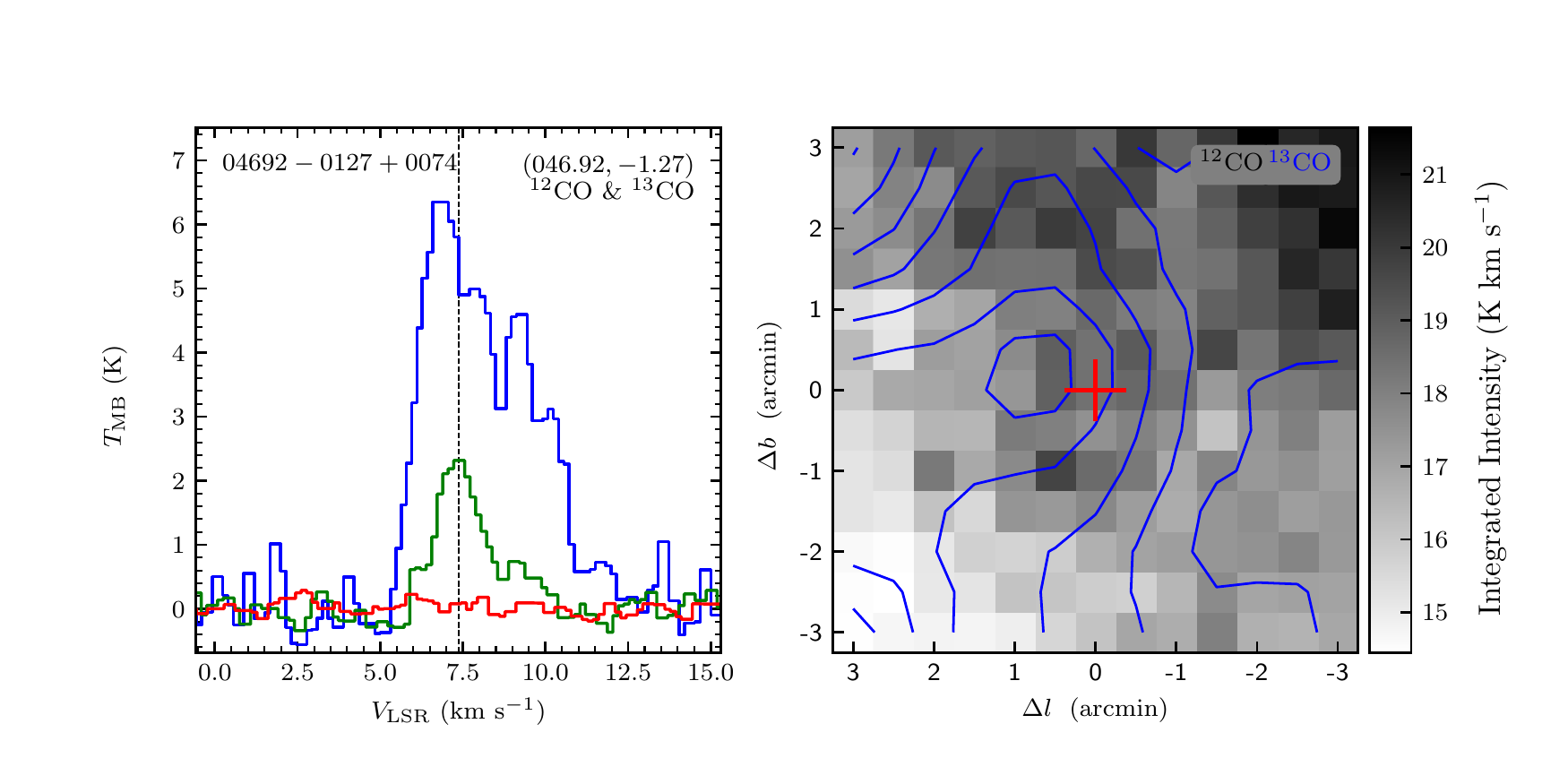}
\vspace{-0.5cm}

\includegraphics[width=9.0cm,angle=0]{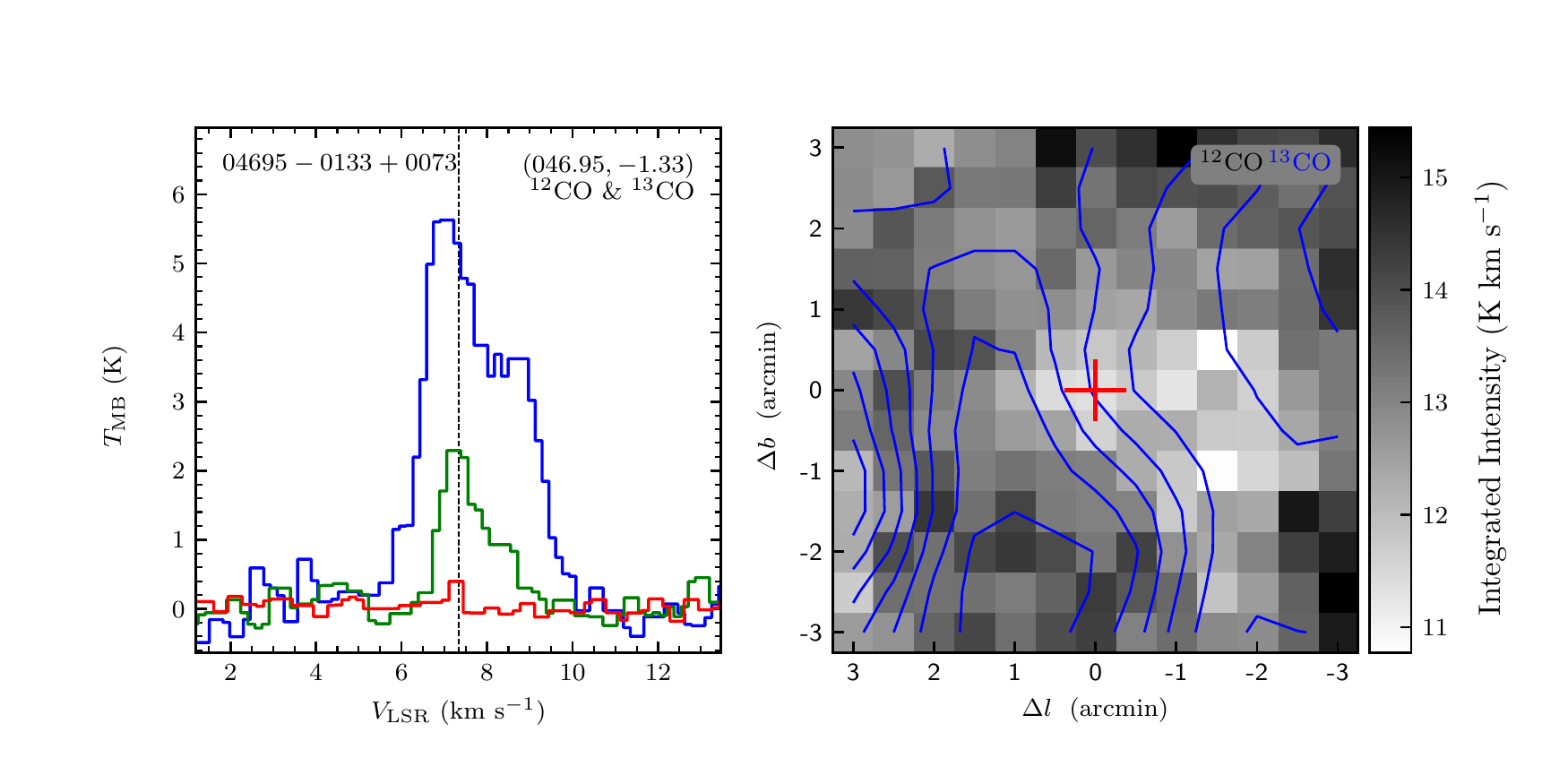}
\includegraphics[width=9.0cm,angle=0]{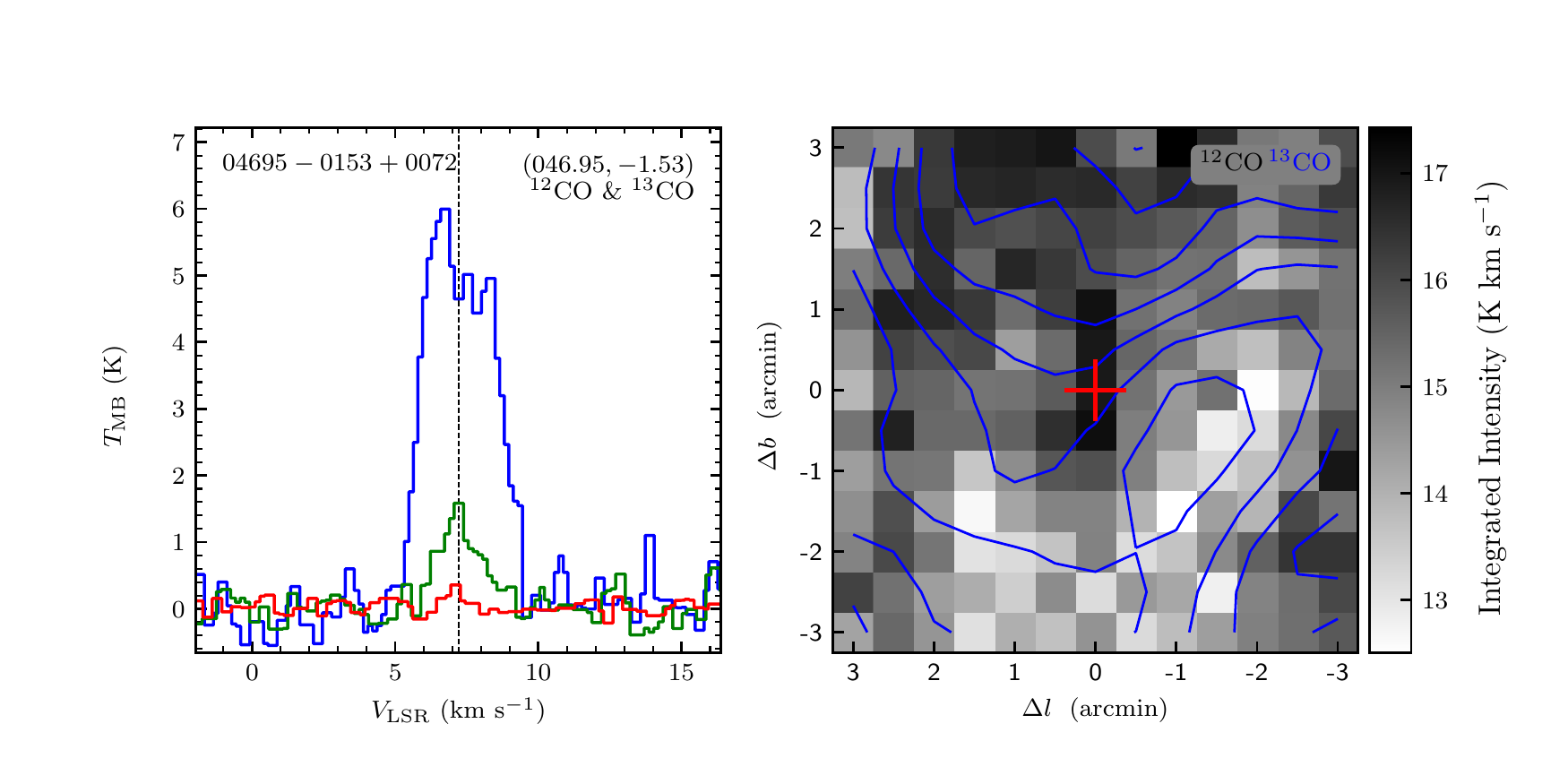}
\vspace{-0.5cm}

\includegraphics[width=9.0cm,angle=0]{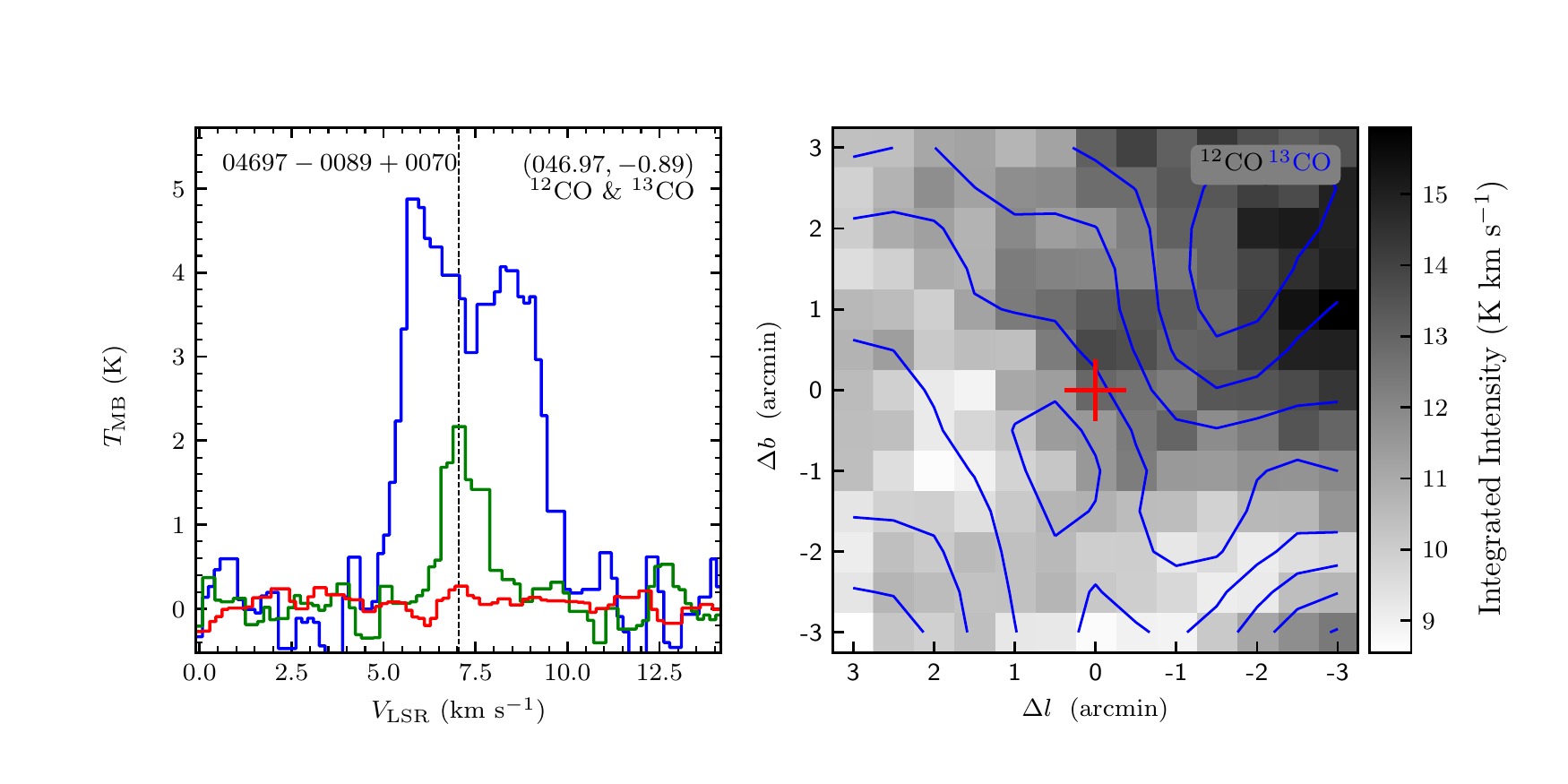}
\includegraphics[width=9.0cm,angle=0]{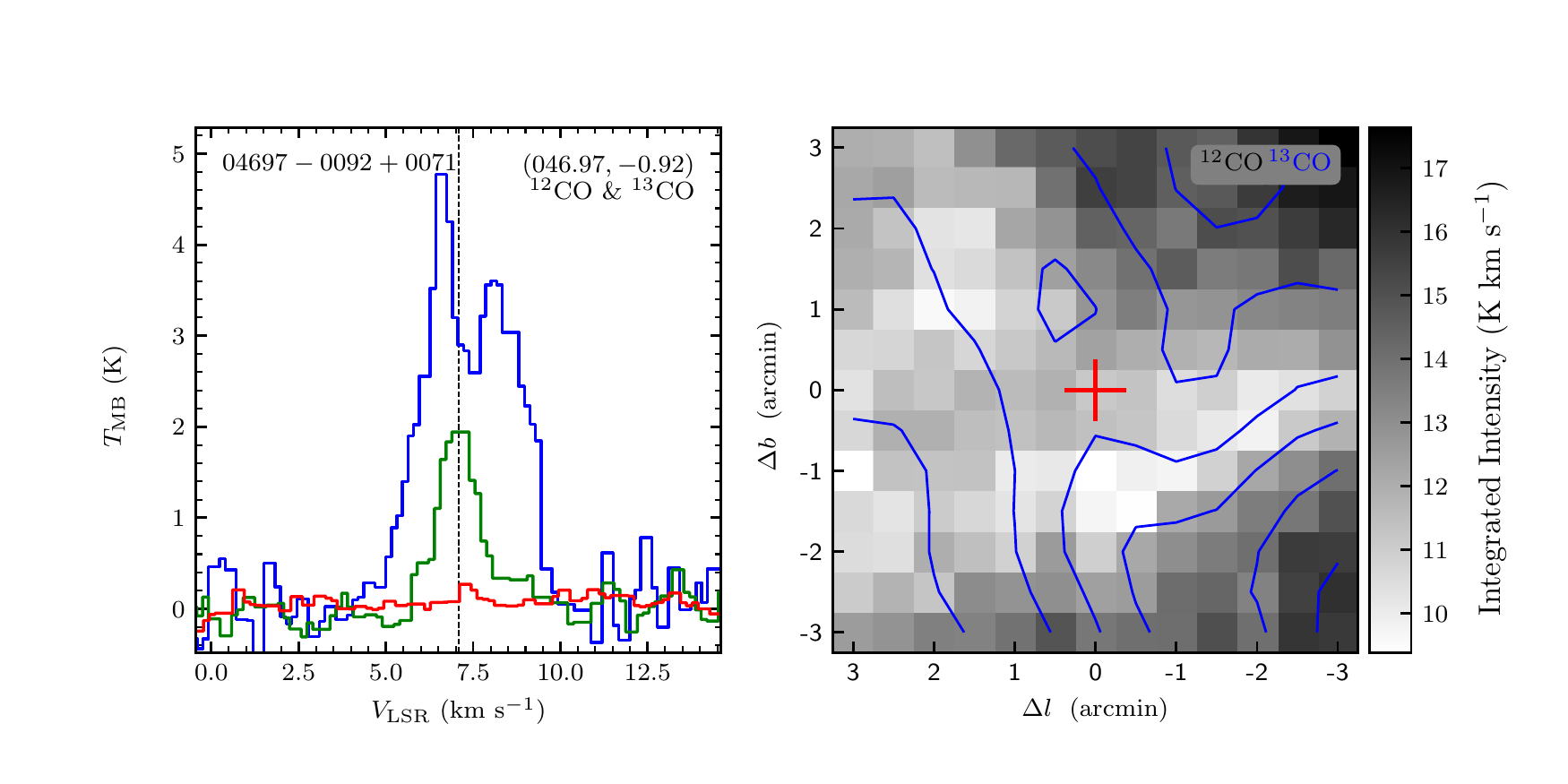}
\vspace{-0.5cm}

\includegraphics[width=9.0cm,angle=0]{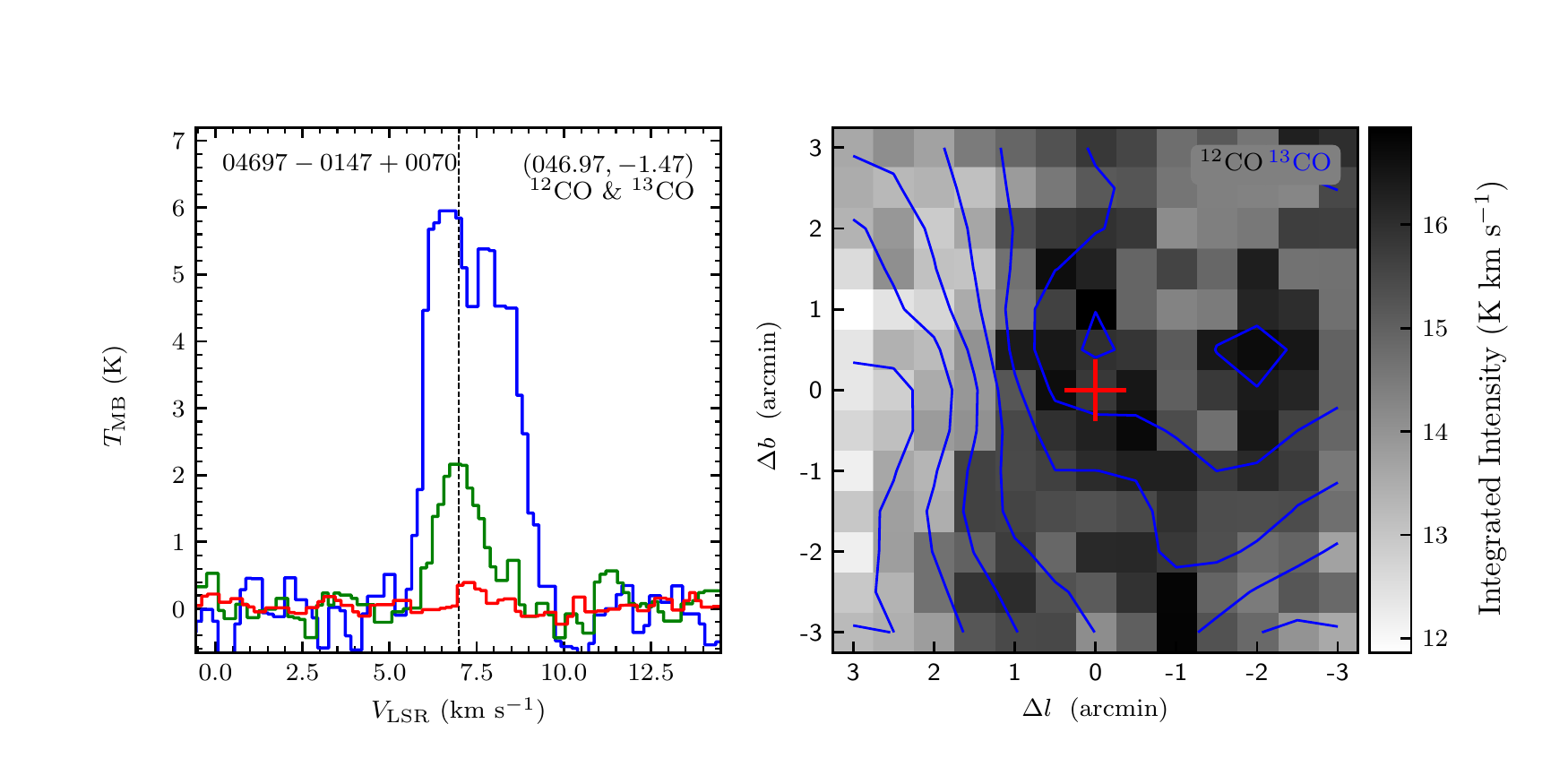}
\includegraphics[width=9.0cm,angle=0]{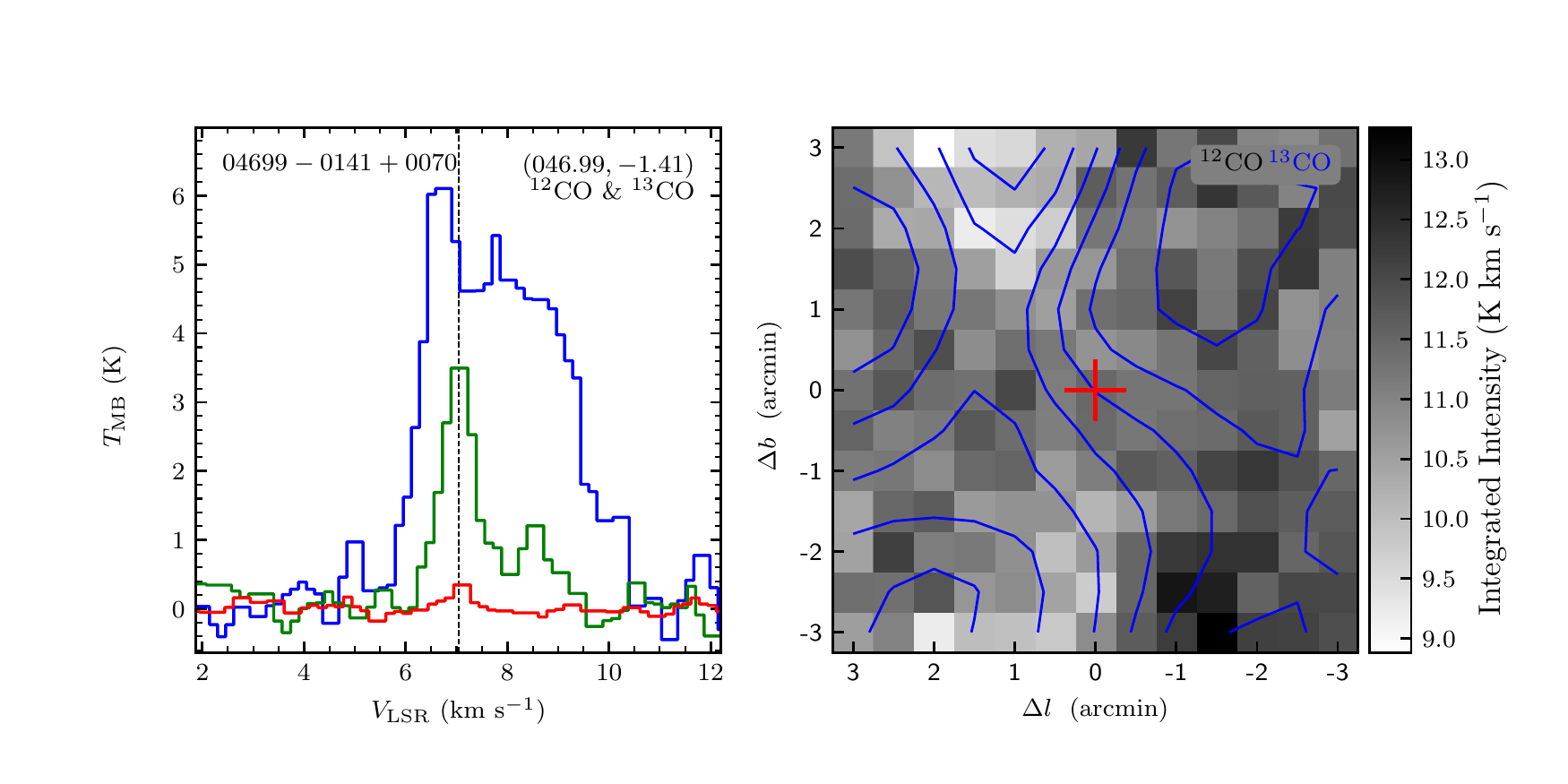}
\end{figure}
\clearpage

\begin{figure}
\includegraphics[width=9.0cm,angle=0]{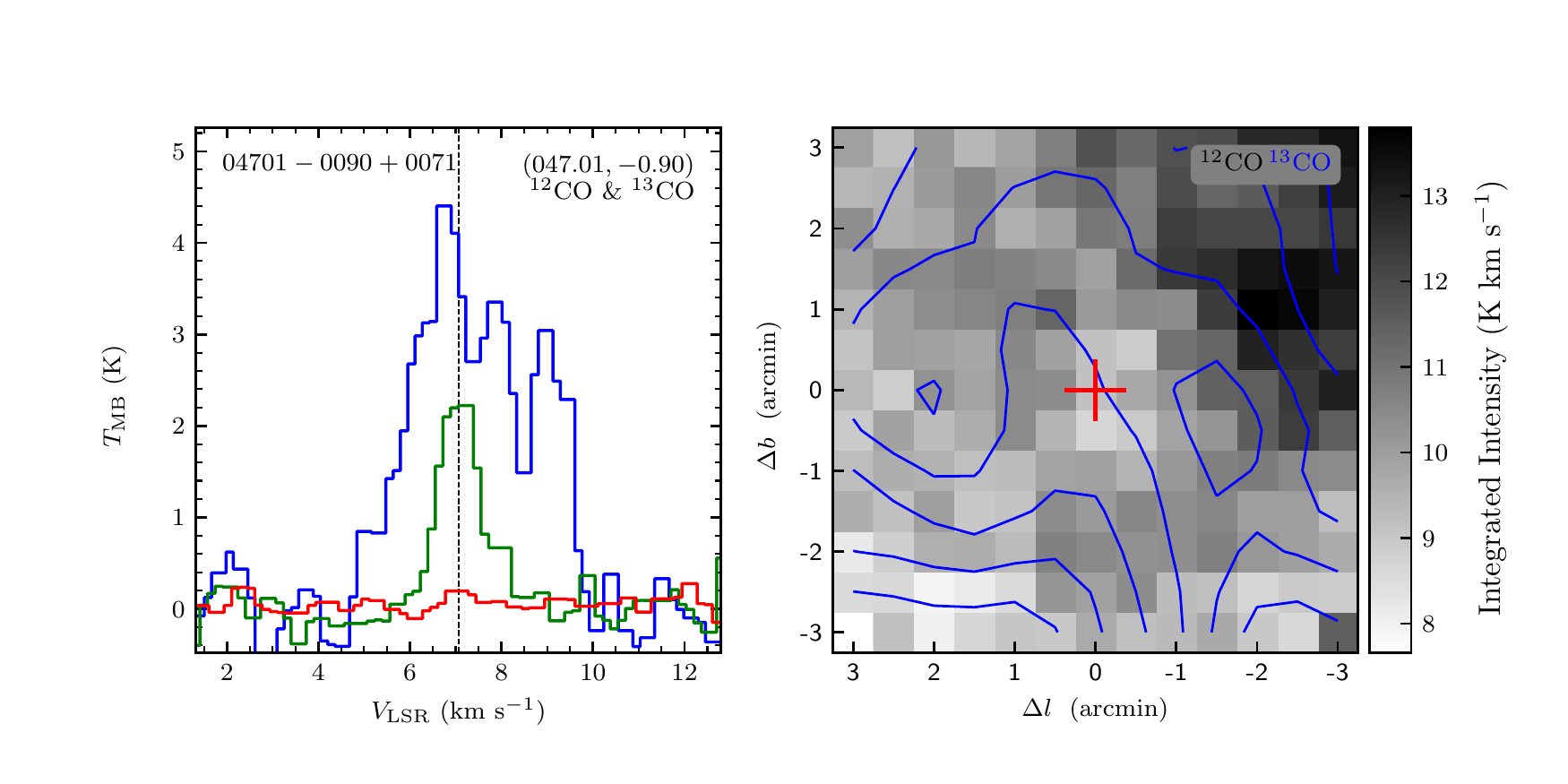}
\includegraphics[width=9.0cm,angle=0]{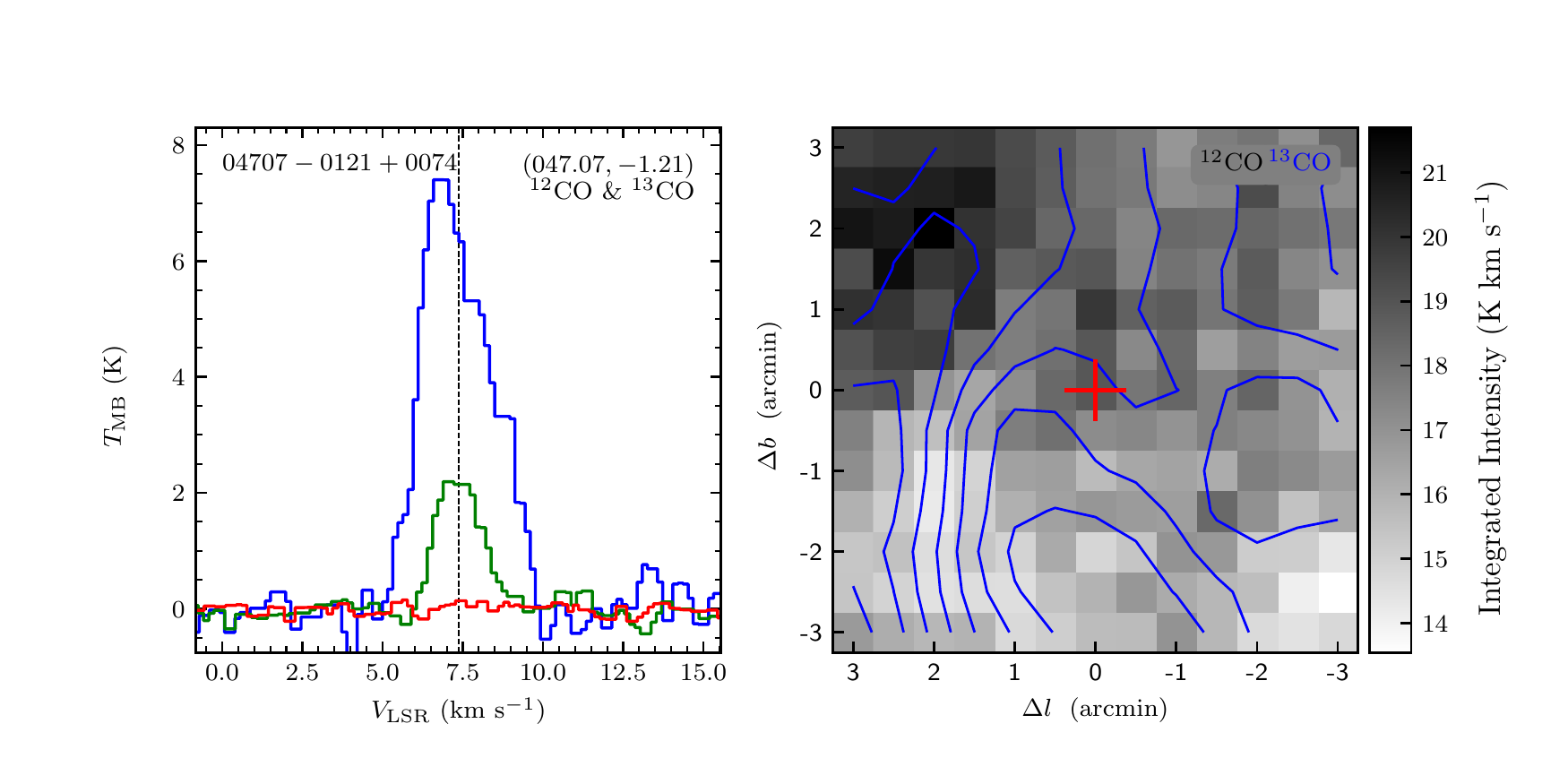}
\vspace{-0.5cm}

\includegraphics[width=9.0cm,angle=0]{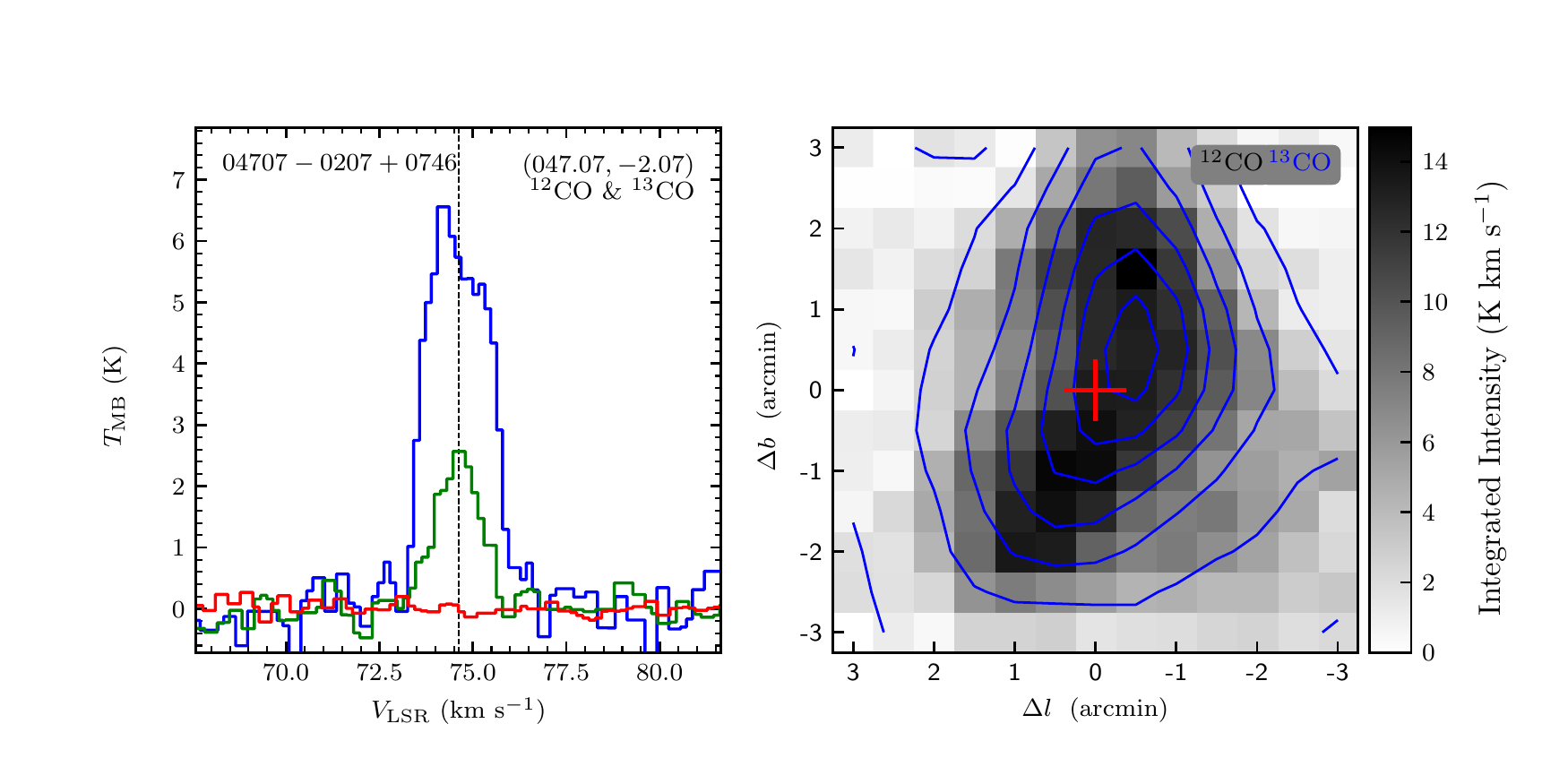}
\includegraphics[width=9.0cm,angle=0]{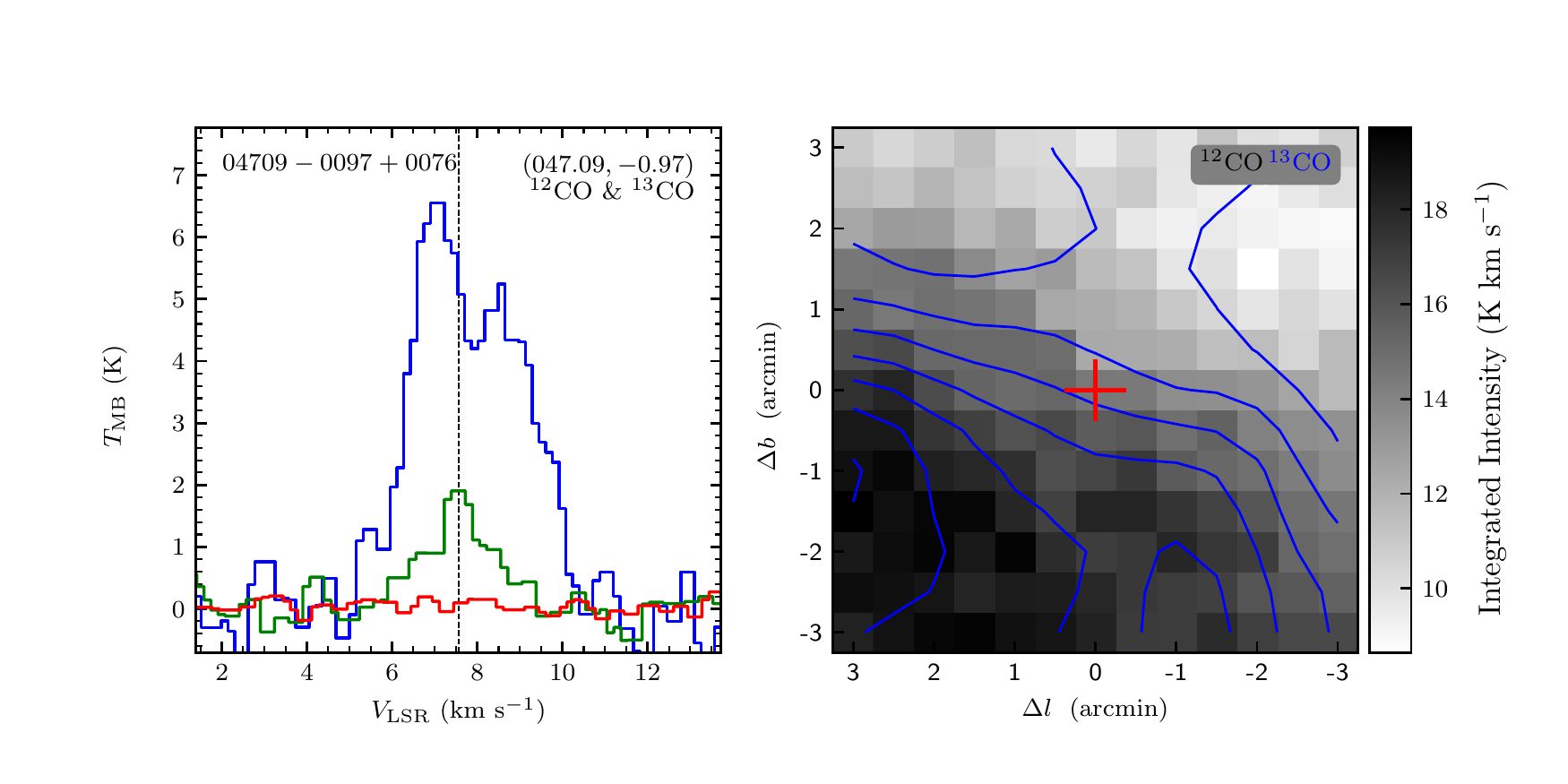}
\vspace{-0.5cm}

\includegraphics[width=9.0cm,angle=0]{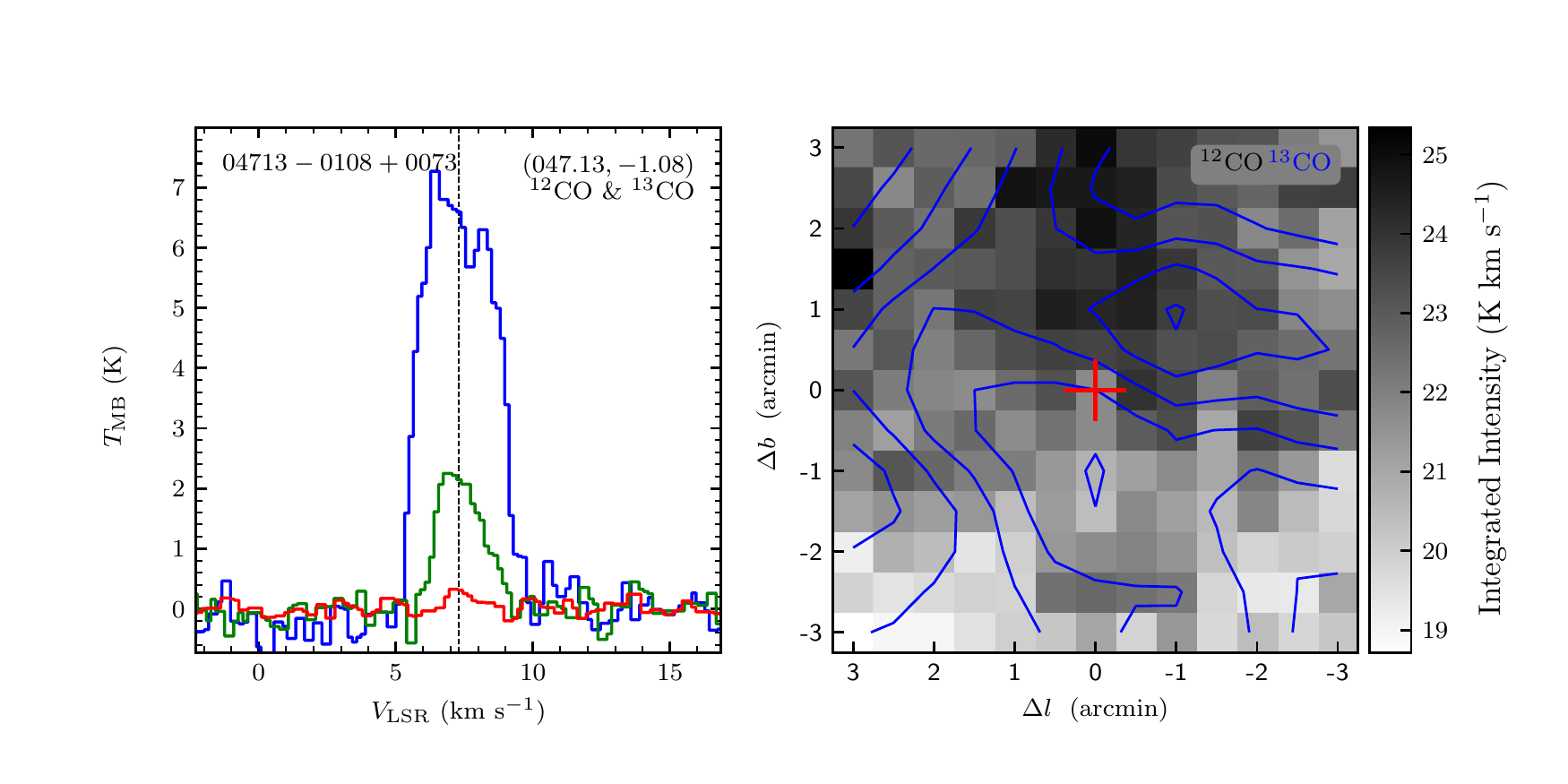}
\includegraphics[width=9.0cm,angle=0]{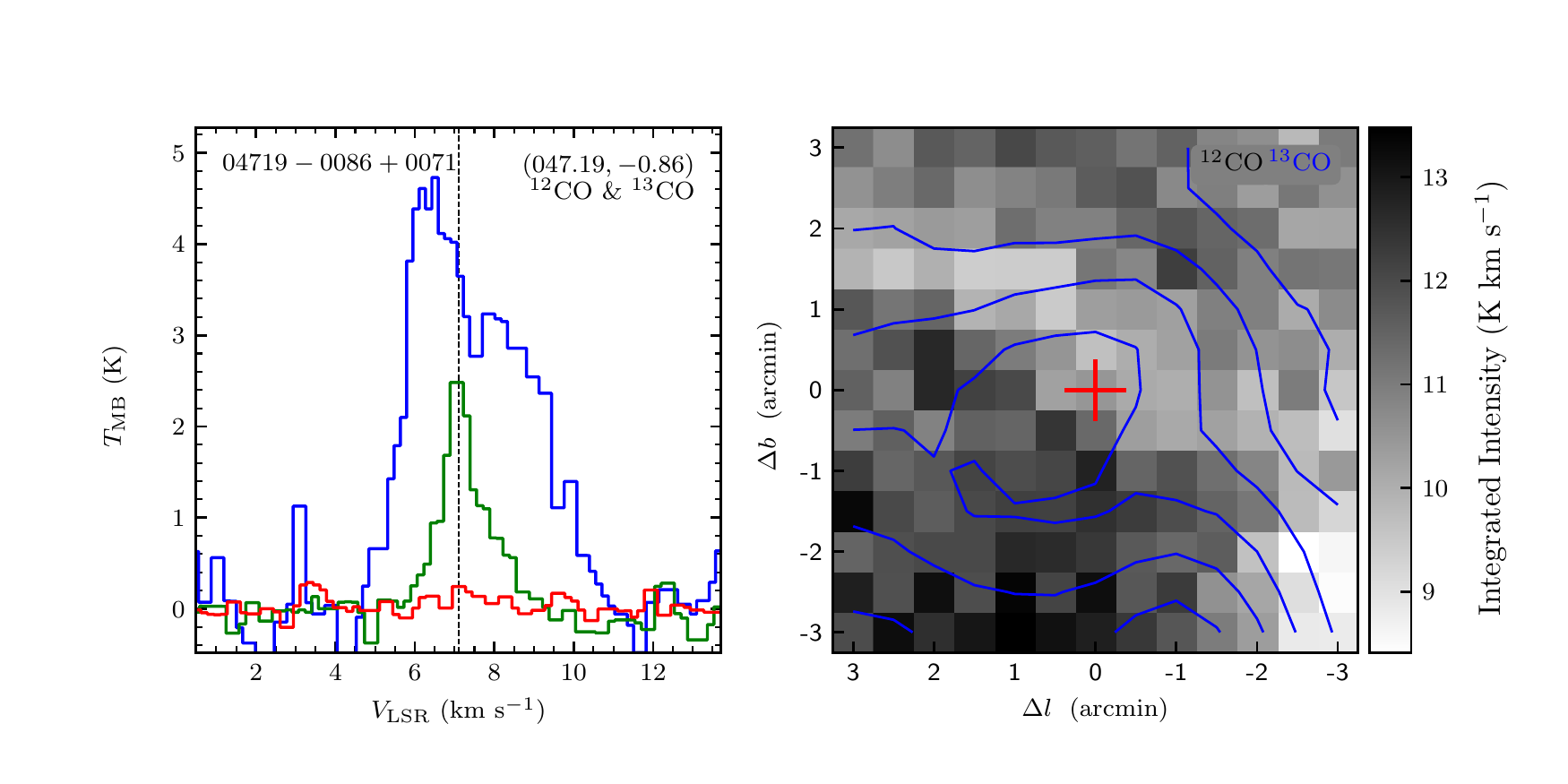}
\vspace{-0.5cm}

\includegraphics[width=9.0cm,angle=0]{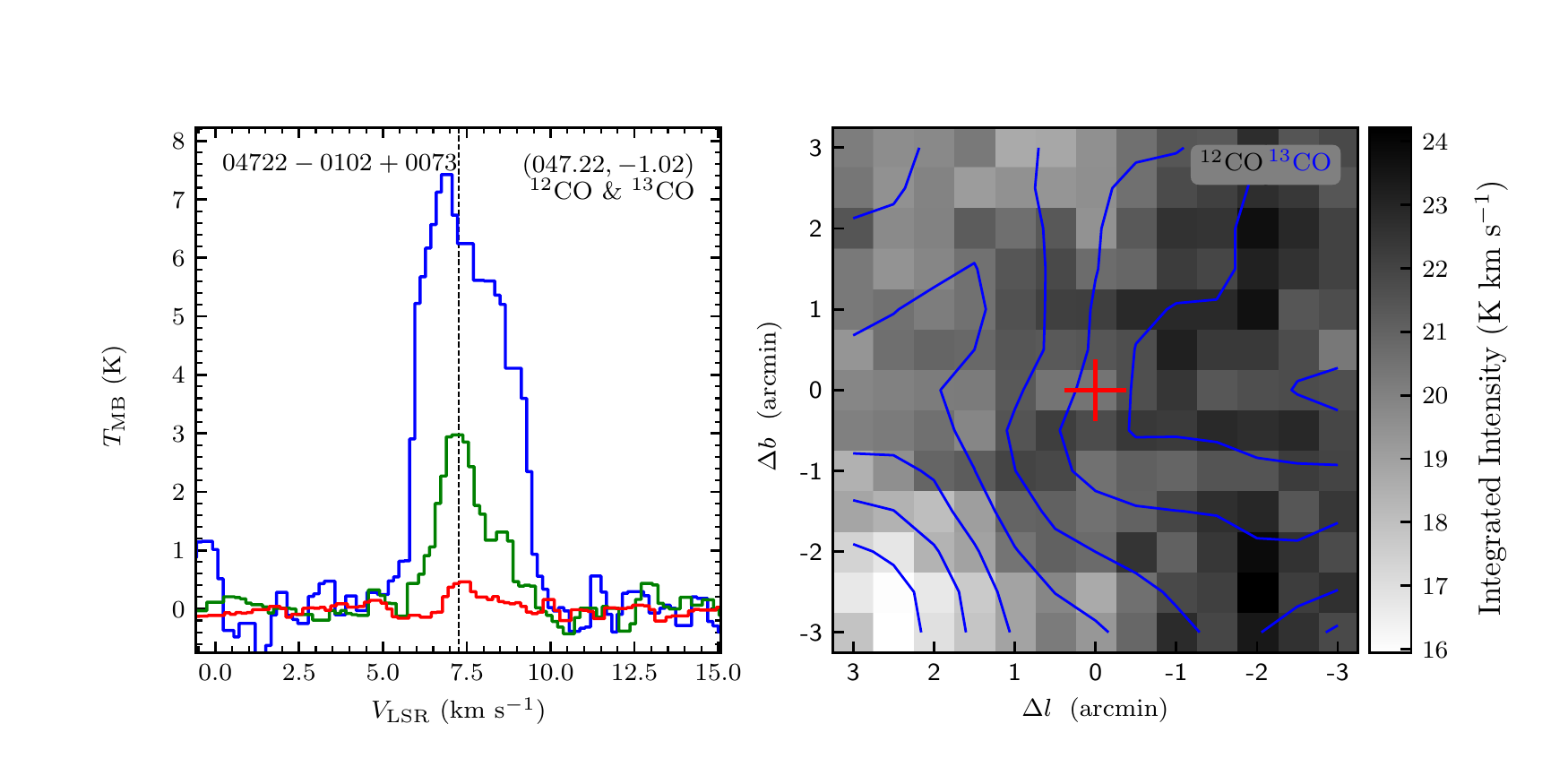}
\includegraphics[width=9.0cm,angle=0]{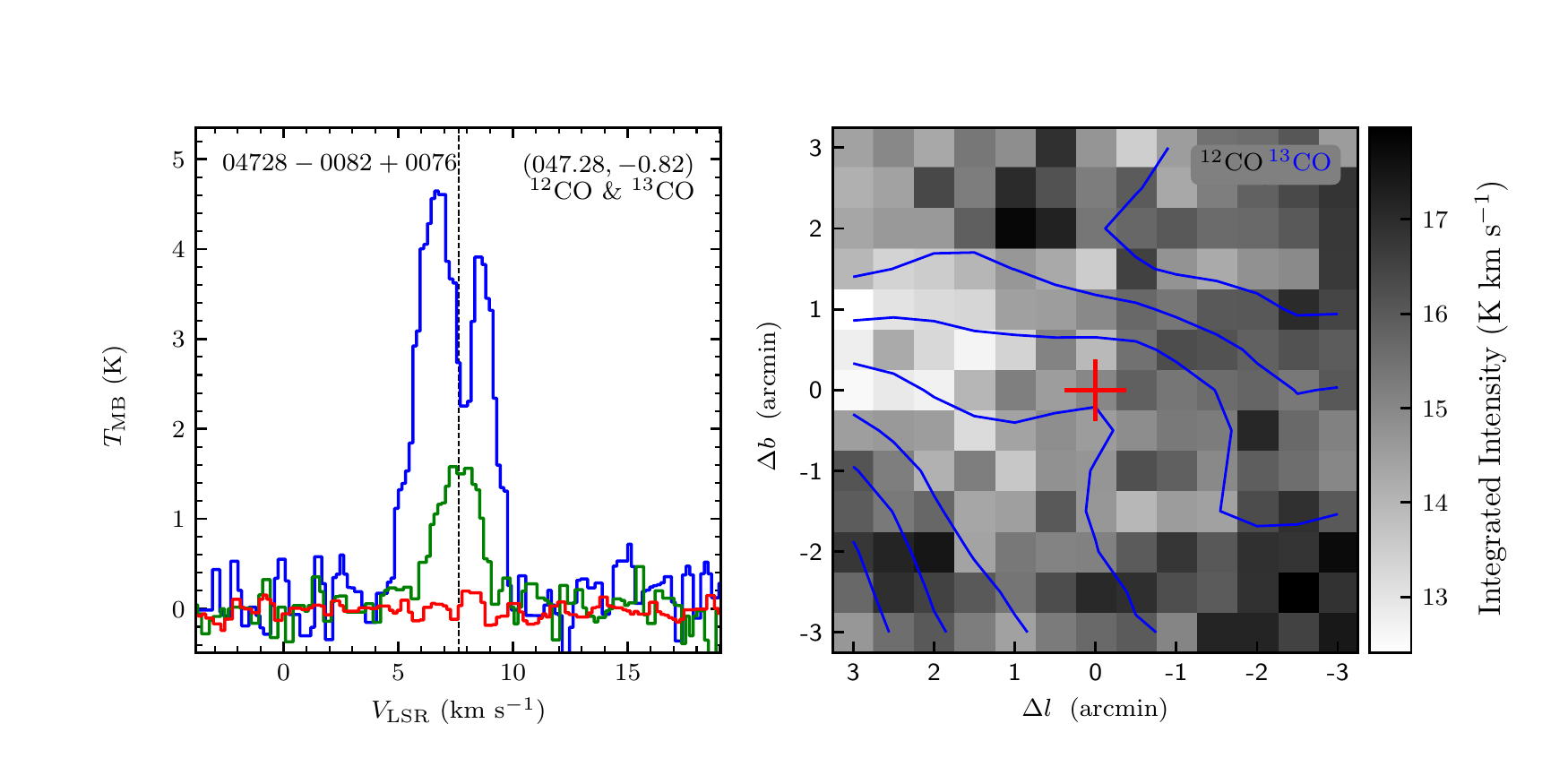}
\vspace{-0.5cm}

\includegraphics[width=9.0cm,angle=0]{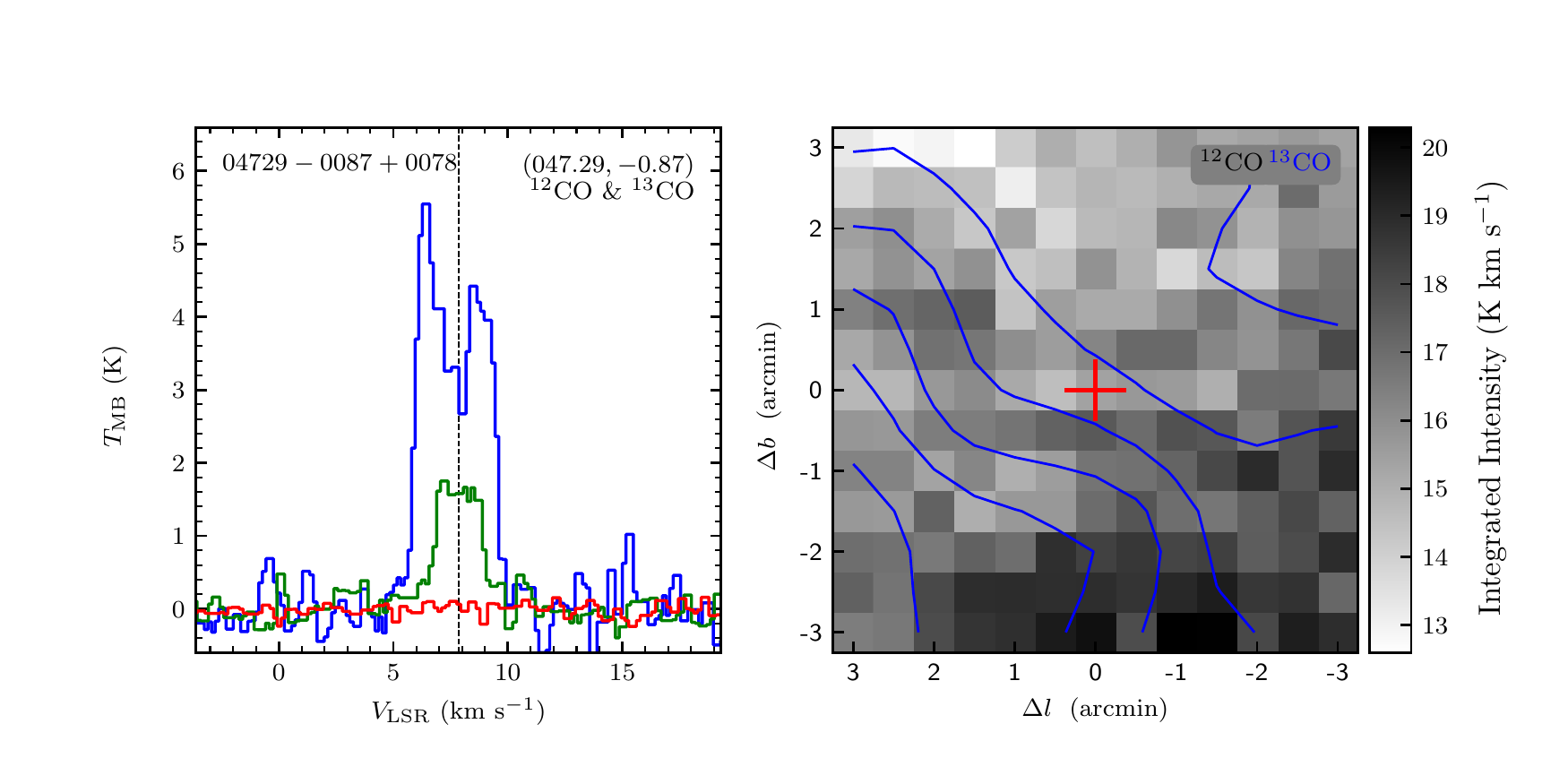}
\includegraphics[width=9.0cm,angle=0]{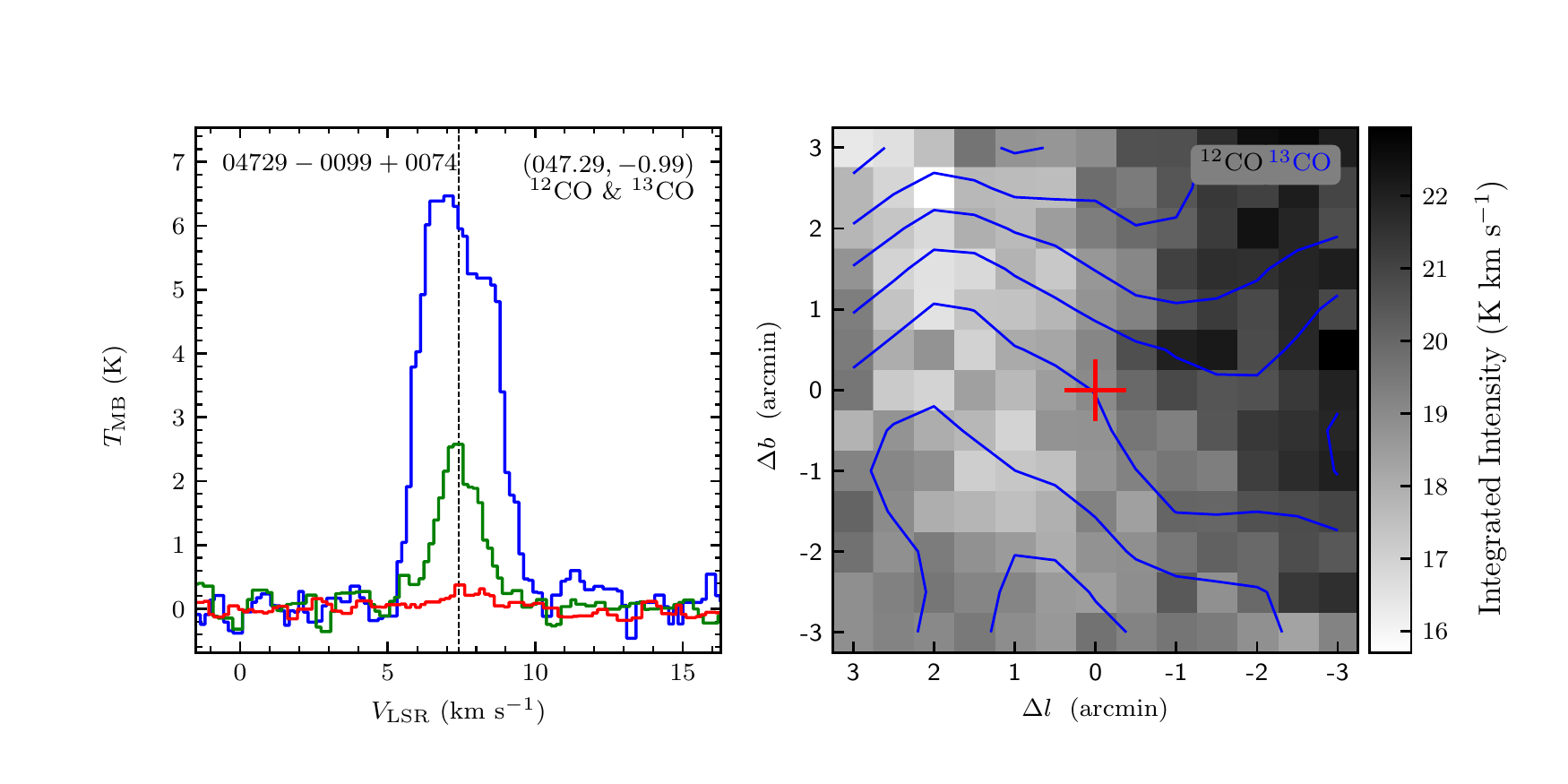}
\end{figure}
\clearpage

\begin{figure}
\includegraphics[width=9.0cm,angle=0]{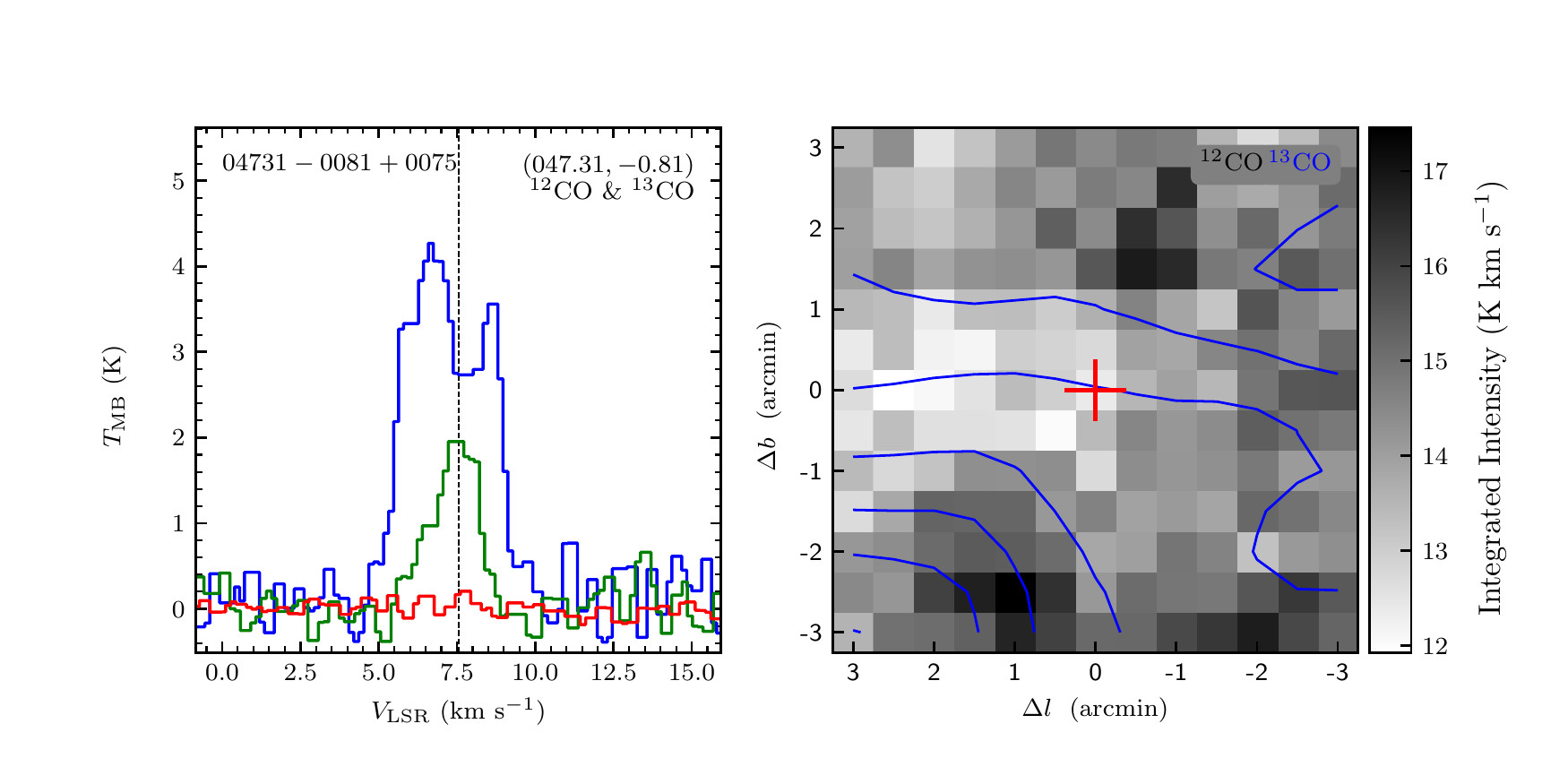}
\includegraphics[width=9.0cm,angle=0]{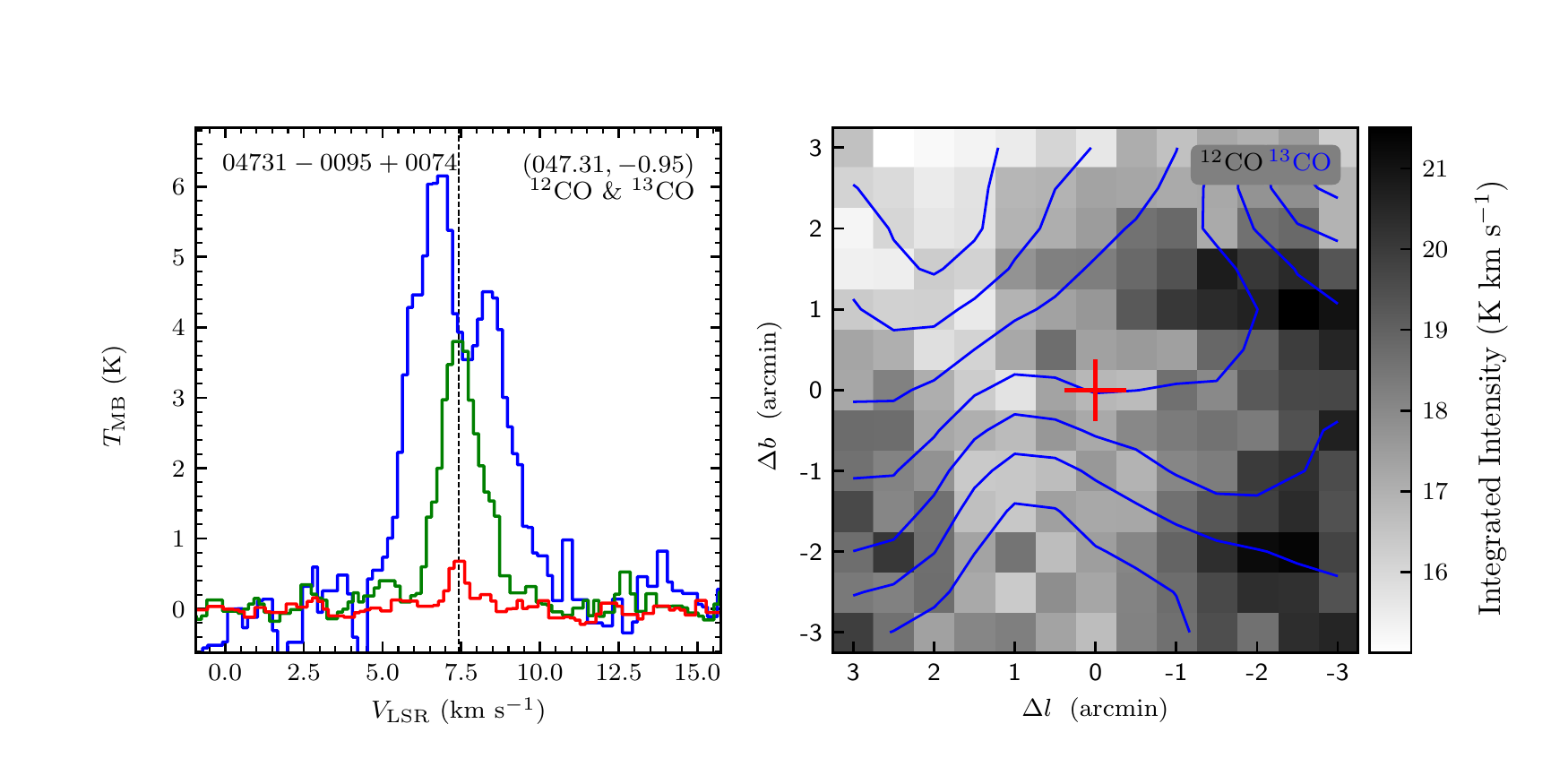}
\vspace{-0.5cm}

\includegraphics[width=9.0cm,angle=0]{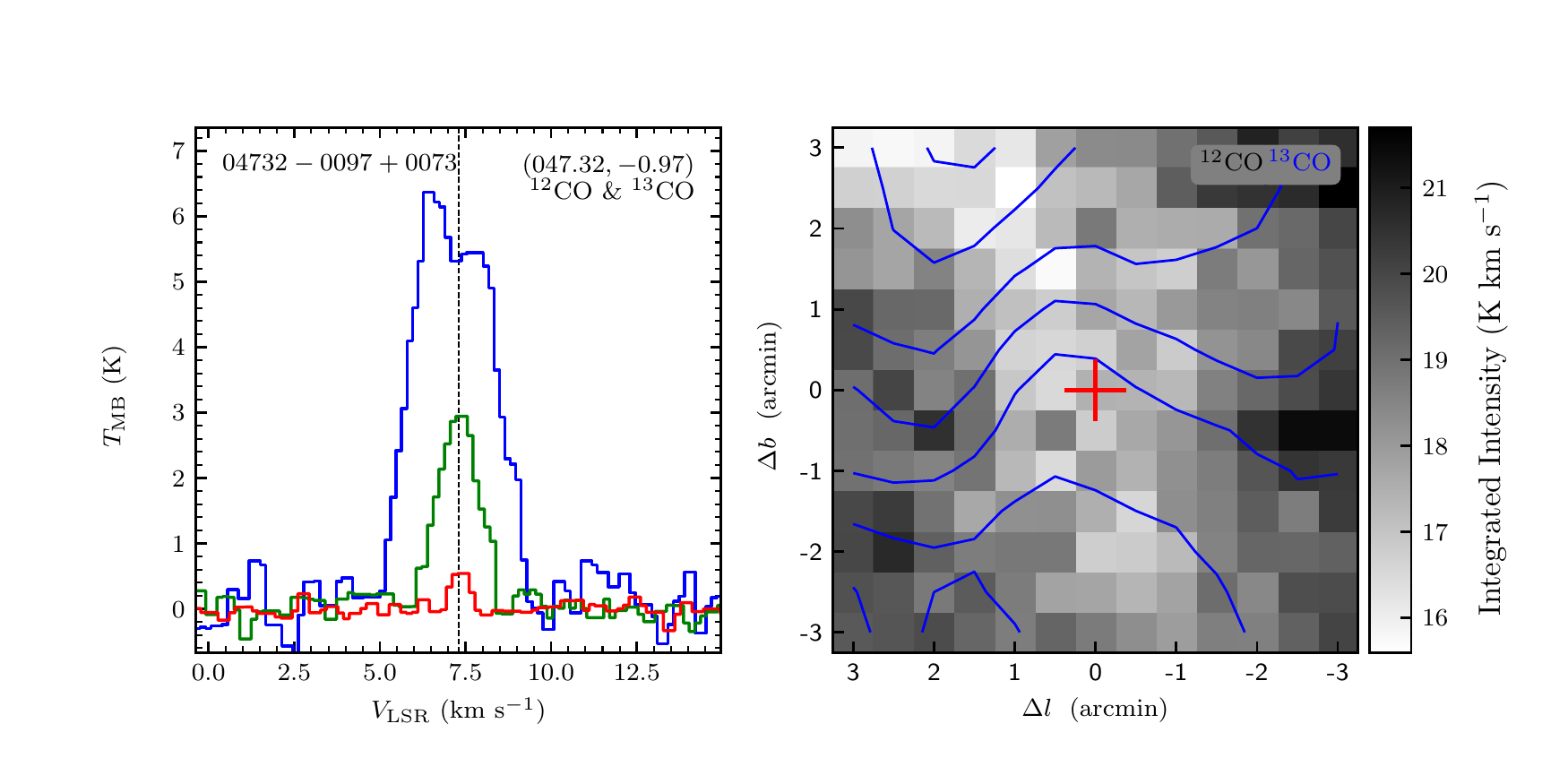}
\includegraphics[width=9.0cm,angle=0]{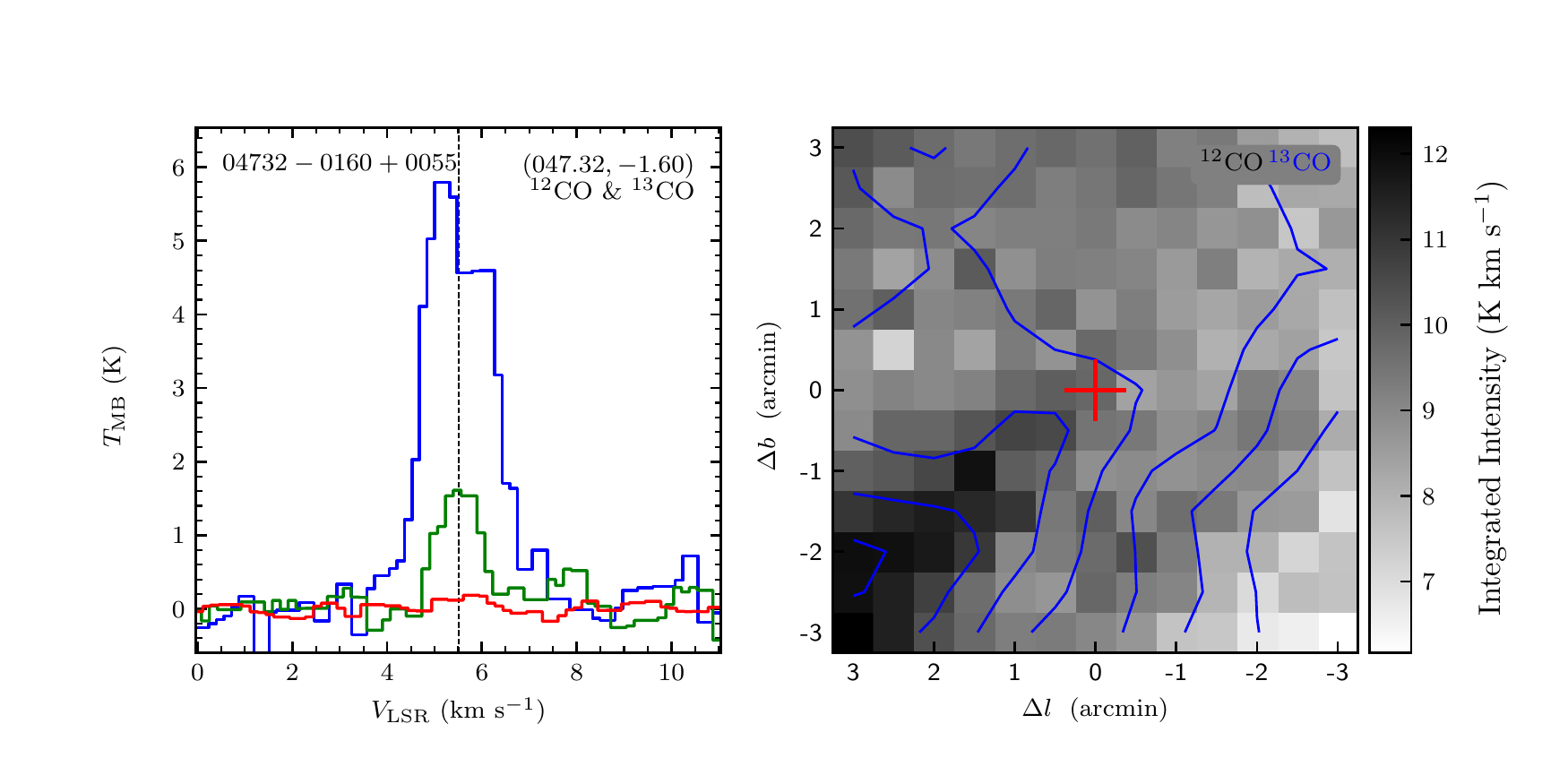}
\vspace{-0.5cm}

\includegraphics[width=9.0cm,angle=0]{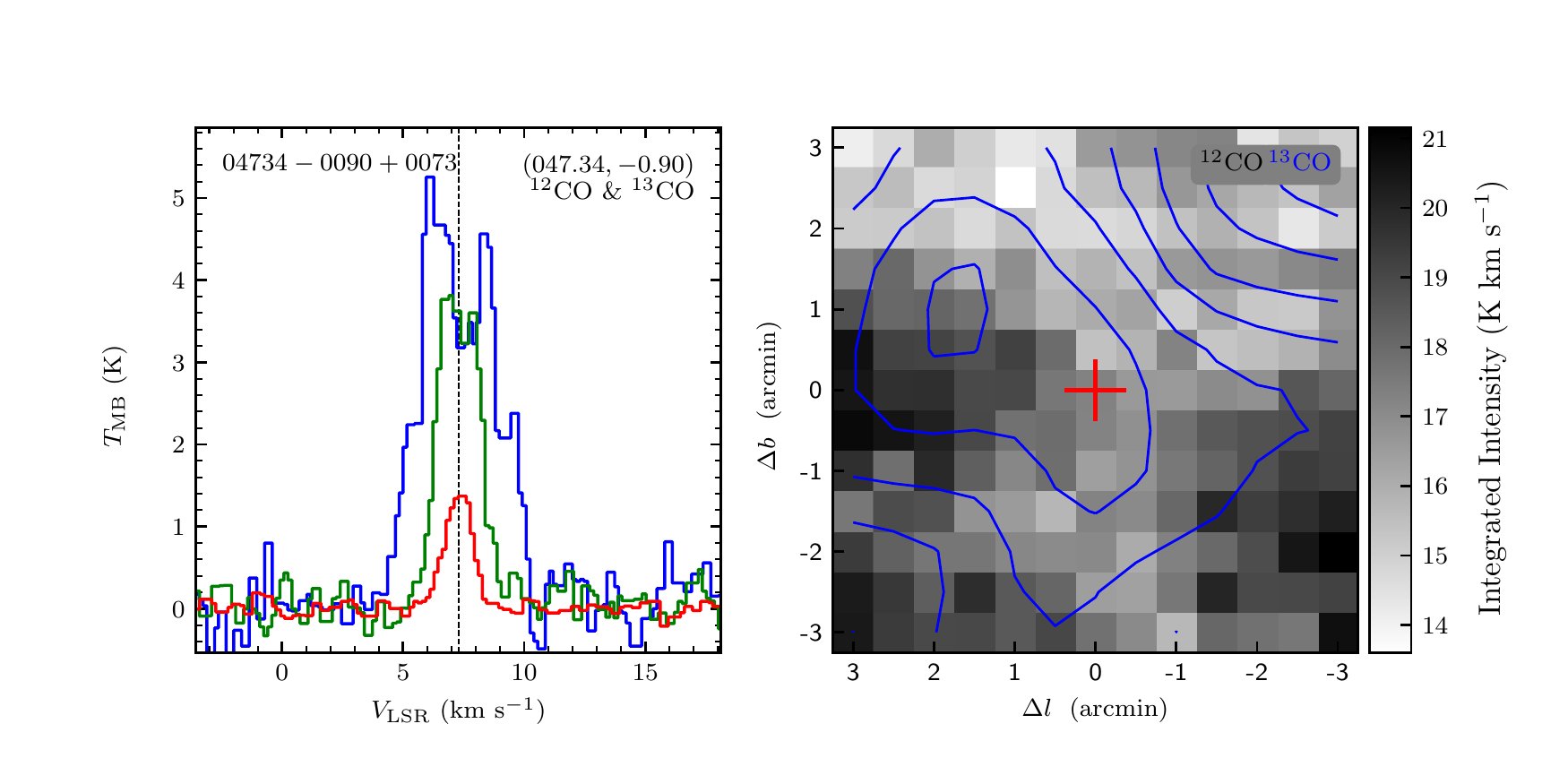}
\includegraphics[width=9.0cm,angle=0]{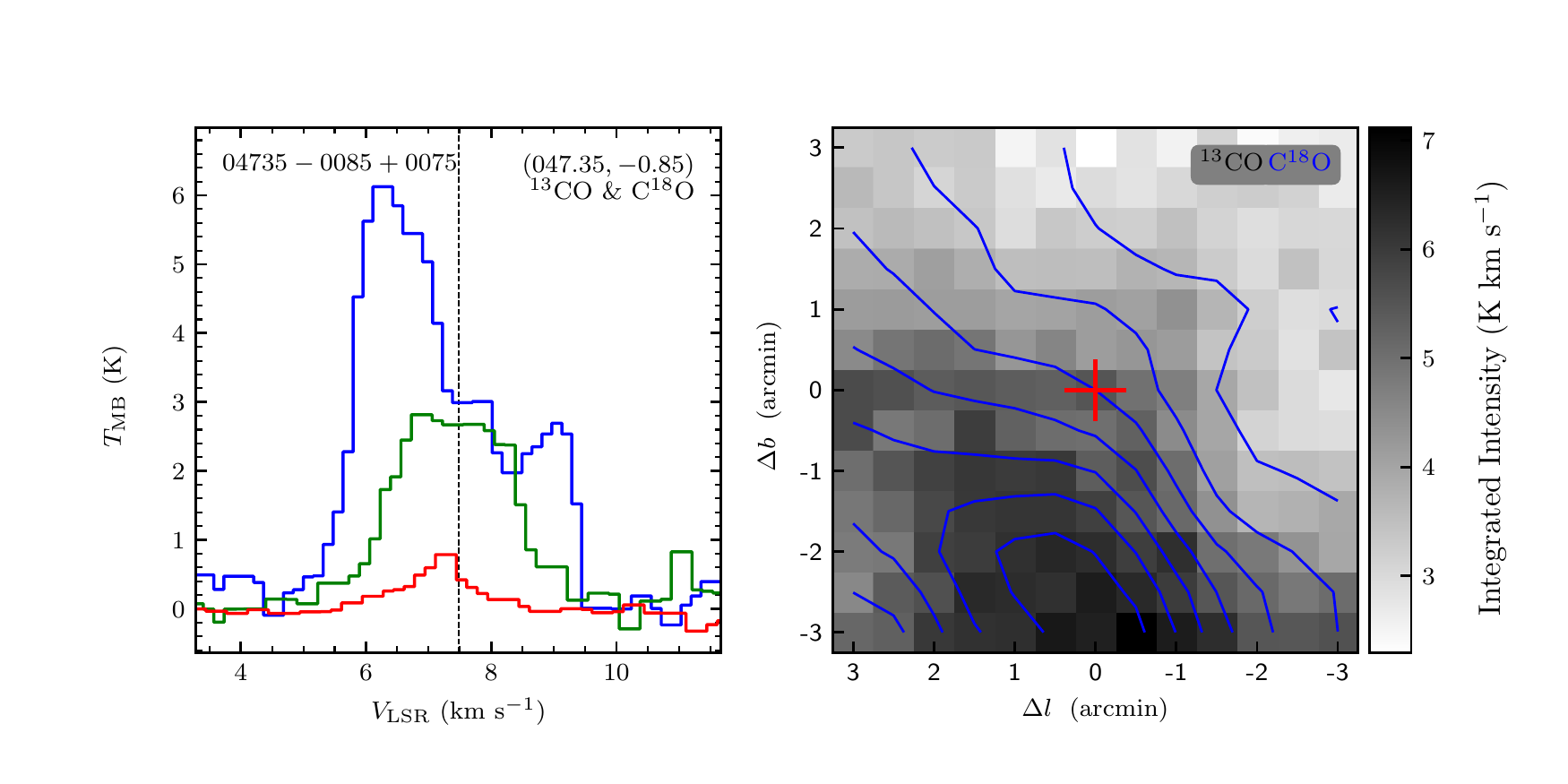}
\vspace{-0.5cm}

\includegraphics[width=9.0cm,angle=0]{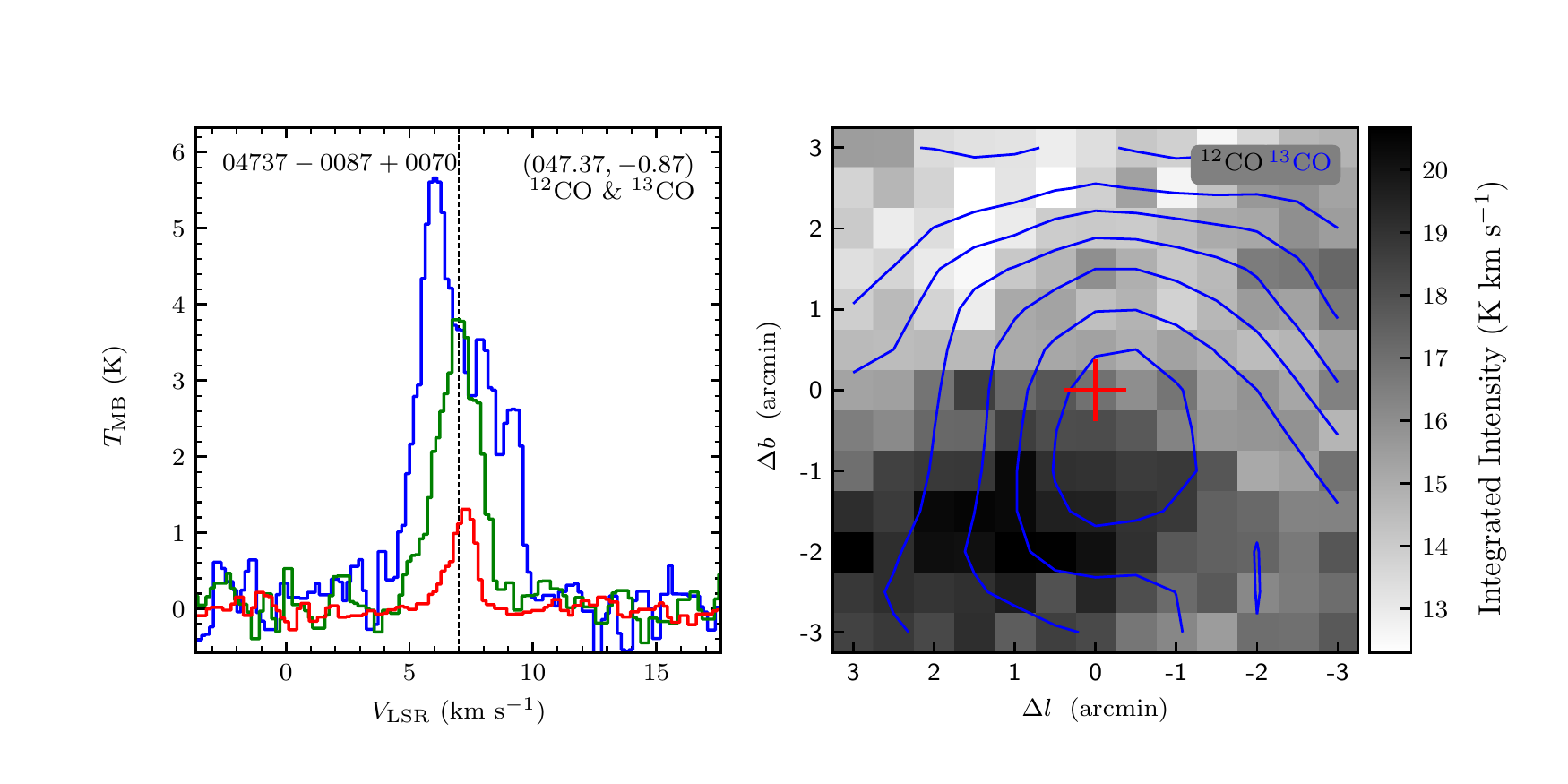}
\includegraphics[width=9.0cm,angle=0]{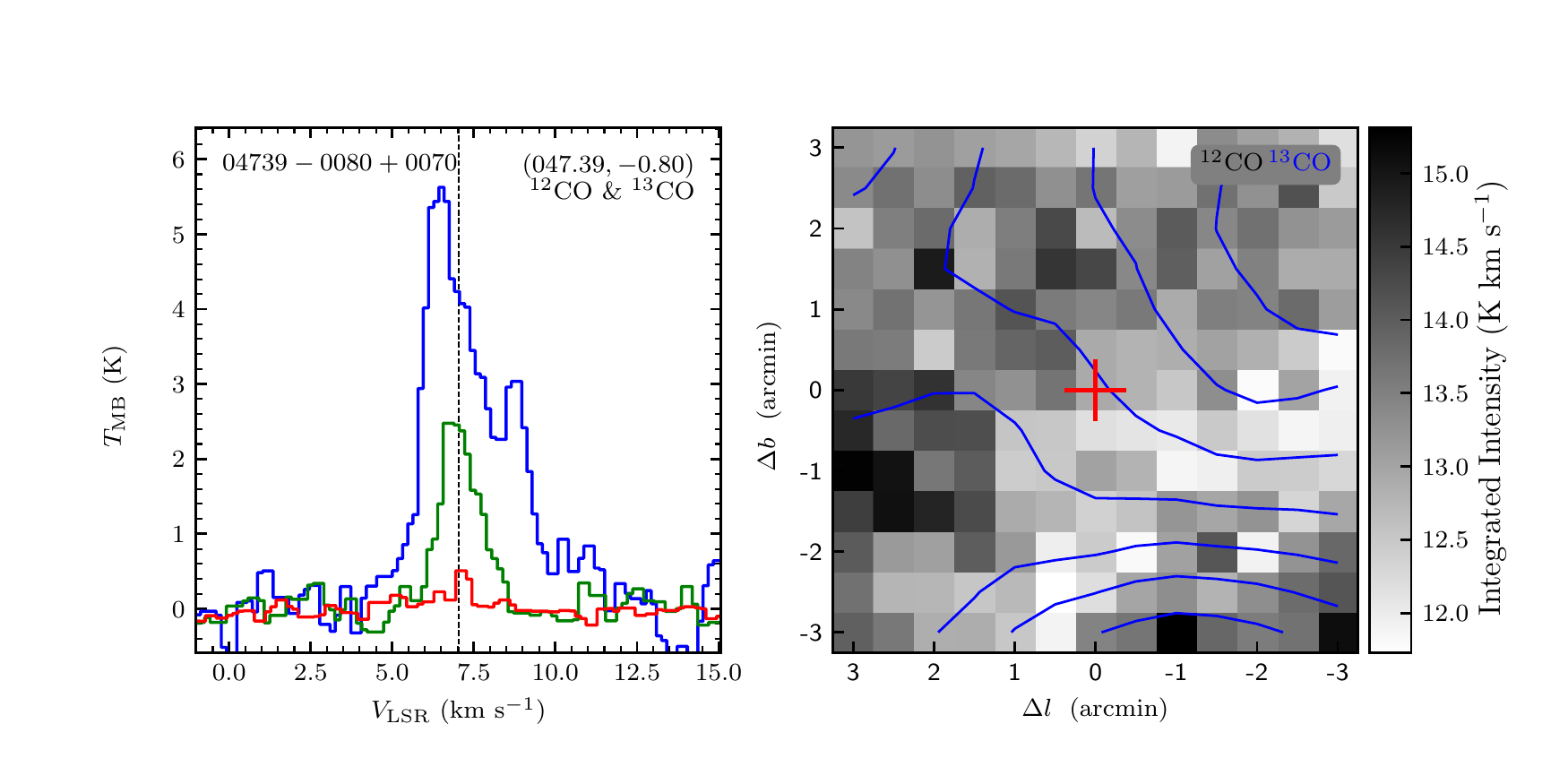}
\vspace{-0.5cm}

\includegraphics[width=9.0cm,angle=0]{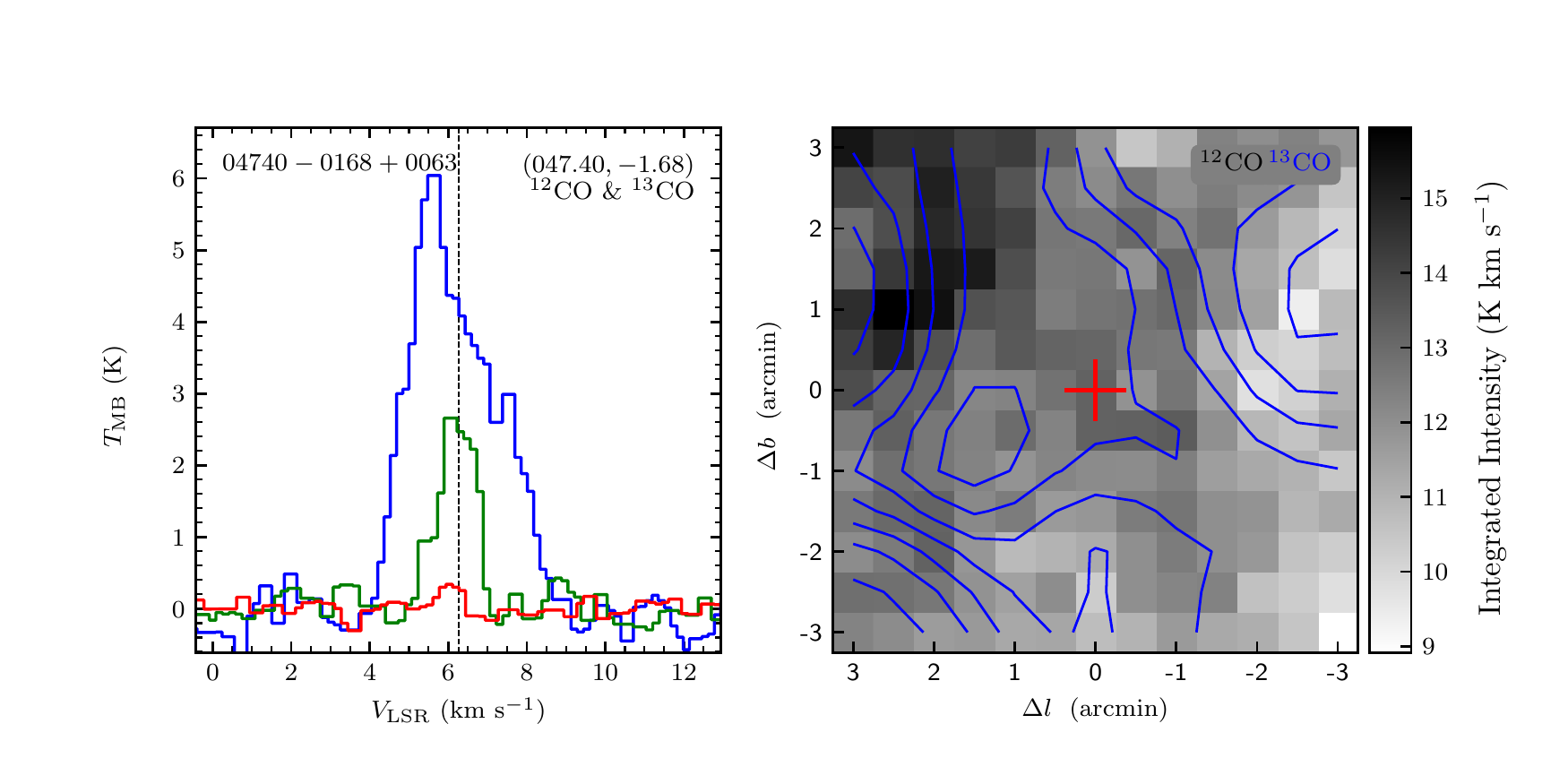}
\includegraphics[width=9.0cm,angle=0]{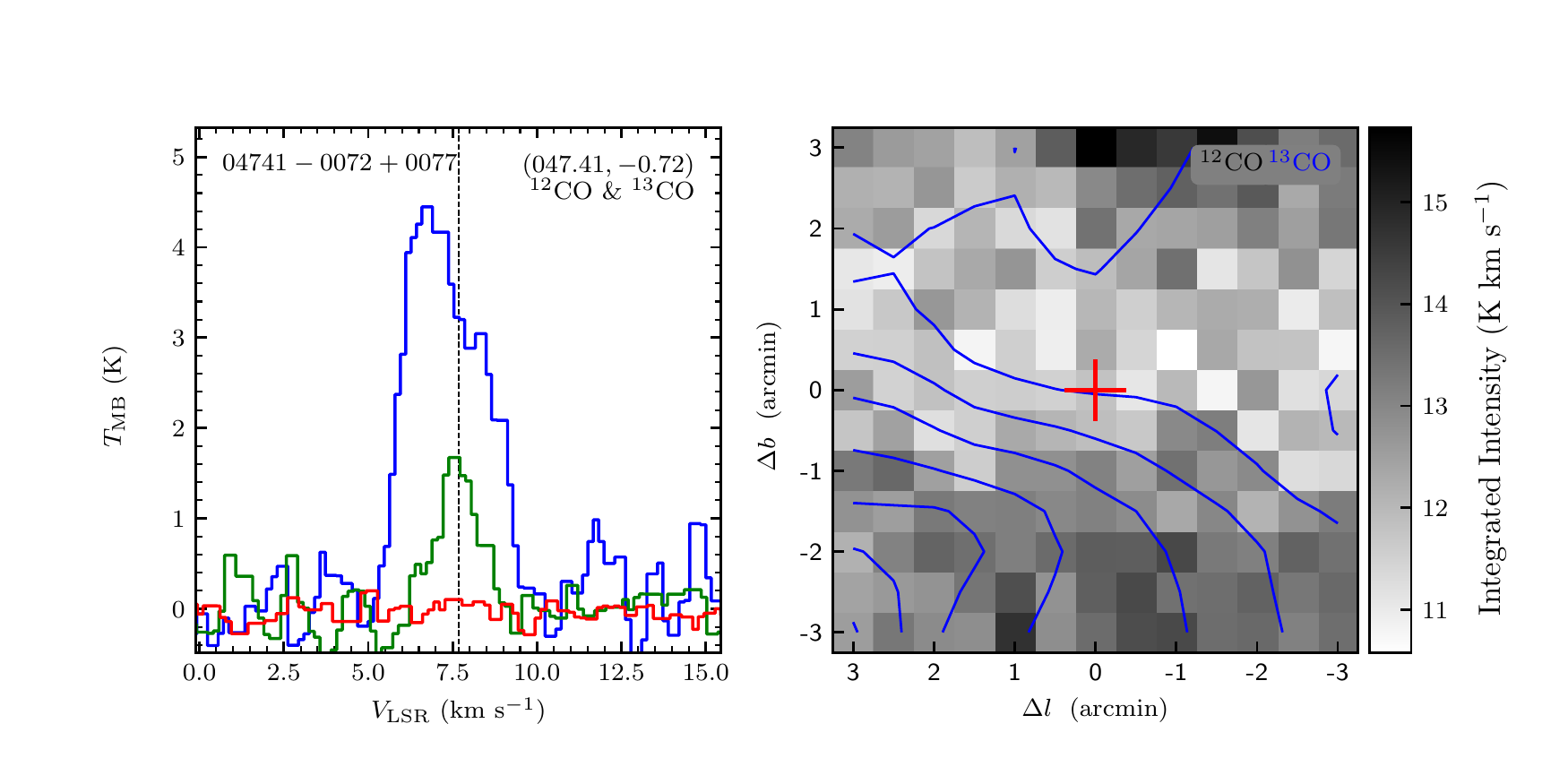}
\end{figure}
\clearpage

\begin{figure}
\includegraphics[width=9.0cm,angle=0]{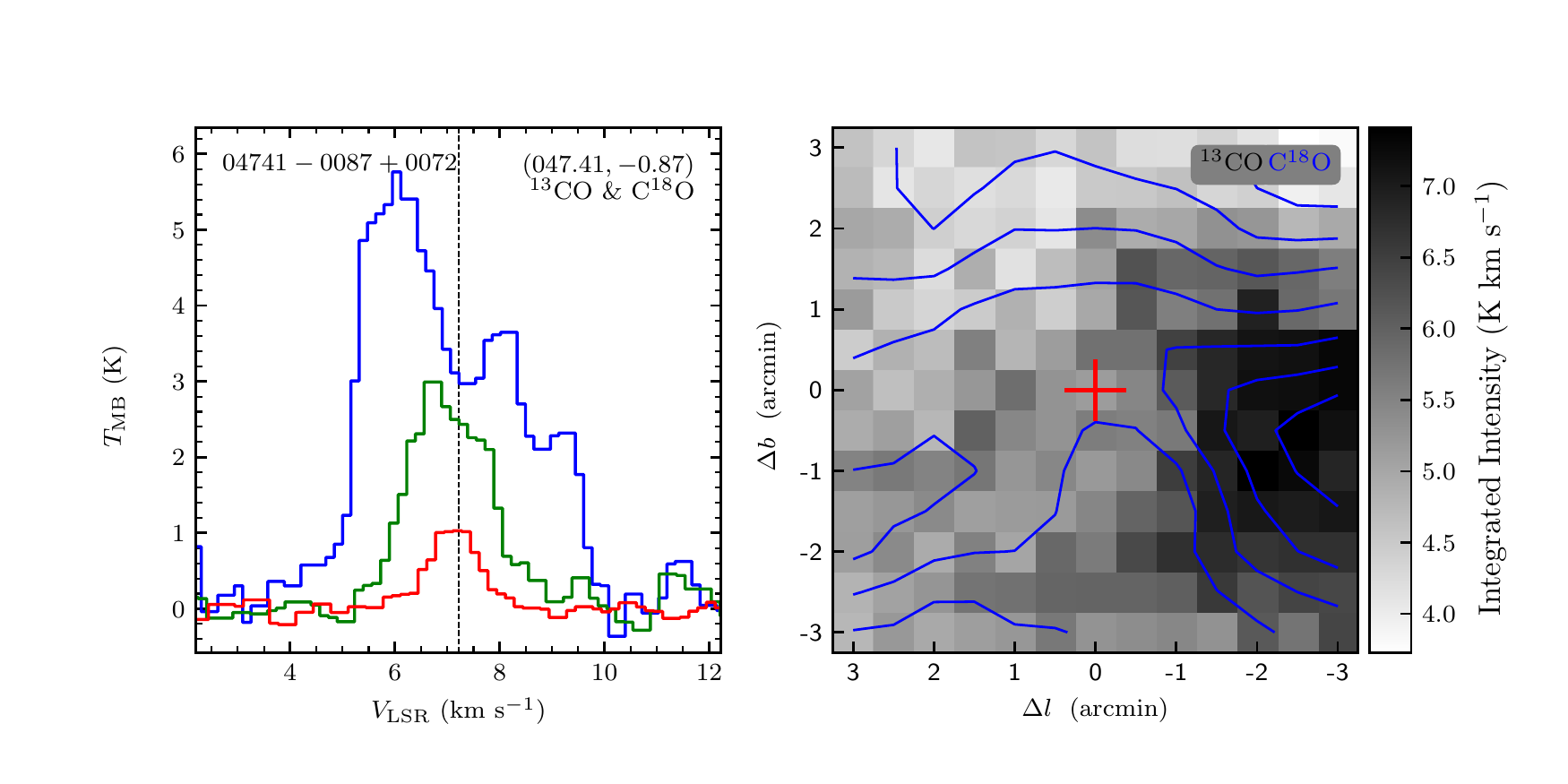}
\includegraphics[width=9.0cm,angle=0]{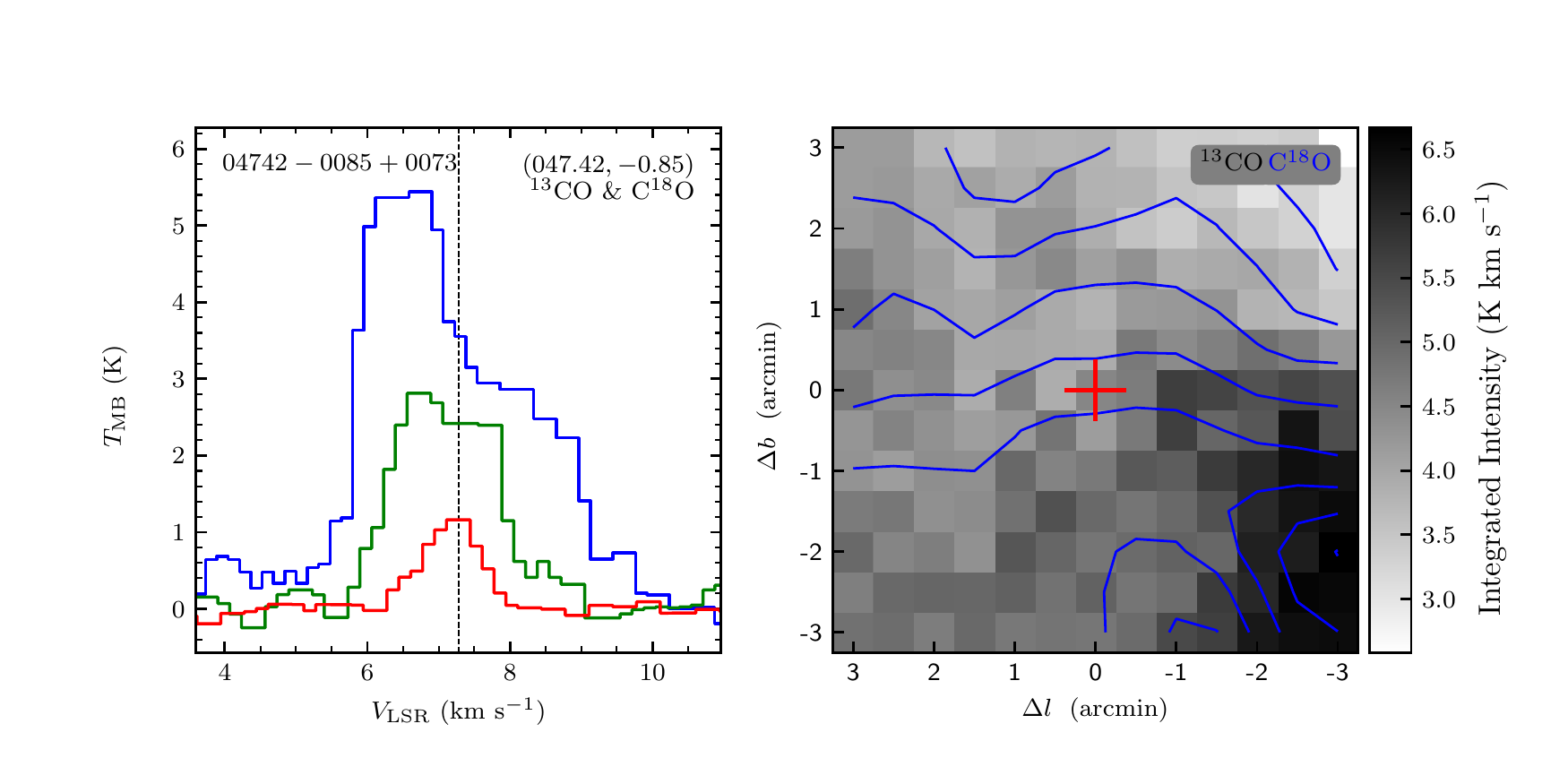}
\vspace{-0.5cm}

\includegraphics[width=9.0cm,angle=0]{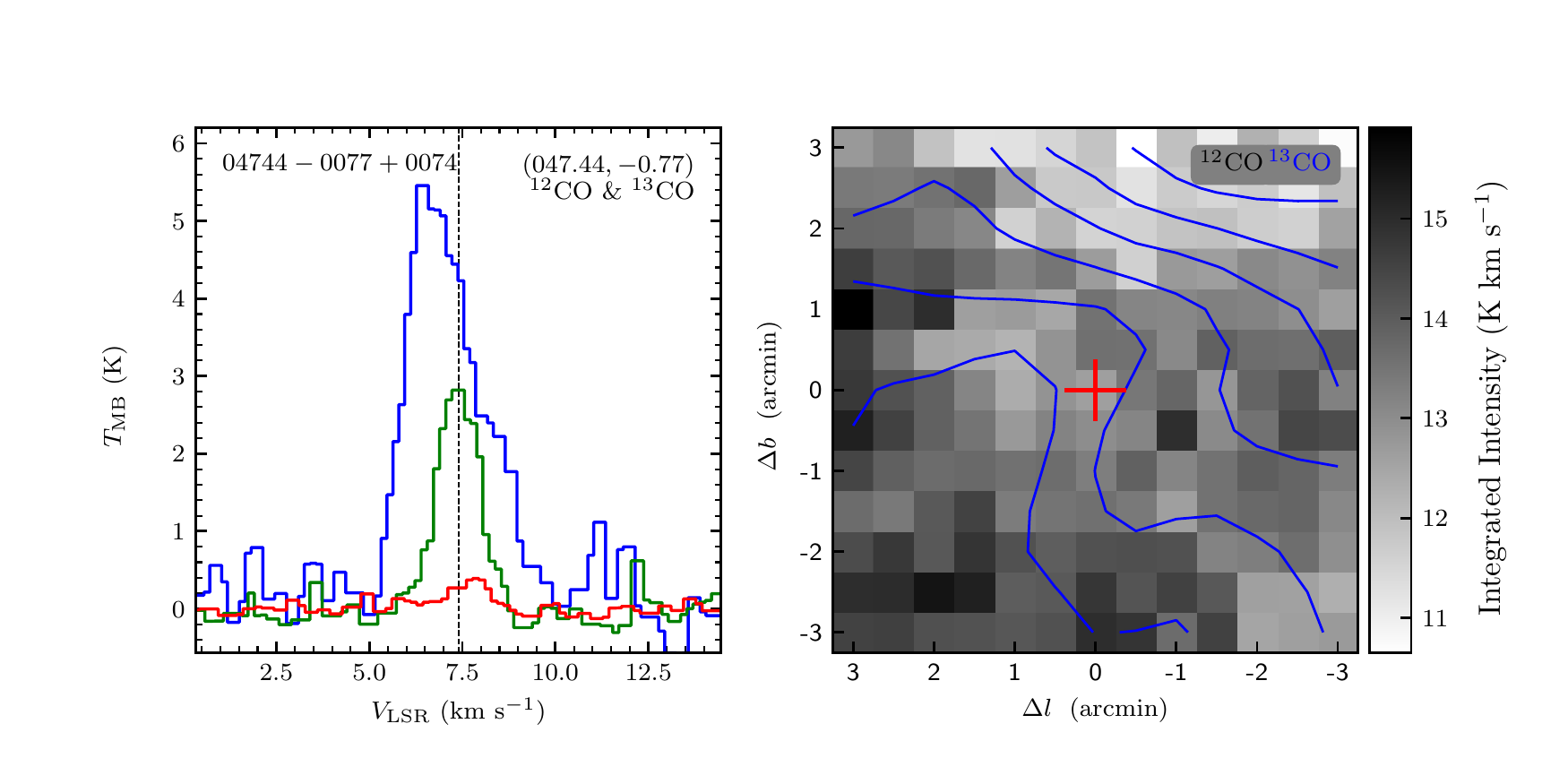}
\includegraphics[width=9.0cm,angle=0]{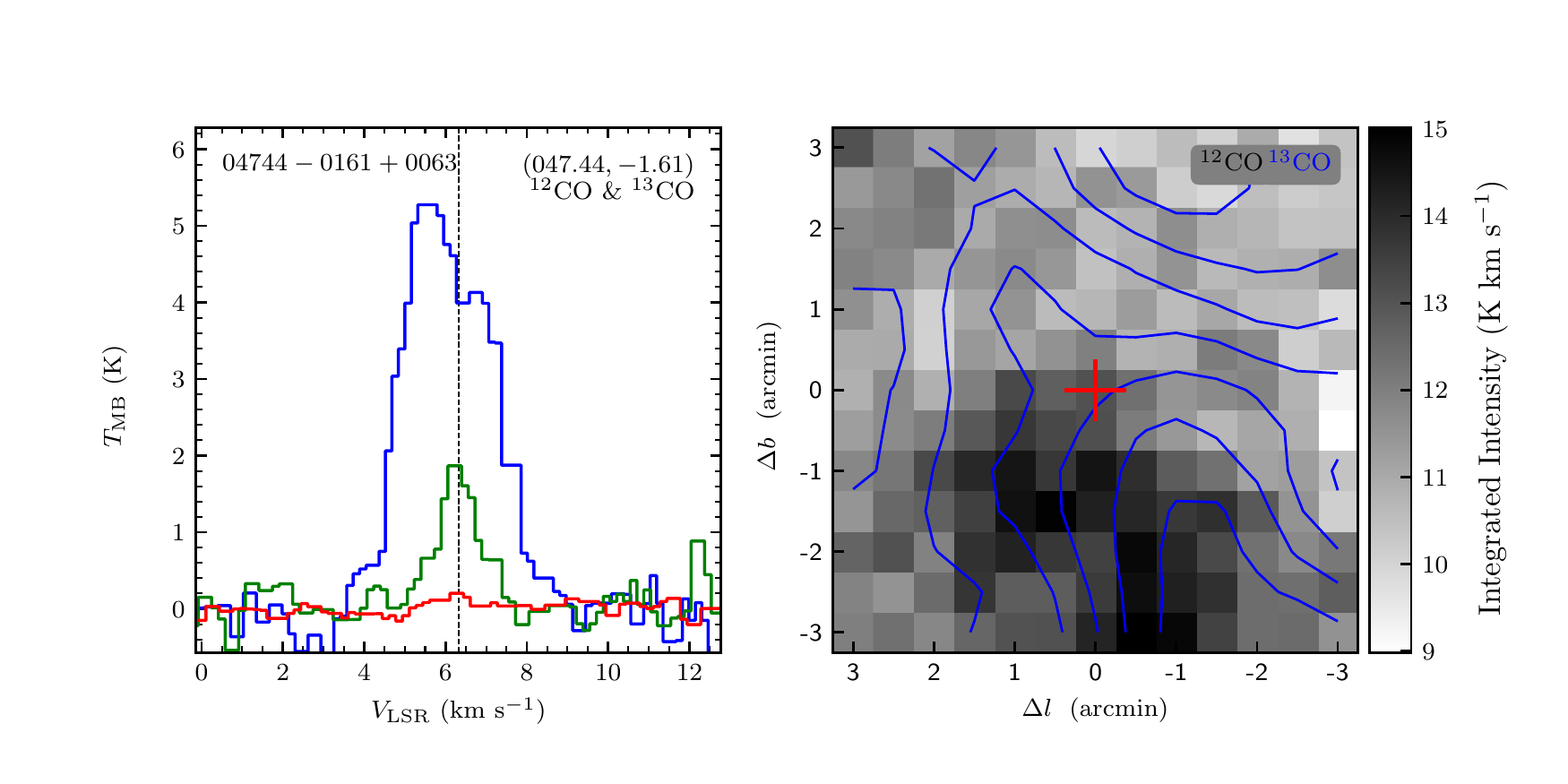}
\vspace{-0.5cm}

\includegraphics[width=9.0cm,angle=0]{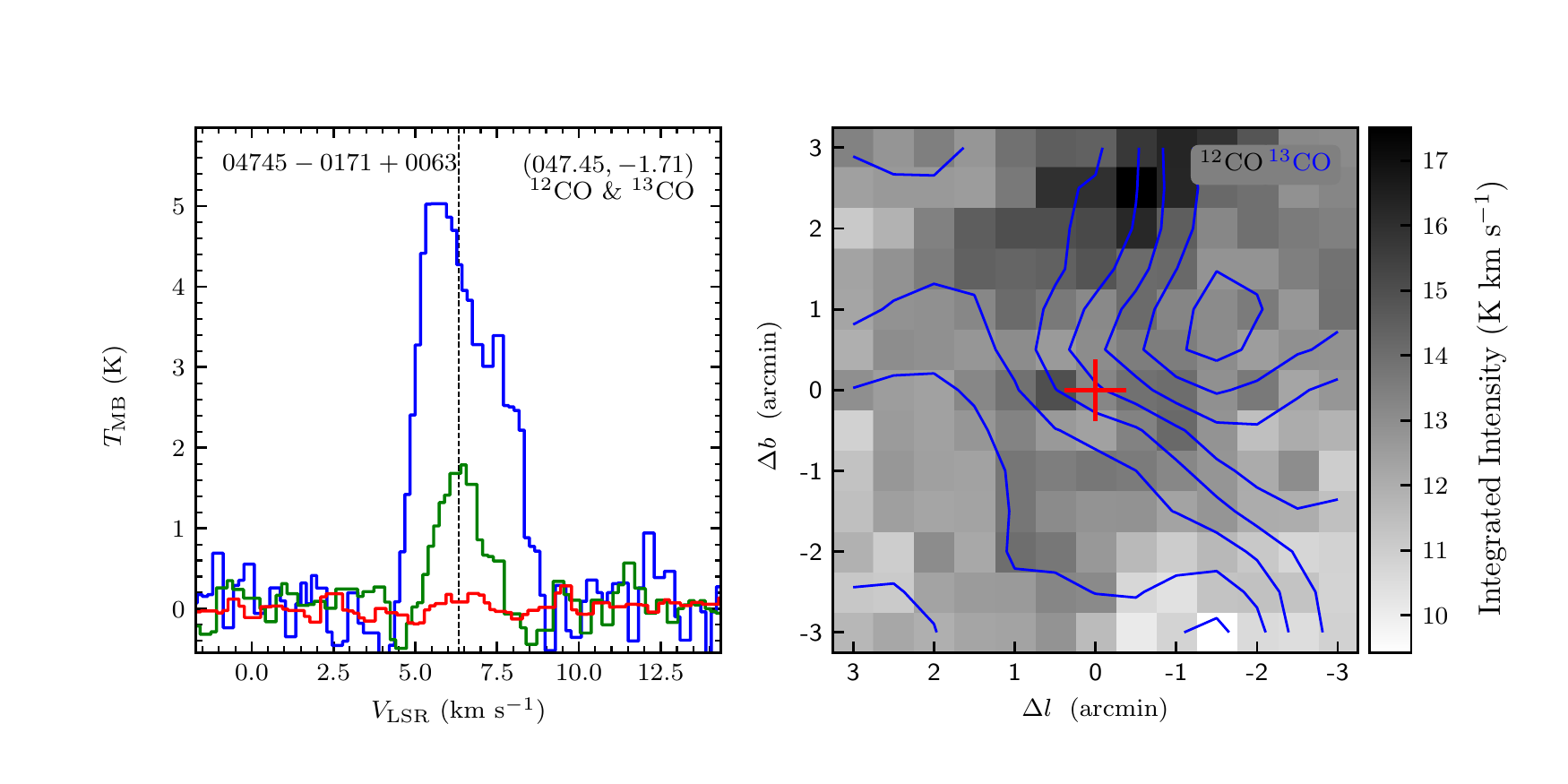}
\includegraphics[width=9.0cm,angle=0]{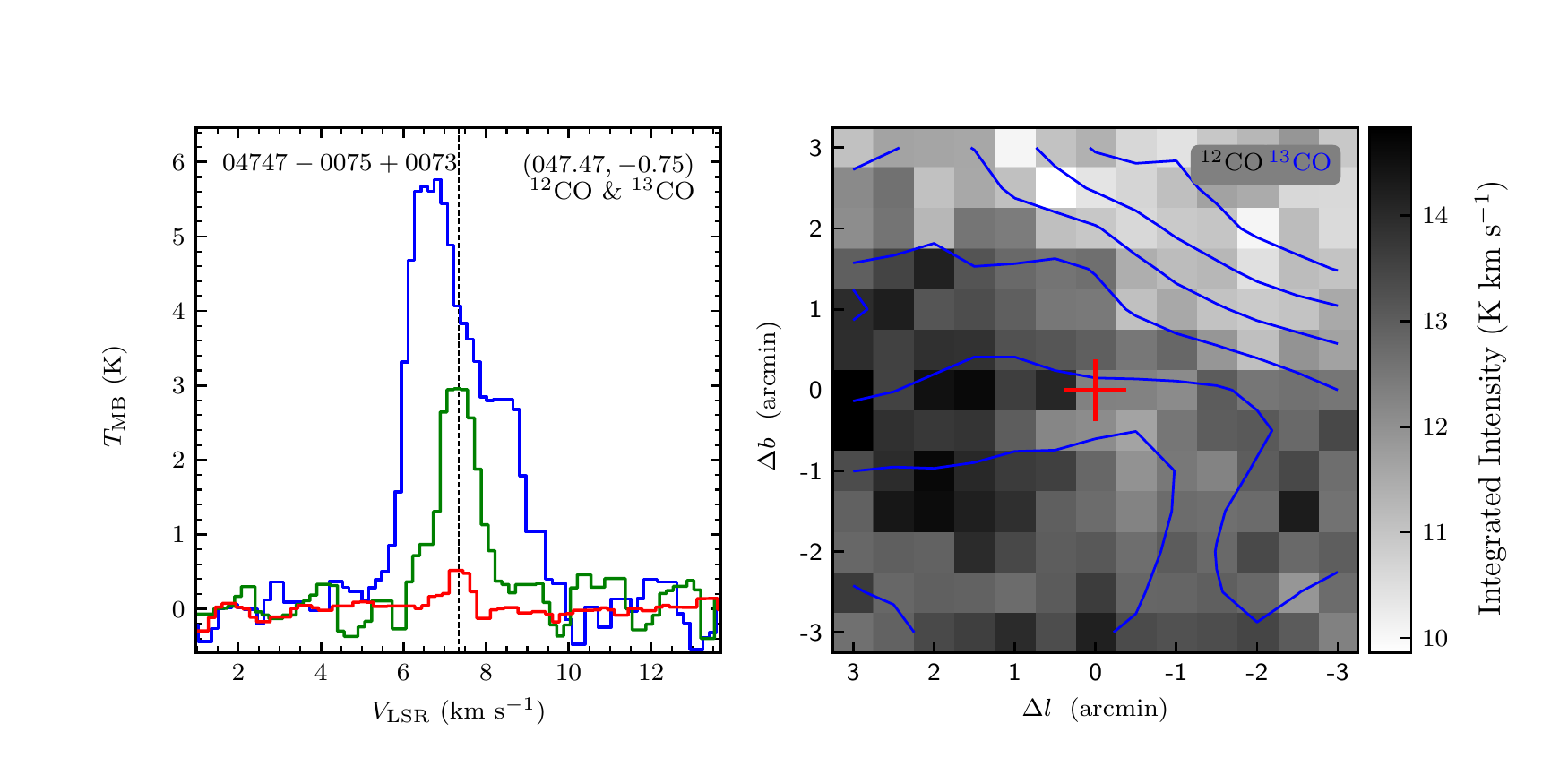}
\vspace{-0.5cm}

\includegraphics[width=9.0cm,angle=0]{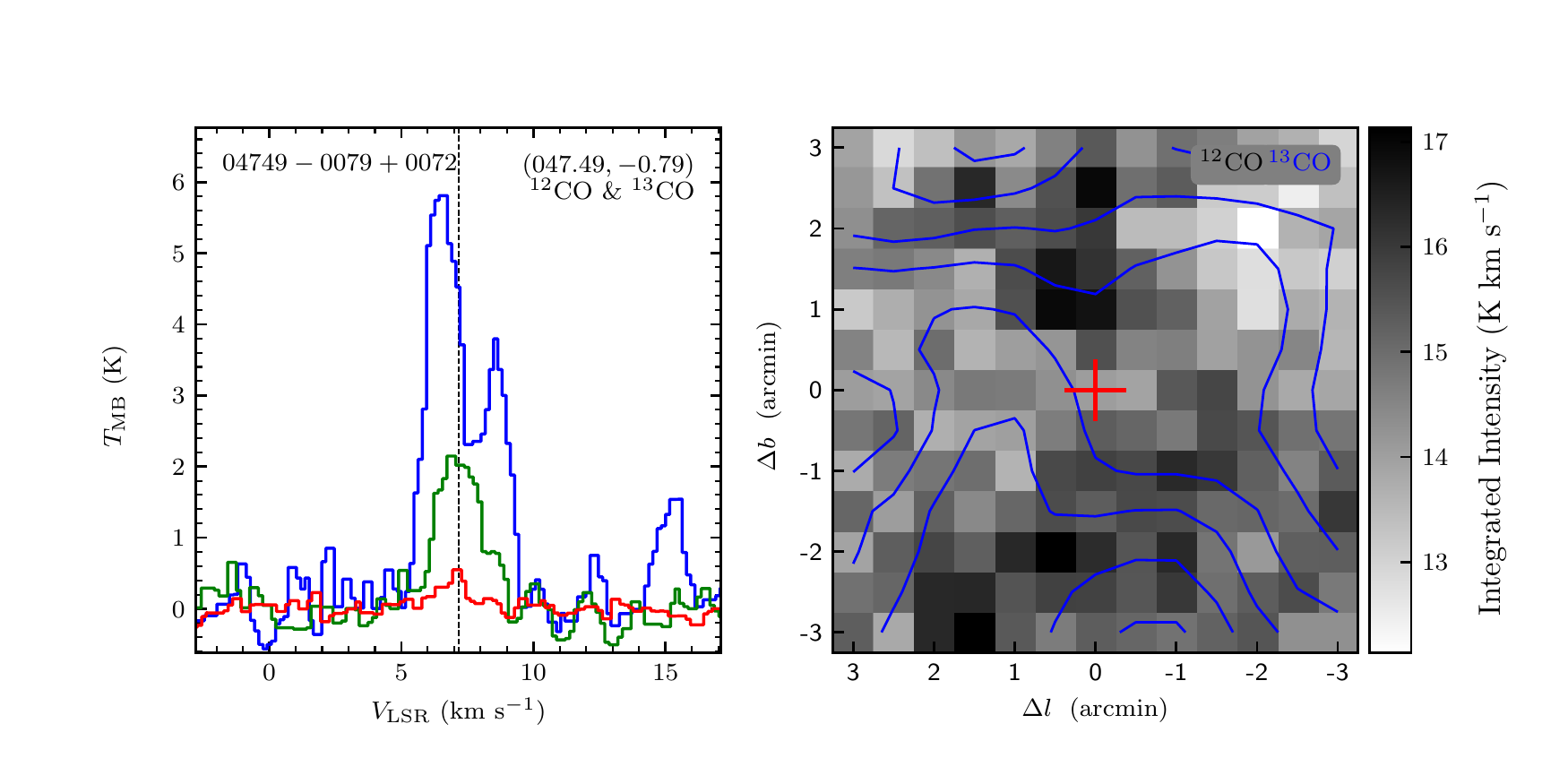}
\includegraphics[width=9.0cm,angle=0]{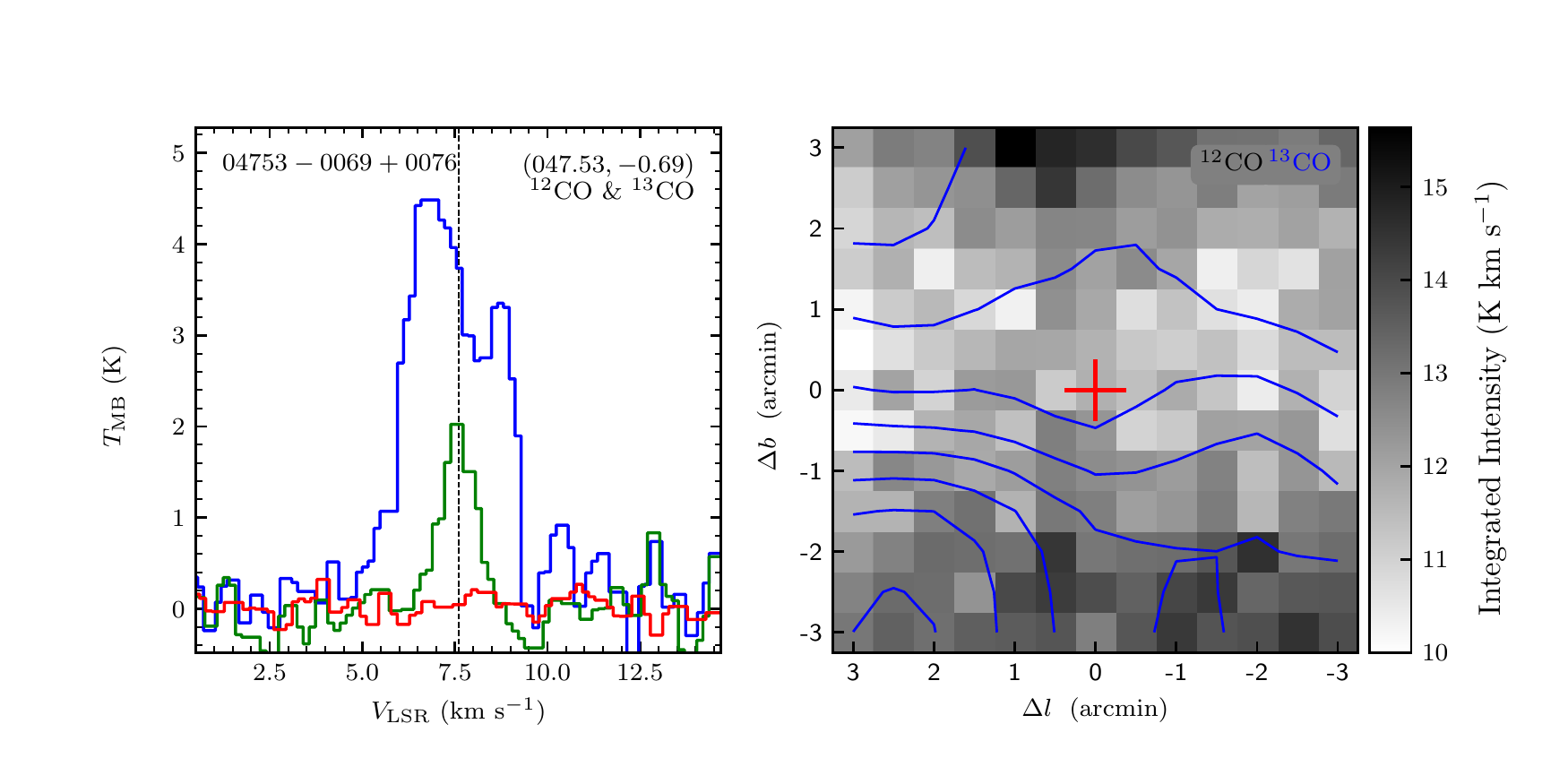}
\vspace{-0.5cm}

\includegraphics[width=9.0cm,angle=0]{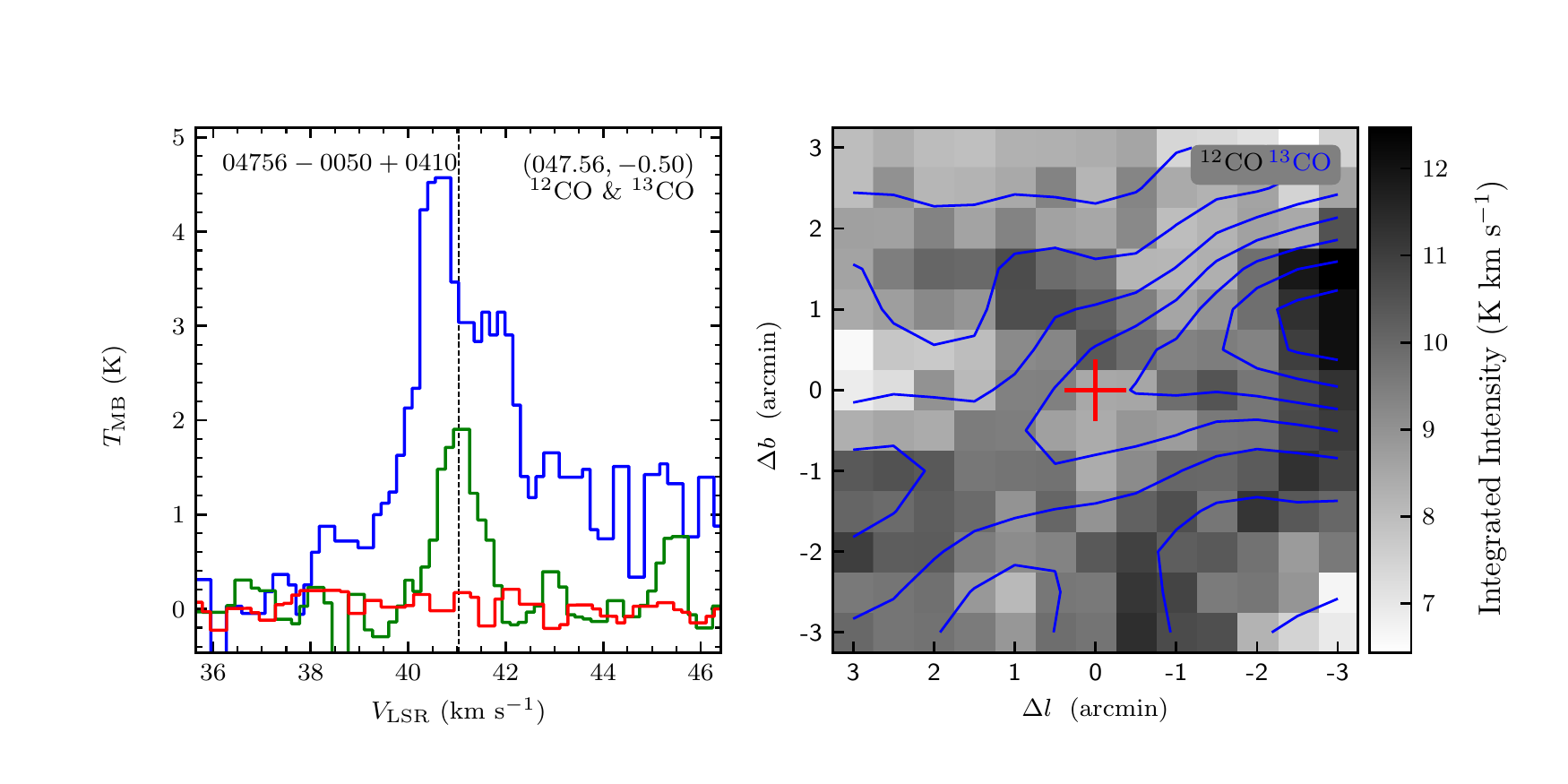}
\includegraphics[width=9.0cm,angle=0]{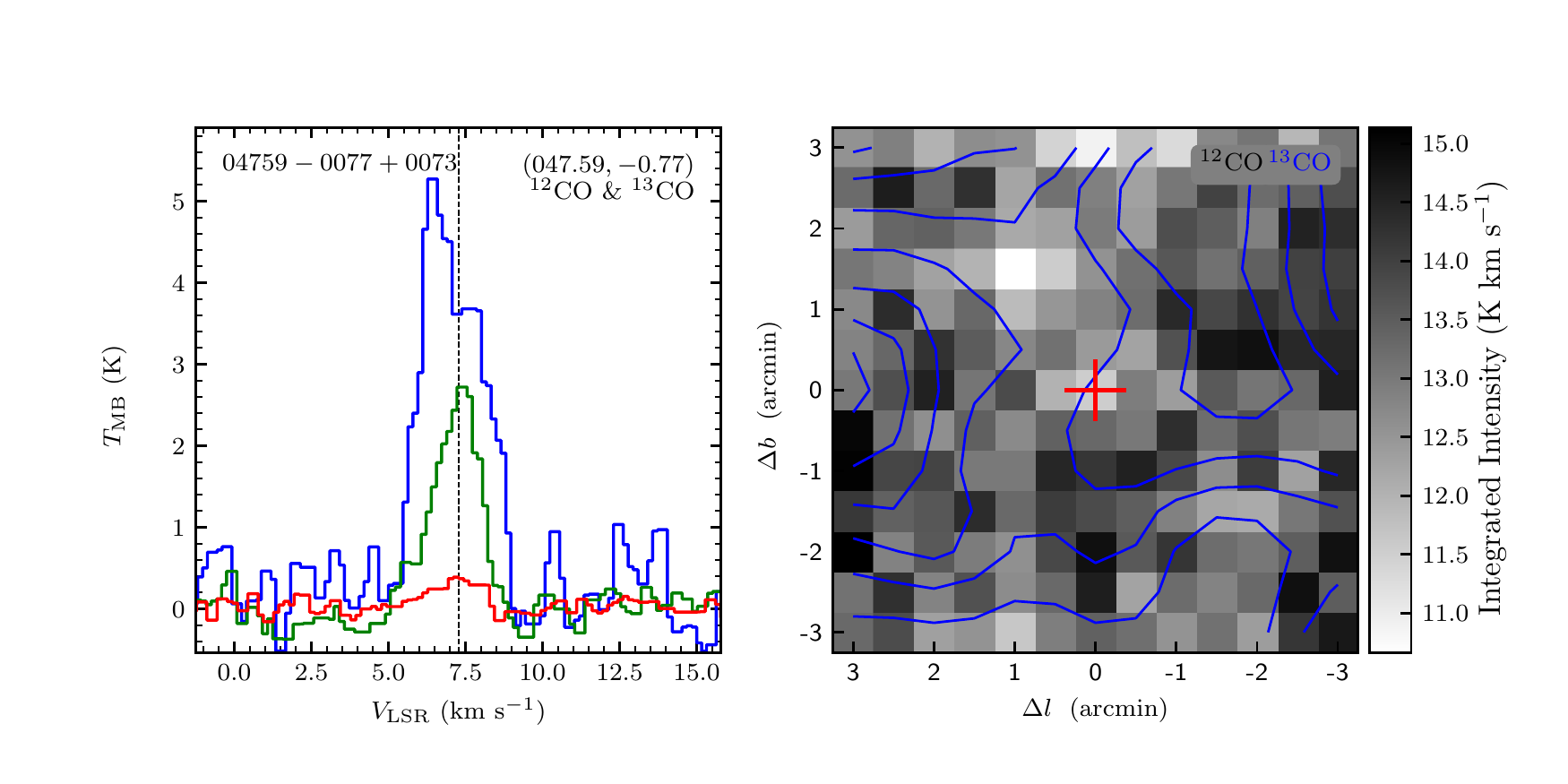}
\end{figure}
\clearpage

\begin{figure}
\includegraphics[width=9.0cm,angle=0]{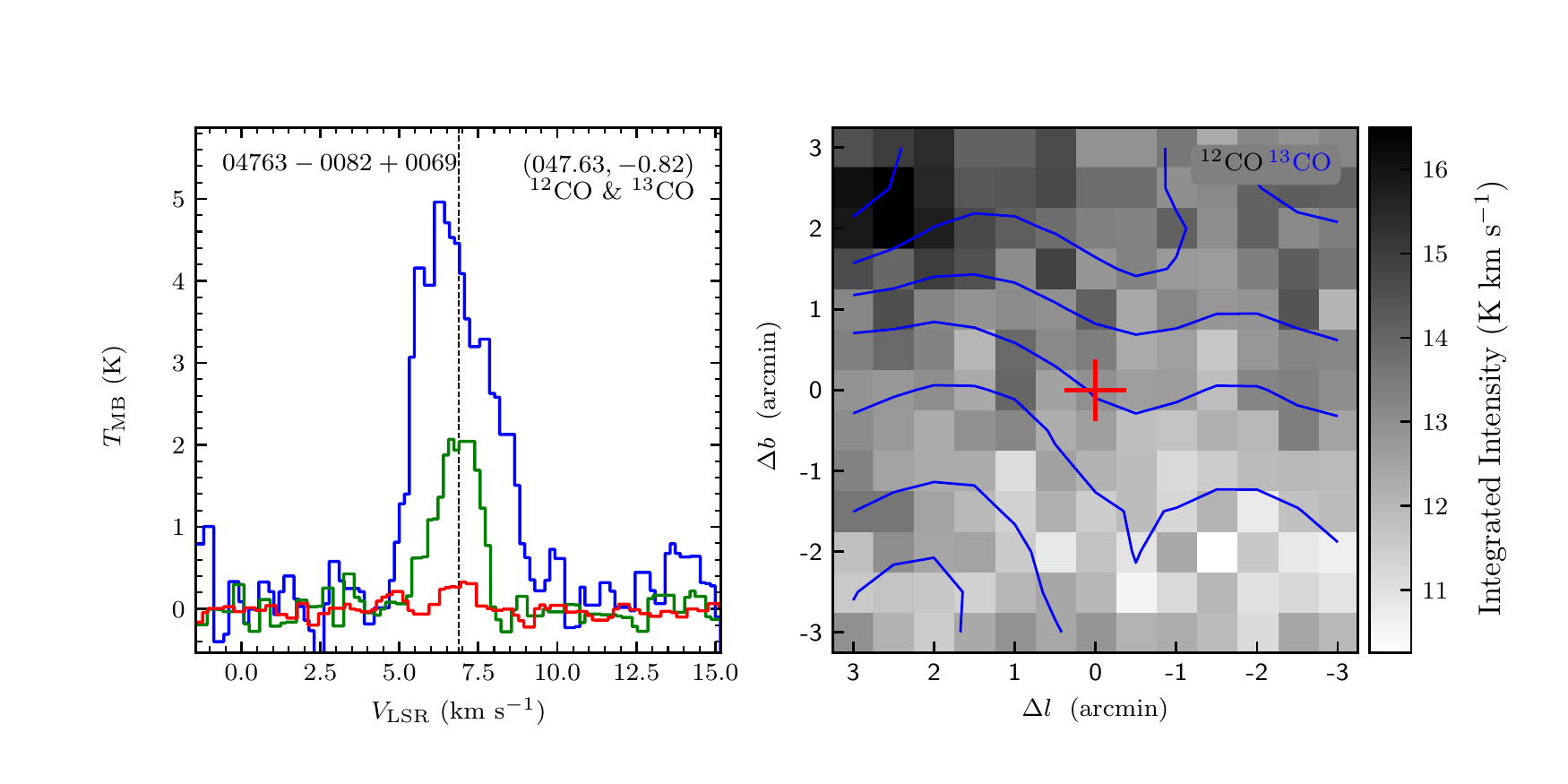}
\includegraphics[width=9.0cm,angle=0]{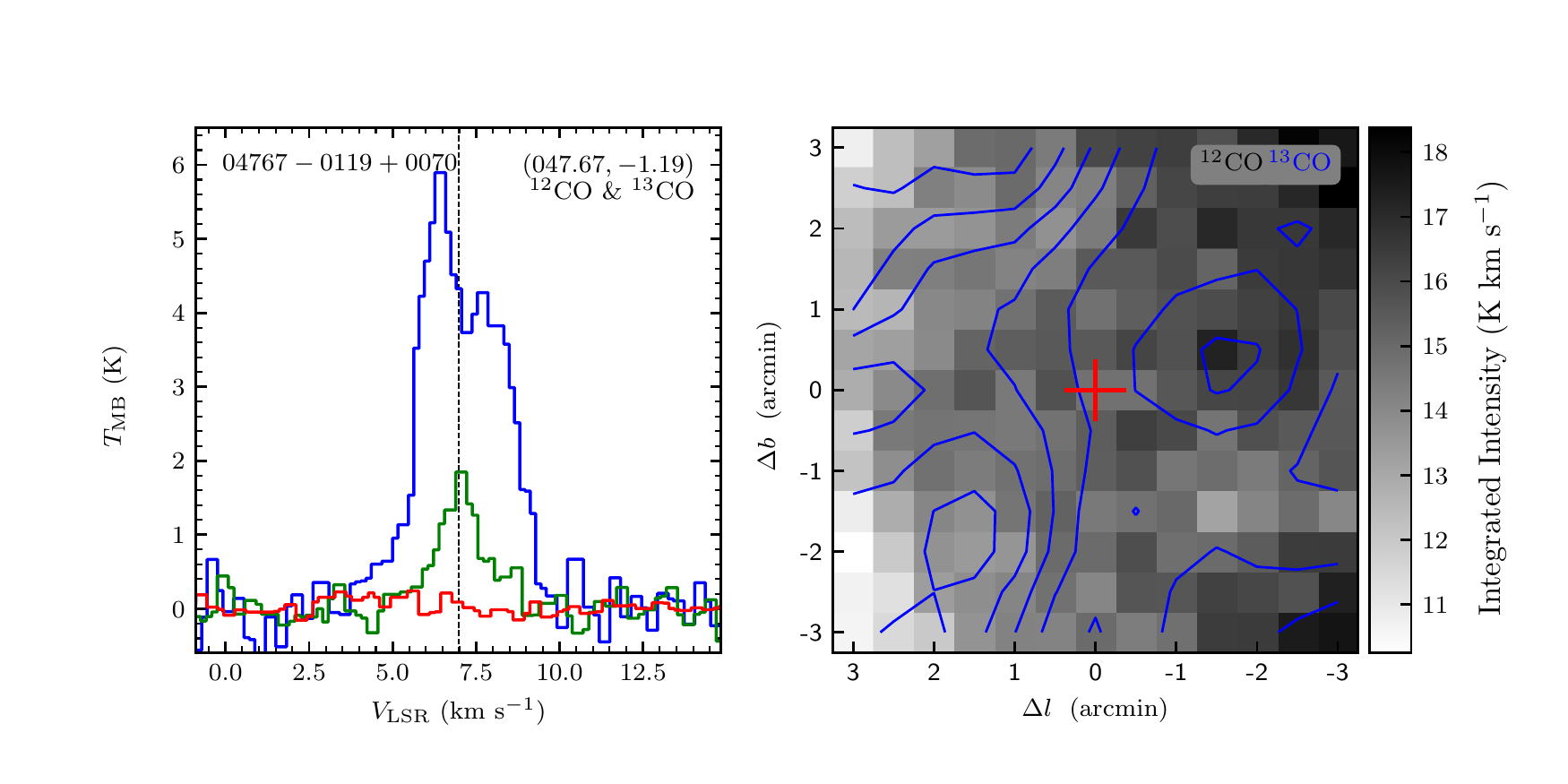}
\vspace{-0.5cm}

\includegraphics[width=9.0cm,angle=0]{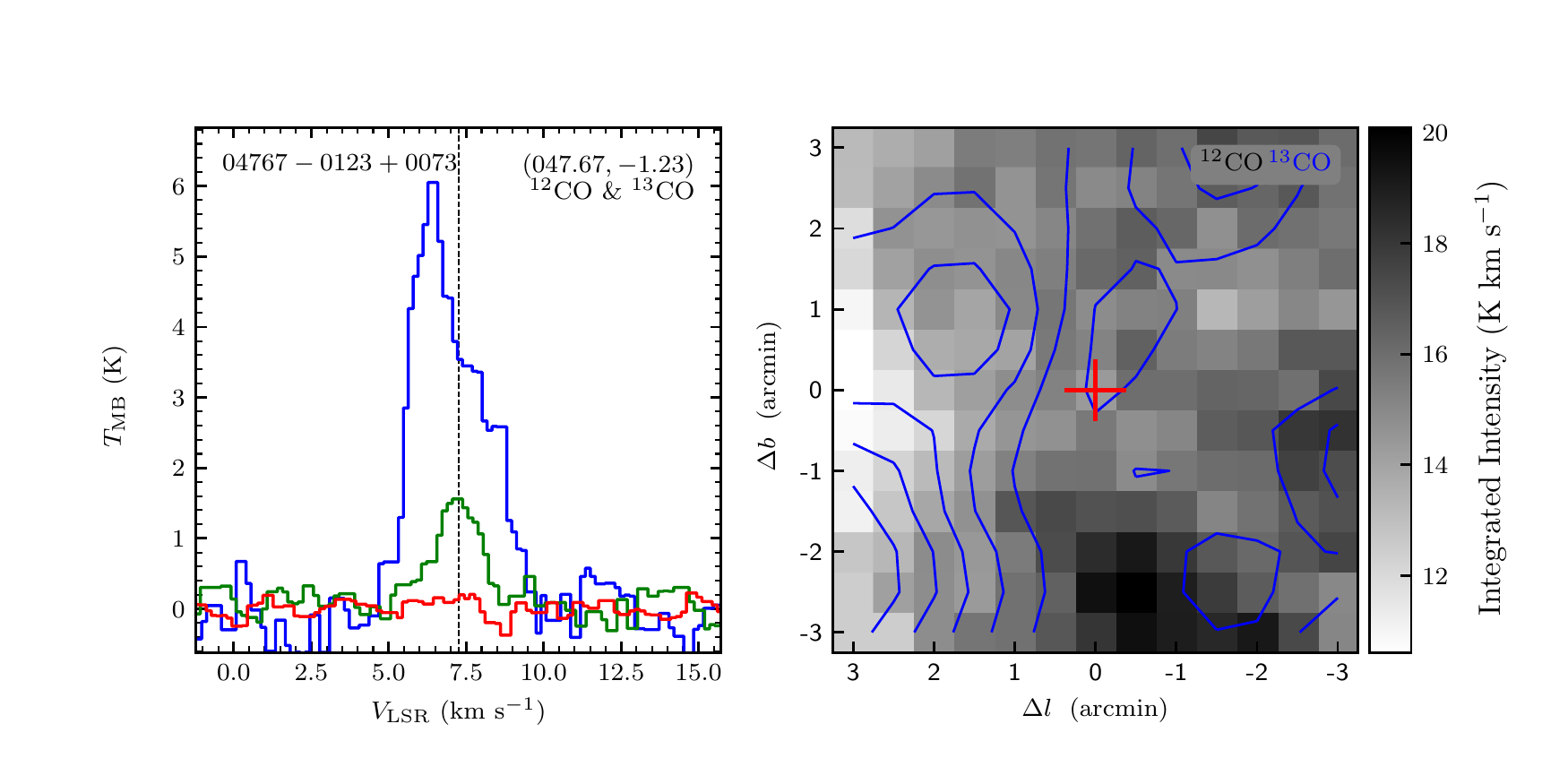}
\includegraphics[width=9.0cm,angle=0]{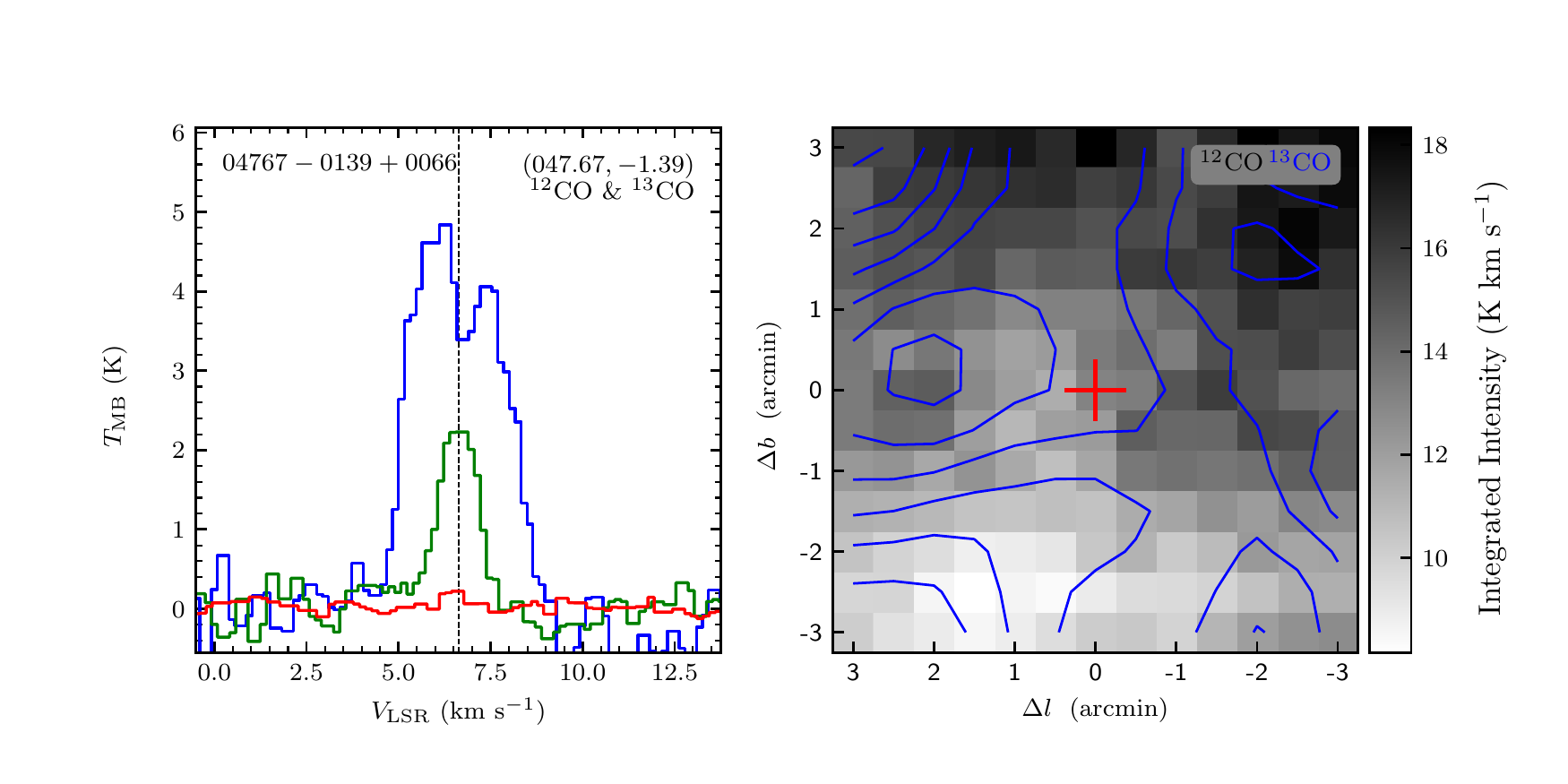}
\vspace{-0.5cm}

\includegraphics[width=9.0cm,angle=0]{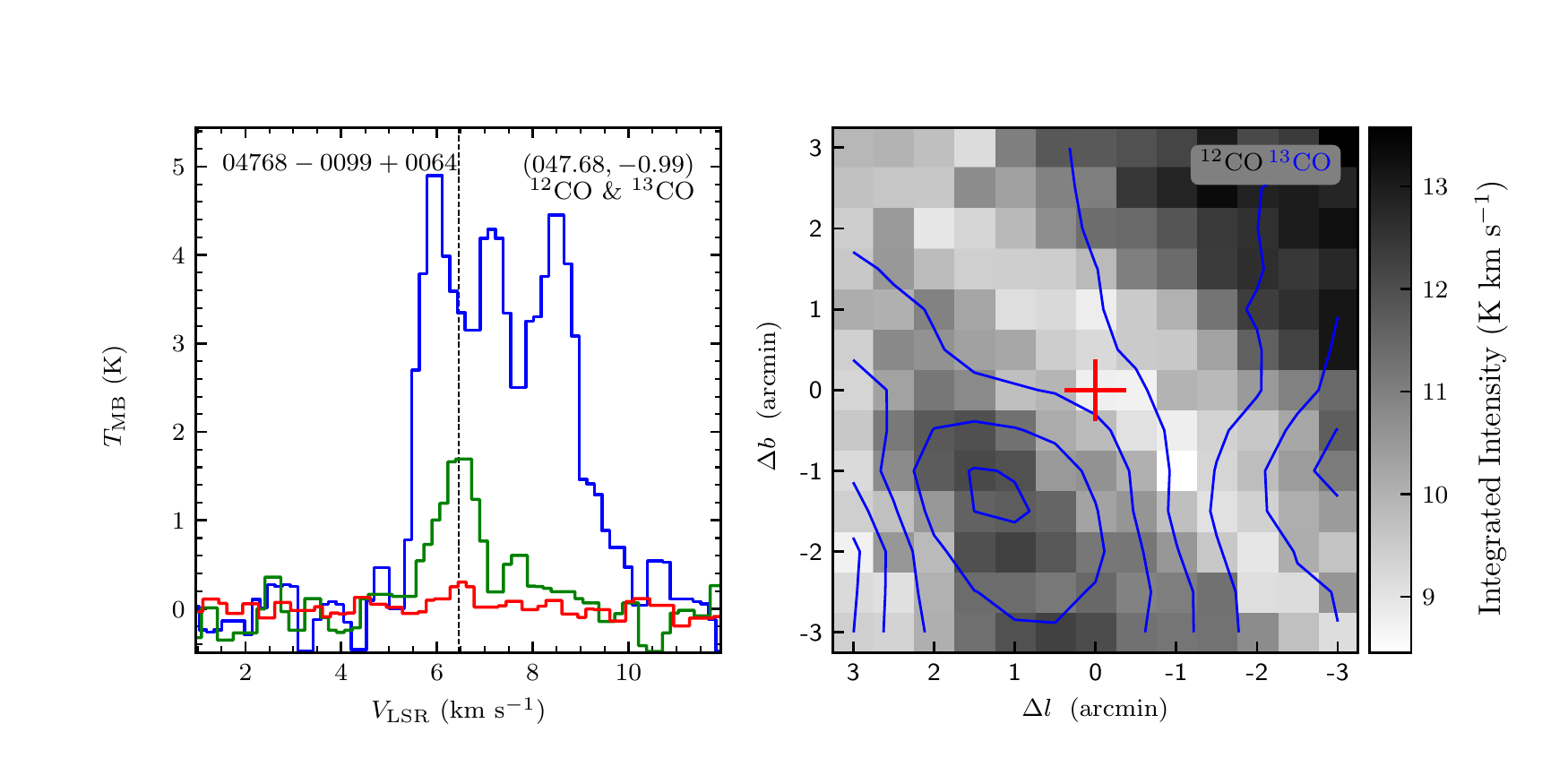}
\includegraphics[width=9.0cm,angle=0]{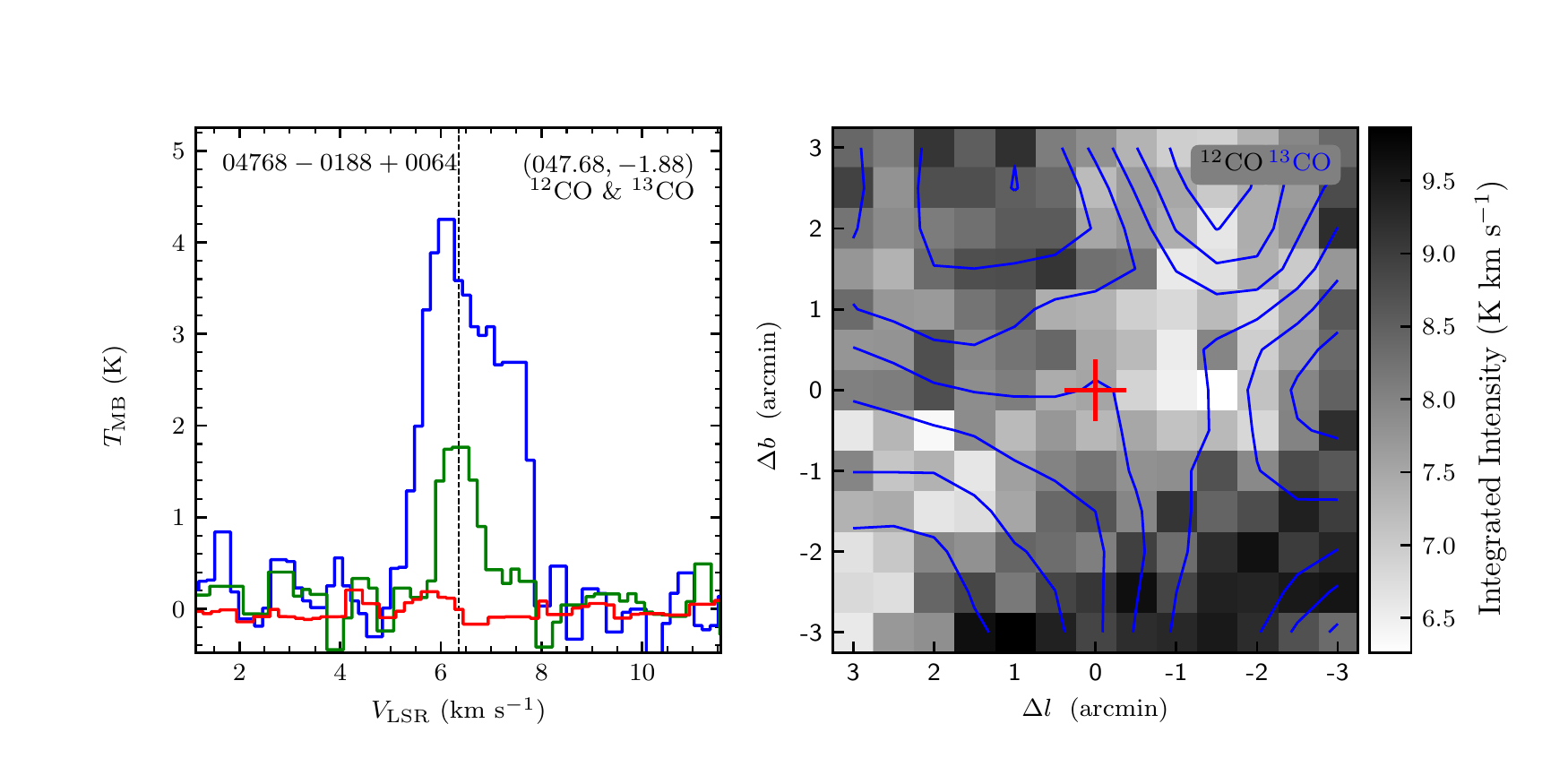}
\vspace{-0.5cm}

\includegraphics[width=9.0cm,angle=0]{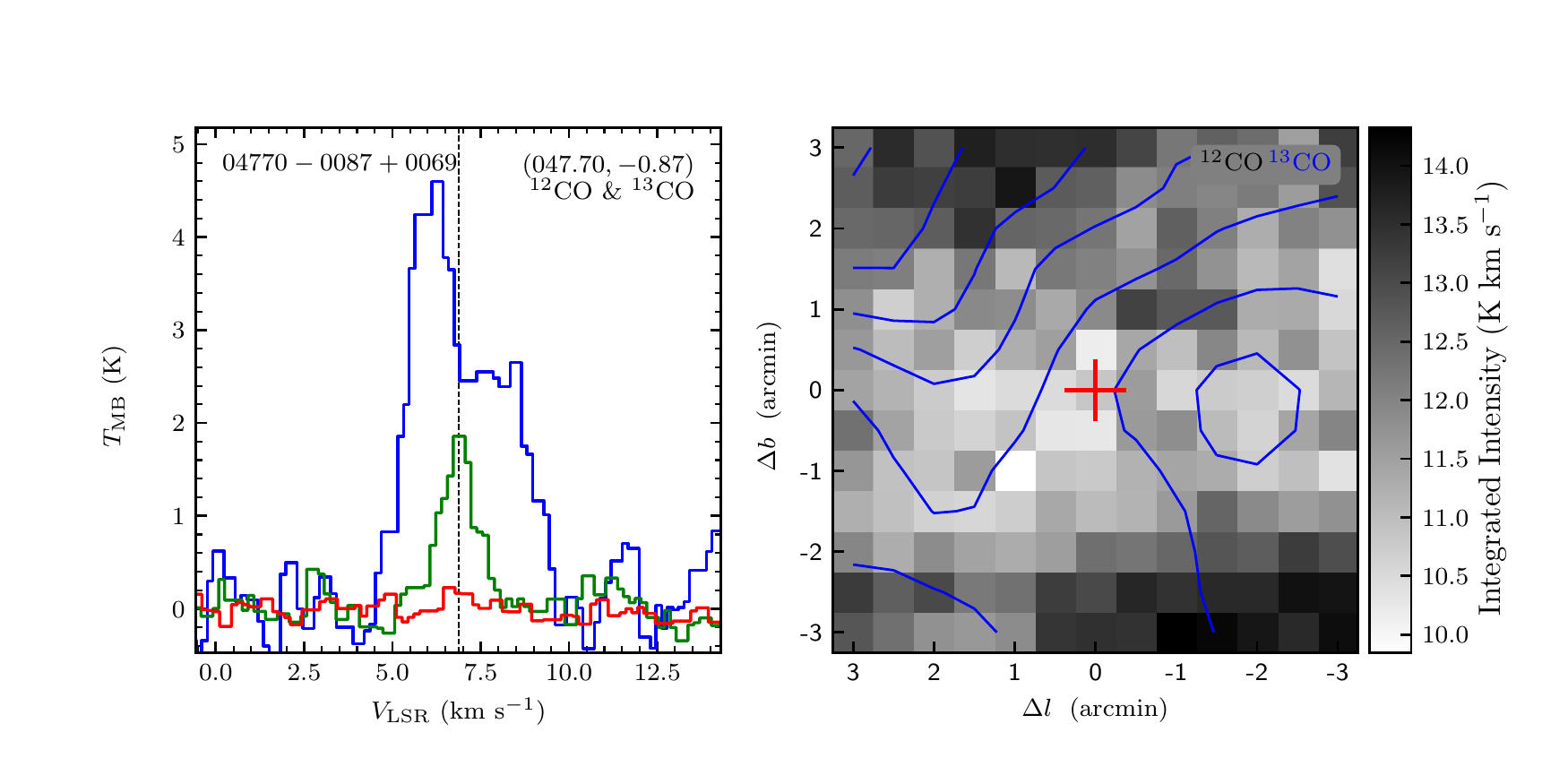}
\includegraphics[width=9.0cm,angle=0]{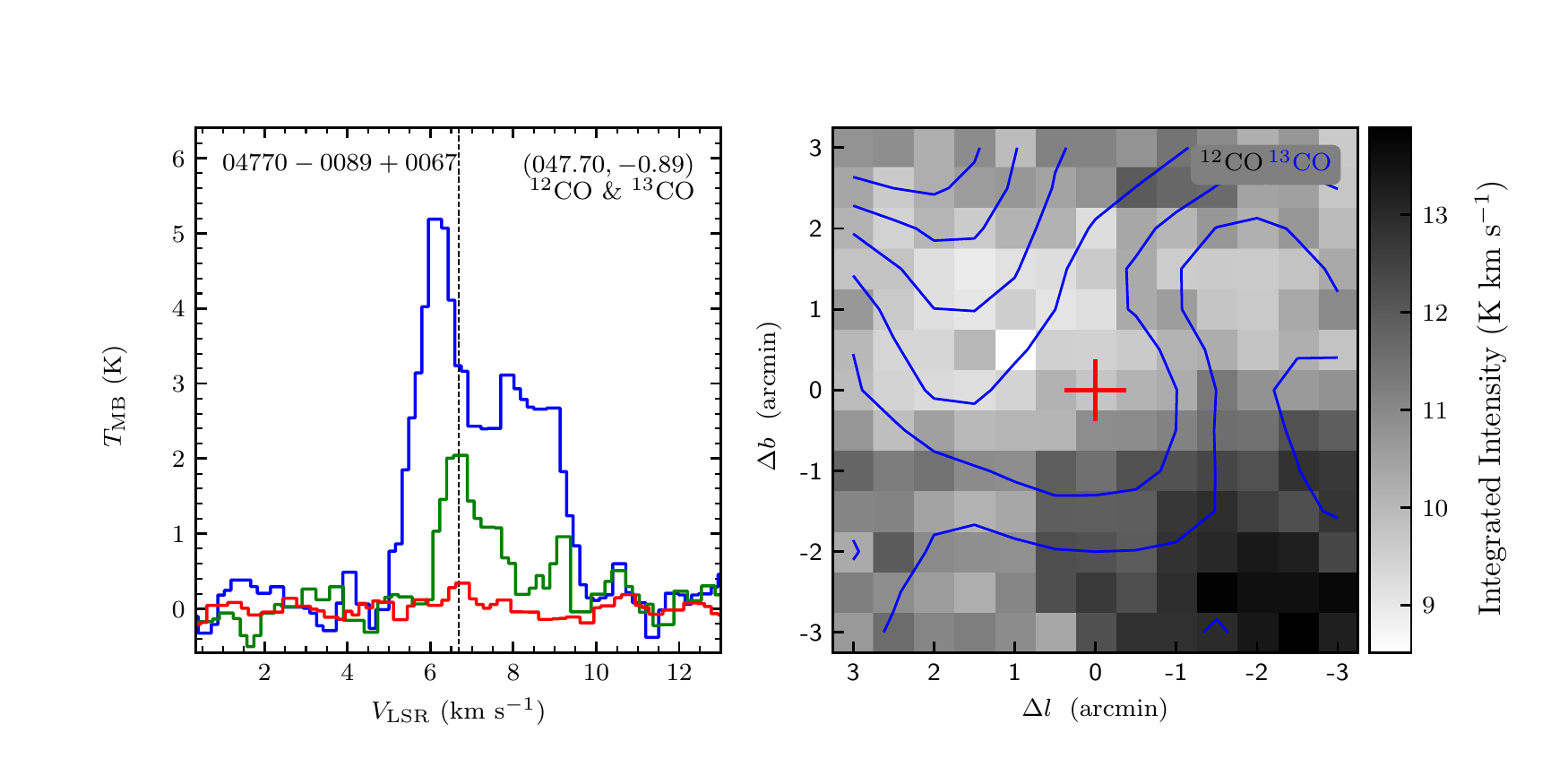}
\vspace{-0.5cm}

\includegraphics[width=9.0cm,angle=0]{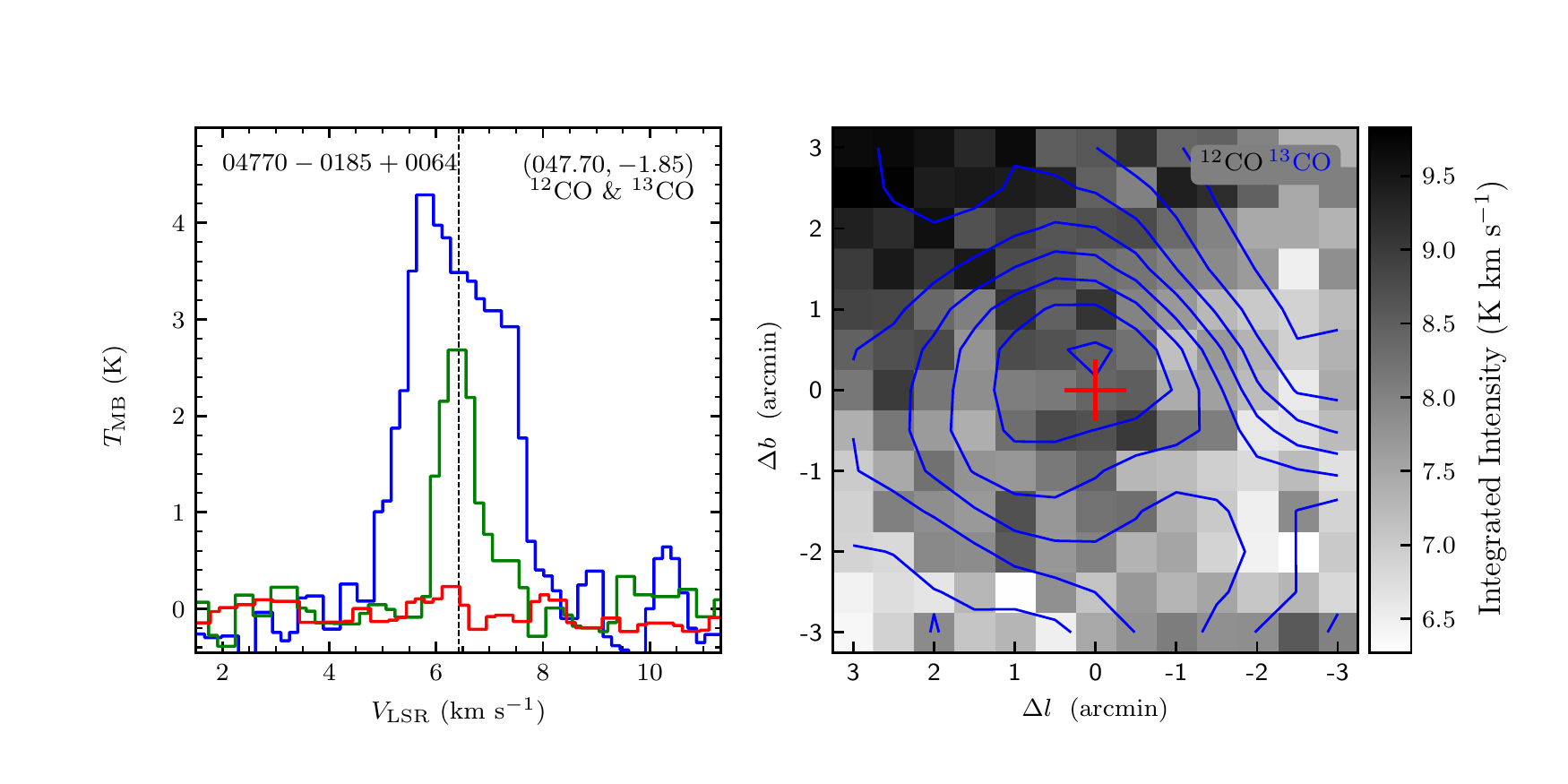}
\includegraphics[width=9.0cm,angle=0]{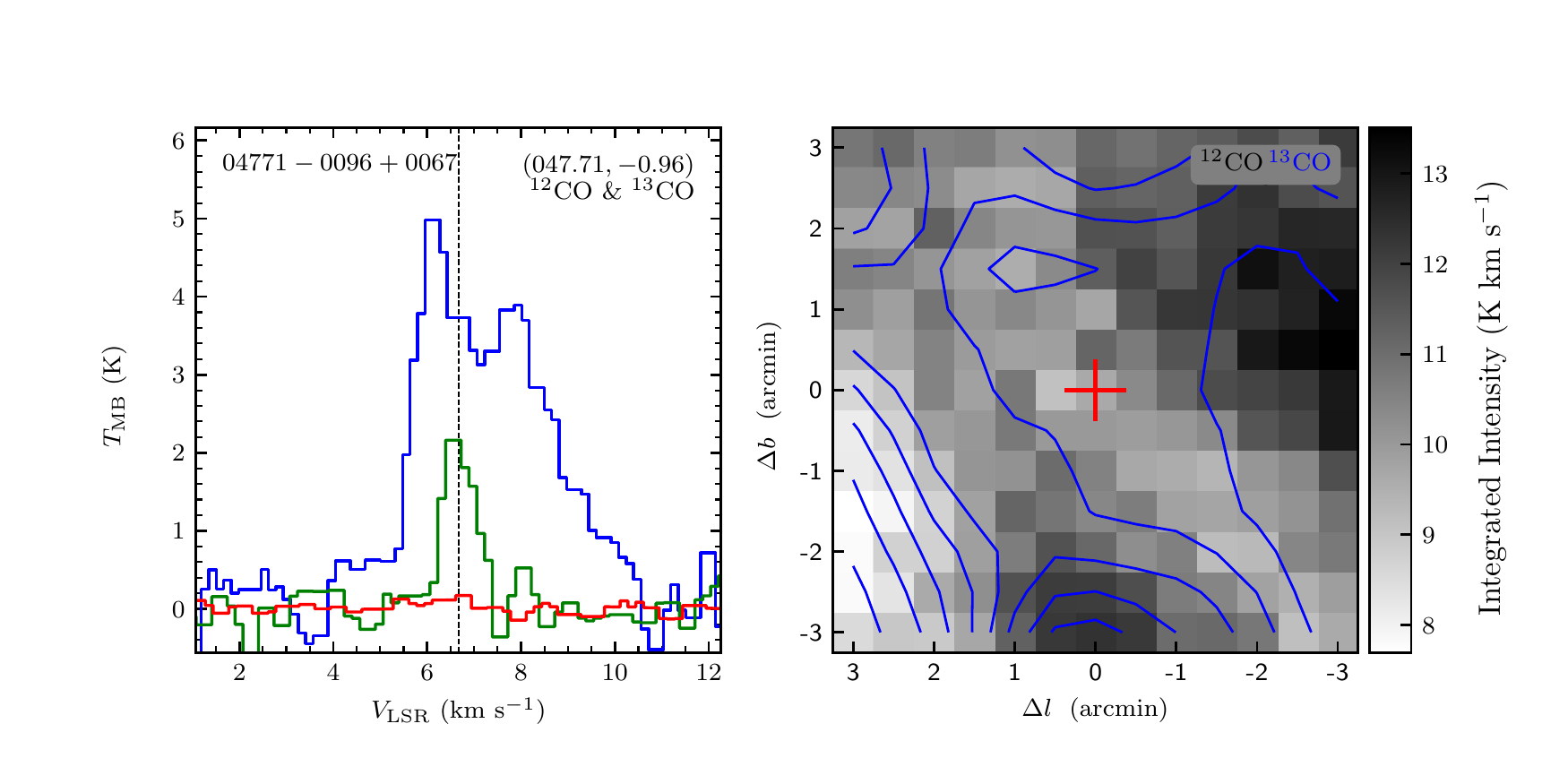}
\end{figure}
\clearpage

\begin{figure}
\includegraphics[width=9.0cm,angle=0]{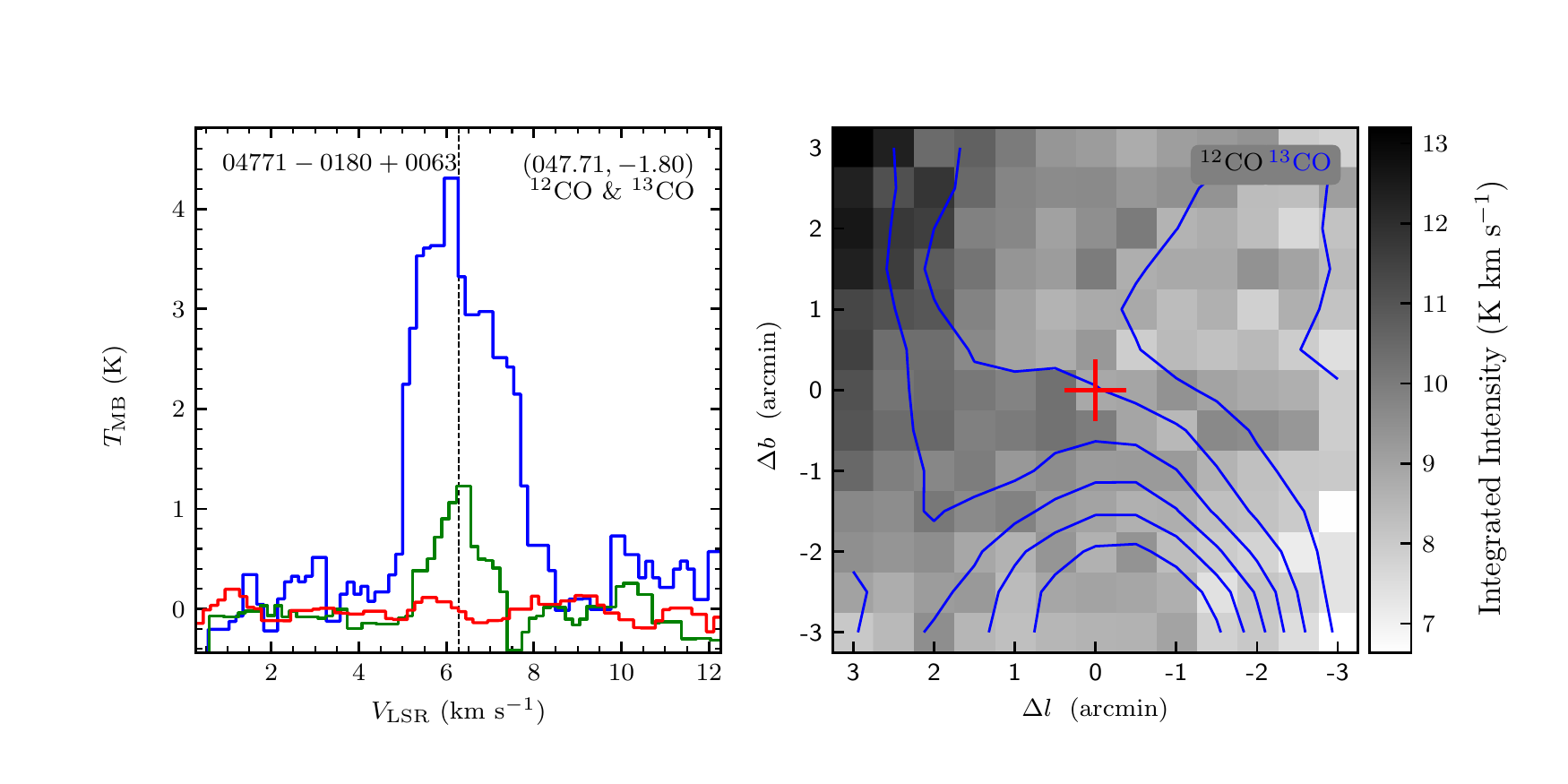}
\includegraphics[width=9.0cm,angle=0]{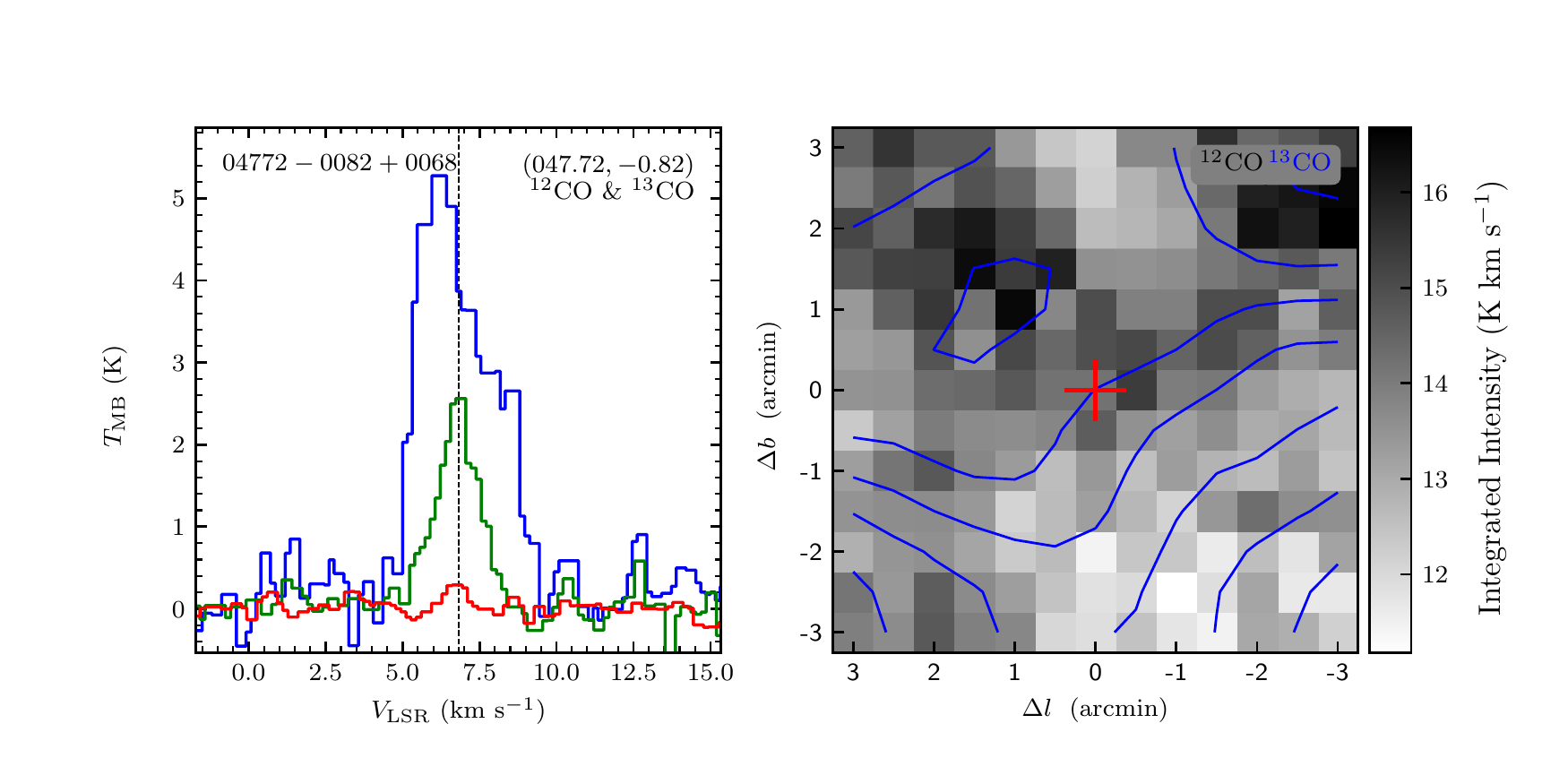}
\vspace{-0.5cm}

\includegraphics[width=9.0cm,angle=0]{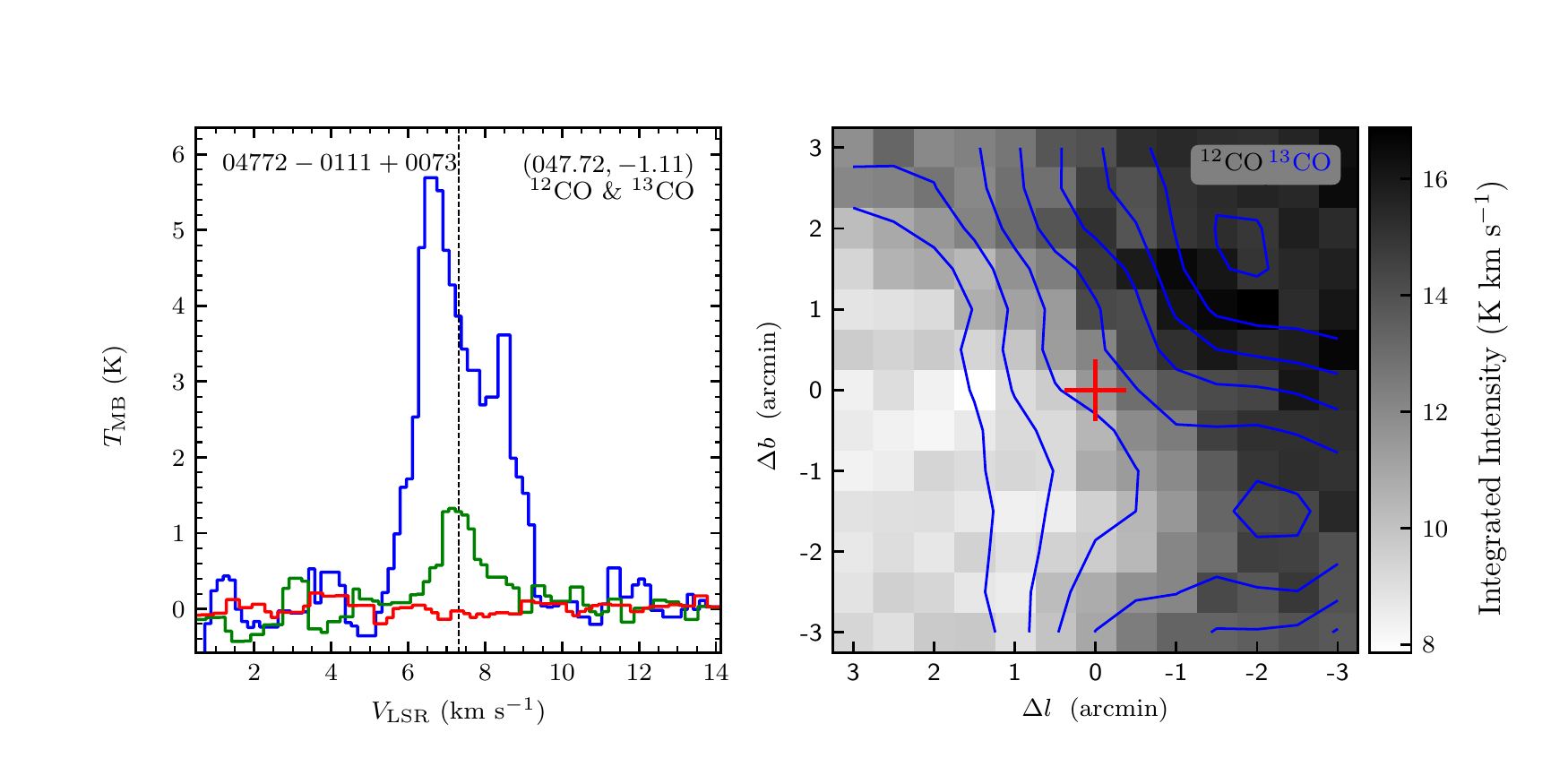}
\includegraphics[width=9.0cm,angle=0]{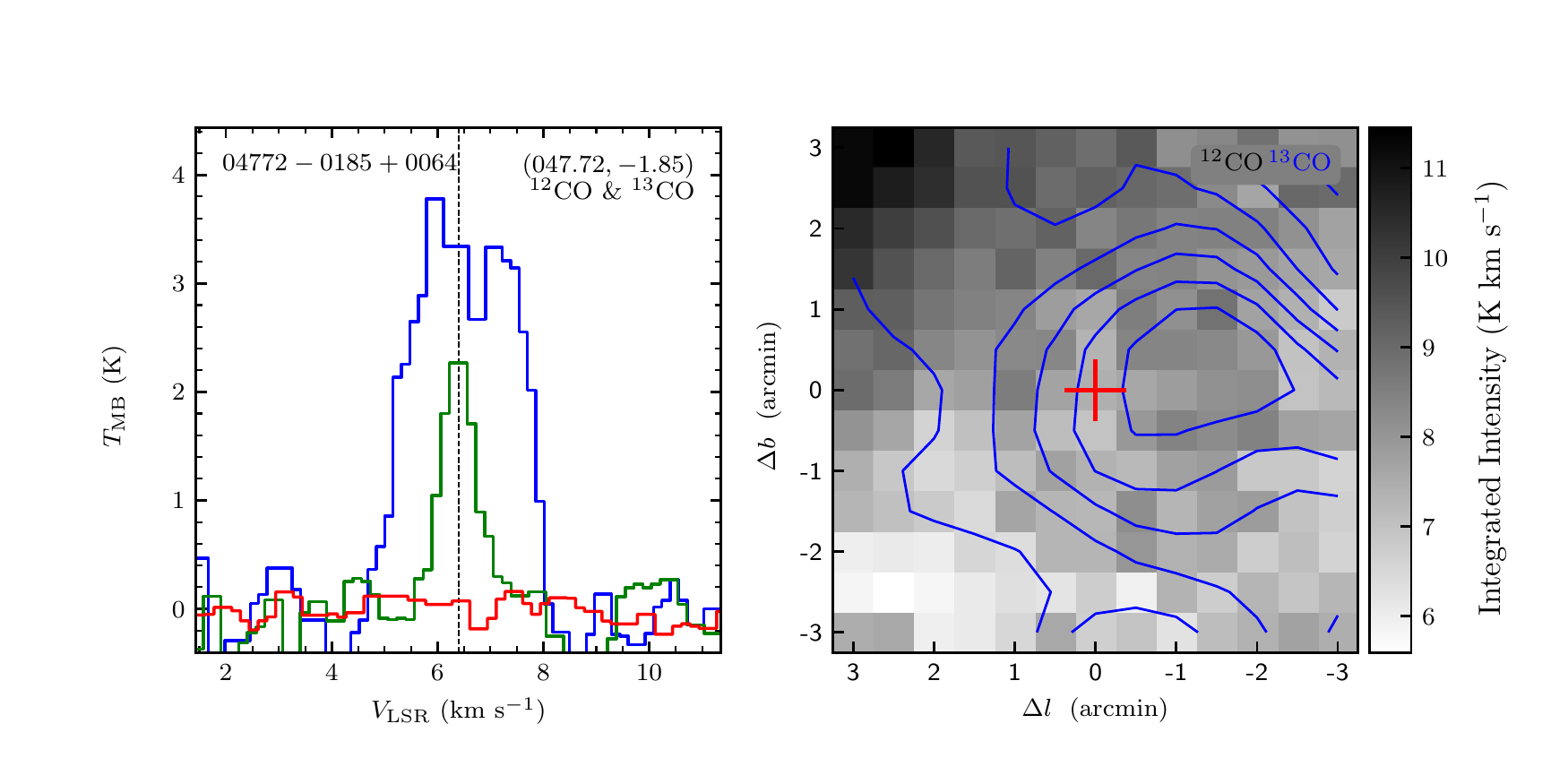}
\vspace{-0.5cm}

\includegraphics[width=9.0cm,angle=0]{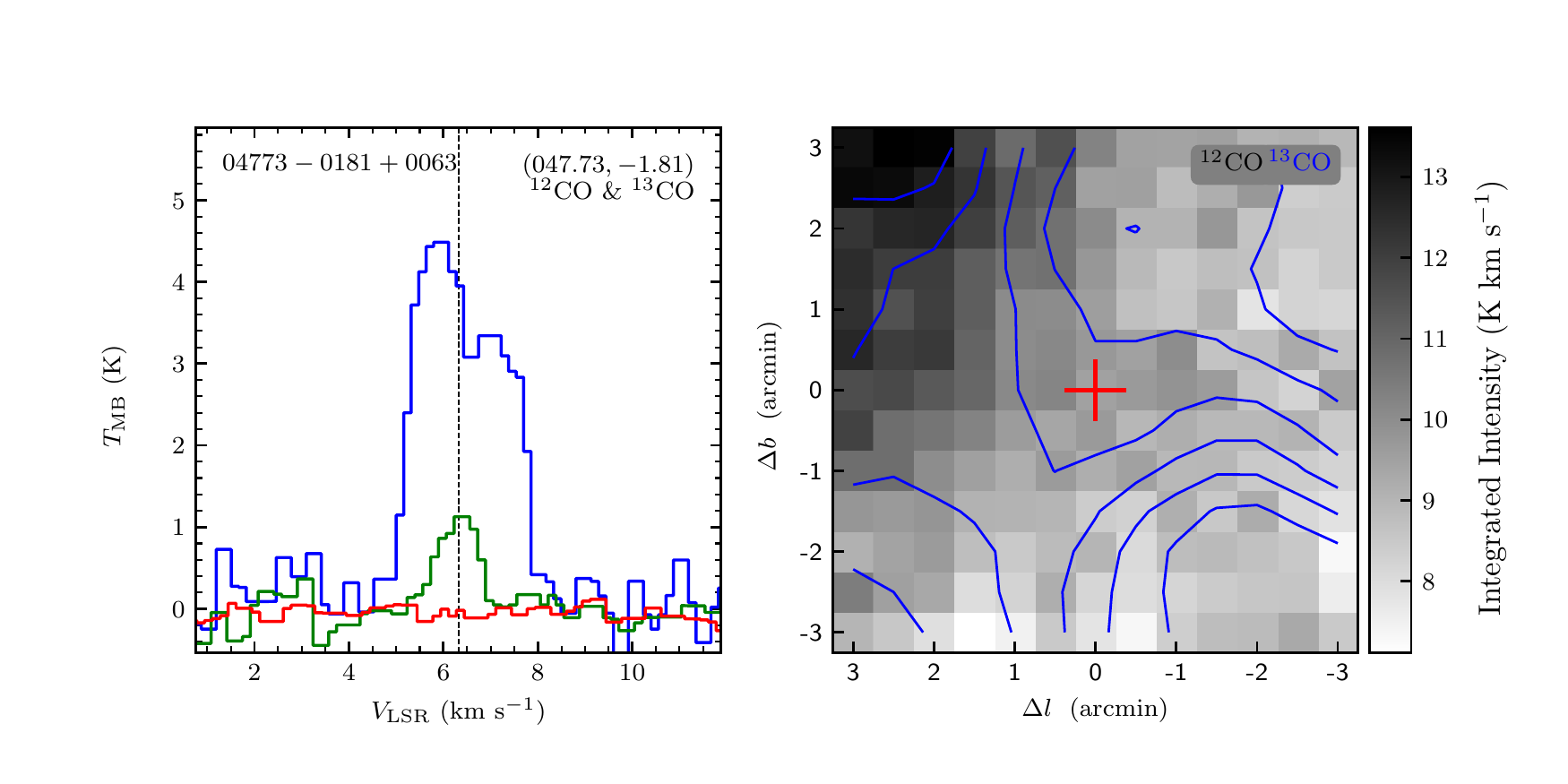}
\includegraphics[width=9.0cm,angle=0]{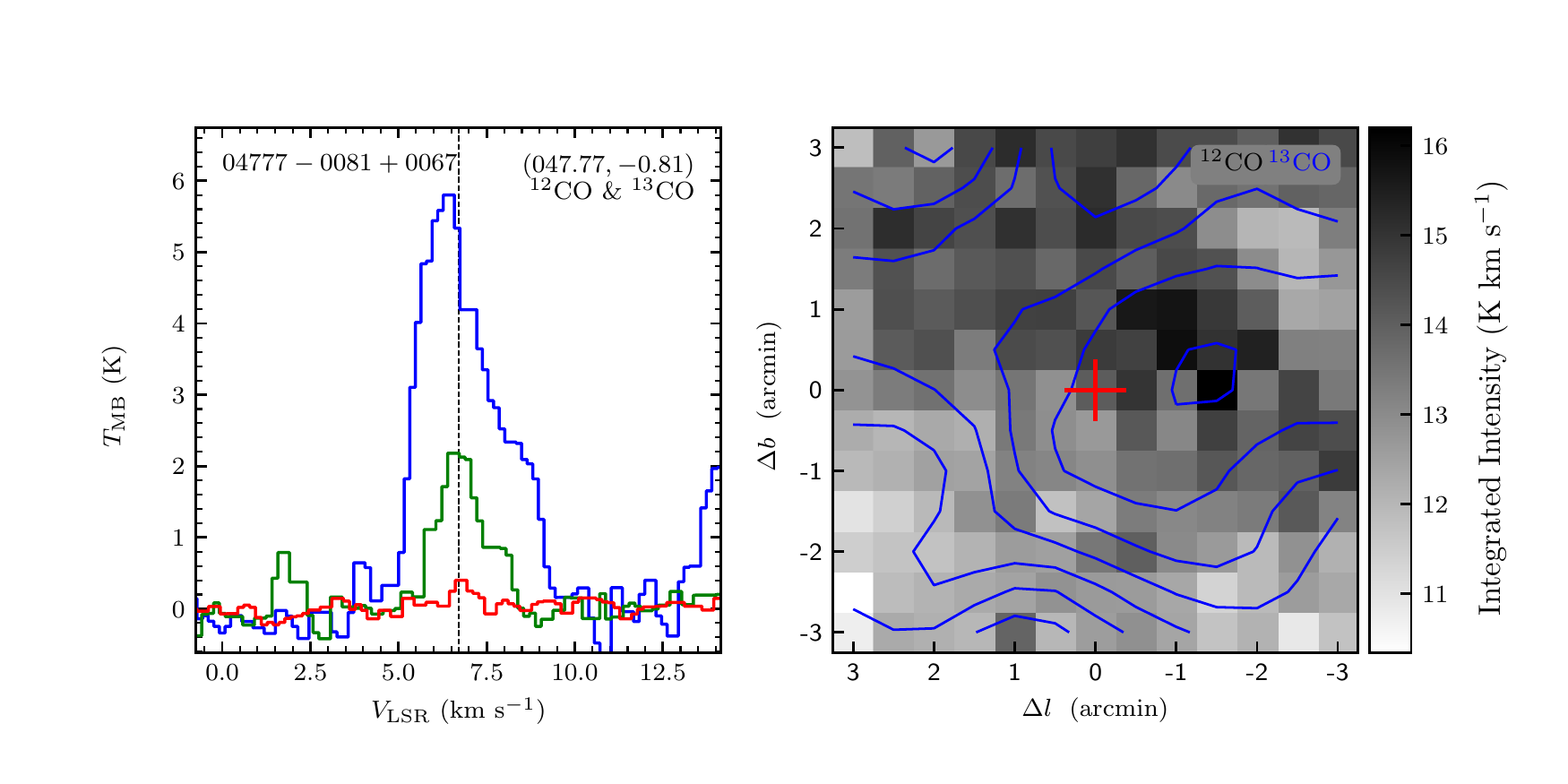}
\vspace{-0.5cm}

\includegraphics[width=9.0cm,angle=0]{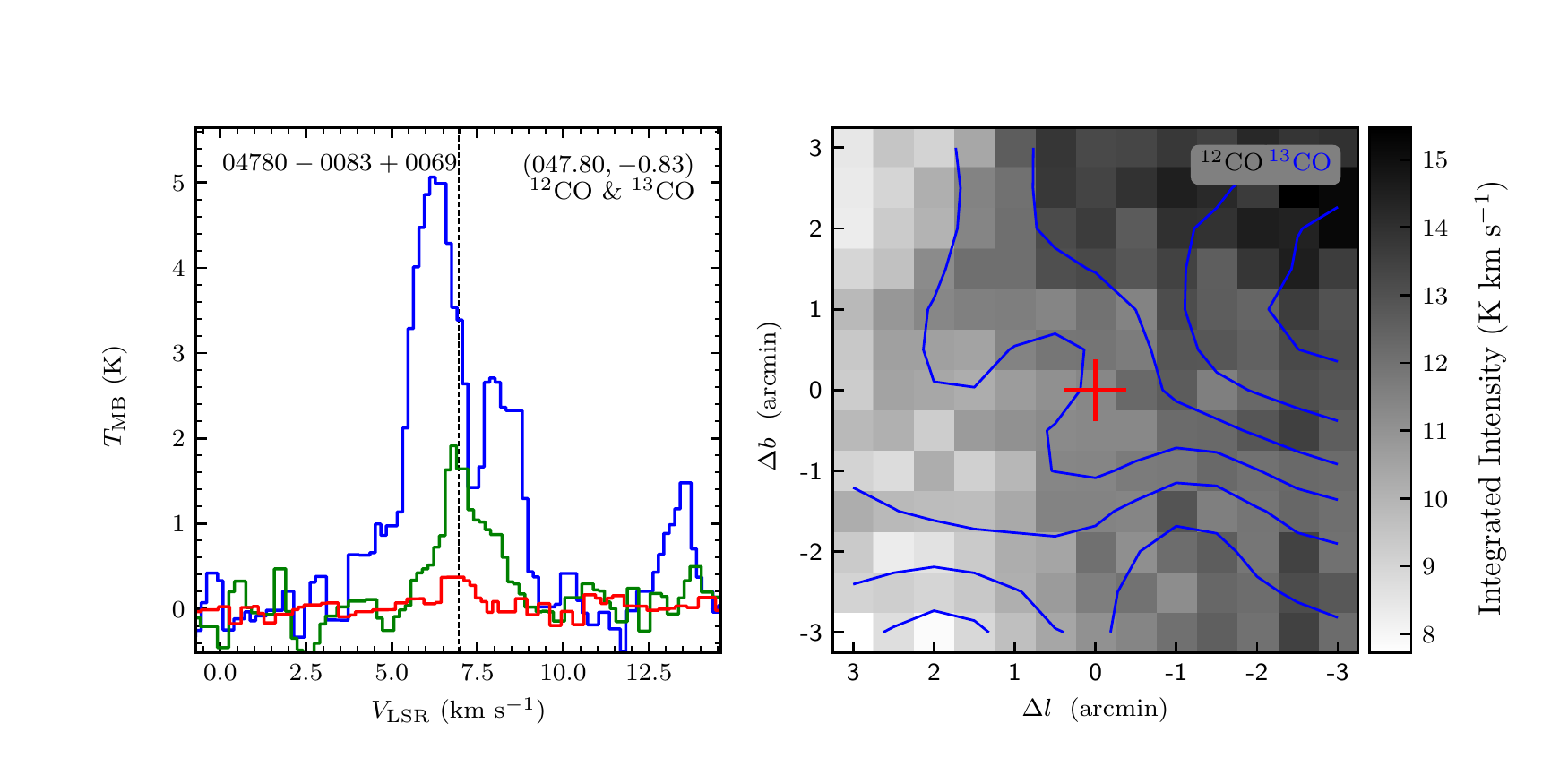}
\includegraphics[width=9.0cm,angle=0]{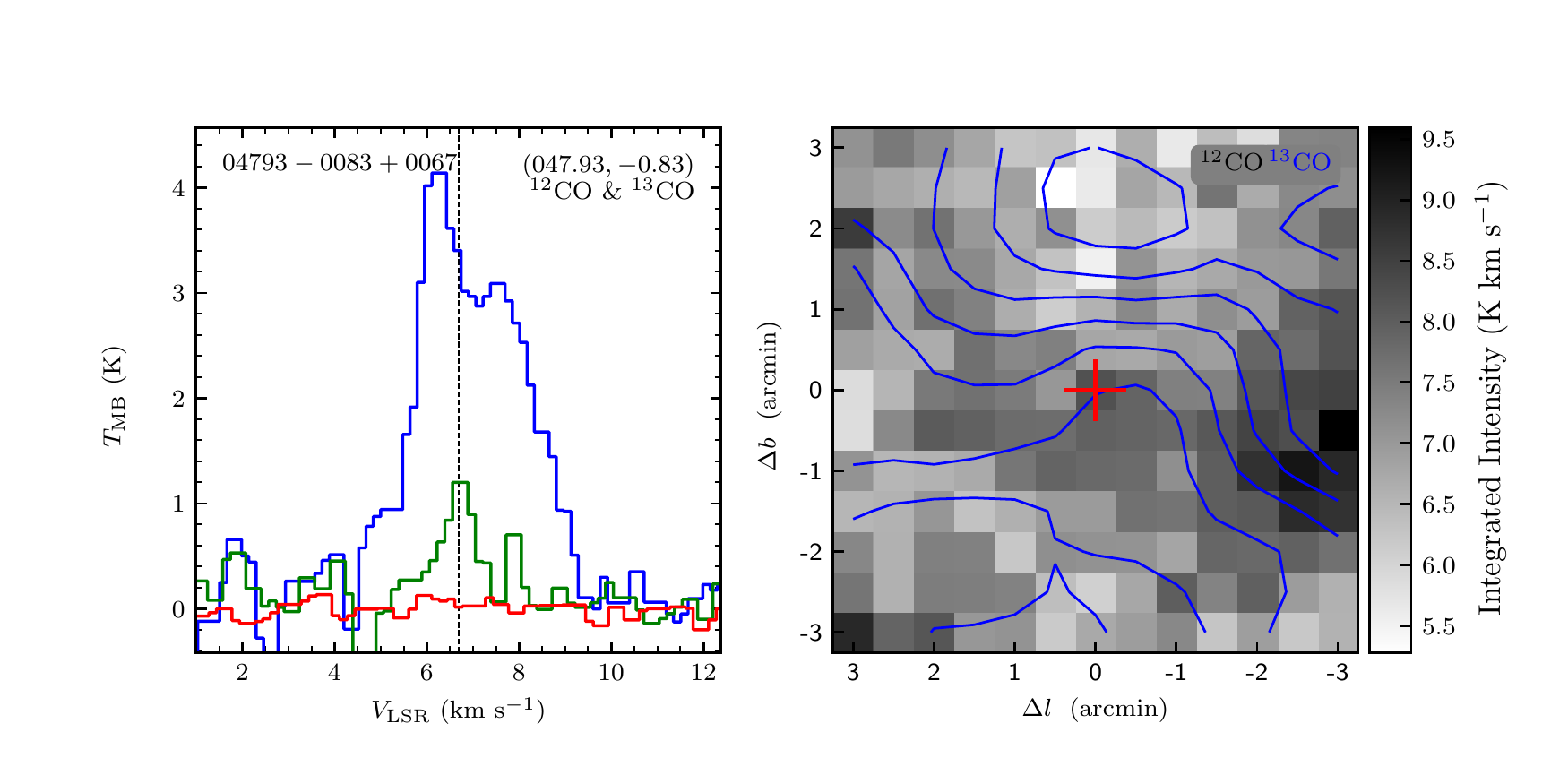}
\vspace{-0.5cm}

\includegraphics[width=9.0cm,angle=0]{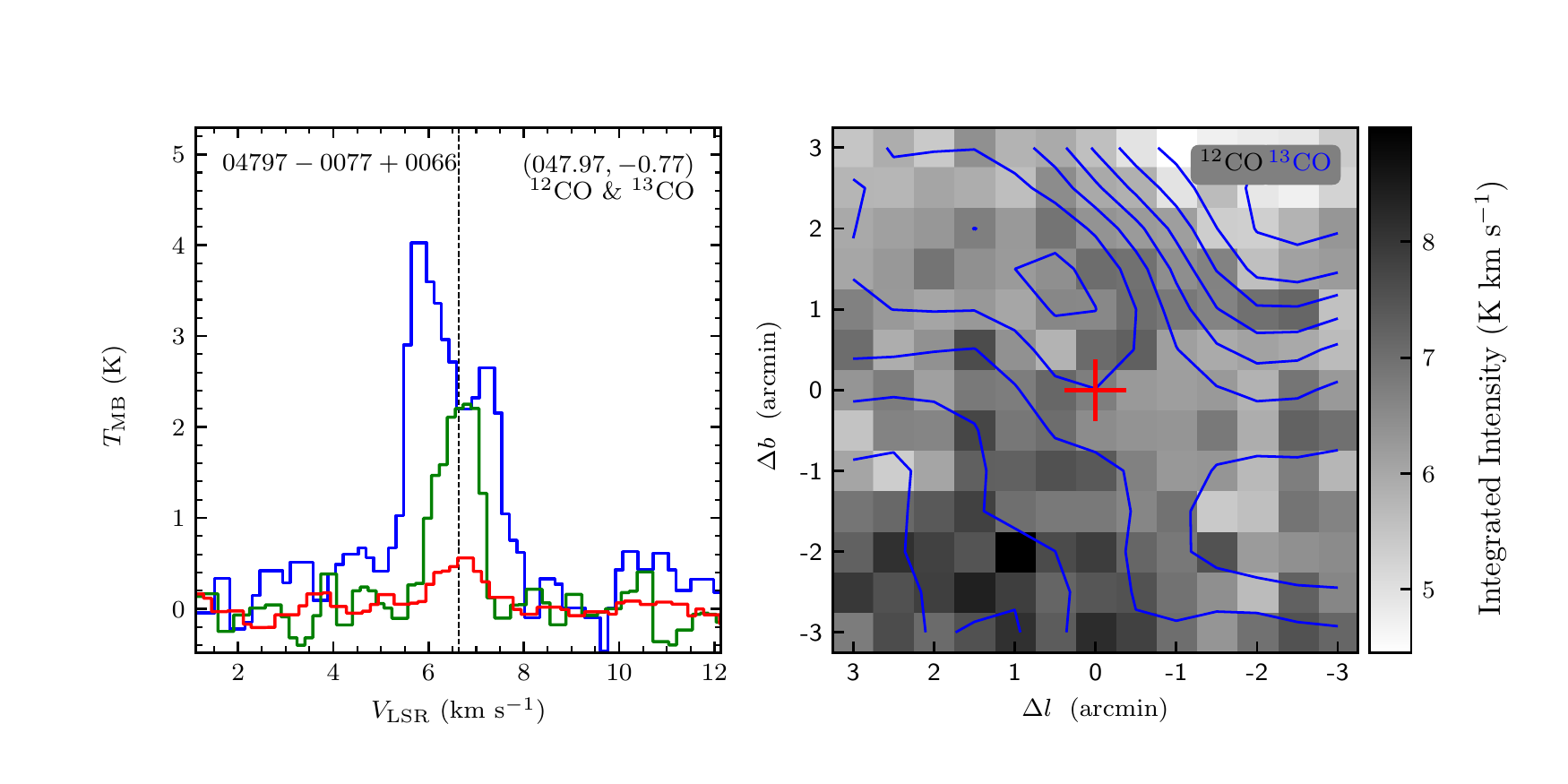}
\includegraphics[width=9.0cm,angle=0]{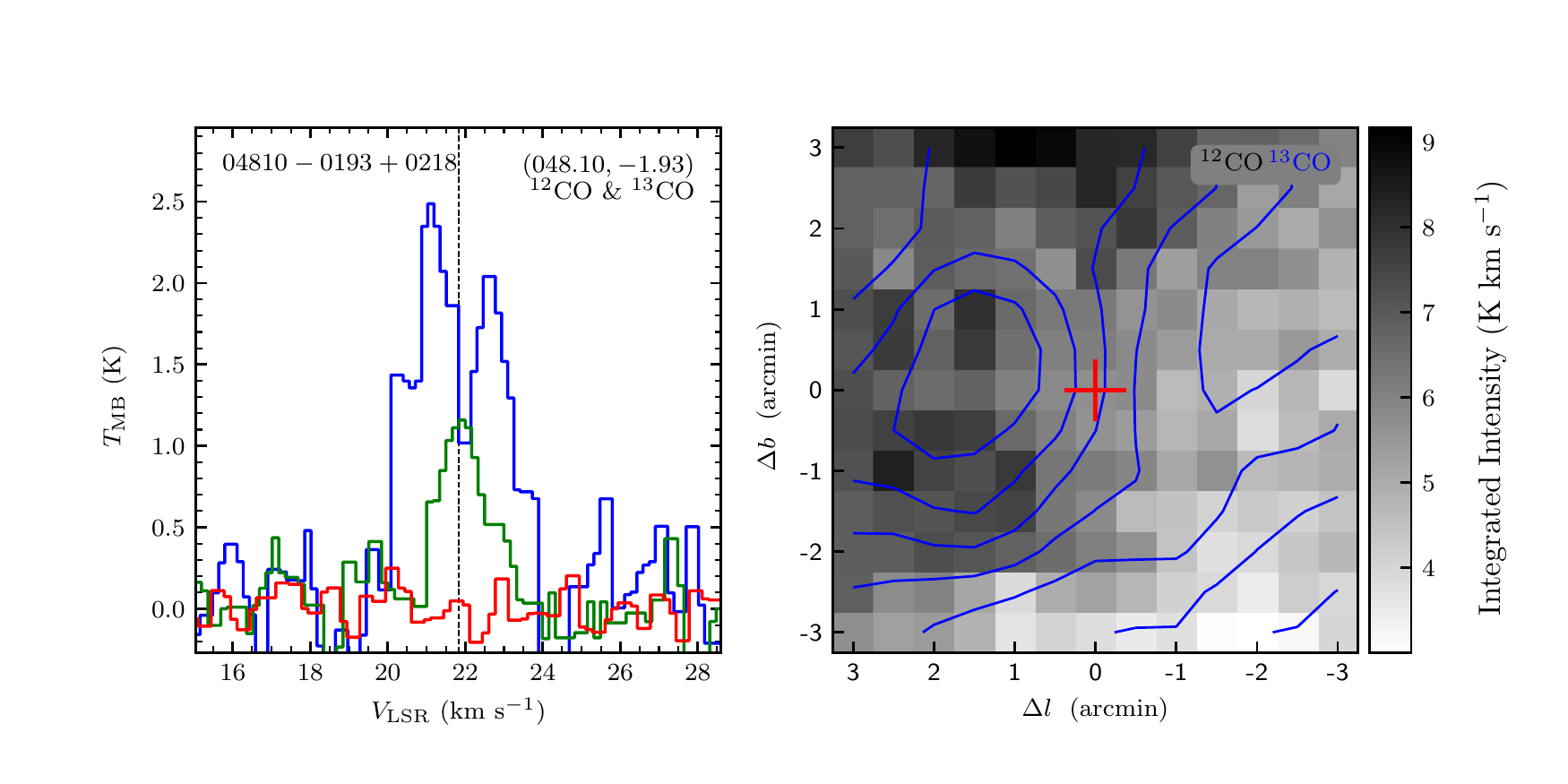}
\end{figure}
\clearpage

\begin{figure}
\includegraphics[width=9.0cm,angle=0]{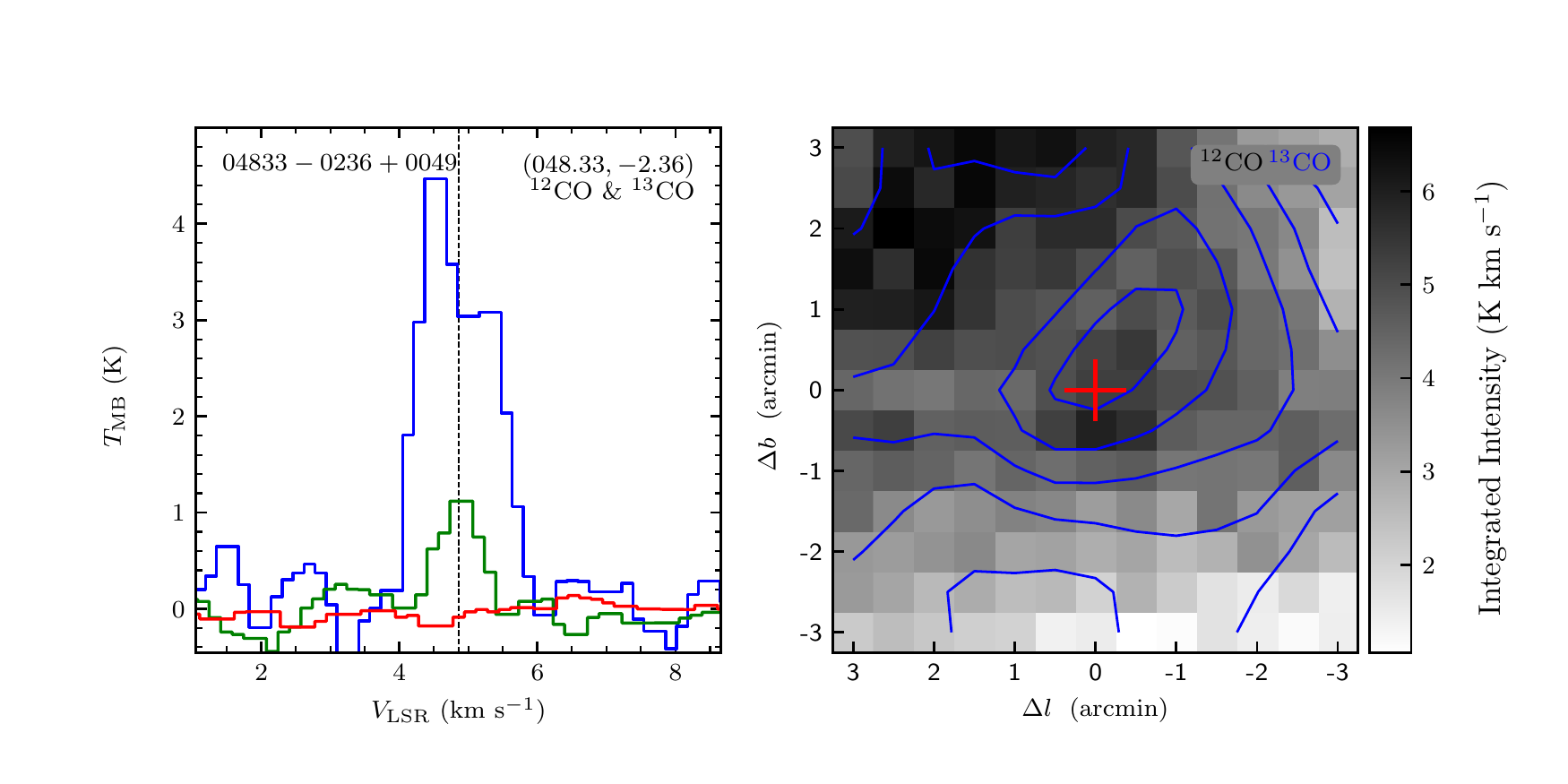}
\includegraphics[width=9.0cm,angle=0]{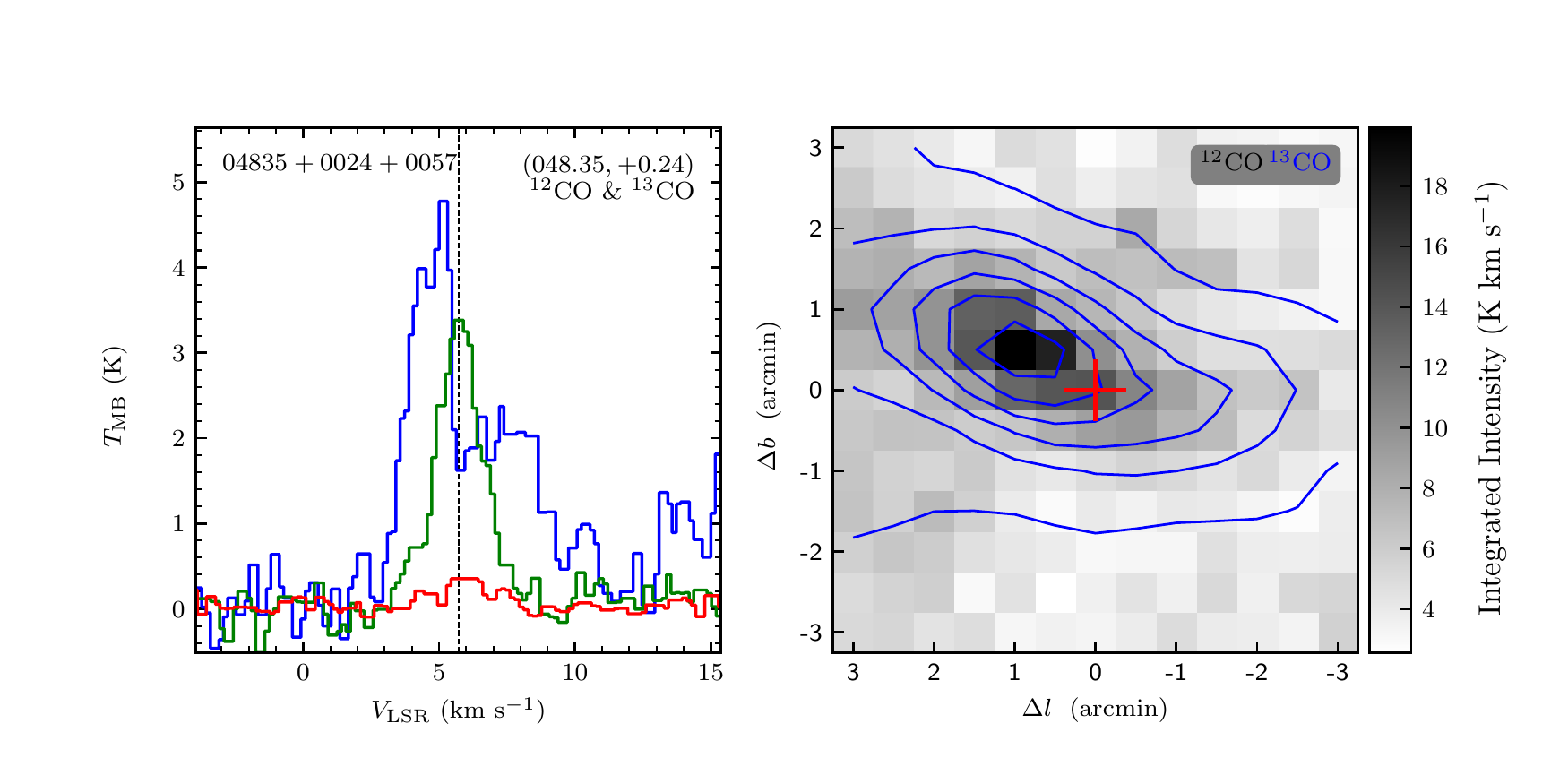}
\vspace{-0.5cm}

\includegraphics[width=9.0cm,angle=0]{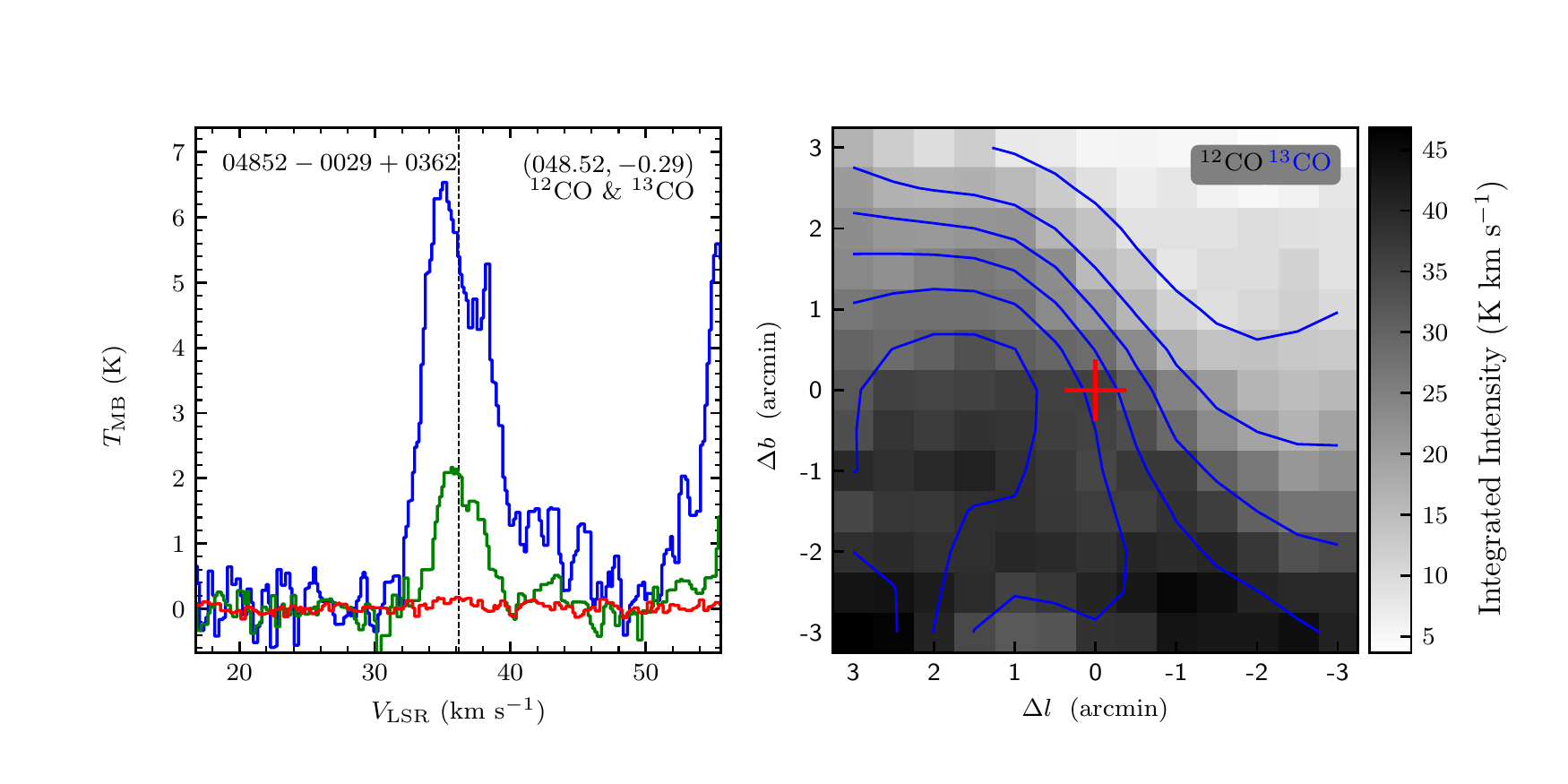}
\includegraphics[width=9.0cm,angle=0]{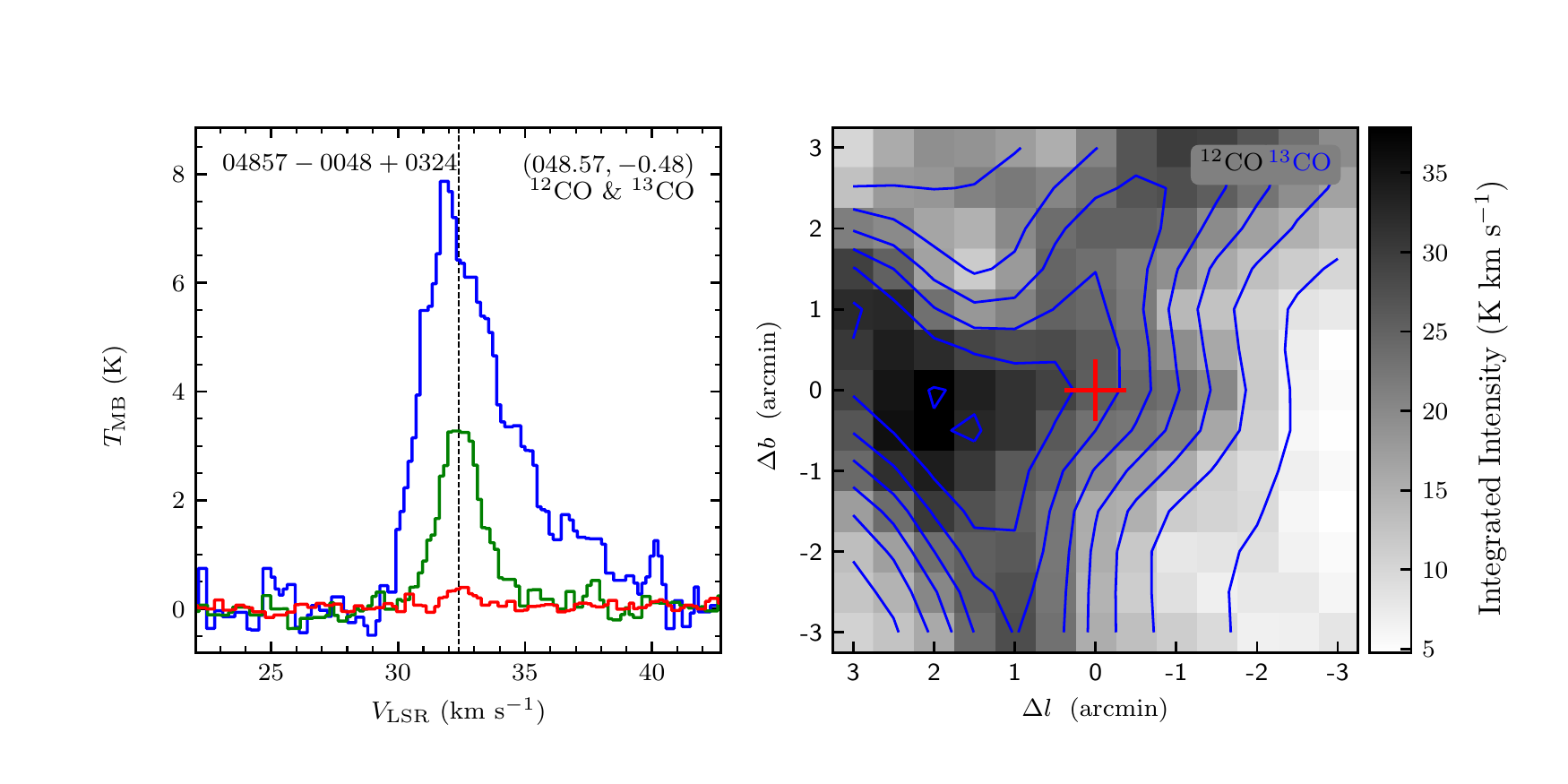}
\vspace{-0.5cm}

\includegraphics[width=9.0cm,angle=0]{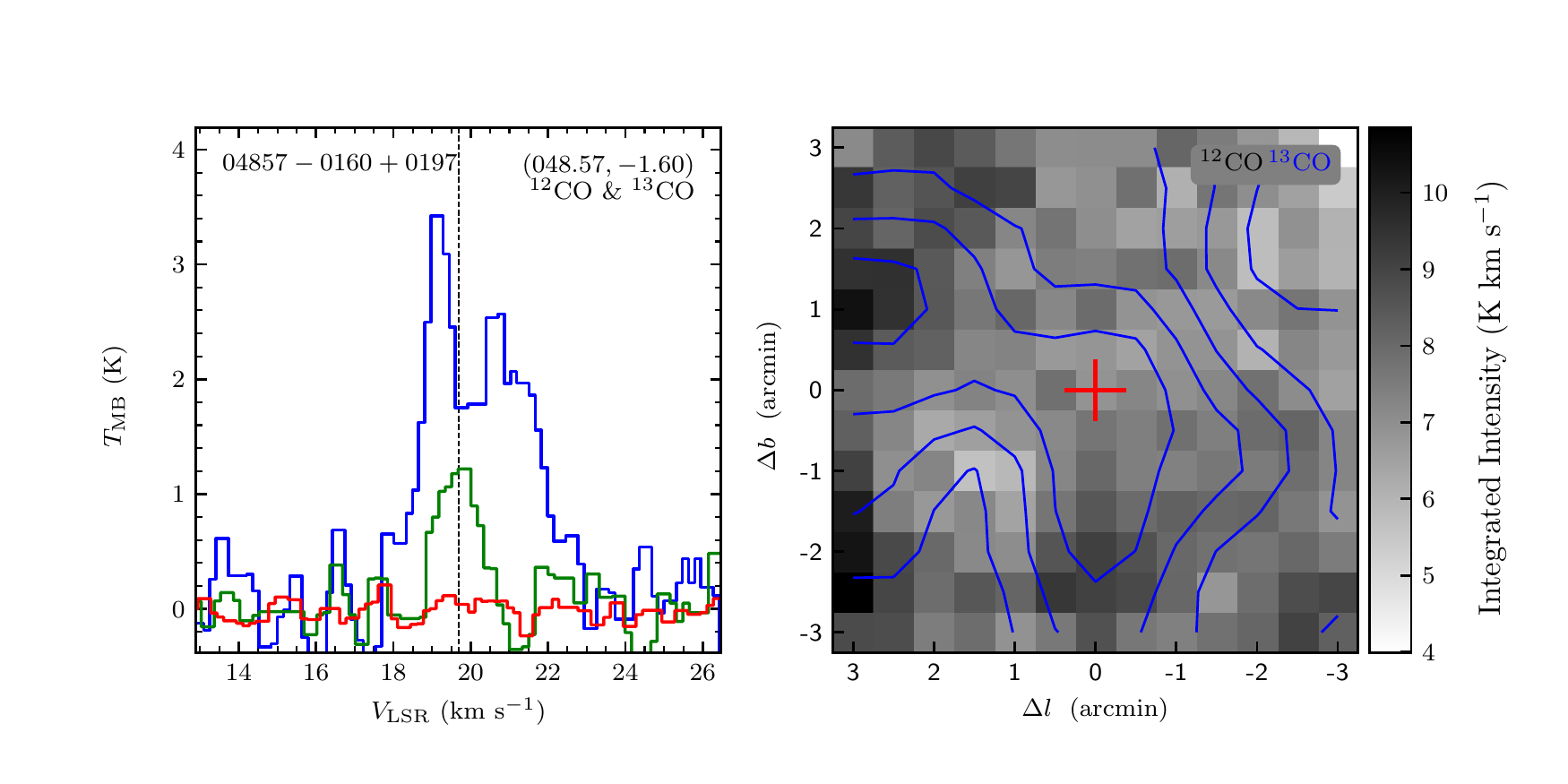}
\includegraphics[width=9.0cm,angle=0]{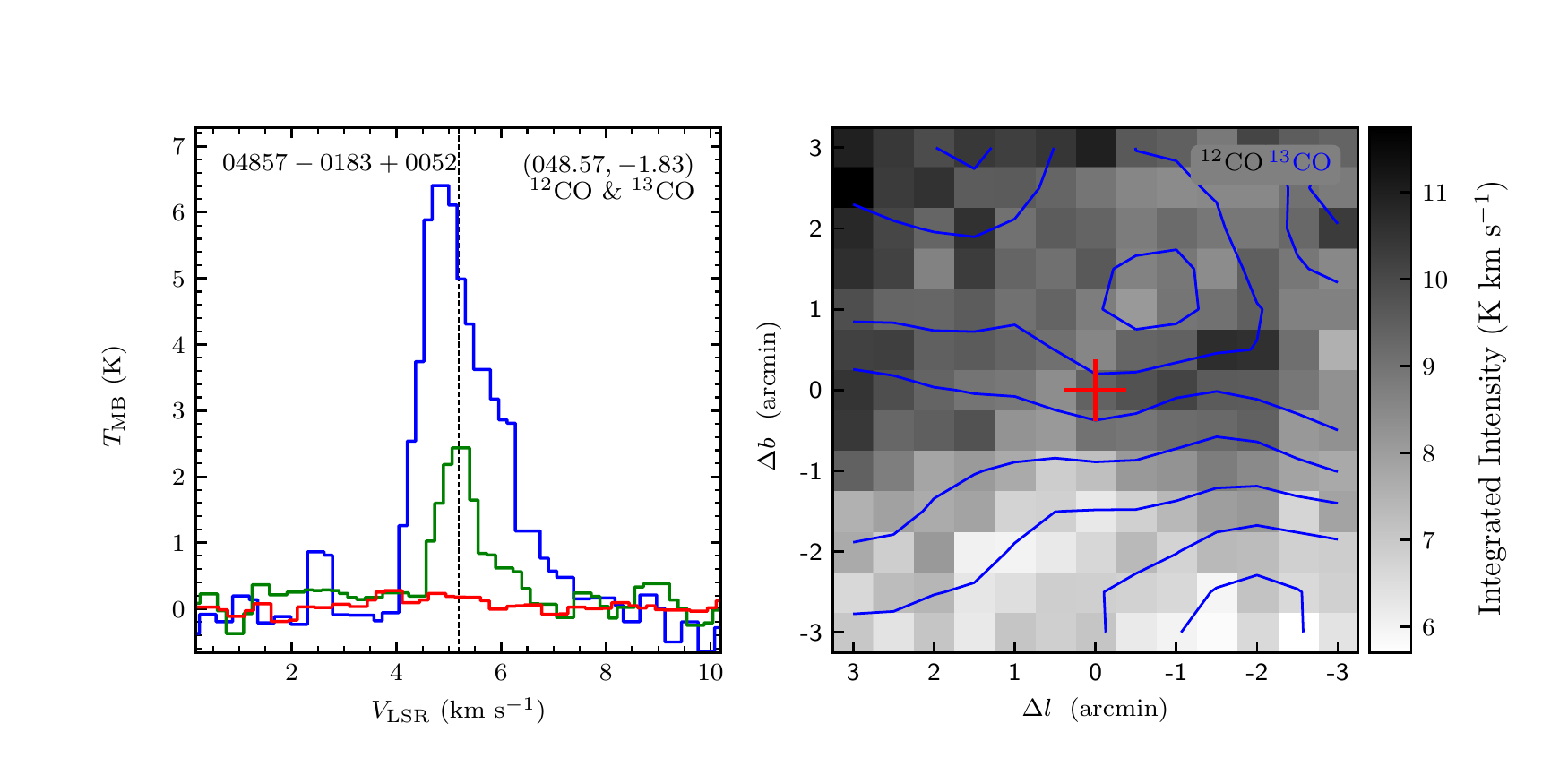}
\vspace{-0.5cm}

\includegraphics[width=9.0cm,angle=0]{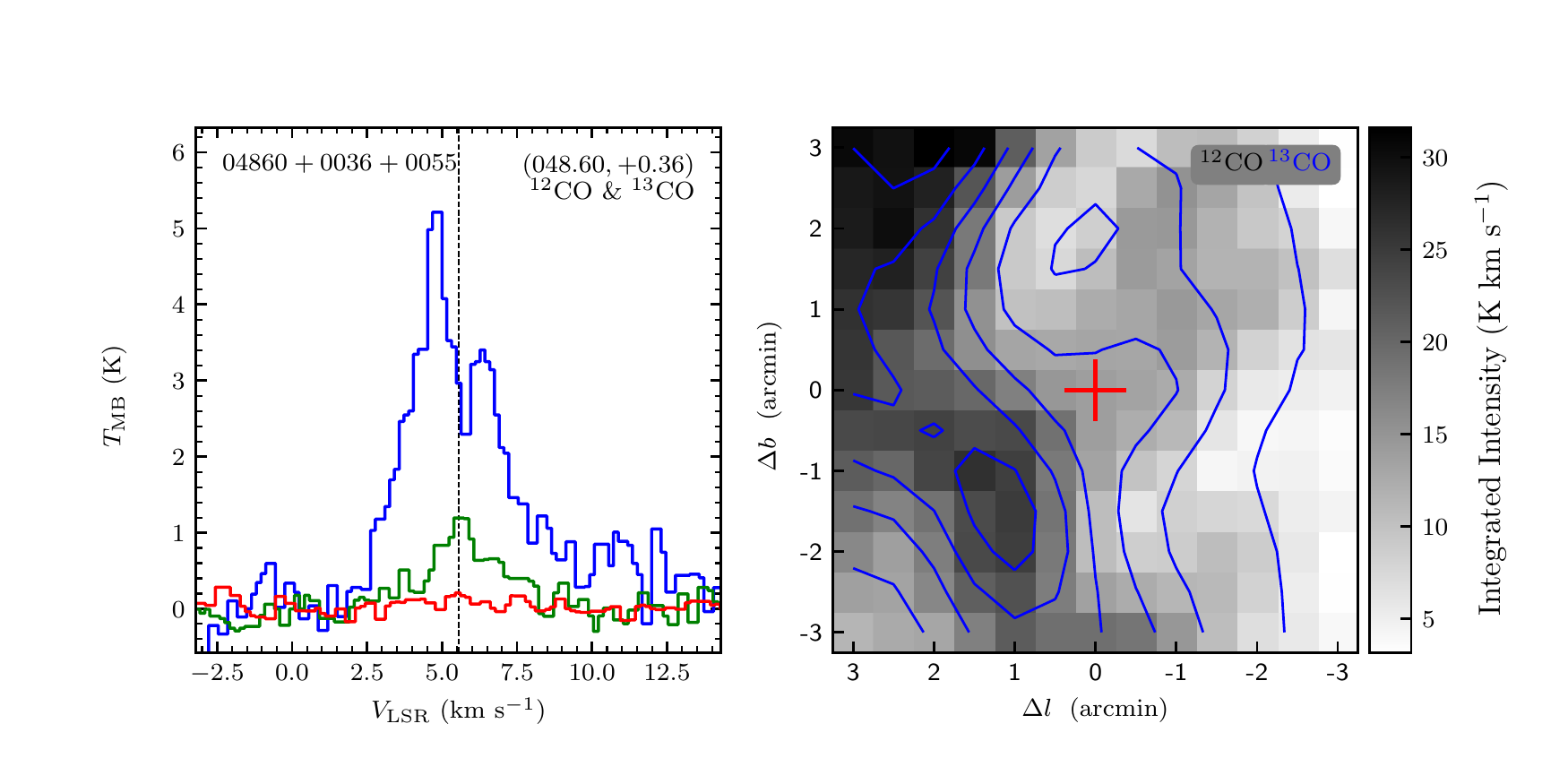}
\includegraphics[width=9.0cm,angle=0]{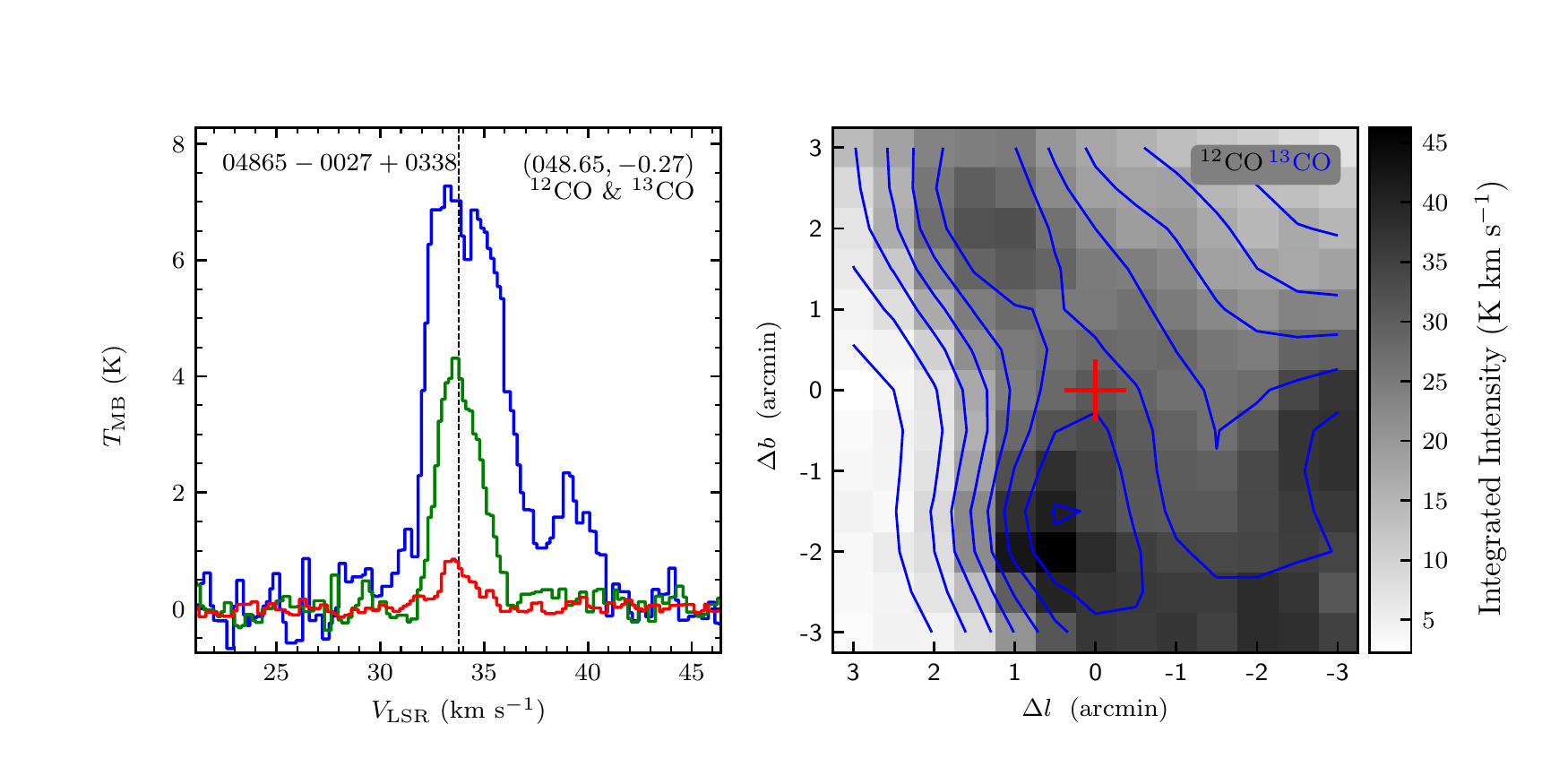}
\vspace{-0.5cm}

\includegraphics[width=9.0cm,angle=0]{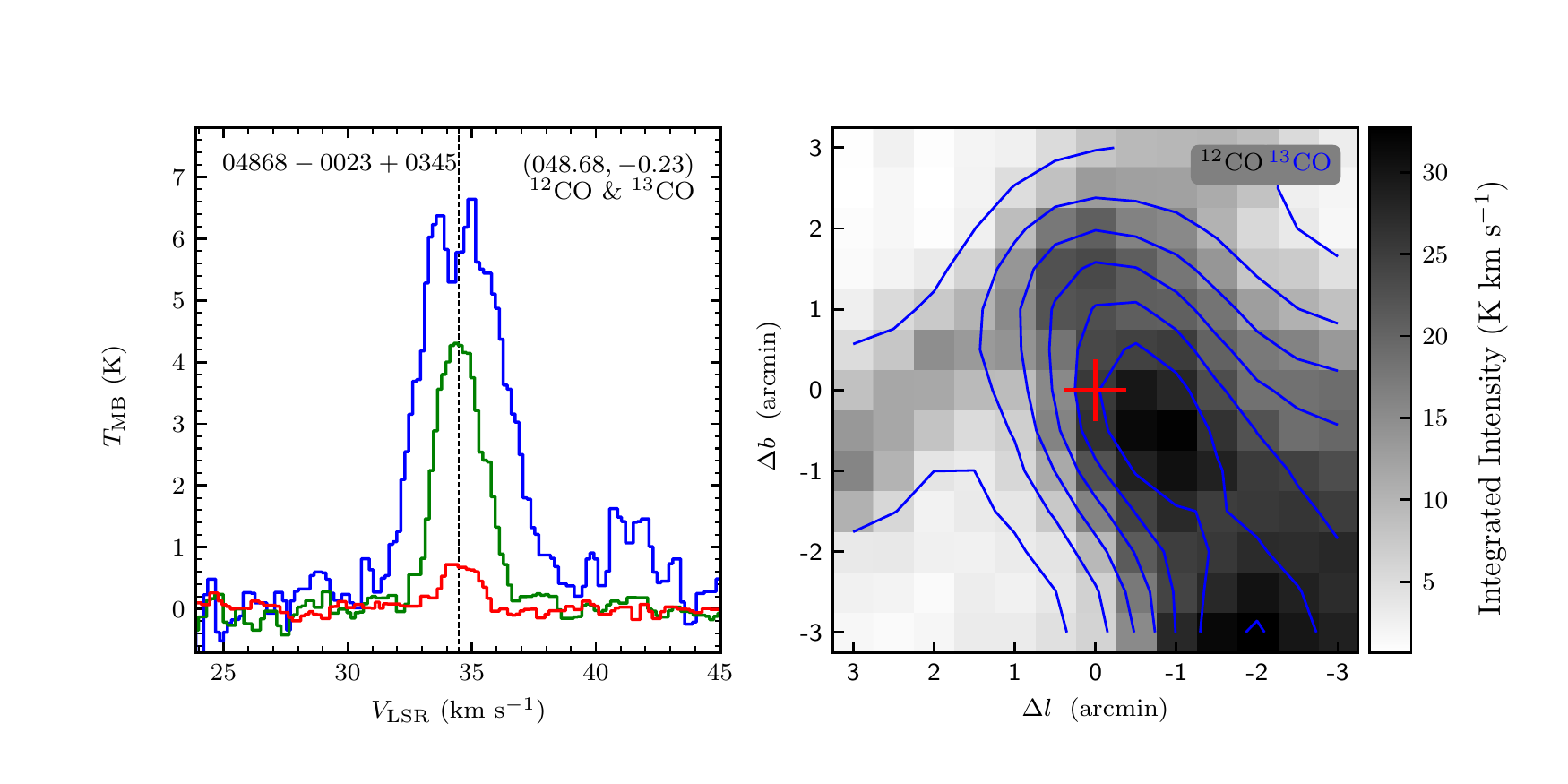}
\includegraphics[width=9.0cm,angle=0]{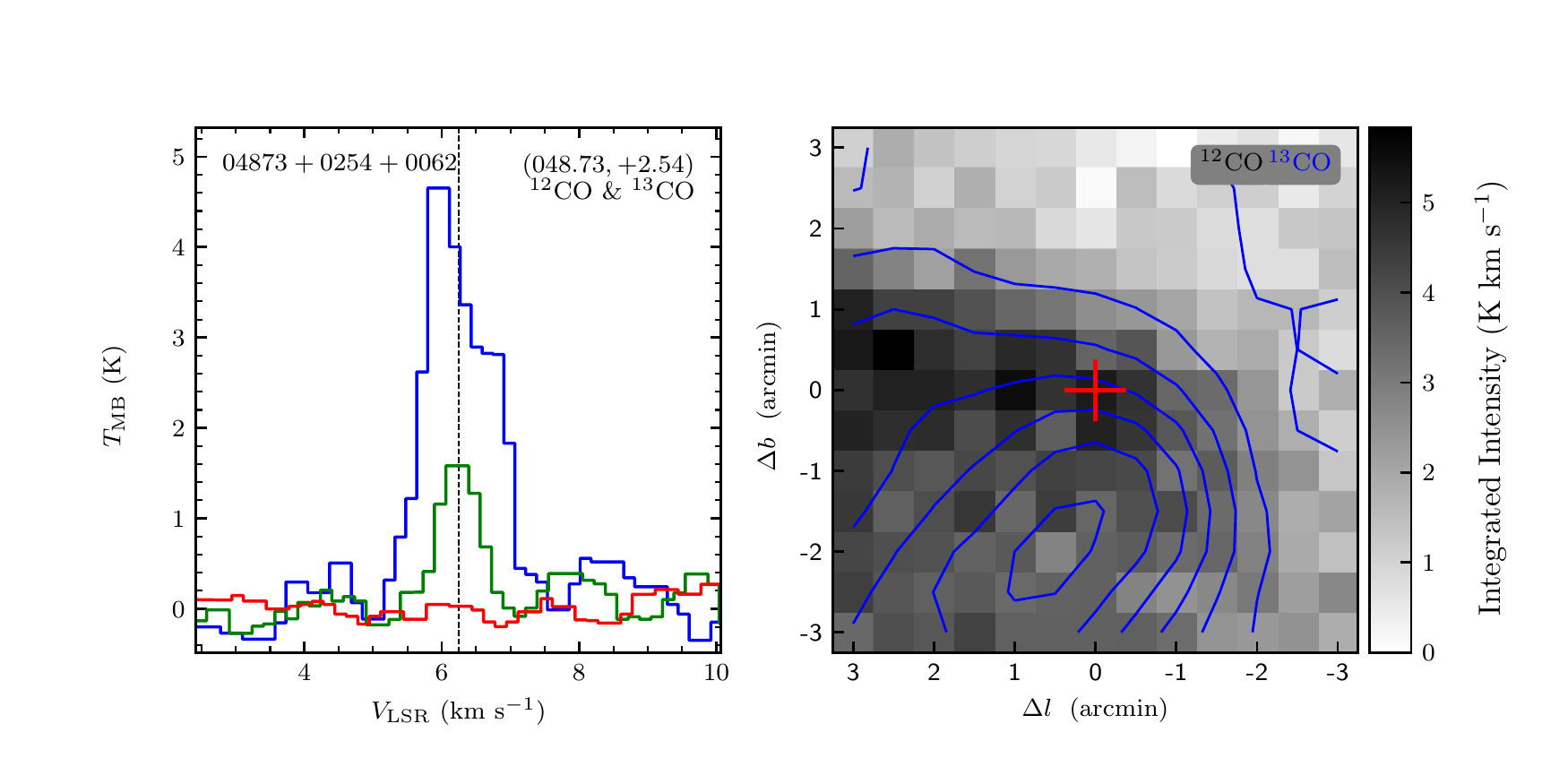}
\end{figure}
\clearpage

\begin{figure}
\includegraphics[width=9.0cm,angle=0]{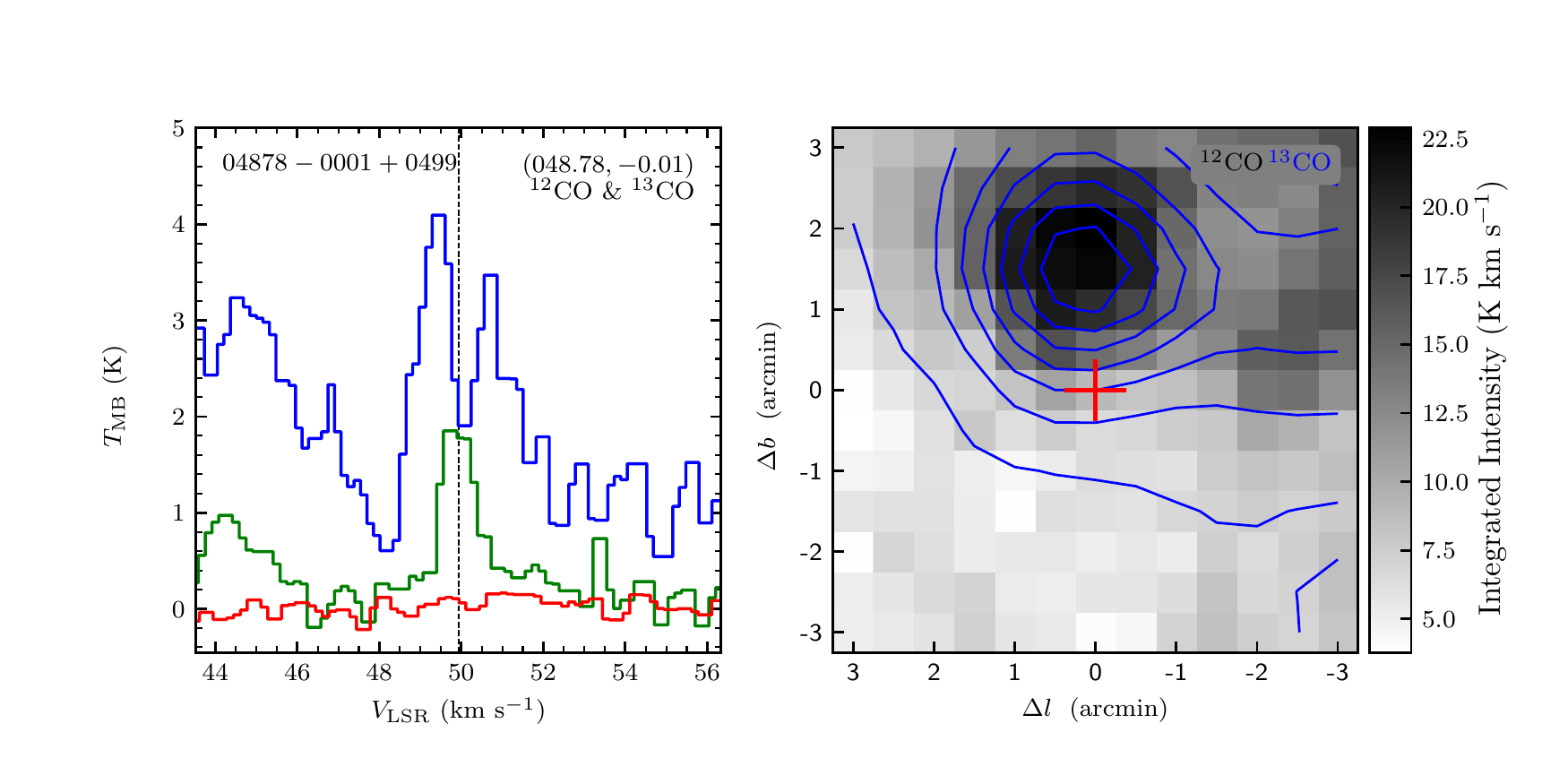}
\includegraphics[width=9.0cm,angle=0]{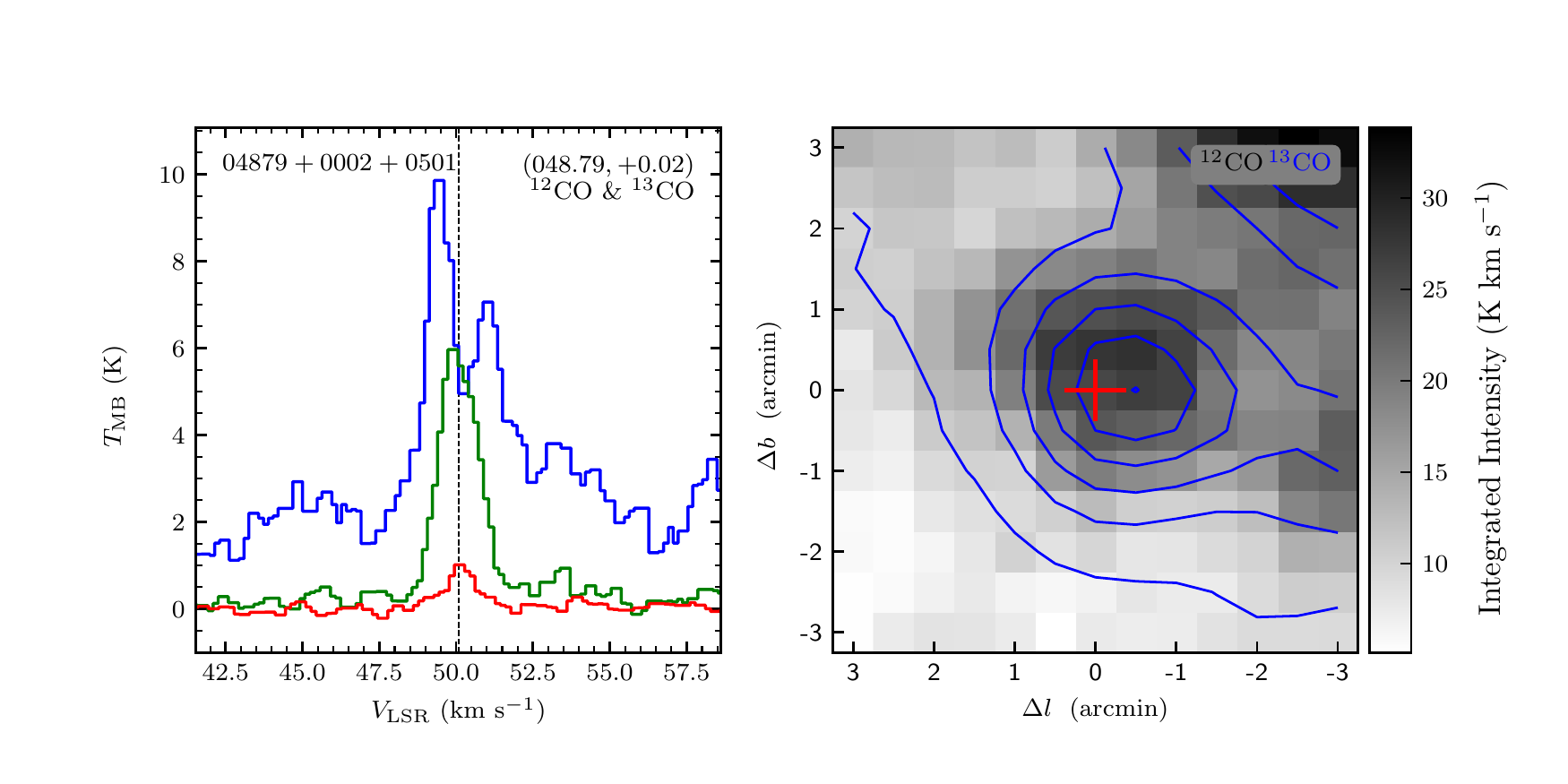}
\vspace{-0.5cm}

\includegraphics[width=9.0cm,angle=0]{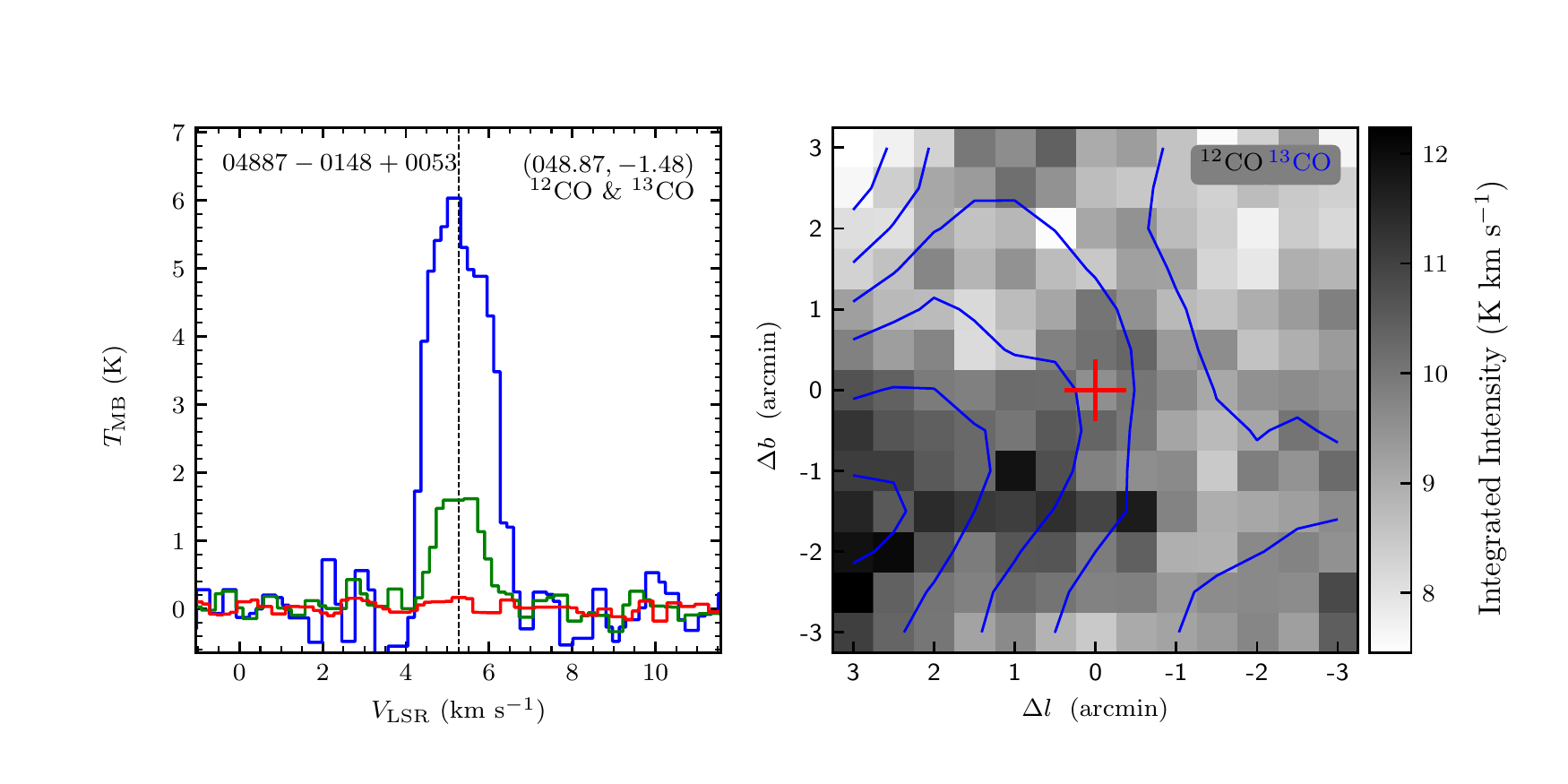}
\includegraphics[width=9.0cm,angle=0]{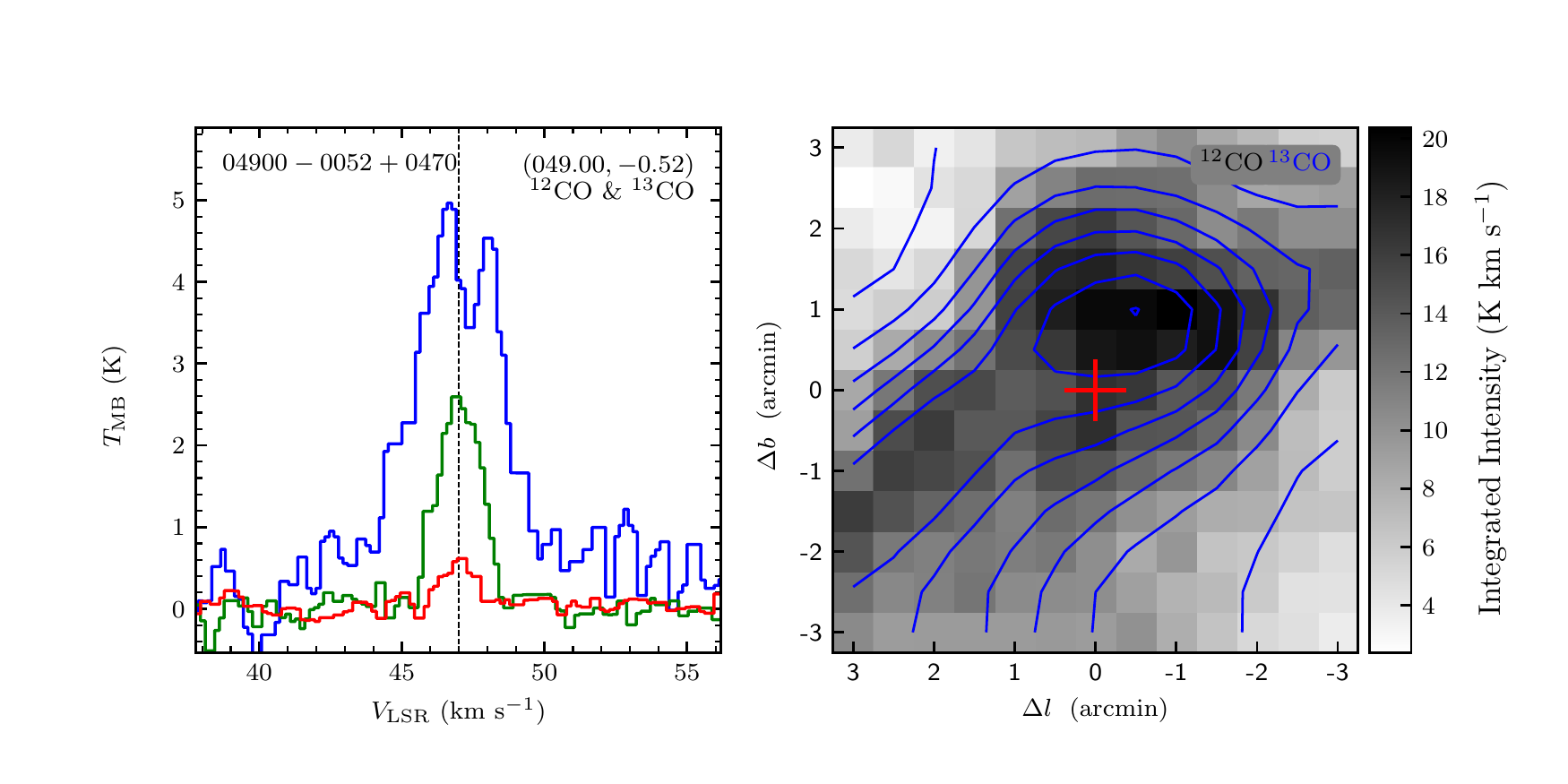}
\vspace{-0.5cm}

\includegraphics[width=9.0cm,angle=0]{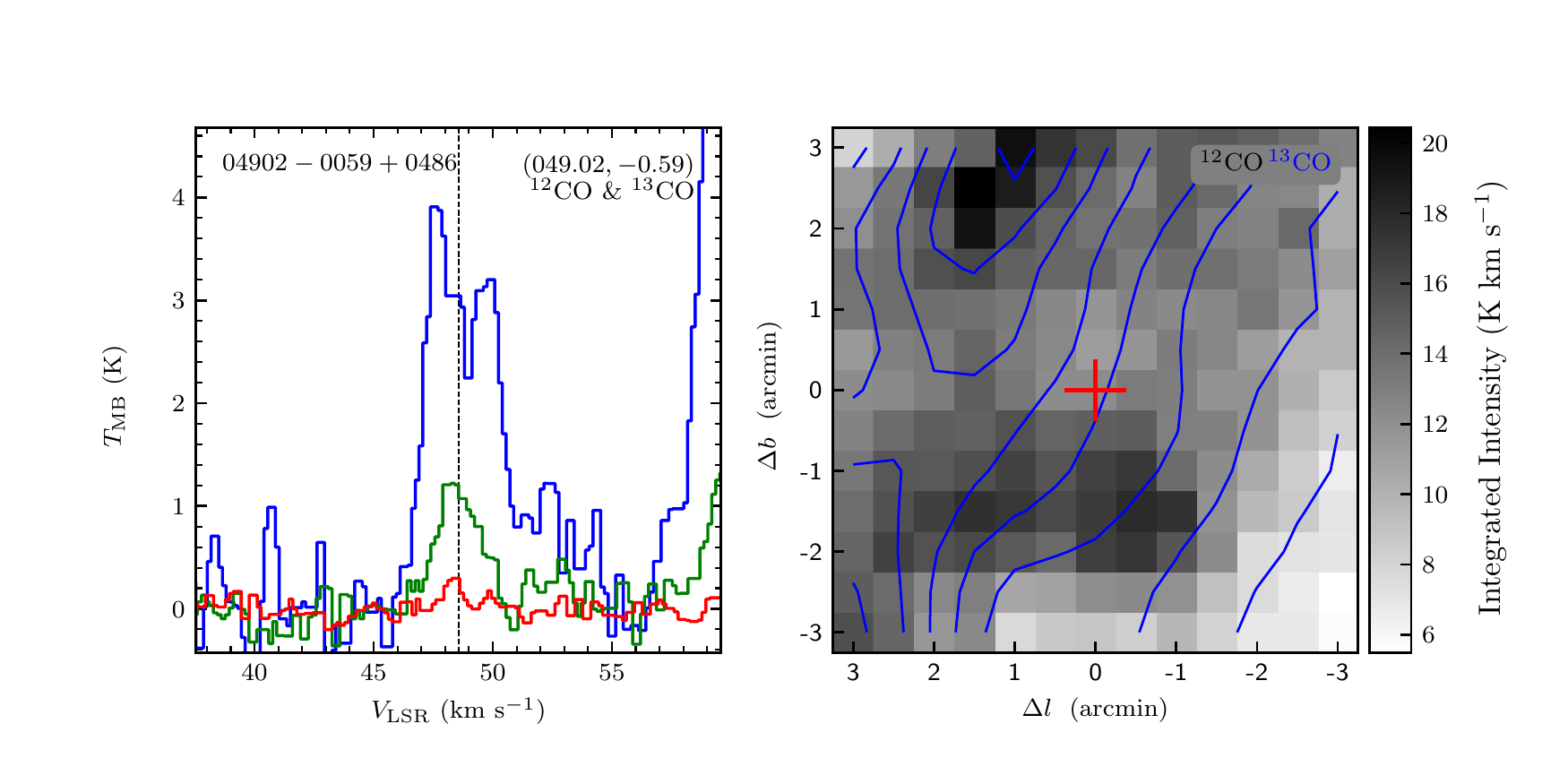}
\includegraphics[width=9.0cm,angle=0]{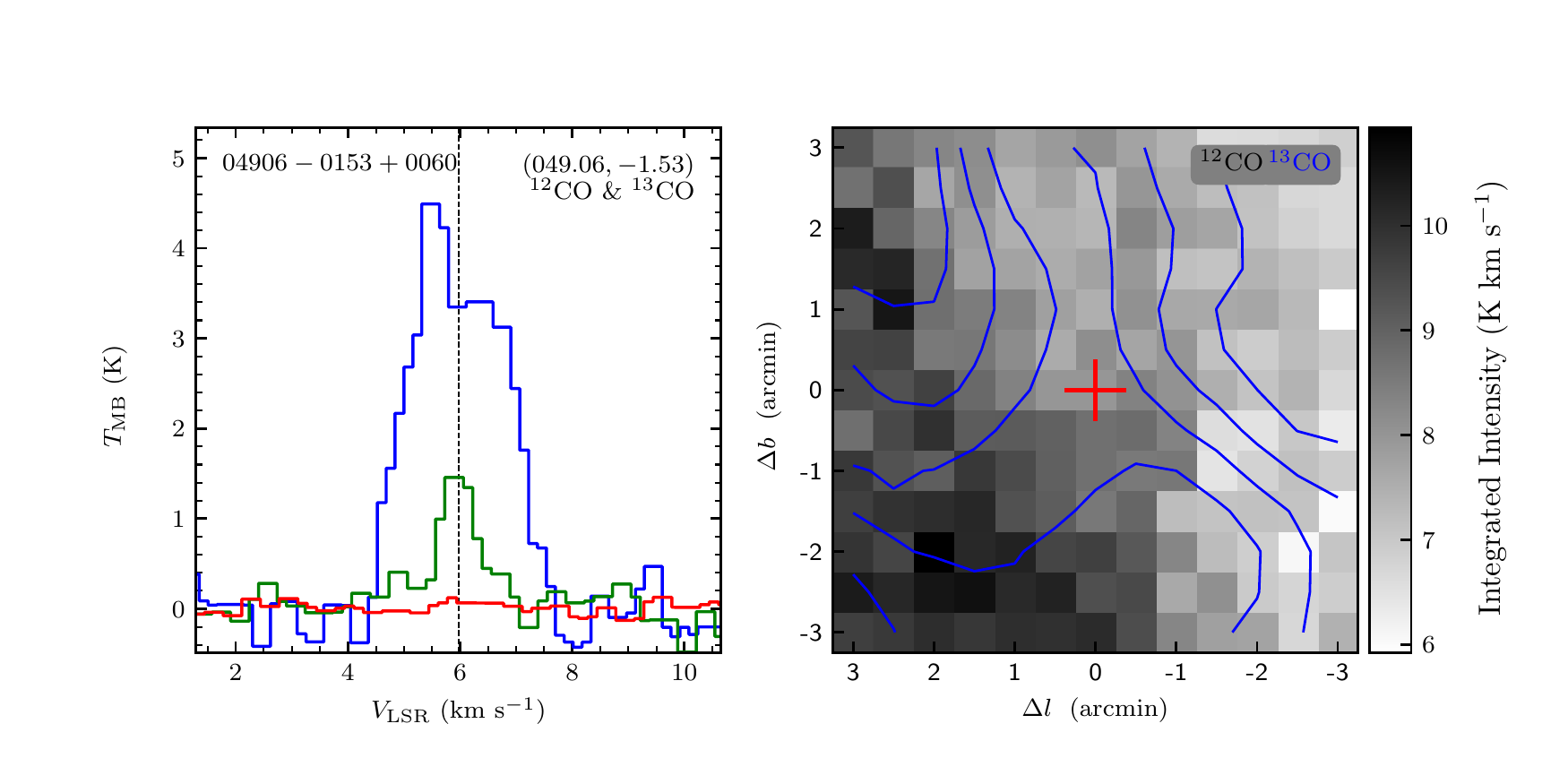}
\vspace{-0.5cm}

\includegraphics[width=9.0cm,angle=0]{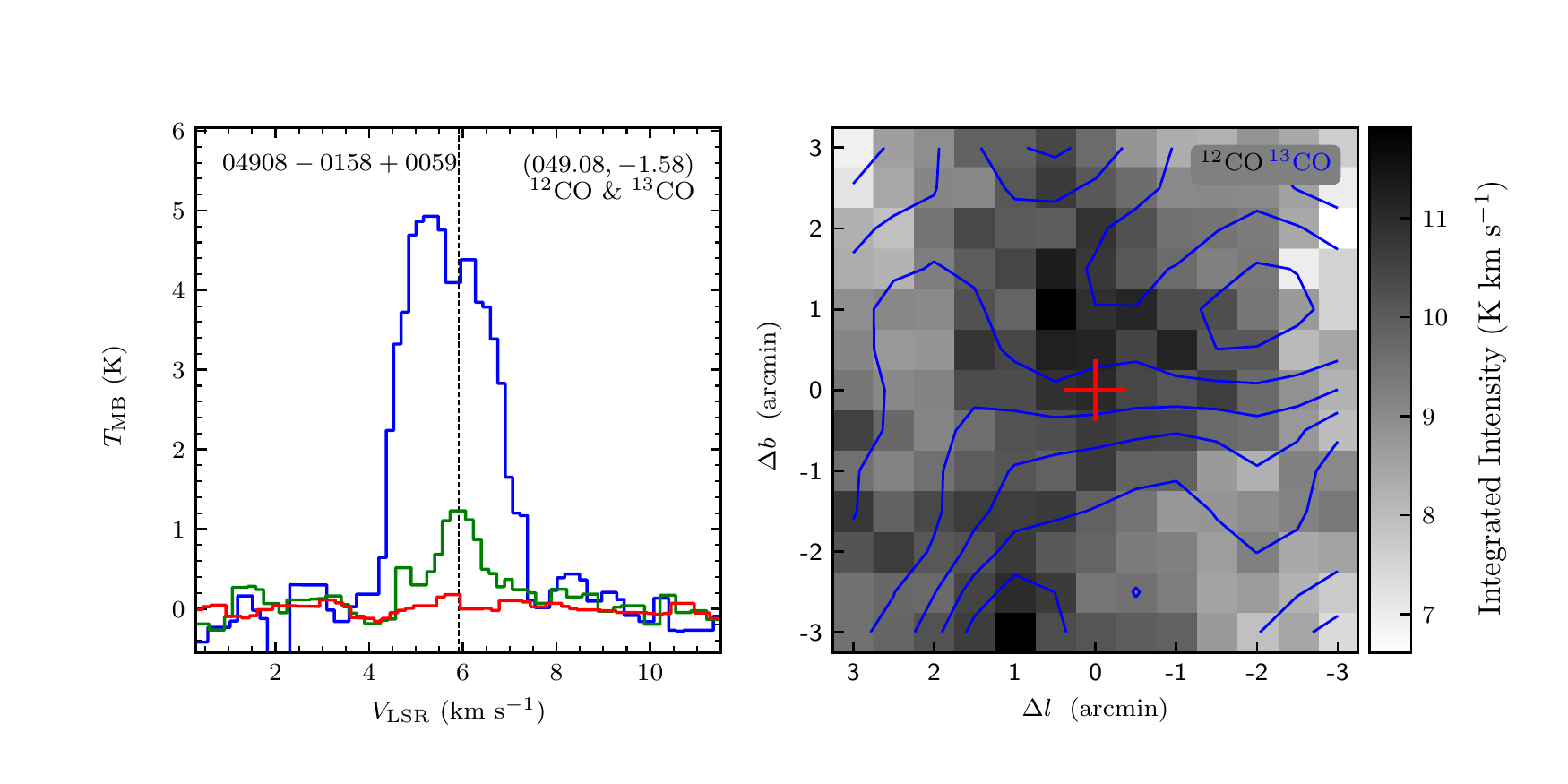}
\includegraphics[width=9.0cm,angle=0]{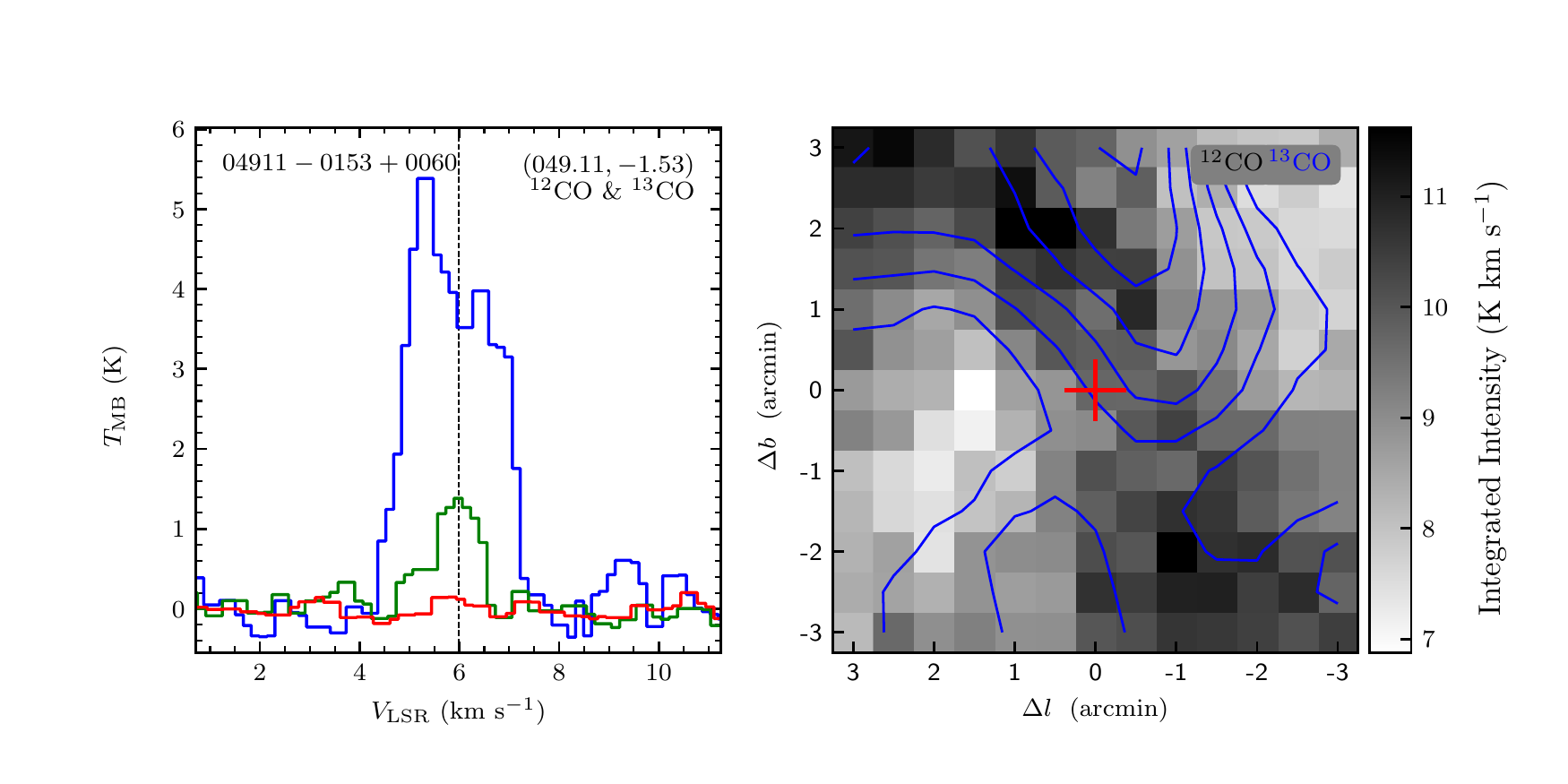}
\vspace{-0.5cm}

\includegraphics[width=9.0cm,angle=0]{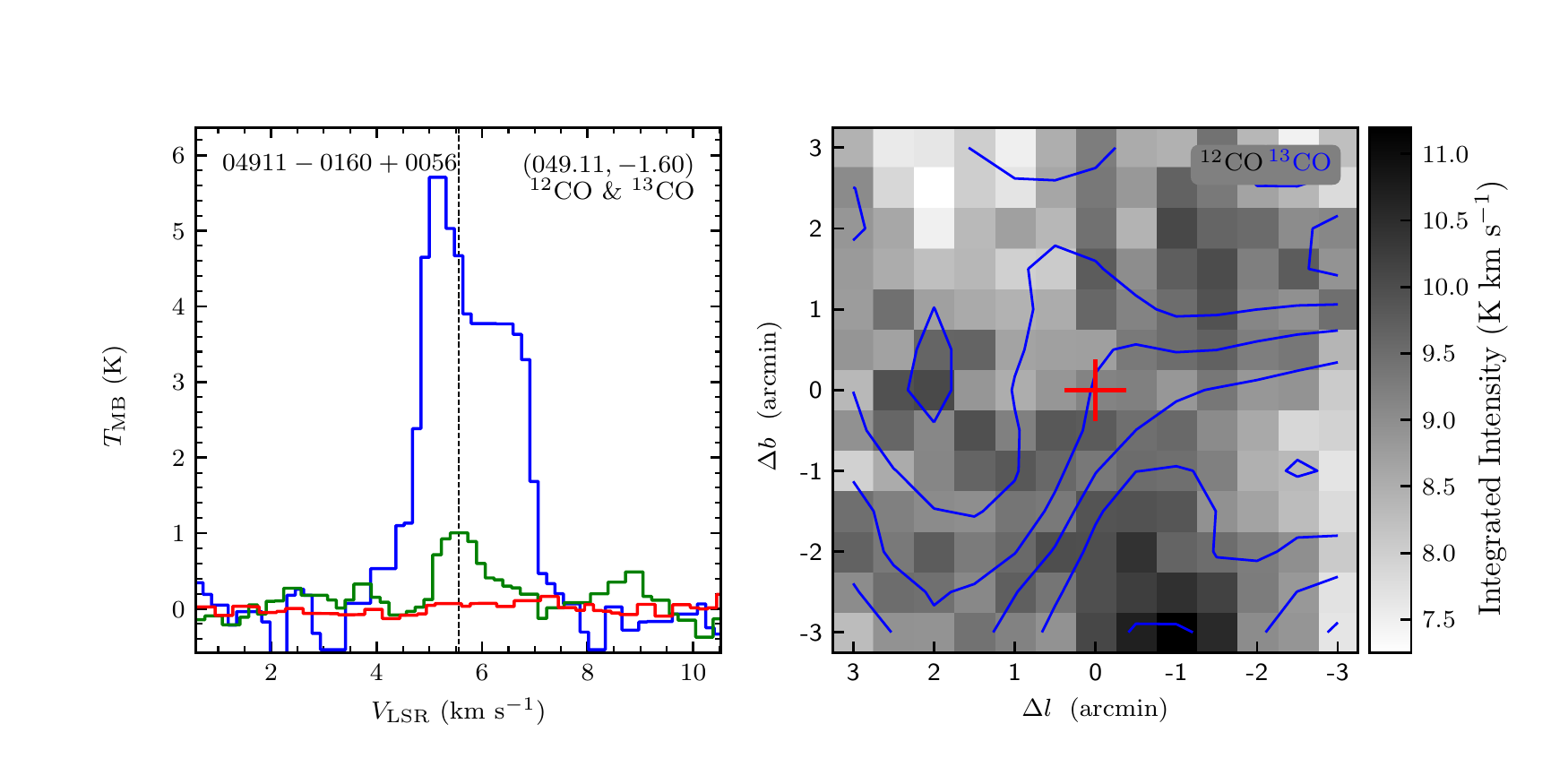}
\includegraphics[width=9.0cm,angle=0]{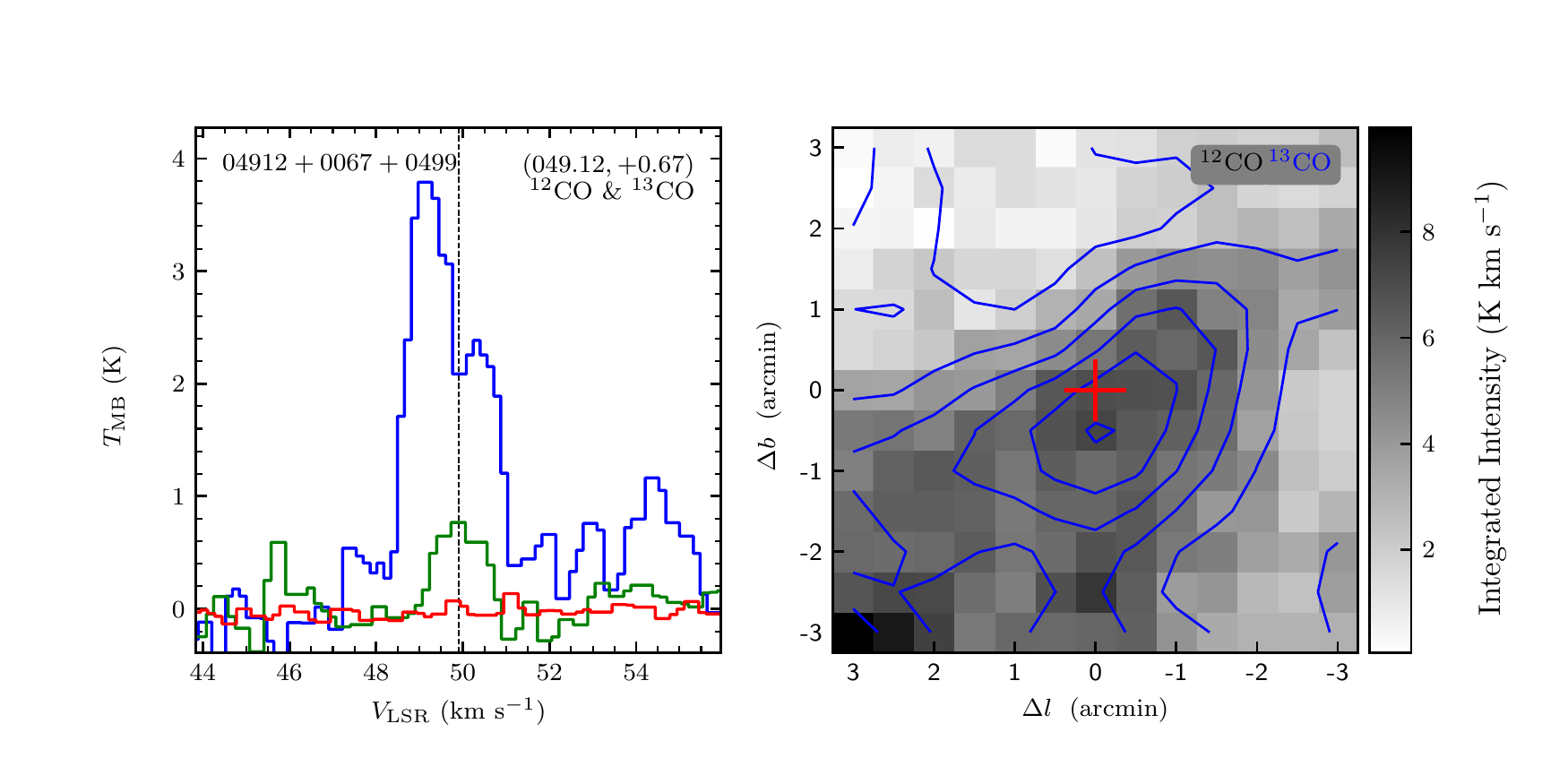}
\end{figure}
\clearpage

\begin{figure}
\includegraphics[width=9.0cm,angle=0]{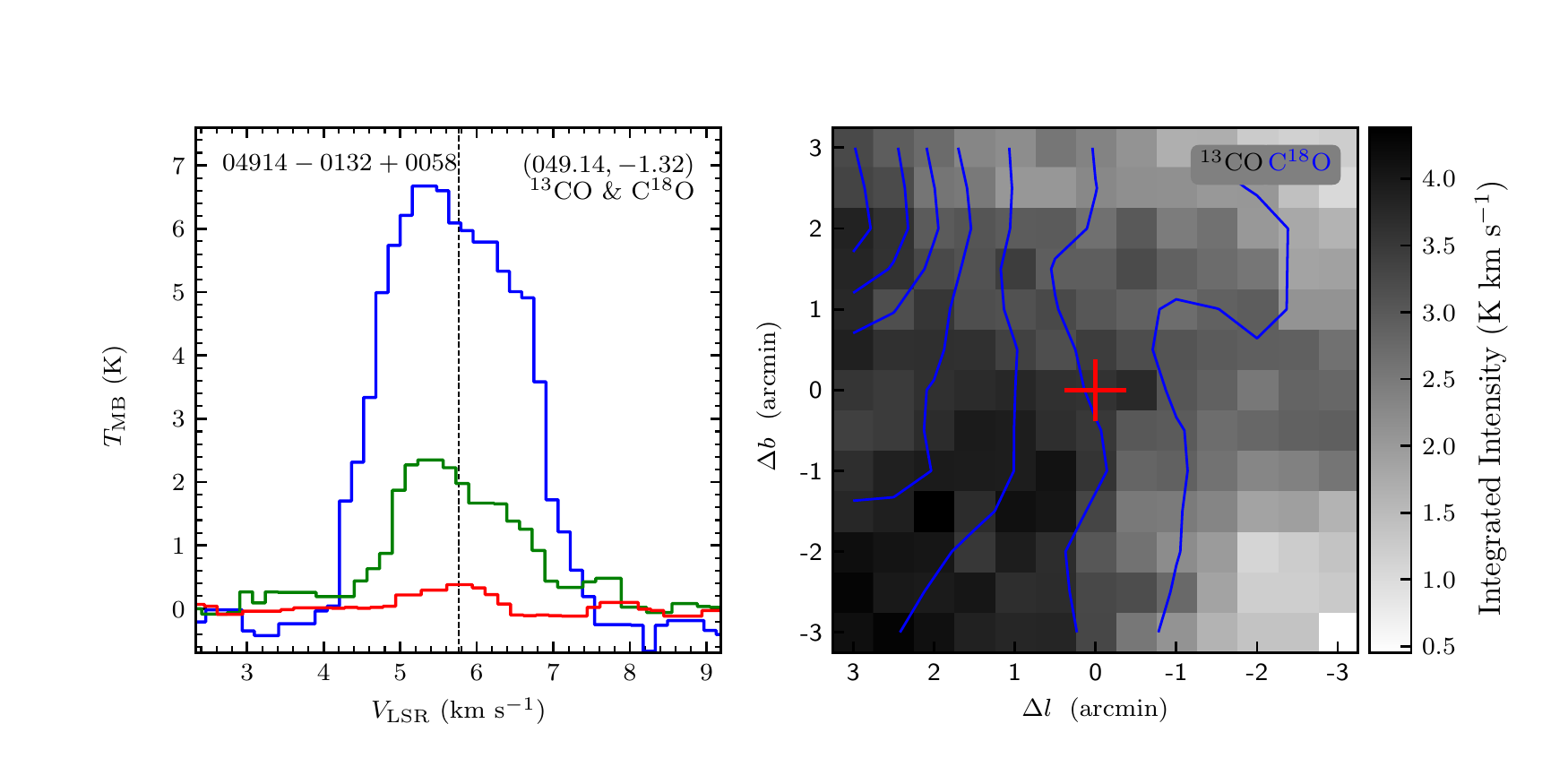}
\includegraphics[width=9.0cm,angle=0]{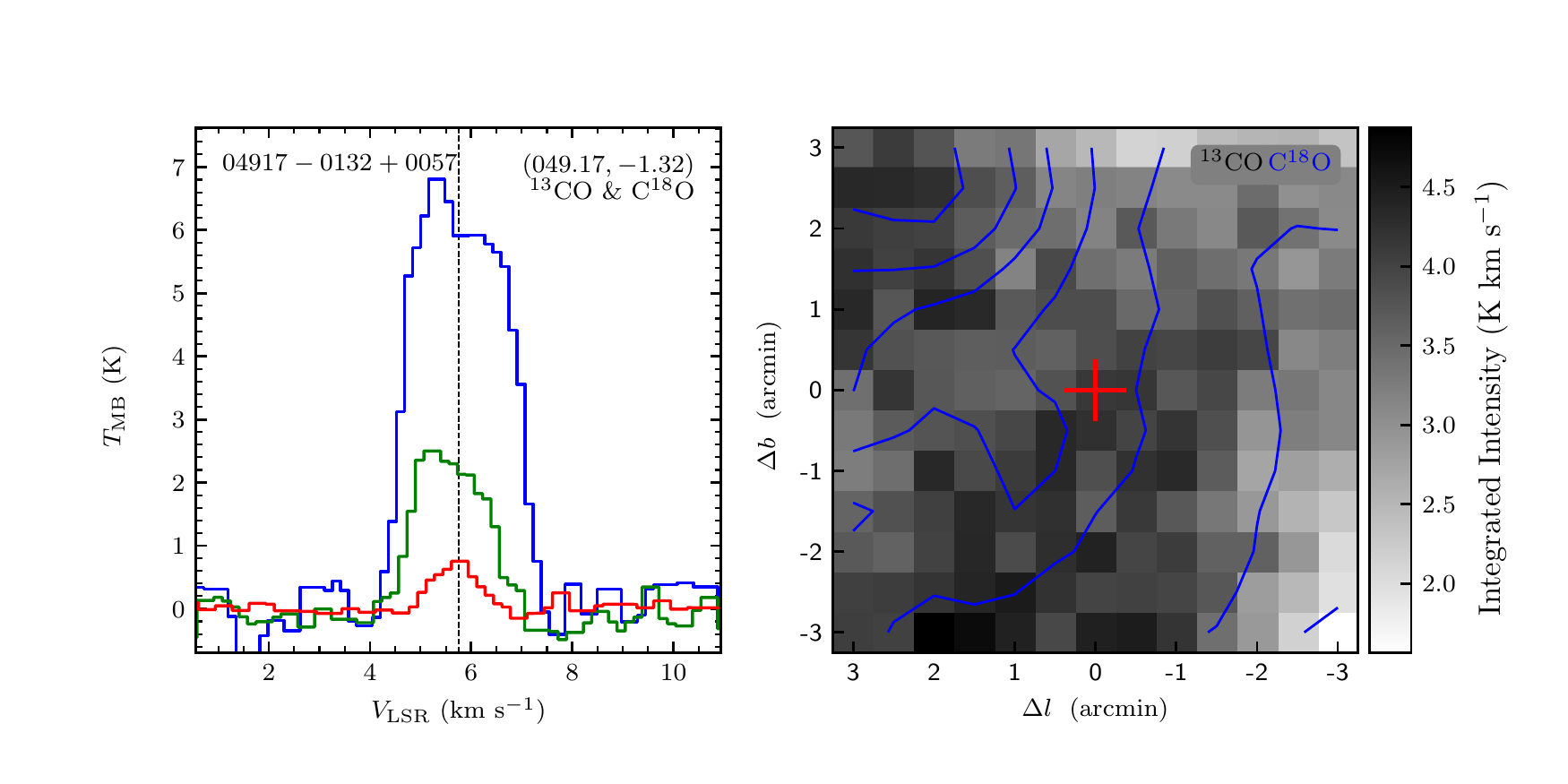}
\vspace{-0.5cm}

\includegraphics[width=9.0cm,angle=0]{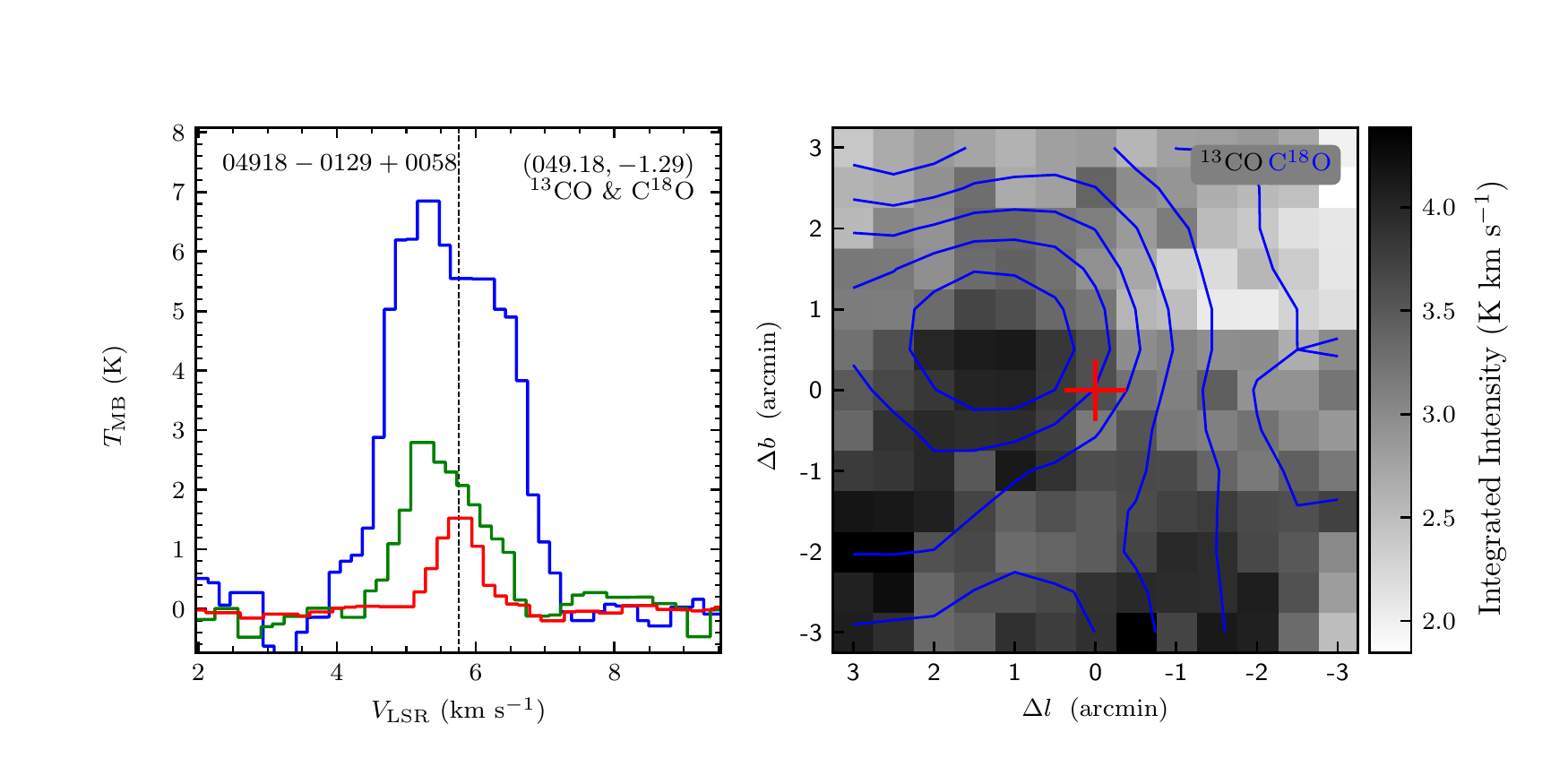}
\includegraphics[width=9.0cm,angle=0]{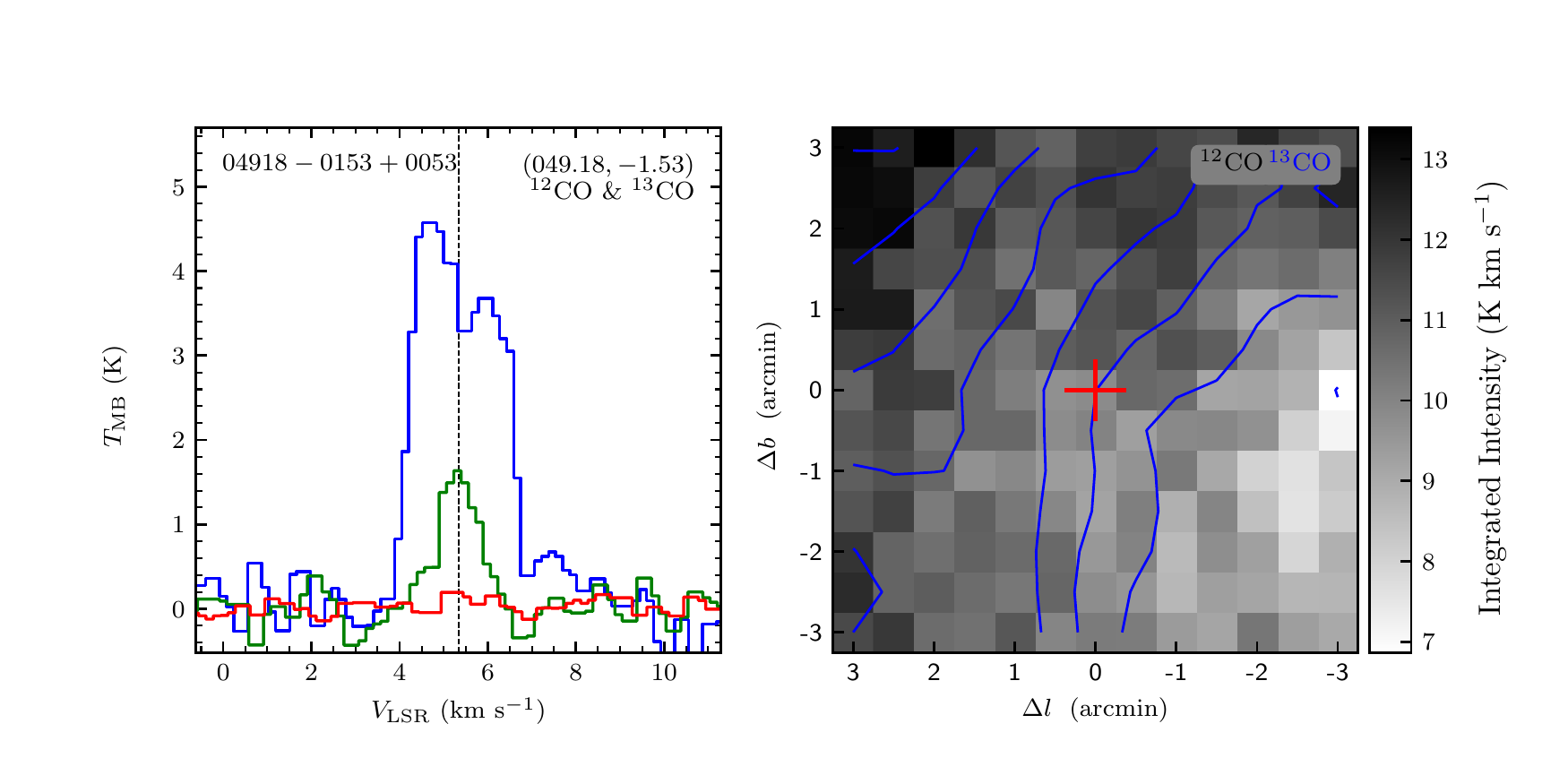}
\vspace{-0.5cm}

\includegraphics[width=9.0cm,angle=0]{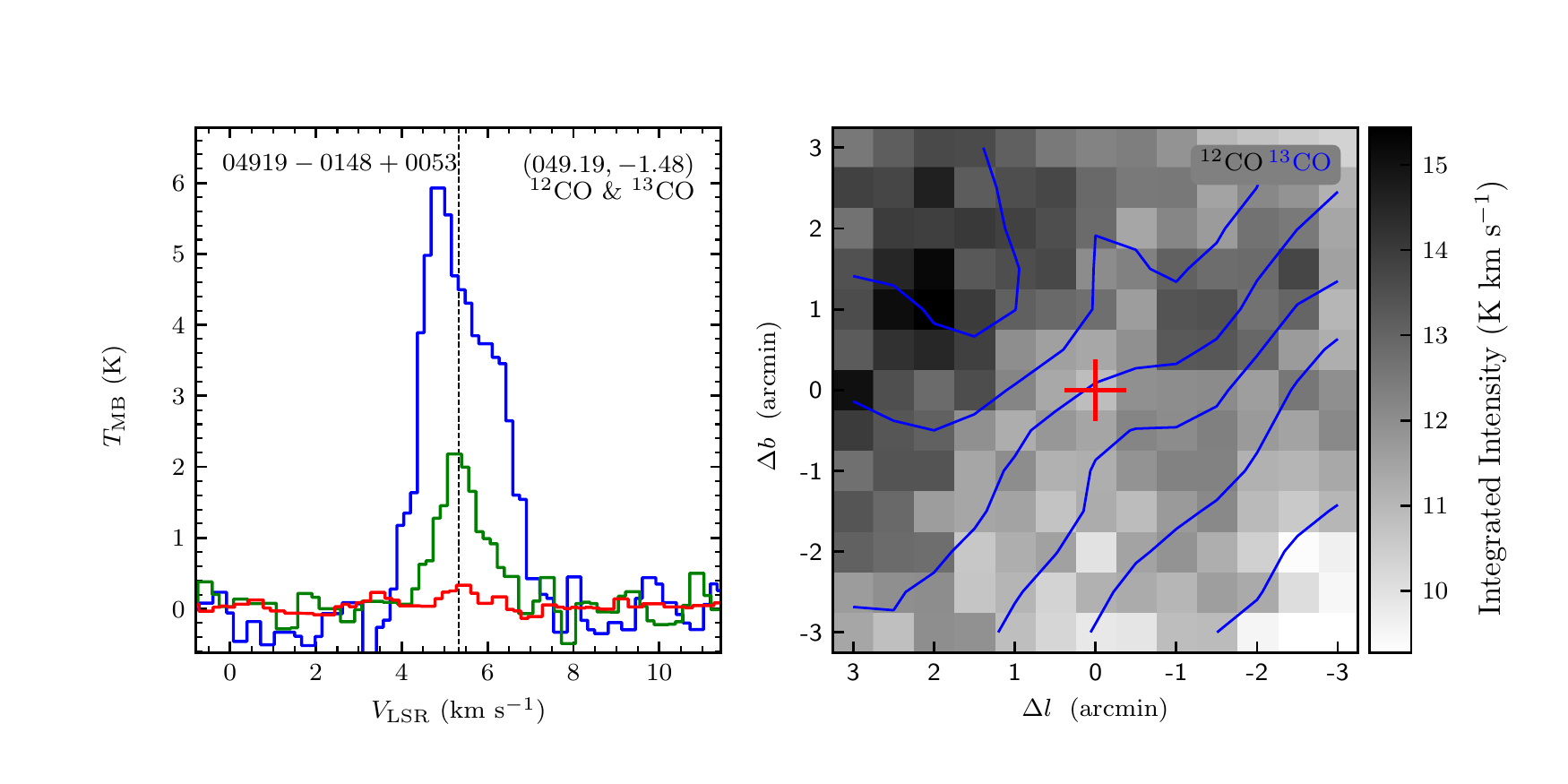}
\includegraphics[width=9.0cm,angle=0]{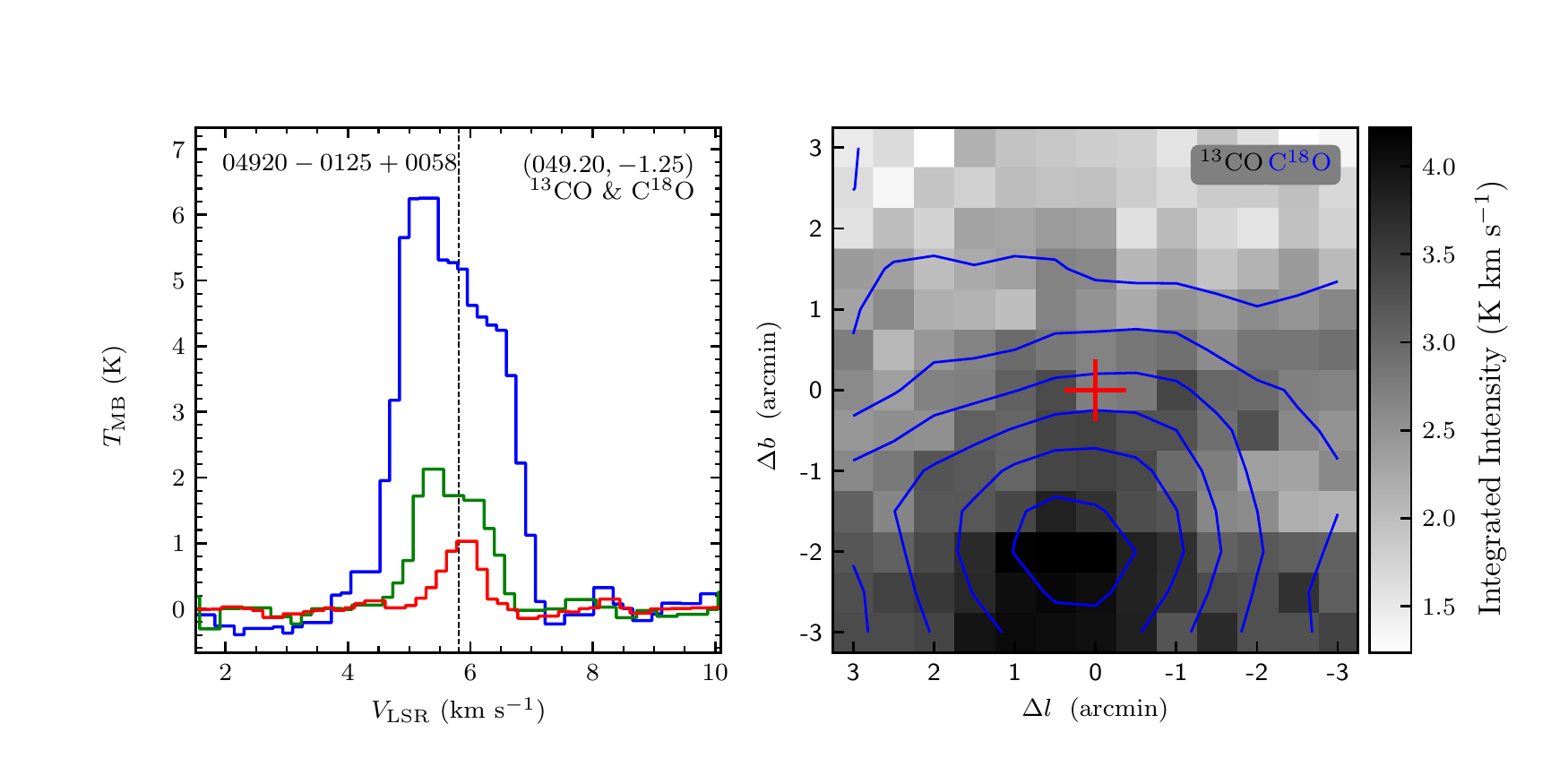}
\vspace{-0.5cm}

\includegraphics[width=9.0cm,angle=0]{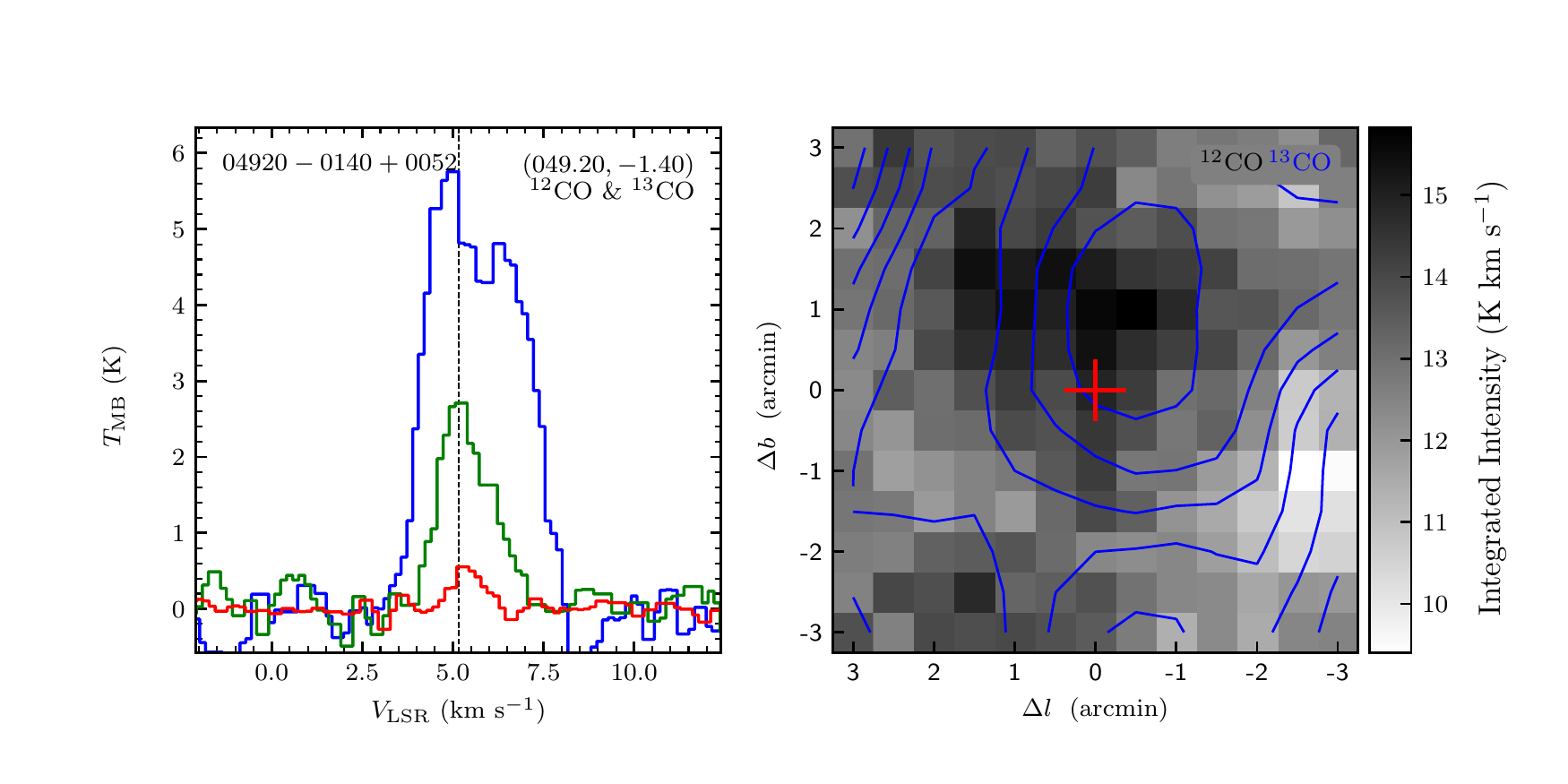}
\includegraphics[width=9.0cm,angle=0]{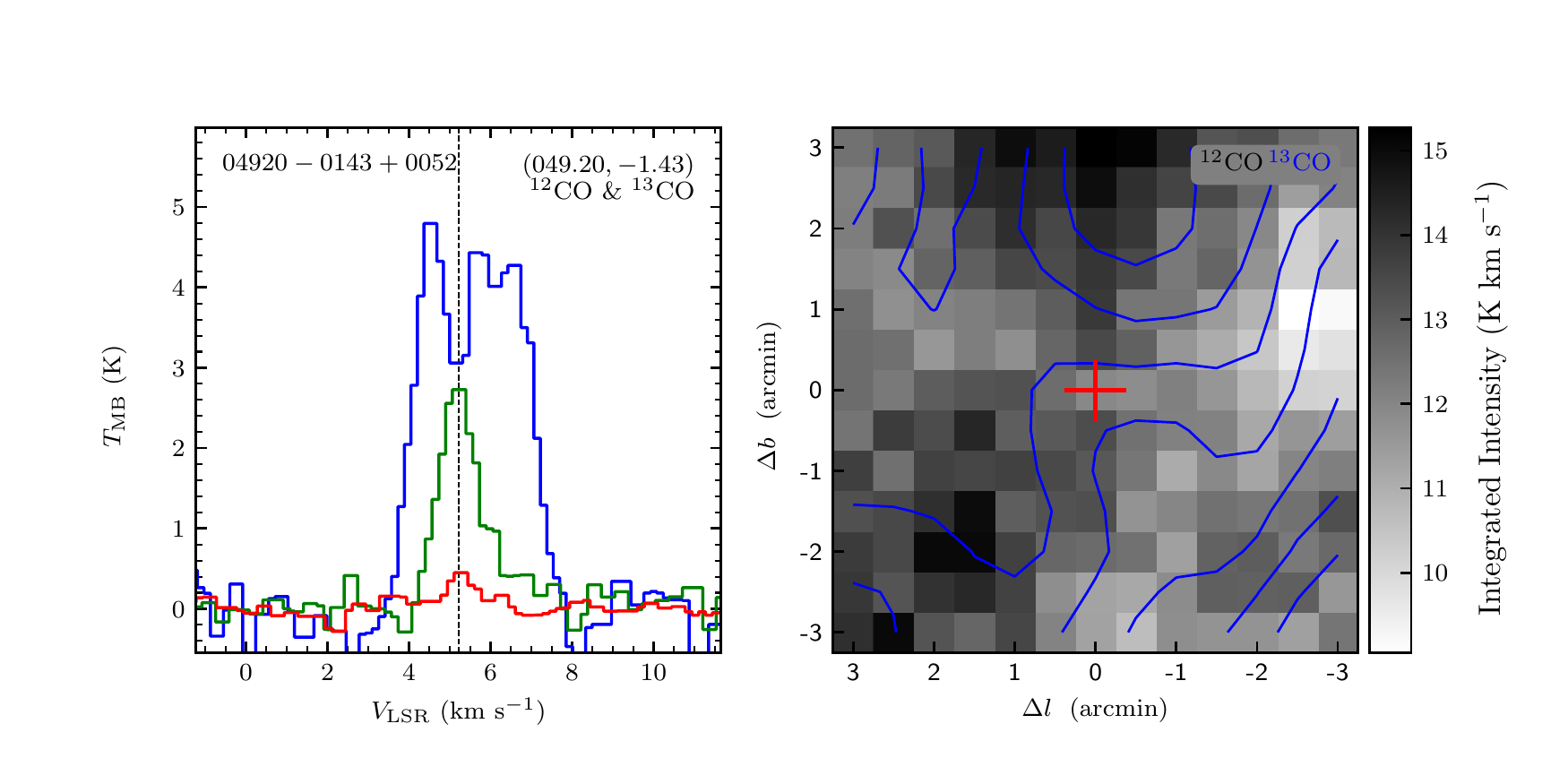}
\vspace{-0.5cm}

\includegraphics[width=9.0cm,angle=0]{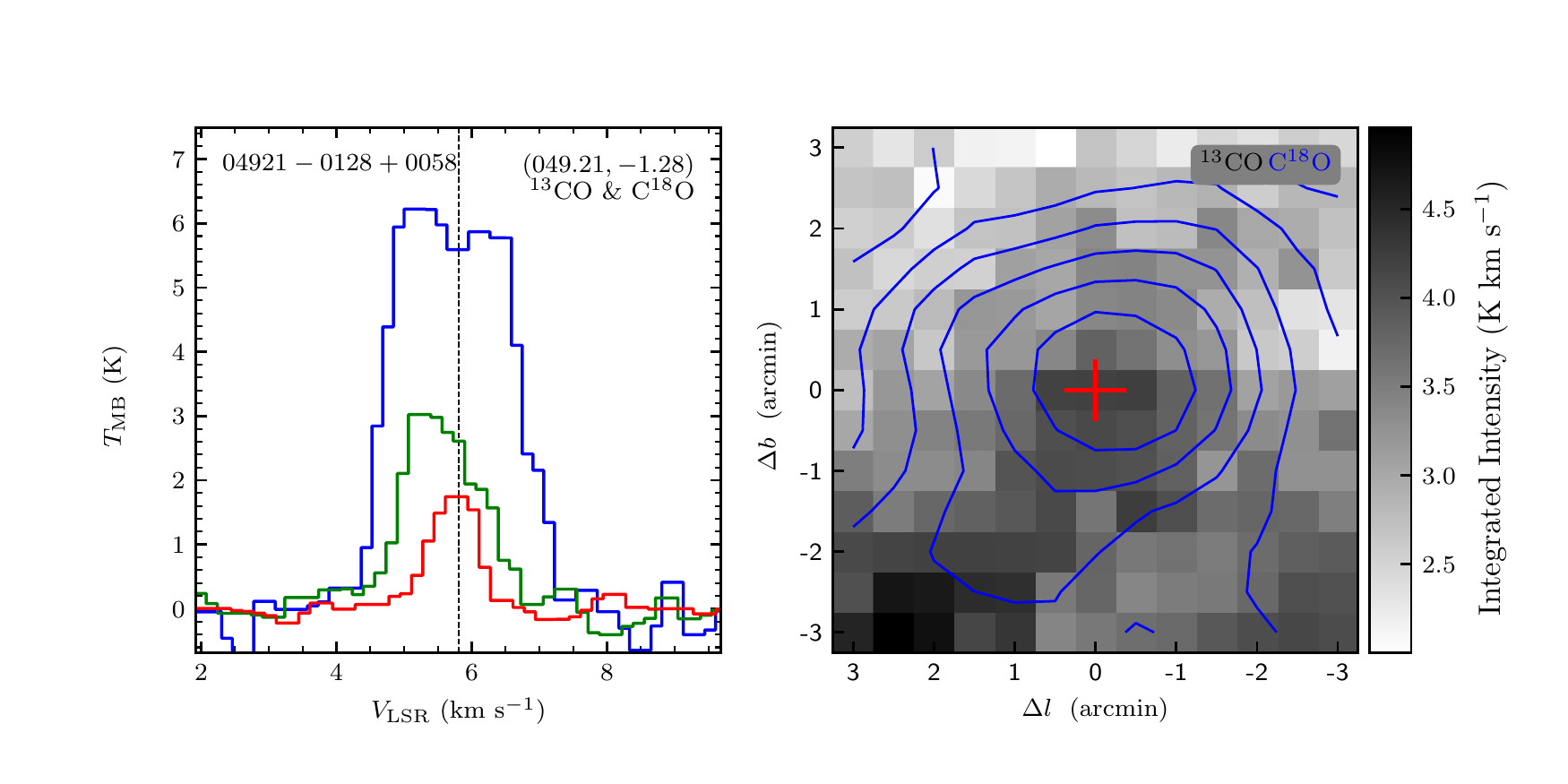}
\includegraphics[width=9.0cm,angle=0]{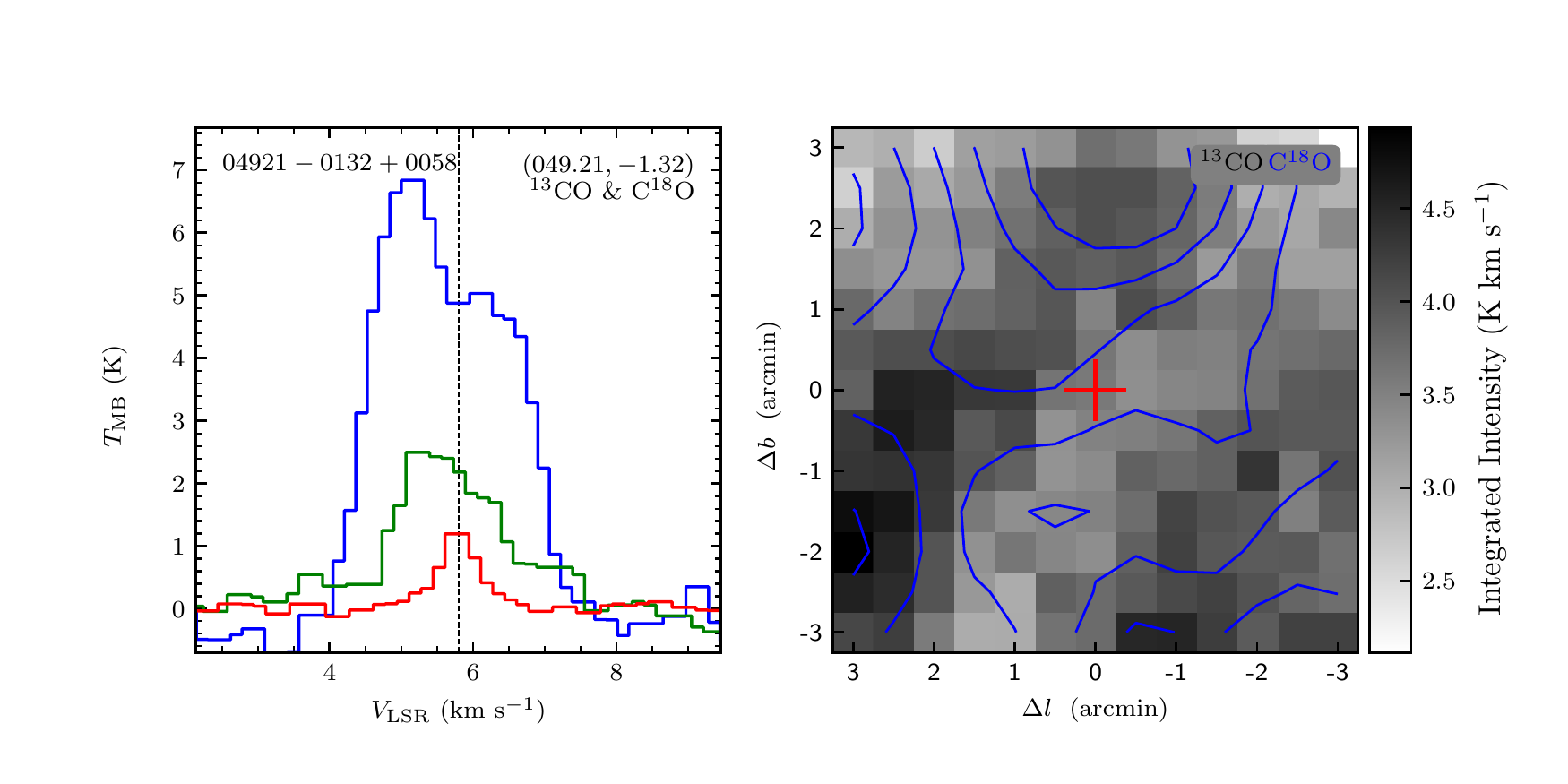}
\end{figure}
\clearpage

\begin{figure}
\includegraphics[width=9.0cm,angle=0]{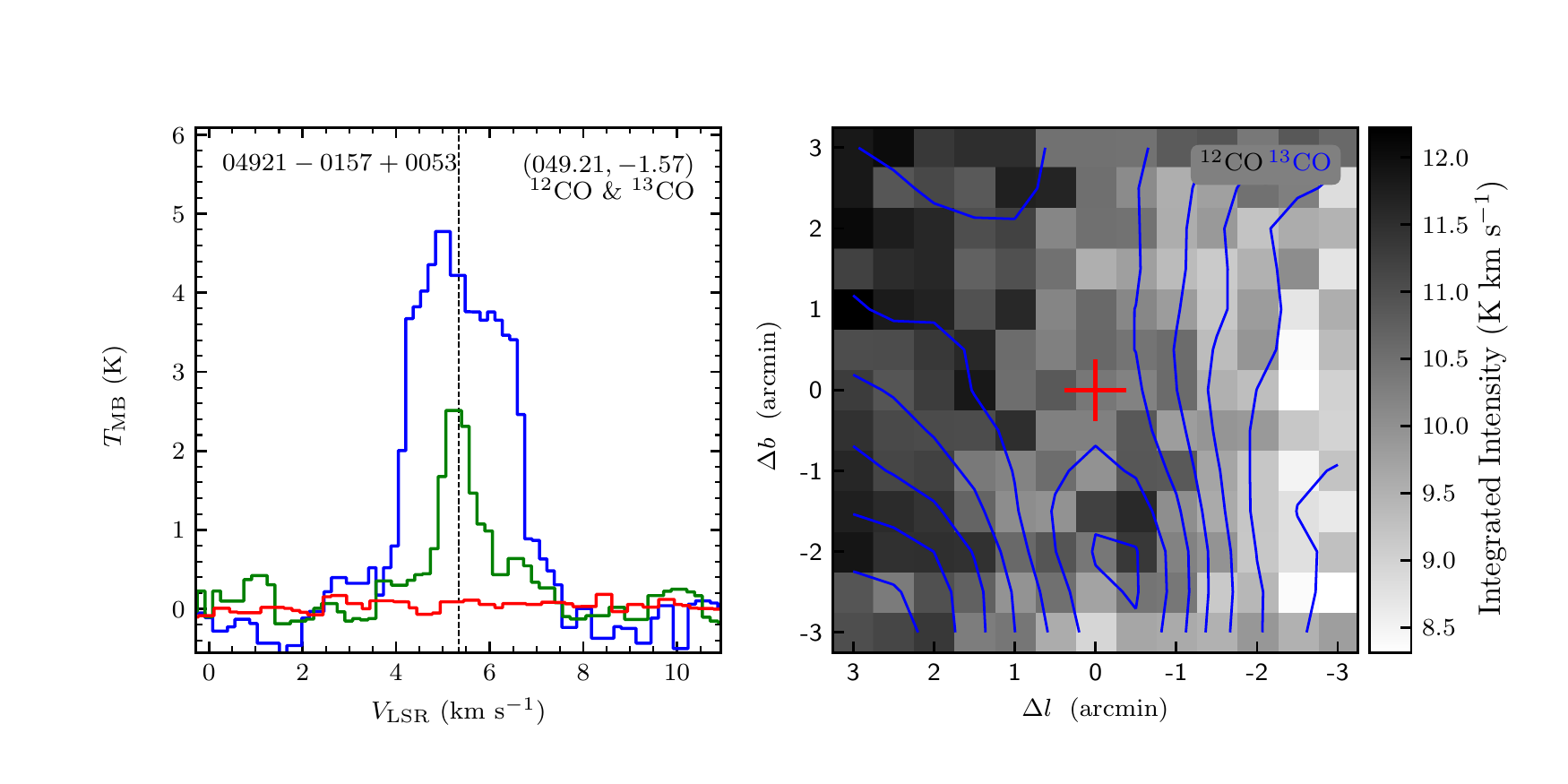}
\includegraphics[width=9.0cm,angle=0]{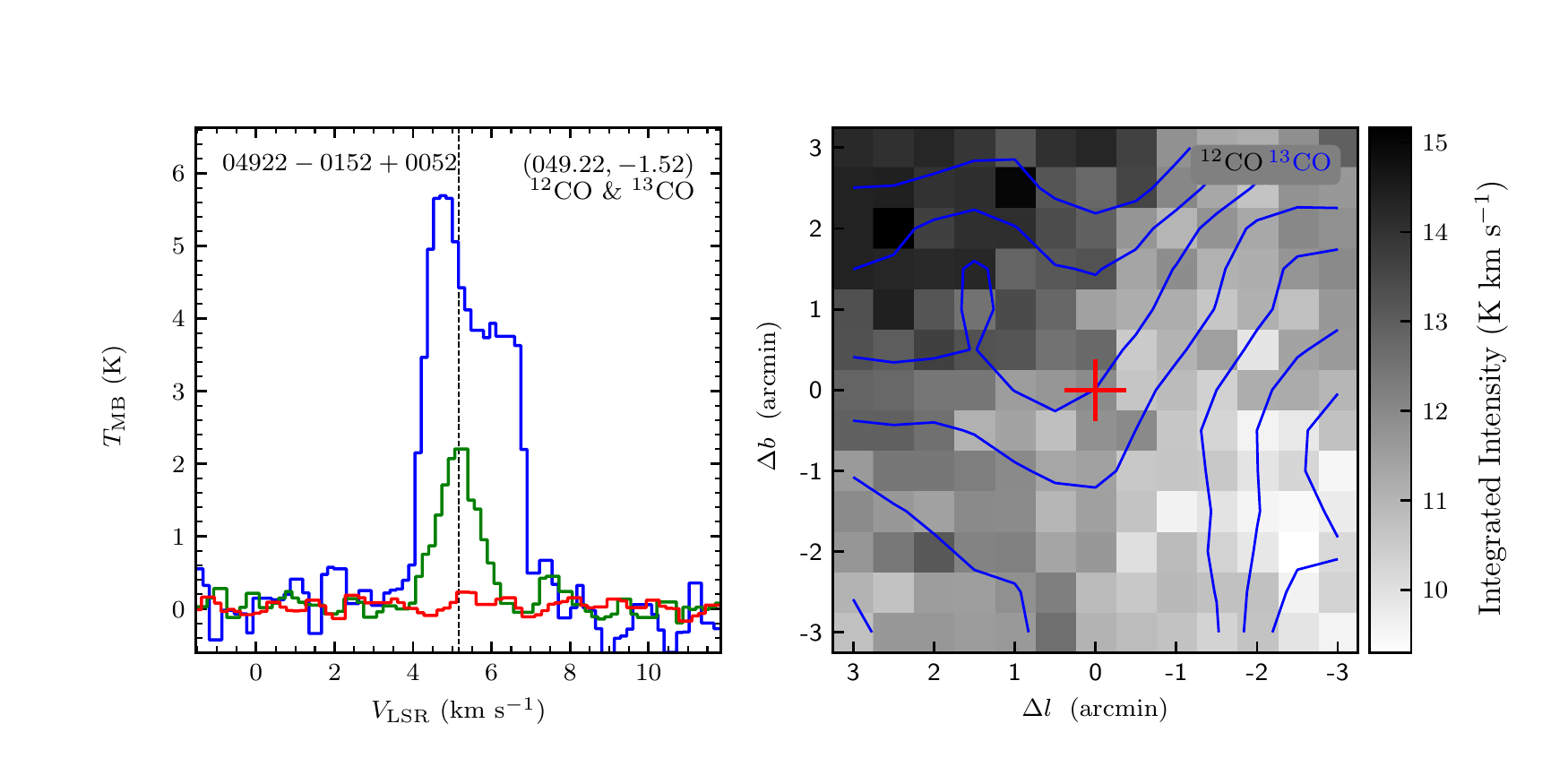}
\vspace{-0.5cm}

\includegraphics[width=9.0cm,angle=0]{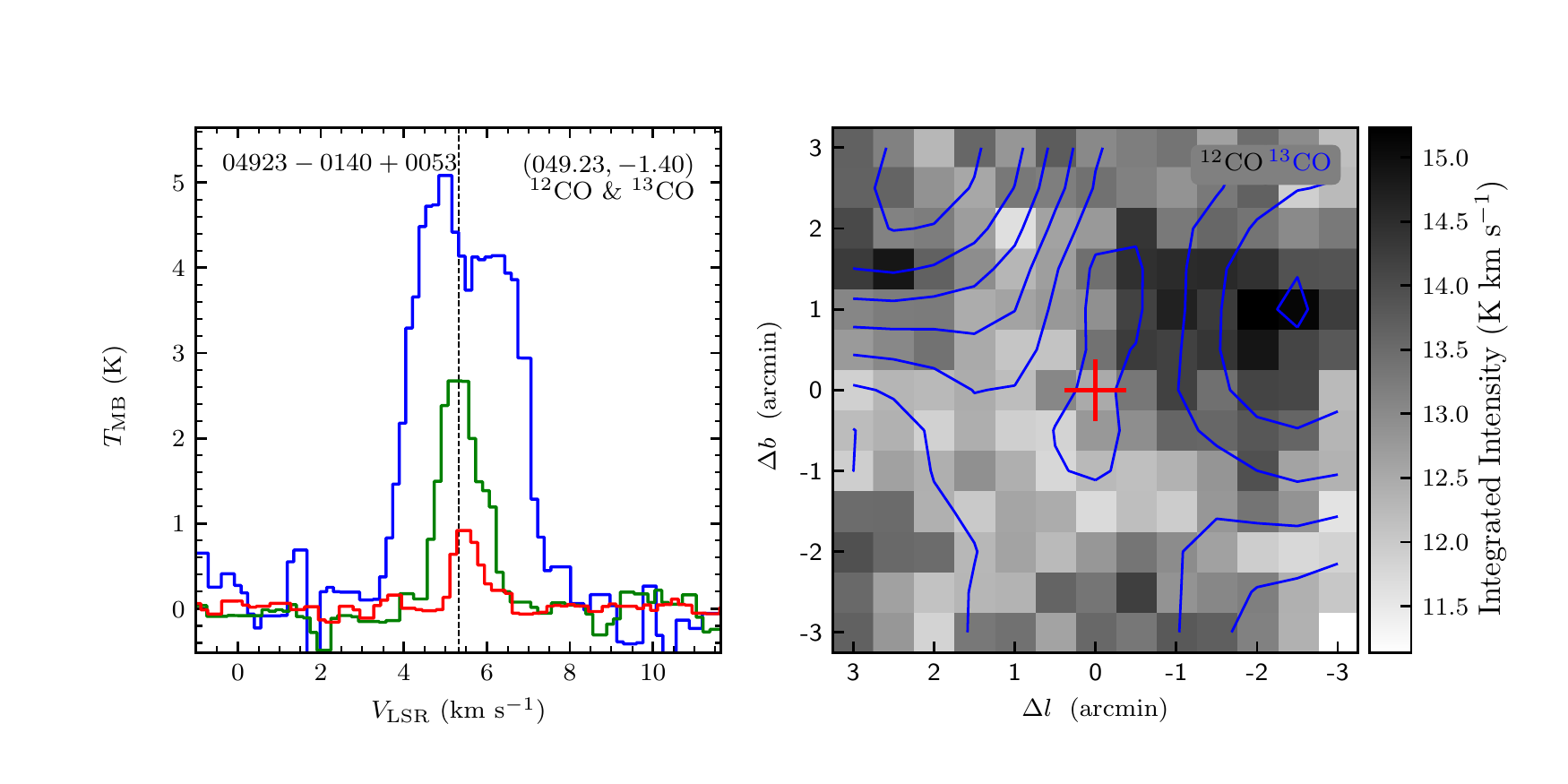}
\includegraphics[width=9.0cm,angle=0]{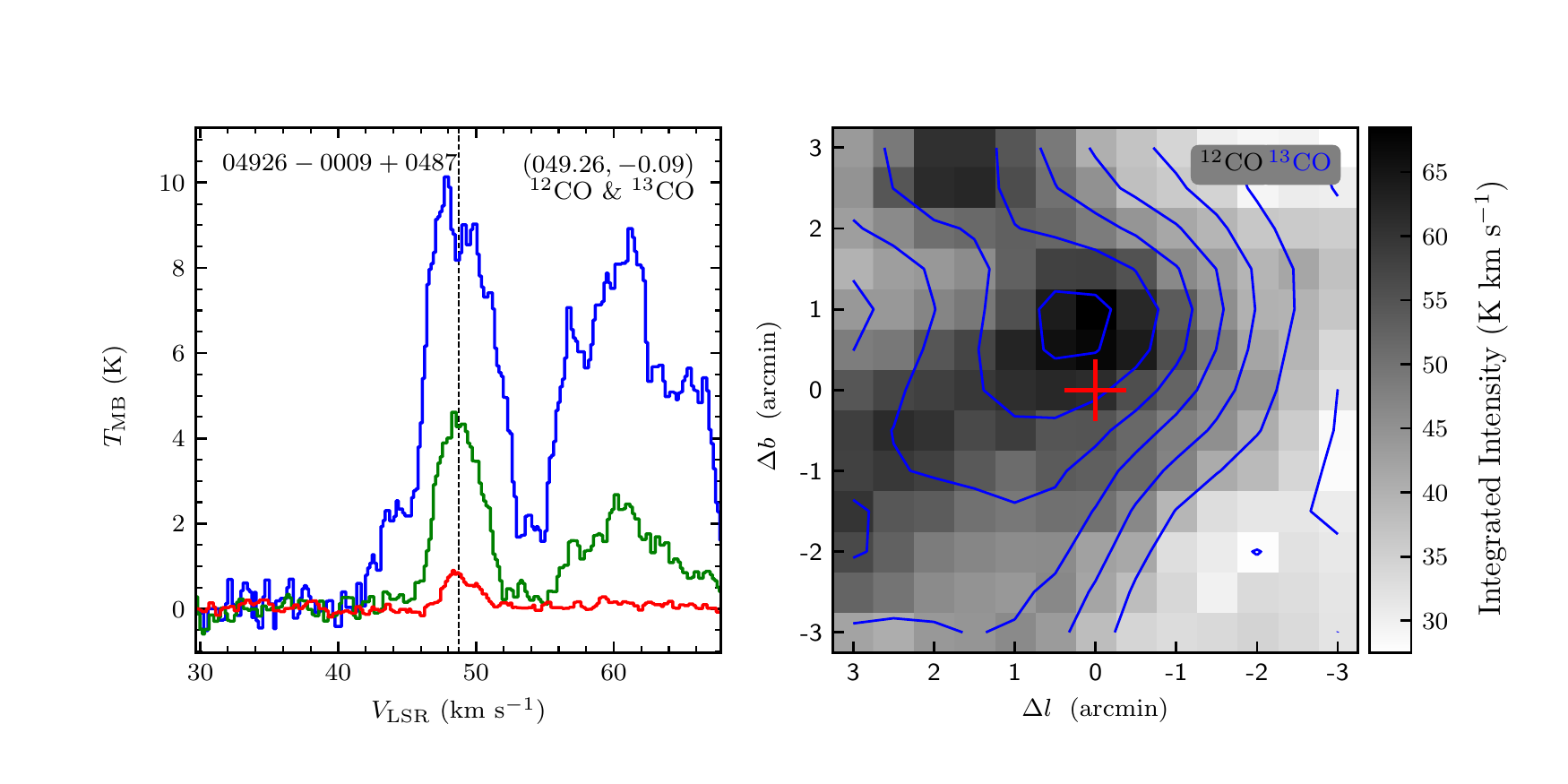}
\vspace{-0.5cm}

\includegraphics[width=9.0cm,angle=0]{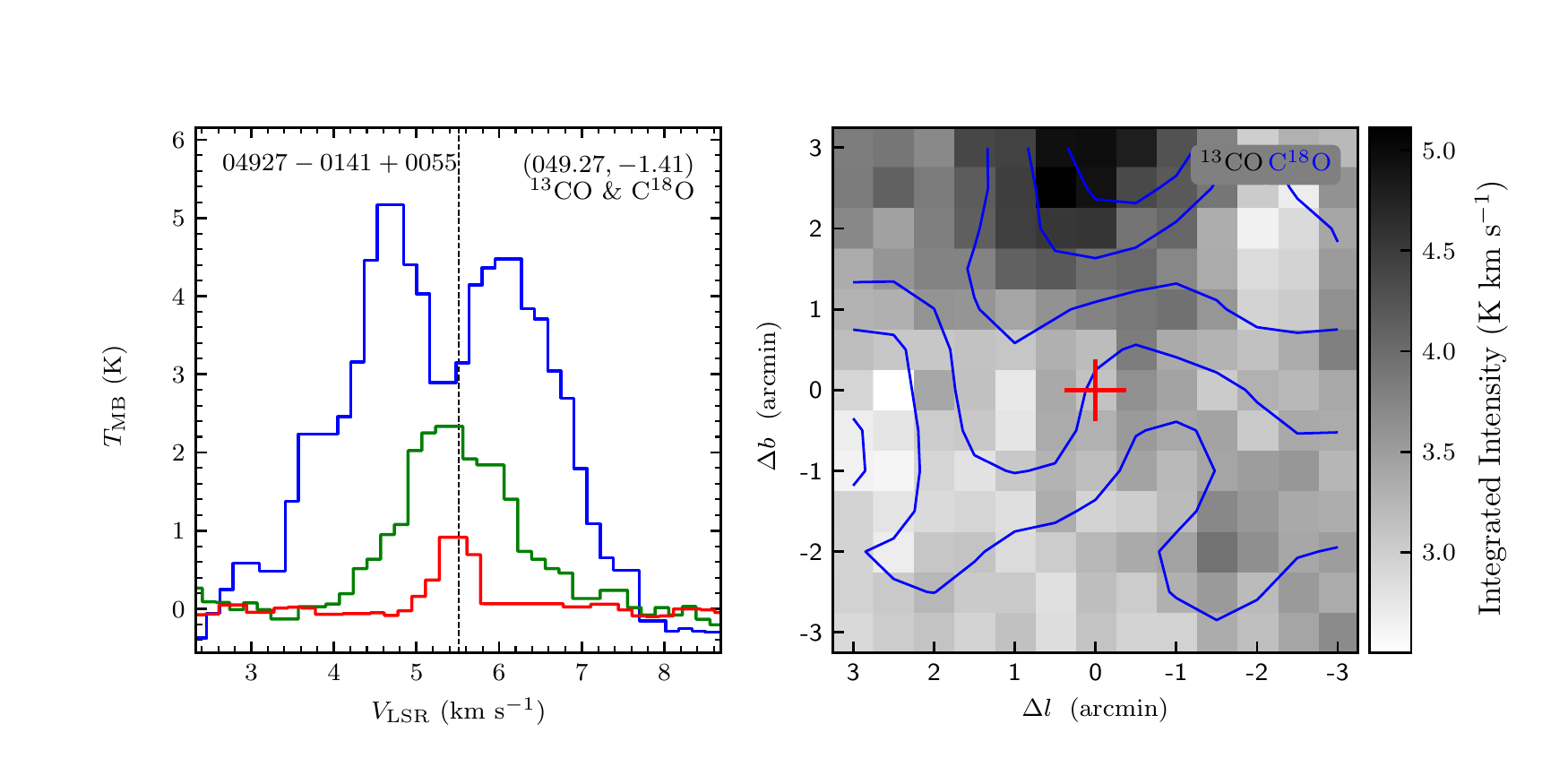}
\includegraphics[width=9.0cm,angle=0]{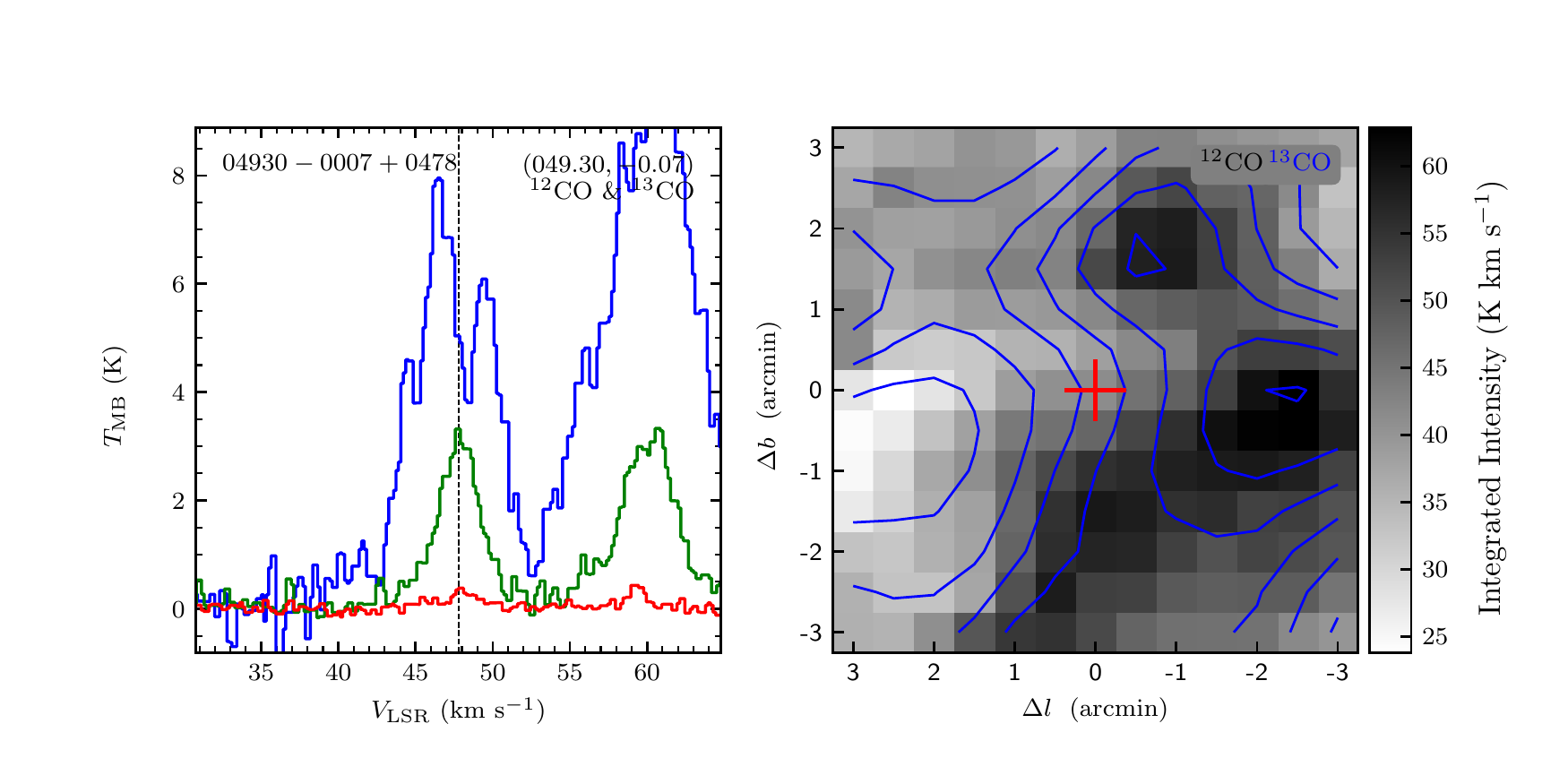}
\vspace{-0.5cm}

\includegraphics[width=9.0cm,angle=0]{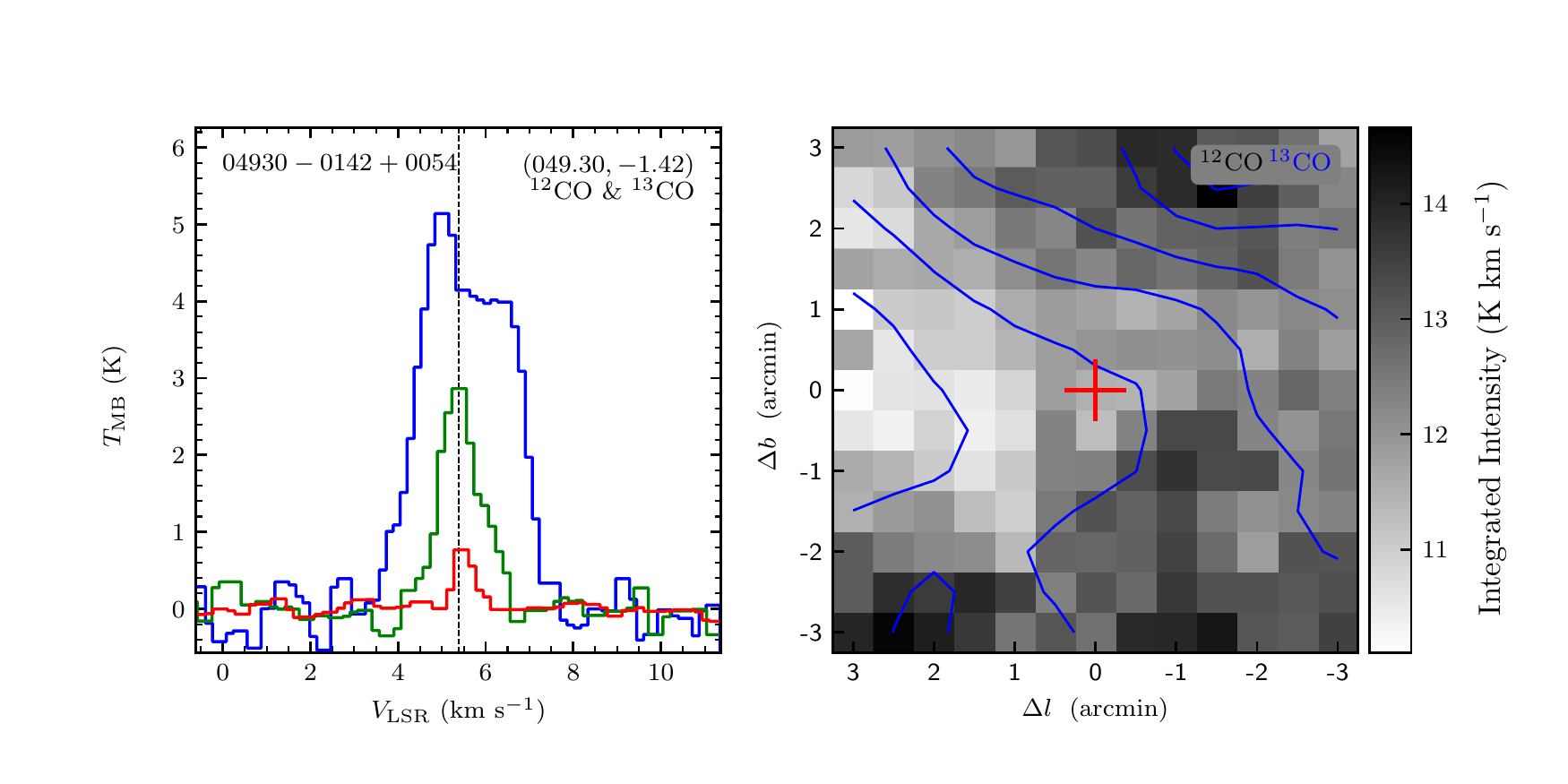}
\includegraphics[width=9.0cm,angle=0]{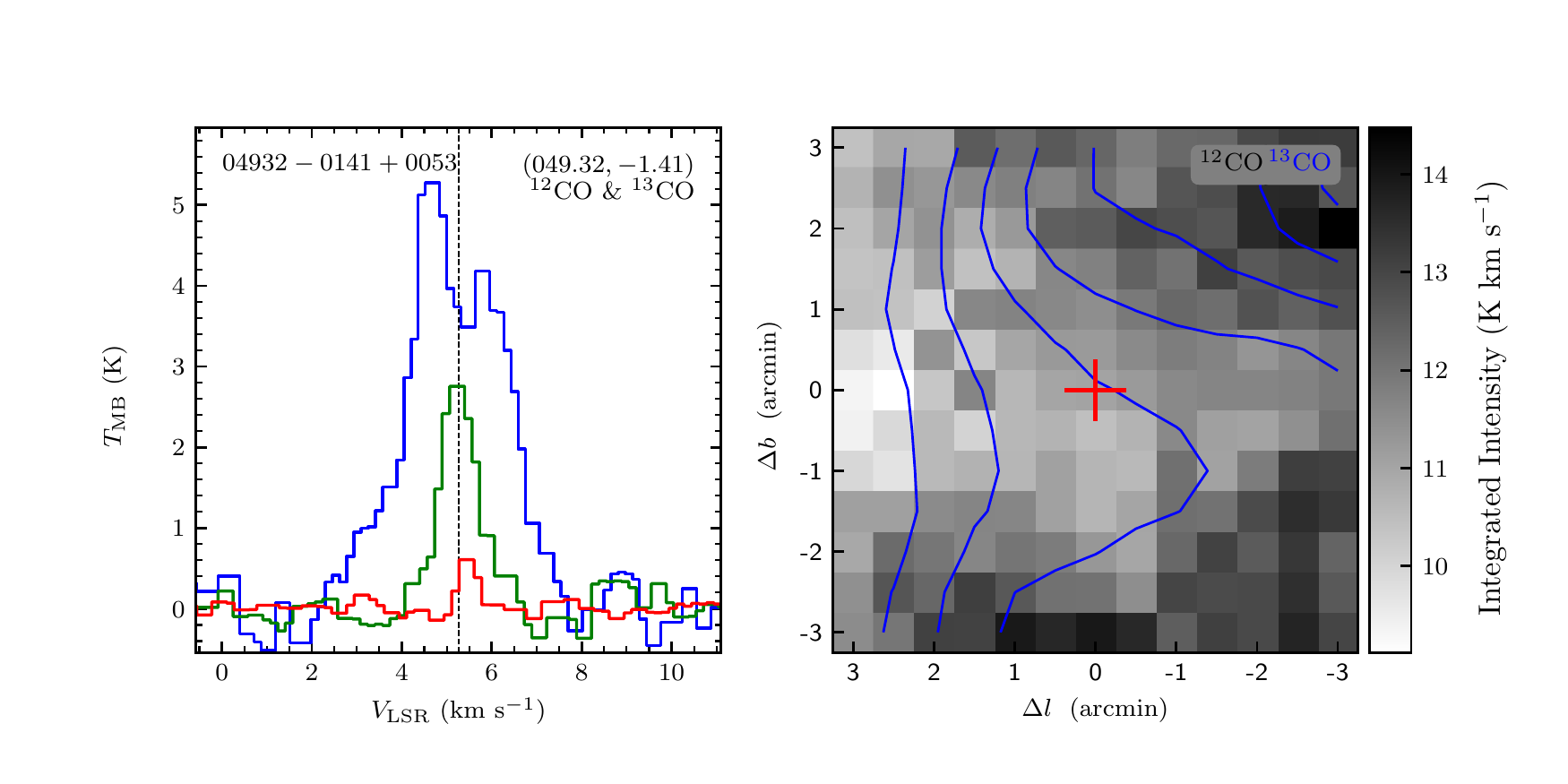}
\vspace{-0.5cm}

\includegraphics[width=9.0cm,angle=0]{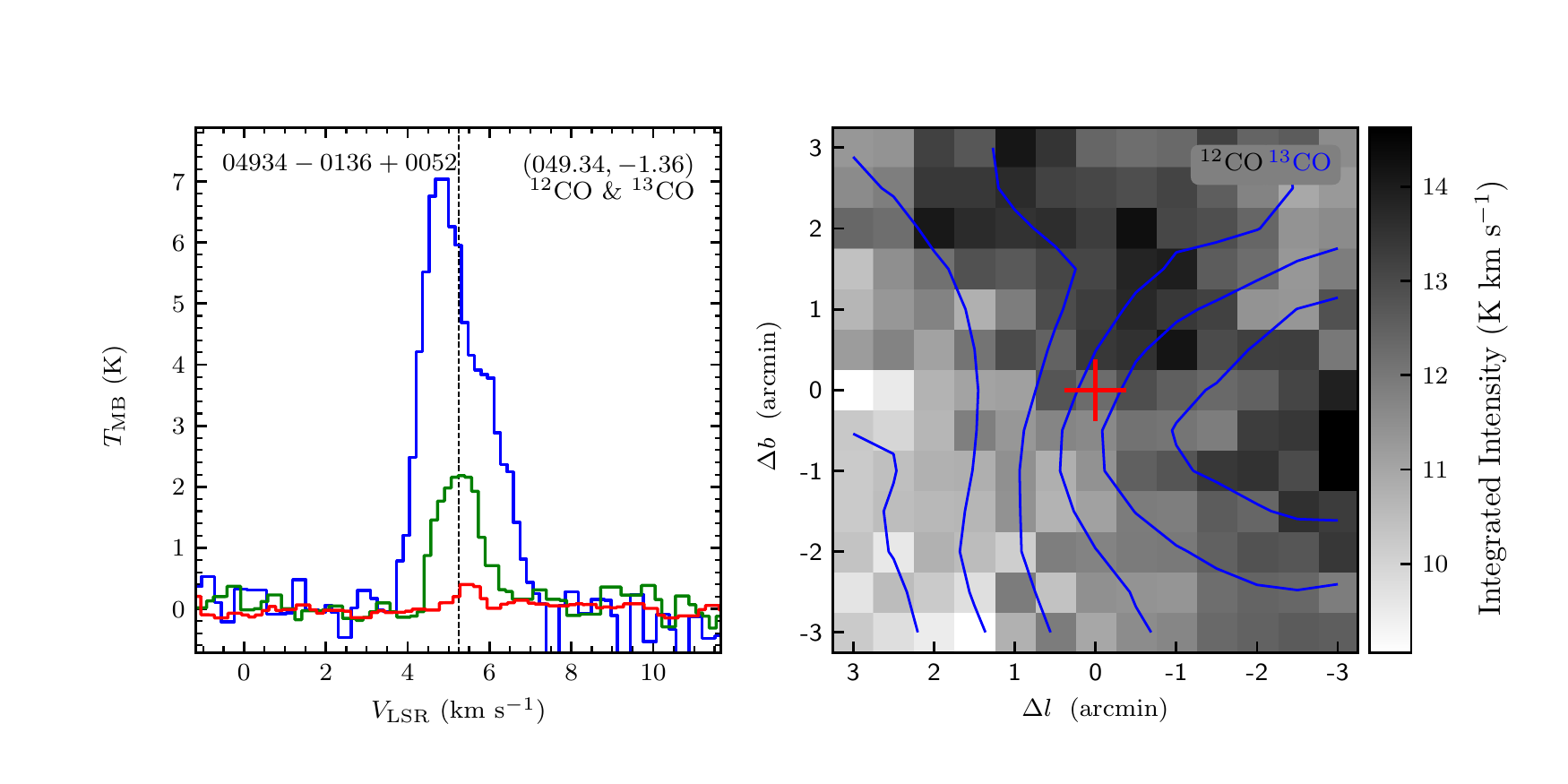}
\includegraphics[width=9.0cm,angle=0]{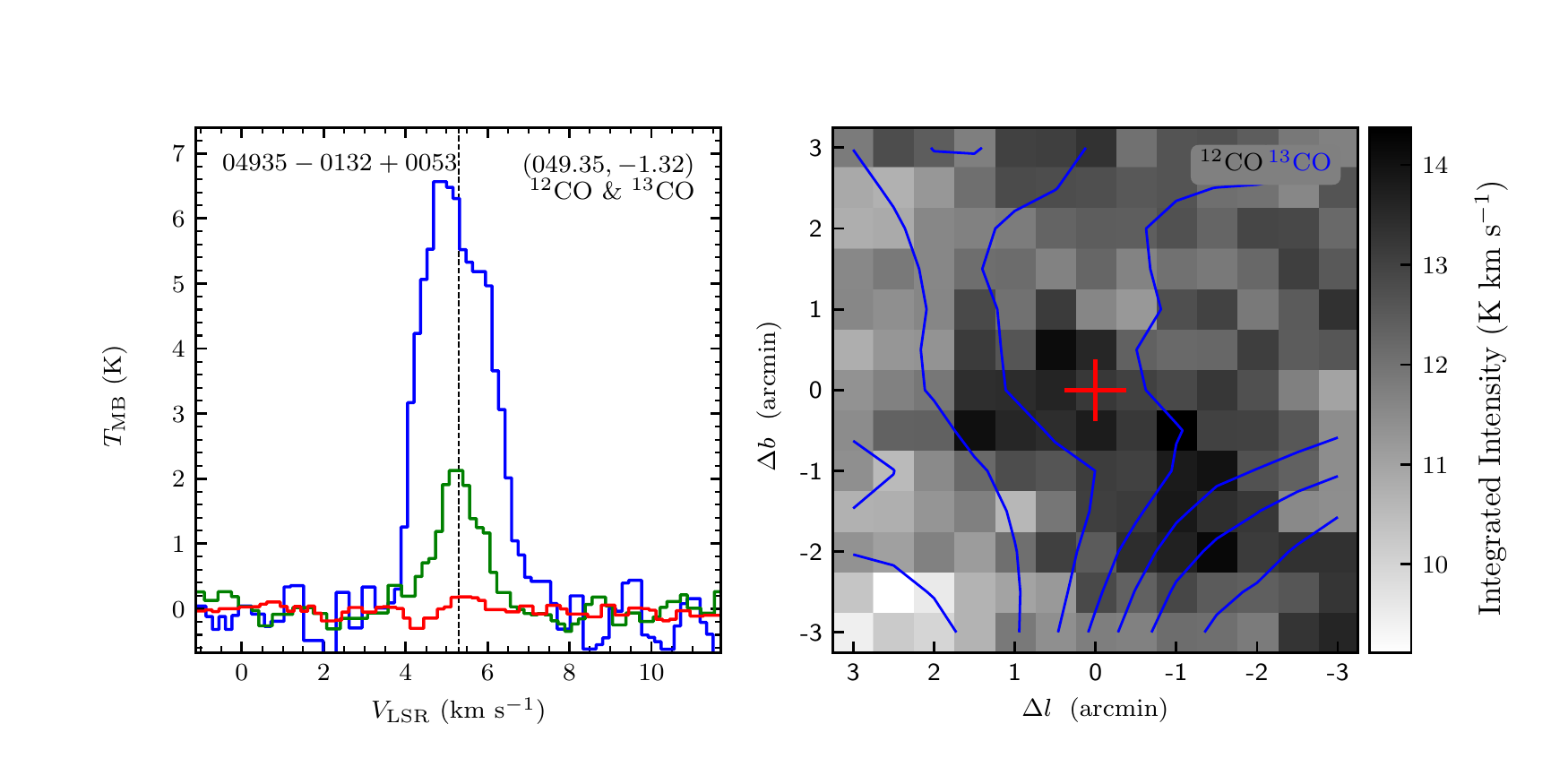}
\end{figure}
\clearpage

\begin{figure}
\includegraphics[width=9.0cm,angle=0]{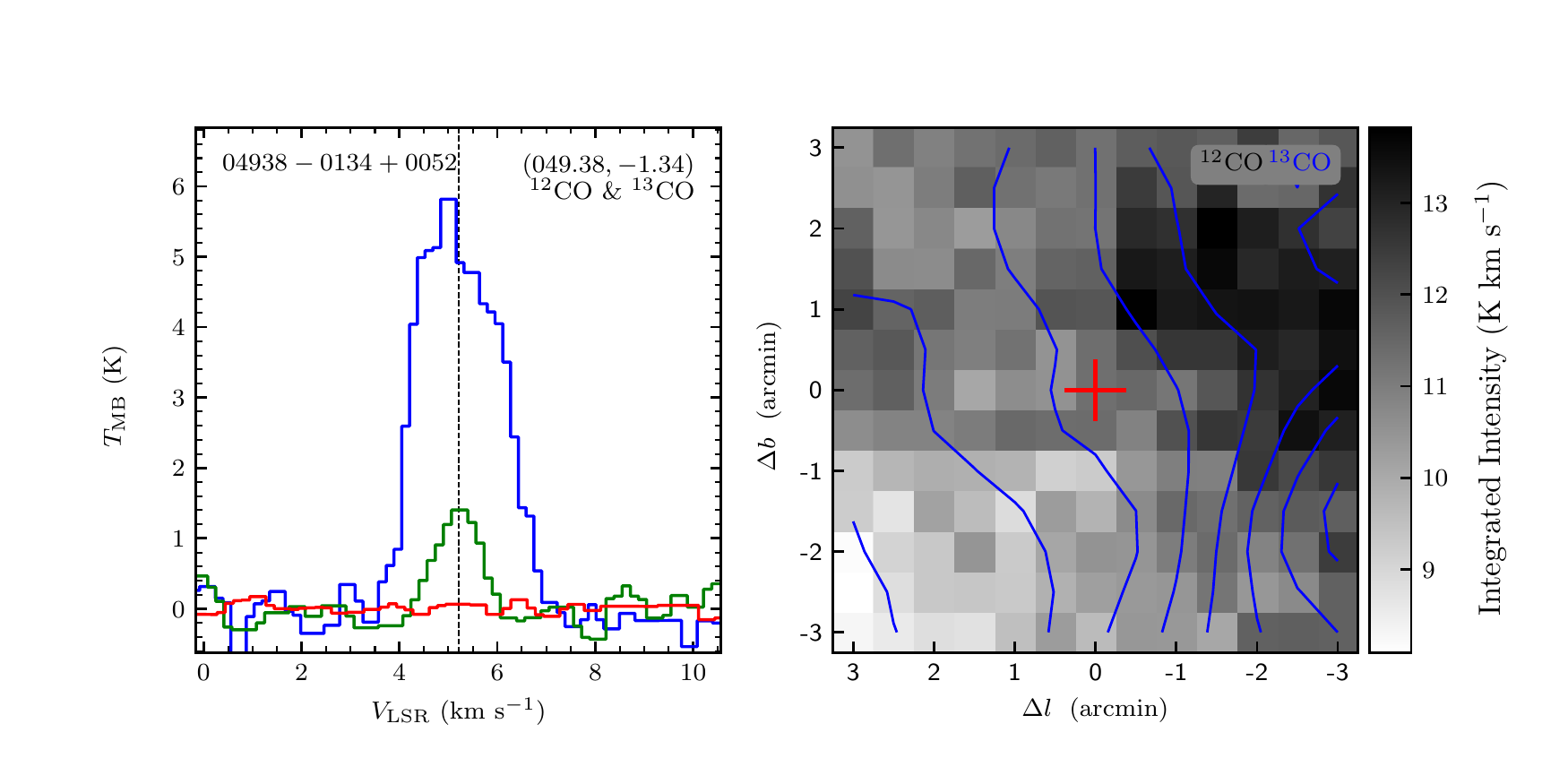}
\includegraphics[width=9.0cm,angle=0]{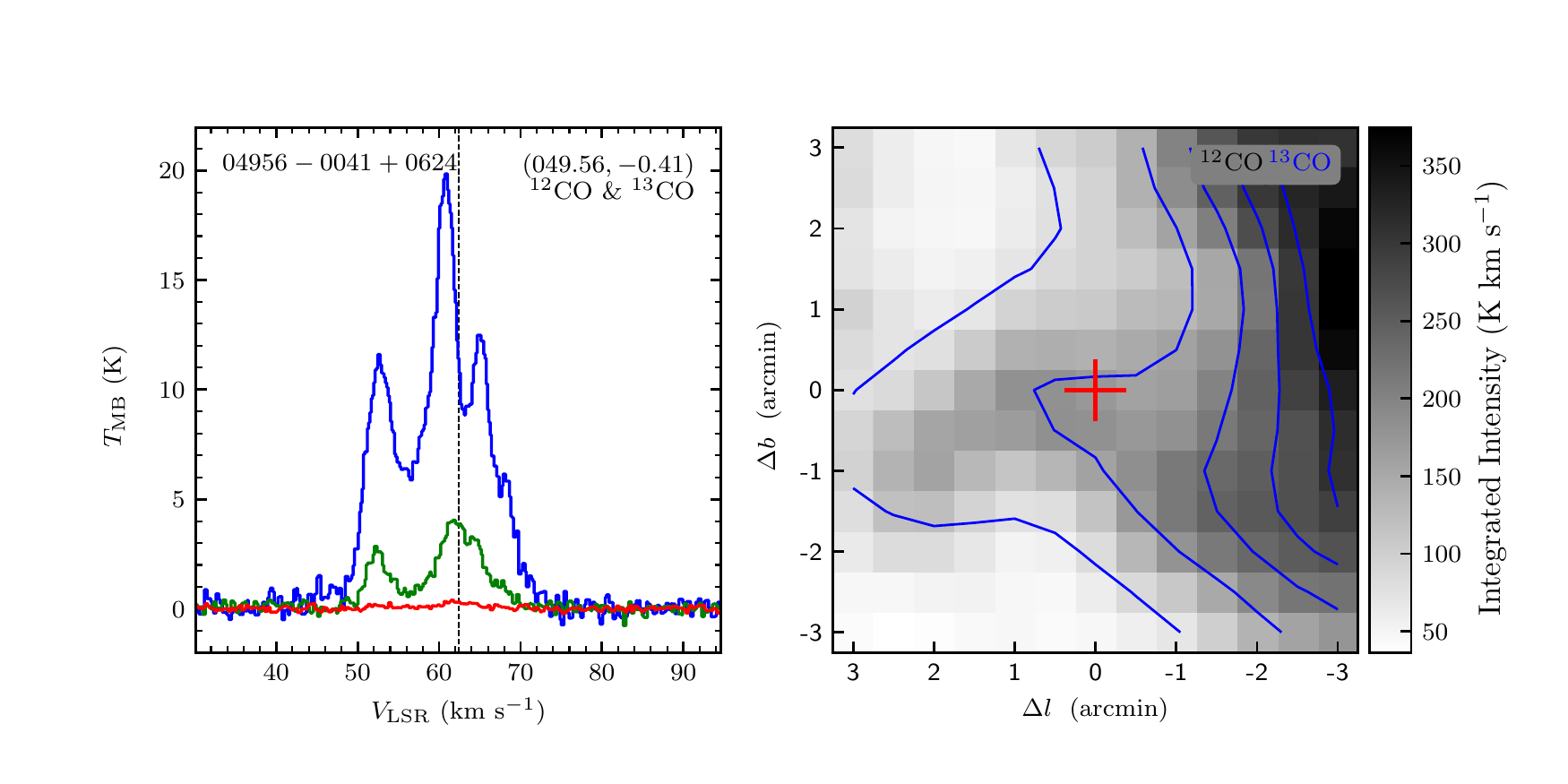}
\vspace{-0.5cm}

\includegraphics[width=9.0cm,angle=0]{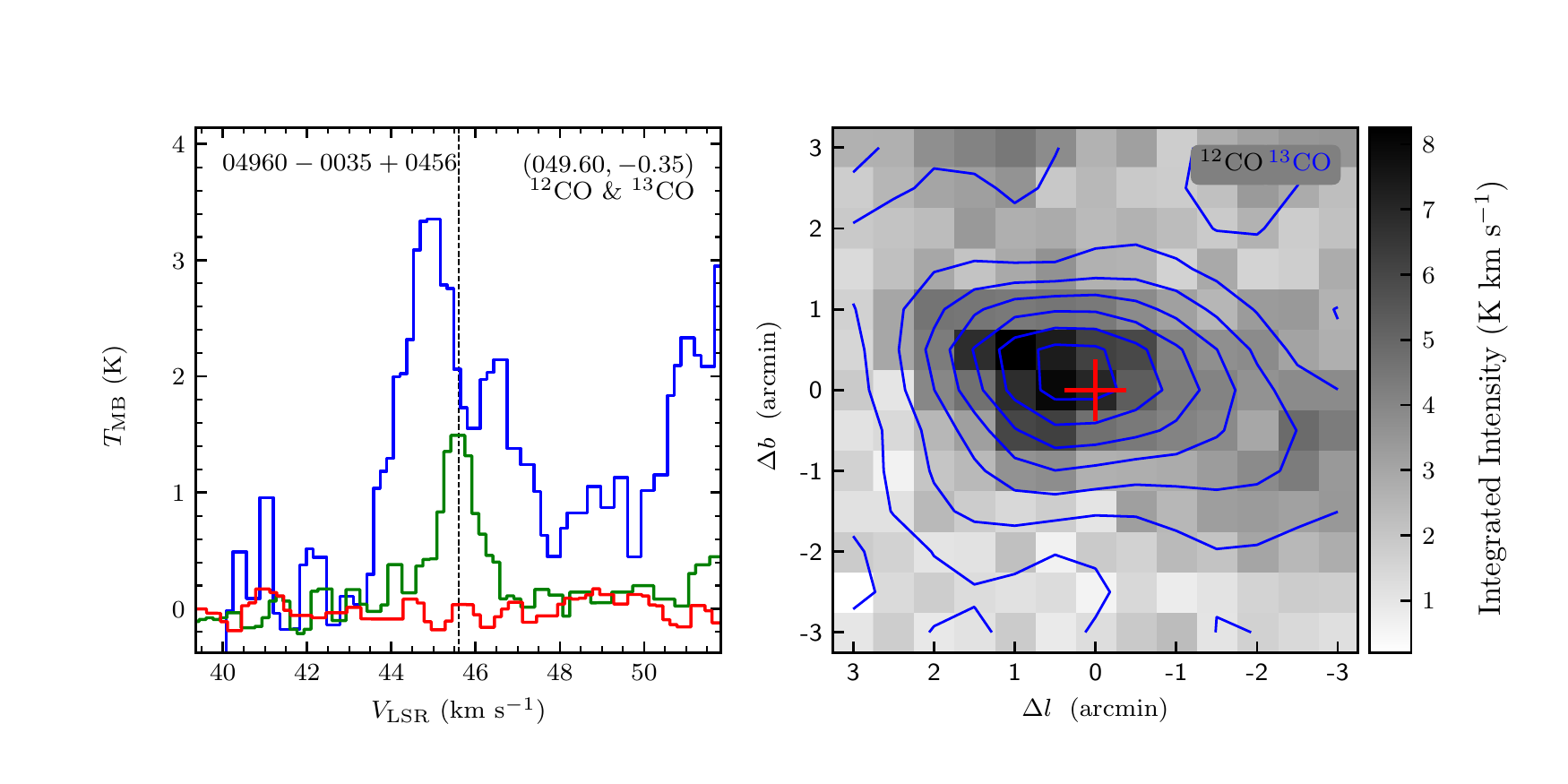}
\includegraphics[width=9.0cm,angle=0]{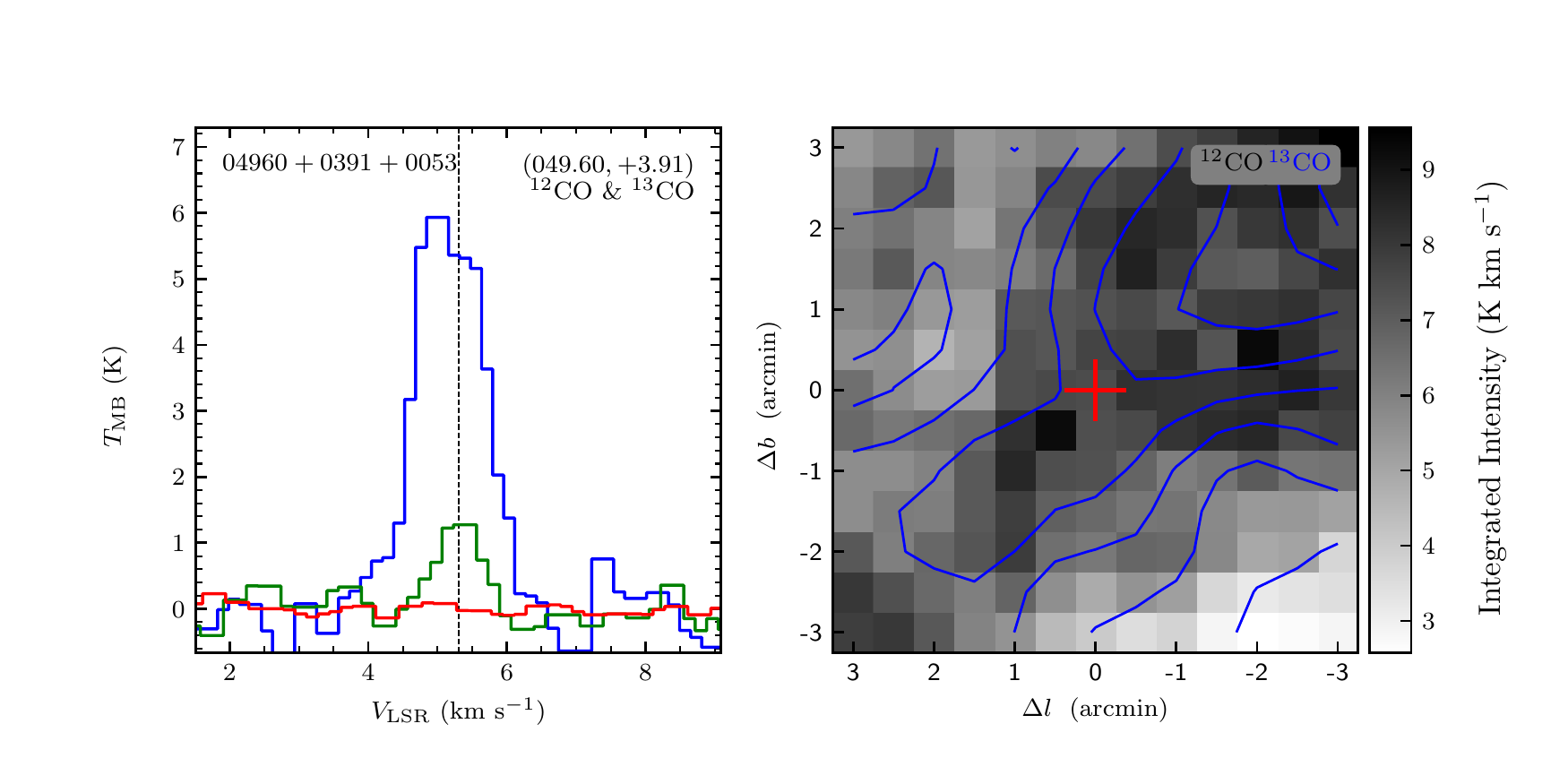}
\vspace{-0.5cm}

\includegraphics[width=9.0cm,angle=0]{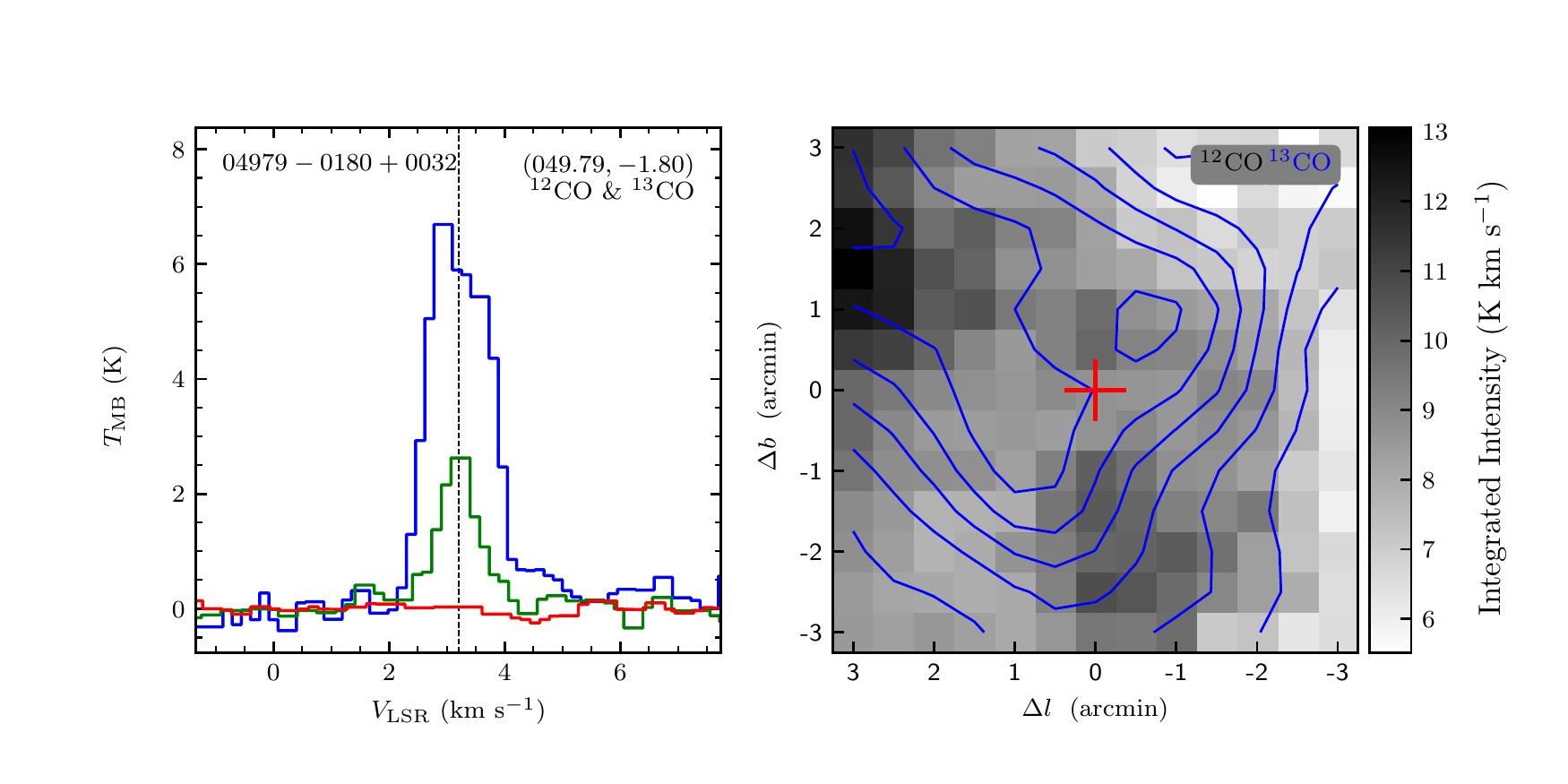}
\includegraphics[width=9.0cm,angle=0]{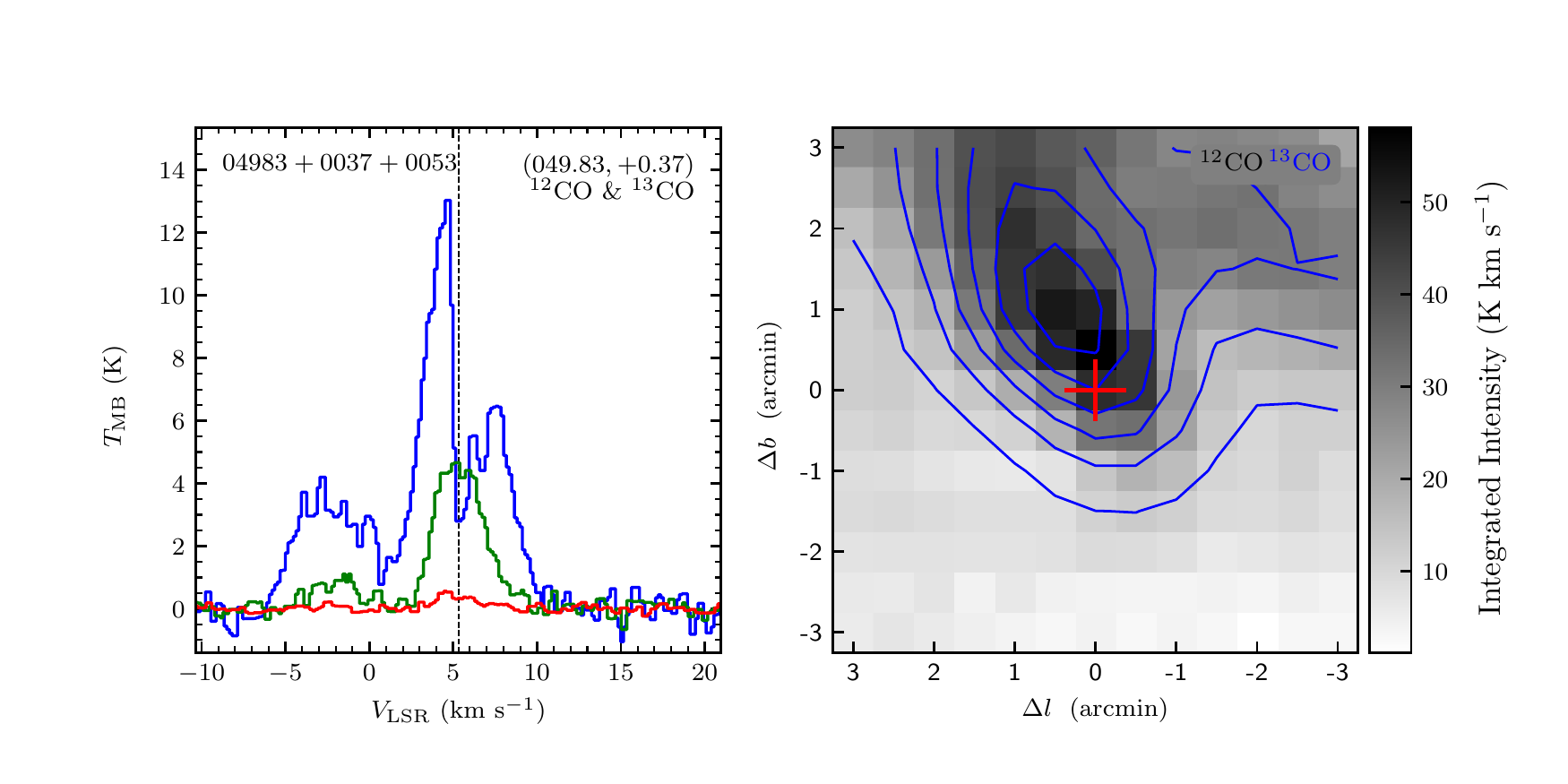}
\vspace{-0.5cm}

\includegraphics[width=9.0cm,angle=0]{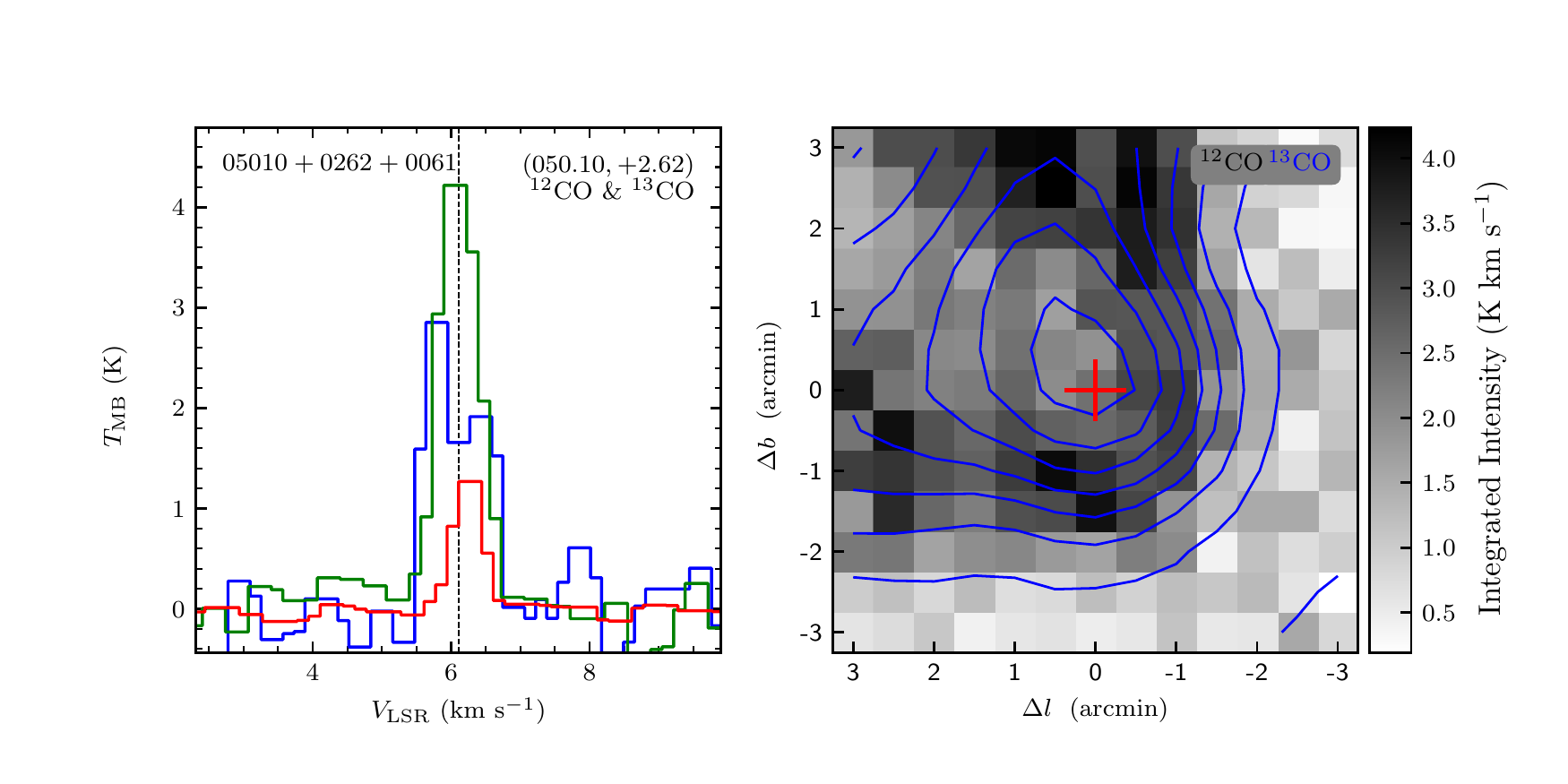}
\includegraphics[width=9.0cm,angle=0]{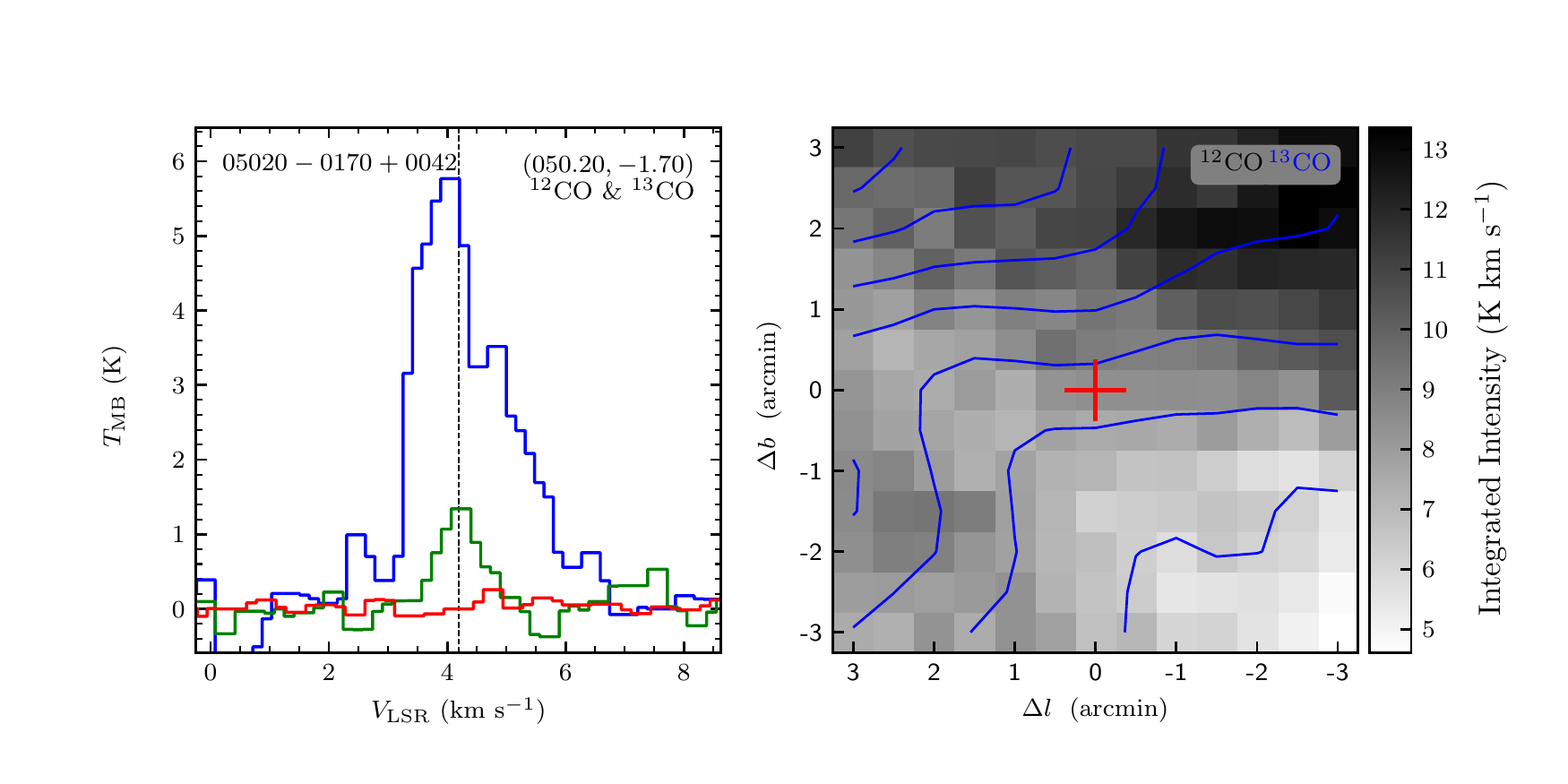}
\vspace{-0.5cm}

\includegraphics[width=9.0cm,angle=0]{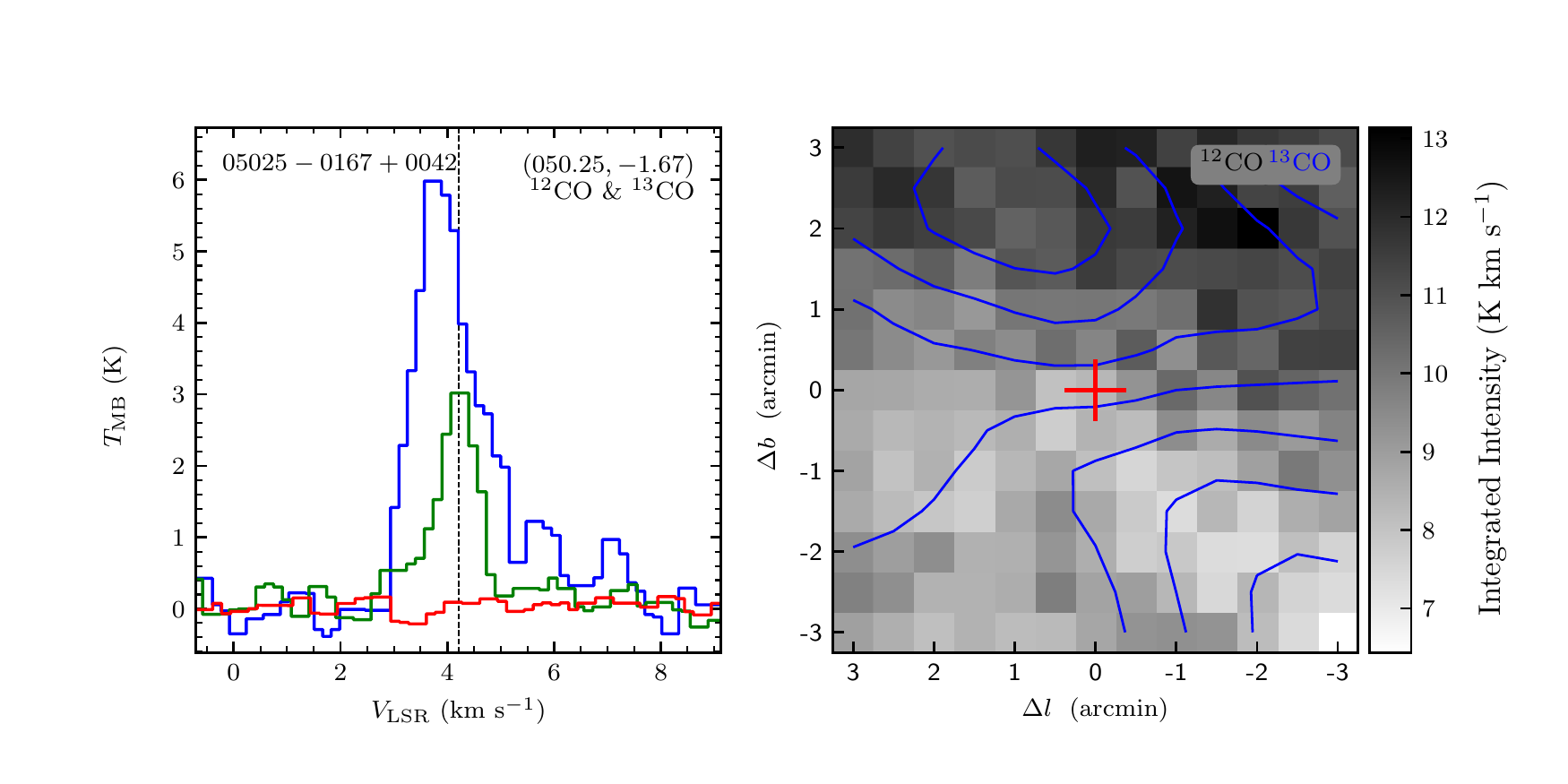}
\includegraphics[width=9.0cm,angle=0]{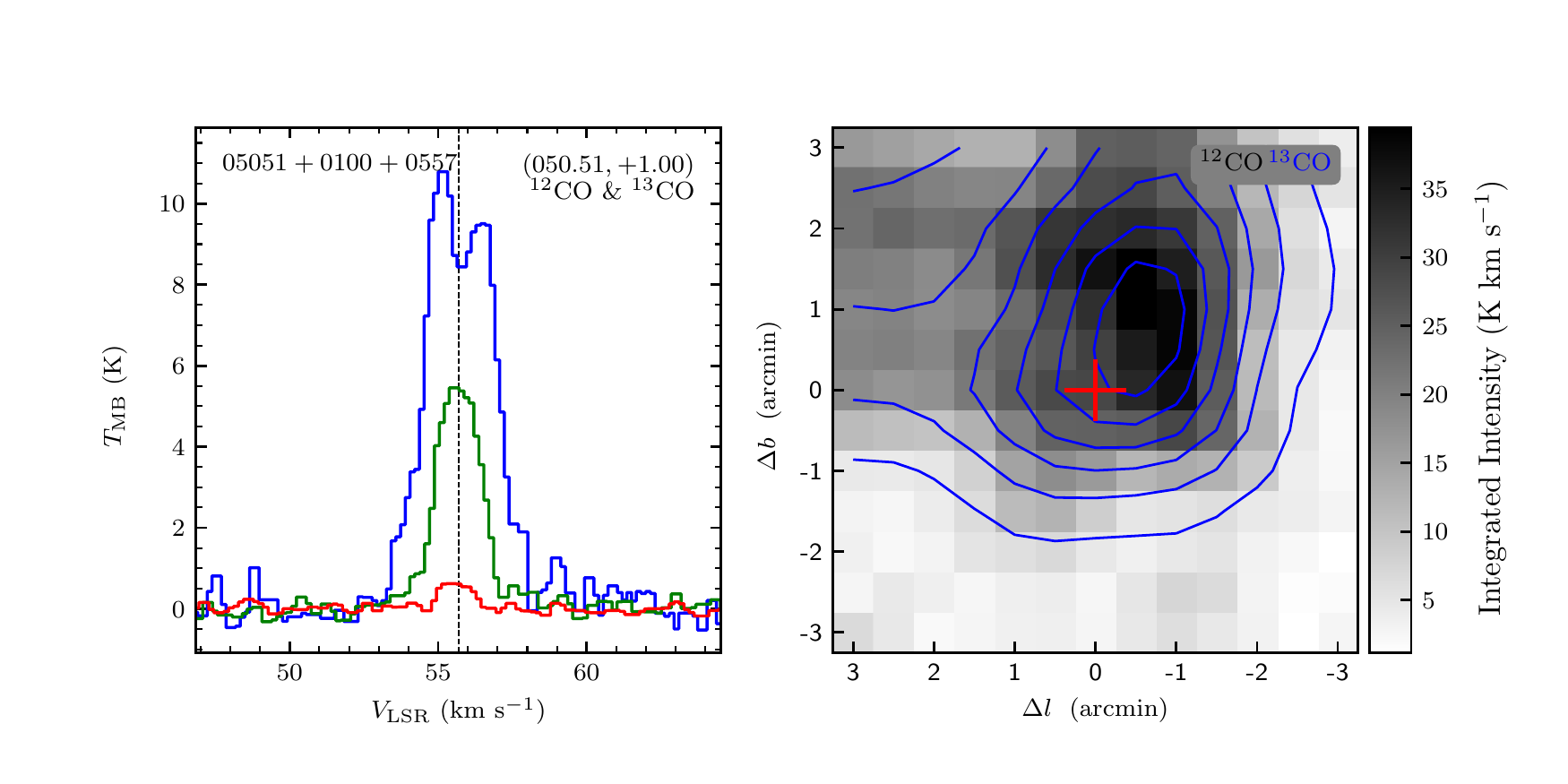}
\end{figure}
\clearpage

\begin{figure}
\includegraphics[width=9.0cm,angle=0]{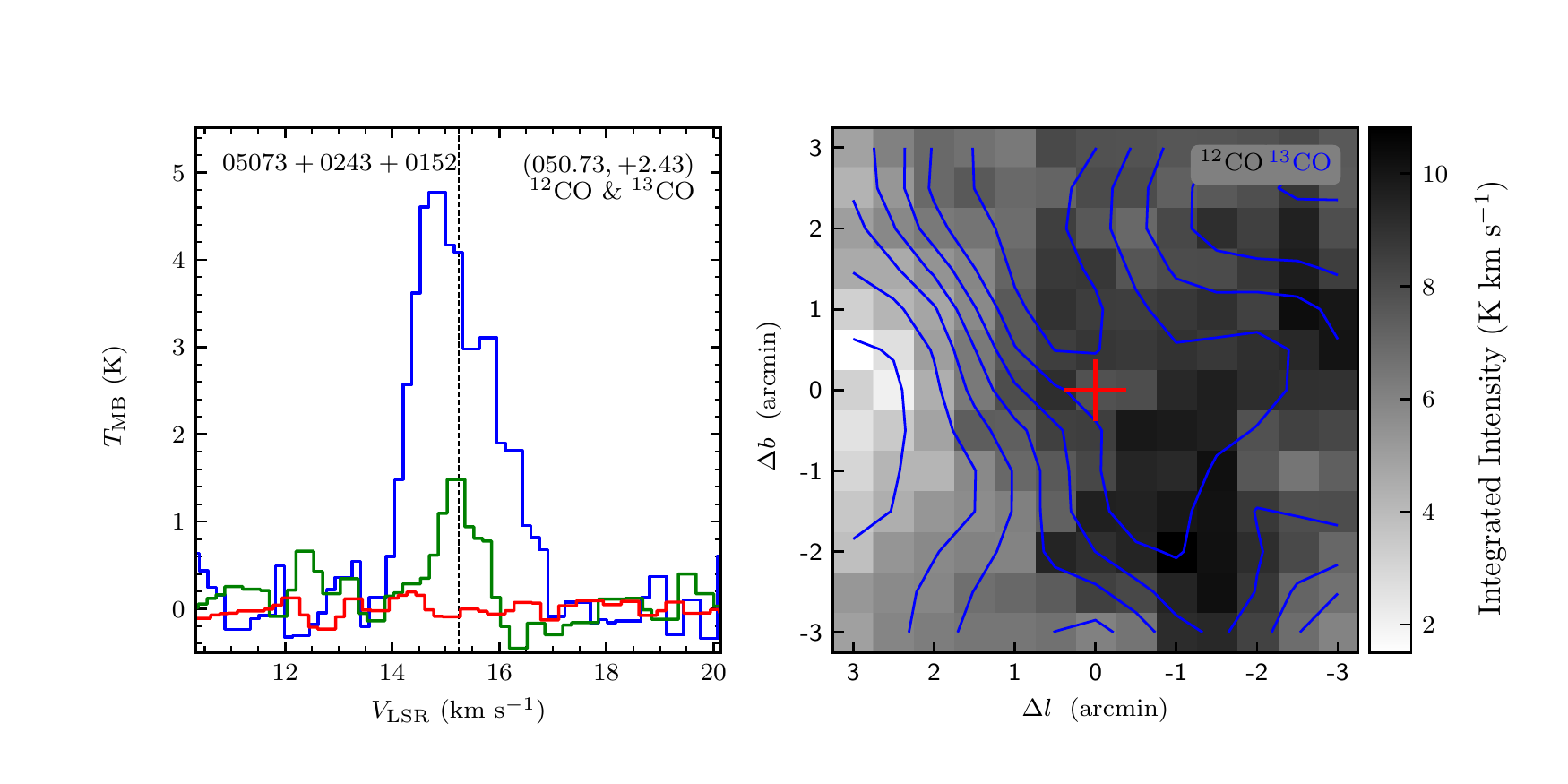}
\includegraphics[width=9.0cm,angle=0]{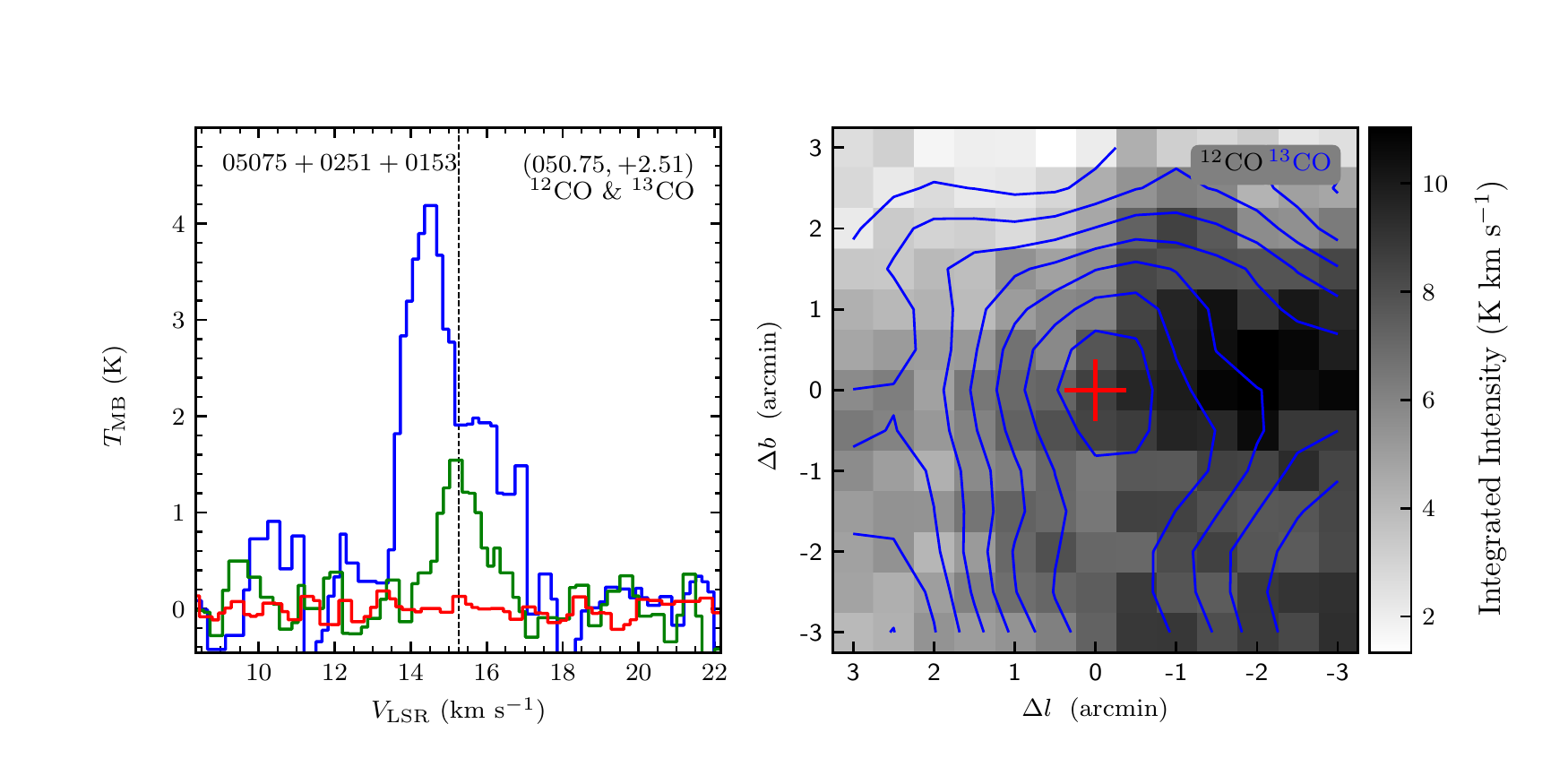}
\vspace{-0.5cm}

\includegraphics[width=9.0cm,angle=0]{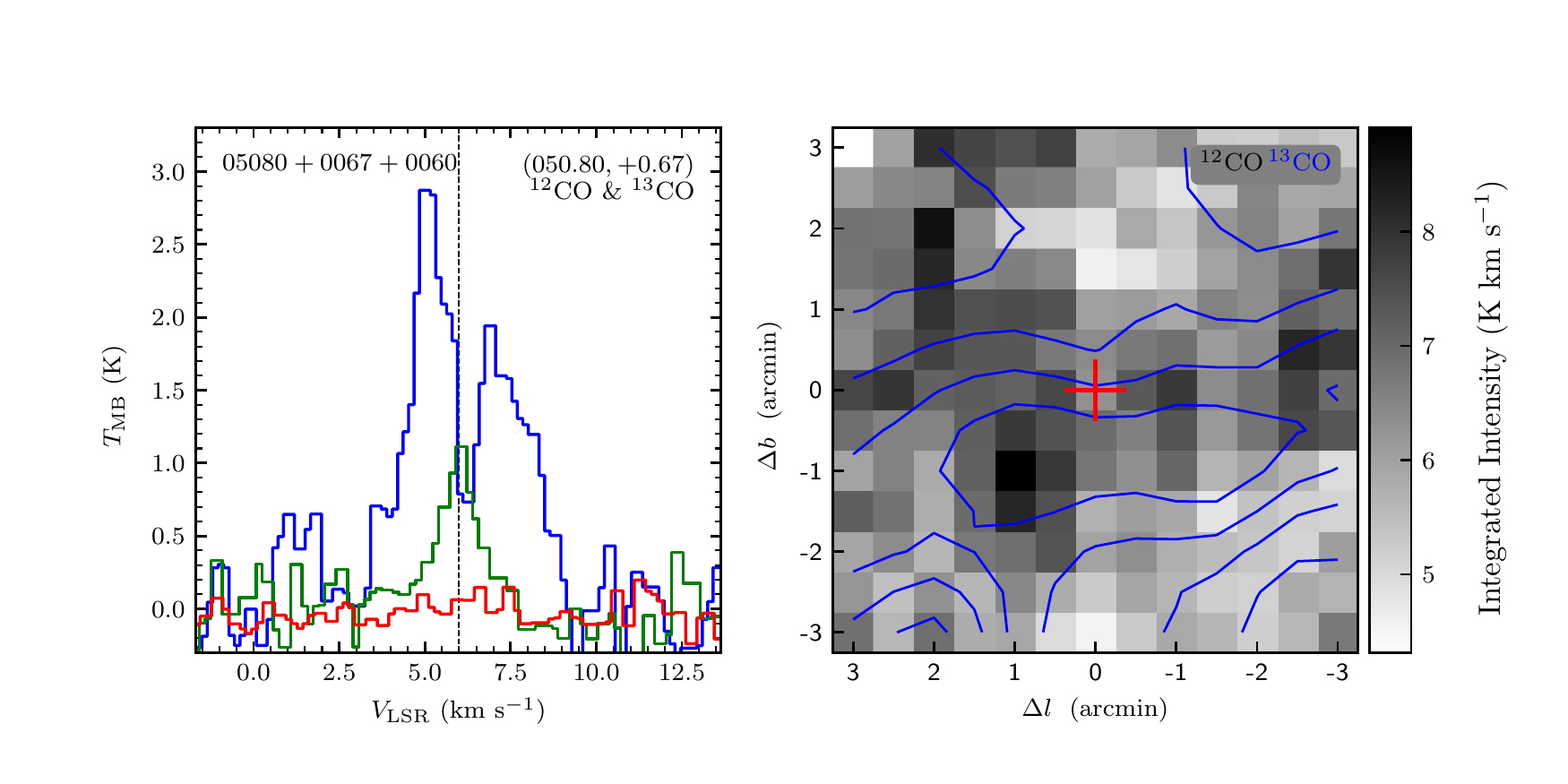}
\includegraphics[width=9.0cm,angle=0]{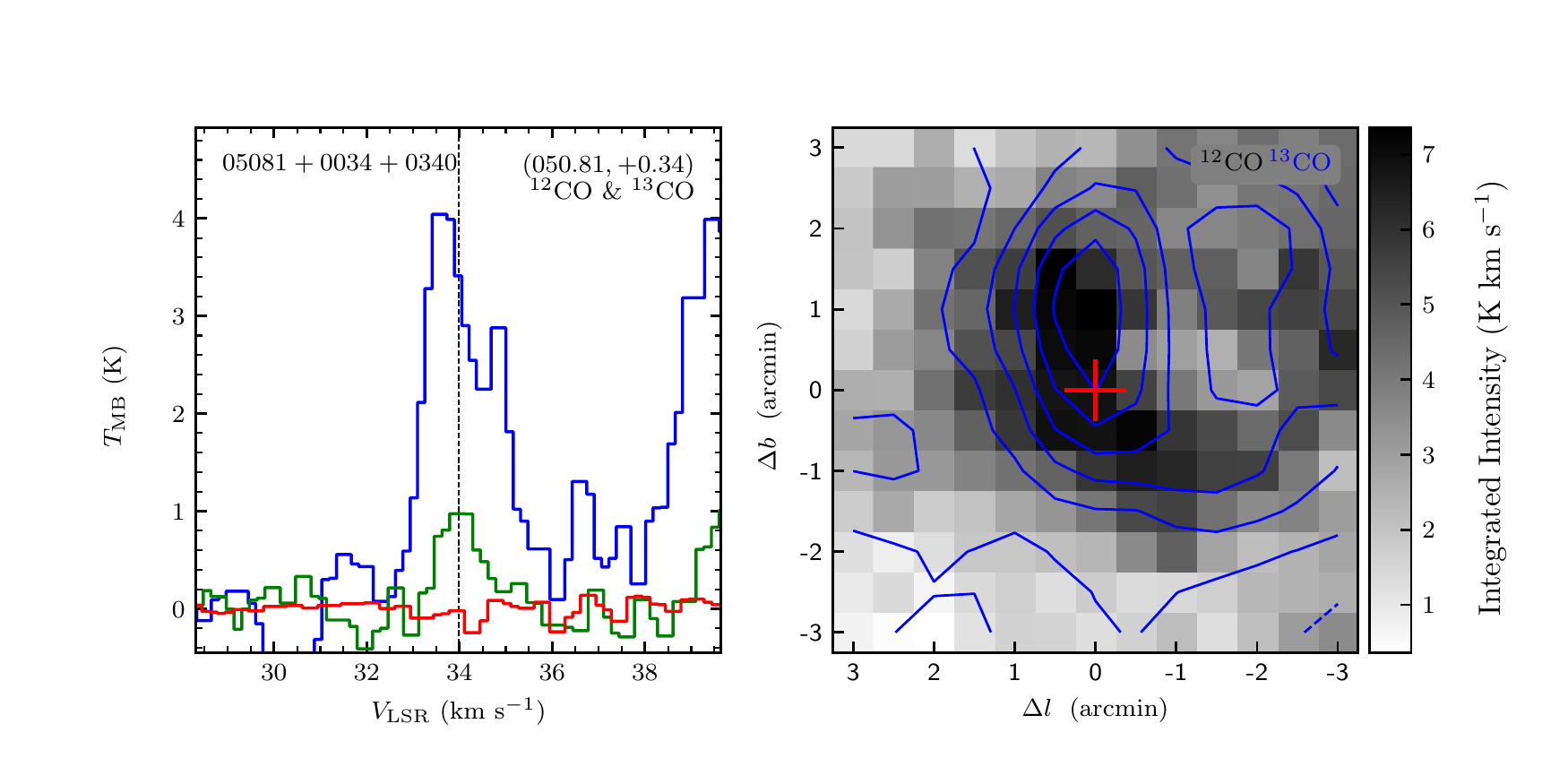}
\vspace{-0.5cm}

\includegraphics[width=9.0cm,angle=0]{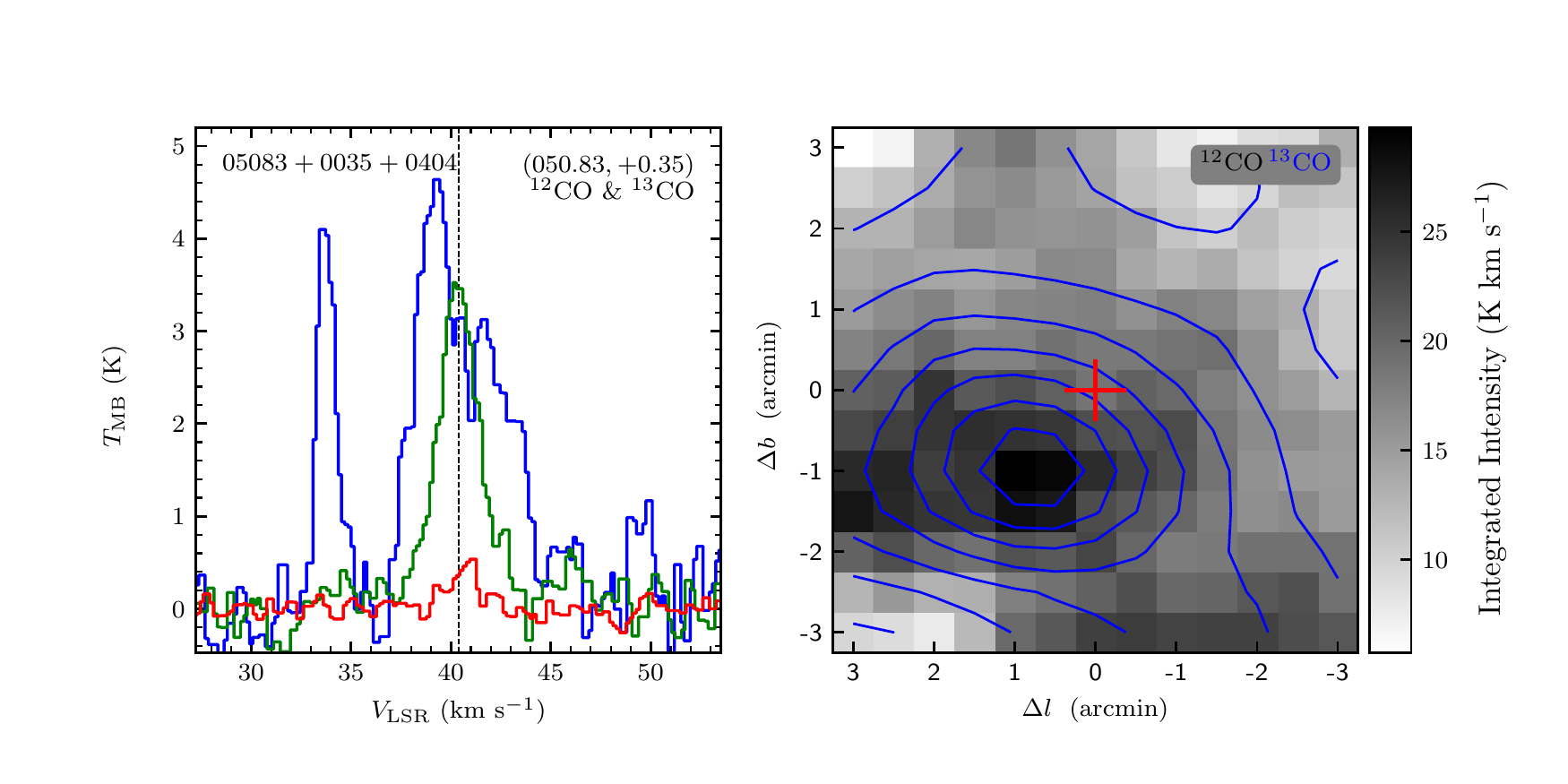}
\includegraphics[width=9.0cm,angle=0]{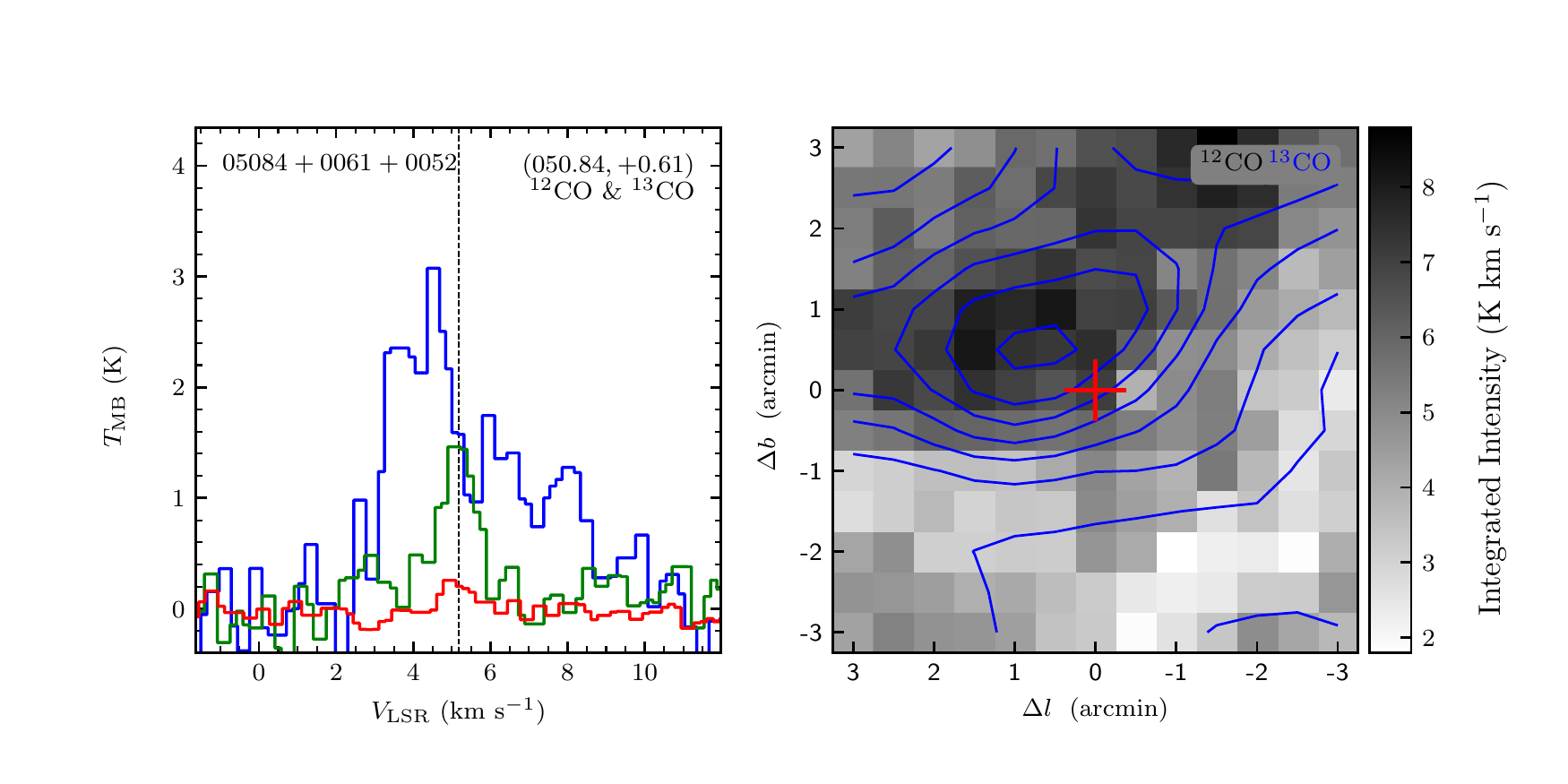}
\vspace{-0.5cm}

\includegraphics[width=9.0cm,angle=0]{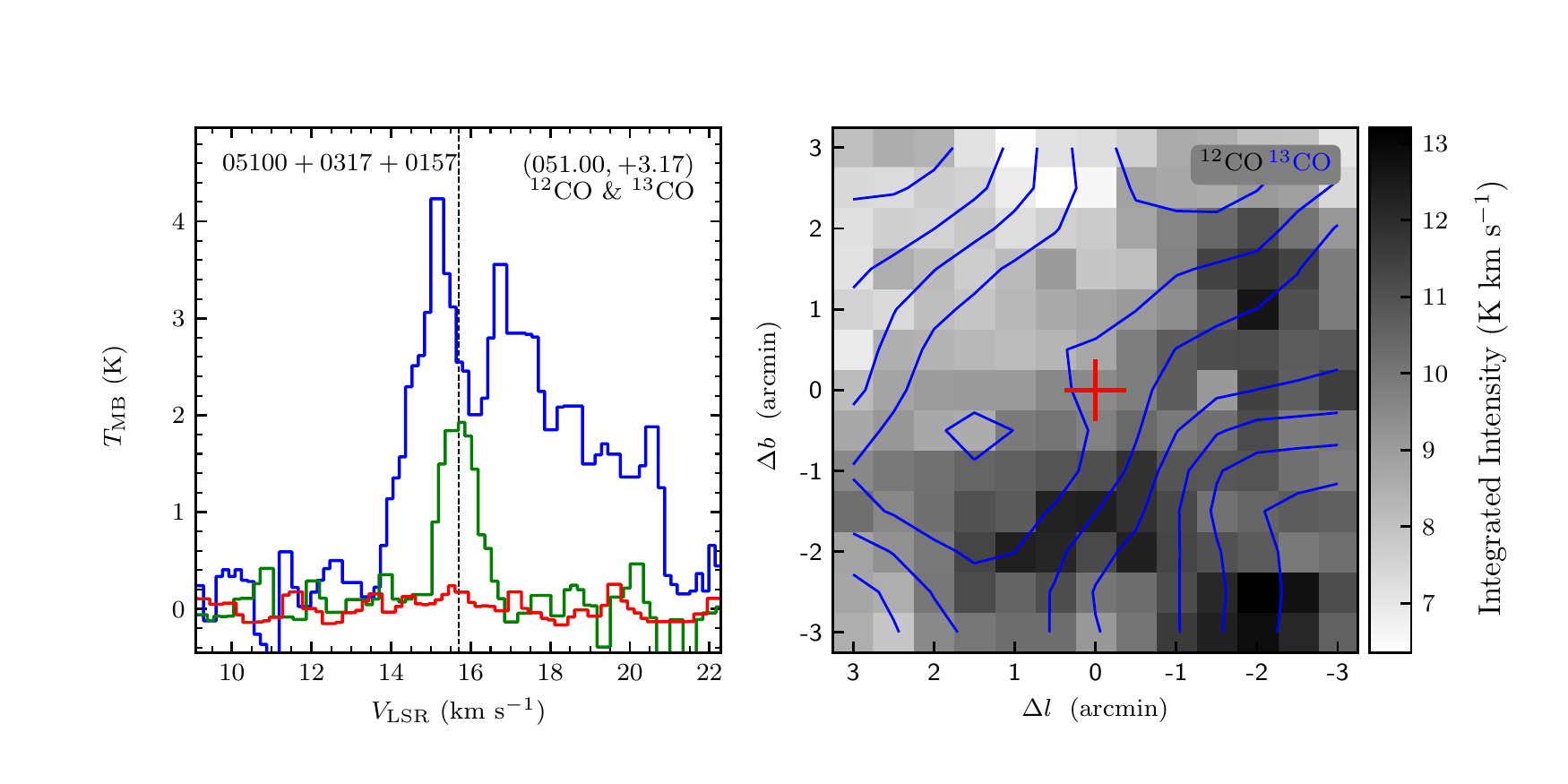}
\includegraphics[width=9.0cm,angle=0]{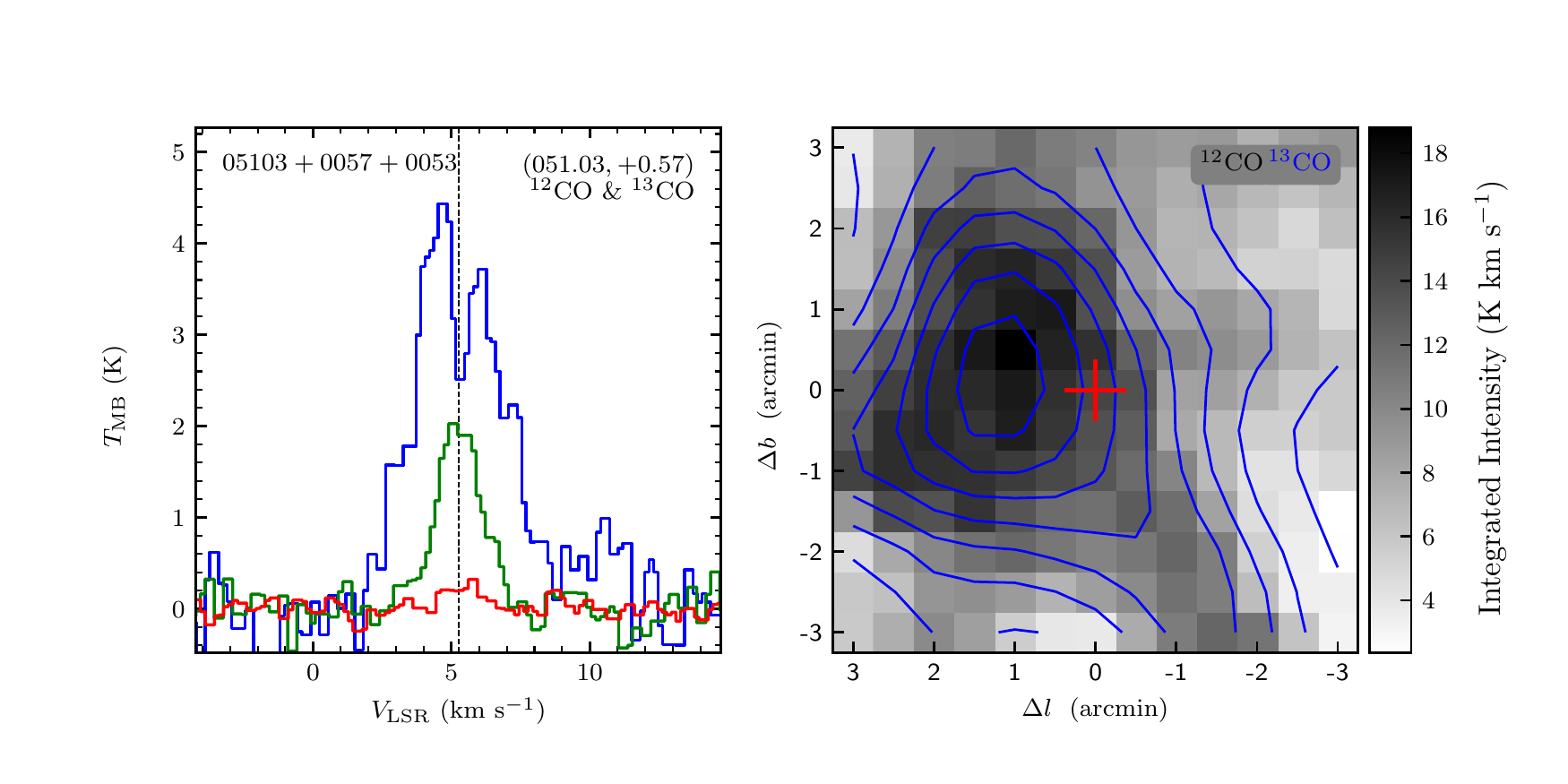}
\vspace{-0.5cm}

\includegraphics[width=9.0cm,angle=0]{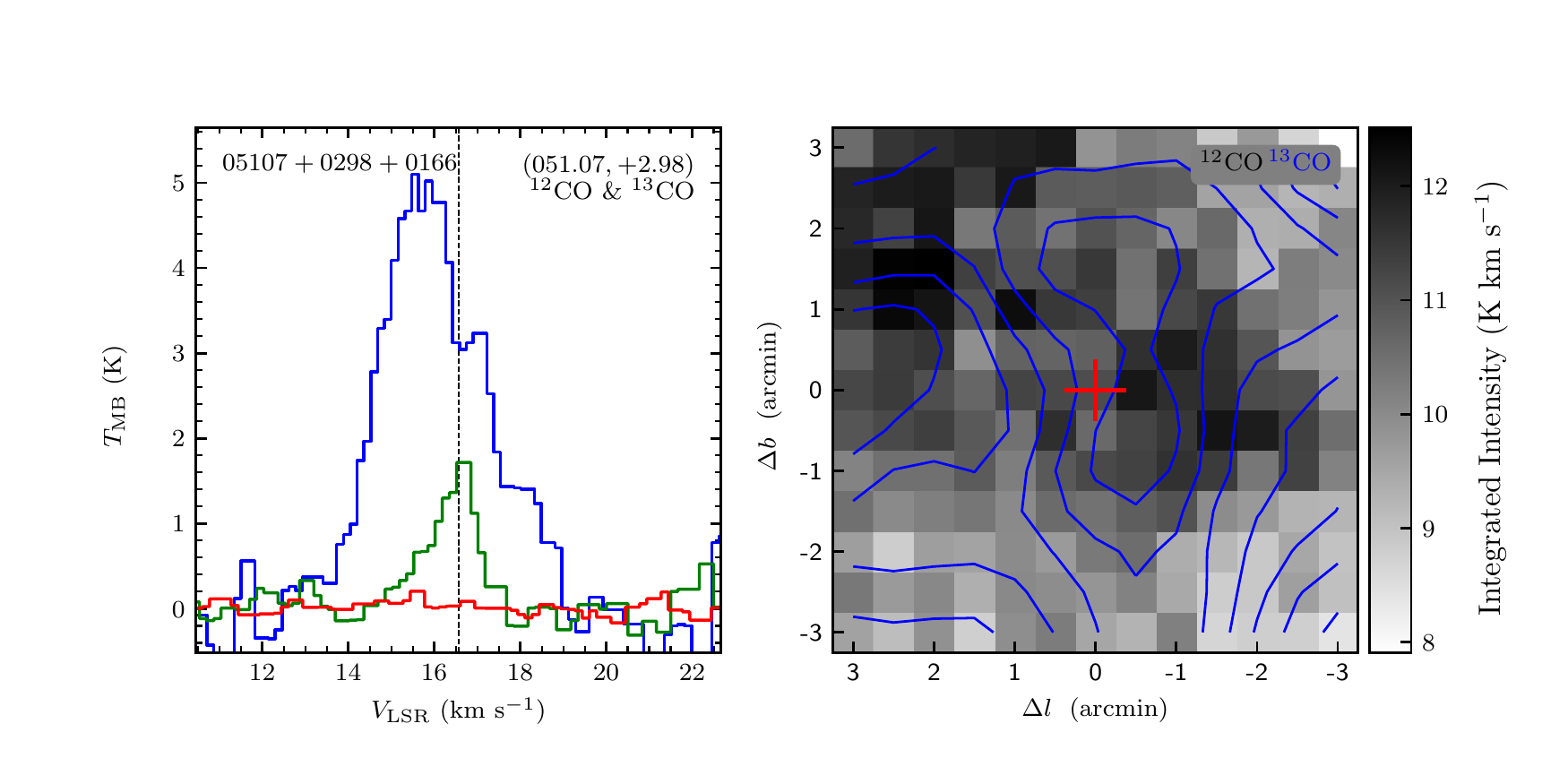}
\includegraphics[width=9.0cm,angle=0]{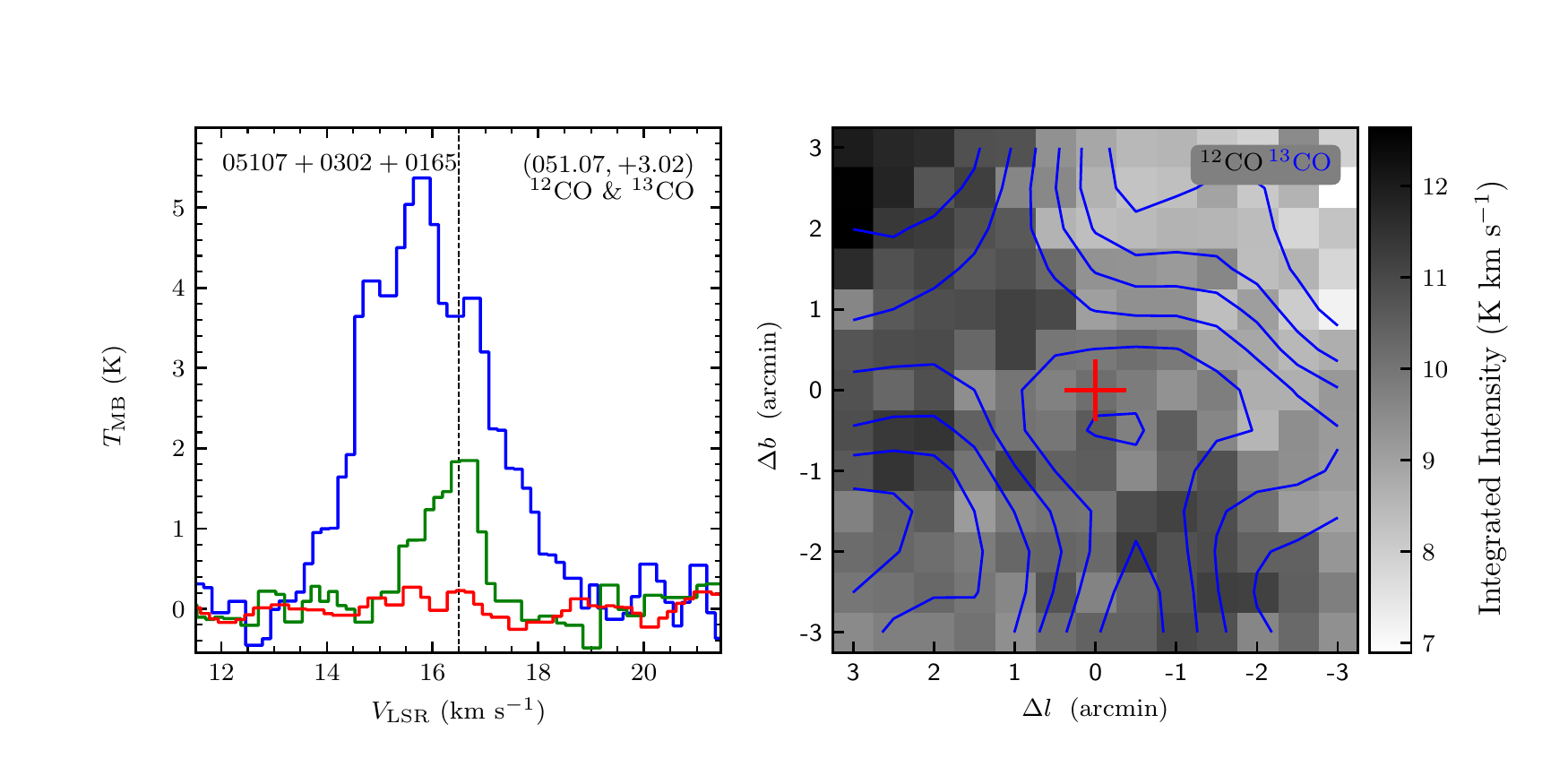}
\end{figure}
\clearpage

\begin{figure}
\includegraphics[width=9.0cm,angle=0]{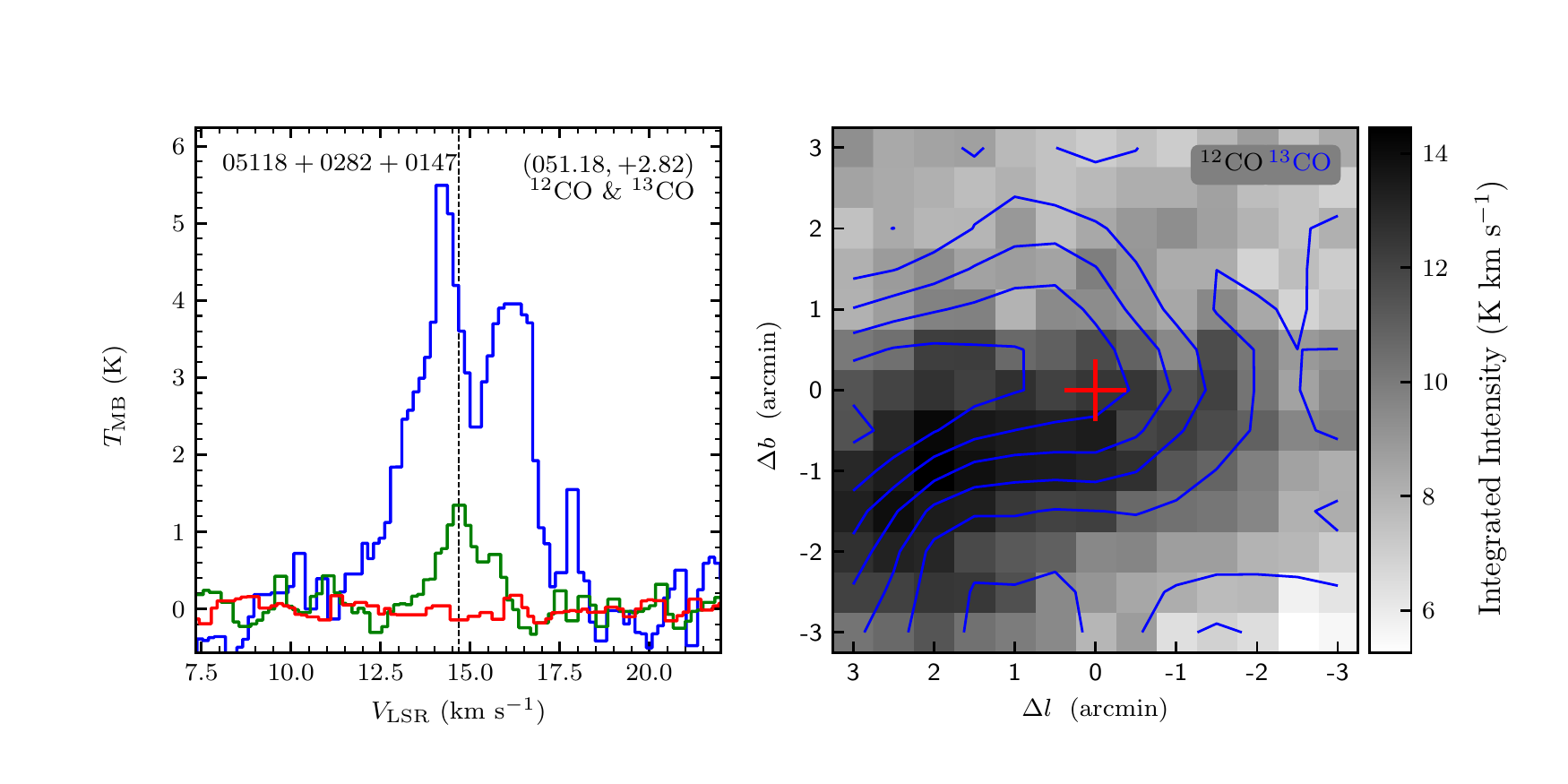}
\includegraphics[width=9.0cm,angle=0]{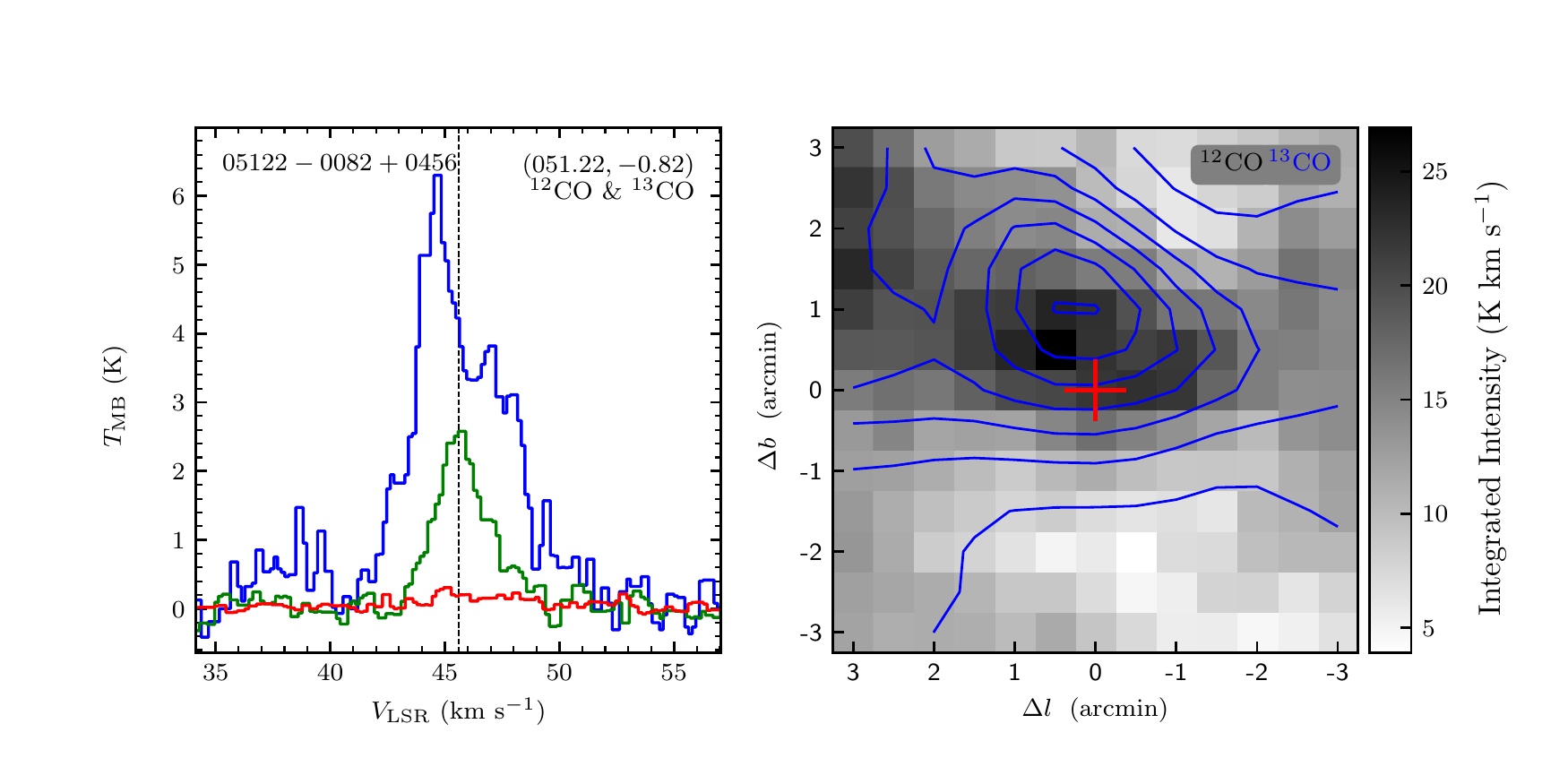}
\vspace{-0.5cm}

\includegraphics[width=9.0cm,angle=0]{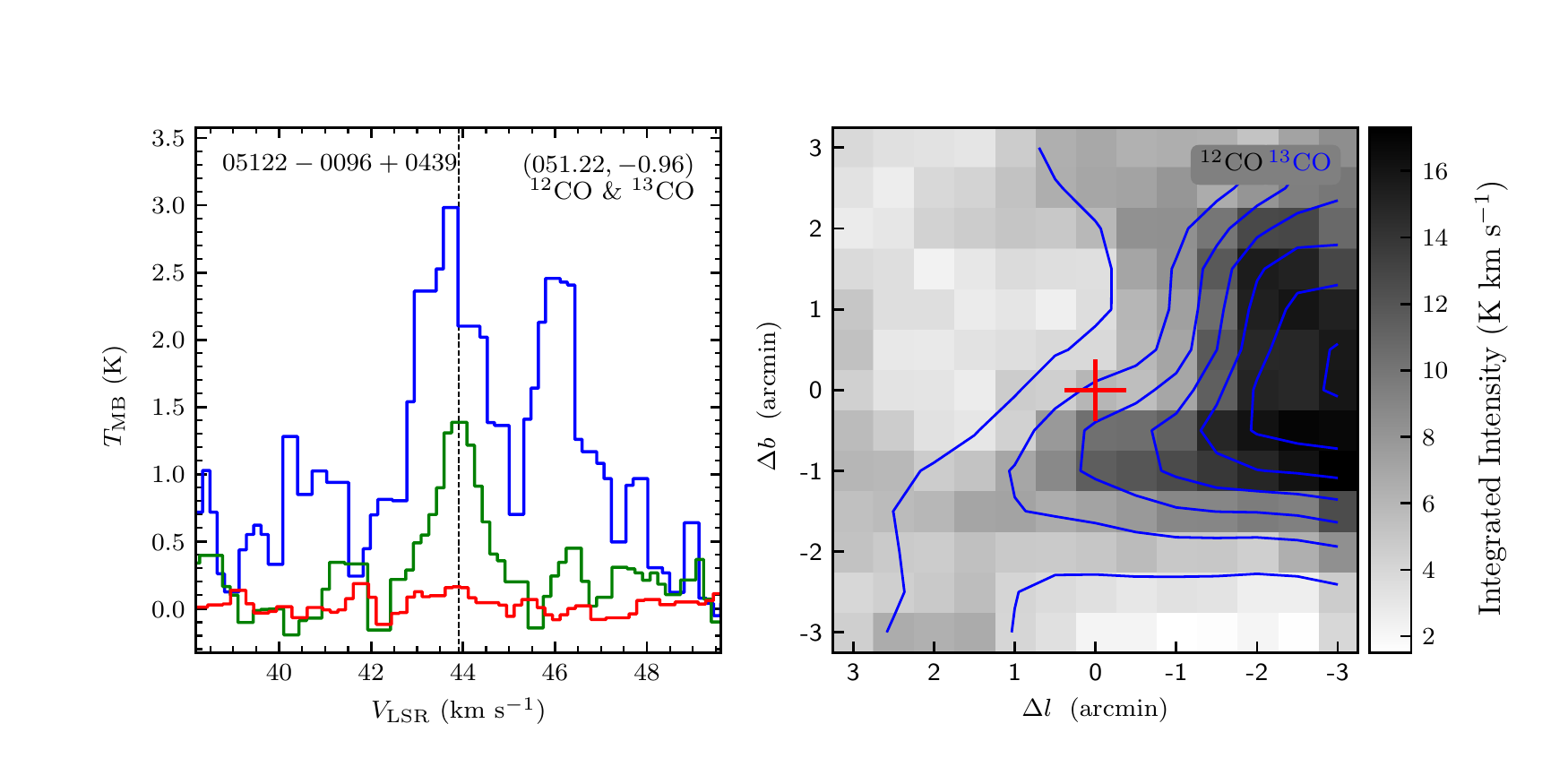}
\includegraphics[width=9.0cm,angle=0]{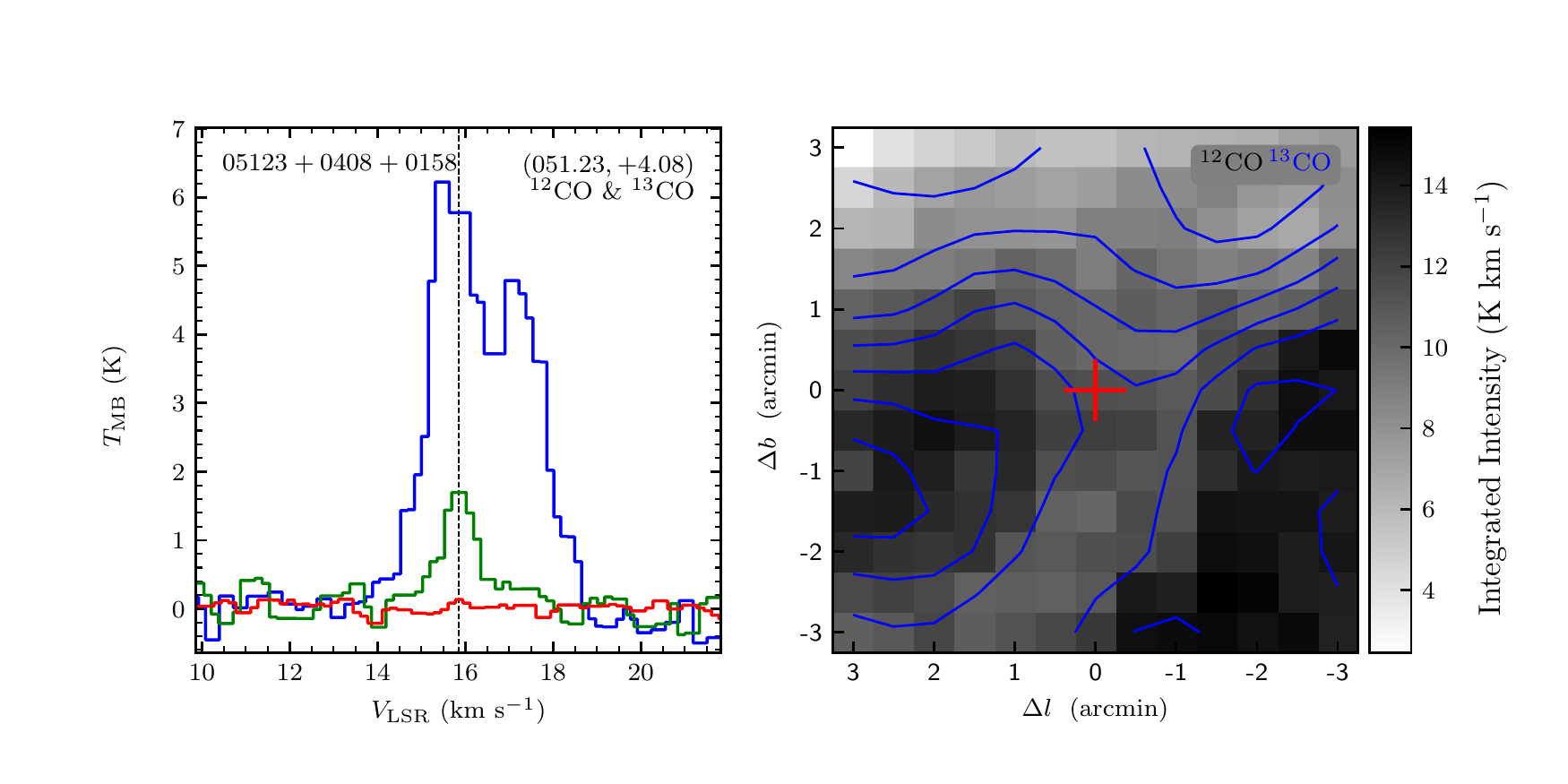}
\vspace{-0.5cm}

\includegraphics[width=9.0cm,angle=0]{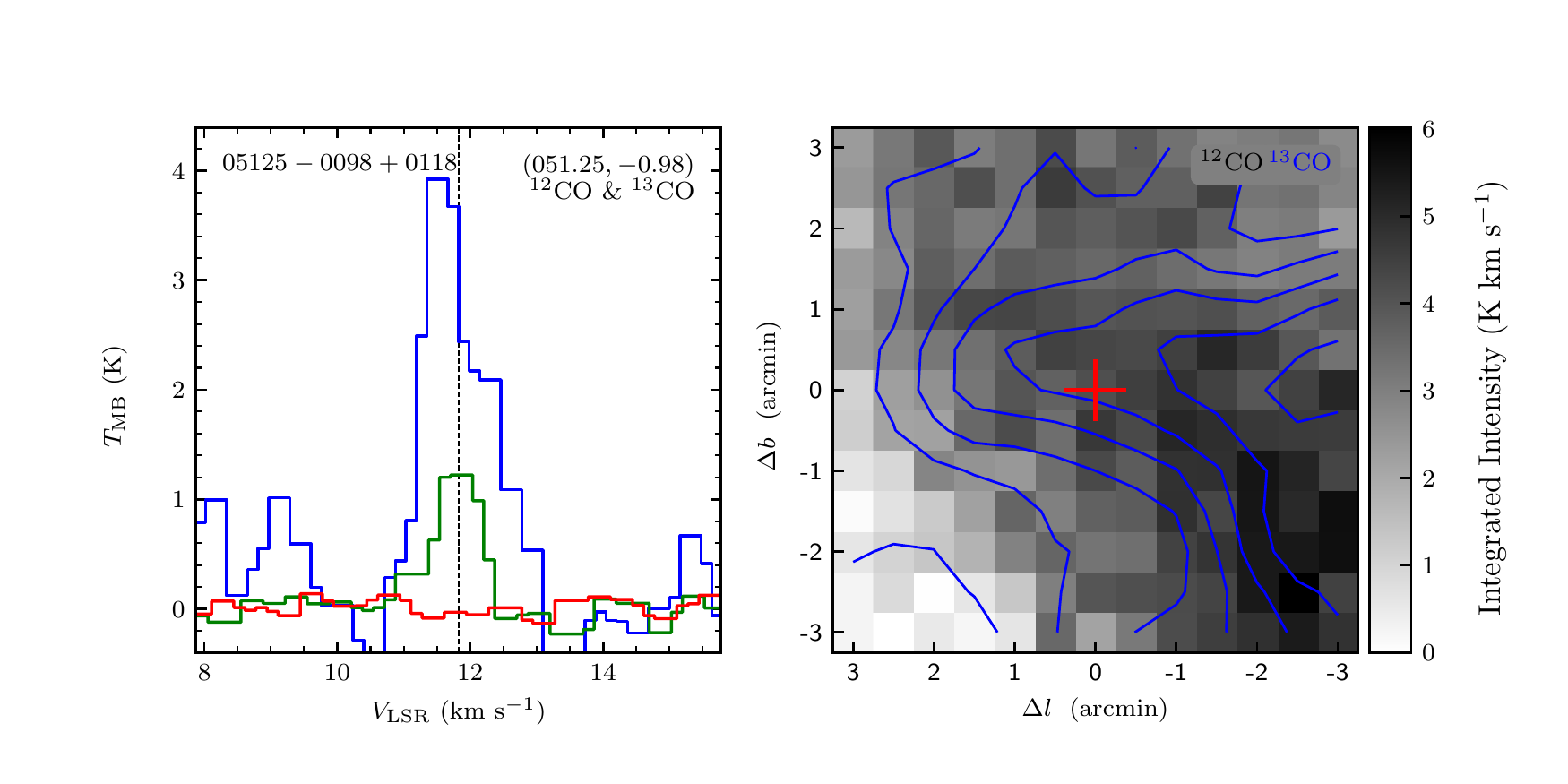}
\includegraphics[width=9.0cm,angle=0]{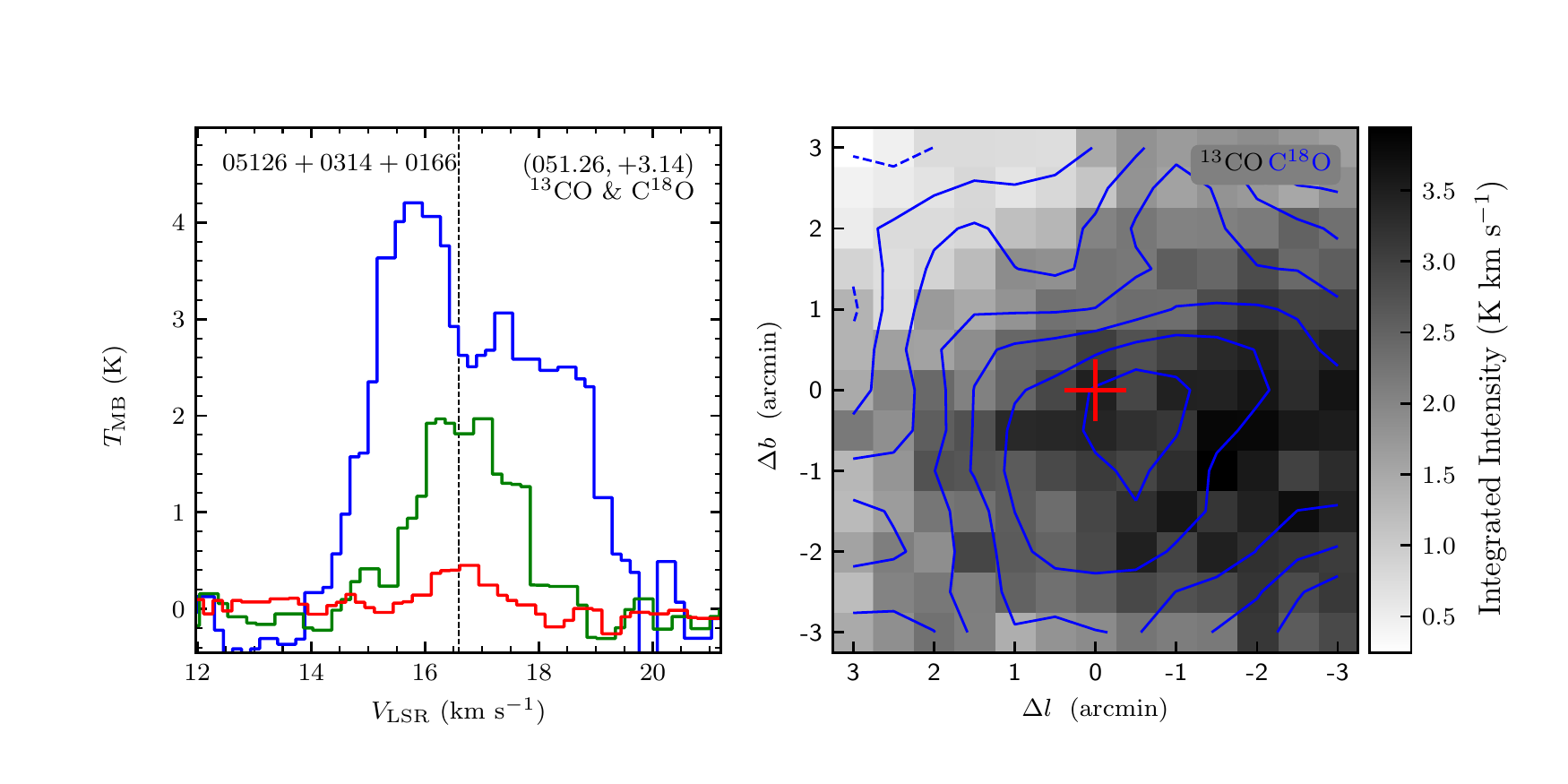}
\vspace{-0.5cm}

\includegraphics[width=9.0cm,angle=0]{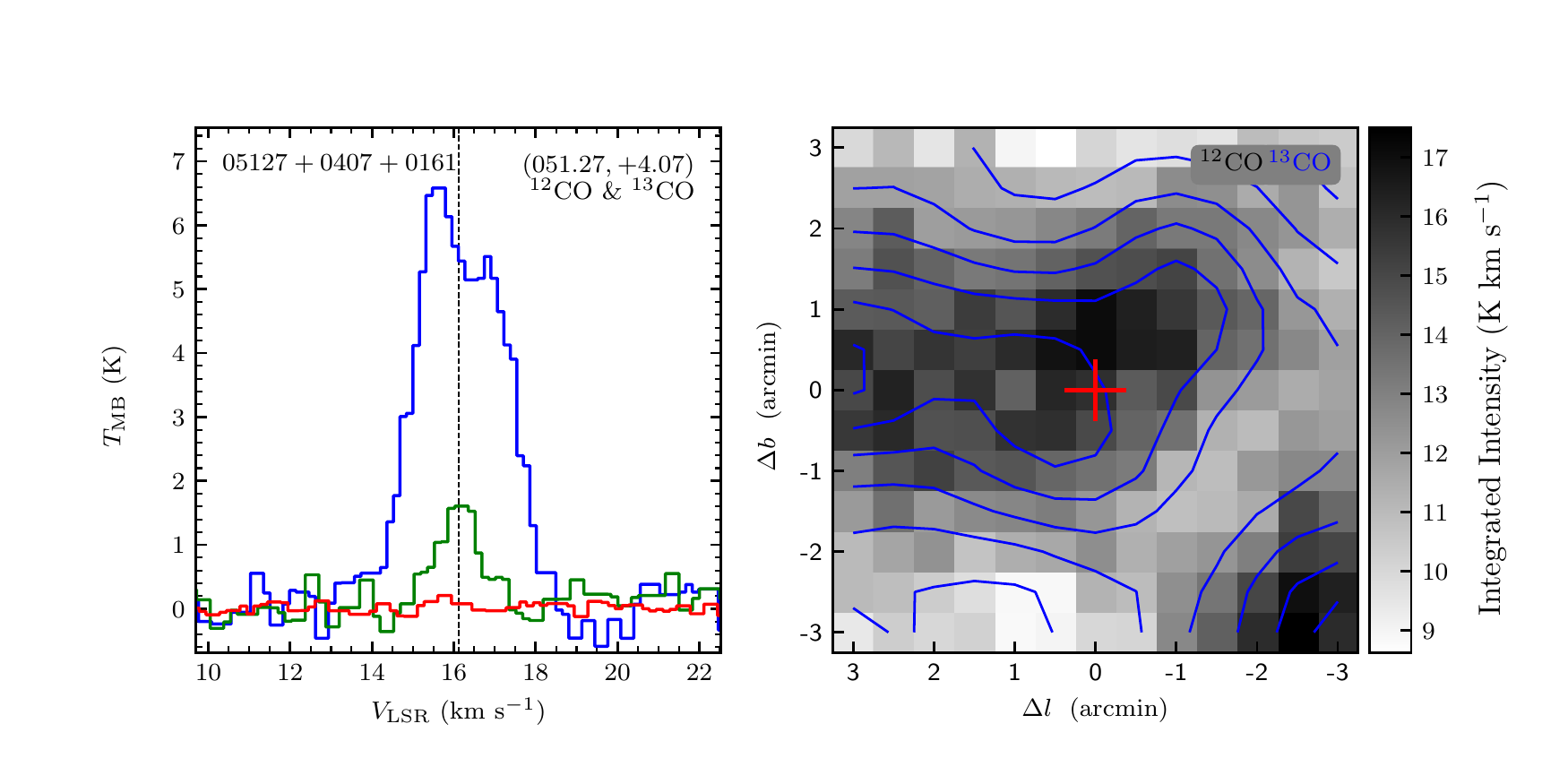}
\includegraphics[width=9.0cm,angle=0]{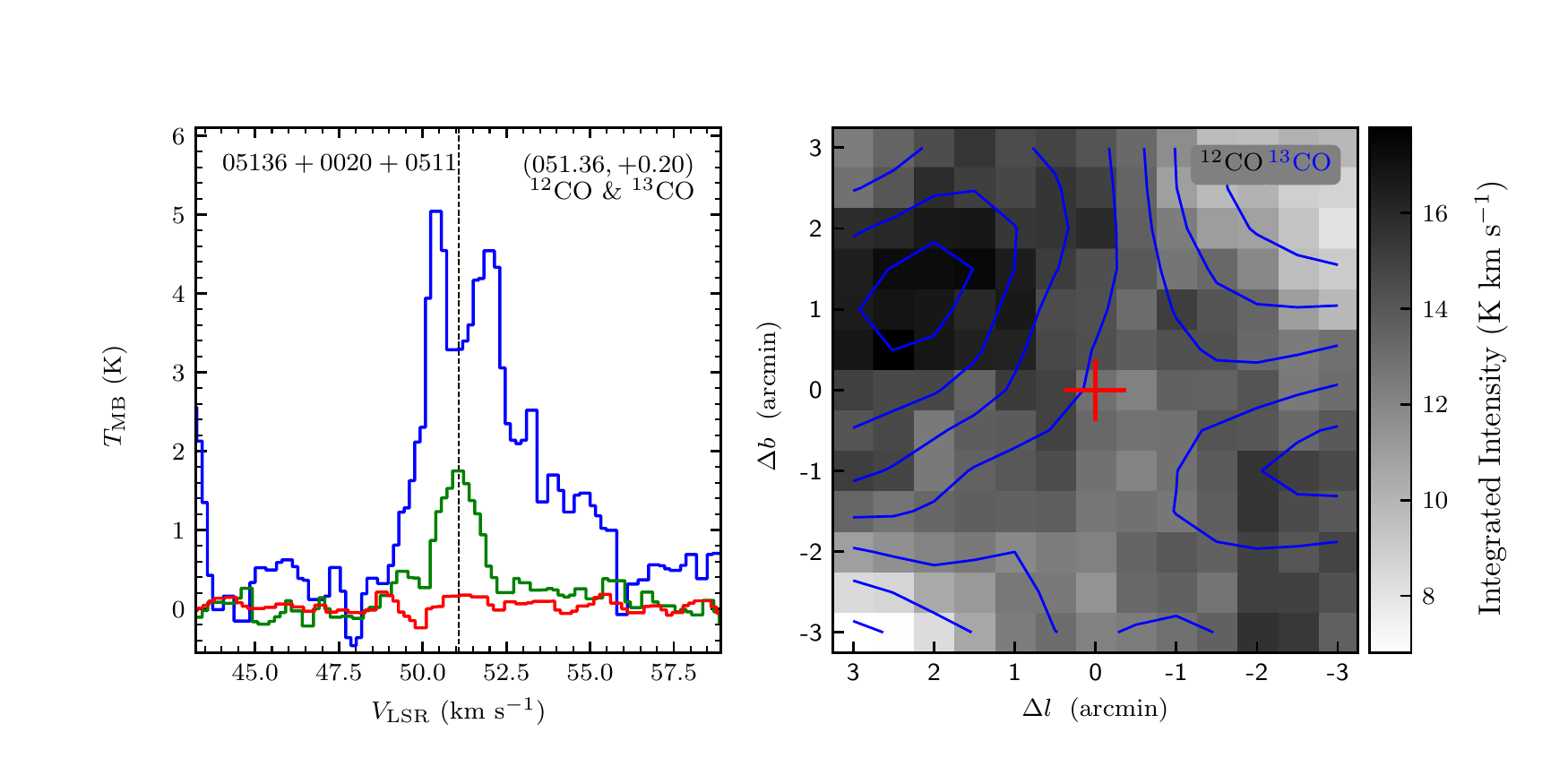}
\vspace{-0.5cm}

\includegraphics[width=9.0cm,angle=0]{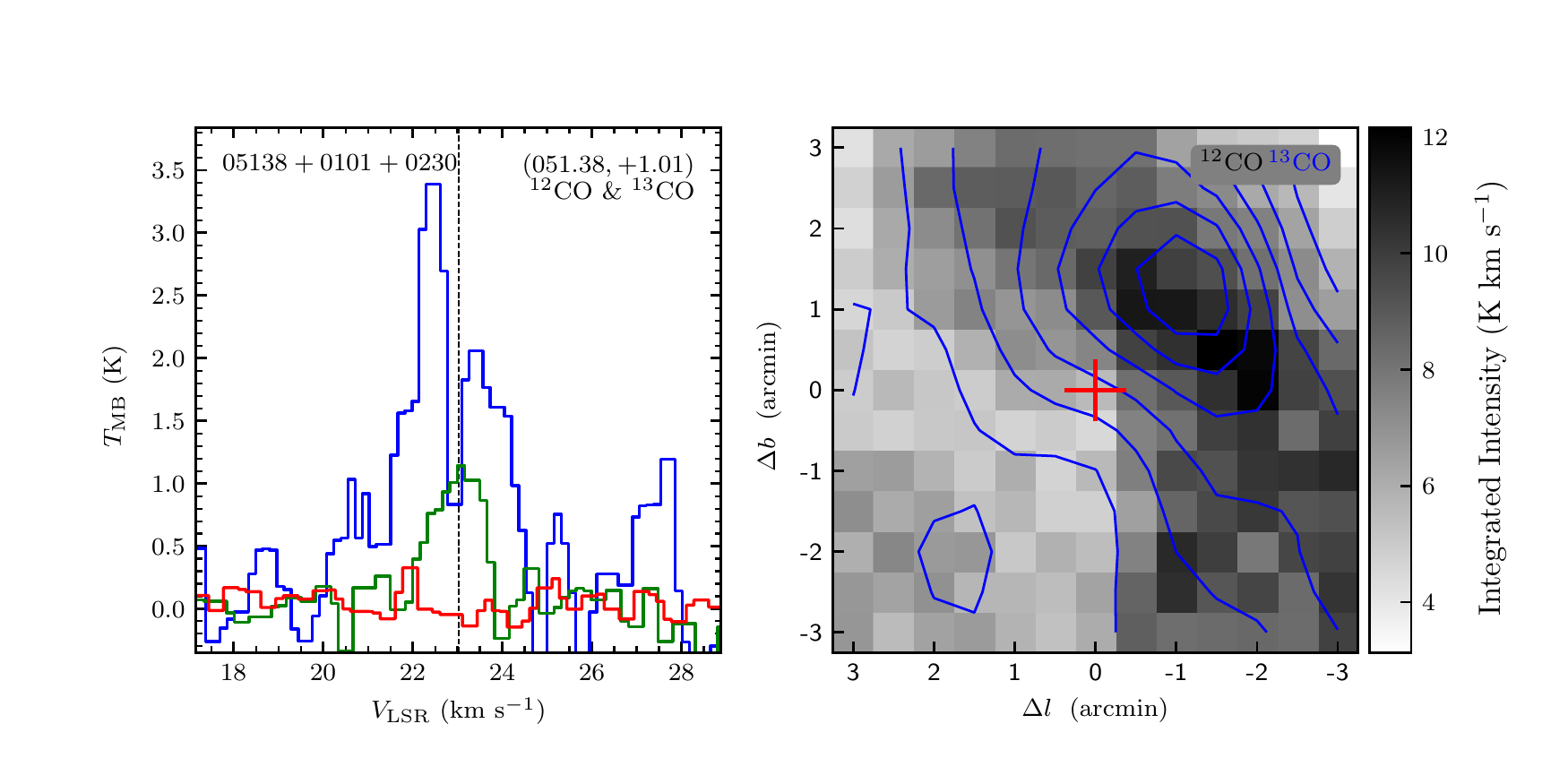}
\includegraphics[width=9.0cm,angle=0]{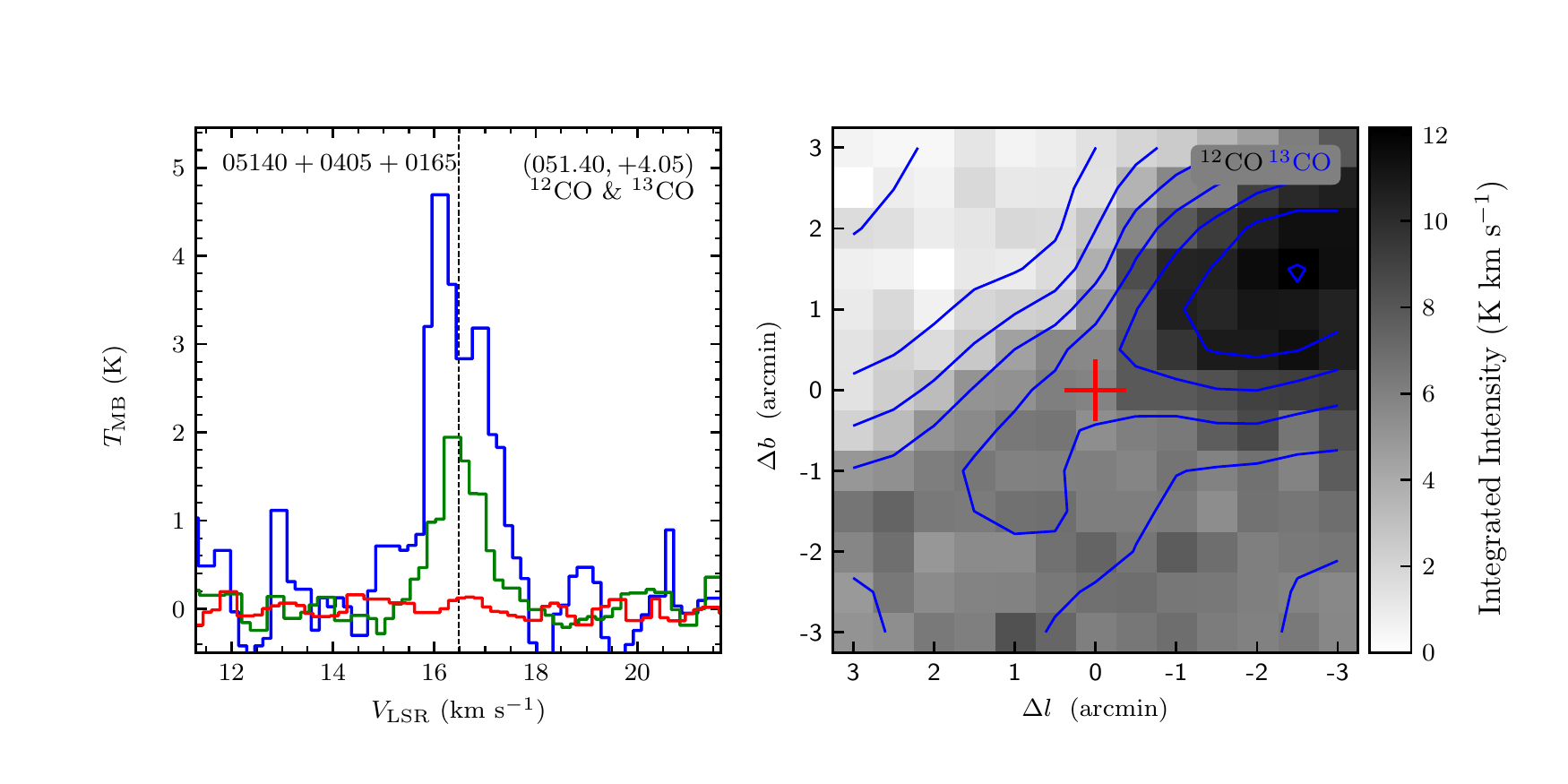}
\end{figure}
\clearpage

\begin{figure}
\includegraphics[width=9.0cm,angle=0]{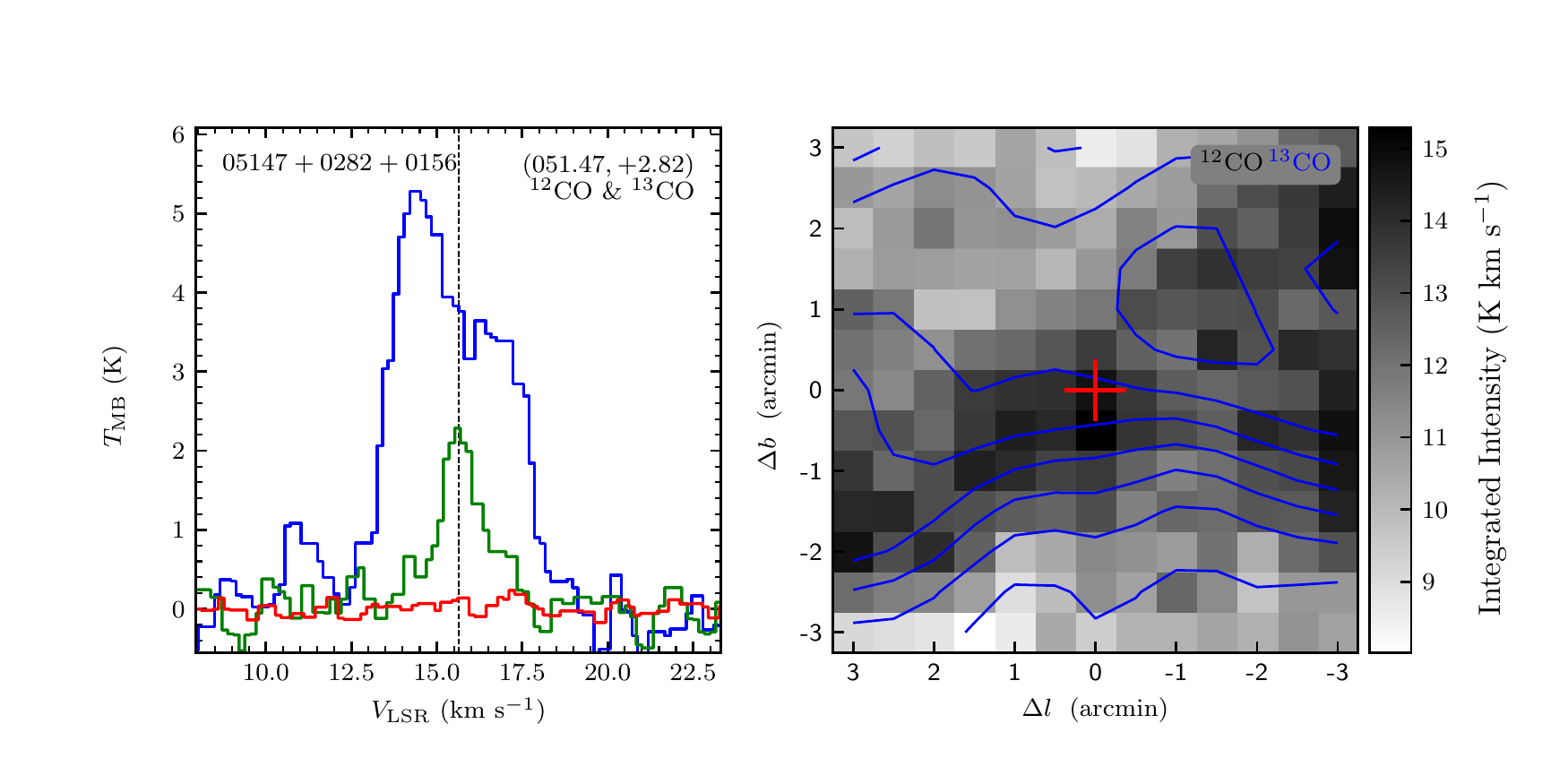}
\includegraphics[width=9.0cm,angle=0]{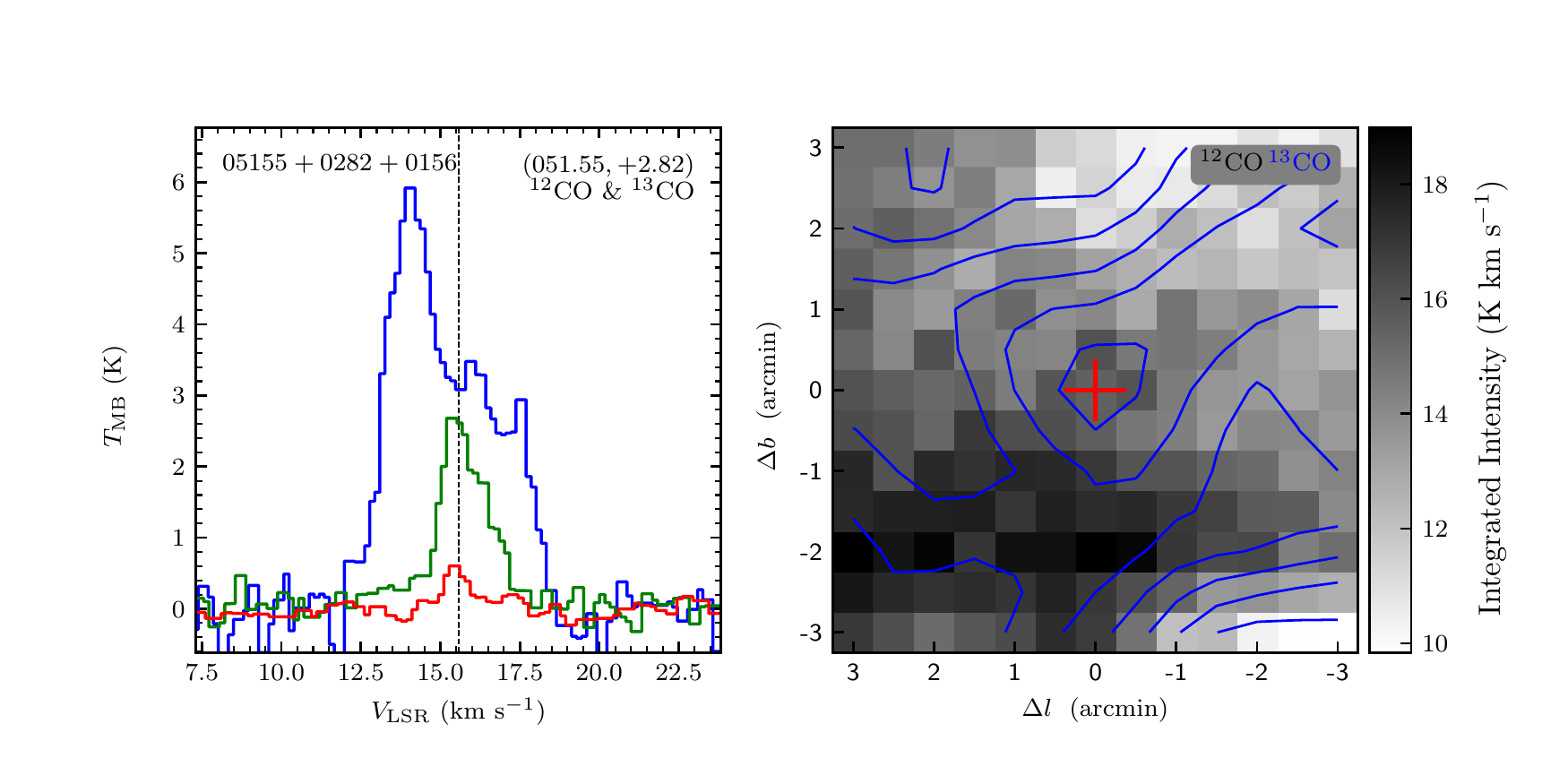}
\vspace{-0.5cm}

\includegraphics[width=9.0cm,angle=0]{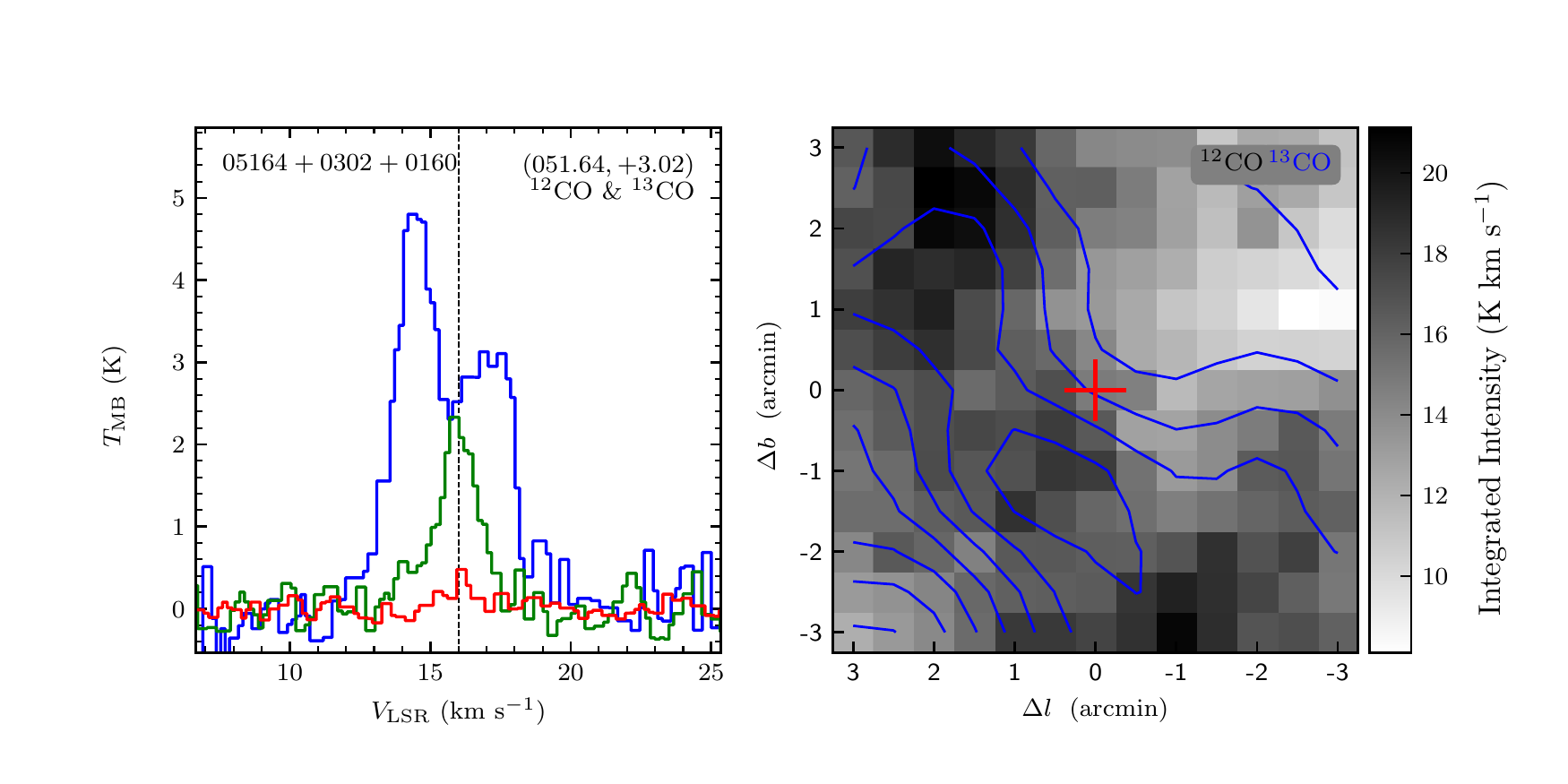}
\includegraphics[width=9.0cm,angle=0]{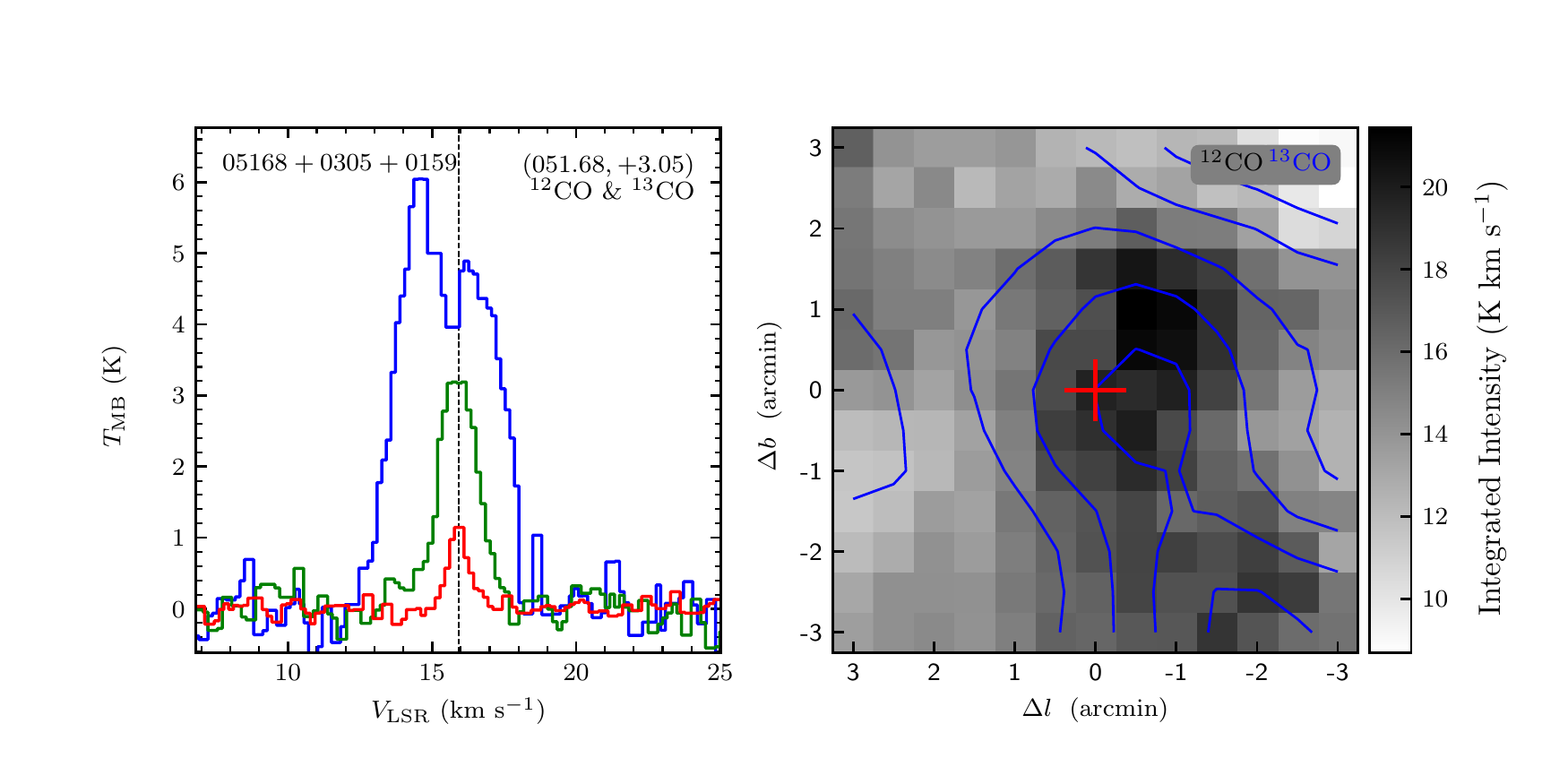}
\vspace{-0.5cm}

\includegraphics[width=9.0cm,angle=0]{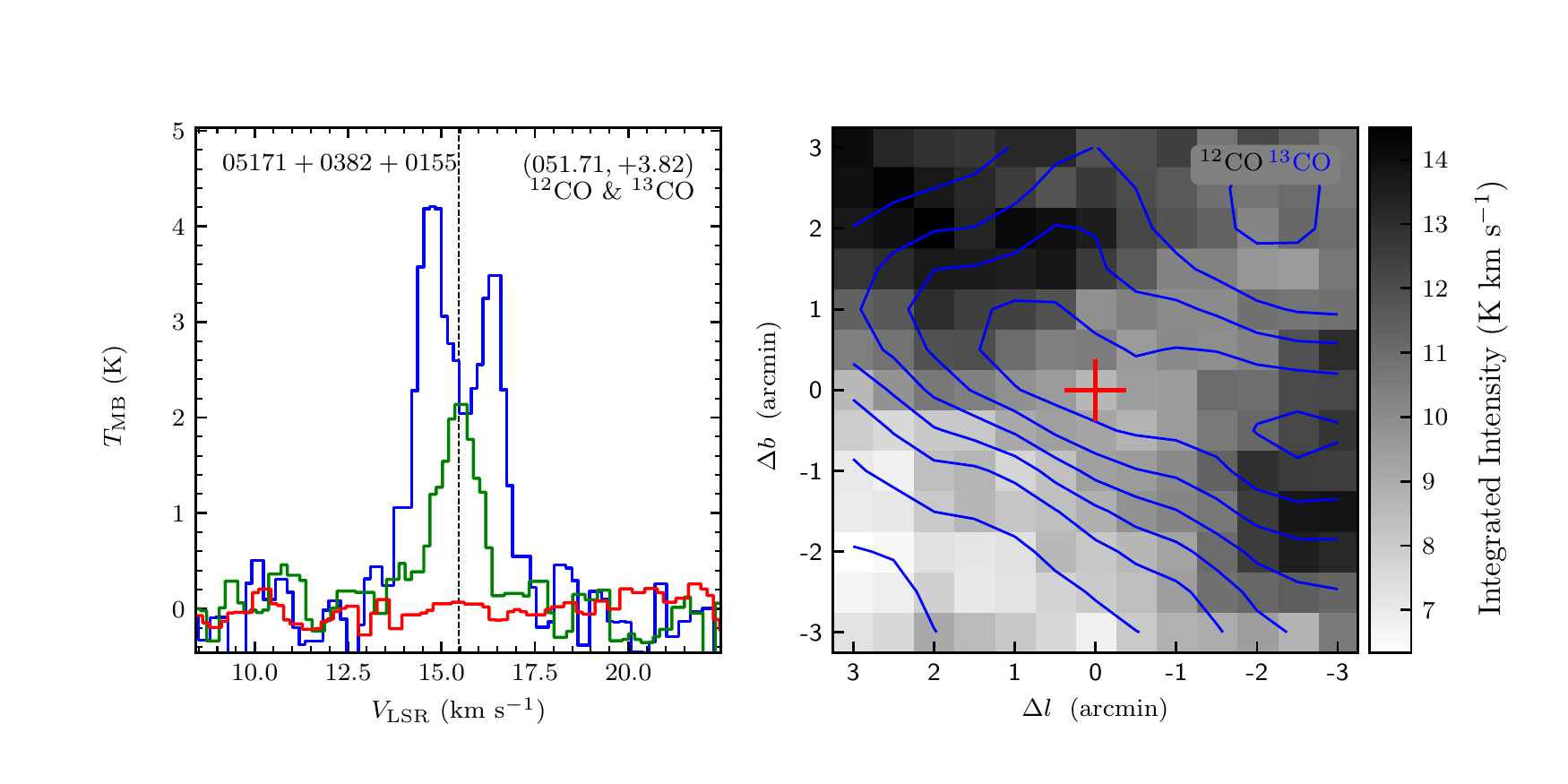}
\includegraphics[width=9.0cm,angle=0]{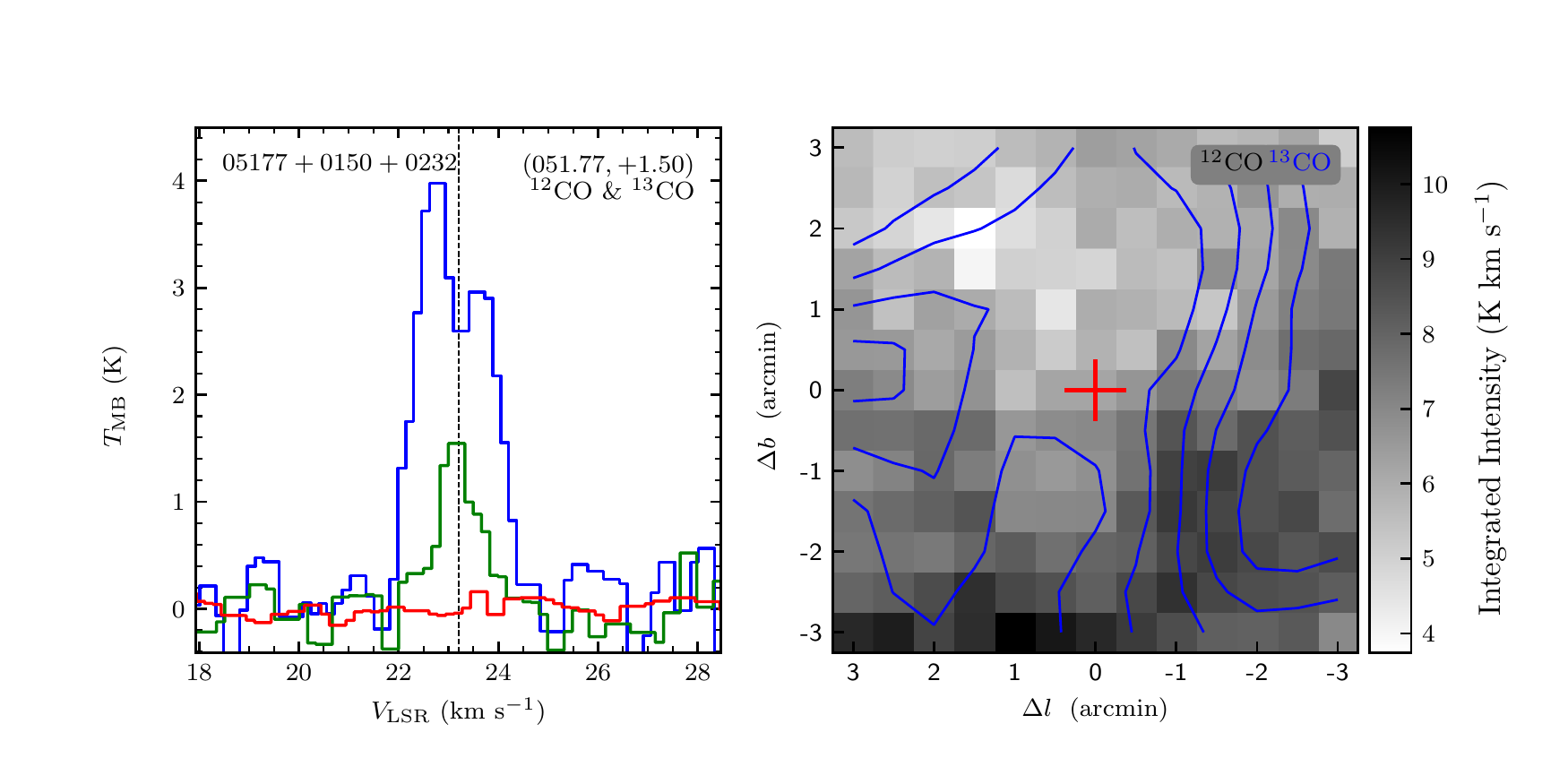}
\vspace{-0.5cm}

\includegraphics[width=9.0cm,angle=0]{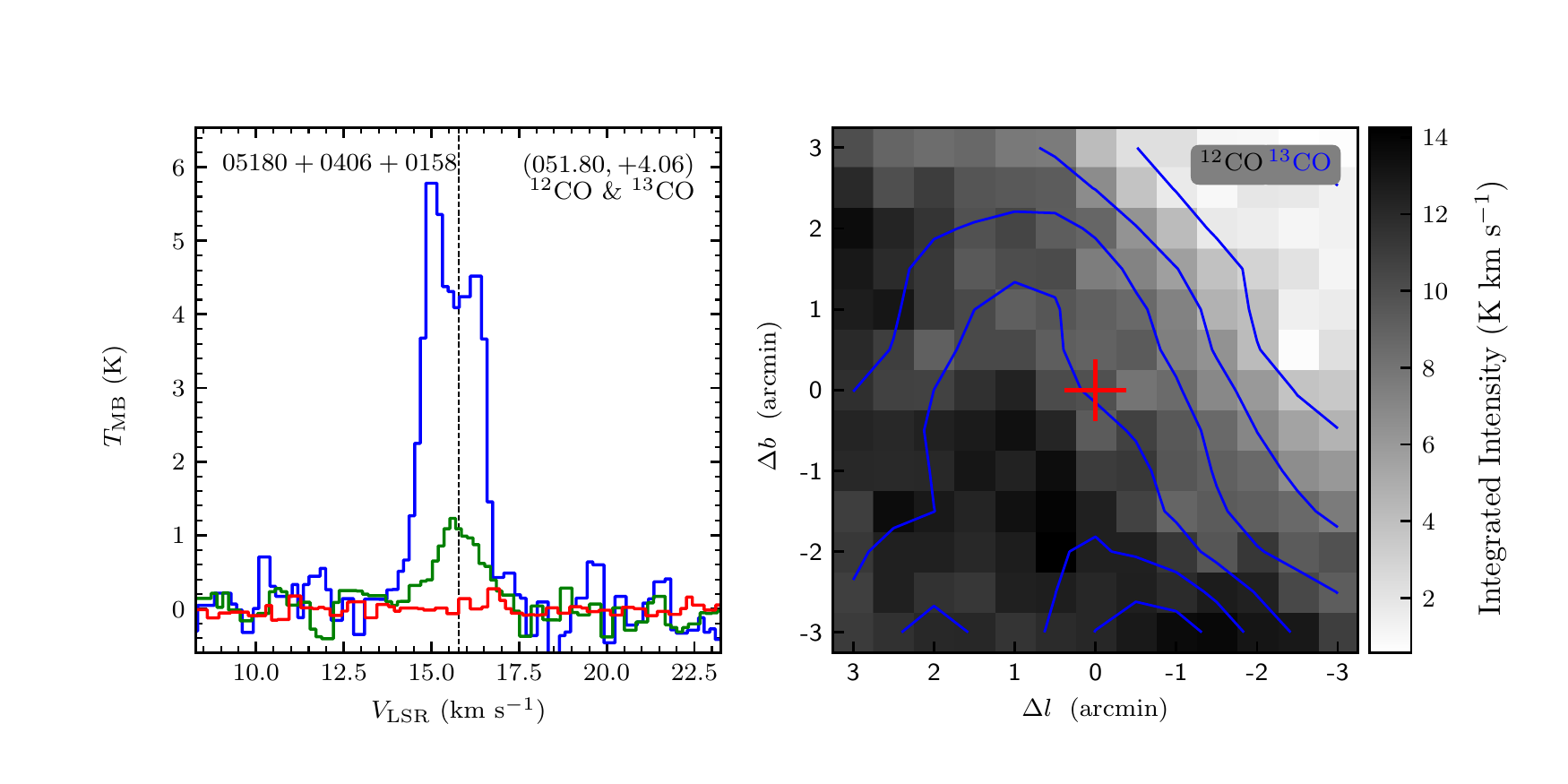}
\includegraphics[width=9.0cm,angle=0]{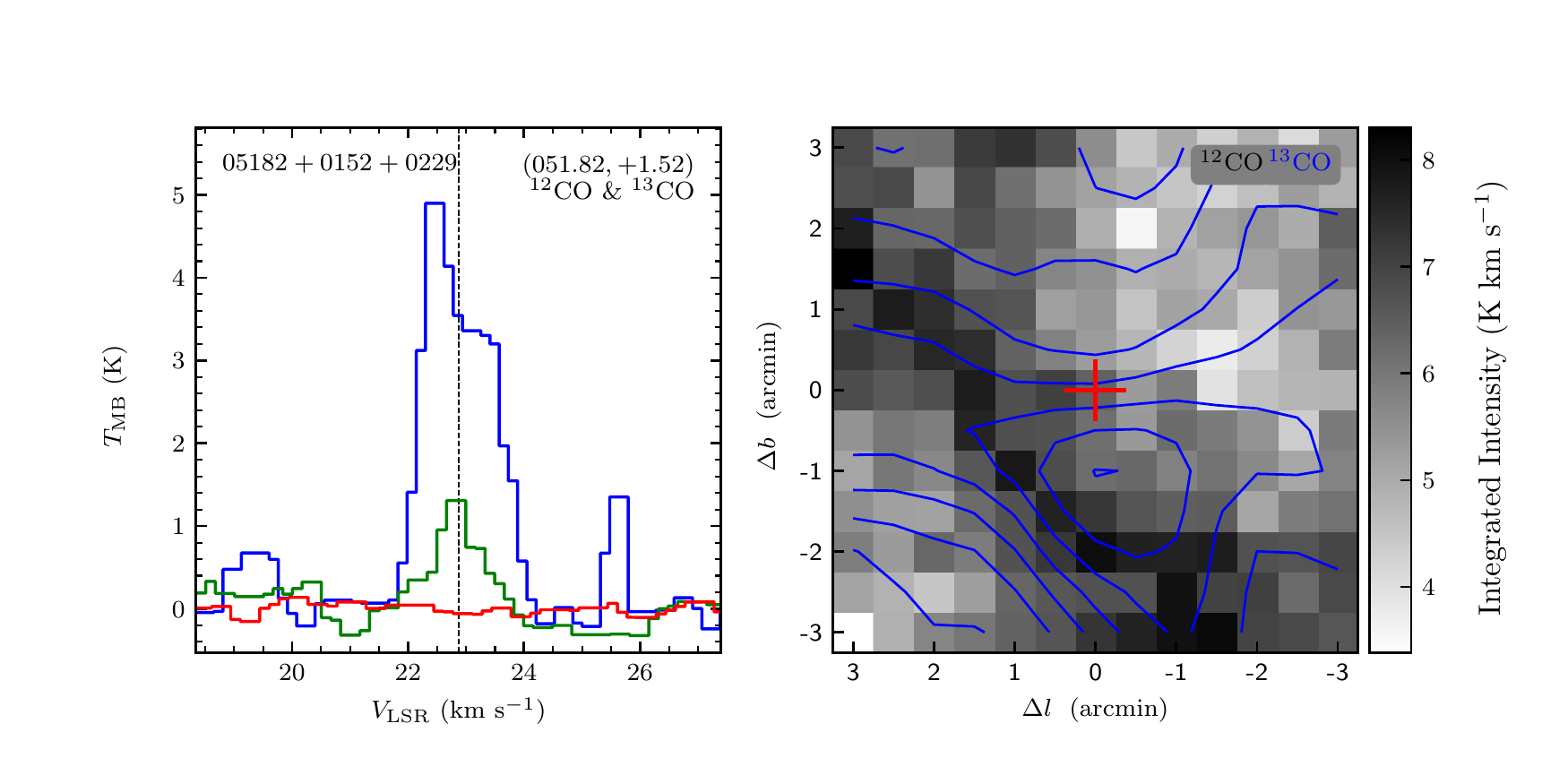}
\vspace{-0.5cm}

\includegraphics[width=9.0cm,angle=0]{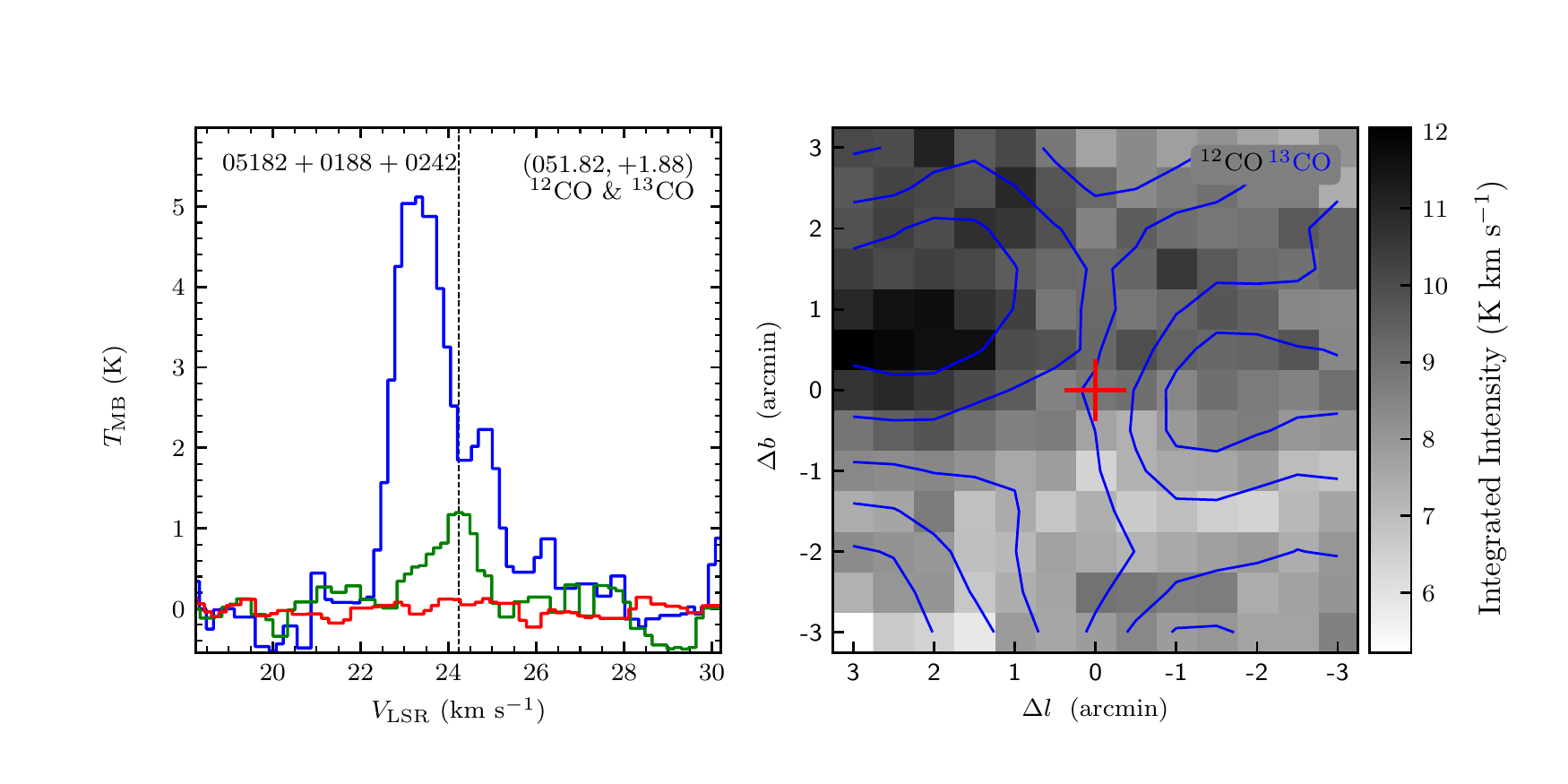}
\includegraphics[width=9.0cm,angle=0]{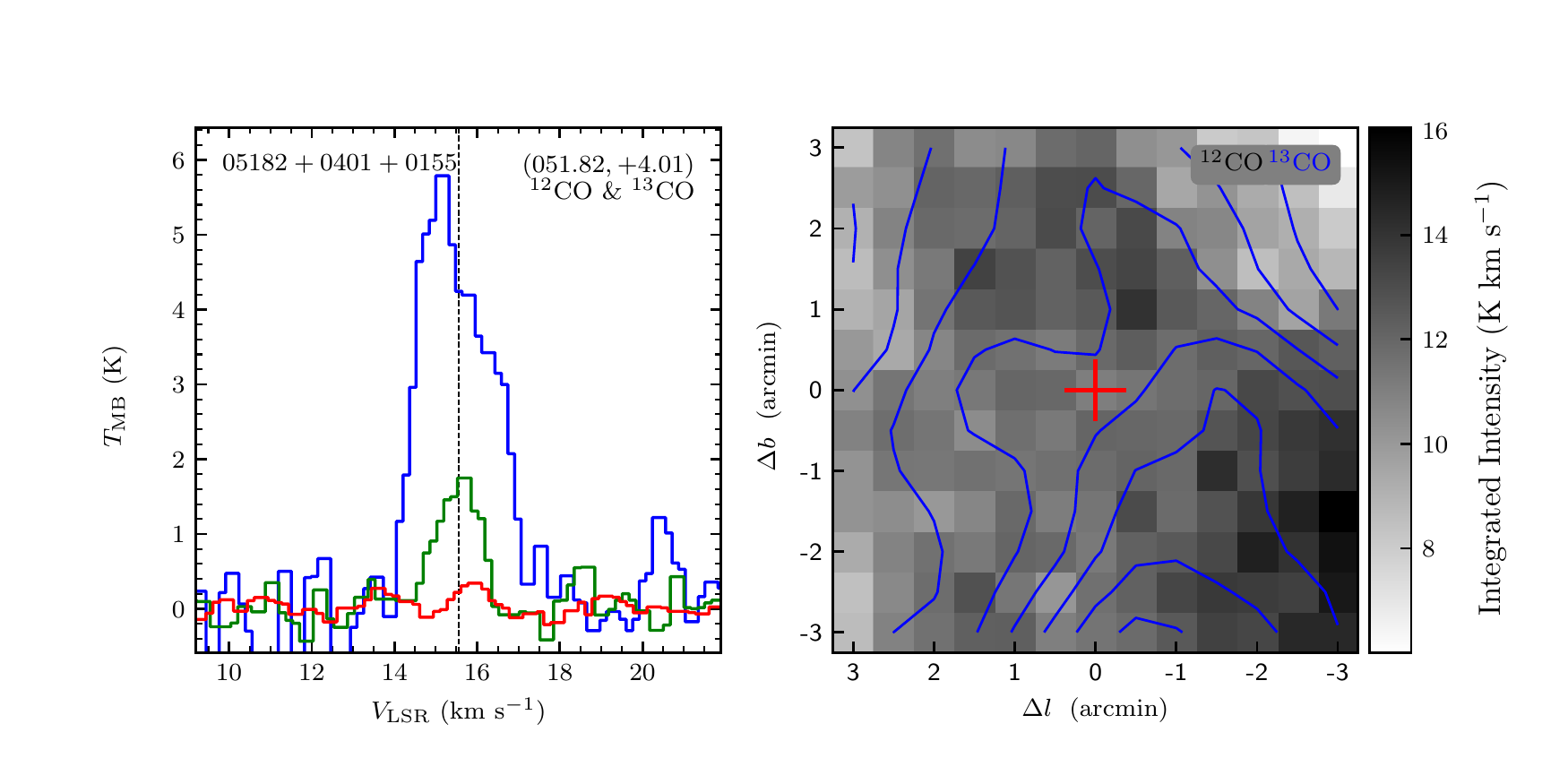}
\end{figure}
\clearpage

\begin{figure}
\includegraphics[width=9.0cm,angle=0]{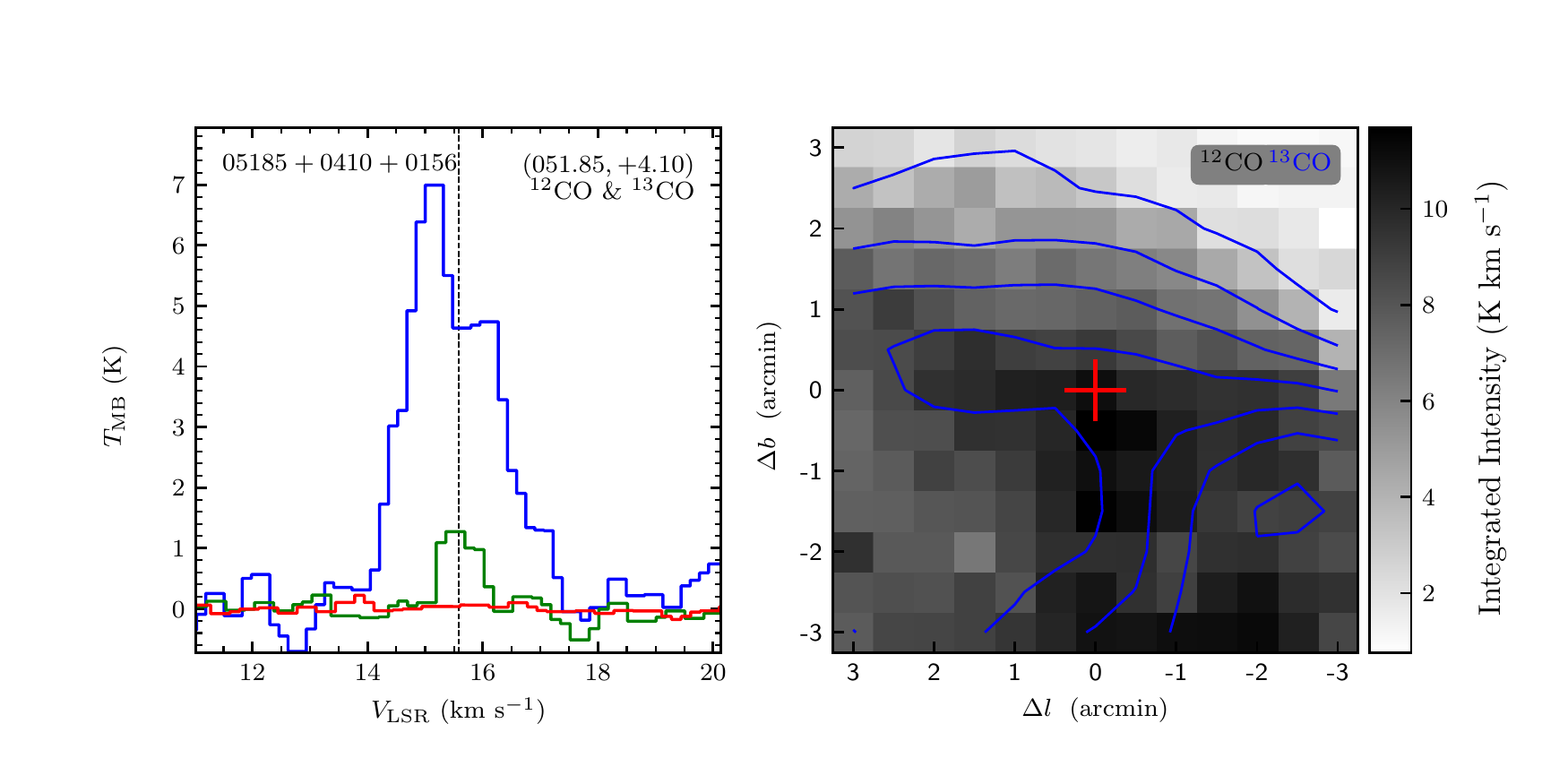}
\includegraphics[width=9.0cm,angle=0]{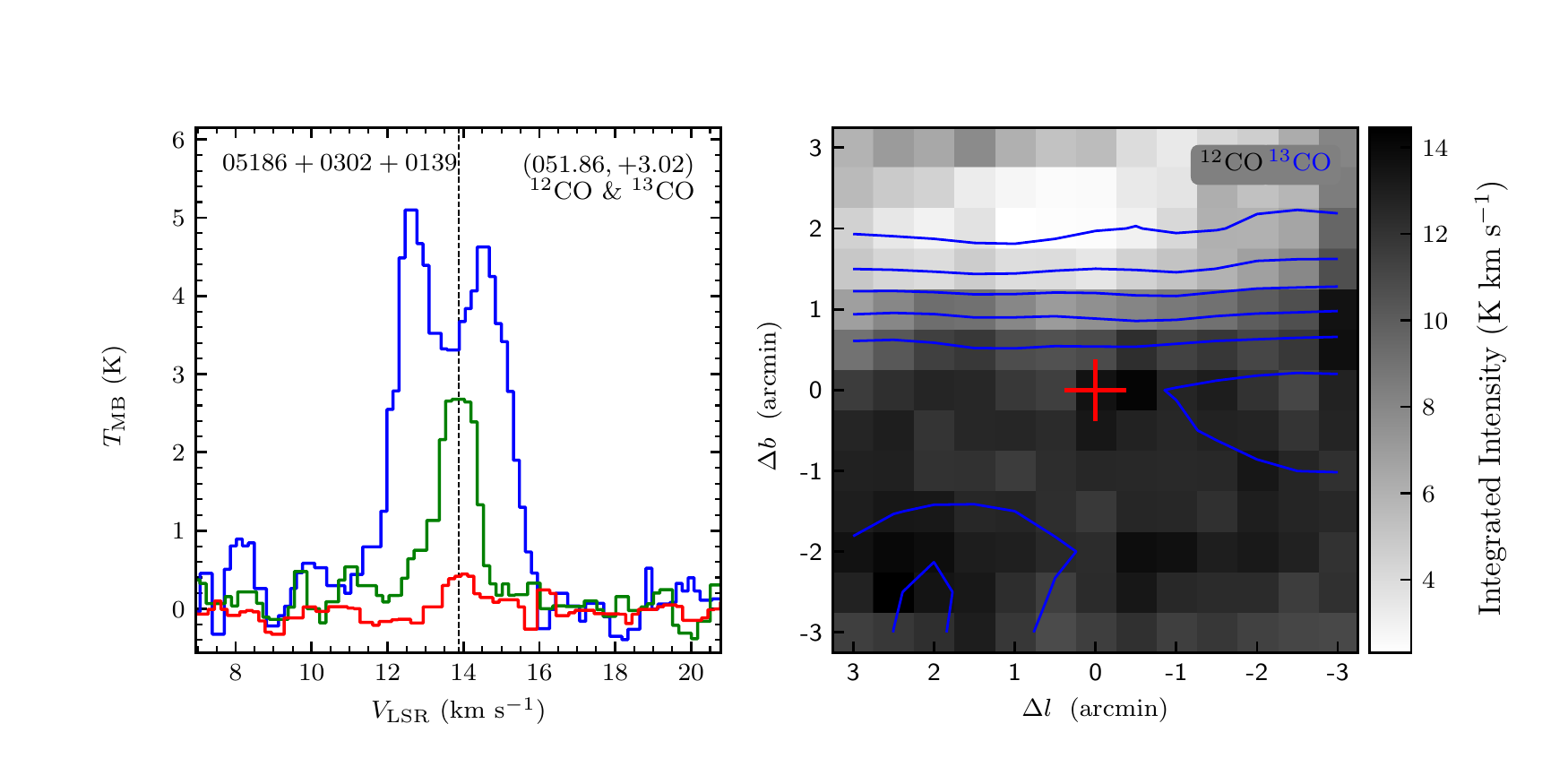}
\vspace{-0.5cm}

\includegraphics[width=9.0cm,angle=0]{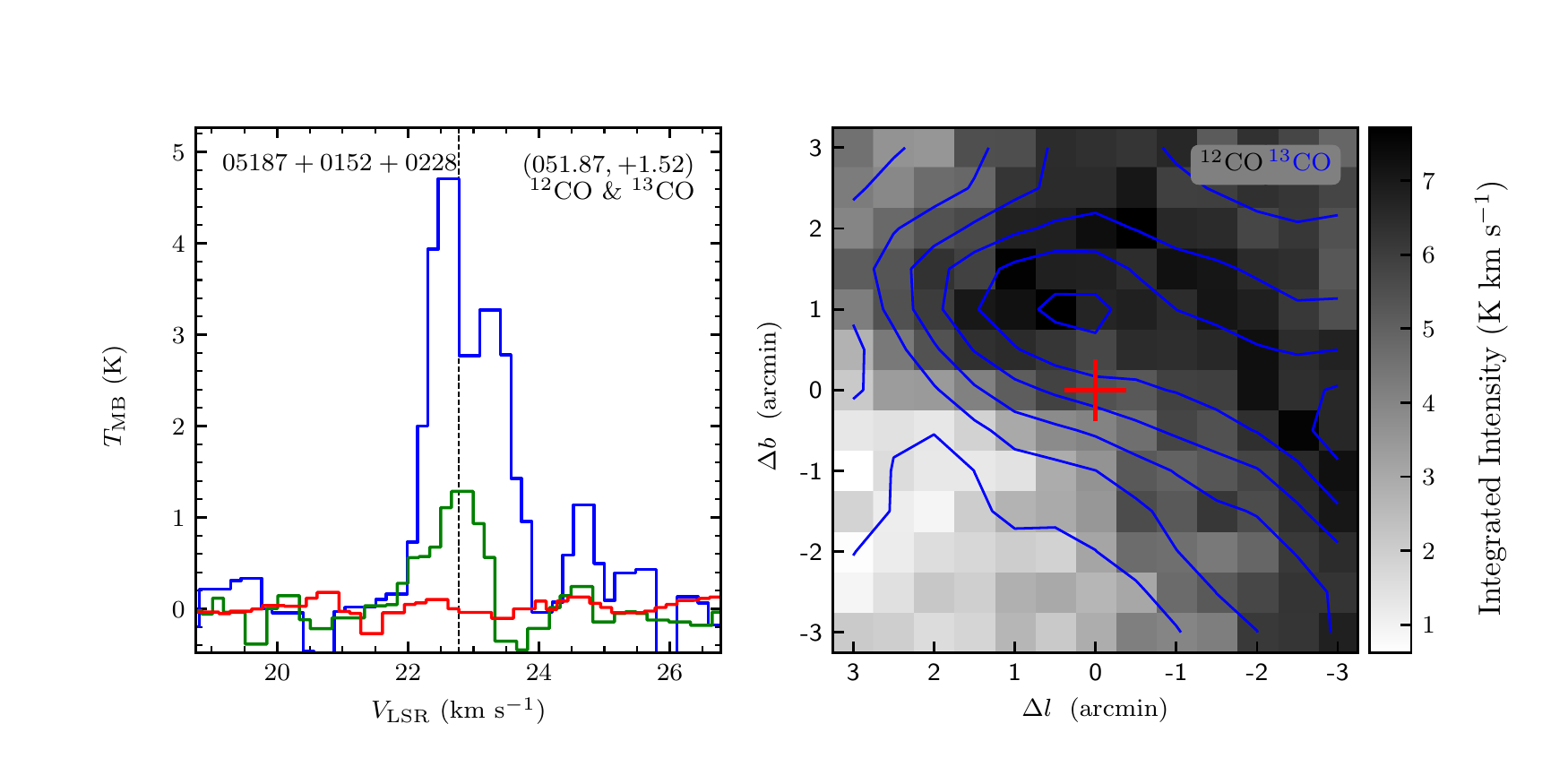}
\includegraphics[width=9.0cm,angle=0]{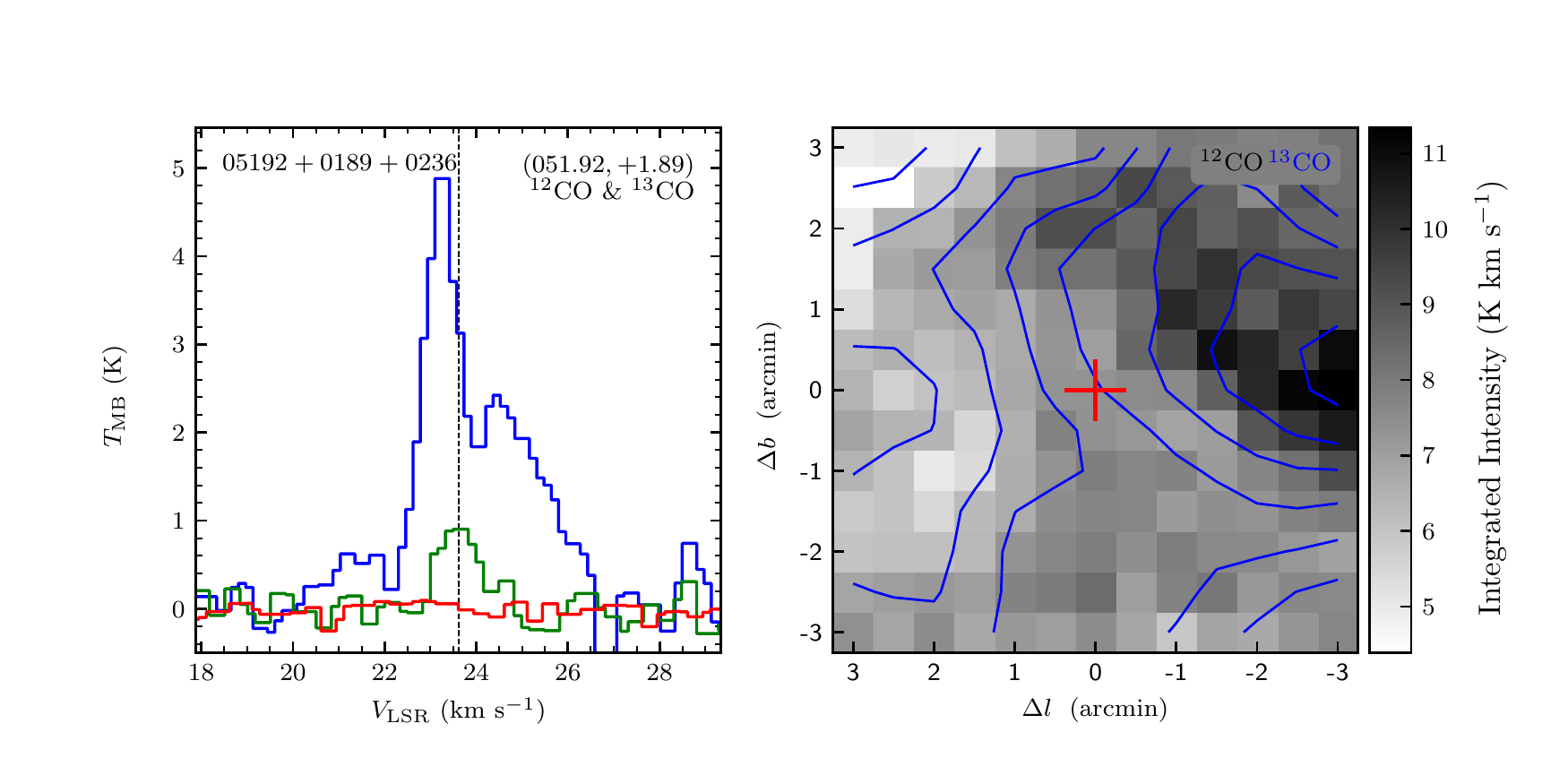}
\vspace{-0.5cm}

\includegraphics[width=9.0cm,angle=0]{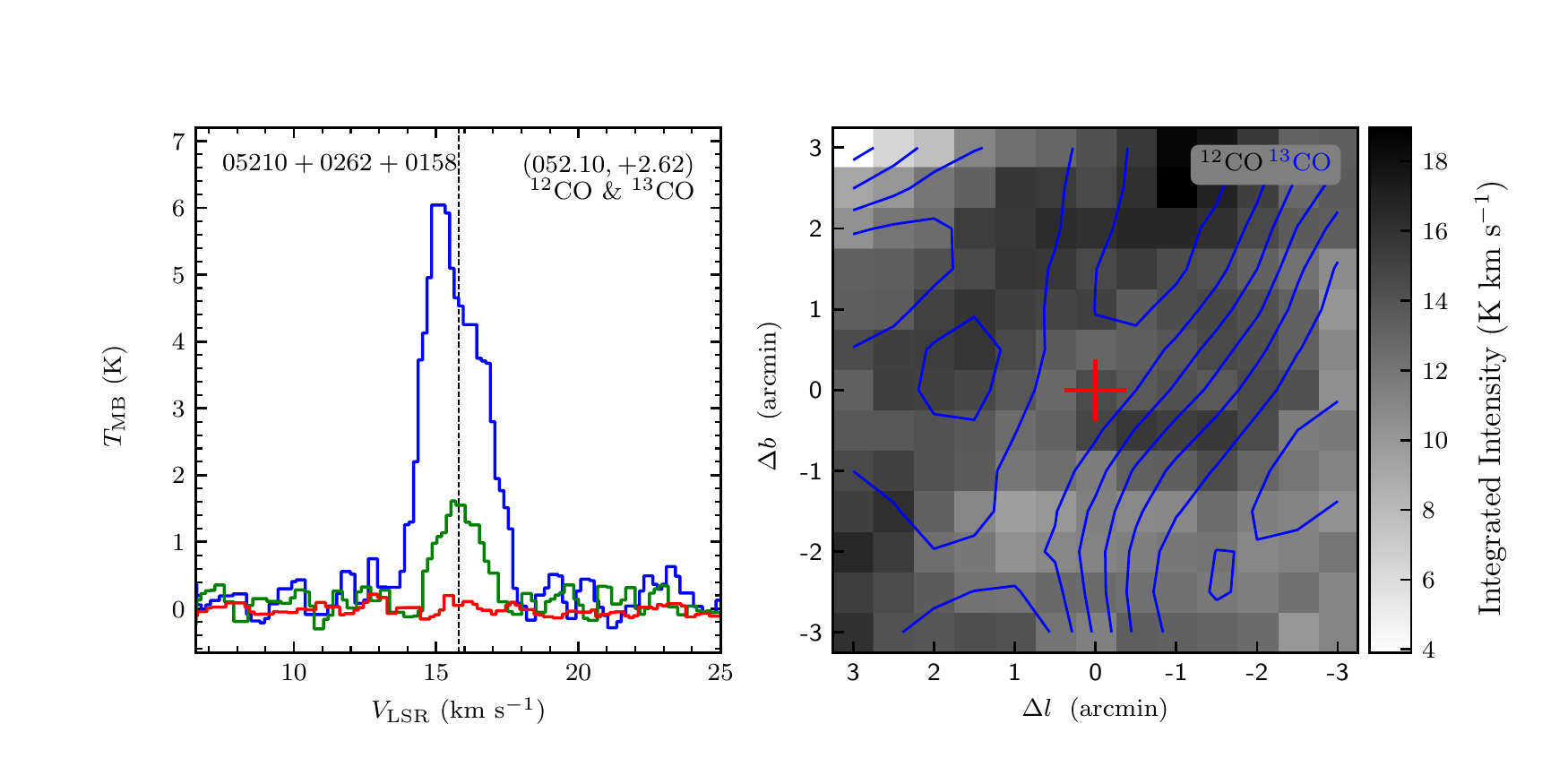}
\includegraphics[width=9.0cm,angle=0]{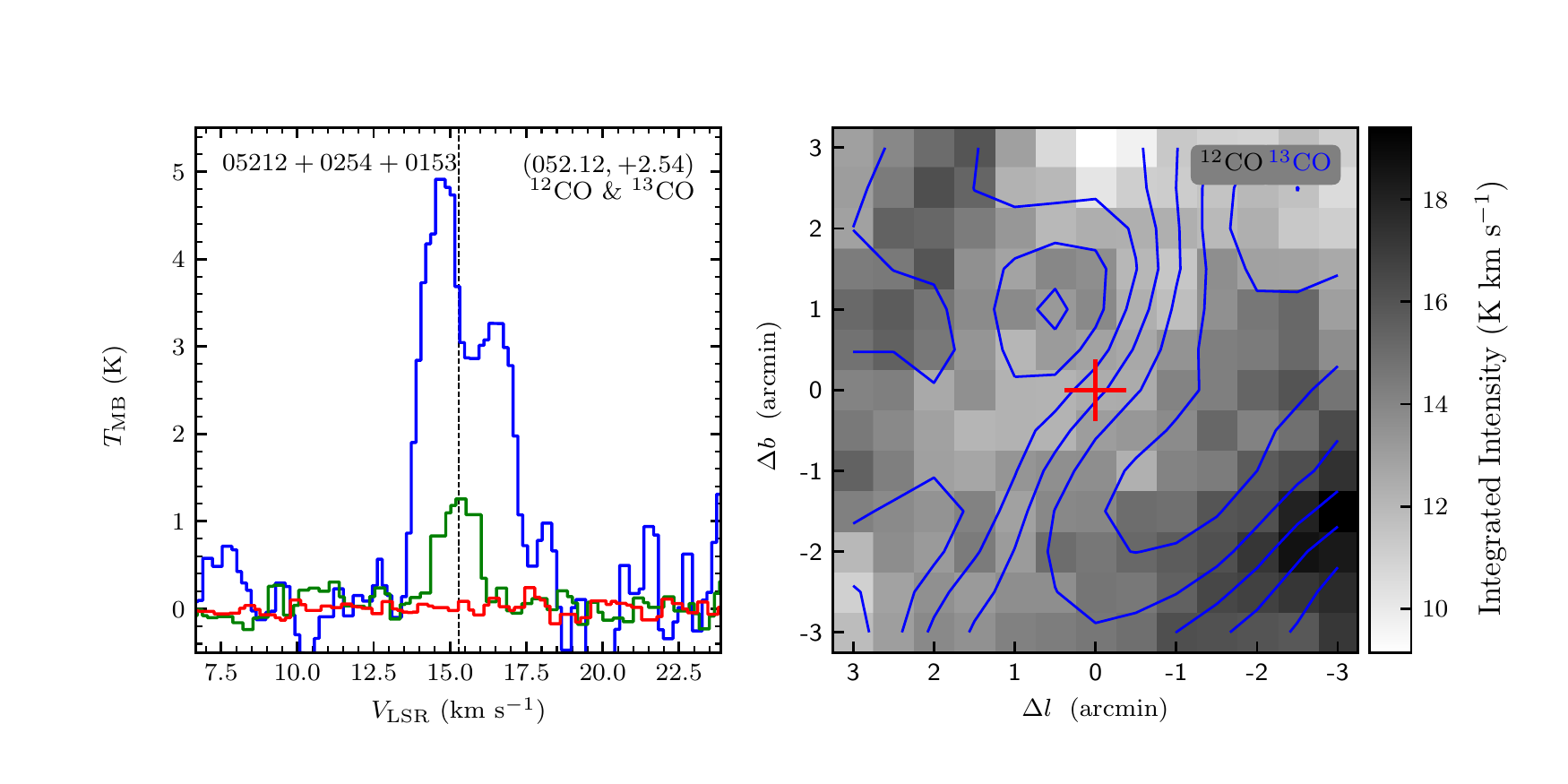}
\vspace{-0.5cm}

\includegraphics[width=9.0cm,angle=0]{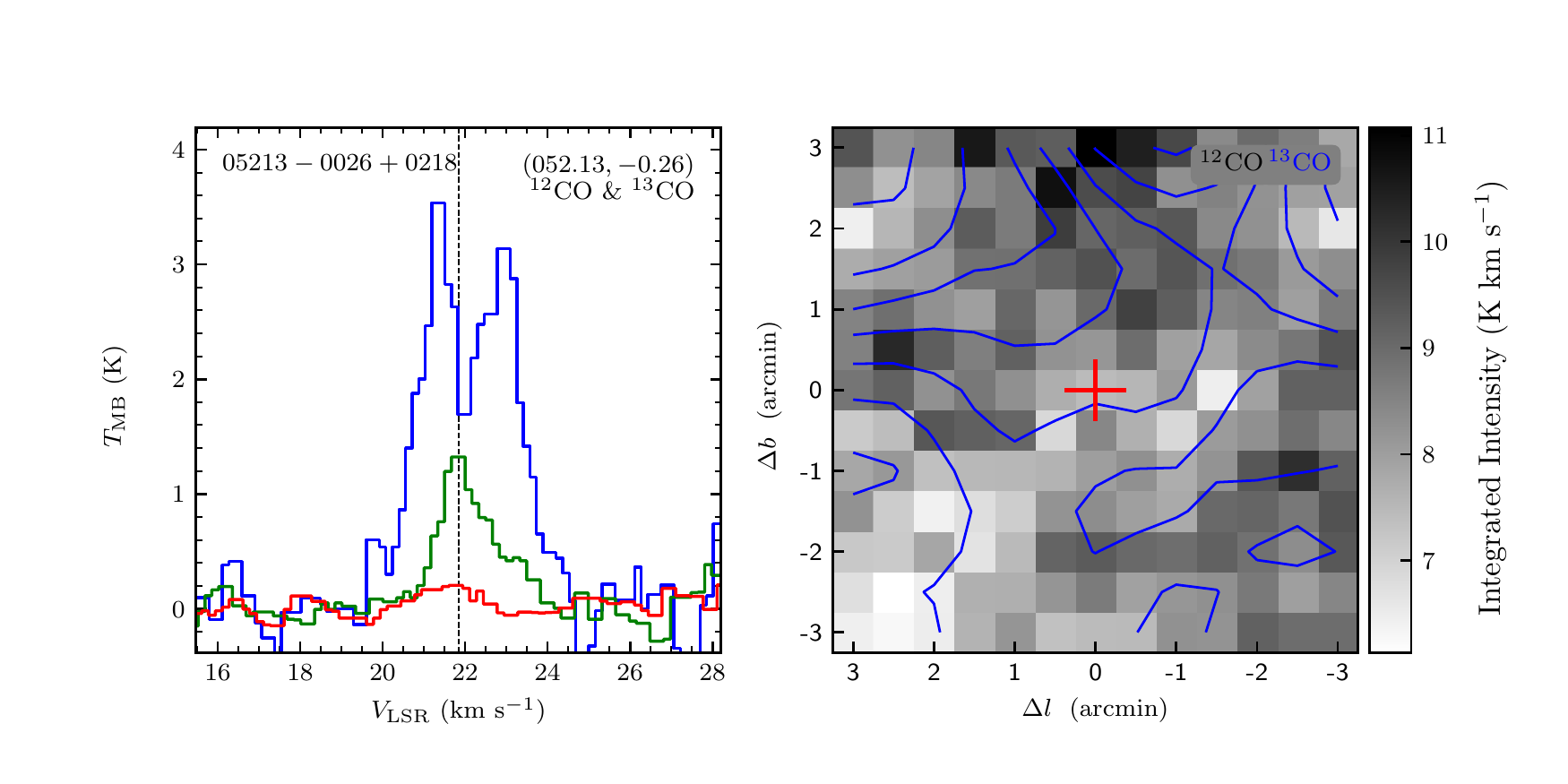}
\includegraphics[width=9.0cm,angle=0]{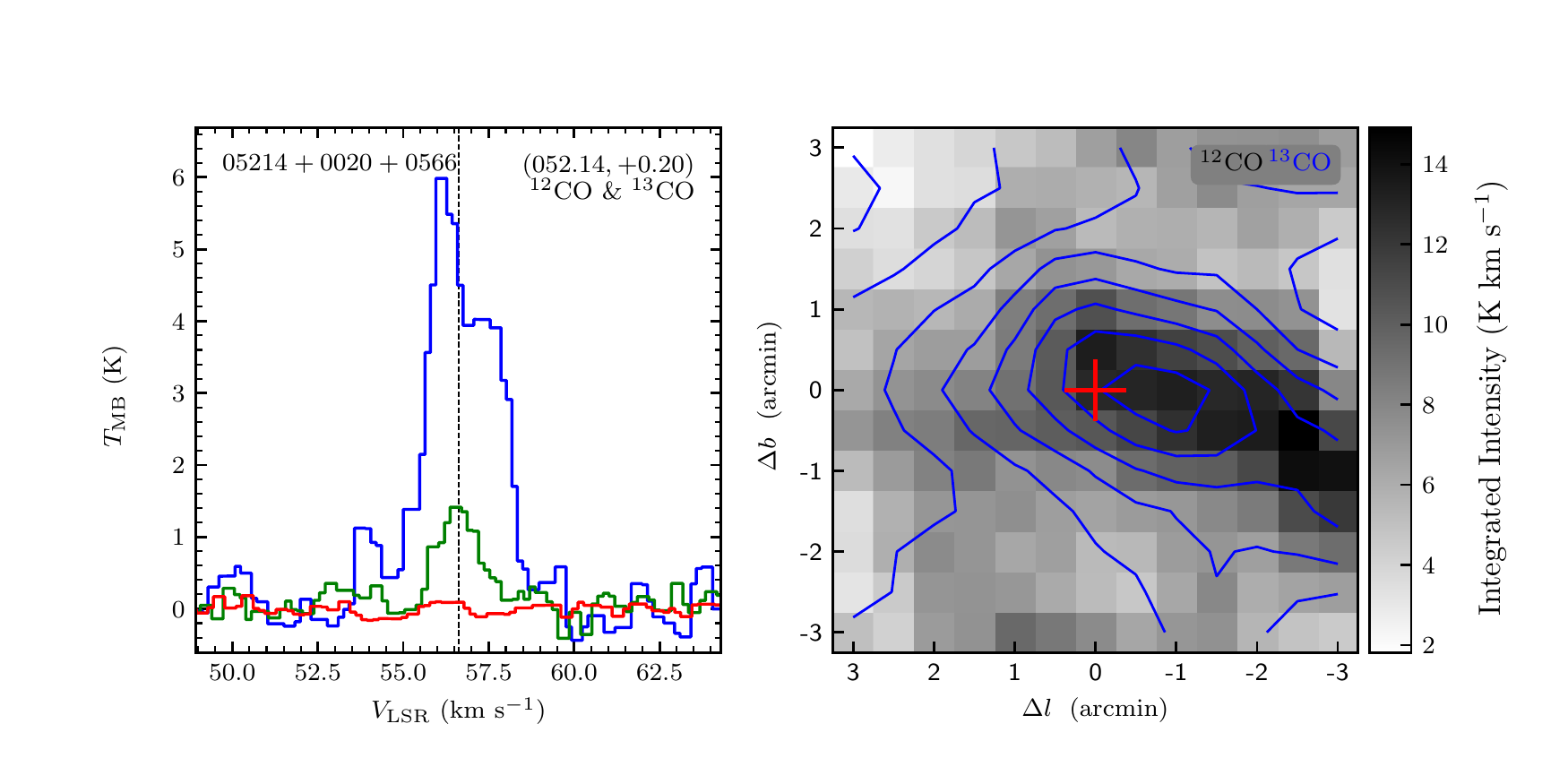}
\vspace{-0.5cm}

\includegraphics[width=9.0cm,angle=0]{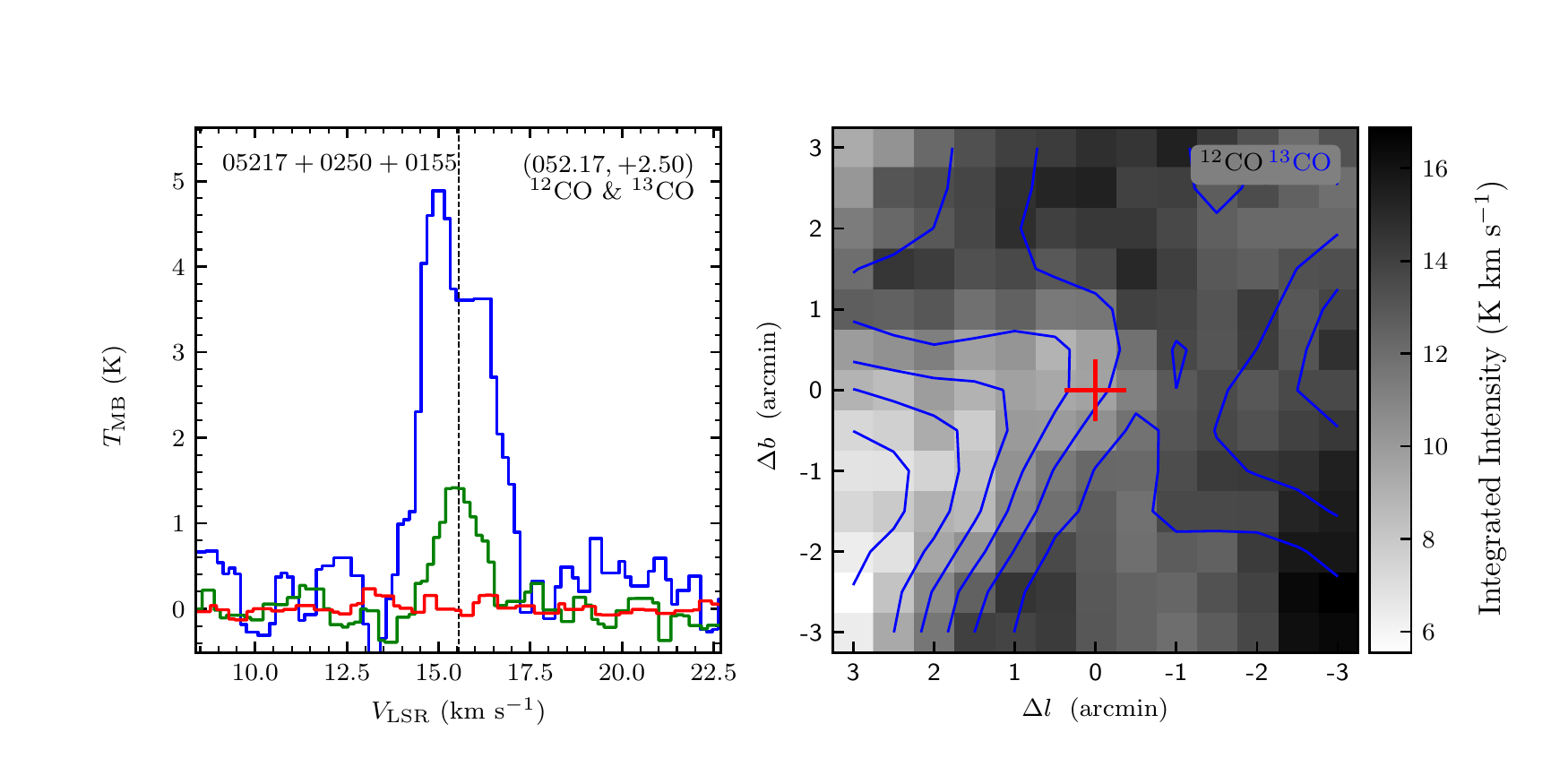}
\includegraphics[width=9.0cm,angle=0]{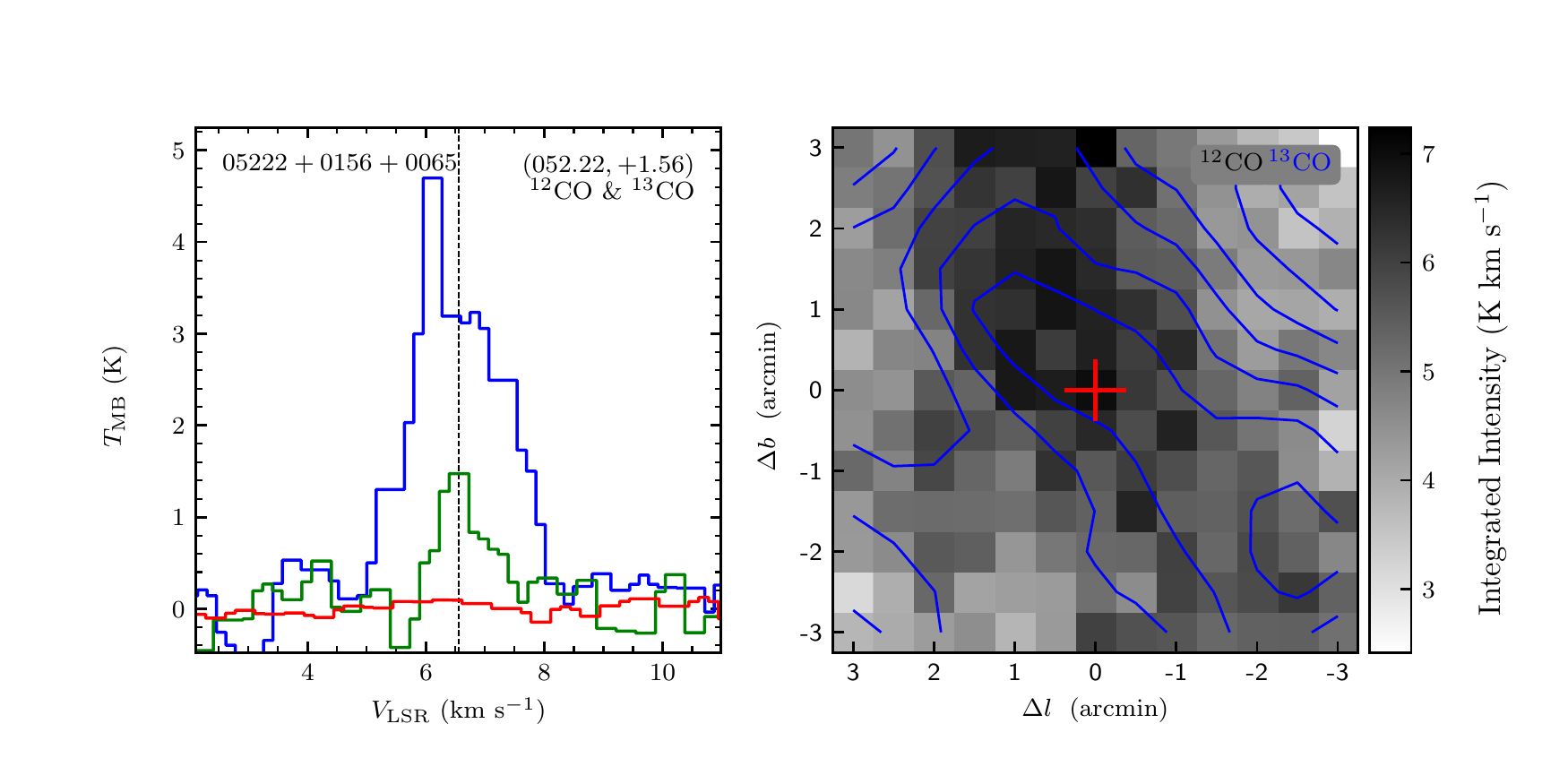}
\end{figure}
\clearpage

\begin{figure}
\includegraphics[width=9.0cm,angle=0]{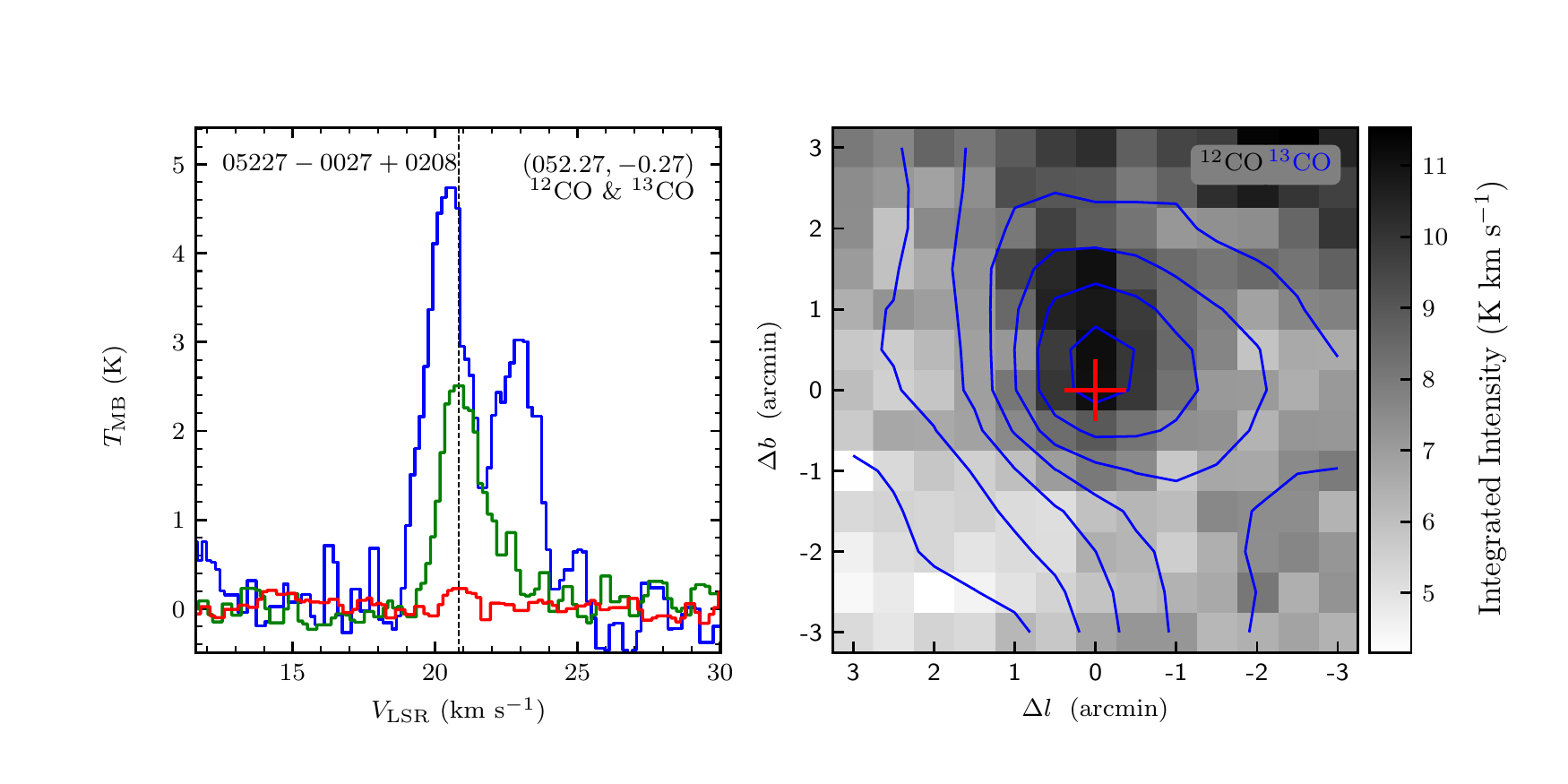}
\includegraphics[width=9.0cm,angle=0]{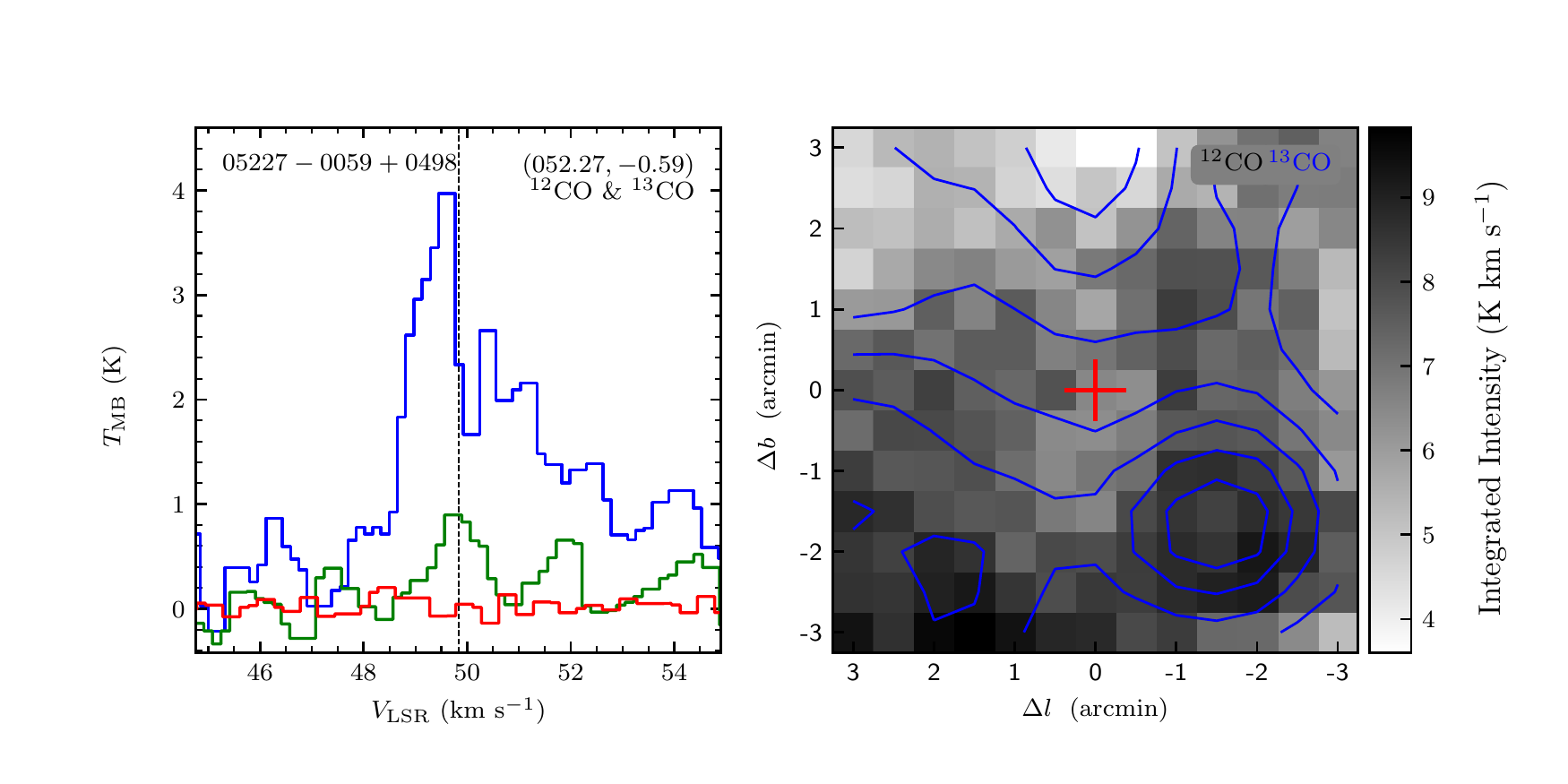}
\vspace{-0.5cm}

\includegraphics[width=9.0cm,angle=0]{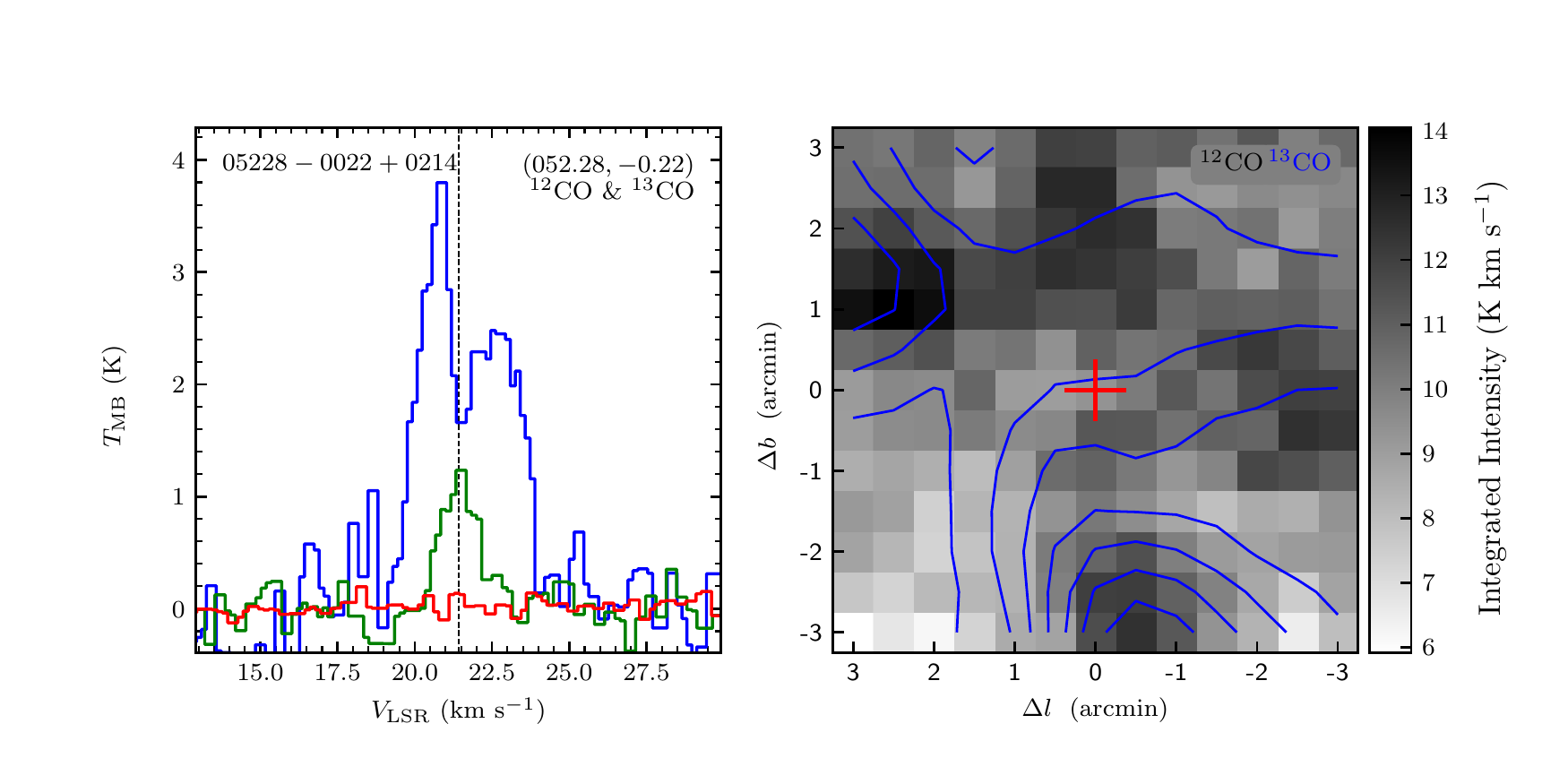}
\includegraphics[width=9.0cm,angle=0]{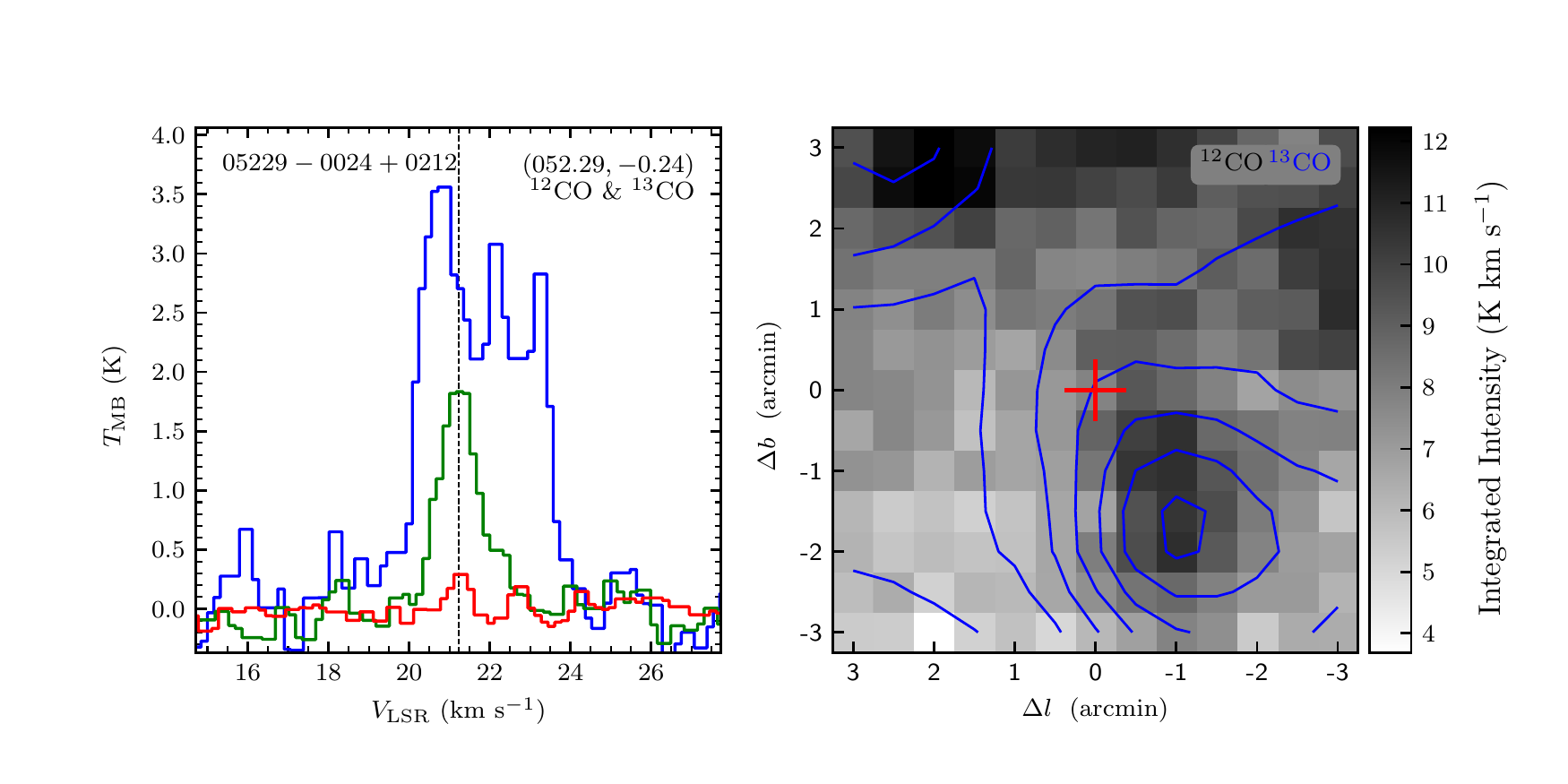}
\vspace{-0.5cm}

\includegraphics[width=9.0cm,angle=0]{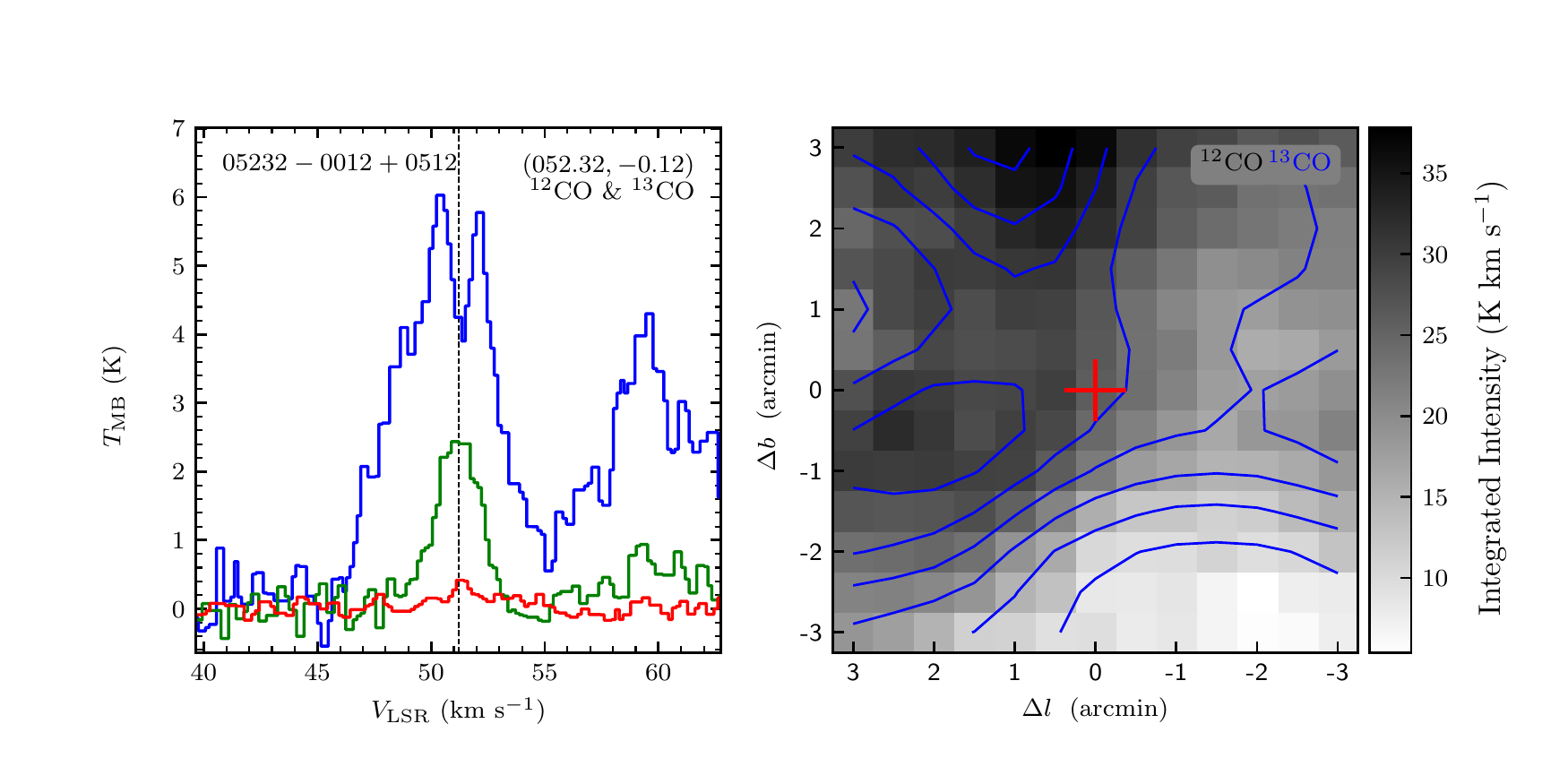}
\includegraphics[width=9.0cm,angle=0]{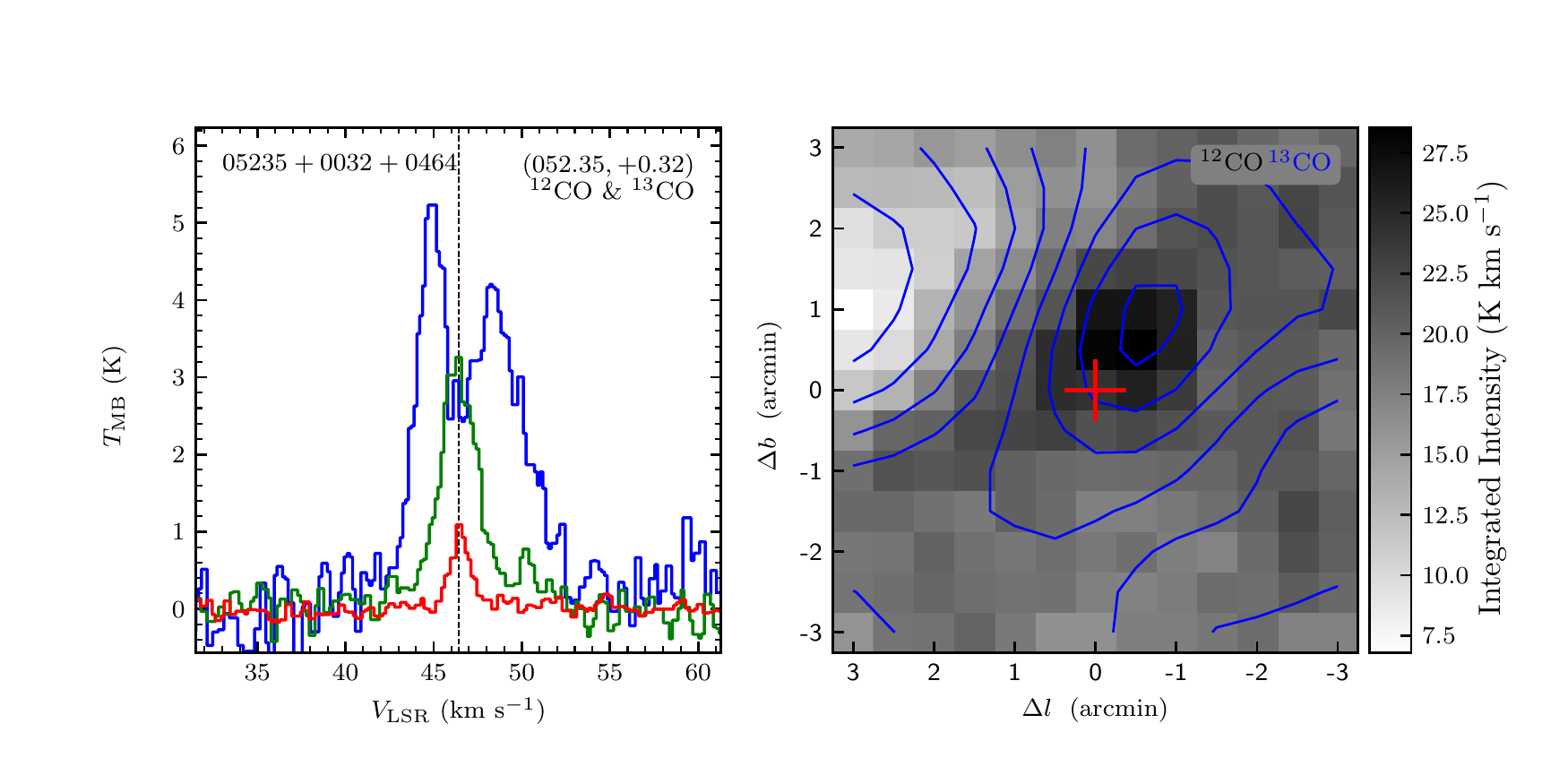}
\vspace{-0.5cm}

\includegraphics[width=9.0cm,angle=0]{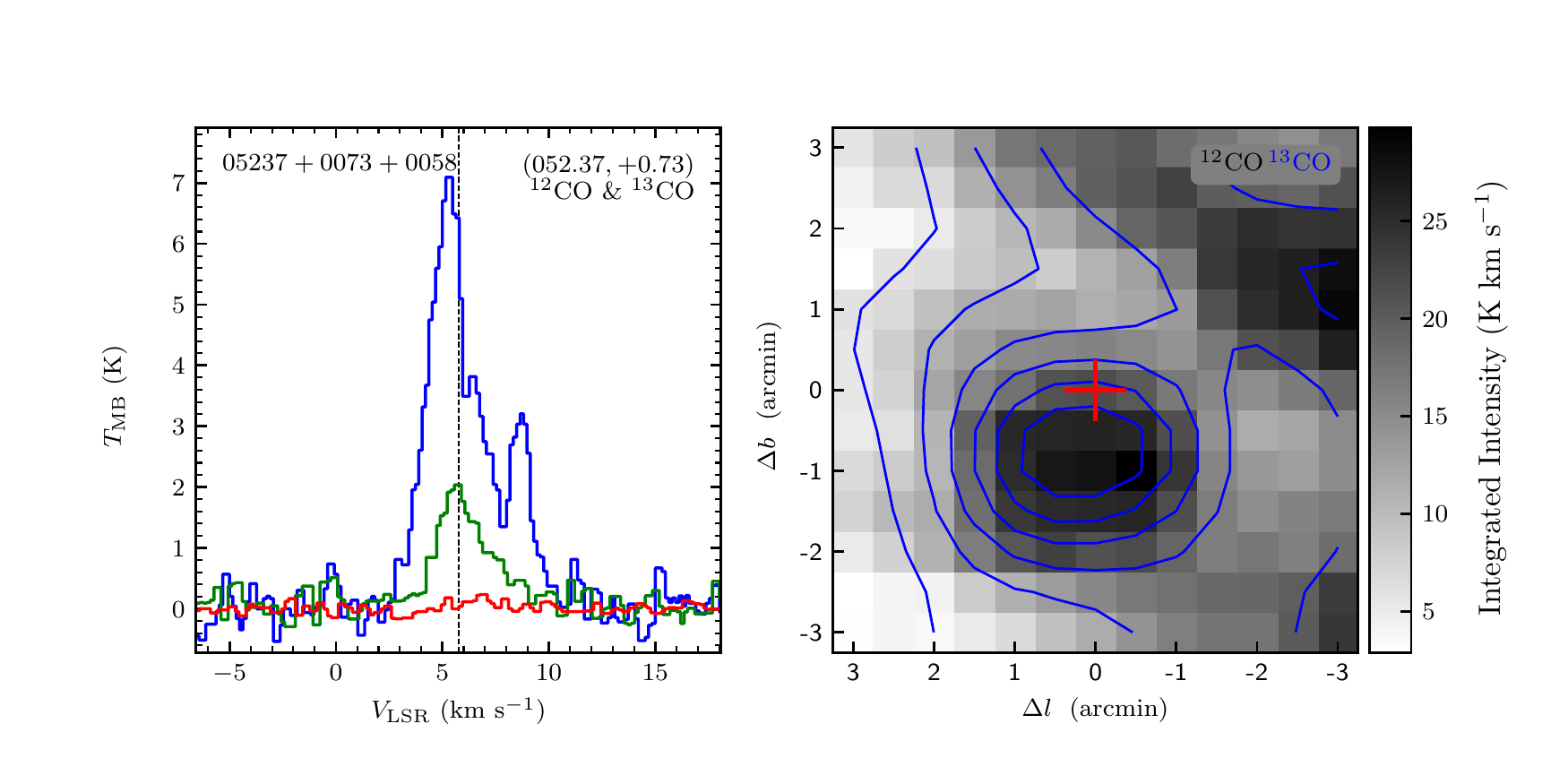}
\includegraphics[width=9.0cm,angle=0]{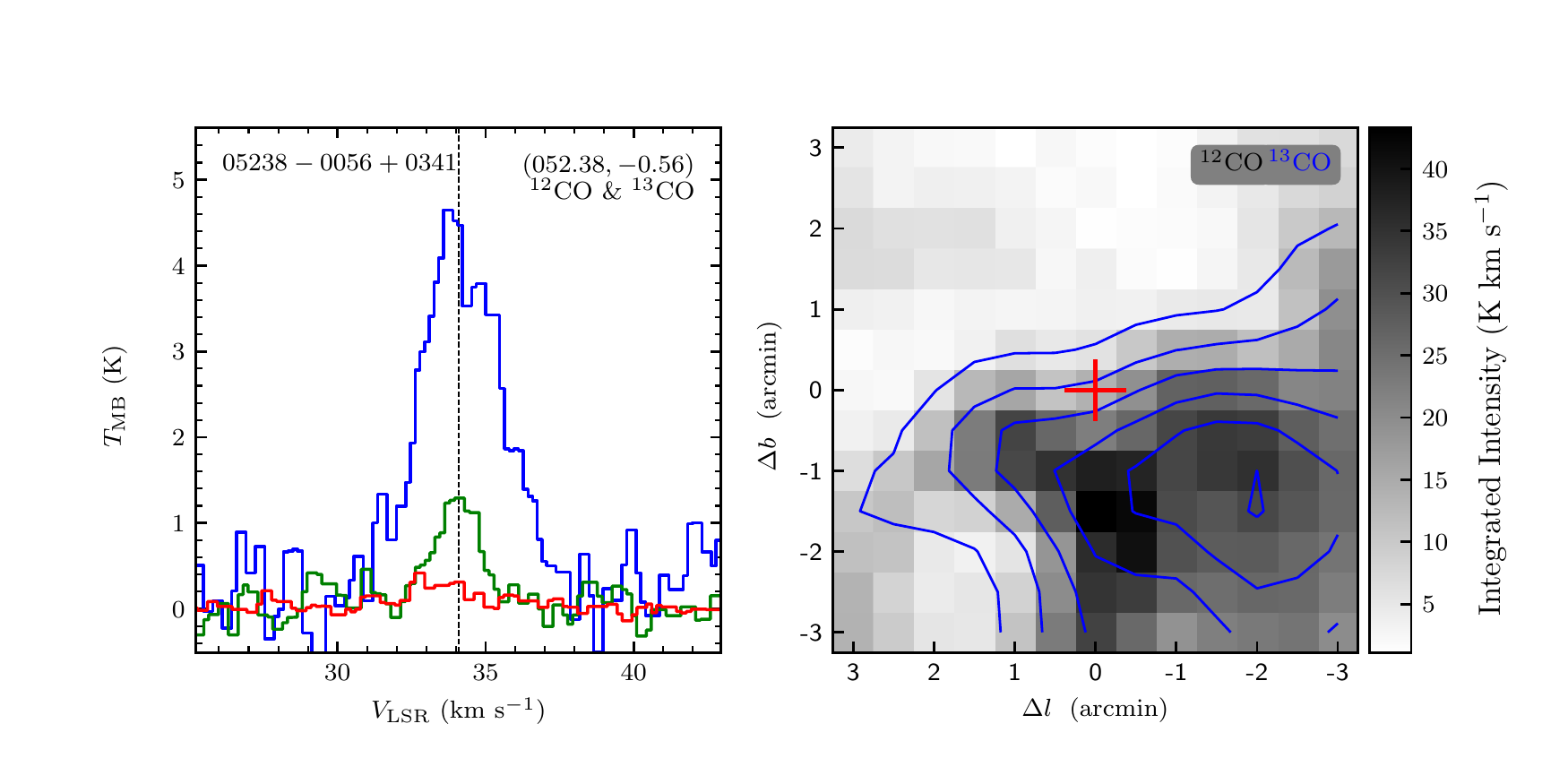}
\vspace{-0.5cm}

\includegraphics[width=9.0cm,angle=0]{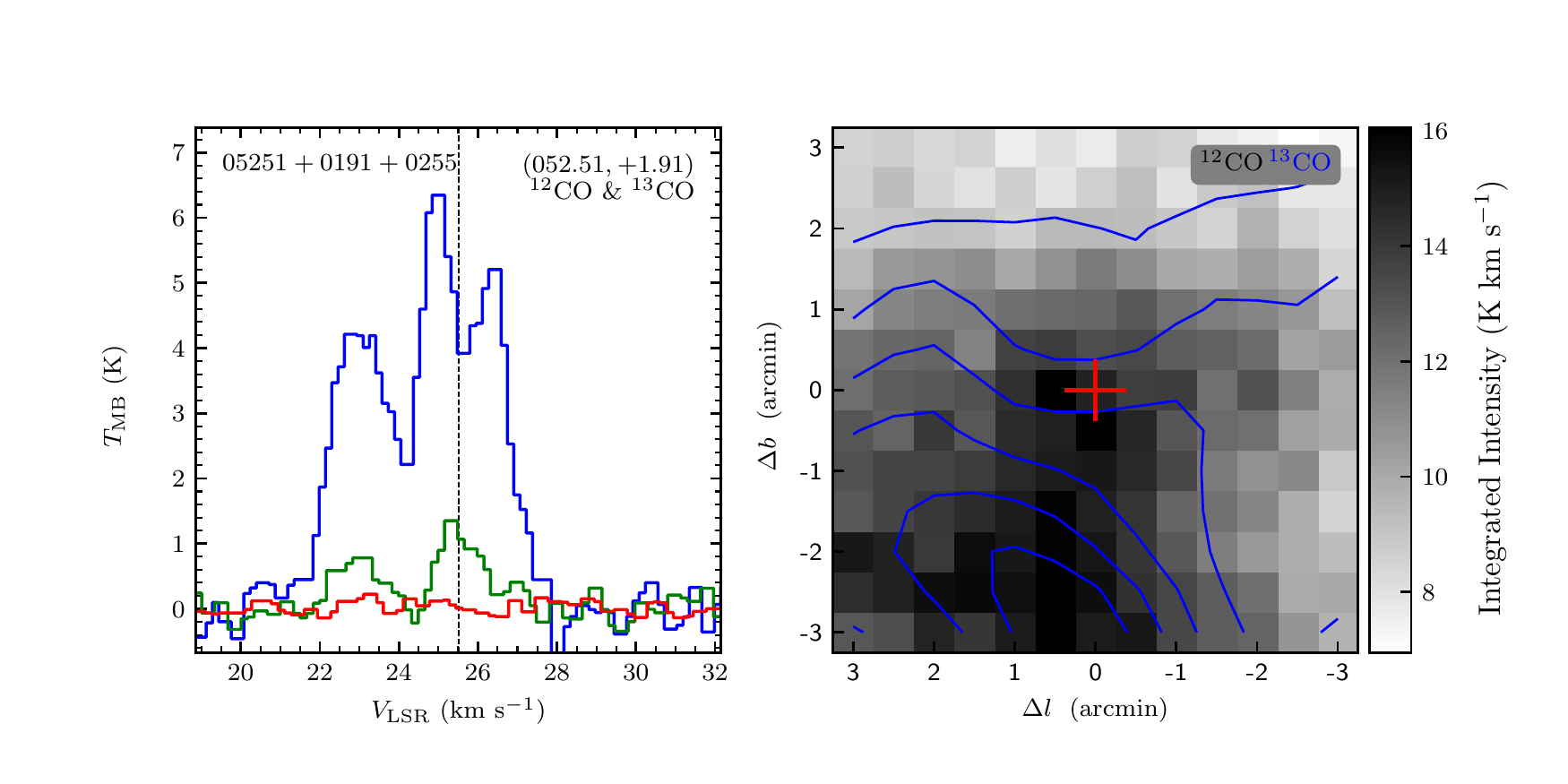}
\includegraphics[width=9.0cm,angle=0]{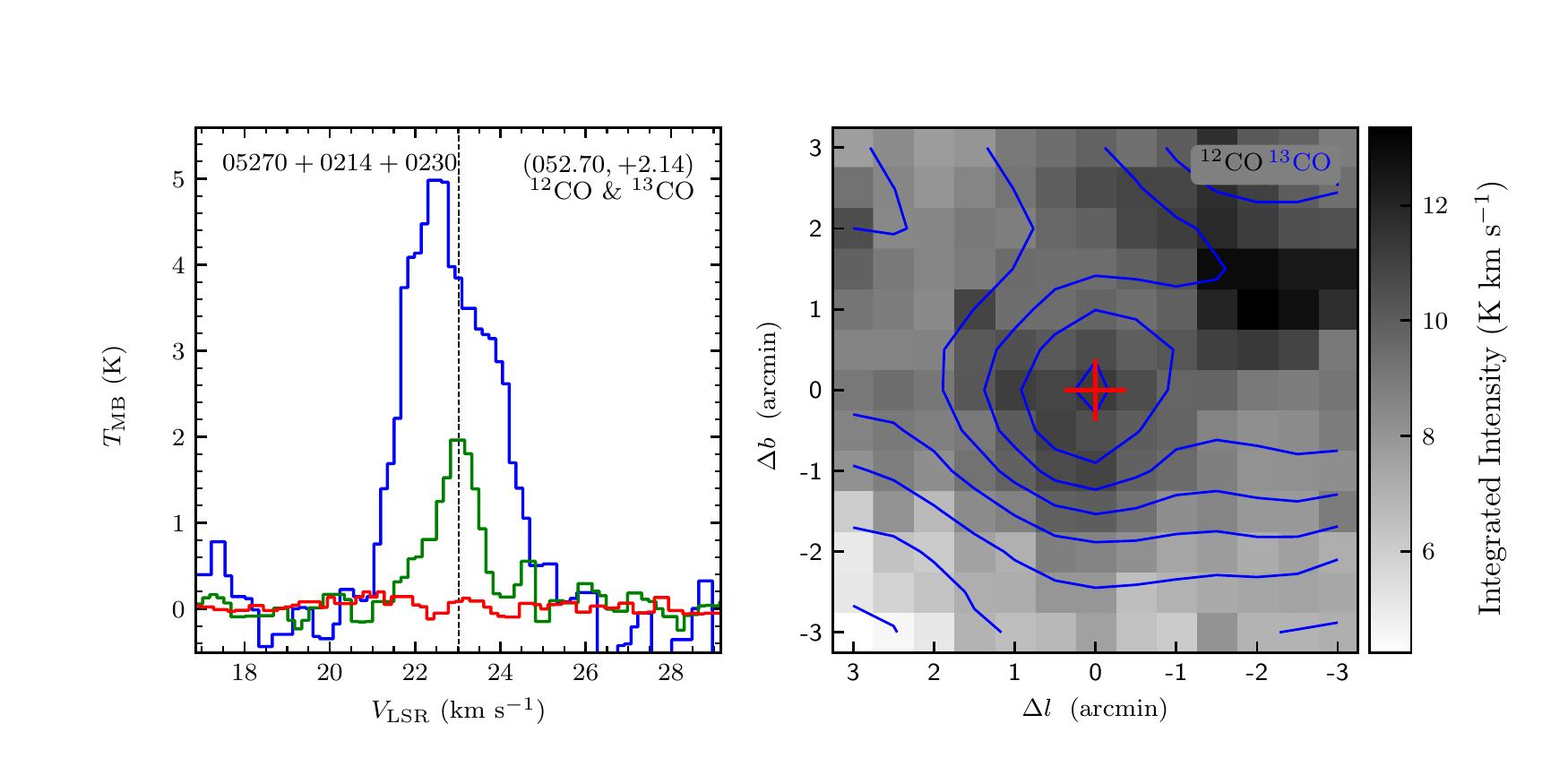}
\end{figure}
\clearpage

\begin{figure}
\includegraphics[width=9.0cm,angle=0]{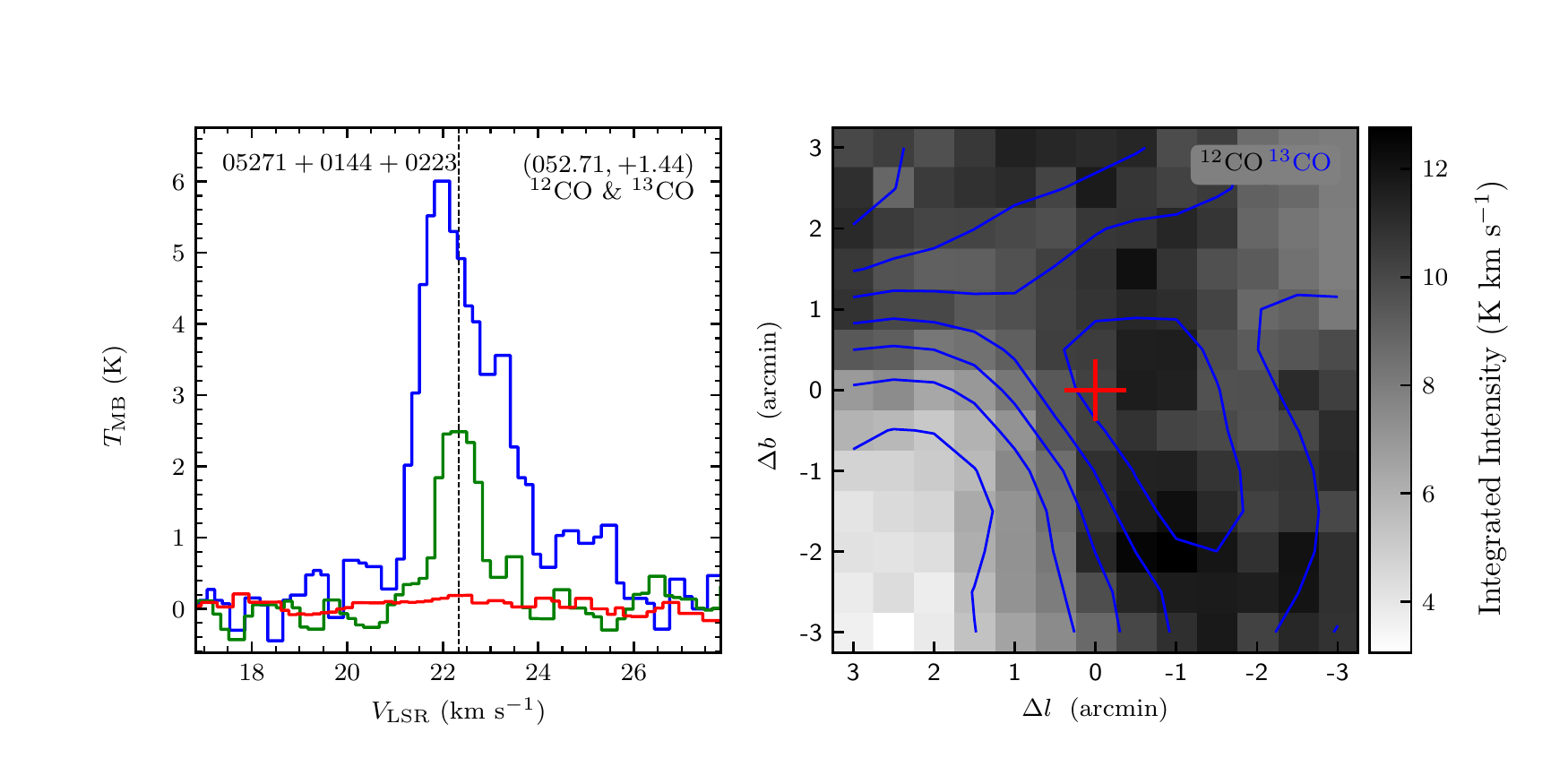}
\includegraphics[width=9.0cm,angle=0]{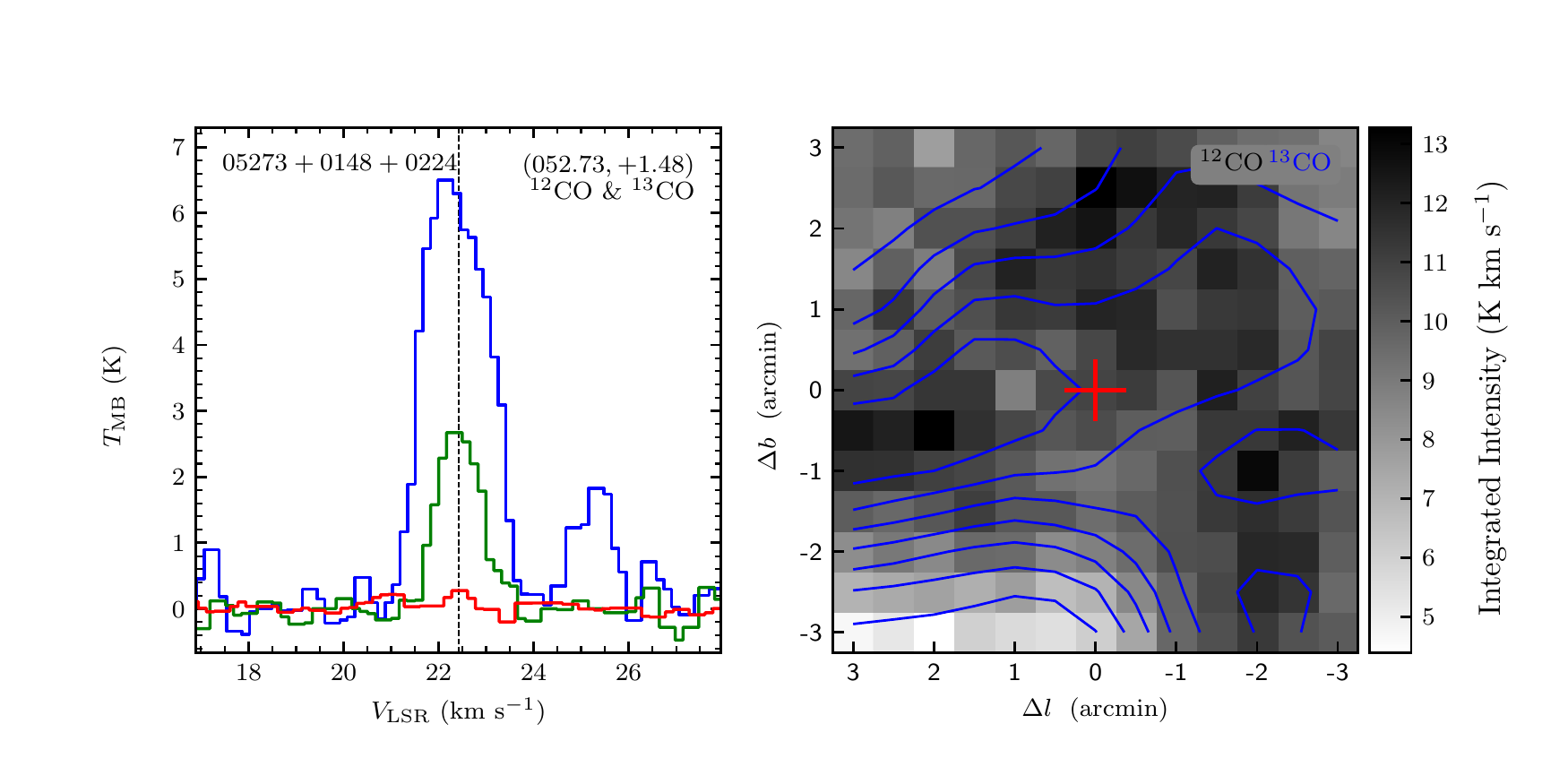}
\vspace{-0.5cm}

\includegraphics[width=9.0cm,angle=0]{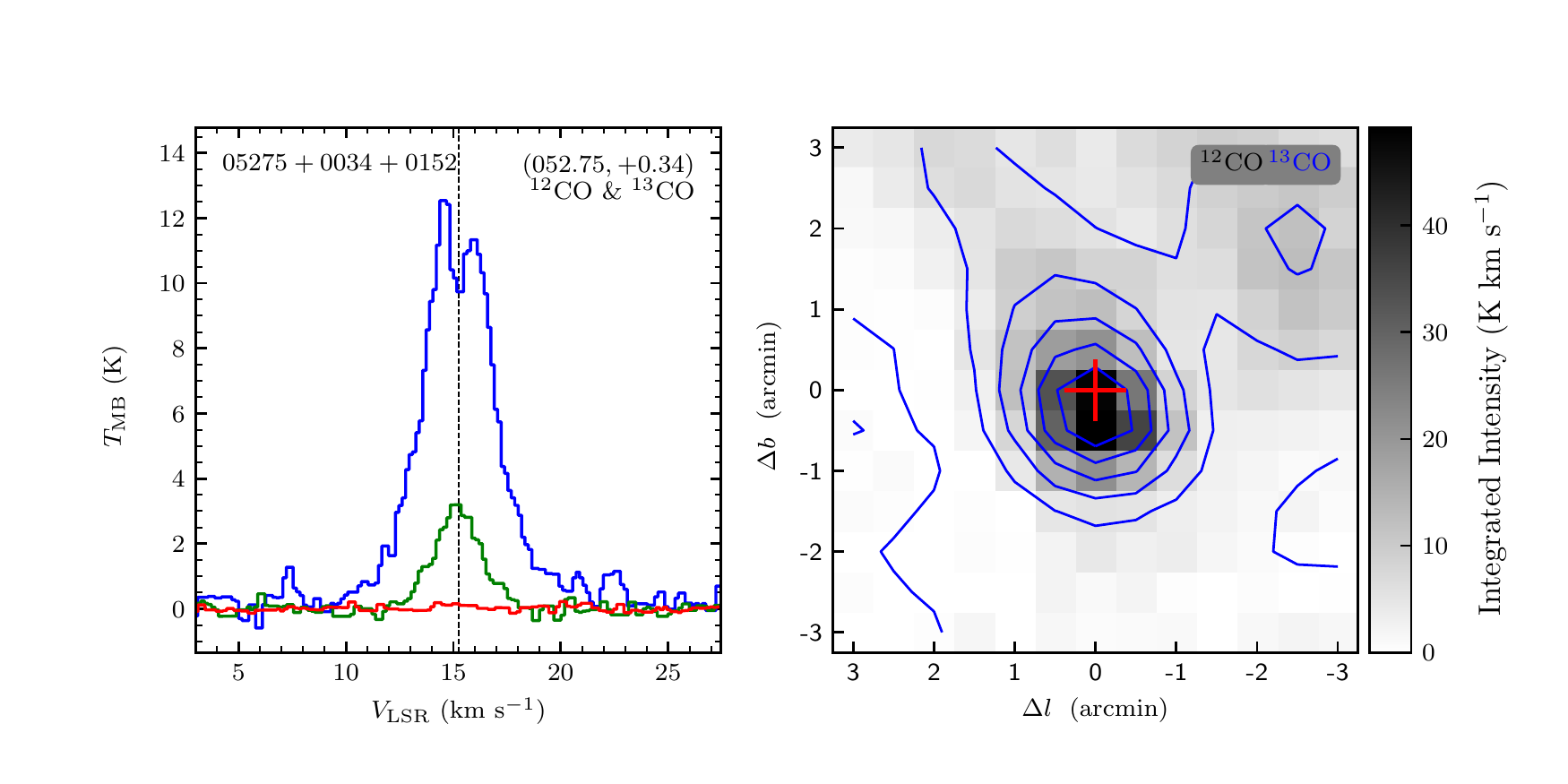}
\includegraphics[width=9.0cm,angle=0]{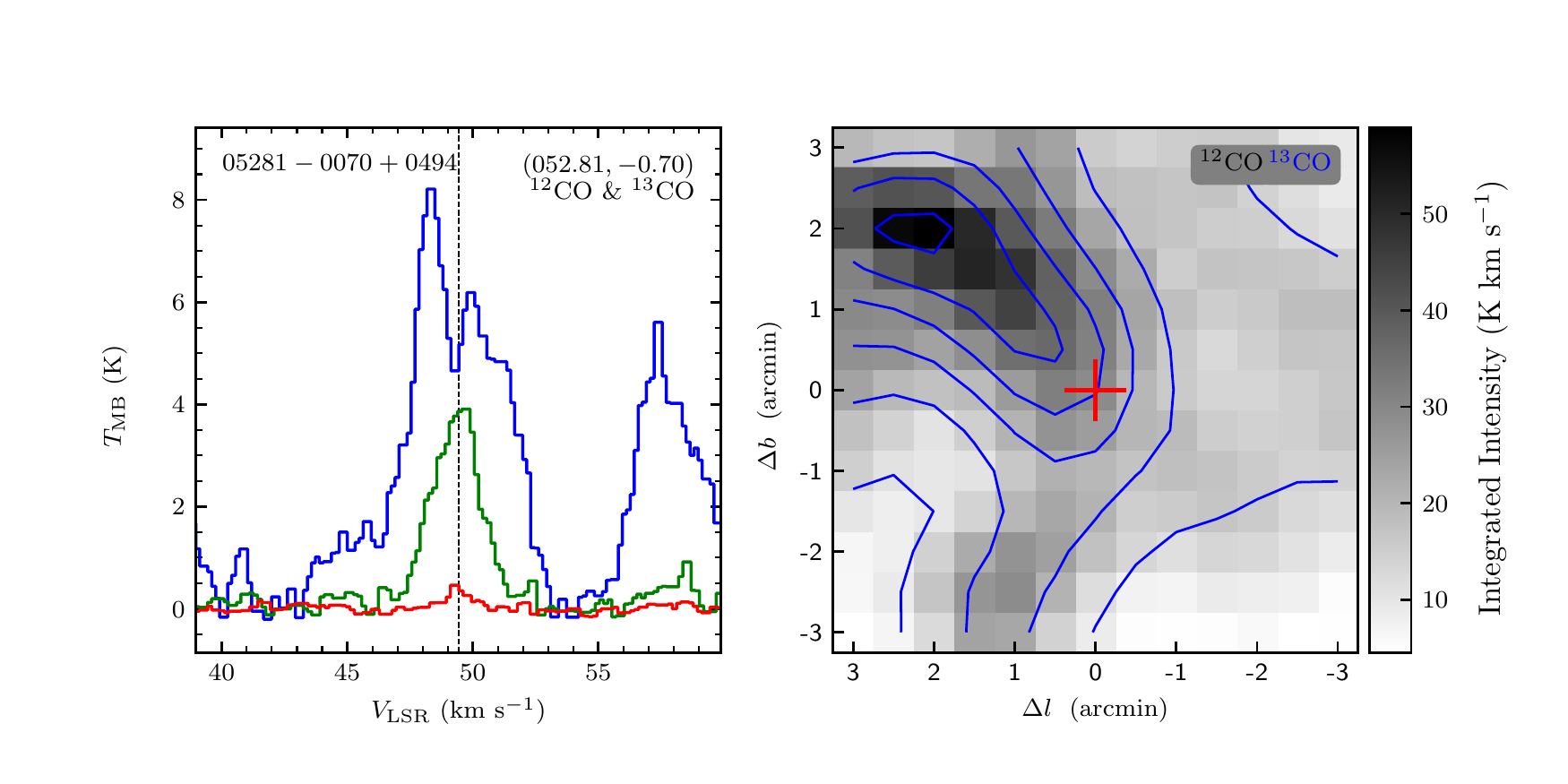}
\vspace{-0.5cm}

\includegraphics[width=9.0cm,angle=0]{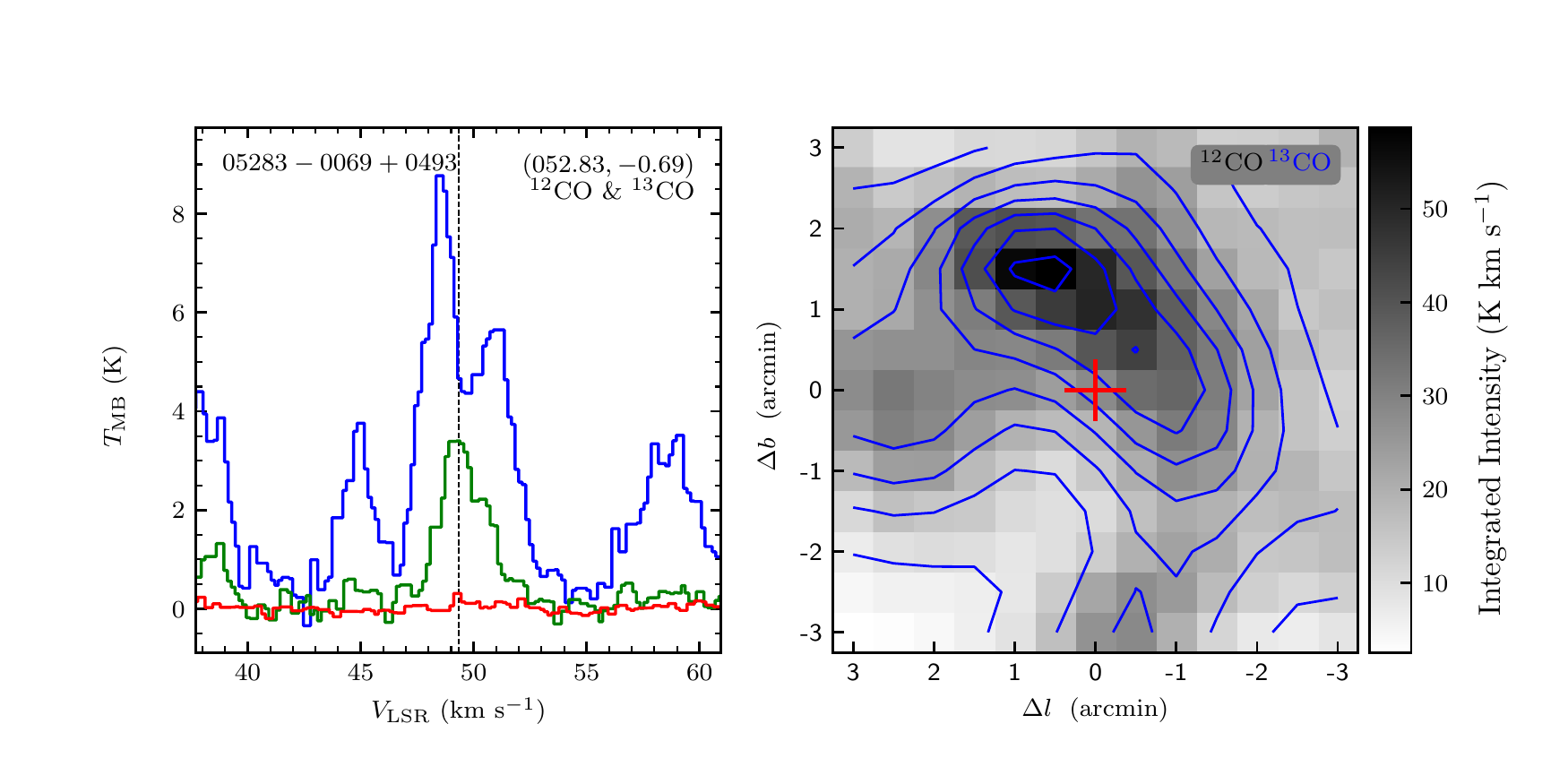}
\includegraphics[width=9.0cm,angle=0]{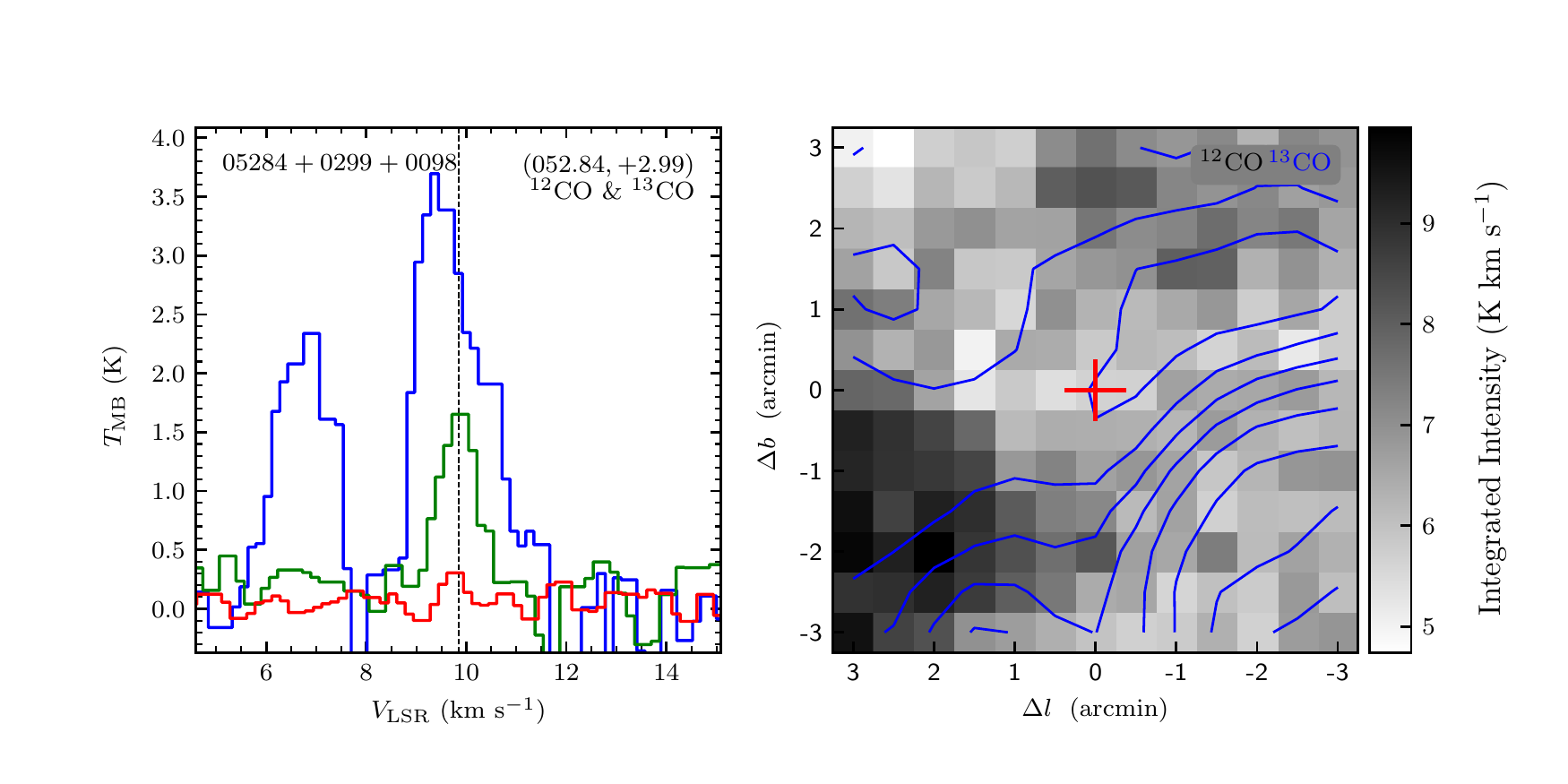}
\vspace{-0.5cm}

\includegraphics[width=9.0cm,angle=0]{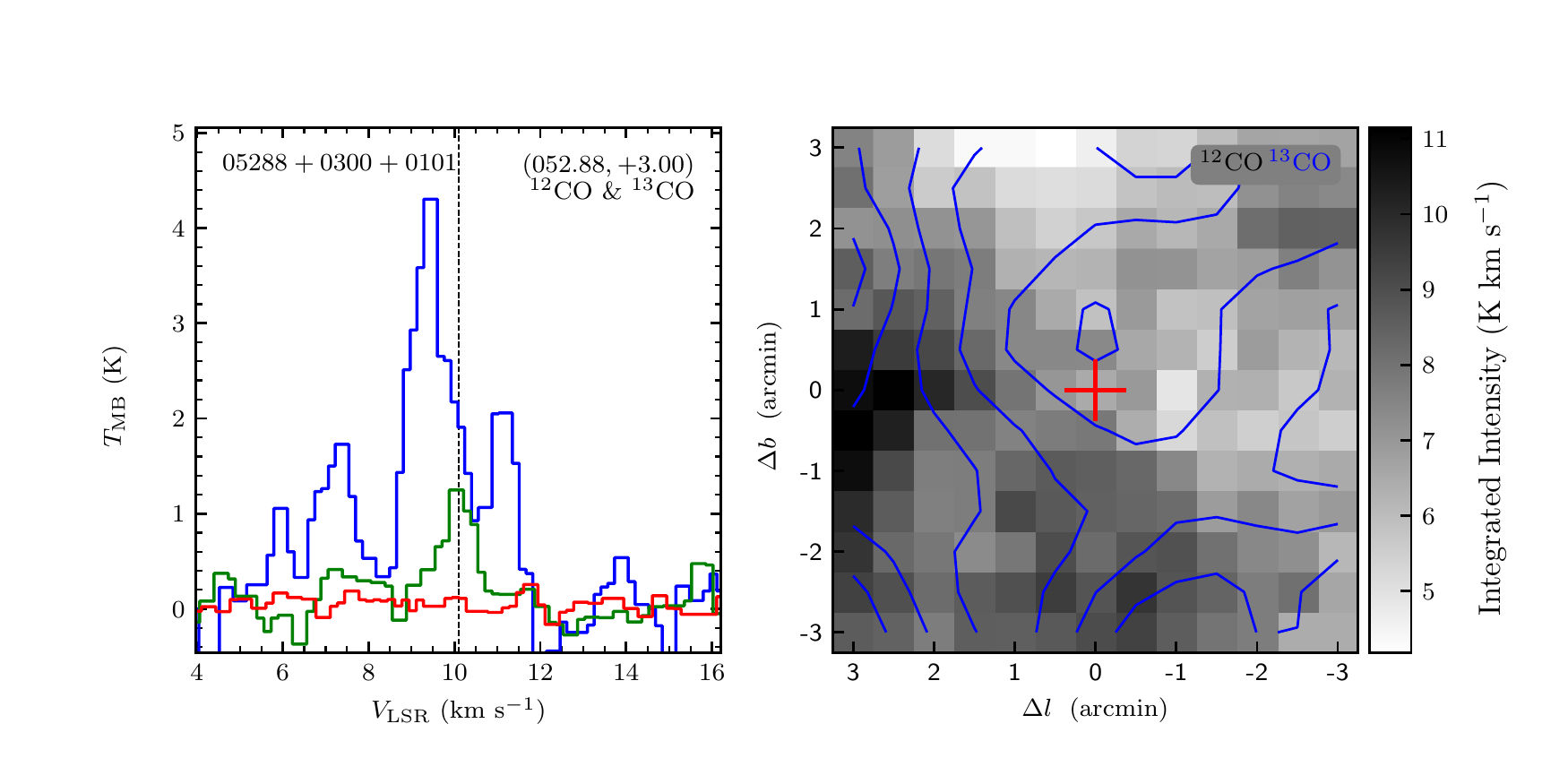}
\includegraphics[width=9.0cm,angle=0]{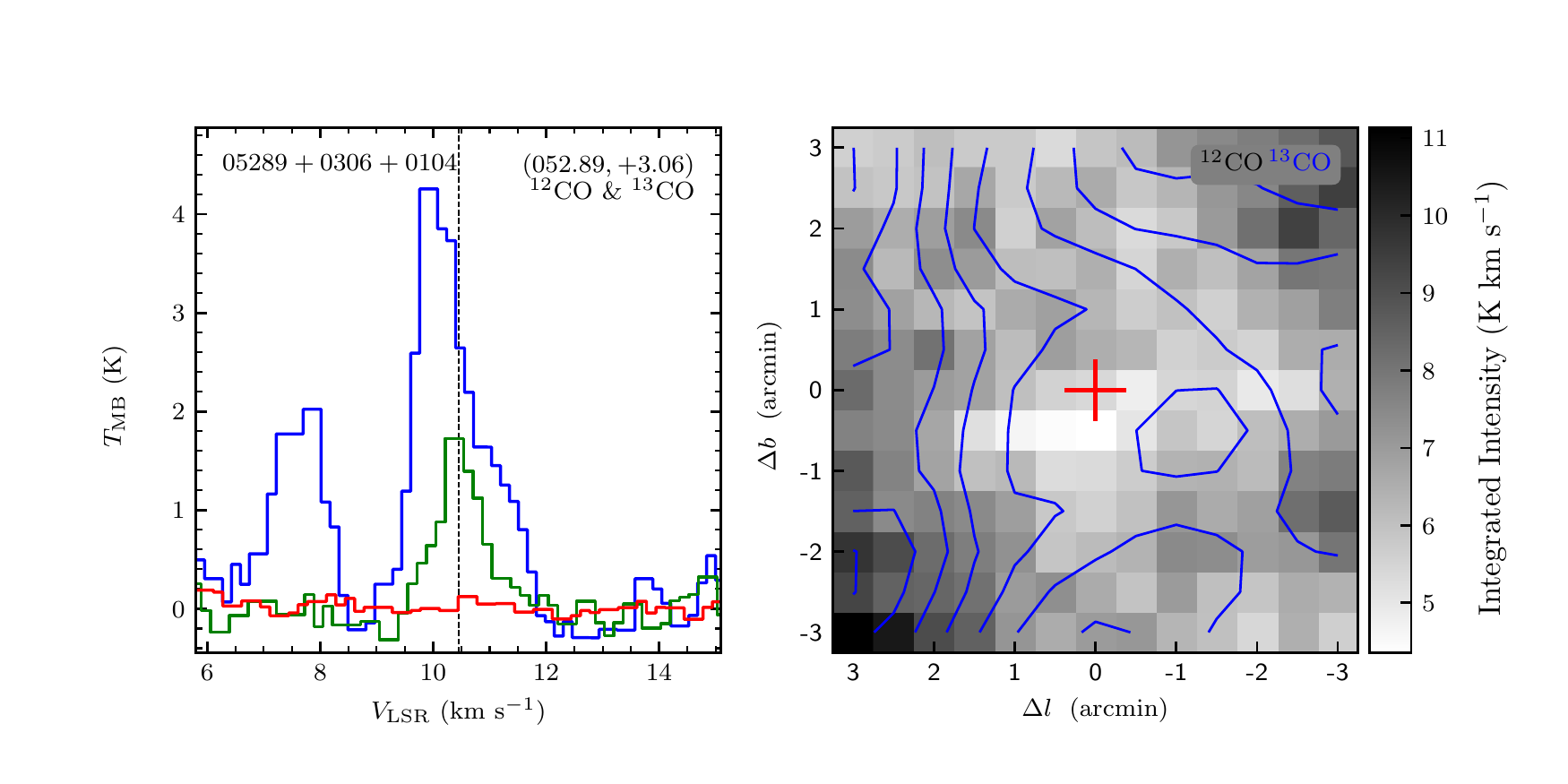}
\vspace{-0.5cm}

\includegraphics[width=9.0cm,angle=0]{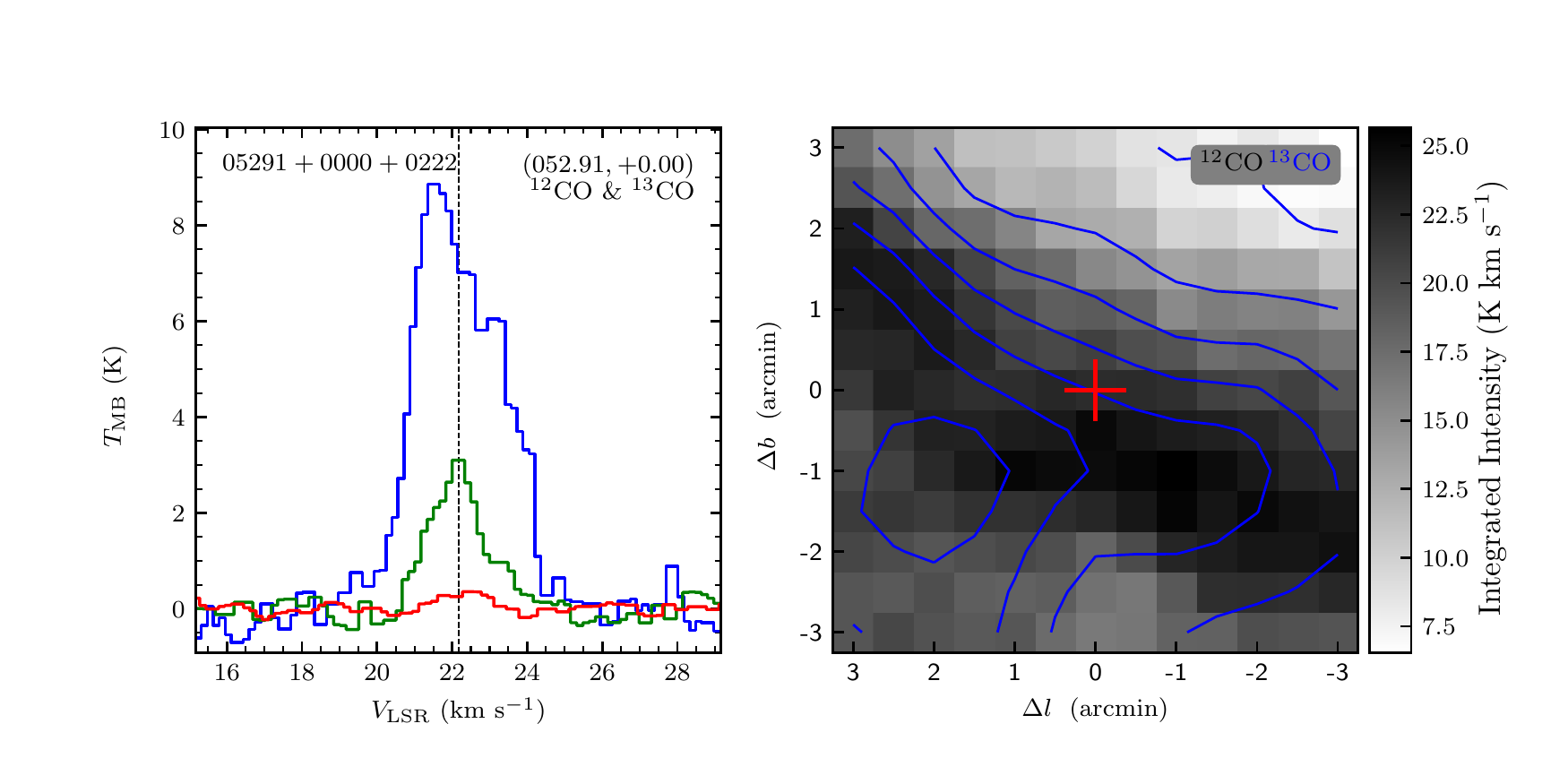}
\includegraphics[width=9.0cm,angle=0]{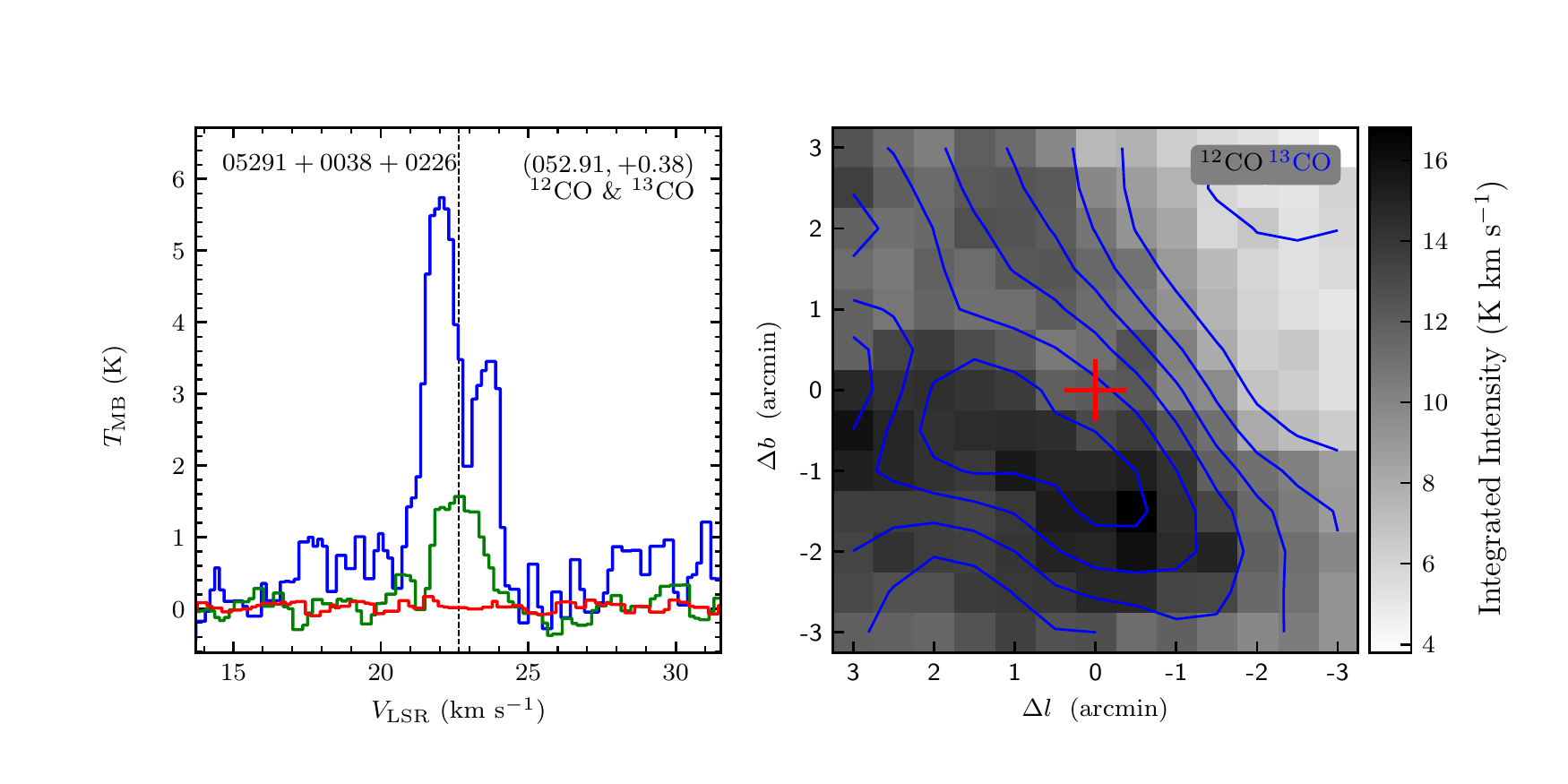}
\end{figure}
\clearpage

\begin{figure}
\includegraphics[width=9.0cm,angle=0]{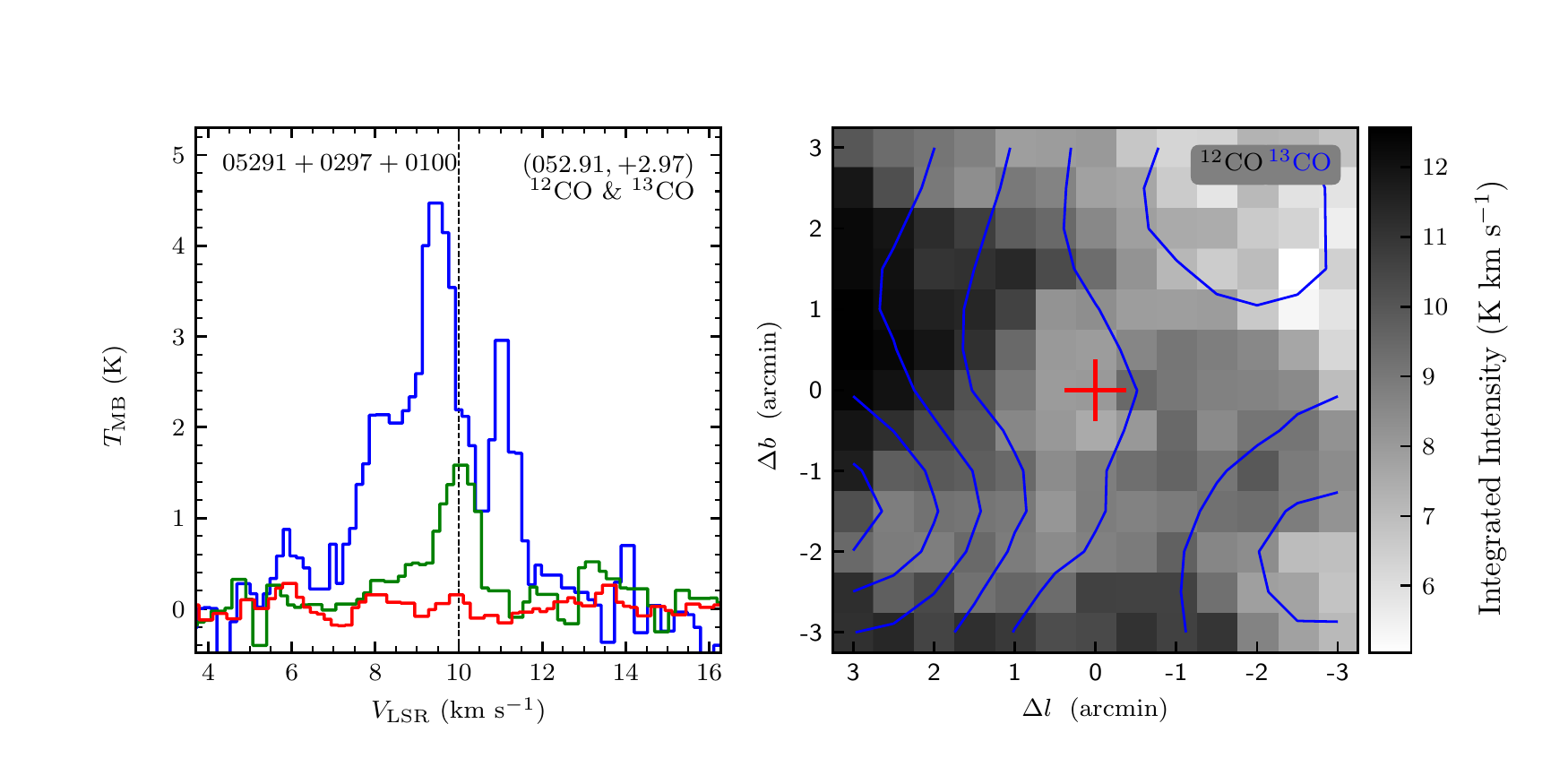}
\includegraphics[width=9.0cm,angle=0]{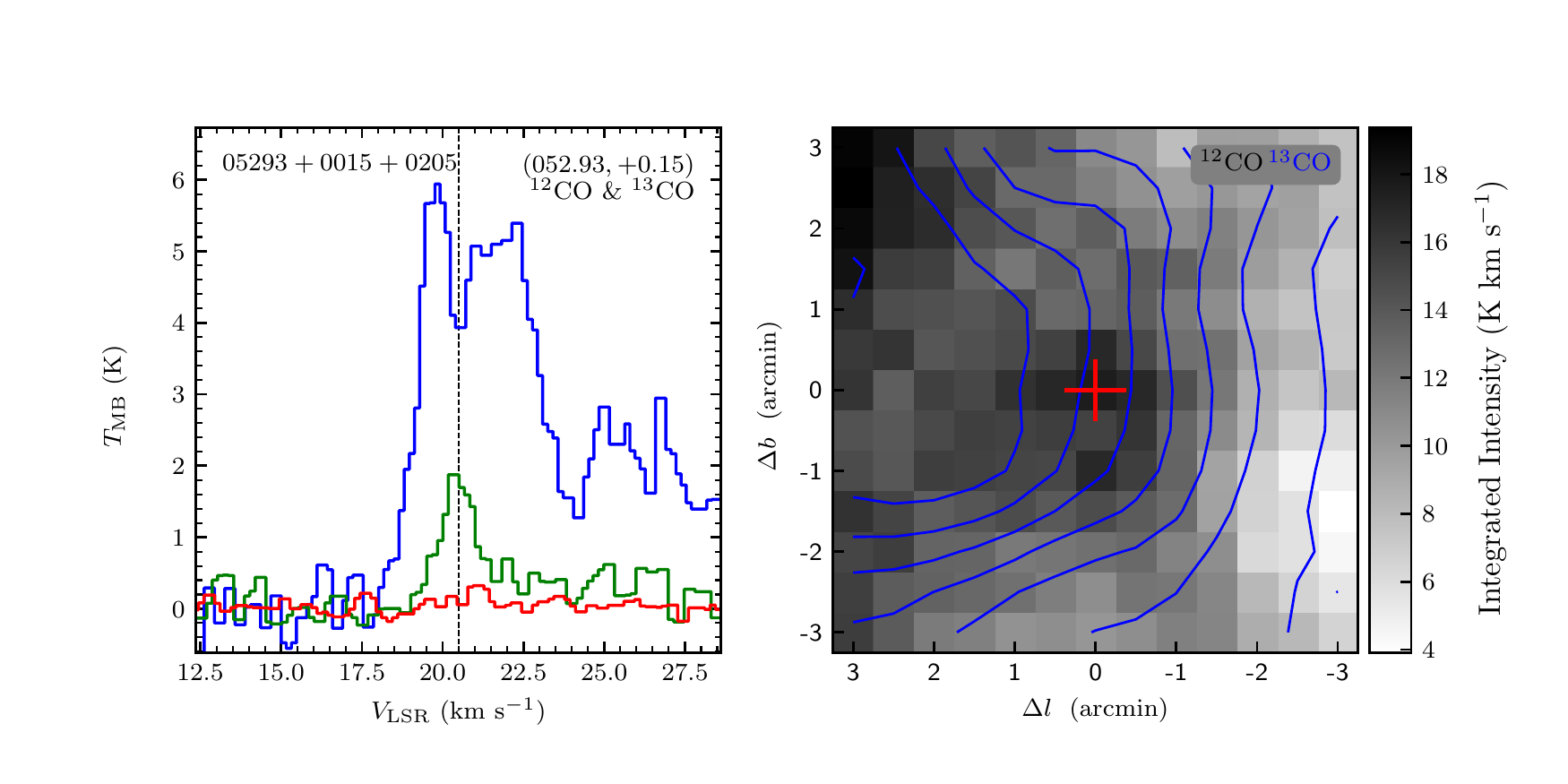}
\vspace{-0.5cm}

\includegraphics[width=9.0cm,angle=0]{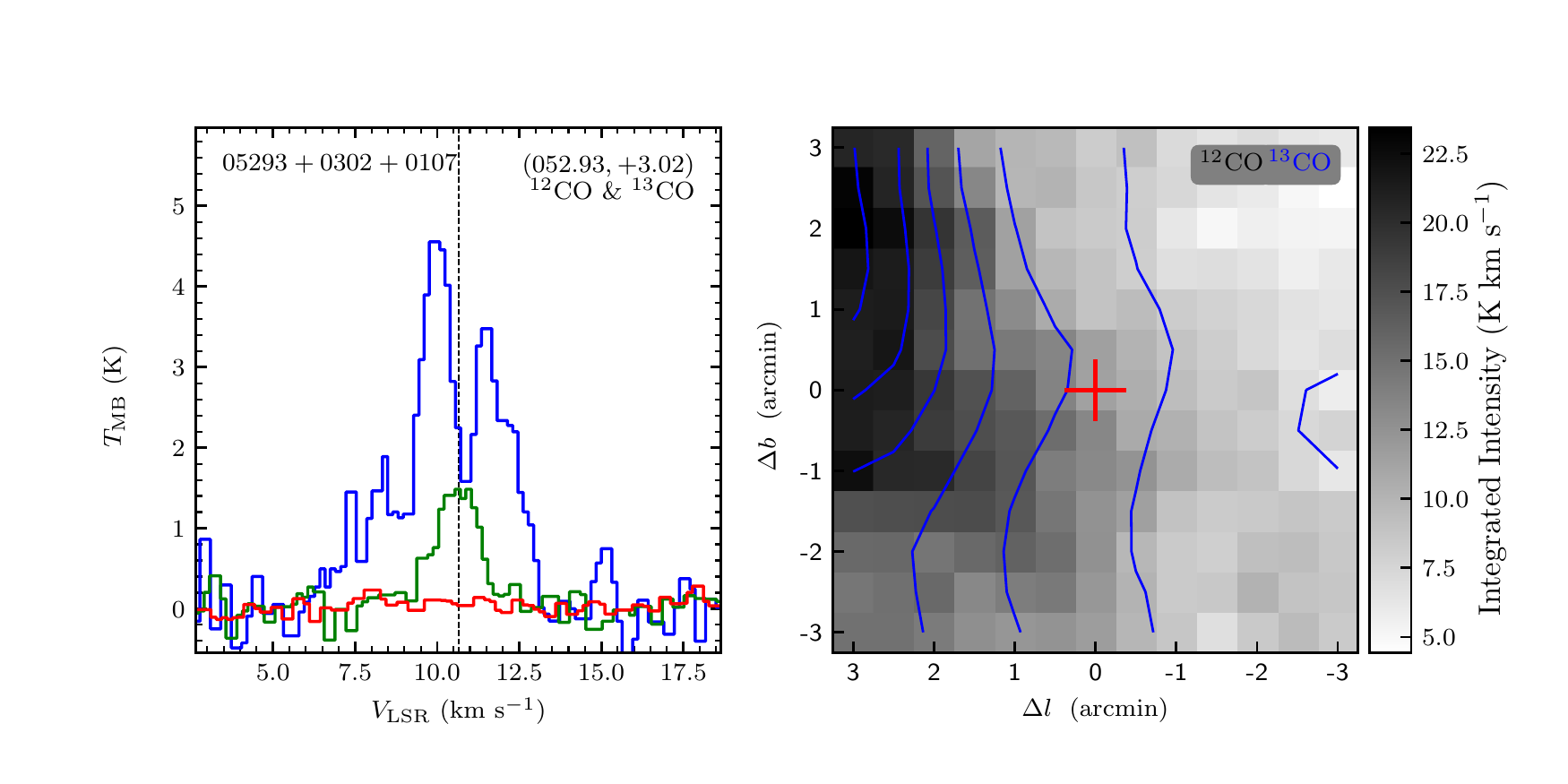}
\includegraphics[width=9.0cm,angle=0]{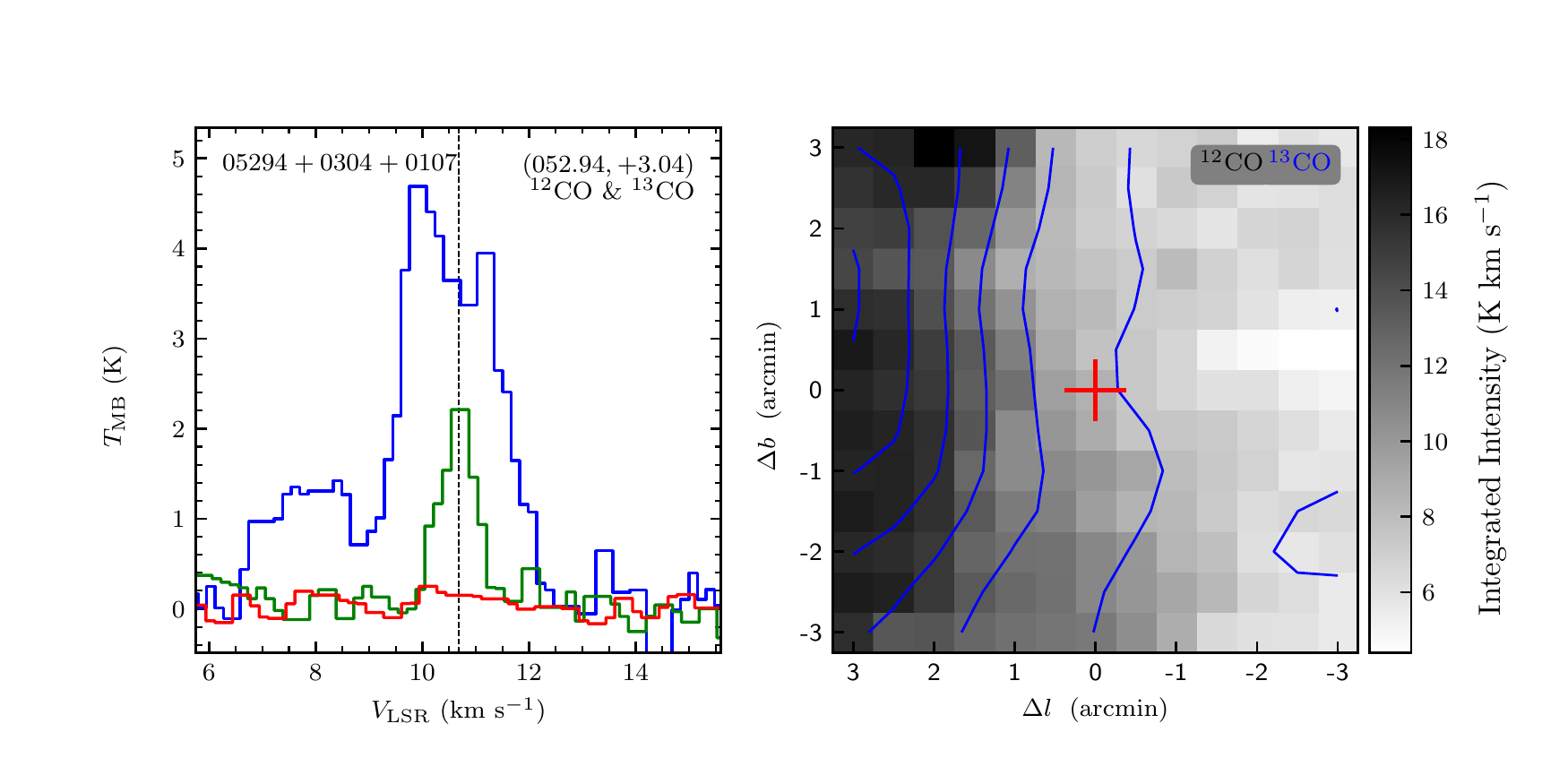}
\vspace{-0.5cm}

\includegraphics[width=9.0cm,angle=0]{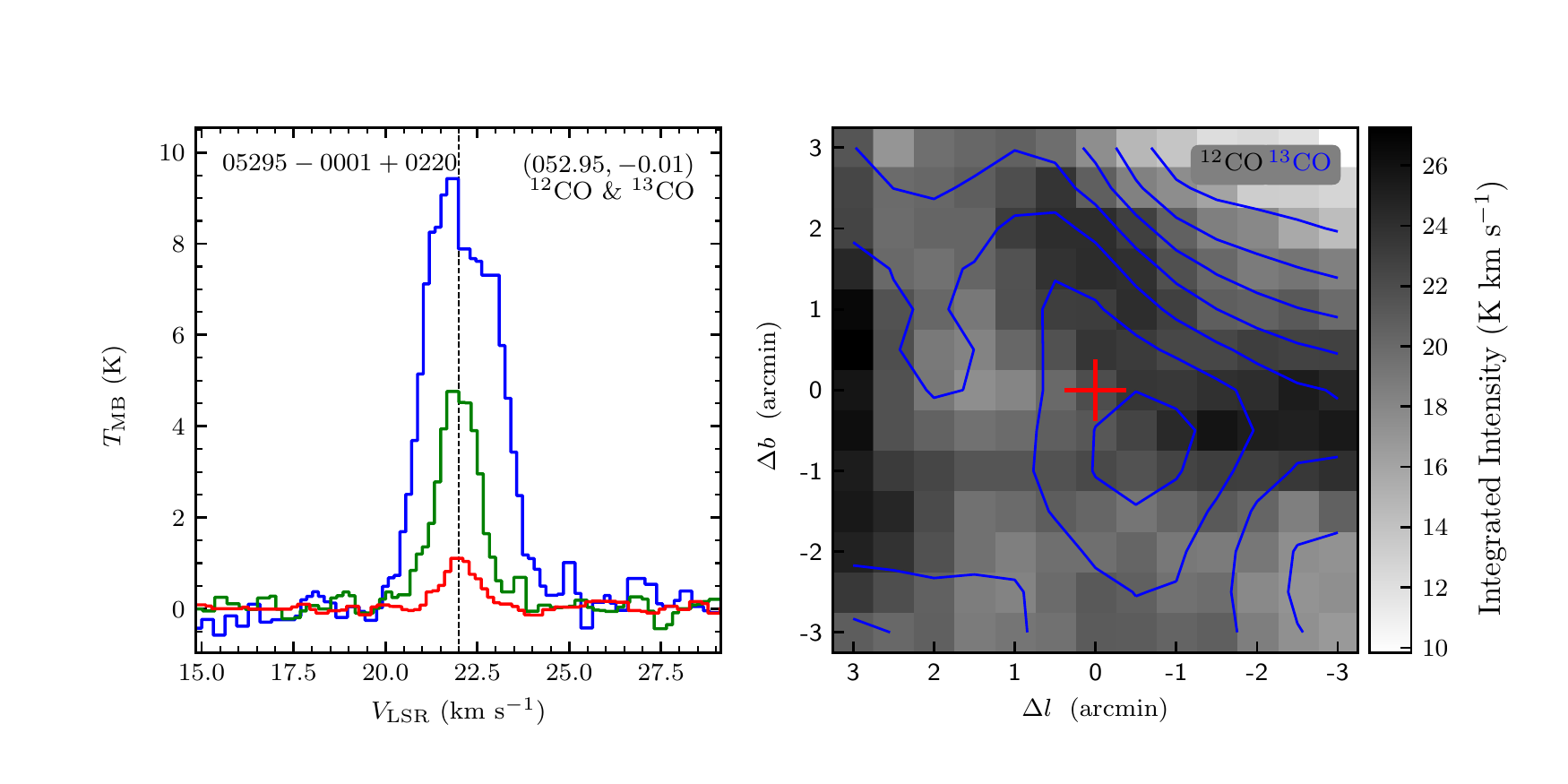}
\includegraphics[width=9.0cm,angle=0]{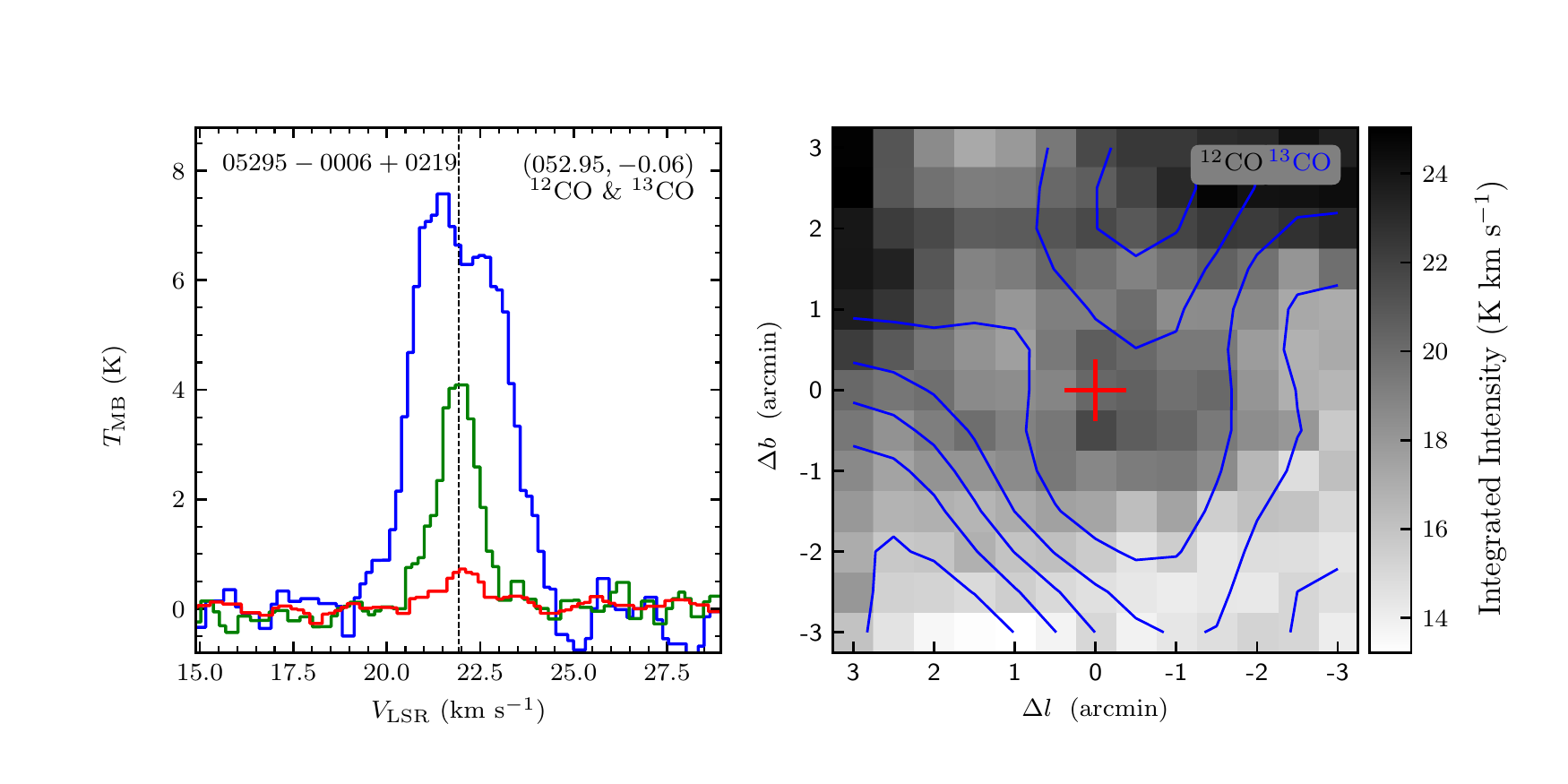}
\vspace{-0.5cm}

\includegraphics[width=9.0cm,angle=0]{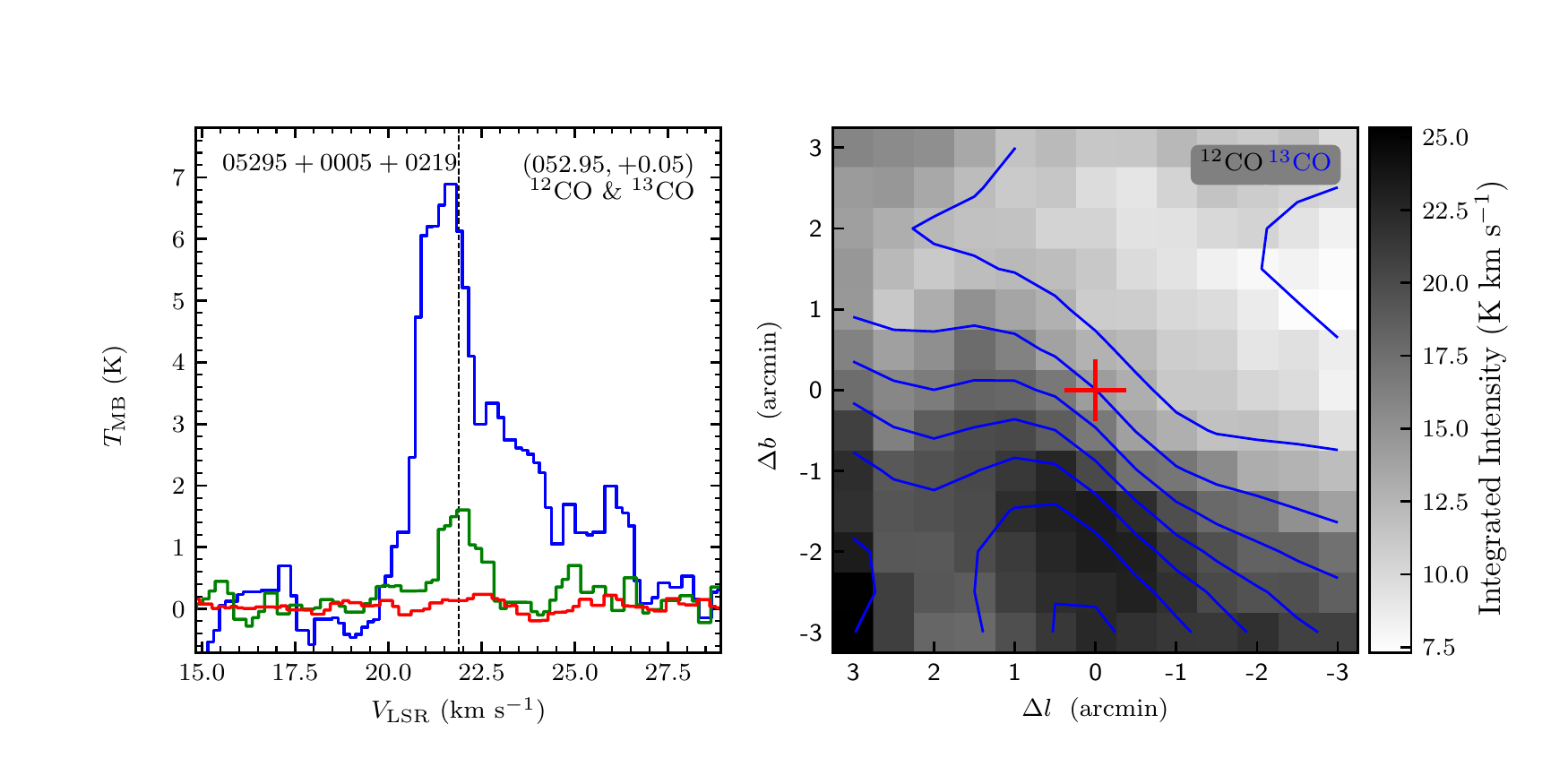}
\includegraphics[width=9.0cm,angle=0]{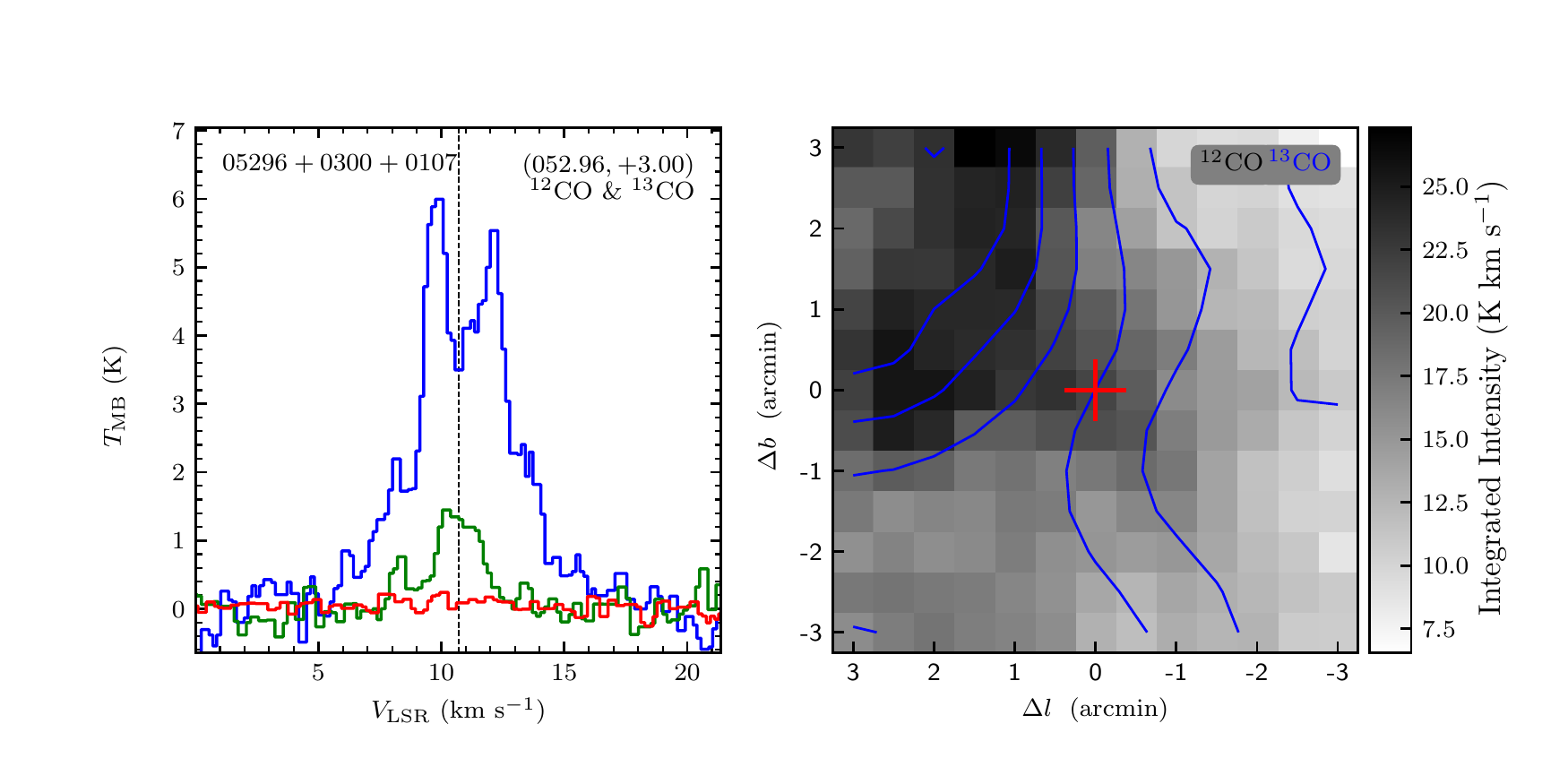}
\vspace{-0.5cm}

\includegraphics[width=9.0cm,angle=0]{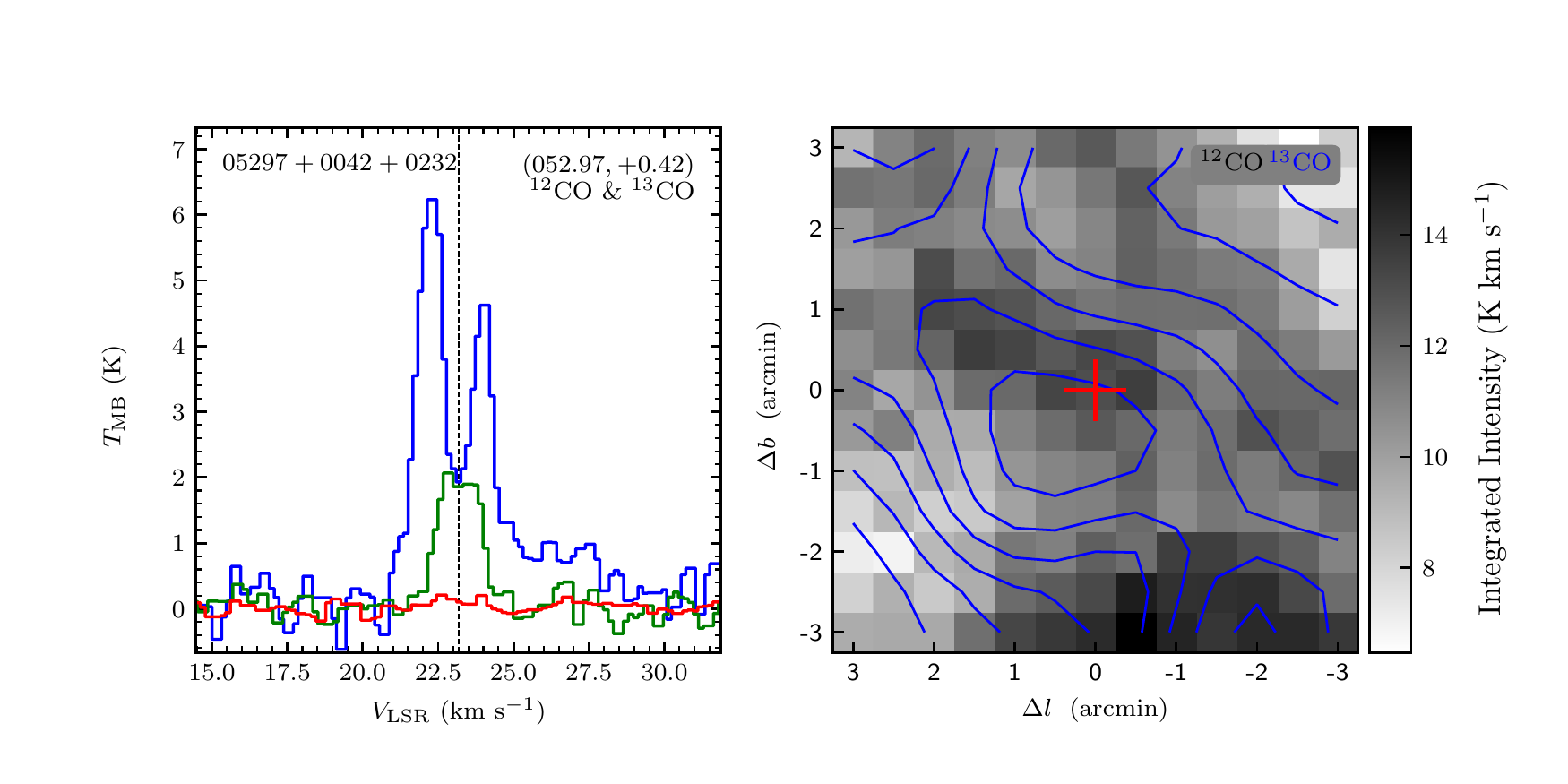}
\includegraphics[width=9.0cm,angle=0]{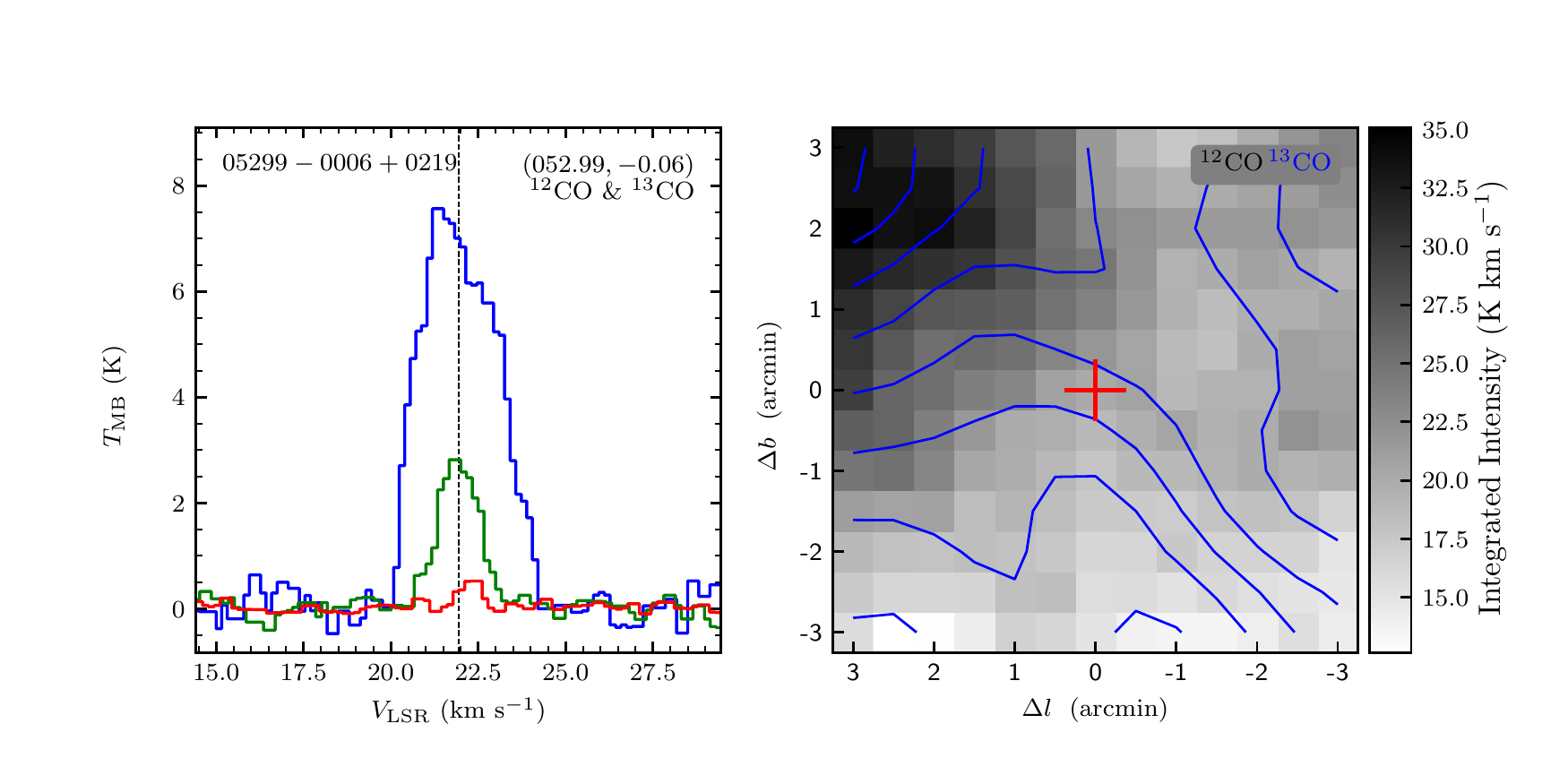}
\end{figure}
\clearpage

\begin{figure}
\includegraphics[width=9.0cm,angle=0]{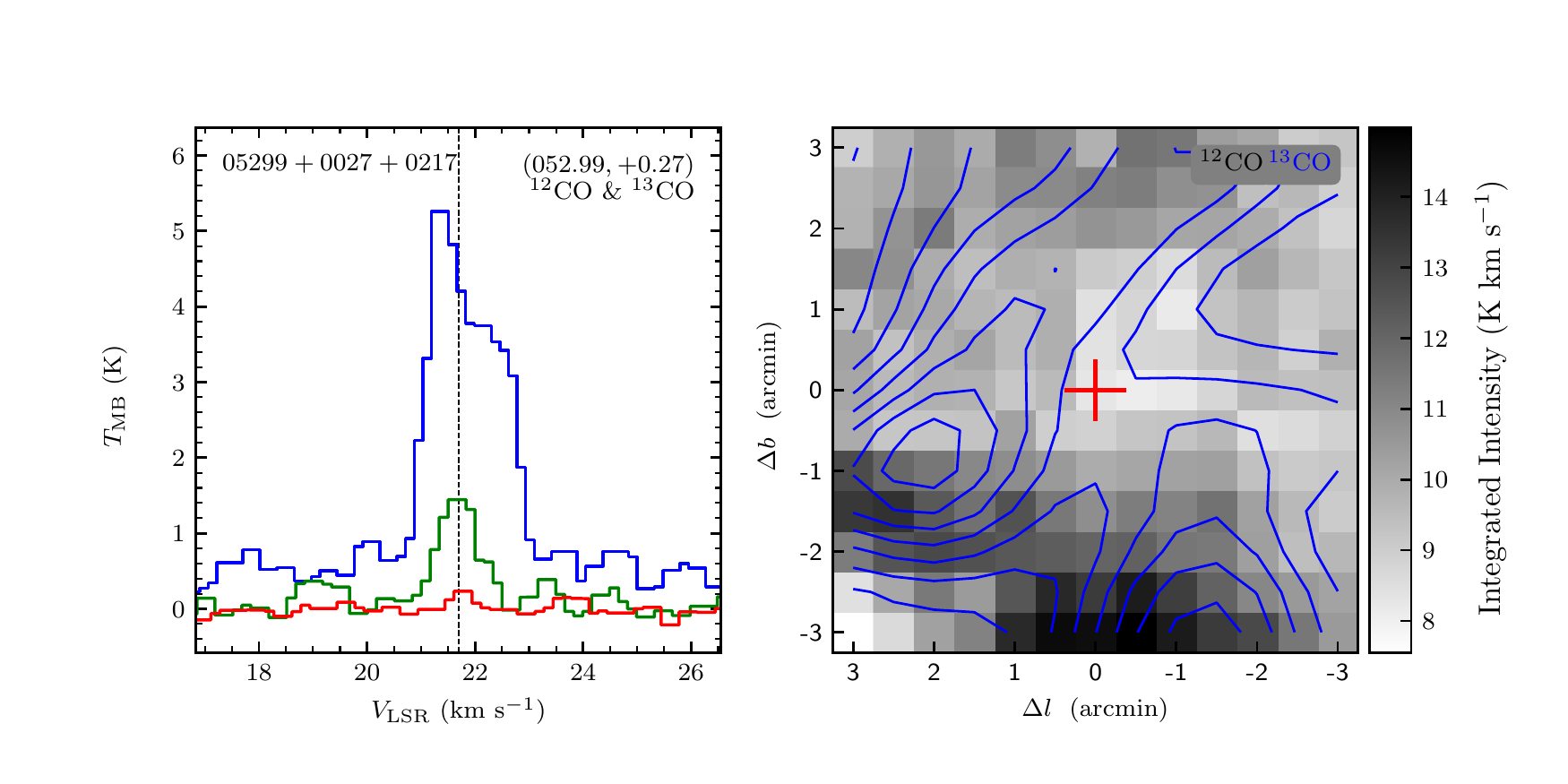}
\includegraphics[width=9.0cm,angle=0]{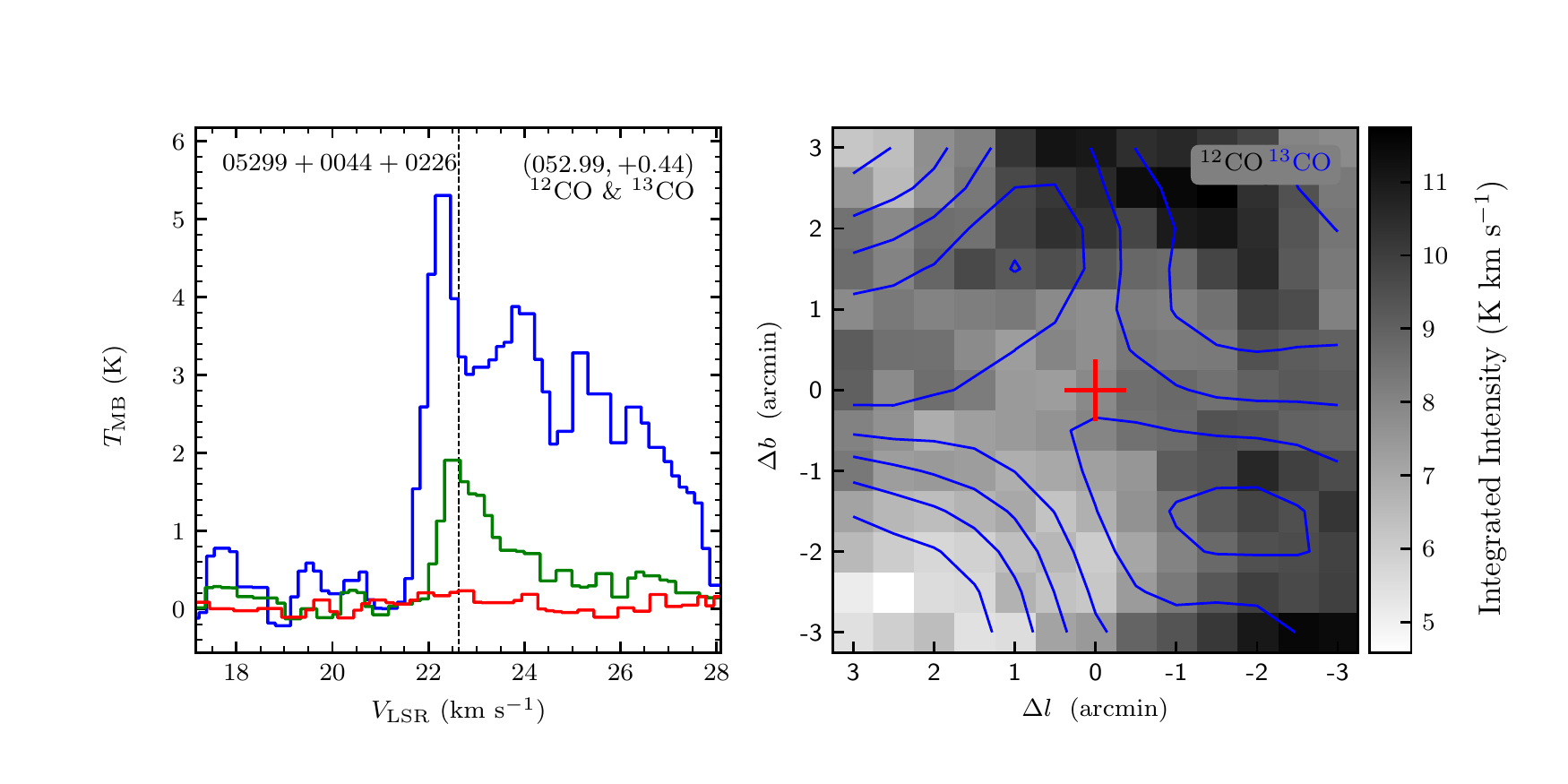}
\vspace{-0.5cm}

\includegraphics[width=9.0cm,angle=0]{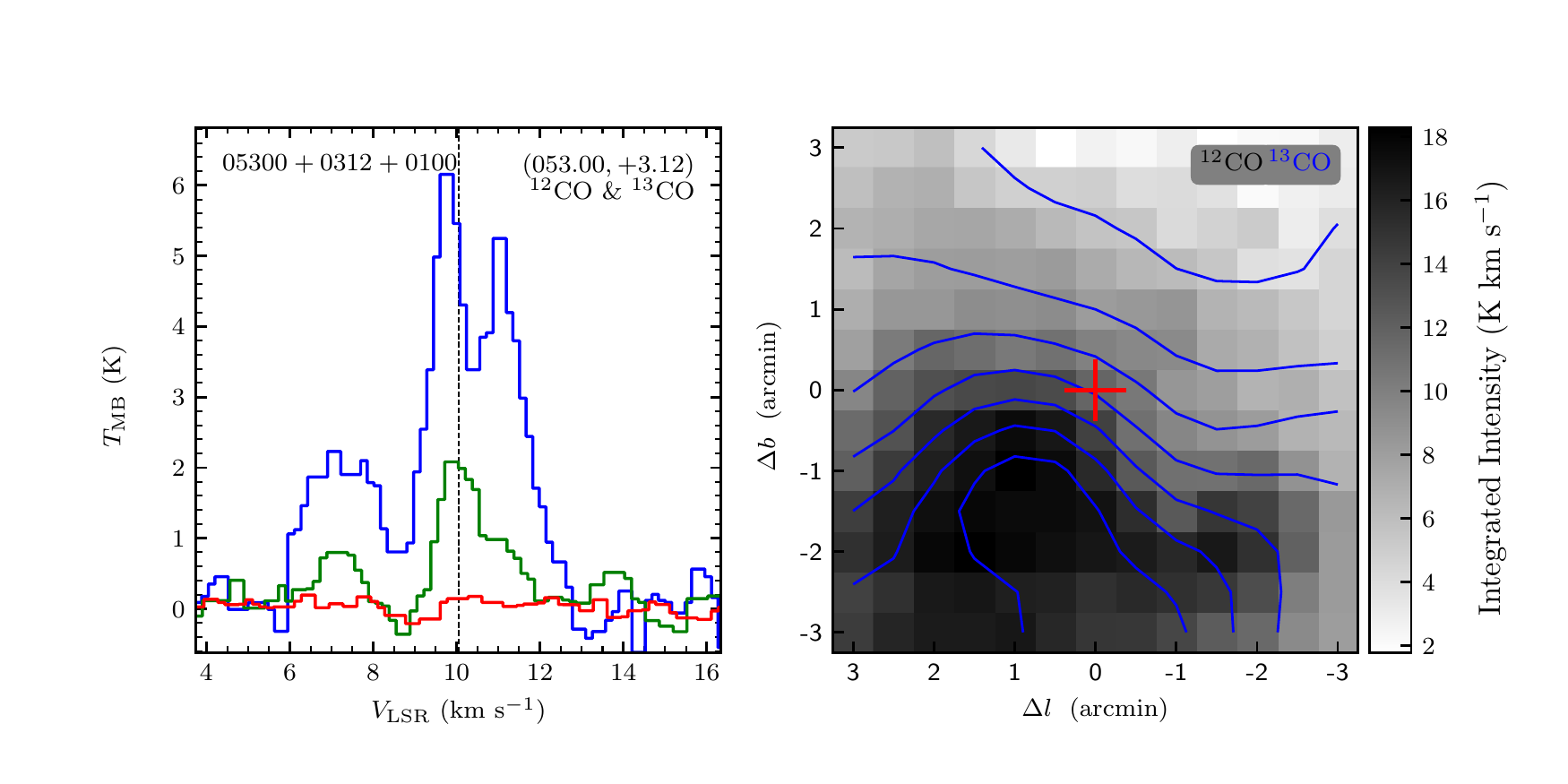}
\includegraphics[width=9.0cm,angle=0]{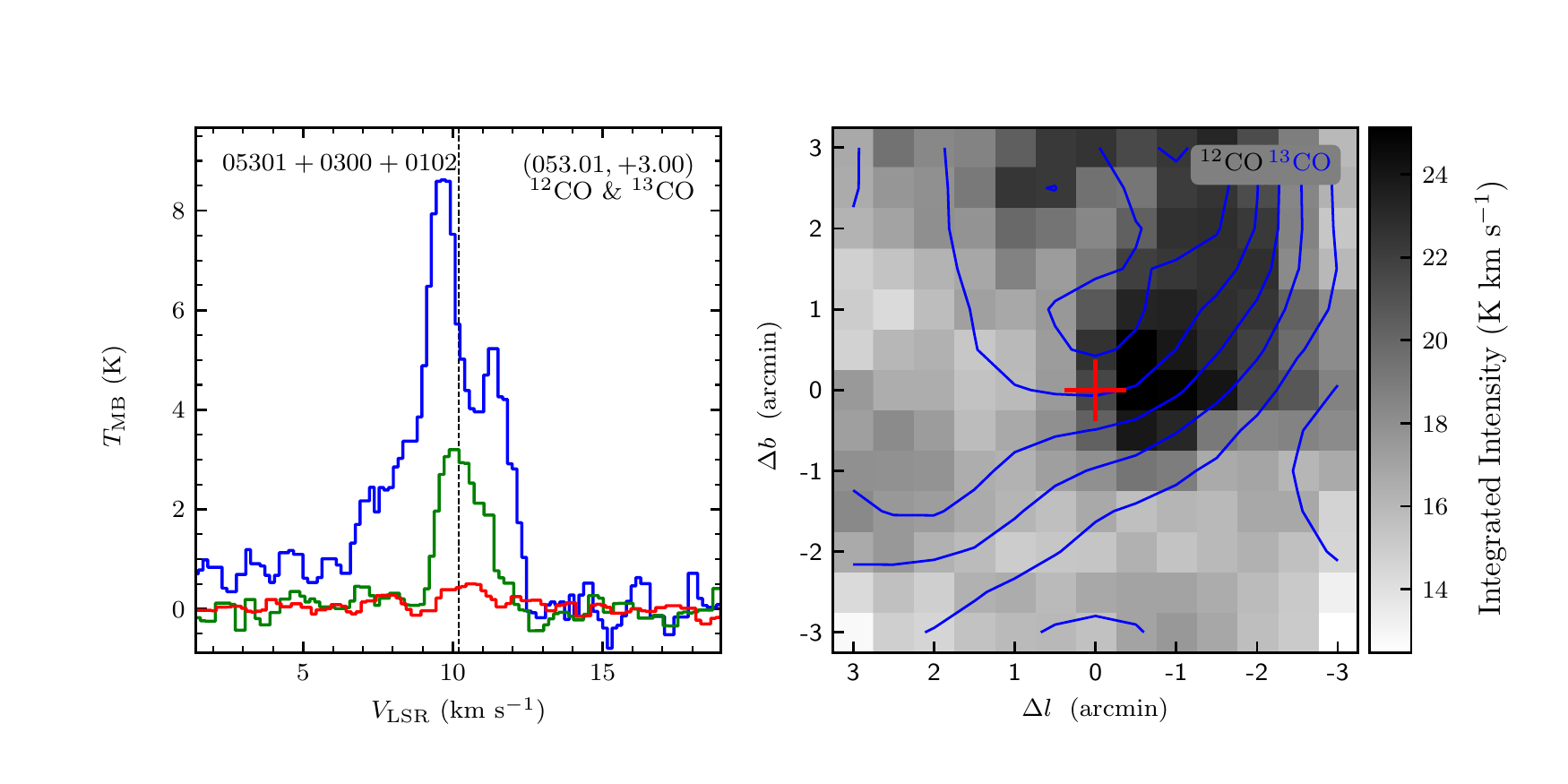}
\vspace{-0.5cm}

\includegraphics[width=9.0cm,angle=0]{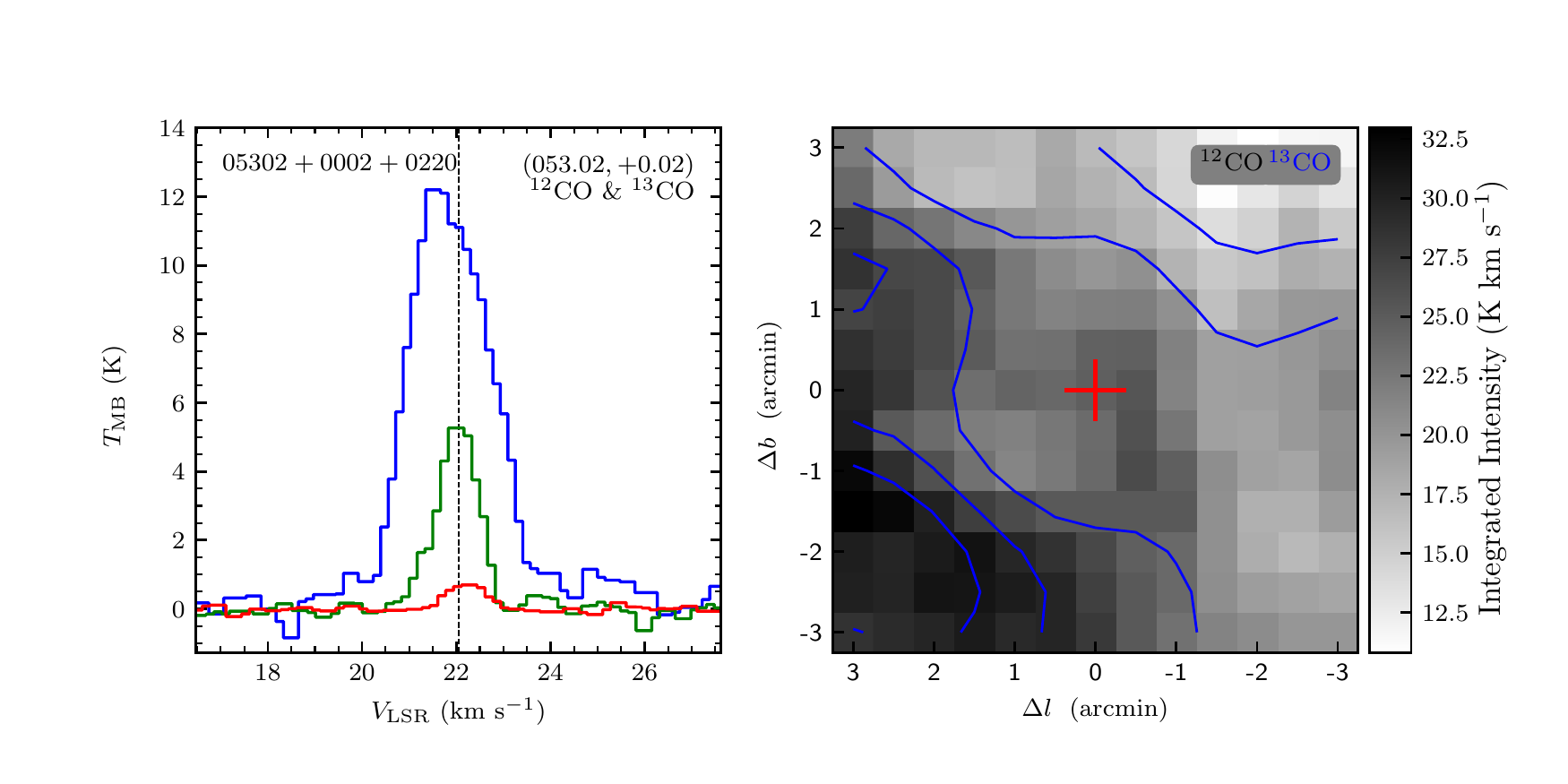}
\includegraphics[width=9.0cm,angle=0]{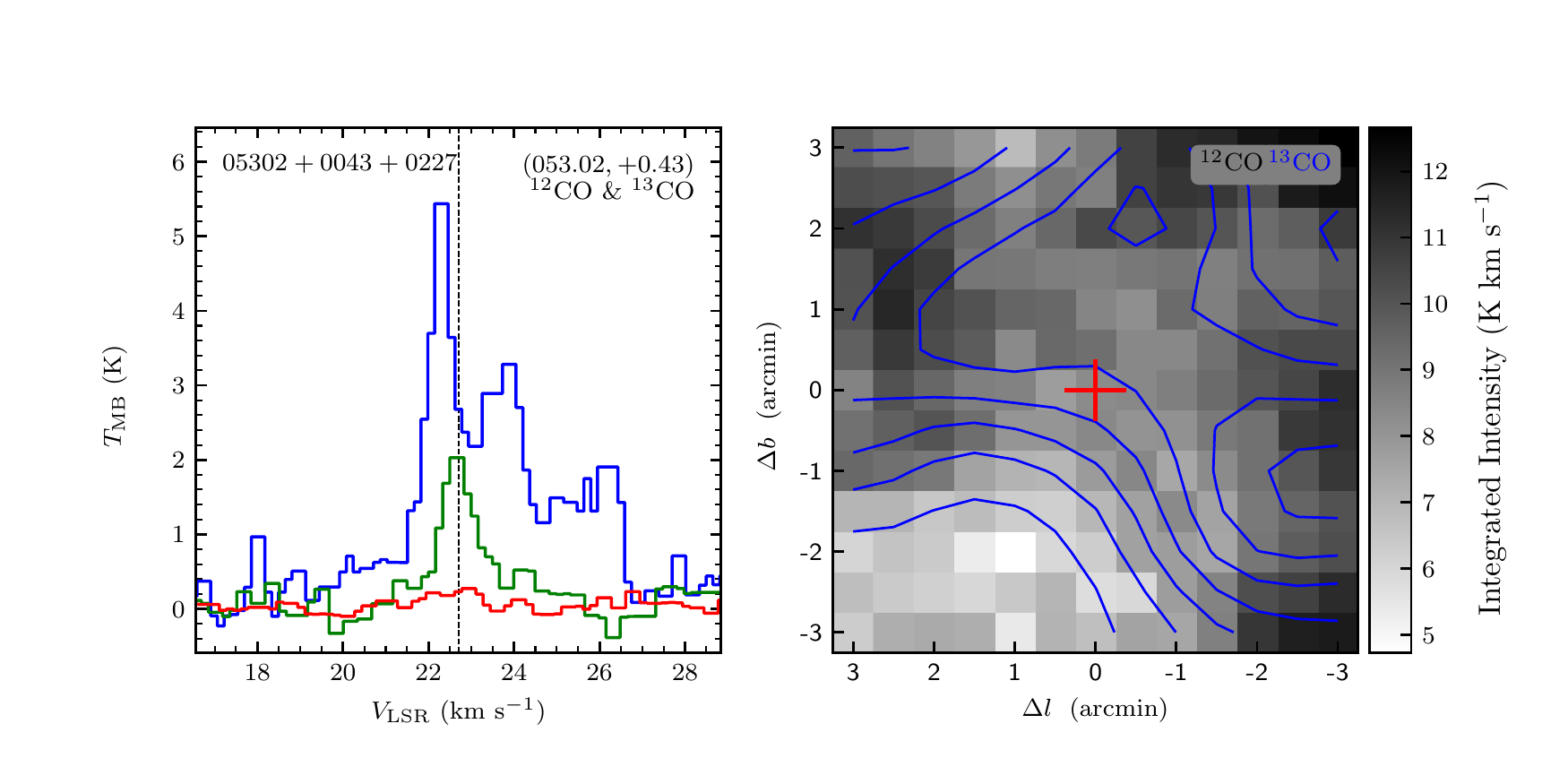}
\vspace{-0.5cm}

\includegraphics[width=9.0cm,angle=0]{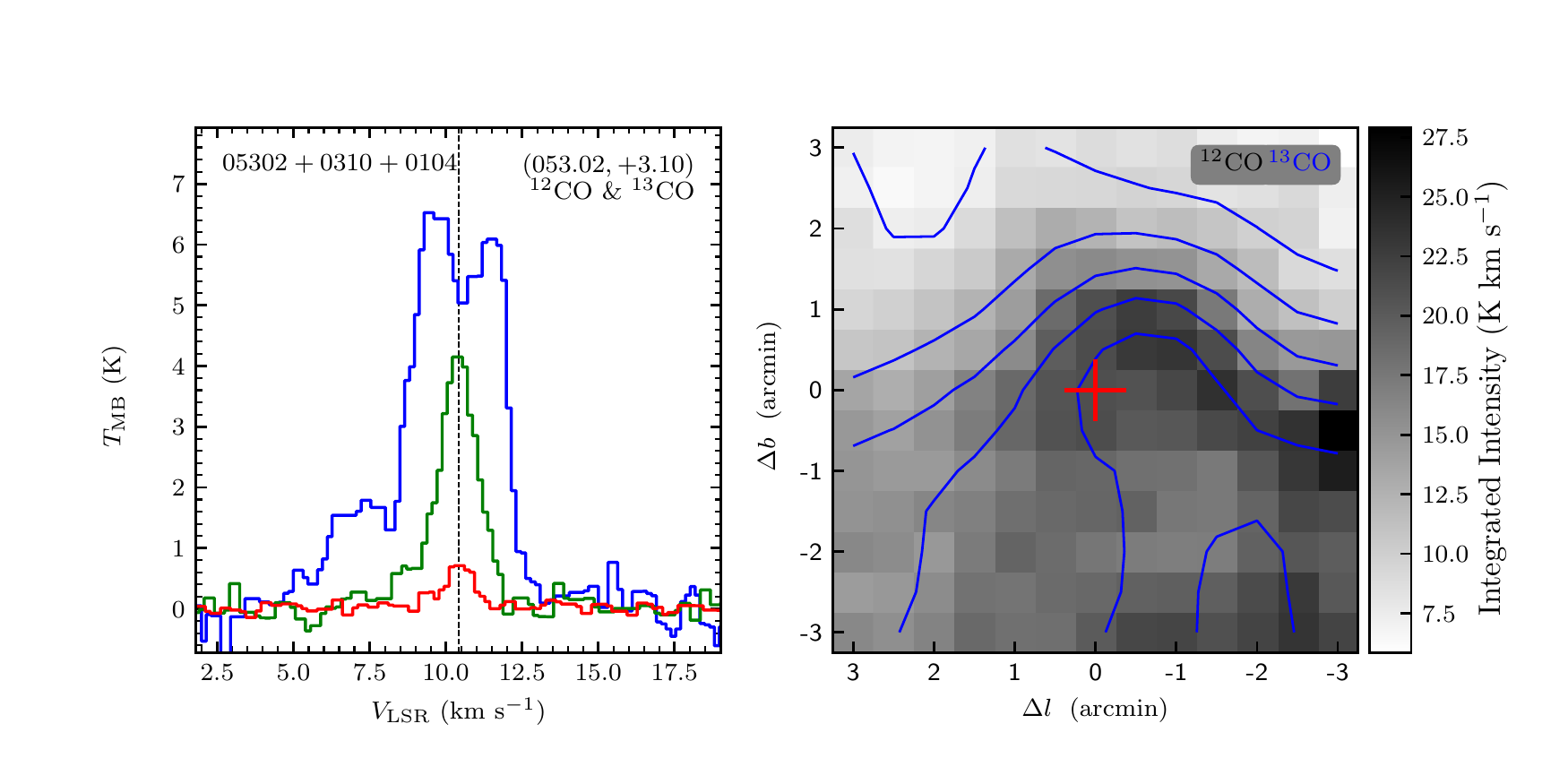}
\includegraphics[width=9.0cm,angle=0]{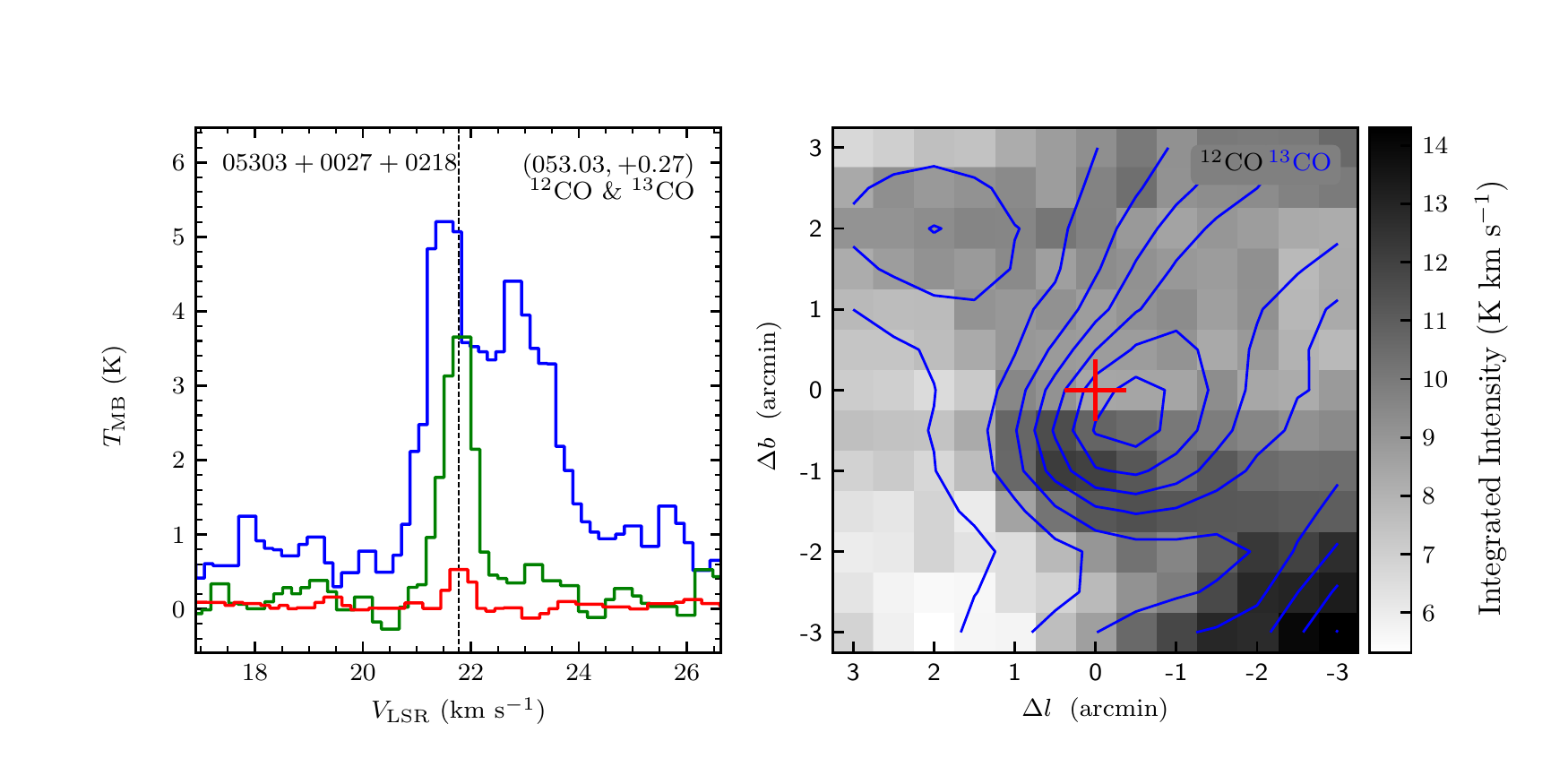}
\vspace{-0.5cm}

\includegraphics[width=9.0cm,angle=0]{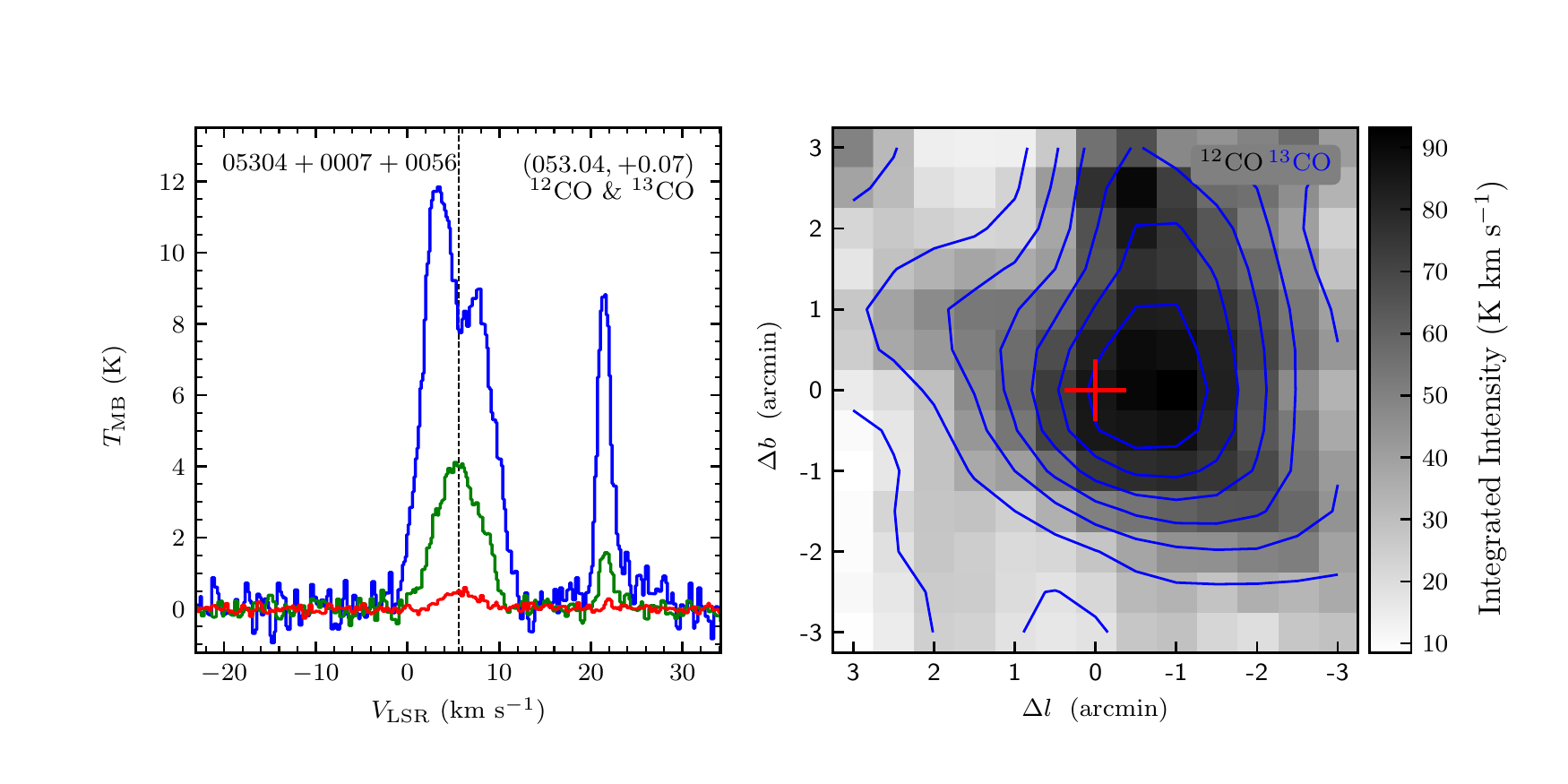}
\includegraphics[width=9.0cm,angle=0]{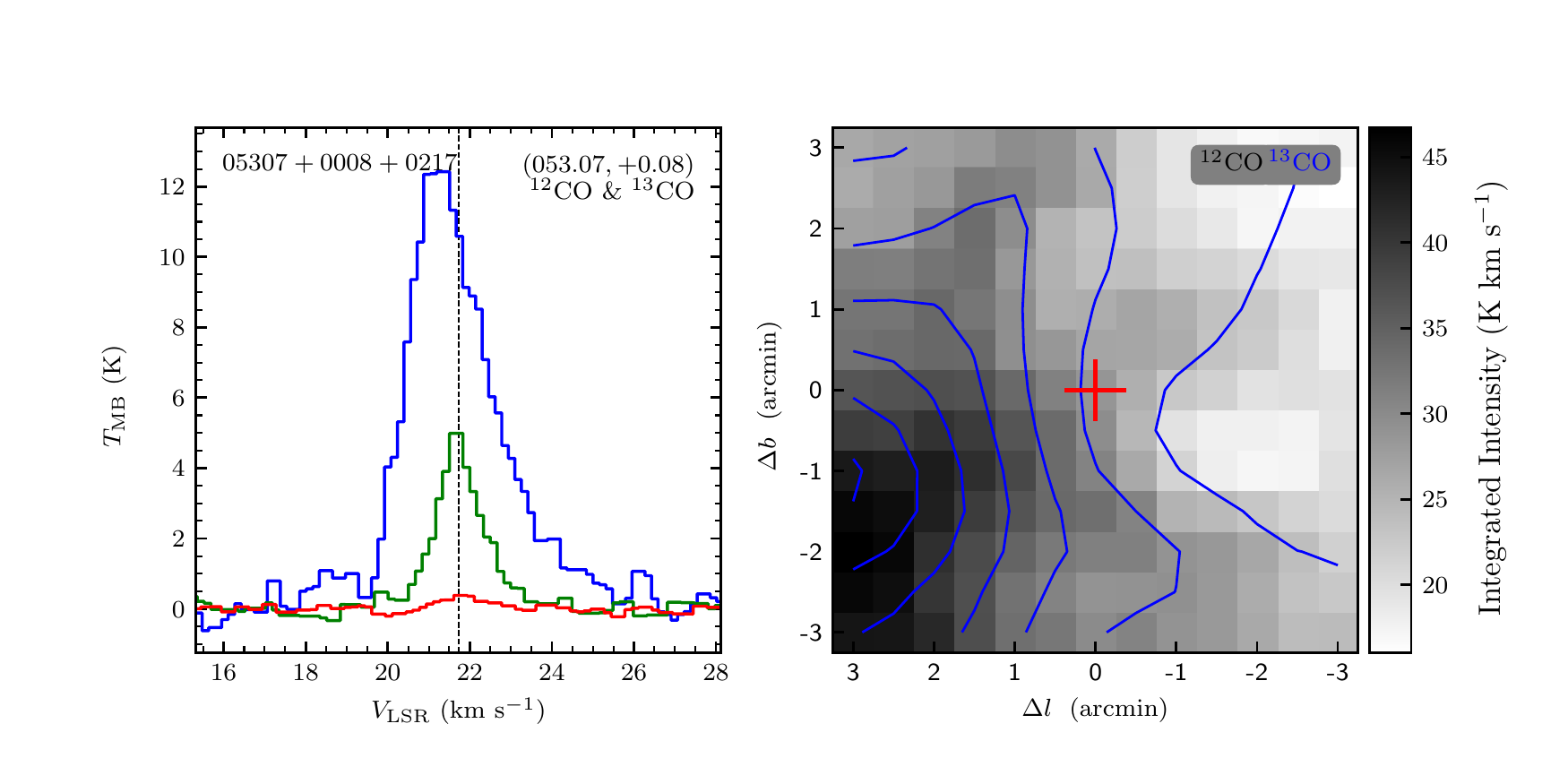}
\end{figure}
\clearpage

\begin{figure}
\includegraphics[width=9.0cm,angle=0]{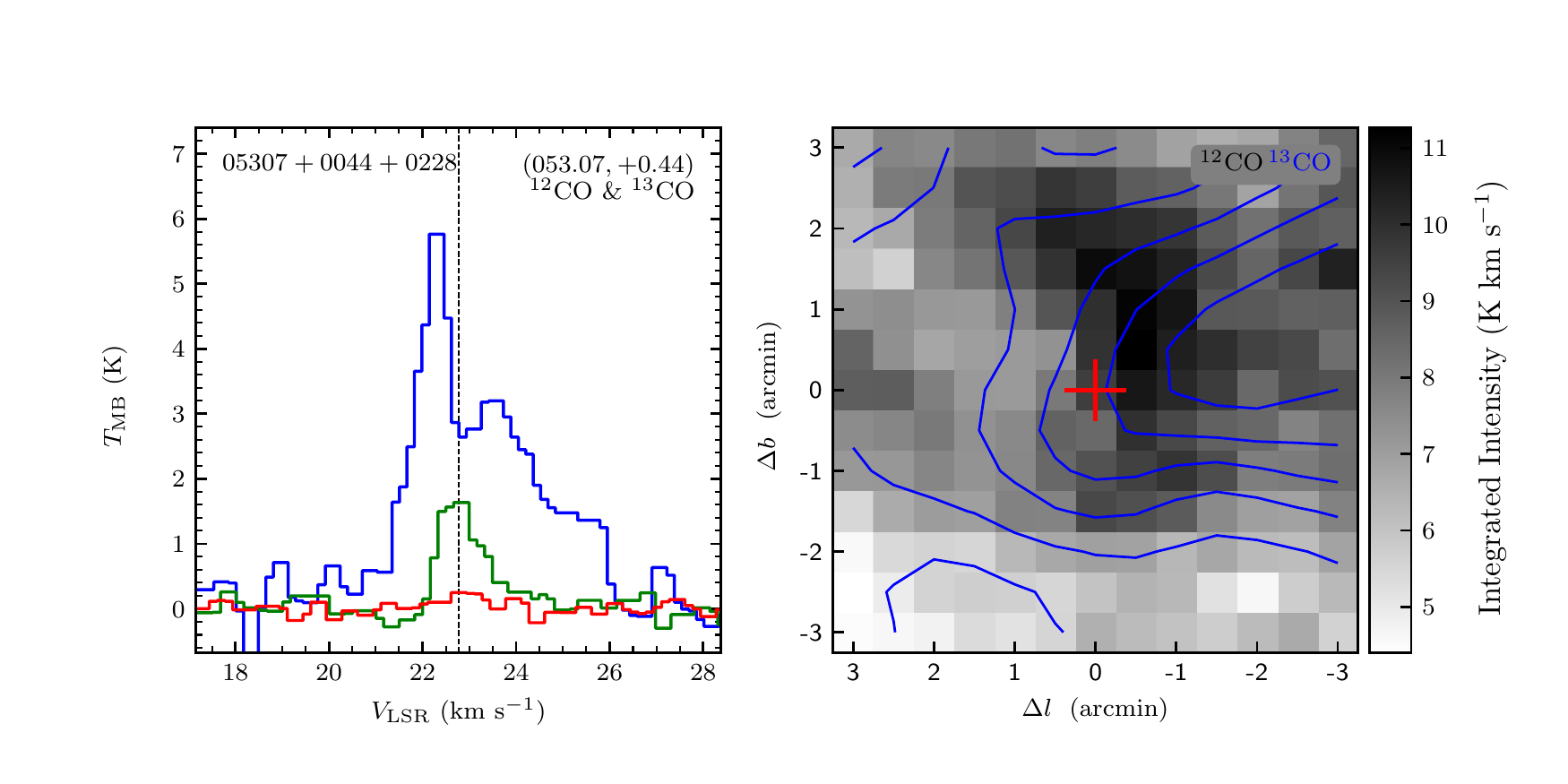}
\includegraphics[width=9.0cm,angle=0]{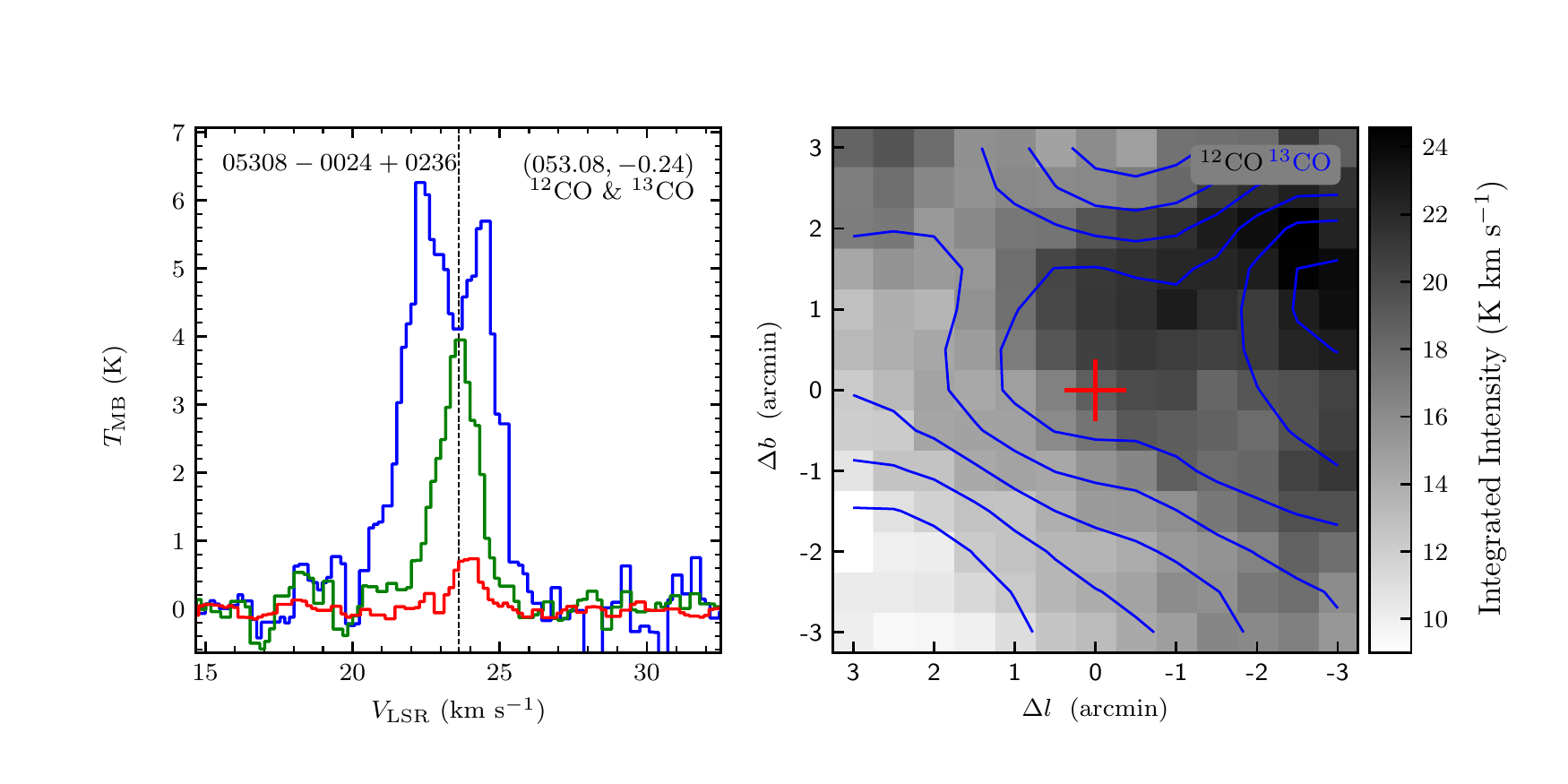}
\vspace{-0.5cm}

\includegraphics[width=9.0cm,angle=0]{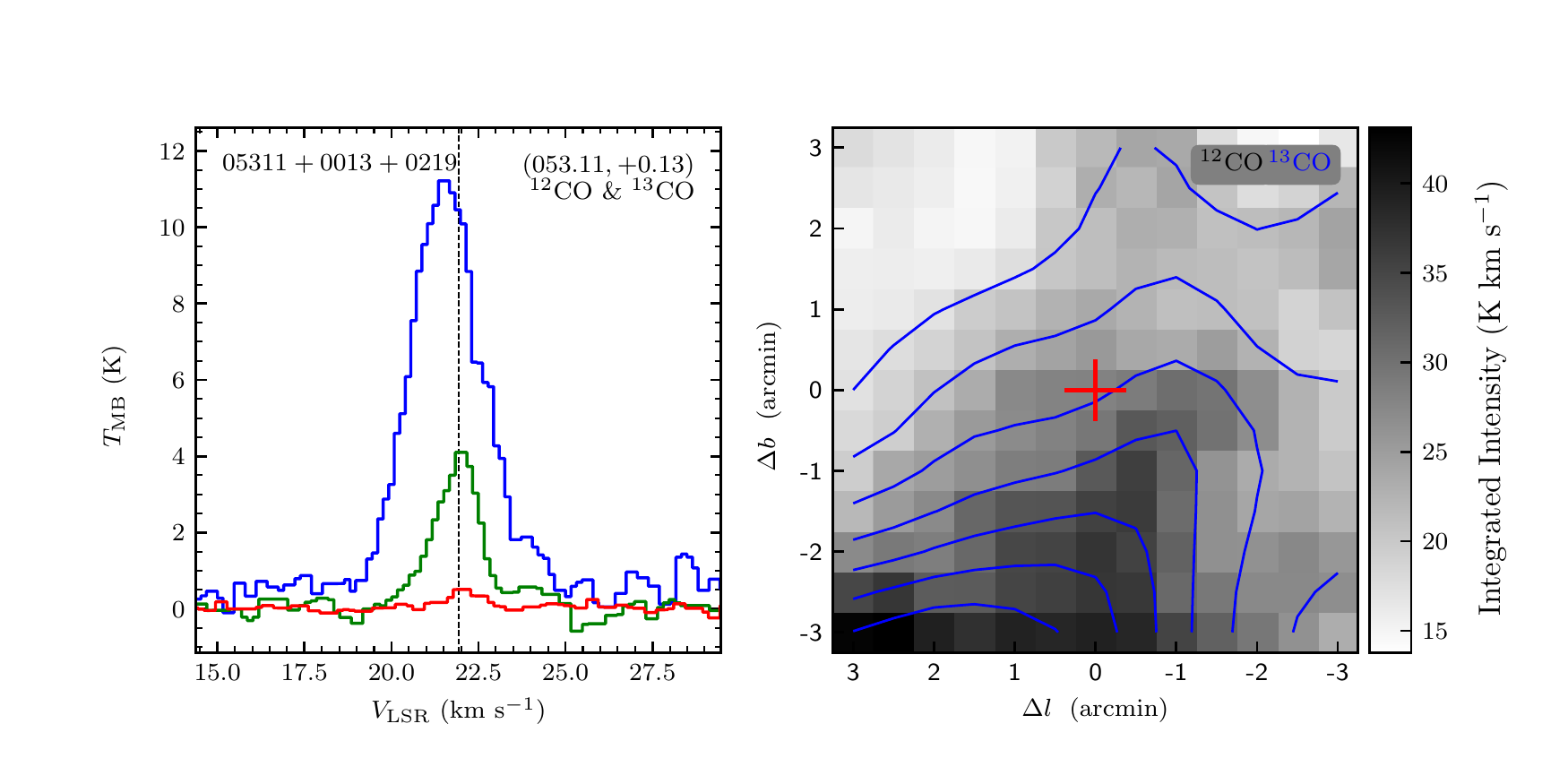}
\includegraphics[width=9.0cm,angle=0]{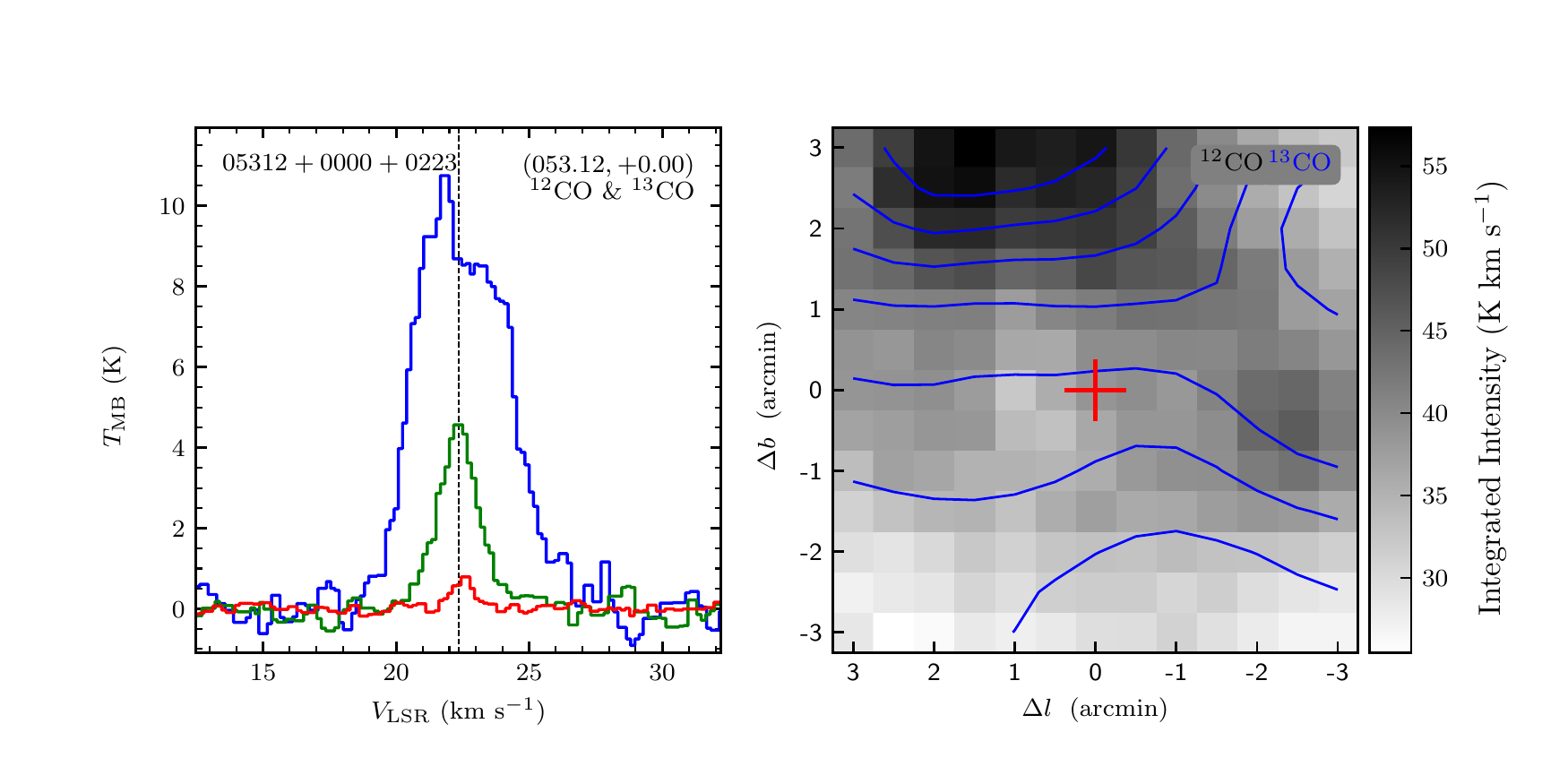}
\vspace{-0.5cm}

\includegraphics[width=9.0cm,angle=0]{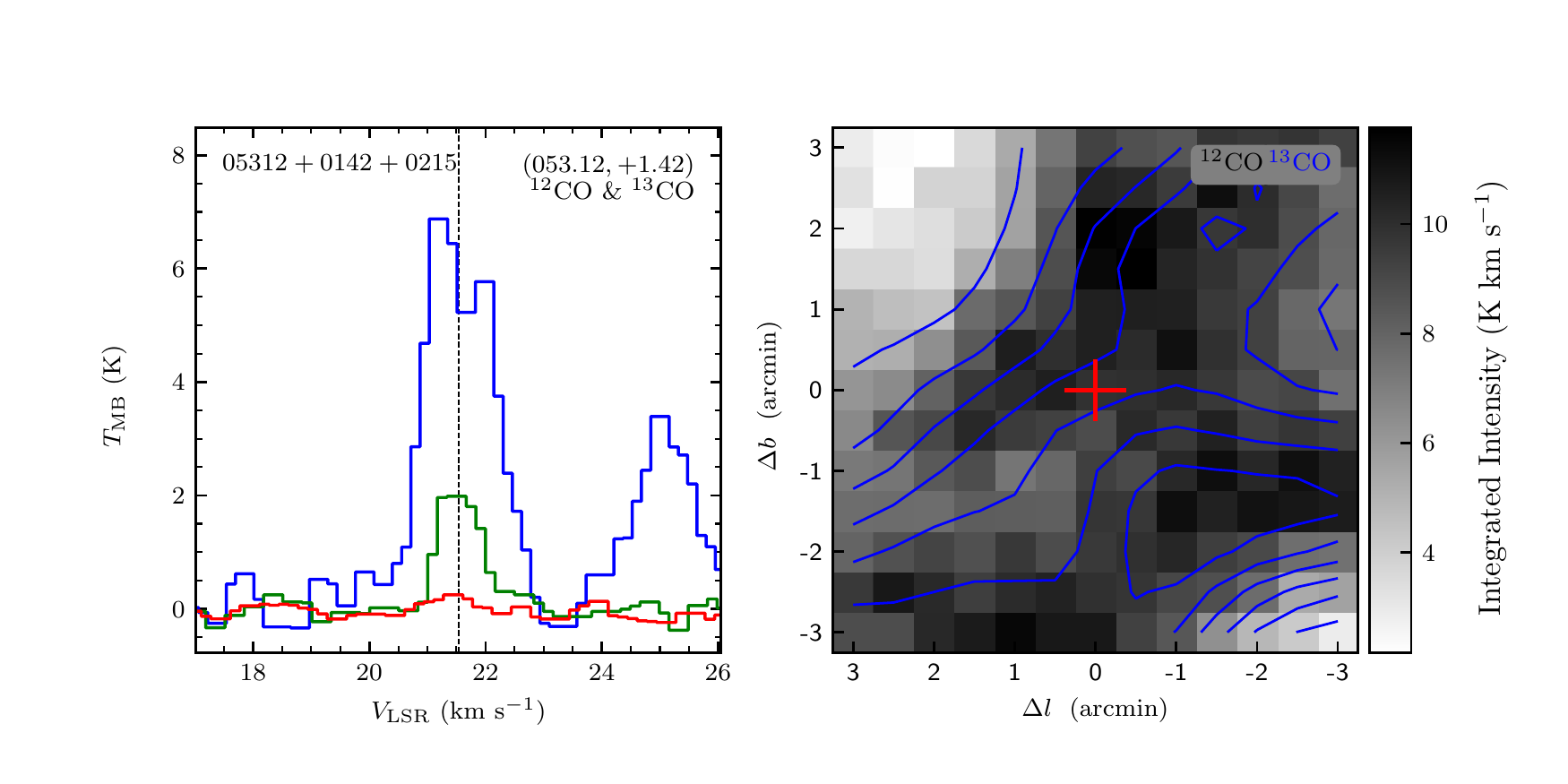}
\includegraphics[width=9.0cm,angle=0]{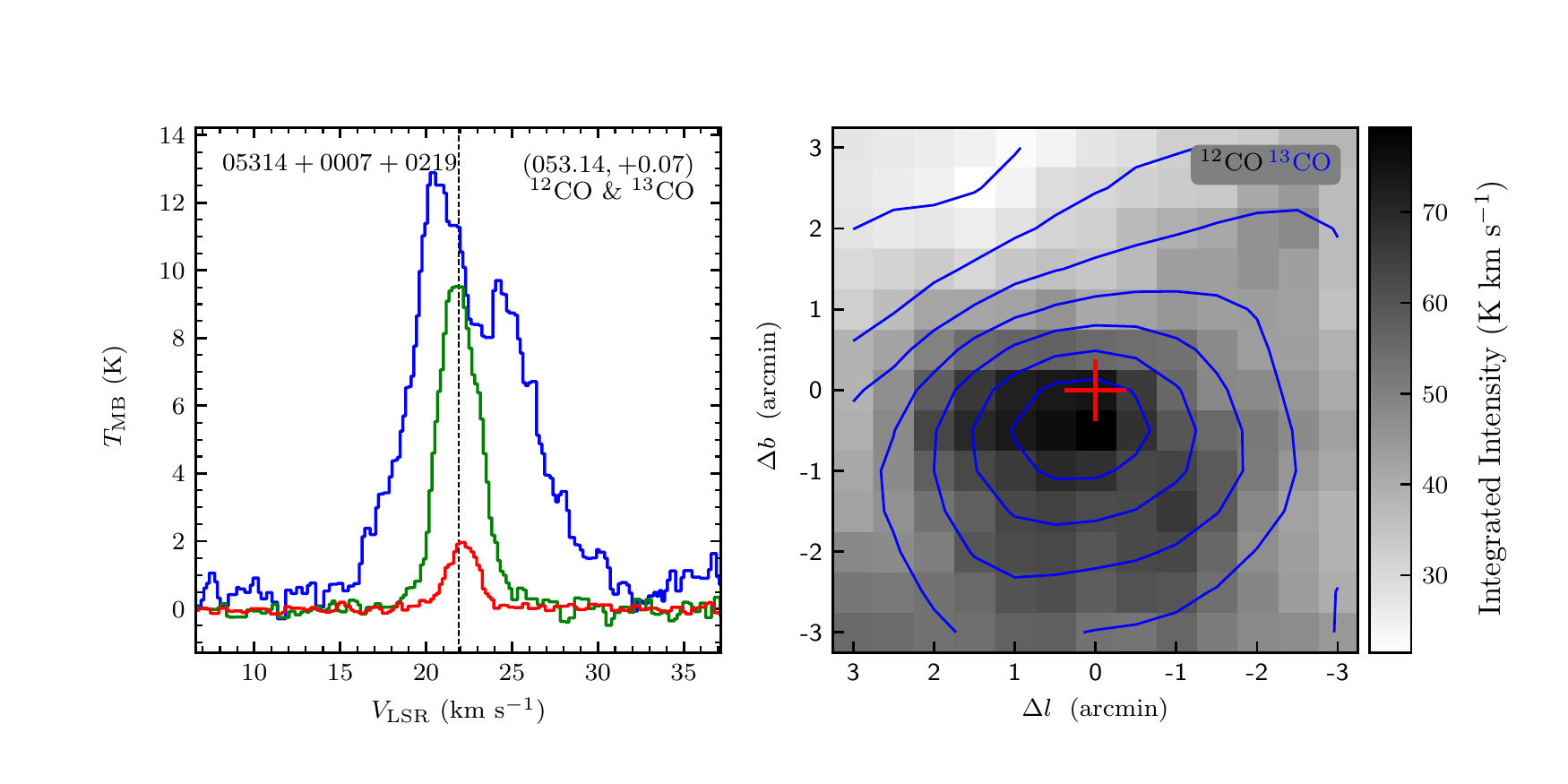}
\vspace{-0.5cm}

\includegraphics[width=9.0cm,angle=0]{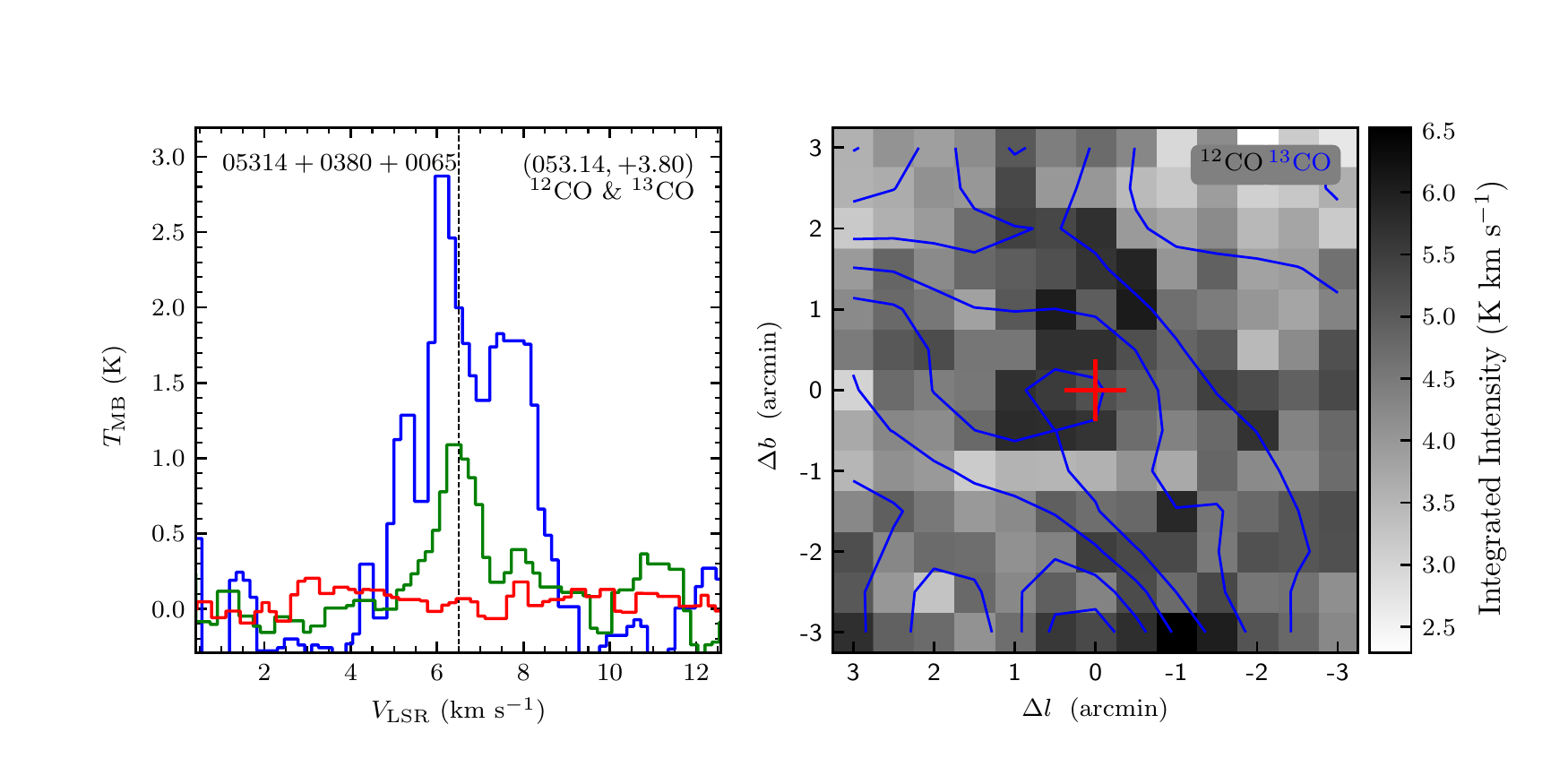}
\includegraphics[width=9.0cm,angle=0]{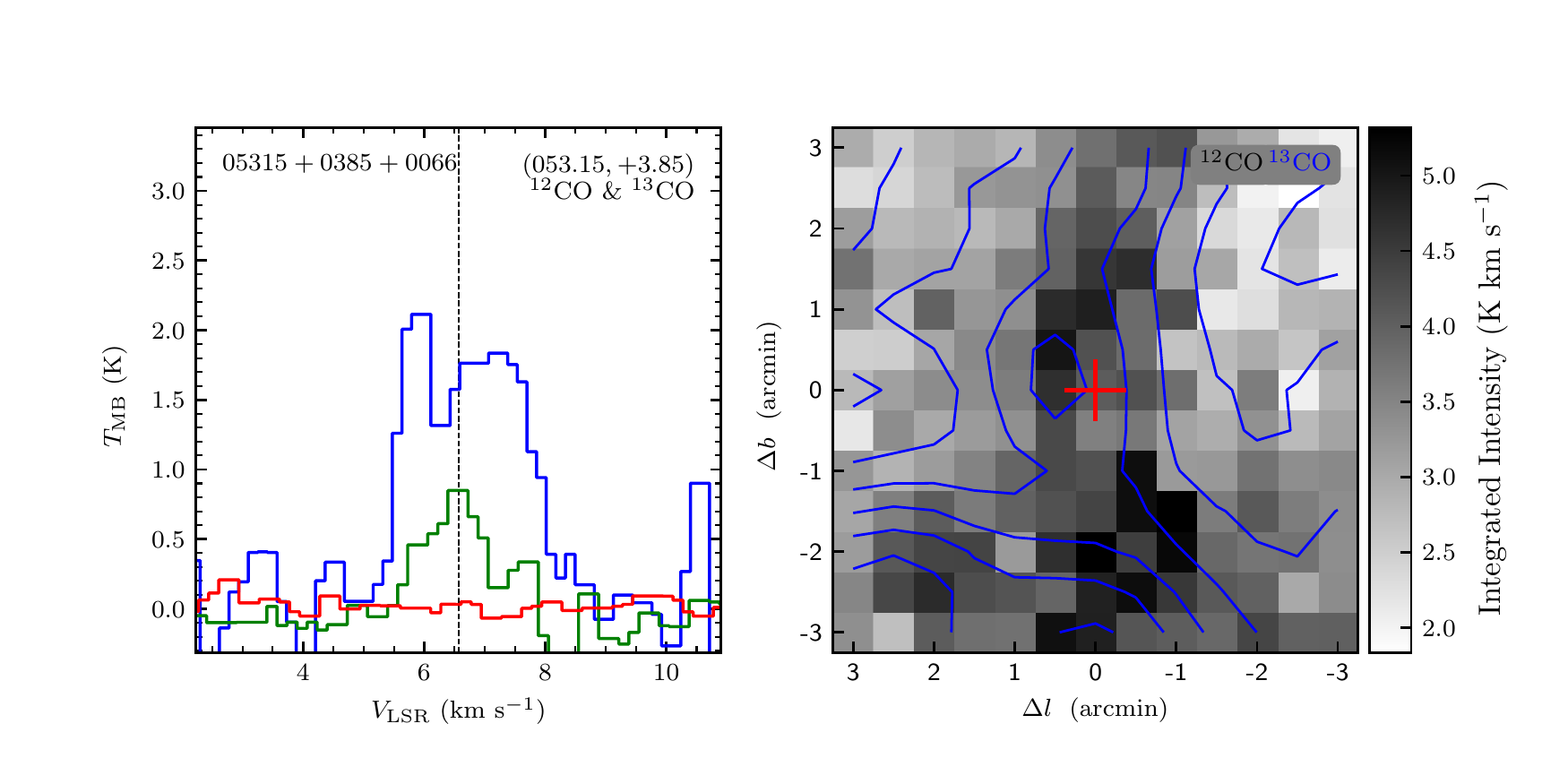}
\vspace{-0.5cm}

\includegraphics[width=9.0cm,angle=0]{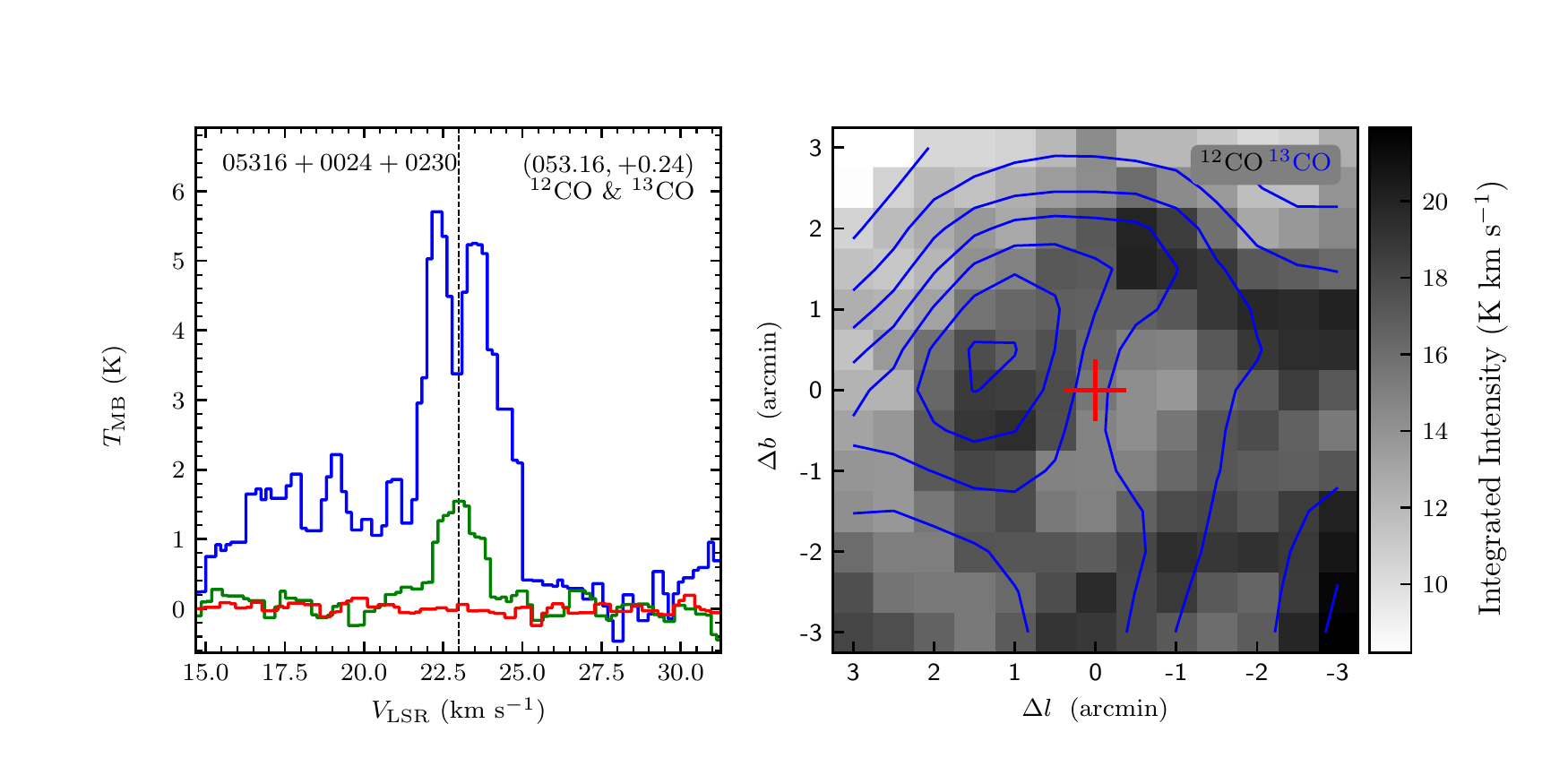}
\includegraphics[width=9.0cm,angle=0]{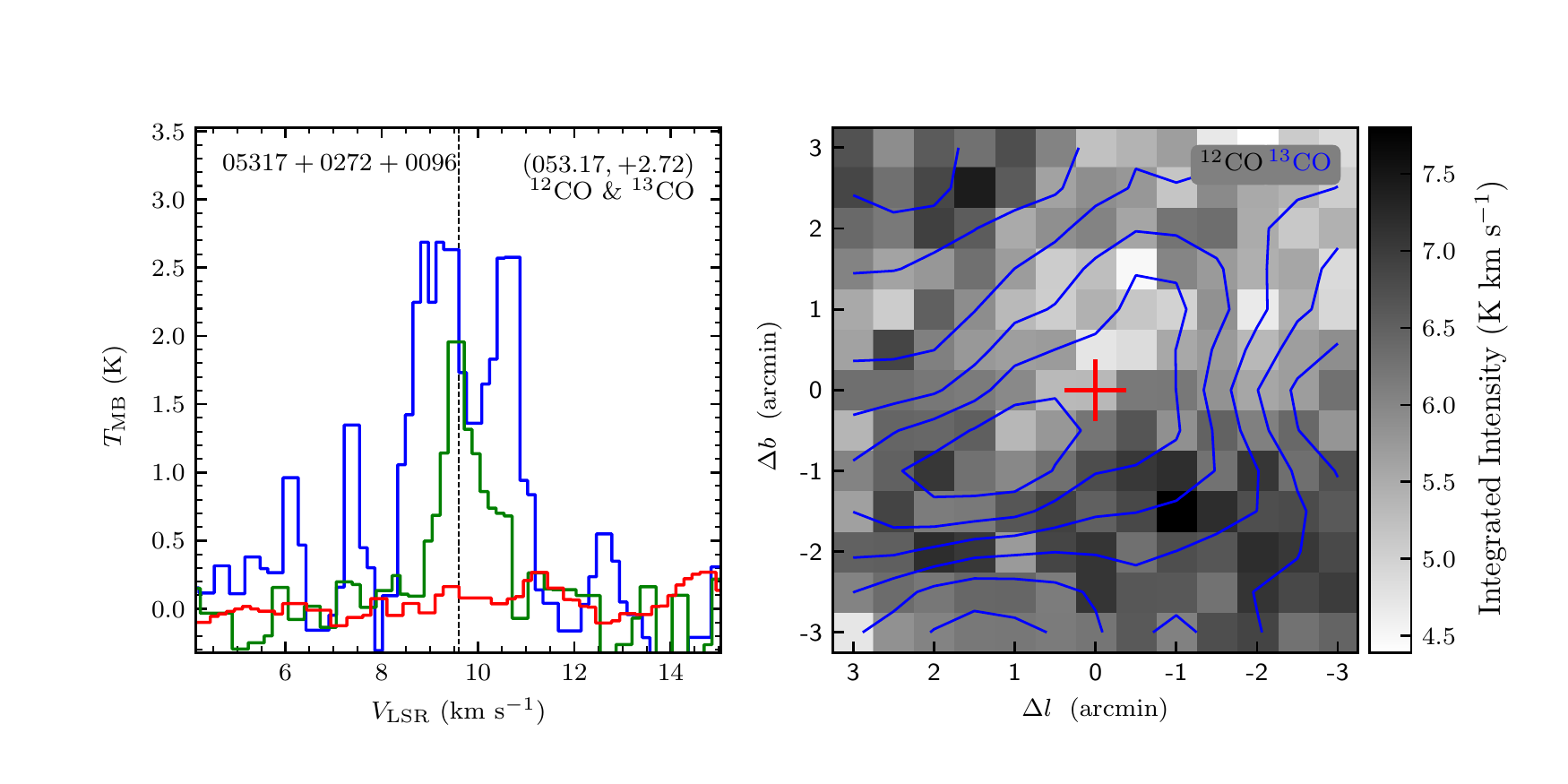}
\end{figure}
\clearpage

\begin{figure}
\includegraphics[width=9.0cm,angle=0]{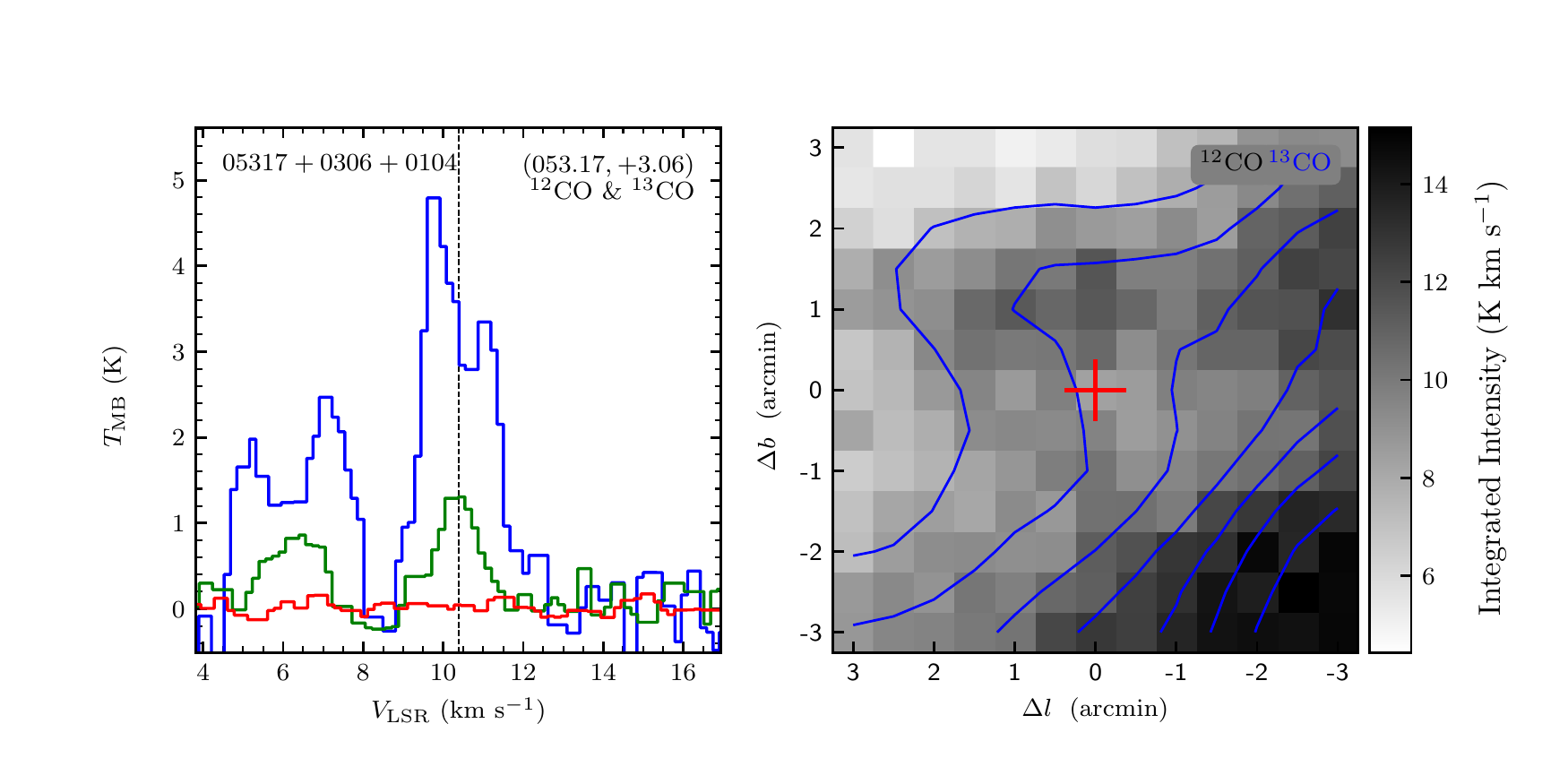}
\includegraphics[width=9.0cm,angle=0]{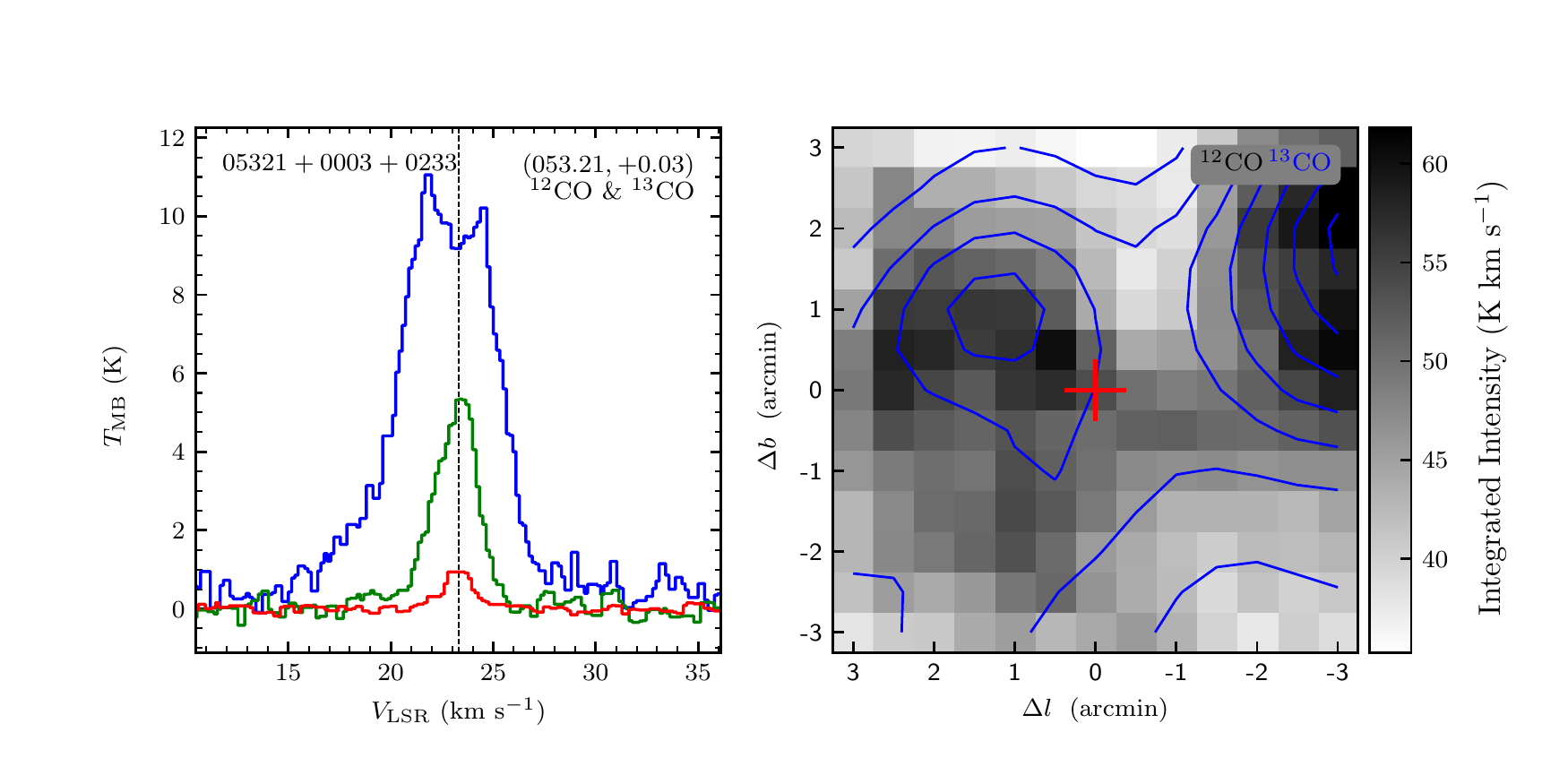}
\vspace{-0.5cm}

\includegraphics[width=9.0cm,angle=0]{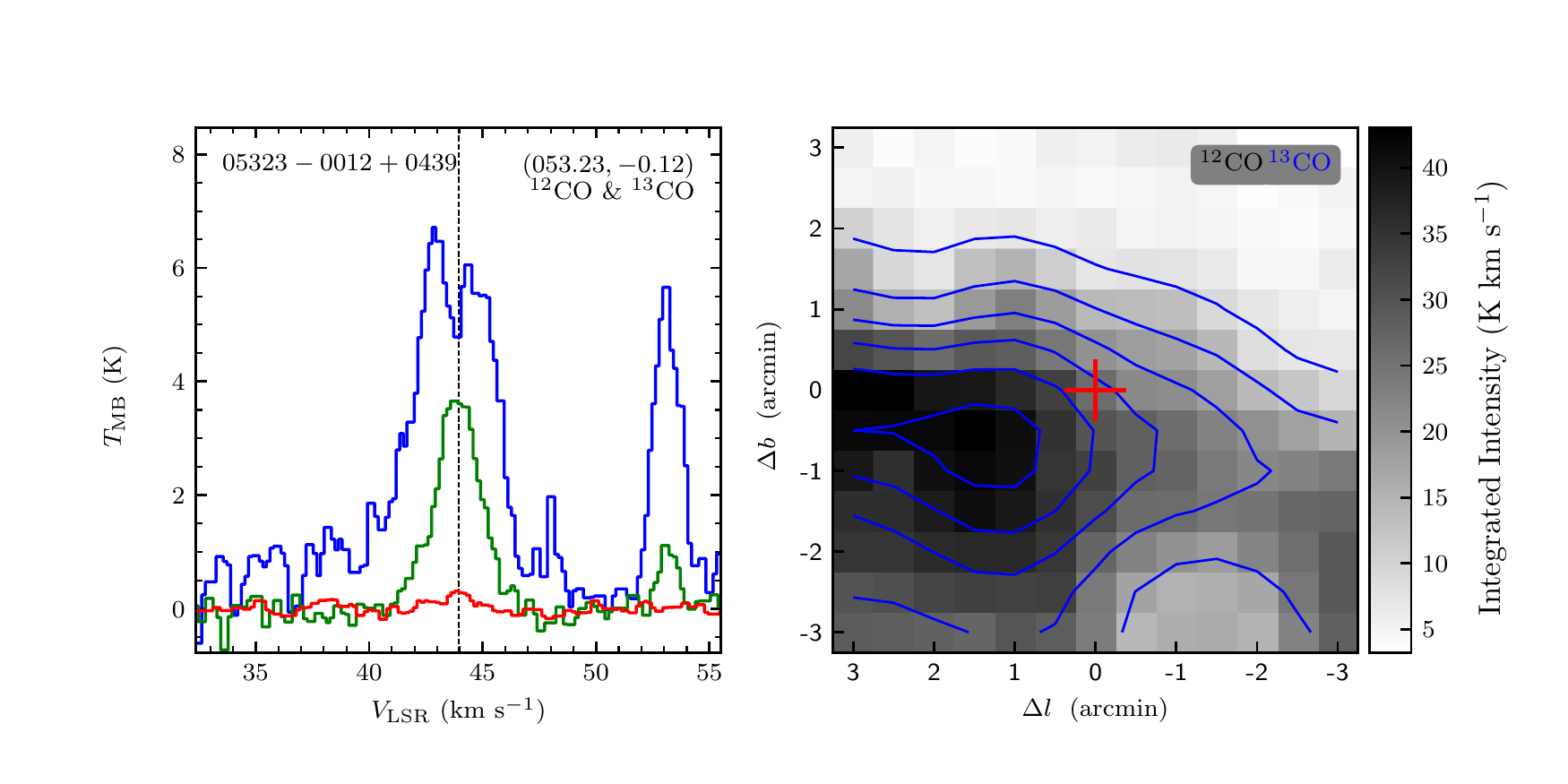}
\includegraphics[width=9.0cm,angle=0]{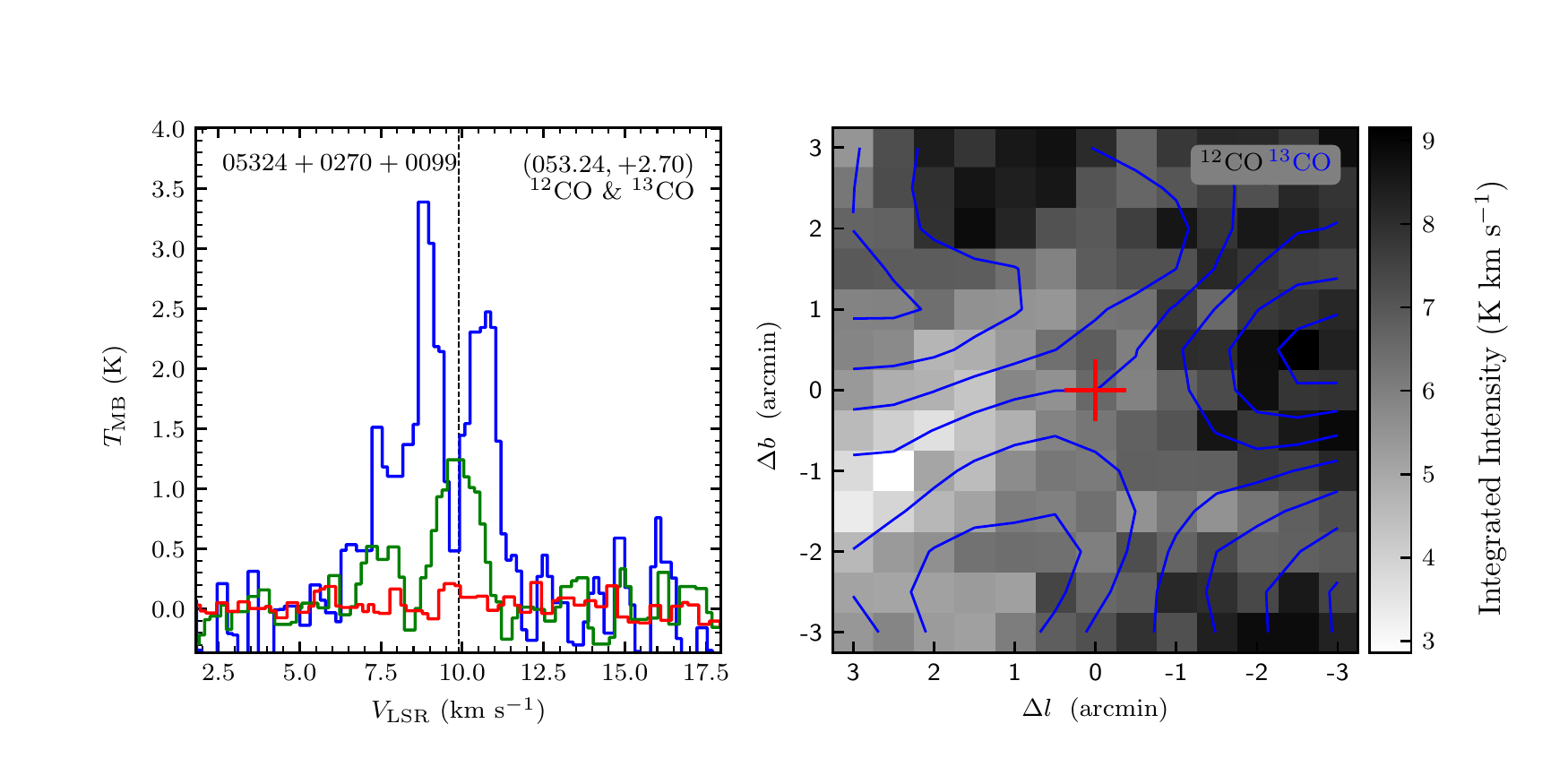}
\vspace{-0.5cm}

\includegraphics[width=9.0cm,angle=0]{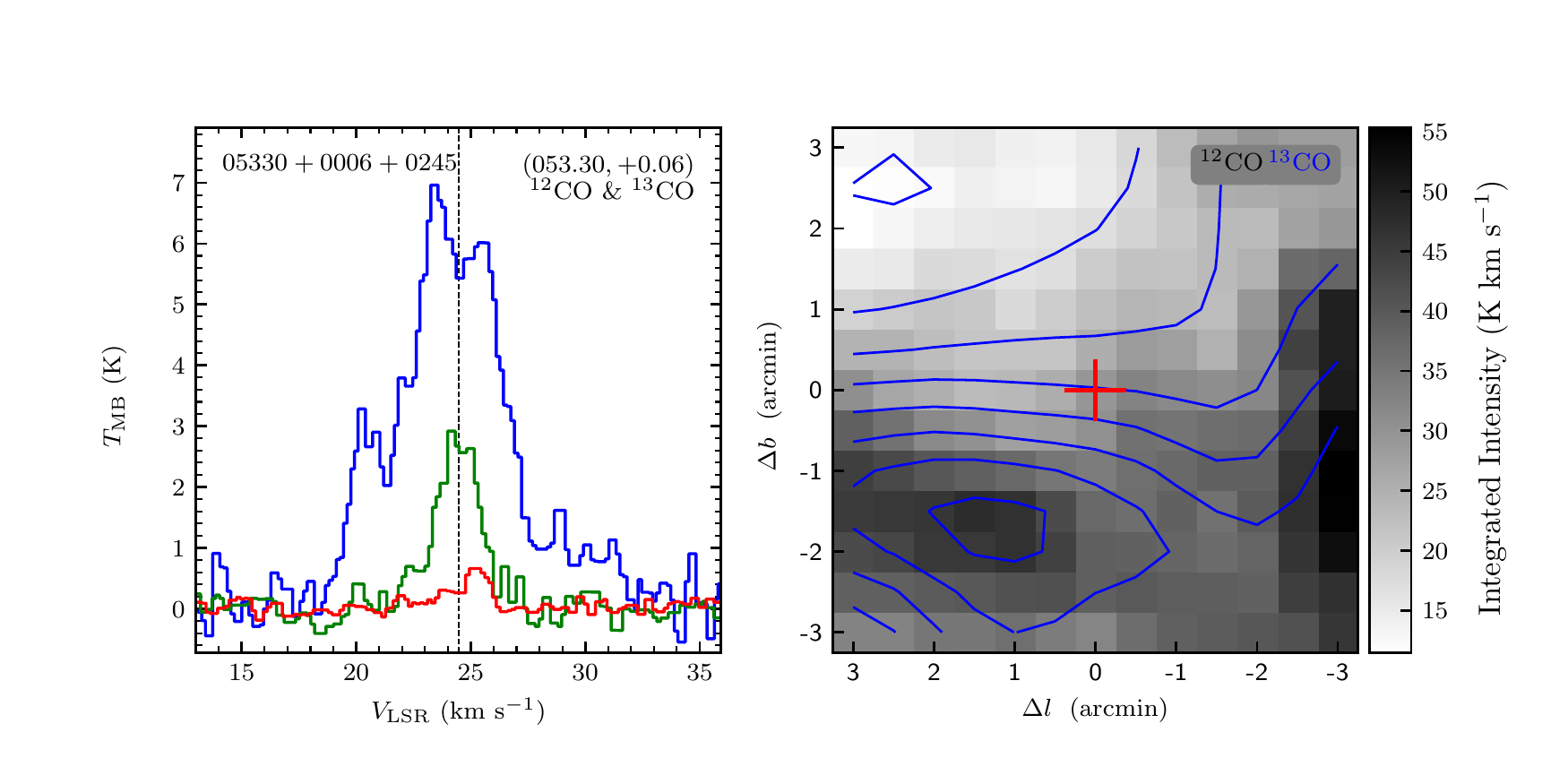}
\includegraphics[width=9.0cm,angle=0]{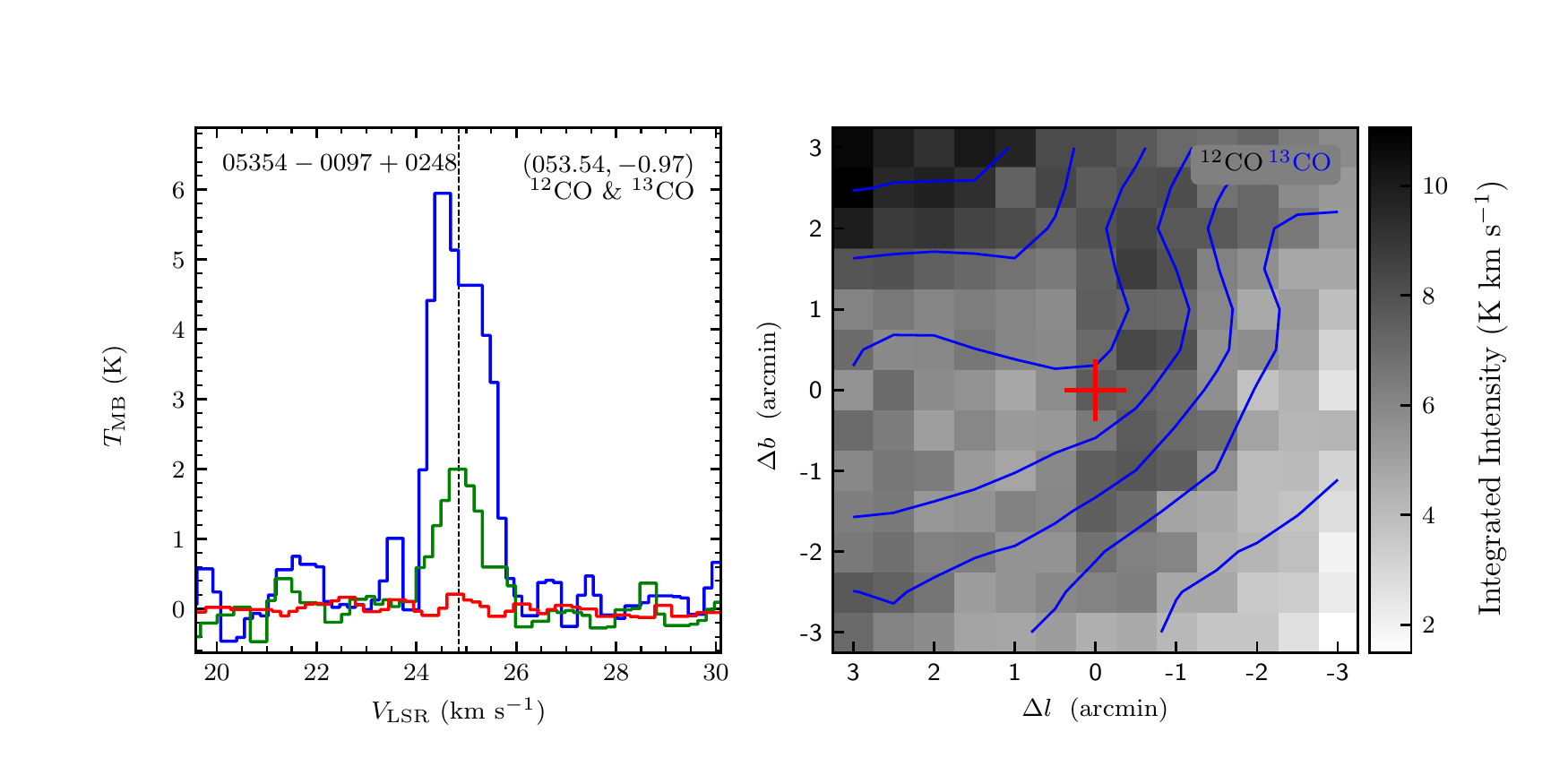}
\vspace{-0.5cm}

\includegraphics[width=9.0cm,angle=0]{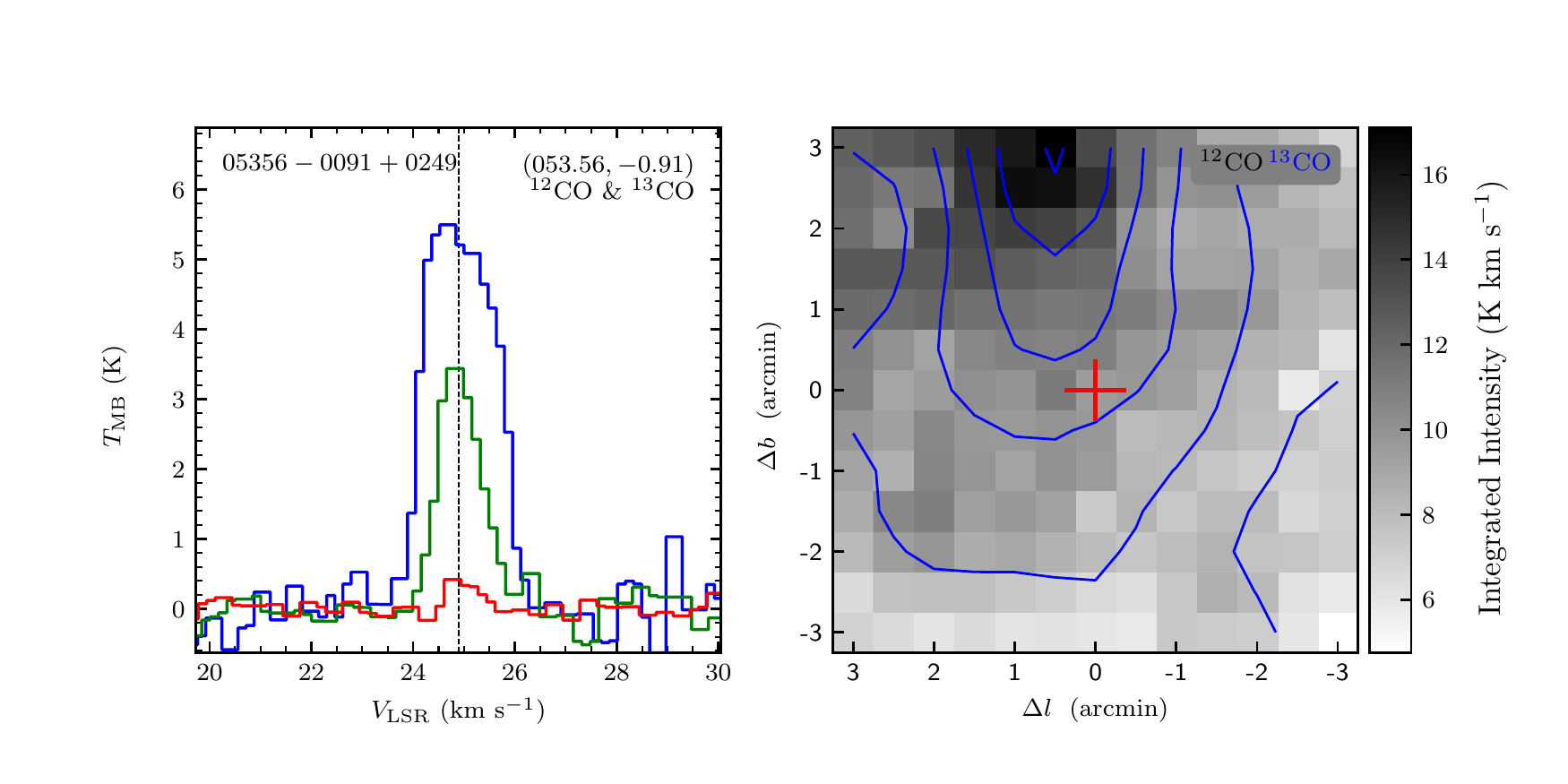}
\includegraphics[width=9.0cm,angle=0]{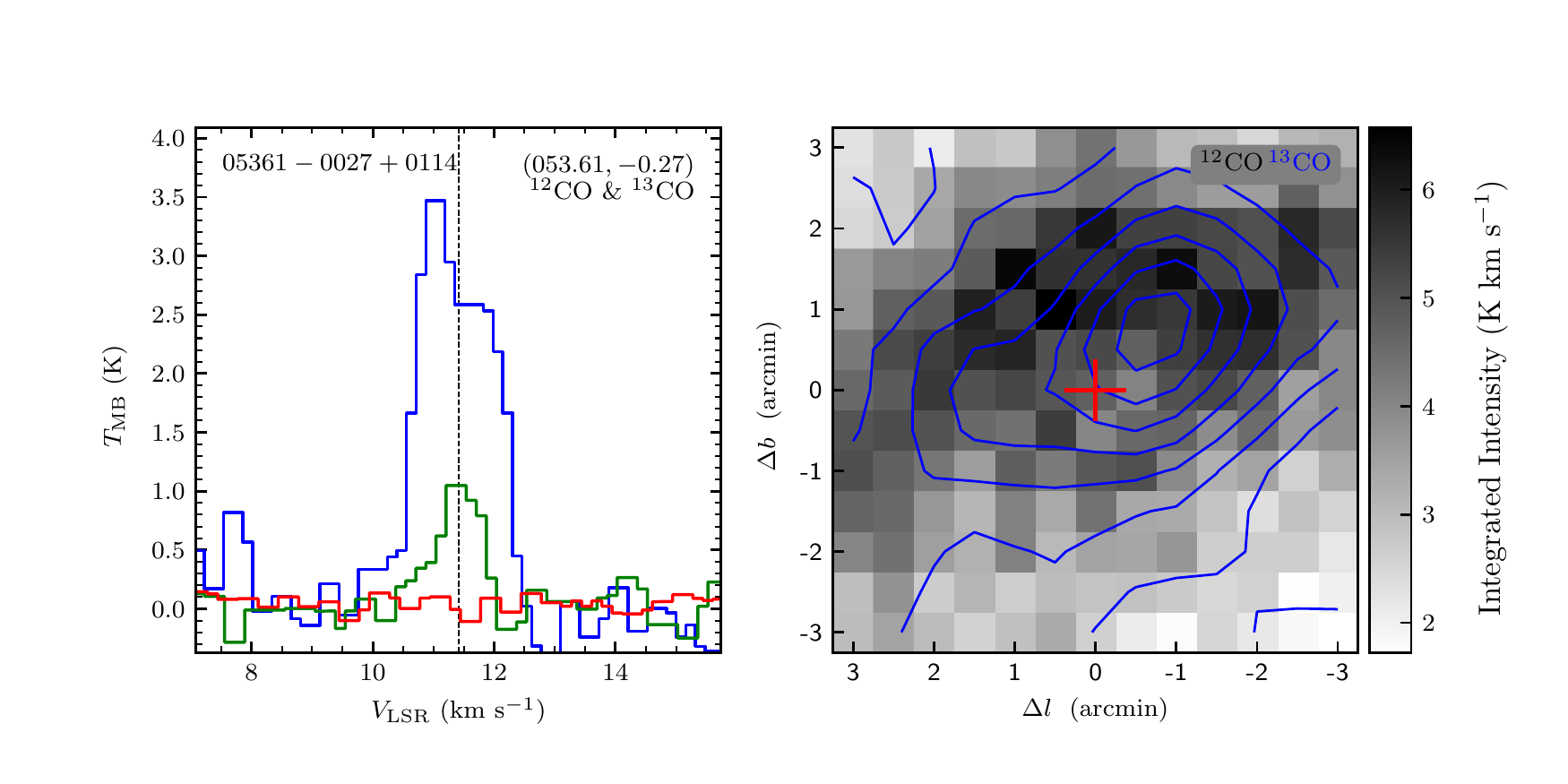}
\vspace{-0.5cm}

\includegraphics[width=9.0cm,angle=0]{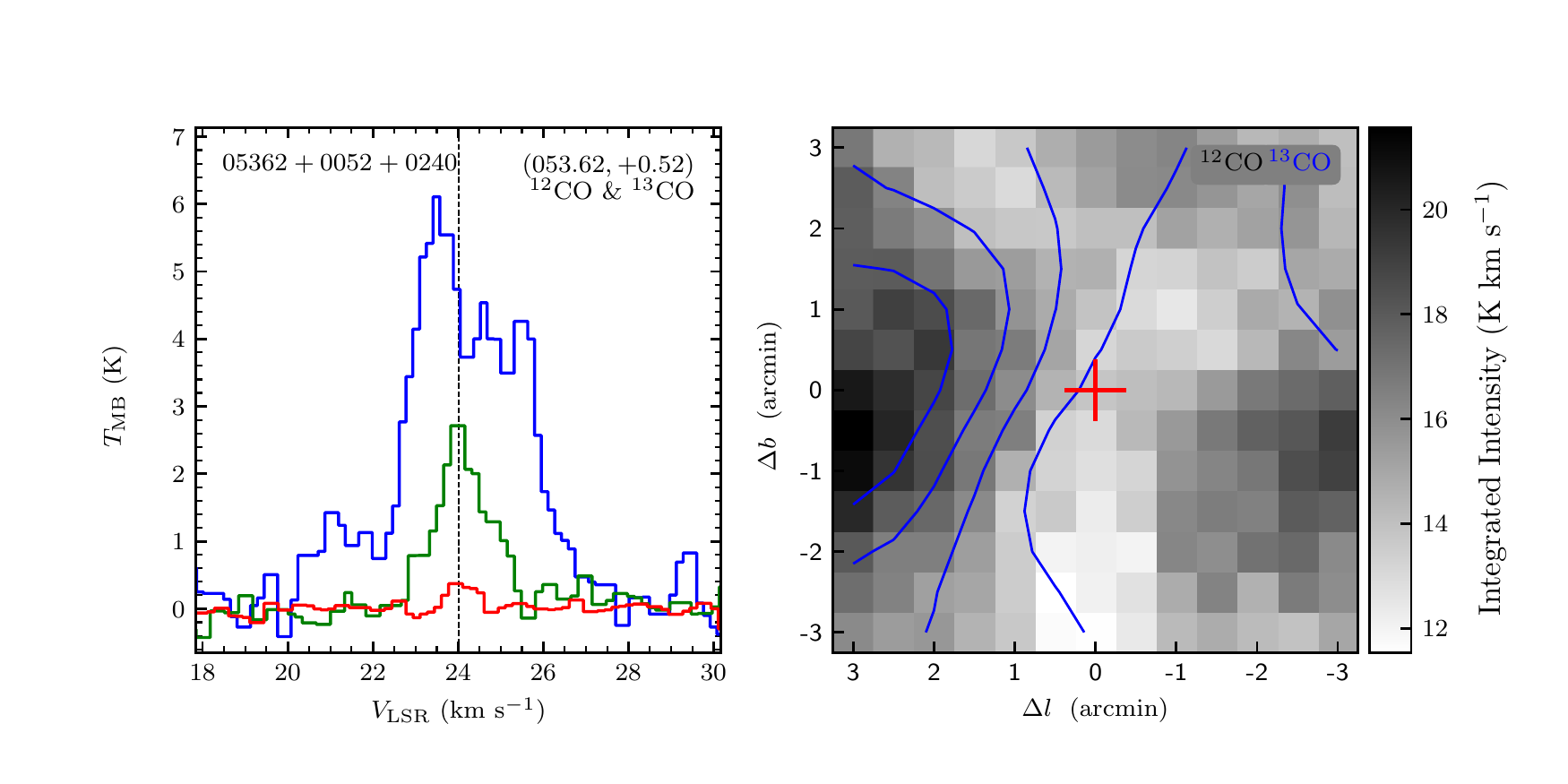}
\includegraphics[width=9.0cm,angle=0]{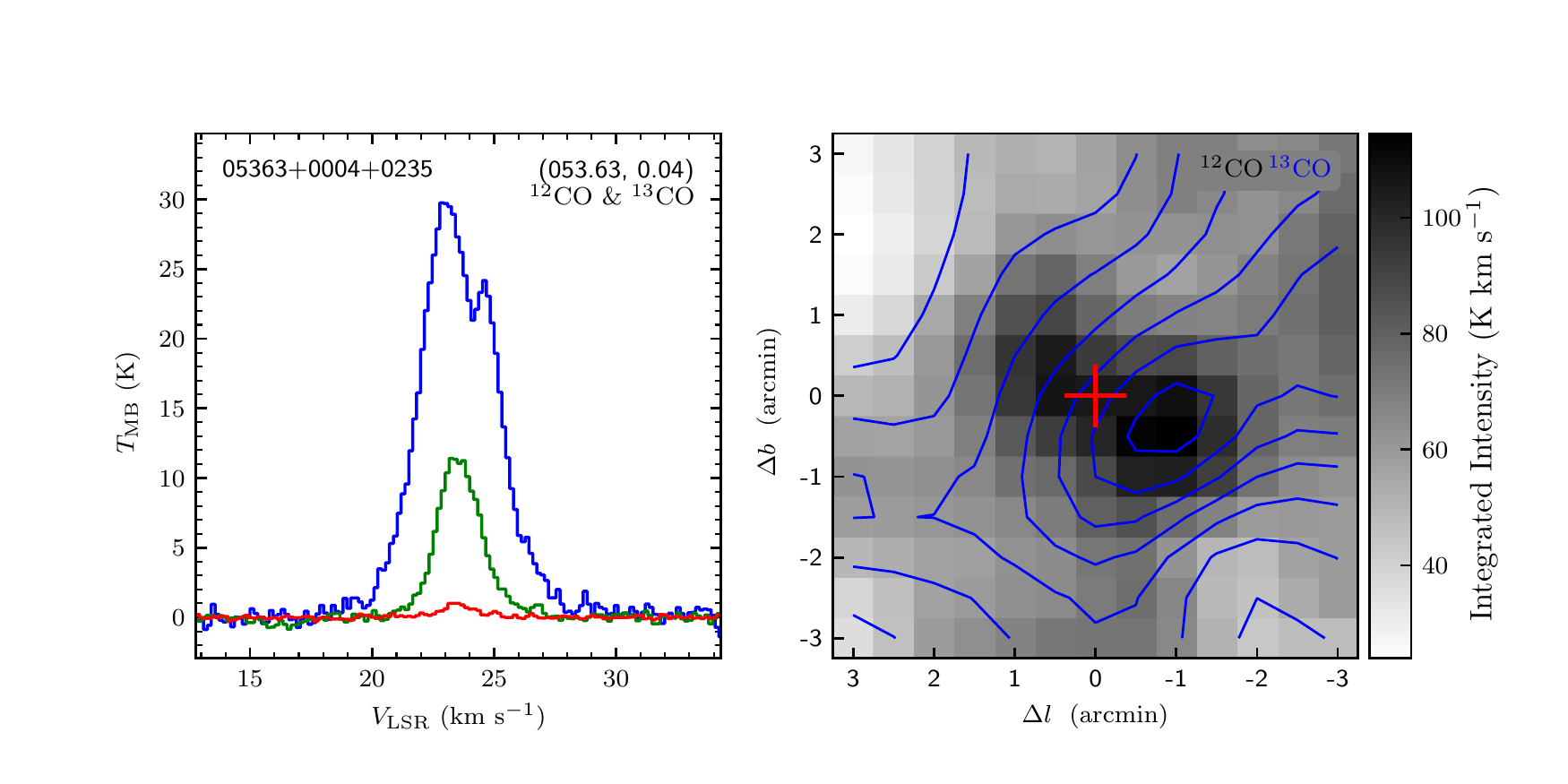}
\end{figure}
\clearpage

\begin{figure}
\includegraphics[width=9.0cm,angle=0]{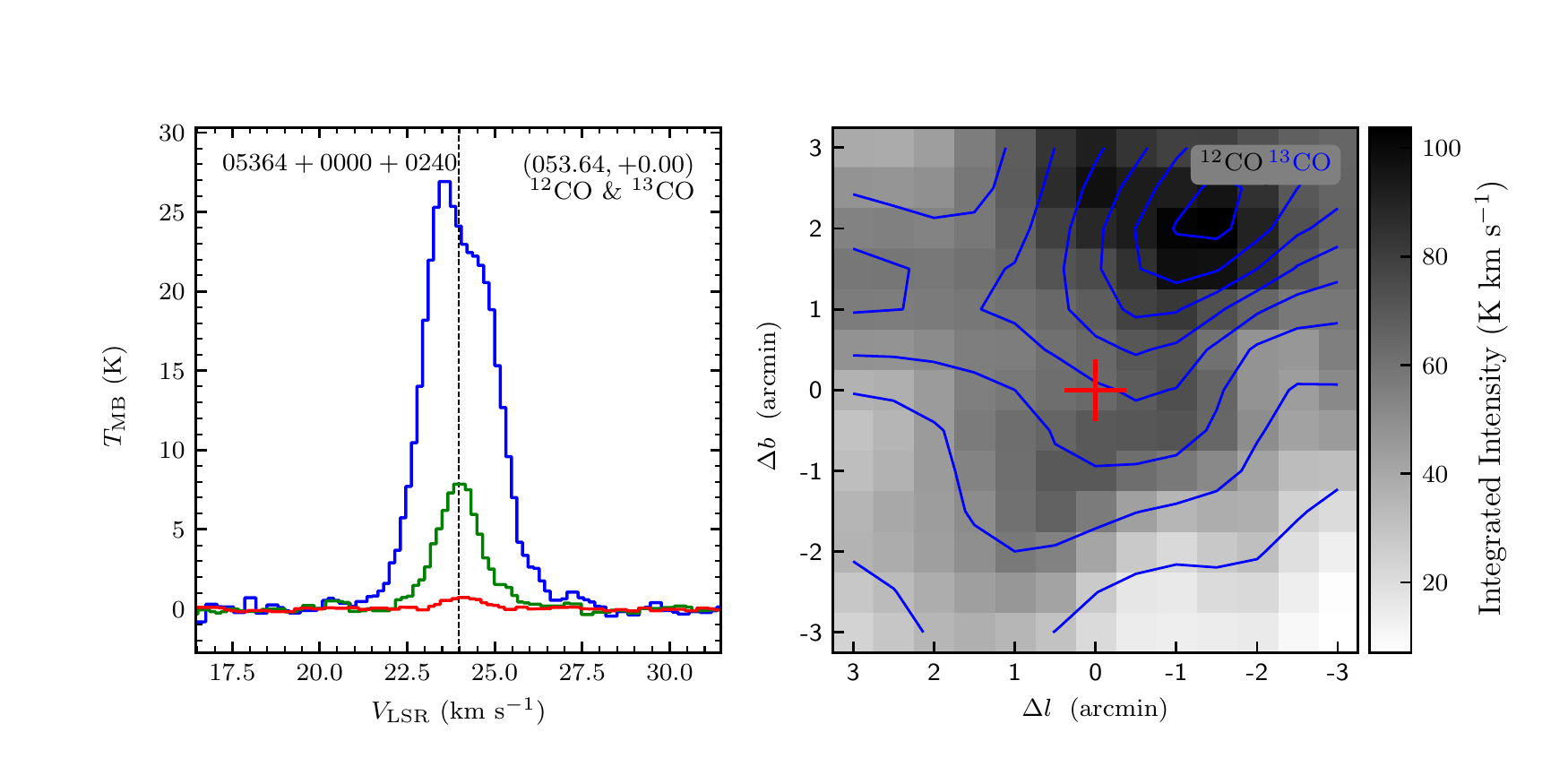}
\includegraphics[width=9.0cm,angle=0]{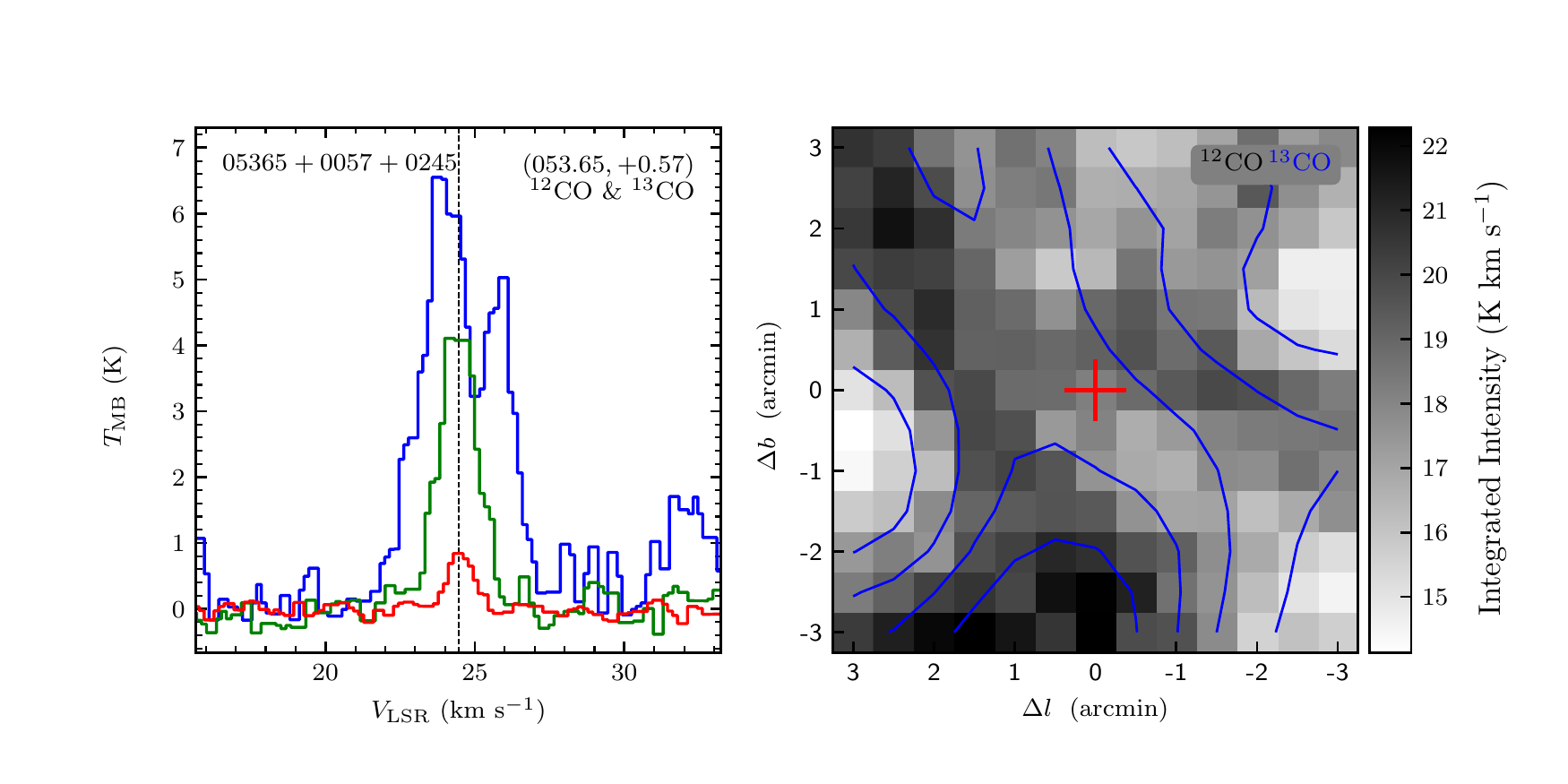}
\vspace{-0.5cm}

\includegraphics[width=9.0cm,angle=0]{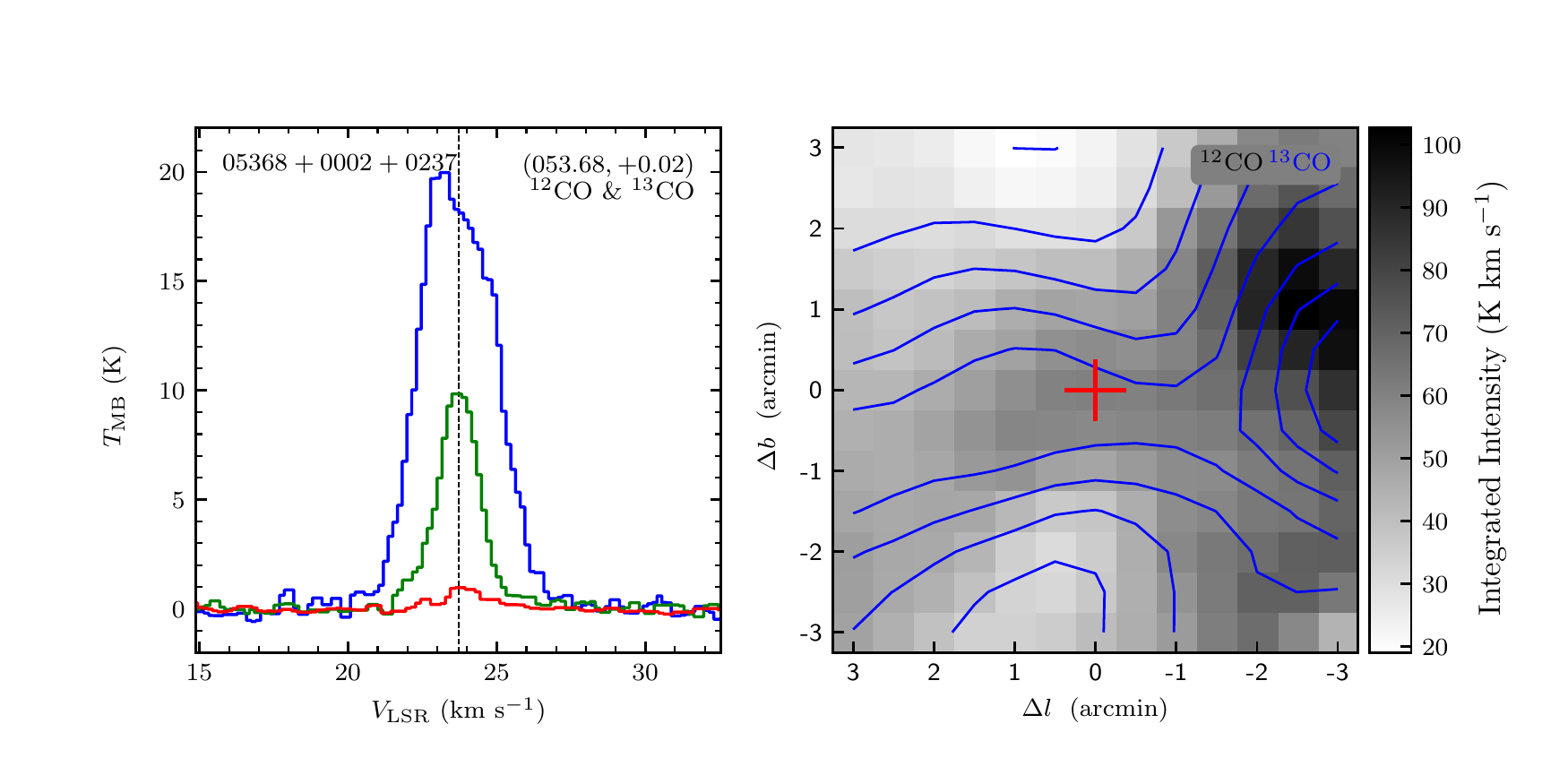}
\includegraphics[width=9.0cm,angle=0]{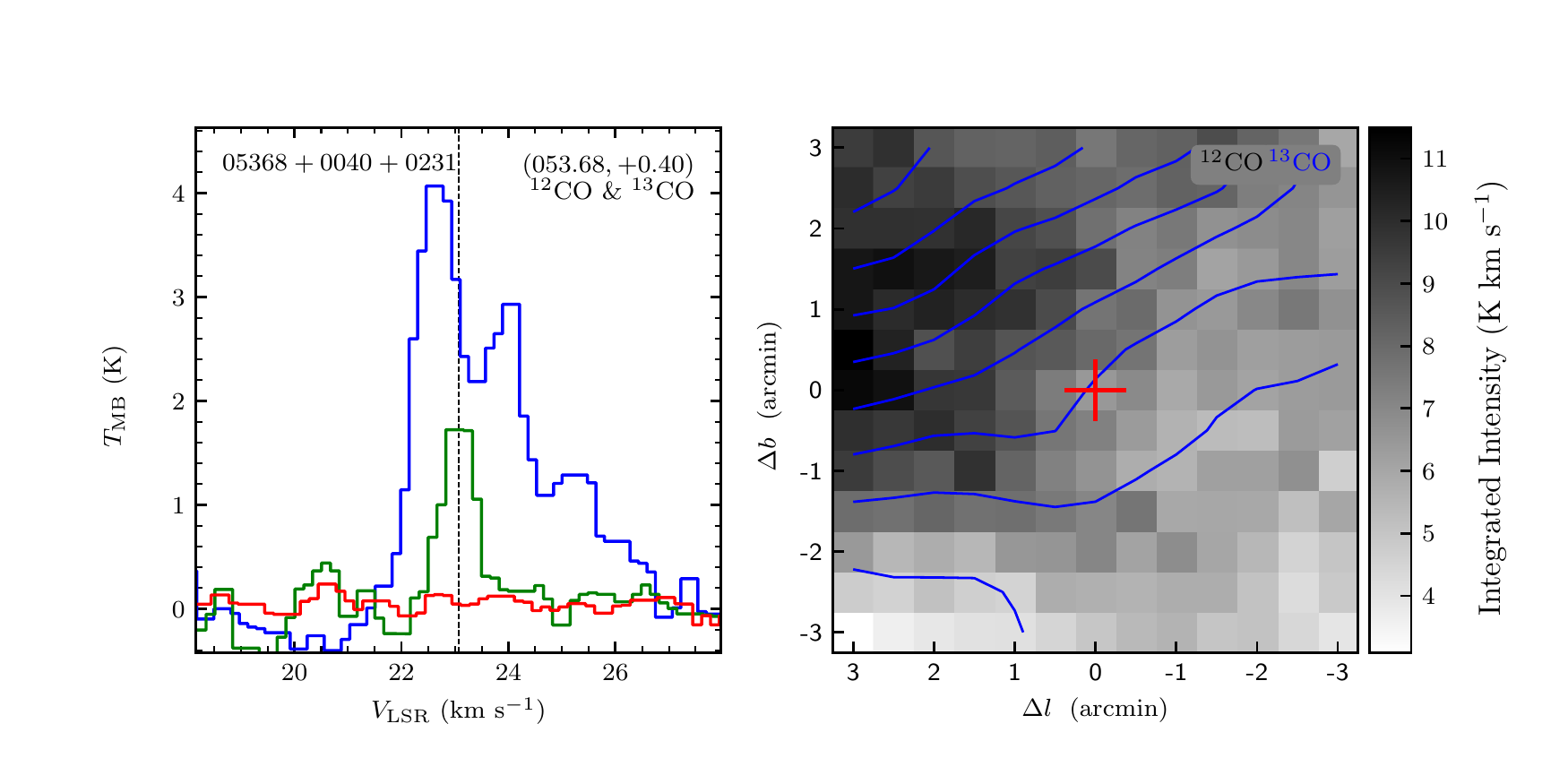}
\vspace{-0.5cm}

\includegraphics[width=9.0cm,angle=0]{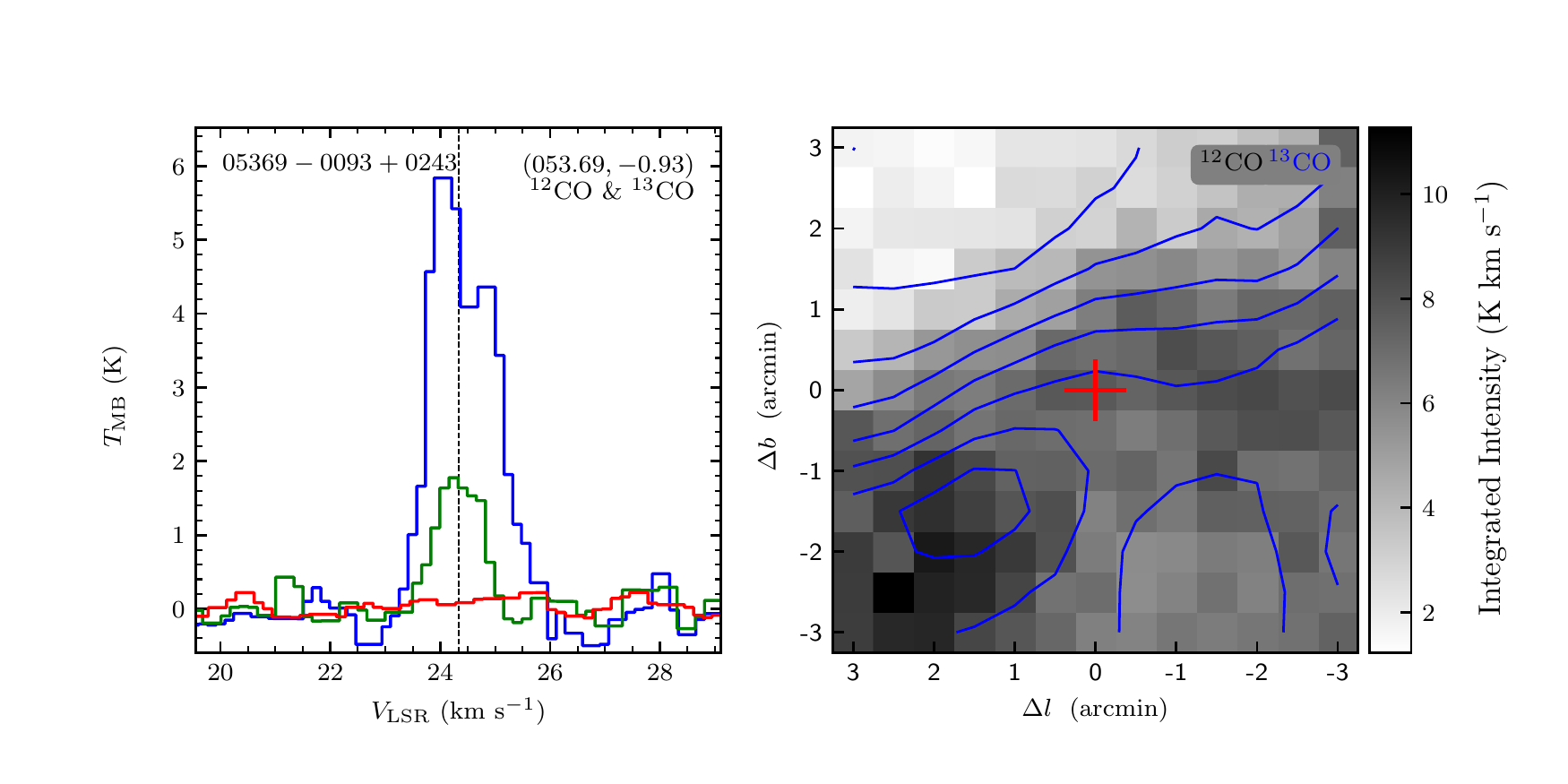}
\includegraphics[width=9.0cm,angle=0]{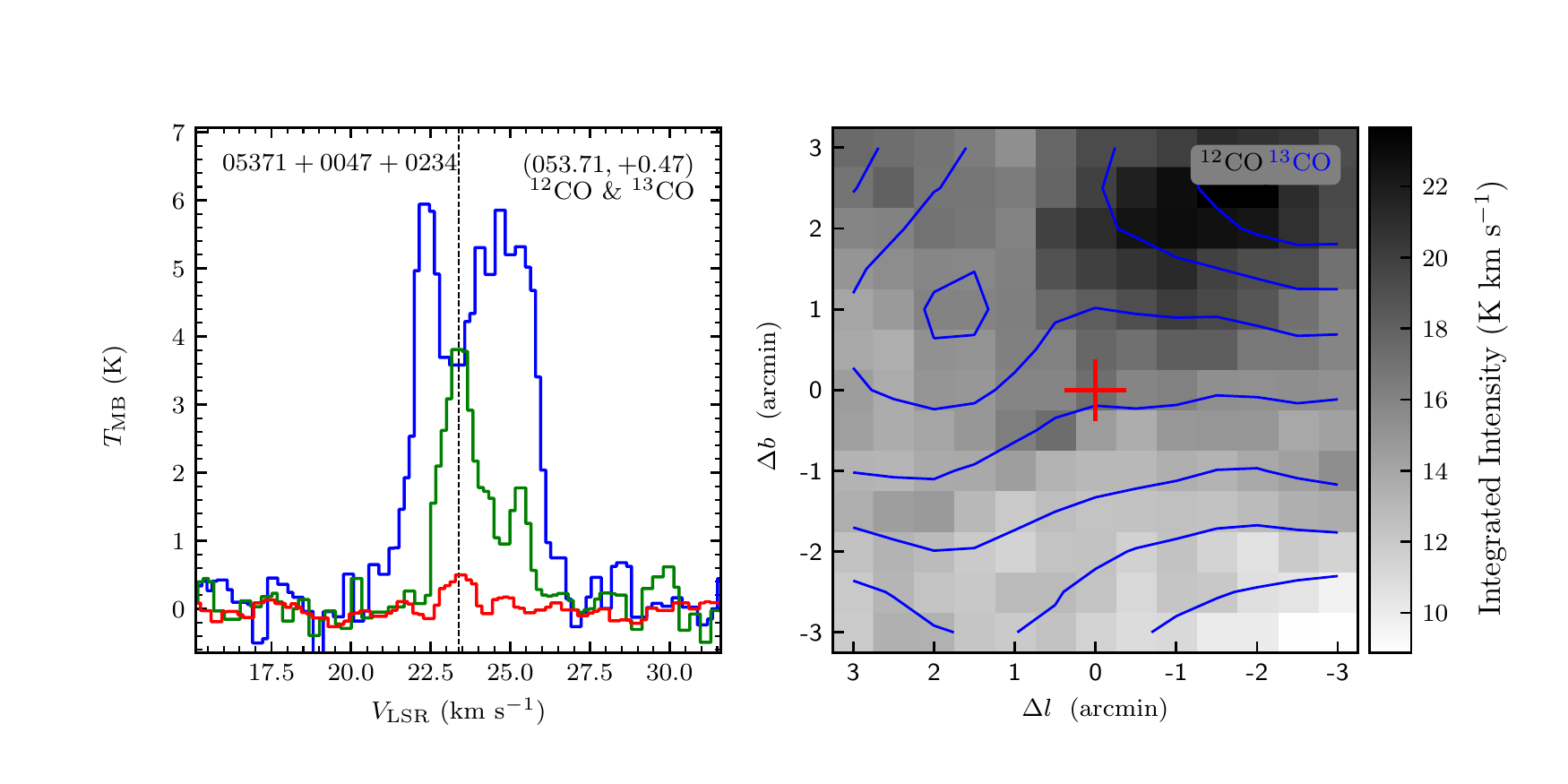}
\vspace{-0.5cm}

\includegraphics[width=9.0cm,angle=0]{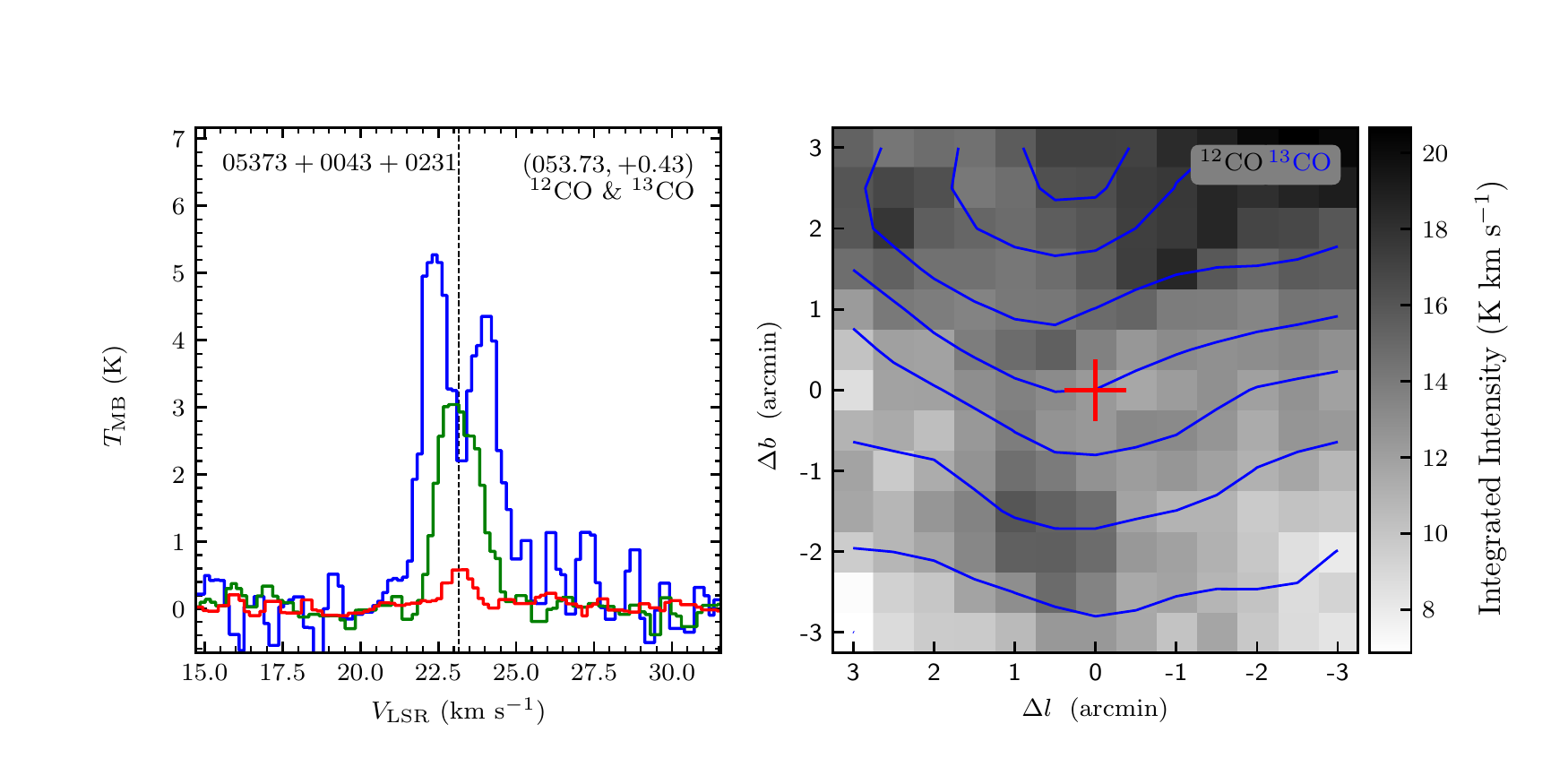}
\includegraphics[width=9.0cm,angle=0]{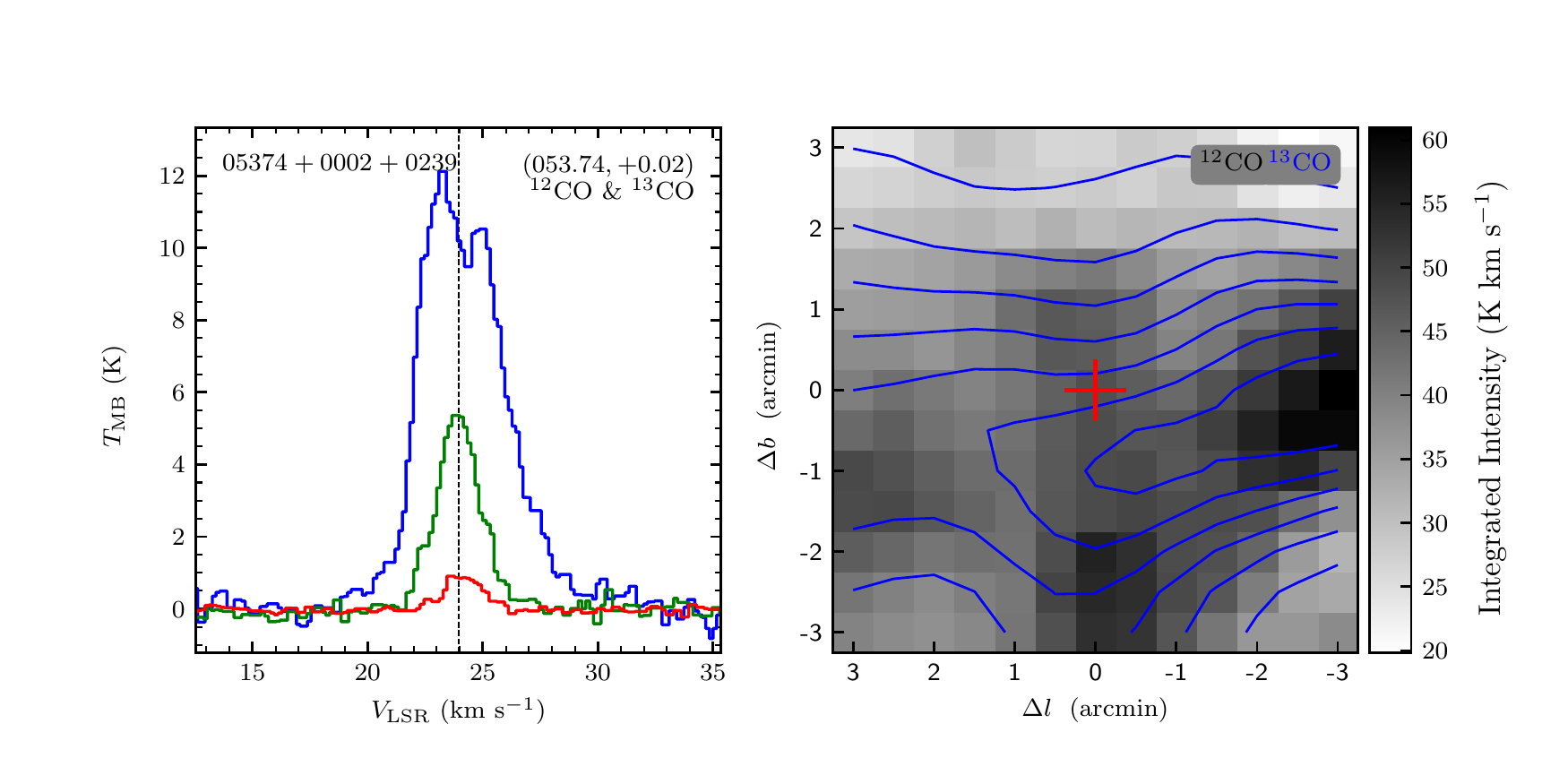}
\vspace{-0.5cm}

\includegraphics[width=9.0cm,angle=0]{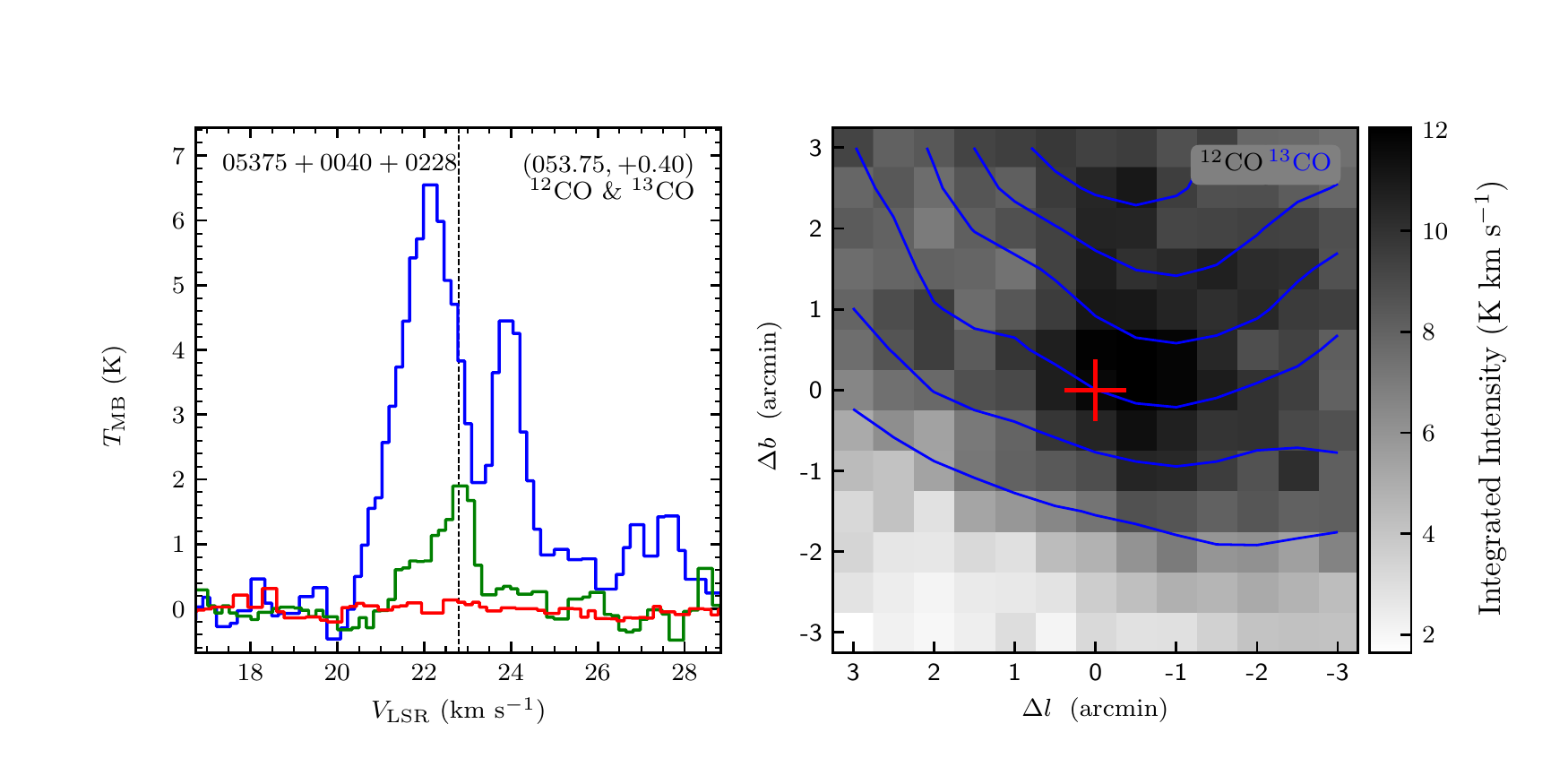}
\includegraphics[width=9.0cm,angle=0]{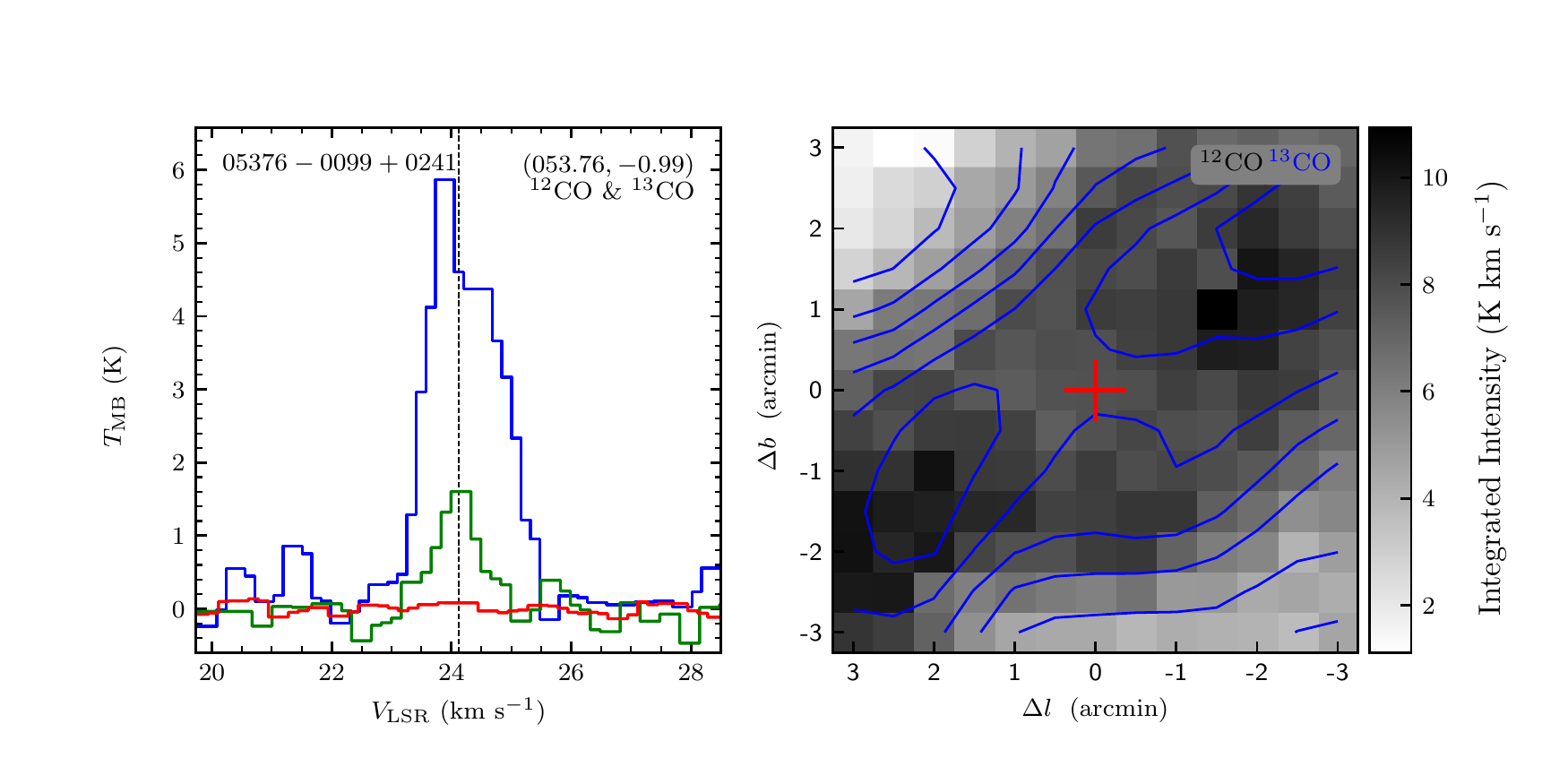}
\end{figure}
\clearpage

\begin{figure}
\includegraphics[width=9.0cm,angle=0]{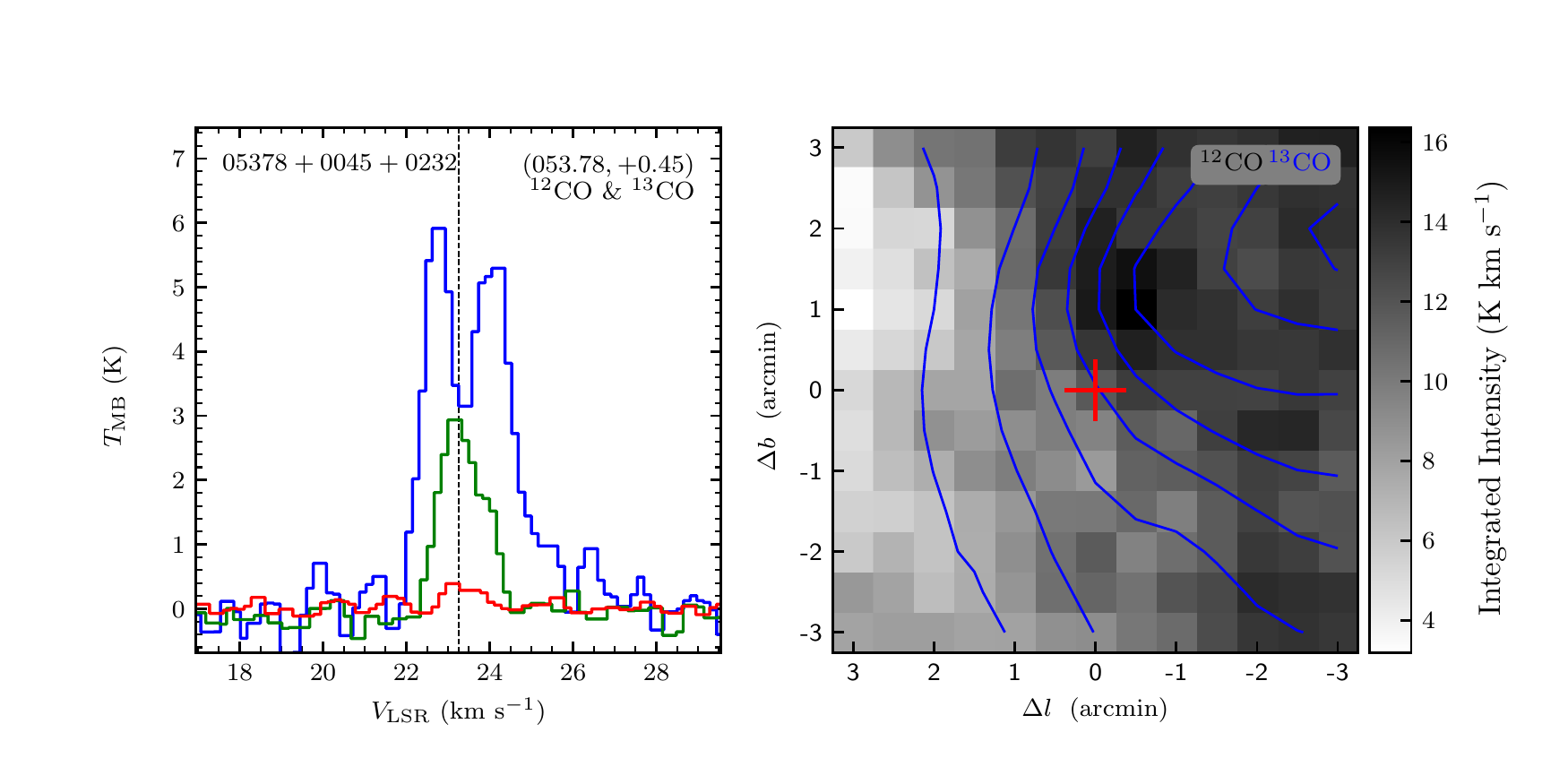}
\includegraphics[width=9.0cm,angle=0]{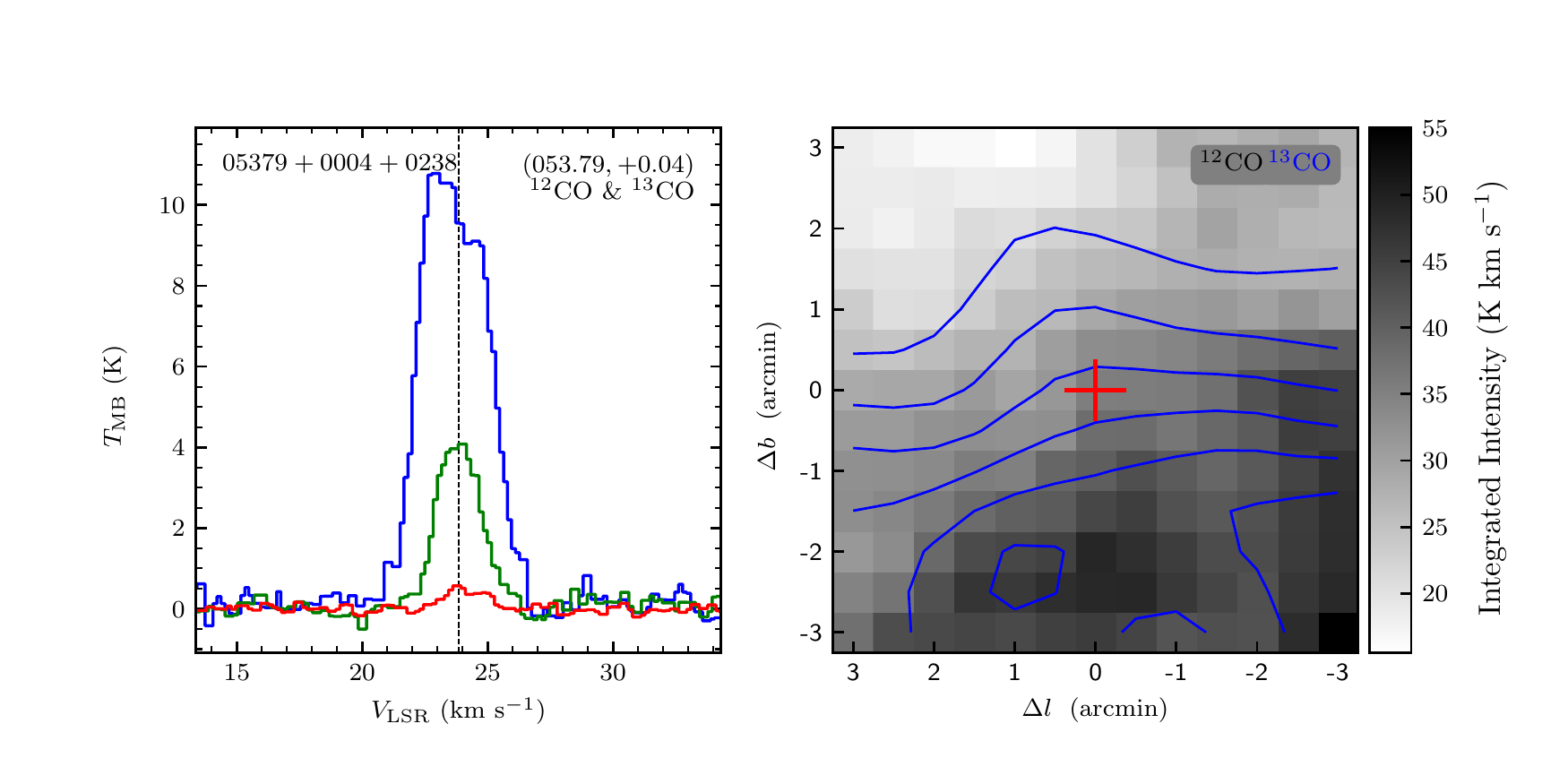}
\vspace{-0.5cm}

\includegraphics[width=9.0cm,angle=0]{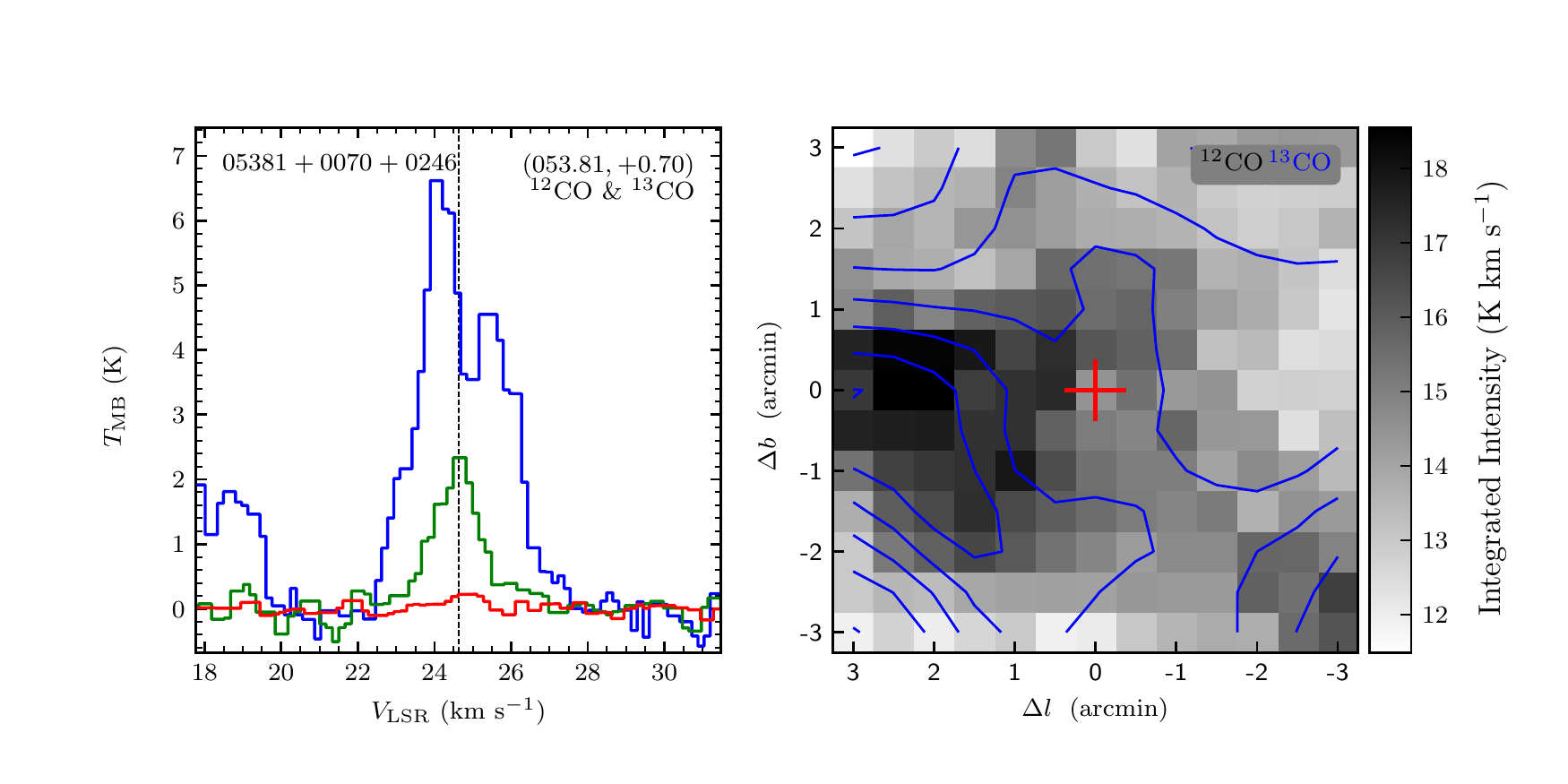}
\includegraphics[width=9.0cm,angle=0]{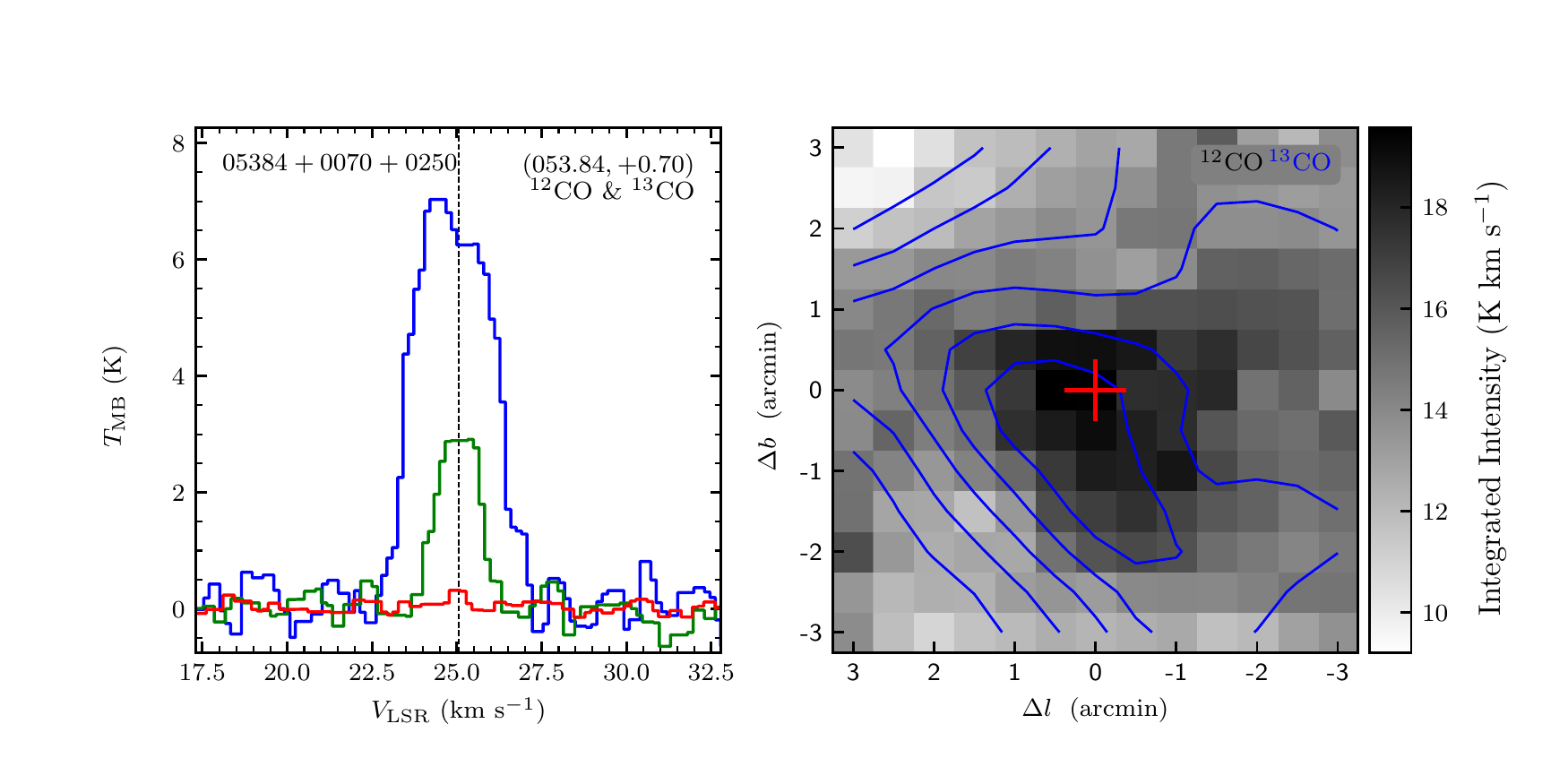}
\vspace{-0.5cm}

\includegraphics[width=9.0cm,angle=0]{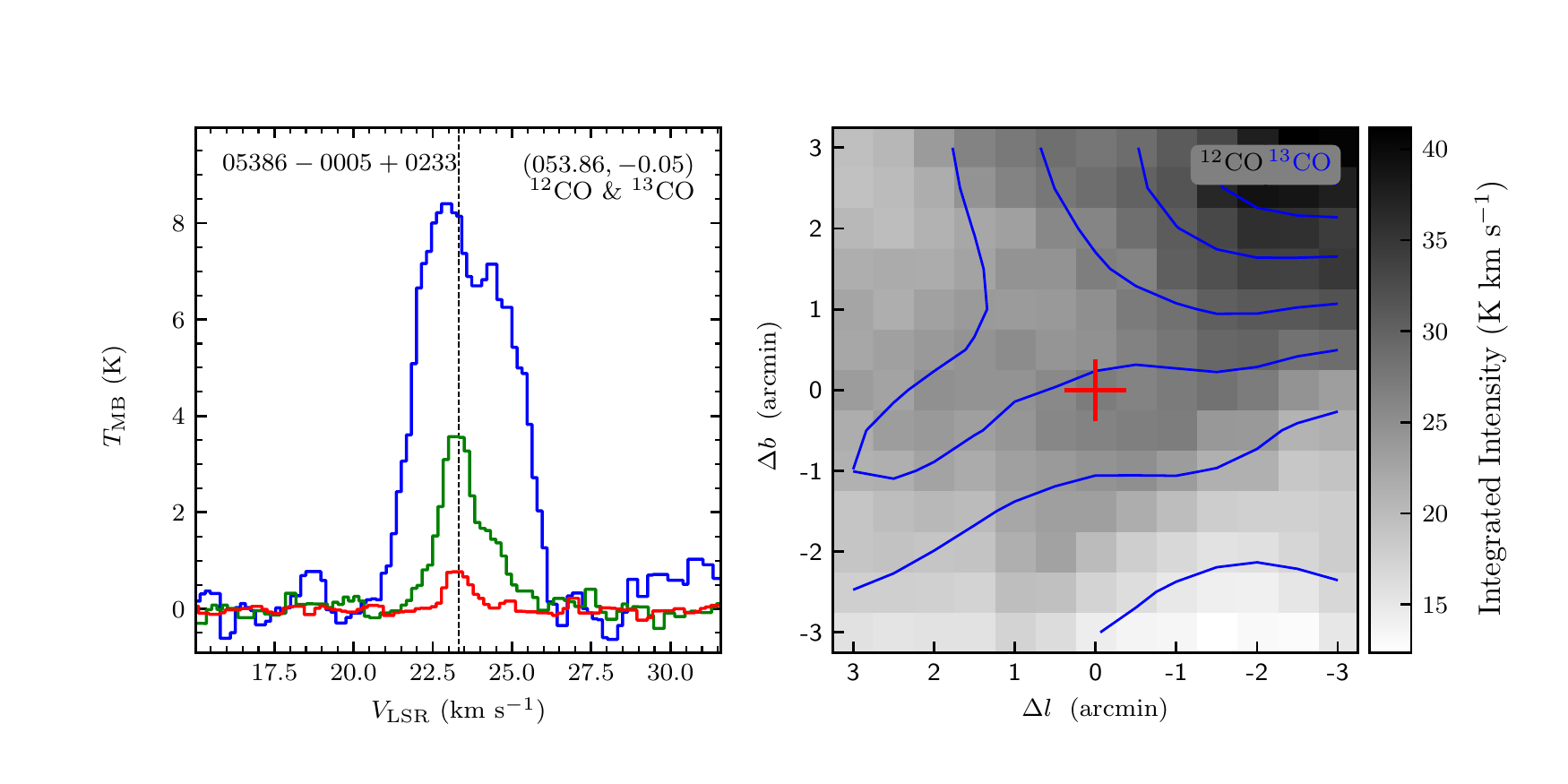}
\includegraphics[width=9.0cm,angle=0]{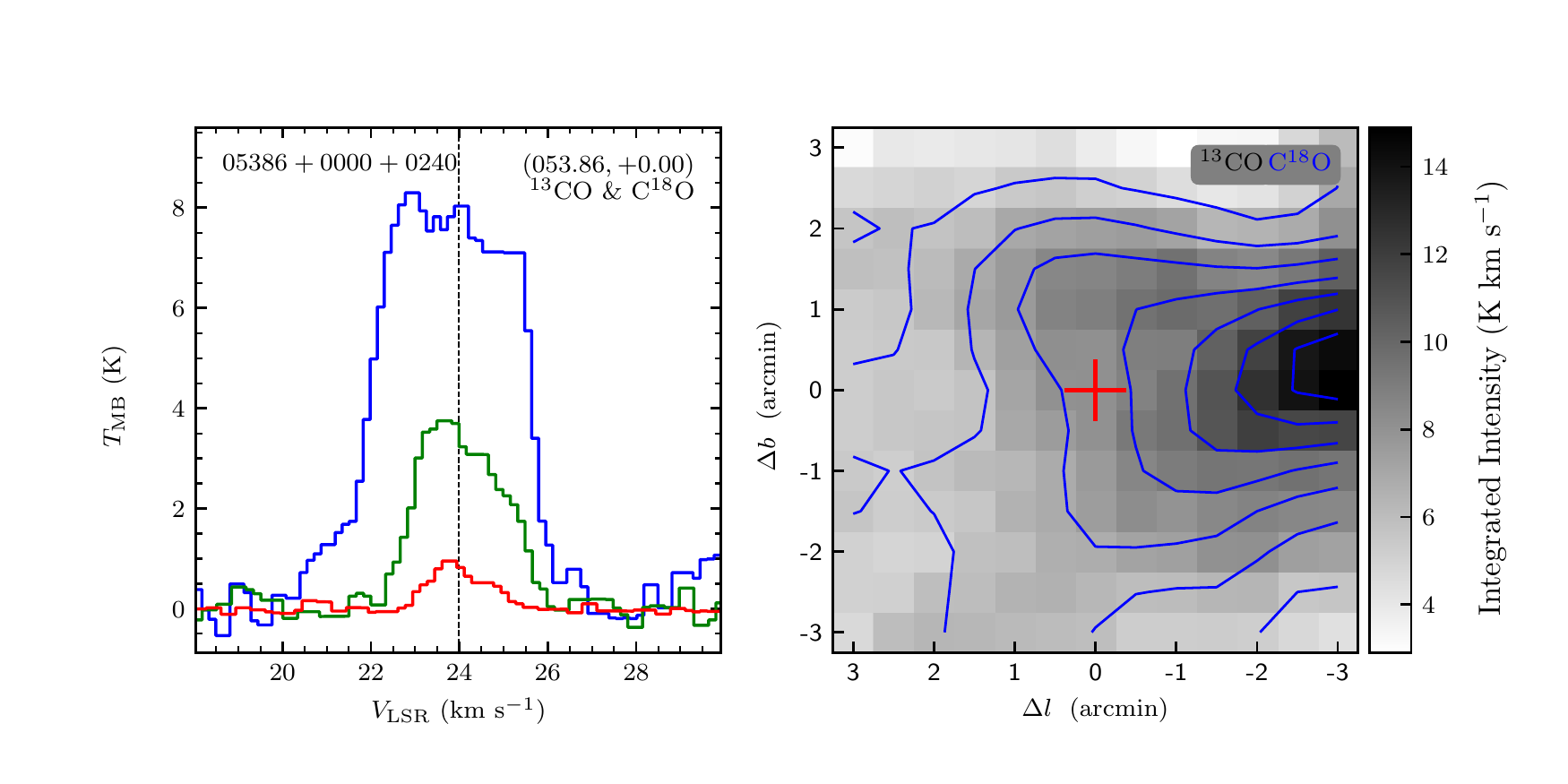}
\vspace{-0.5cm}

\includegraphics[width=9.0cm,angle=0]{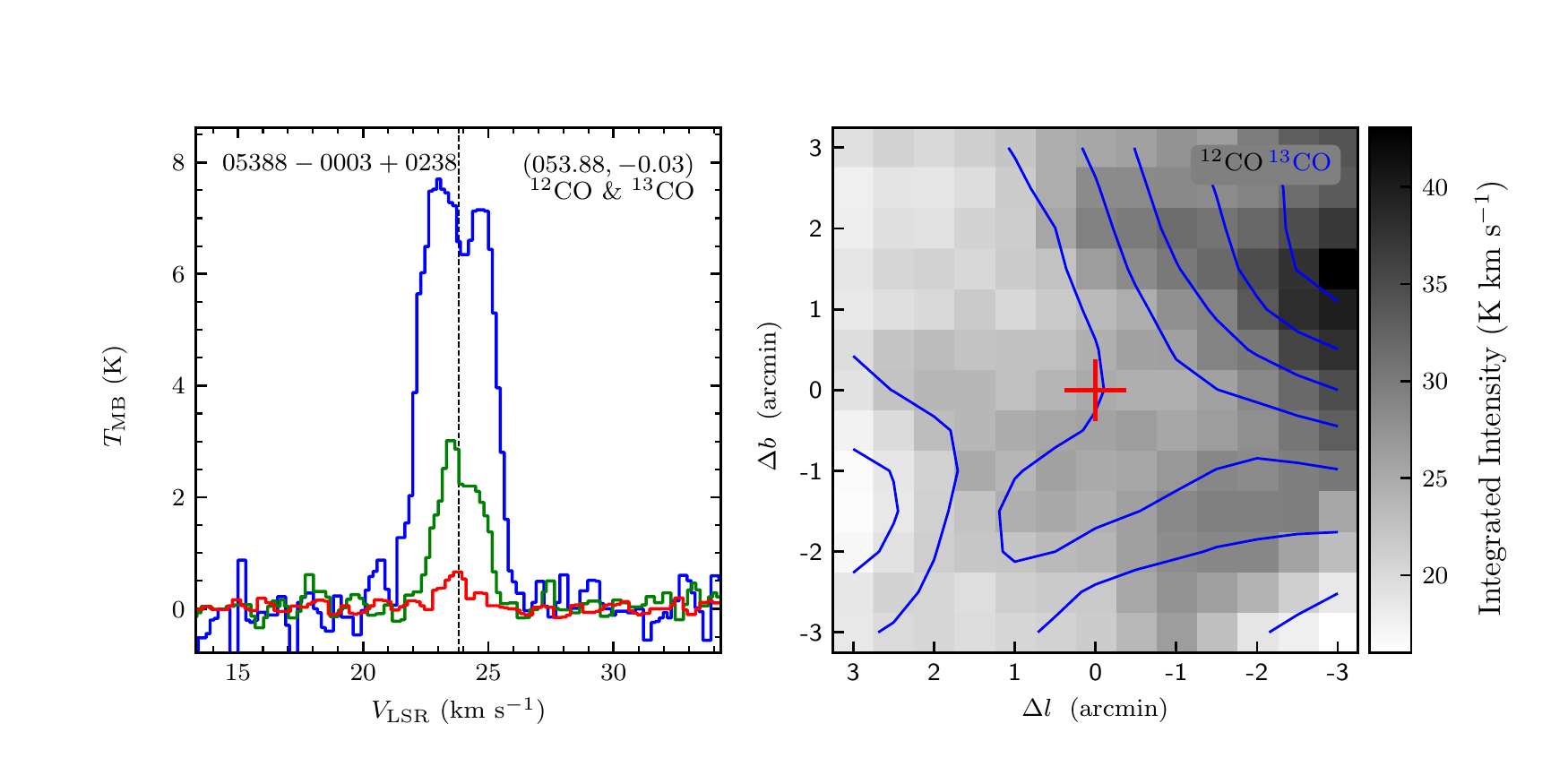}
\includegraphics[width=9.0cm,angle=0]{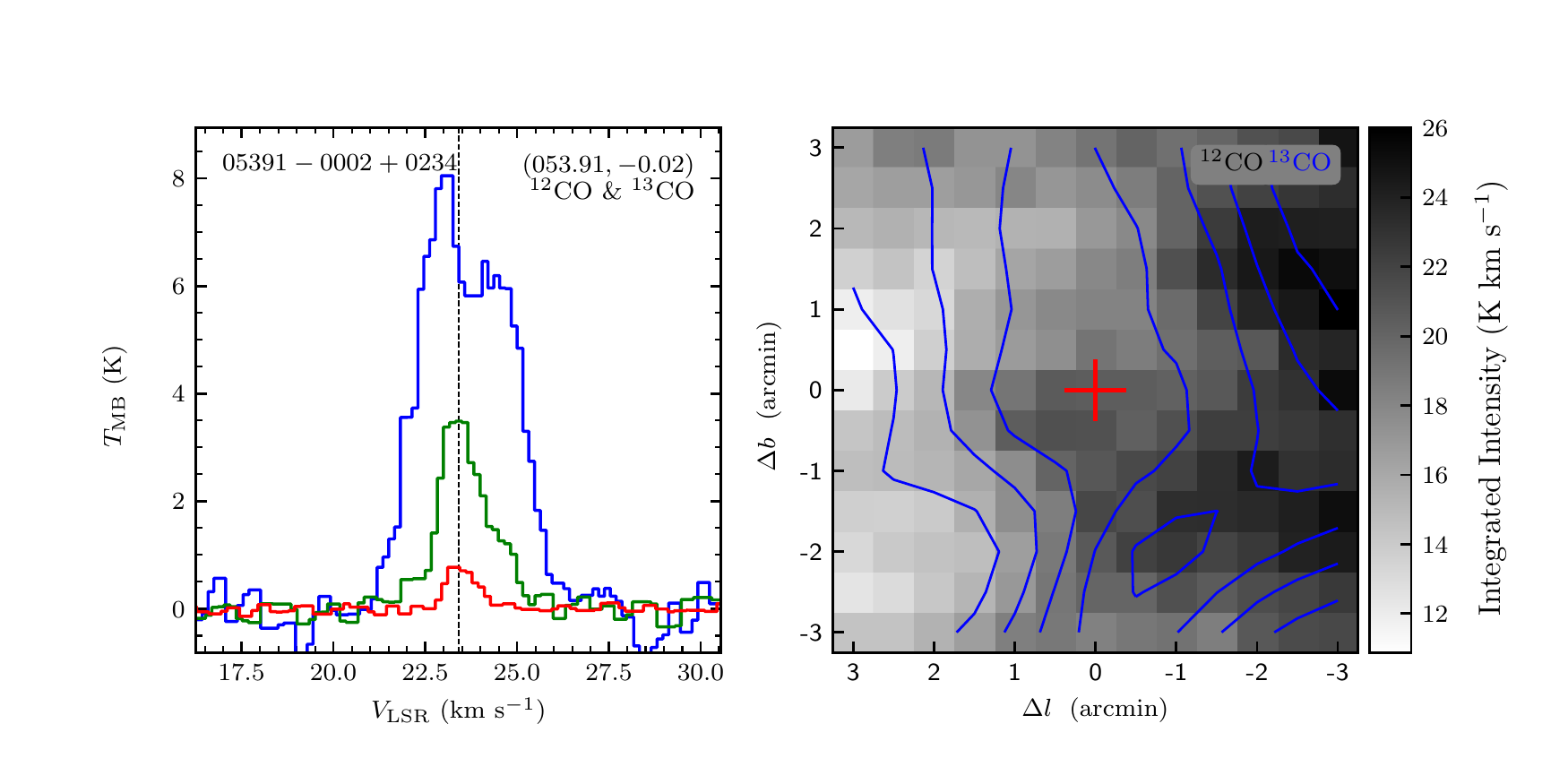}
\vspace{-0.5cm}

\includegraphics[width=9.0cm,angle=0]{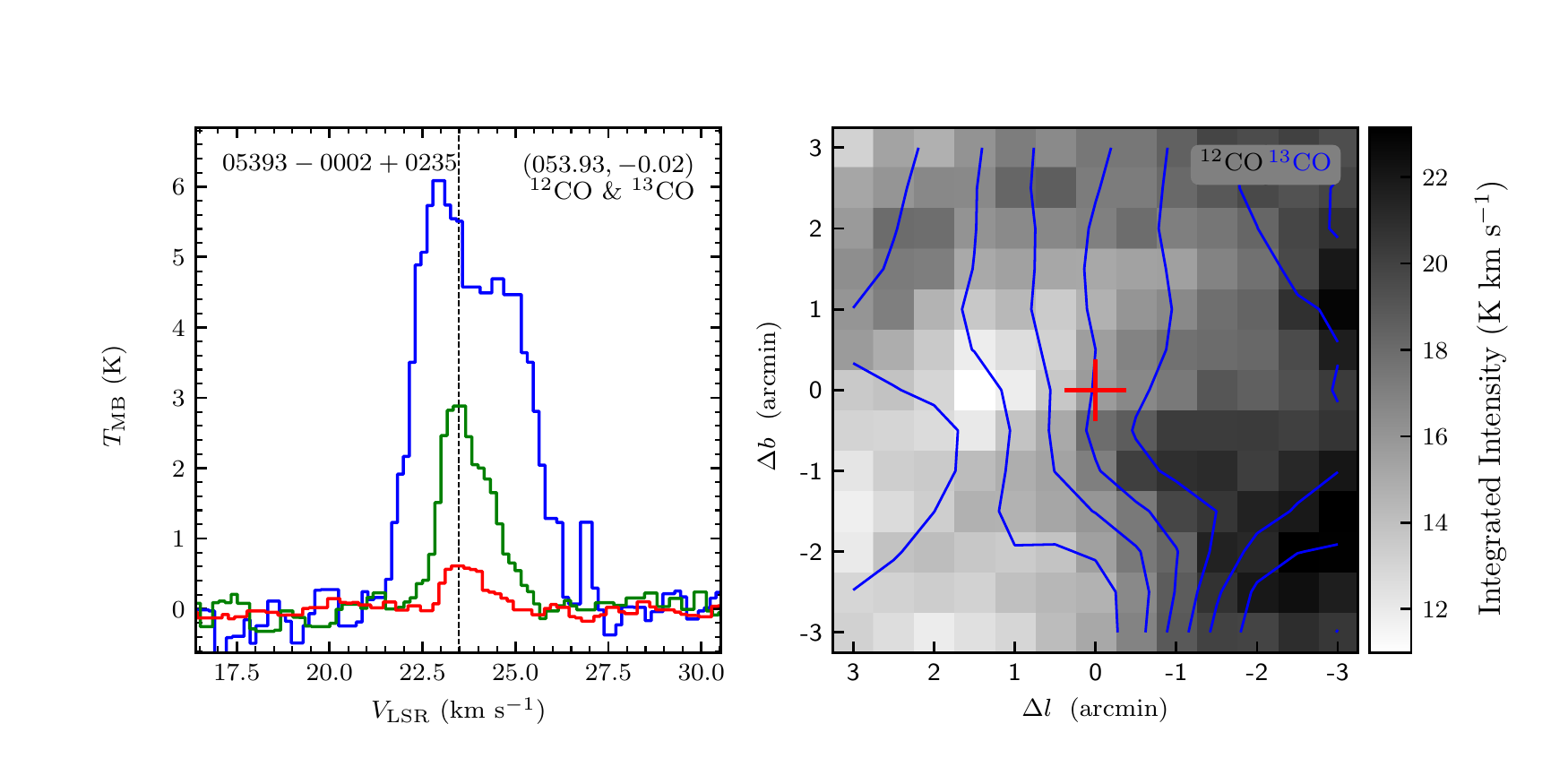}
\includegraphics[width=9.0cm,angle=0]{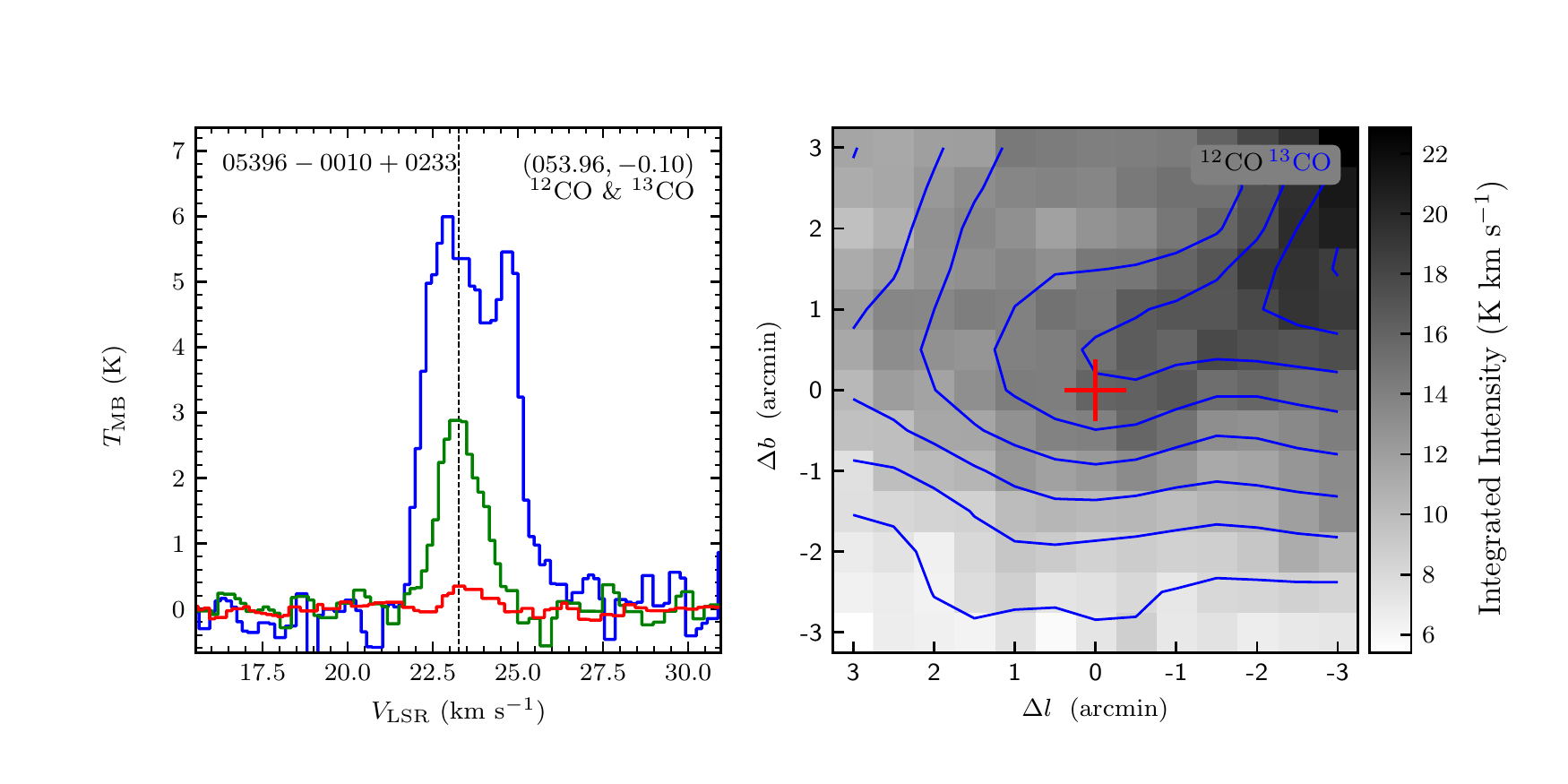}
\end{figure}
\clearpage

\begin{figure}
\includegraphics[width=9.0cm,angle=0]{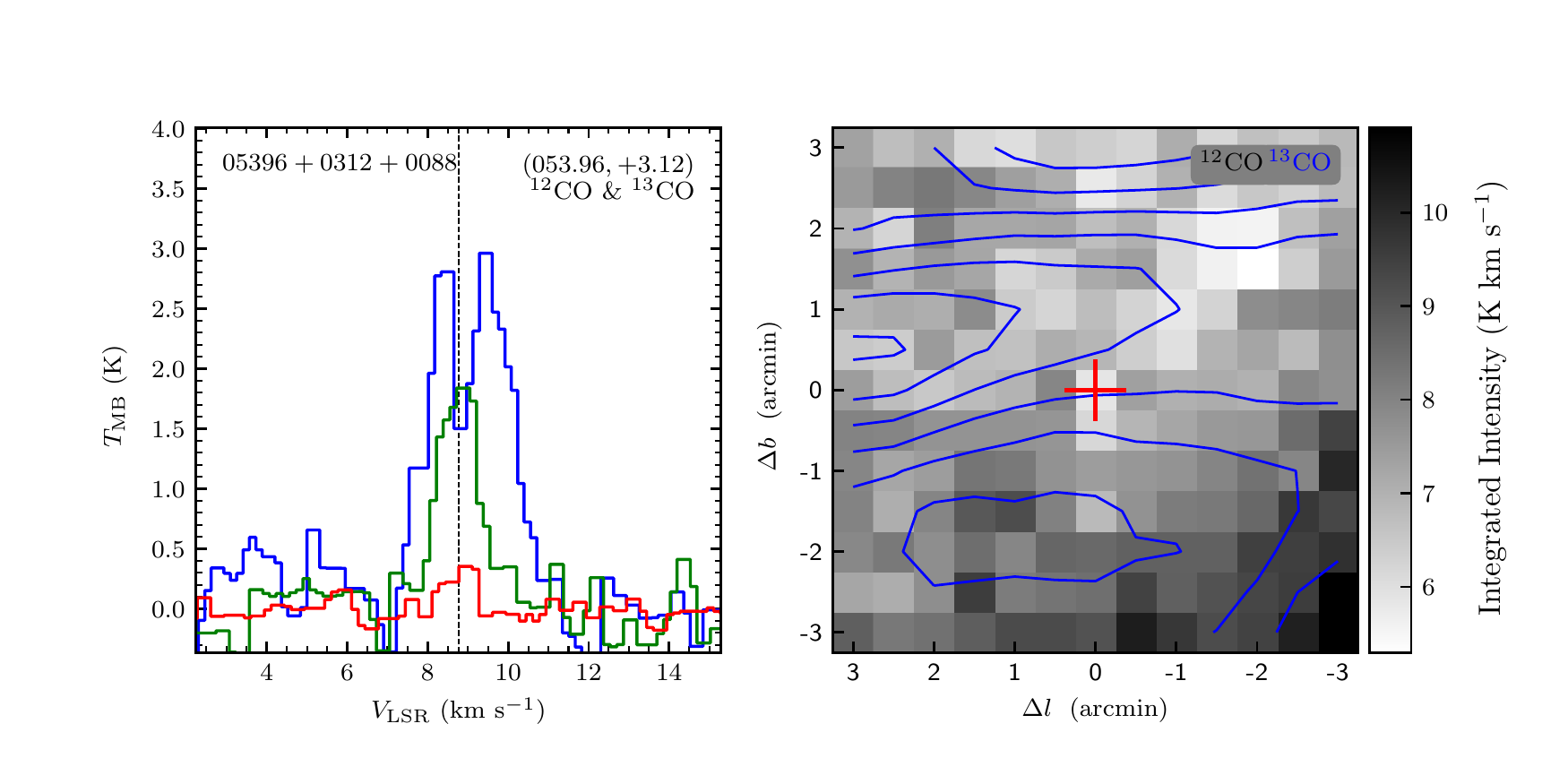}
\includegraphics[width=9.0cm,angle=0]{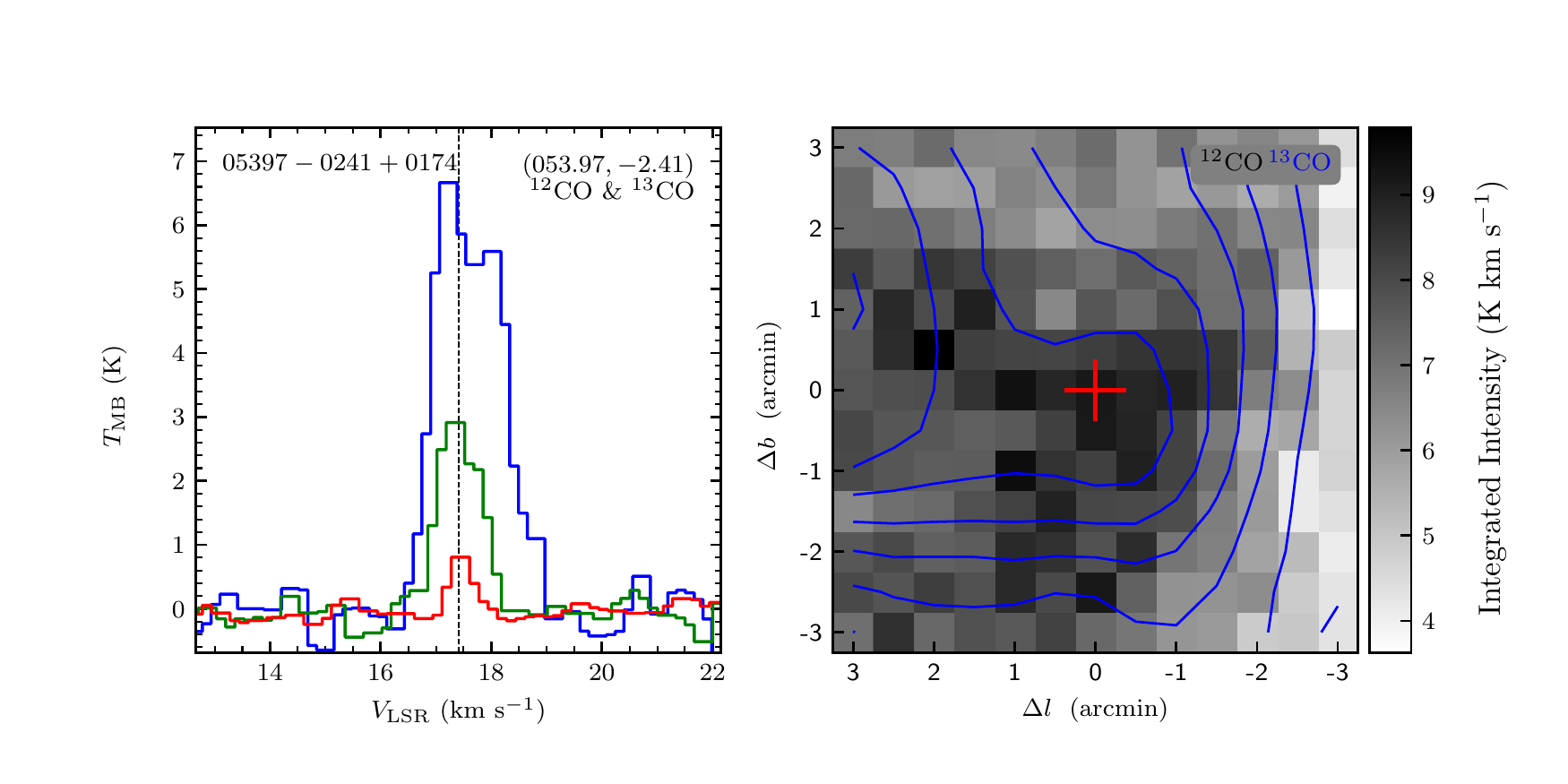}
\vspace{-0.5cm}

\includegraphics[width=9.0cm,angle=0]{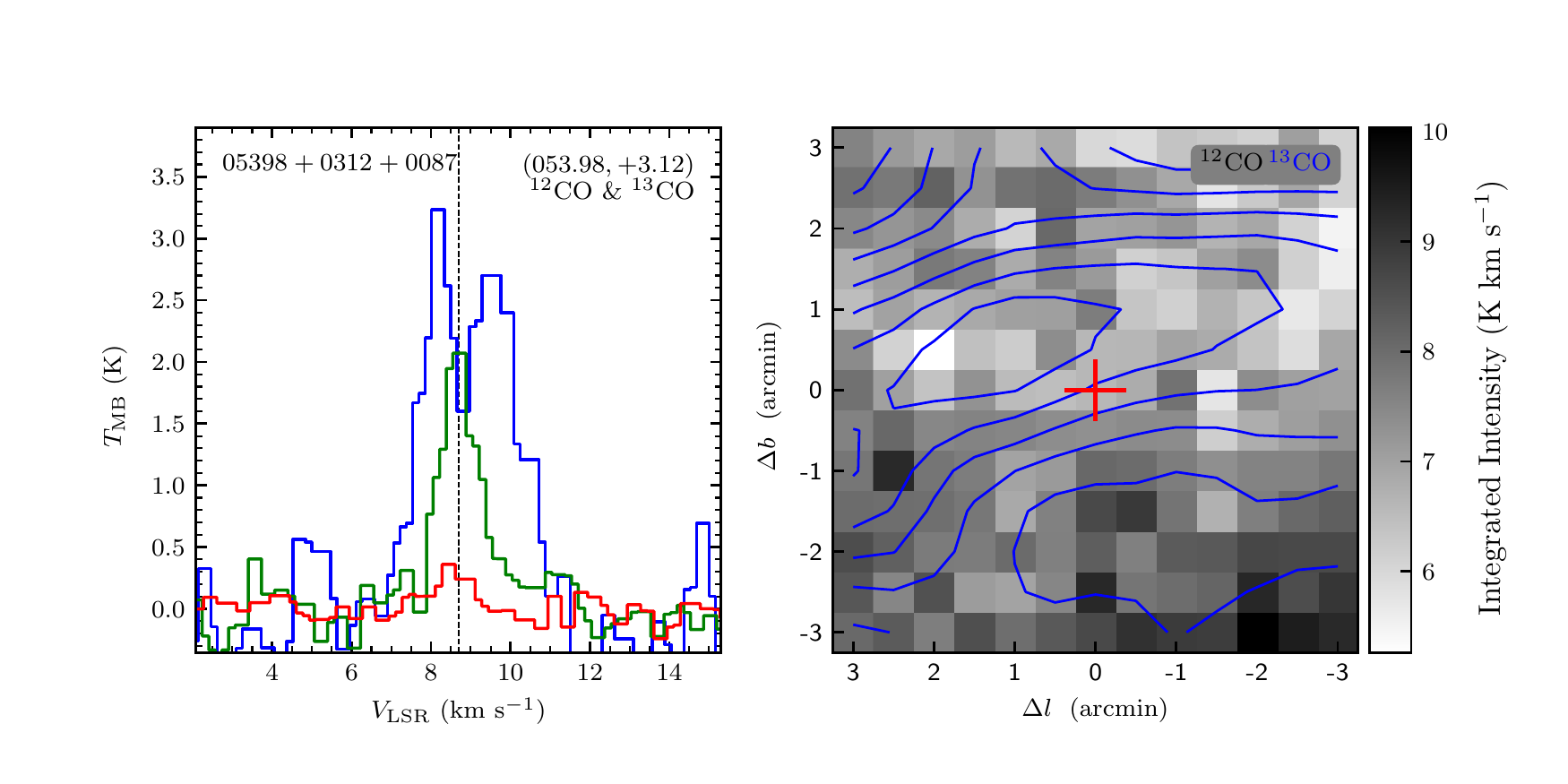}
\includegraphics[width=9.0cm,angle=0]{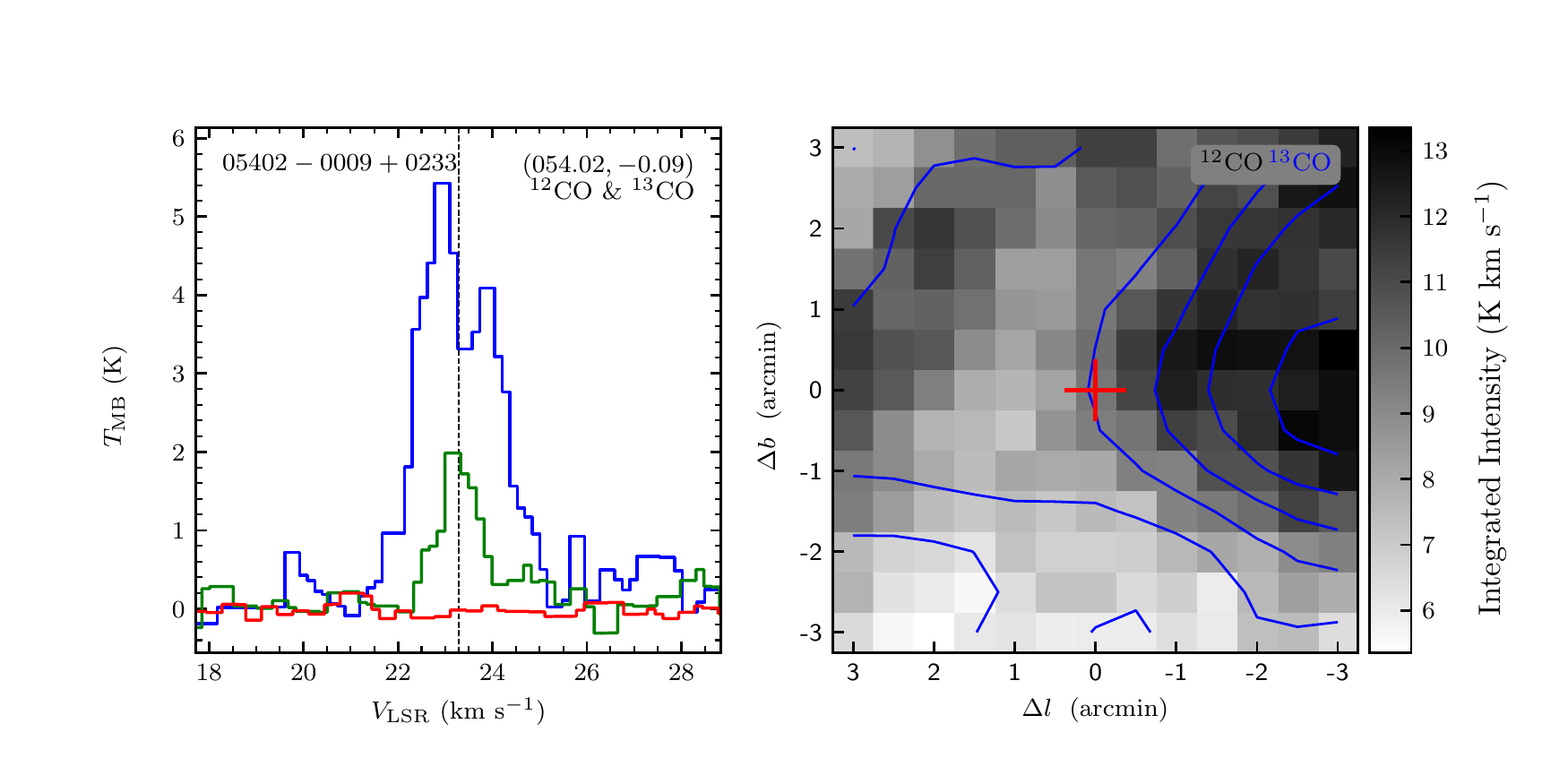}
\vspace{-0.5cm}

\includegraphics[width=9.0cm,angle=0]{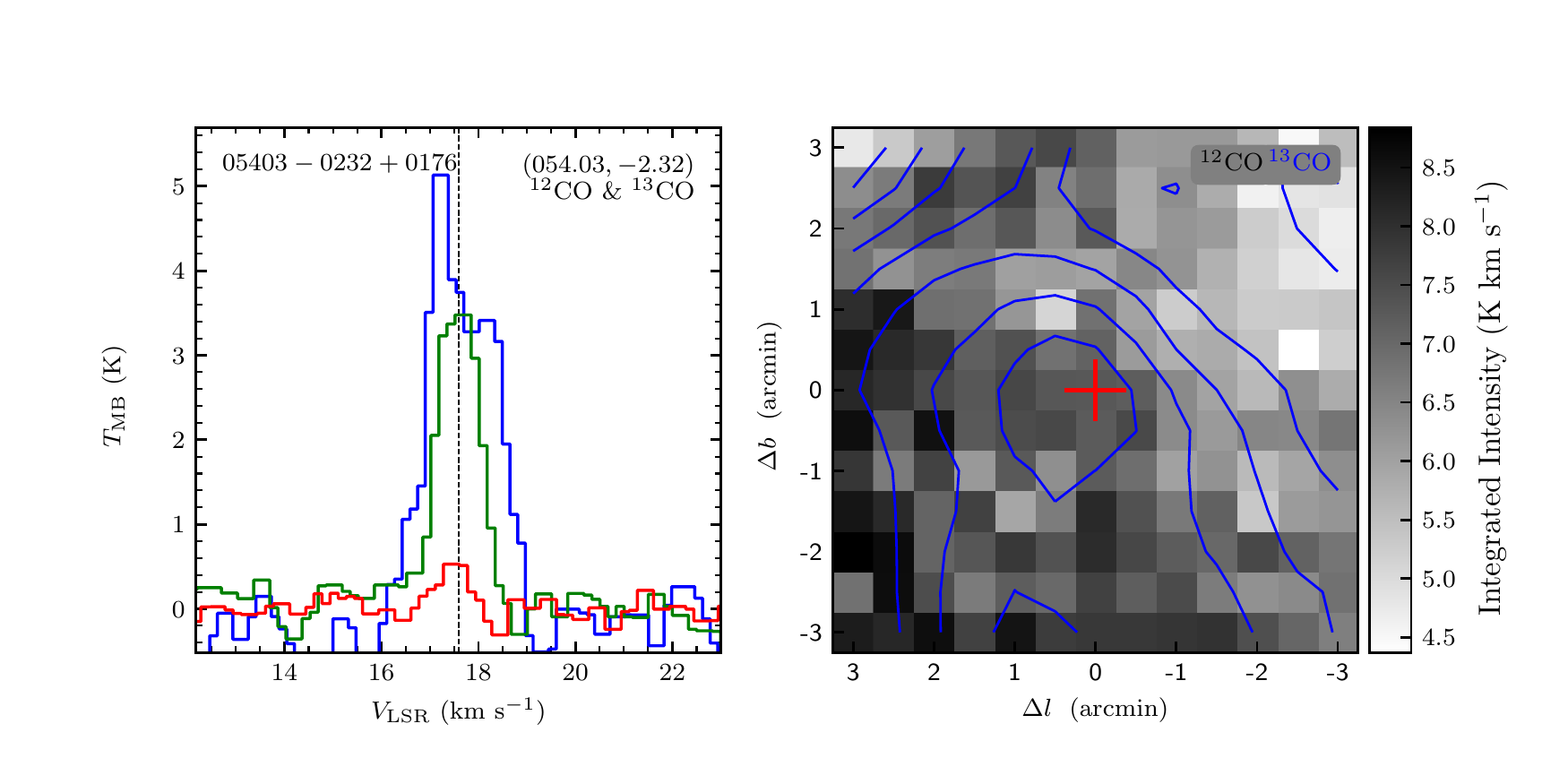}
\includegraphics[width=9.0cm,angle=0]{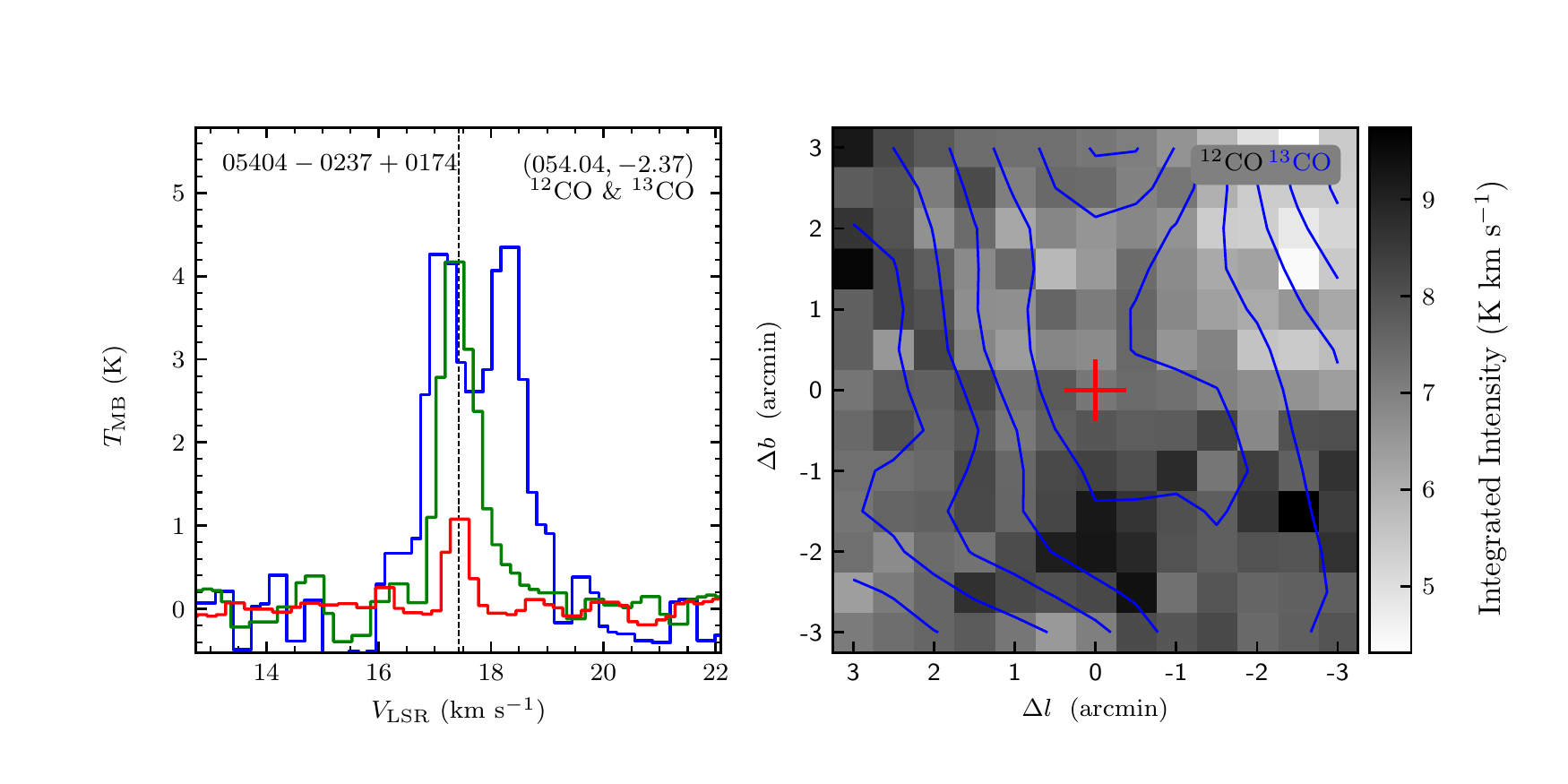}
\vspace{-0.5cm}

\includegraphics[width=9.0cm,angle=0]{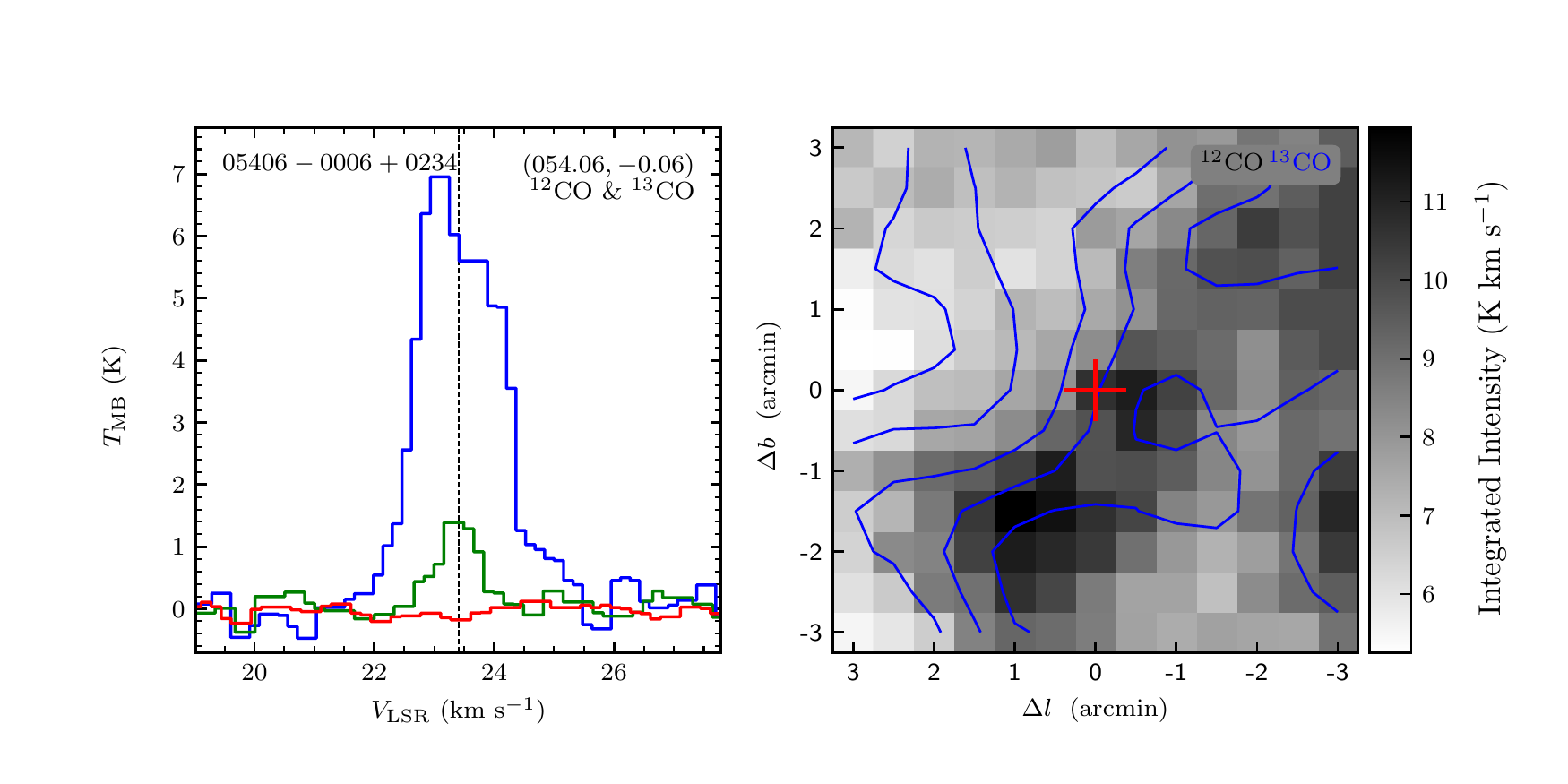}
\includegraphics[width=9.0cm,angle=0]{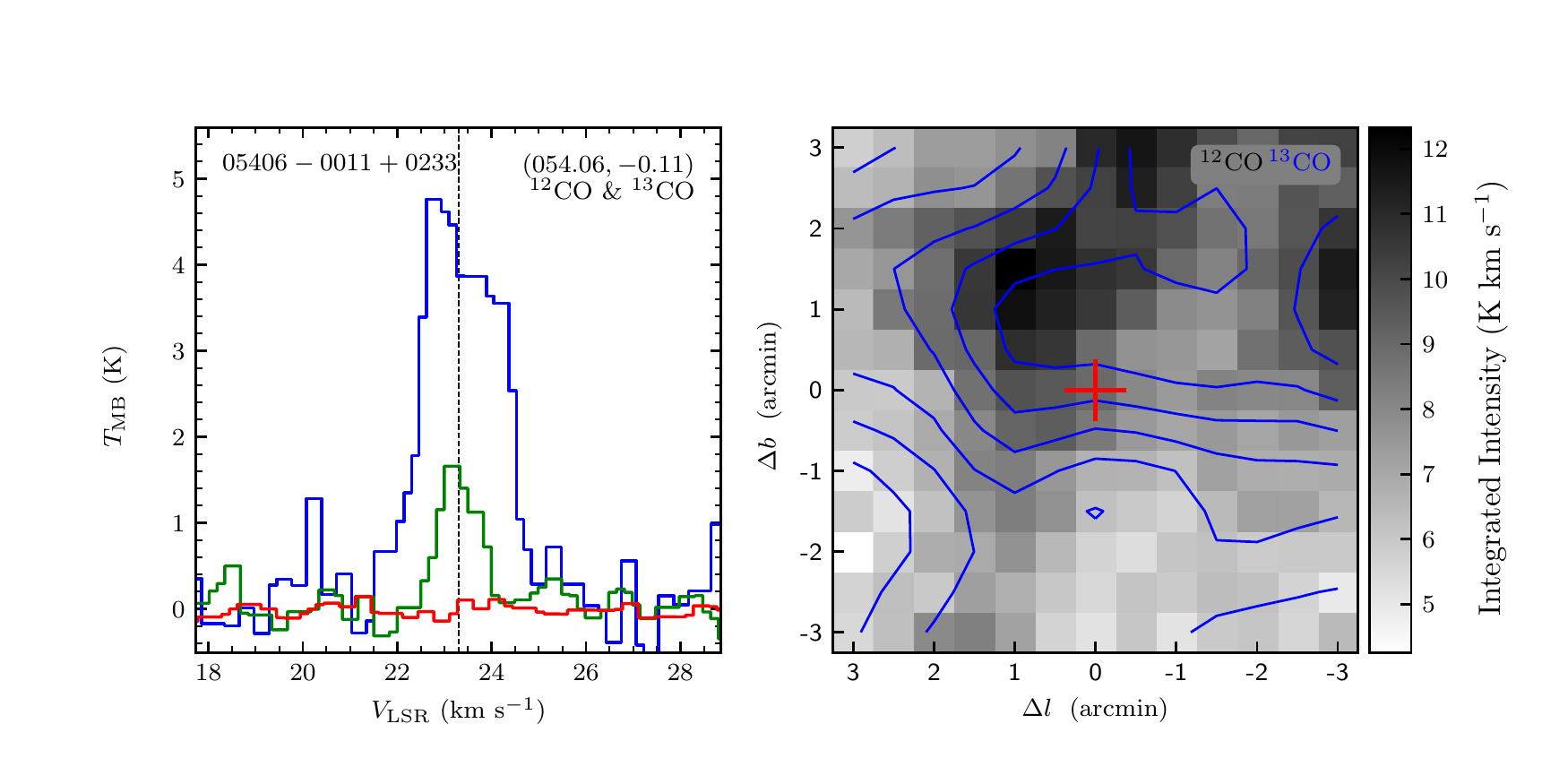}
\vspace{-0.5cm}

\includegraphics[width=9.0cm,angle=0]{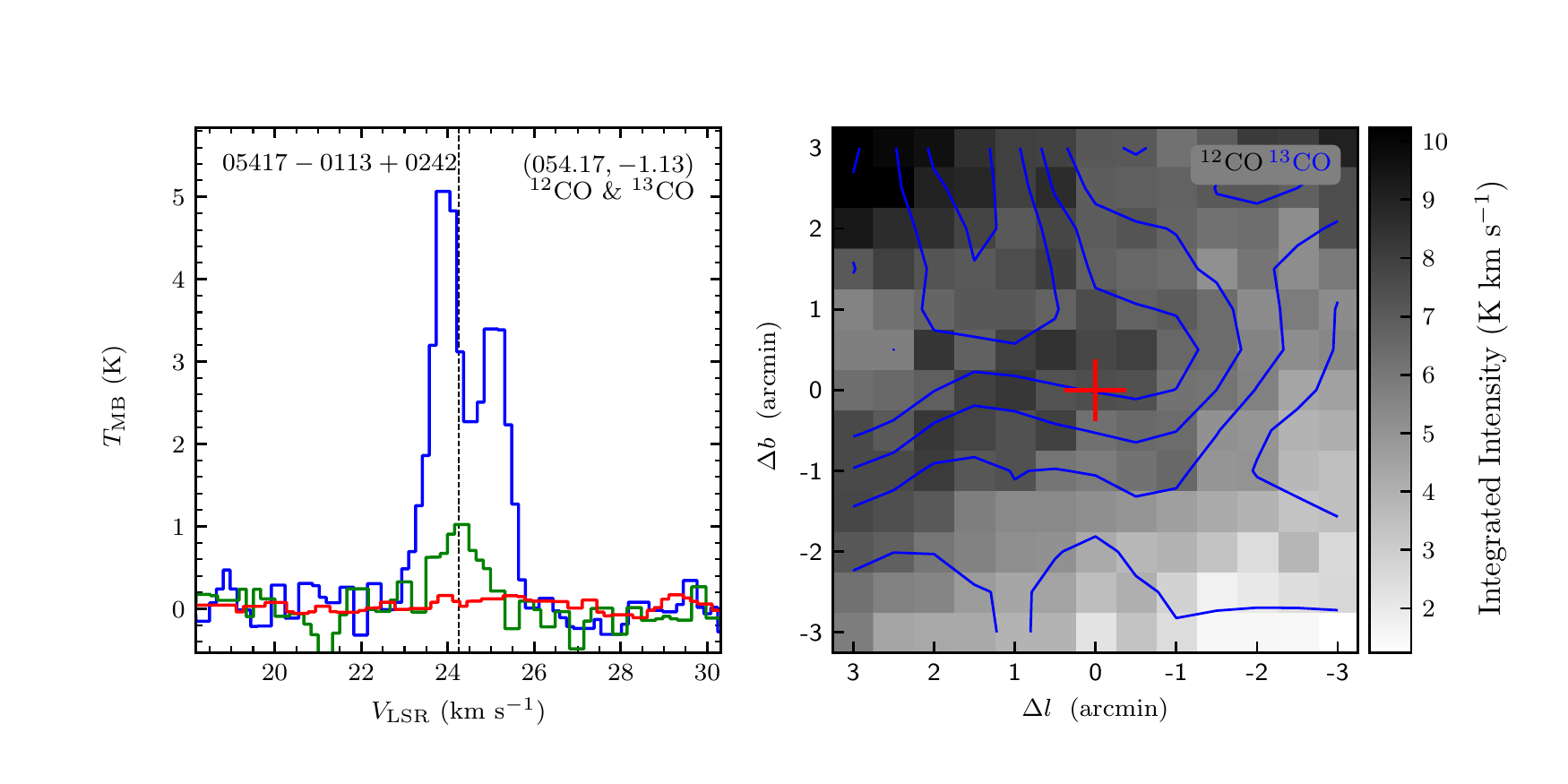}
\includegraphics[width=9.0cm,angle=0]{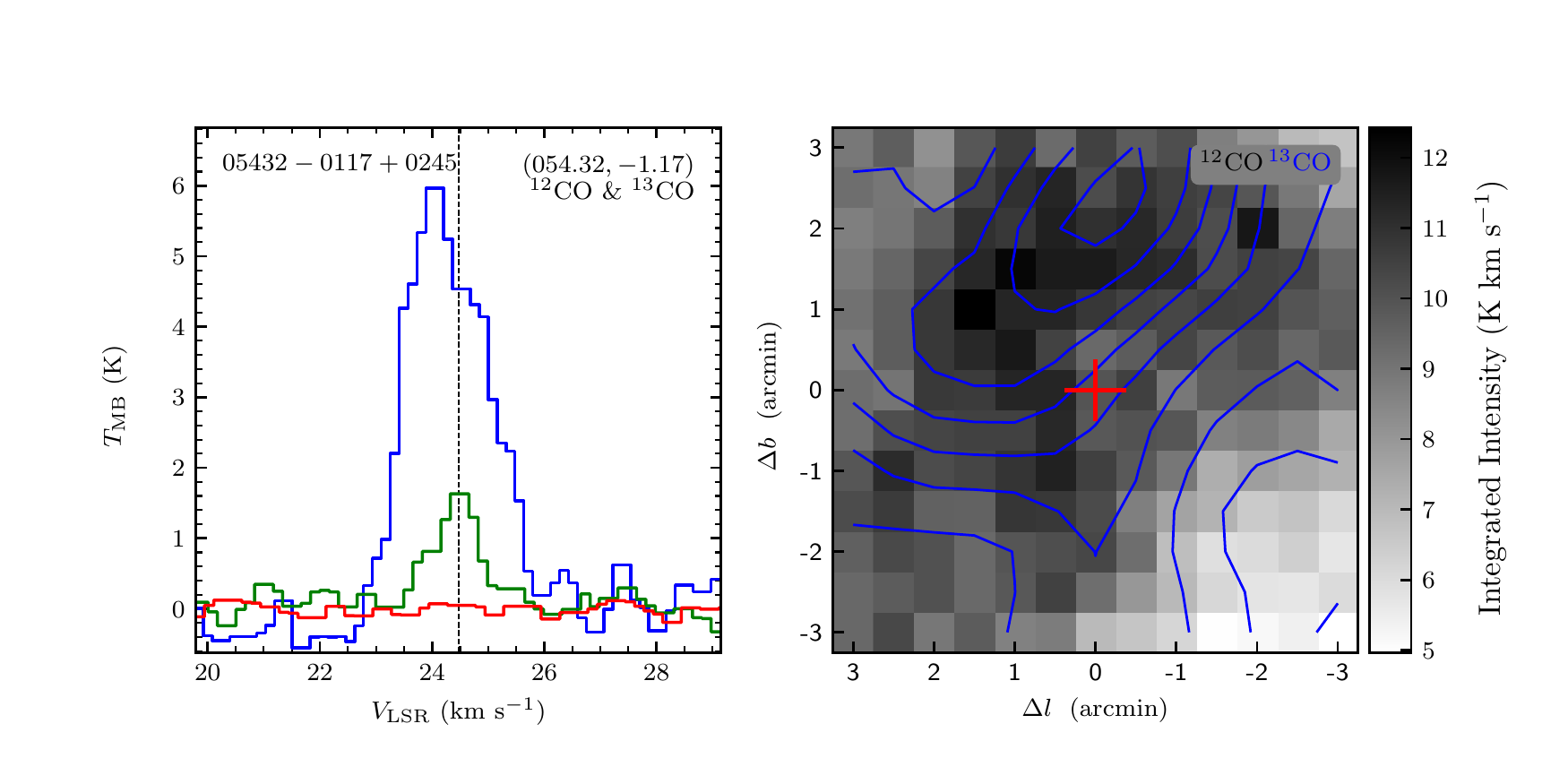}
\end{figure}
\clearpage

\begin{figure}
\includegraphics[width=9.0cm,angle=0]{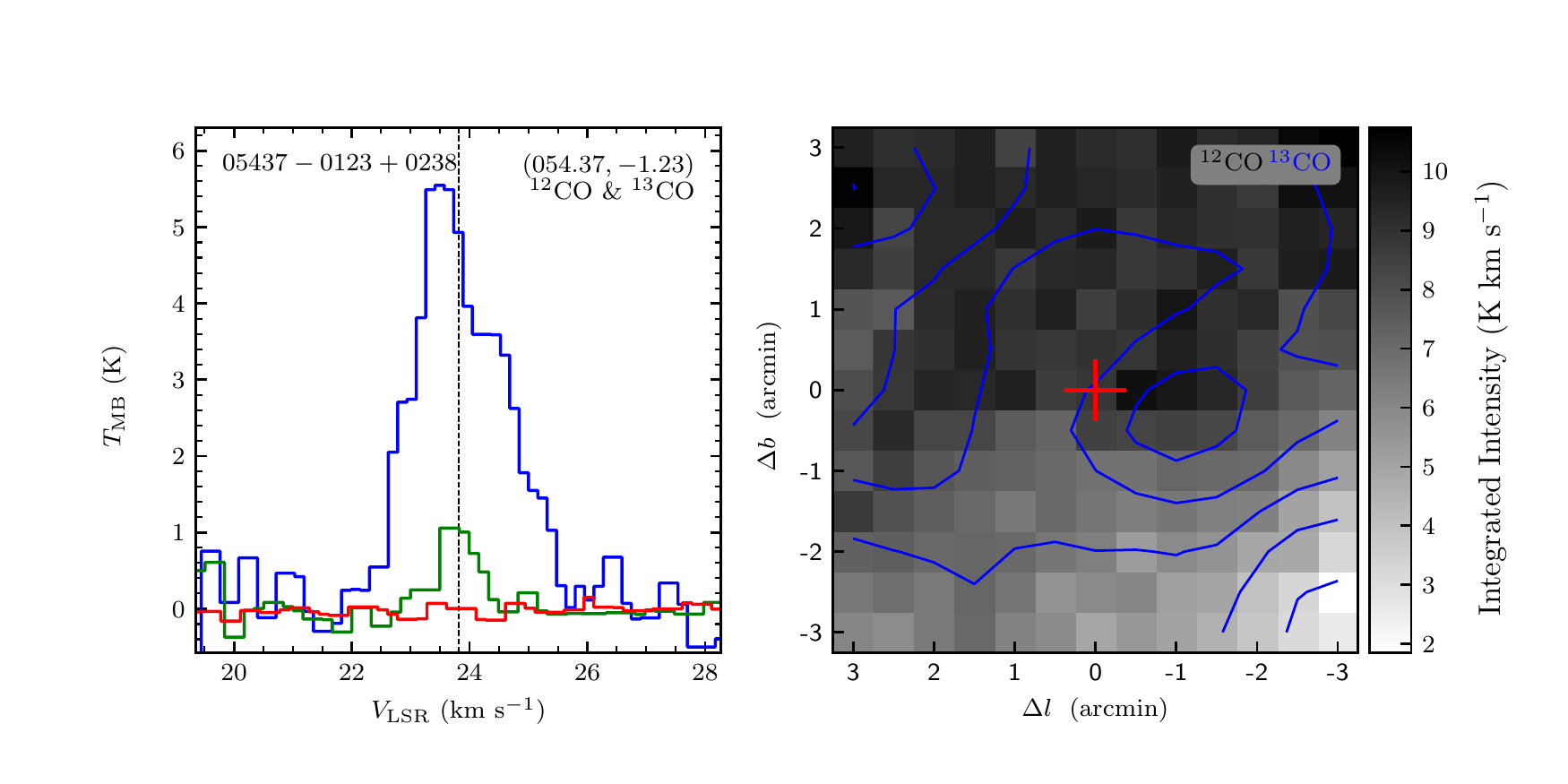}
\includegraphics[width=9.0cm,angle=0]{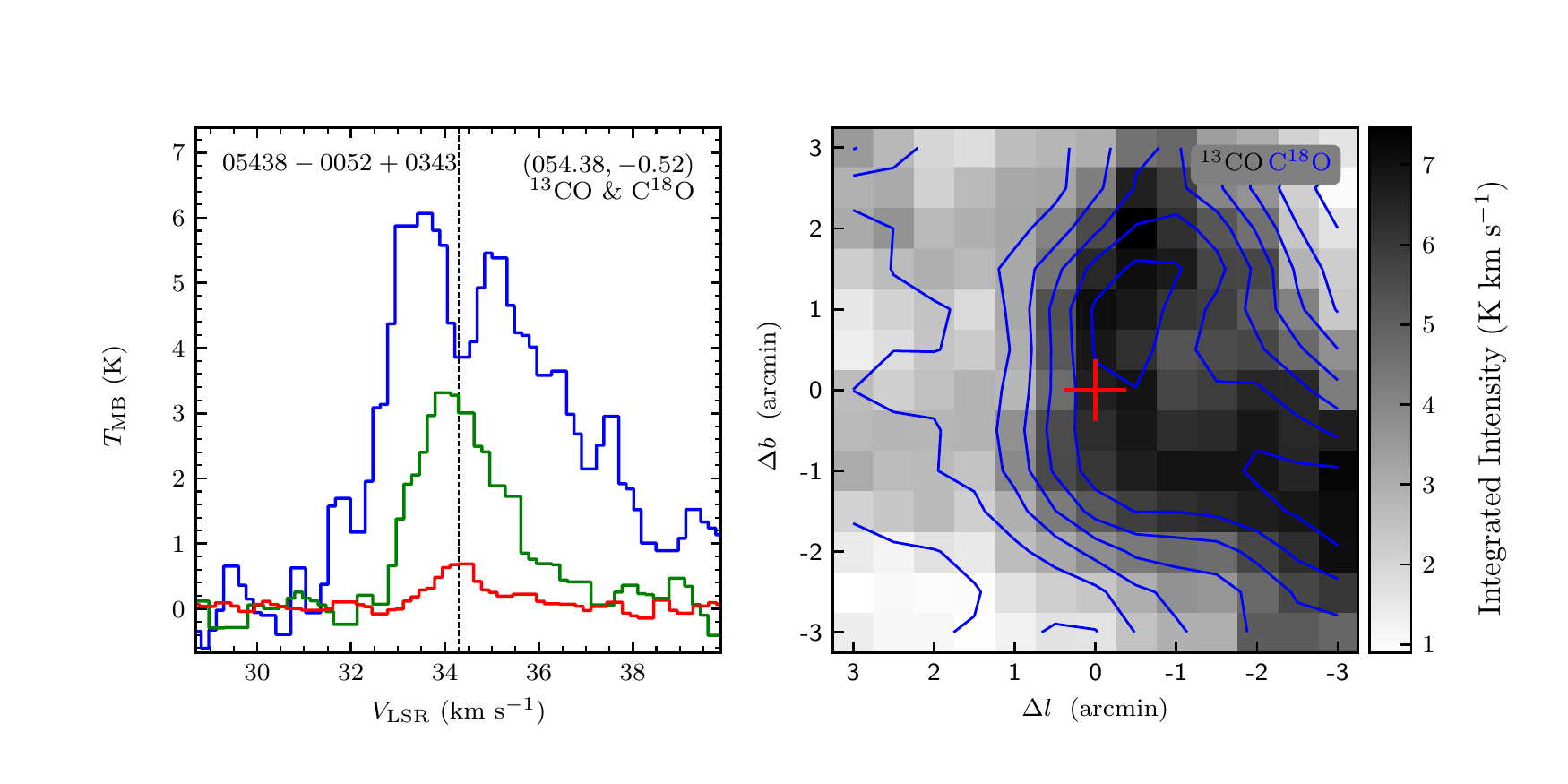}
\vspace{-0.5cm}

\includegraphics[width=9.0cm,angle=0]{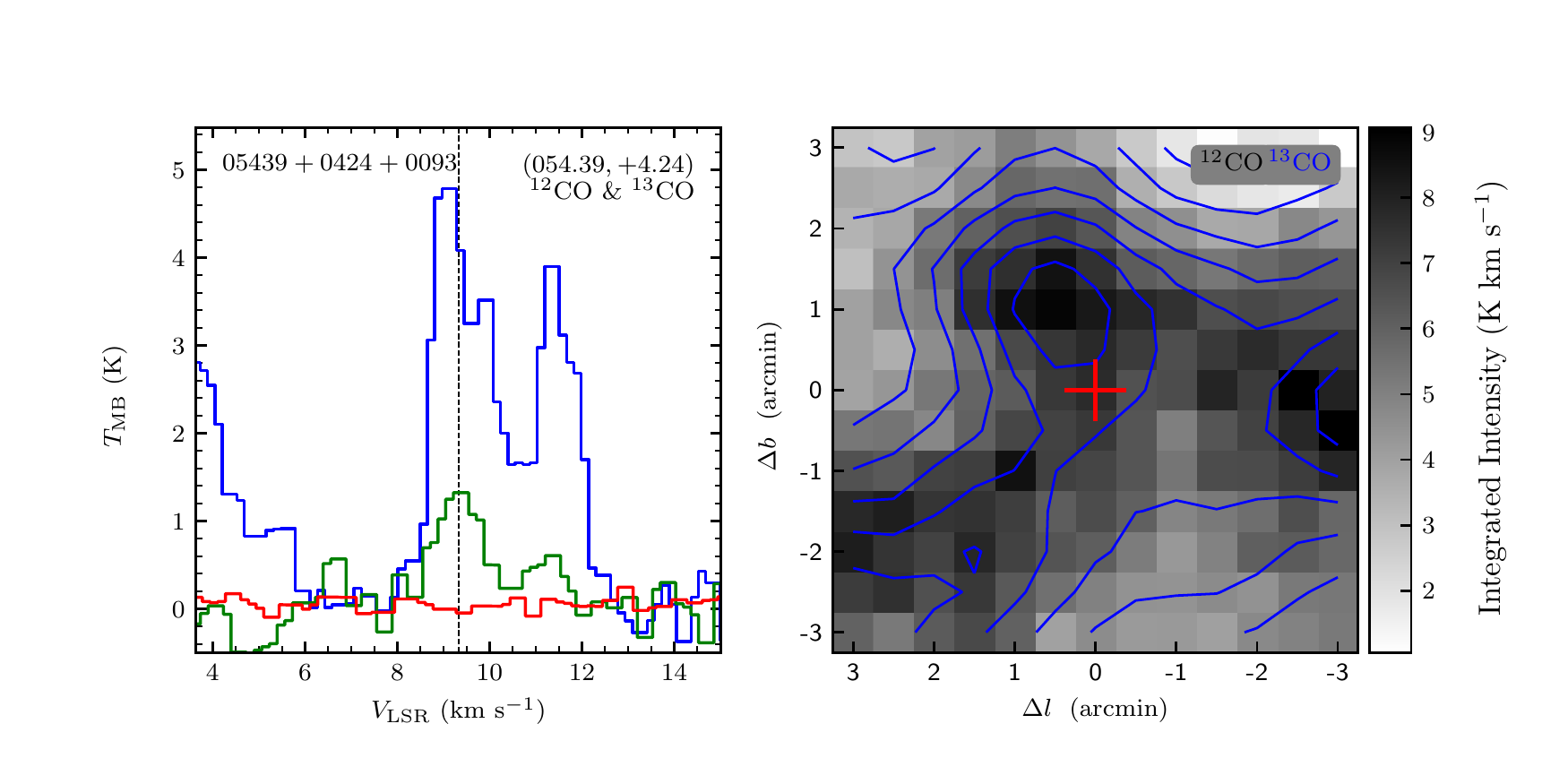}
\includegraphics[width=9.0cm,angle=0]{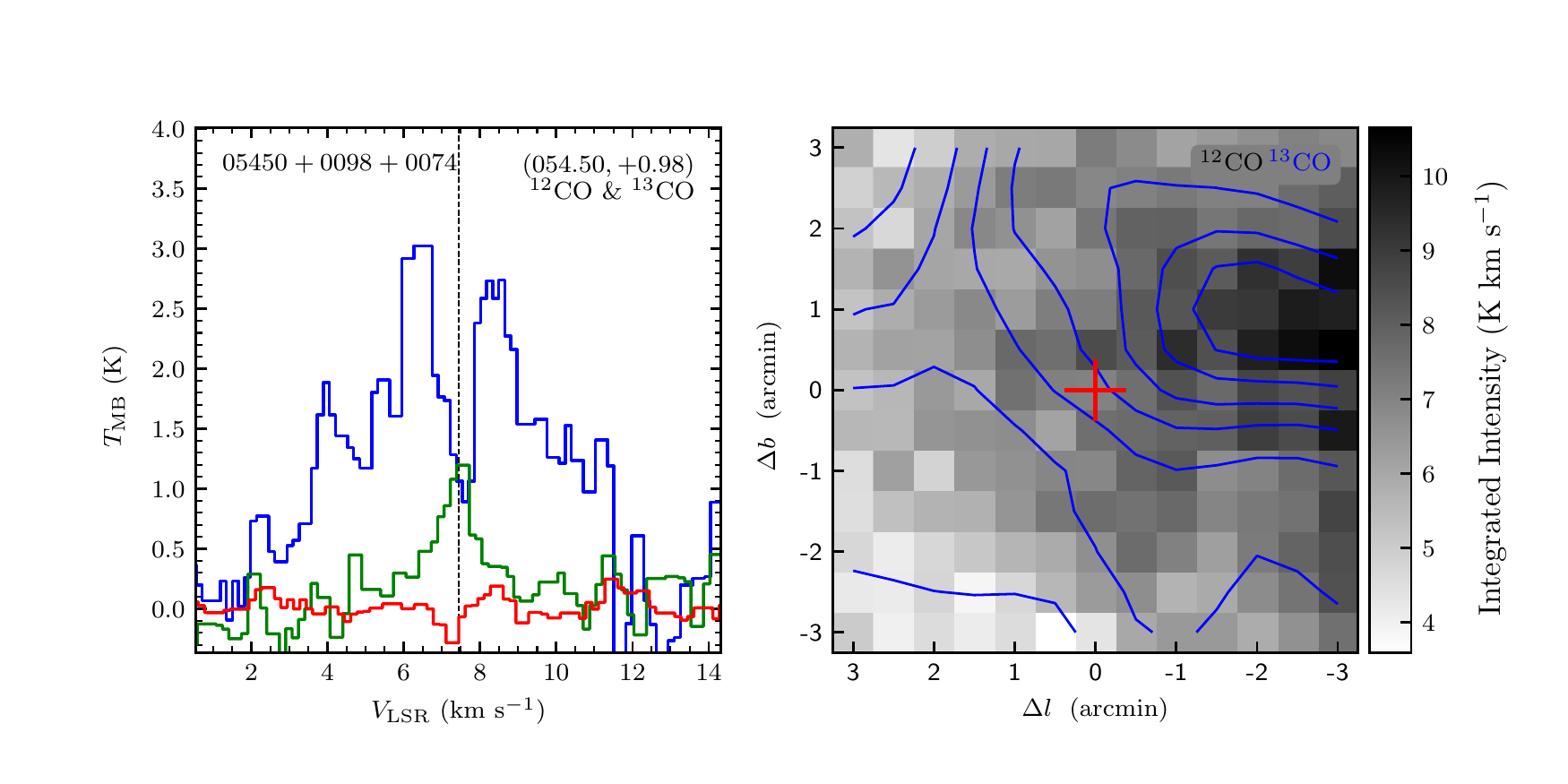}
\vspace{-0.5cm}

\includegraphics[width=9.0cm,angle=0]{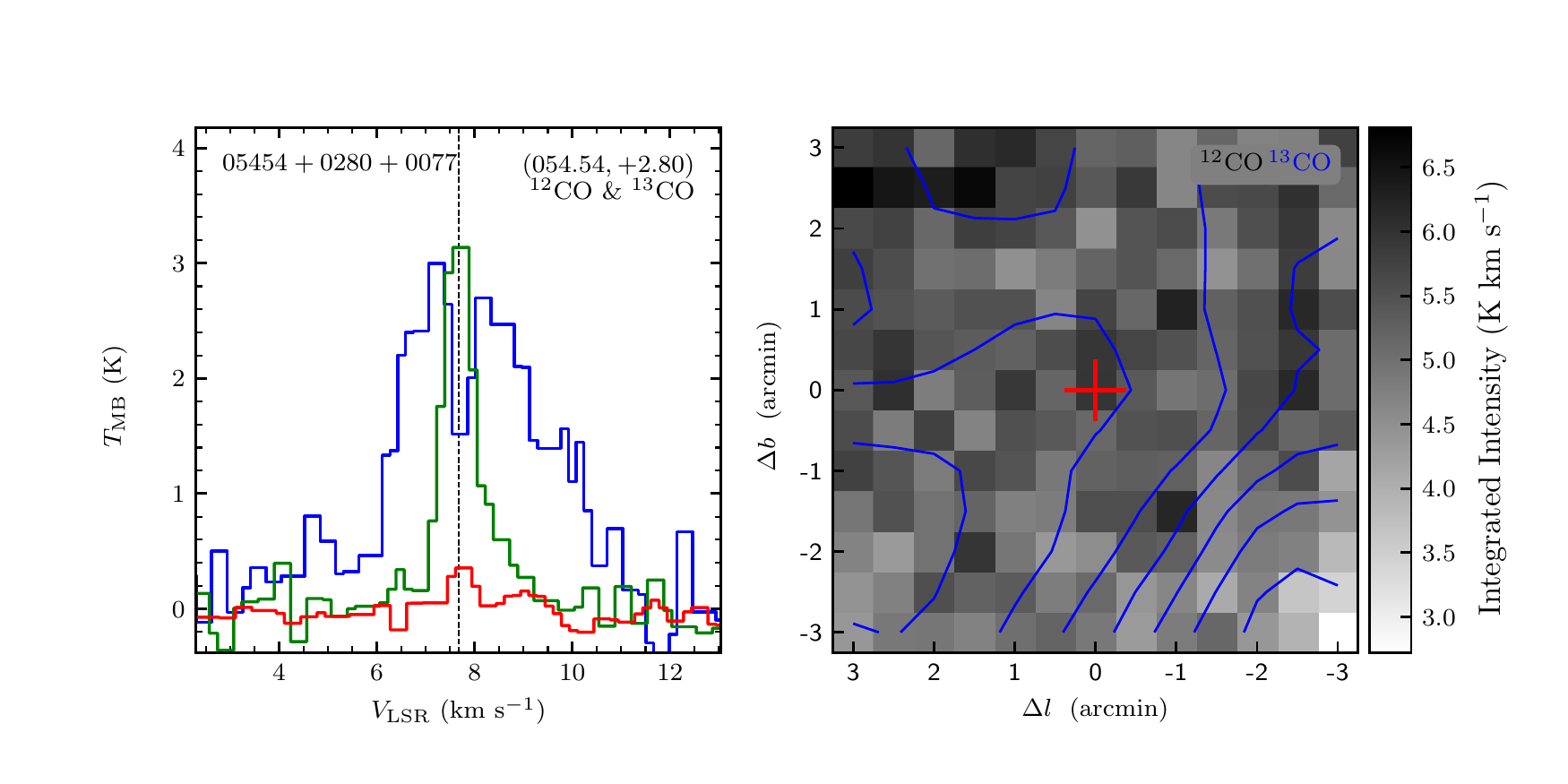}
\includegraphics[width=9.0cm,angle=0]{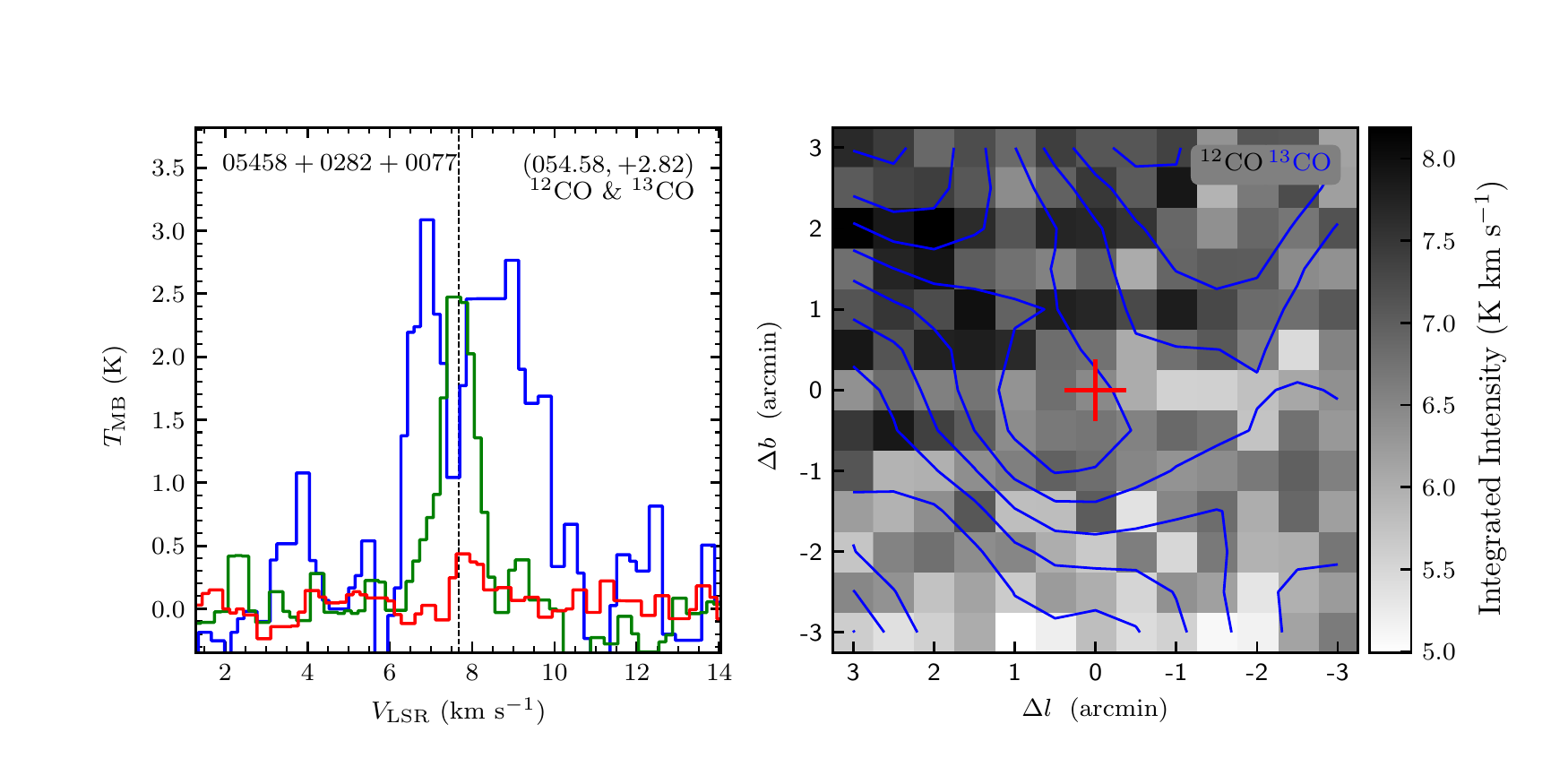}
\vspace{-0.5cm}

\includegraphics[width=9.0cm,angle=0]{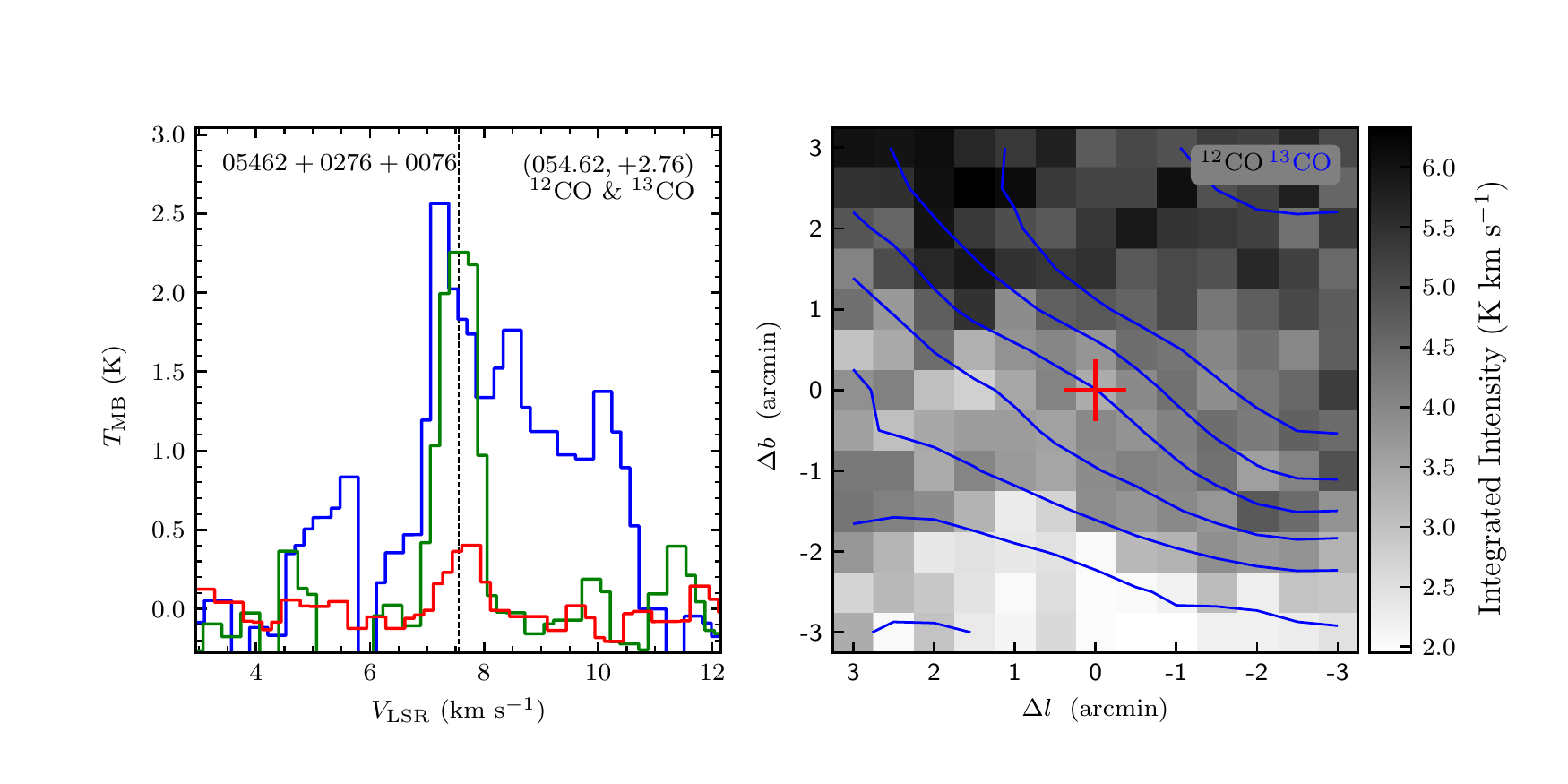}
\includegraphics[width=9.0cm,angle=0]{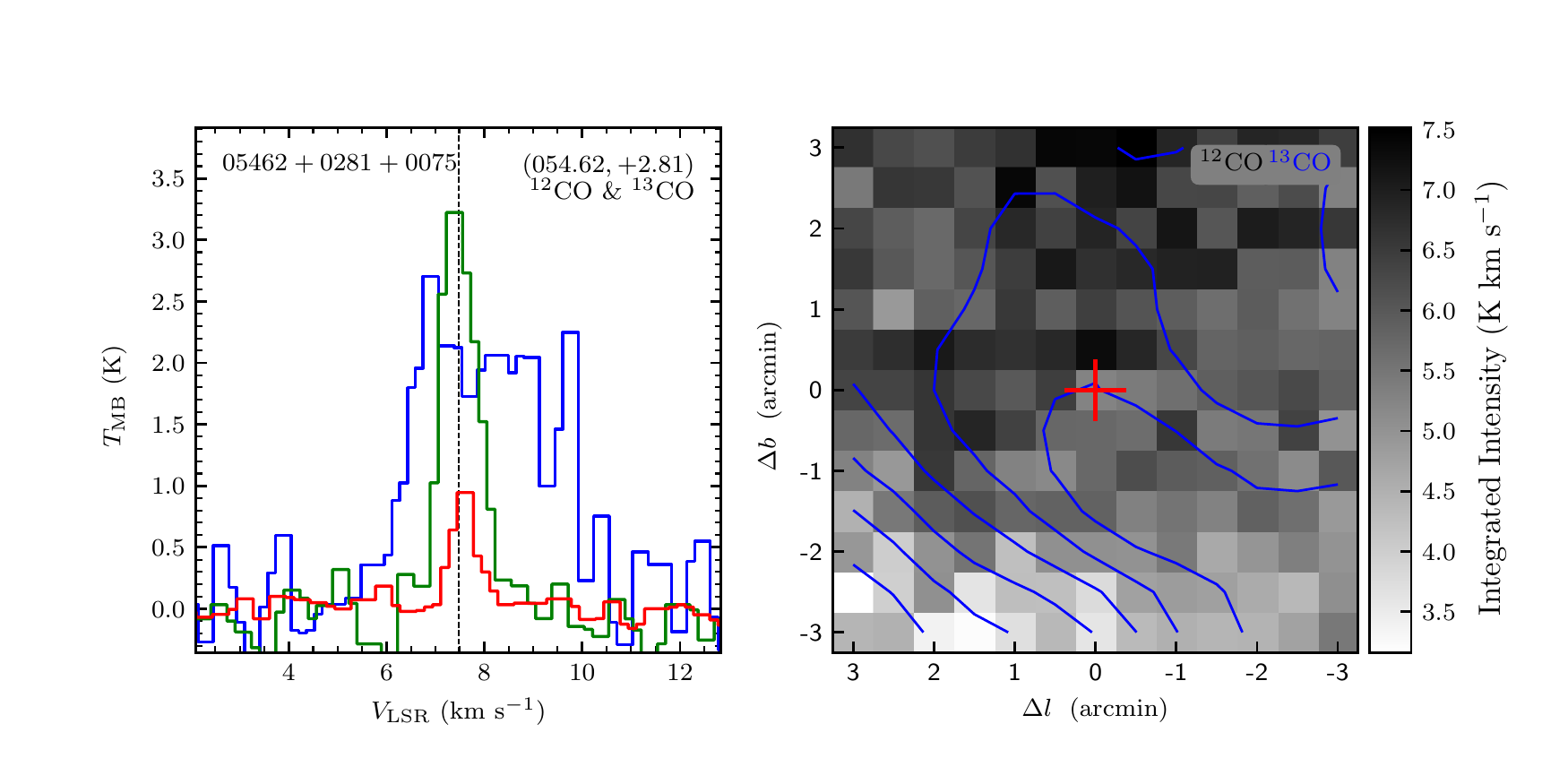}
\vspace{-0.5cm}

\includegraphics[width=9.0cm,angle=0]{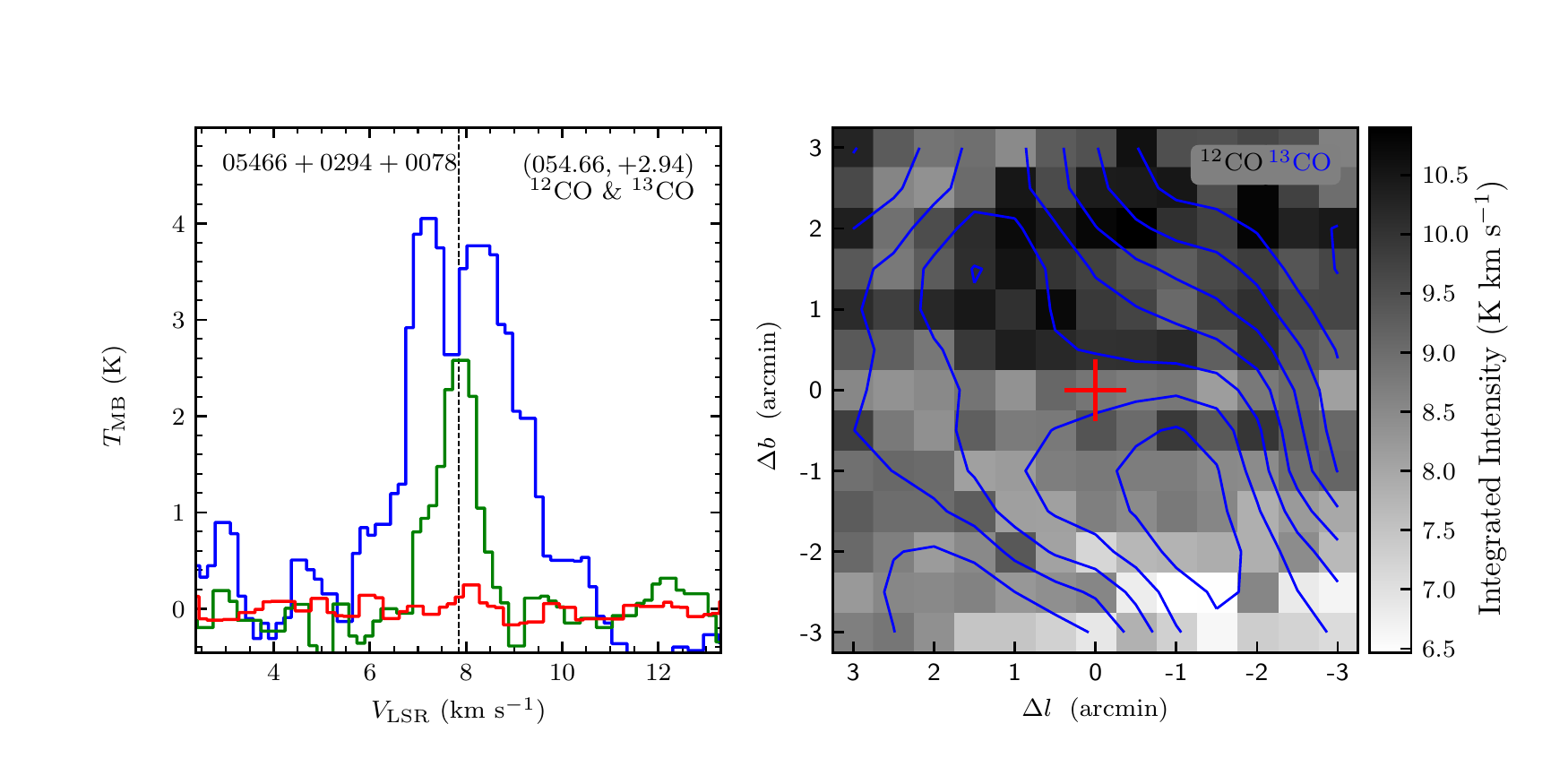}
\includegraphics[width=9.0cm,angle=0]{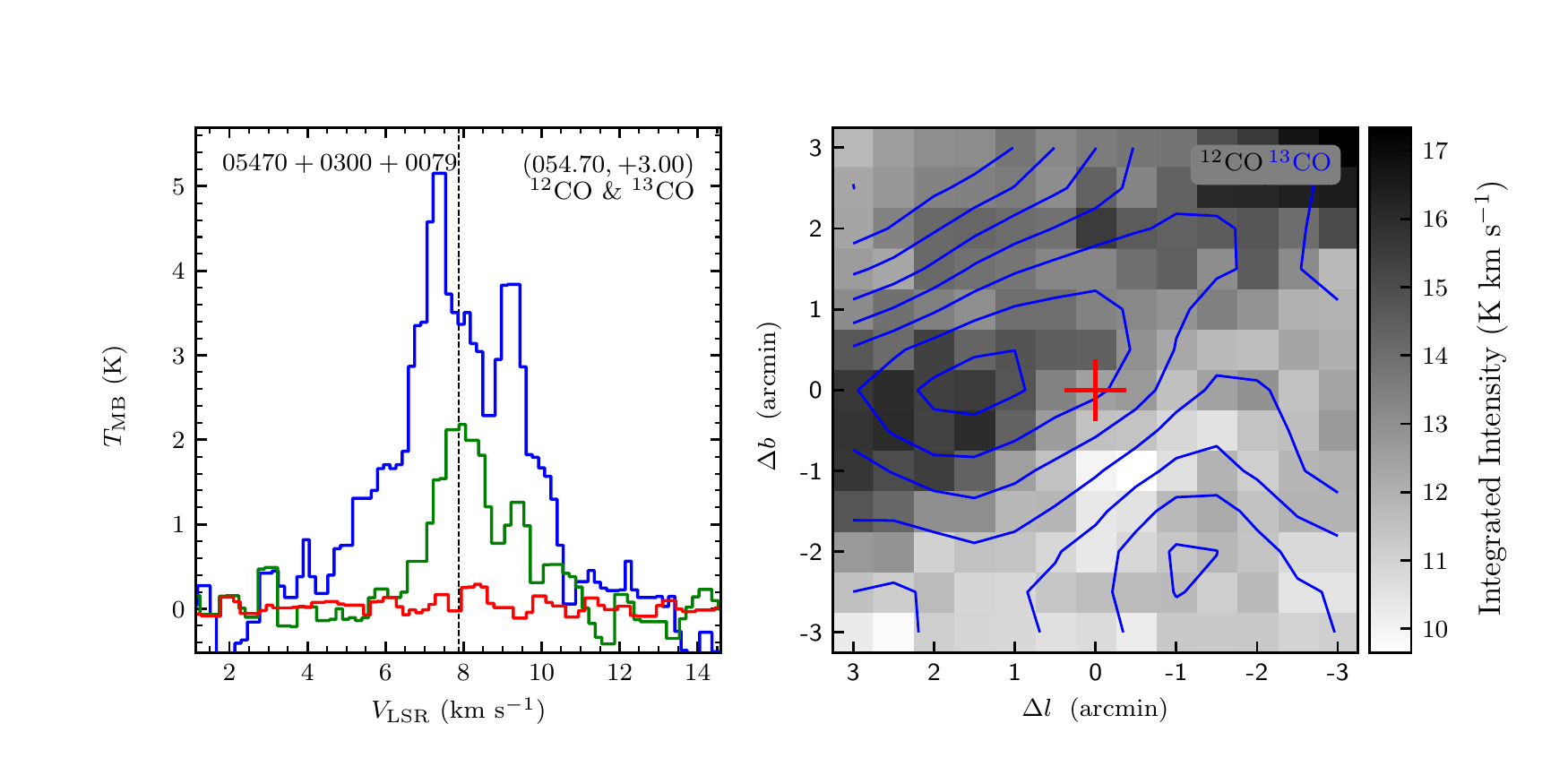}
\end{figure}
\clearpage

\begin{figure}
\includegraphics[width=9.0cm,angle=0]{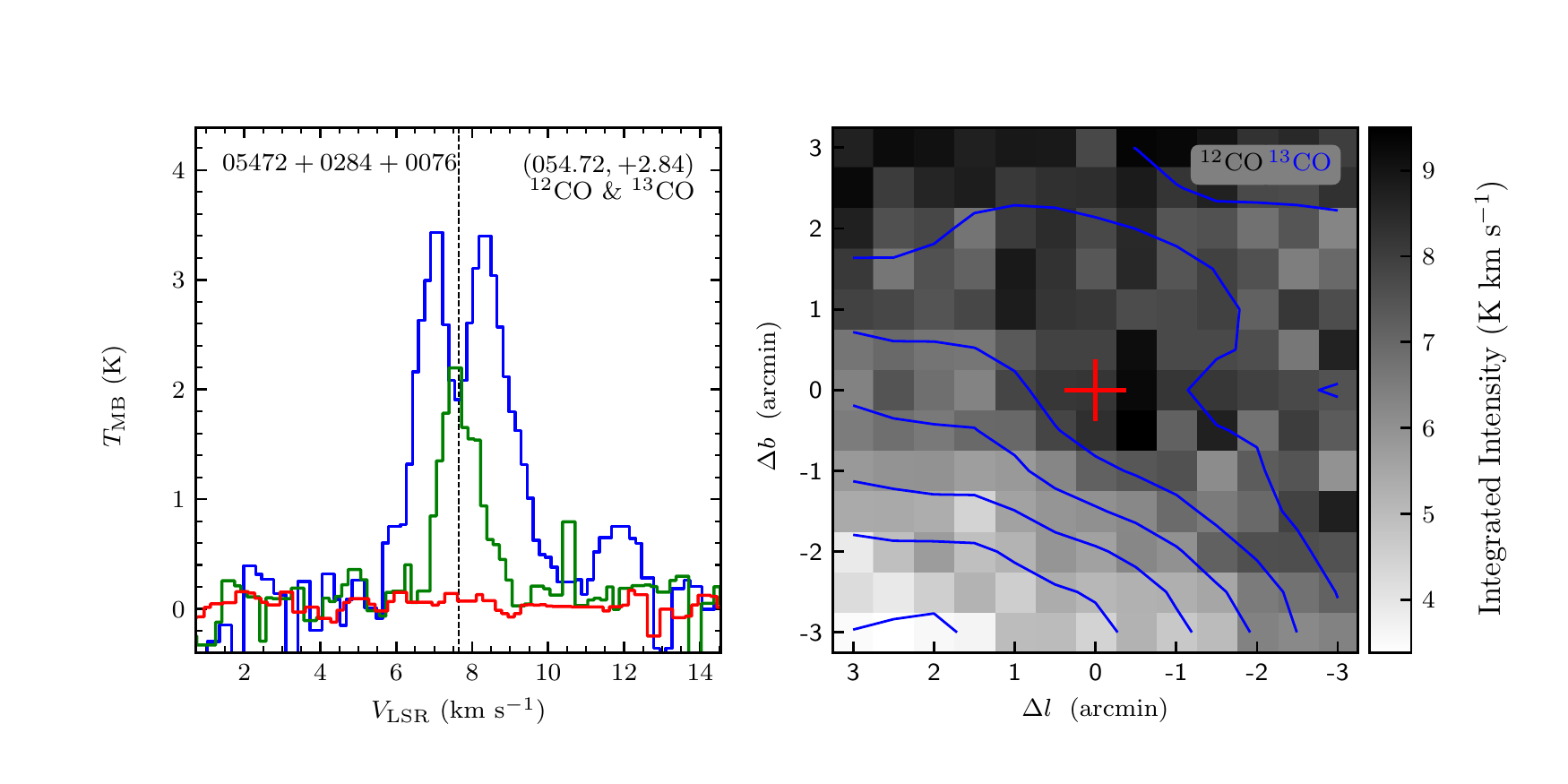}
\includegraphics[width=9.0cm,angle=0]{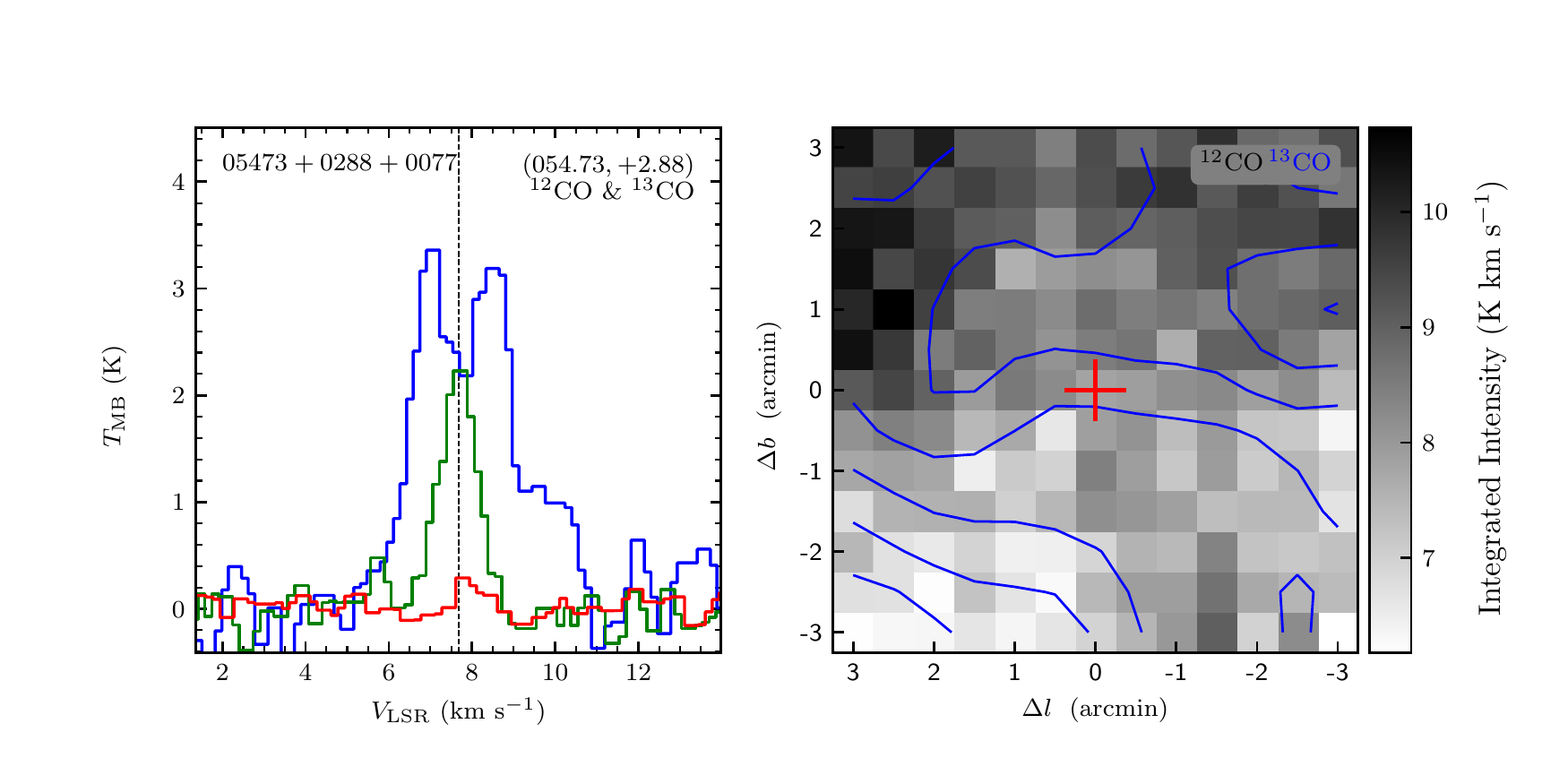}
\vspace{-0.5cm}

\includegraphics[width=9.0cm,angle=0]{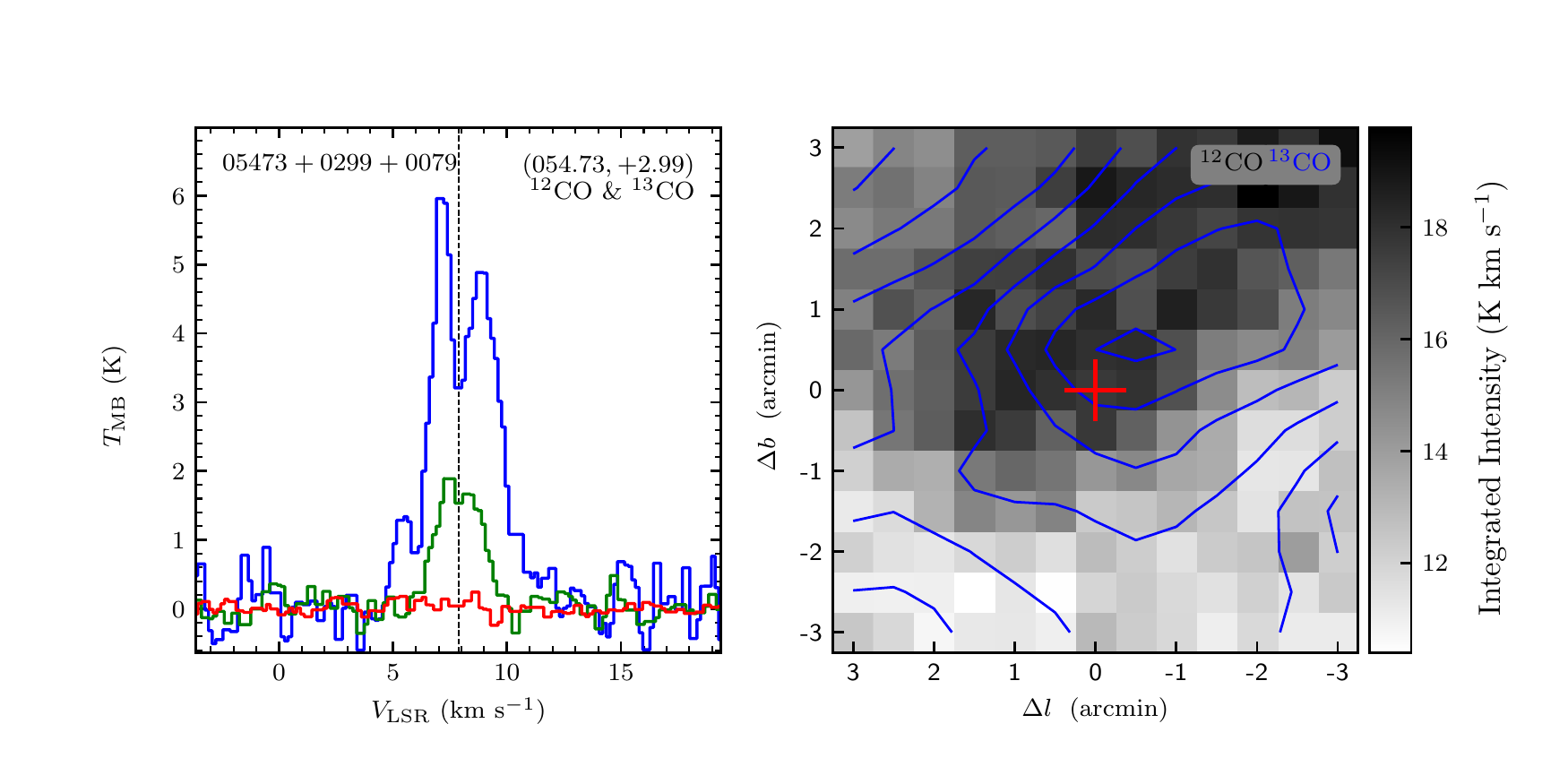}
\includegraphics[width=9.0cm,angle=0]{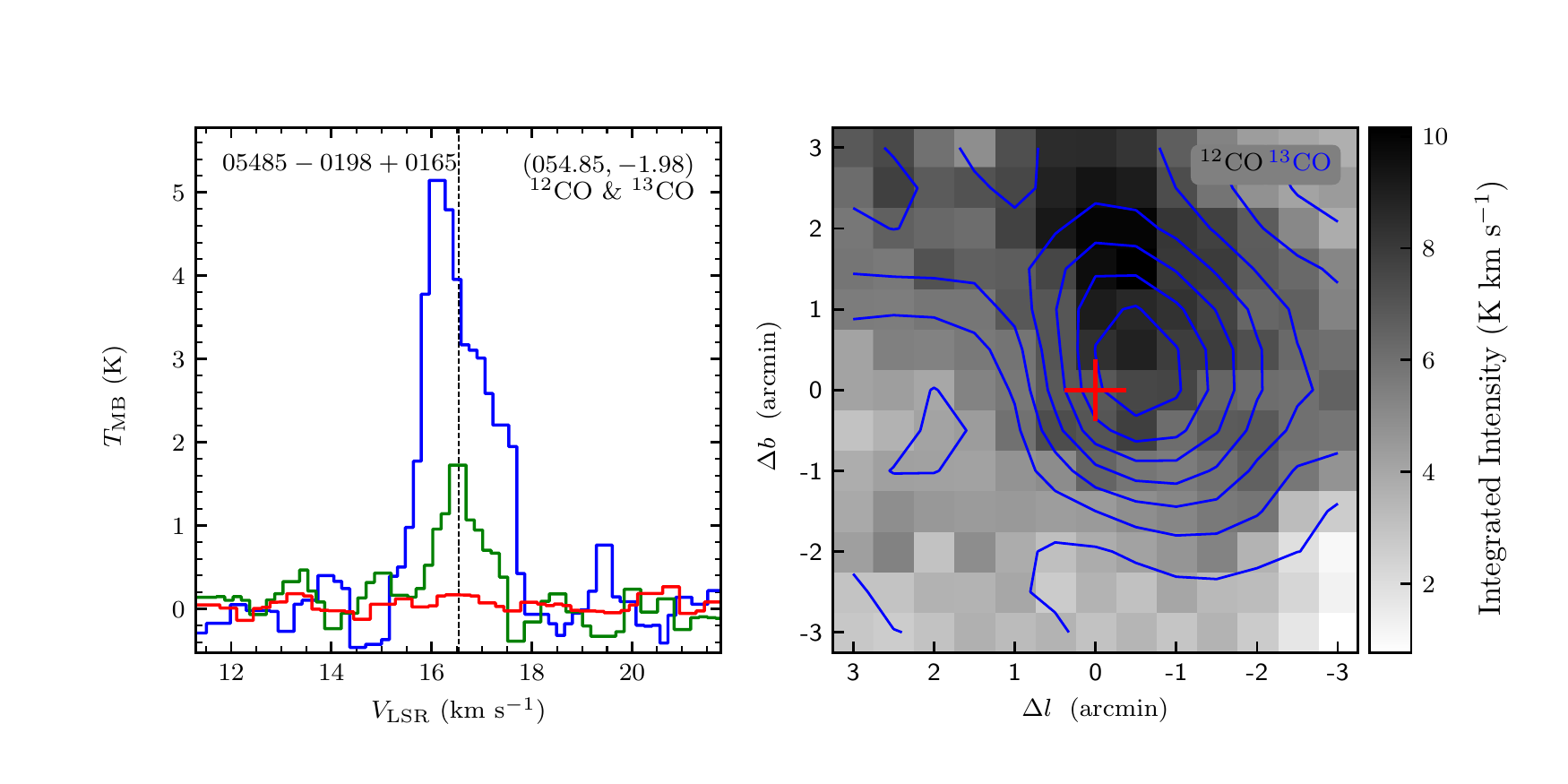}
\vspace{-0.5cm}

\includegraphics[width=9.0cm,angle=0]{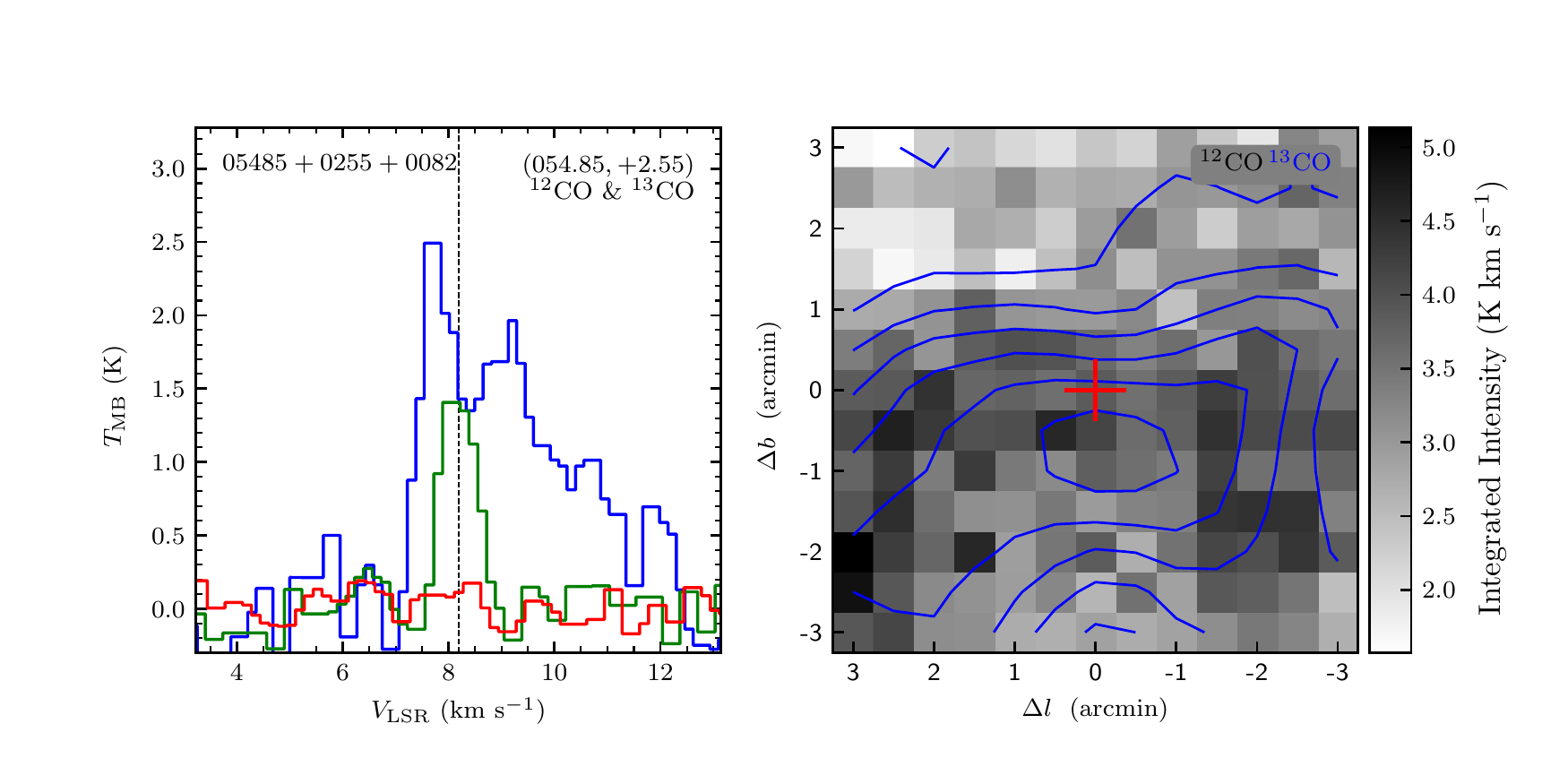}
\includegraphics[width=9.0cm,angle=0]{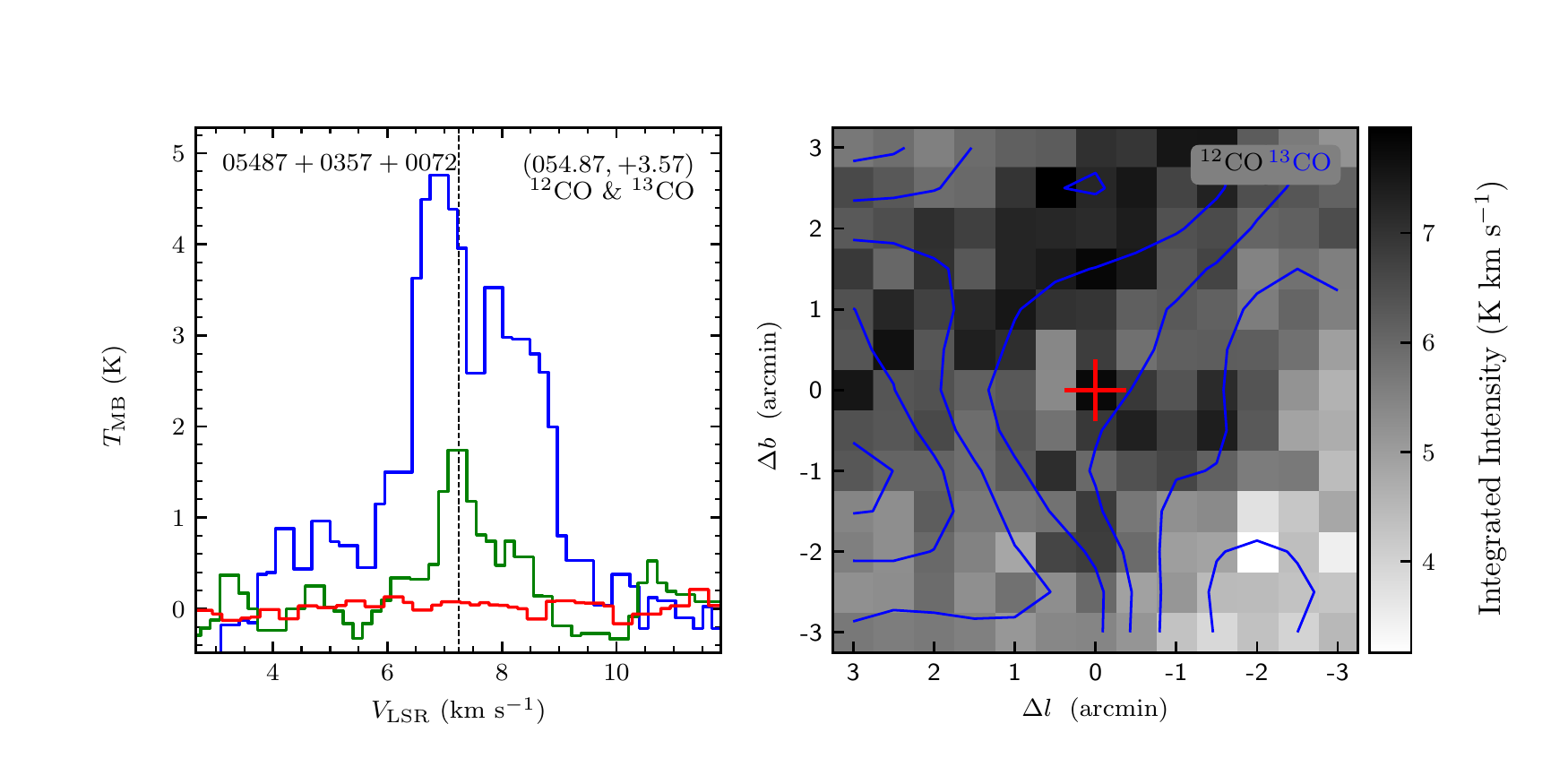}
\vspace{-0.5cm}

\includegraphics[width=9.0cm,angle=0]{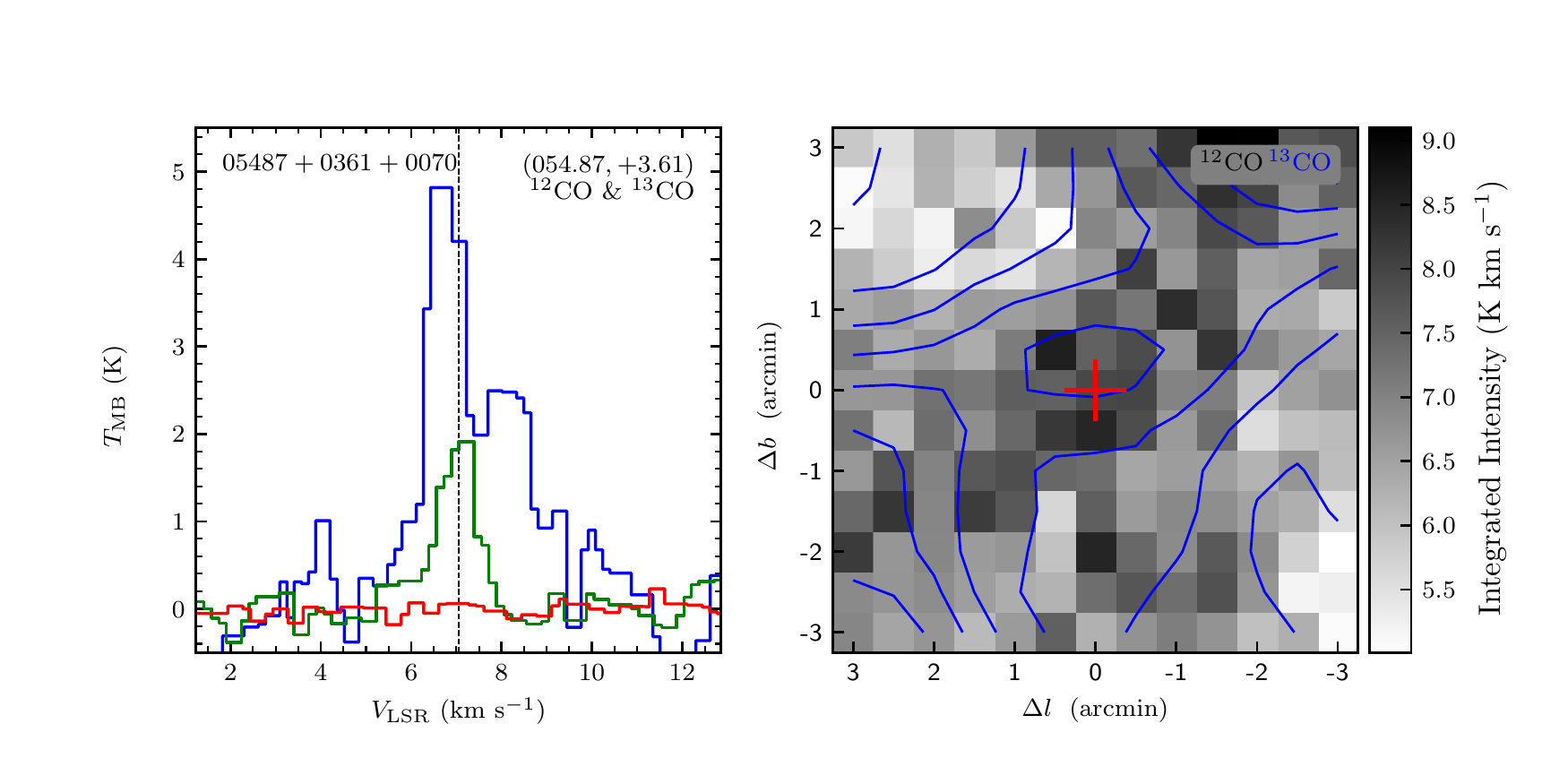}
\includegraphics[width=9.0cm,angle=0]{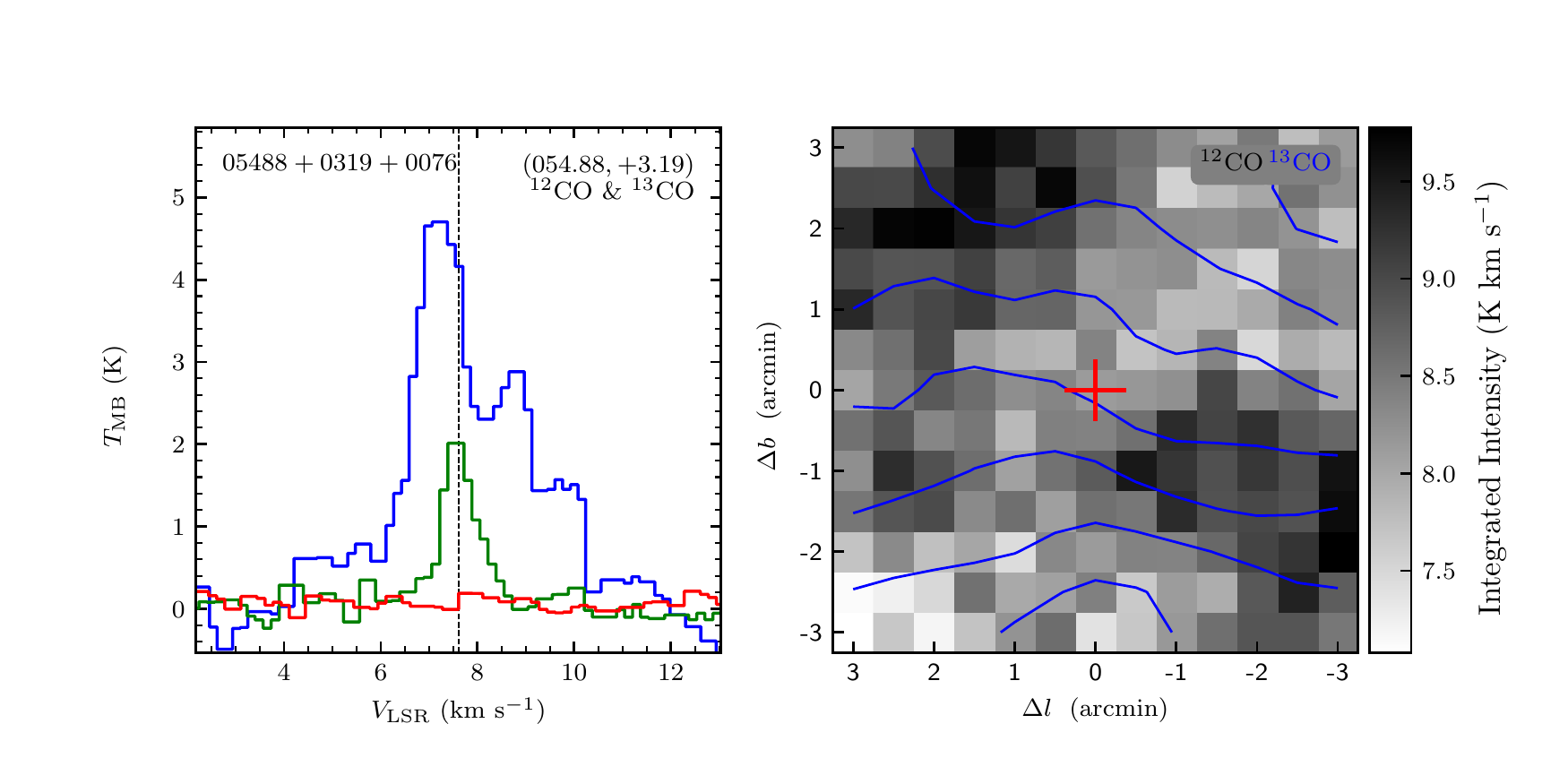}
\vspace{-0.5cm}

\includegraphics[width=9.0cm,angle=0]{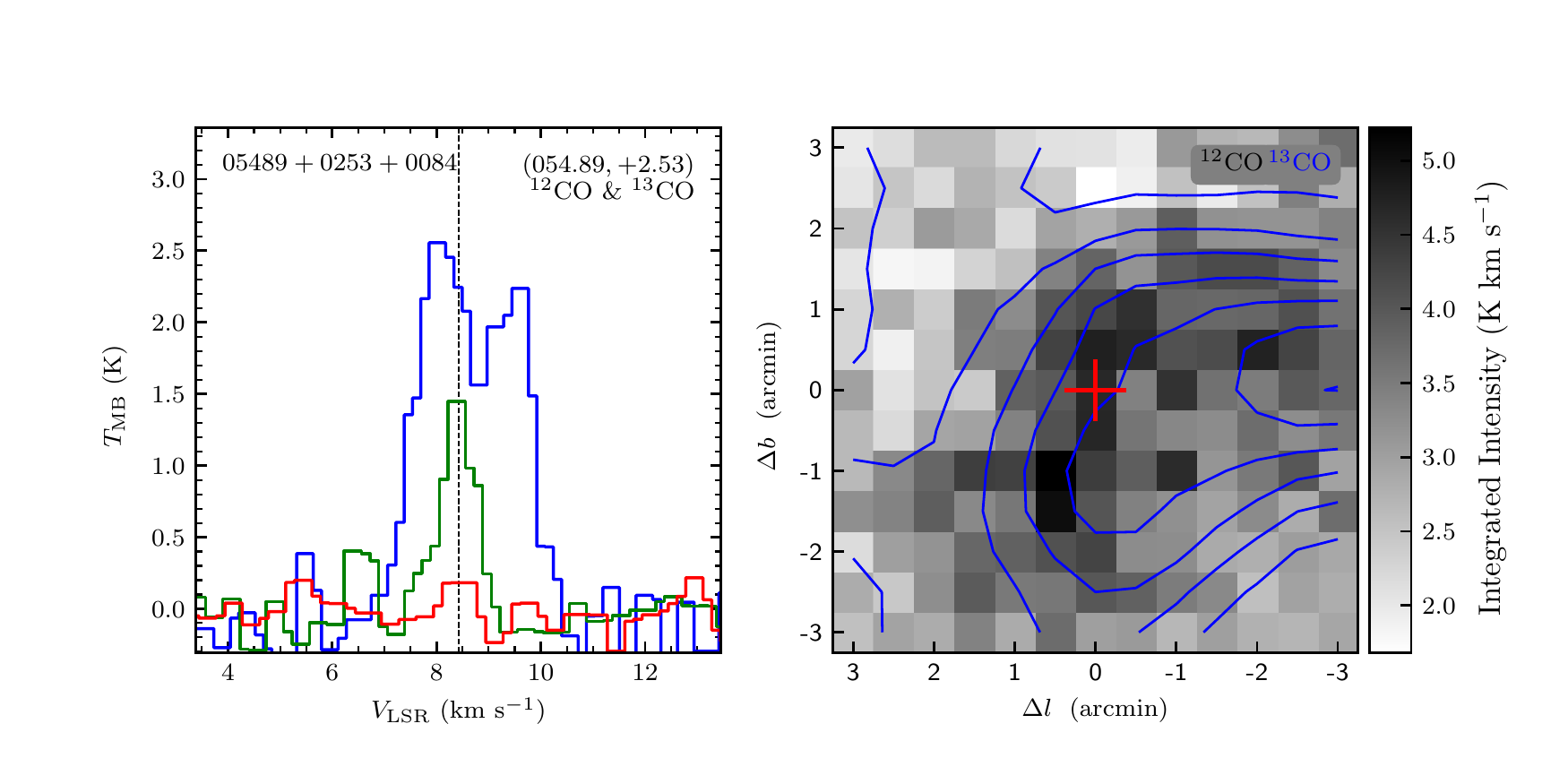}
\includegraphics[width=9.0cm,angle=0]{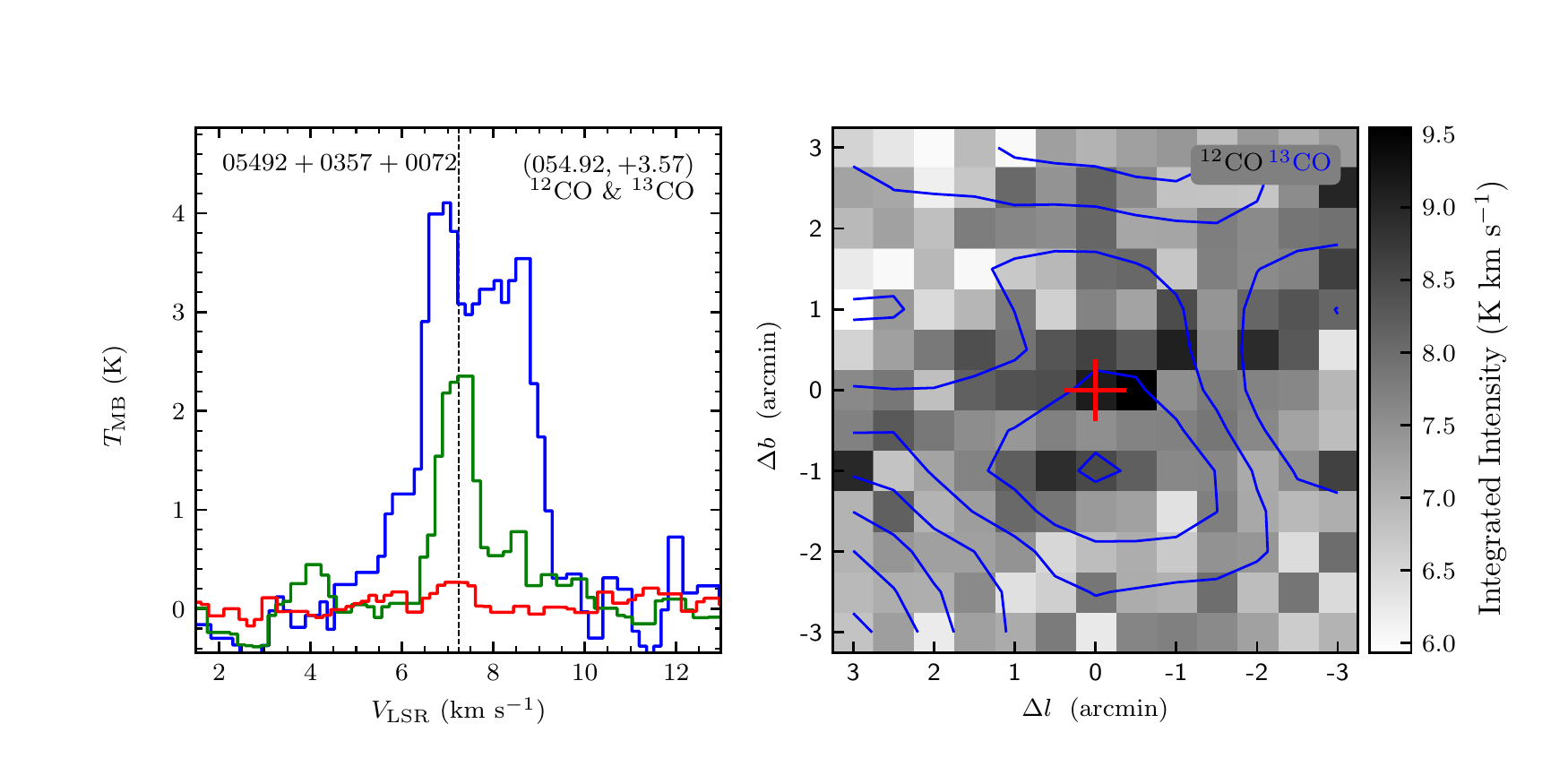}
\end{figure}
\clearpage

\begin{figure}
\includegraphics[width=9.0cm,angle=0]{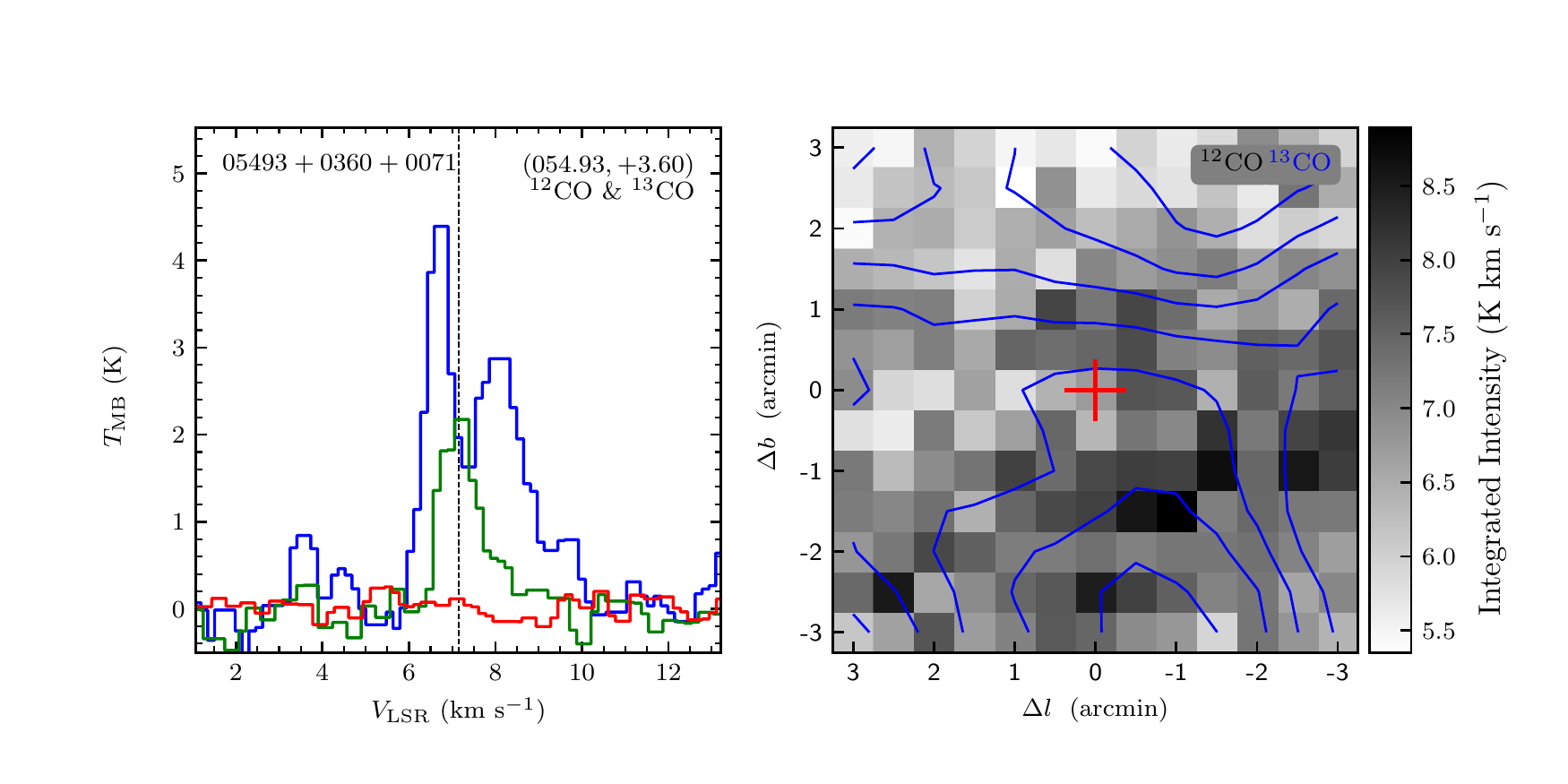}
\includegraphics[width=9.0cm,angle=0]{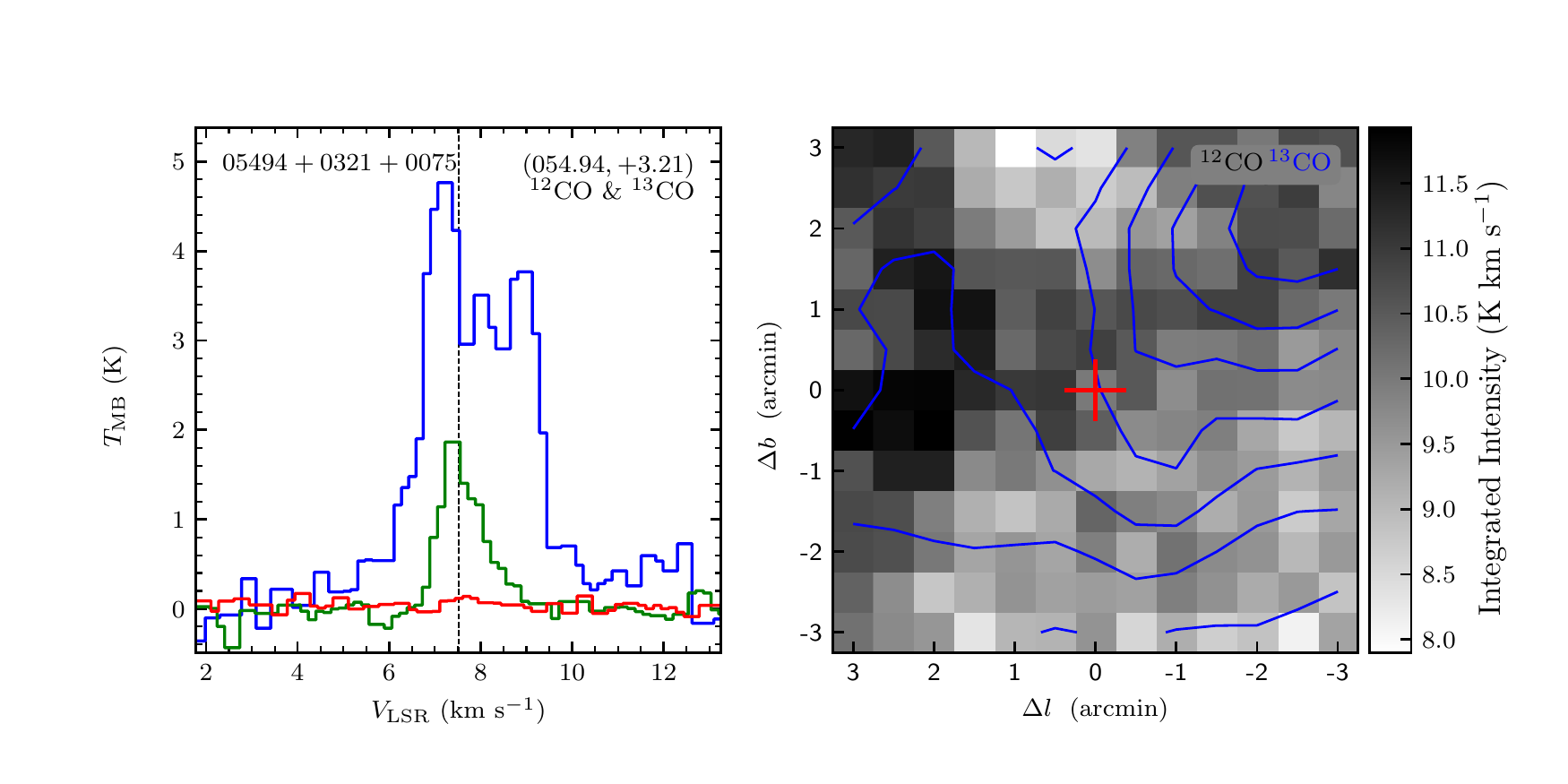}
\vspace{-0.5cm}

\includegraphics[width=9.0cm,angle=0]{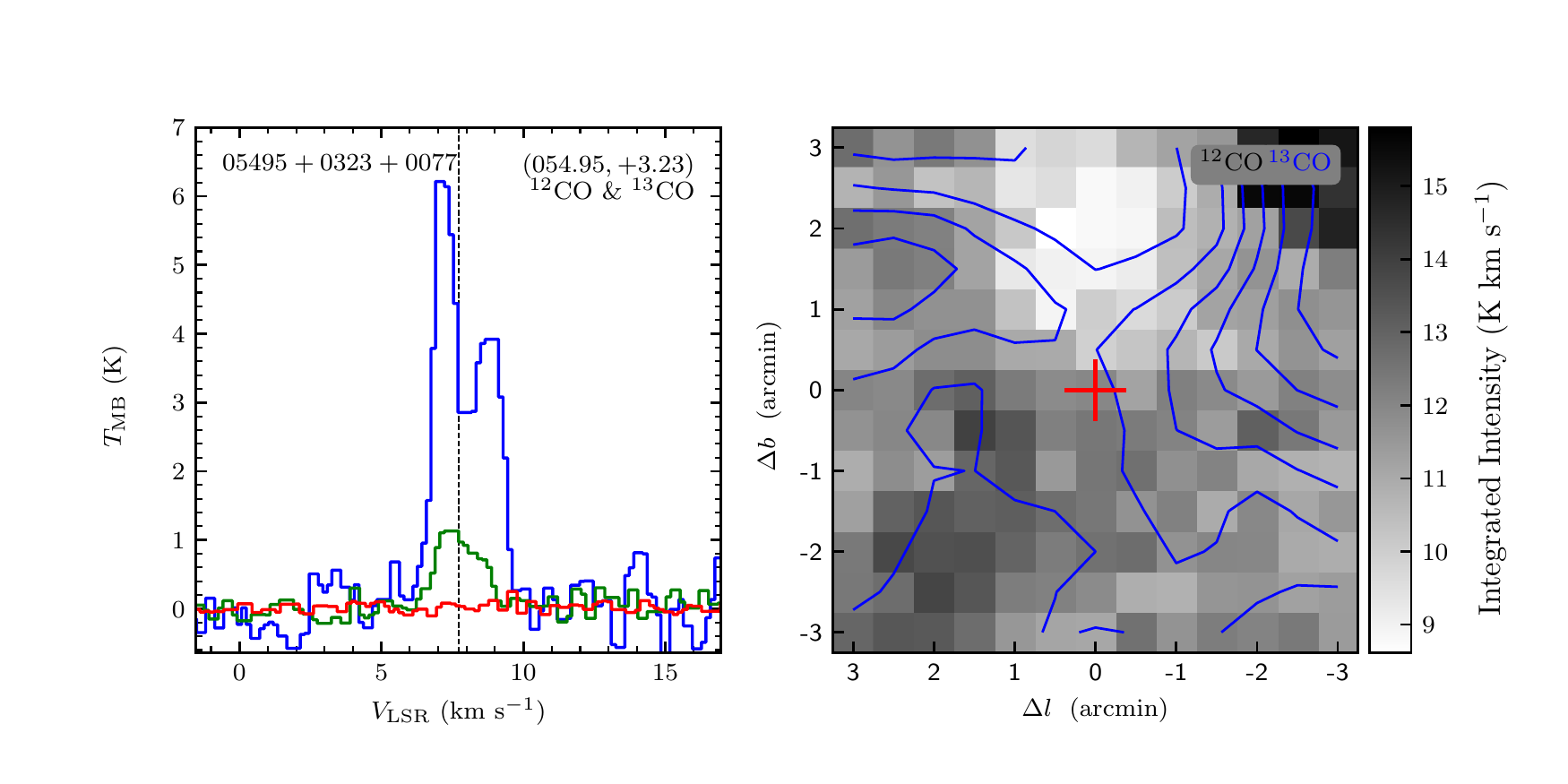}
\includegraphics[width=9.0cm,angle=0]{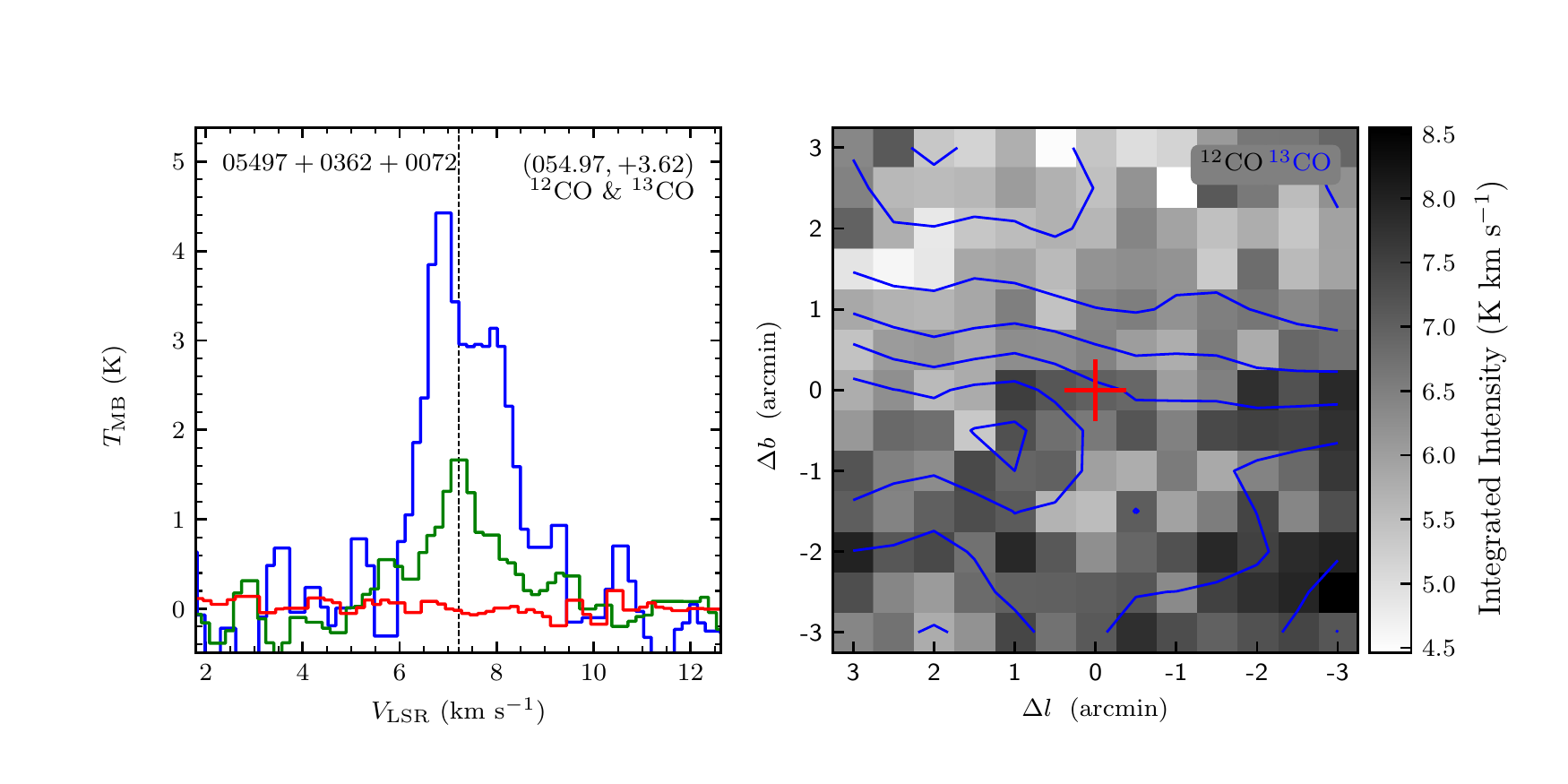}
\vspace{-0.5cm}

\includegraphics[width=9.0cm,angle=0]{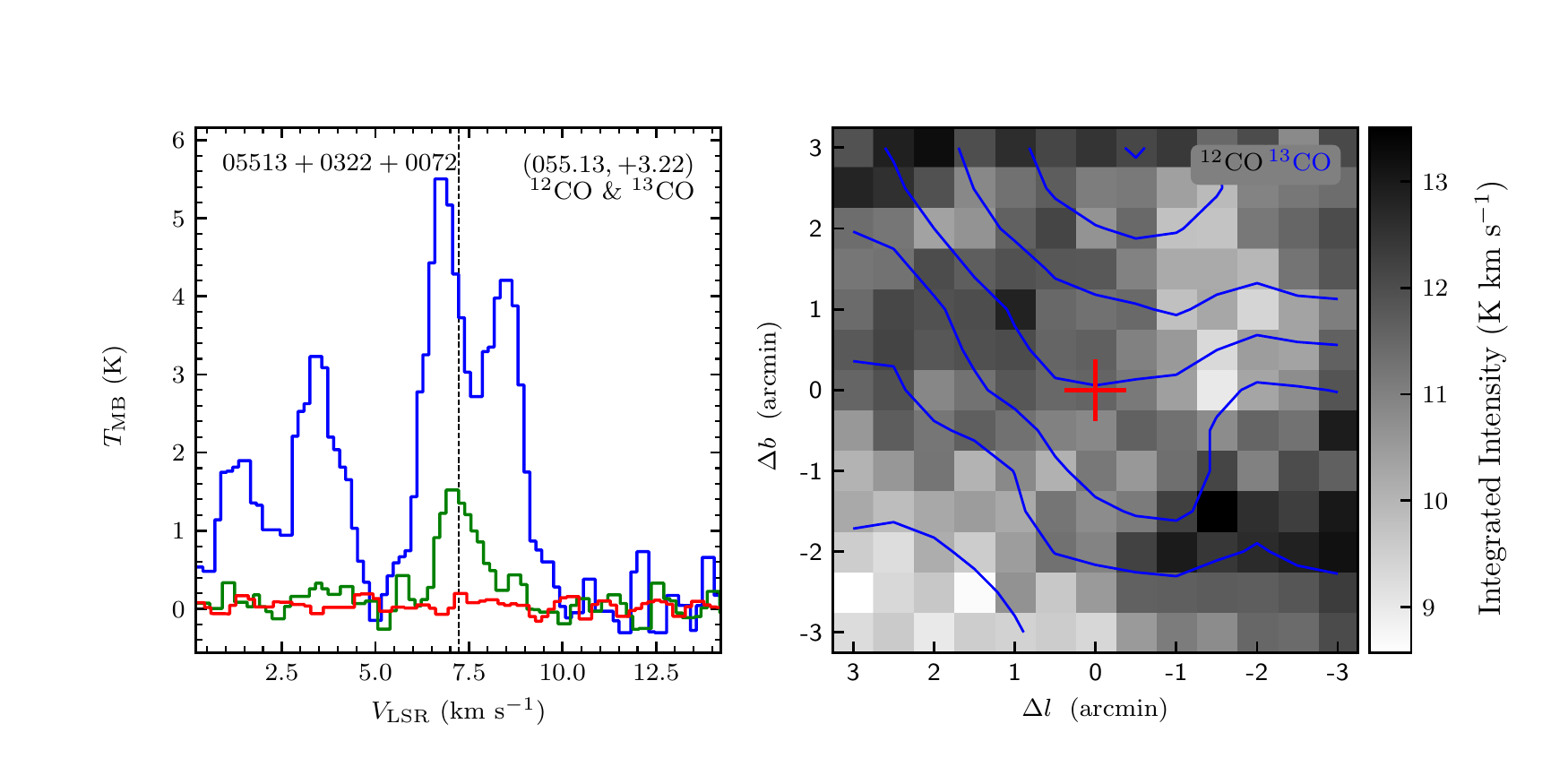}
\includegraphics[width=9.0cm,angle=0]{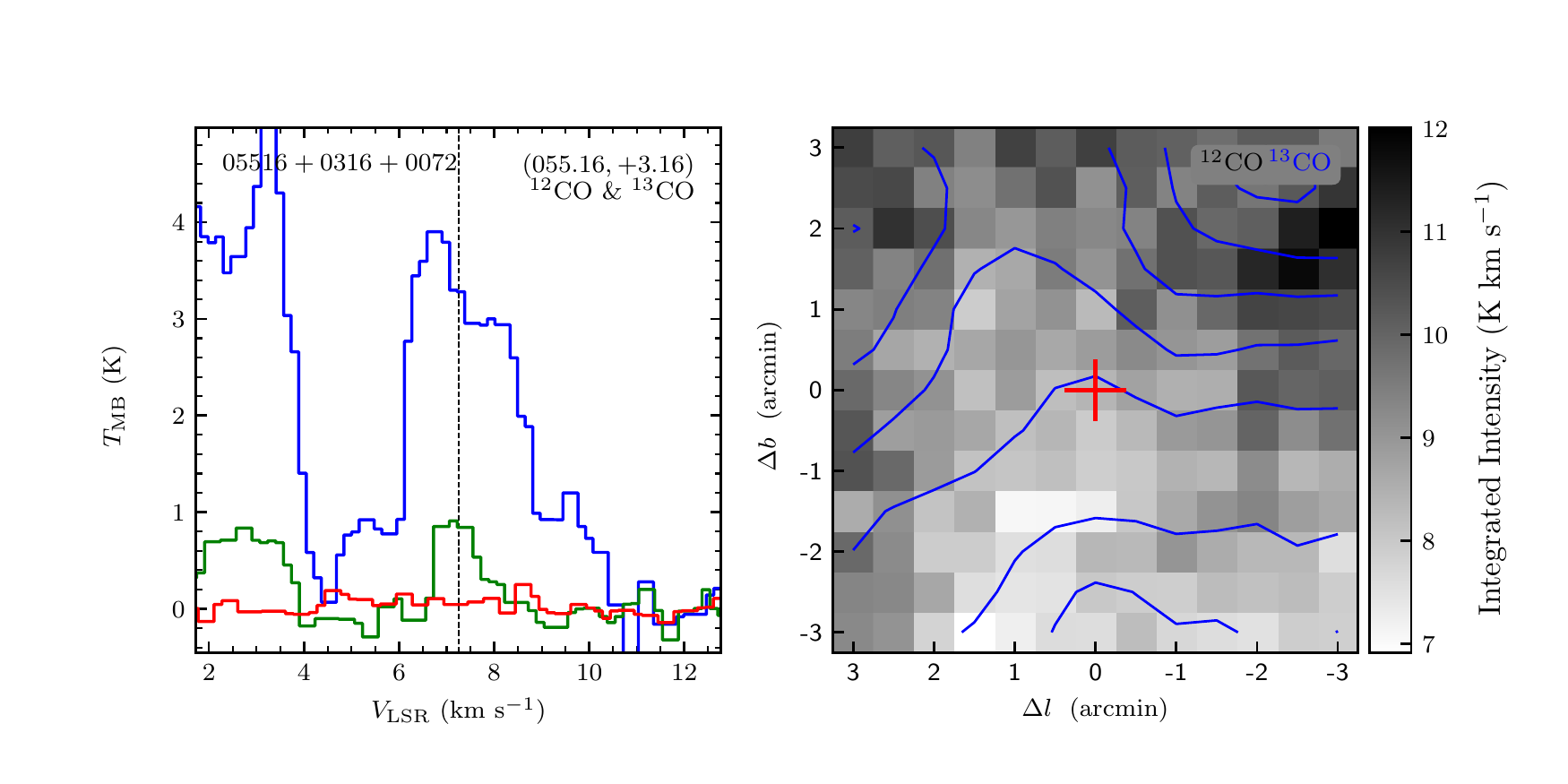}
\vspace{-0.5cm}

\includegraphics[width=9.0cm,angle=0]{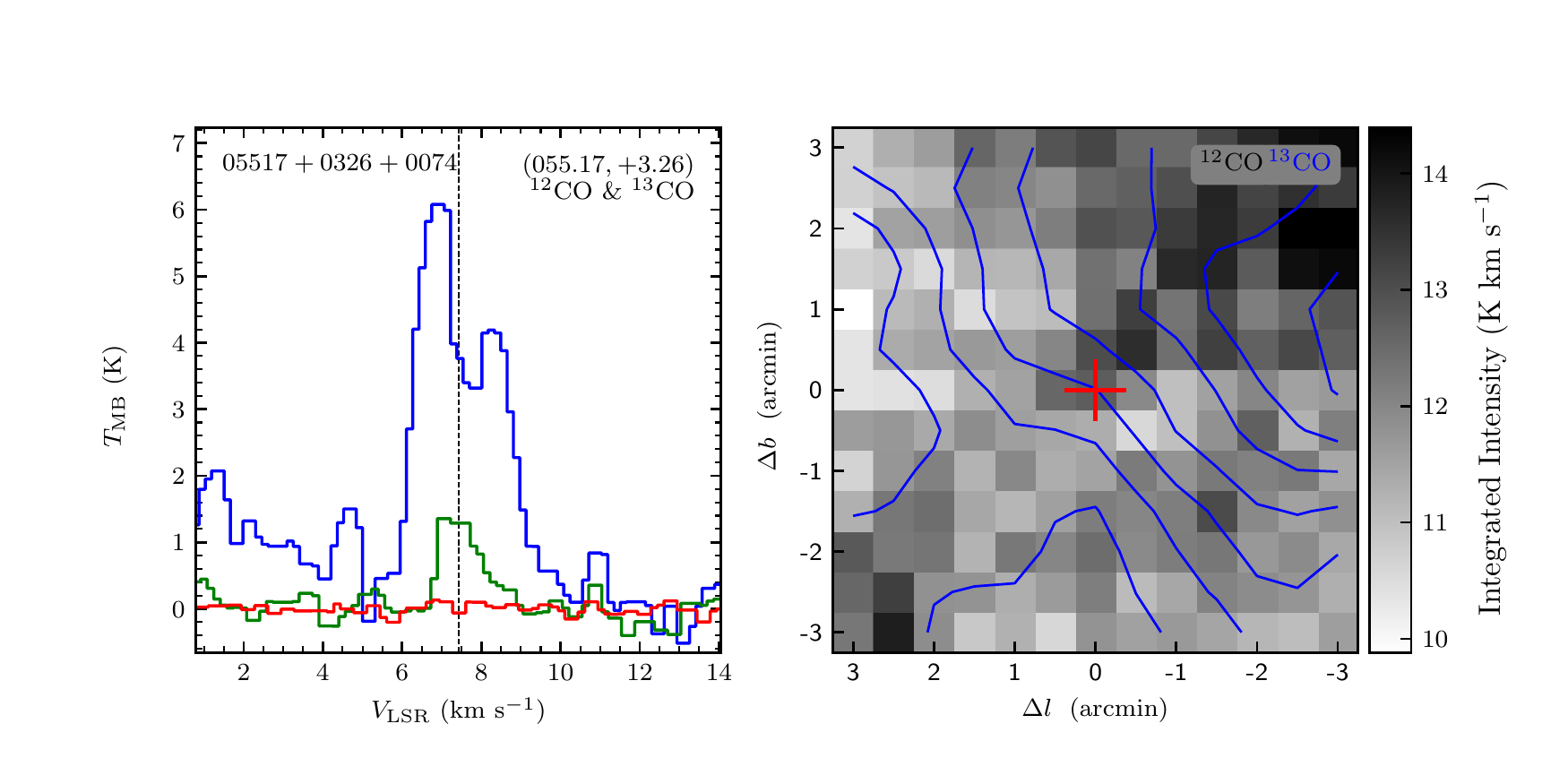}
\includegraphics[width=9.0cm,angle=0]{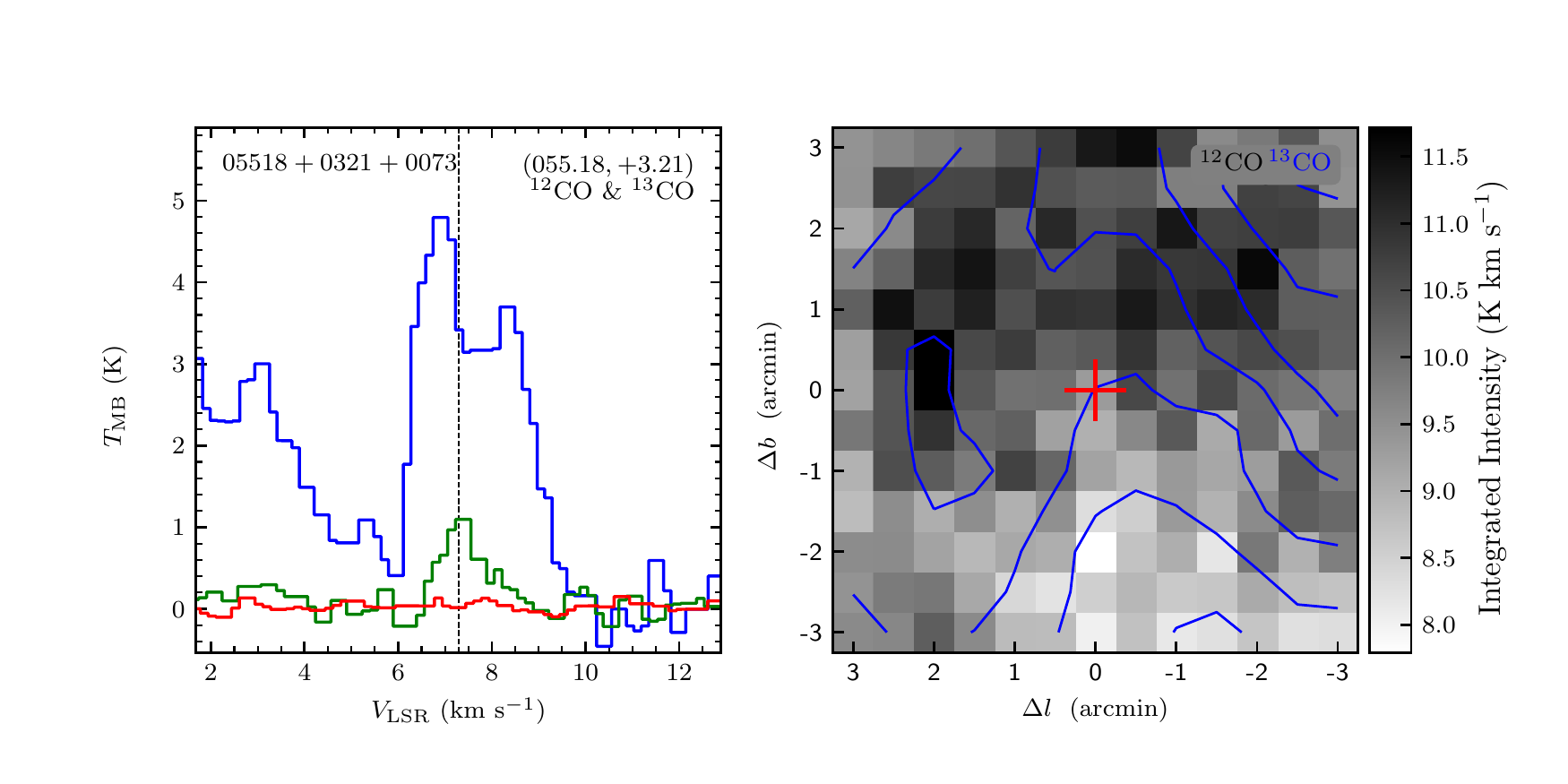}
\vspace{-0.5cm}

\includegraphics[width=9.0cm,angle=0]{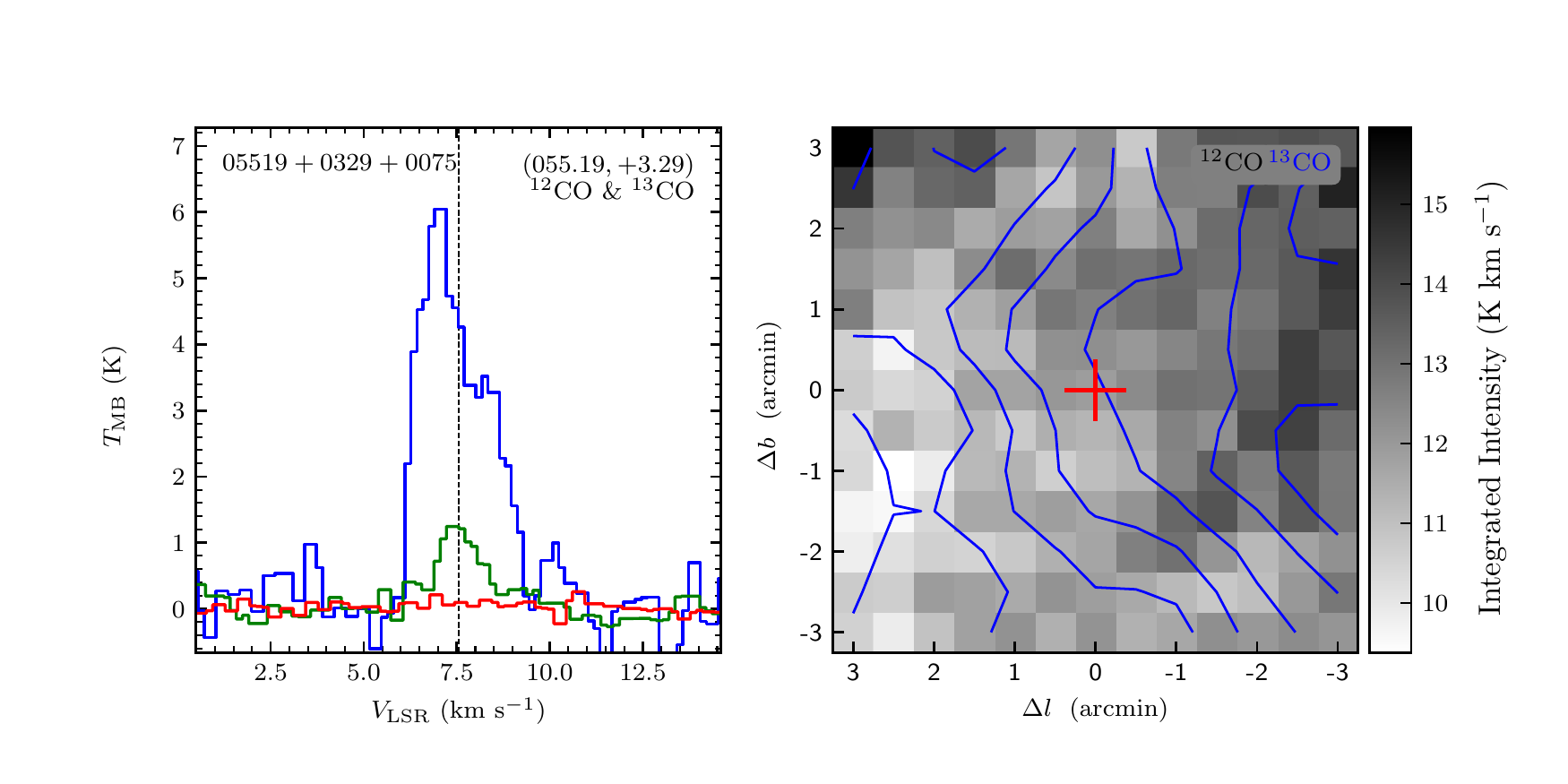}
\includegraphics[width=9.0cm,angle=0]{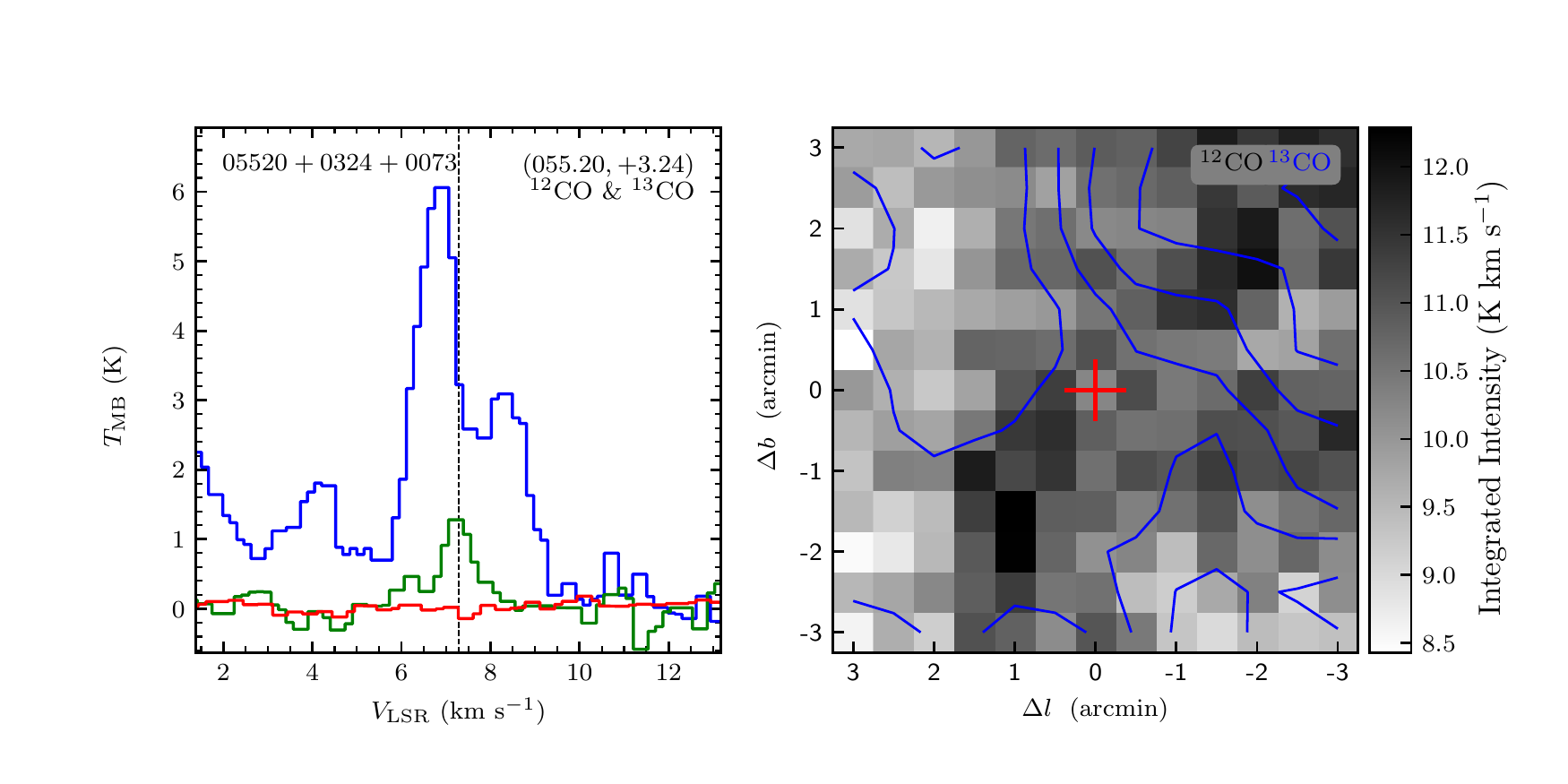}
\end{figure}
\clearpage

\begin{figure}
\includegraphics[width=9.0cm,angle=0]{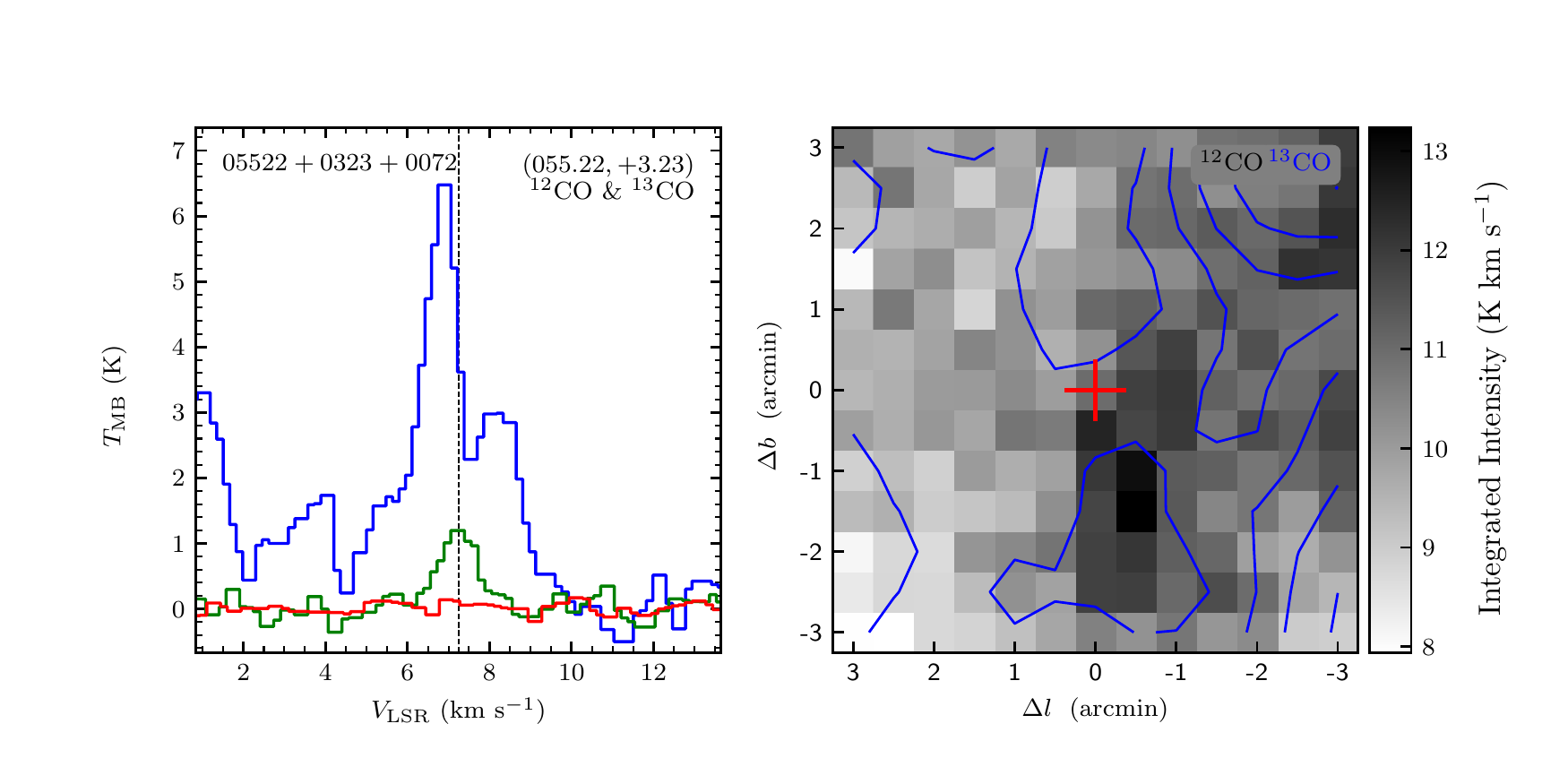}
\includegraphics[width=9.0cm,angle=0]{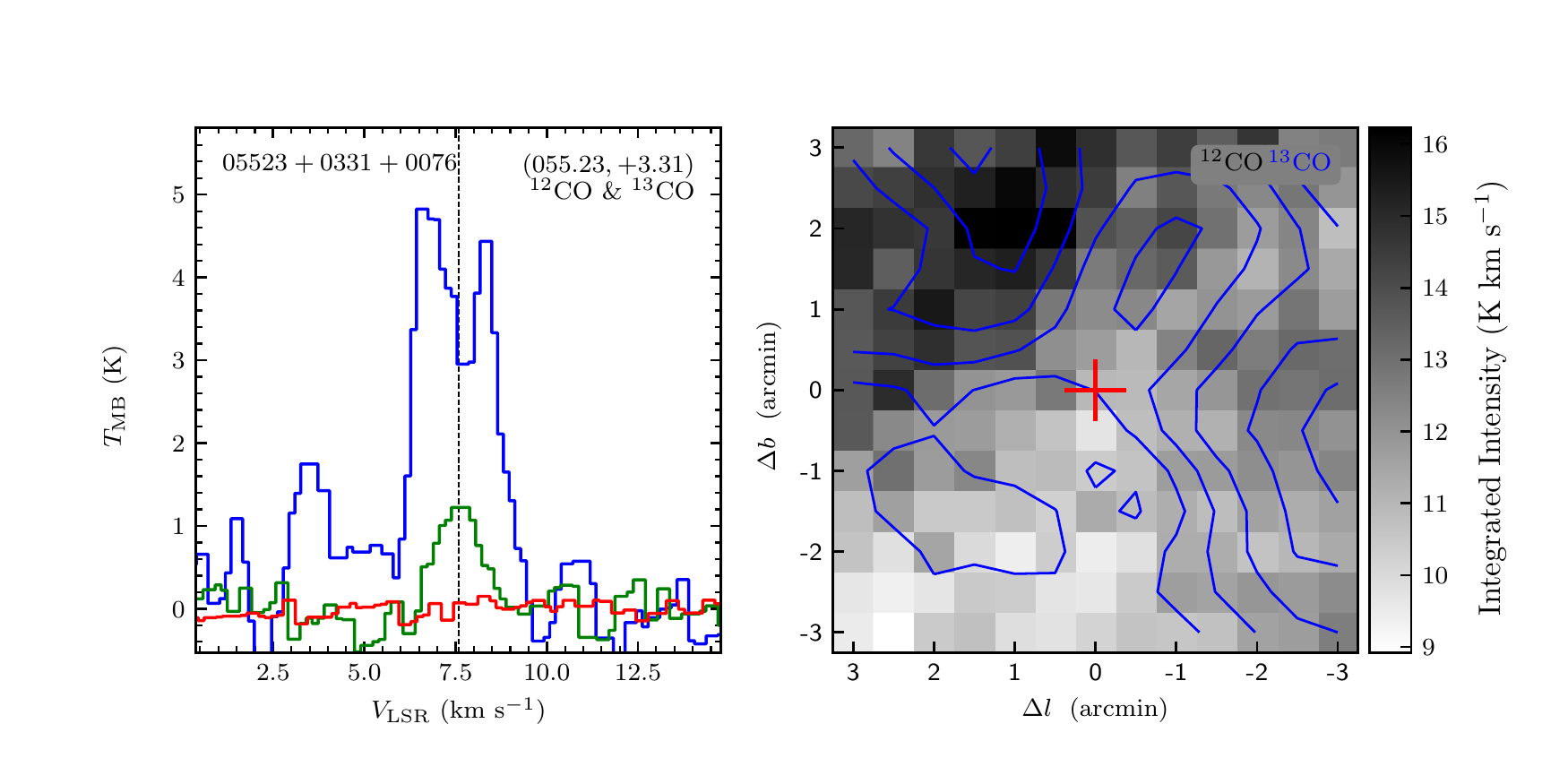}
\vspace{-0.5cm}

\includegraphics[width=9.0cm,angle=0]{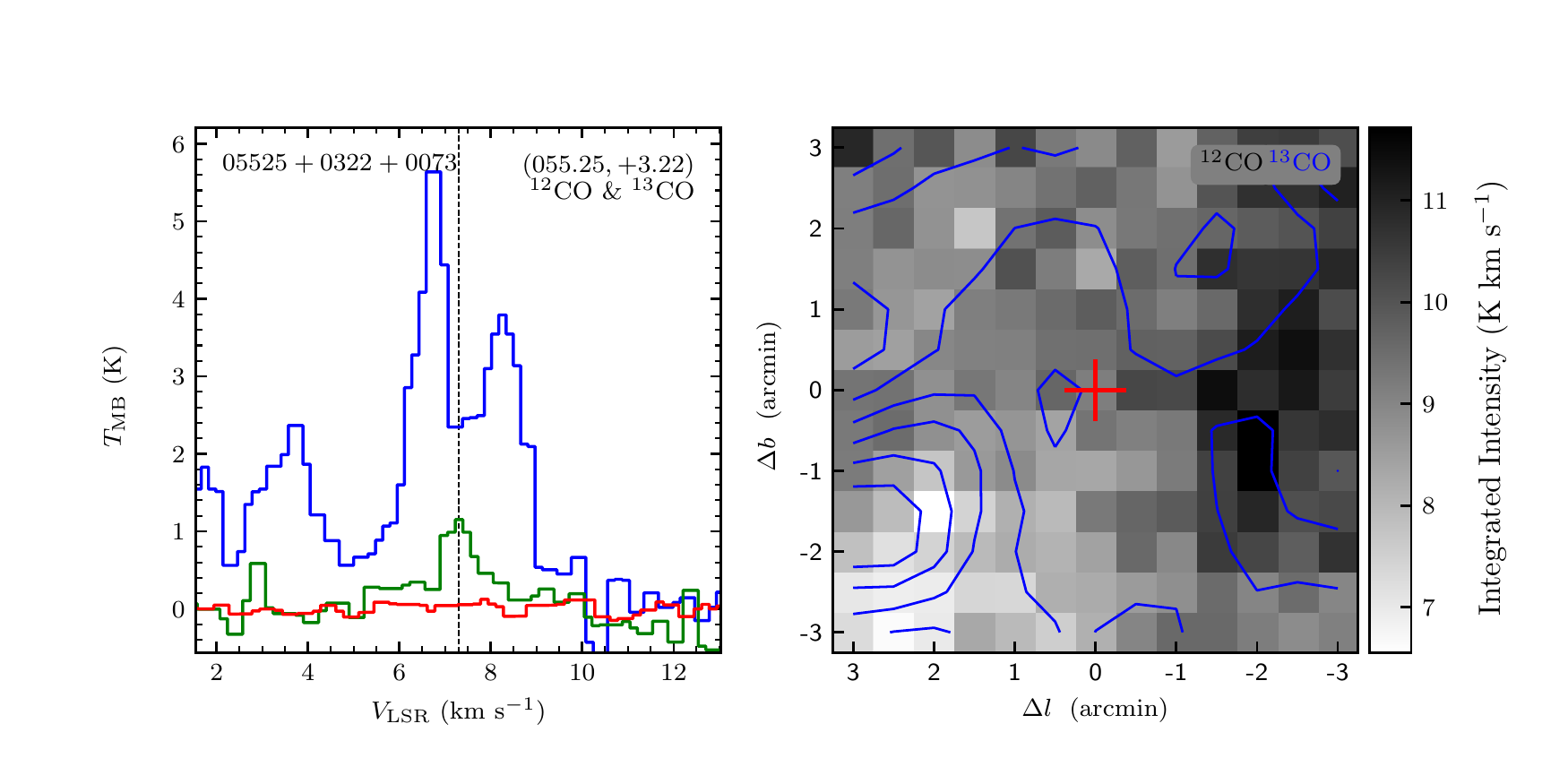}
\includegraphics[width=9.0cm,angle=0]{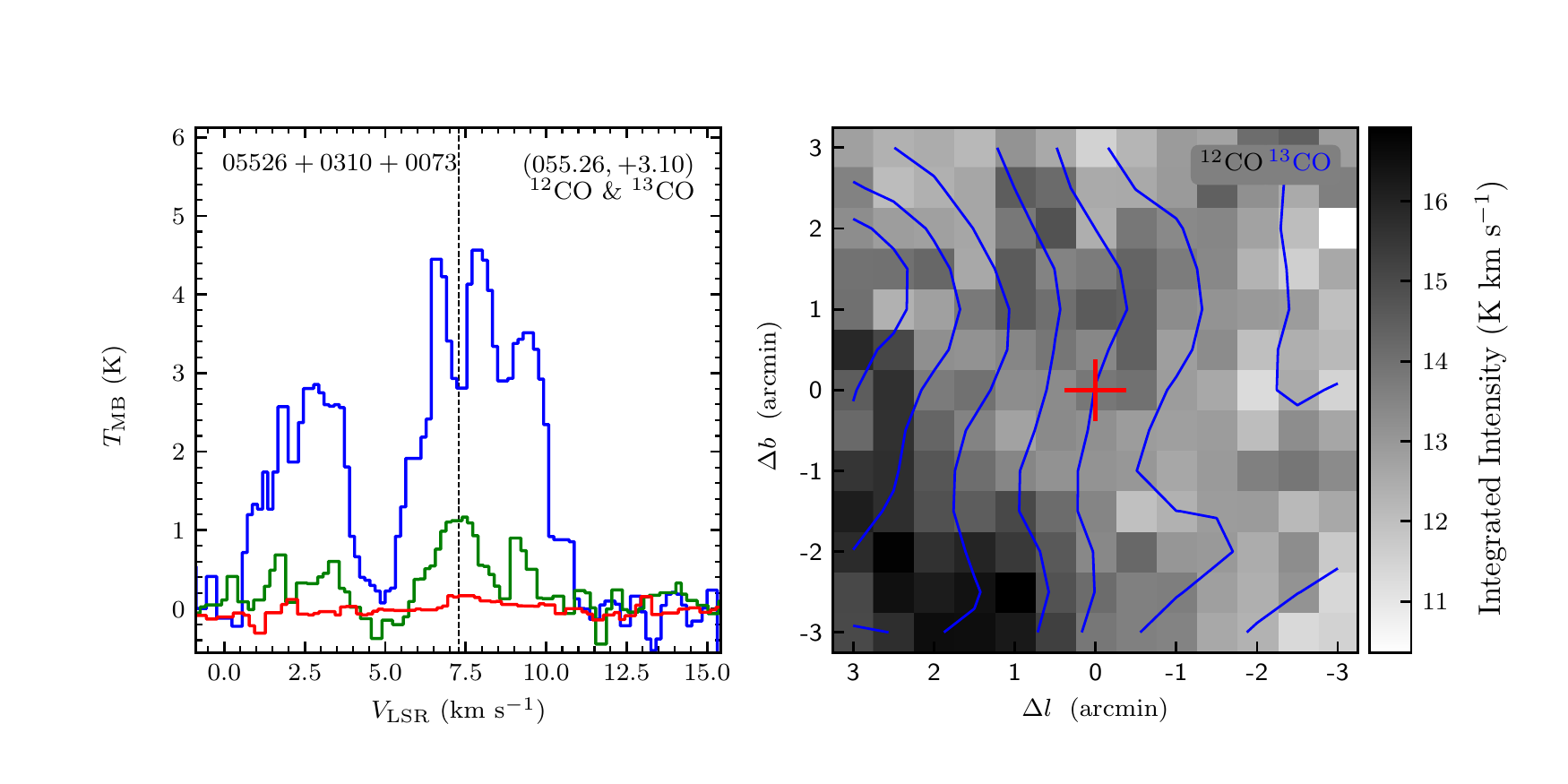}
\vspace{-0.5cm}

\includegraphics[width=9.0cm,angle=0]{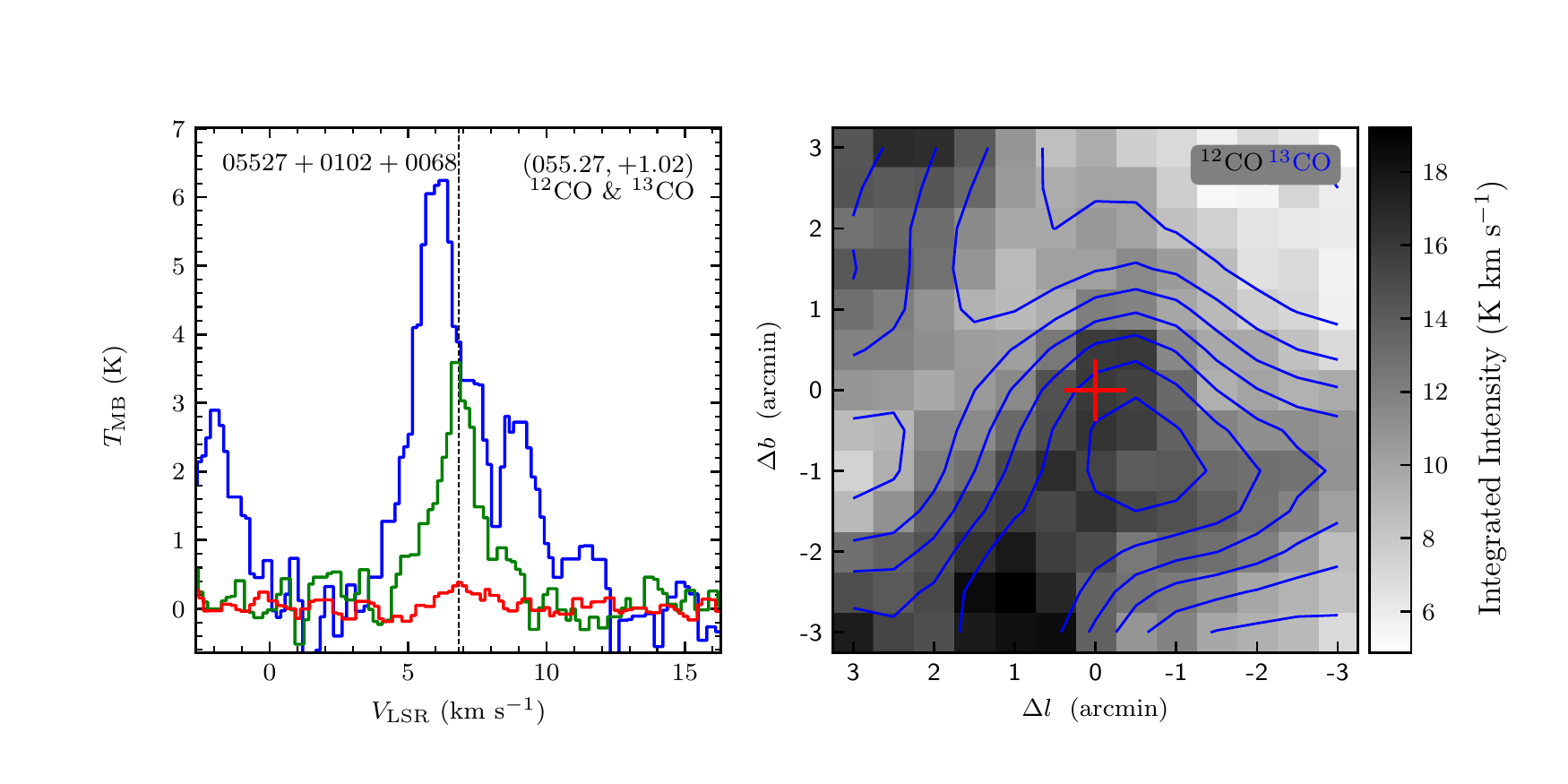}
\includegraphics[width=9.0cm,angle=0]{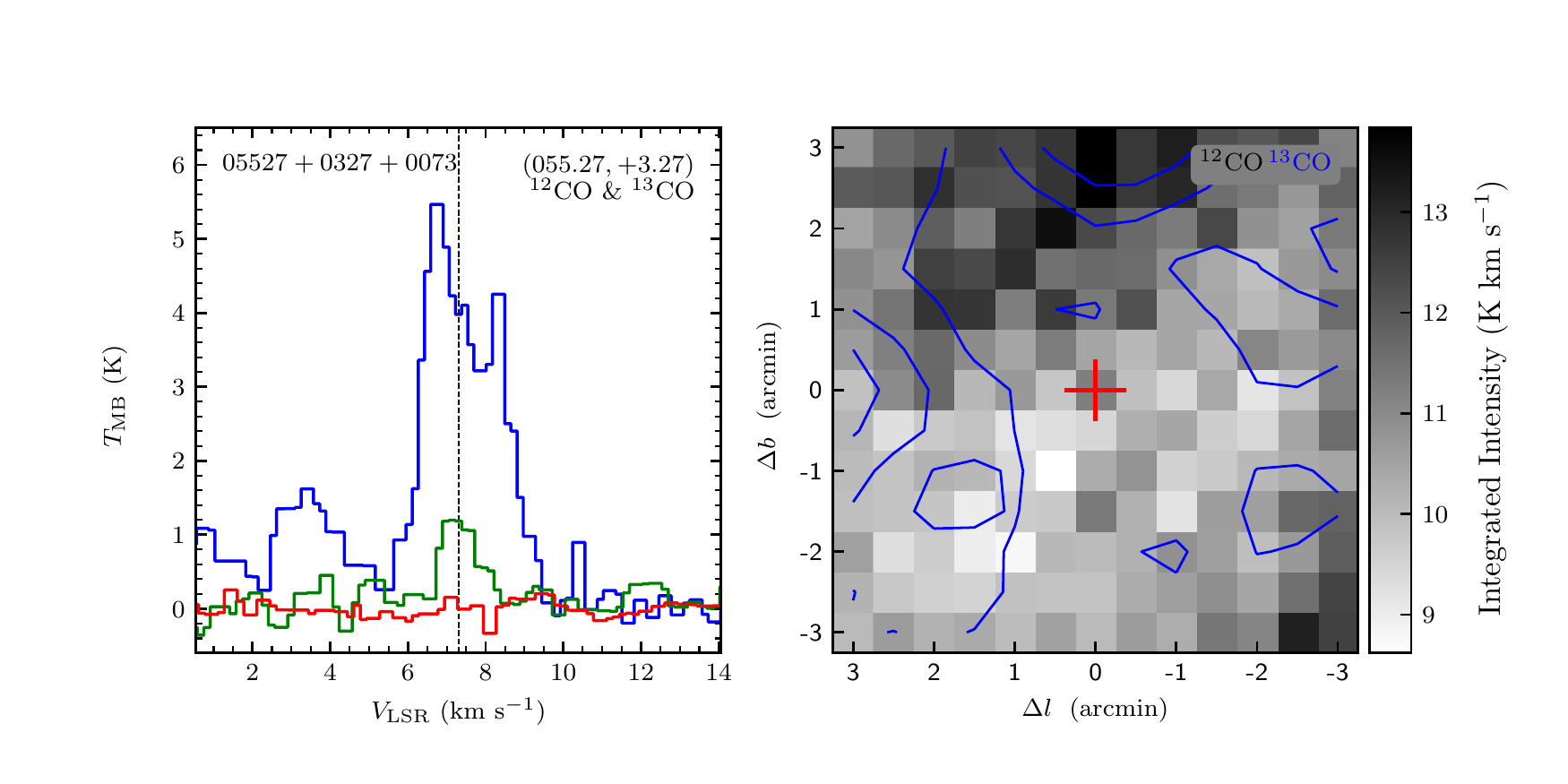}
\vspace{-0.5cm}

\includegraphics[width=9.0cm,angle=0]{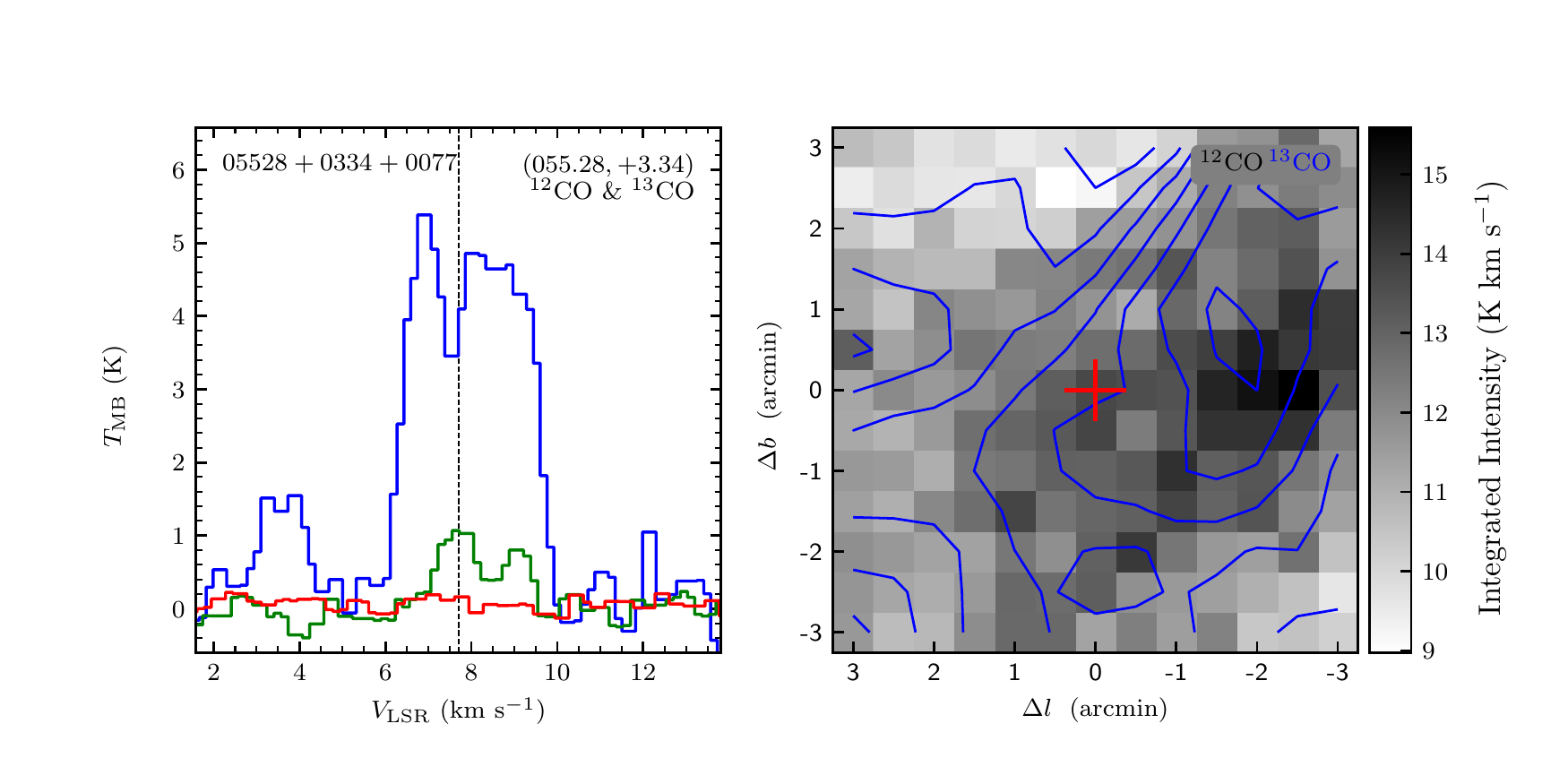}
\includegraphics[width=9.0cm,angle=0]{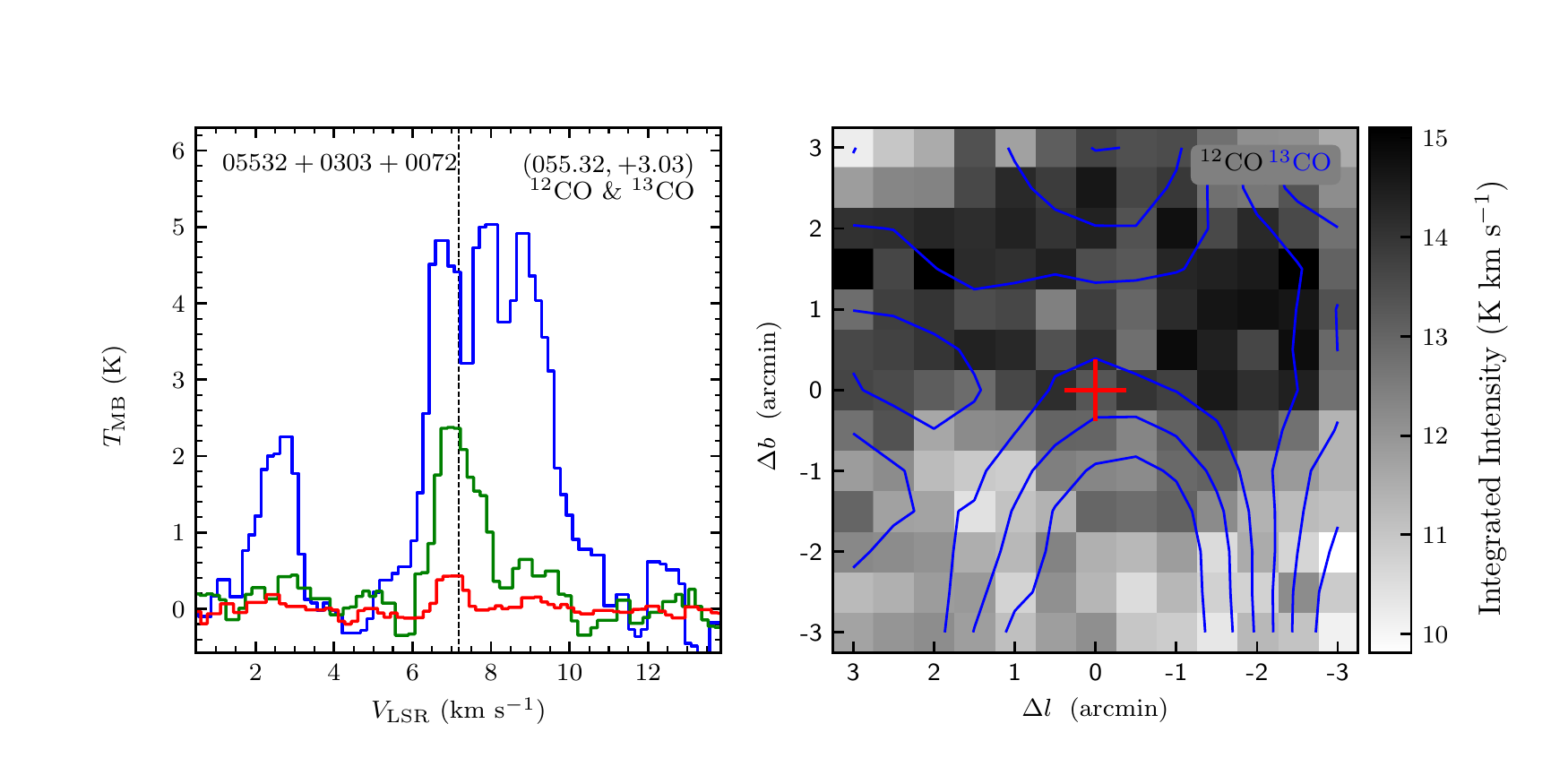}
\vspace{-0.5cm}

\includegraphics[width=9.0cm,angle=0]{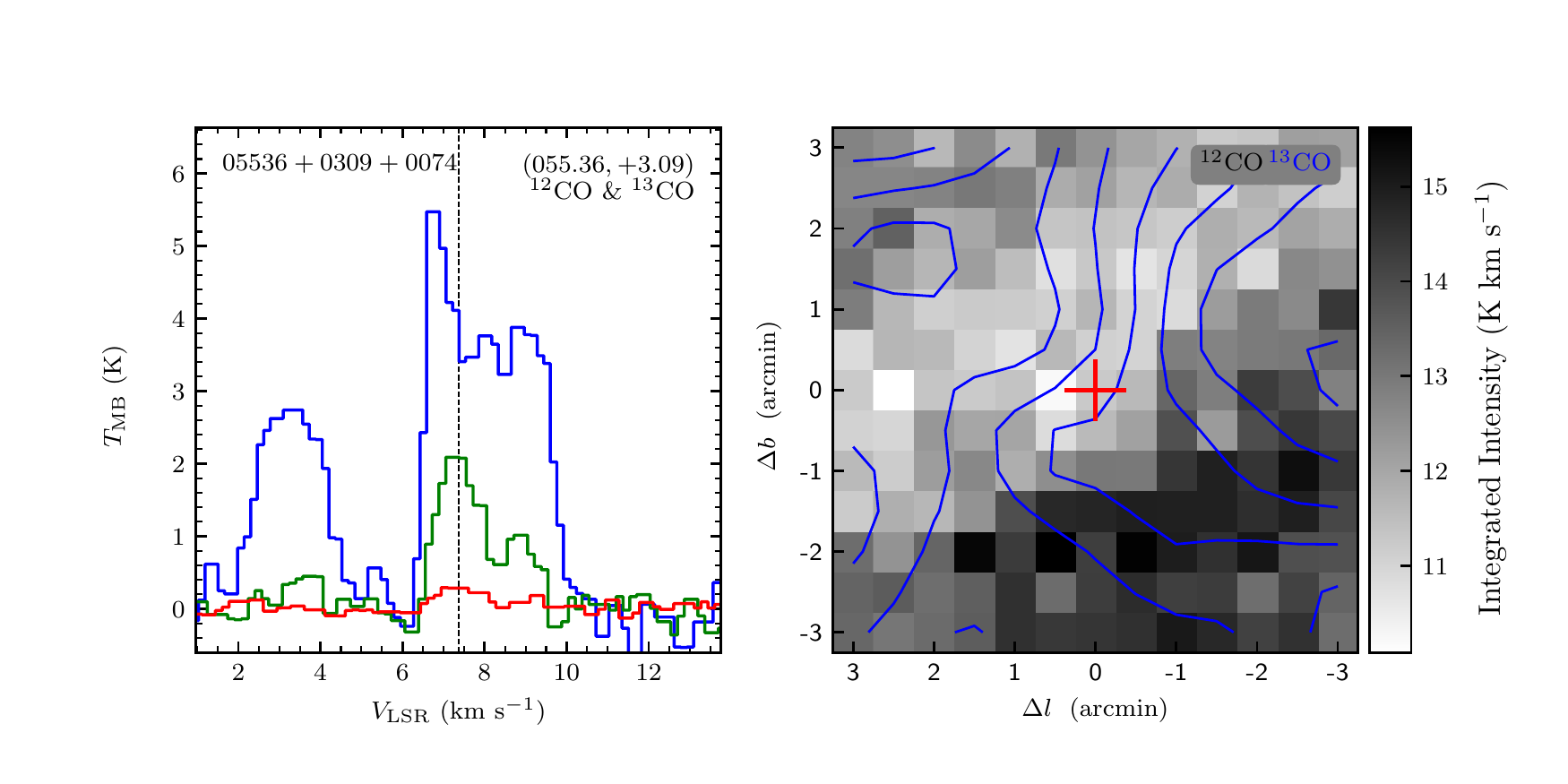}
\includegraphics[width=9.0cm,angle=0]{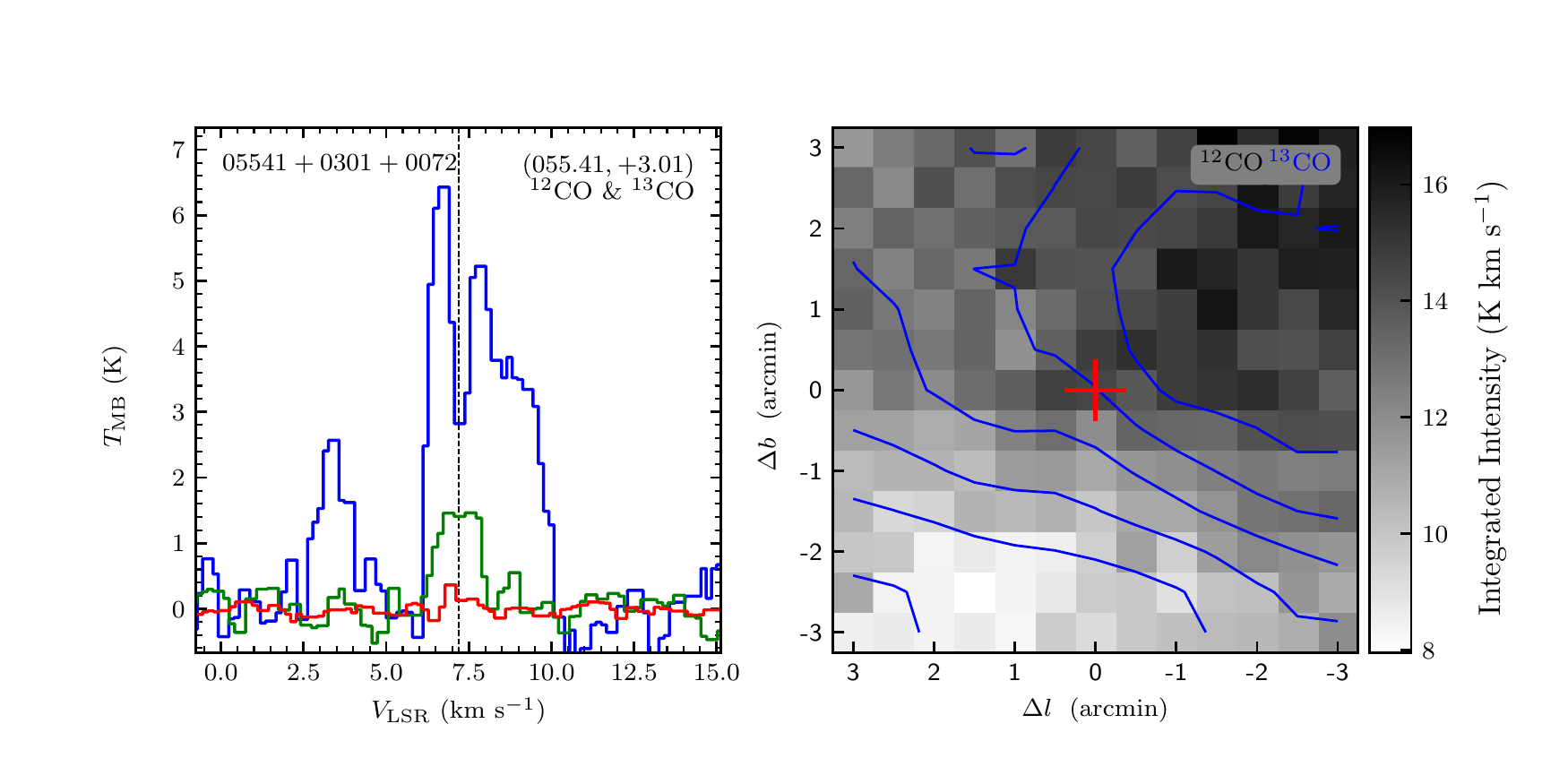}
\end{figure}
\clearpage

\begin{figure}
\includegraphics[width=9.0cm,angle=0]{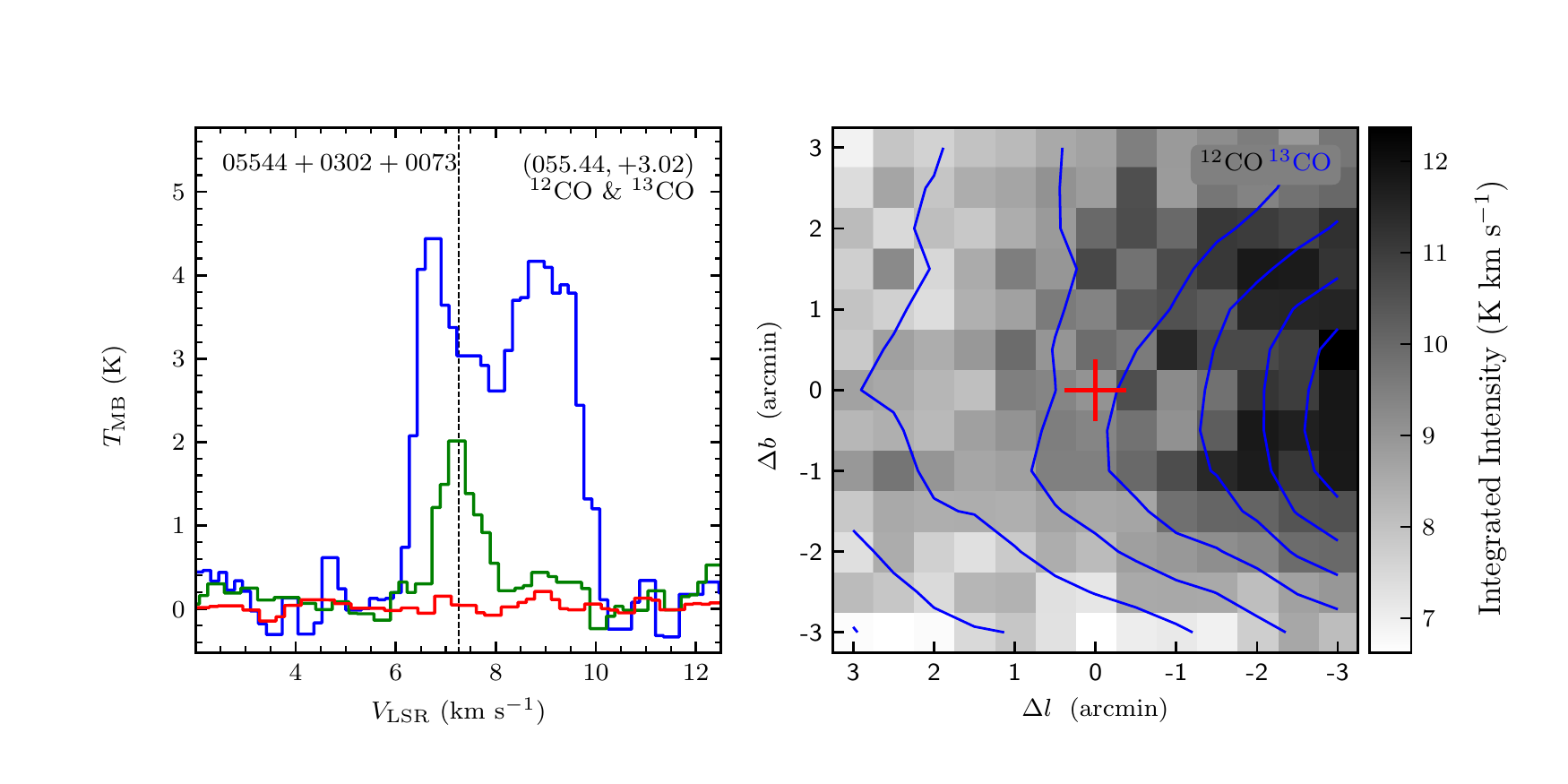}
\includegraphics[width=9.0cm,angle=0]{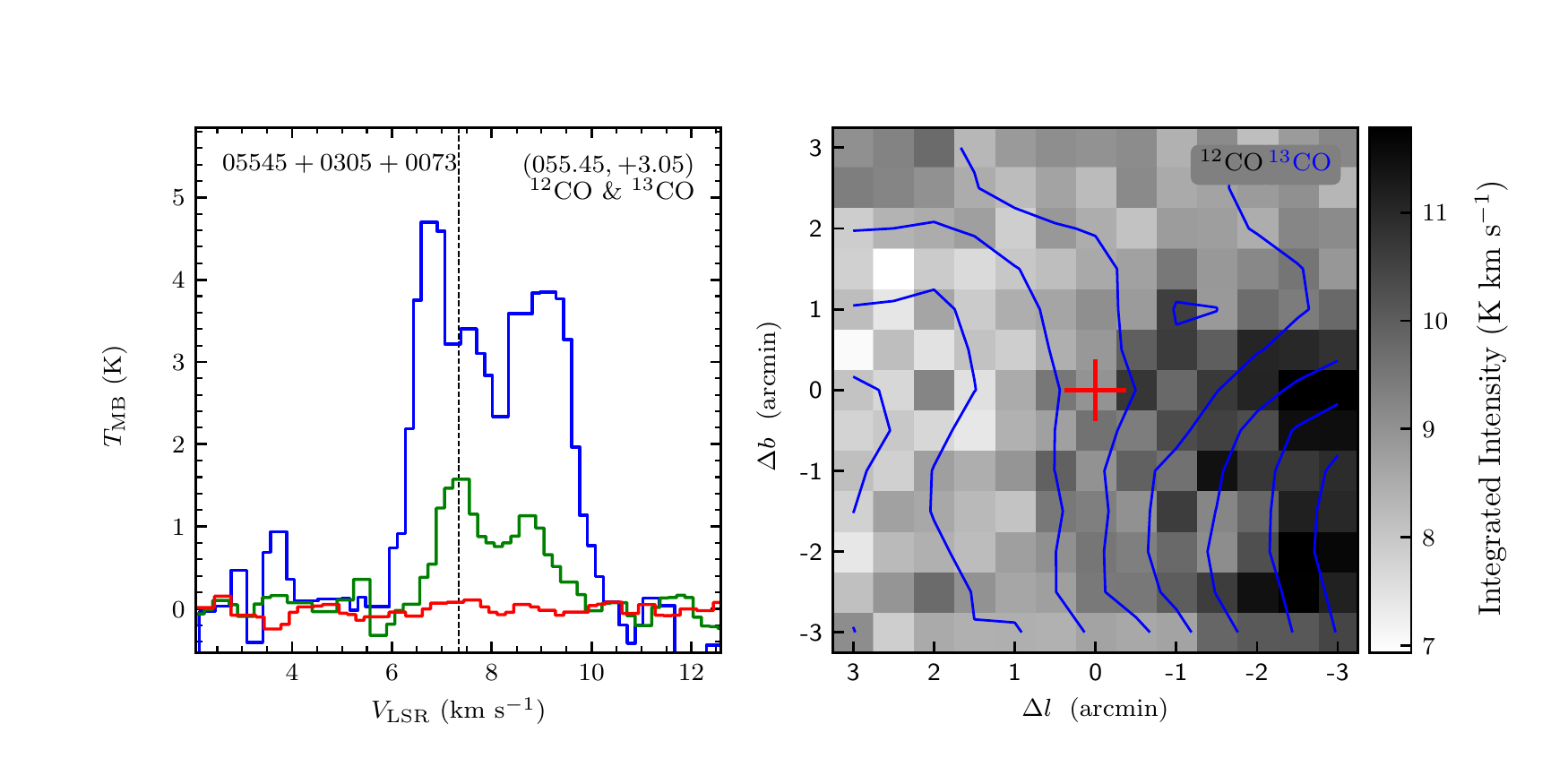}
\vspace{-0.5cm}

\includegraphics[width=9.0cm,angle=0]{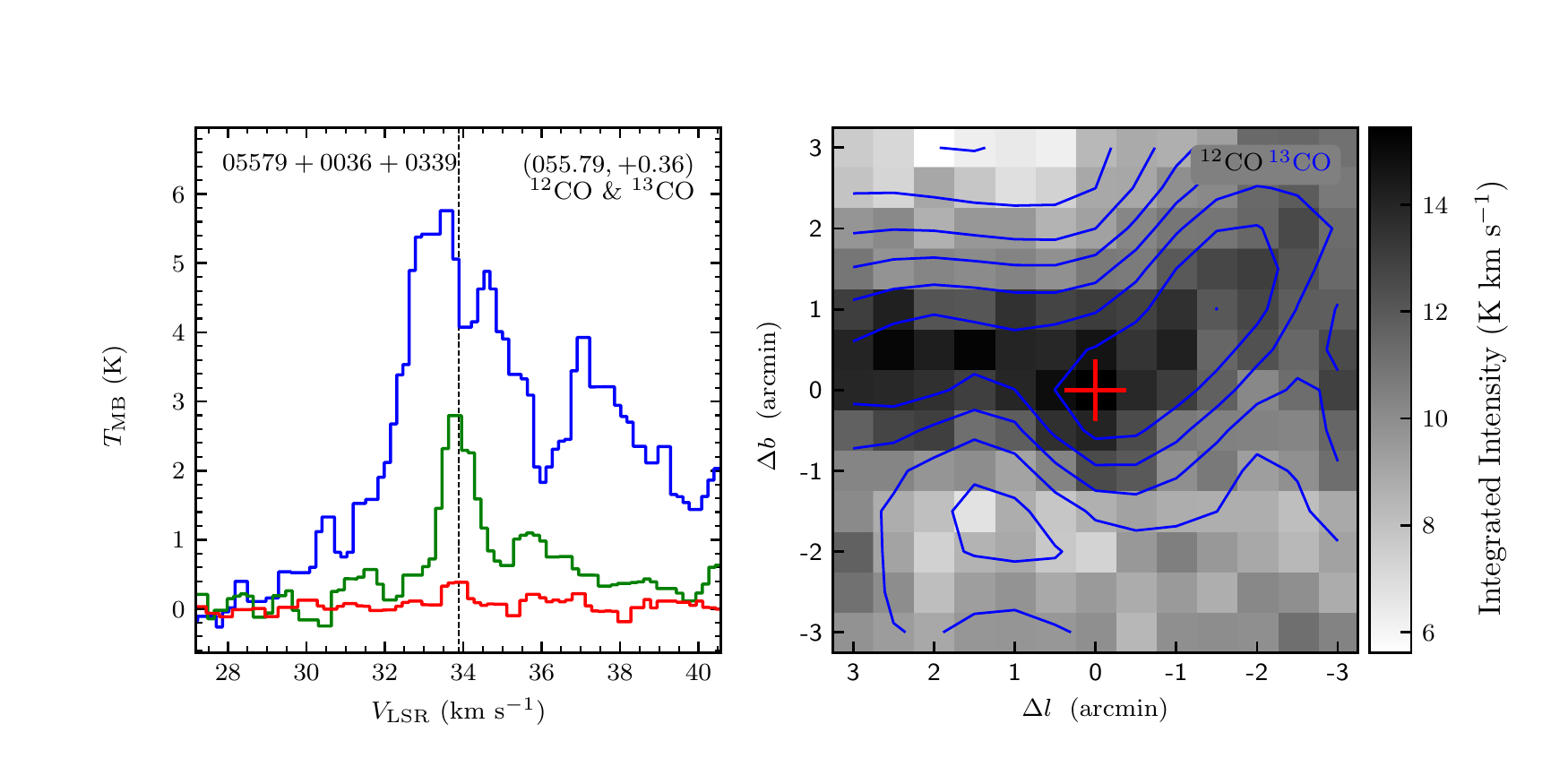}
\includegraphics[width=9.0cm,angle=0]{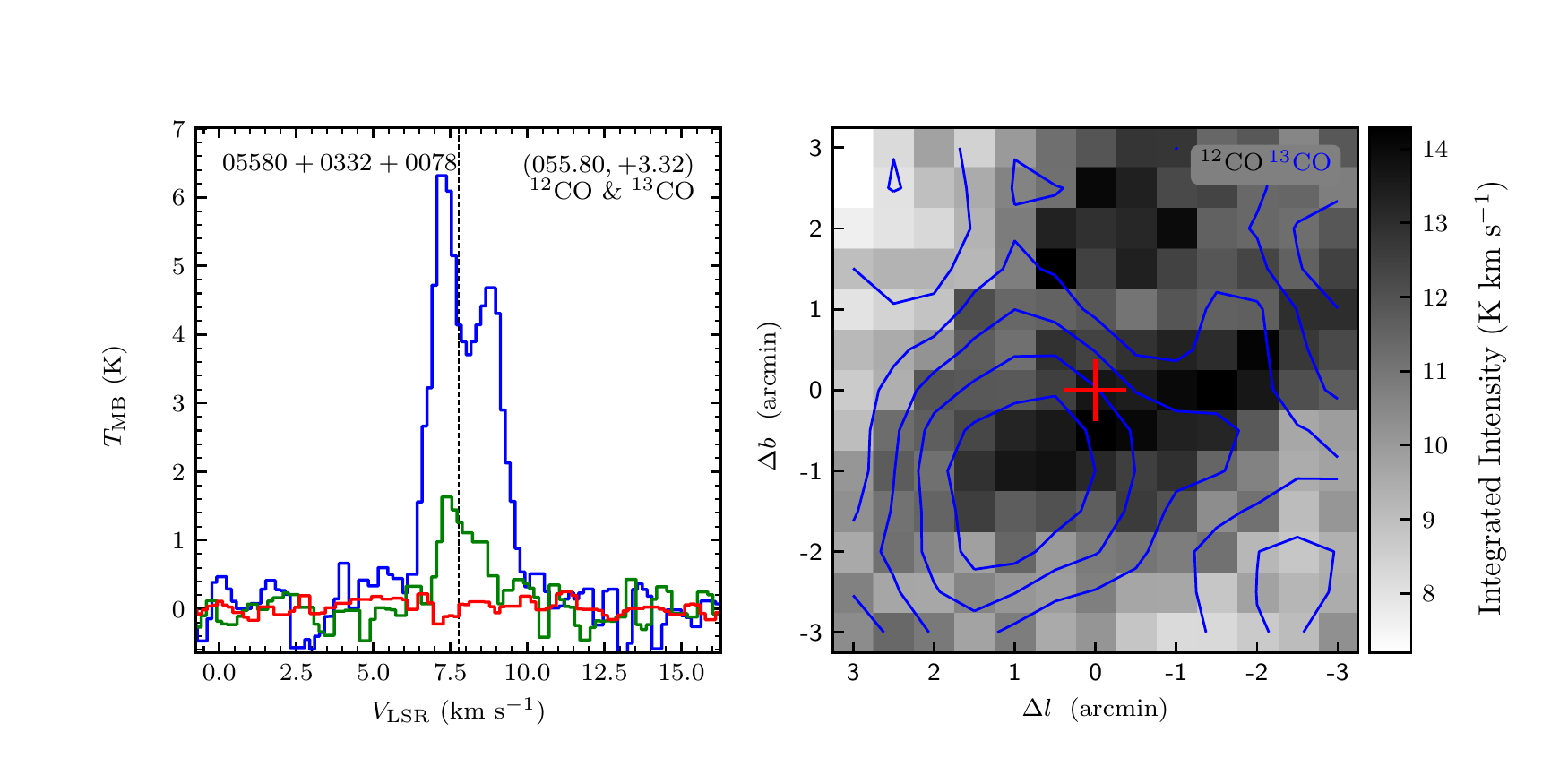}
\vspace{-0.5cm}

\includegraphics[width=9.0cm,angle=0]{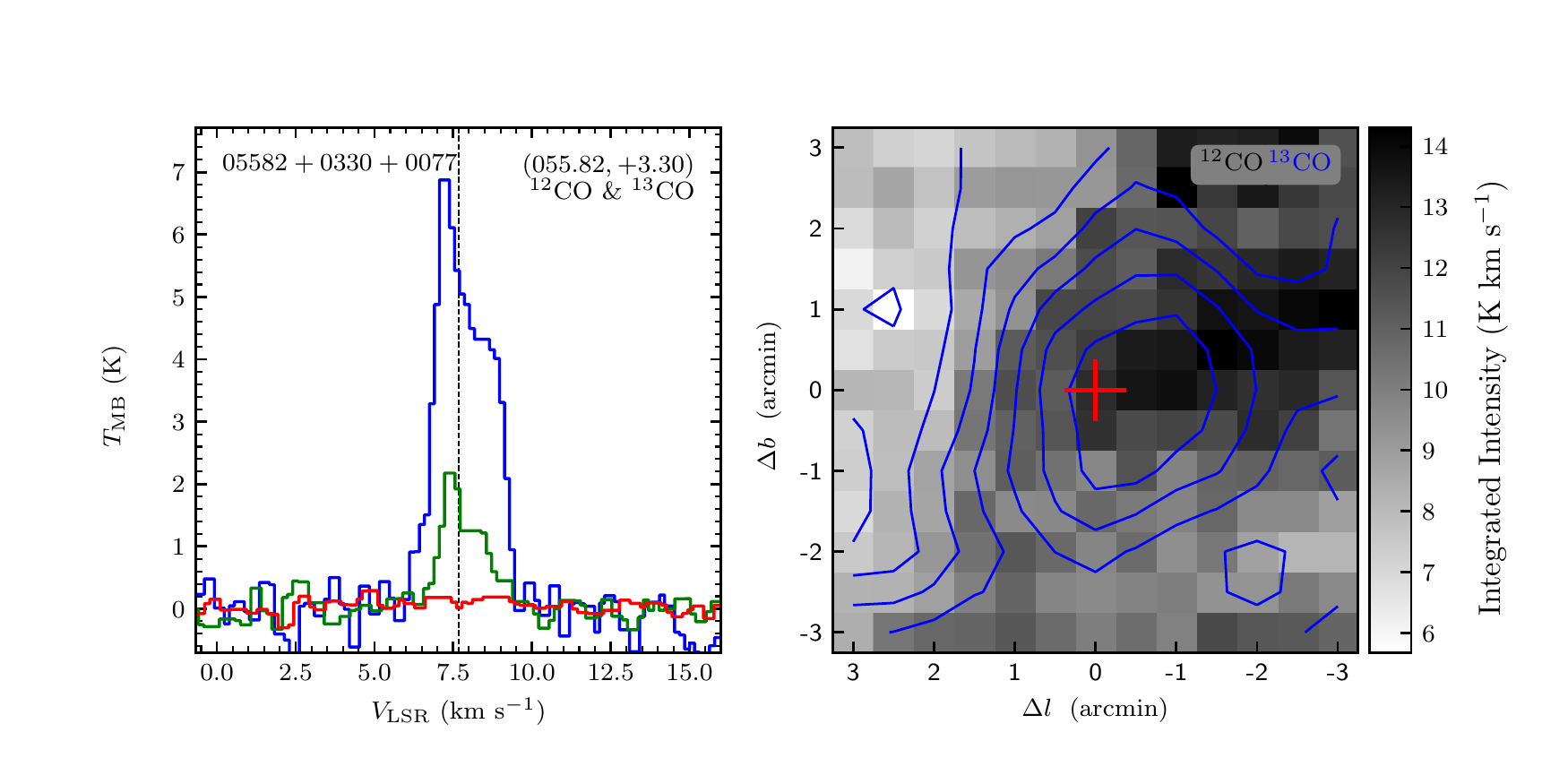}
\includegraphics[width=9.0cm,angle=0]{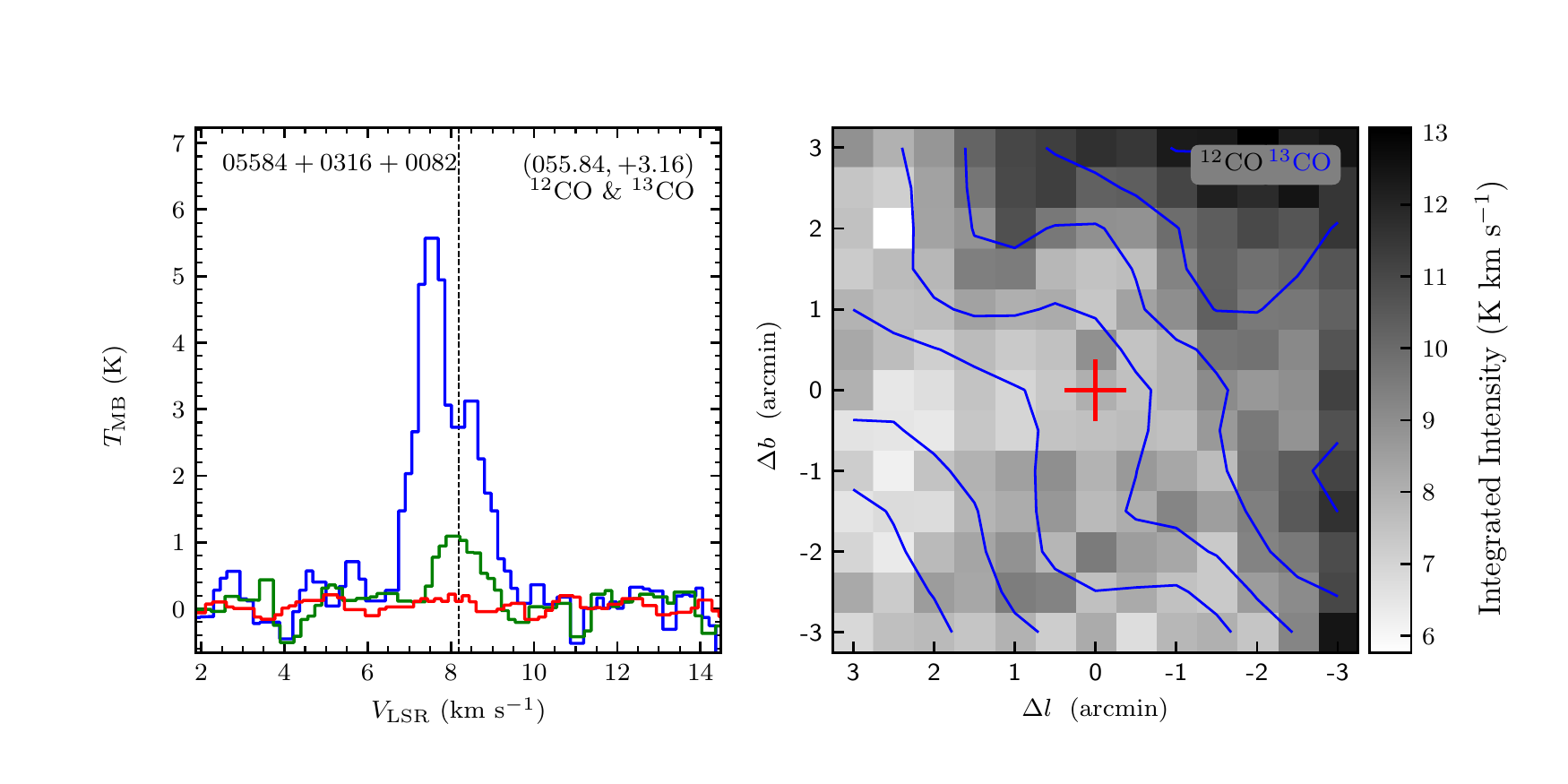}
\vspace{-0.5cm}

\includegraphics[width=9.0cm,angle=0]{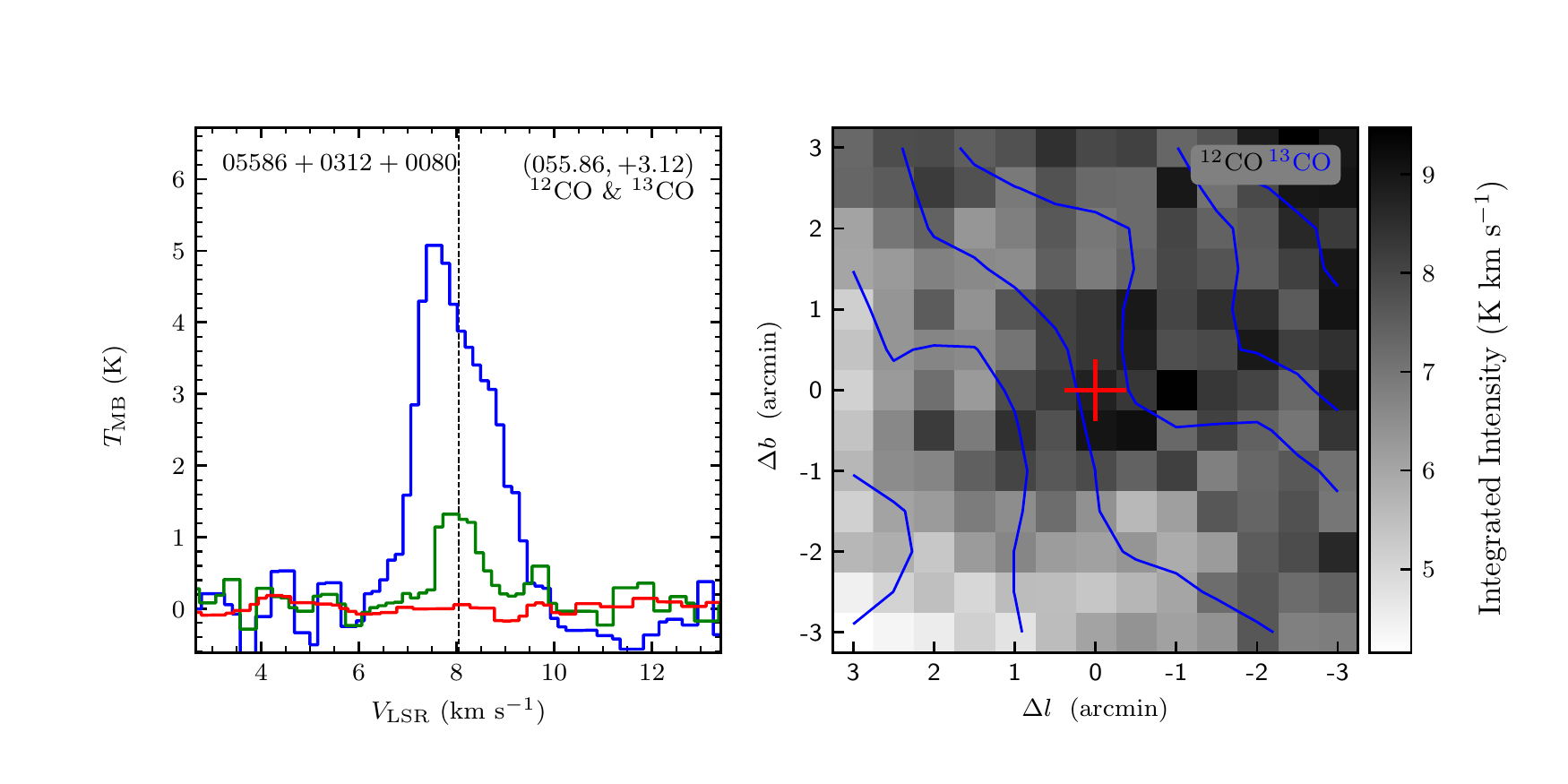}
\includegraphics[width=9.0cm,angle=0]{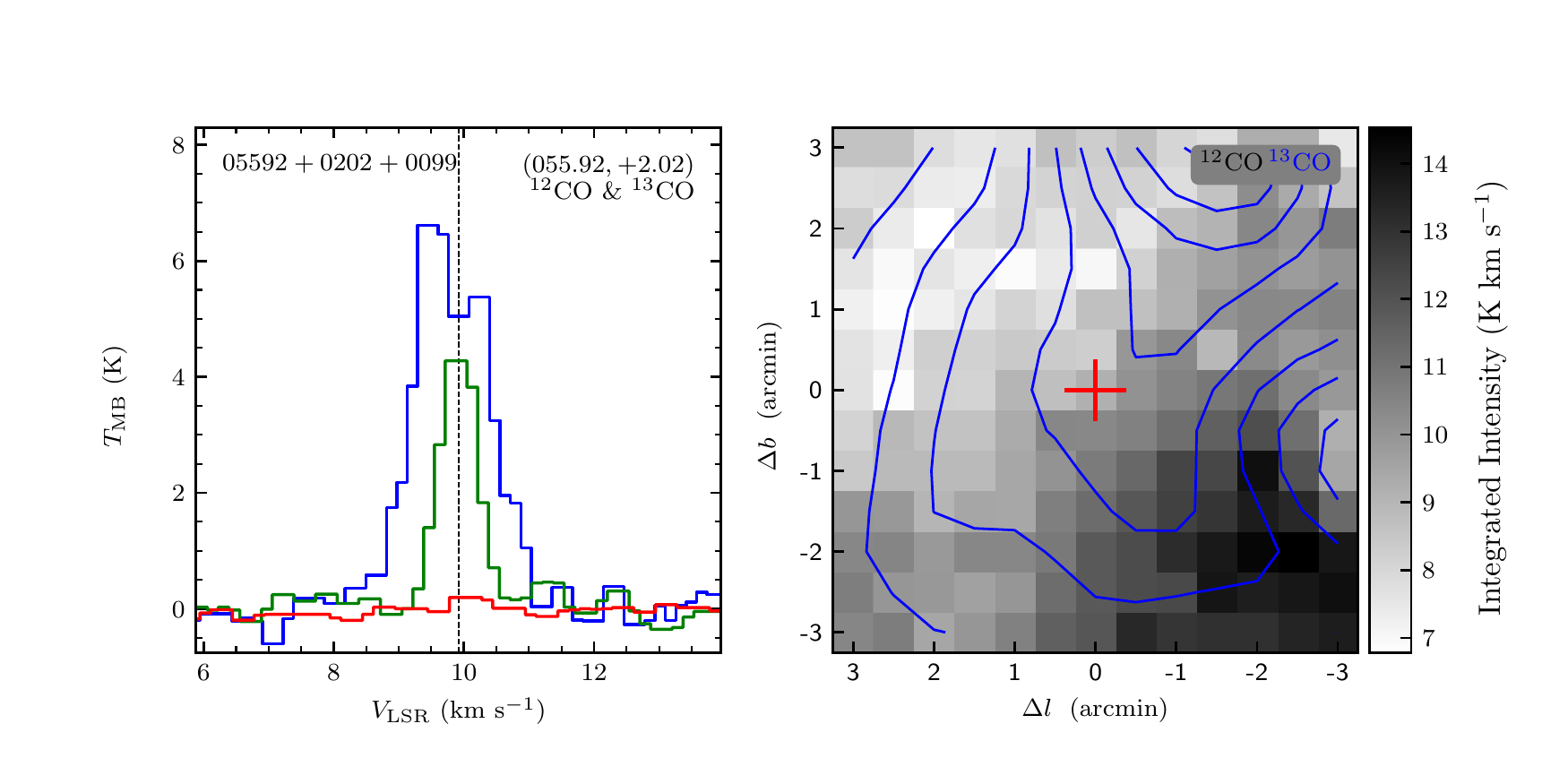}
\vspace{-0.5cm}

\includegraphics[width=9.0cm,angle=0]{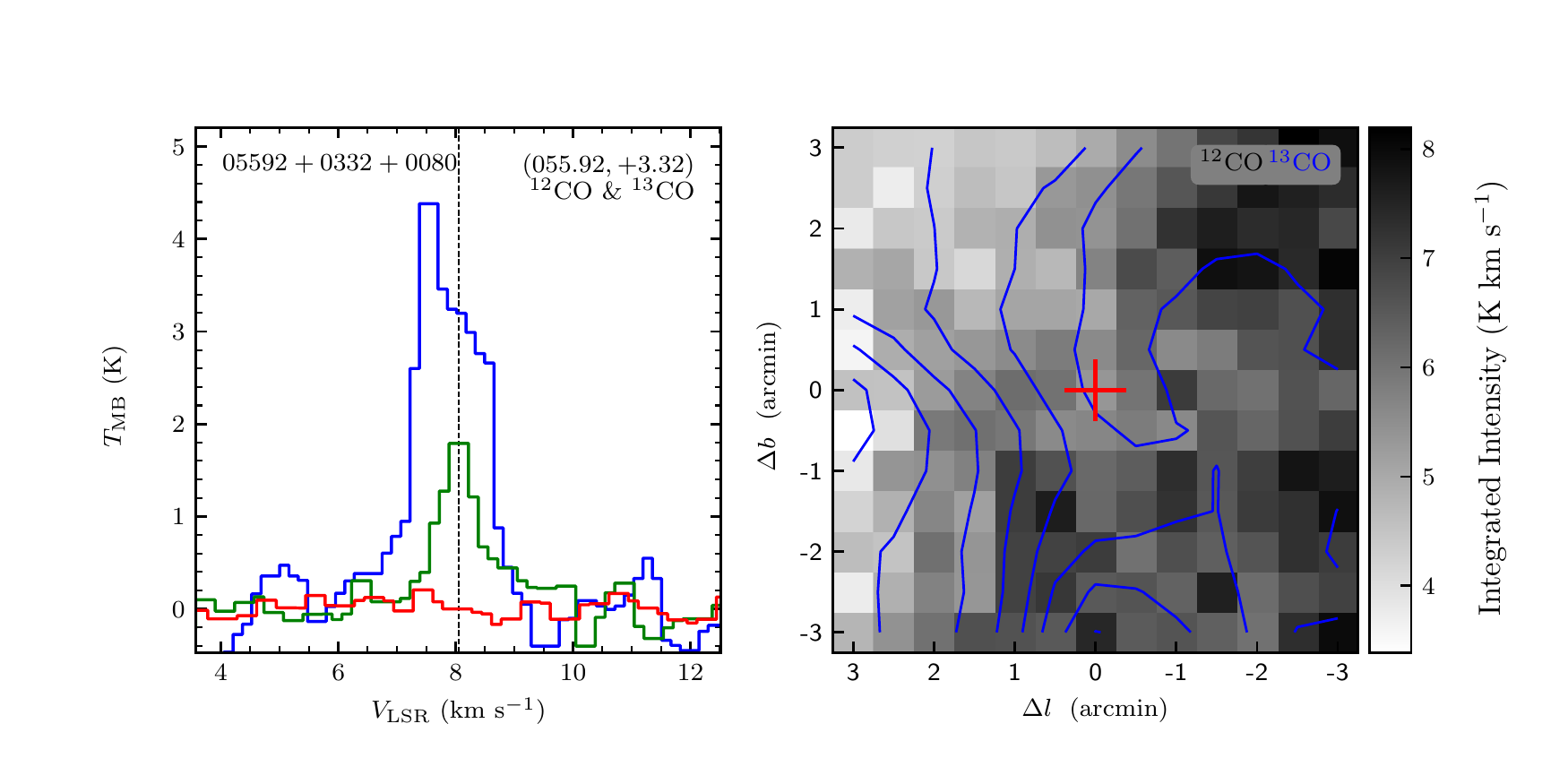}
\includegraphics[width=9.0cm,angle=0]{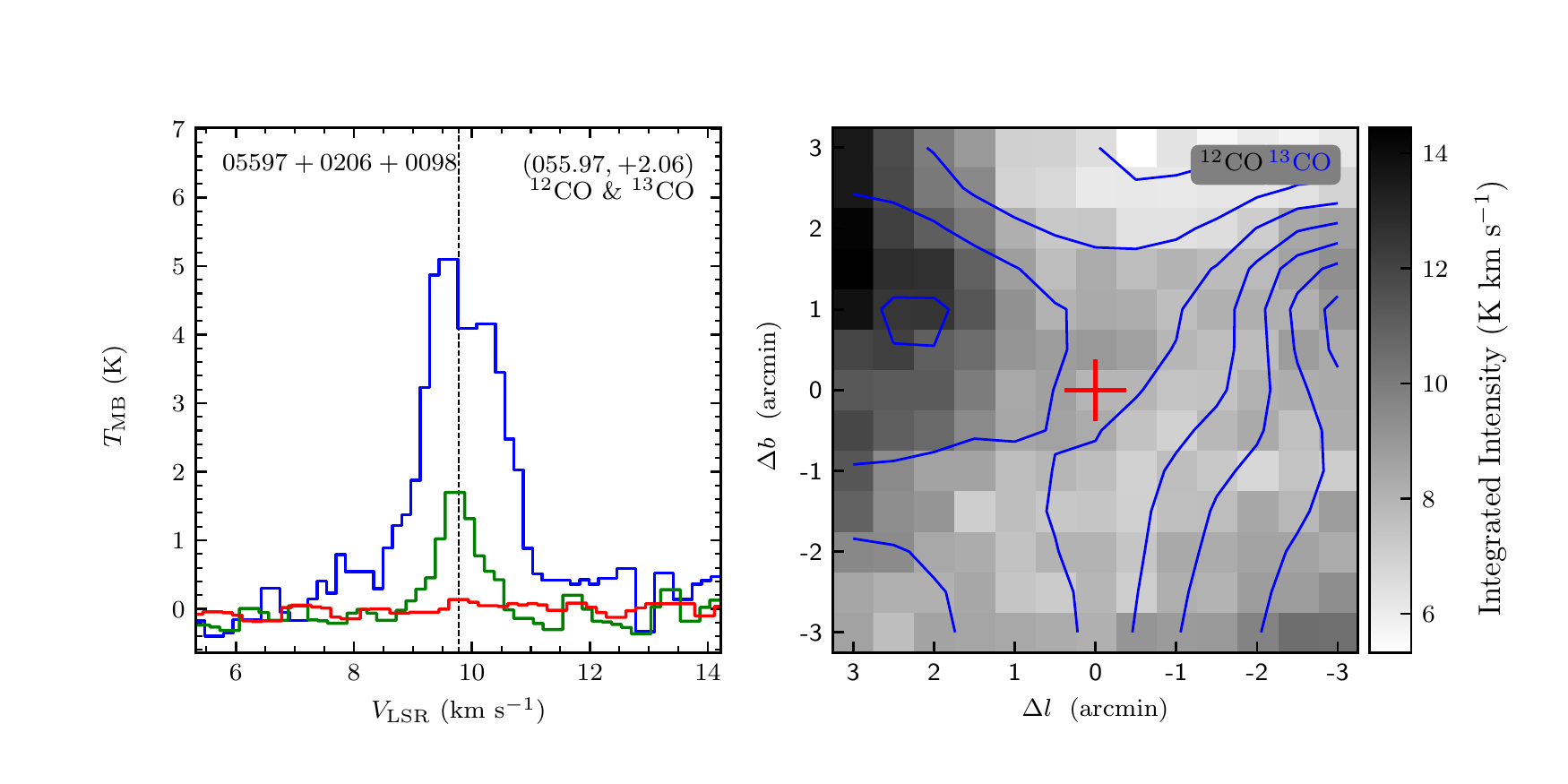}
\end{figure}
\clearpage

\begin{figure}
\includegraphics[width=9.0cm,angle=0]{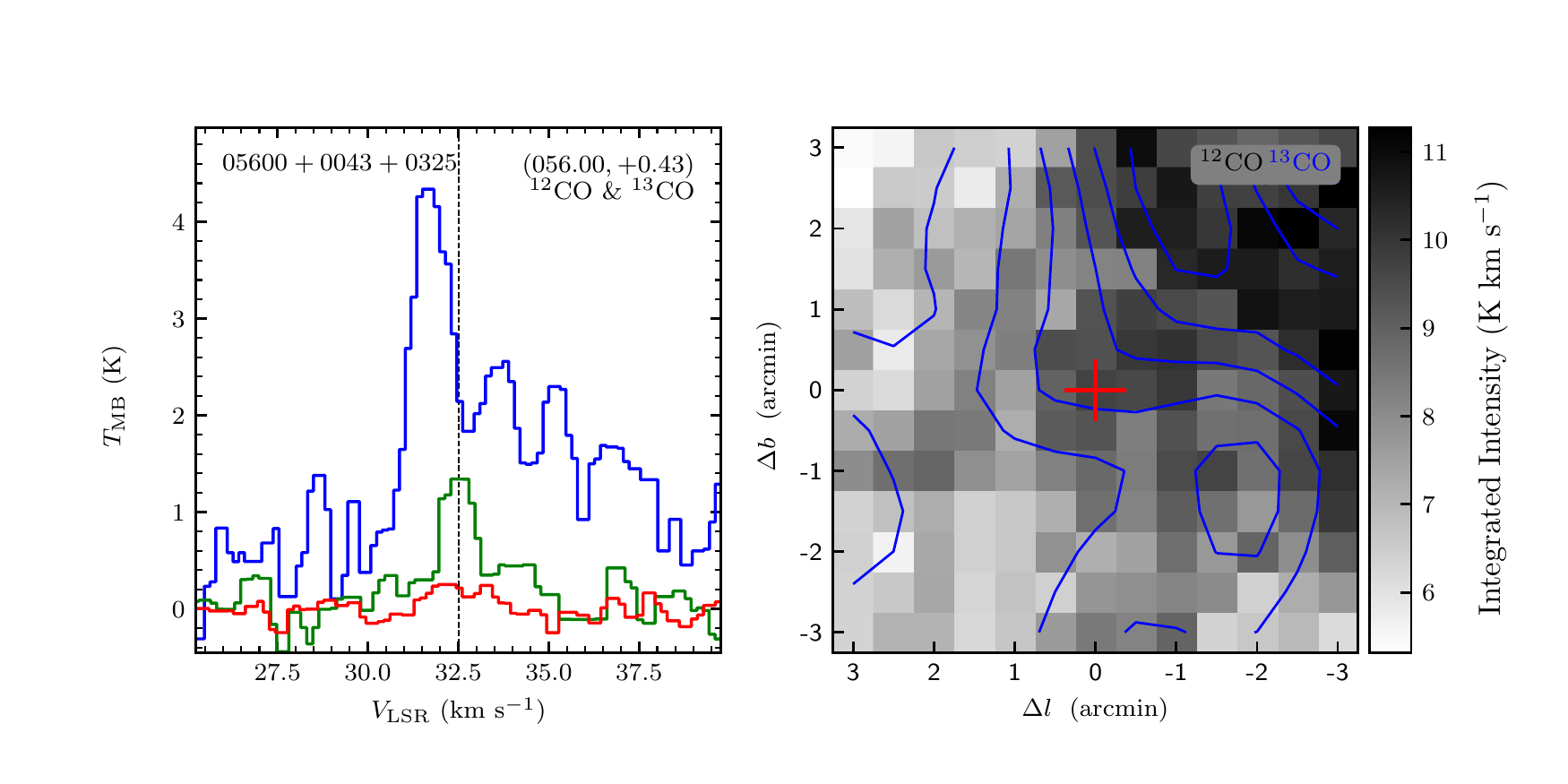}
\includegraphics[width=9.0cm,angle=0]{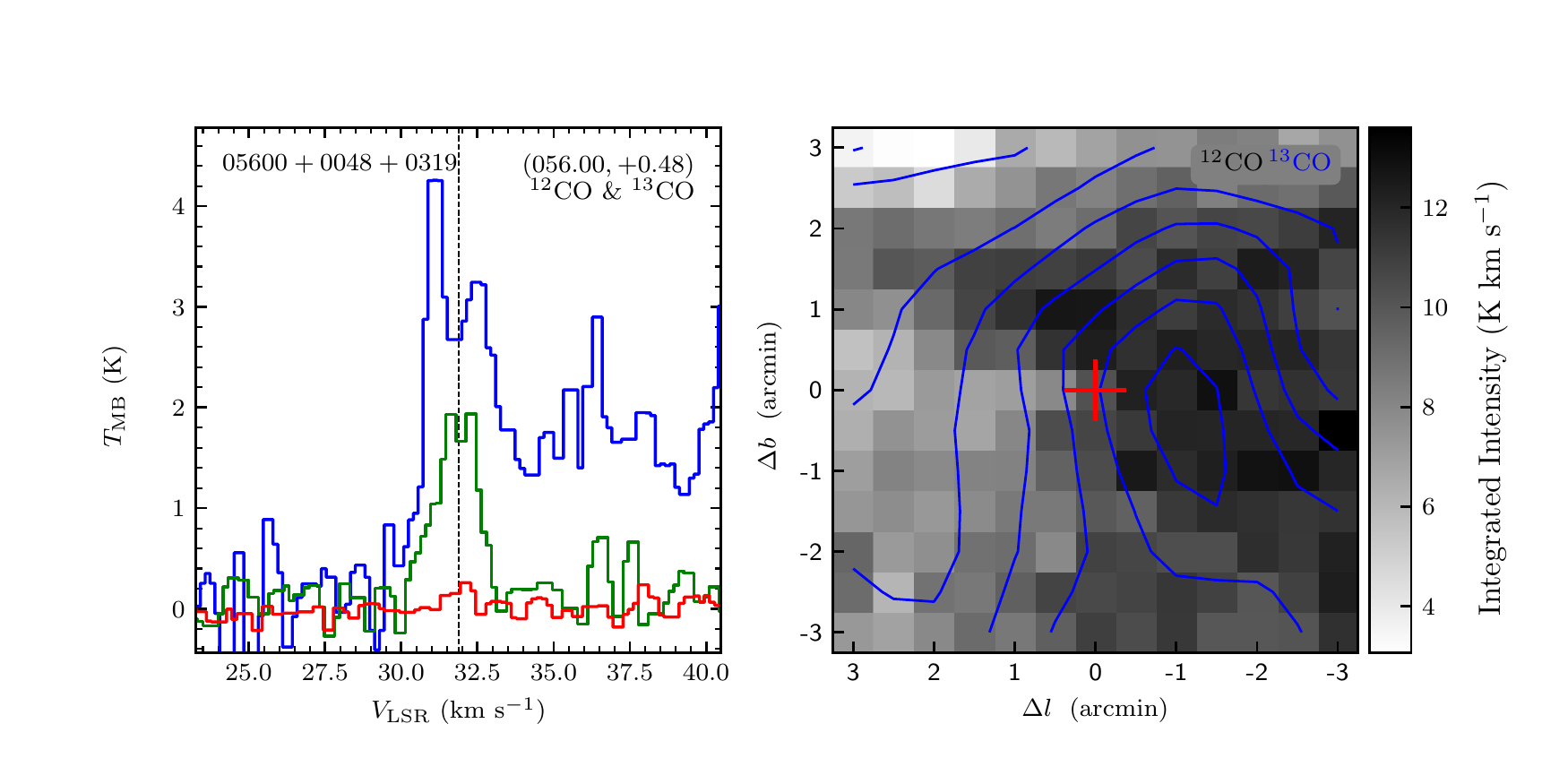}
\vspace{-0.5cm}

\includegraphics[width=9.0cm,angle=0]{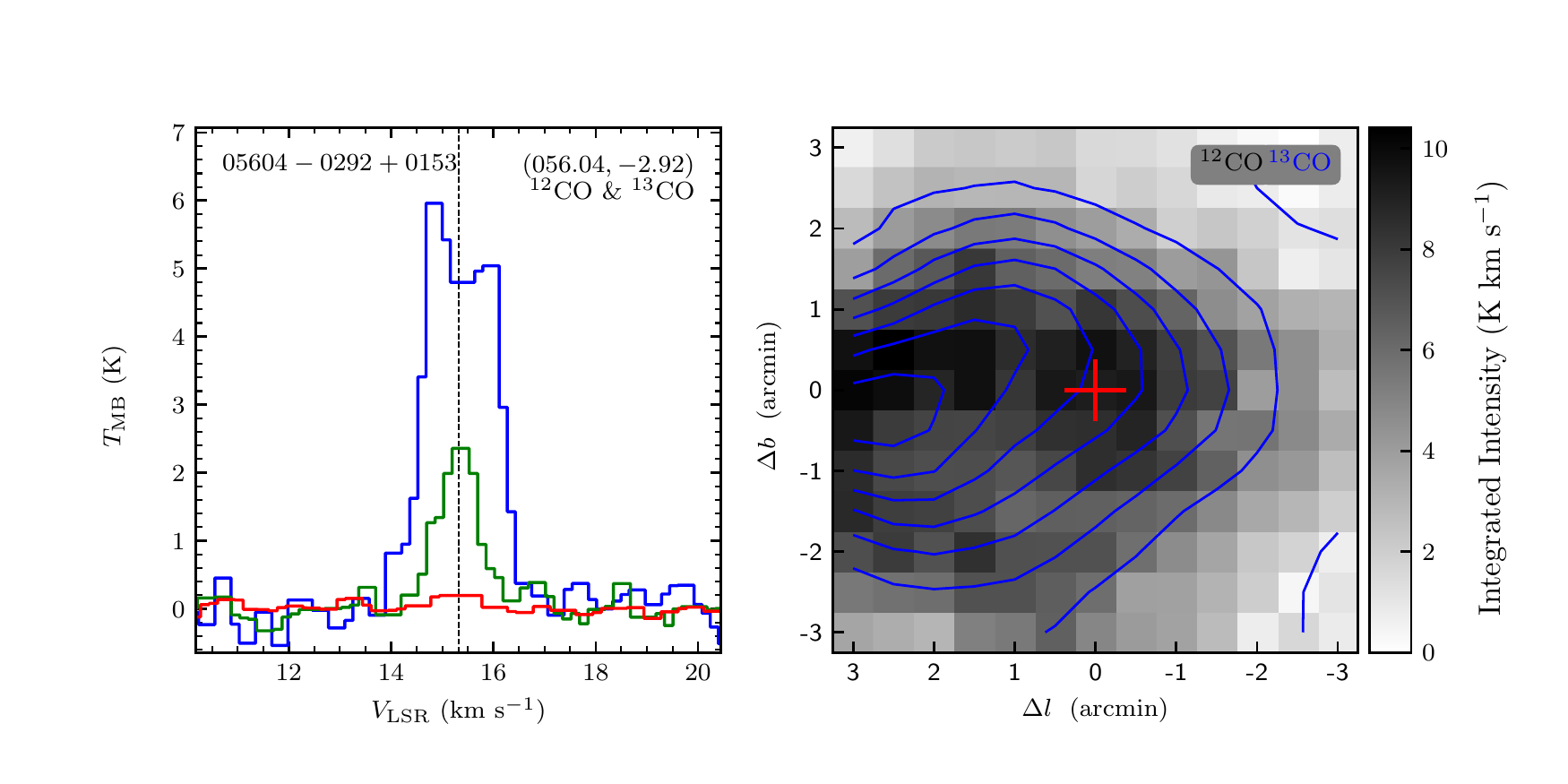}
\includegraphics[width=9.0cm,angle=0]{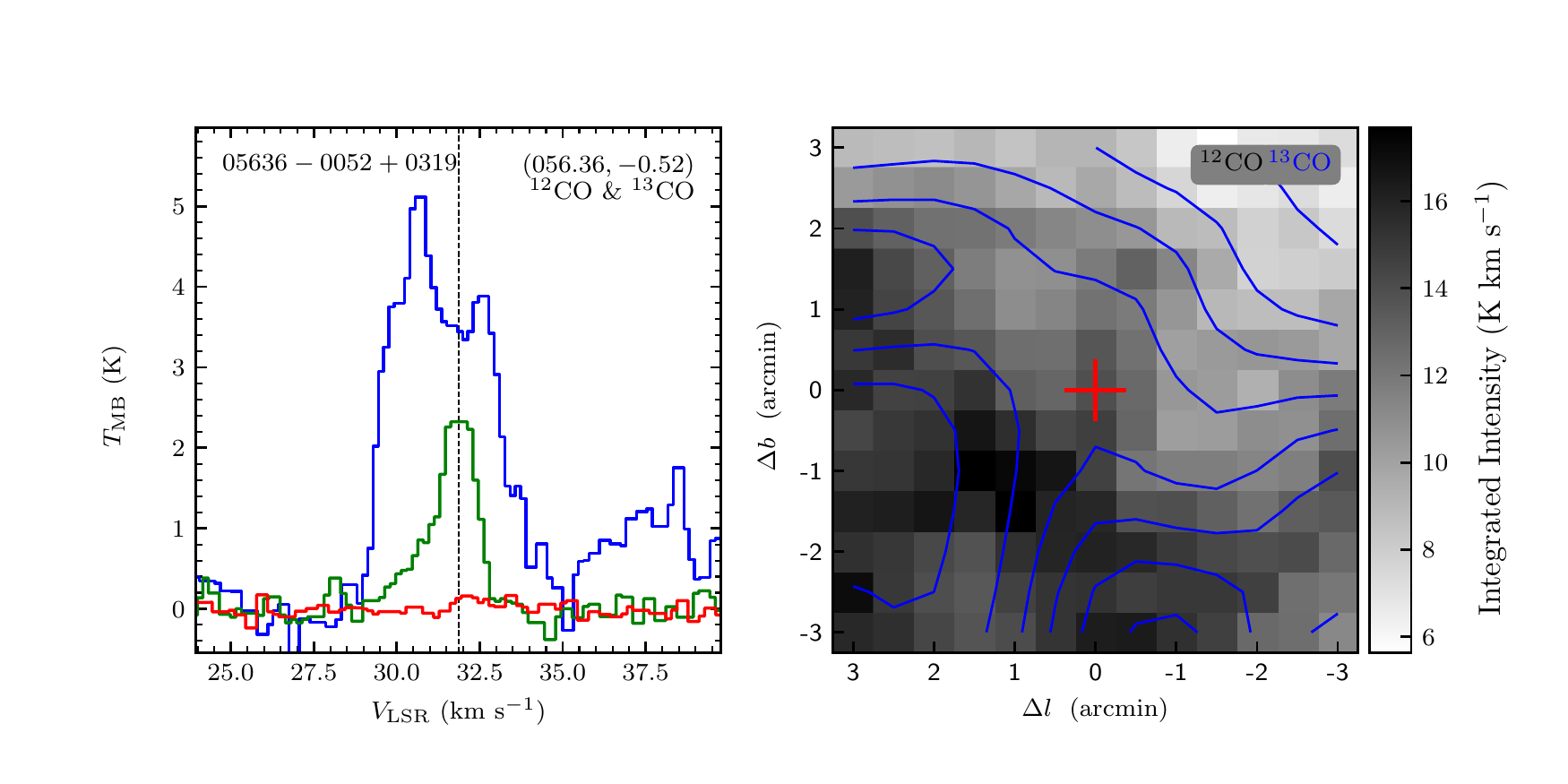}
\vspace{-0.5cm}

\includegraphics[width=9.0cm,angle=0]{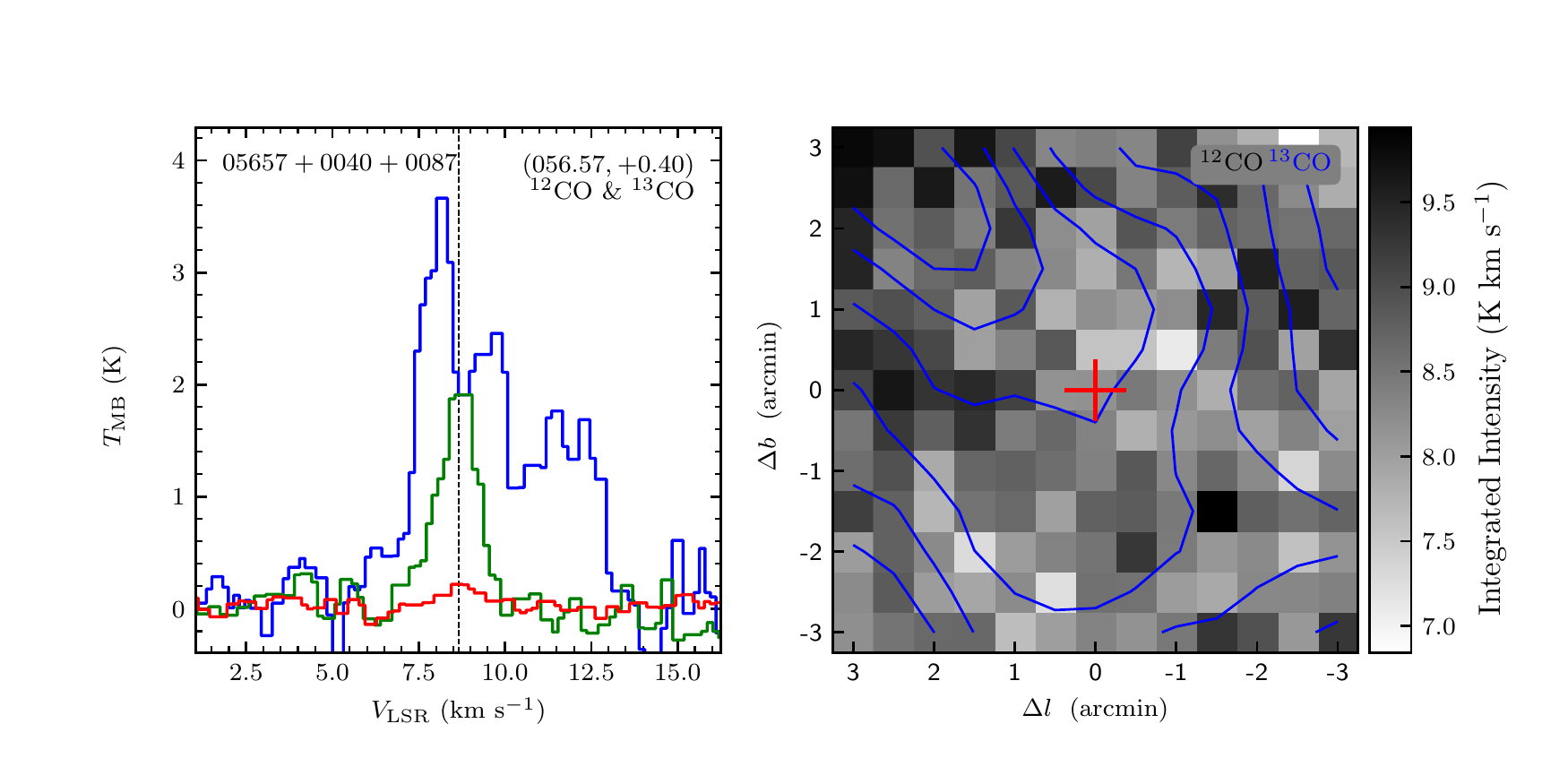}
\includegraphics[width=9.0cm,angle=0]{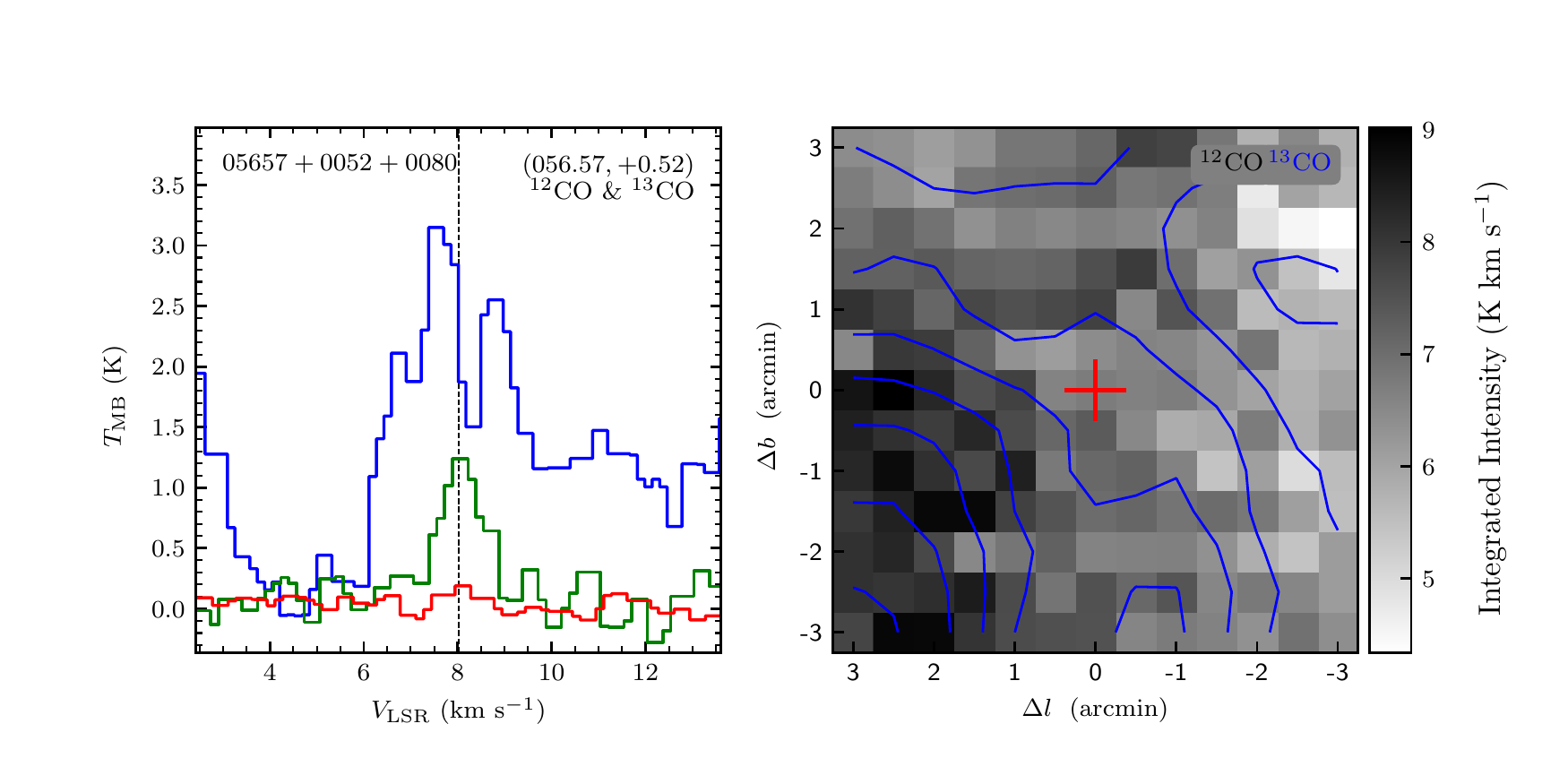}
\vspace{-0.5cm}

\includegraphics[width=9.0cm,angle=0]{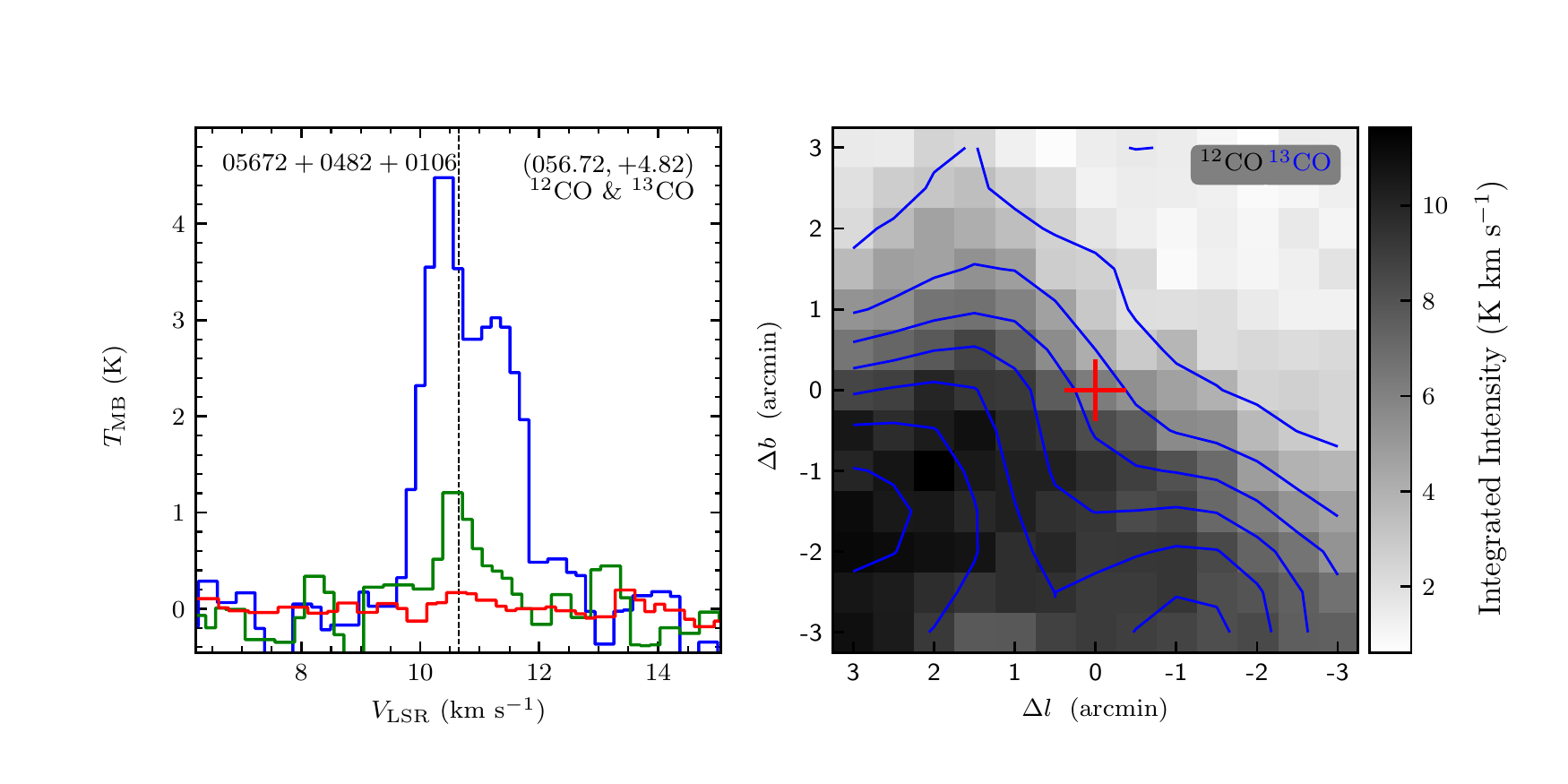}
\includegraphics[width=9.0cm,angle=0]{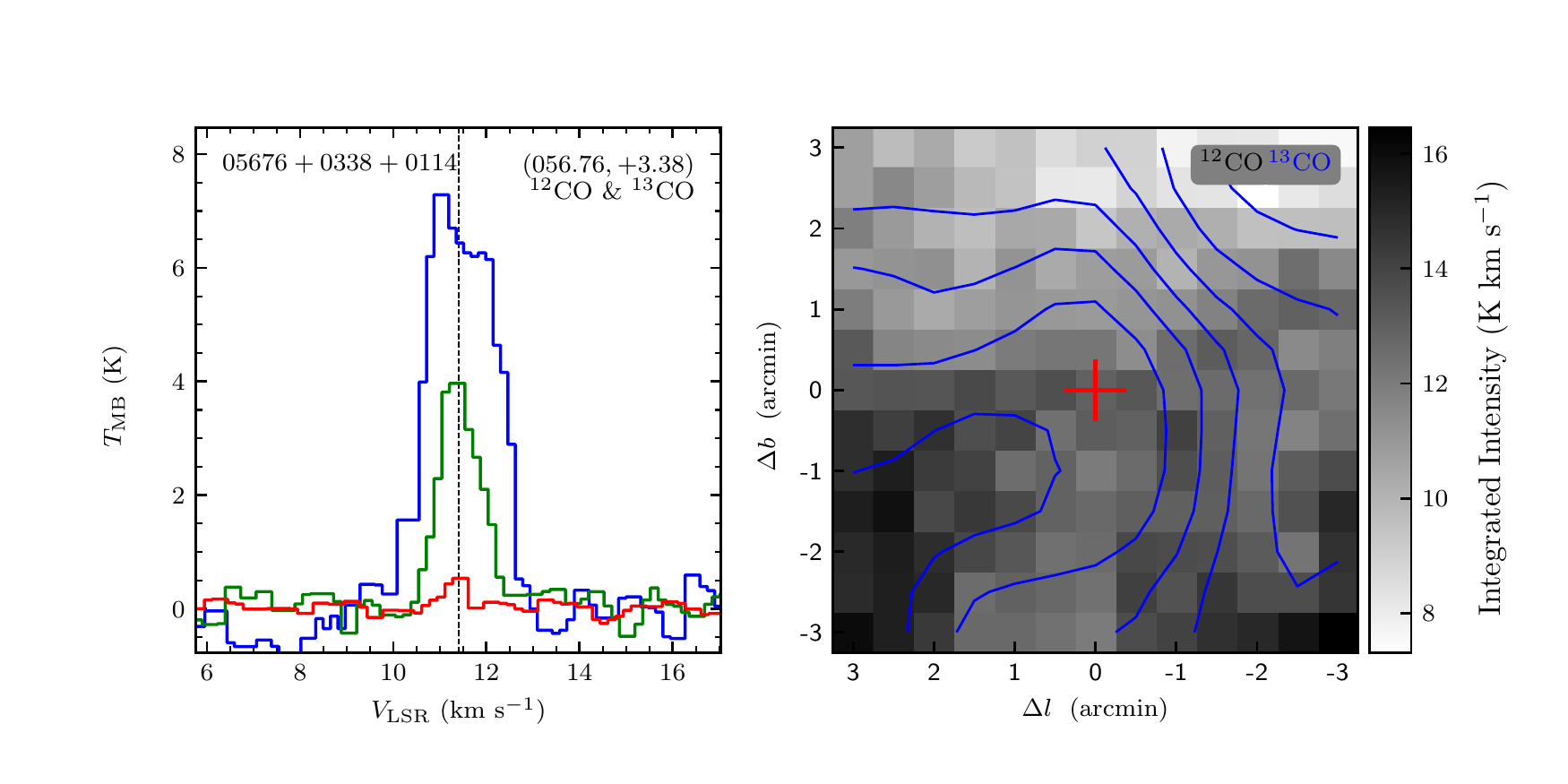}
\vspace{-0.5cm}

\includegraphics[width=9.0cm,angle=0]{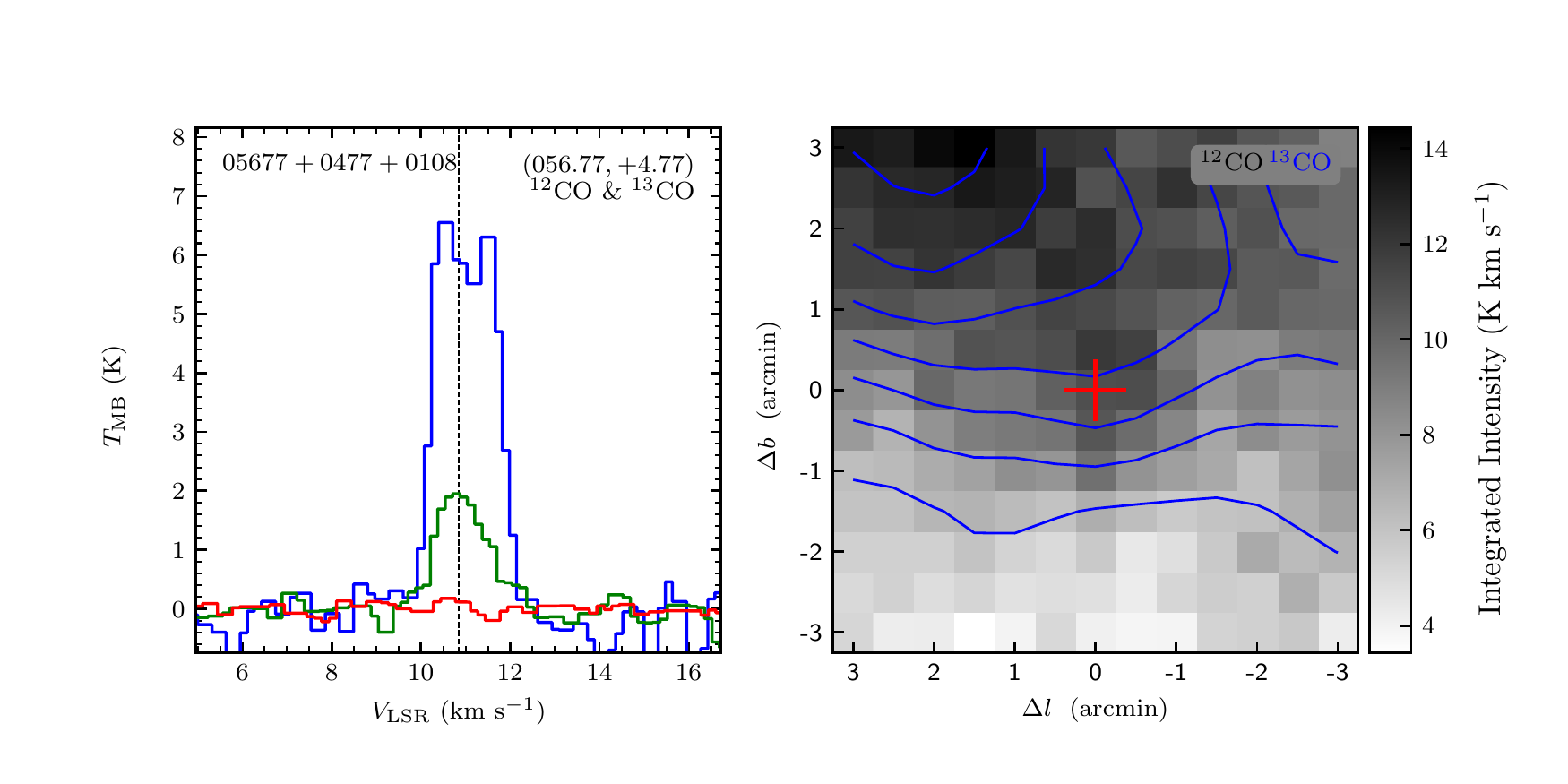}
\includegraphics[width=9.0cm,angle=0]{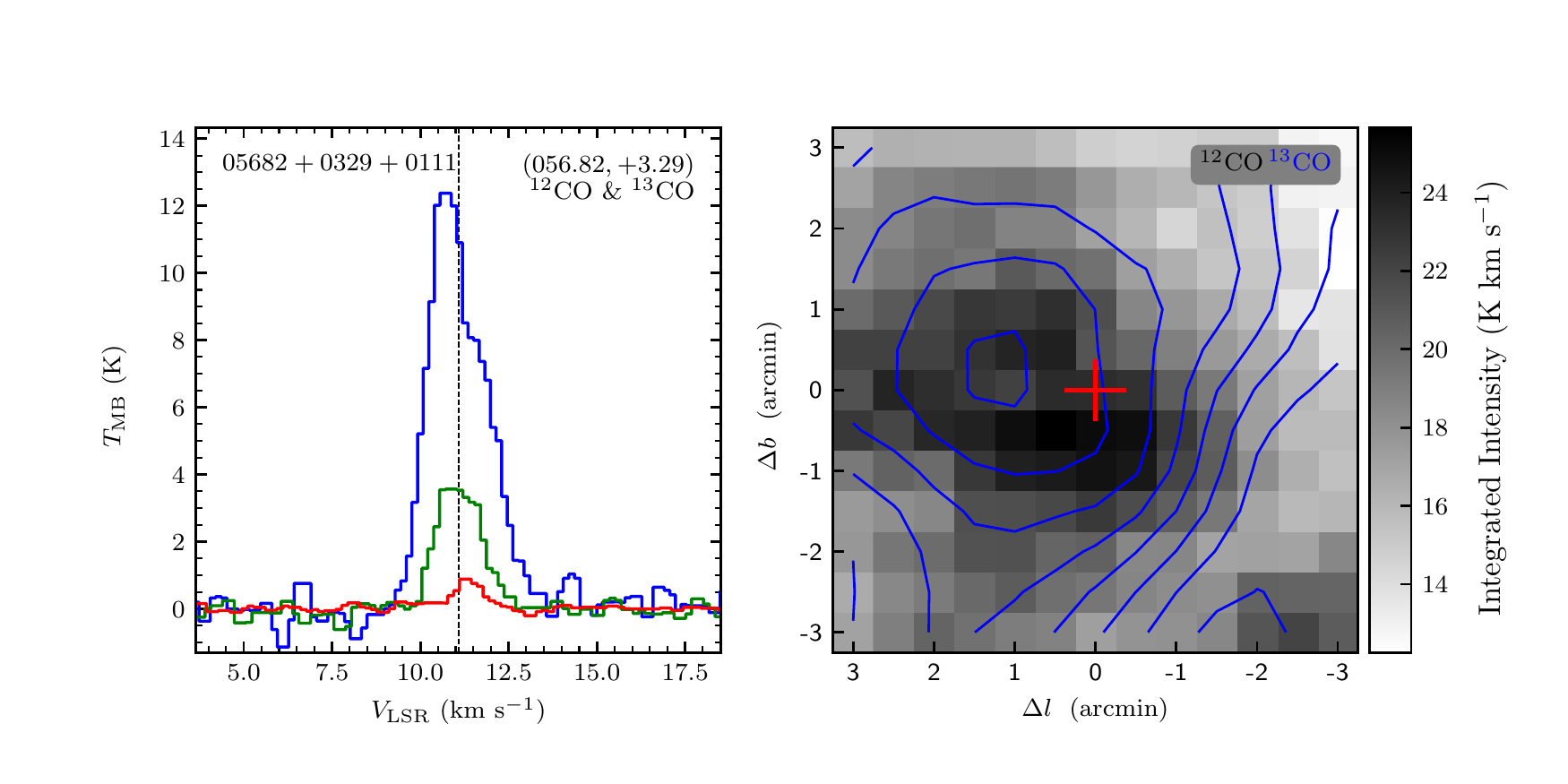}
\end{figure}
\clearpage

\begin{figure}
\includegraphics[width=9.0cm,angle=0]{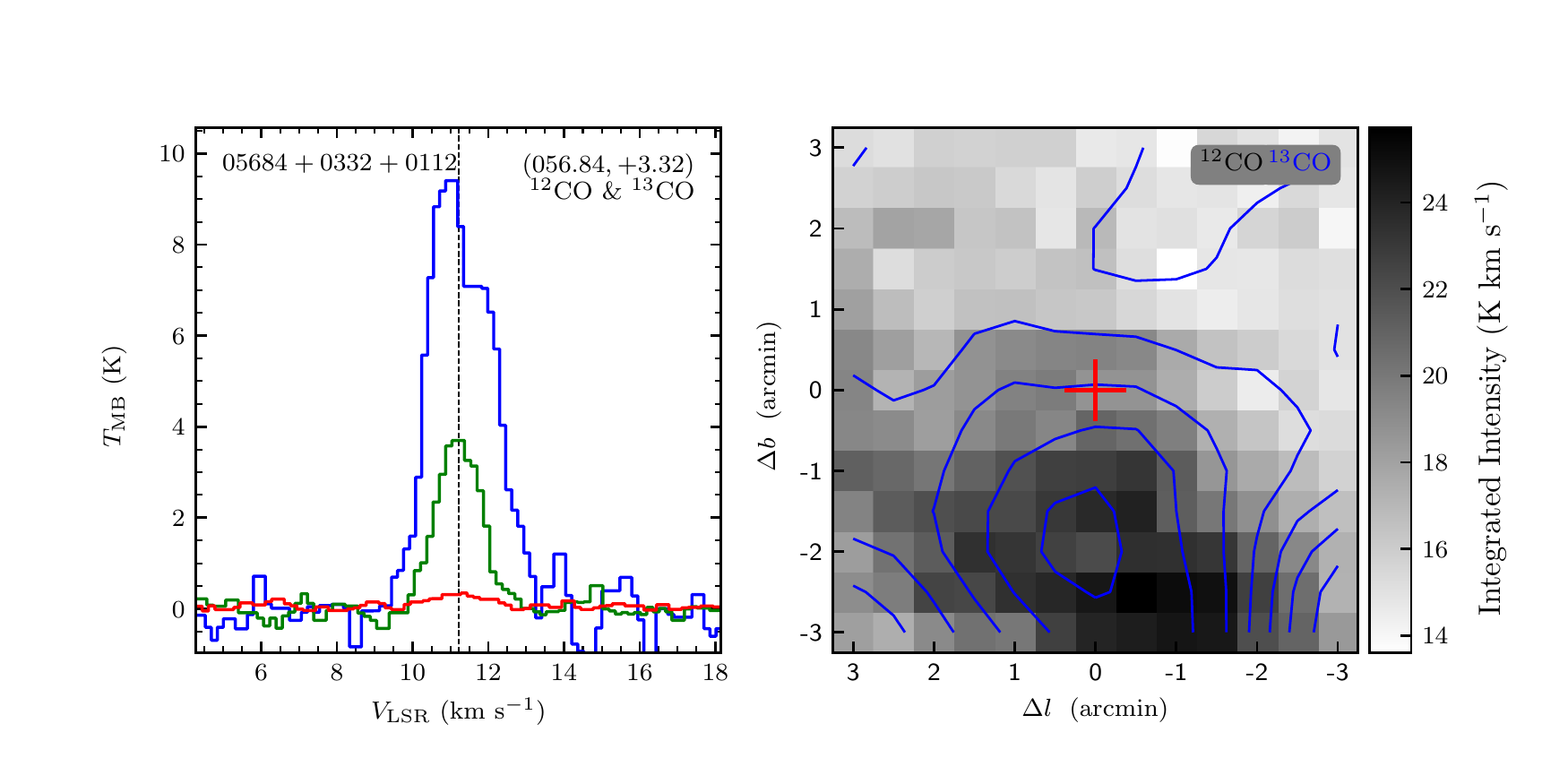}
\includegraphics[width=9.0cm,angle=0]{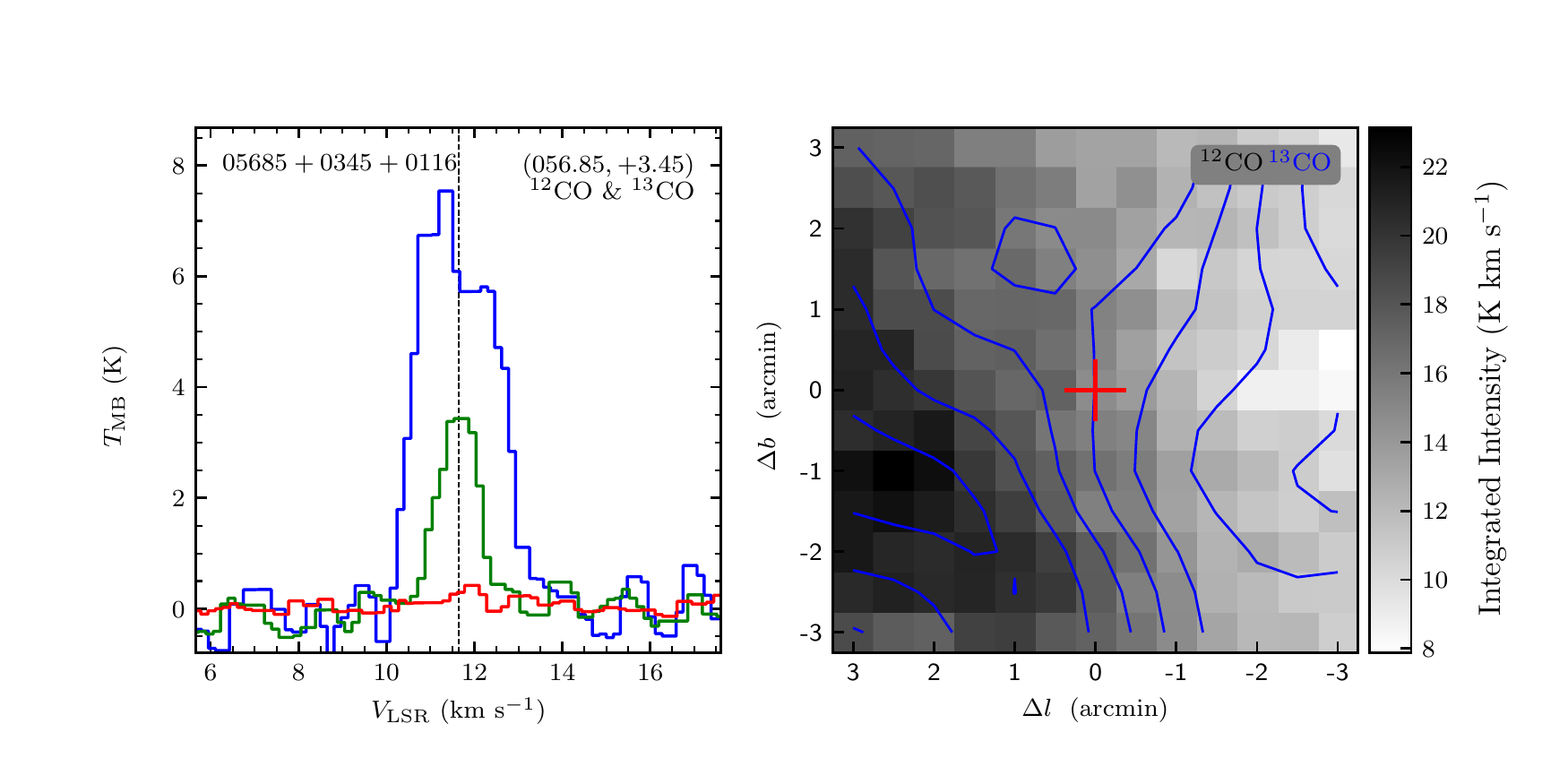}
\vspace{-0.5cm}

\includegraphics[width=9.0cm,angle=0]{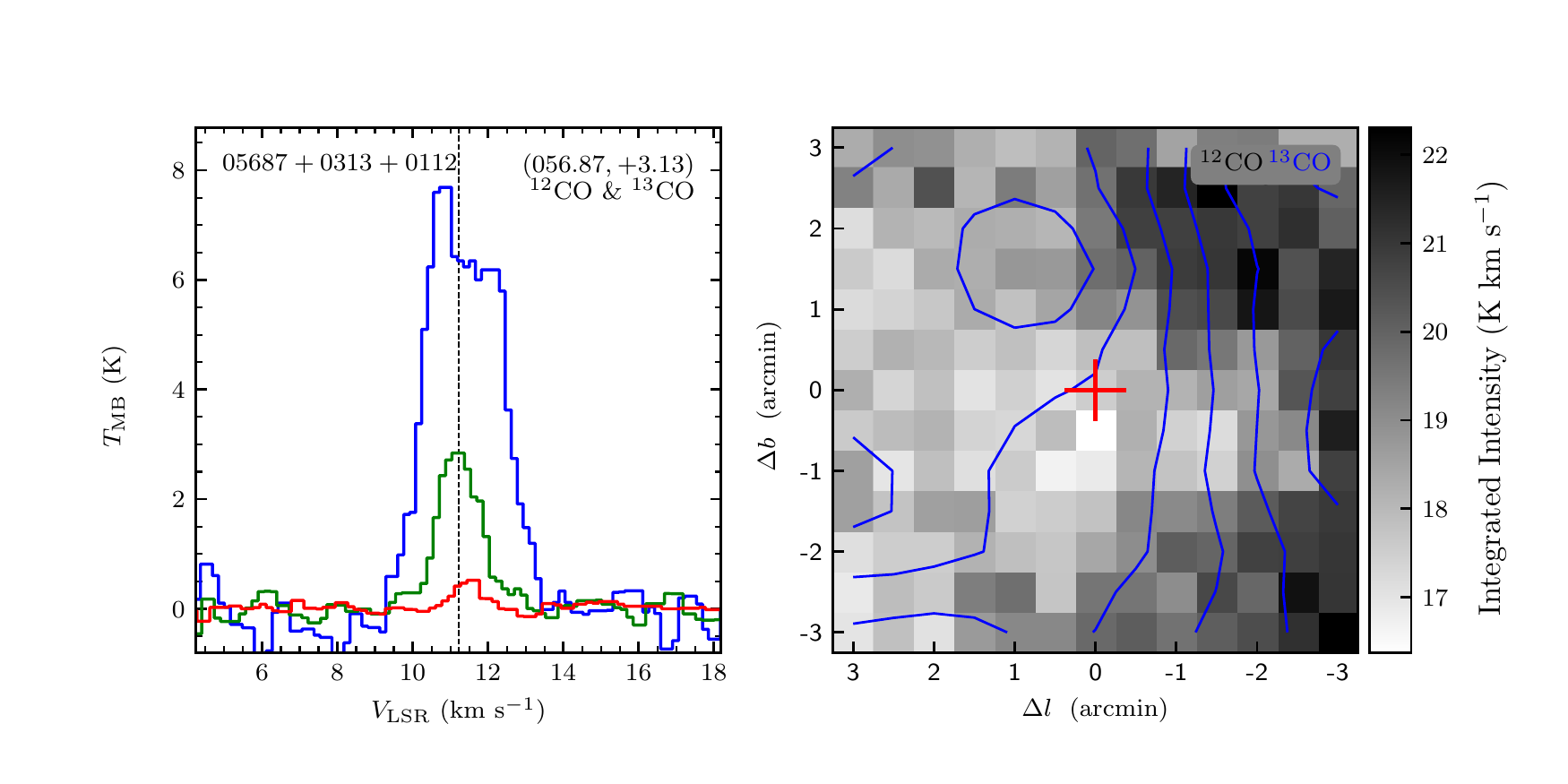}
\includegraphics[width=9.0cm,angle=0]{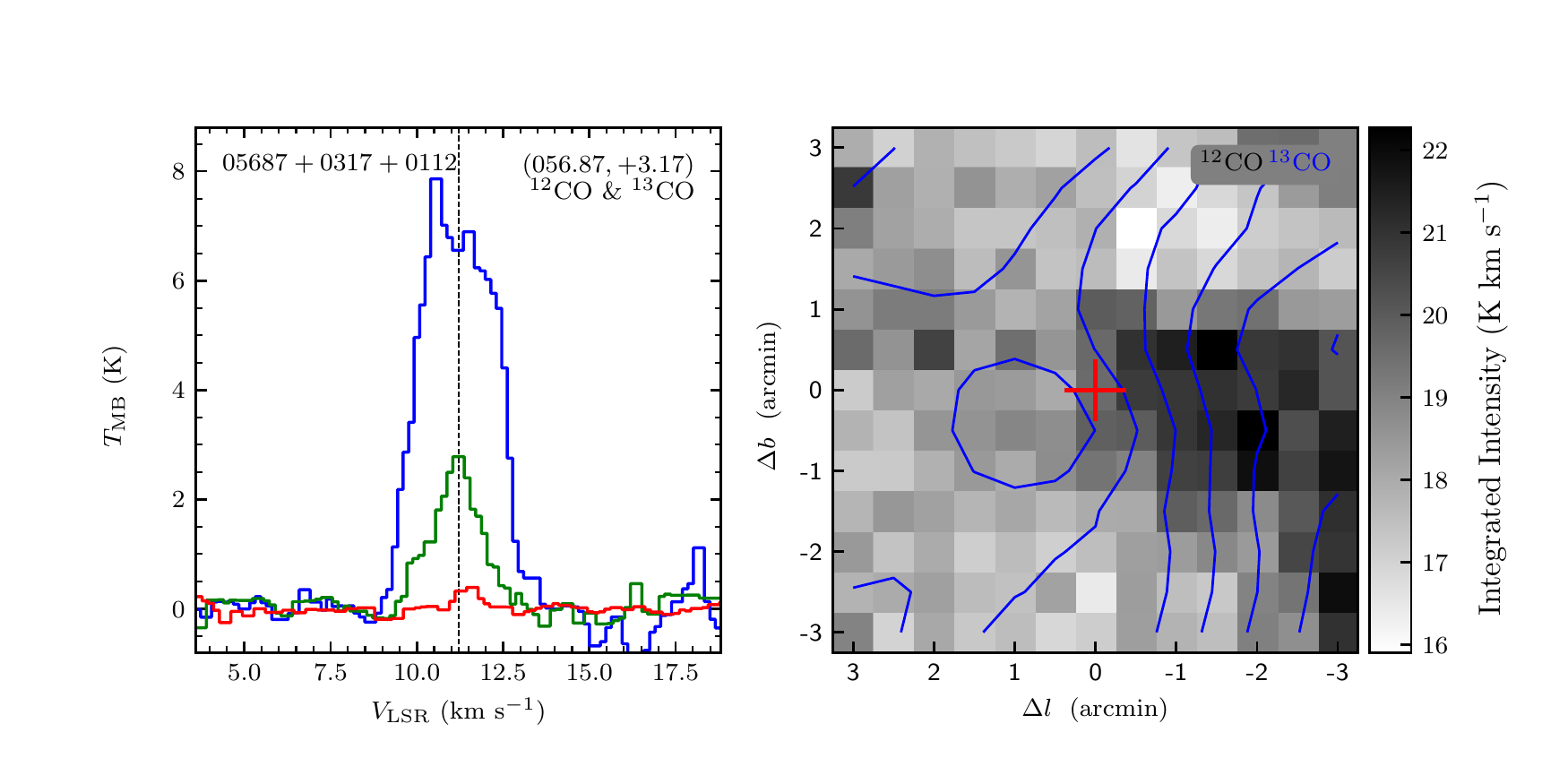}
\vspace{-0.5cm}

\includegraphics[width=9.0cm,angle=0]{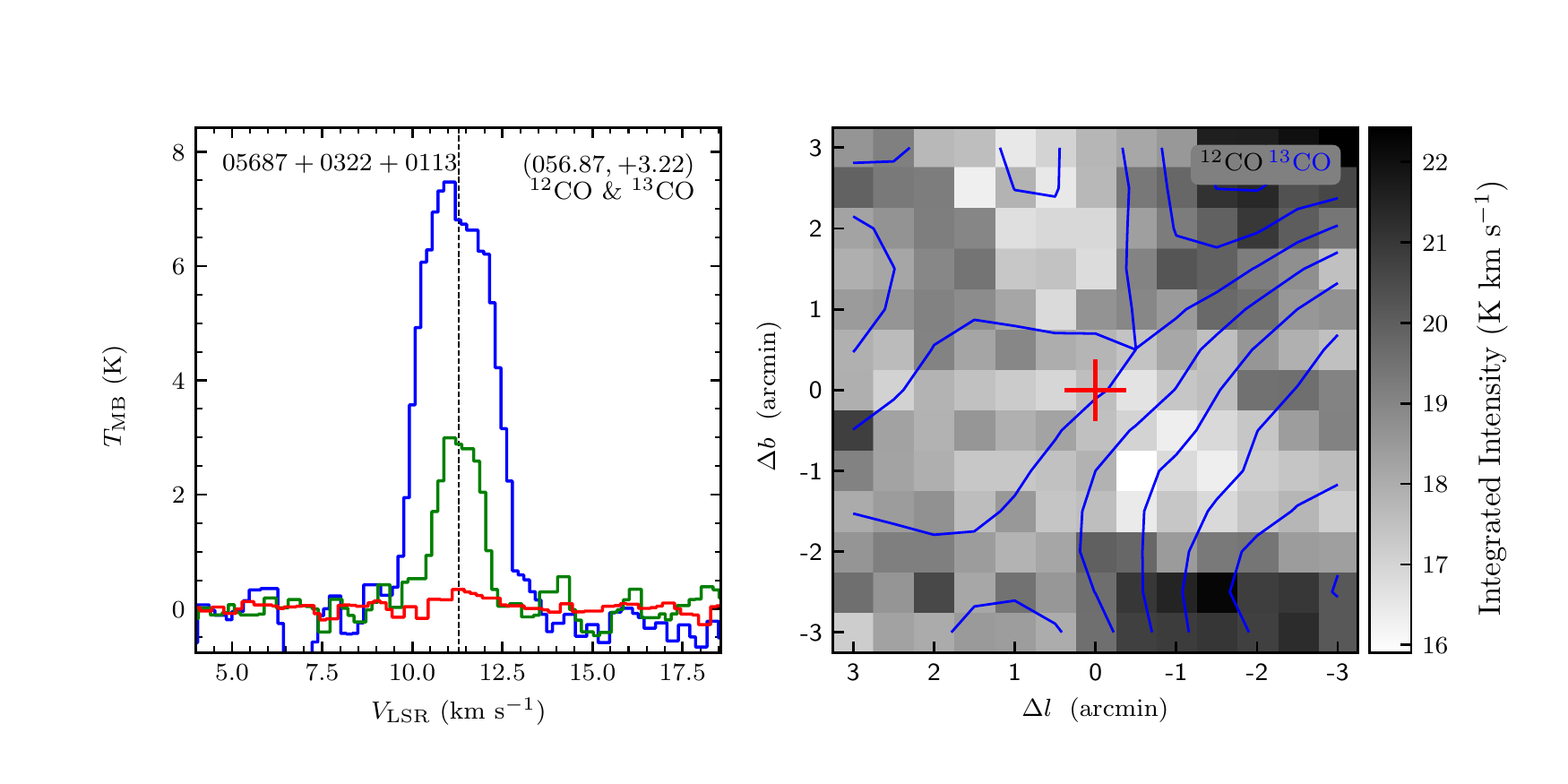}
\includegraphics[width=9.0cm,angle=0]{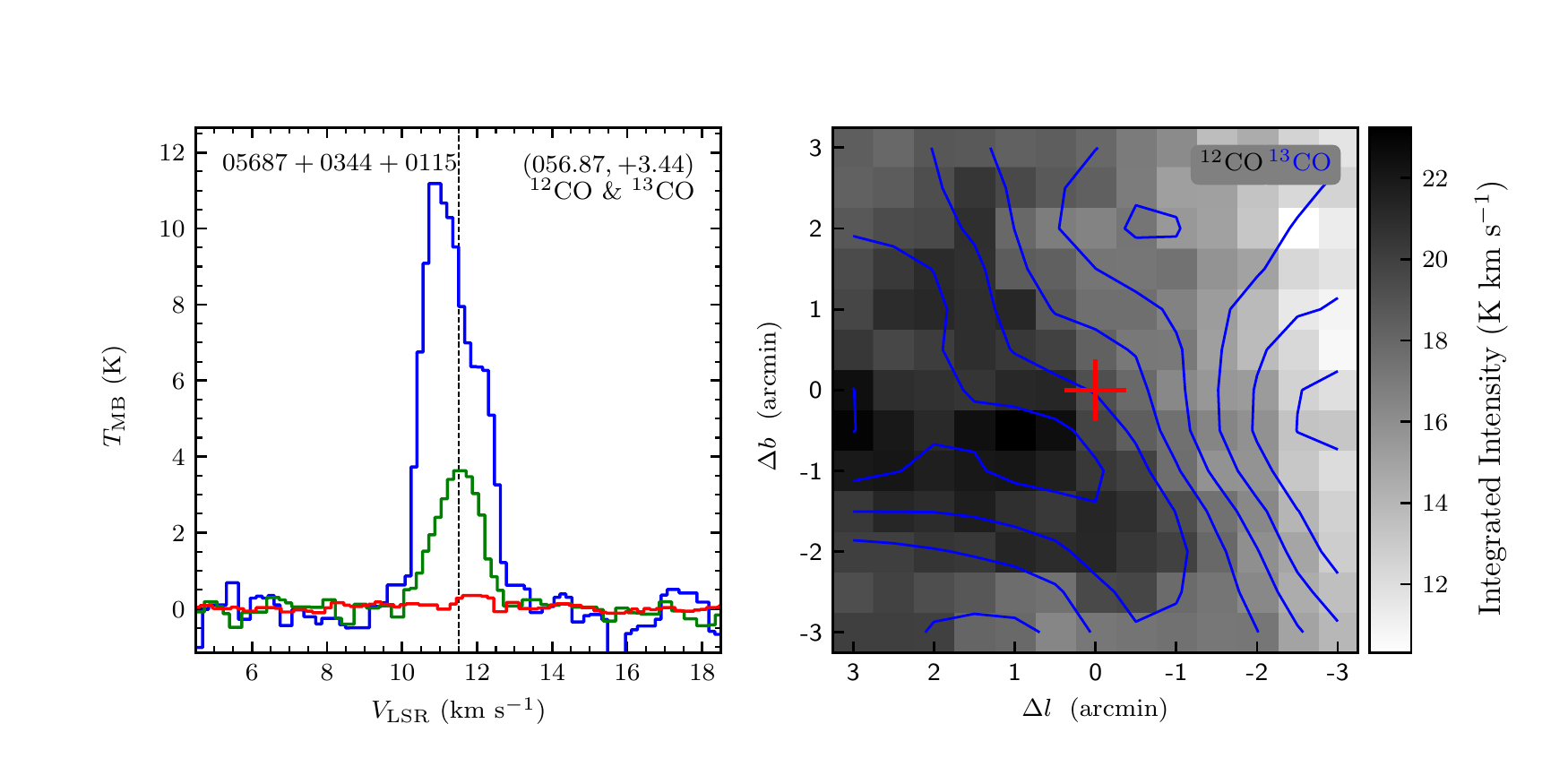}
\vspace{-0.5cm}

\includegraphics[width=9.0cm,angle=0]{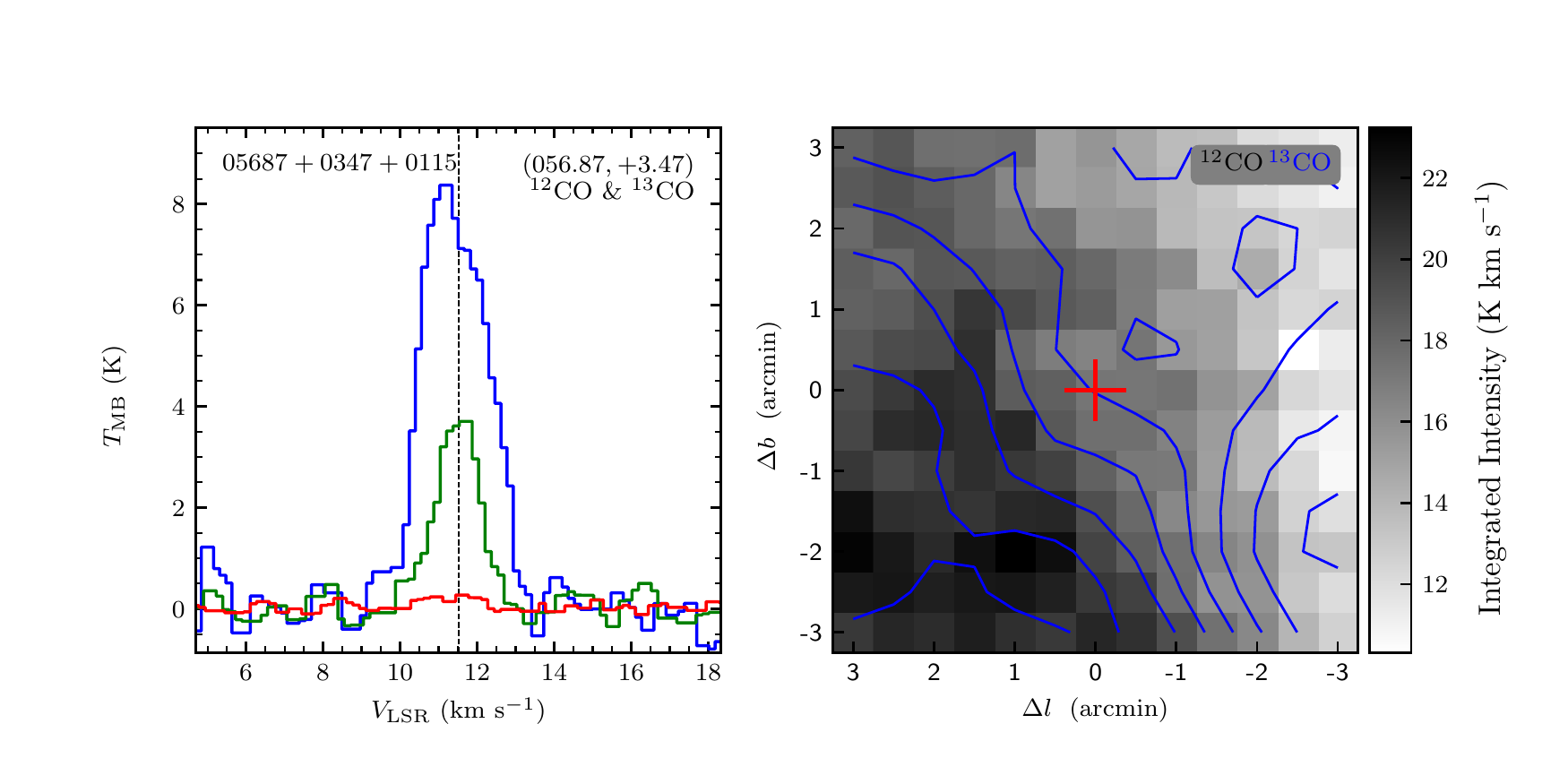}
\includegraphics[width=9.0cm,angle=0]{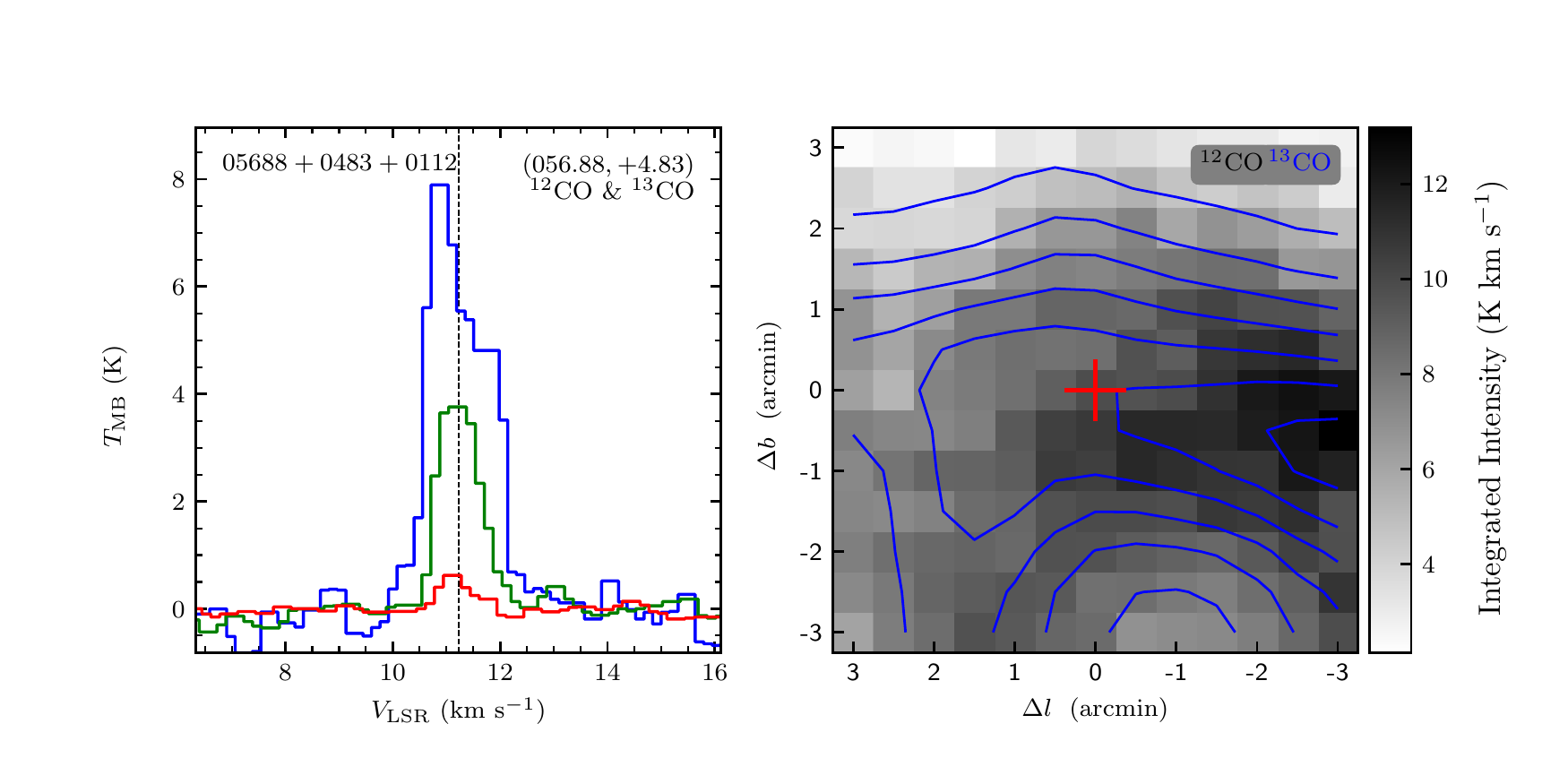}
\vspace{-0.5cm}

\includegraphics[width=9.0cm,angle=0]{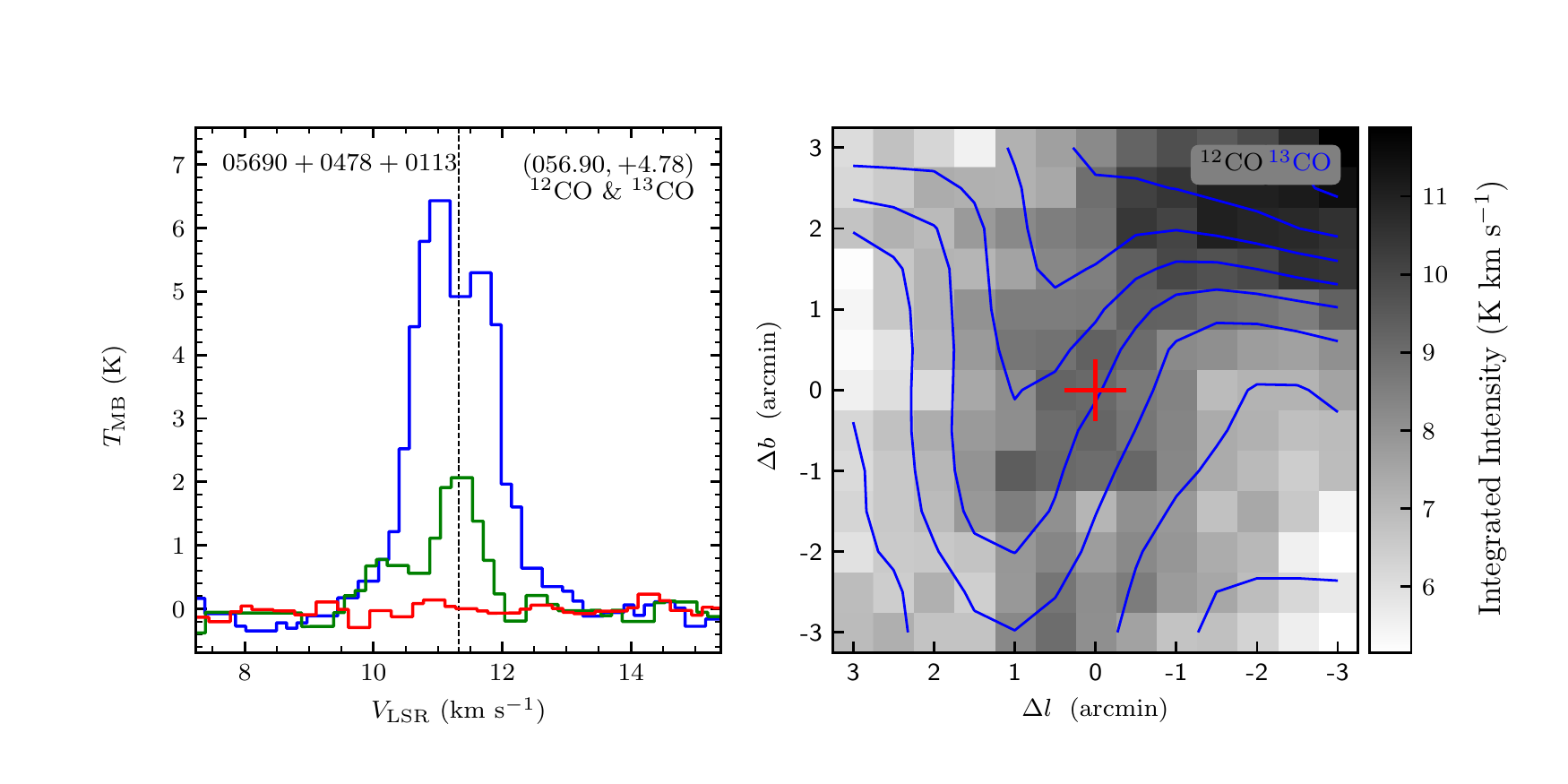}
\includegraphics[width=9.0cm,angle=0]{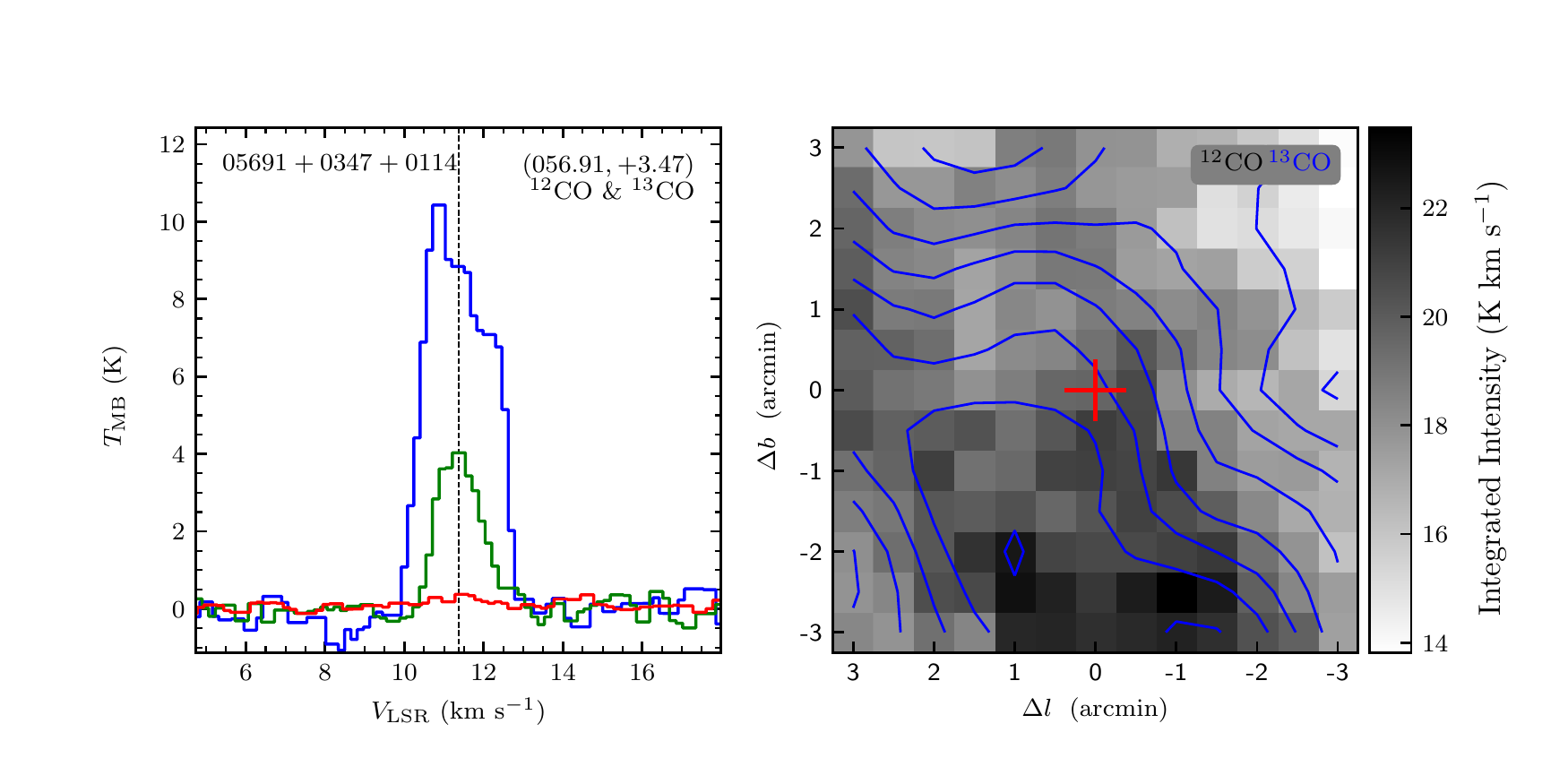}
\end{figure}
\clearpage

\begin{figure}
\includegraphics[width=9.0cm,angle=0]{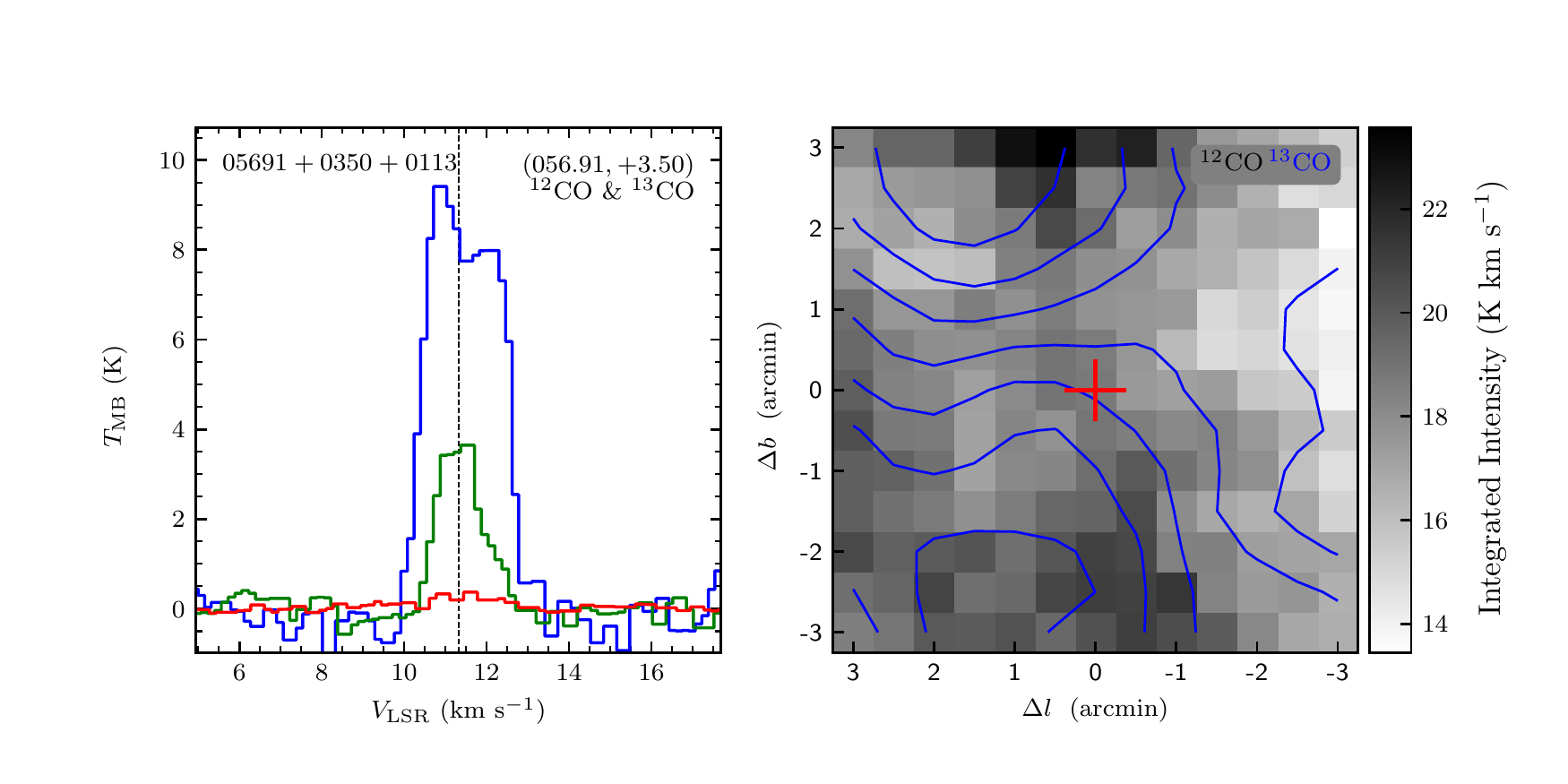}
\includegraphics[width=9.0cm,angle=0]{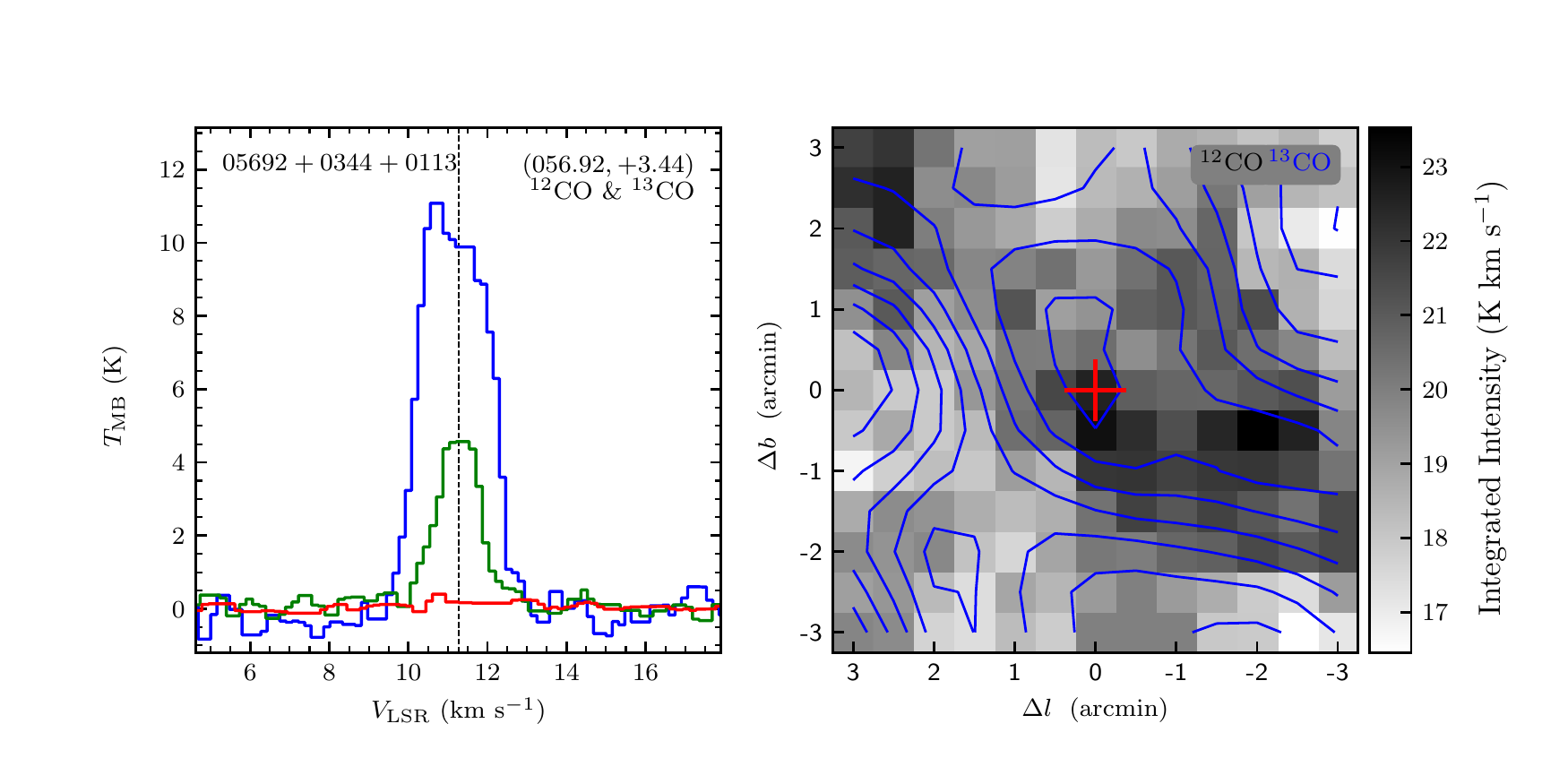}
\vspace{-0.5cm}

\includegraphics[width=9.0cm,angle=0]{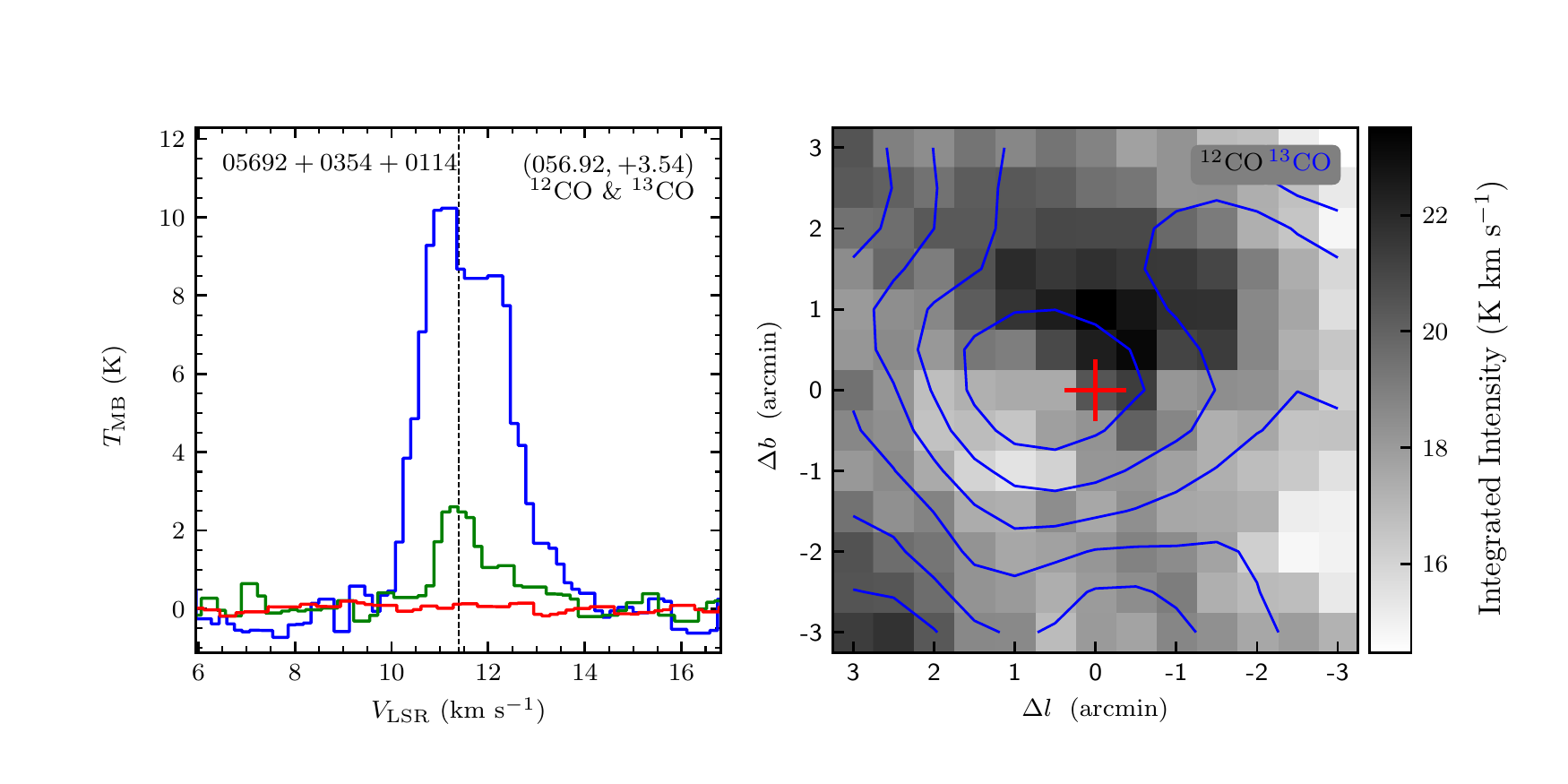}
\includegraphics[width=9.0cm,angle=0]{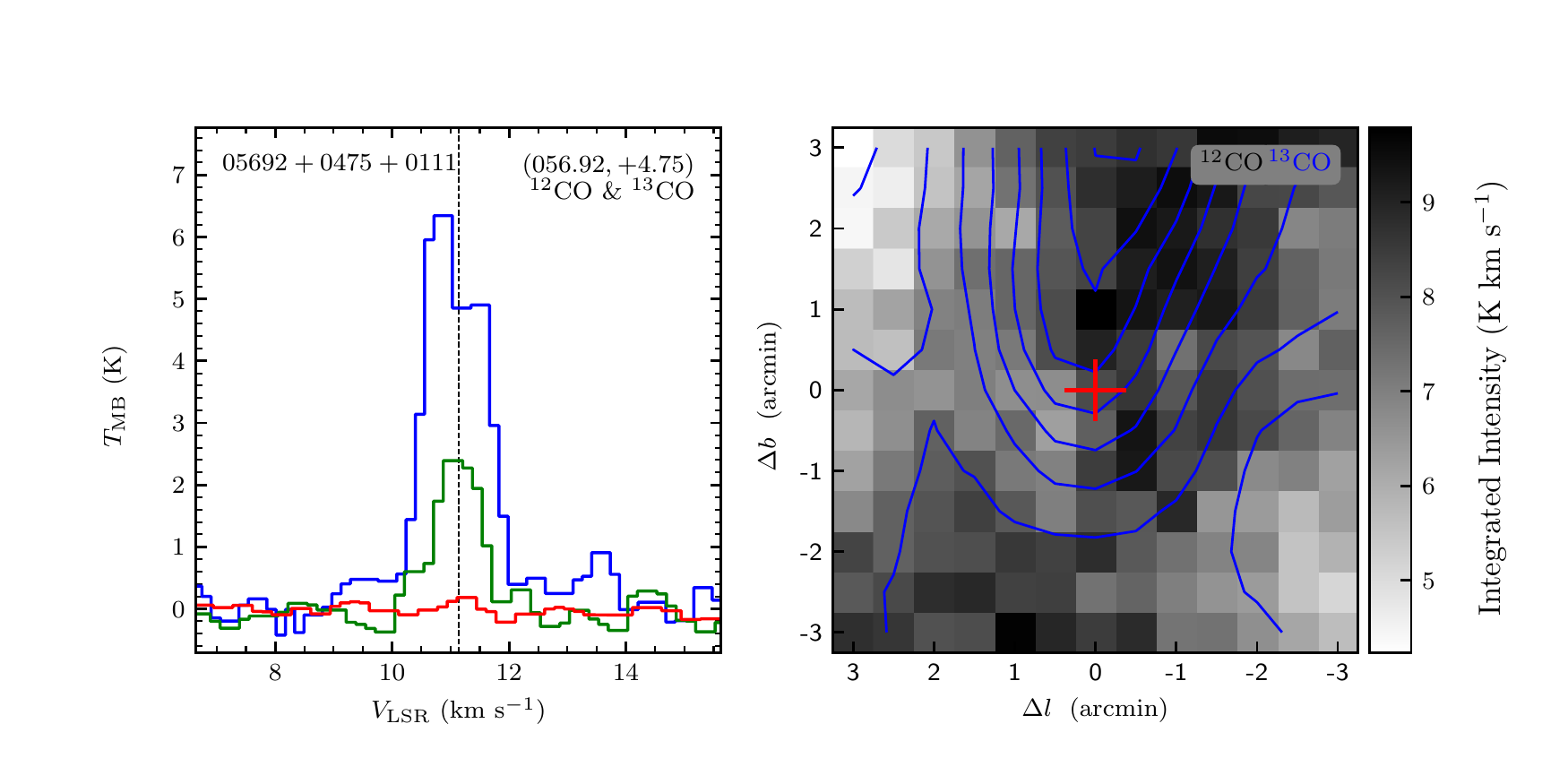}
\vspace{-0.5cm}

\includegraphics[width=9.0cm,angle=0]{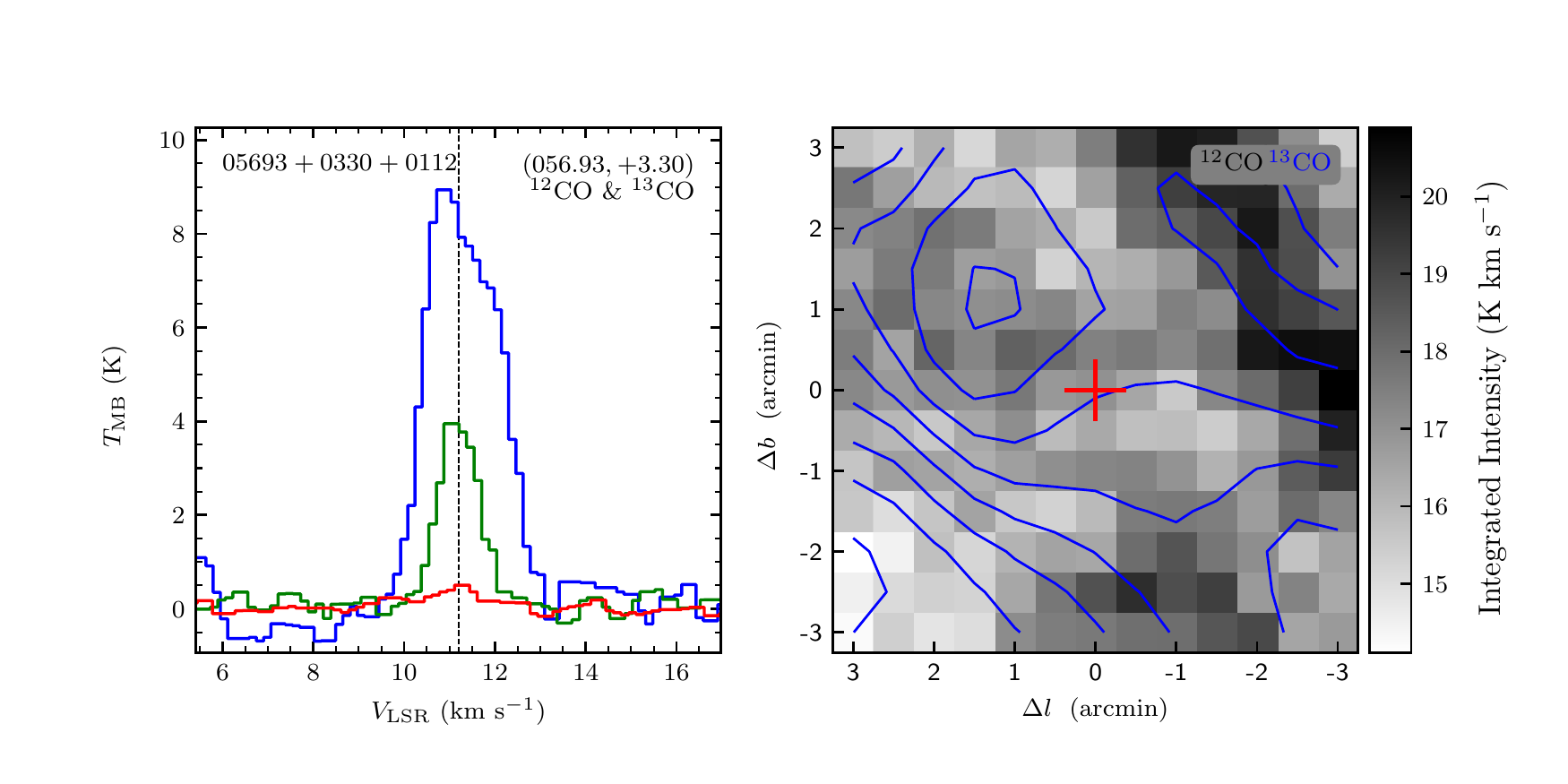}
\includegraphics[width=9.0cm,angle=0]{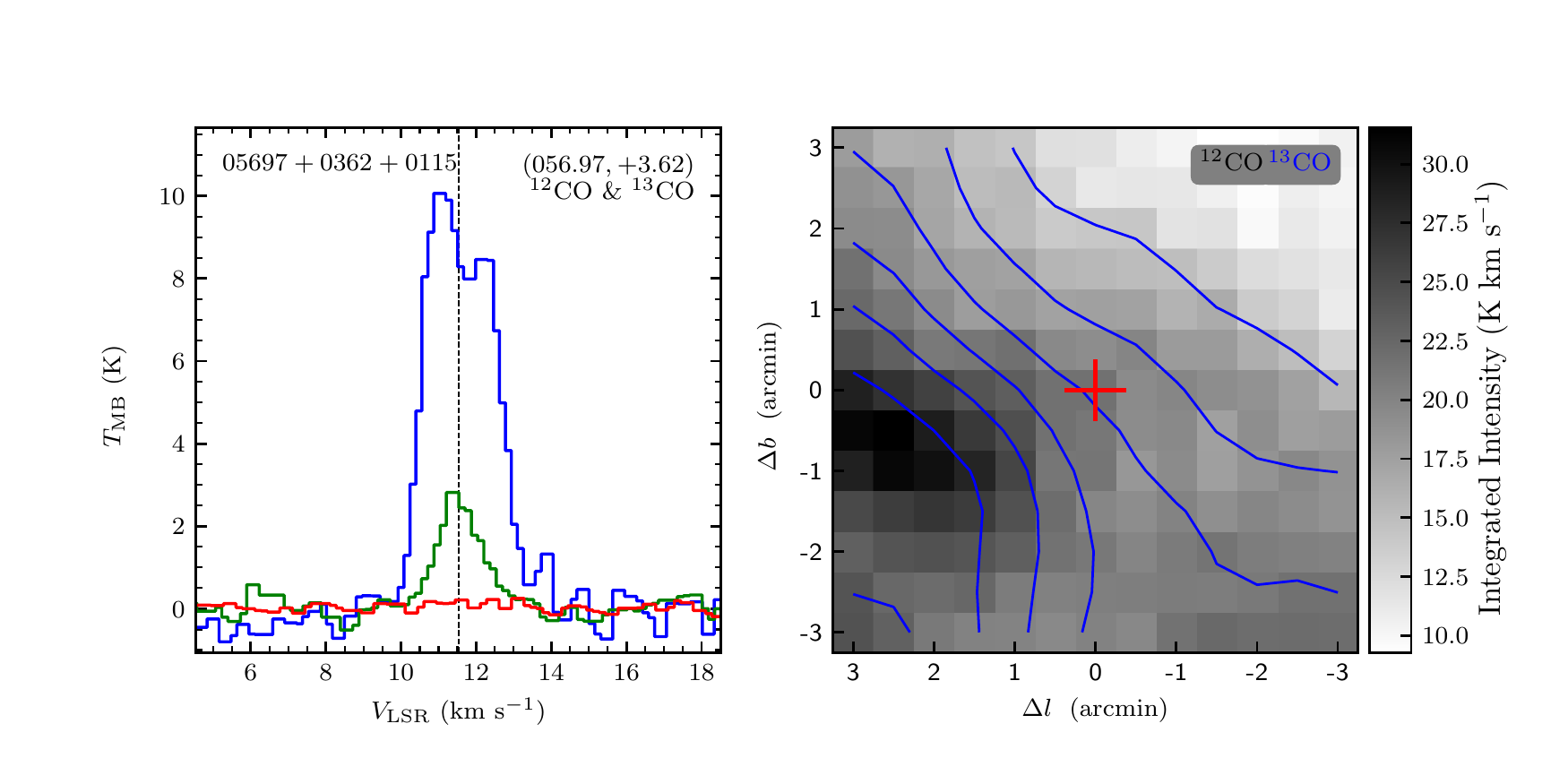}
\vspace{-0.5cm}

\includegraphics[width=9.0cm,angle=0]{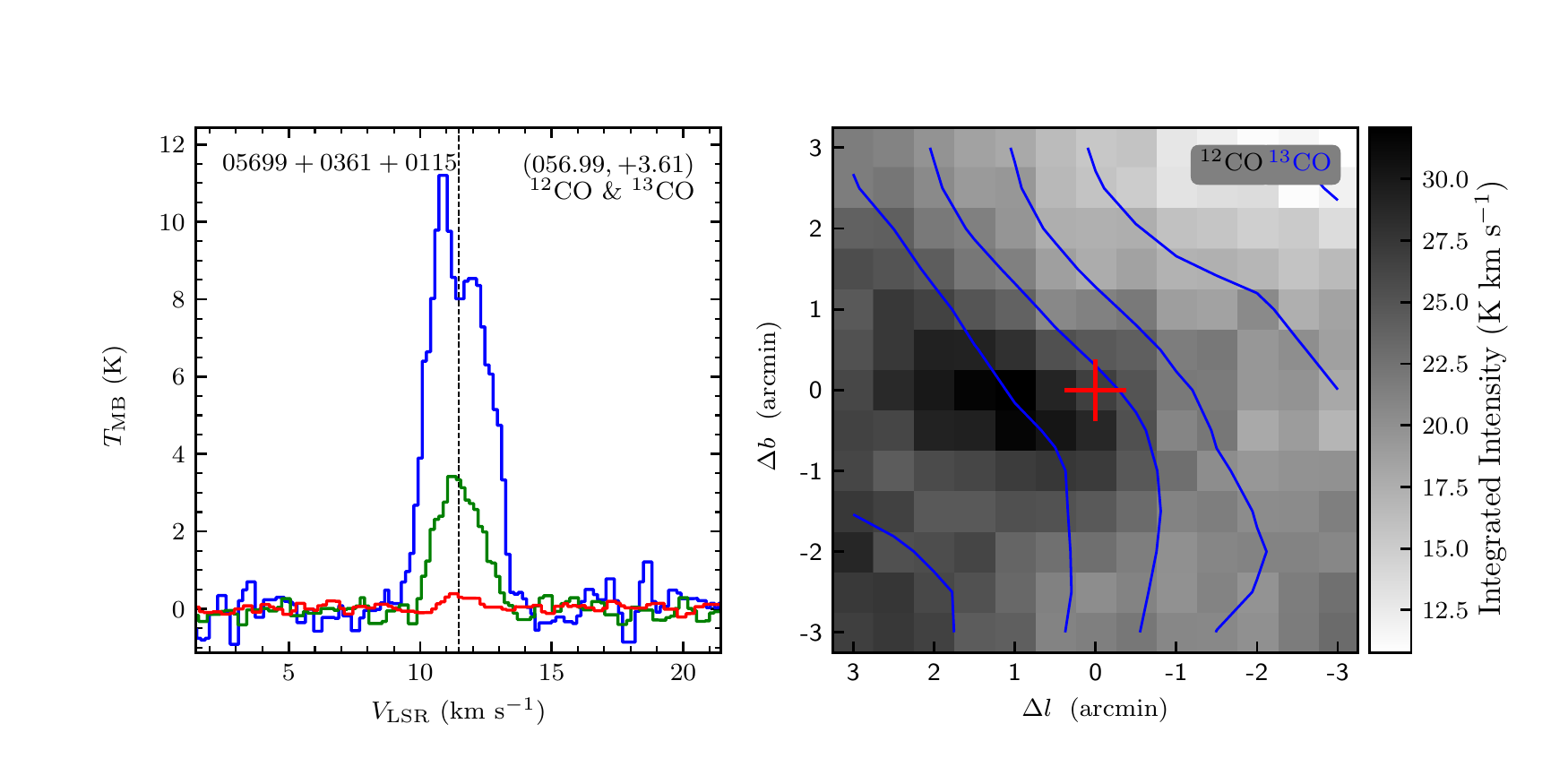}
\includegraphics[width=9.0cm,angle=0]{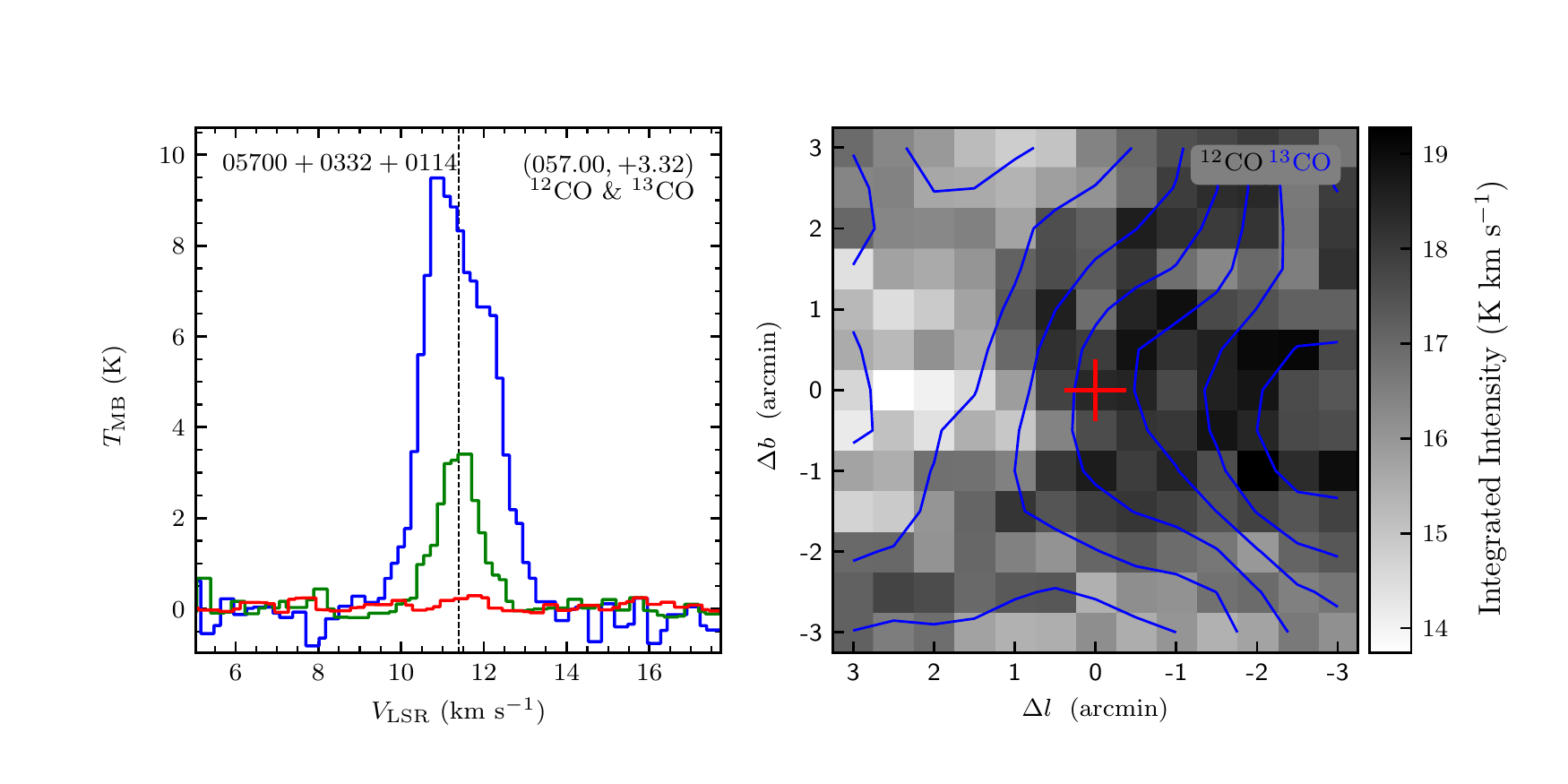}
\vspace{-0.5cm}

\includegraphics[width=9.0cm,angle=0]{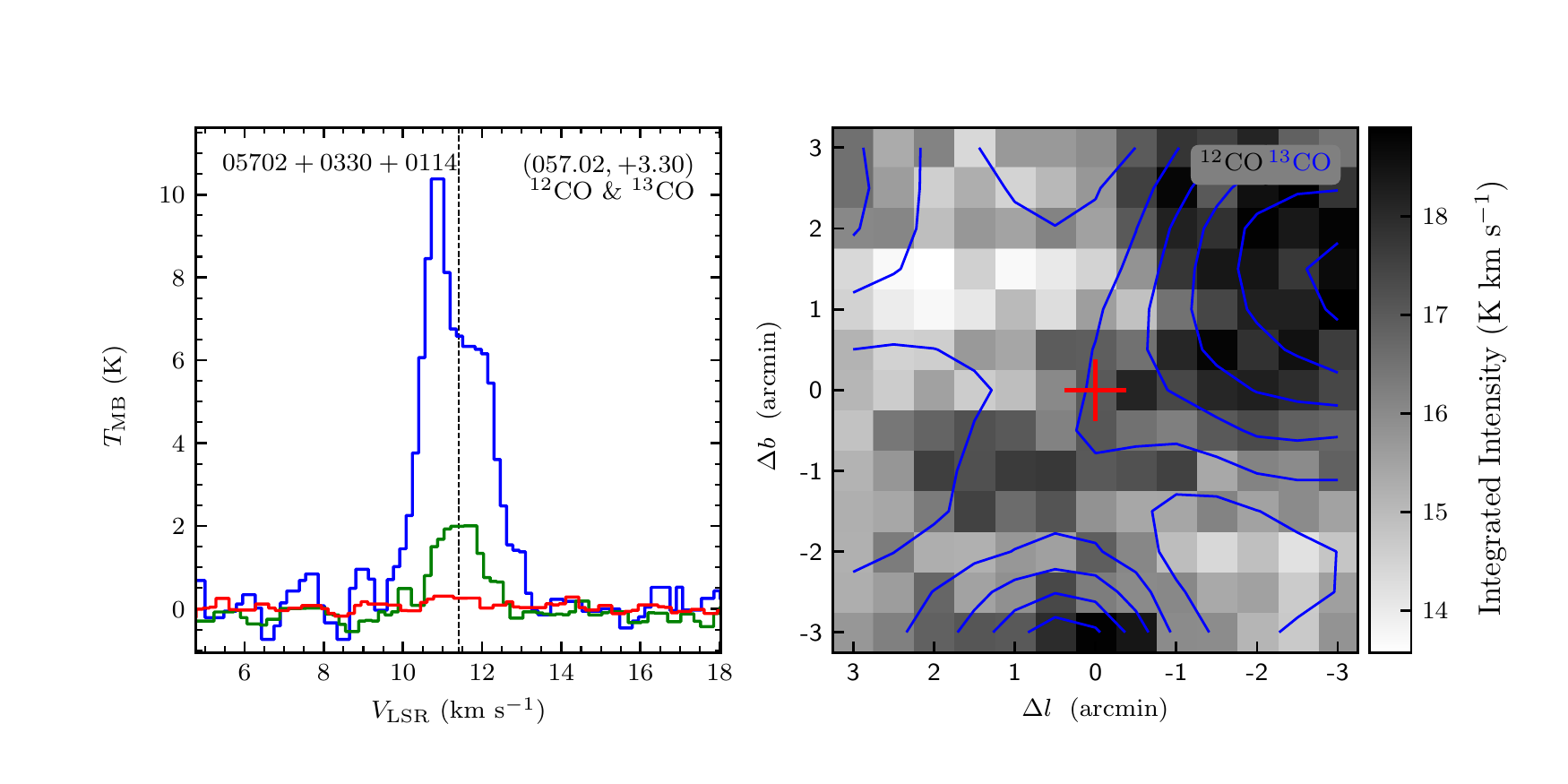}
\includegraphics[width=9.0cm,angle=0]{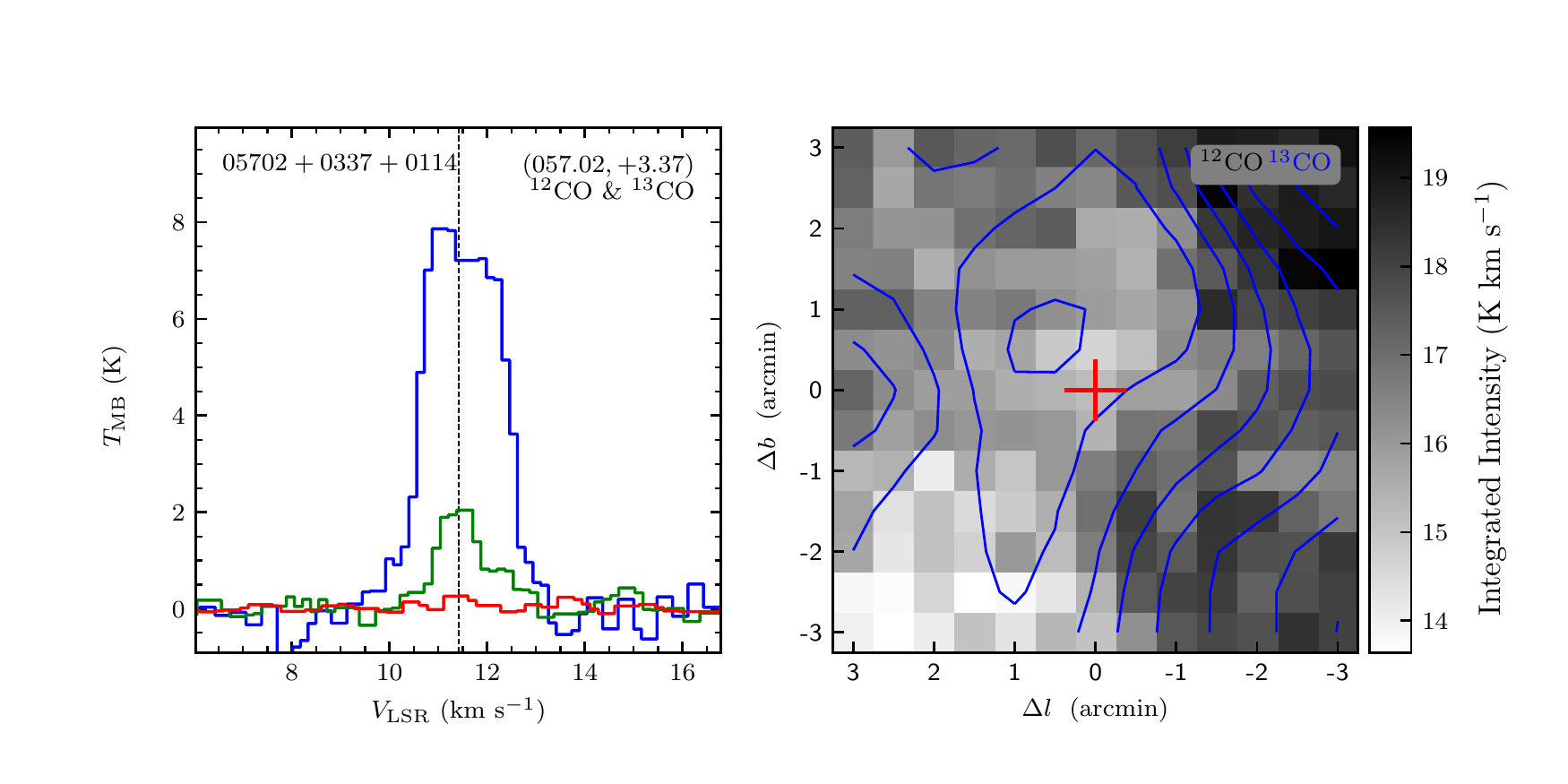}
\end{figure}
\clearpage

\begin{figure}
\includegraphics[width=9.0cm,angle=0]{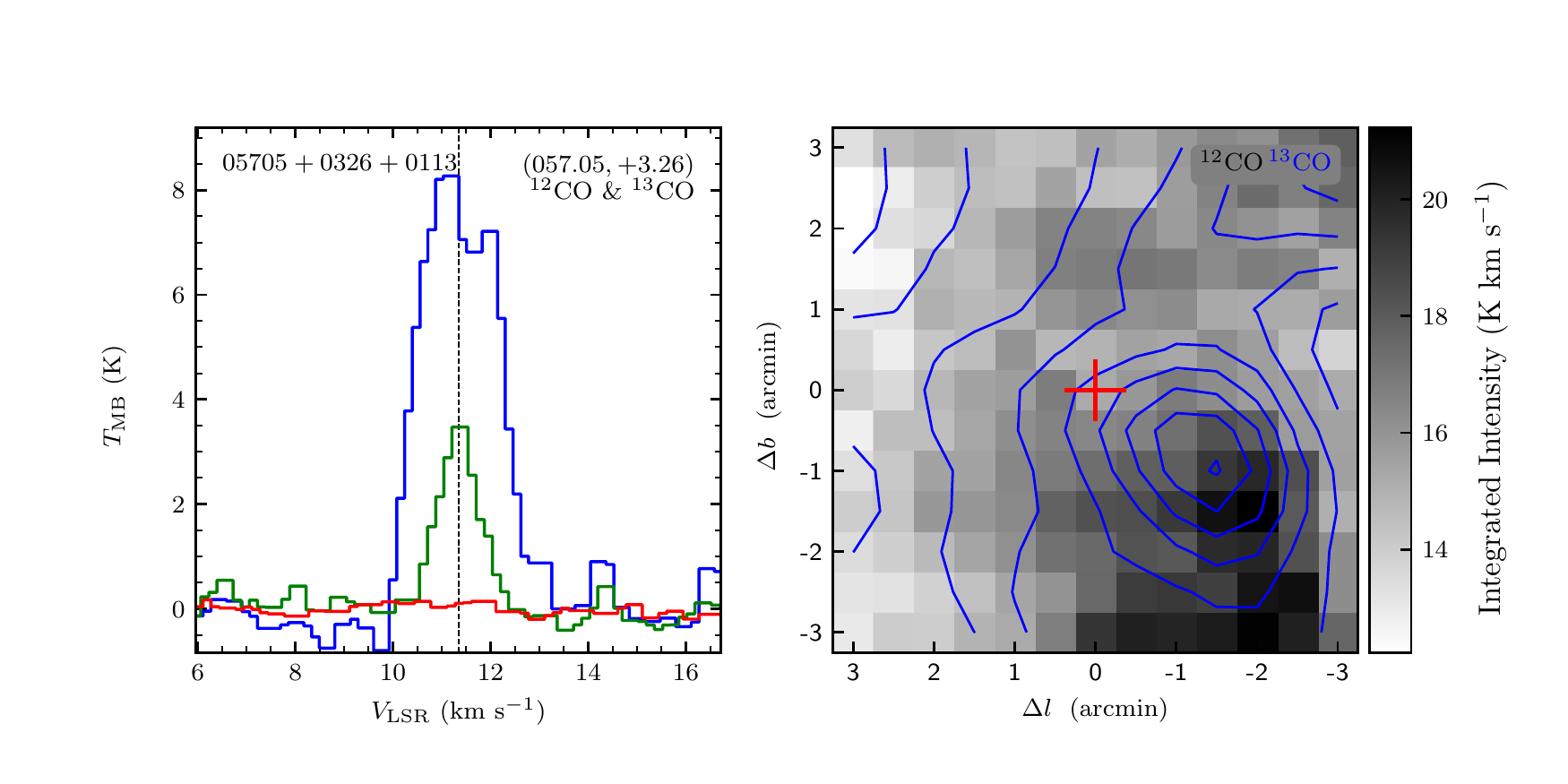}
\includegraphics[width=9.0cm,angle=0]{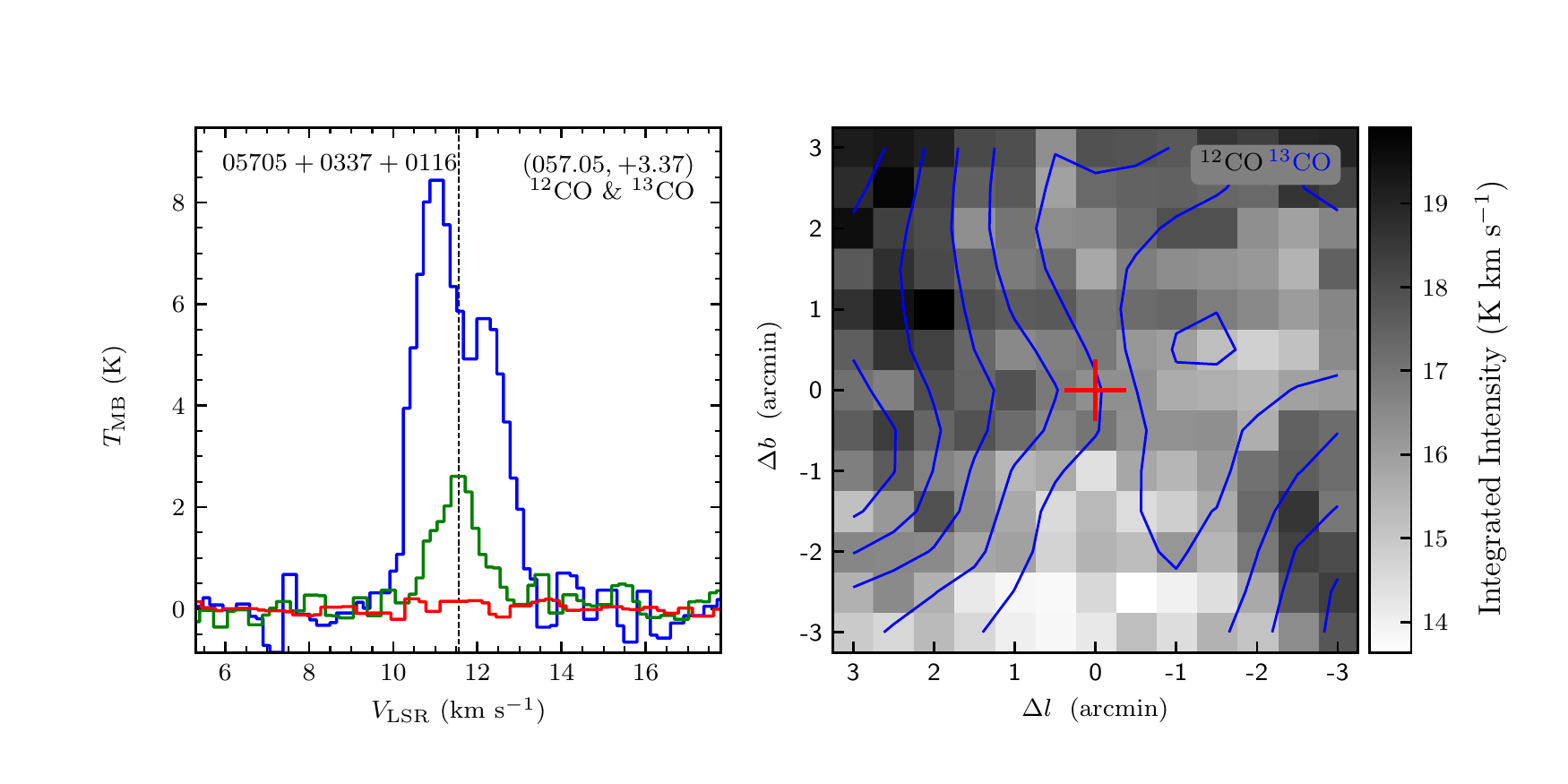}
\vspace{-0.5cm}

\includegraphics[width=9.0cm,angle=0]{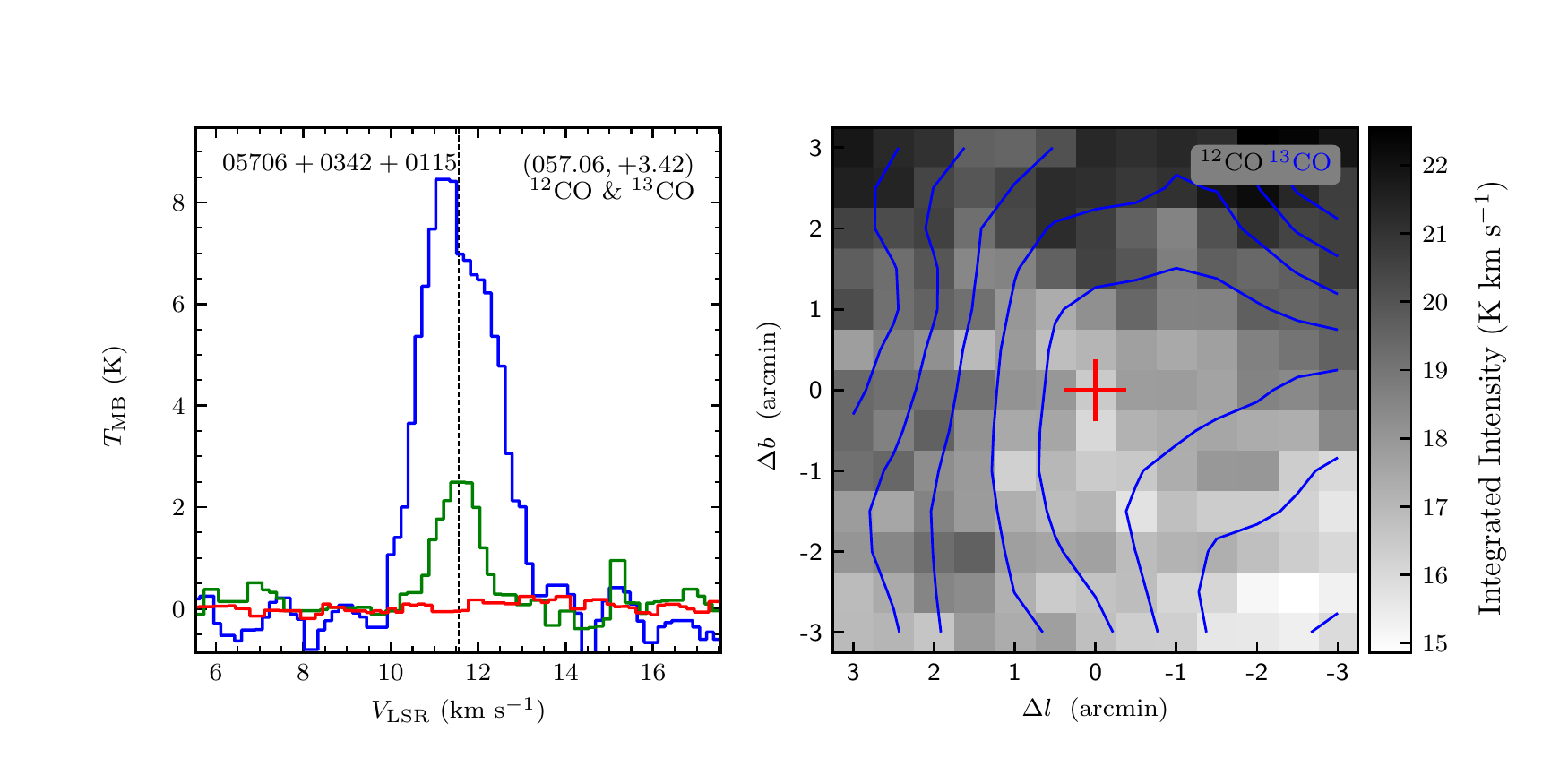}
\includegraphics[width=9.0cm,angle=0]{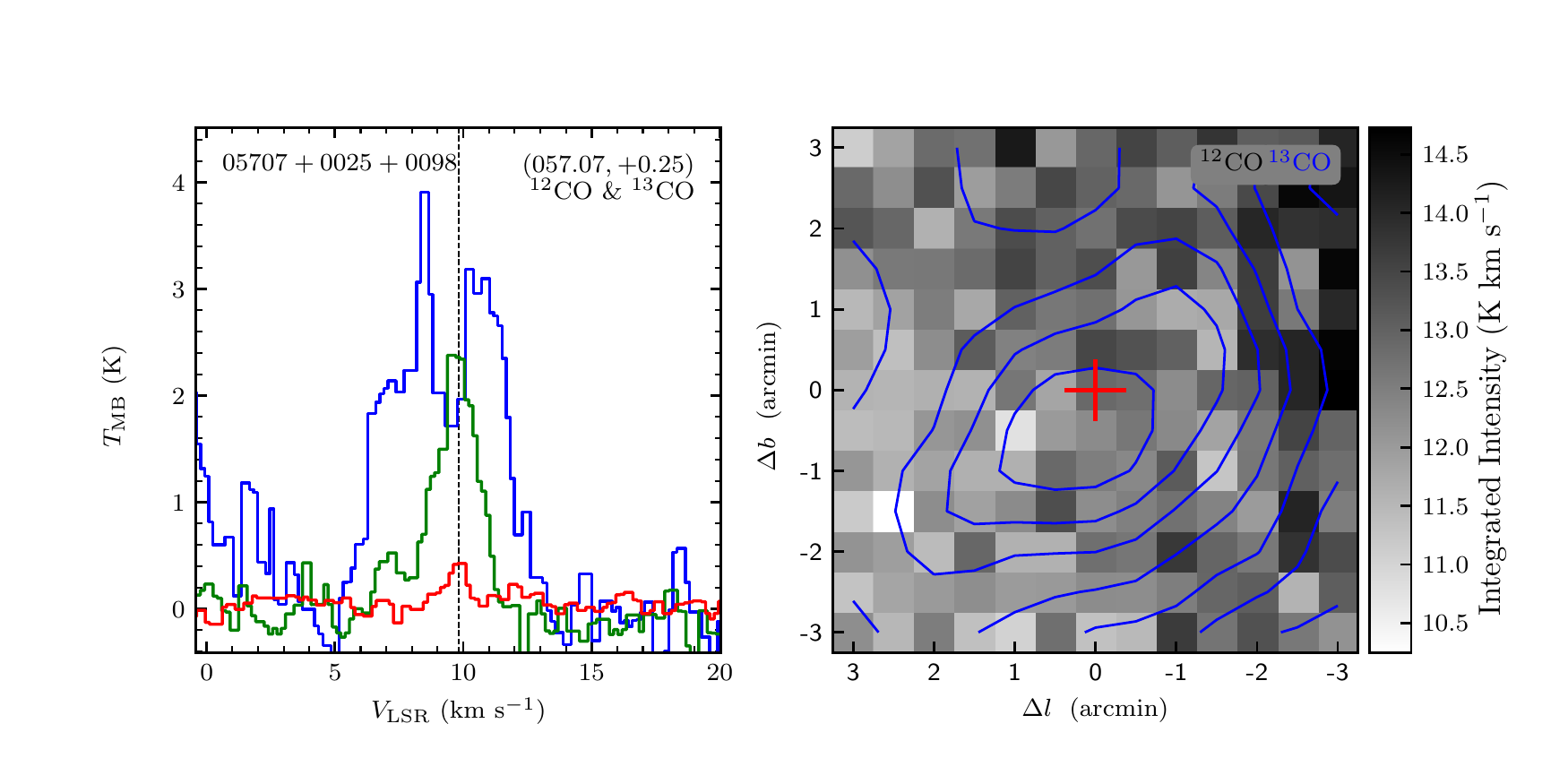}
\vspace{-0.5cm}

\includegraphics[width=9.0cm,angle=0]{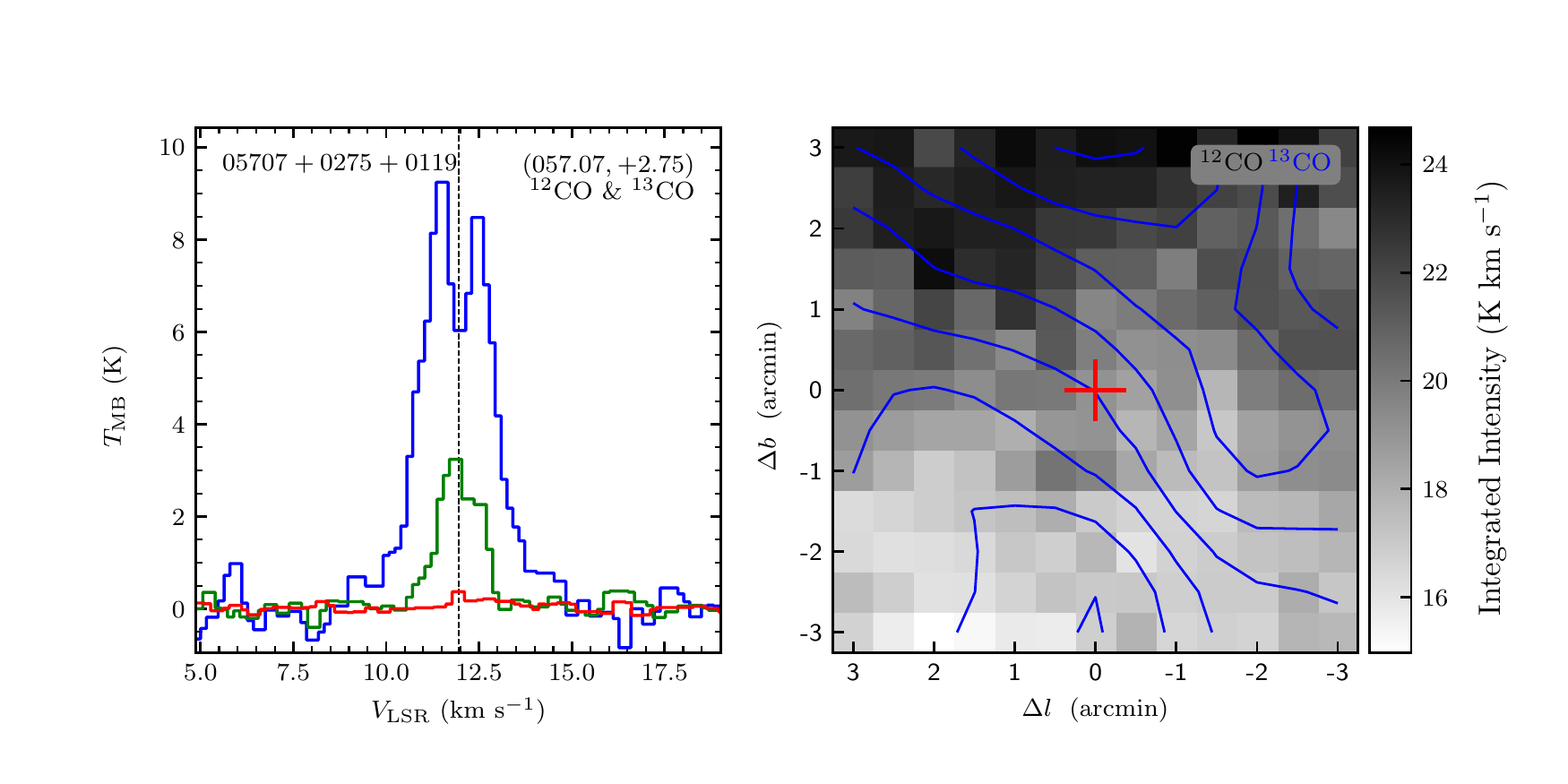}
\includegraphics[width=9.0cm,angle=0]{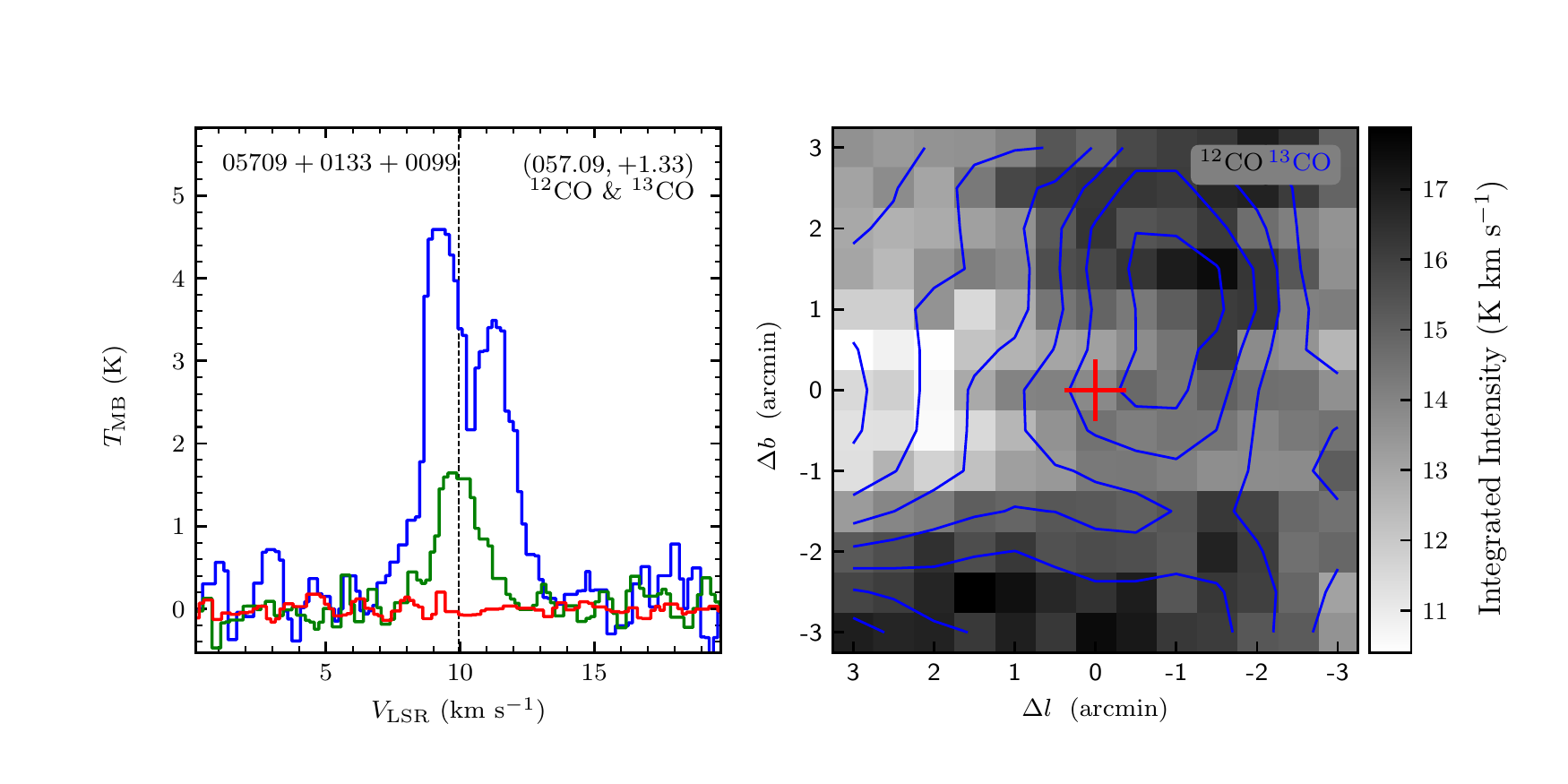}
\vspace{-0.5cm}

\includegraphics[width=9.0cm,angle=0]{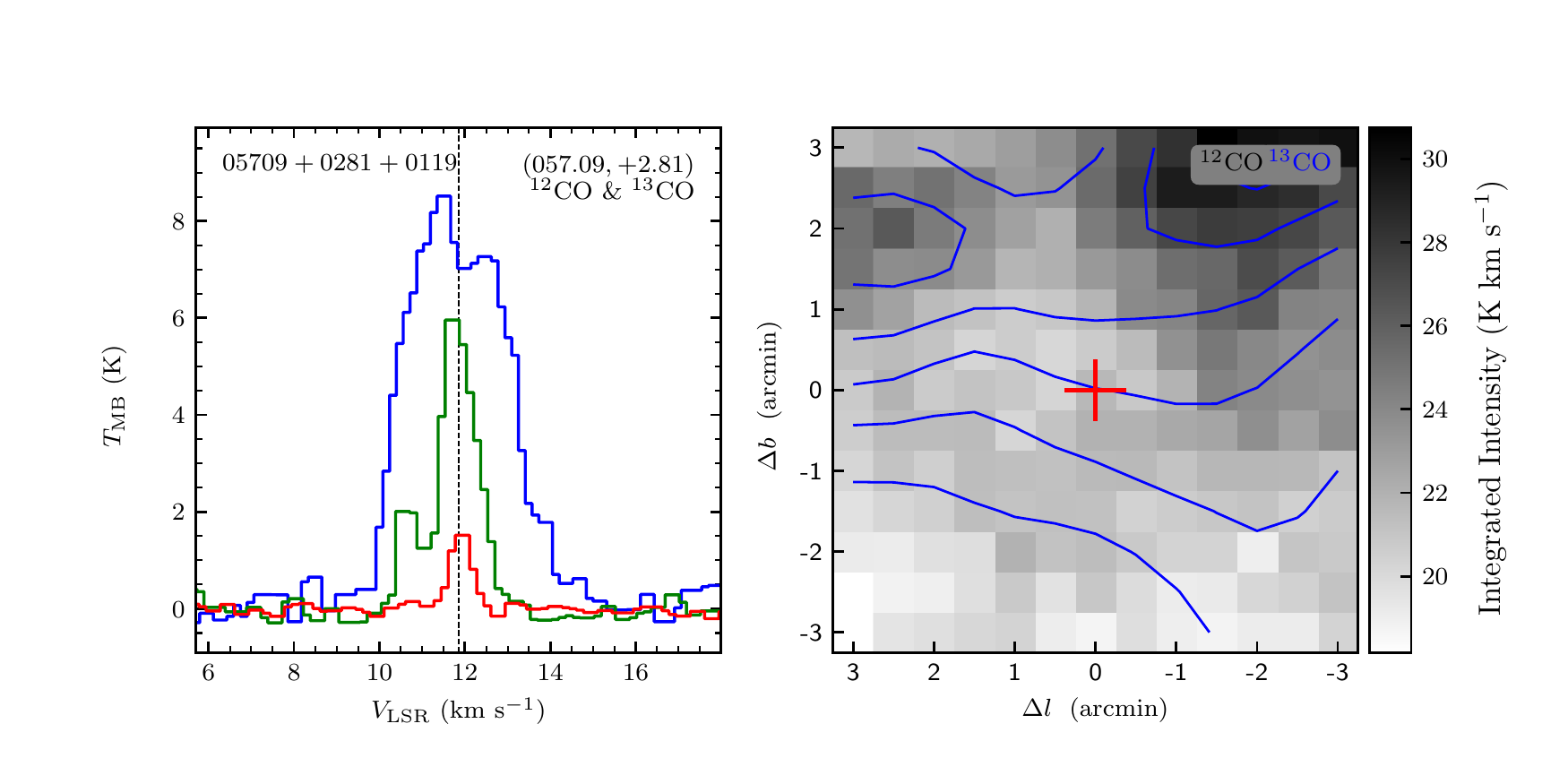}
\includegraphics[width=9.0cm,angle=0]{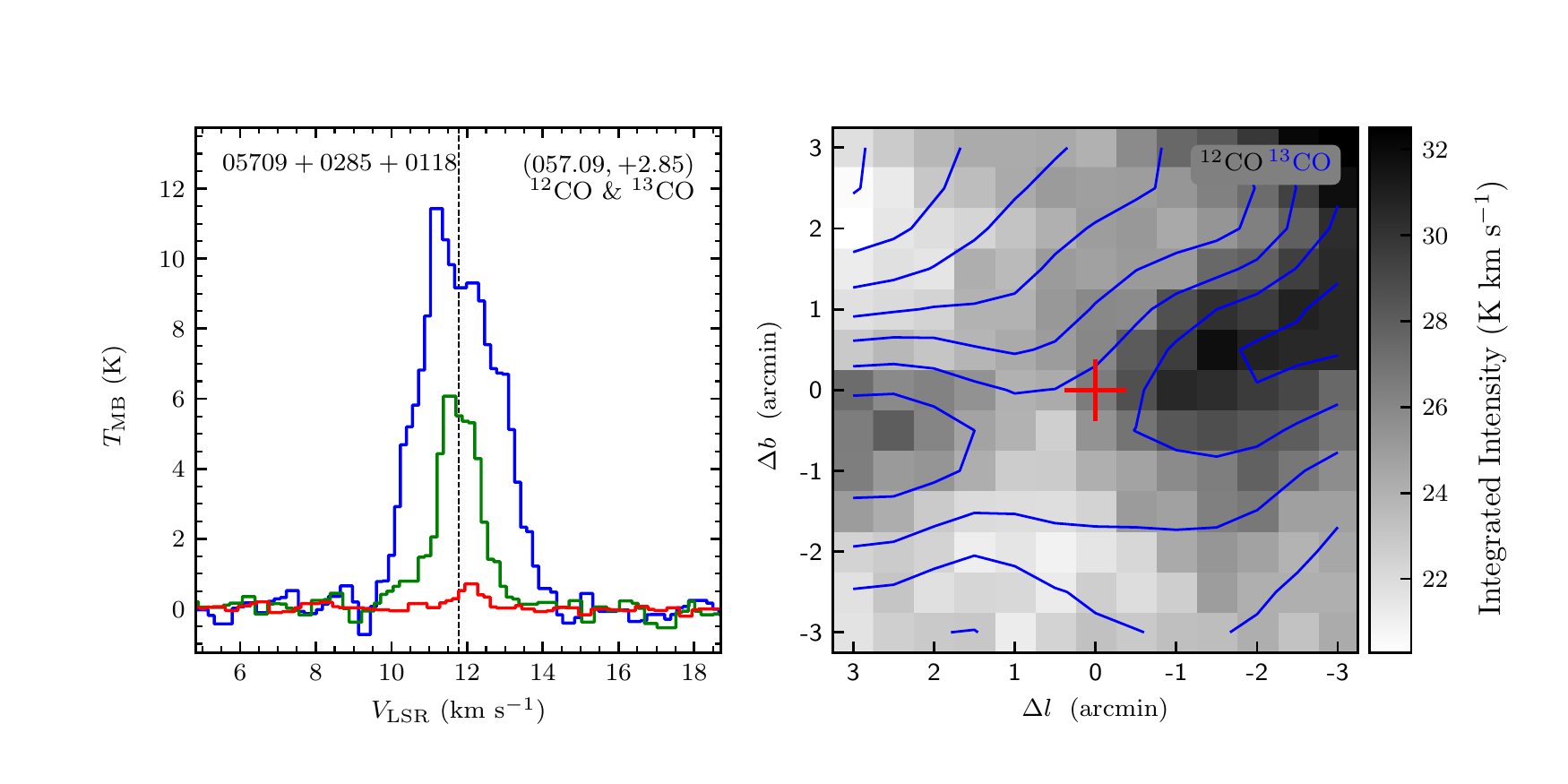}
\vspace{-0.5cm}

\includegraphics[width=9.0cm,angle=0]{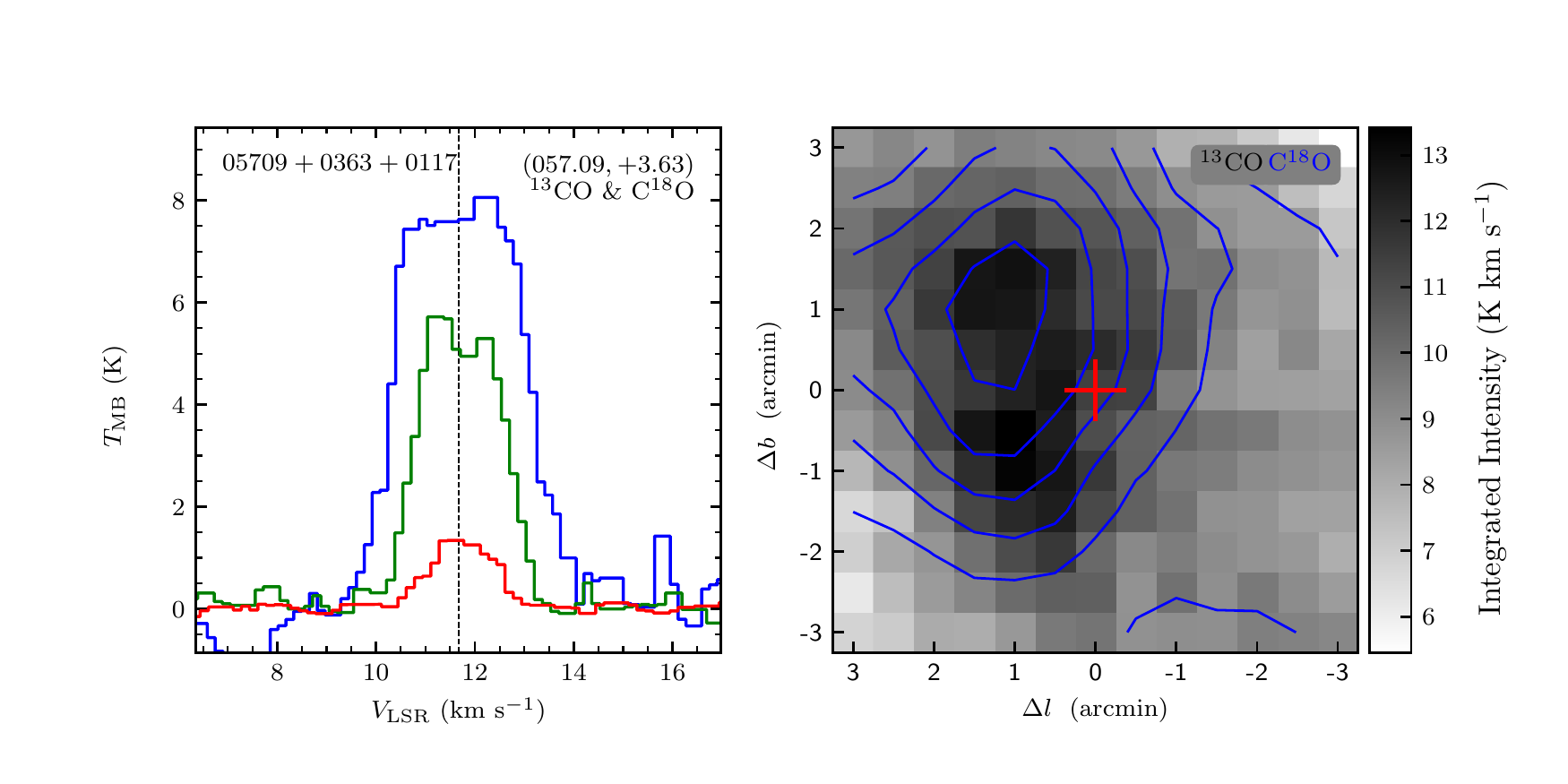}
\includegraphics[width=9.0cm,angle=0]{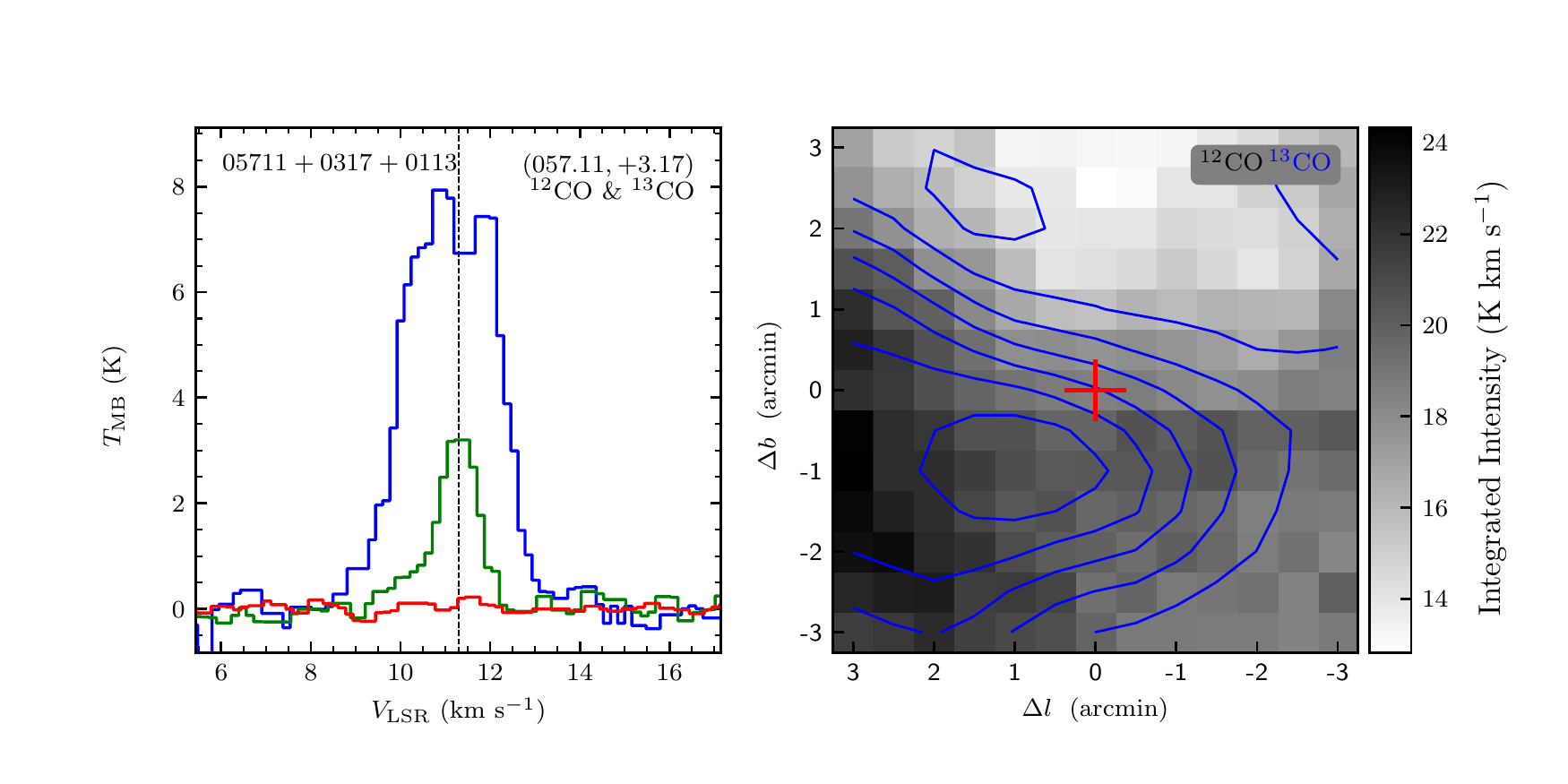}
\end{figure}
\clearpage

\begin{figure}
\includegraphics[width=9.0cm,angle=0]{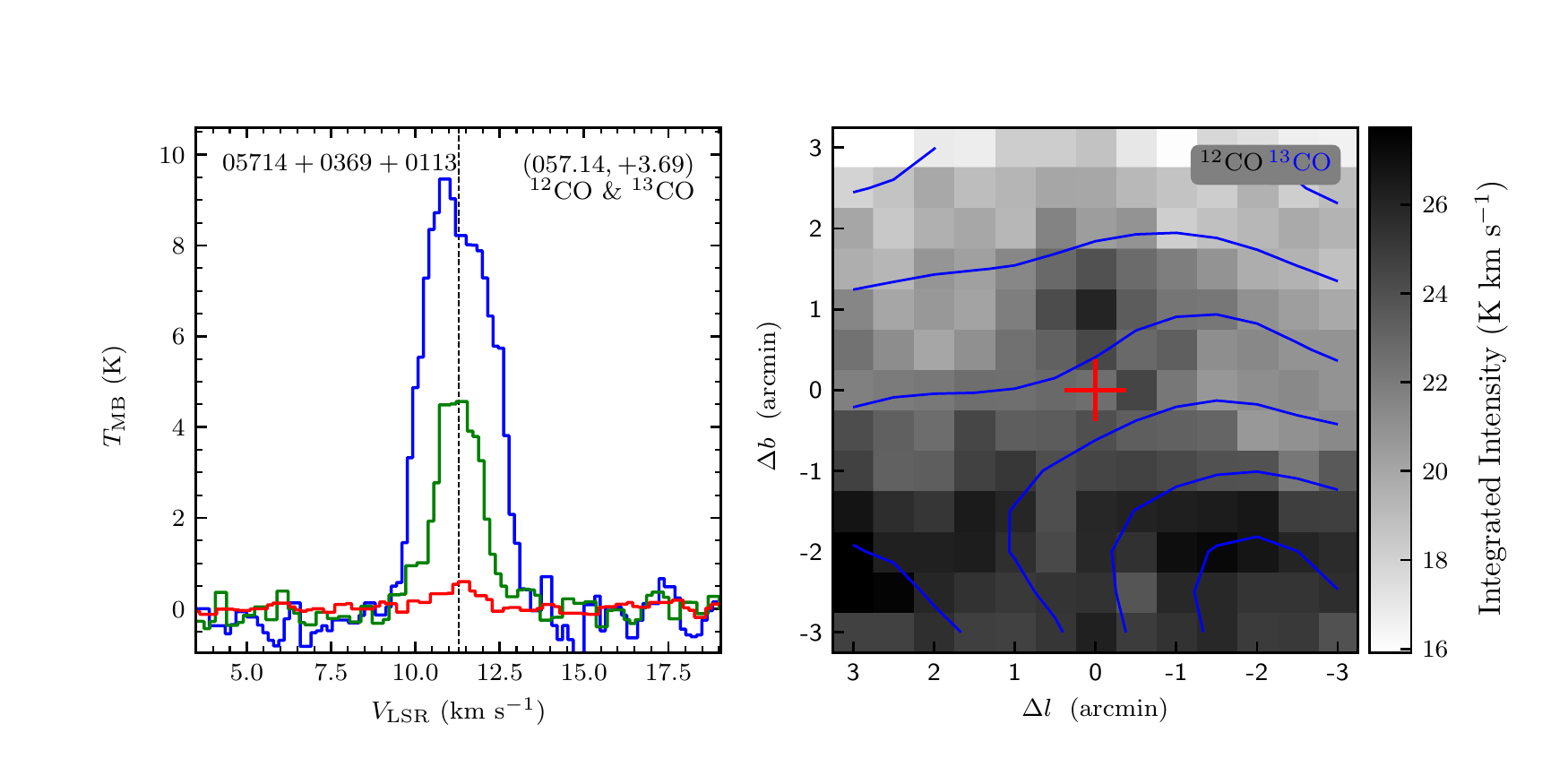}
\includegraphics[width=9.0cm,angle=0]{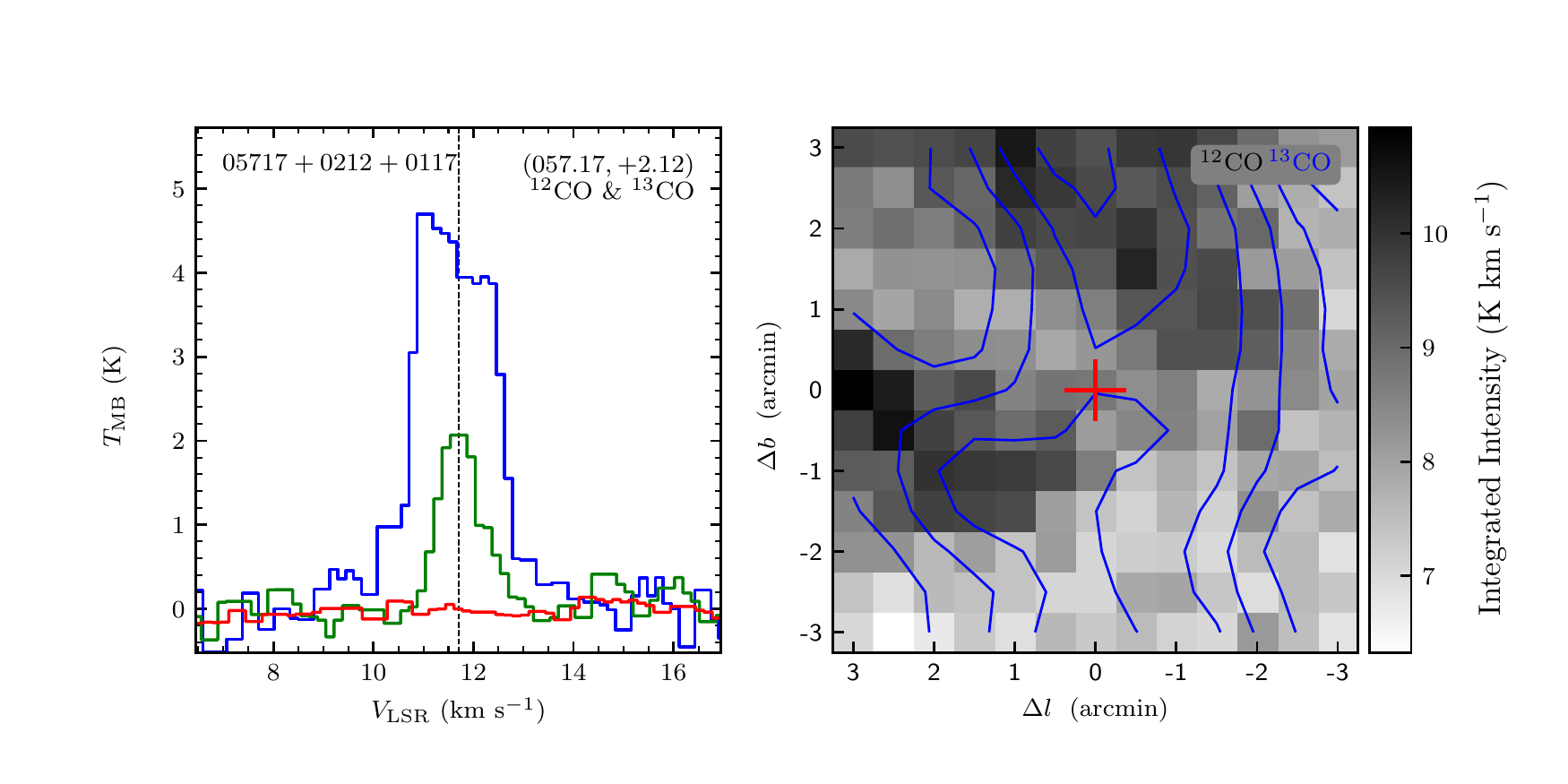}
\vspace{-0.5cm}

\includegraphics[width=9.0cm,angle=0]{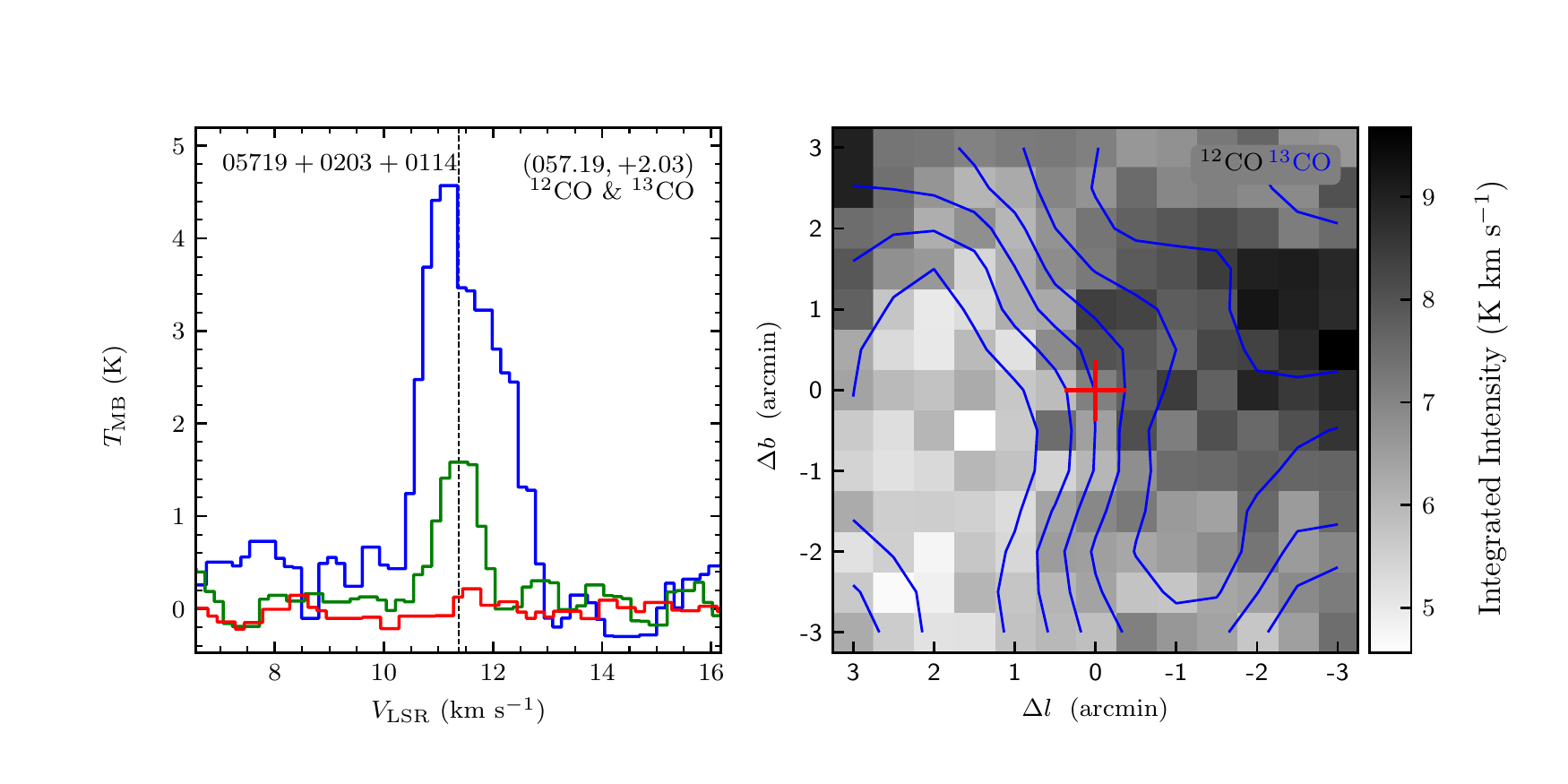}
\includegraphics[width=9.0cm,angle=0]{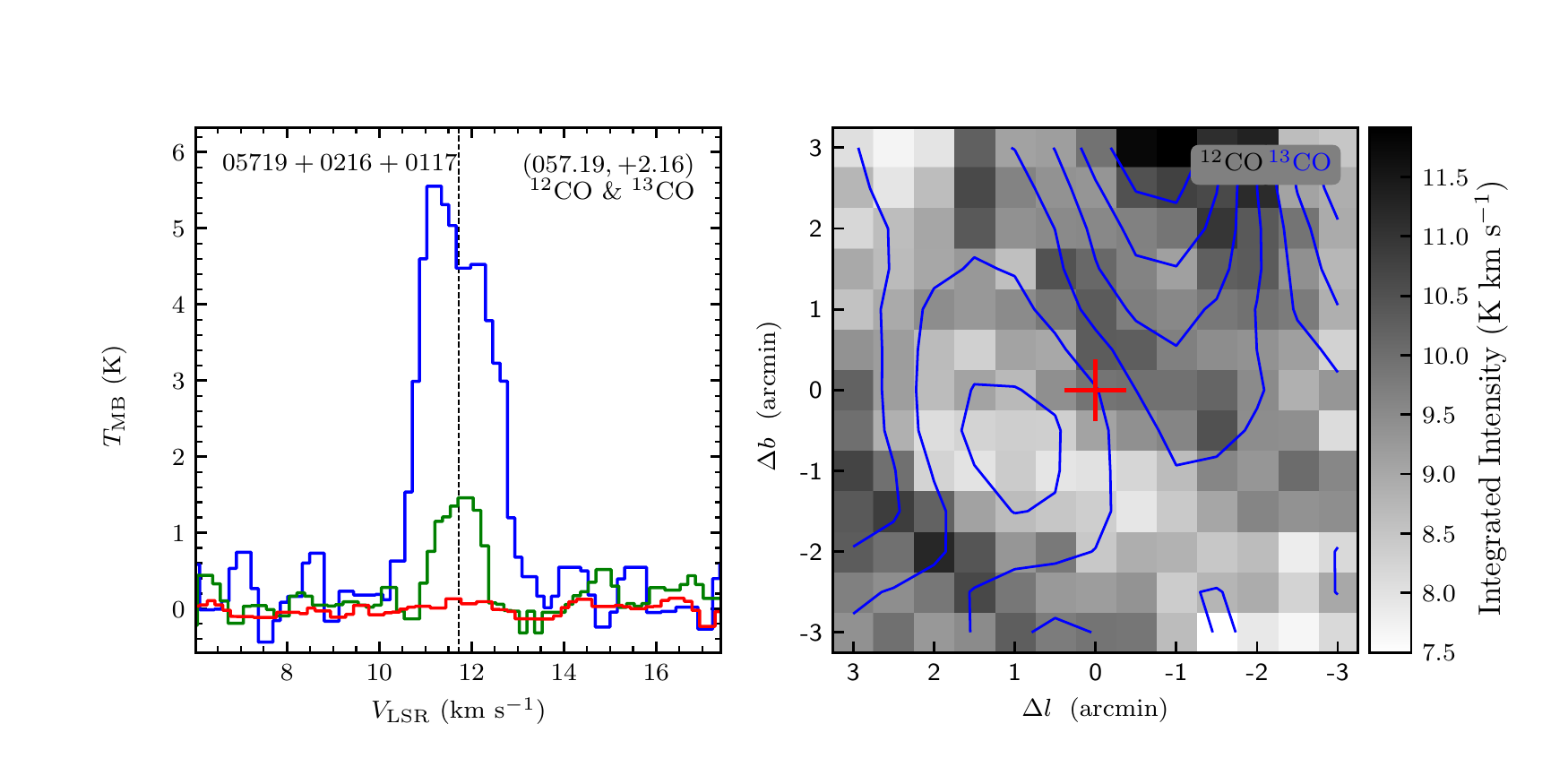}
\vspace{-0.5cm}

\includegraphics[width=9.0cm,angle=0]{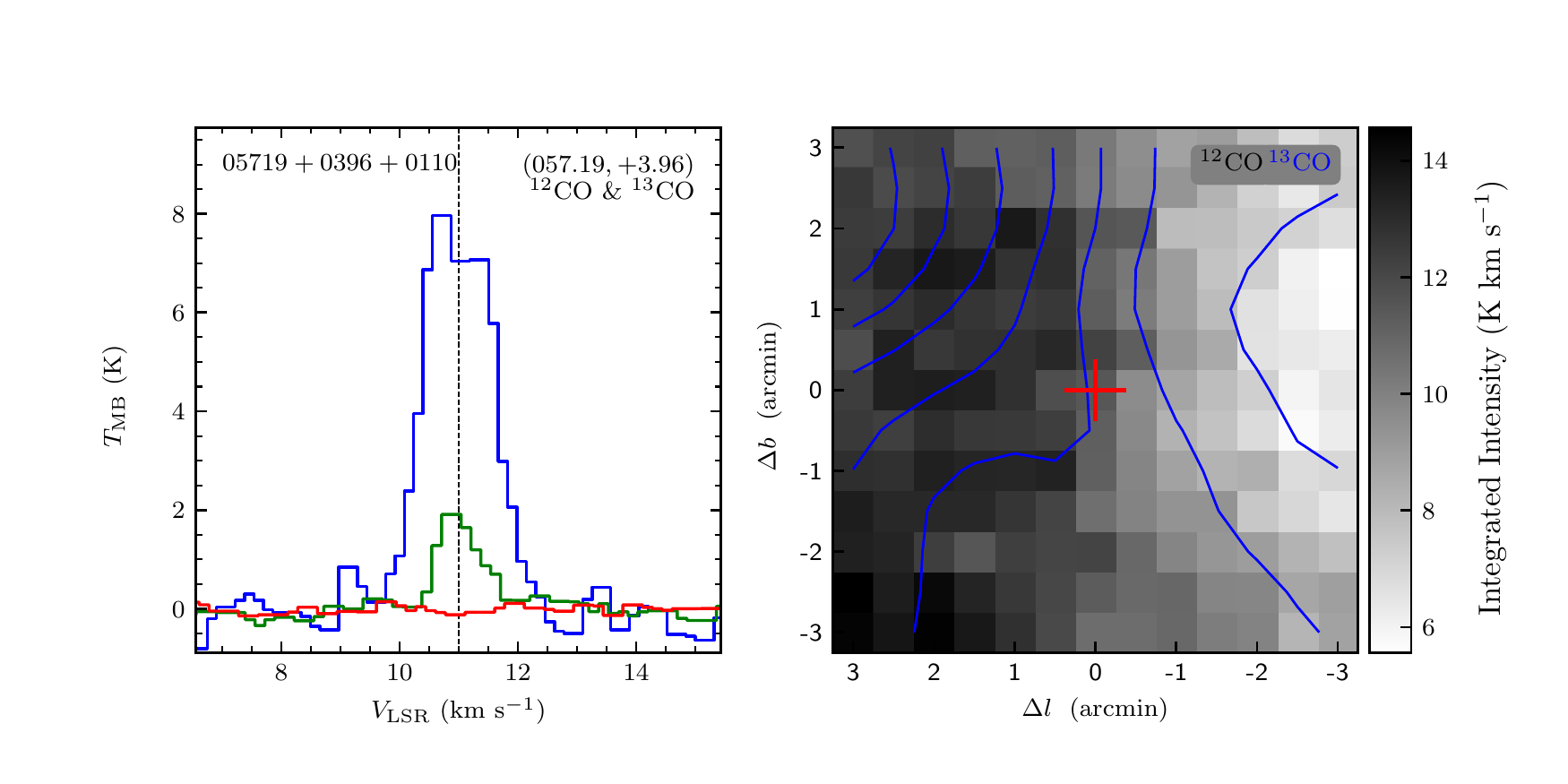}
\includegraphics[width=9.0cm,angle=0]{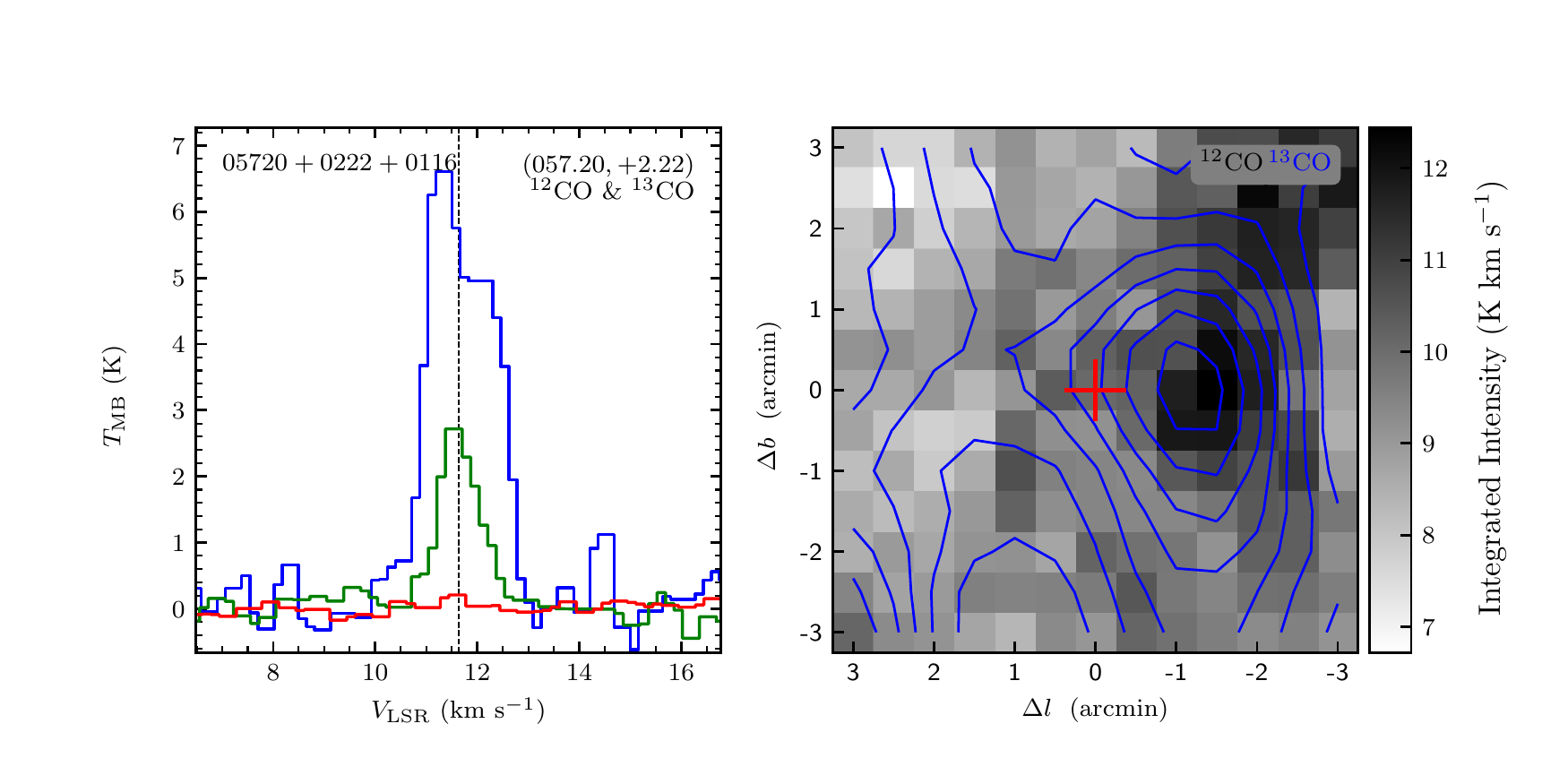}
\vspace{-0.5cm}

\includegraphics[width=9.0cm,angle=0]{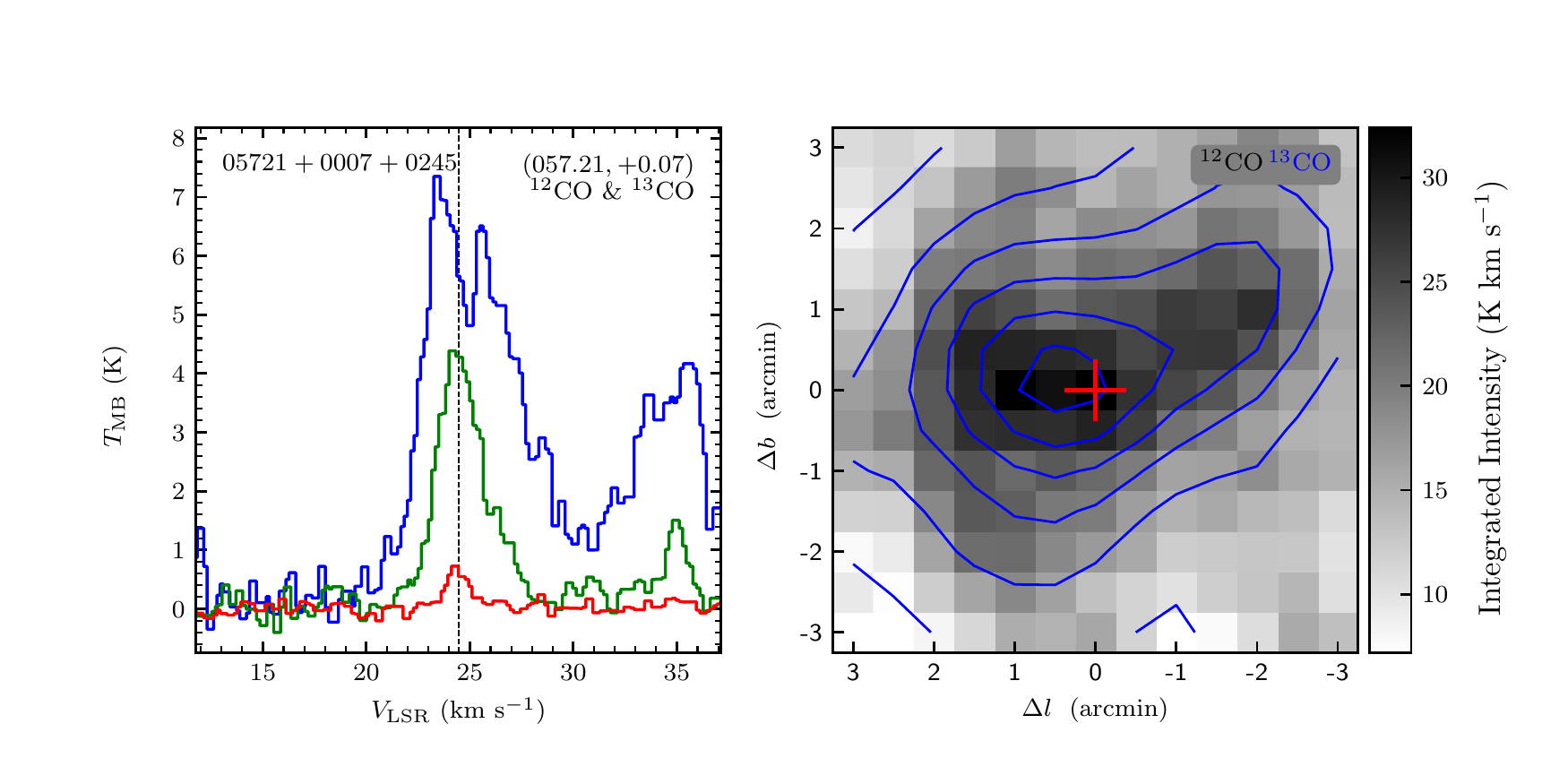}
\includegraphics[width=9.0cm,angle=0]{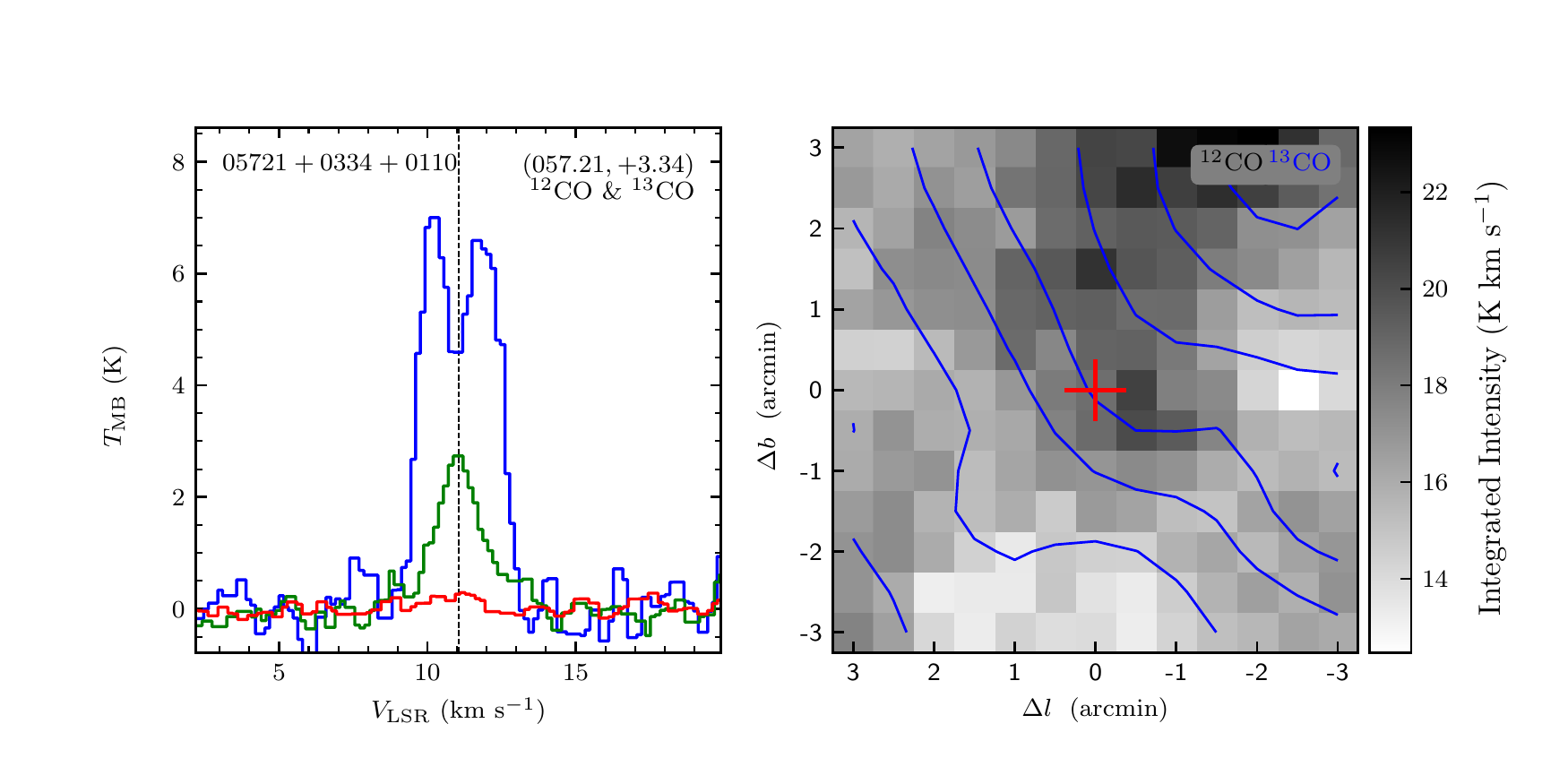}
\vspace{-0.5cm}

\includegraphics[width=9.0cm,angle=0]{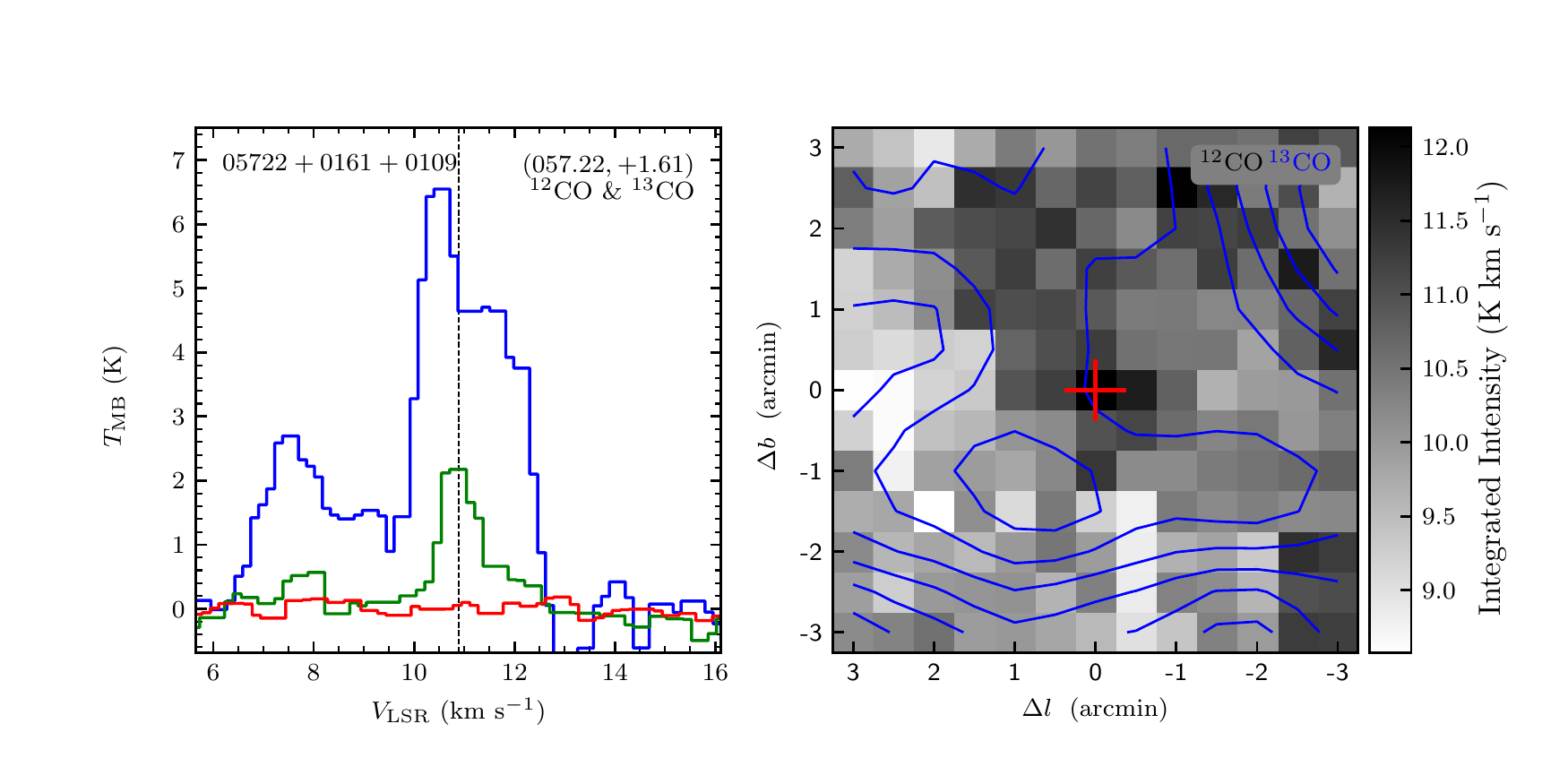}
\includegraphics[width=9.0cm,angle=0]{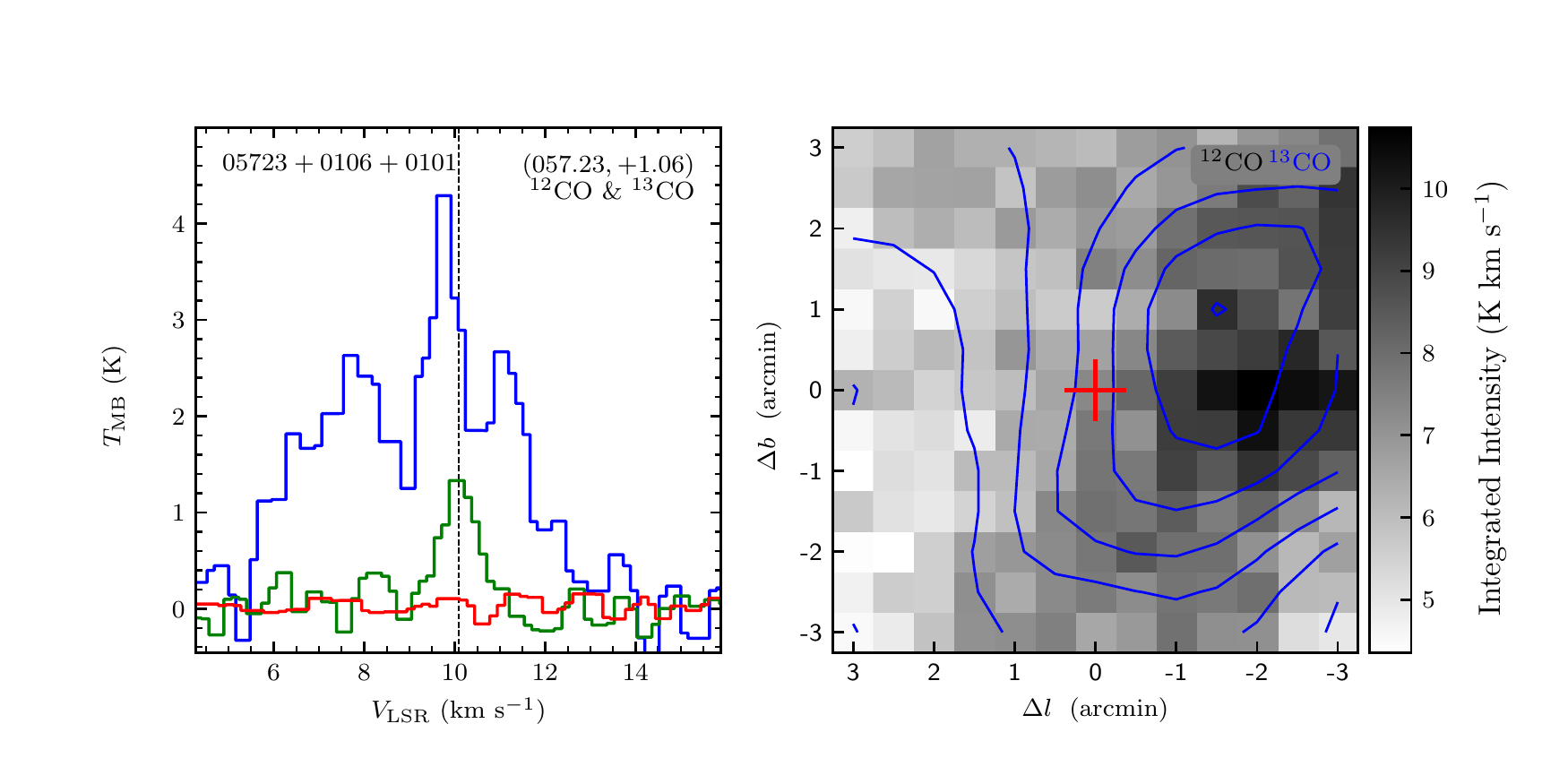}
\end{figure}
\clearpage

\begin{figure}
\includegraphics[width=9.0cm,angle=0]{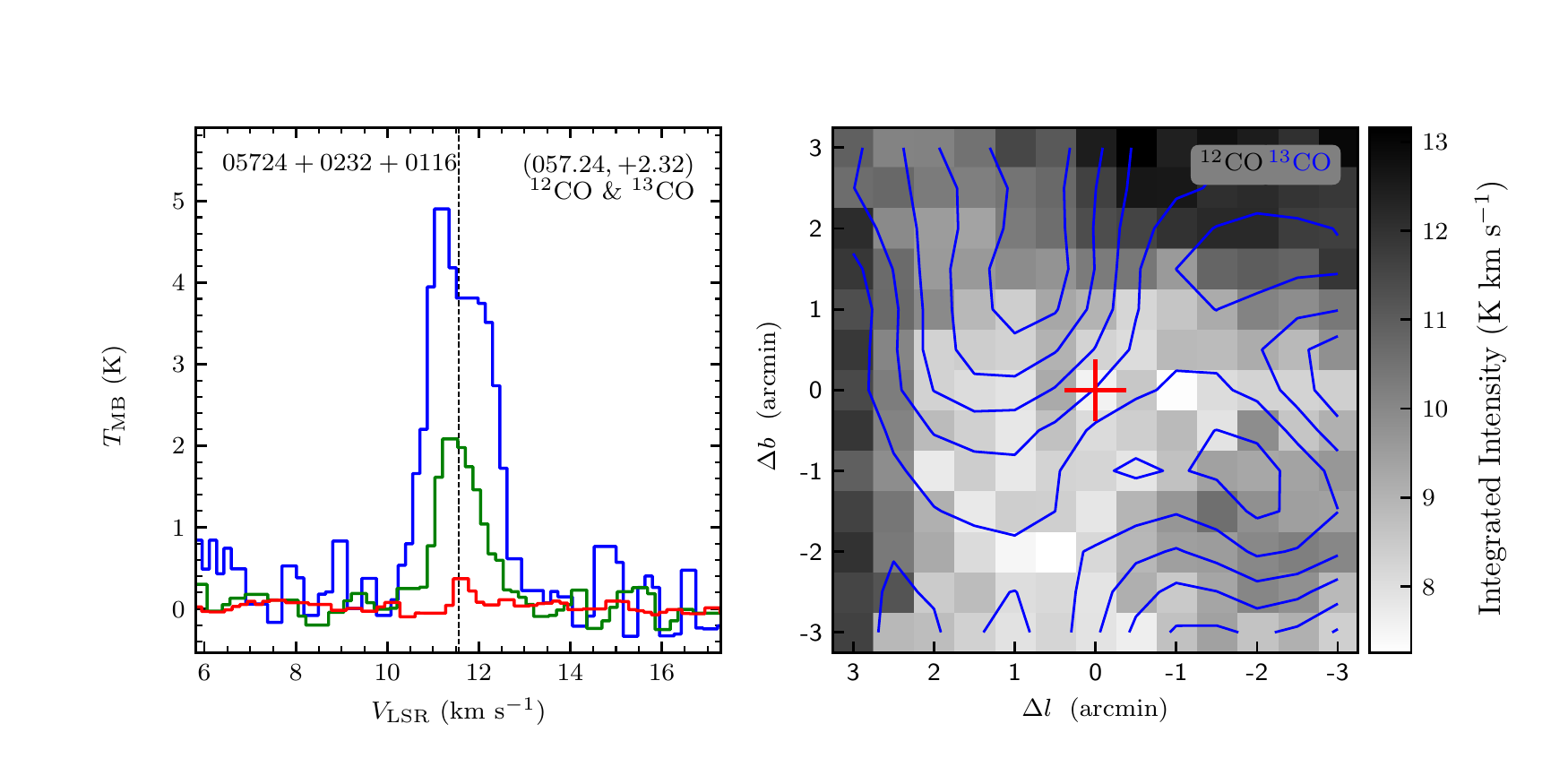}
\includegraphics[width=9.0cm,angle=0]{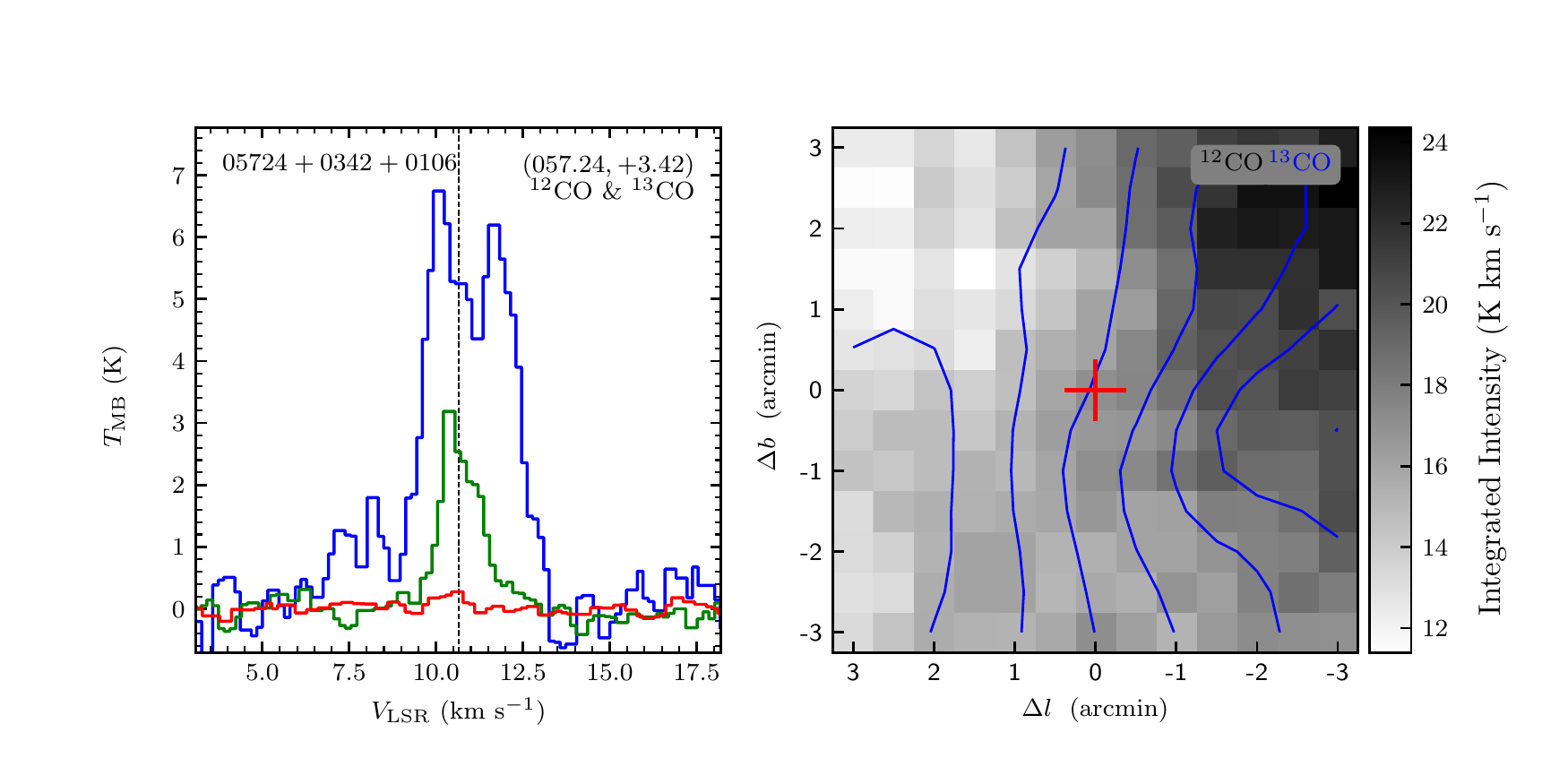}
\vspace{-0.5cm}

\includegraphics[width=9.0cm,angle=0]{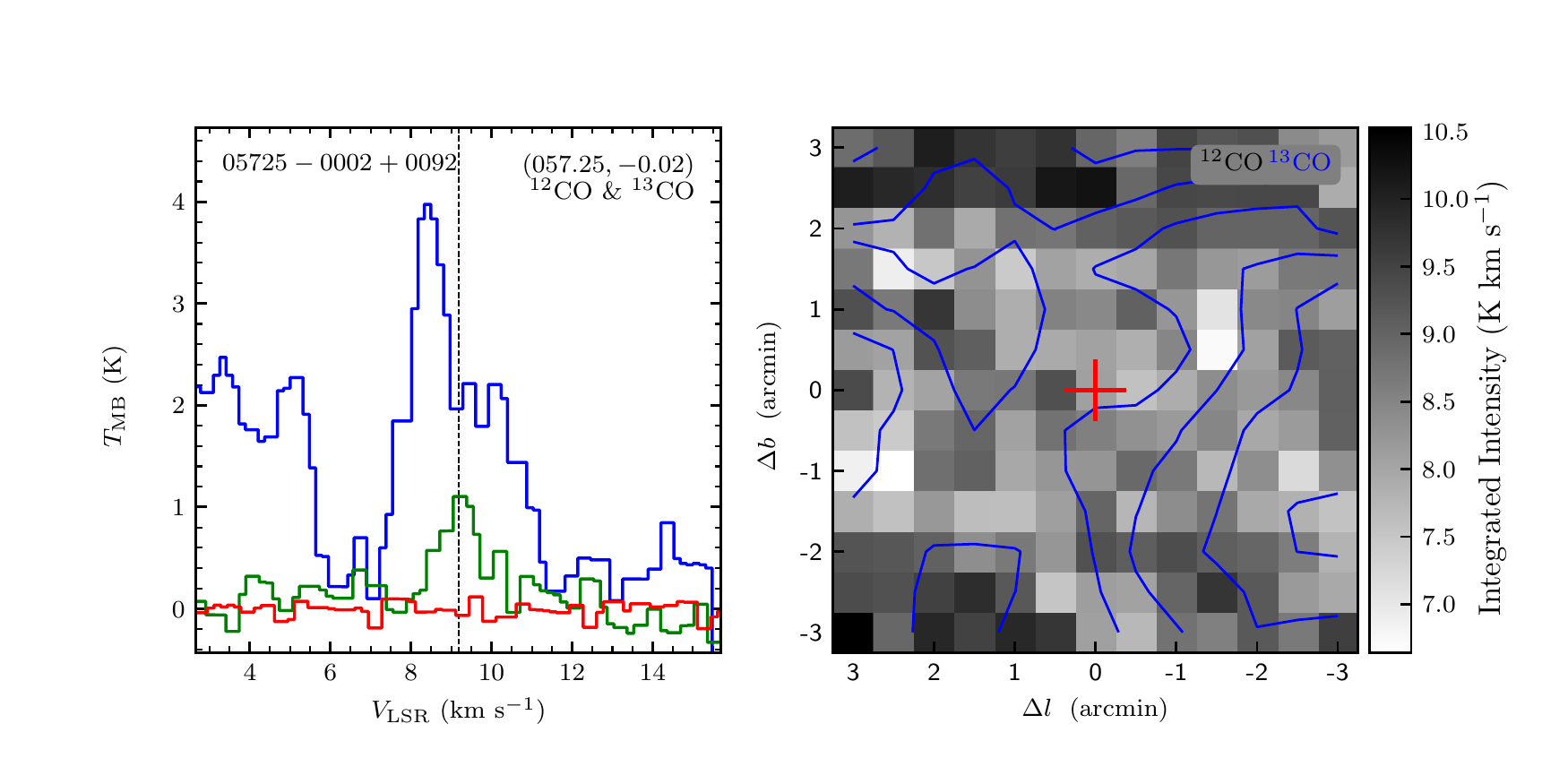}
\includegraphics[width=9.0cm,angle=0]{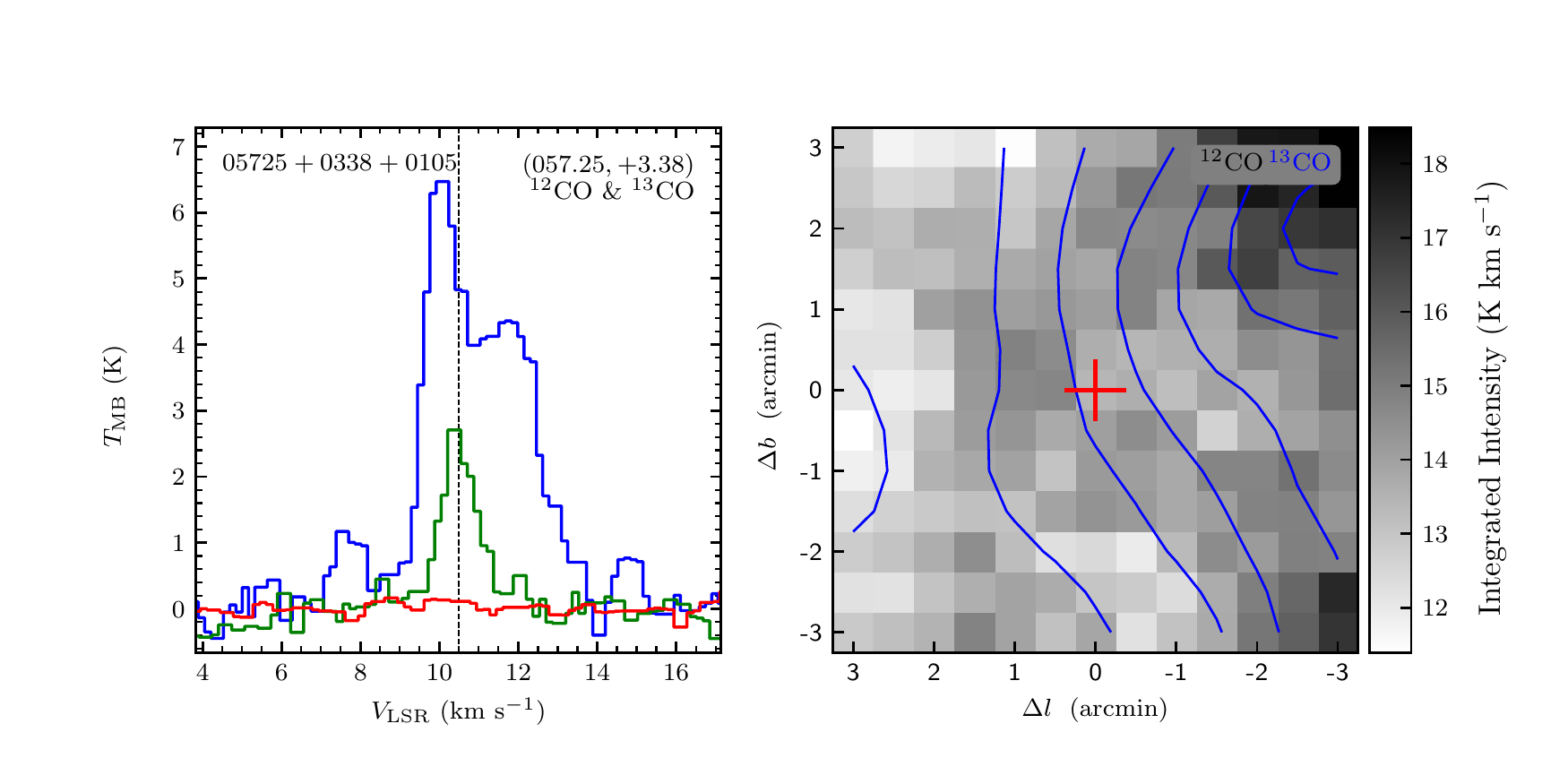}
\vspace{-0.5cm}

\includegraphics[width=9.0cm,angle=0]{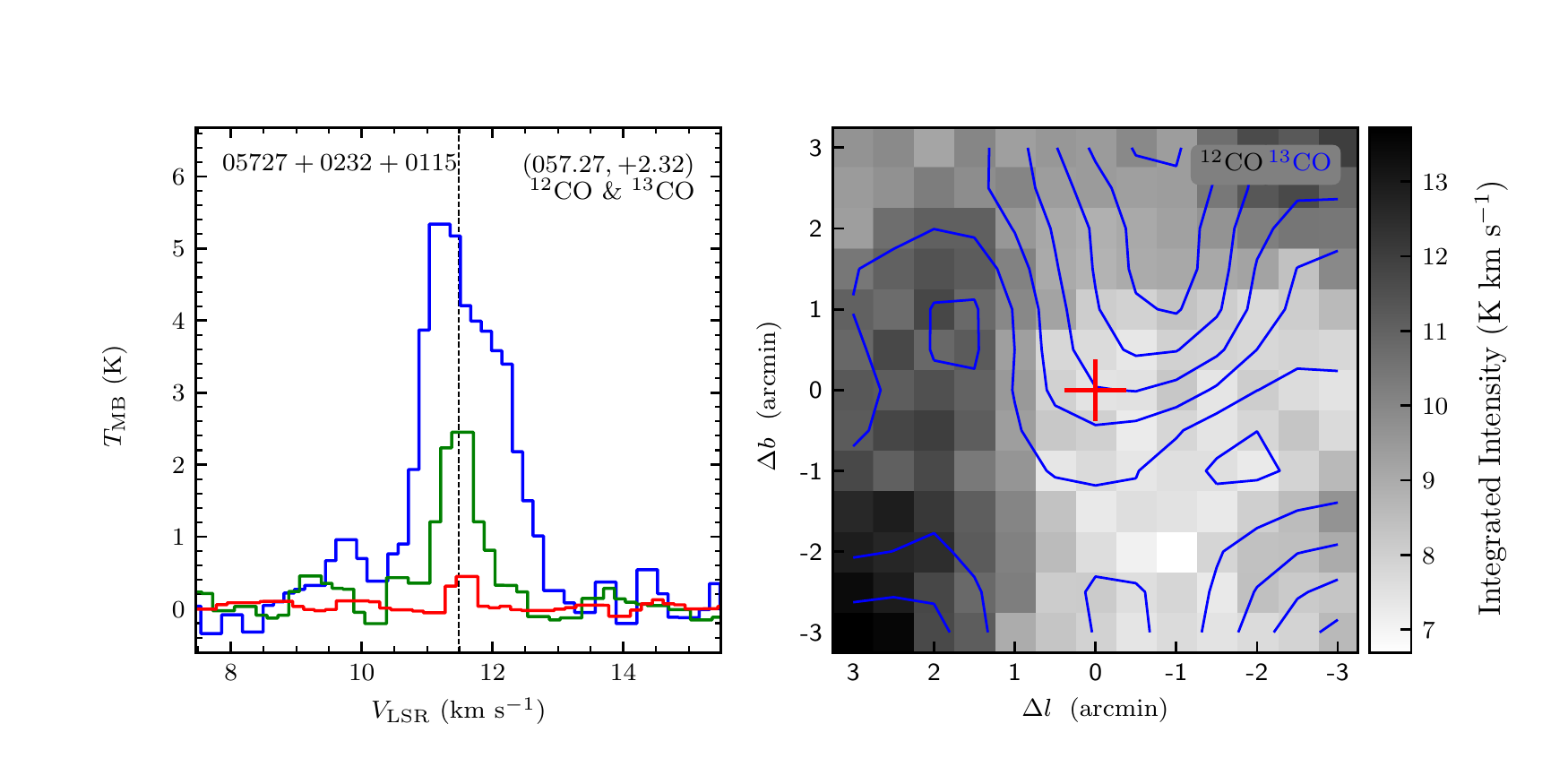}
\includegraphics[width=9.0cm,angle=0]{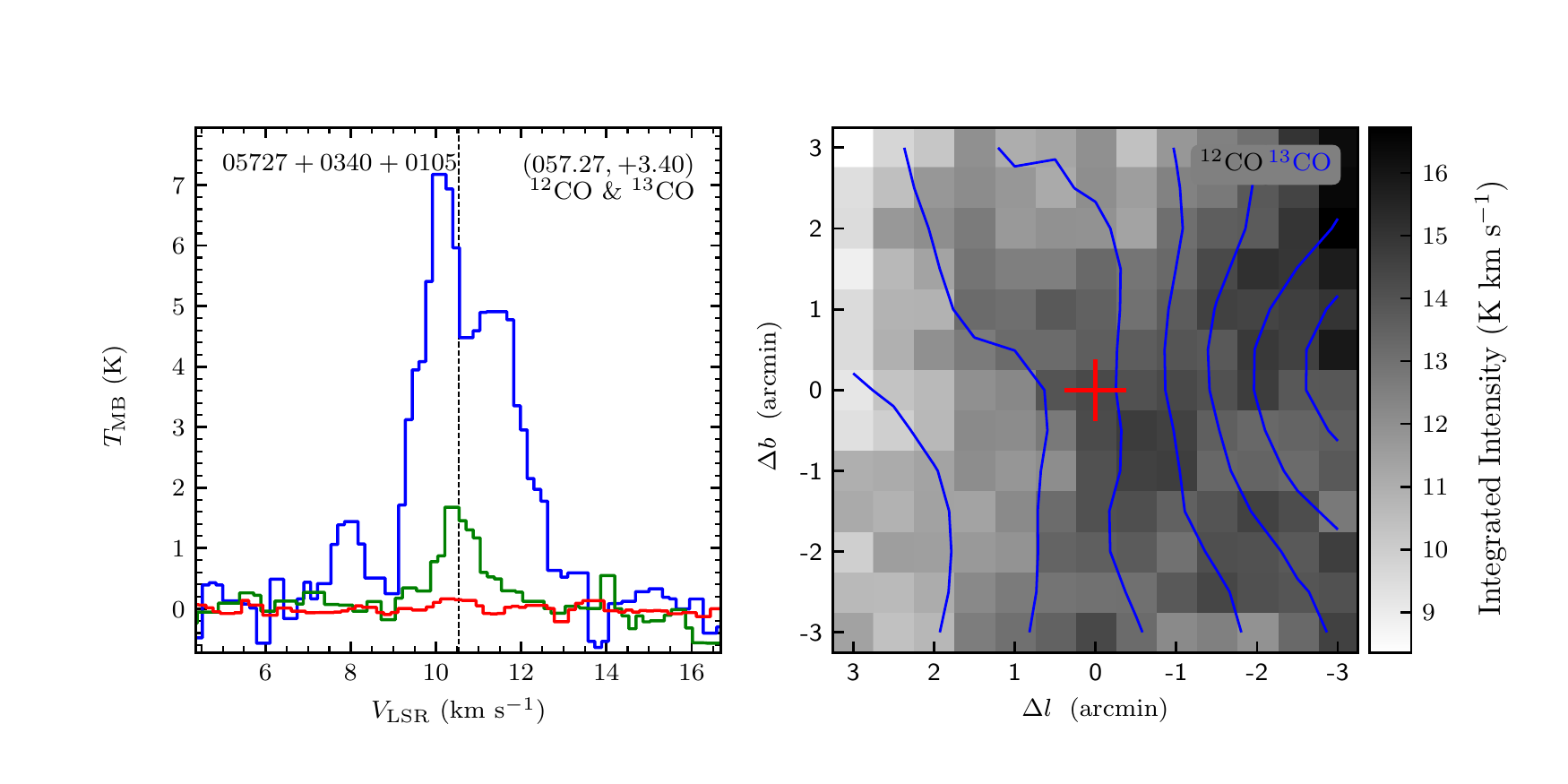}
\vspace{-0.5cm}

\includegraphics[width=9.0cm,angle=0]{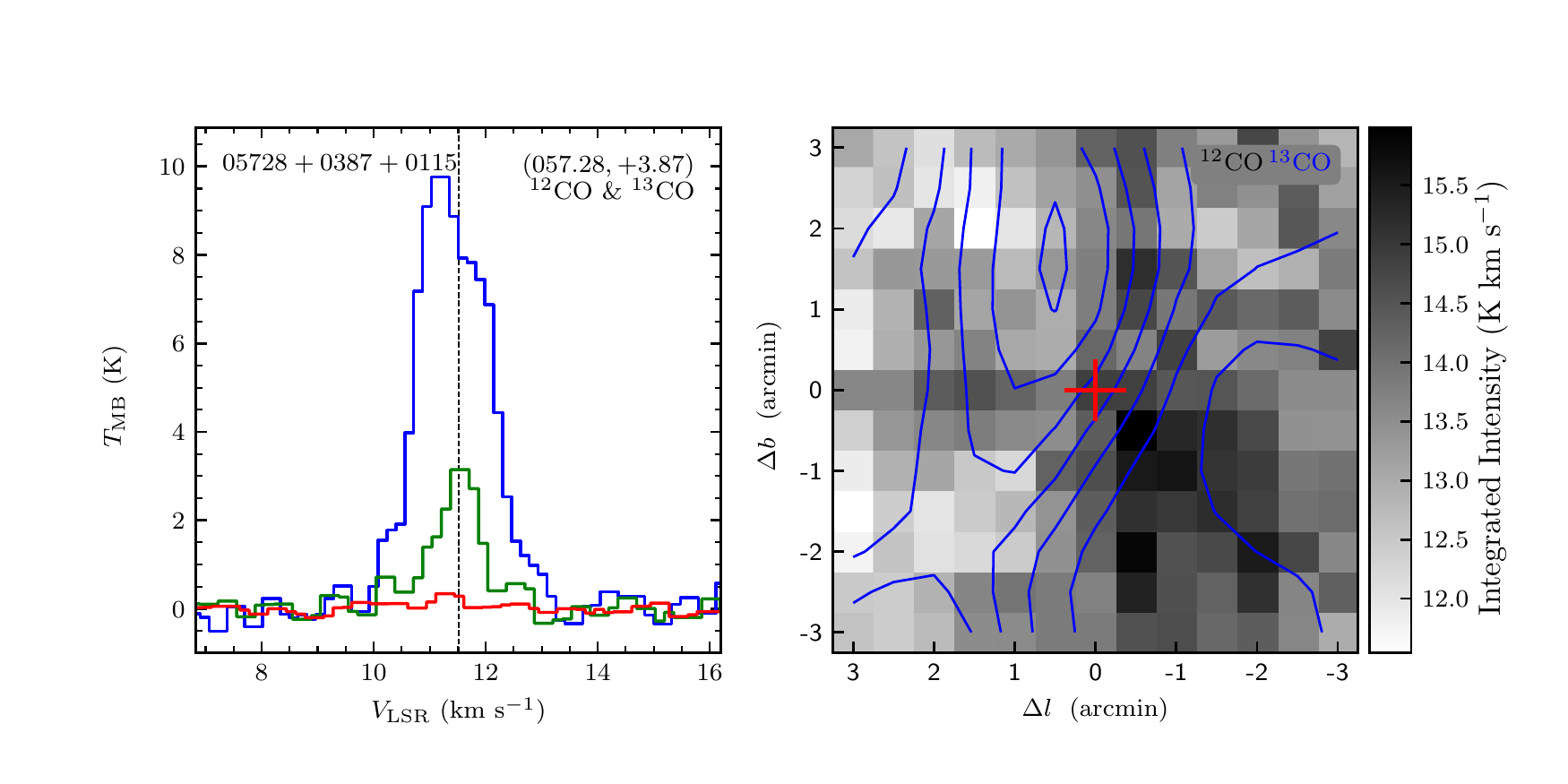}
\includegraphics[width=9.0cm,angle=0]{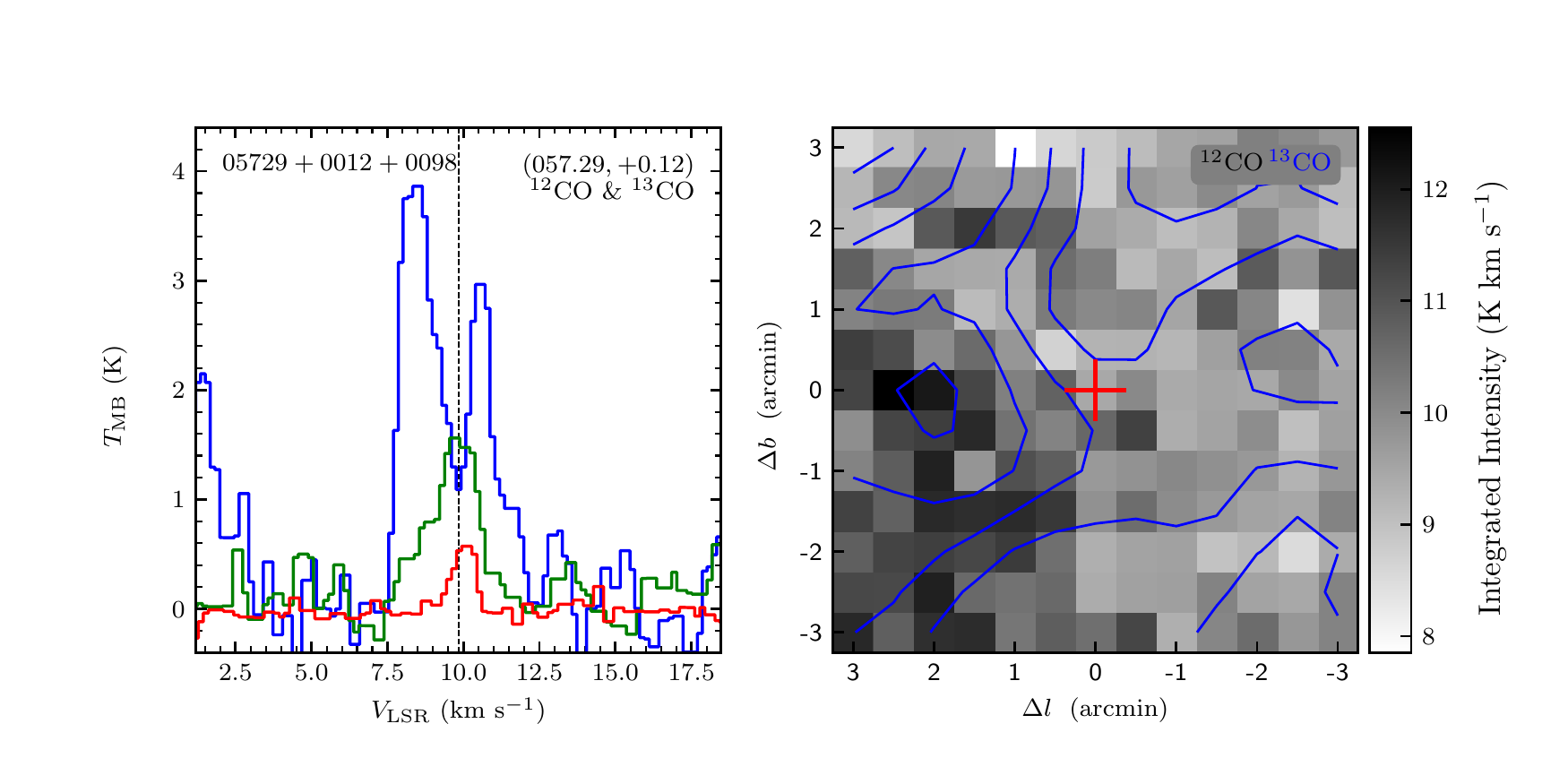}
\vspace{-0.5cm}

\includegraphics[width=9.0cm,angle=0]{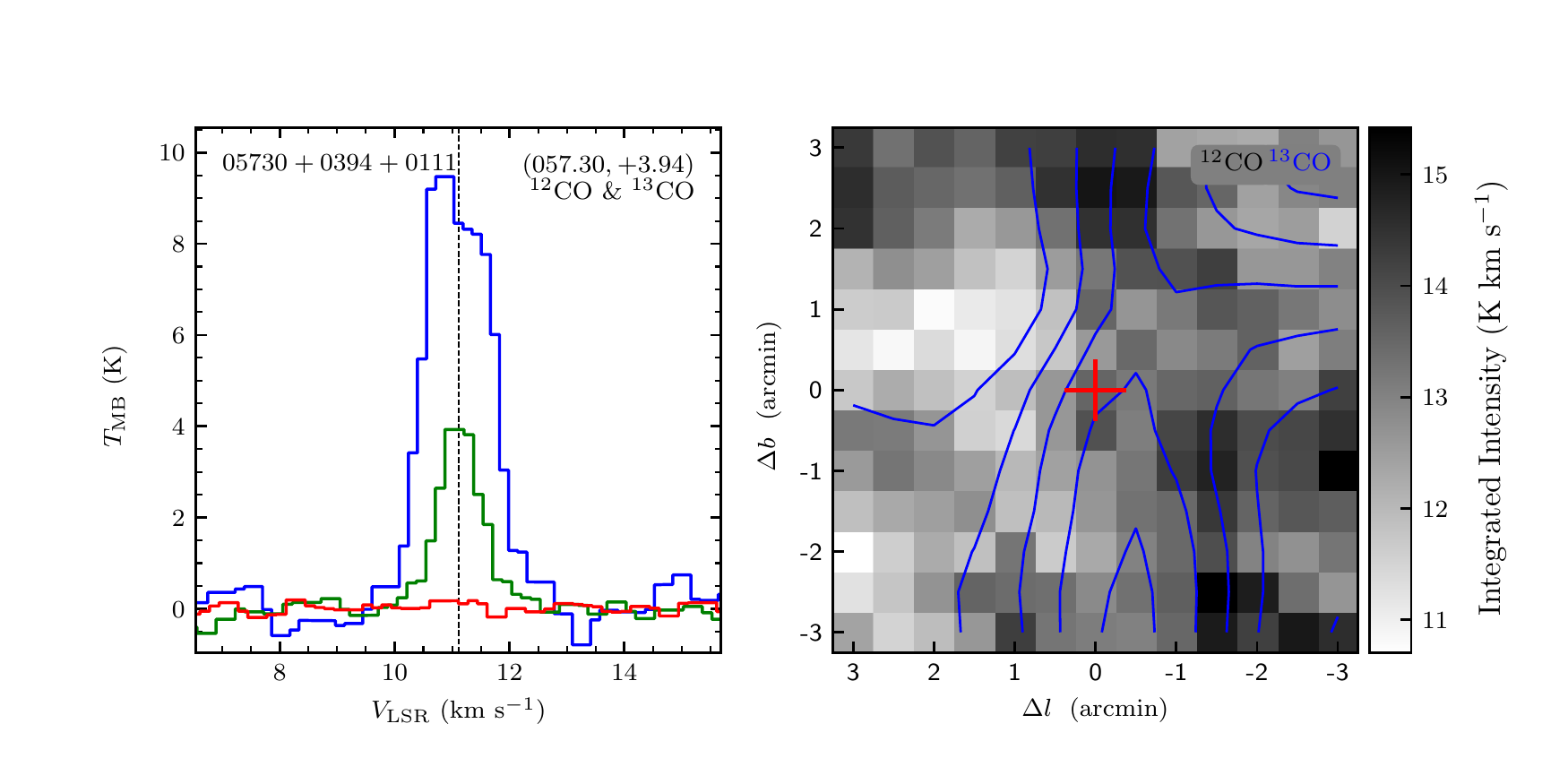}
\includegraphics[width=9.0cm,angle=0]{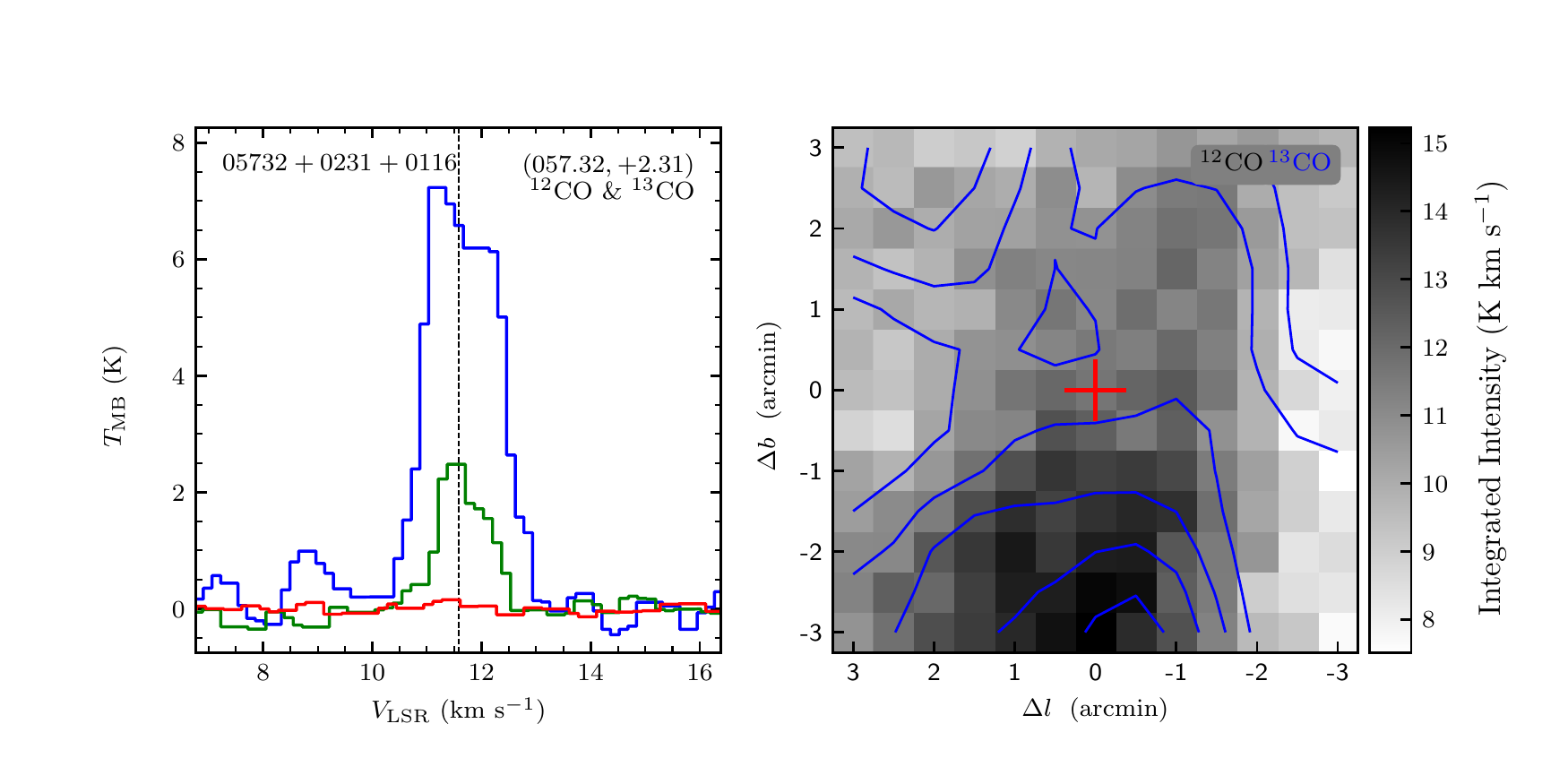}
\end{figure}
\clearpage

\begin{figure}
\includegraphics[width=9.0cm,angle=0]{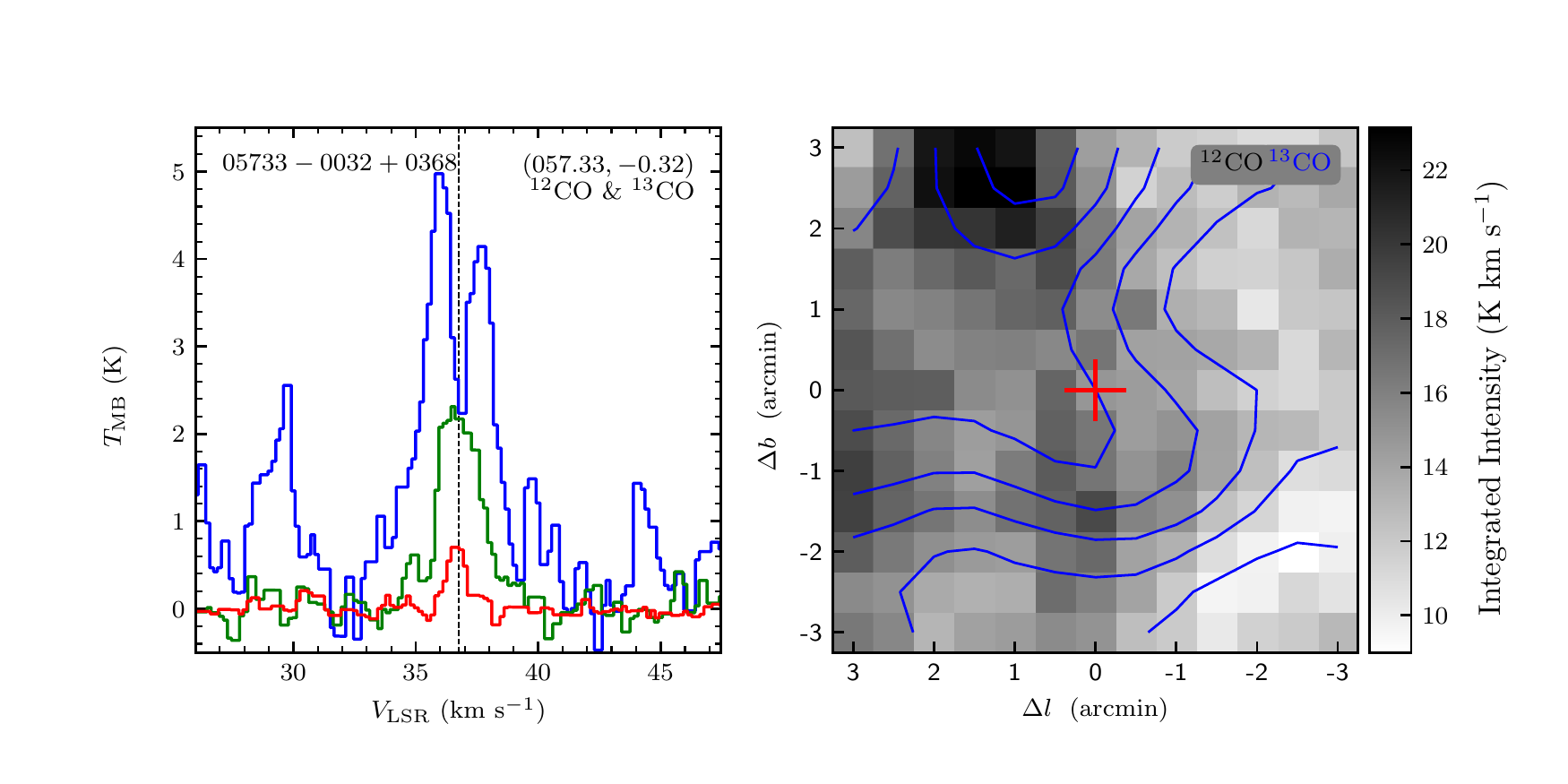}
\includegraphics[width=9.0cm,angle=0]{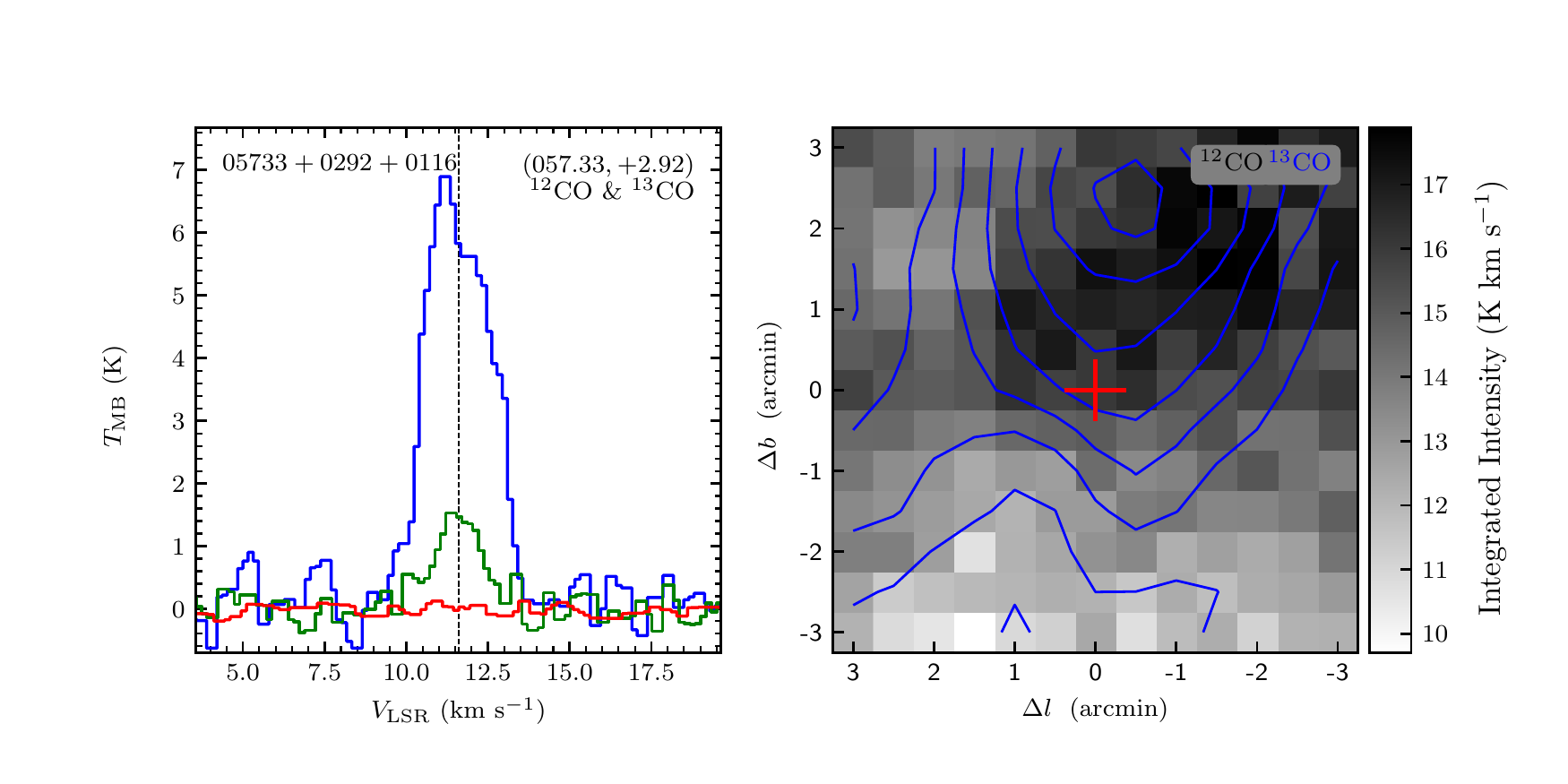}
\vspace{-0.5cm}

\includegraphics[width=9.0cm,angle=0]{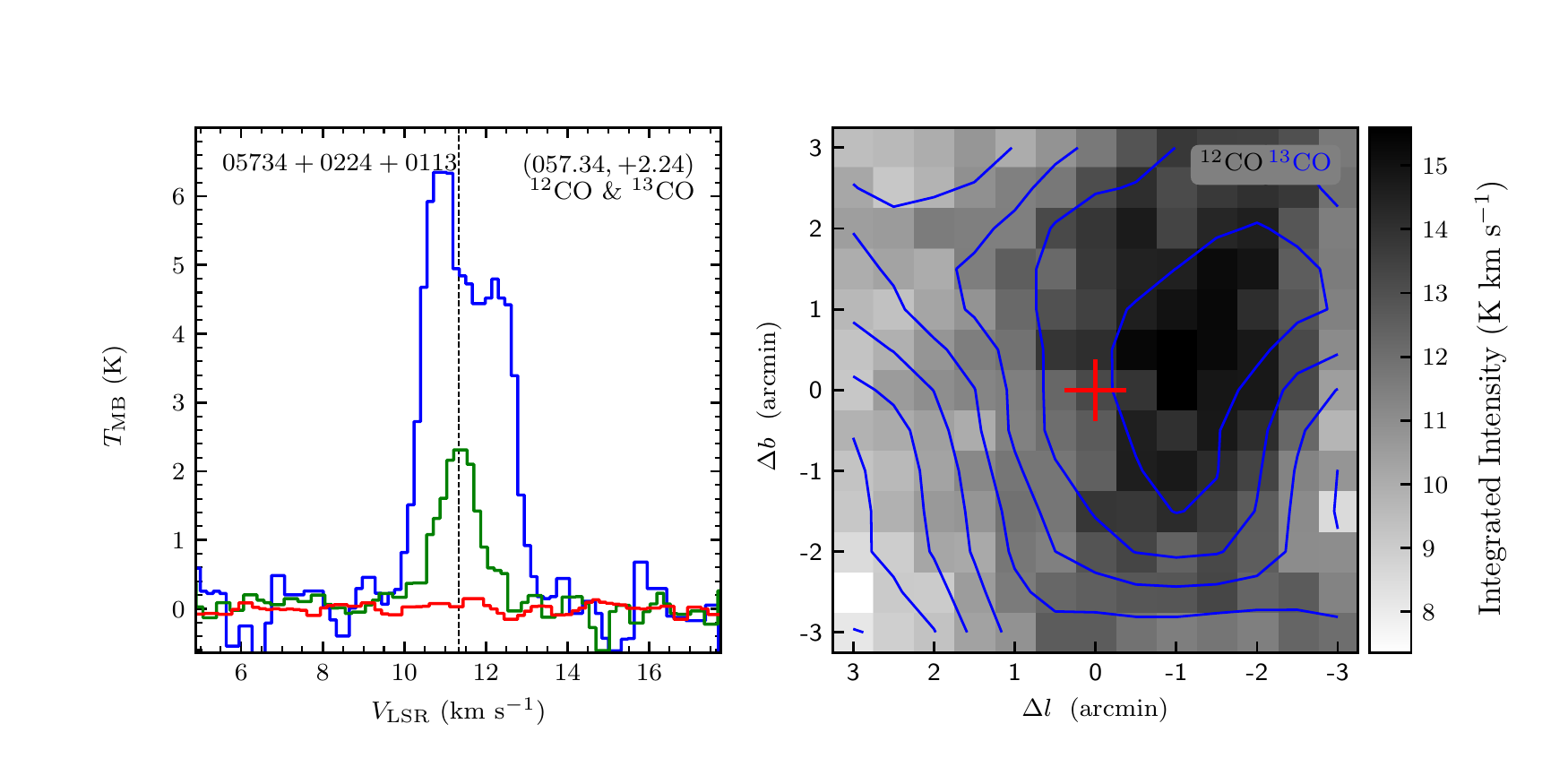}
\includegraphics[width=9.0cm,angle=0]{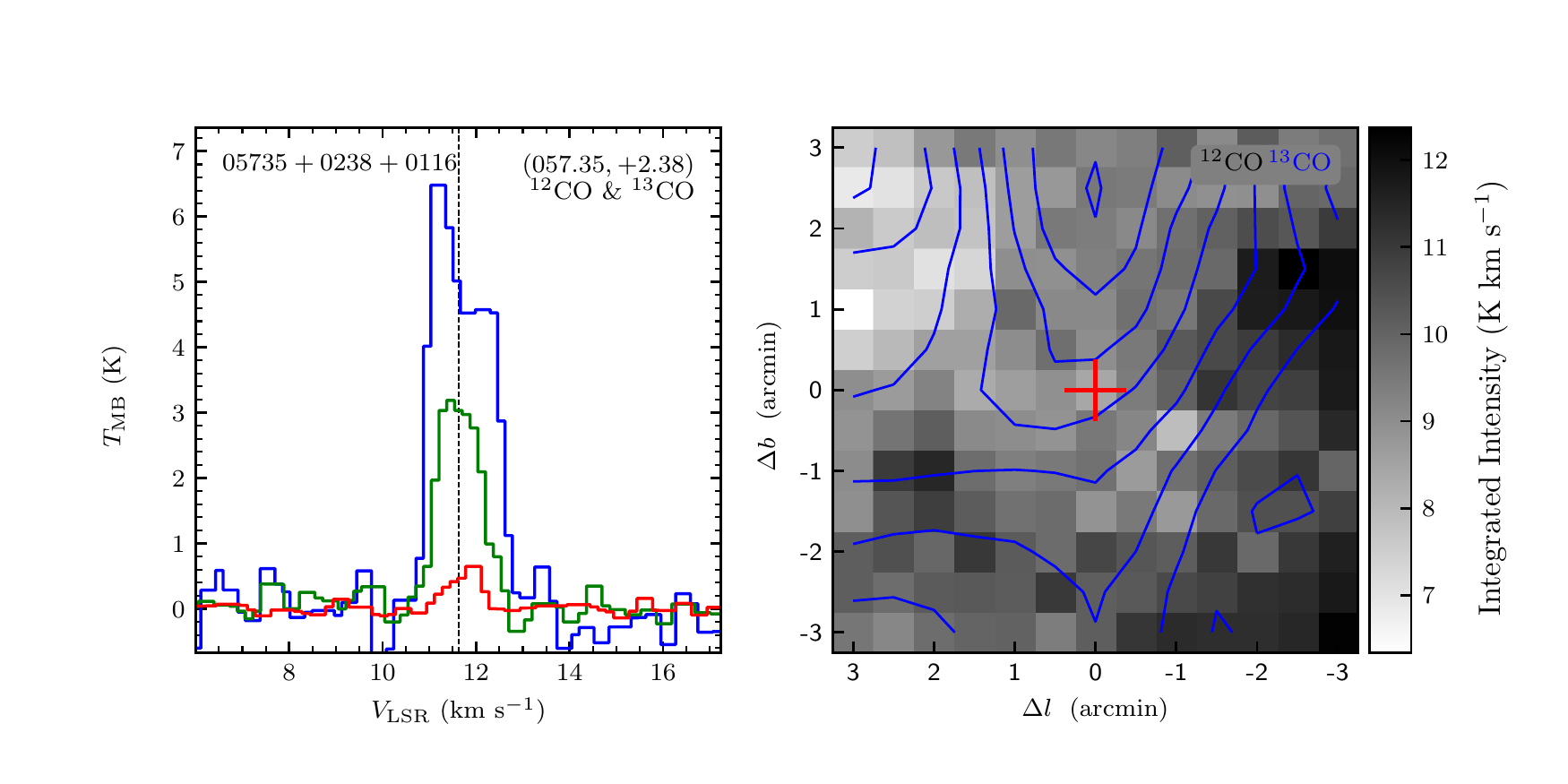}
\vspace{-0.5cm}

\includegraphics[width=9.0cm,angle=0]{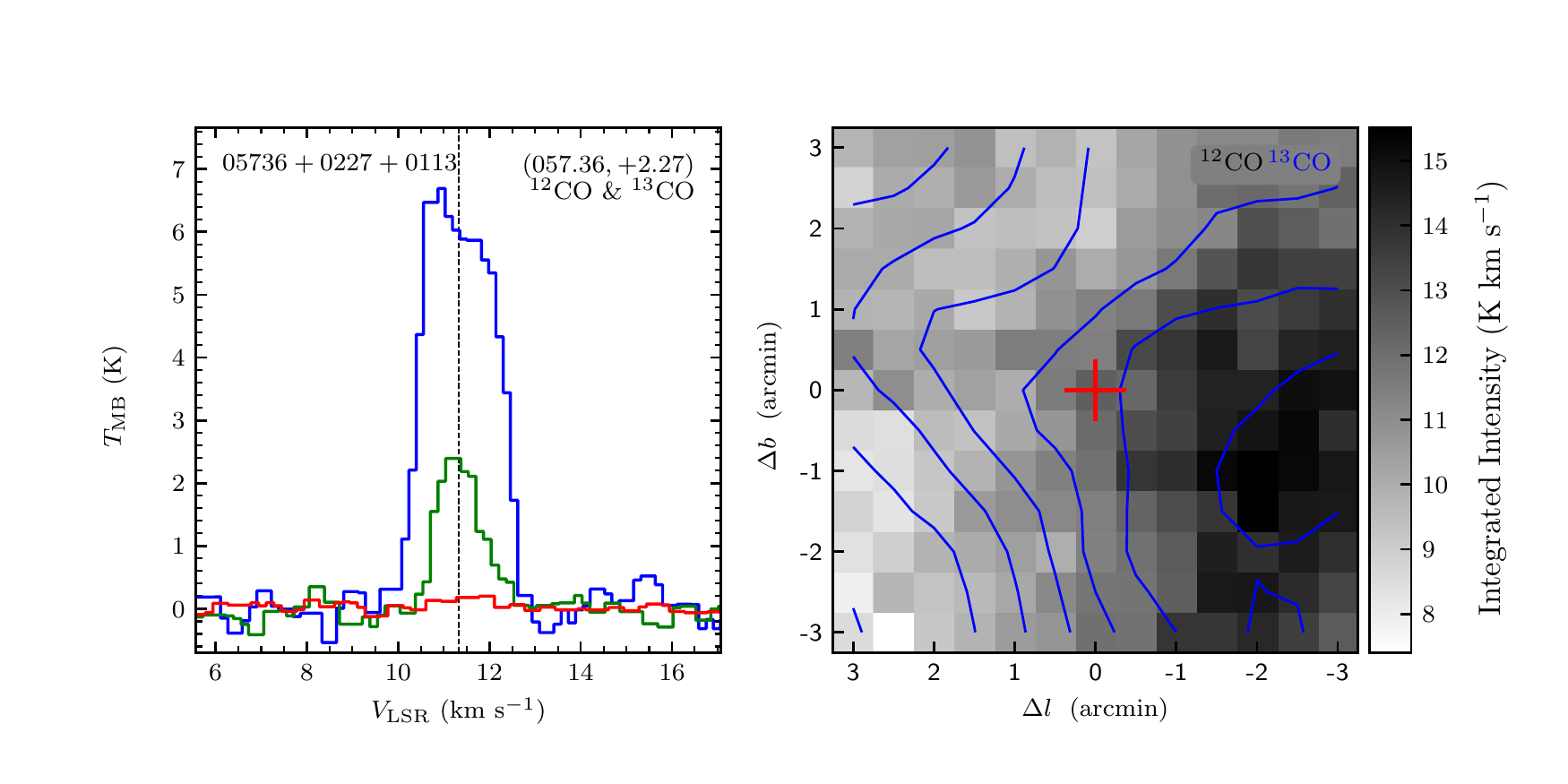}
\includegraphics[width=9.0cm,angle=0]{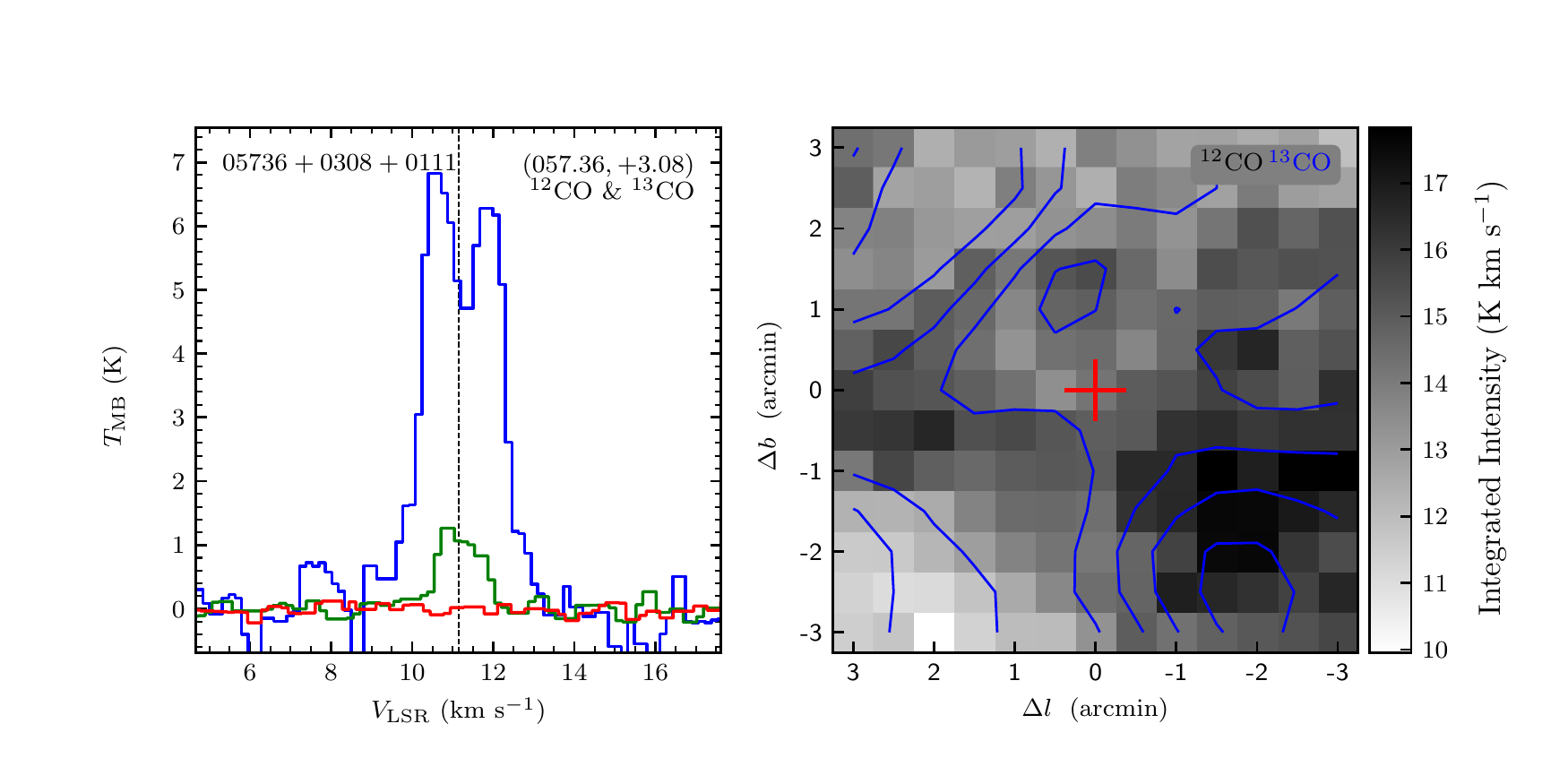}
\vspace{-0.5cm}

\includegraphics[width=9.0cm,angle=0]{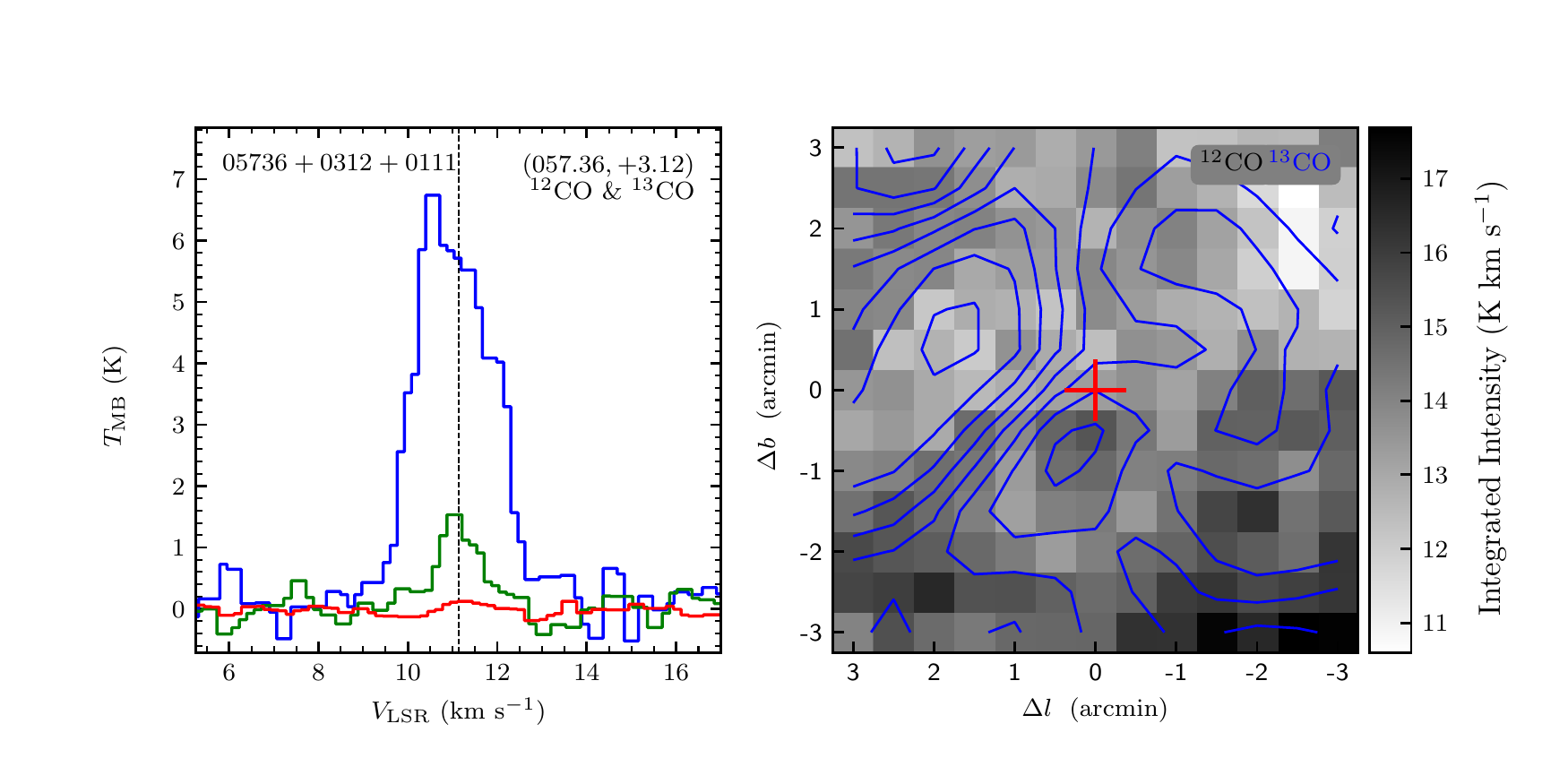}
\includegraphics[width=9.0cm,angle=0]{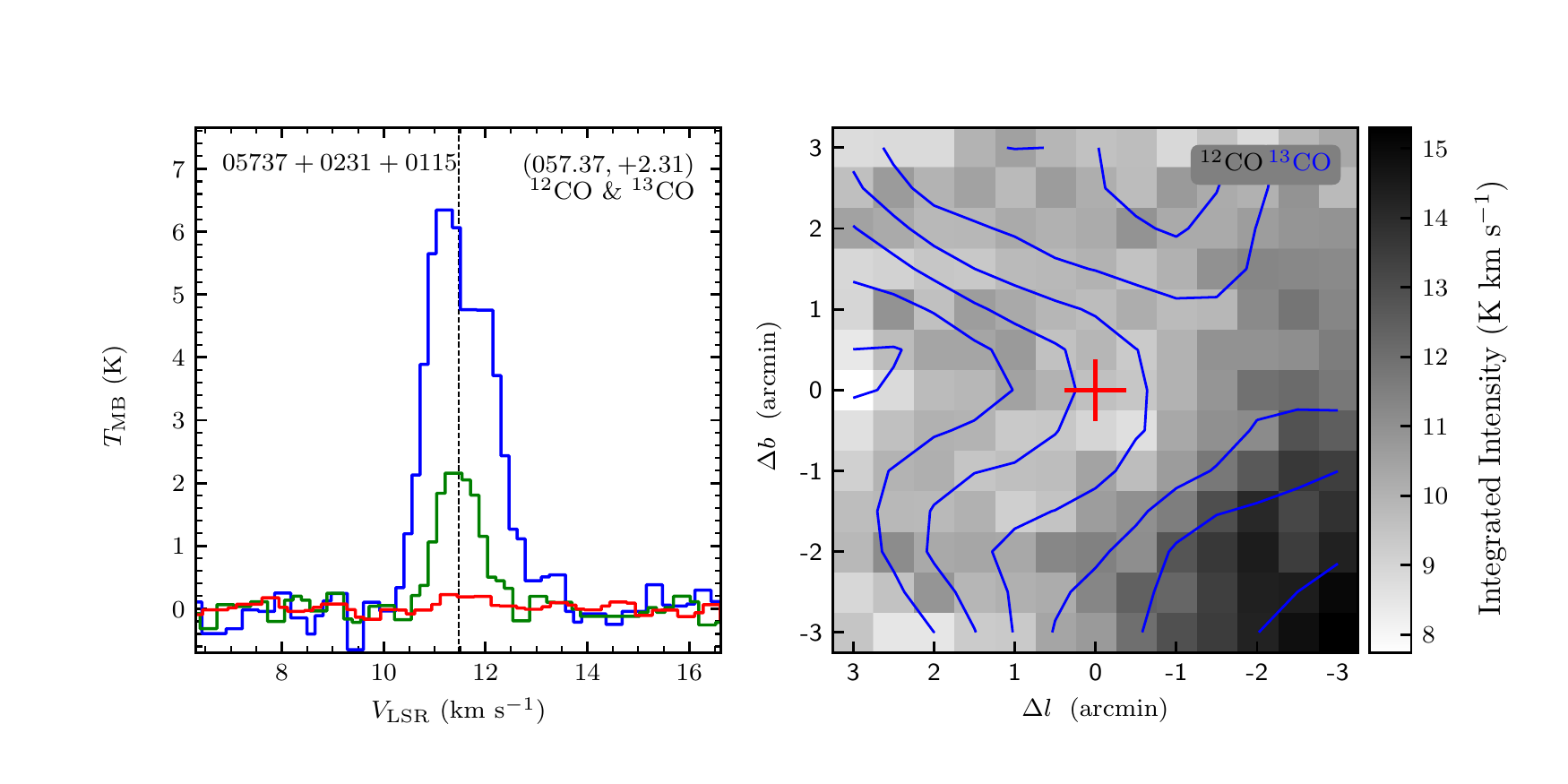}
\vspace{-0.5cm}

\includegraphics[width=9.0cm,angle=0]{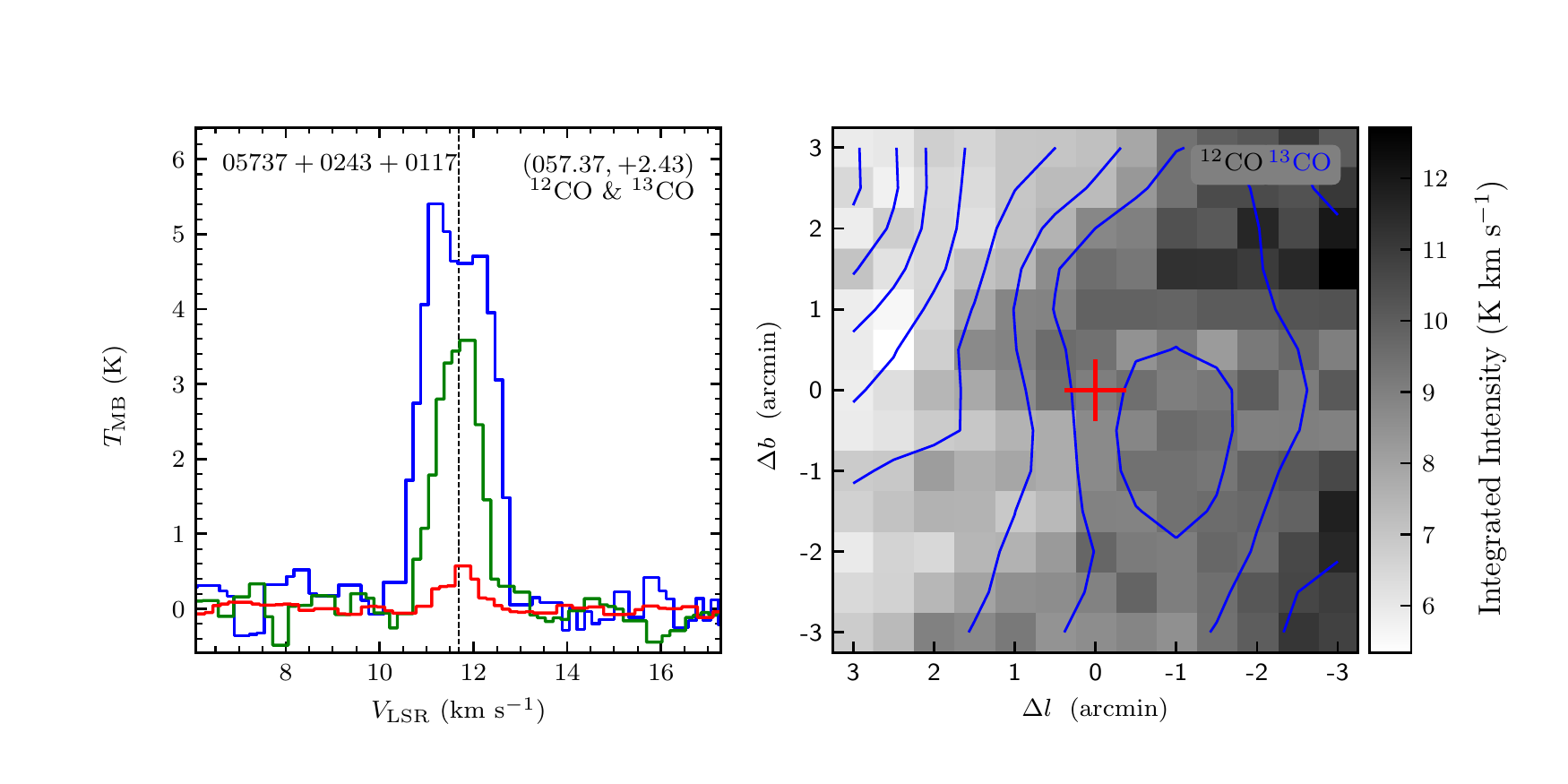}
\includegraphics[width=9.0cm,angle=0]{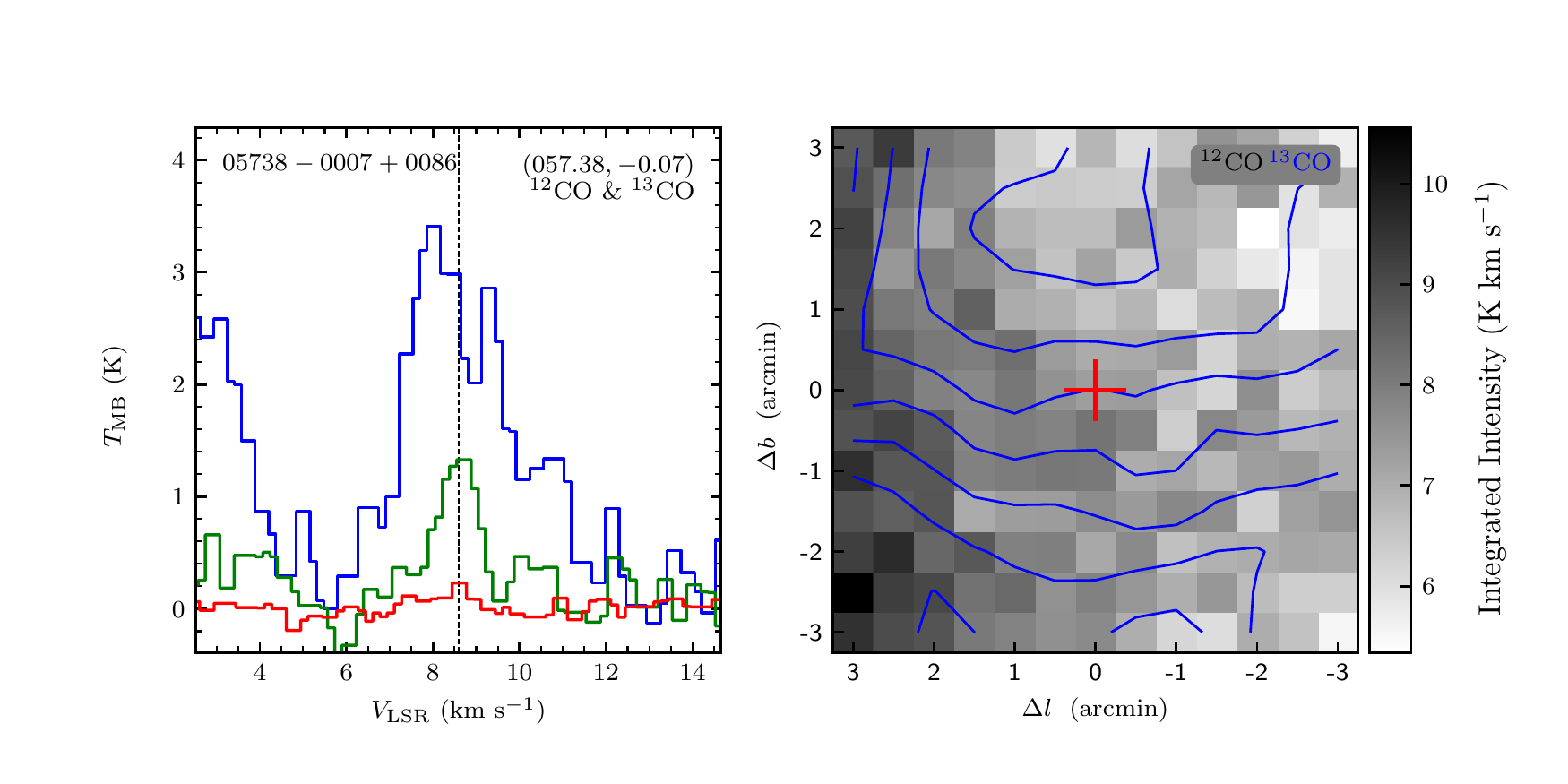}
\end{figure}
\clearpage

\begin{figure}
\includegraphics[width=9.0cm,angle=0]{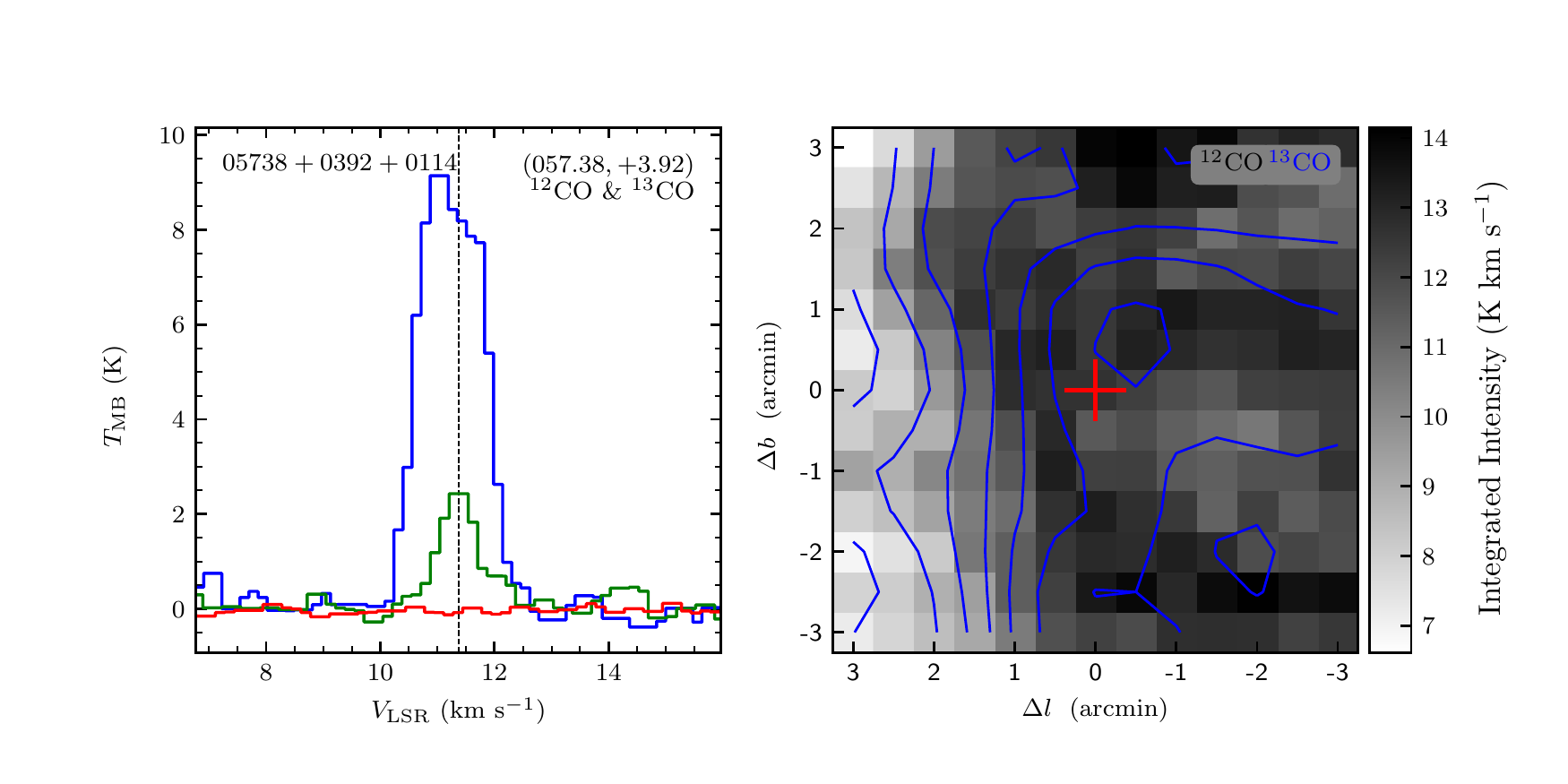}
\includegraphics[width=9.0cm,angle=0]{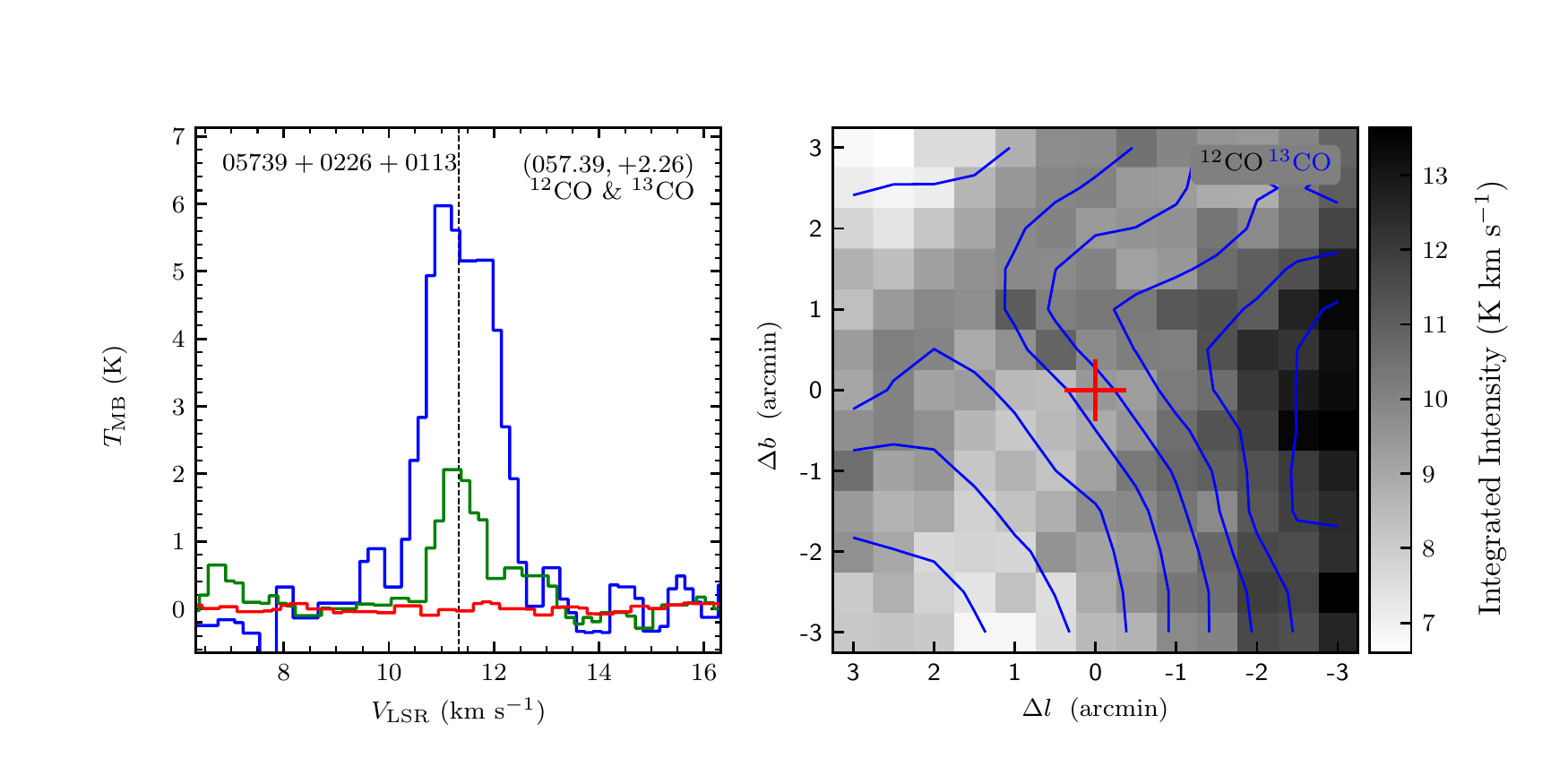}
\vspace{-0.5cm}

\includegraphics[width=9.0cm,angle=0]{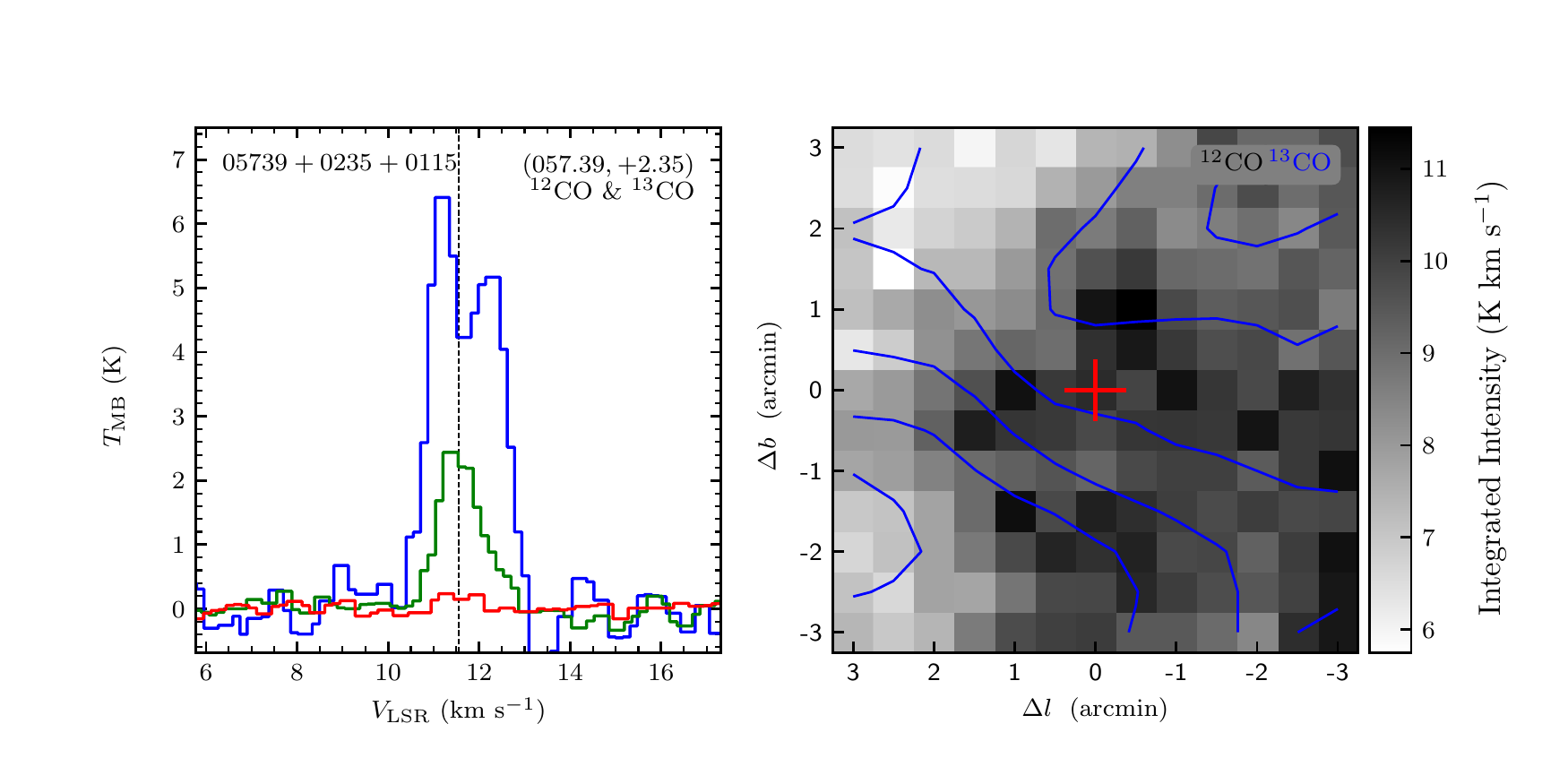}
\includegraphics[width=9.0cm,angle=0]{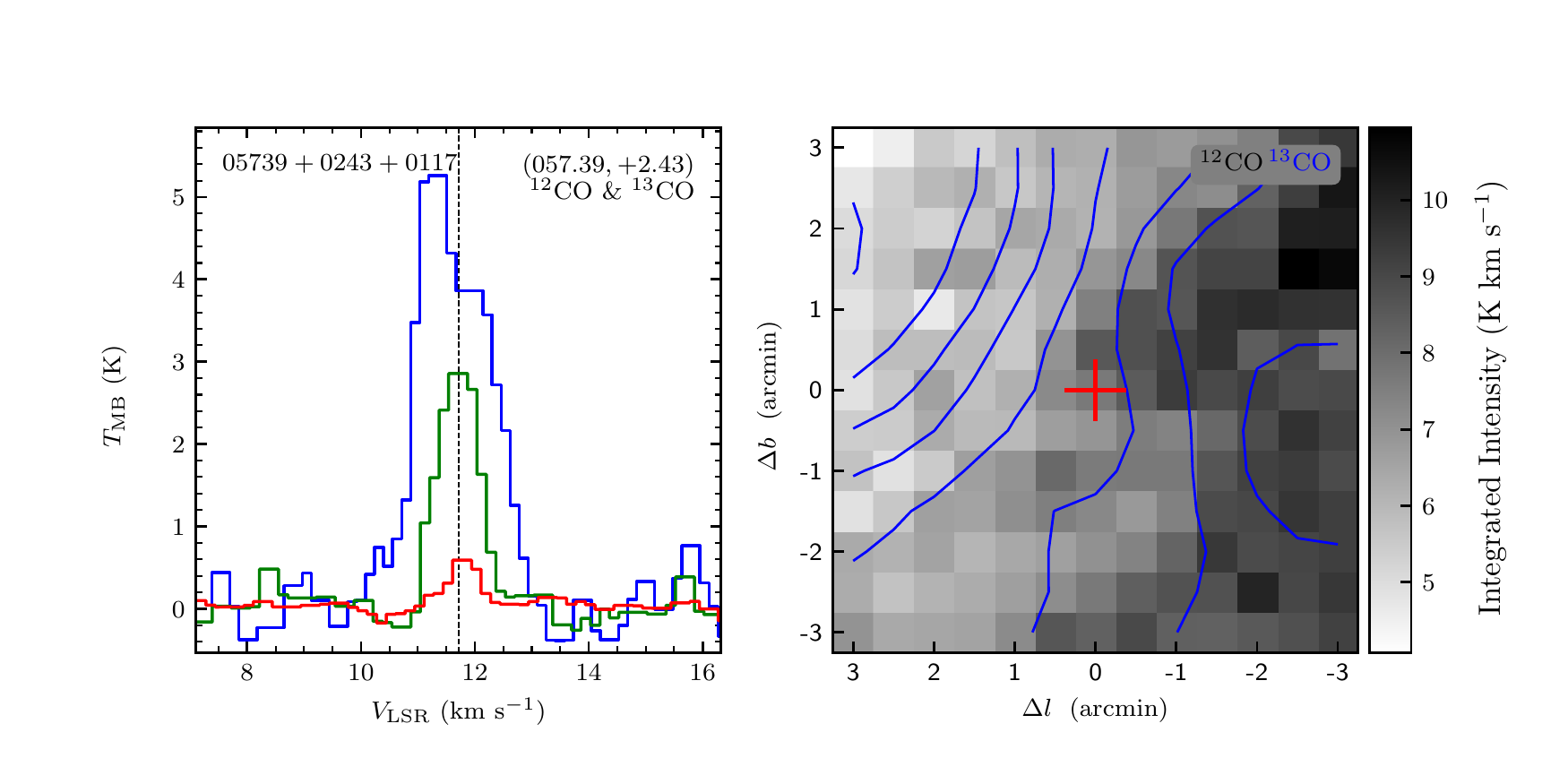}
\vspace{-0.5cm}

\includegraphics[width=9.0cm,angle=0]{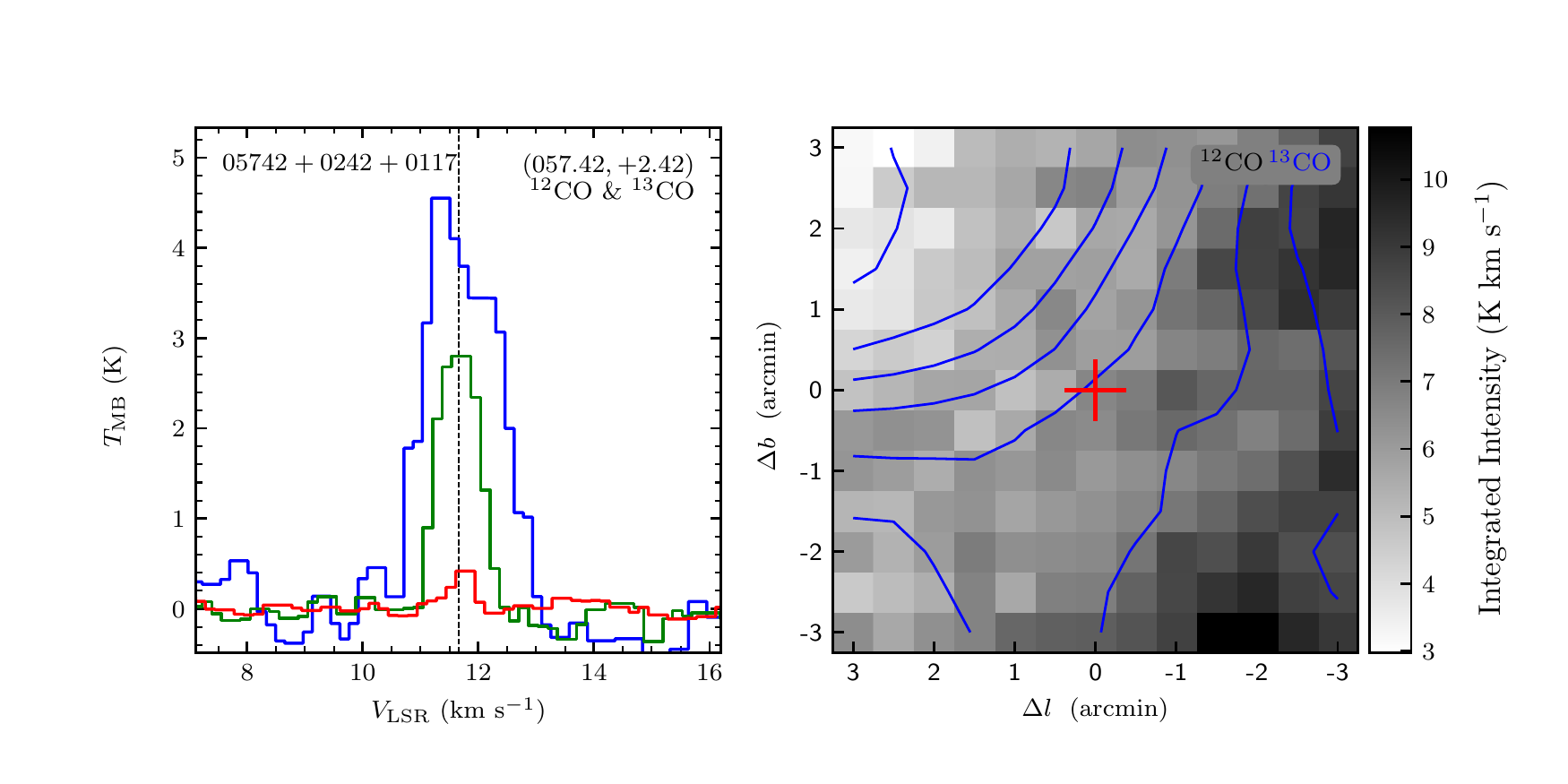}
\includegraphics[width=9.0cm,angle=0]{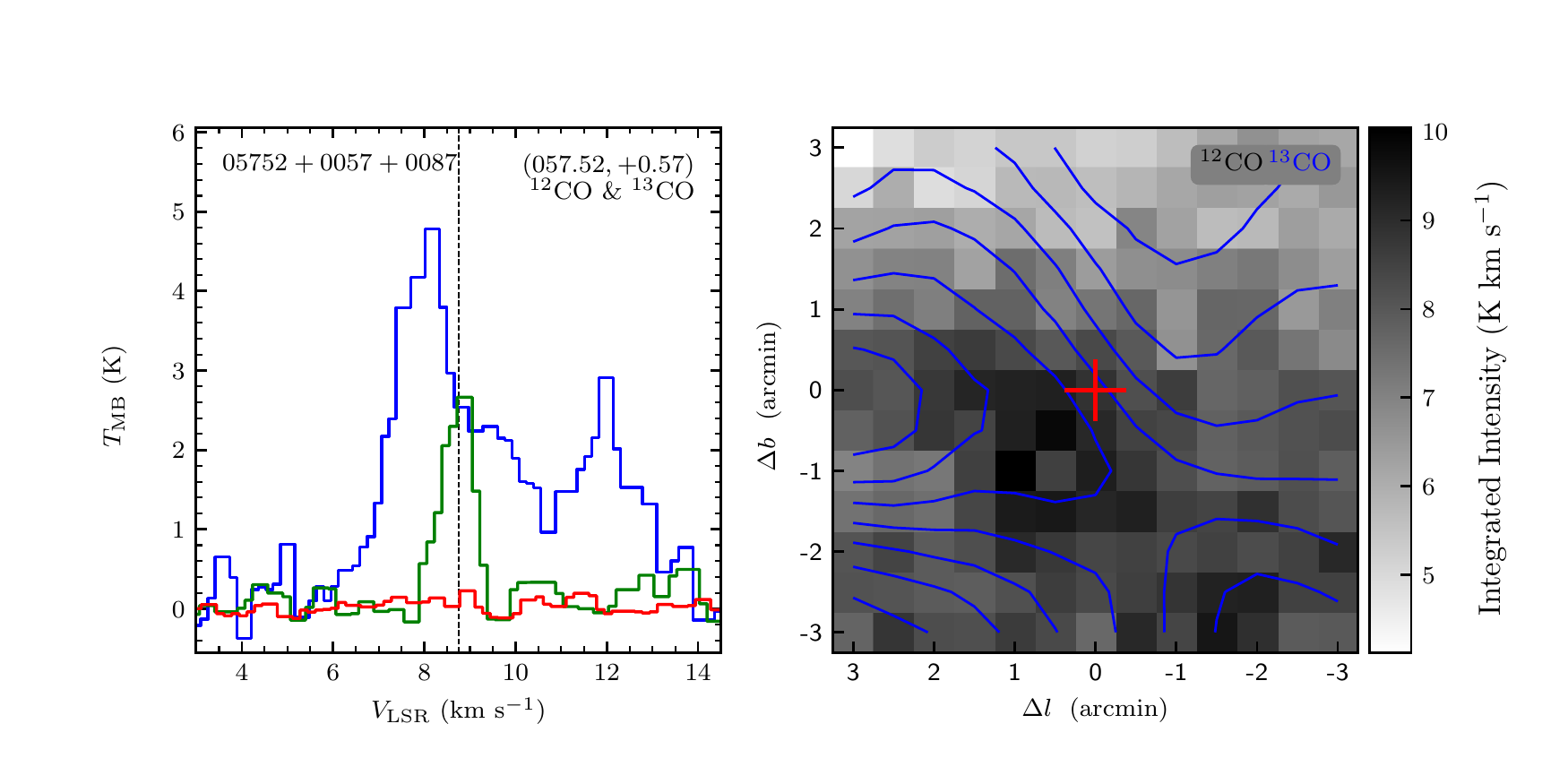}
\vspace{-0.5cm}

\includegraphics[width=9.0cm,angle=0]{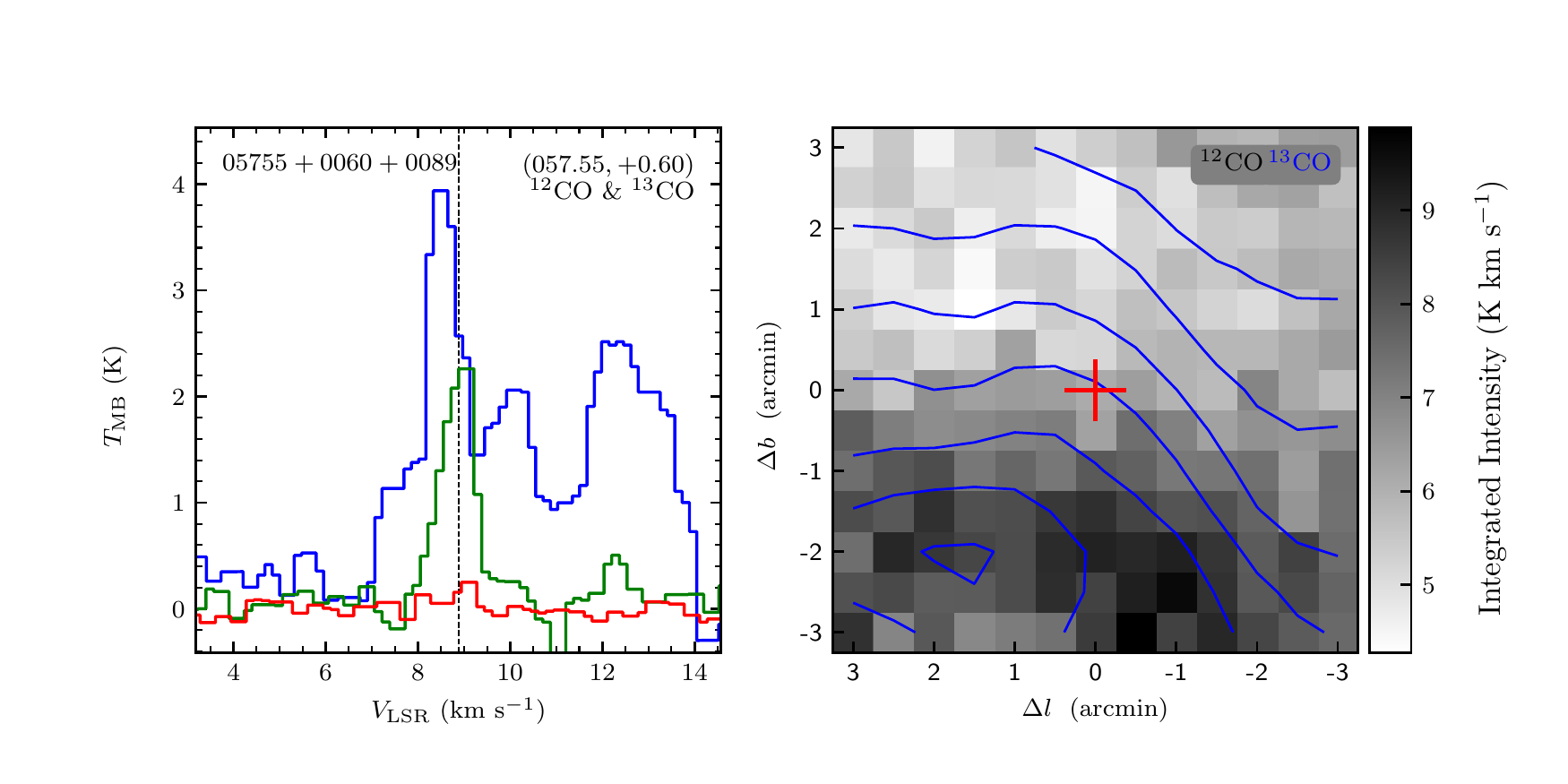}
\includegraphics[width=9.0cm,angle=0]{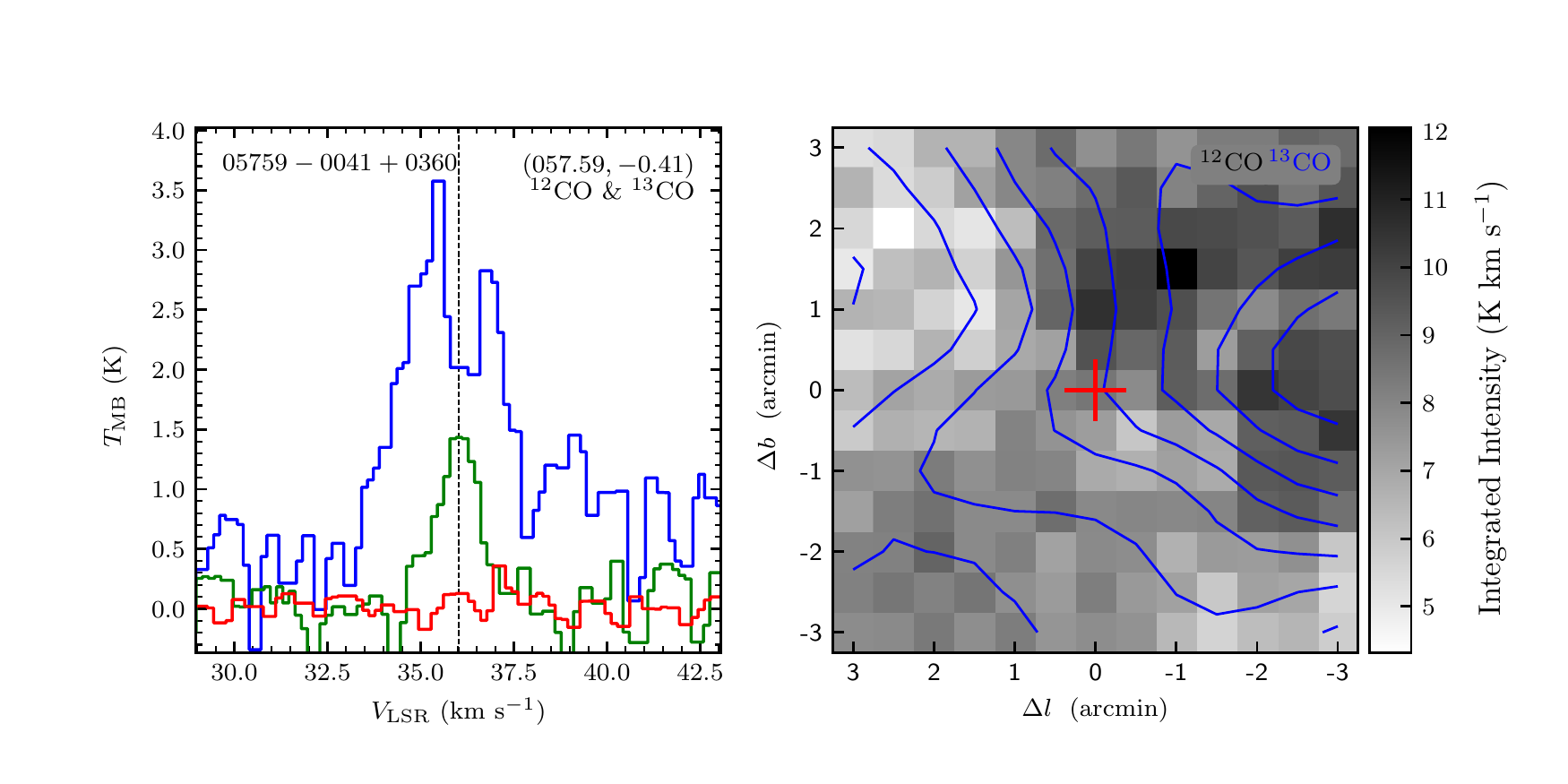}
\vspace{-0.5cm}

\includegraphics[width=9.0cm,angle=0]{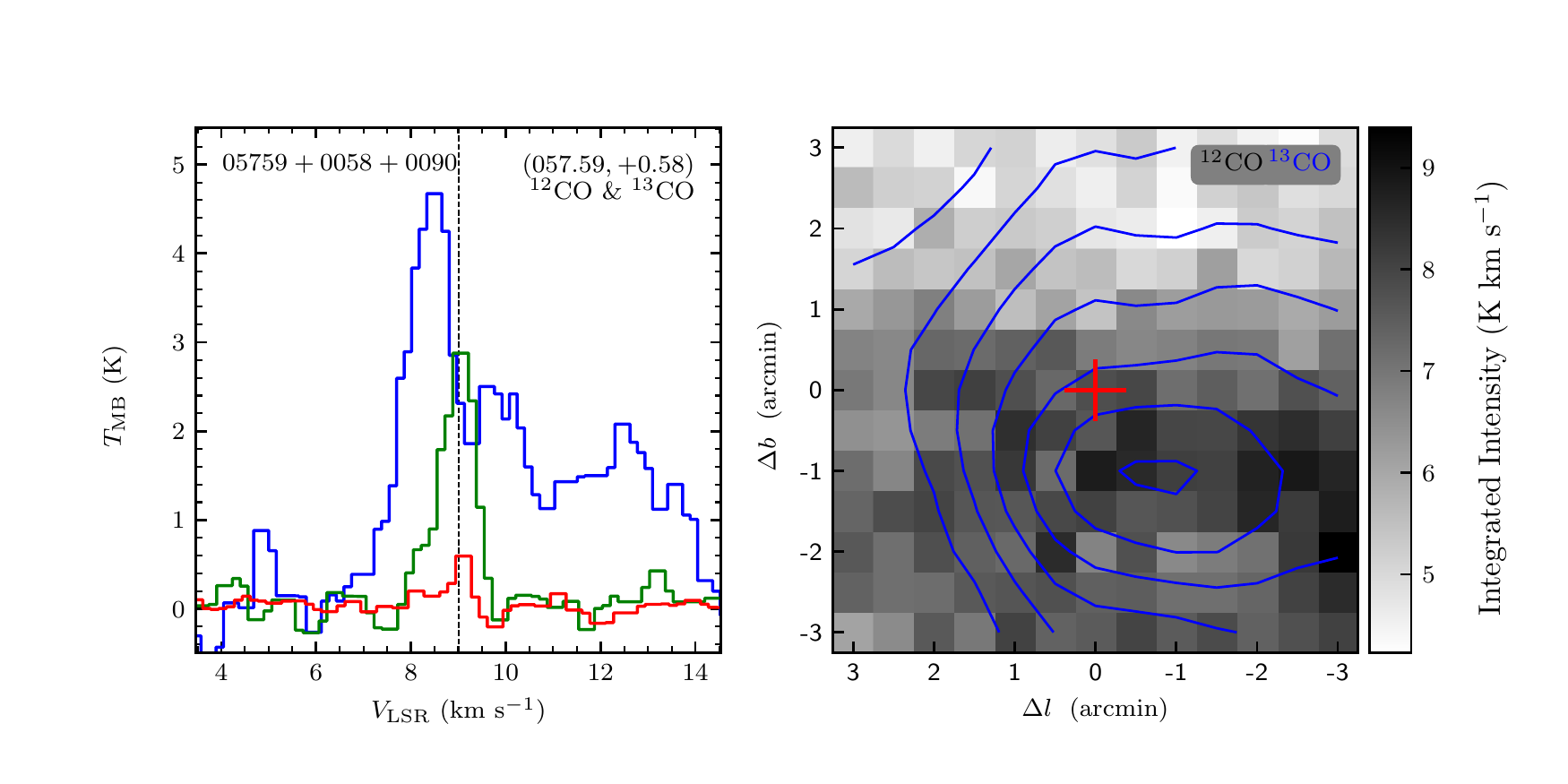}
\includegraphics[width=9.0cm,angle=0]{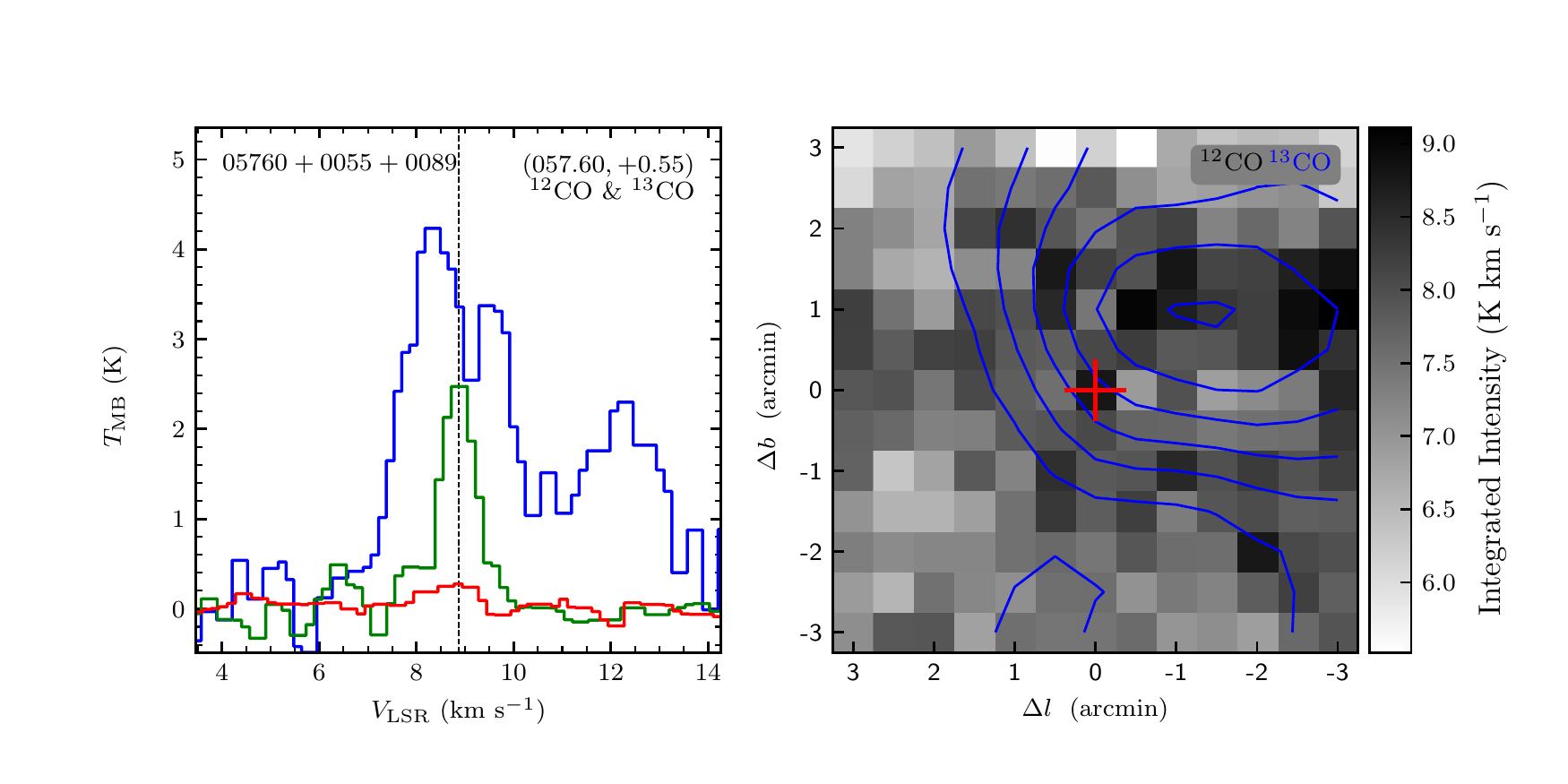}
\end{figure}
\clearpage

\begin{figure}
\includegraphics[width=9.0cm,angle=0]{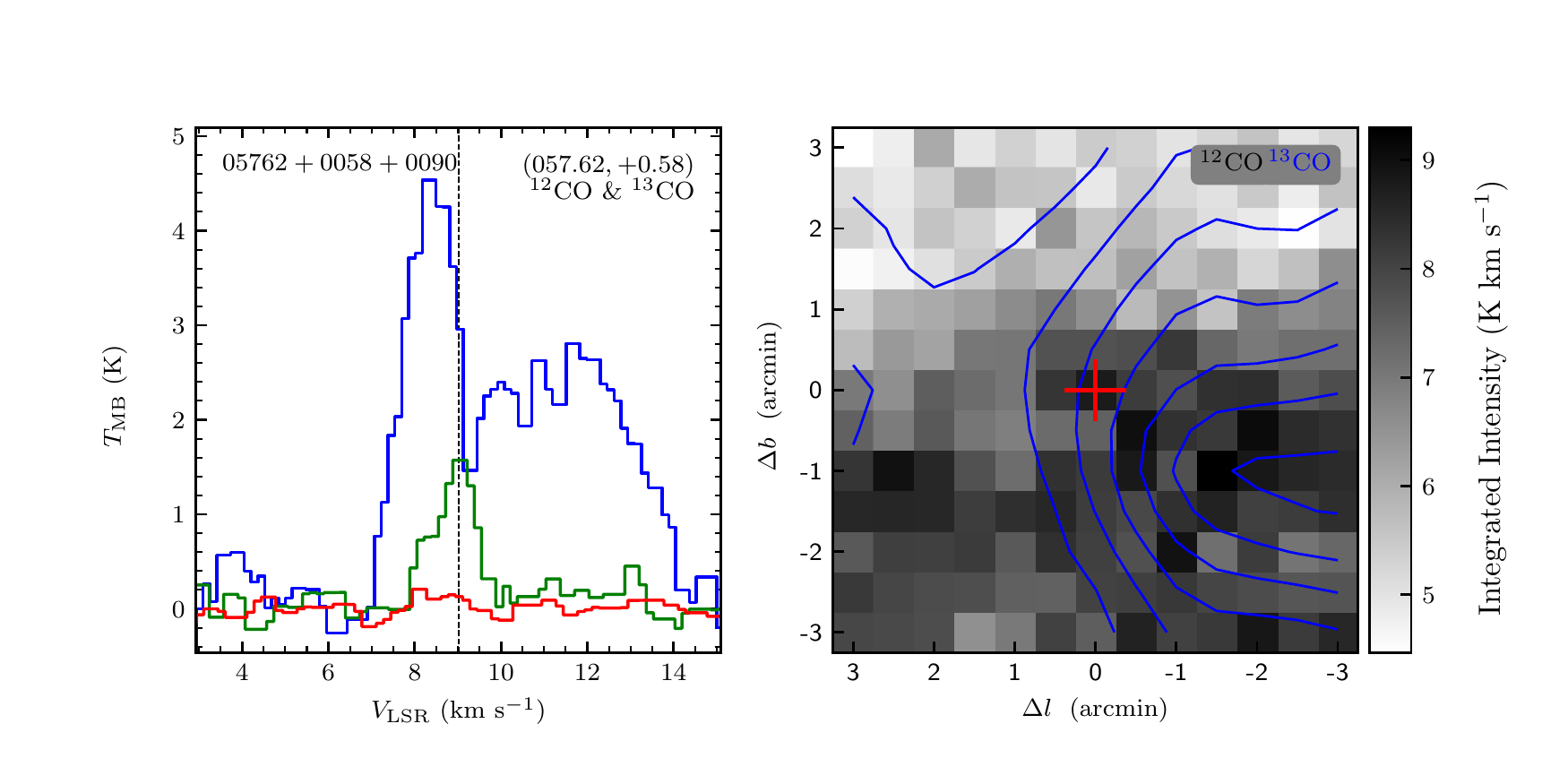}
\includegraphics[width=9.0cm,angle=0]{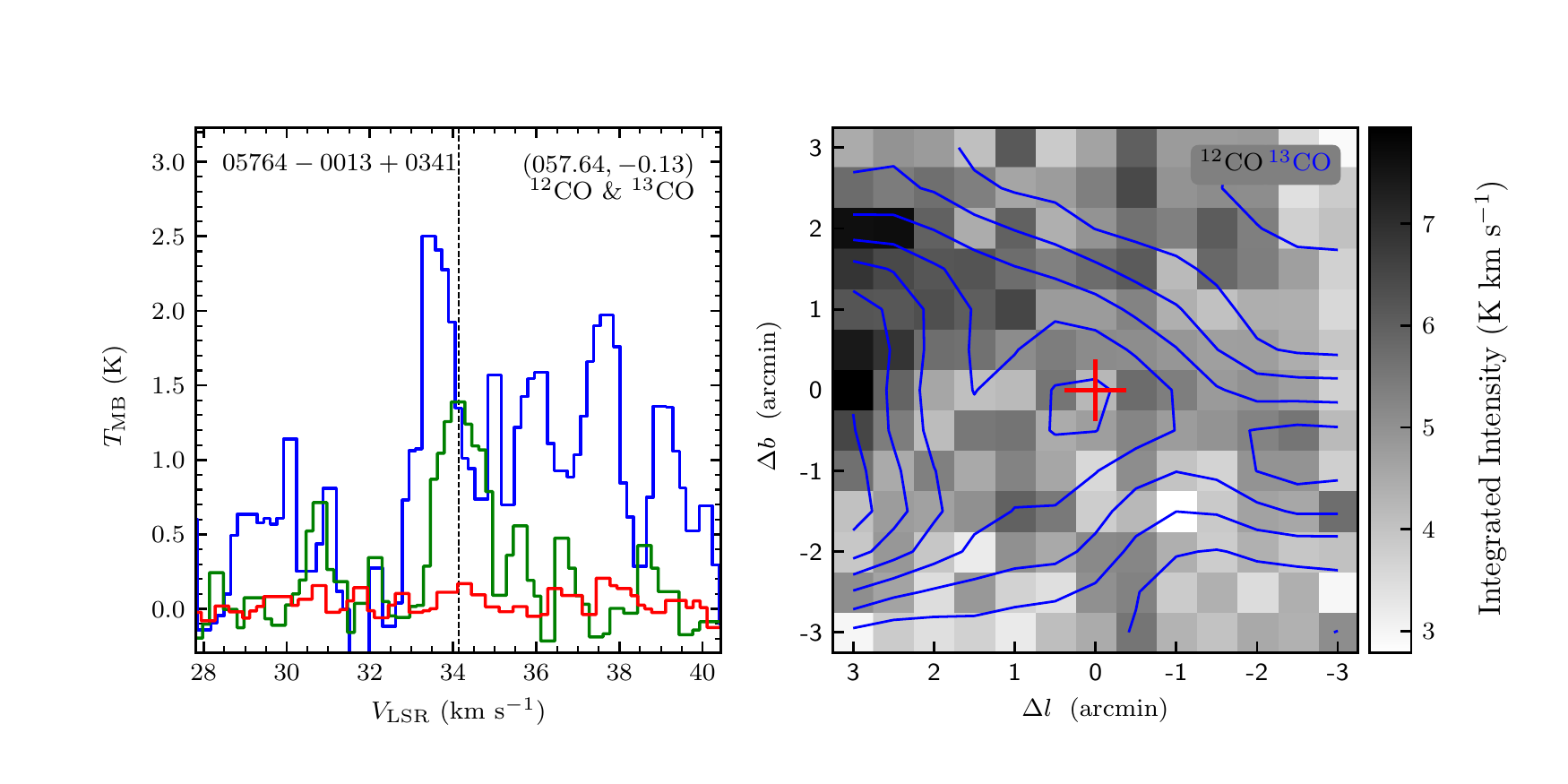}
\vspace{-0.5cm}

\includegraphics[width=9.0cm,angle=0]{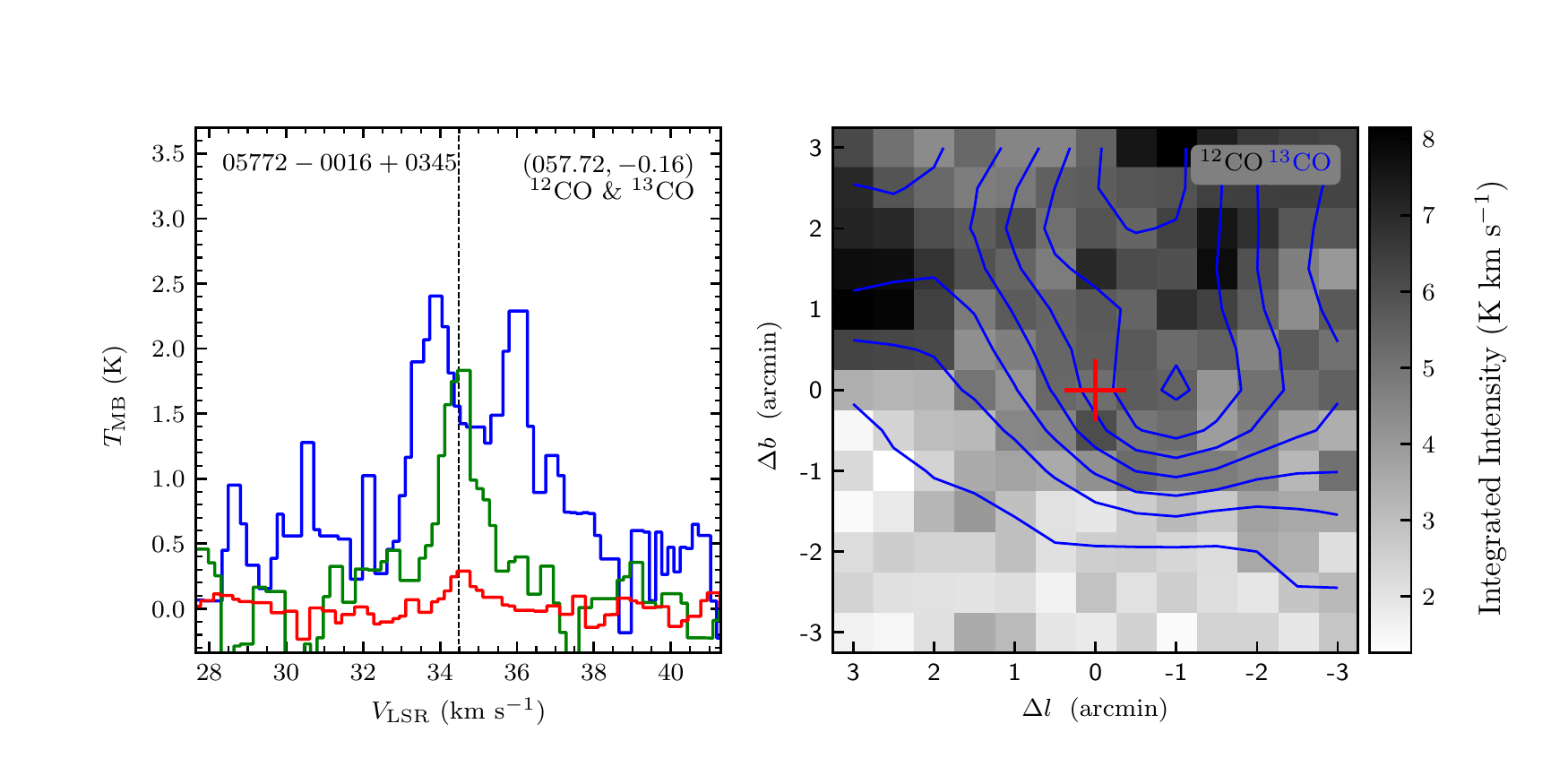}
\includegraphics[width=9.0cm,angle=0]{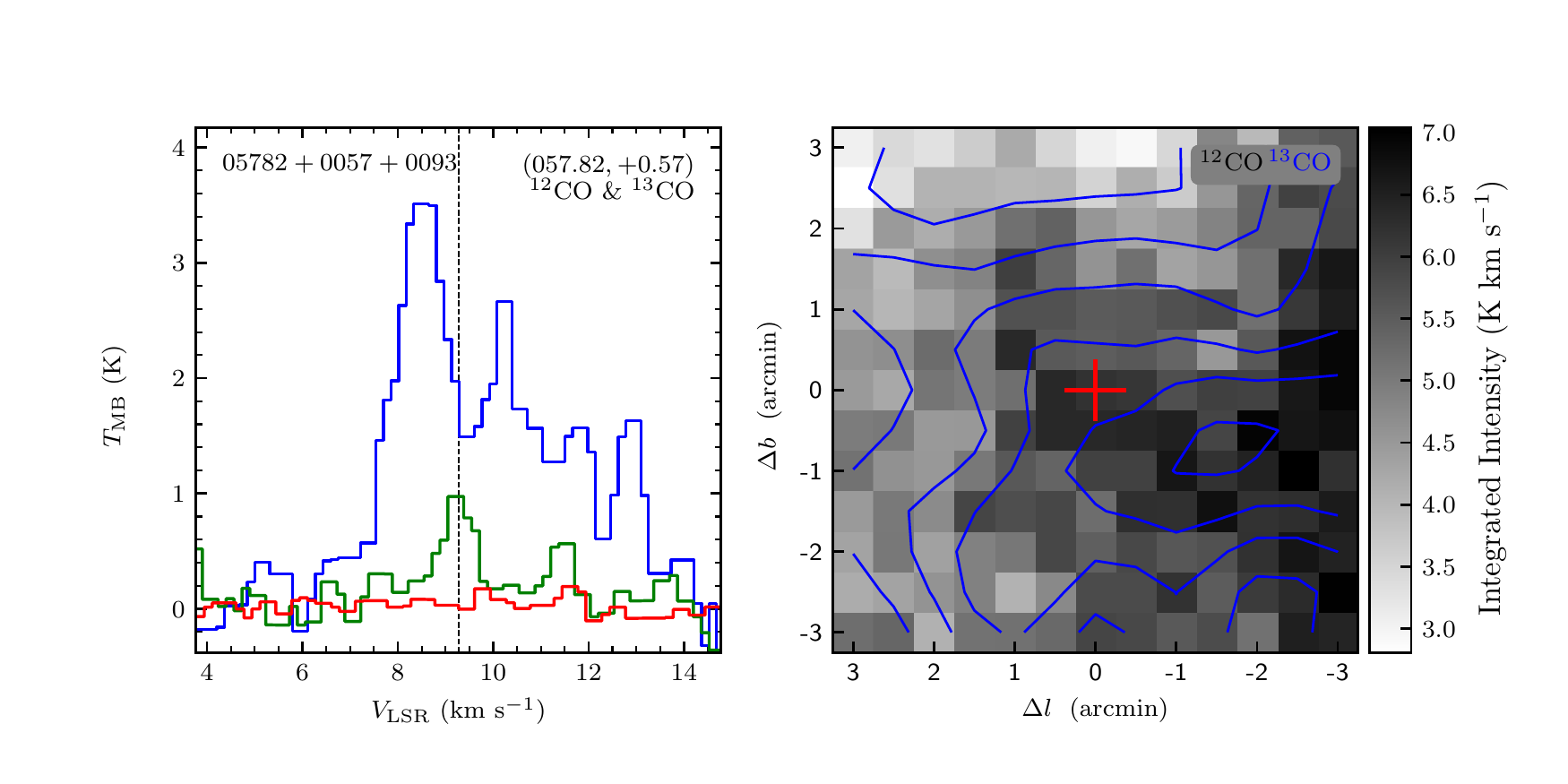}
\vspace{-0.5cm}

\includegraphics[width=9.0cm,angle=0]{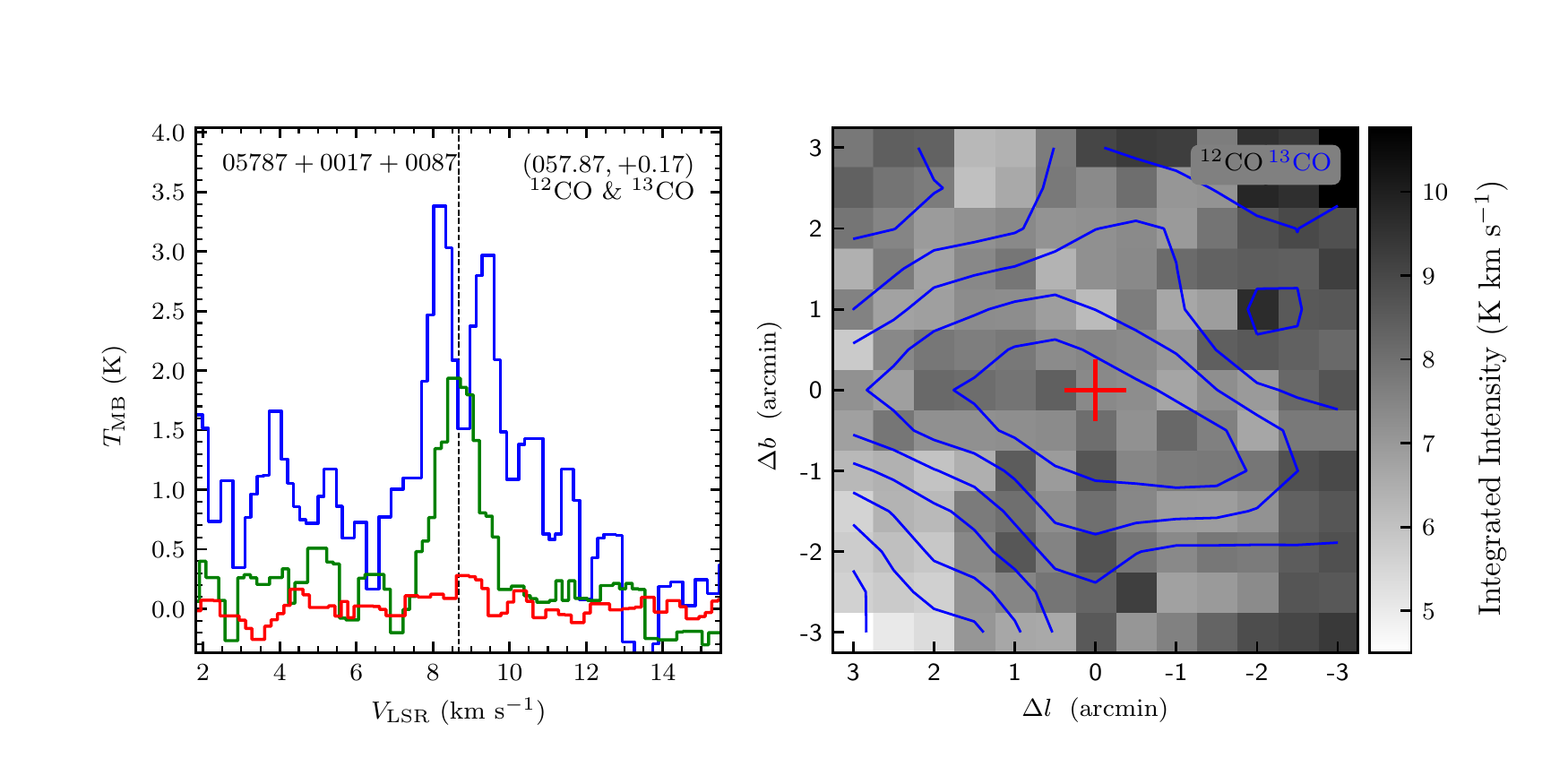}
\includegraphics[width=9.0cm,angle=0]{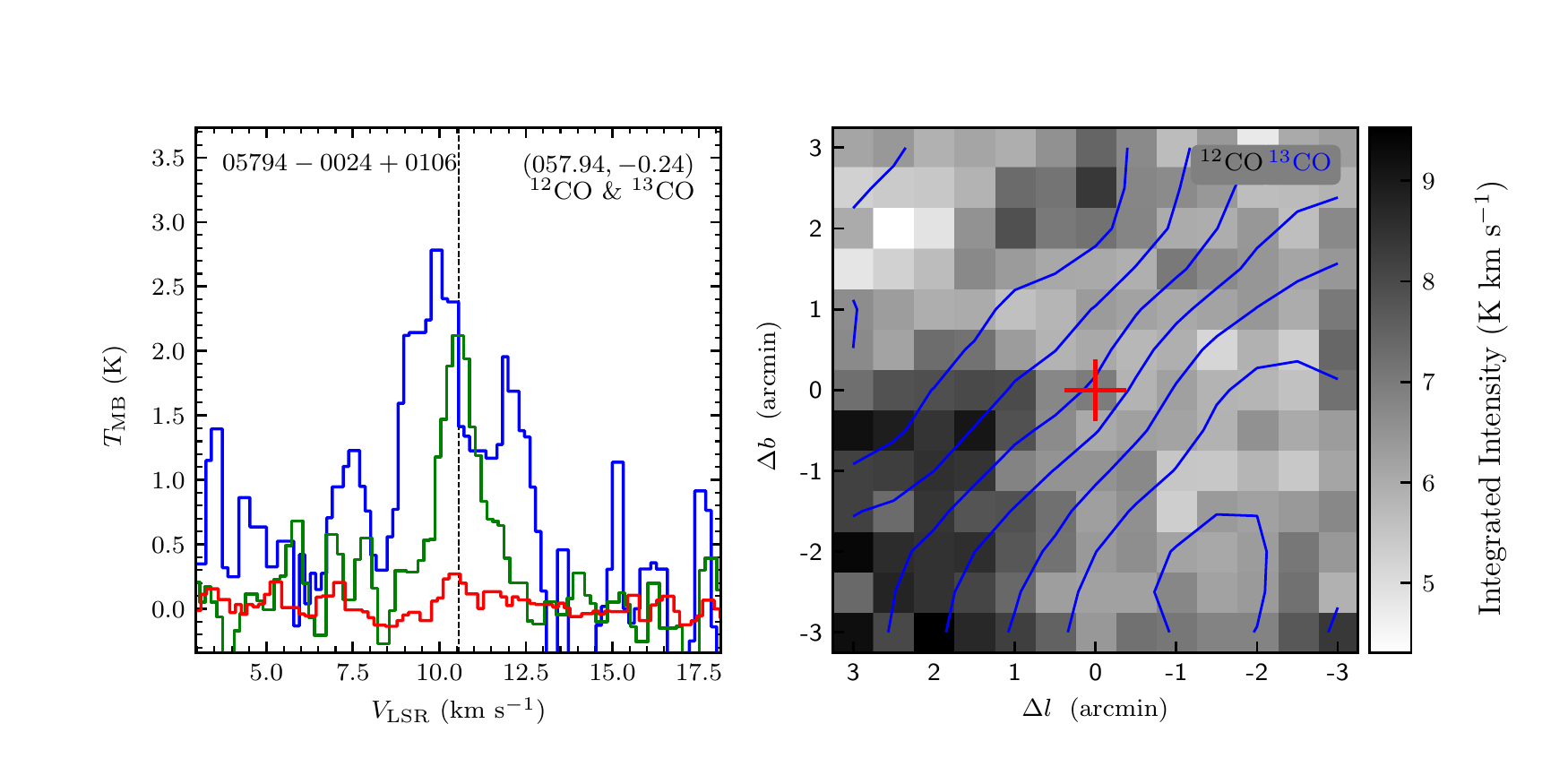}
\vspace{-0.5cm}

\includegraphics[width=9.0cm,angle=0]{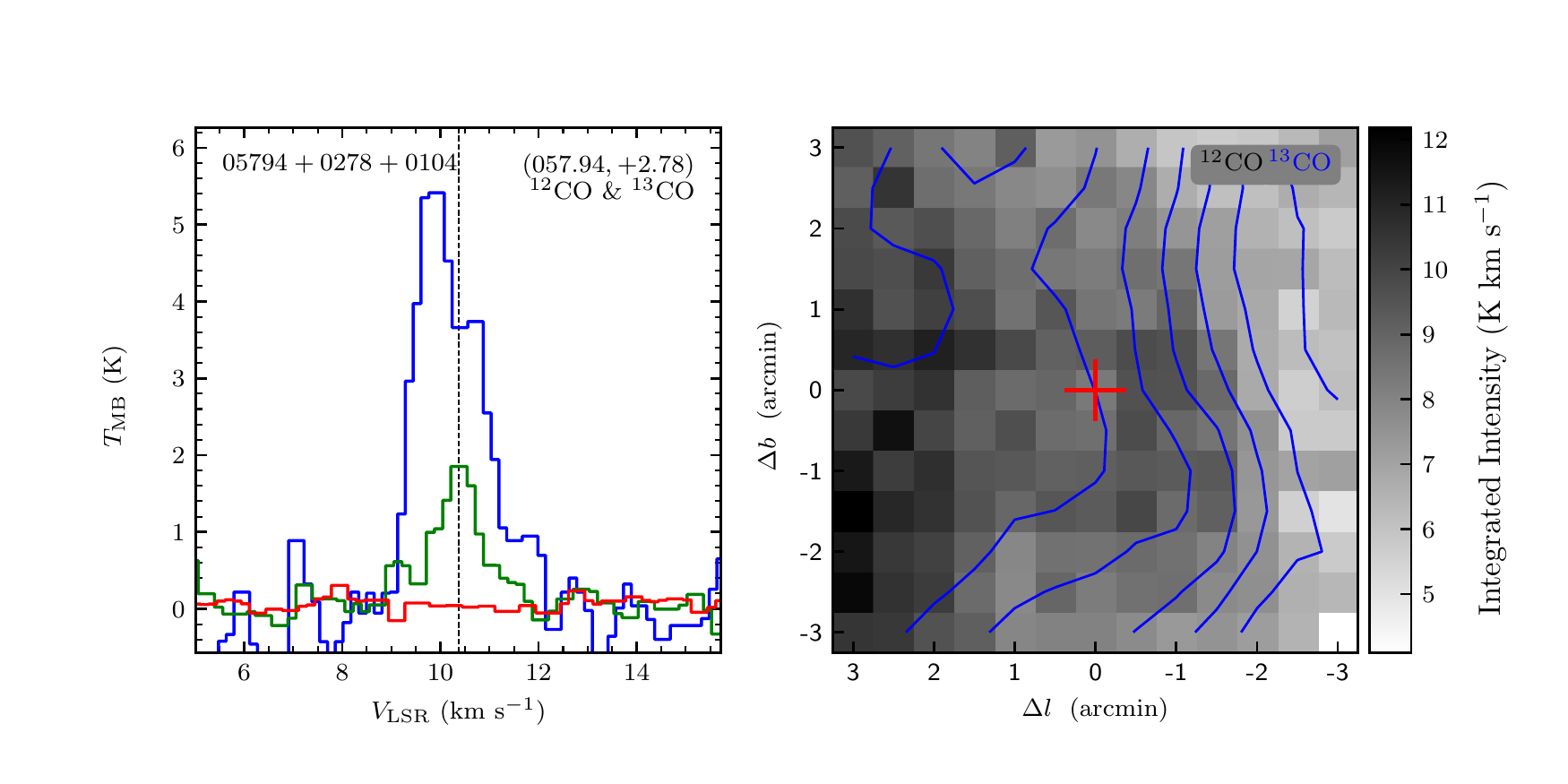}
\includegraphics[width=9.0cm,angle=0]{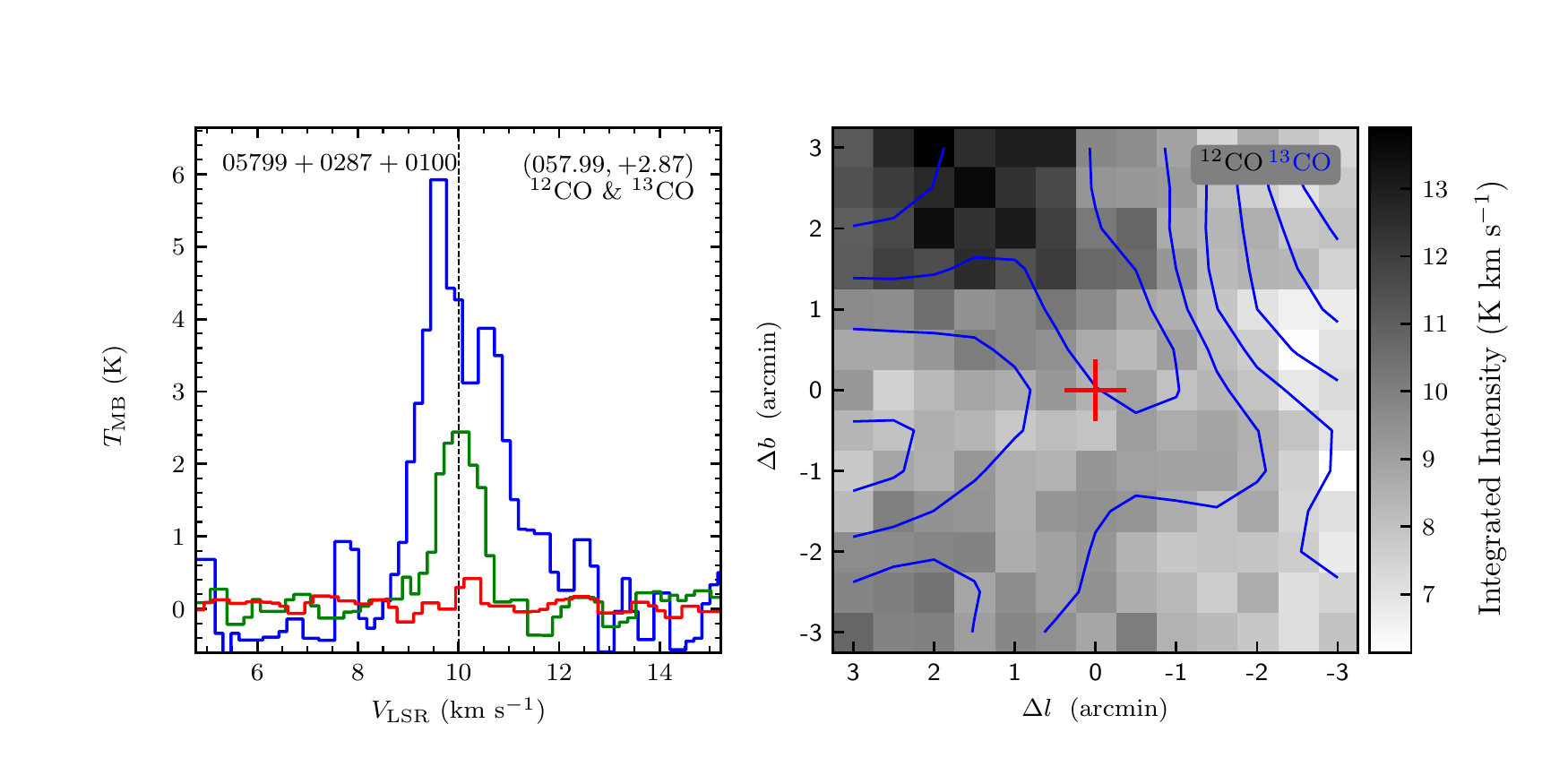}
\vspace{-0.5cm}

\includegraphics[width=9.0cm,angle=0]{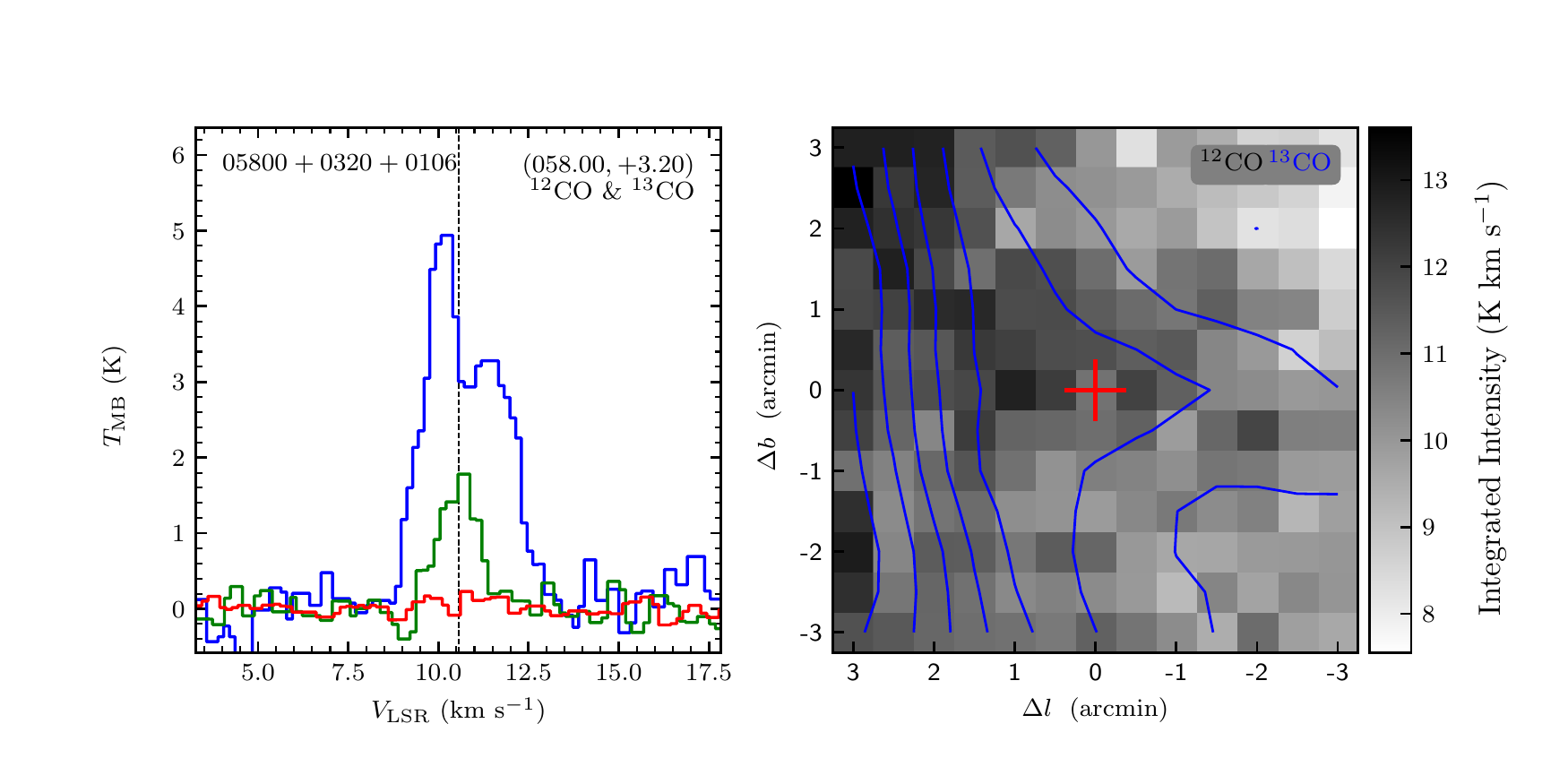}
\includegraphics[width=9.0cm,angle=0]{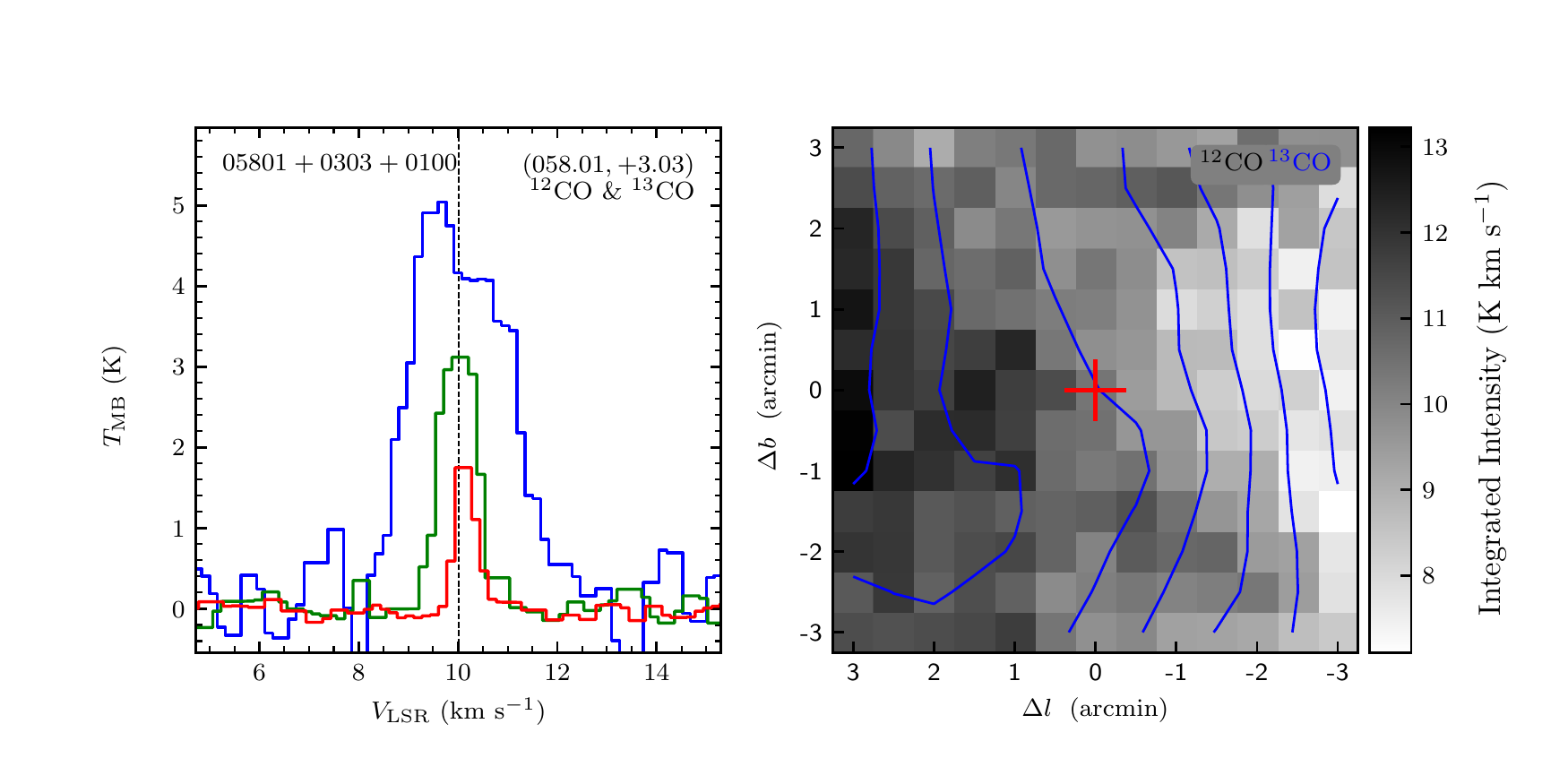}
\end{figure}
\clearpage

\begin{figure}
\includegraphics[width=9.0cm,angle=0]{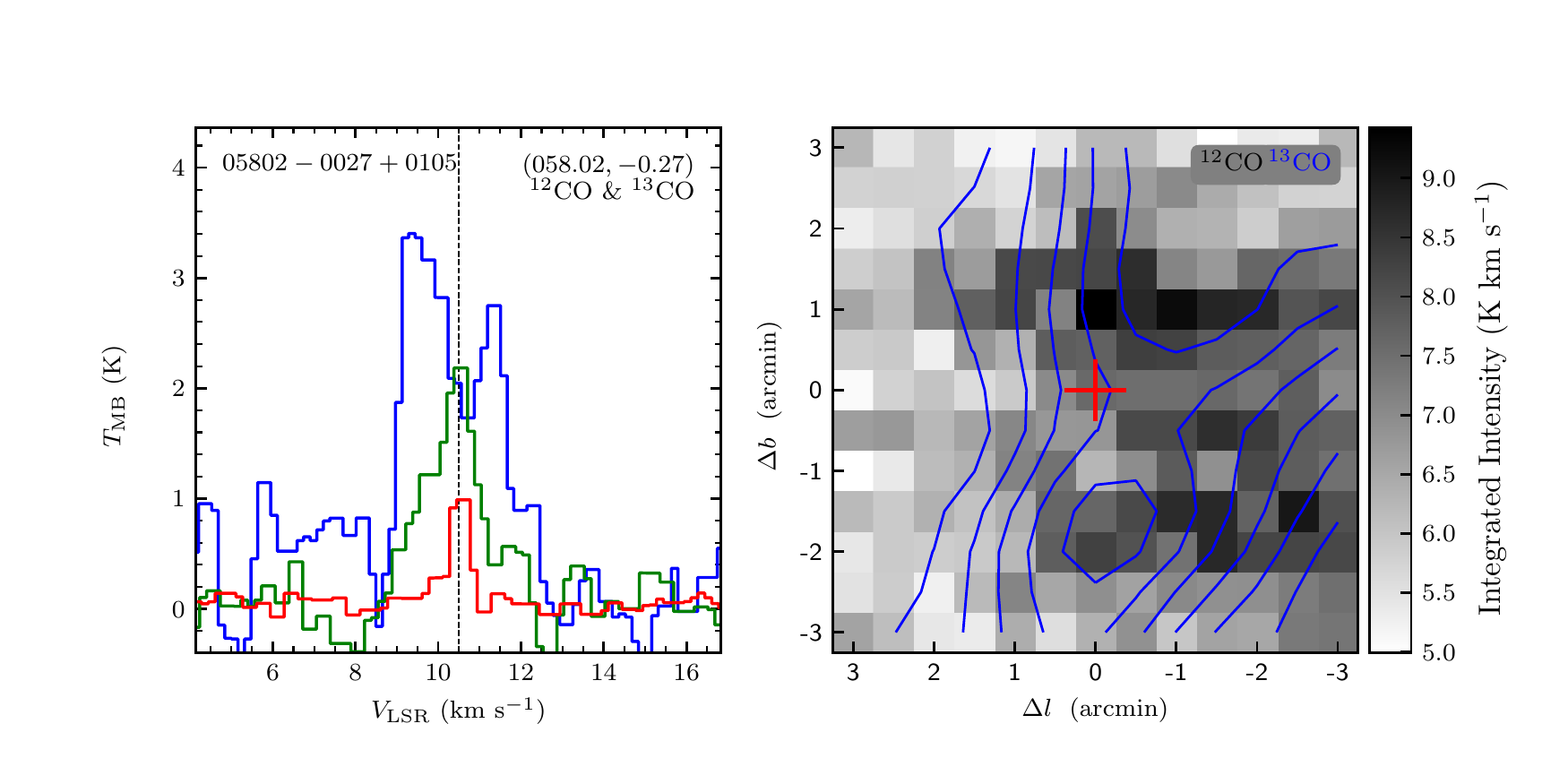}
\includegraphics[width=9.0cm,angle=0]{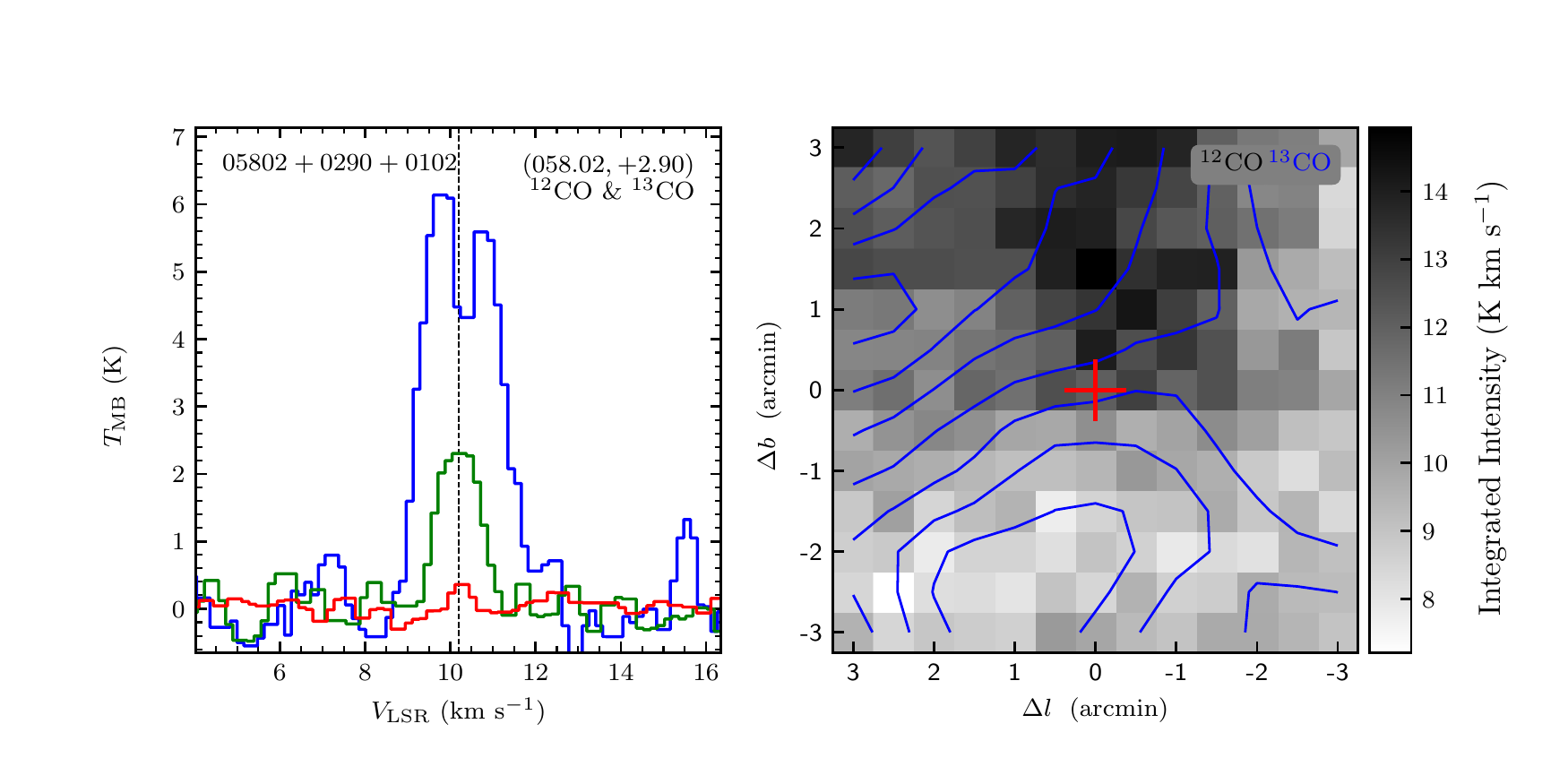}
\vspace{-0.5cm}

\includegraphics[width=9.0cm,angle=0]{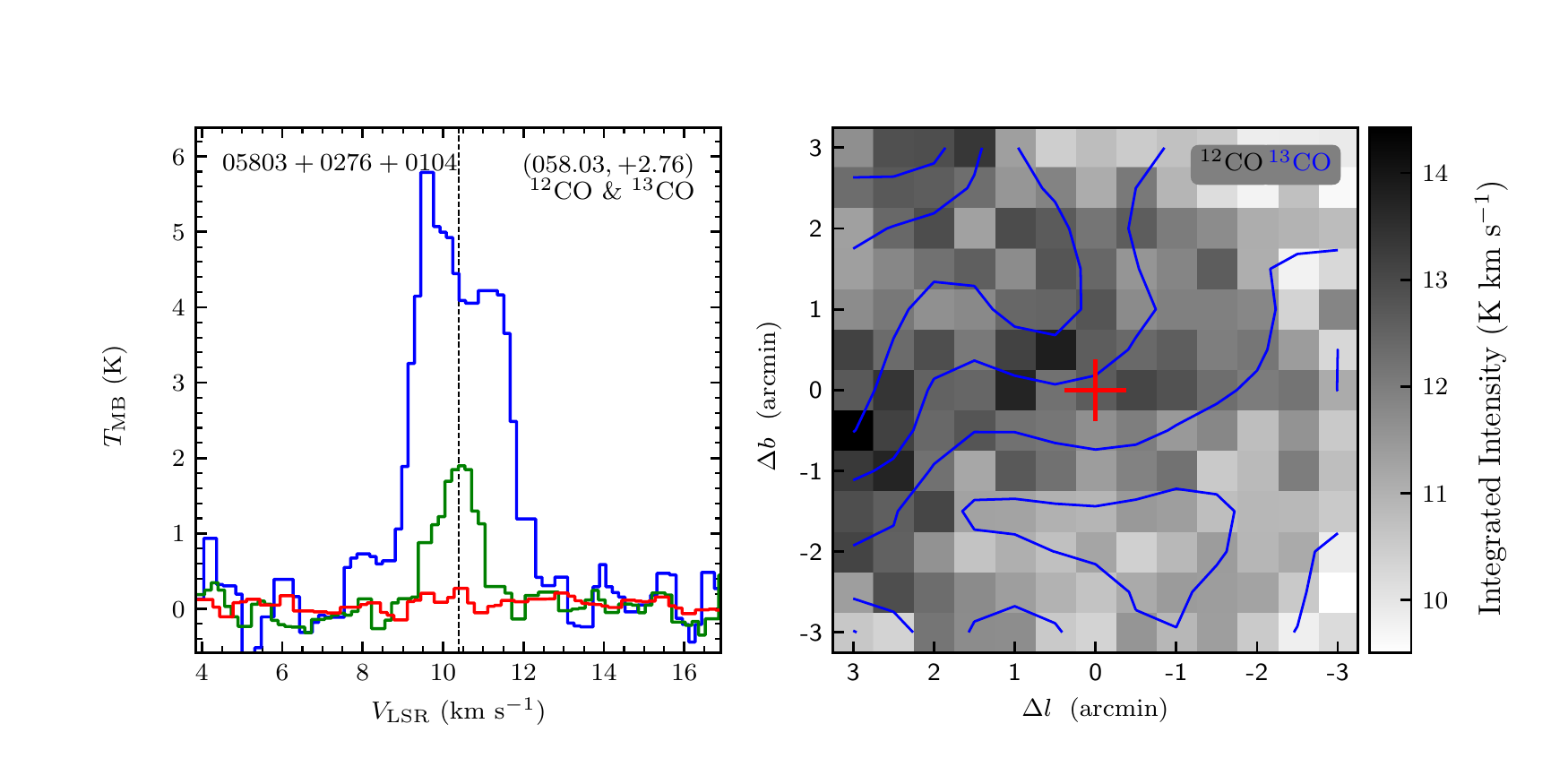}
\includegraphics[width=9.0cm,angle=0]{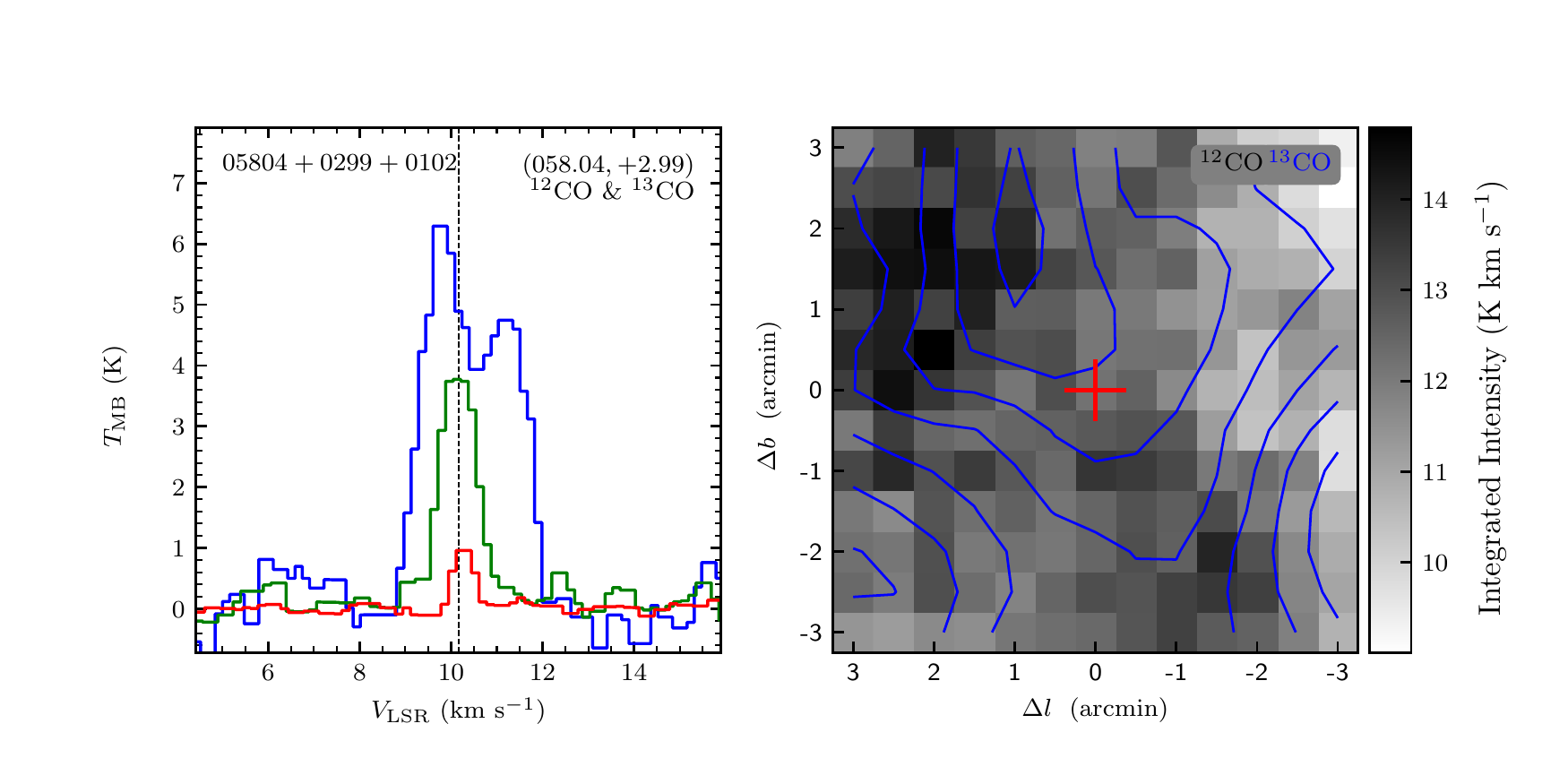}
\vspace{-0.5cm}

\includegraphics[width=9.0cm,angle=0]{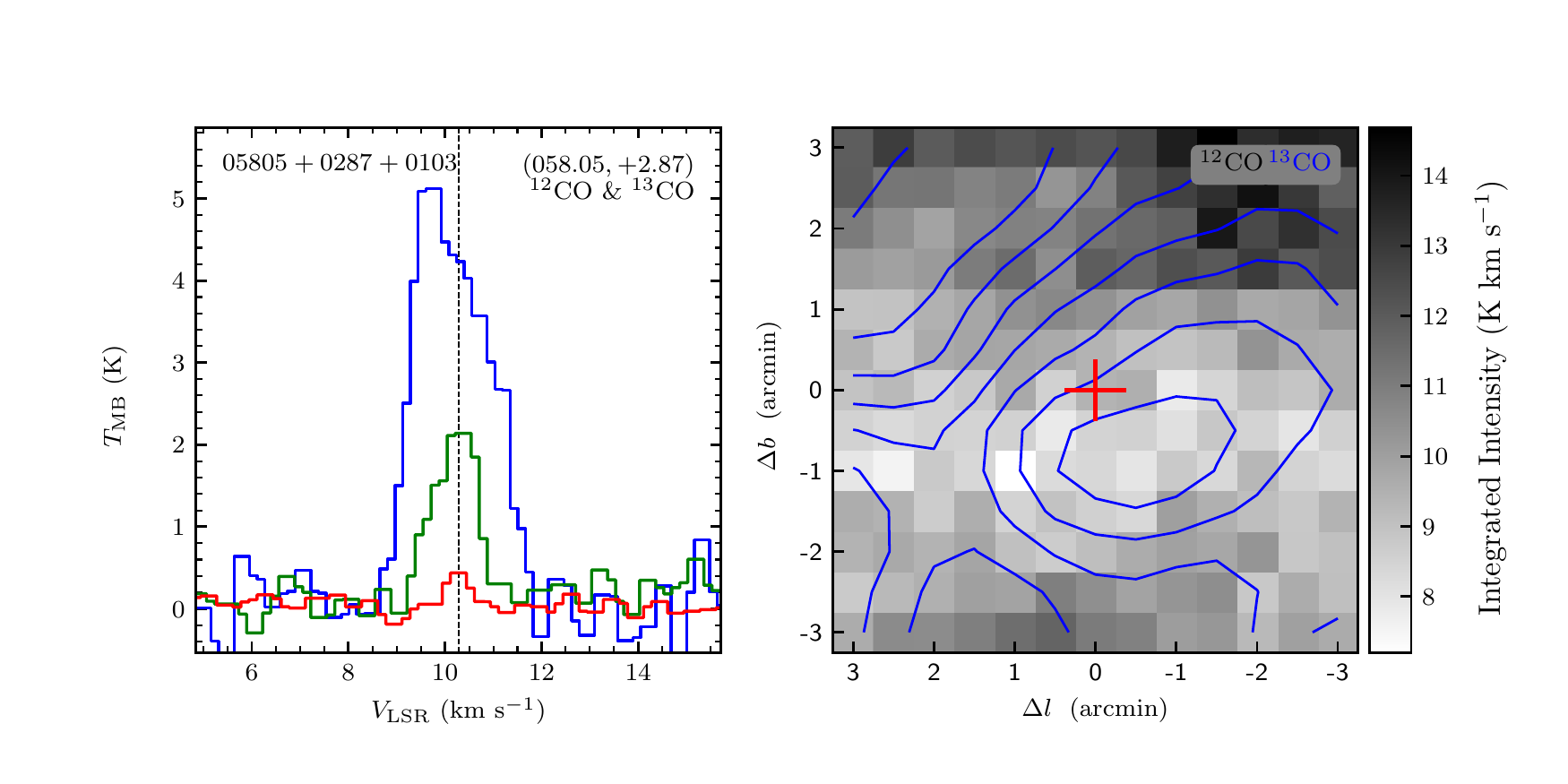}
\includegraphics[width=9.0cm,angle=0]{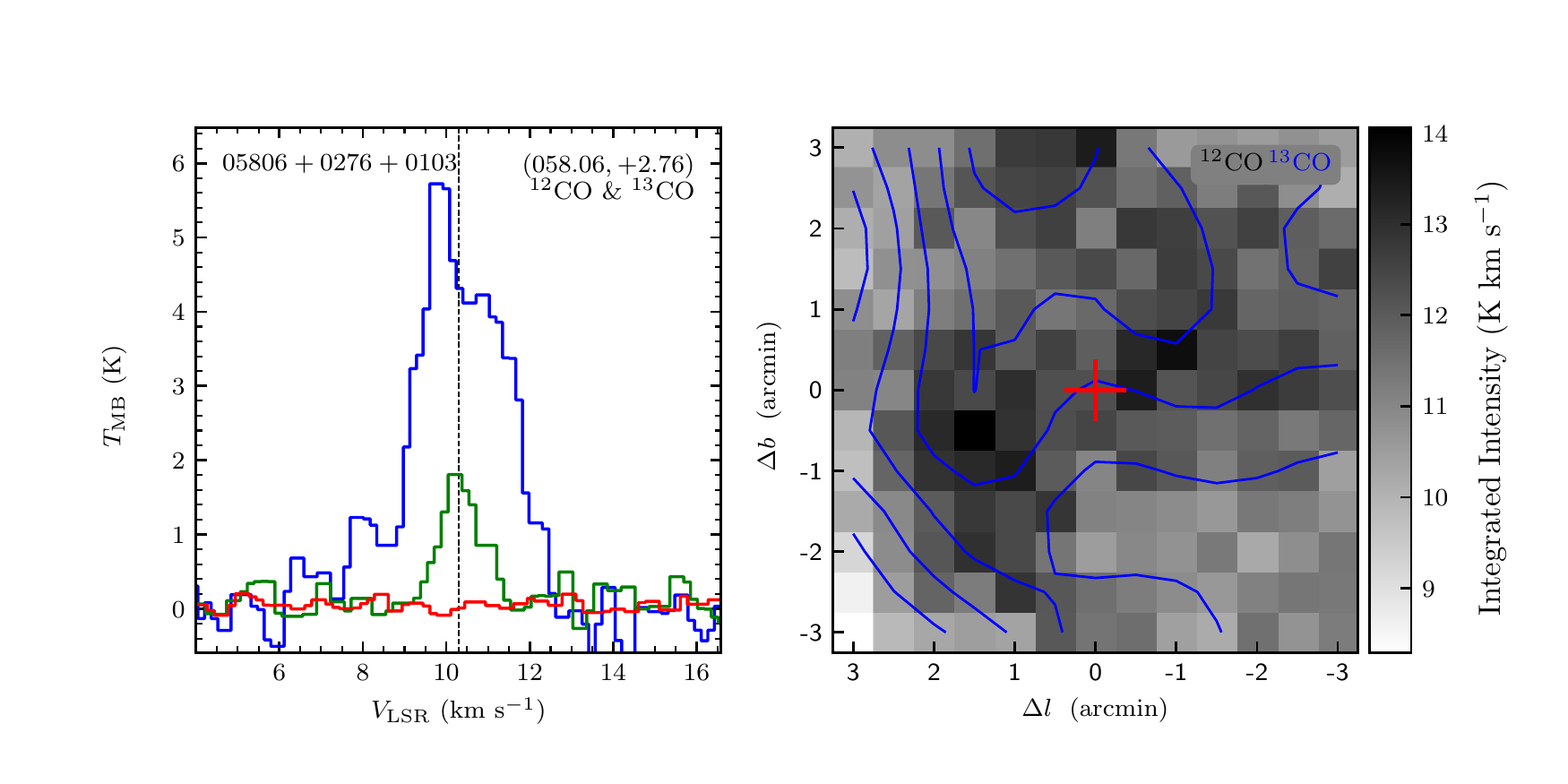}
\vspace{-0.5cm}

\includegraphics[width=9.0cm,angle=0]{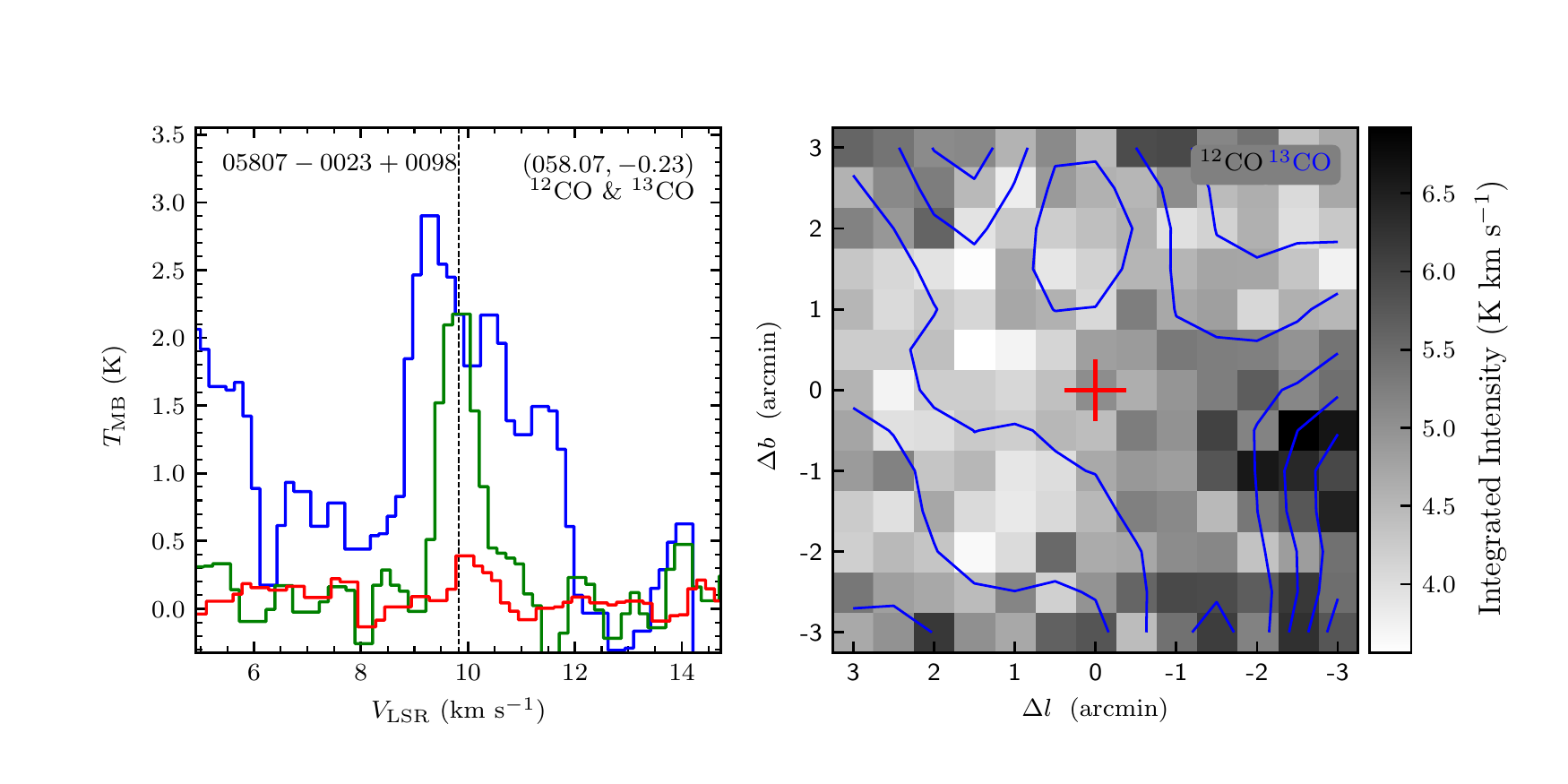}
\includegraphics[width=9.0cm,angle=0]{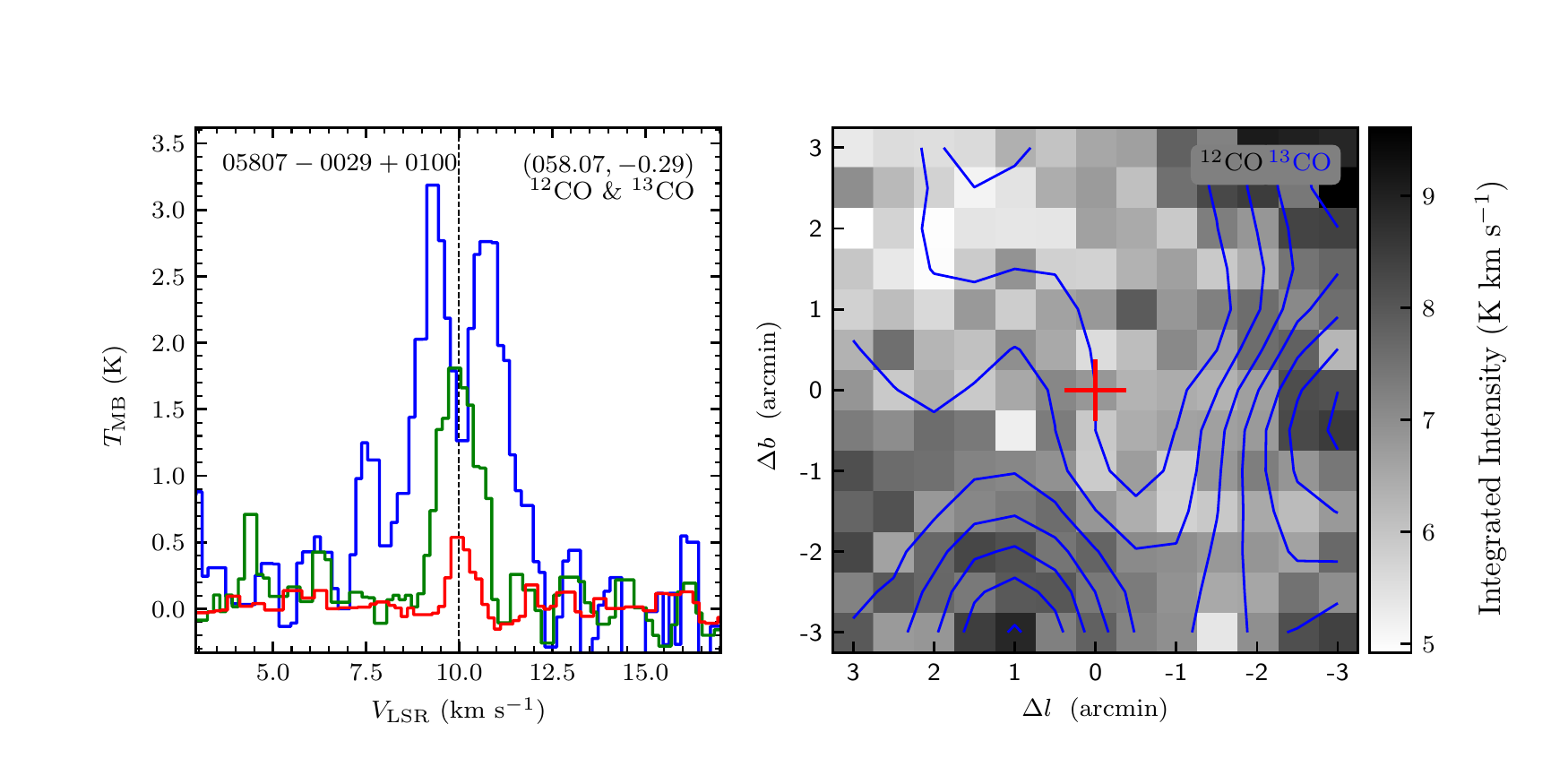}
\vspace{-0.5cm}

\includegraphics[width=9.0cm,angle=0]{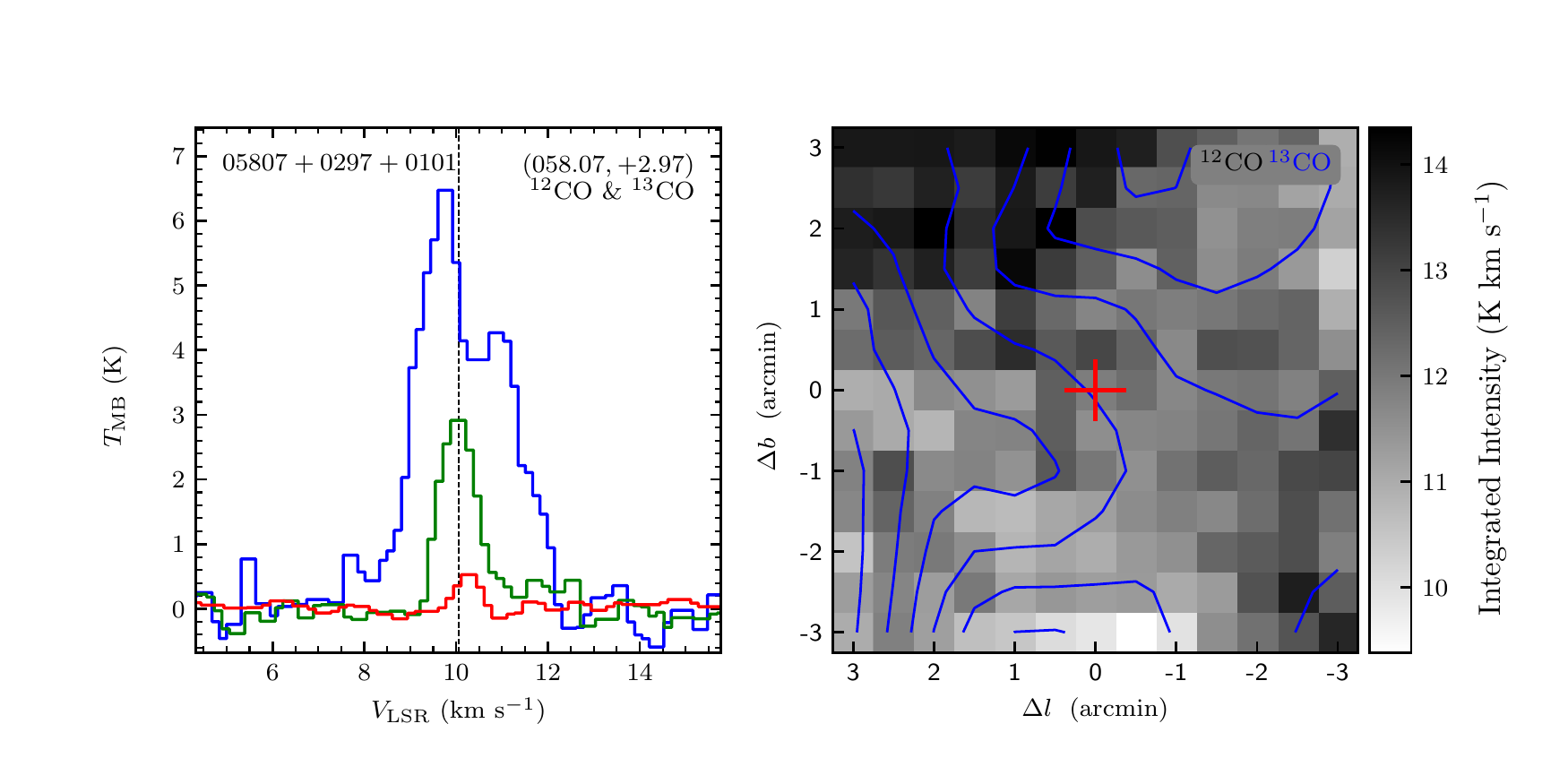}
\includegraphics[width=9.0cm,angle=0]{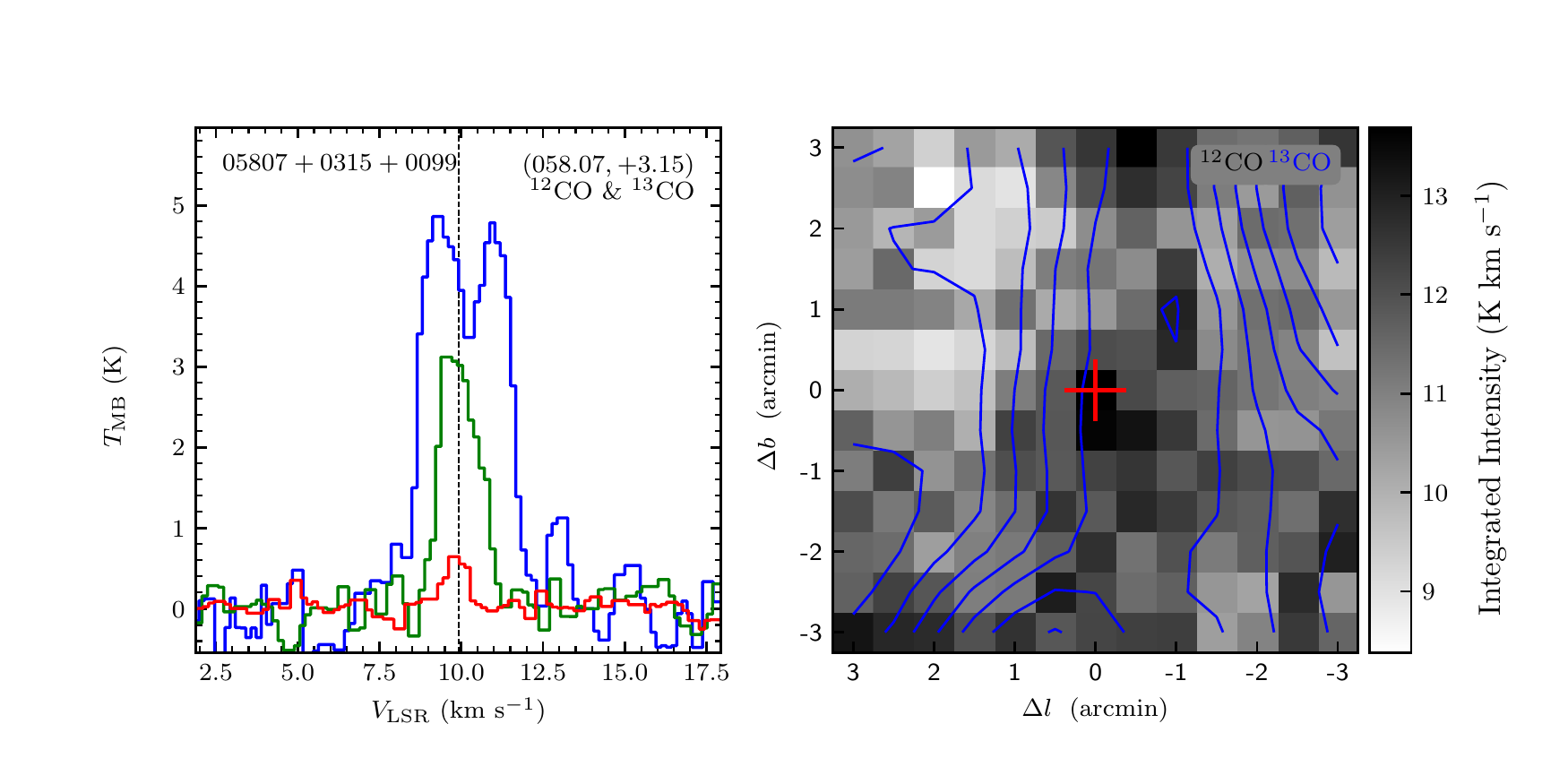}
\end{figure}
\clearpage

\begin{figure}
\includegraphics[width=9.0cm,angle=0]{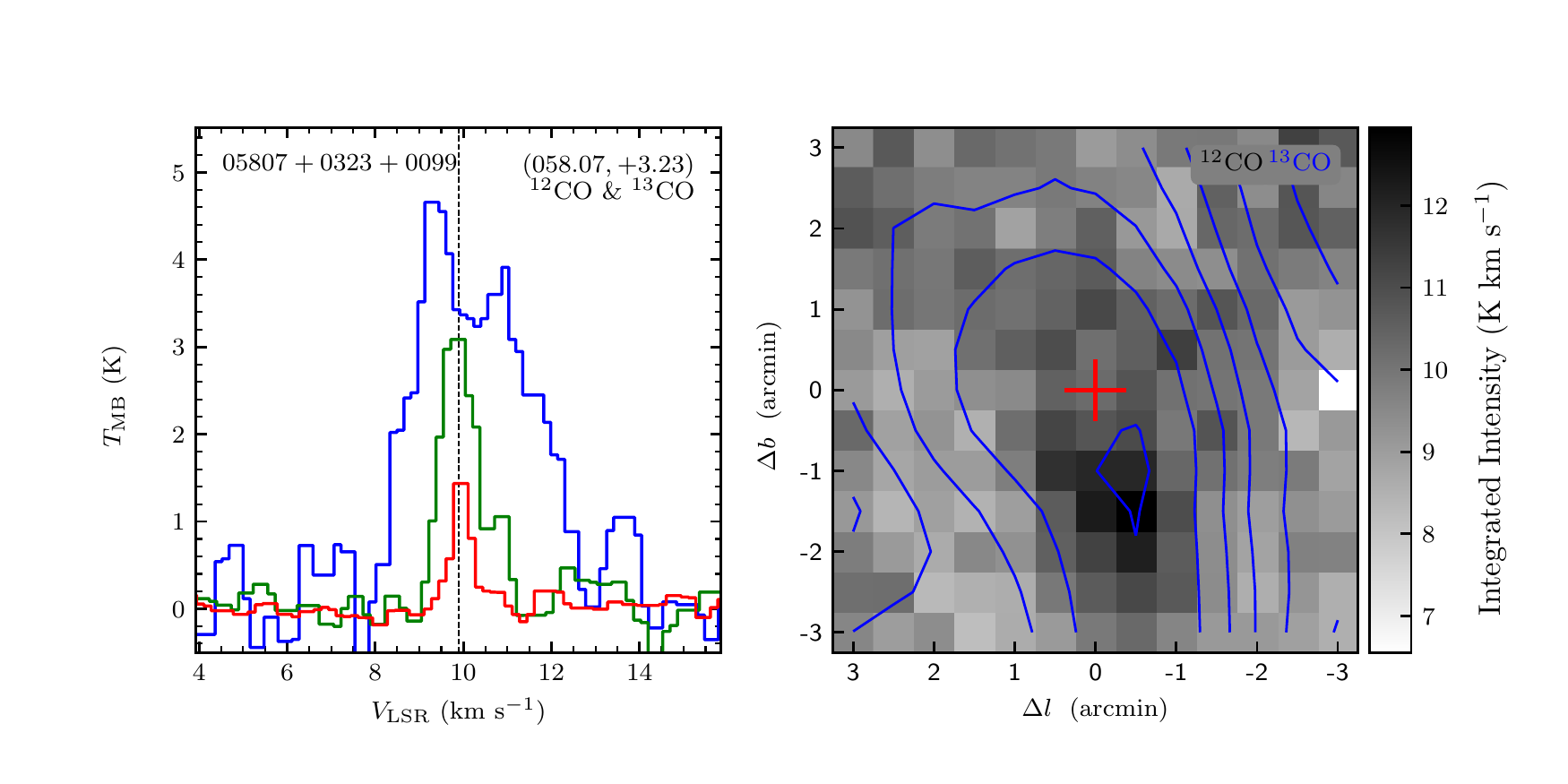}
\includegraphics[width=9.0cm,angle=0]{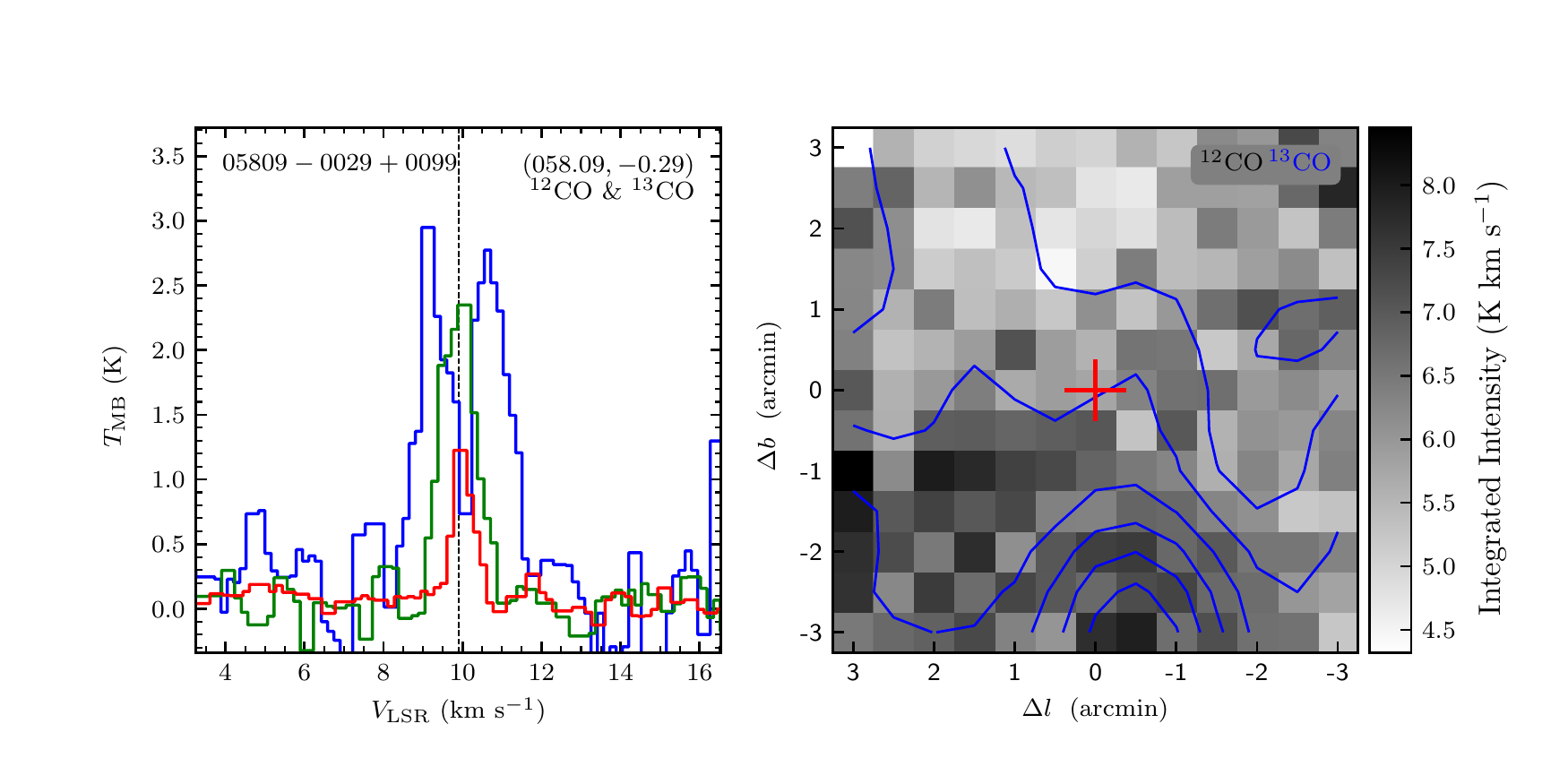}
\vspace{-0.5cm}

\includegraphics[width=9.0cm,angle=0]{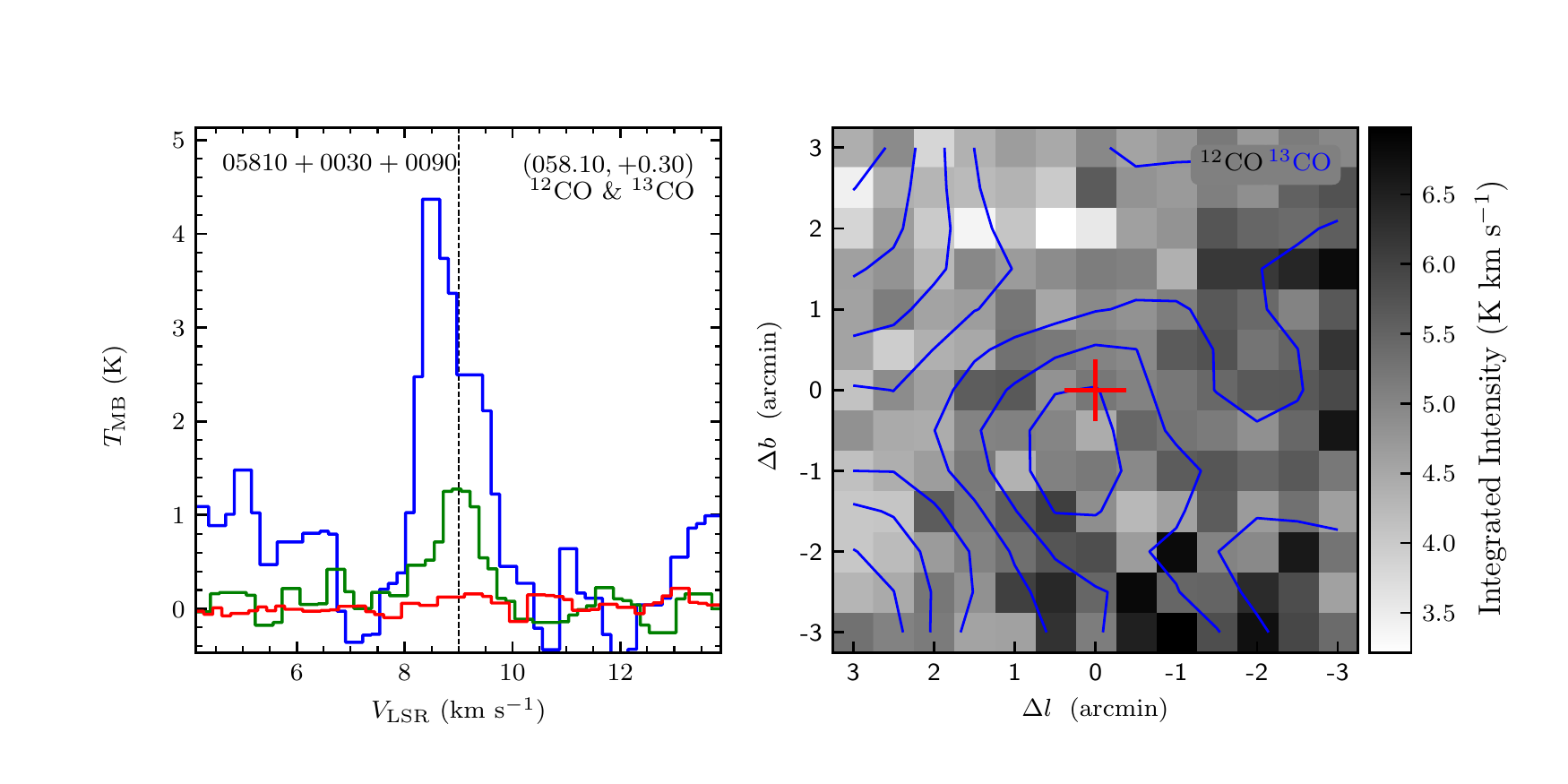}
\includegraphics[width=9.0cm,angle=0]{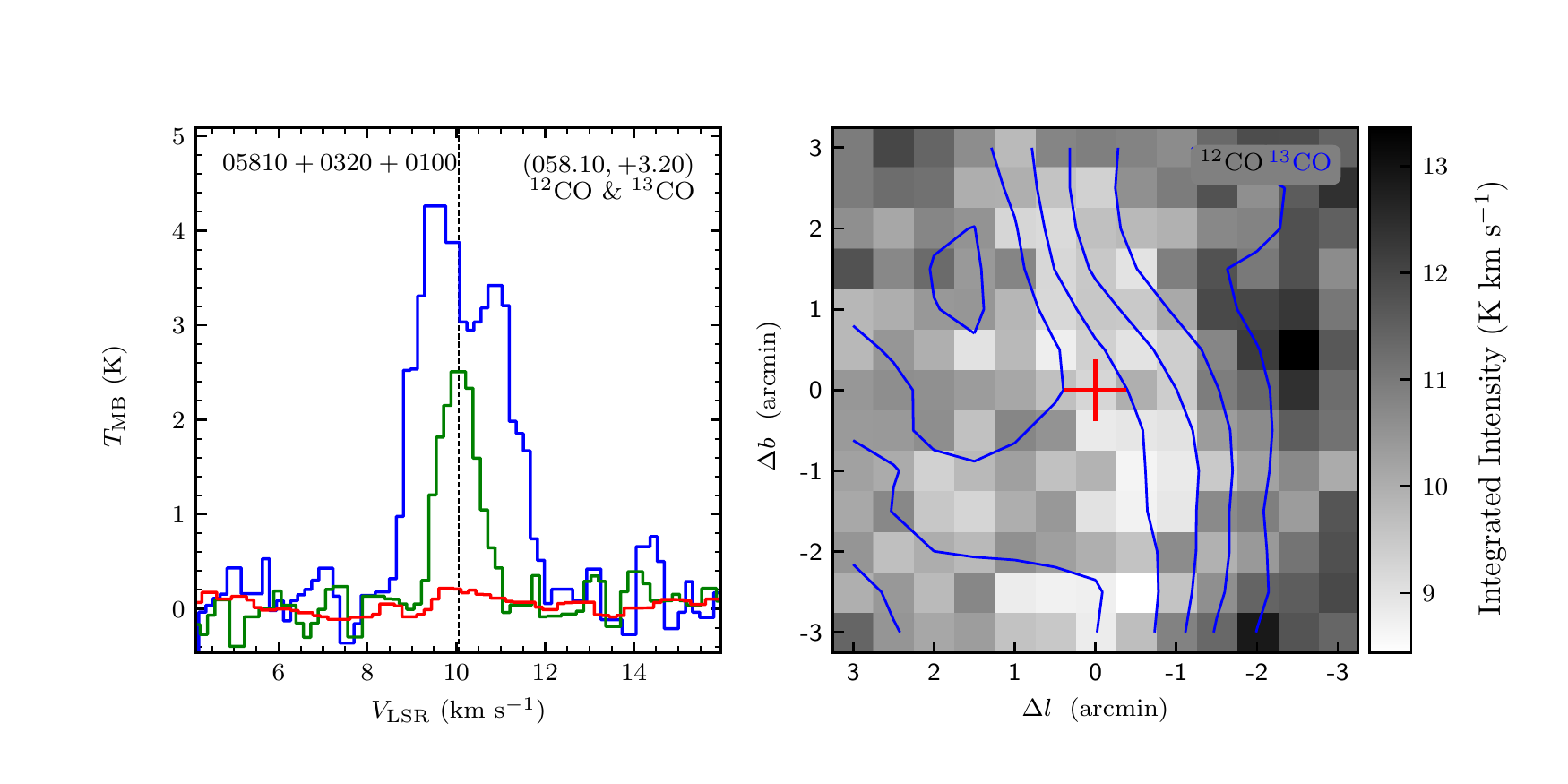}
\vspace{-0.5cm}

\includegraphics[width=9.0cm,angle=0]{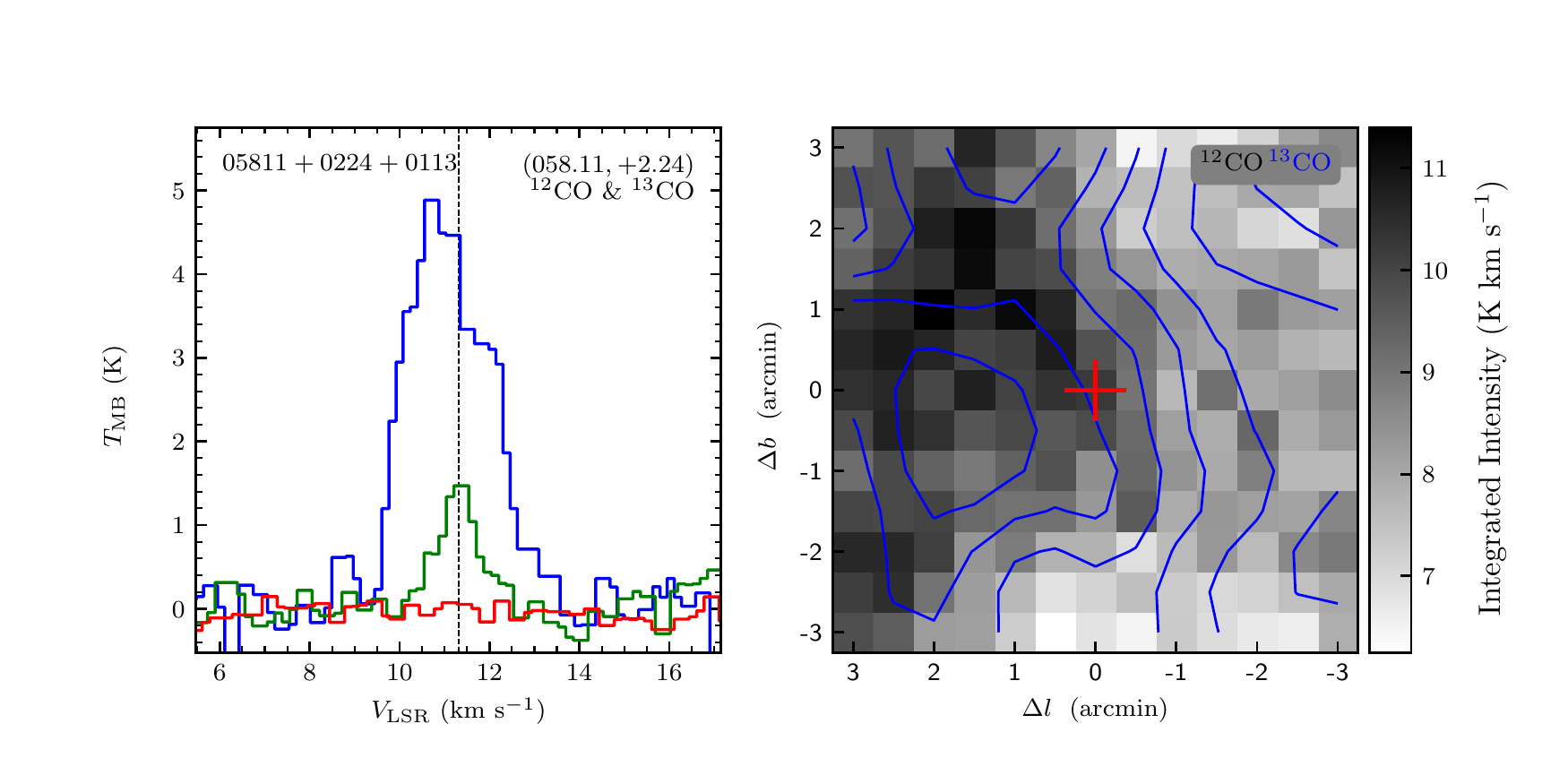}
\includegraphics[width=9.0cm,angle=0]{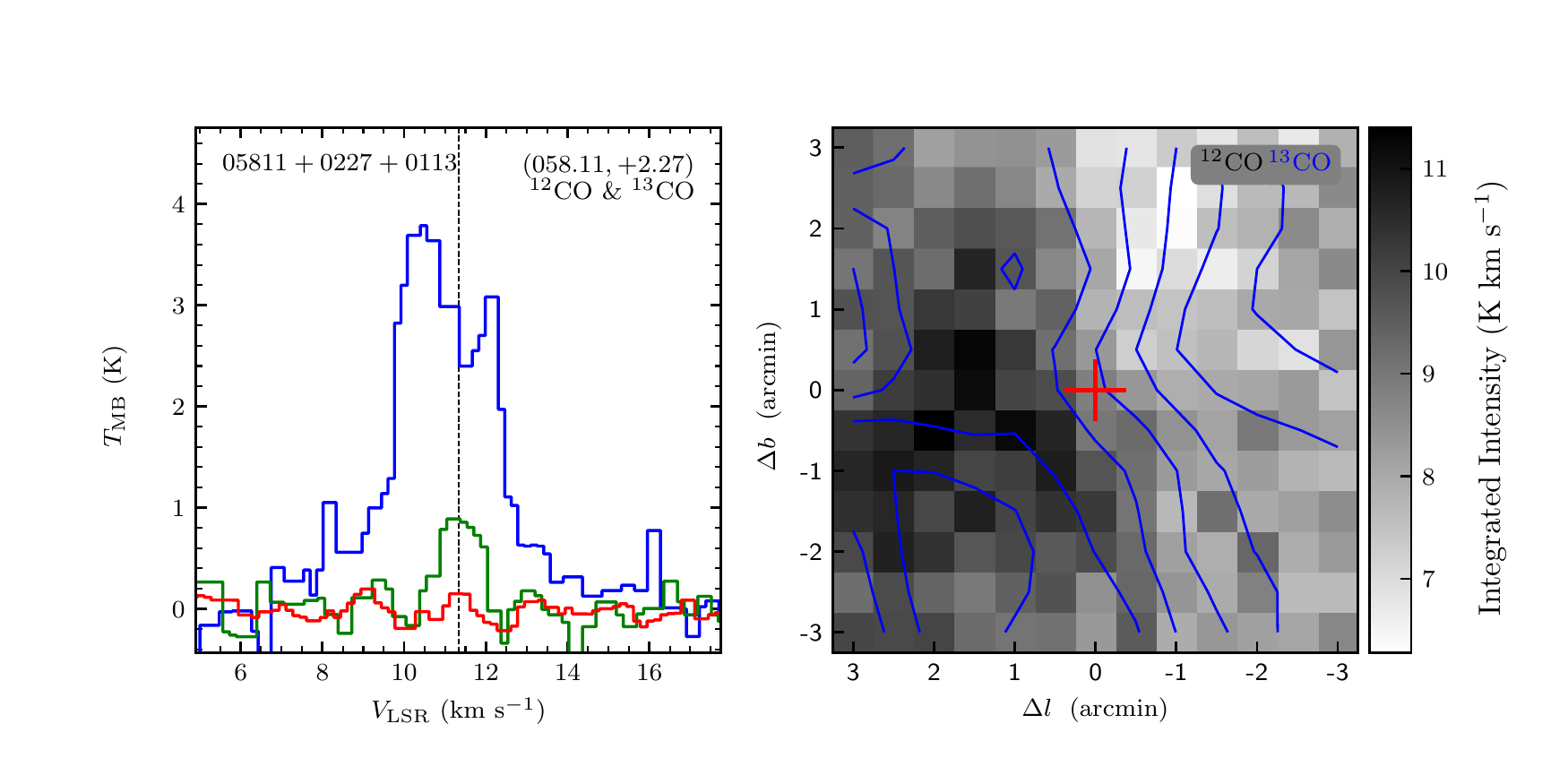}
\vspace{-0.5cm}

\includegraphics[width=9.0cm,angle=0]{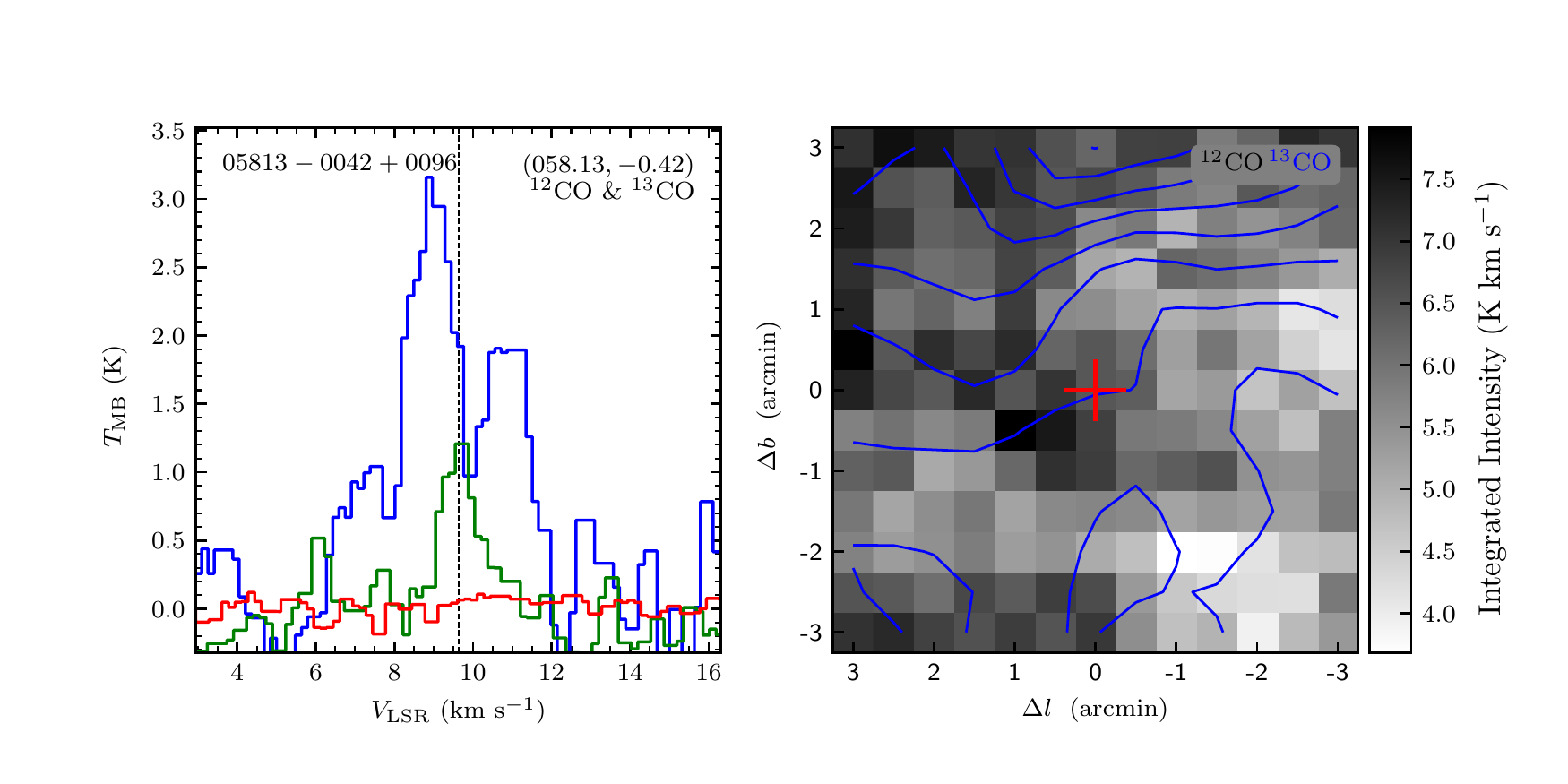}
\includegraphics[width=9.0cm,angle=0]{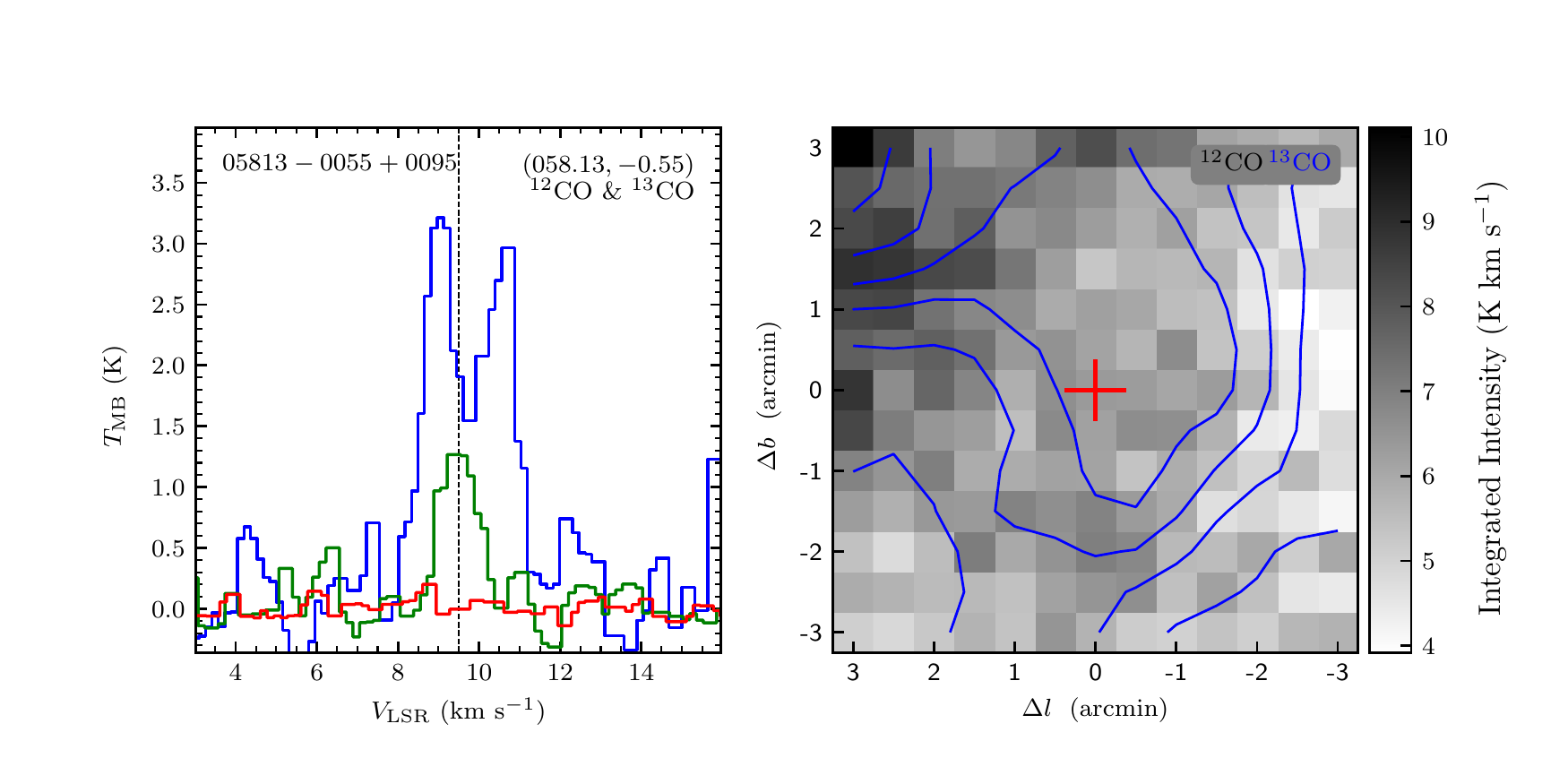}
\vspace{-0.5cm}

\includegraphics[width=9.0cm,angle=0]{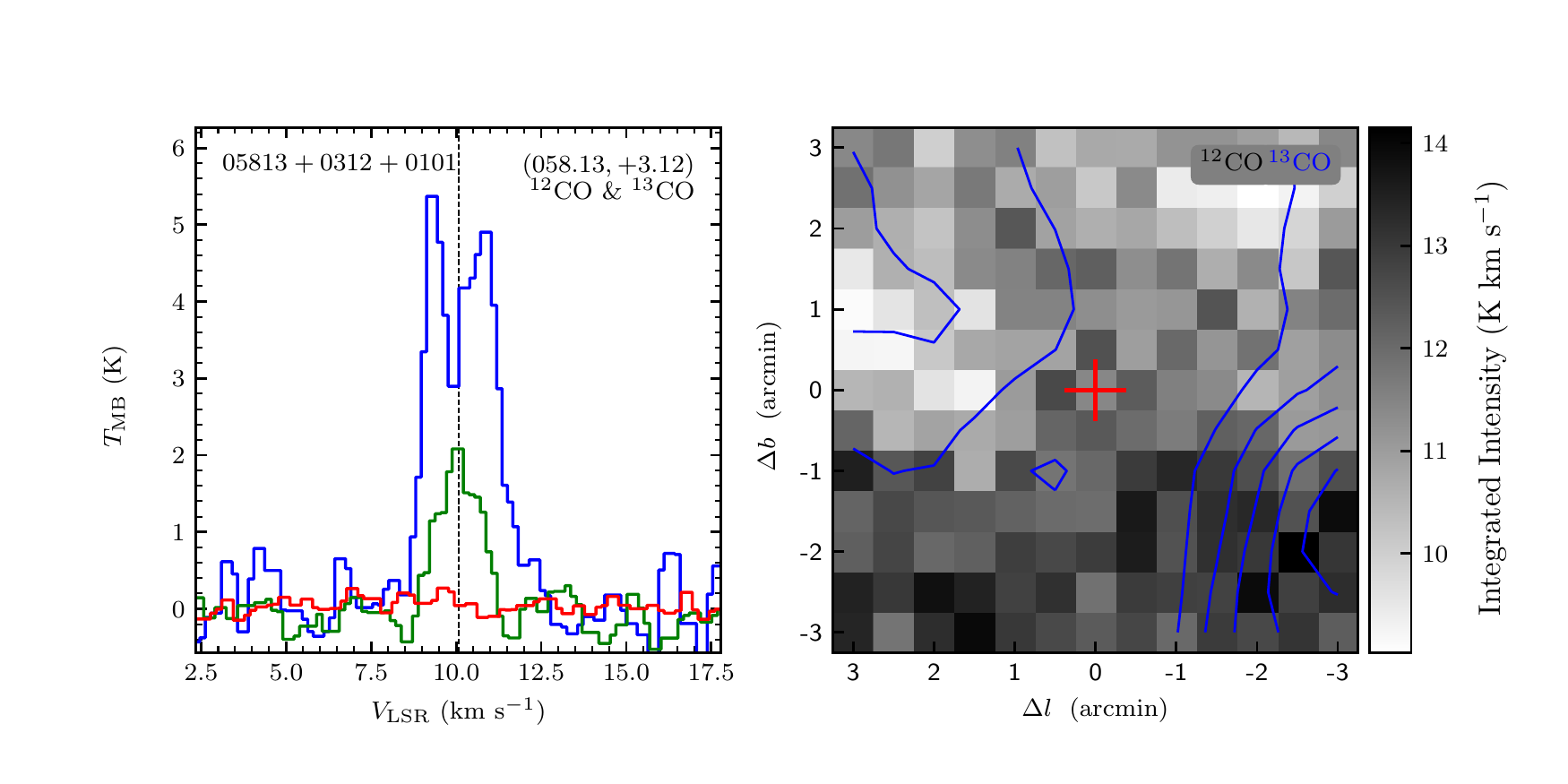}
\includegraphics[width=9.0cm,angle=0]{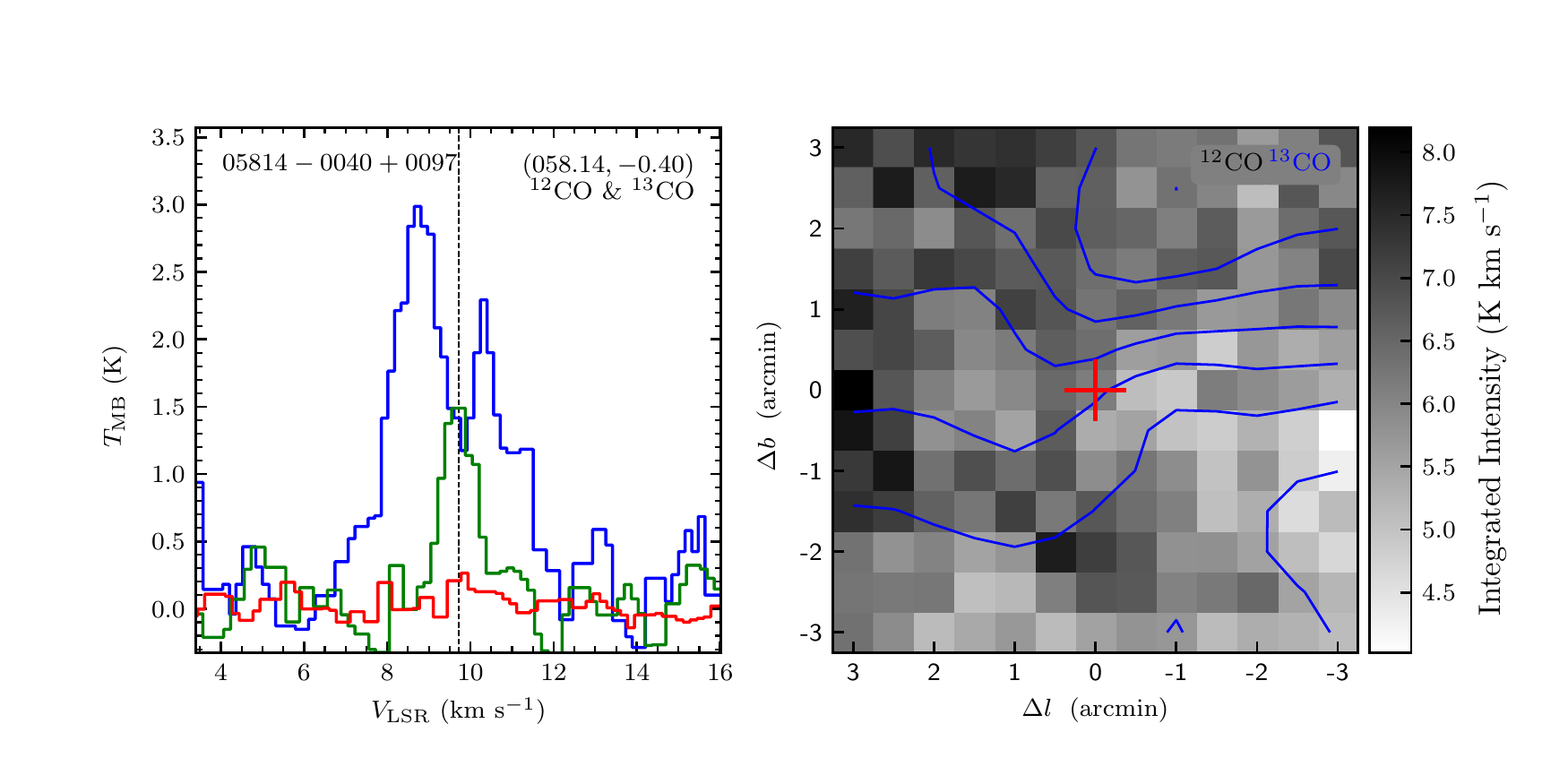}
\end{figure}
\clearpage

\begin{figure}
\includegraphics[width=9.0cm,angle=0]{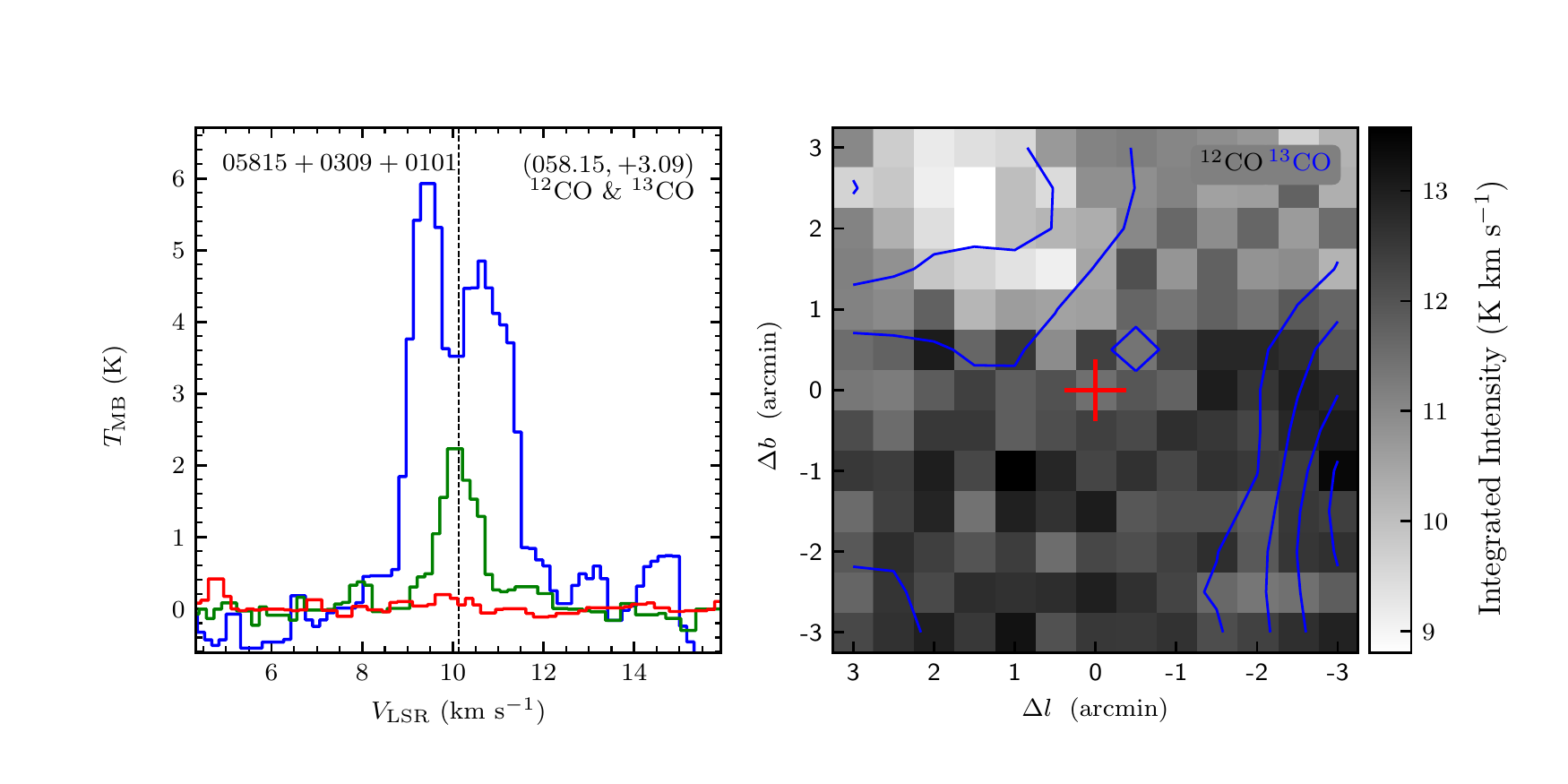}
\includegraphics[width=9.0cm,angle=0]{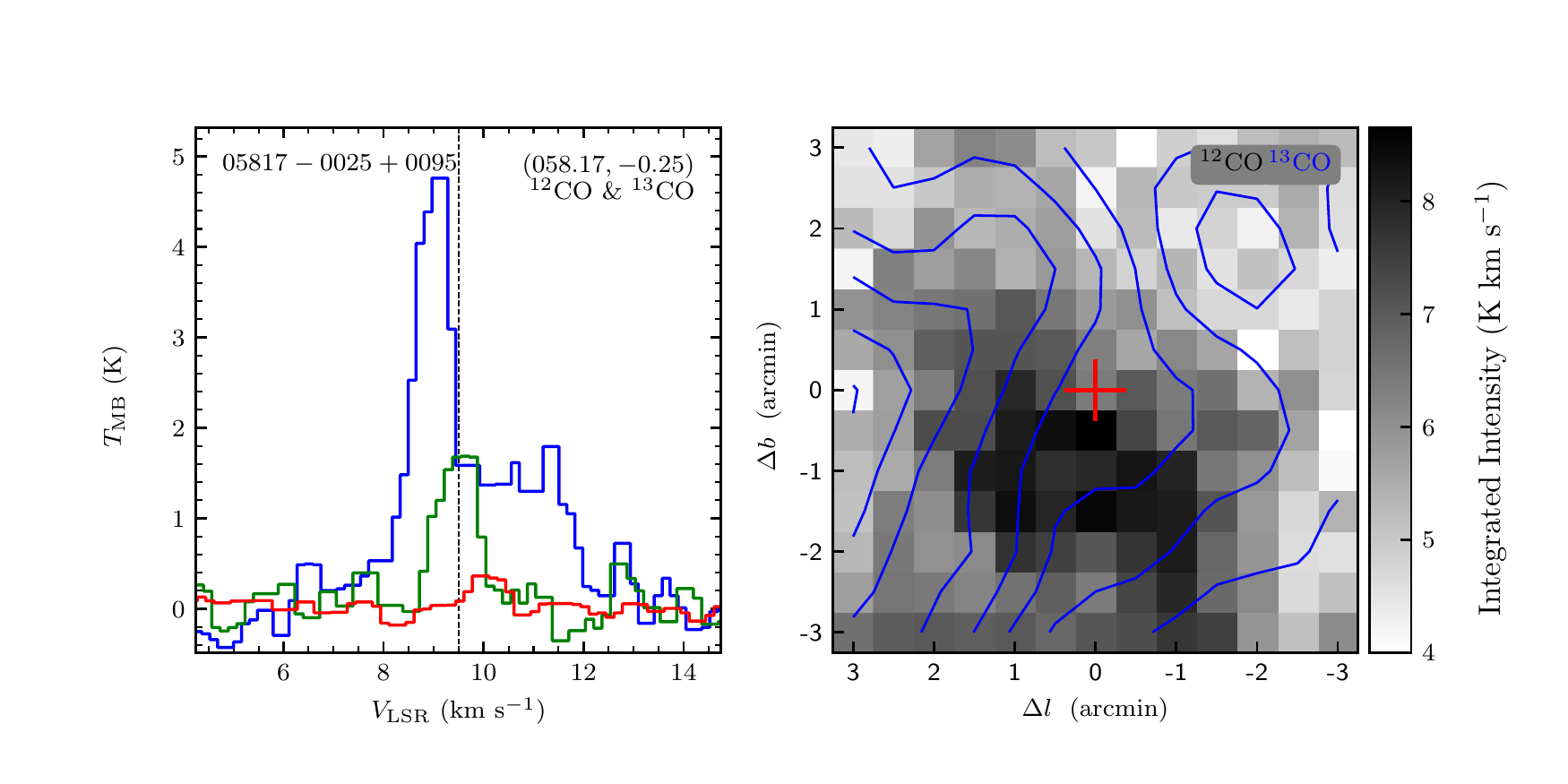}
\vspace{-0.5cm}

\includegraphics[width=9.0cm,angle=0]{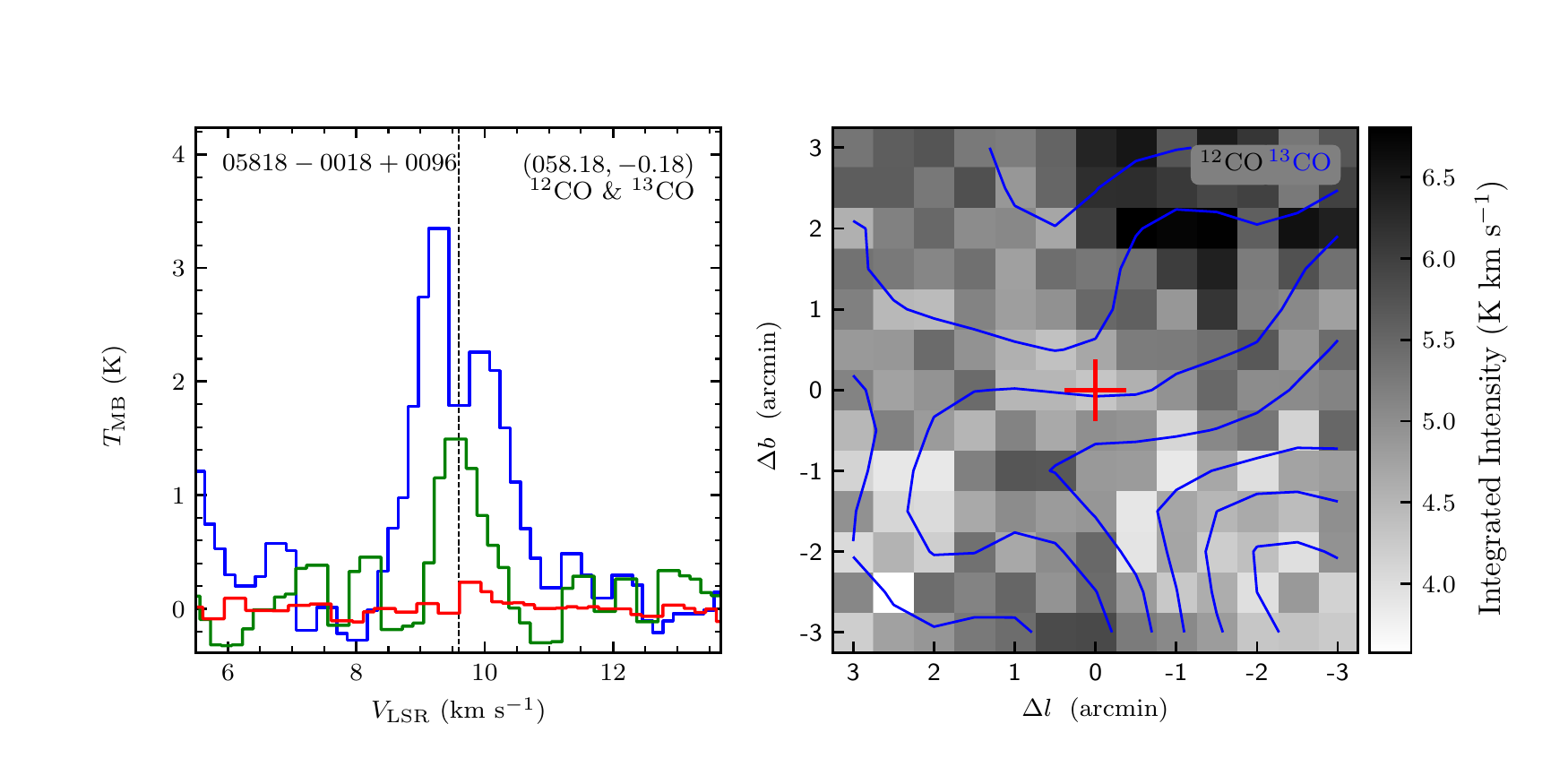}
\includegraphics[width=9.0cm,angle=0]{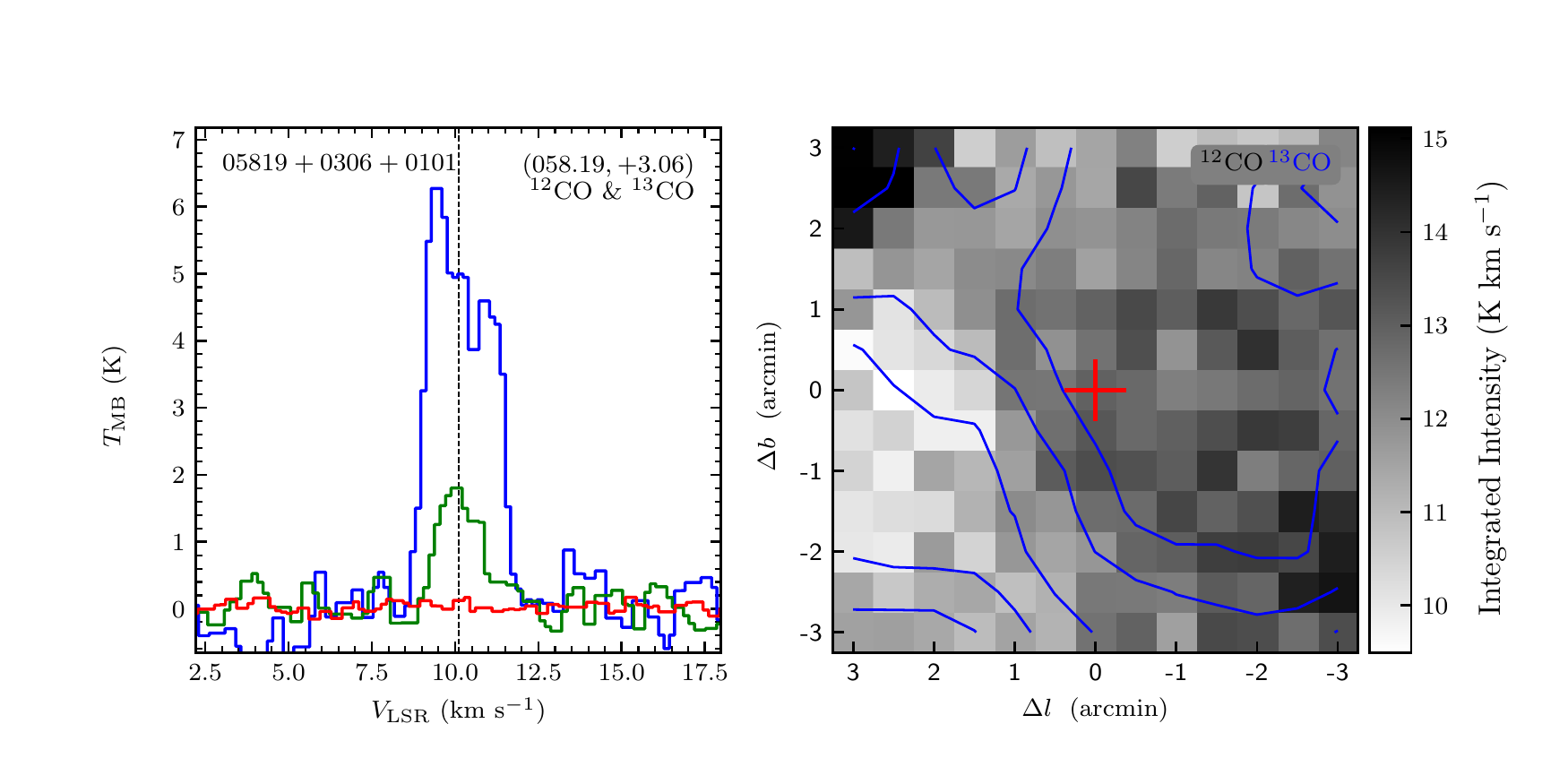}
\vspace{-0.5cm}

\includegraphics[width=9.0cm,angle=0]{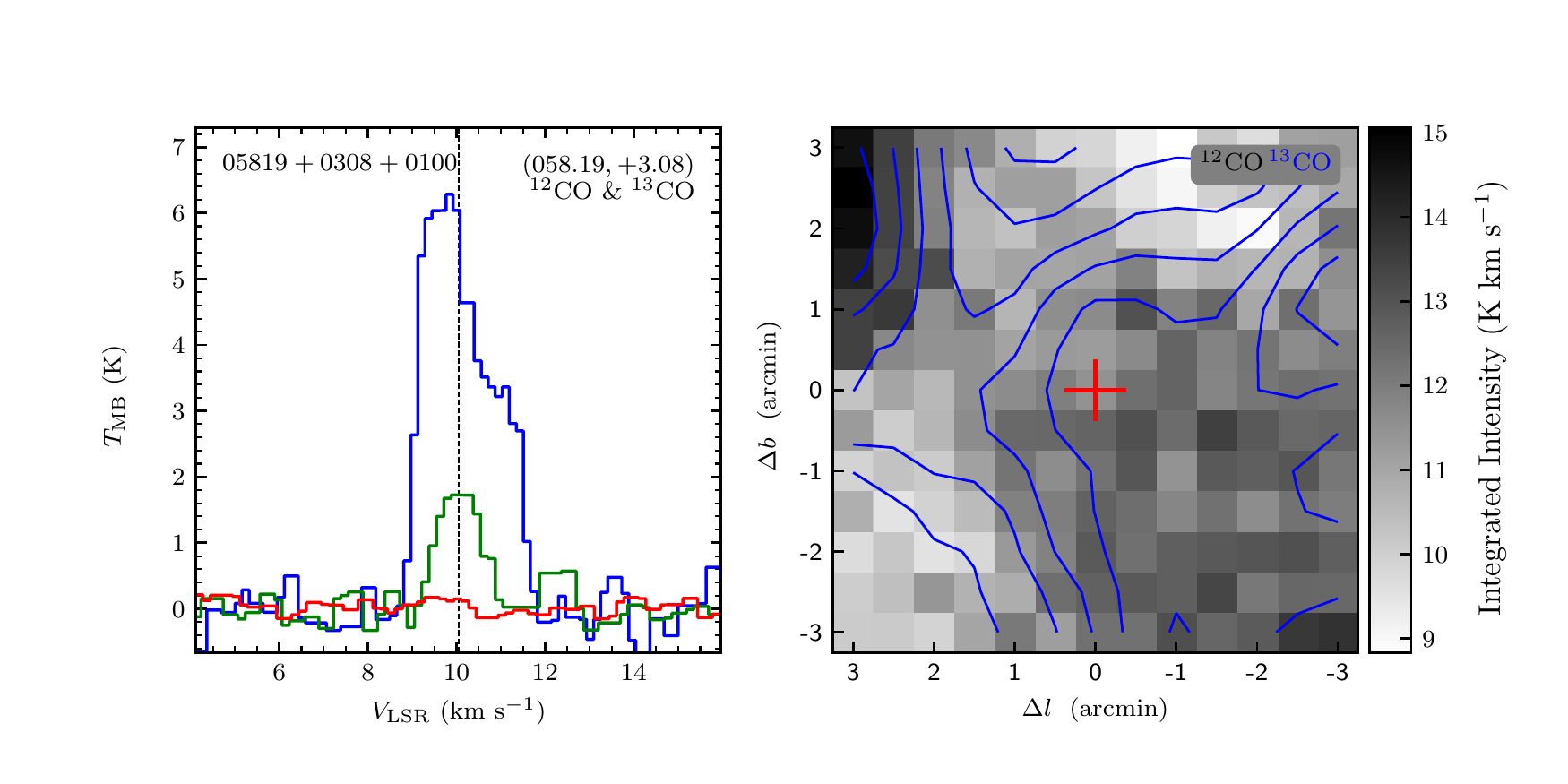}
\includegraphics[width=9.0cm,angle=0]{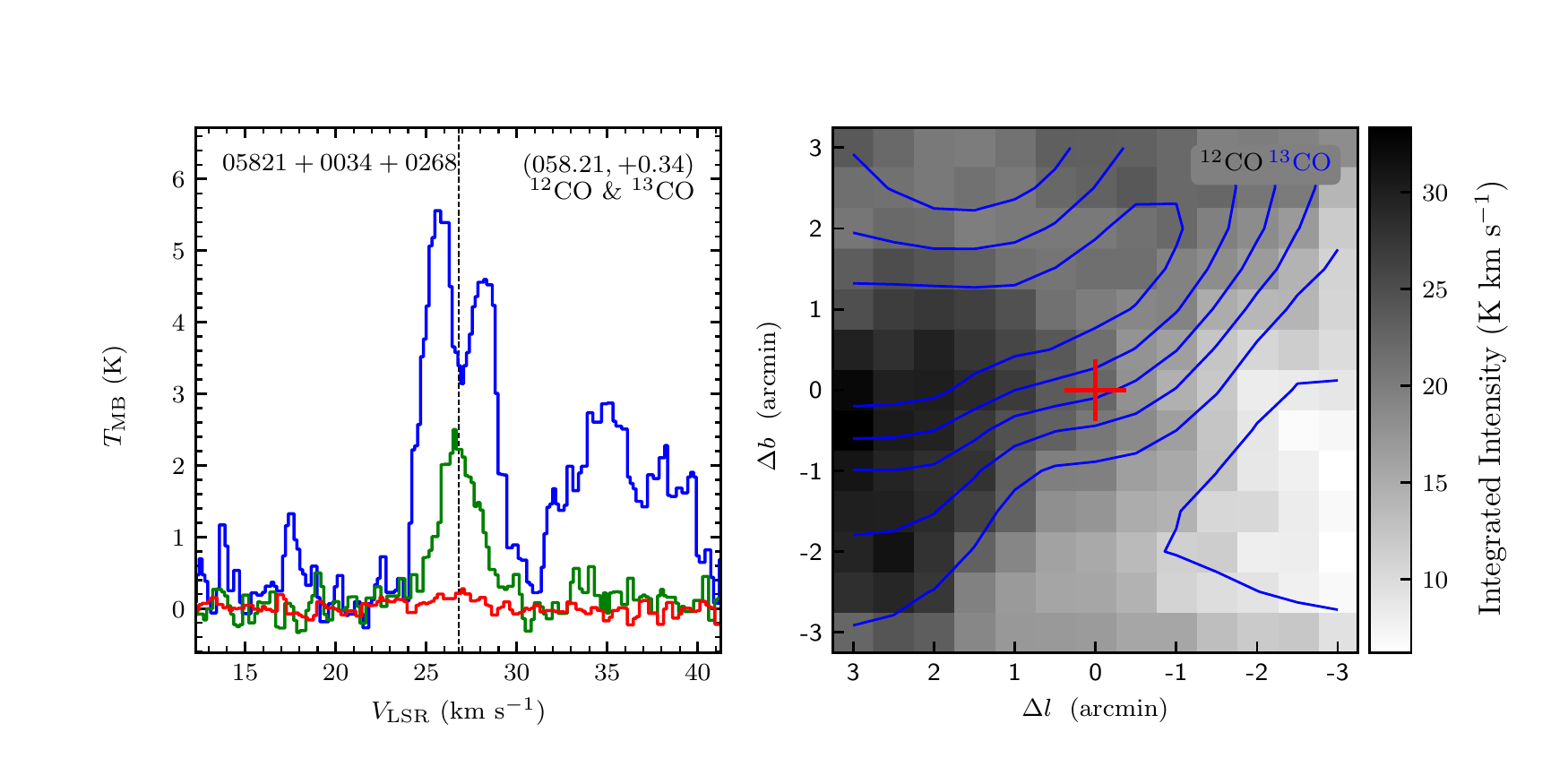}
\vspace{-0.5cm}

\includegraphics[width=9.0cm,angle=0]{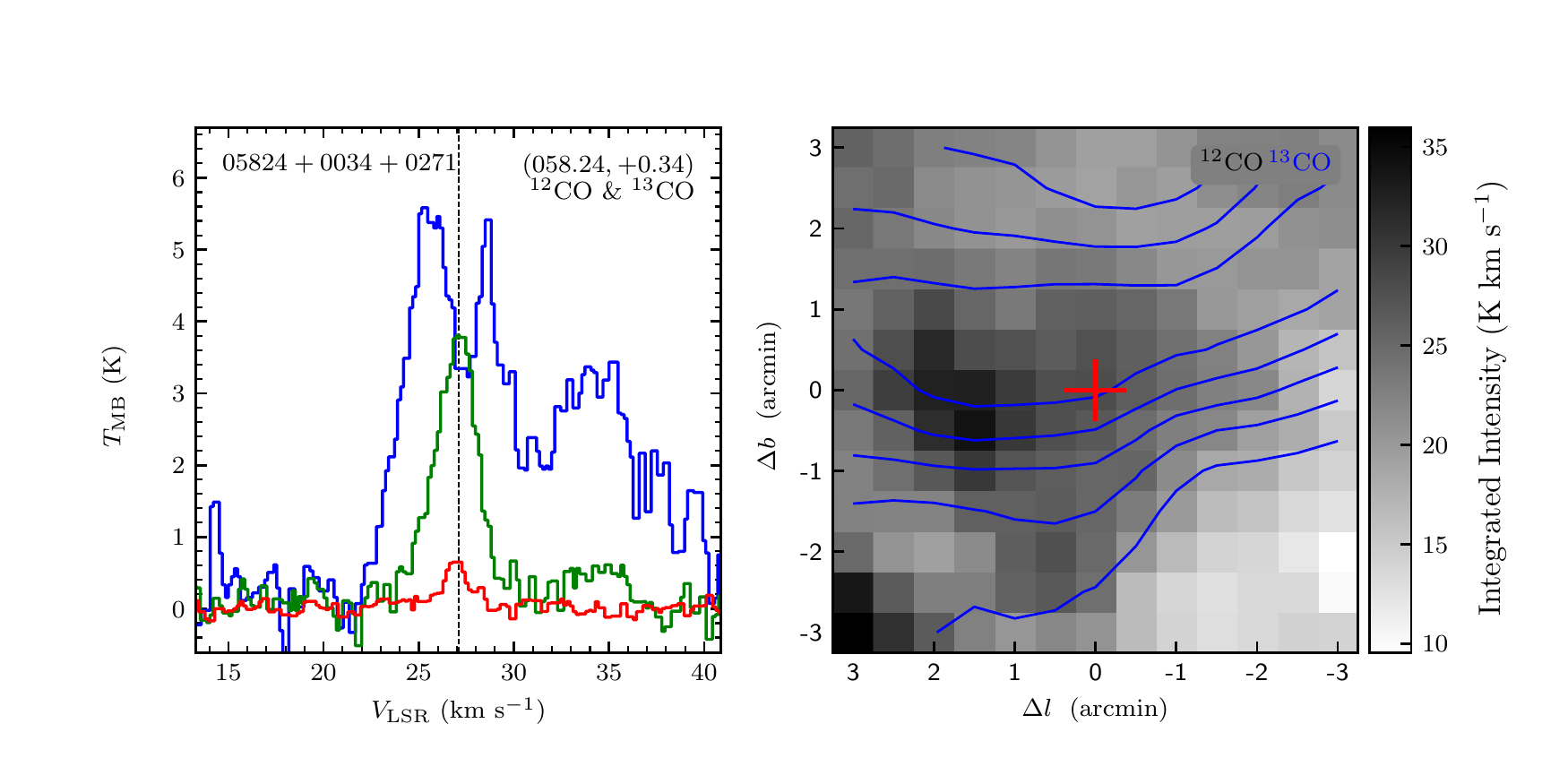}
\includegraphics[width=9.0cm,angle=0]{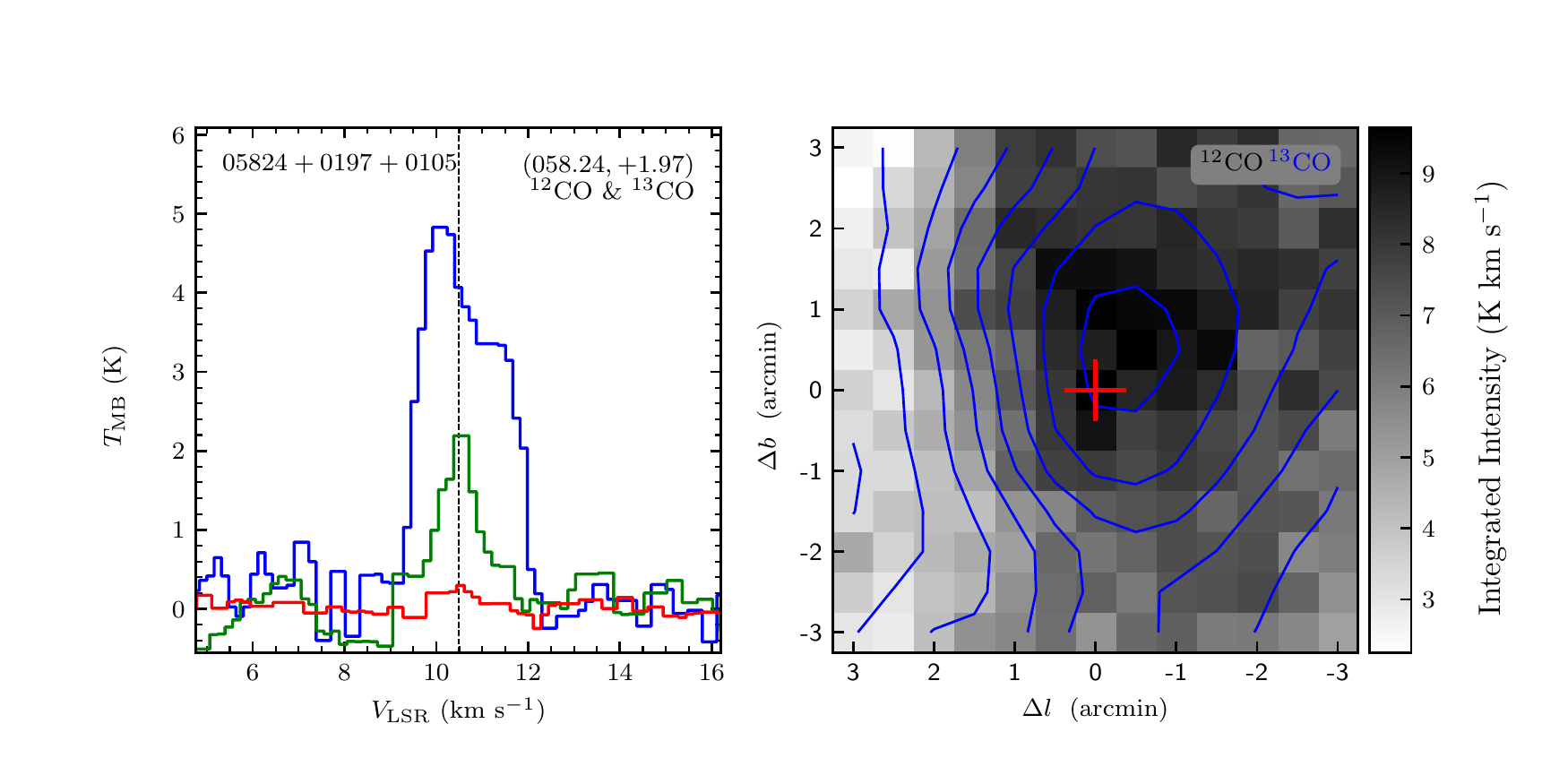}
\vspace{-0.5cm}

\includegraphics[width=9.0cm,angle=0]{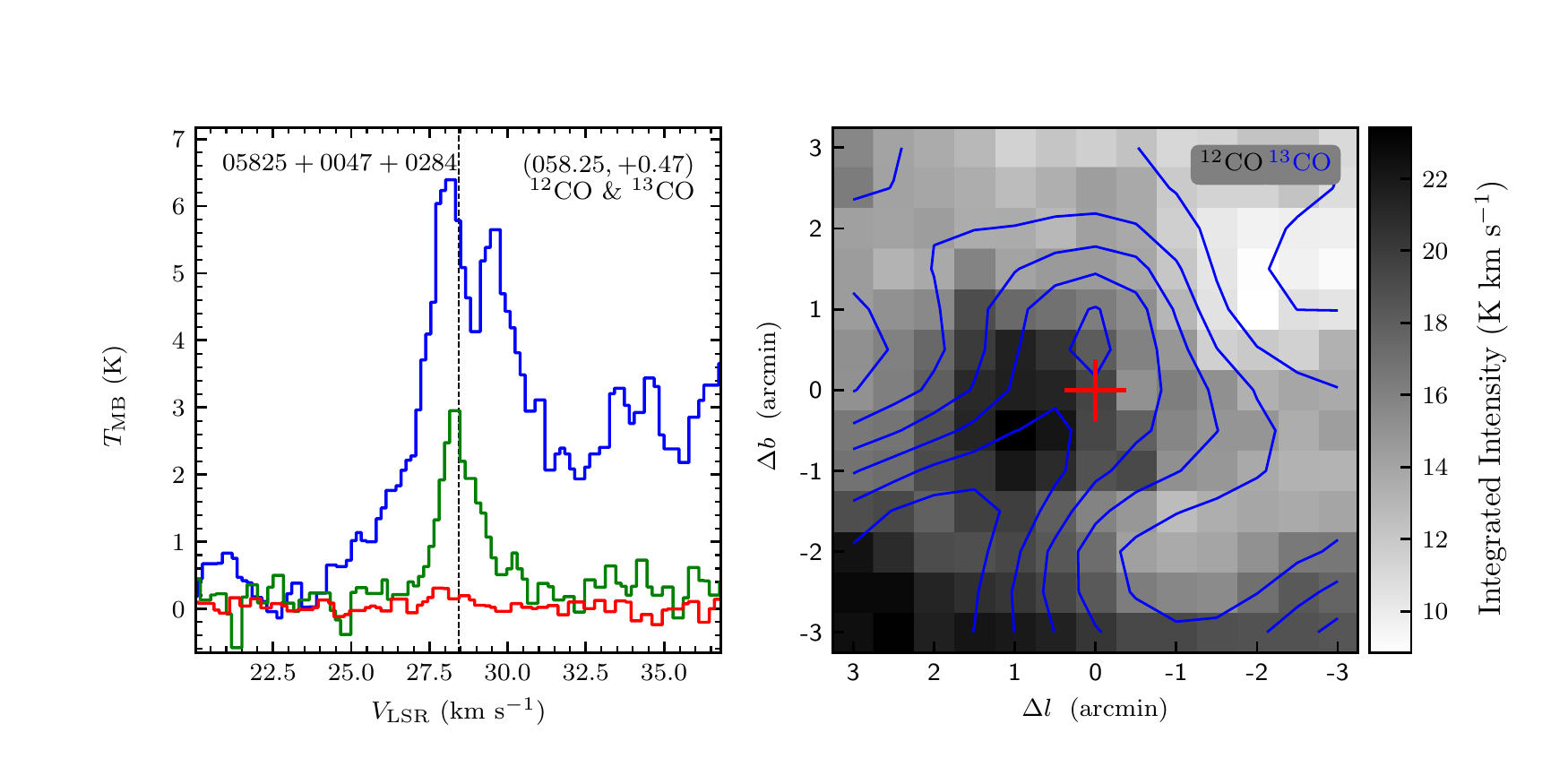}
\includegraphics[width=9.0cm,angle=0]{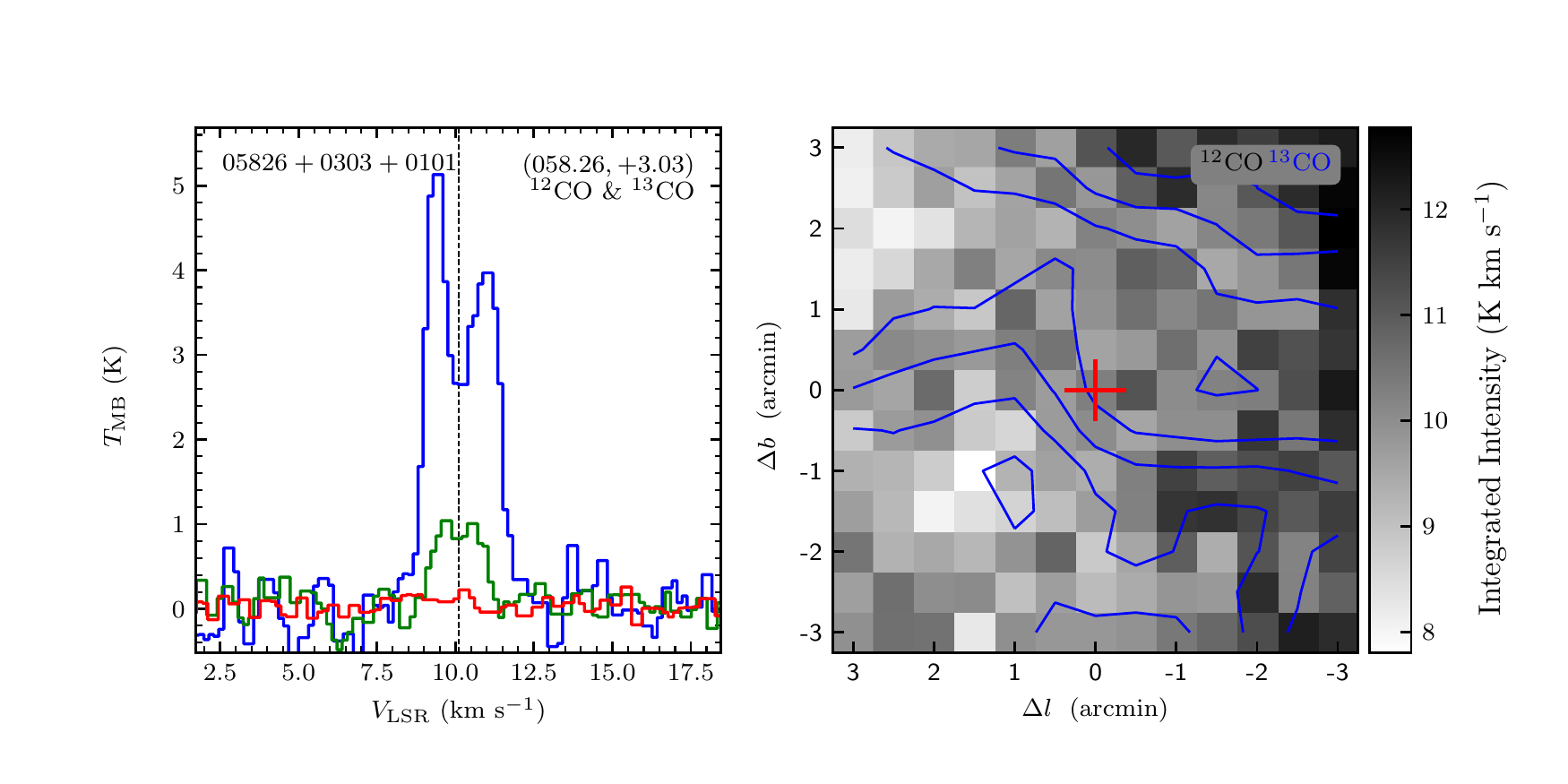}
\end{figure}
\clearpage

\begin{figure}
\includegraphics[width=9.0cm,angle=0]{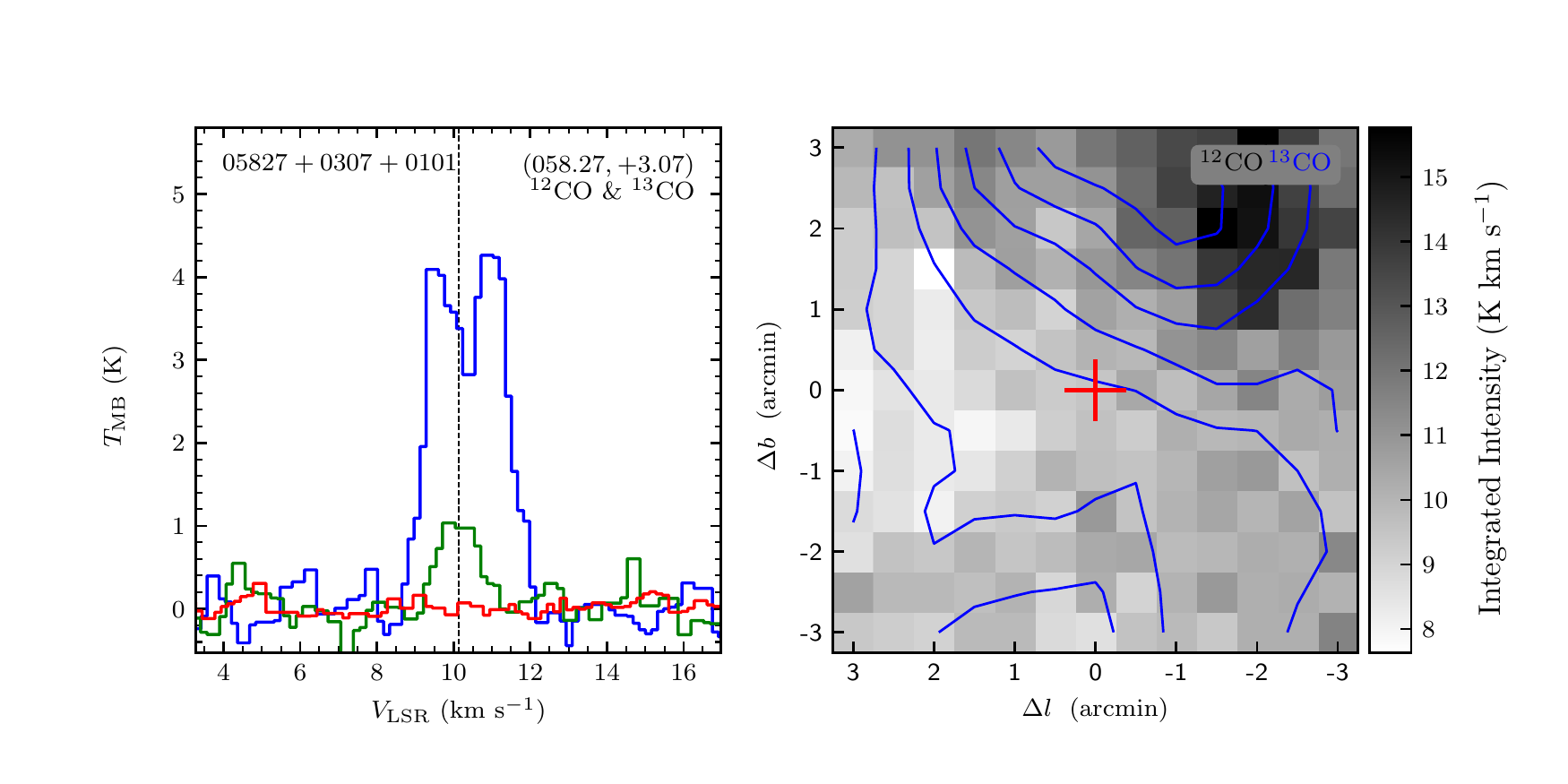}
\includegraphics[width=9.0cm,angle=0]{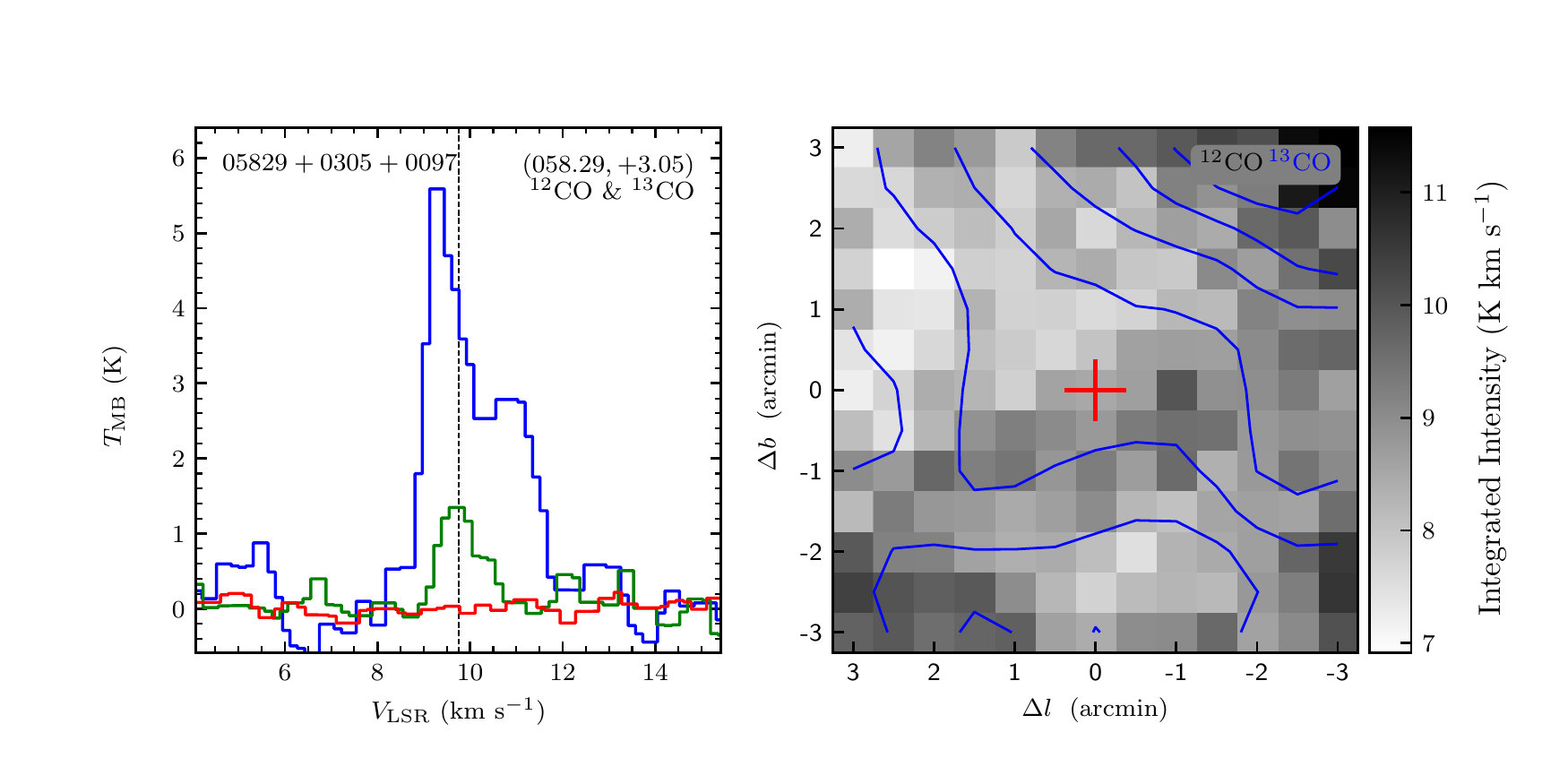}
\vspace{-0.5cm}

\includegraphics[width=9.0cm,angle=0]{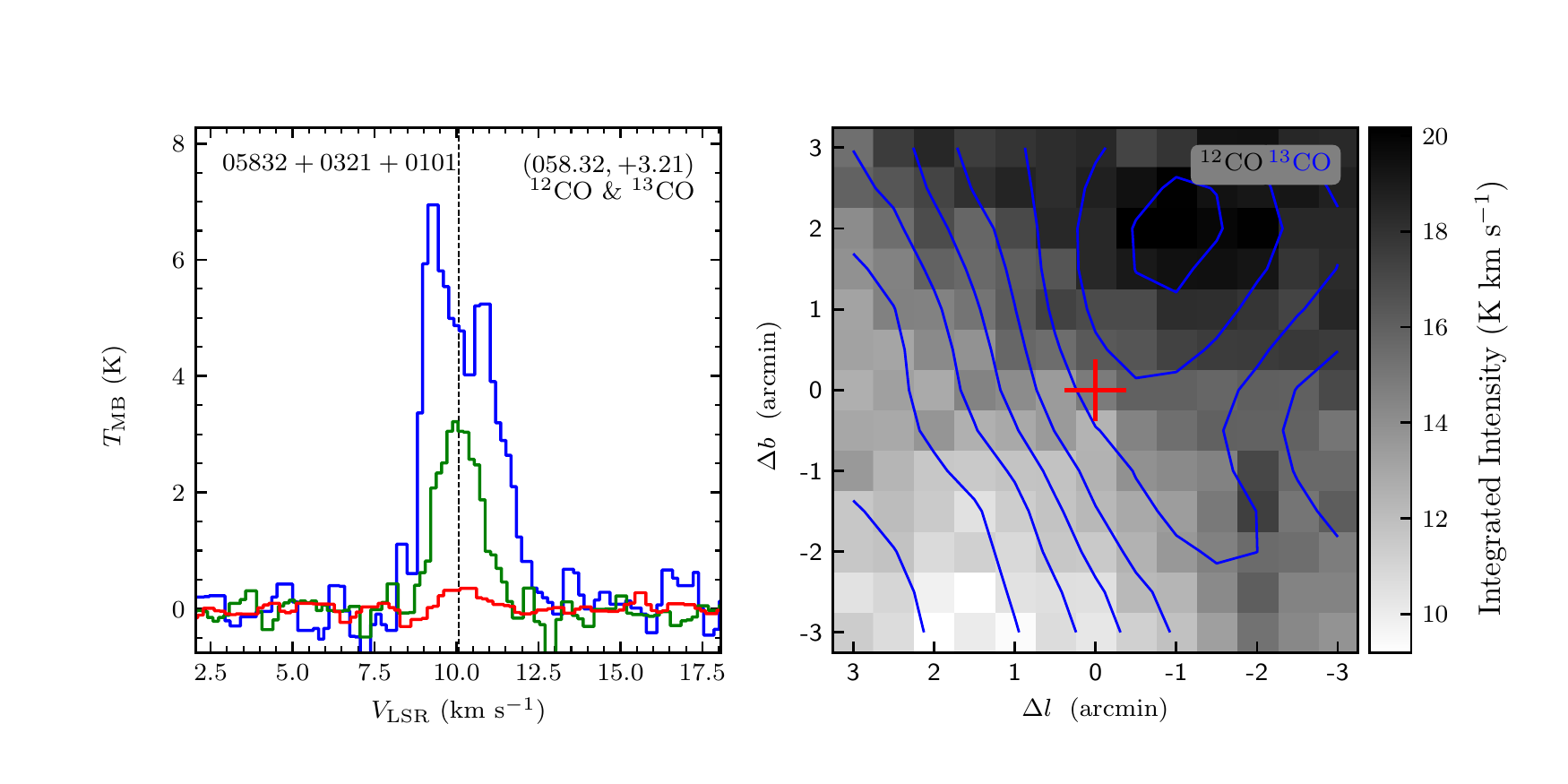}
\includegraphics[width=9.0cm,angle=0]{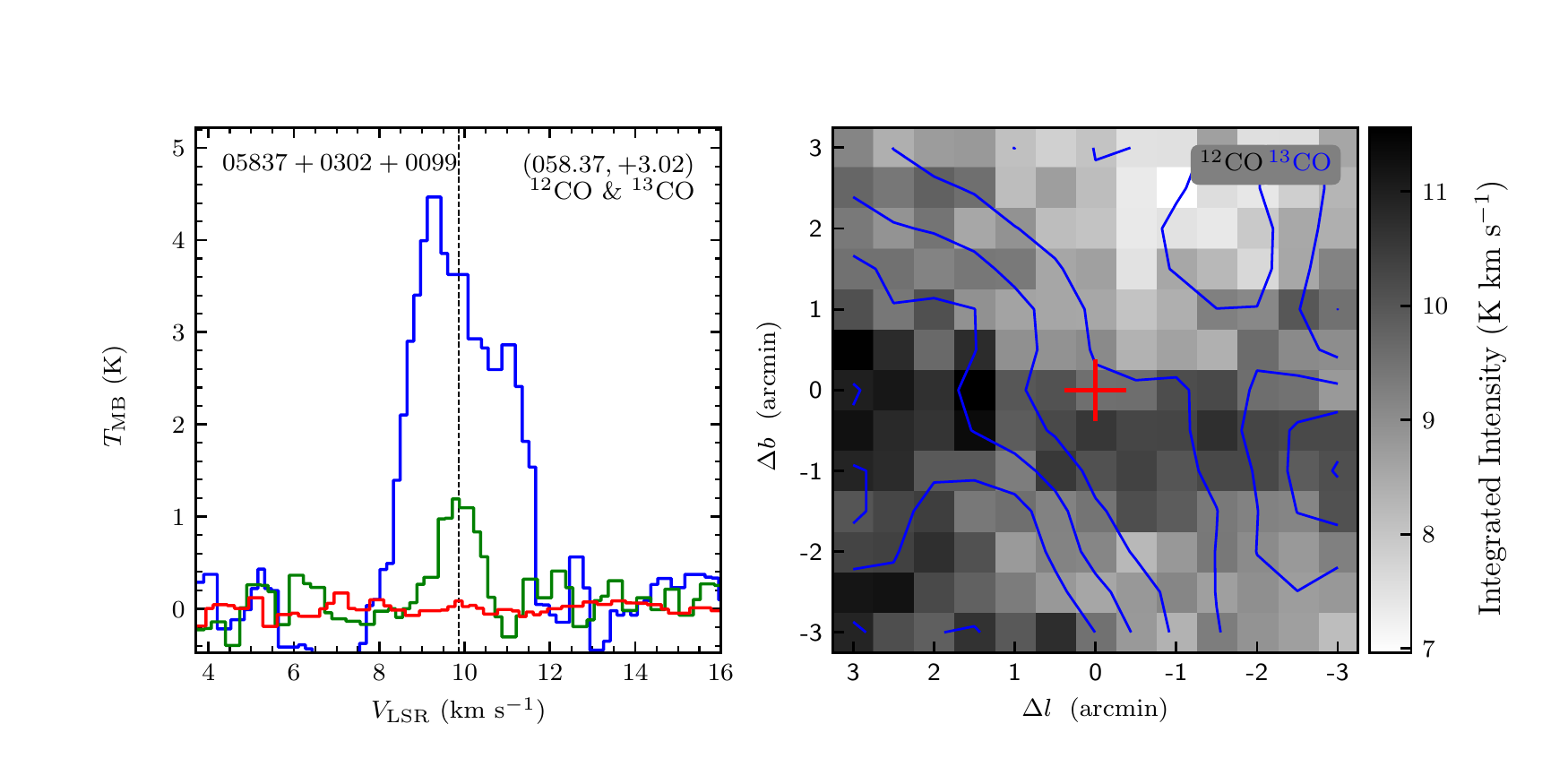}
\vspace{-0.5cm}

\includegraphics[width=9.0cm,angle=0]{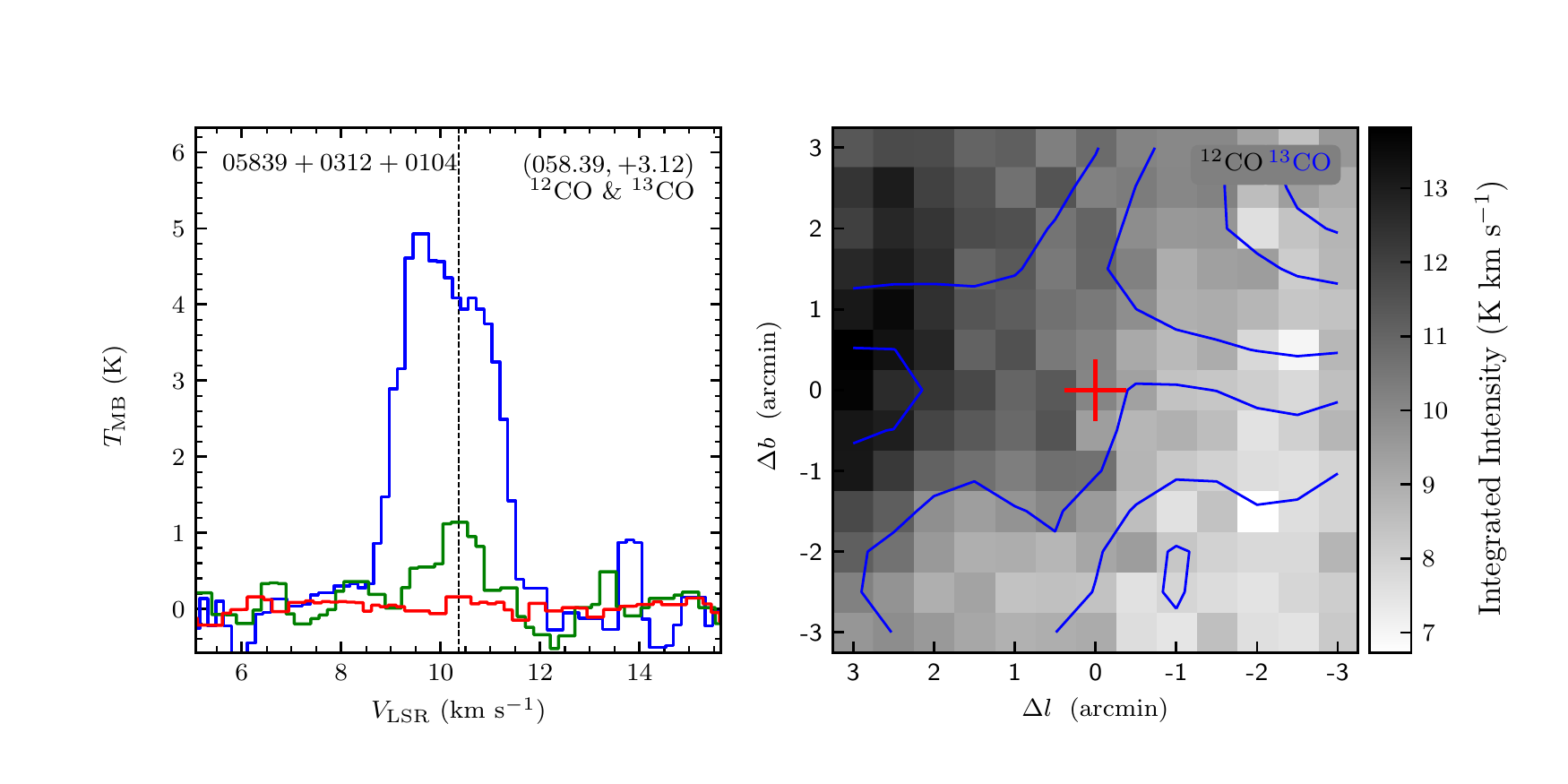}
\includegraphics[width=9.0cm,angle=0]{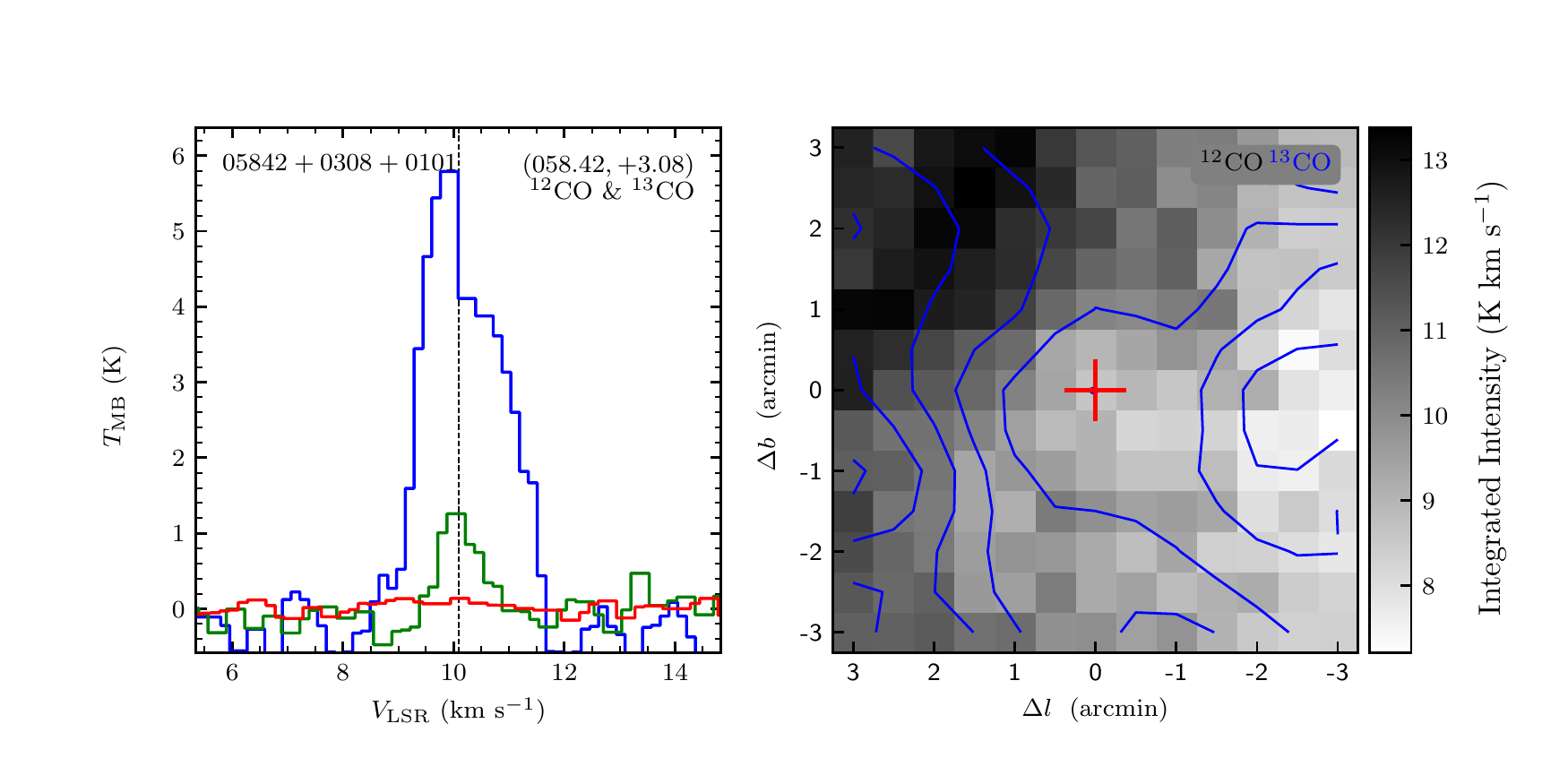}
\vspace{-0.5cm}

\includegraphics[width=9.0cm,angle=0]{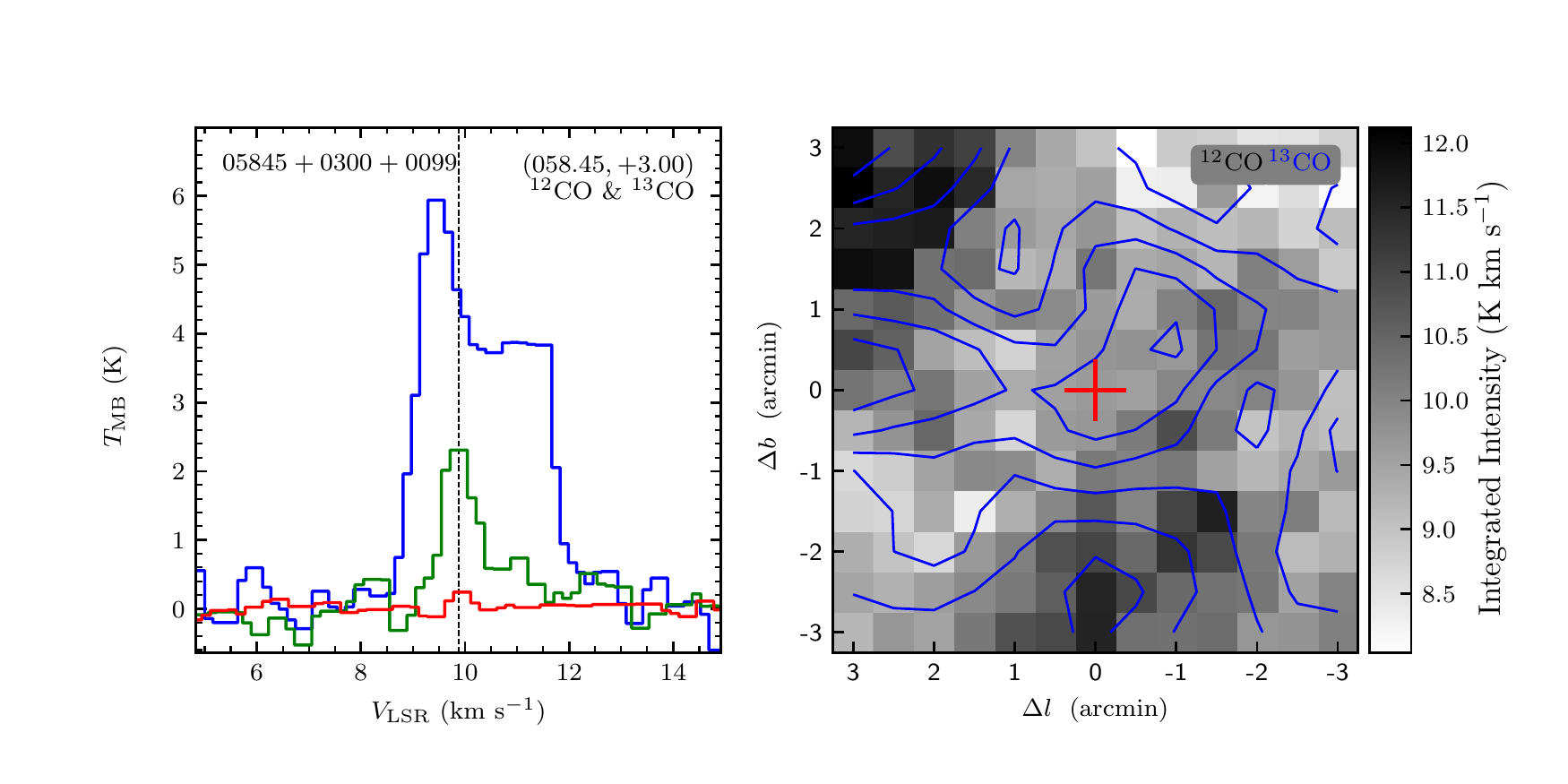}
\includegraphics[width=9.0cm,angle=0]{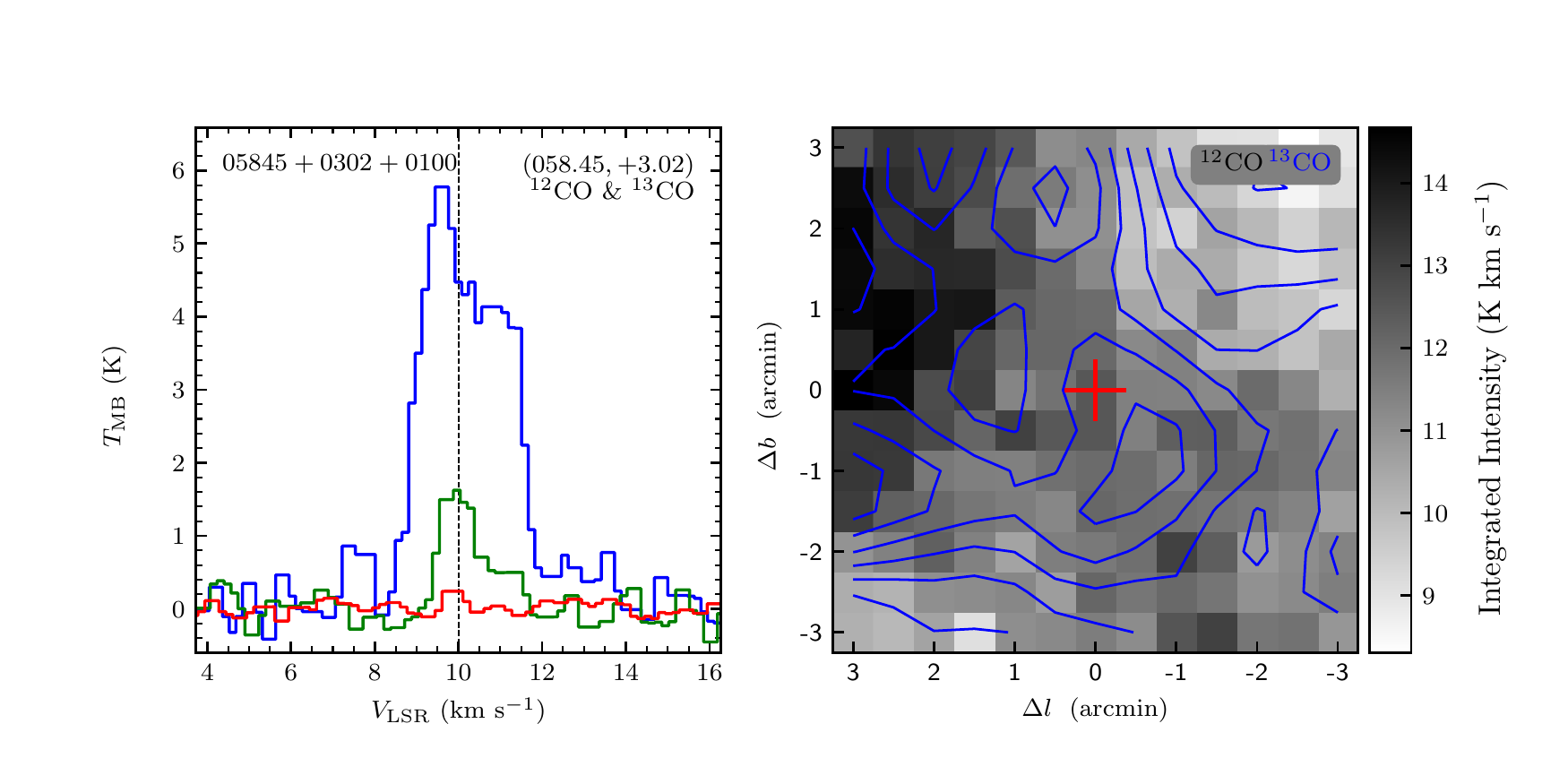}
\vspace{-0.5cm}

\includegraphics[width=9.0cm,angle=0]{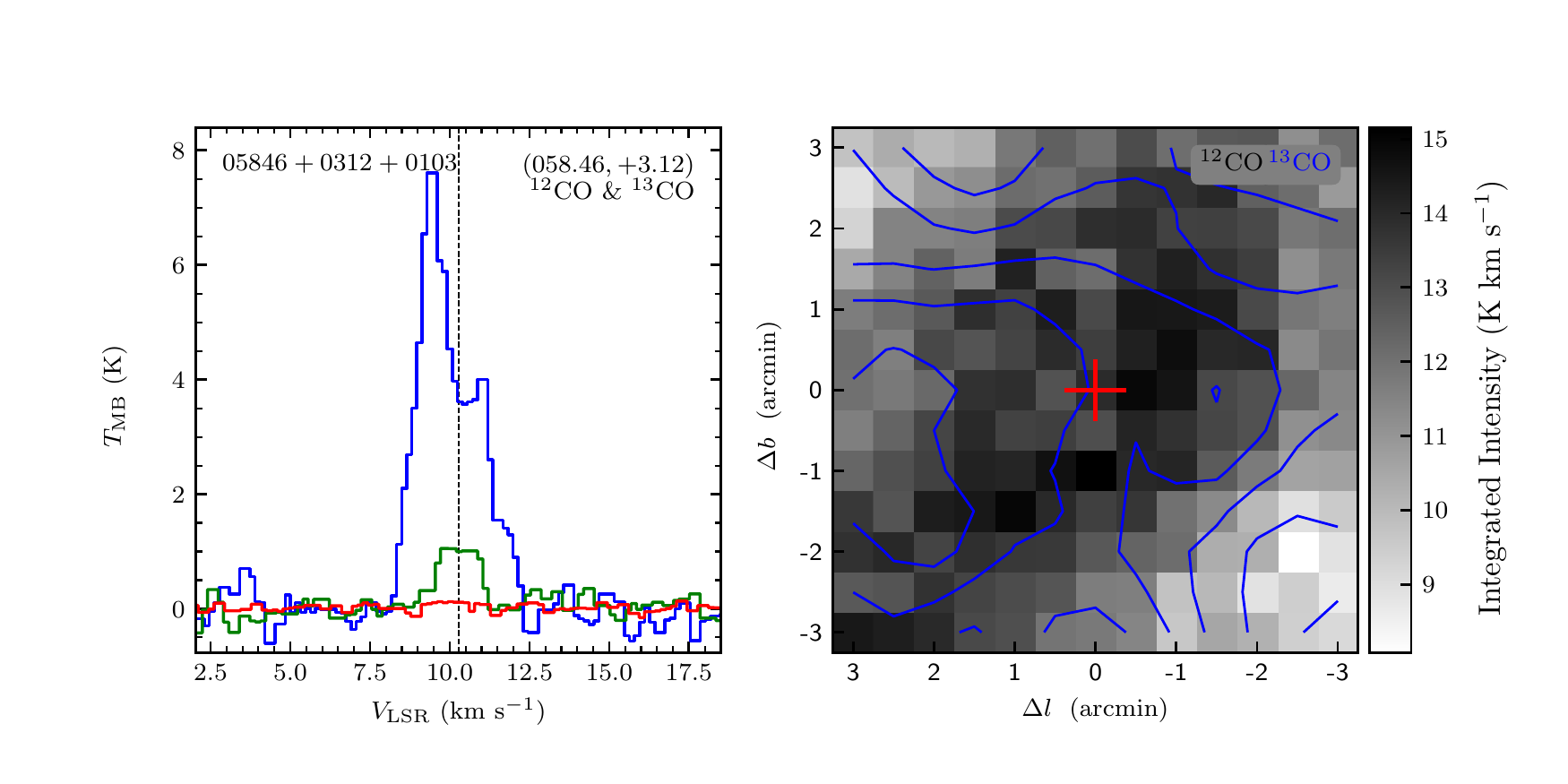}
\includegraphics[width=9.0cm,angle=0]{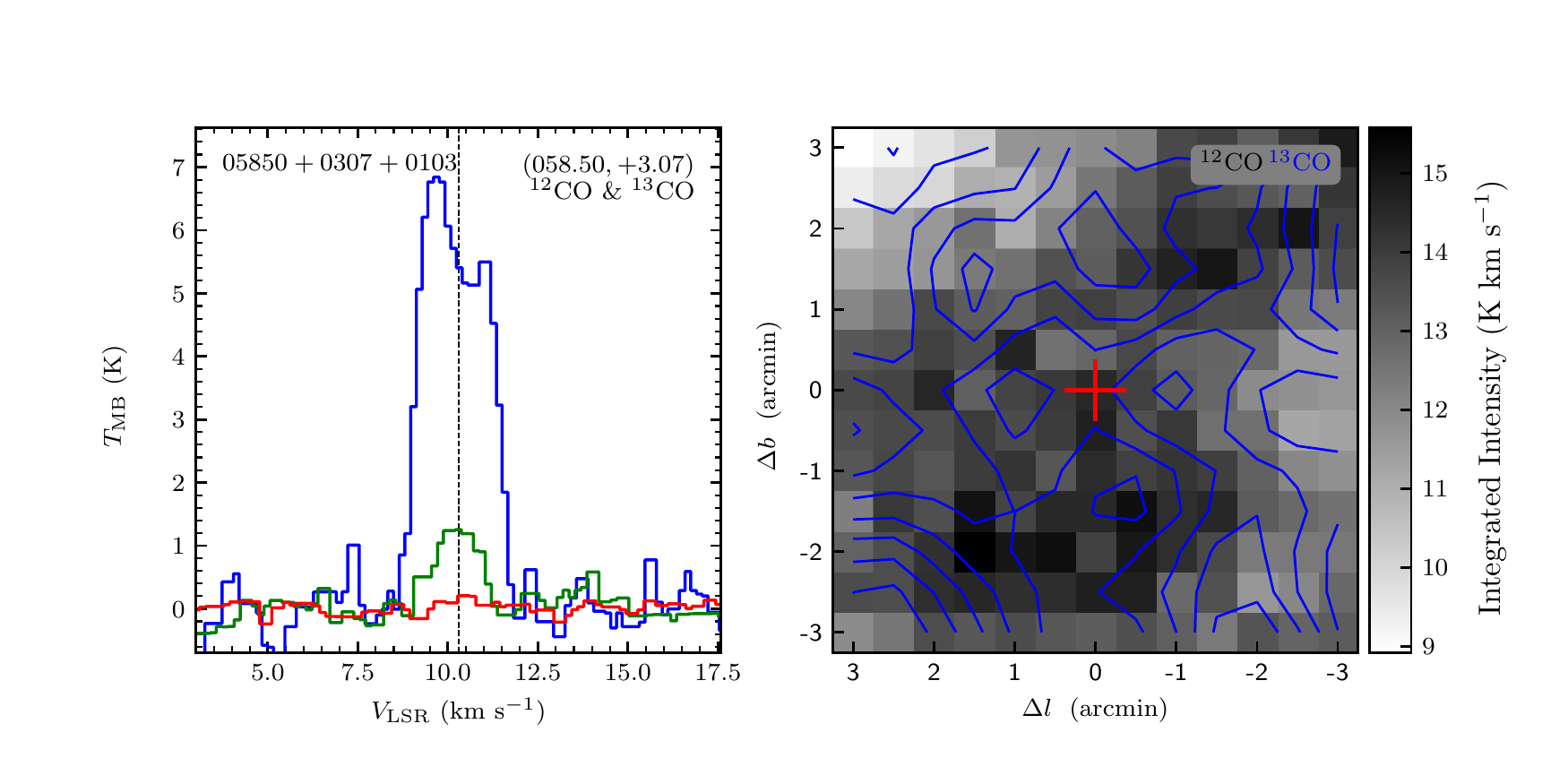}
\end{figure}
\clearpage

\begin{figure}
\includegraphics[width=9.0cm,angle=0]{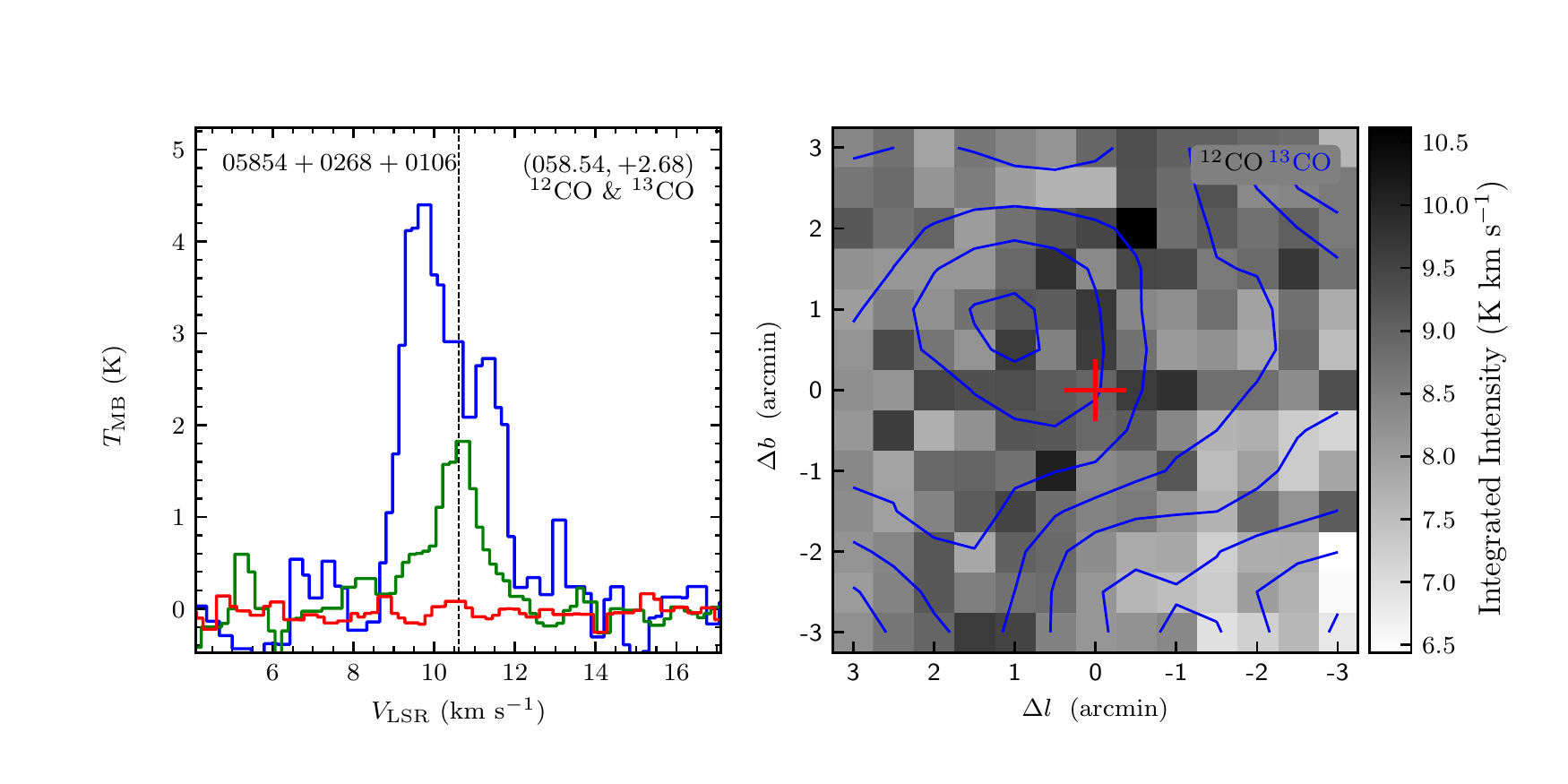}
\includegraphics[width=9.0cm,angle=0]{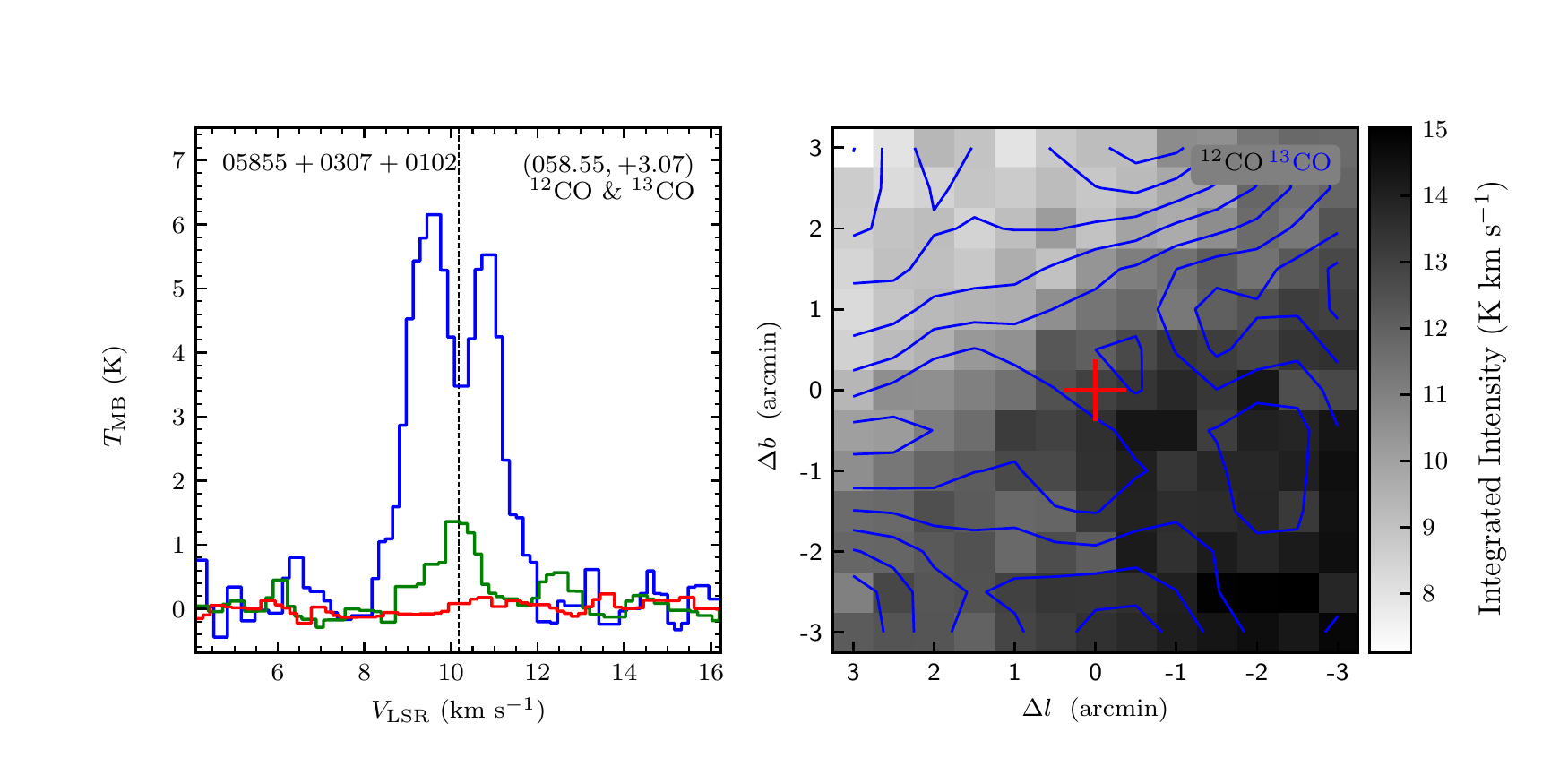}
\vspace{-0.5cm}

\includegraphics[width=9.0cm,angle=0]{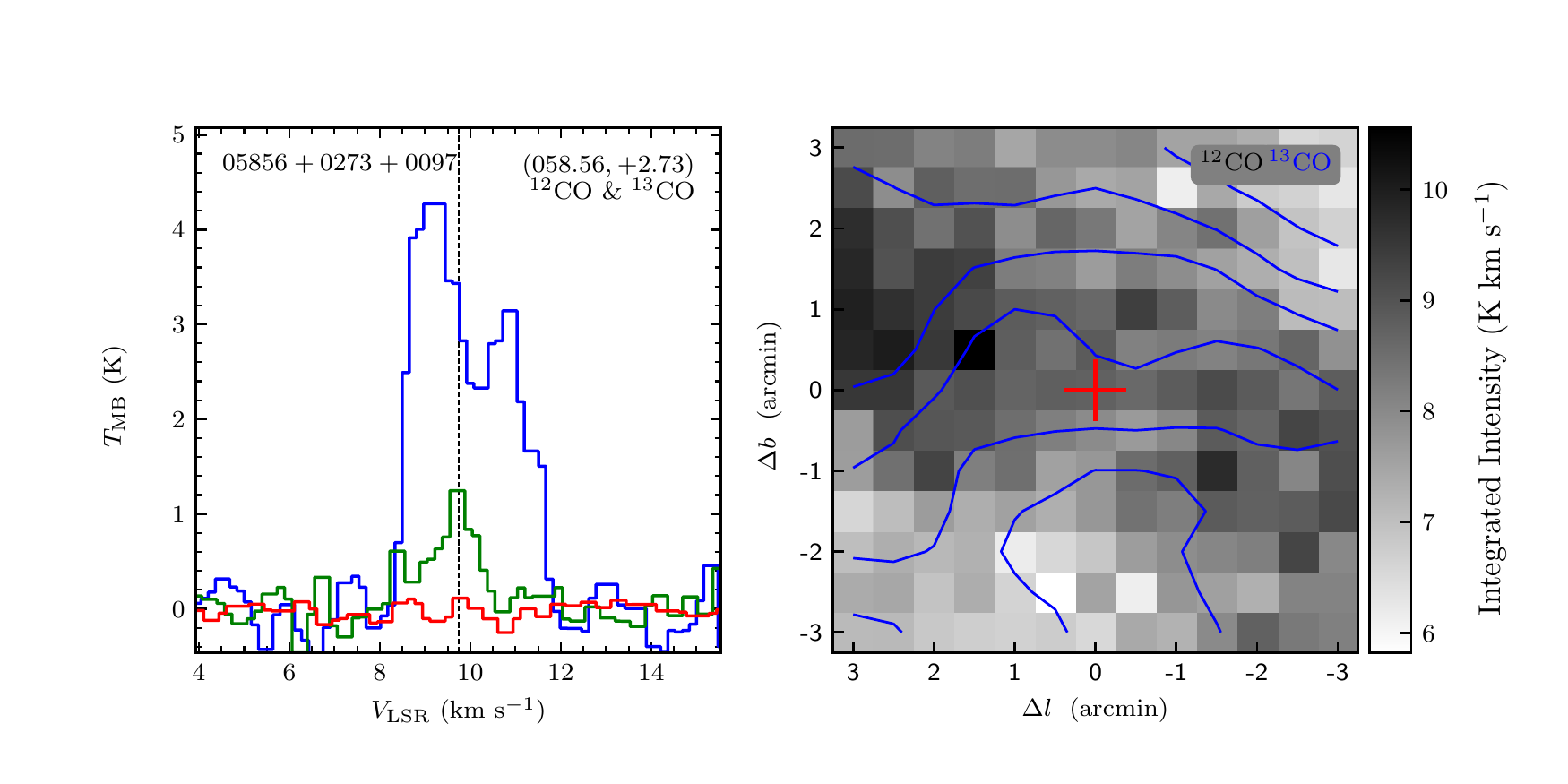}
\includegraphics[width=9.0cm,angle=0]{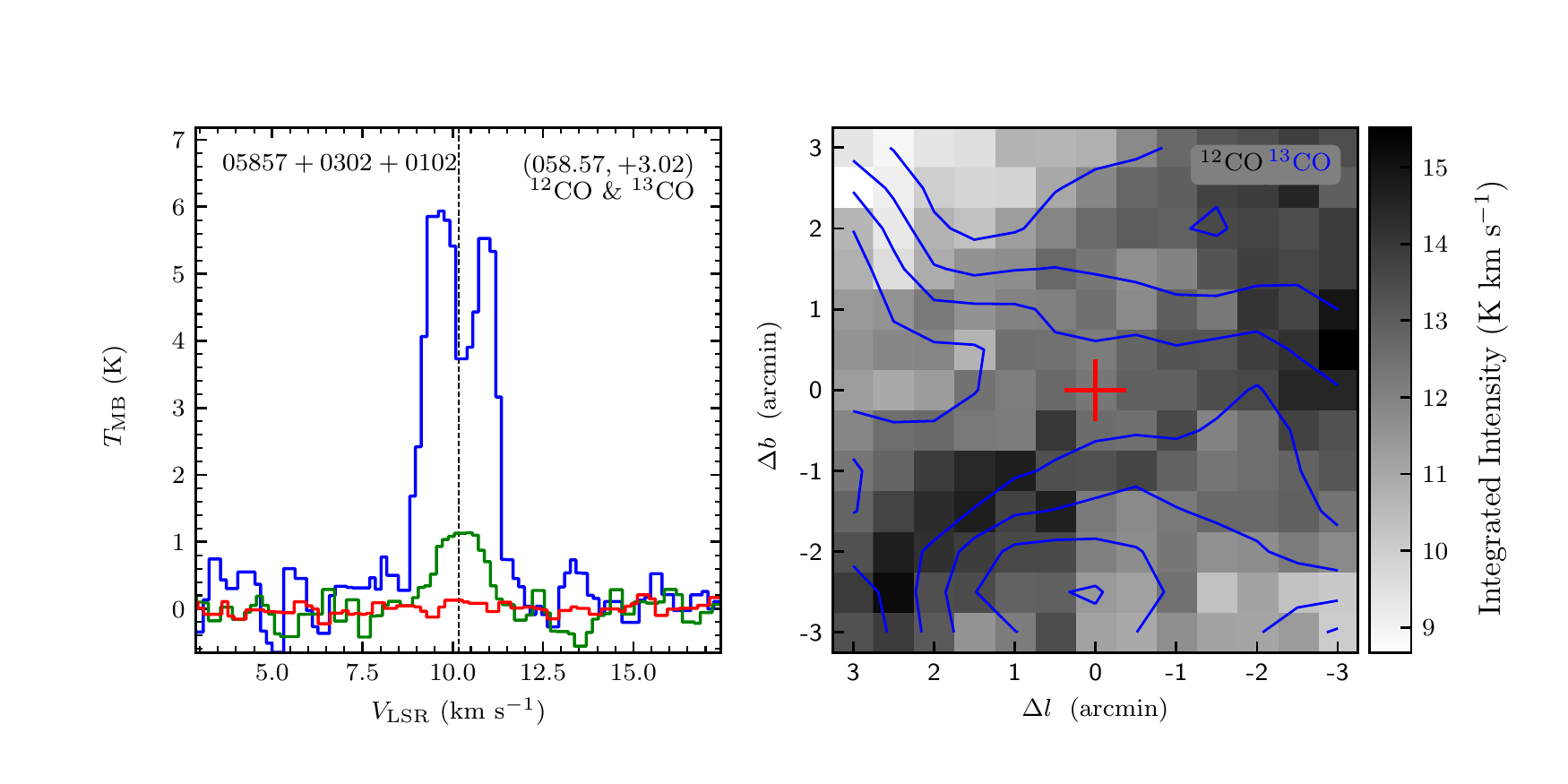}
\vspace{-0.5cm}

\includegraphics[width=9.0cm,angle=0]{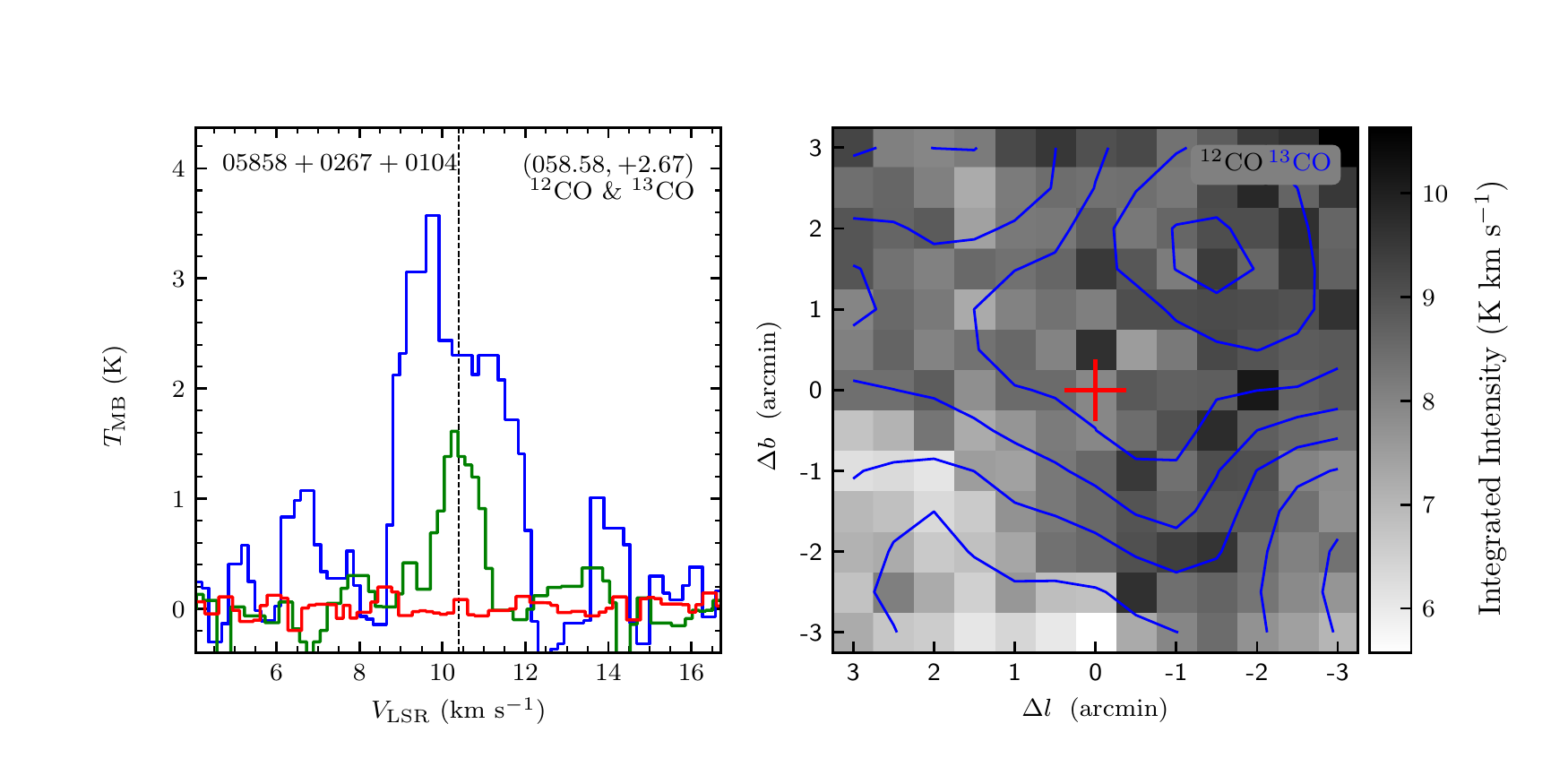}
\includegraphics[width=9.0cm,angle=0]{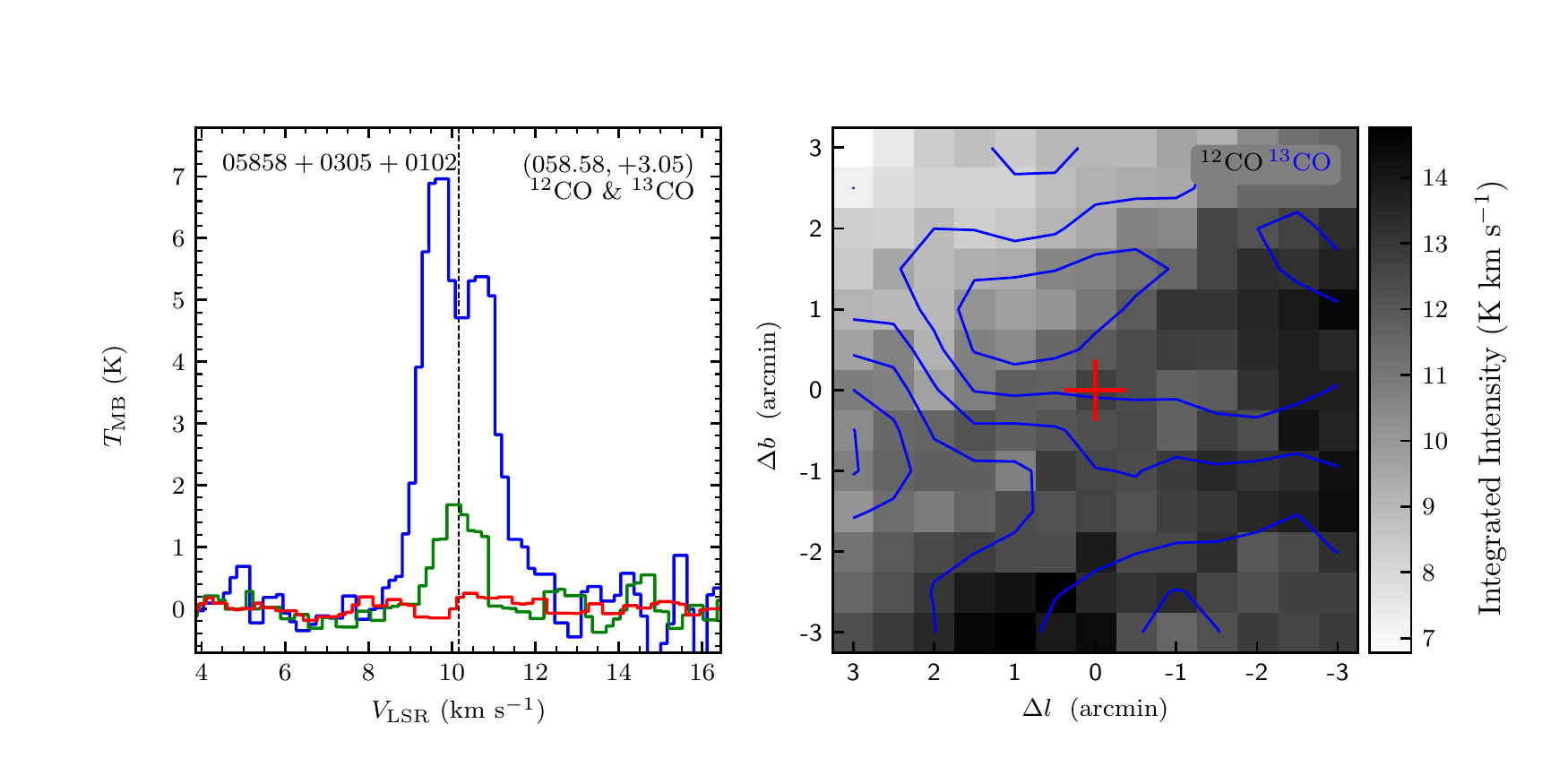}
\vspace{-0.5cm}

\includegraphics[width=9.0cm,angle=0]{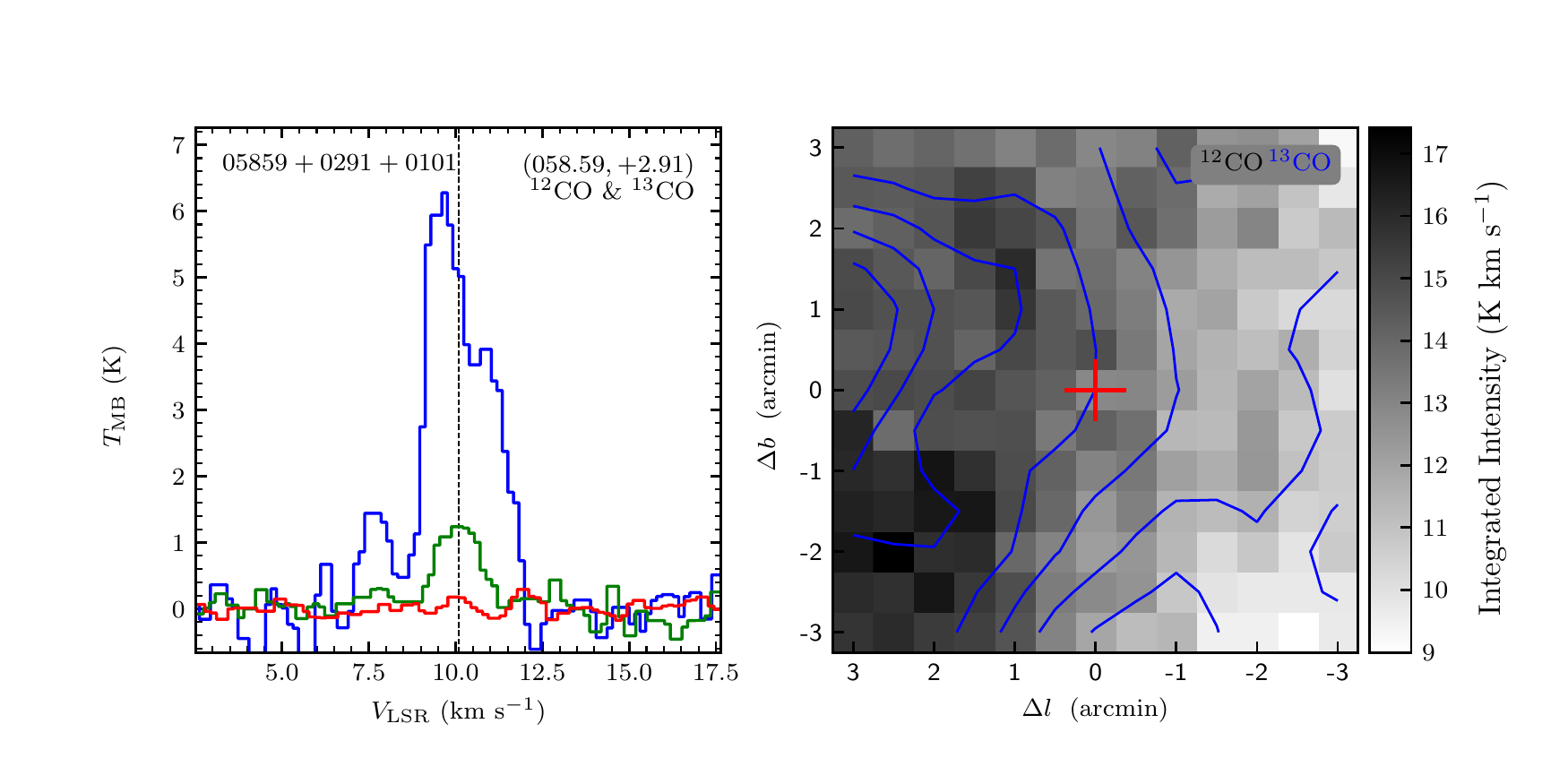}
\includegraphics[width=9.0cm,angle=0]{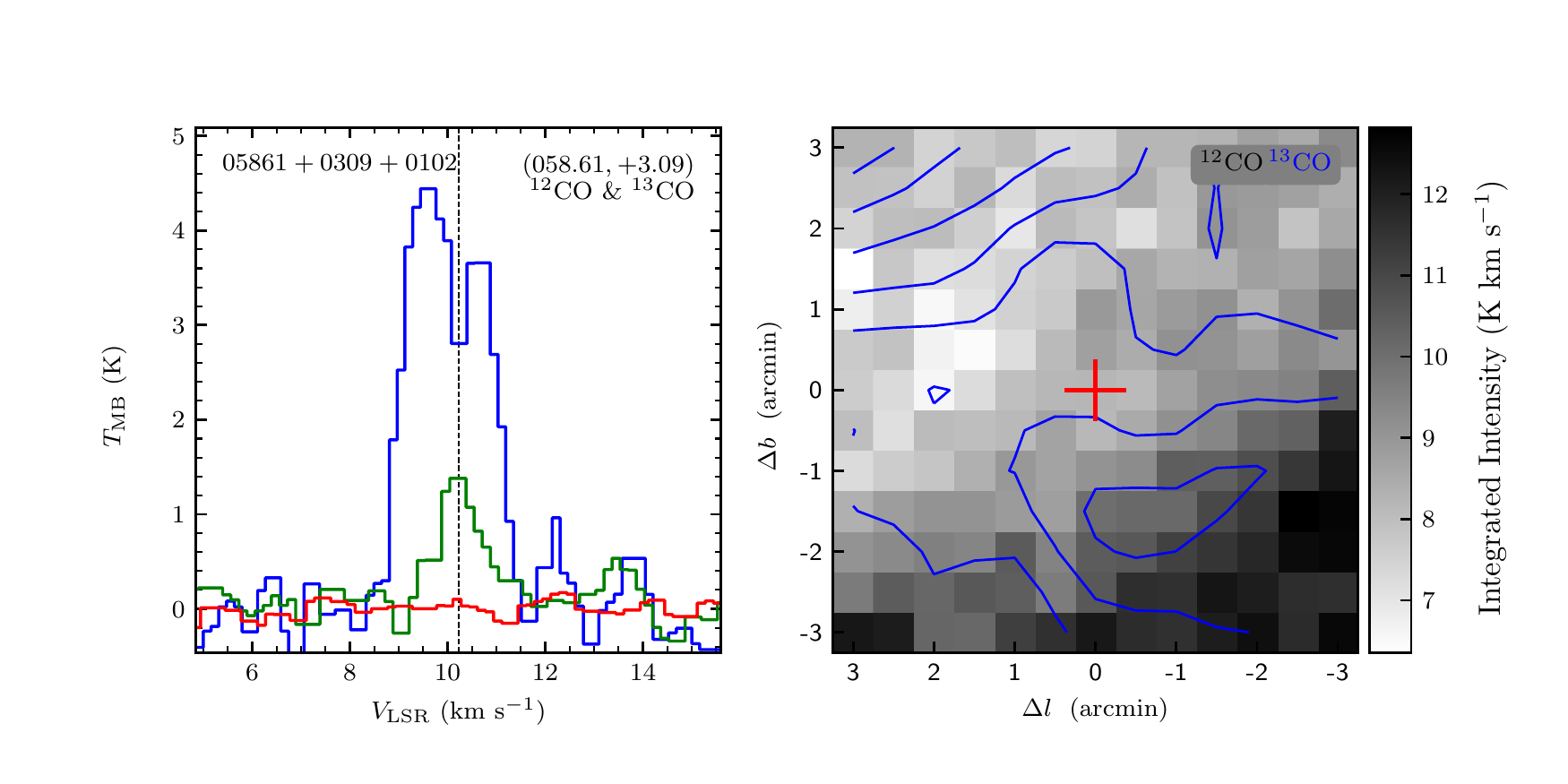}
\vspace{-0.5cm}

\includegraphics[width=9.0cm,angle=0]{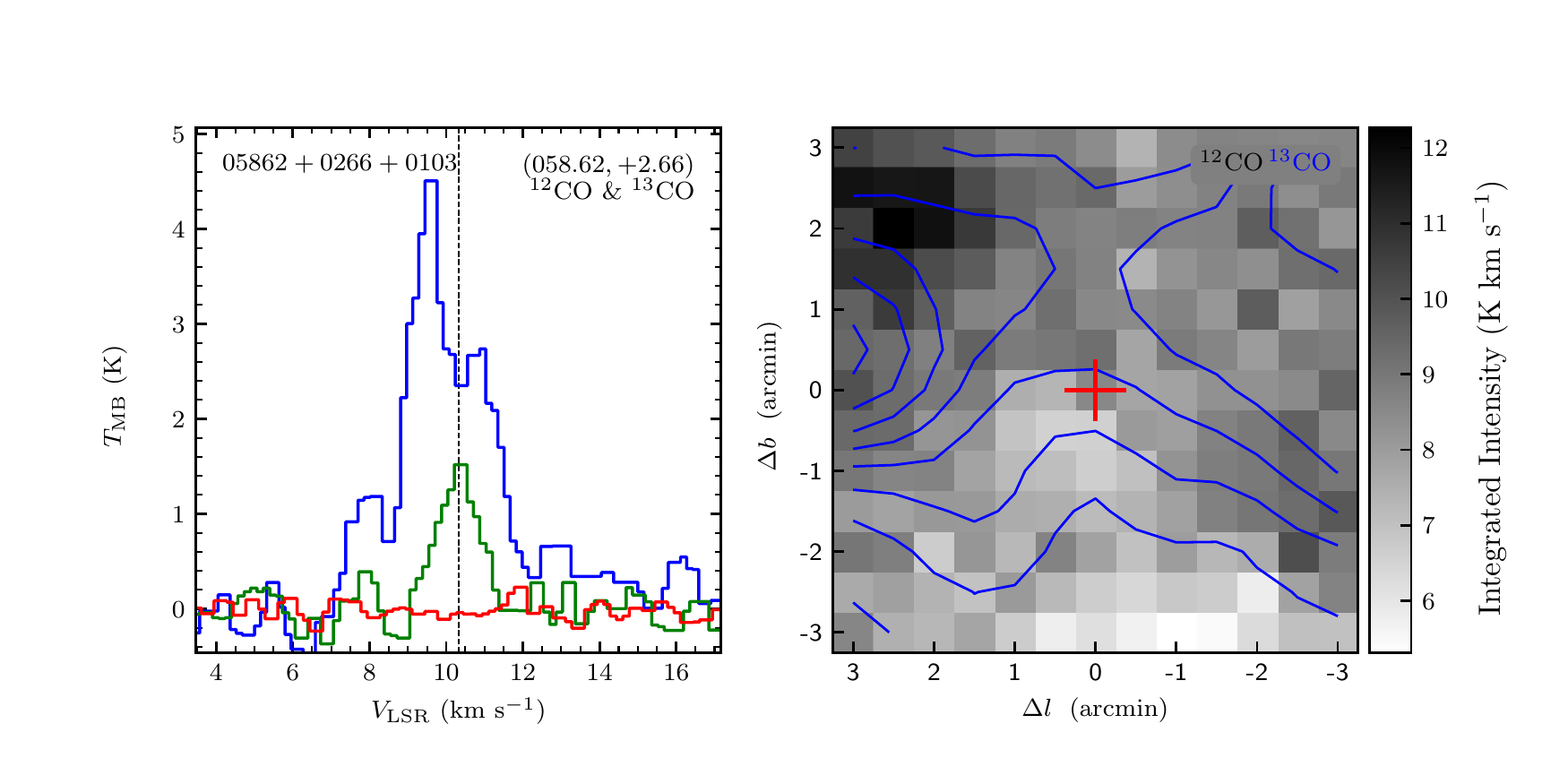}
\includegraphics[width=9.0cm,angle=0]{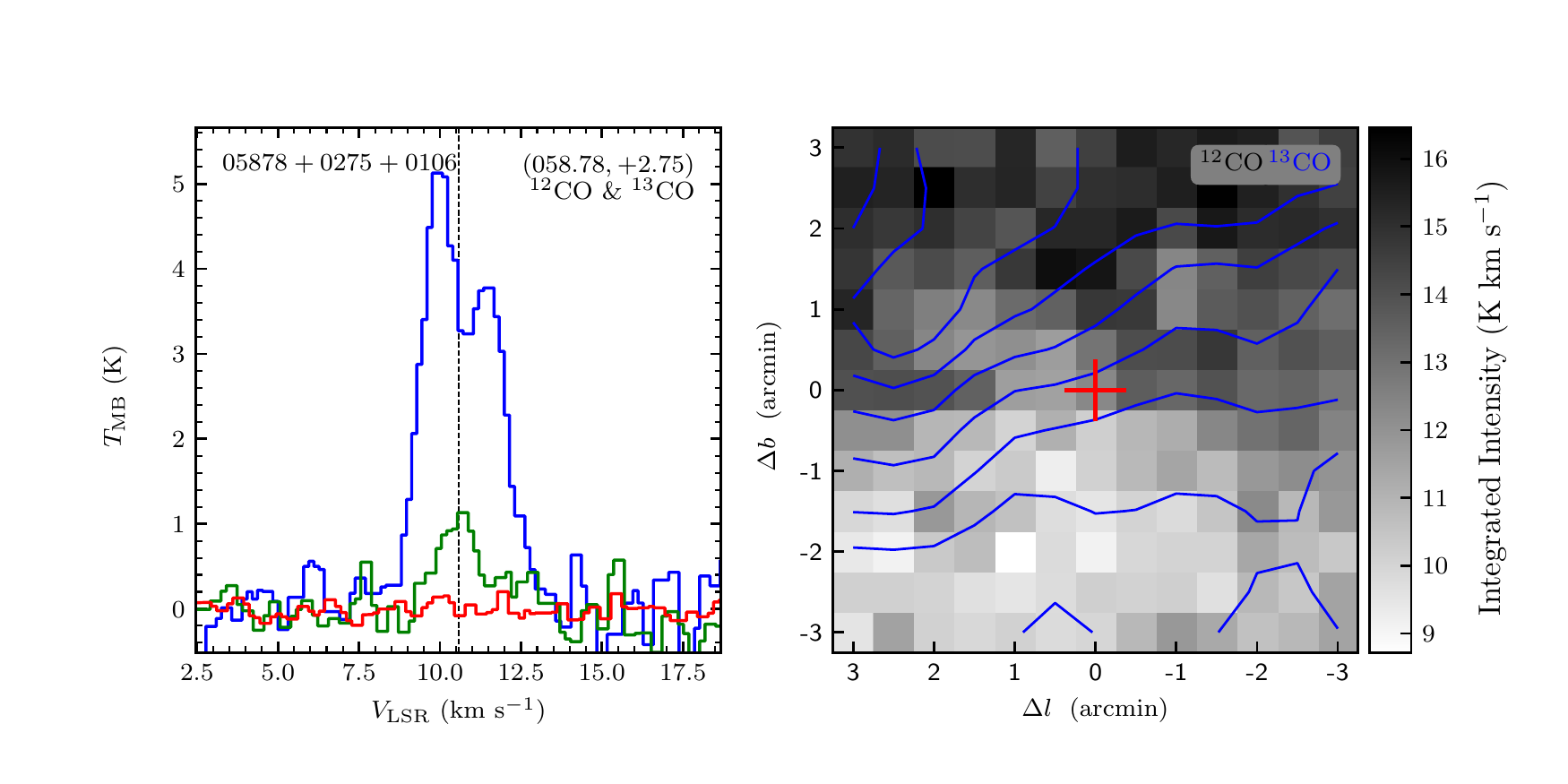}
\end{figure}
\clearpage

\begin{figure}
\includegraphics[width=9.0cm,angle=0]{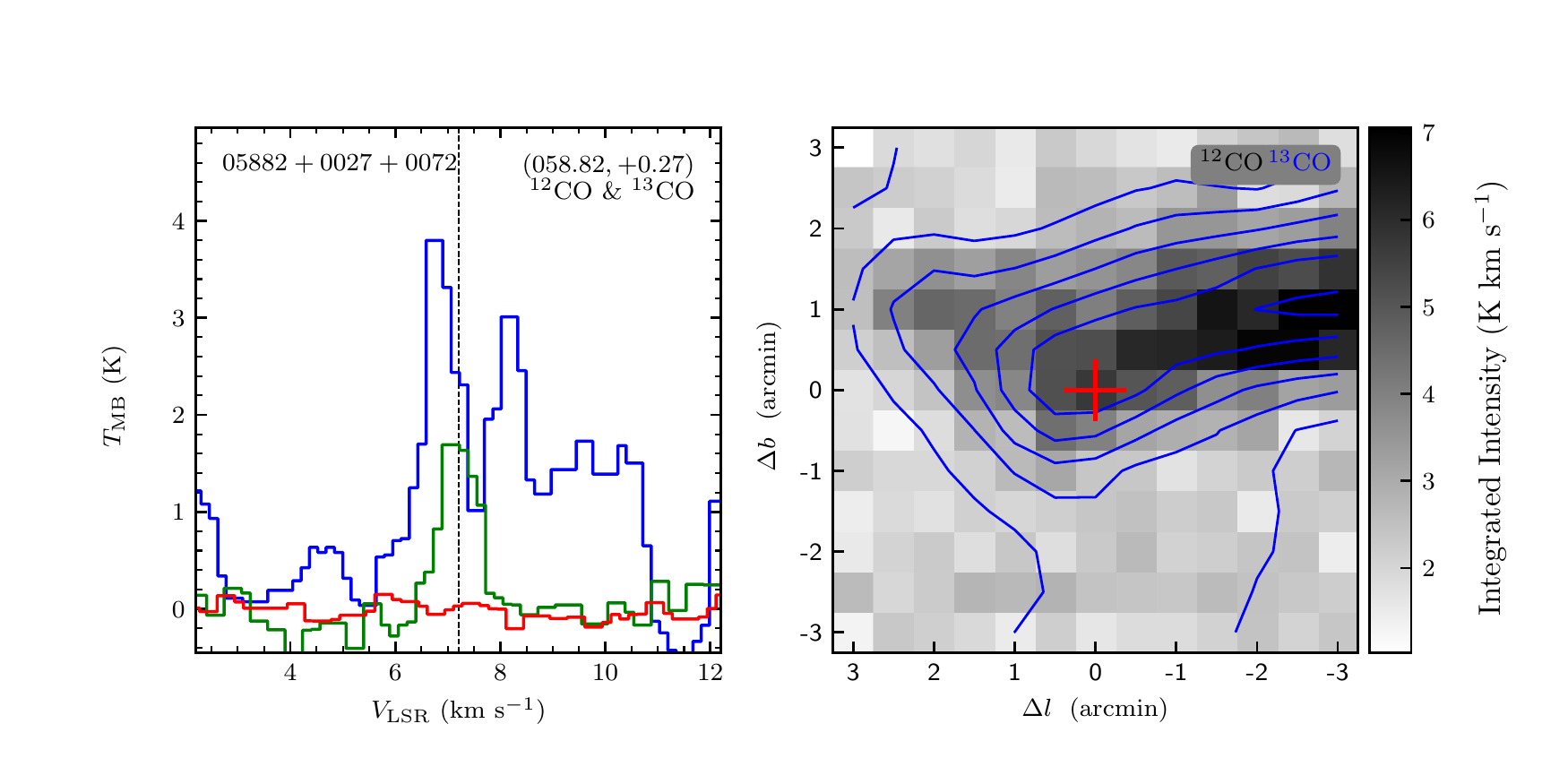}
\includegraphics[width=9.0cm,angle=0]{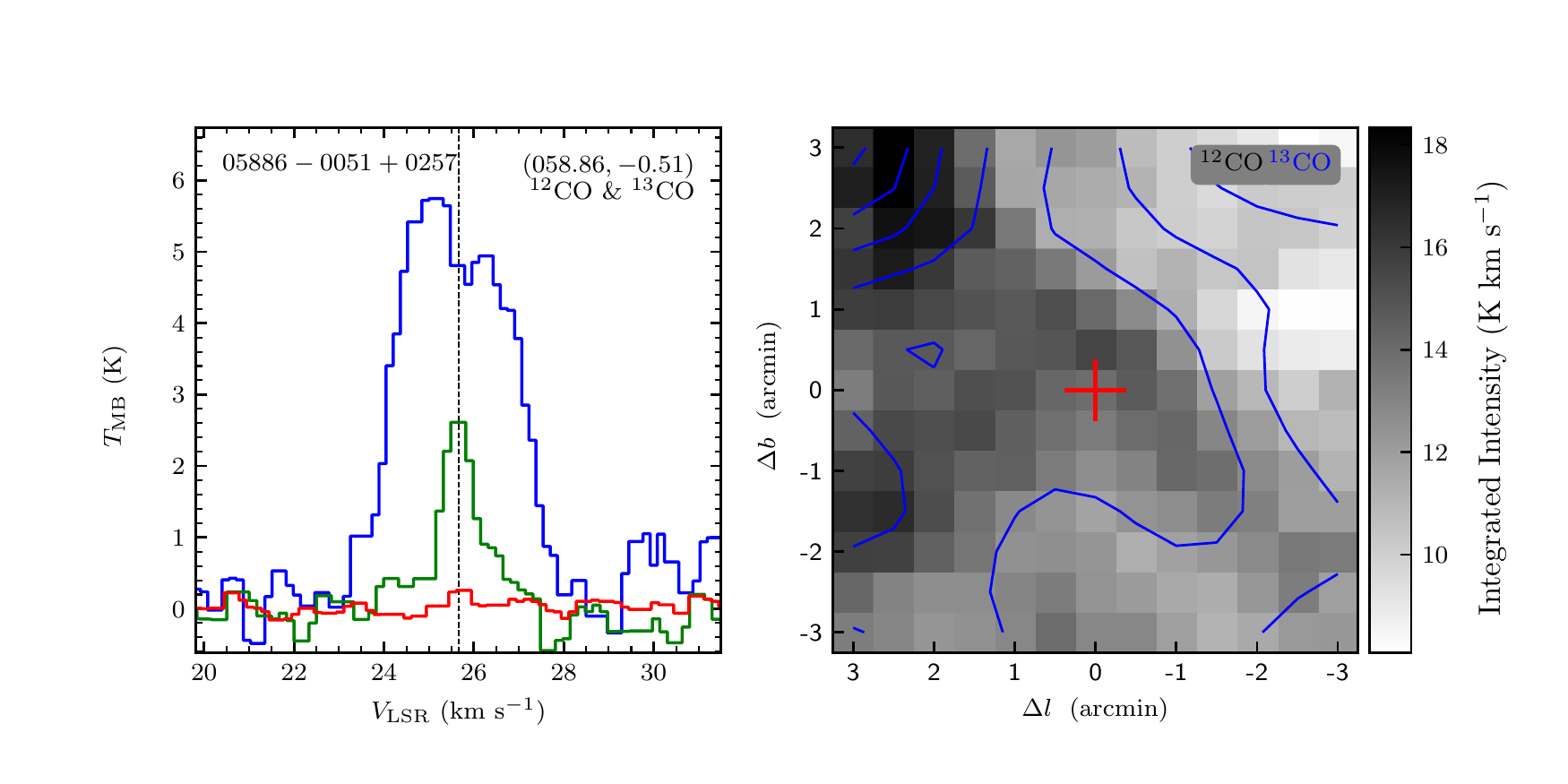}
\vspace{-0.5cm}

\includegraphics[width=9.0cm,angle=0]{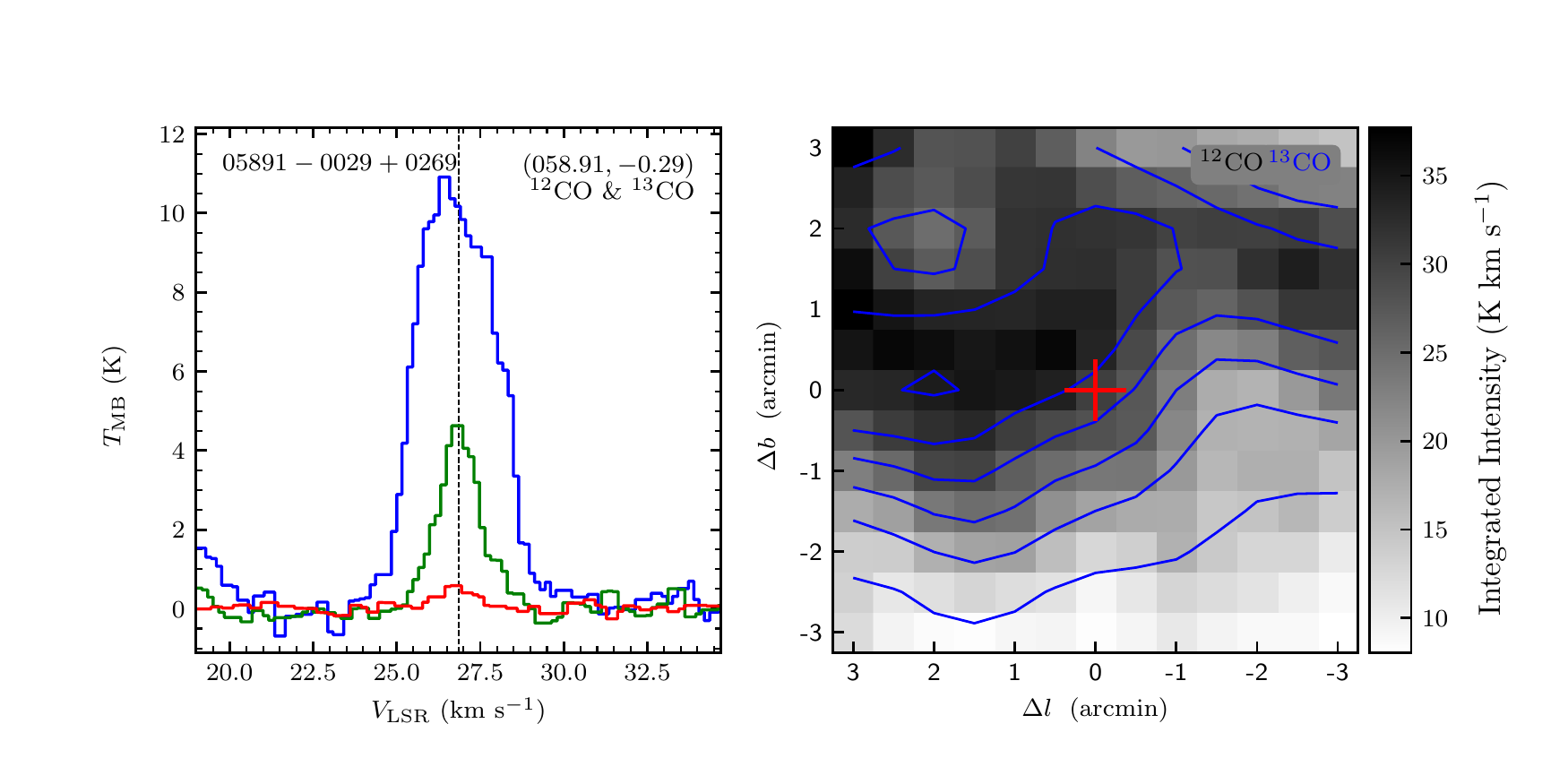}
\includegraphics[width=9.0cm,angle=0]{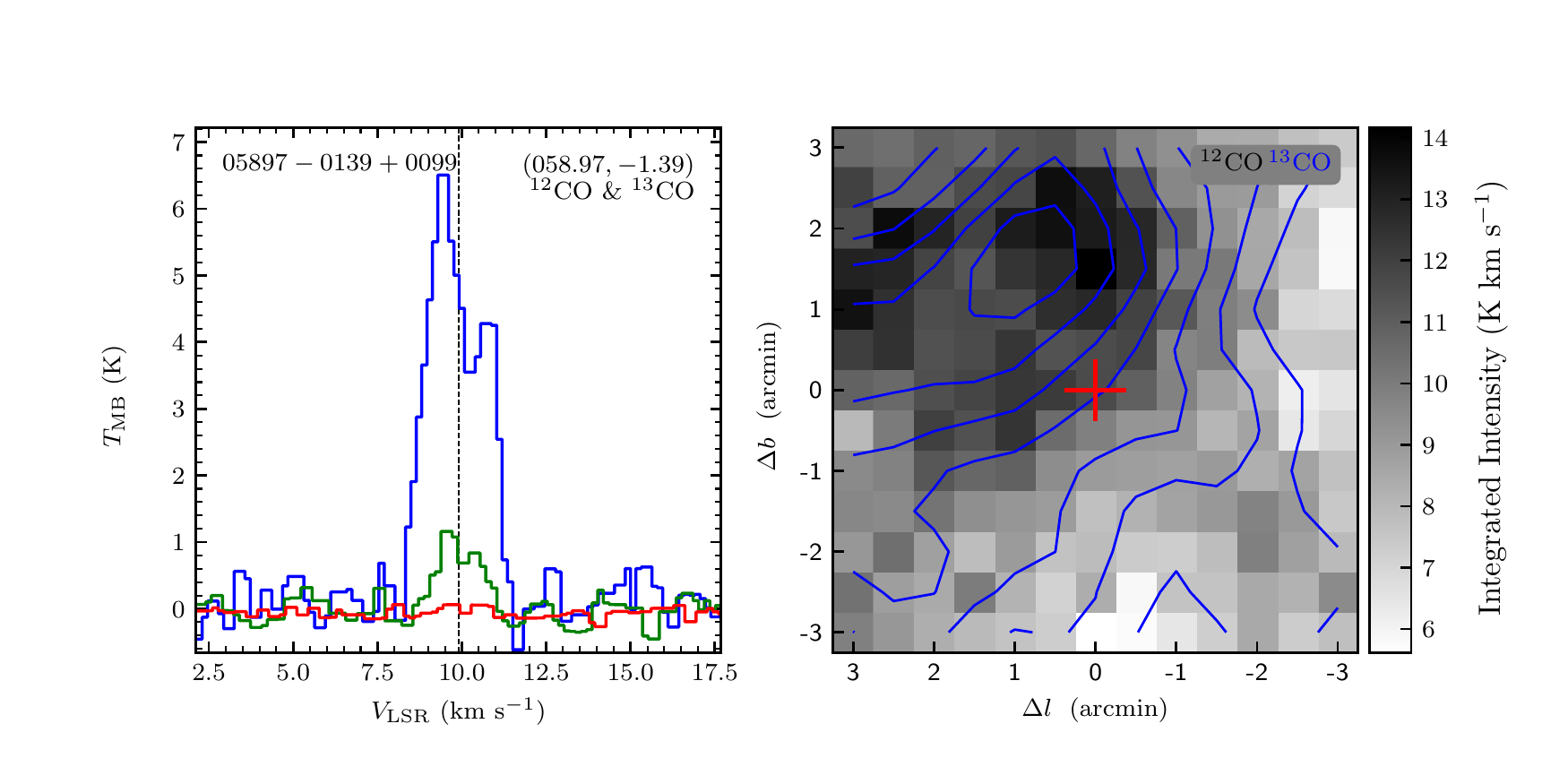}
\vspace{-0.5cm}

\includegraphics[width=9.0cm,angle=0]{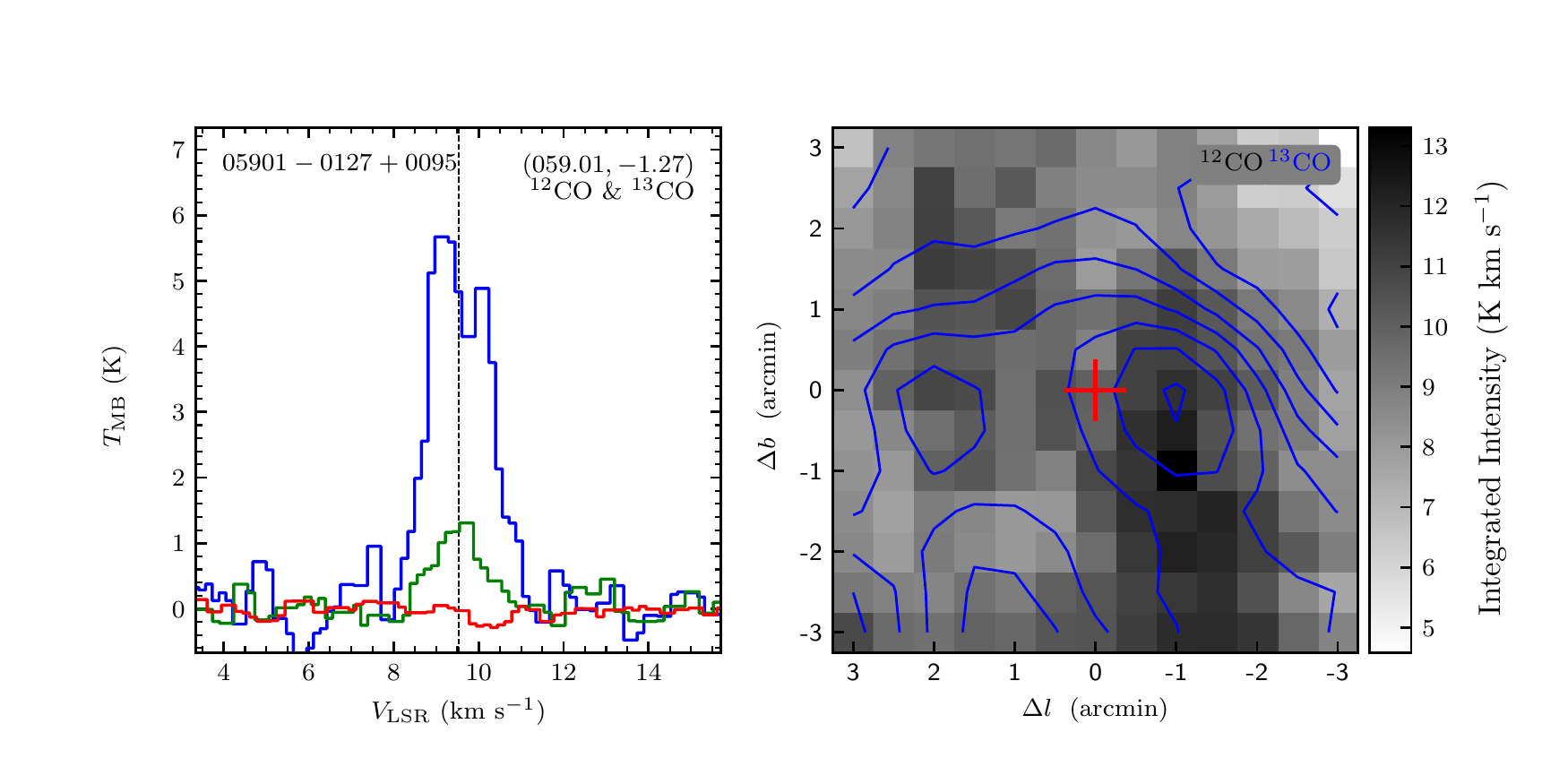}
\includegraphics[width=9.0cm,angle=0]{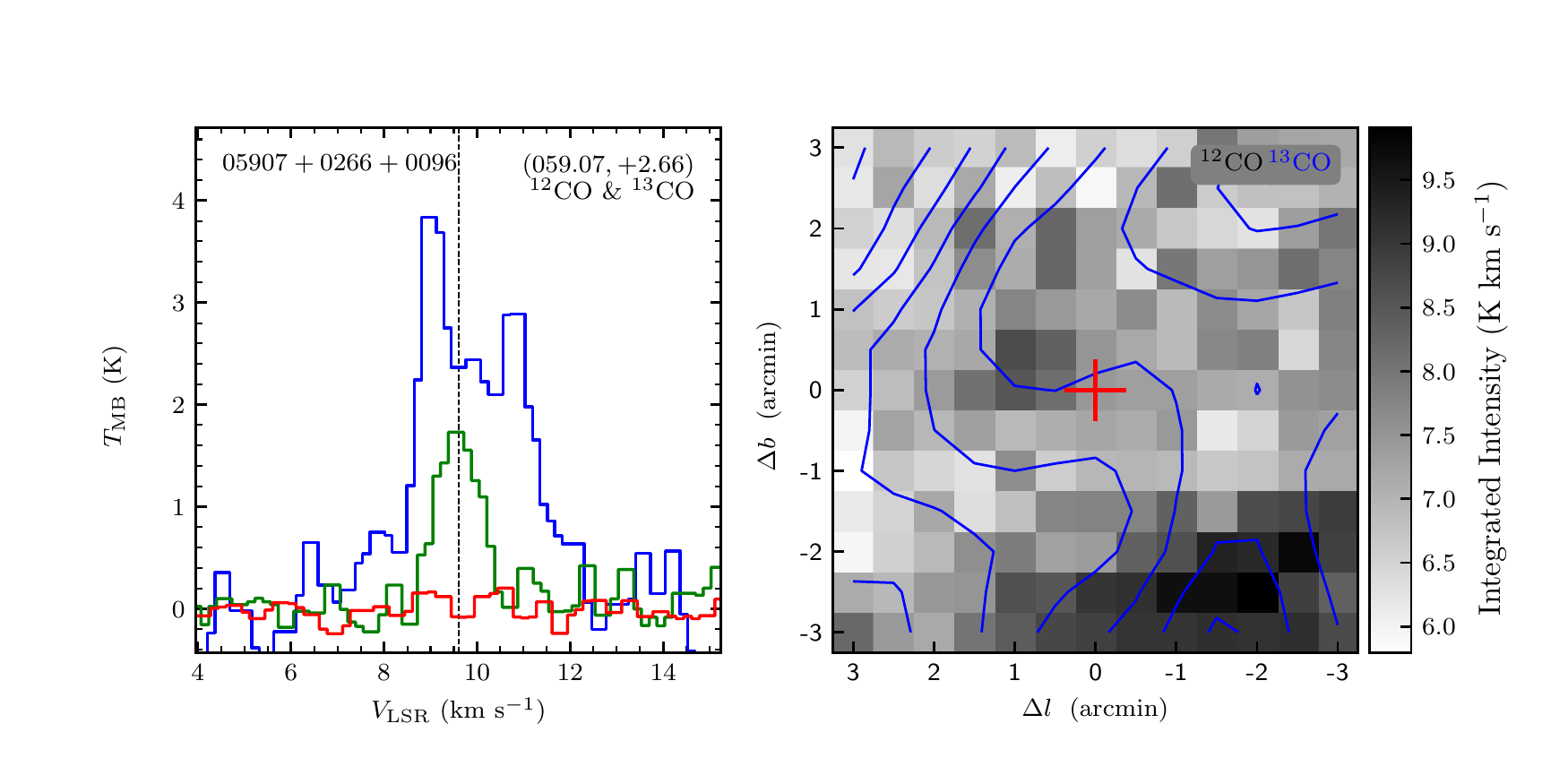}
\vspace{-0.5cm}

\includegraphics[width=9.0cm,angle=0]{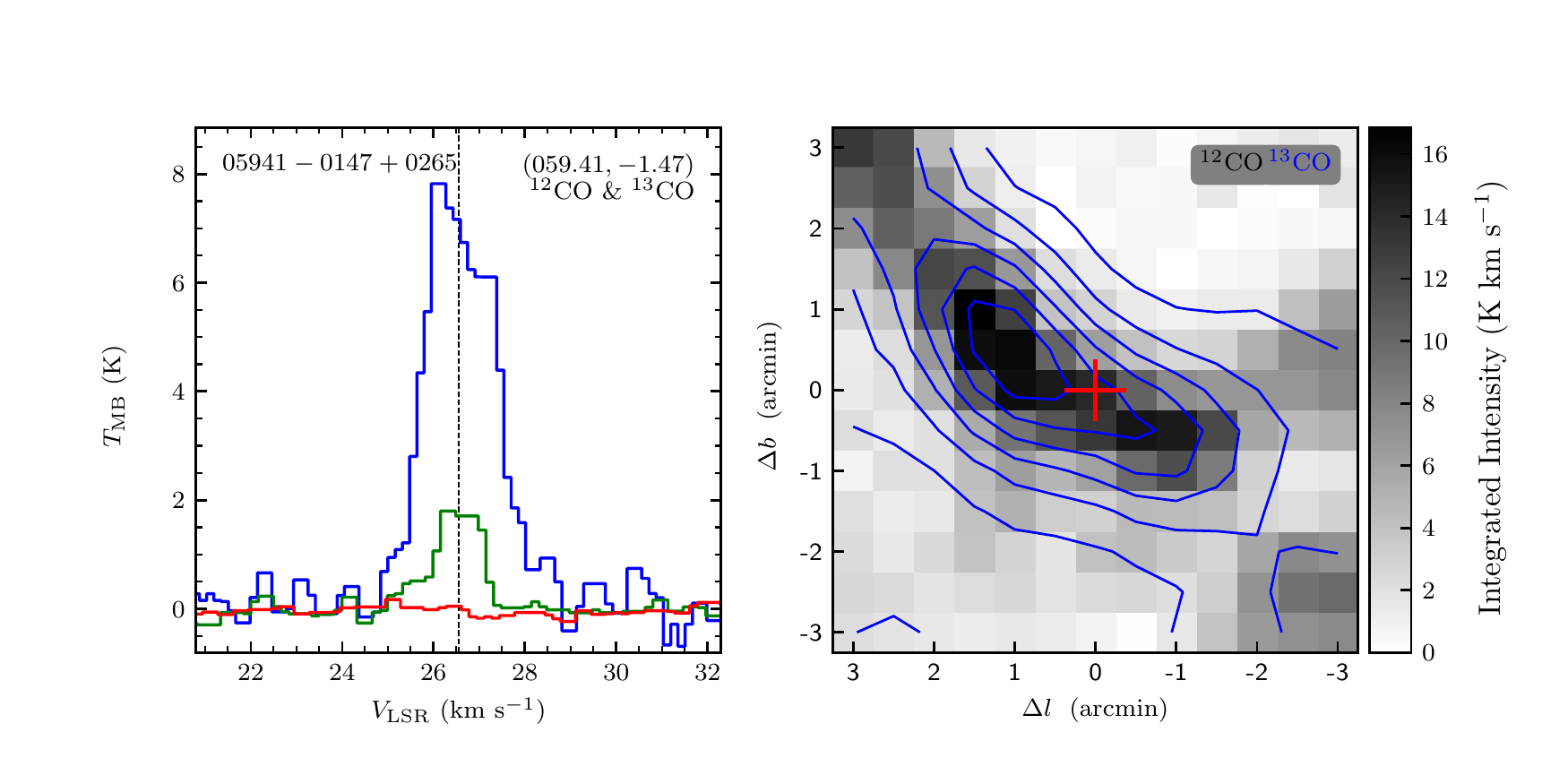}
\includegraphics[width=9.0cm,angle=0]{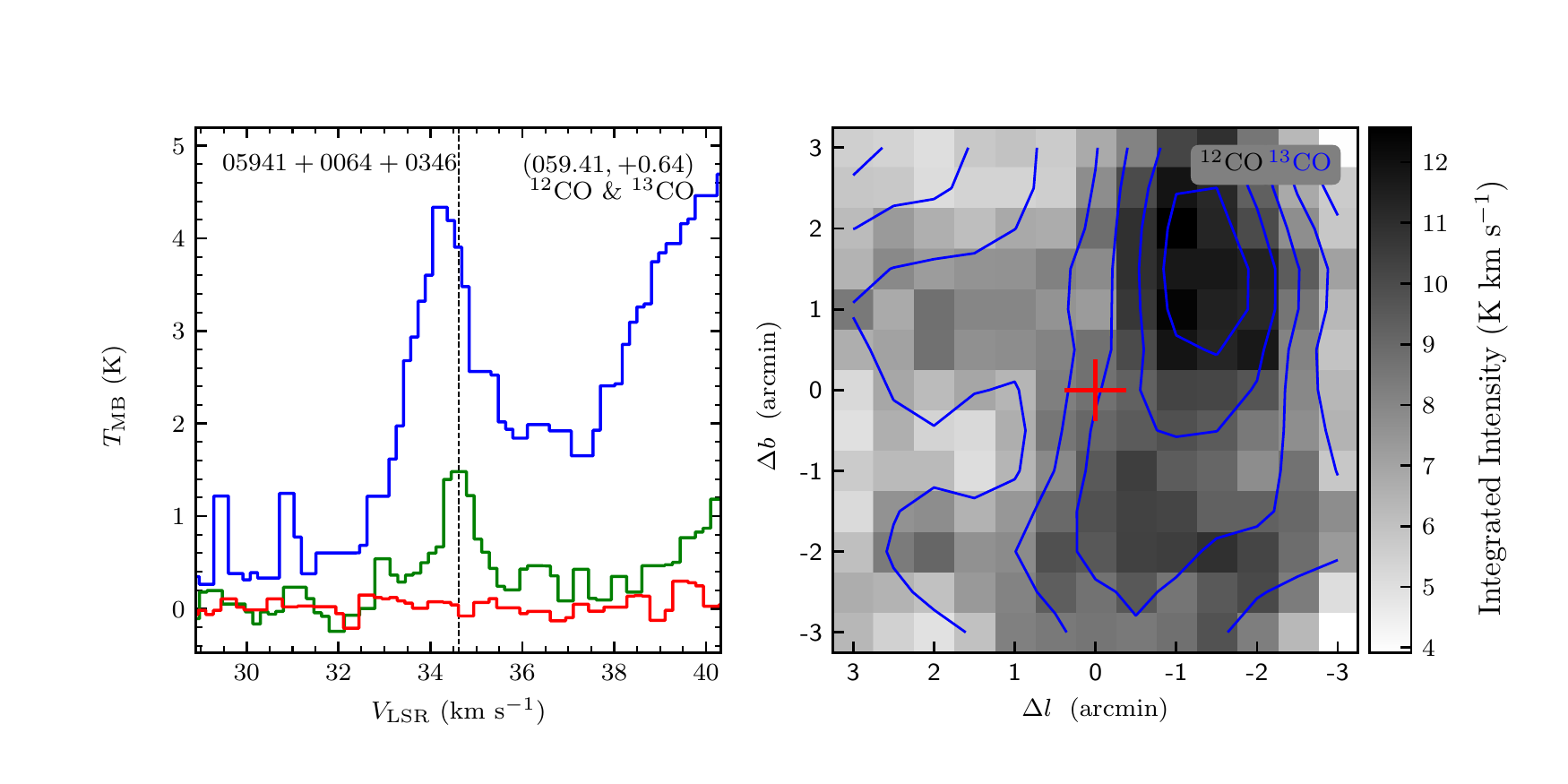}
\vspace{-0.5cm}

\includegraphics[width=9.0cm,angle=0]{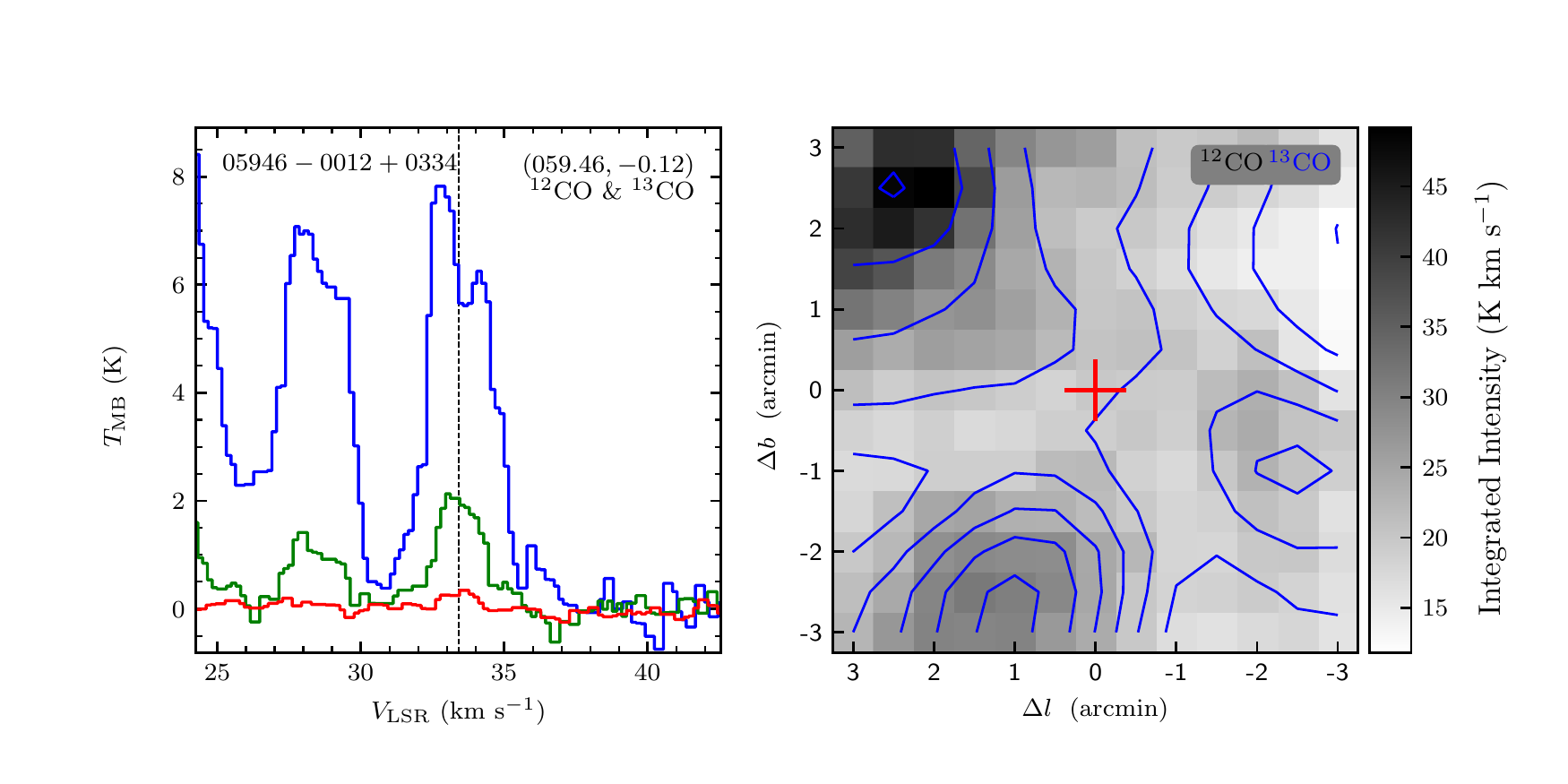}
\includegraphics[width=9.0cm,angle=0]{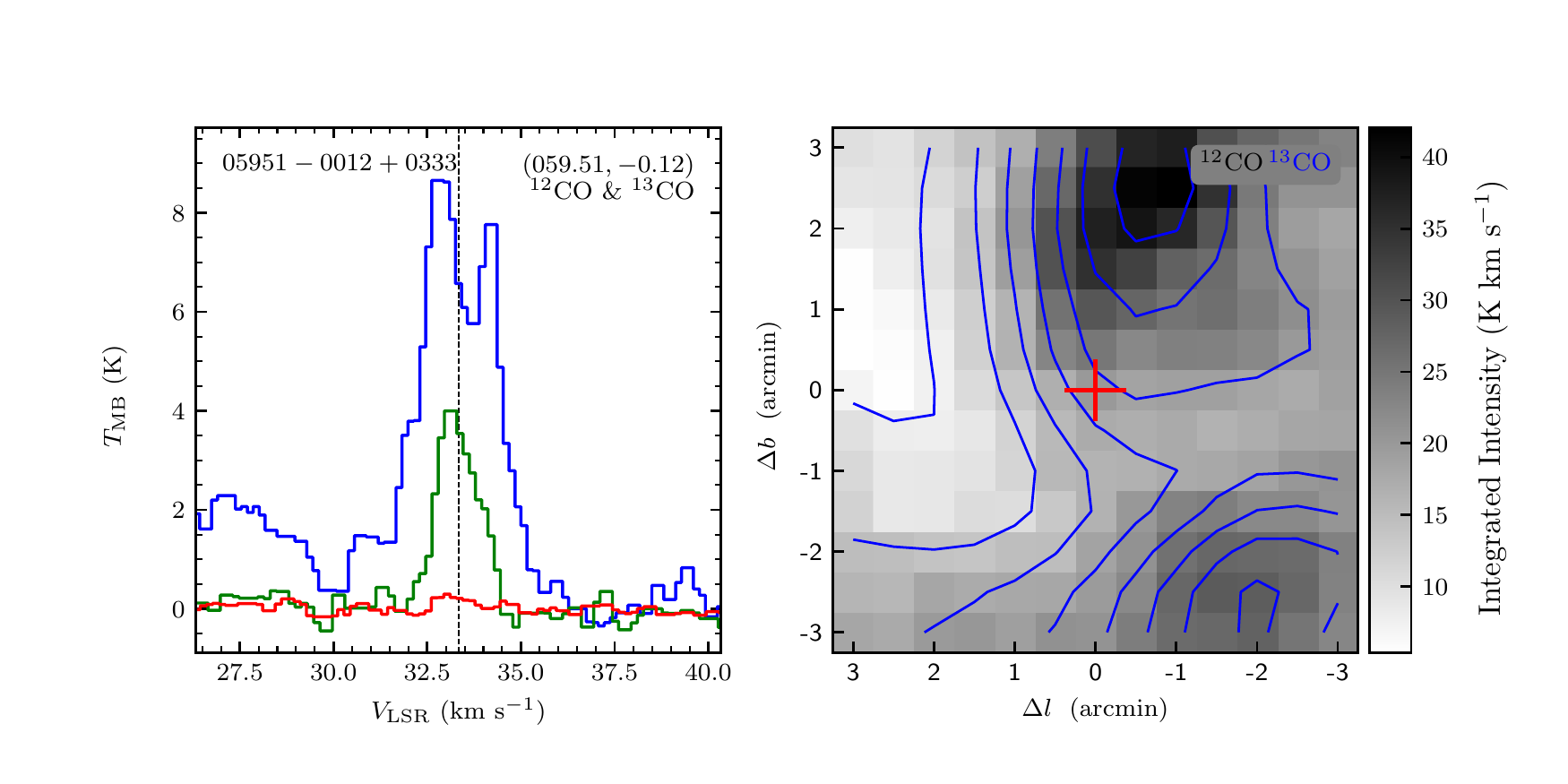}
\end{figure}
\clearpage

\begin{figure}
\includegraphics[width=9.0cm,angle=0]{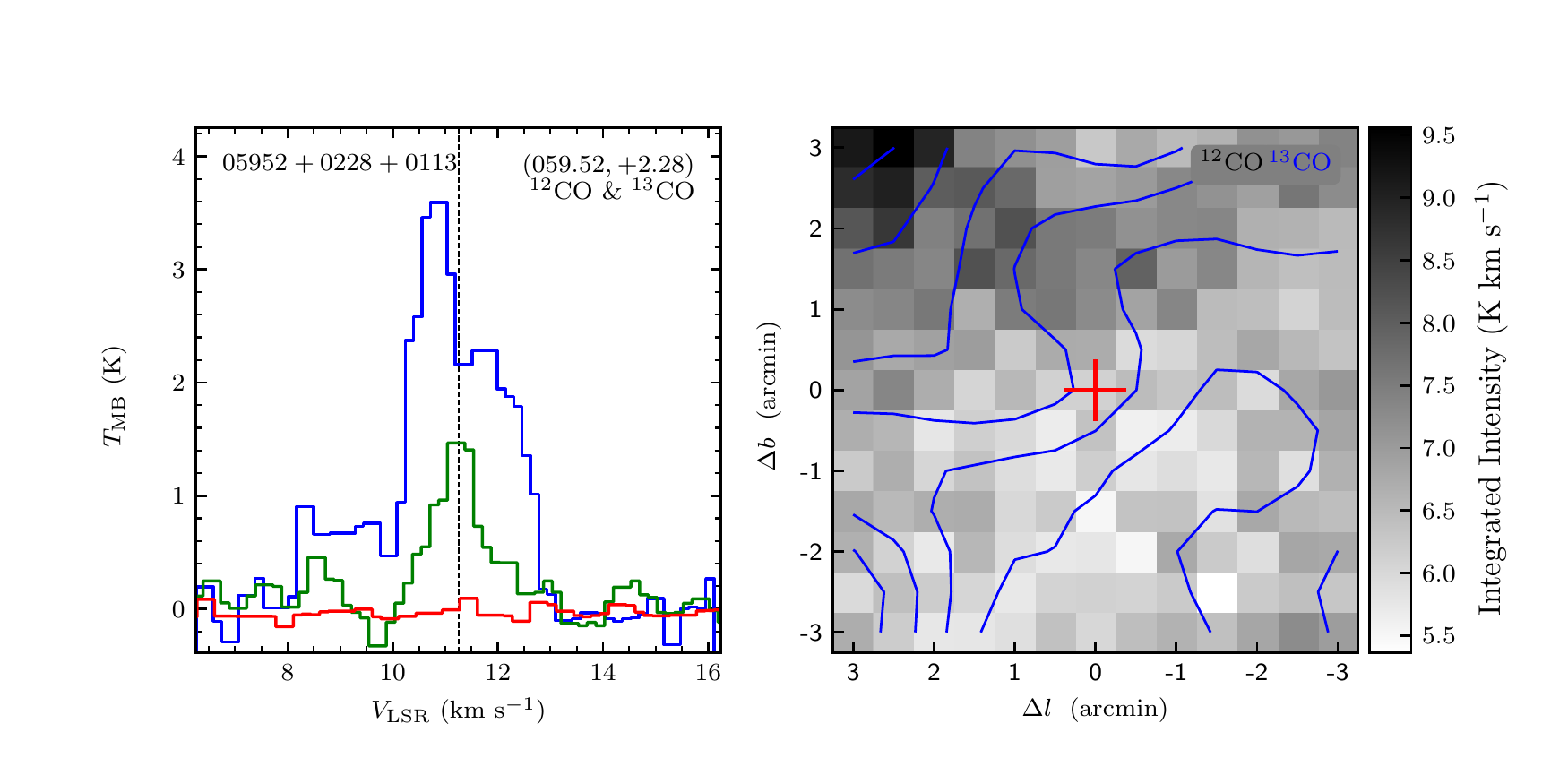}
\includegraphics[width=9.0cm,angle=0]{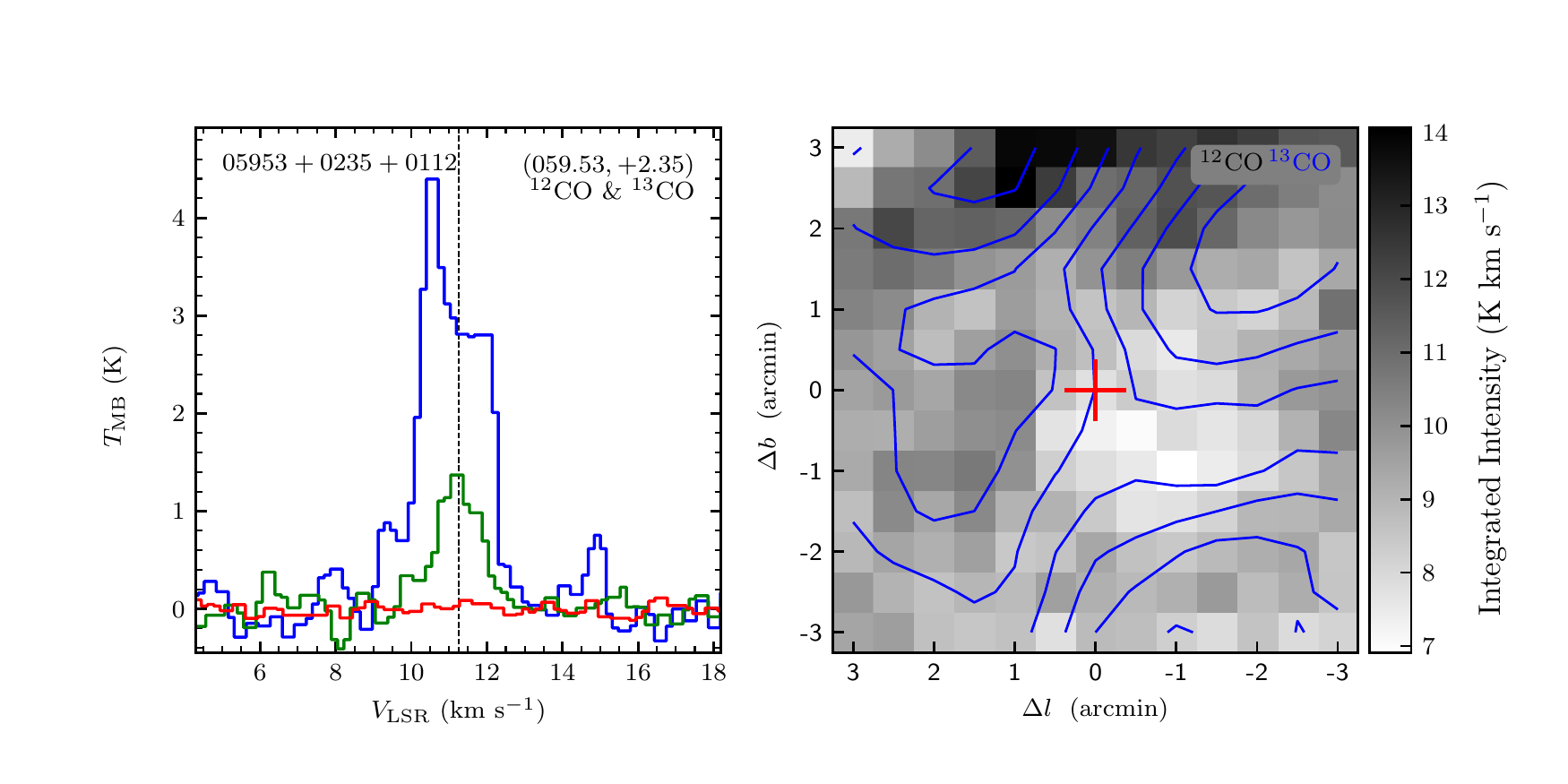}
\vspace{-0.5cm}

\includegraphics[width=9.0cm,angle=0]{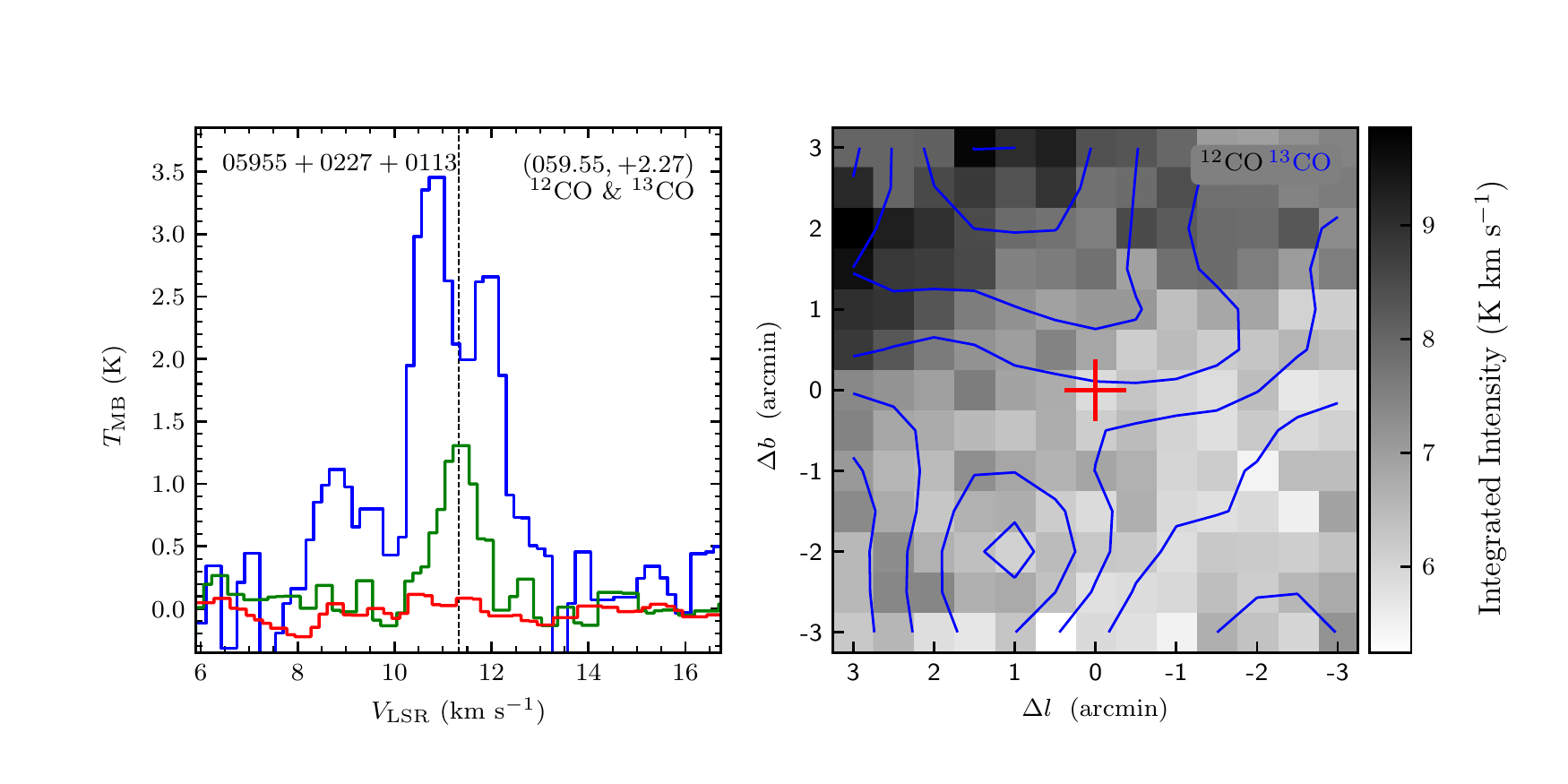}
\includegraphics[width=9.0cm,angle=0]{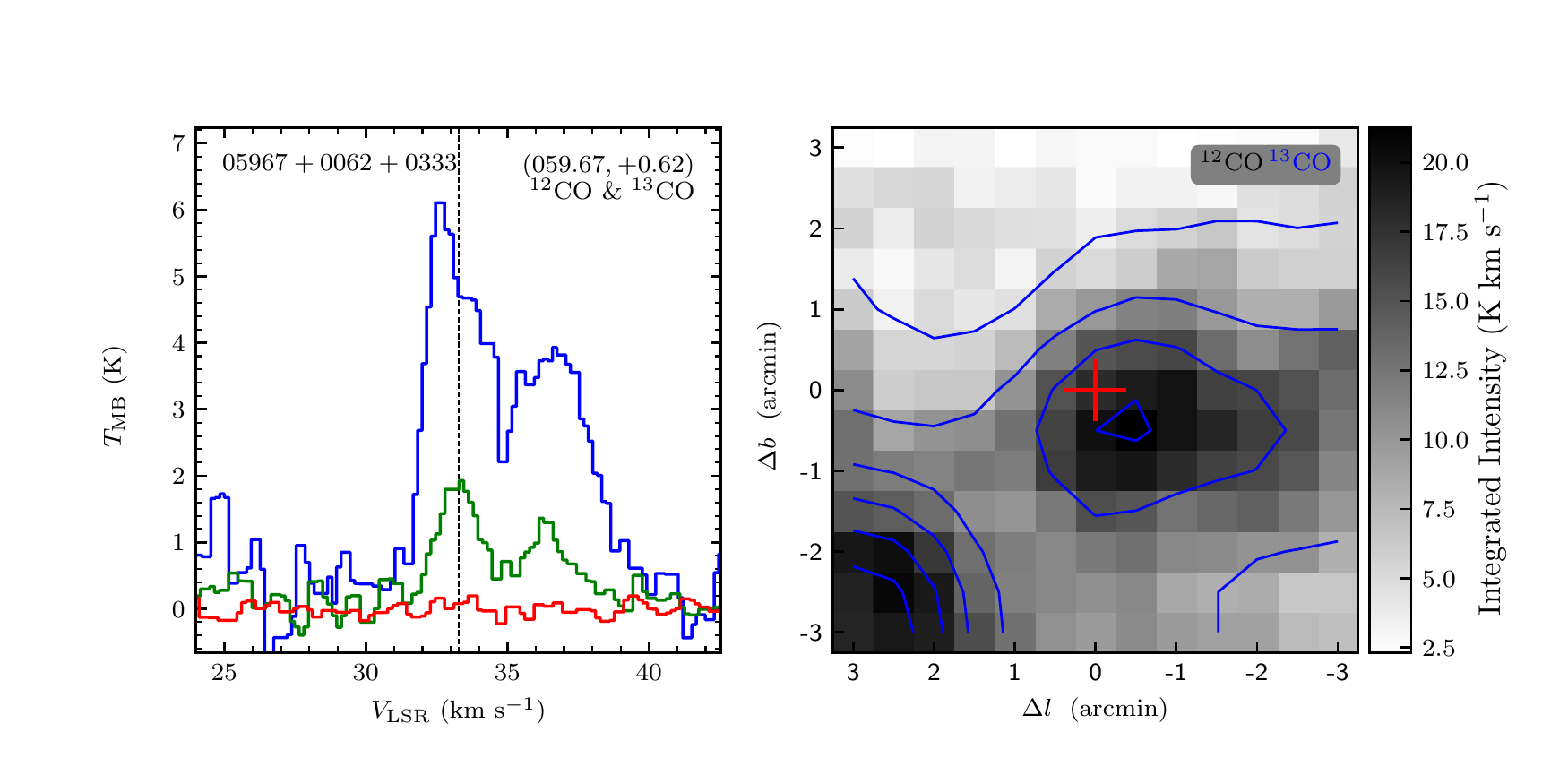}
\vspace{-0.5cm}

\includegraphics[width=9.0cm,angle=0]{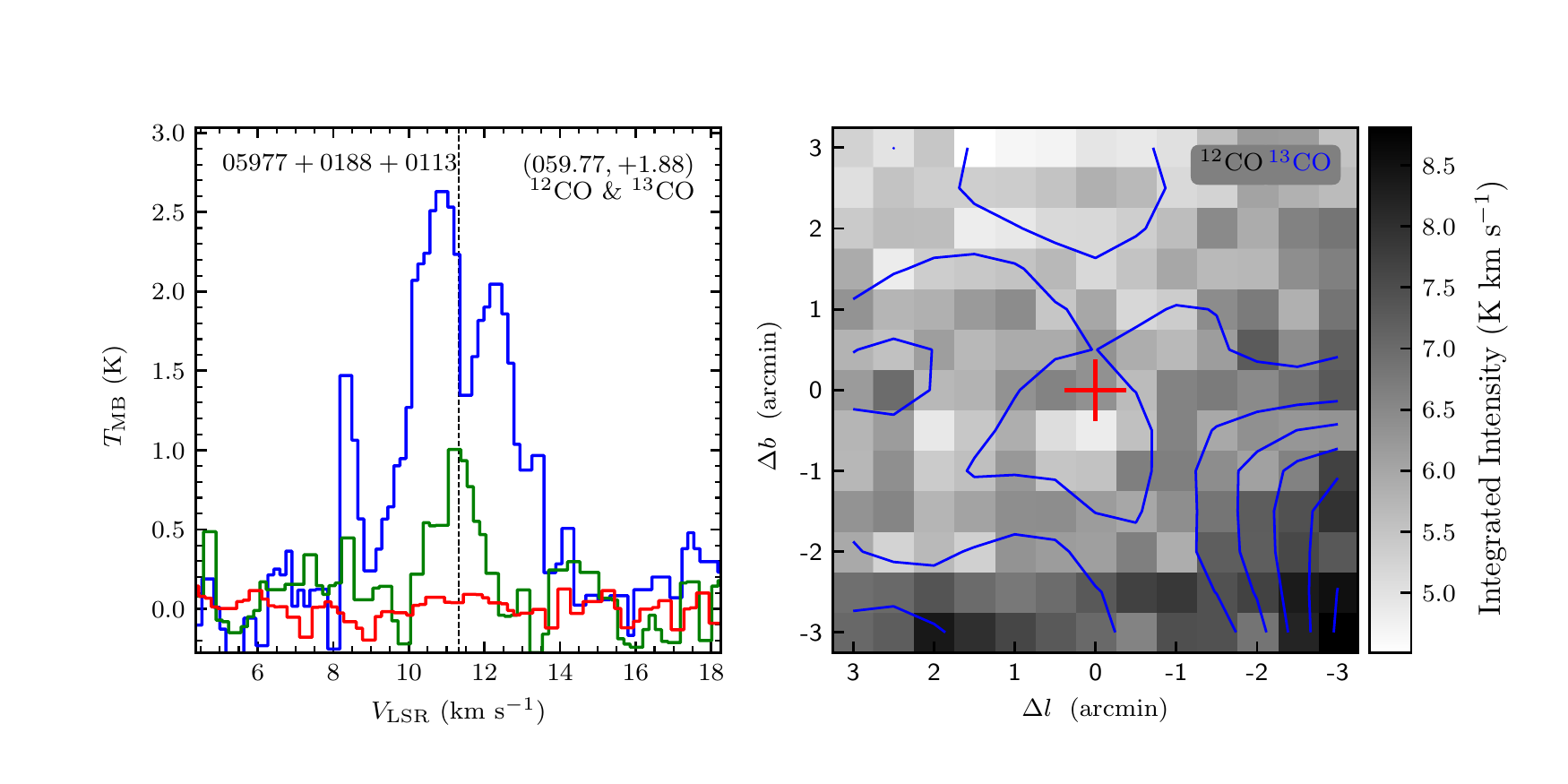}
\includegraphics[width=9.0cm,angle=0]{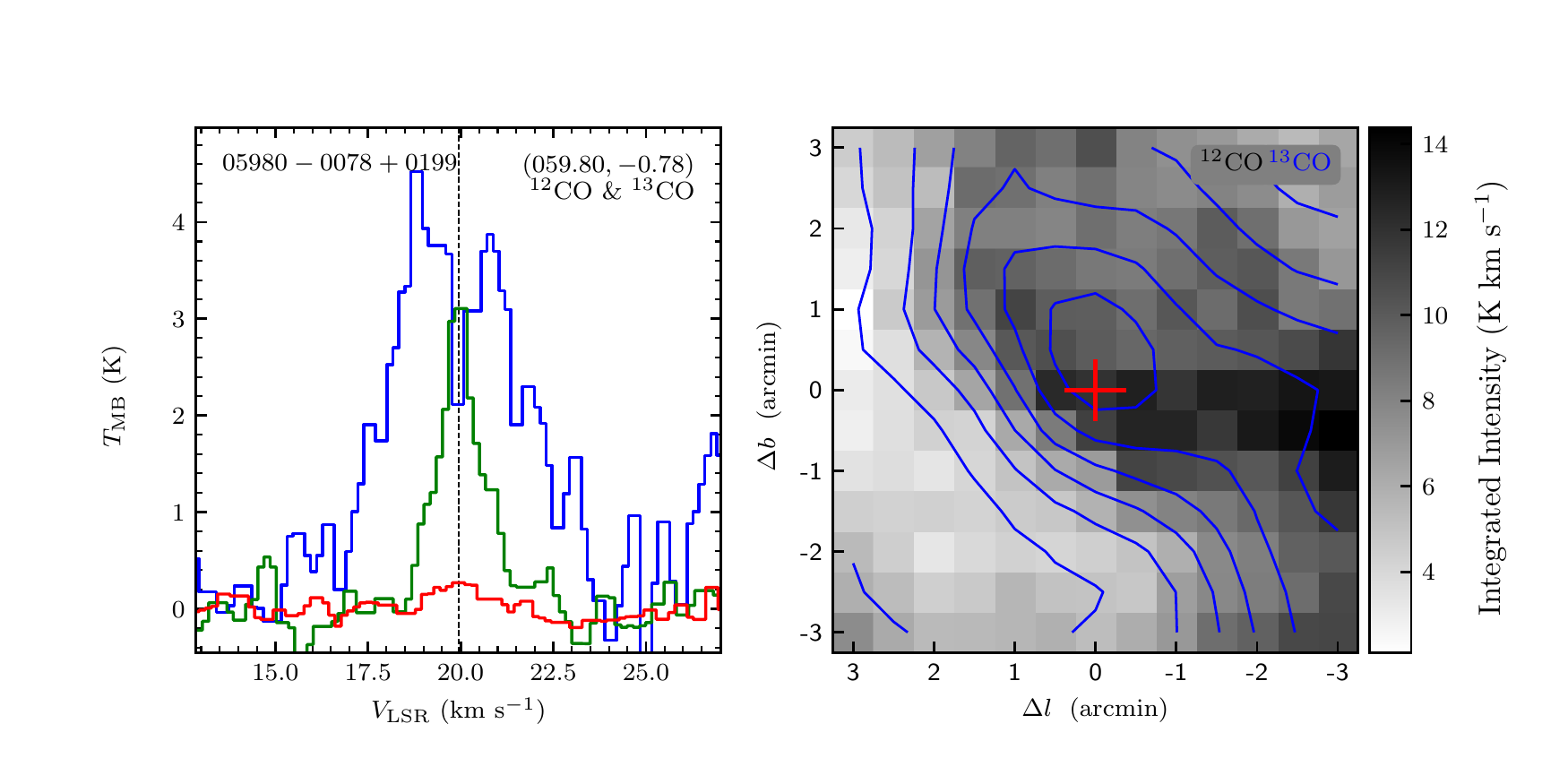}
\vspace{-0.5cm}

\includegraphics[width=9.0cm,angle=0]{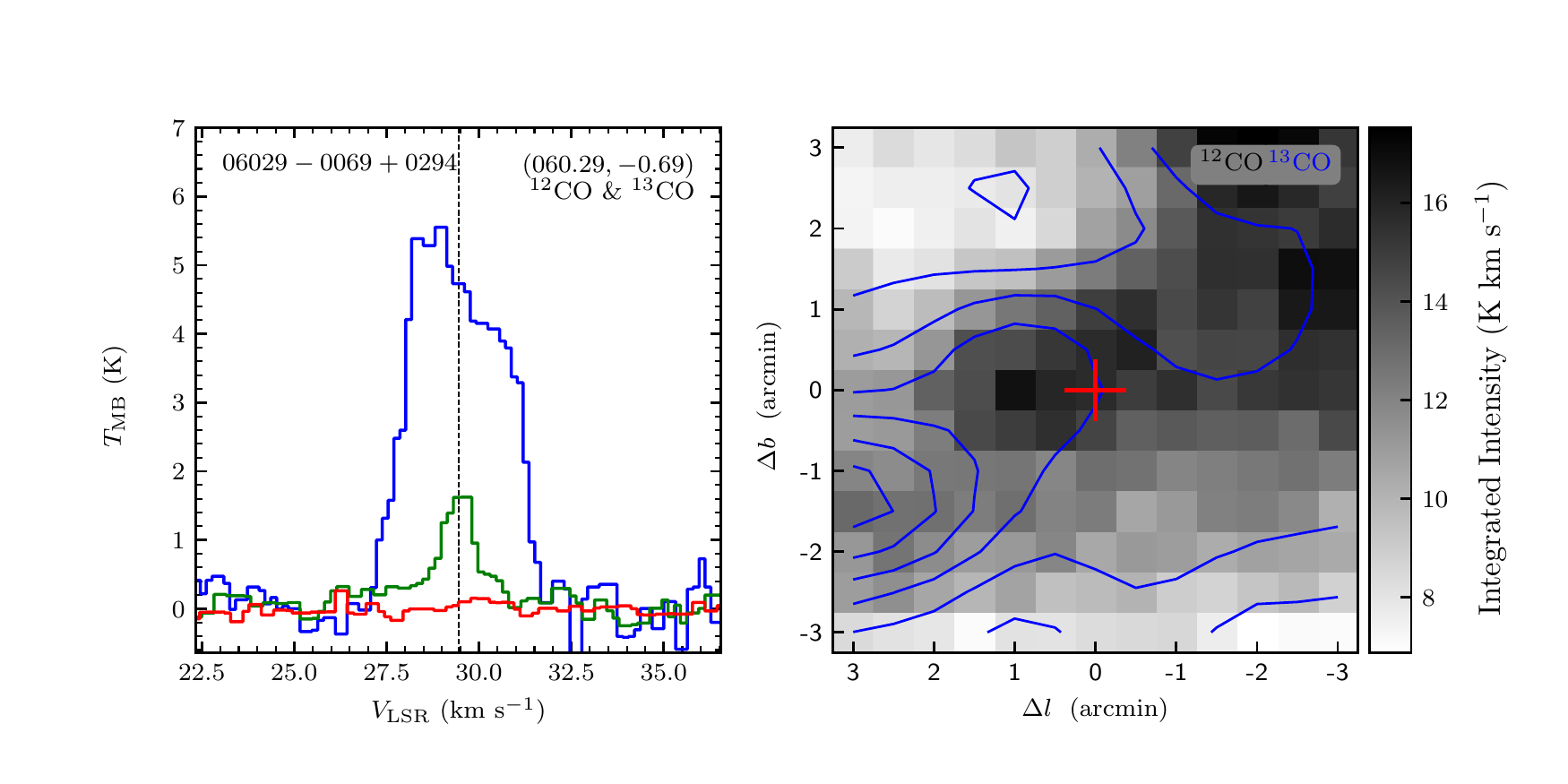}
\includegraphics[width=9.0cm,angle=0]{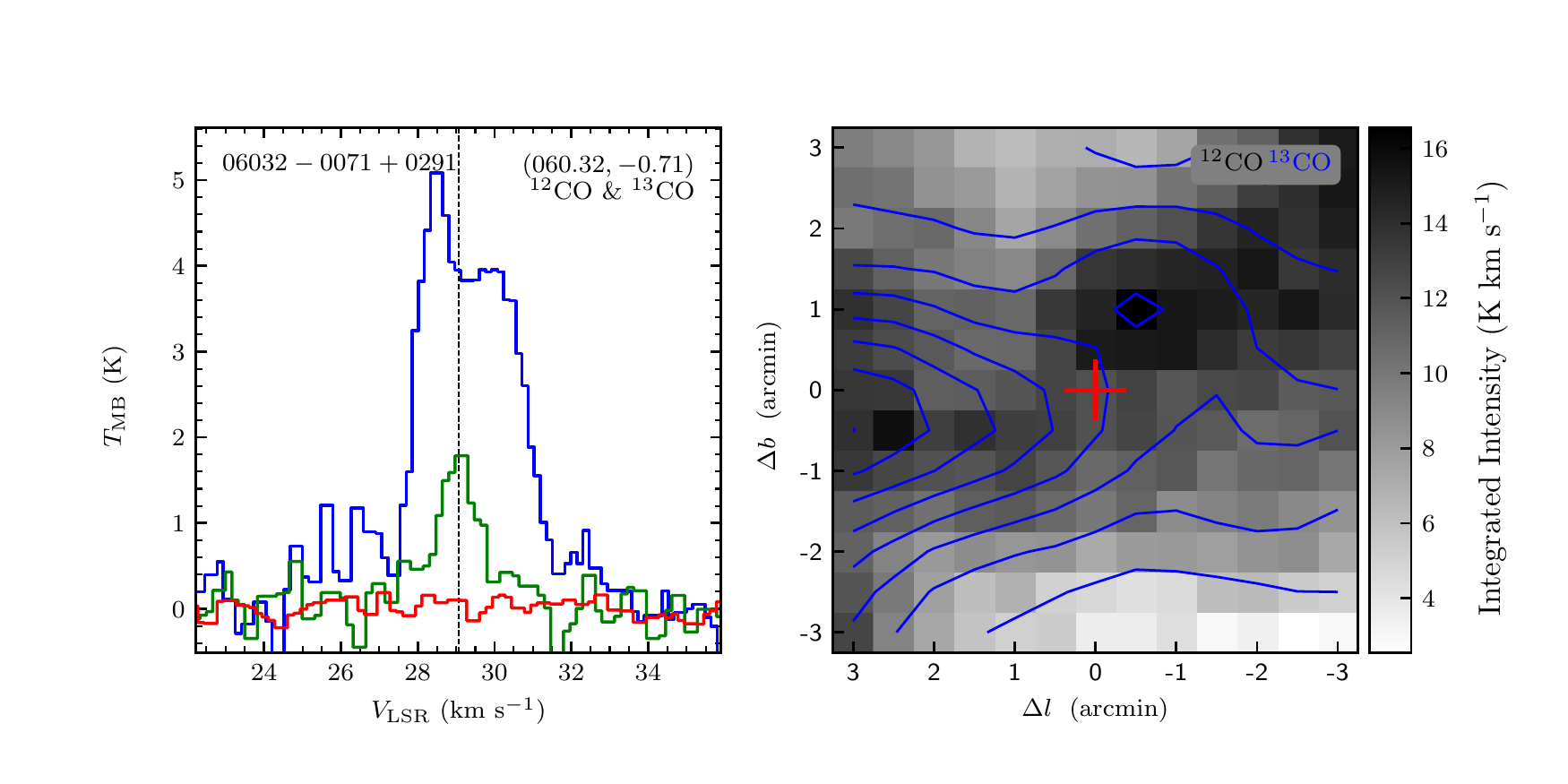}
\vspace{-0.5cm}

\includegraphics[width=9.0cm,angle=0]{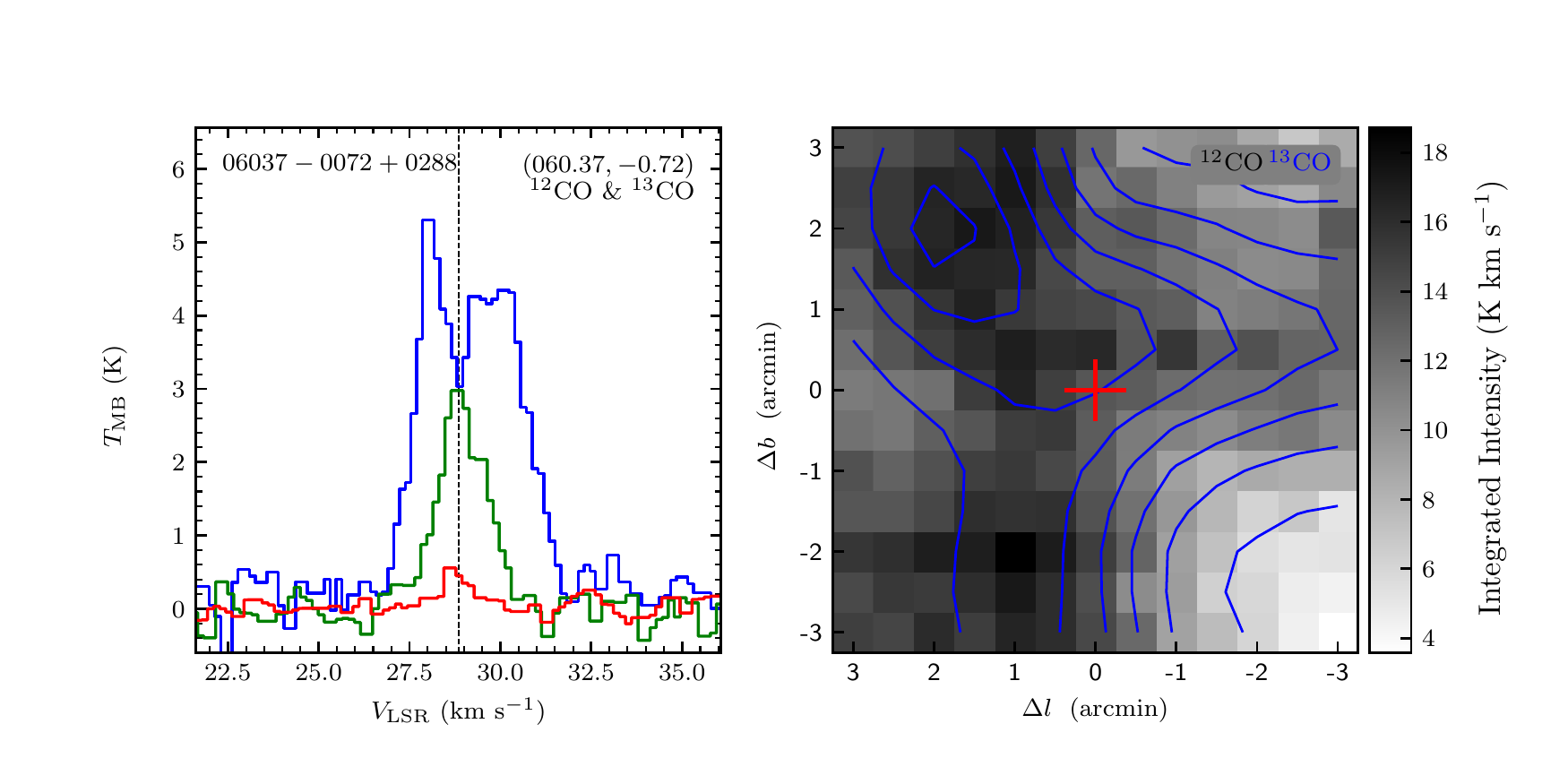}
\includegraphics[width=9.0cm,angle=0]{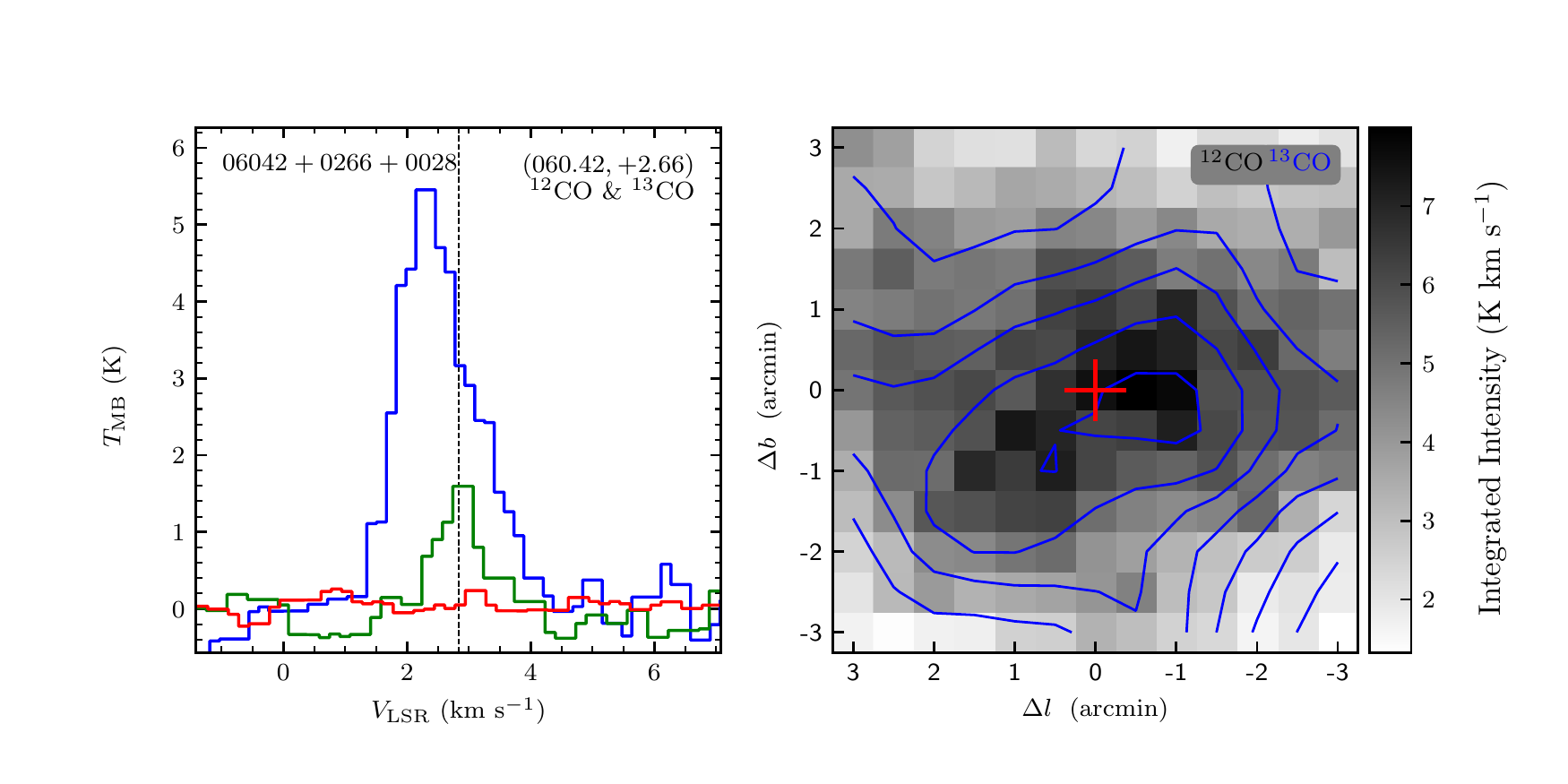}
\end{figure}
\clearpage

\begin{figure}
\includegraphics[width=9.0cm,angle=0]{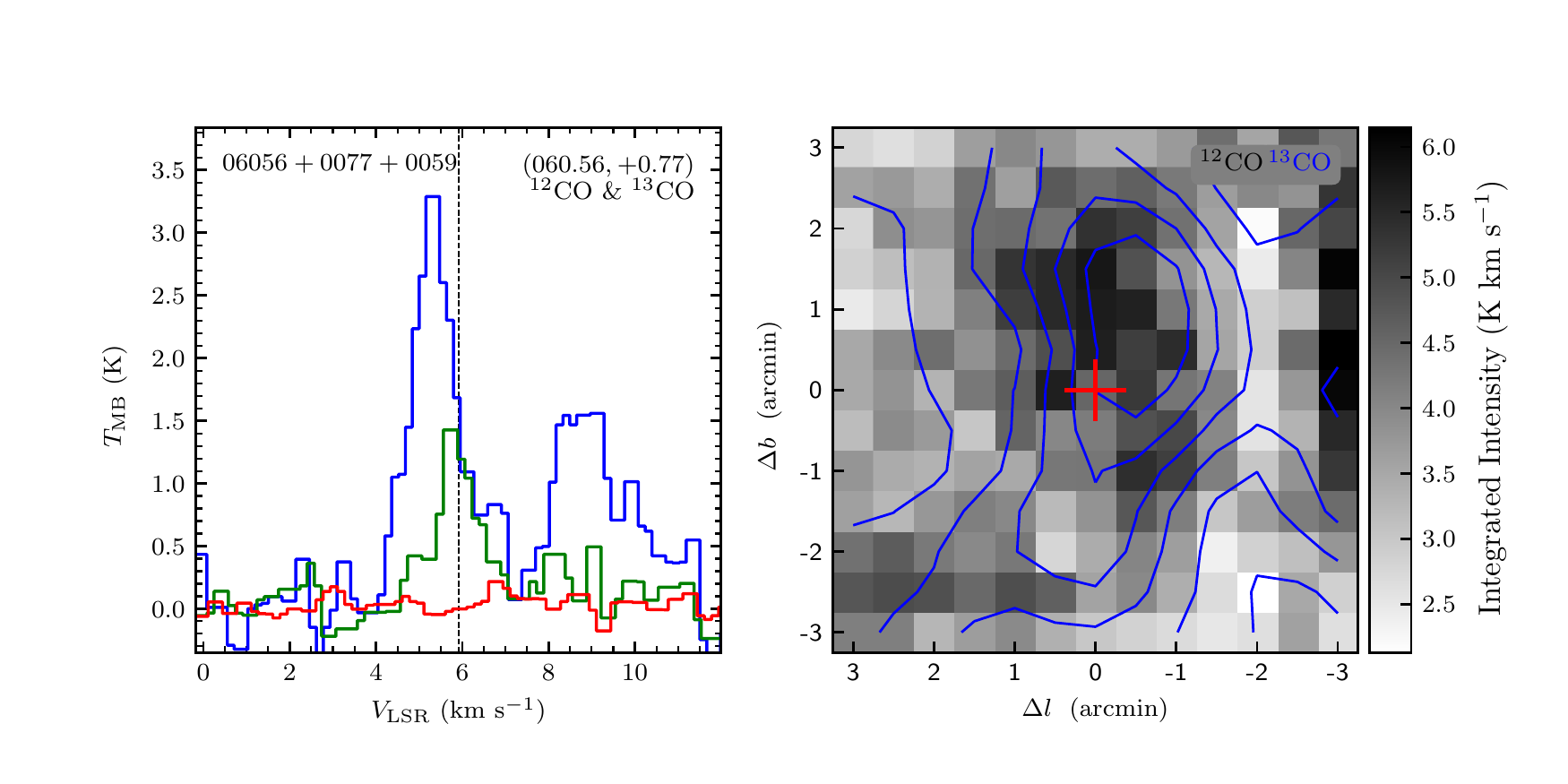}
\includegraphics[width=9.0cm,angle=0]{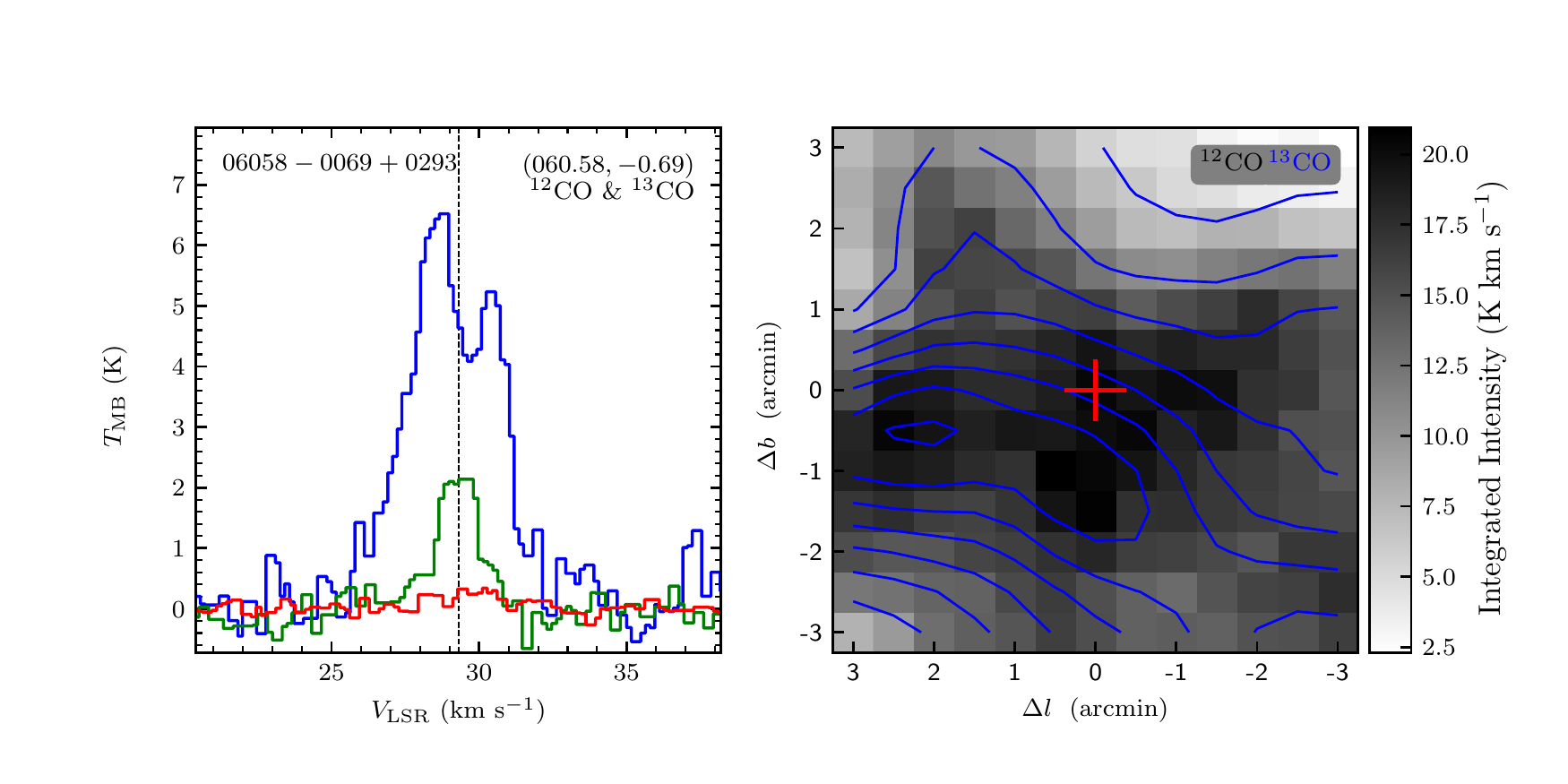}
\vspace{-0.5cm}

\includegraphics[width=9.0cm,angle=0]{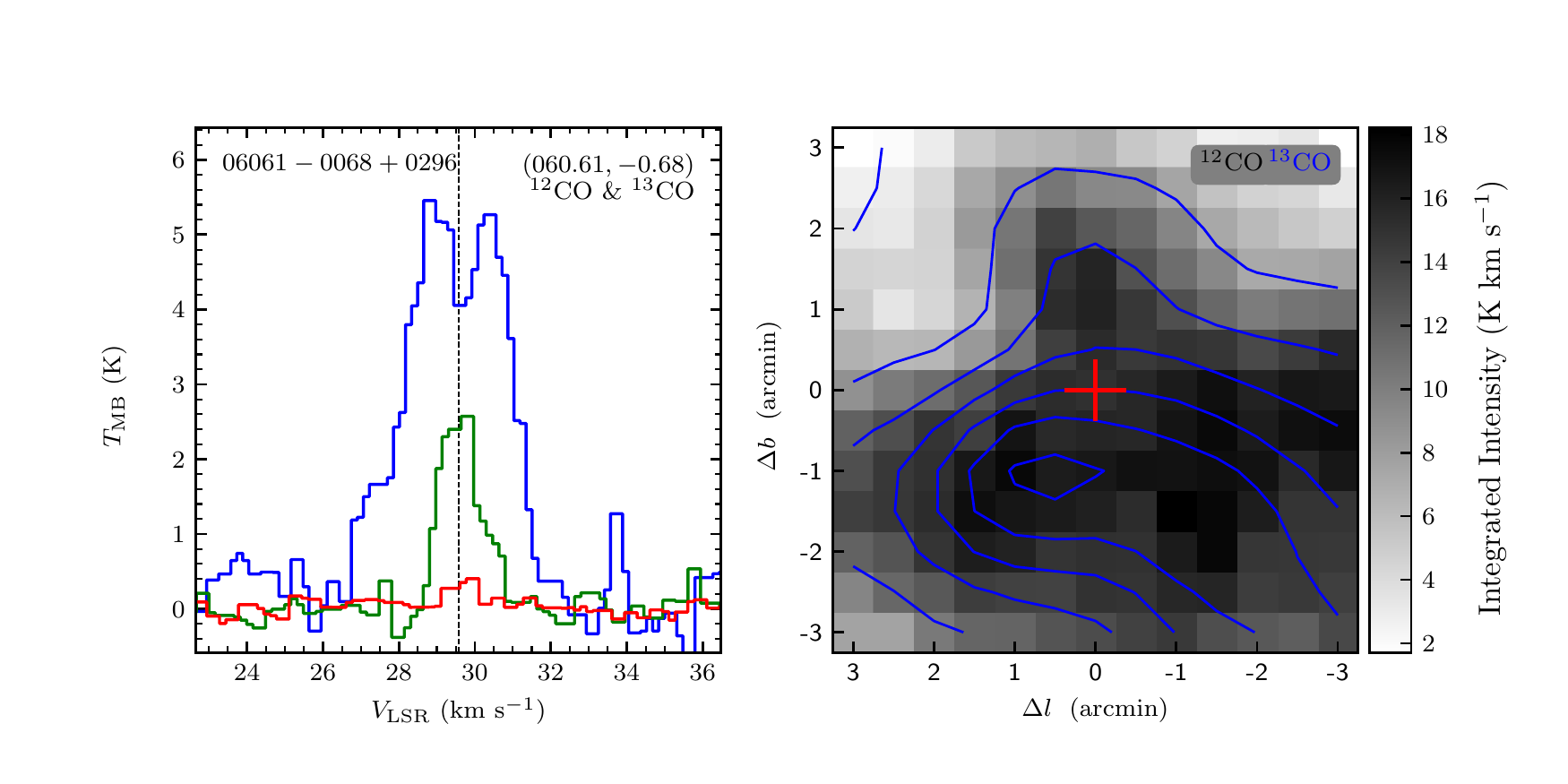}
\includegraphics[width=9.0cm,angle=0]{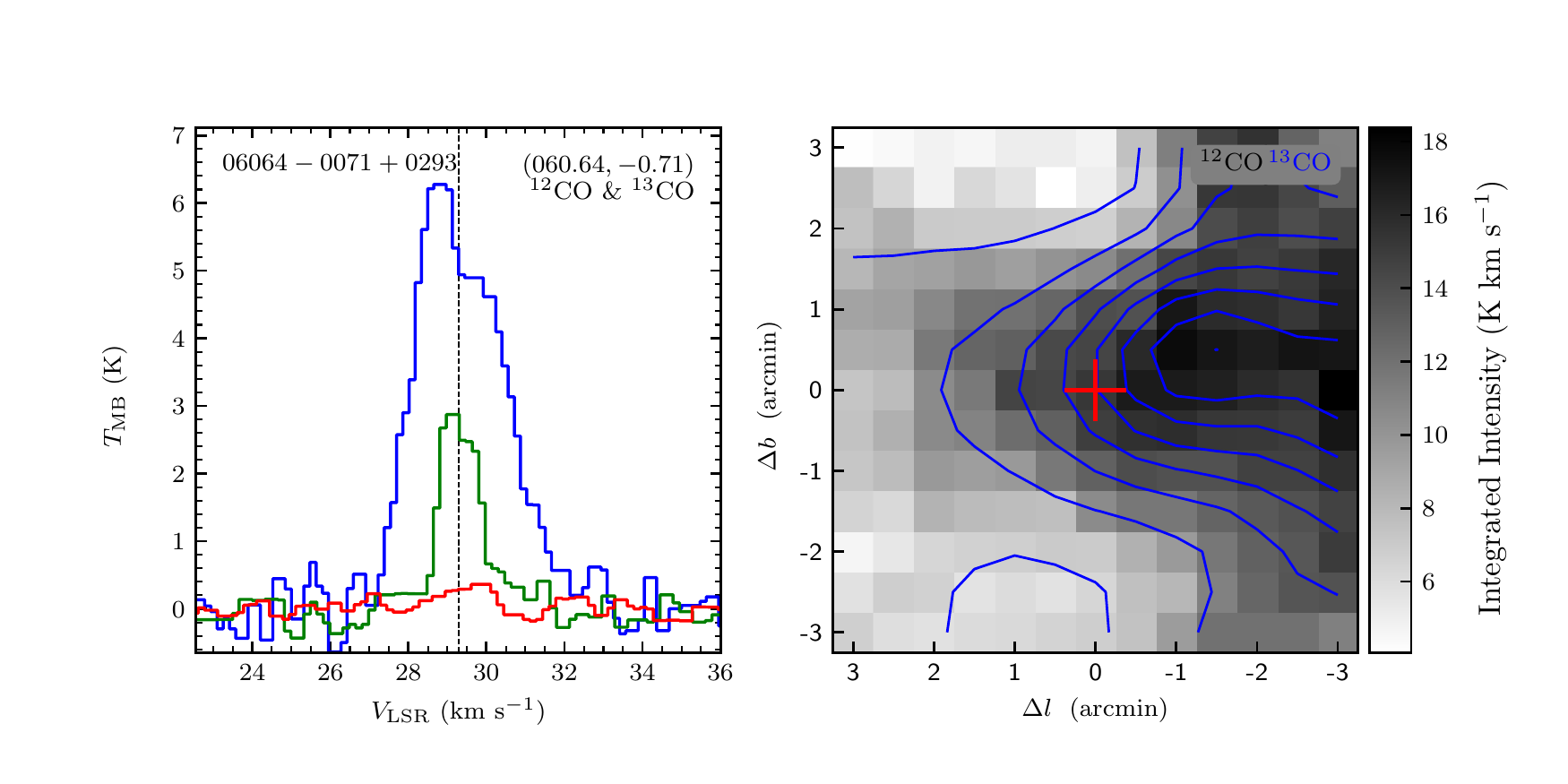}
\vspace{-0.5cm}

\includegraphics[width=9.0cm,angle=0]{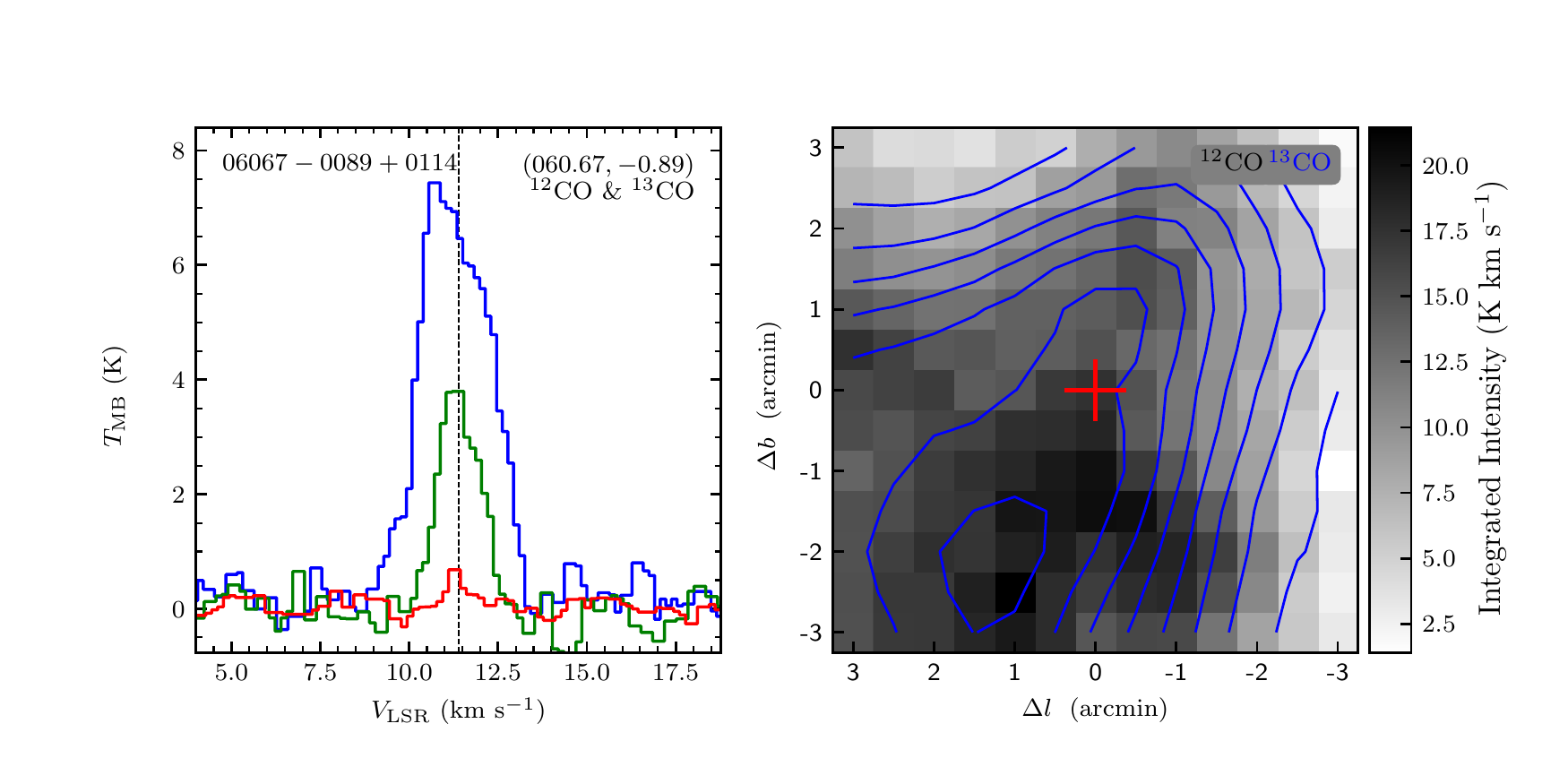}
\includegraphics[width=9.0cm,angle=0]{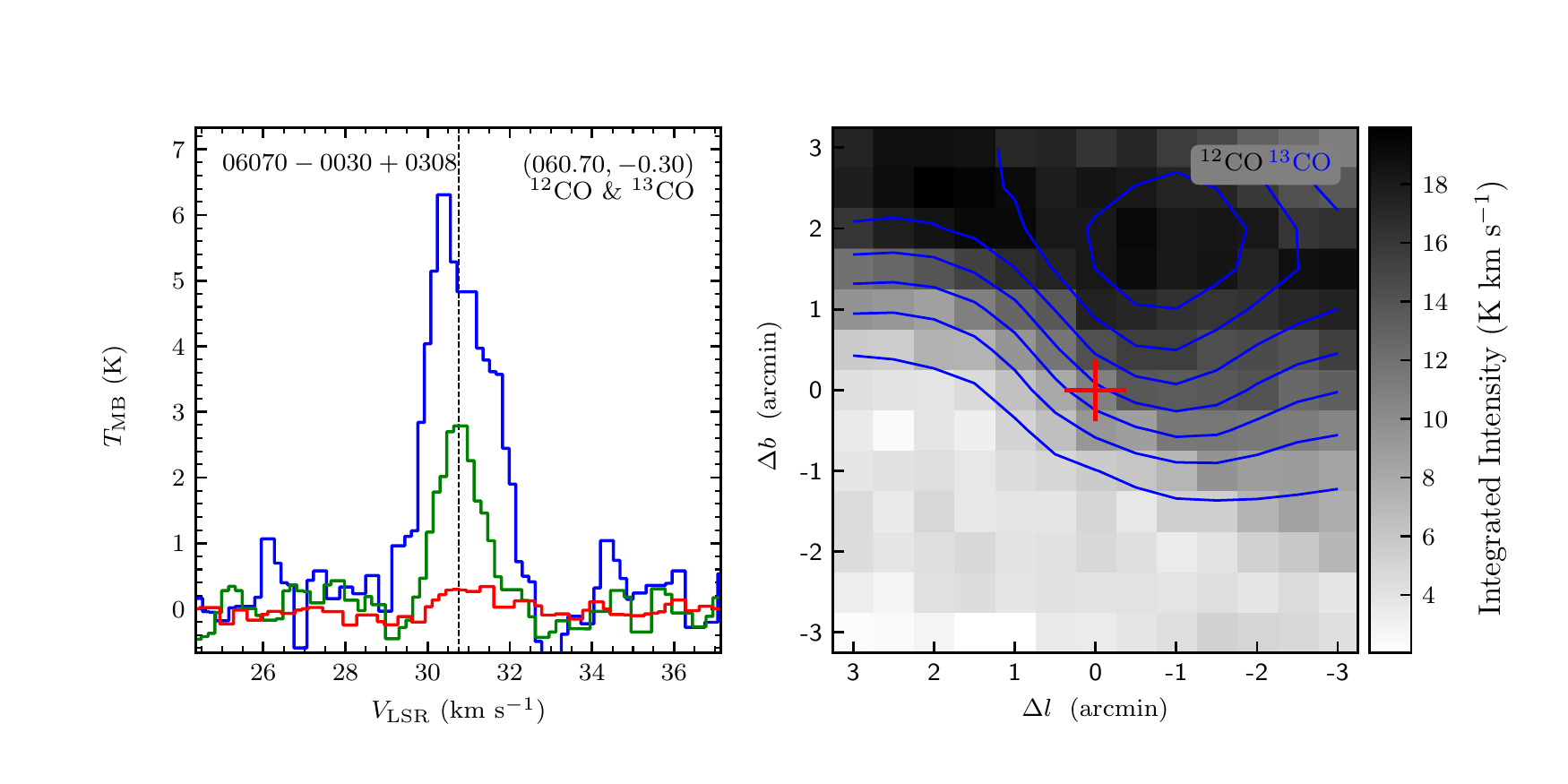}
\vspace{-0.5cm}

\includegraphics[width=9.0cm,angle=0]{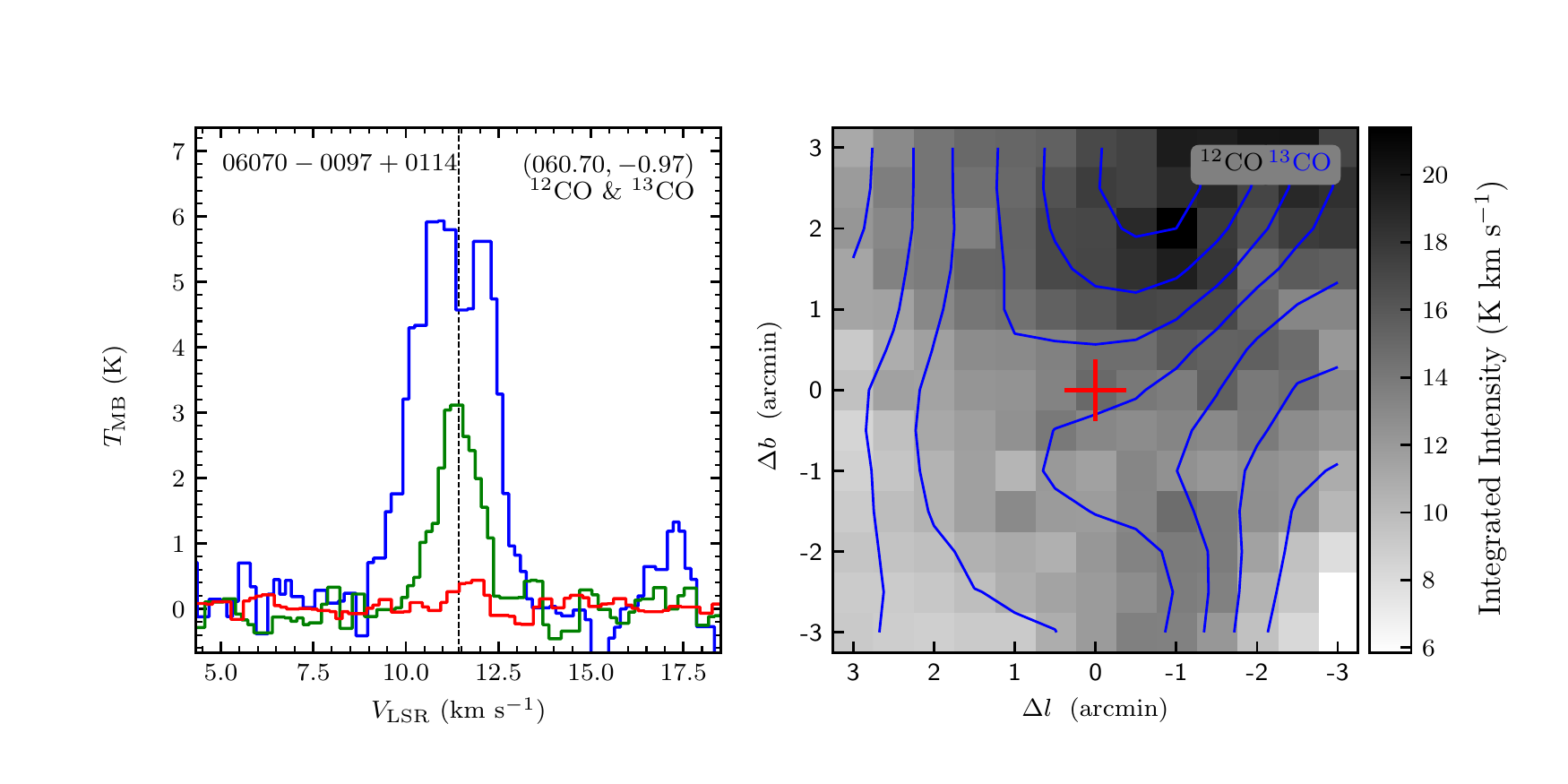}
\includegraphics[width=9.0cm,angle=0]{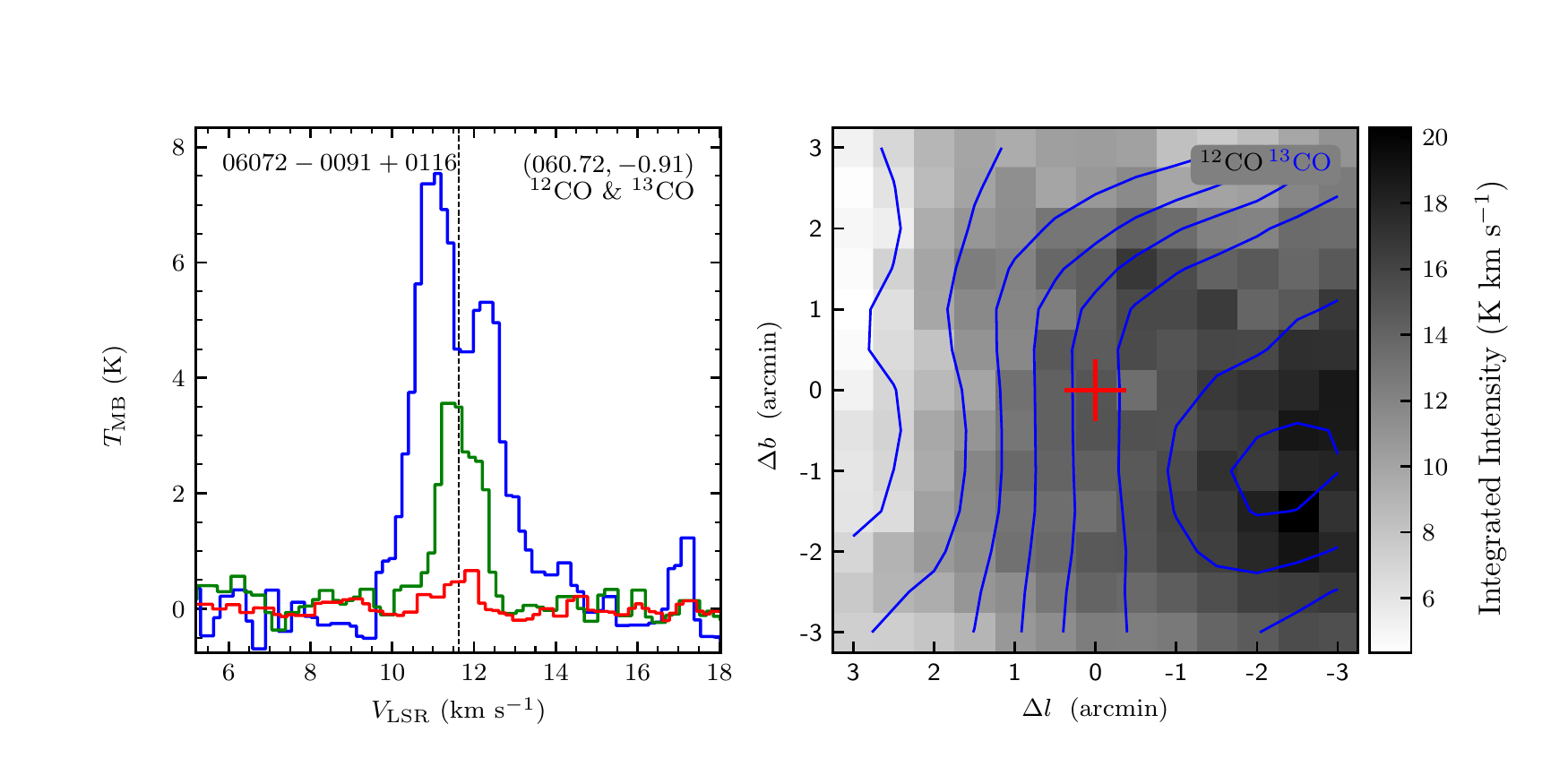}
\vspace{-0.5cm}

\includegraphics[width=9.0cm,angle=0]{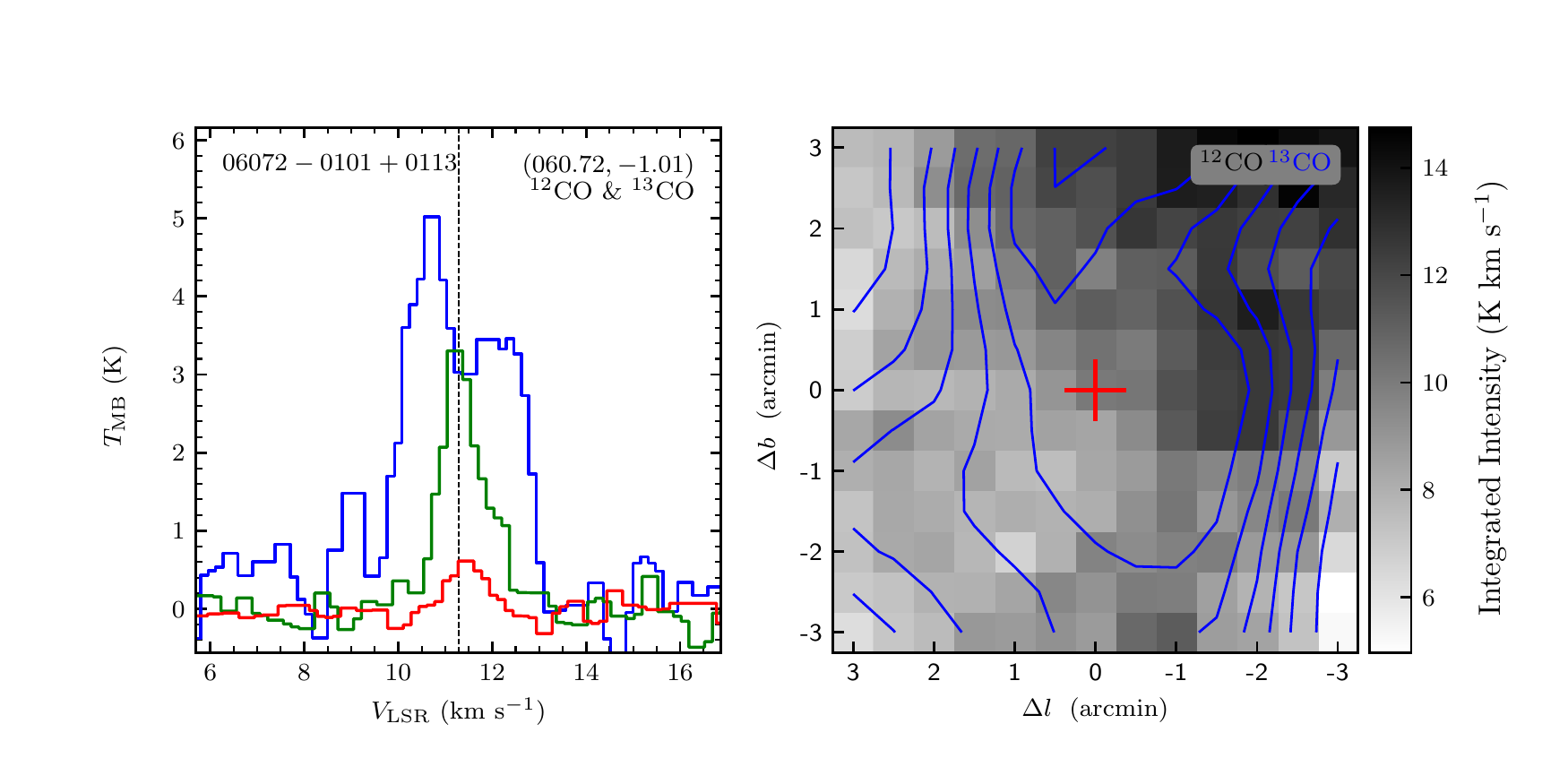}
\includegraphics[width=9.0cm,angle=0]{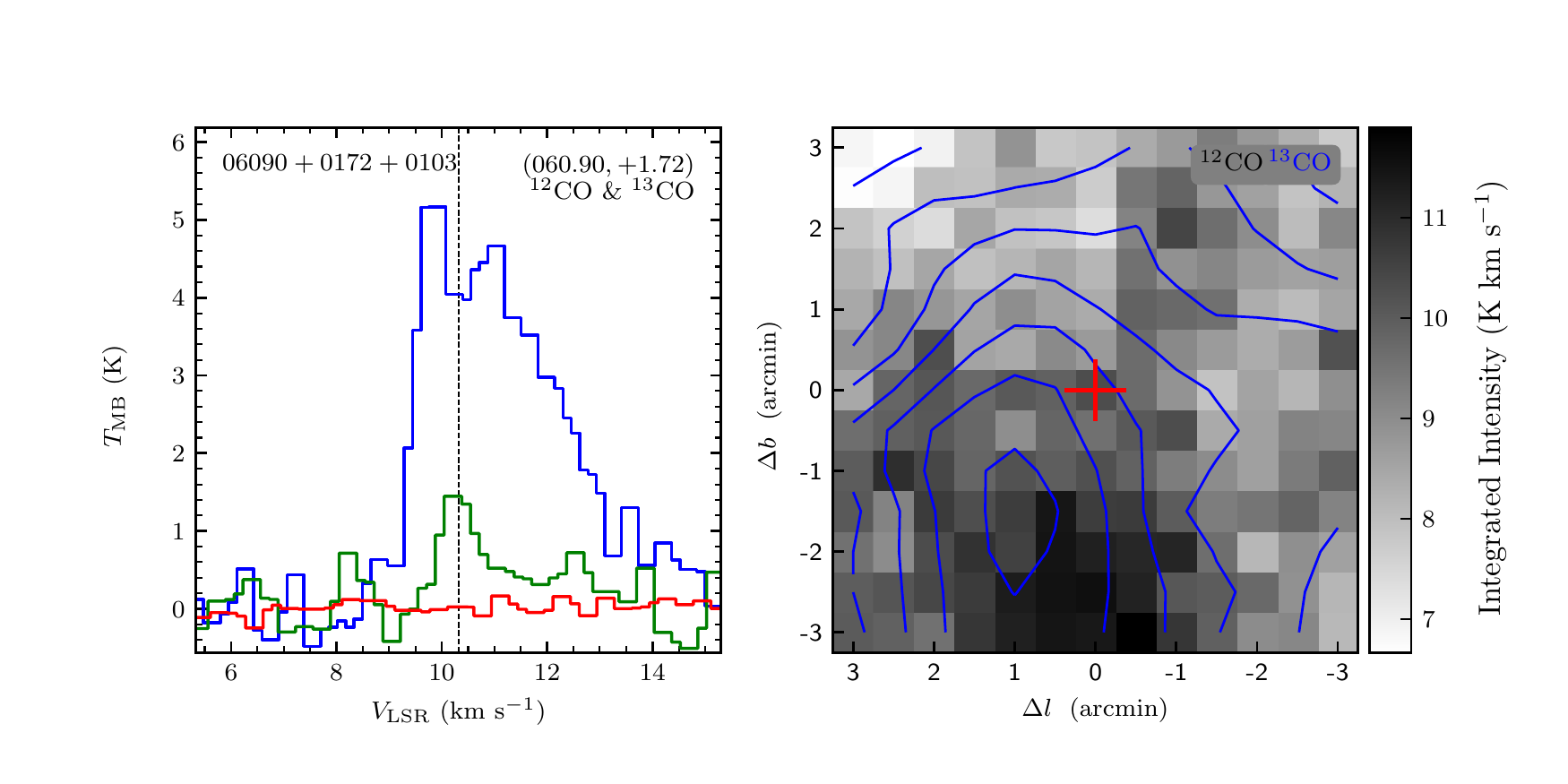}
\end{figure}
\clearpage

\begin{figure}
\includegraphics[width=9.0cm,angle=0]{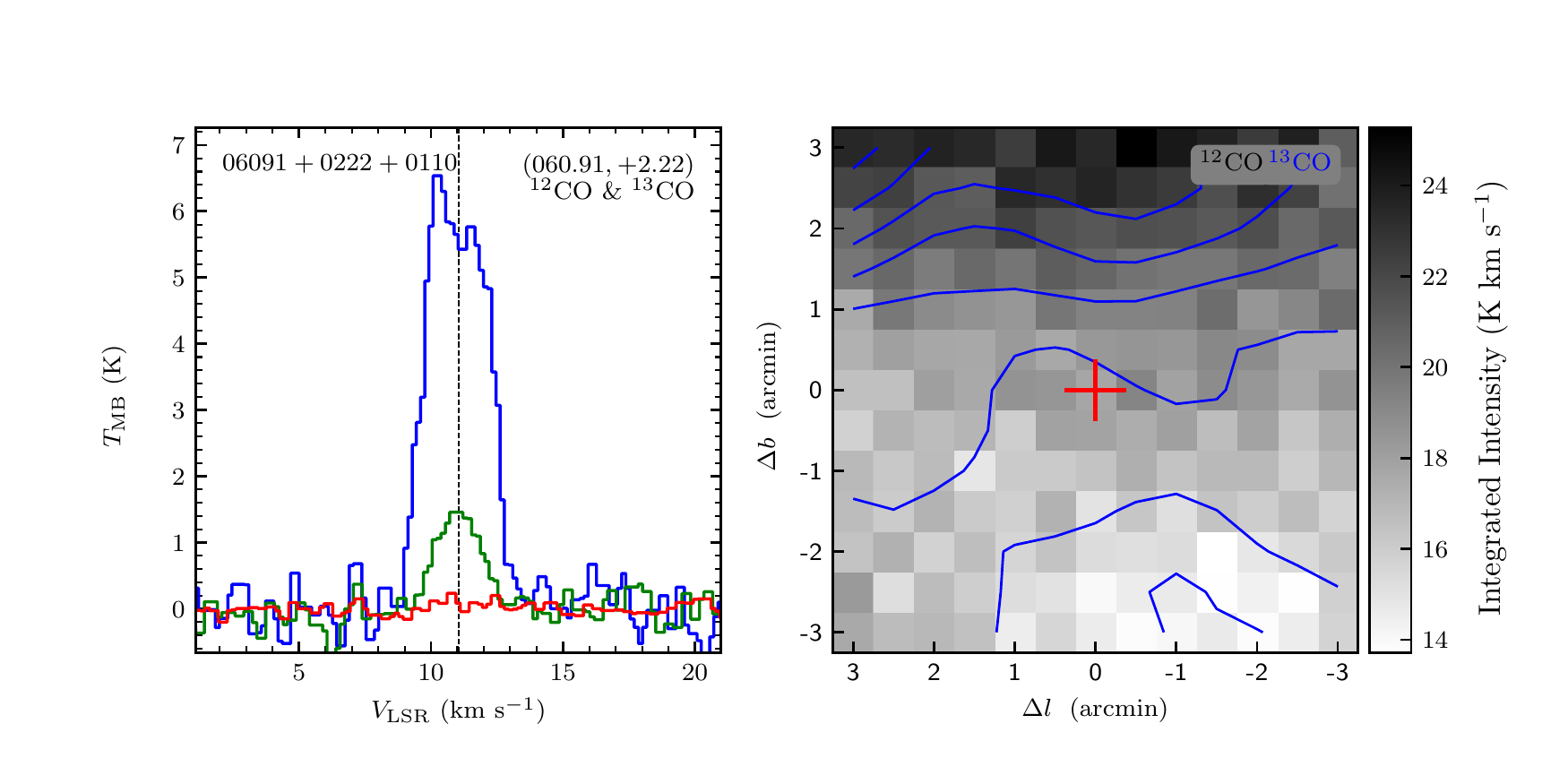}
\includegraphics[width=9.0cm,angle=0]{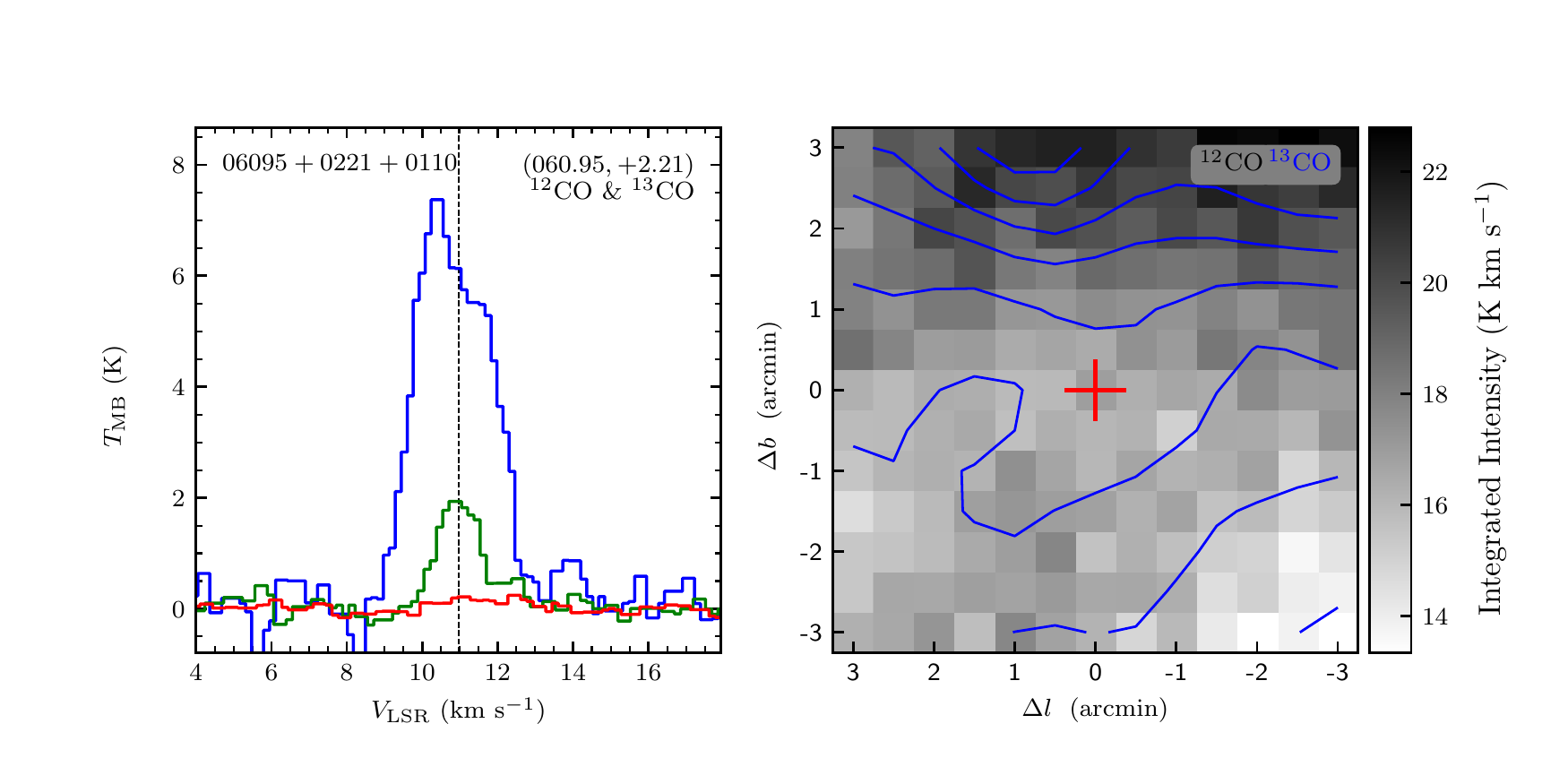}
\vspace{-0.5cm}

\includegraphics[width=9.0cm,angle=0]{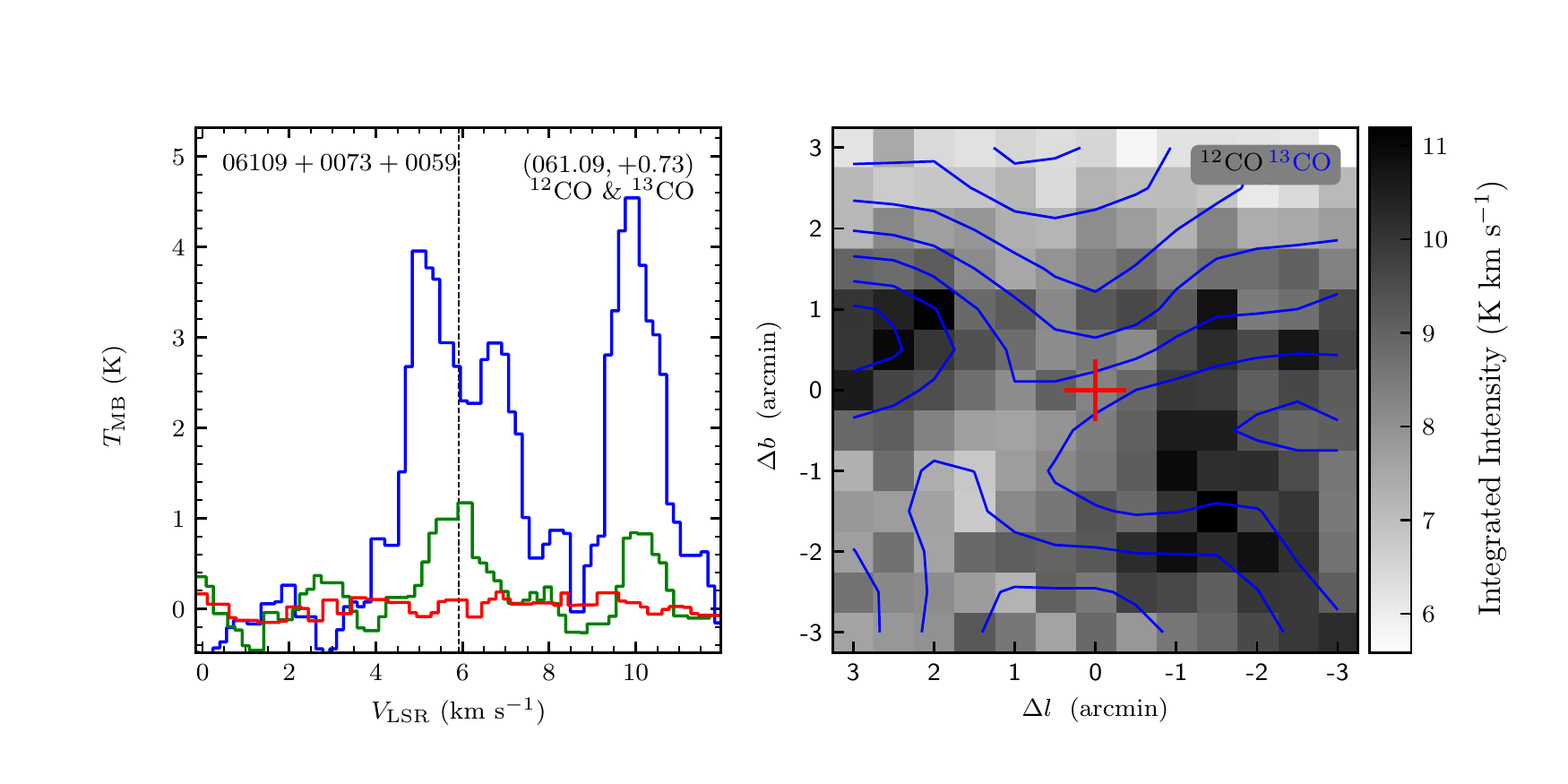}
\includegraphics[width=9.0cm,angle=0]{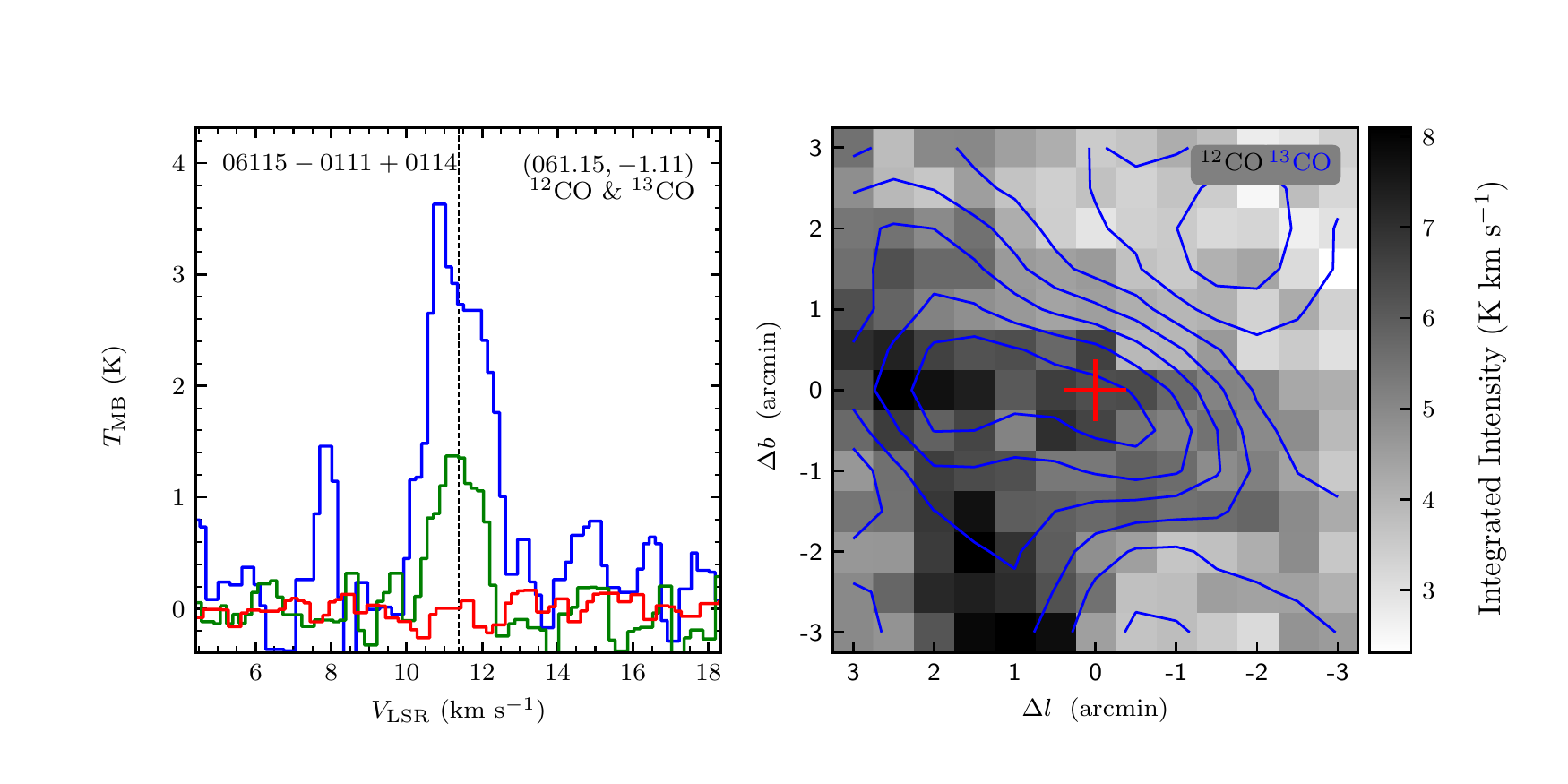}
\vspace{-0.5cm}

\includegraphics[width=9.0cm,angle=0]{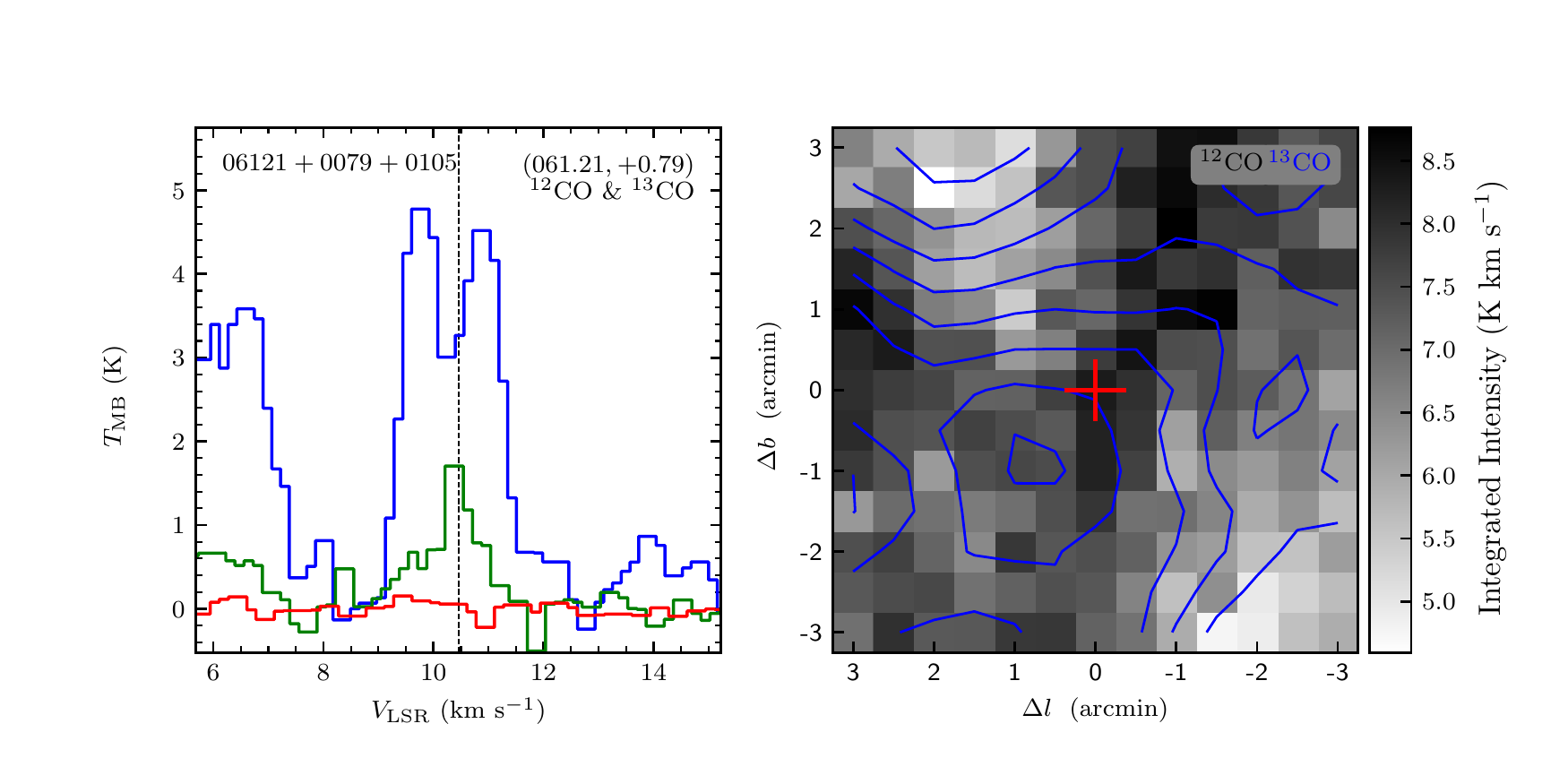}
\includegraphics[width=9.0cm,angle=0]{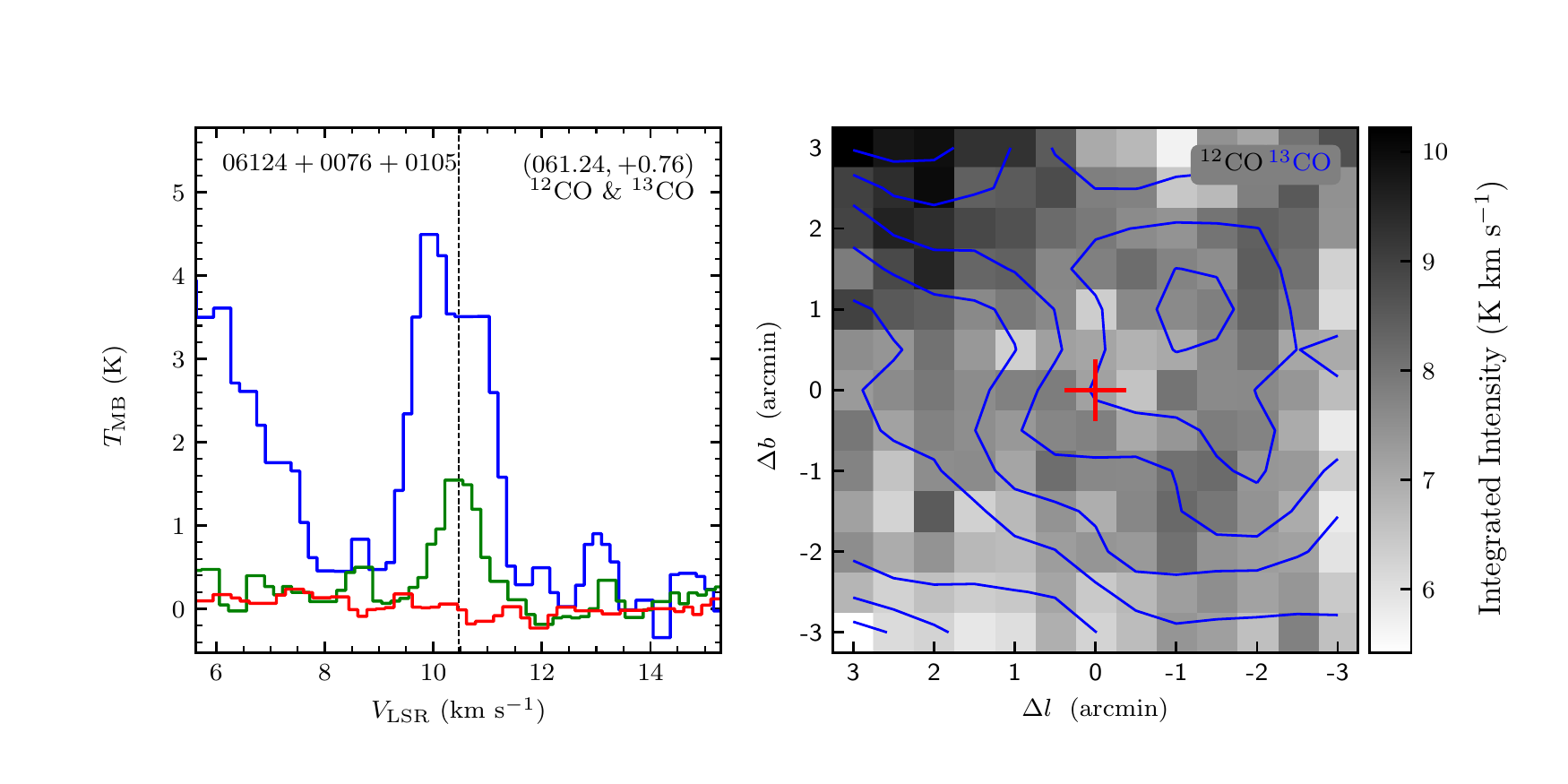}
\vspace{-0.5cm}

\includegraphics[width=9.0cm,angle=0]{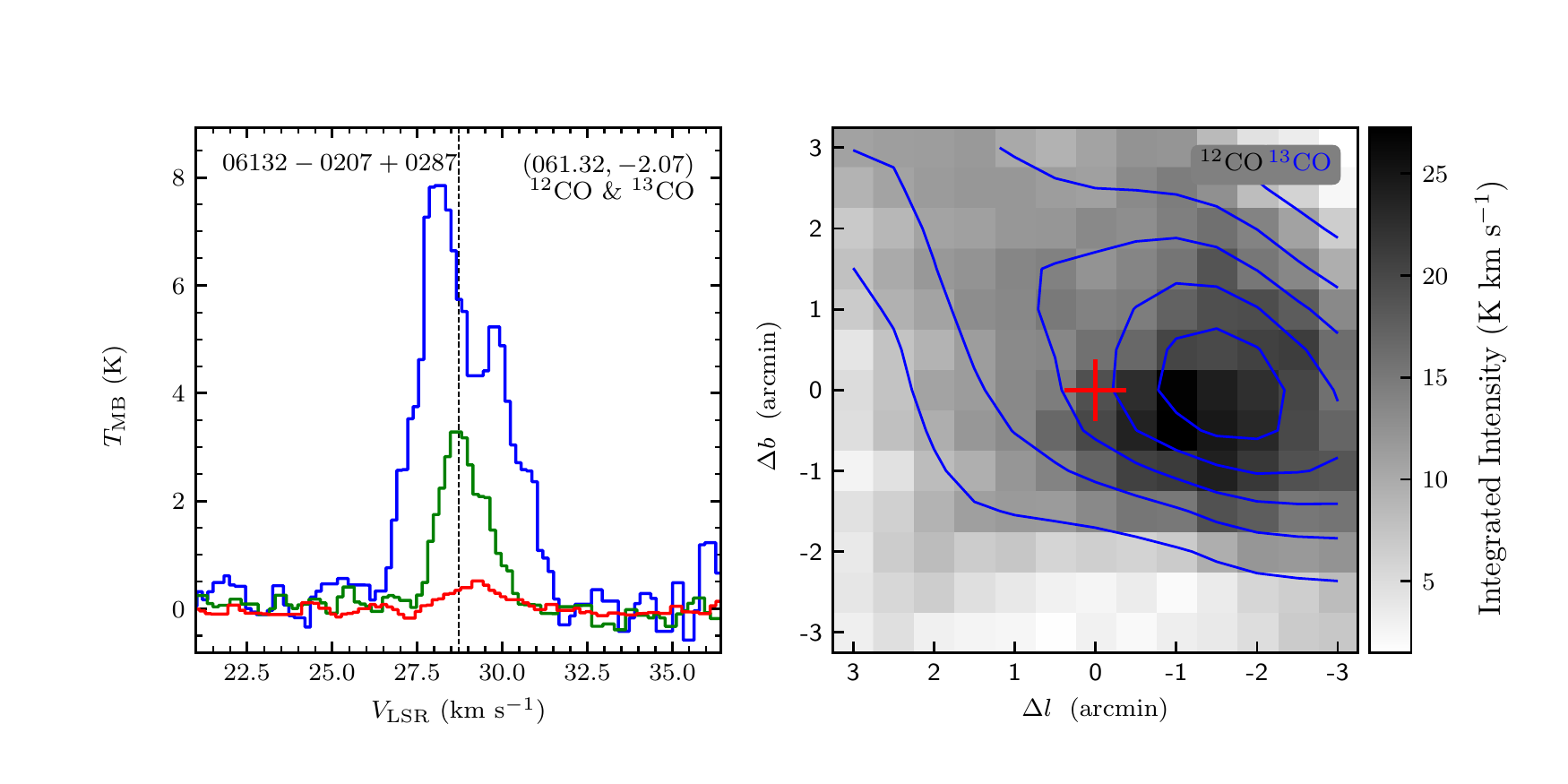}
\includegraphics[width=9.0cm,angle=0]{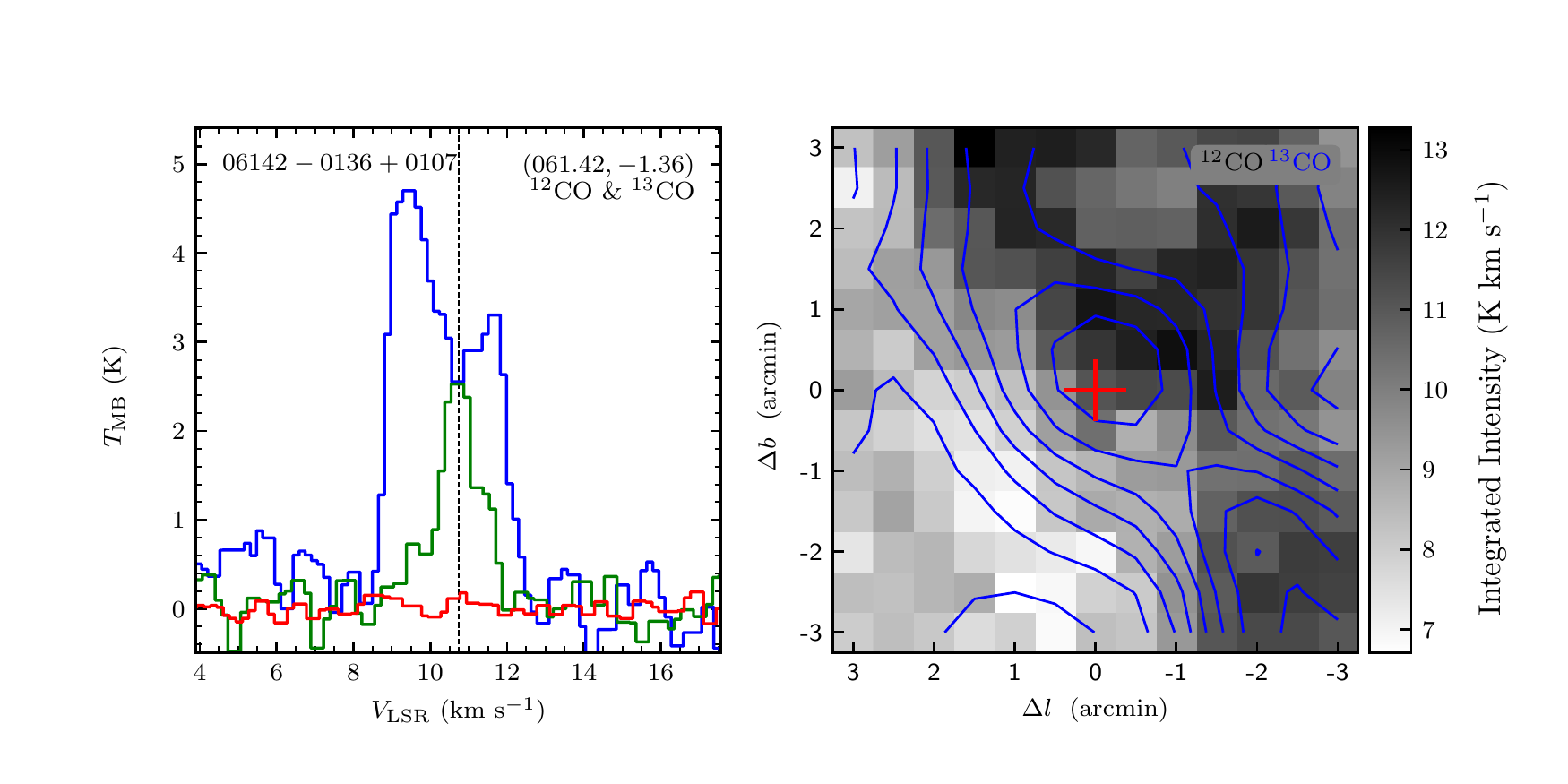}
\vspace{-0.5cm}

\includegraphics[width=9.0cm,angle=0]{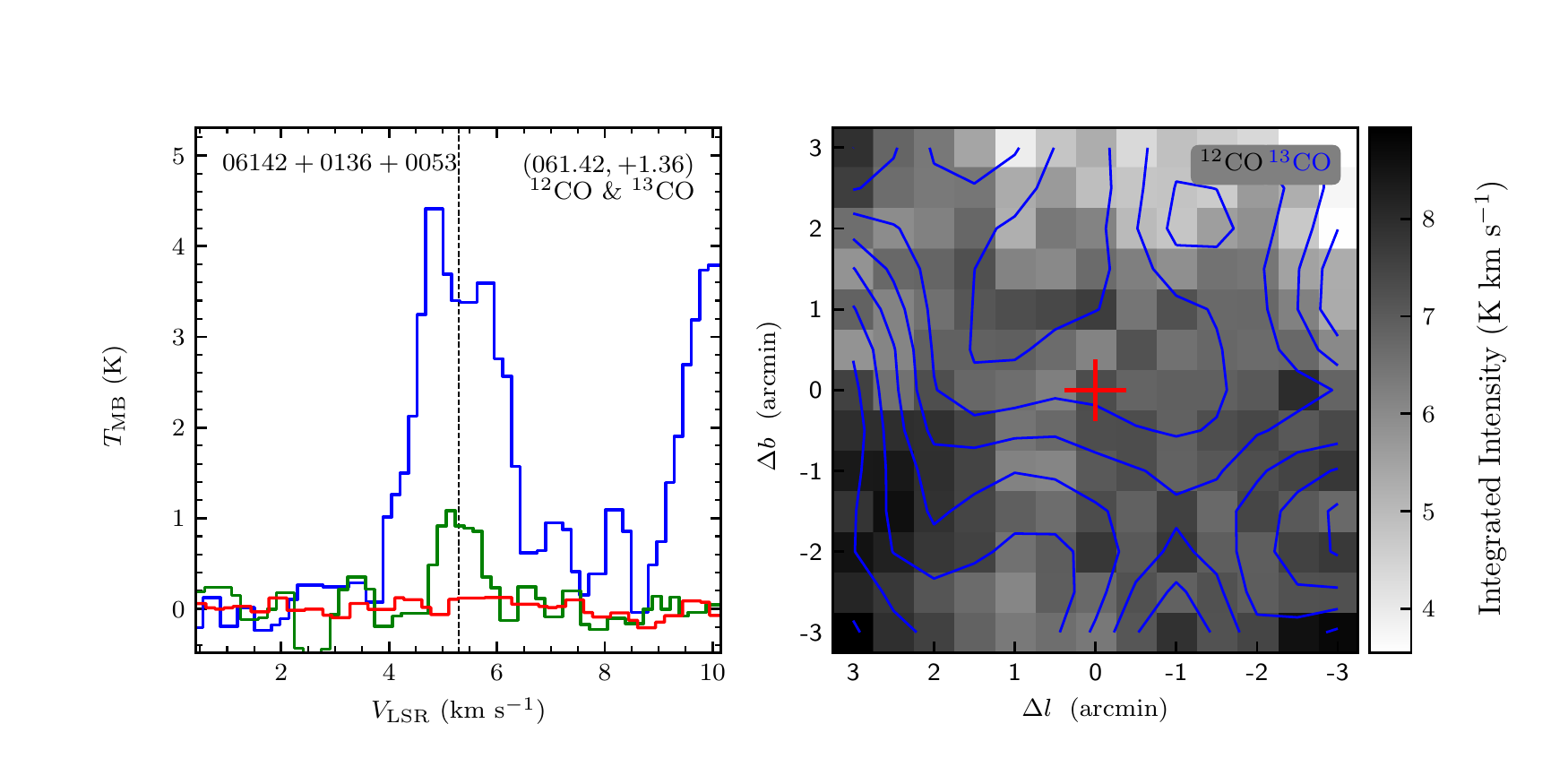}
\includegraphics[width=9.0cm,angle=0]{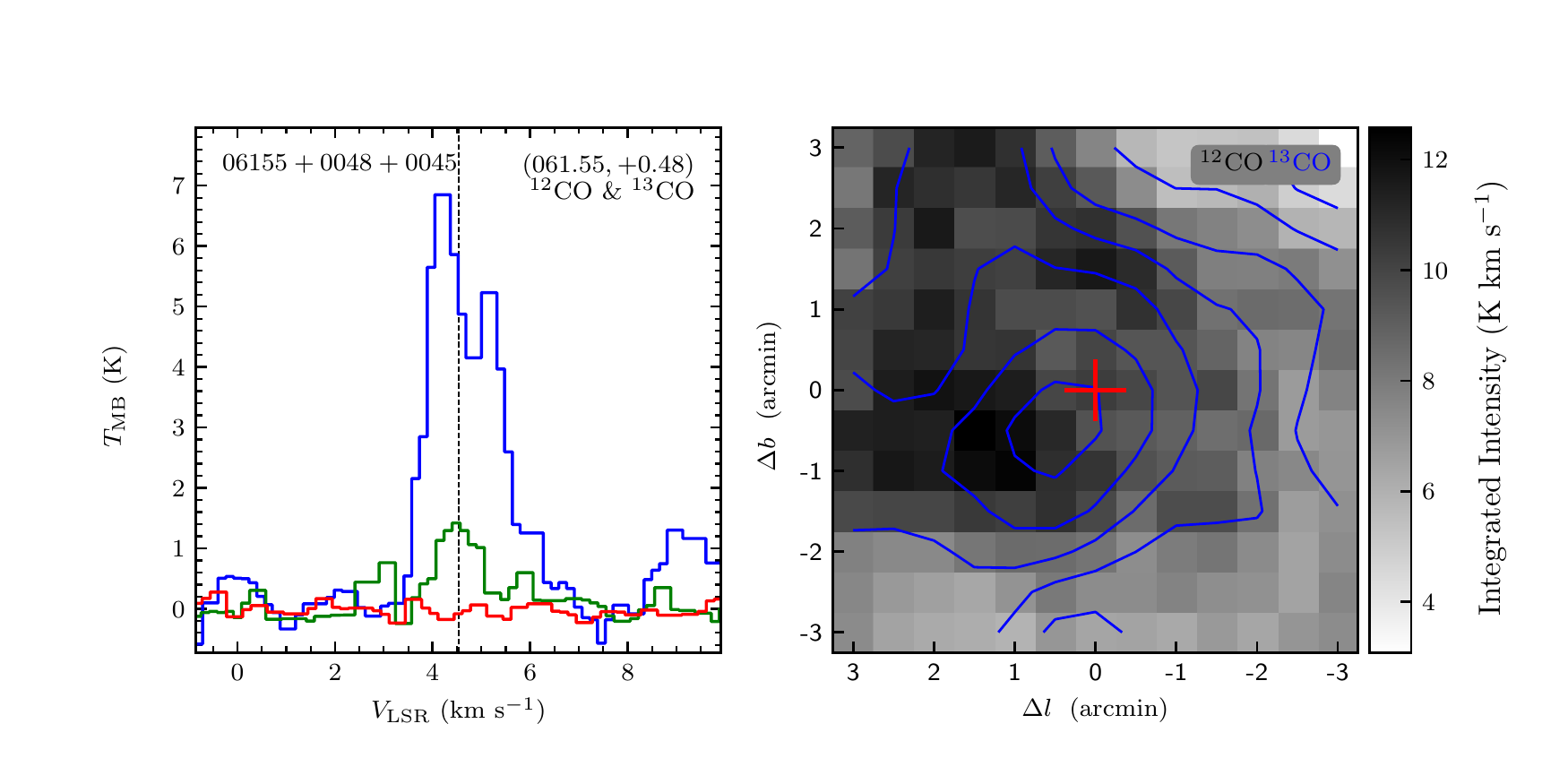}
\end{figure}
\clearpage

\begin{figure}
\includegraphics[width=9.0cm,angle=0]{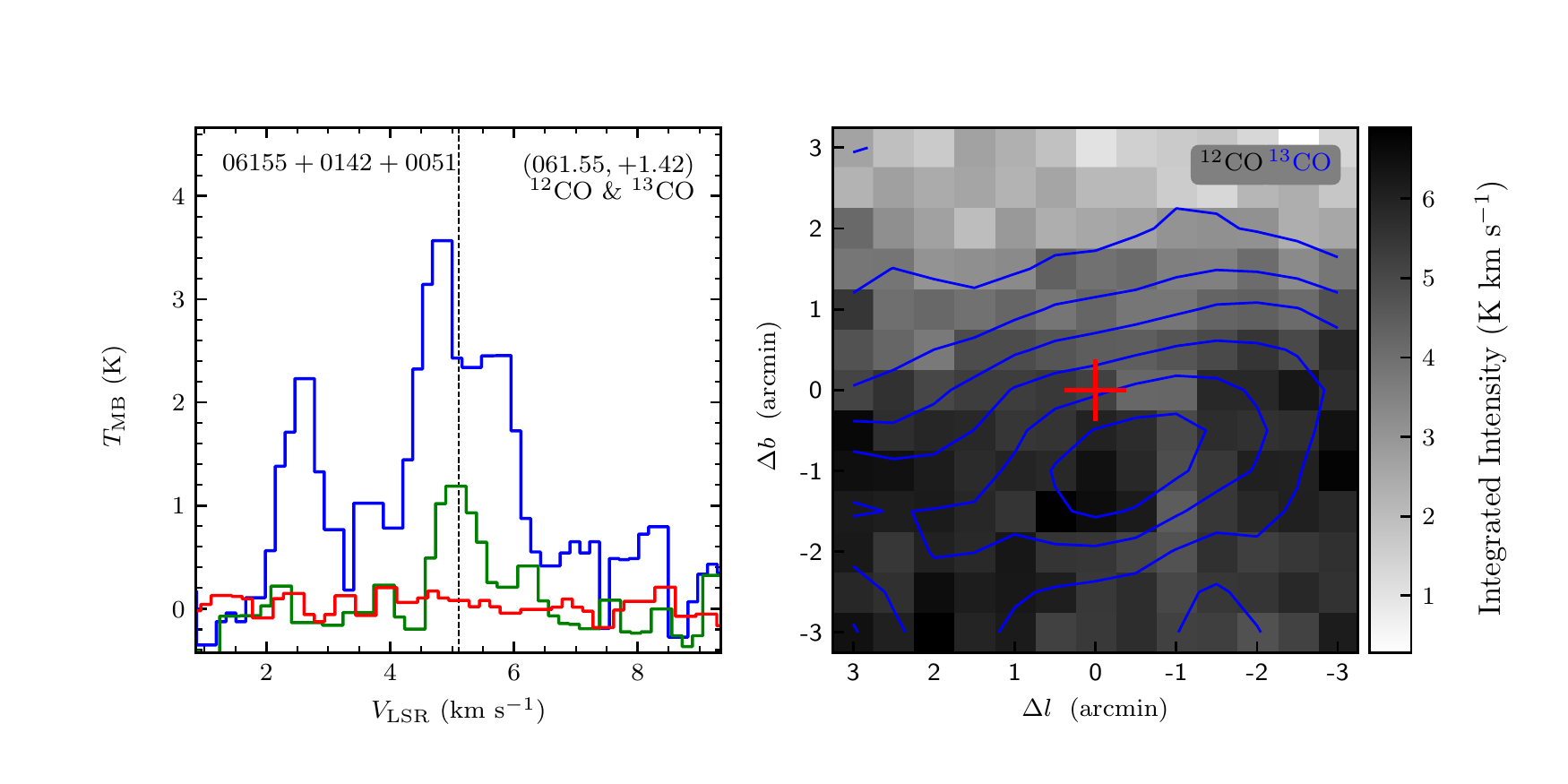}
\includegraphics[width=9.0cm,angle=0]{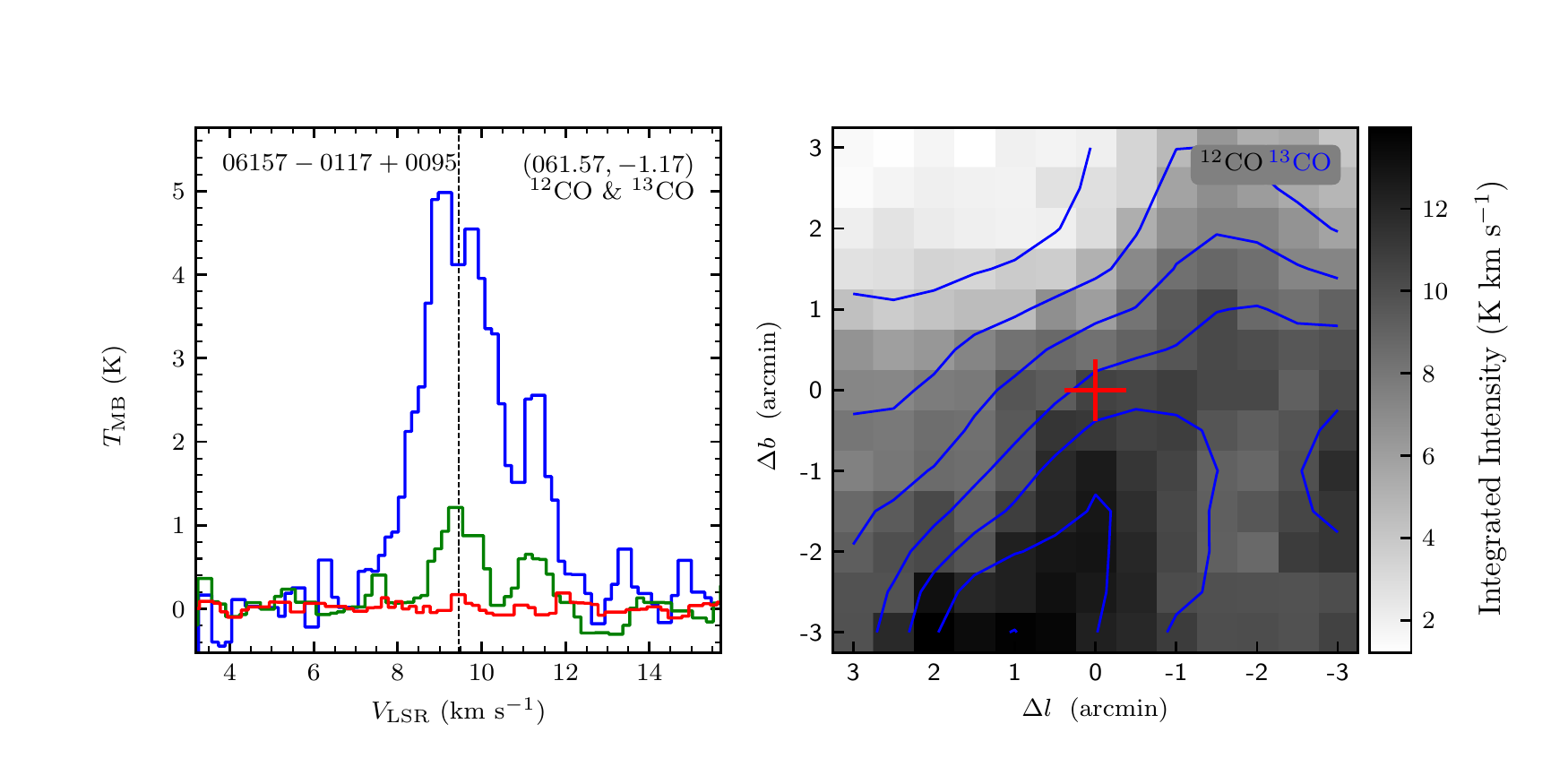}
\vspace{-0.5cm}

\includegraphics[width=9.0cm,angle=0]{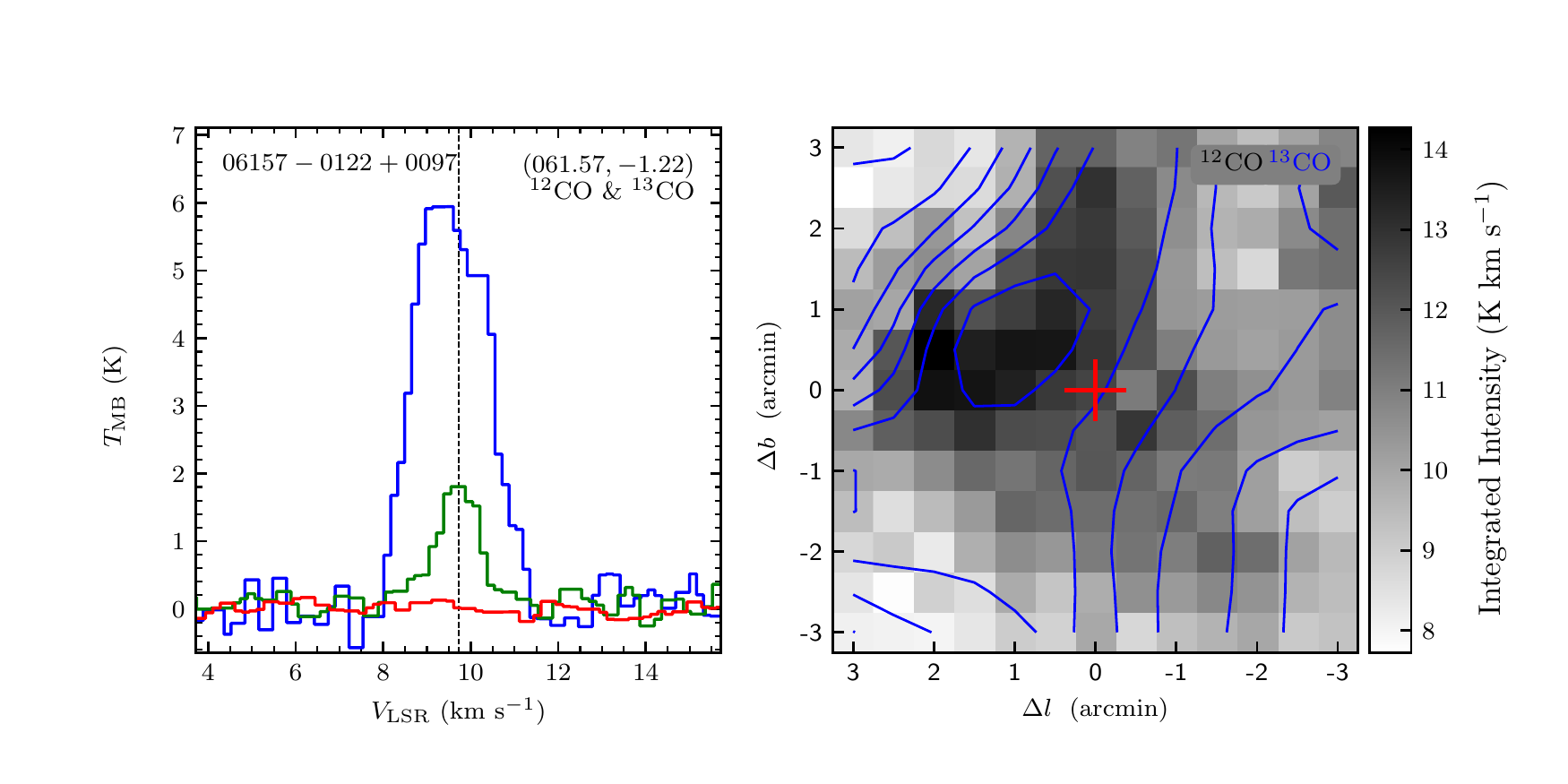}
\includegraphics[width=9.0cm,angle=0]{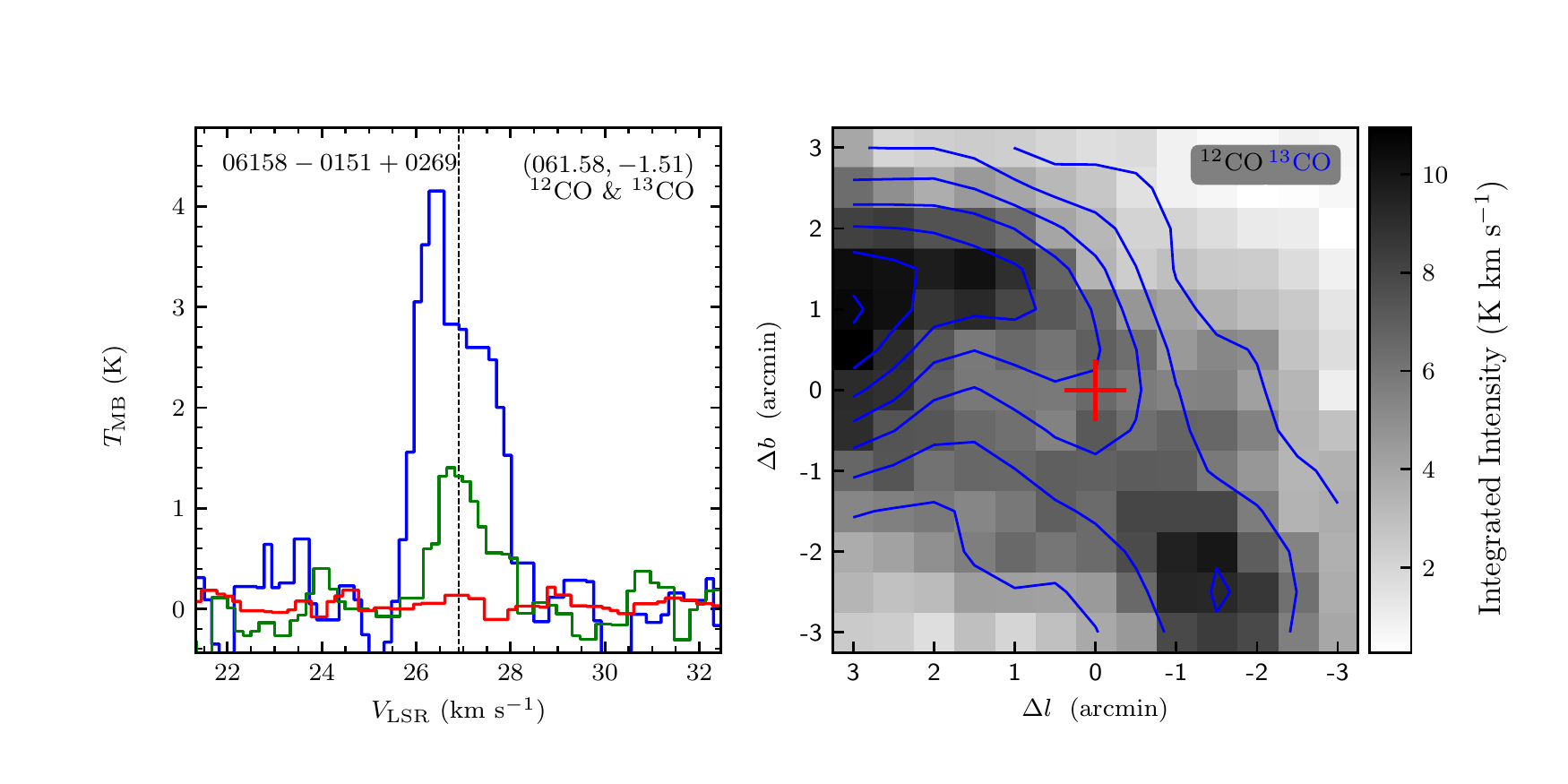}
\vspace{-0.5cm}

\includegraphics[width=9.0cm,angle=0]{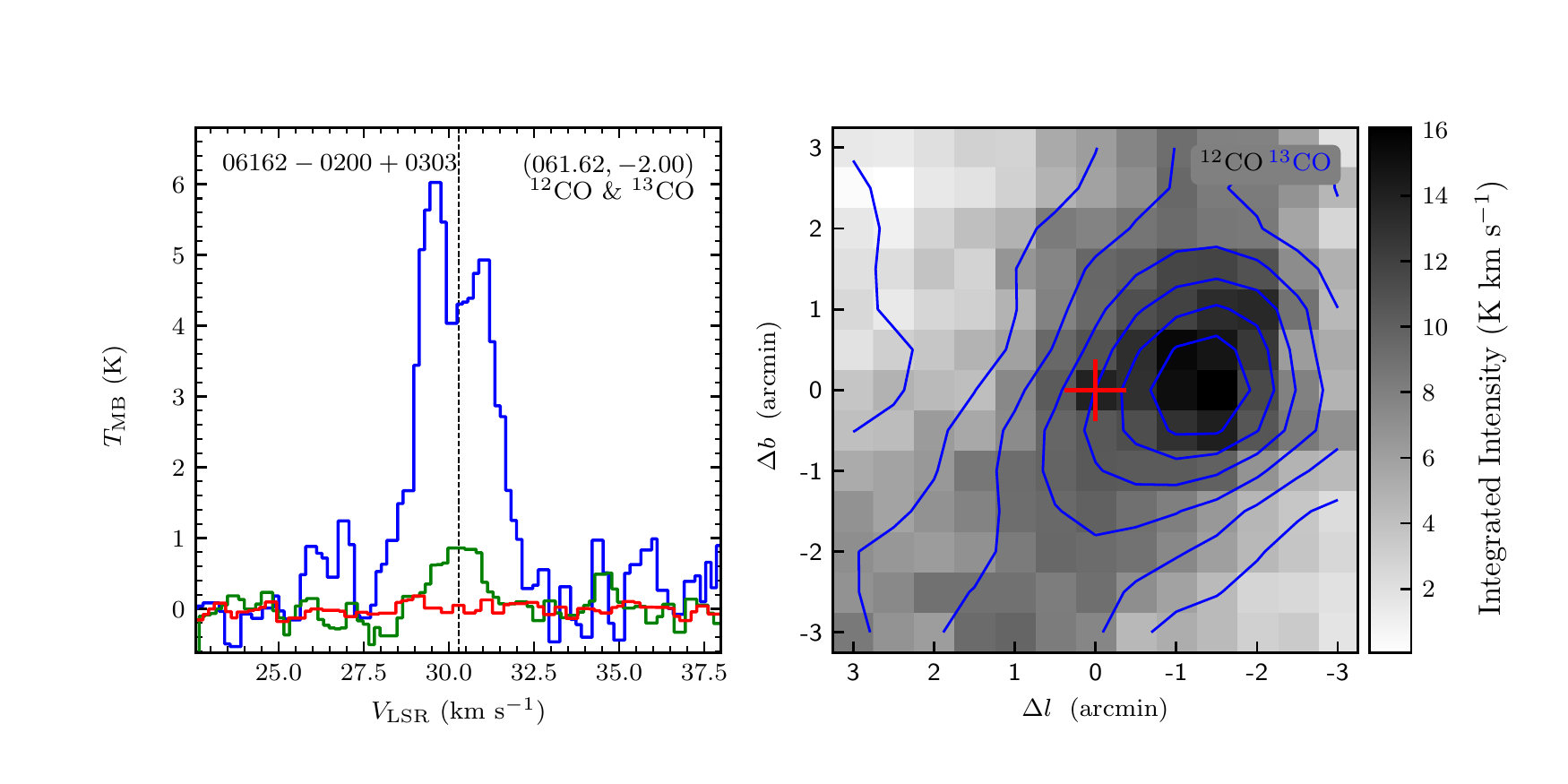}
\includegraphics[width=9.0cm,angle=0]{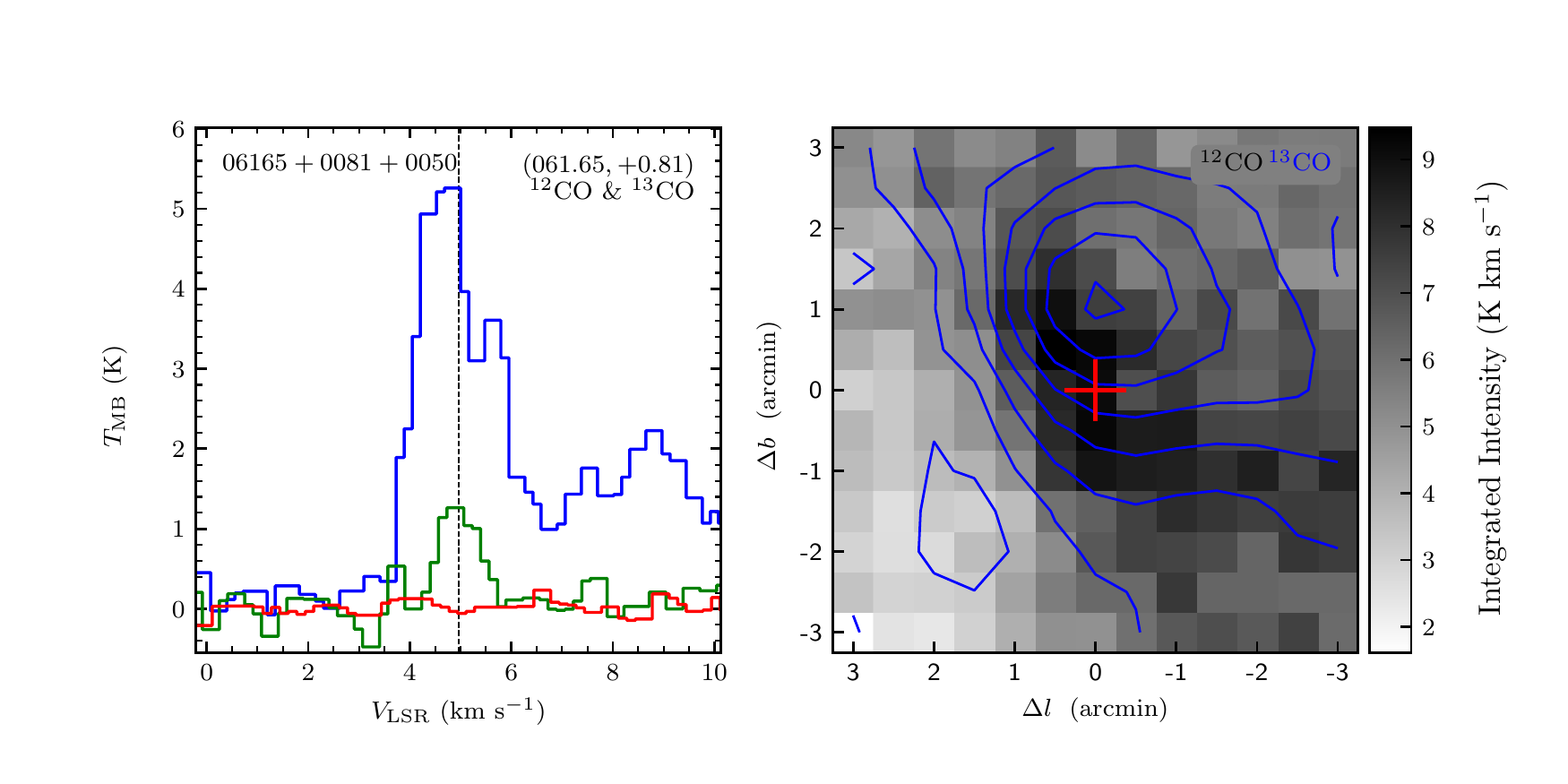}
\vspace{-0.5cm}

\includegraphics[width=9.0cm,angle=0]{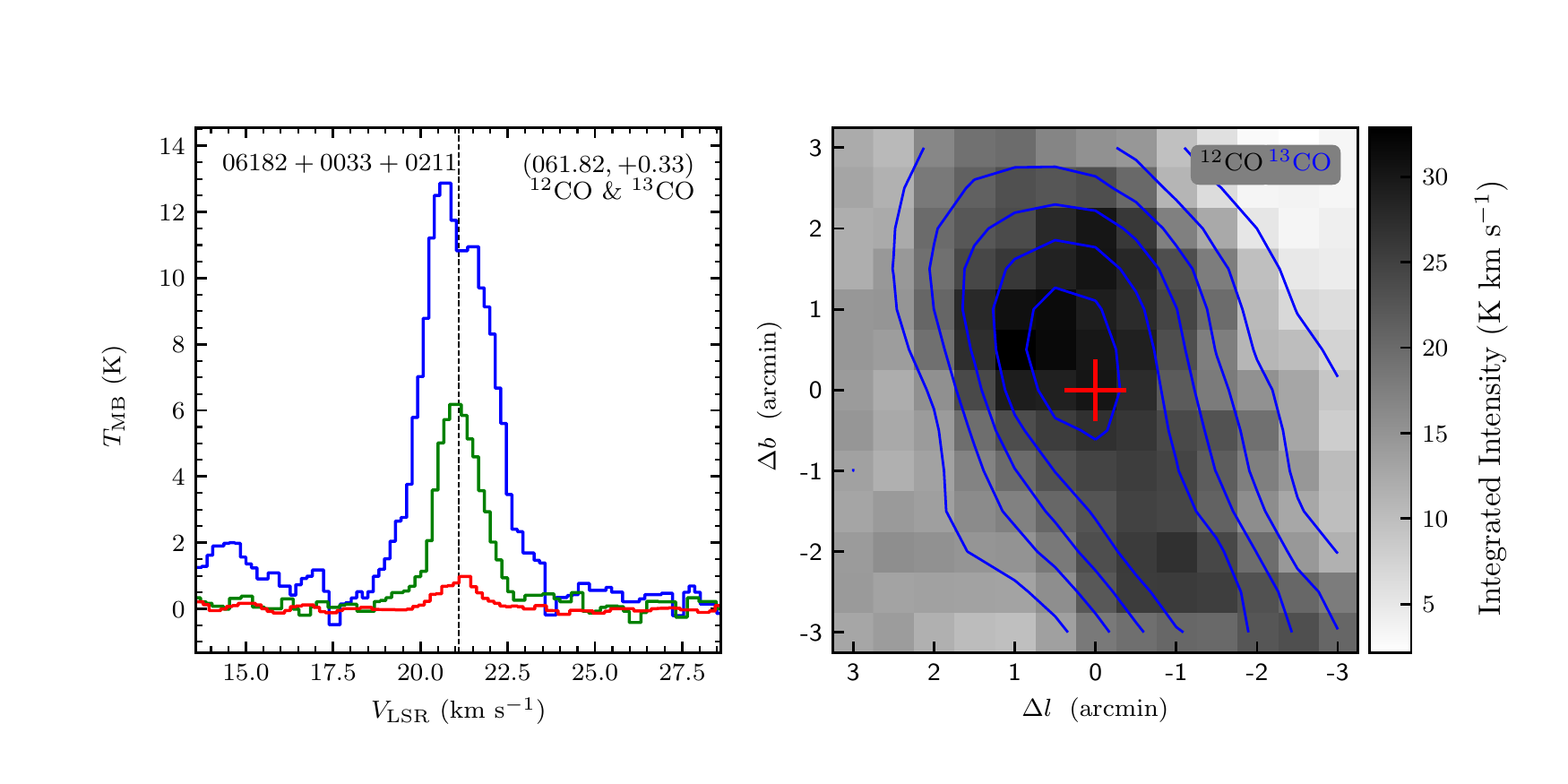}
\includegraphics[width=9.0cm,angle=0]{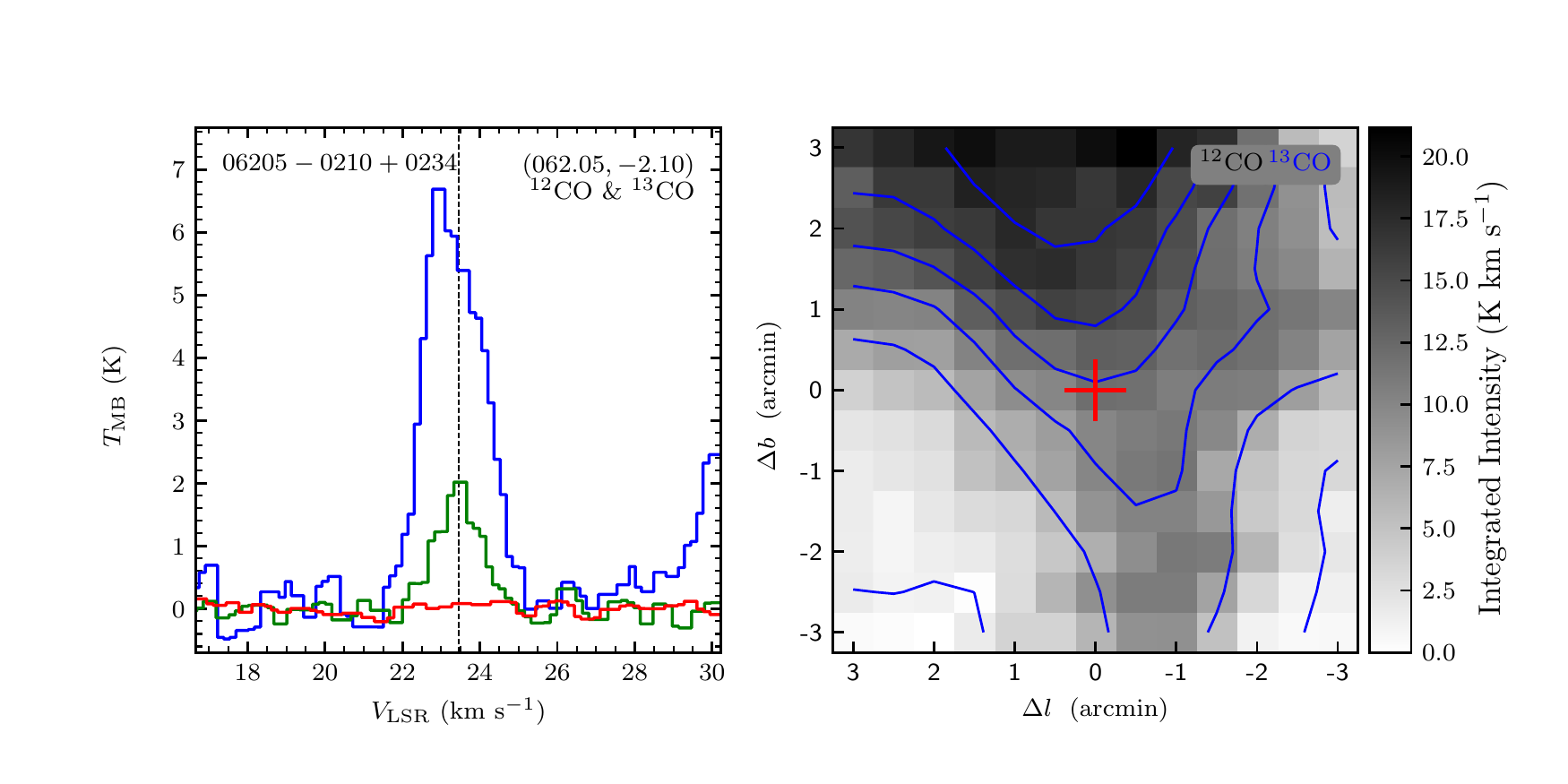}
\vspace{-0.5cm}

\includegraphics[width=9.0cm,angle=0]{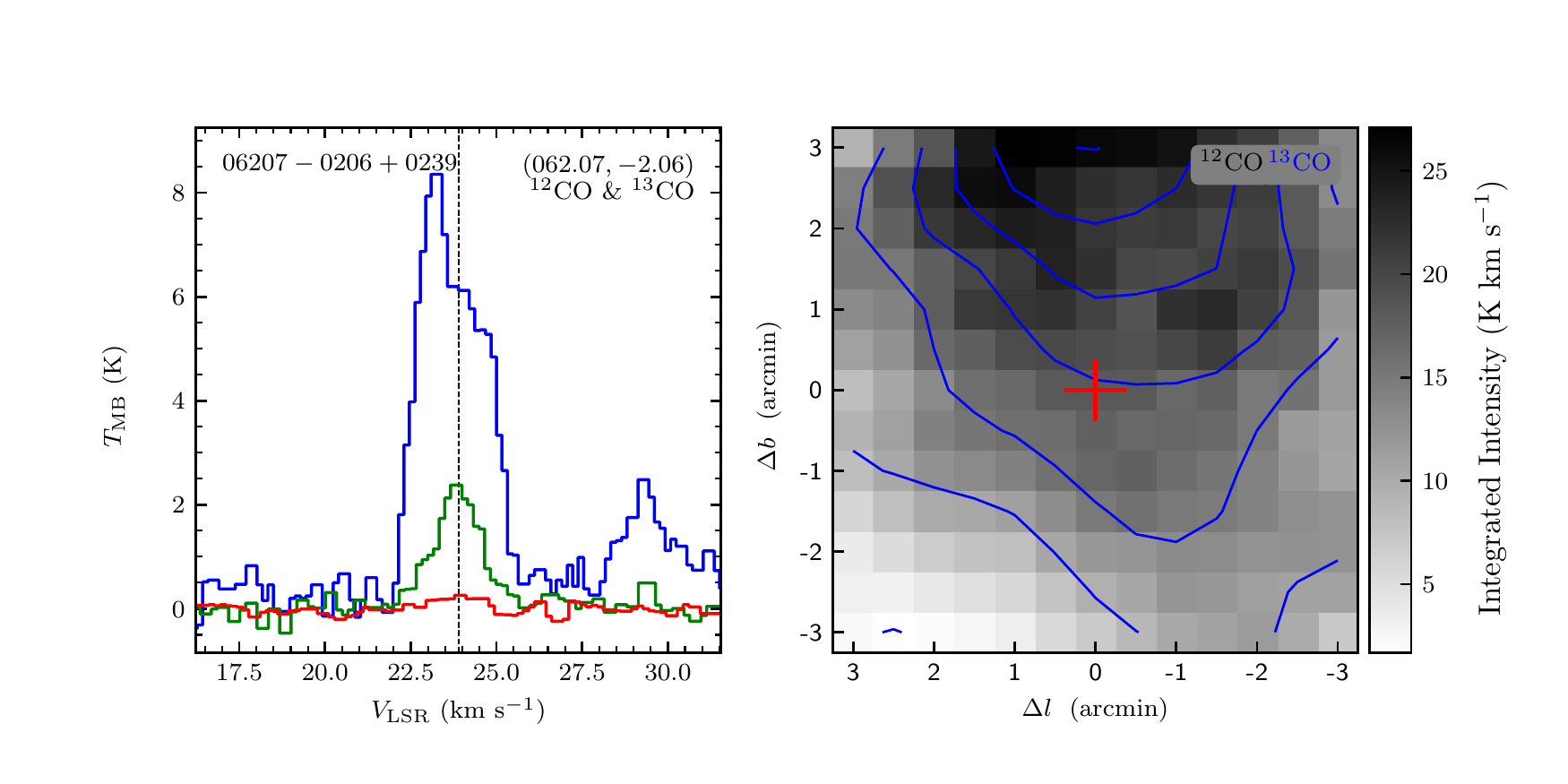}
\includegraphics[width=9.0cm,angle=0]{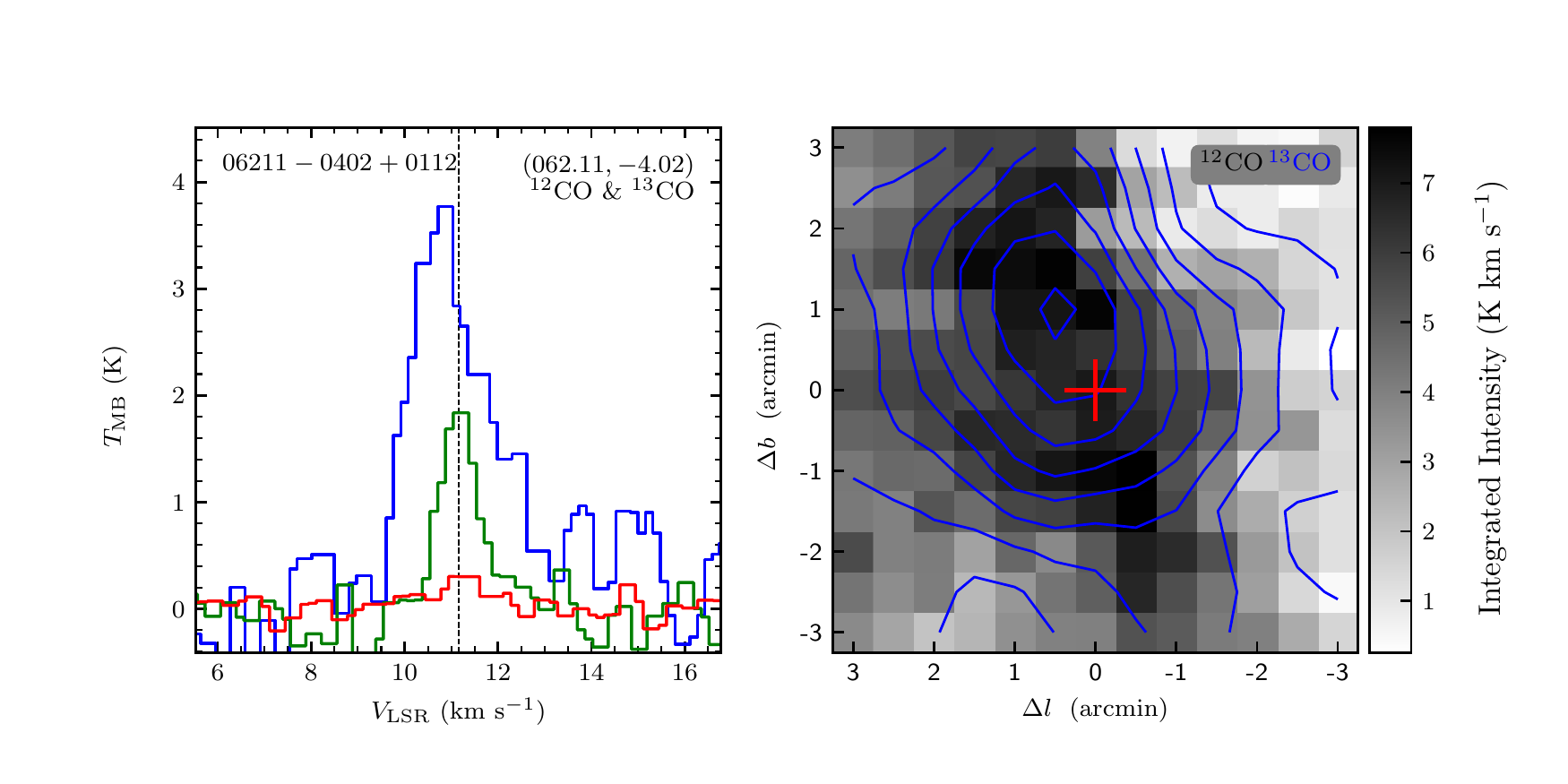}
\end{figure}
\clearpage

\begin{figure}
\includegraphics[width=9.0cm,angle=0]{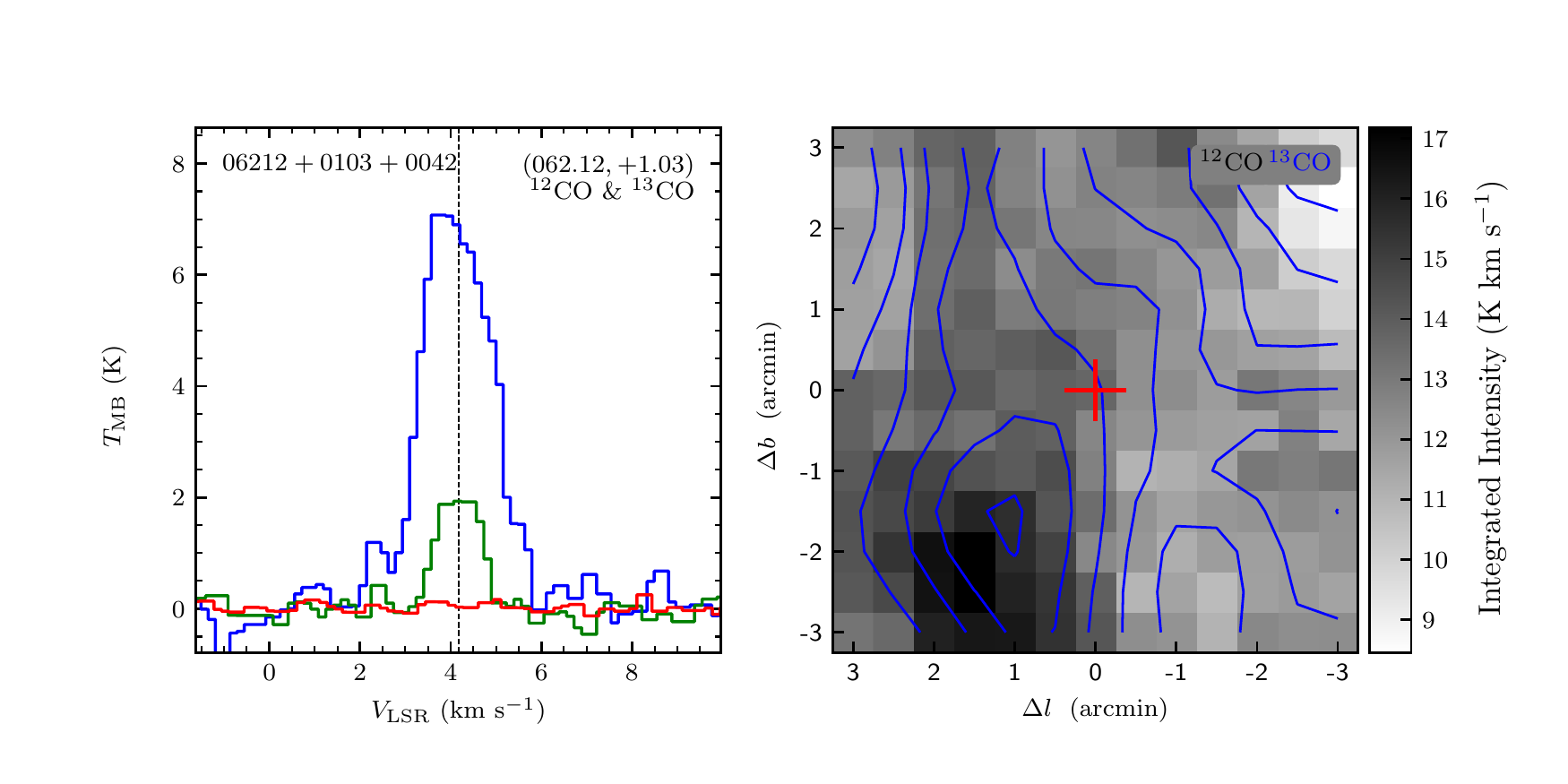}
\includegraphics[width=9.0cm,angle=0]{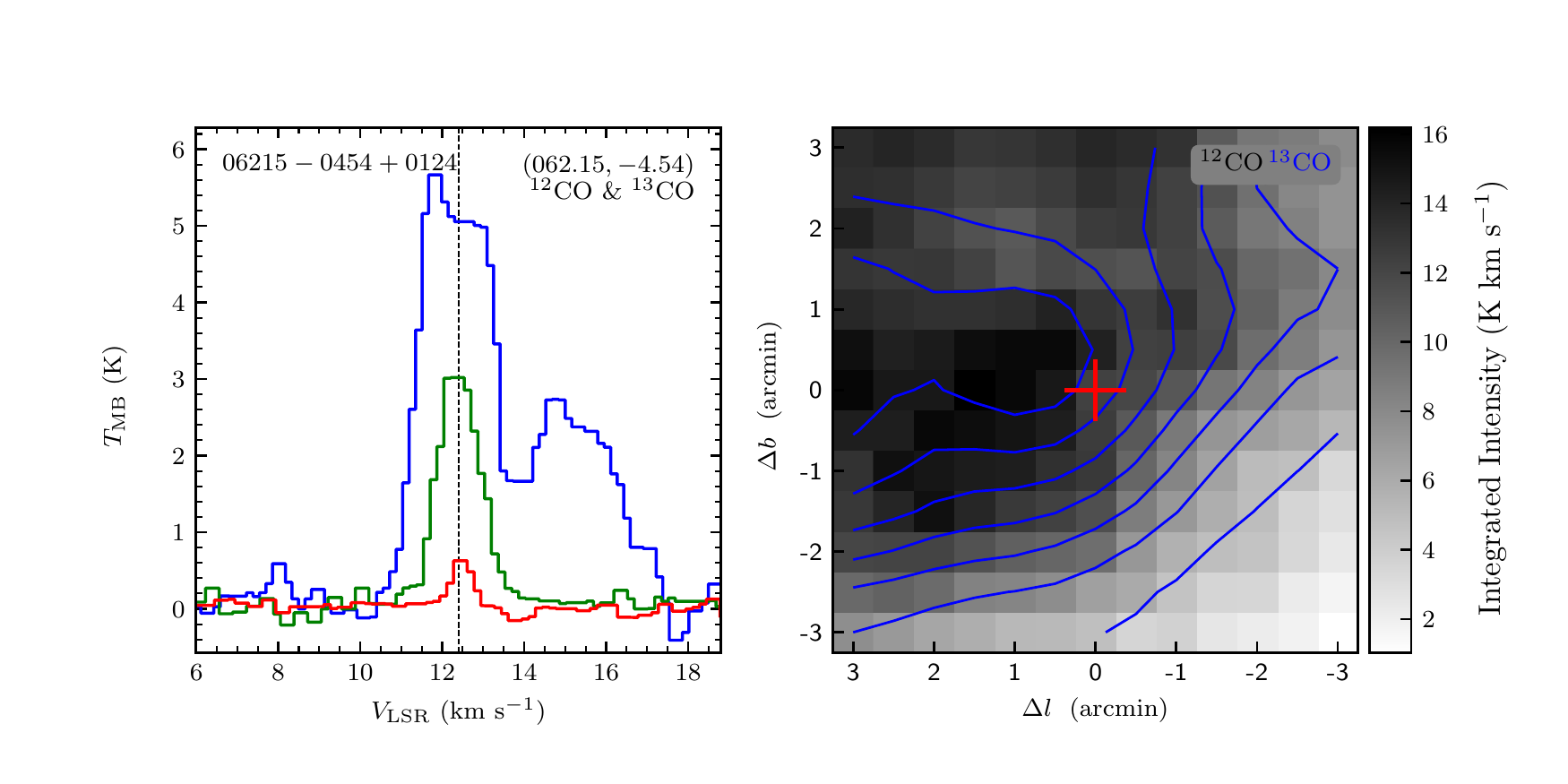}
\vspace{-0.5cm}

\includegraphics[width=9.0cm,angle=0]{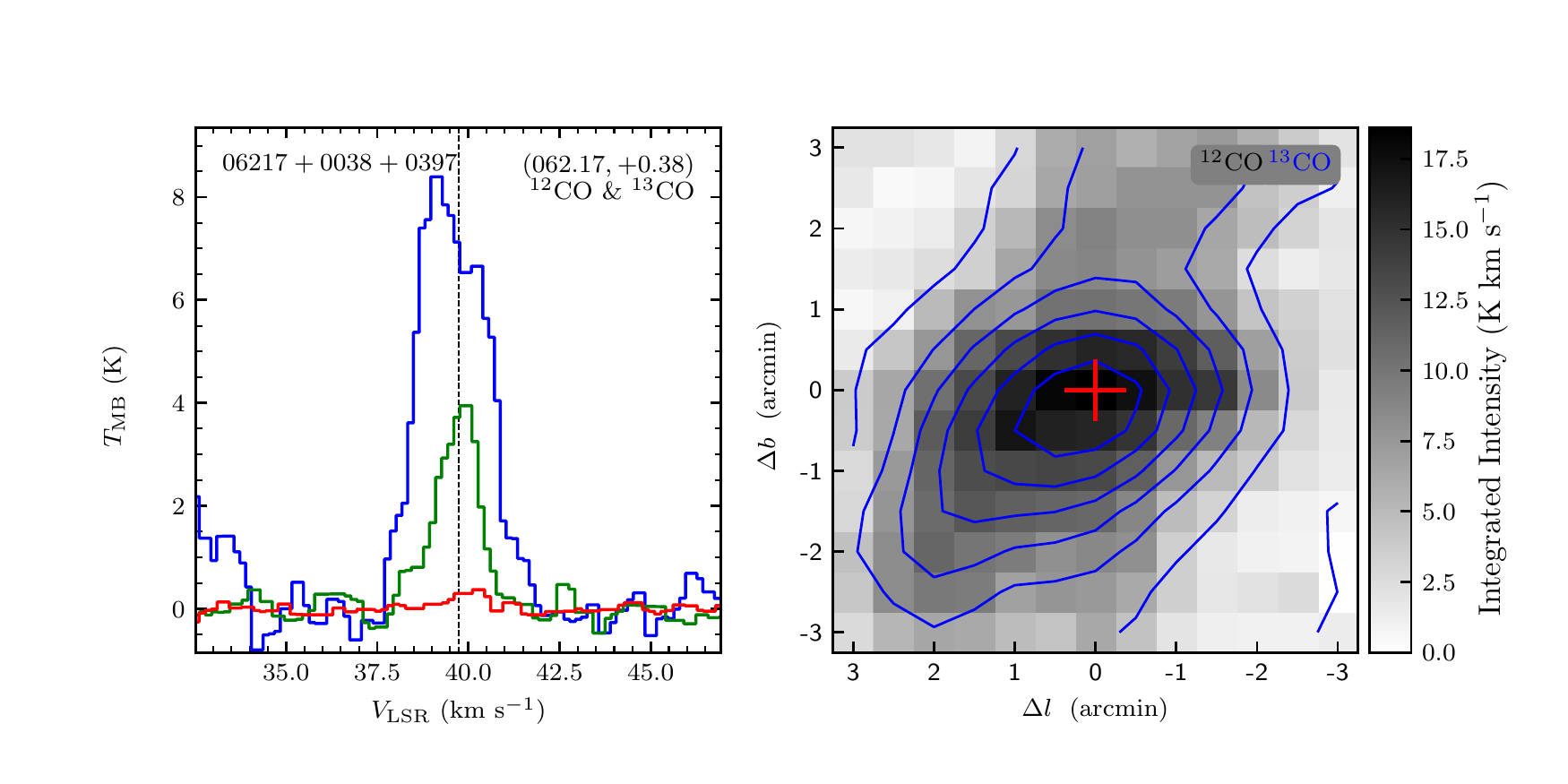}
\includegraphics[width=9.0cm,angle=0]{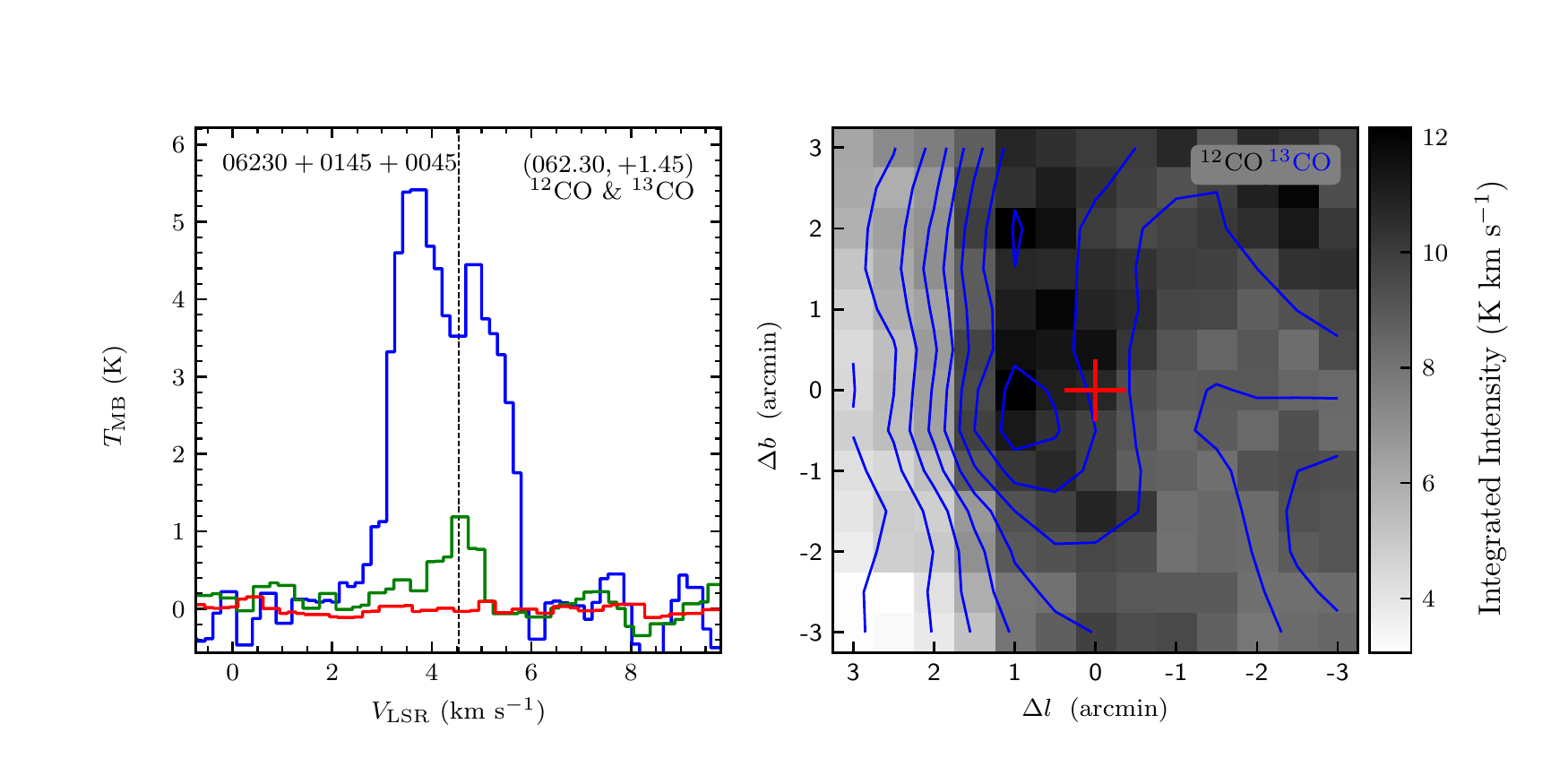}
\vspace{-0.5cm}

\includegraphics[width=9.0cm,angle=0]{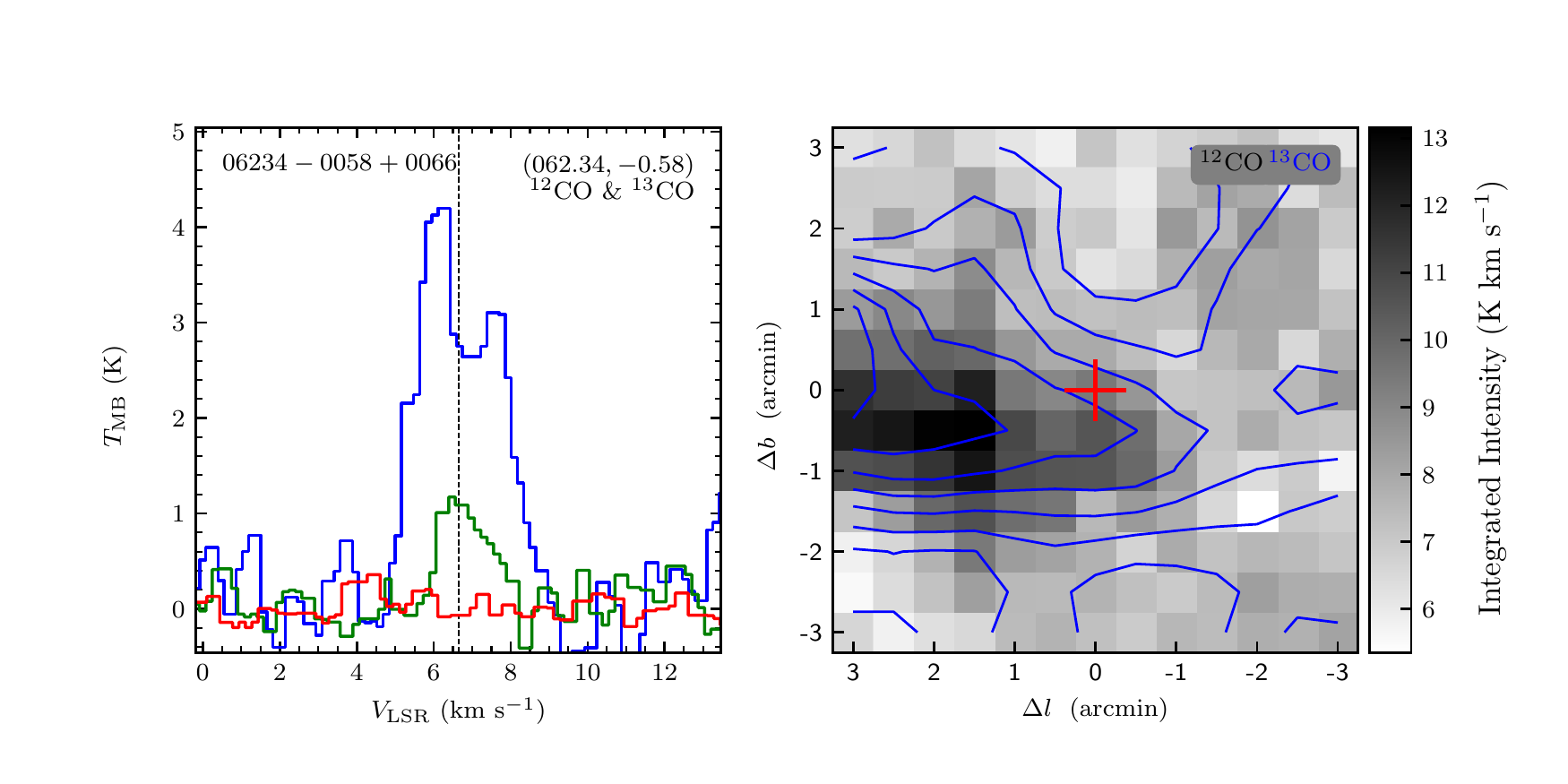}
\includegraphics[width=9.0cm,angle=0]{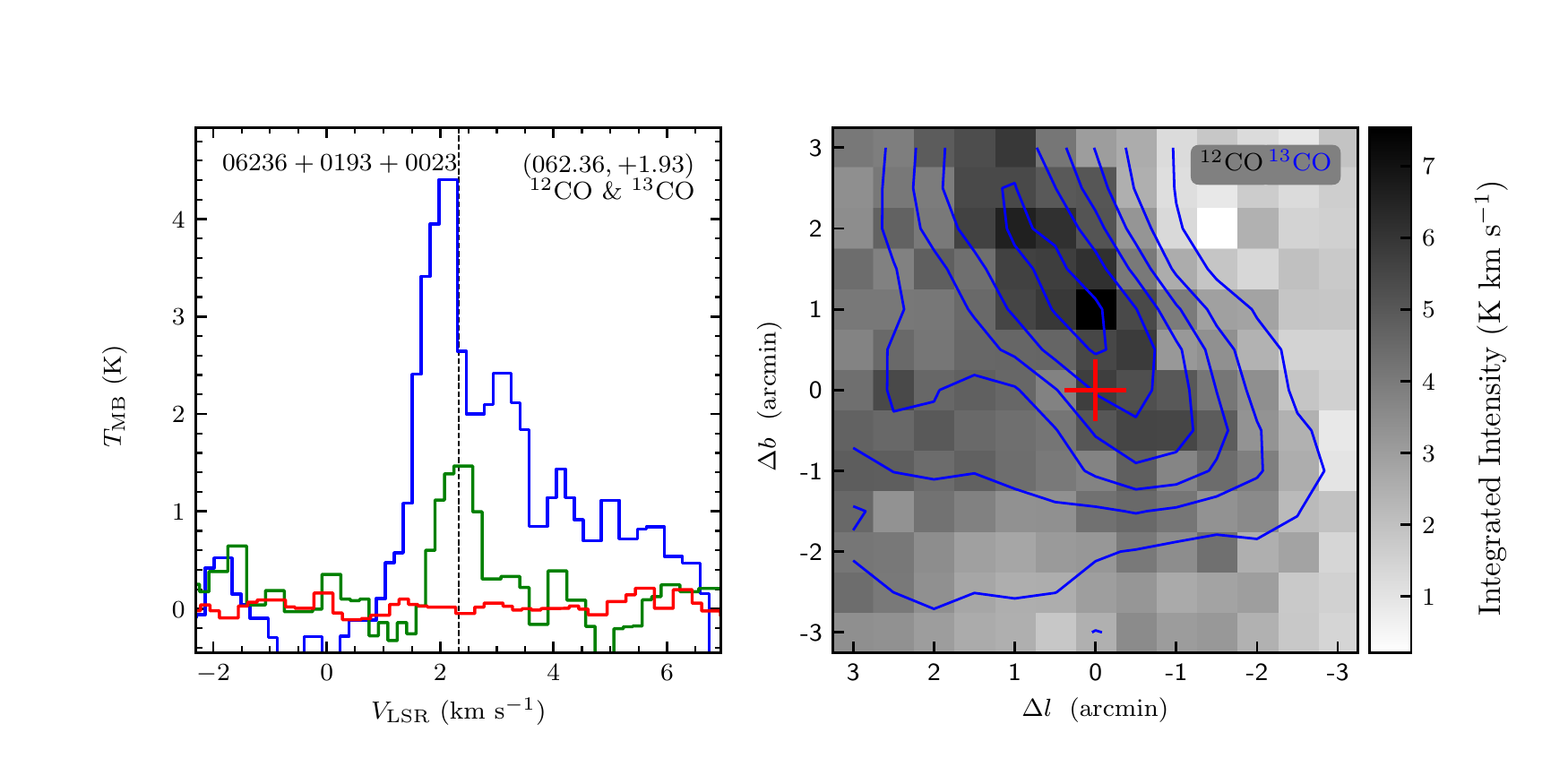}
\vspace{-0.5cm}

\includegraphics[width=9.0cm,angle=0]{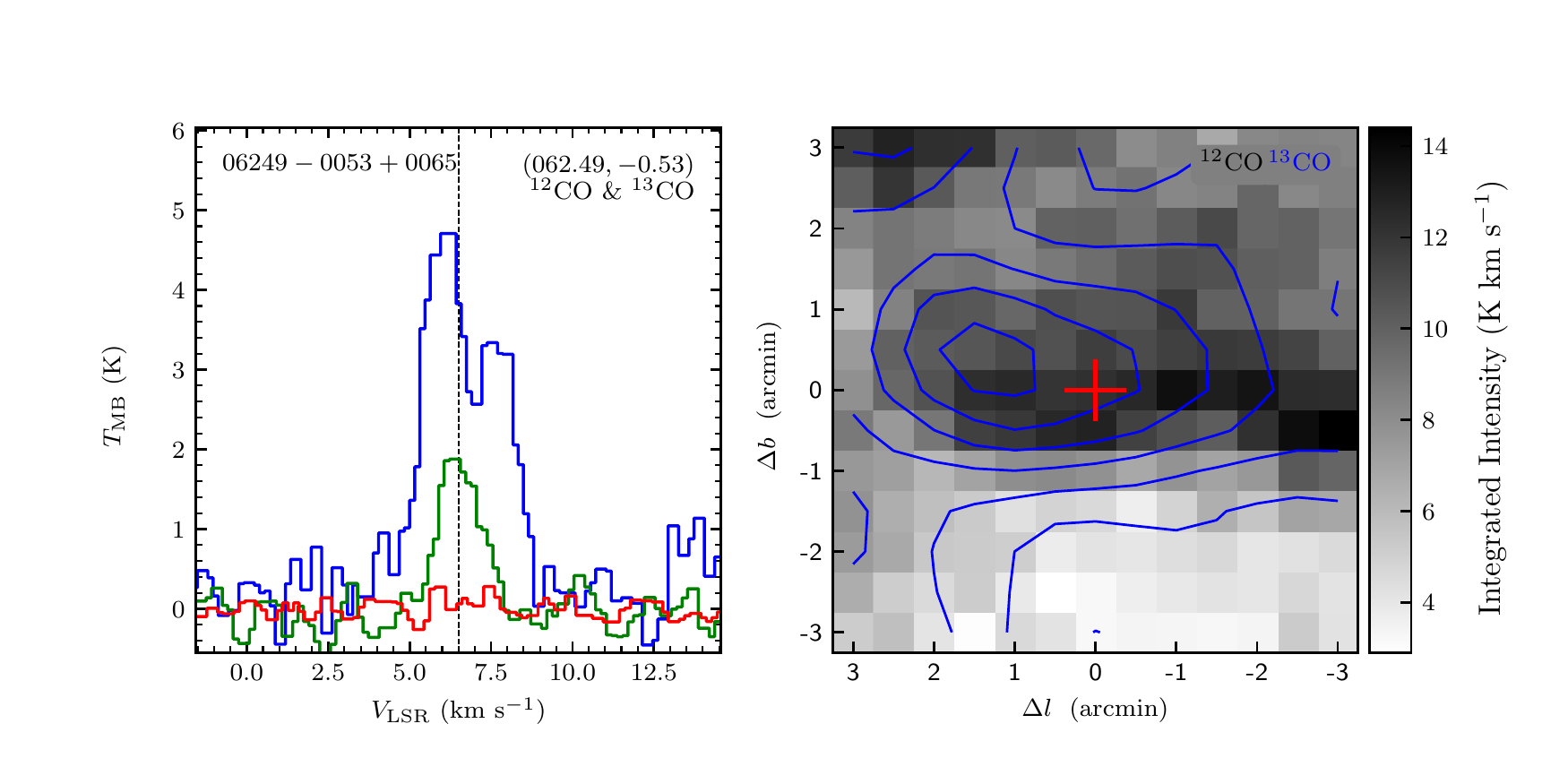}
\includegraphics[width=9.0cm,angle=0]{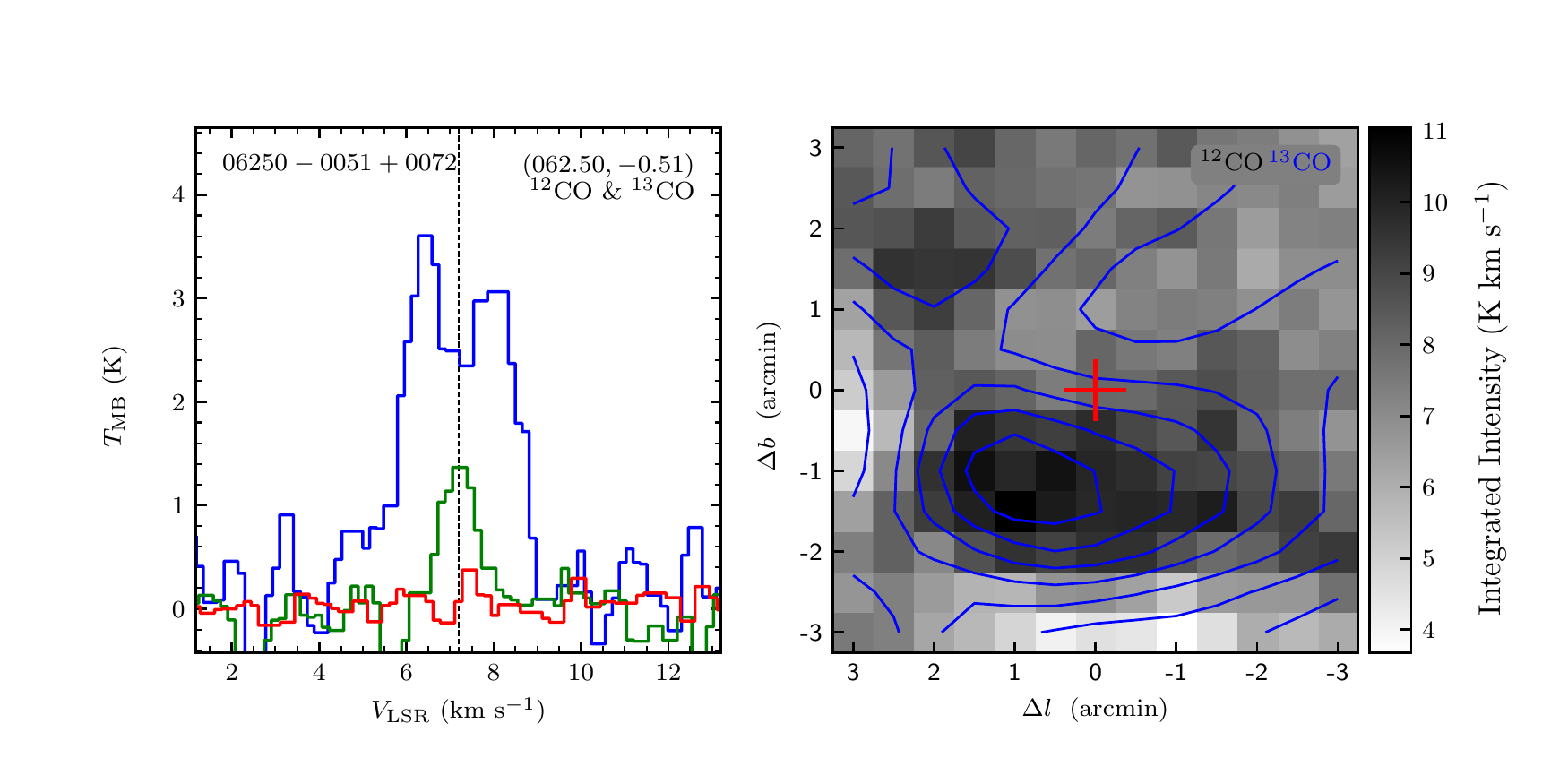}
\vspace{-0.5cm}

\includegraphics[width=9.0cm,angle=0]{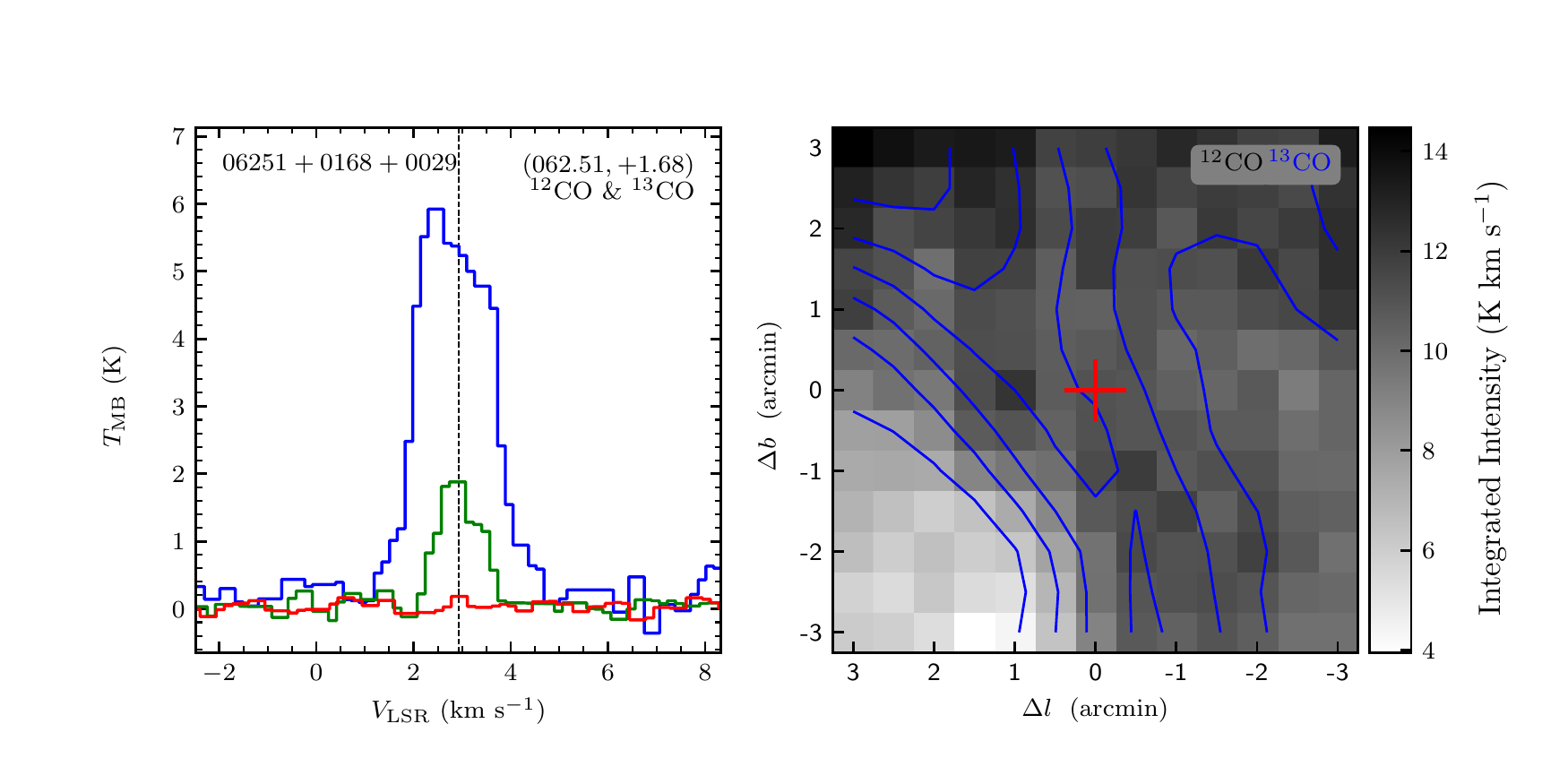}
\includegraphics[width=9.0cm,angle=0]{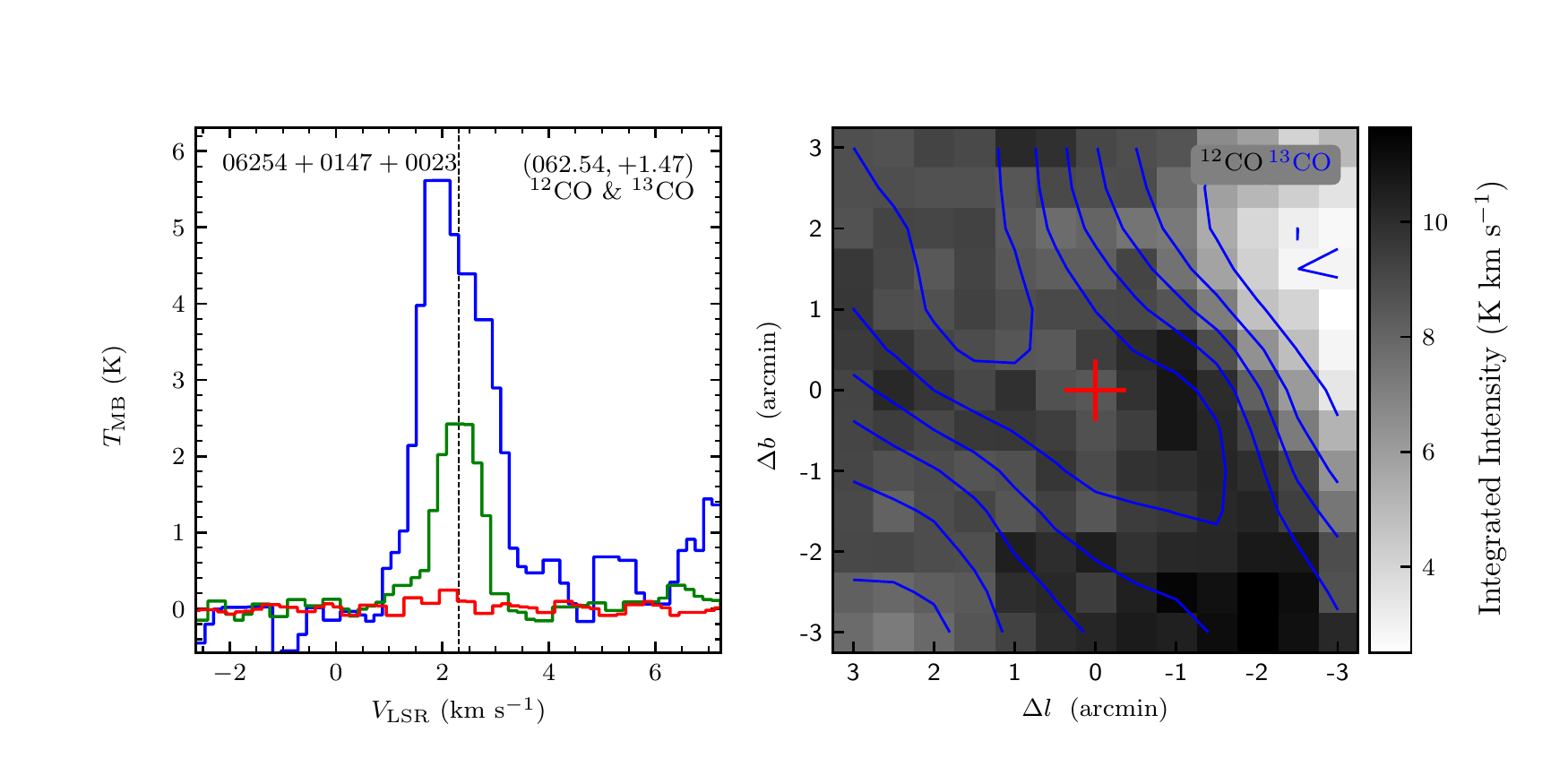}
\end{figure}
\clearpage

\begin{figure}
\includegraphics[width=9.0cm,angle=0]{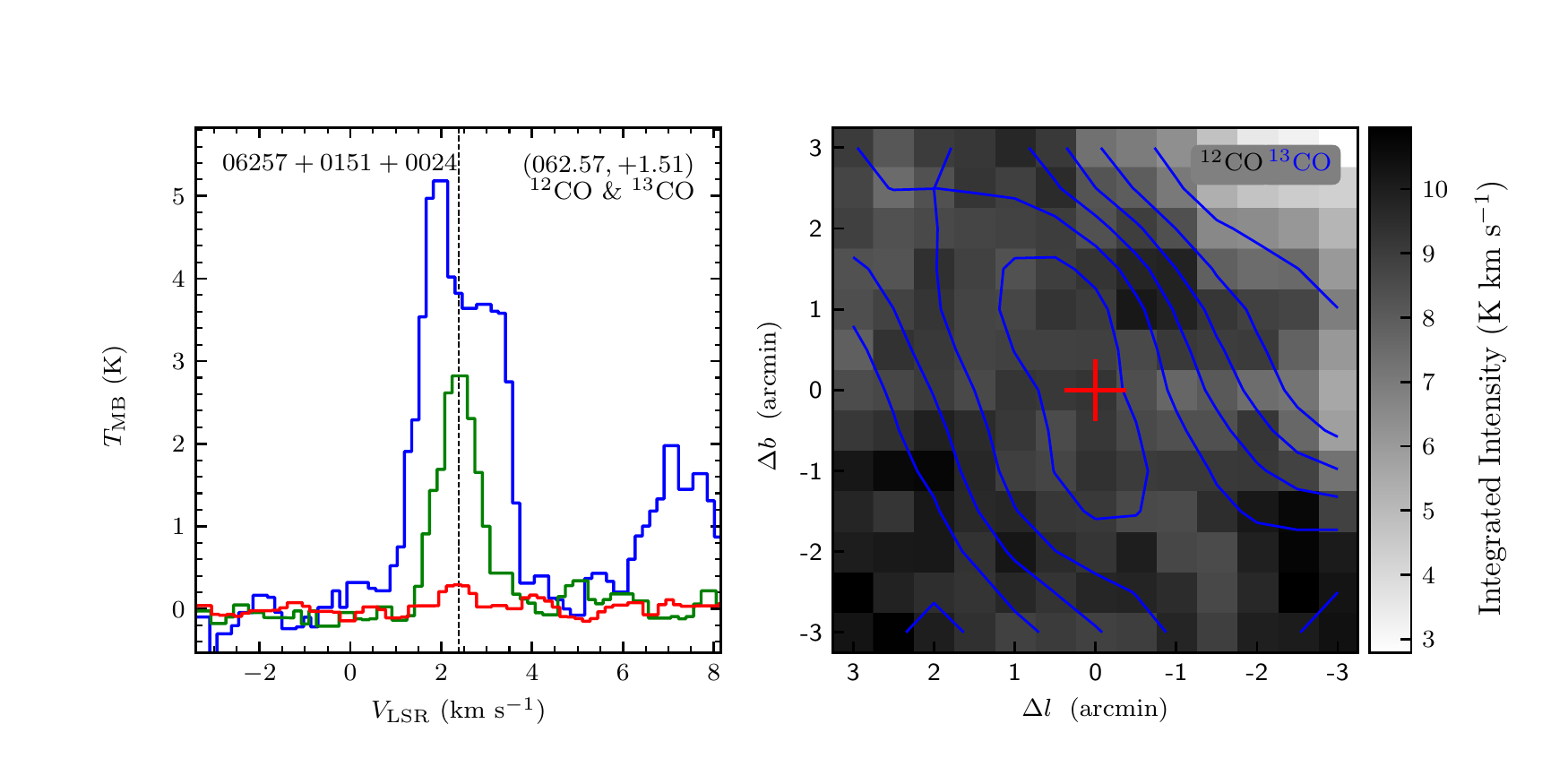}
\includegraphics[width=9.0cm,angle=0]{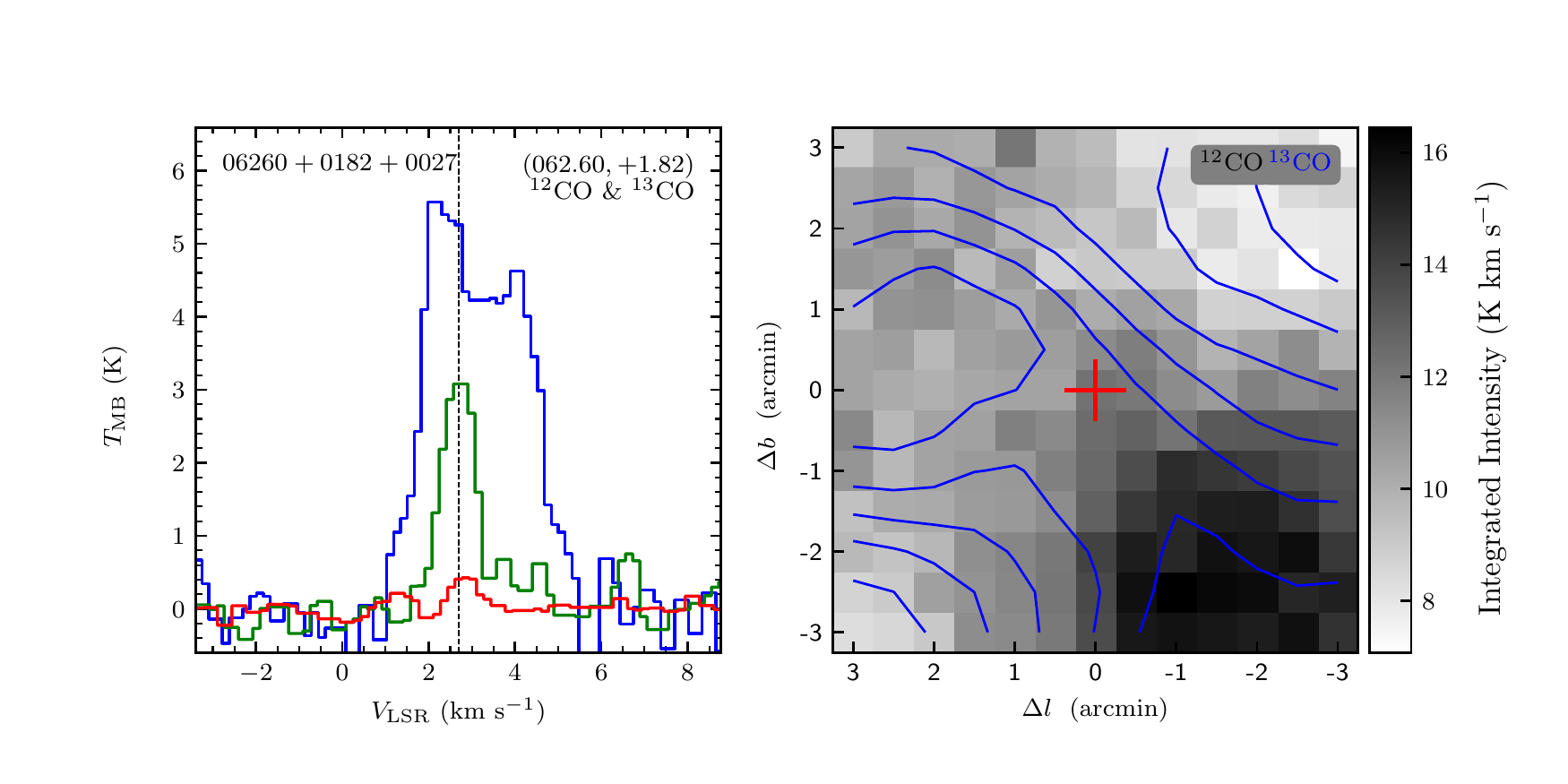}
\vspace{-0.5cm}

\includegraphics[width=9.0cm,angle=0]{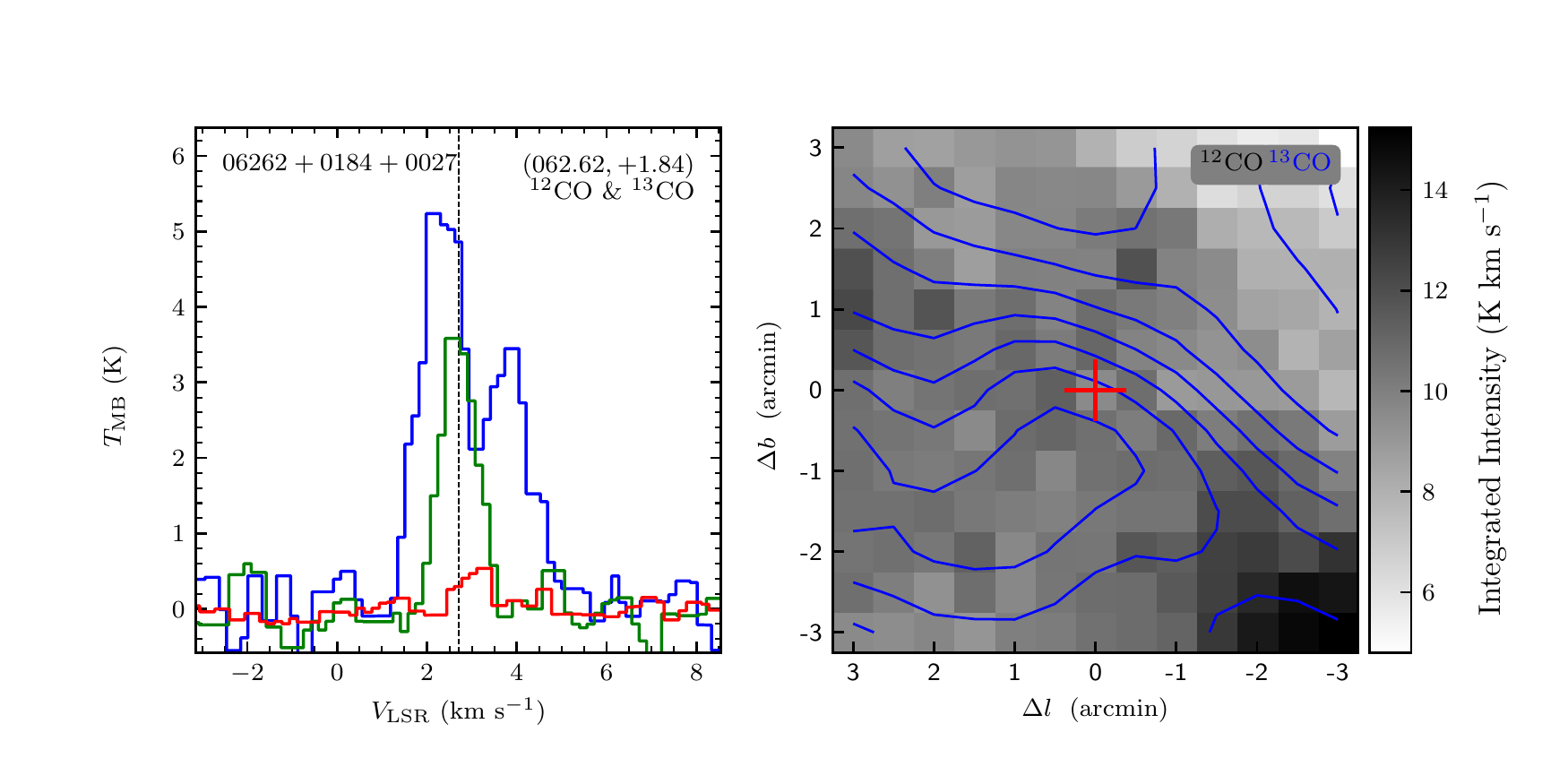}
\includegraphics[width=9.0cm,angle=0]{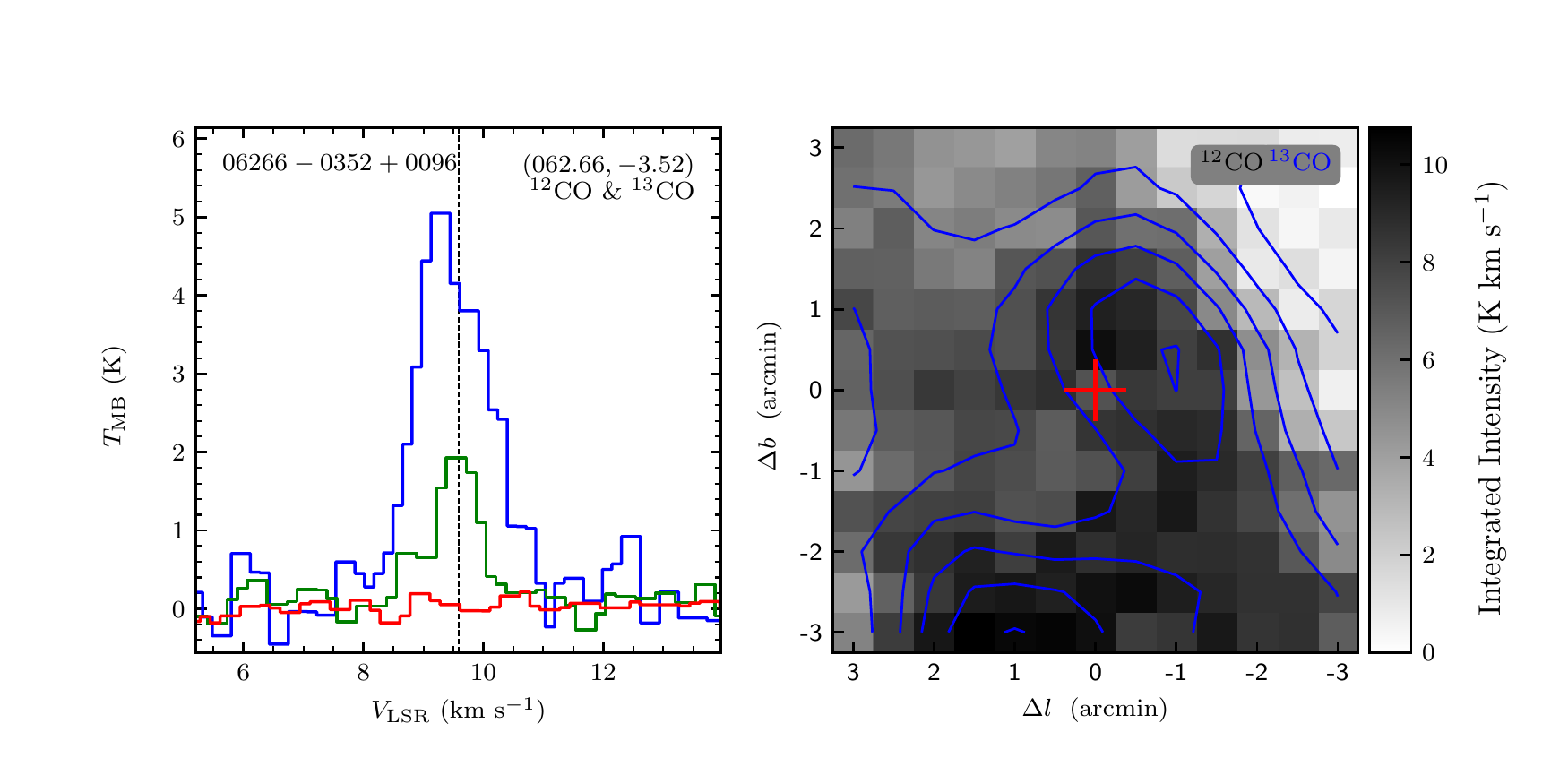}
\vspace{-0.5cm}

\includegraphics[width=9.0cm,angle=0]{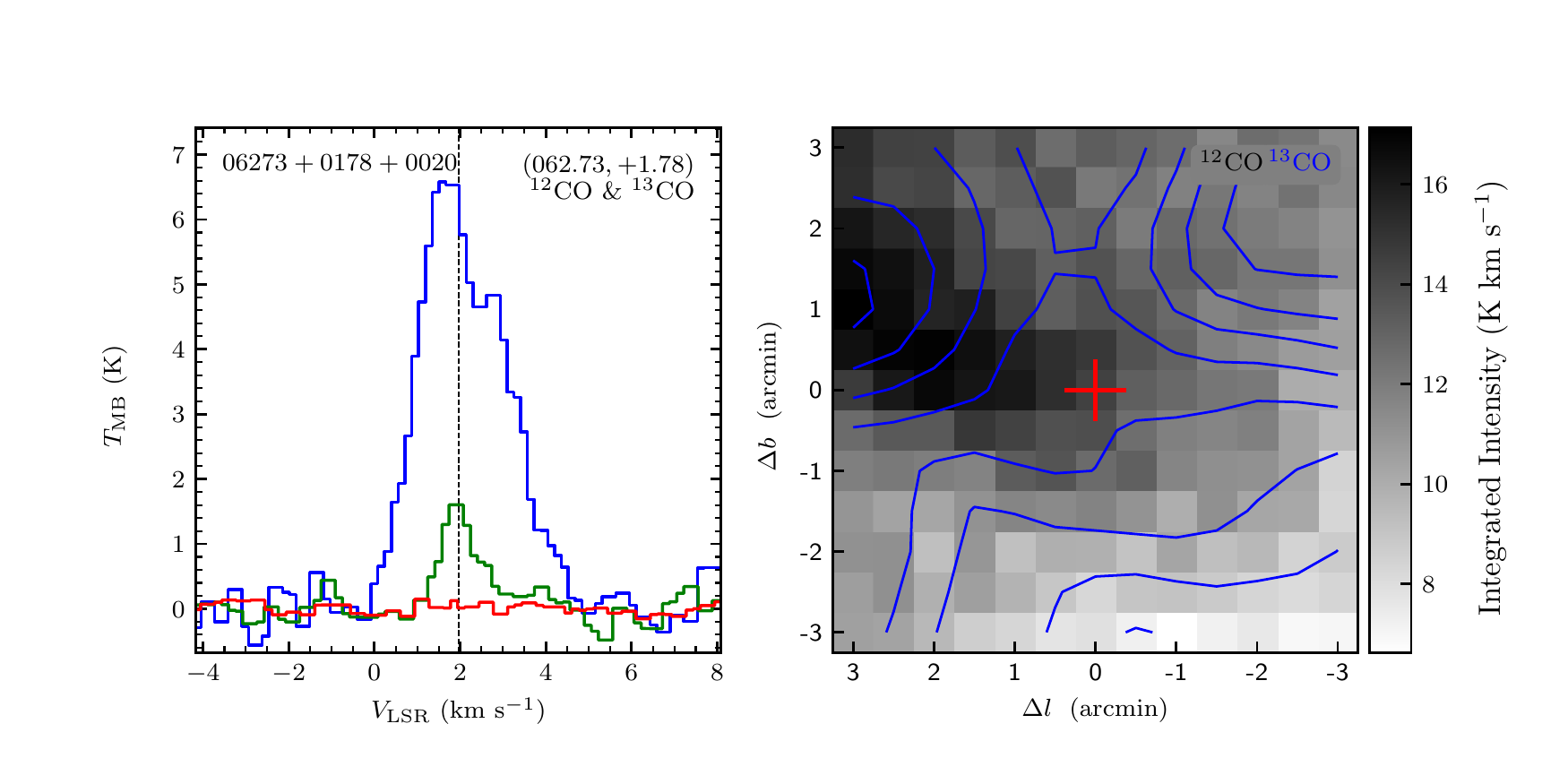}
\includegraphics[width=9.0cm,angle=0]{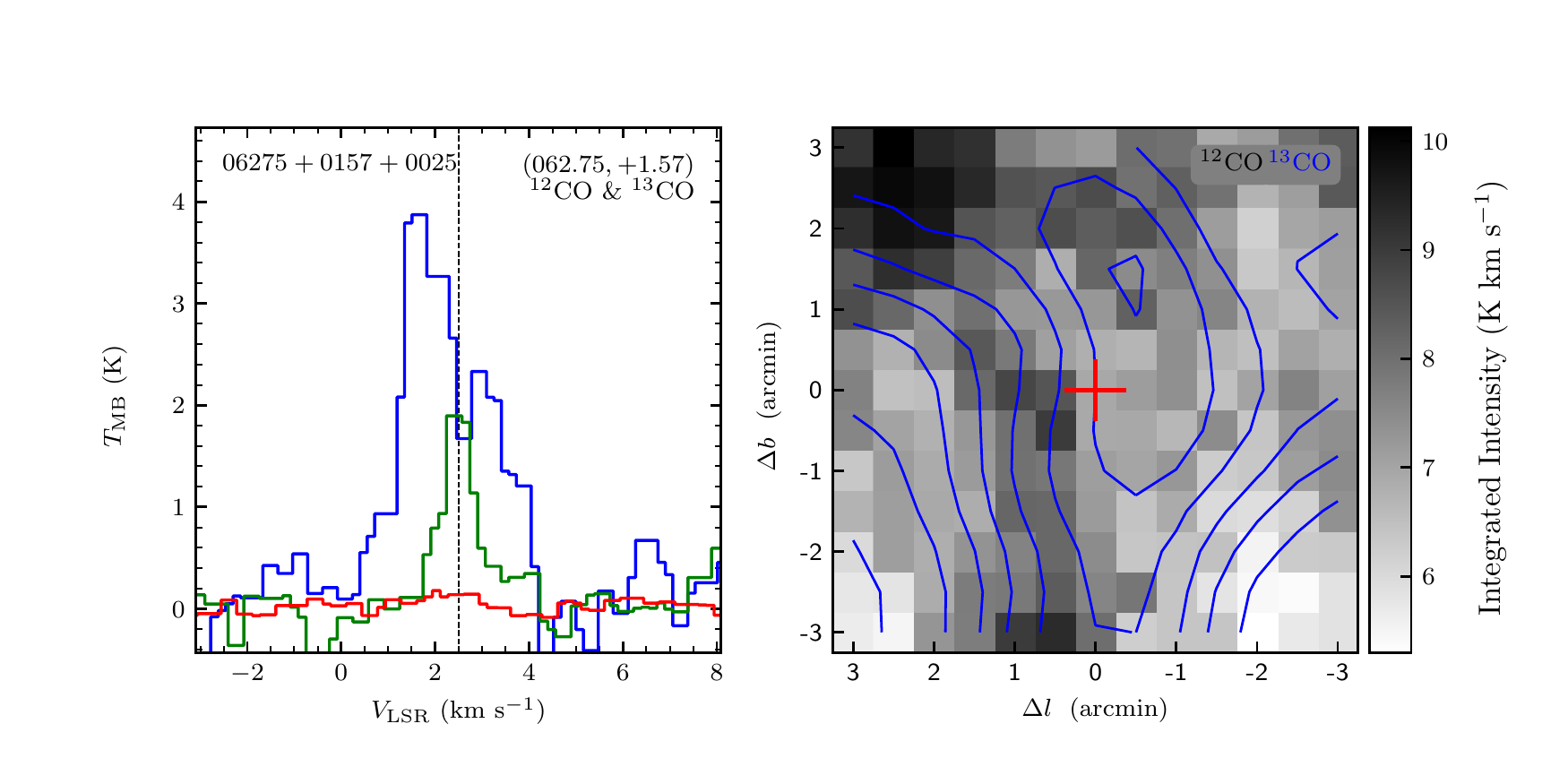}
\vspace{-0.5cm}

\includegraphics[width=9.0cm,angle=0]{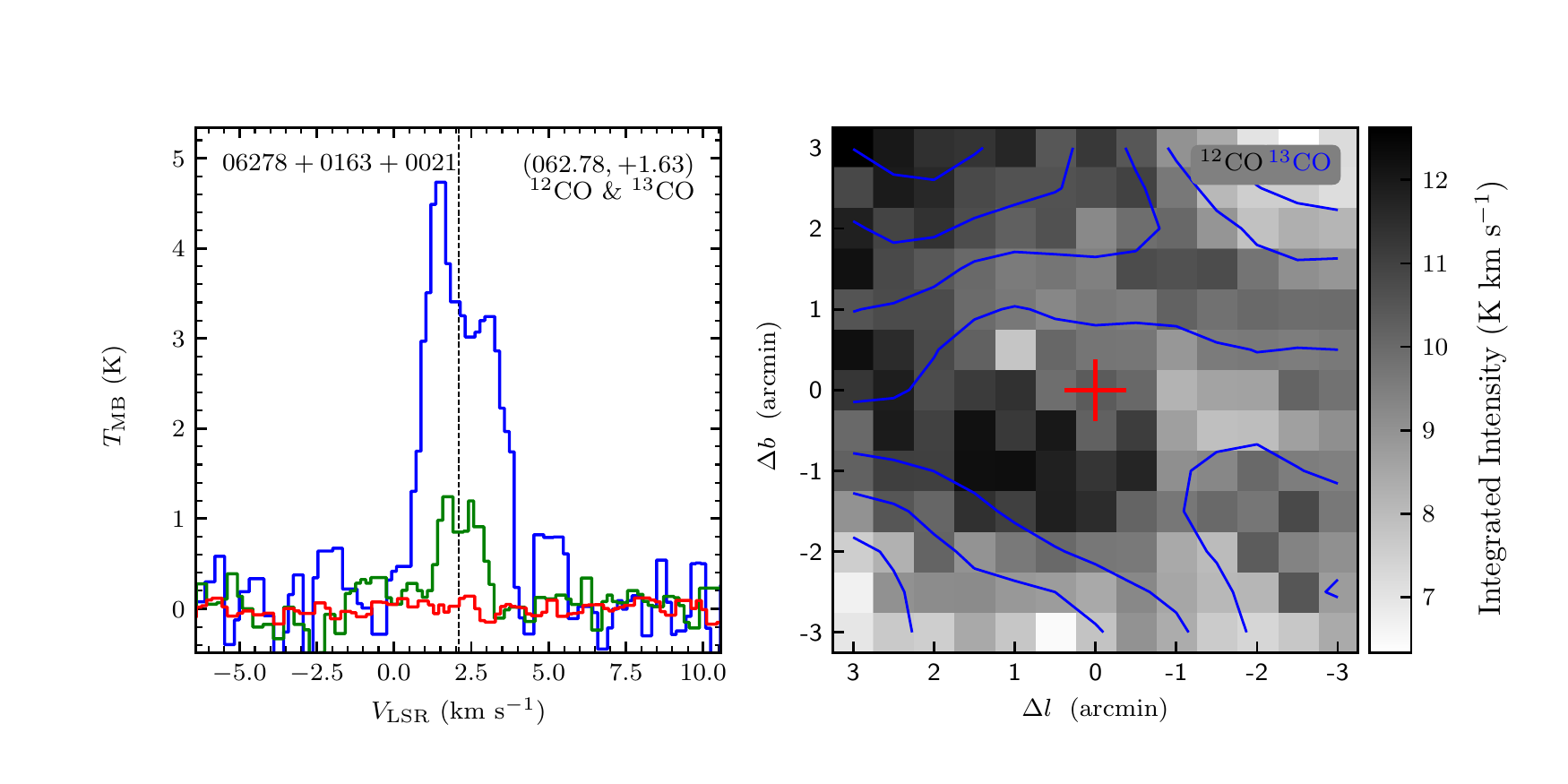}
\includegraphics[width=9.0cm,angle=0]{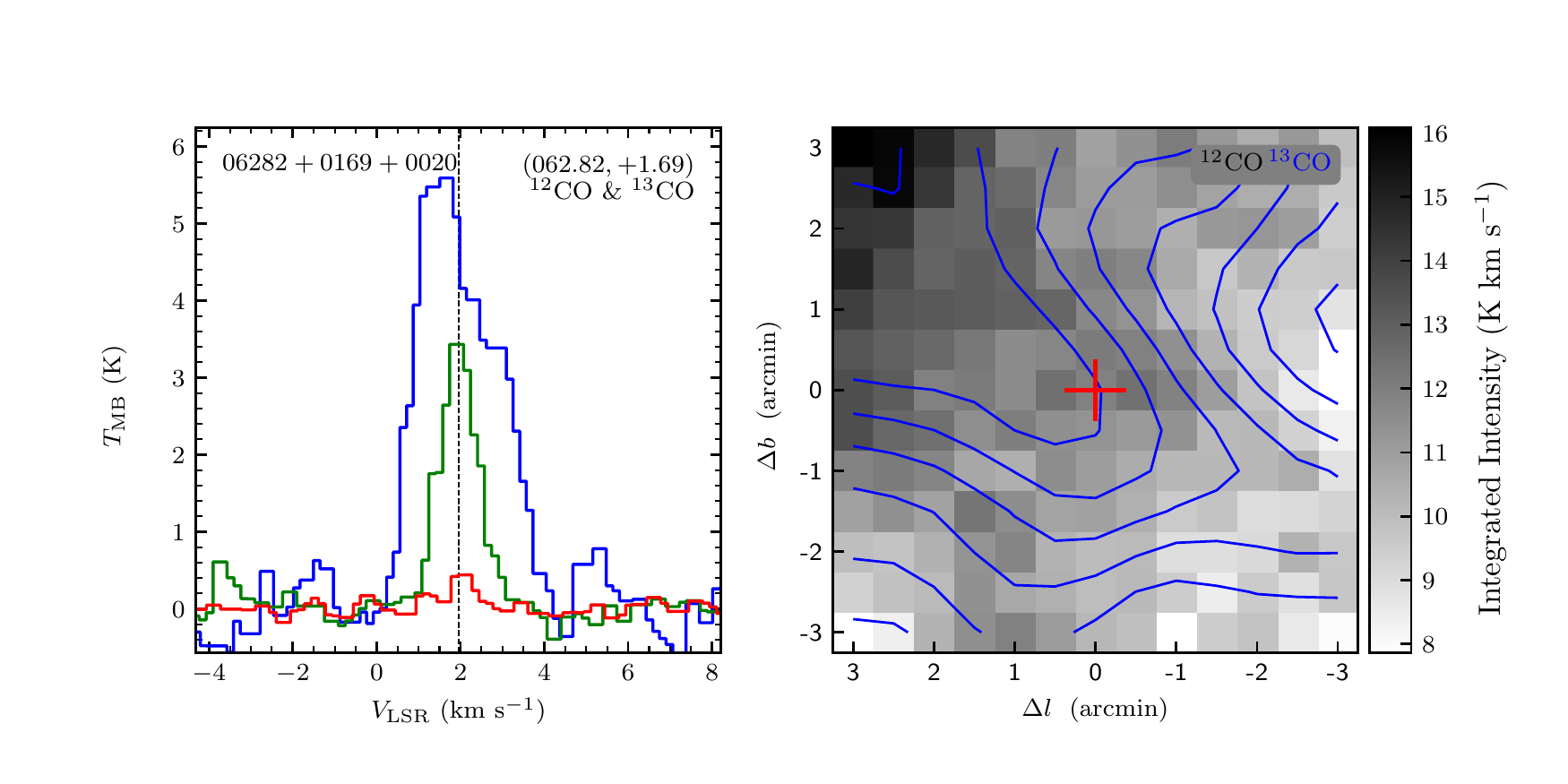}
\vspace{-0.5cm}

\includegraphics[width=9.0cm,angle=0]{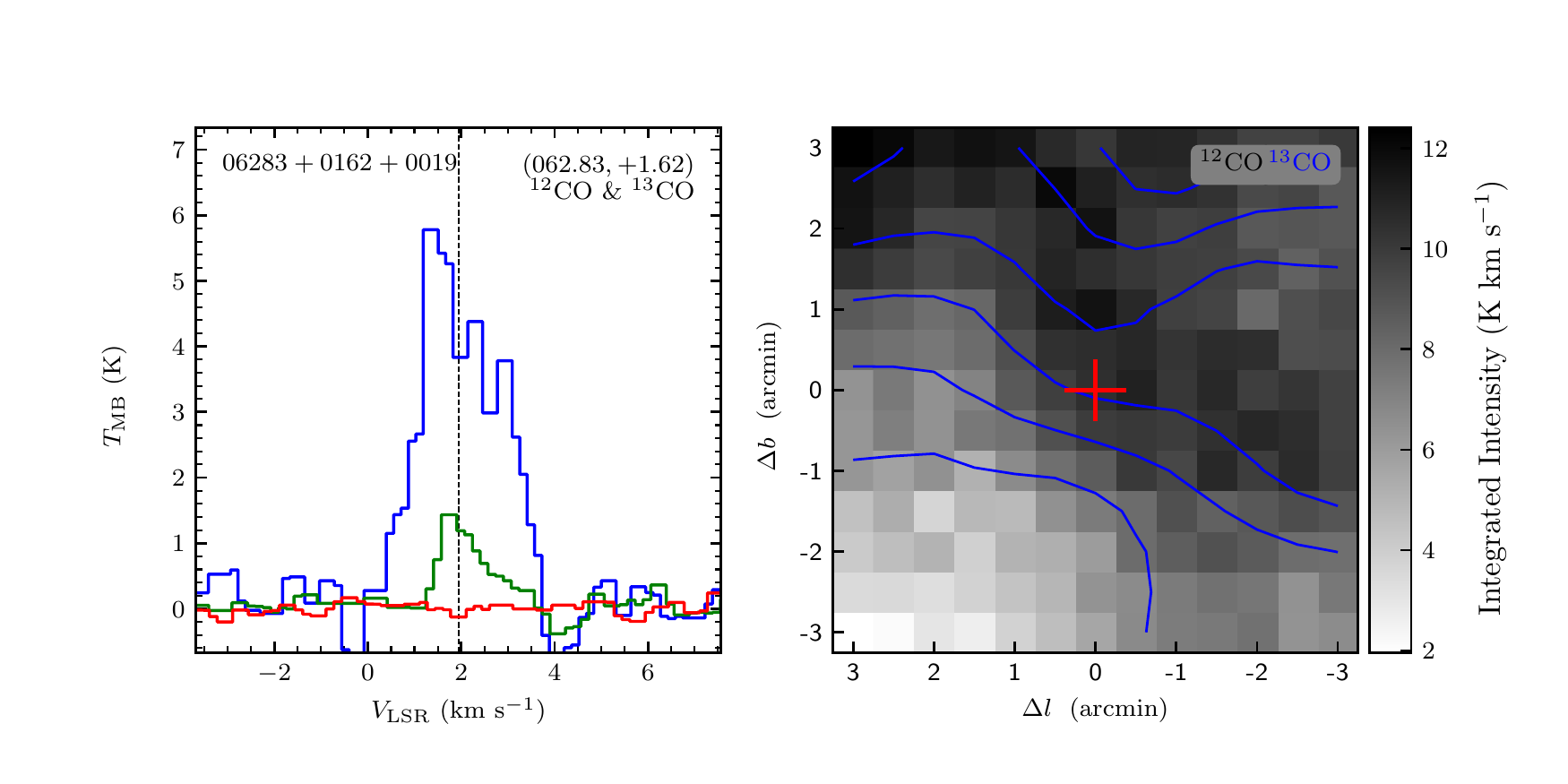}
\includegraphics[width=9.0cm,angle=0]{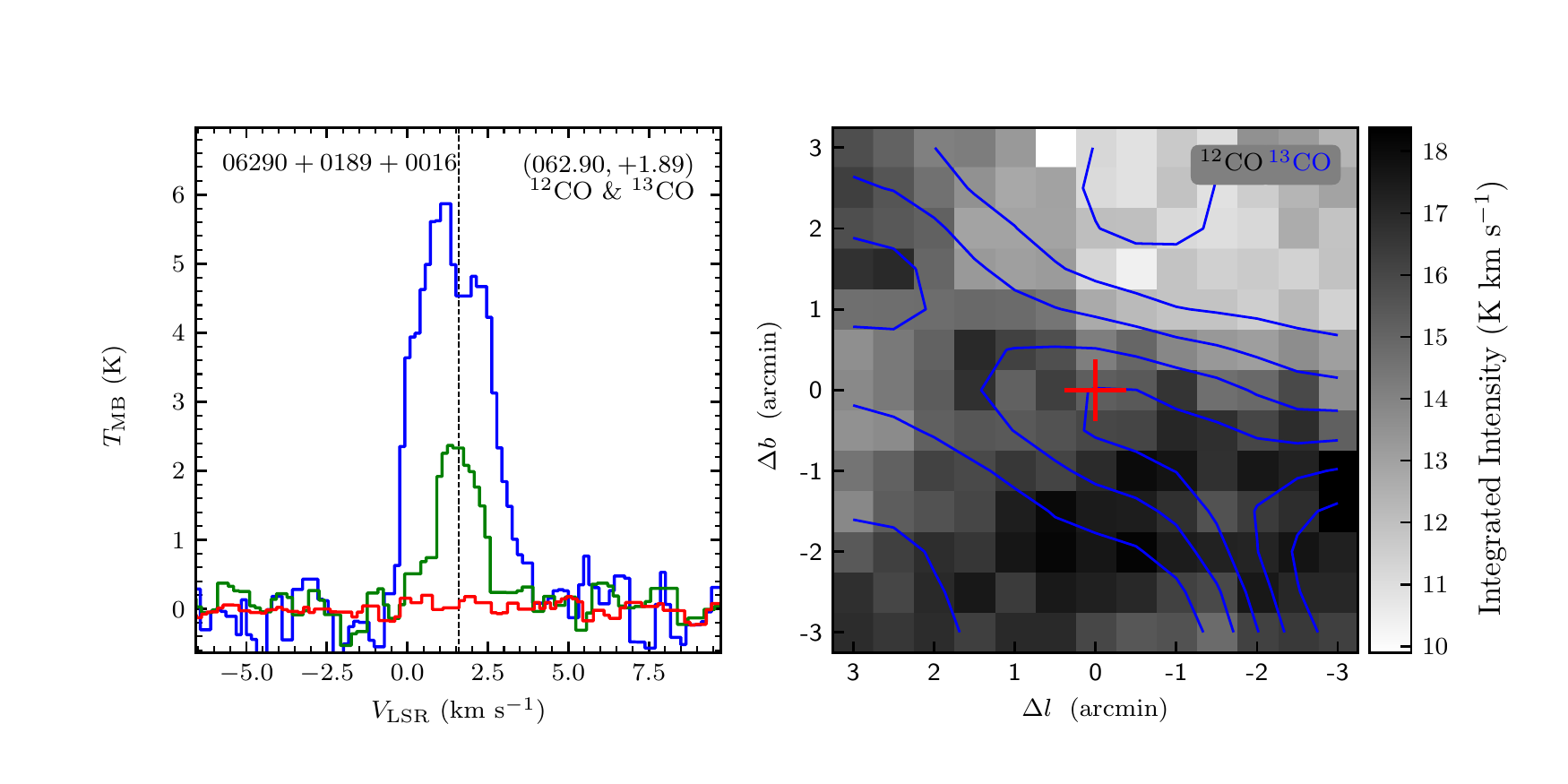}
\end{figure}
\clearpage

\begin{figure}
\includegraphics[width=9.0cm,angle=0]{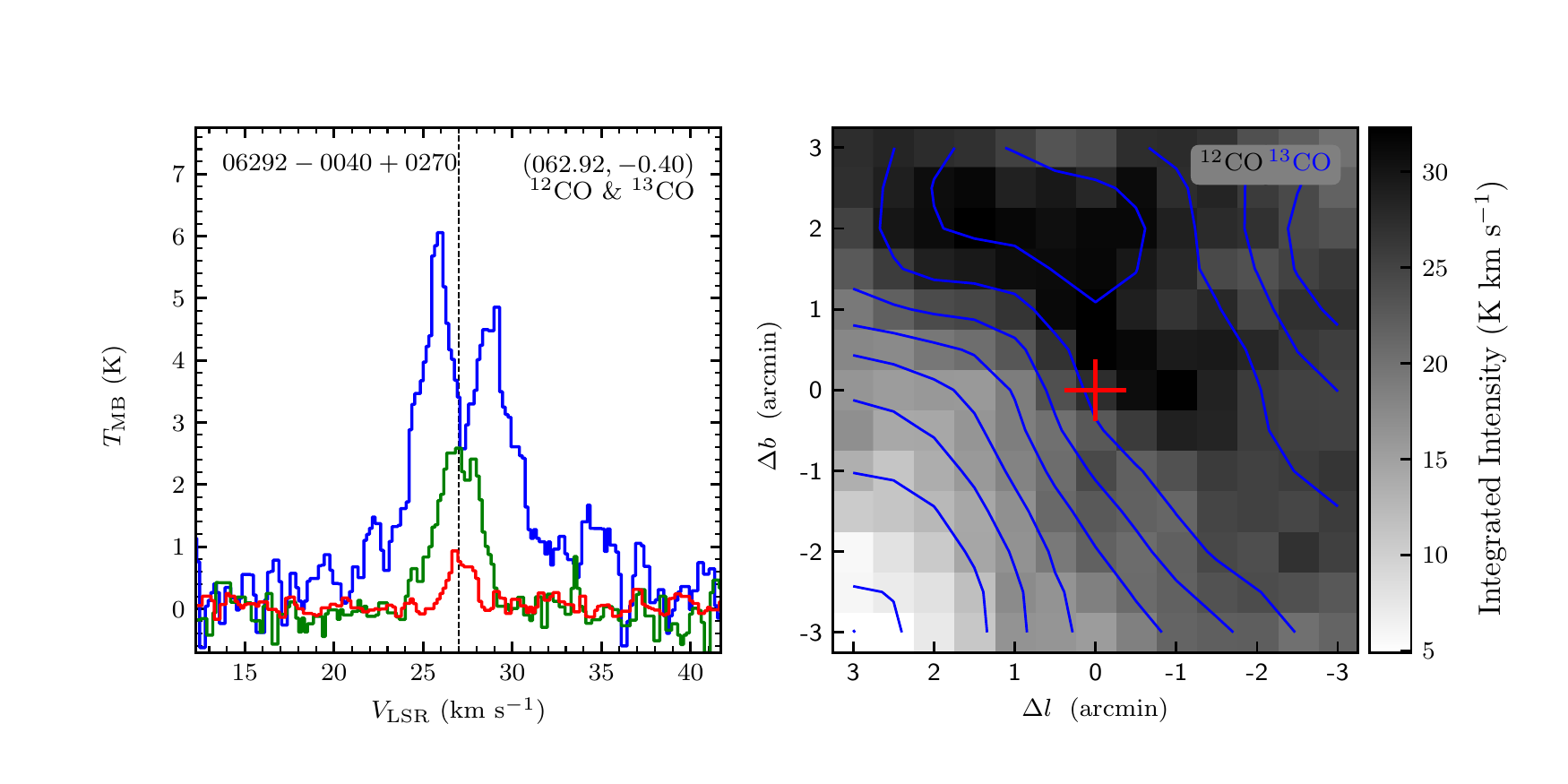}
\includegraphics[width=9.0cm,angle=0]{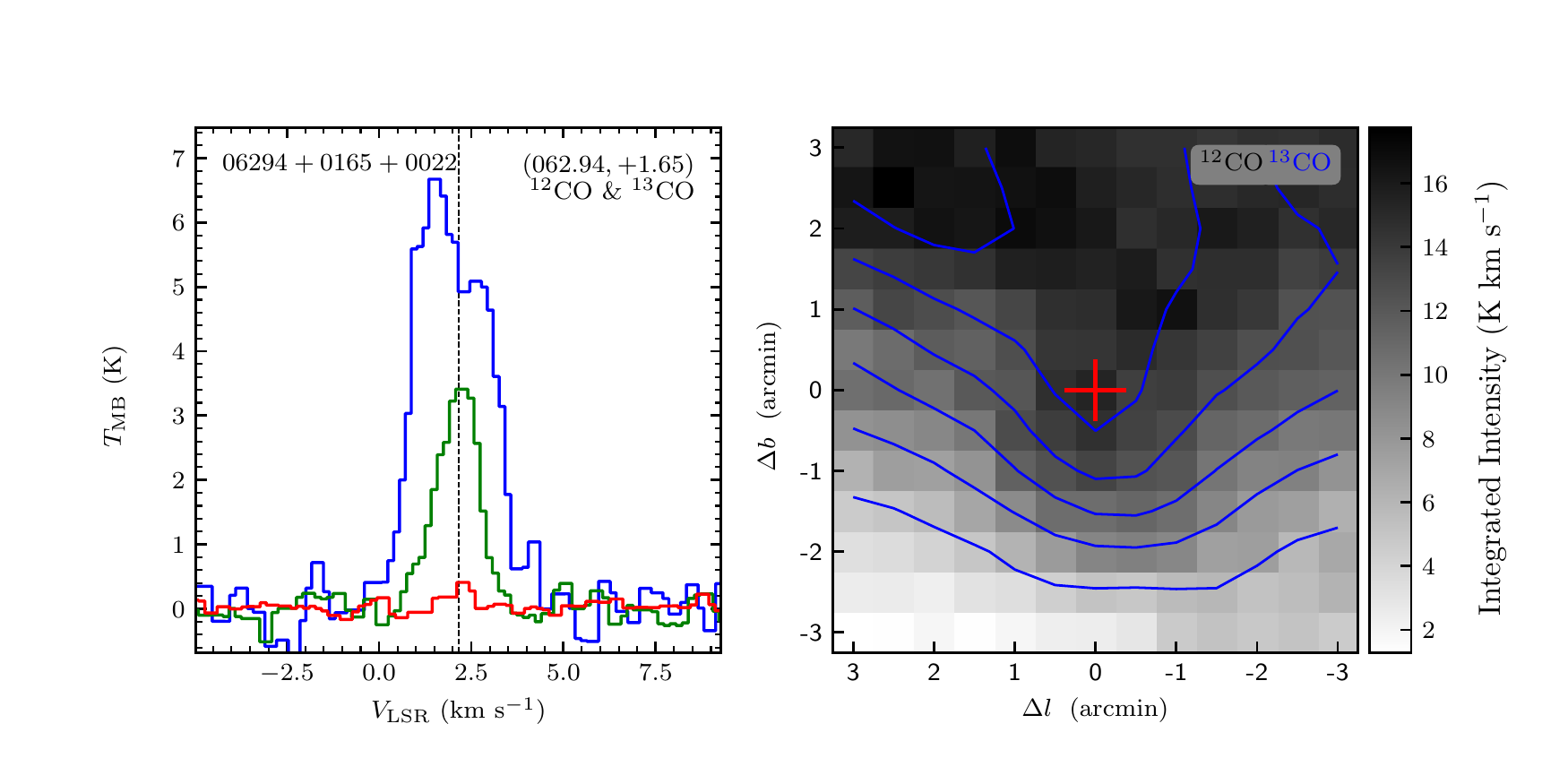}
\vspace{-0.5cm}

\includegraphics[width=9.0cm,angle=0]{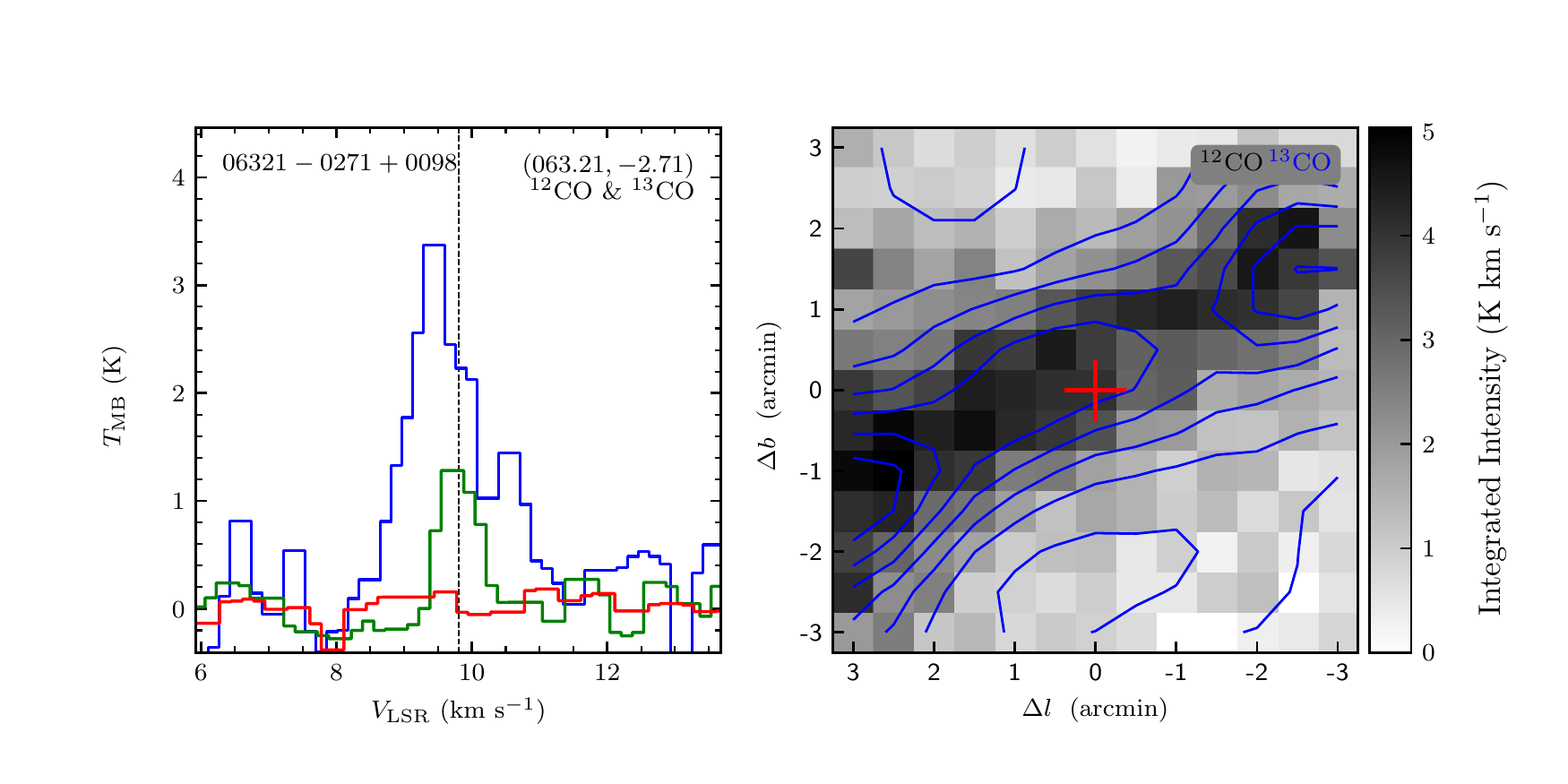}
\includegraphics[width=9.0cm,angle=0]{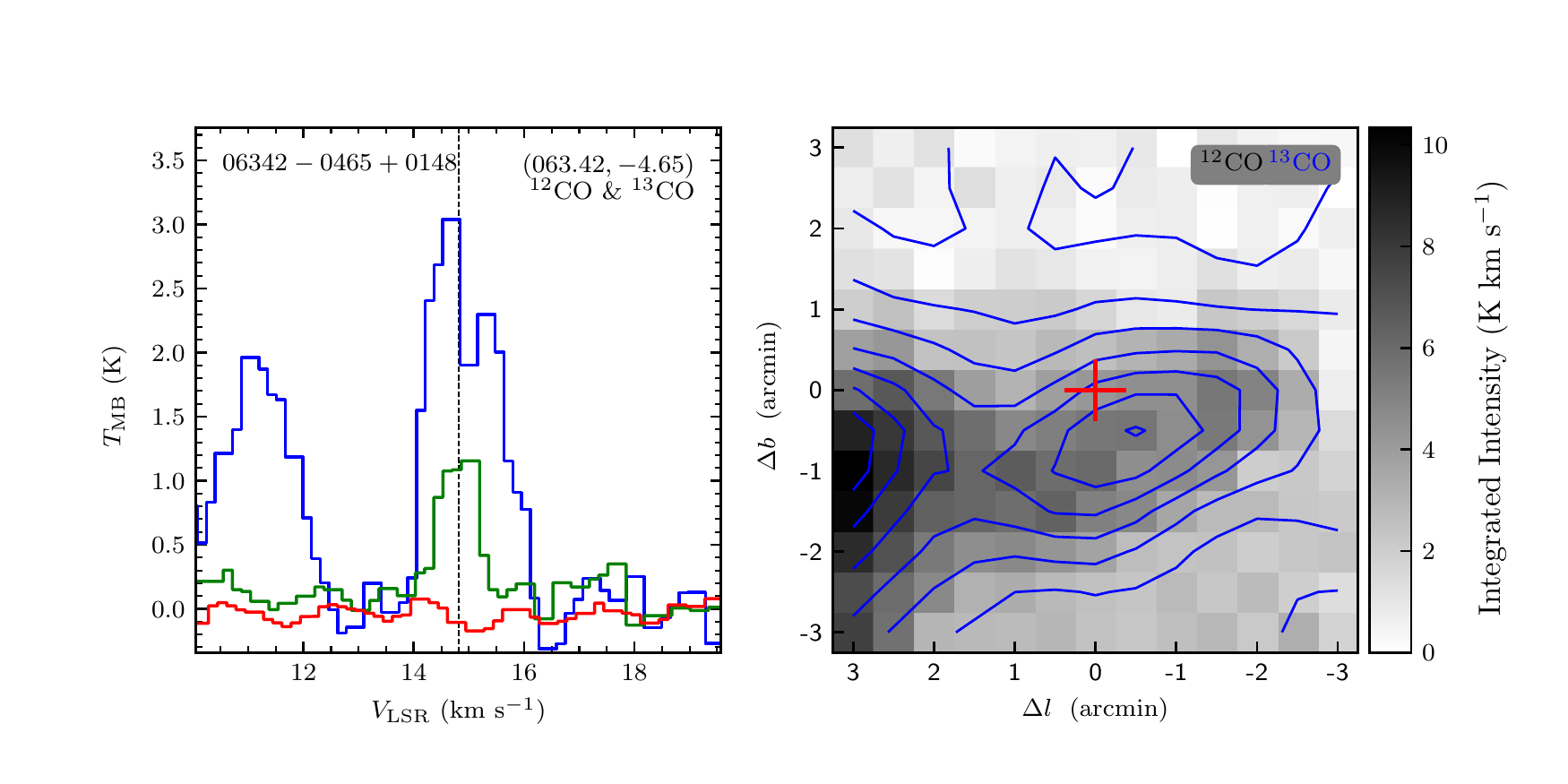}
\vspace{-0.5cm}

\includegraphics[width=9.0cm,angle=0]{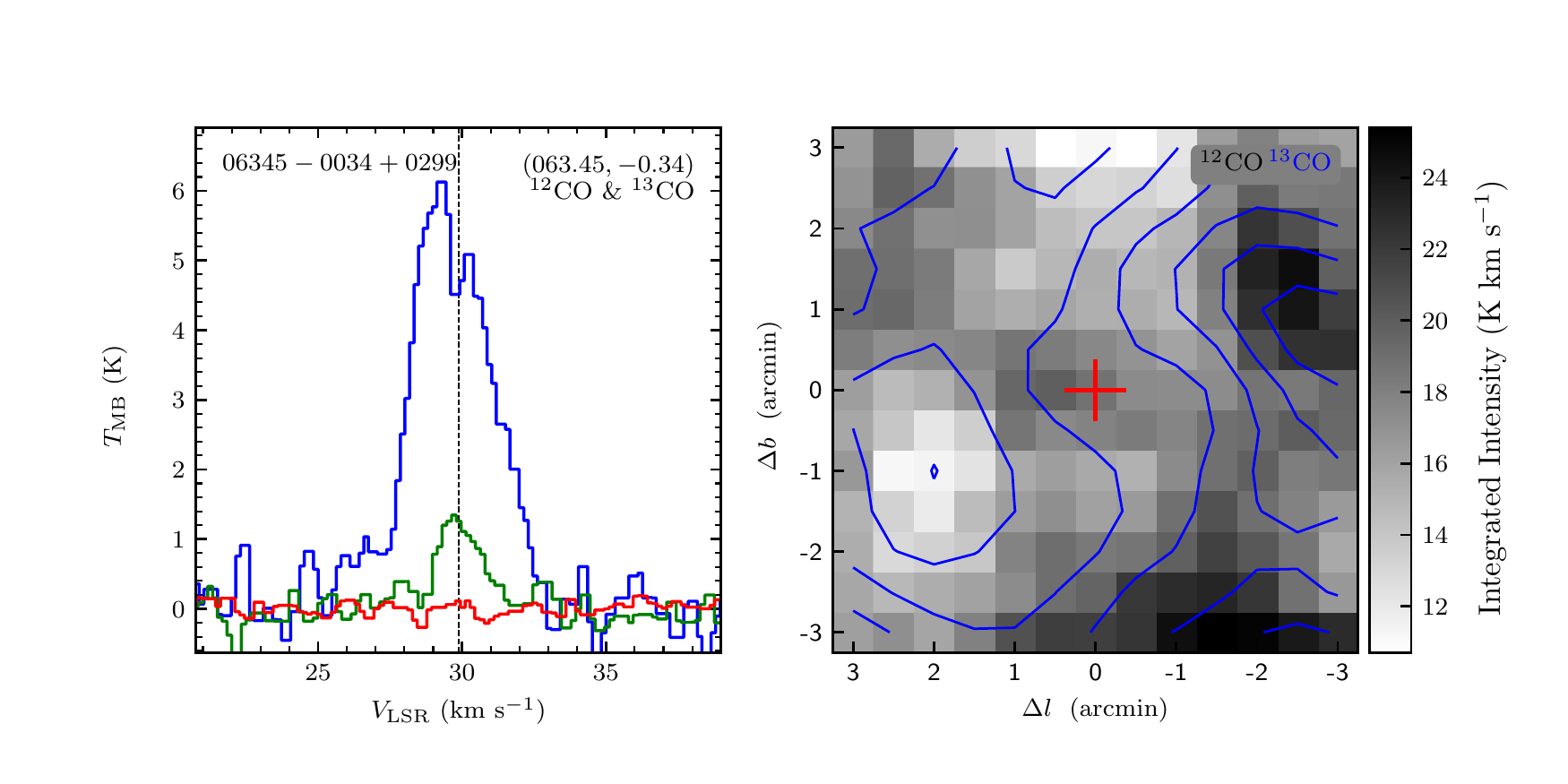}
\includegraphics[width=9.0cm,angle=0]{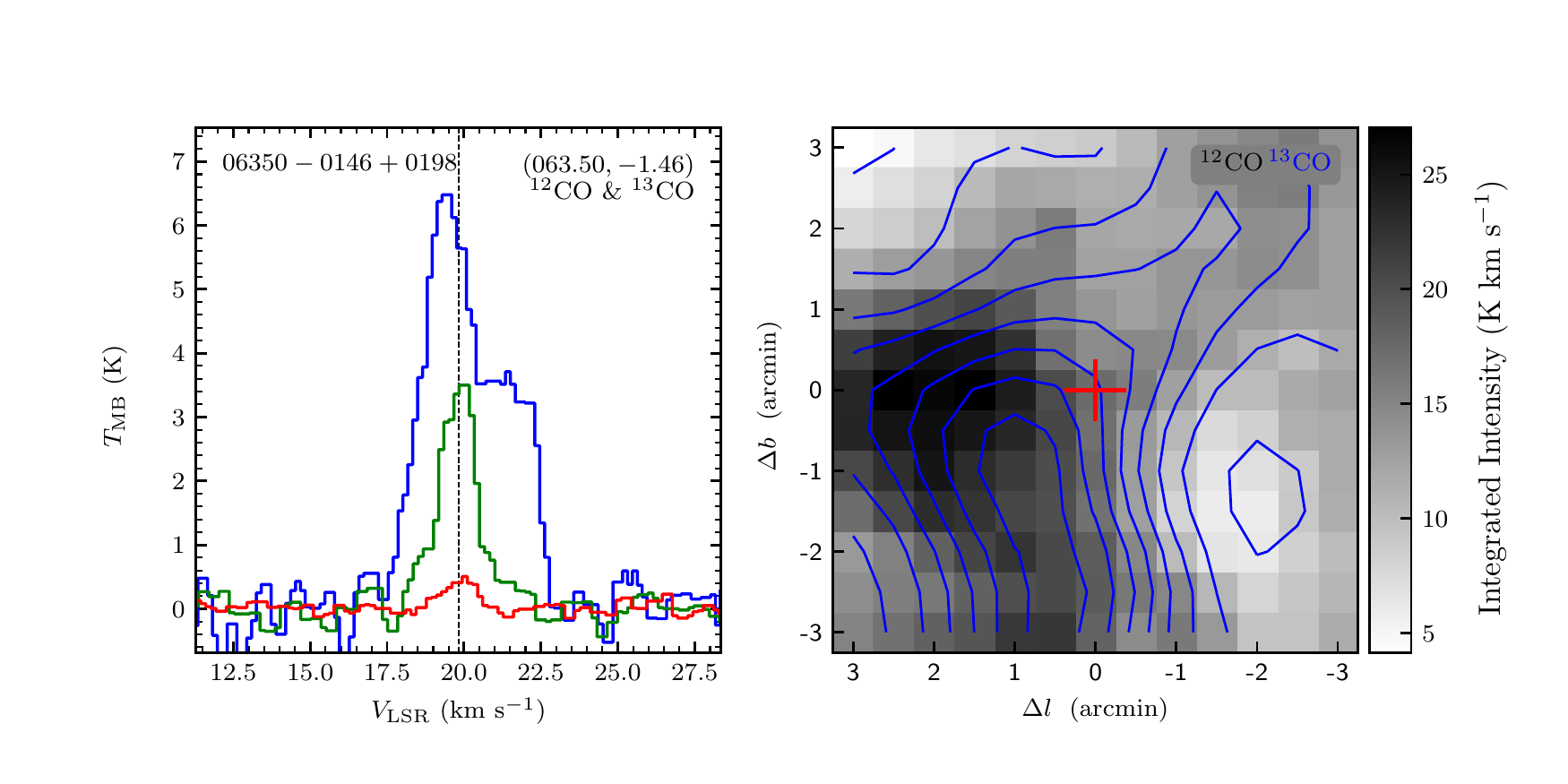}
\vspace{-0.5cm}

\includegraphics[width=9.0cm,angle=0]{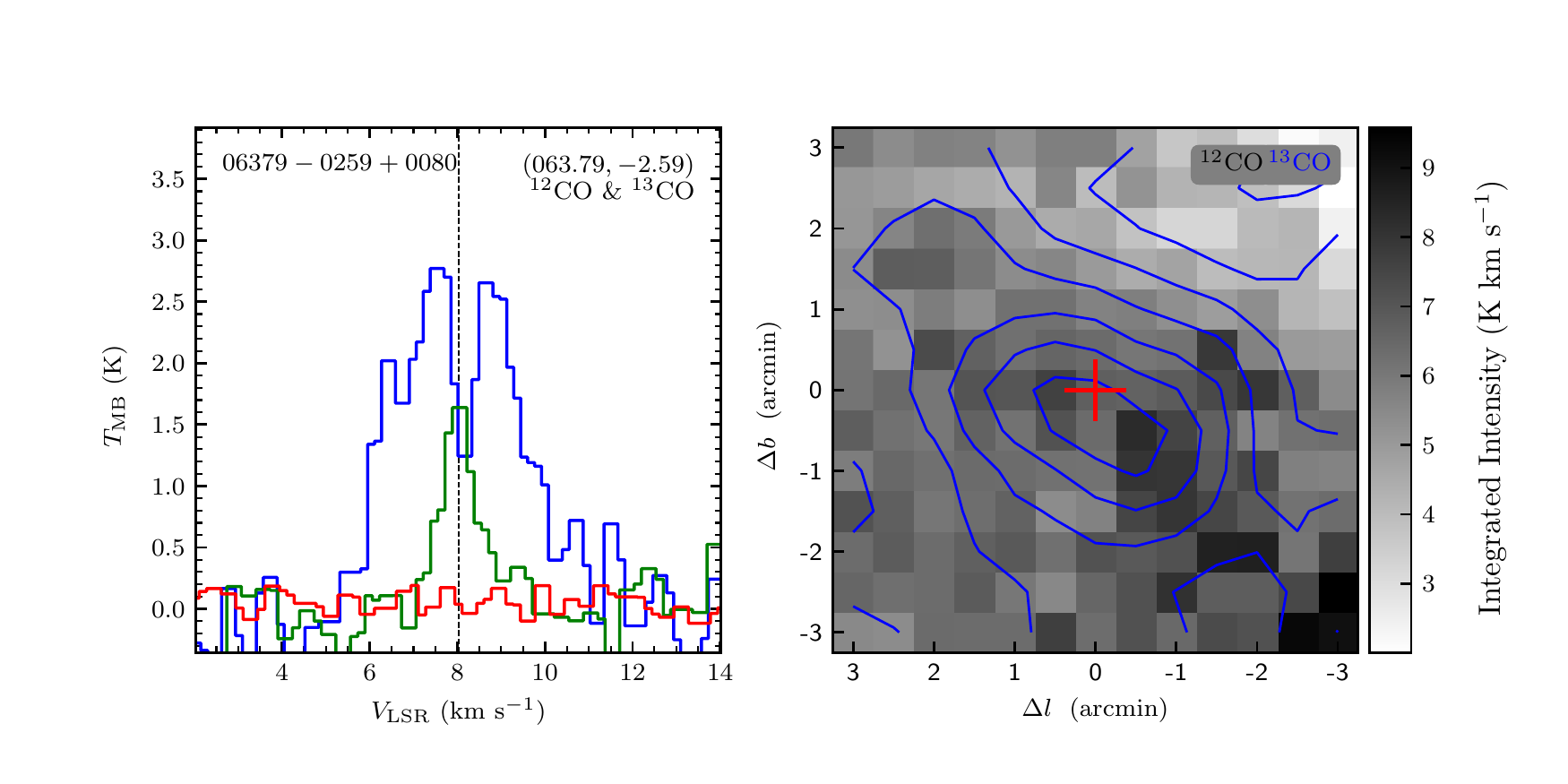}
\includegraphics[width=9.0cm,angle=0]{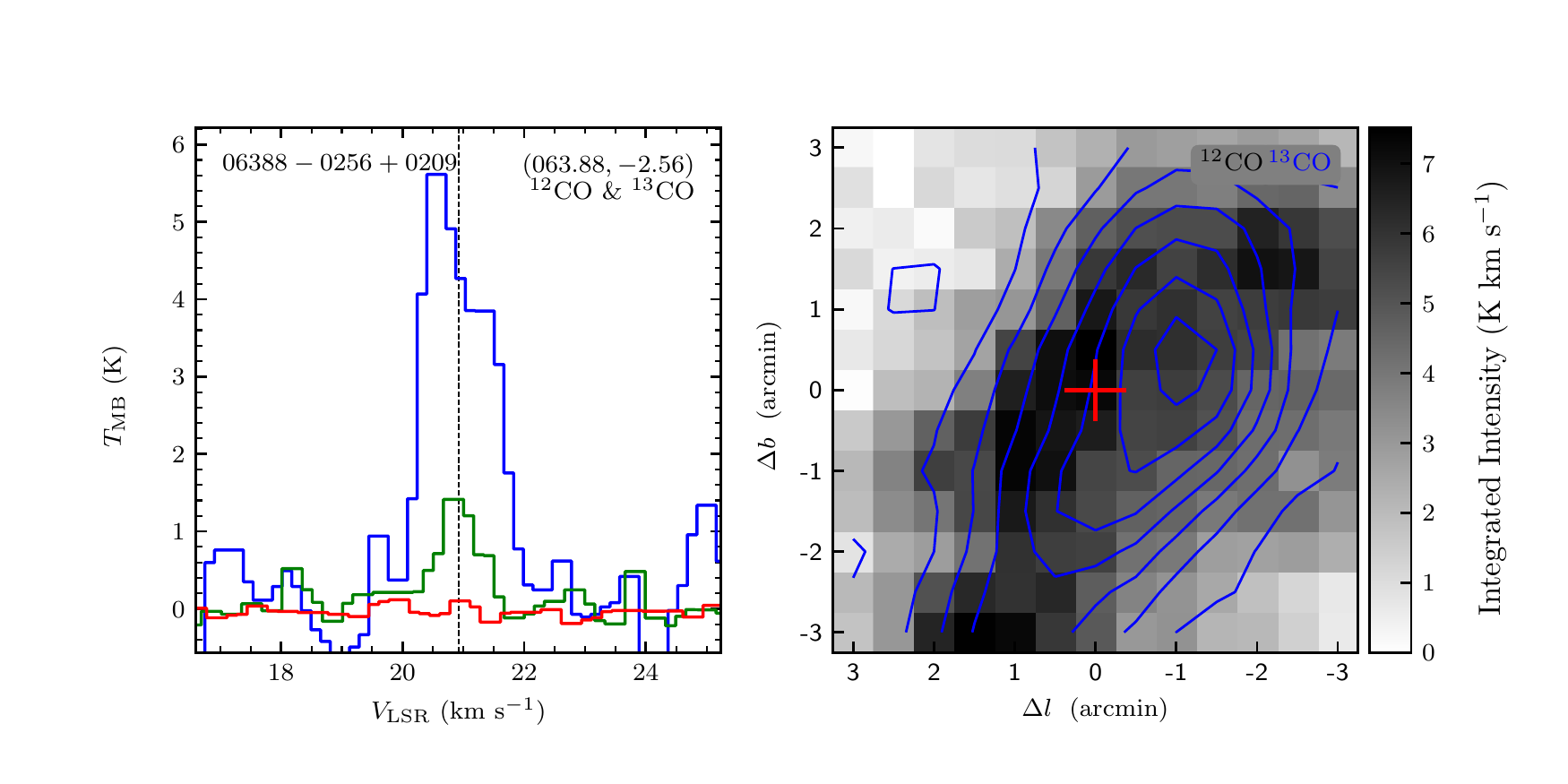}
\vspace{-0.5cm}

\includegraphics[width=9.0cm,angle=0]{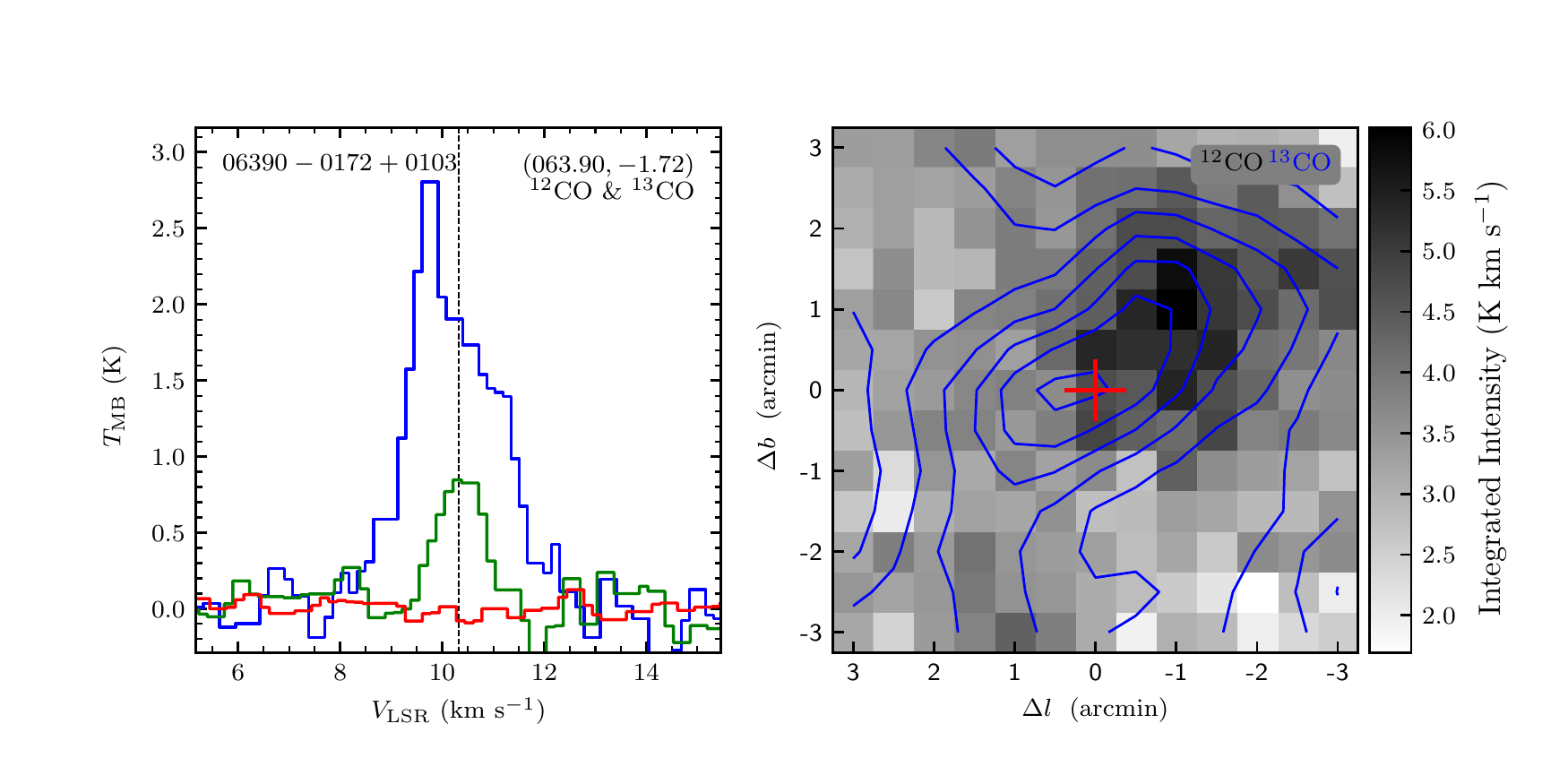}
\includegraphics[width=9.0cm,angle=0]{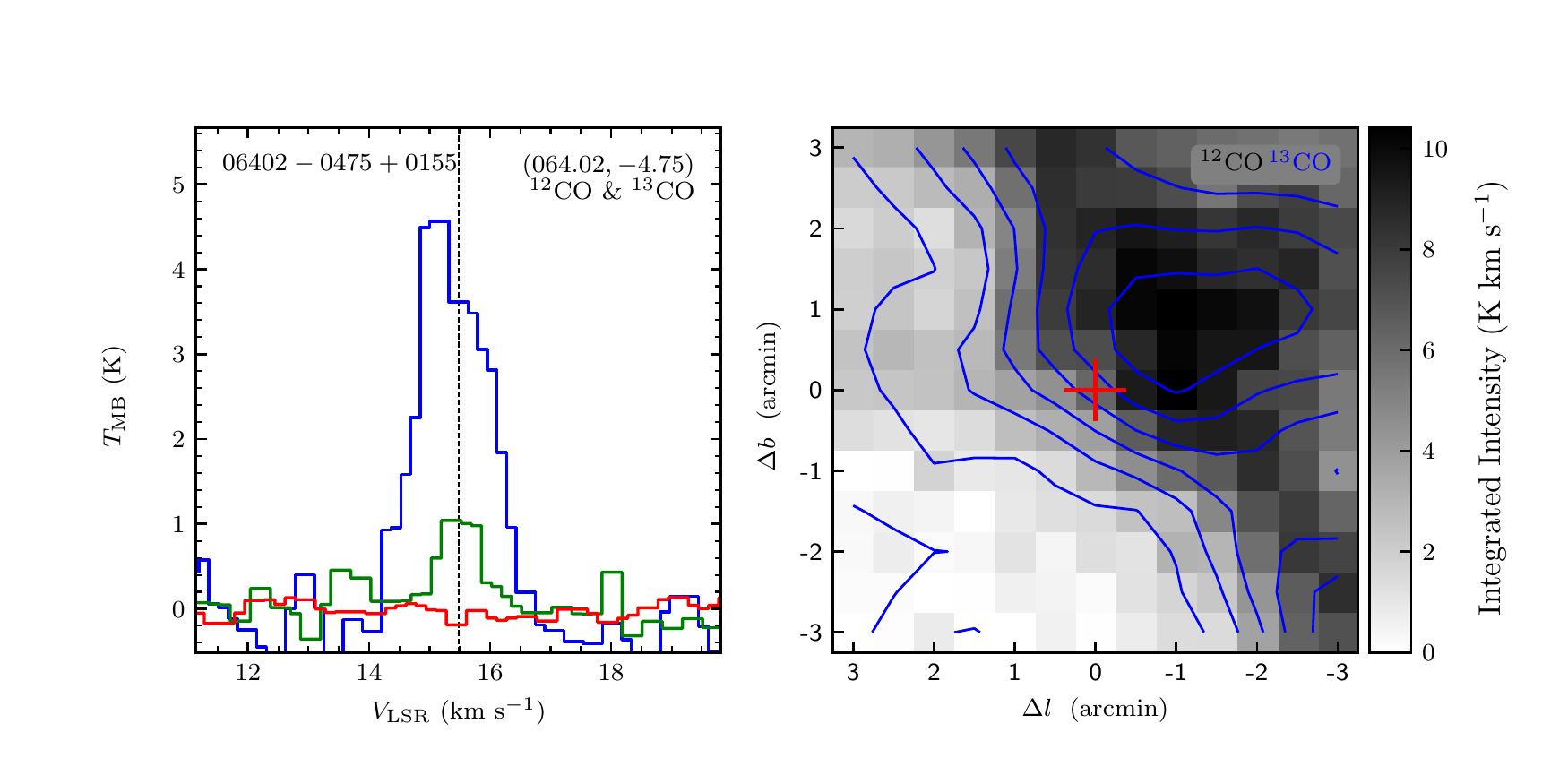}
\end{figure}
\clearpage

\begin{figure}
\includegraphics[width=9.0cm,angle=0]{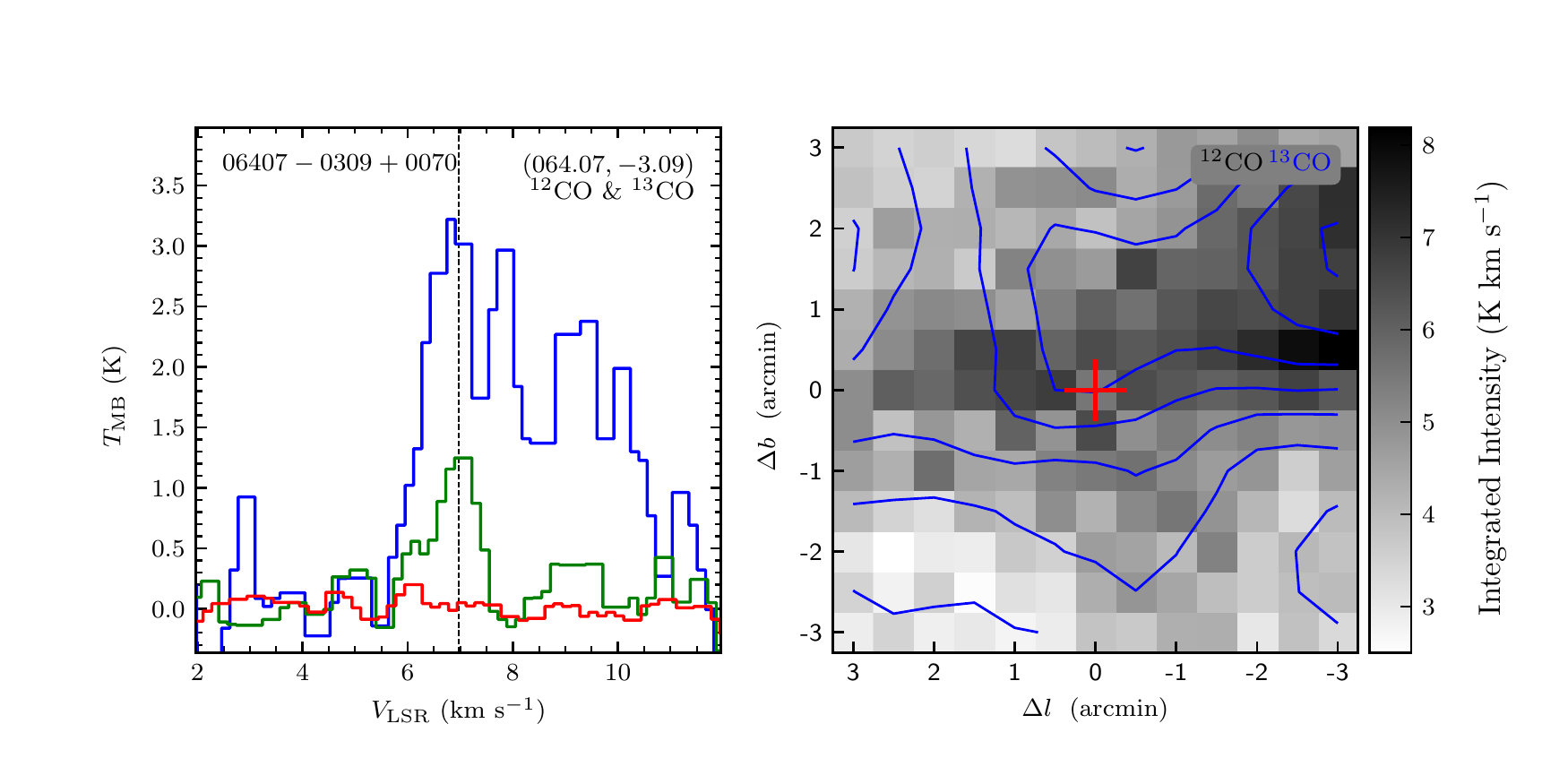}
\includegraphics[width=9.0cm,angle=0]{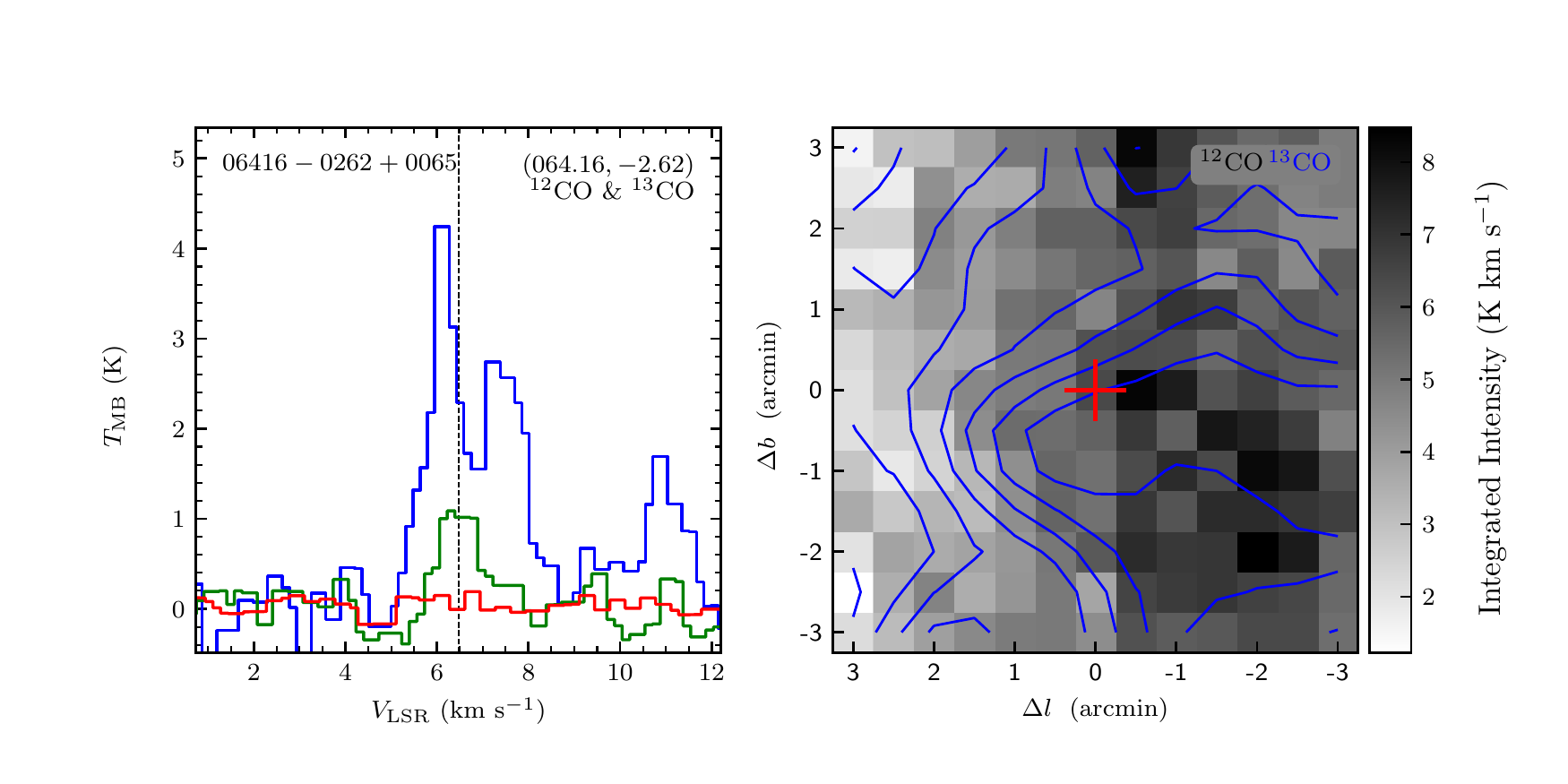}
\vspace{-0.5cm}

\includegraphics[width=9.0cm,angle=0]{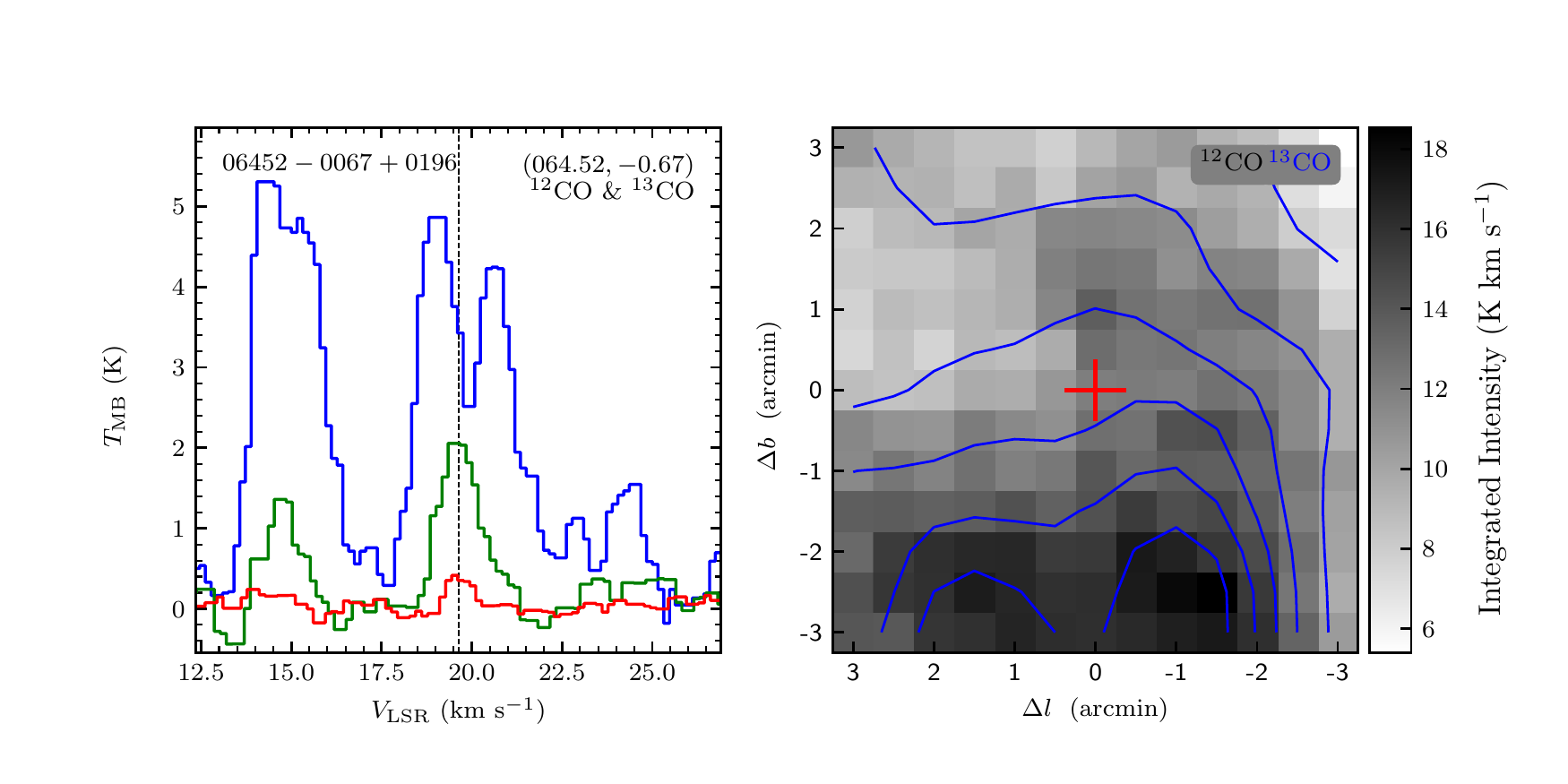}
\includegraphics[width=9.0cm,angle=0]{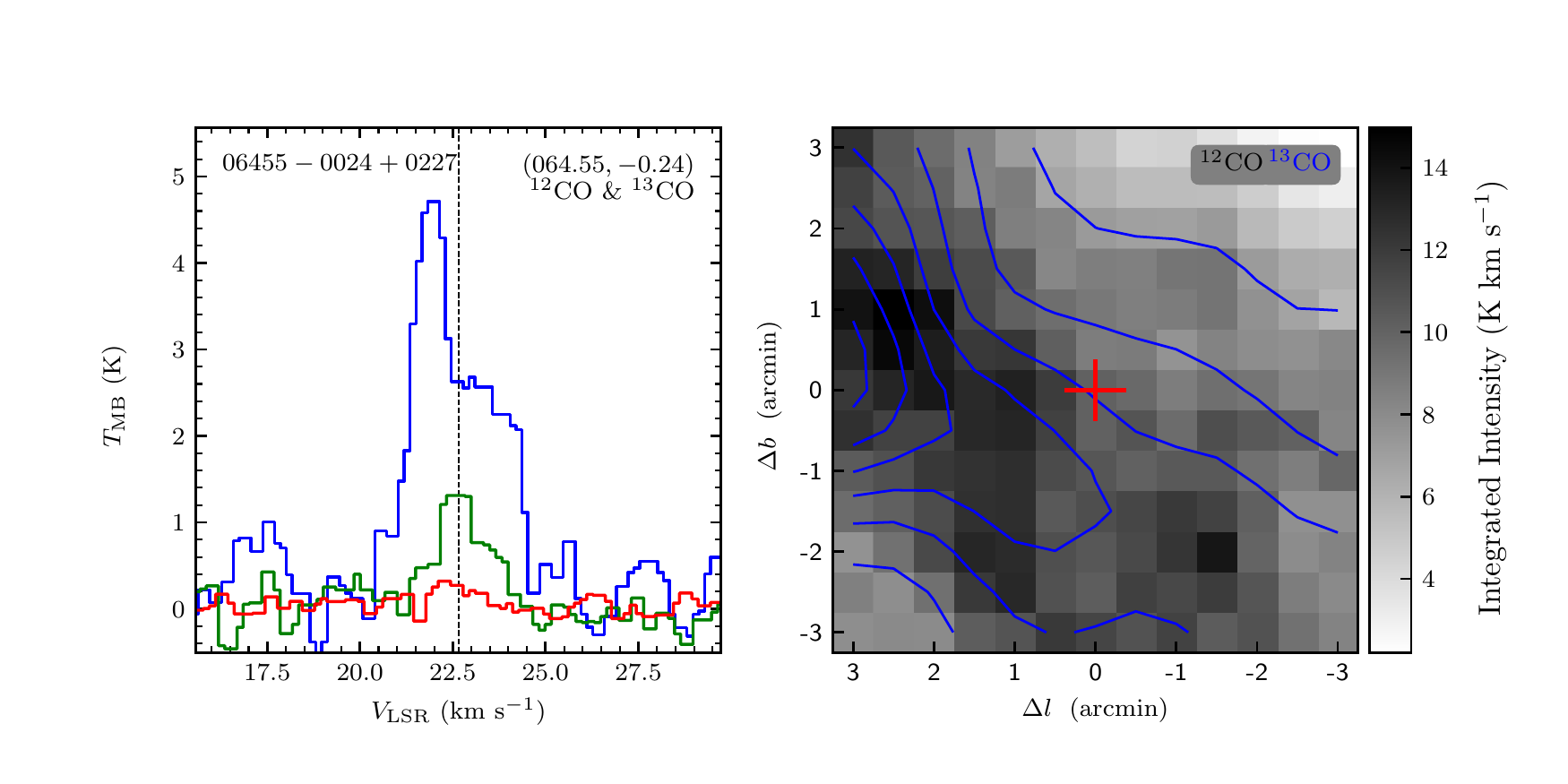}
\vspace{-0.5cm}

\includegraphics[width=9.0cm,angle=0]{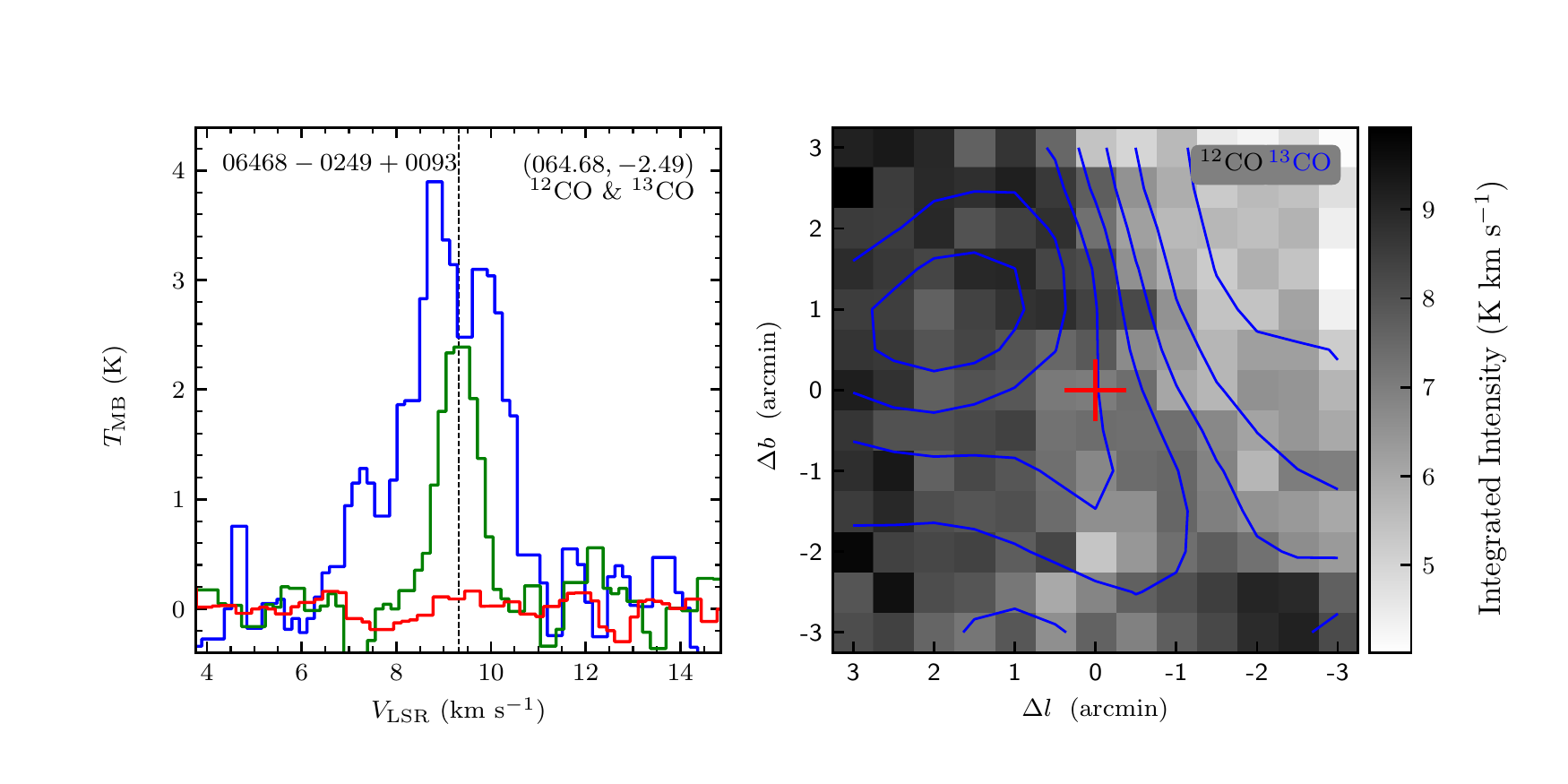}
\includegraphics[width=9.0cm,angle=0]{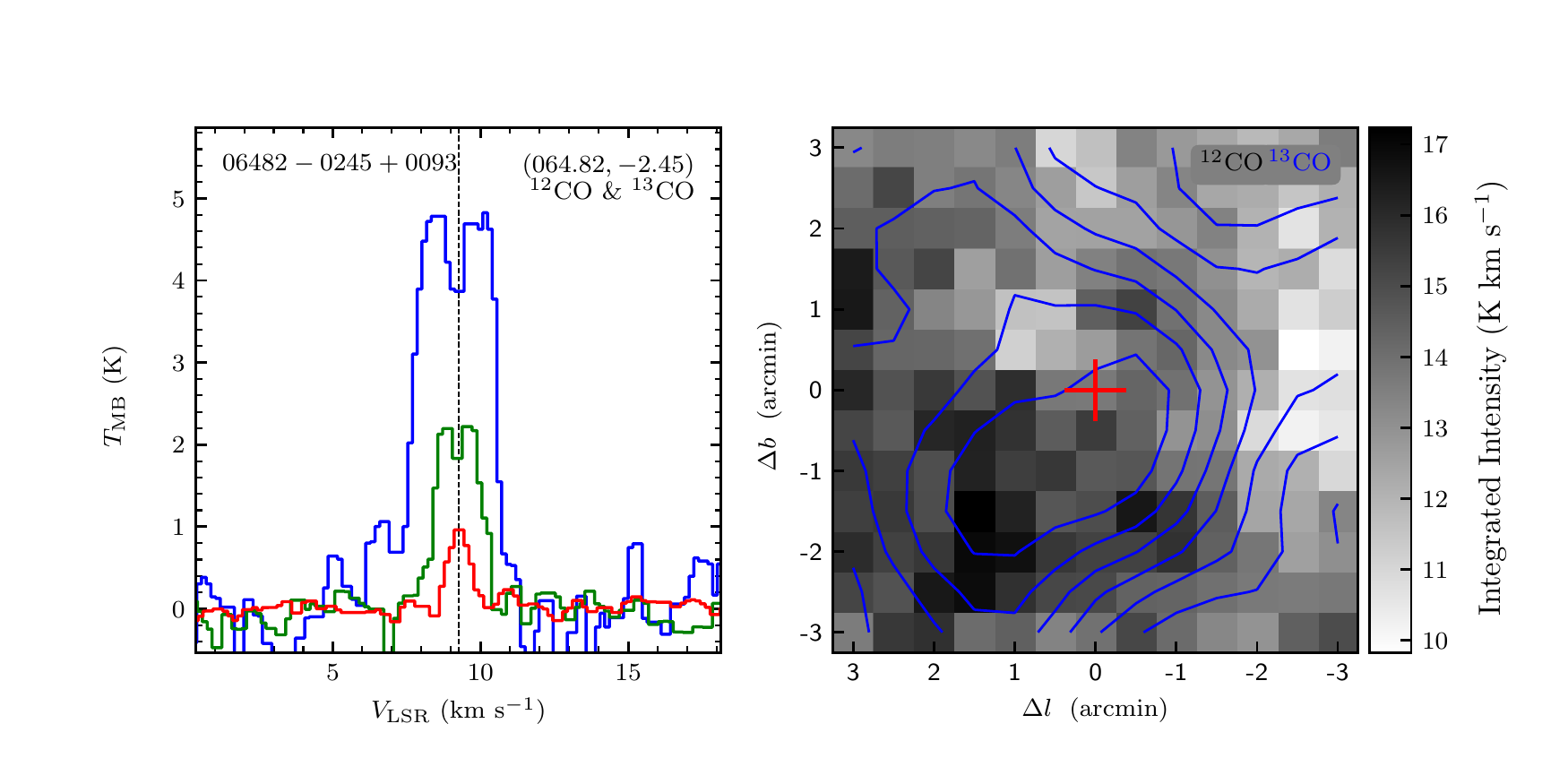}
\vspace{-0.5cm}

\includegraphics[width=9.0cm,angle=0]{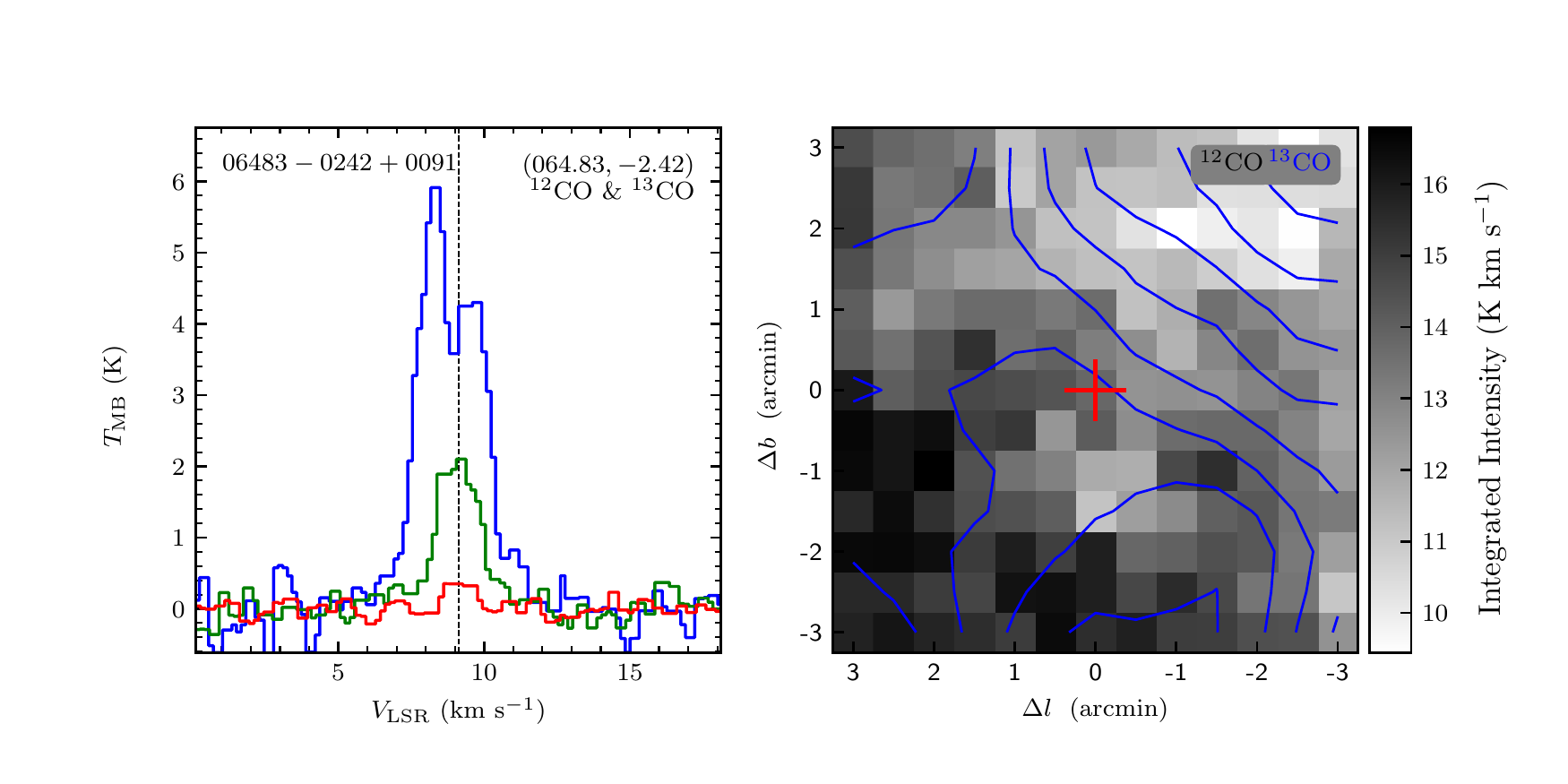}
\includegraphics[width=9.0cm,angle=0]{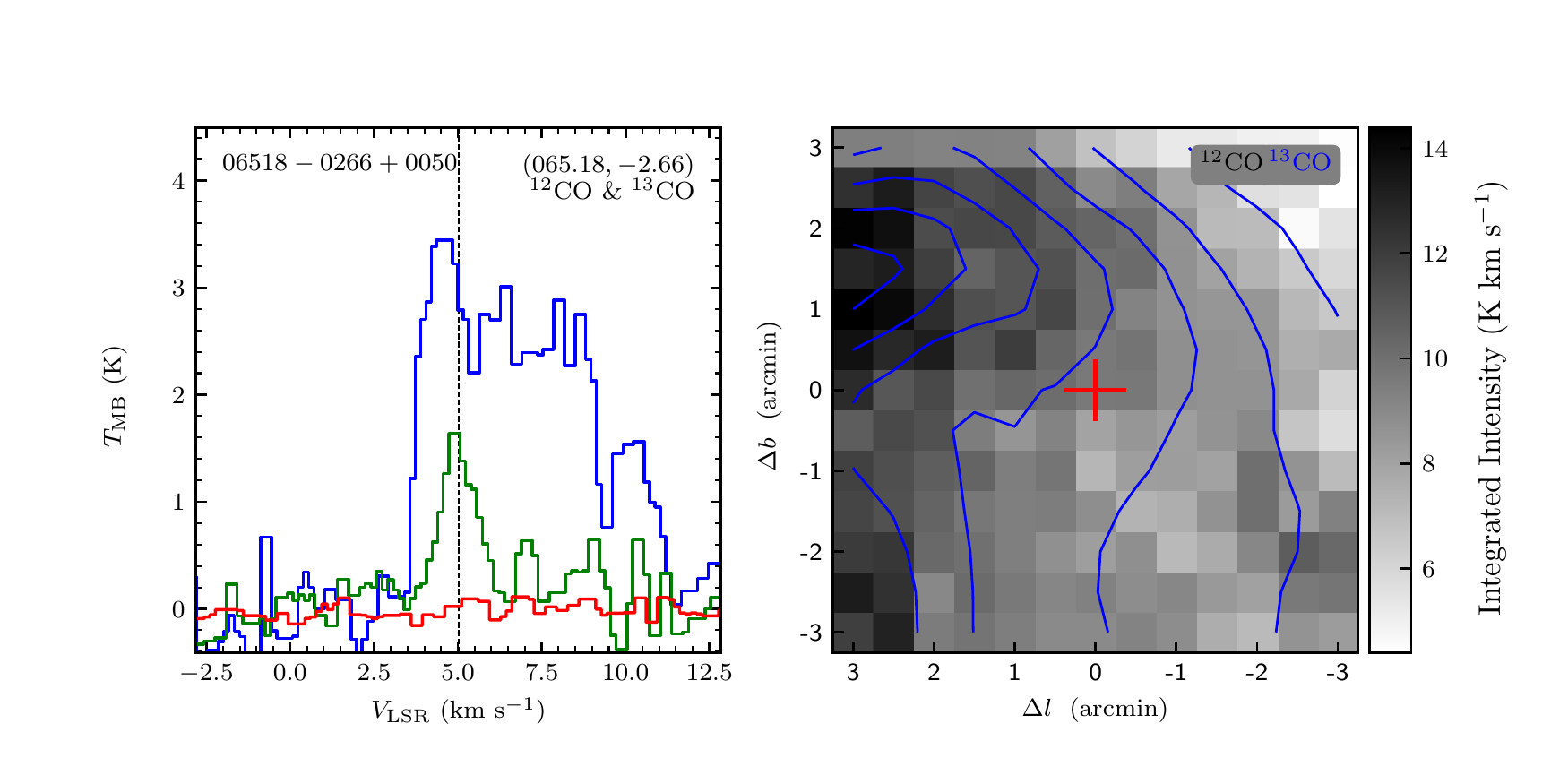}
\vspace{-0.5cm}

\includegraphics[width=9.0cm,angle=0]{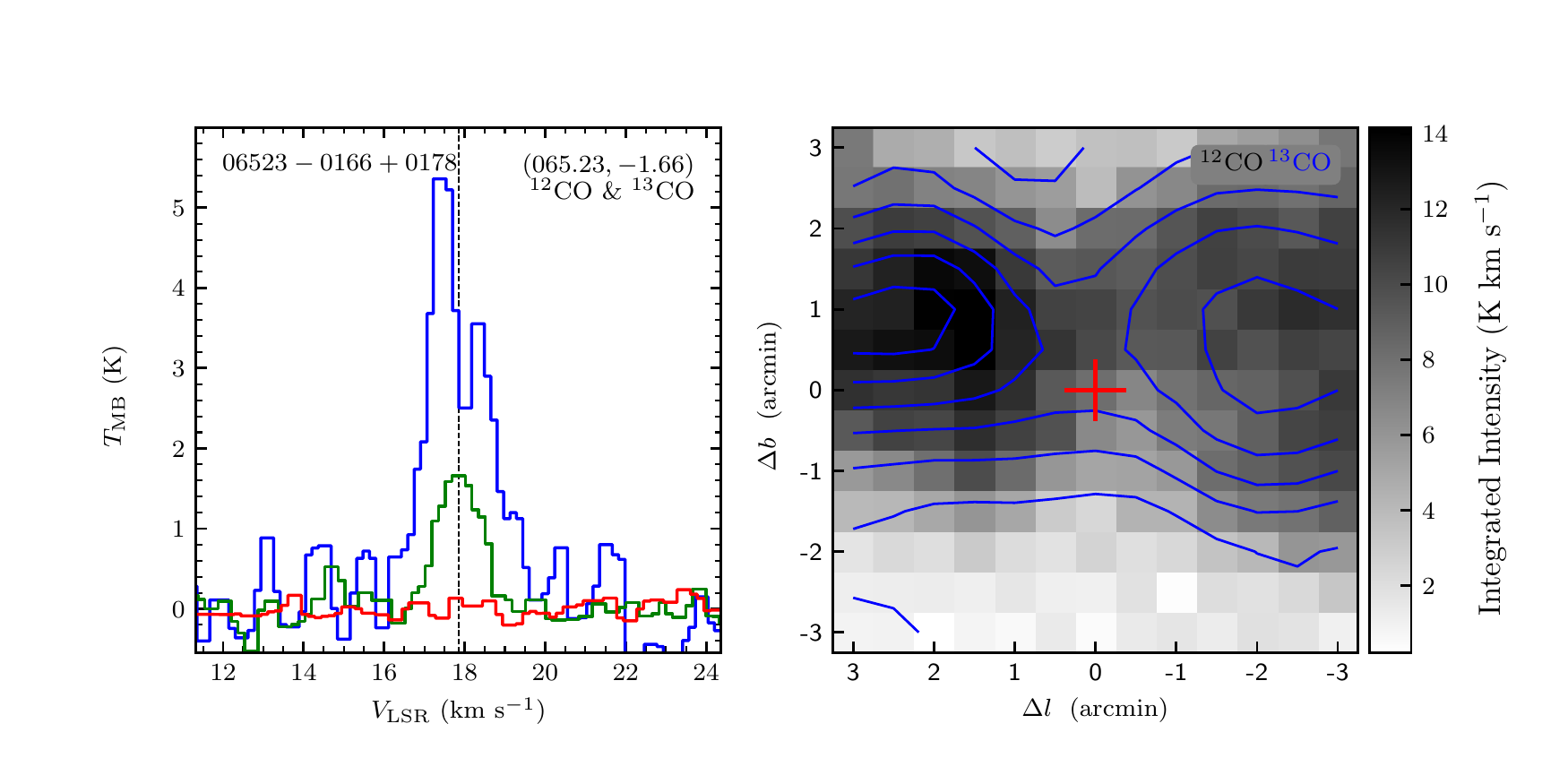}
\includegraphics[width=9.0cm,angle=0]{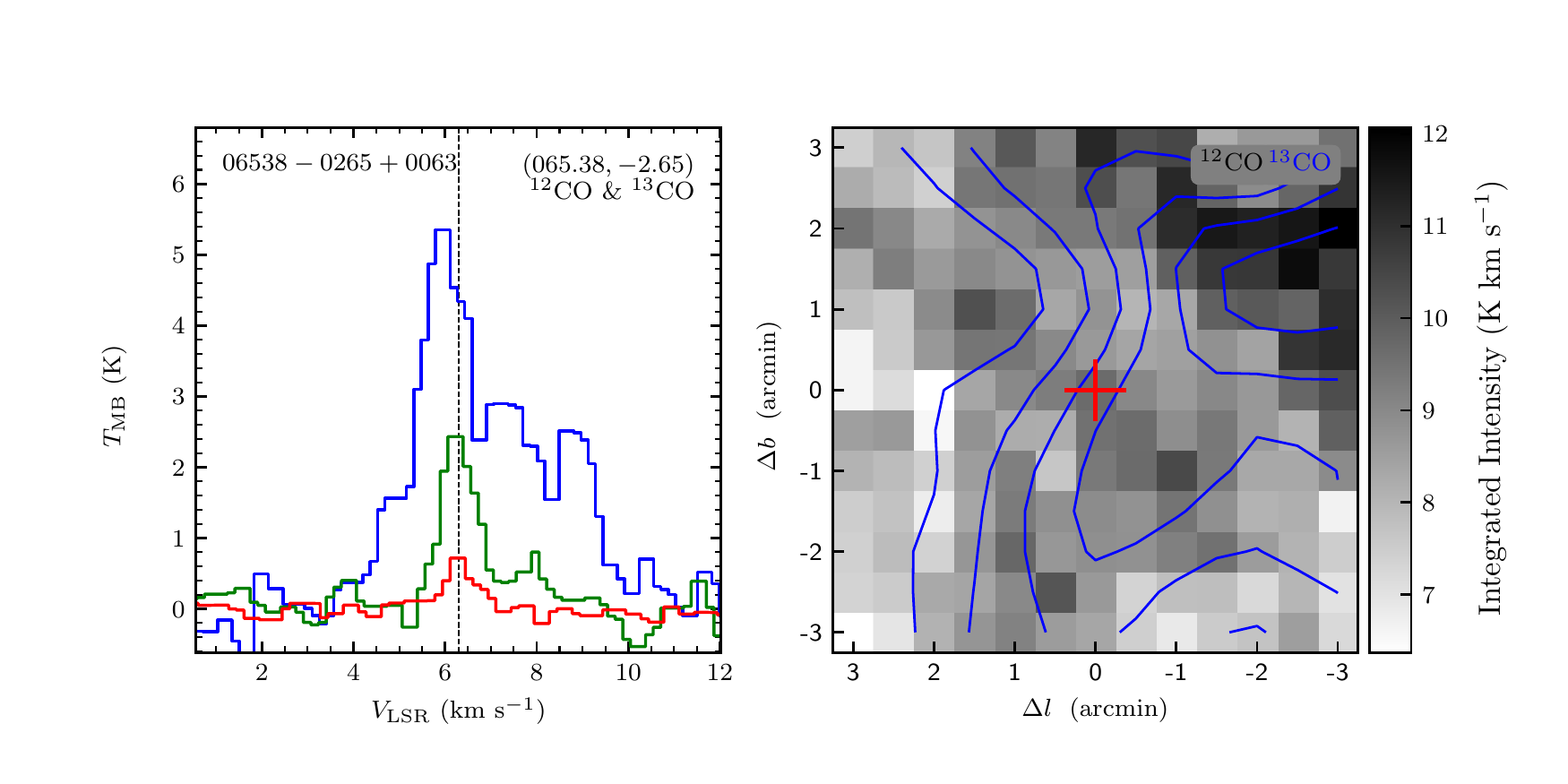}
\end{figure}
\clearpage

\begin{figure}
\includegraphics[width=9.0cm,angle=0]{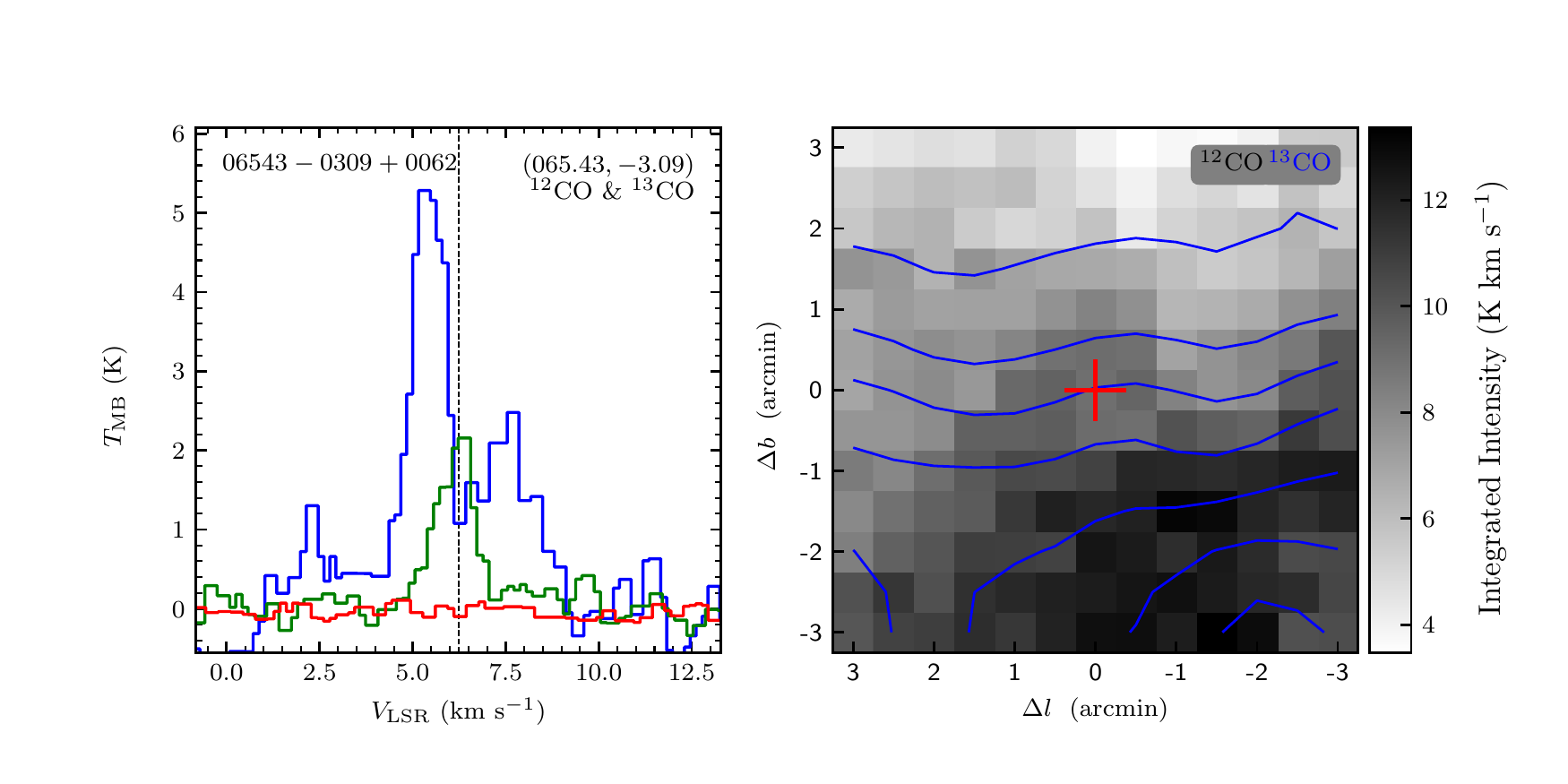}
\includegraphics[width=9.0cm,angle=0]{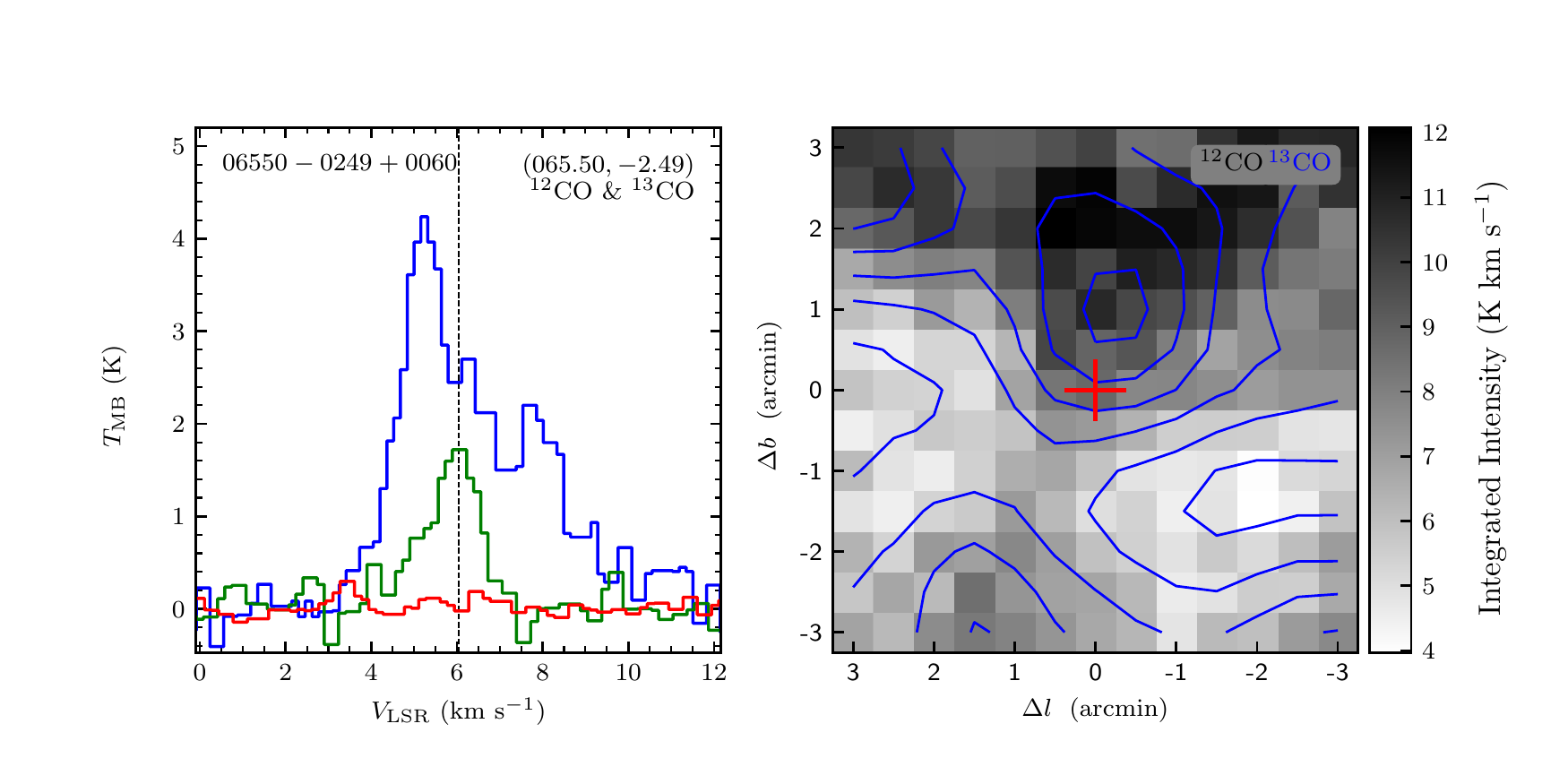}
\vspace{-0.5cm}

\includegraphics[width=9.0cm,angle=0]{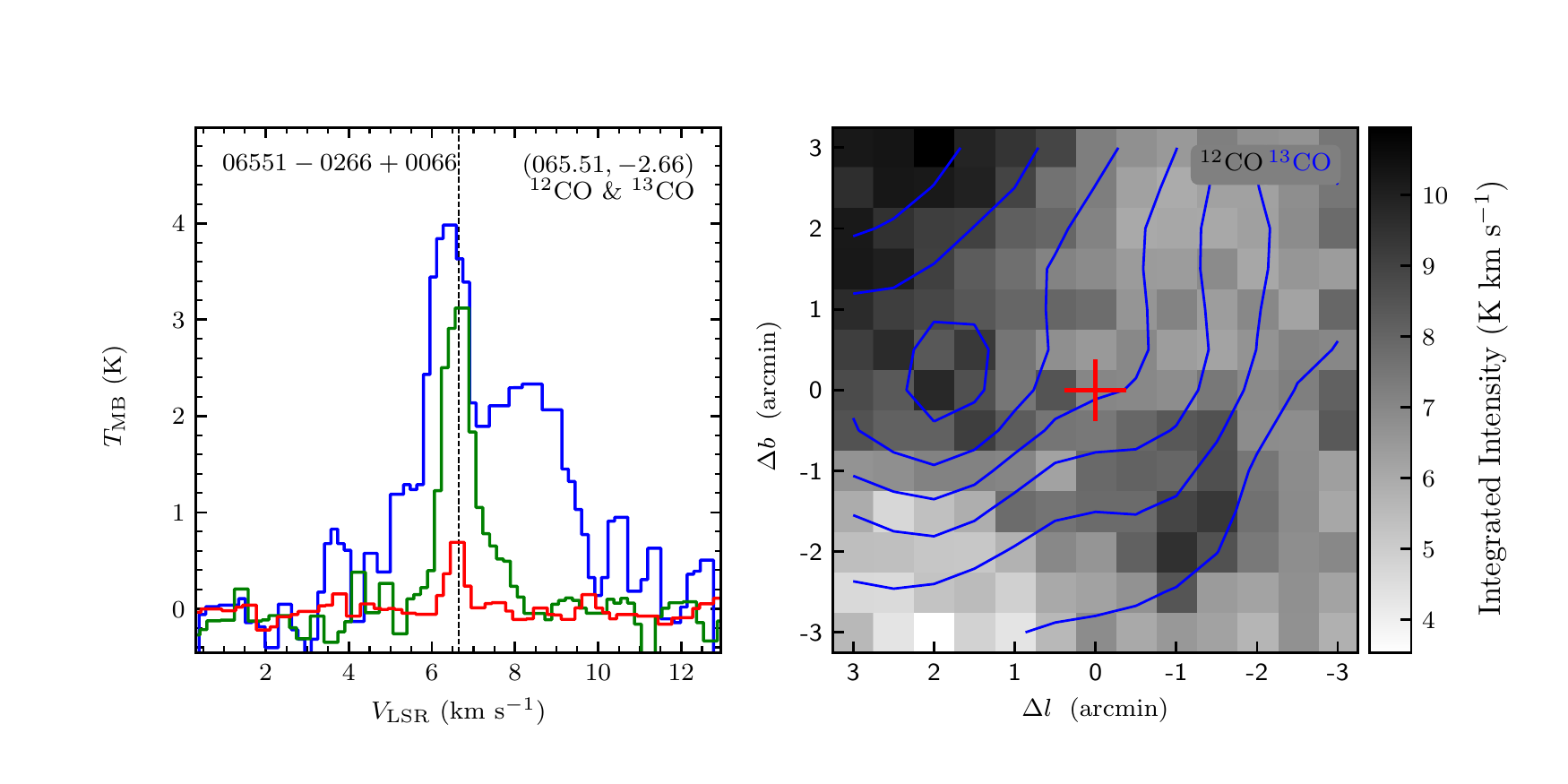}
\includegraphics[width=9.0cm,angle=0]{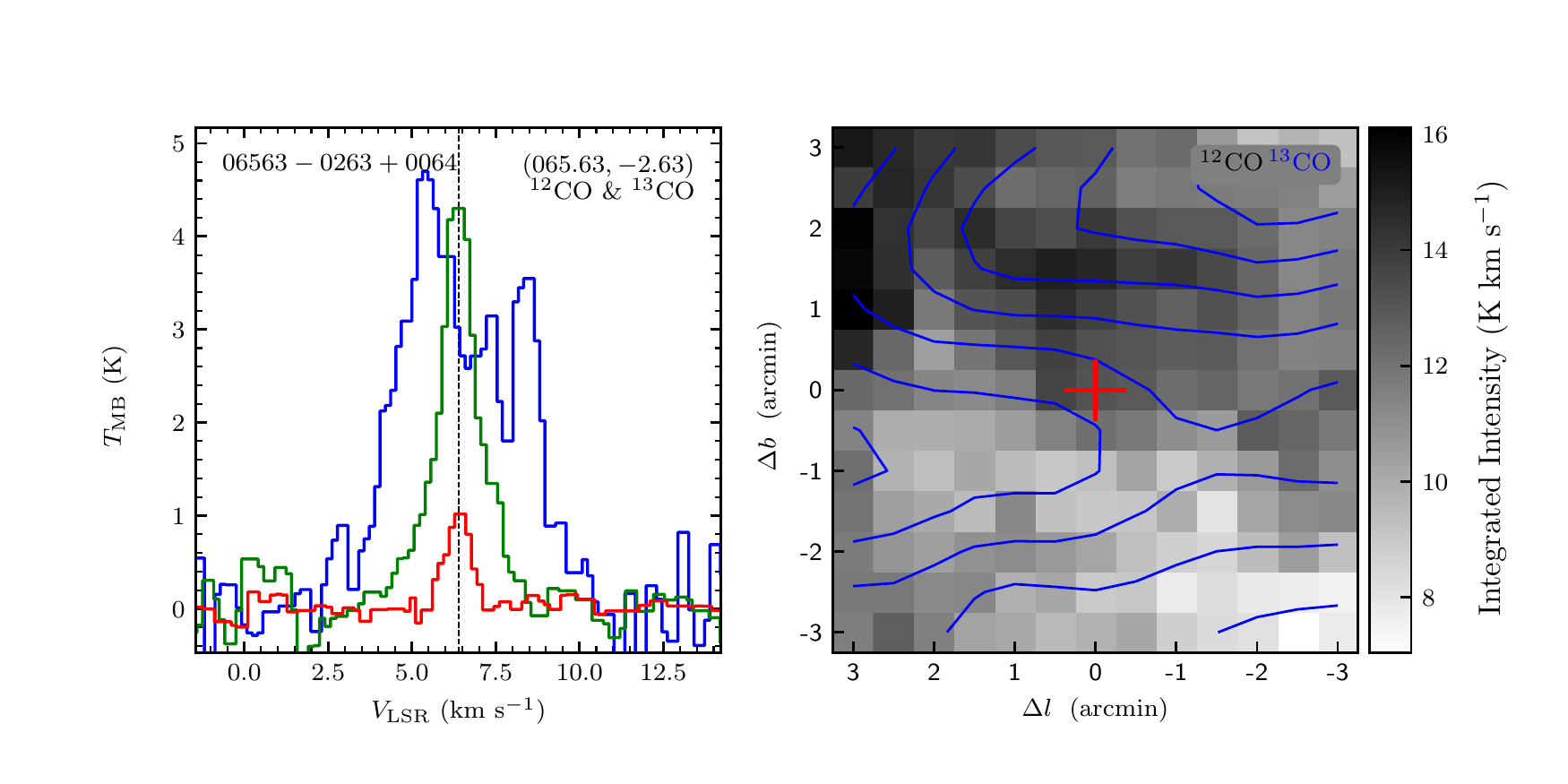}
\vspace{-0.5cm}

\includegraphics[width=9.0cm,angle=0]{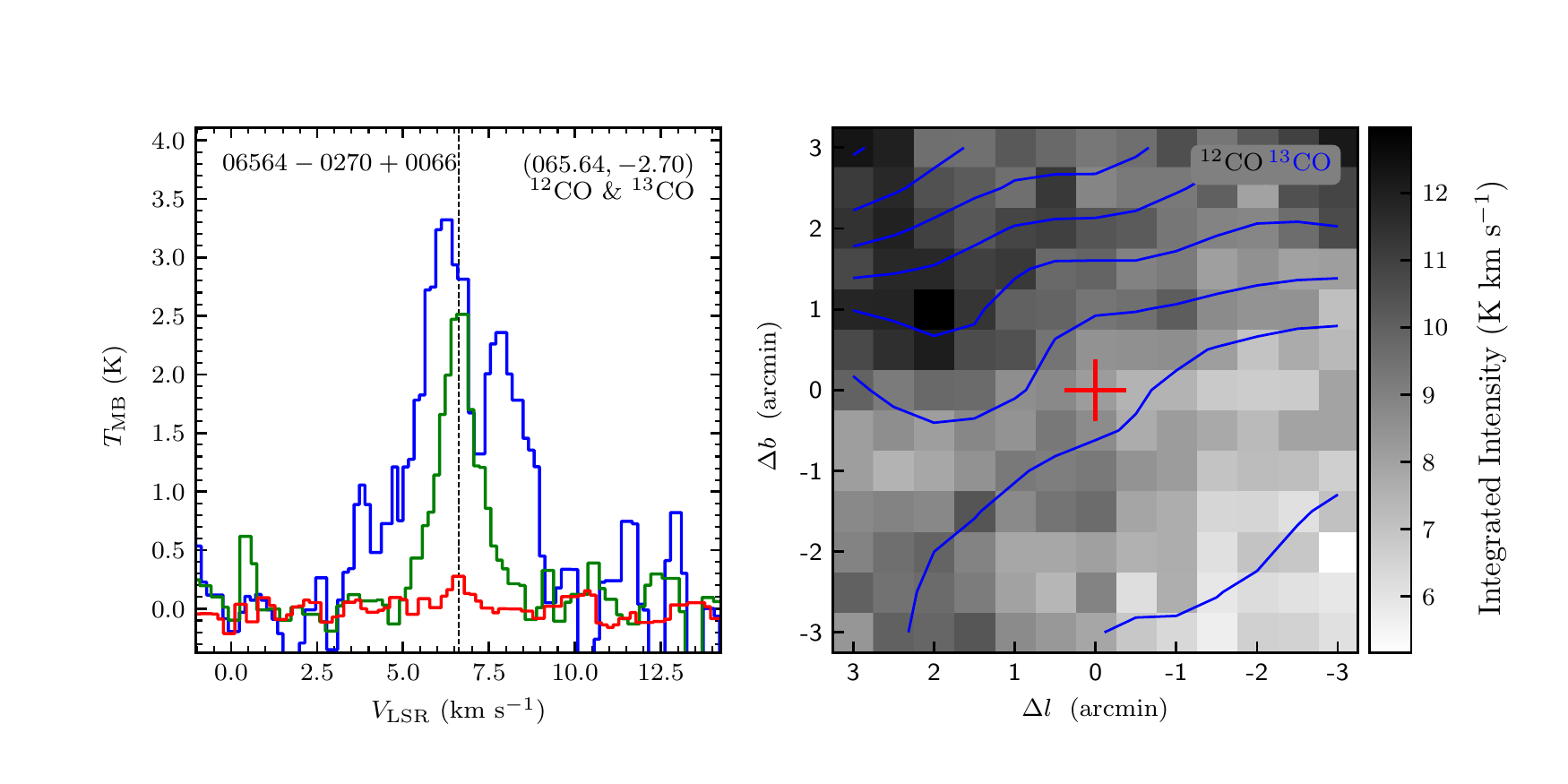}
\includegraphics[width=9.0cm,angle=0]{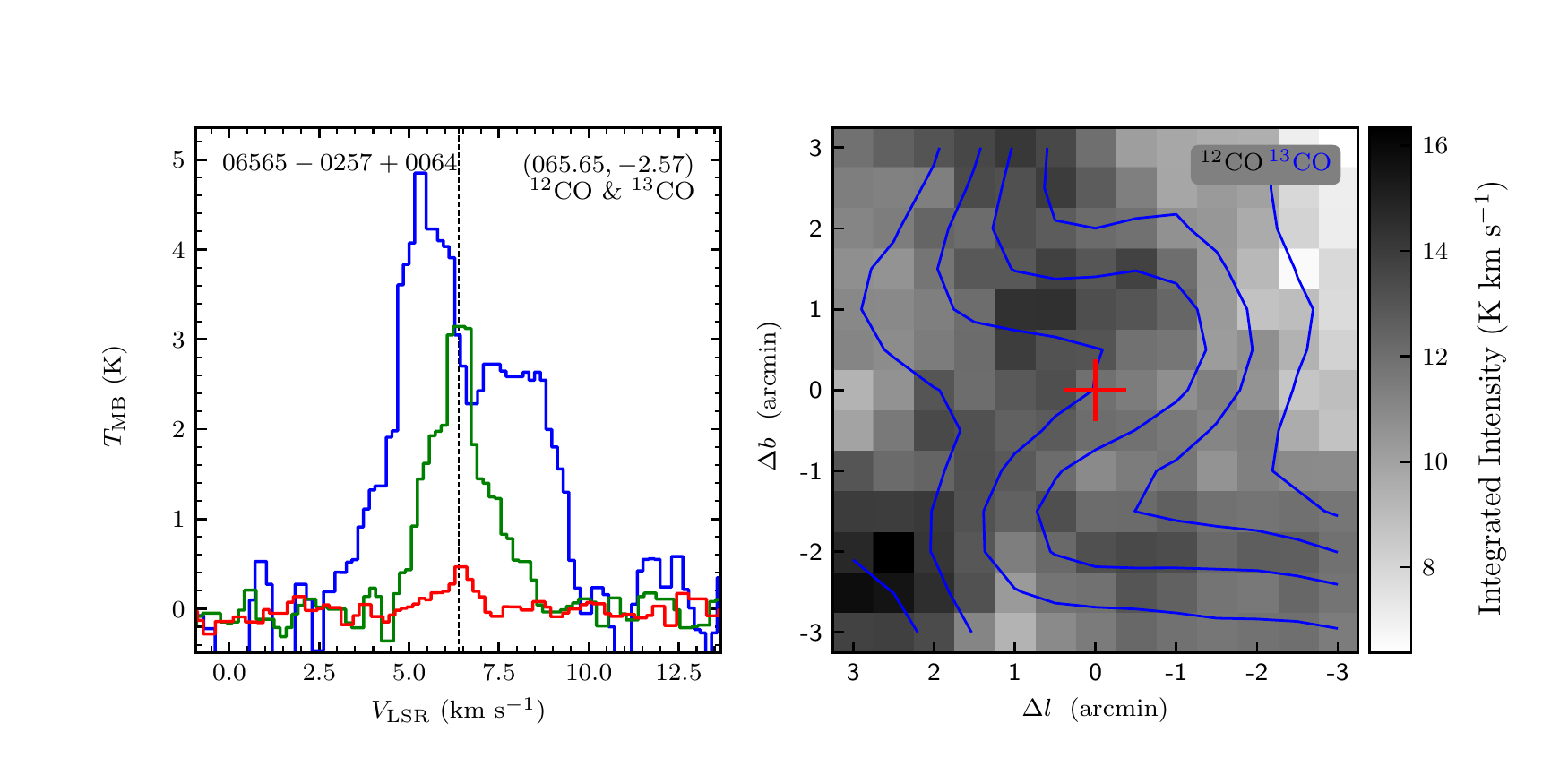}
\vspace{-0.5cm}

\includegraphics[width=9.0cm,angle=0]{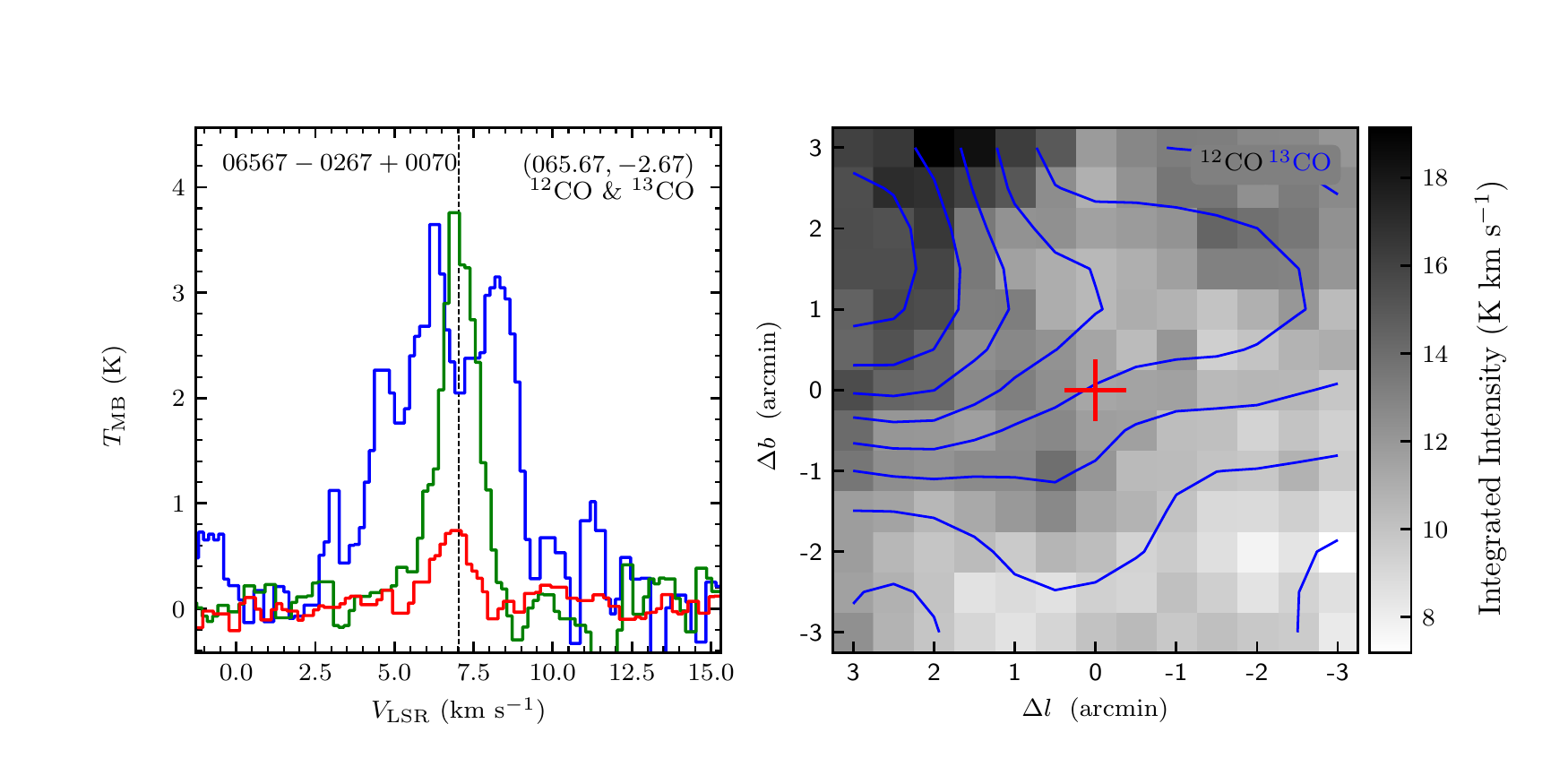}
\includegraphics[width=9.0cm,angle=0]{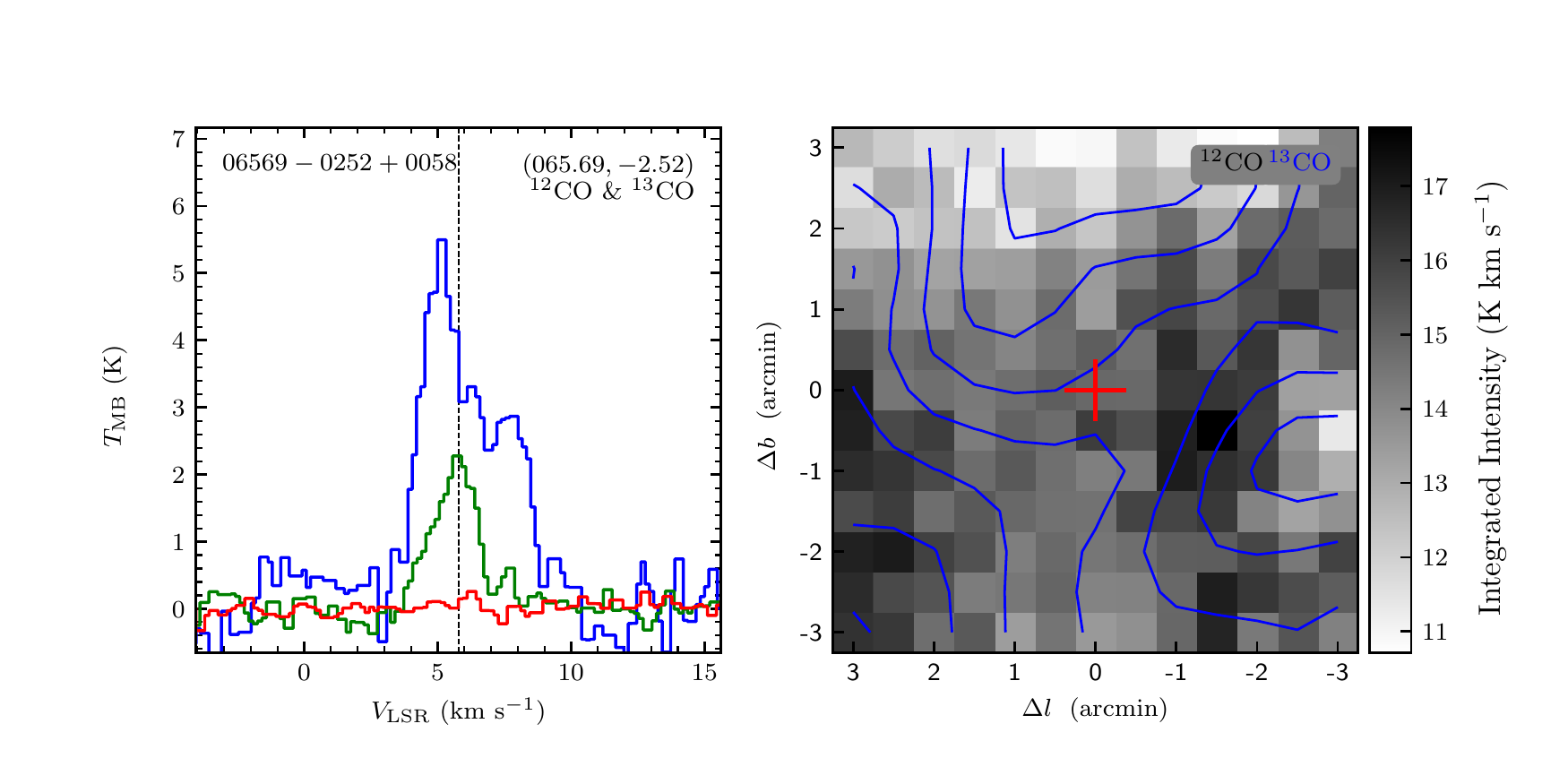}
\vspace{-0.5cm}

\includegraphics[width=9.0cm,angle=0]{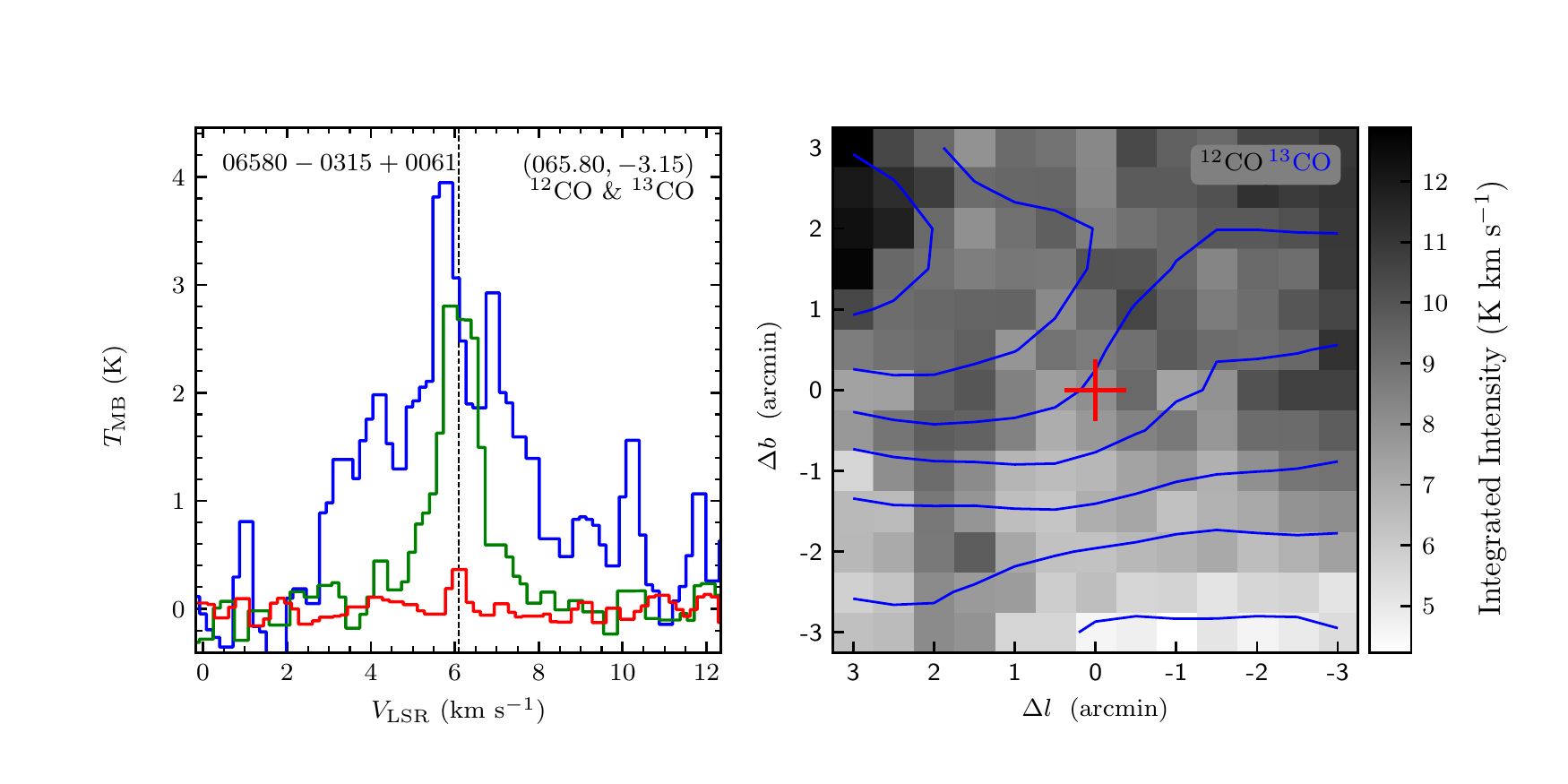}
\includegraphics[width=9.0cm,angle=0]{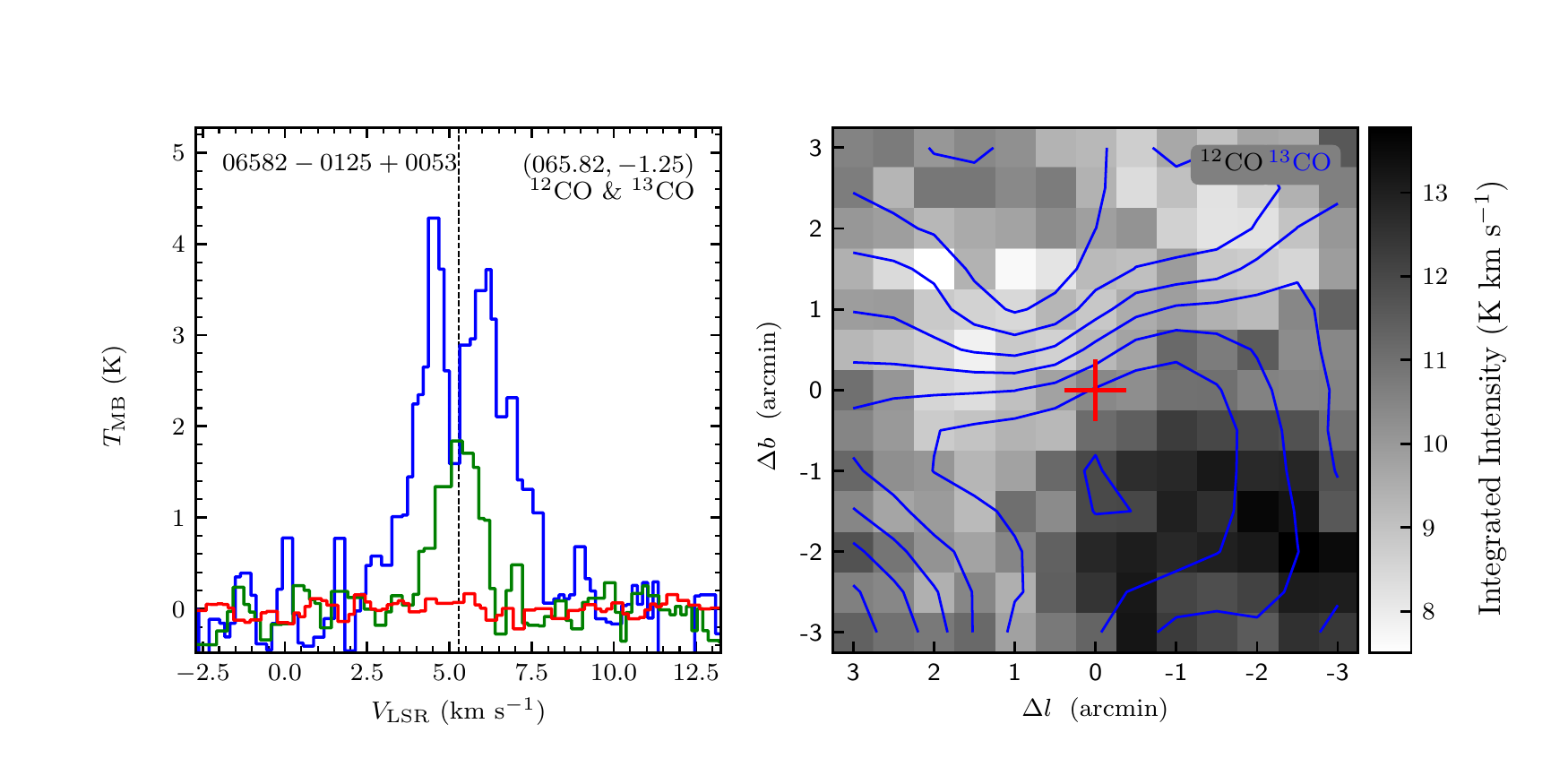}
\end{figure}
\clearpage

\begin{figure}
\includegraphics[width=9.0cm,angle=0]{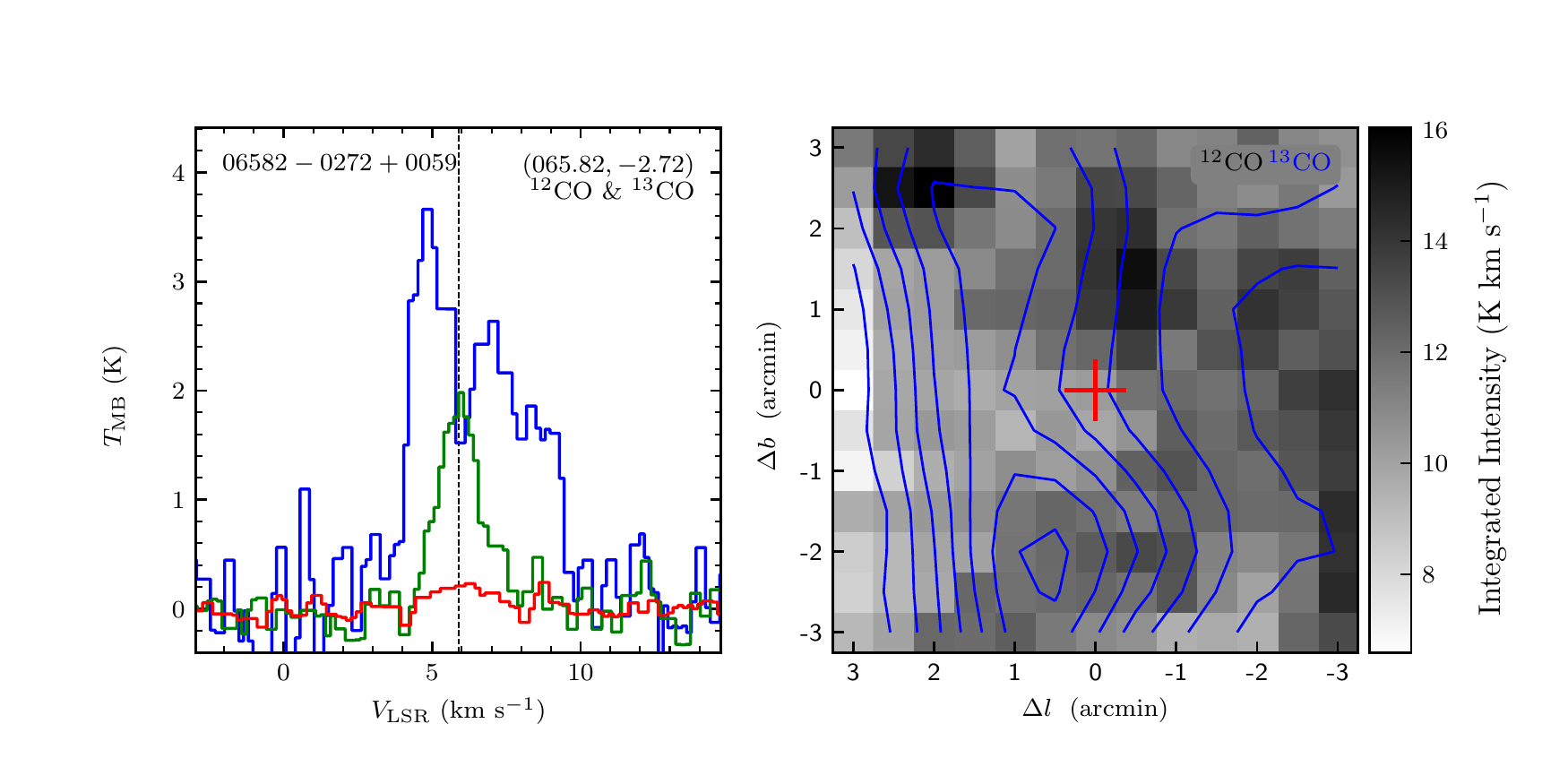}
\includegraphics[width=9.0cm,angle=0]{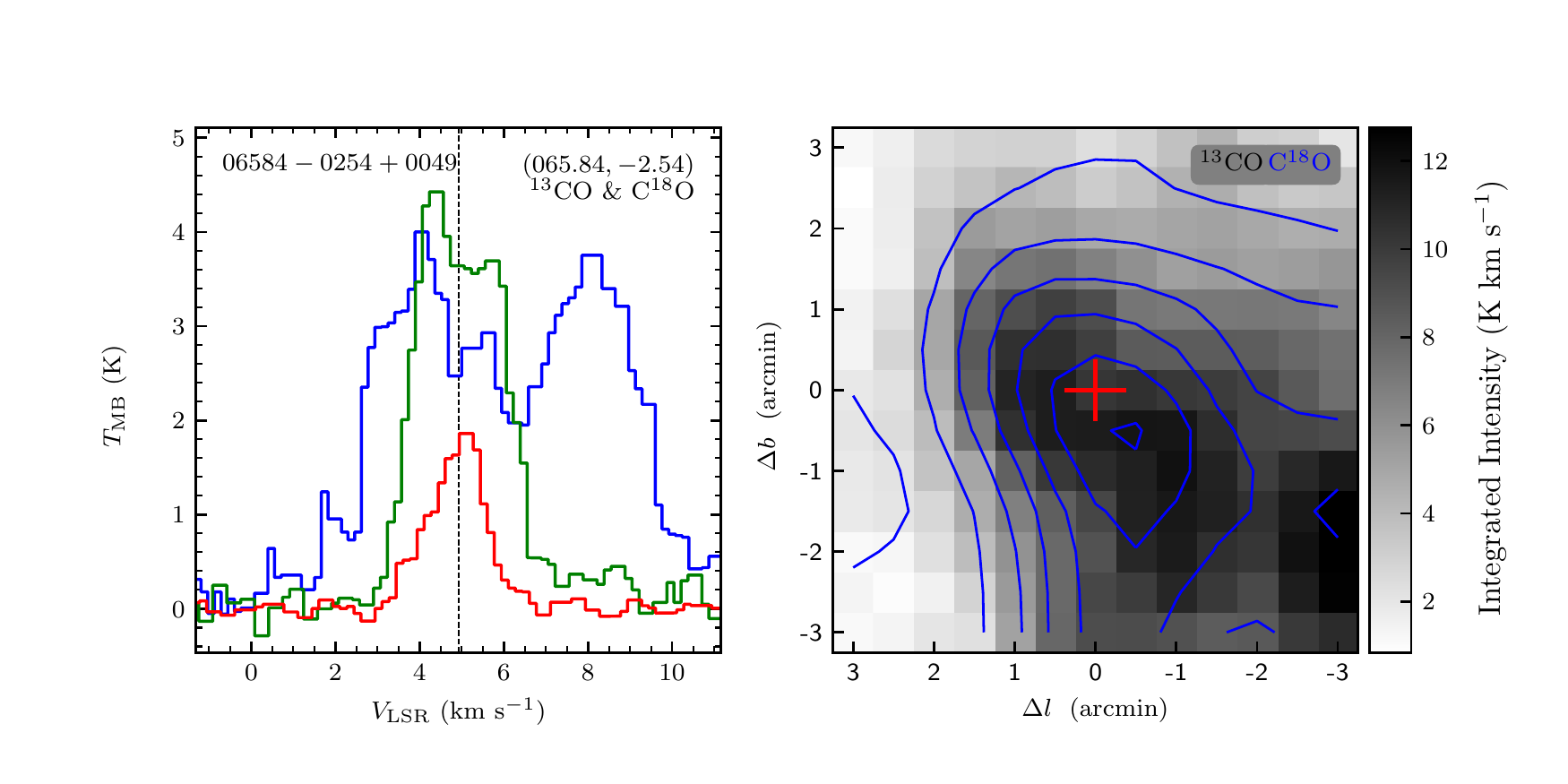}
\vspace{-0.5cm}

\includegraphics[width=9.0cm,angle=0]{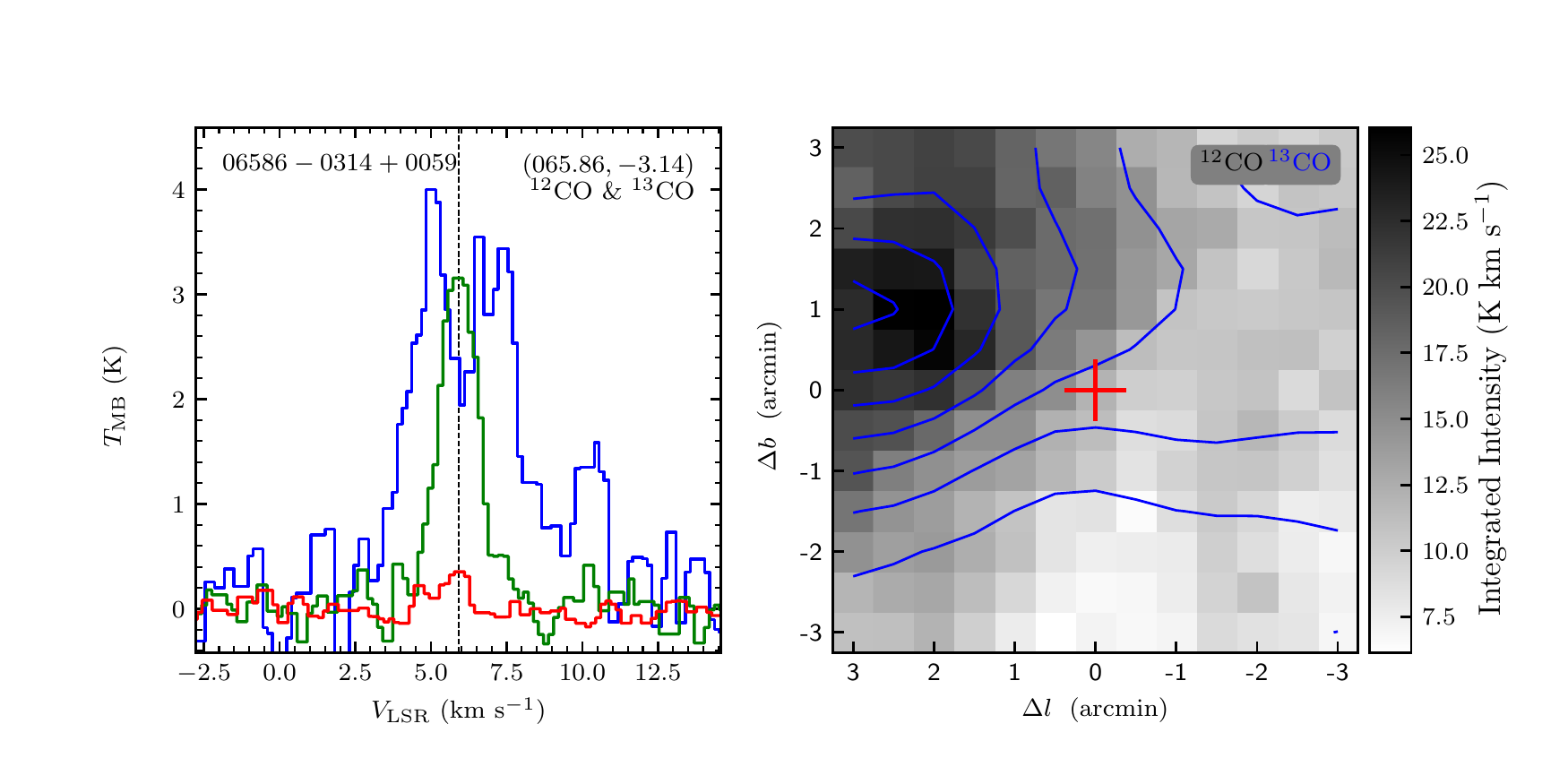}
\includegraphics[width=9.0cm,angle=0]{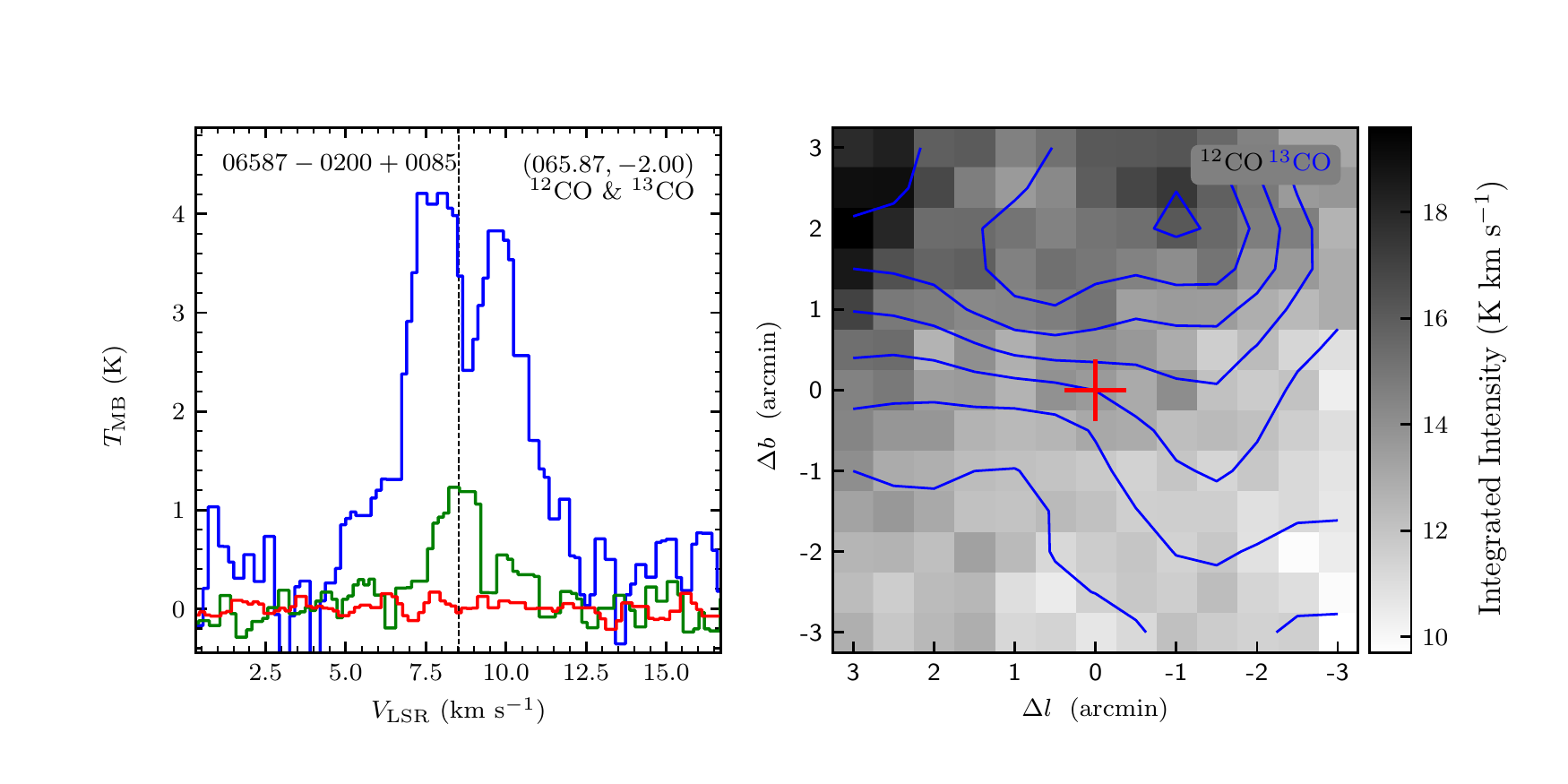}
\vspace{-0.5cm}

\includegraphics[width=9.0cm,angle=0]{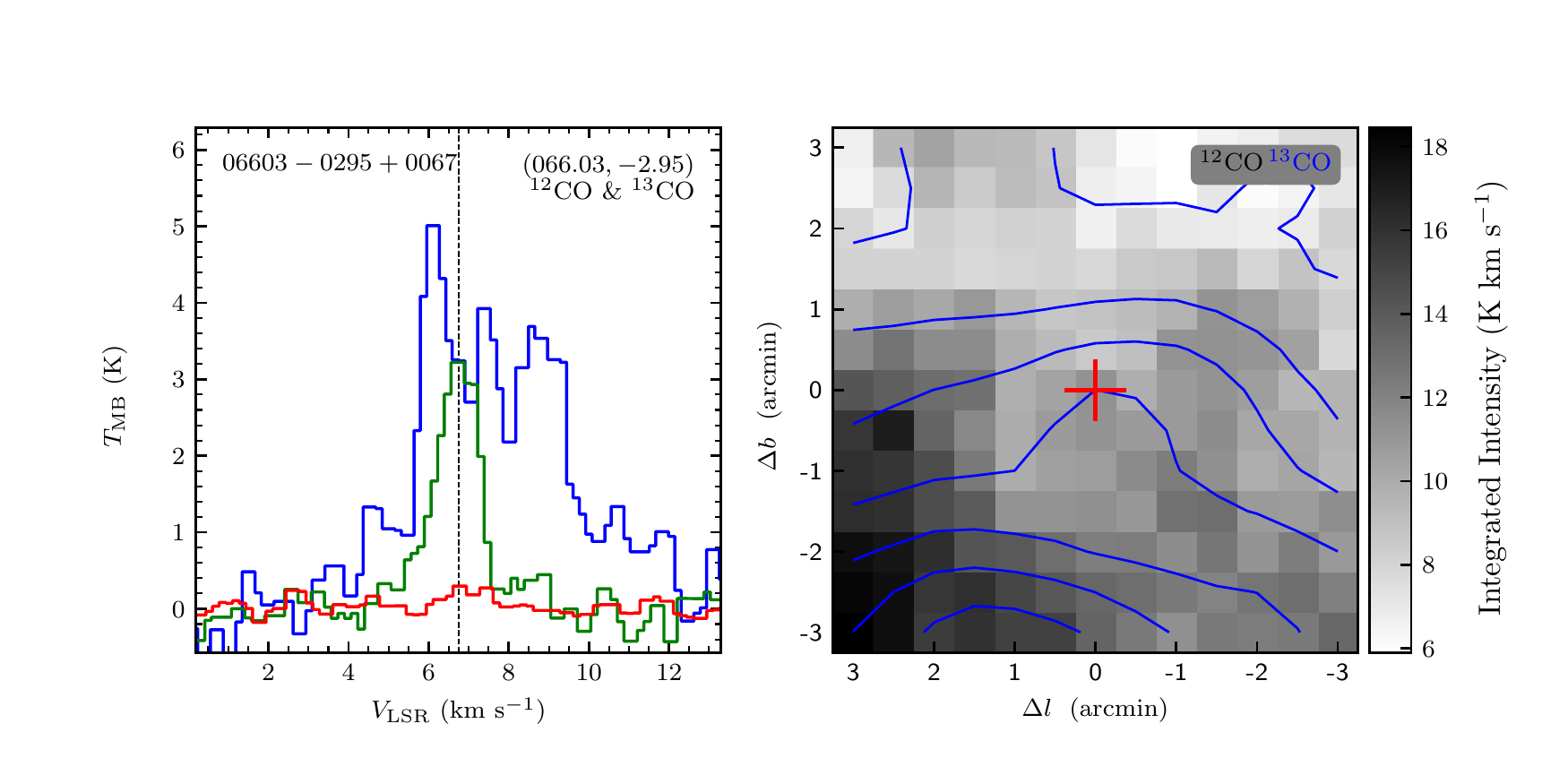}
\includegraphics[width=9.0cm,angle=0]{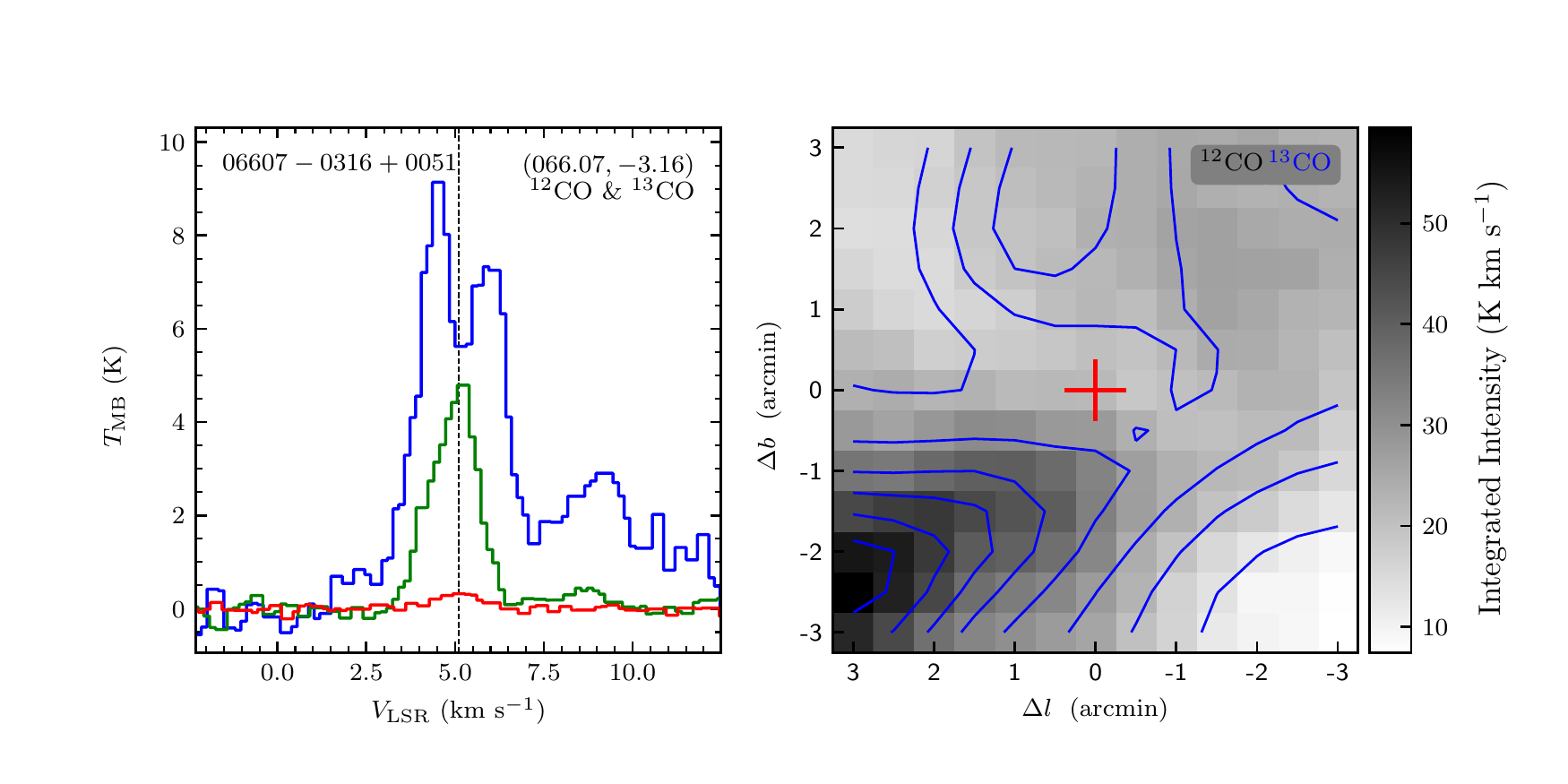}
\vspace{-0.5cm}

\includegraphics[width=9.0cm,angle=0]{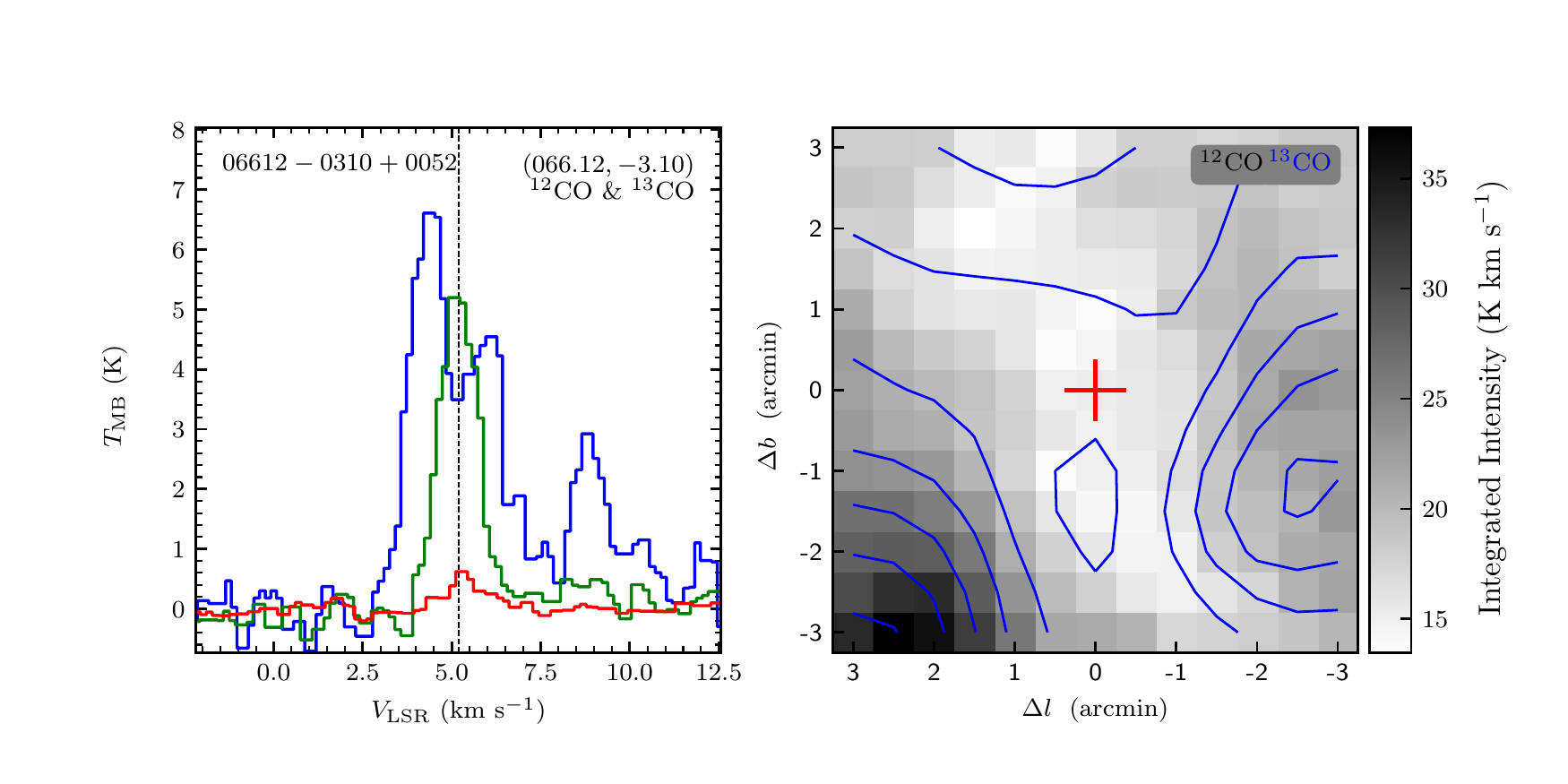}
\includegraphics[width=9.0cm,angle=0]{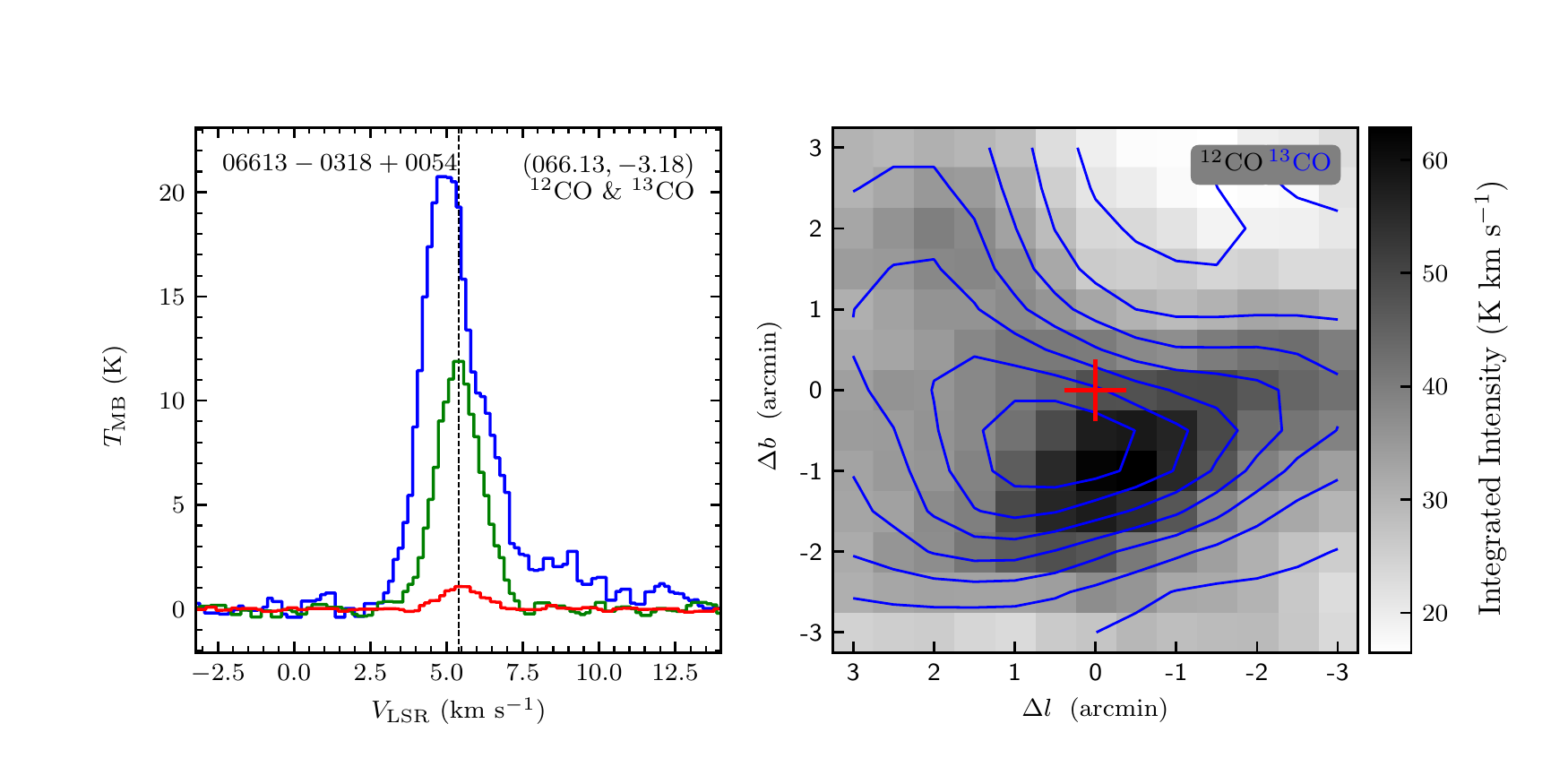}
\vspace{-0.5cm}

\includegraphics[width=9.0cm,angle=0]{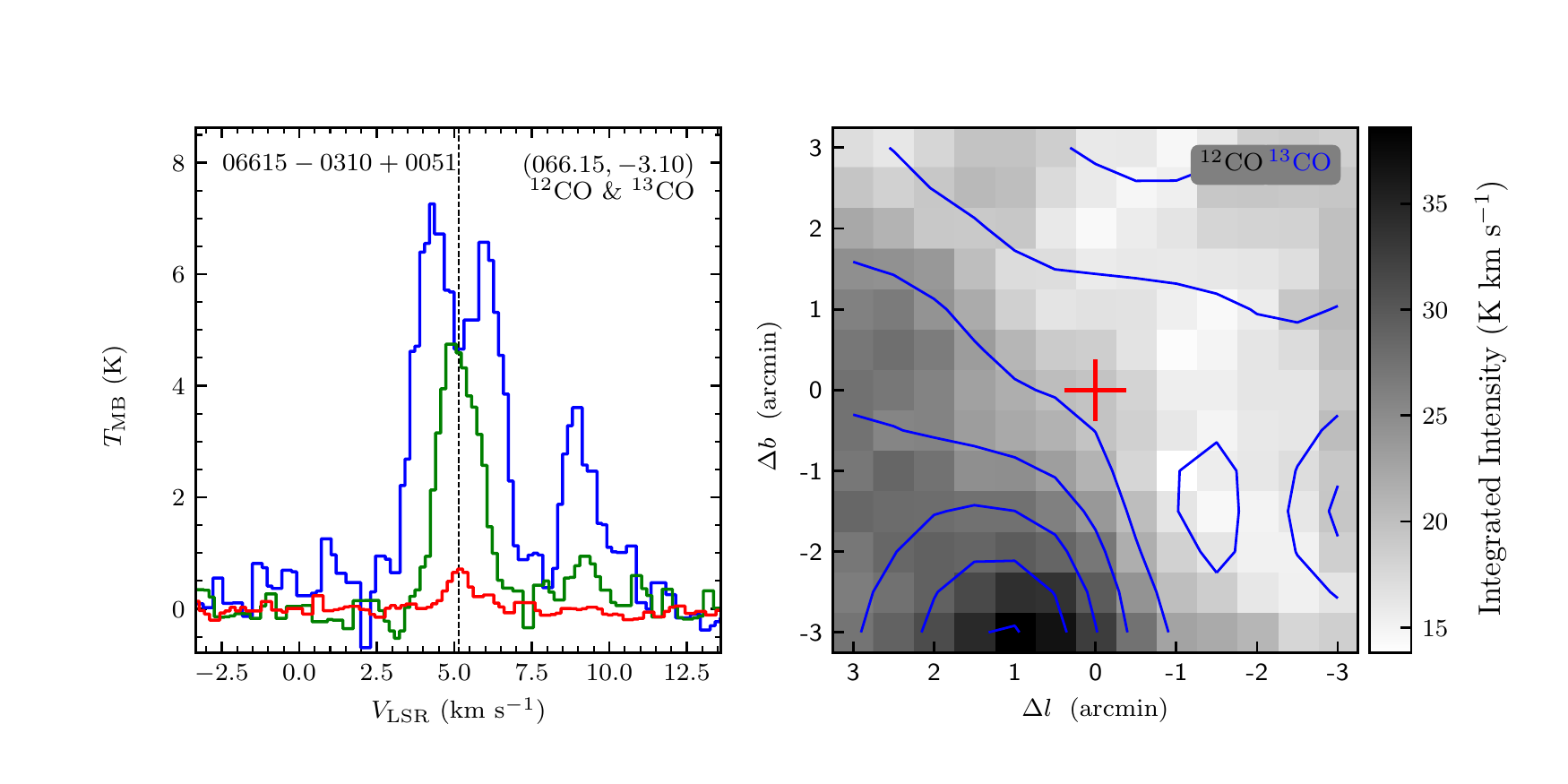}
\includegraphics[width=9.0cm,angle=0]{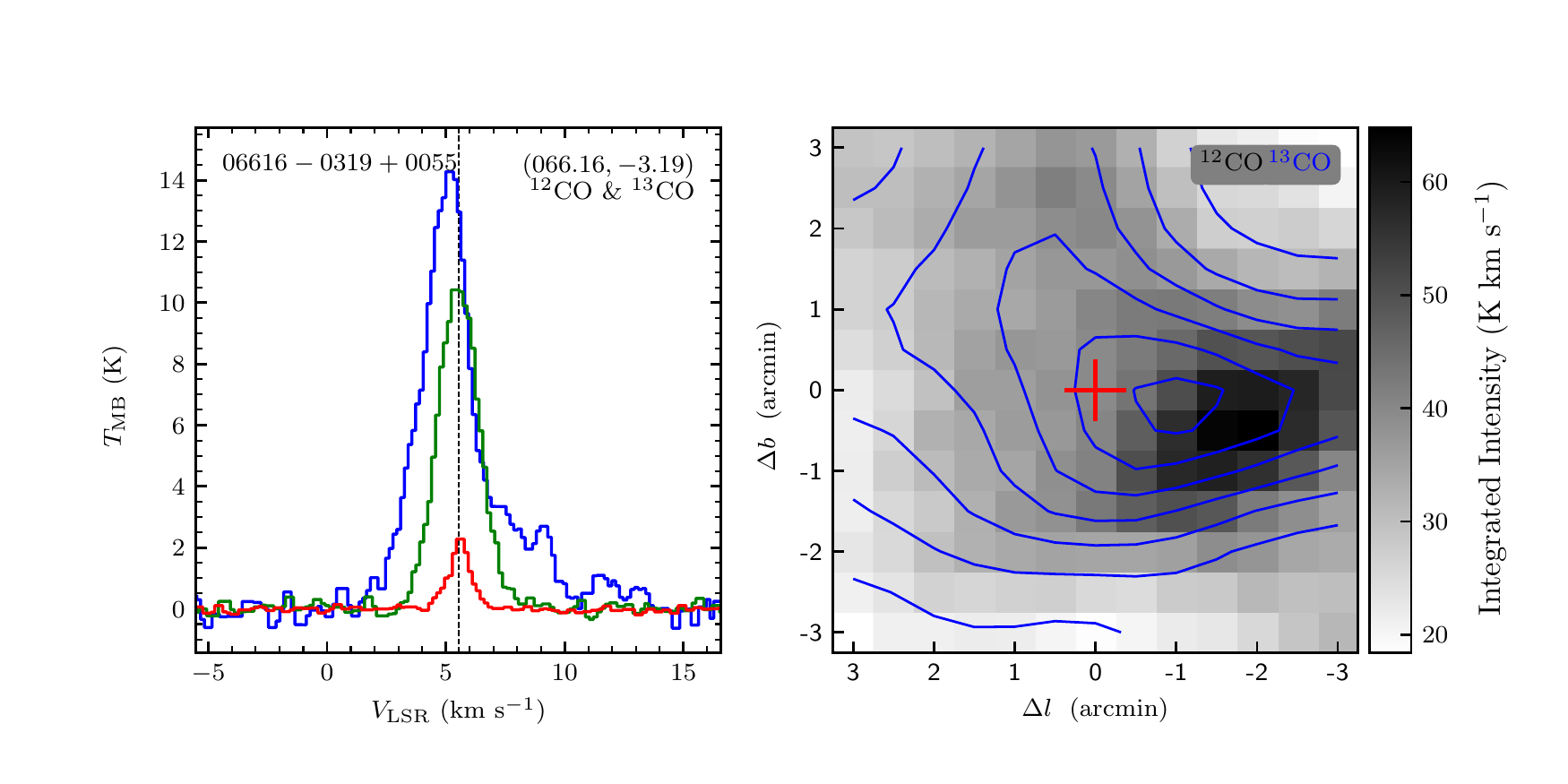}
\end{figure}
\clearpage

\begin{figure}
\includegraphics[width=9.0cm,angle=0]{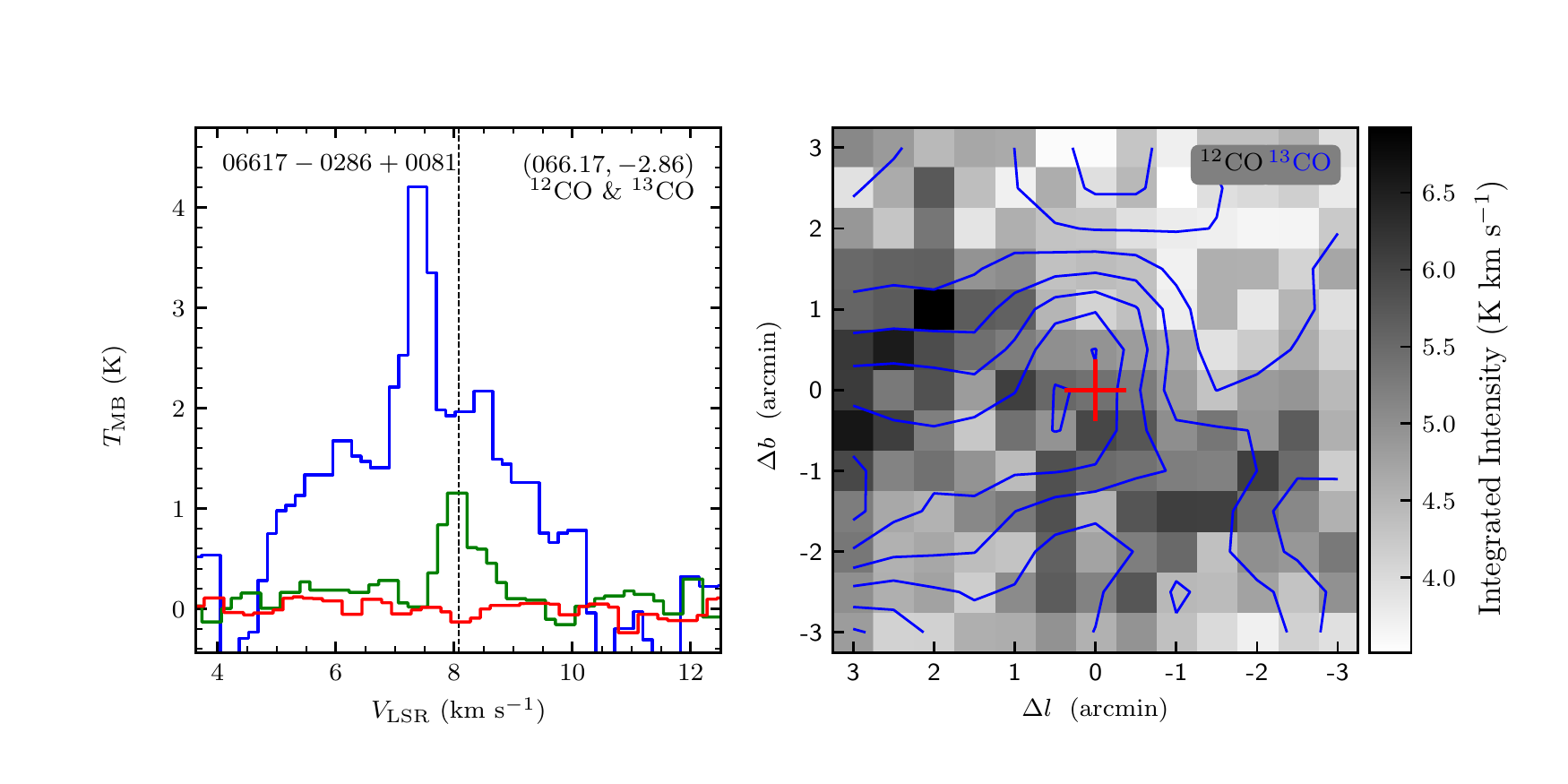}
\includegraphics[width=9.0cm,angle=0]{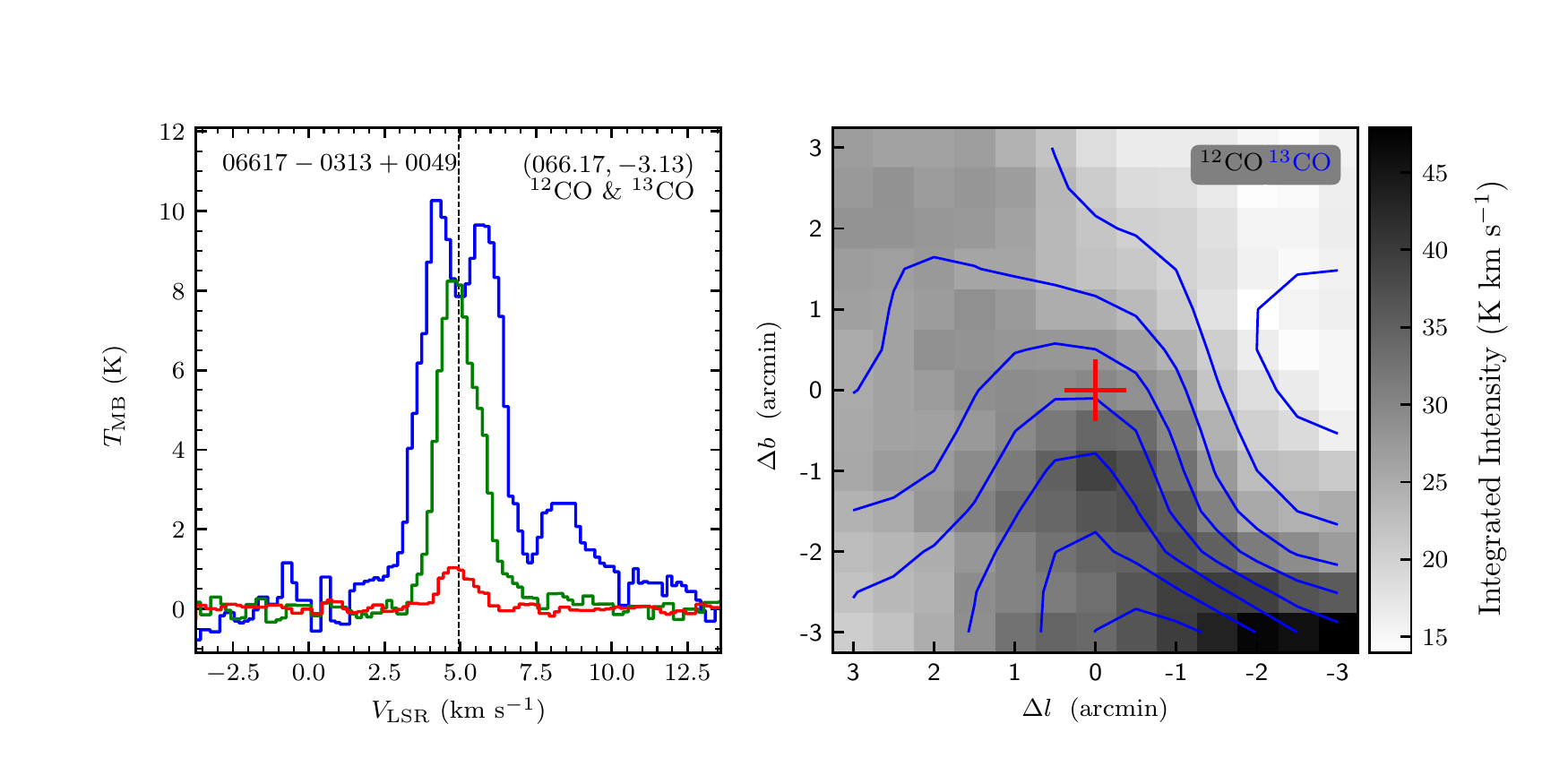}
\vspace{-0.5cm}

\includegraphics[width=9.0cm,angle=0]{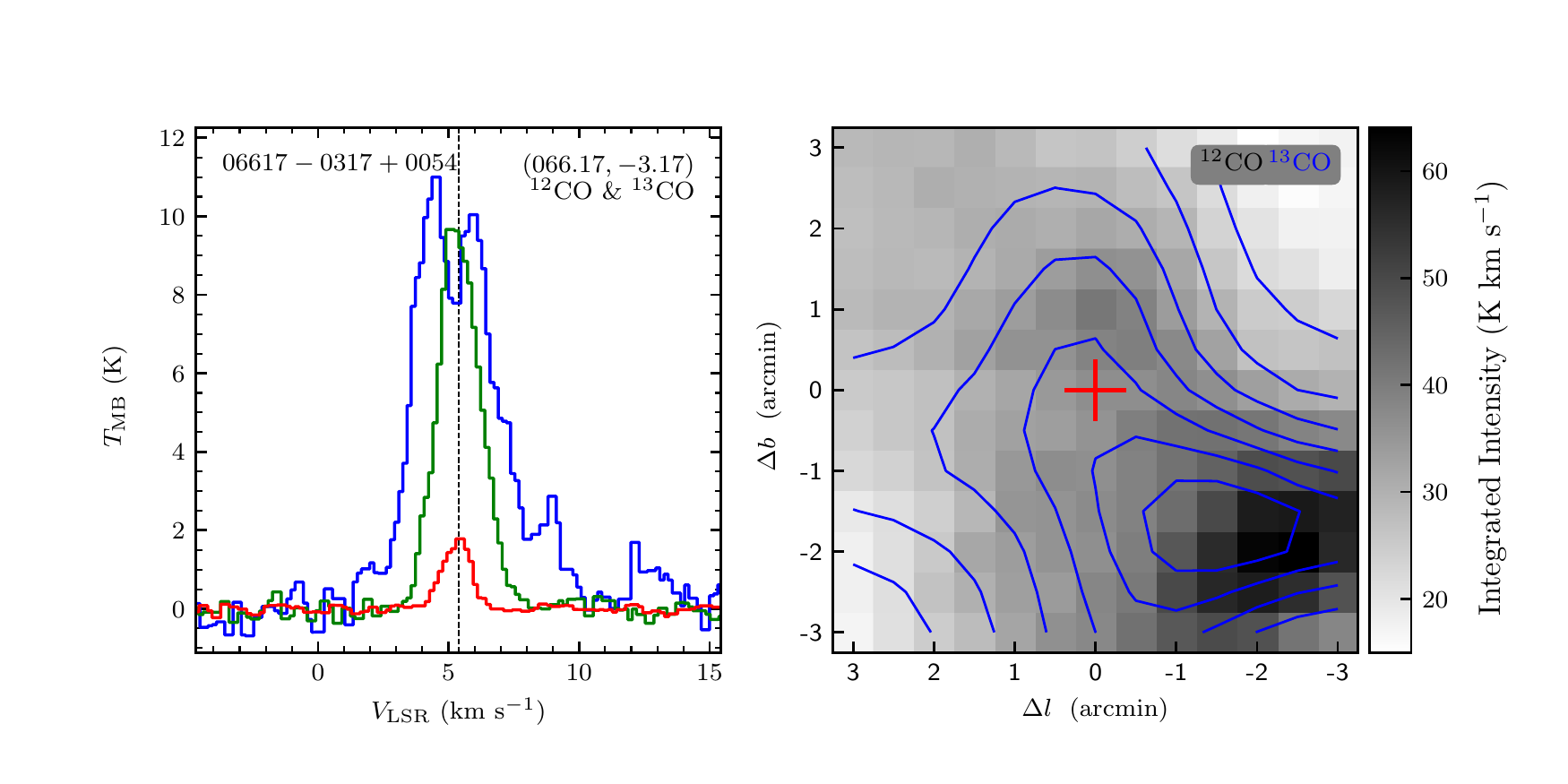}
\includegraphics[width=9.0cm,angle=0]{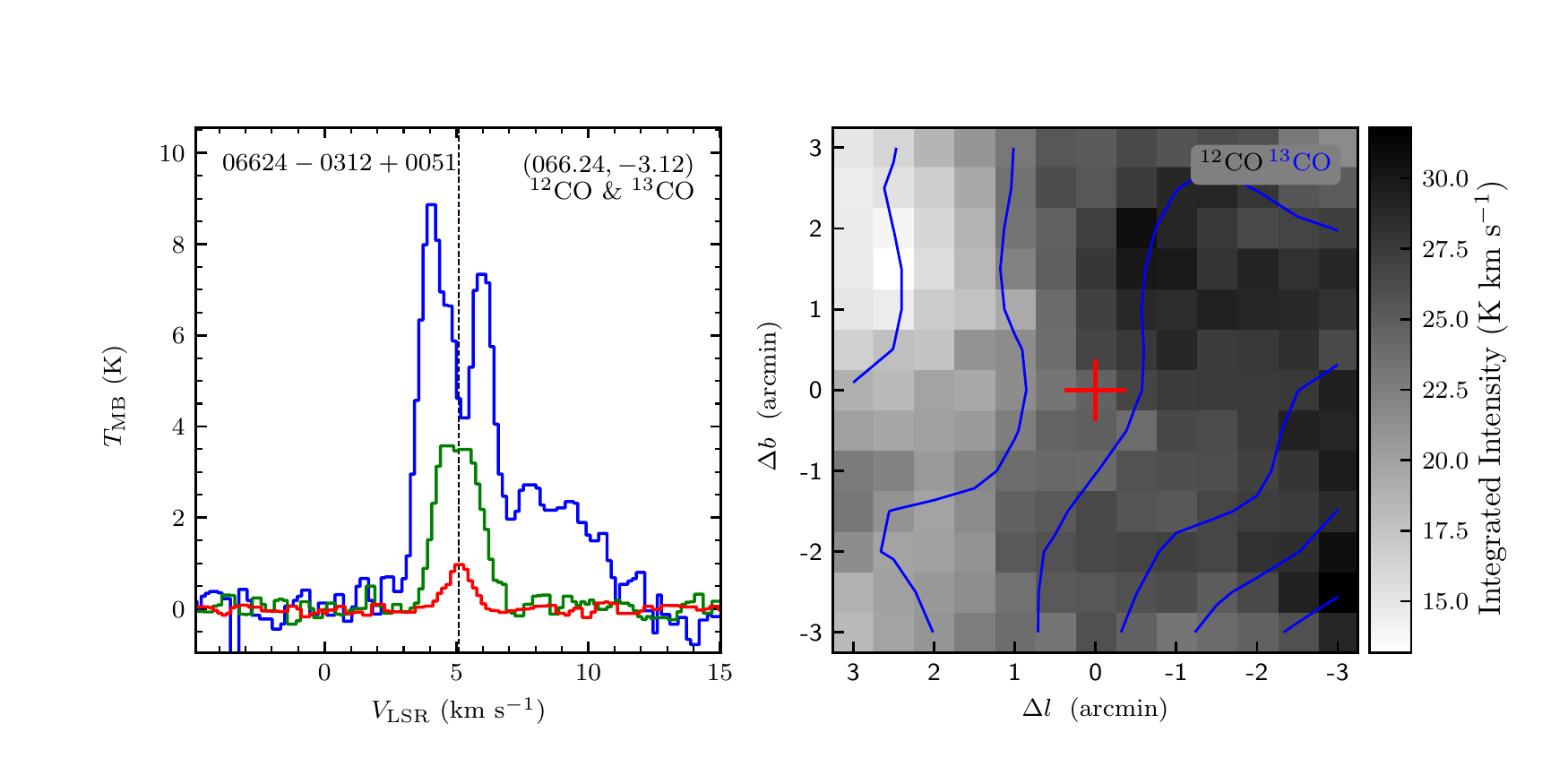}
\vspace{-0.5cm}

\includegraphics[width=9.0cm,angle=0]{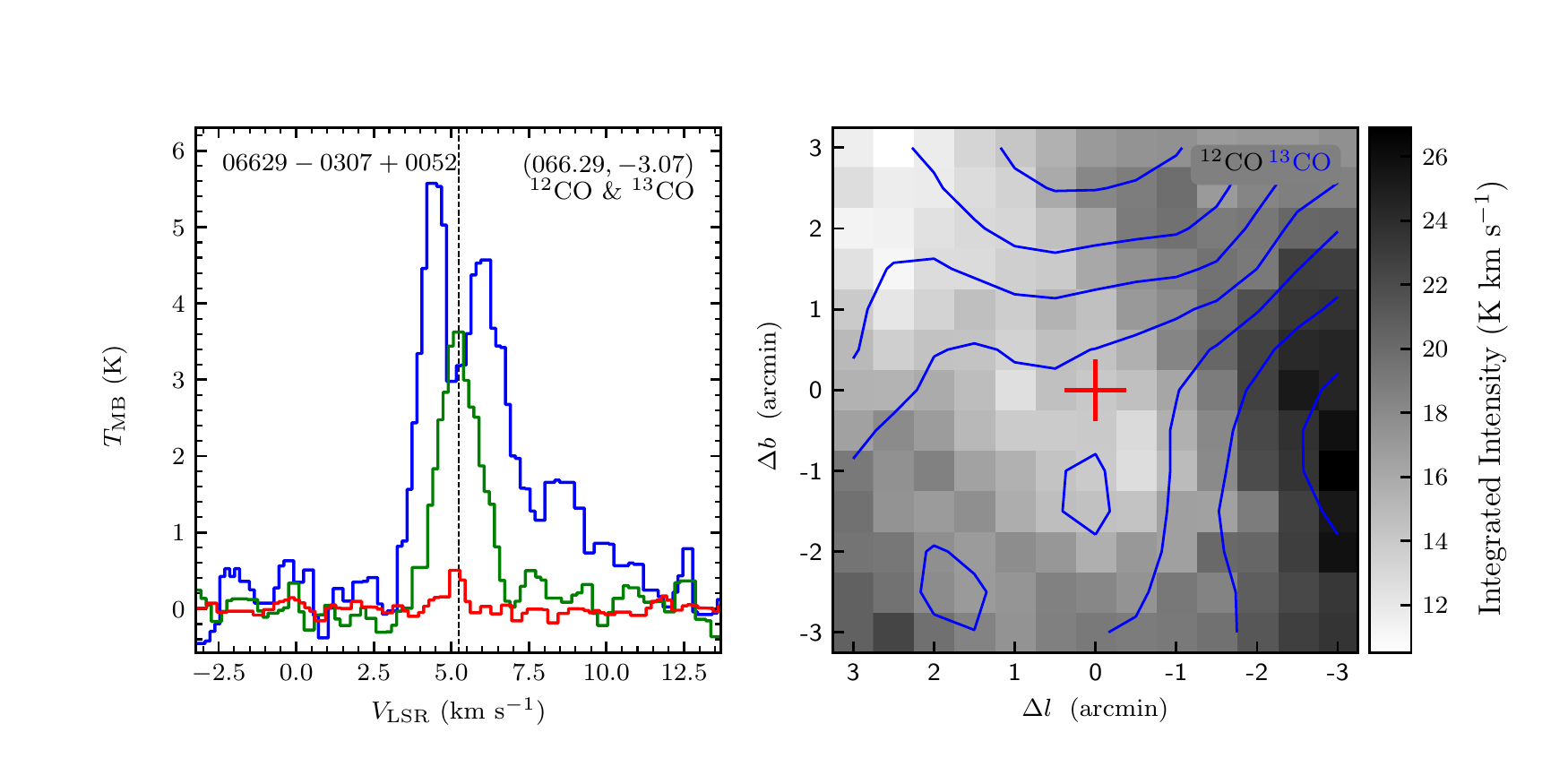}
\includegraphics[width=9.0cm,angle=0]{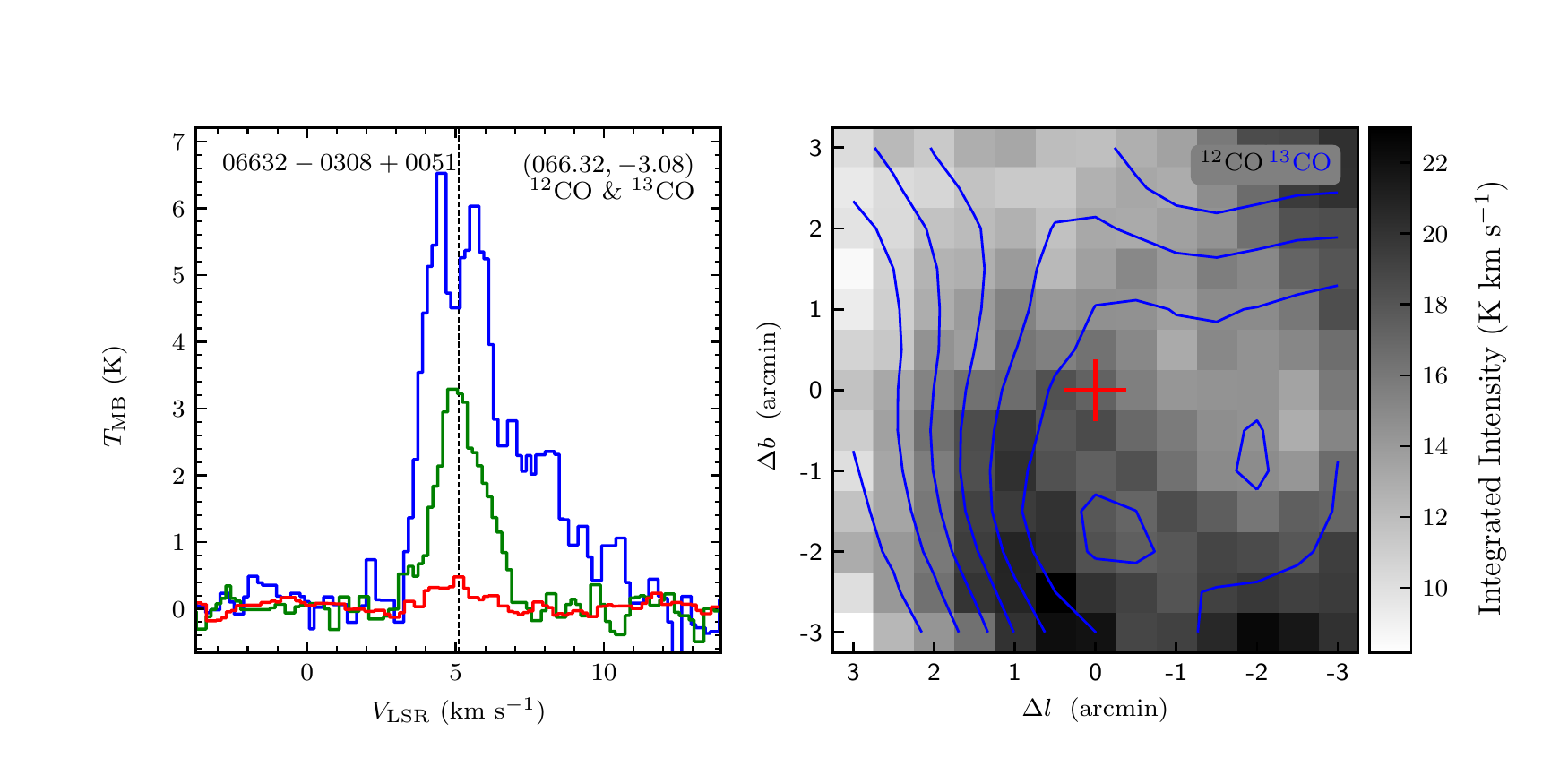}
\vspace{-0.5cm}

\includegraphics[width=9.0cm,angle=0]{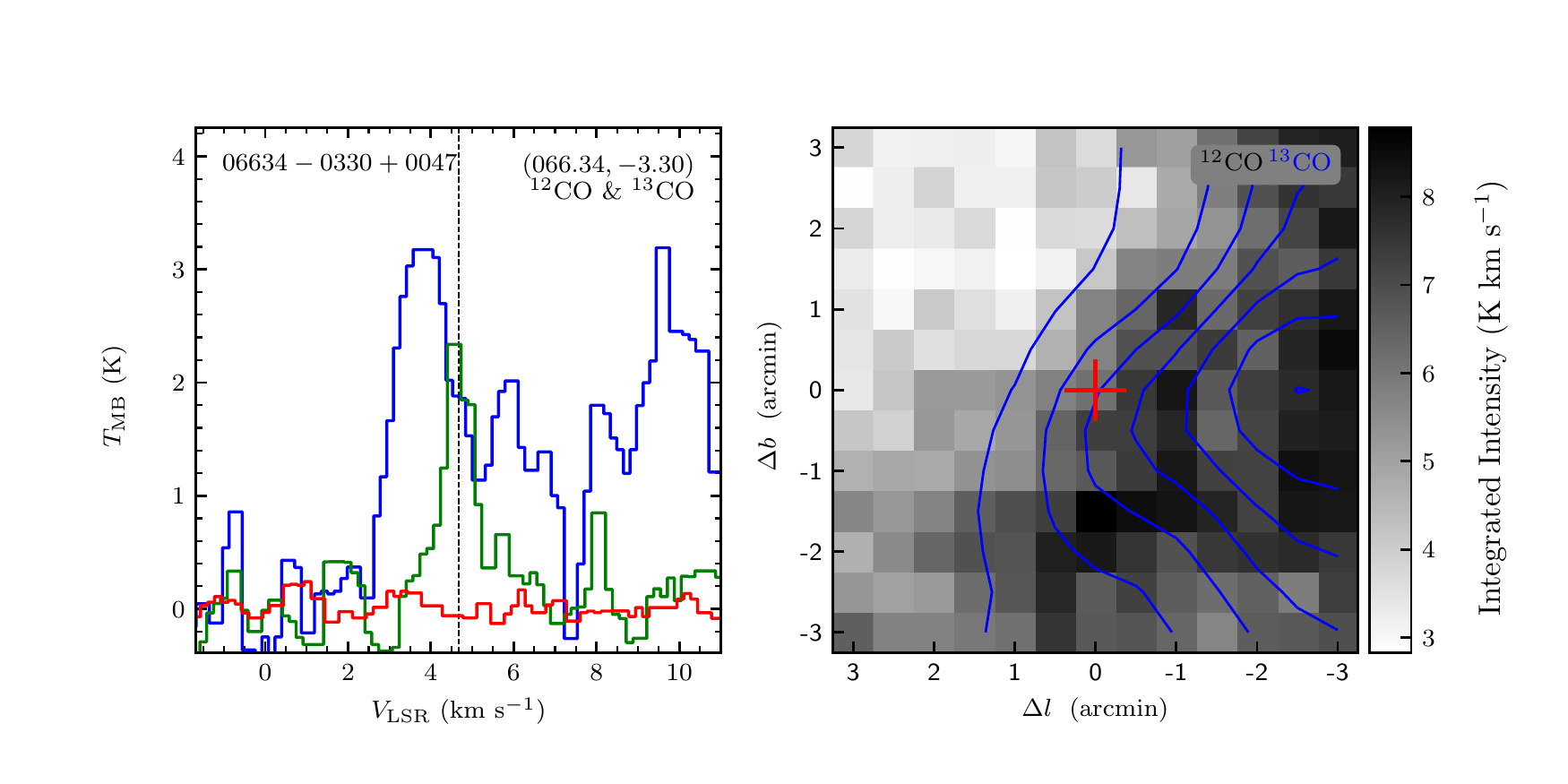}
\includegraphics[width=9.0cm,angle=0]{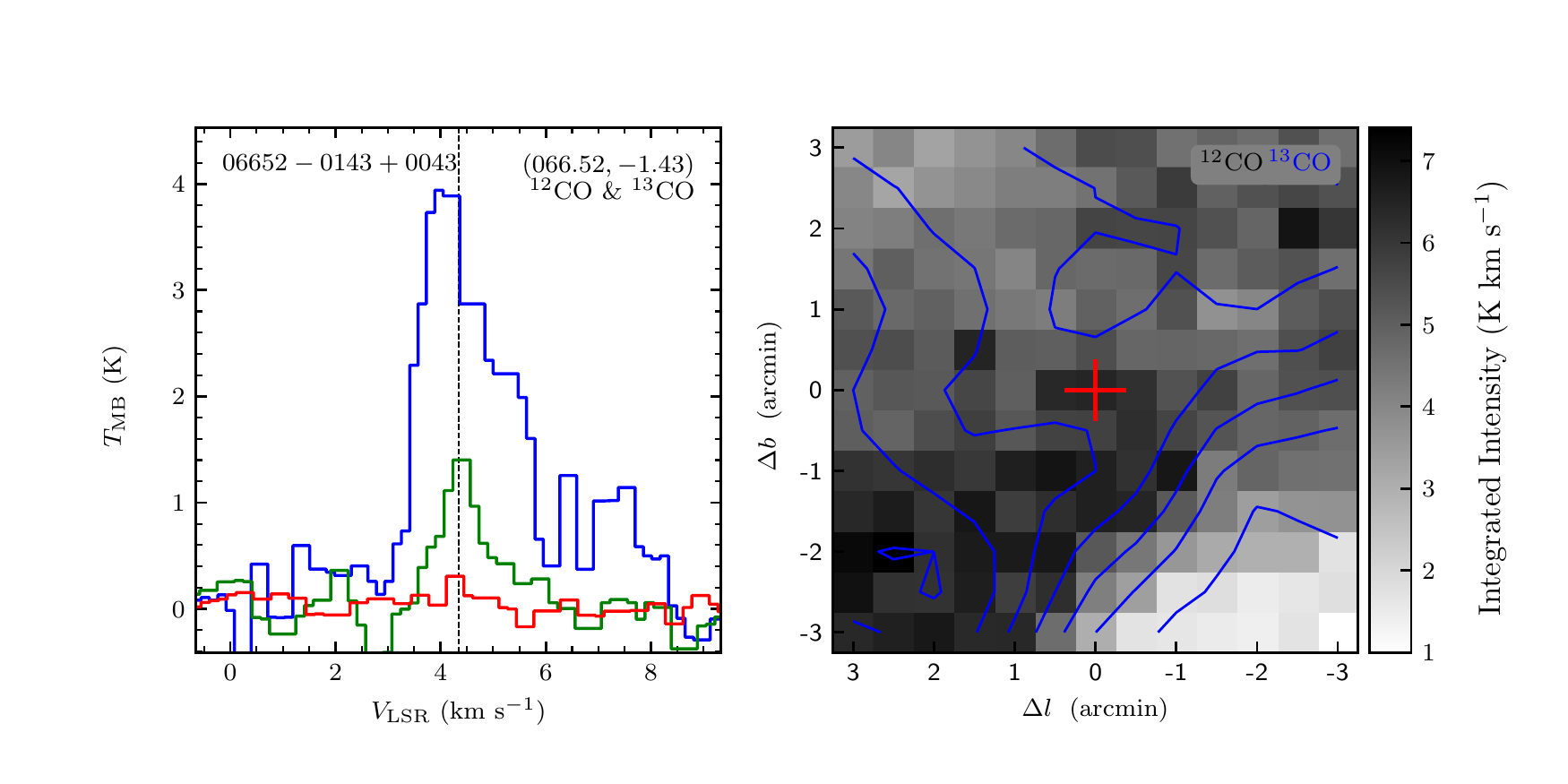}
\vspace{-0.5cm}

\includegraphics[width=9.0cm,angle=0]{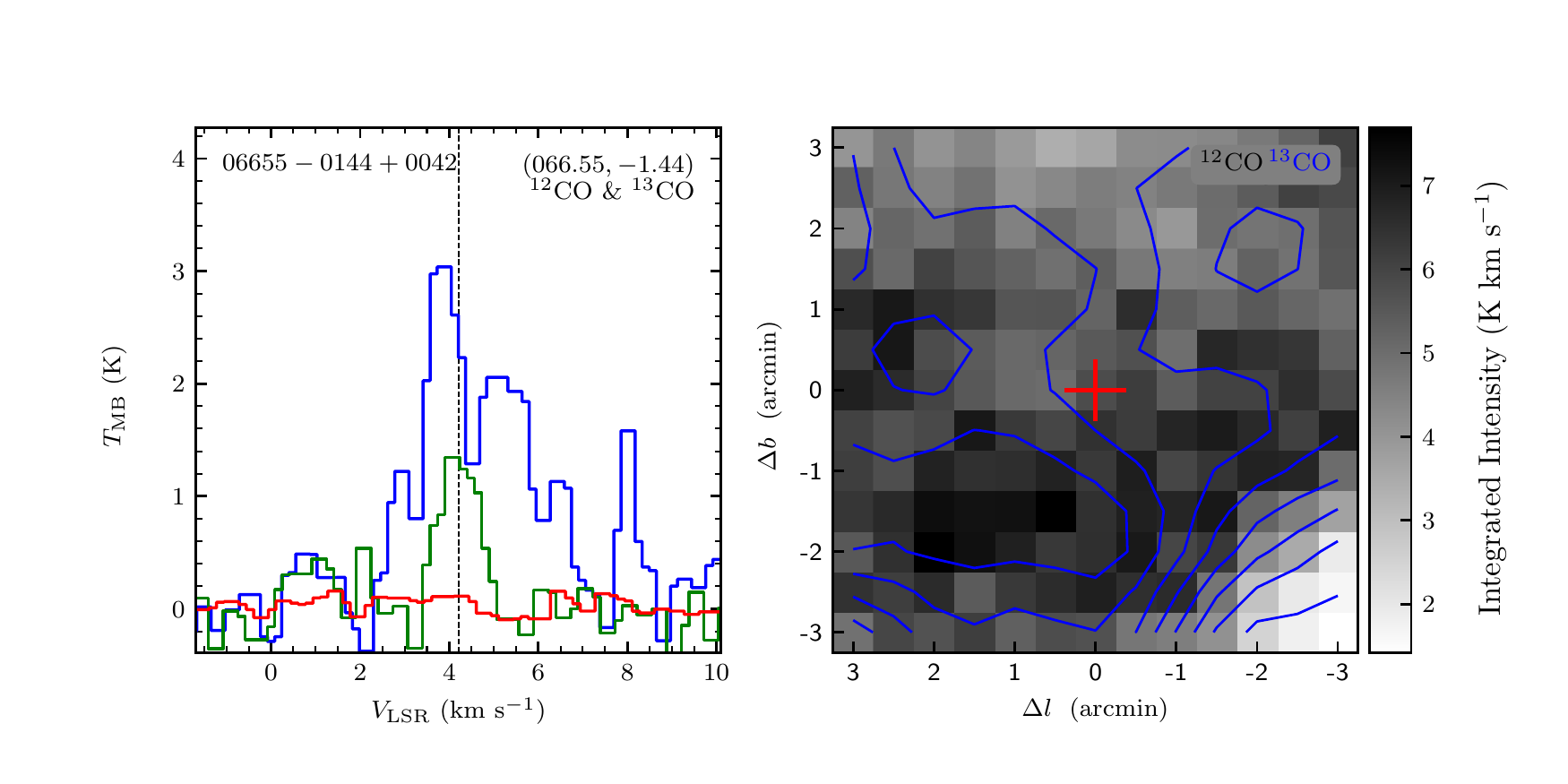}
\includegraphics[width=9.0cm,angle=0]{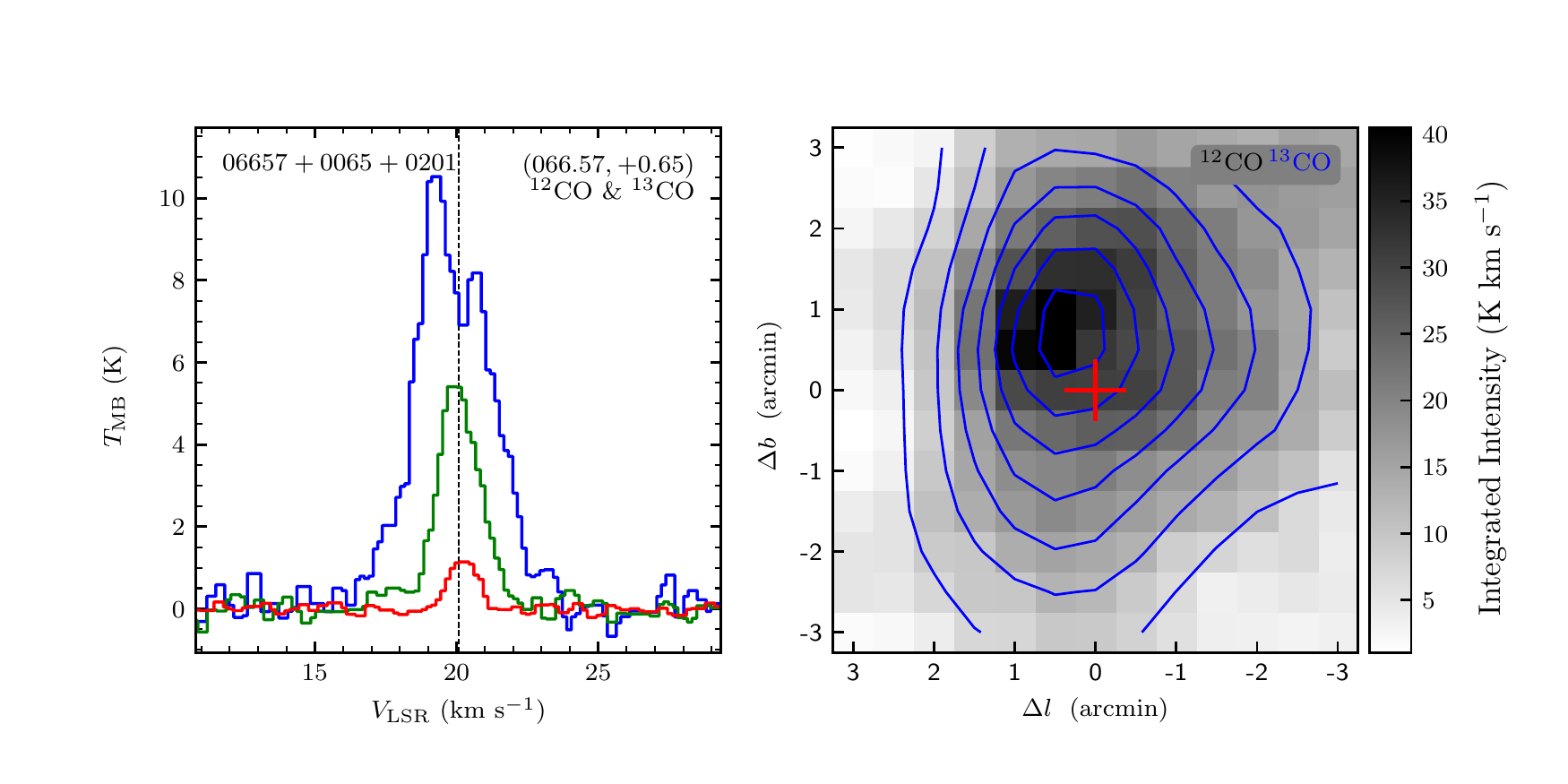}
\end{figure}
\clearpage

\begin{figure}
\includegraphics[width=9.0cm,angle=0]{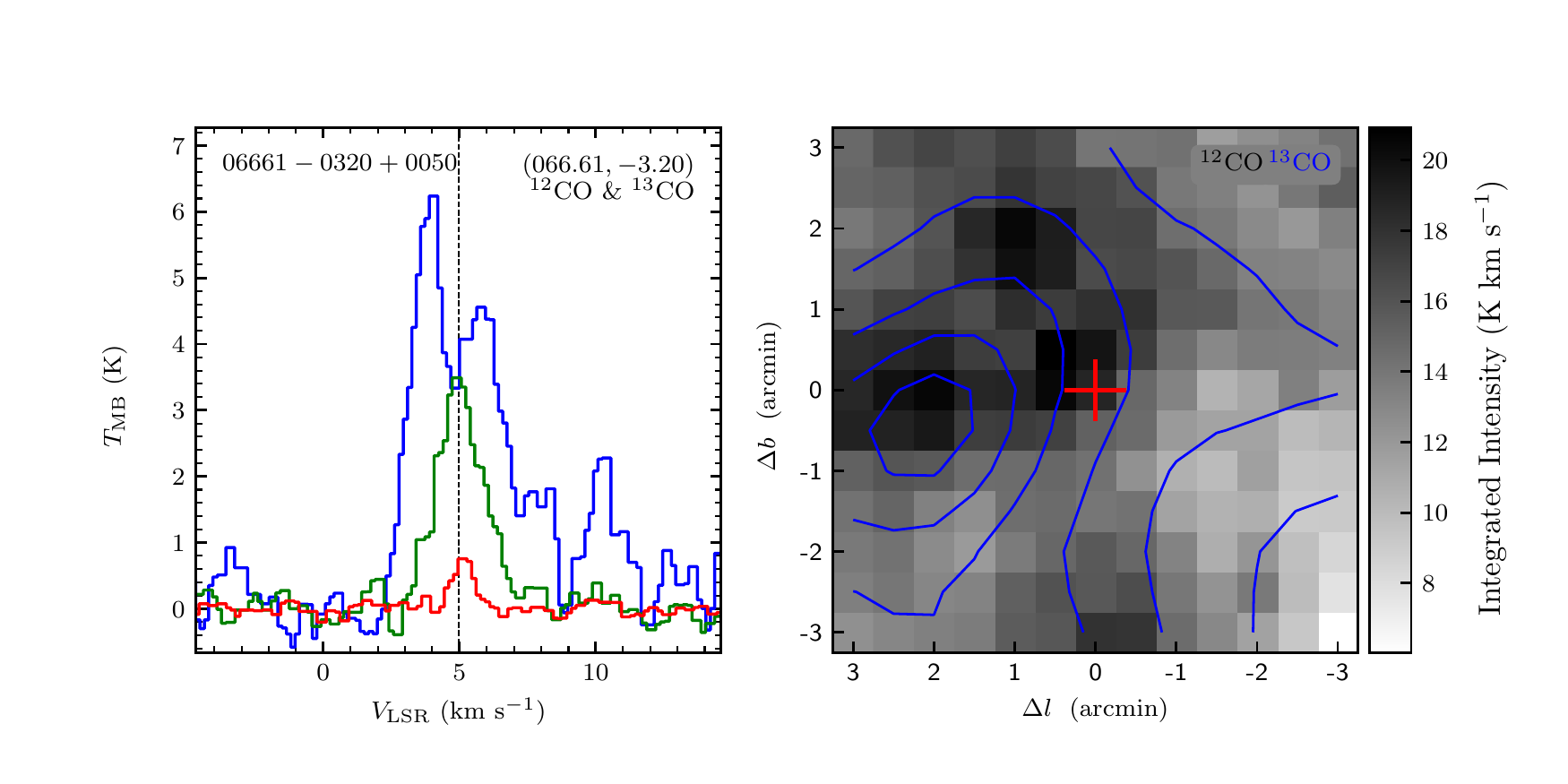}
\includegraphics[width=9.0cm,angle=0]{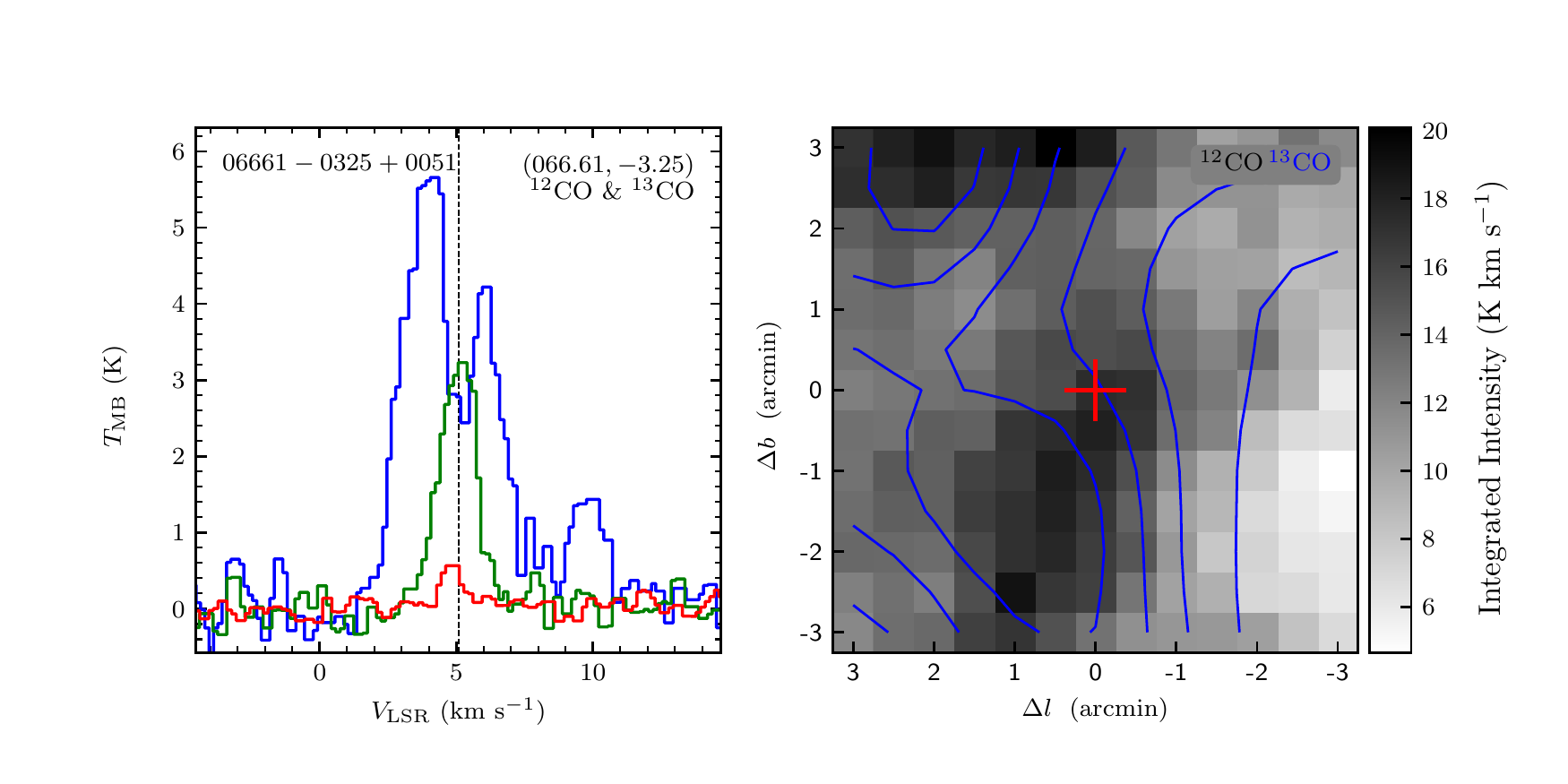}
\vspace{-0.5cm}

\includegraphics[width=9.0cm,angle=0]{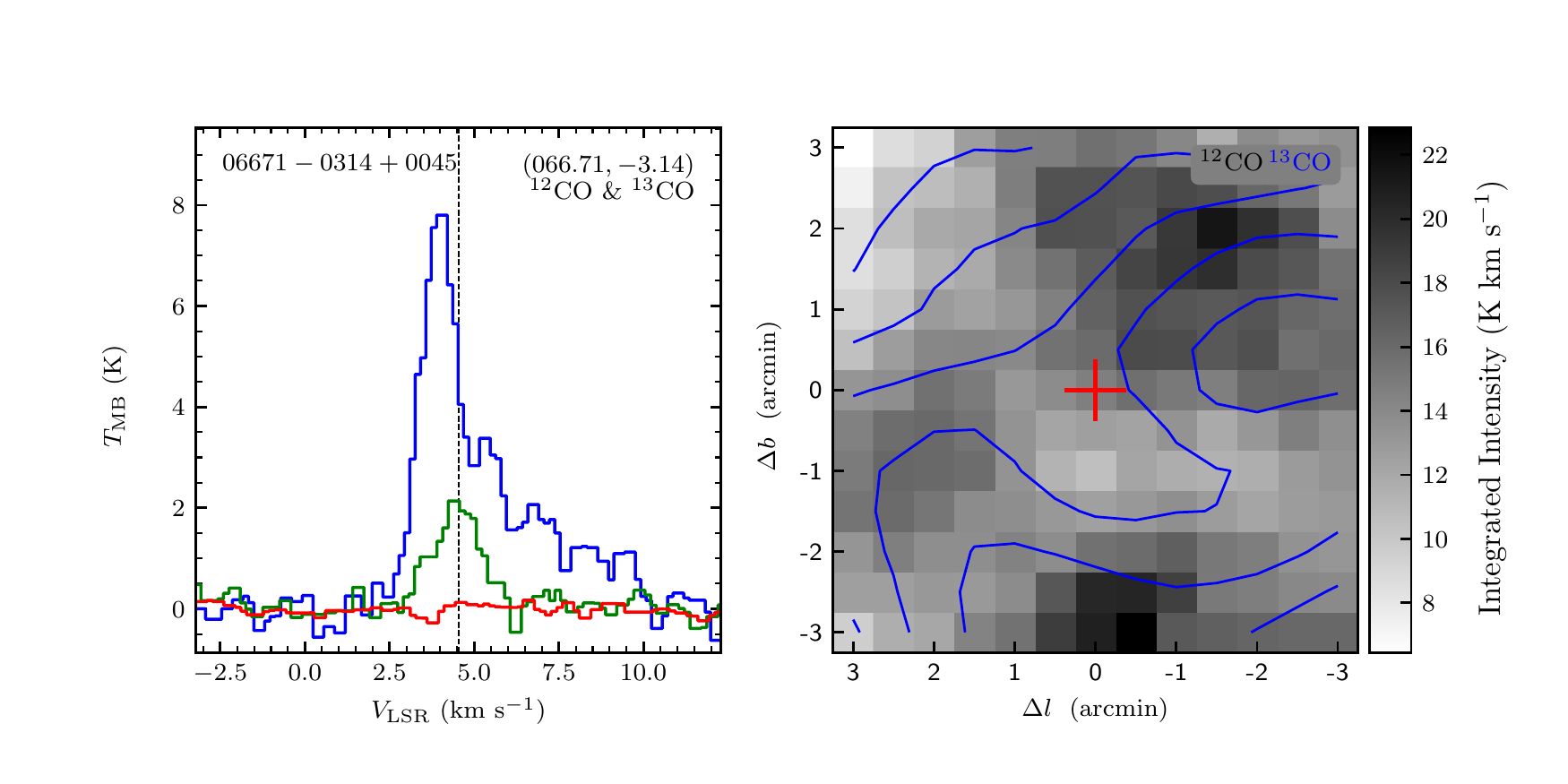}
\includegraphics[width=9.0cm,angle=0]{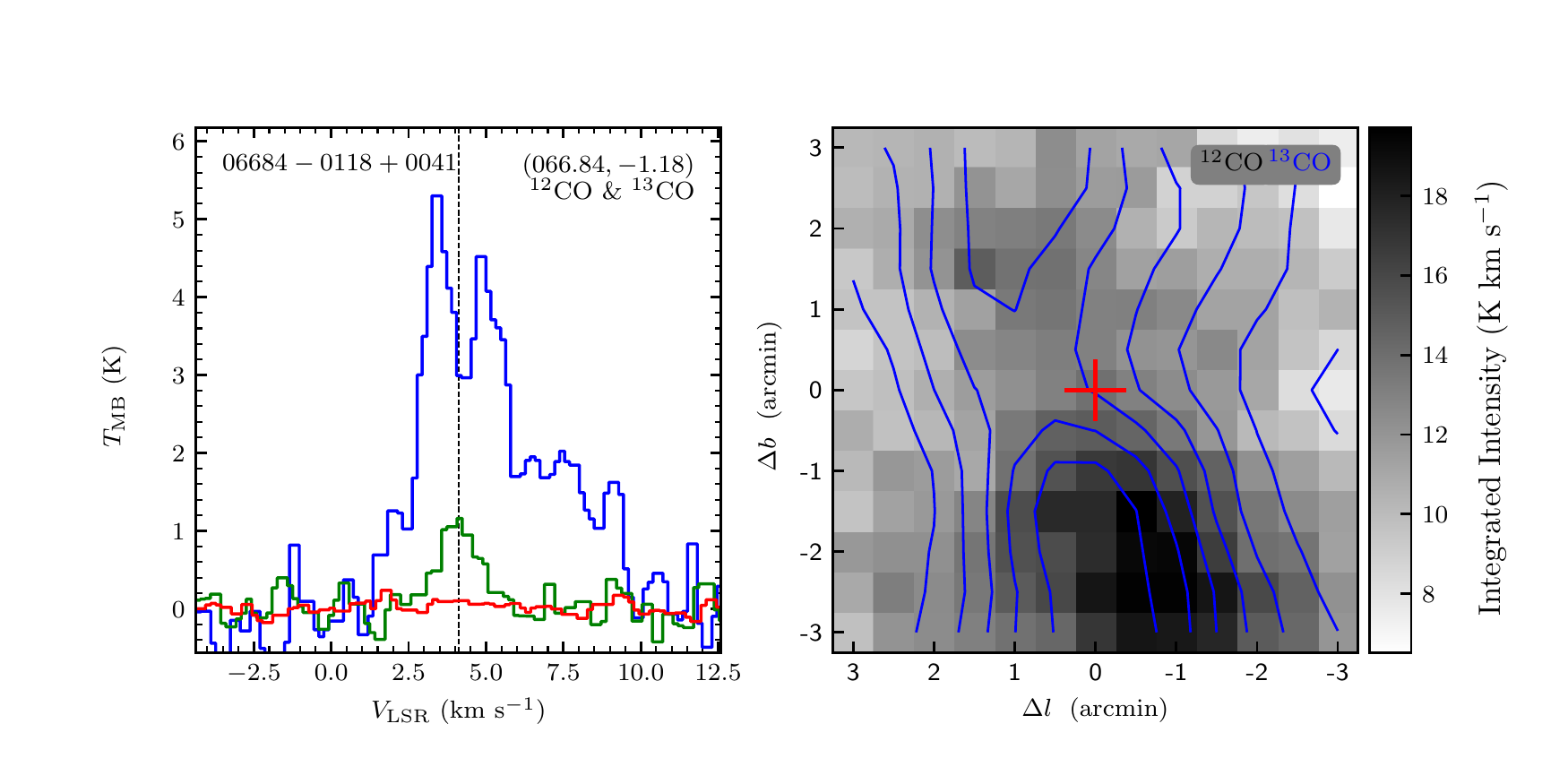}
\vspace{-0.5cm}

\includegraphics[width=9.0cm,angle=0]{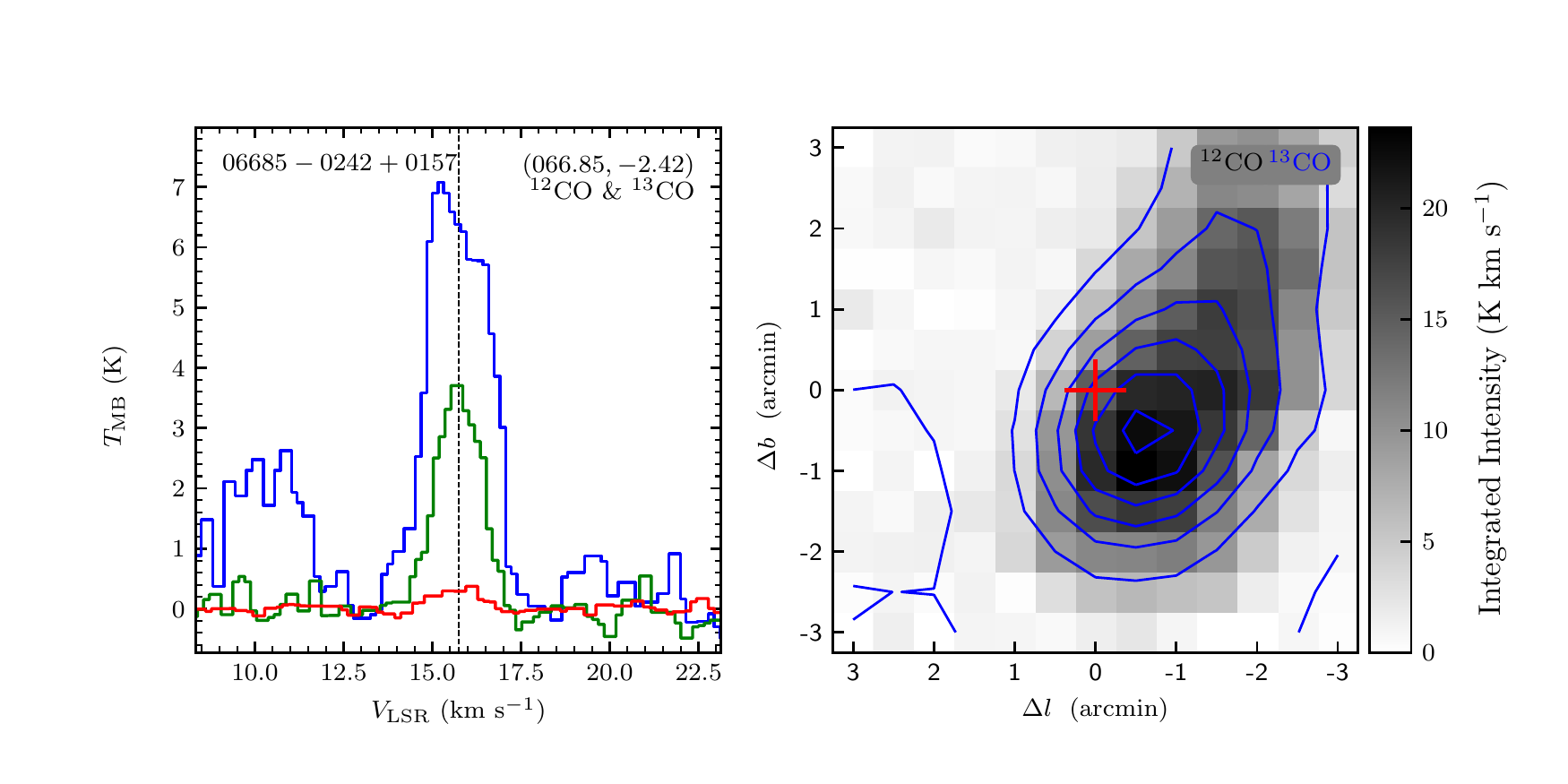}
\includegraphics[width=9.0cm,angle=0]{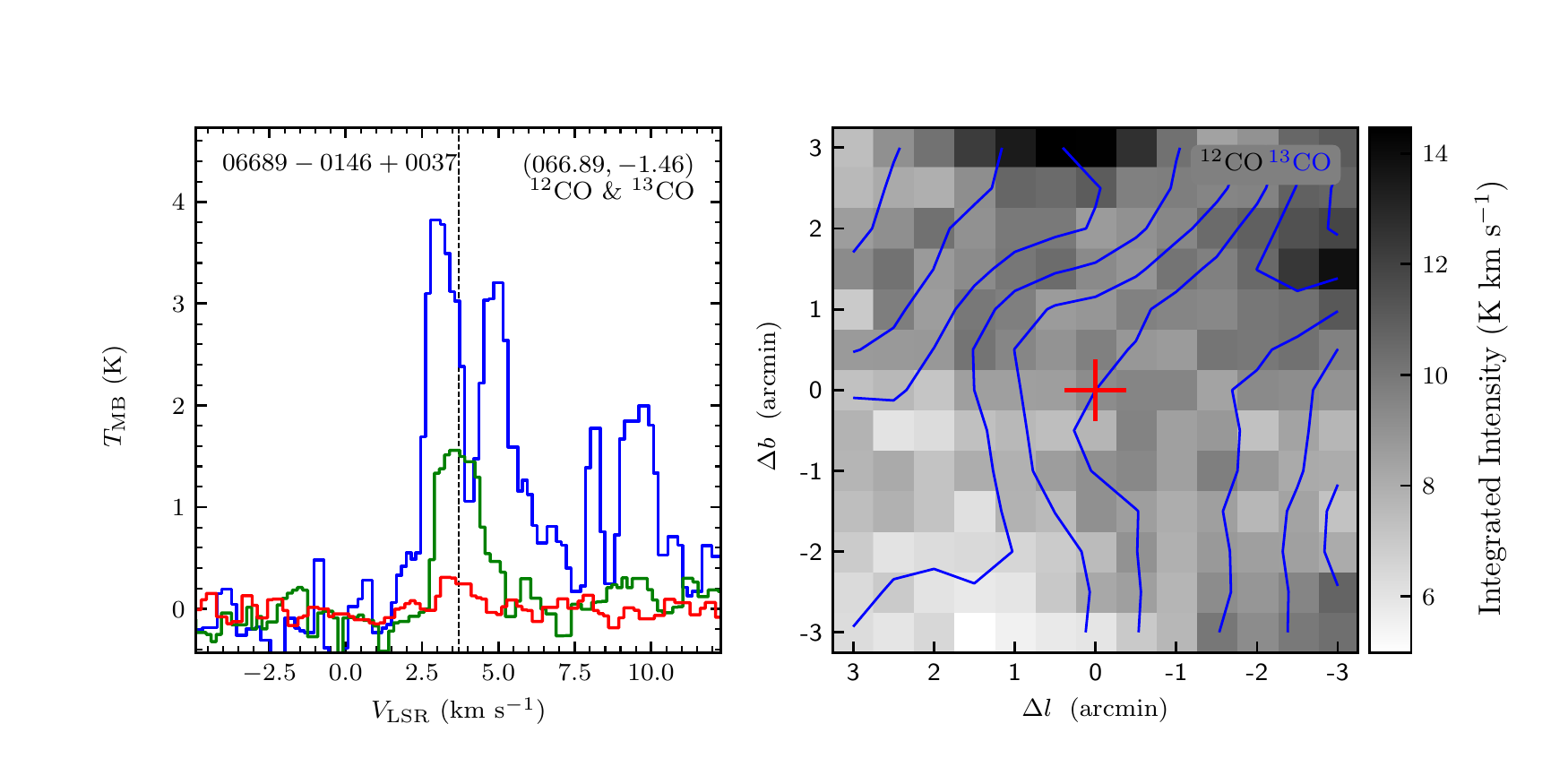}
\vspace{-0.5cm}

\includegraphics[width=9.0cm,angle=0]{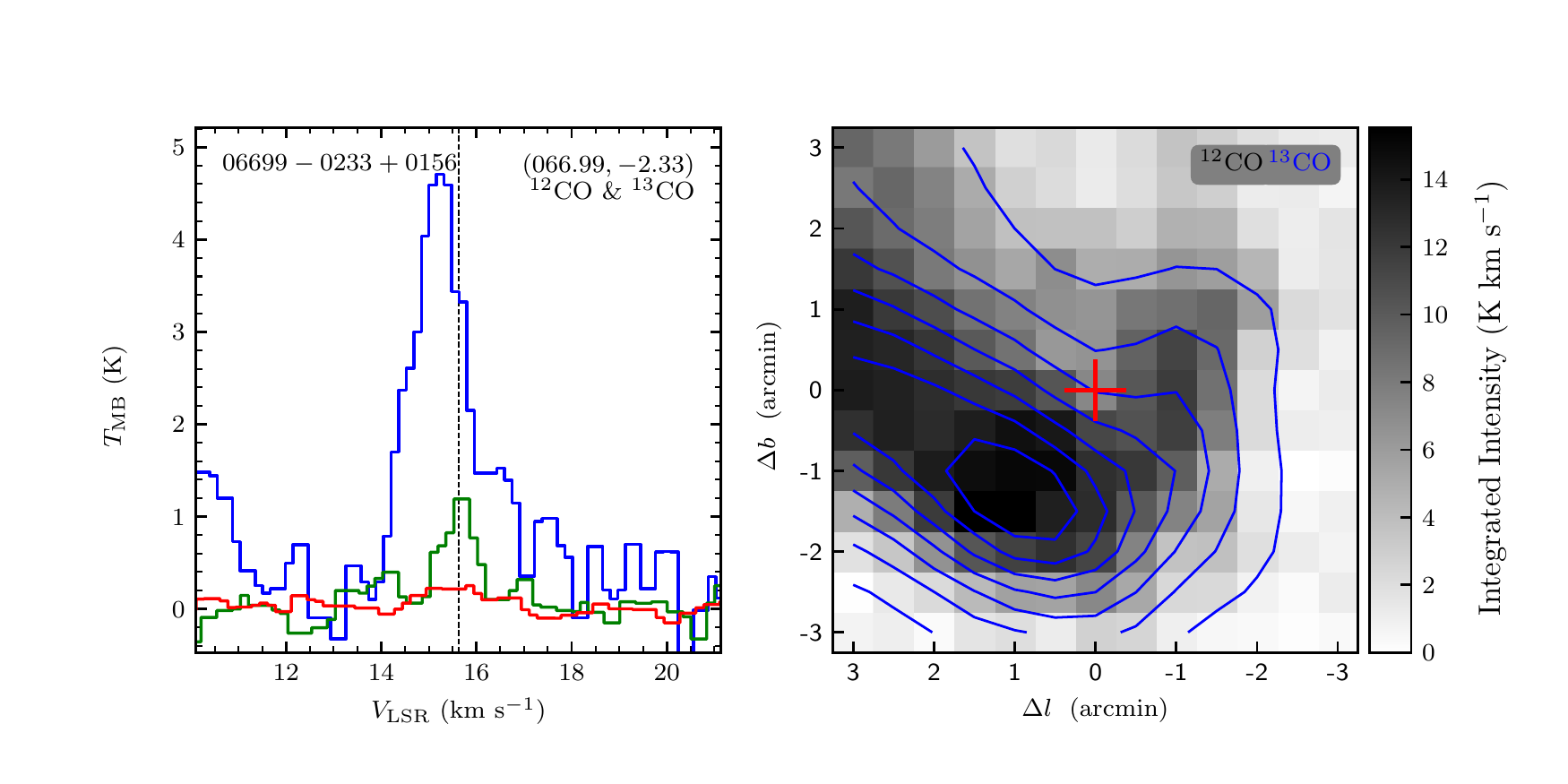}
\includegraphics[width=9.0cm,angle=0]{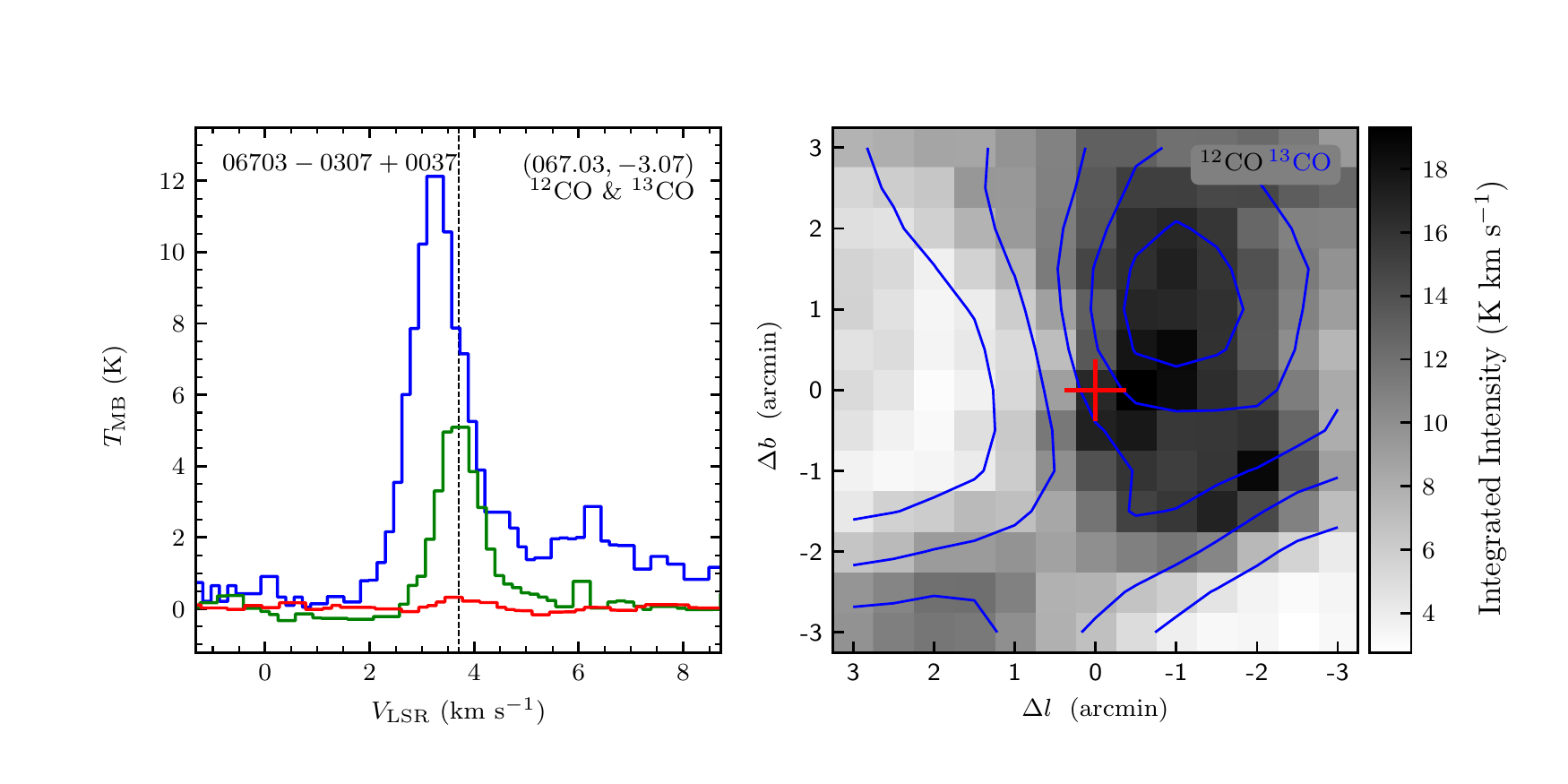}
\vspace{-0.5cm}

\includegraphics[width=9.0cm,angle=0]{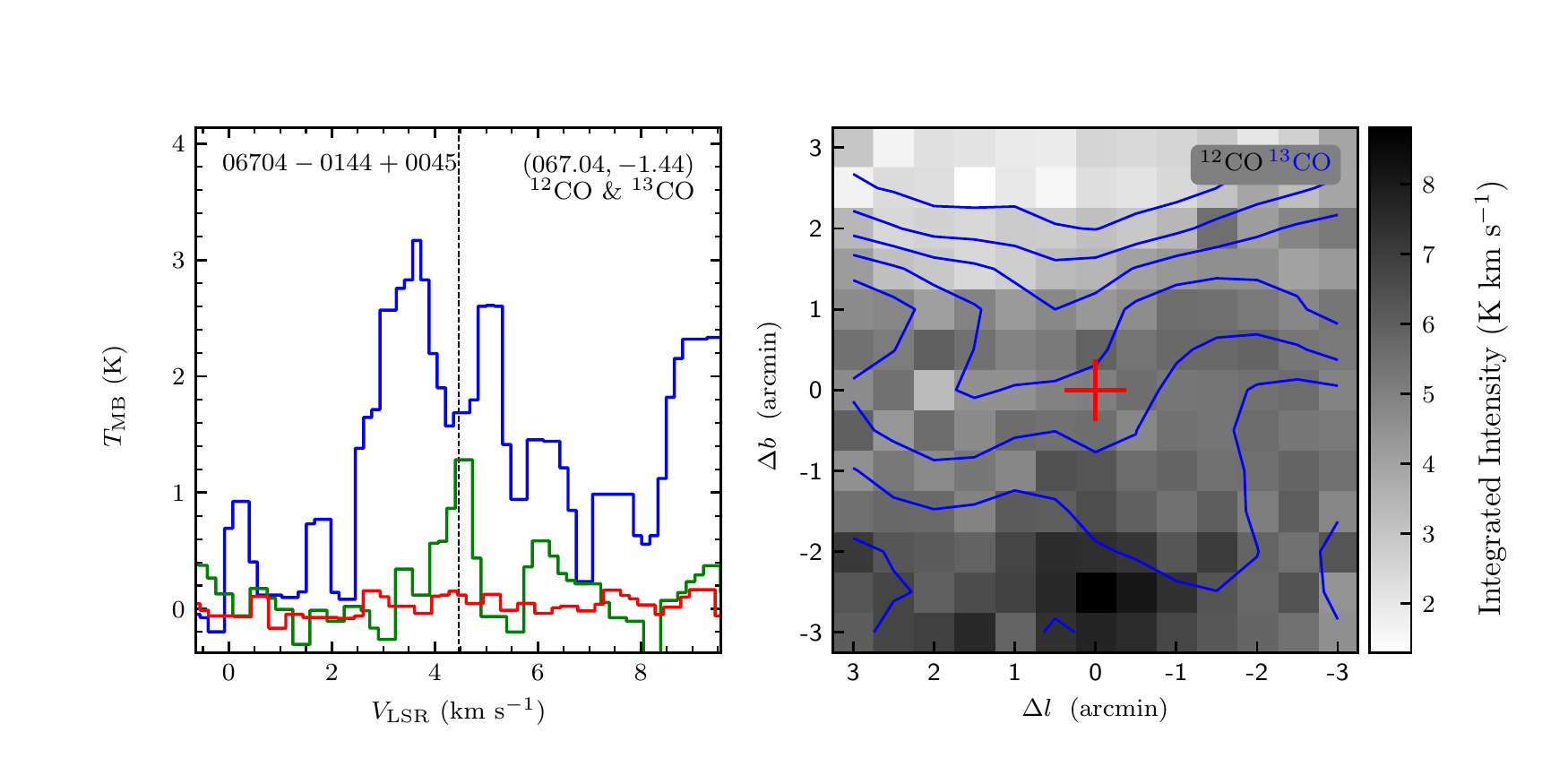}
\includegraphics[width=9.0cm,angle=0]{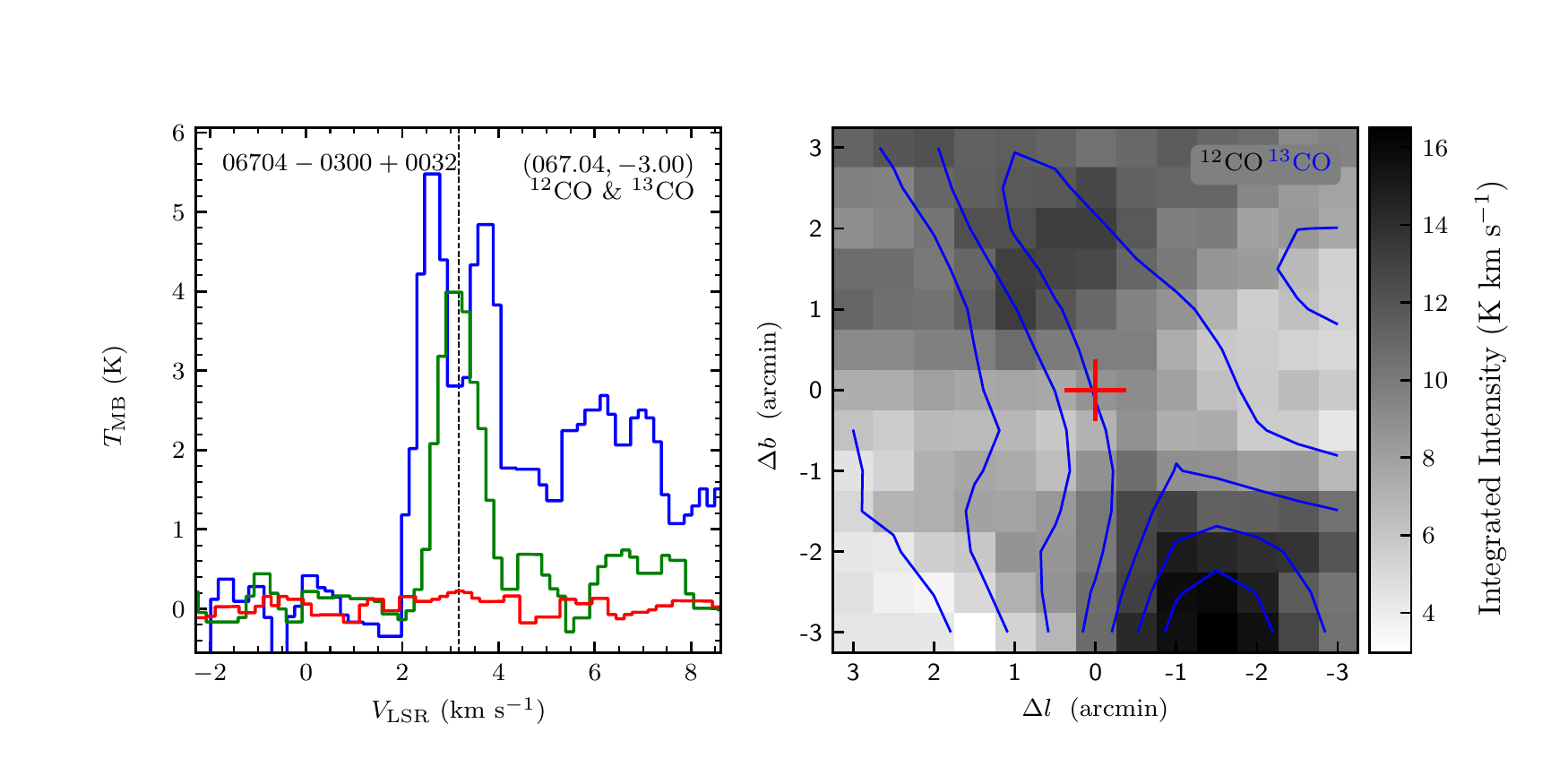}
\end{figure}
\clearpage

\begin{figure}
\includegraphics[width=9.0cm,angle=0]{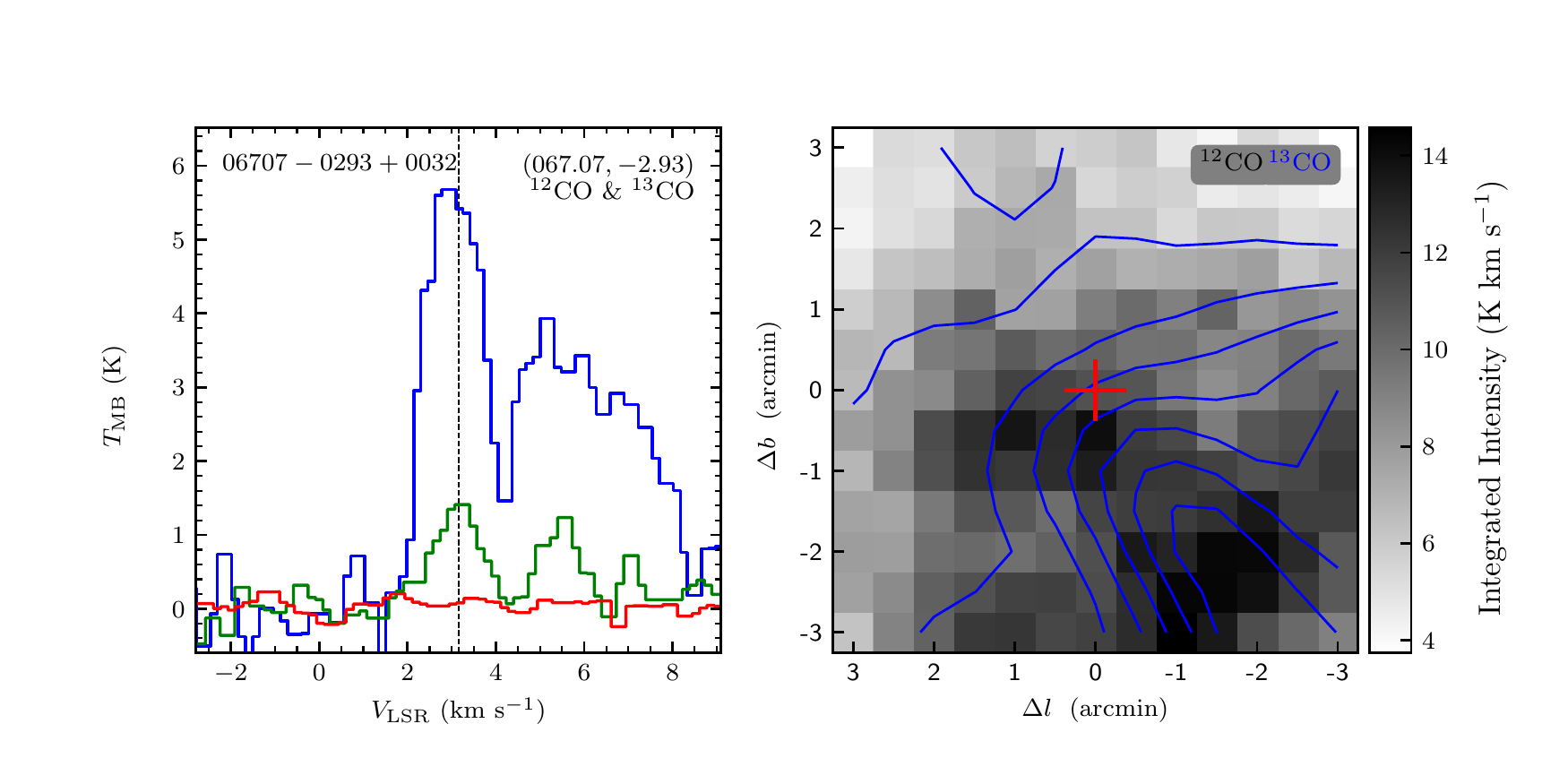}
\includegraphics[width=9.0cm,angle=0]{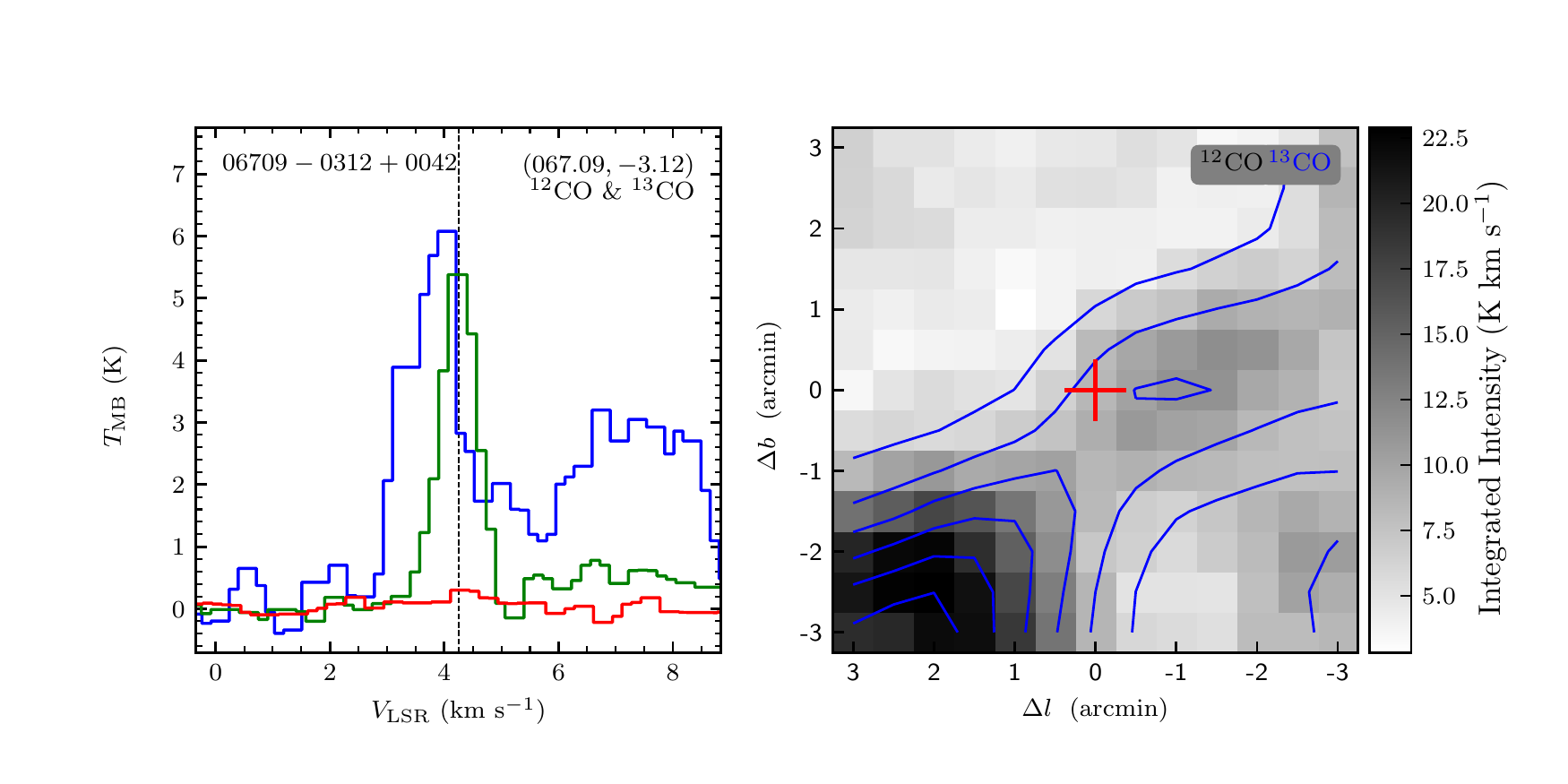}
\vspace{-0.5cm}

\includegraphics[width=9.0cm,angle=0]{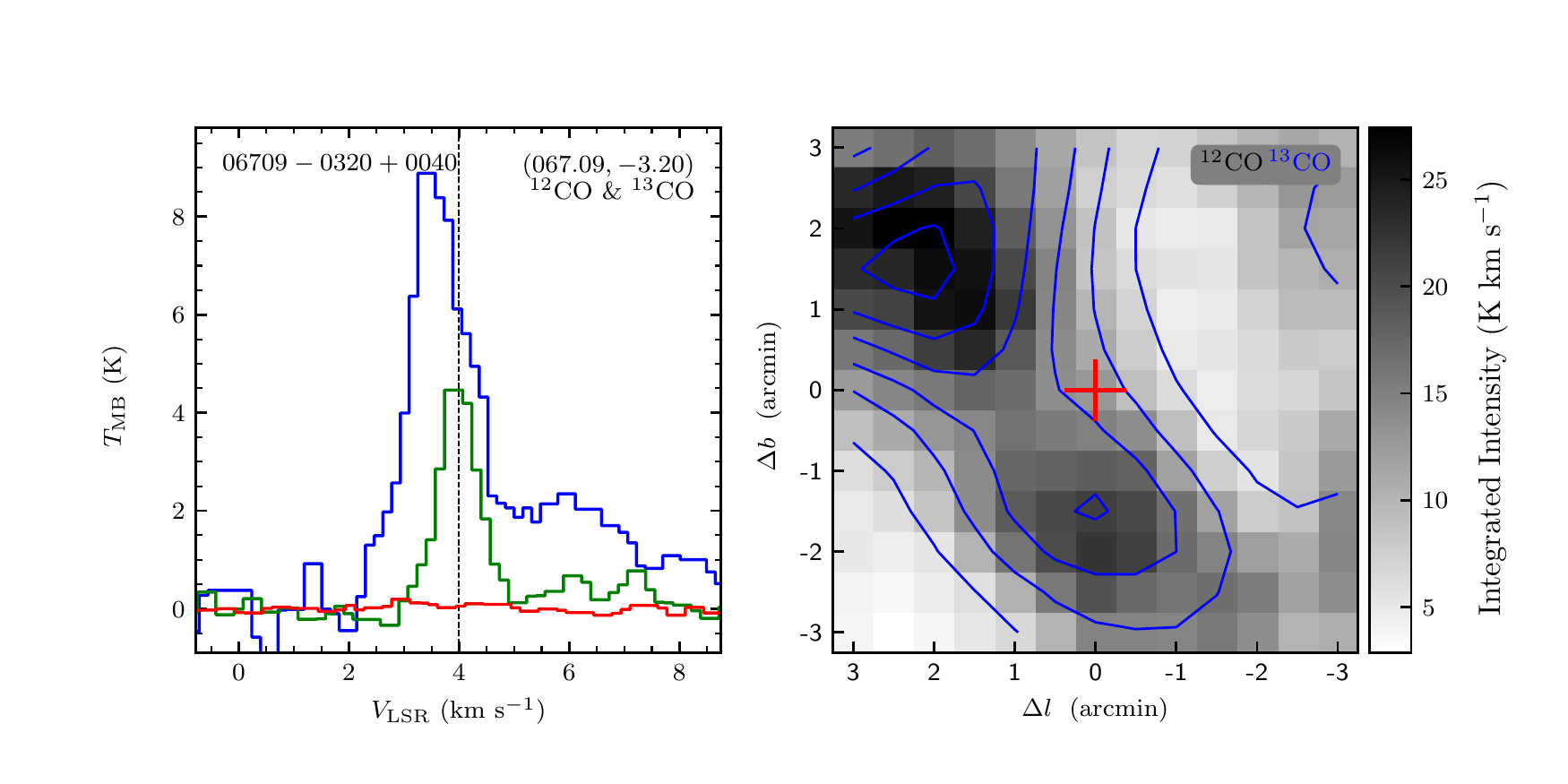}
\includegraphics[width=9.0cm,angle=0]{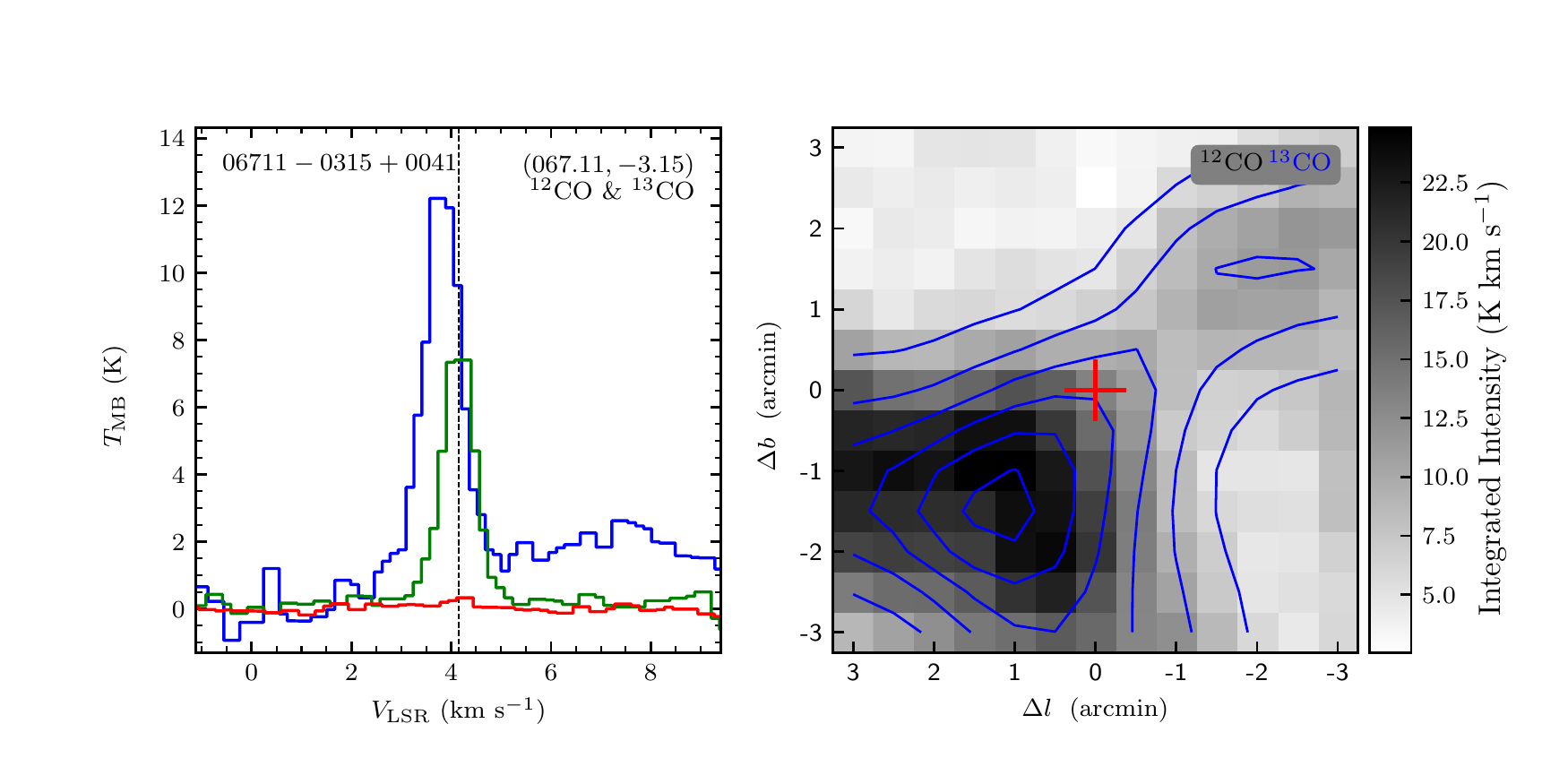}
\vspace{-0.5cm}

\includegraphics[width=9.0cm,angle=0]{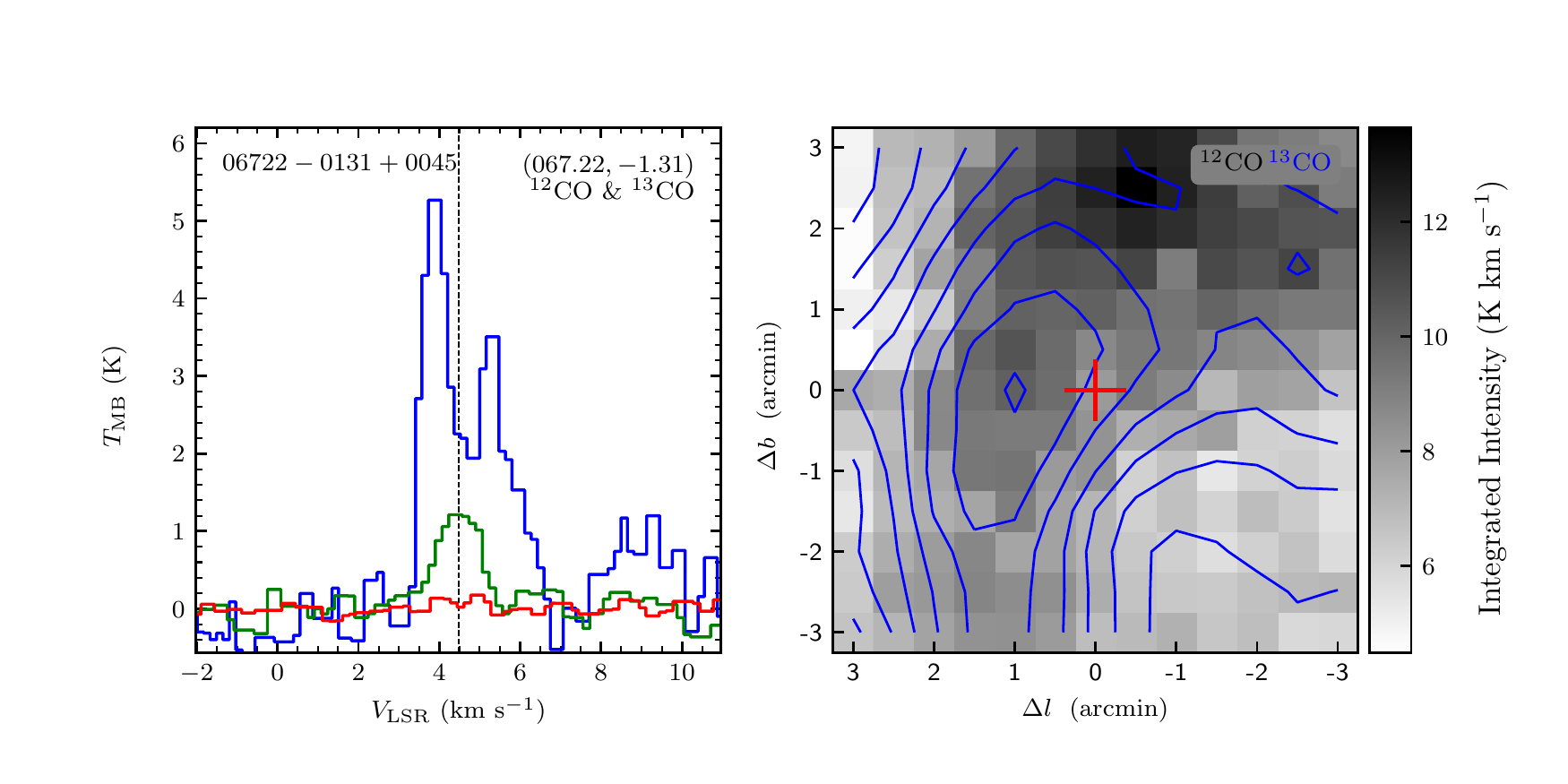}
\includegraphics[width=9.0cm,angle=0]{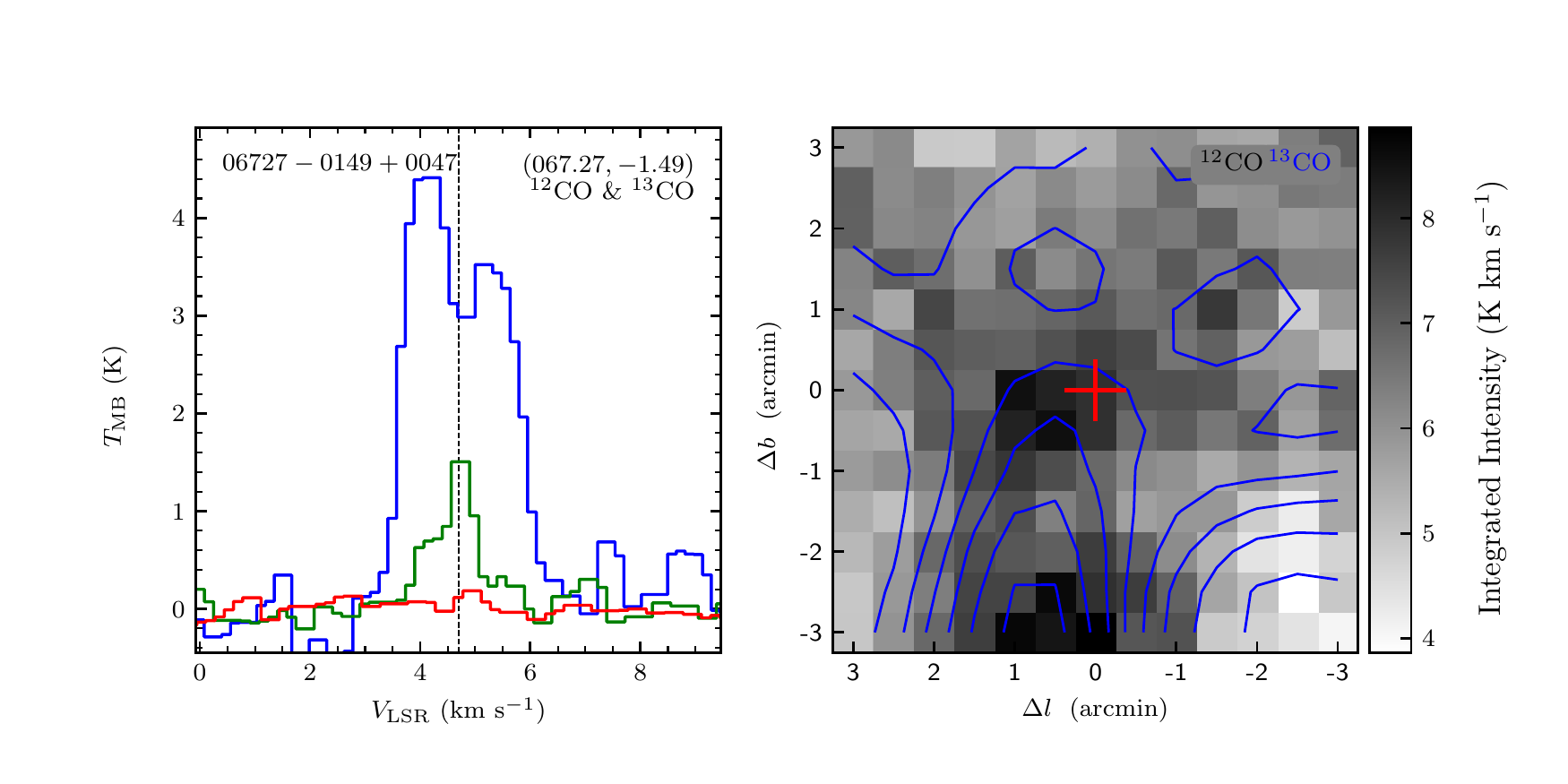}
\vspace{-0.5cm}

\includegraphics[width=9.0cm,angle=0]{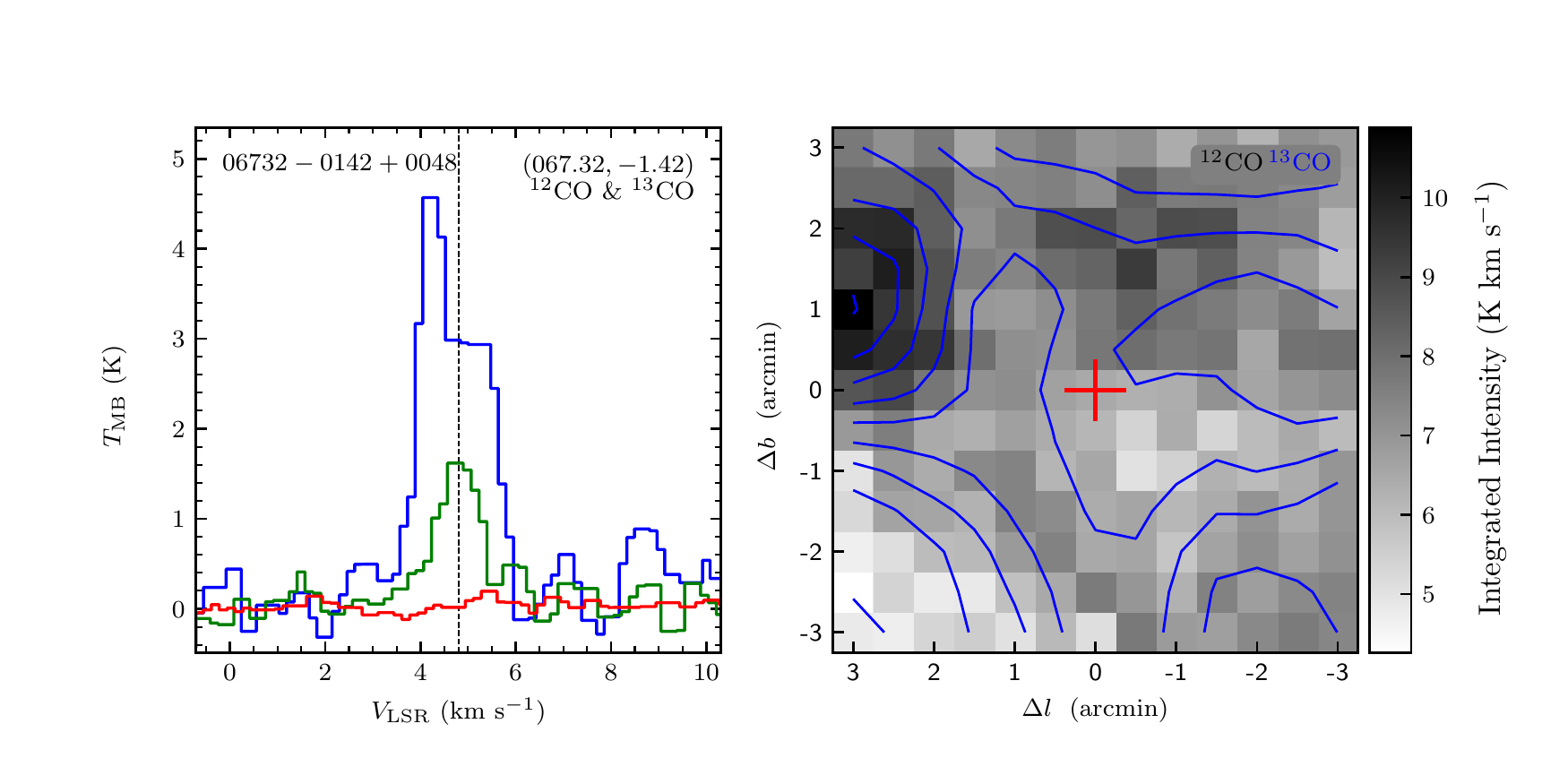}
\includegraphics[width=9.0cm,angle=0]{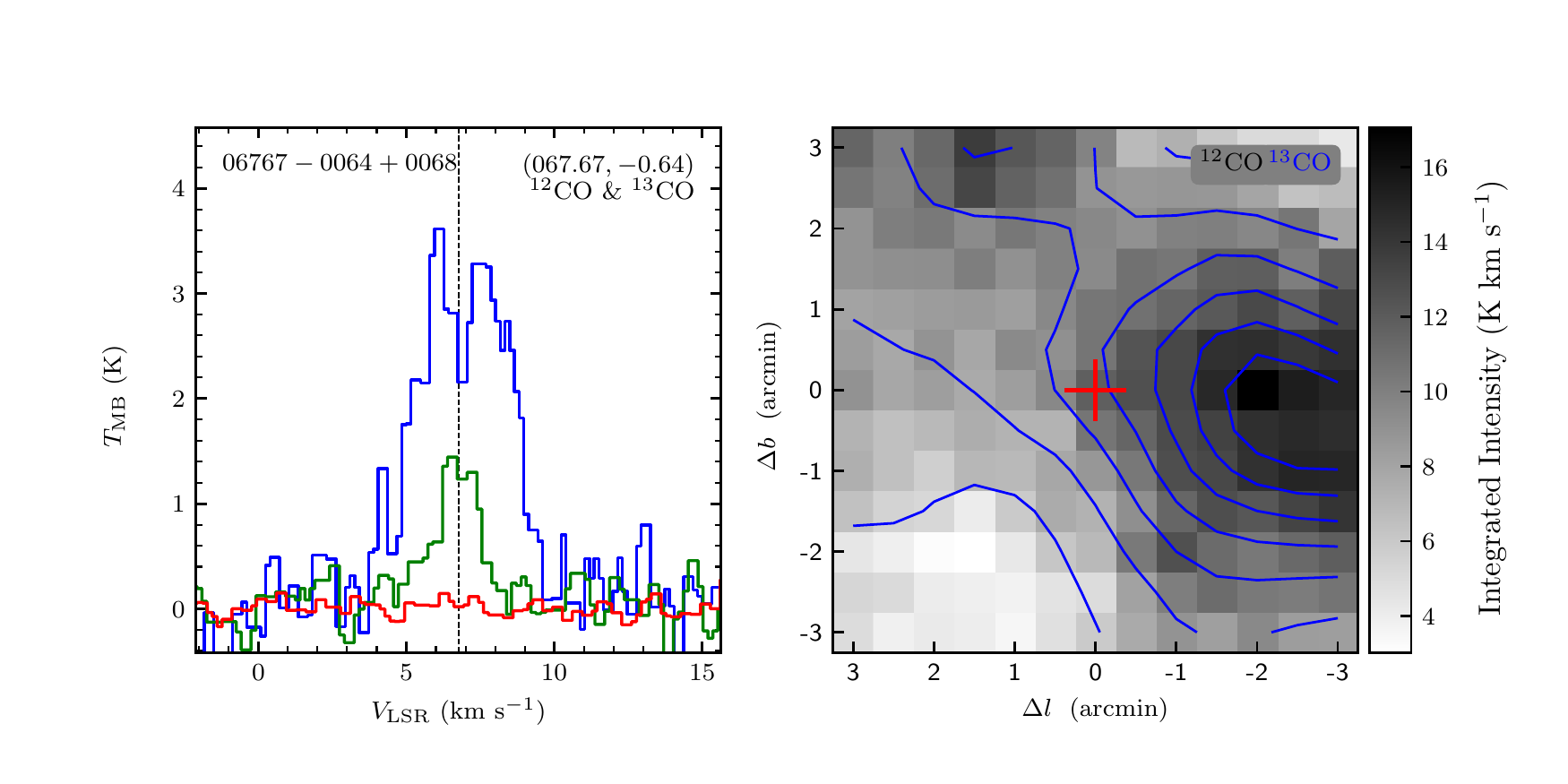}
\vspace{-0.5cm}

\includegraphics[width=9.0cm,angle=0]{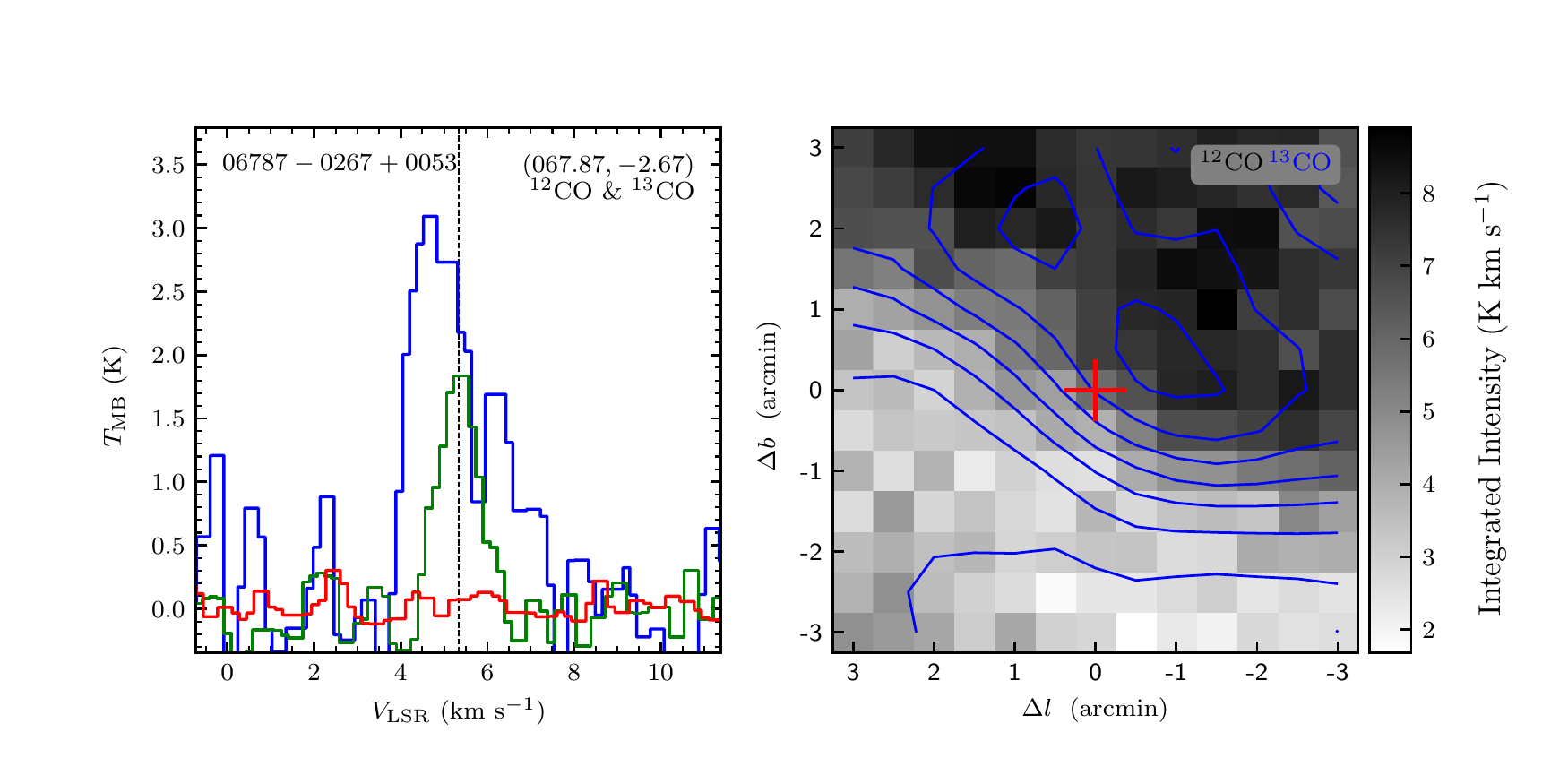}
\includegraphics[width=9.0cm,angle=0]{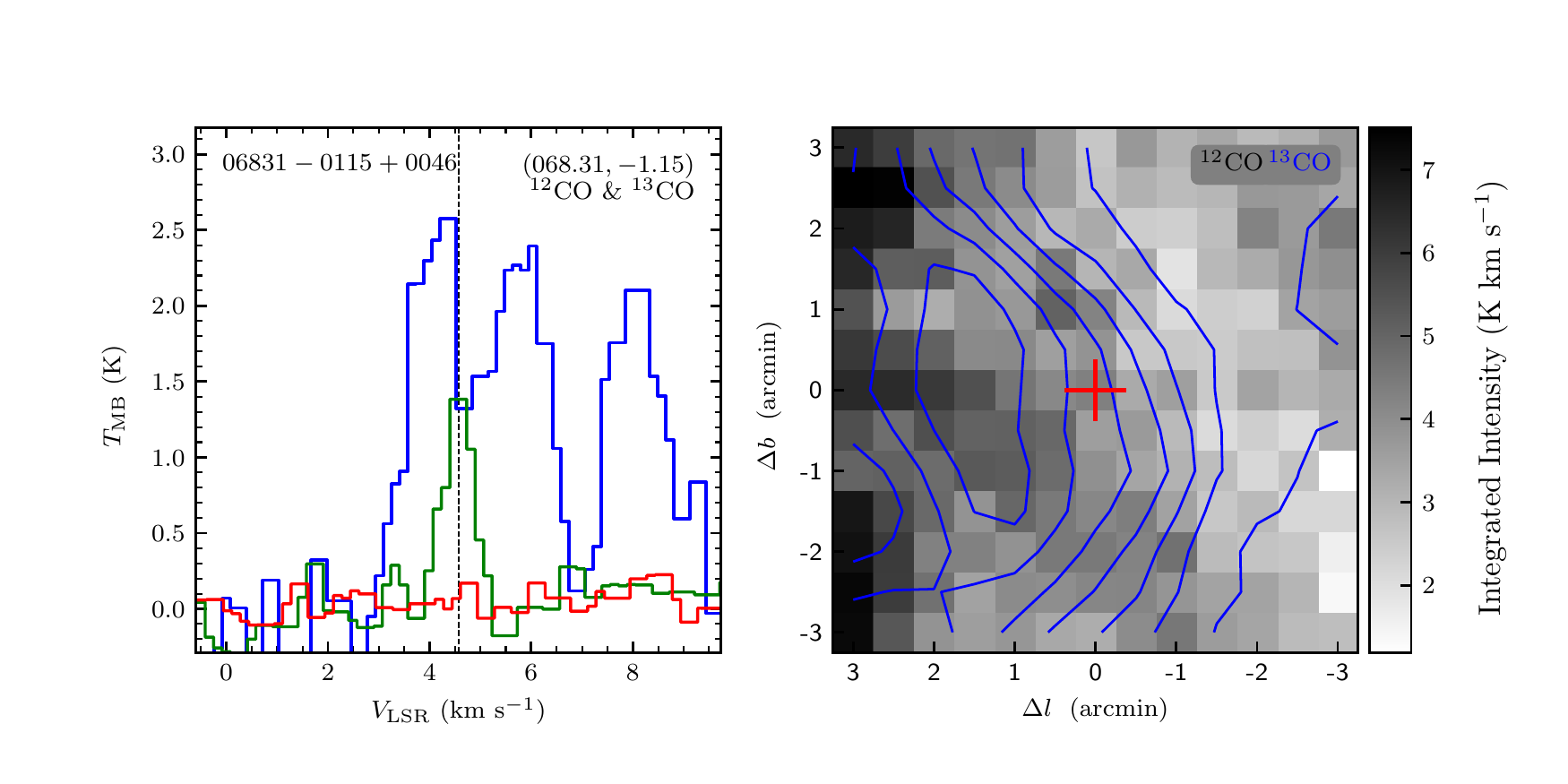}
\end{figure}
\clearpage

\begin{figure}
\includegraphics[width=9.0cm,angle=0]{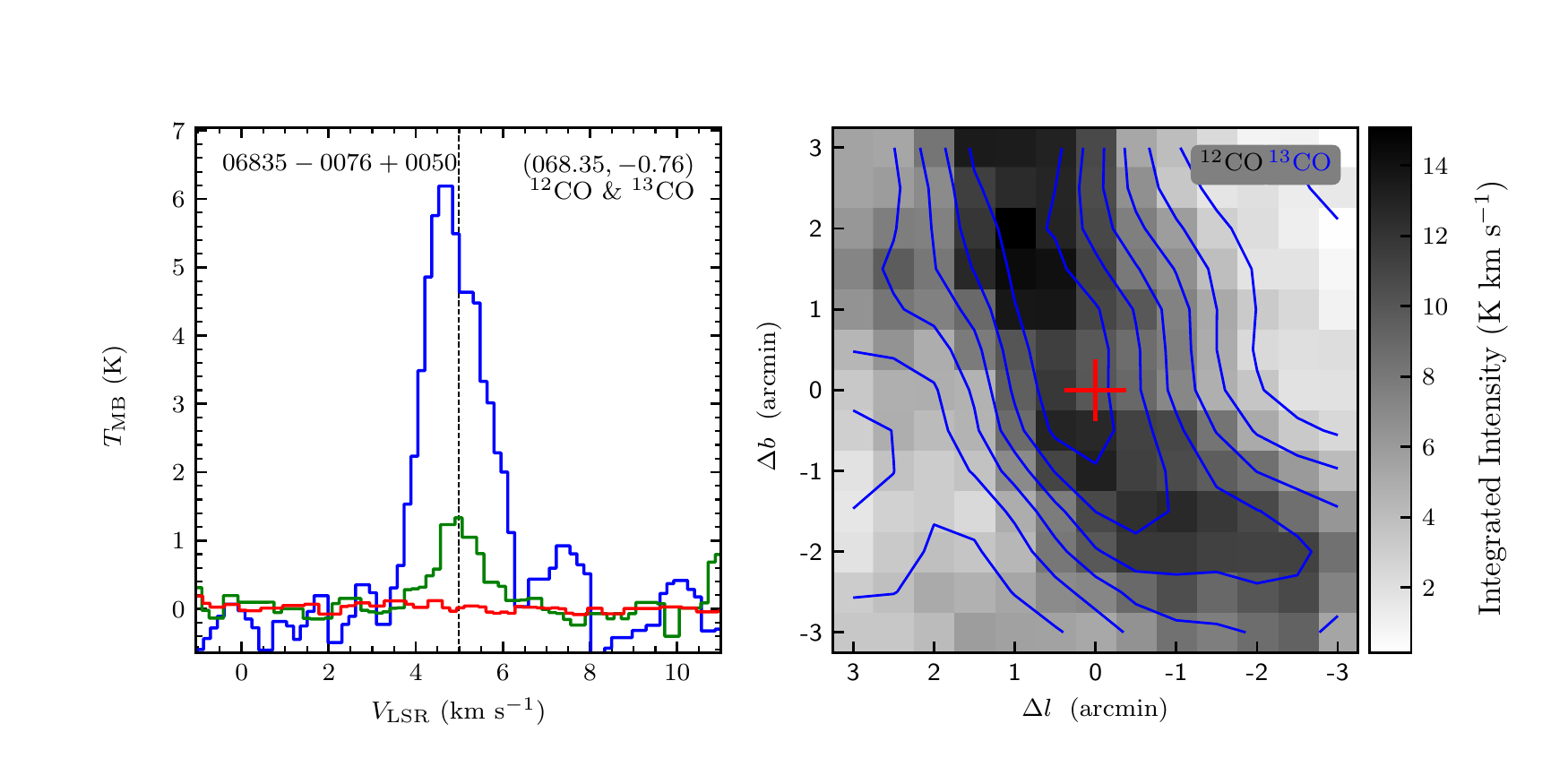}
\includegraphics[width=9.0cm,angle=0]{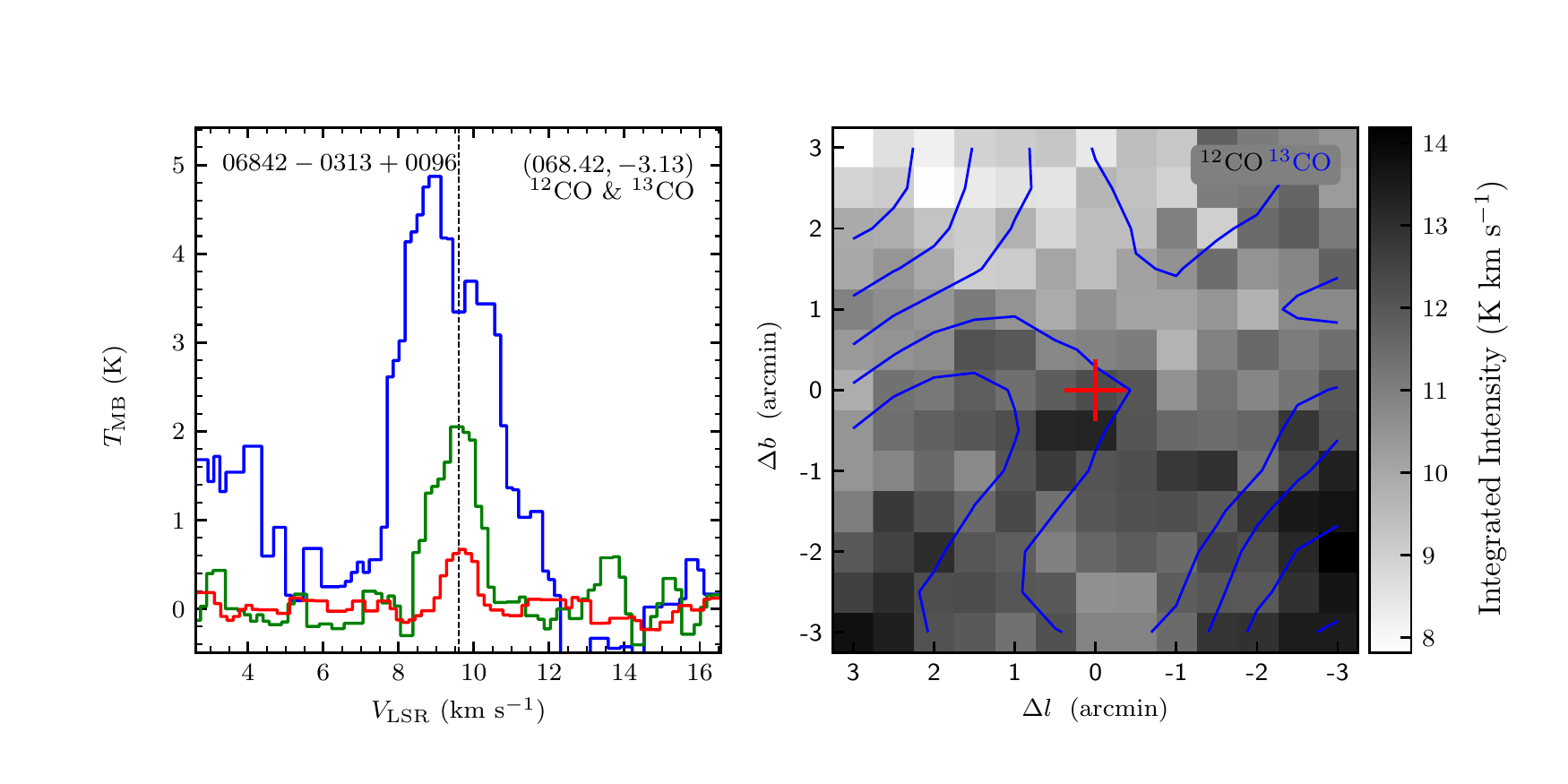}
\vspace{-0.5cm}

\includegraphics[width=9.0cm,angle=0]{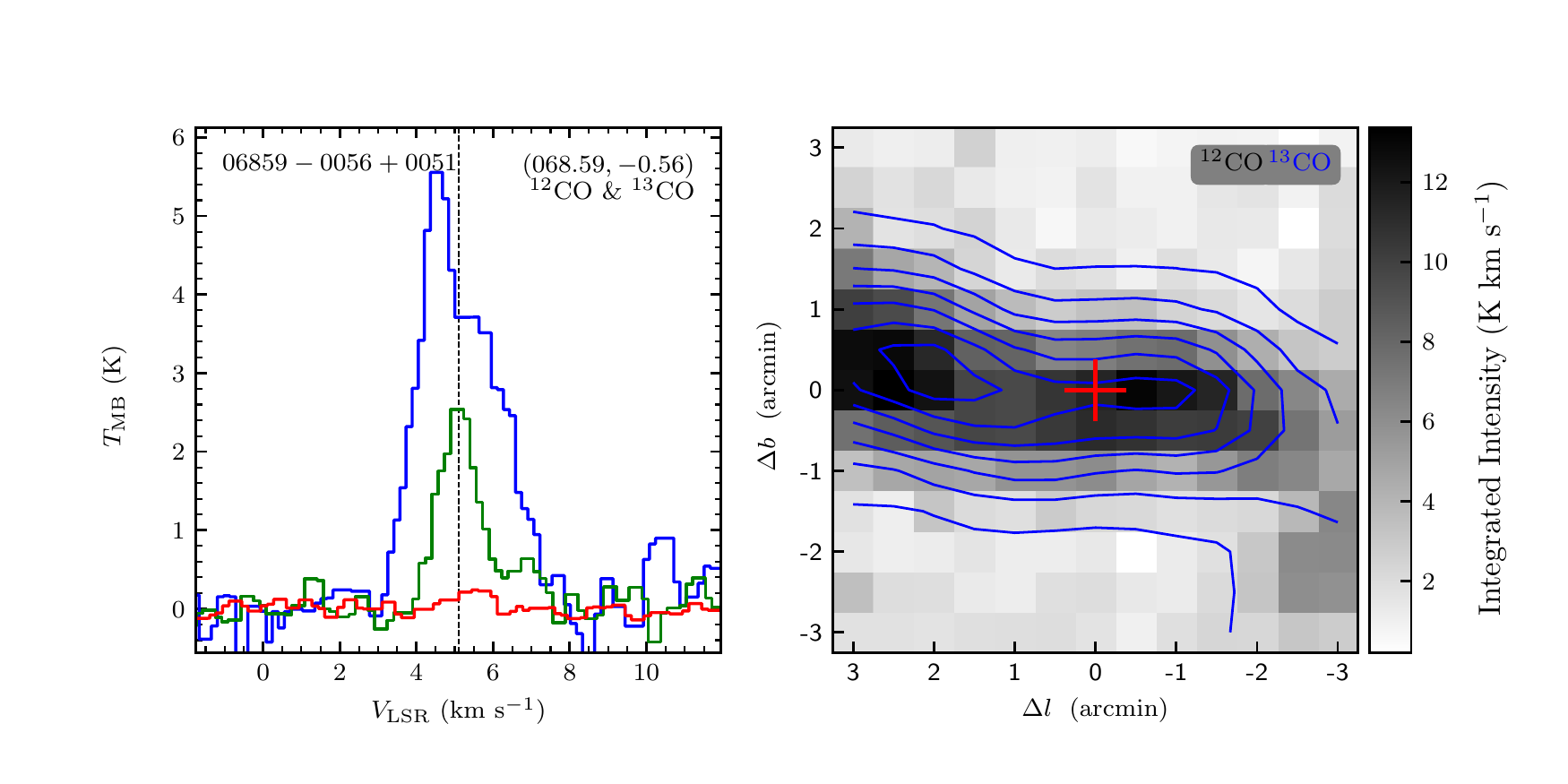}
\includegraphics[width=9.0cm,angle=0]{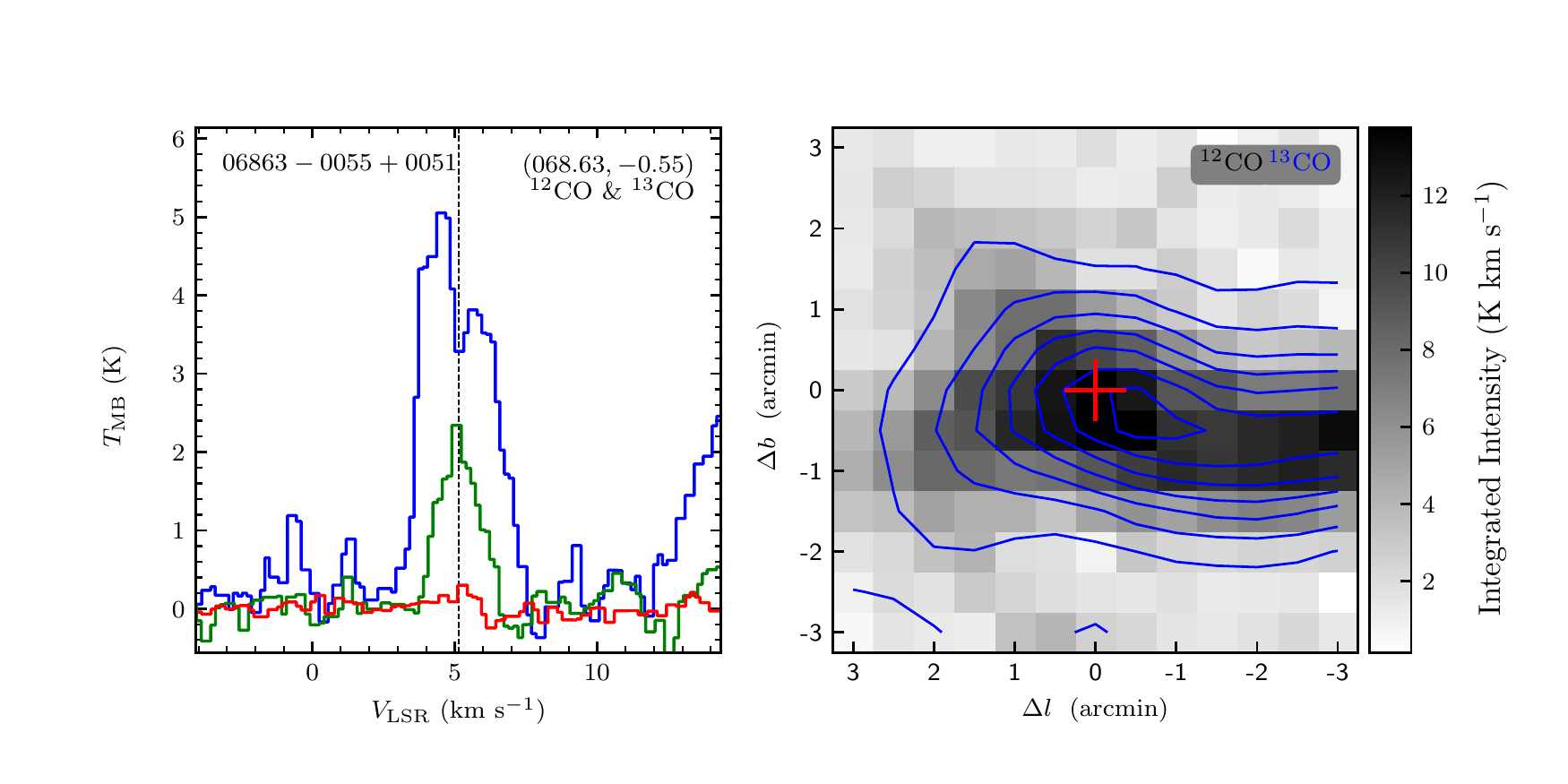}
\vspace{-0.5cm}

\includegraphics[width=9.0cm,angle=0]{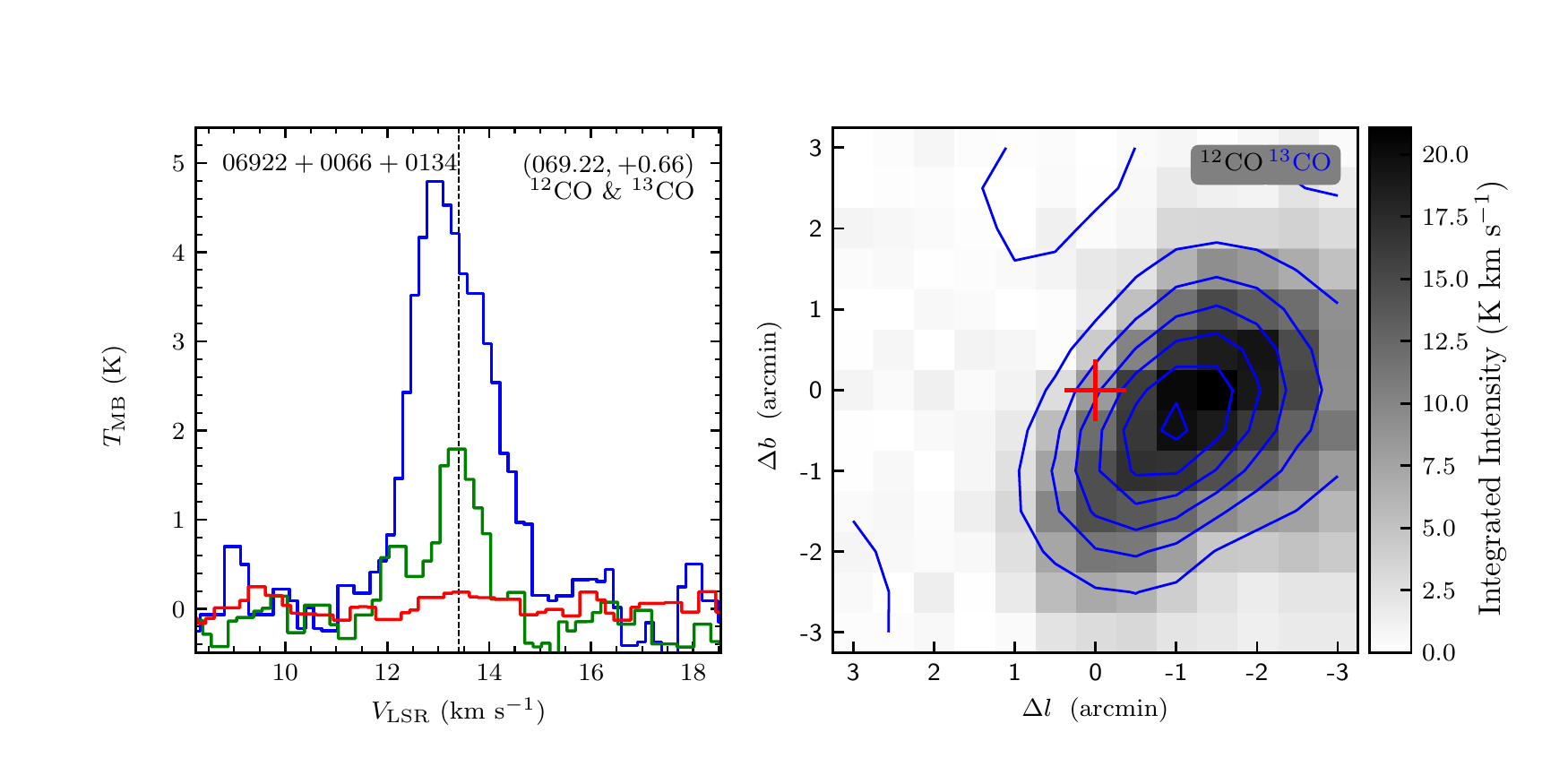}
\includegraphics[width=9.0cm,angle=0]{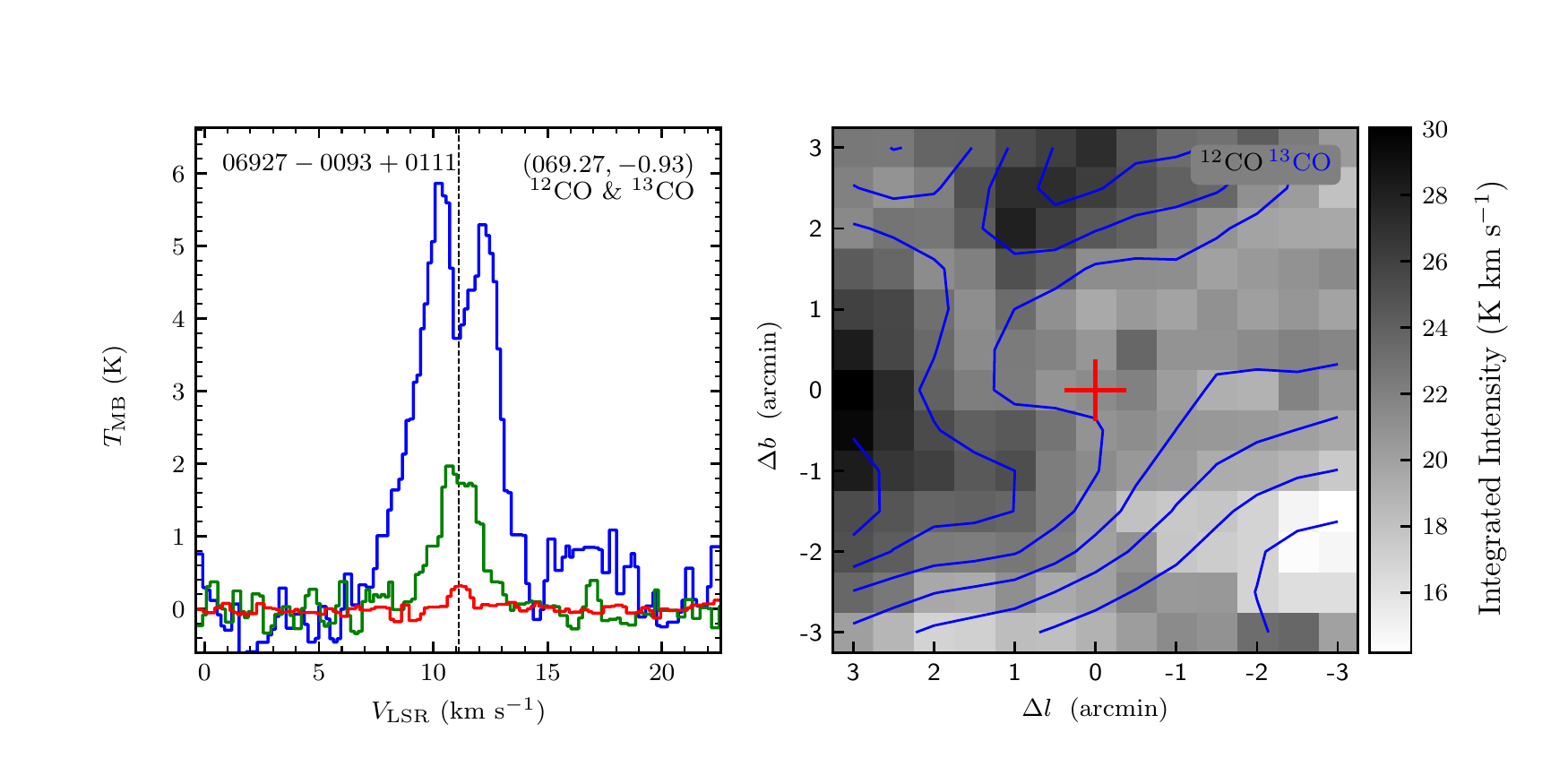}
\vspace{-0.5cm}

\includegraphics[width=9.0cm,angle=0]{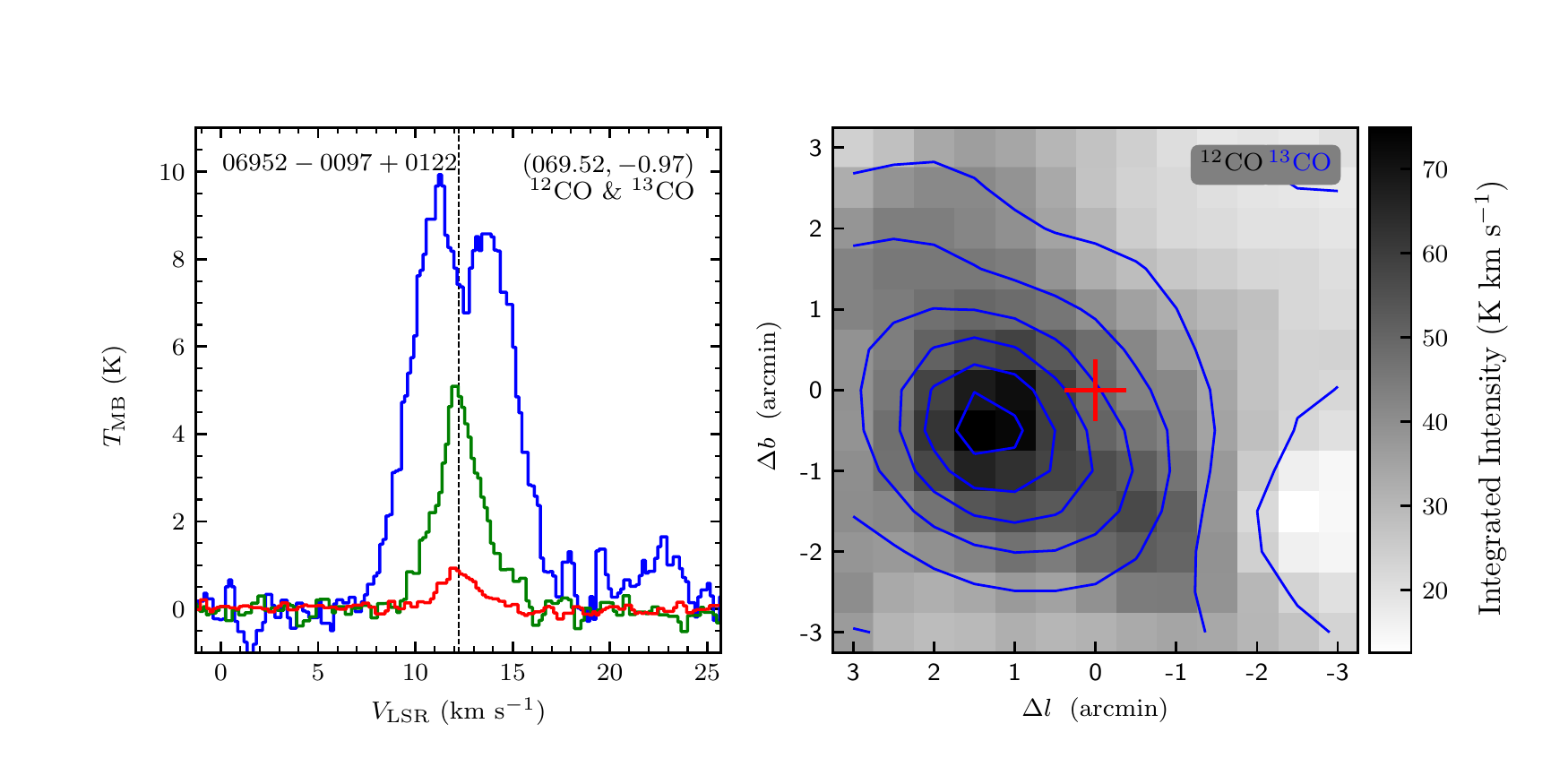}
\includegraphics[width=9.0cm,angle=0]{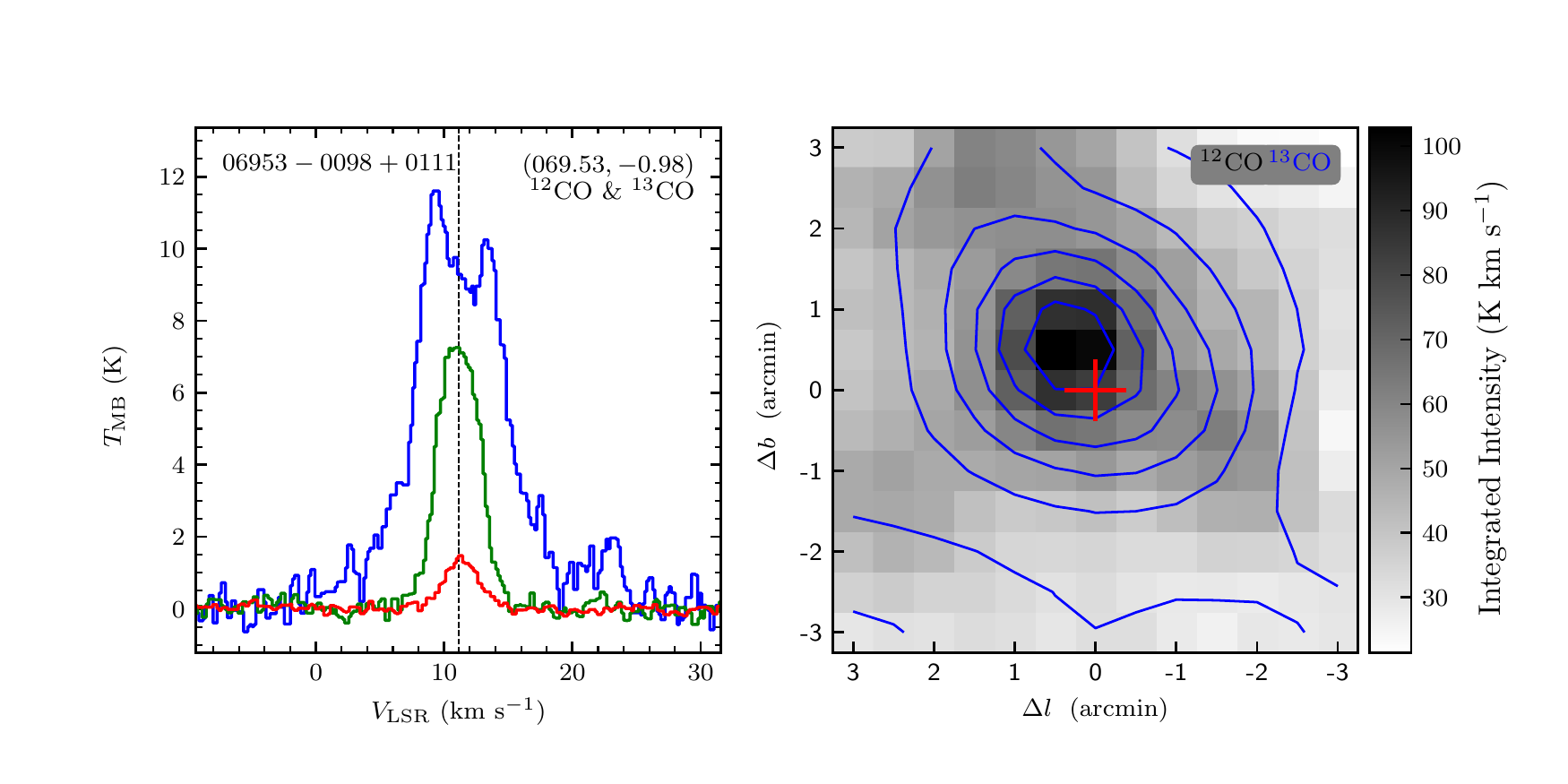}
\vspace{-0.5cm}

\includegraphics[width=9.0cm,angle=0]{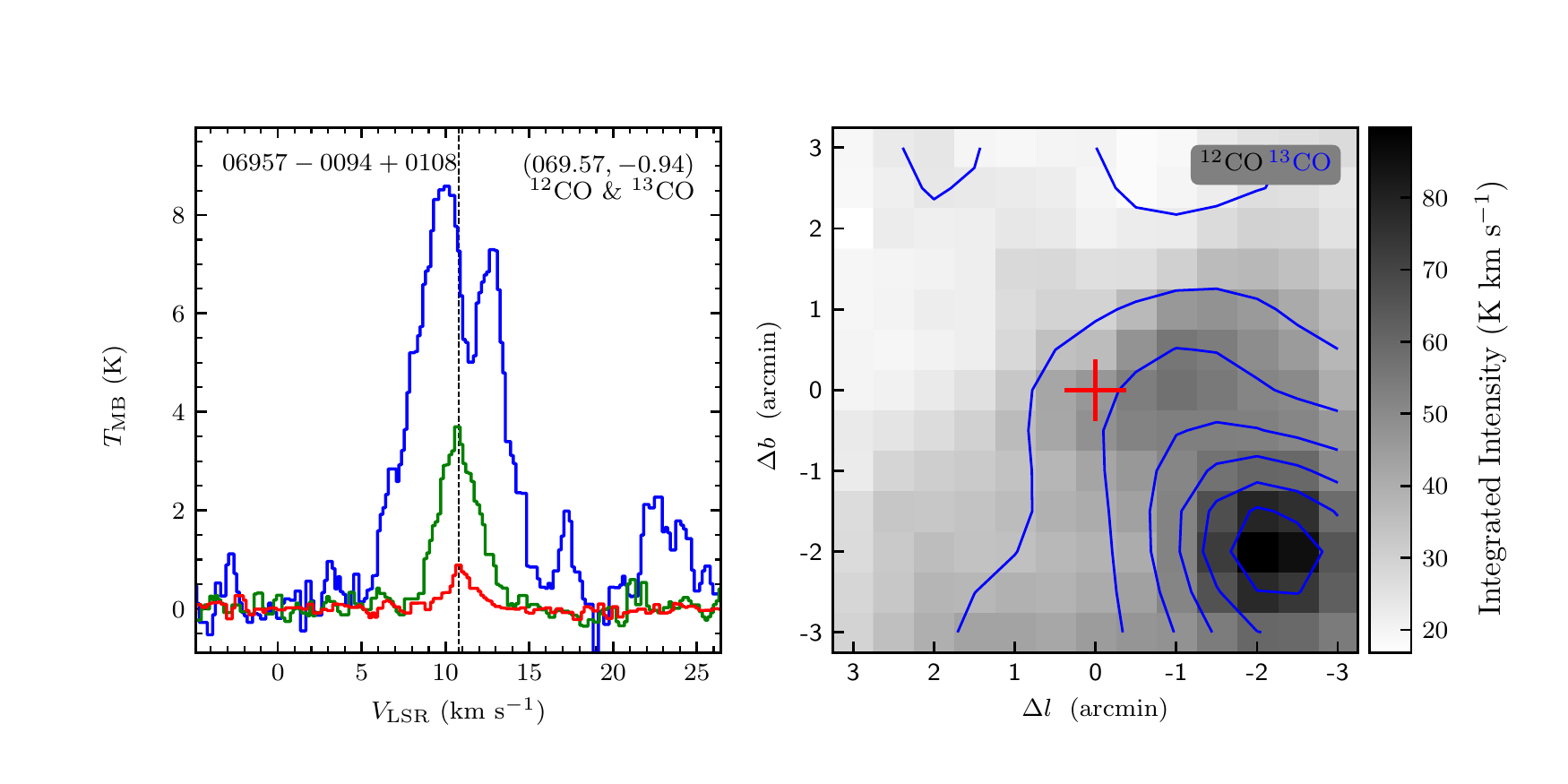}
\includegraphics[width=9.0cm,angle=0]{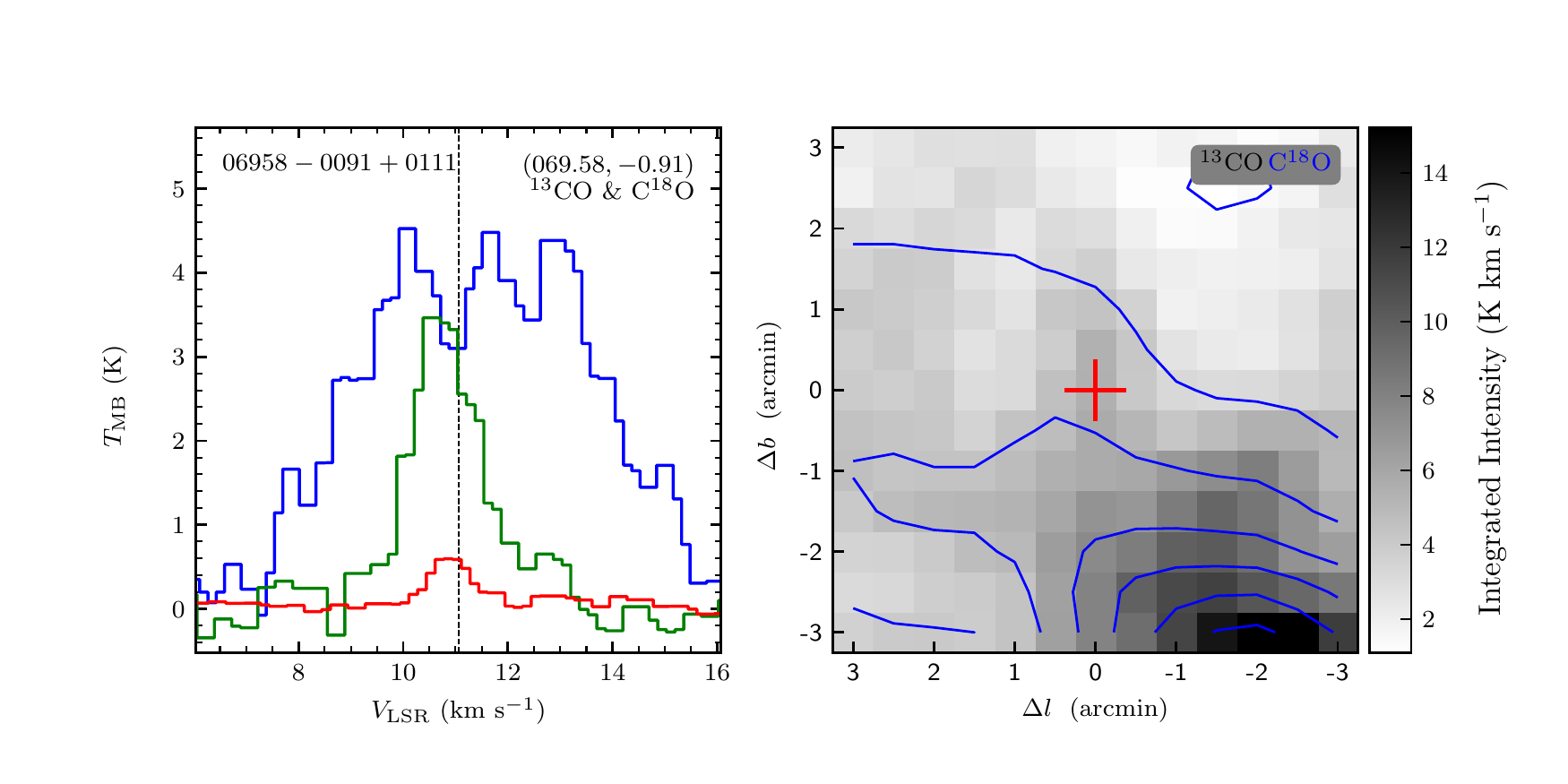}
\end{figure}
\clearpage

\begin{figure}
\includegraphics[width=9.0cm,angle=0]{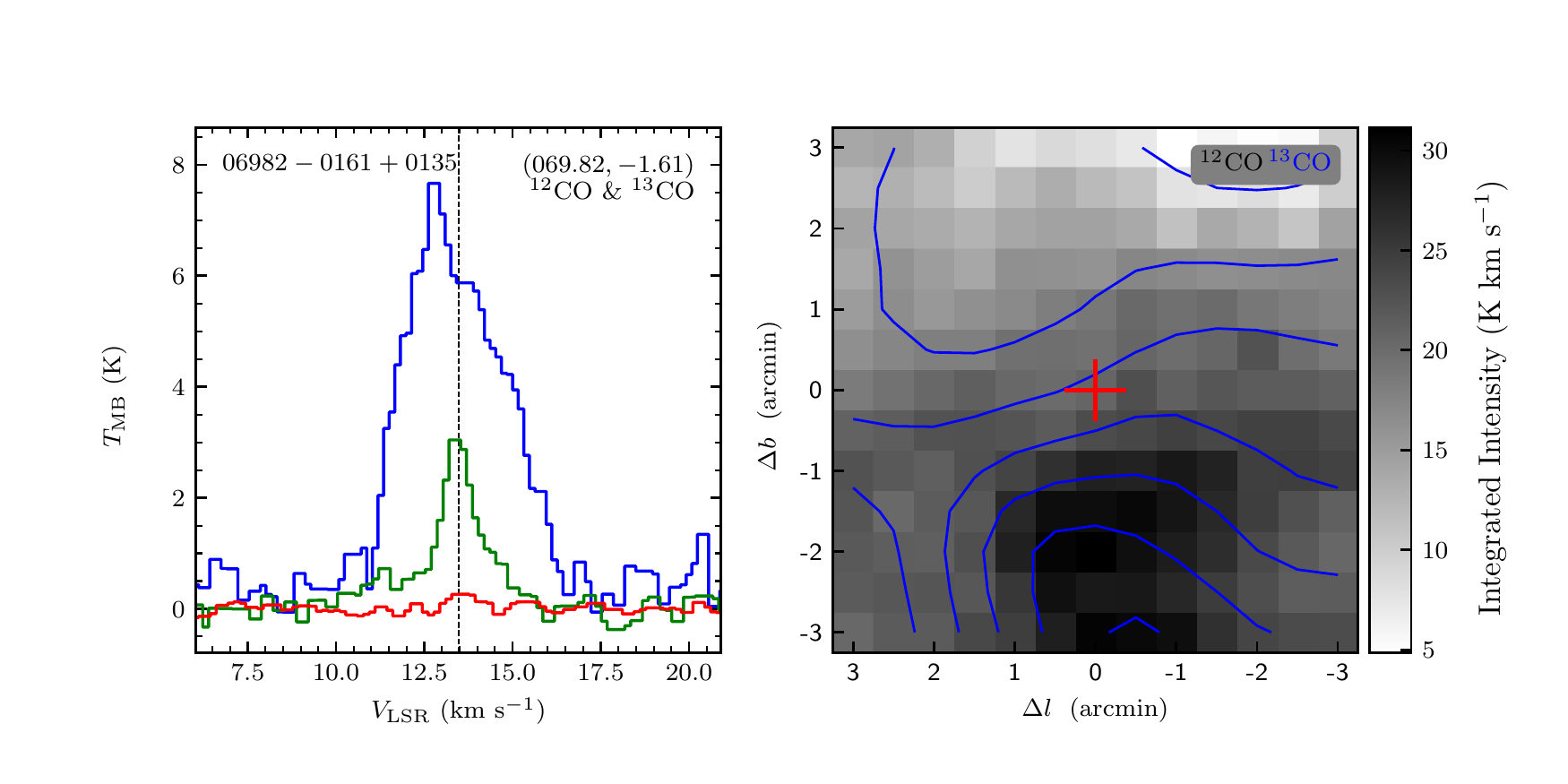}
\includegraphics[width=9.0cm,angle=0]{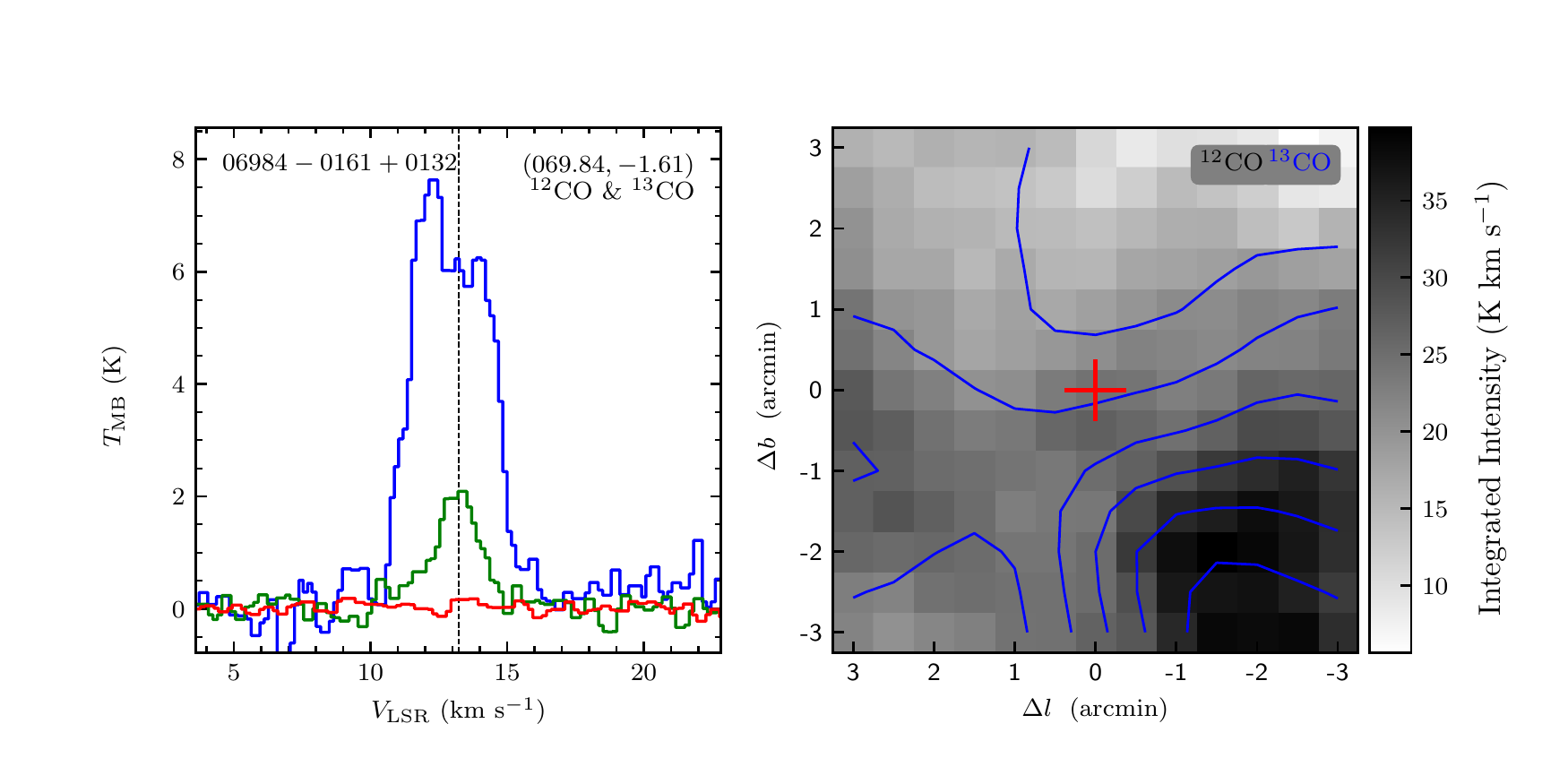}
\vspace{-0.5cm}

\includegraphics[width=9.0cm,angle=0]{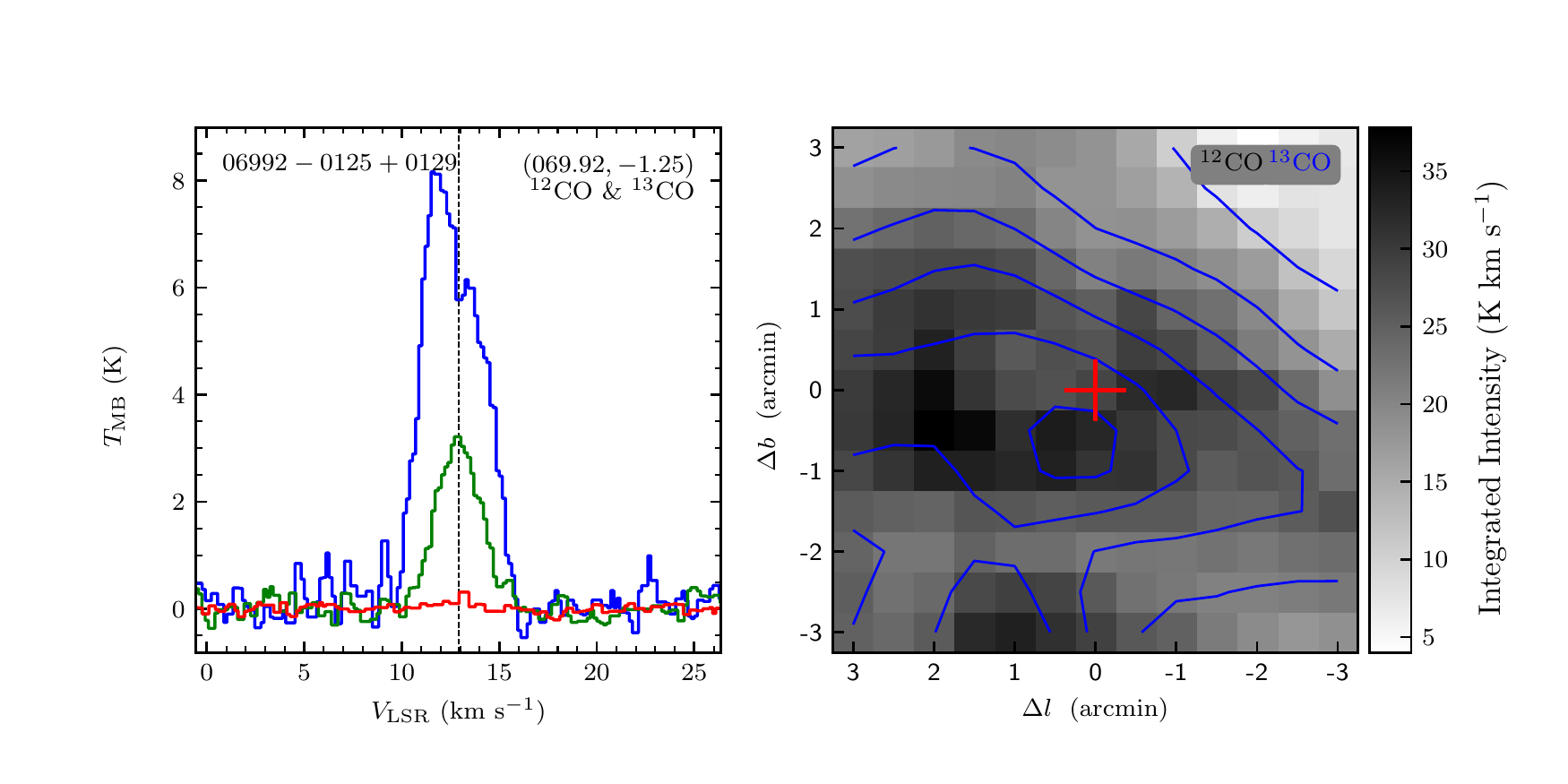}
\includegraphics[width=9.0cm,angle=0]{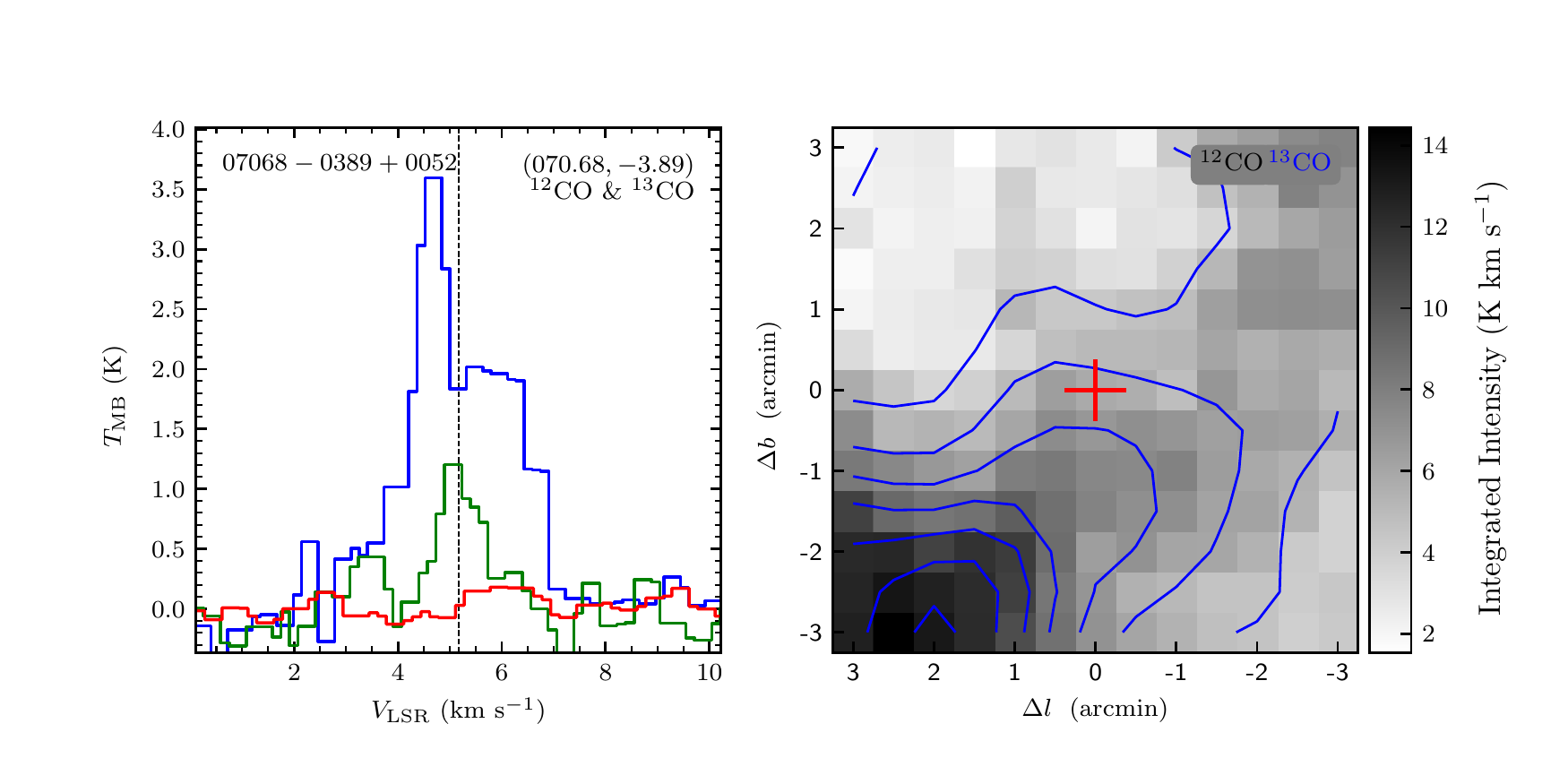}
\vspace{-0.5cm}

\includegraphics[width=9.0cm,angle=0]{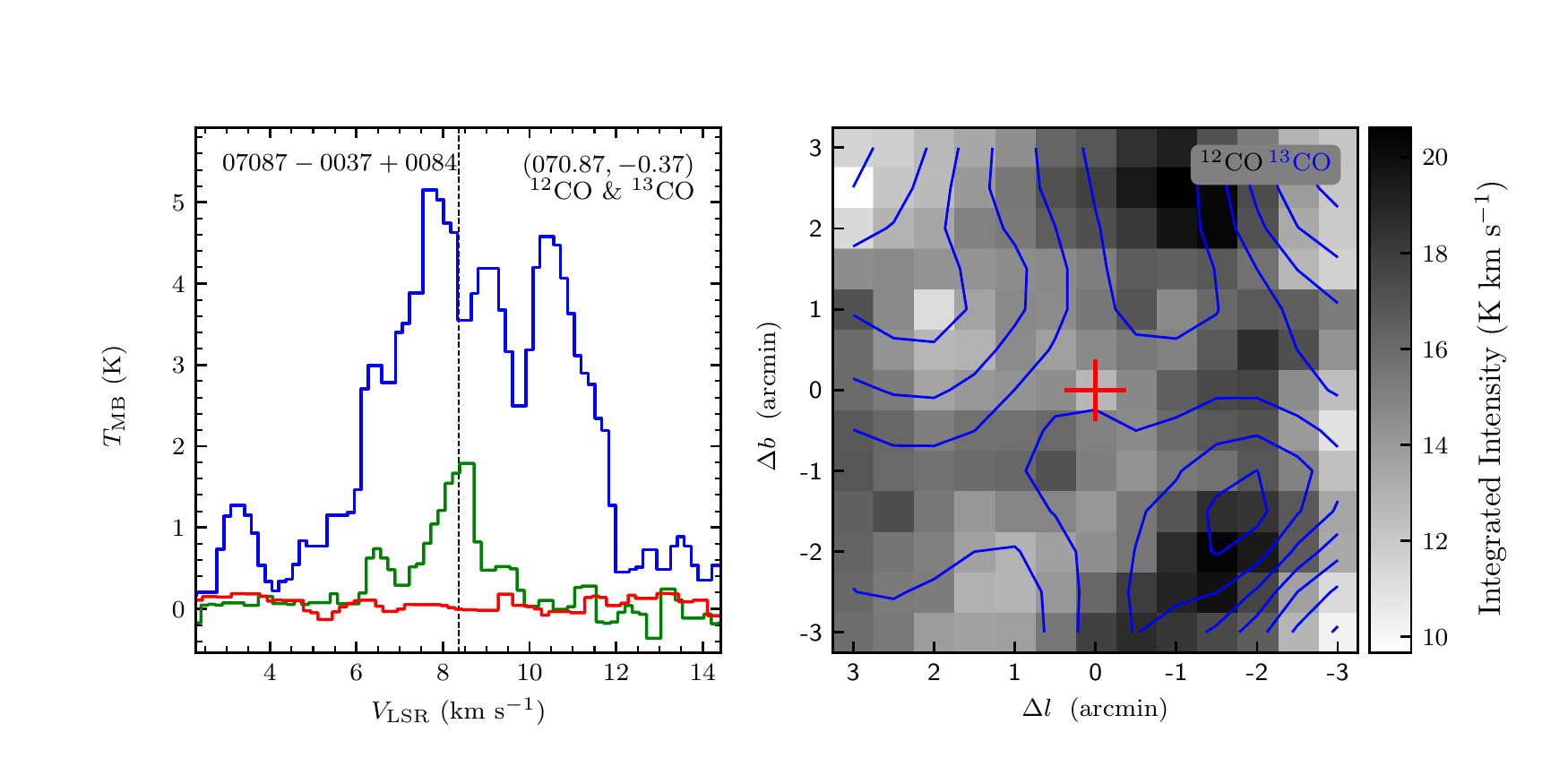}
\includegraphics[width=9.0cm,angle=0]{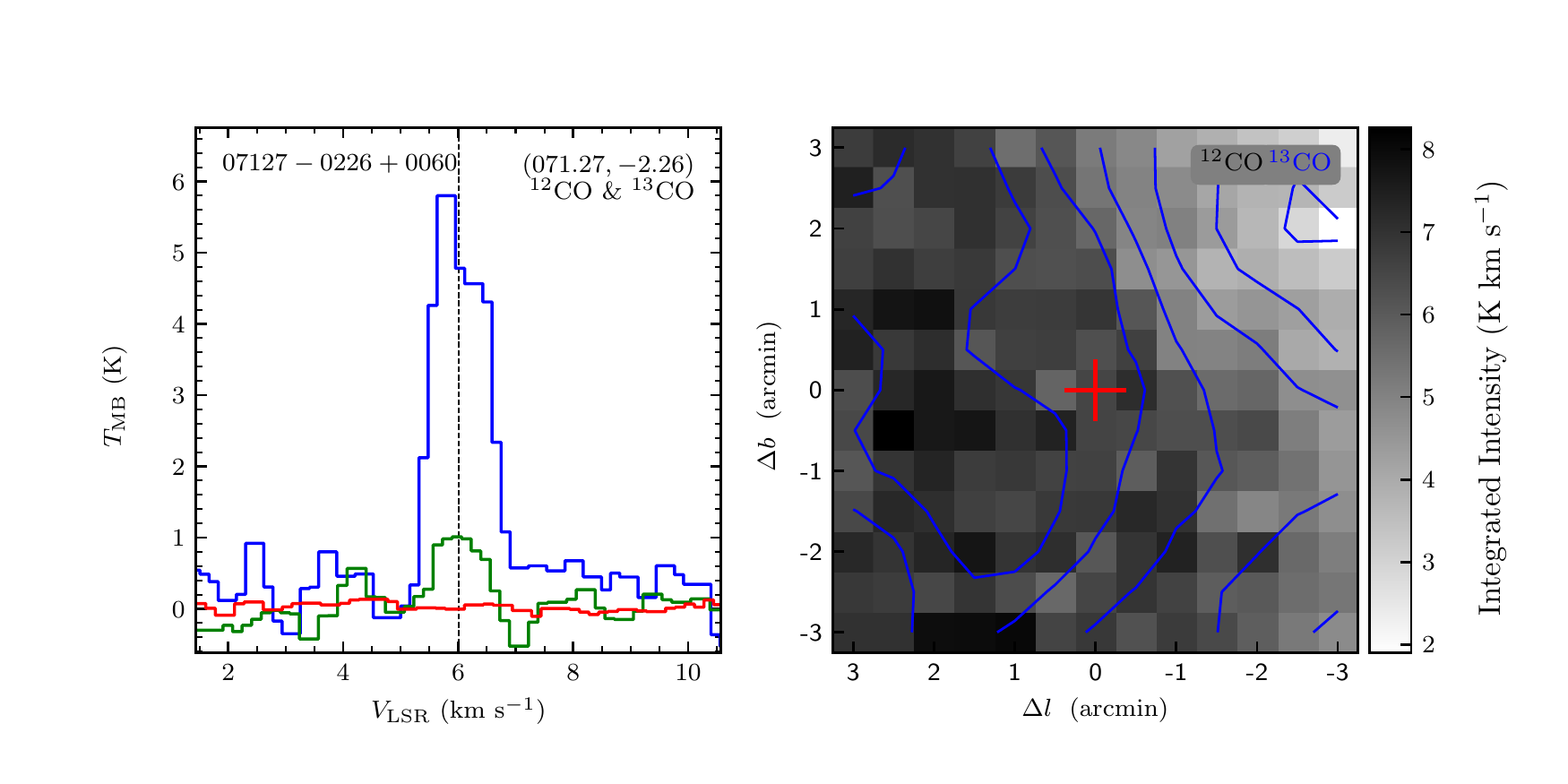}
\vspace{-0.5cm}

\includegraphics[width=9.0cm,angle=0]{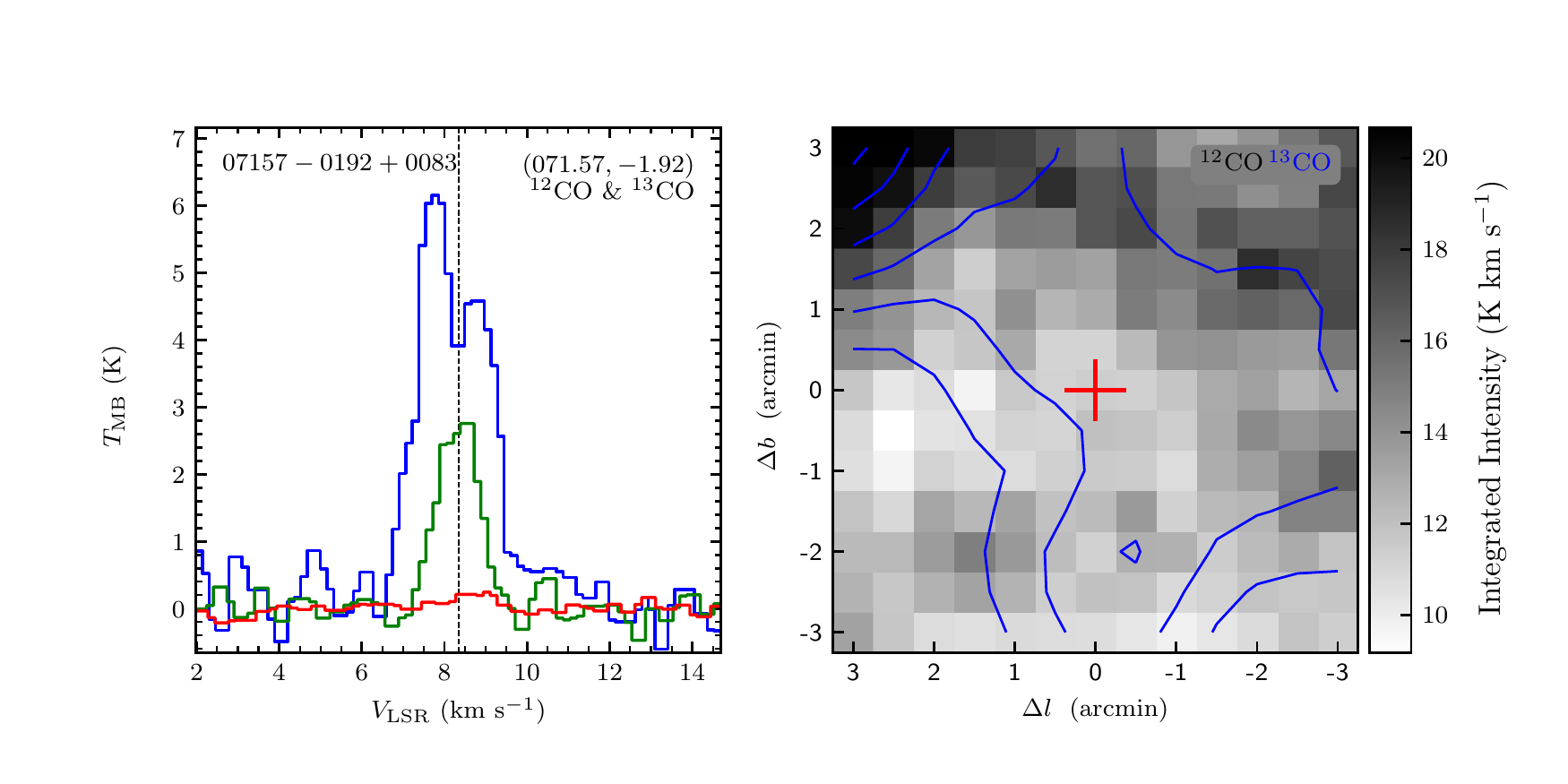}
\includegraphics[width=9.0cm,angle=0]{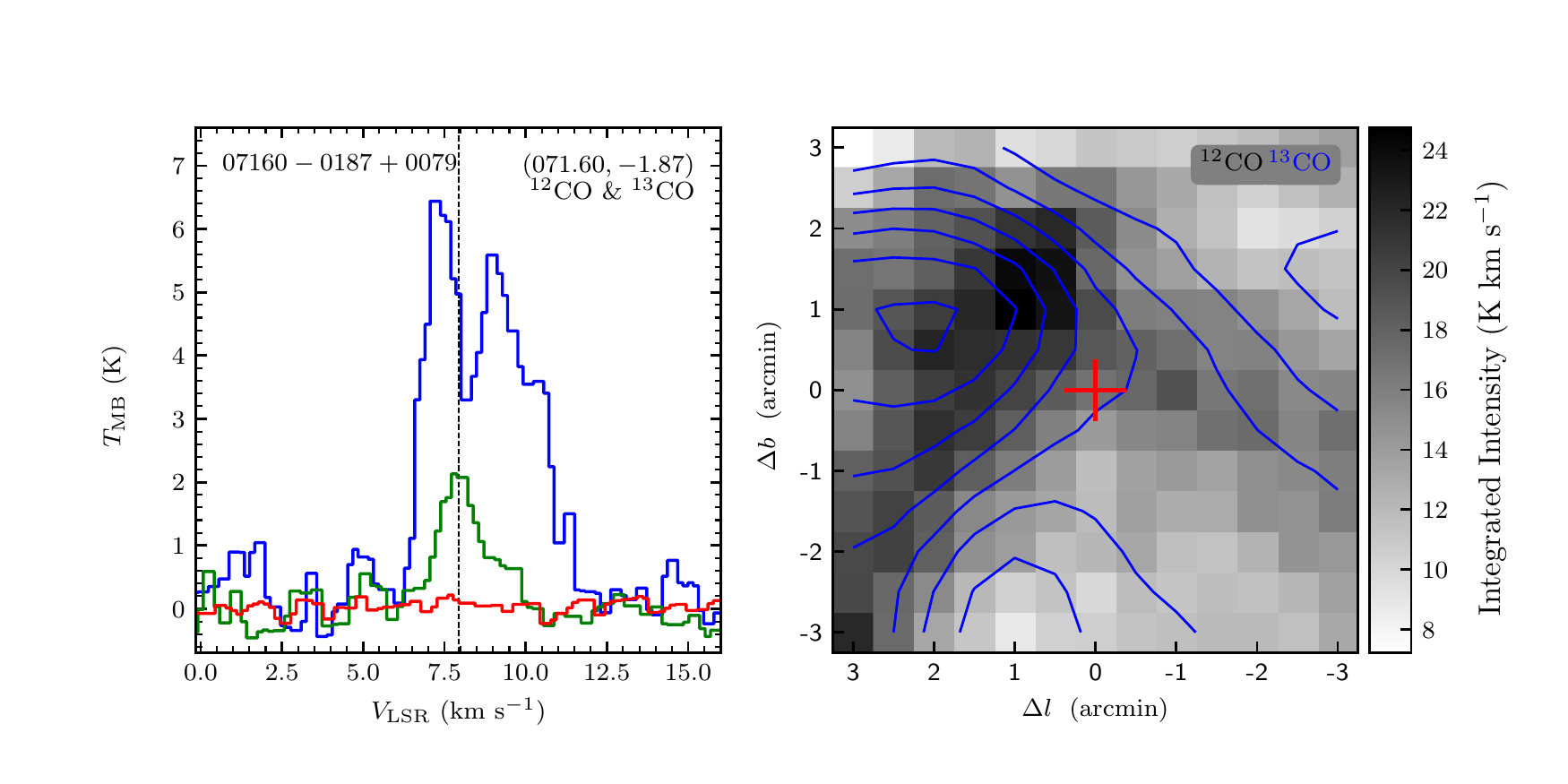}
\vspace{-0.5cm}

\includegraphics[width=9.0cm,angle=0]{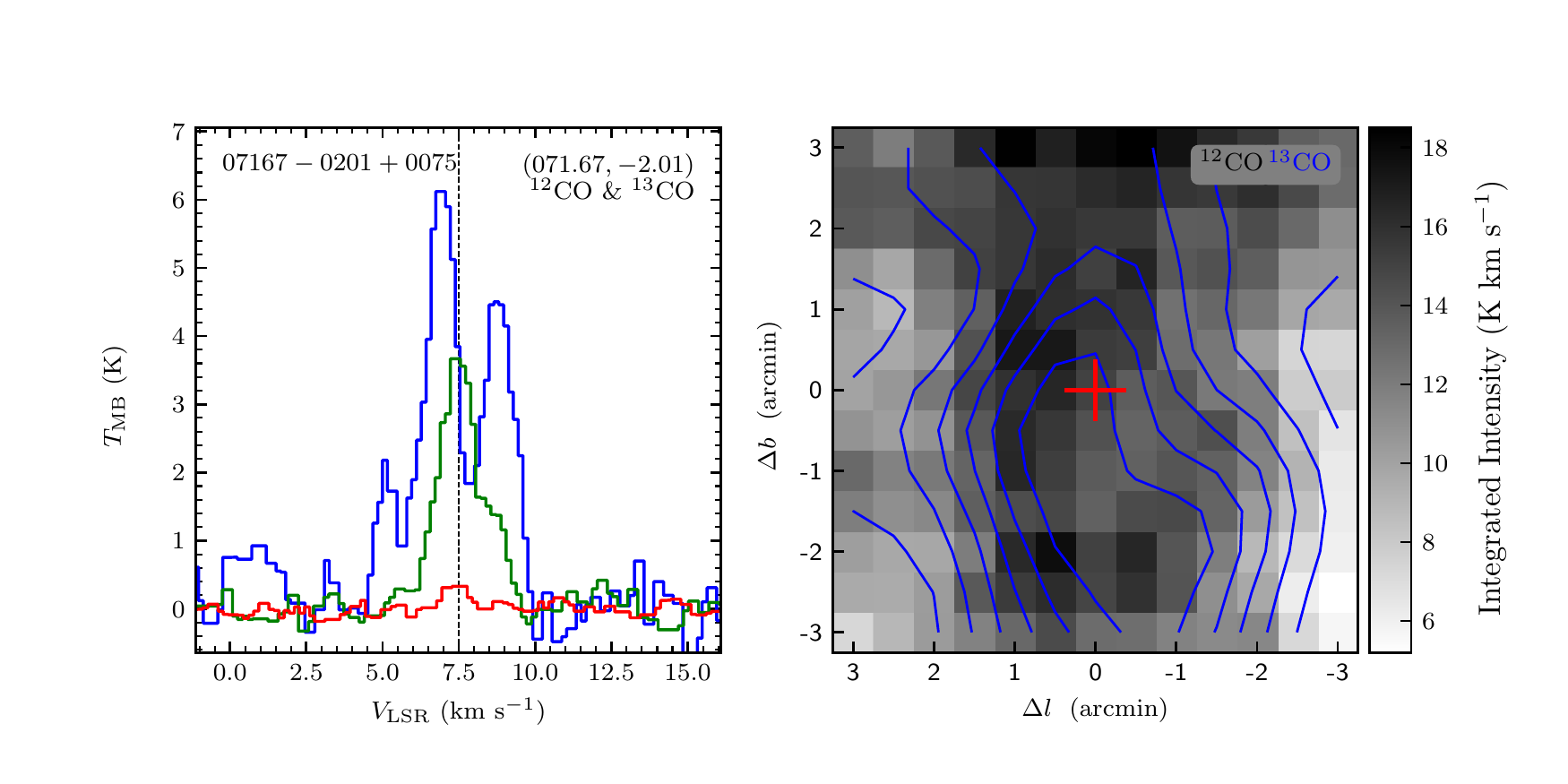}
\includegraphics[width=9.0cm,angle=0]{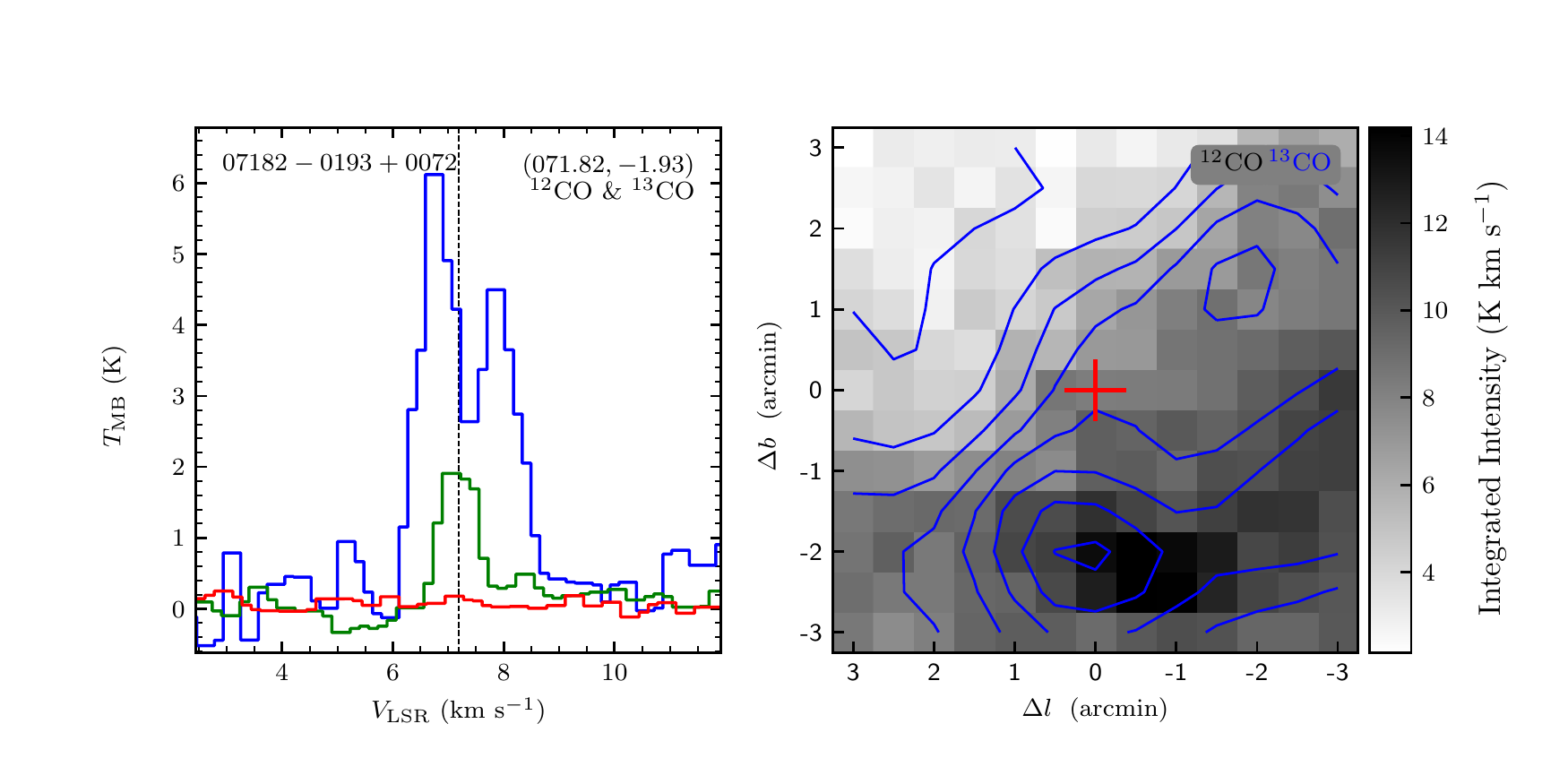}
\end{figure}
\clearpage

\begin{figure}
\includegraphics[width=9.0cm,angle=0]{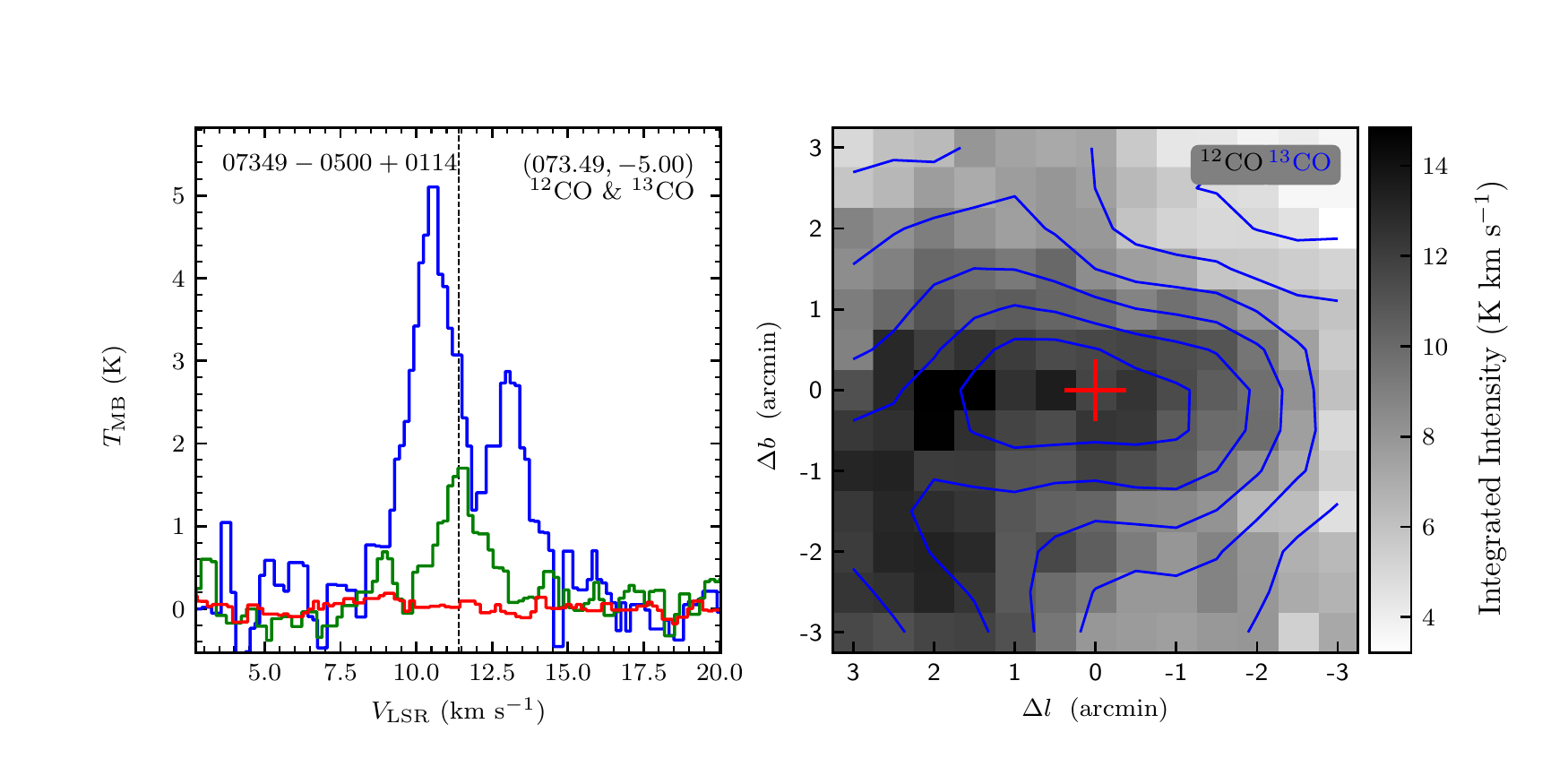}
\includegraphics[width=9.0cm,angle=0]{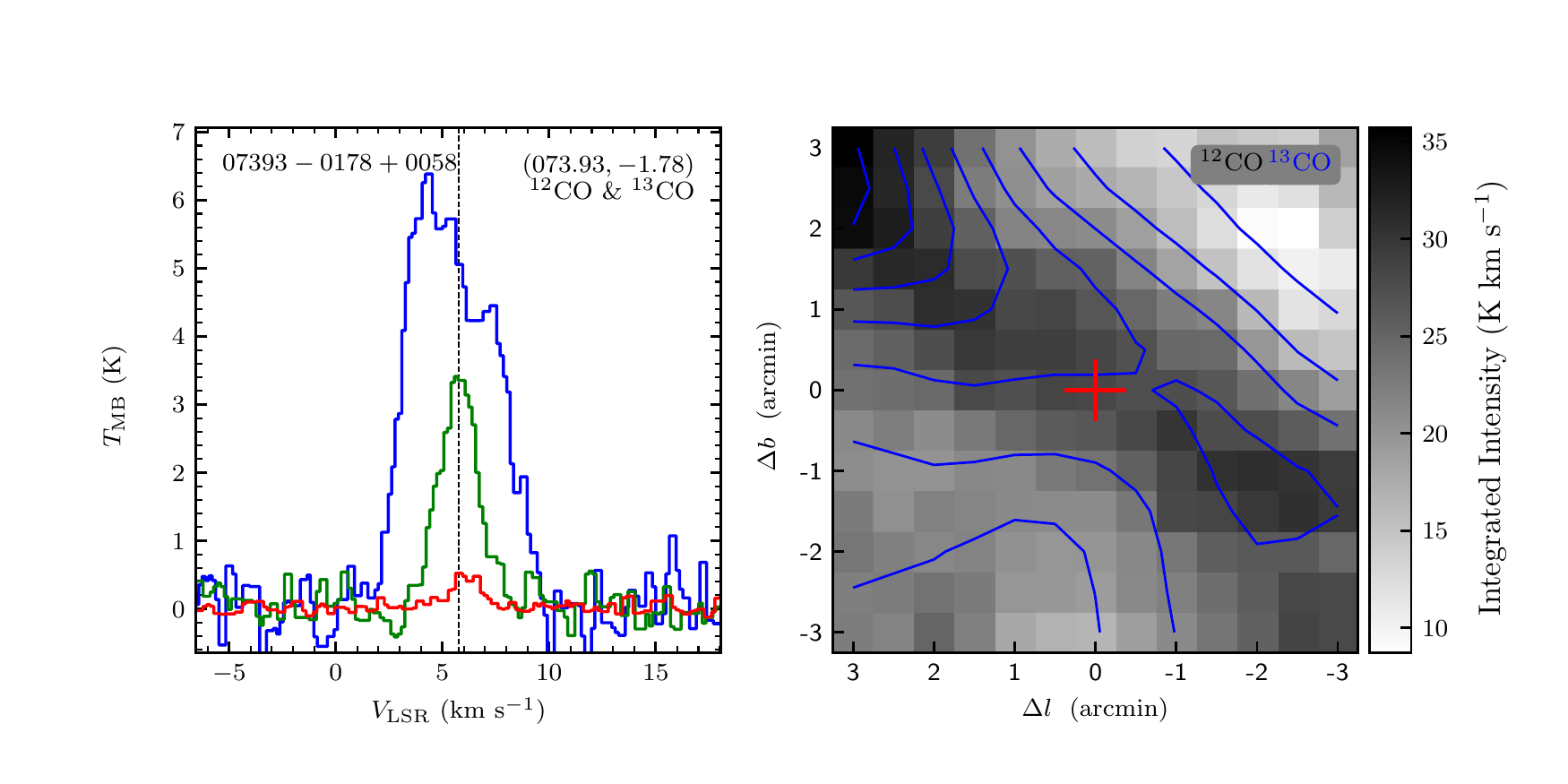}
\vspace{-0.5cm}

\includegraphics[width=9.0cm,angle=0]{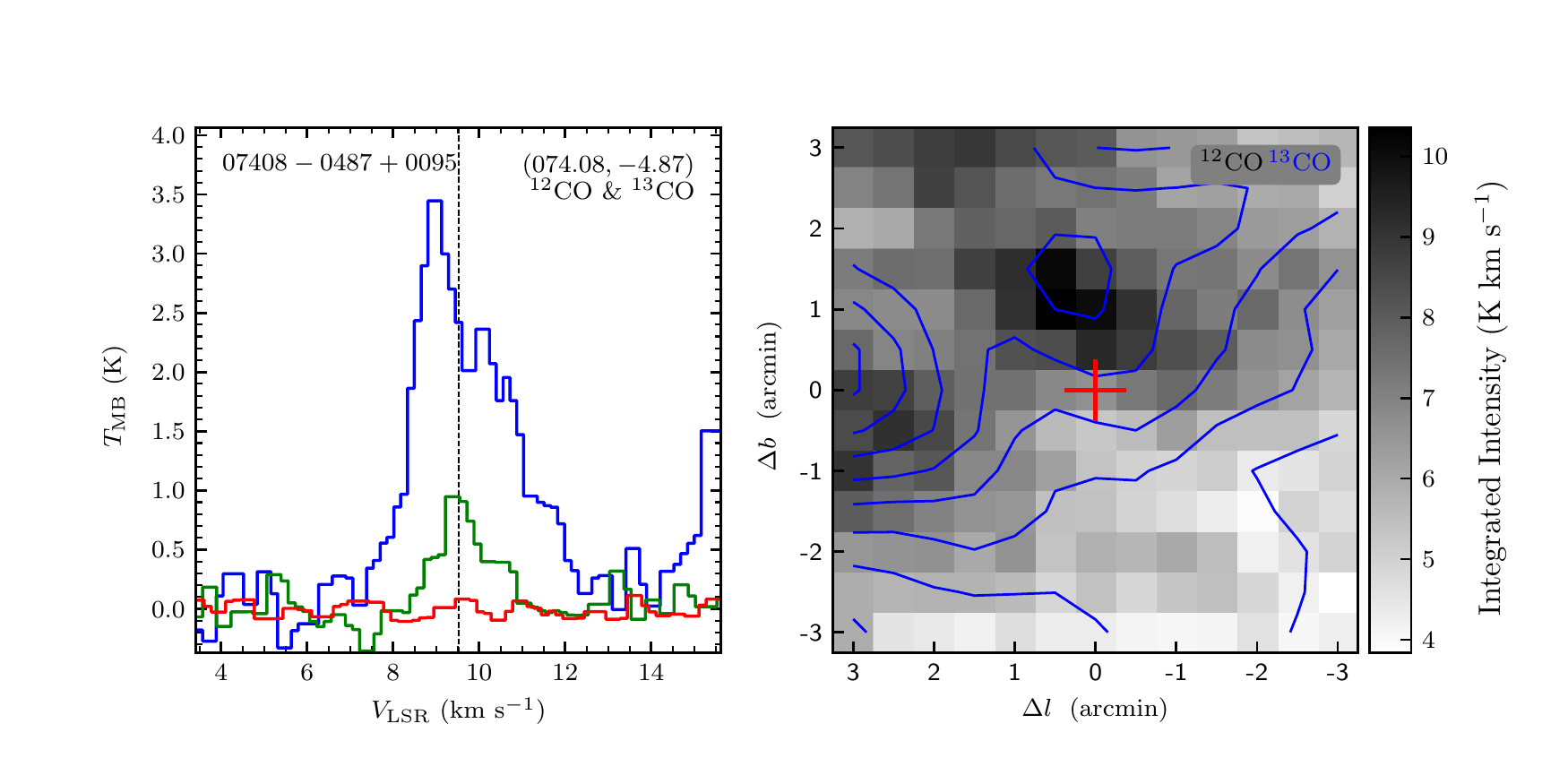}
\includegraphics[width=9.0cm,angle=0]{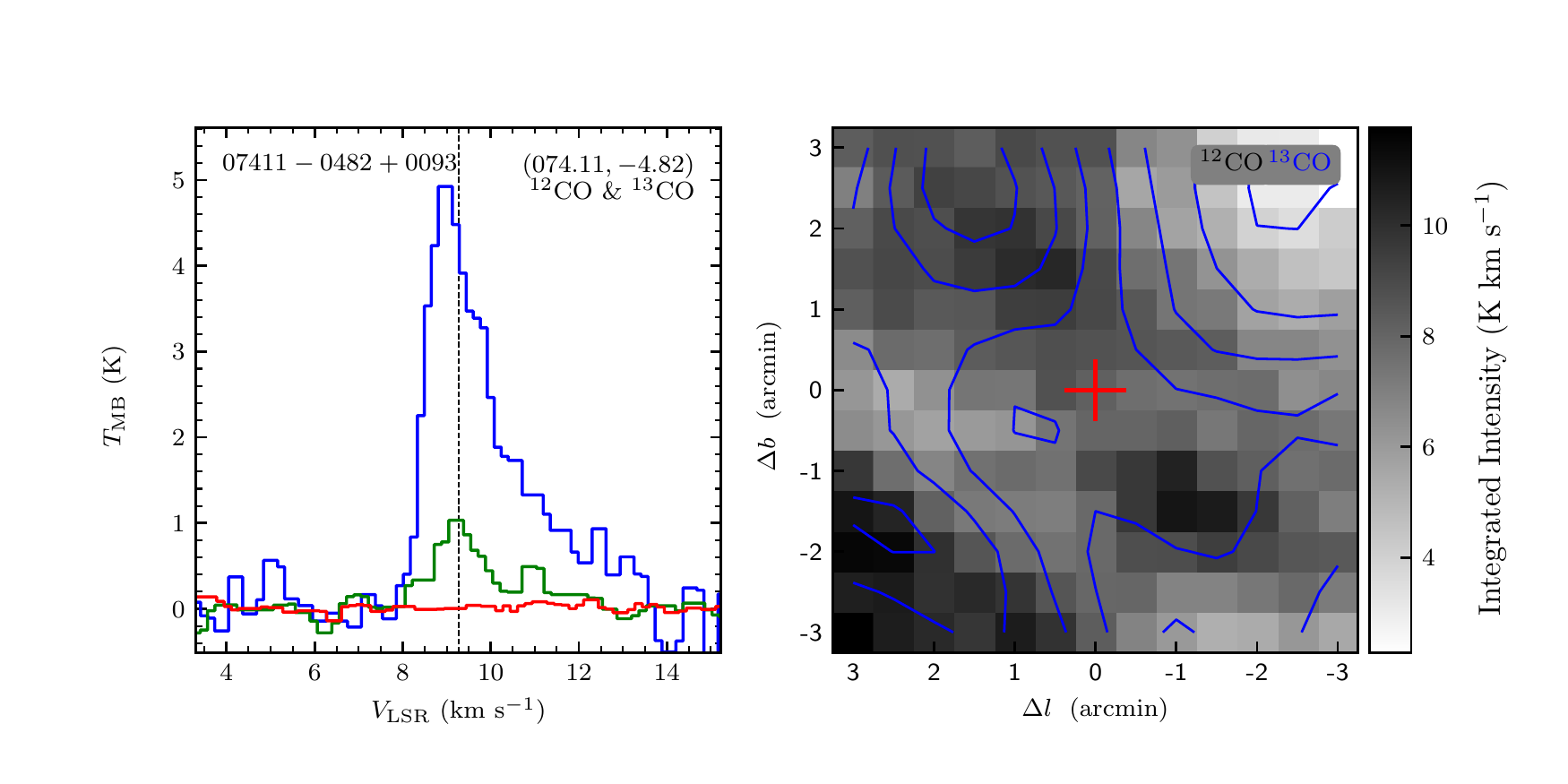}
\vspace{-0.5cm}

\includegraphics[width=9.0cm,angle=0]{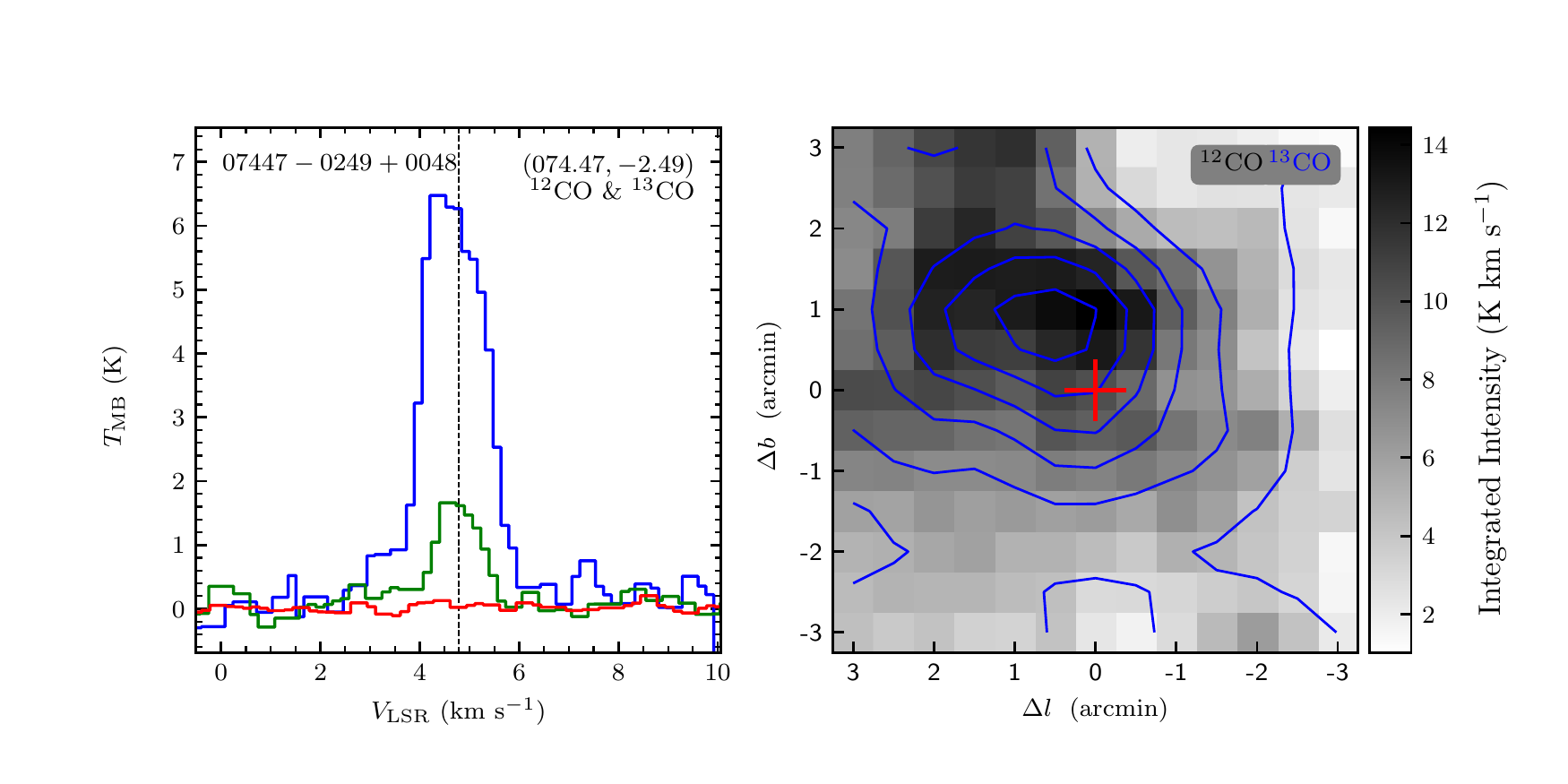}
\includegraphics[width=9.0cm,angle=0]{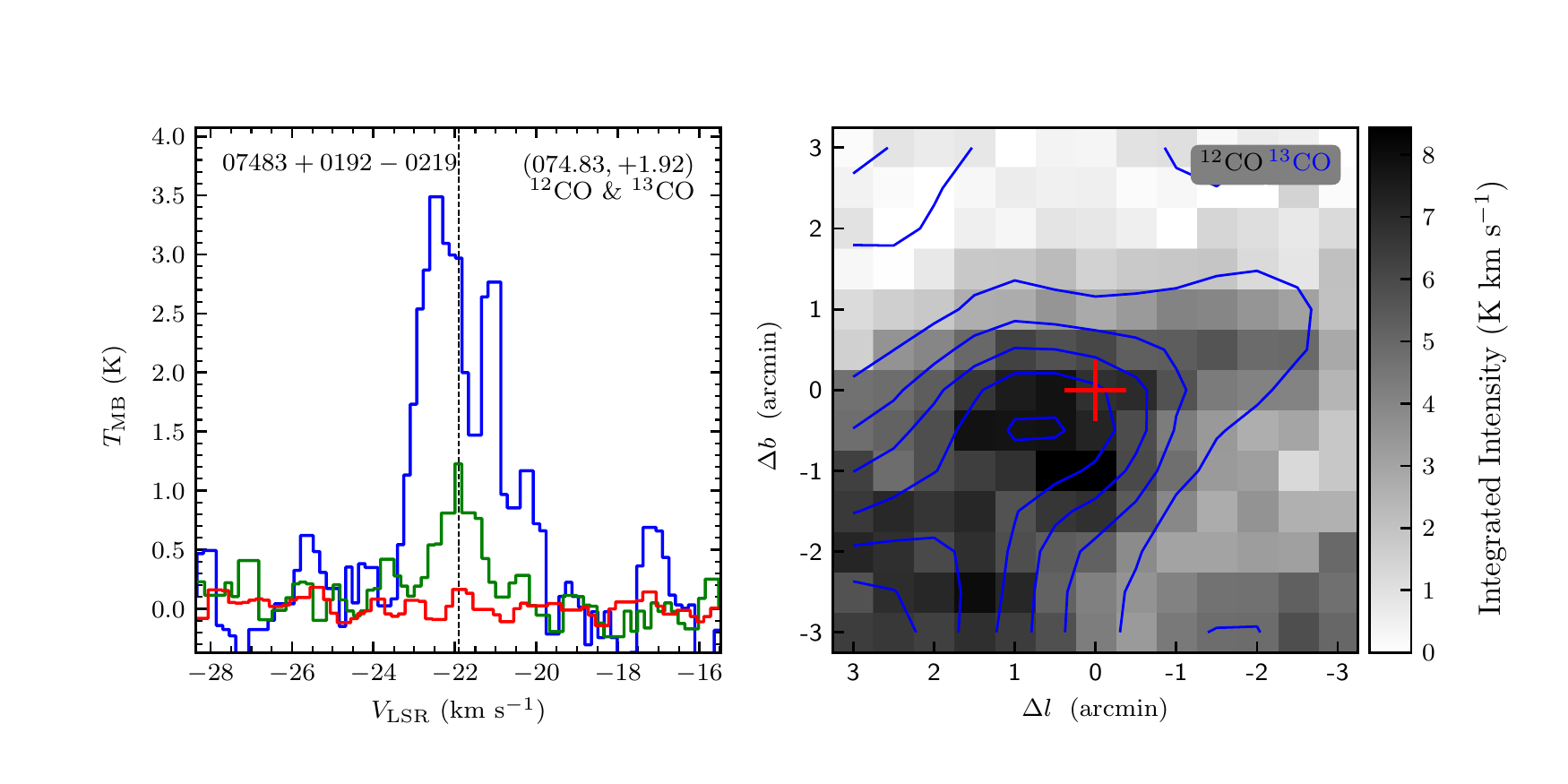}
\vspace{-0.5cm}

\includegraphics[width=9.0cm,angle=0]{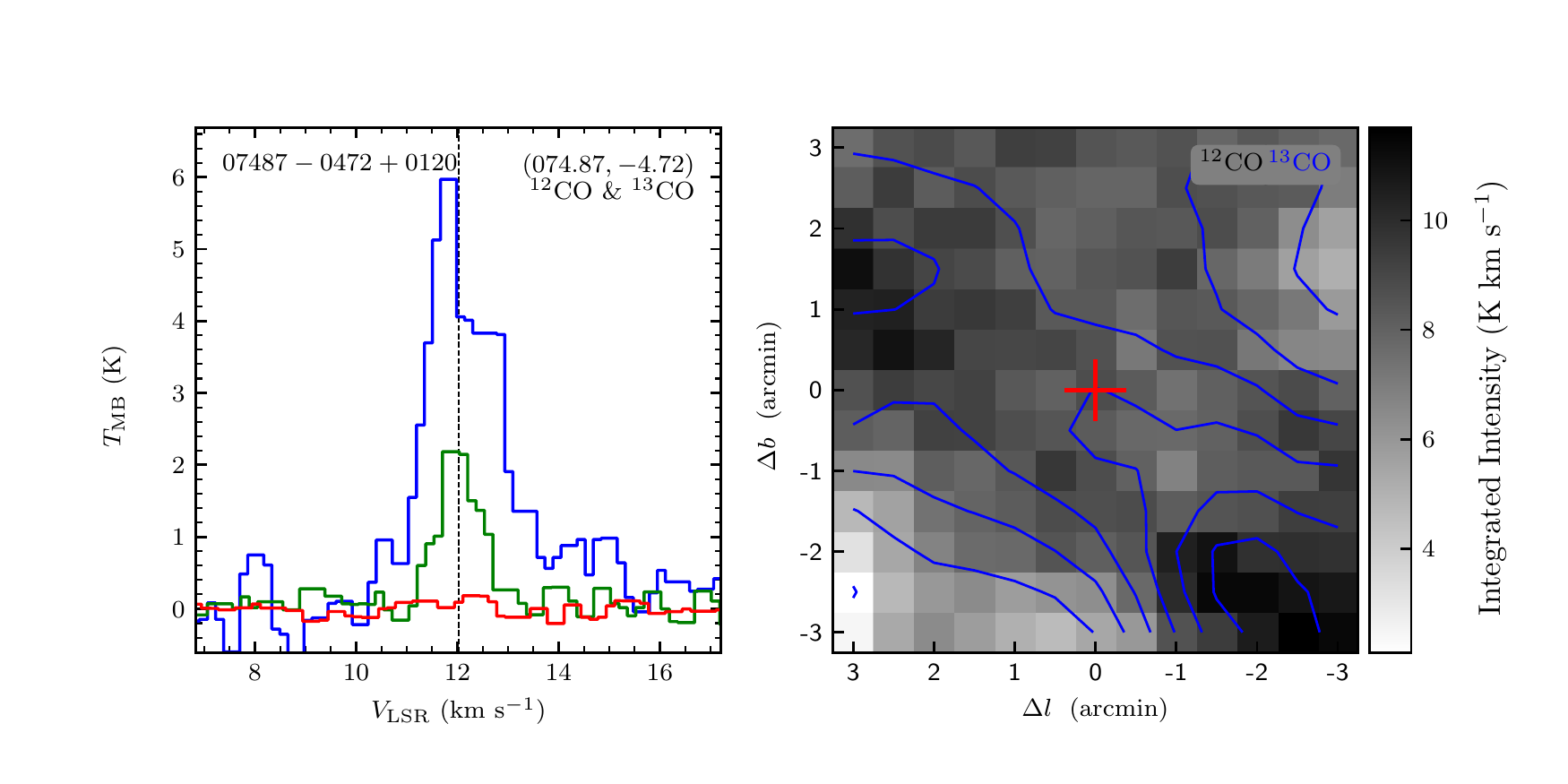}
\includegraphics[width=9.0cm,angle=0]{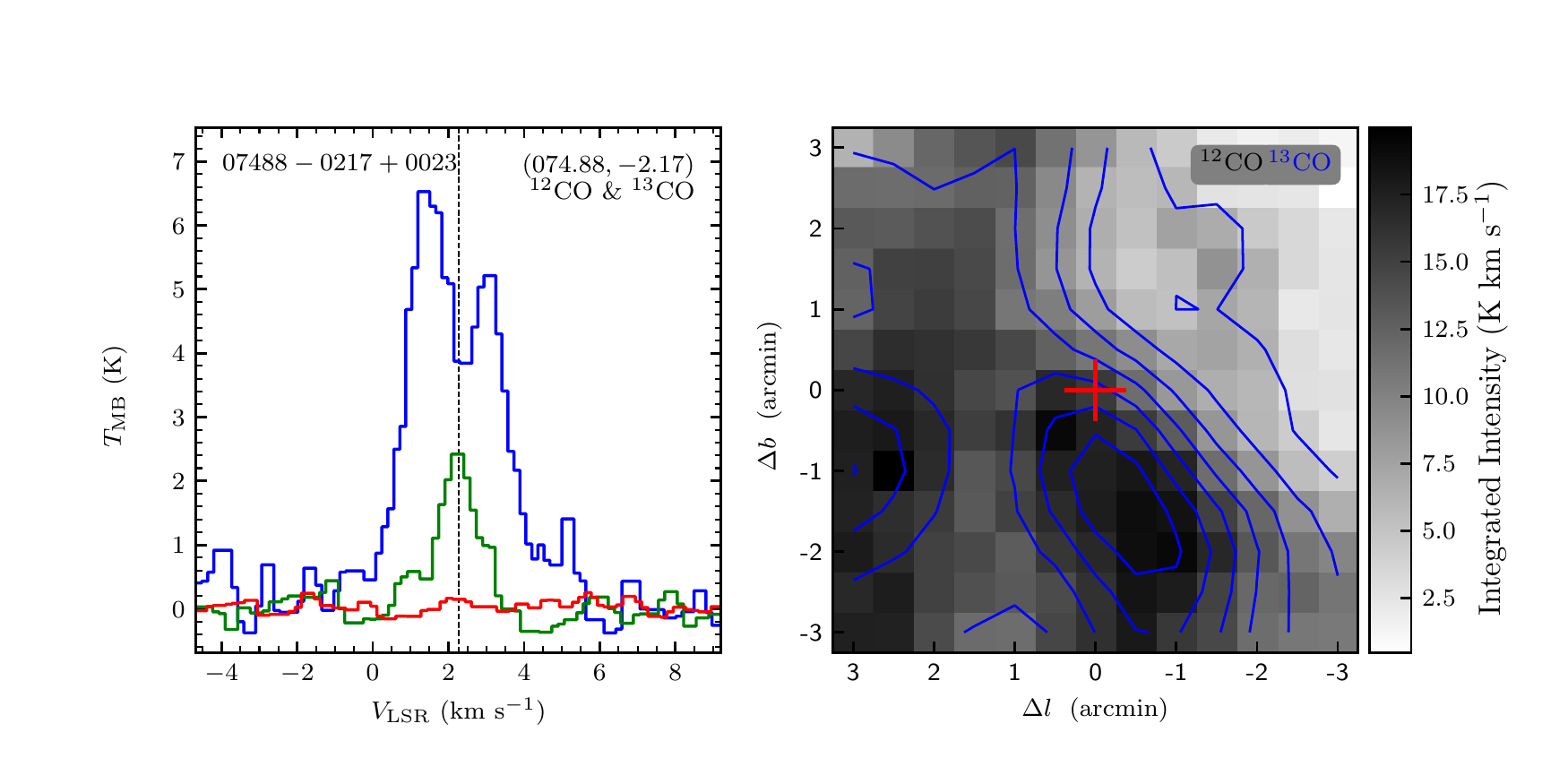}
\vspace{-0.5cm}

\includegraphics[width=9.0cm,angle=0]{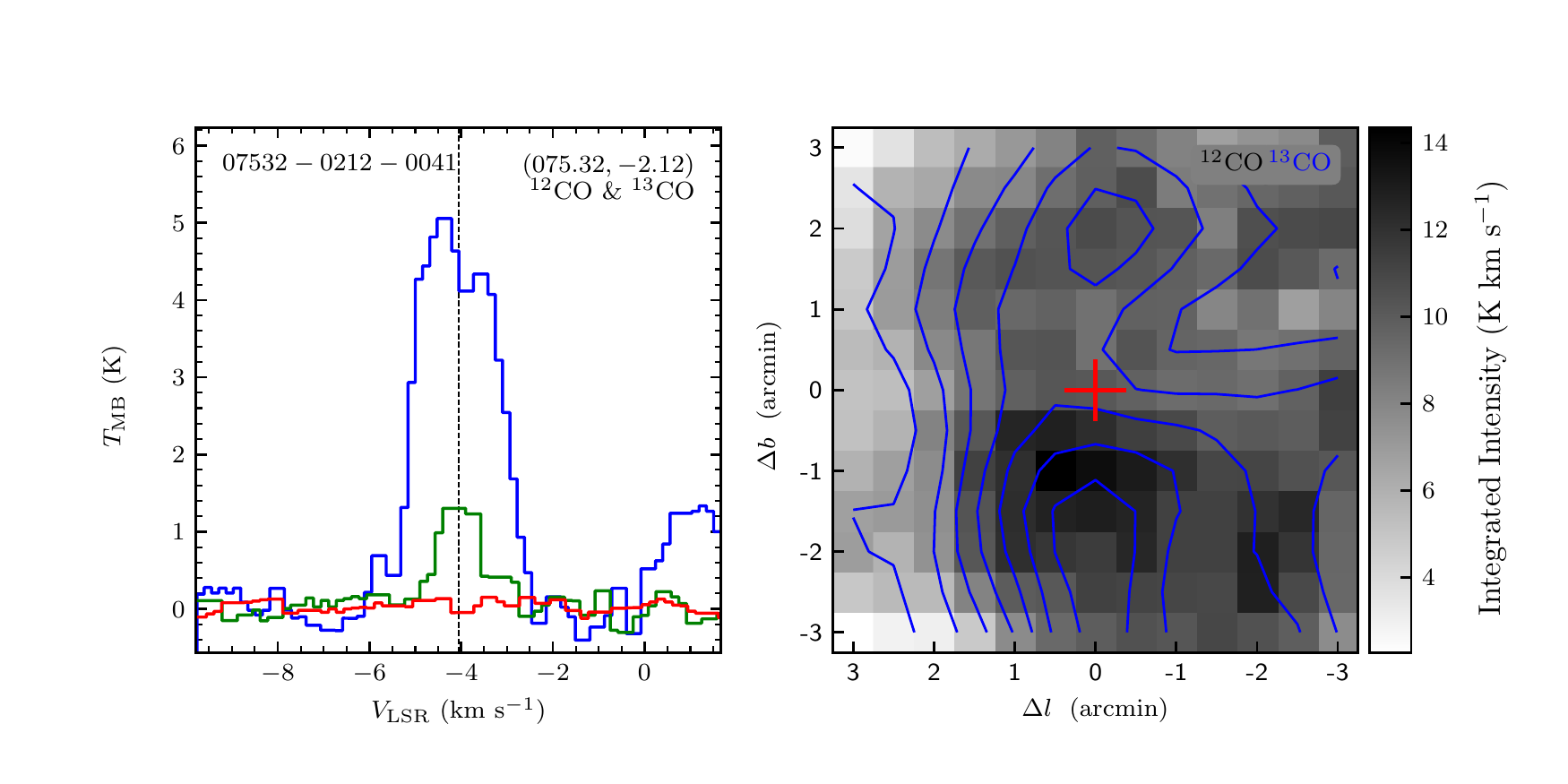}
\includegraphics[width=9.0cm,angle=0]{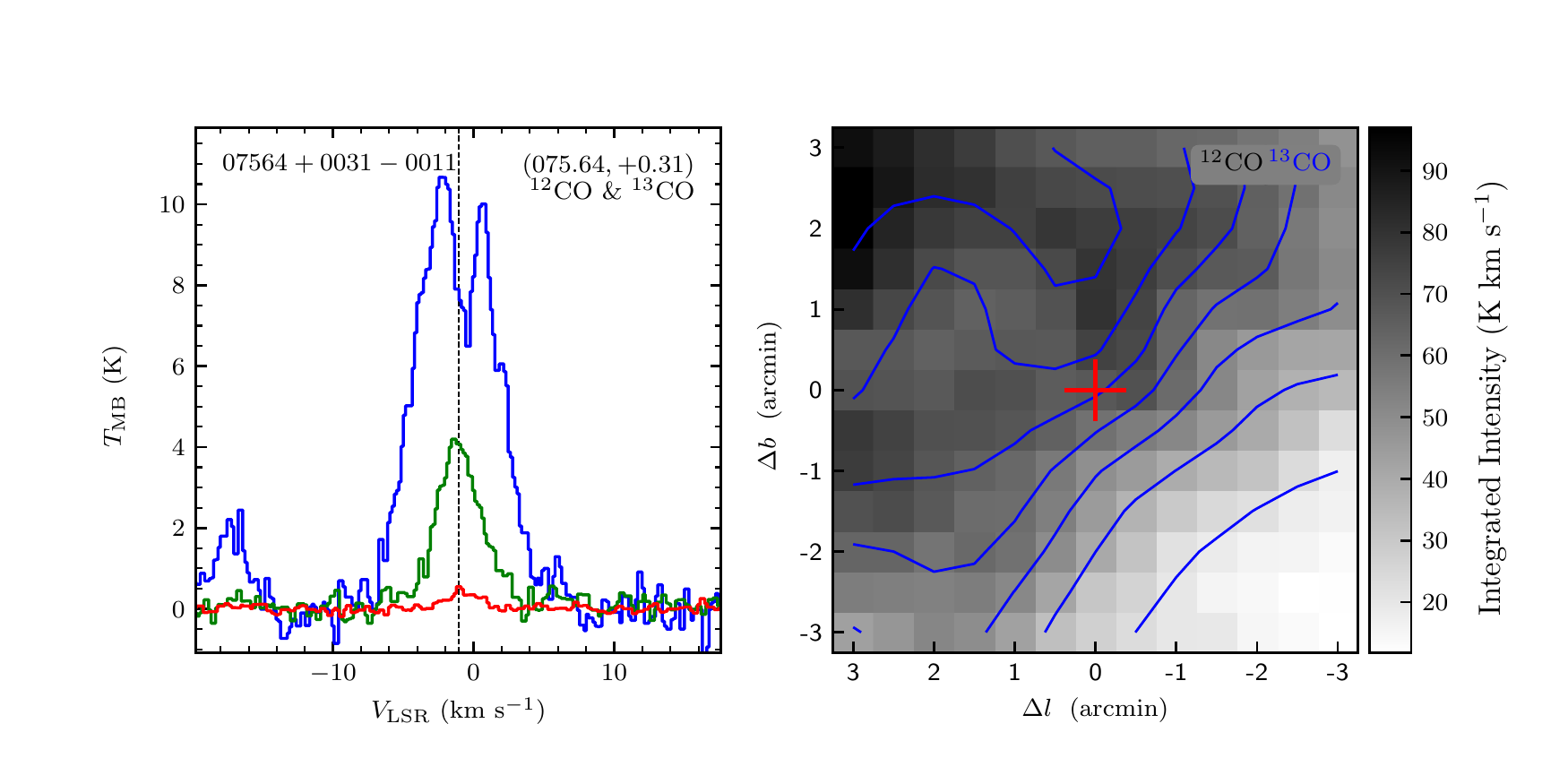}
\end{figure}
\clearpage

\begin{figure}
\includegraphics[width=9.0cm,angle=0]{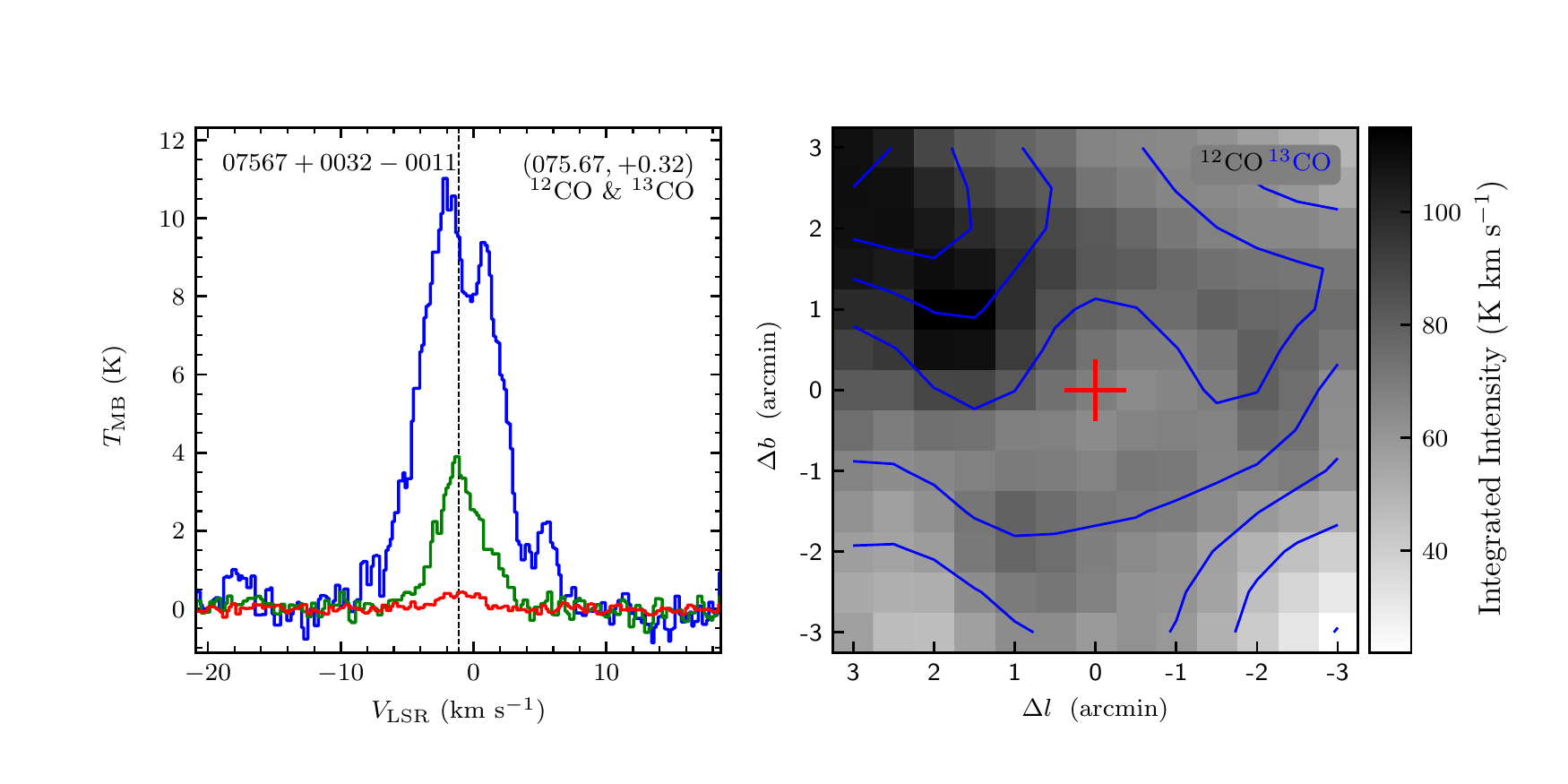}
\includegraphics[width=9.0cm,angle=0]{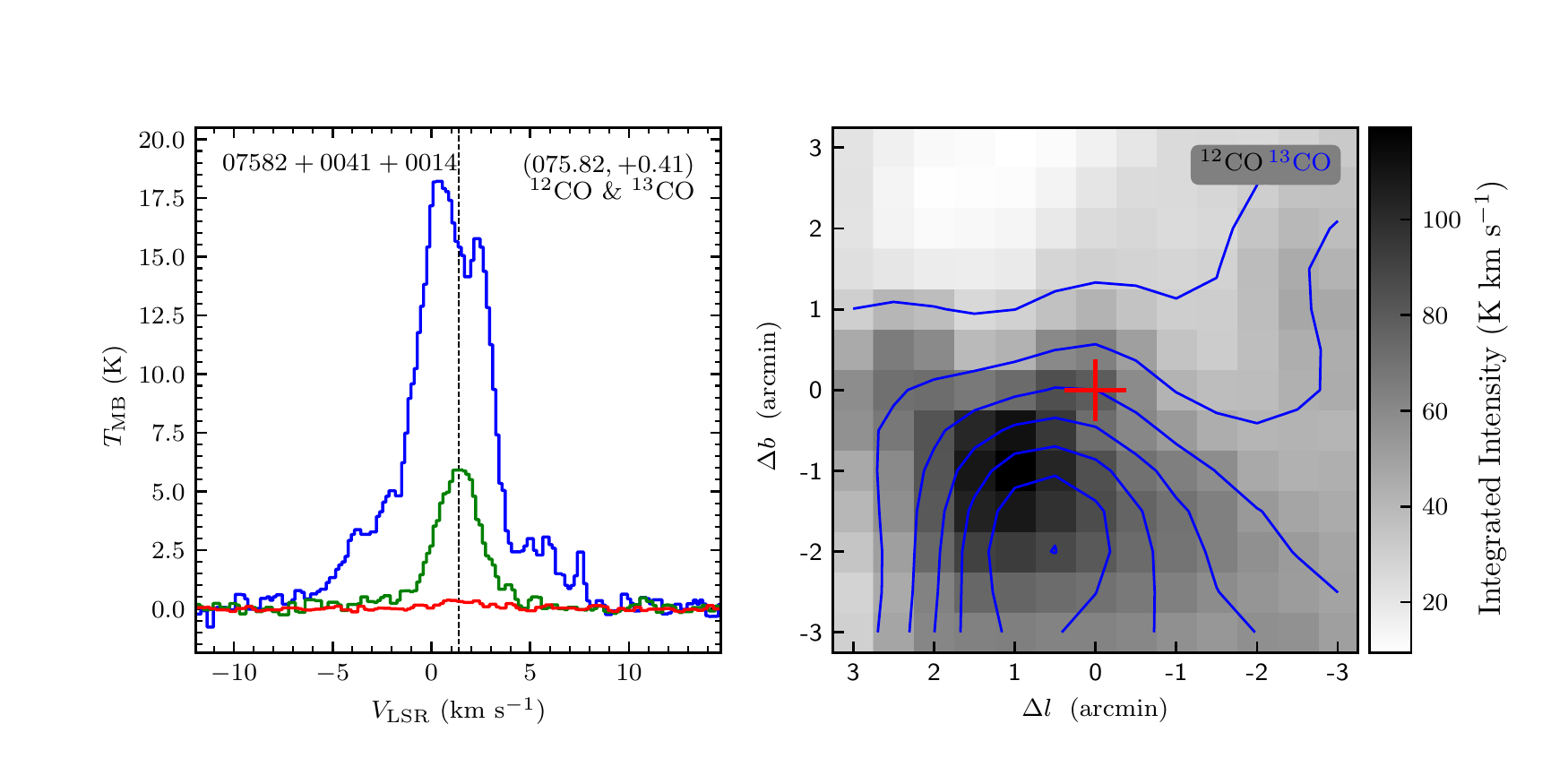}
\vspace{-0.5cm}

\includegraphics[width=9.0cm,angle=0]{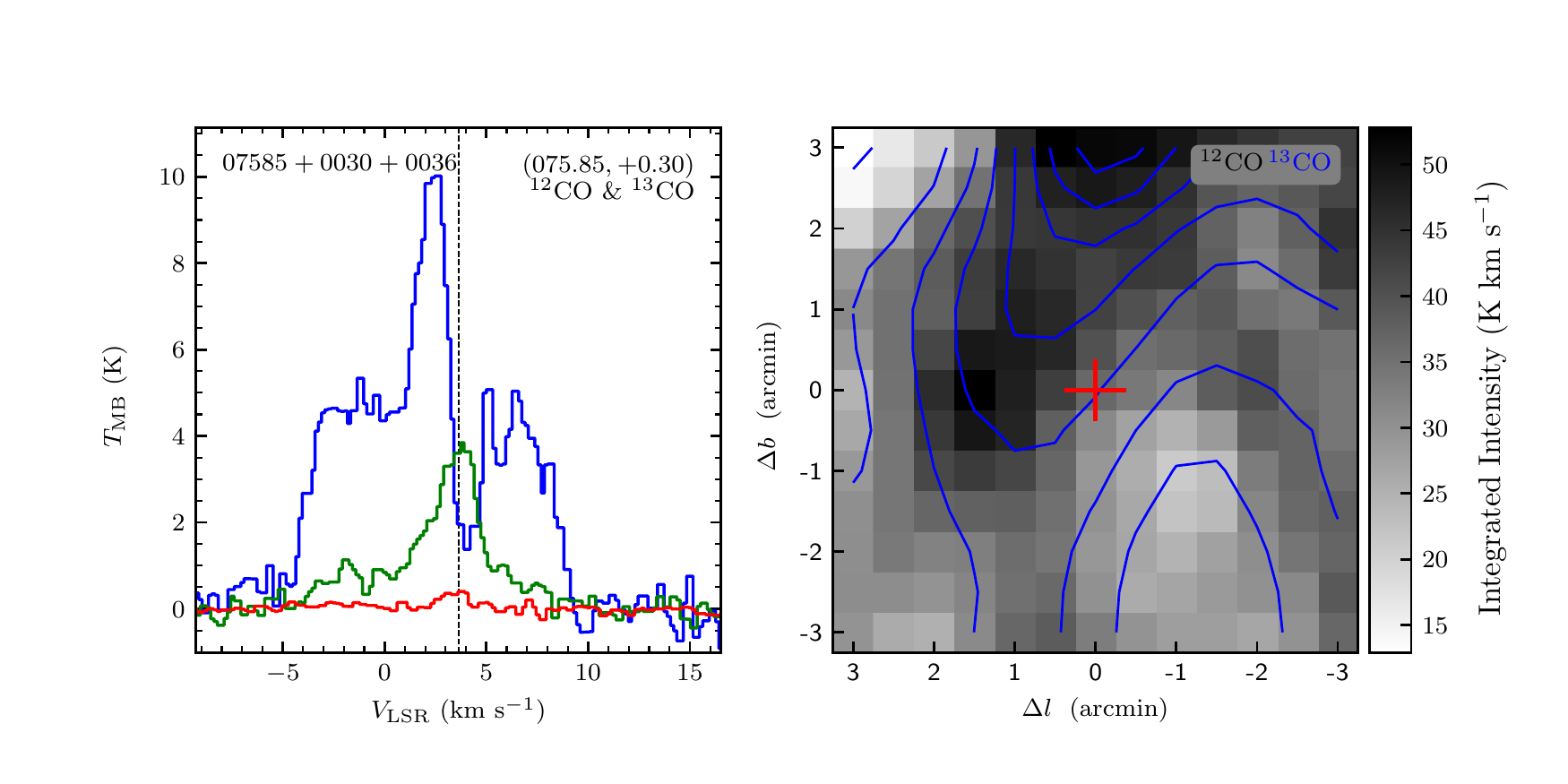}
\includegraphics[width=9.0cm,angle=0]{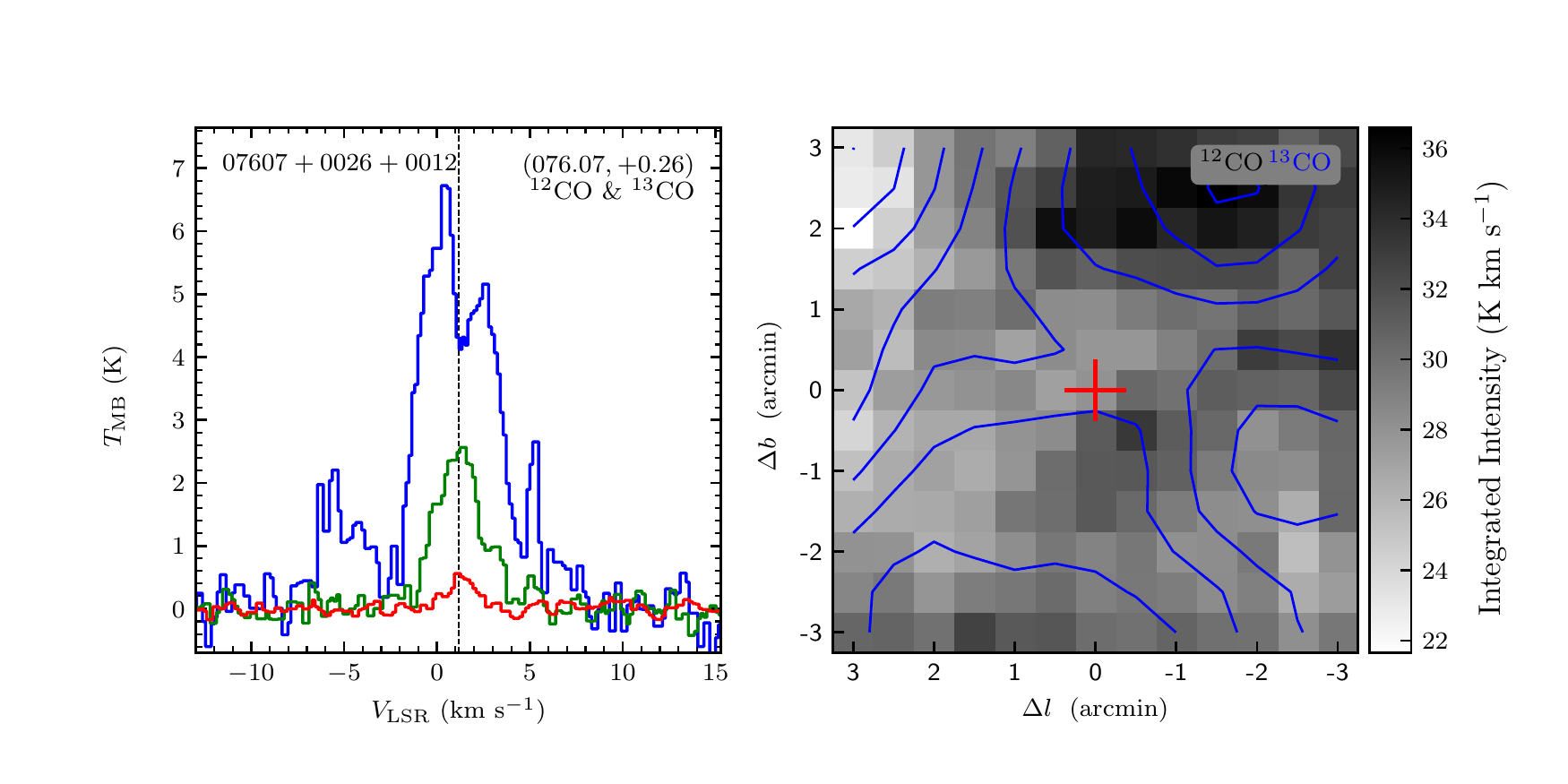}
\vspace{-0.5cm}

\includegraphics[width=9.0cm,angle=0]{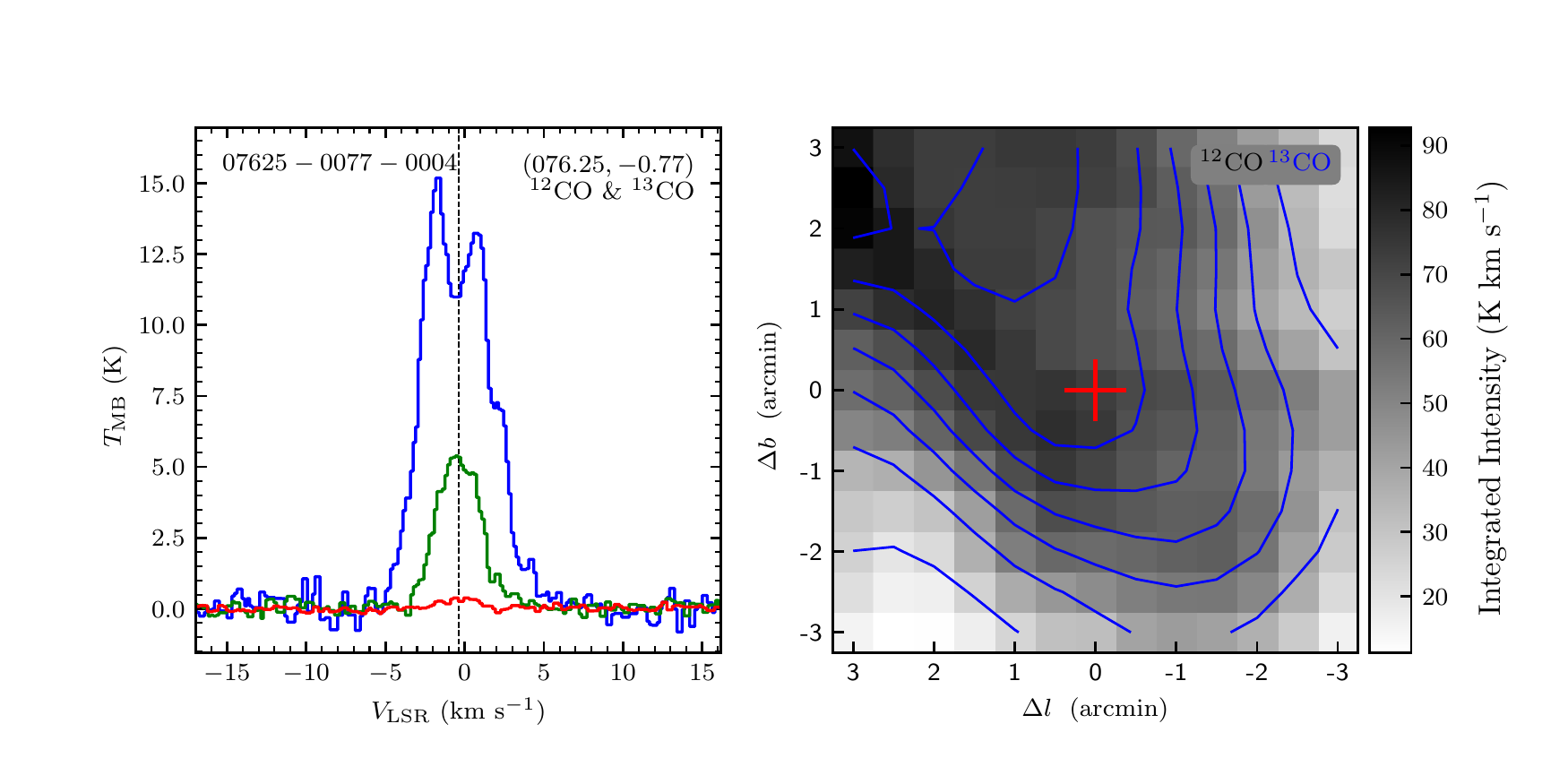}
\includegraphics[width=9.0cm,angle=0]{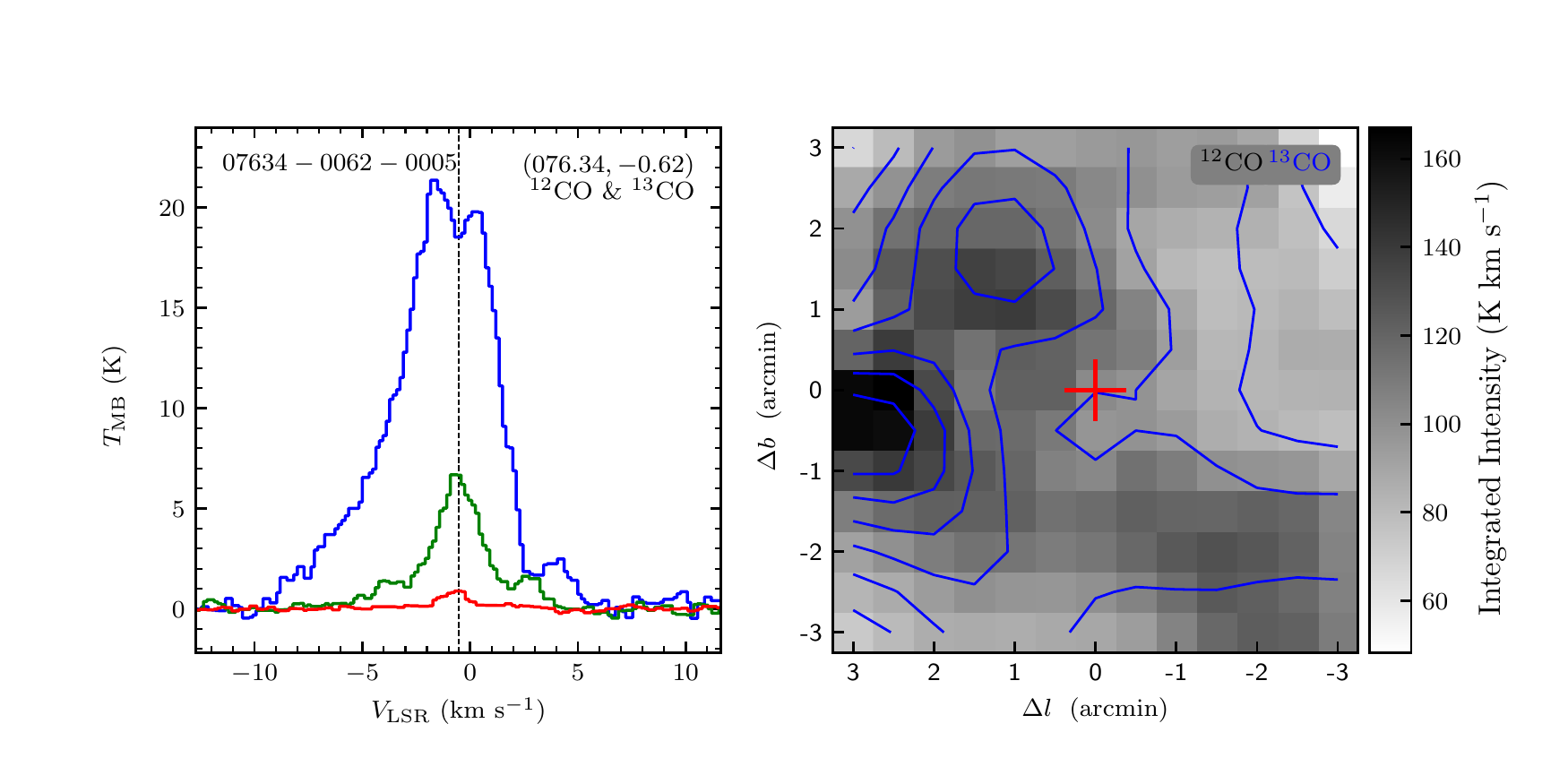}
\vspace{-0.5cm}

\includegraphics[width=9.0cm,angle=0]{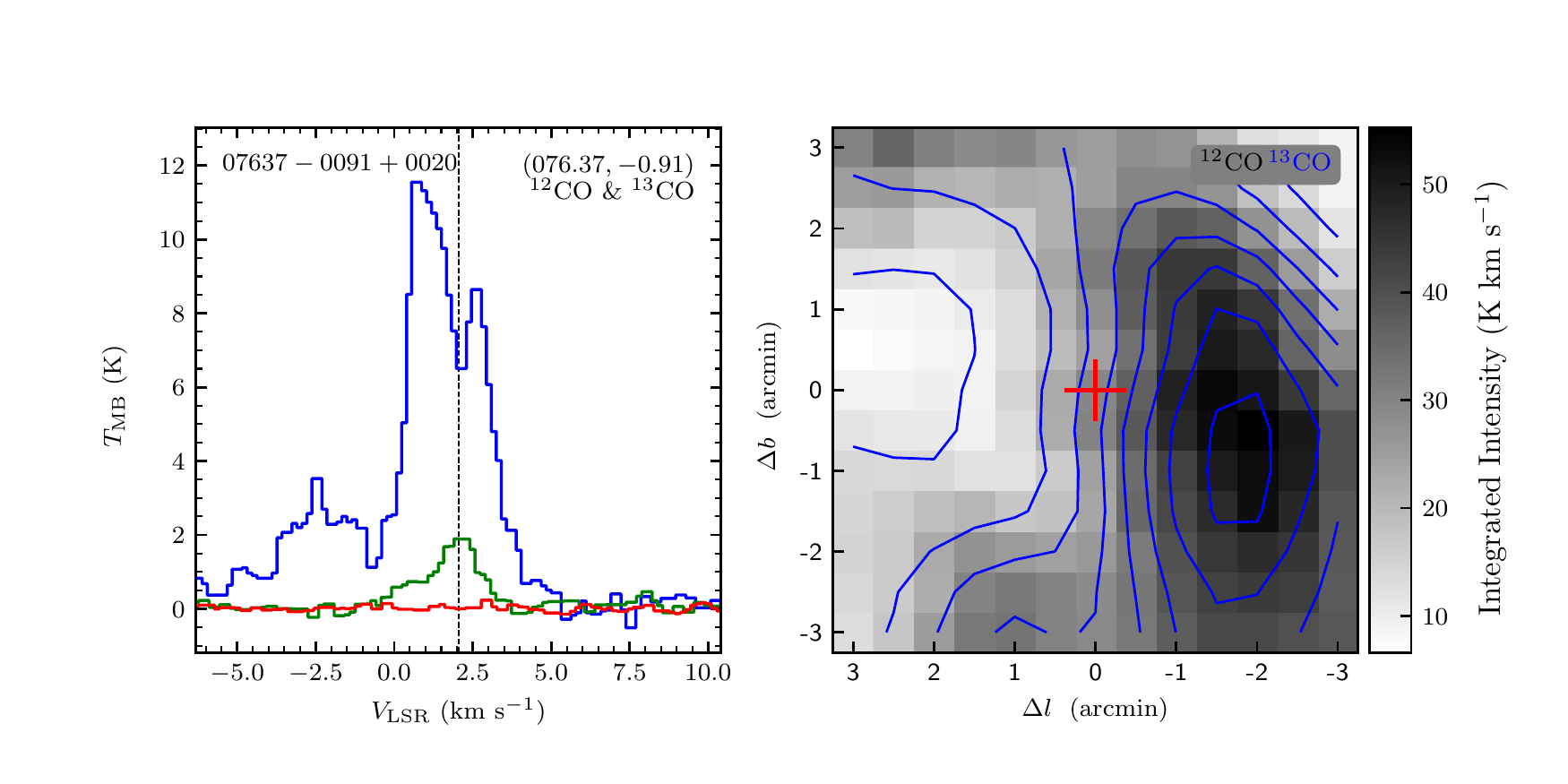}
\includegraphics[width=9.0cm,angle=0]{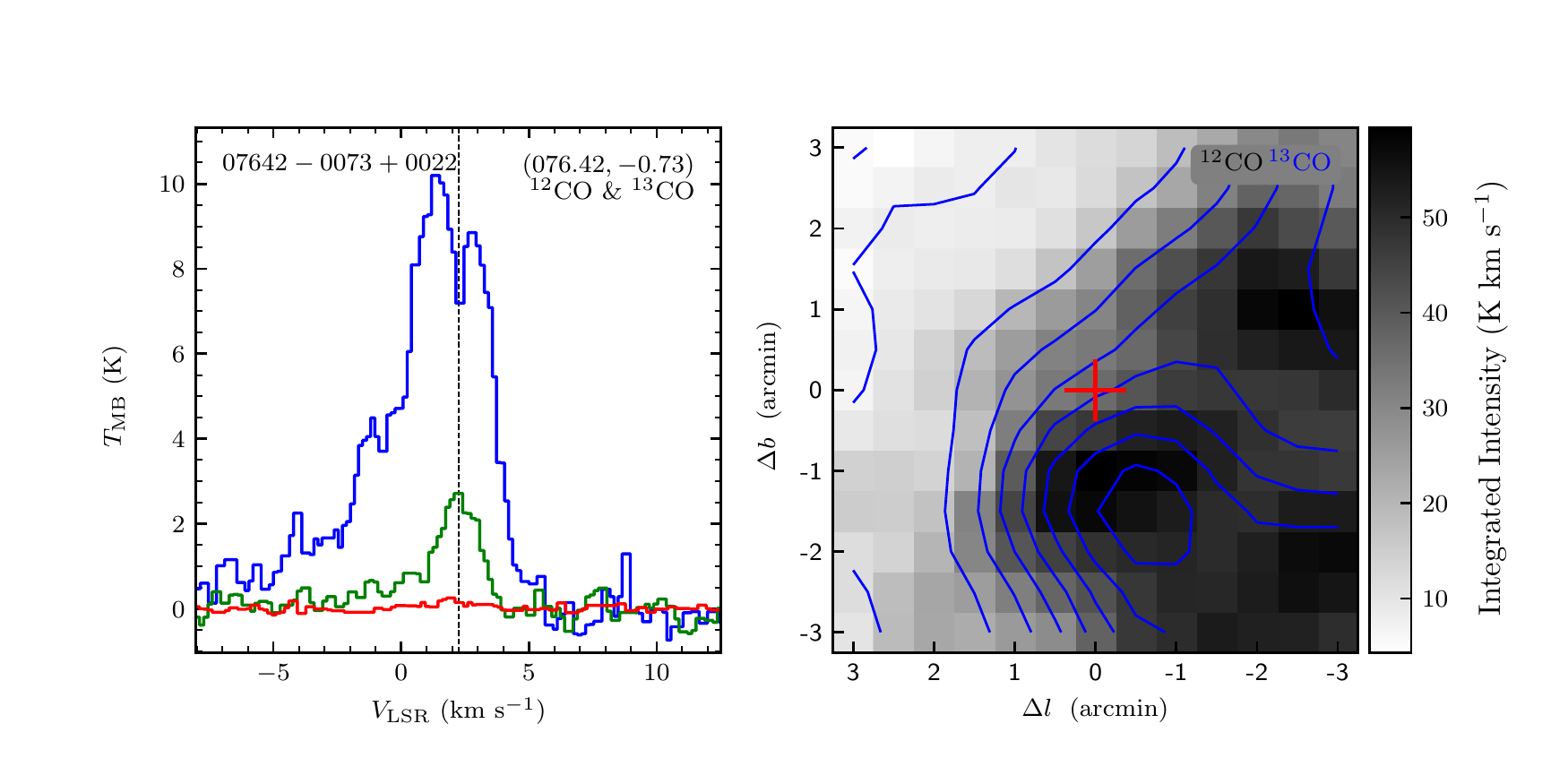}
\vspace{-0.5cm}

\includegraphics[width=9.0cm,angle=0]{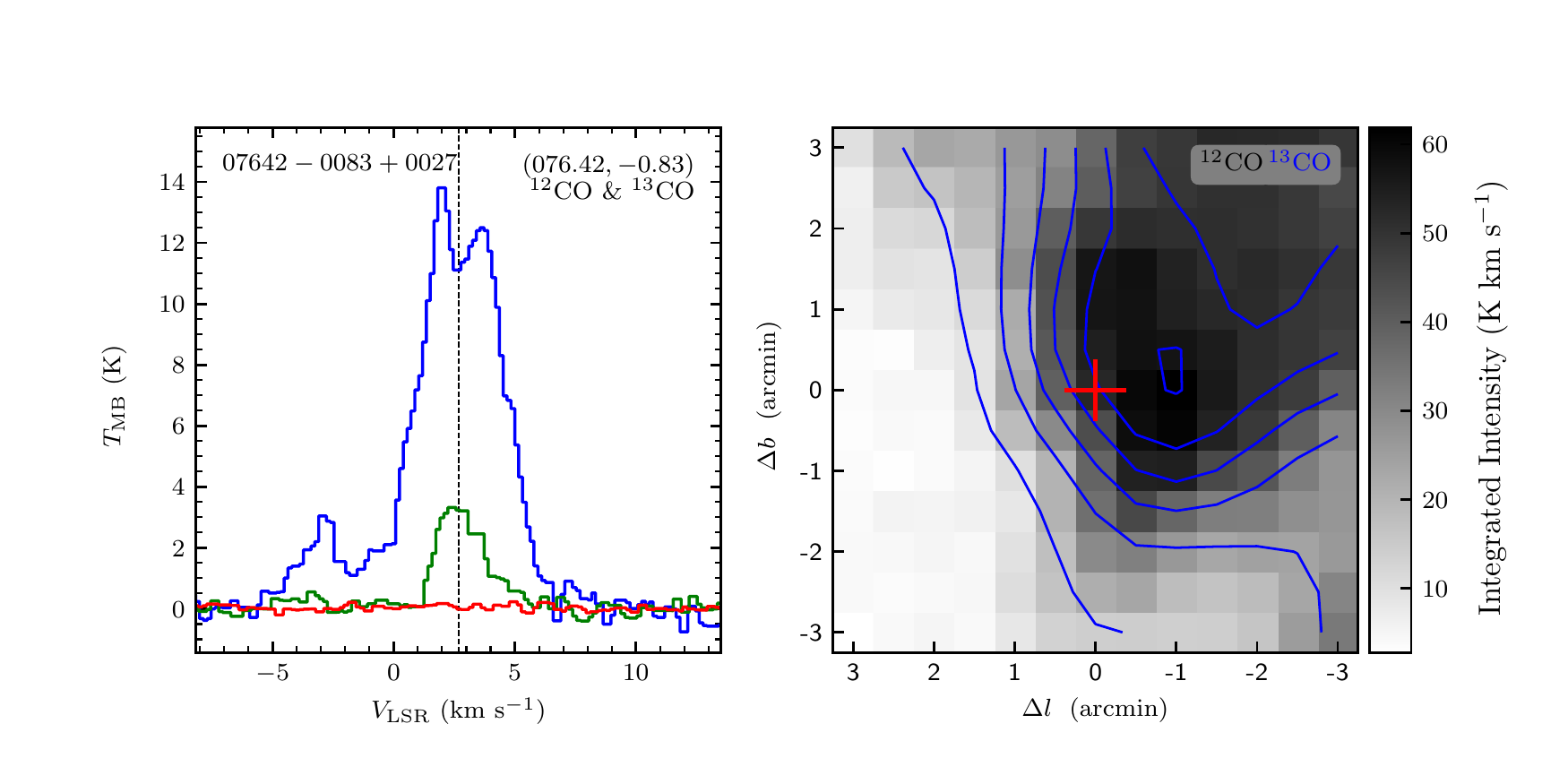}
\includegraphics[width=9.0cm,angle=0]{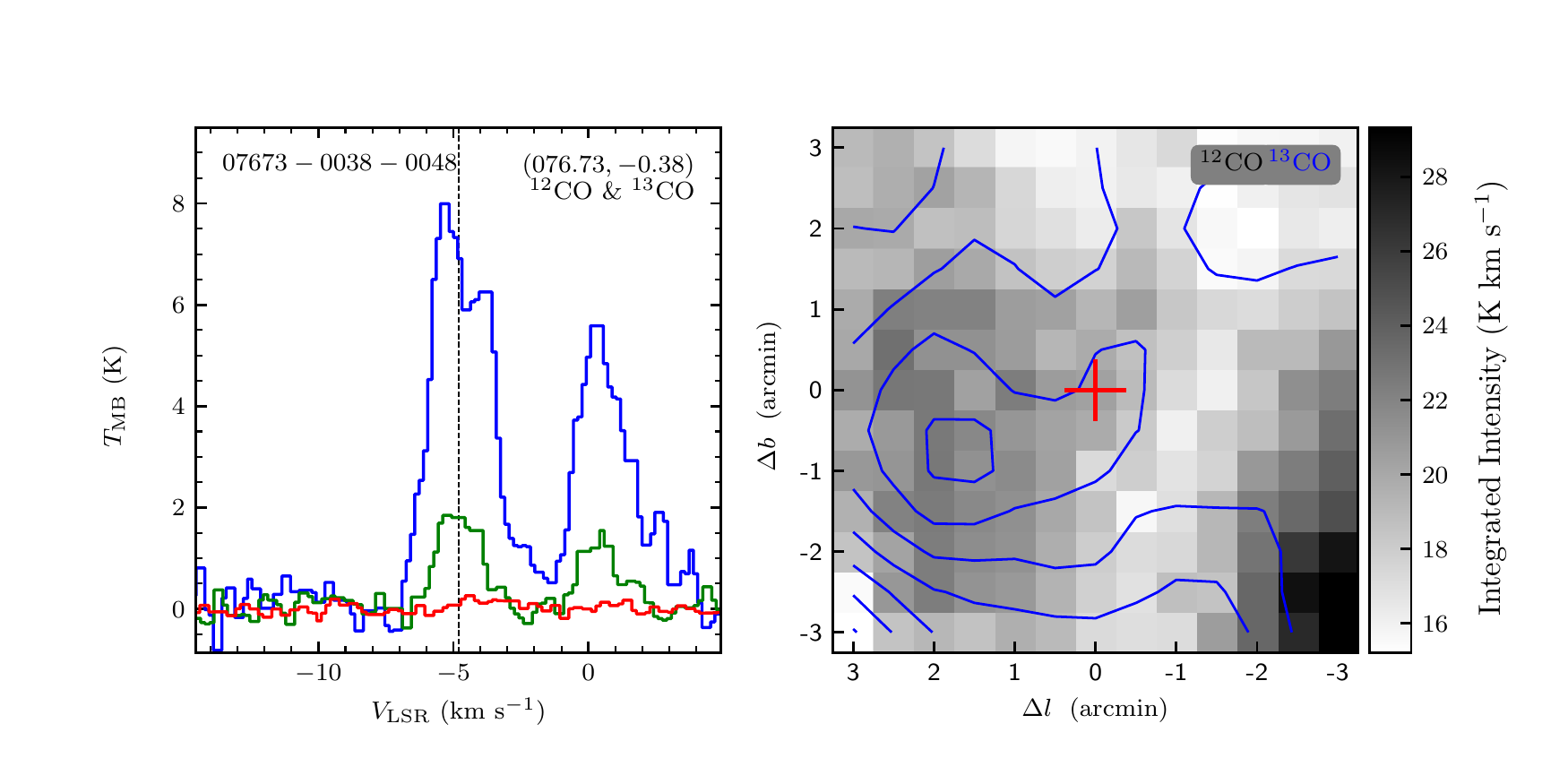}
\end{figure}
\clearpage

\begin{figure}
\includegraphics[width=9.0cm,angle=0]{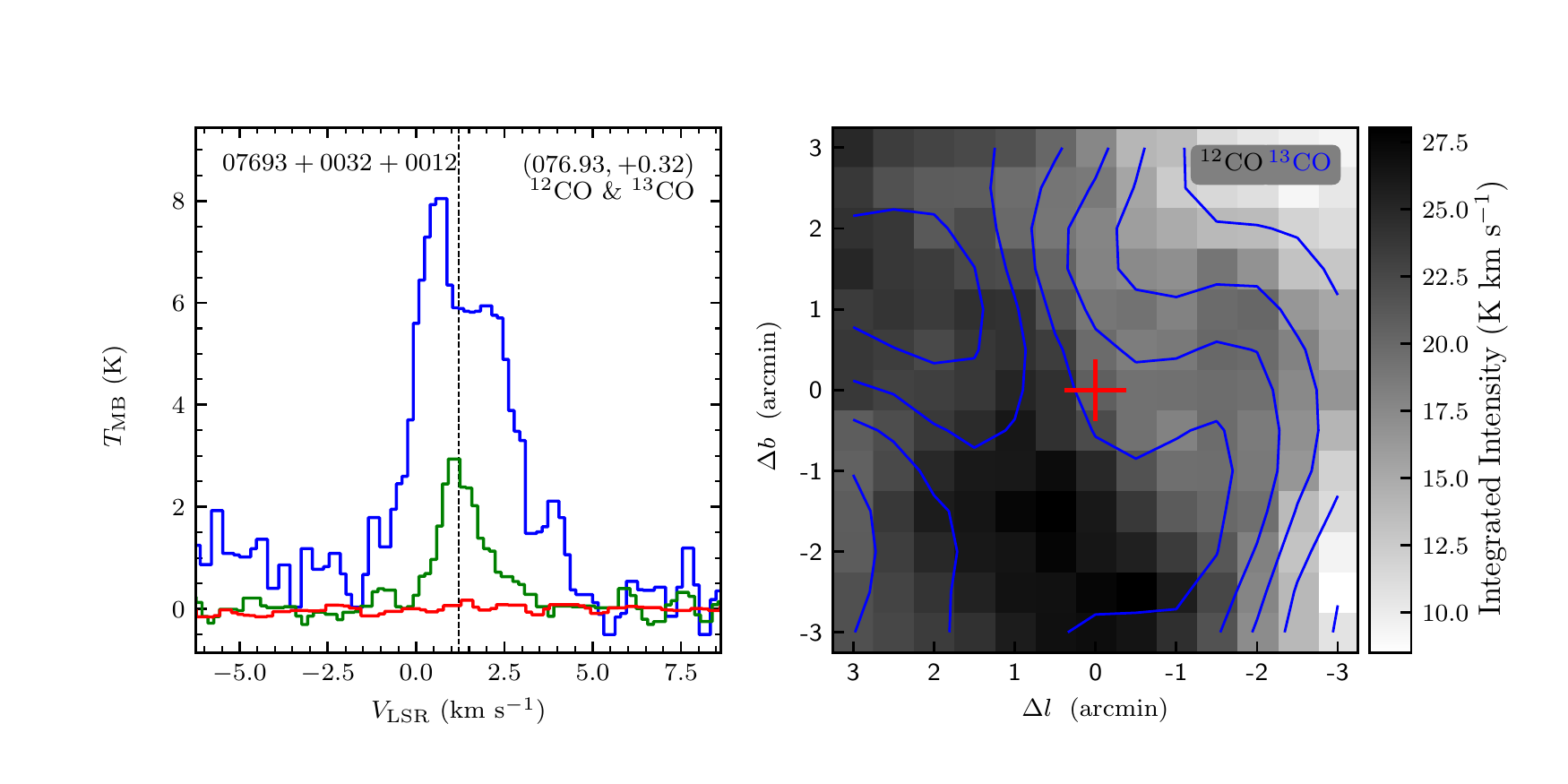}
\includegraphics[width=9.0cm,angle=0]{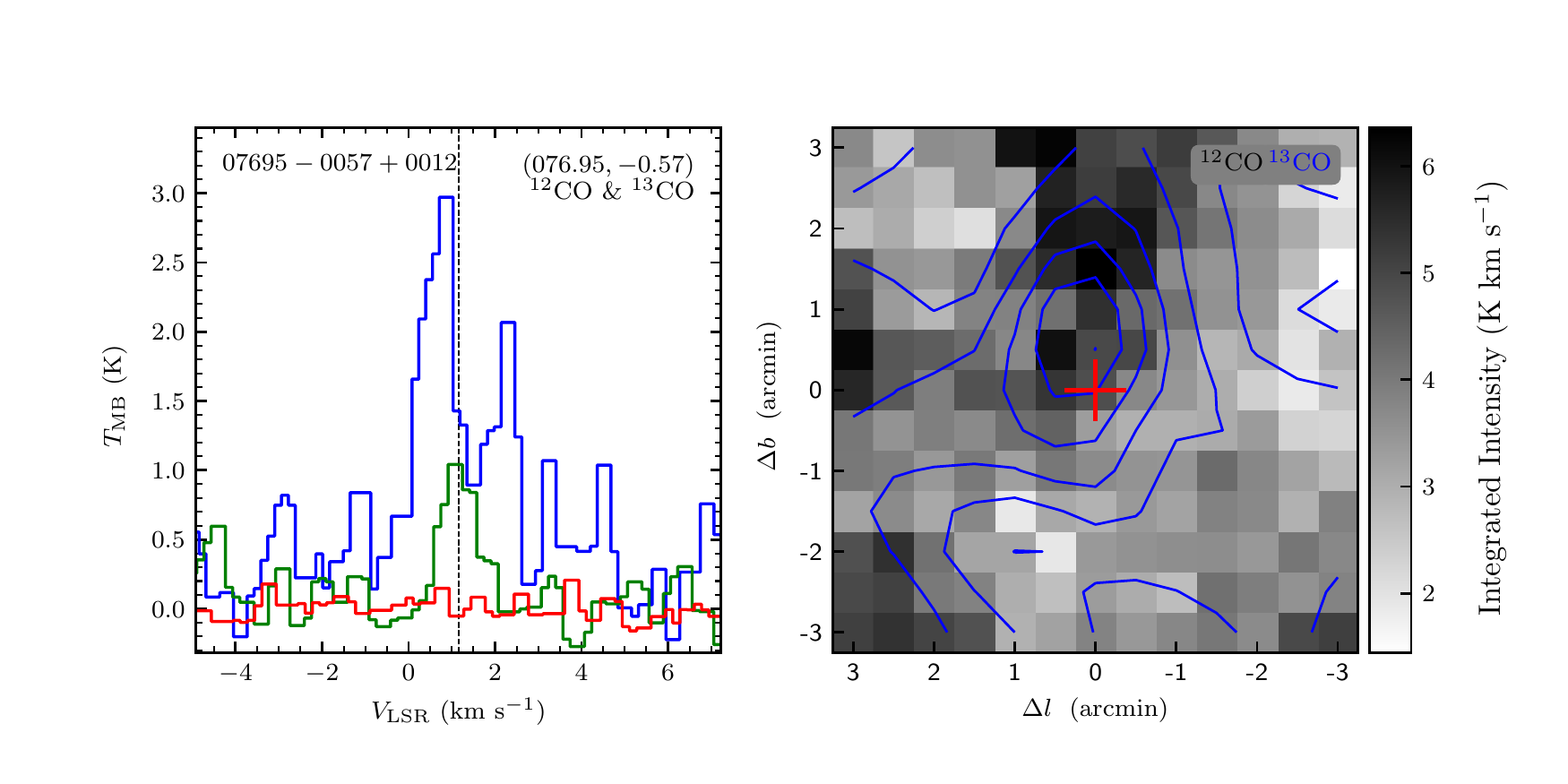}
\vspace{-0.5cm}

\includegraphics[width=9.0cm,angle=0]{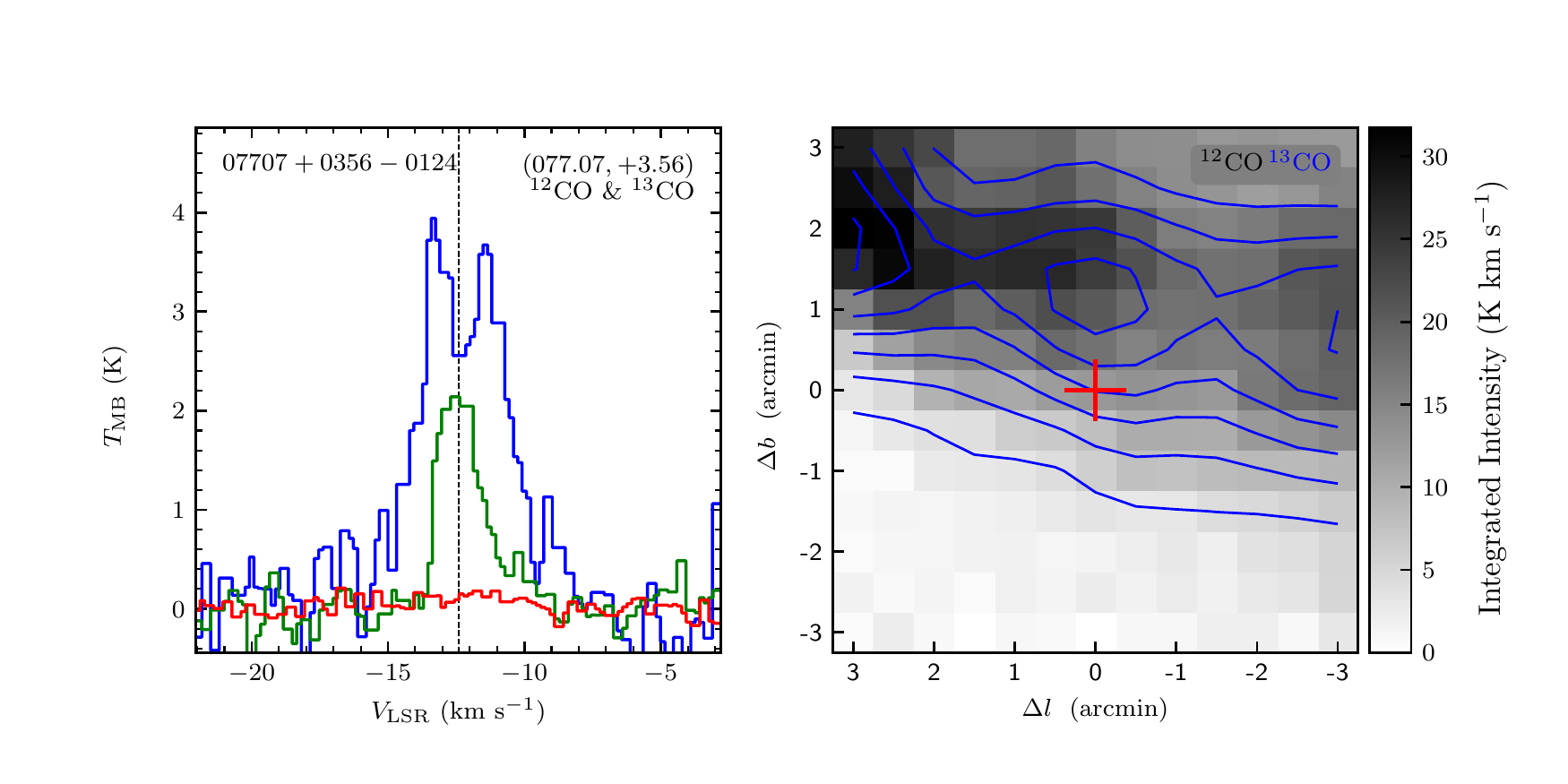}
\includegraphics[width=9.0cm,angle=0]{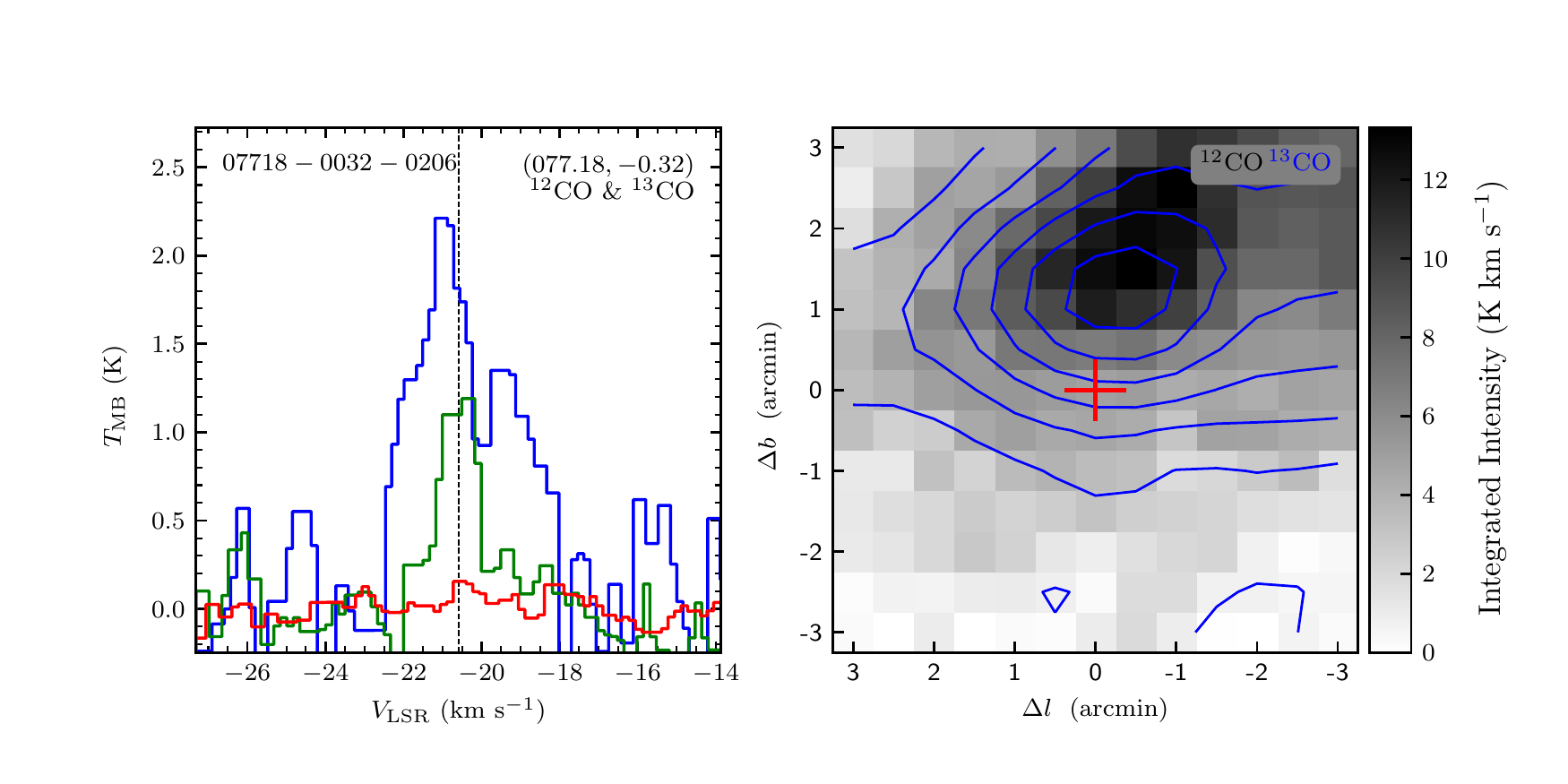}
\vspace{-0.5cm}

\includegraphics[width=9.0cm,angle=0]{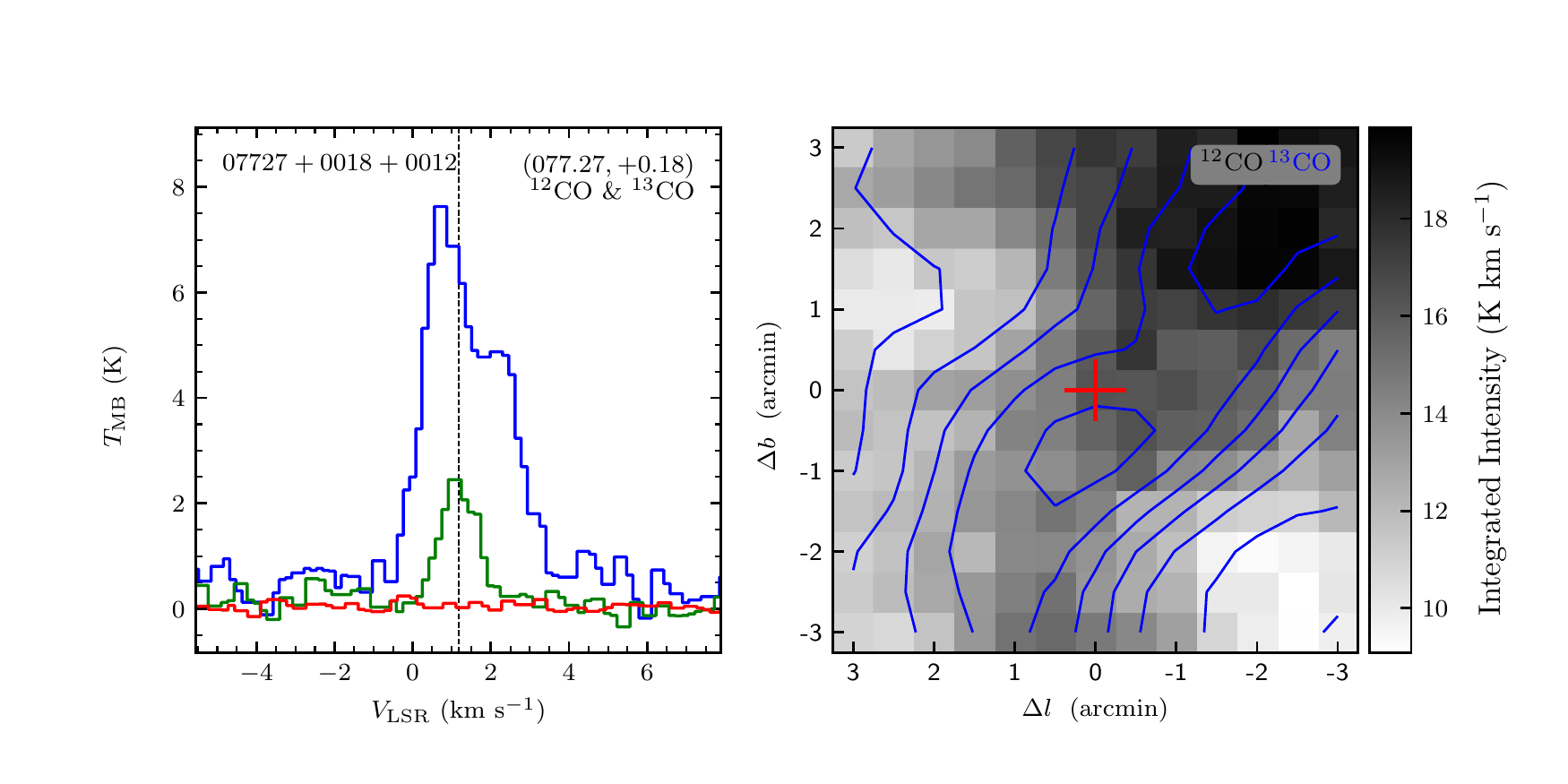}
\includegraphics[width=9.0cm,angle=0]{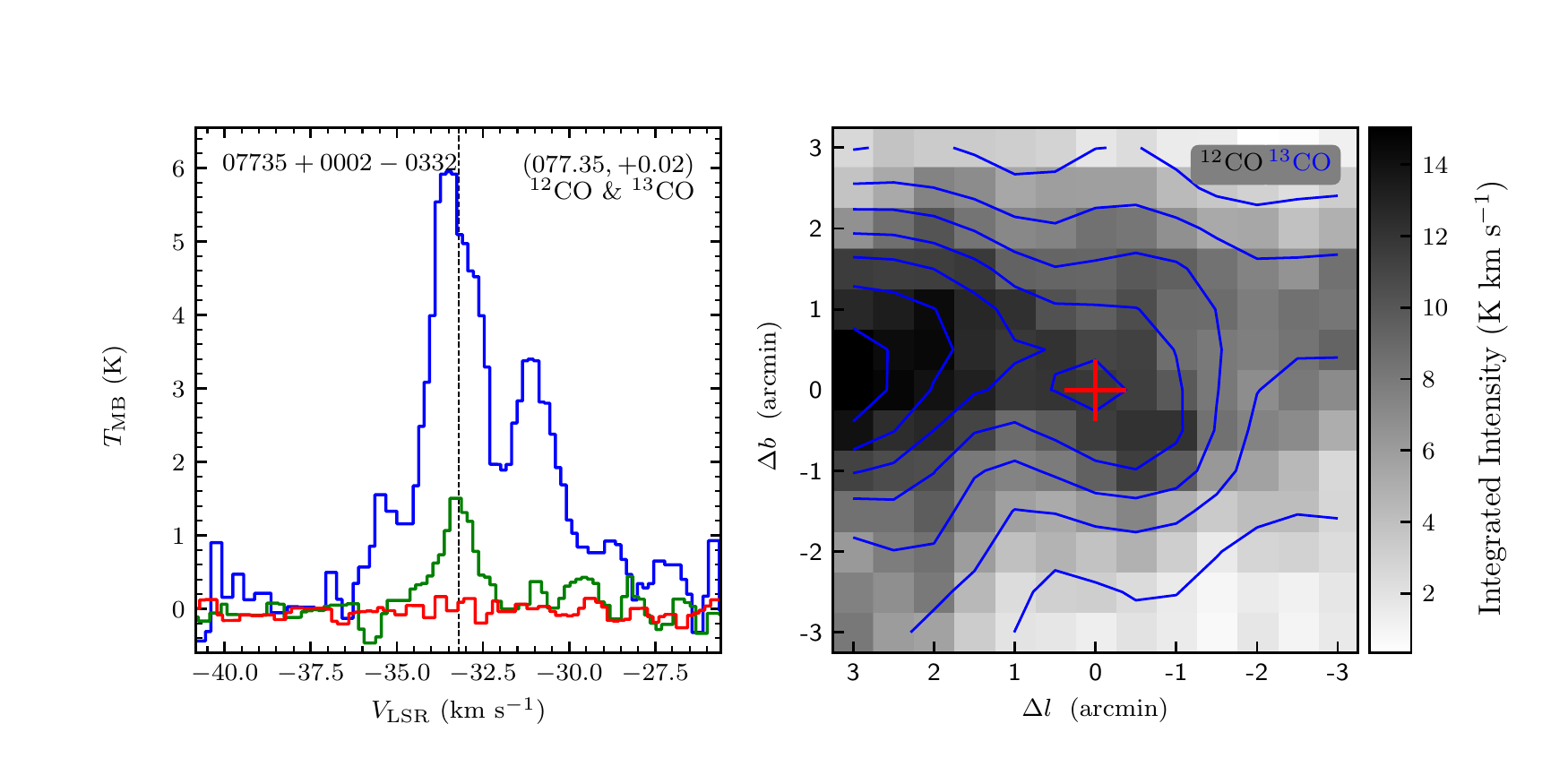}
\vspace{-0.5cm}

\includegraphics[width=9.0cm,angle=0]{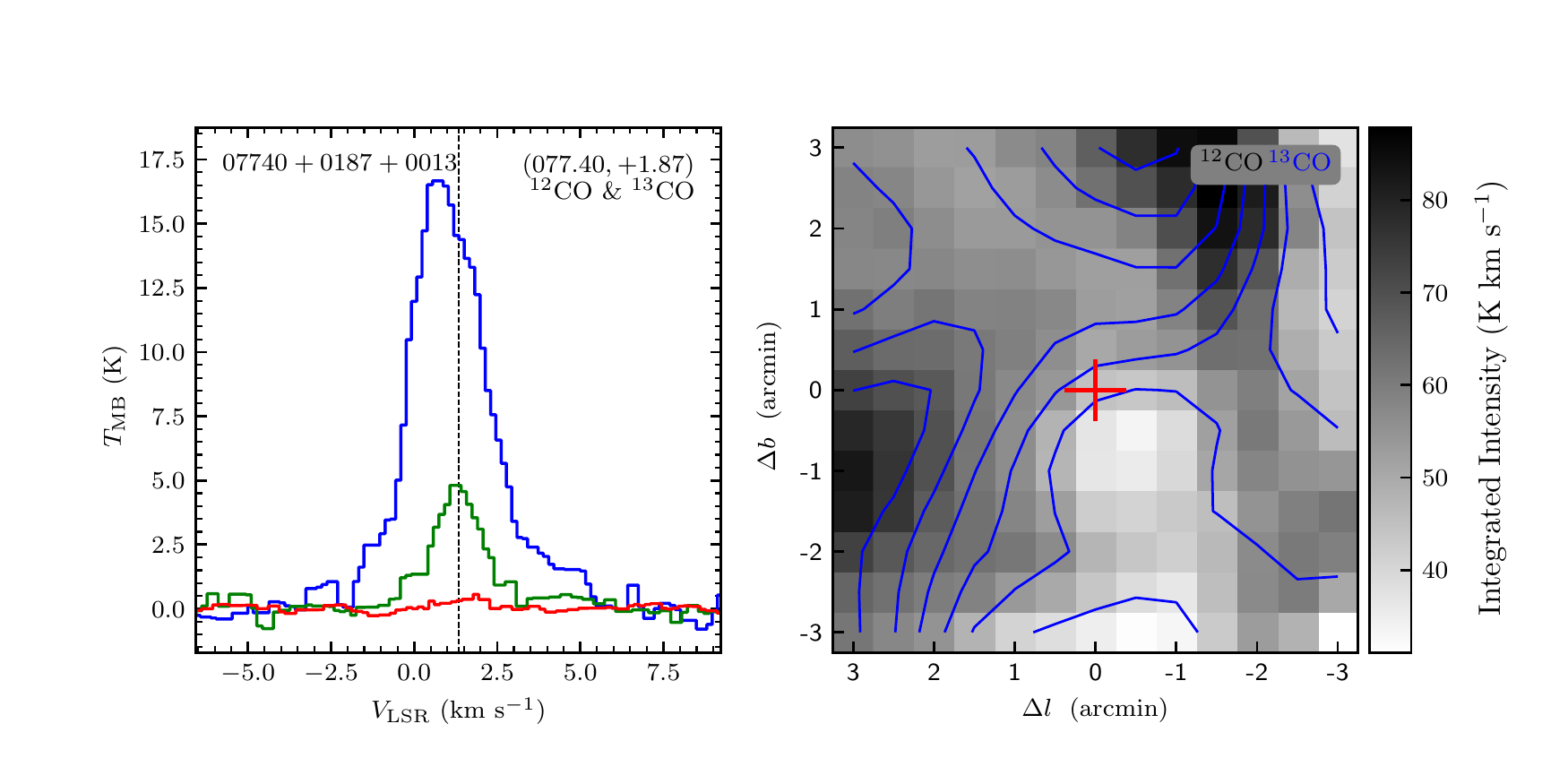}
\includegraphics[width=9.0cm,angle=0]{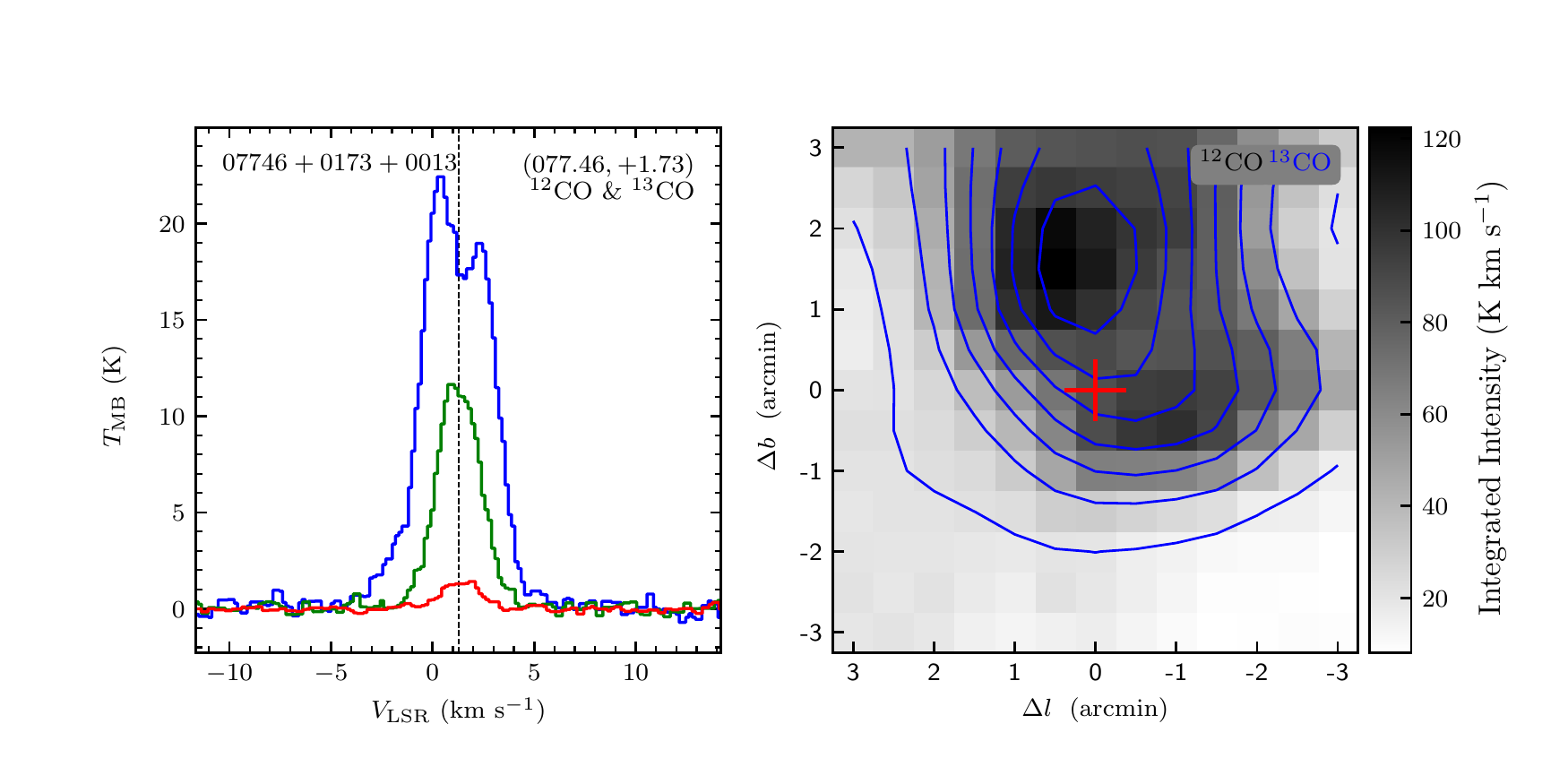}
\vspace{-0.5cm}

\includegraphics[width=9.0cm,angle=0]{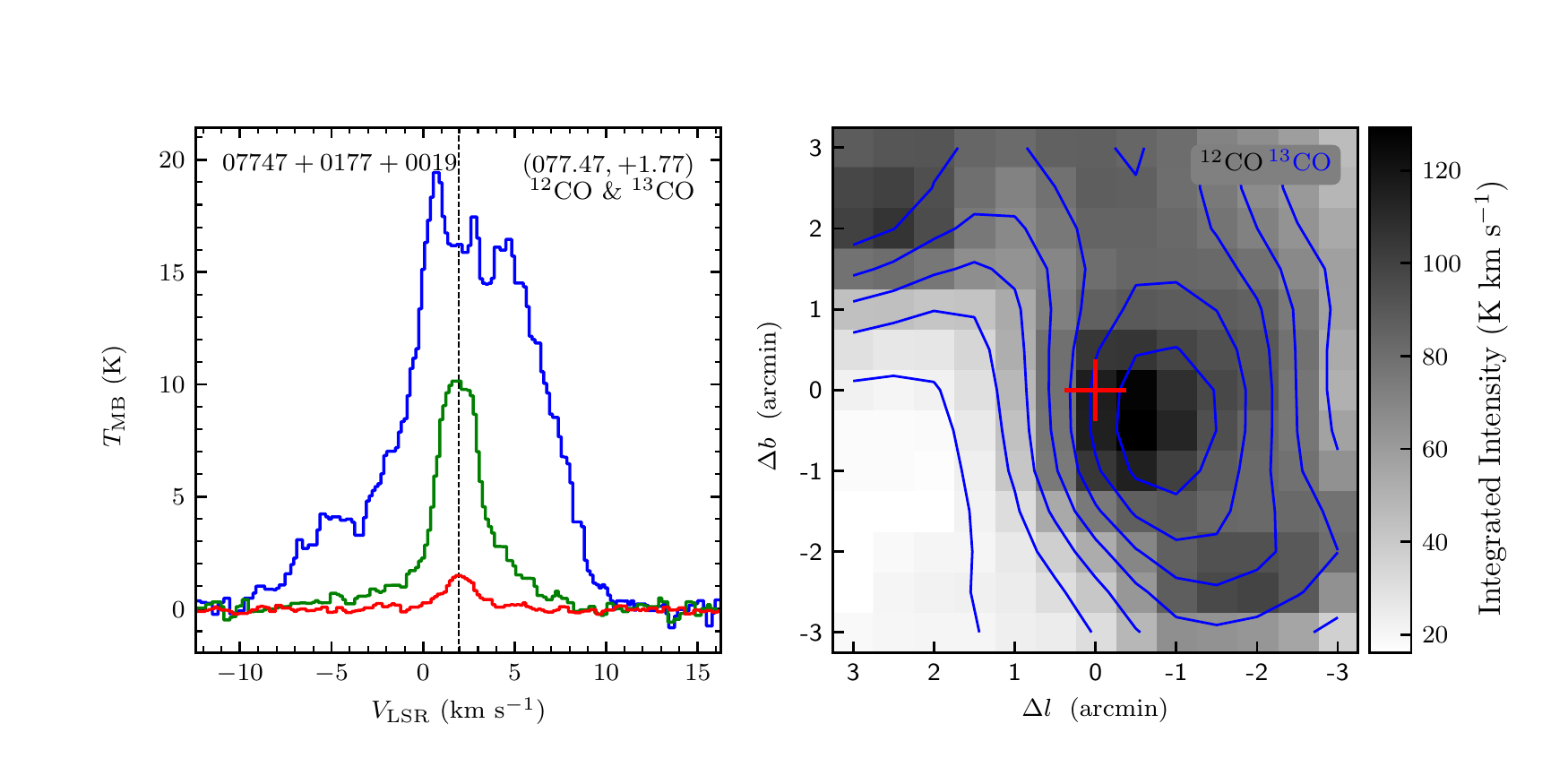}
\includegraphics[width=9.0cm,angle=0]{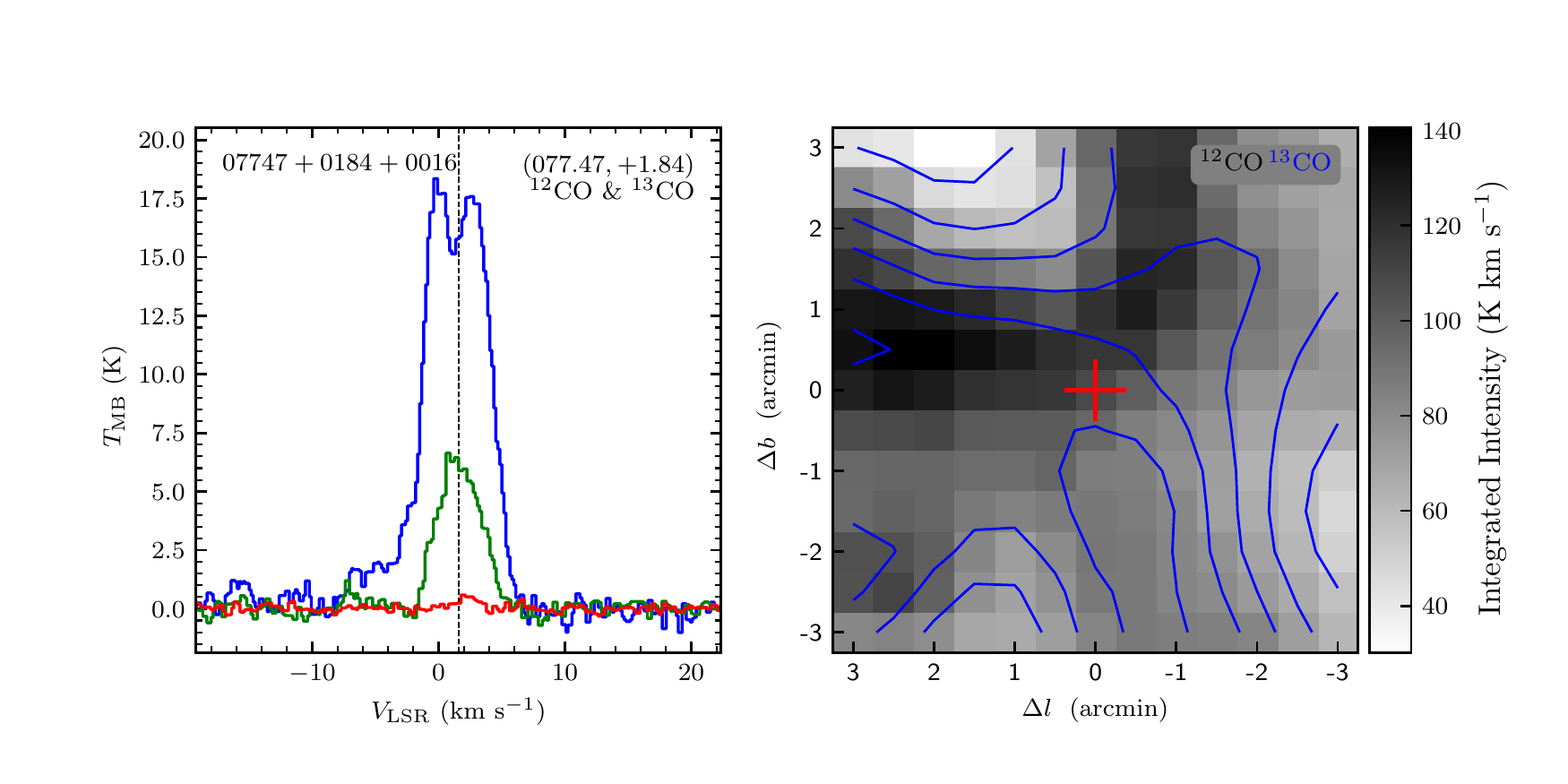}
\end{figure}
\clearpage

\begin{figure}
\includegraphics[width=9.0cm,angle=0]{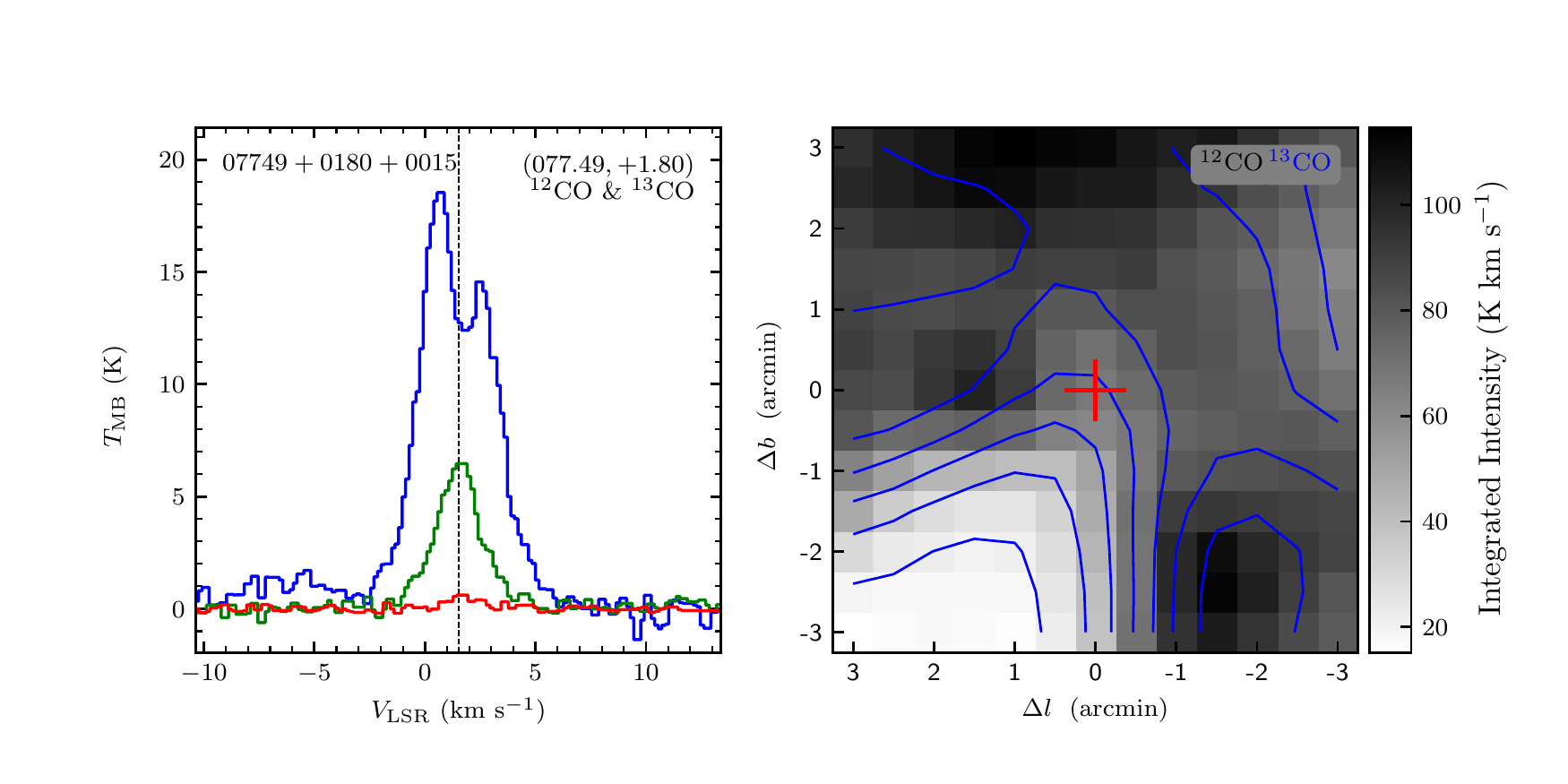}
\includegraphics[width=9.0cm,angle=0]{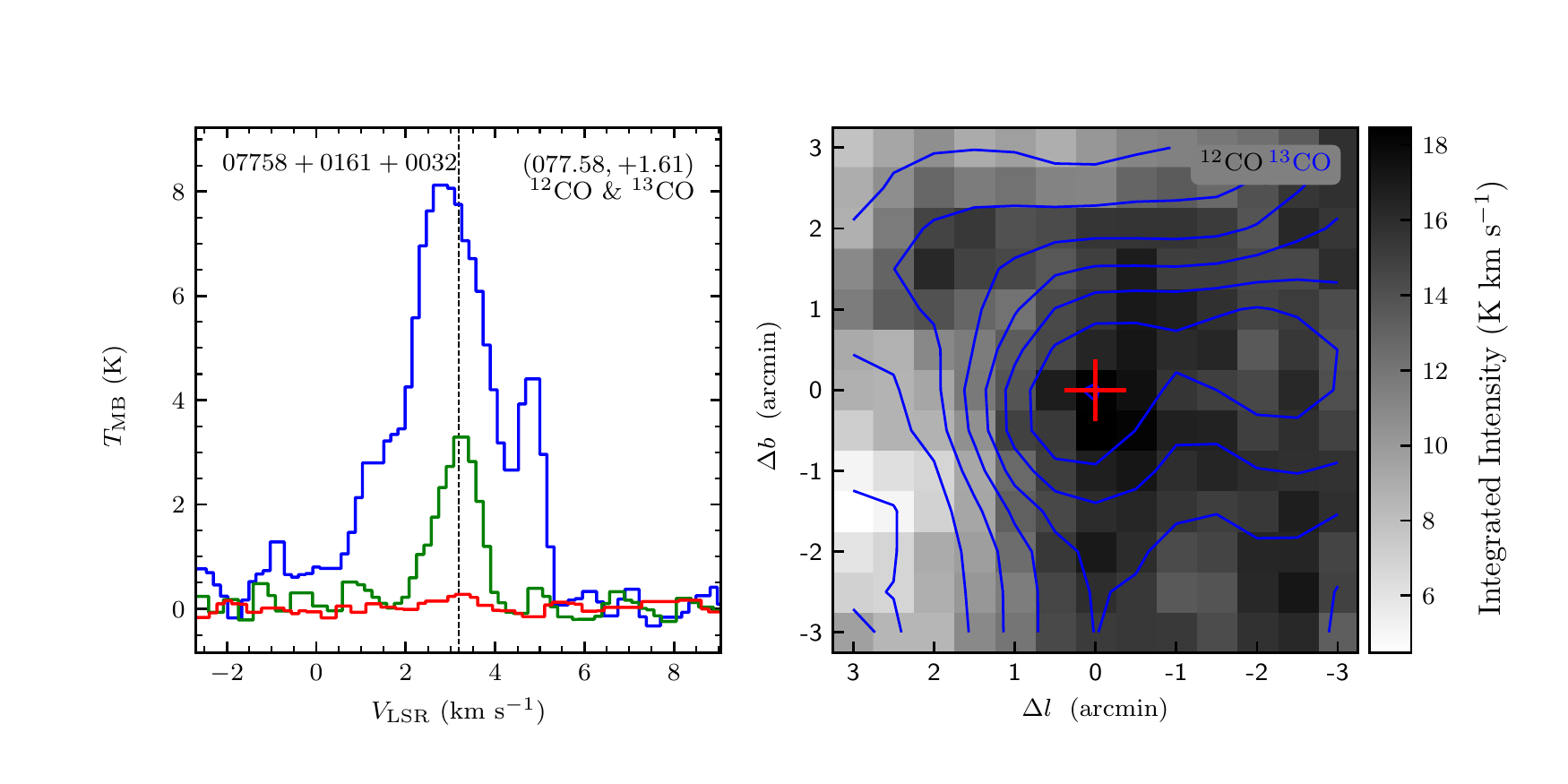}
\vspace{-0.5cm}

\includegraphics[width=9.0cm,angle=0]{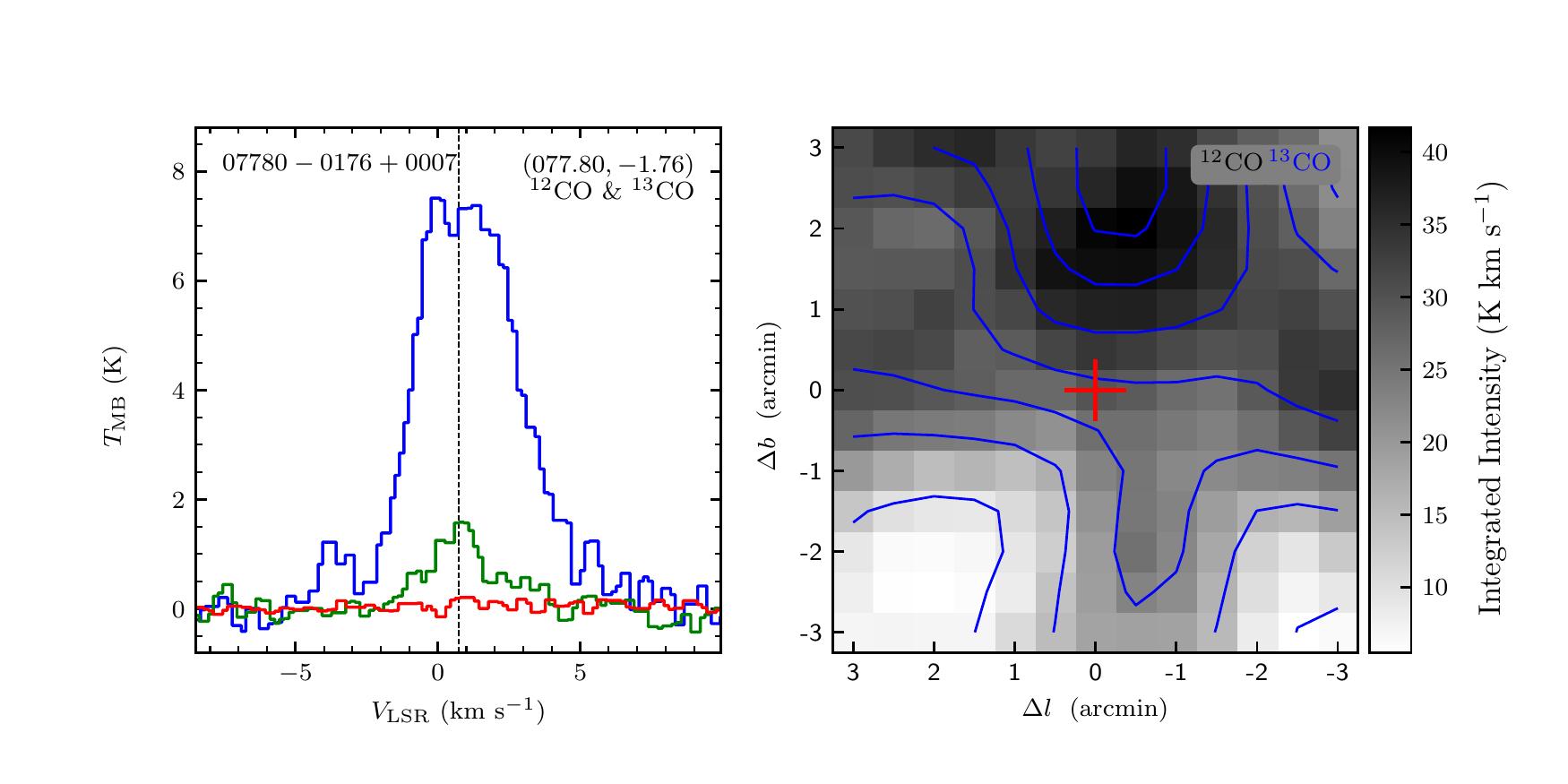}
\includegraphics[width=9.0cm,angle=0]{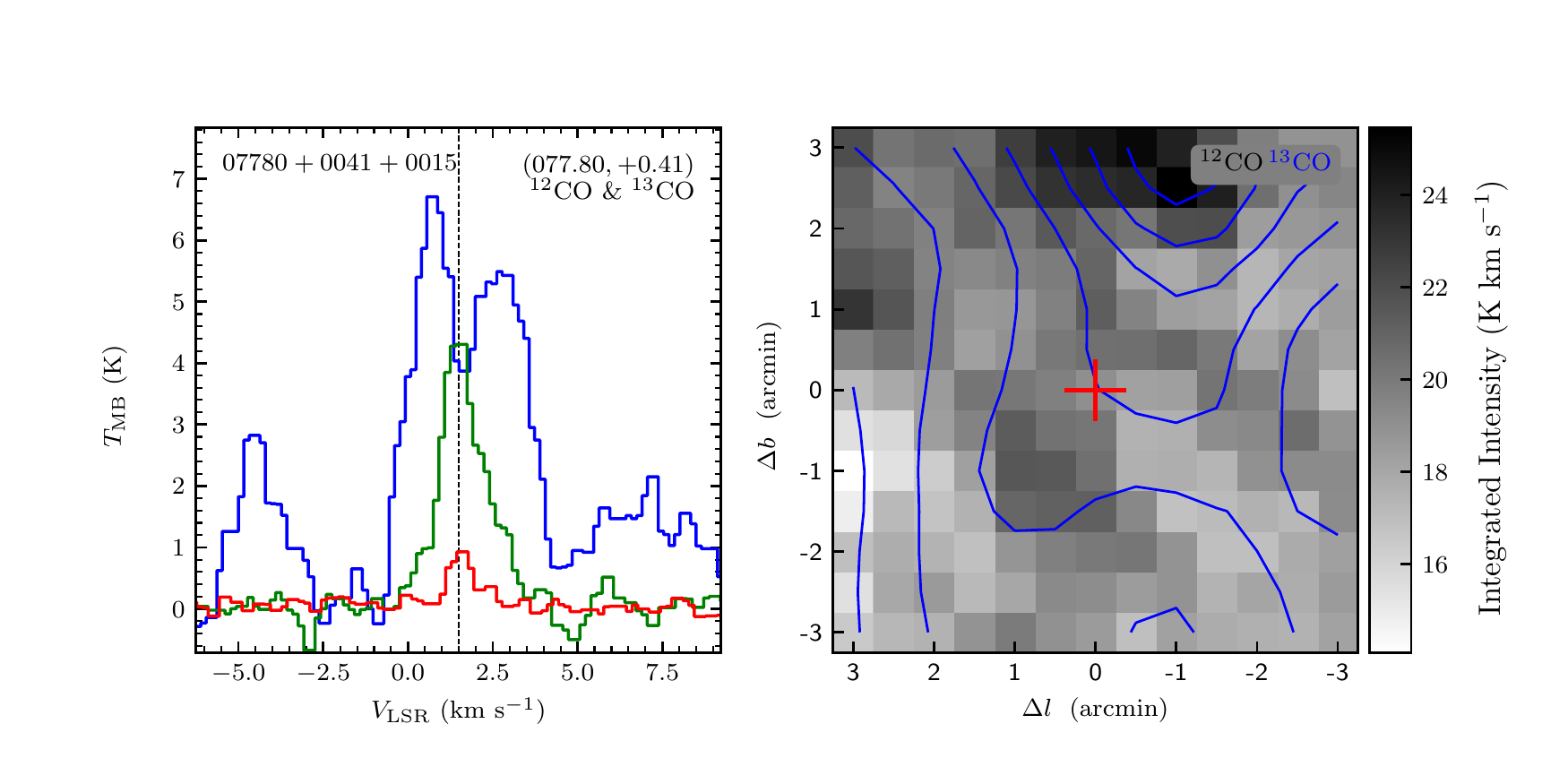}
\vspace{-0.5cm}

\includegraphics[width=9.0cm,angle=0]{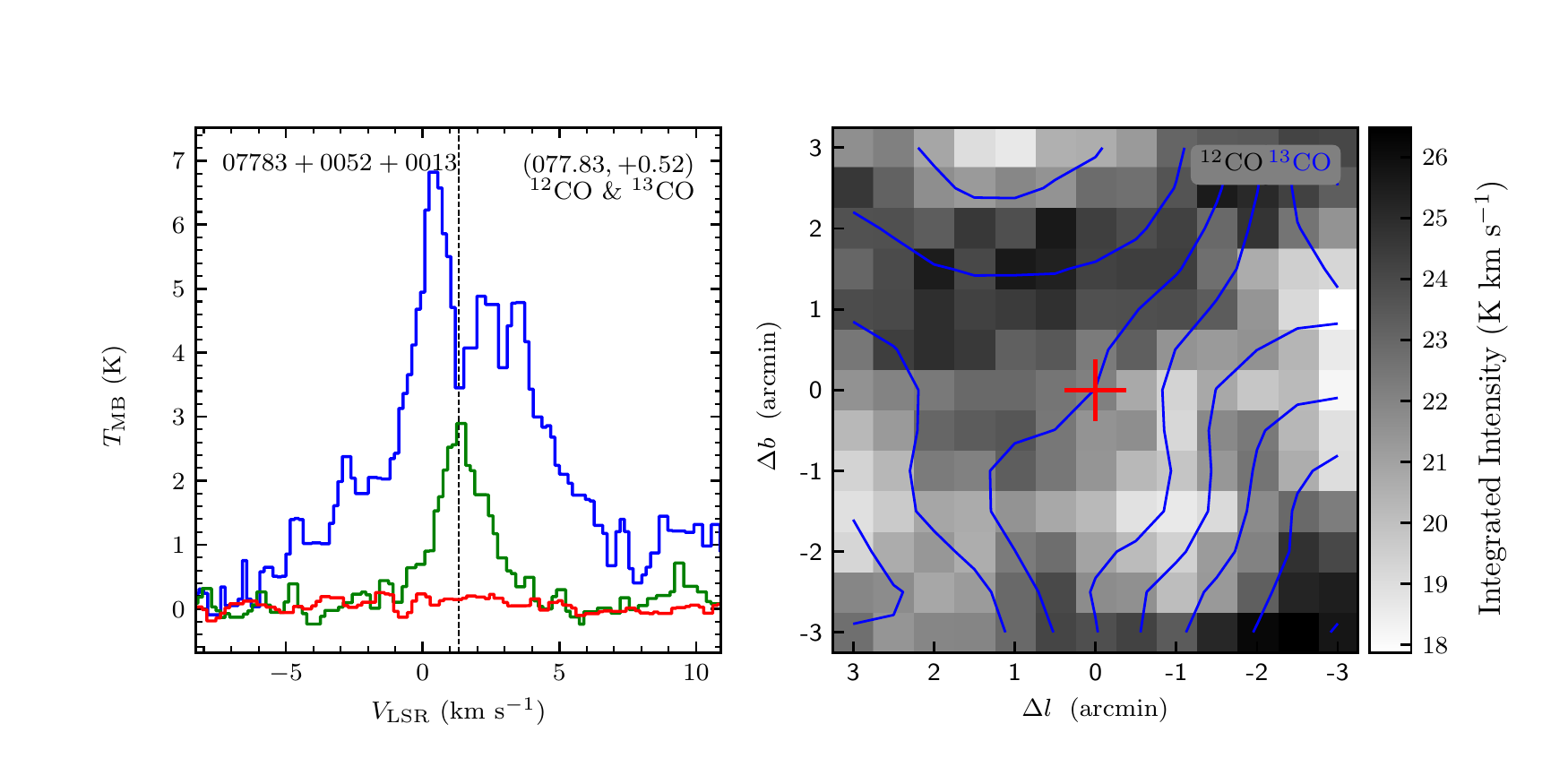}
\includegraphics[width=9.0cm,angle=0]{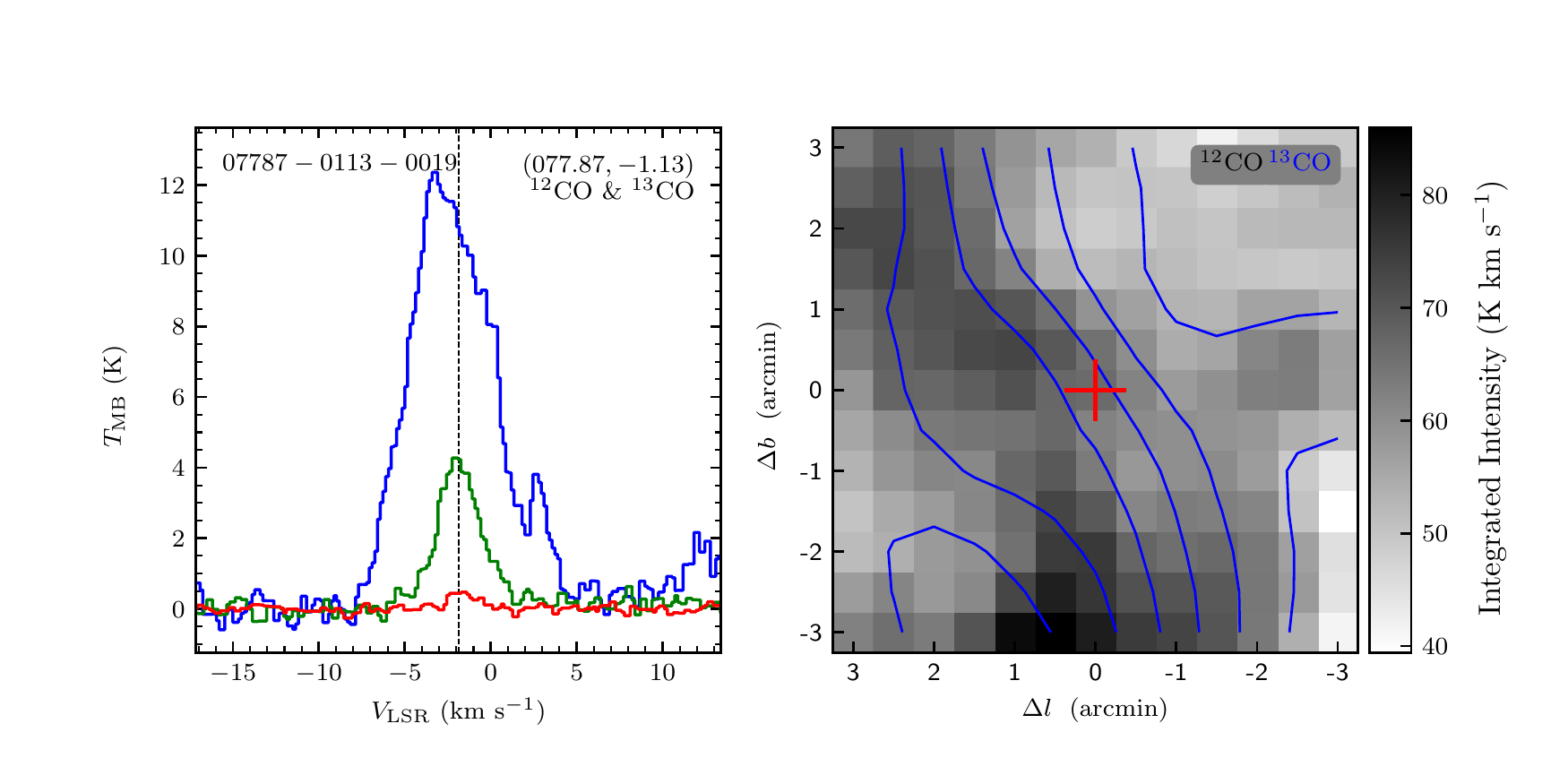}
\vspace{-0.5cm}

\includegraphics[width=9.0cm,angle=0]{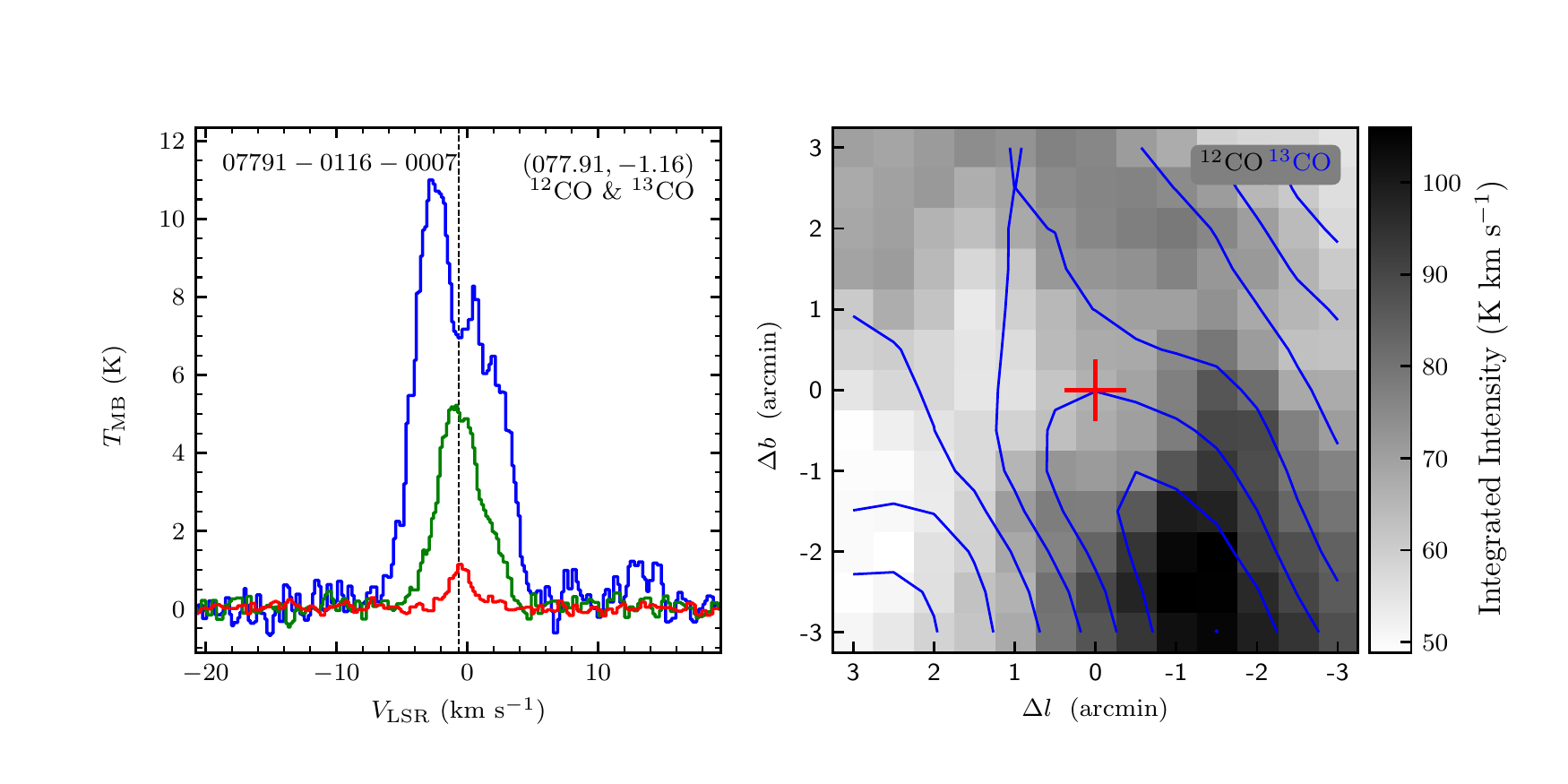}
\includegraphics[width=9.0cm,angle=0]{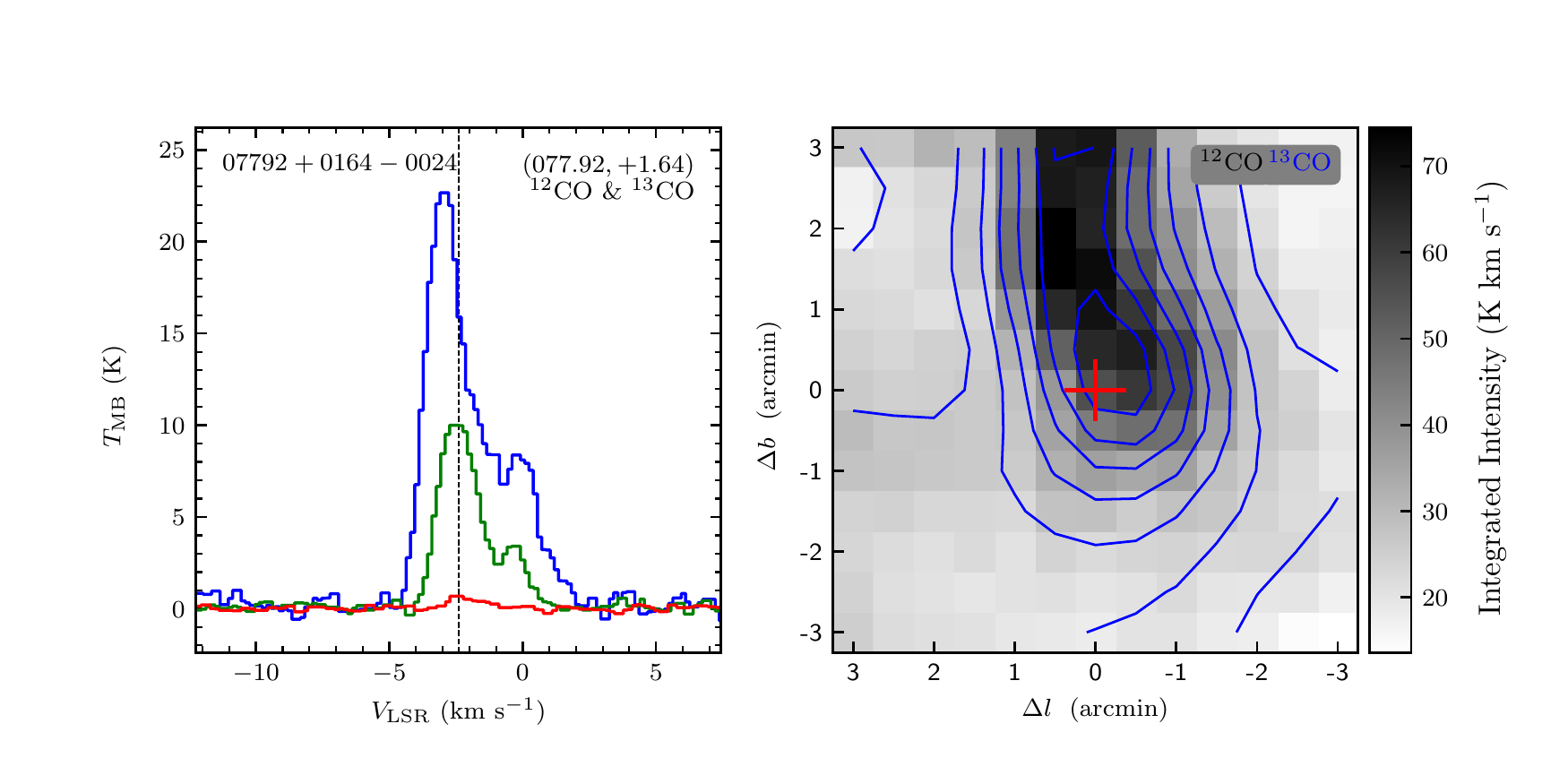}
\vspace{-0.5cm}

\includegraphics[width=9.0cm,angle=0]{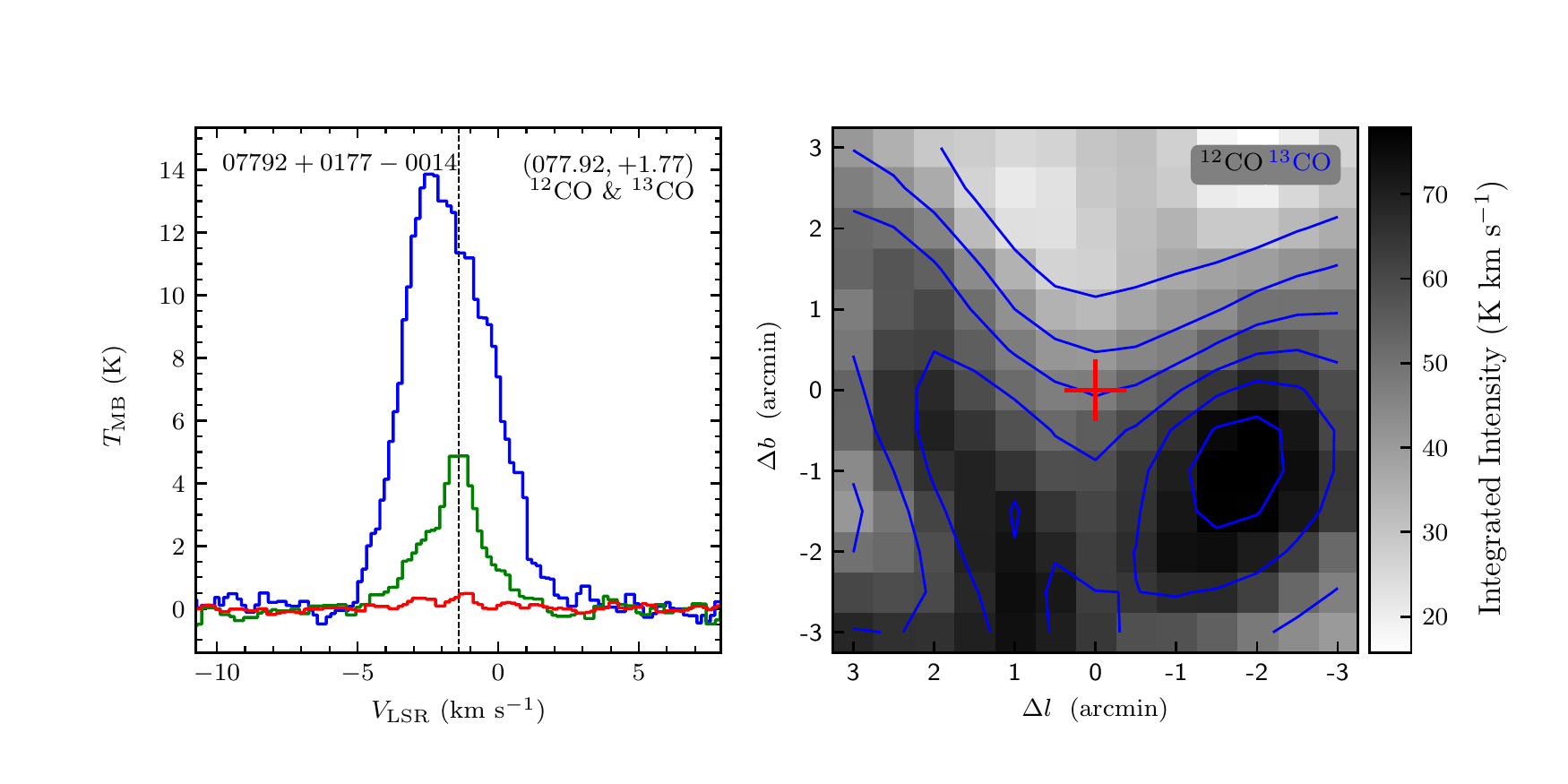}
\includegraphics[width=9.0cm,angle=0]{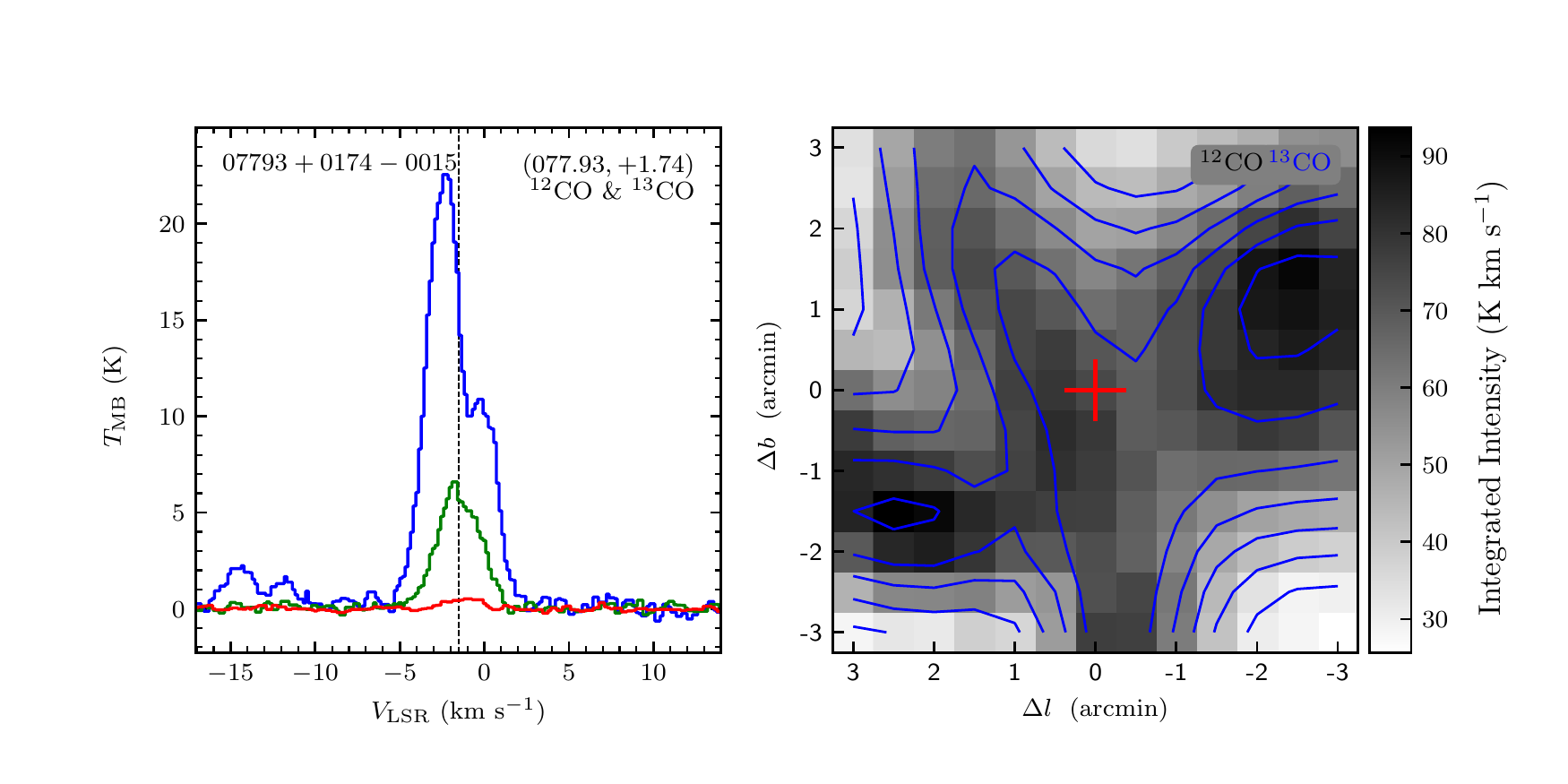}
\end{figure}
\clearpage

\begin{figure}
\includegraphics[width=9.0cm,angle=0]{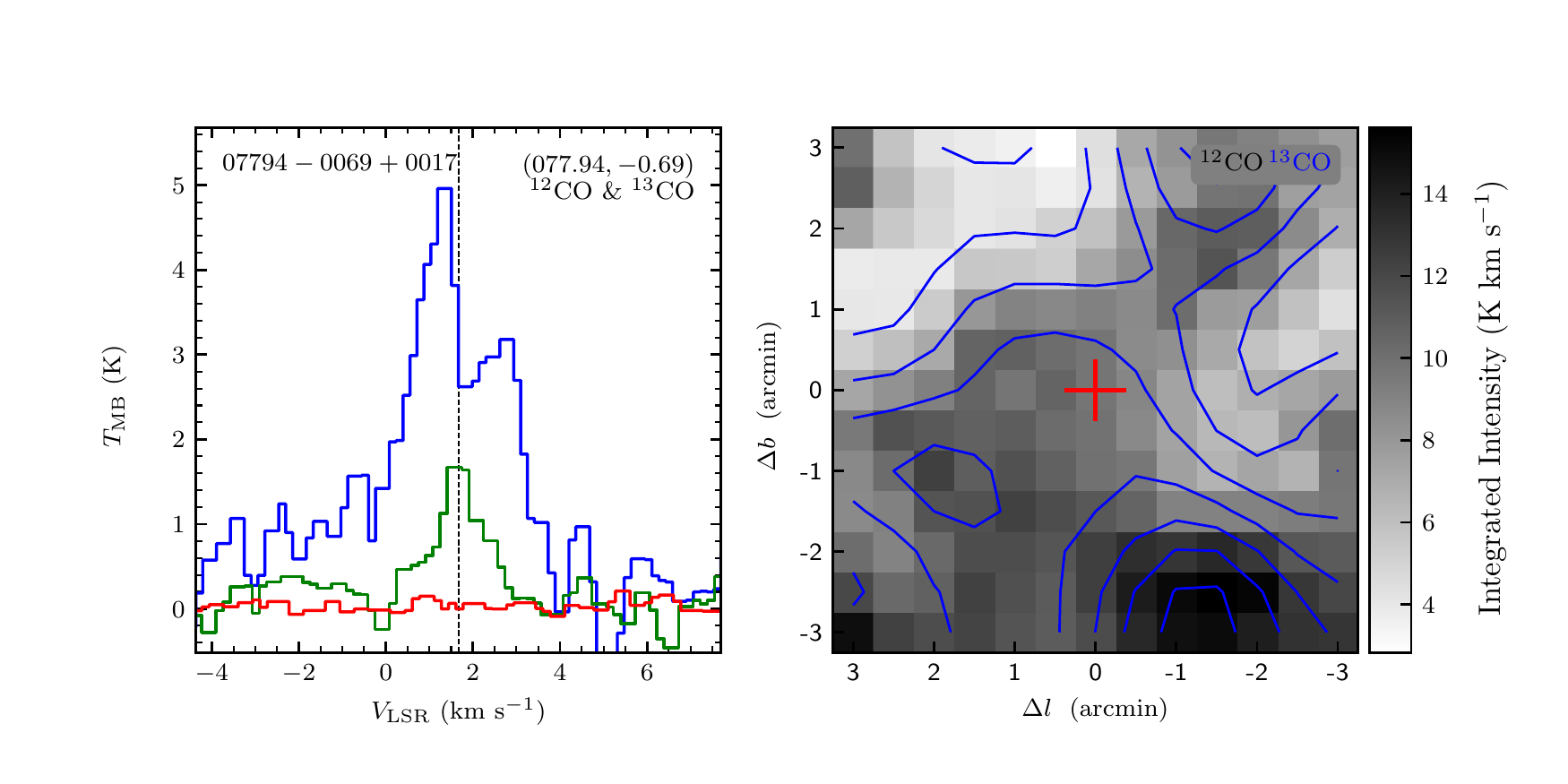}
\includegraphics[width=9.0cm,angle=0]{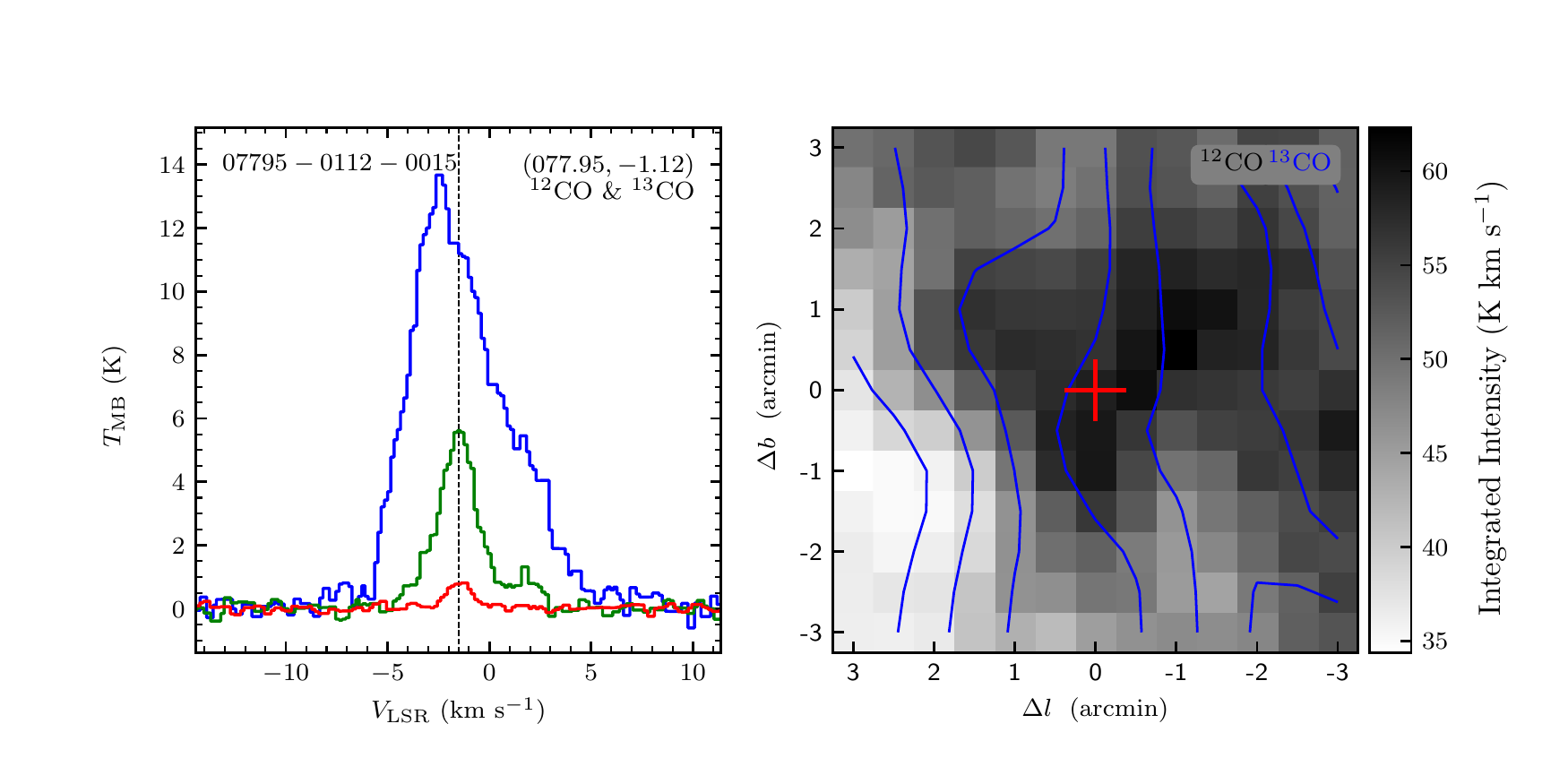}
\vspace{-0.5cm}

\includegraphics[width=9.0cm,angle=0]{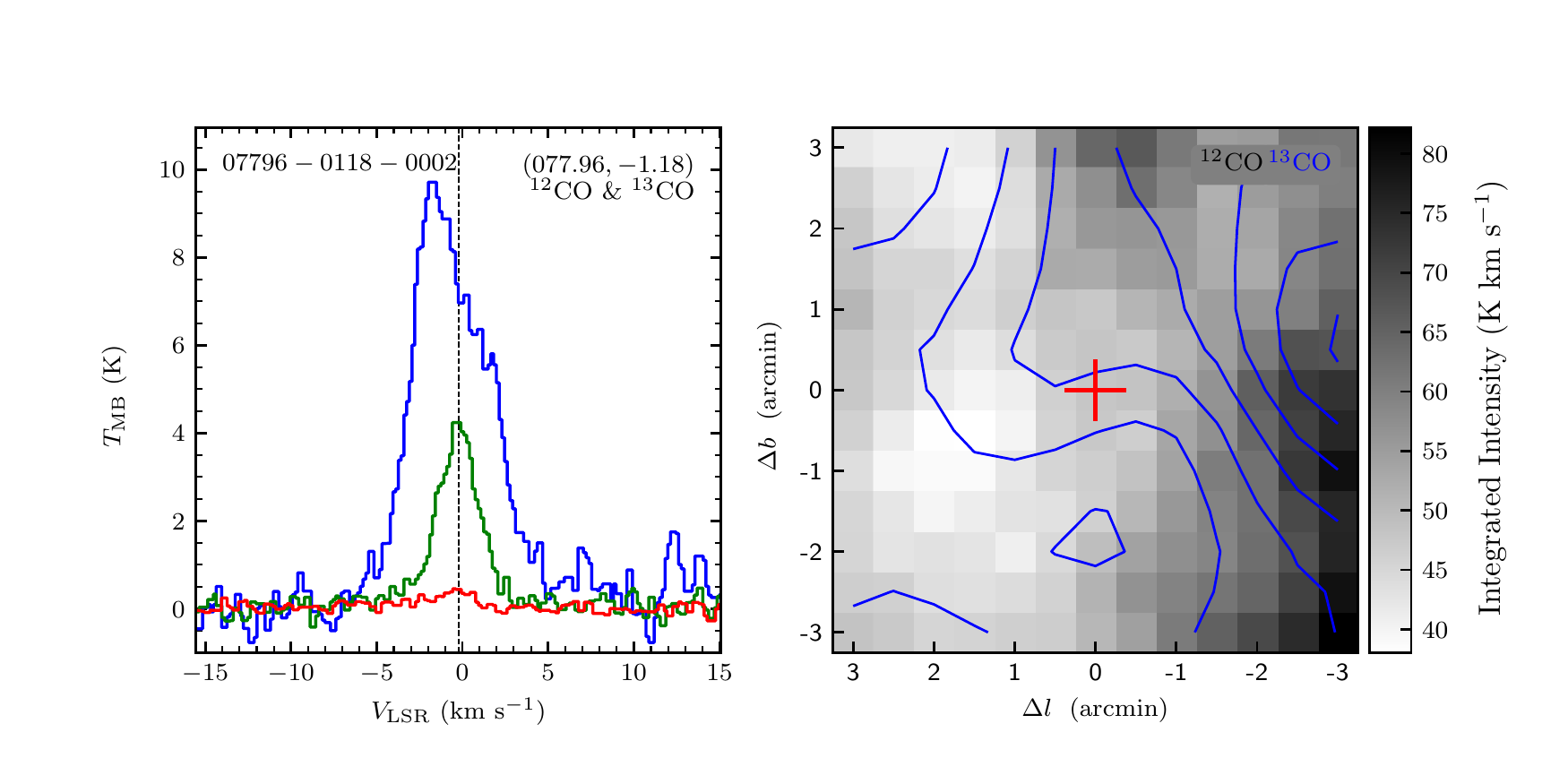}
\includegraphics[width=9.0cm,angle=0]{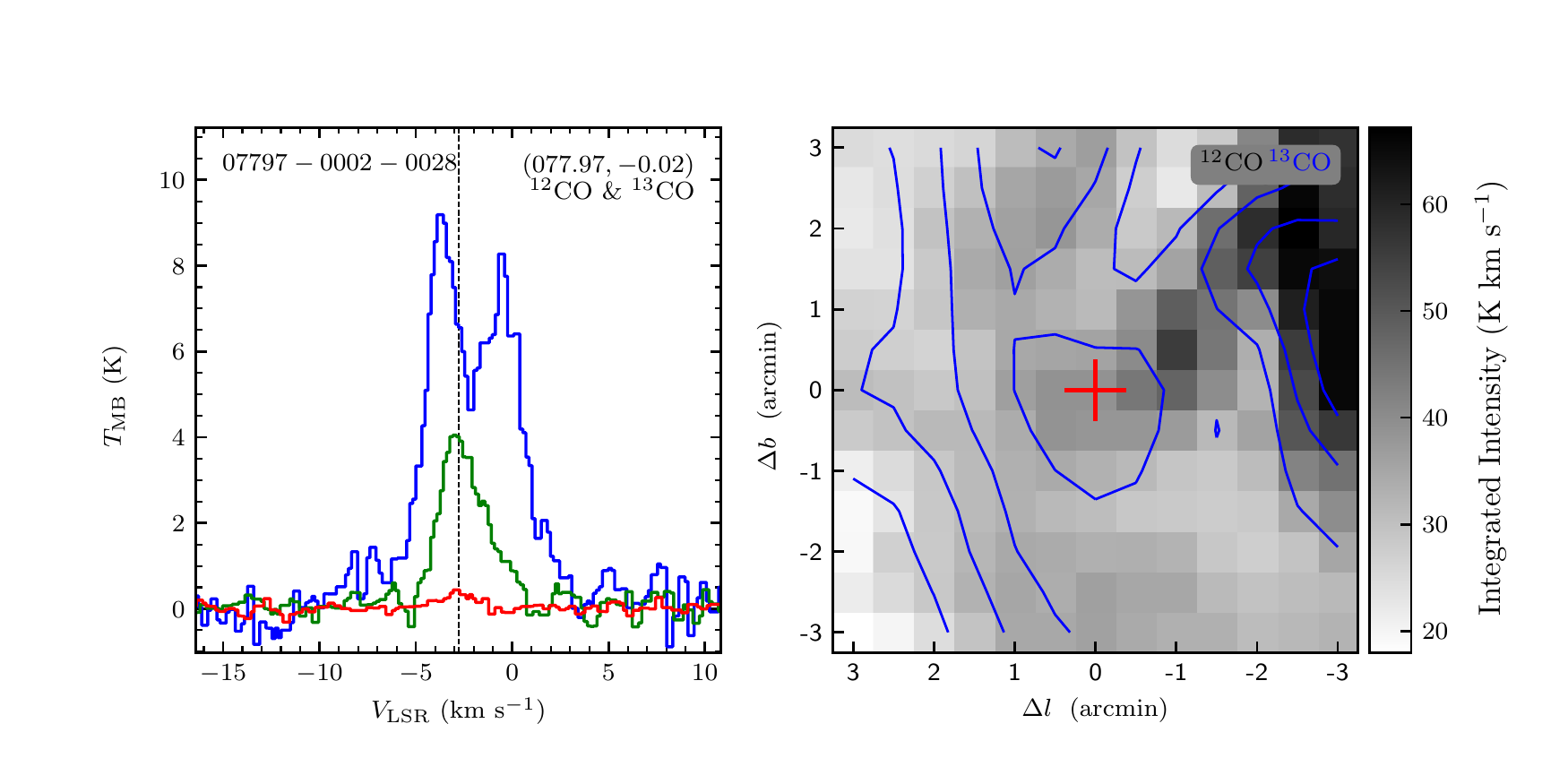}
\vspace{-0.5cm}

\includegraphics[width=9.0cm,angle=0]{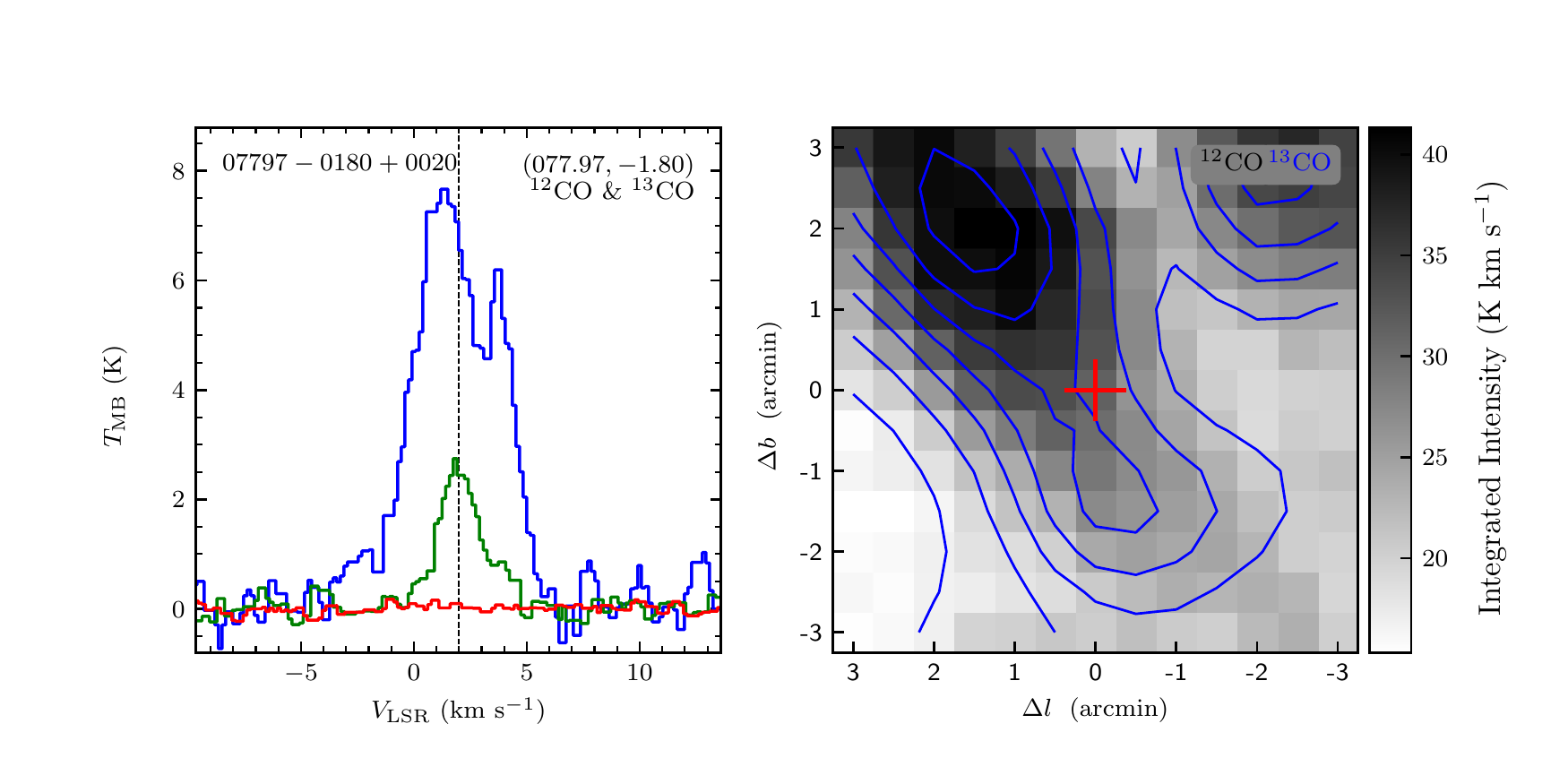}
\includegraphics[width=9.0cm,angle=0]{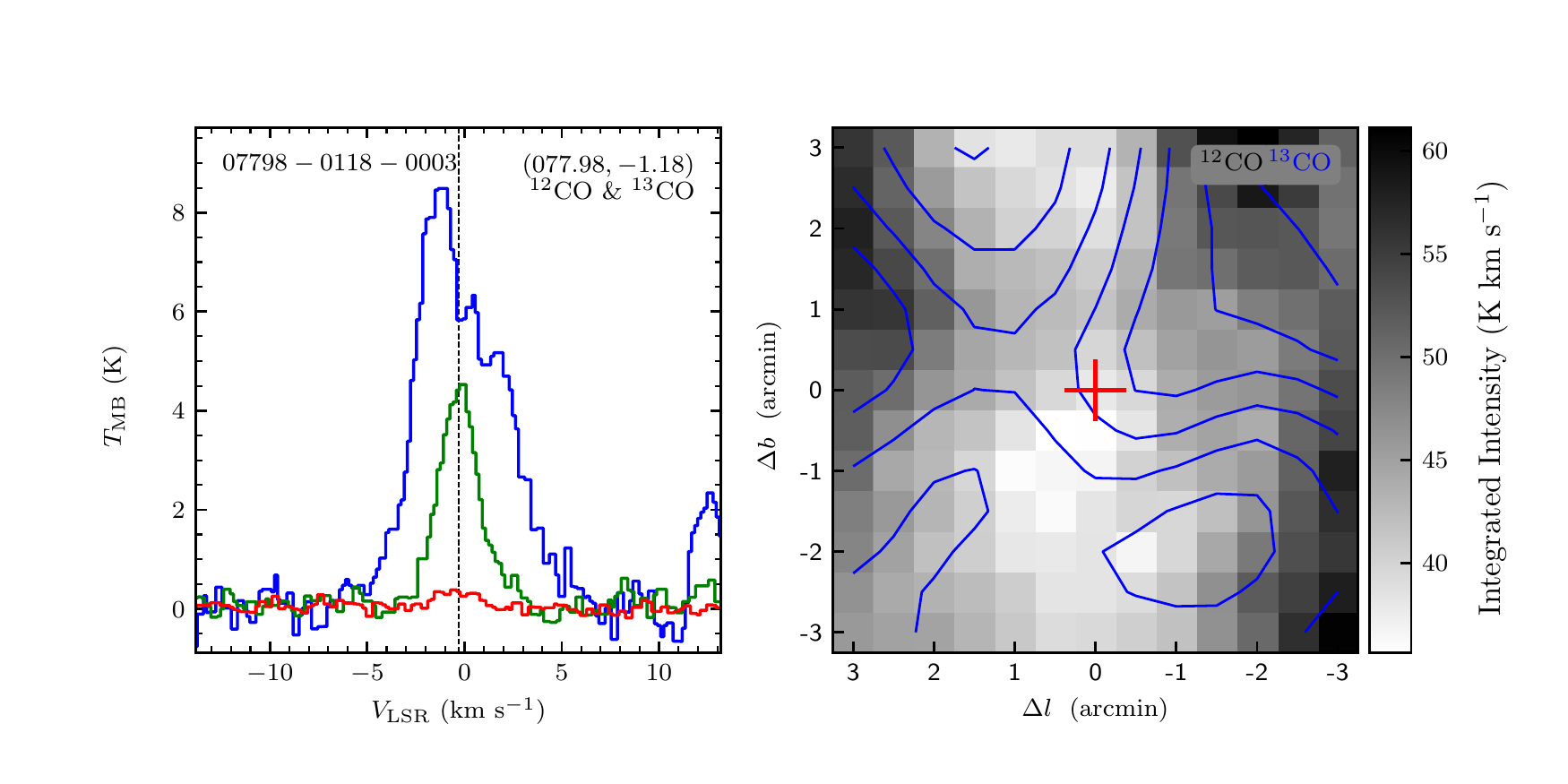}
\vspace{-0.5cm}

\includegraphics[width=9.0cm,angle=0]{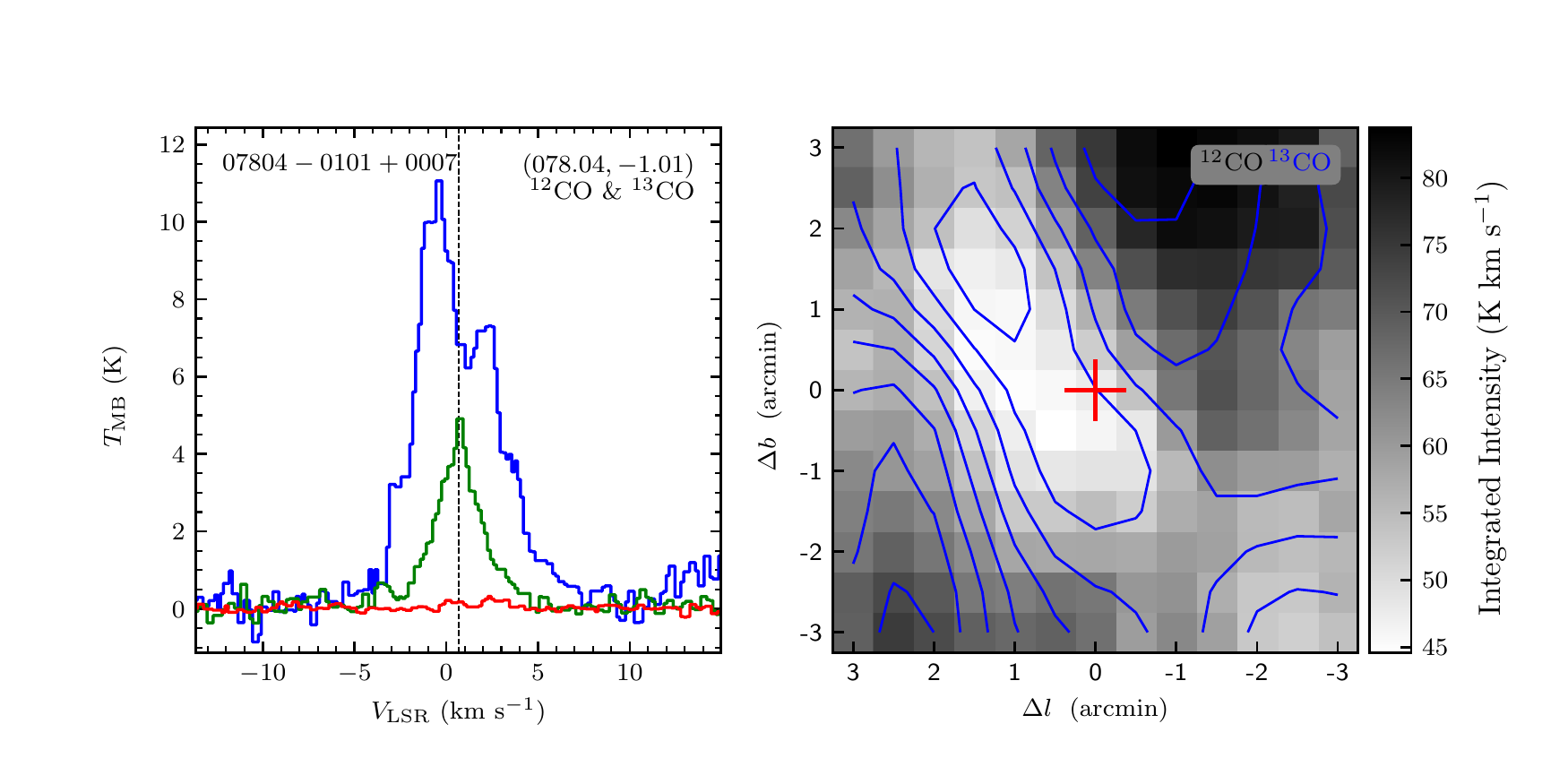}
\includegraphics[width=9.0cm,angle=0]{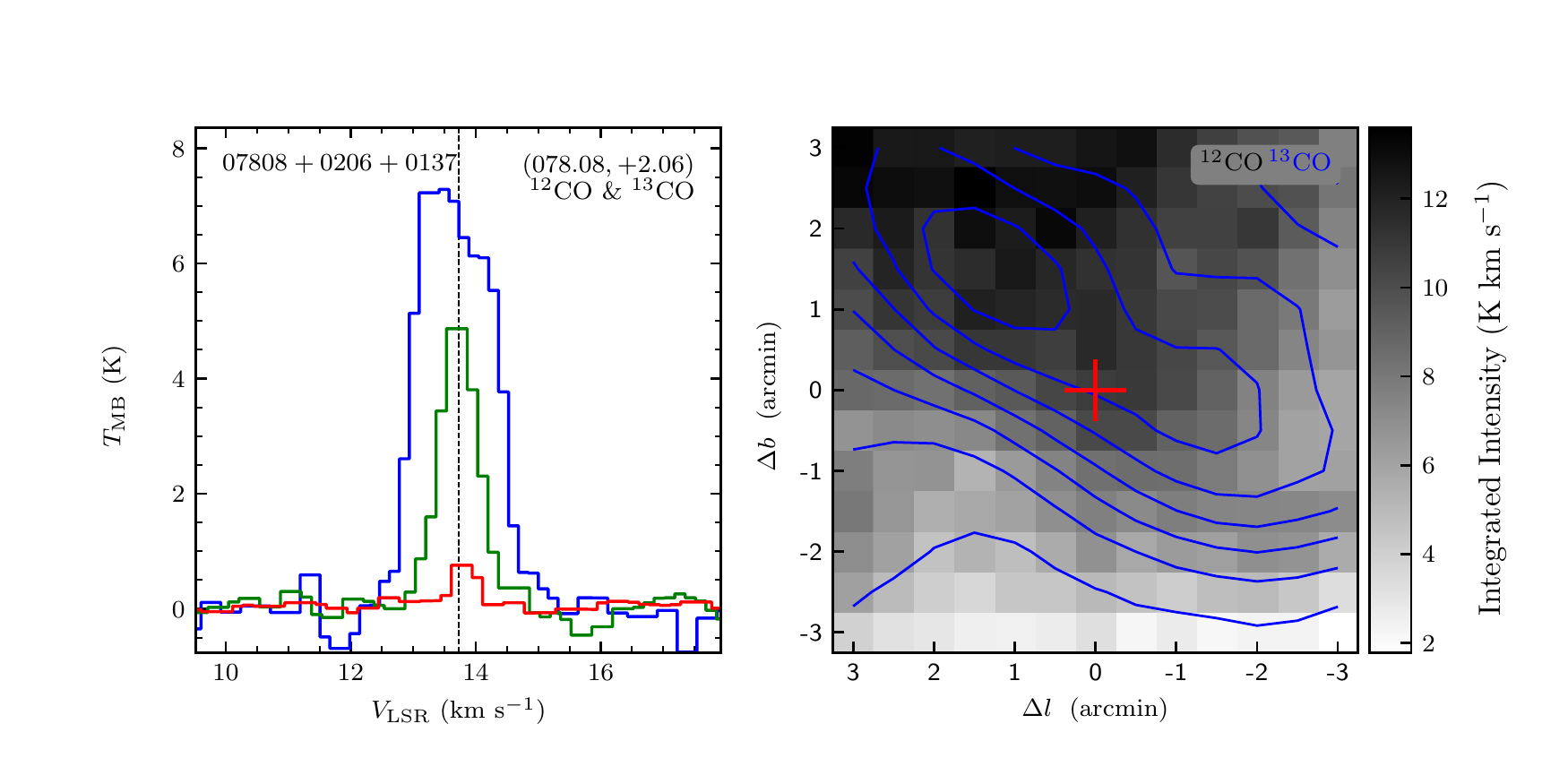}
\vspace{-0.5cm}

\includegraphics[width=9.0cm,angle=0]{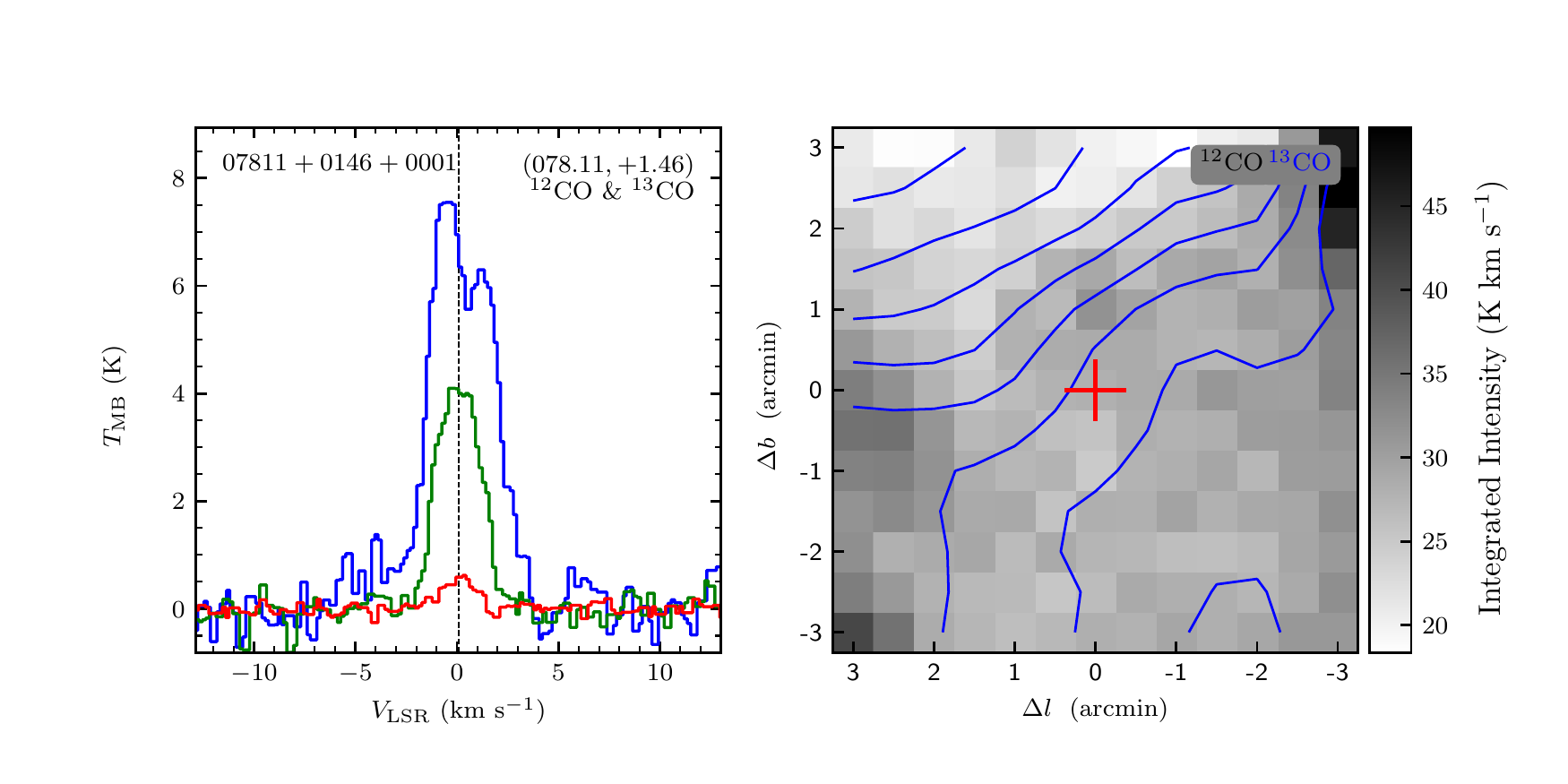}
\includegraphics[width=9.0cm,angle=0]{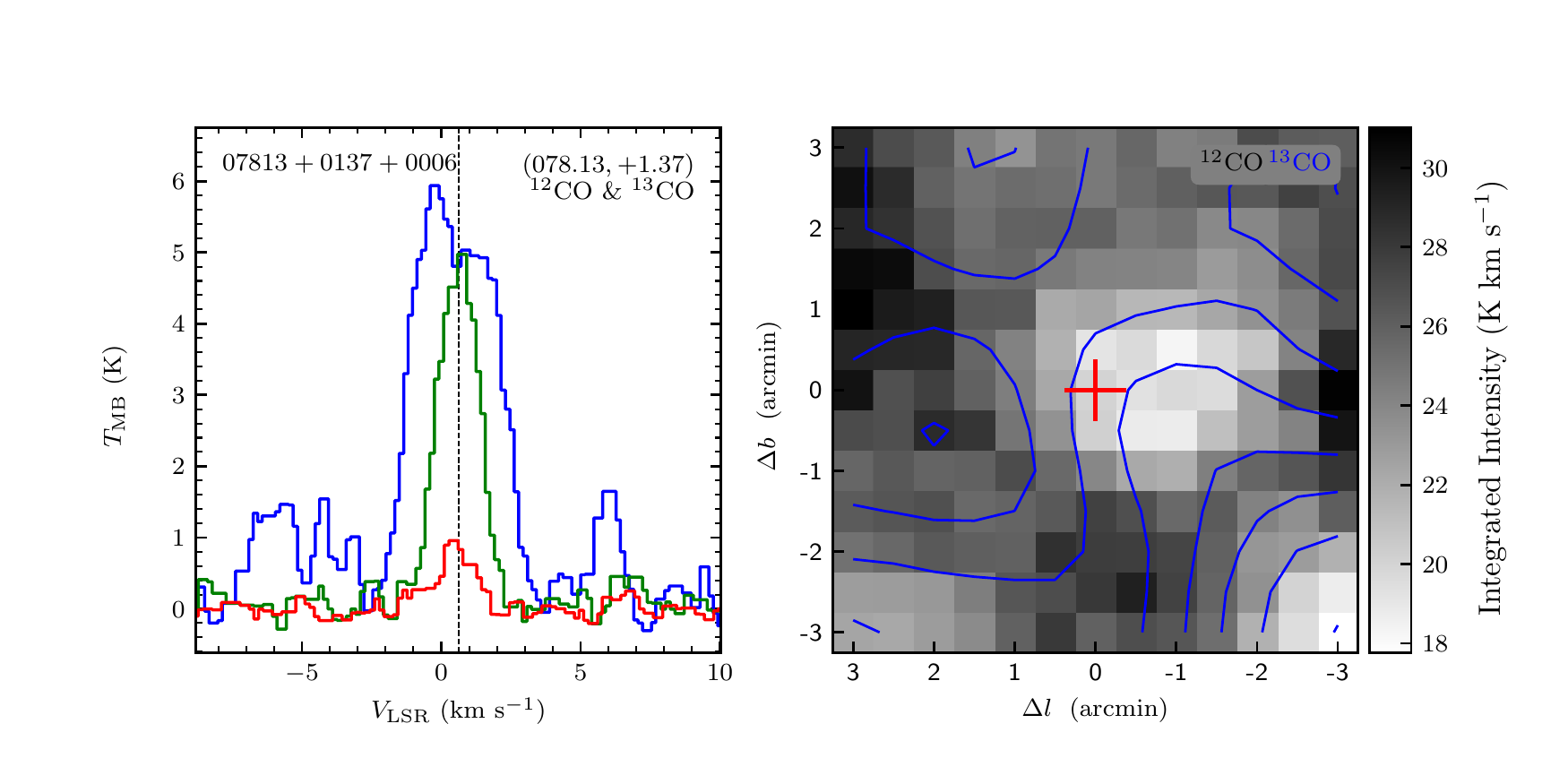}
\end{figure}
\clearpage

\begin{figure}
\includegraphics[width=9.0cm,angle=0]{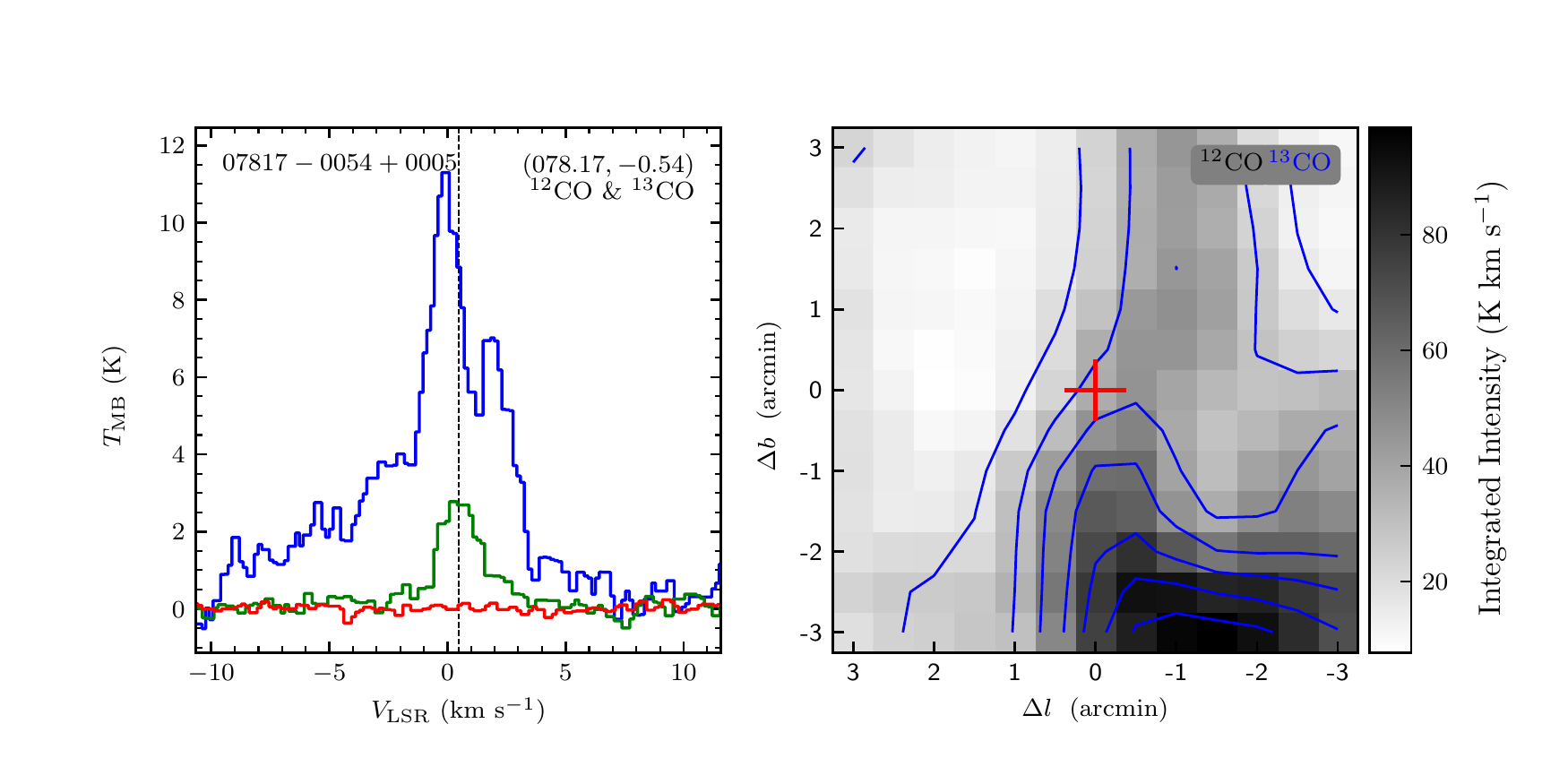}
\includegraphics[width=9.0cm,angle=0]{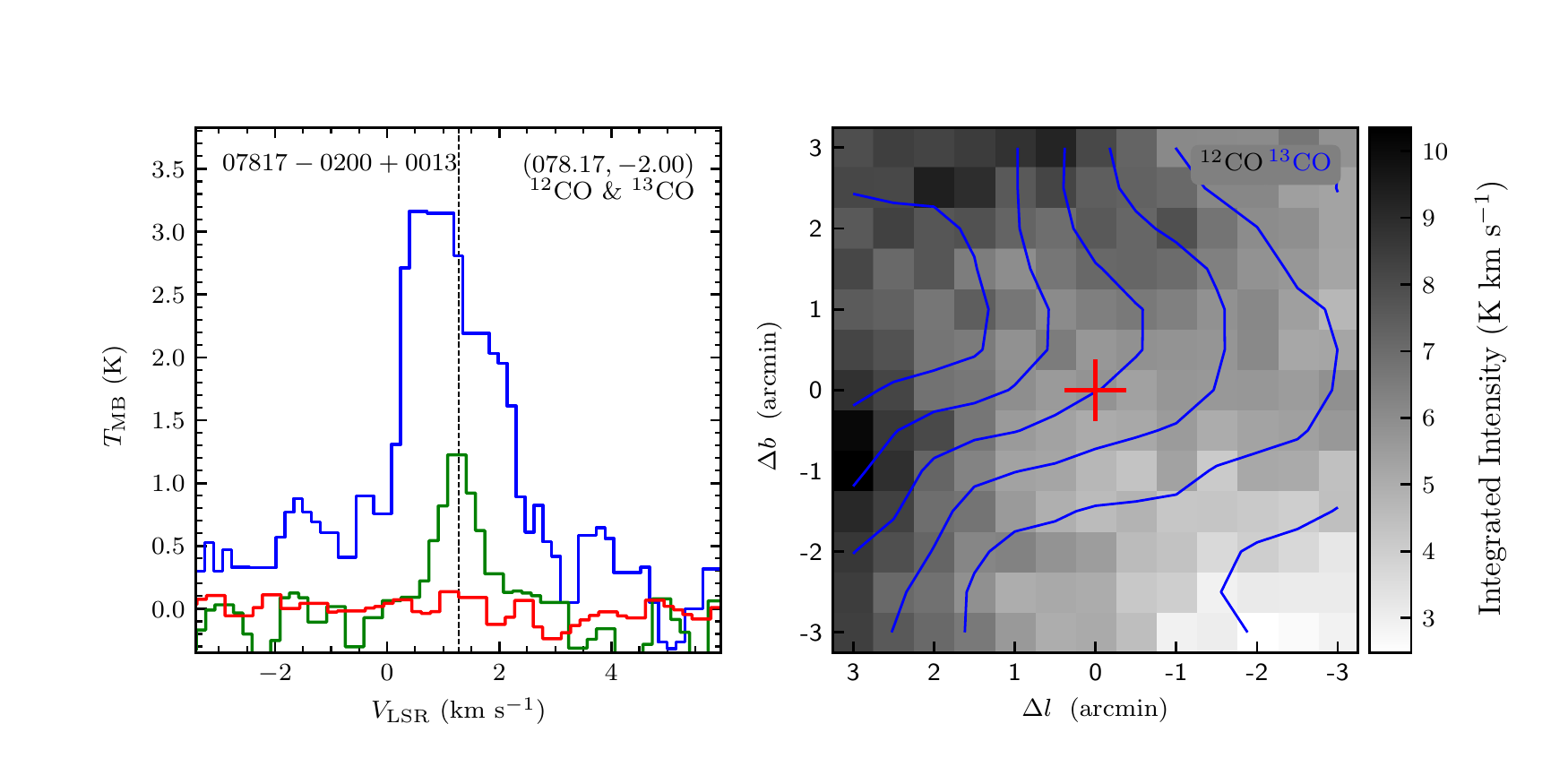}
\vspace{-0.5cm}

\includegraphics[width=9.0cm,angle=0]{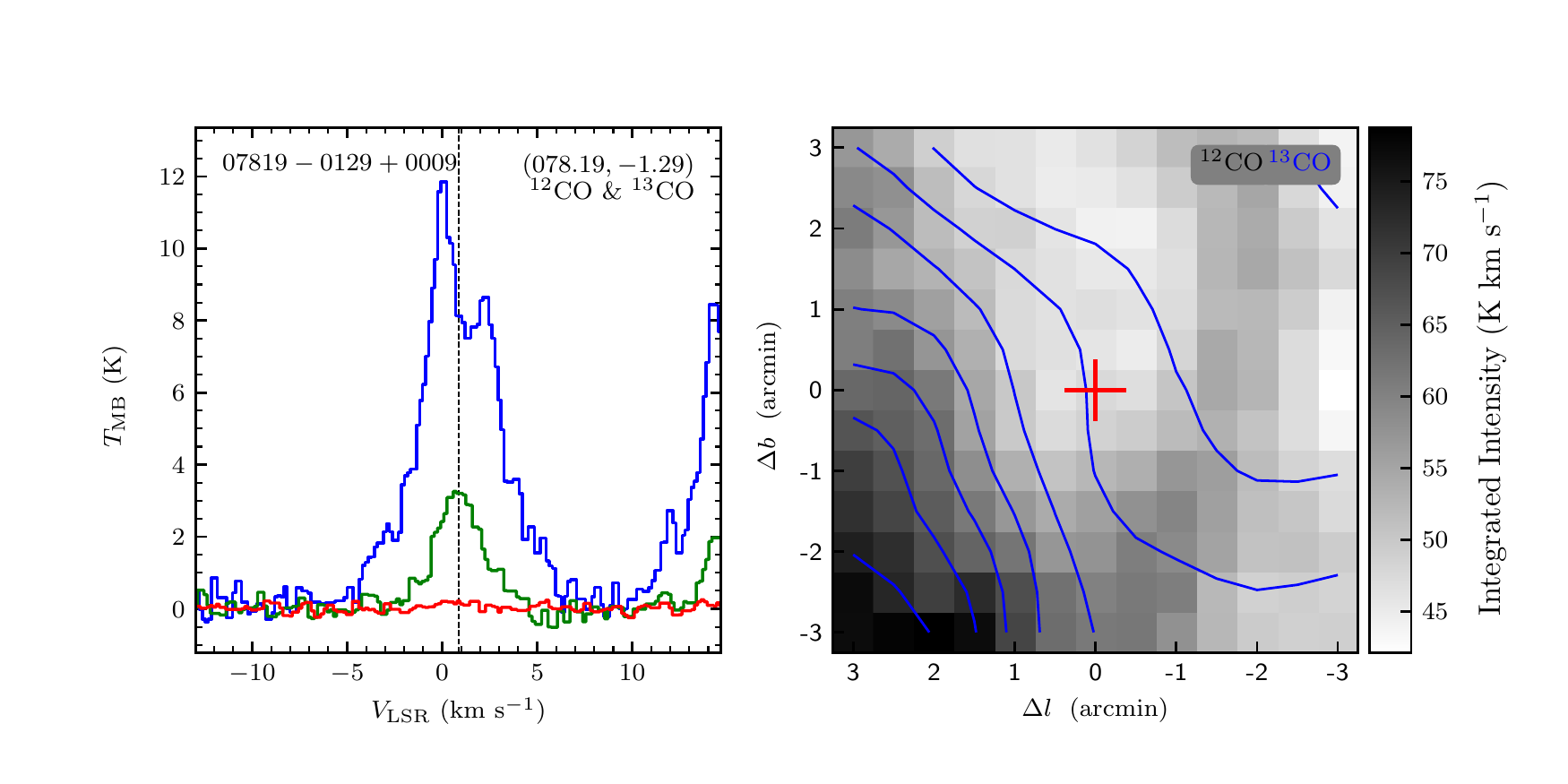}
\includegraphics[width=9.0cm,angle=0]{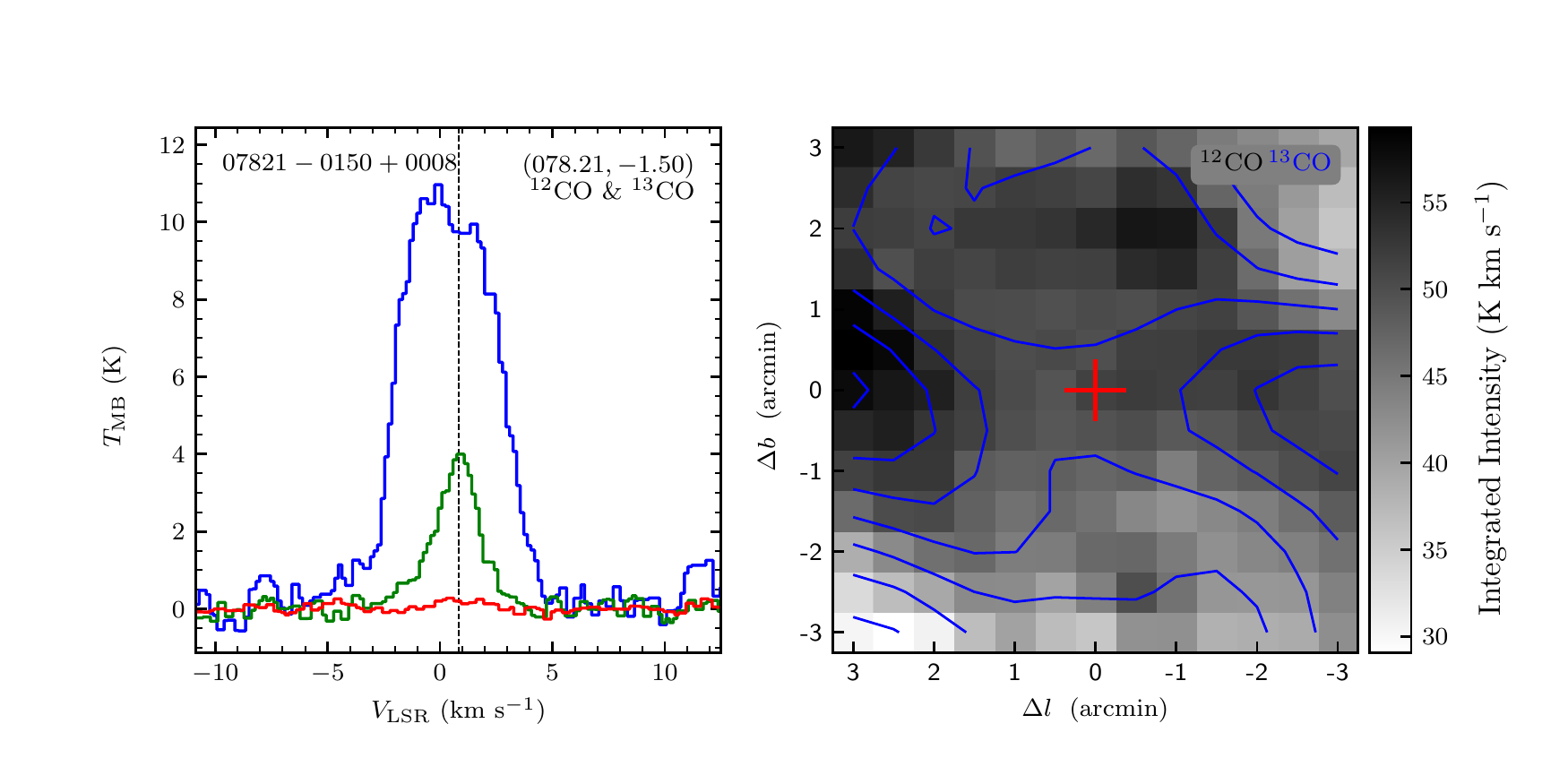}
\vspace{-0.5cm}

\includegraphics[width=9.0cm,angle=0]{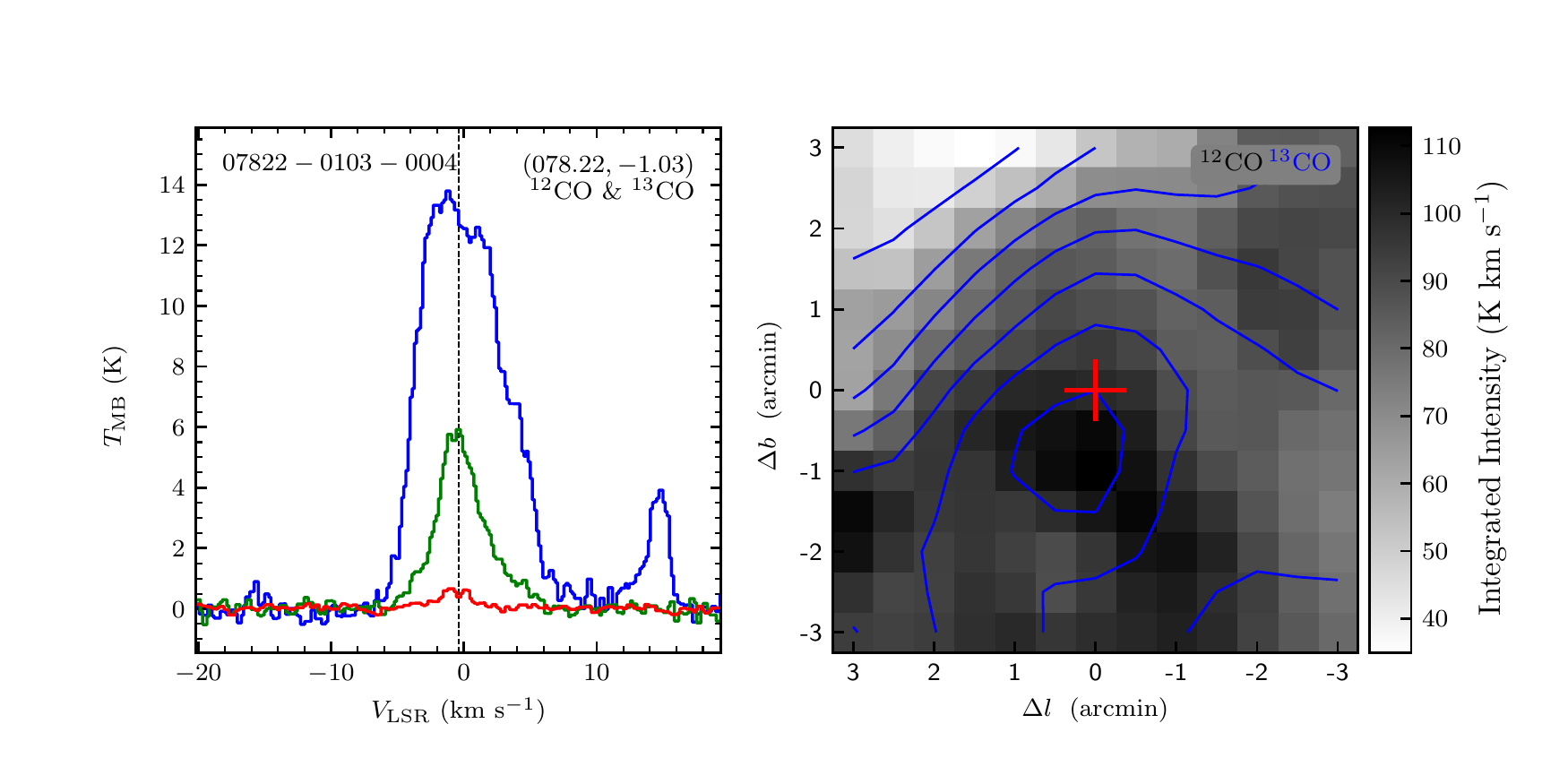}
\includegraphics[width=9.0cm,angle=0]{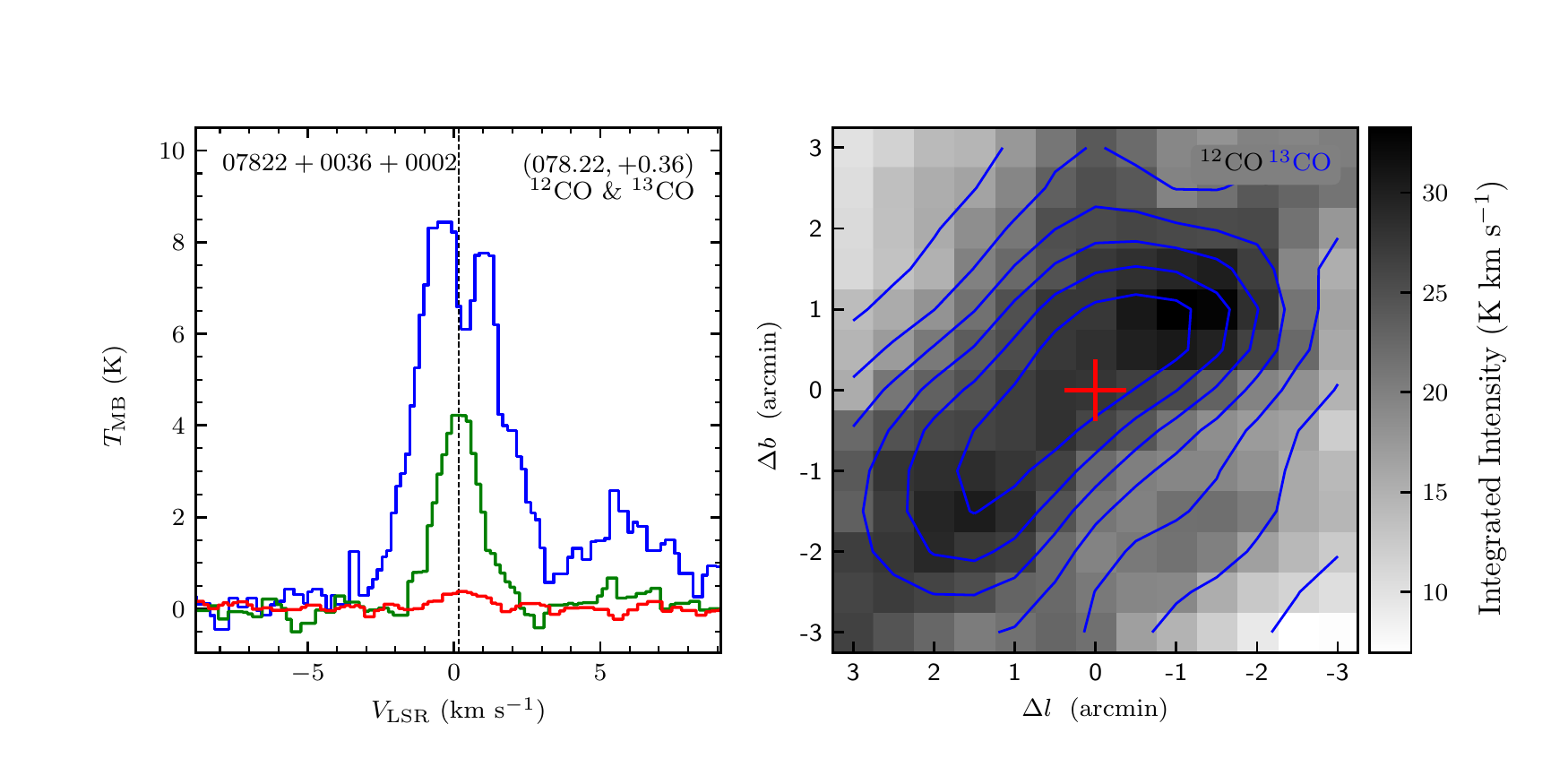}
\vspace{-0.5cm}

\includegraphics[width=9.0cm,angle=0]{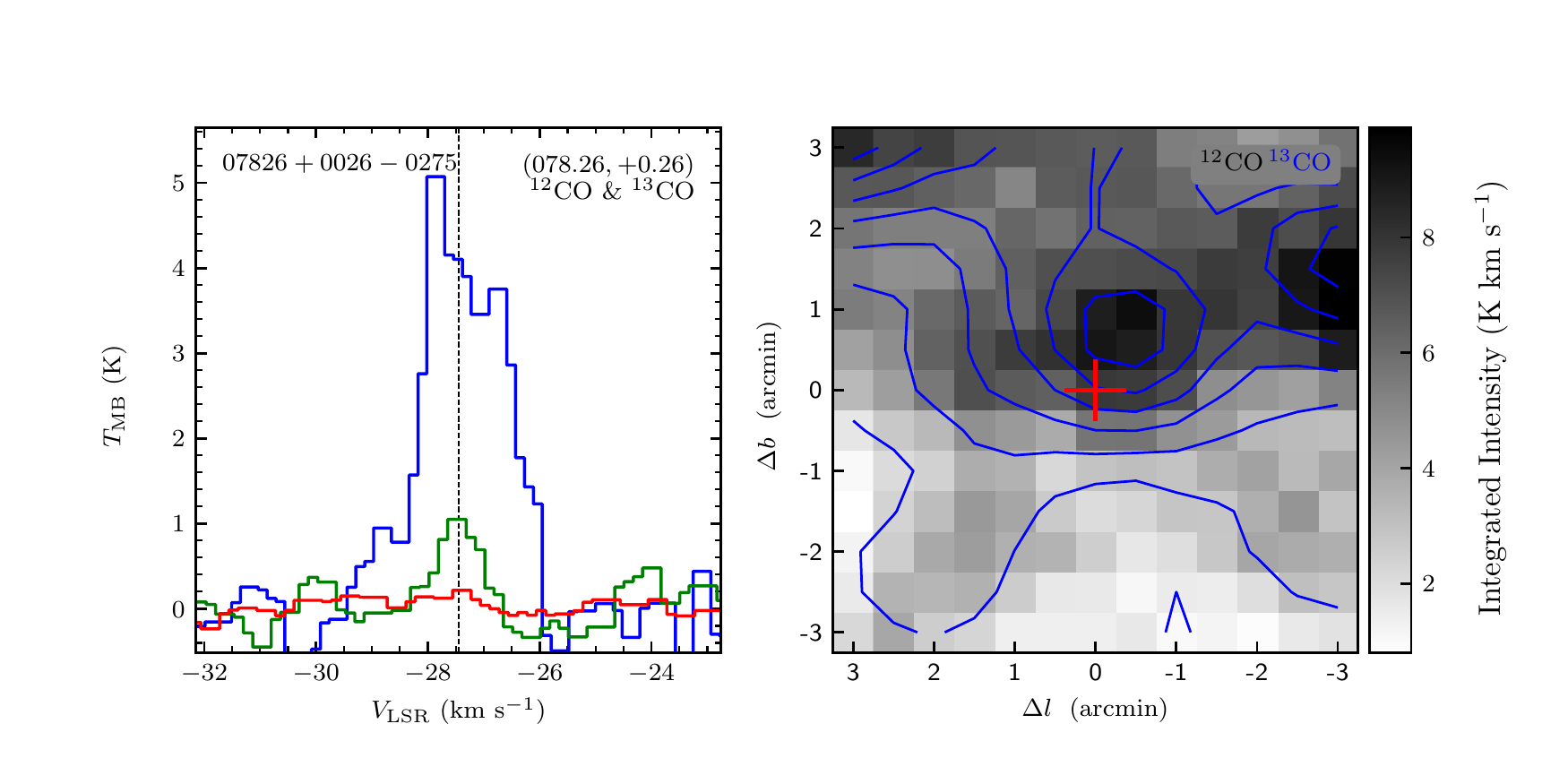}
\includegraphics[width=9.0cm,angle=0]{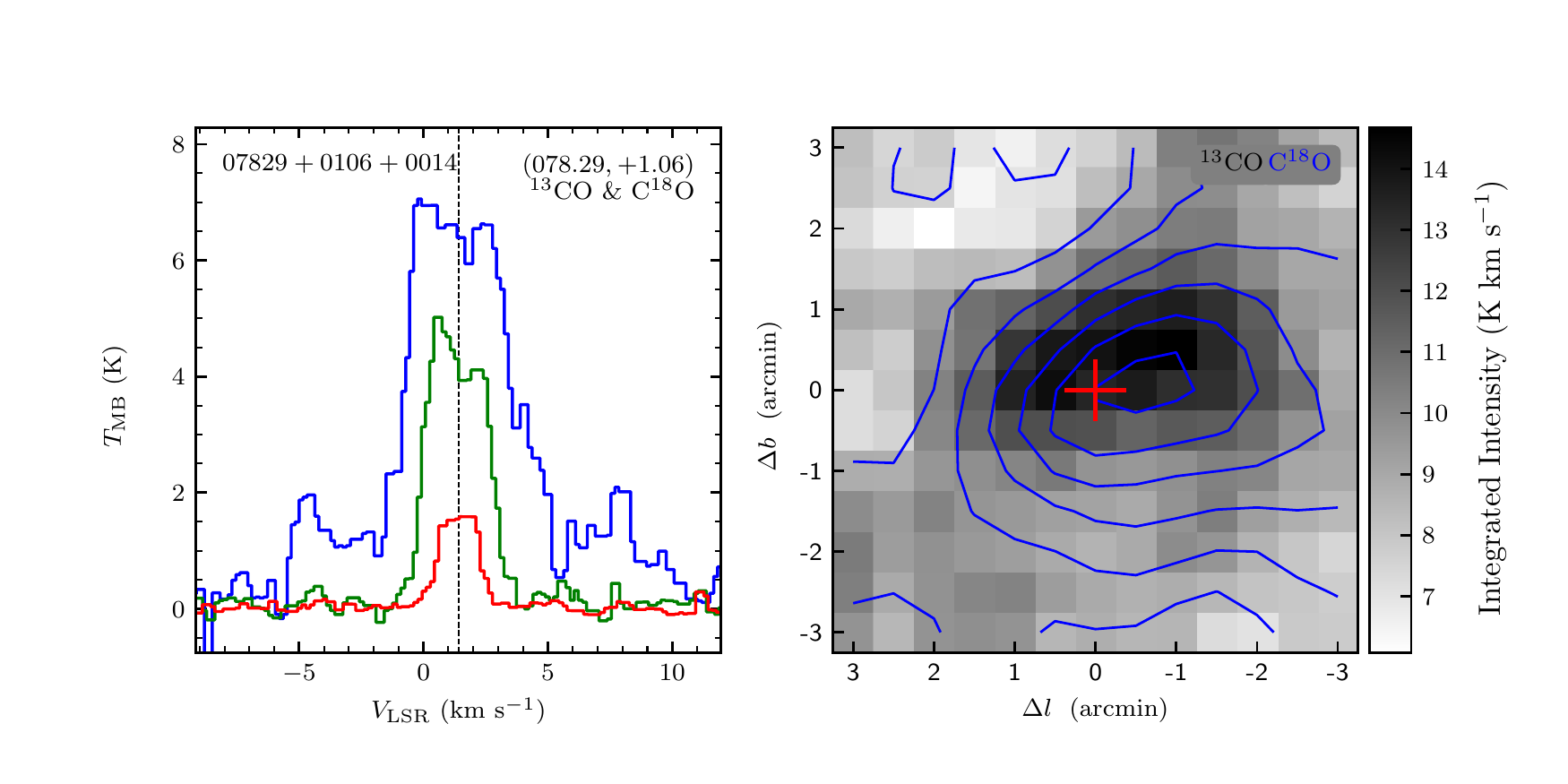}
\vspace{-0.5cm}

\includegraphics[width=9.0cm,angle=0]{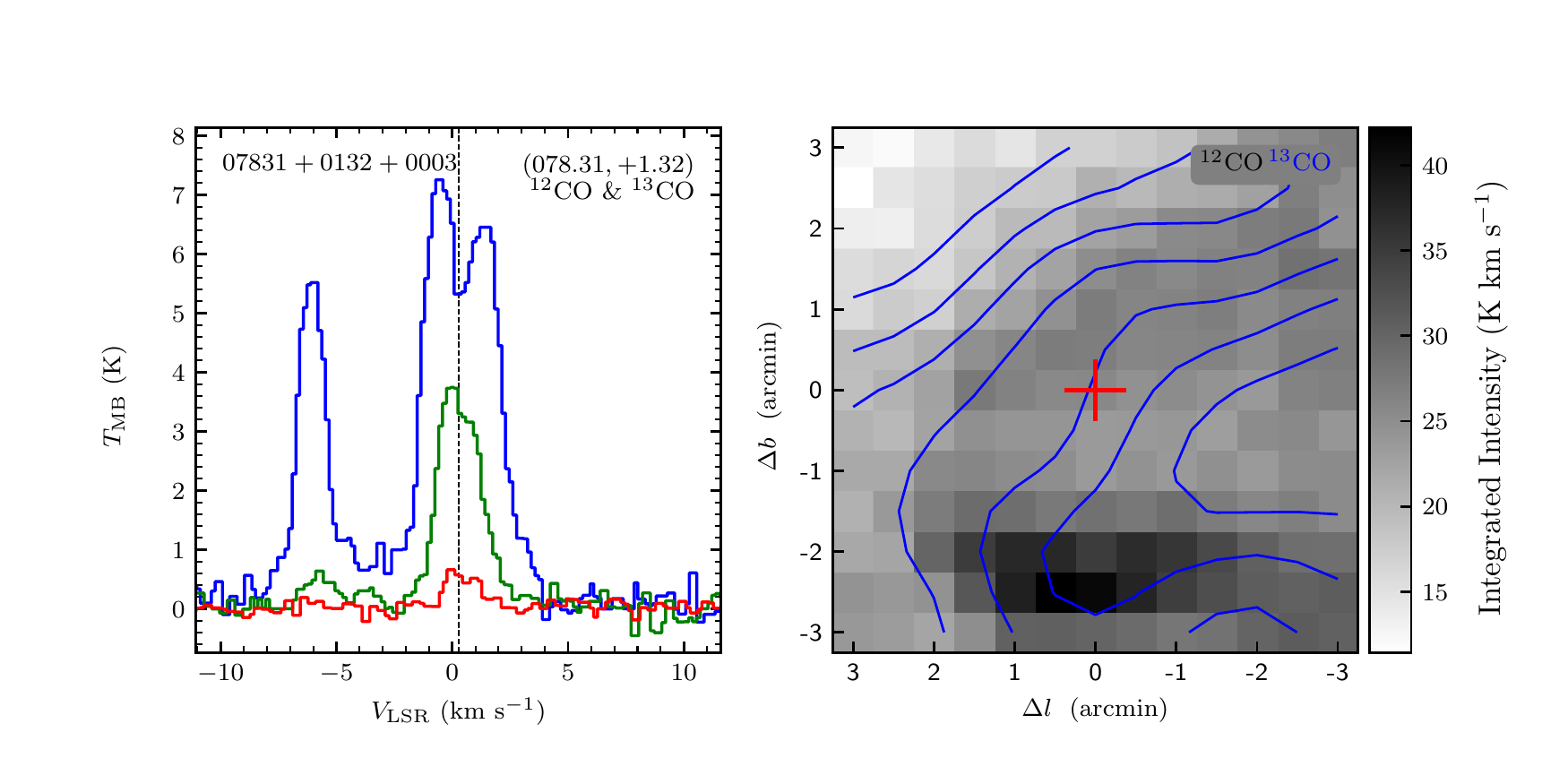}
\includegraphics[width=9.0cm,angle=0]{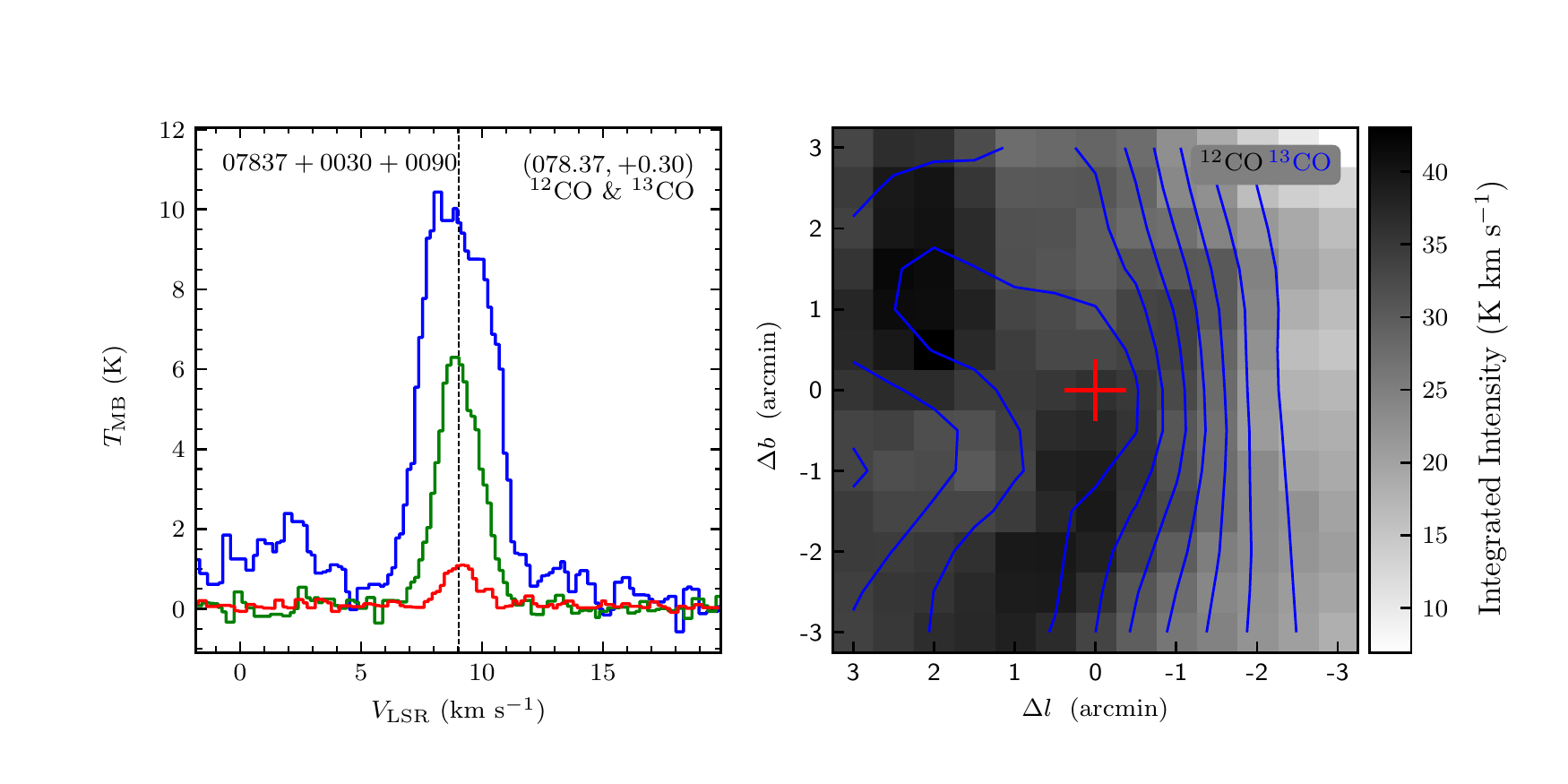}
\end{figure}
\clearpage

\begin{figure}
\includegraphics[width=9.0cm,angle=0]{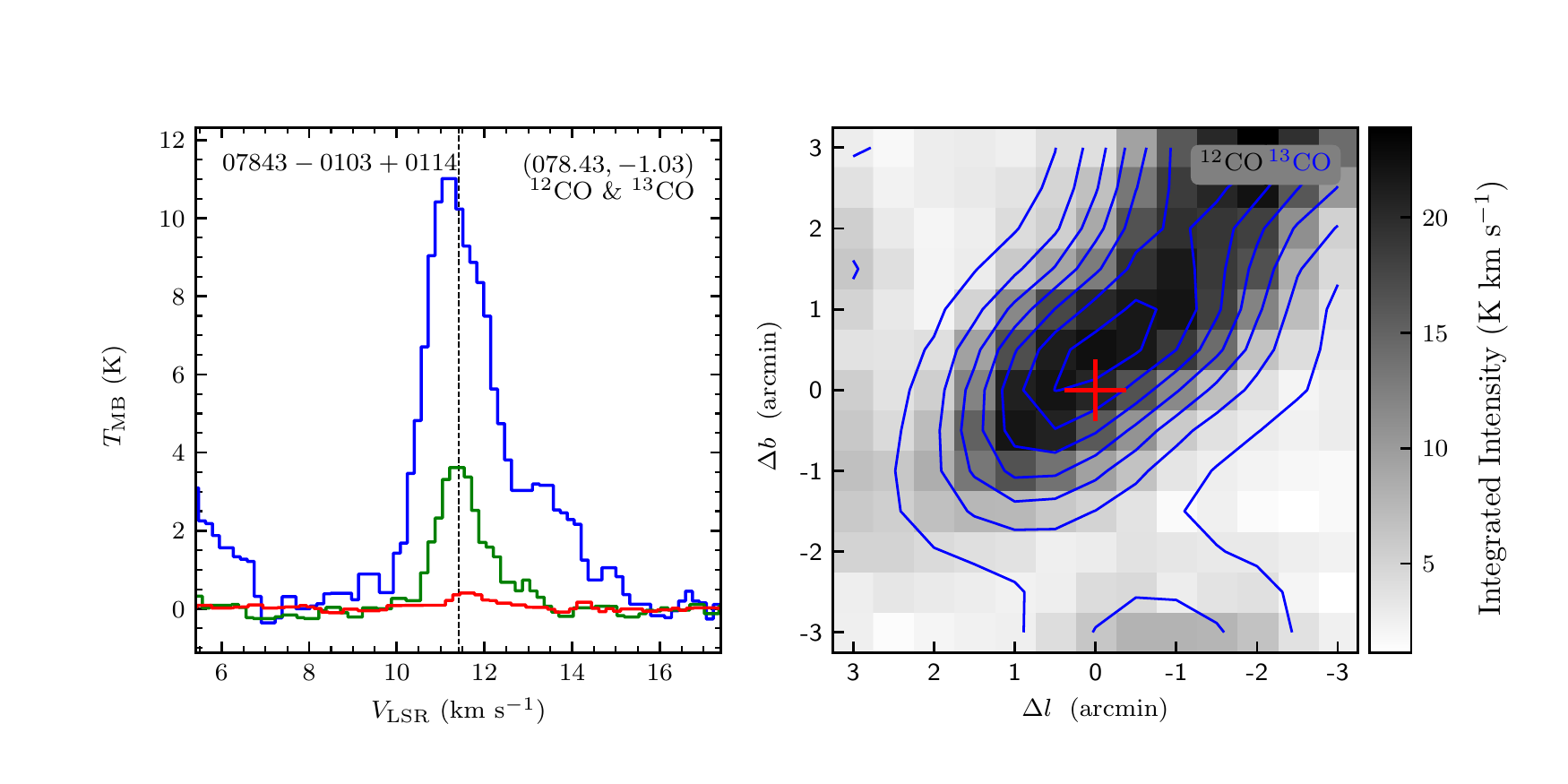}
\includegraphics[width=9.0cm,angle=0]{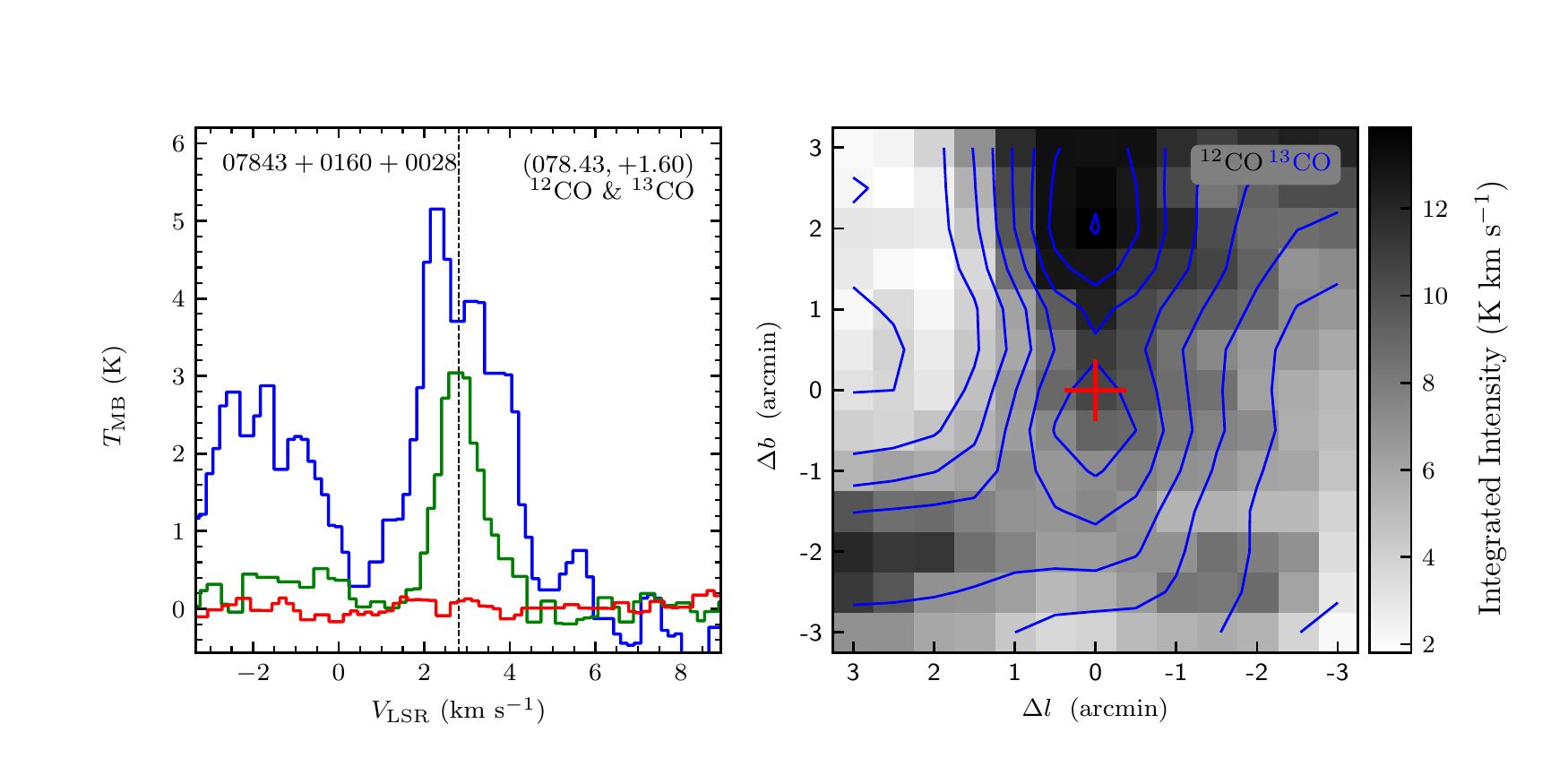}
\vspace{-0.5cm}

\includegraphics[width=9.0cm,angle=0]{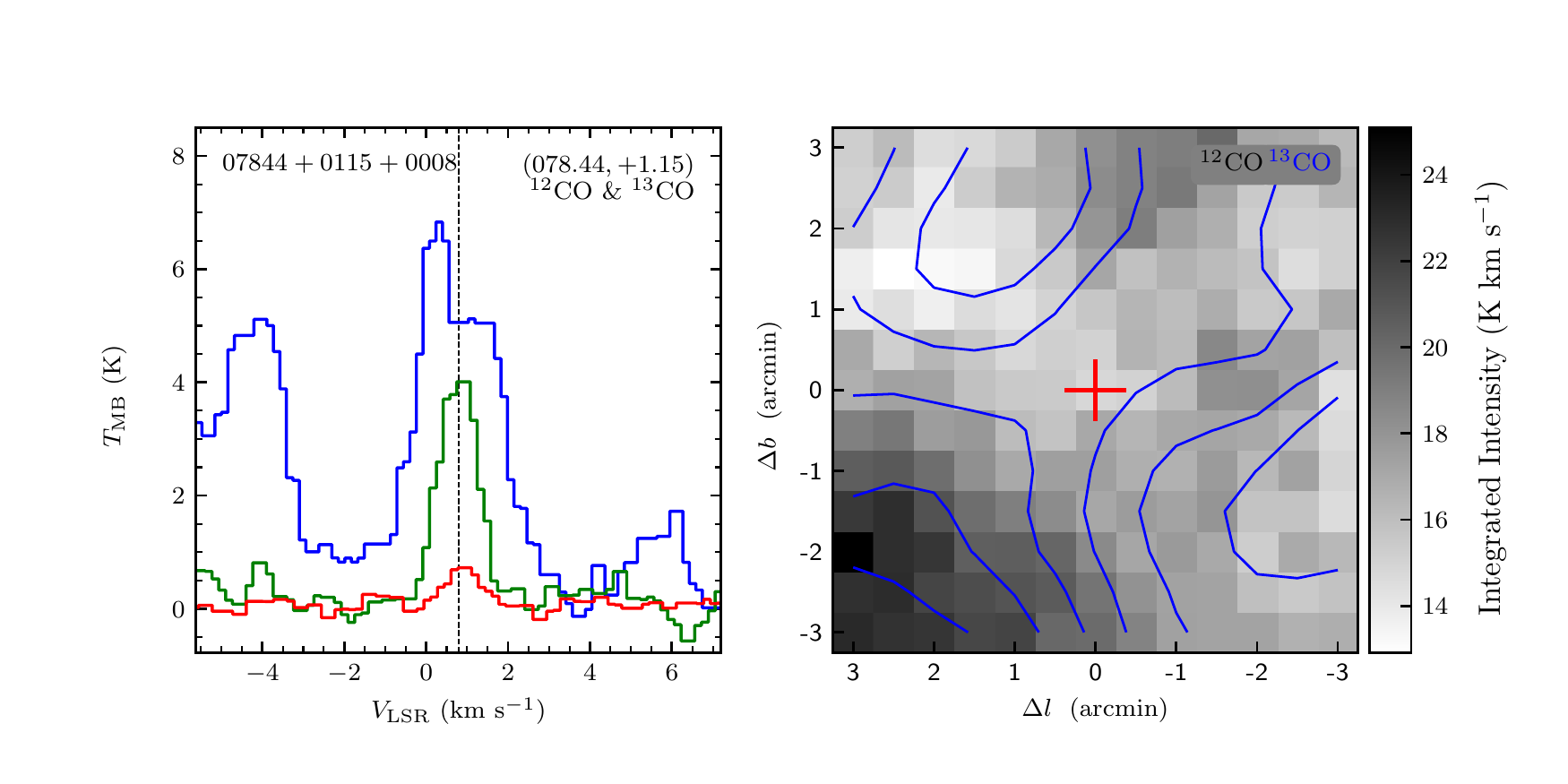}
\includegraphics[width=9.0cm,angle=0]{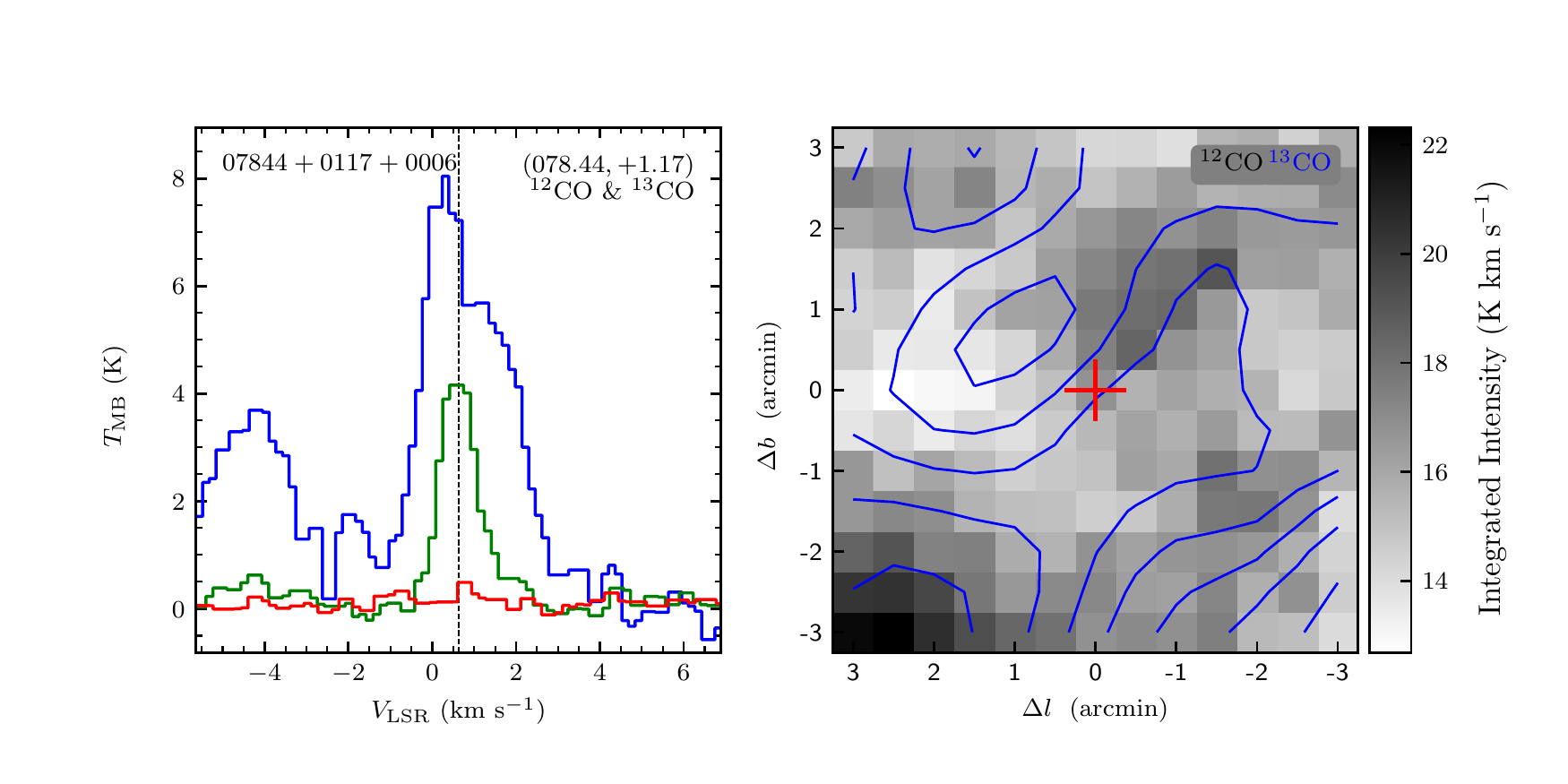}
\vspace{-0.5cm}

\includegraphics[width=9.0cm,angle=0]{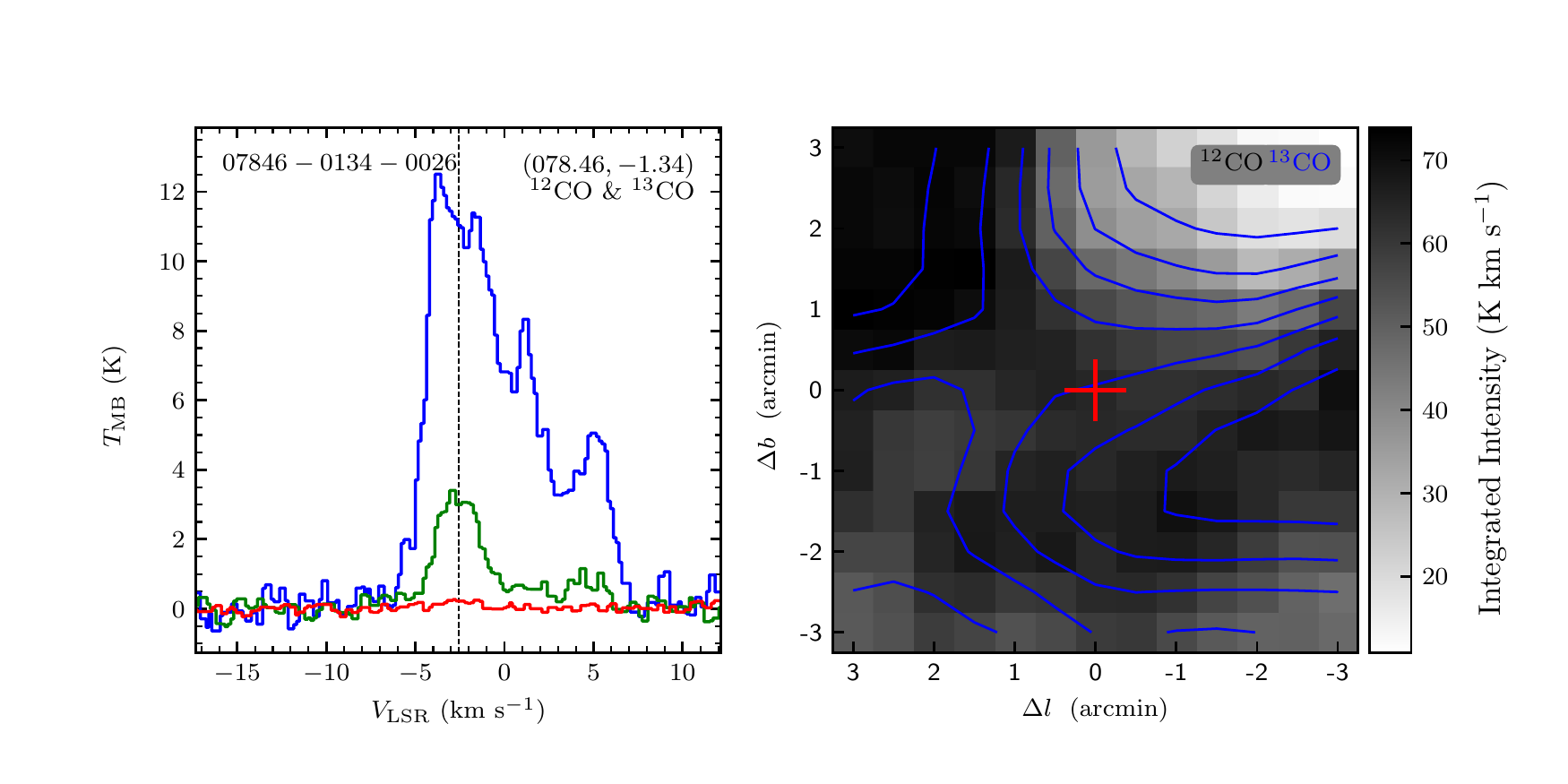}
\includegraphics[width=9.0cm,angle=0]{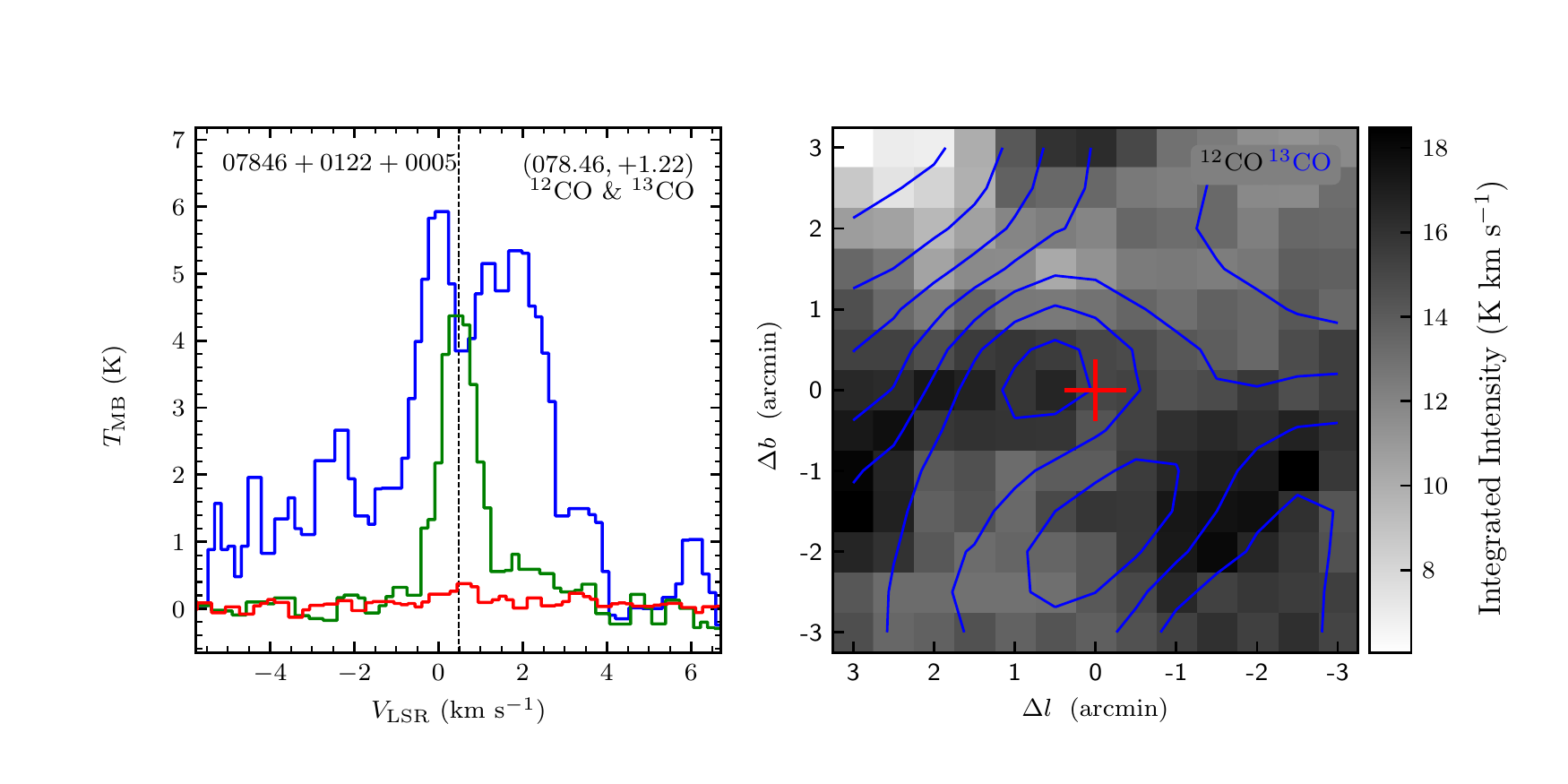}
\vspace{-0.5cm}

\includegraphics[width=9.0cm,angle=0]{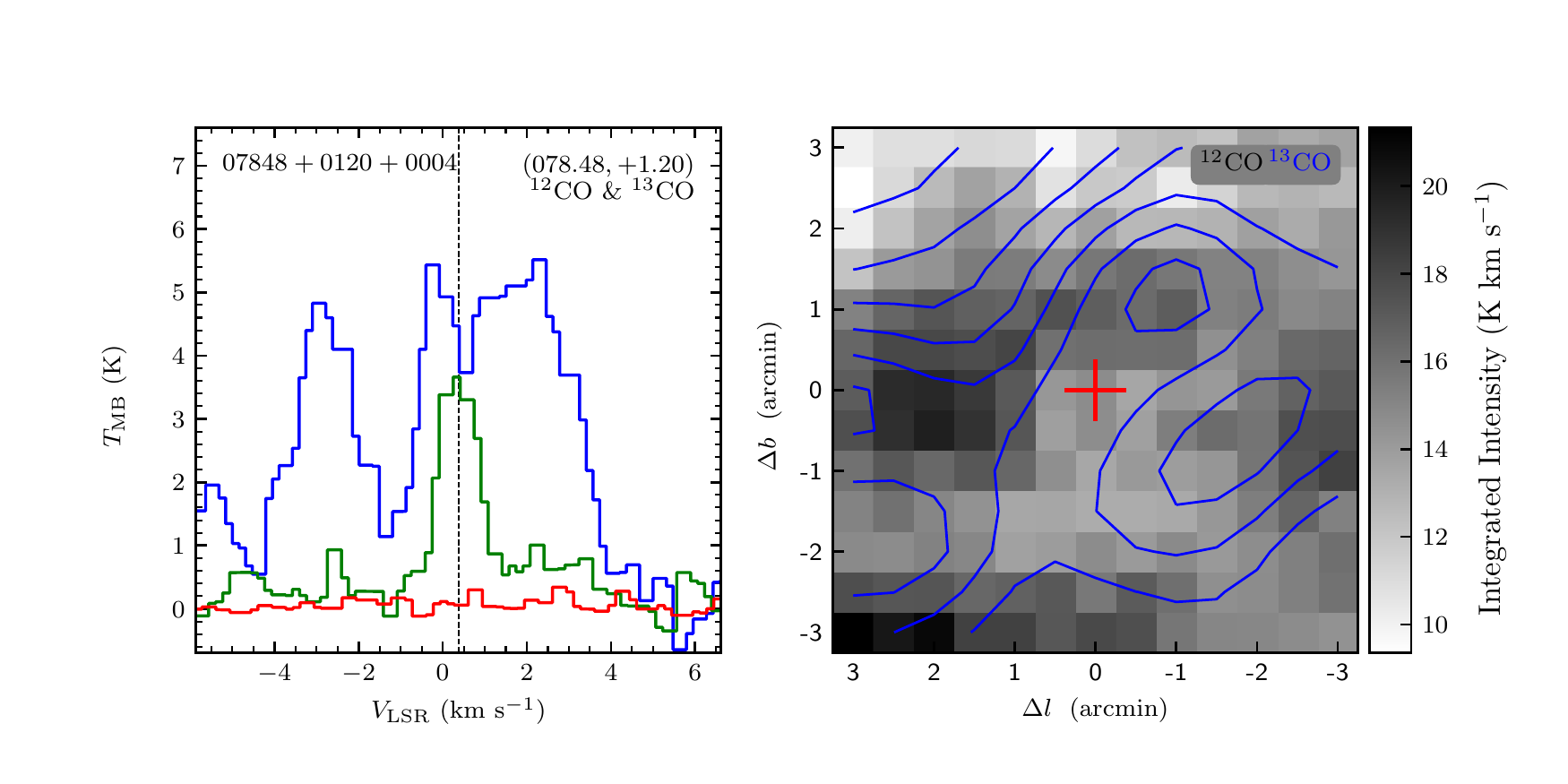}
\includegraphics[width=9.0cm,angle=0]{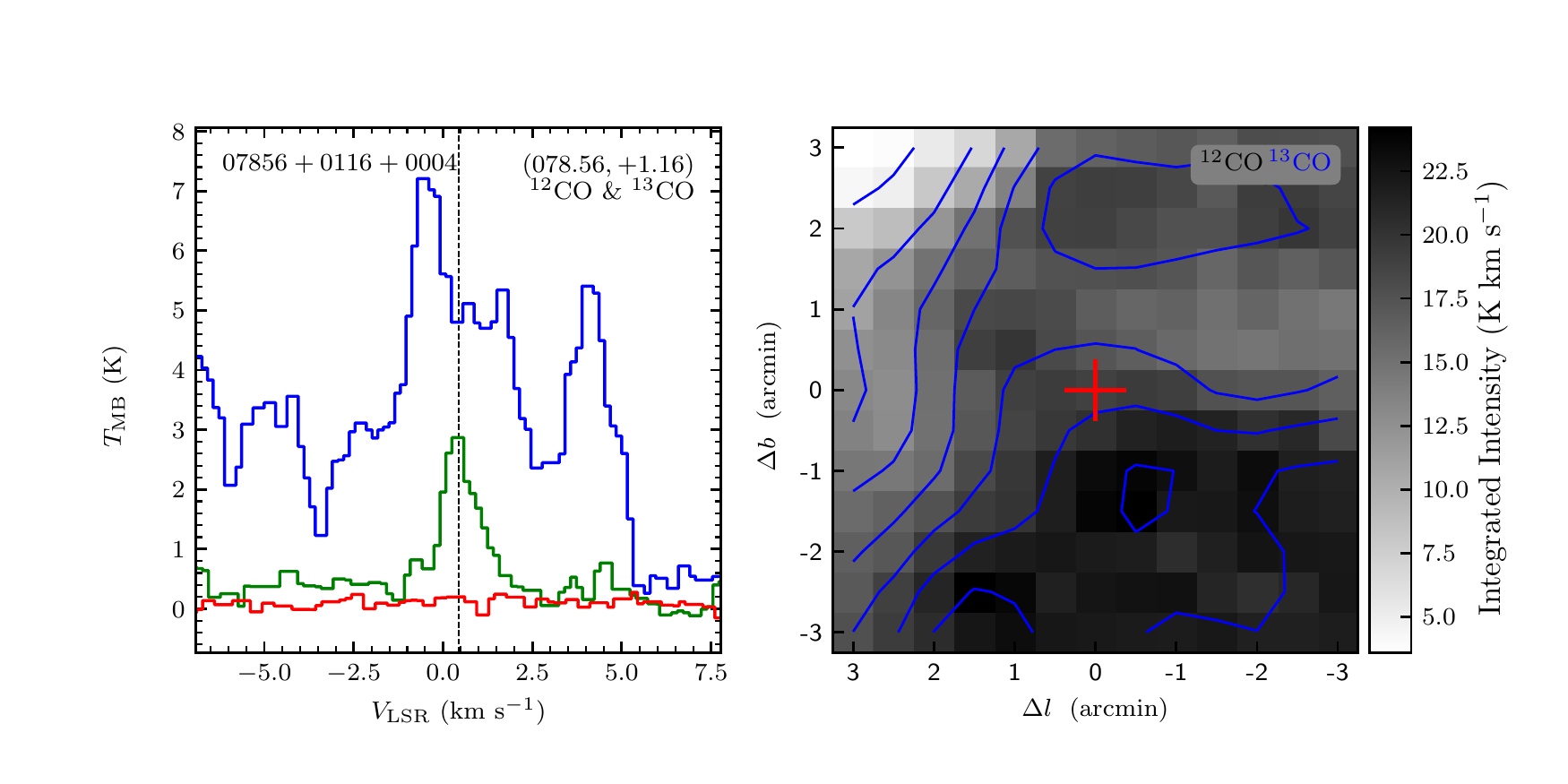}
\vspace{-0.5cm}

\includegraphics[width=9.0cm,angle=0]{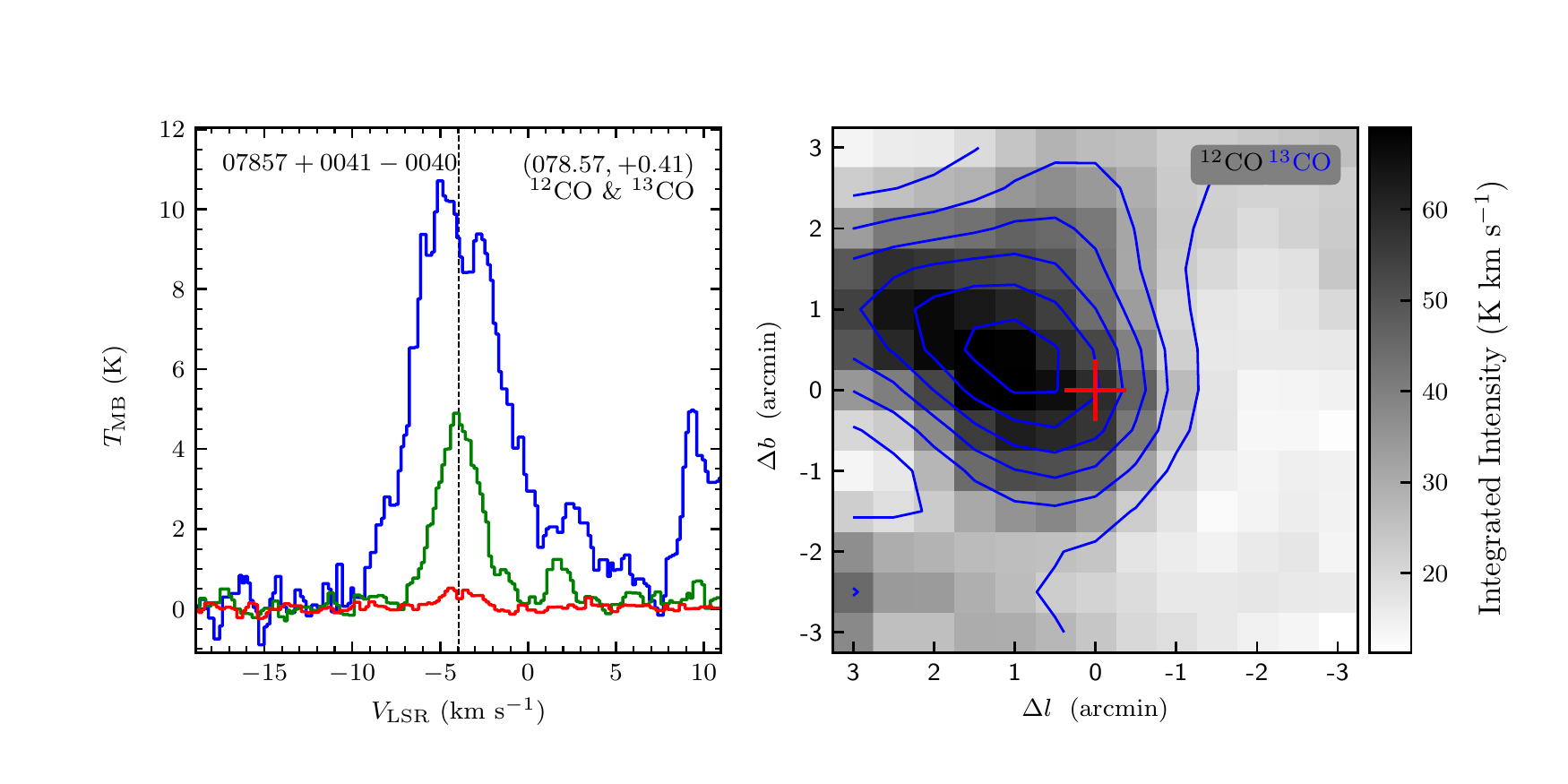}
\includegraphics[width=9.0cm,angle=0]{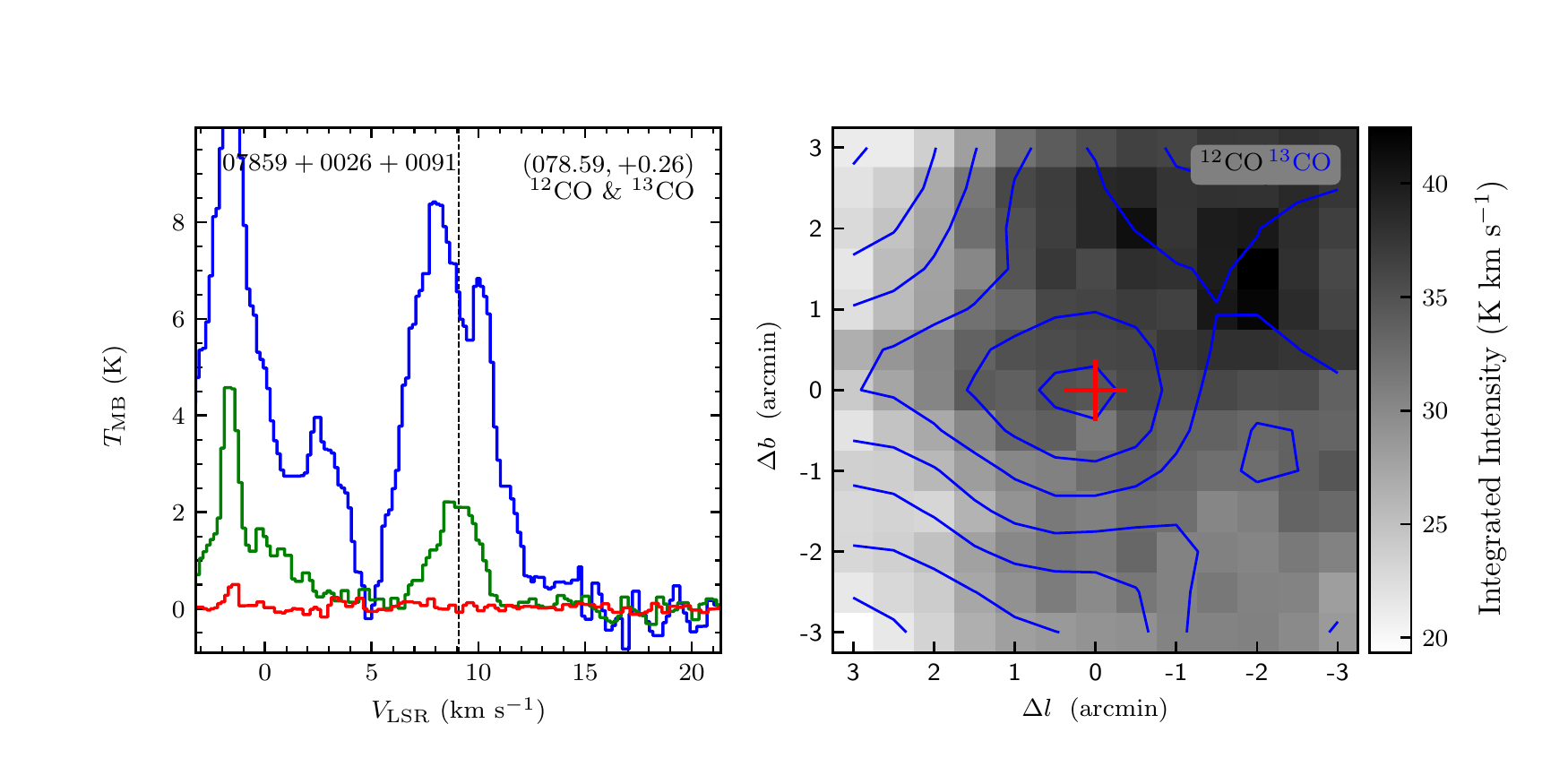}
\end{figure}
\clearpage

\begin{figure}
\includegraphics[width=9.0cm,angle=0]{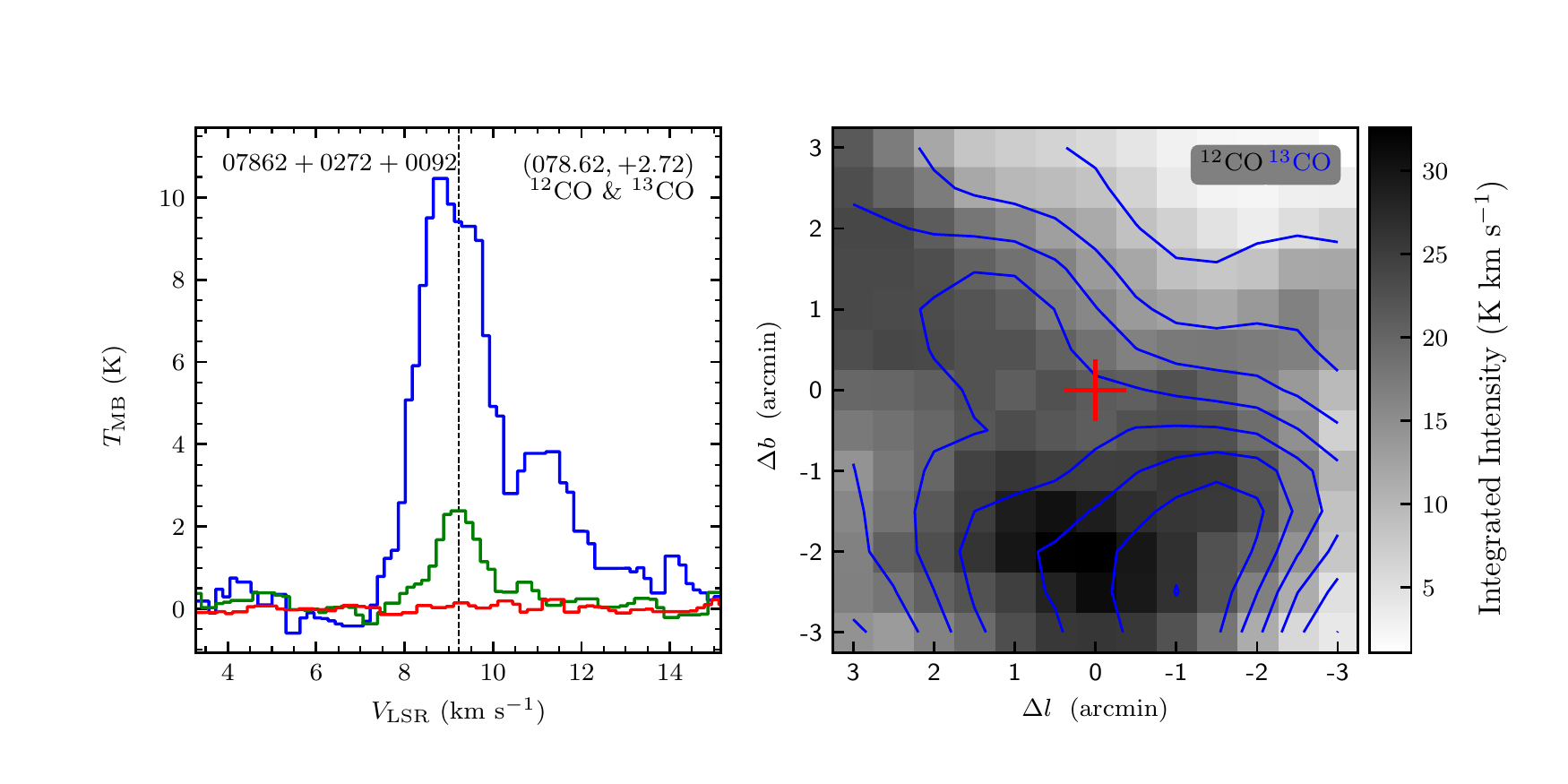}
\includegraphics[width=9.0cm,angle=0]{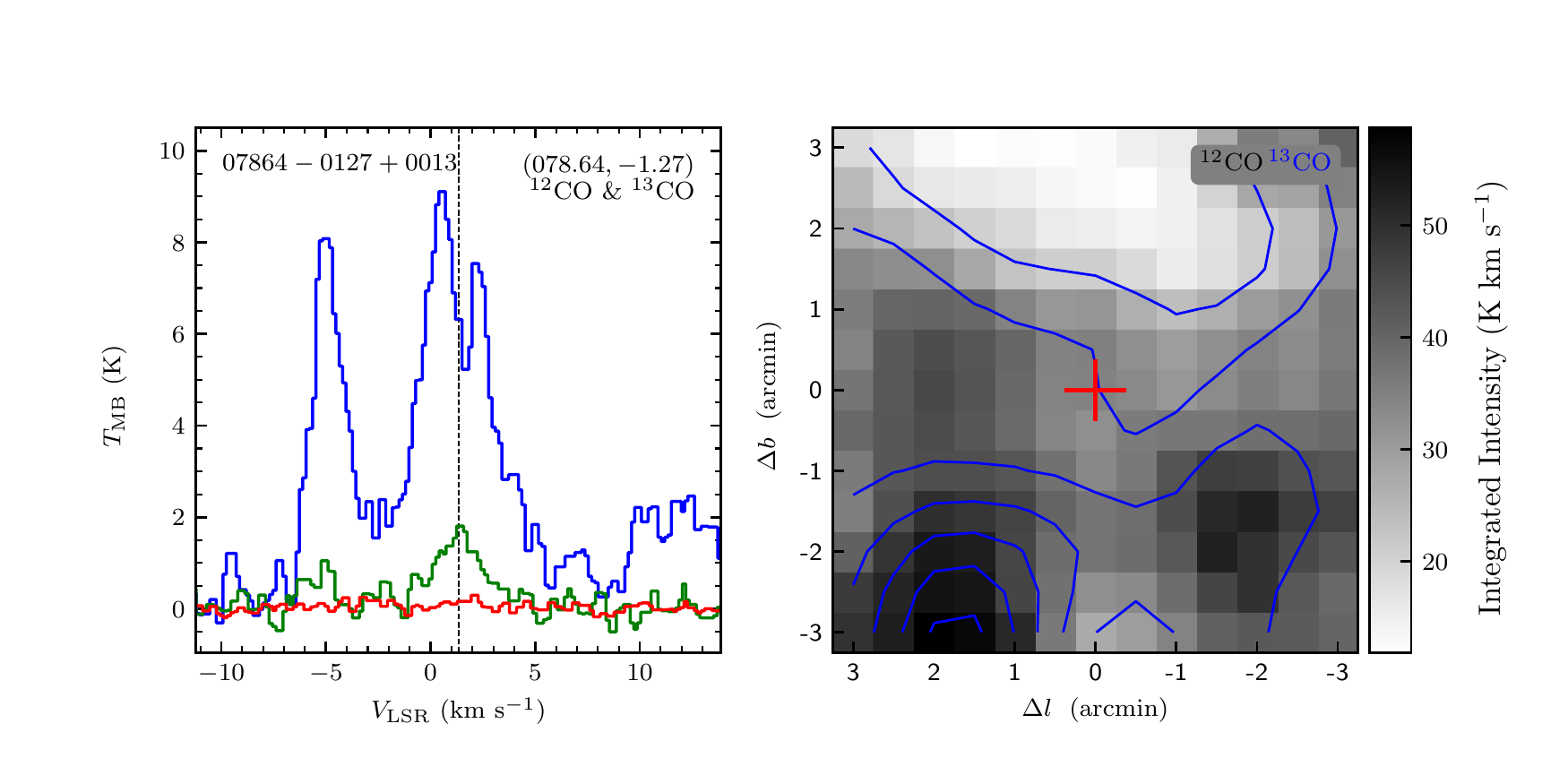}
\vspace{-0.5cm}

\includegraphics[width=9.0cm,angle=0]{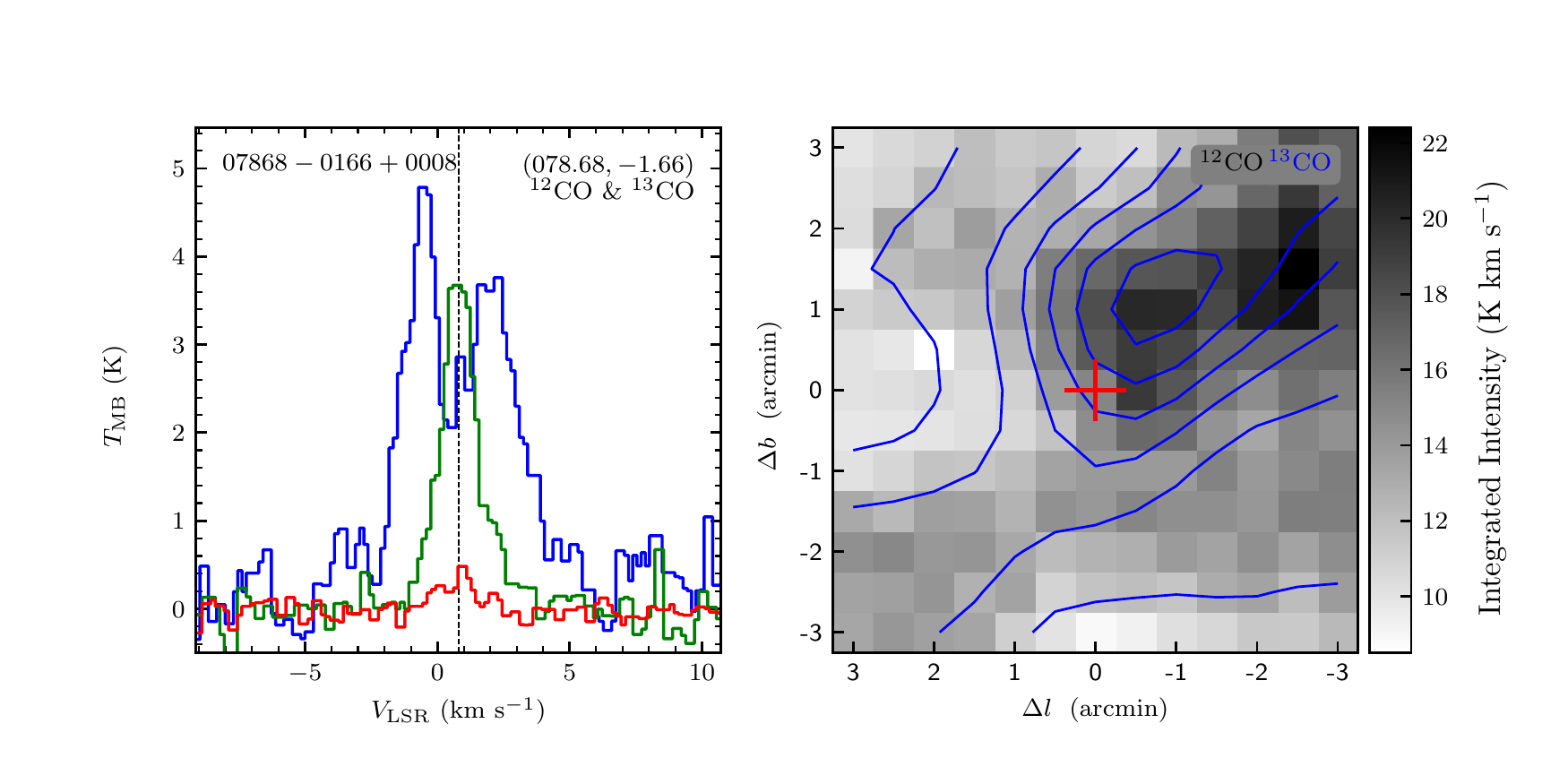}
\includegraphics[width=9.0cm,angle=0]{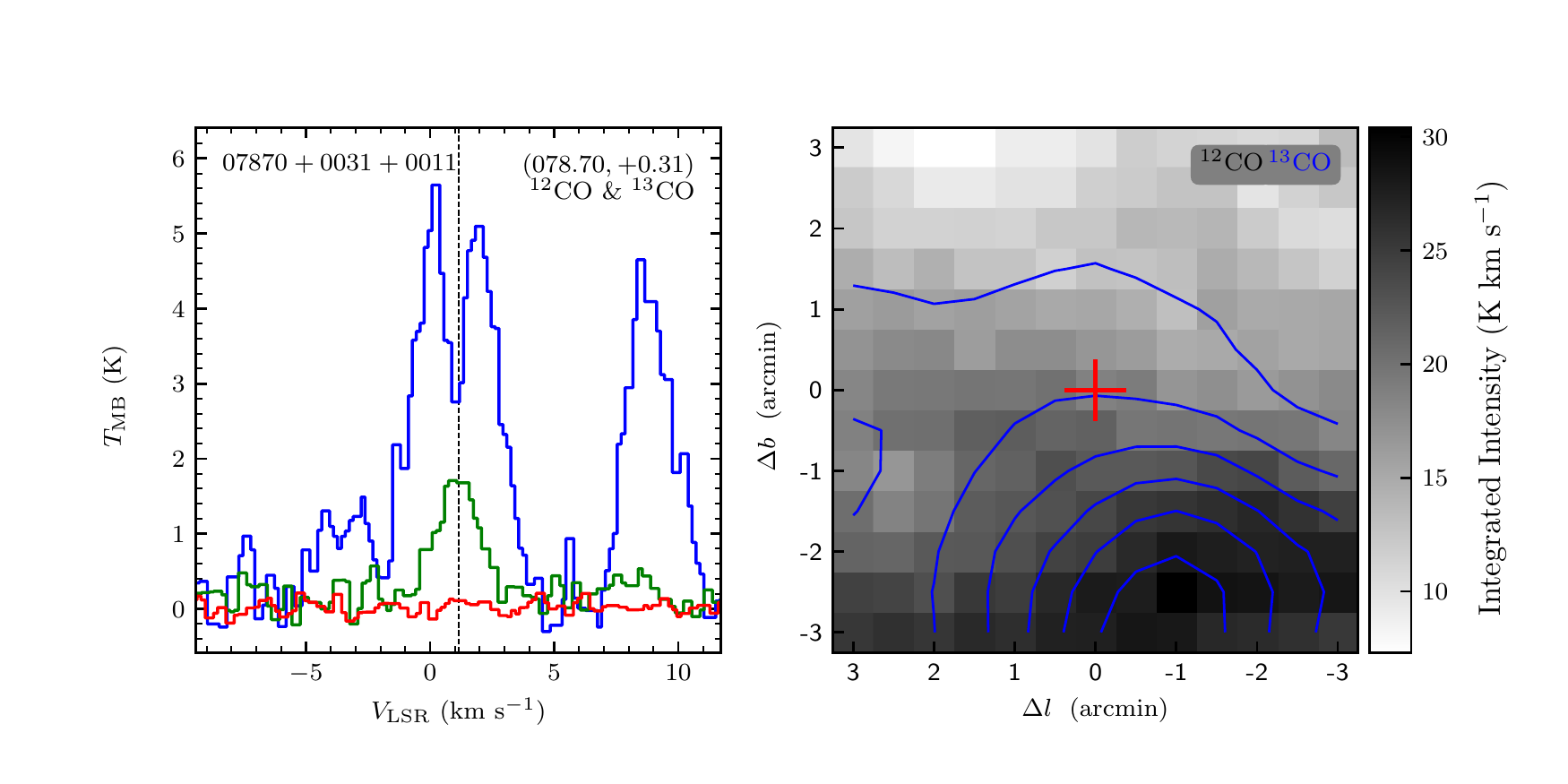}
\vspace{-0.5cm}

\includegraphics[width=9.0cm,angle=0]{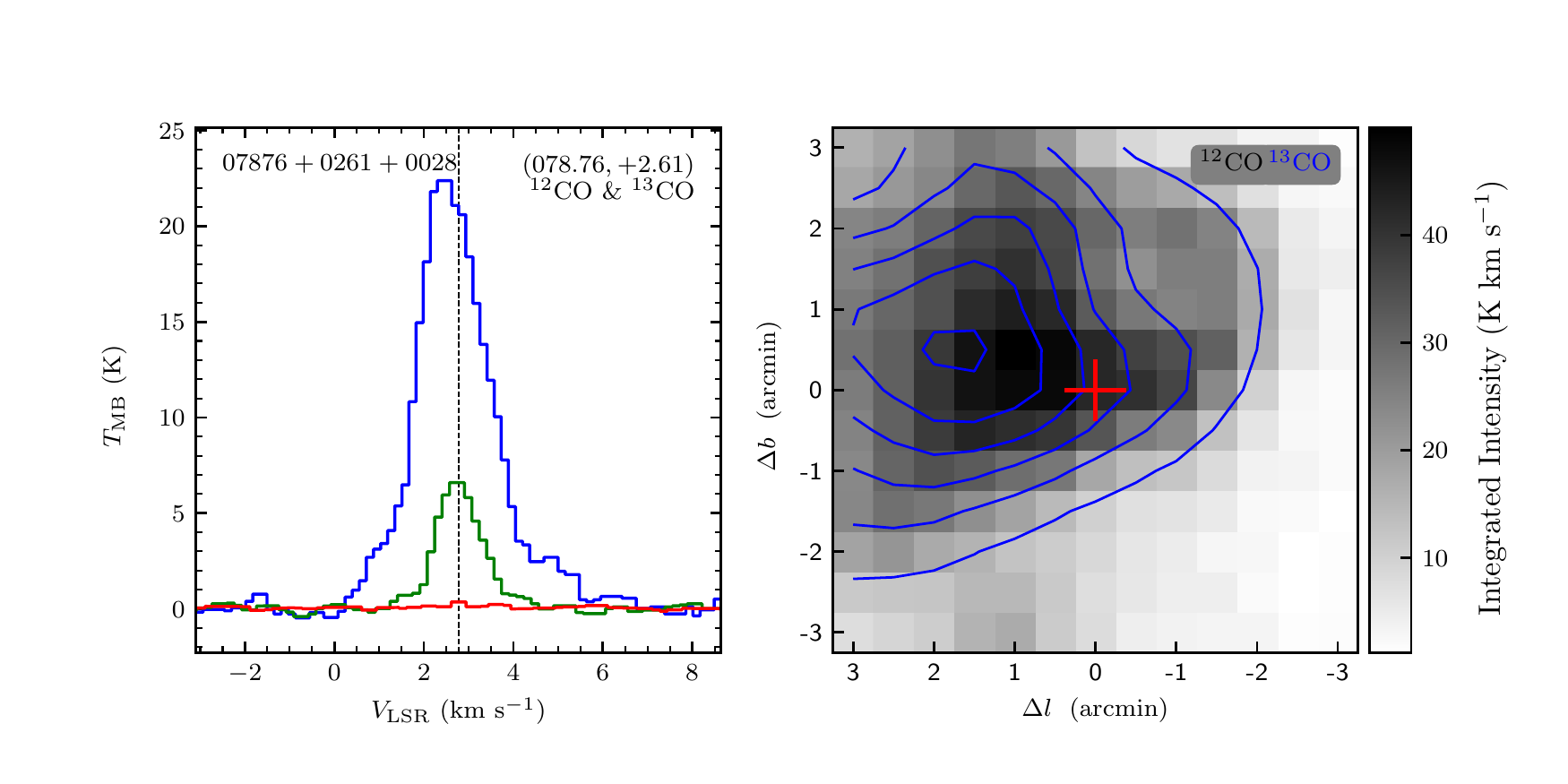}
\includegraphics[width=9.0cm,angle=0]{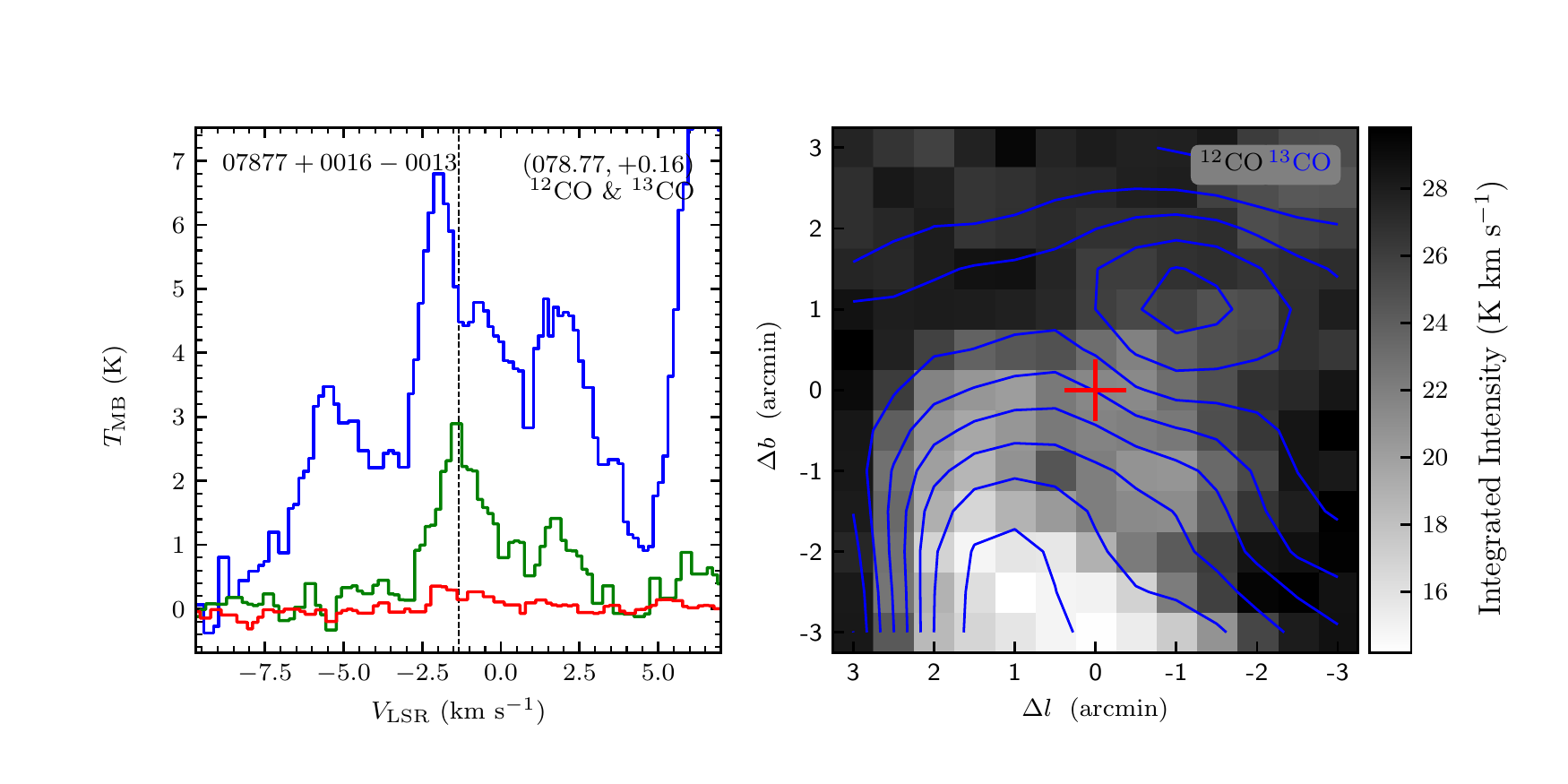}
\vspace{-0.5cm}

\includegraphics[width=9.0cm,angle=0]{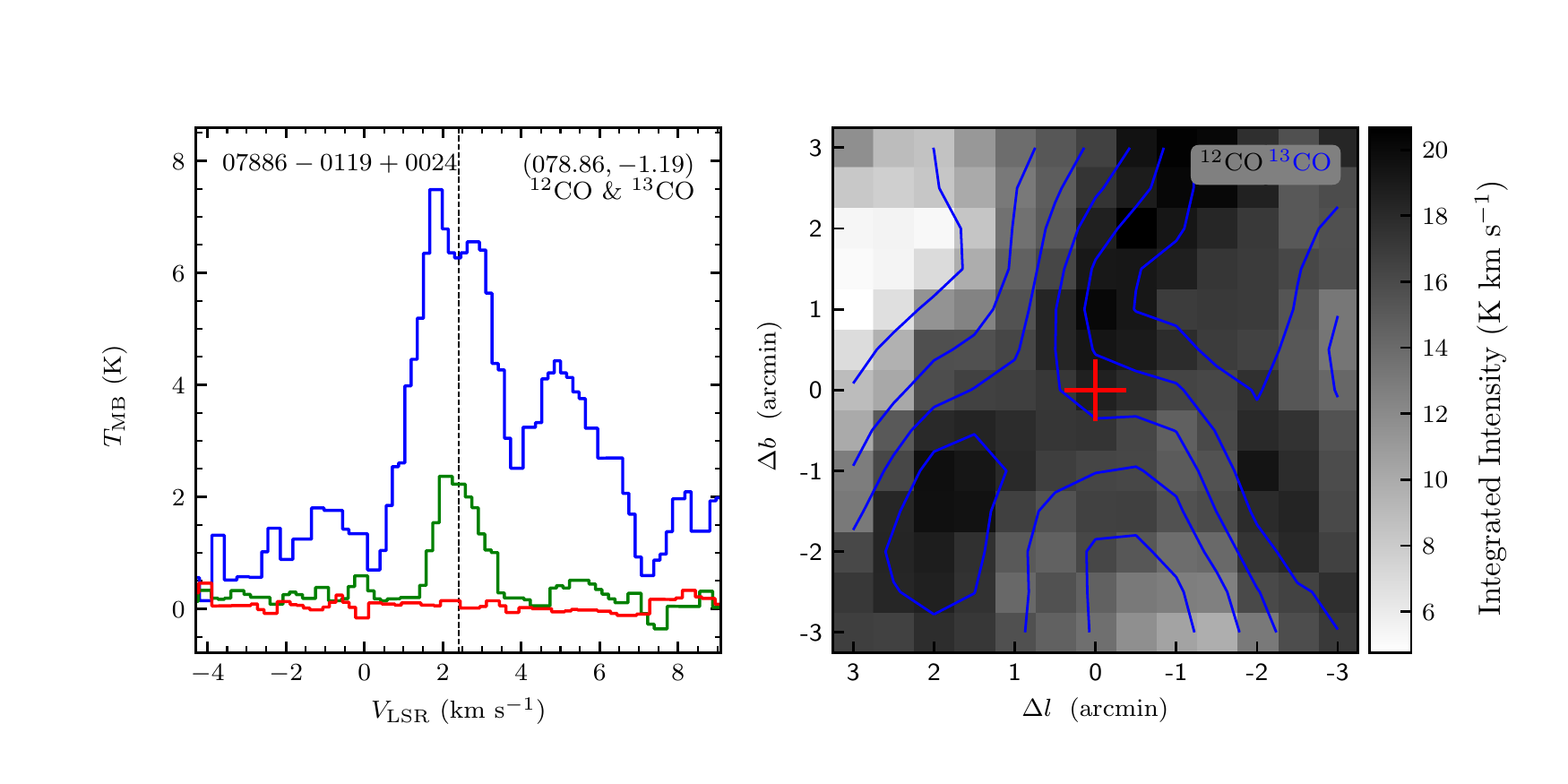}
\includegraphics[width=9.0cm,angle=0]{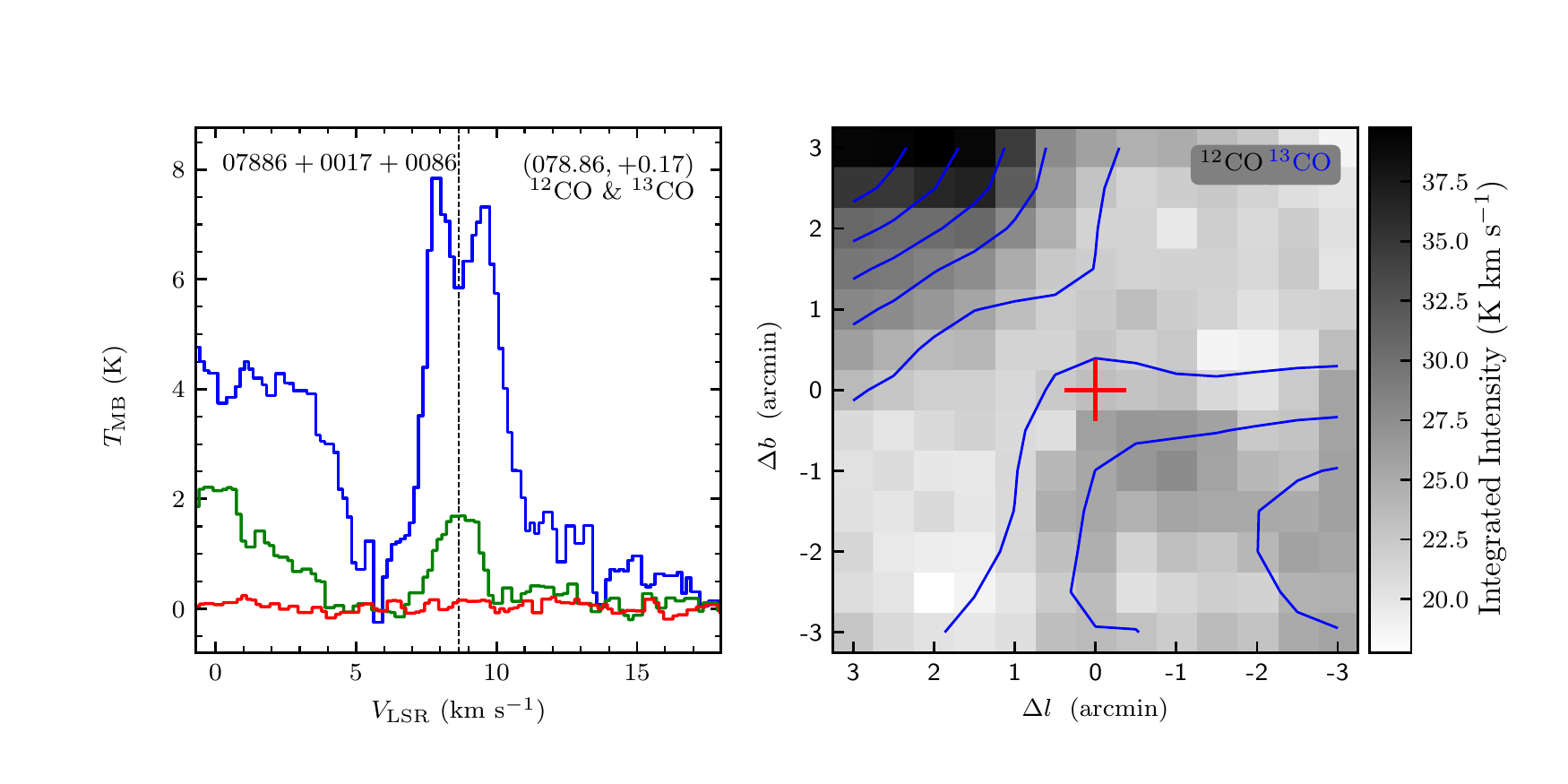}
\vspace{-0.5cm}

\includegraphics[width=9.0cm,angle=0]{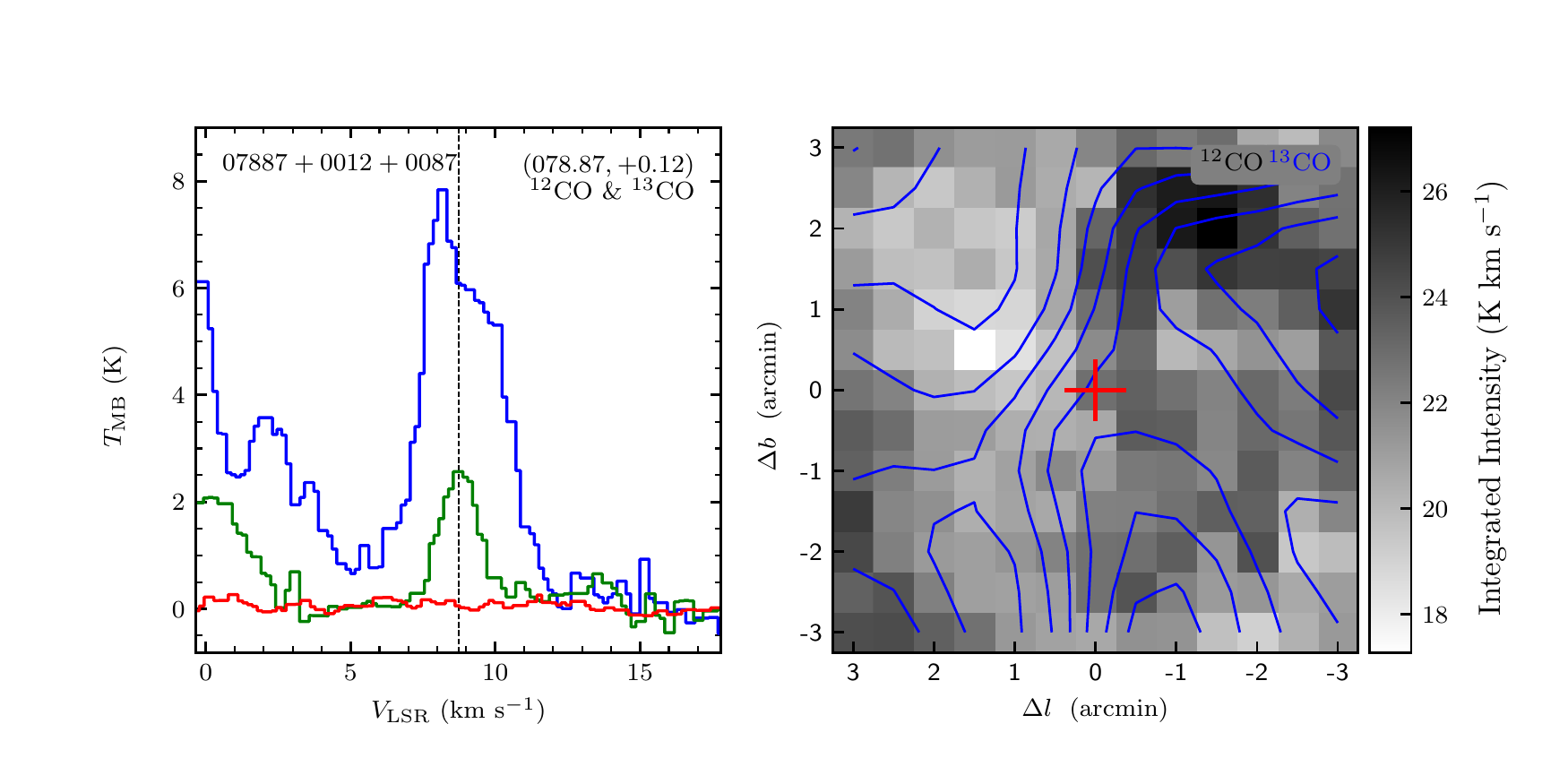}
\includegraphics[width=9.0cm,angle=0]{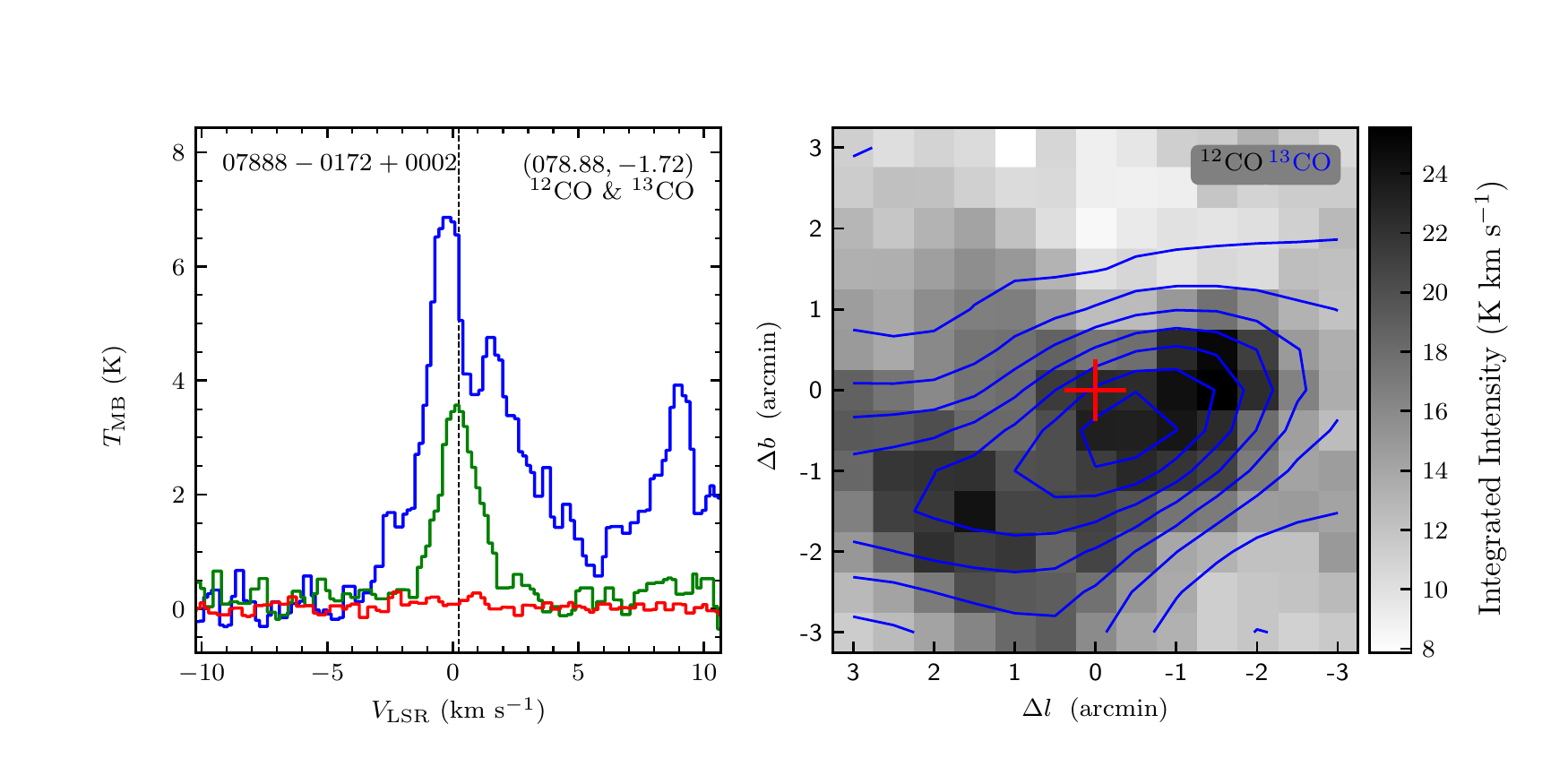}
\end{figure}
\clearpage

\begin{figure}
\includegraphics[width=9.0cm,angle=0]{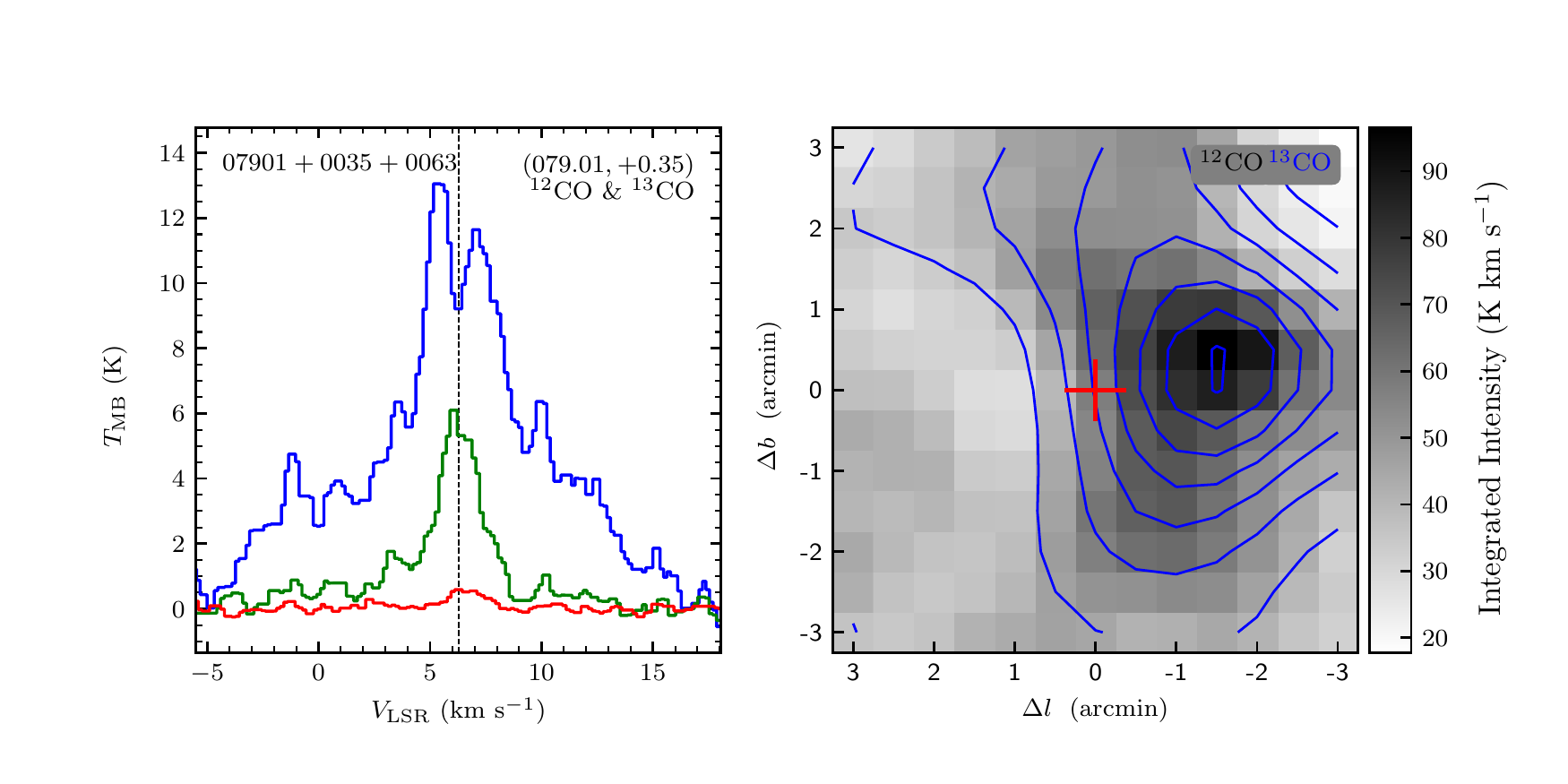}
\includegraphics[width=9.0cm,angle=0]{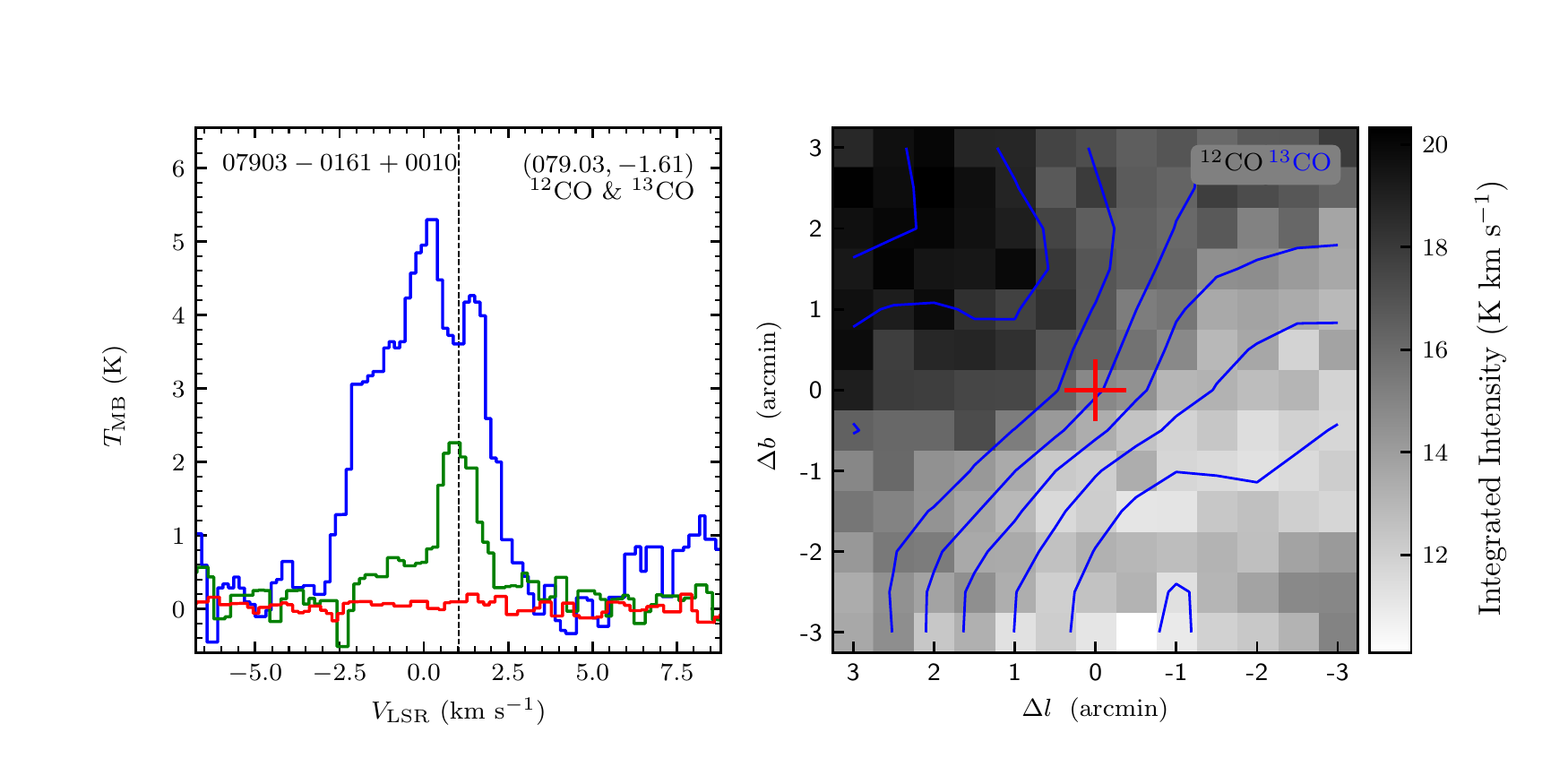}
\vspace{-0.5cm}

\includegraphics[width=9.0cm,angle=0]{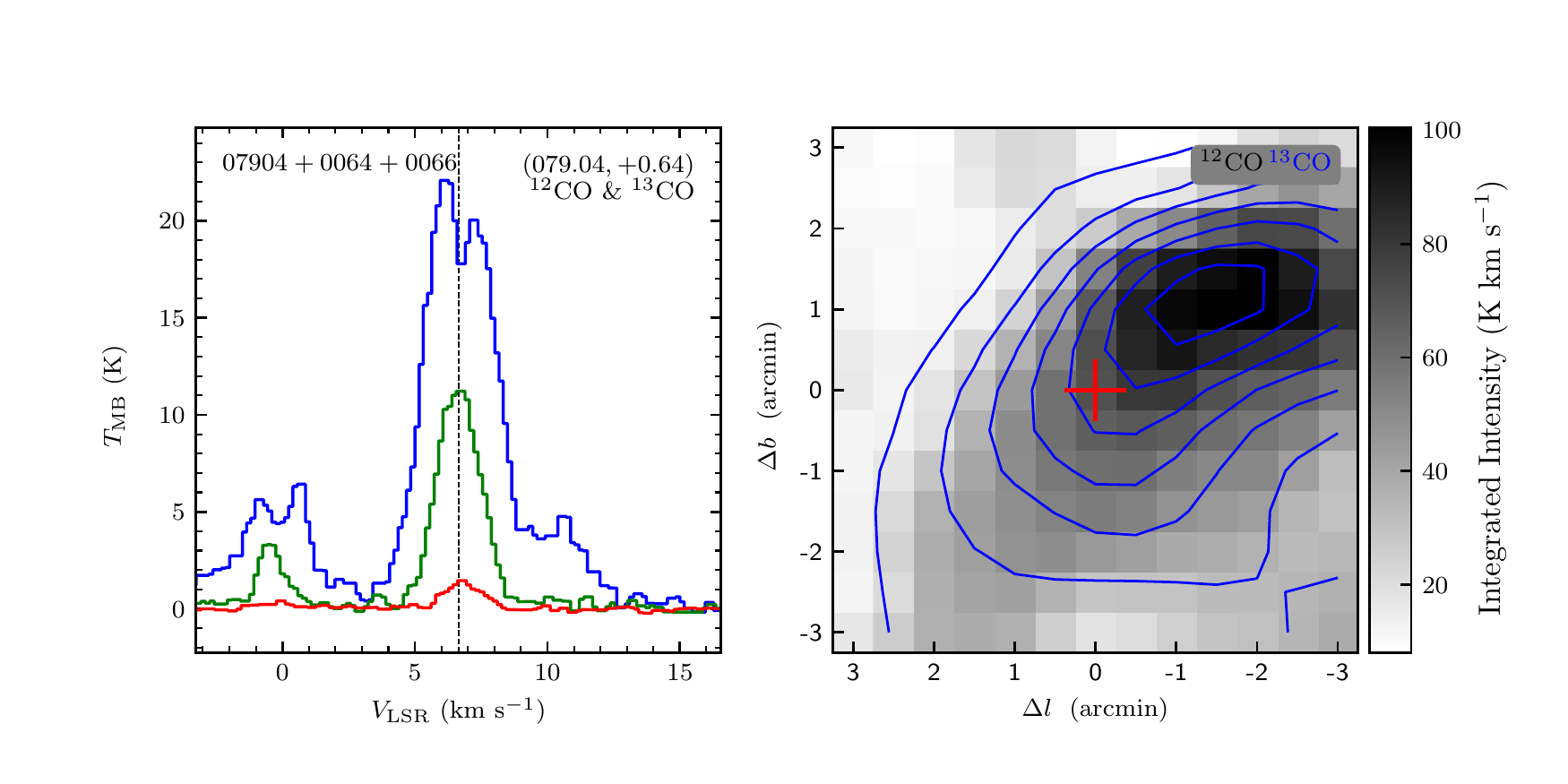}
\includegraphics[width=9.0cm,angle=0]{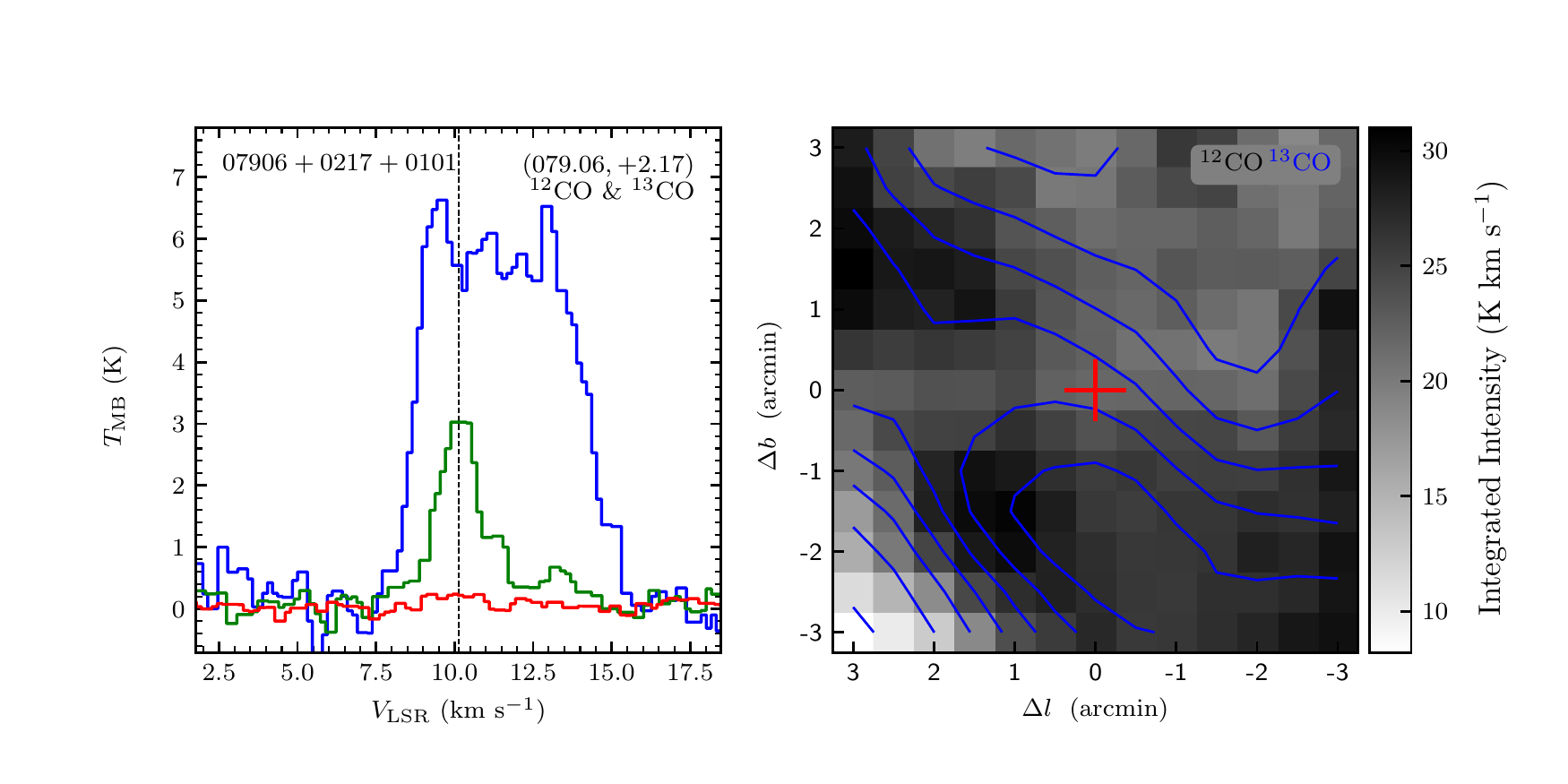}
\vspace{-0.5cm}

\includegraphics[width=9.0cm,angle=0]{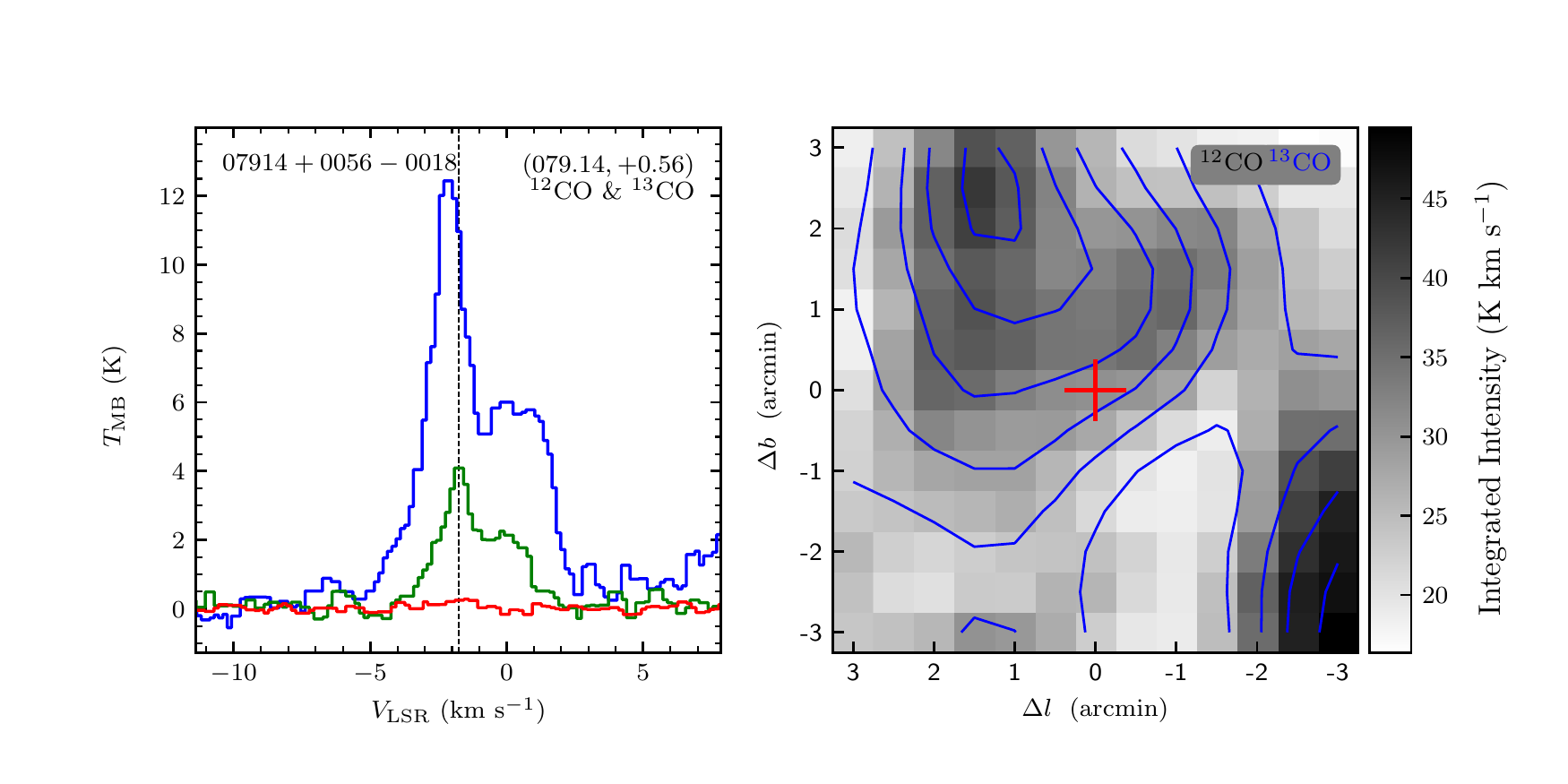}
\includegraphics[width=9.0cm,angle=0]{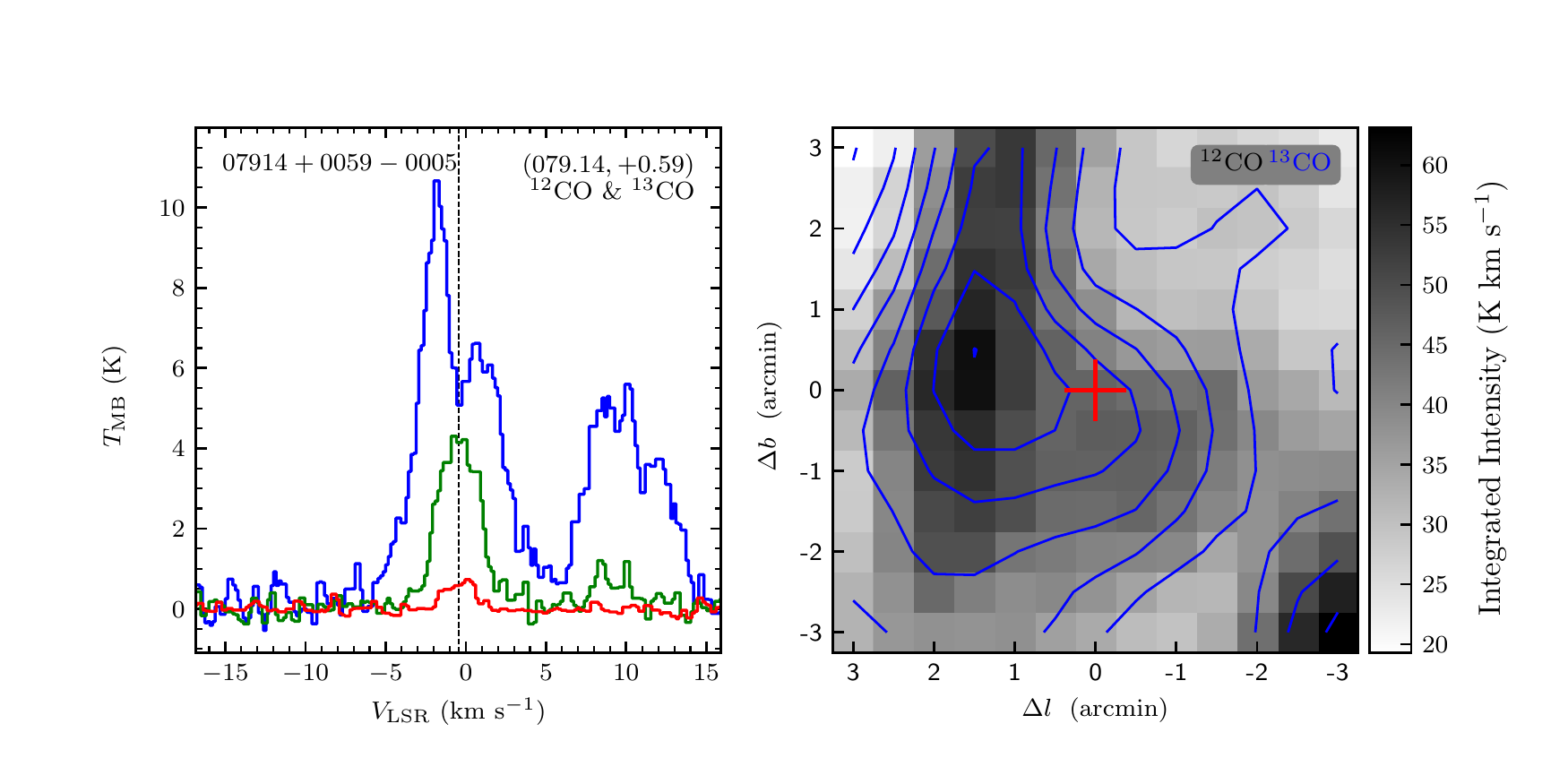}
\vspace{-0.5cm}

\includegraphics[width=9.0cm,angle=0]{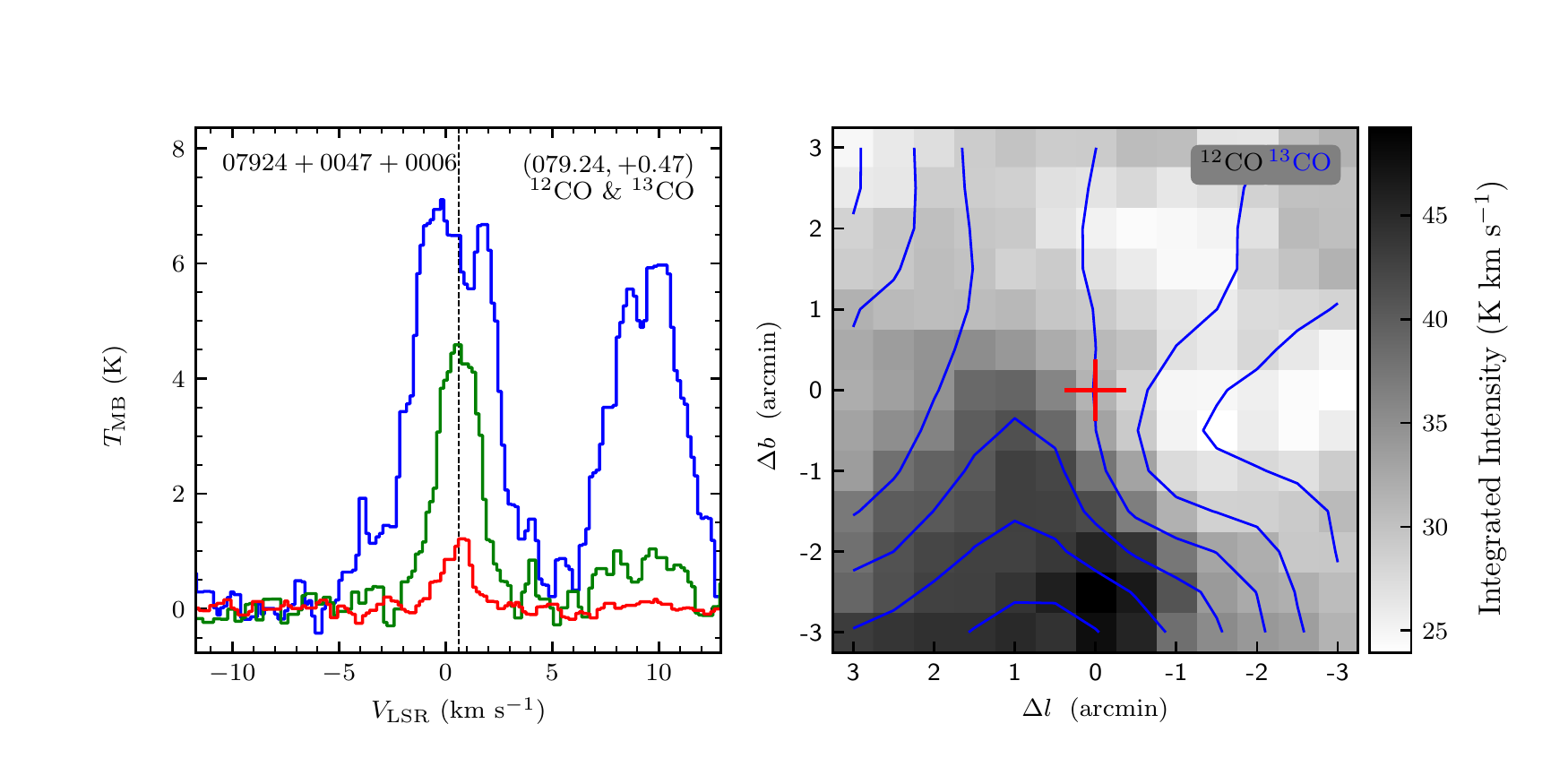}
\includegraphics[width=9.0cm,angle=0]{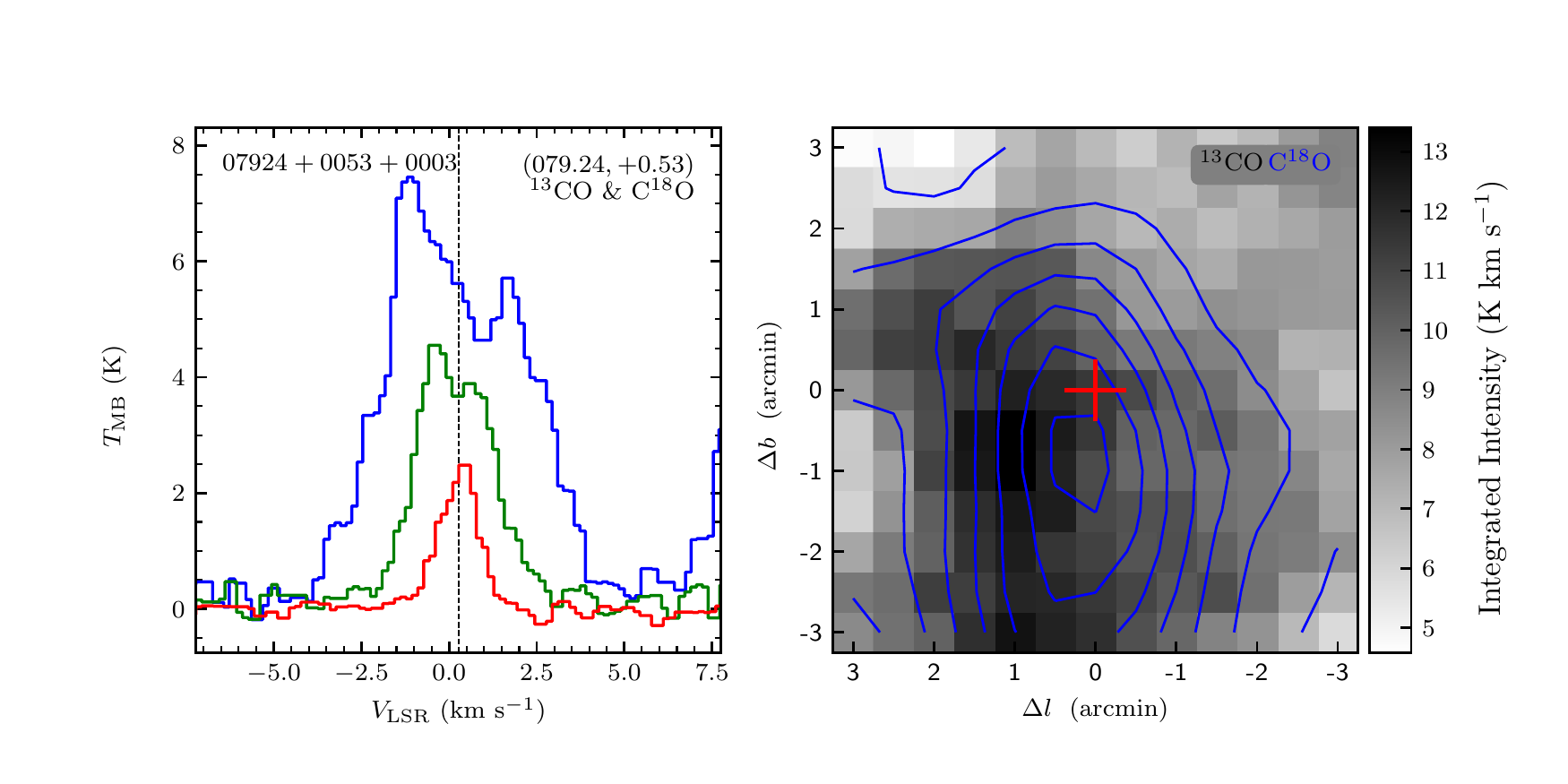}
\vspace{-0.5cm}

\includegraphics[width=9.0cm,angle=0]{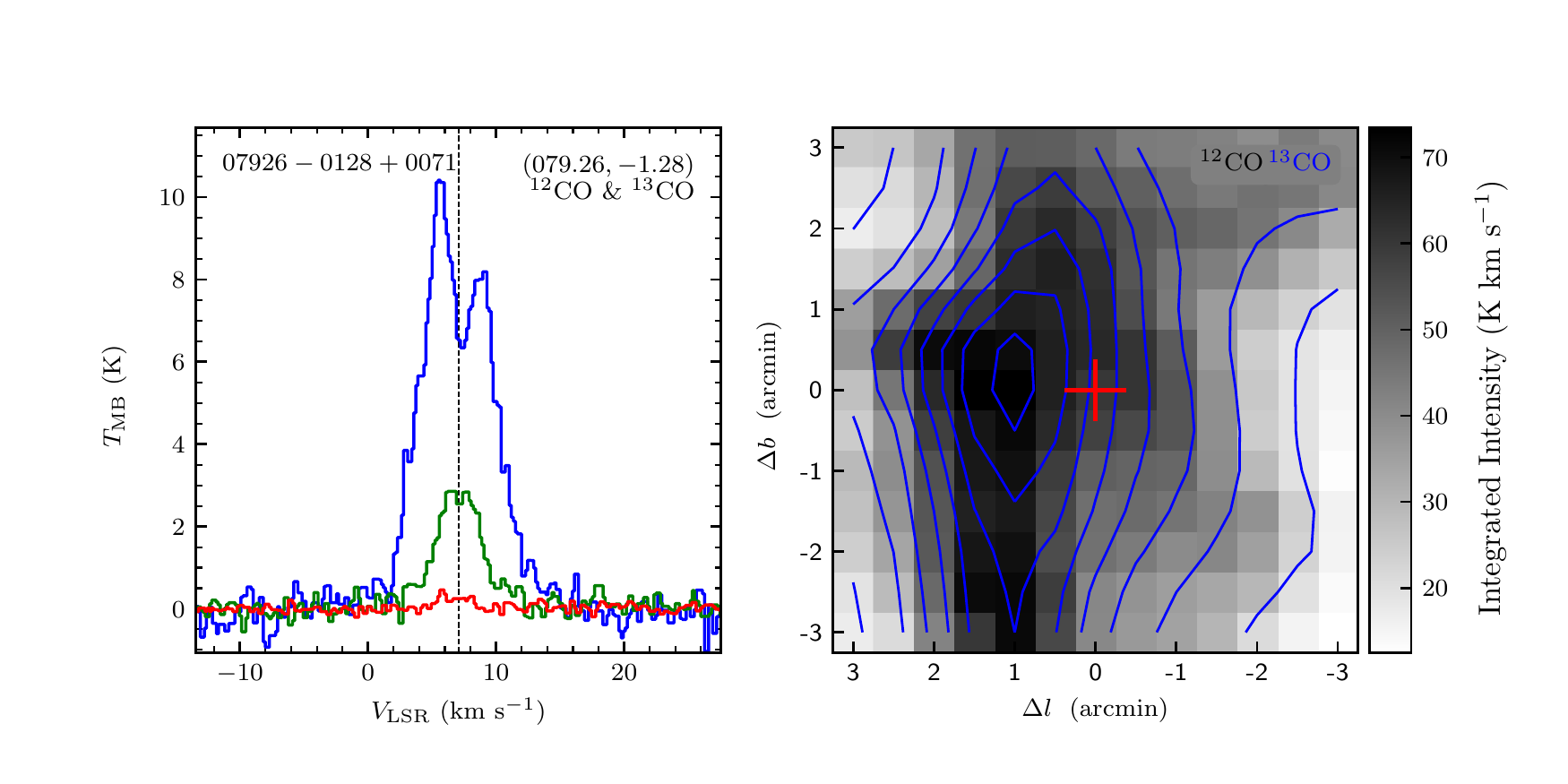}
\includegraphics[width=9.0cm,angle=0]{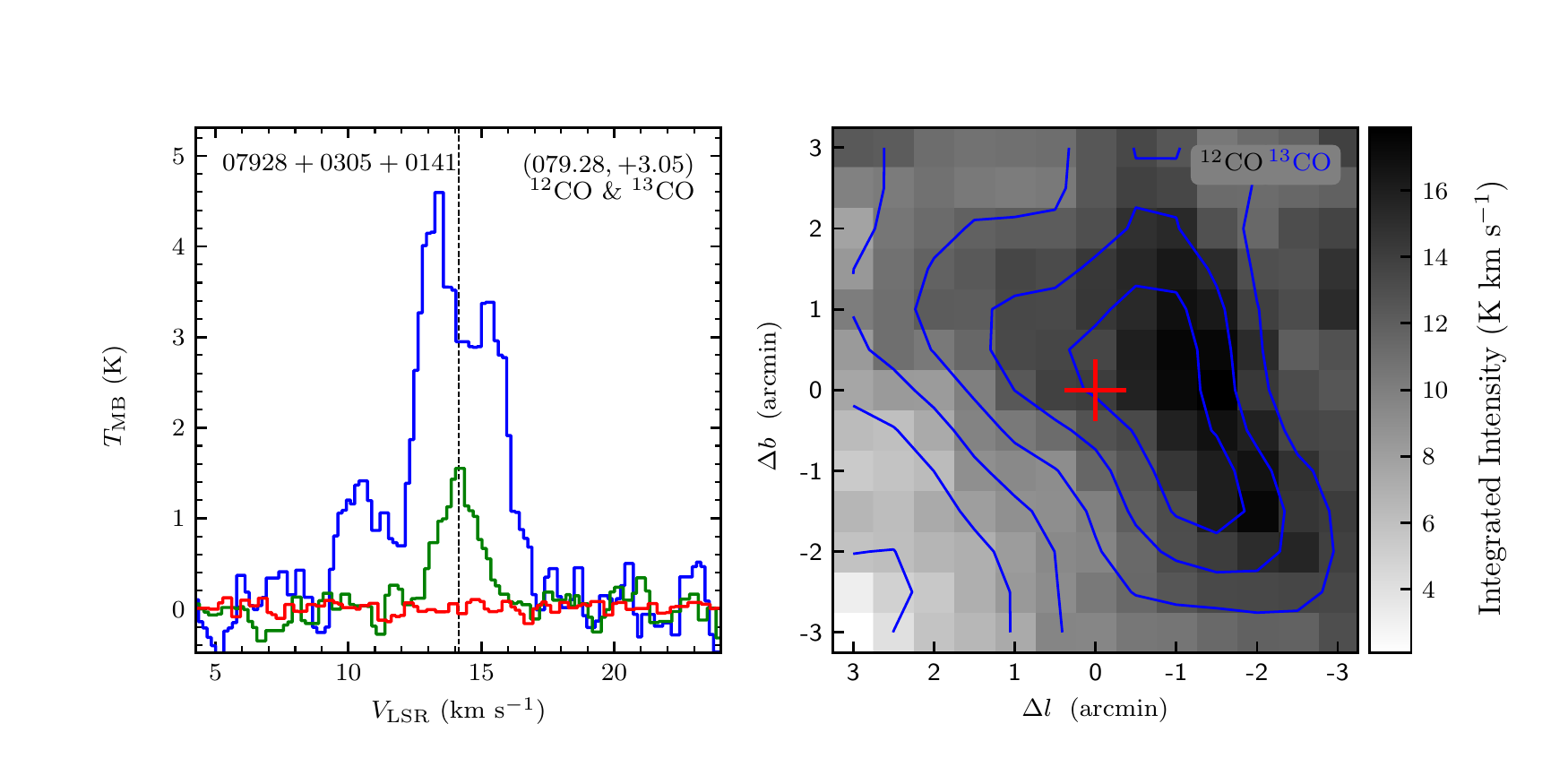}
\end{figure}
\clearpage

\begin{figure}
\includegraphics[width=9.0cm,angle=0]{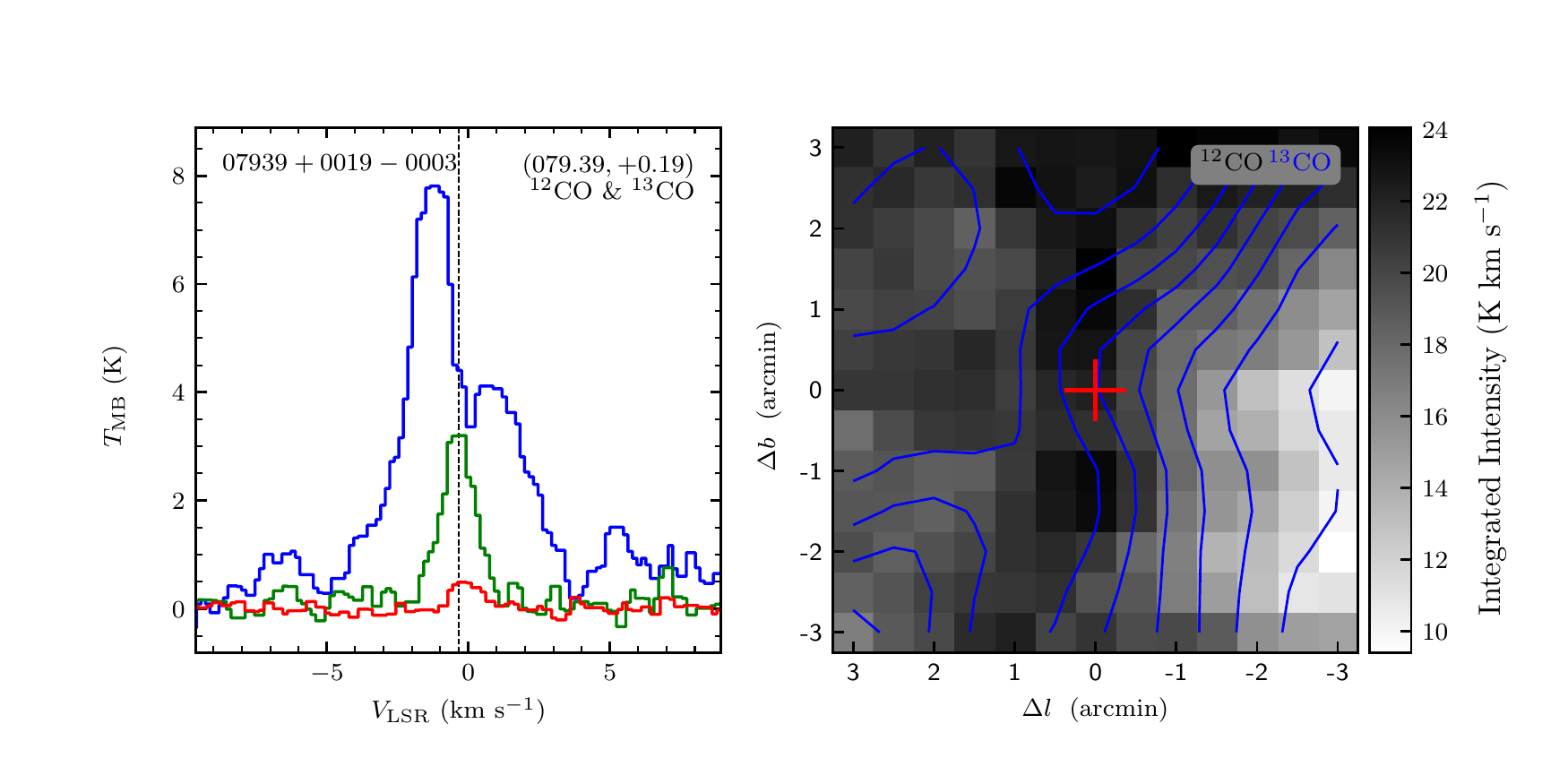}
\includegraphics[width=9.0cm,angle=0]{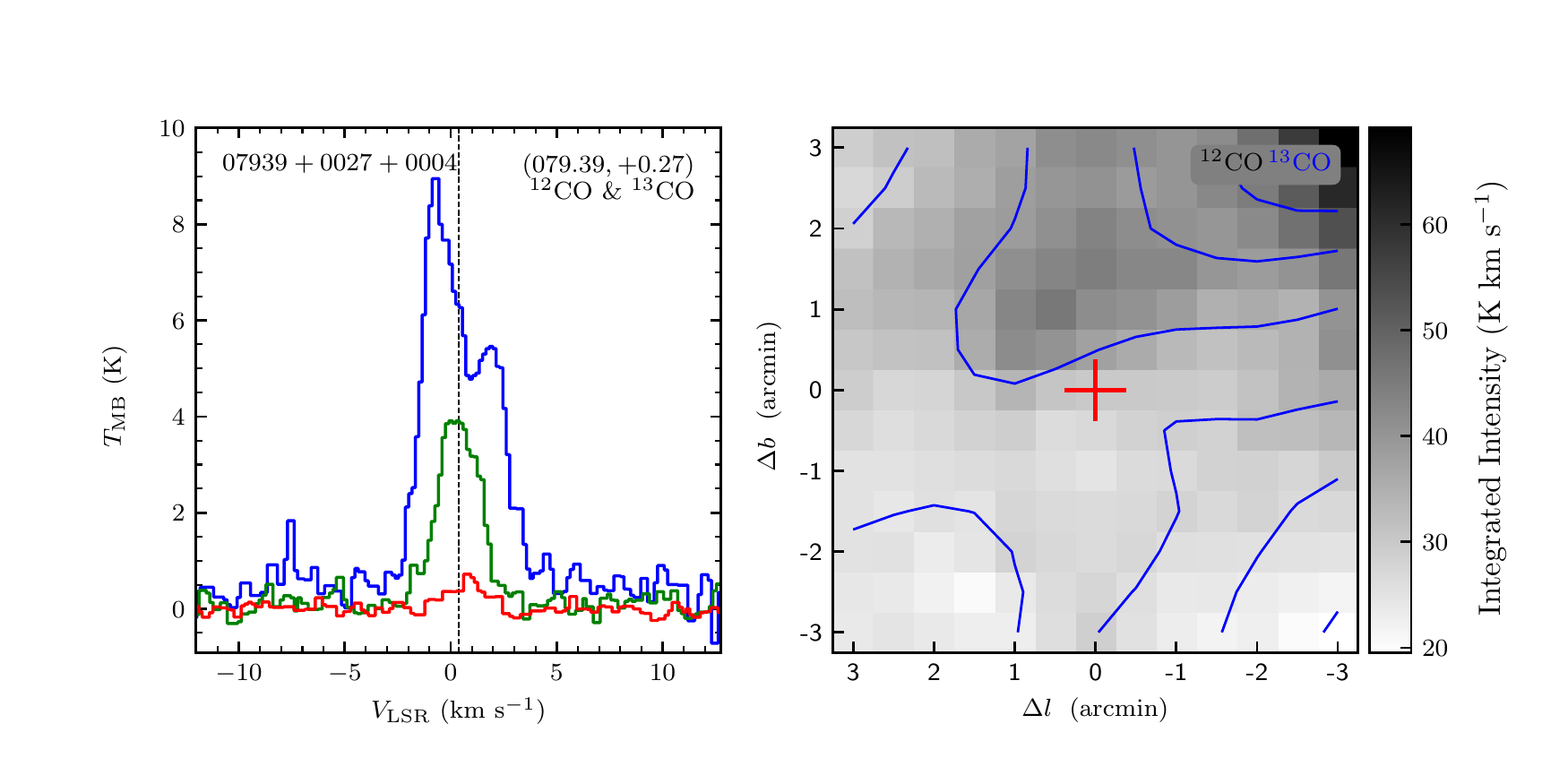}
\vspace{-0.5cm}

\includegraphics[width=9.0cm,angle=0]{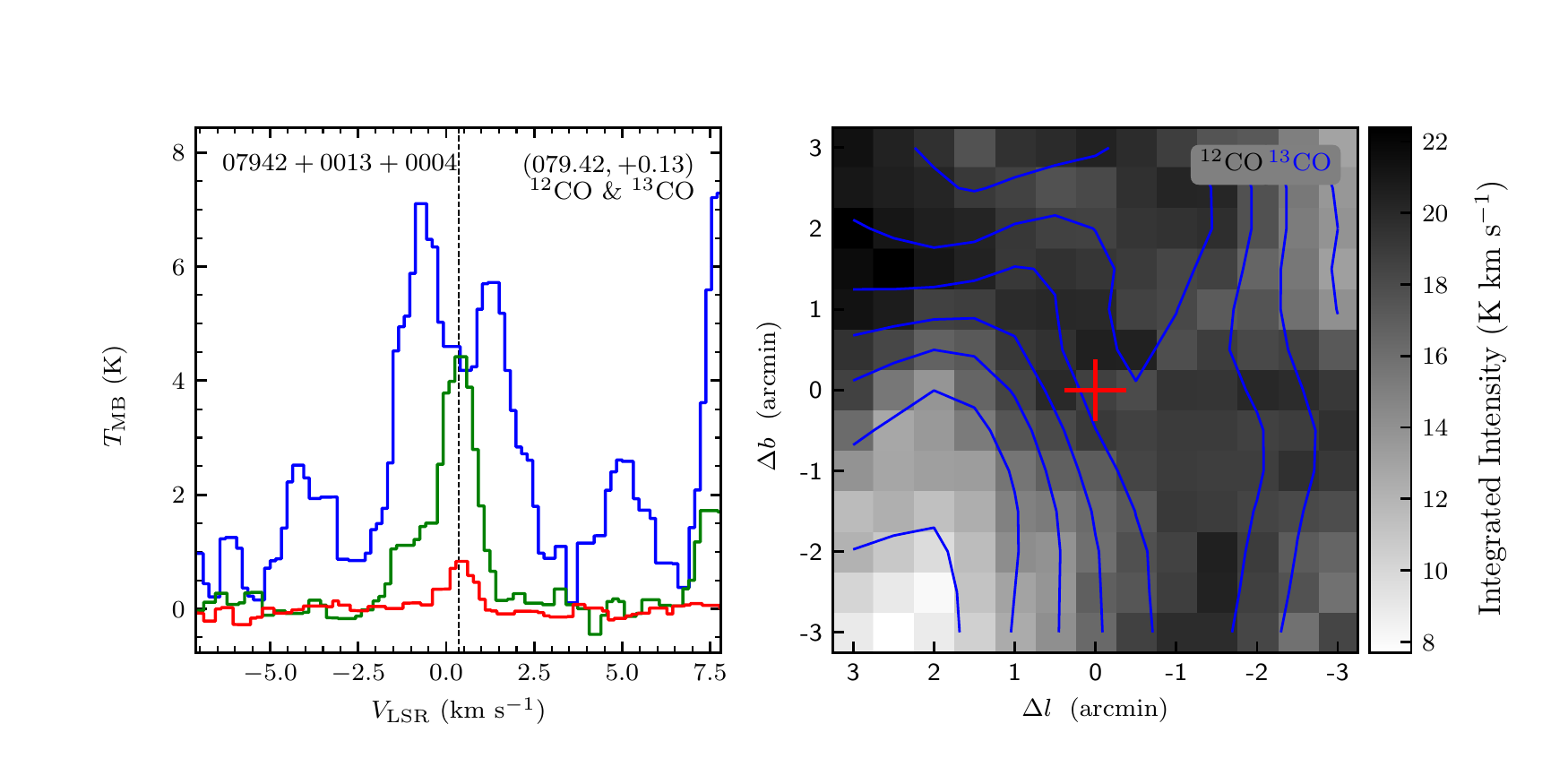}
\includegraphics[width=9.0cm,angle=0]{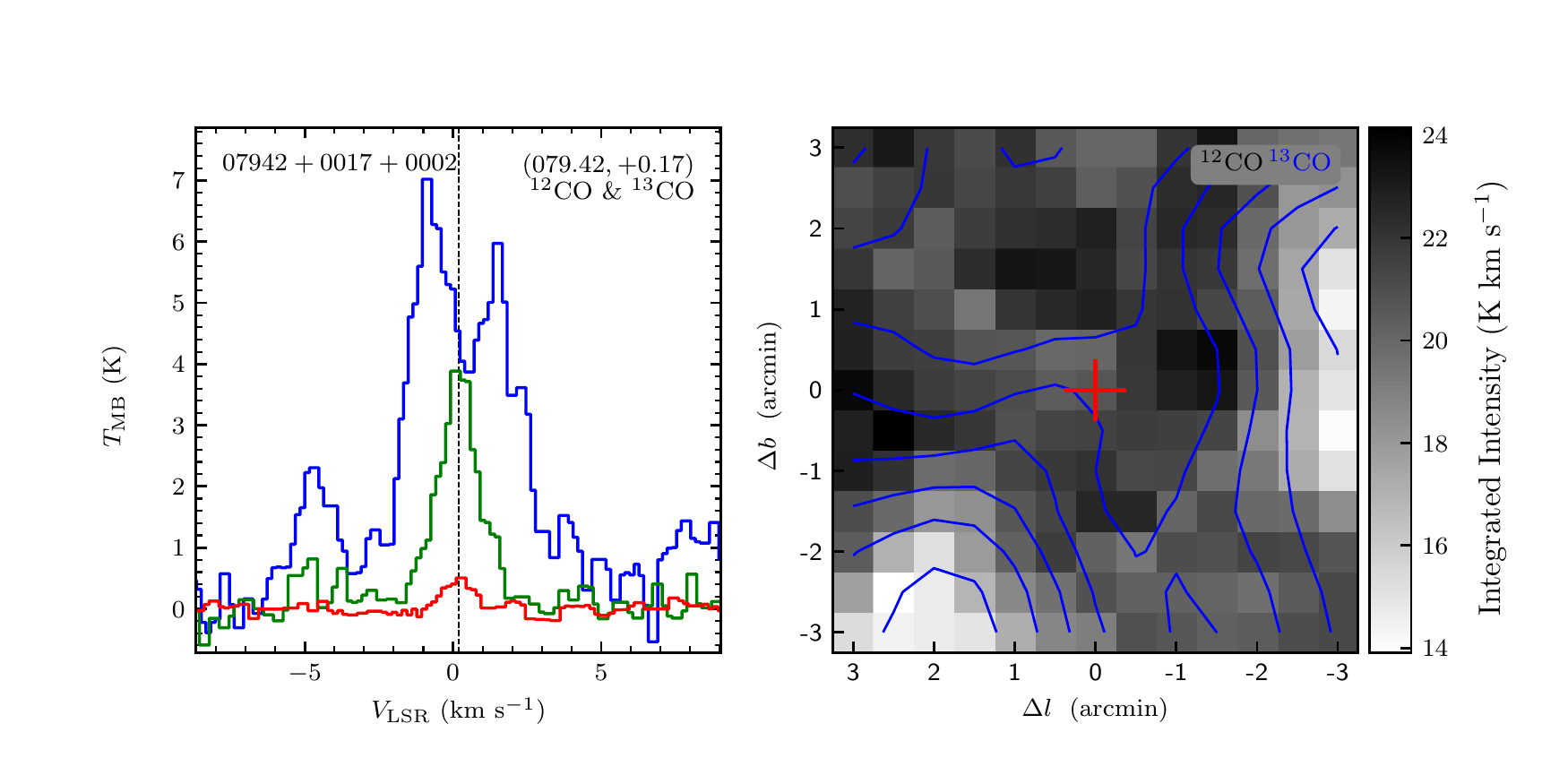}
\vspace{-0.5cm}

\includegraphics[width=9.0cm,angle=0]{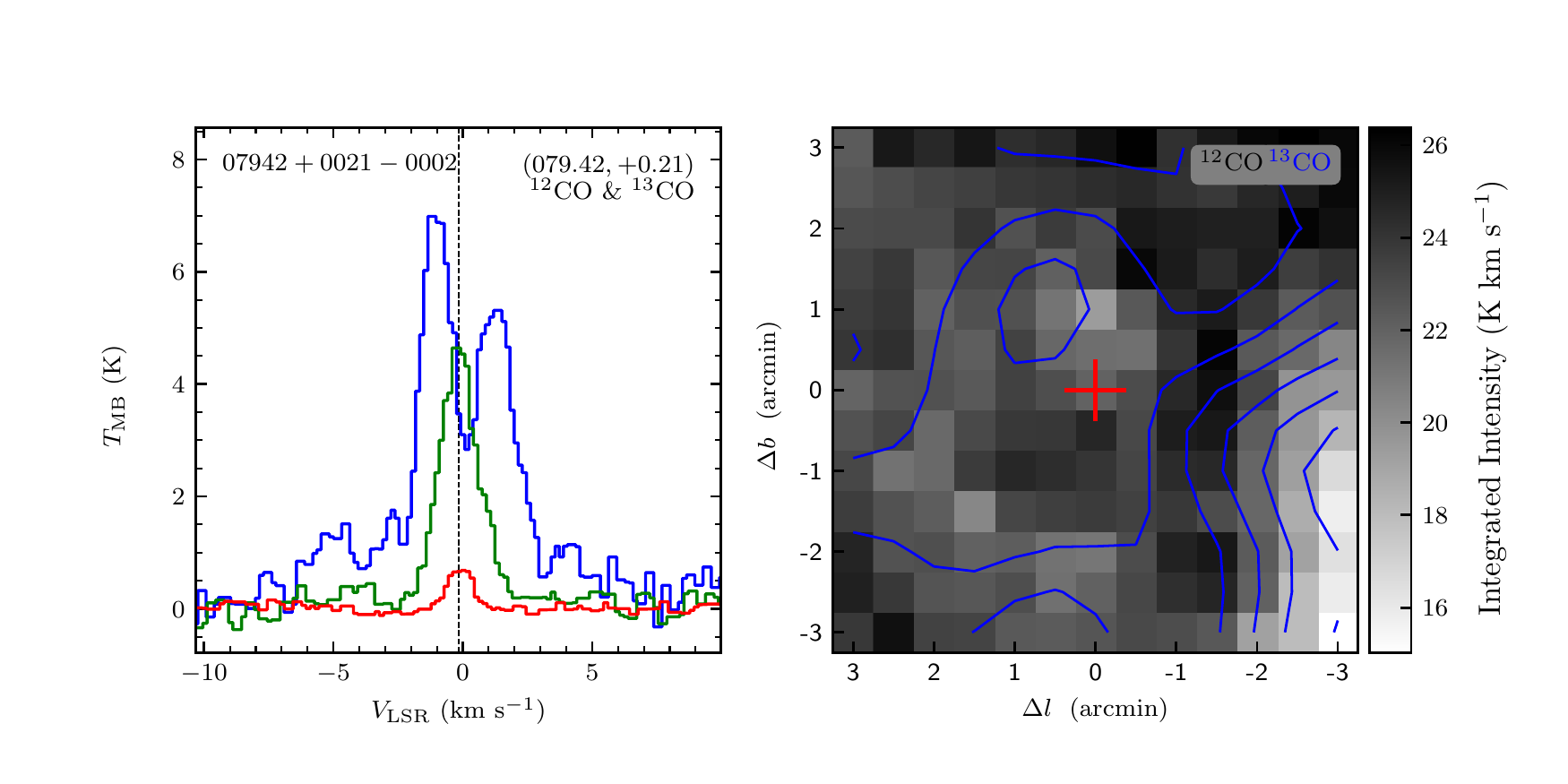}
\includegraphics[width=9.0cm,angle=0]{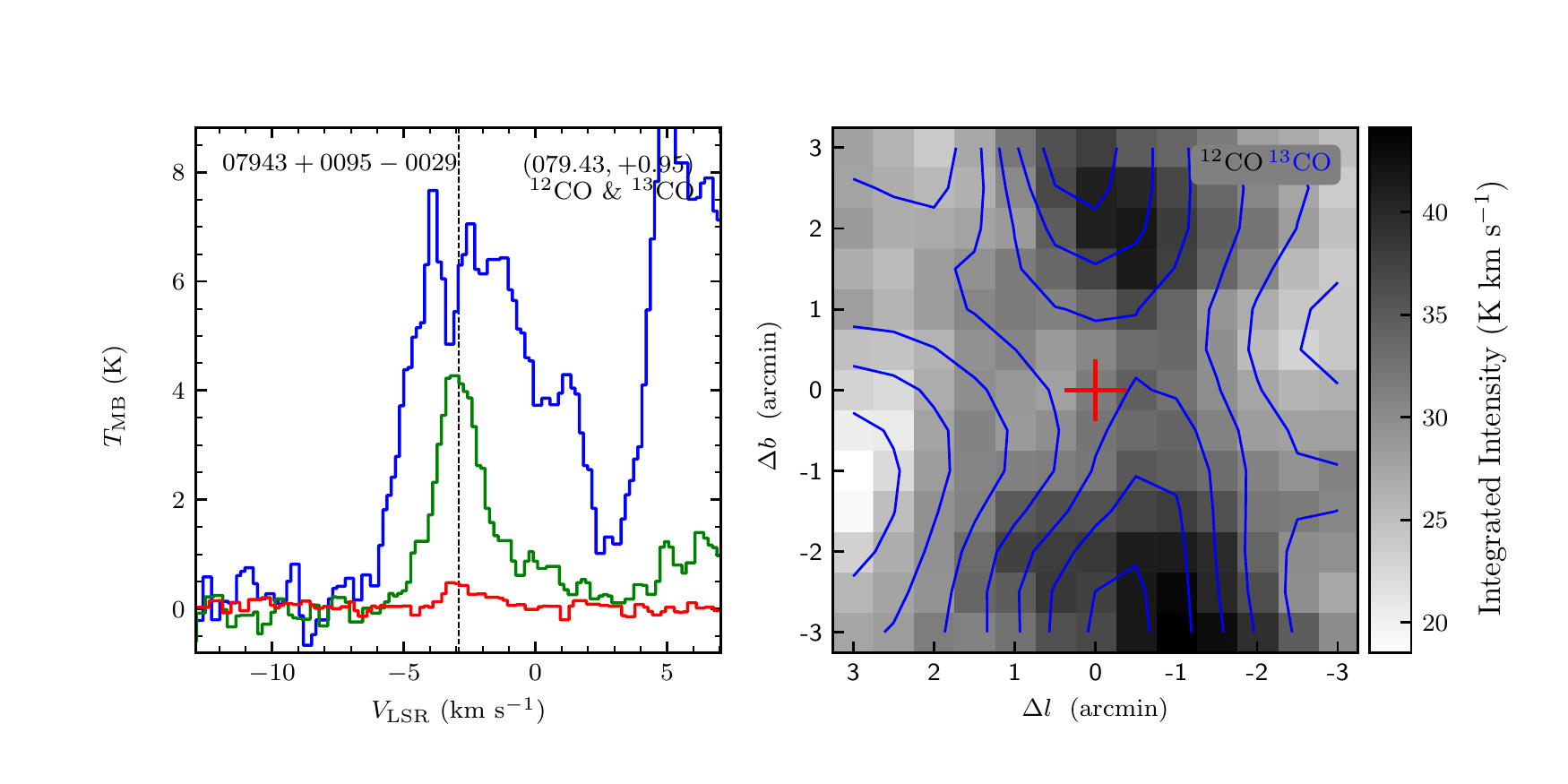}
\vspace{-0.5cm}

\includegraphics[width=9.0cm,angle=0]{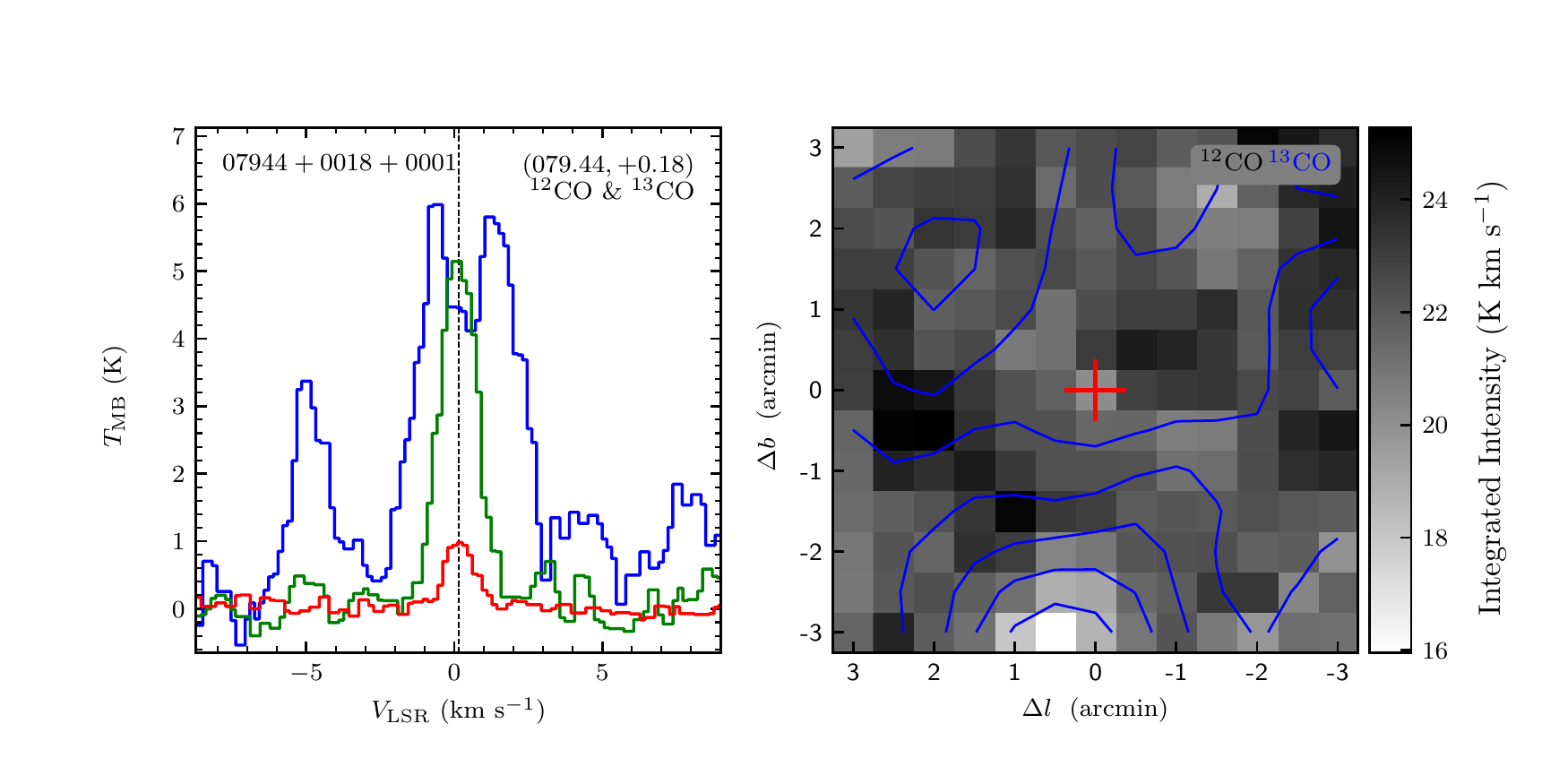}
\includegraphics[width=9.0cm,angle=0]{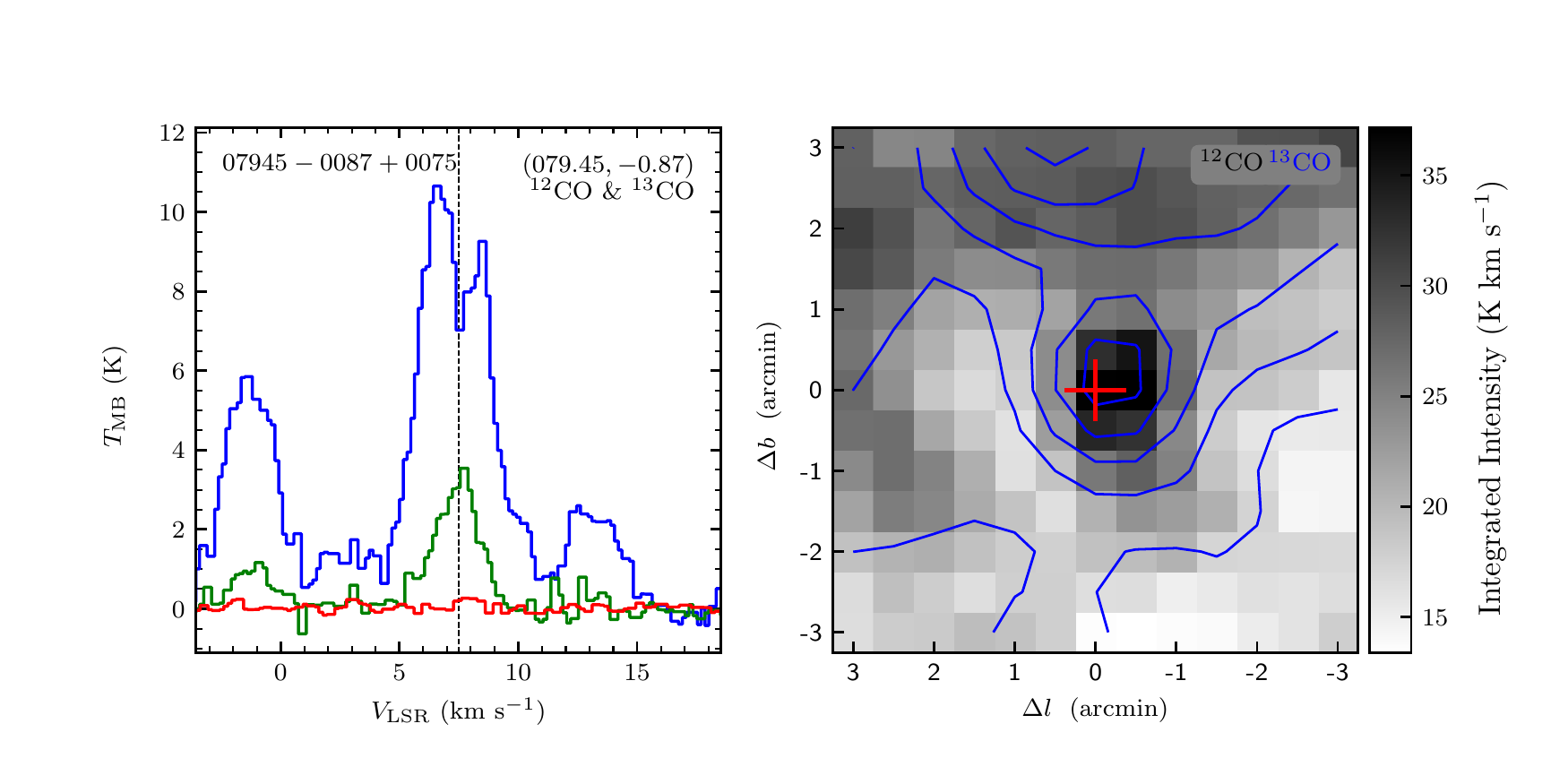}
\vspace{-0.5cm}

\includegraphics[width=9.0cm,angle=0]{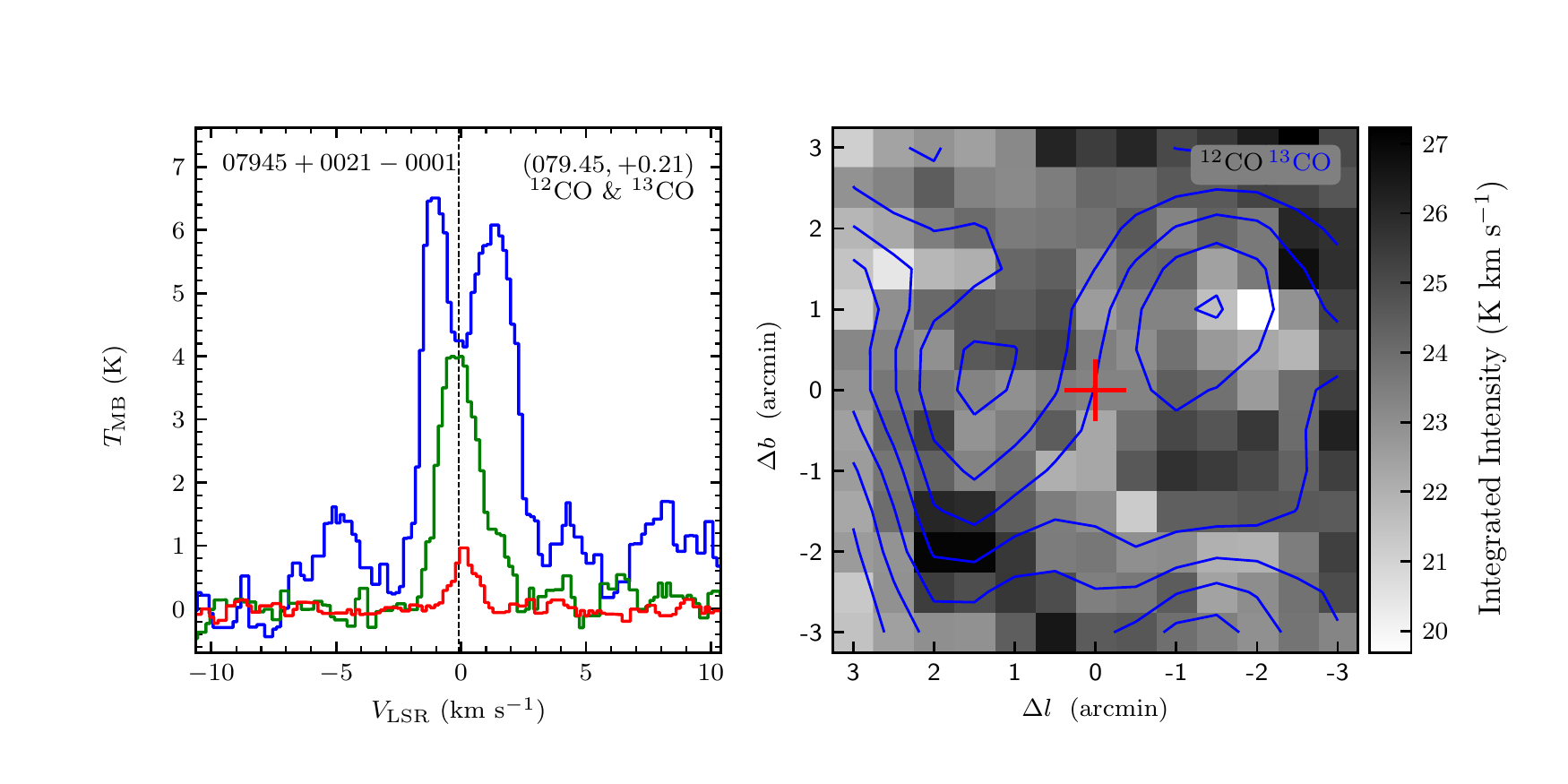}
\includegraphics[width=9.0cm,angle=0]{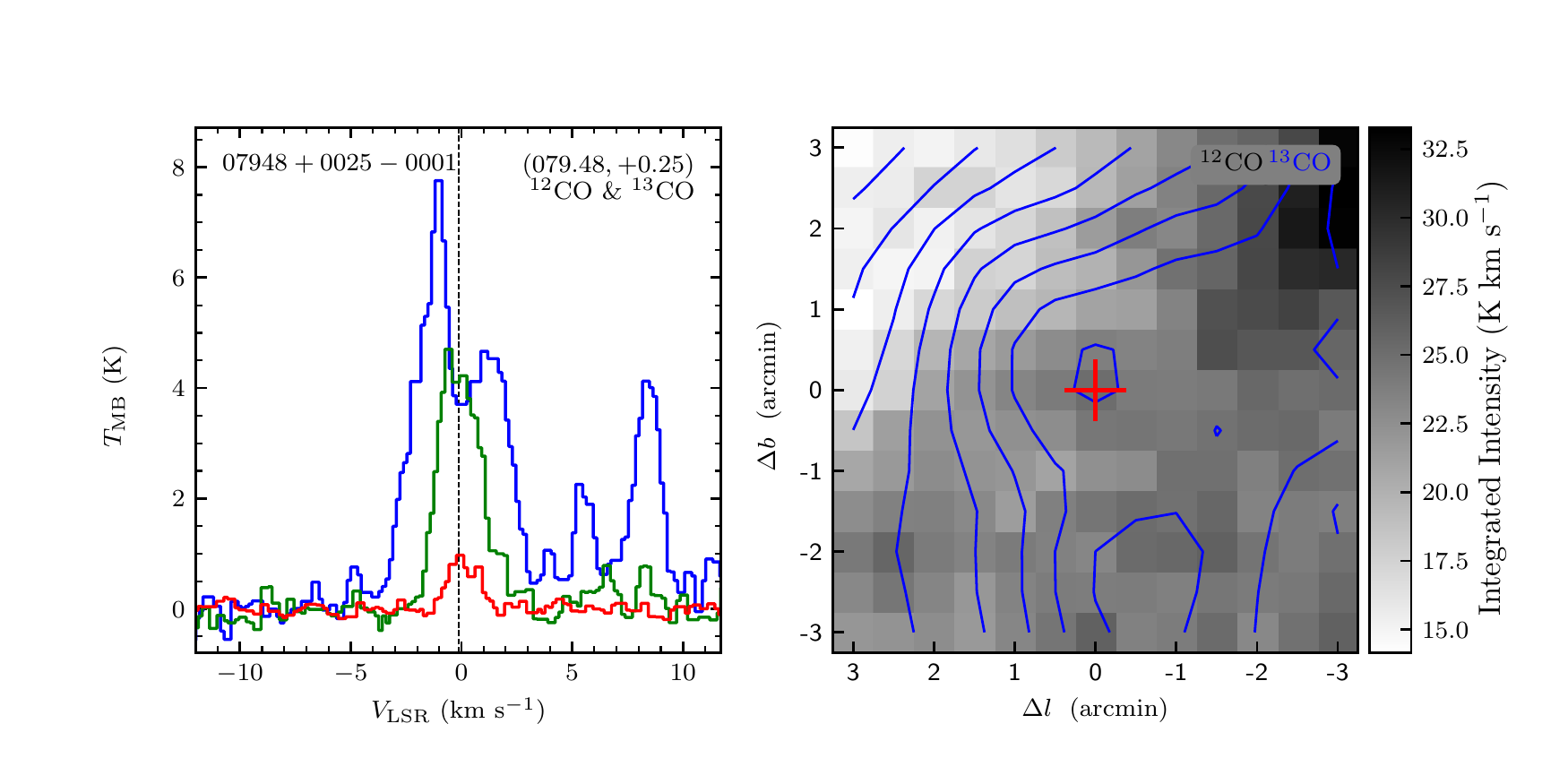}
\end{figure}
\clearpage

\begin{figure}
\includegraphics[width=9.0cm,angle=0]{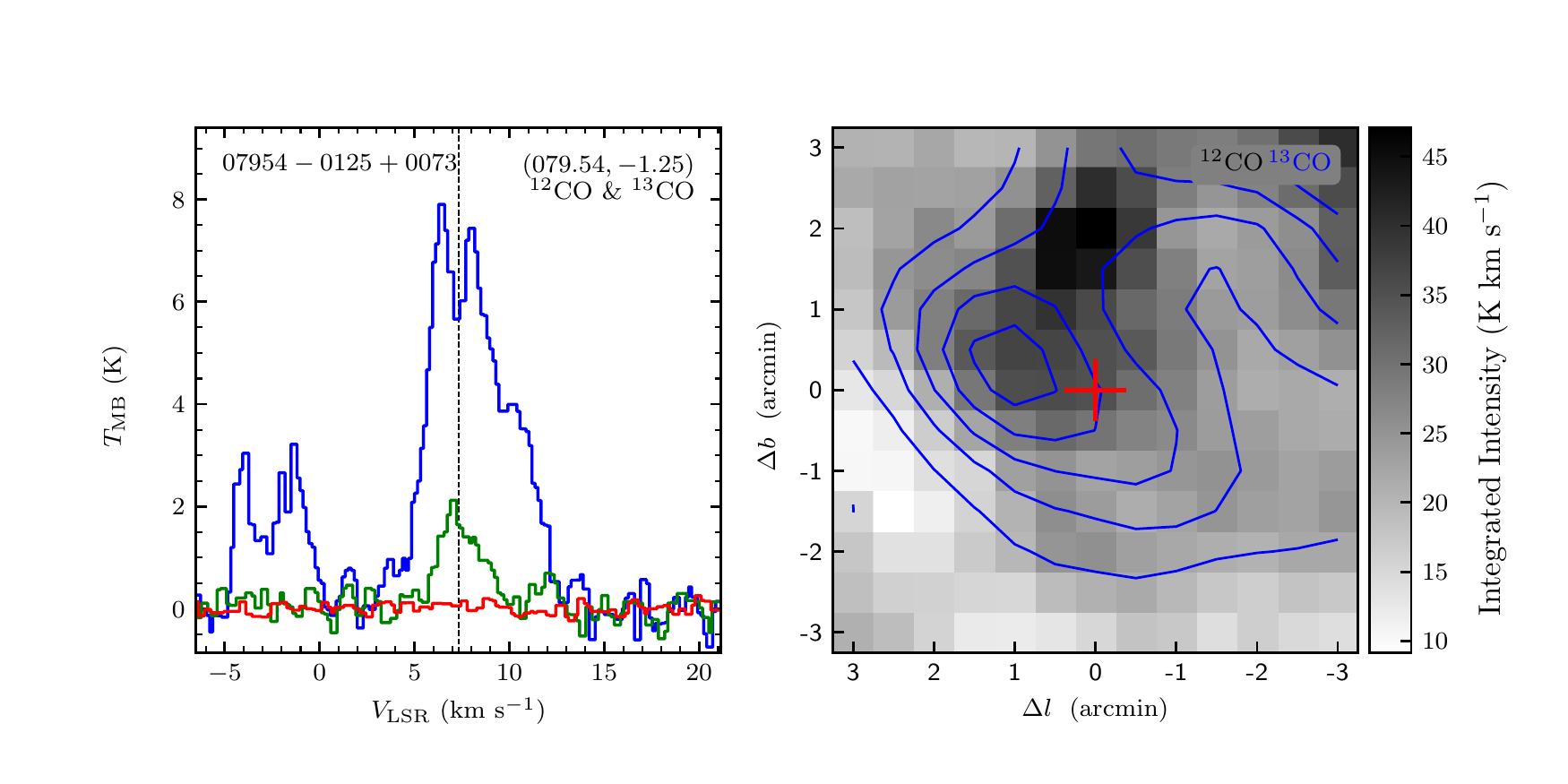}
\includegraphics[width=9.0cm,angle=0]{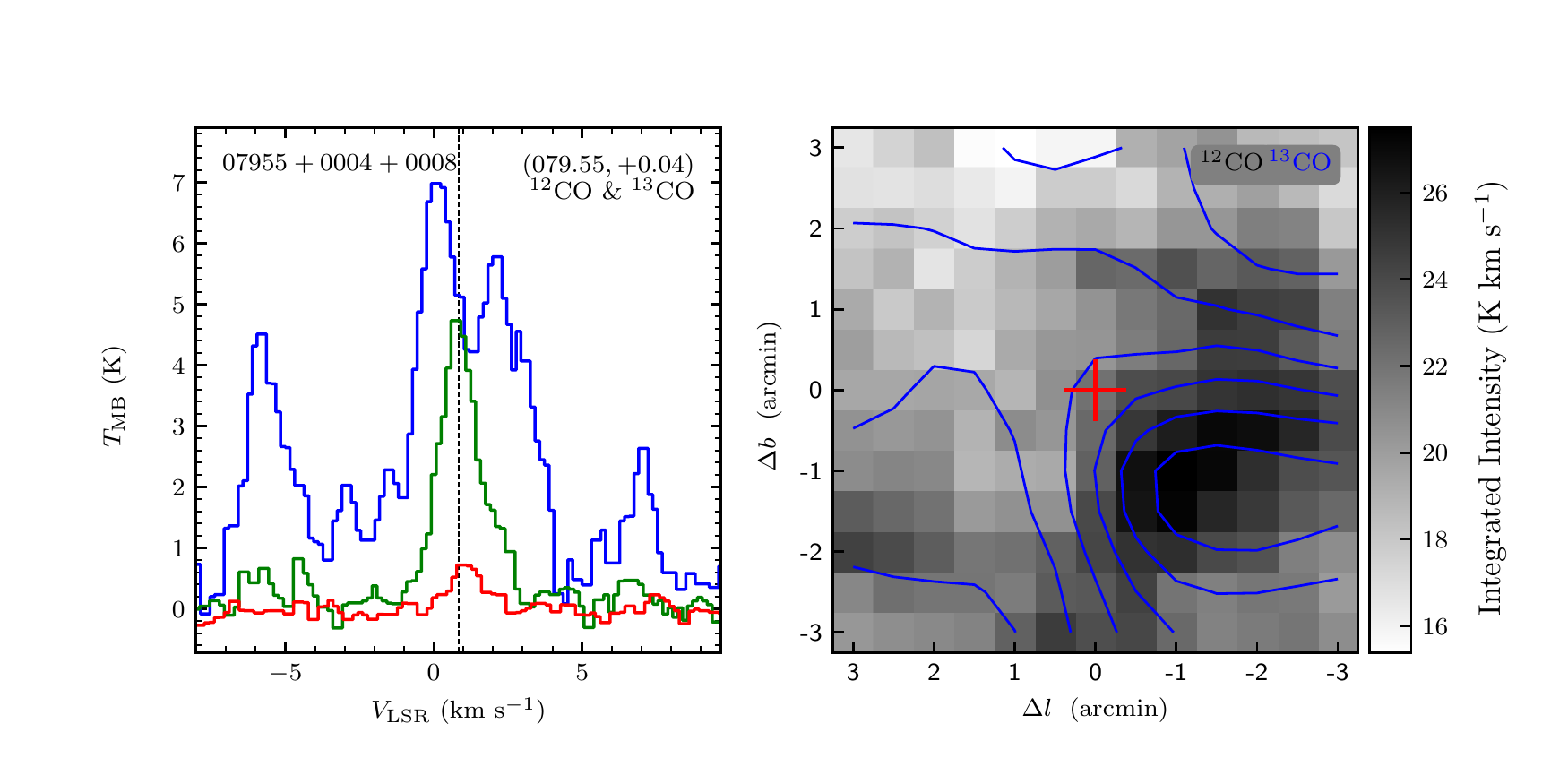}
\vspace{-0.5cm}

\includegraphics[width=9.0cm,angle=0]{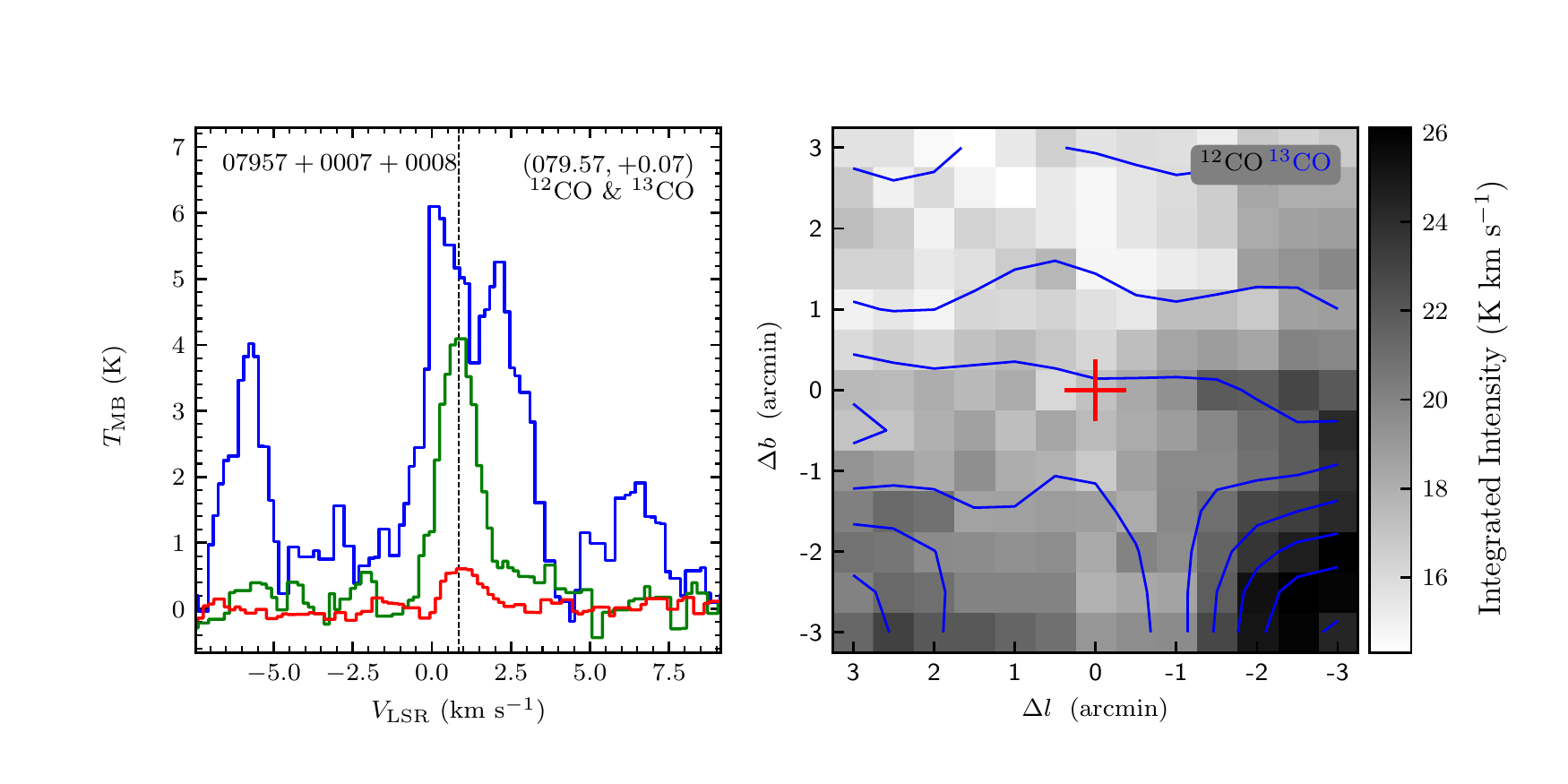}
\includegraphics[width=9.0cm,angle=0]{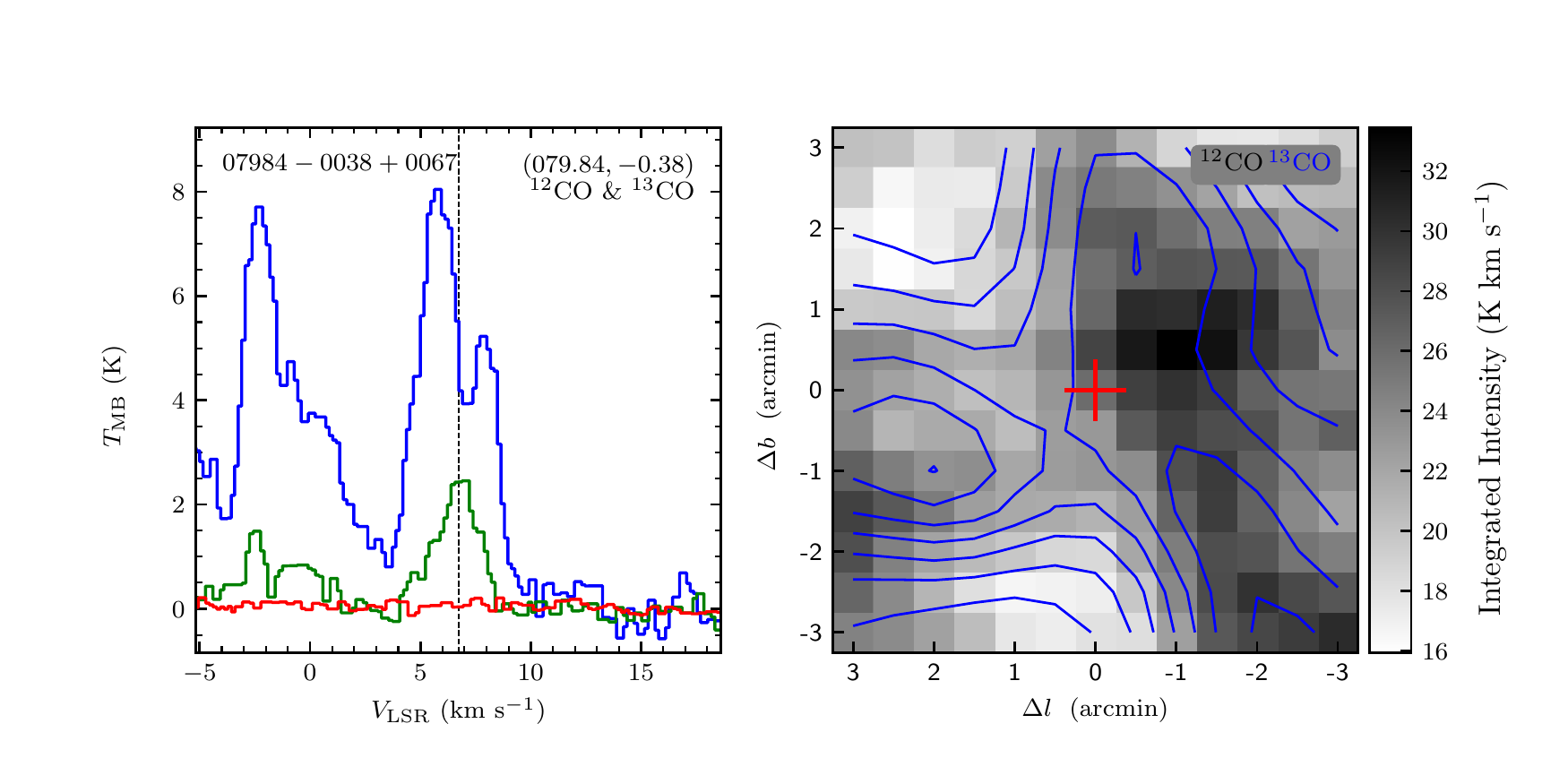}
\vspace{-0.5cm}

\includegraphics[width=9.0cm,angle=0]{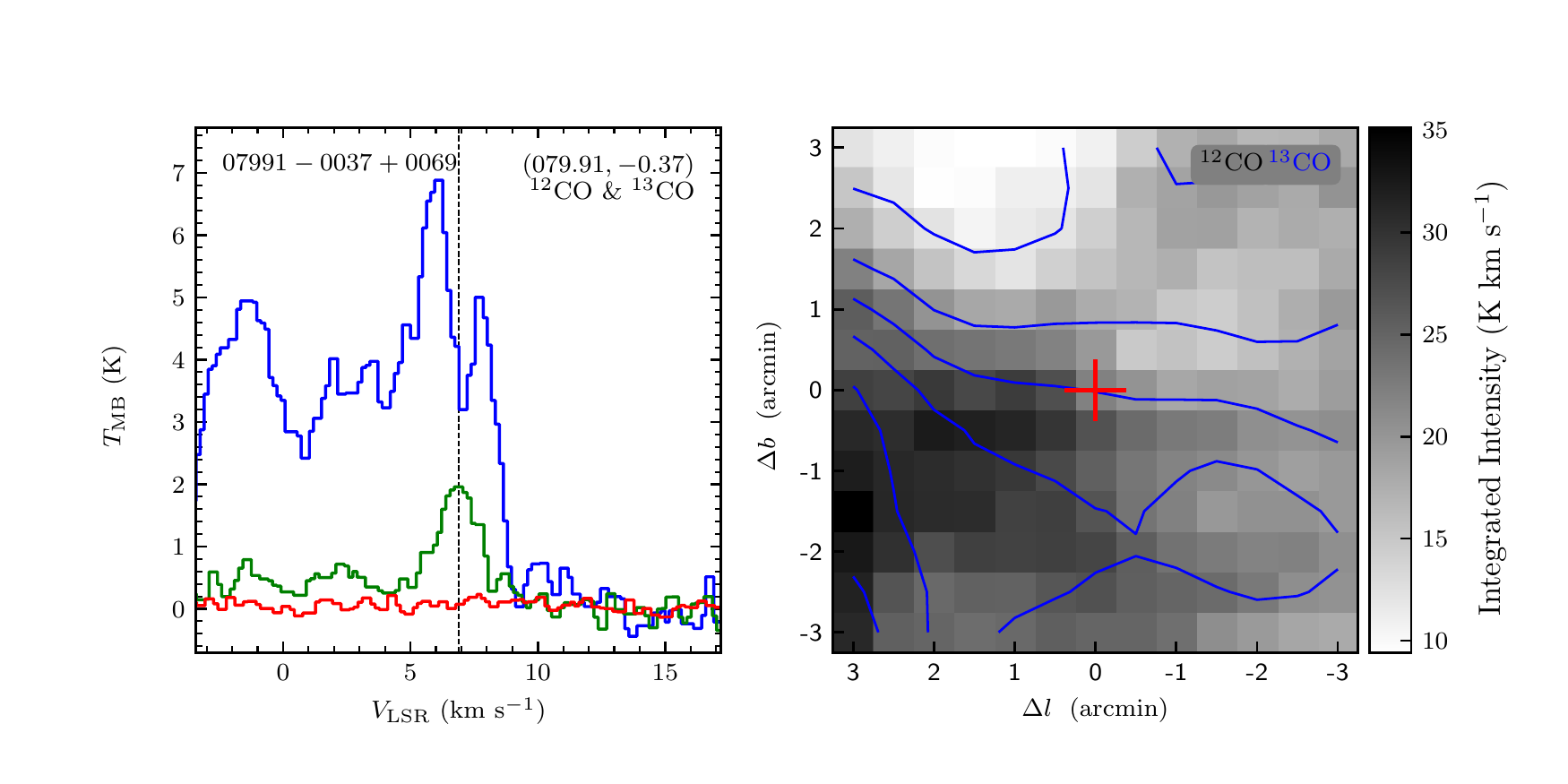}
\includegraphics[width=9.0cm,angle=0]{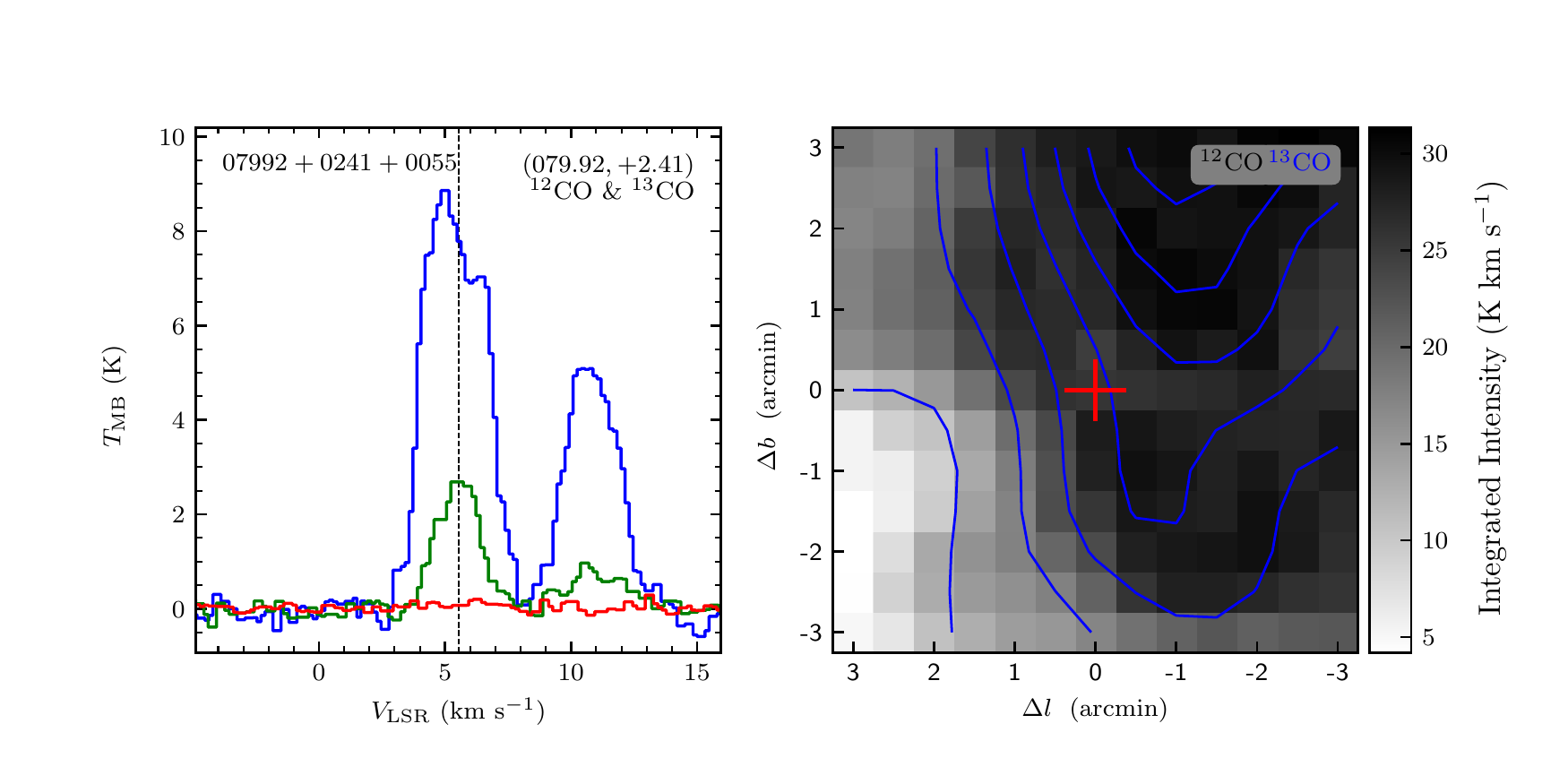}
\vspace{-0.5cm}

\includegraphics[width=9.0cm,angle=0]{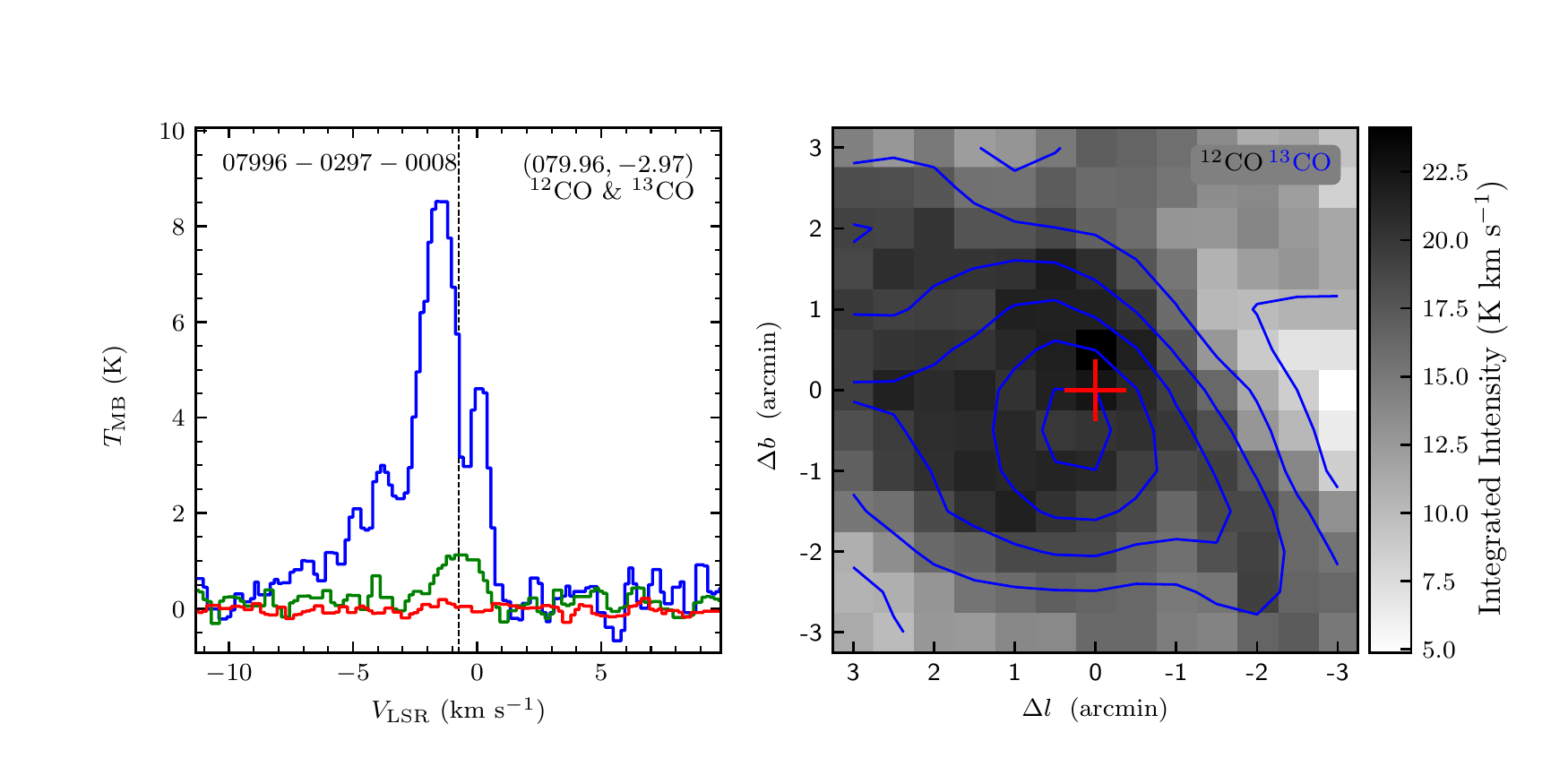}
\includegraphics[width=9.0cm,angle=0]{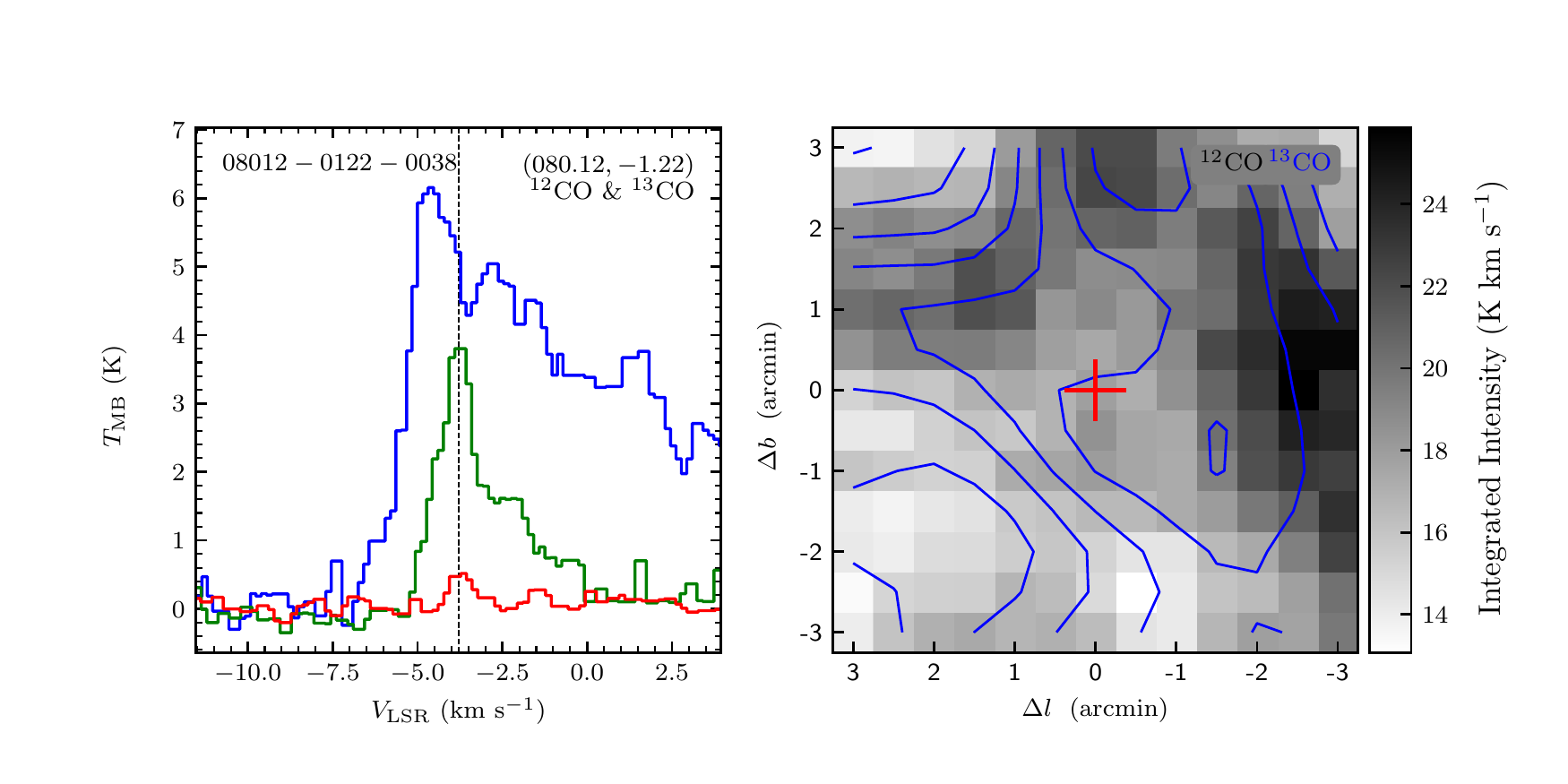}
\vspace{-0.5cm}

\includegraphics[width=9.0cm,angle=0]{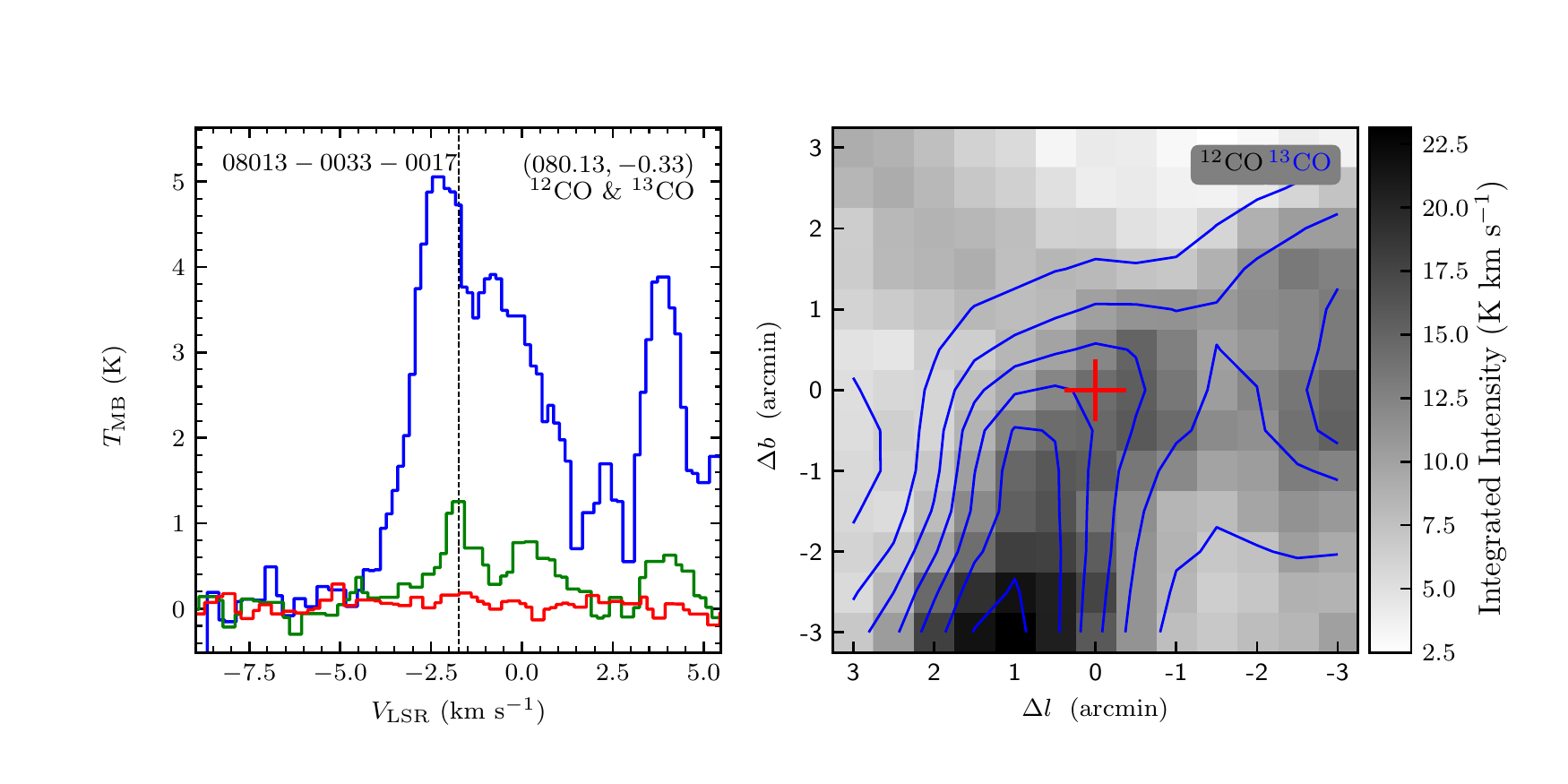}
\includegraphics[width=9.0cm,angle=0]{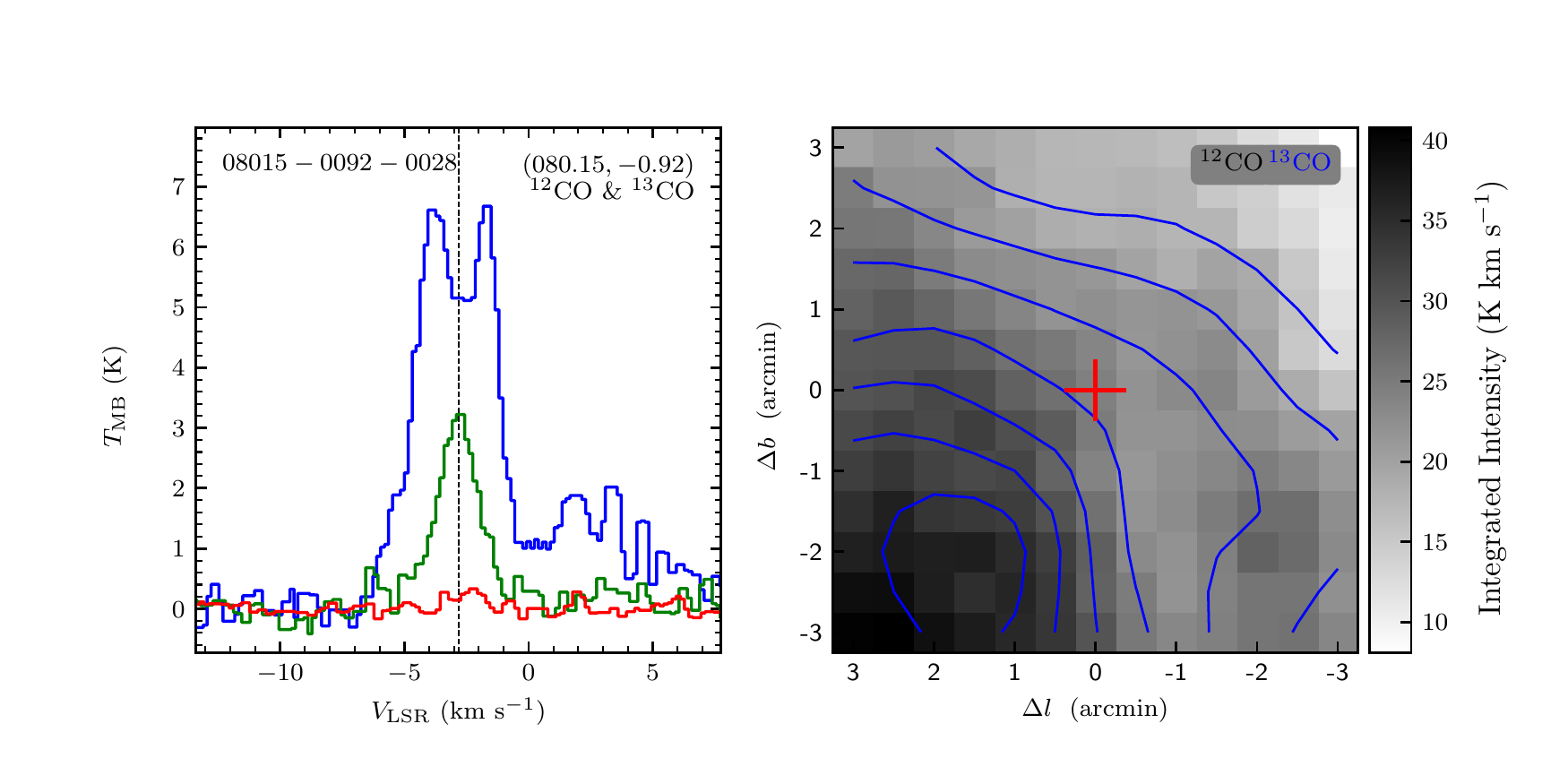}
\end{figure}
\clearpage

\begin{figure}
\includegraphics[width=9.0cm,angle=0]{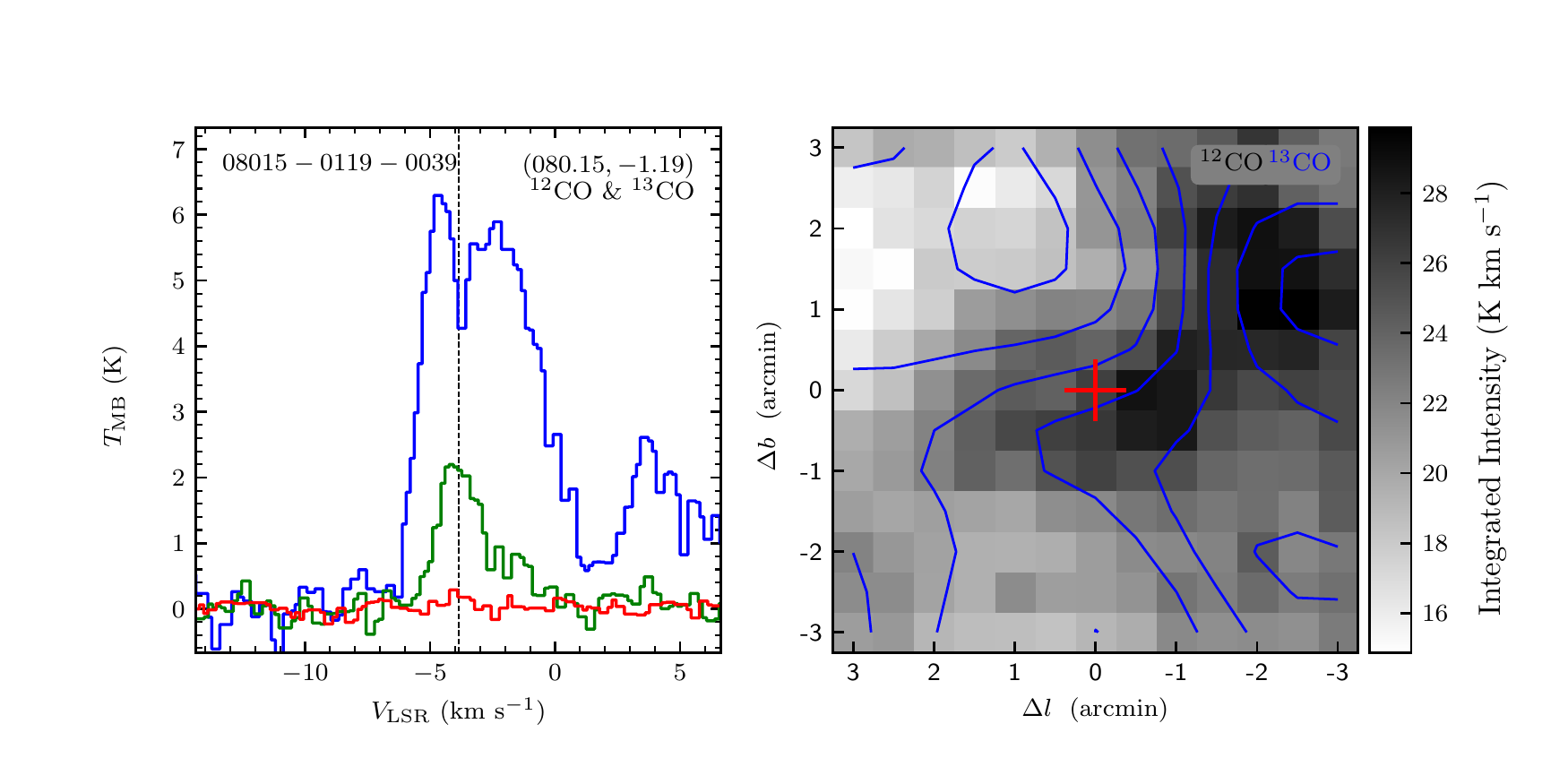}
\includegraphics[width=9.0cm,angle=0]{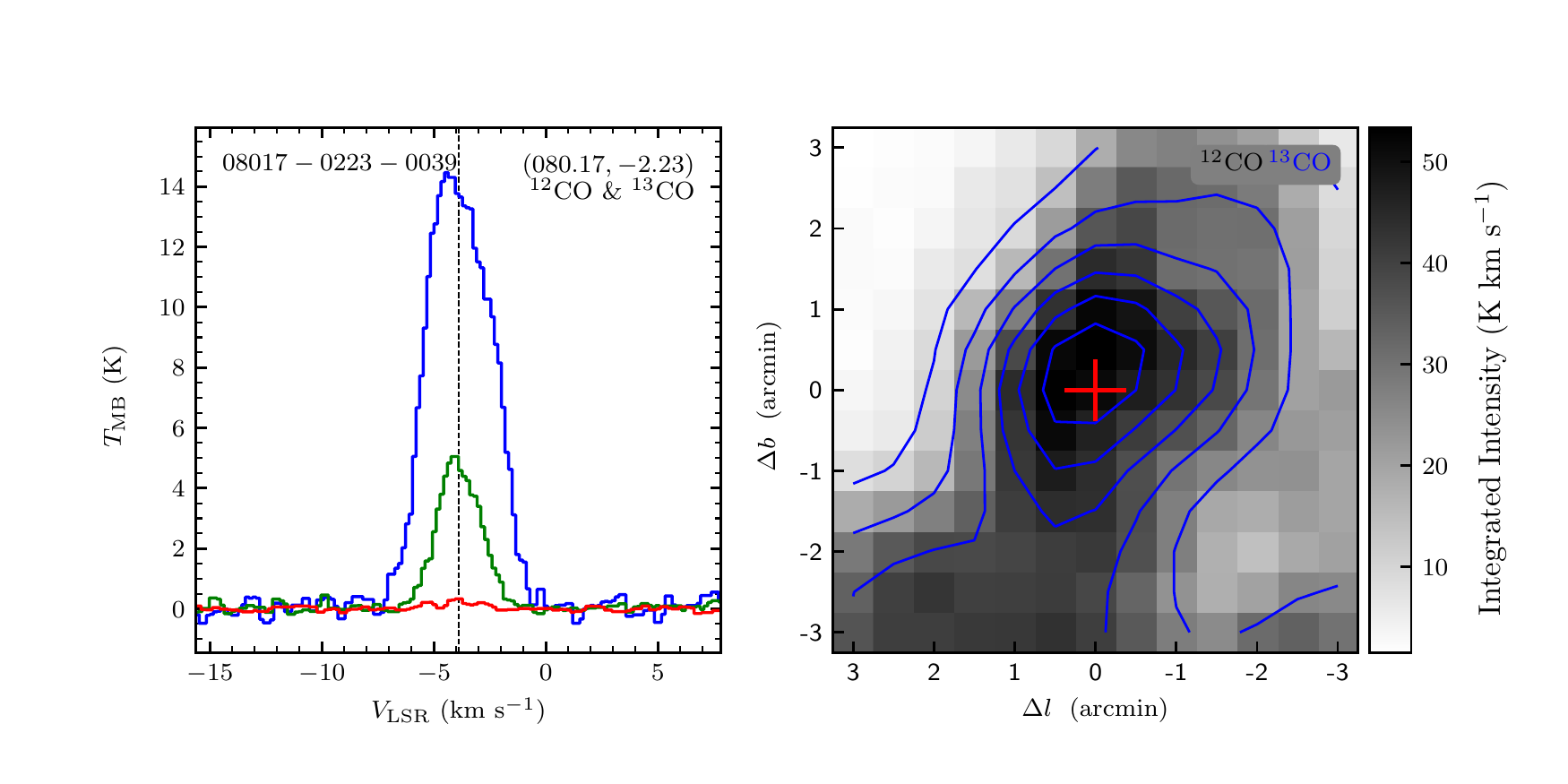}
\vspace{-0.5cm}

\includegraphics[width=9.0cm,angle=0]{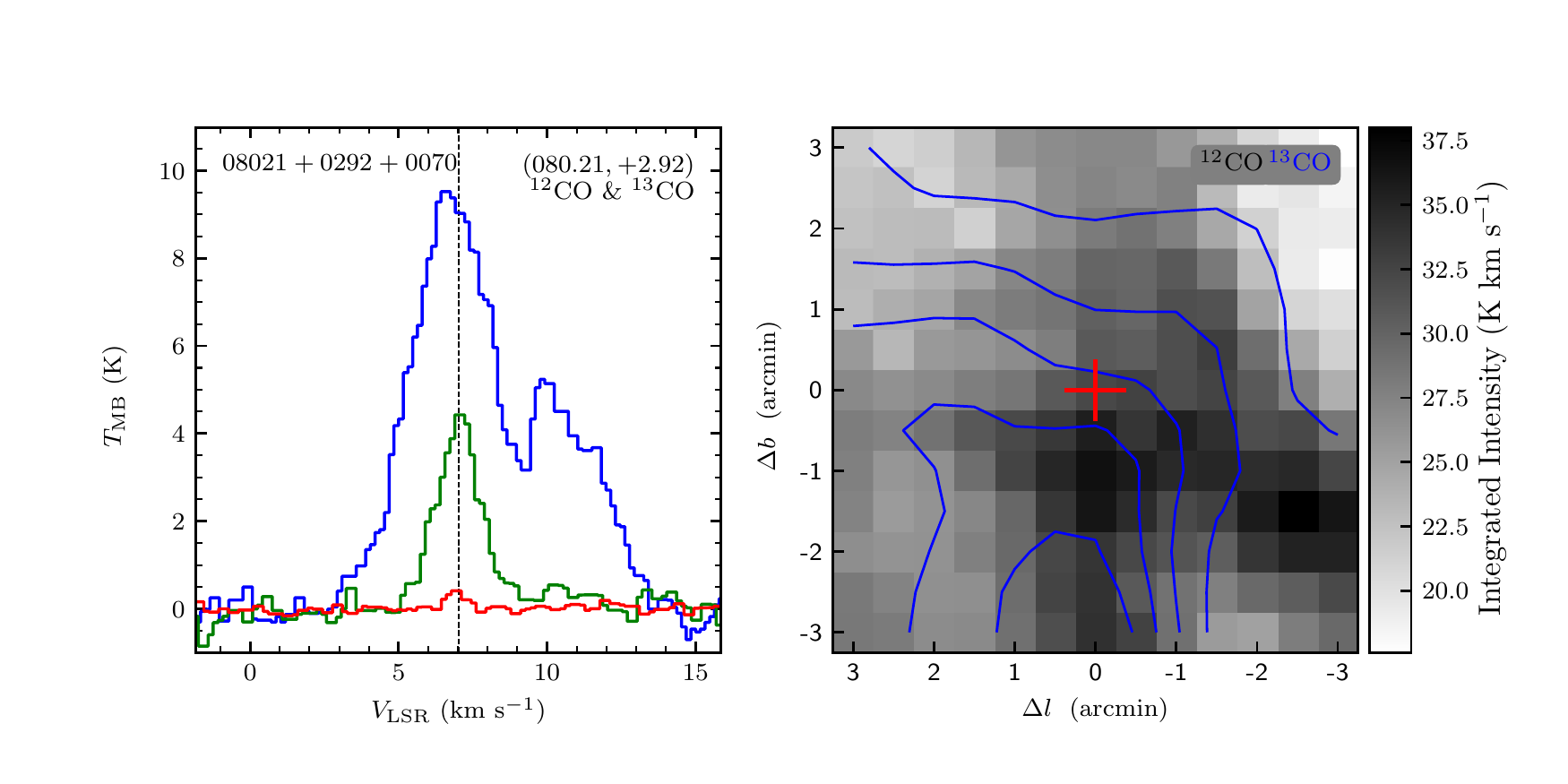}
\includegraphics[width=9.0cm,angle=0]{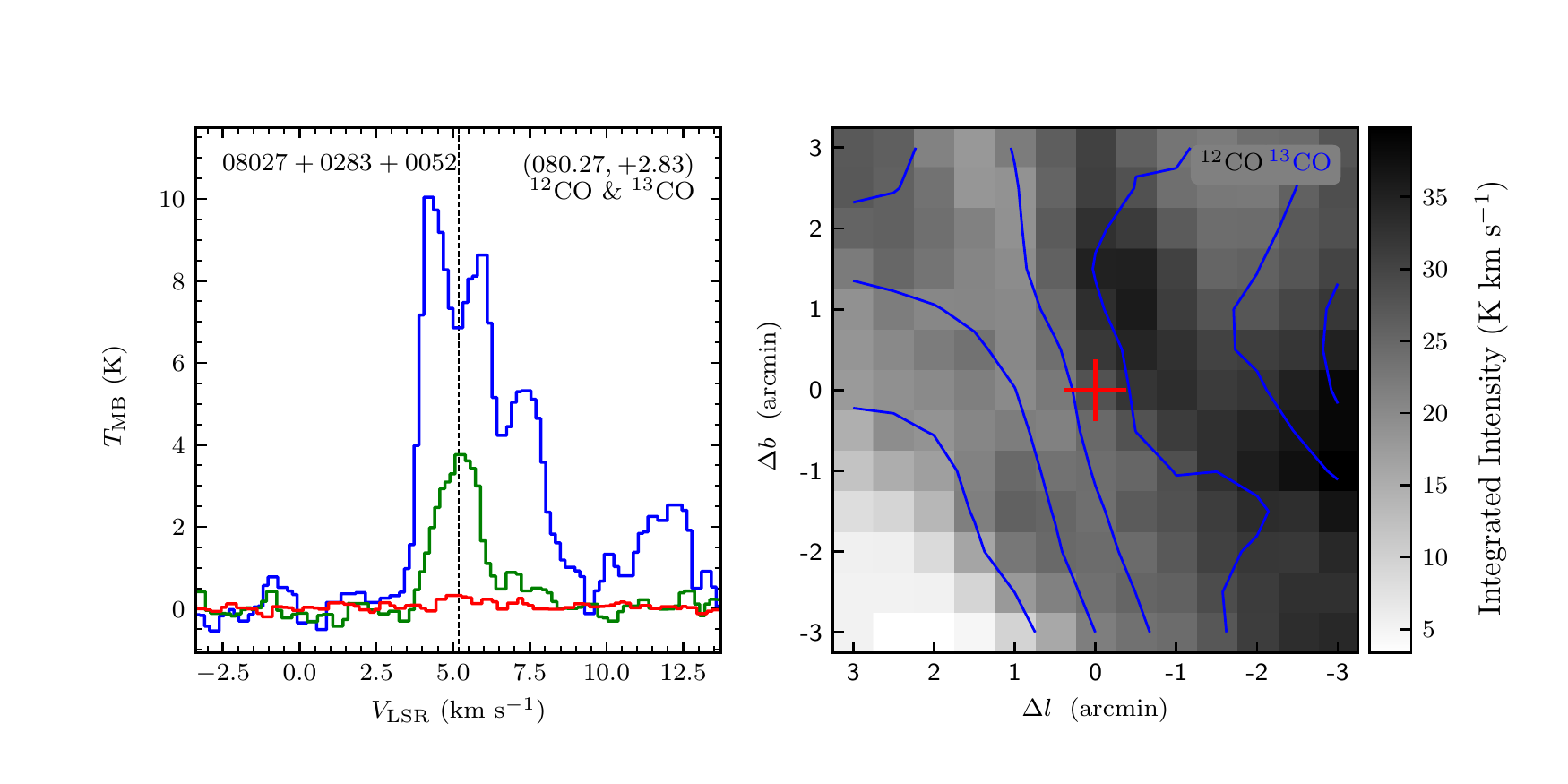}
\vspace{-0.5cm}

\includegraphics[width=9.0cm,angle=0]{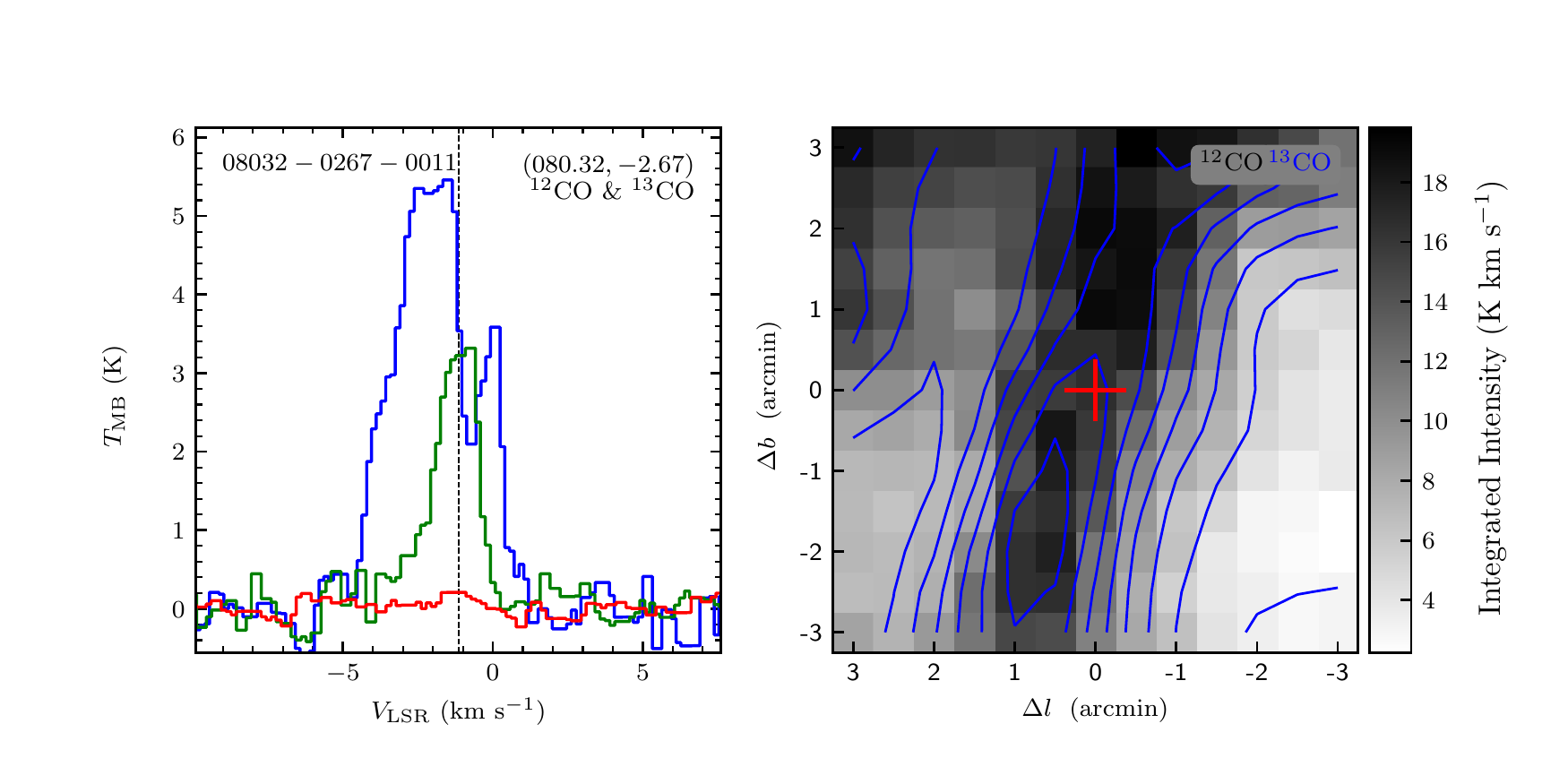}
\includegraphics[width=9.0cm,angle=0]{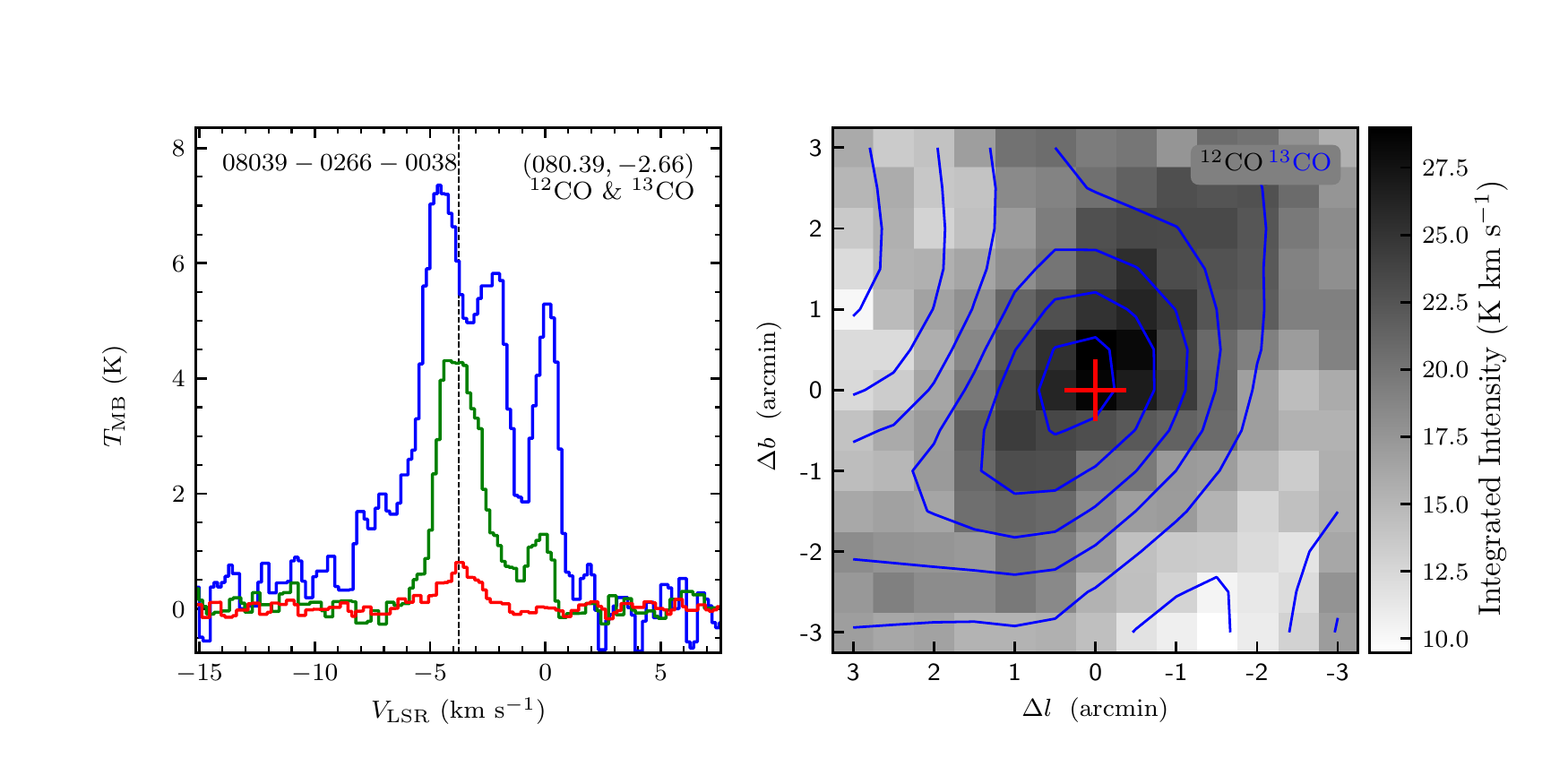}
\vspace{-0.5cm}

\includegraphics[width=9.0cm,angle=0]{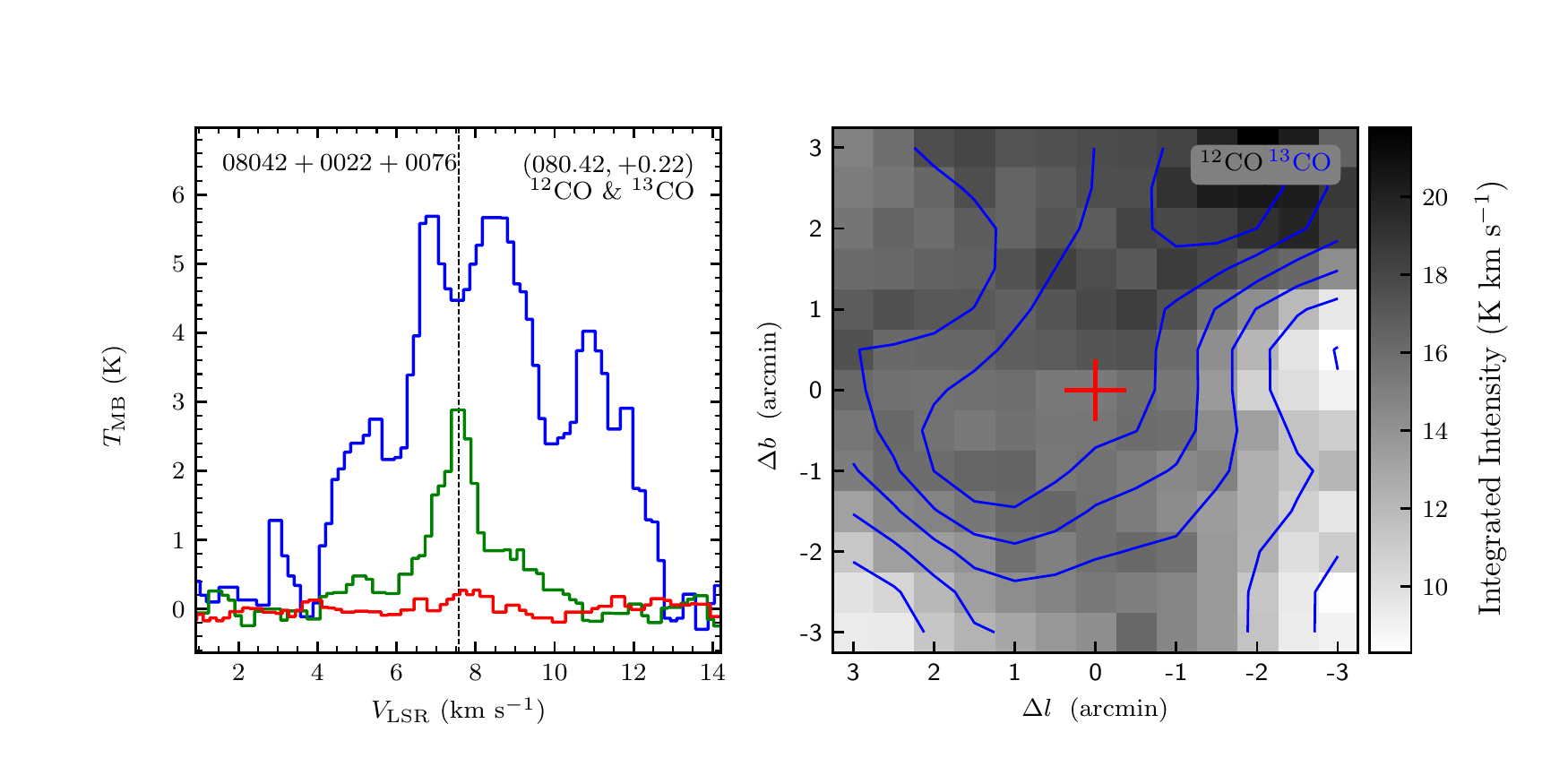}
\includegraphics[width=9.0cm,angle=0]{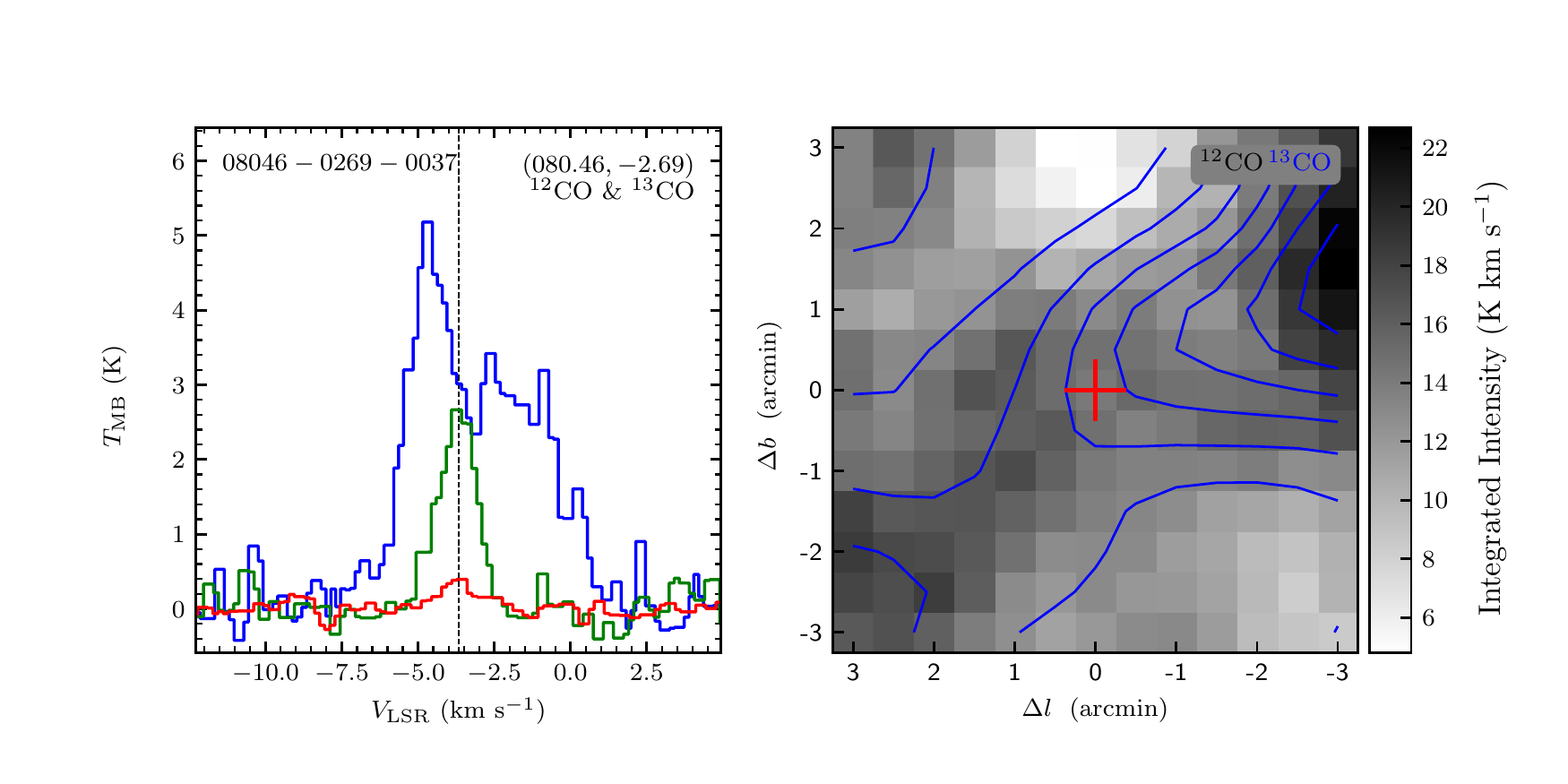}
\vspace{-0.5cm}

\includegraphics[width=9.0cm,angle=0]{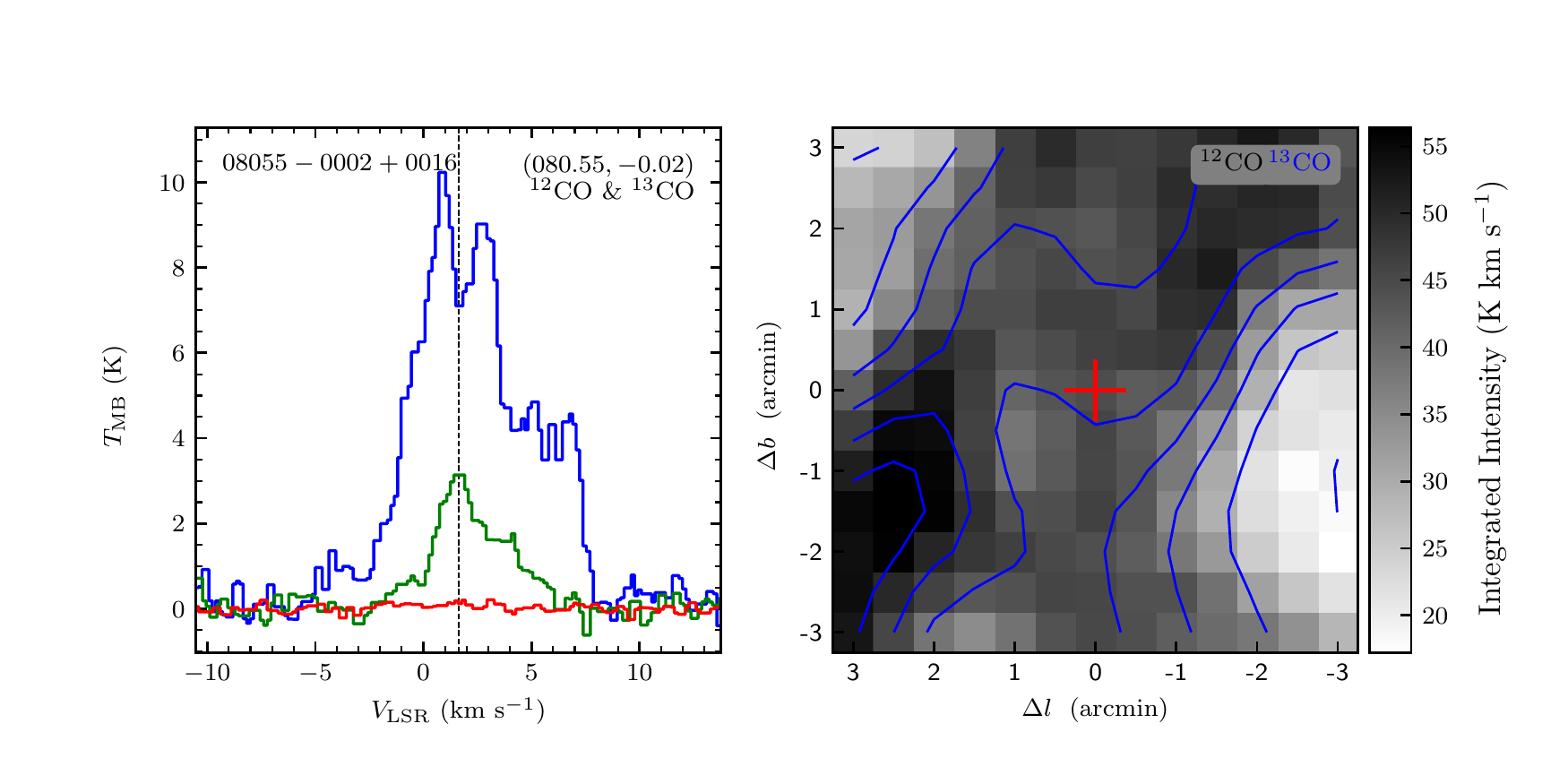}
\includegraphics[width=9.0cm,angle=0]{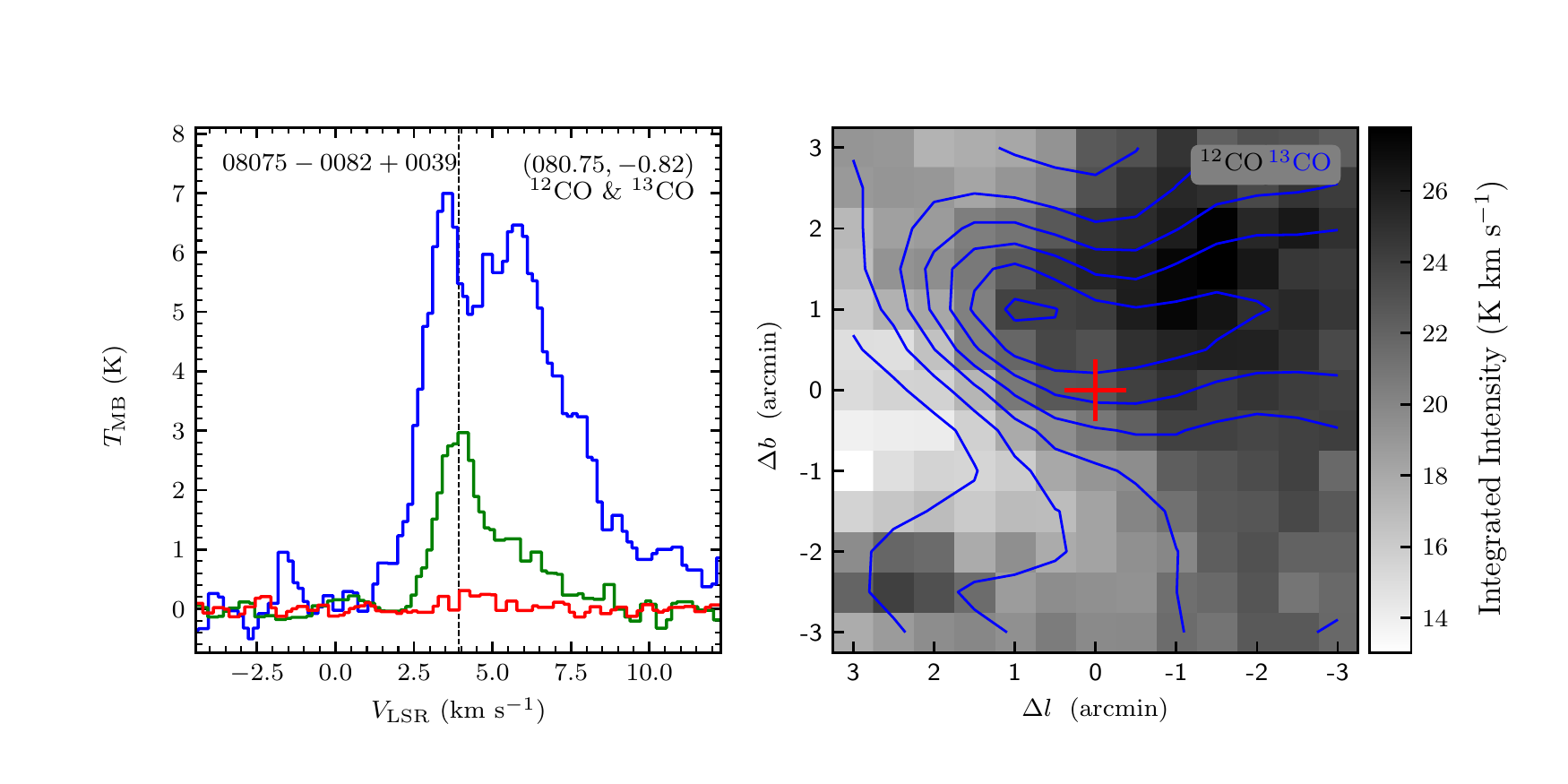}
\end{figure}
\clearpage

\begin{figure}
\includegraphics[width=9.0cm,angle=0]{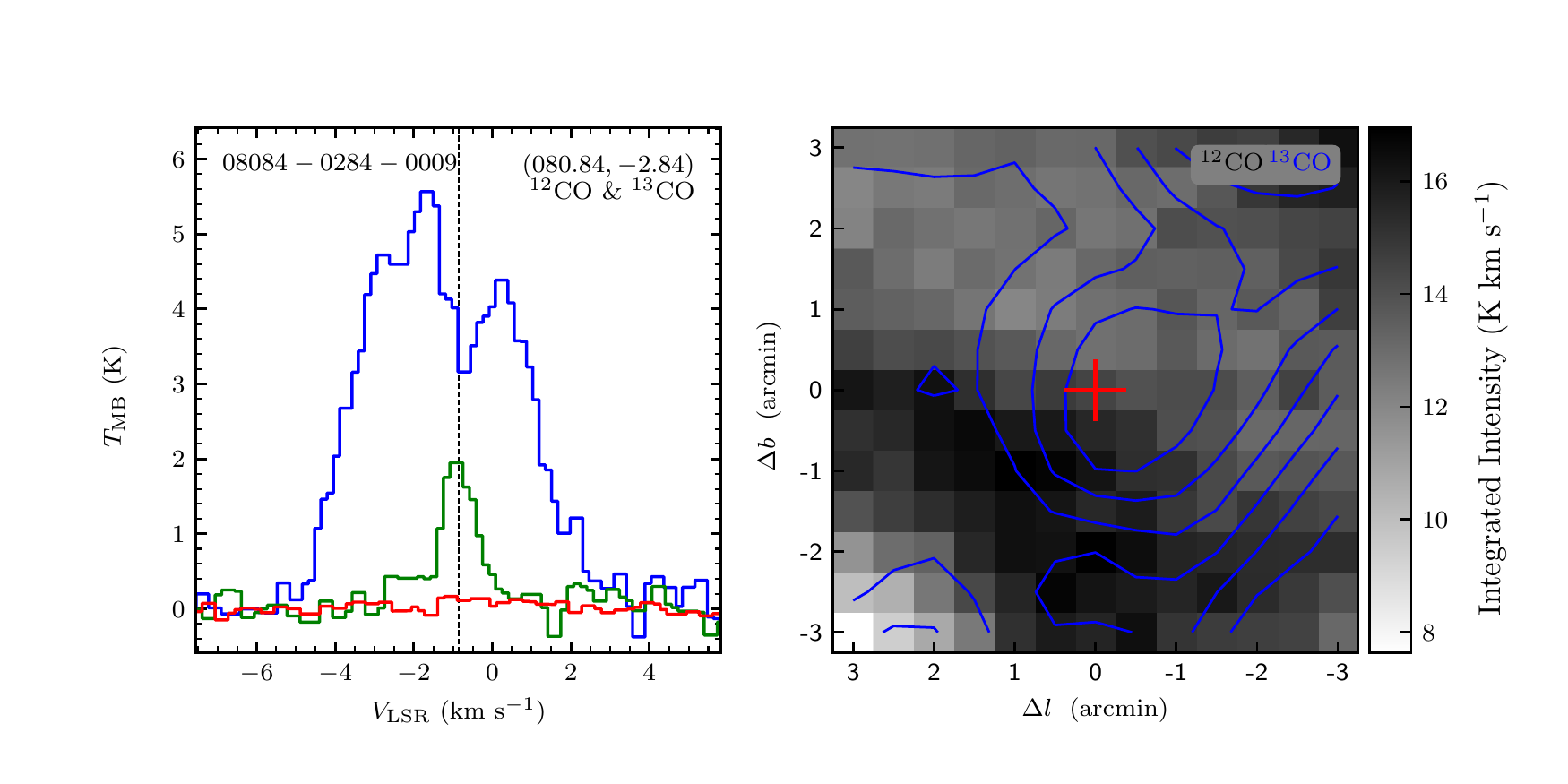}
\includegraphics[width=9.0cm,angle=0]{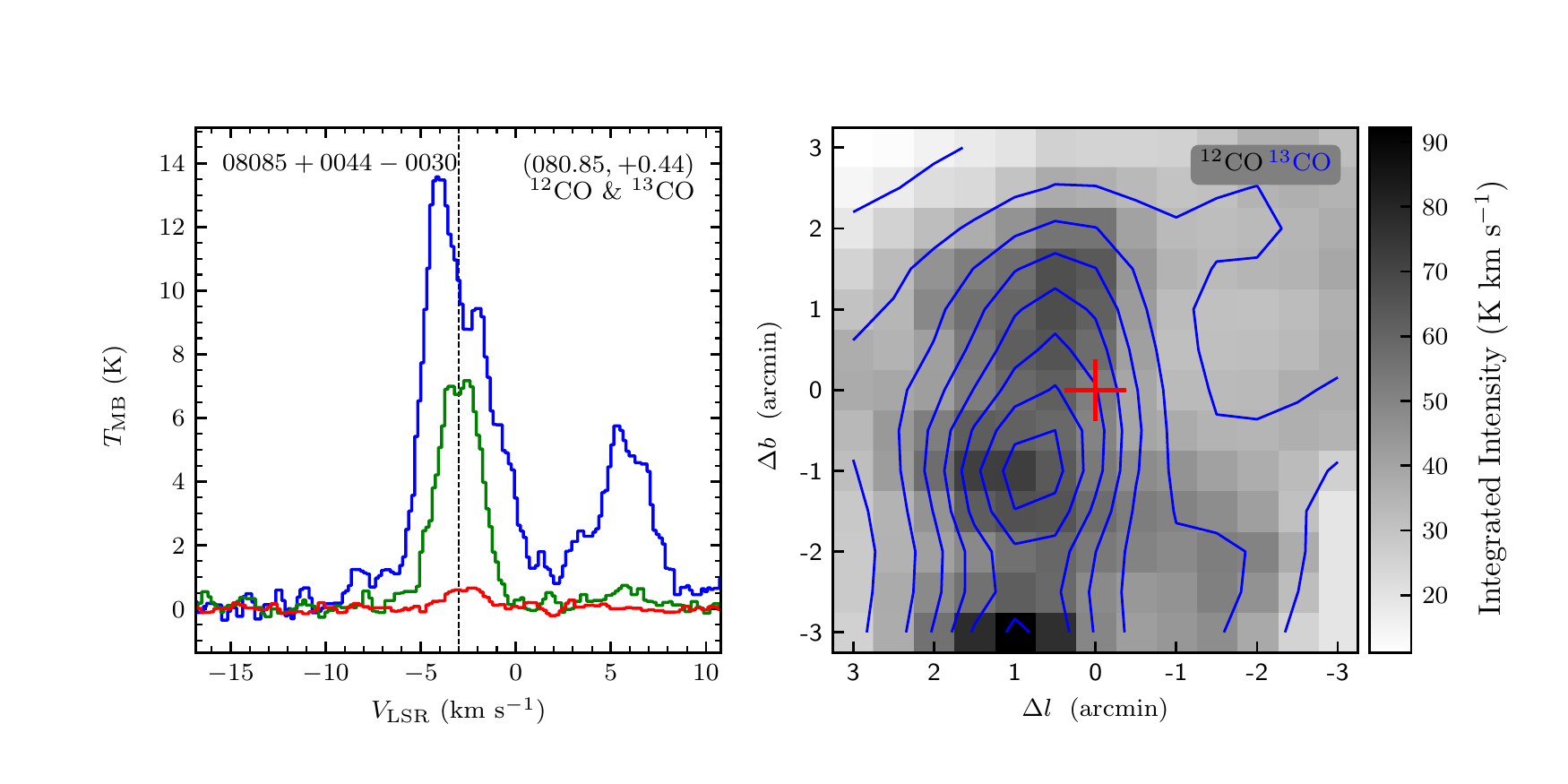}
\vspace{-0.5cm}

\includegraphics[width=9.0cm,angle=0]{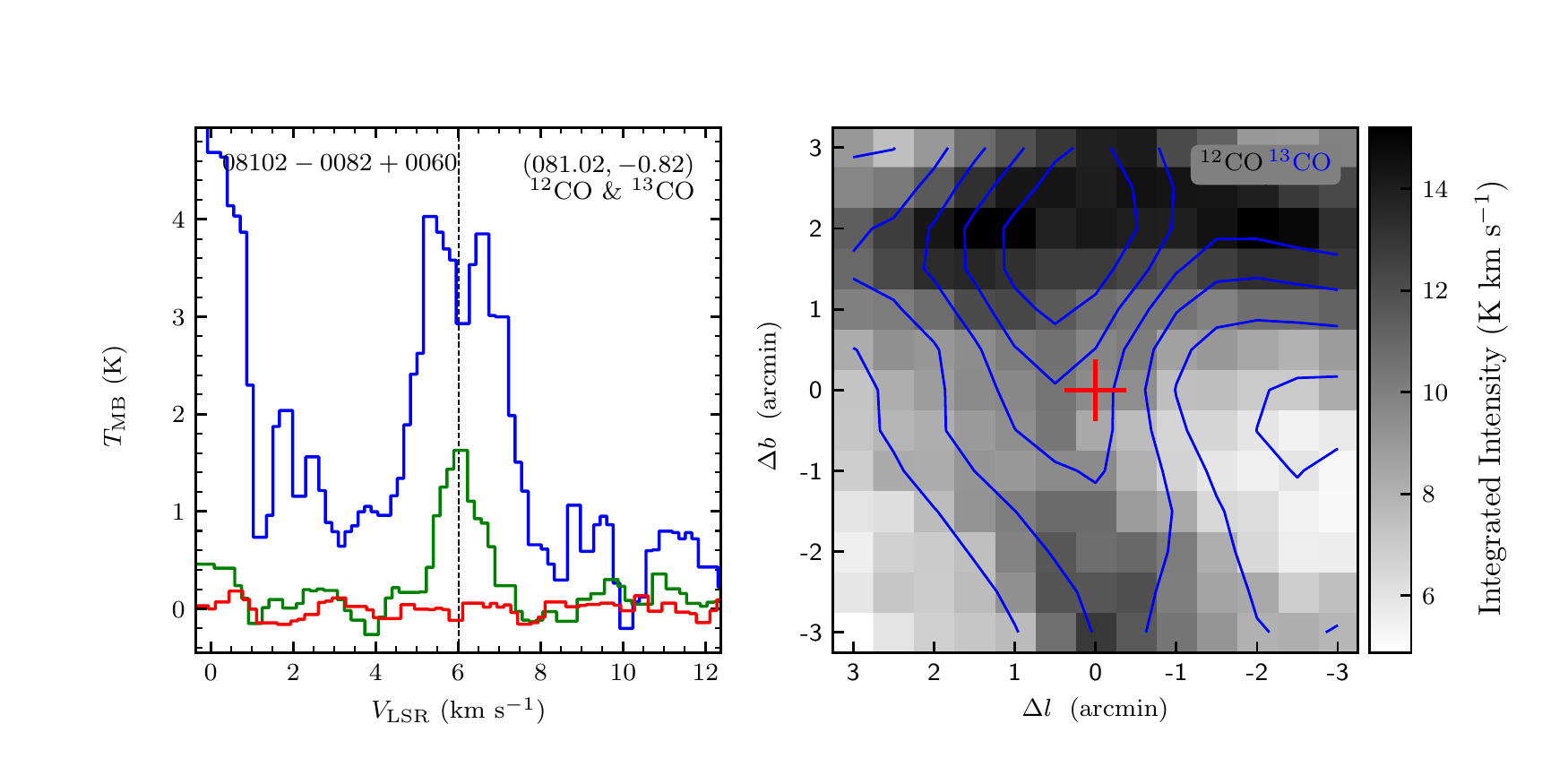}
\includegraphics[width=9.0cm,angle=0]{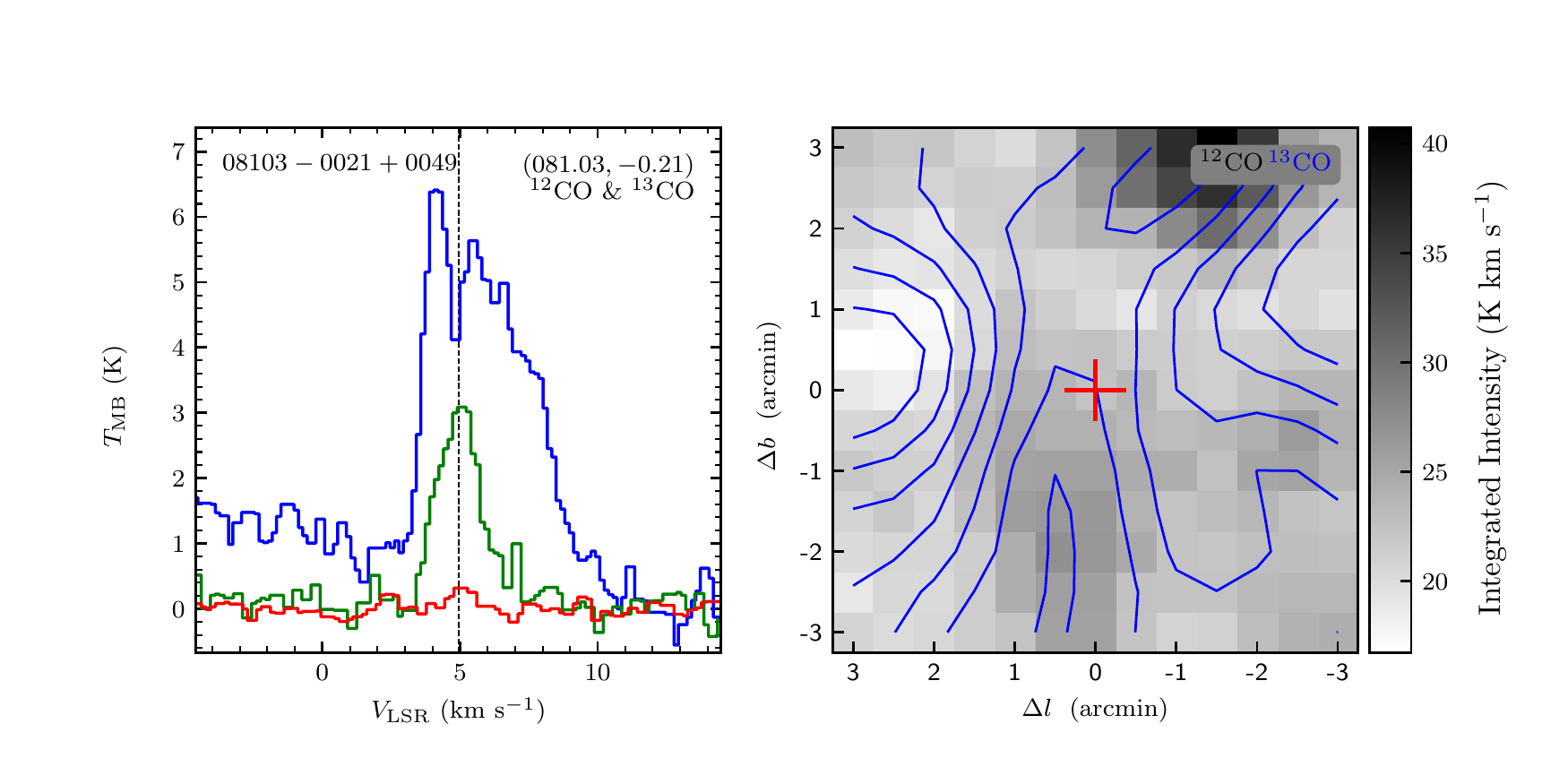}
\vspace{-0.5cm}

\includegraphics[width=9.0cm,angle=0]{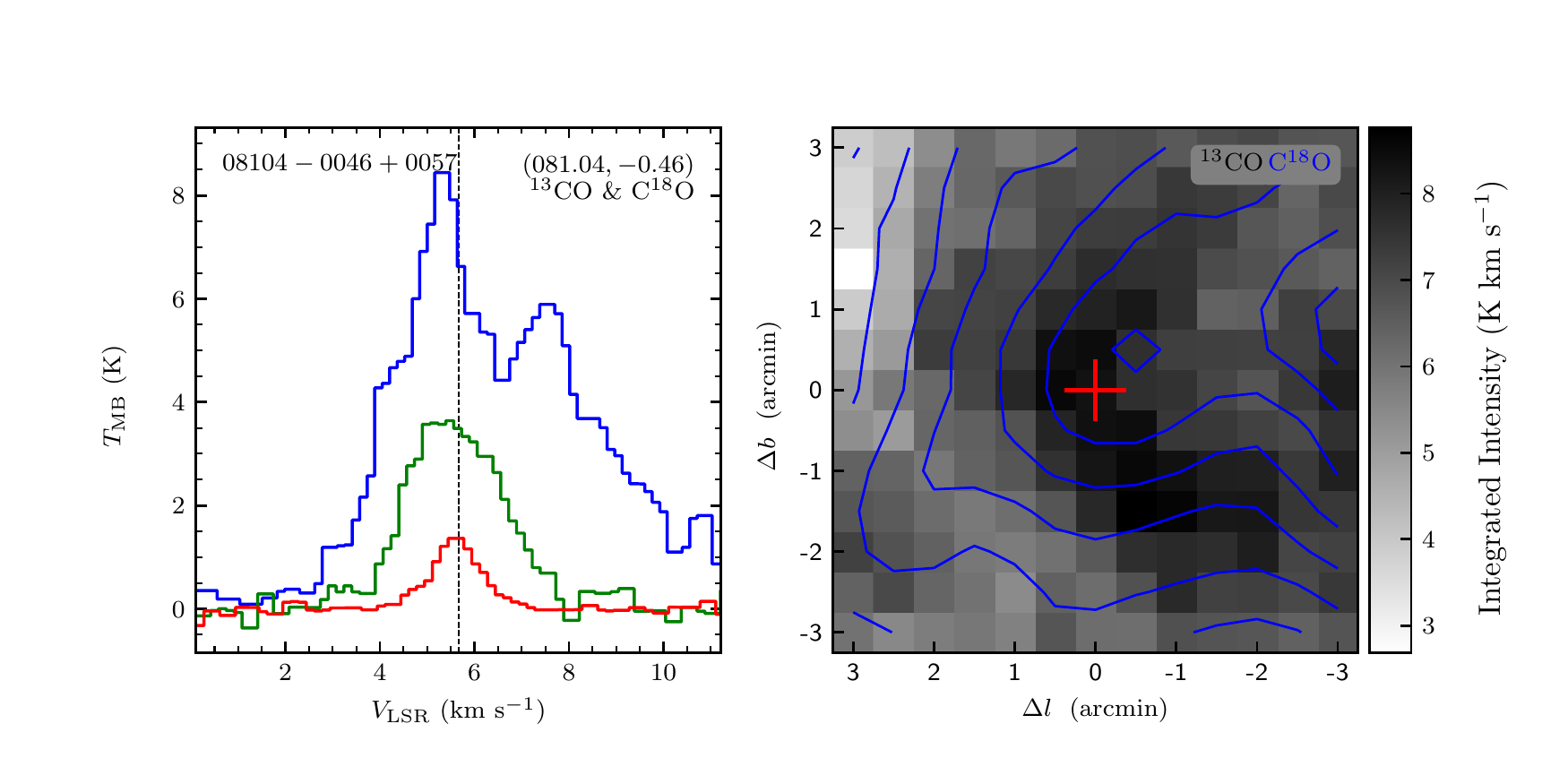}
\includegraphics[width=9.0cm,angle=0]{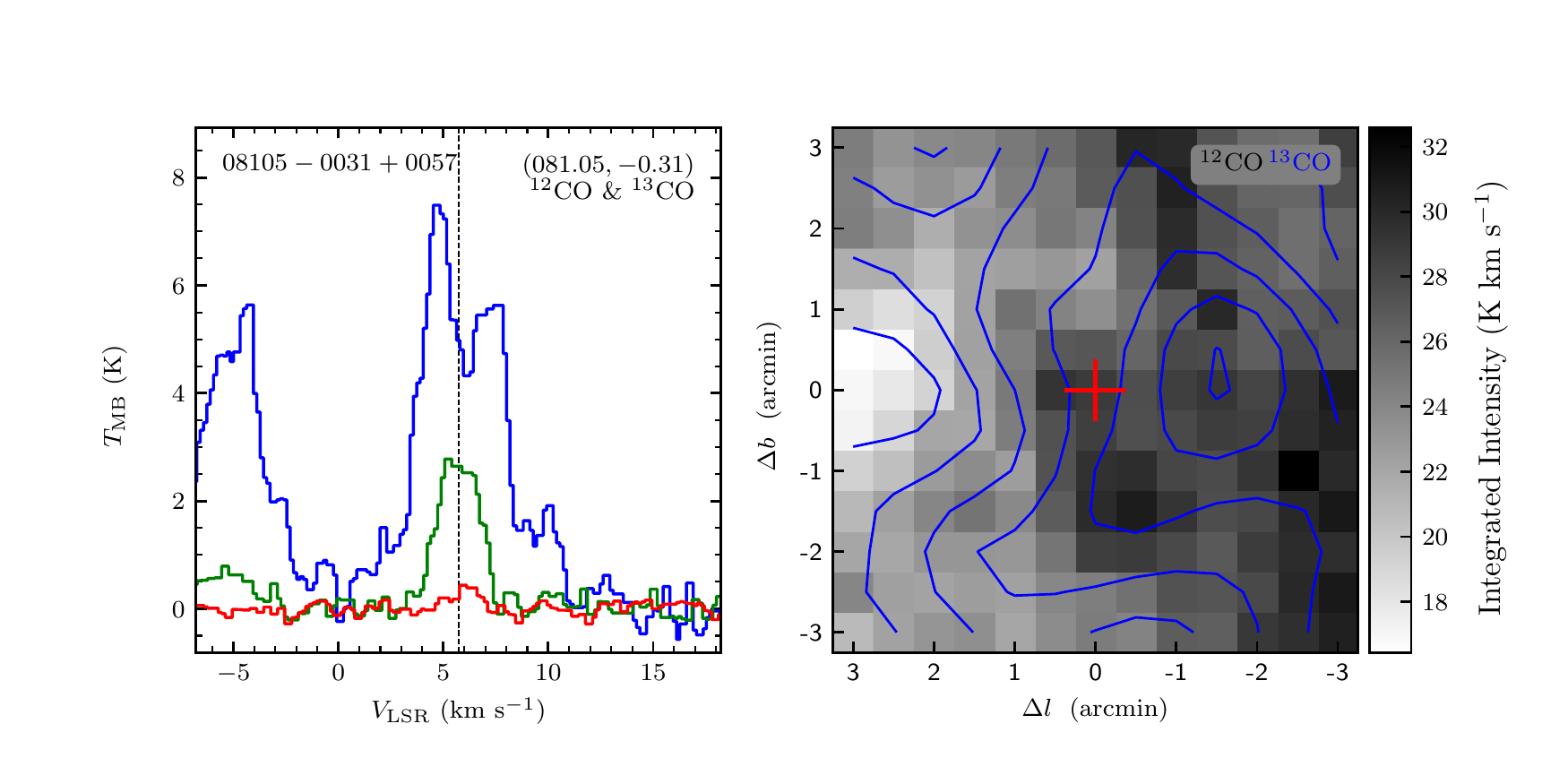}
\vspace{-0.5cm}

\includegraphics[width=9.0cm,angle=0]{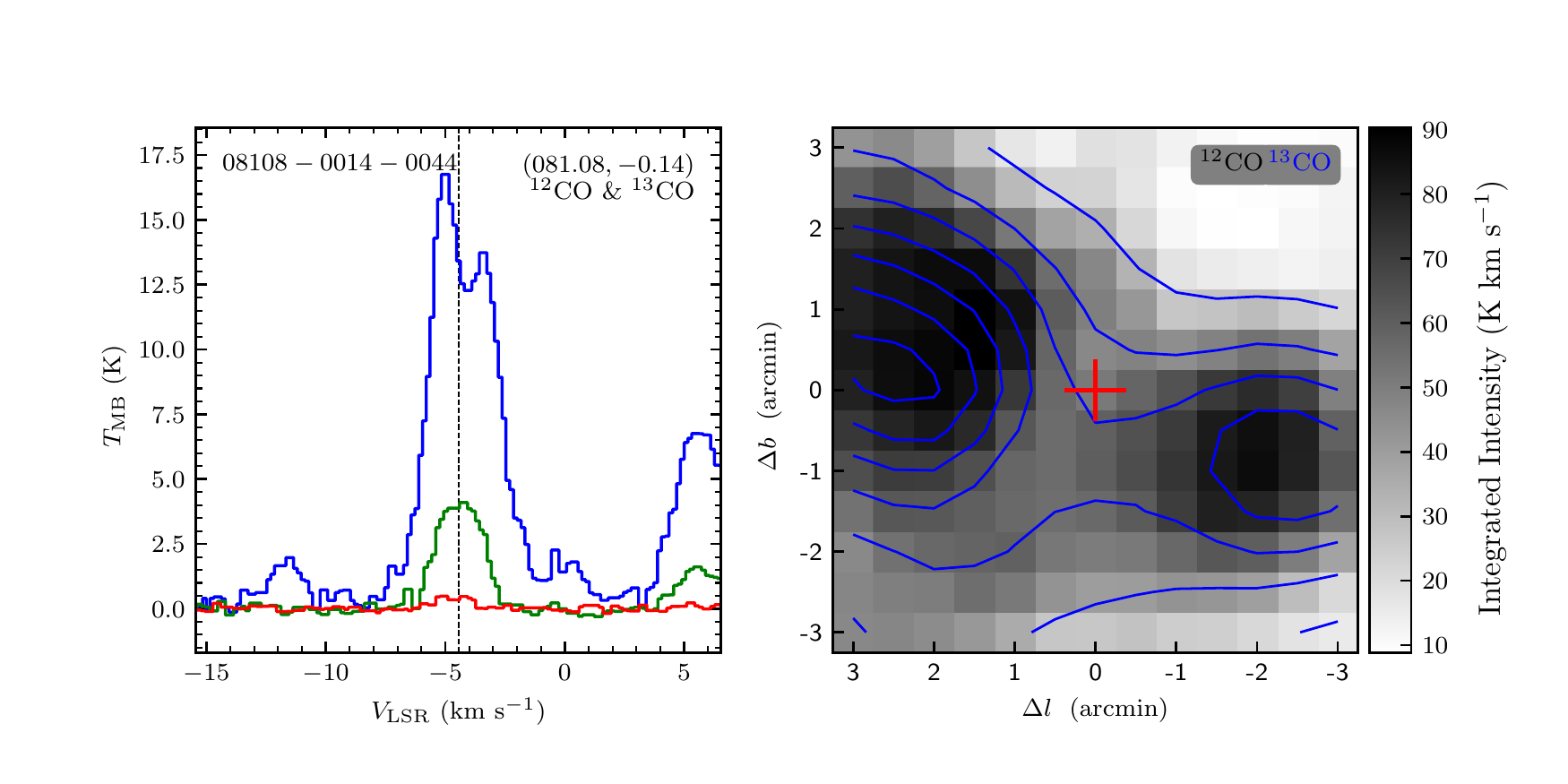}
\includegraphics[width=9.0cm,angle=0]{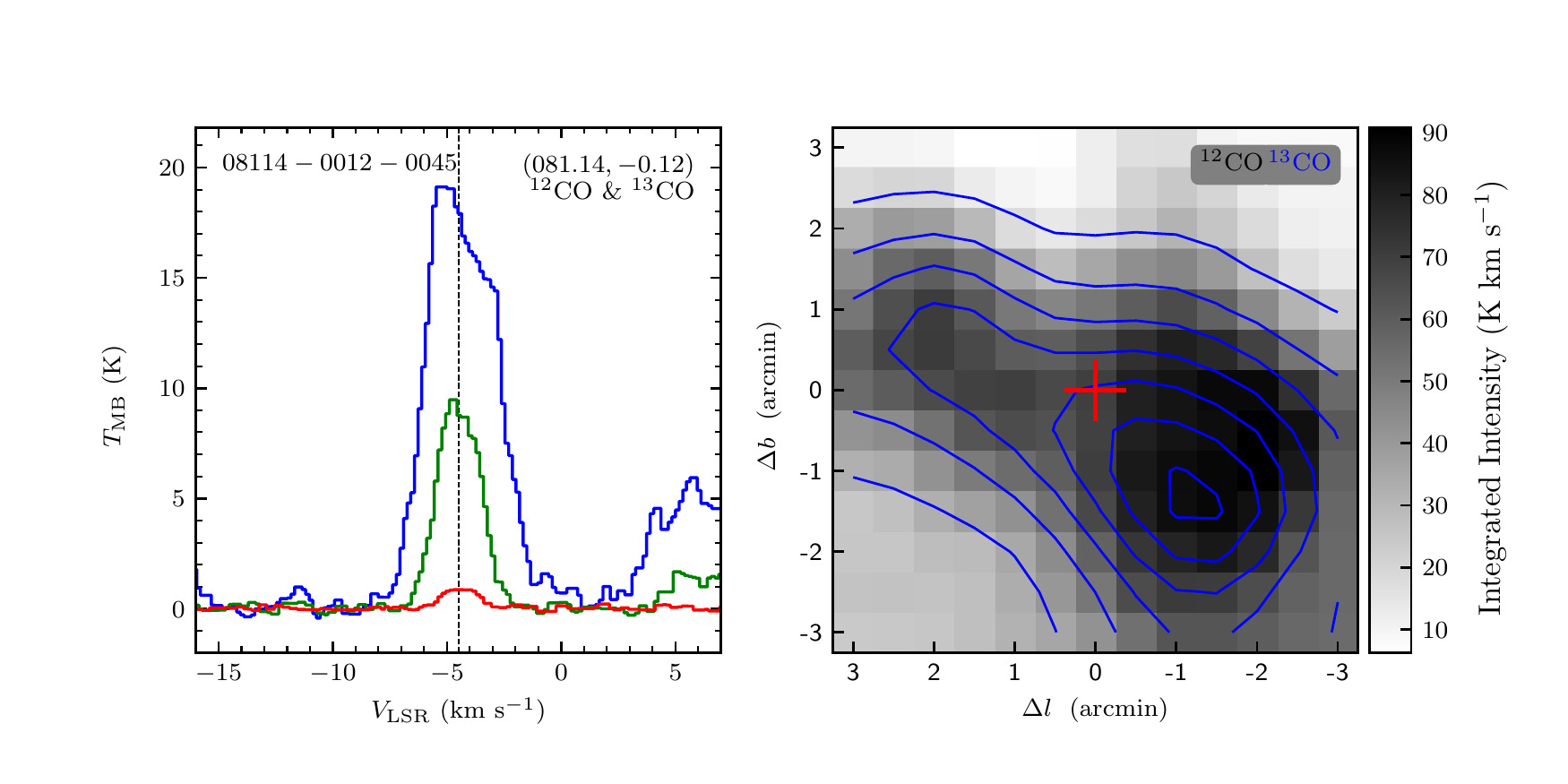}
\vspace{-0.5cm}

\includegraphics[width=9.0cm,angle=0]{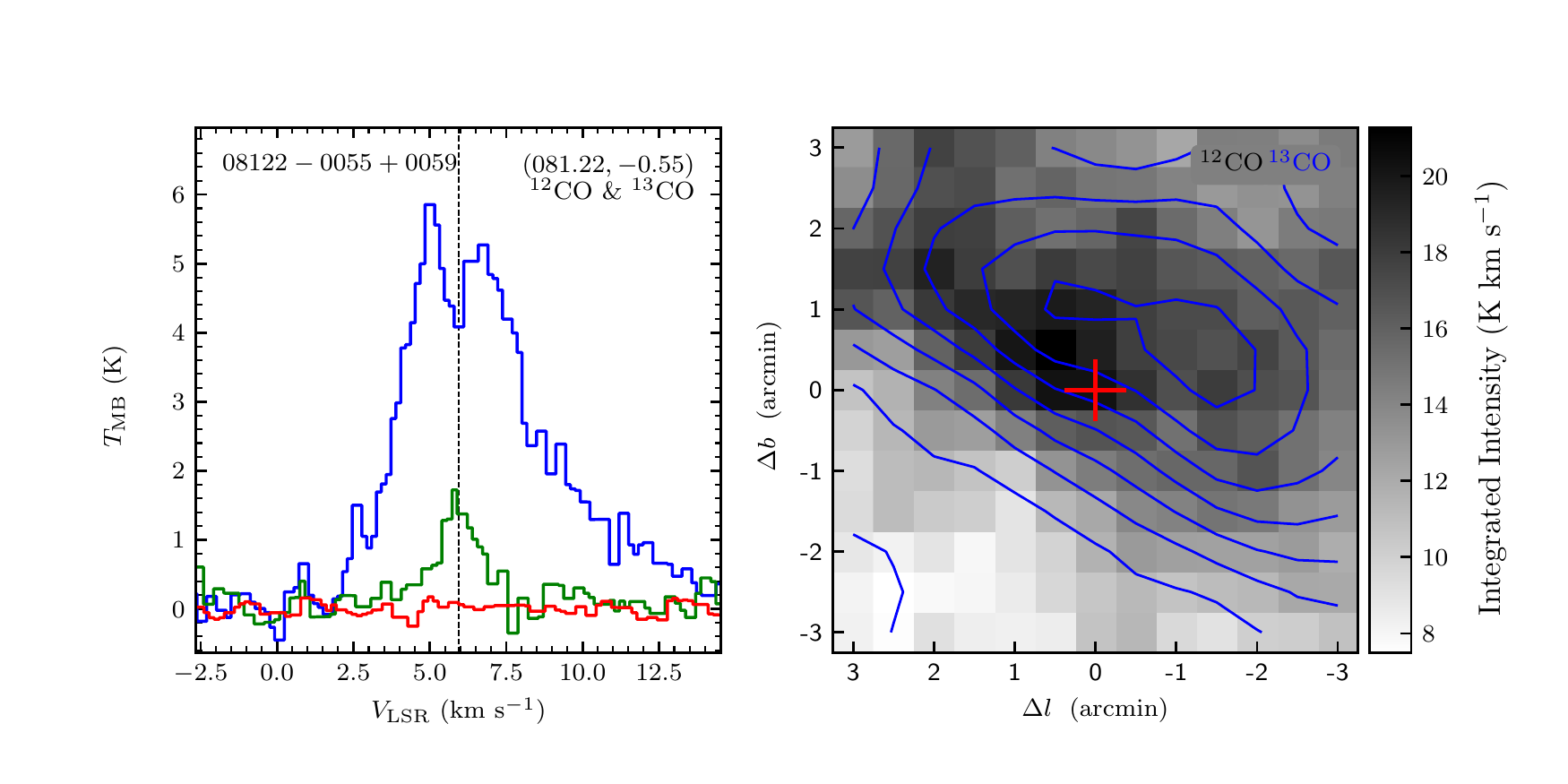}
\includegraphics[width=9.0cm,angle=0]{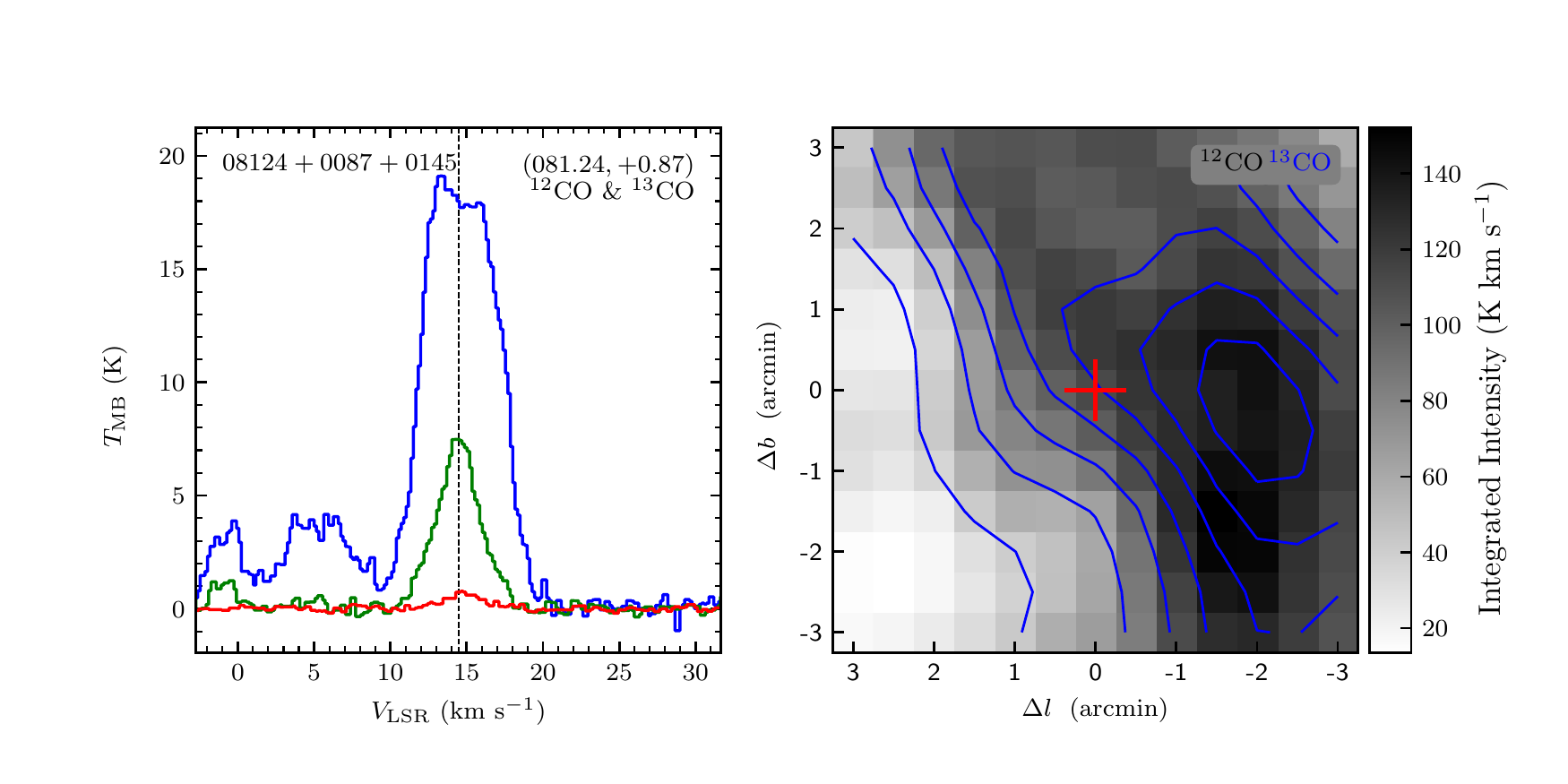}
\end{figure}
\clearpage

\begin{figure}
\includegraphics[width=9.0cm,angle=0]{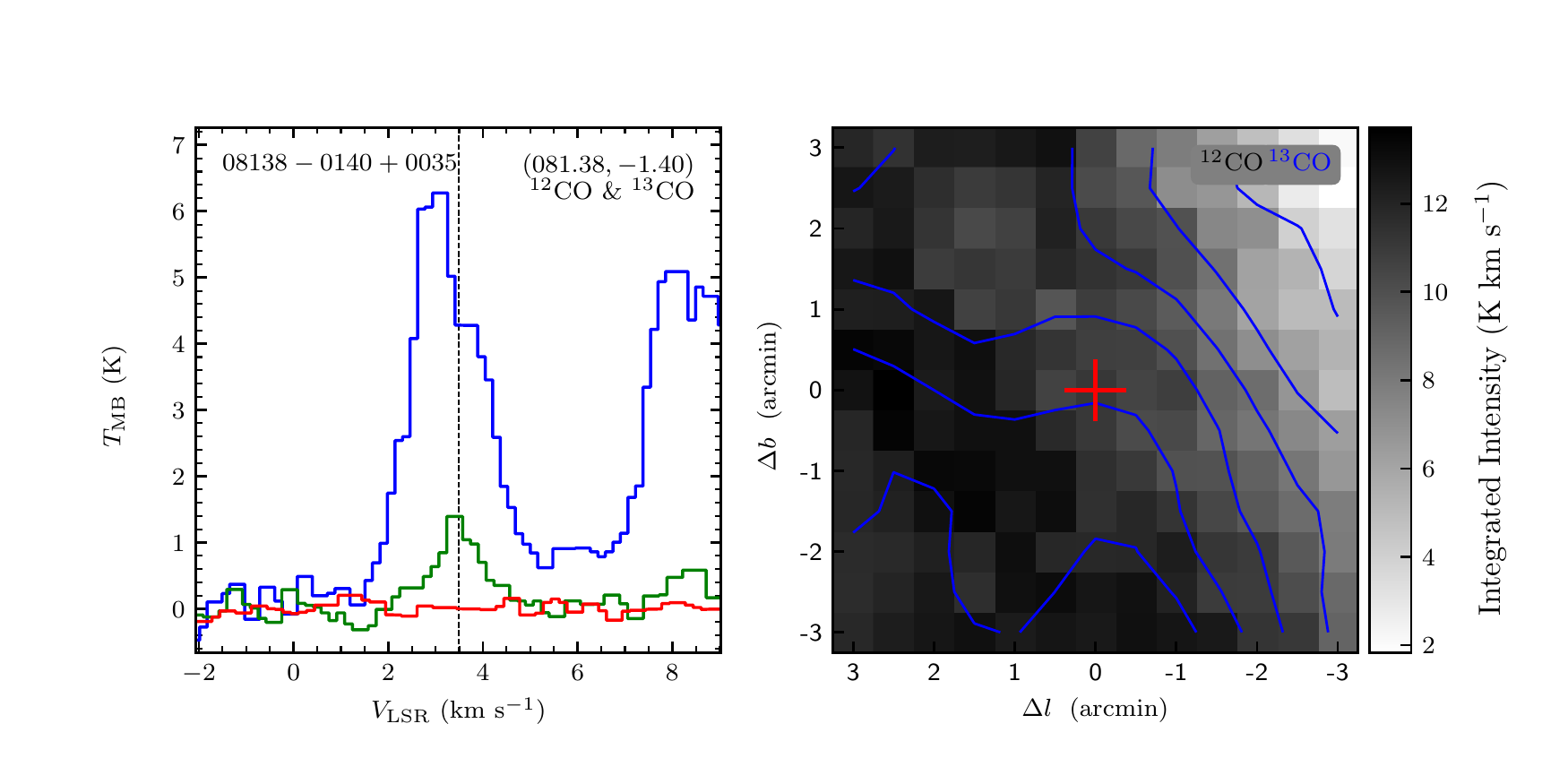}
\includegraphics[width=9.0cm,angle=0]{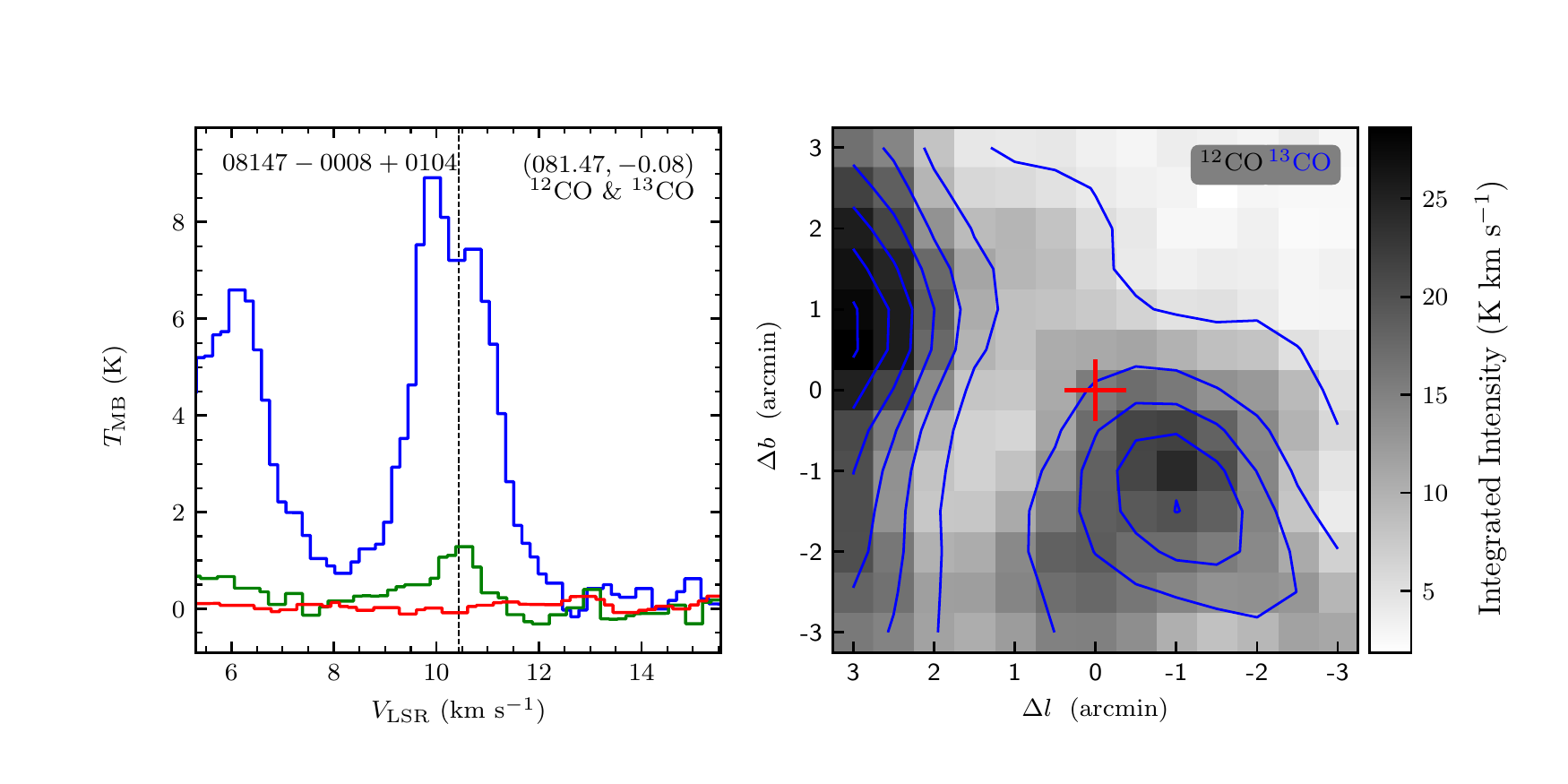}
\vspace{-0.5cm}

\includegraphics[width=9.0cm,angle=0]{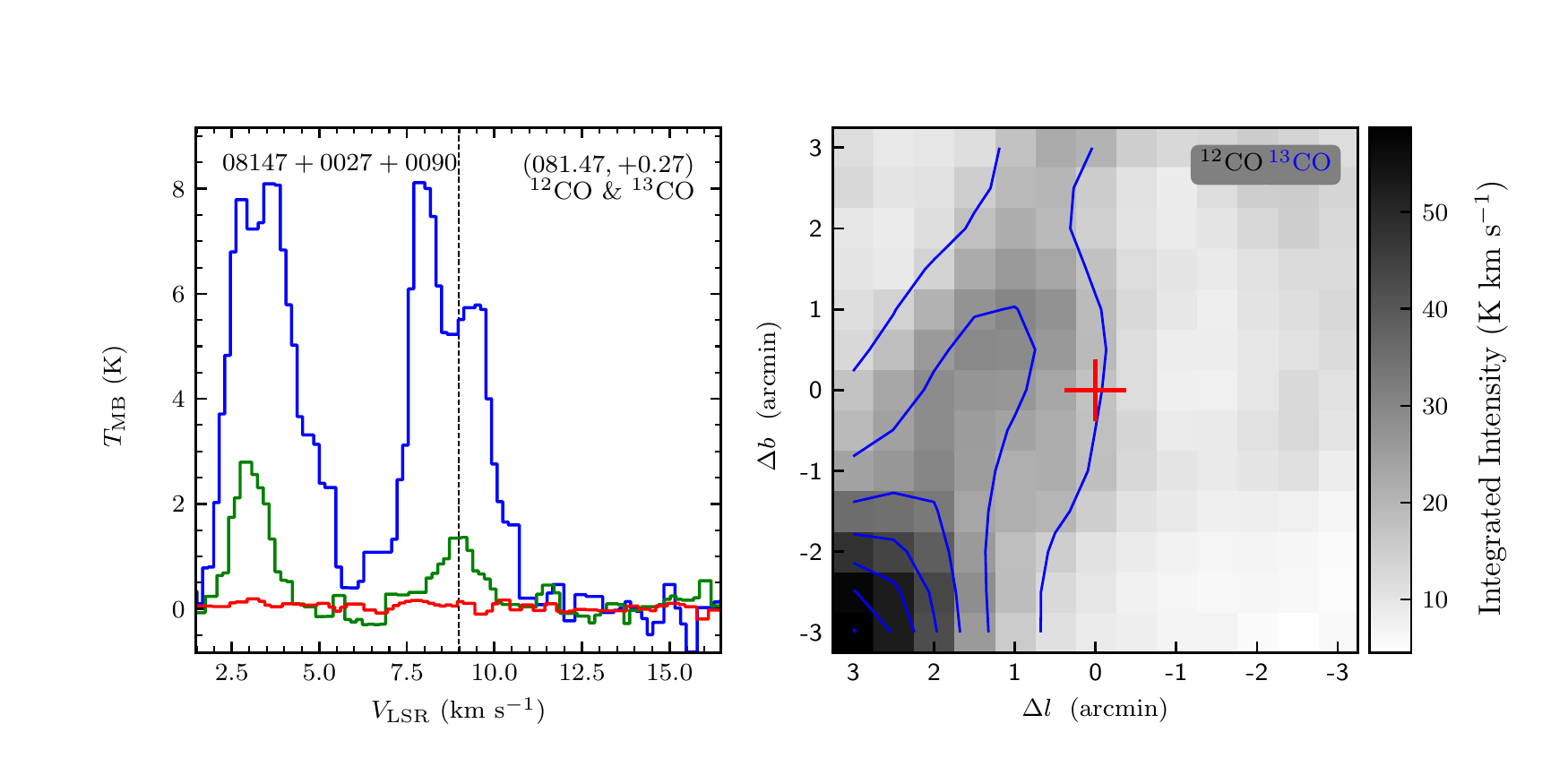}
\includegraphics[width=9.0cm,angle=0]{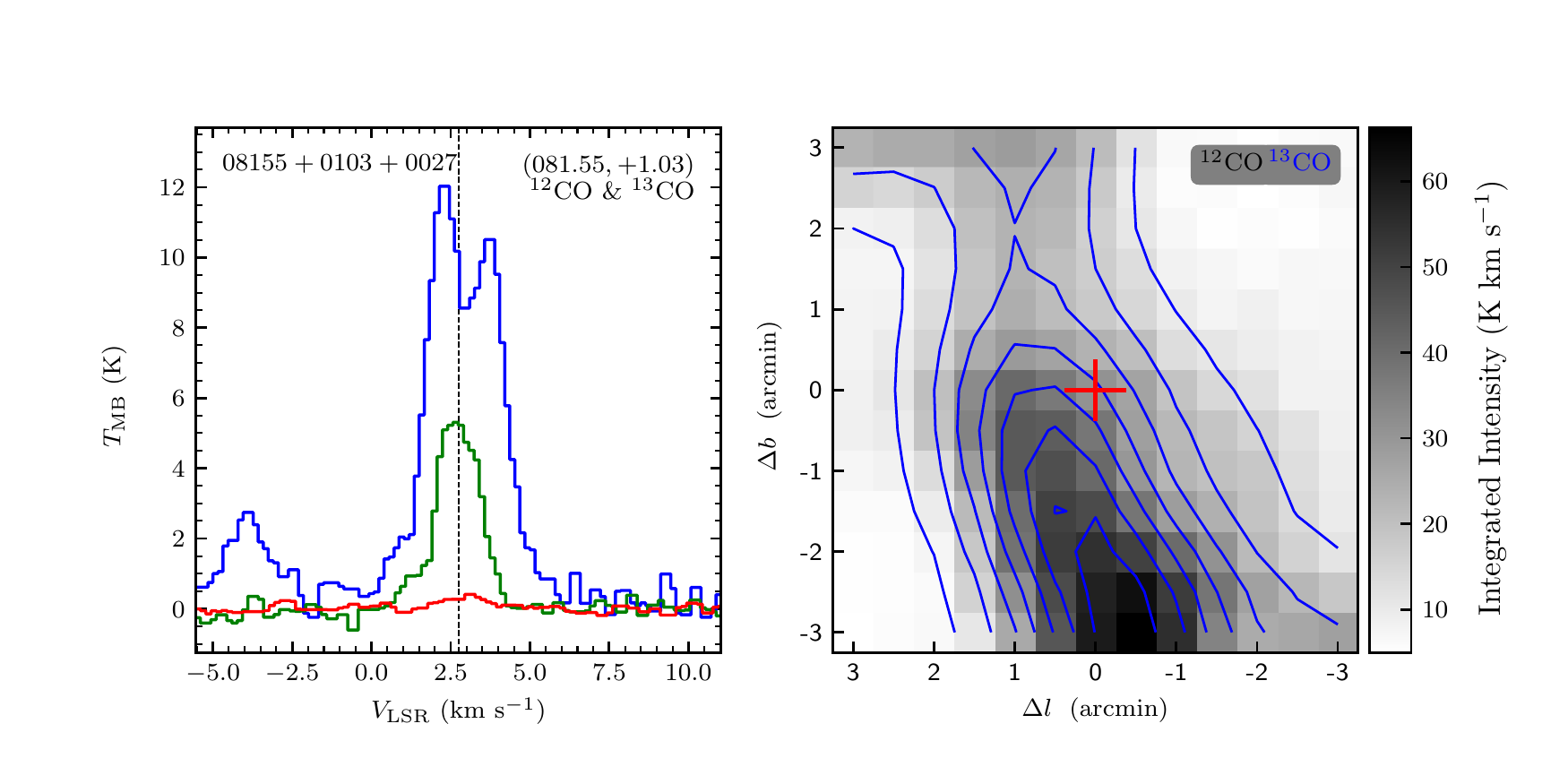}
\vspace{-0.5cm}

\includegraphics[width=9.0cm,angle=0]{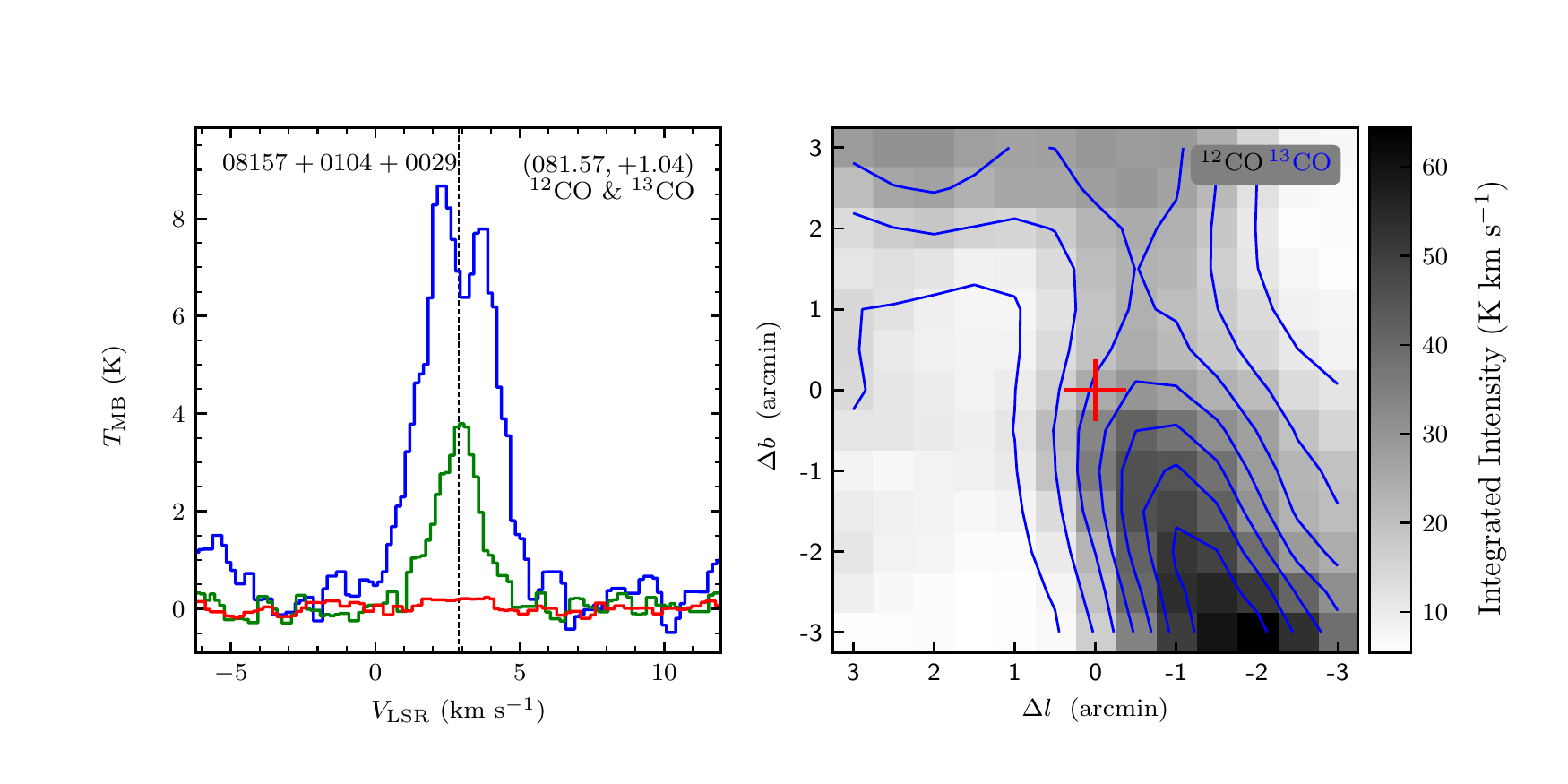}
\includegraphics[width=9.0cm,angle=0]{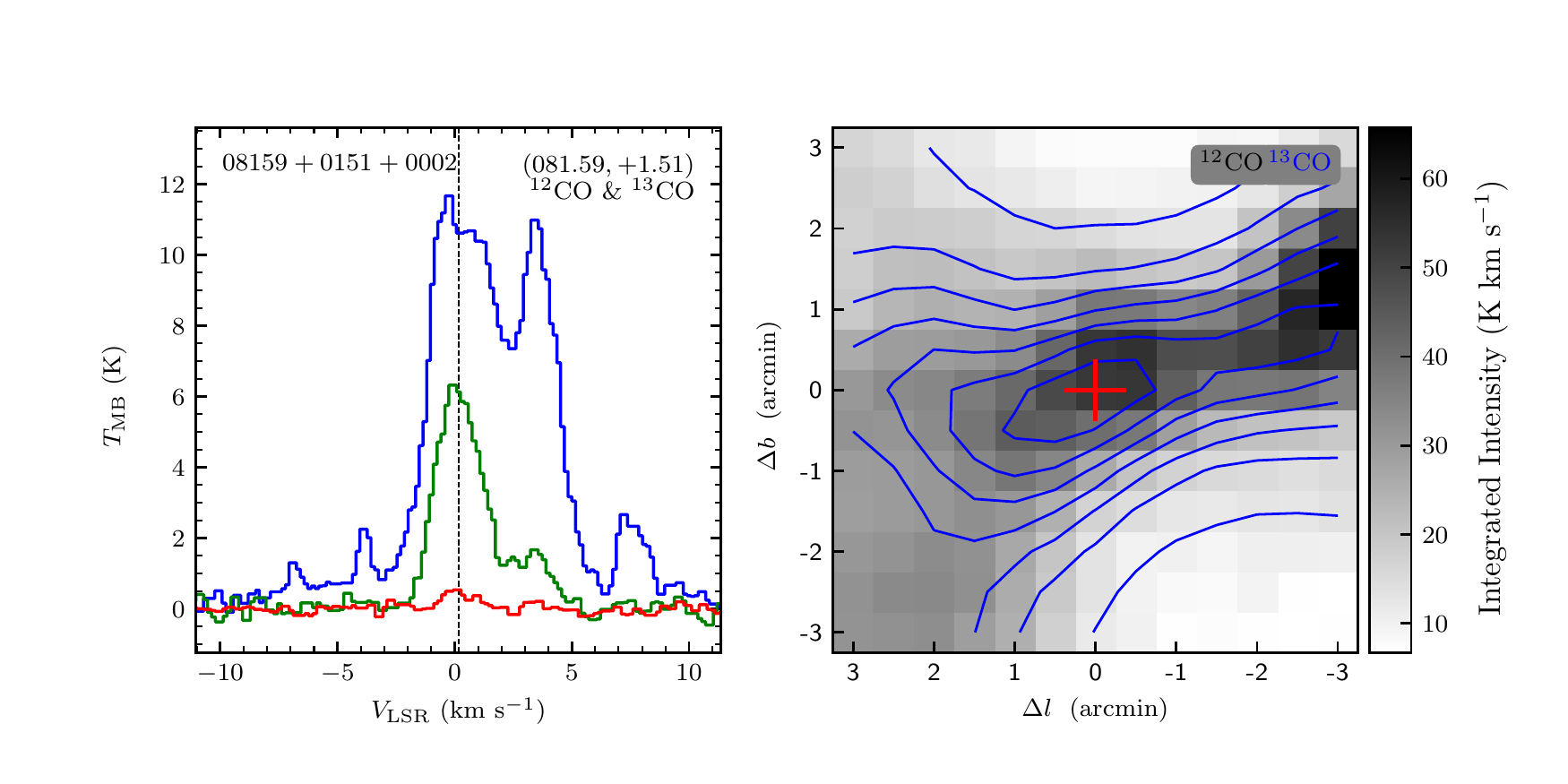}
\vspace{-0.5cm}

\includegraphics[width=9.0cm,angle=0]{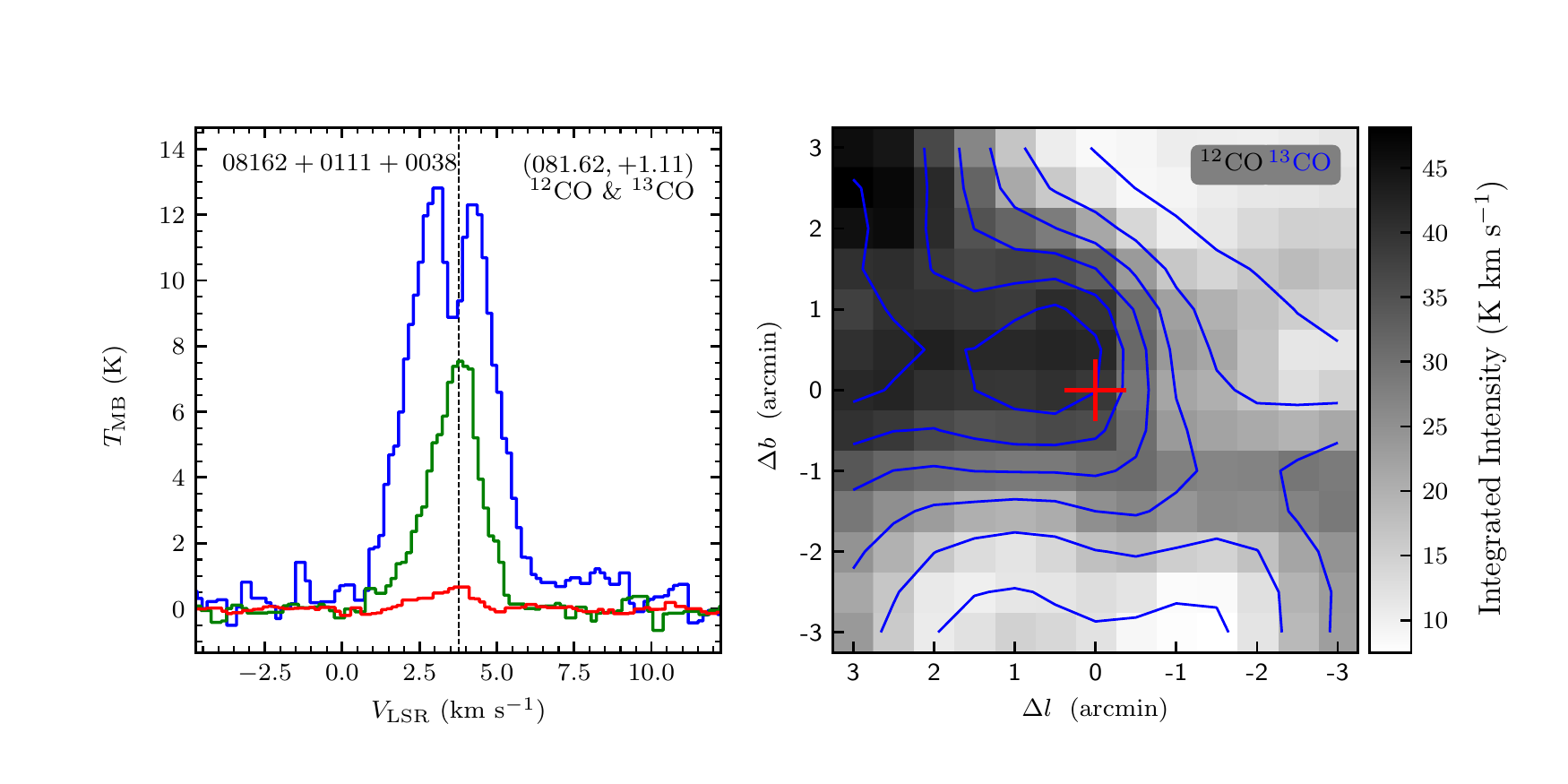}
\includegraphics[width=9.0cm,angle=0]{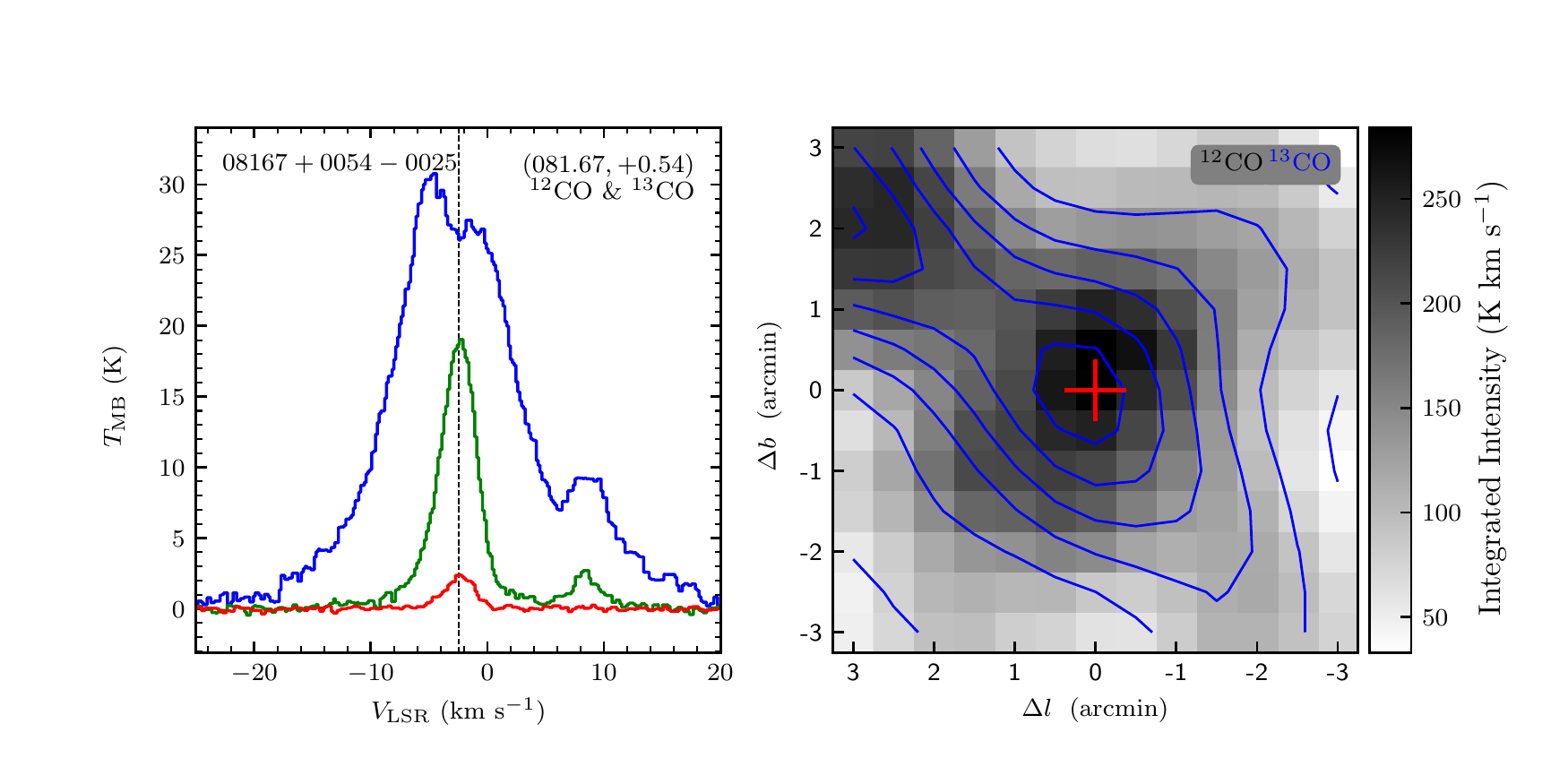}
\vspace{-0.5cm}

\includegraphics[width=9.0cm,angle=0]{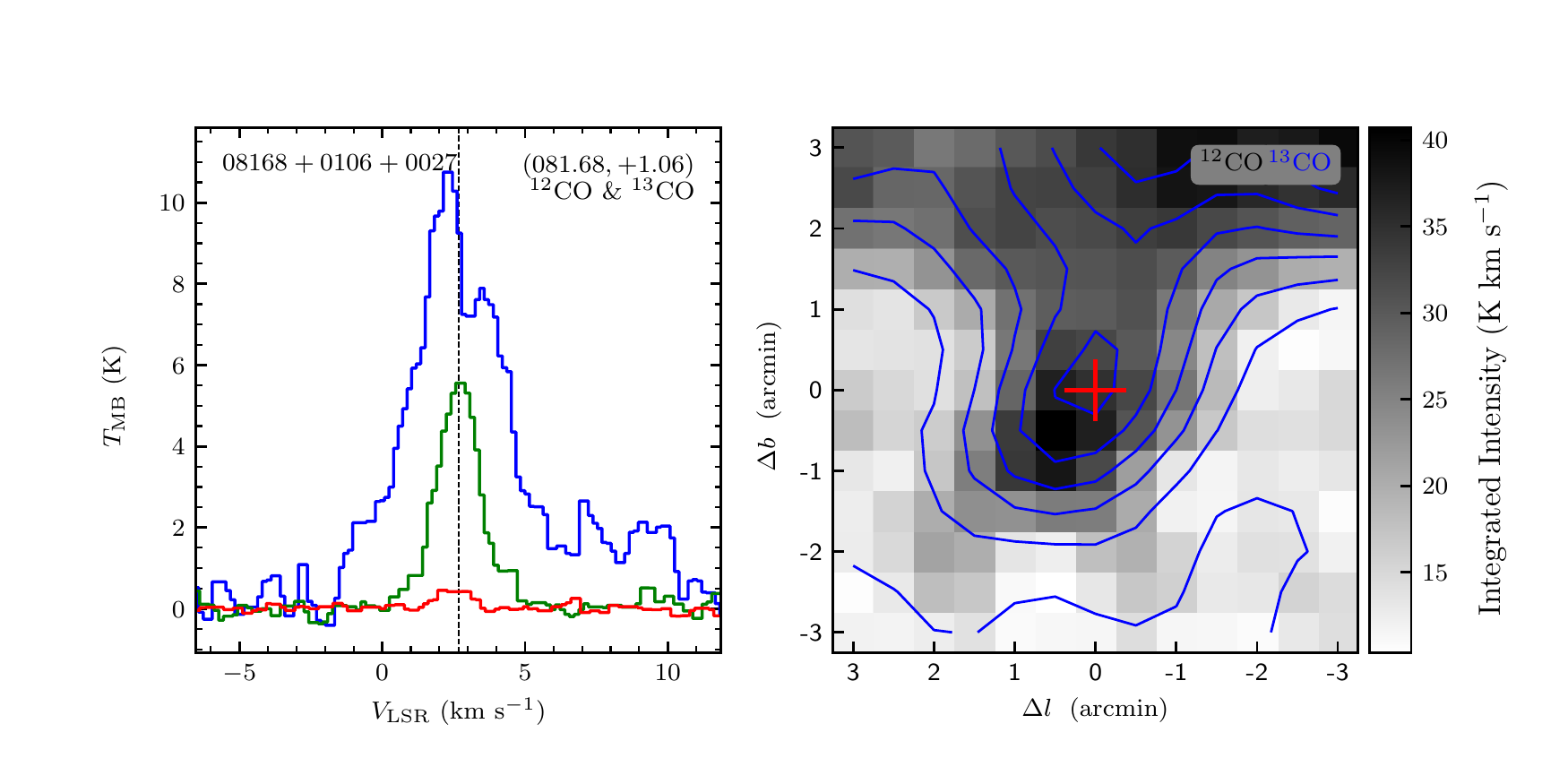}
\includegraphics[width=9.0cm,angle=0]{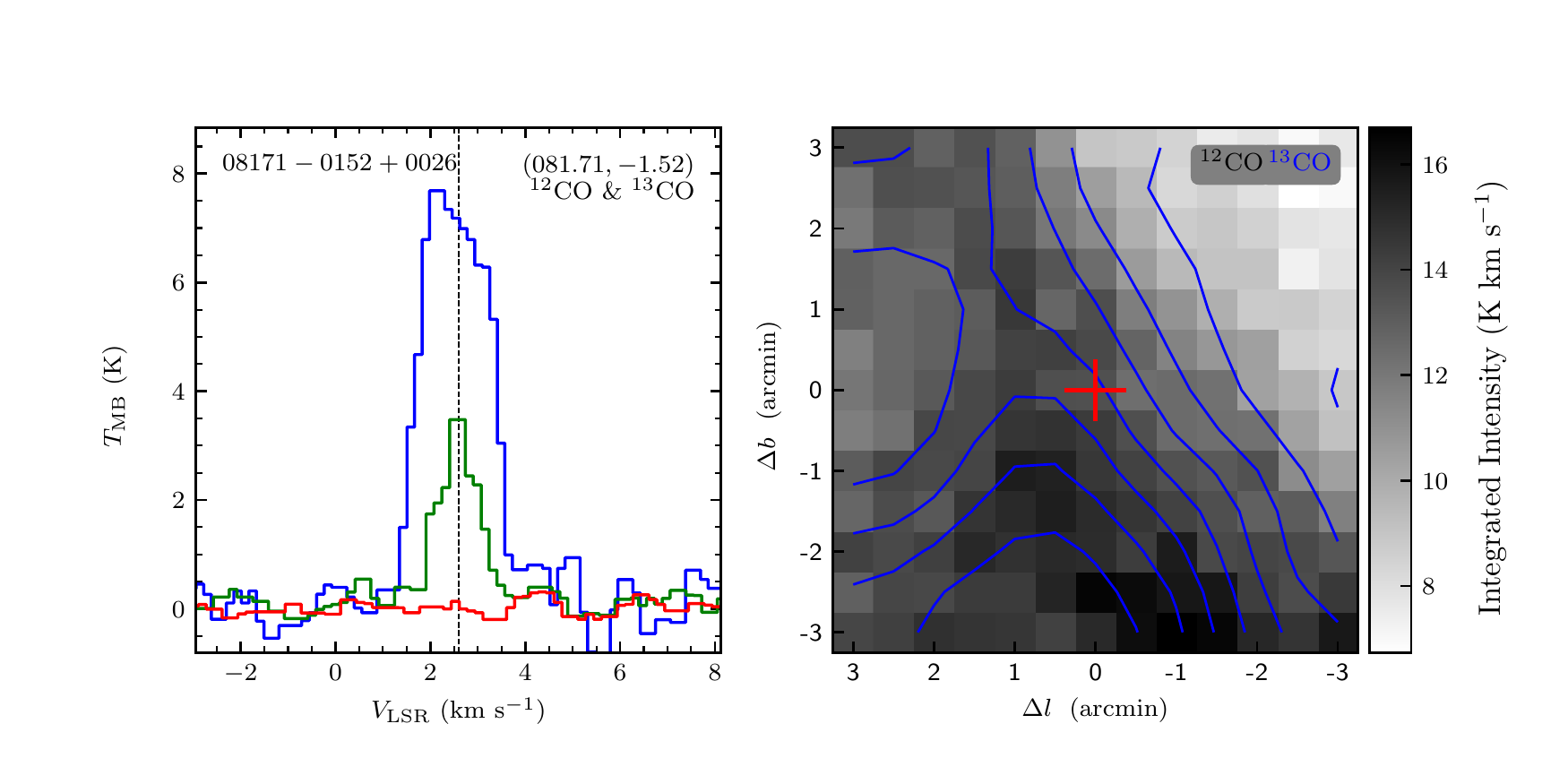}
\end{figure}
\clearpage

\begin{figure}
\includegraphics[width=9.0cm,angle=0]{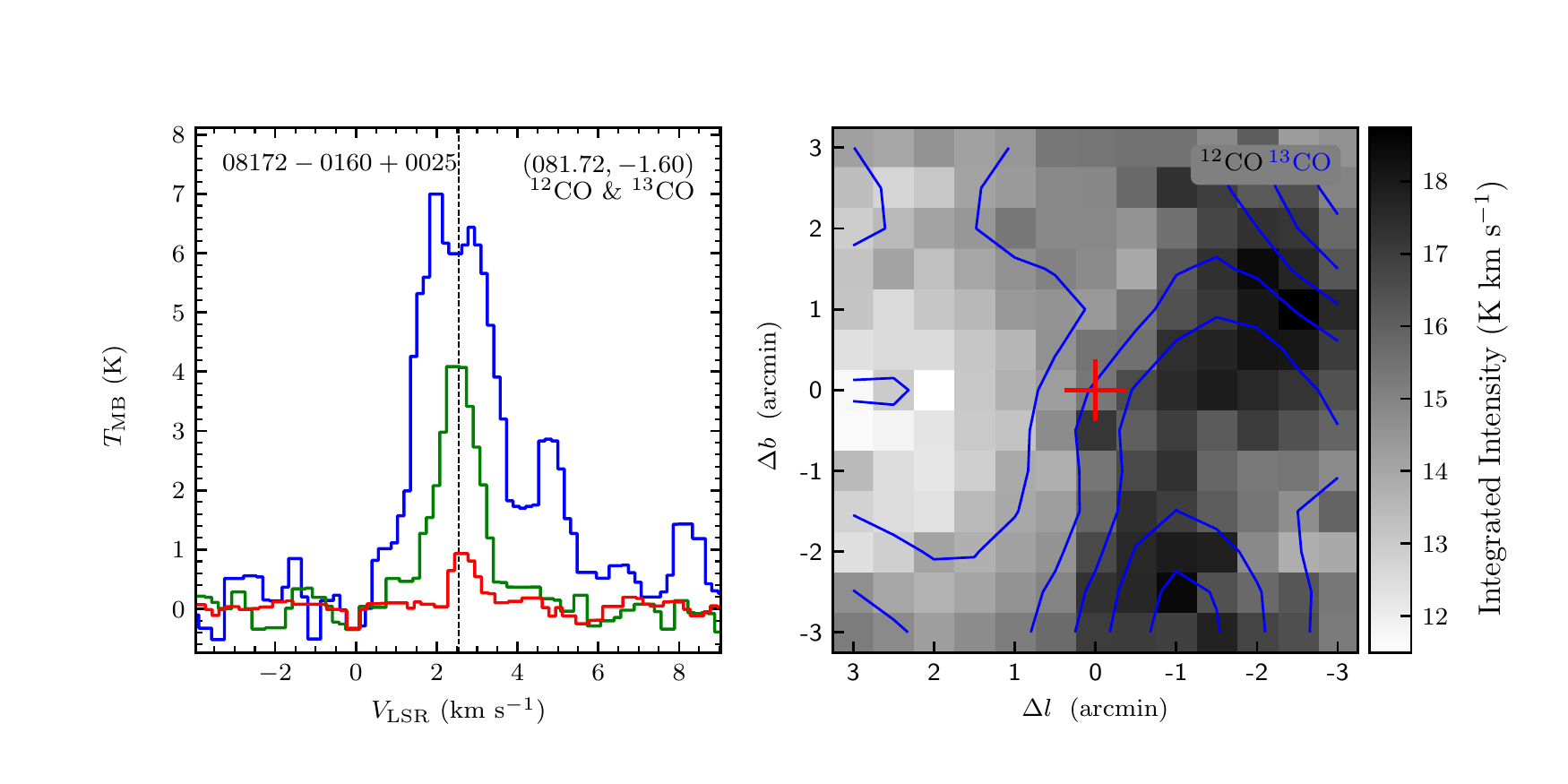}
\includegraphics[width=9.0cm,angle=0]{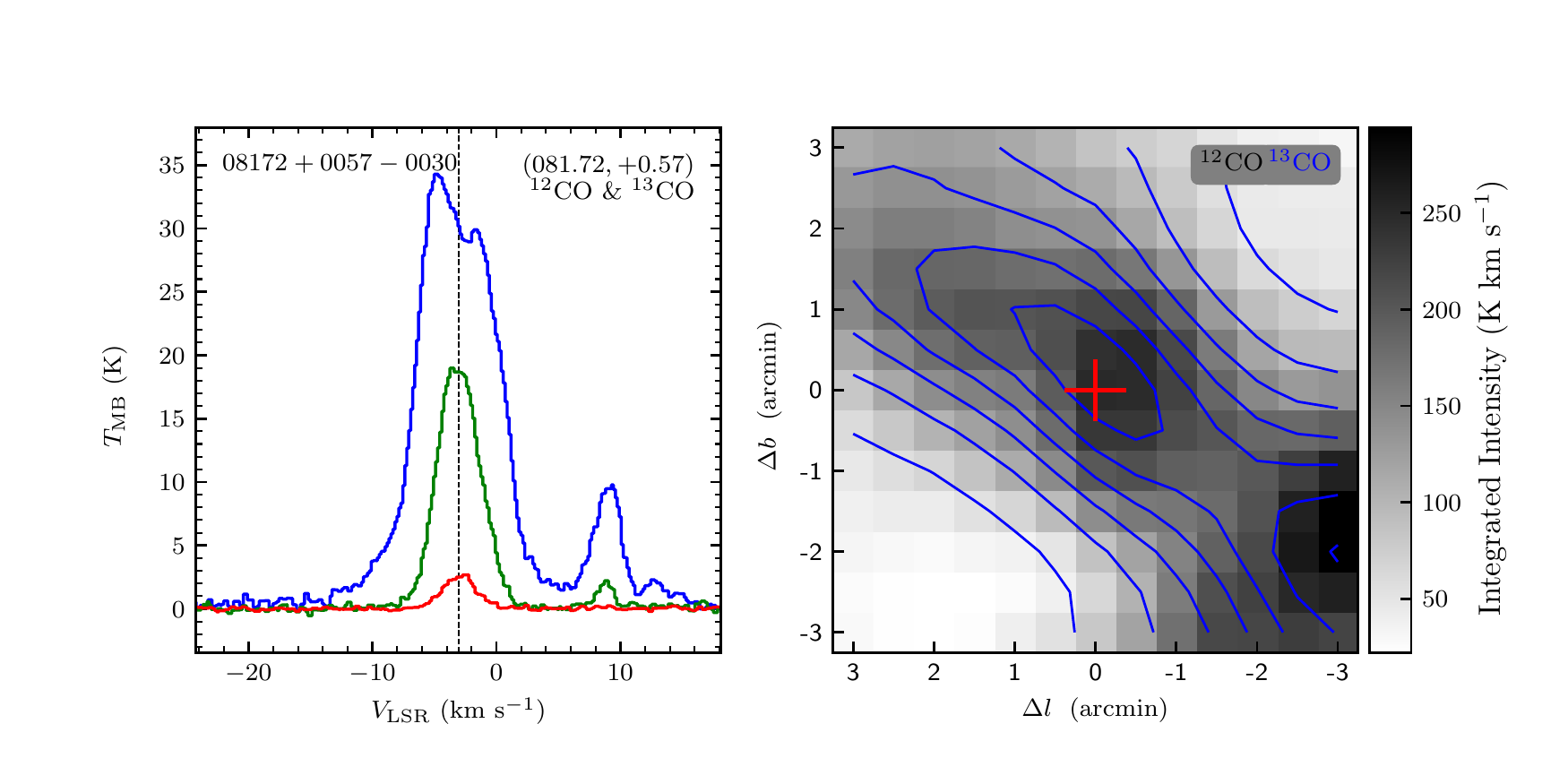}
\vspace{-0.5cm}

\includegraphics[width=9.0cm,angle=0]{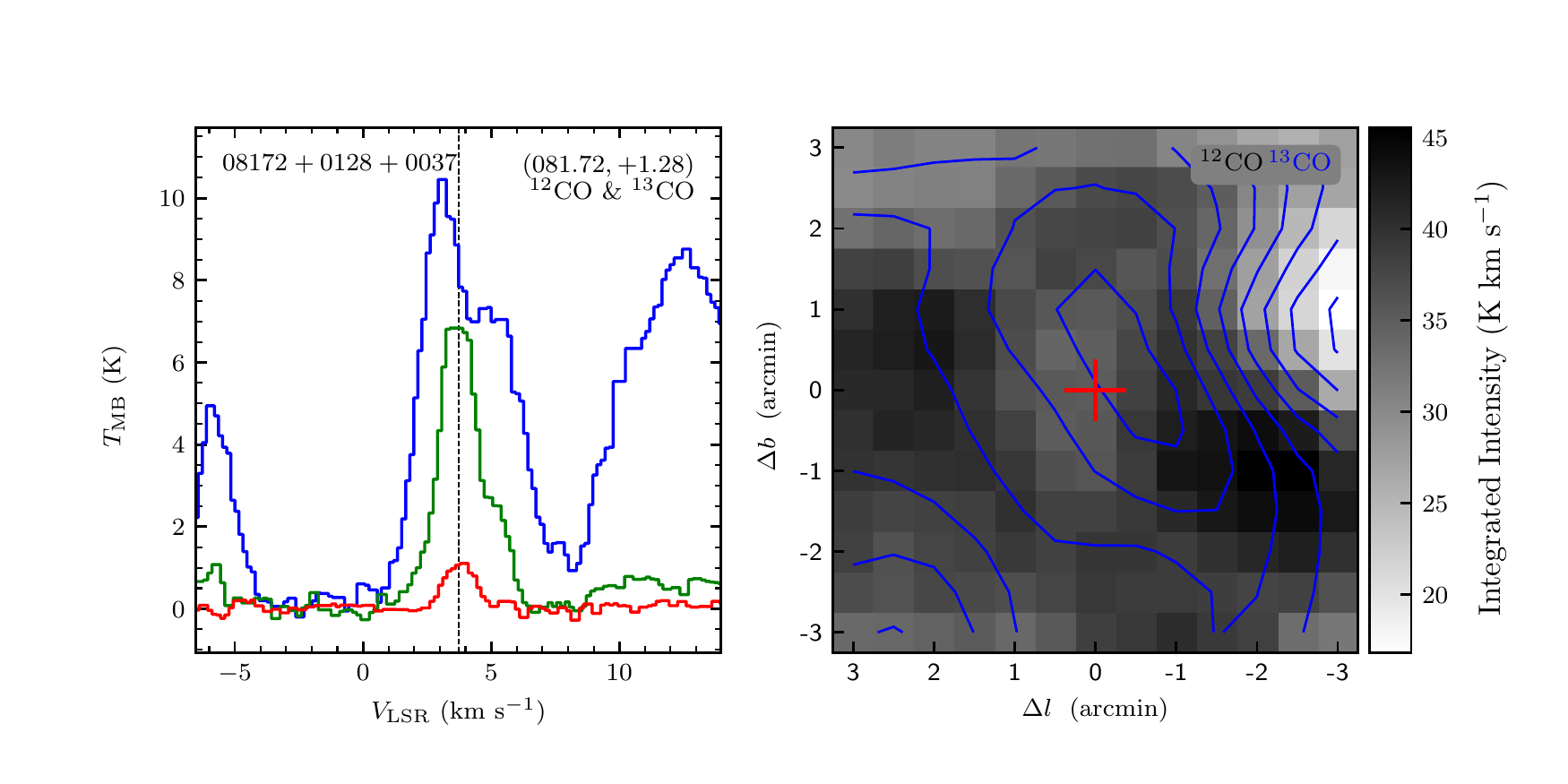}
\includegraphics[width=9.0cm,angle=0]{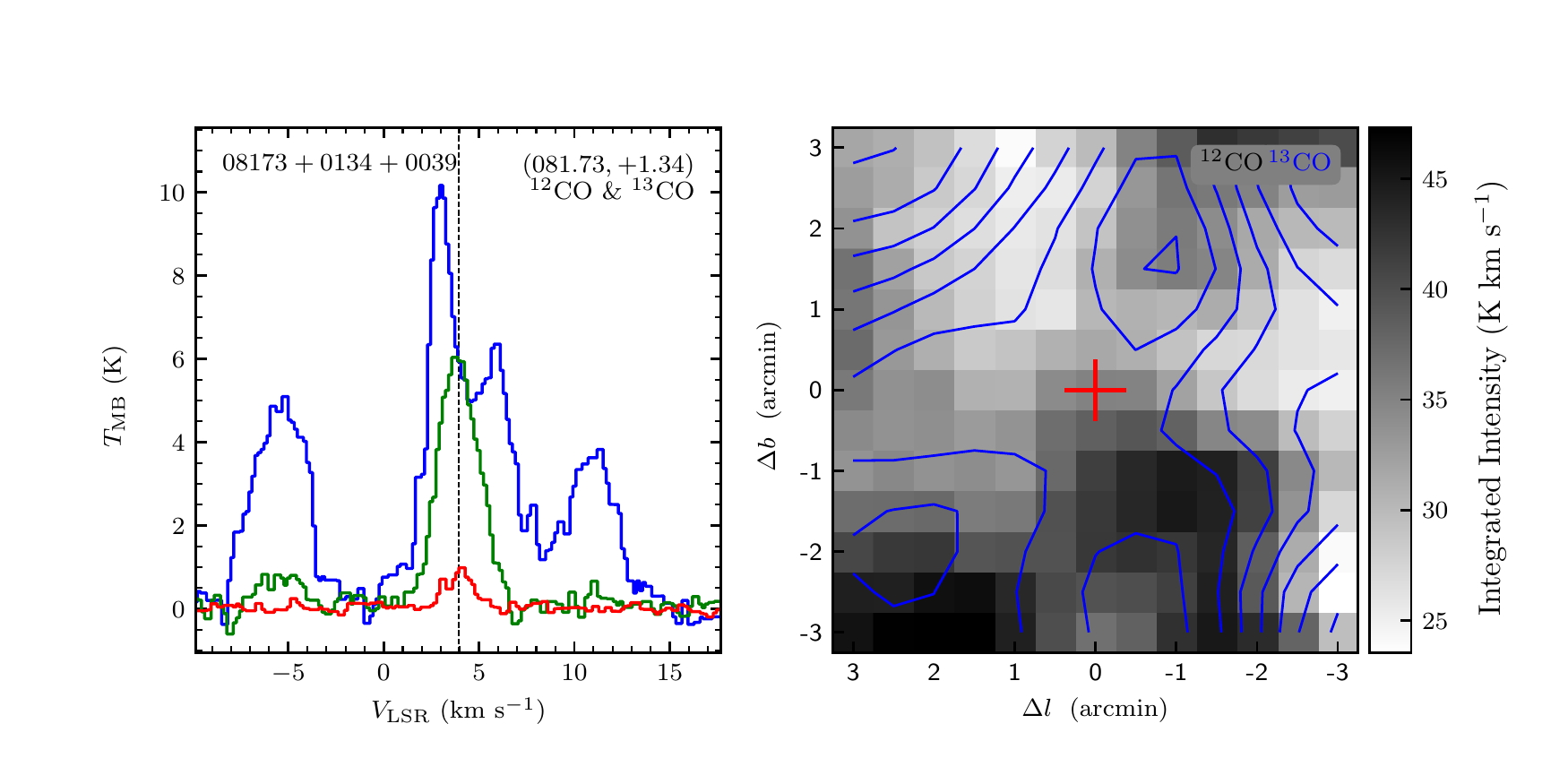}
\vspace{-0.5cm}

\includegraphics[width=9.0cm,angle=0]{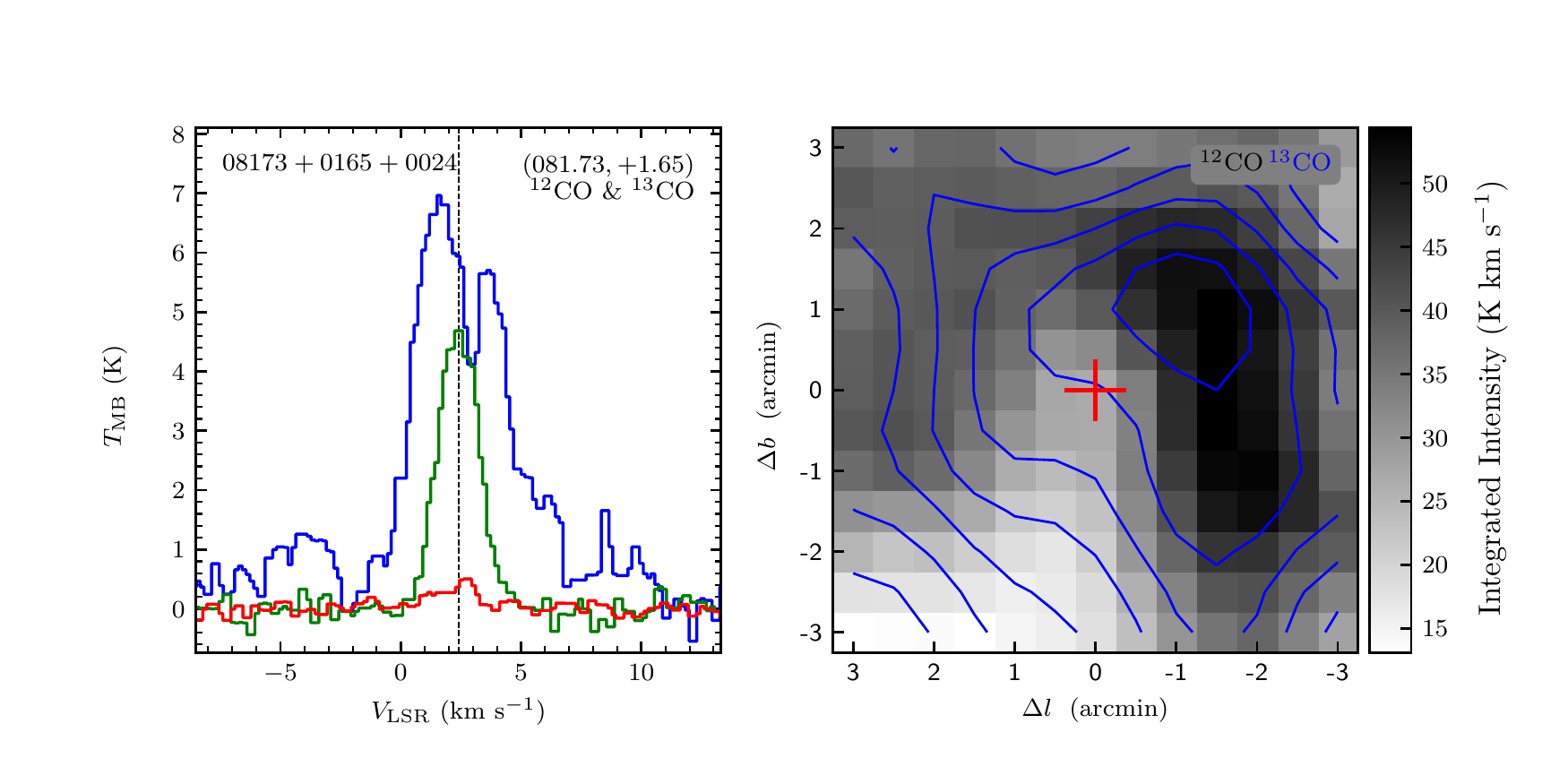}
\includegraphics[width=9.0cm,angle=0]{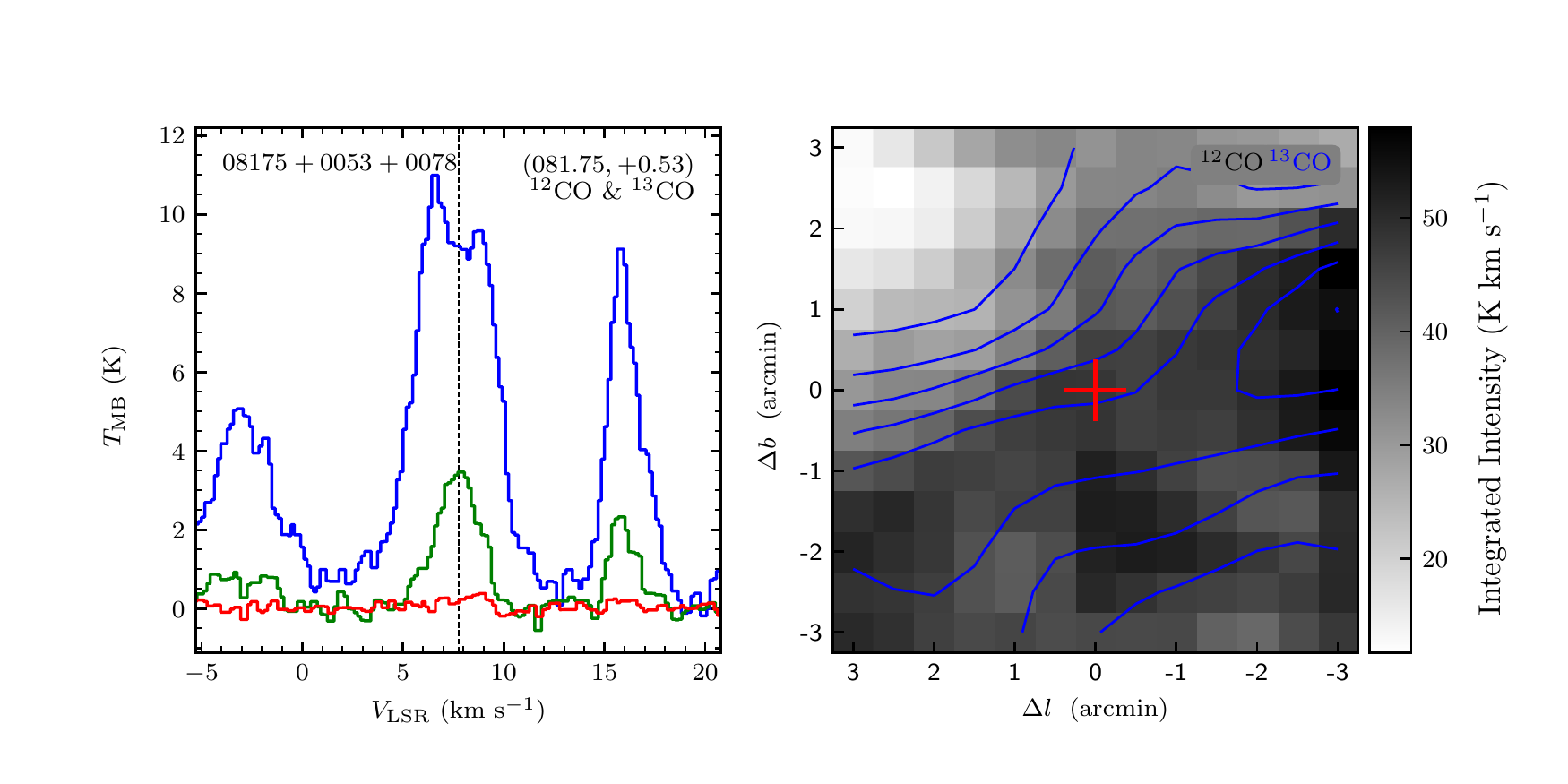}
\vspace{-0.5cm}

\includegraphics[width=9.0cm,angle=0]{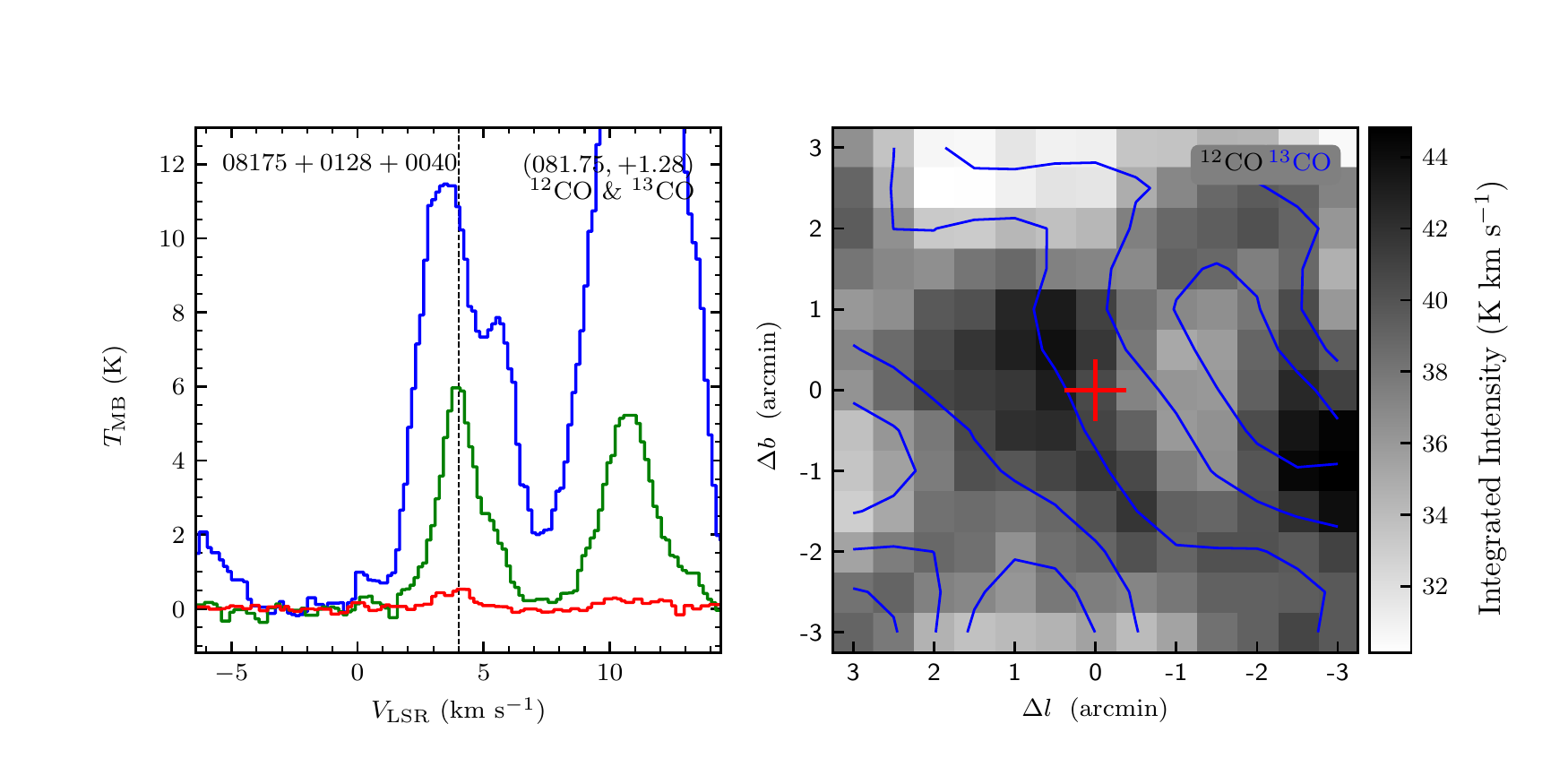}
\includegraphics[width=9.0cm,angle=0]{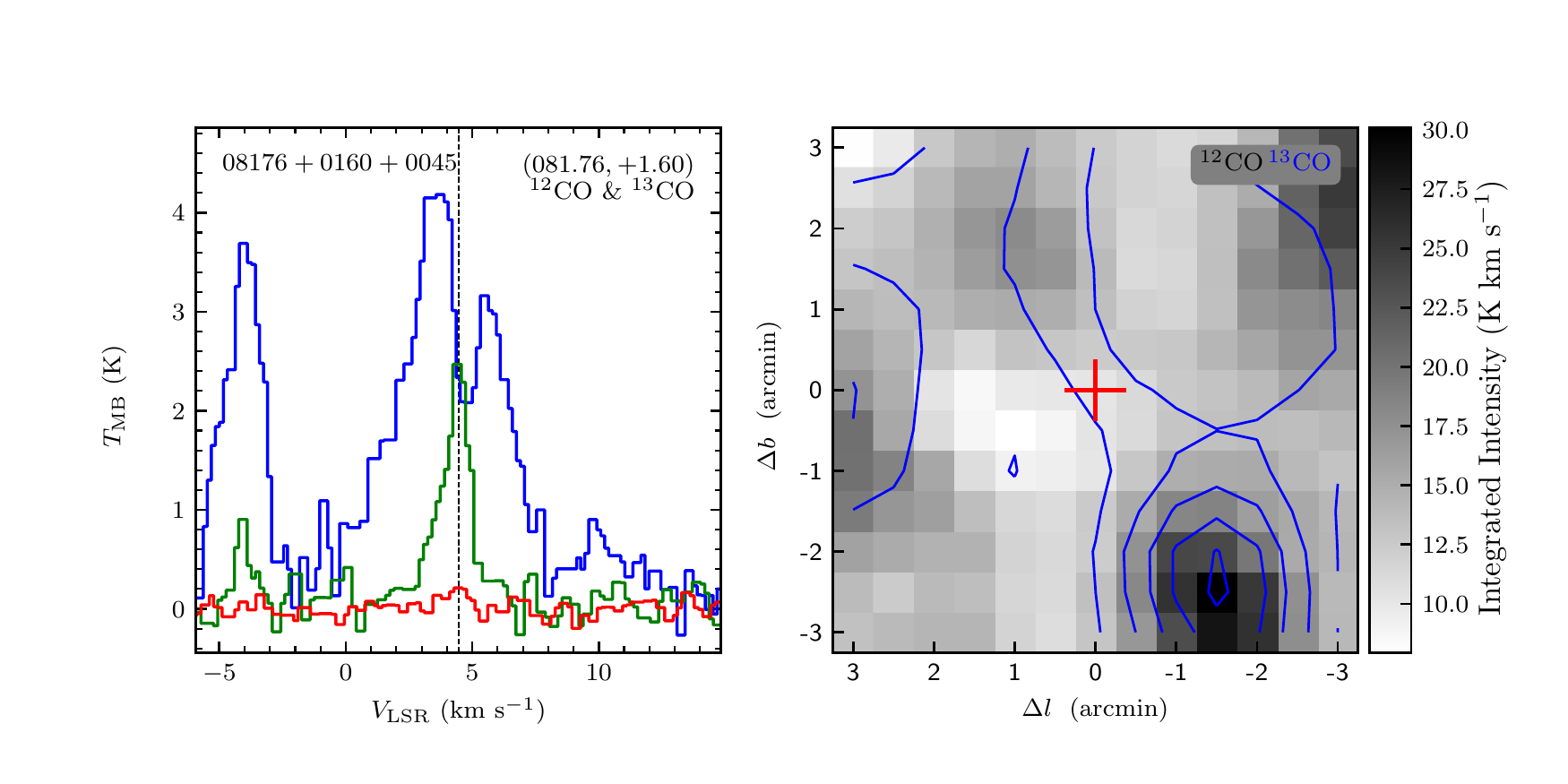}
\vspace{-0.5cm}

\includegraphics[width=9.0cm,angle=0]{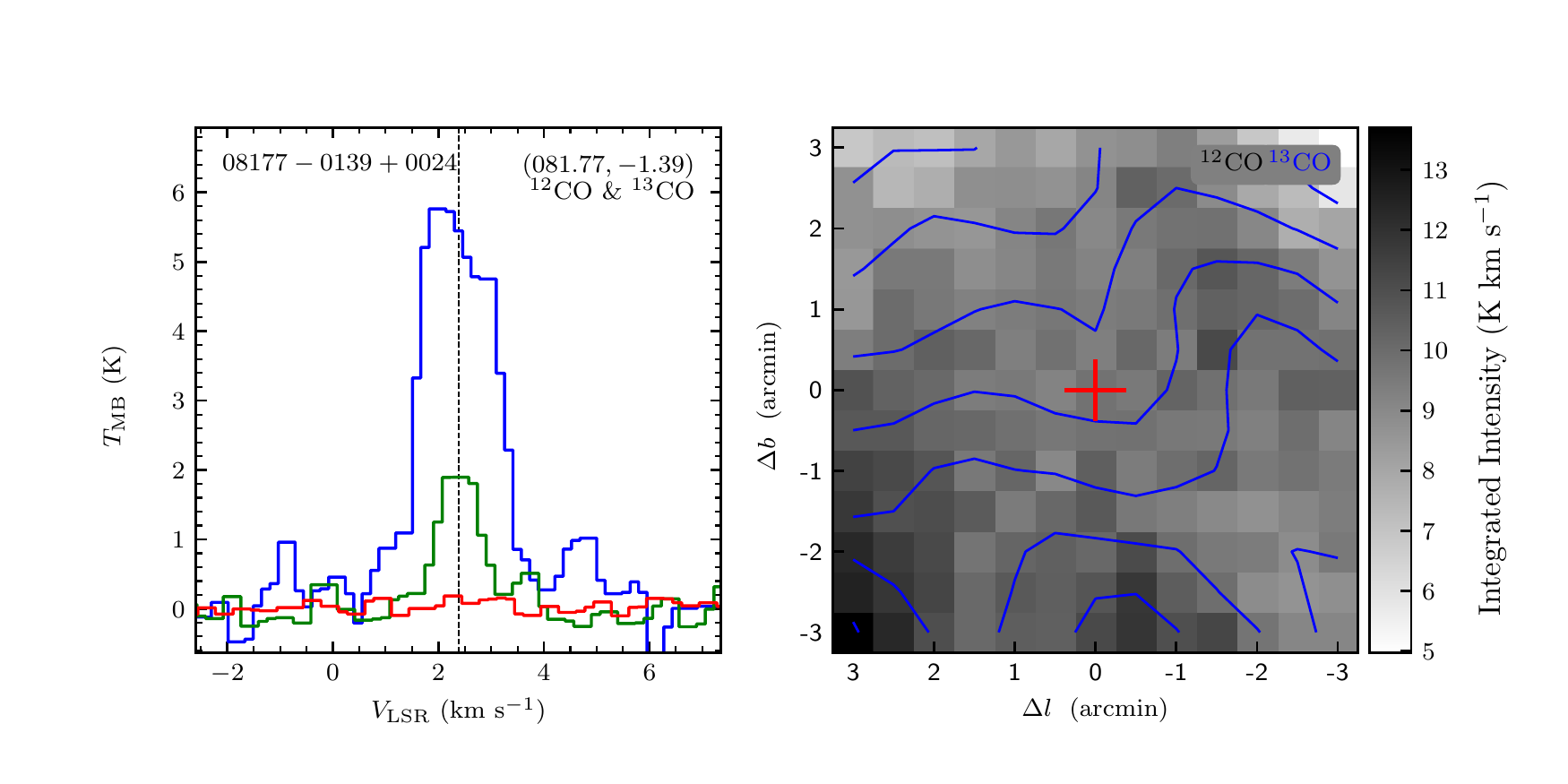}
\includegraphics[width=9.0cm,angle=0]{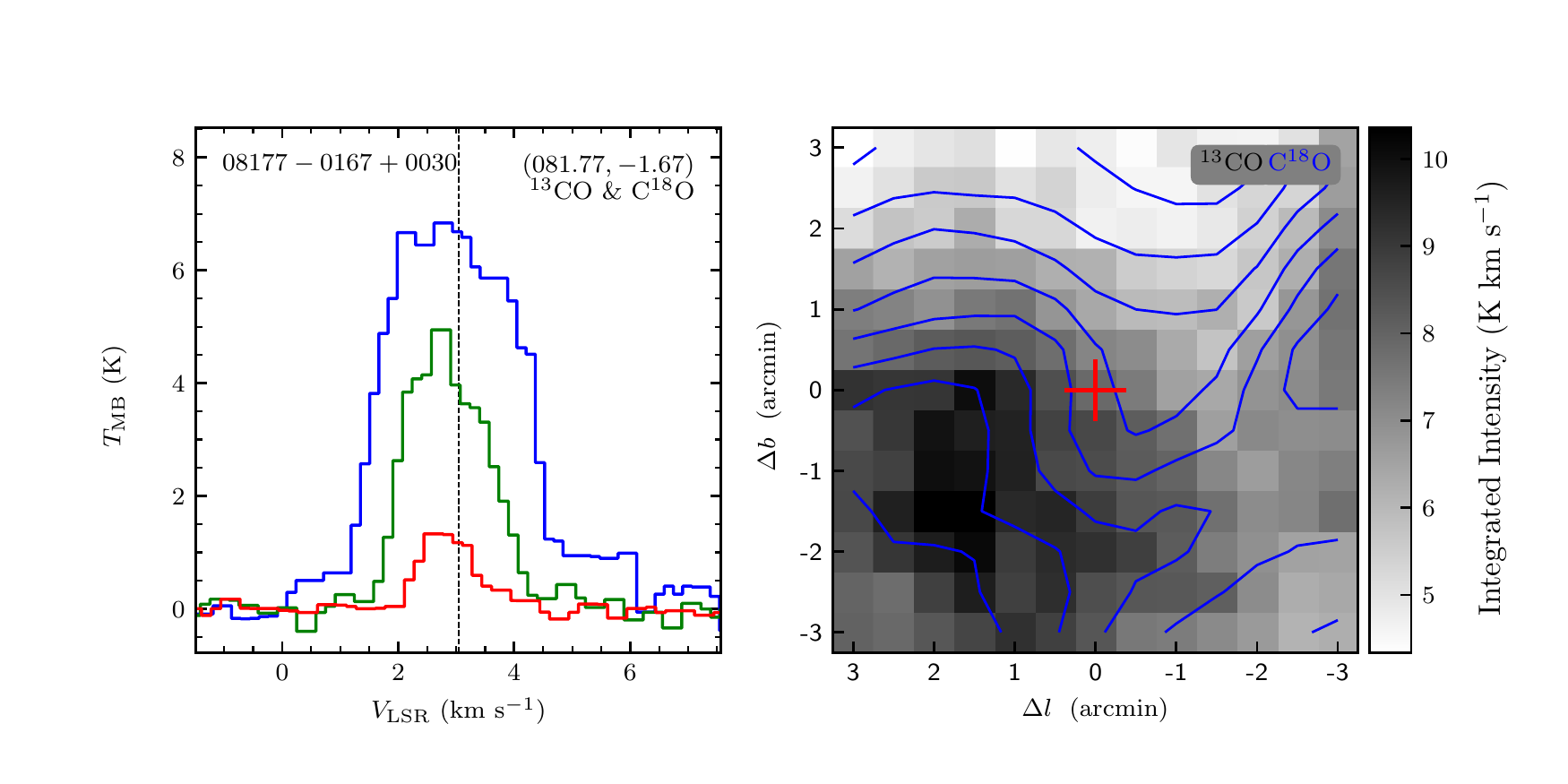}
\end{figure}
\clearpage

\begin{figure}
\includegraphics[width=9.0cm,angle=0]{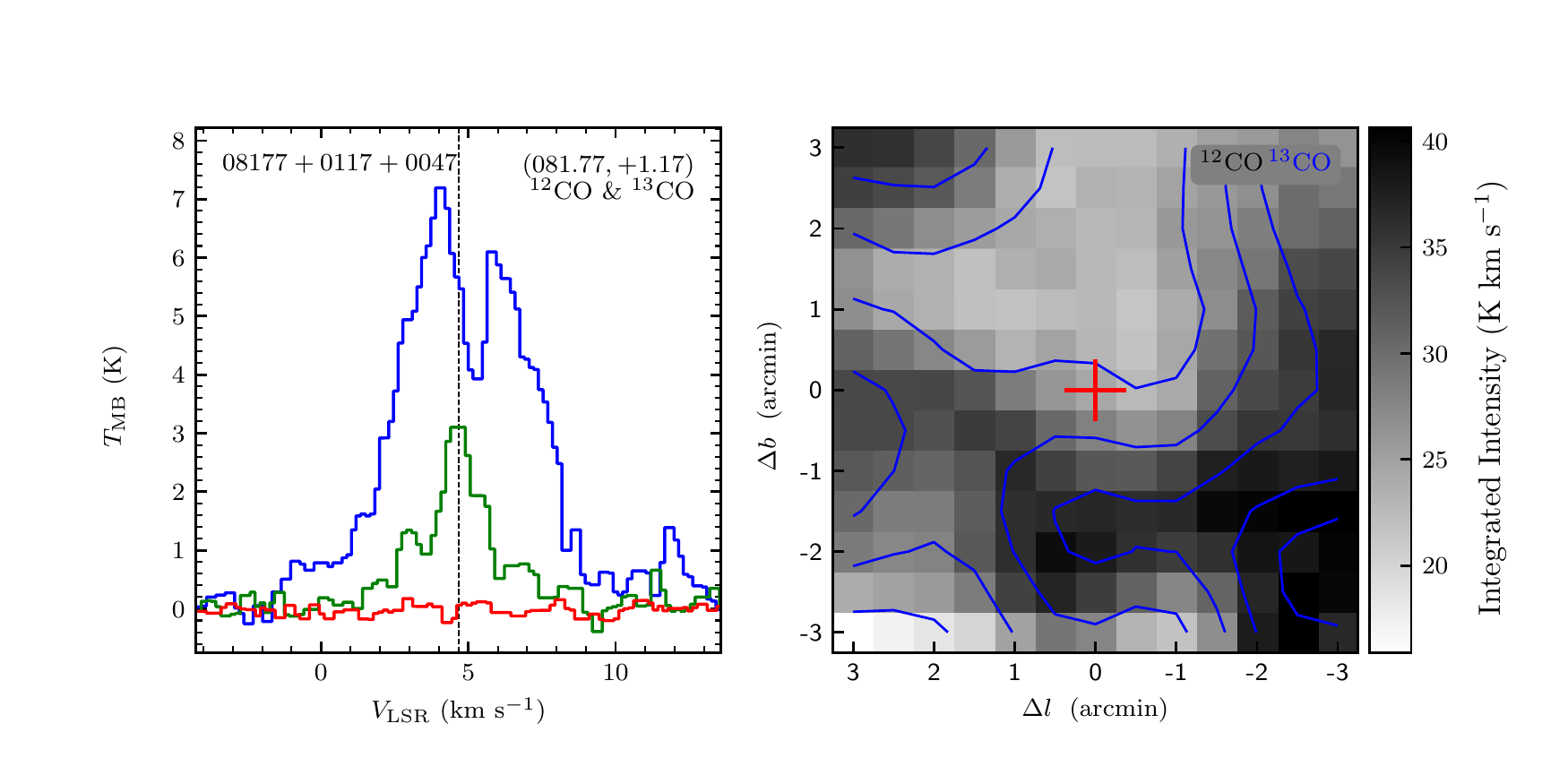}
\includegraphics[width=9.0cm,angle=0]{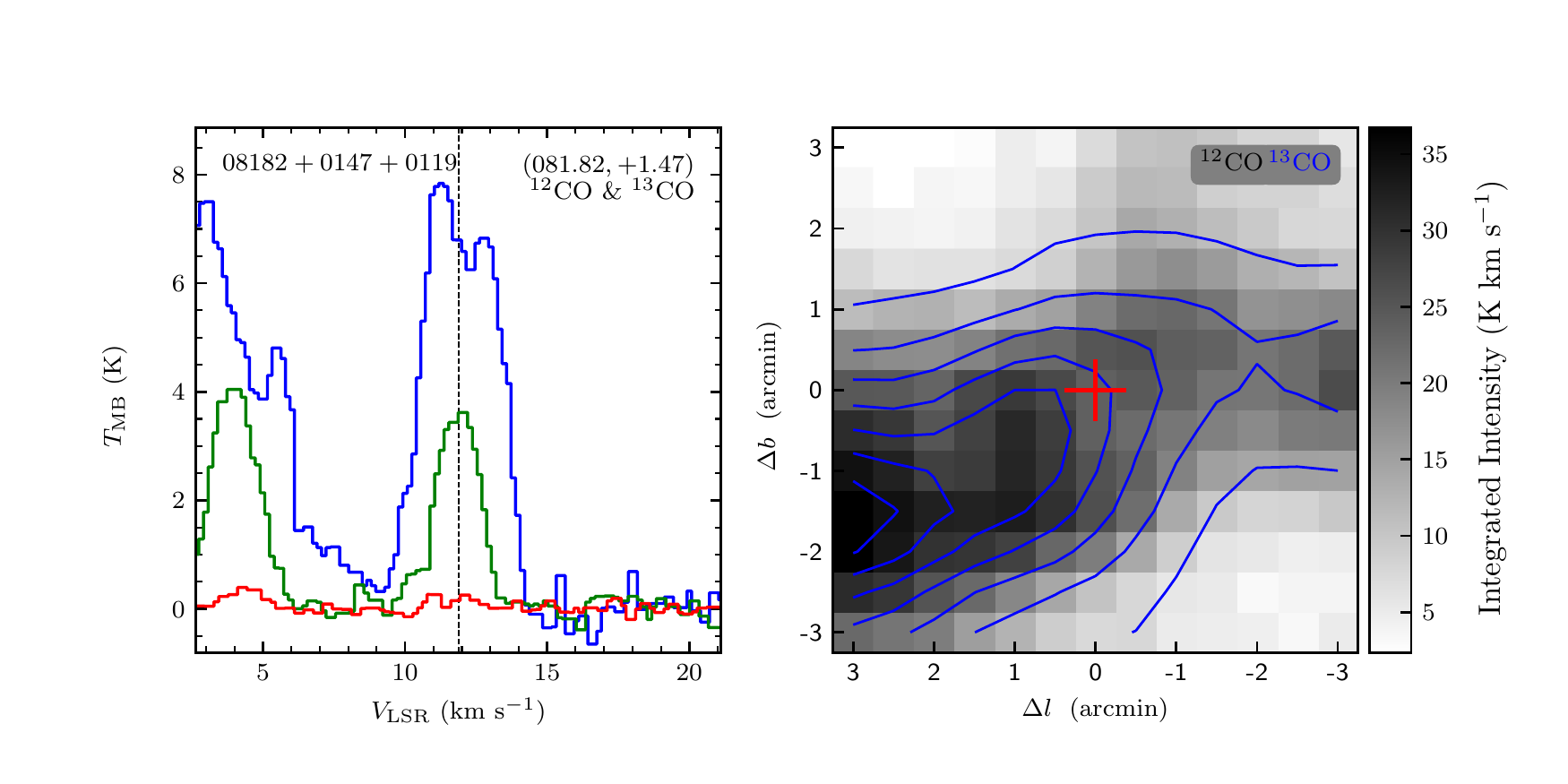}
\vspace{-0.5cm}

\includegraphics[width=9.0cm,angle=0]{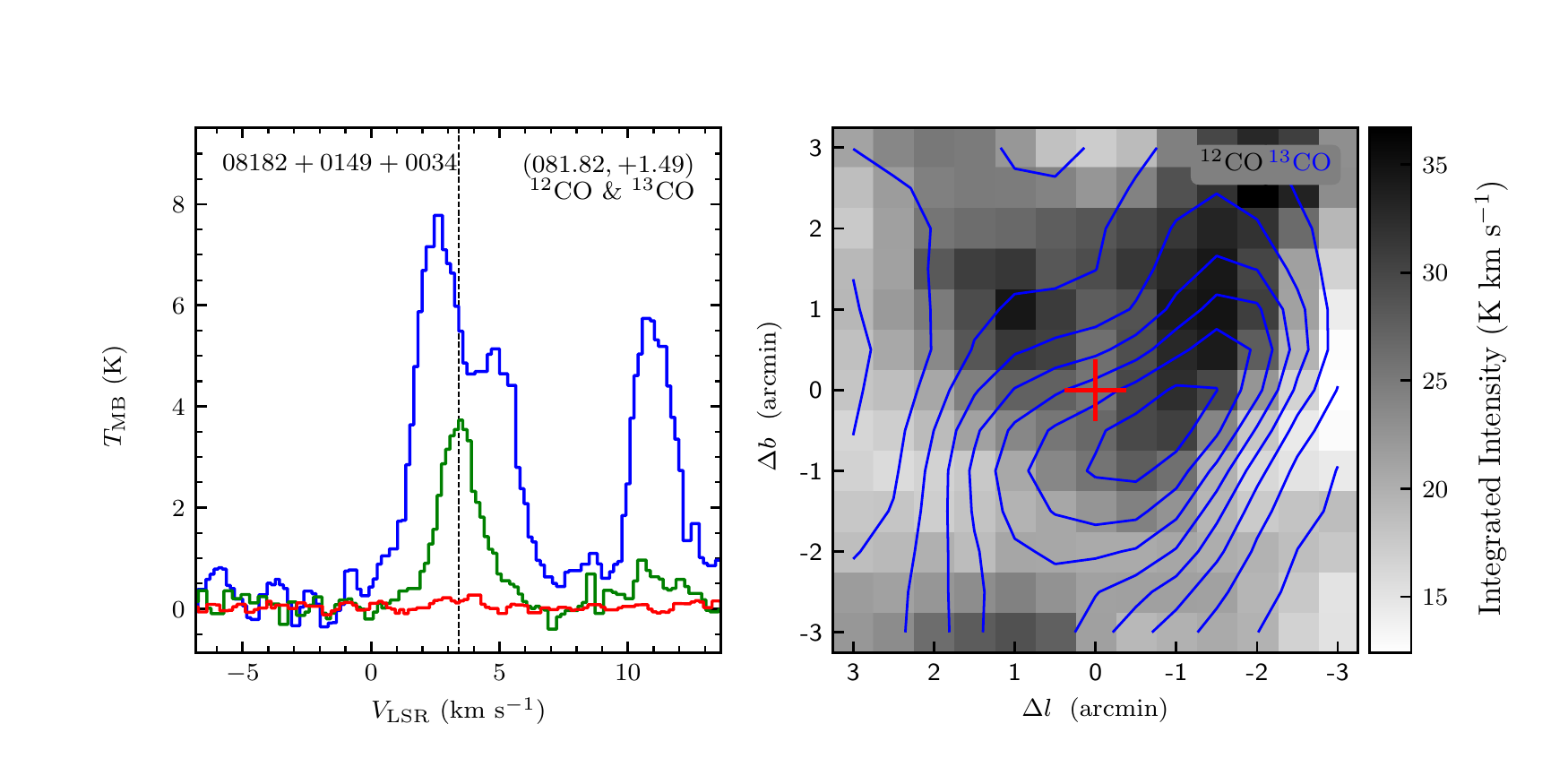}
\includegraphics[width=9.0cm,angle=0]{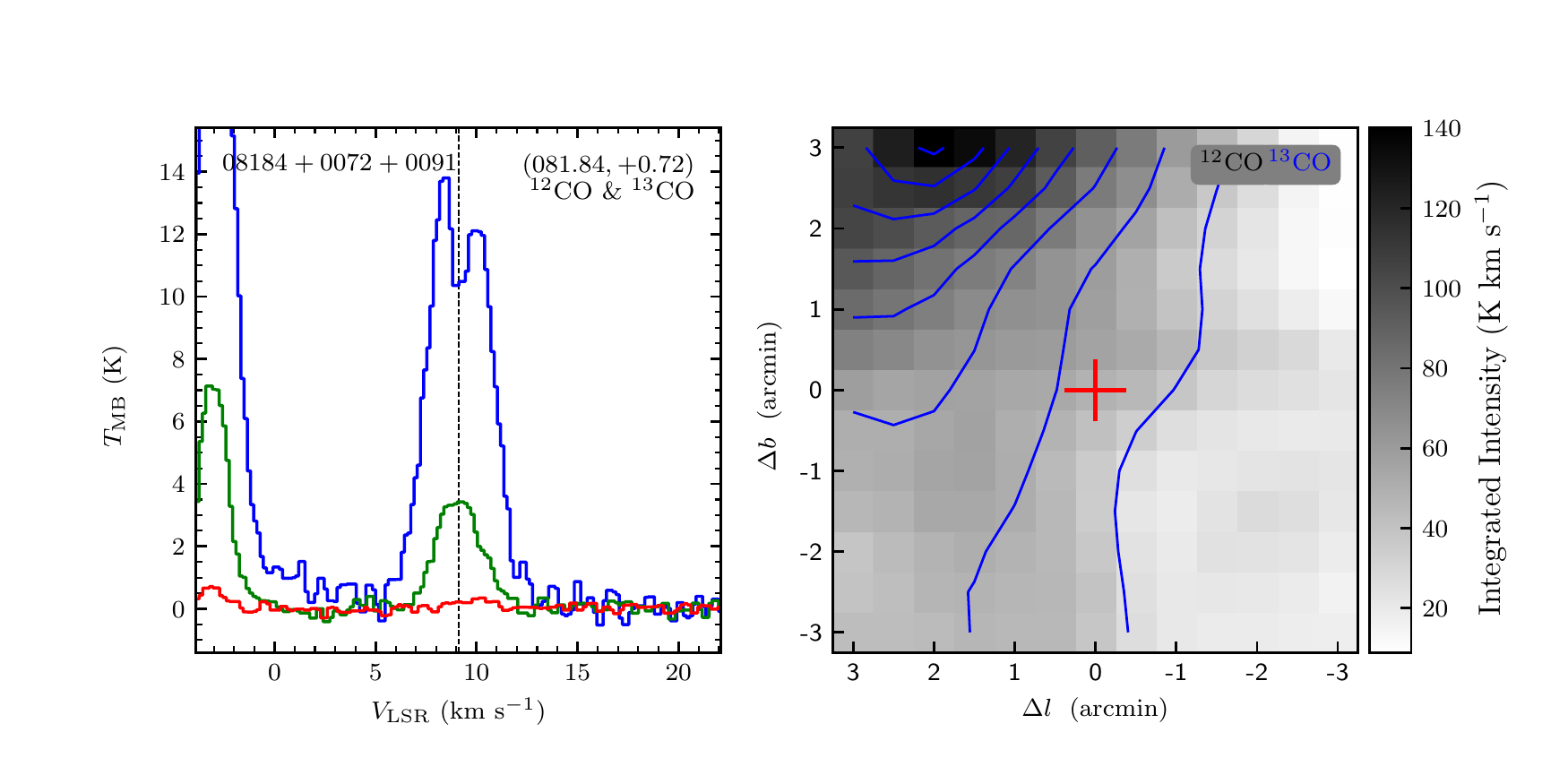}
\vspace{-0.5cm}

\includegraphics[width=9.0cm,angle=0]{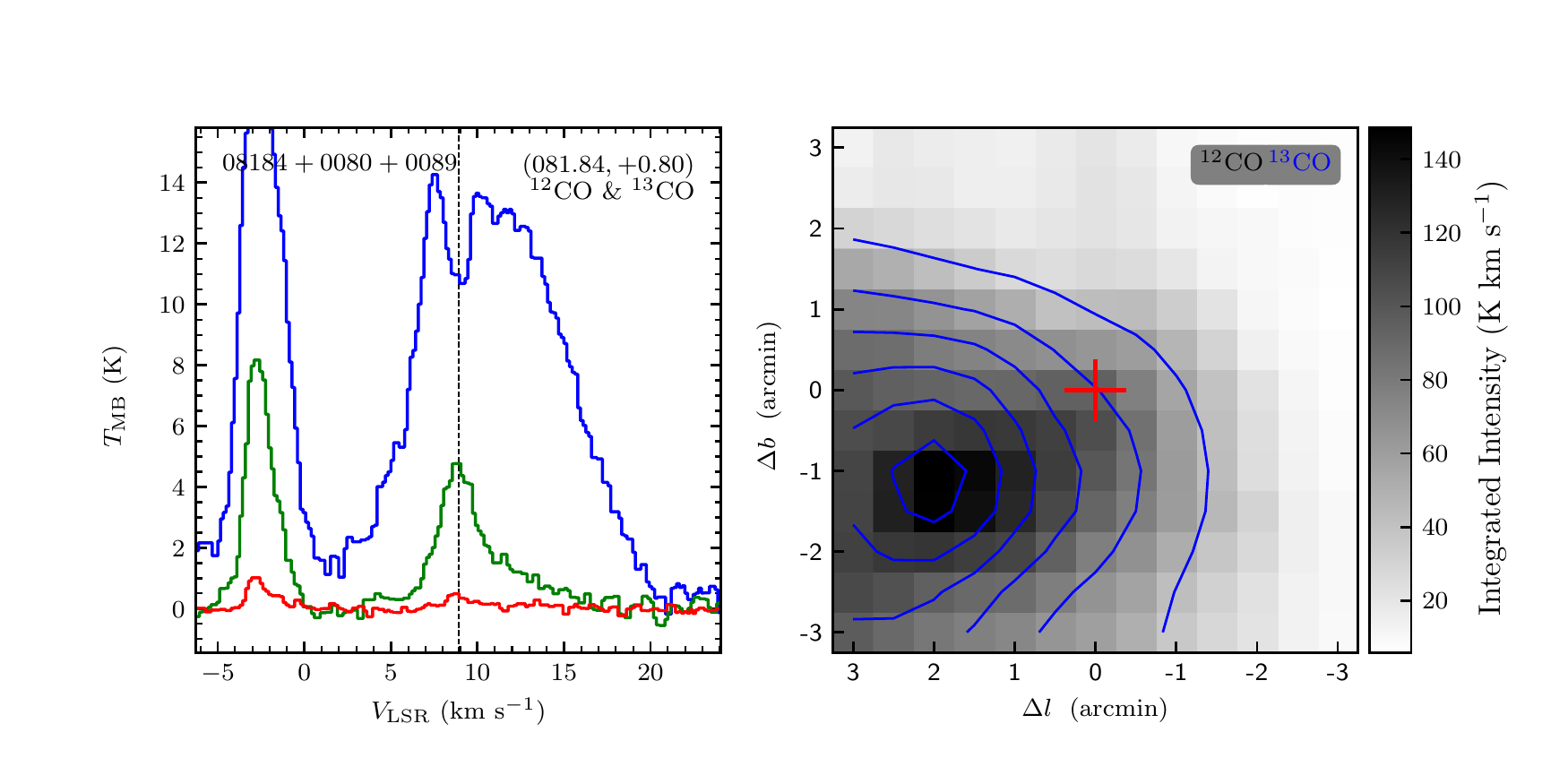}
\includegraphics[width=9.0cm,angle=0]{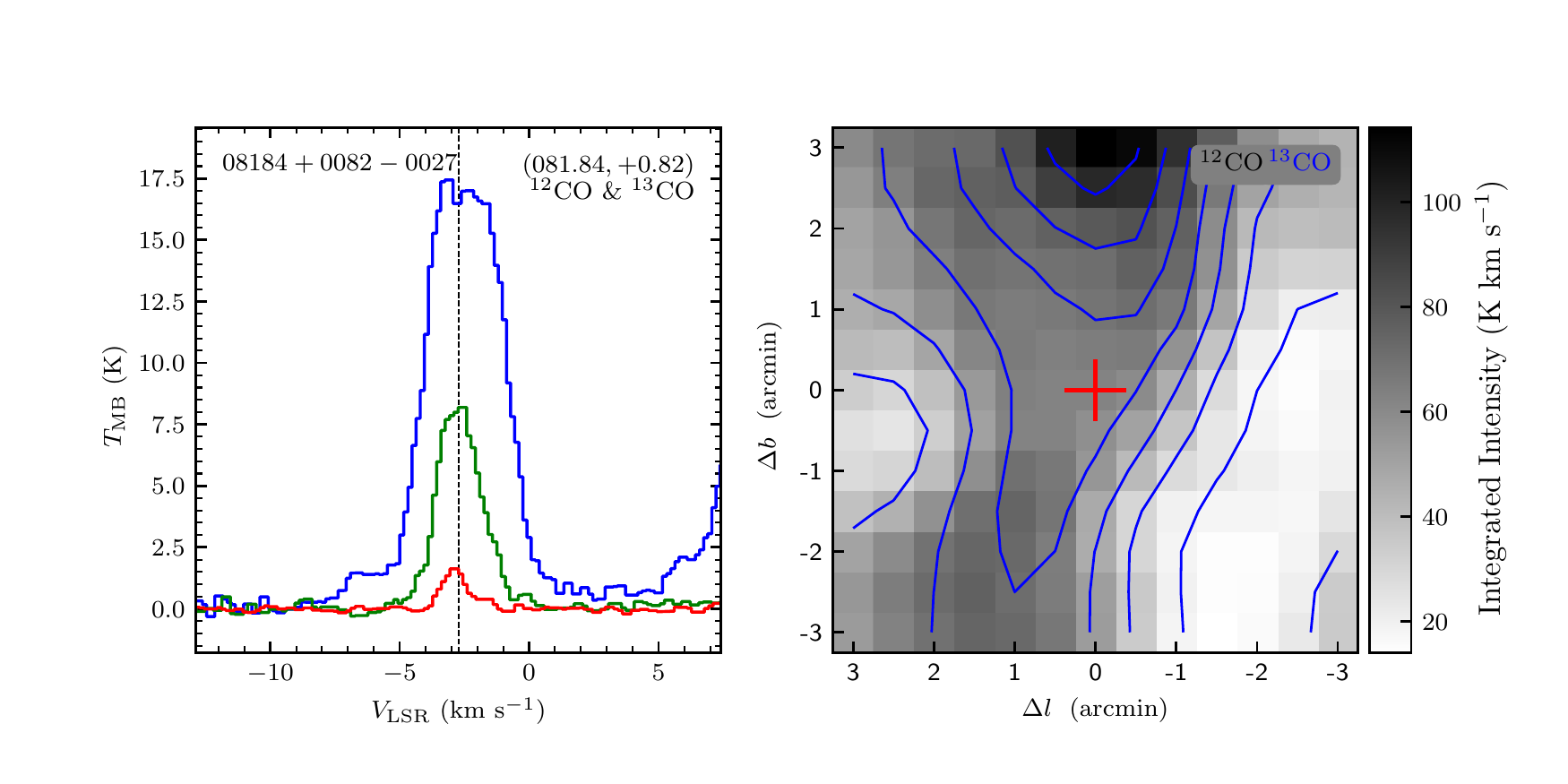}
\vspace{-0.5cm}

\includegraphics[width=9.0cm,angle=0]{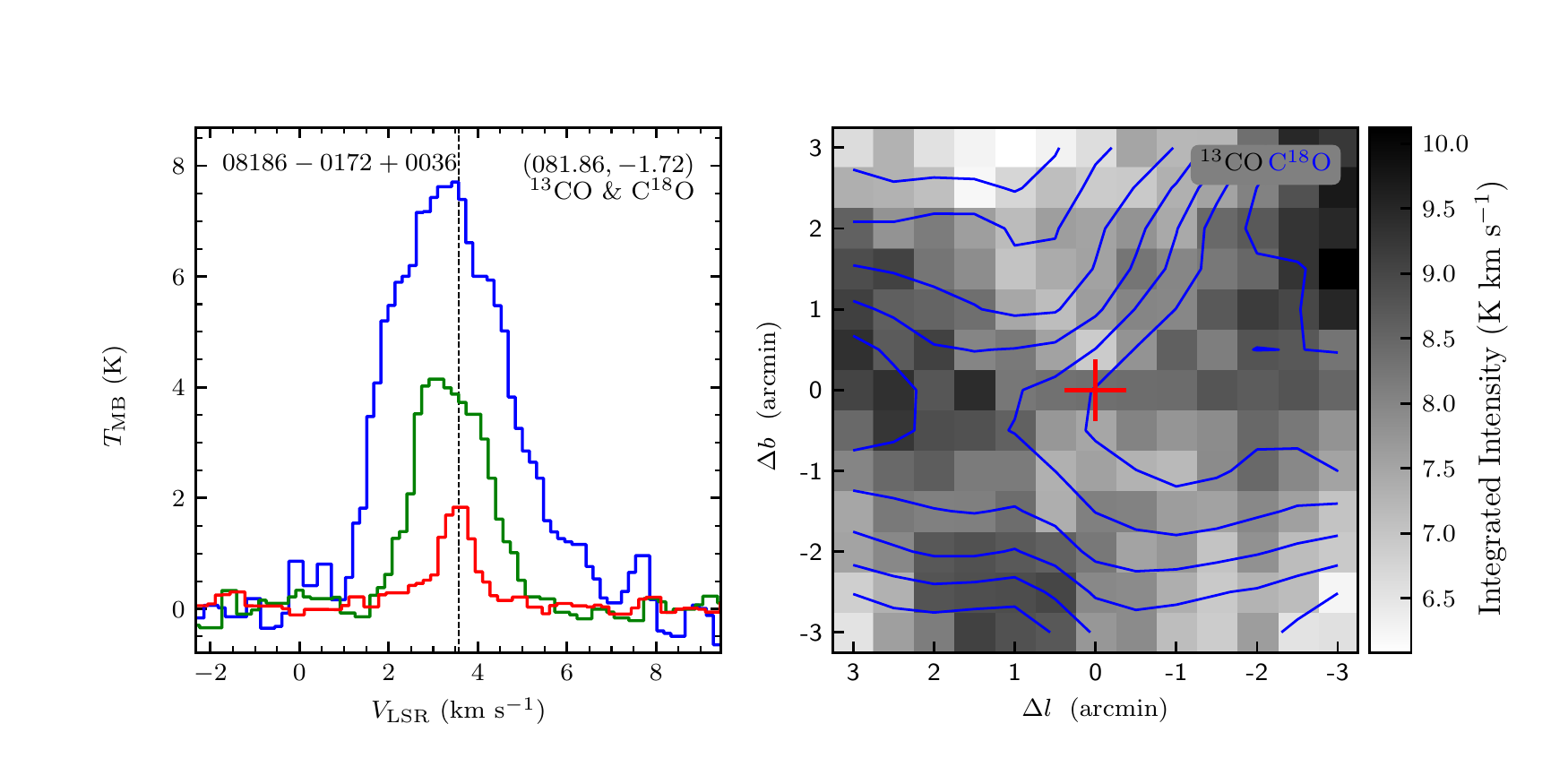}
\includegraphics[width=9.0cm,angle=0]{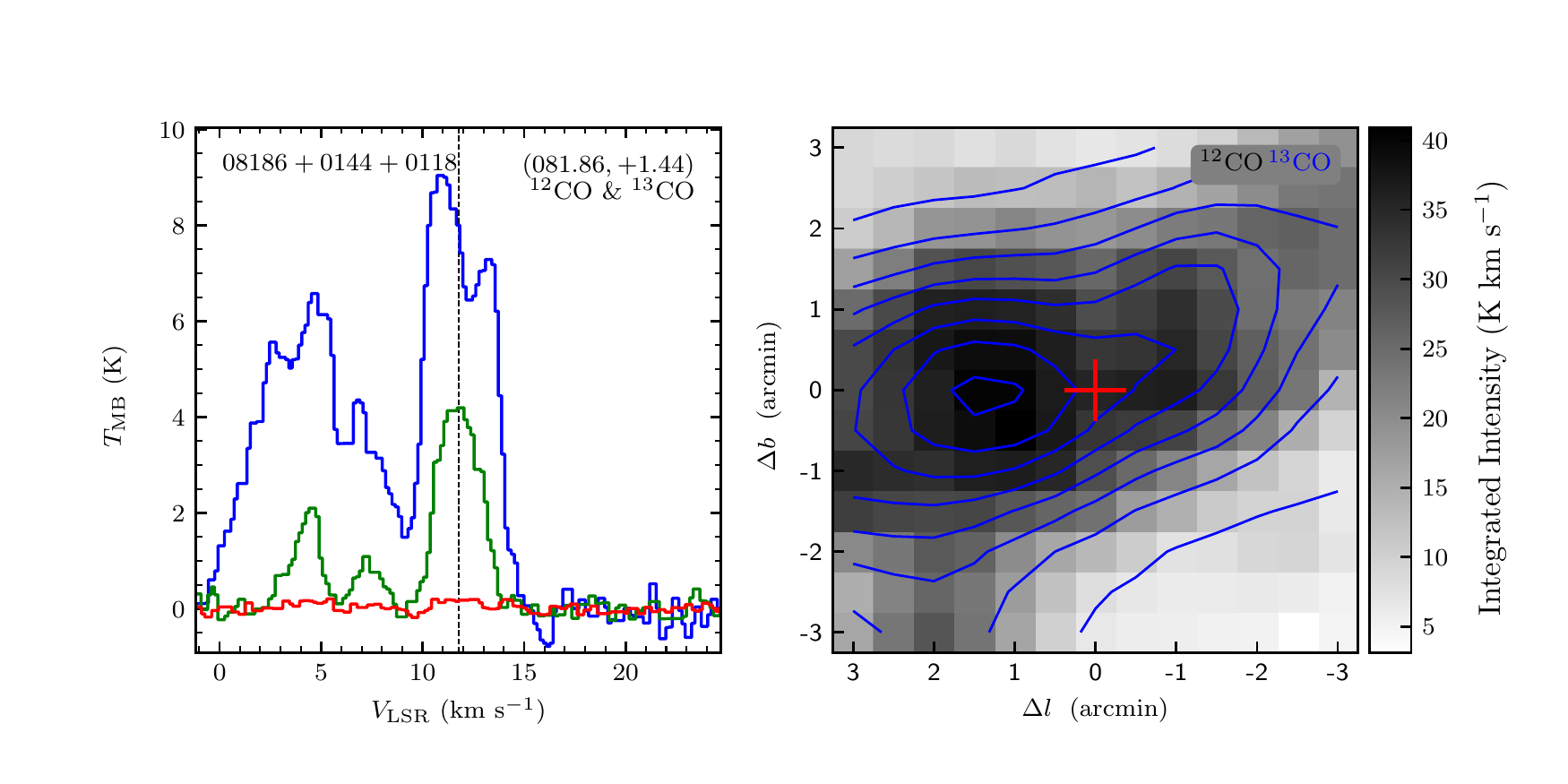}
\vspace{-0.5cm}

\includegraphics[width=9.0cm,angle=0]{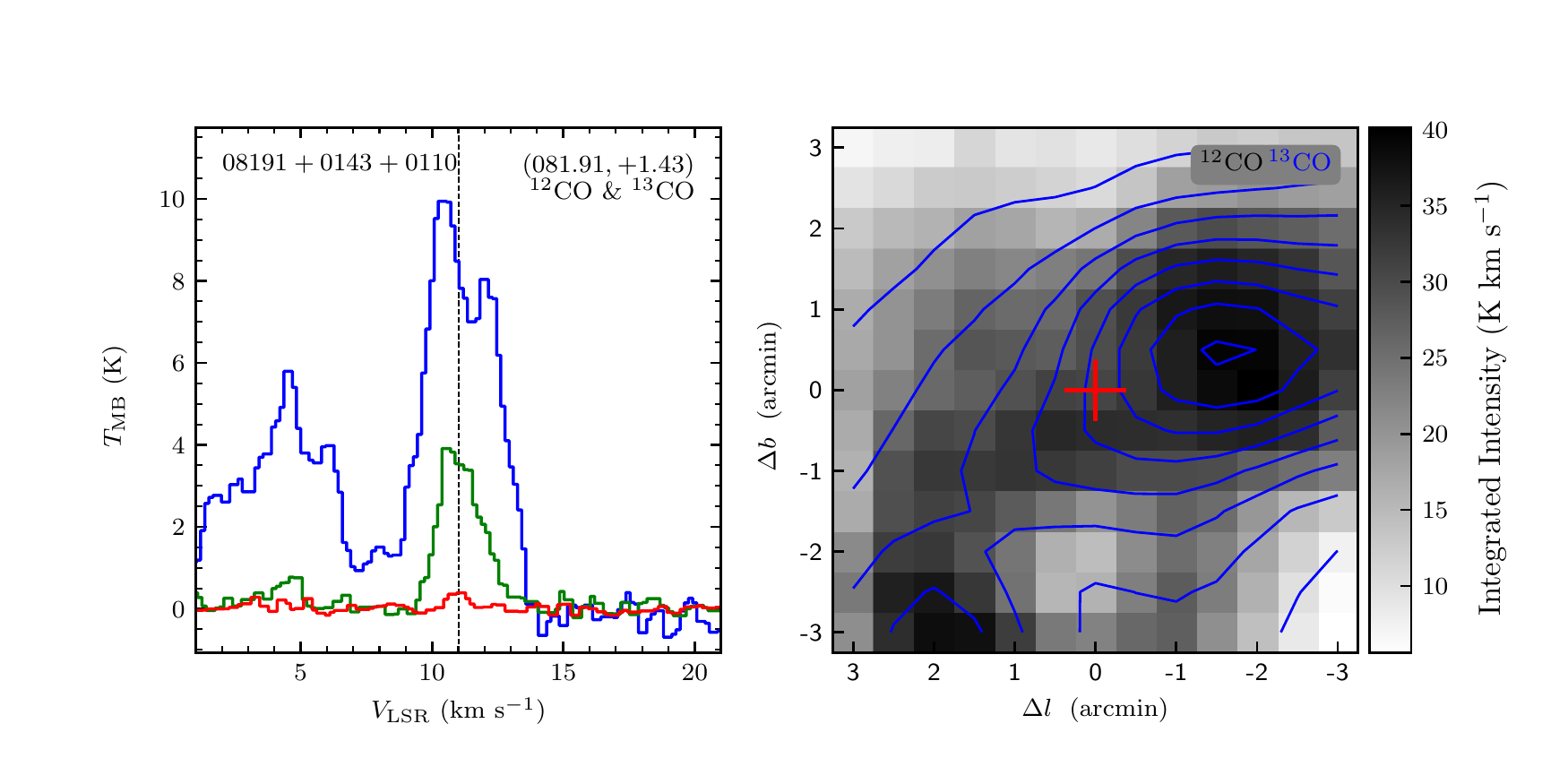}
\includegraphics[width=9.0cm,angle=0]{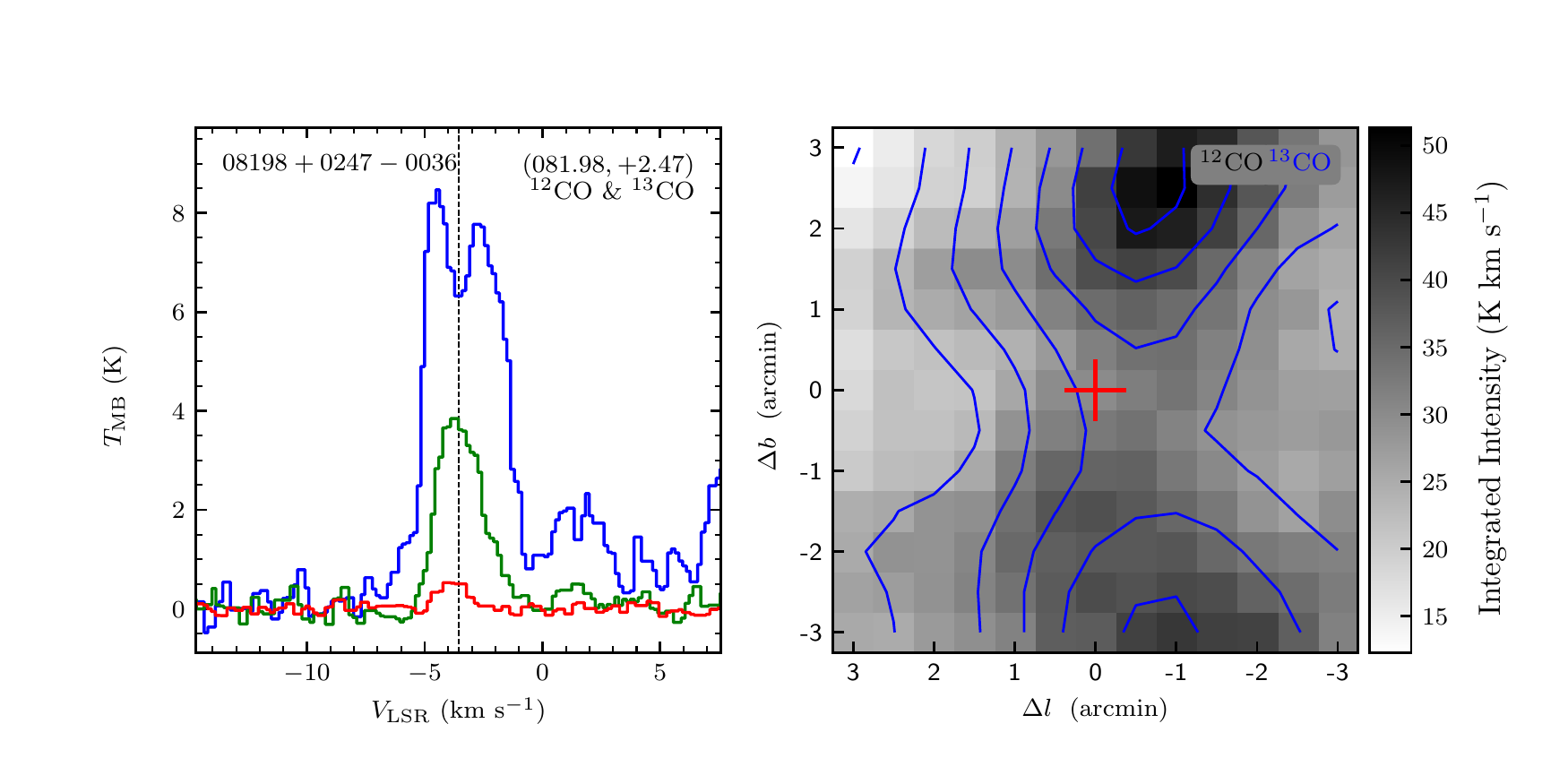}
\end{figure}
\clearpage

\begin{figure}
\includegraphics[width=9.0cm,angle=0]{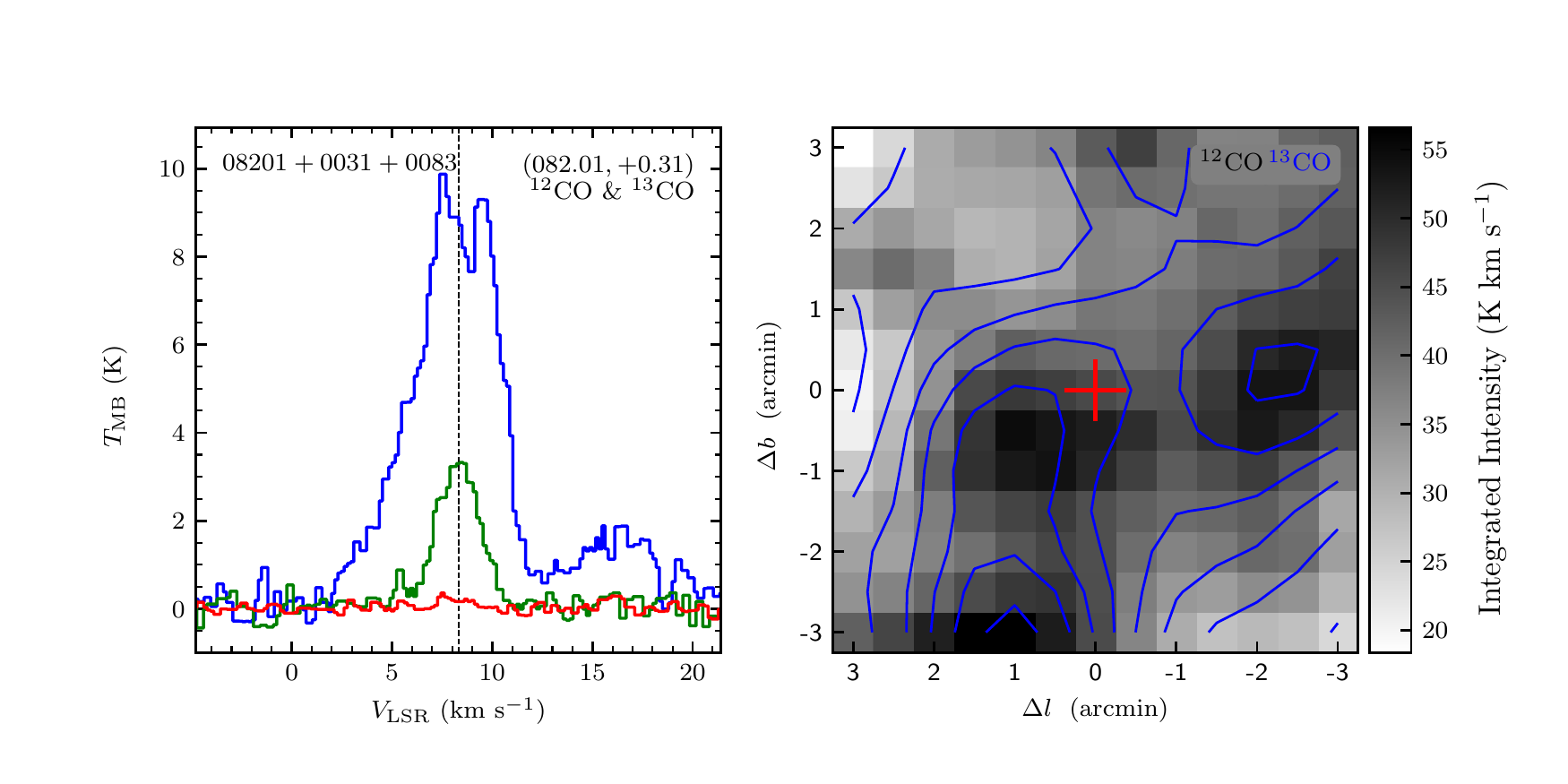}
\includegraphics[width=9.0cm,angle=0]{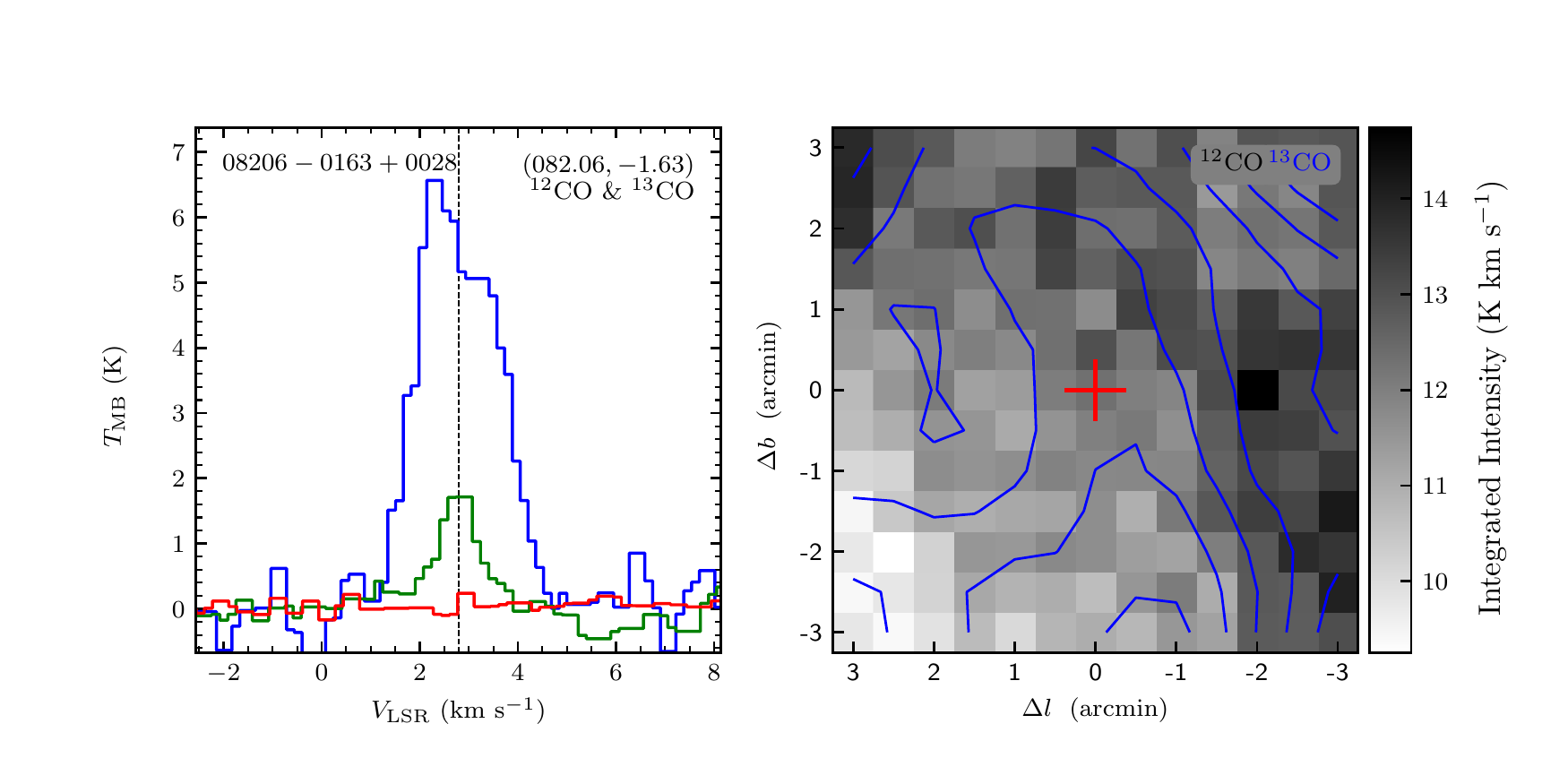}
\vspace{-0.5cm}

\includegraphics[width=9.0cm,angle=0]{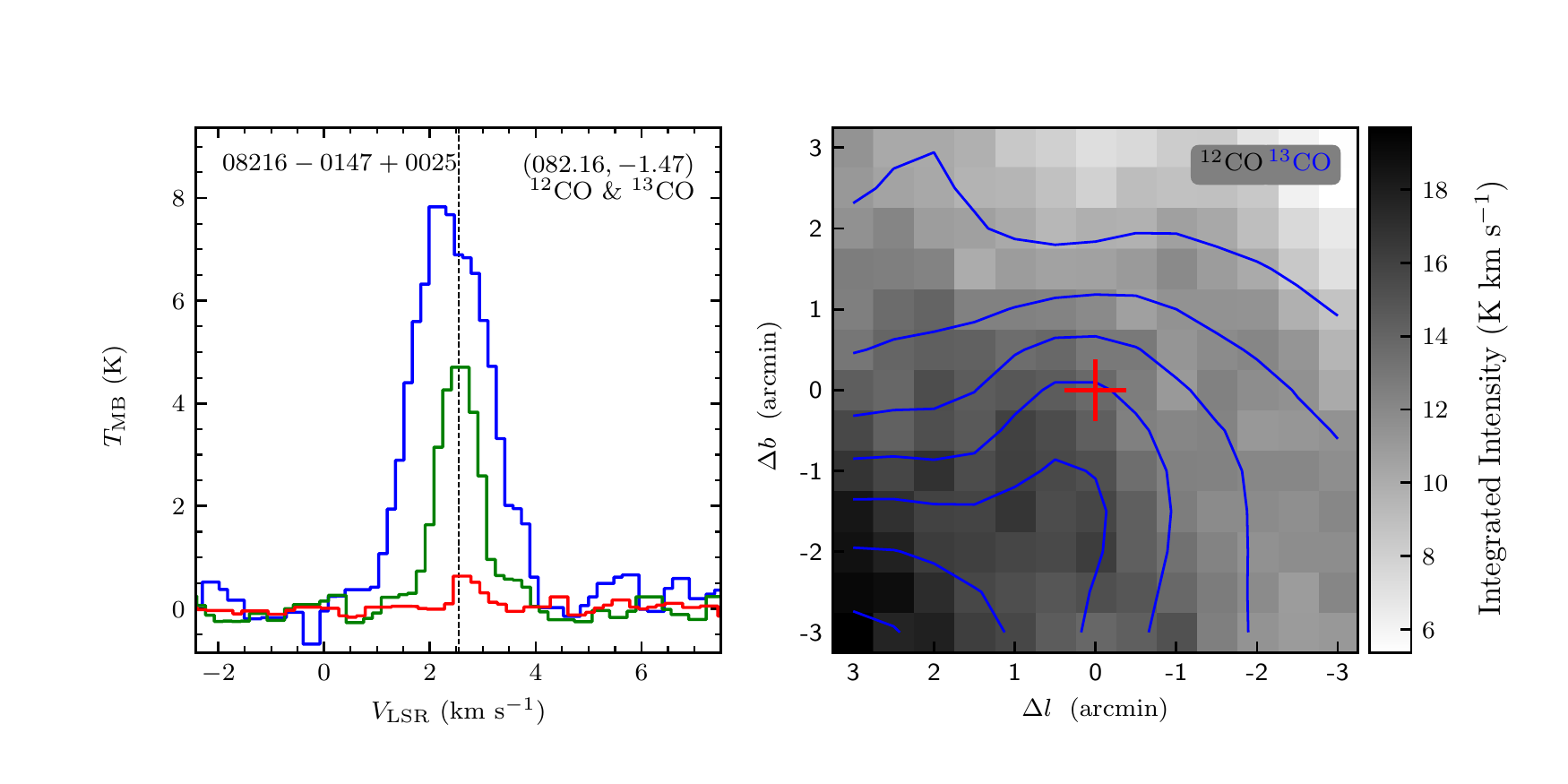}
\includegraphics[width=9.0cm,angle=0]{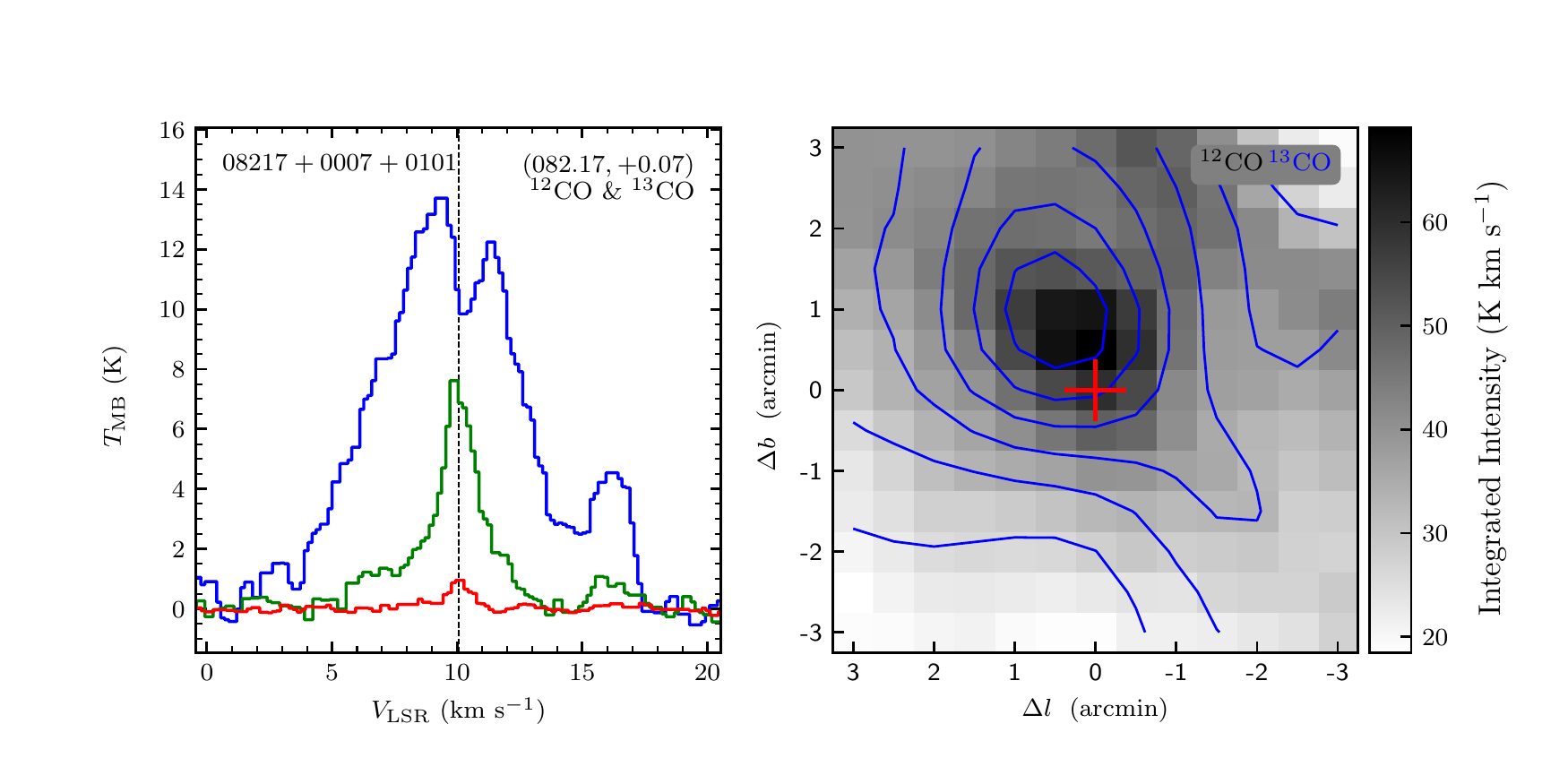}
\vspace{-0.5cm}

\includegraphics[width=9.0cm,angle=0]{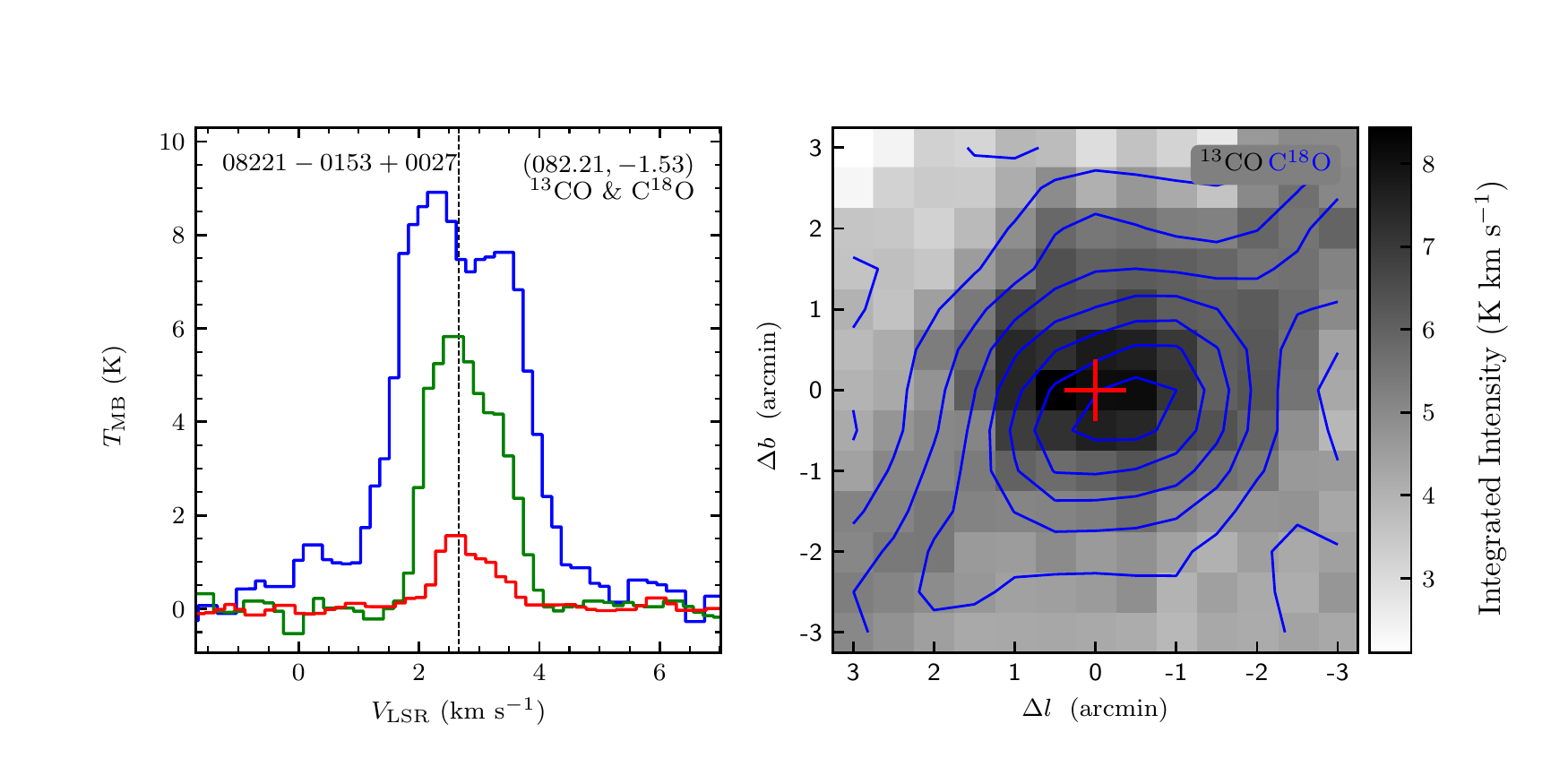}
\includegraphics[width=9.0cm,angle=0]{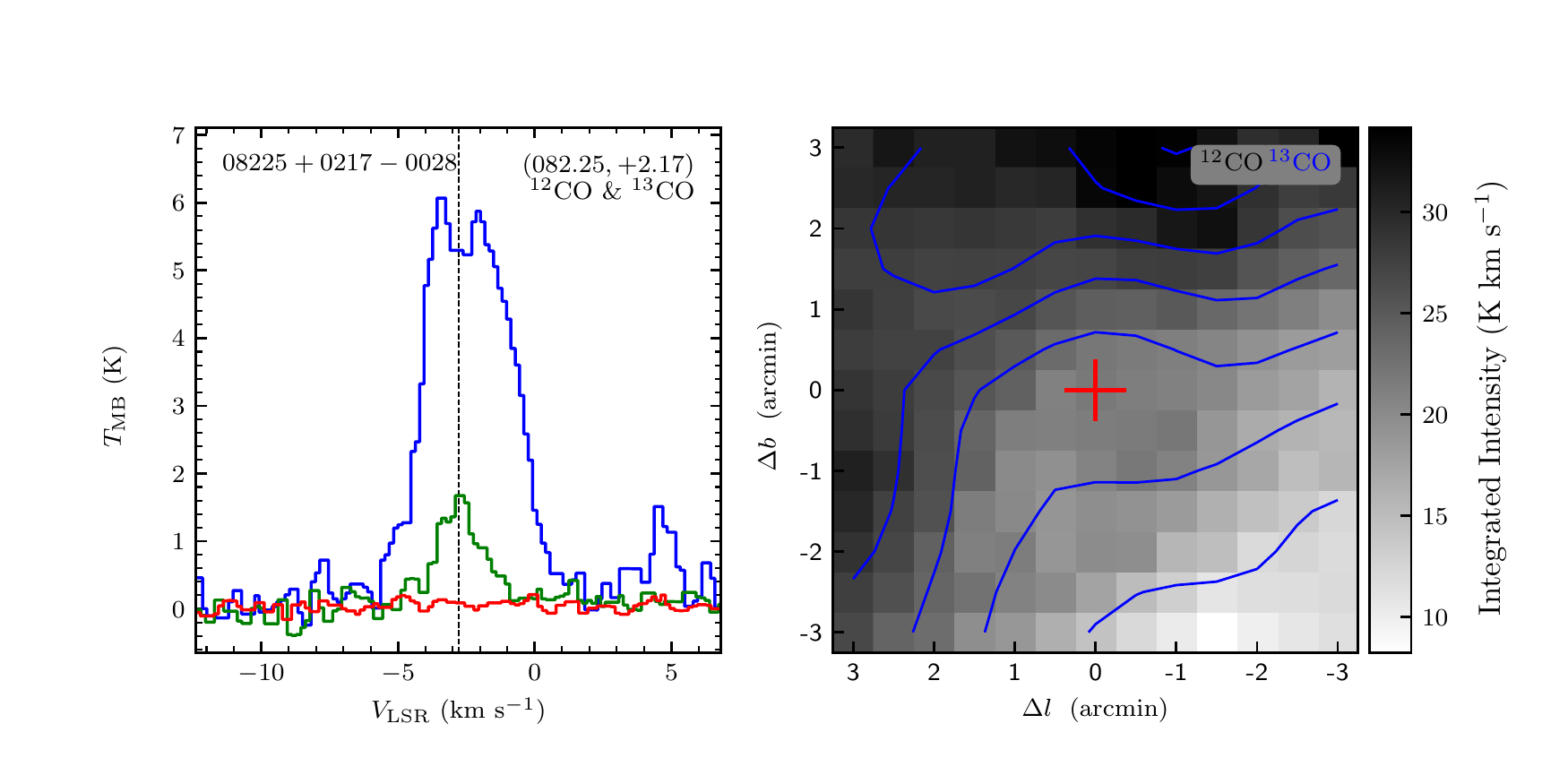}
\vspace{-0.5cm}

\includegraphics[width=9.0cm,angle=0]{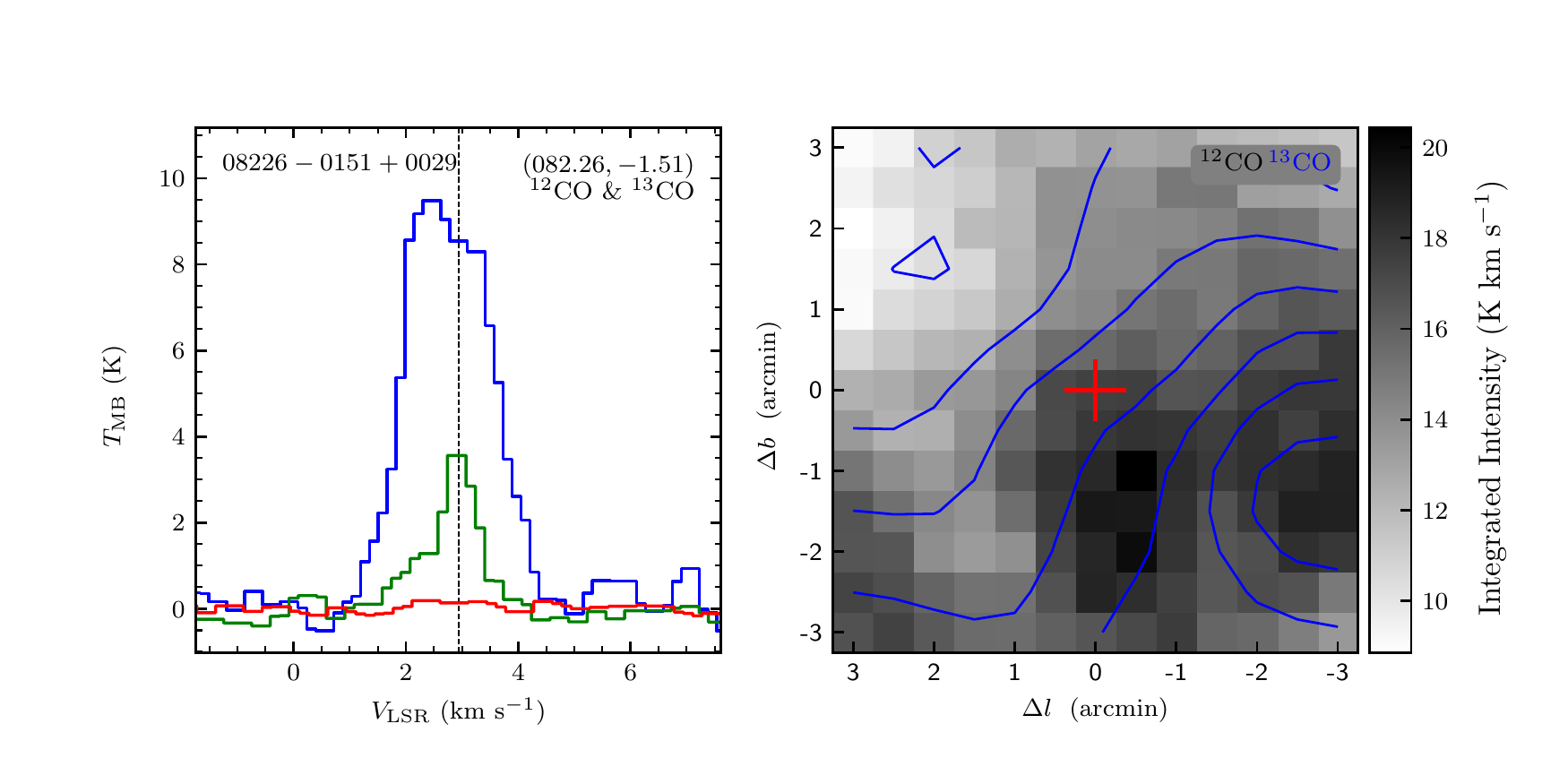}
\includegraphics[width=9.0cm,angle=0]{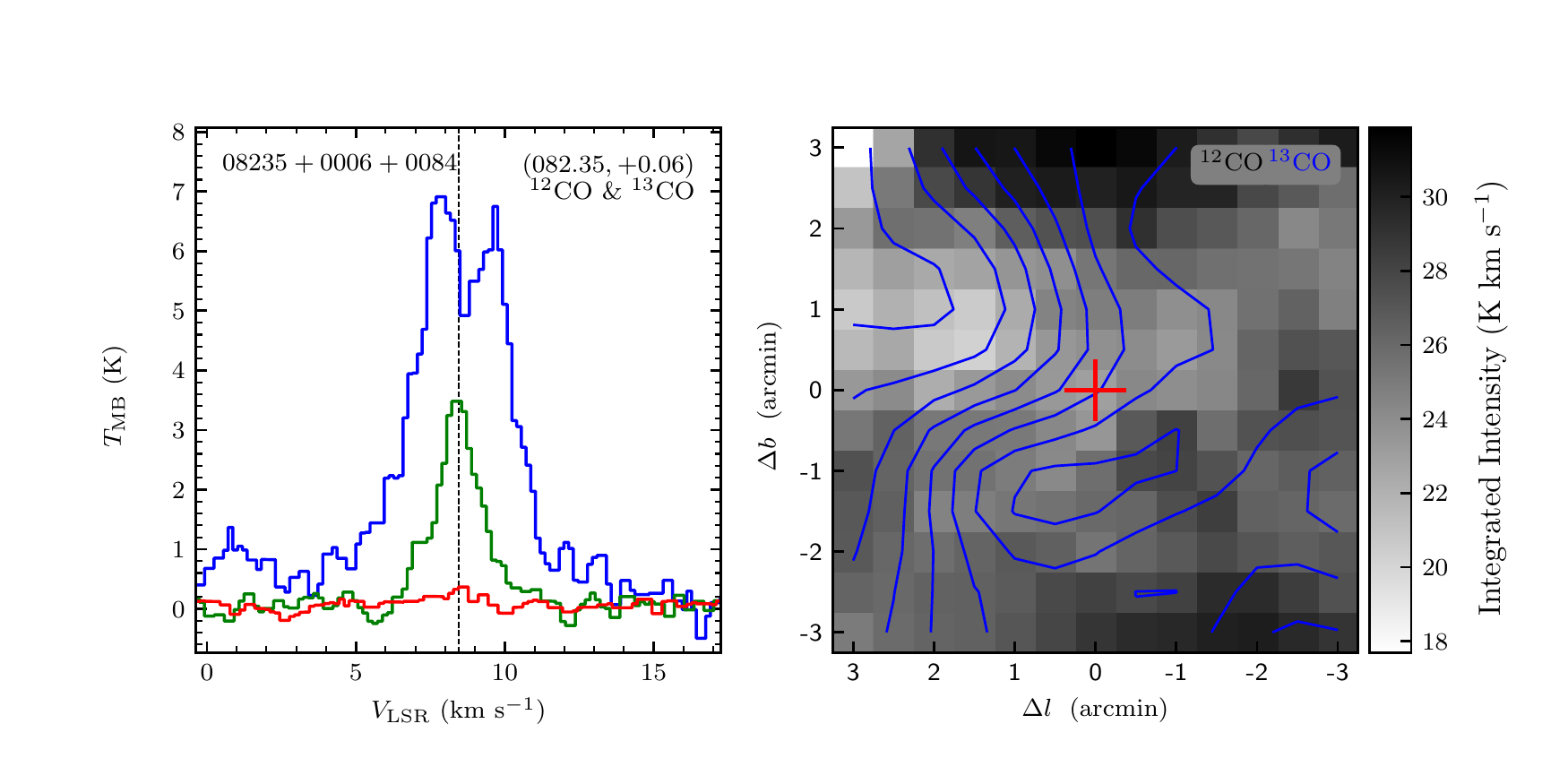}
\vspace{-0.5cm}

\includegraphics[width=9.0cm,angle=0]{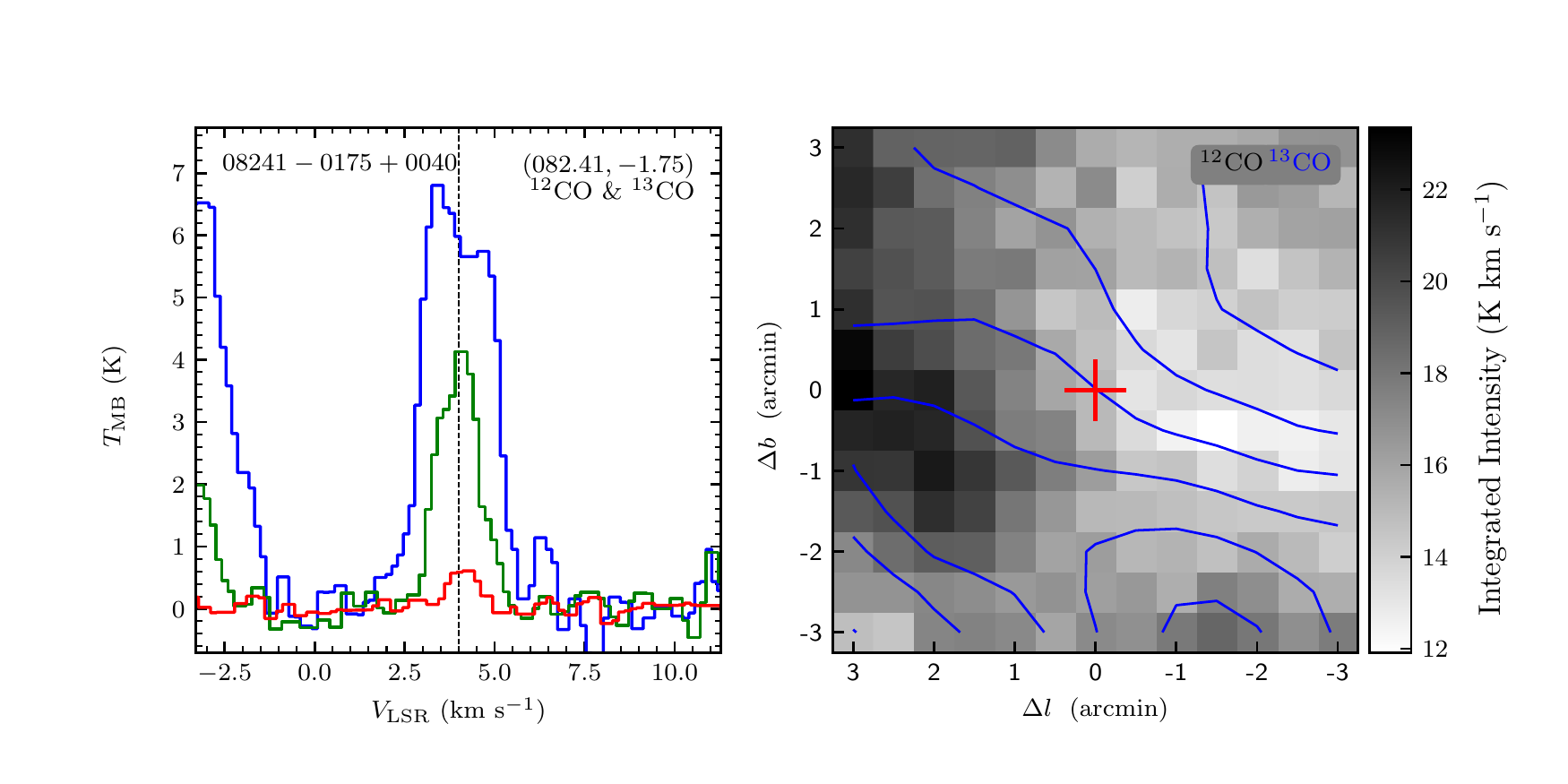}
\includegraphics[width=9.0cm,angle=0]{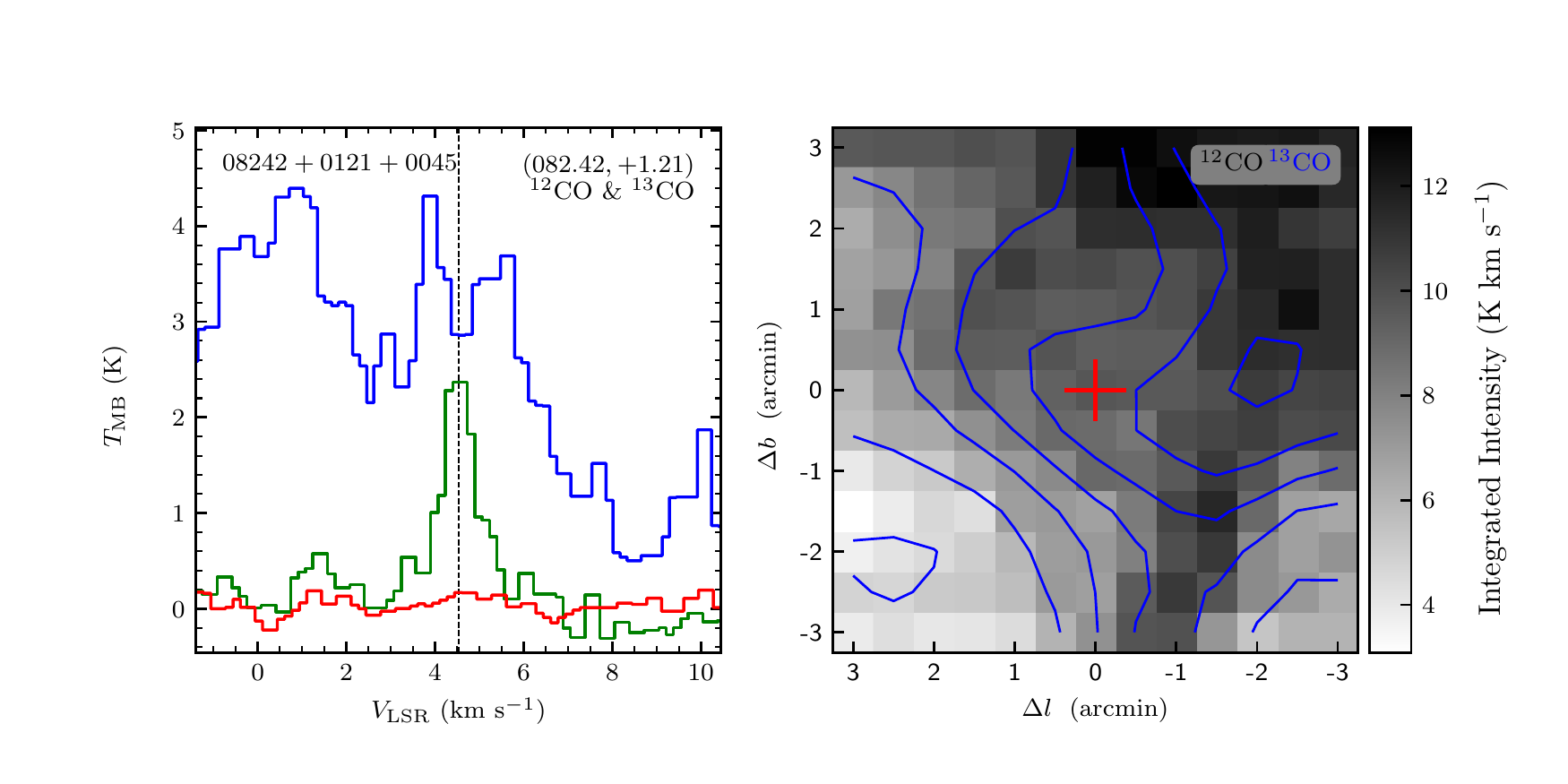}
\end{figure}
\clearpage

\begin{figure}
\includegraphics[width=9.0cm,angle=0]{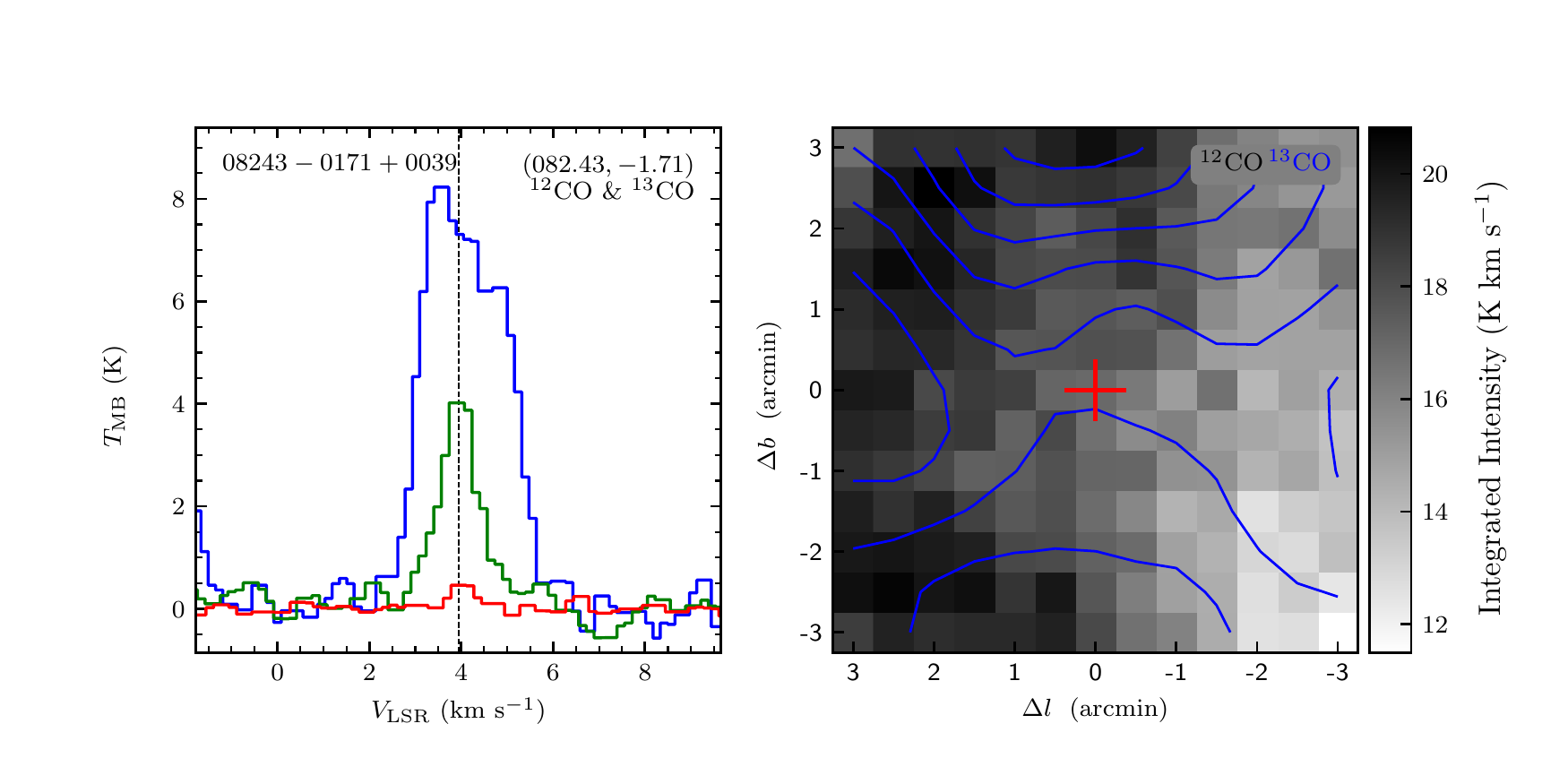}
\includegraphics[width=9.0cm,angle=0]{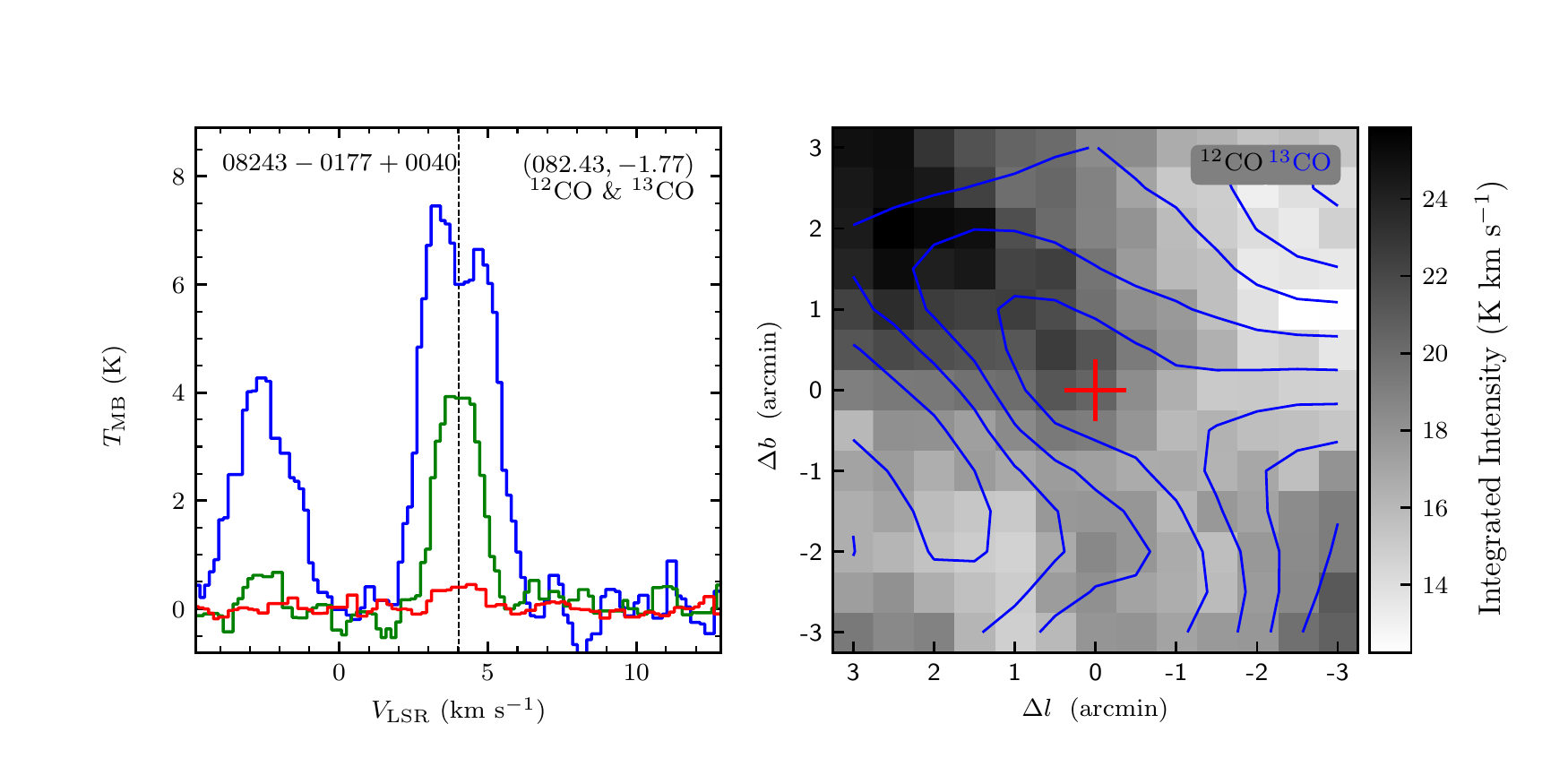}
\vspace{-0.5cm}

\includegraphics[width=9.0cm,angle=0]{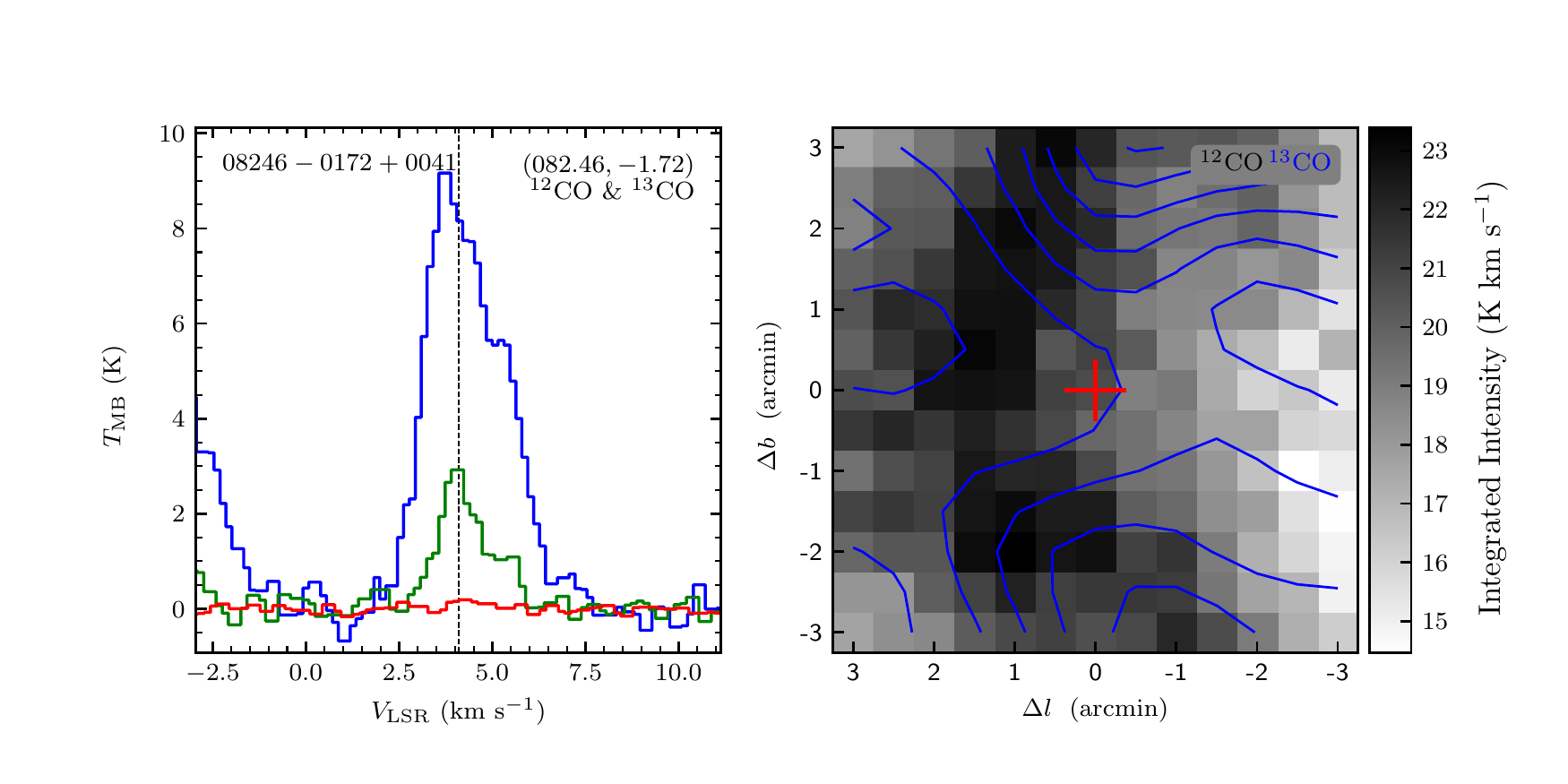}
\includegraphics[width=9.0cm,angle=0]{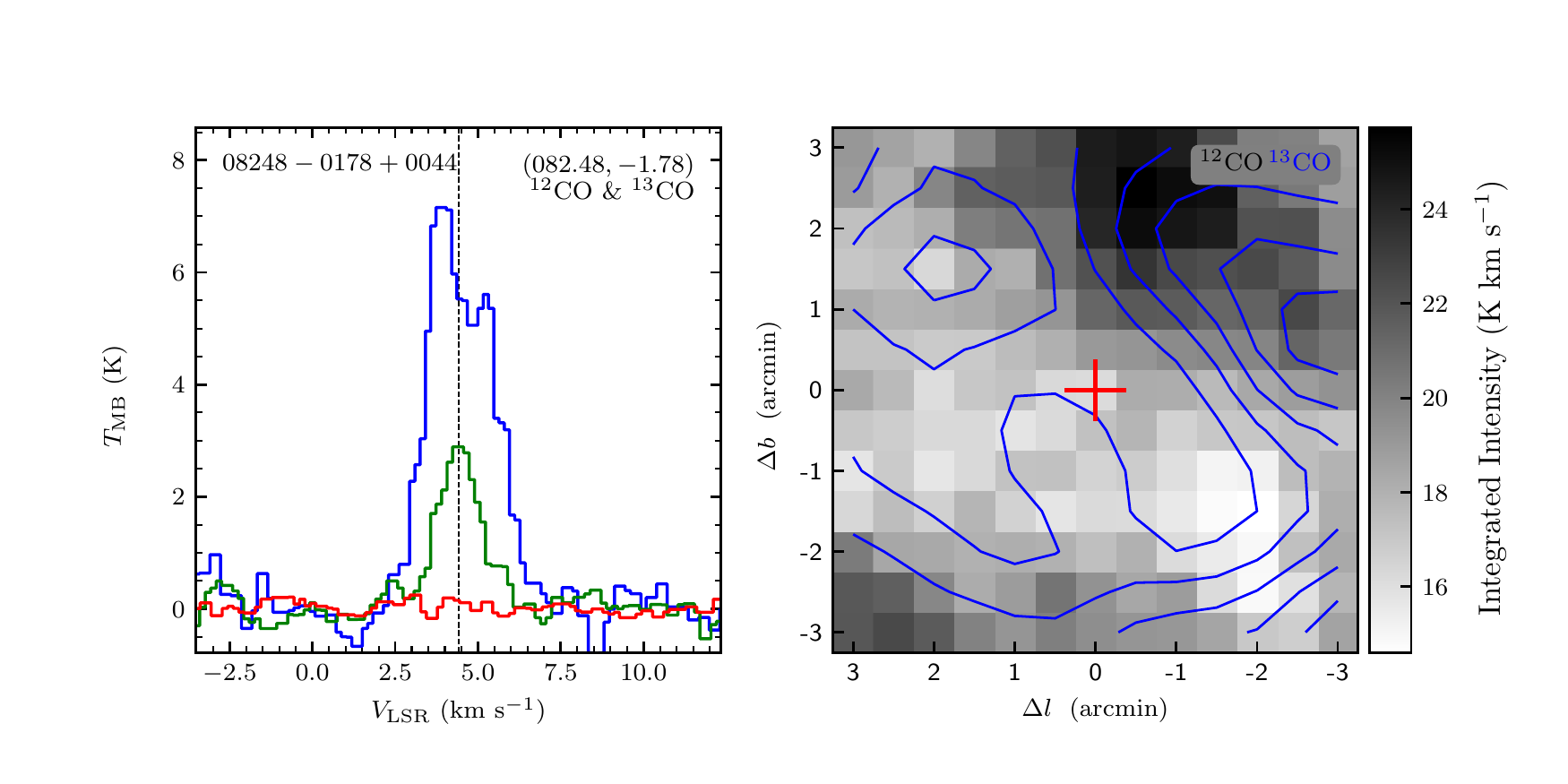}
\vspace{-0.5cm}

\includegraphics[width=9.0cm,angle=0]{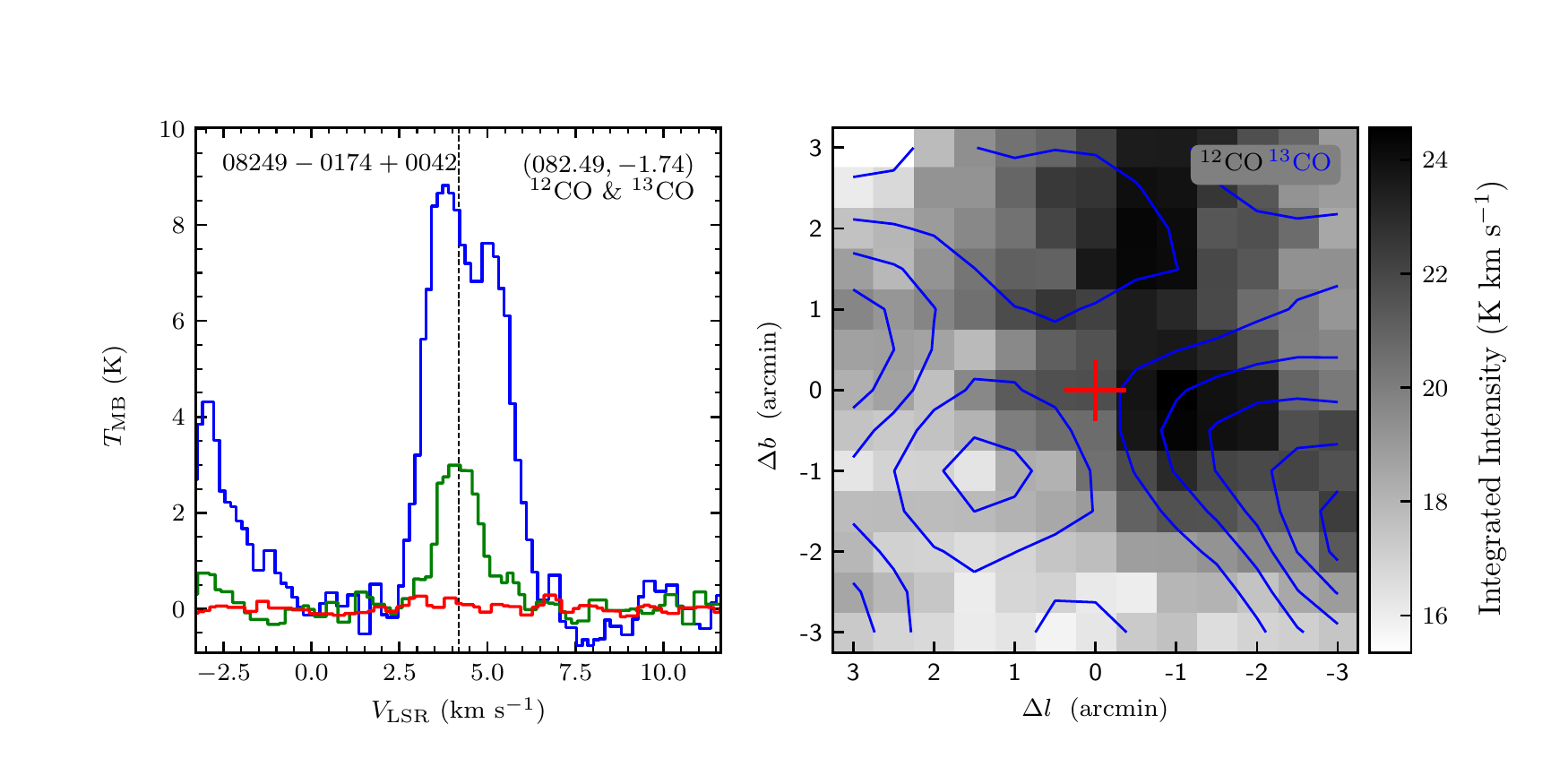}
\includegraphics[width=9.0cm,angle=0]{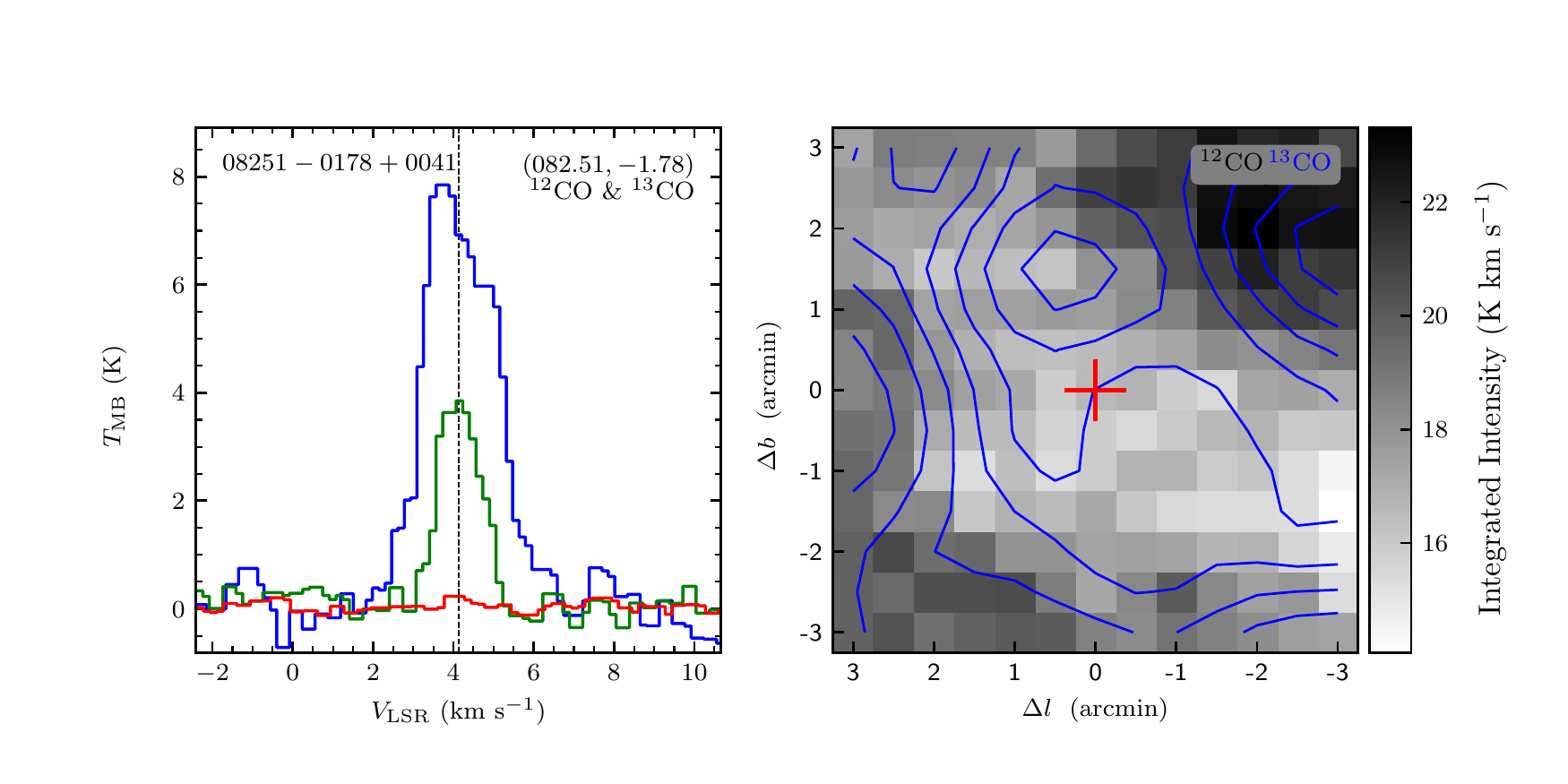}
\vspace{-0.5cm}

\includegraphics[width=9.0cm,angle=0]{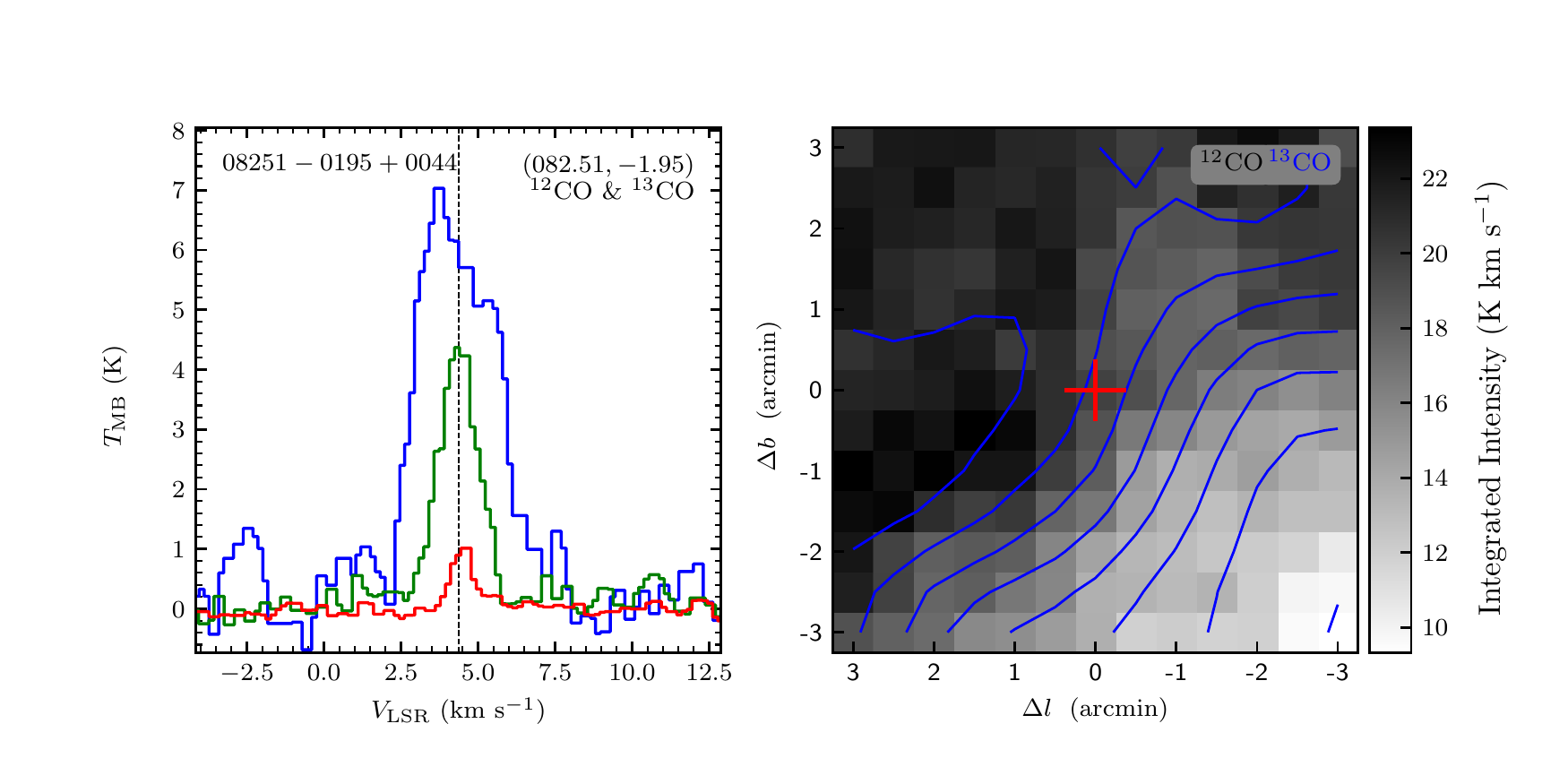}
\includegraphics[width=9.0cm,angle=0]{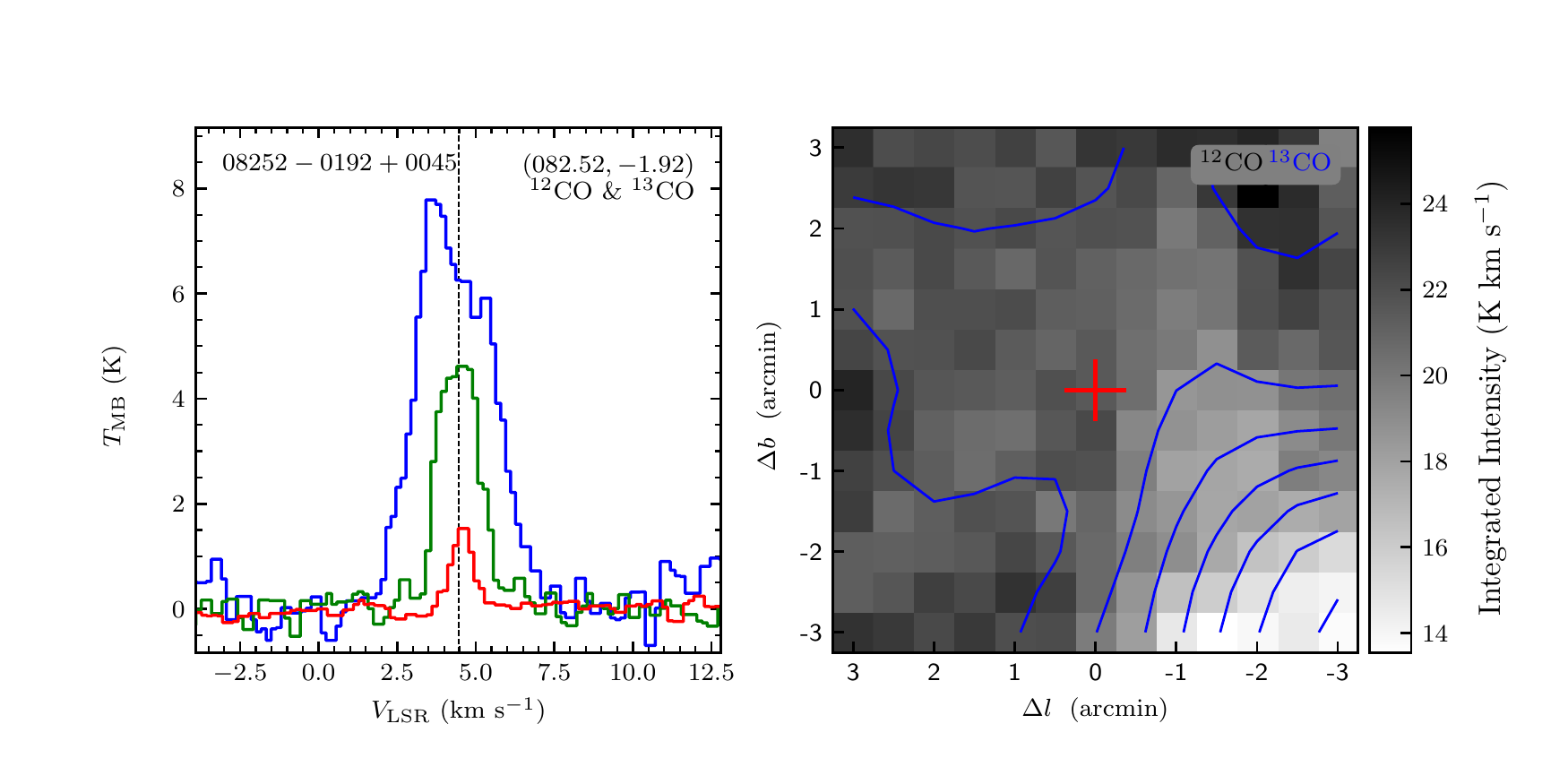}
\vspace{-0.5cm}

\includegraphics[width=9.0cm,angle=0]{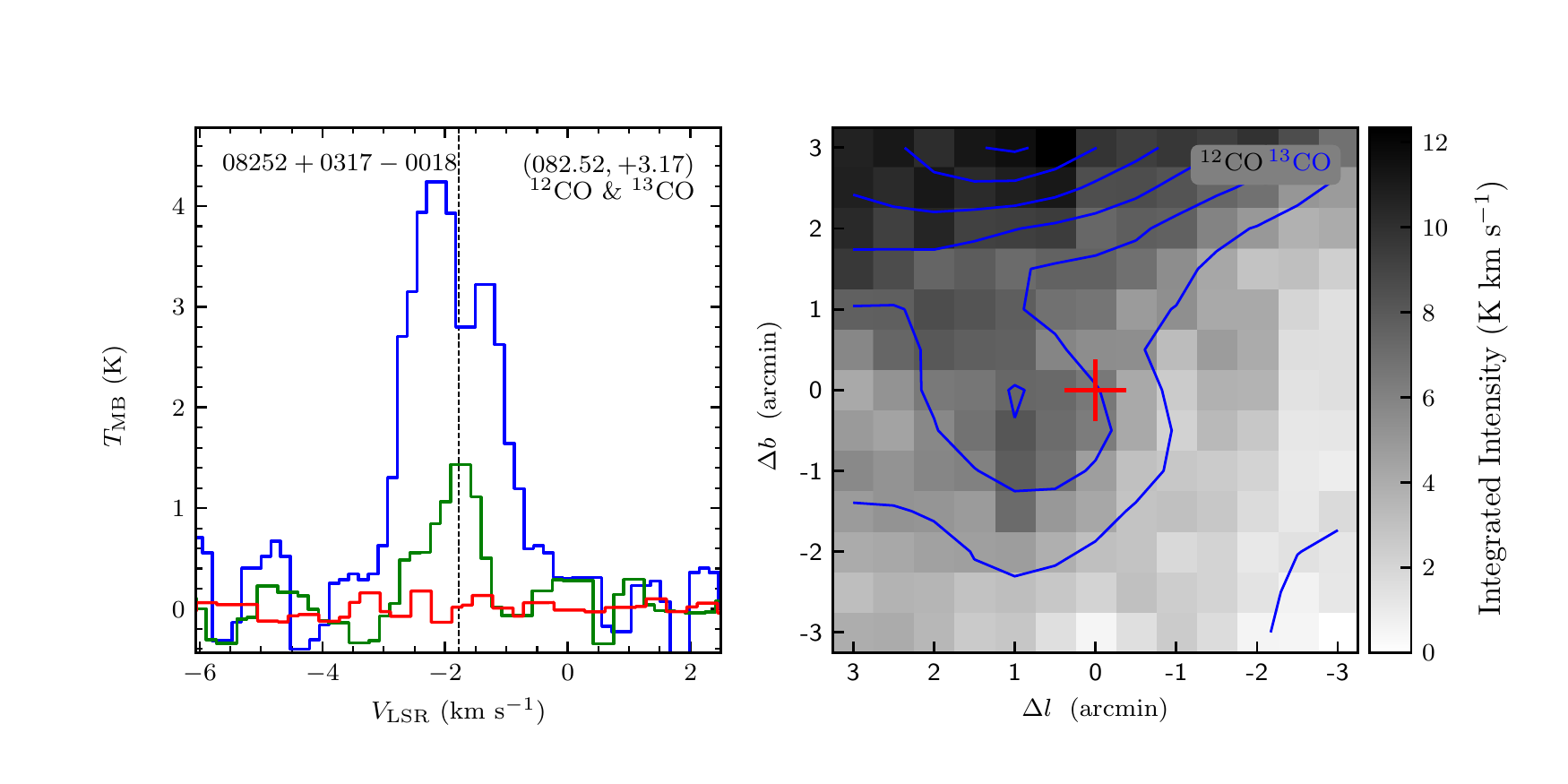}
\includegraphics[width=9.0cm,angle=0]{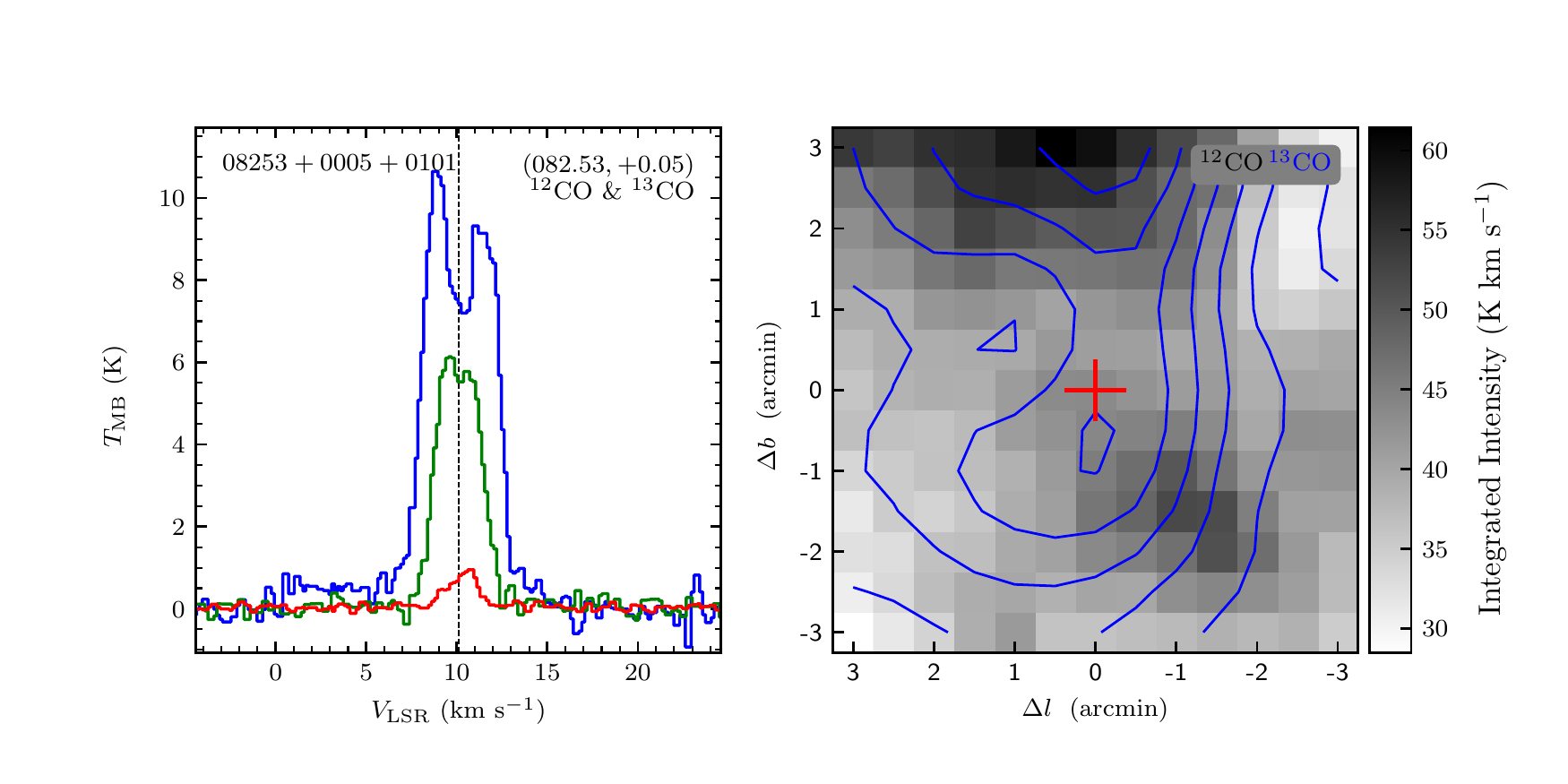}
\end{figure}
\clearpage

\begin{figure}
\includegraphics[width=9.0cm,angle=0]{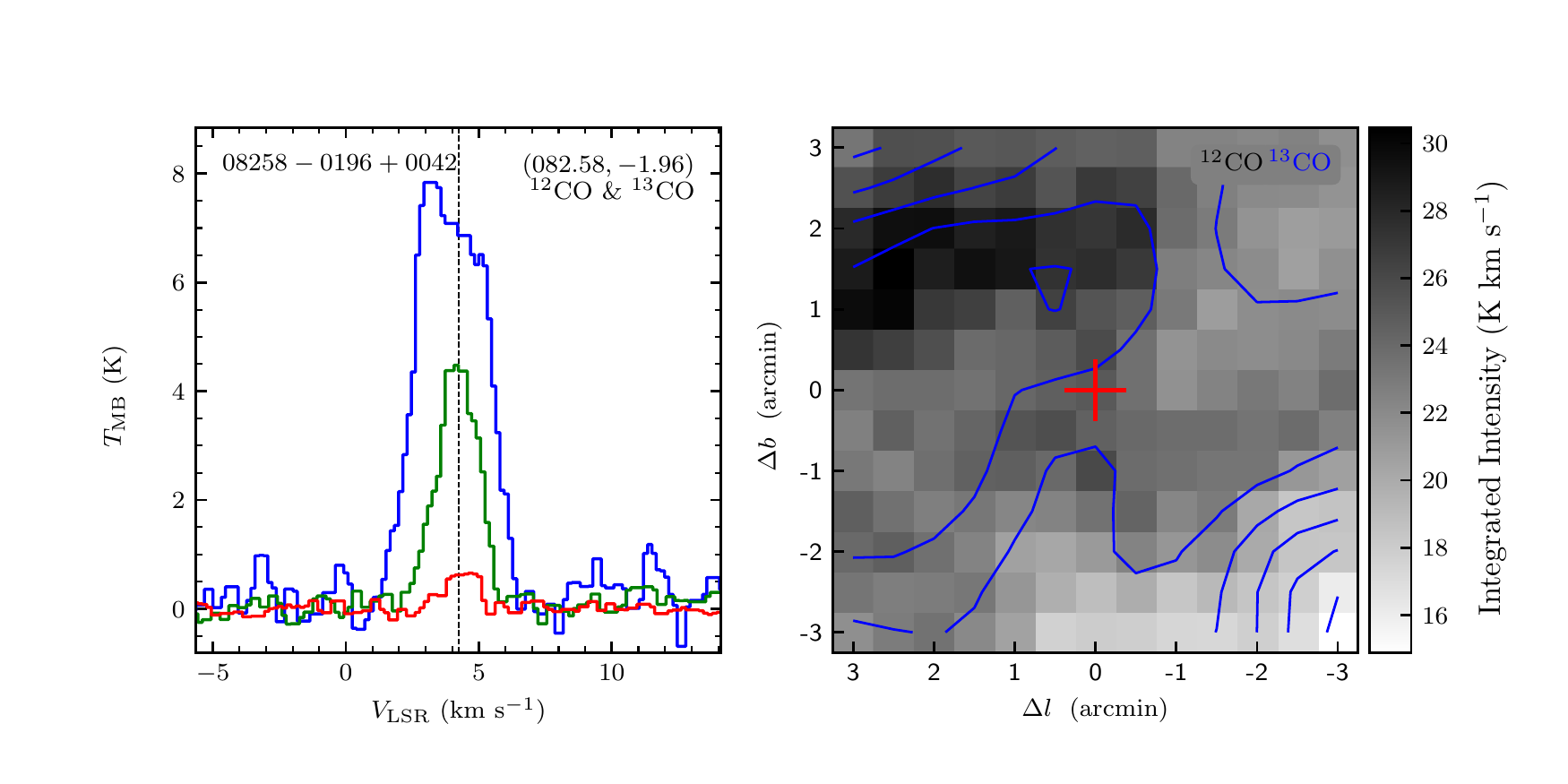}
\includegraphics[width=9.0cm,angle=0]{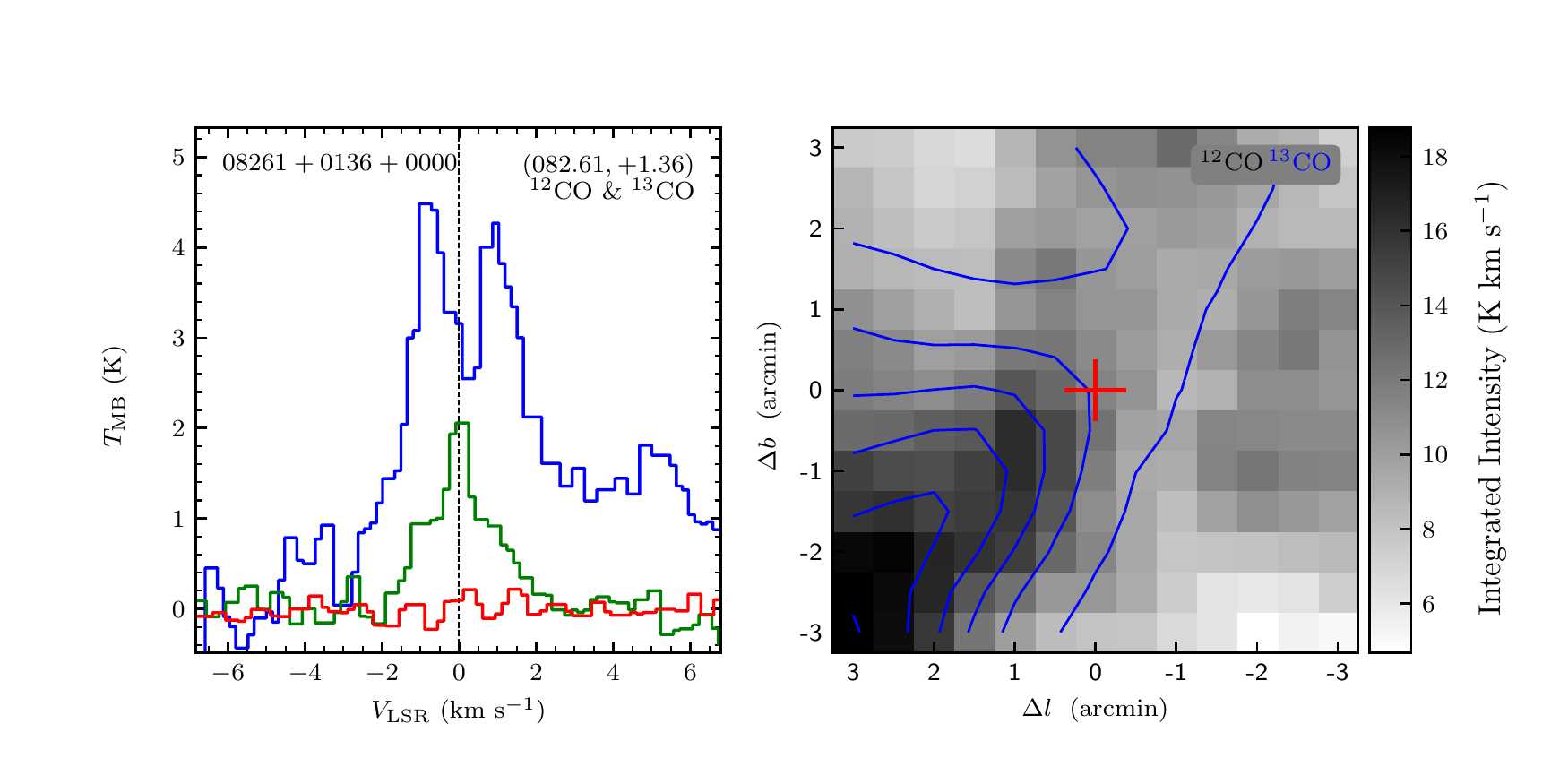}
\vspace{-0.5cm}

\includegraphics[width=9.0cm,angle=0]{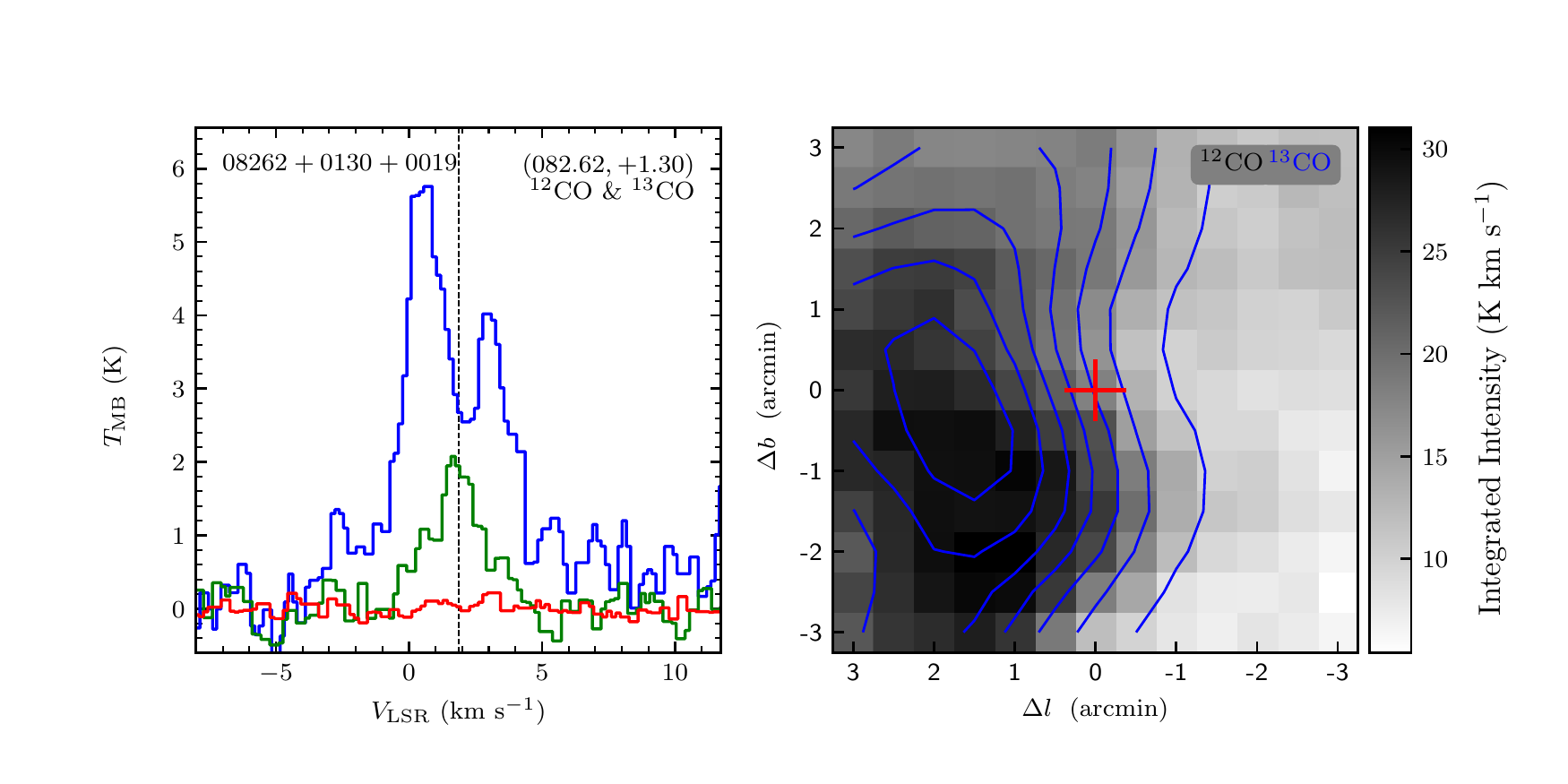}
\includegraphics[width=9.0cm,angle=0]{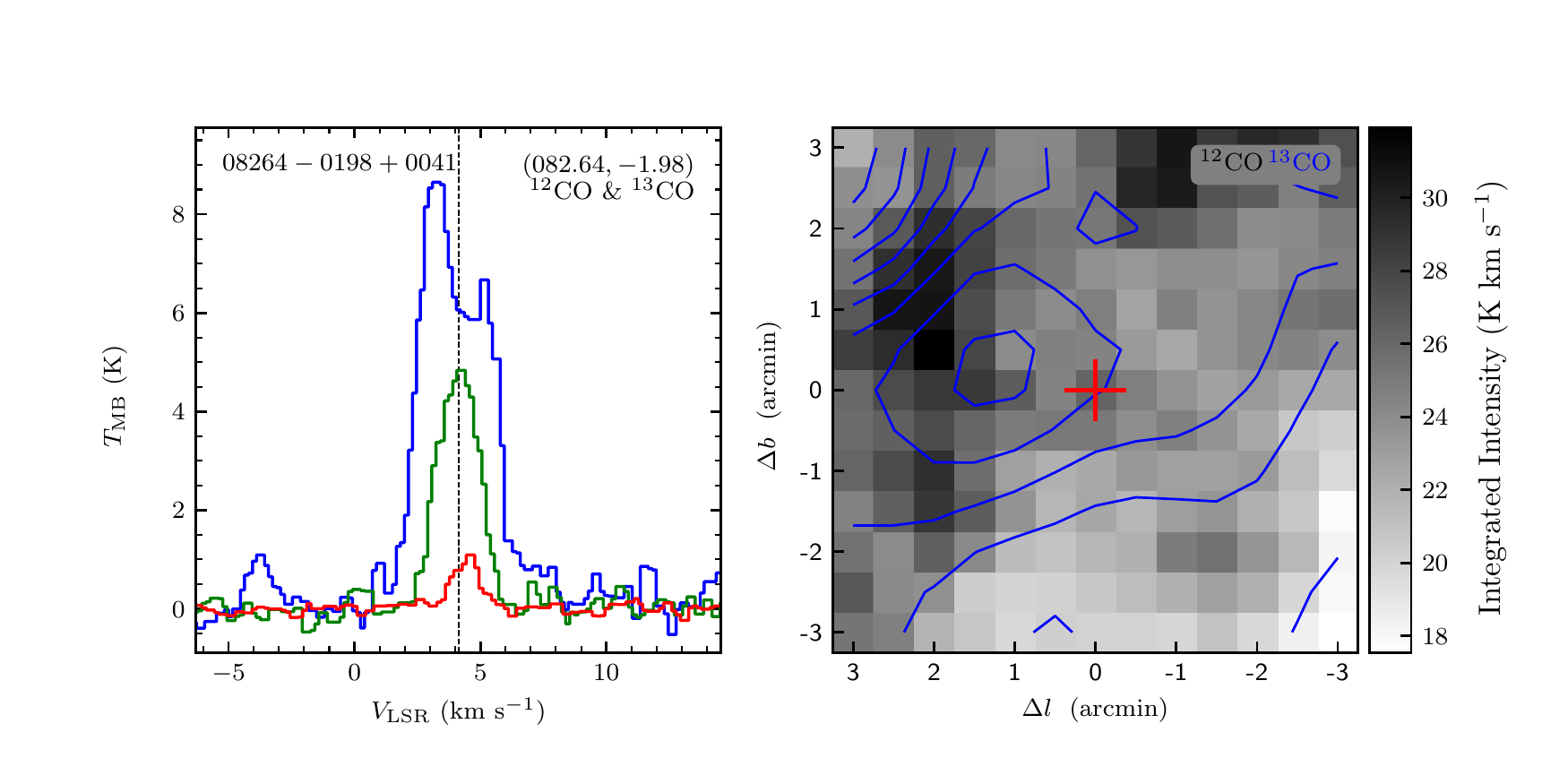}
\vspace{-0.5cm}

\includegraphics[width=9.0cm,angle=0]{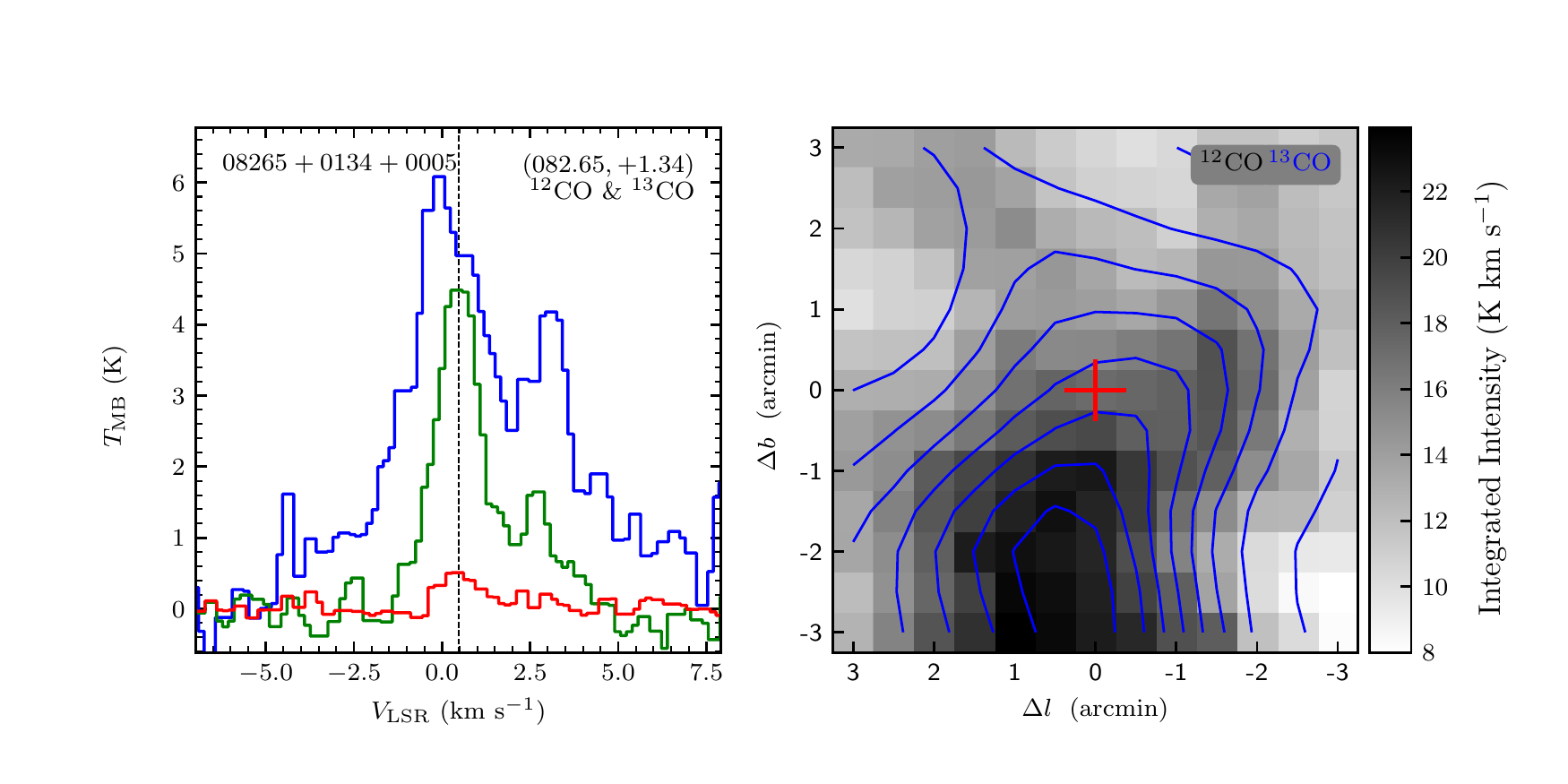}
\includegraphics[width=9.0cm,angle=0]{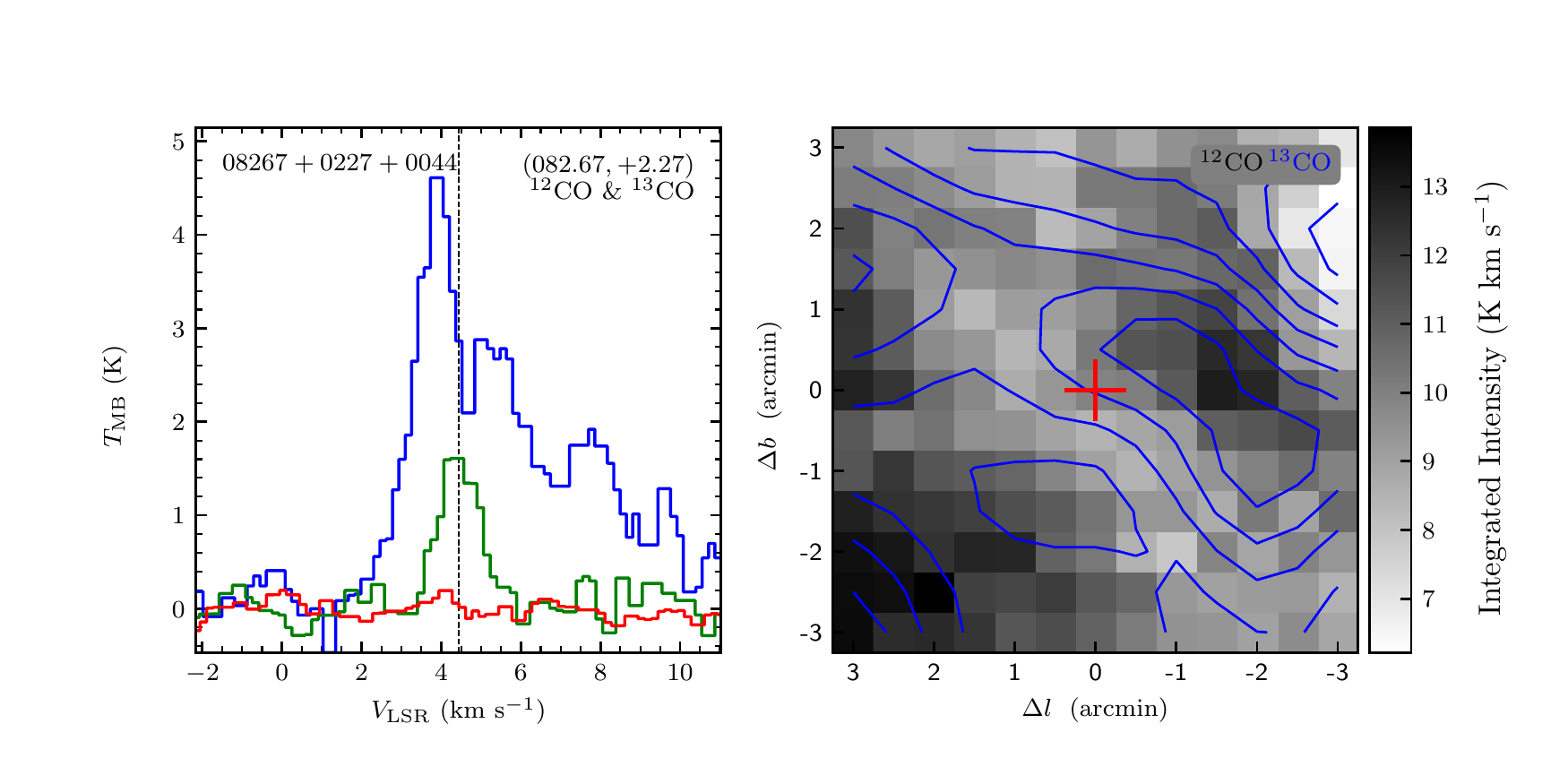}
\vspace{-0.5cm}

\includegraphics[width=9.0cm,angle=0]{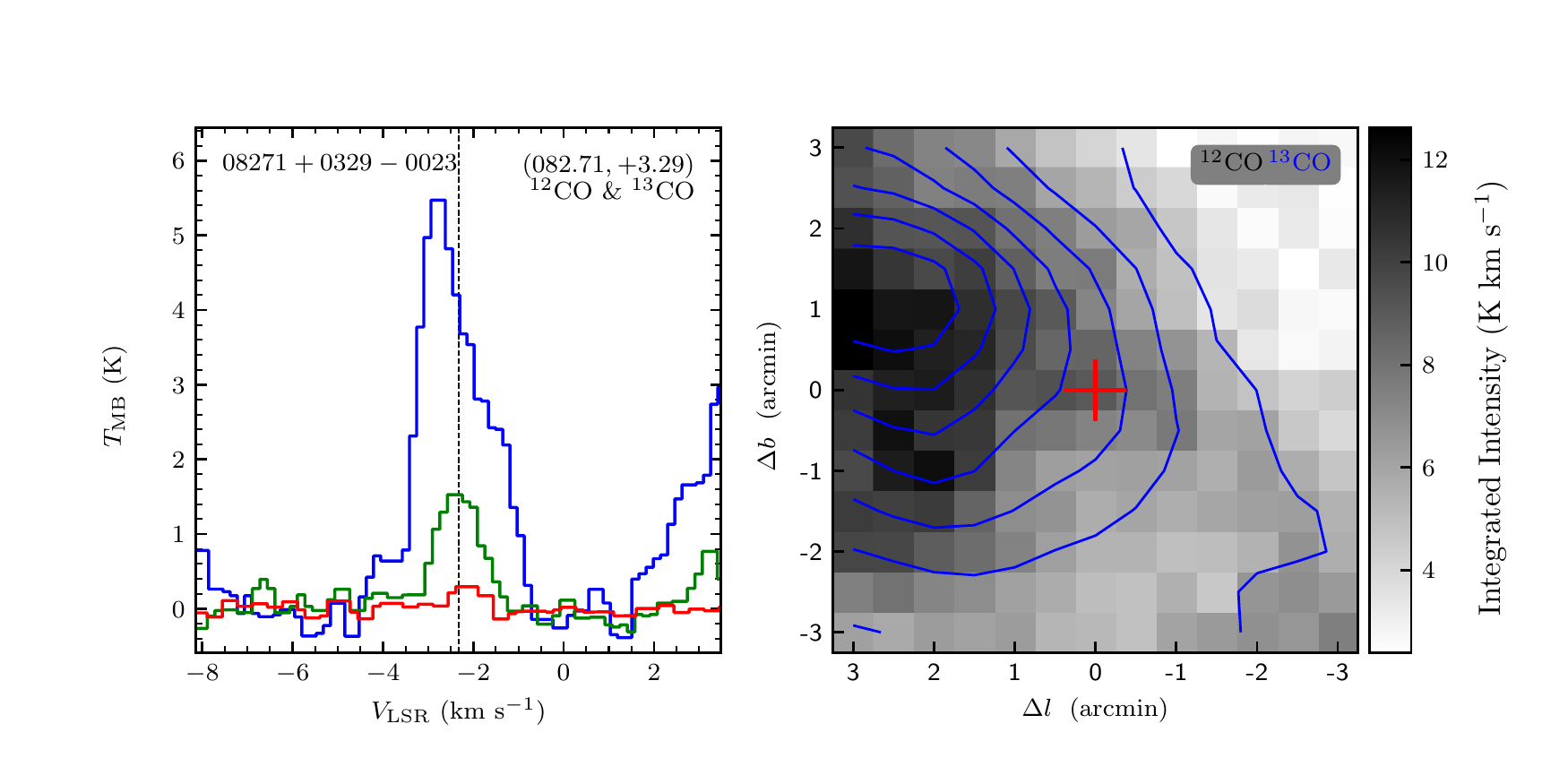}
\includegraphics[width=9.0cm,angle=0]{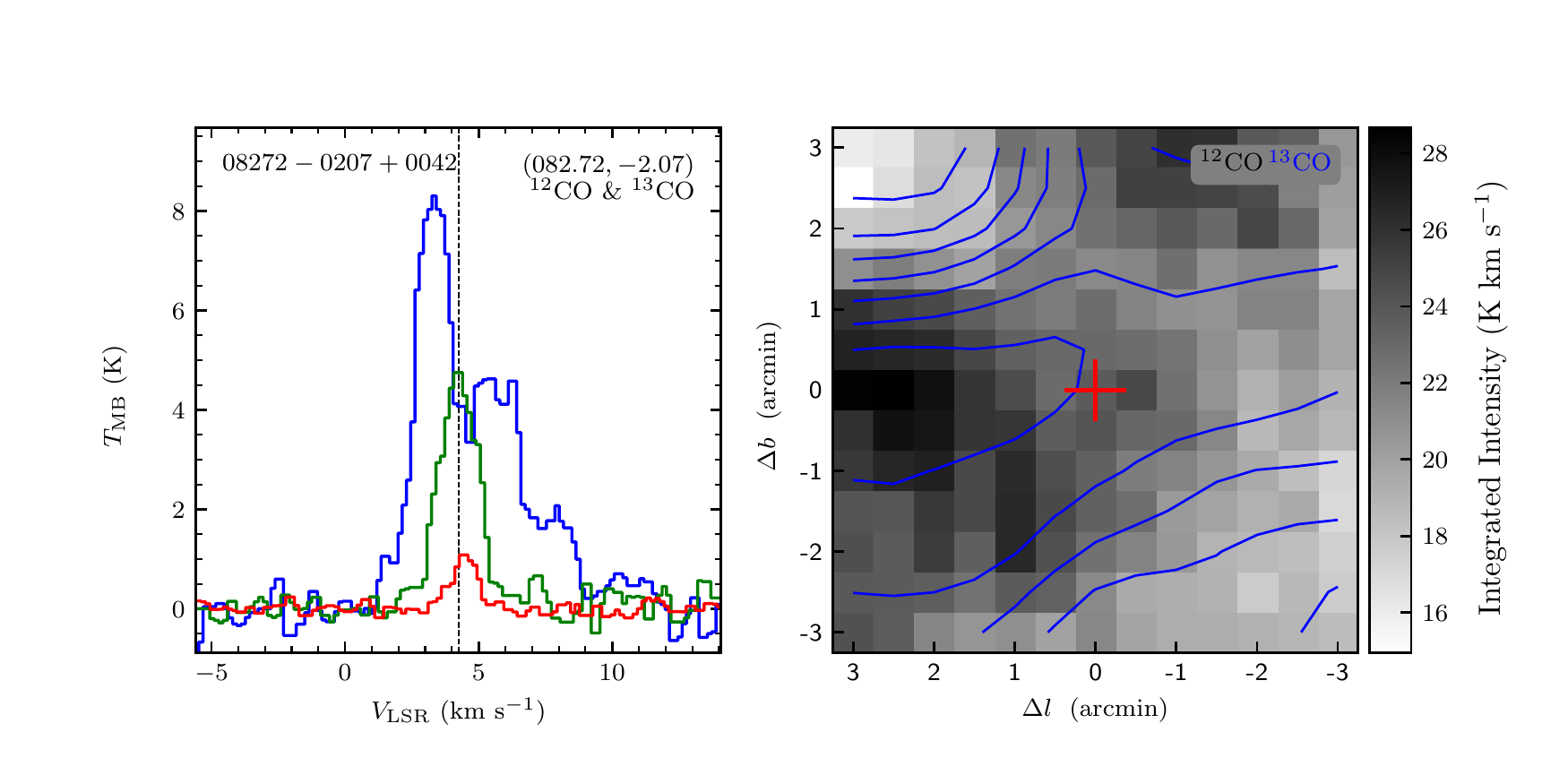}
\vspace{-0.5cm}

\includegraphics[width=9.0cm,angle=0]{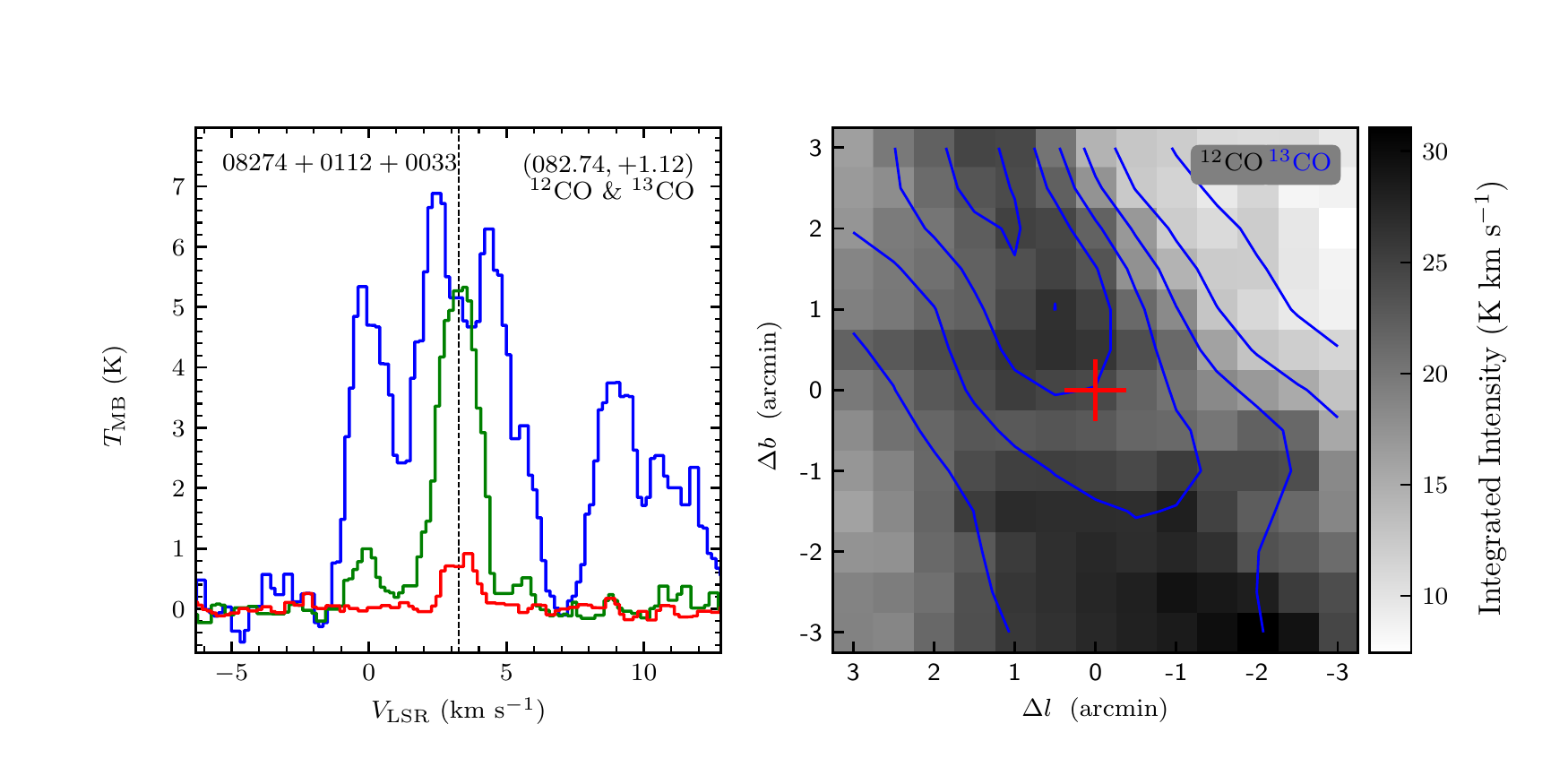}
\includegraphics[width=9.0cm,angle=0]{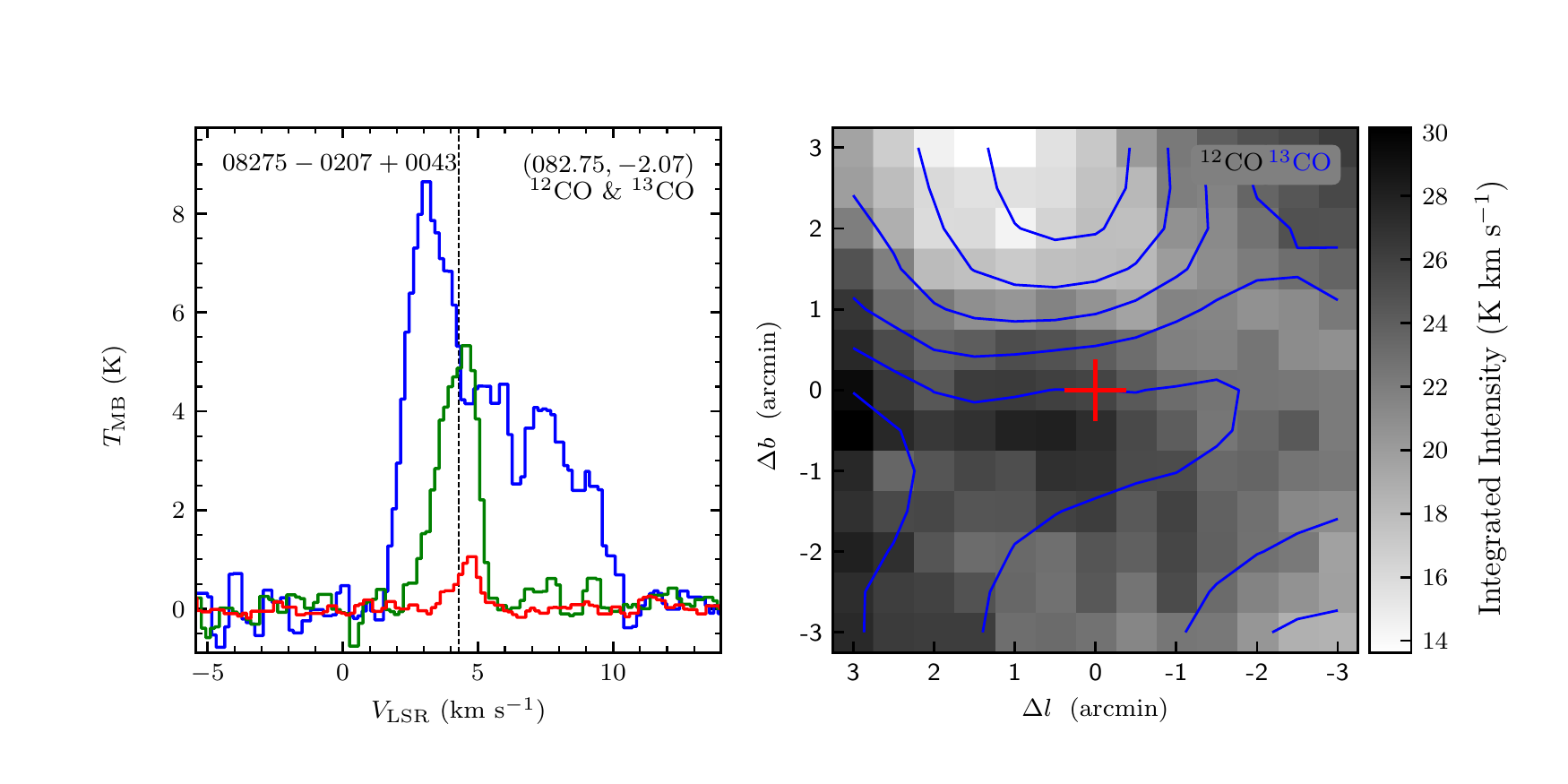}
\end{figure}
\clearpage

\begin{figure}
\includegraphics[width=9.0cm,angle=0]{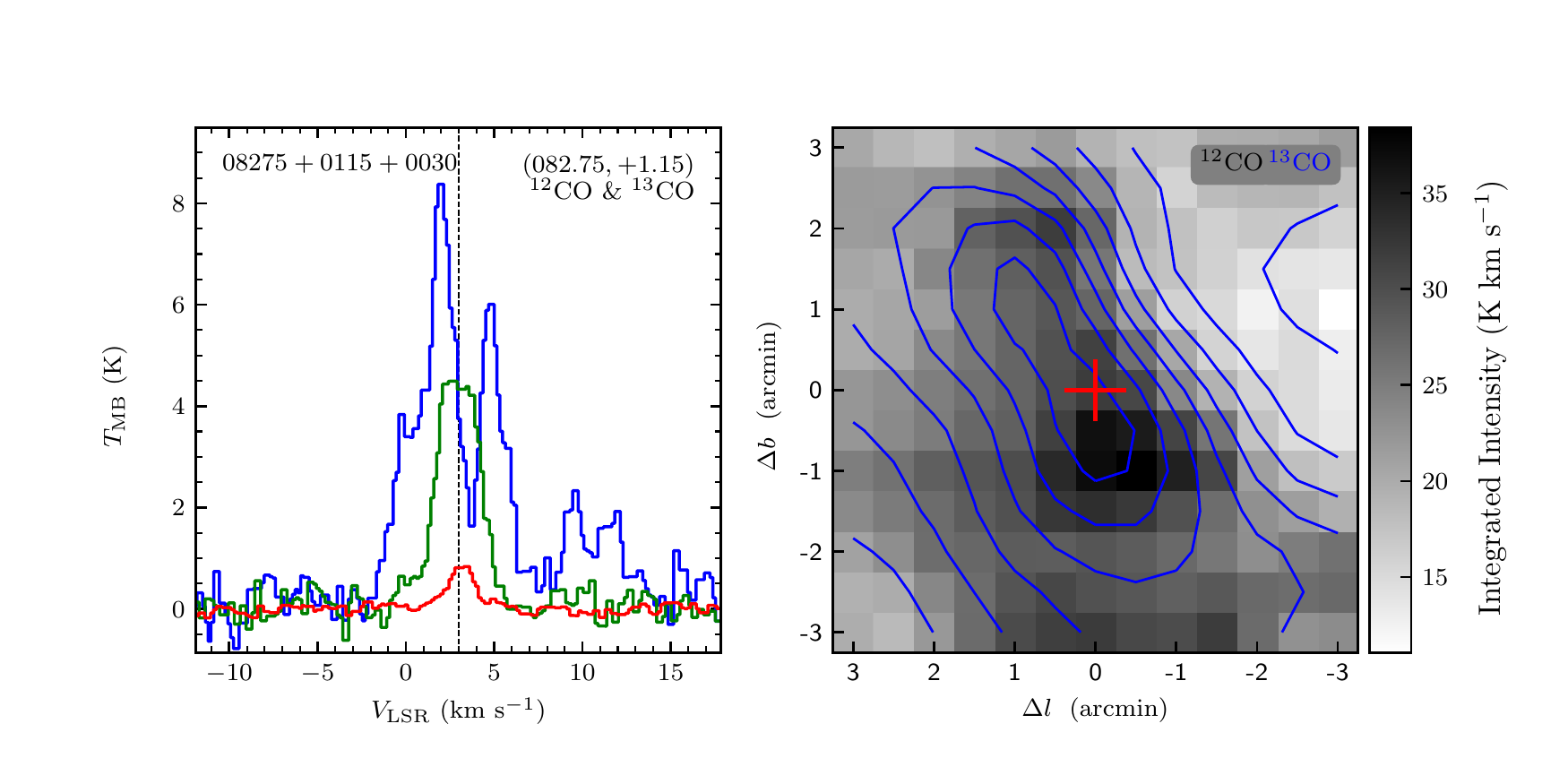}
\includegraphics[width=9.0cm,angle=0]{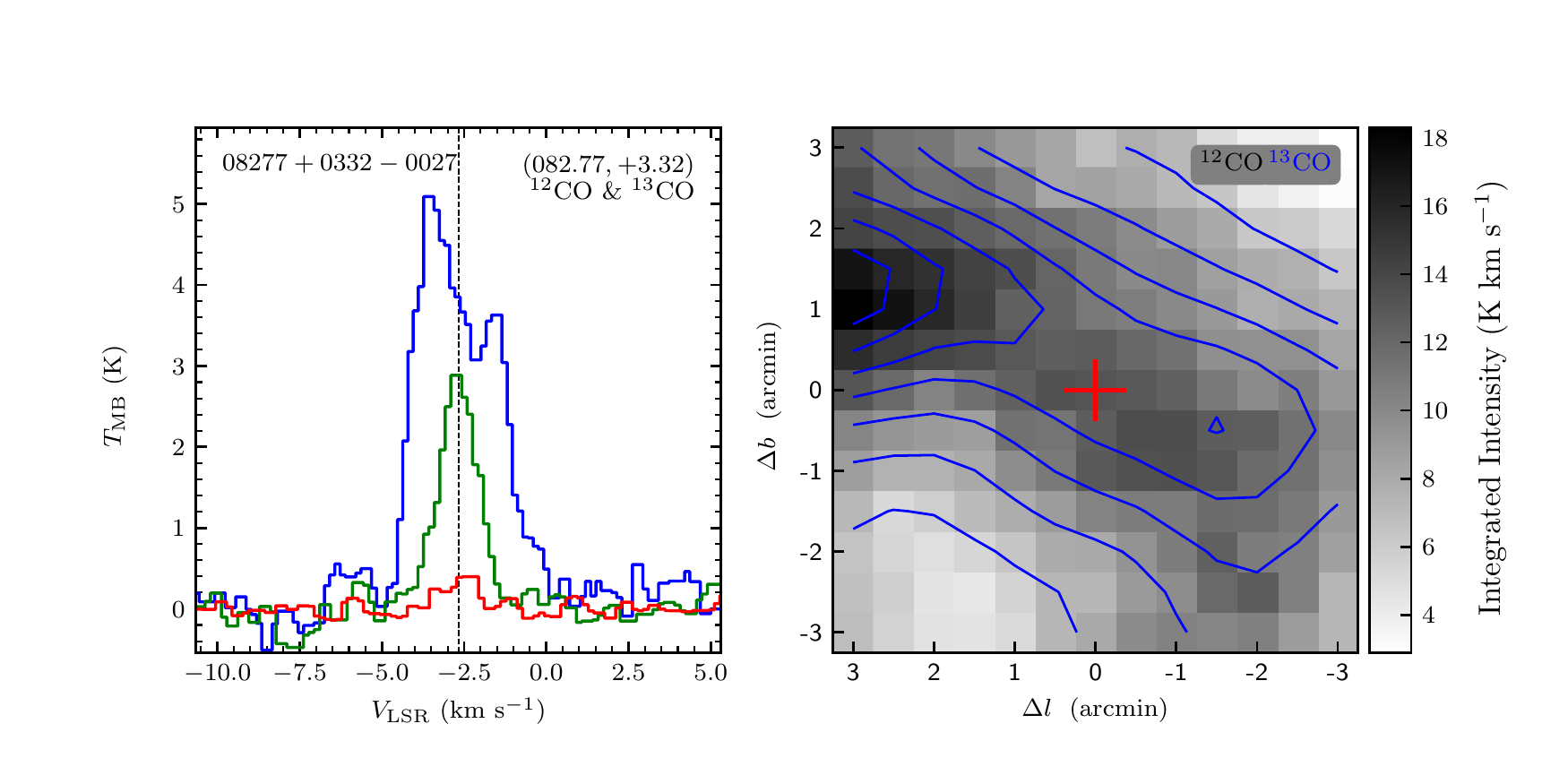}
\vspace{-0.5cm}

\includegraphics[width=9.0cm,angle=0]{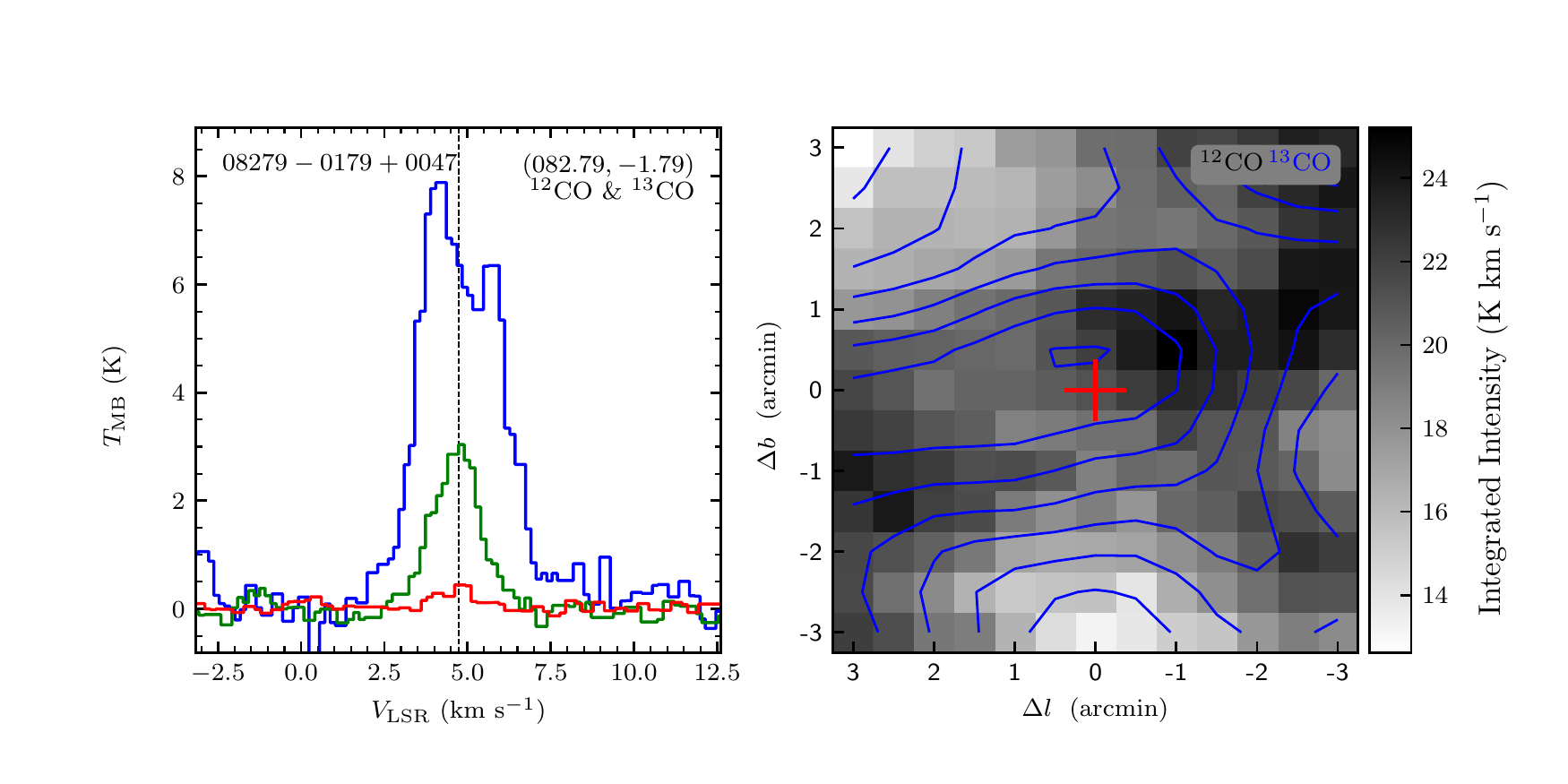}
\includegraphics[width=9.0cm,angle=0]{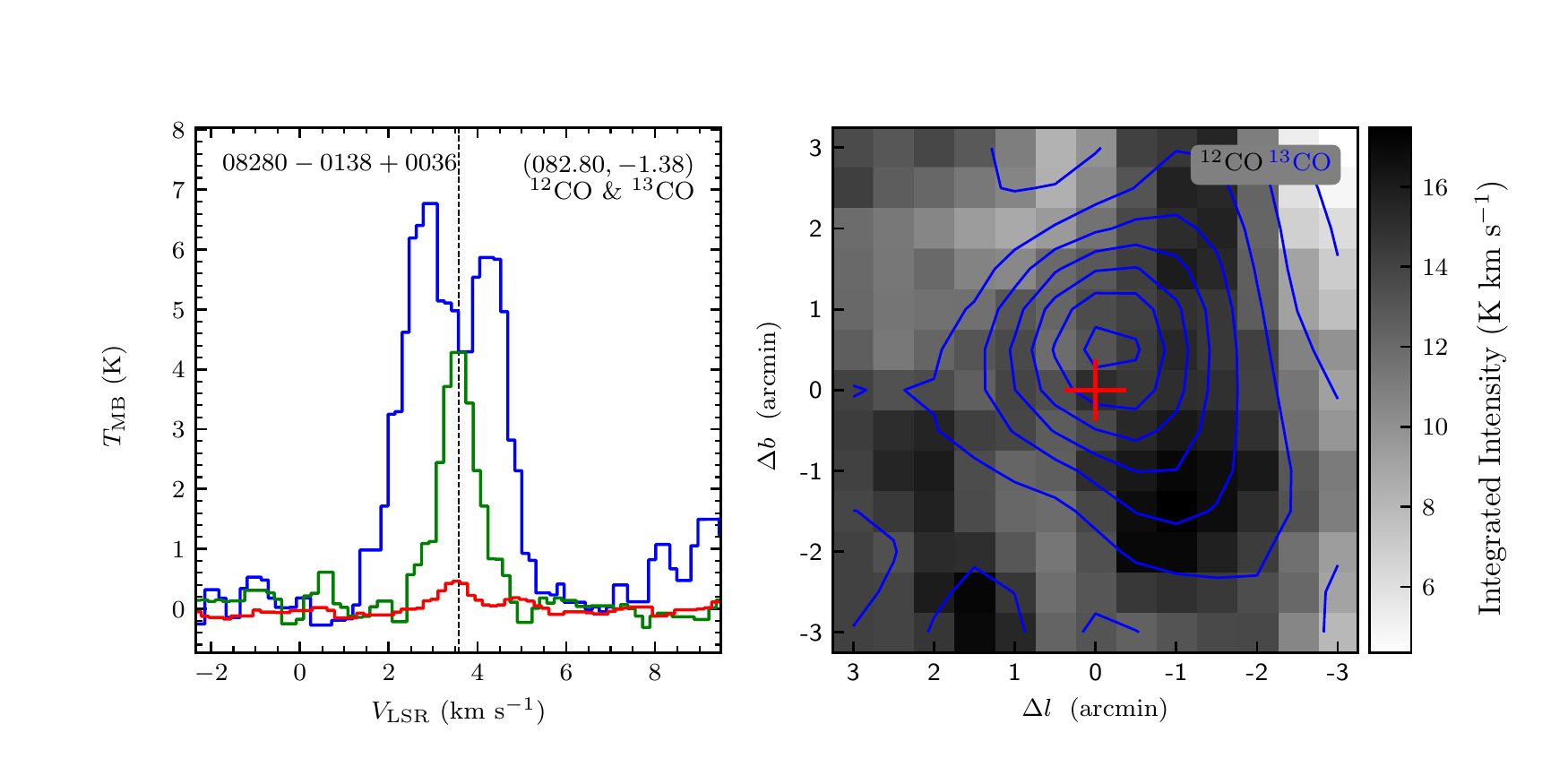}
\vspace{-0.5cm}

\includegraphics[width=9.0cm,angle=0]{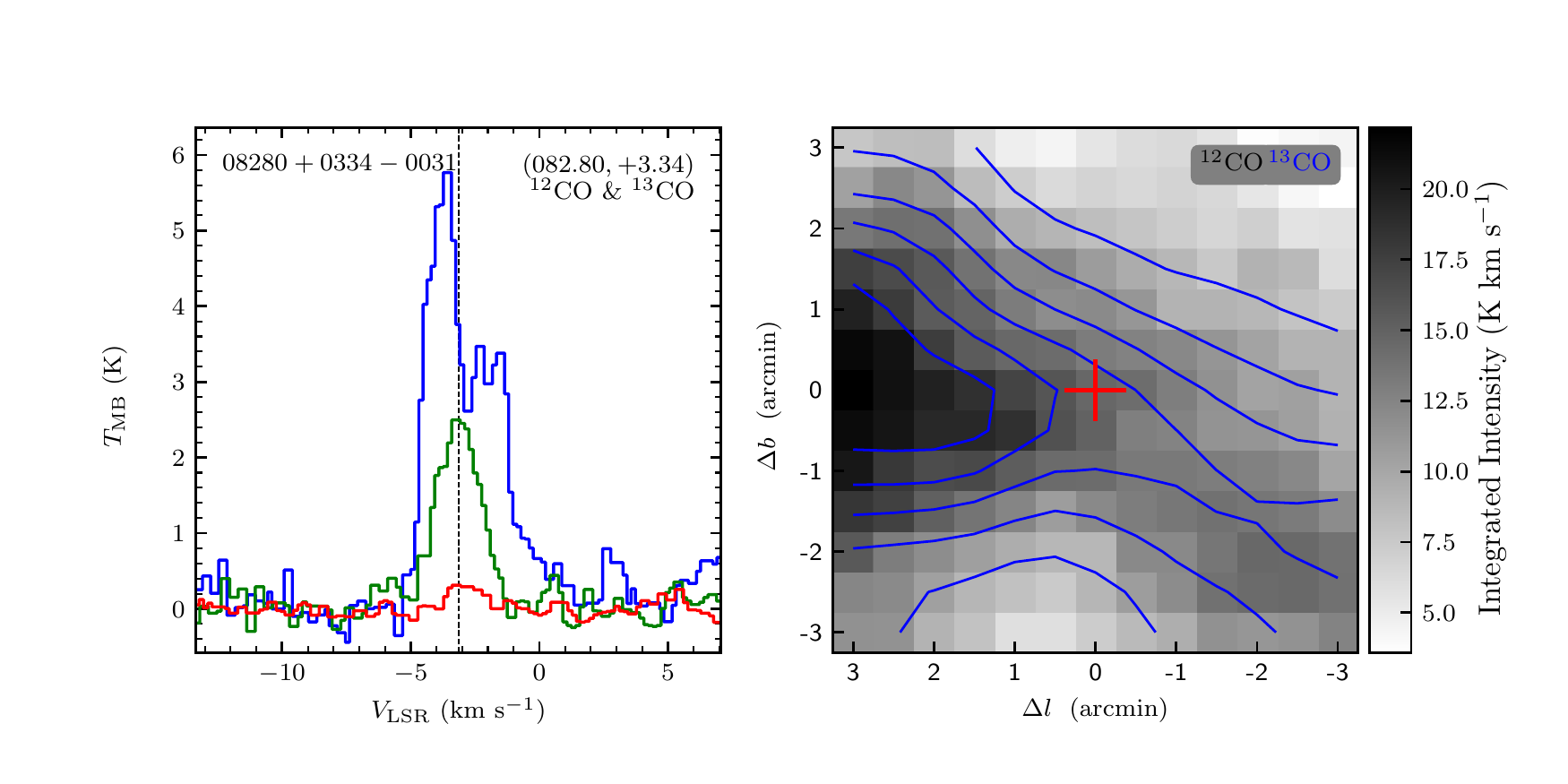}
\includegraphics[width=9.0cm,angle=0]{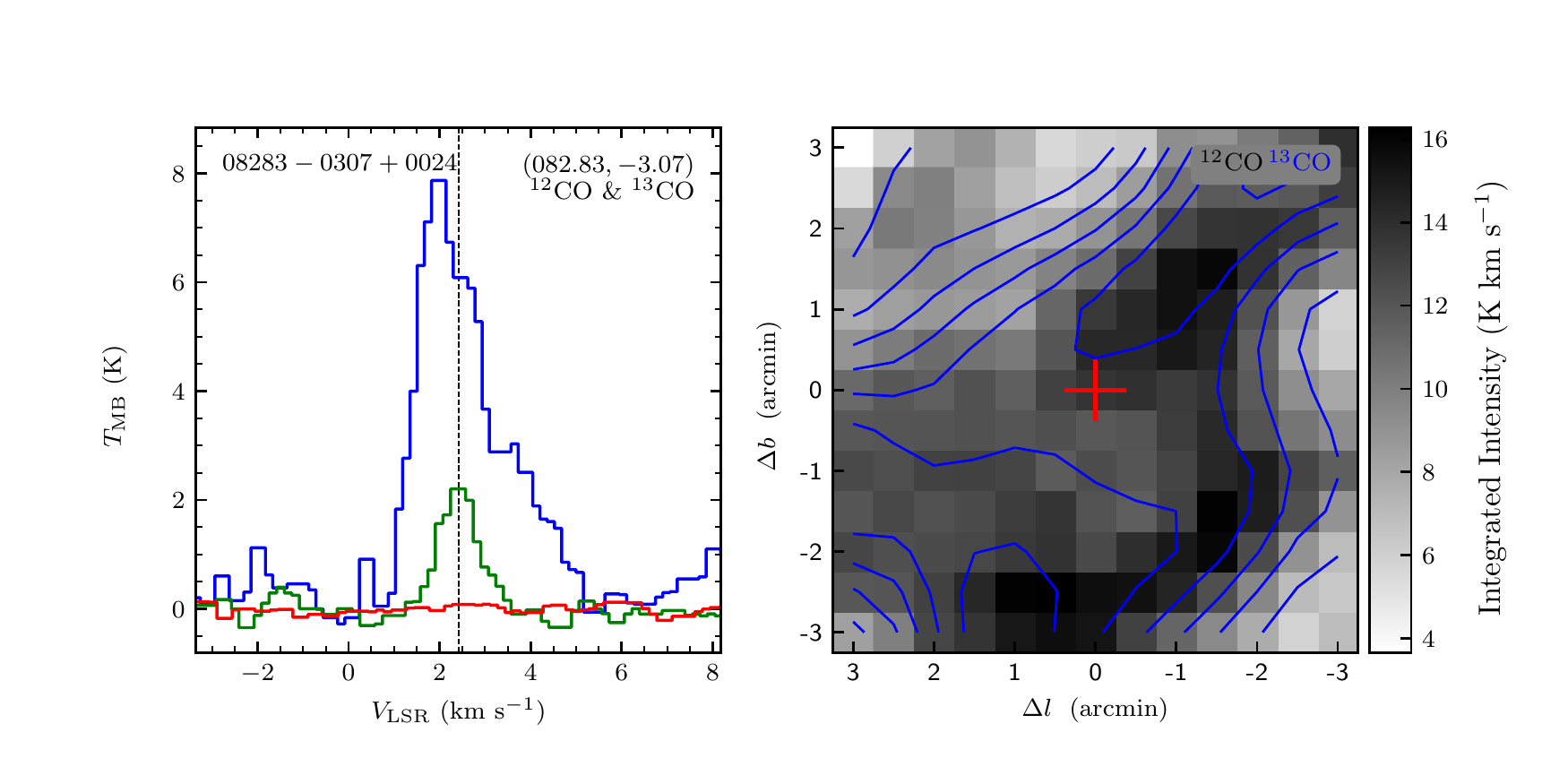}
\vspace{-0.5cm}

\includegraphics[width=9.0cm,angle=0]{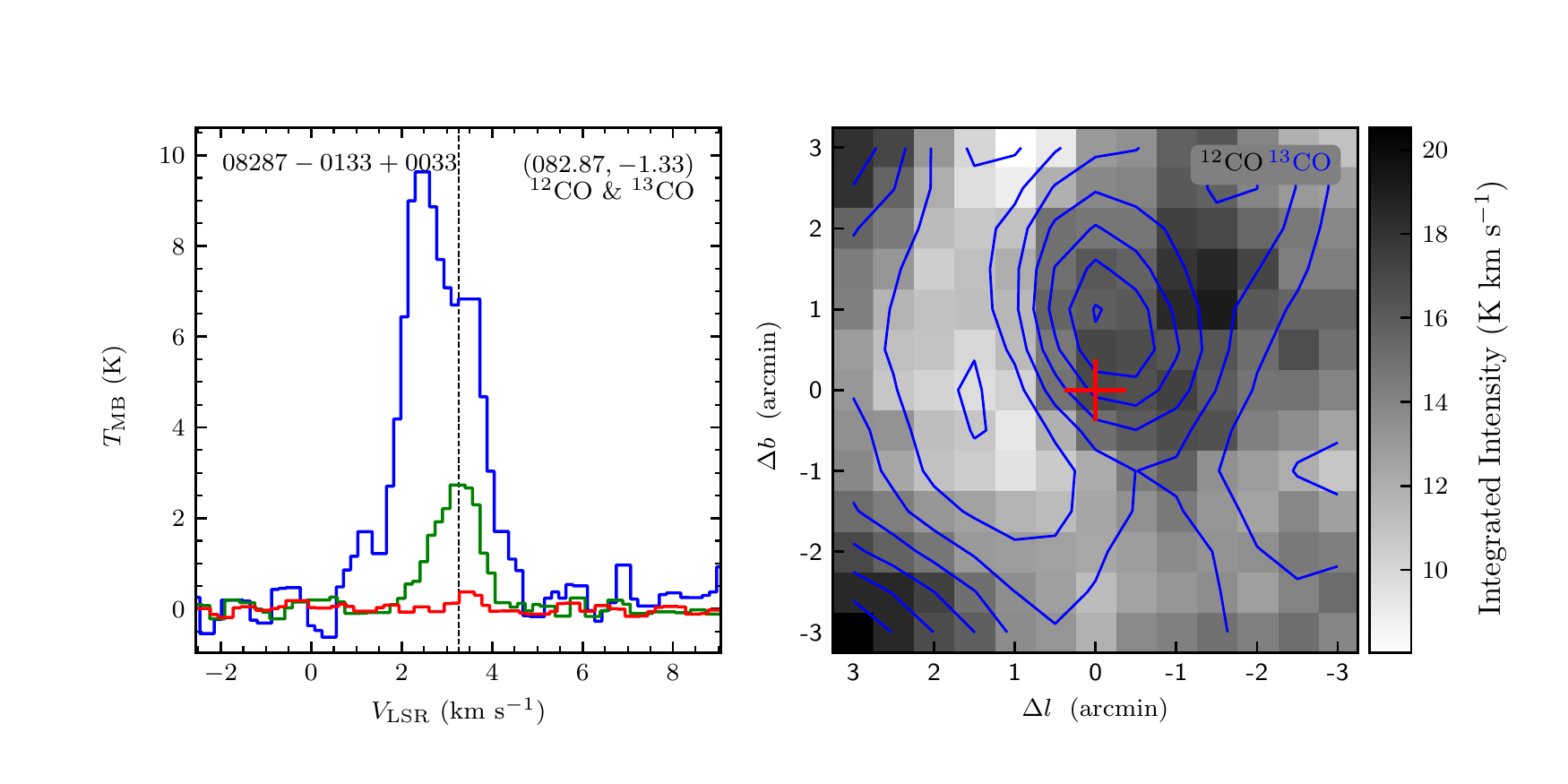}
\includegraphics[width=9.0cm,angle=0]{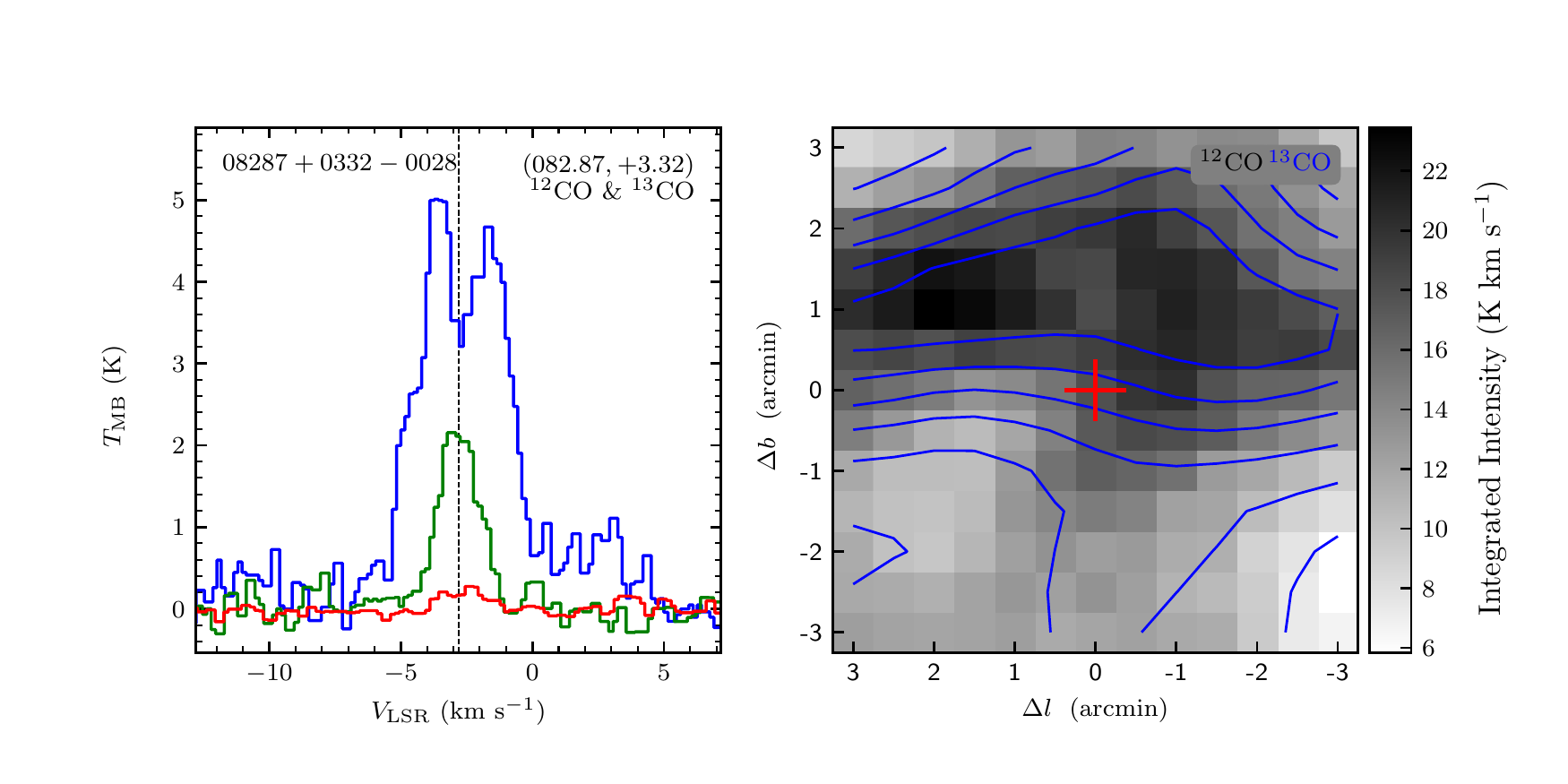}
\vspace{-0.5cm}

\includegraphics[width=9.0cm,angle=0]{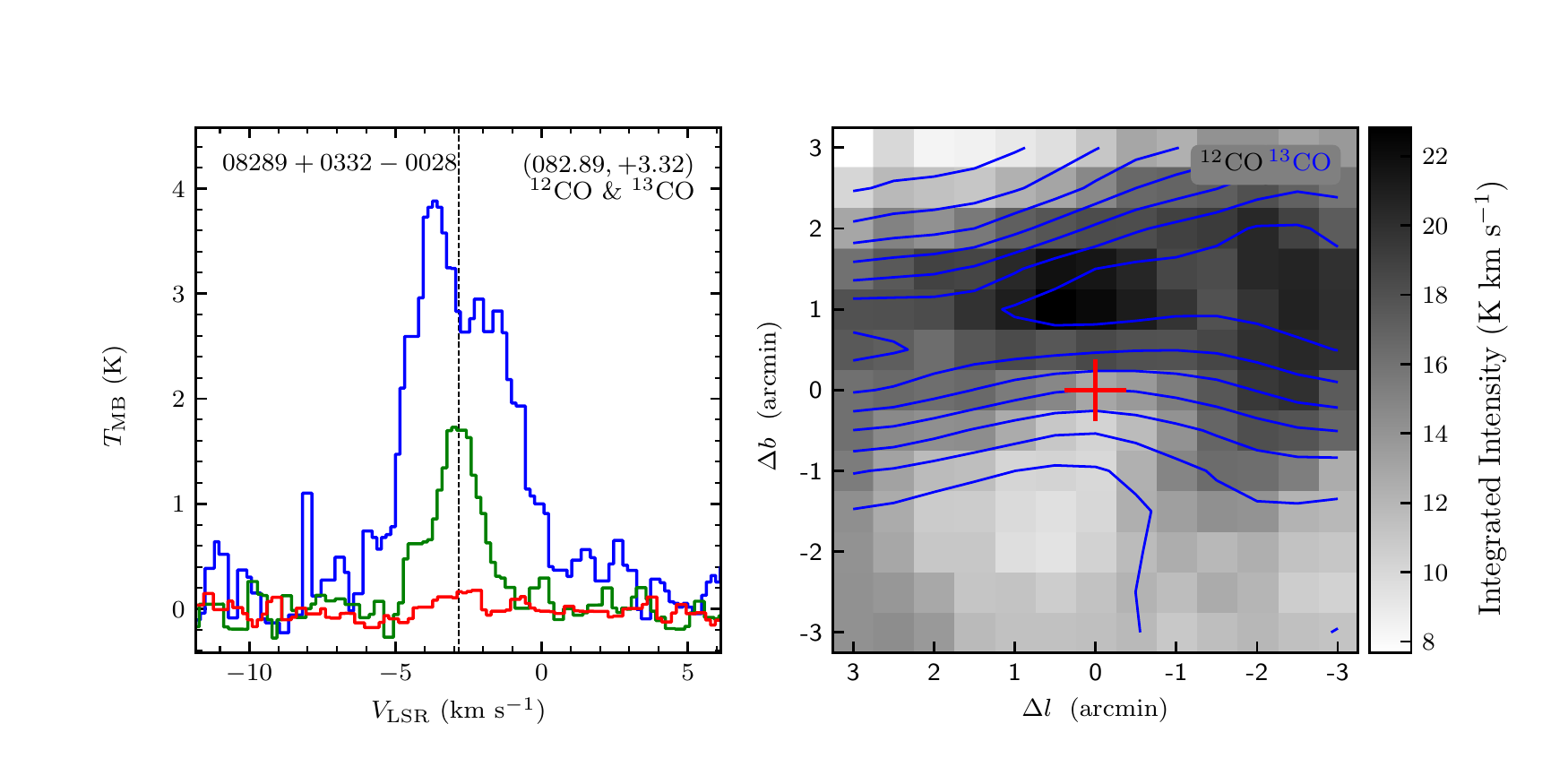}
\includegraphics[width=9.0cm,angle=0]{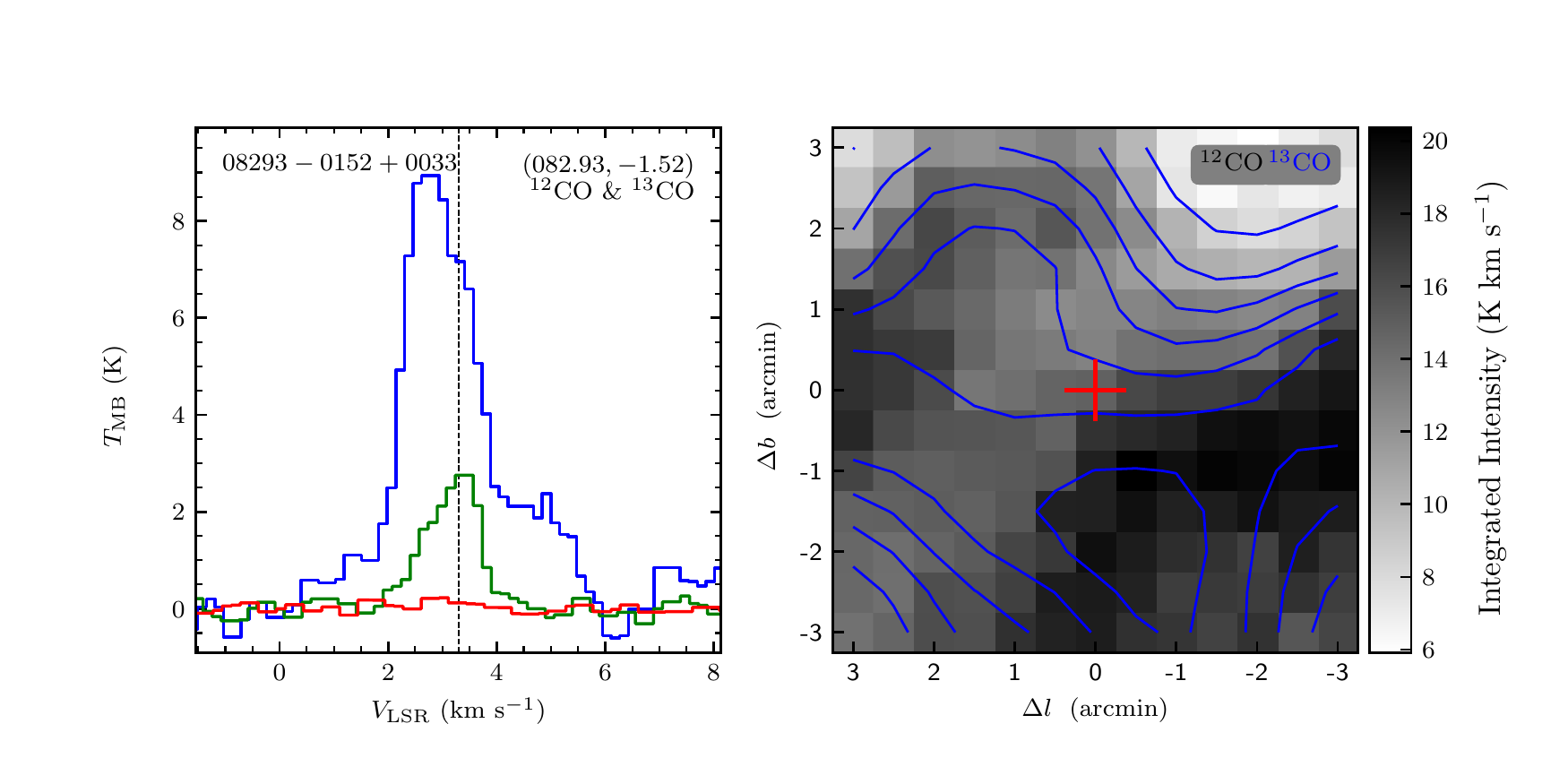}
\end{figure}
\clearpage

\begin{figure}
\includegraphics[width=9.0cm,angle=0]{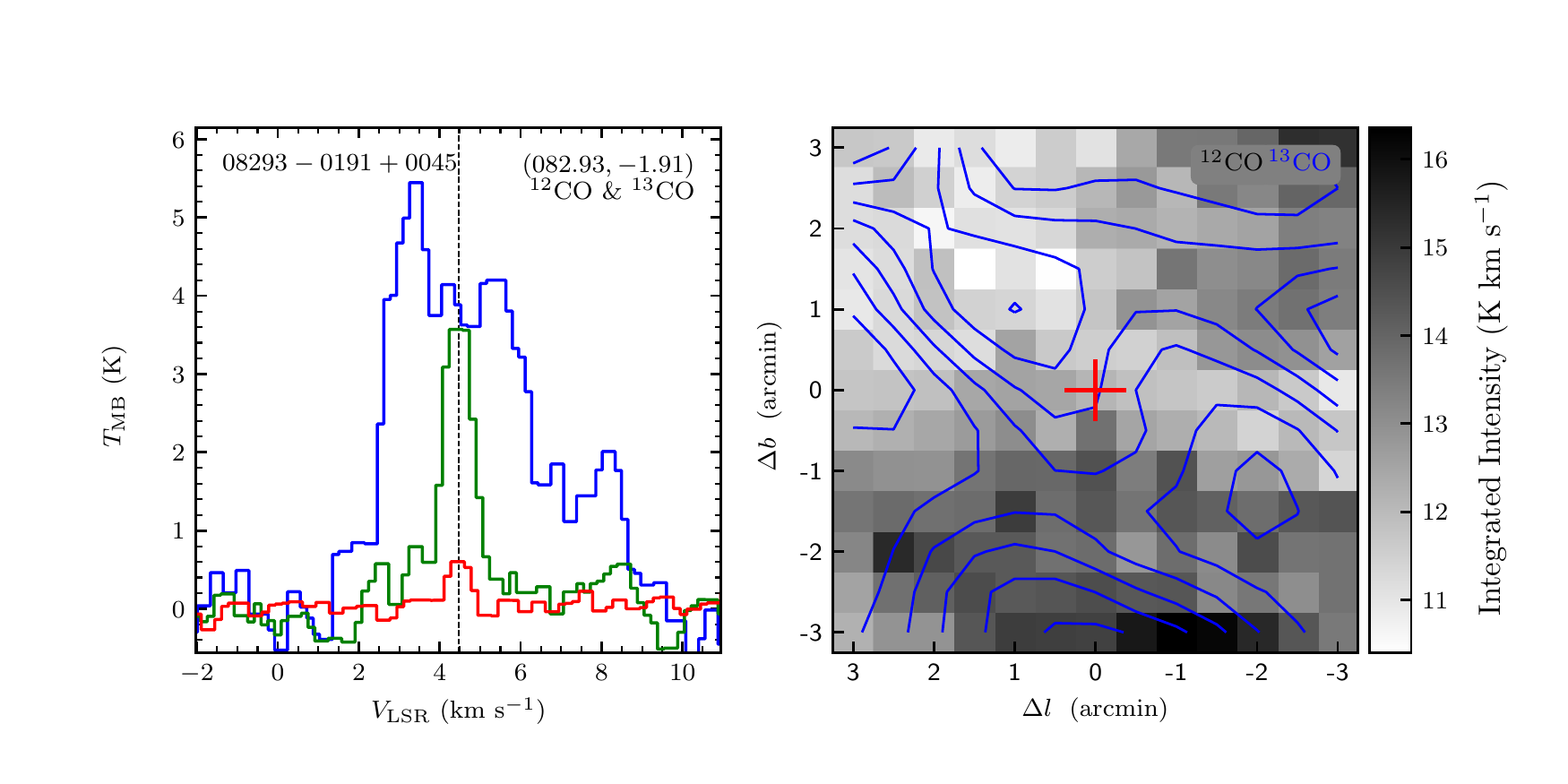}
\includegraphics[width=9.0cm,angle=0]{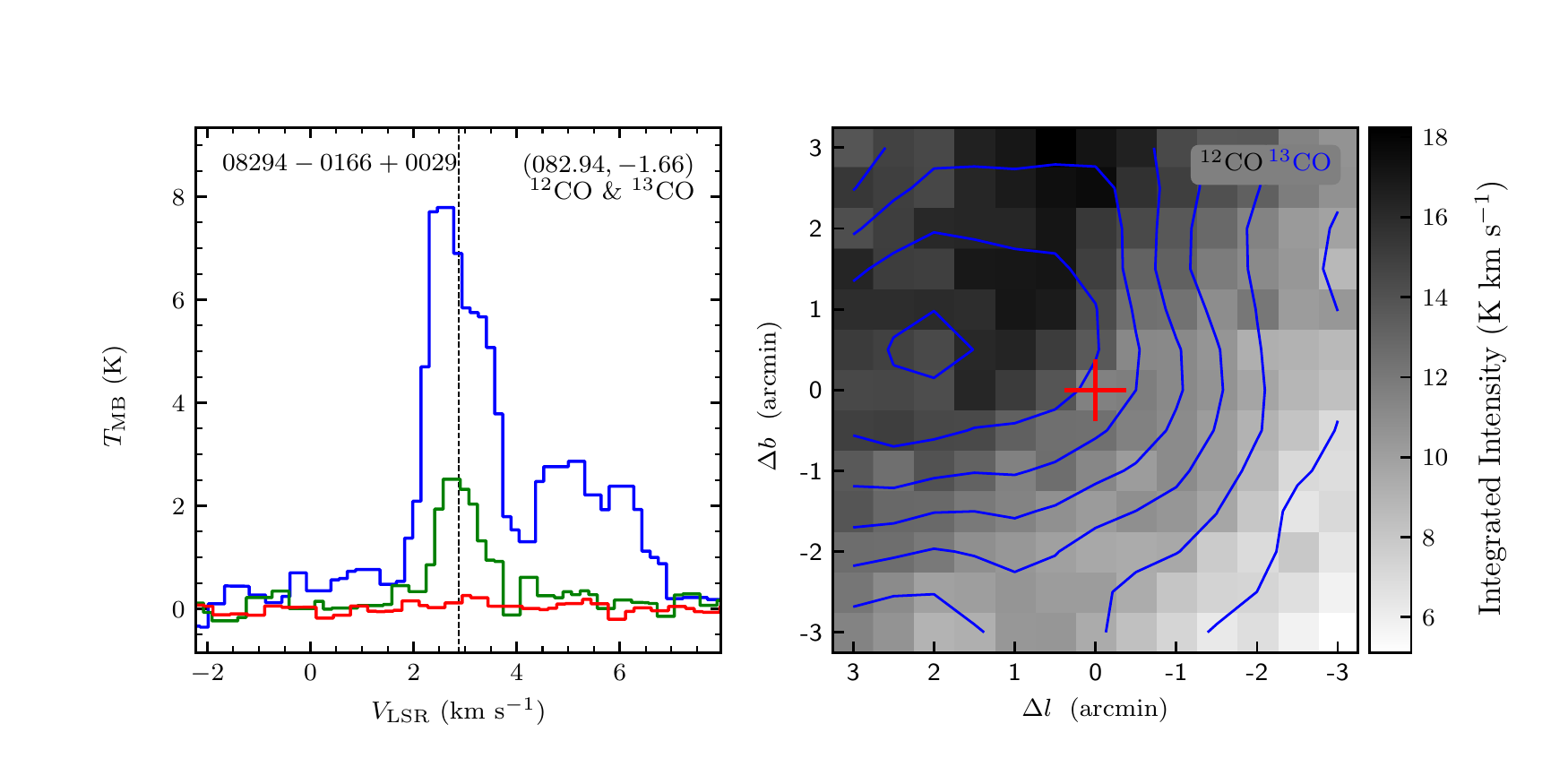}
\vspace{-0.5cm}

\includegraphics[width=9.0cm,angle=0]{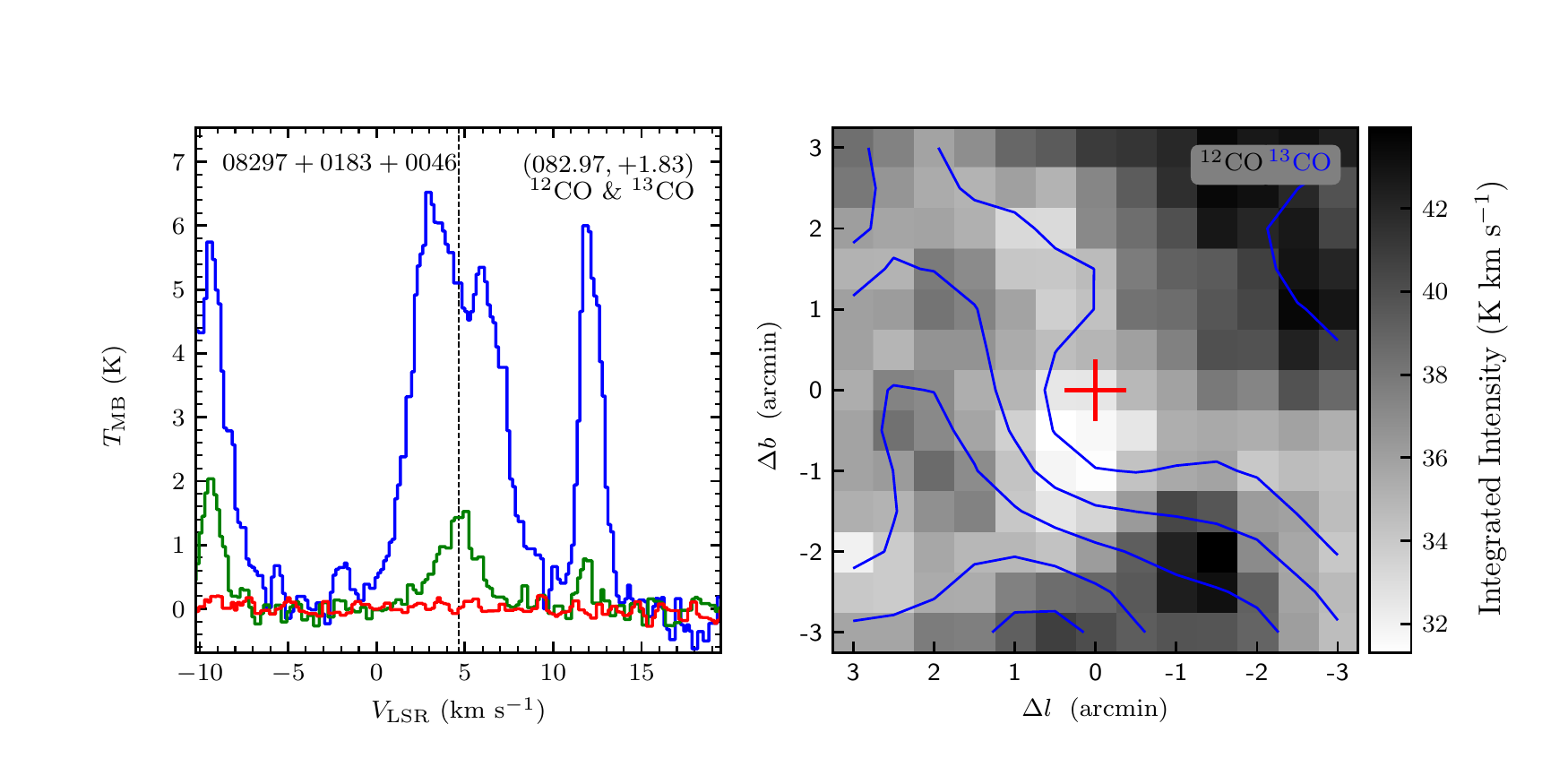}
\includegraphics[width=9.0cm,angle=0]{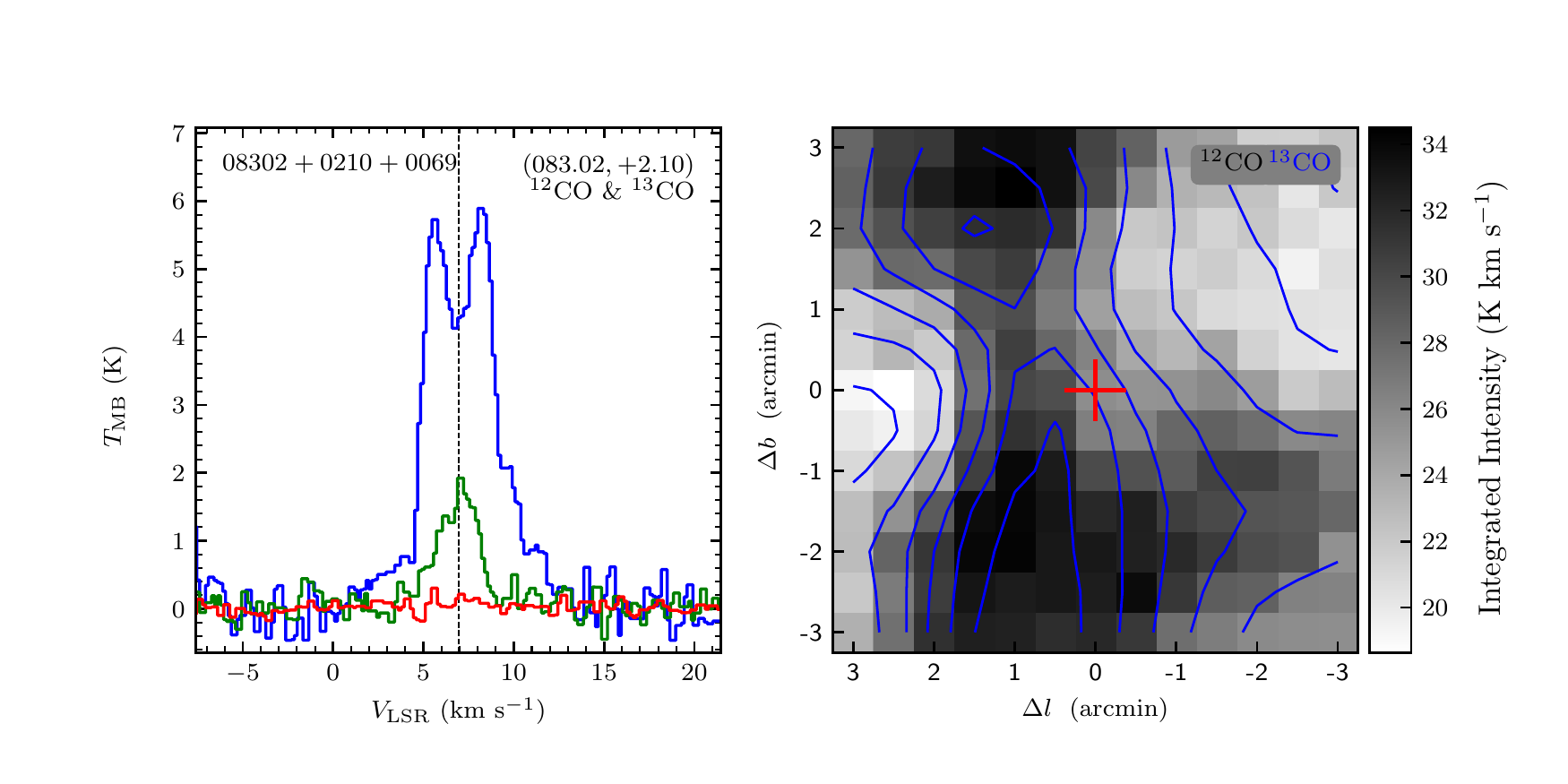}
\vspace{-0.5cm}

\includegraphics[width=9.0cm,angle=0]{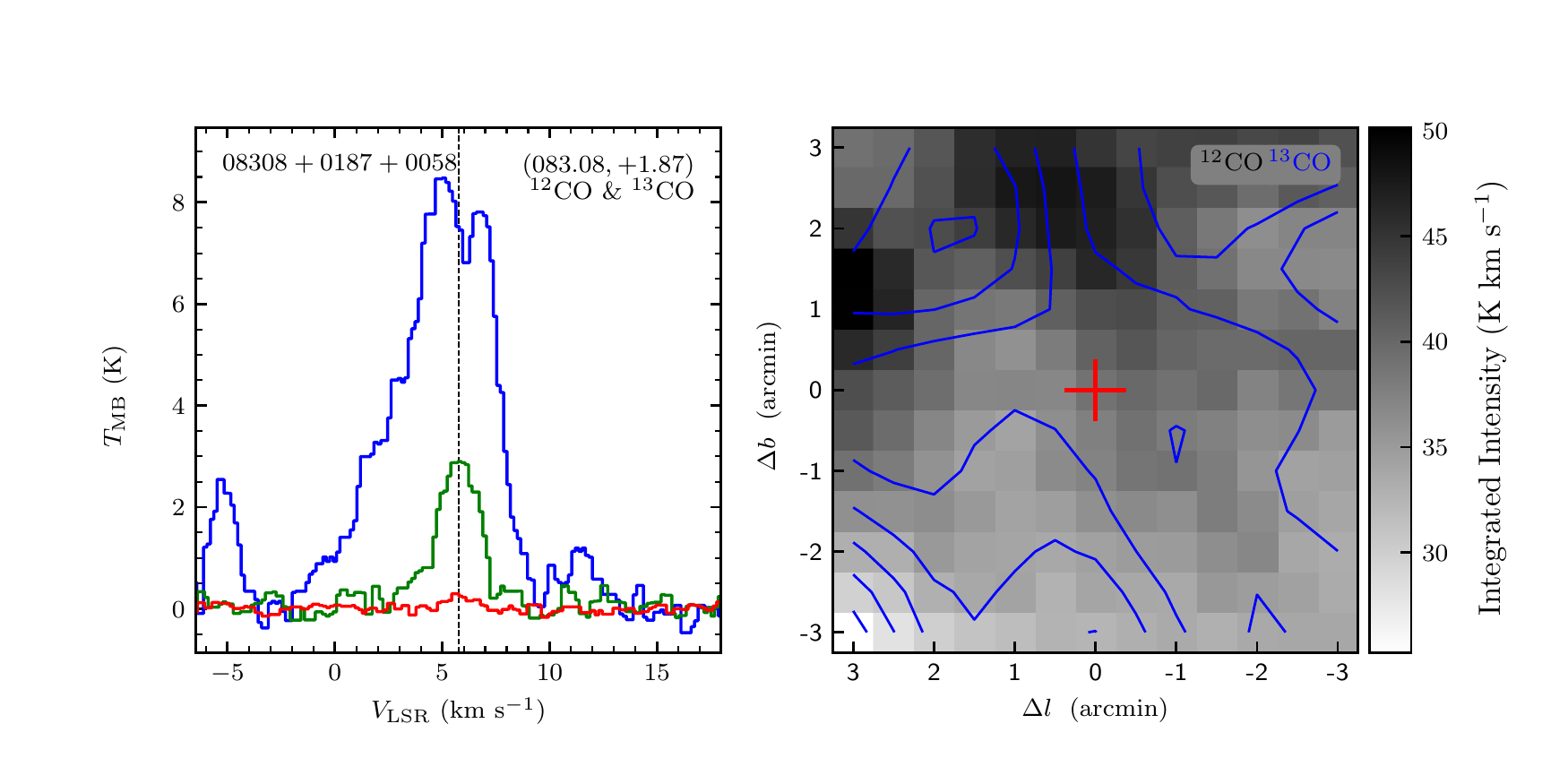}
\includegraphics[width=9.0cm,angle=0]{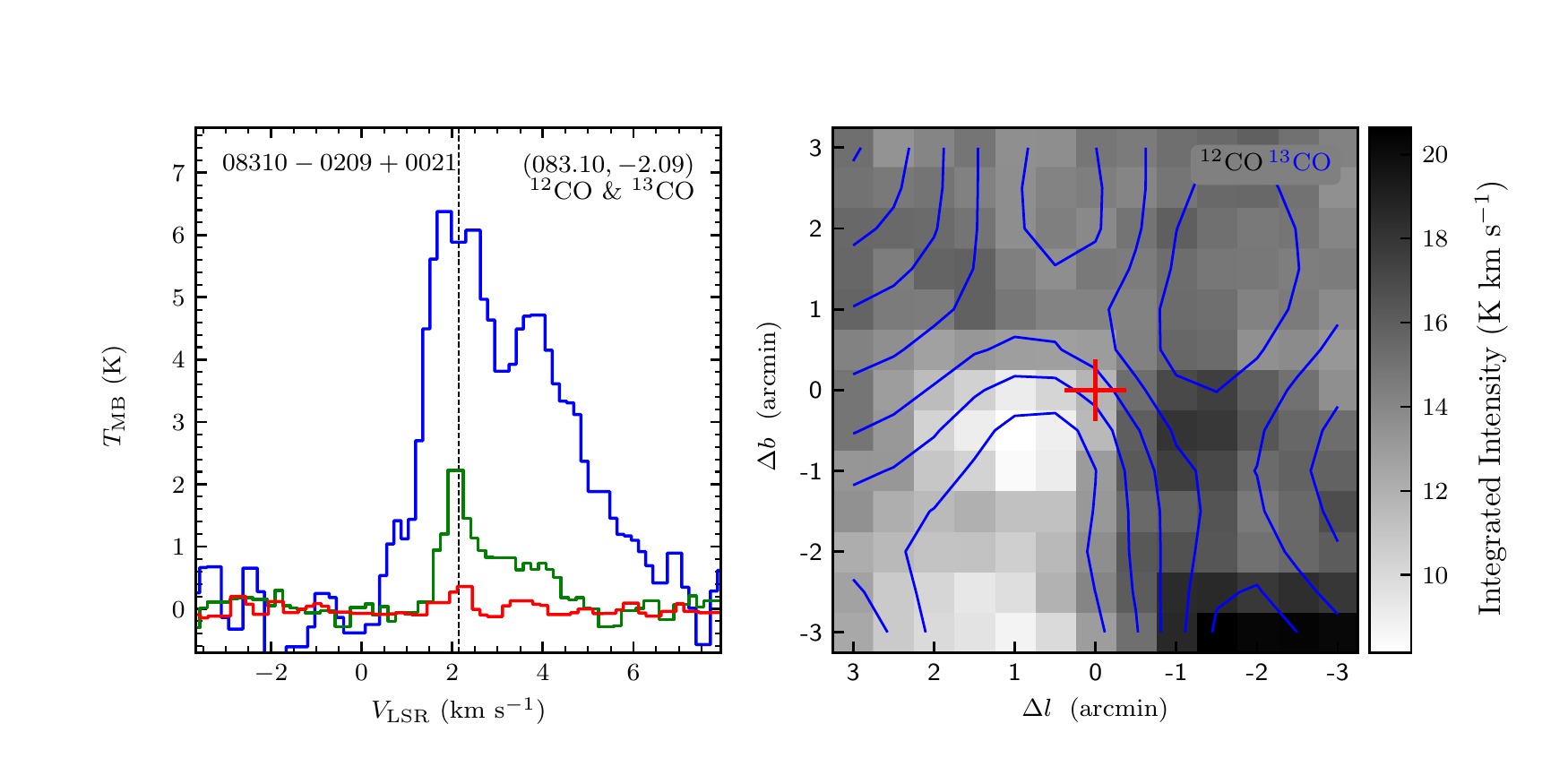}
\vspace{-0.5cm}

\includegraphics[width=9.0cm,angle=0]{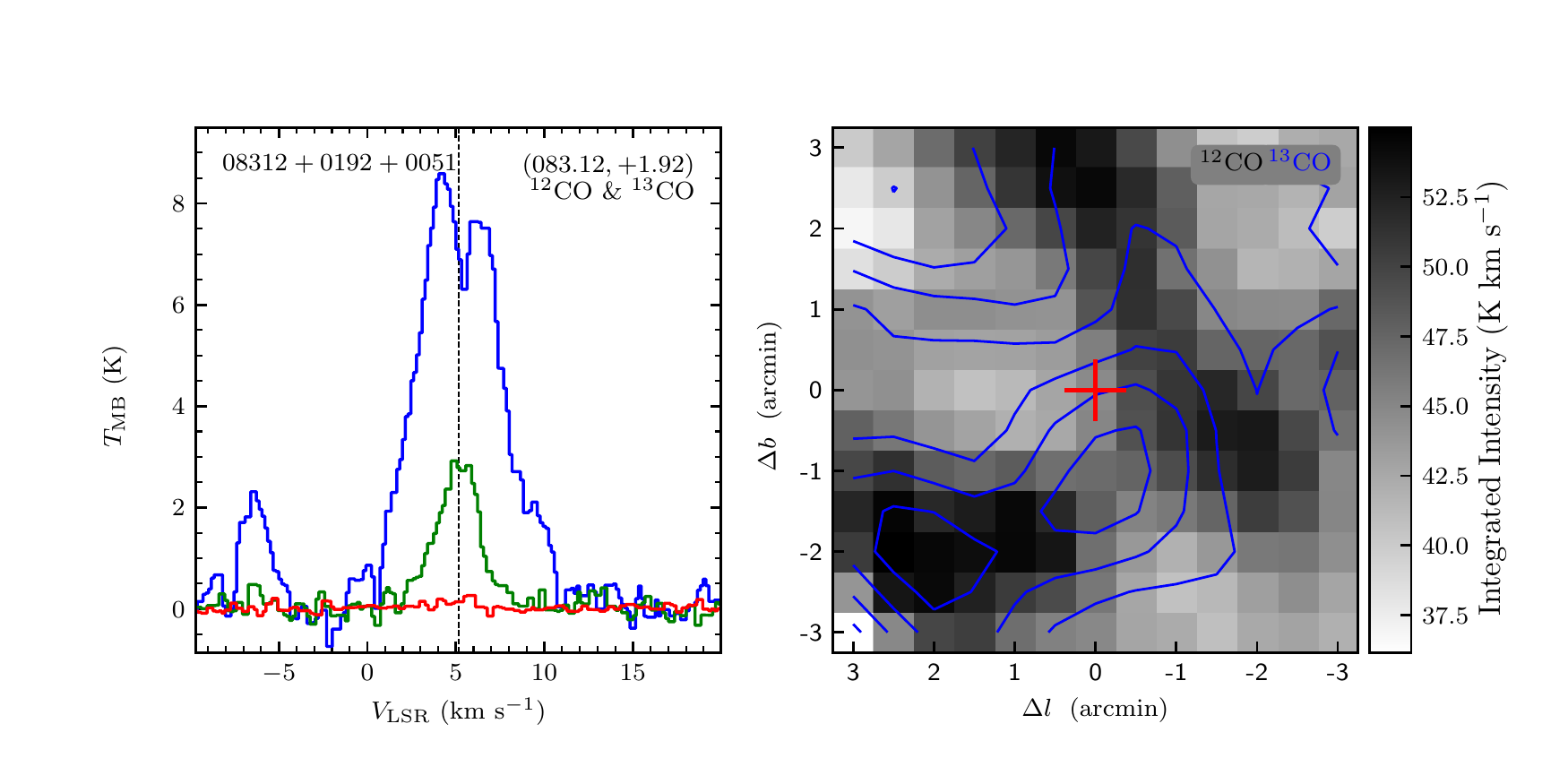}
\includegraphics[width=9.0cm,angle=0]{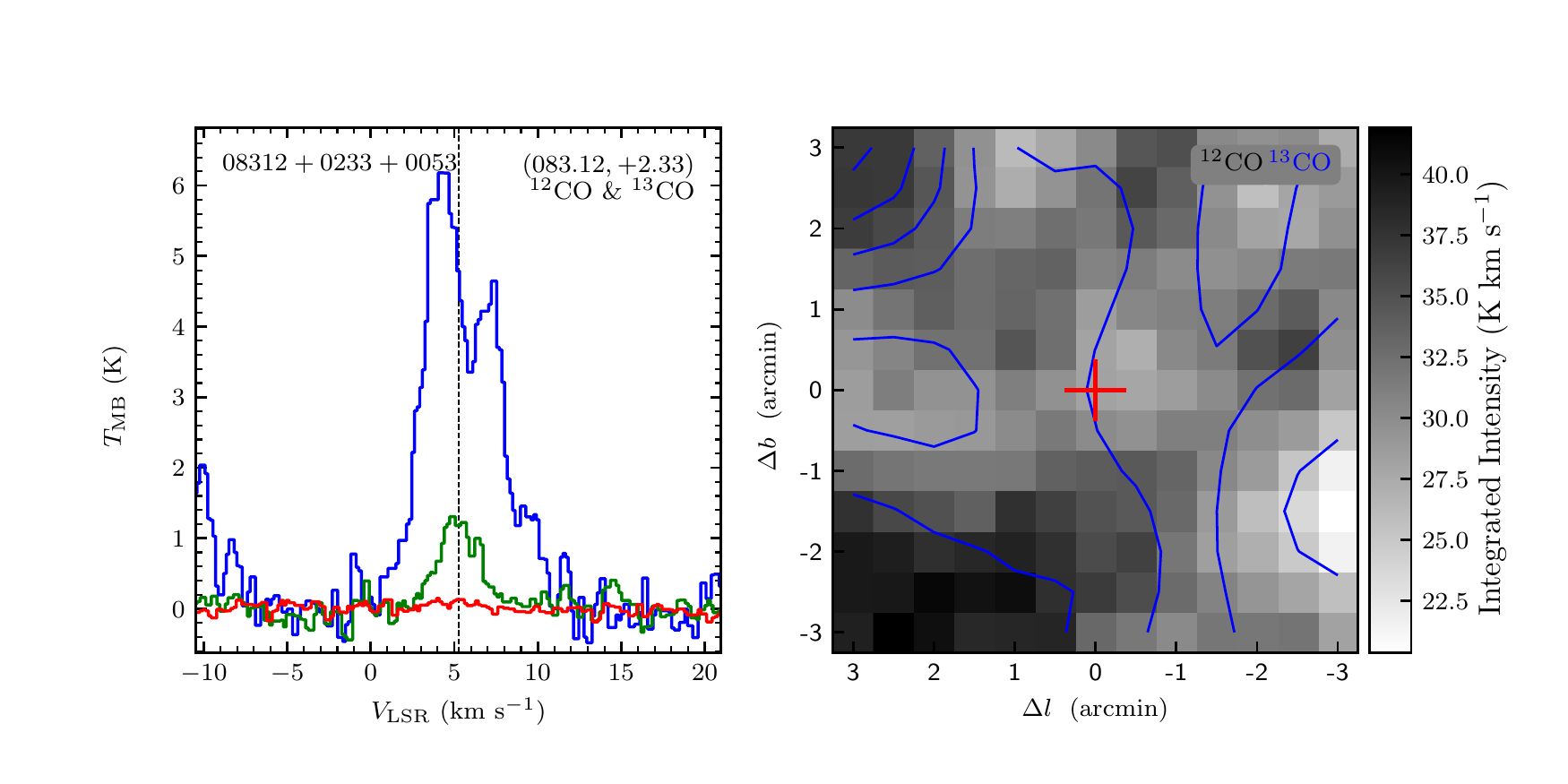}
\vspace{-0.5cm}

\includegraphics[width=9.0cm,angle=0]{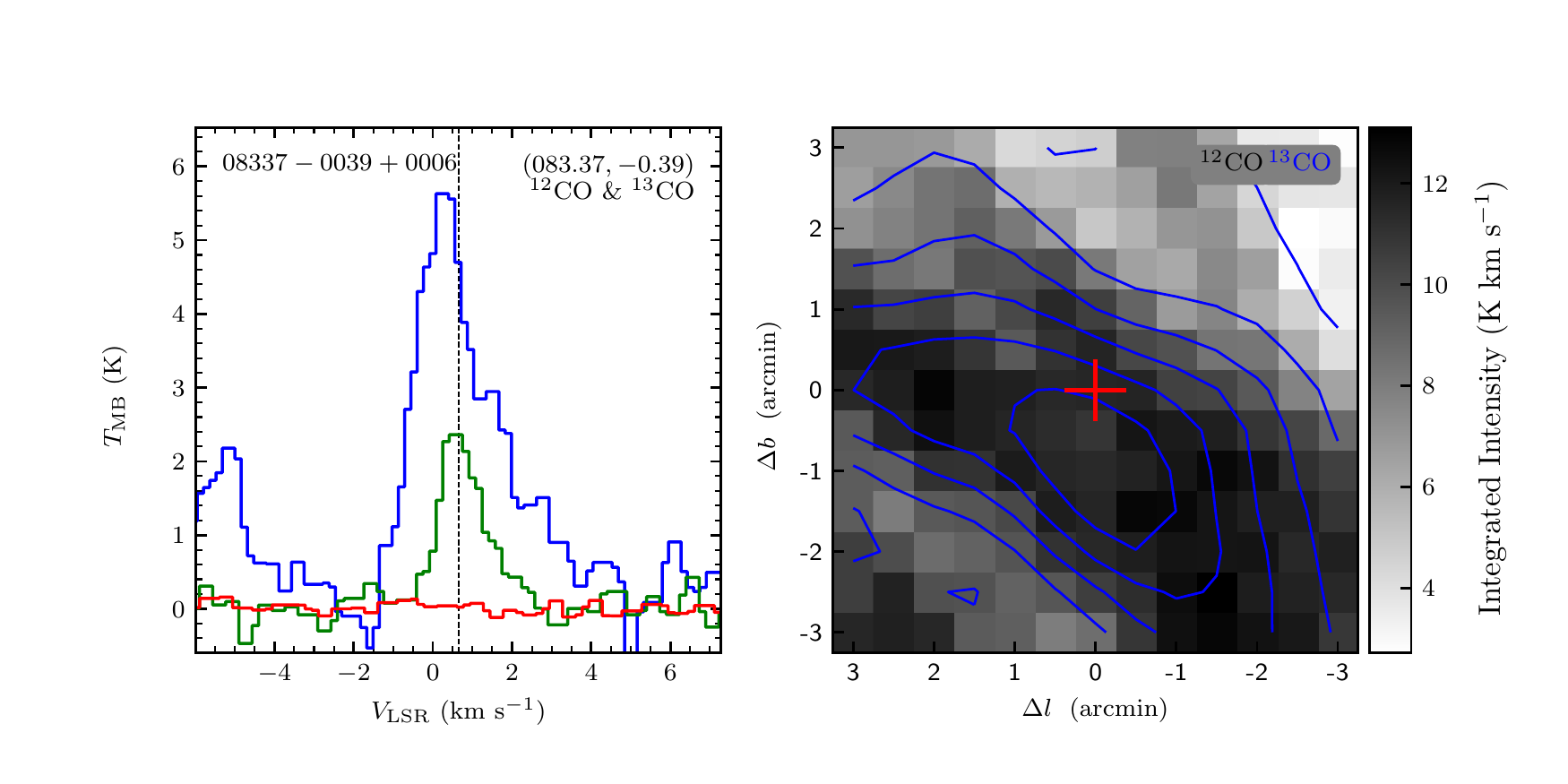}
\includegraphics[width=9.0cm,angle=0]{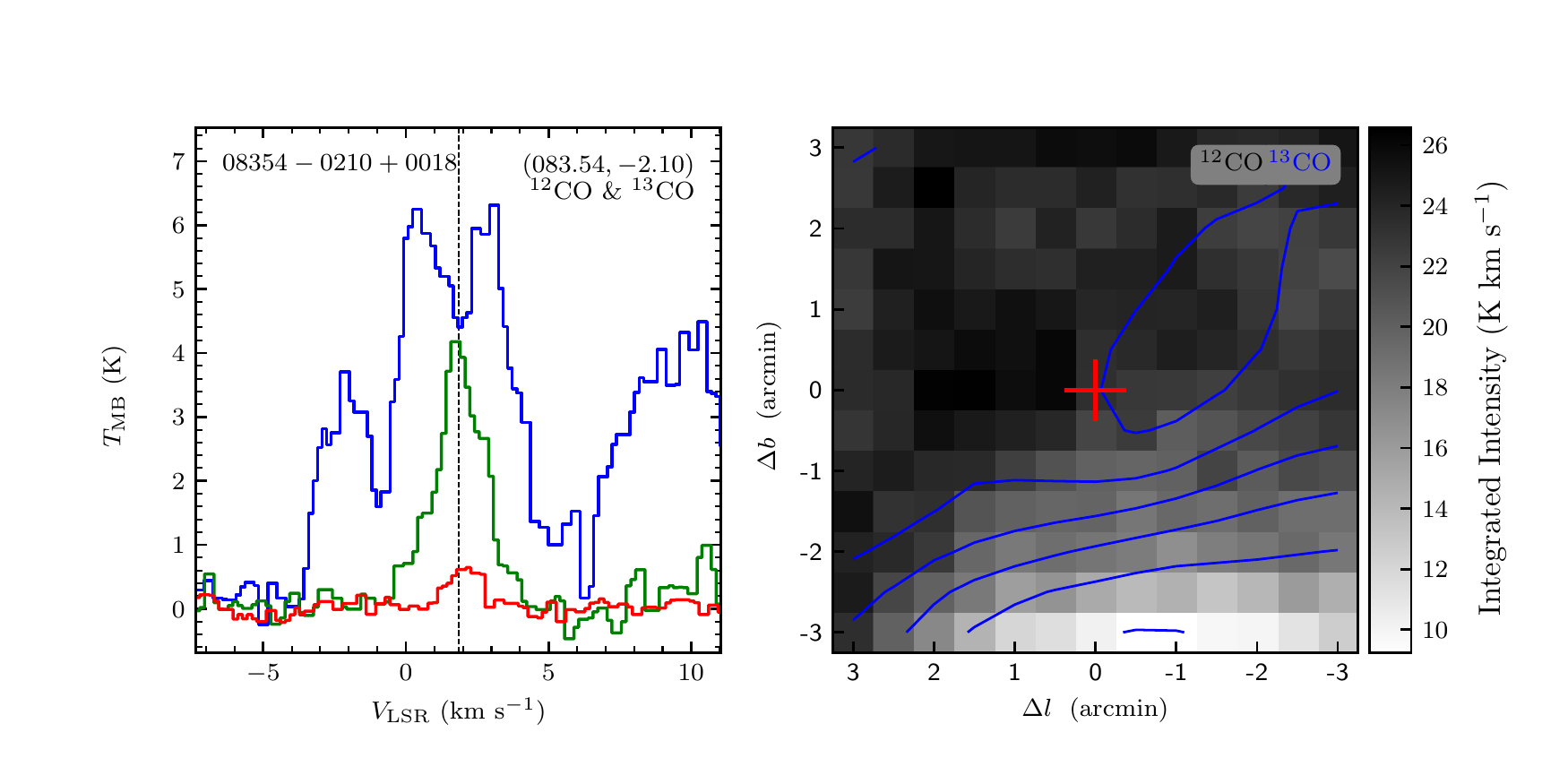}
\end{figure}
\clearpage

\begin{figure}
\includegraphics[width=9.0cm,angle=0]{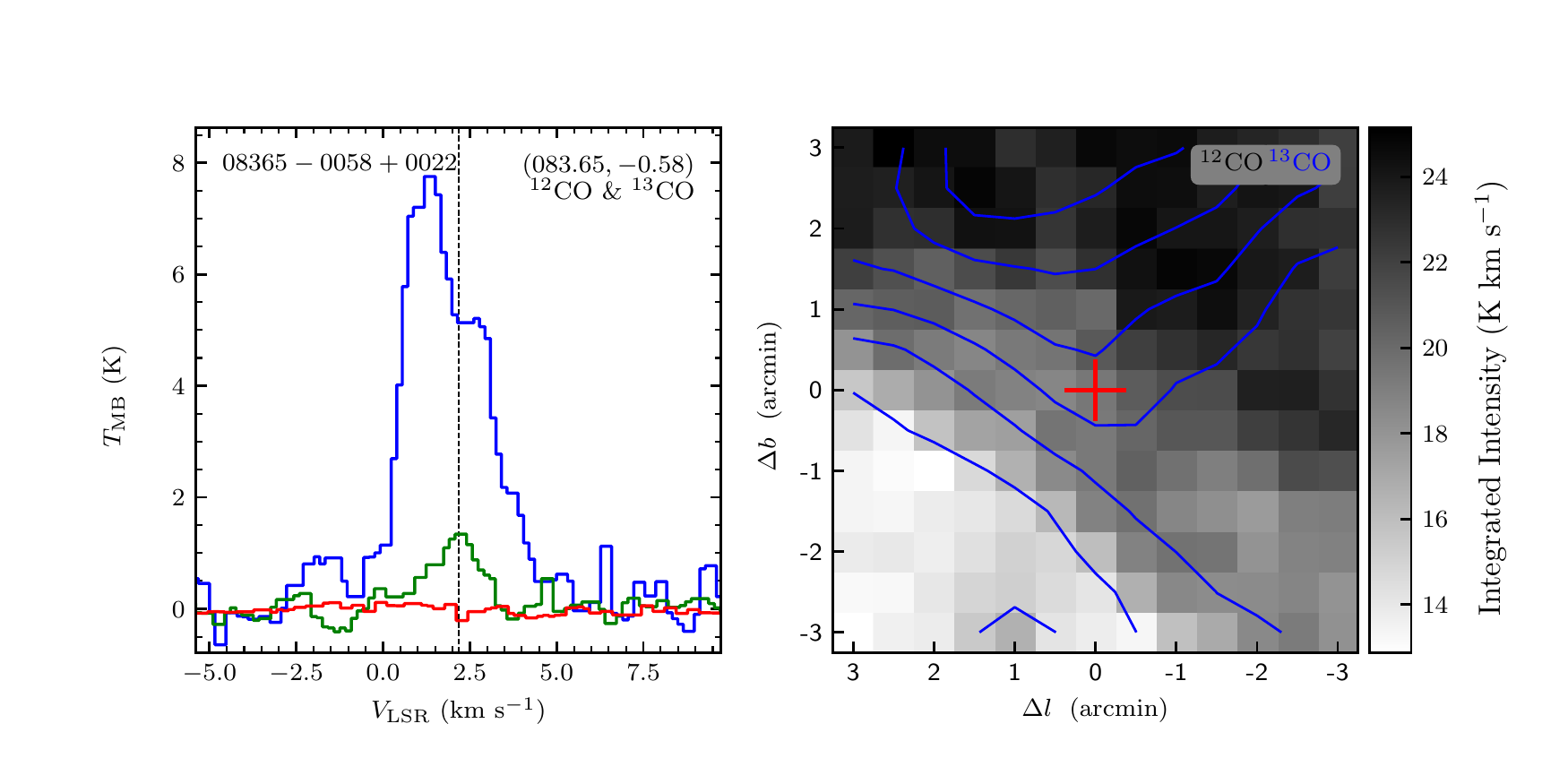}
\includegraphics[width=9.0cm,angle=0]{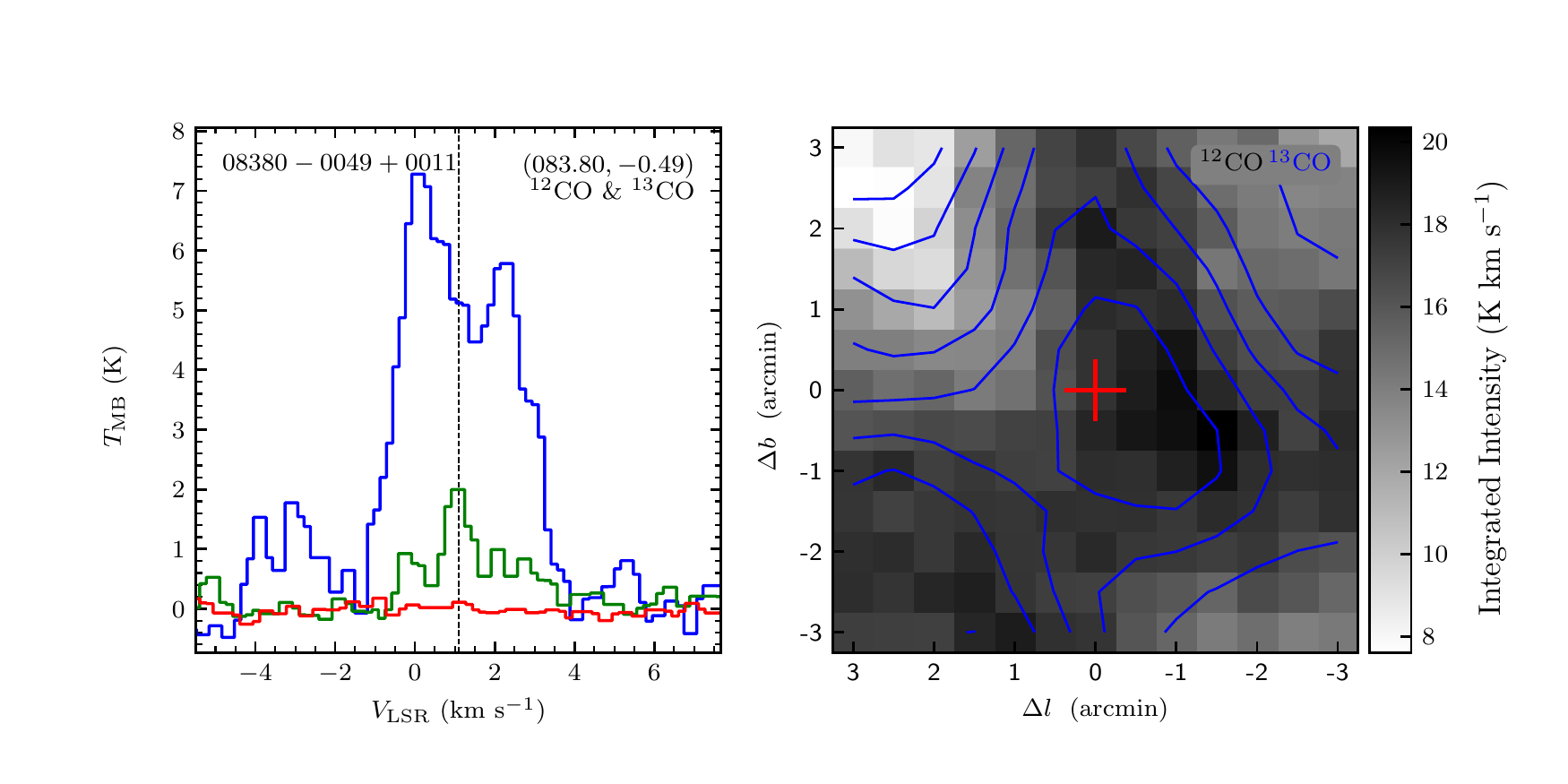}
\vspace{-0.5cm}

\includegraphics[width=9.0cm,angle=0]{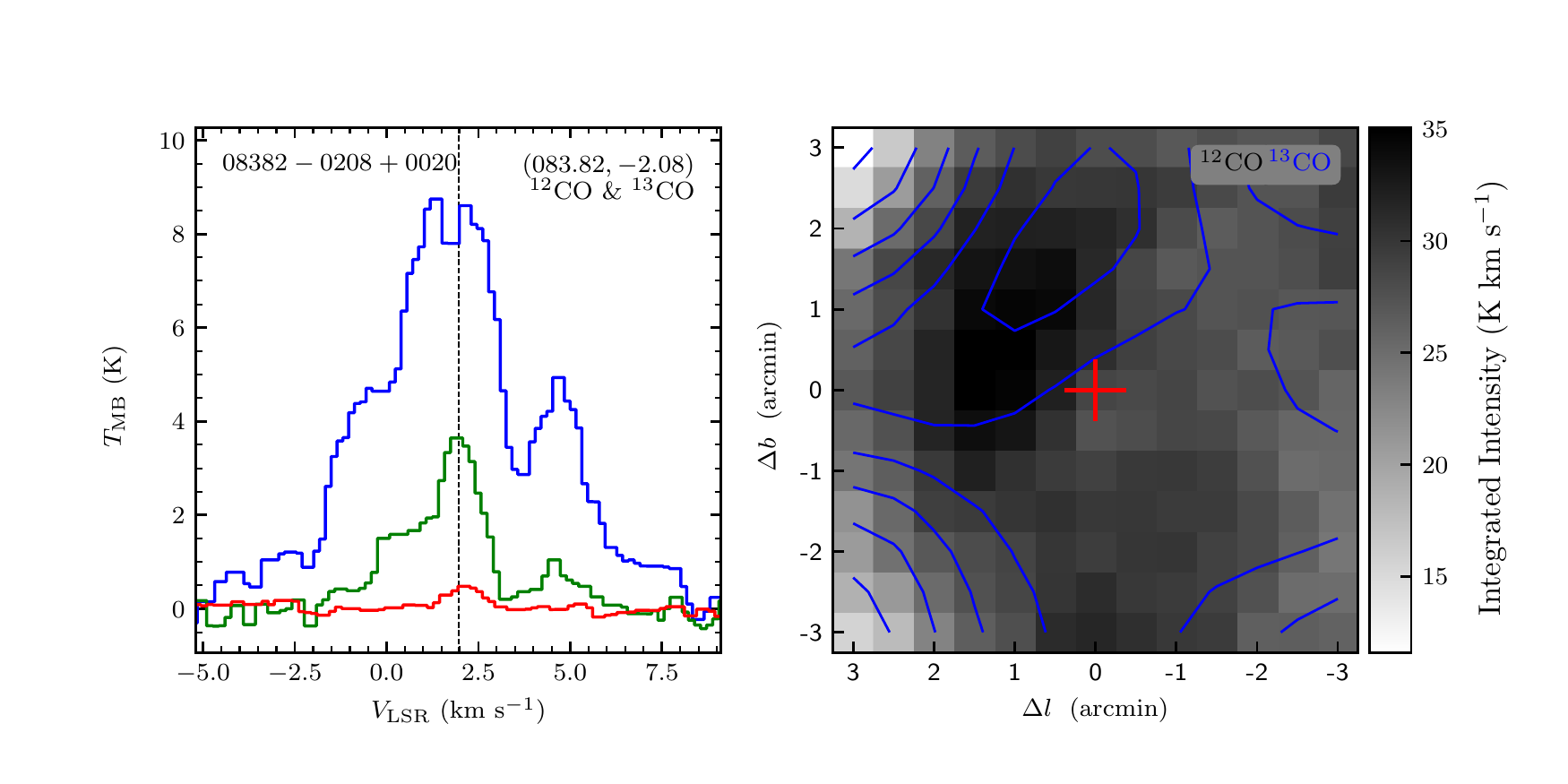}
\includegraphics[width=9.0cm,angle=0]{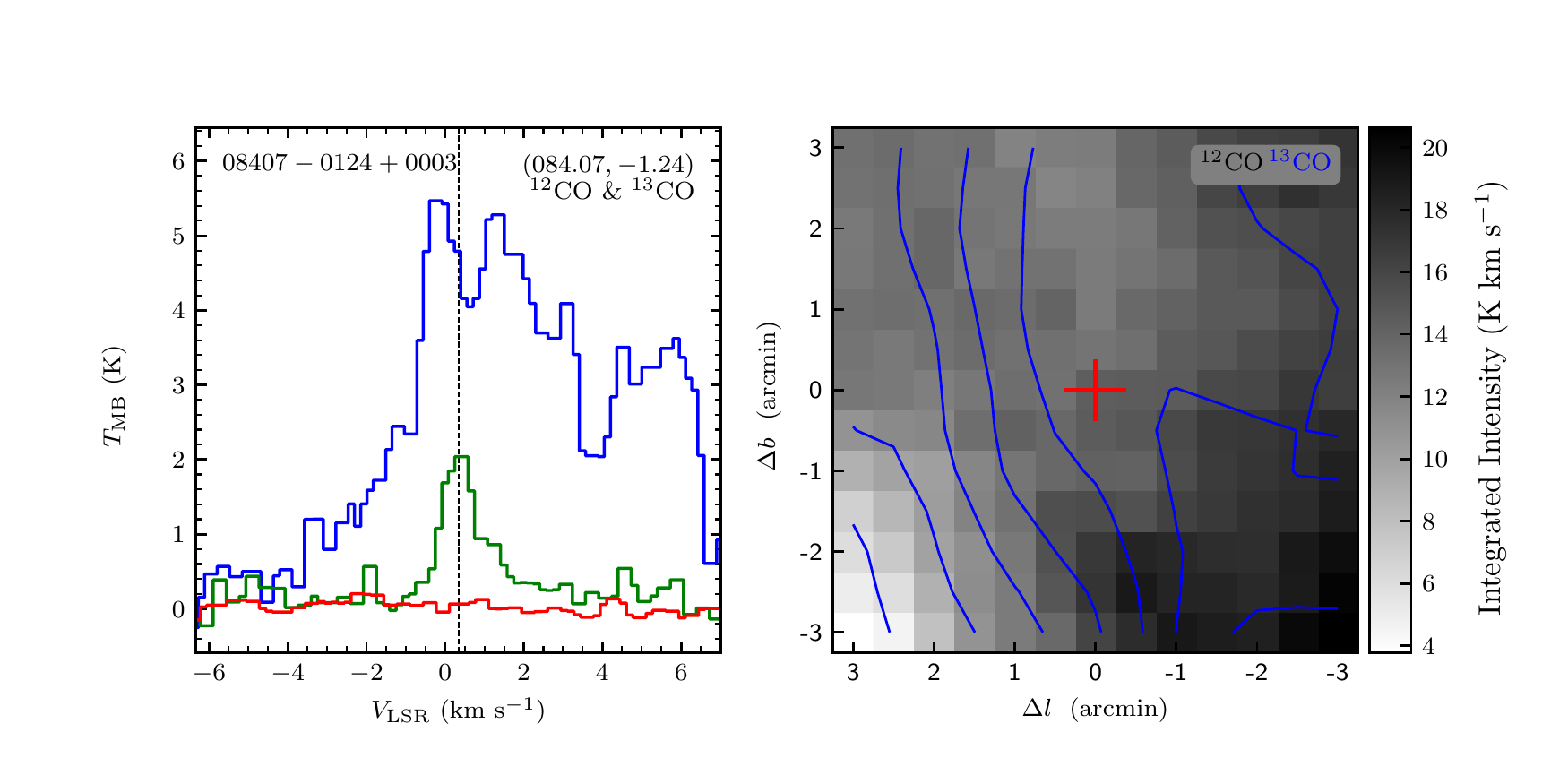}
\vspace{-0.5cm}

\includegraphics[width=9.0cm,angle=0]{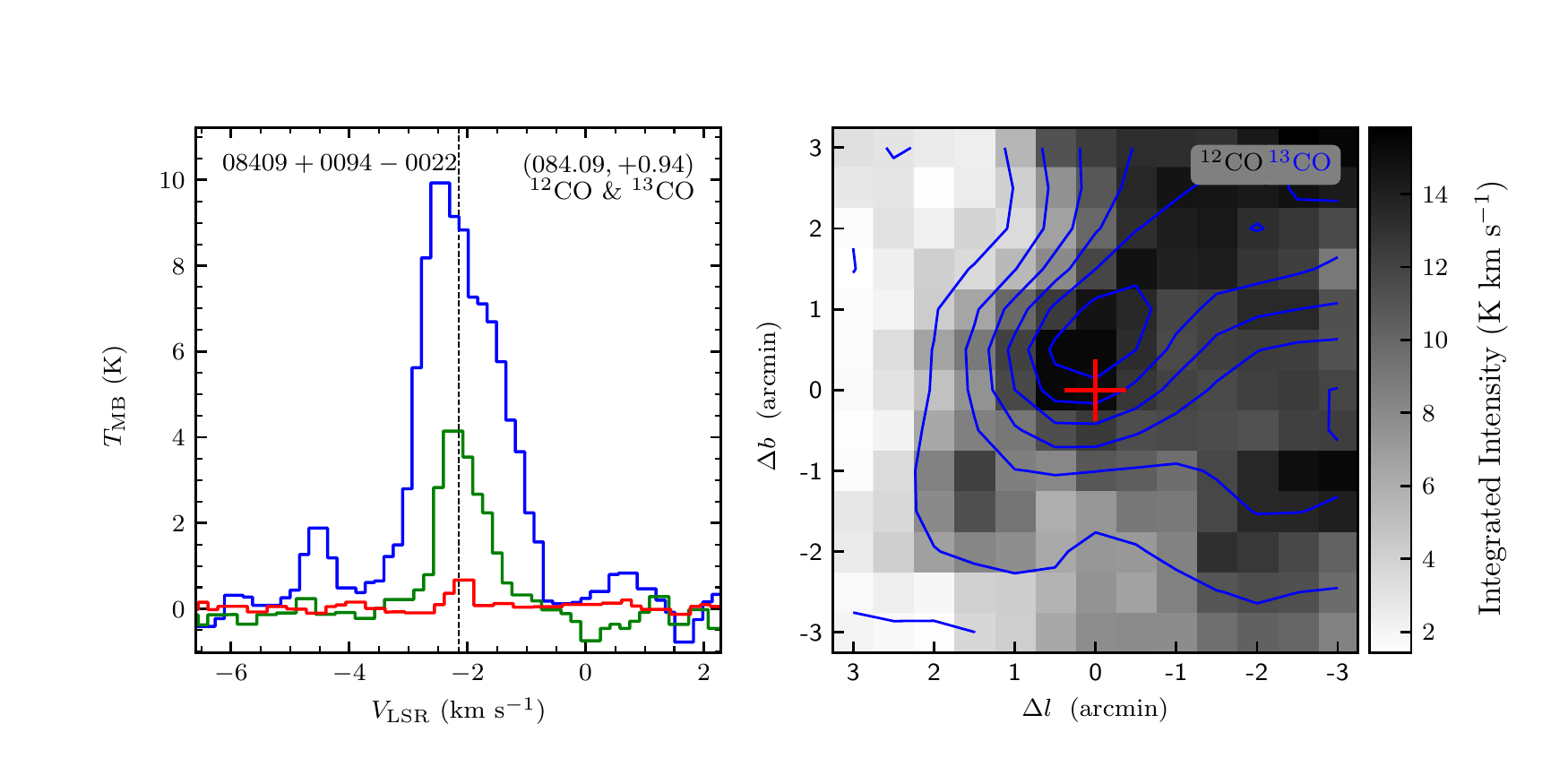}
\includegraphics[width=9.0cm,angle=0]{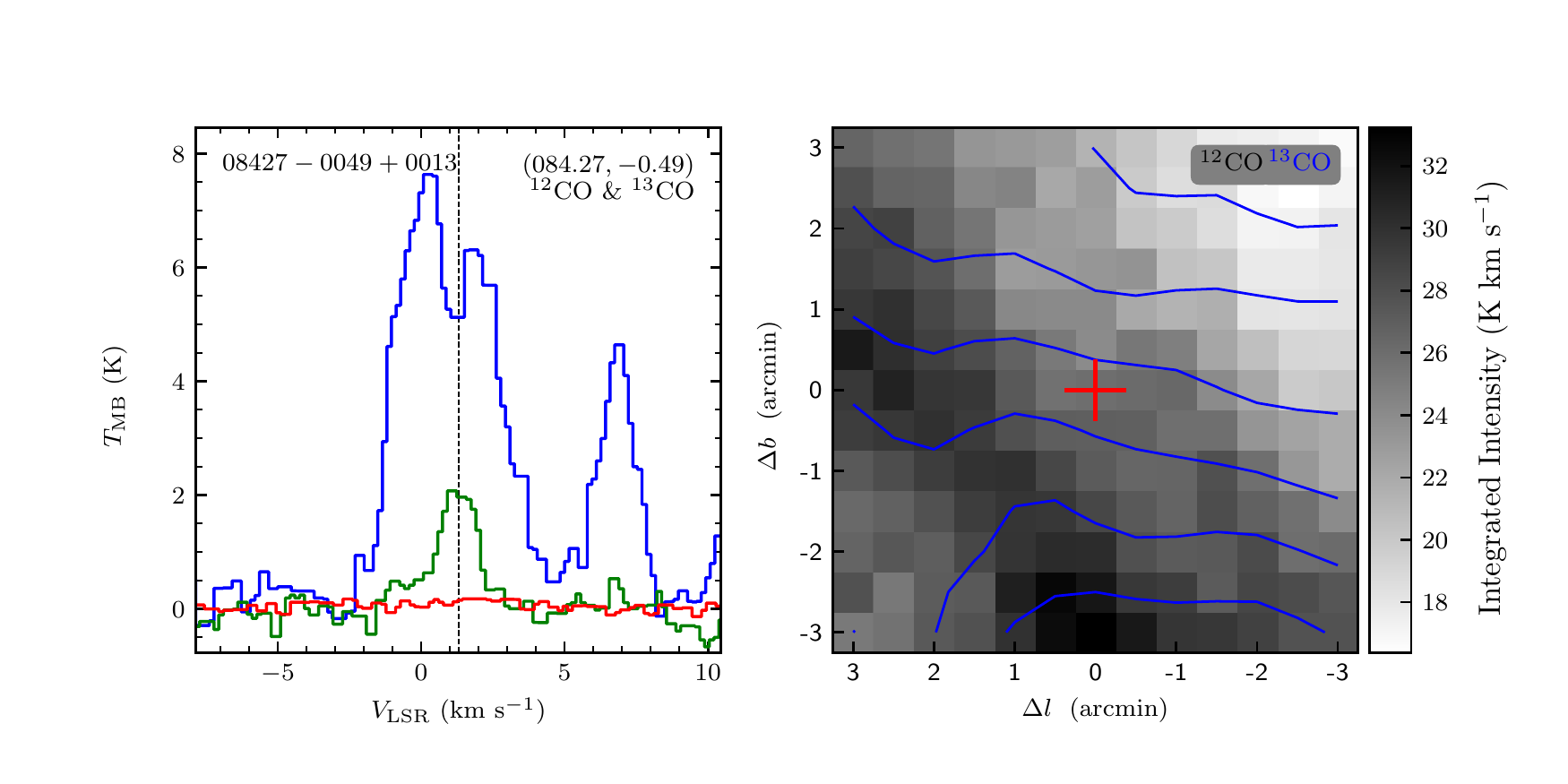}
\vspace{-0.5cm}

\includegraphics[width=9.0cm,angle=0]{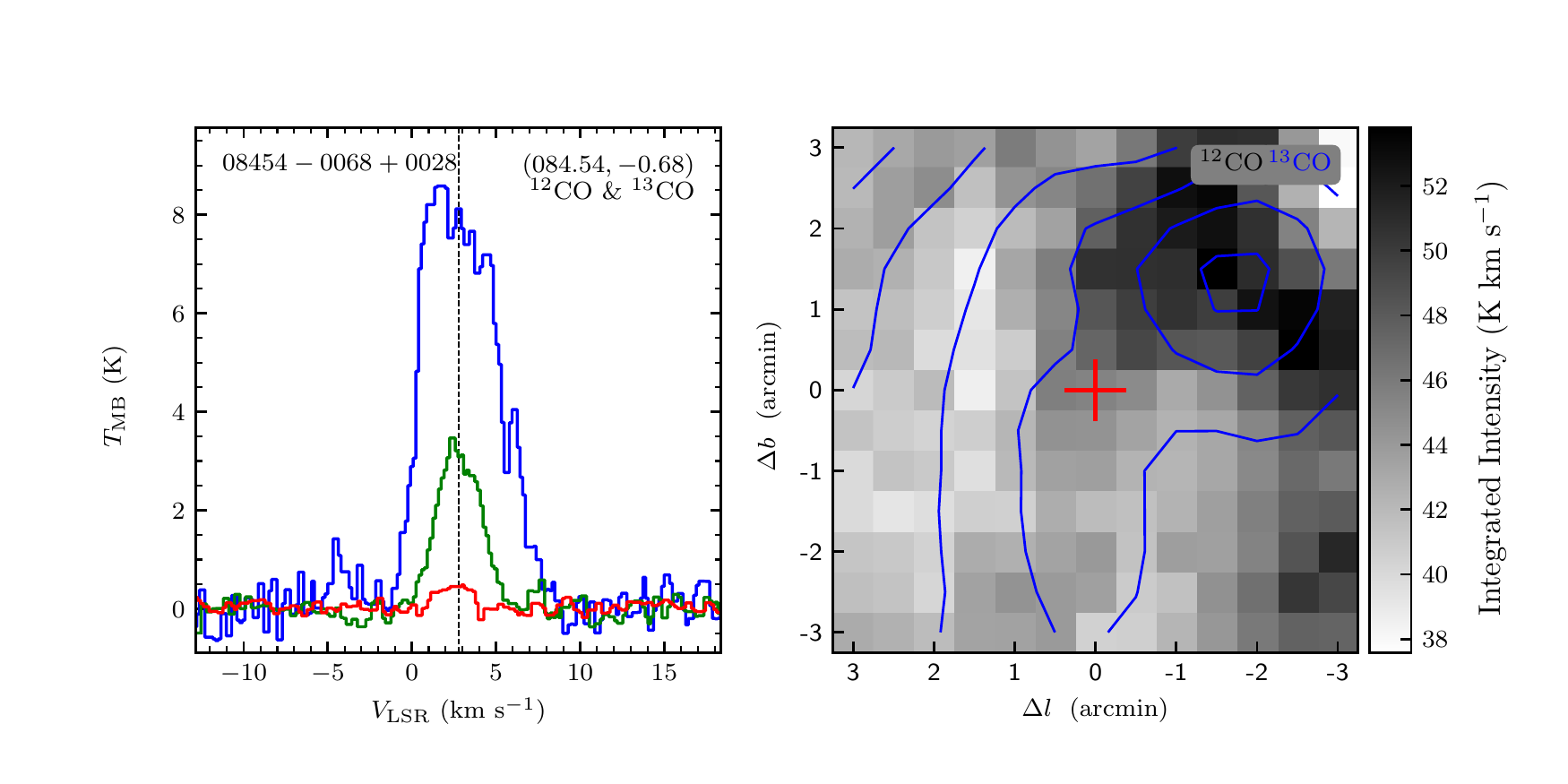}
\includegraphics[width=9.0cm,angle=0]{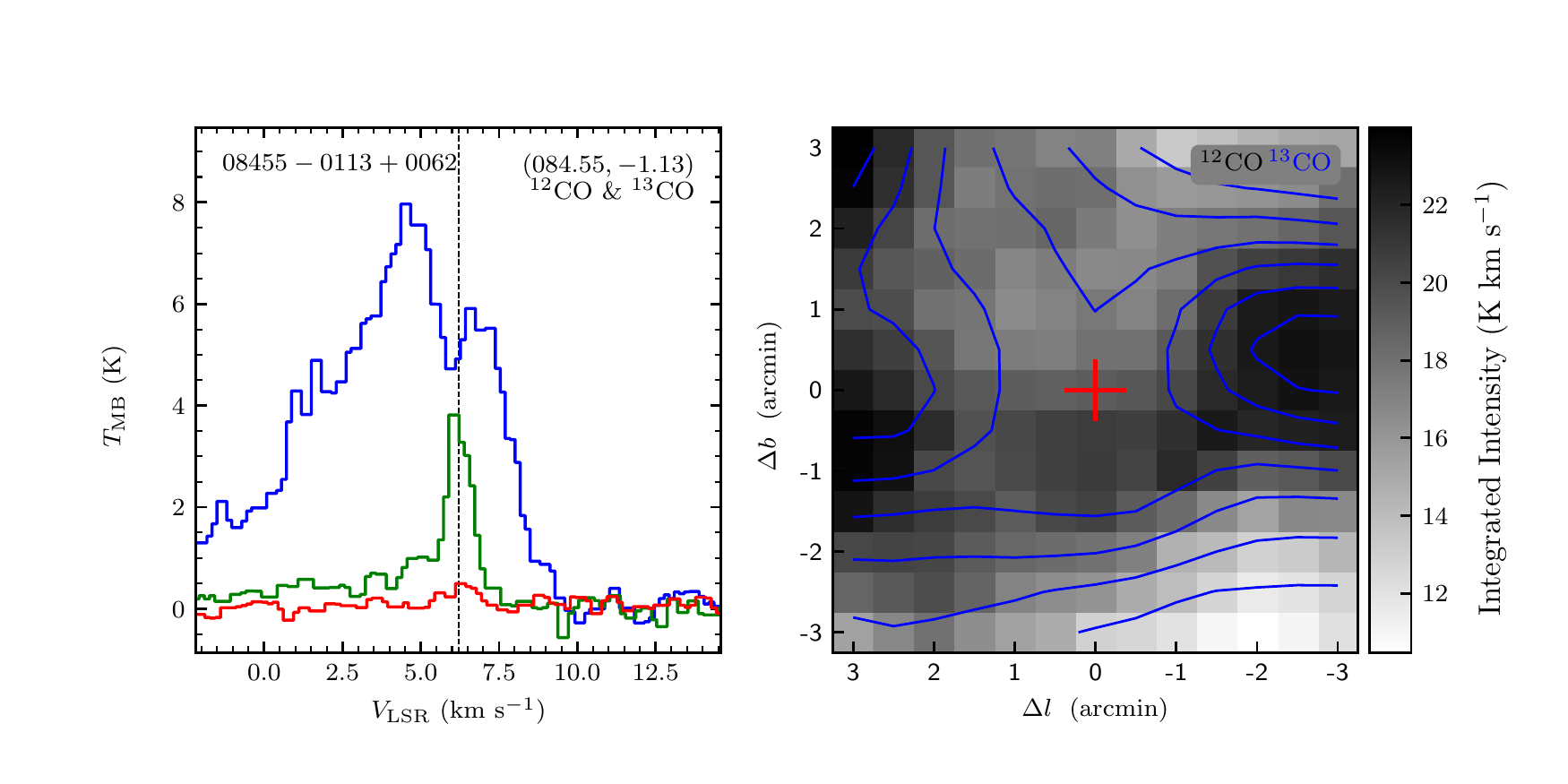}
\vspace{-0.5cm}

\includegraphics[width=9.0cm,angle=0]{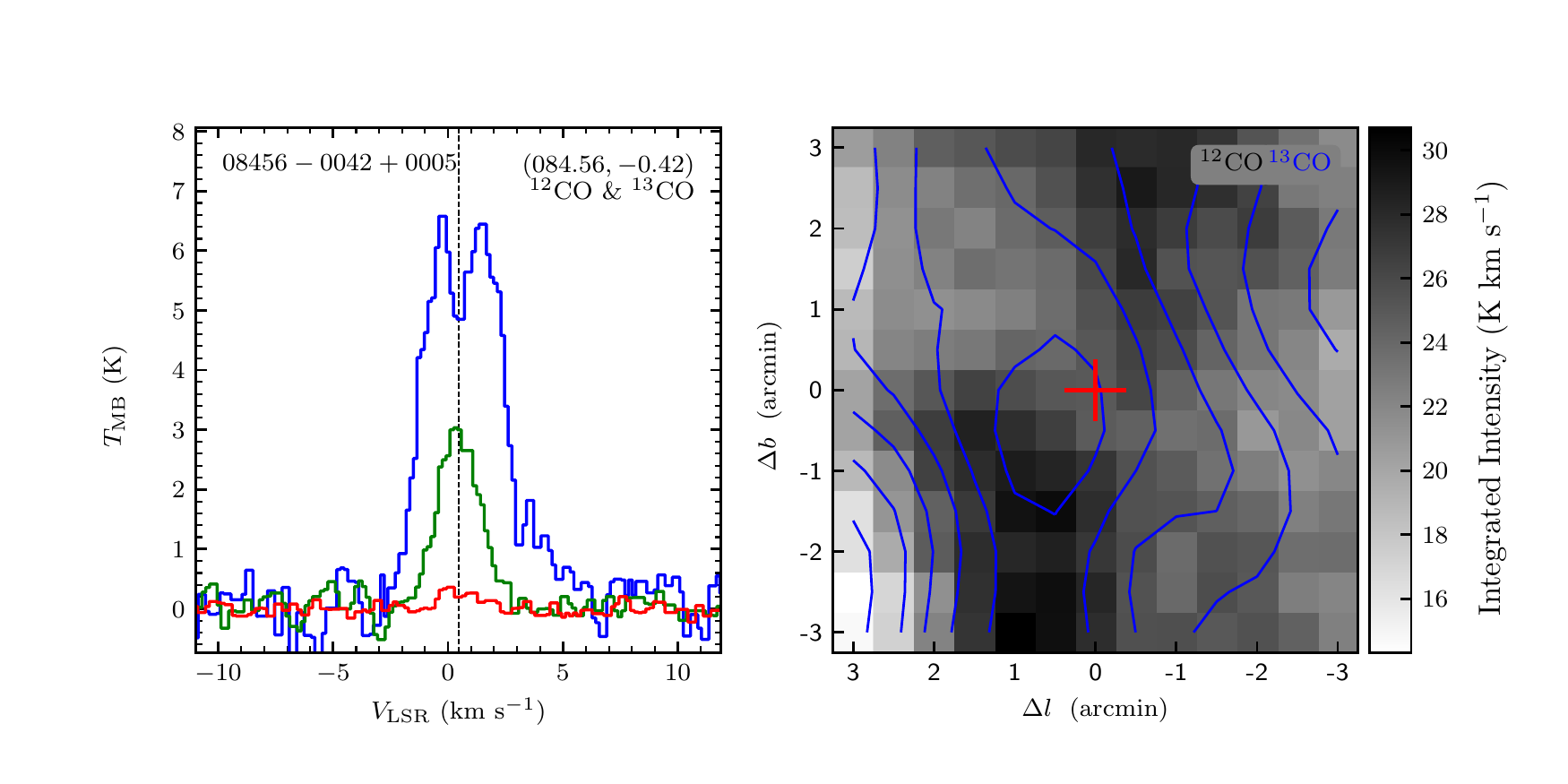}
\includegraphics[width=9.0cm,angle=0]{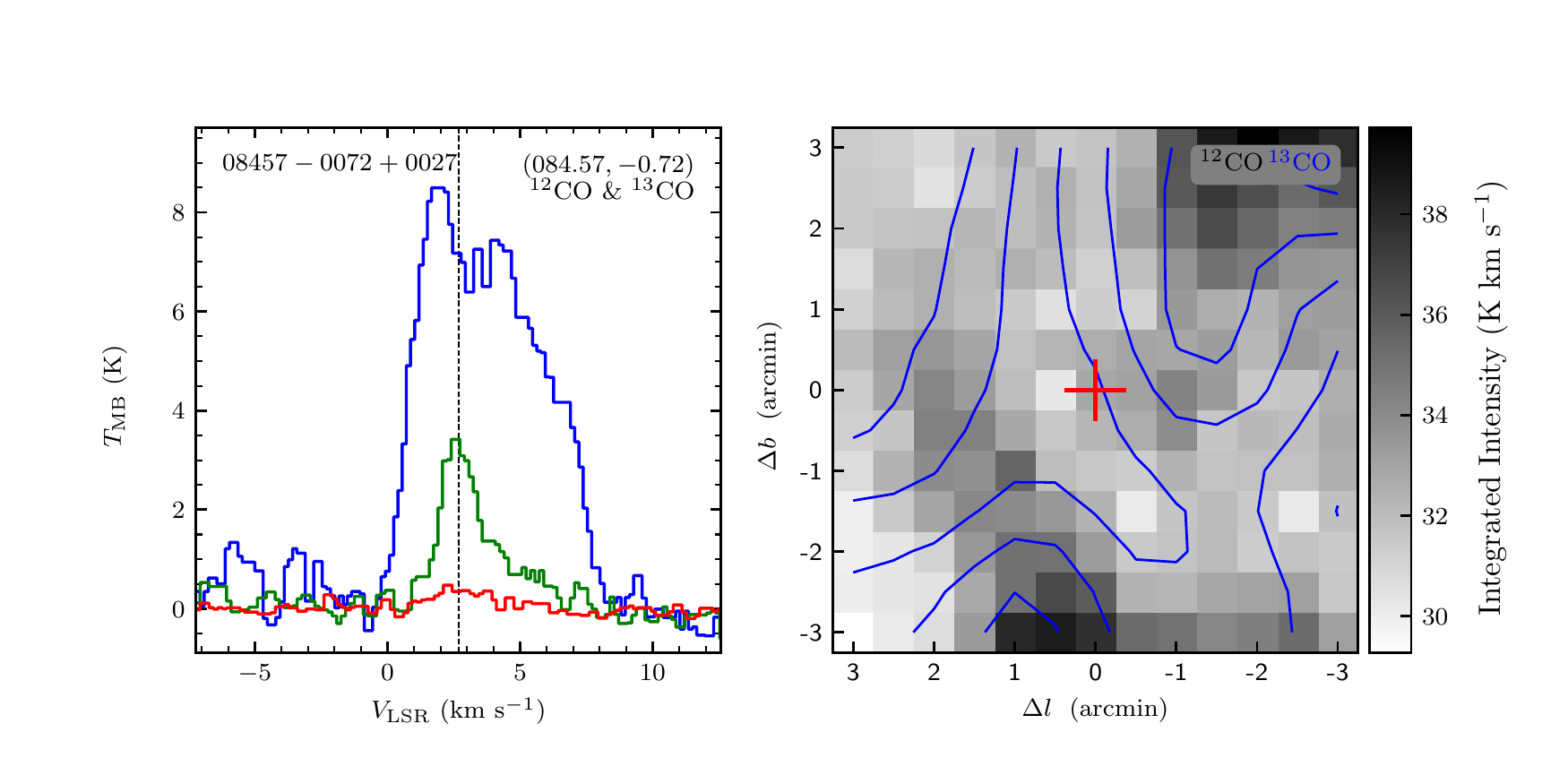}
\end{figure}
\clearpage

\begin{figure}
\includegraphics[width=9.0cm,angle=0]{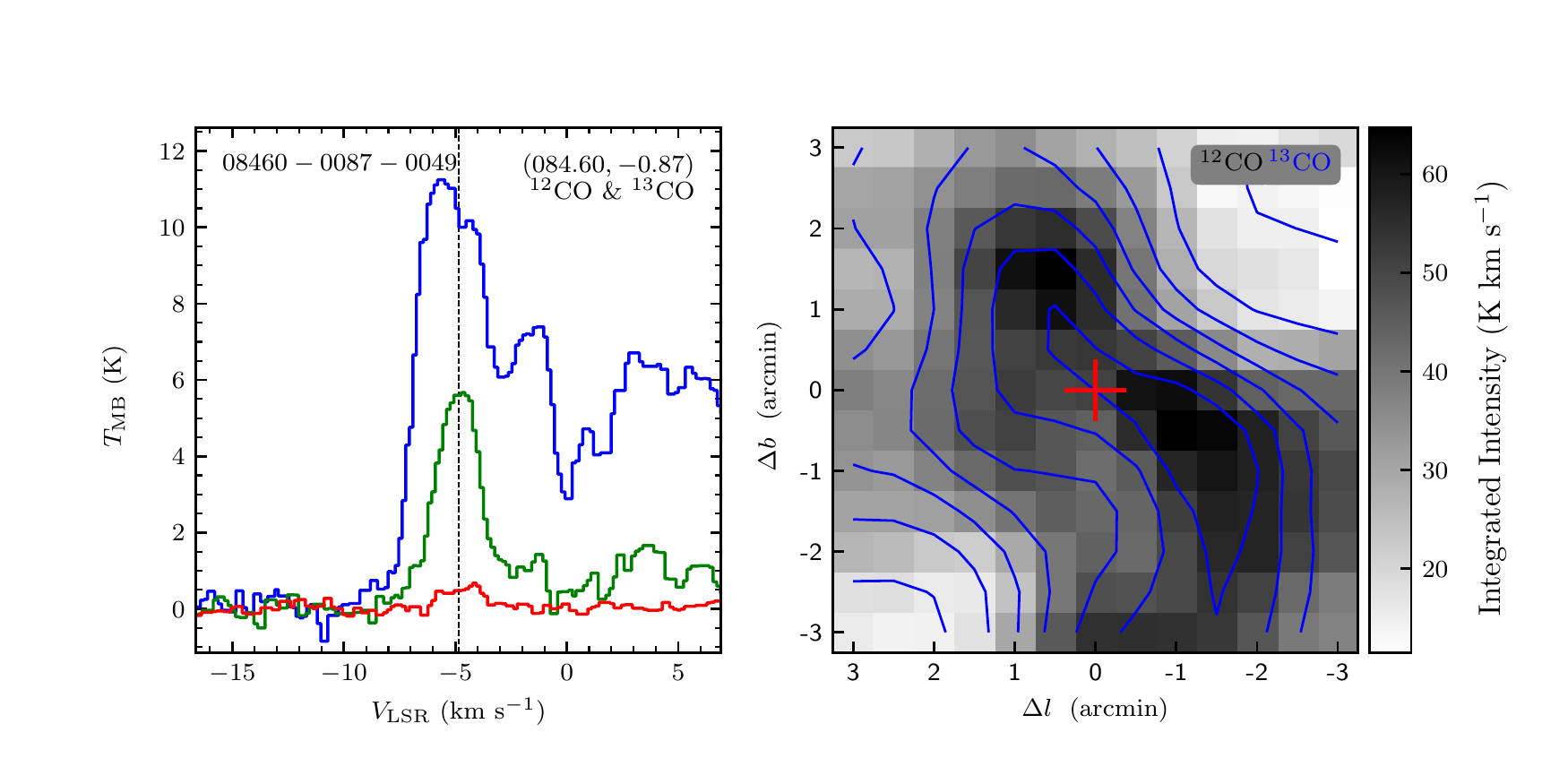}
\includegraphics[width=9.0cm,angle=0]{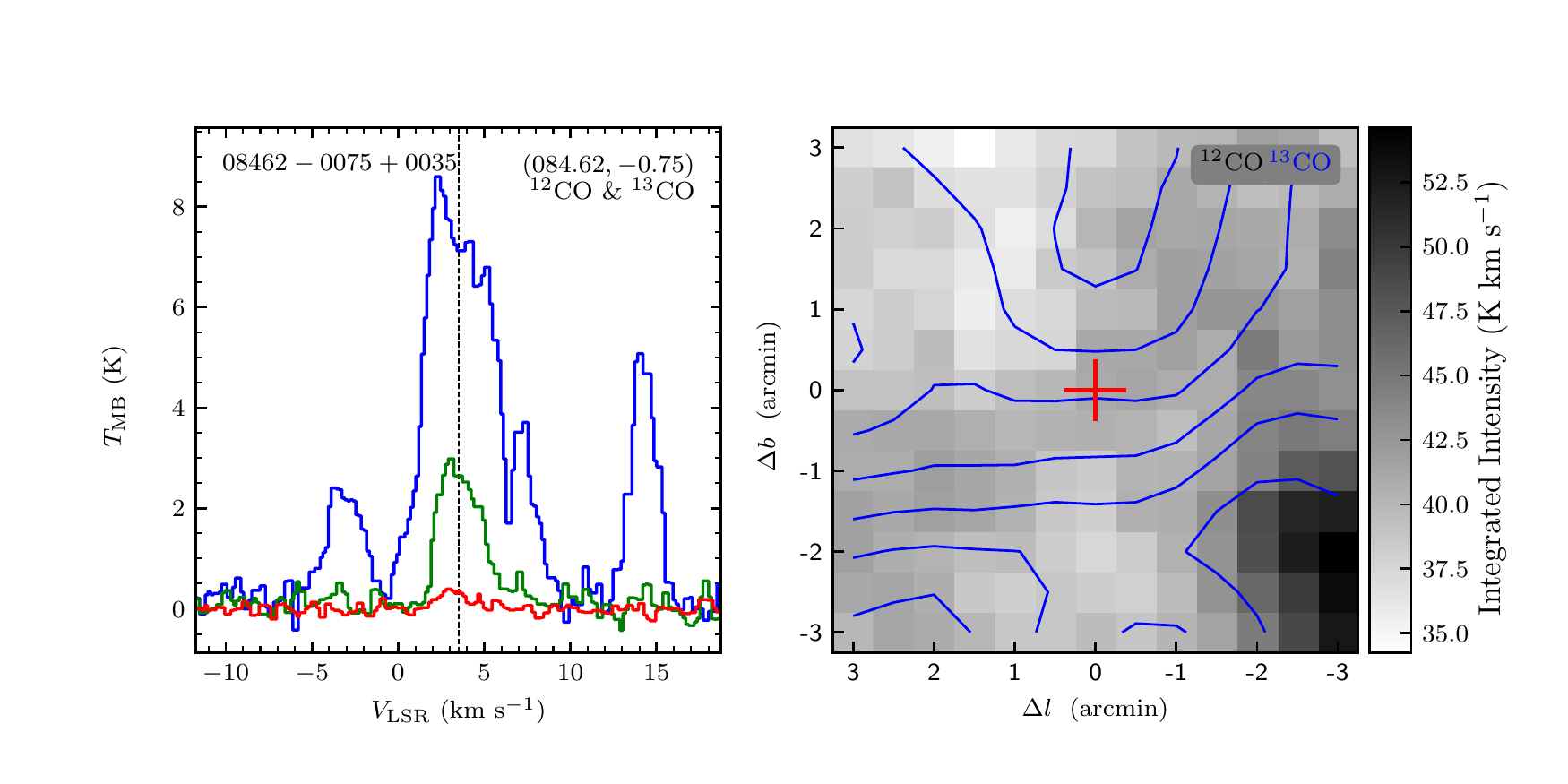}
\vspace{-0.5cm}

\includegraphics[width=9.0cm,angle=0]{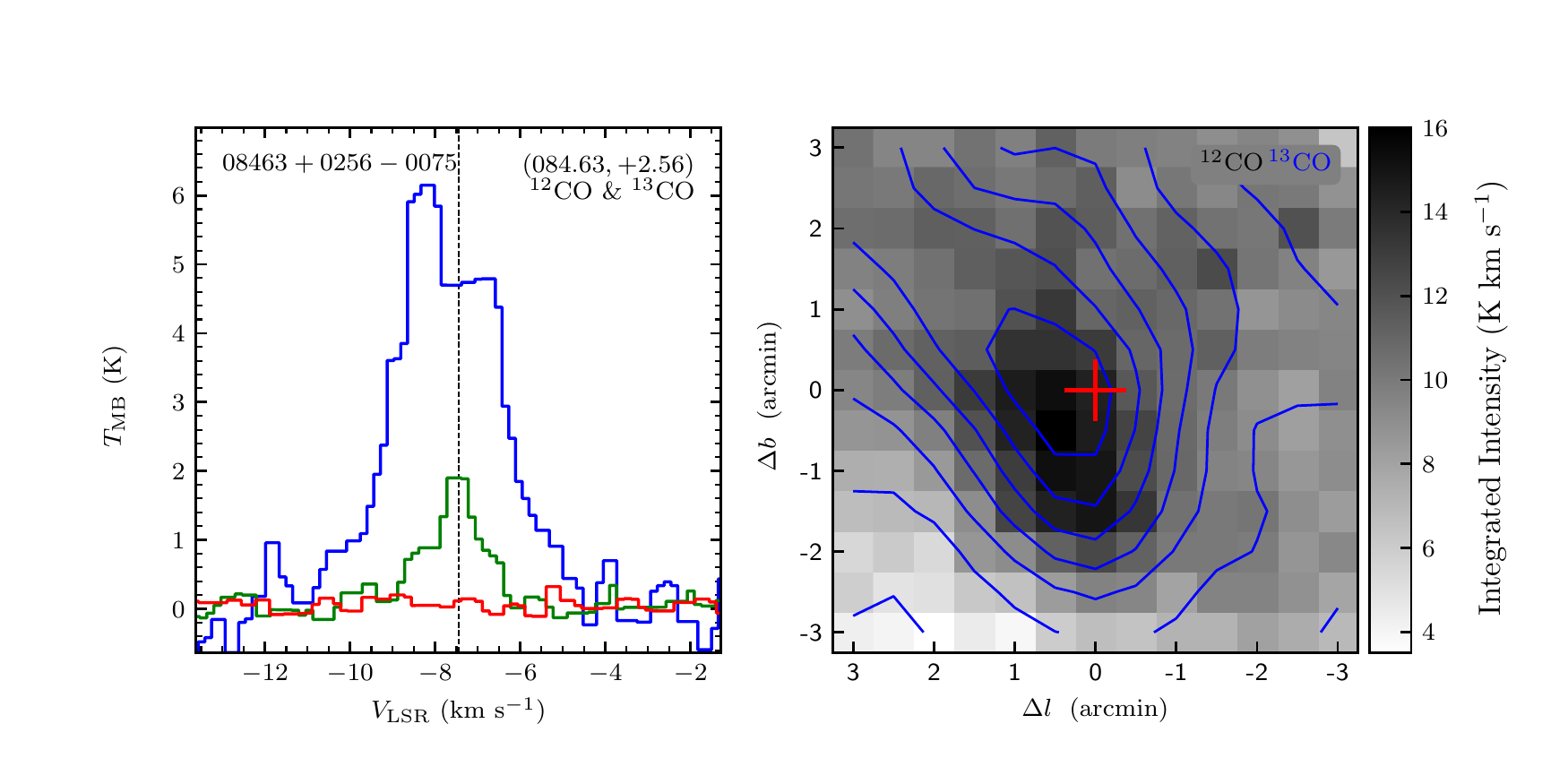}
\includegraphics[width=9.0cm,angle=0]{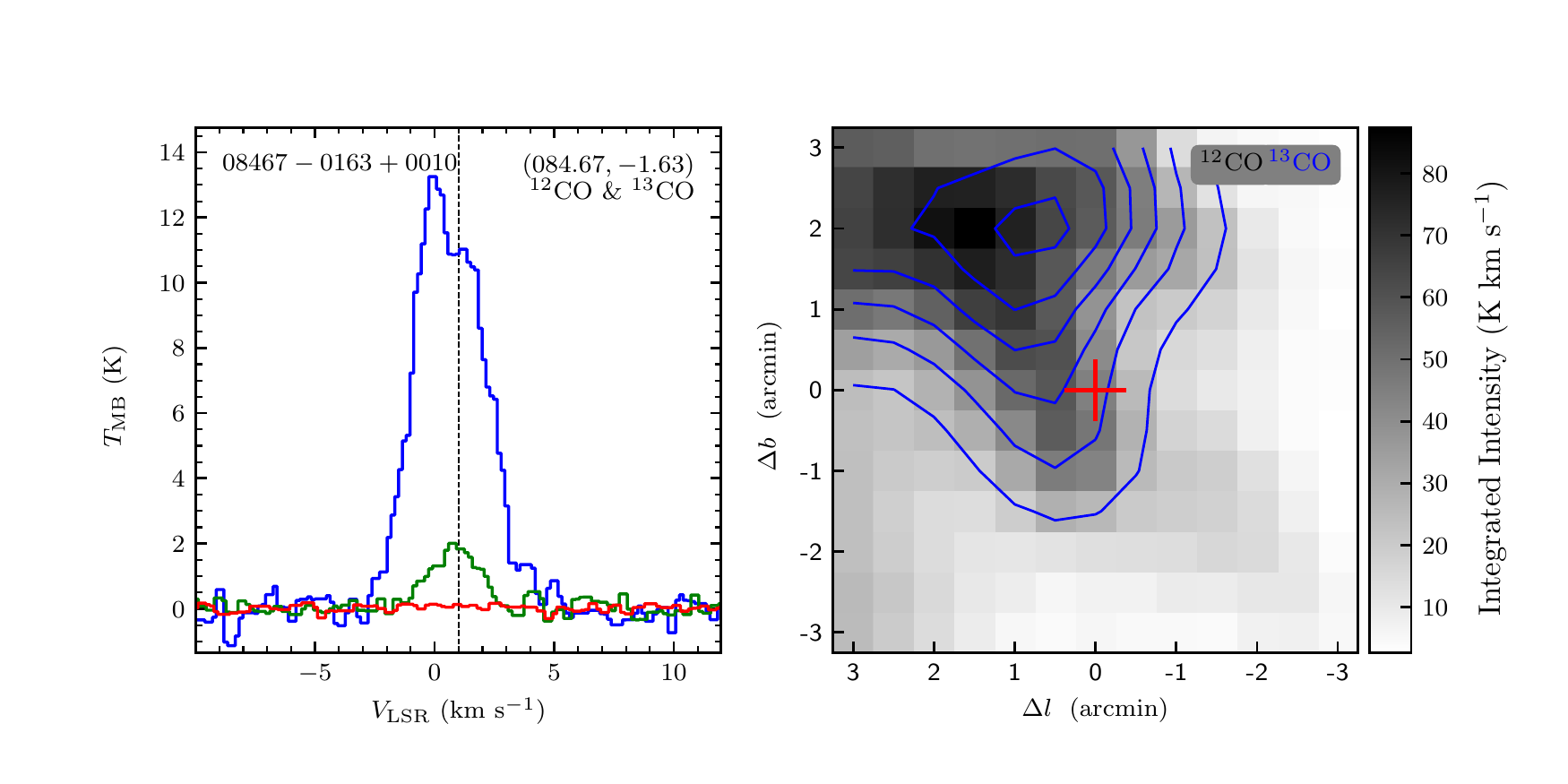}
\vspace{-0.5cm}

\includegraphics[width=9.0cm,angle=0]{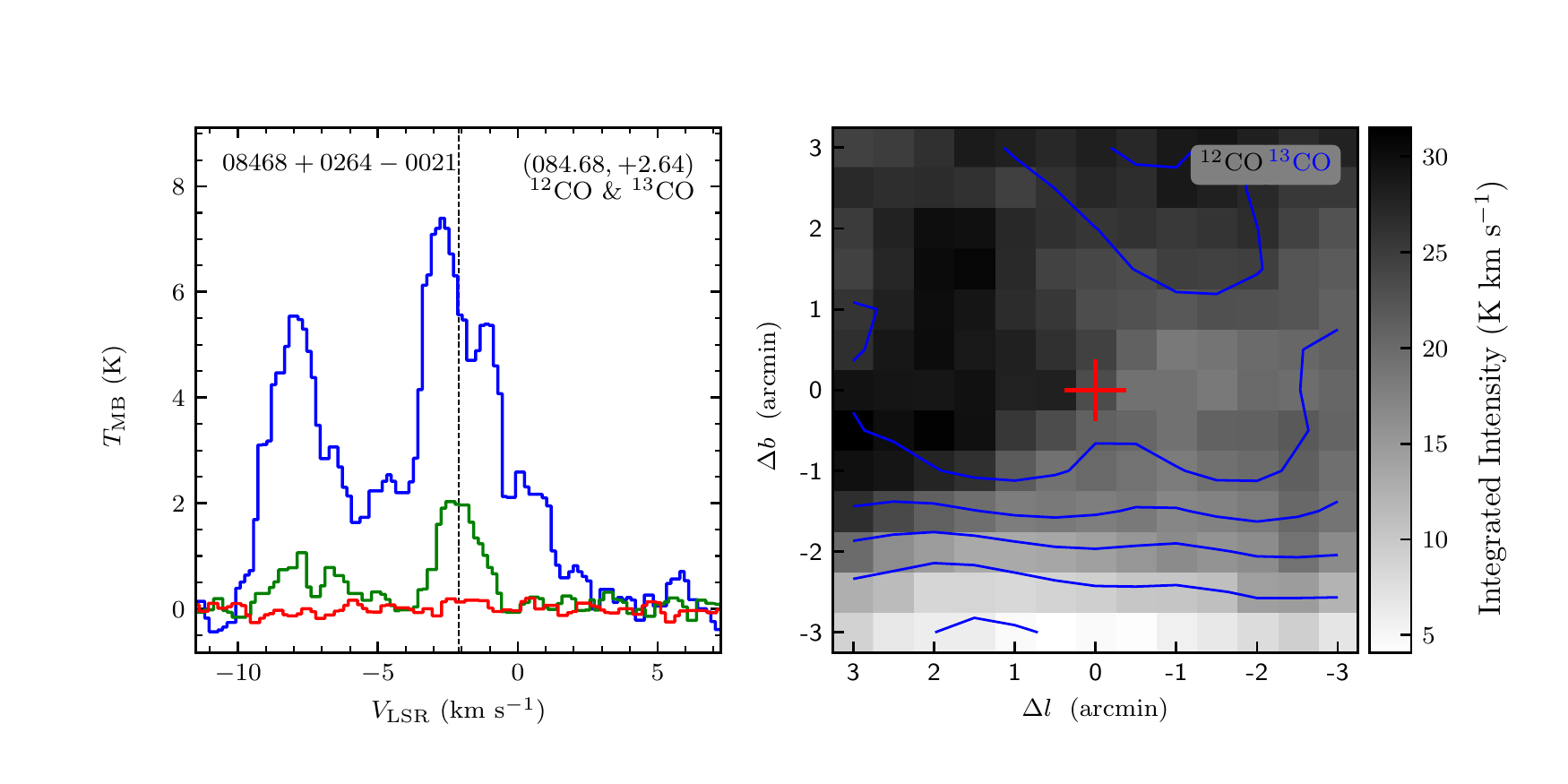}
\includegraphics[width=9.0cm,angle=0]{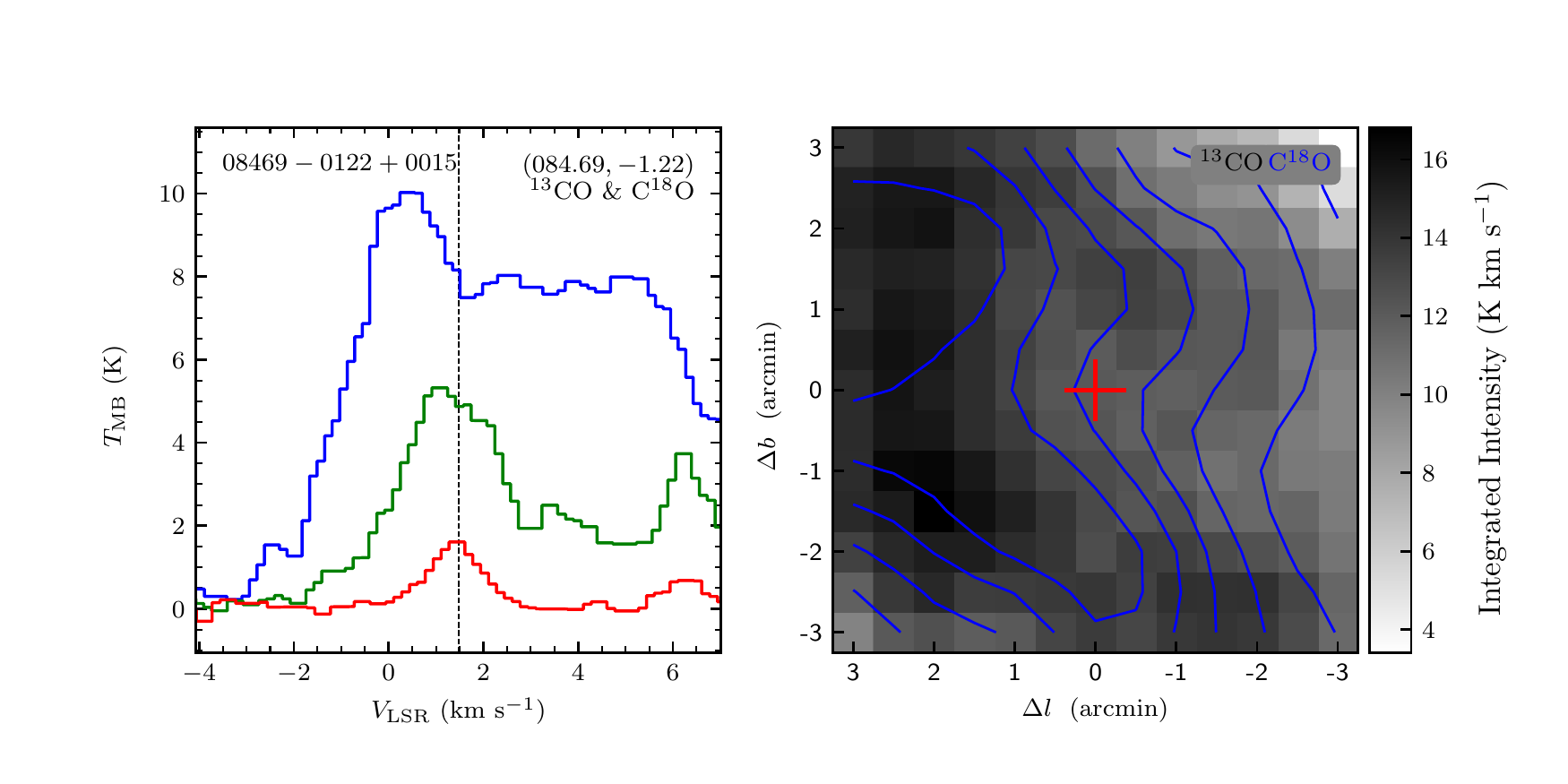}
\vspace{-0.5cm}

\includegraphics[width=9.0cm,angle=0]{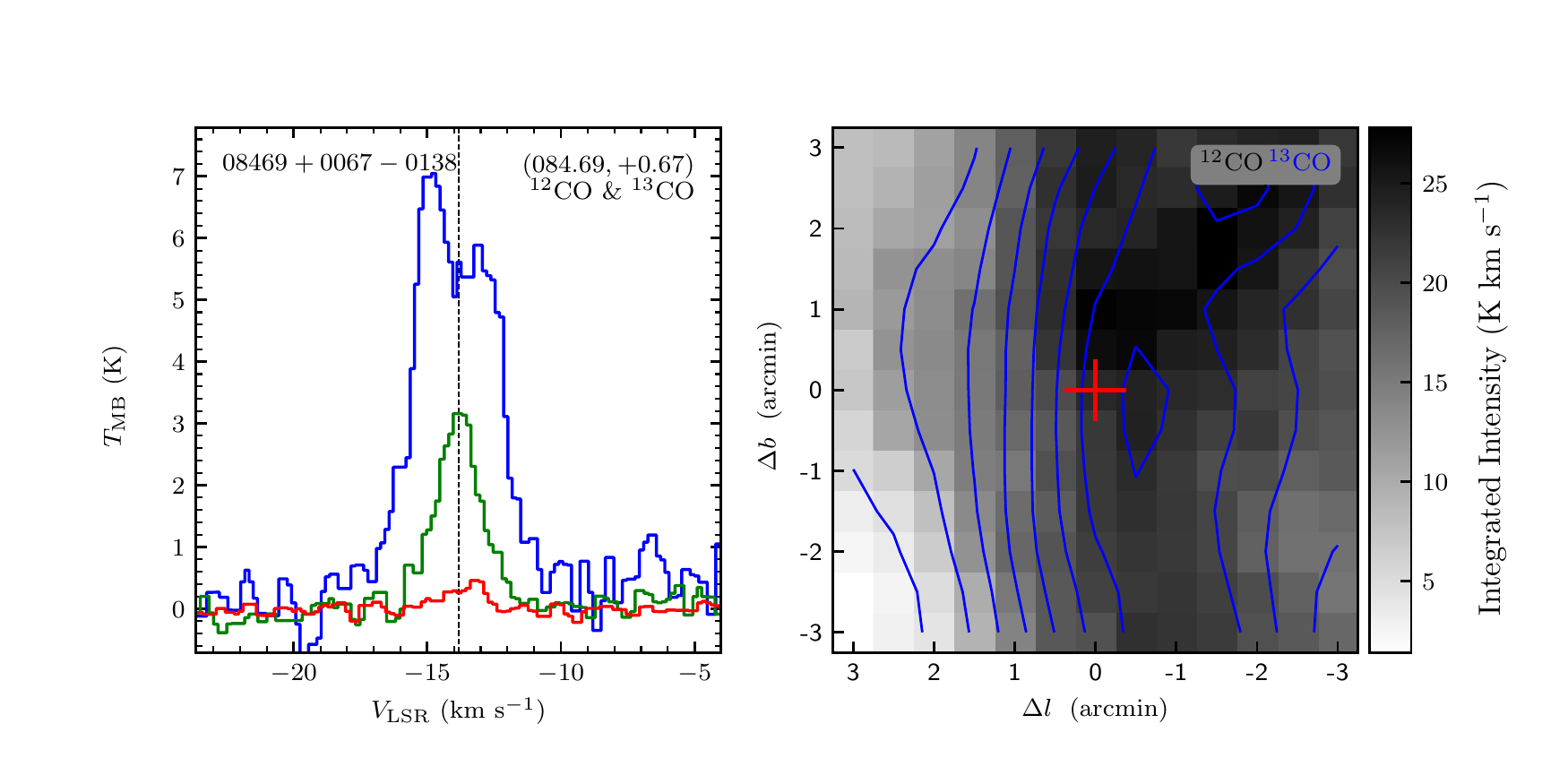}
\includegraphics[width=9.0cm,angle=0]{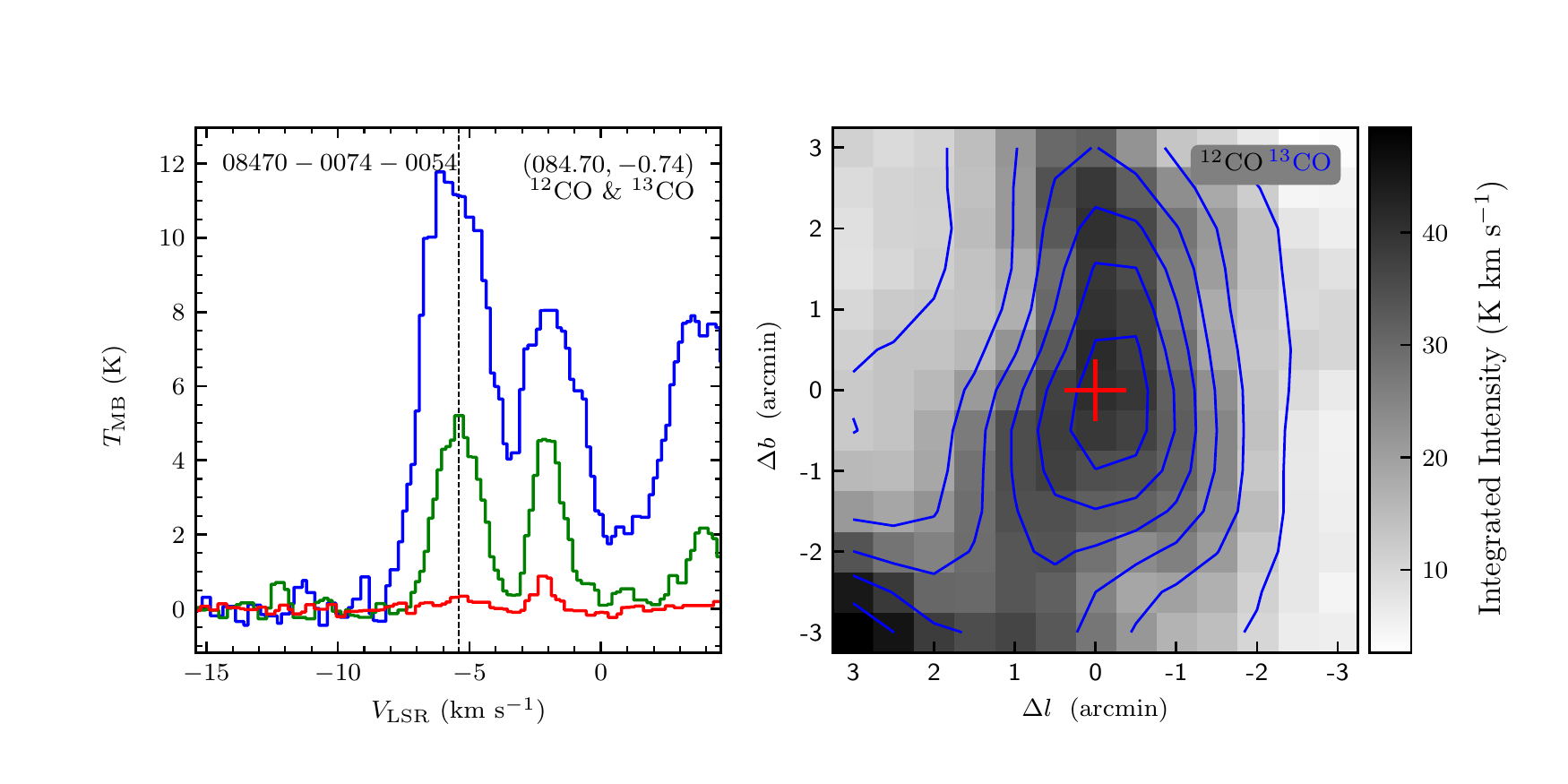}
\vspace{-0.5cm}

\includegraphics[width=9.0cm,angle=0]{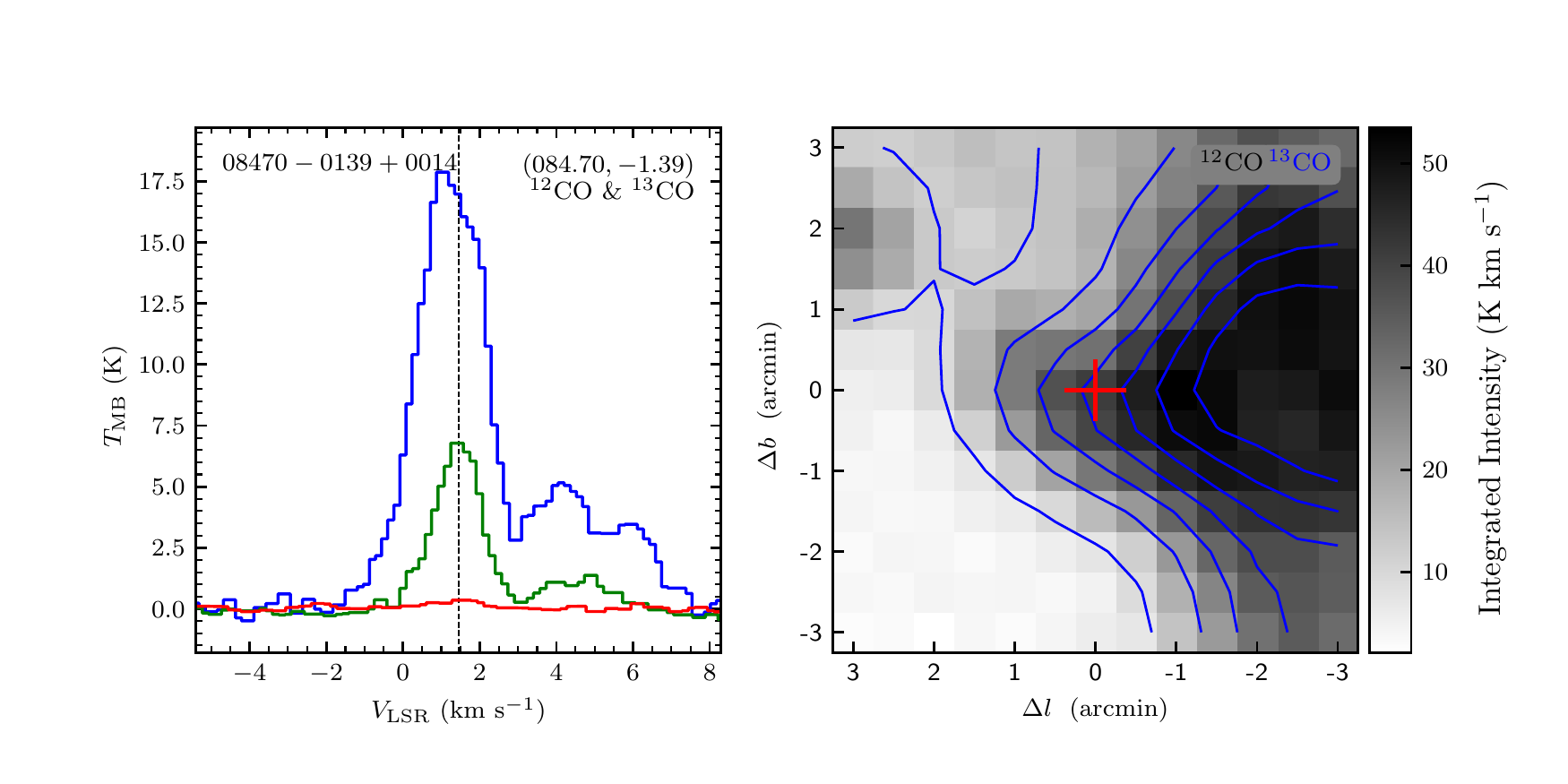}
\includegraphics[width=9.0cm,angle=0]{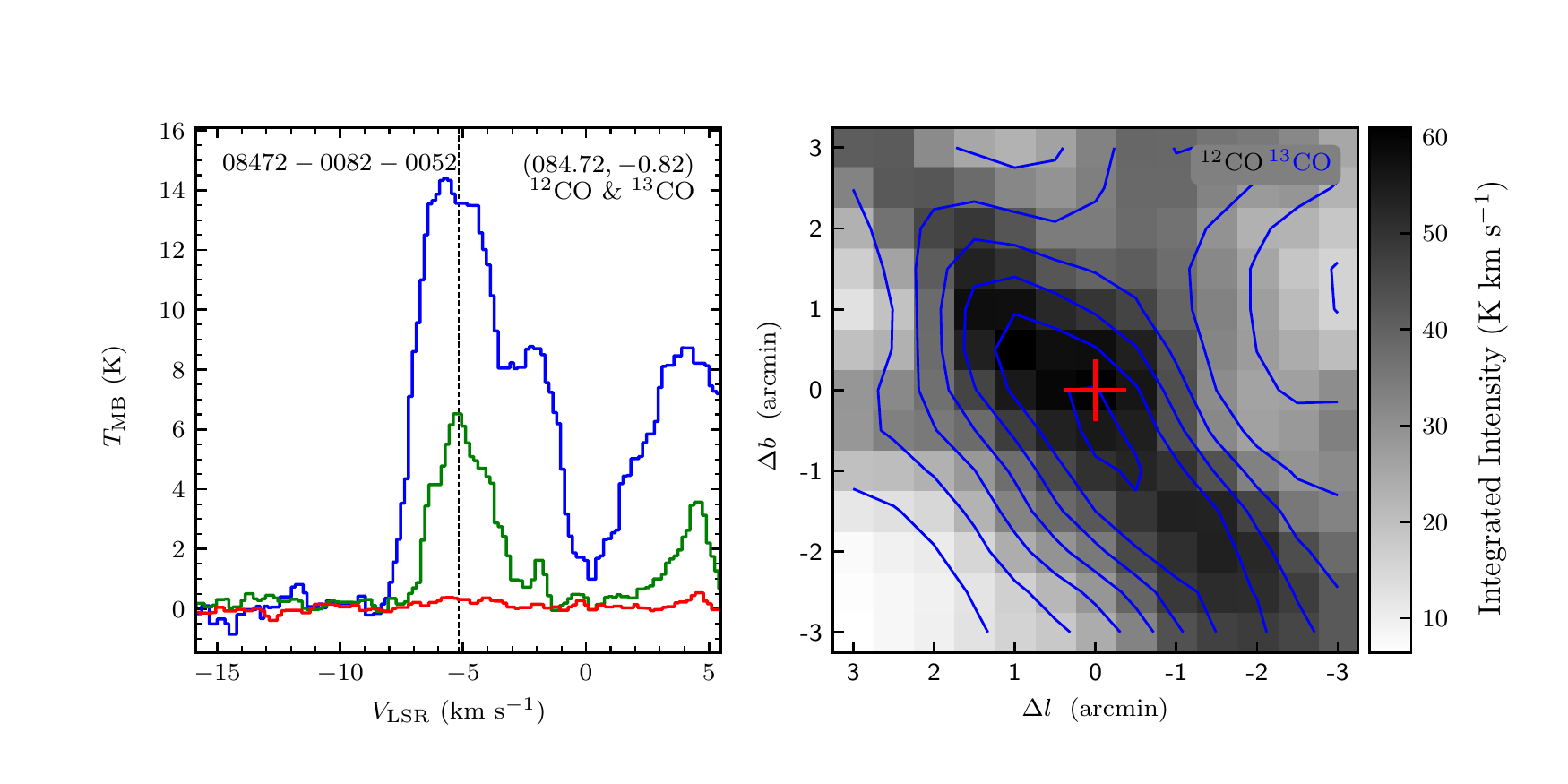}
\end{figure}
\clearpage

\begin{figure}
\includegraphics[width=9.0cm,angle=0]{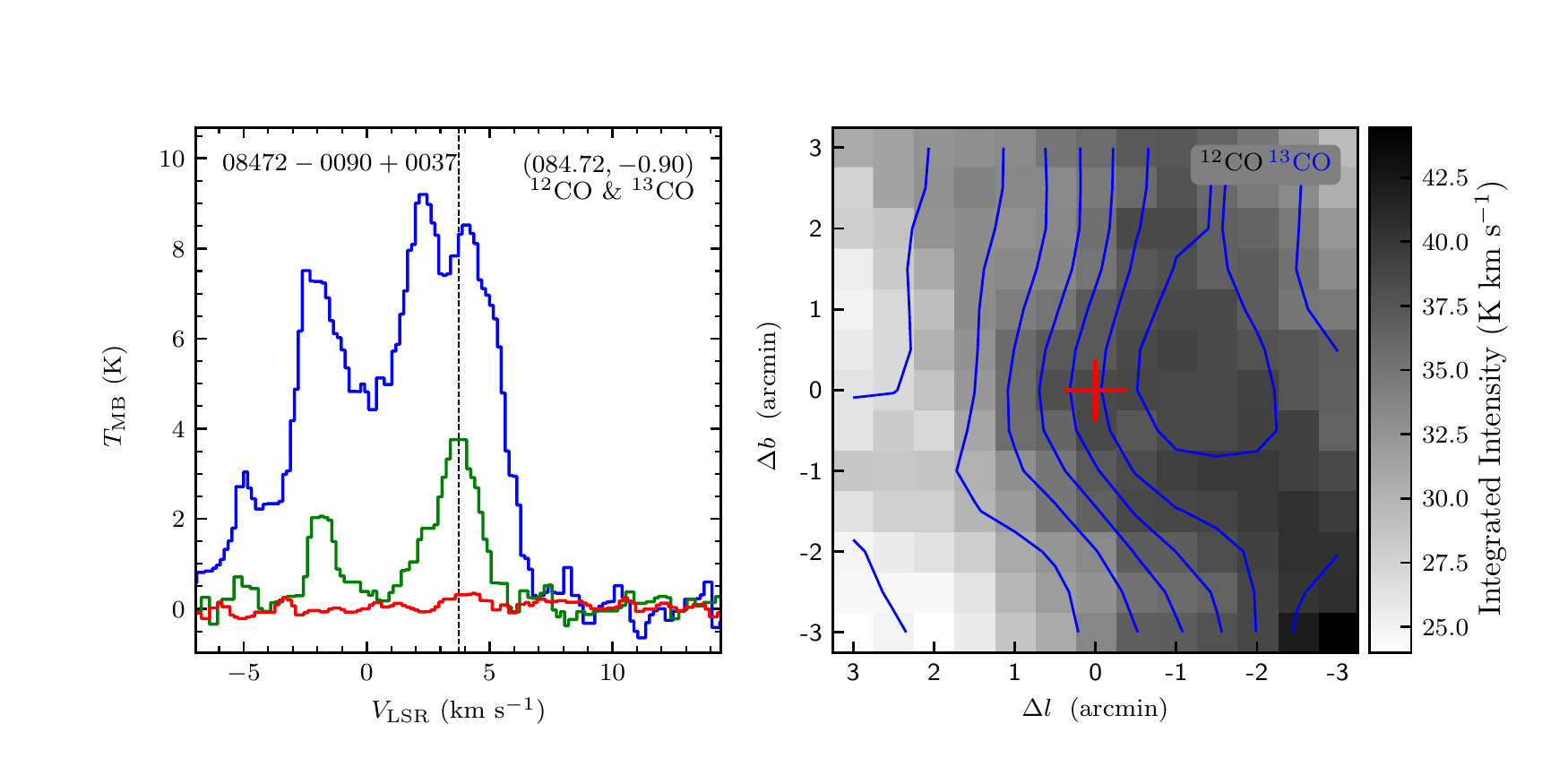}
\includegraphics[width=9.0cm,angle=0]{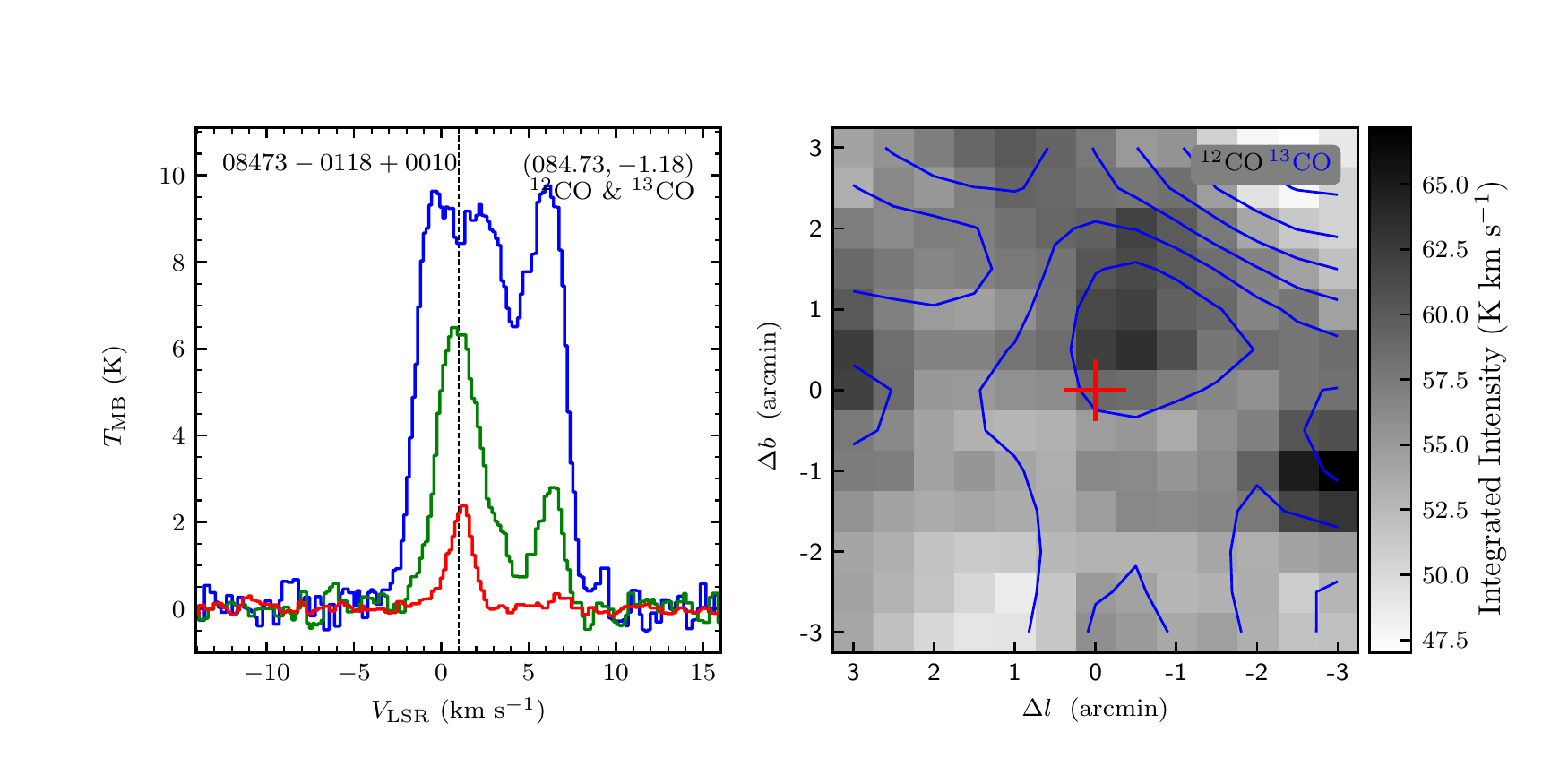}
\vspace{-0.5cm}

\includegraphics[width=9.0cm,angle=0]{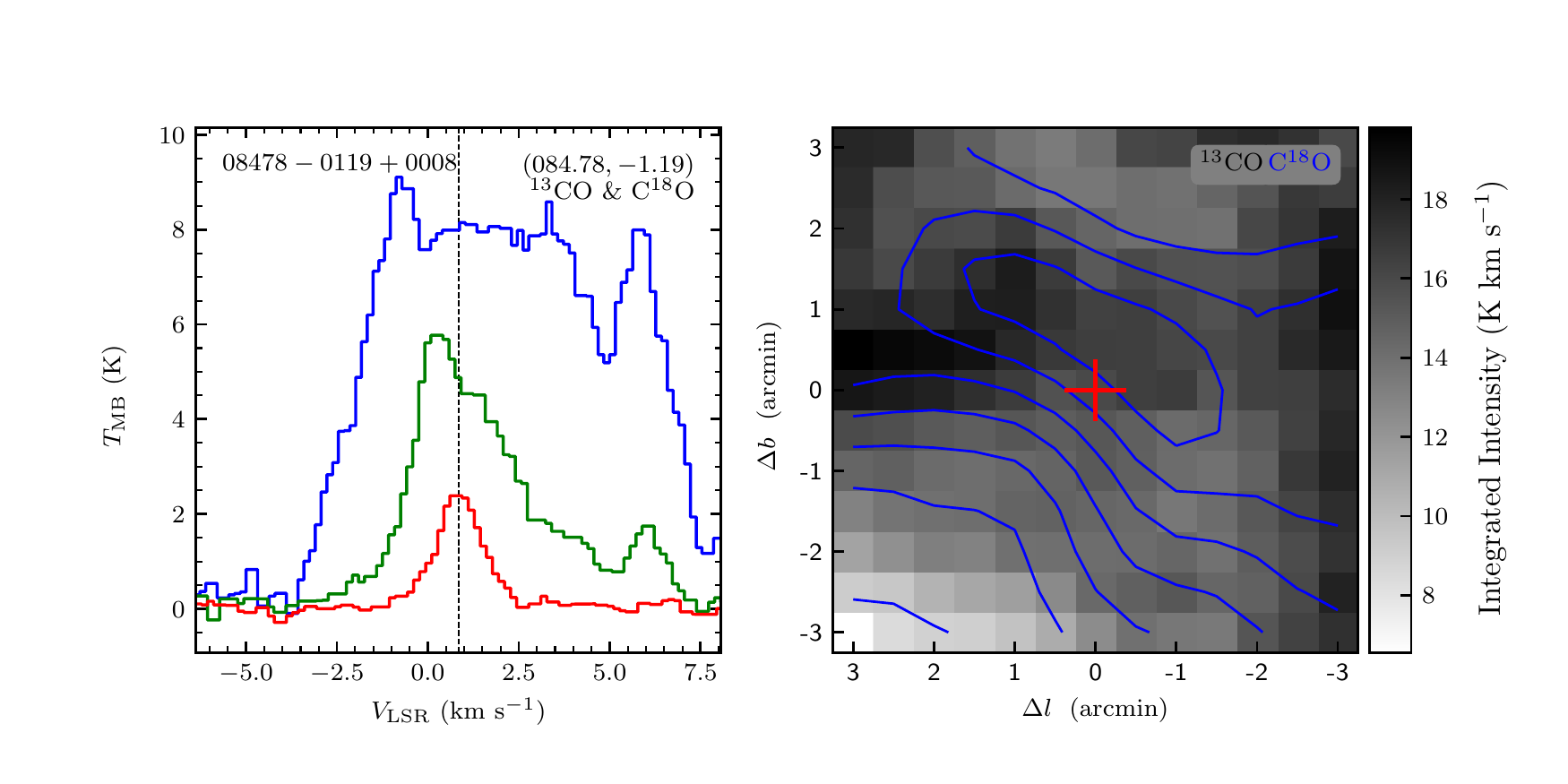}
\includegraphics[width=9.0cm,angle=0]{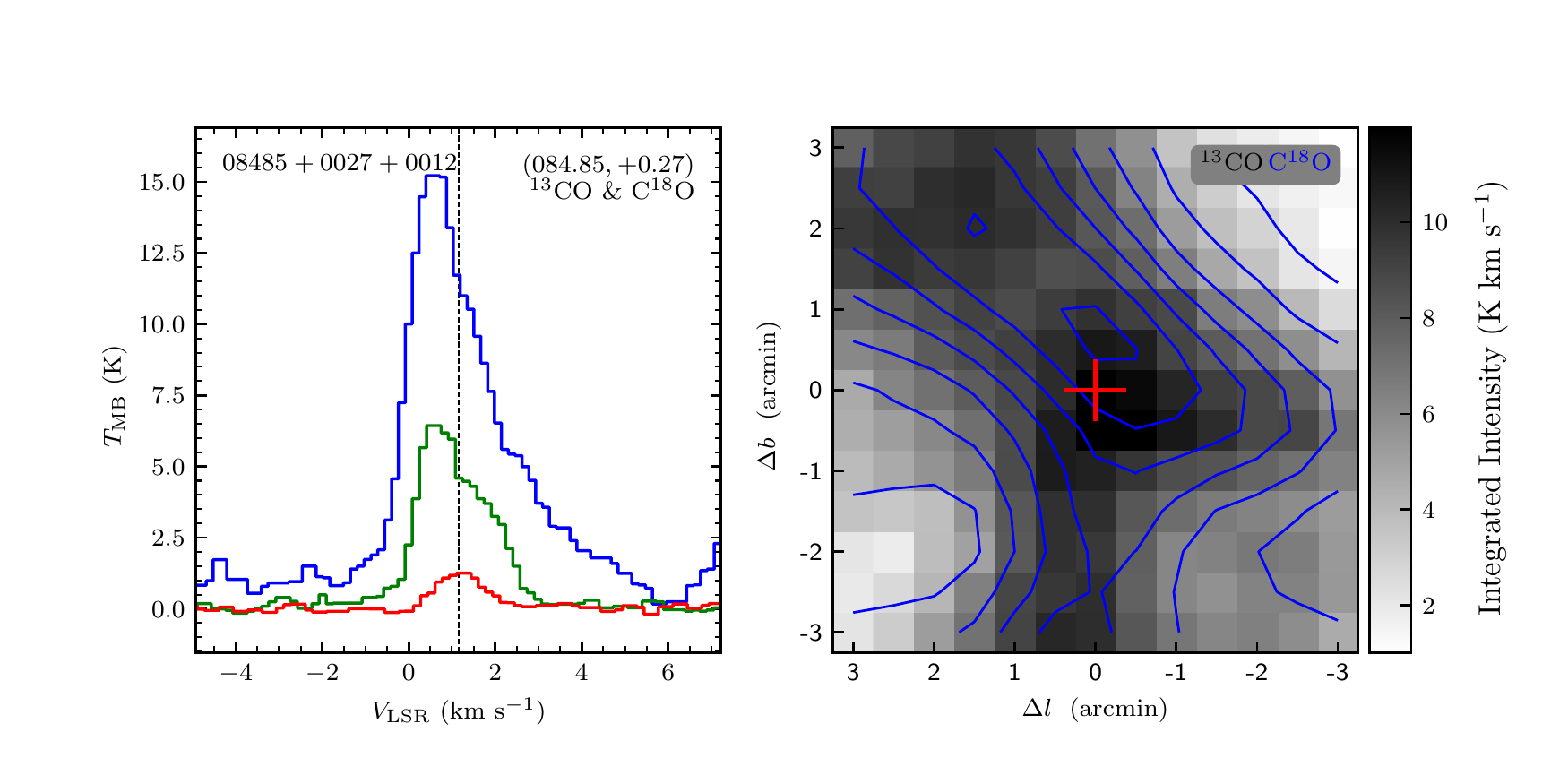}
\vspace{-0.5cm}

\includegraphics[width=9.0cm,angle=0]{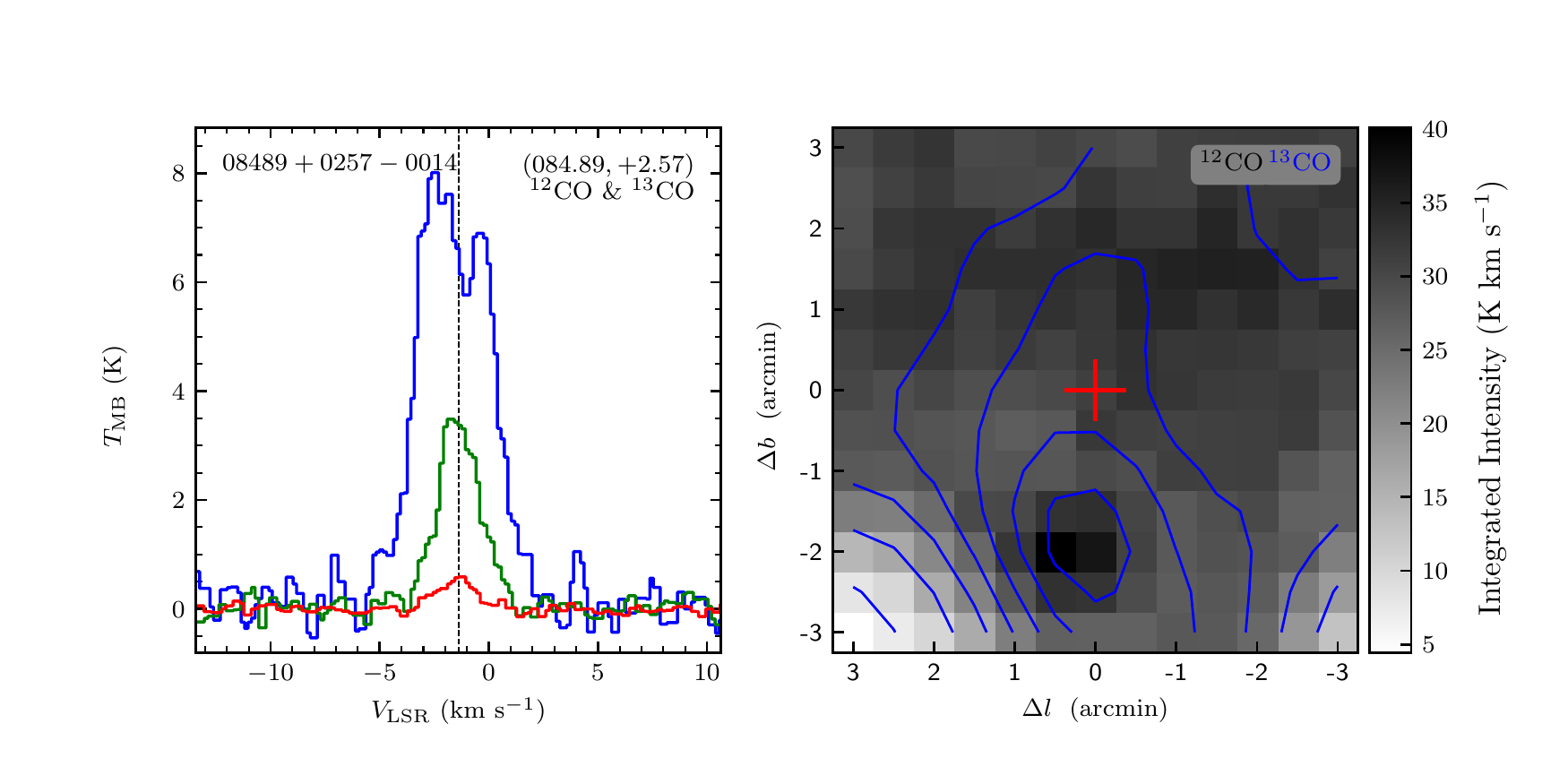}
\includegraphics[width=9.0cm,angle=0]{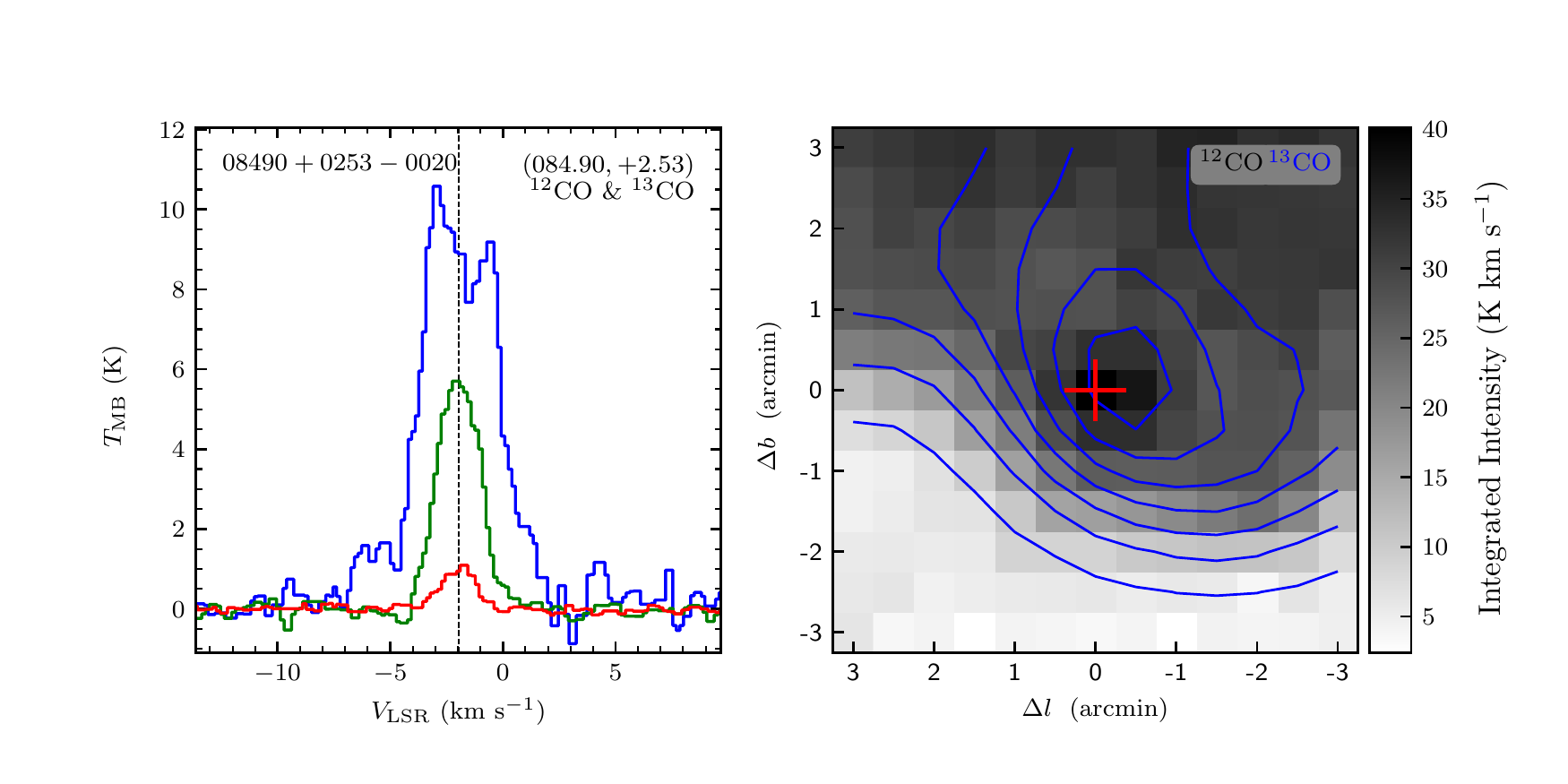}
\vspace{-0.5cm}

\includegraphics[width=9.0cm,angle=0]{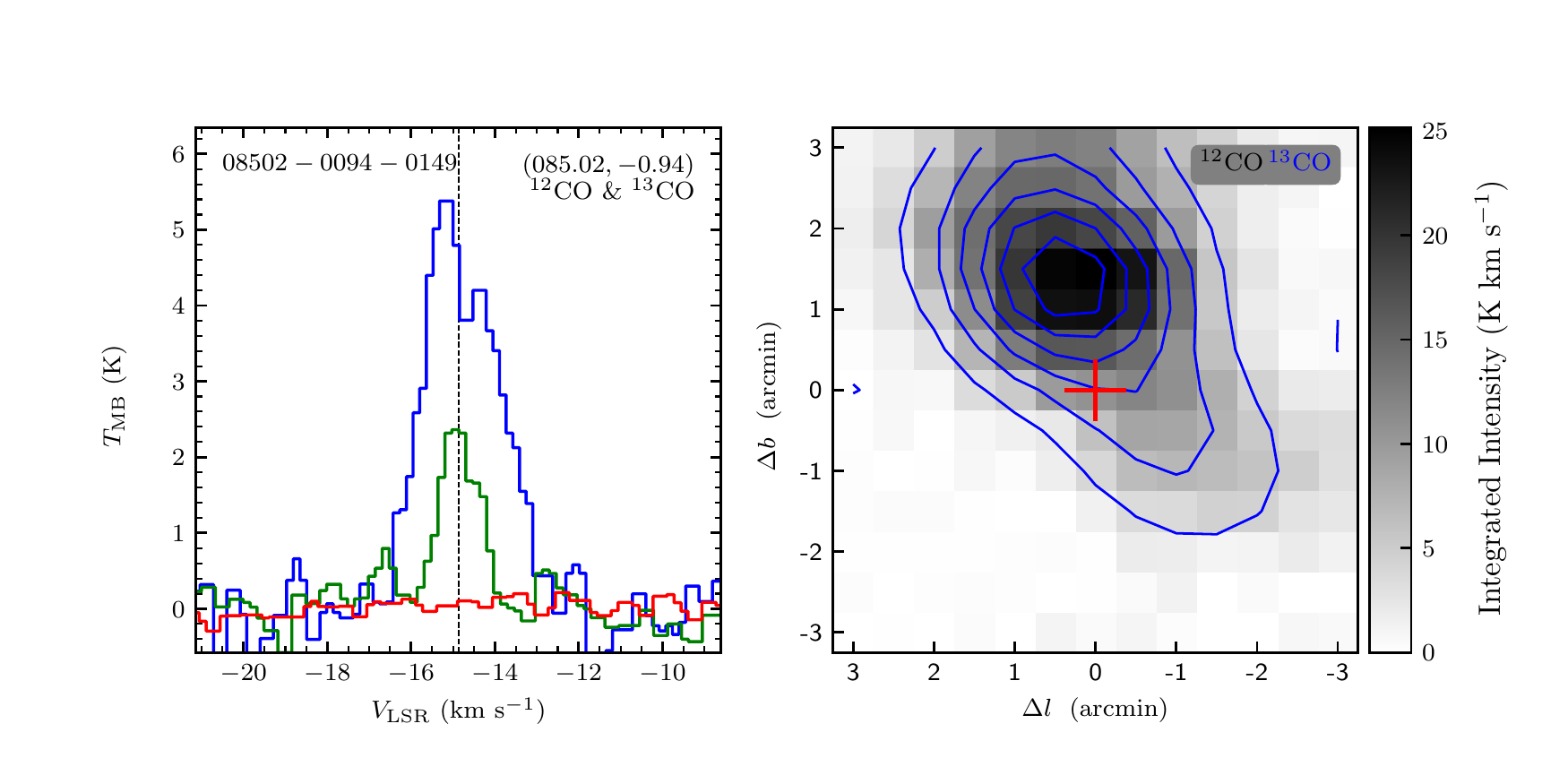}
\includegraphics[width=9.0cm,angle=0]{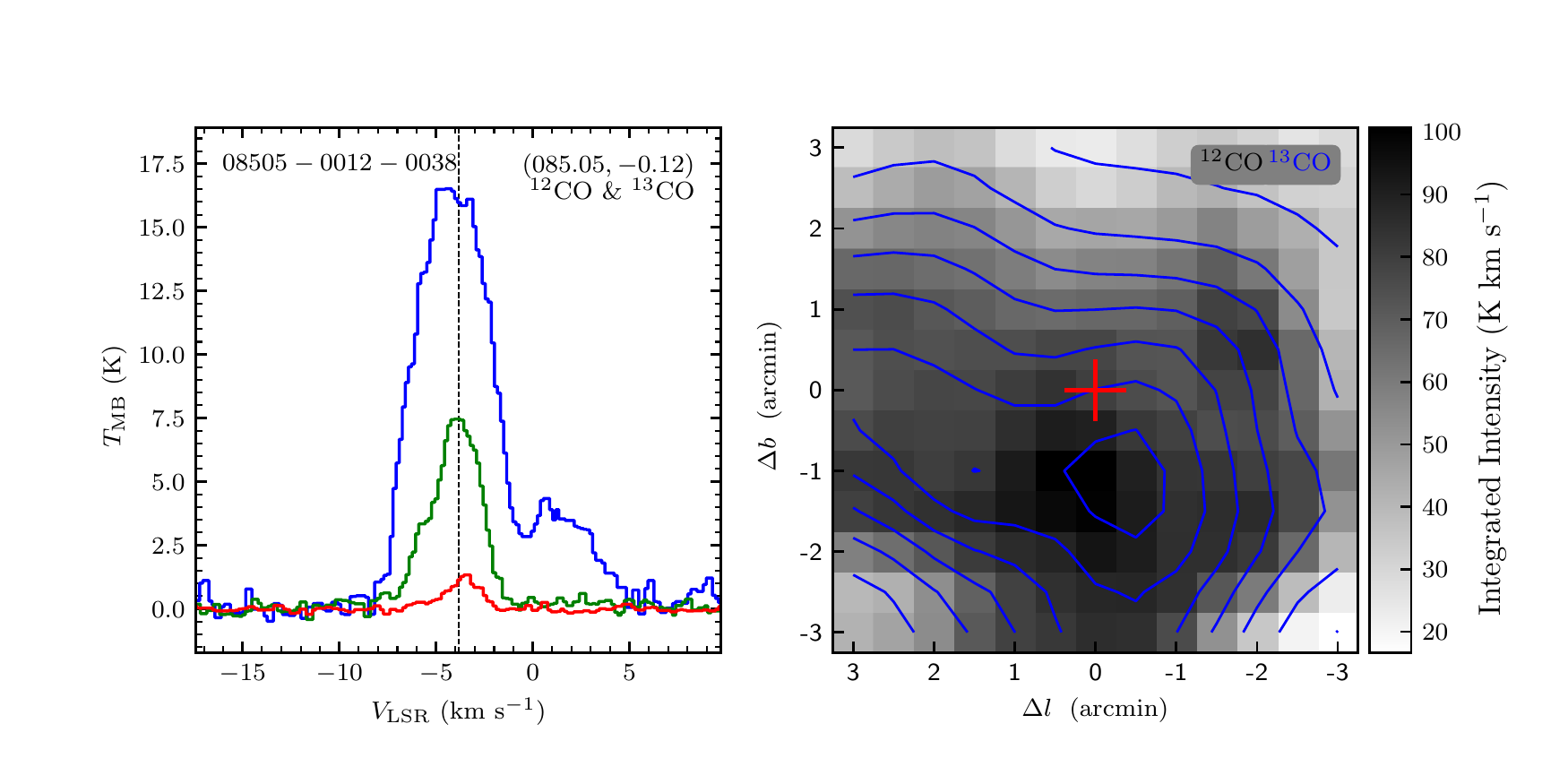}
\vspace{-0.5cm}

\includegraphics[width=9.0cm,angle=0]{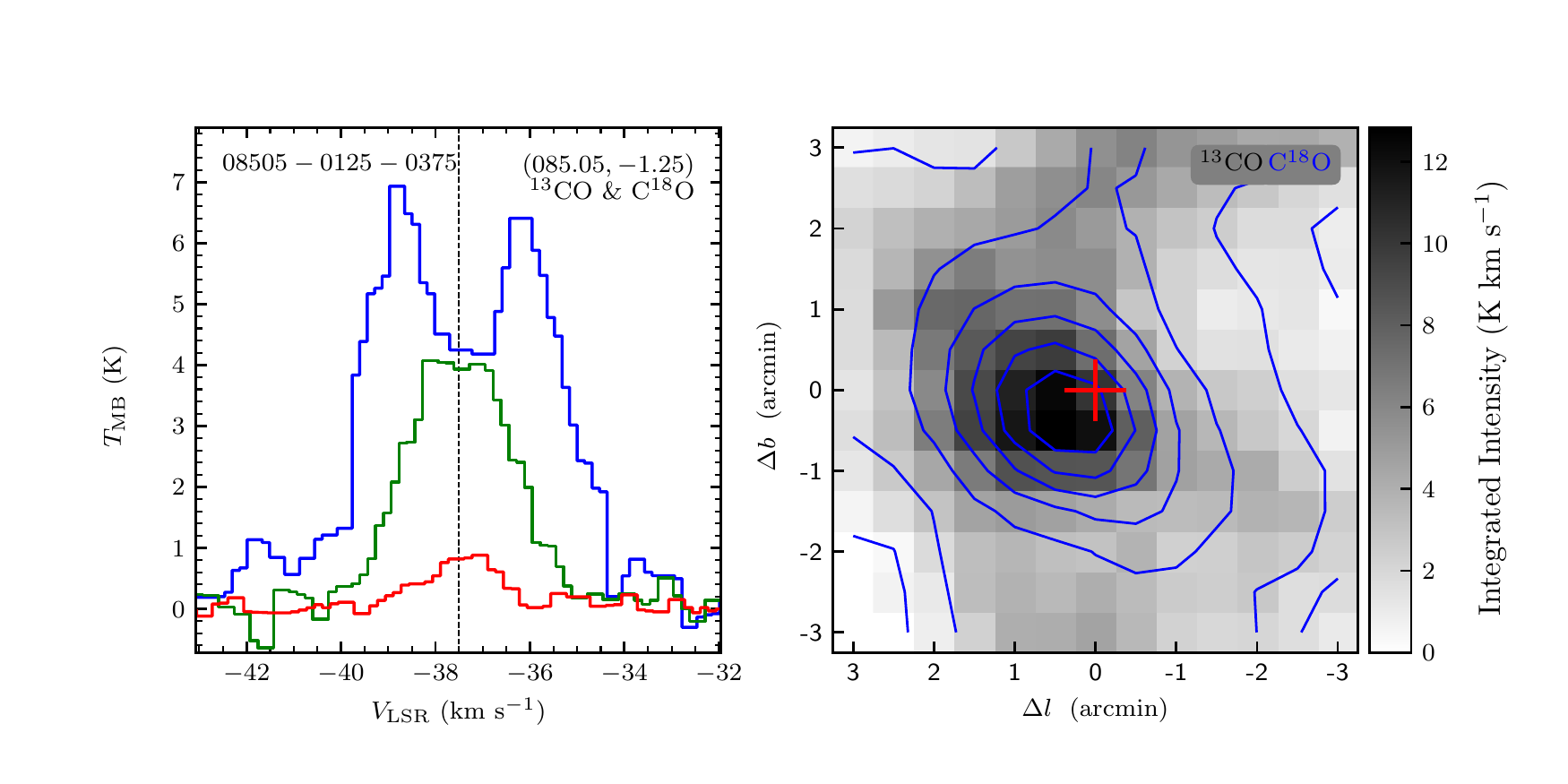}
\includegraphics[width=9.0cm,angle=0]{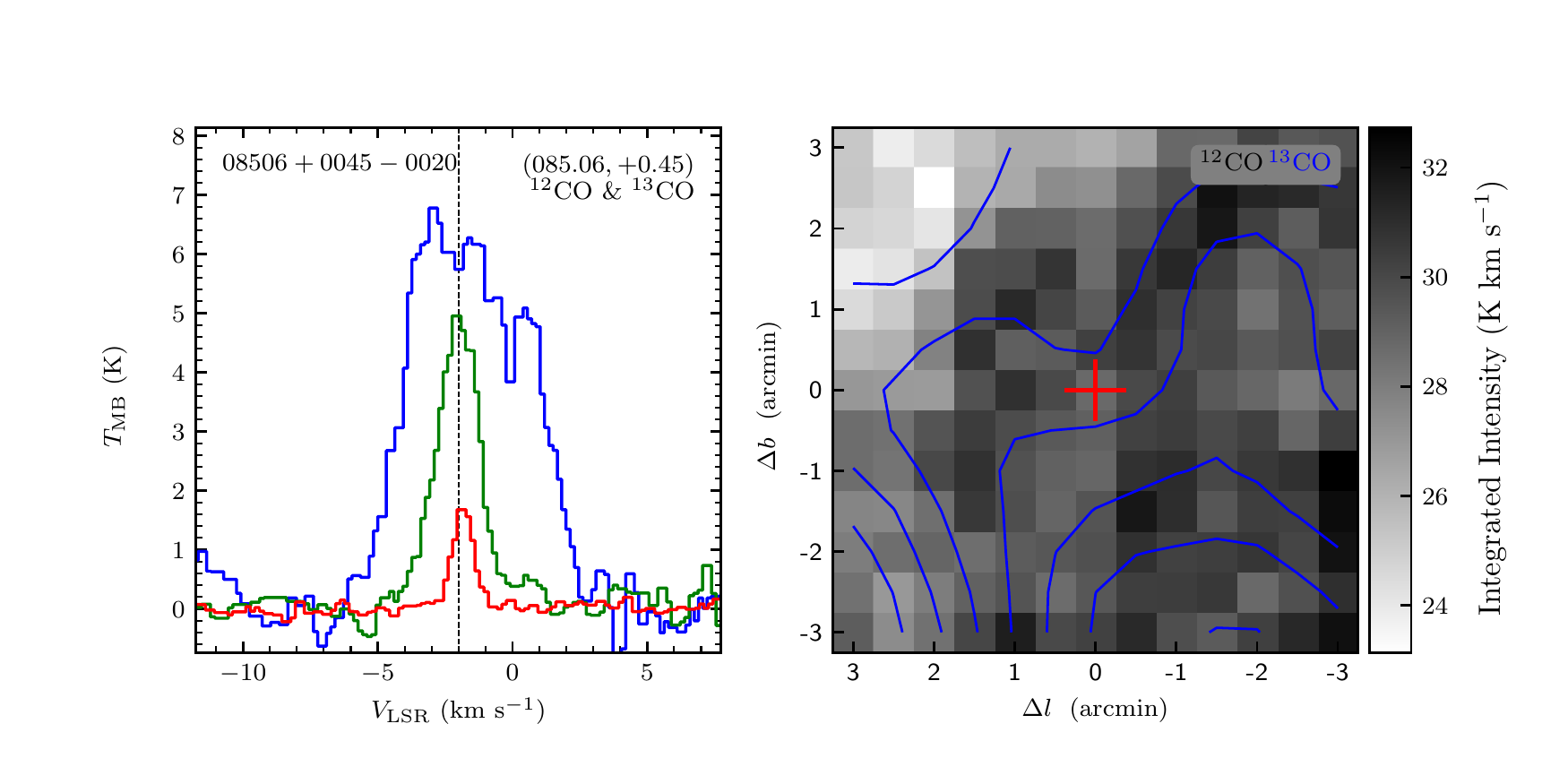}
\end{figure}
\clearpage

\begin{figure}
\includegraphics[width=9.0cm,angle=0]{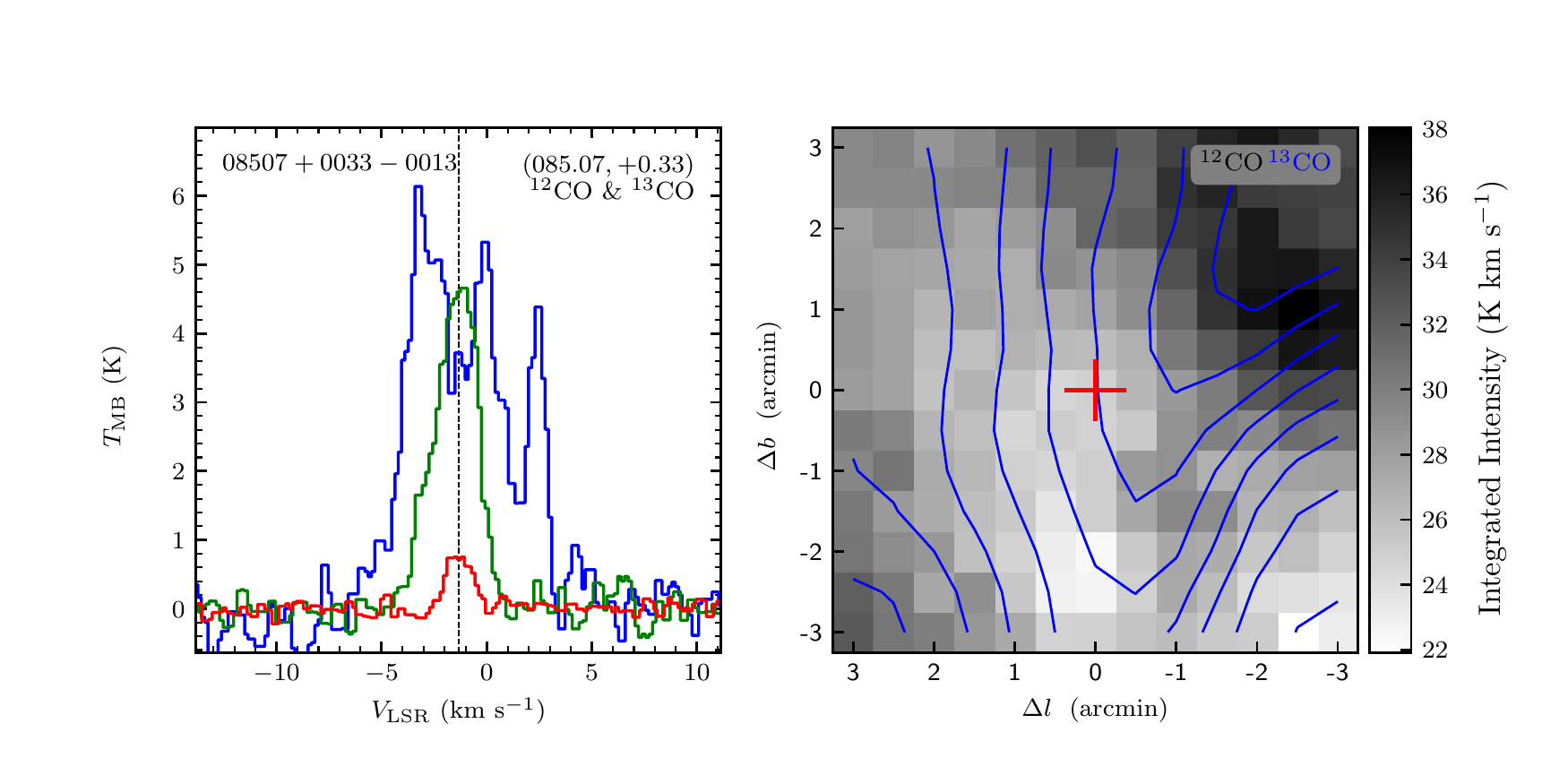}
\includegraphics[width=9.0cm,angle=0]{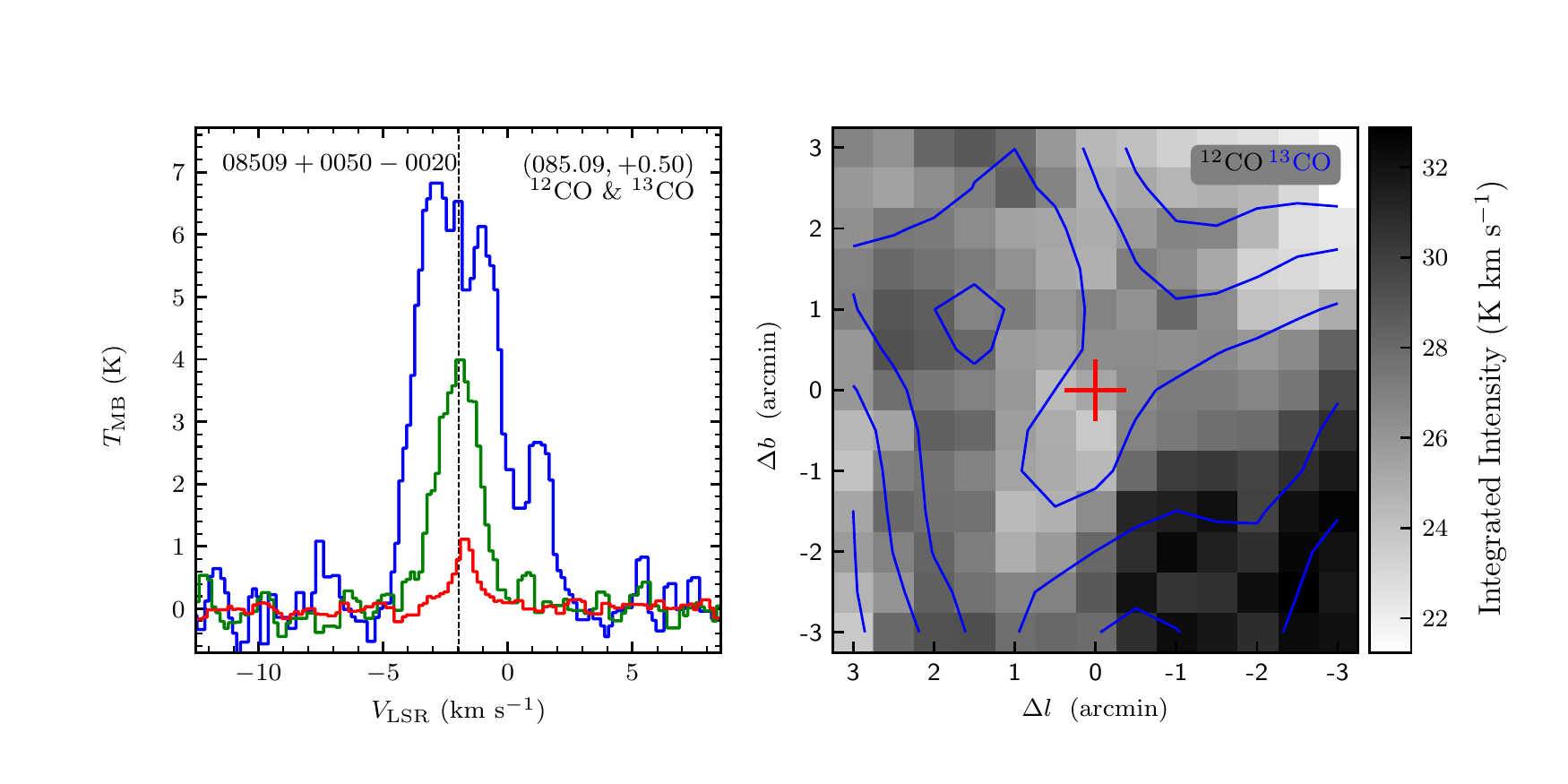}
\vspace{-0.5cm}

\includegraphics[width=9.0cm,angle=0]{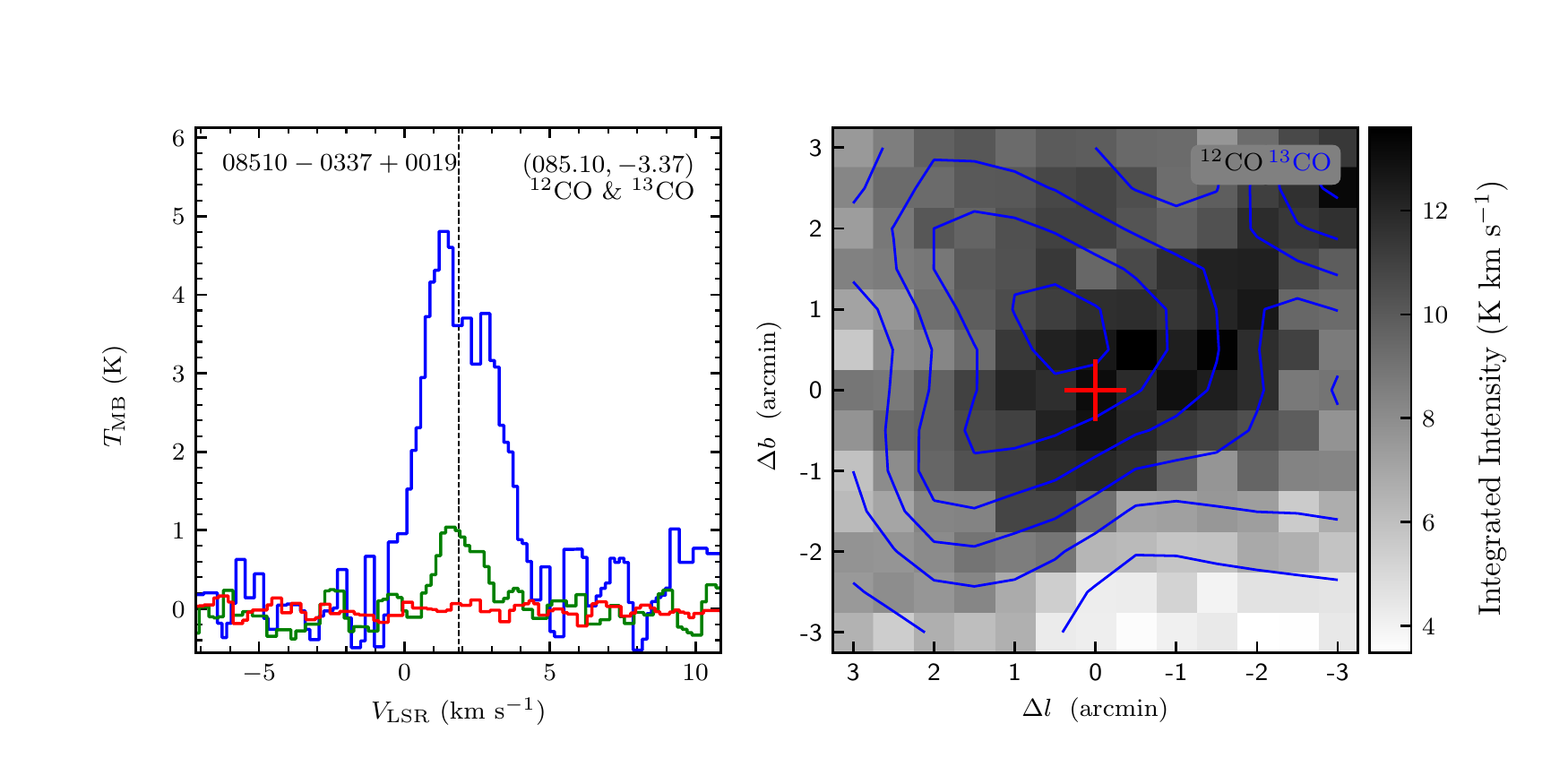}
\includegraphics[width=9.0cm,angle=0]{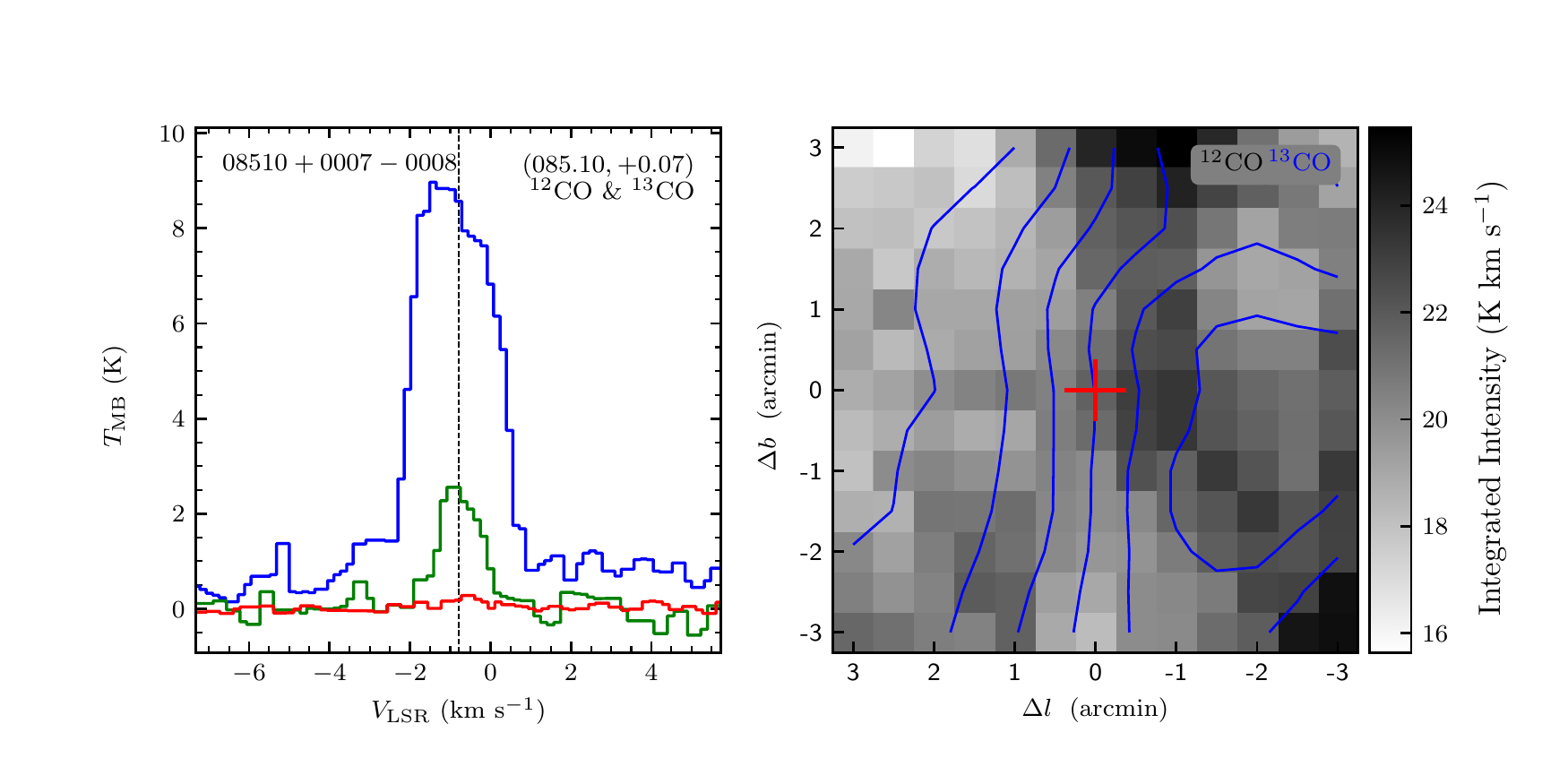}
\vspace{-0.5cm}

\includegraphics[width=9.0cm,angle=0]{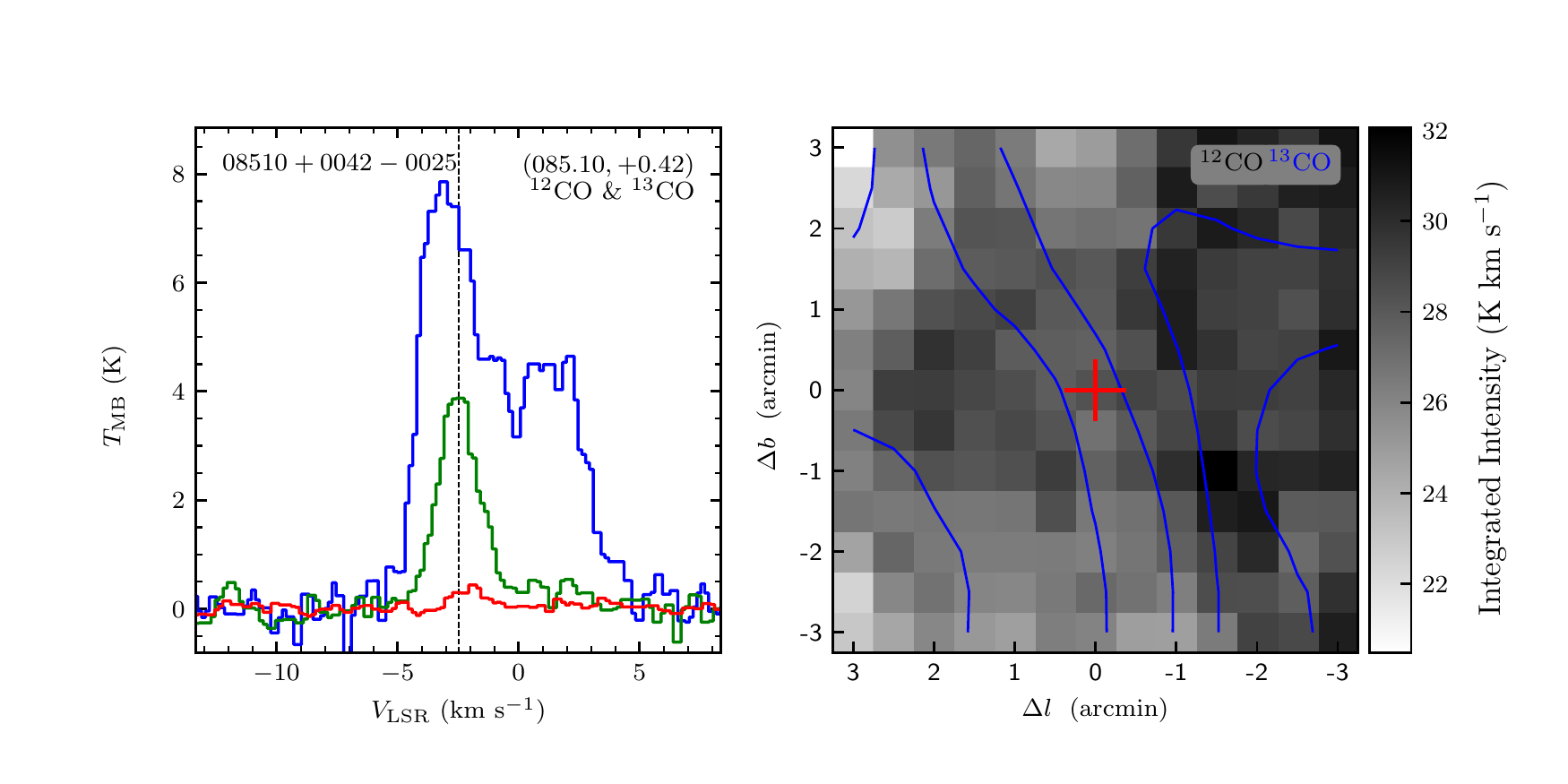}
\includegraphics[width=9.0cm,angle=0]{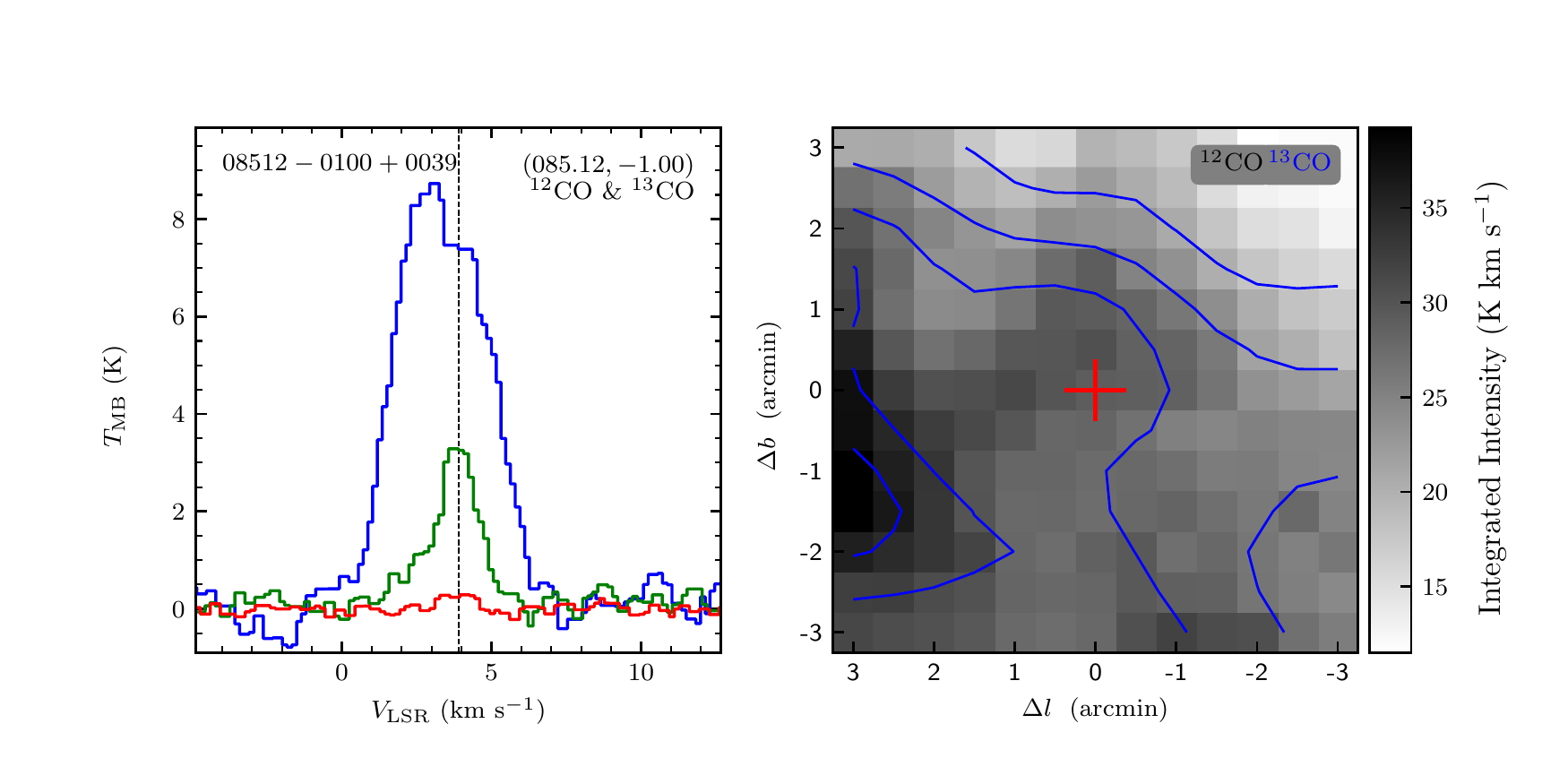}
\vspace{-0.5cm}

\includegraphics[width=9.0cm,angle=0]{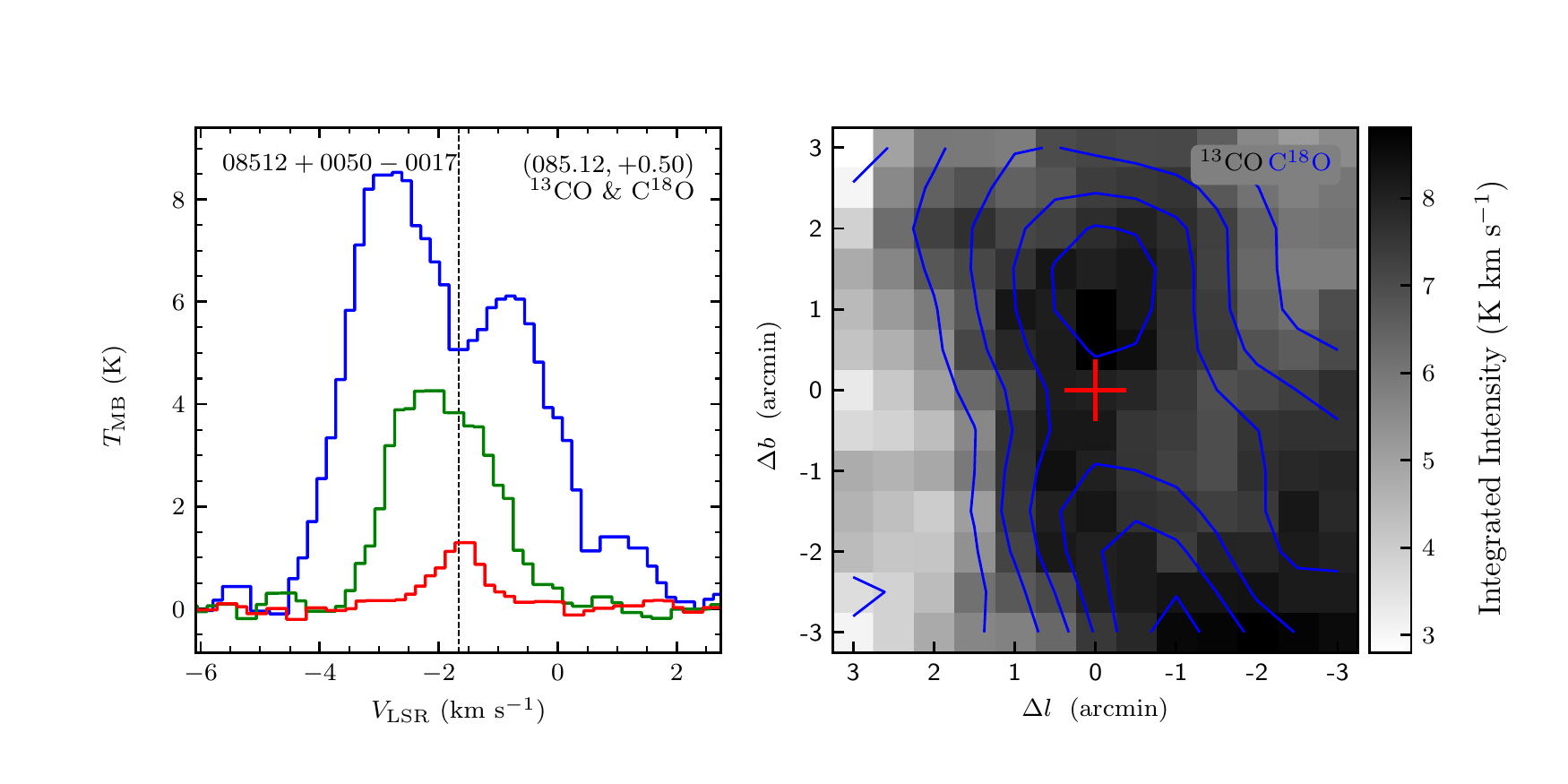}
\includegraphics[width=9.0cm,angle=0]{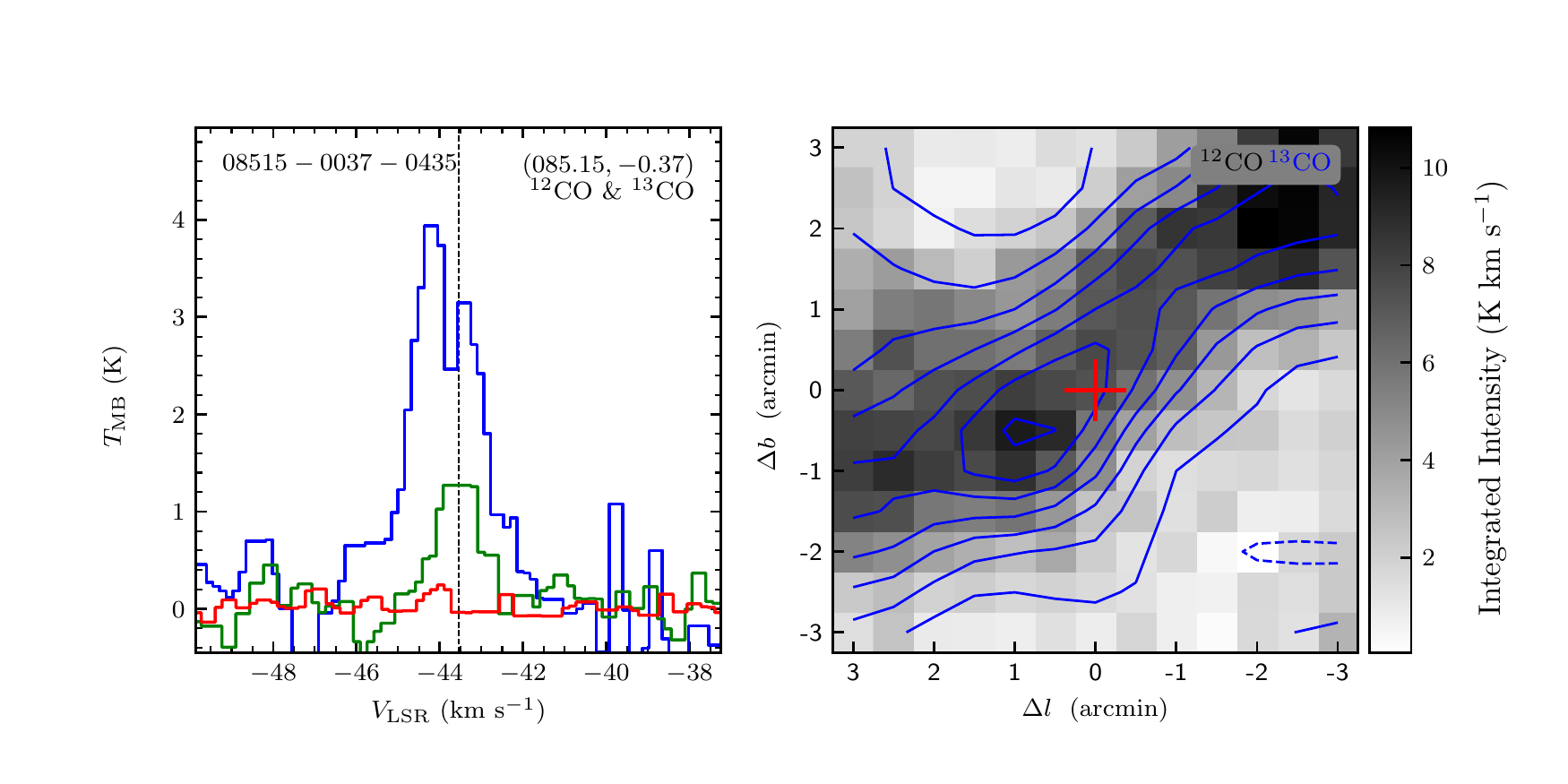}
\vspace{-0.5cm}

\includegraphics[width=9.0cm,angle=0]{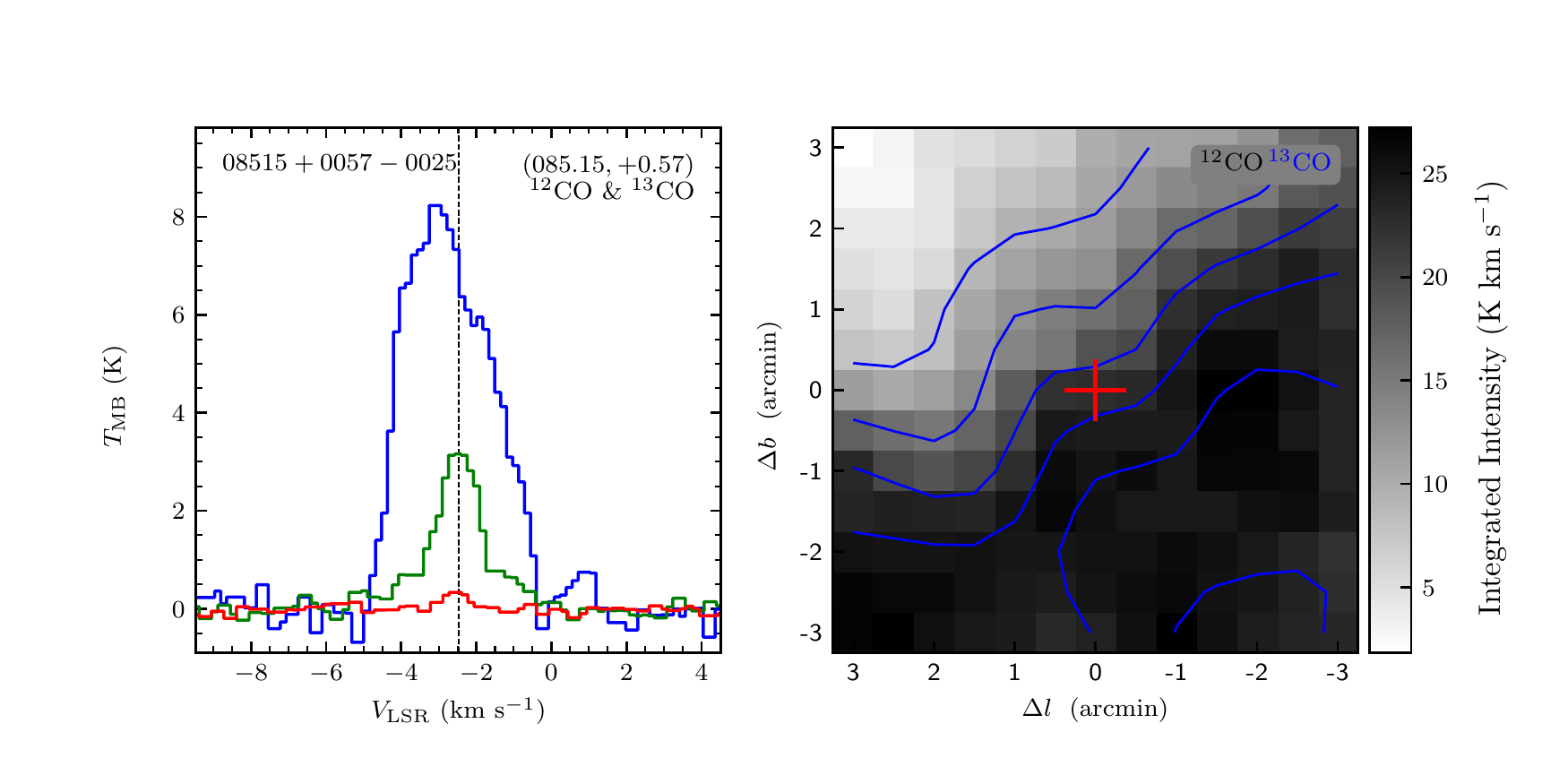}
\includegraphics[width=9.0cm,angle=0]{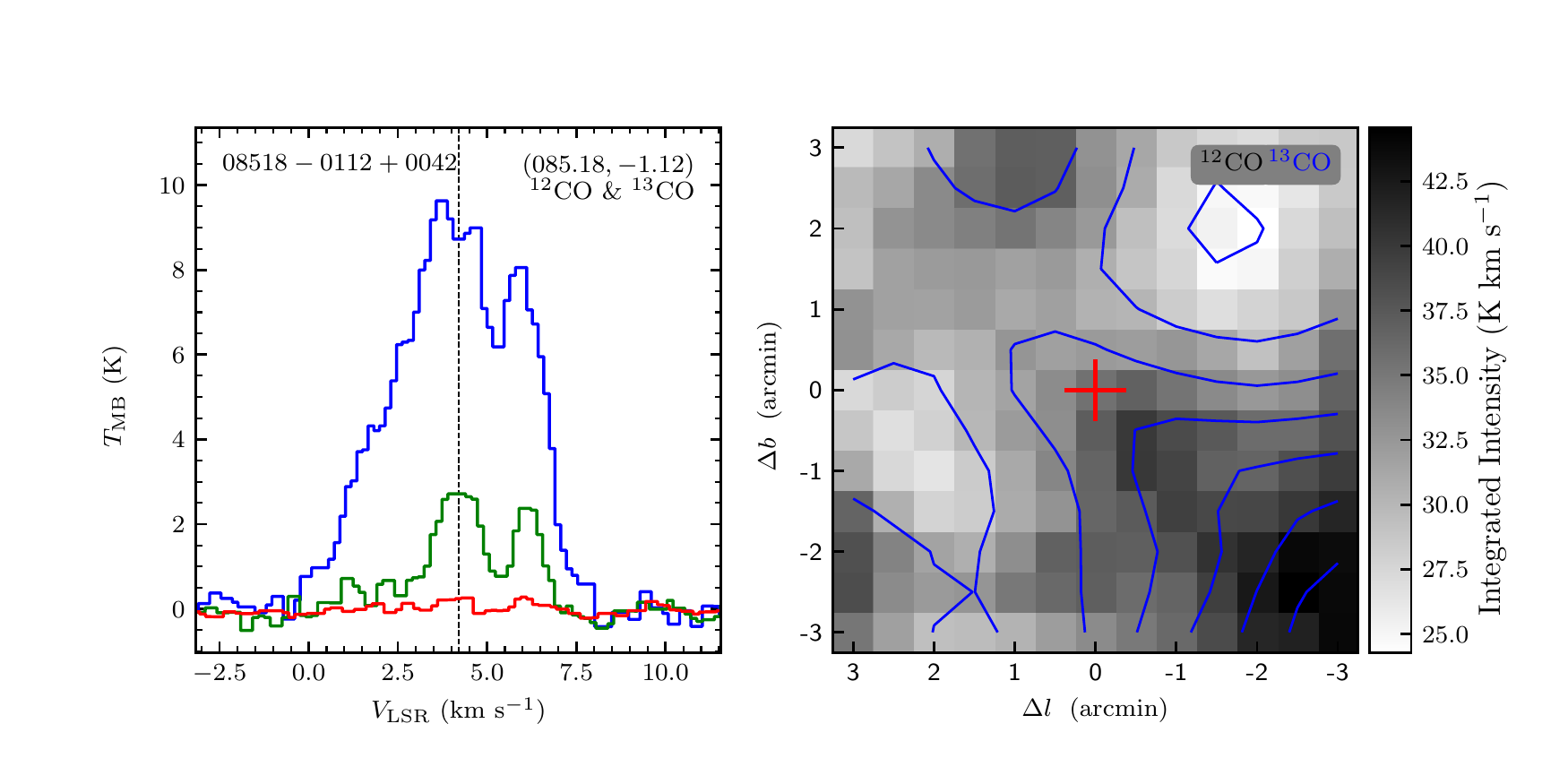}
\end{figure}
\clearpage

\begin{figure}
\includegraphics[width=9.0cm,angle=0]{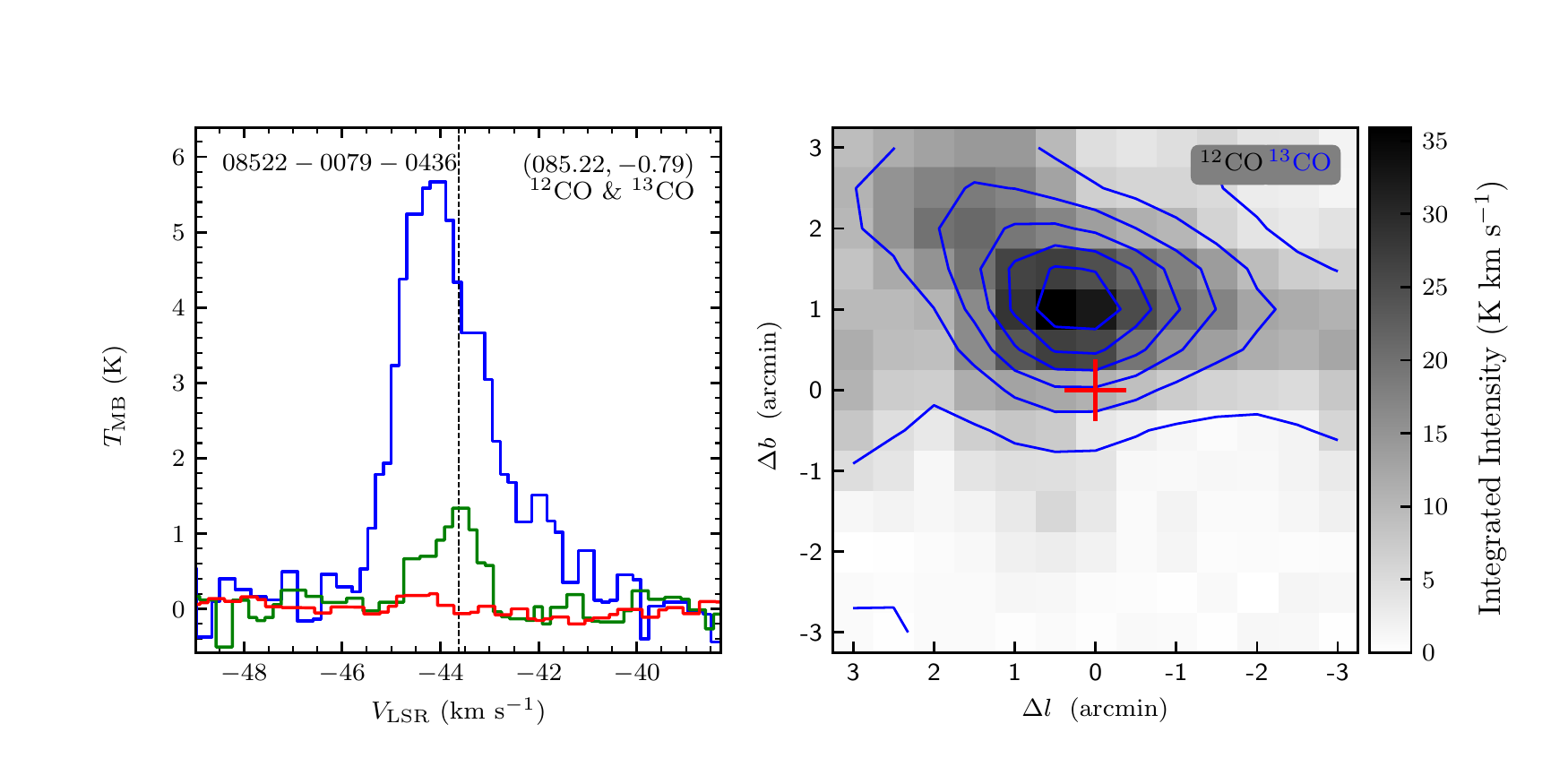}
\includegraphics[width=9.0cm,angle=0]{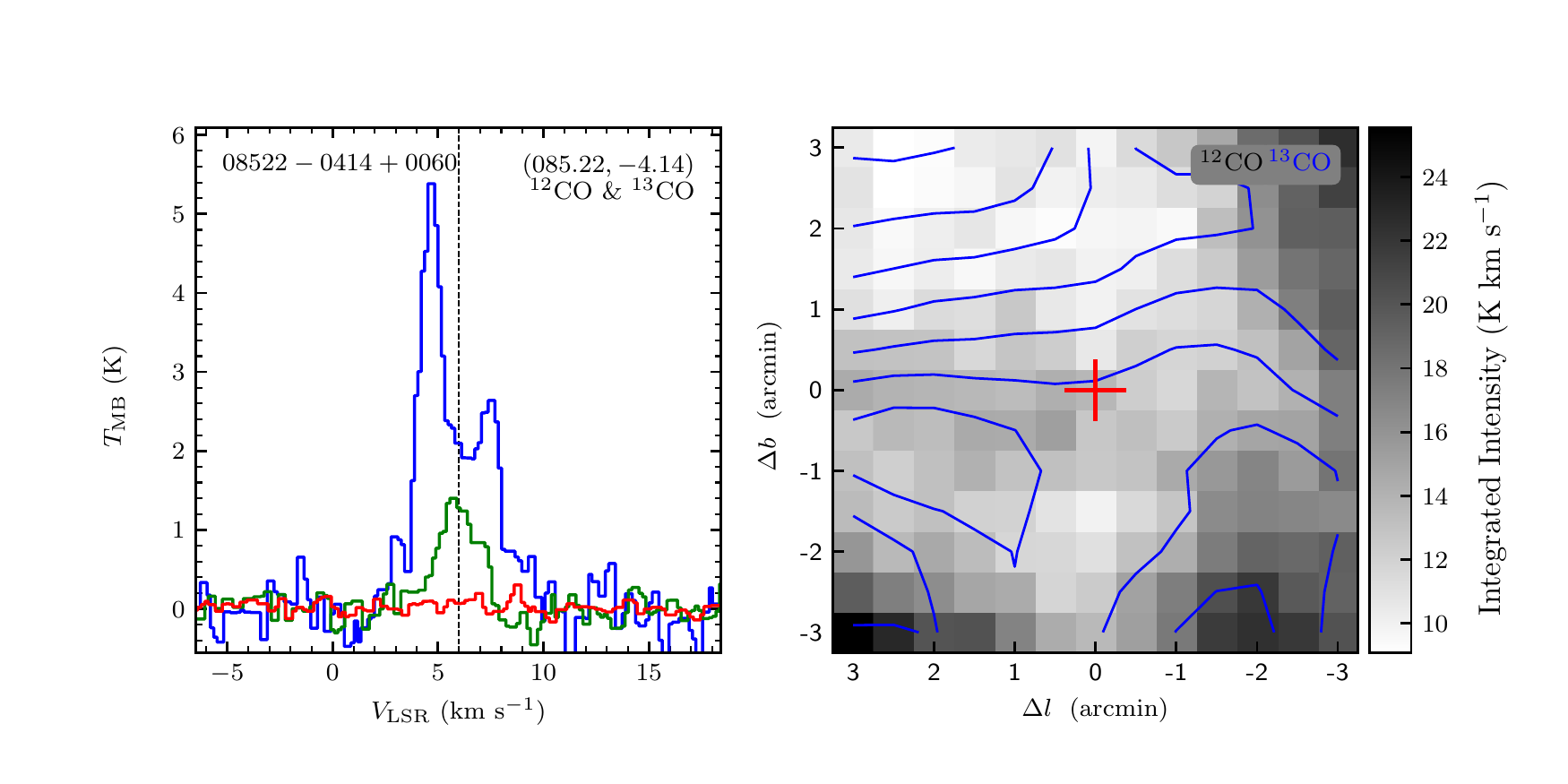}
\vspace{-0.5cm}

\includegraphics[width=9.0cm,angle=0]{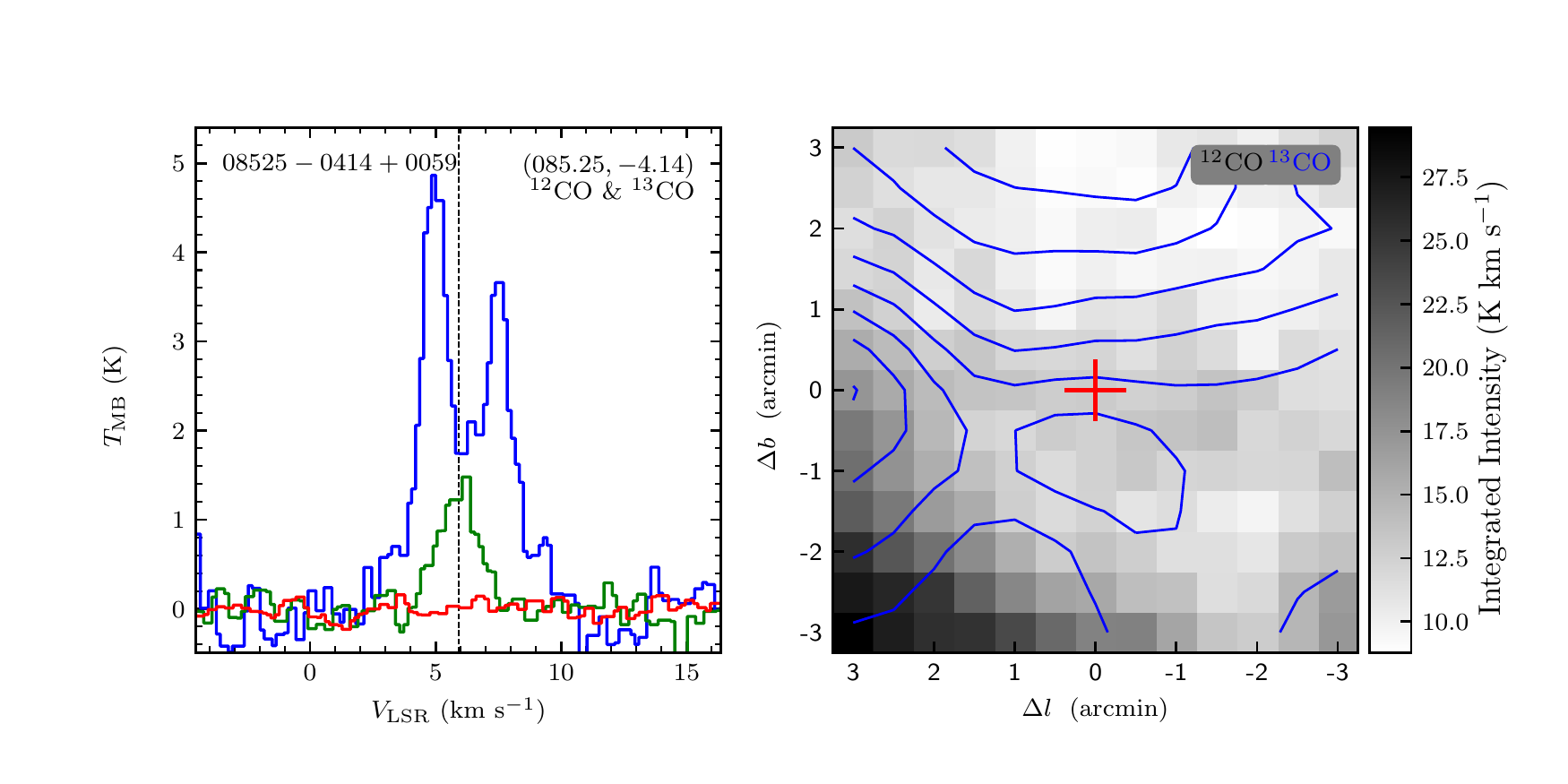}
\includegraphics[width=9.0cm,angle=0]{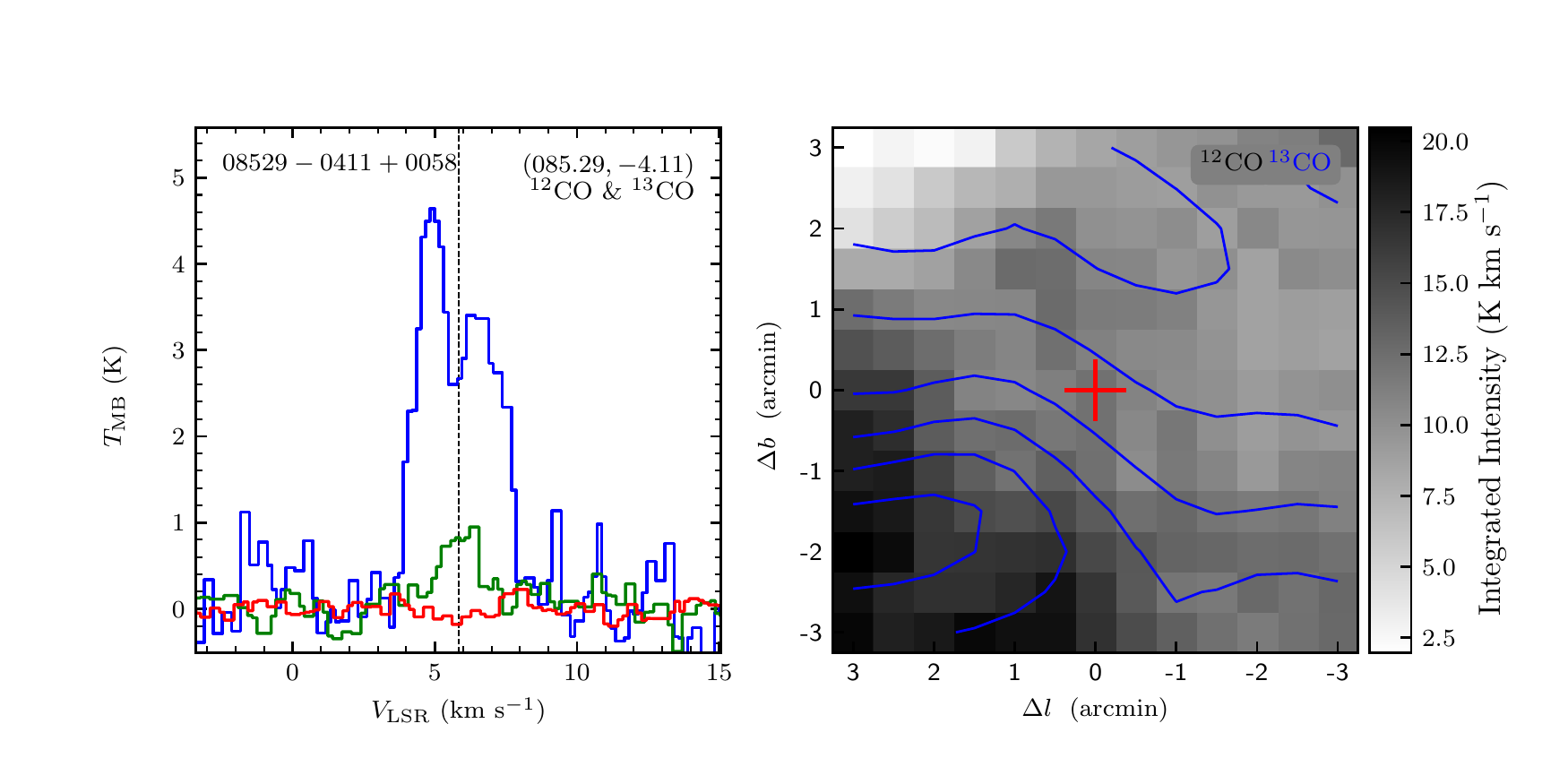}
\vspace{-0.5cm}

\includegraphics[width=9.0cm,angle=0]{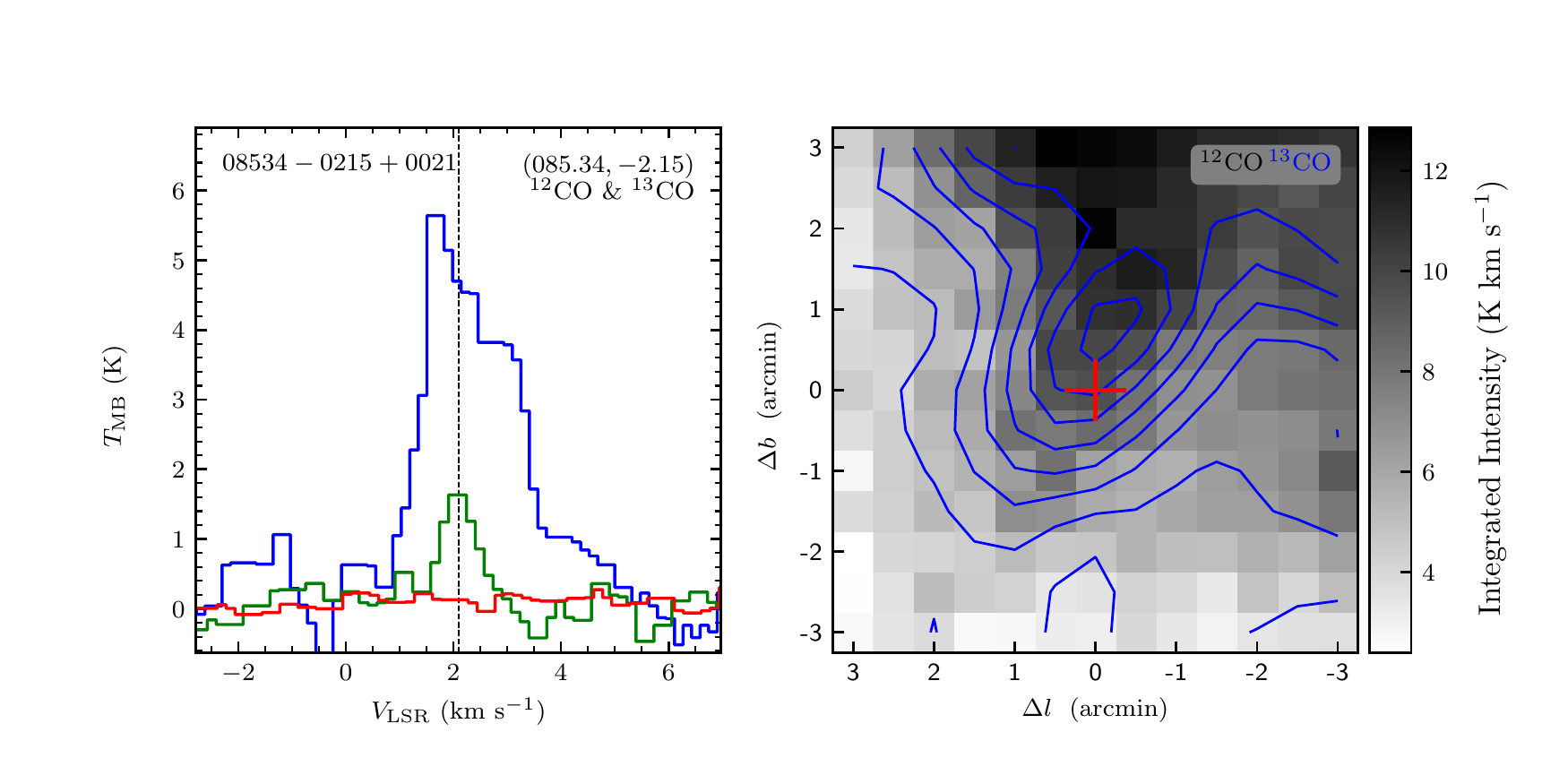}
\includegraphics[width=9.0cm,angle=0]{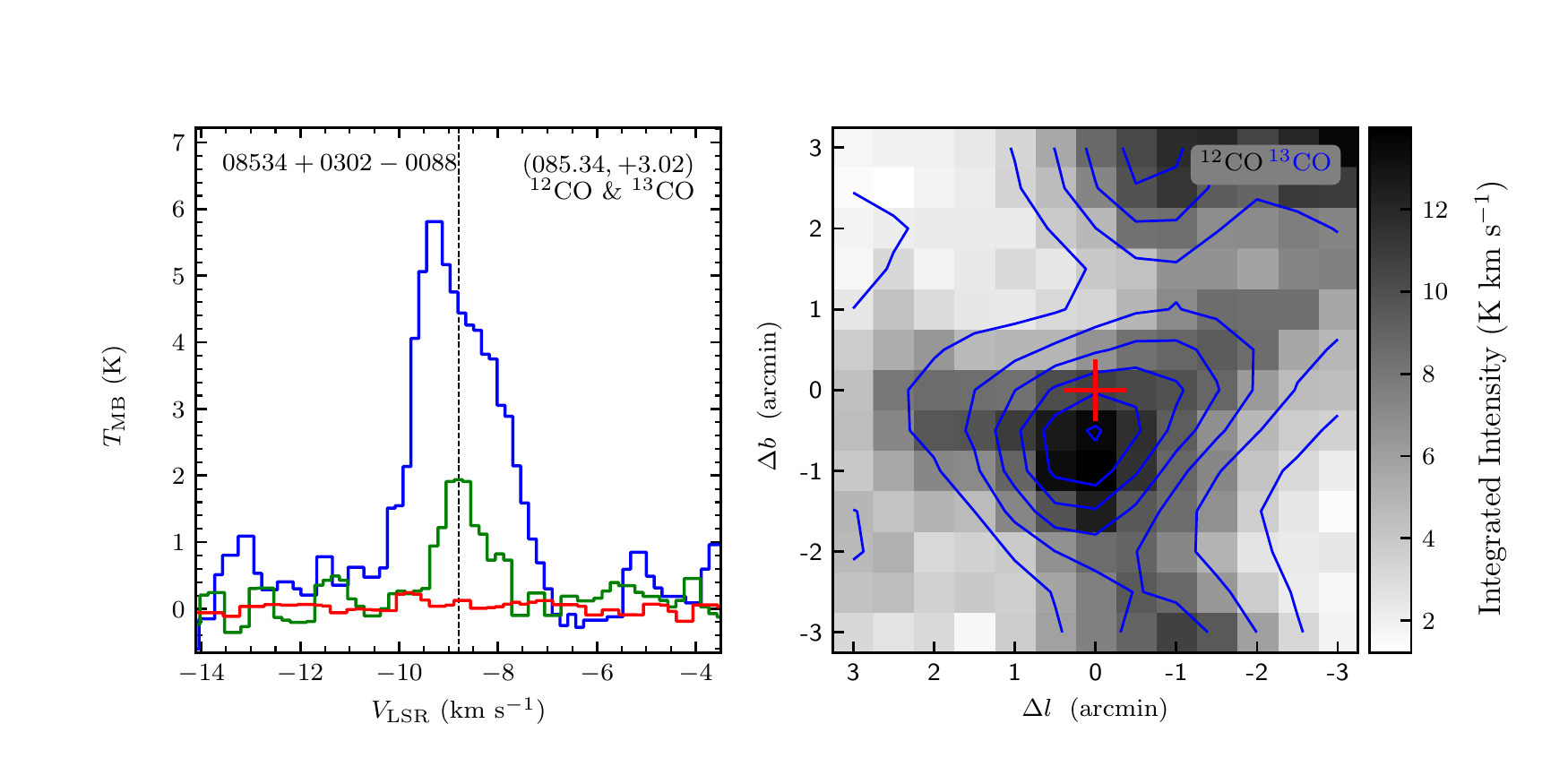}
\vspace{-0.5cm}

\includegraphics[width=9.0cm,angle=0]{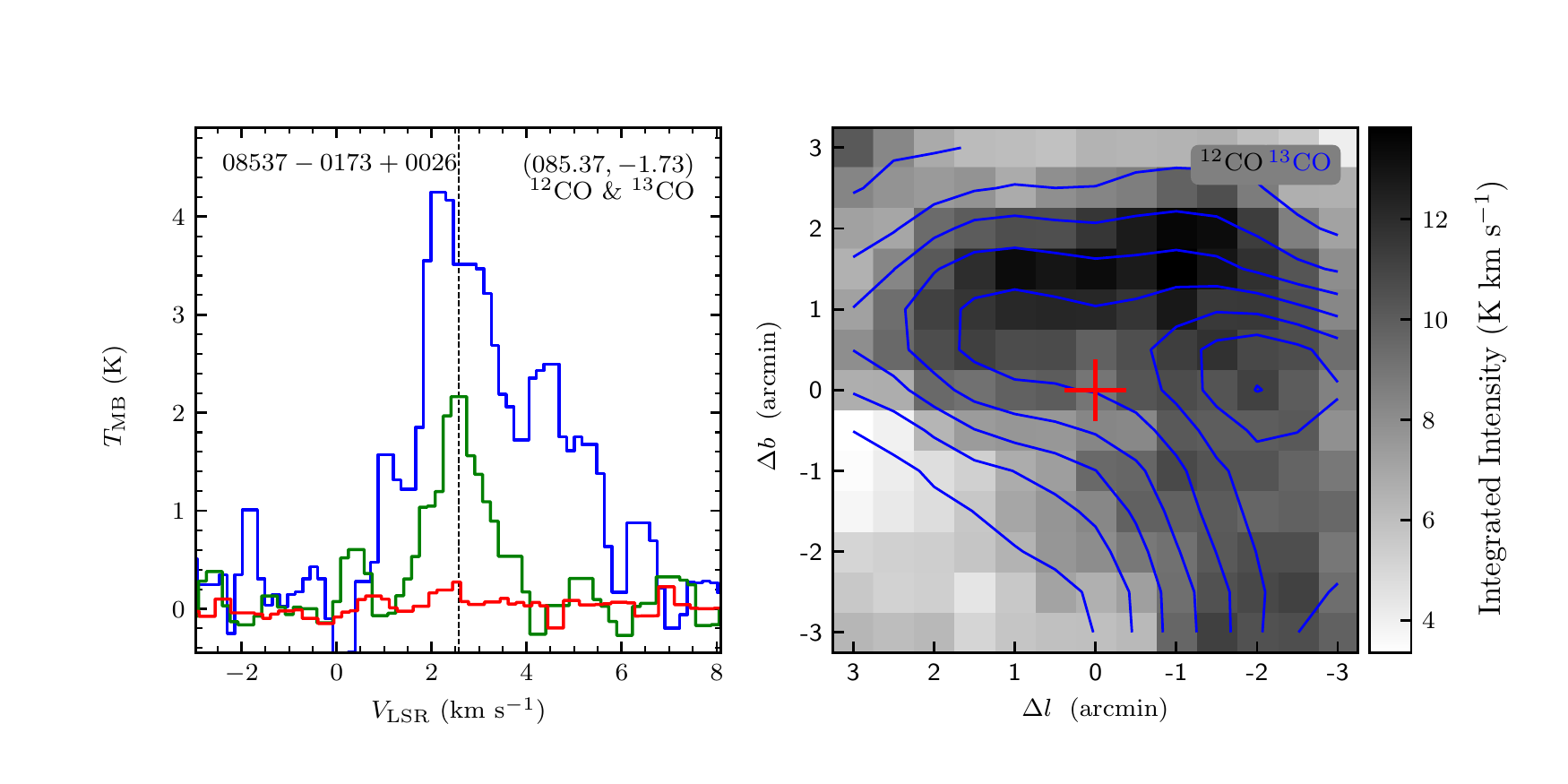}
\includegraphics[width=9.0cm,angle=0]{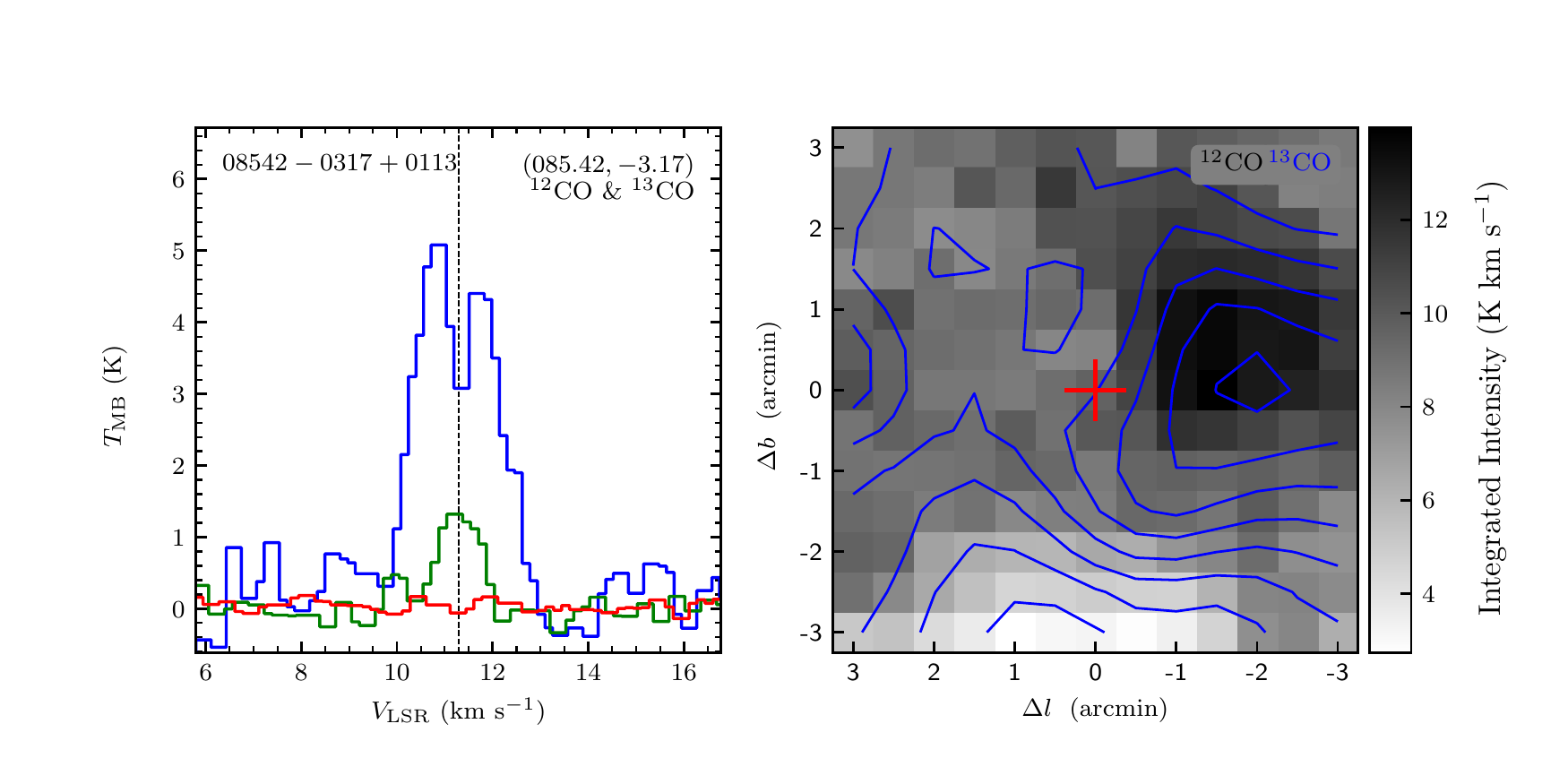}
\vspace{-0.5cm}

\includegraphics[width=9.0cm,angle=0]{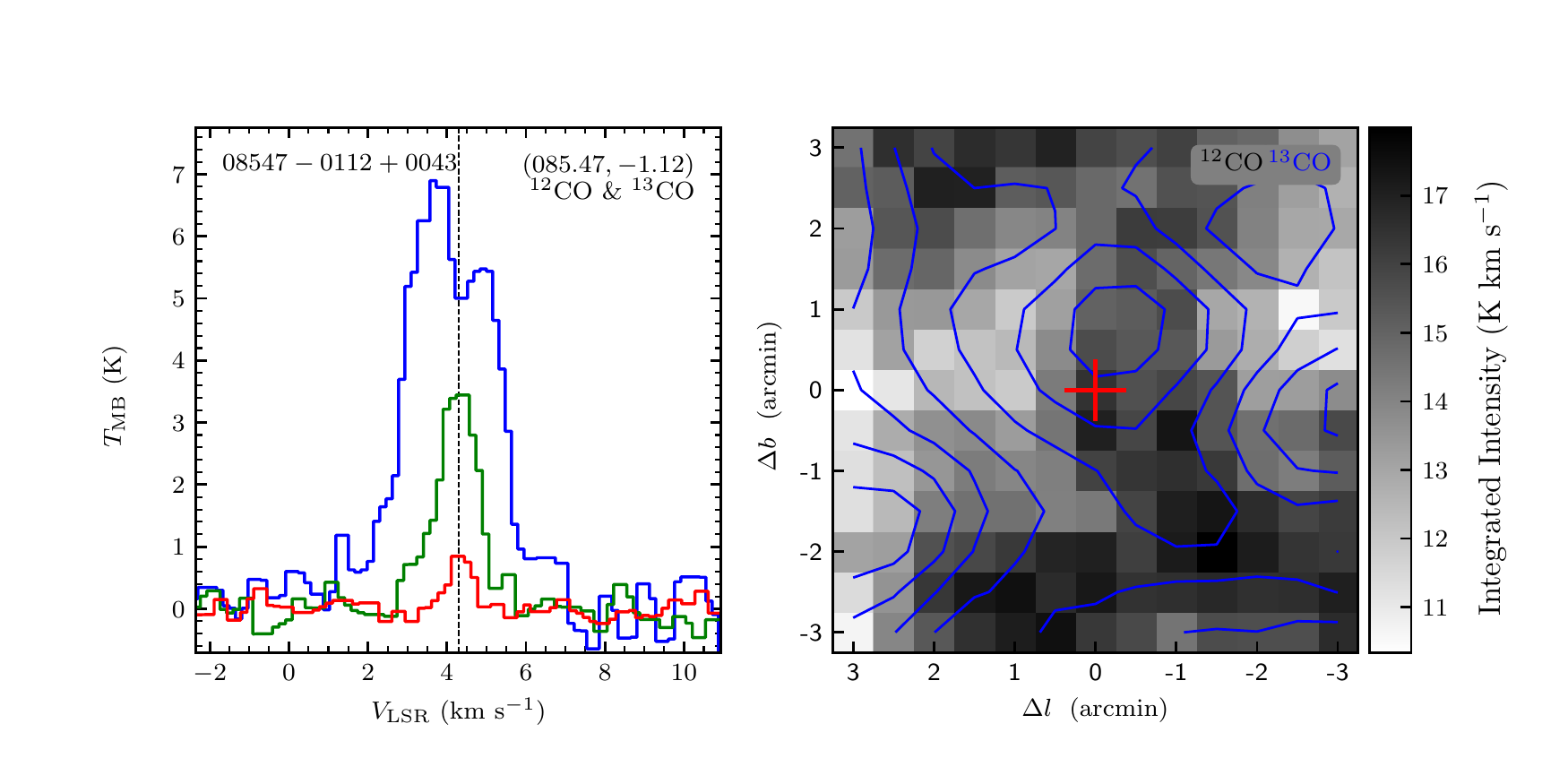}
\includegraphics[width=9.0cm,angle=0]{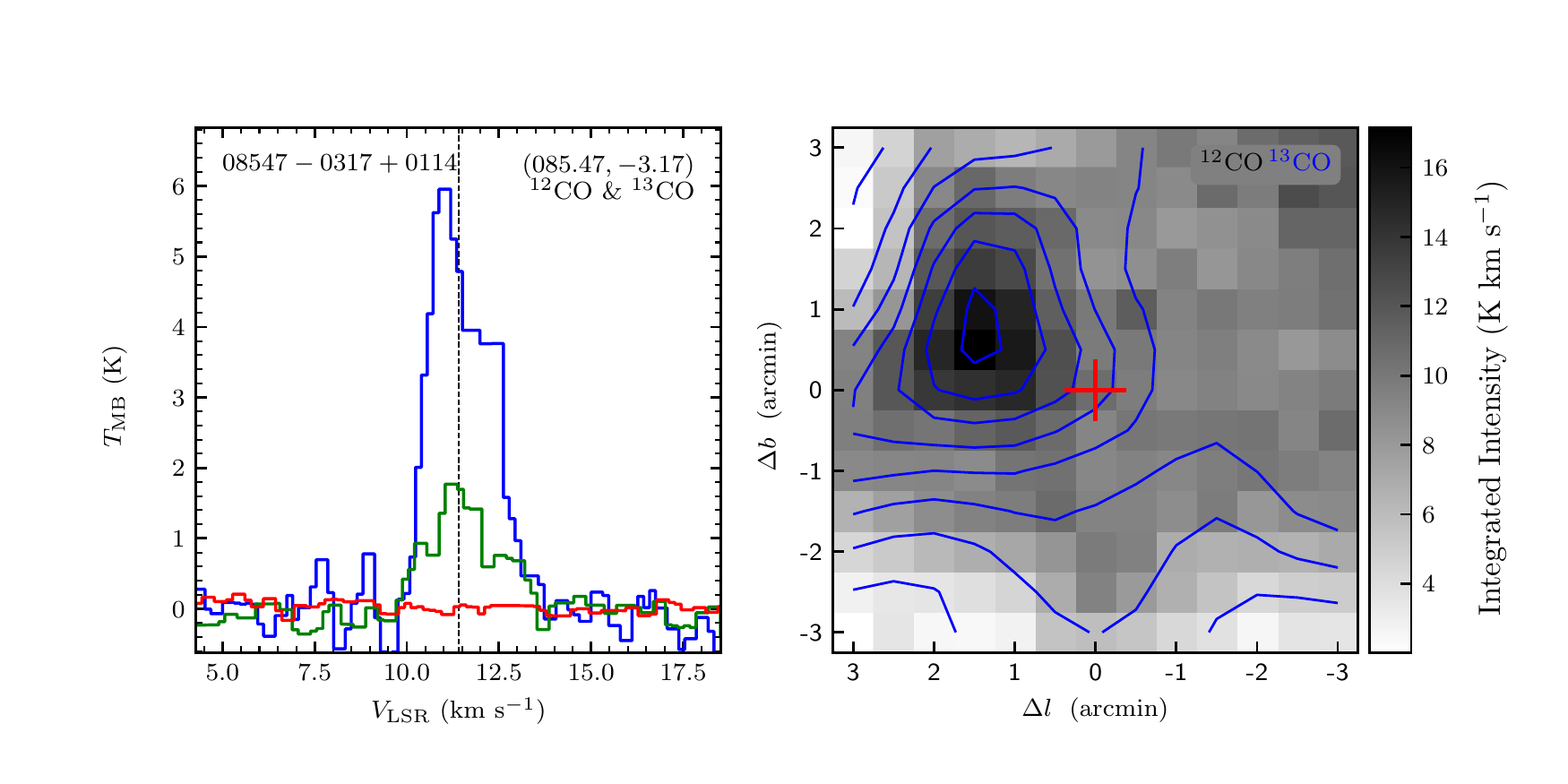}
\end{figure}
\clearpage

\begin{figure}
\includegraphics[width=9.0cm,angle=0]{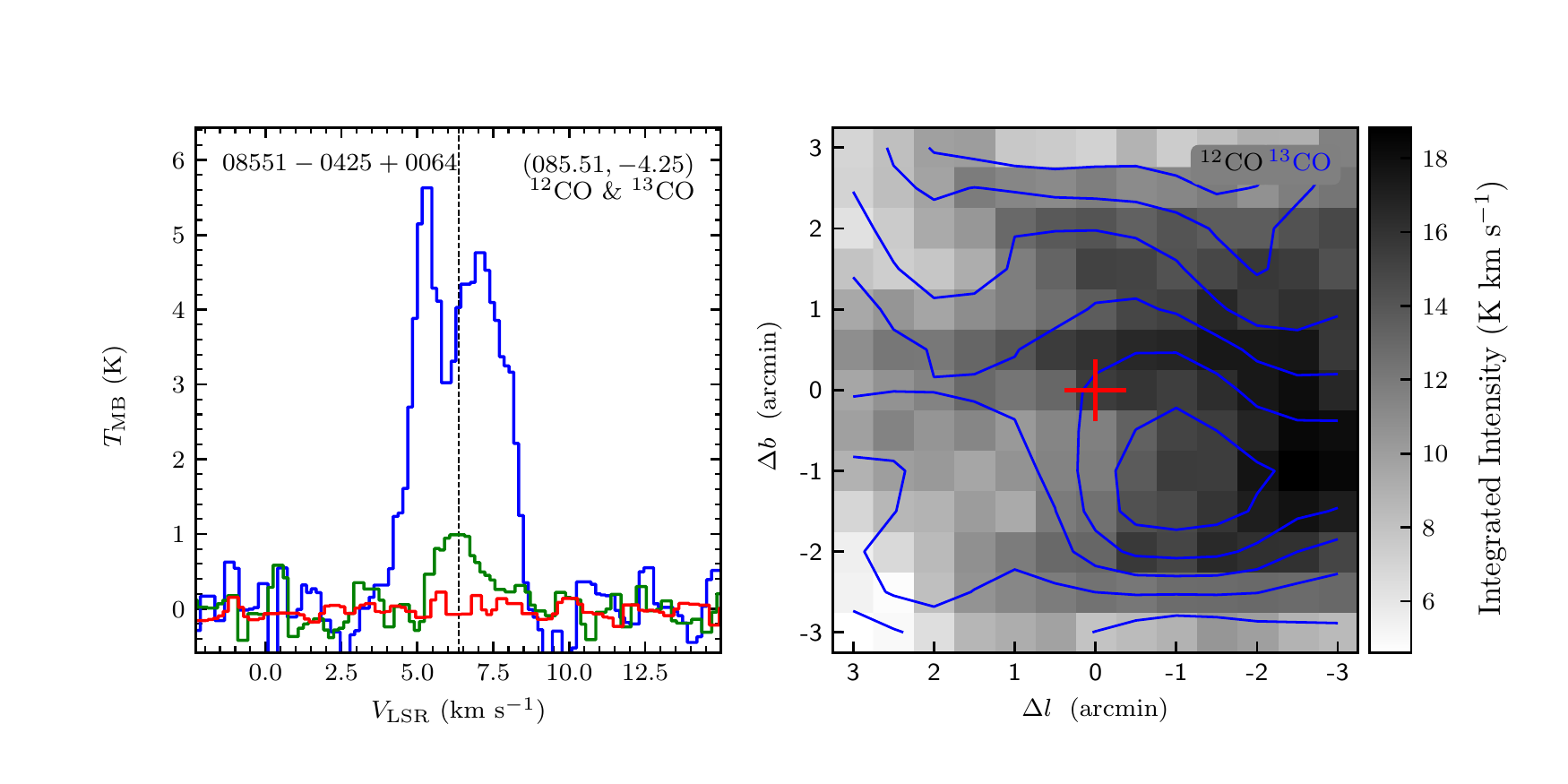}
\includegraphics[width=9.0cm,angle=0]{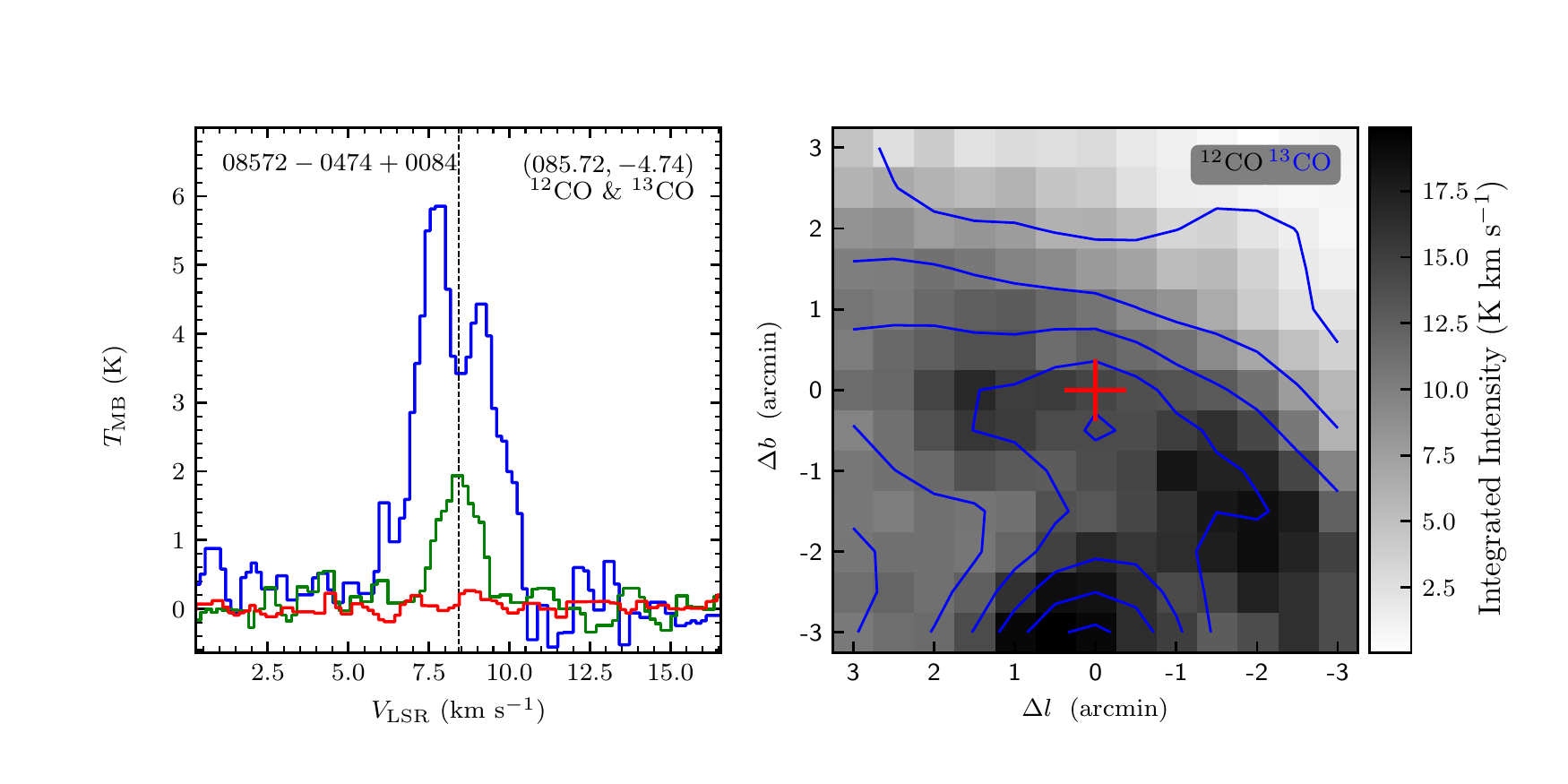}
\vspace{-0.5cm}

\includegraphics[width=9.0cm,angle=0]{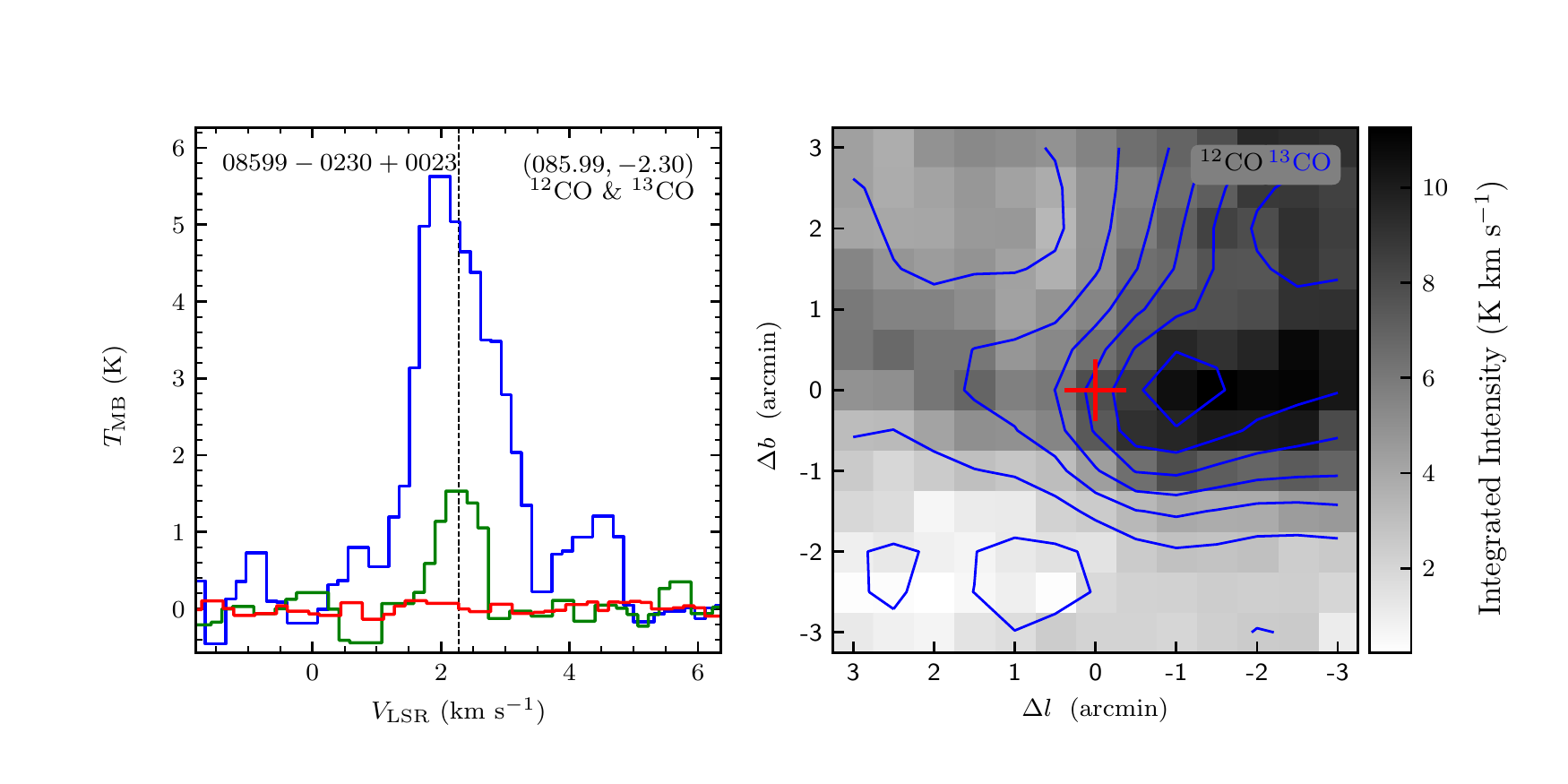}
\includegraphics[width=9.0cm,angle=0]{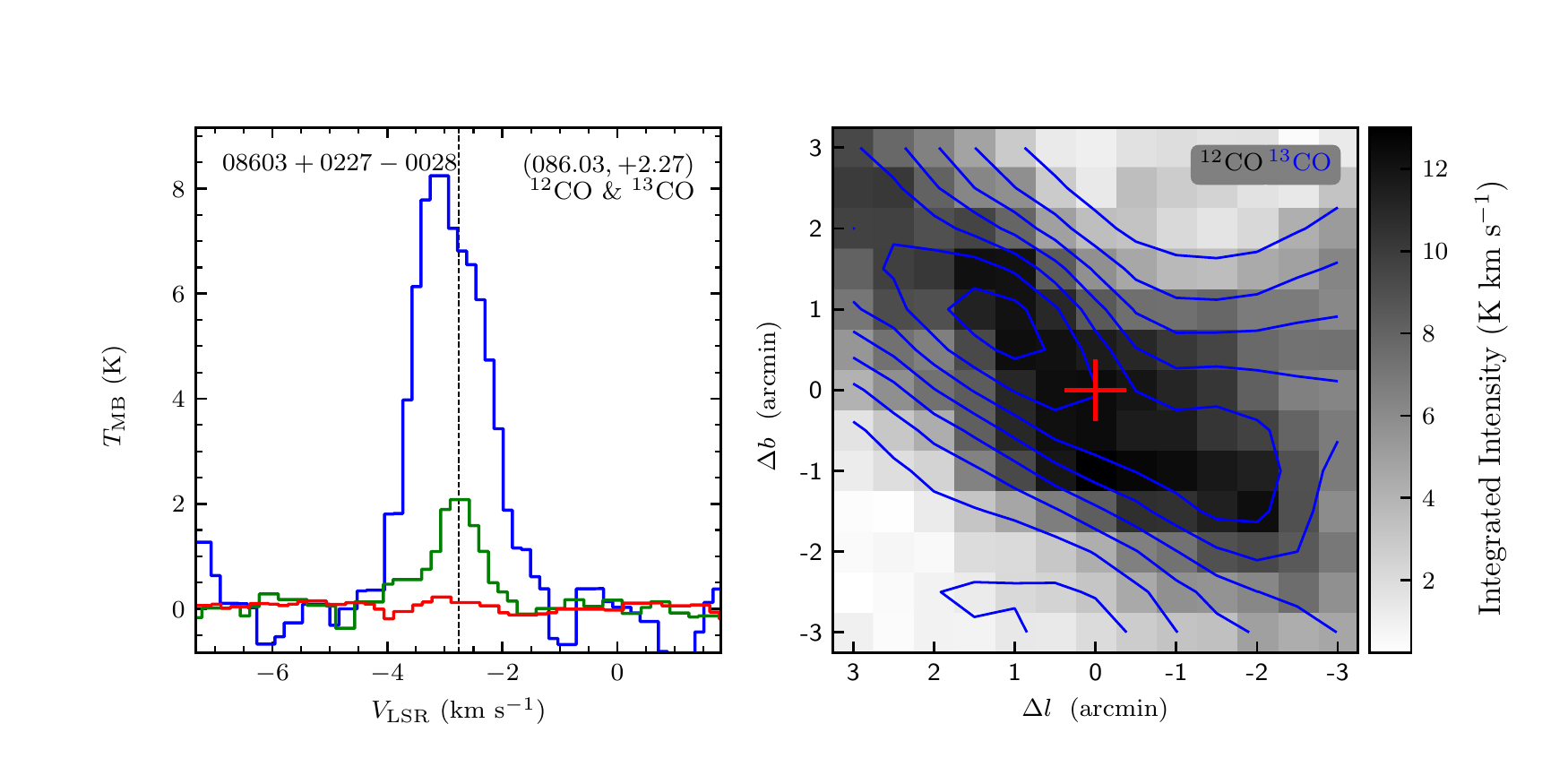}
\vspace{-0.5cm}

\includegraphics[width=9.0cm,angle=0]{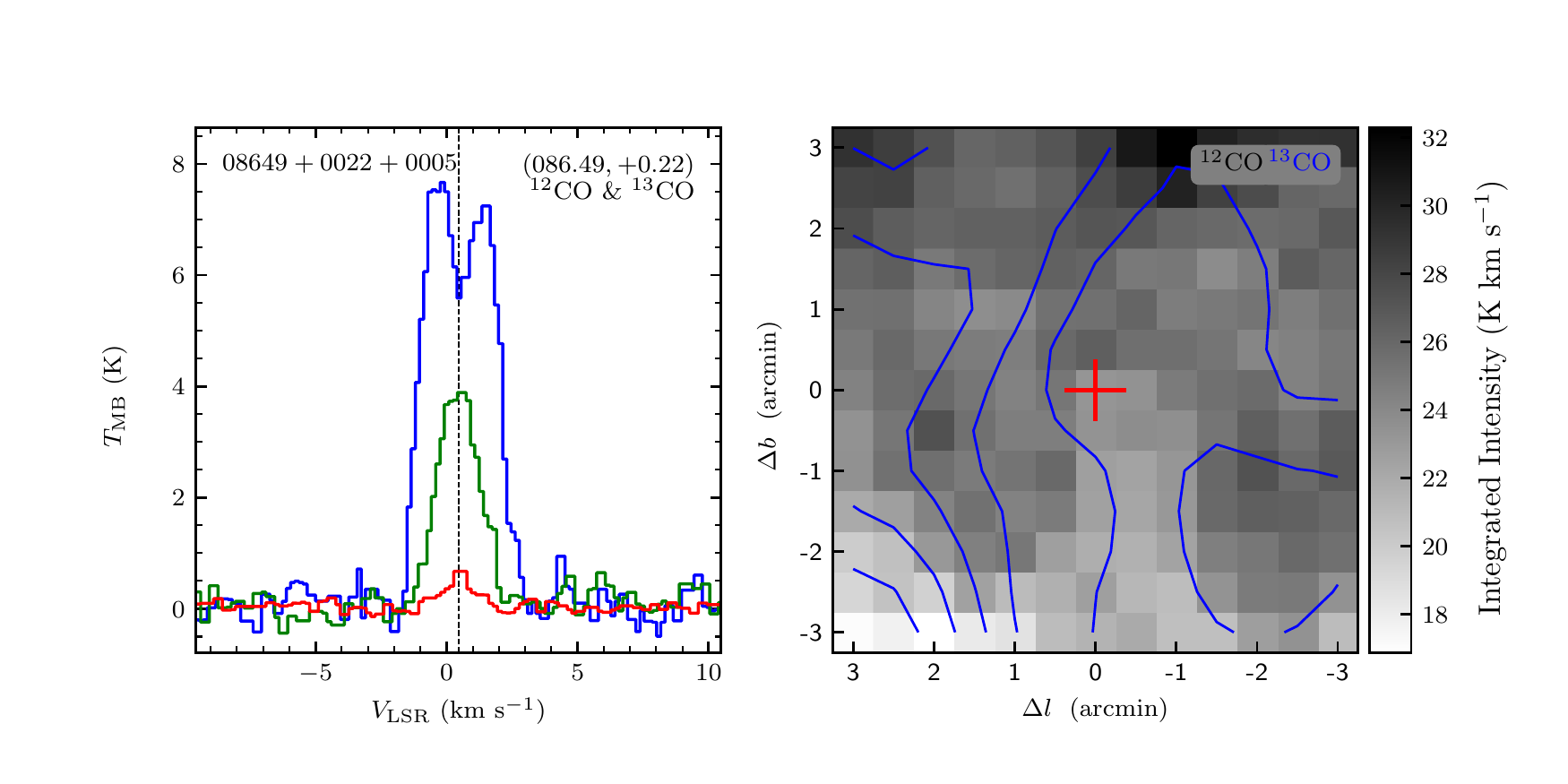}
\includegraphics[width=9.0cm,angle=0]{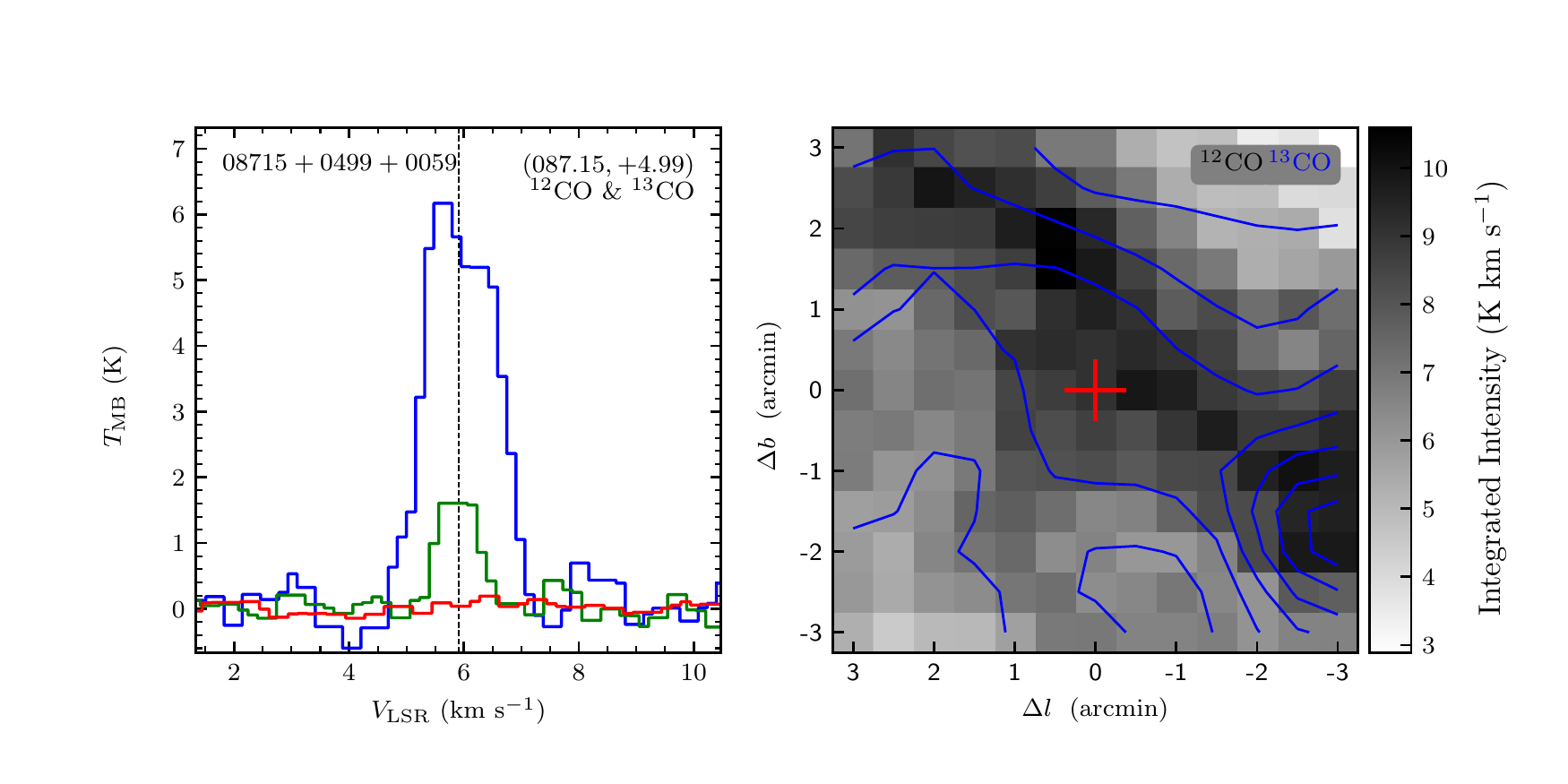}
\vspace{-0.5cm}

\includegraphics[width=9.0cm,angle=0]{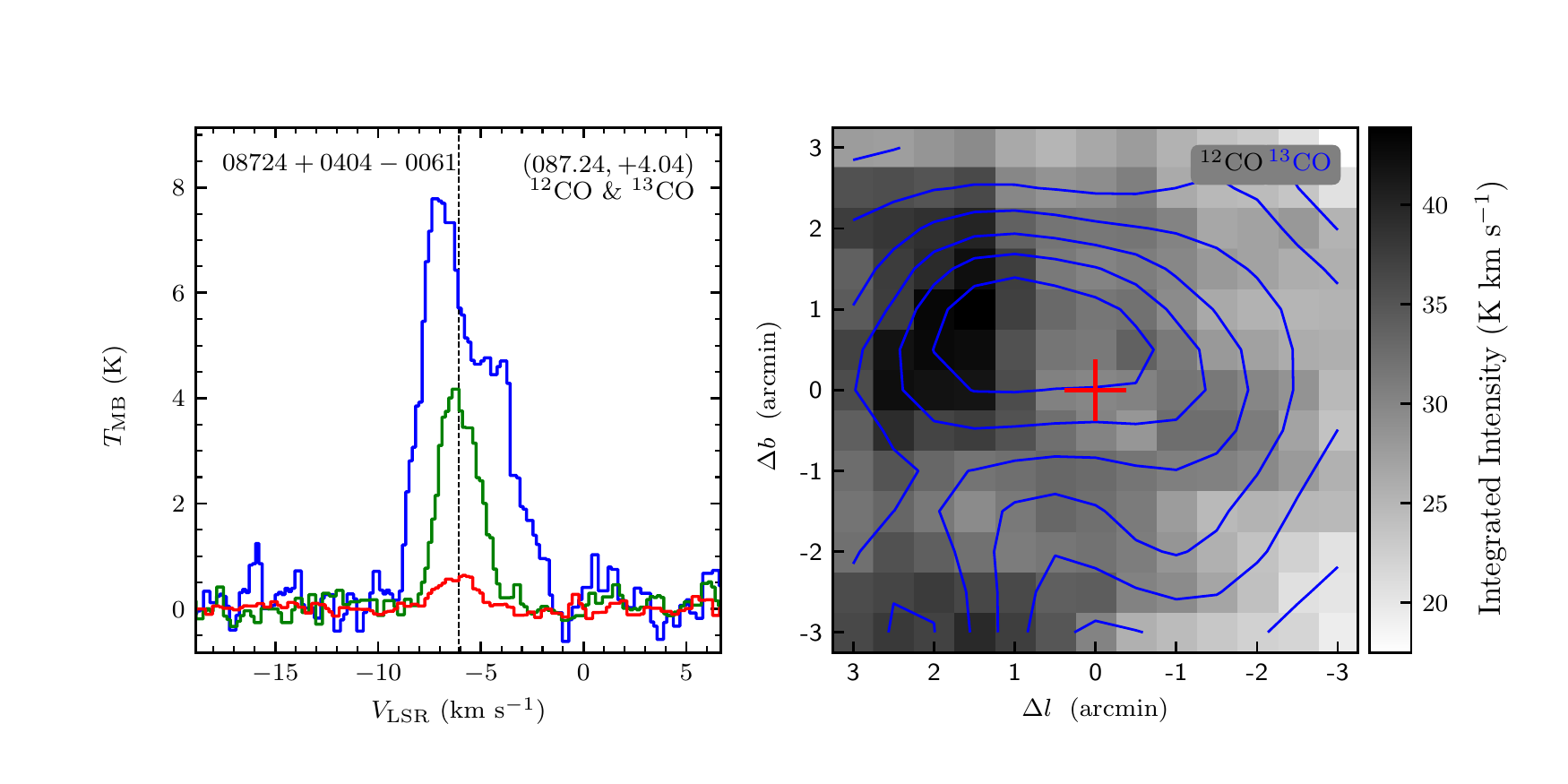}
\includegraphics[width=9.0cm,angle=0]{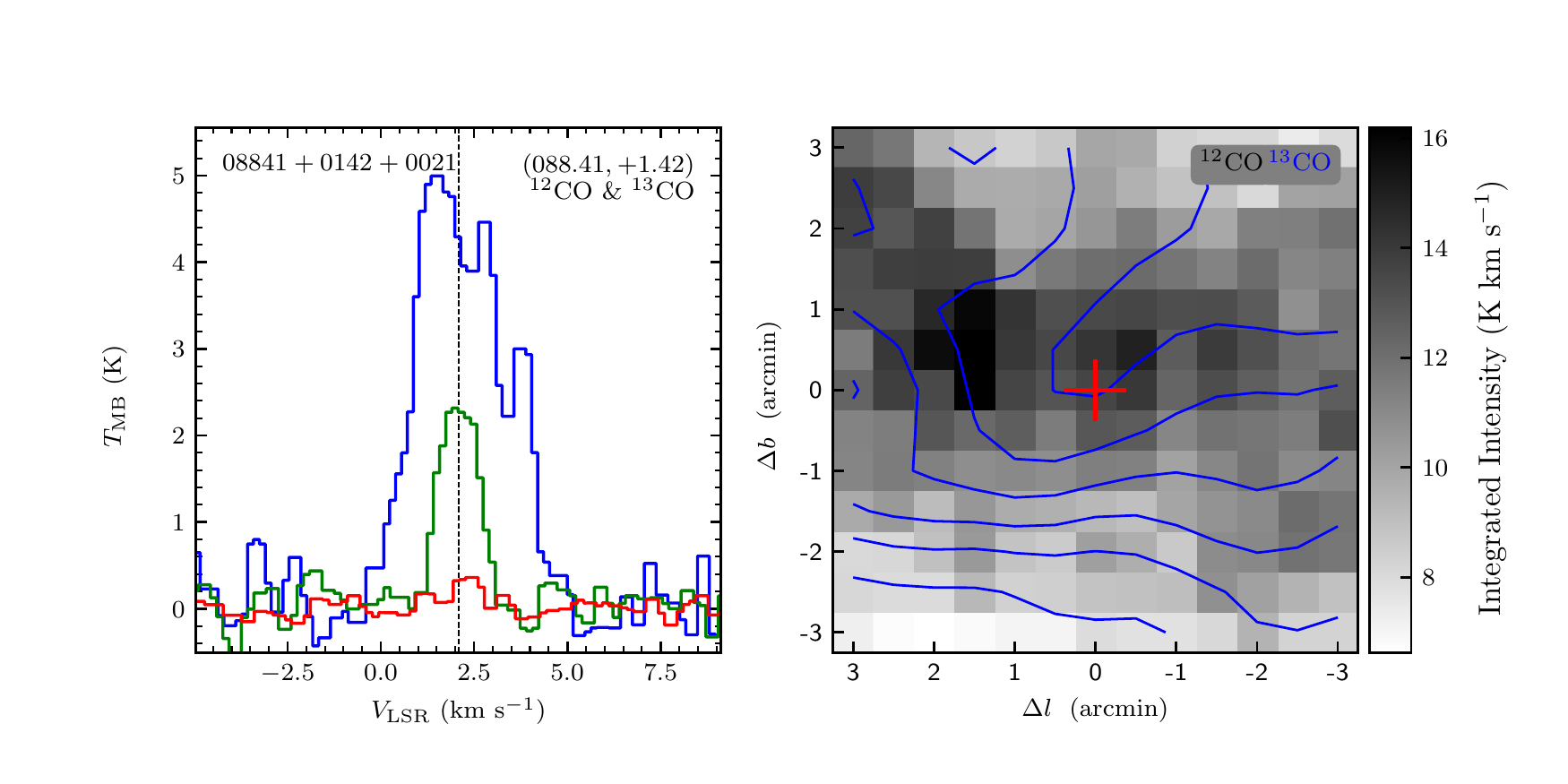}
\vspace{-0.5cm}

\includegraphics[width=9.0cm,angle=0]{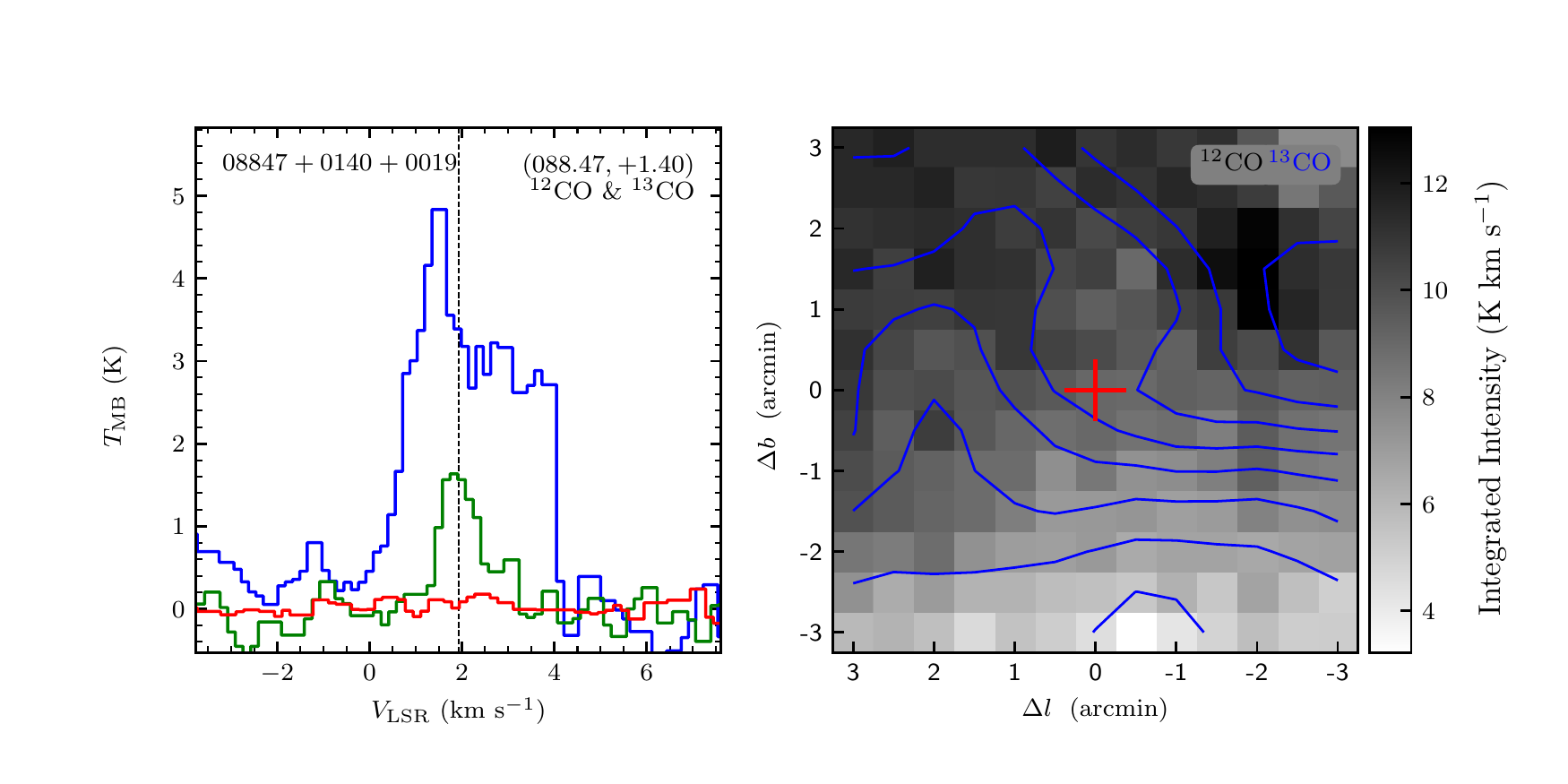}
\includegraphics[width=9.0cm,angle=0]{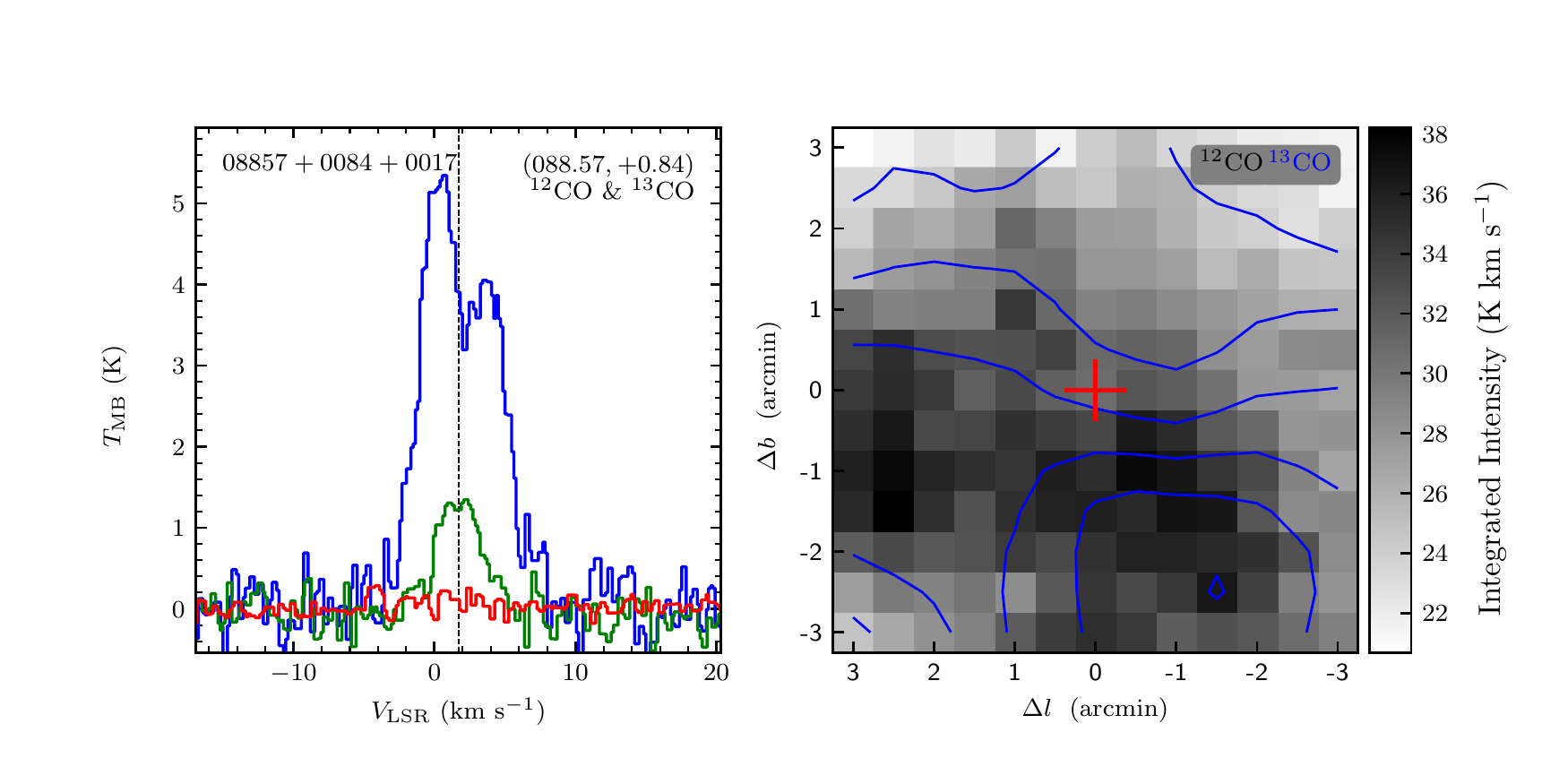}
\end{figure}
\clearpage

\begin{figure}
\includegraphics[width=9.0cm,angle=0]{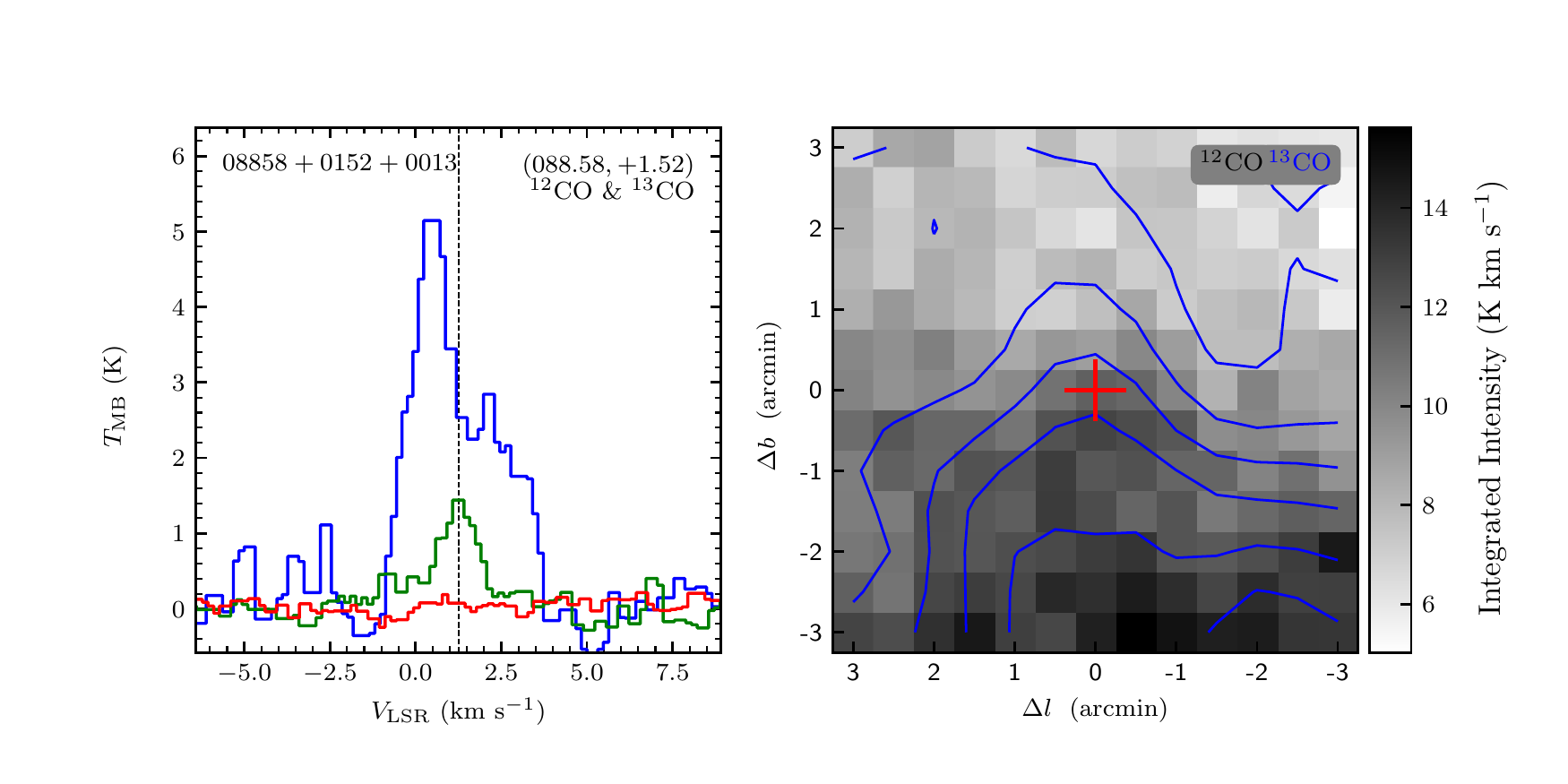}
\includegraphics[width=9.0cm,angle=0]{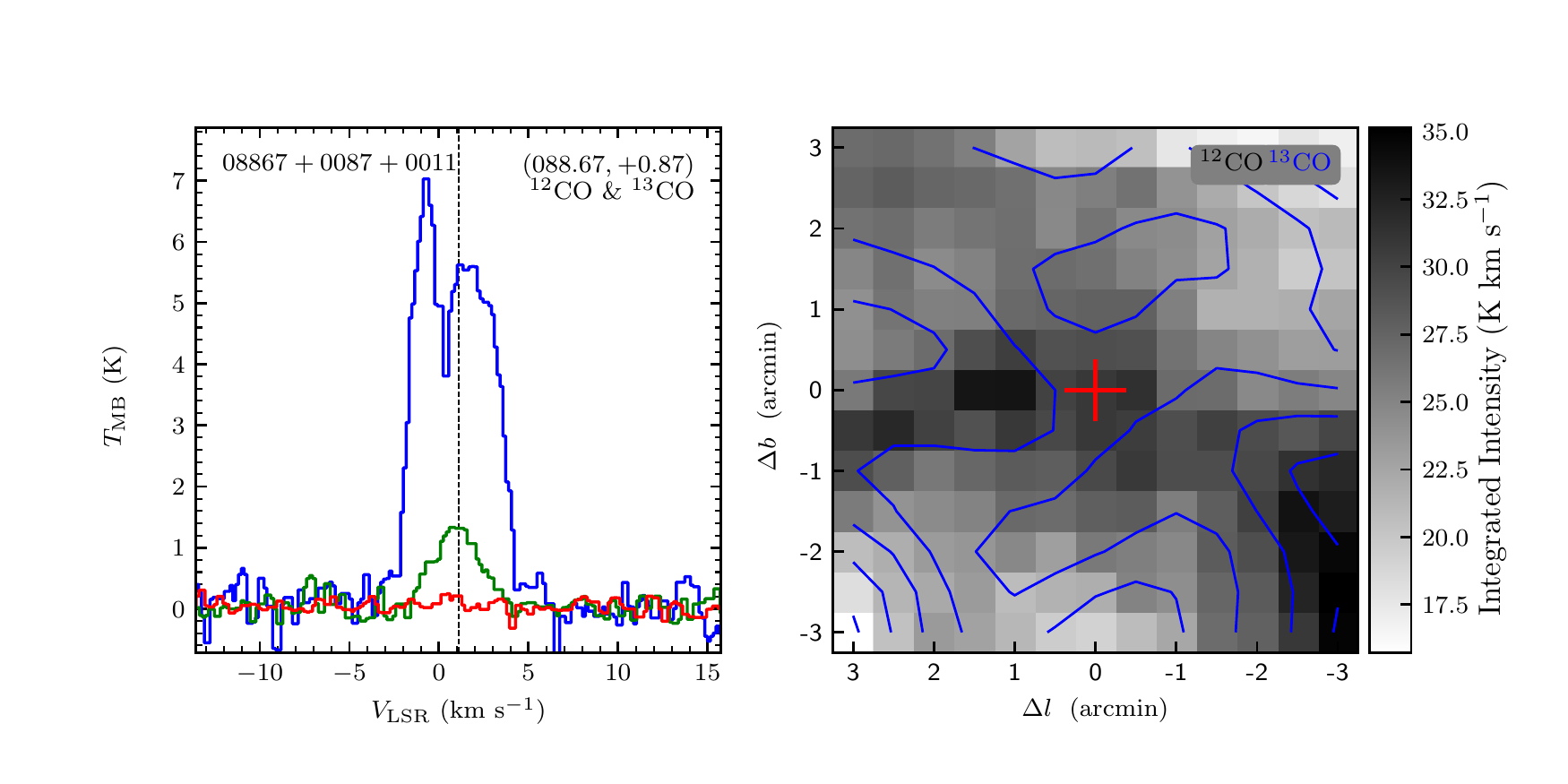}
\vspace{-0.5cm}

\includegraphics[width=9.0cm,angle=0]{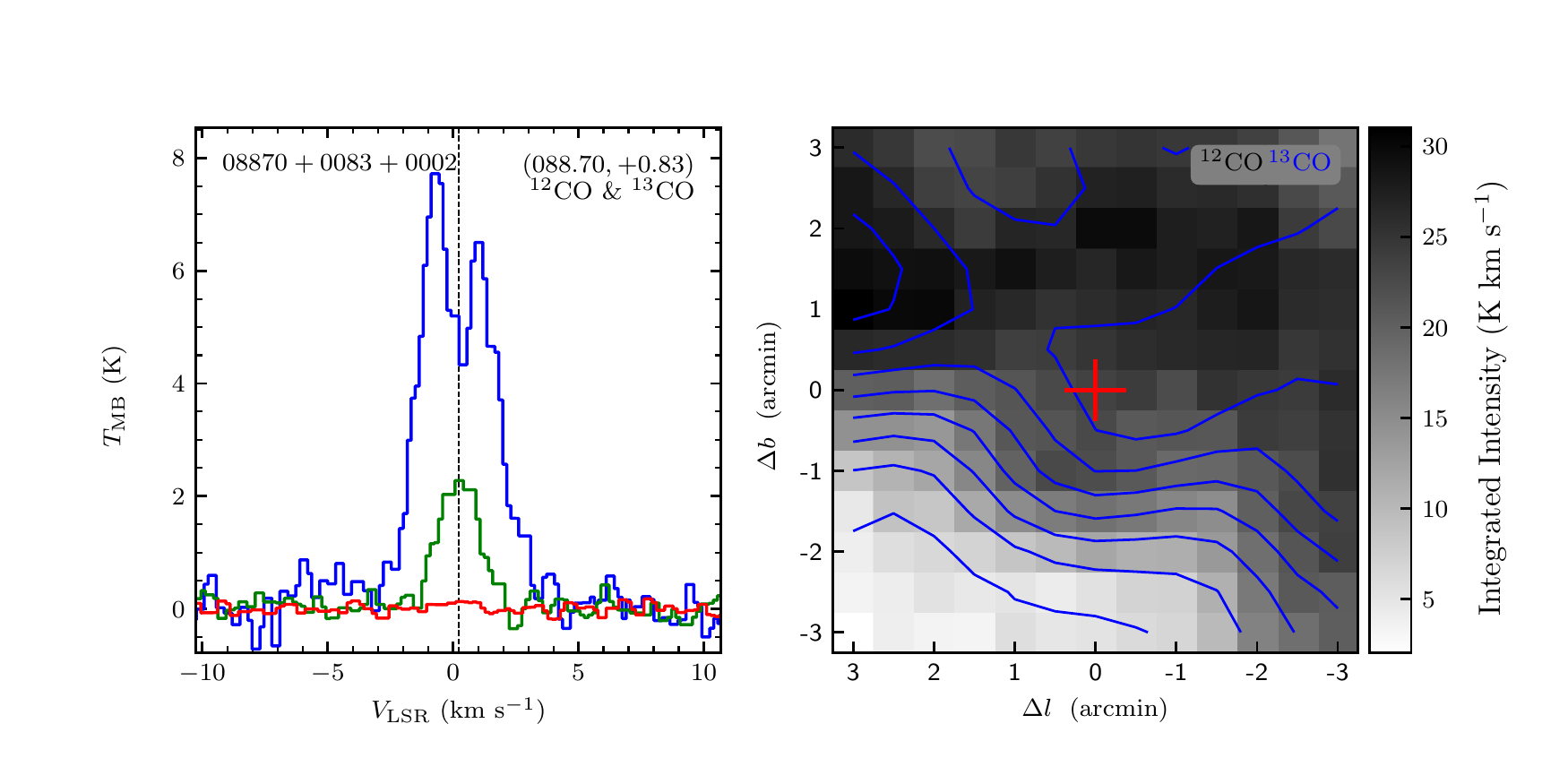}
\includegraphics[width=9.0cm,angle=0]{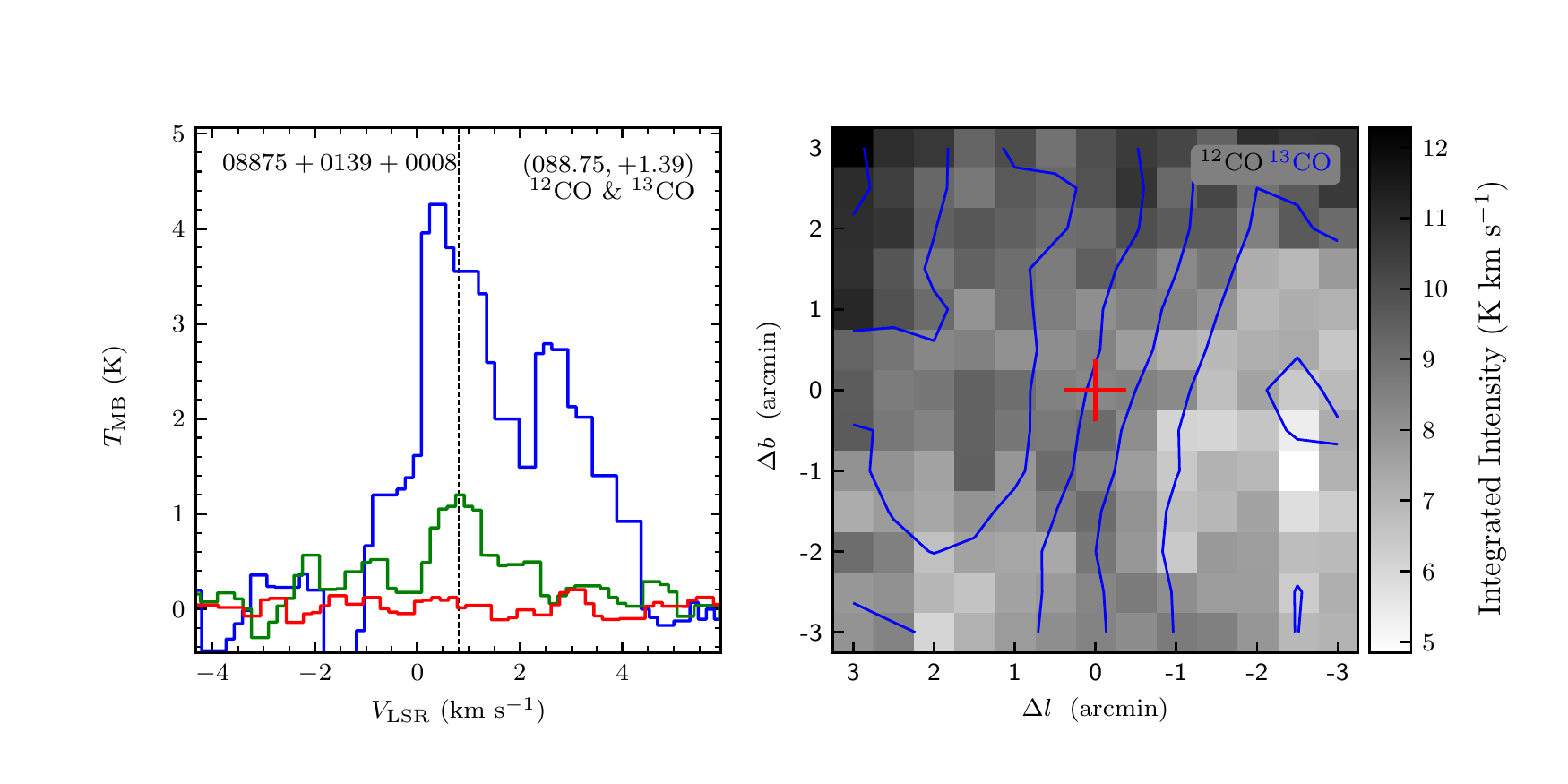}
\vspace{-0.5cm}

\includegraphics[width=9.0cm,angle=0]{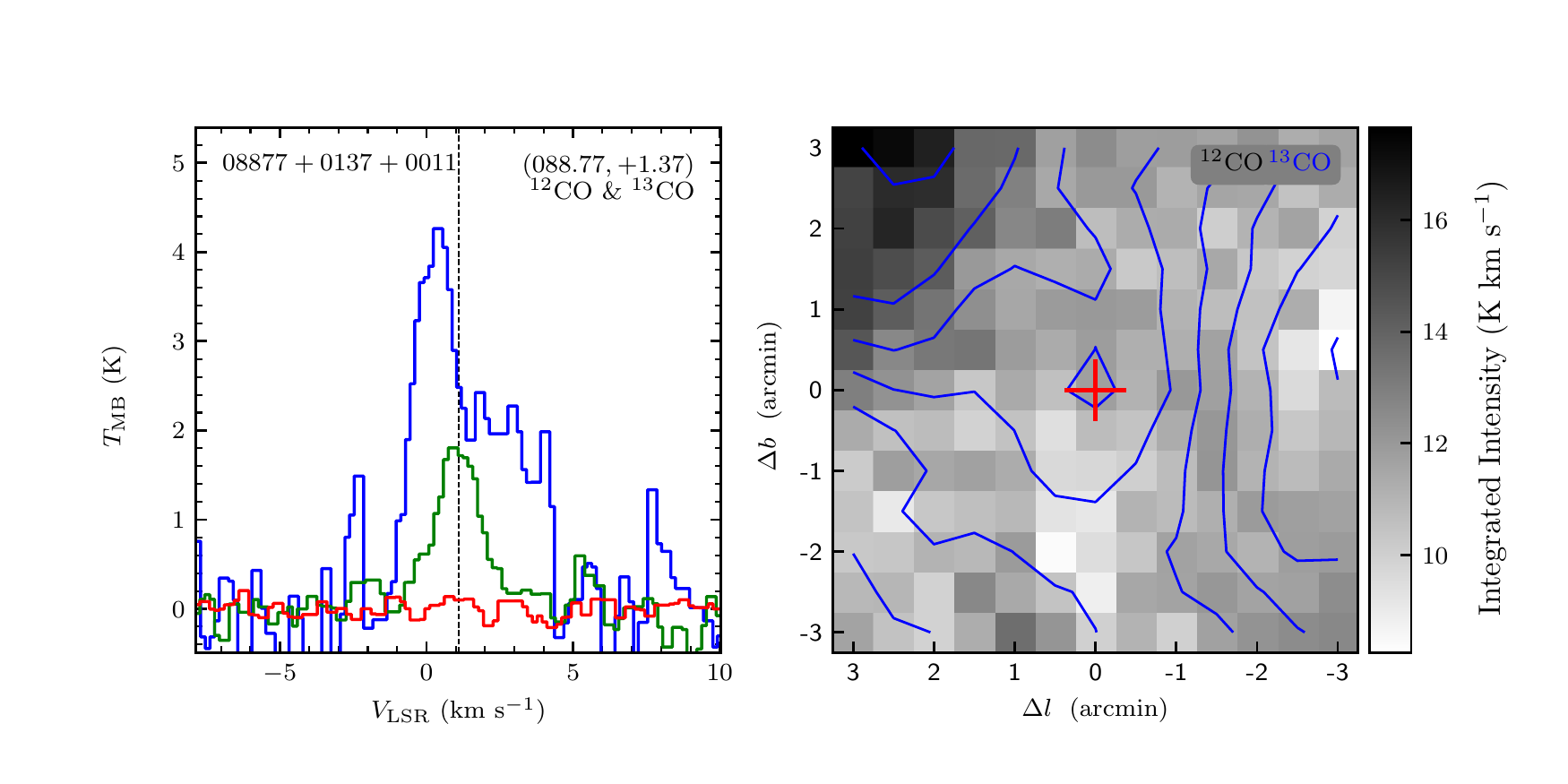}
\includegraphics[width=9.0cm,angle=0]{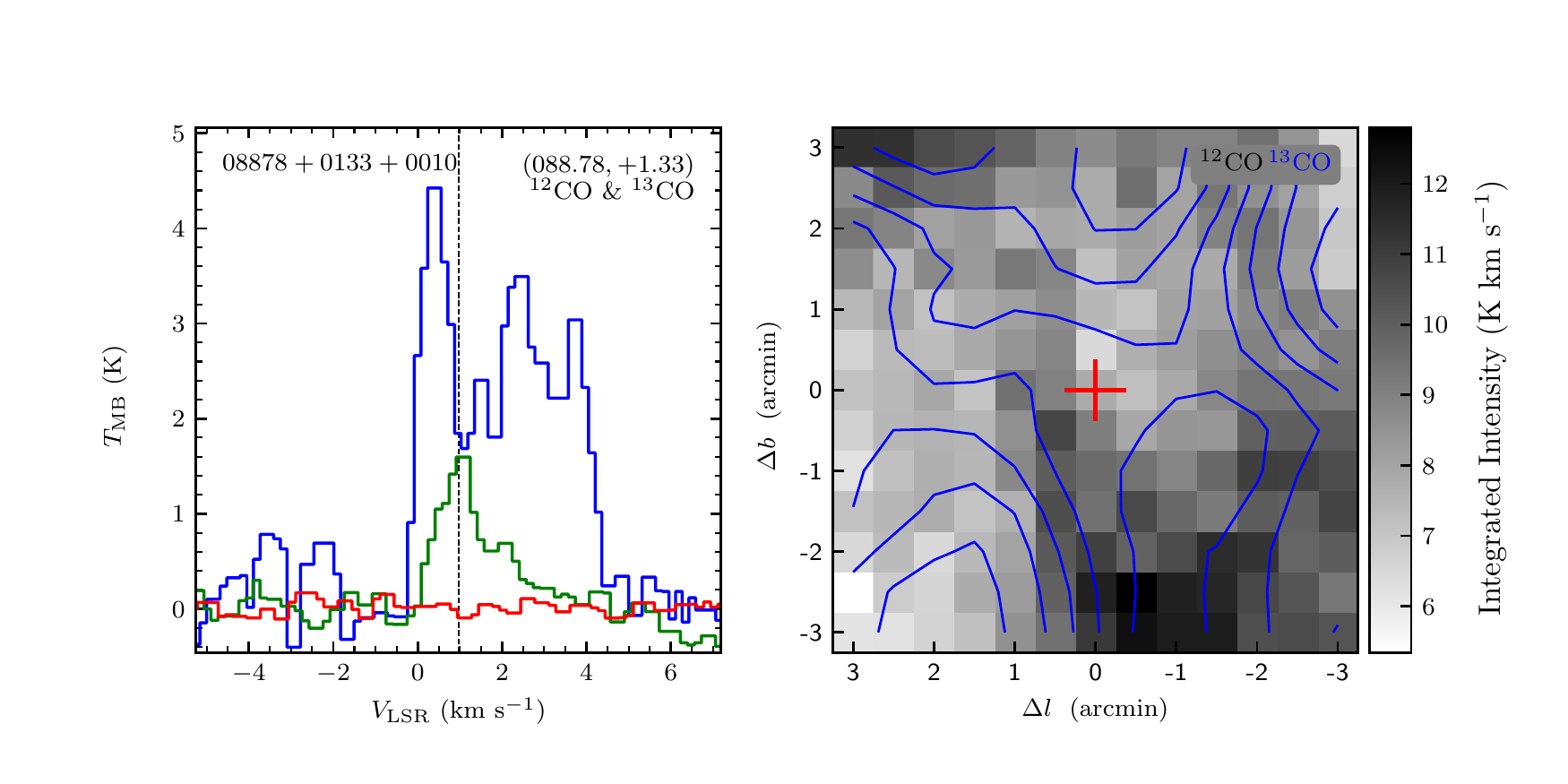}
\vspace{-0.5cm}

\includegraphics[width=9.0cm,angle=0]{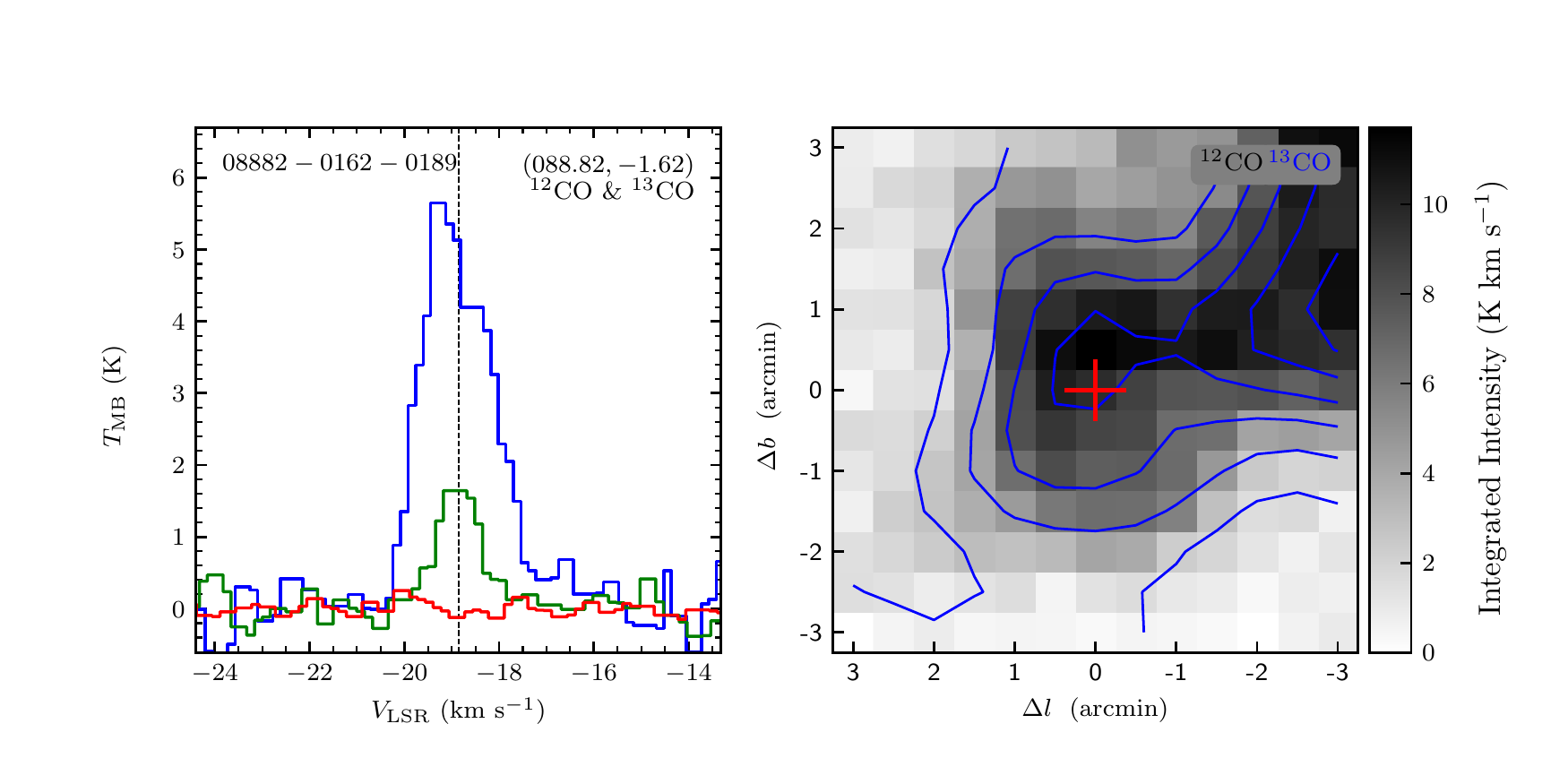}
\includegraphics[width=9.0cm,angle=0]{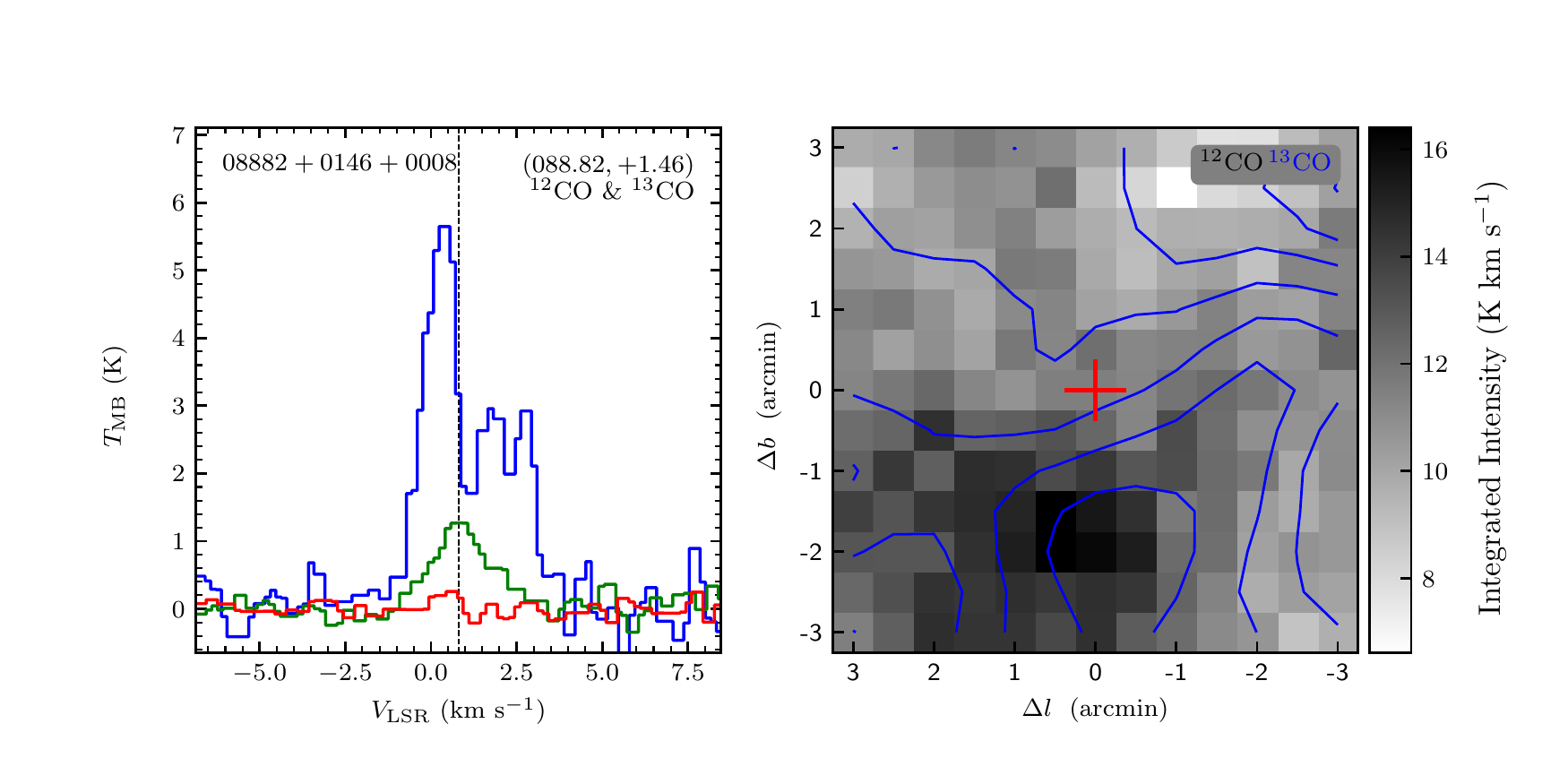}
\vspace{-0.5cm}

\includegraphics[width=9.0cm,angle=0]{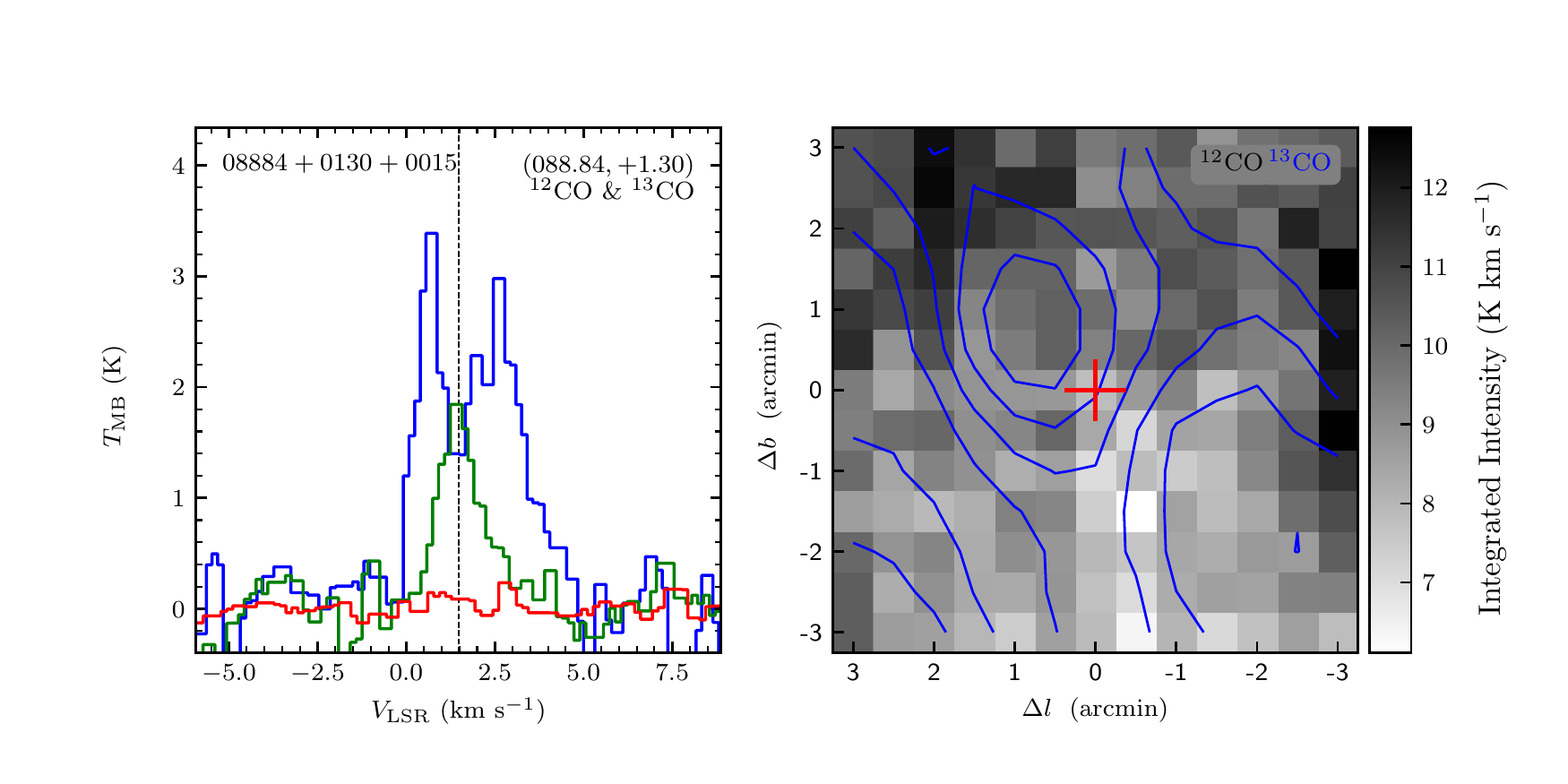}
\includegraphics[width=9.0cm,angle=0]{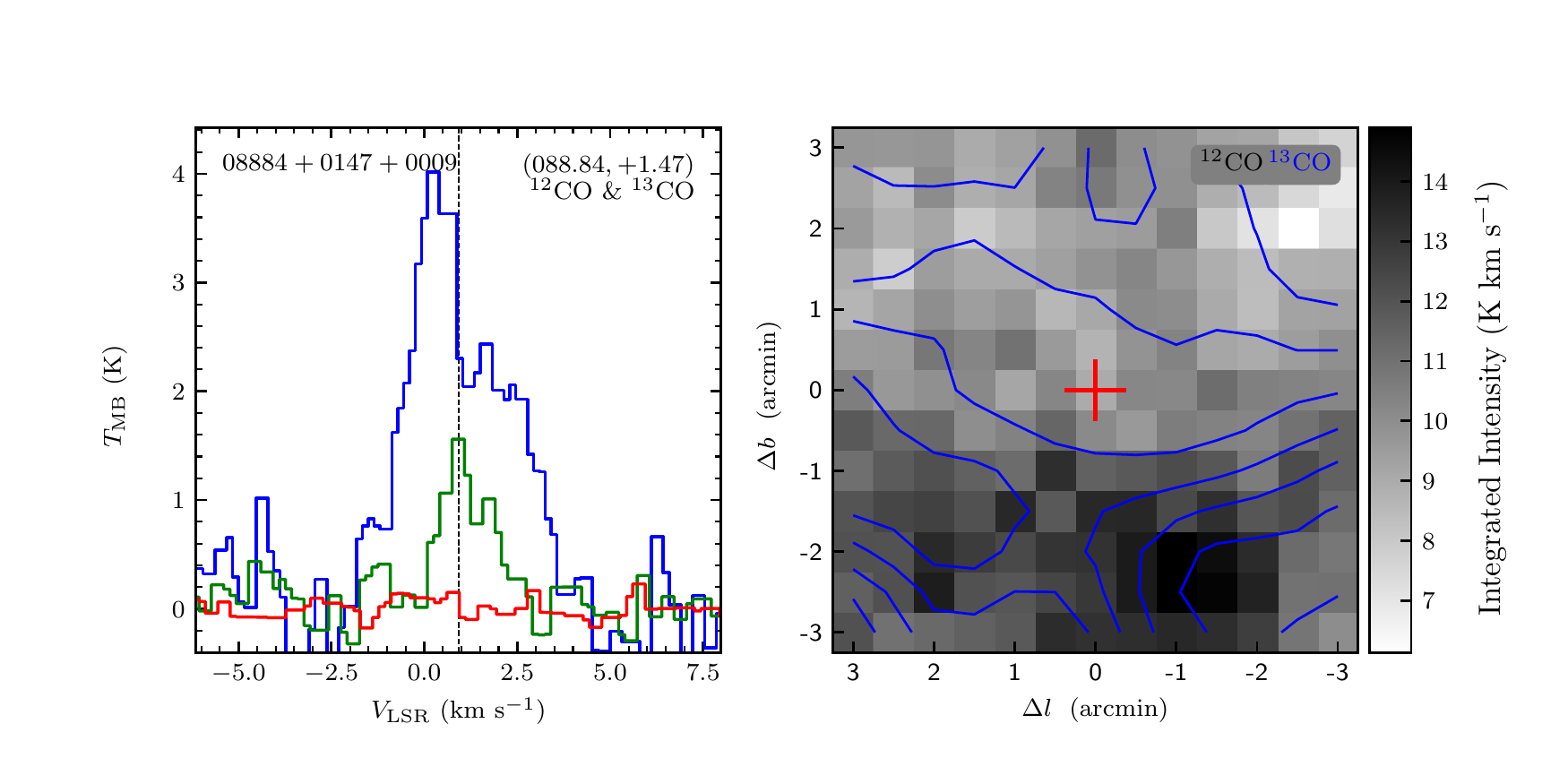}
\end{figure}
\clearpage

\begin{figure}
\includegraphics[width=9.0cm,angle=0]{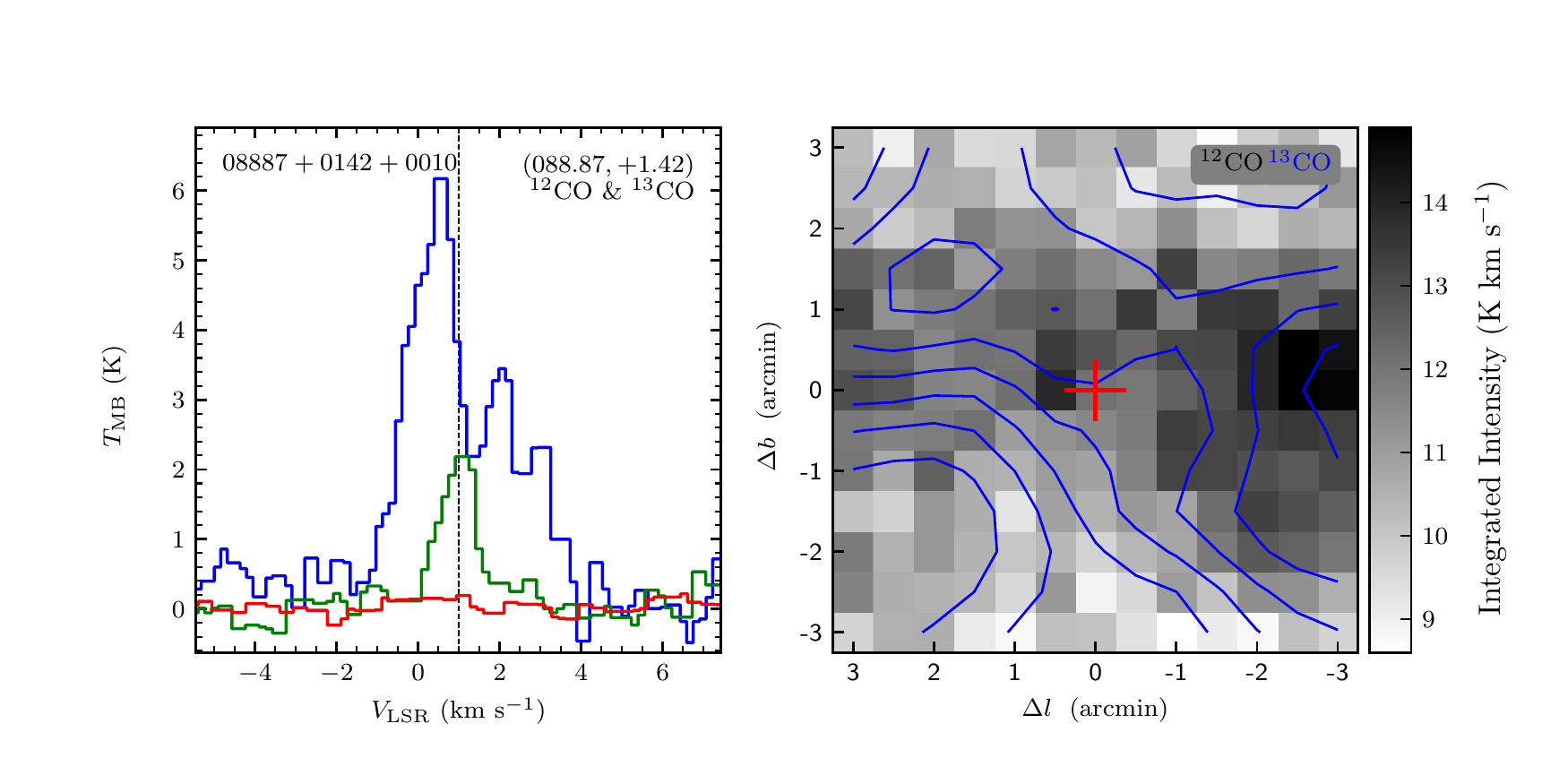}
\includegraphics[width=9.0cm,angle=0]{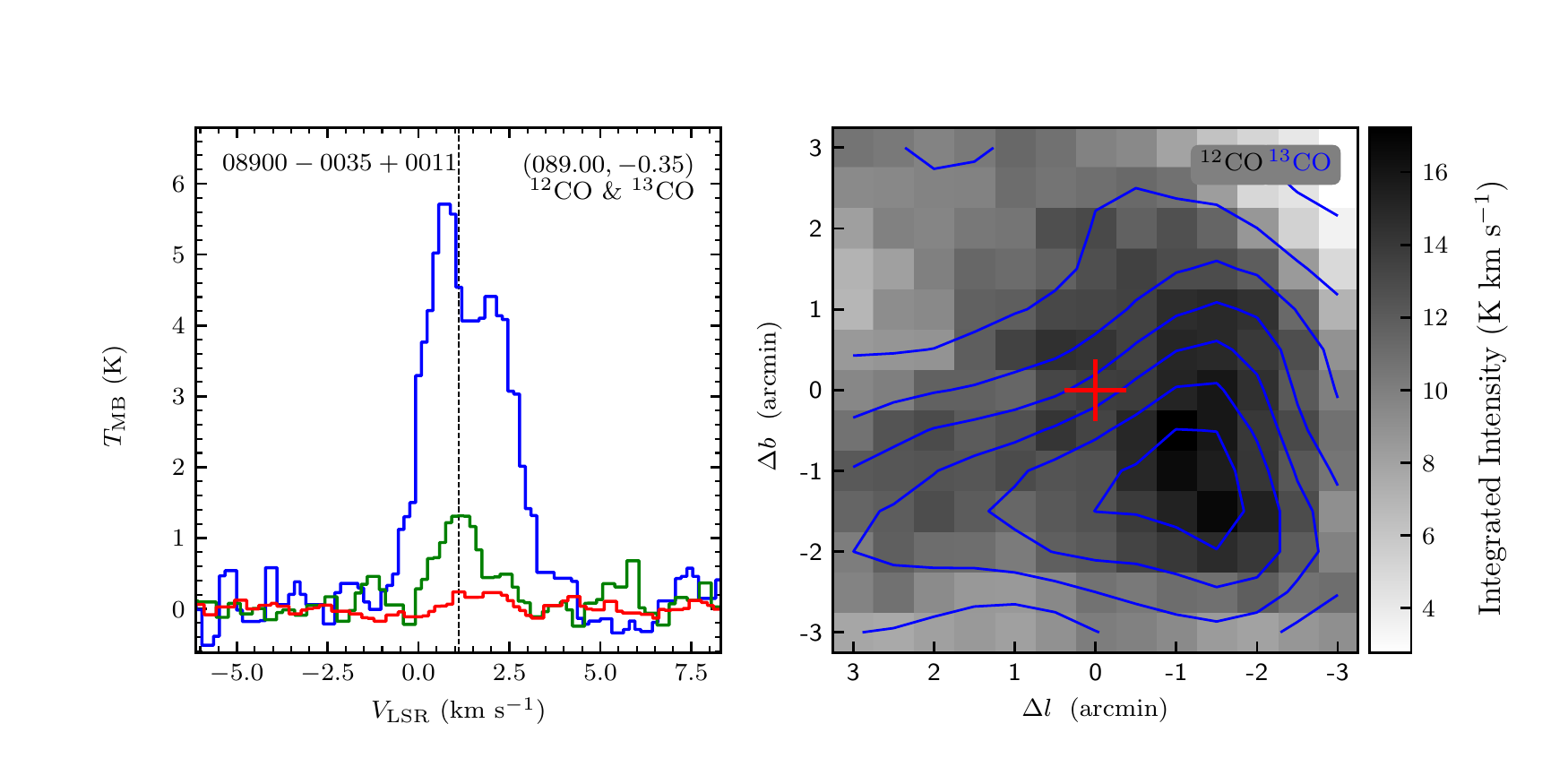}
\vspace{-0.5cm}

\includegraphics[width=9.0cm,angle=0]{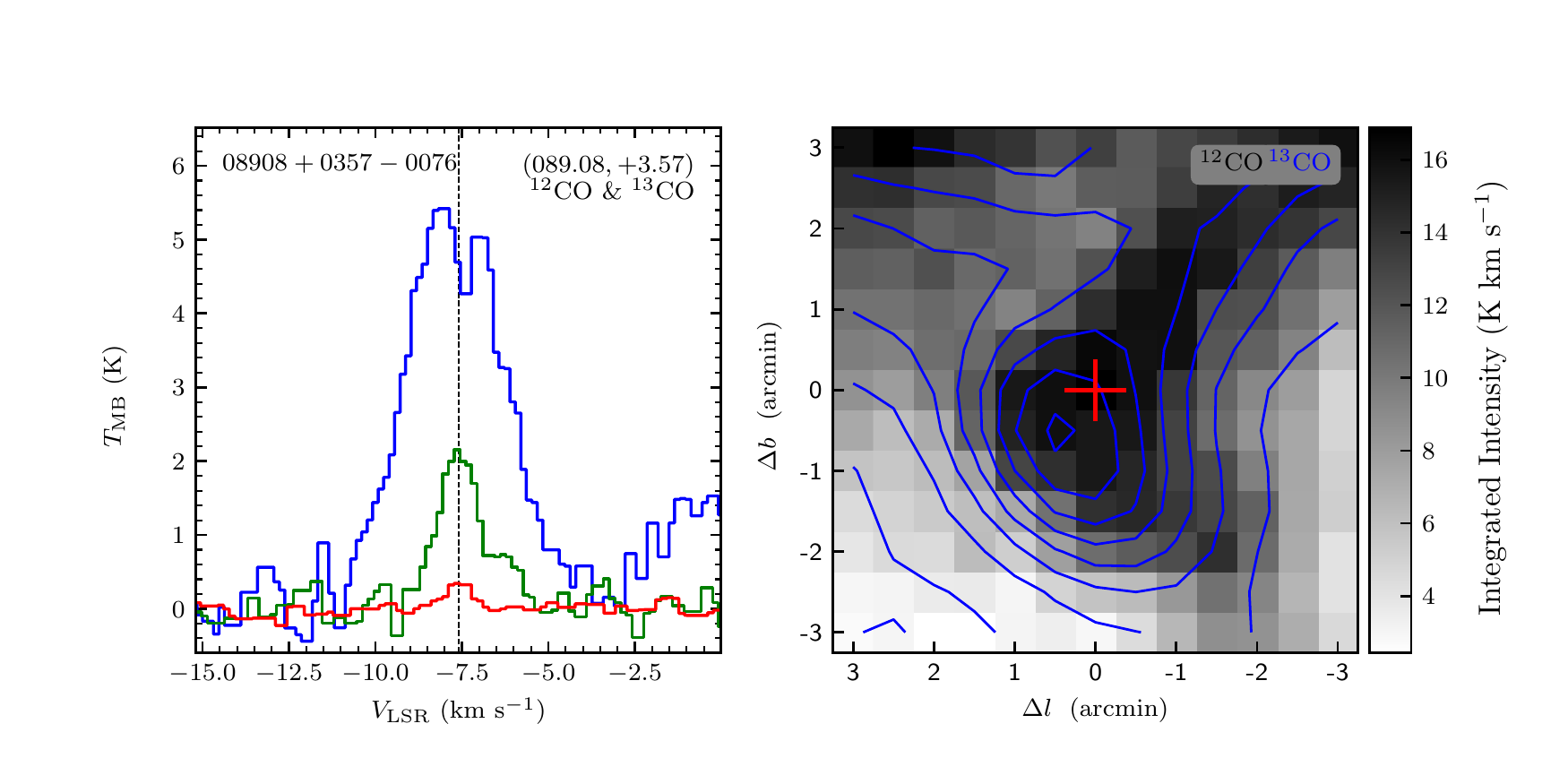}
\includegraphics[width=9.0cm,angle=0]{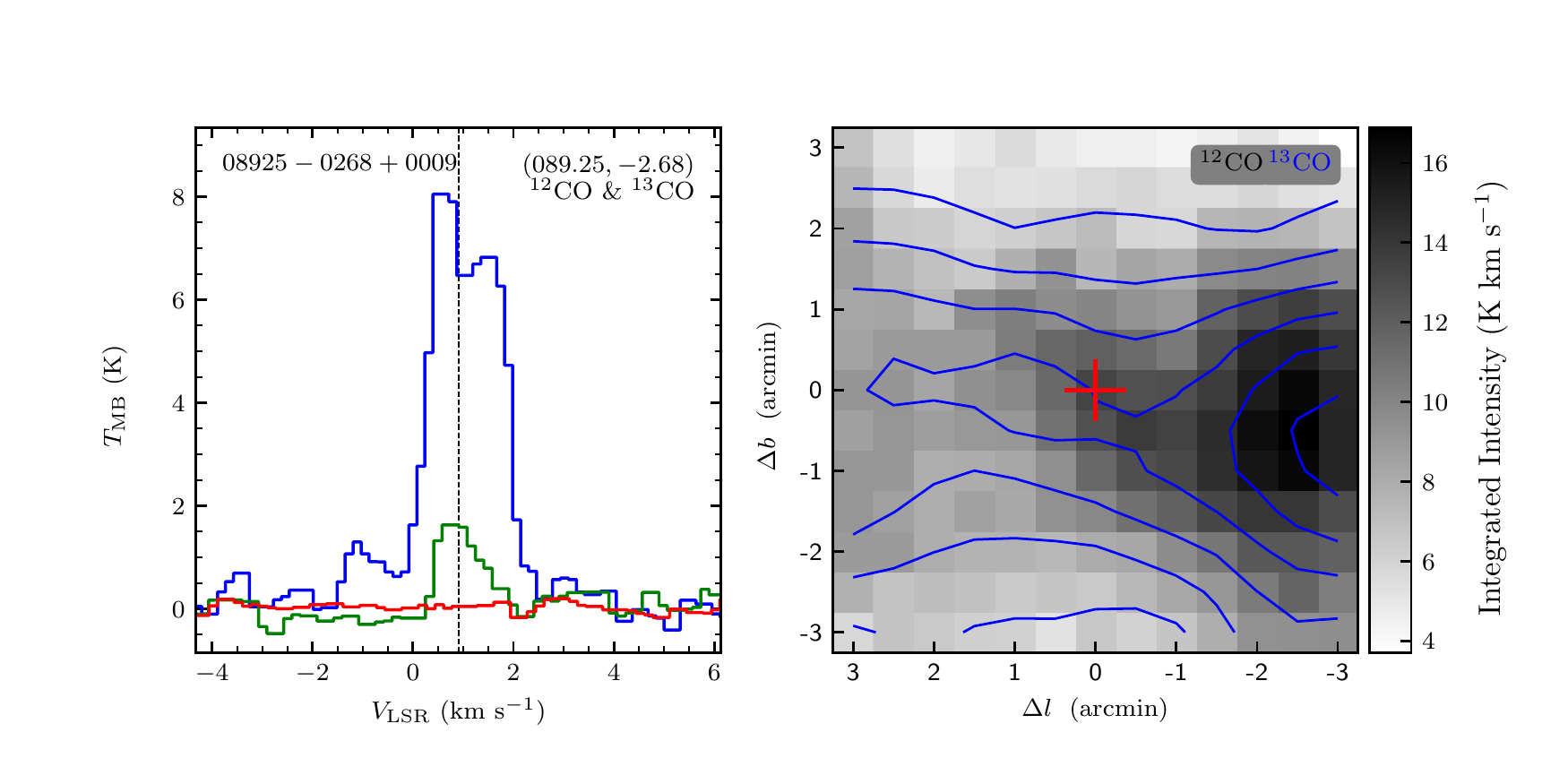}
\vspace{-0.5cm}

\includegraphics[width=9.0cm,angle=0]{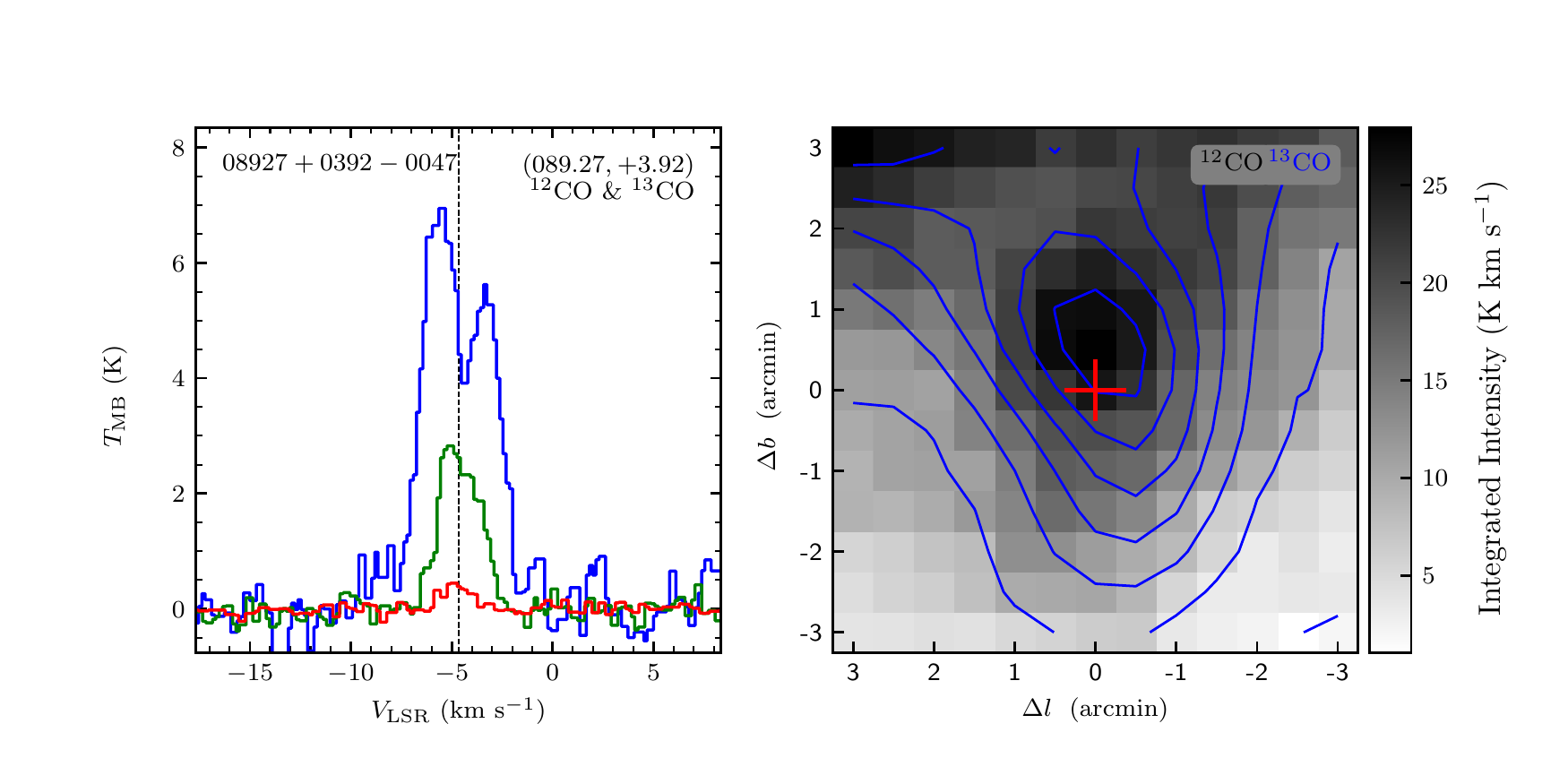}
\includegraphics[width=9.0cm,angle=0]{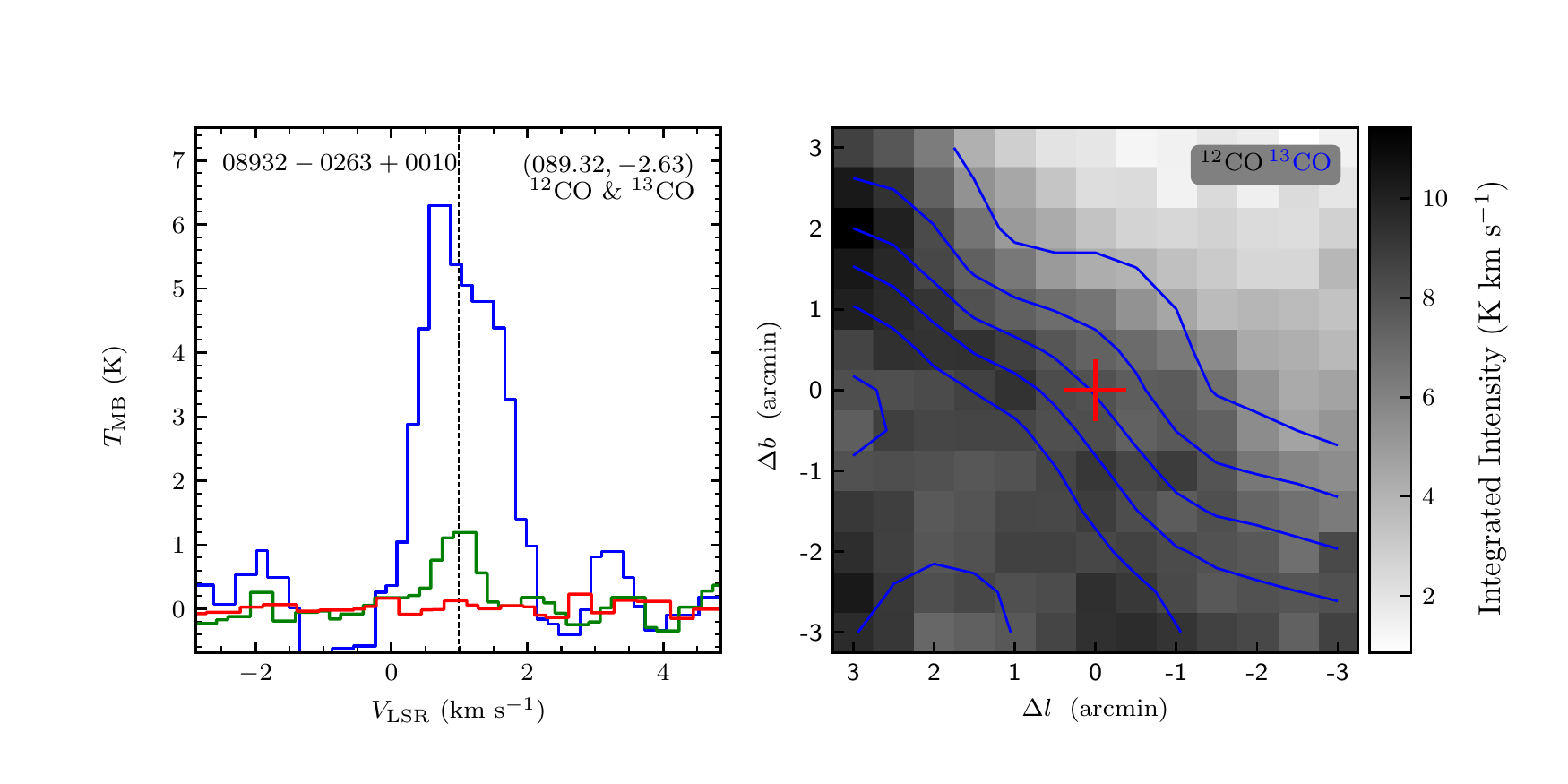}
\vspace{-0.5cm}

\includegraphics[width=9.0cm,angle=0]{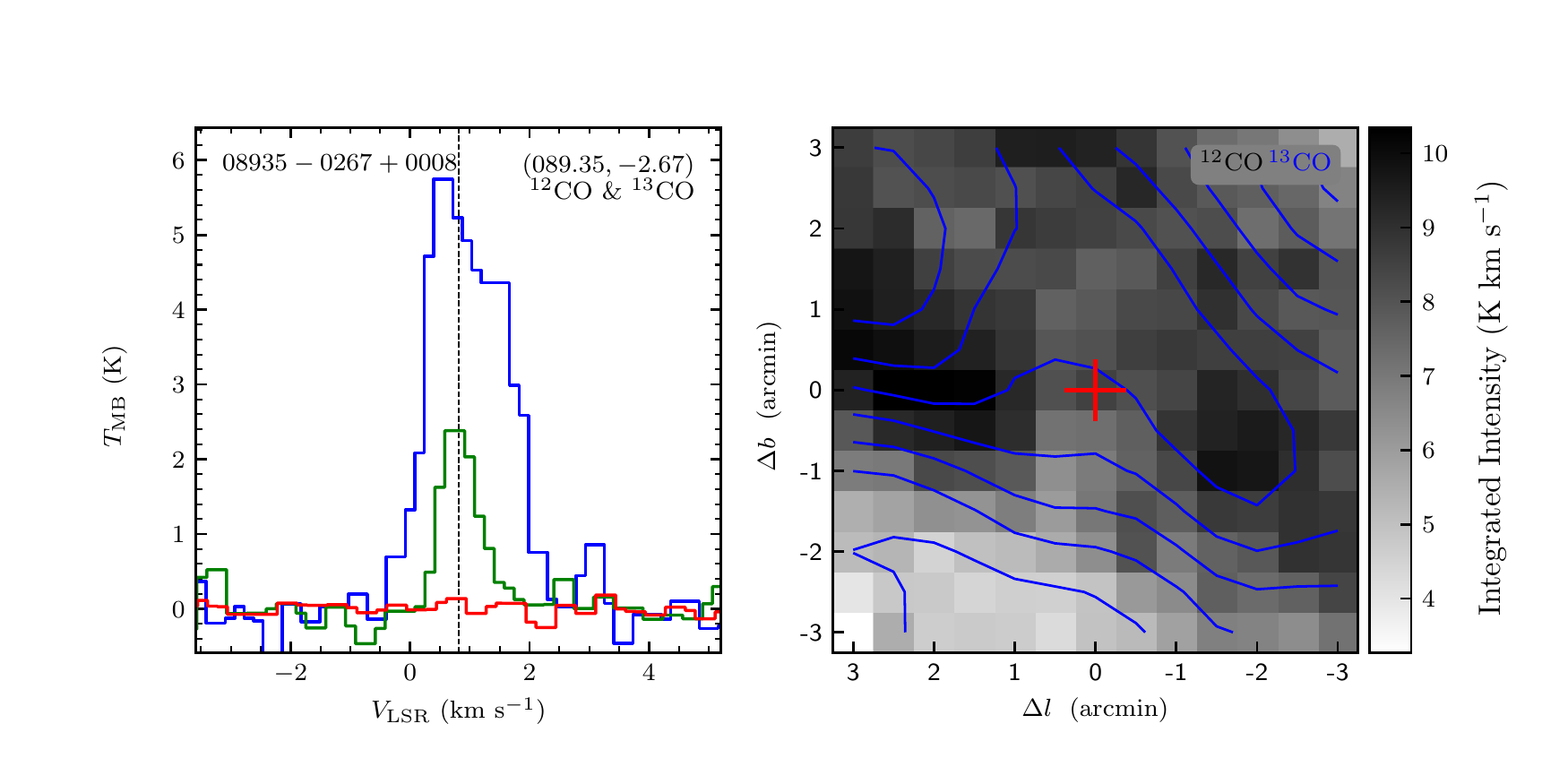}
\includegraphics[width=9.0cm,angle=0]{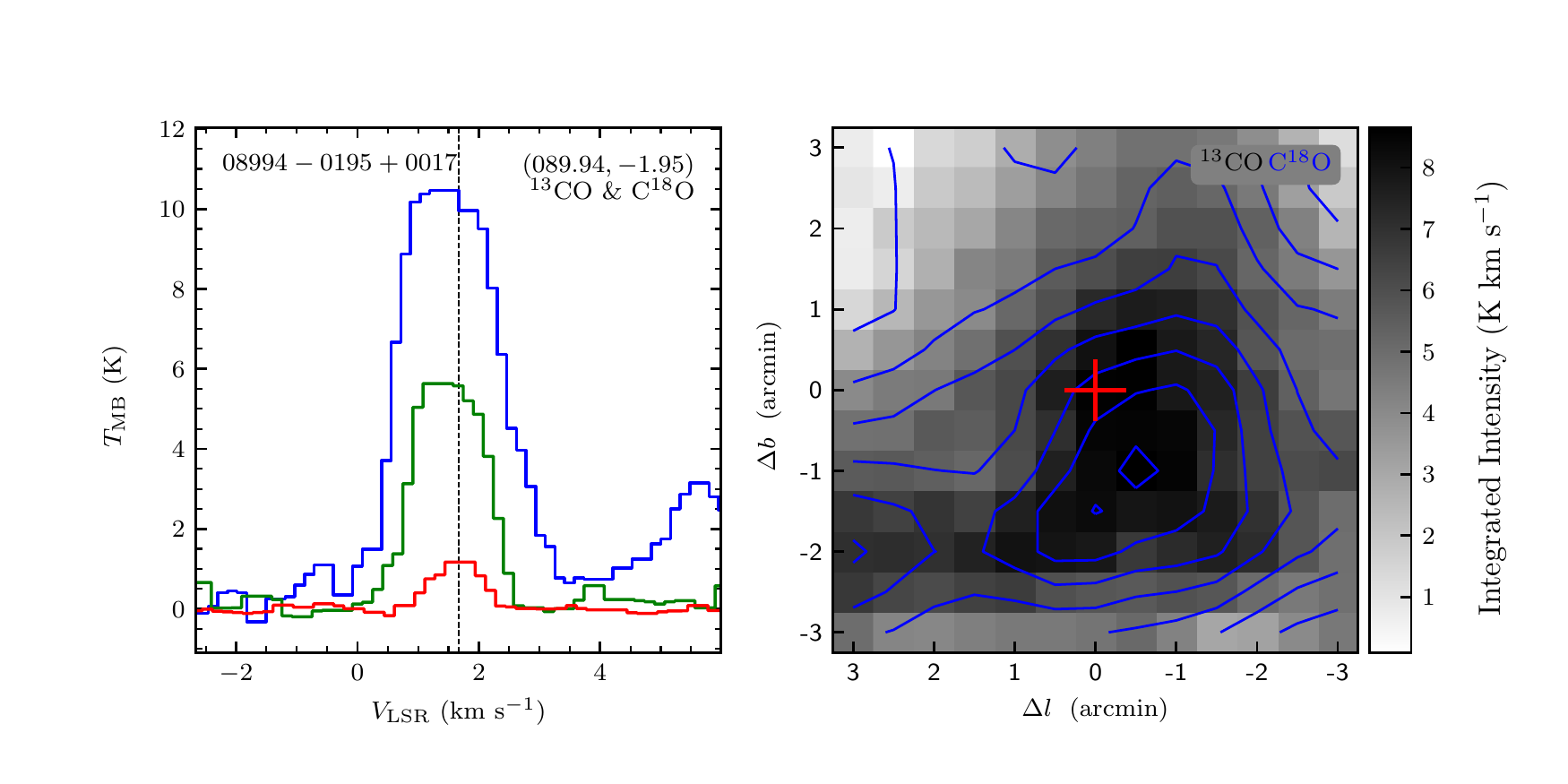}
\vspace{-0.5cm}

\includegraphics[width=9.0cm,angle=0]{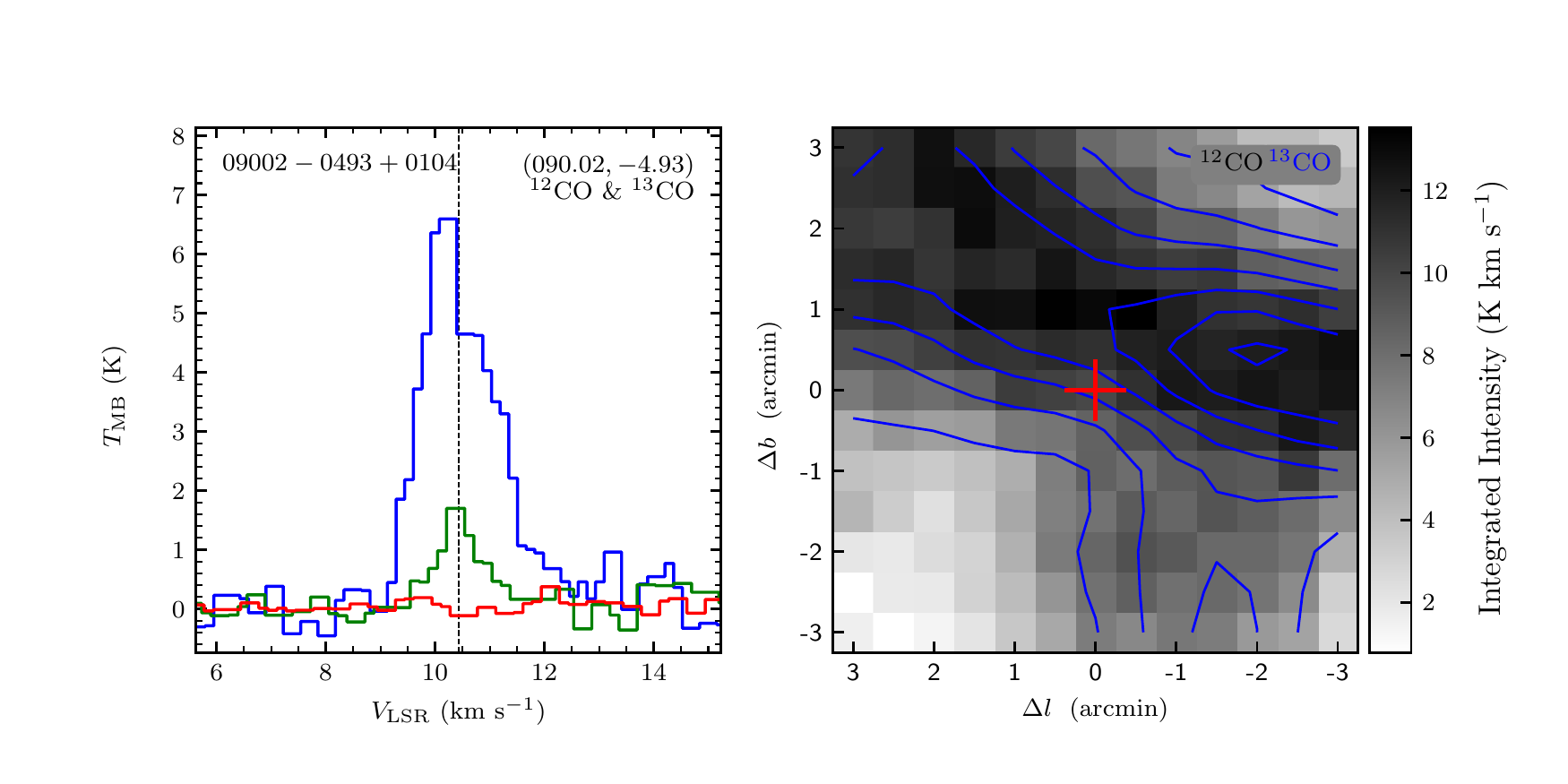}
\includegraphics[width=9.0cm,angle=0]{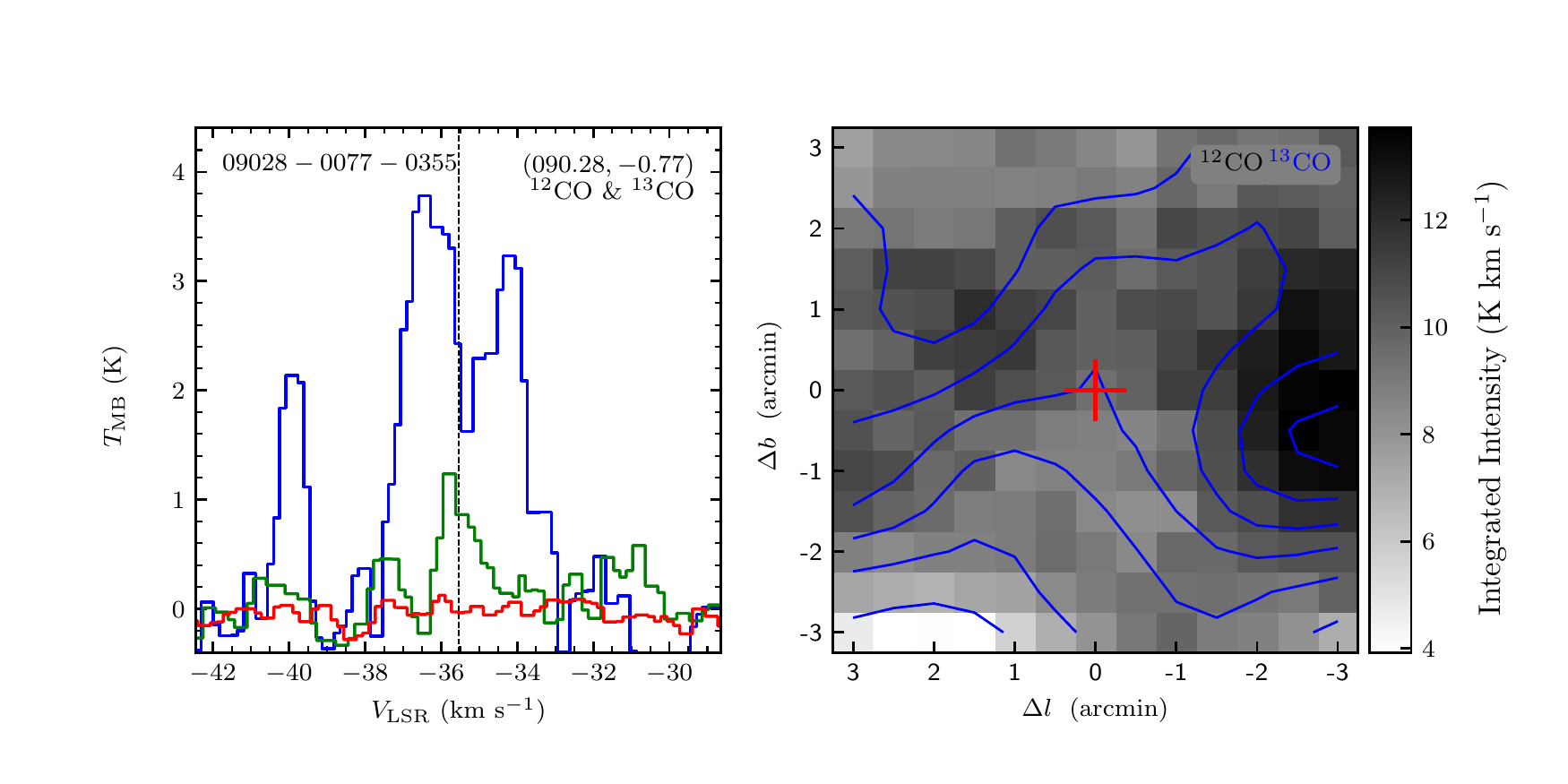}
\end{figure}
\clearpage

\begin{figure}
\includegraphics[width=9.0cm,angle=0]{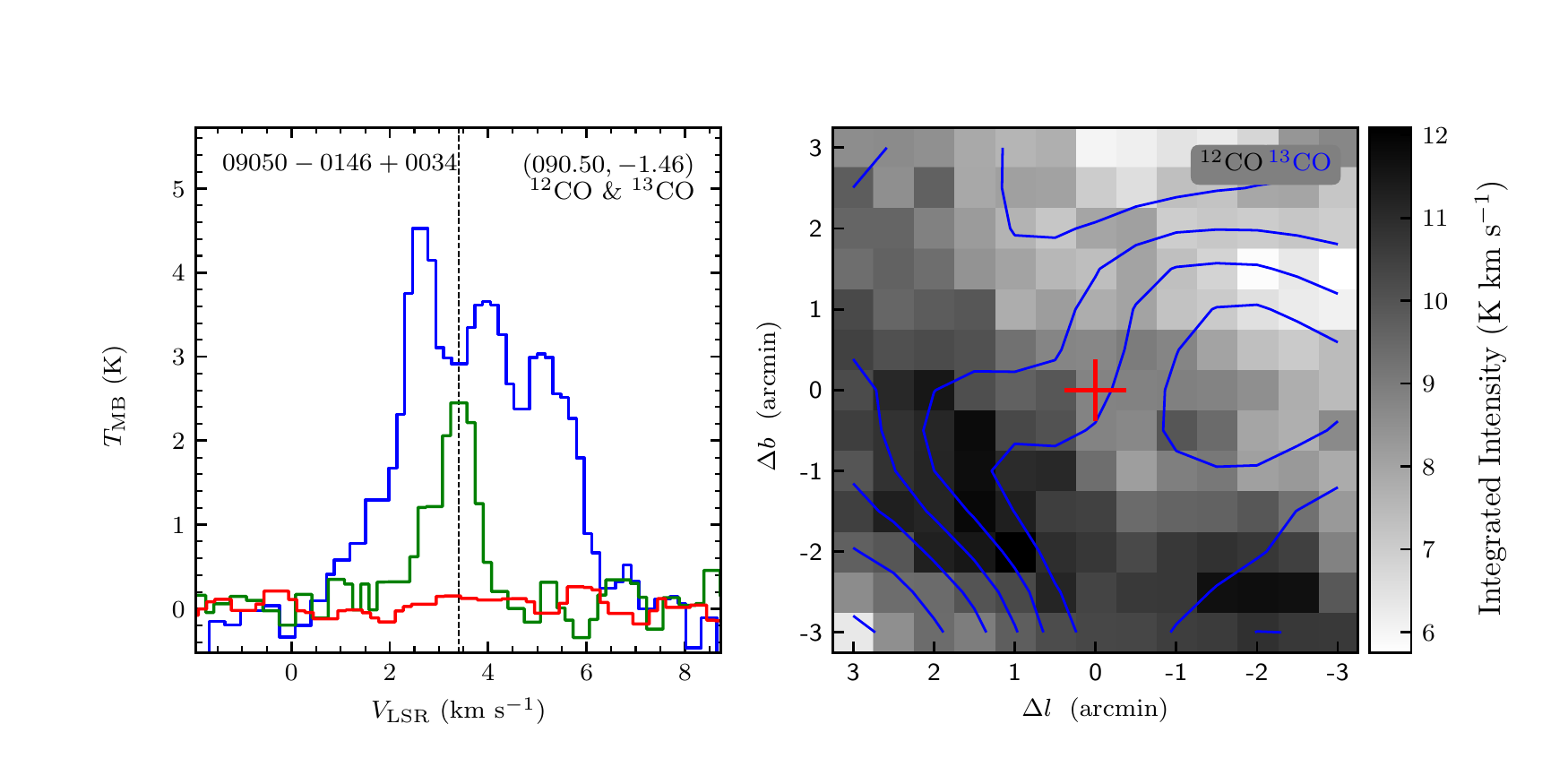}
\includegraphics[width=9.0cm,angle=0]{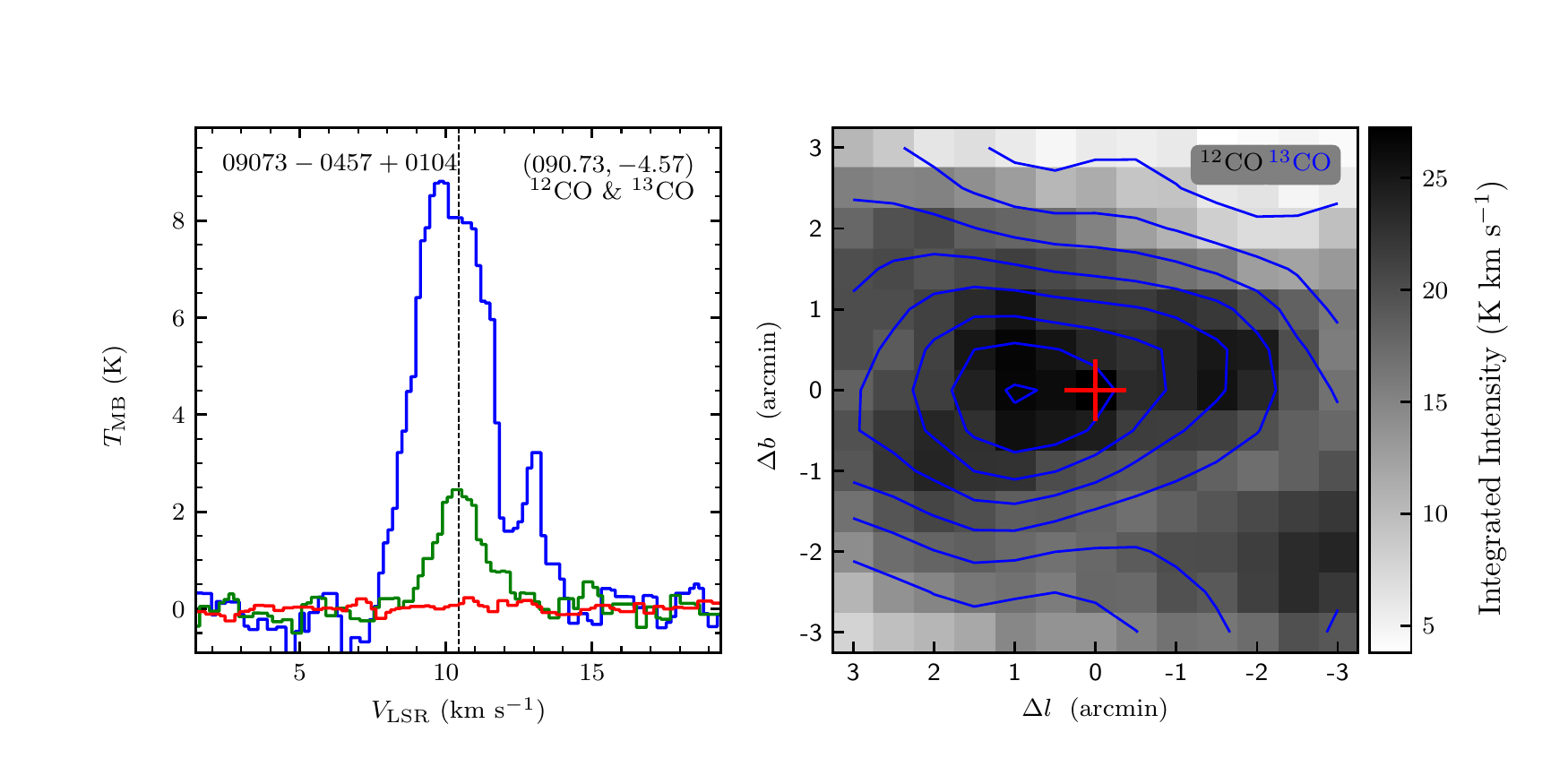}
\vspace{-0.5cm}

\includegraphics[width=9.0cm,angle=0]{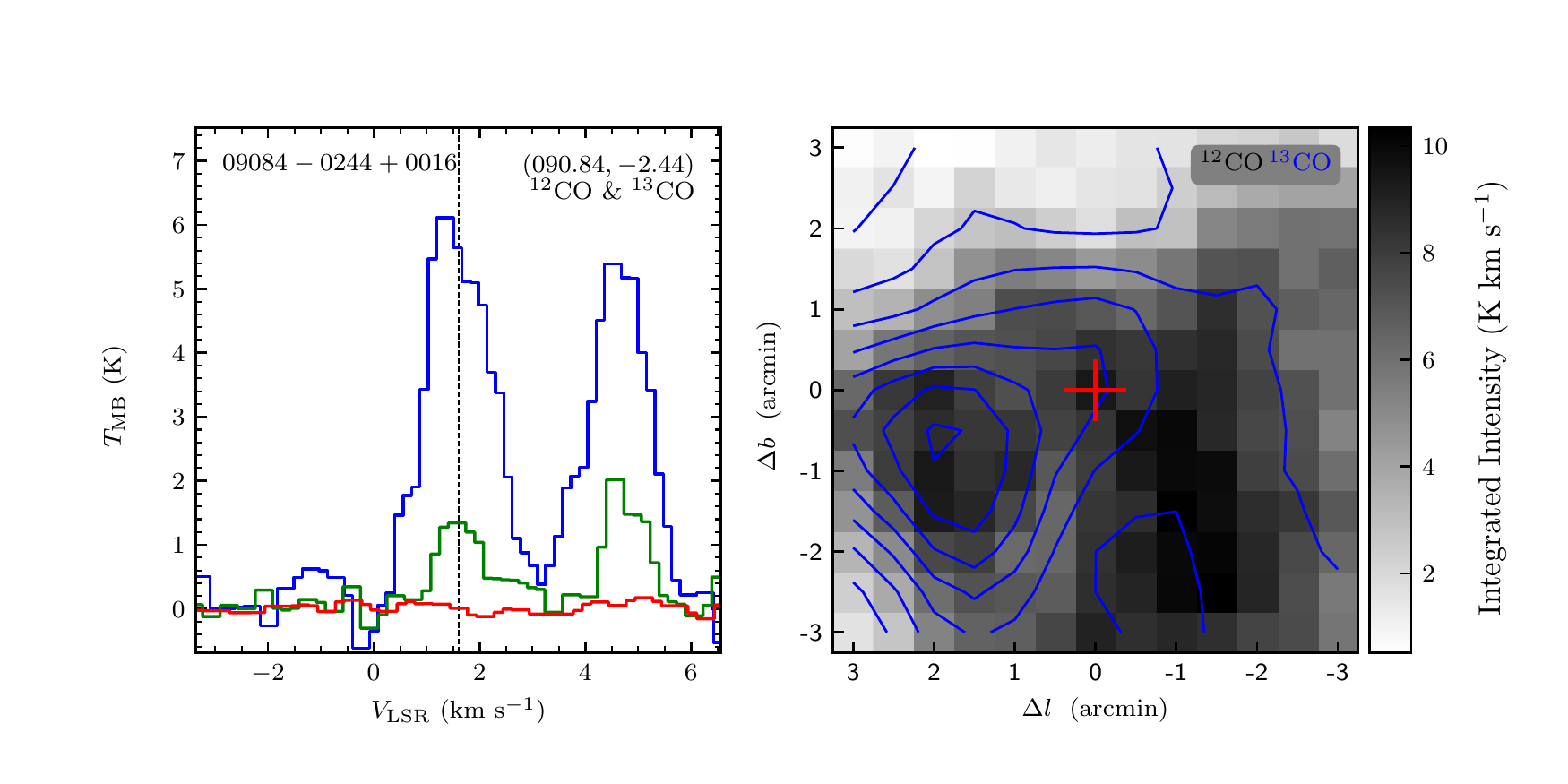}
\includegraphics[width=9.0cm,angle=0]{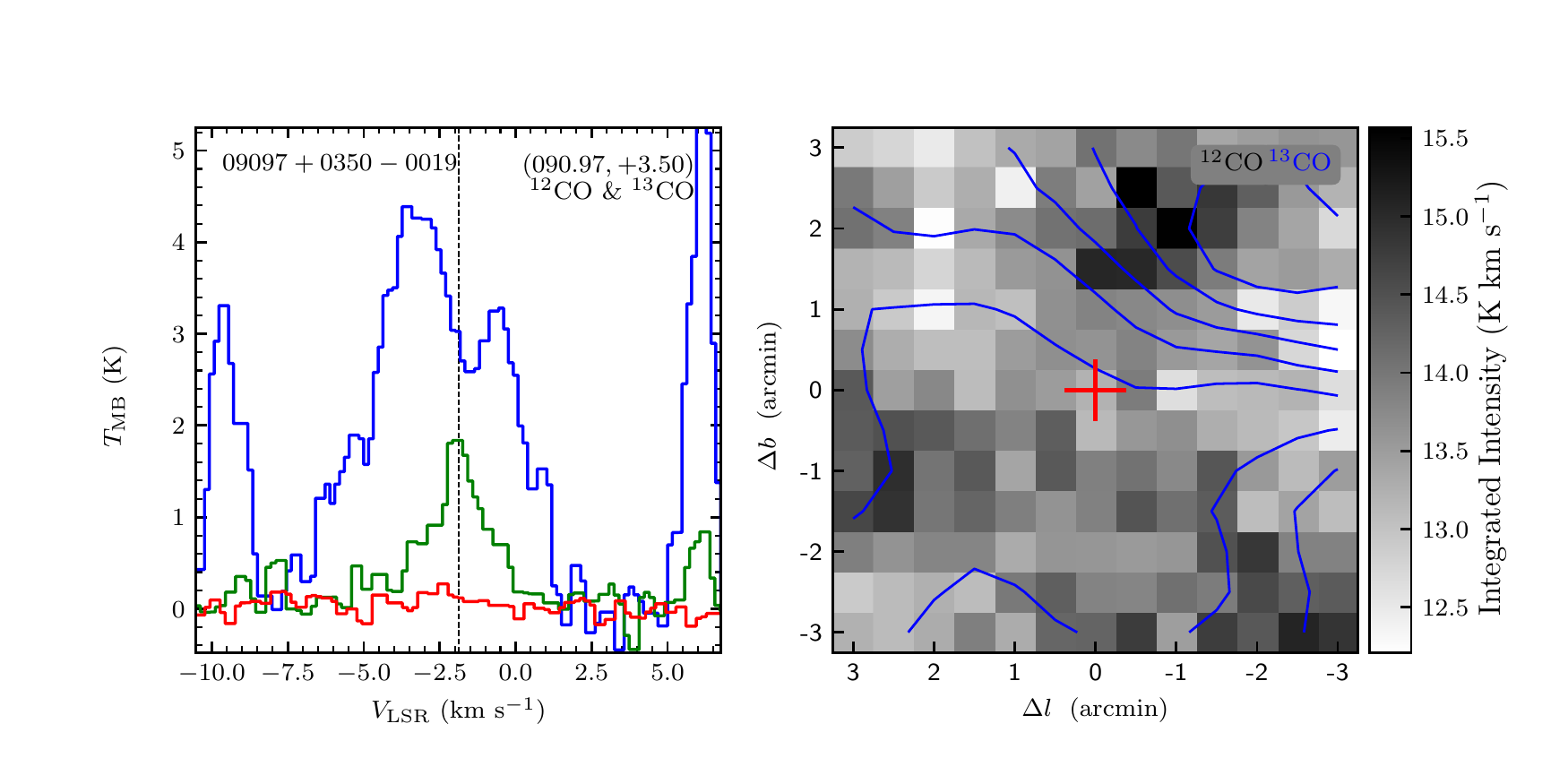}
\vspace{-0.5cm}

\includegraphics[width=9.0cm,angle=0]{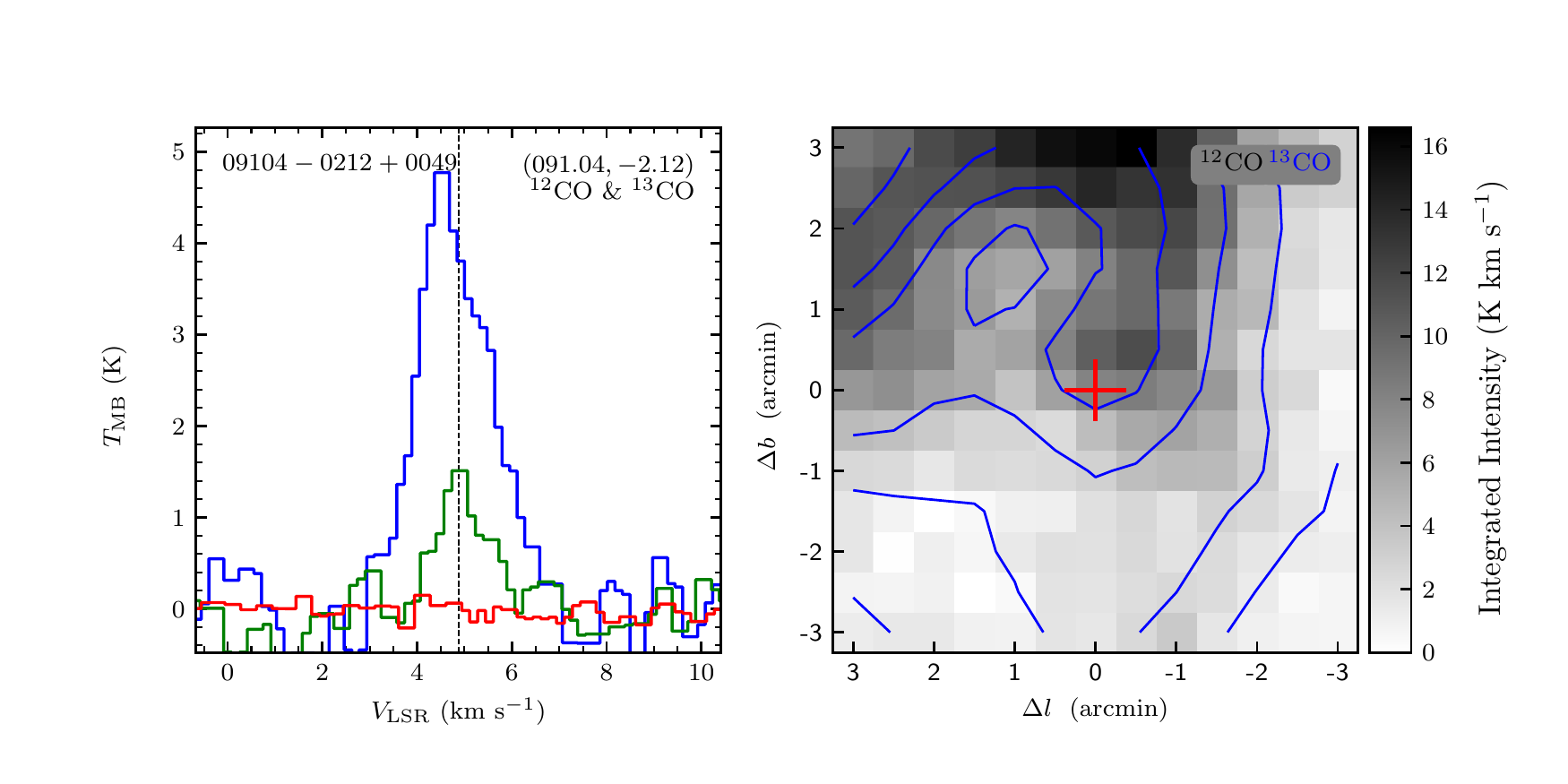}
\includegraphics[width=9.0cm,angle=0]{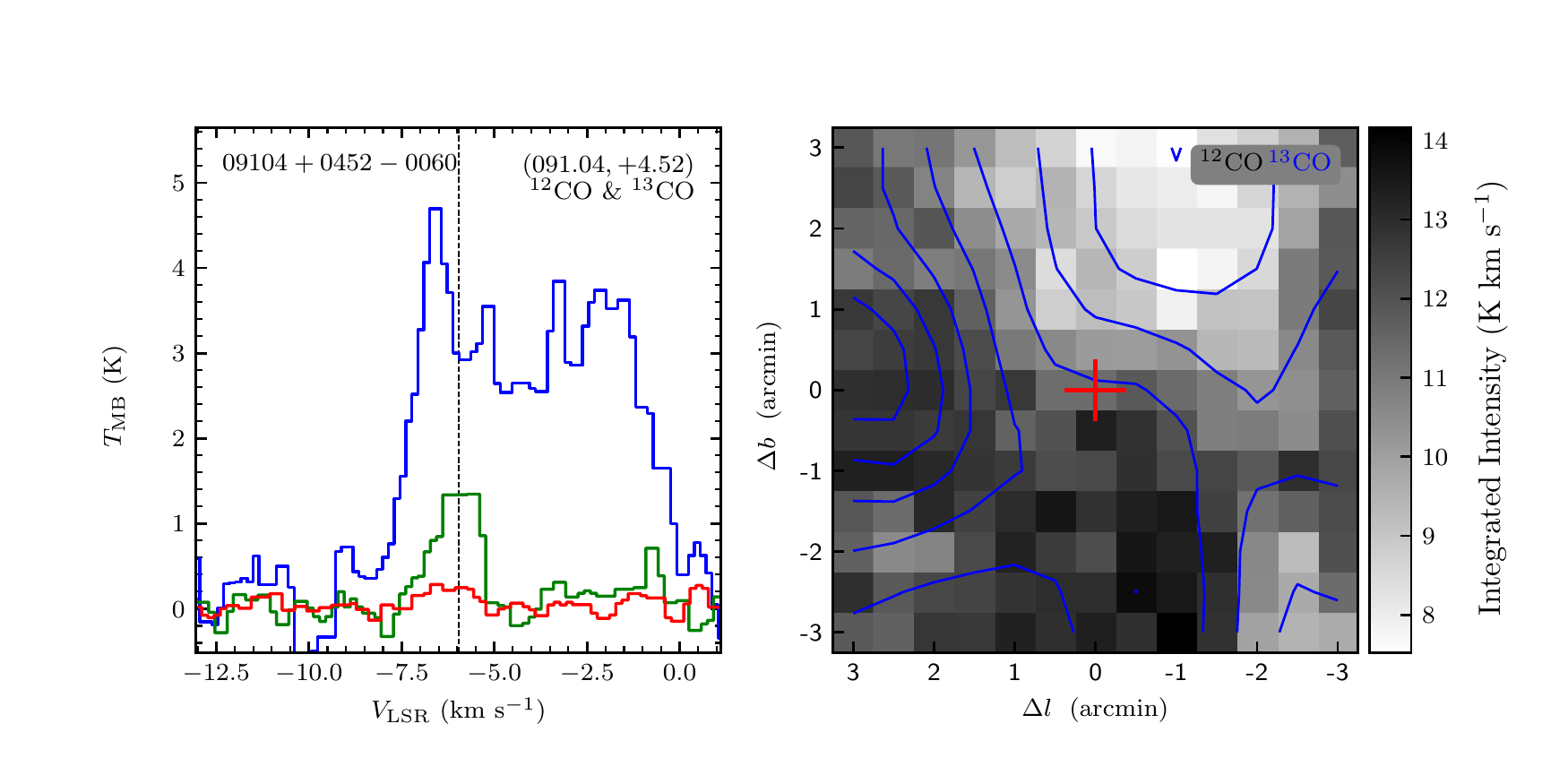}
\vspace{-0.5cm}

\includegraphics[width=9.0cm,angle=0]{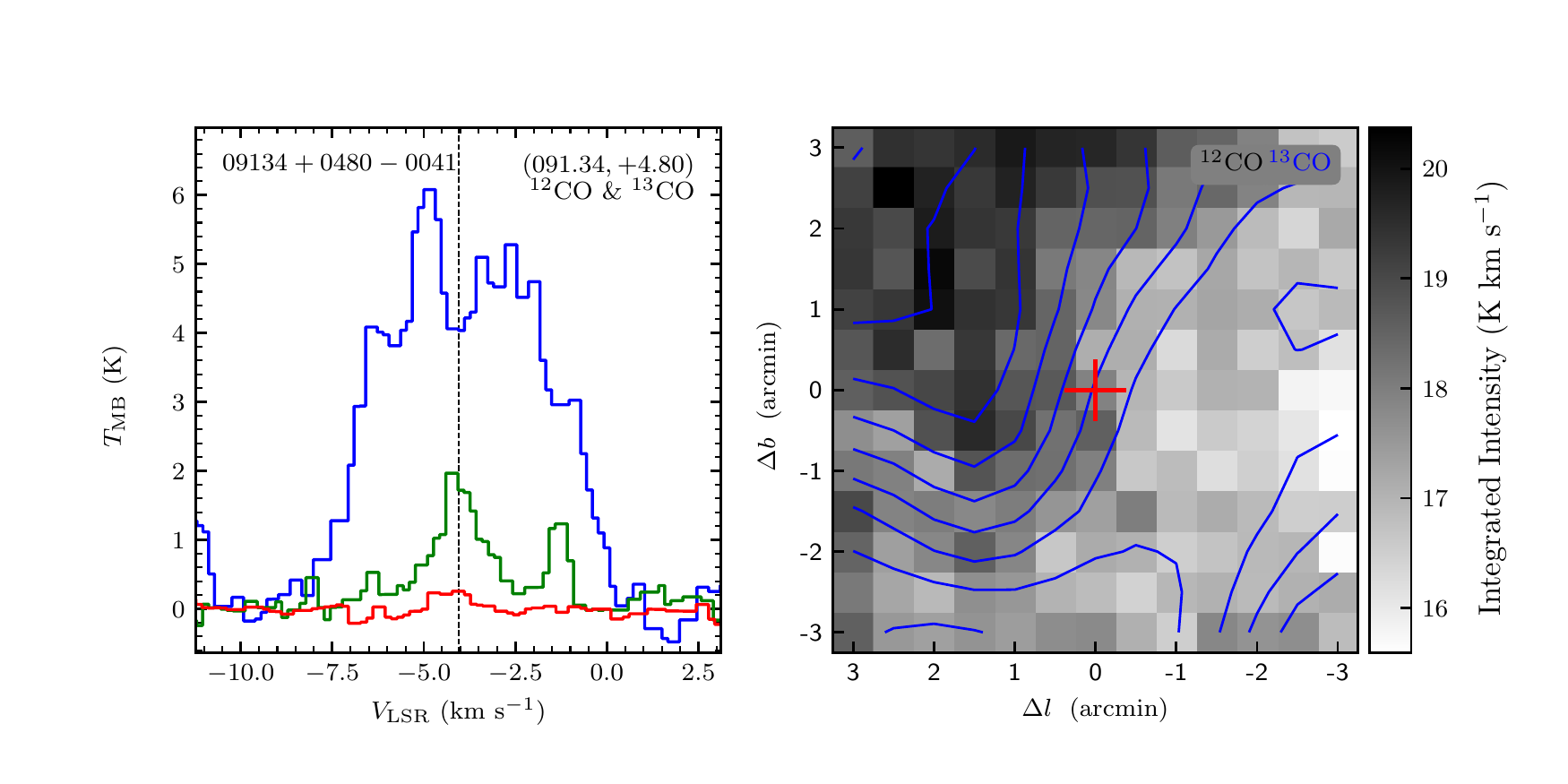}
\includegraphics[width=9.0cm,angle=0]{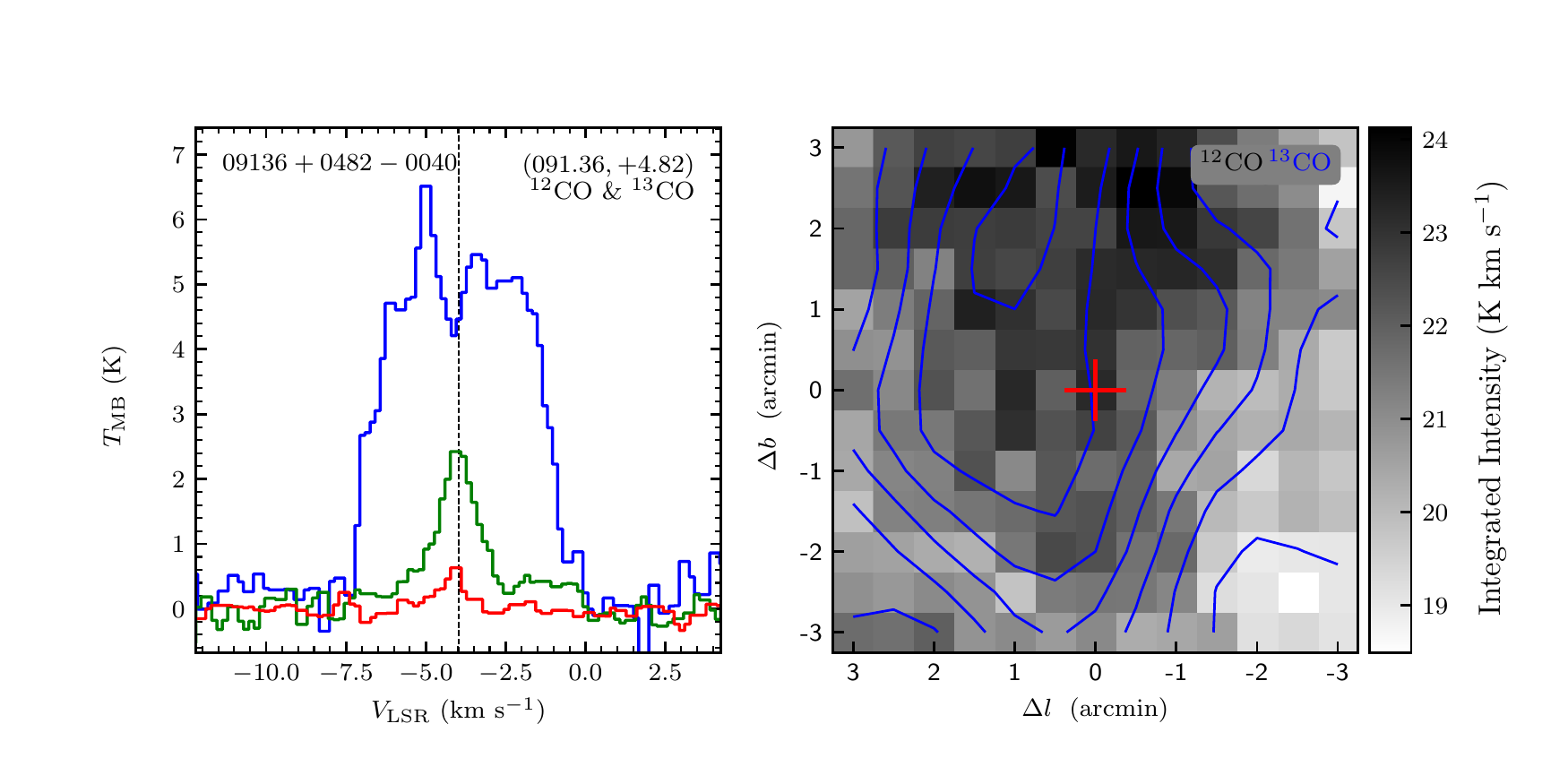}
\vspace{-0.5cm}

\includegraphics[width=9.0cm,angle=0]{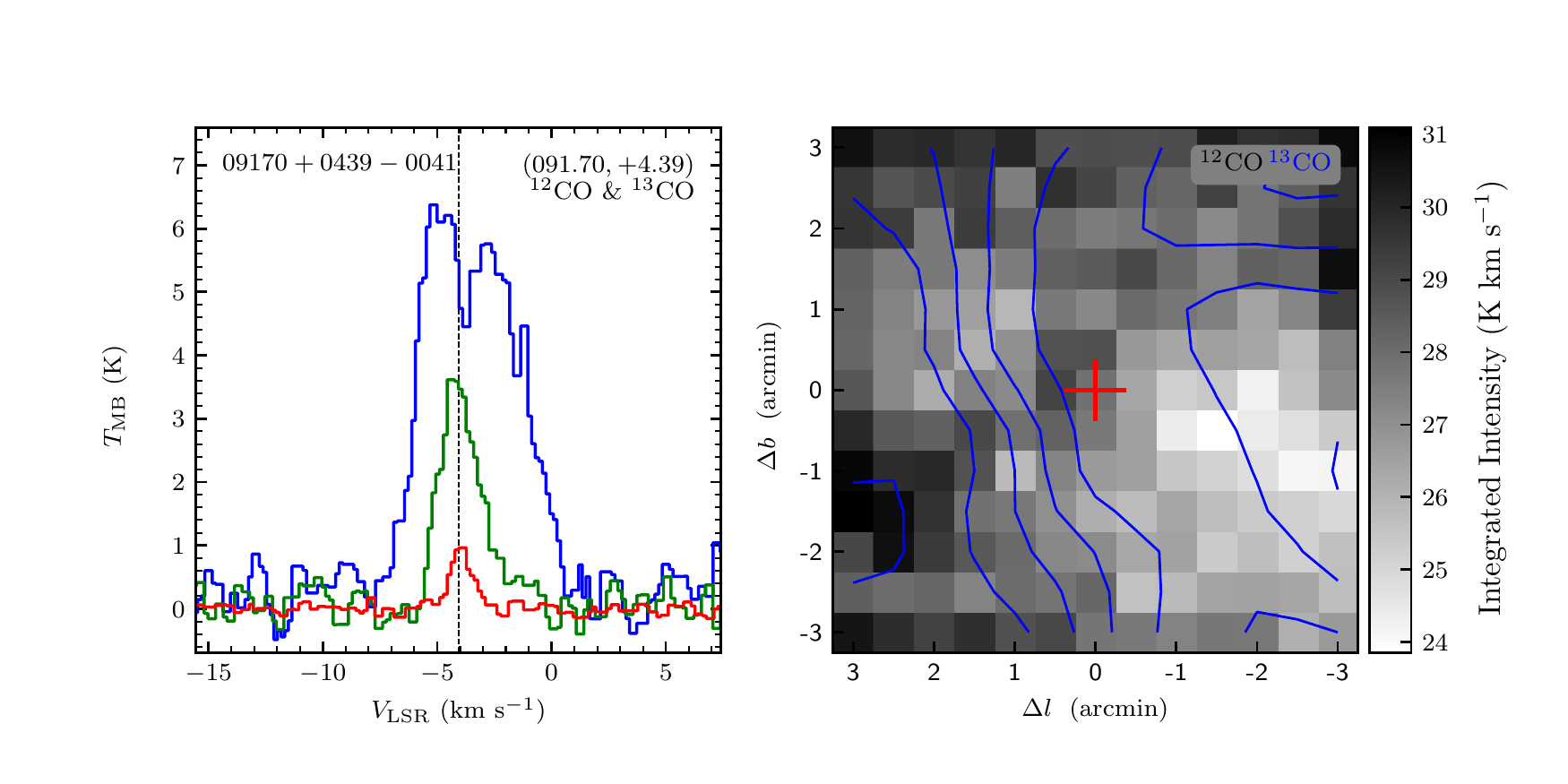}
\includegraphics[width=9.0cm,angle=0]{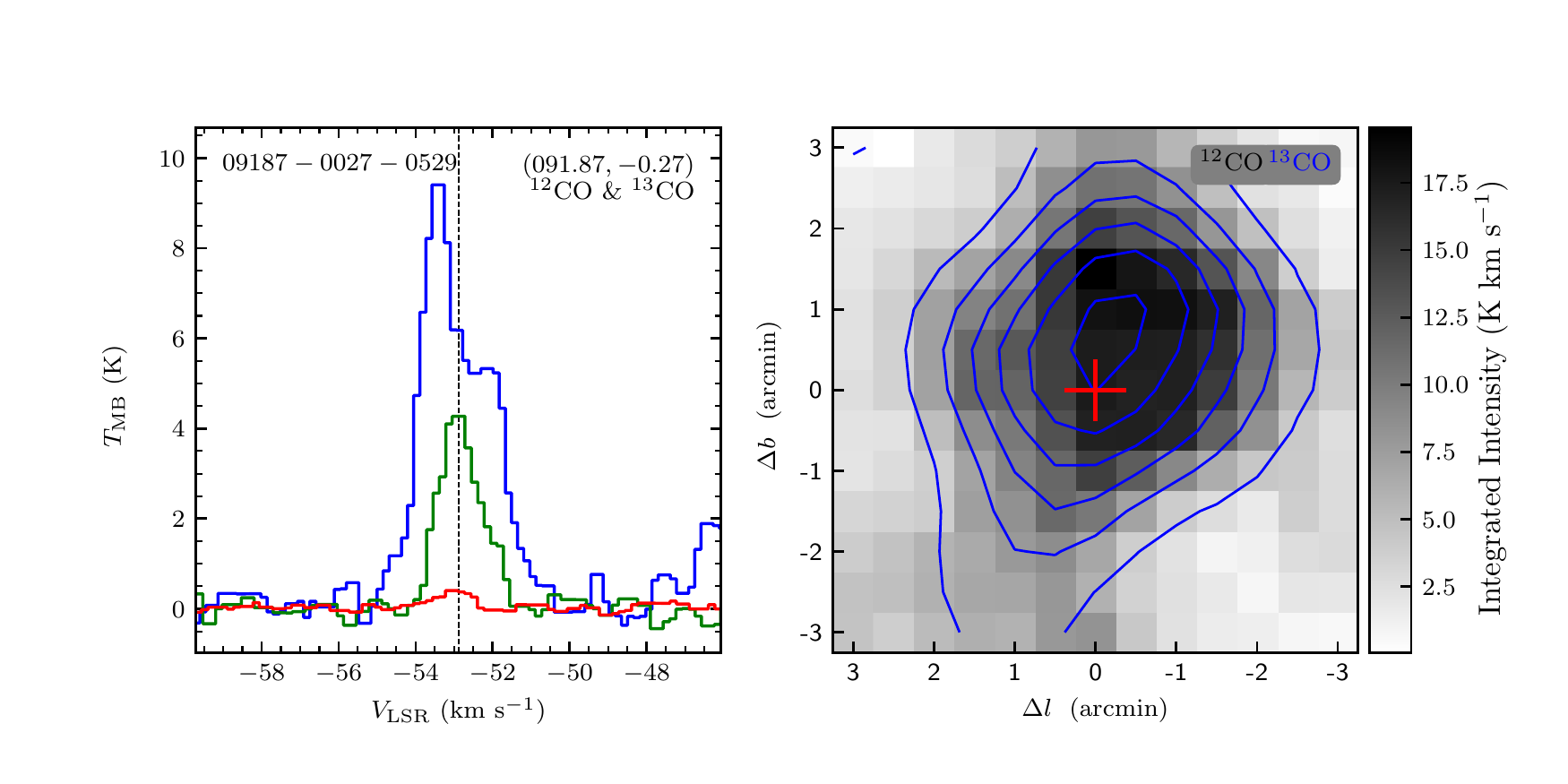}
\end{figure}
\clearpage

\begin{figure}
\includegraphics[width=9.0cm,angle=0]{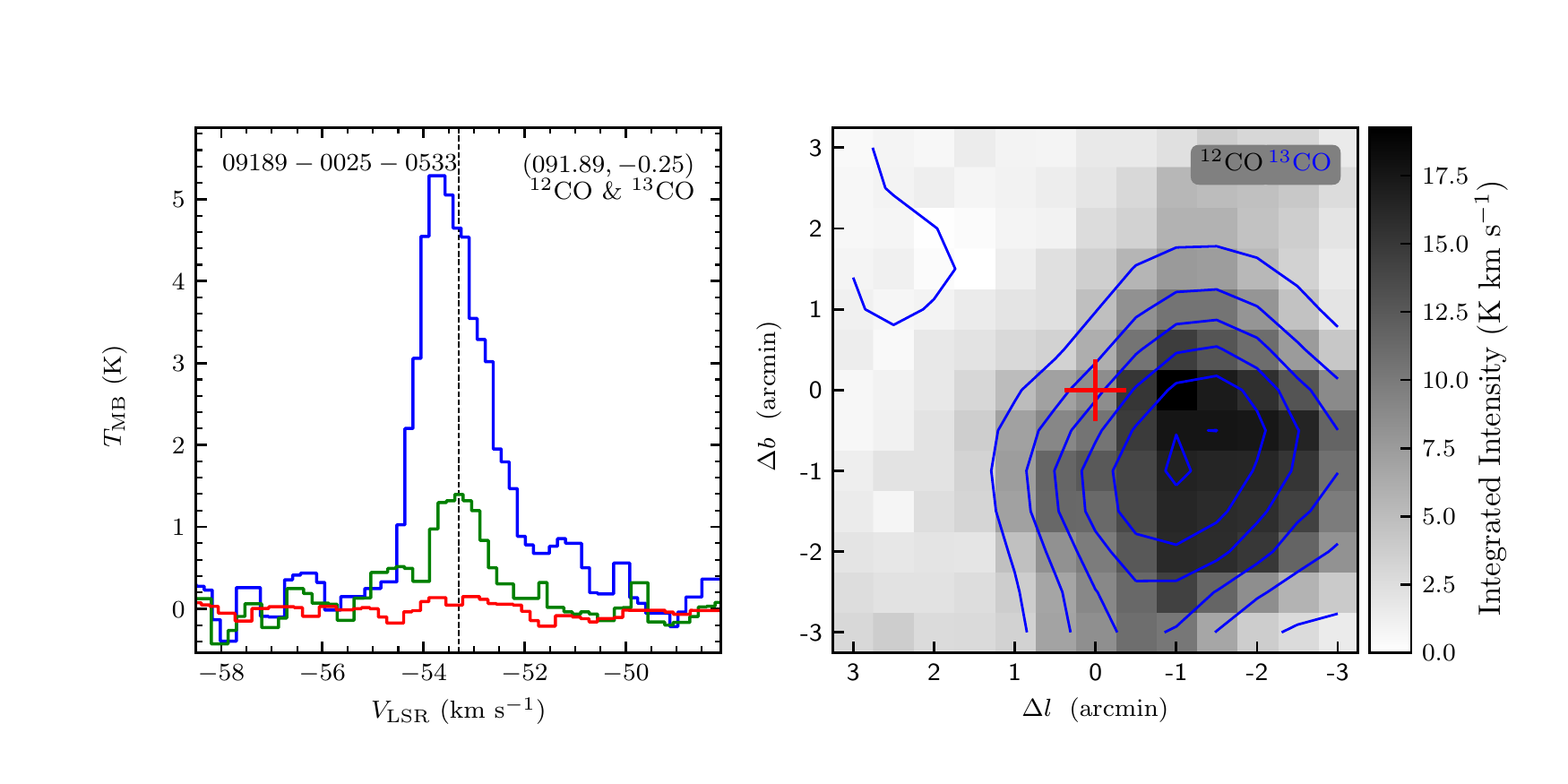}
\includegraphics[width=9.0cm,angle=0]{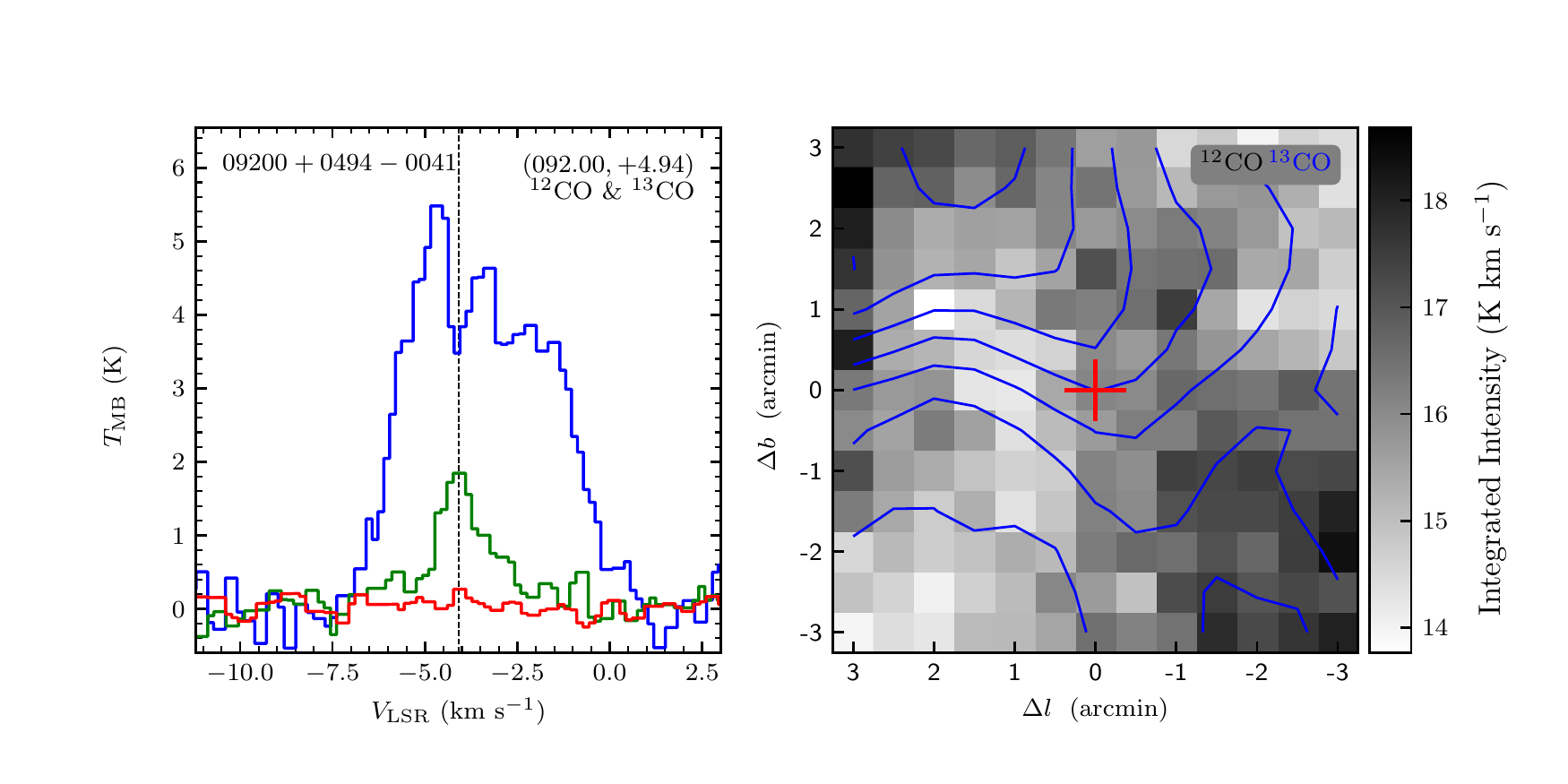}
\vspace{-0.5cm}

\includegraphics[width=9.0cm,angle=0]{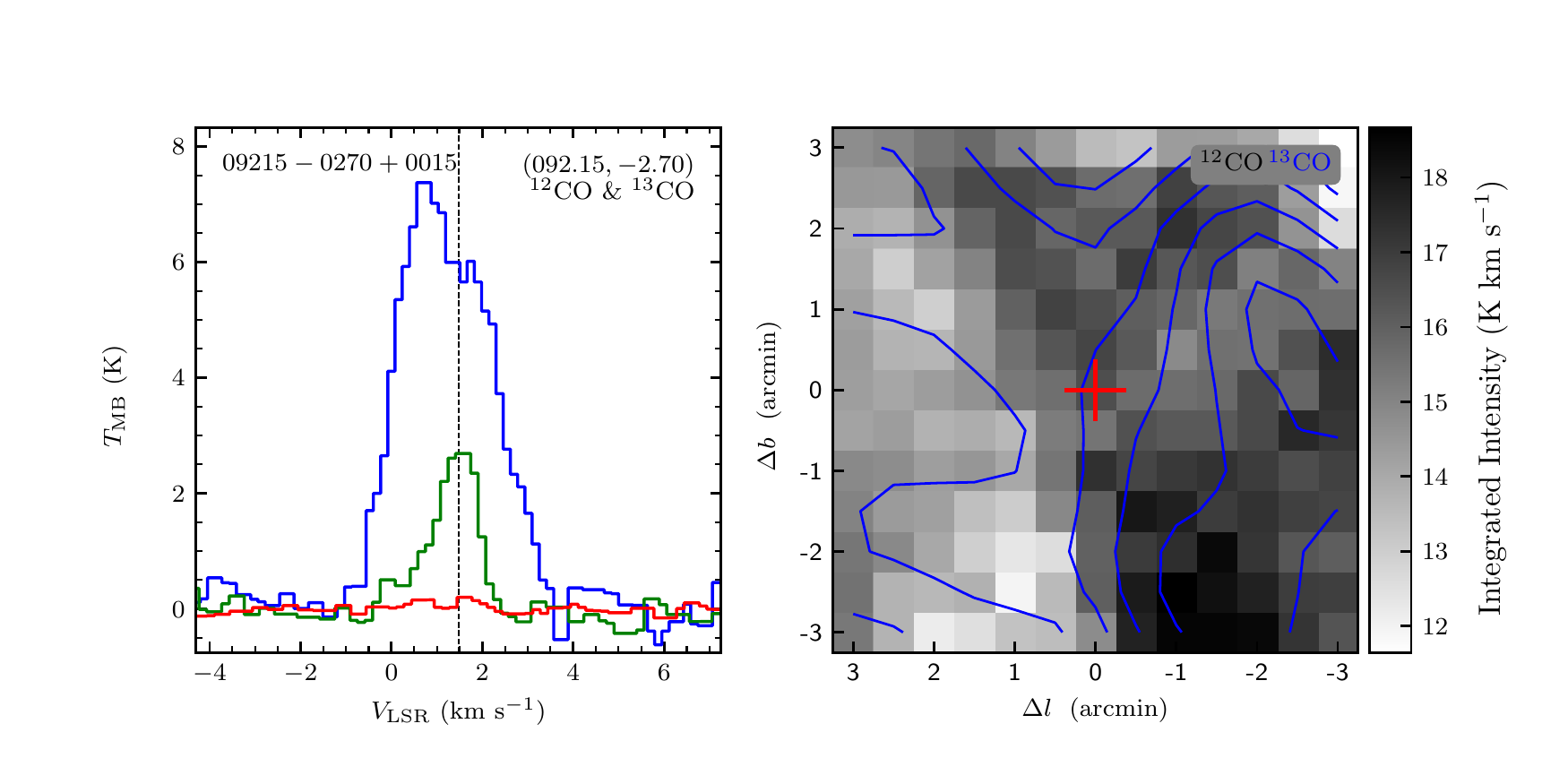}
\includegraphics[width=9.0cm,angle=0]{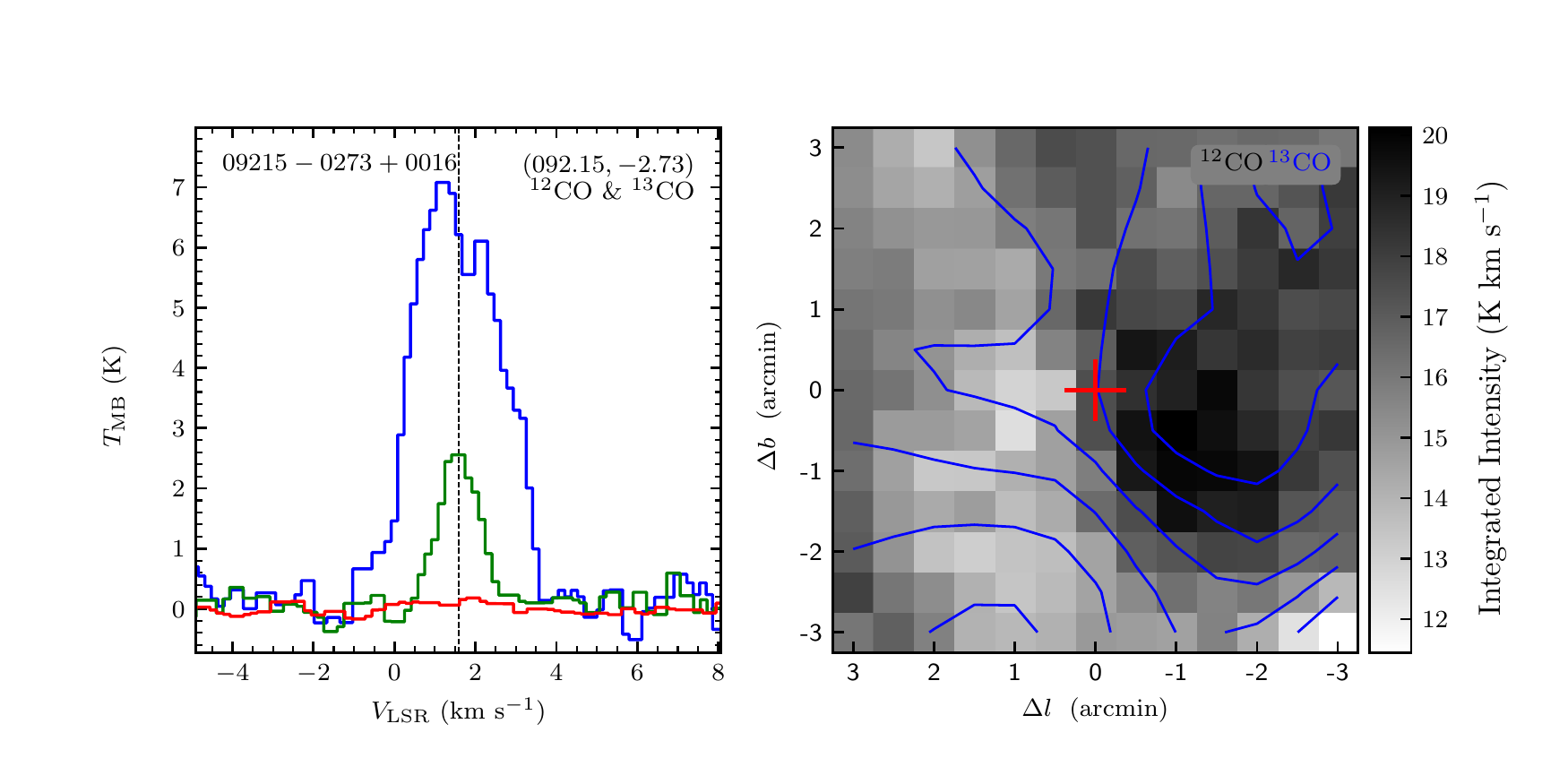}
\vspace{-0.5cm}

\includegraphics[width=9.0cm,angle=0]{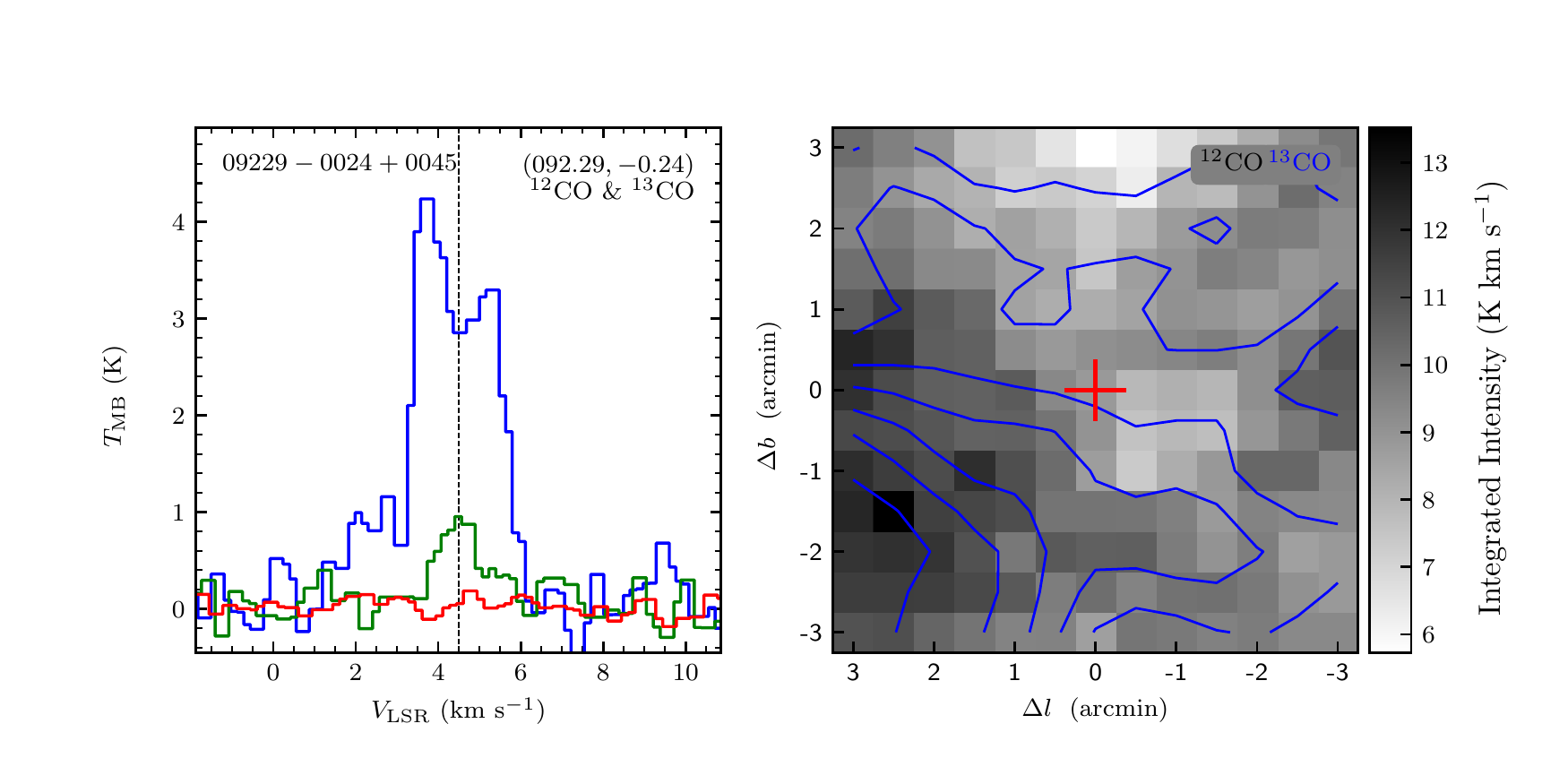}
\includegraphics[width=9.0cm,angle=0]{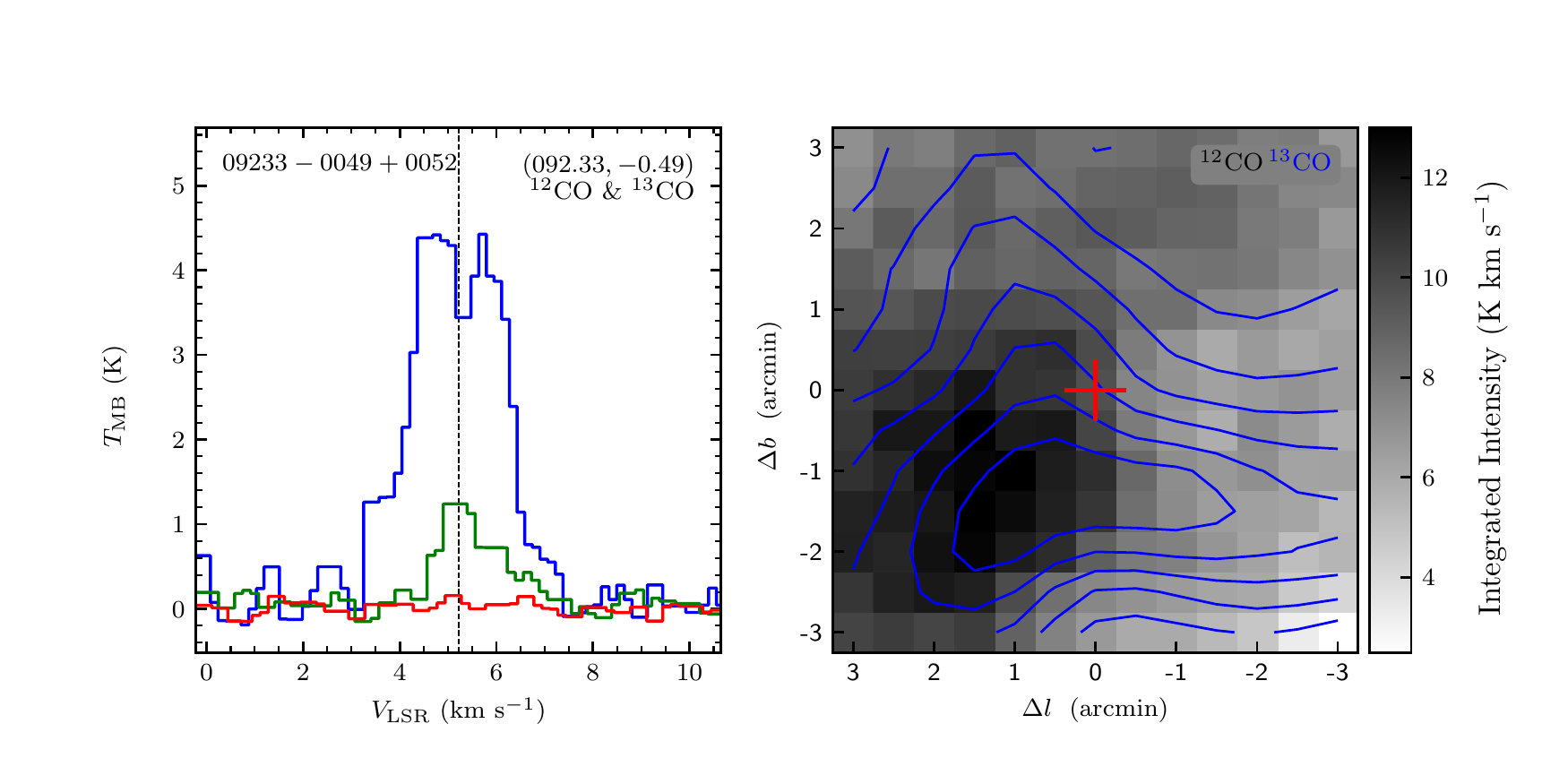}
\vspace{-0.5cm}

\includegraphics[width=9.0cm,angle=0]{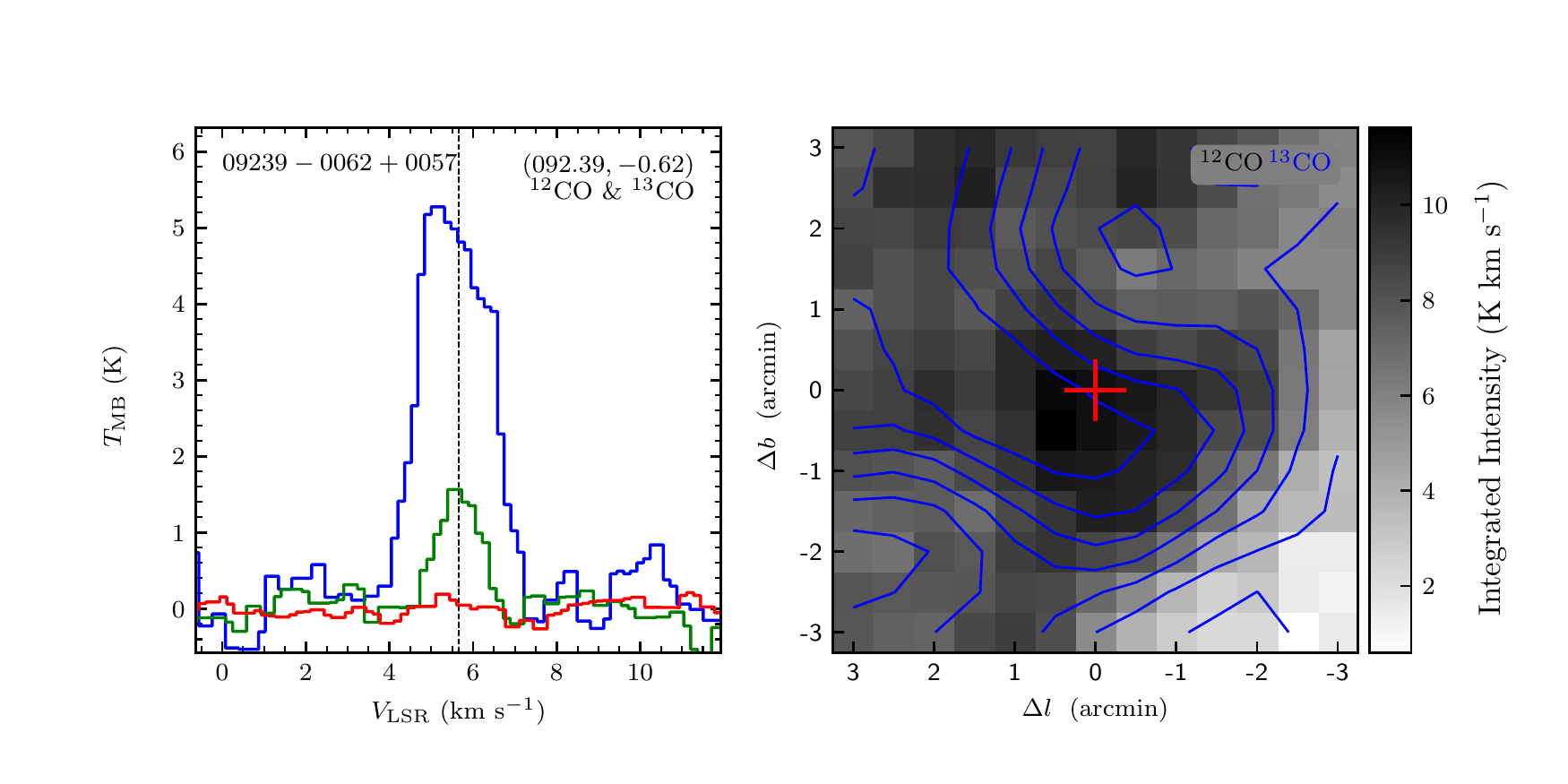}
\includegraphics[width=9.0cm,angle=0]{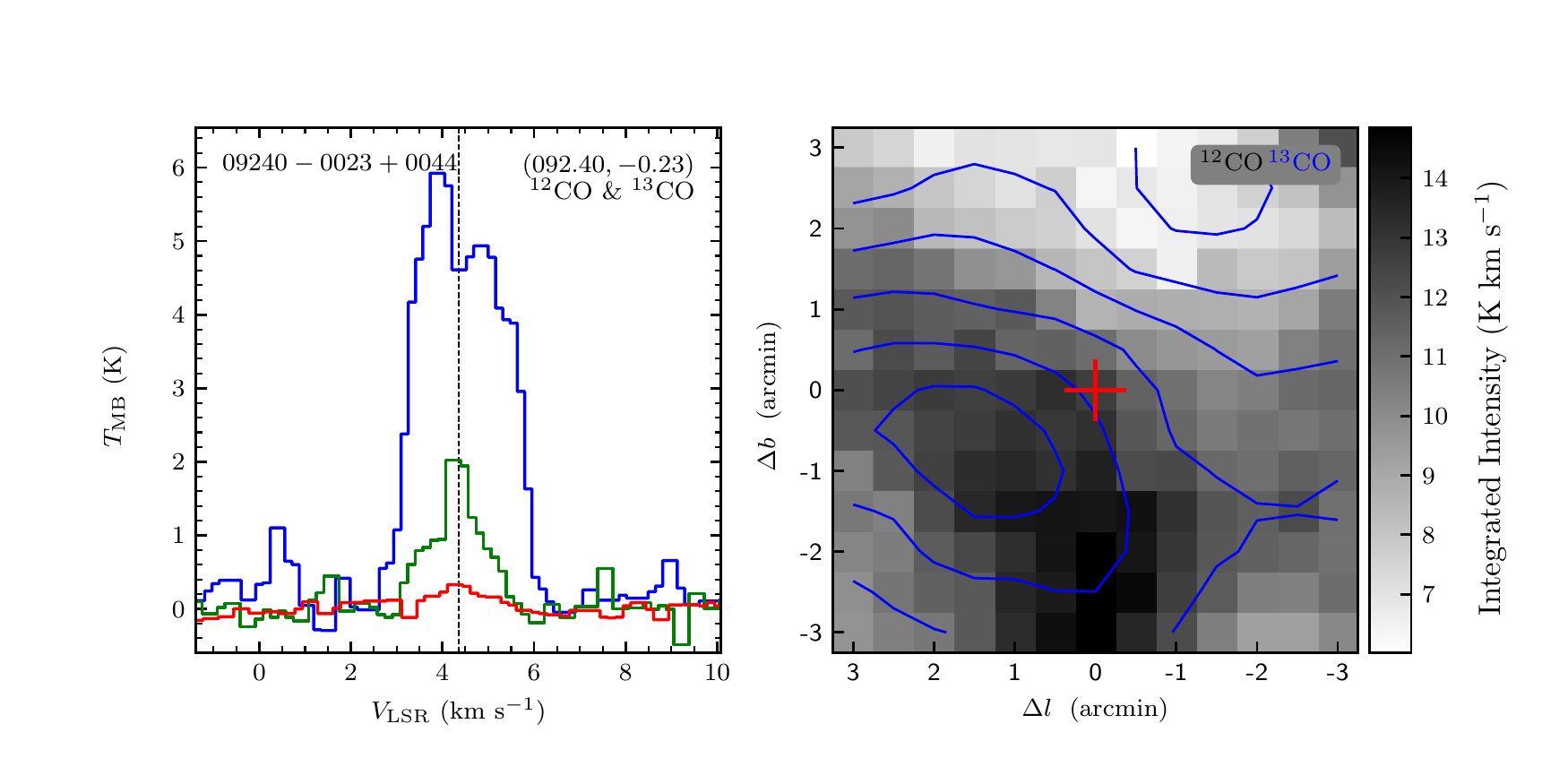}
\vspace{-0.5cm}

\includegraphics[width=9.0cm,angle=0]{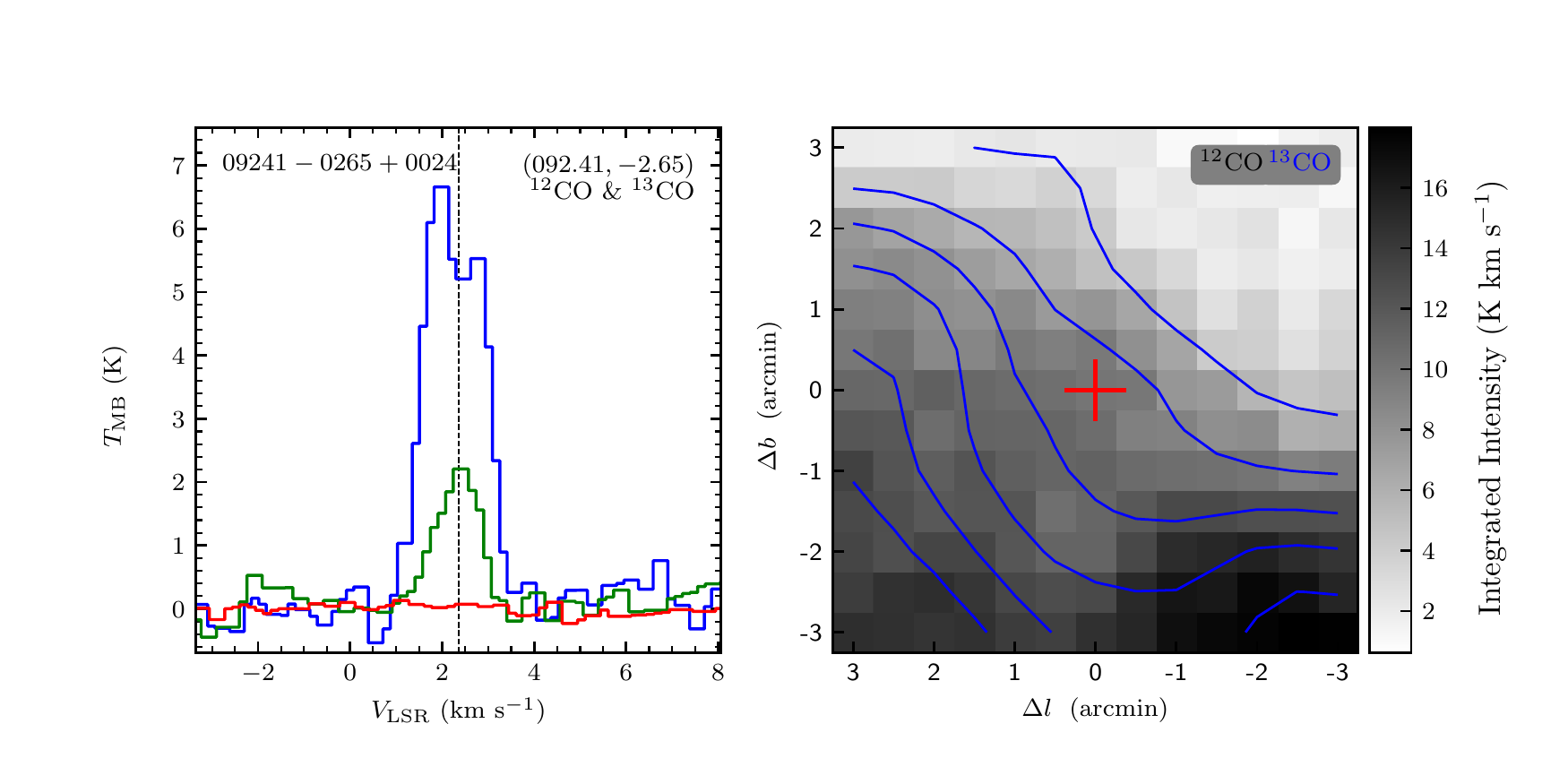}
\includegraphics[width=9.0cm,angle=0]{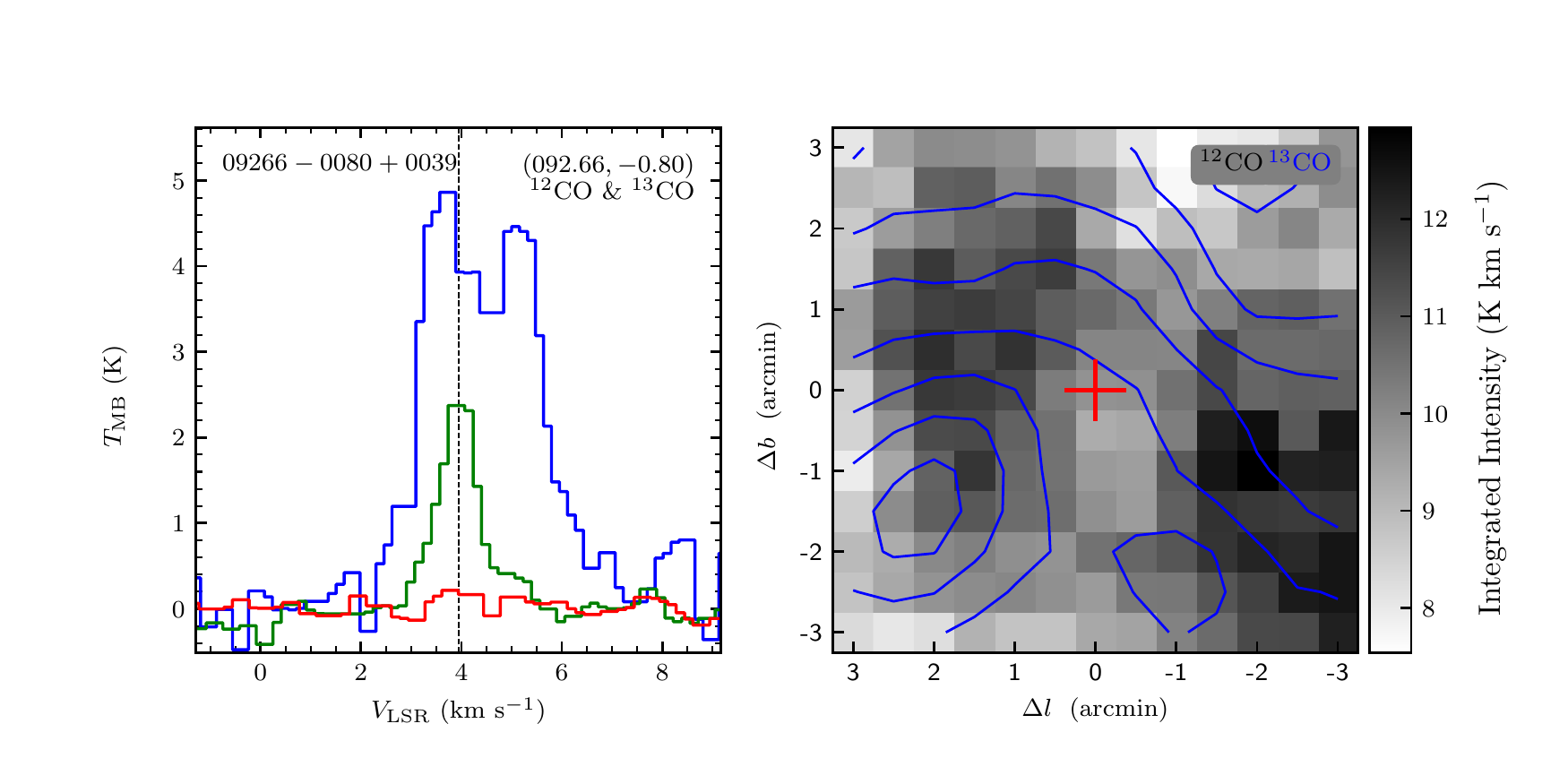}
\end{figure}
\clearpage

\begin{figure}
\includegraphics[width=9.0cm,angle=0]{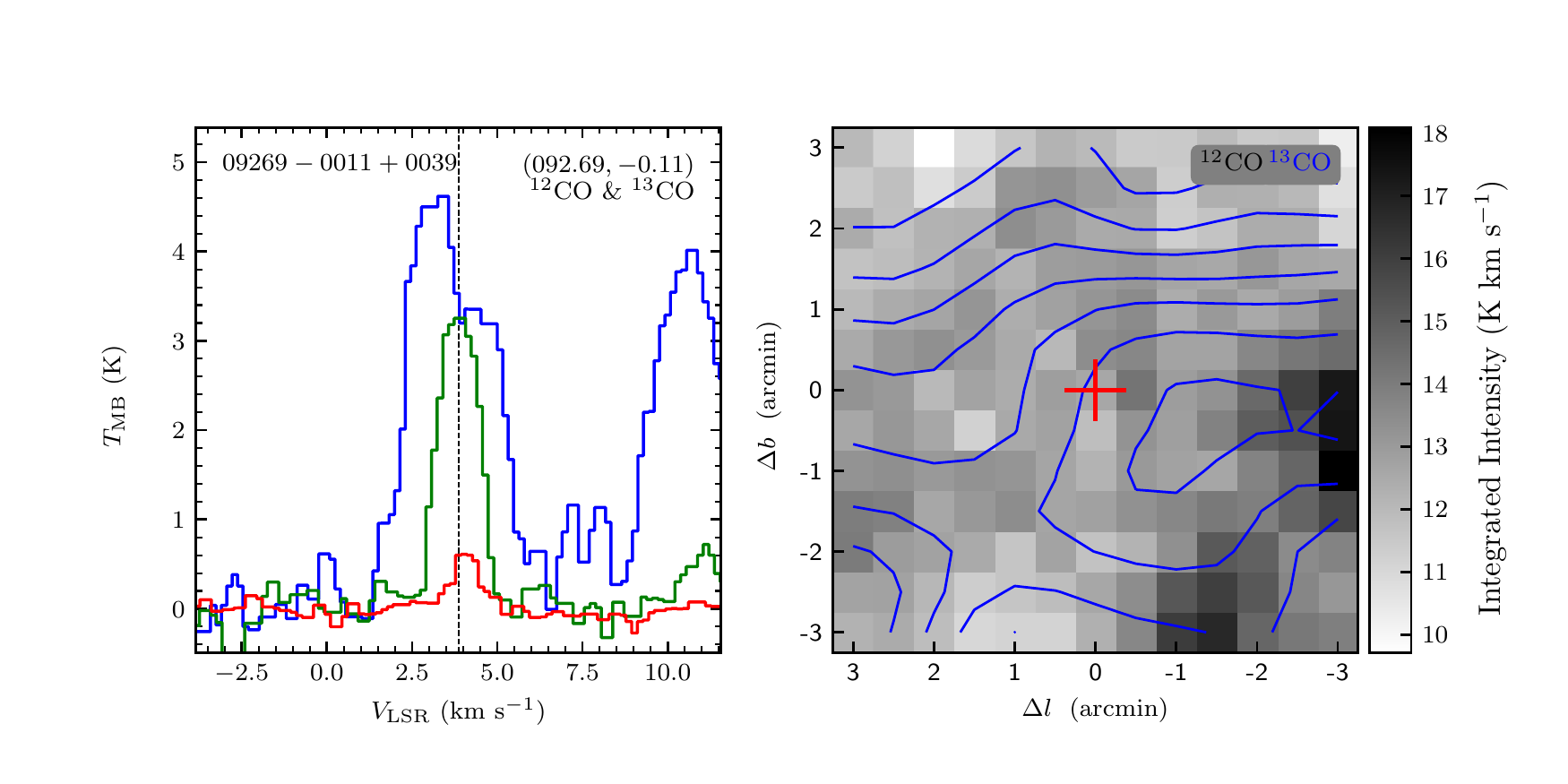}
\includegraphics[width=9.0cm,angle=0]{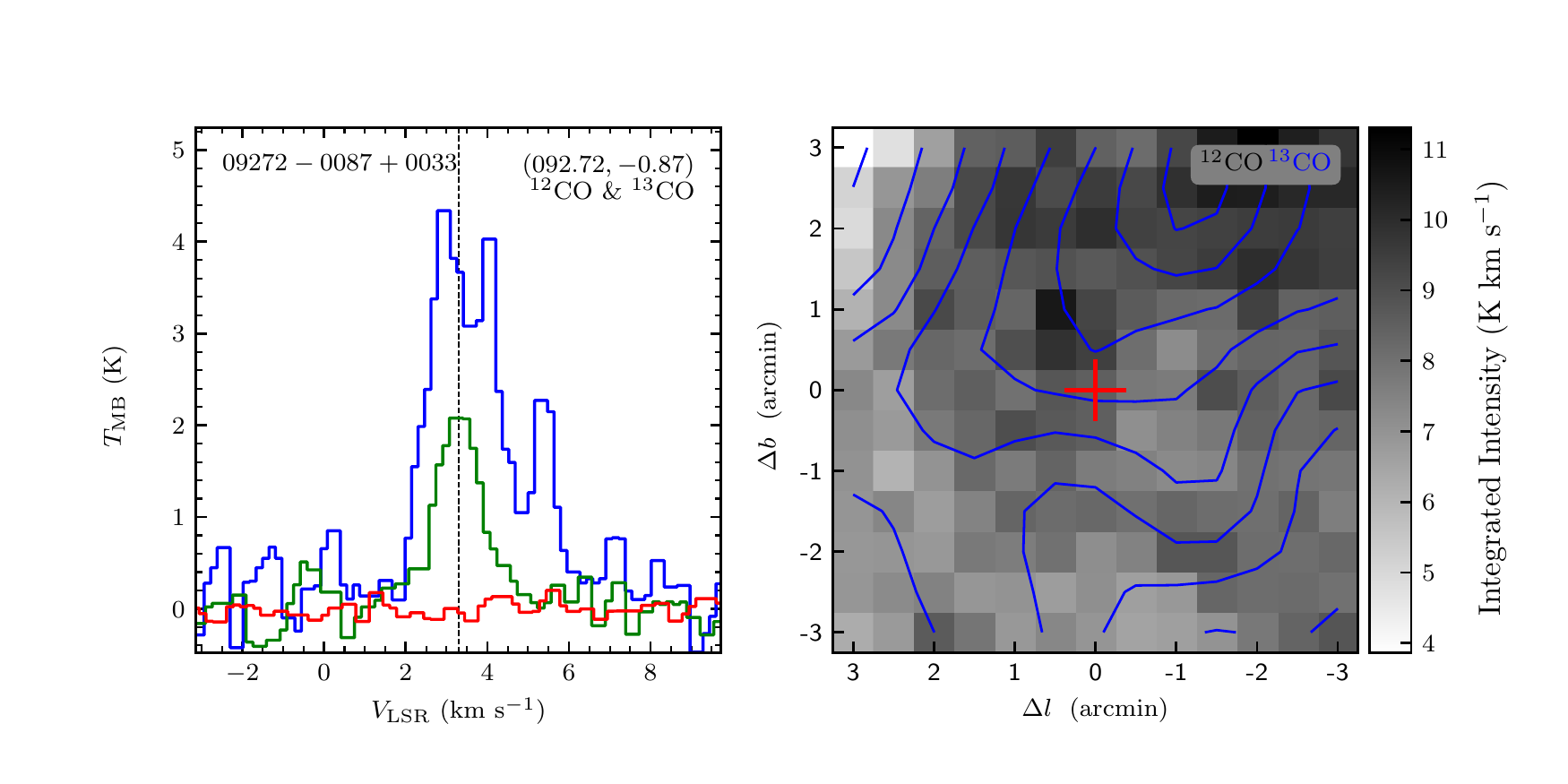}
\vspace{-0.5cm}

\includegraphics[width=9.0cm,angle=0]{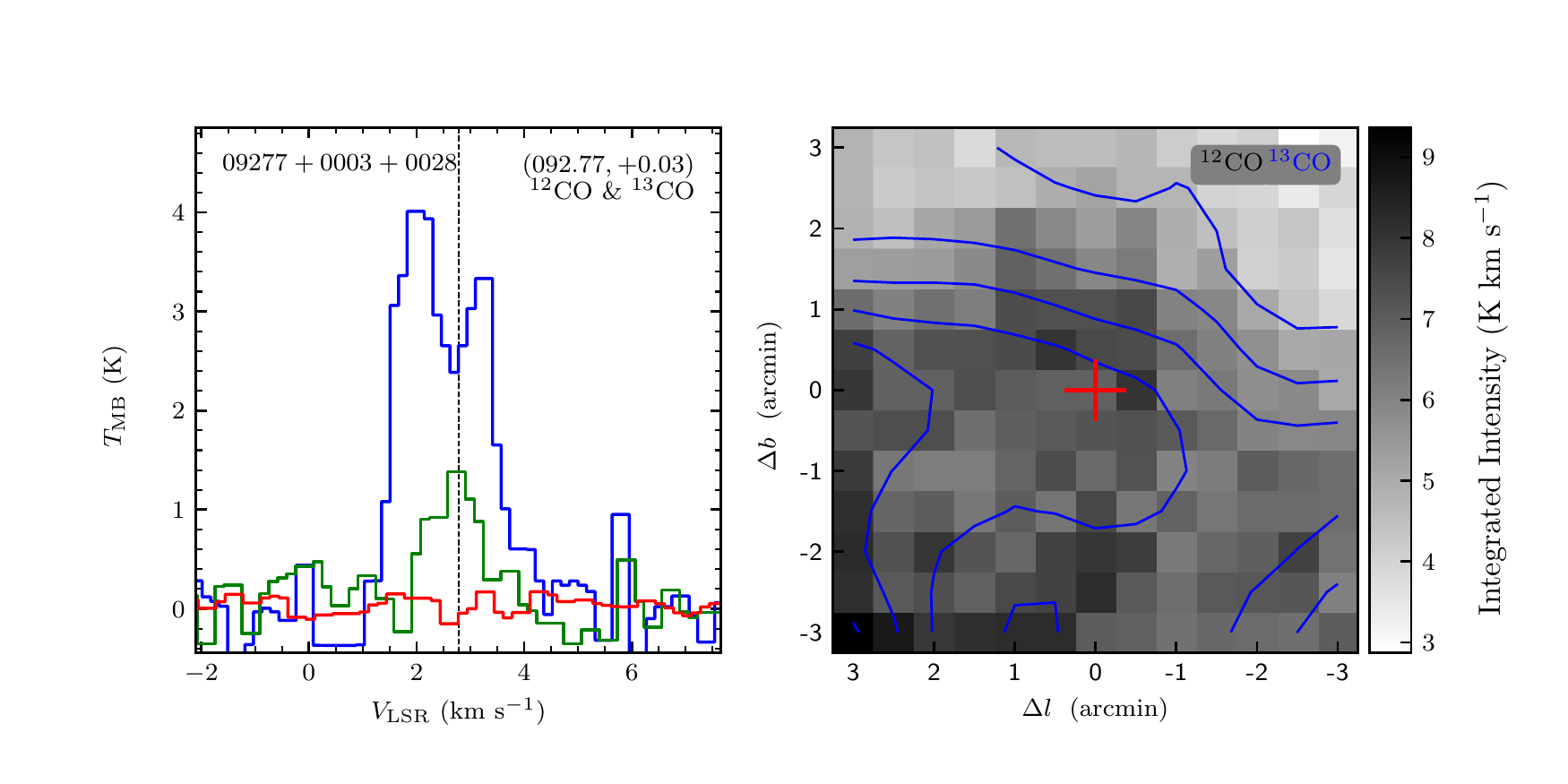}
\includegraphics[width=9.0cm,angle=0]{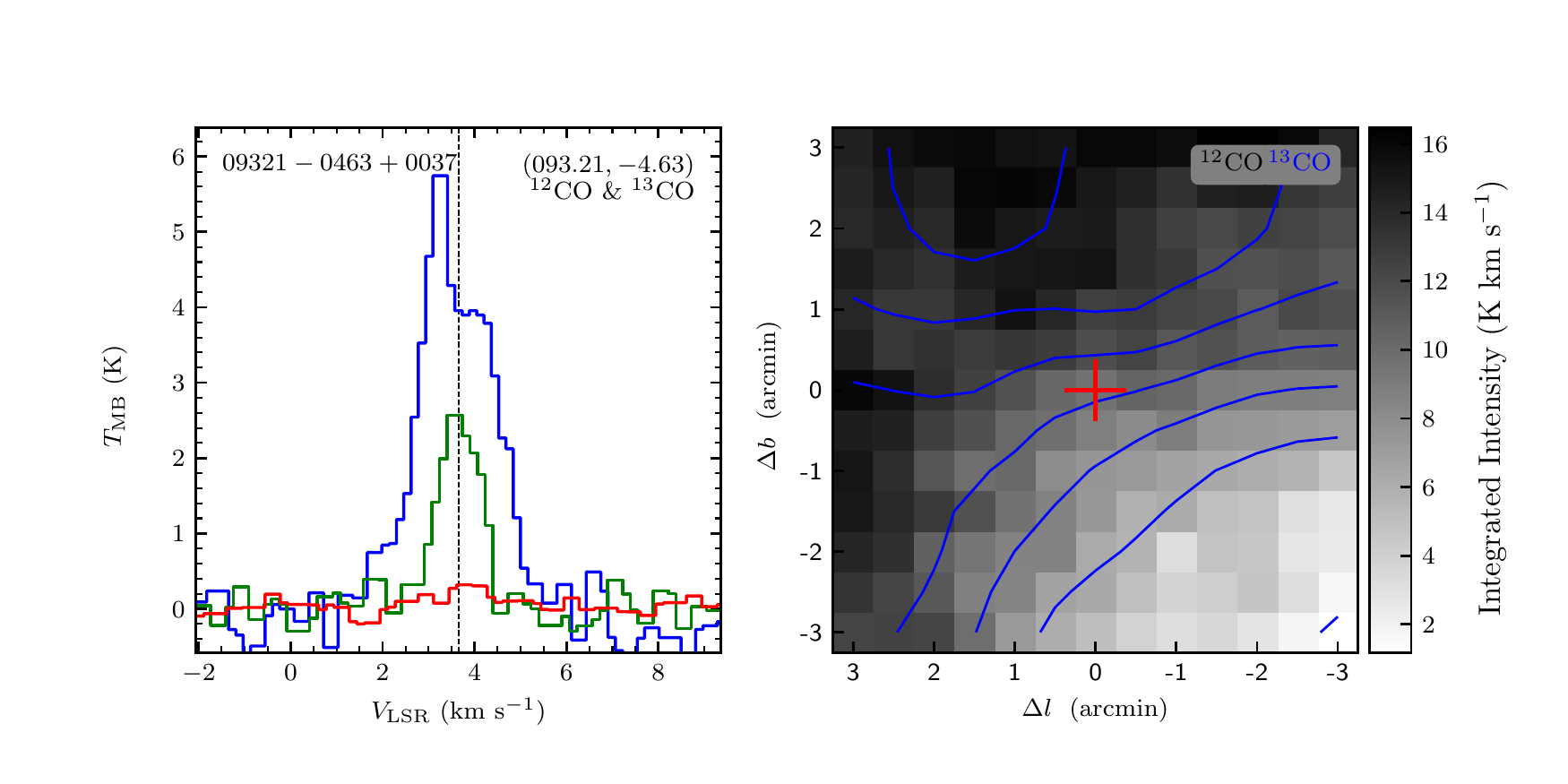}
\vspace{-0.5cm}

\includegraphics[width=9.0cm,angle=0]{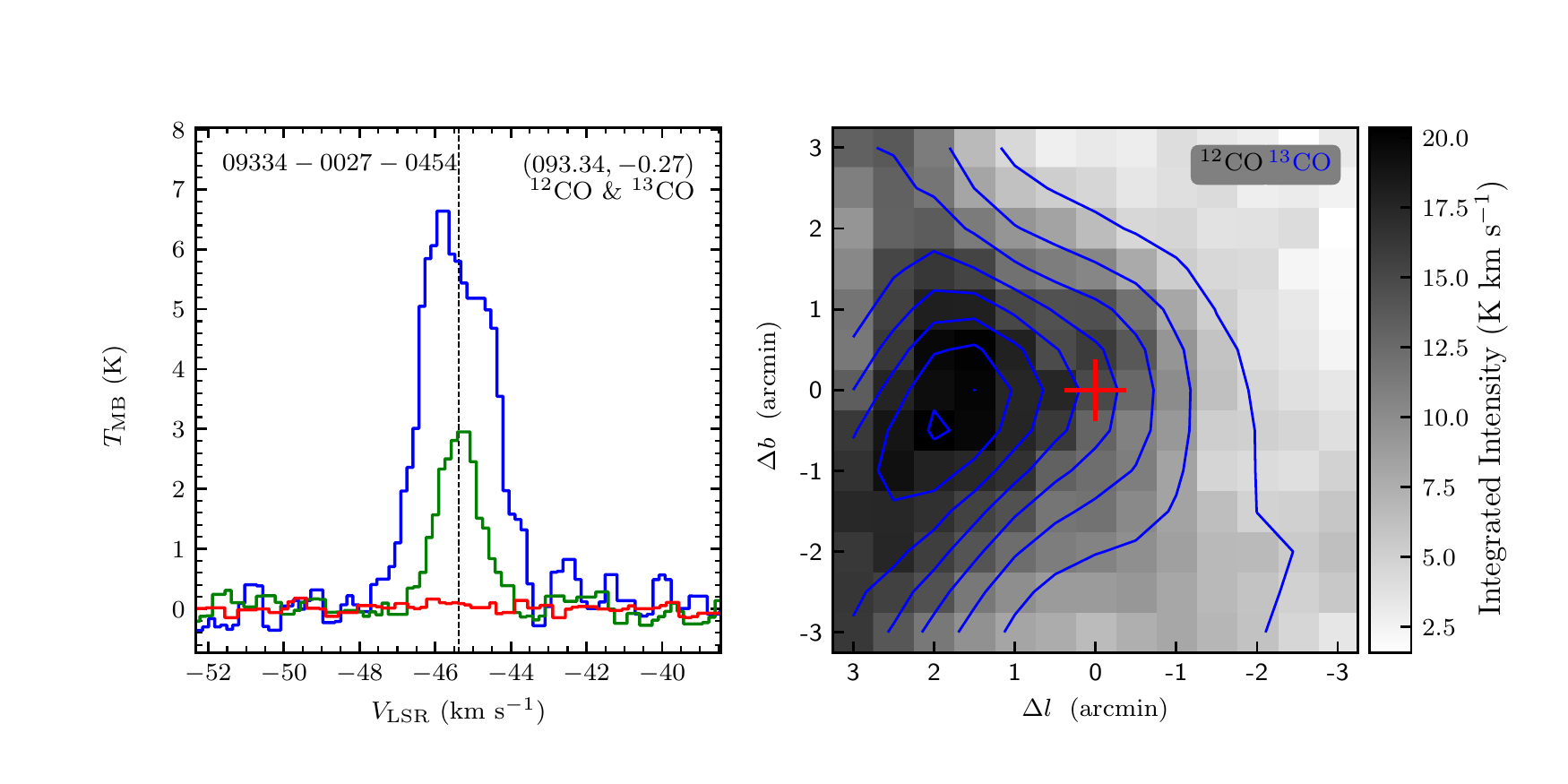}
\includegraphics[width=9.0cm,angle=0]{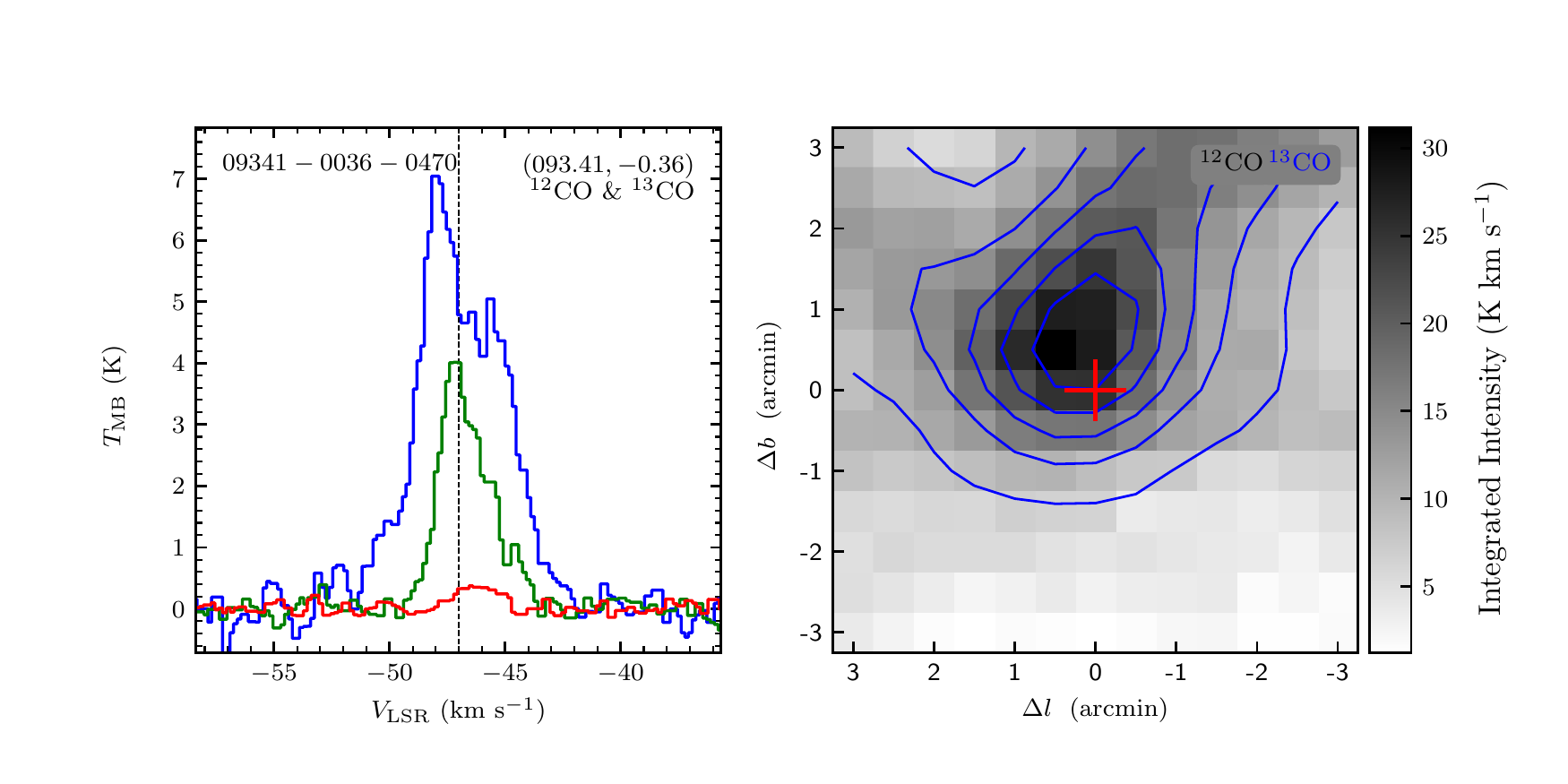}
\vspace{-0.5cm}

\includegraphics[width=9.0cm,angle=0]{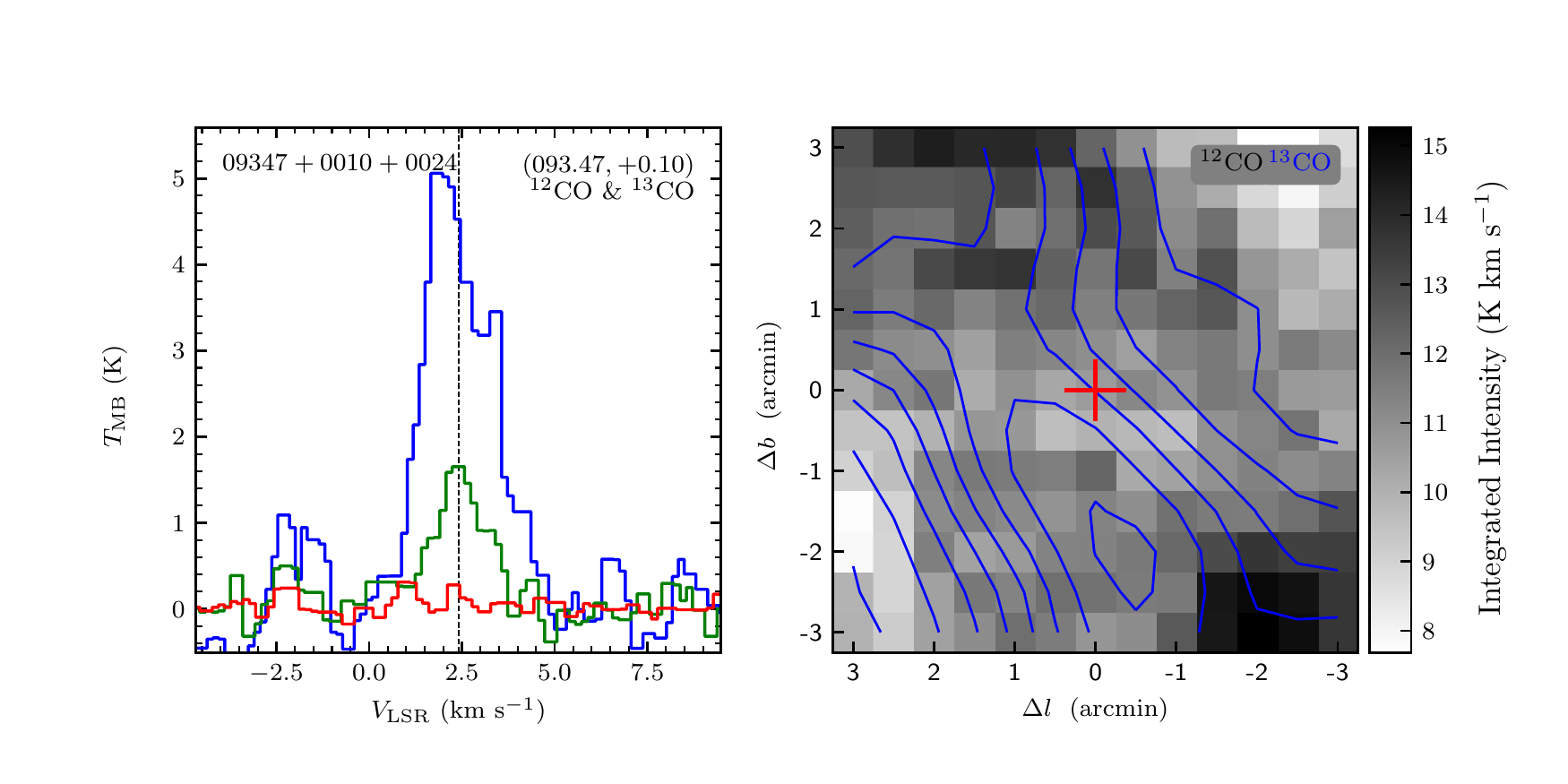}
\includegraphics[width=9.0cm,angle=0]{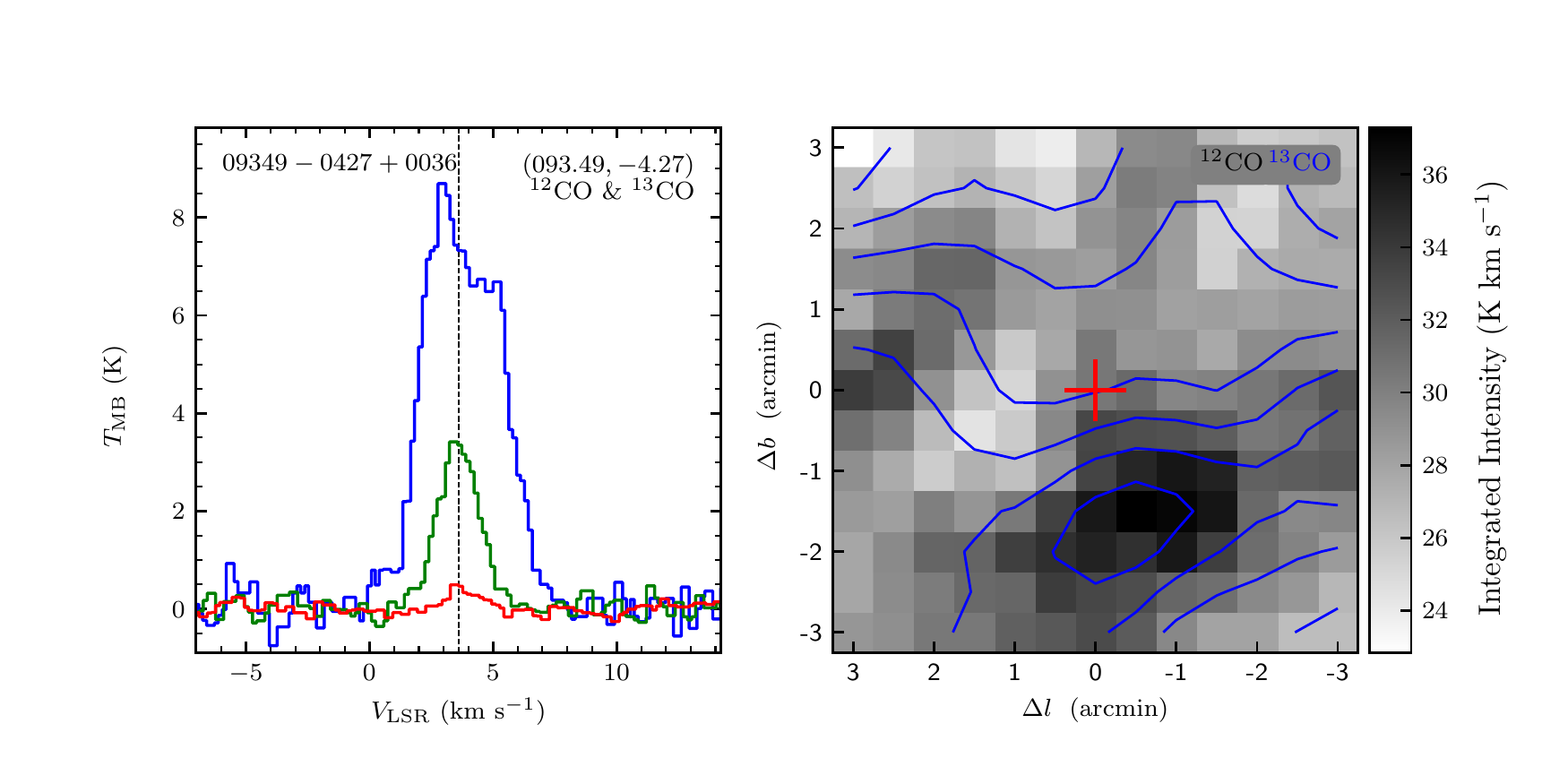}
\vspace{-0.5cm}

\includegraphics[width=9.0cm,angle=0]{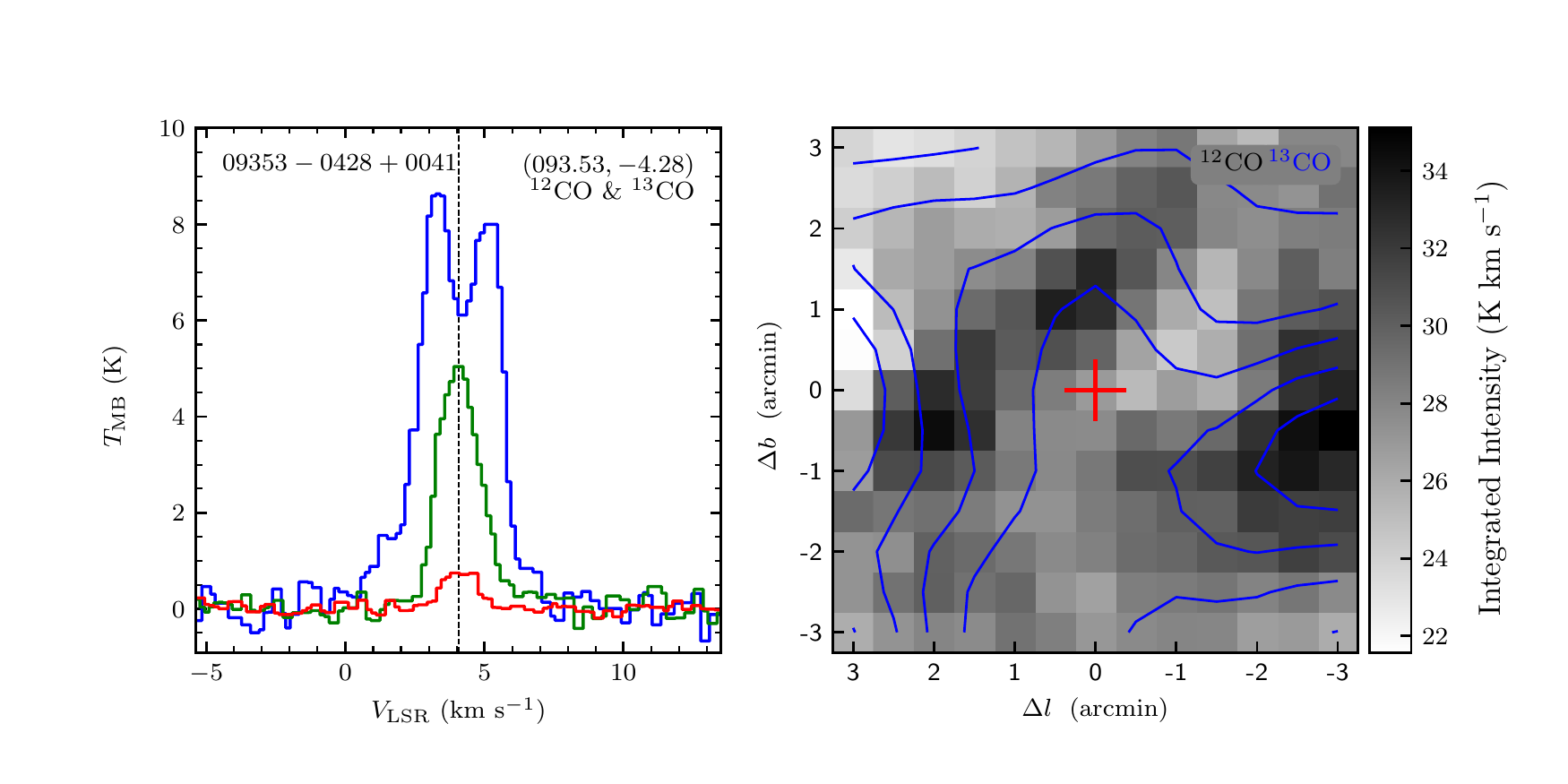}
\includegraphics[width=9.0cm,angle=0]{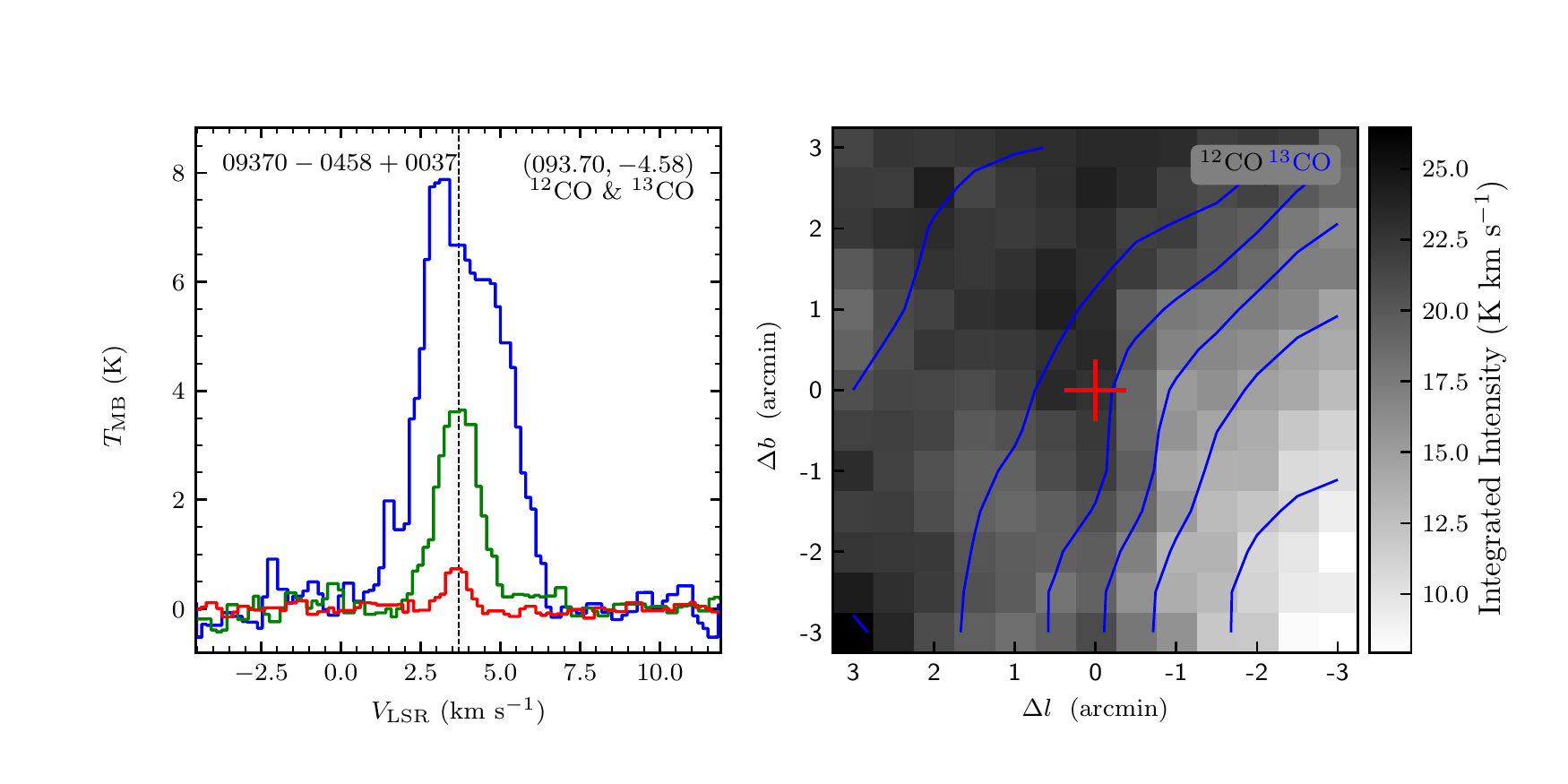}
\end{figure}
\clearpage

\begin{figure}
\includegraphics[width=9.0cm,angle=0]{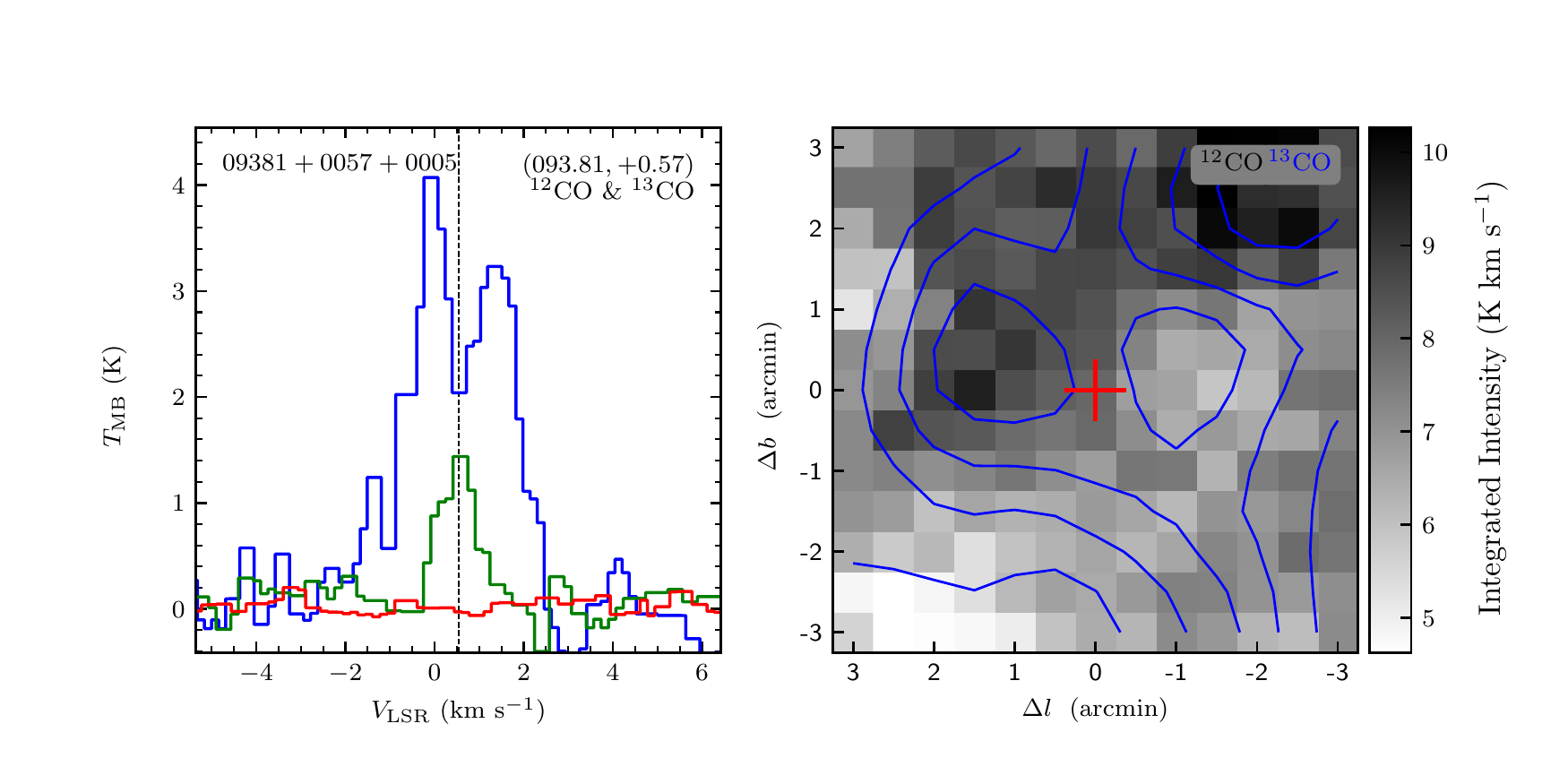}
\includegraphics[width=9.0cm,angle=0]{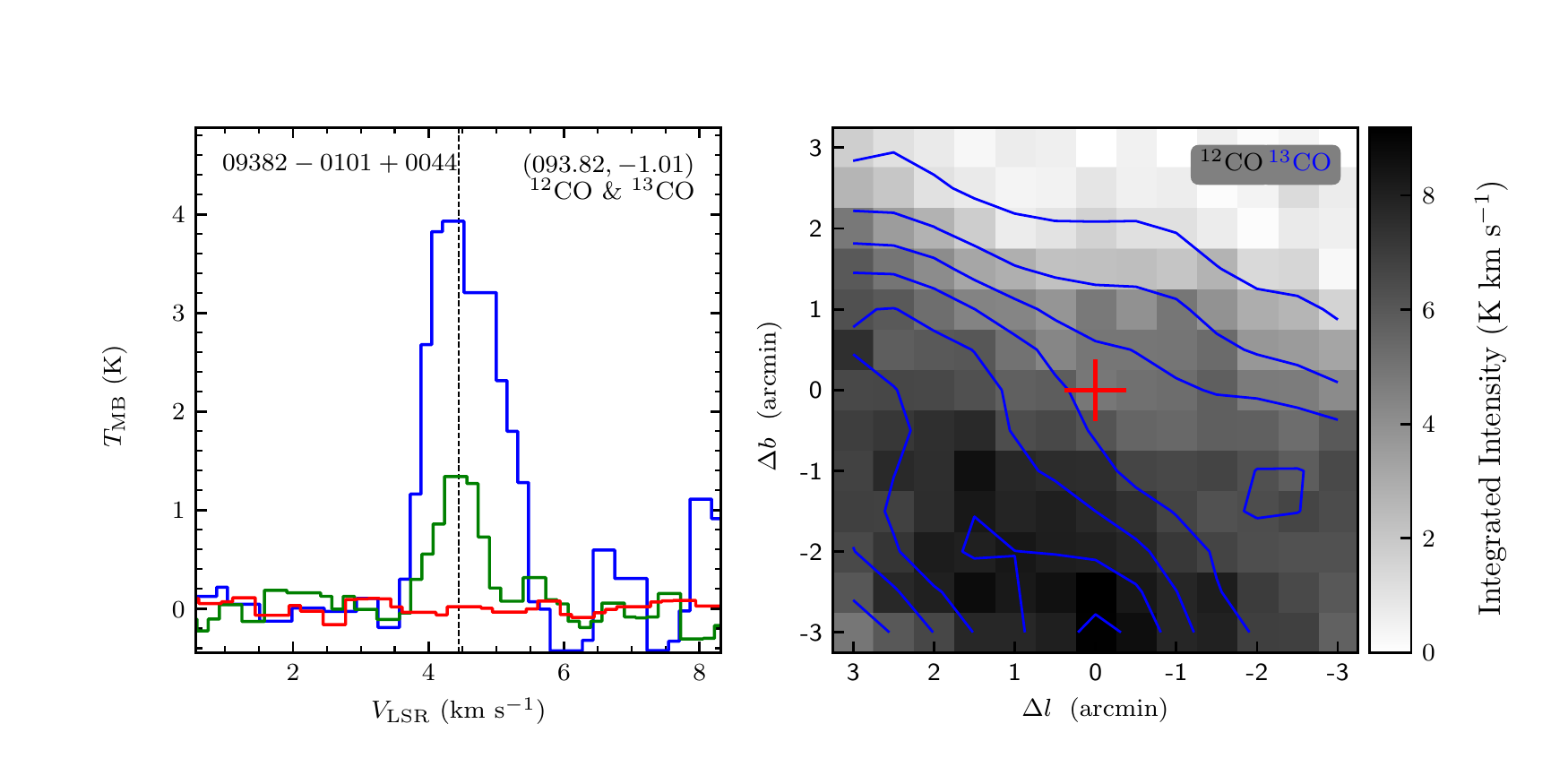}
\vspace{-0.5cm}

\includegraphics[width=9.0cm,angle=0]{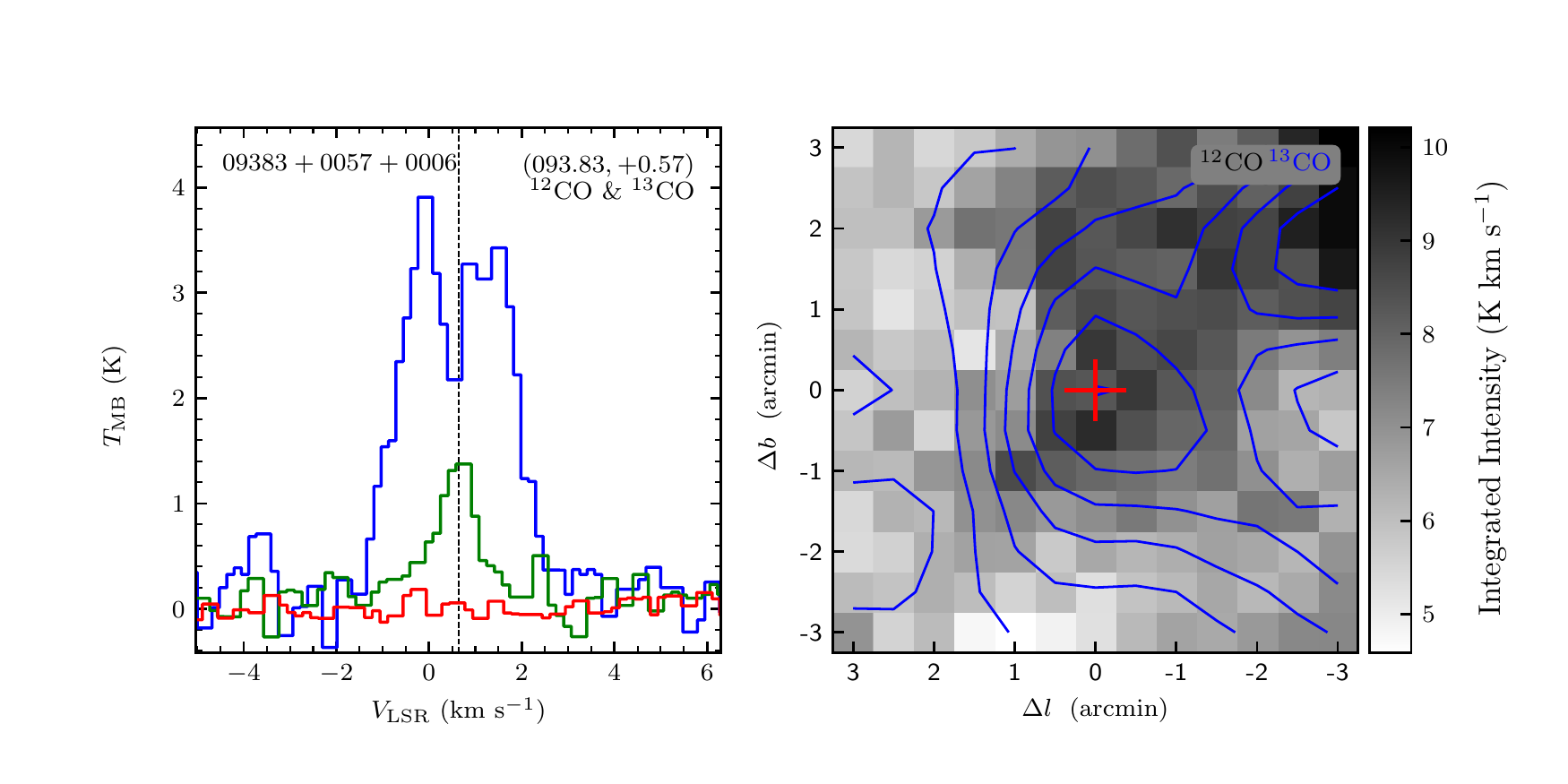}
\includegraphics[width=9.0cm,angle=0]{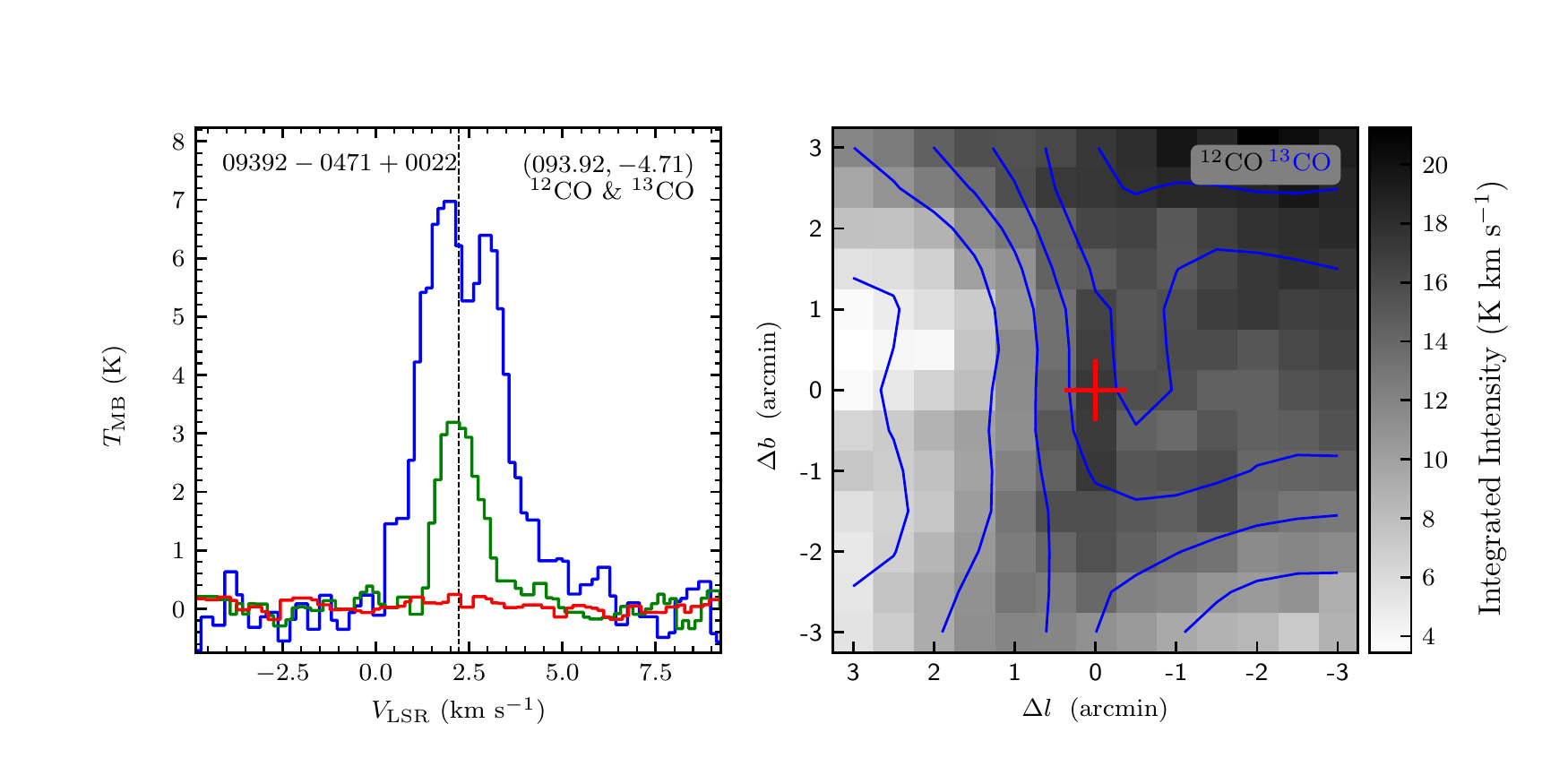}
\vspace{-0.5cm}

\includegraphics[width=9.0cm,angle=0]{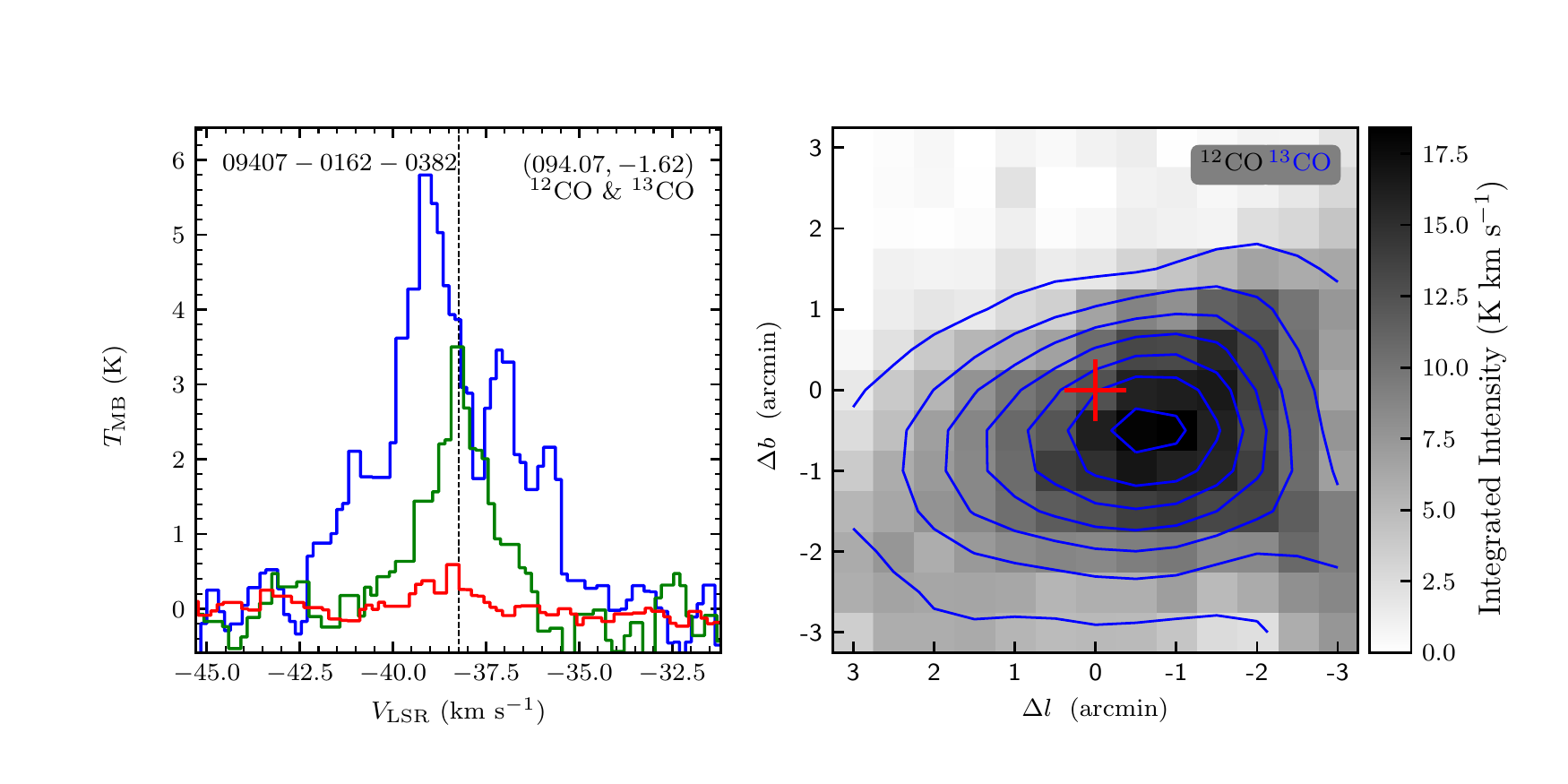}
\includegraphics[width=9.0cm,angle=0]{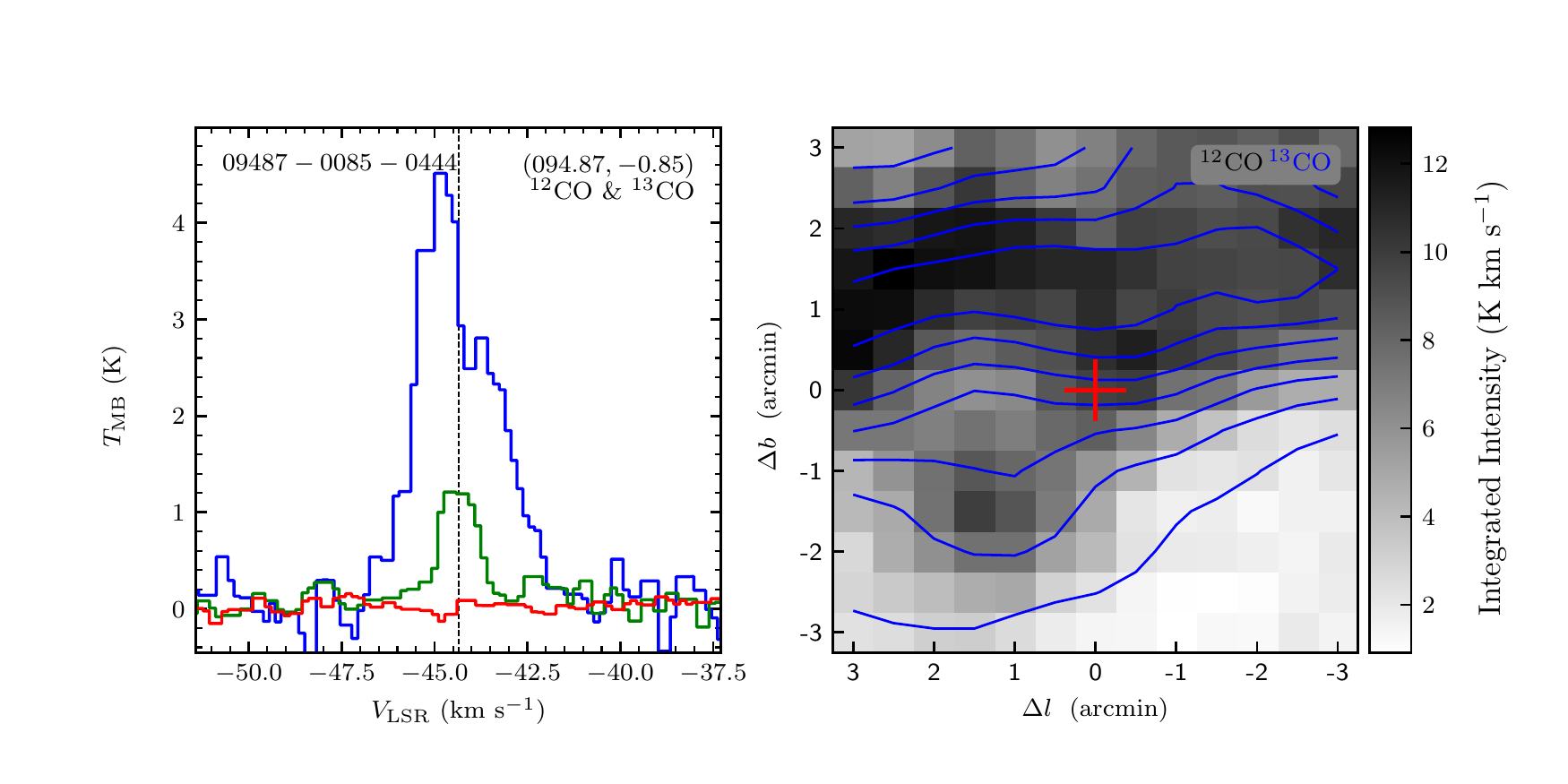}
\vspace{-0.5cm}

\includegraphics[width=9.0cm,angle=0]{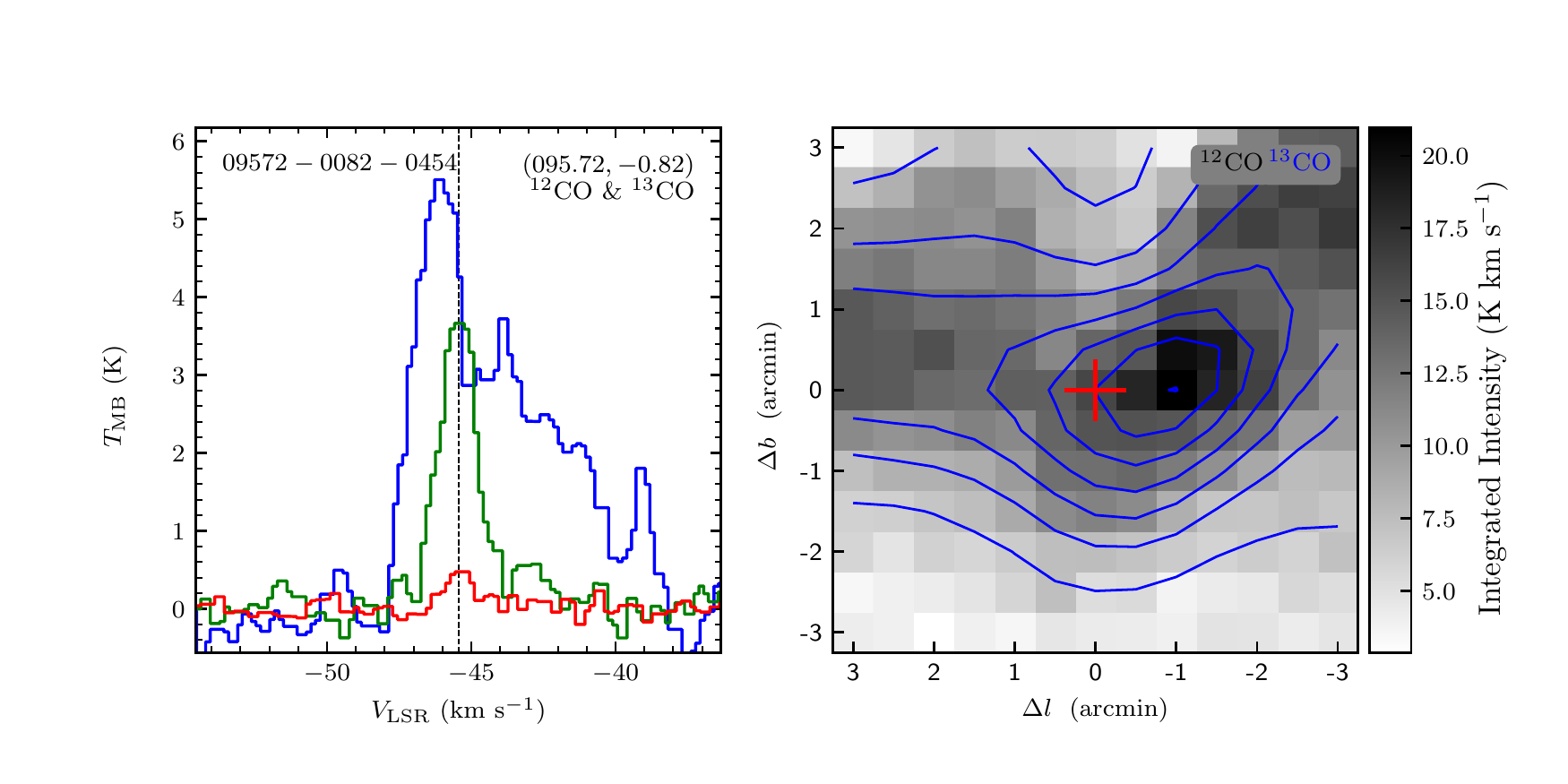}
\includegraphics[width=9.0cm,angle=0]{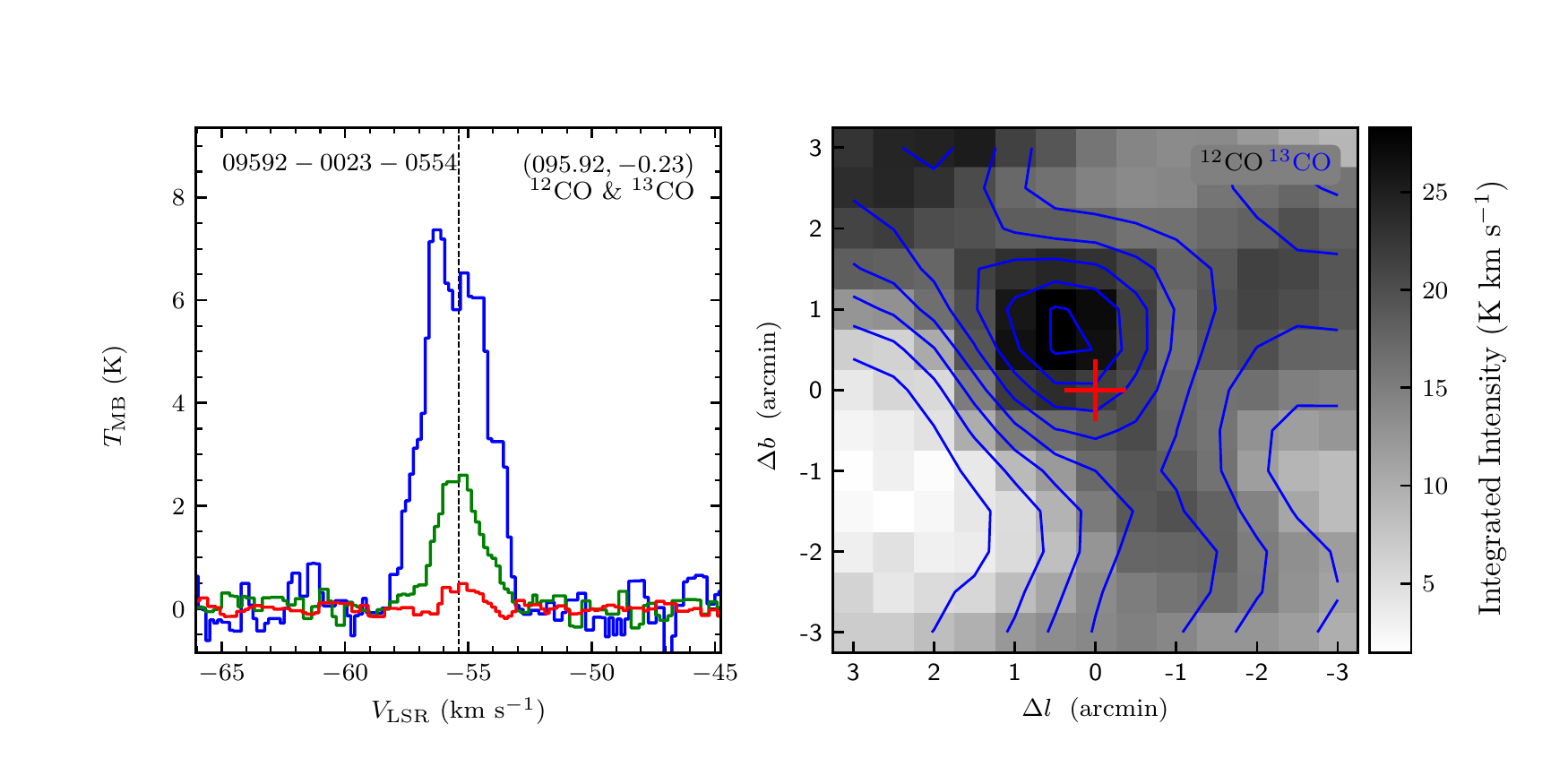}
\vspace{-0.5cm}

\includegraphics[width=9.0cm,angle=0]{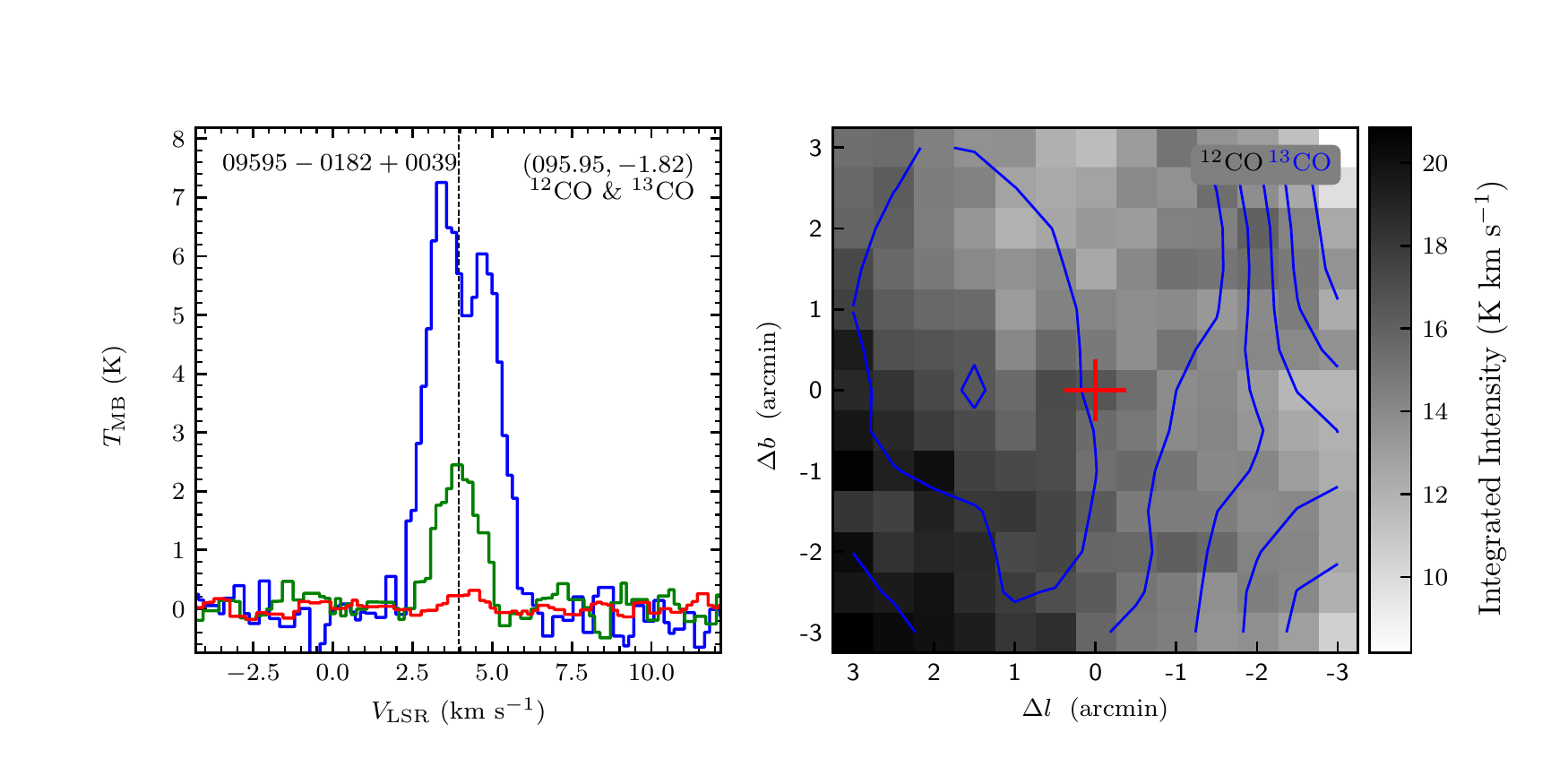}
\includegraphics[width=9.0cm,angle=0]{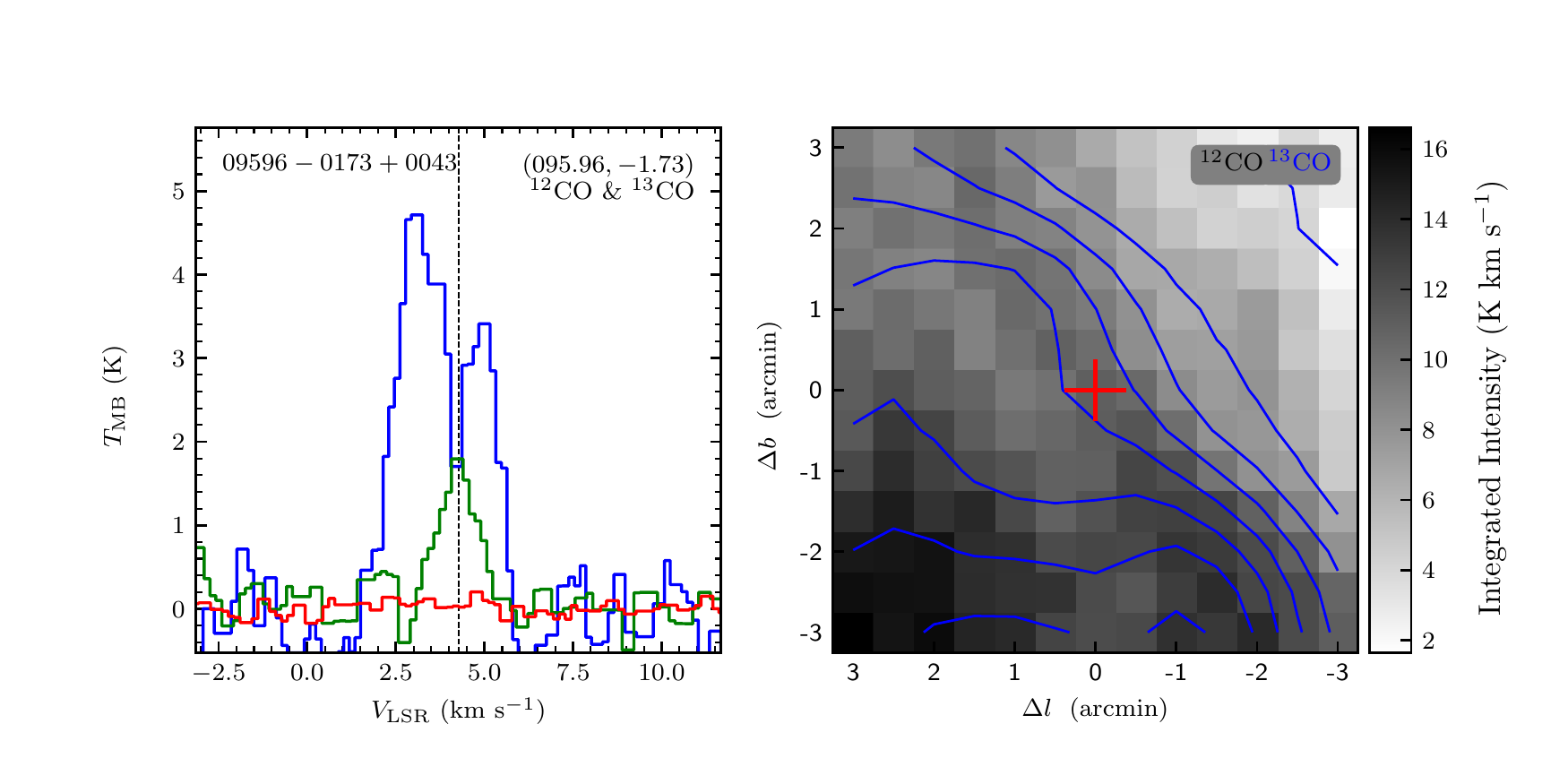}
\end{figure}
\clearpage

\begin{figure}
\includegraphics[width=9.0cm,angle=0]{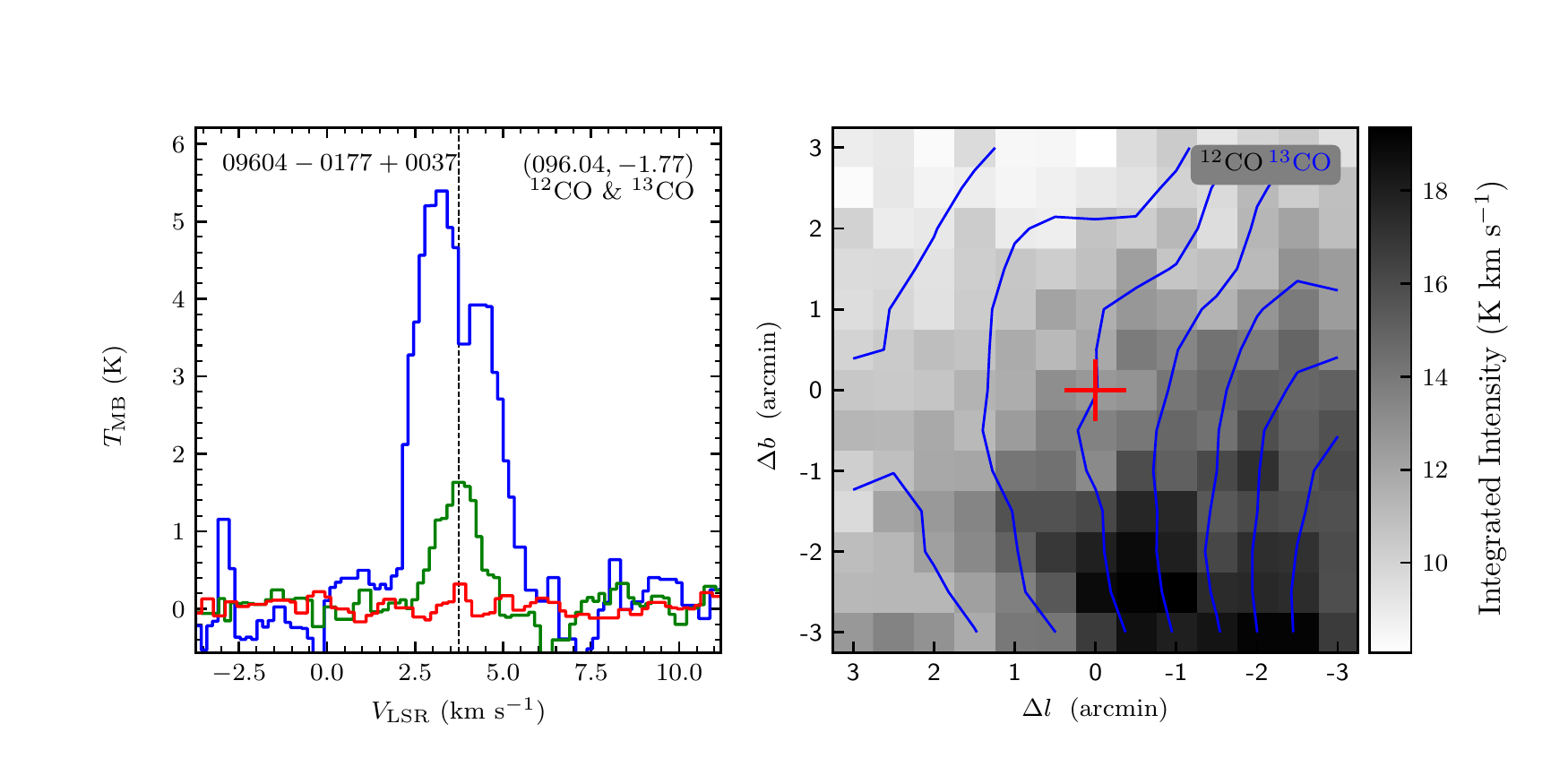}
\includegraphics[width=9.0cm,angle=0]{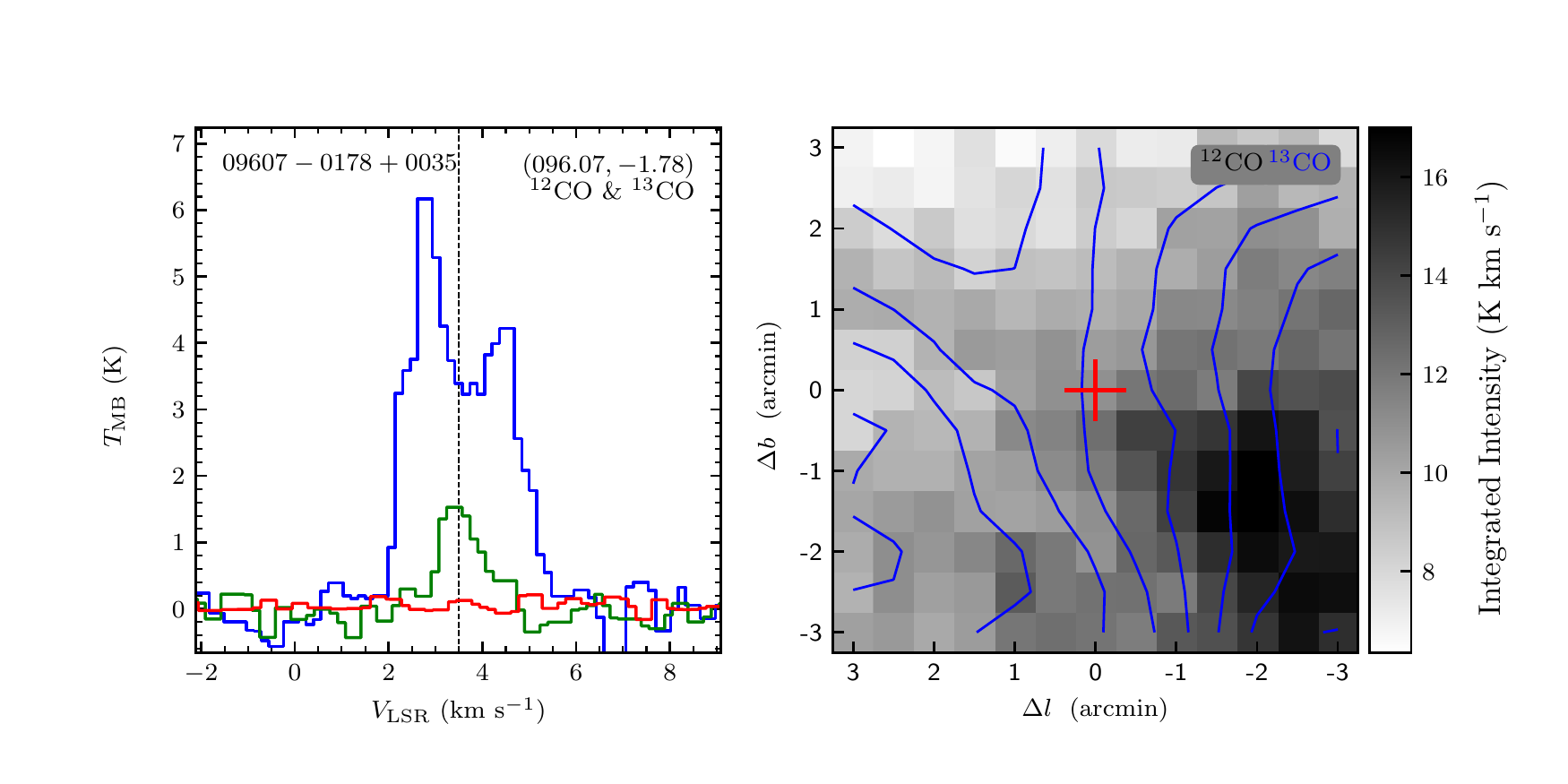}
\vspace{-0.5cm}

\includegraphics[width=9.0cm,angle=0]{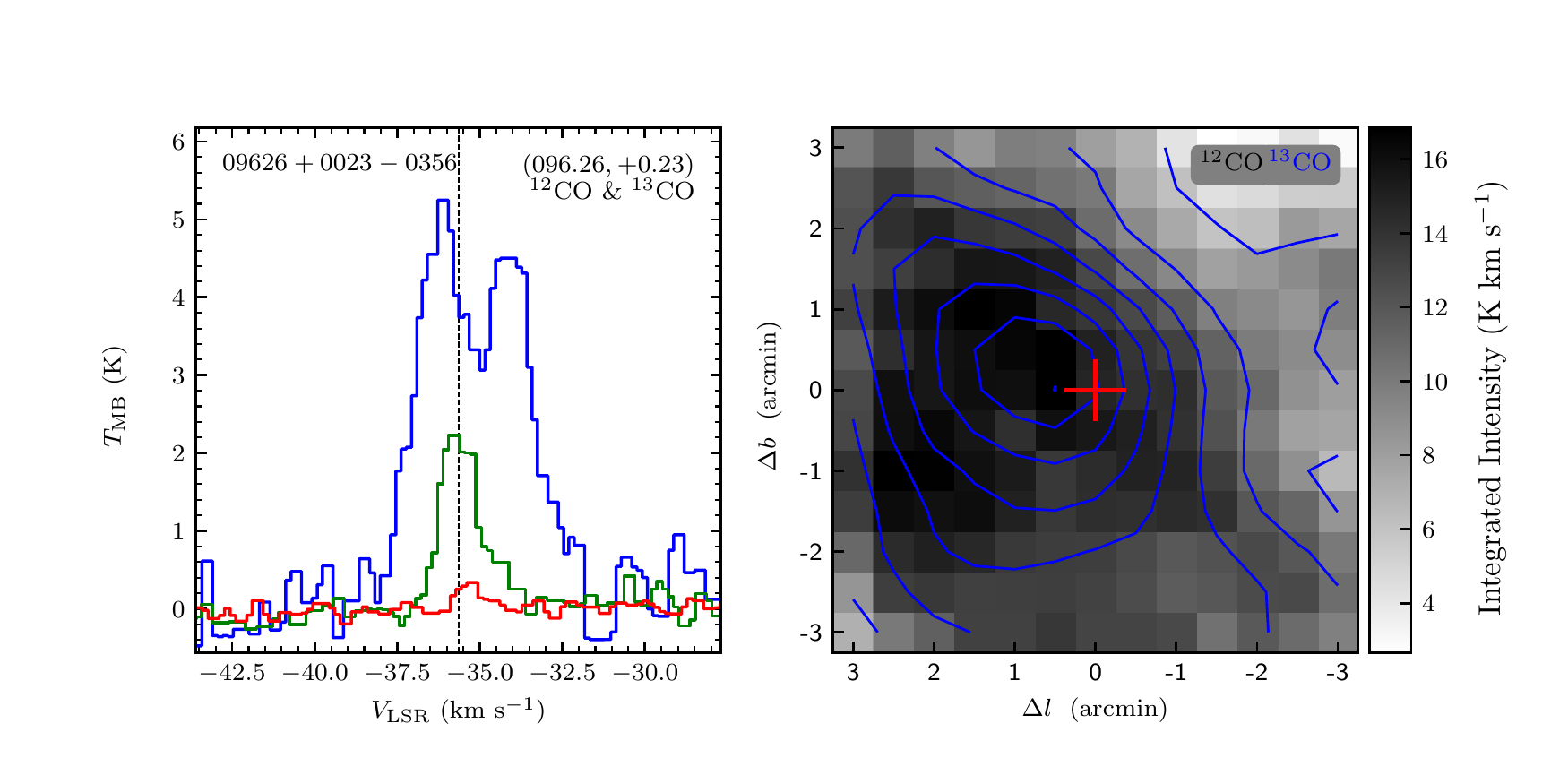}
\includegraphics[width=9.0cm,angle=0]{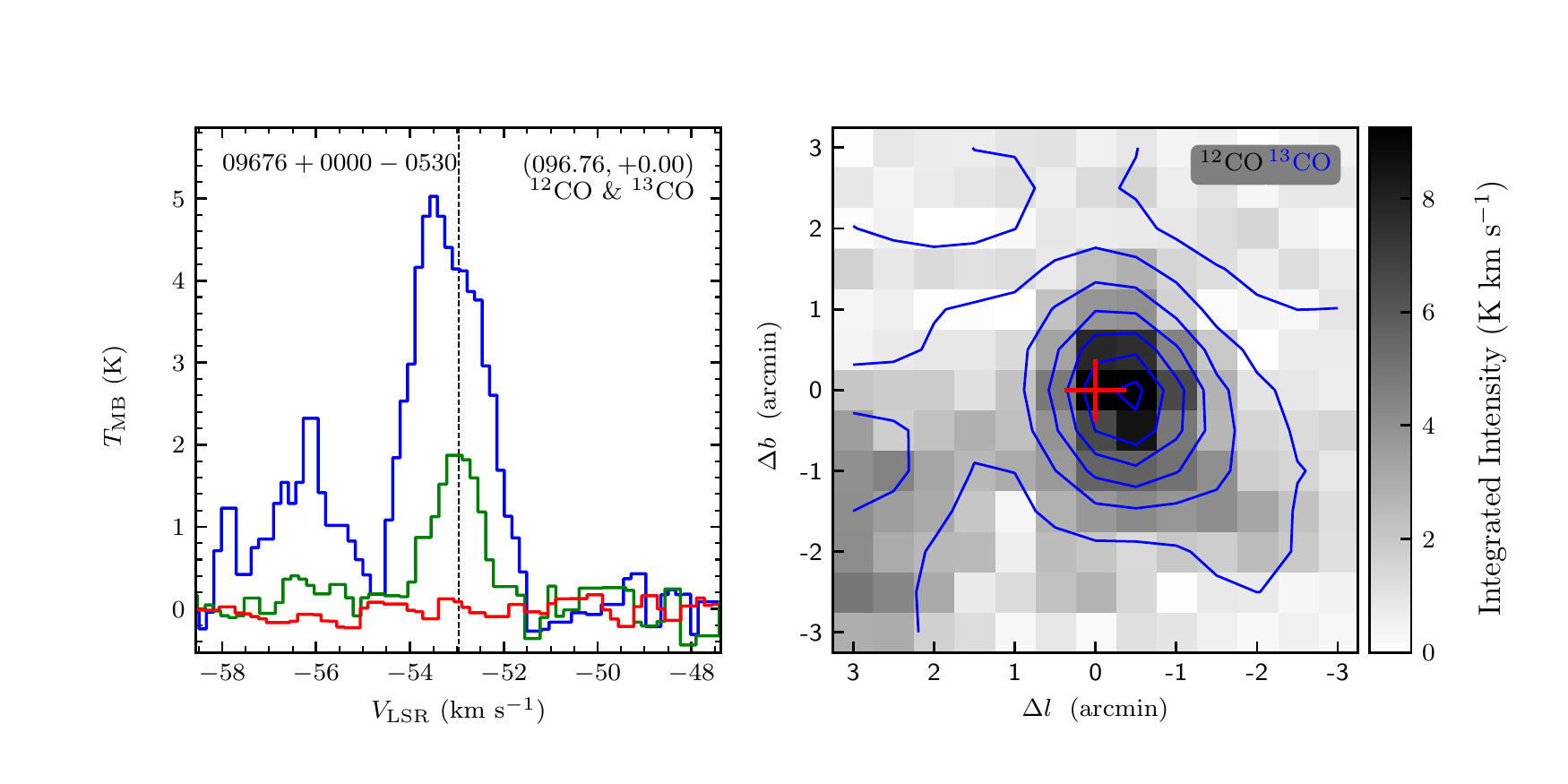}
\vspace{-0.5cm}

\includegraphics[width=9.0cm,angle=0]{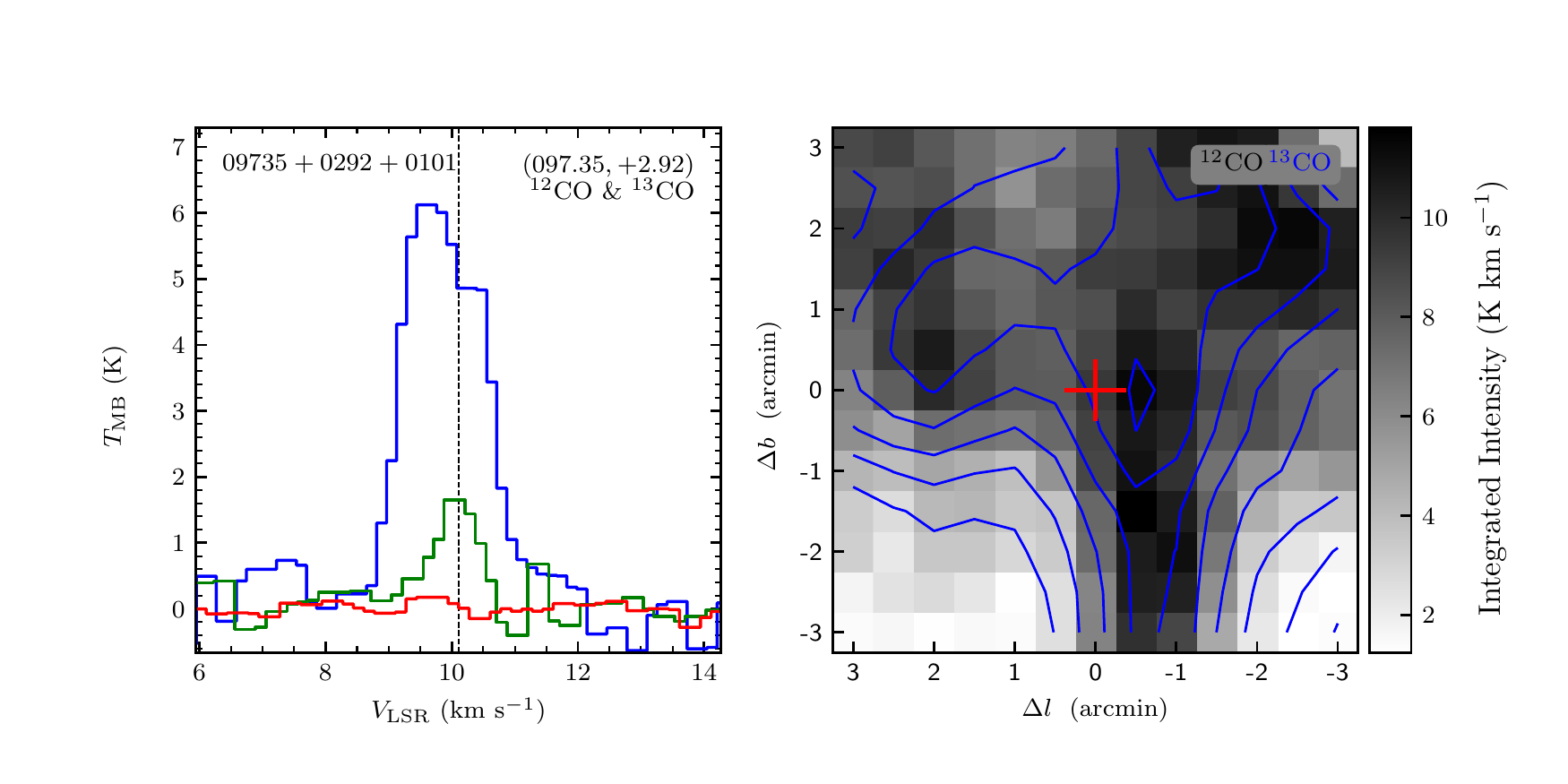}
\includegraphics[width=9.0cm,angle=0]{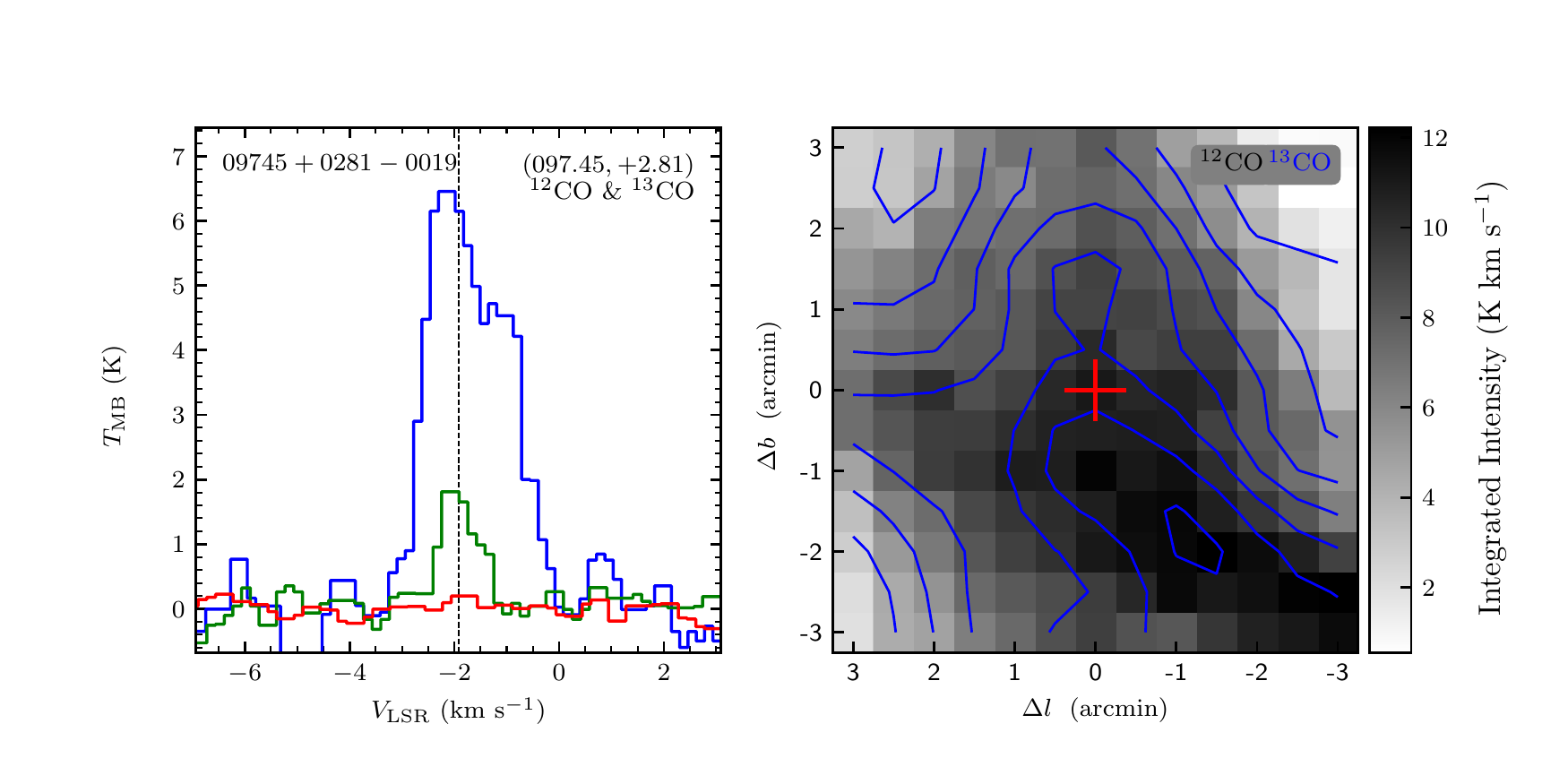}
\vspace{-0.5cm}

\includegraphics[width=9.0cm,angle=0]{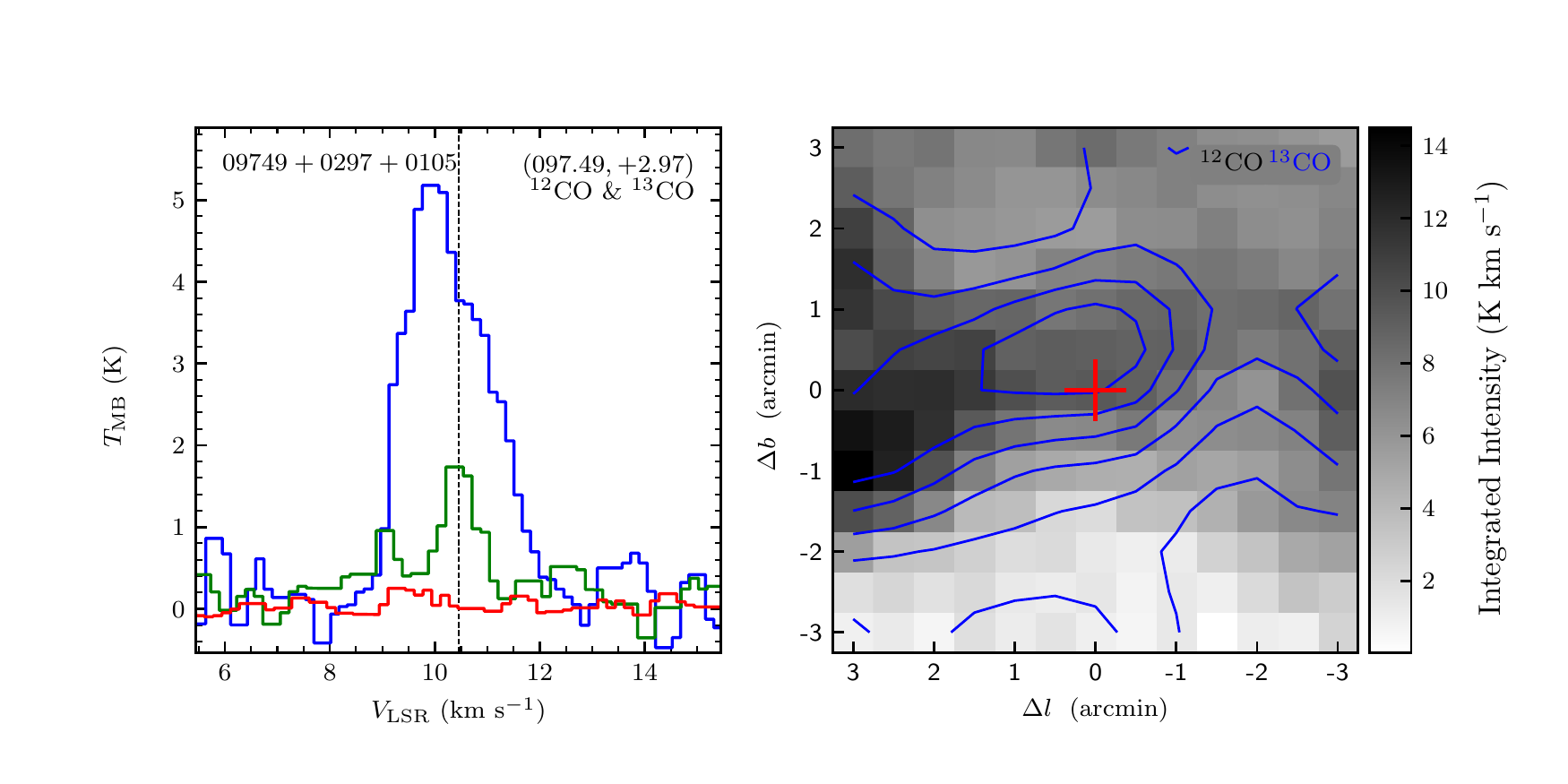}
\includegraphics[width=9.0cm,angle=0]{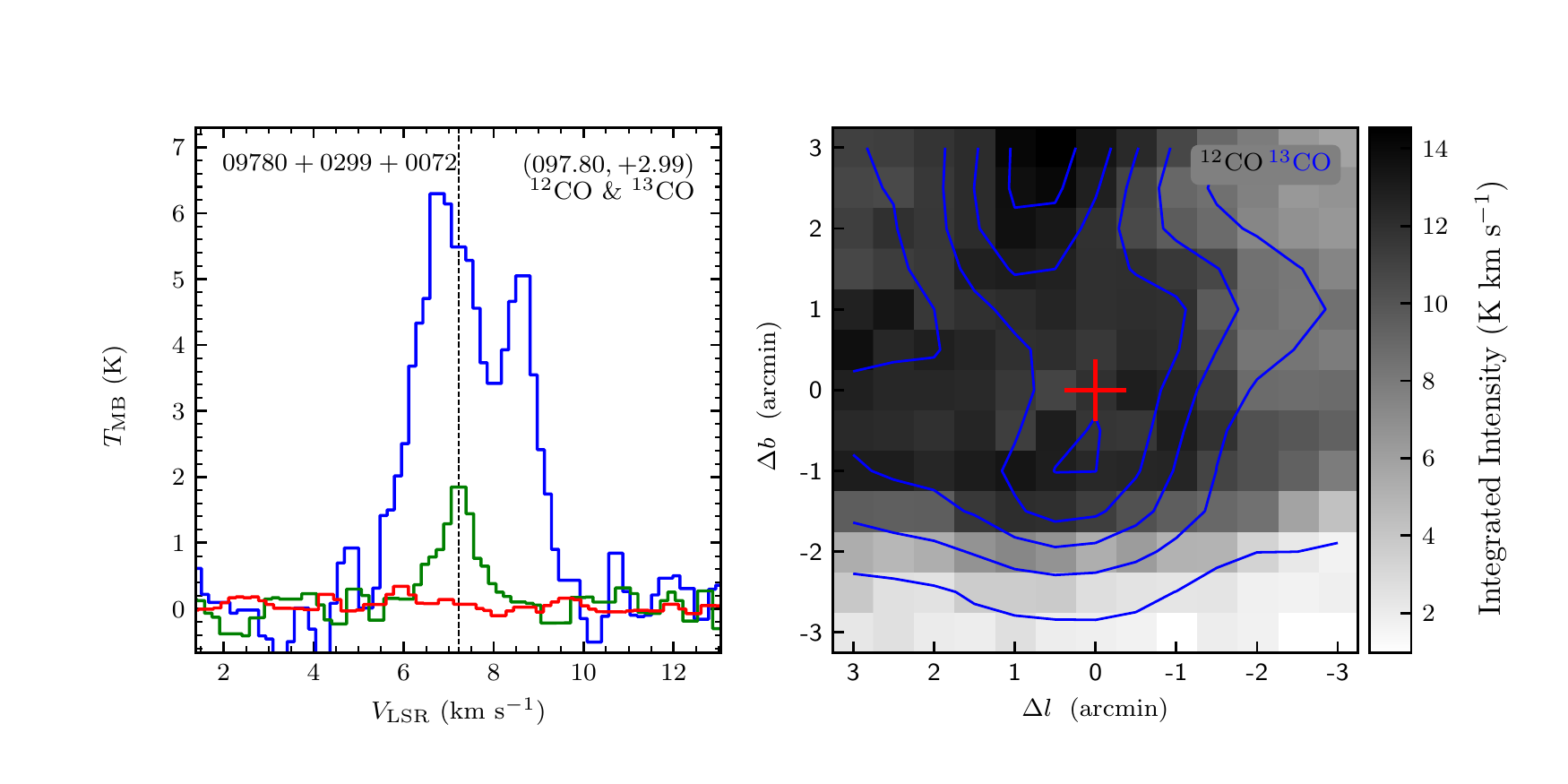}
\vspace{-0.5cm}

\includegraphics[width=9.0cm,angle=0]{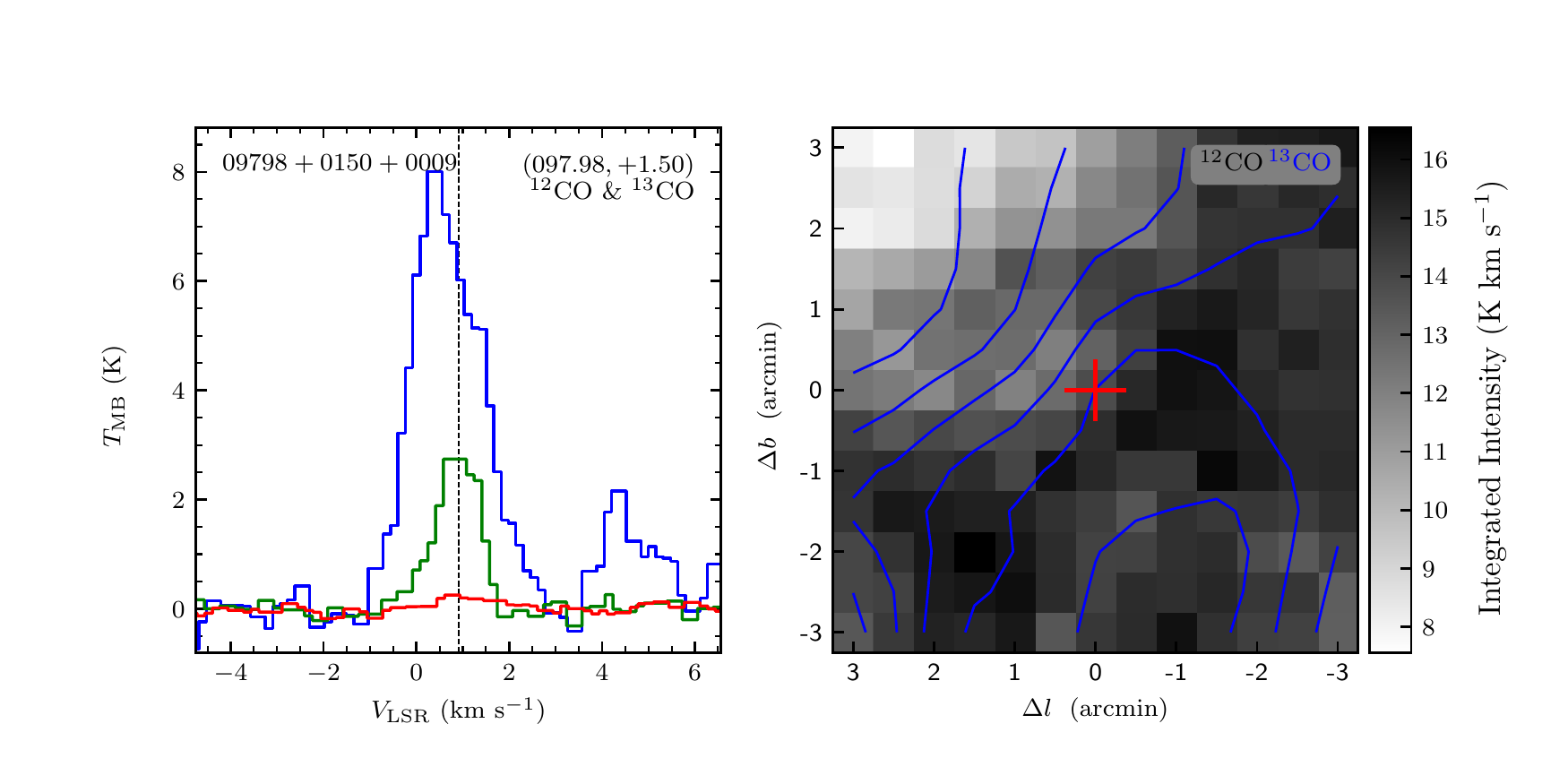}
\includegraphics[width=9.0cm,angle=0]{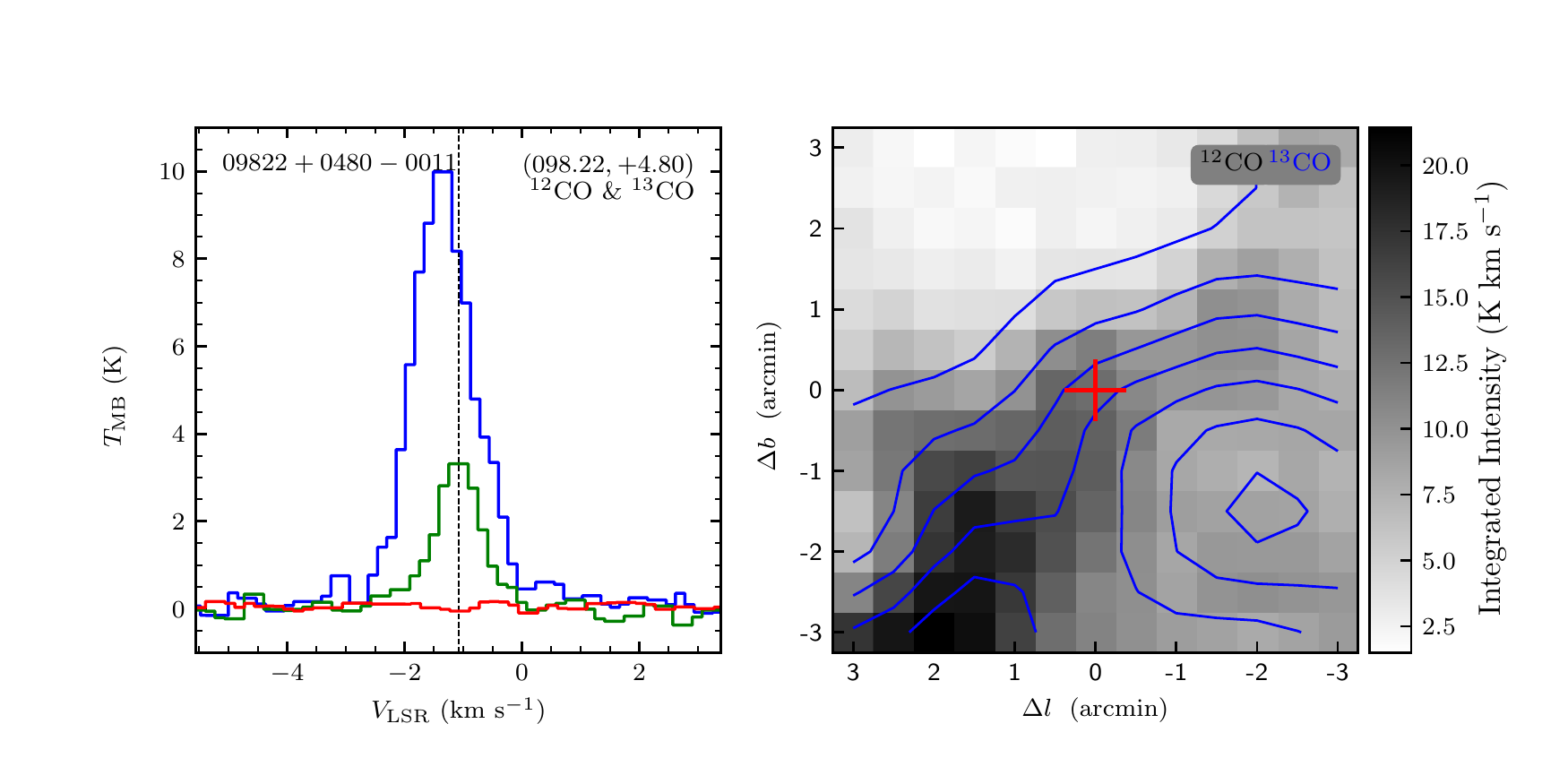}
\end{figure}
\clearpage

\begin{figure}
\includegraphics[width=9.0cm,angle=0]{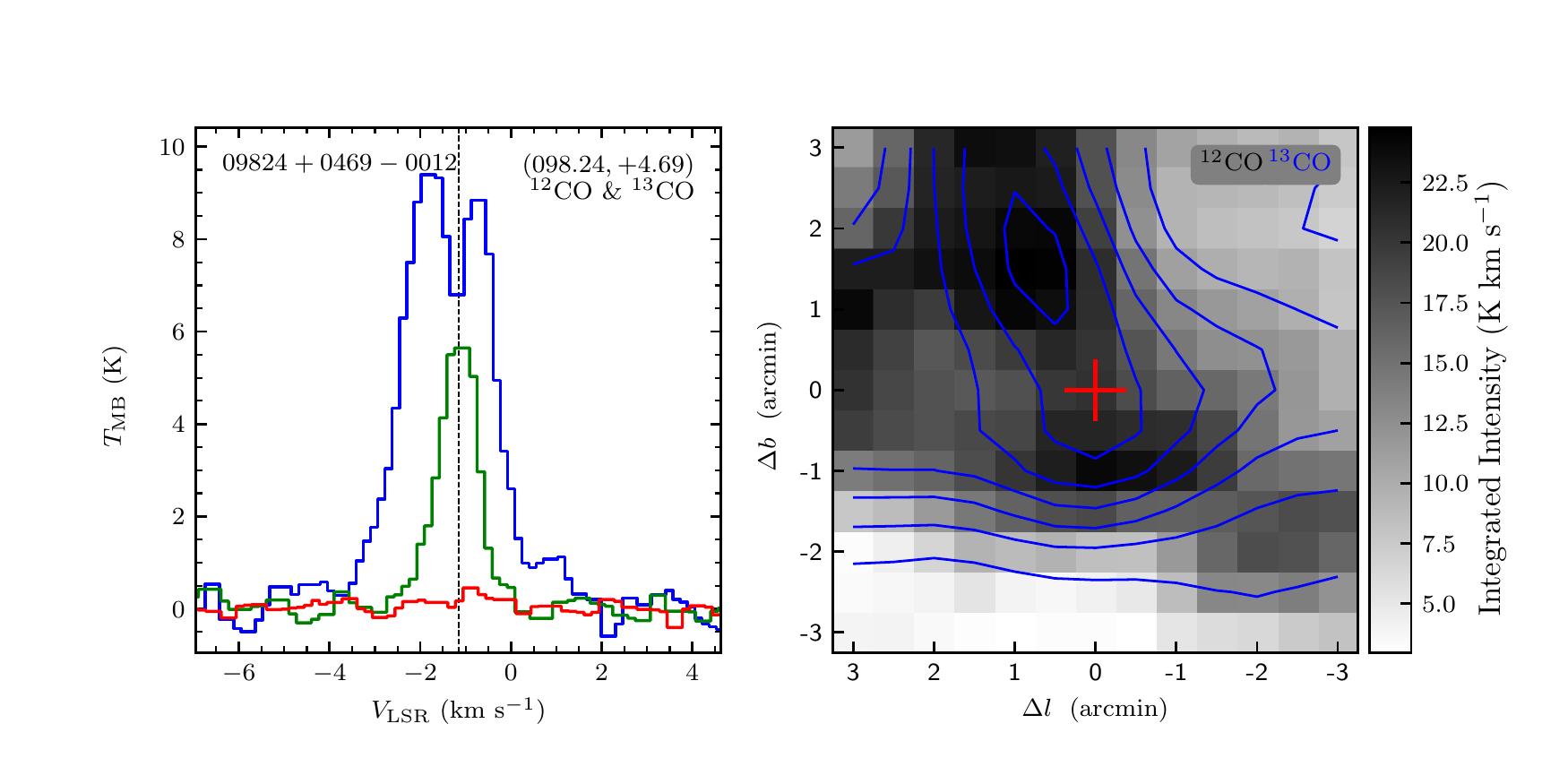}
\includegraphics[width=9.0cm,angle=0]{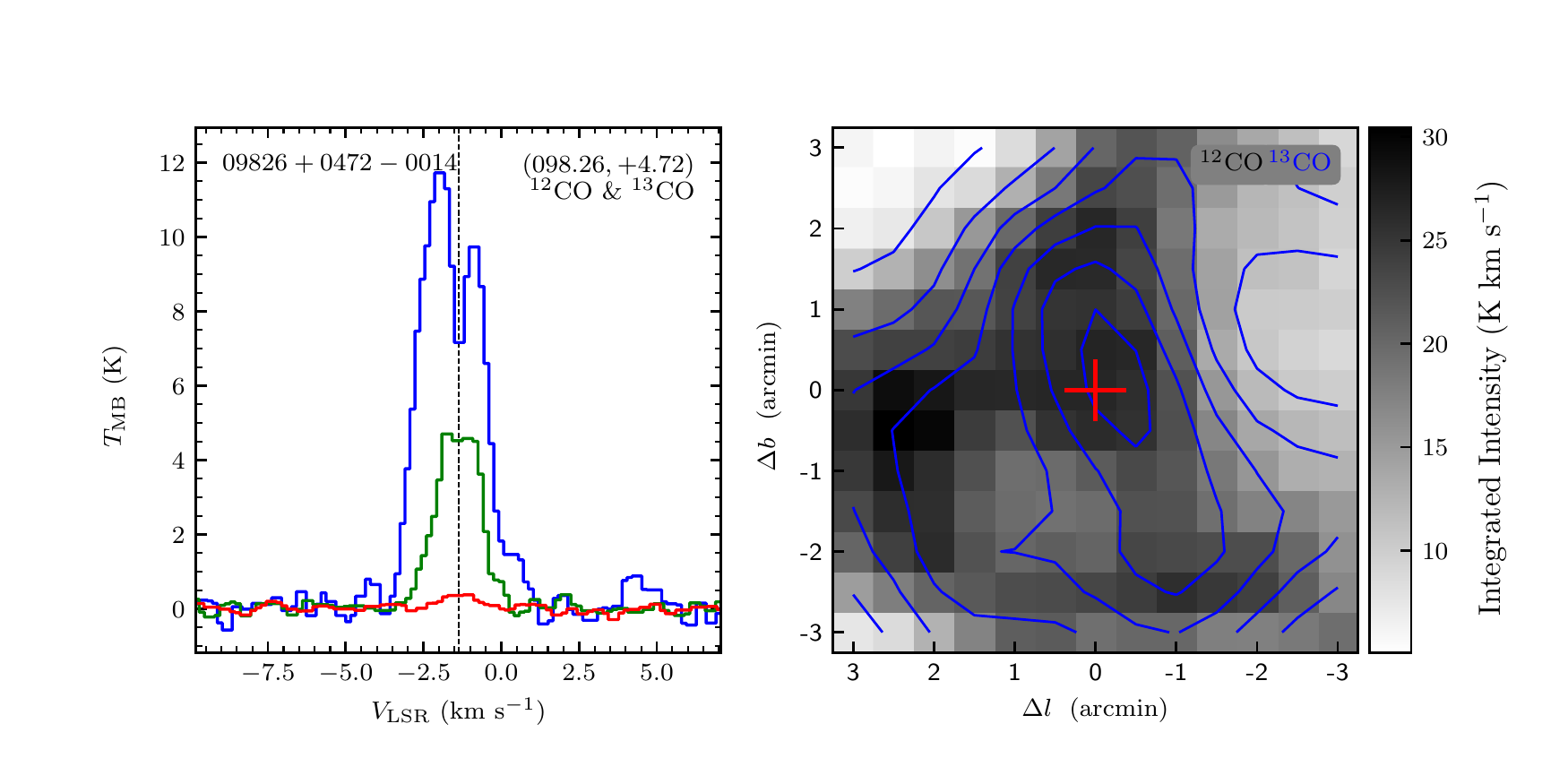}
\vspace{-0.5cm}

\includegraphics[width=9.0cm,angle=0]{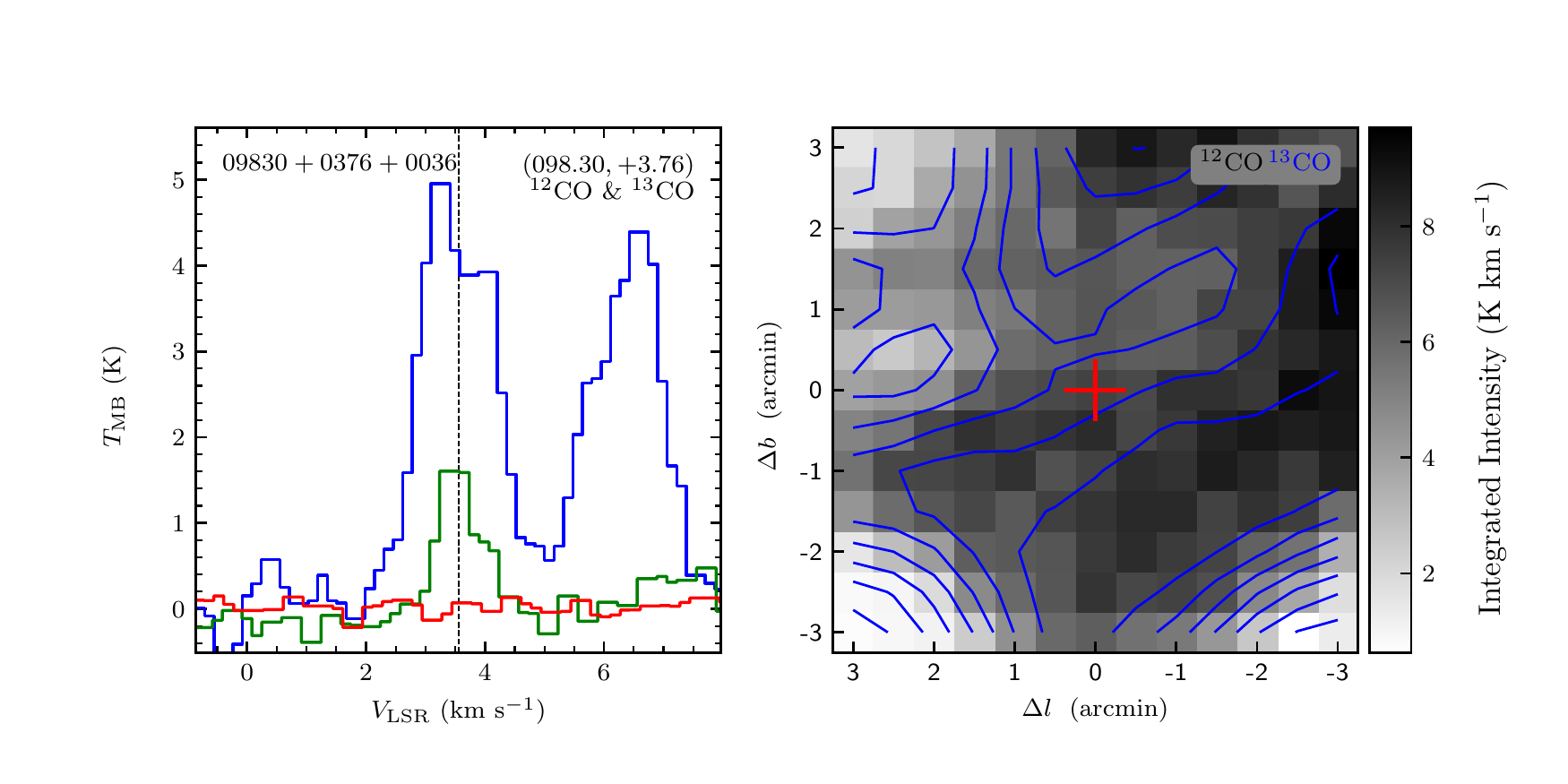}
\includegraphics[width=9.0cm,angle=0]{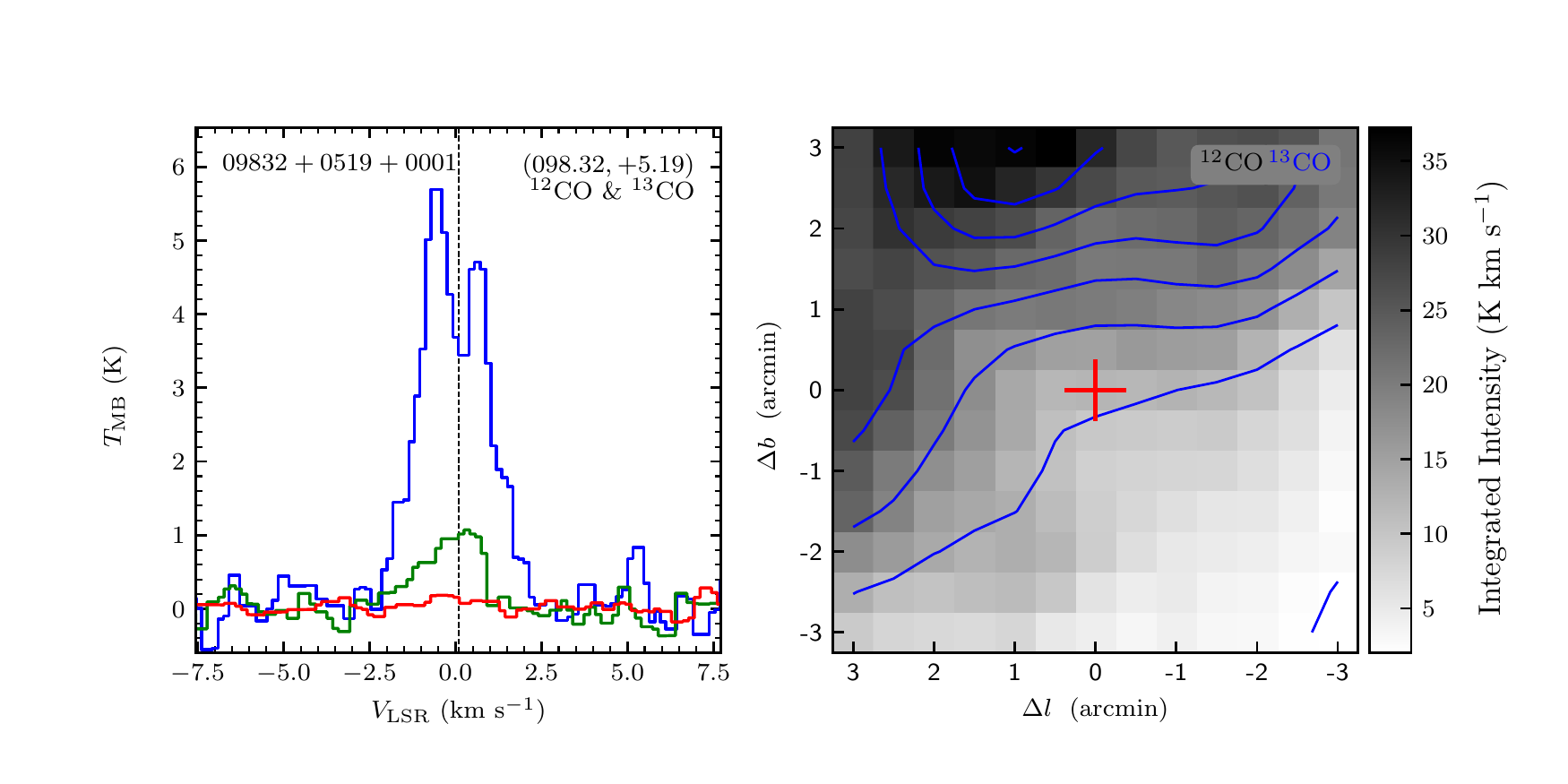}
\vspace{-0.5cm}

\includegraphics[width=9.0cm,angle=0]{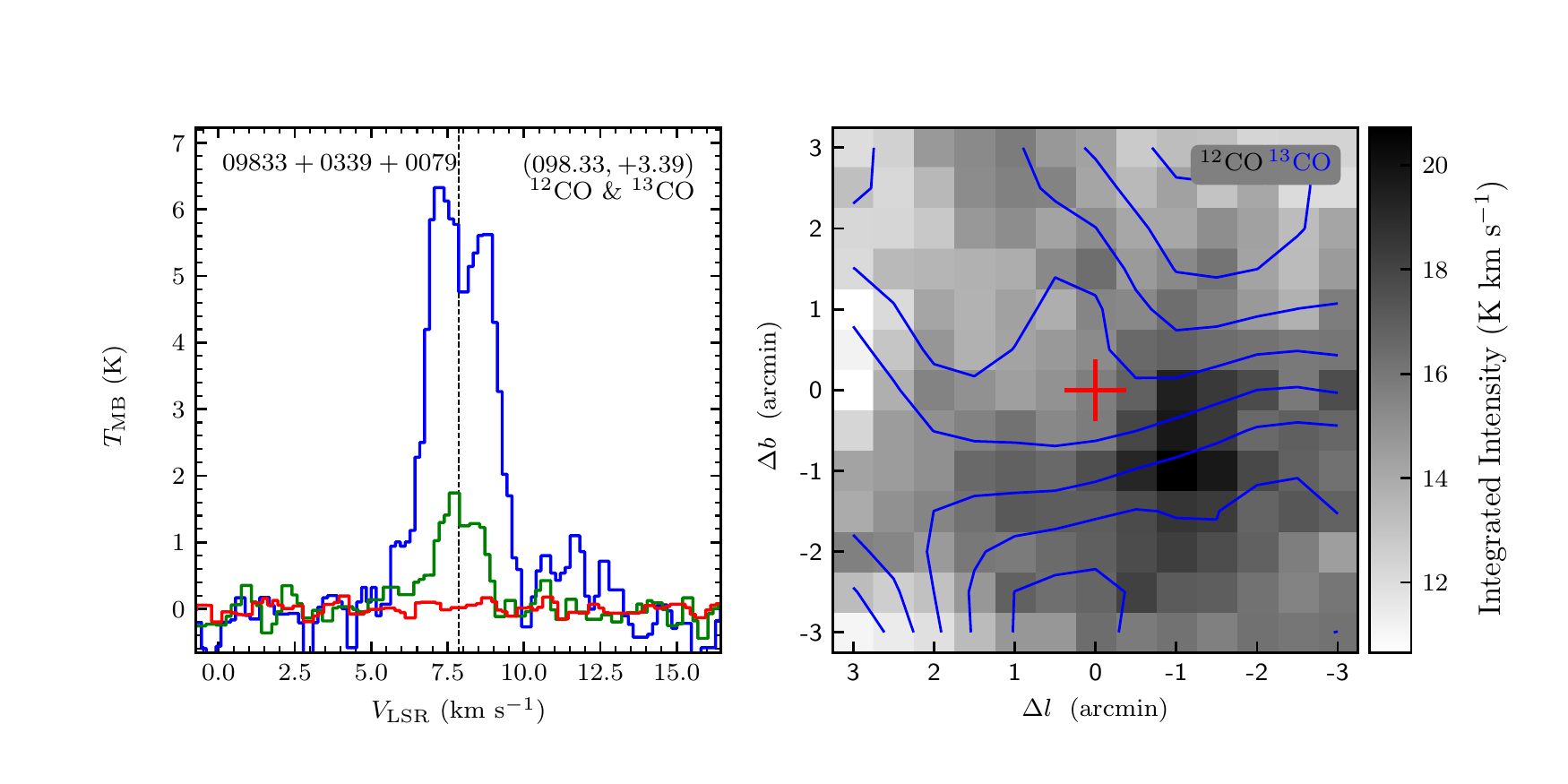}
\includegraphics[width=9.0cm,angle=0]{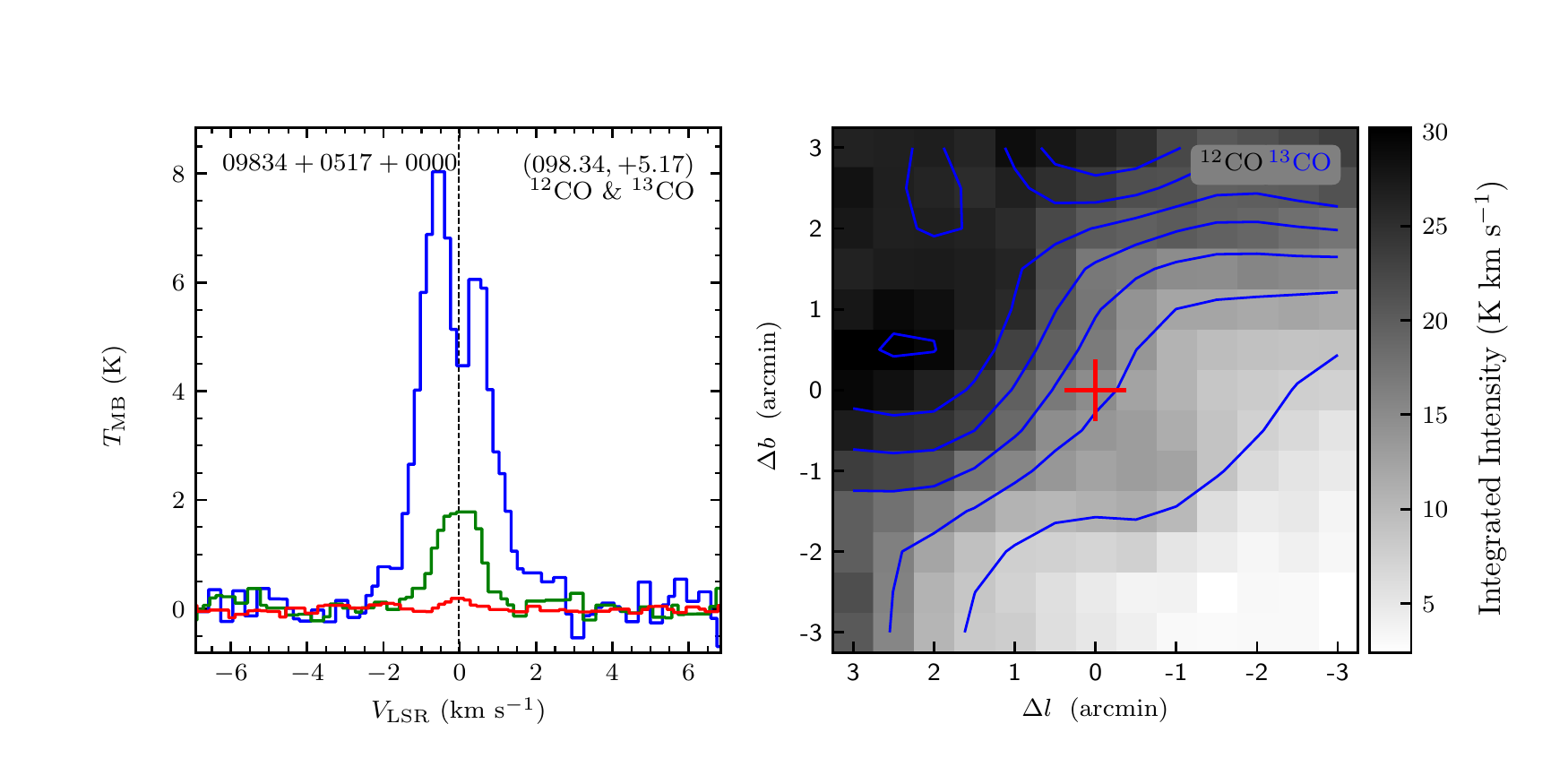}
\vspace{-0.5cm}

\includegraphics[width=9.0cm,angle=0]{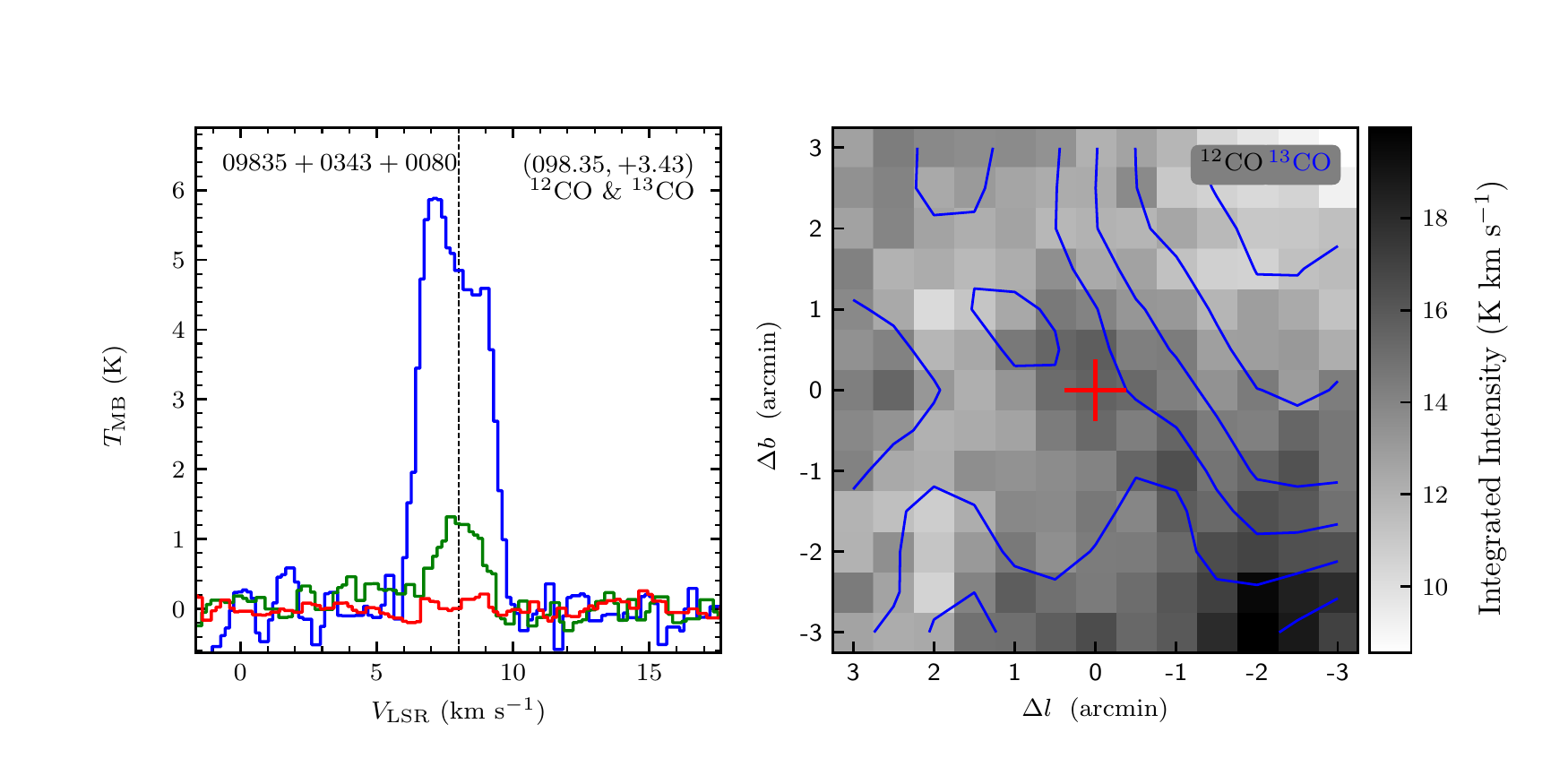}
\includegraphics[width=9.0cm,angle=0]{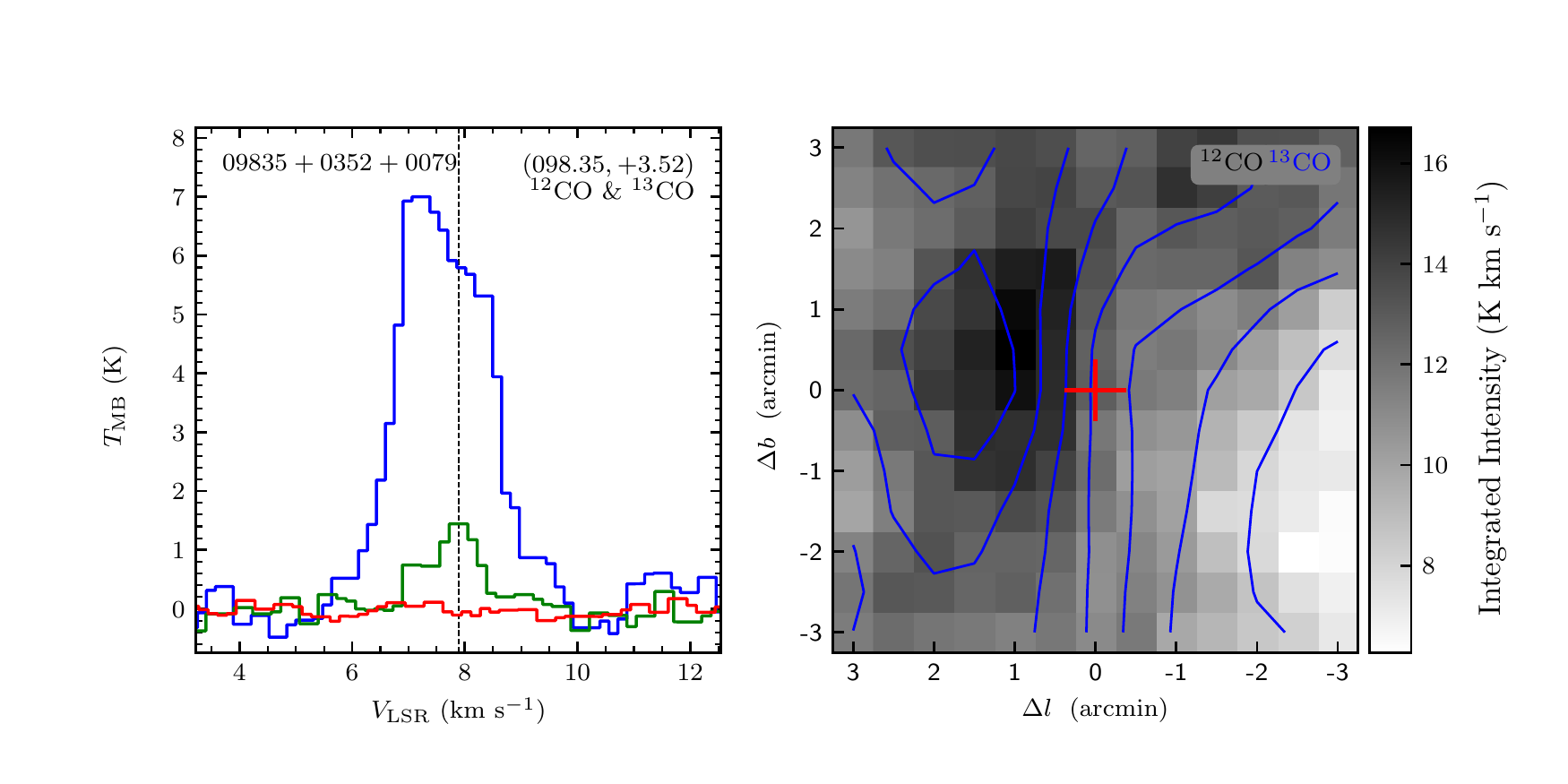}
\vspace{-0.5cm}

\includegraphics[width=9.0cm,angle=0]{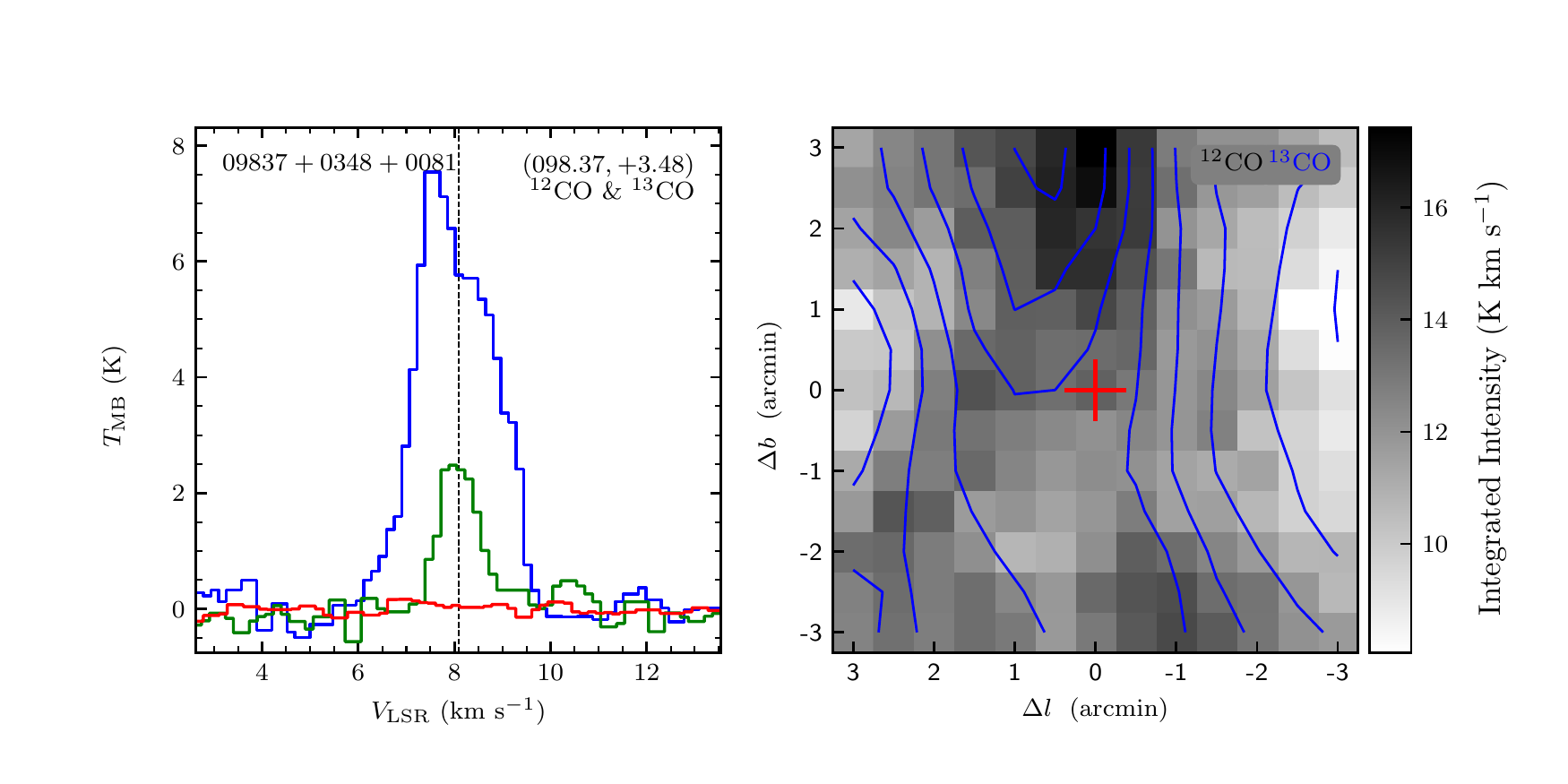}
\includegraphics[width=9.0cm,angle=0]{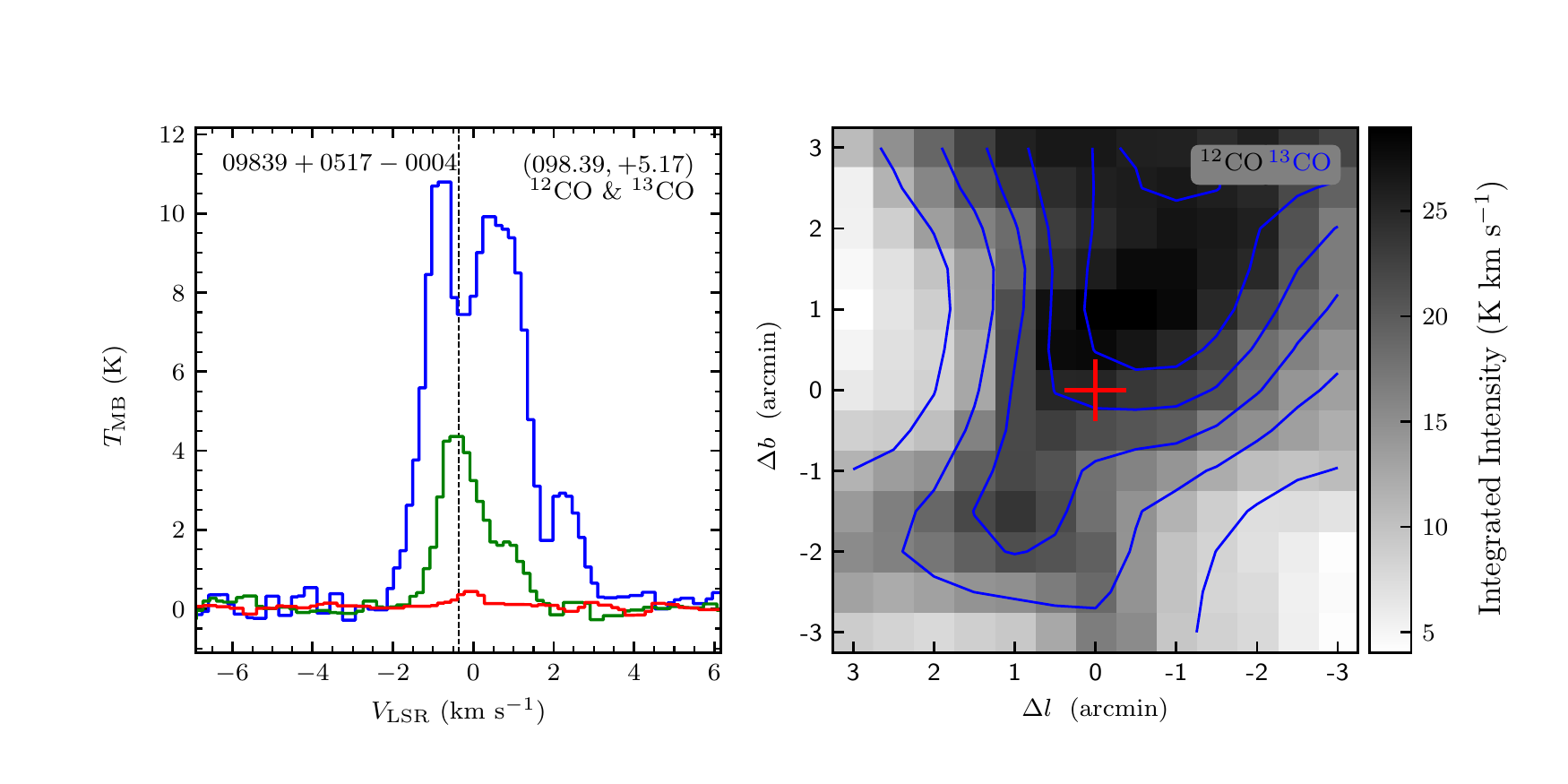}
\end{figure}
\clearpage

\begin{figure}
\includegraphics[width=9.0cm,angle=0]{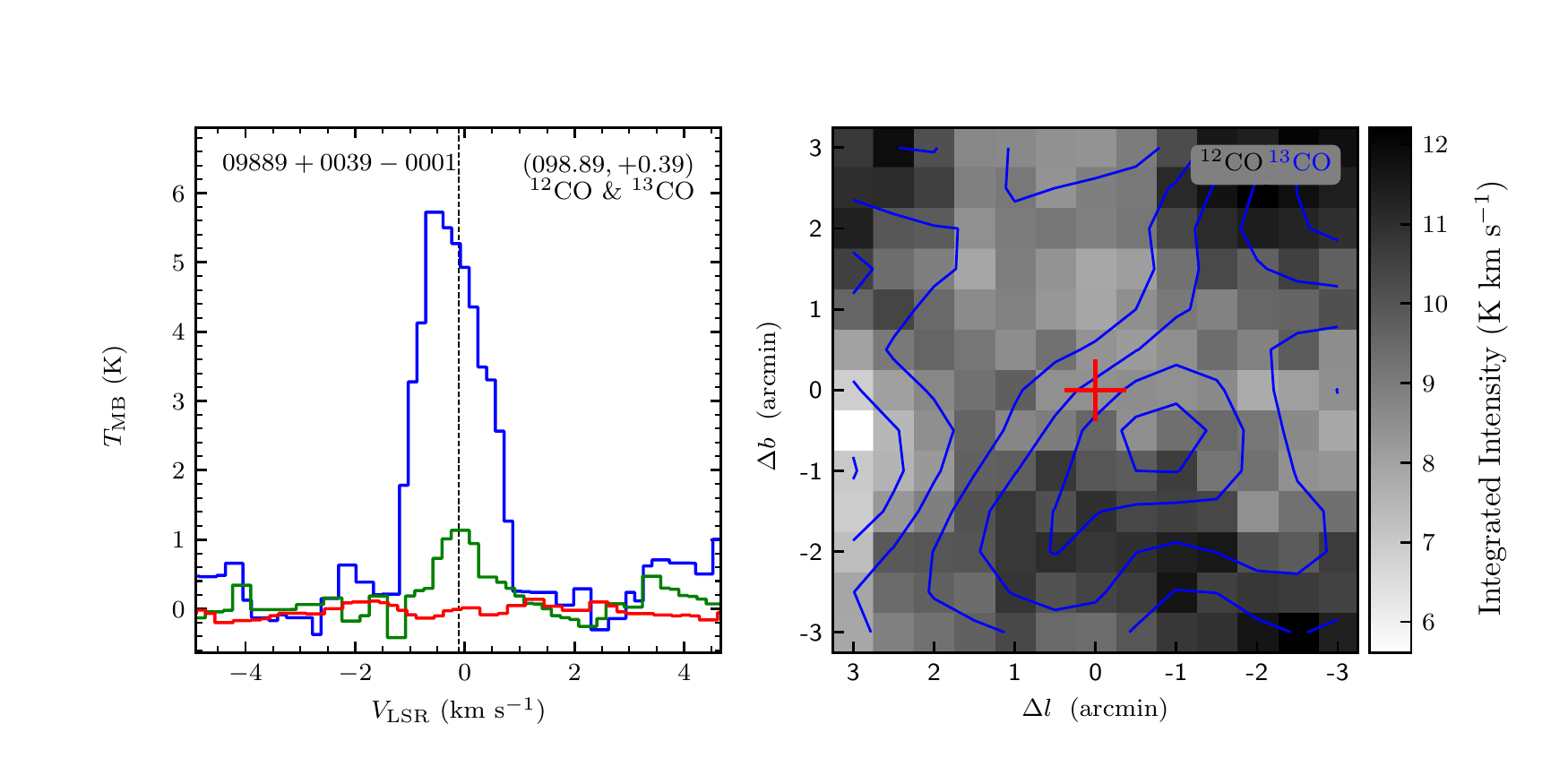}
\includegraphics[width=9.0cm,angle=0]{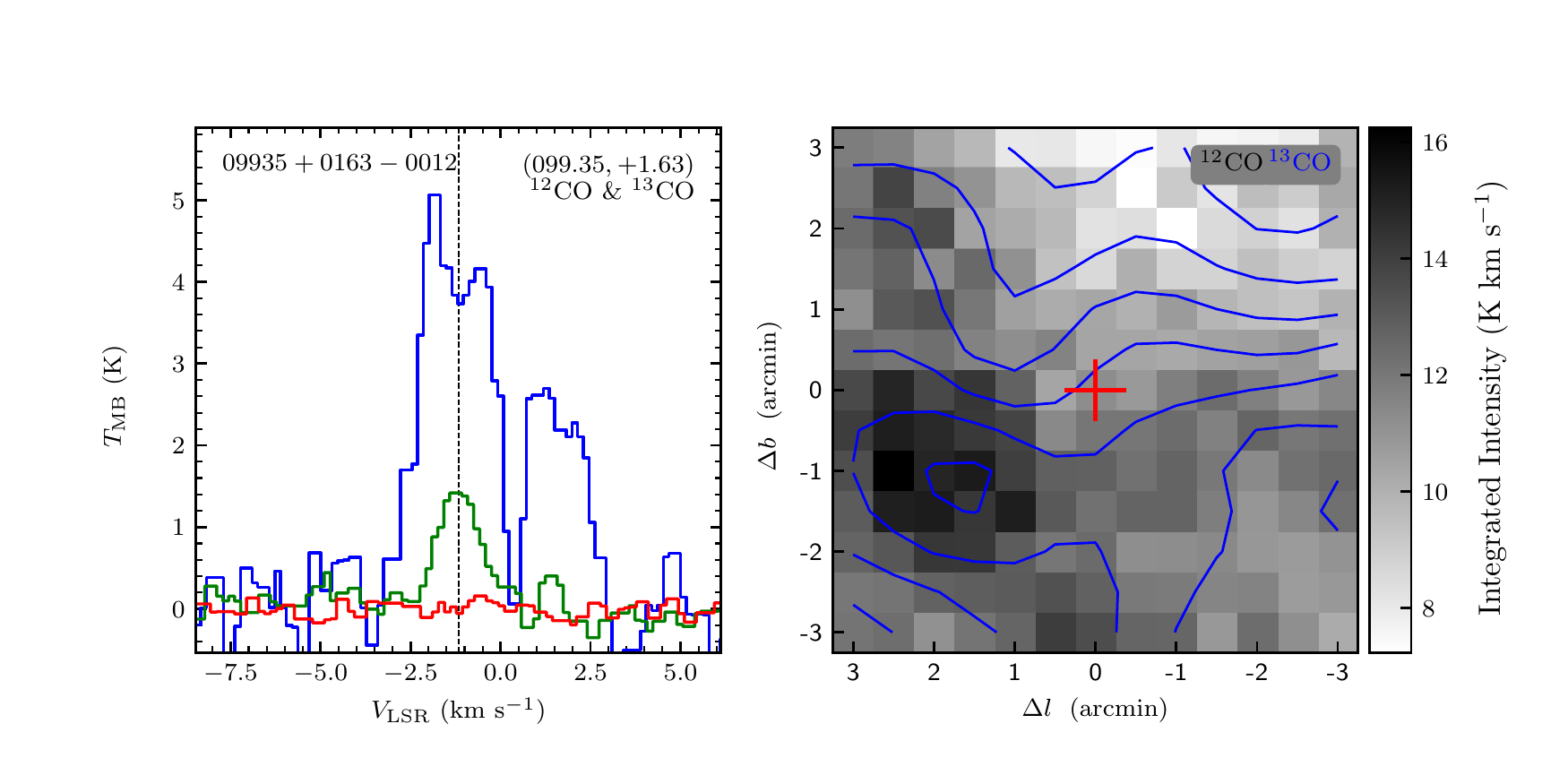}
\vspace{-0.5cm}

\includegraphics[width=9.0cm,angle=0]{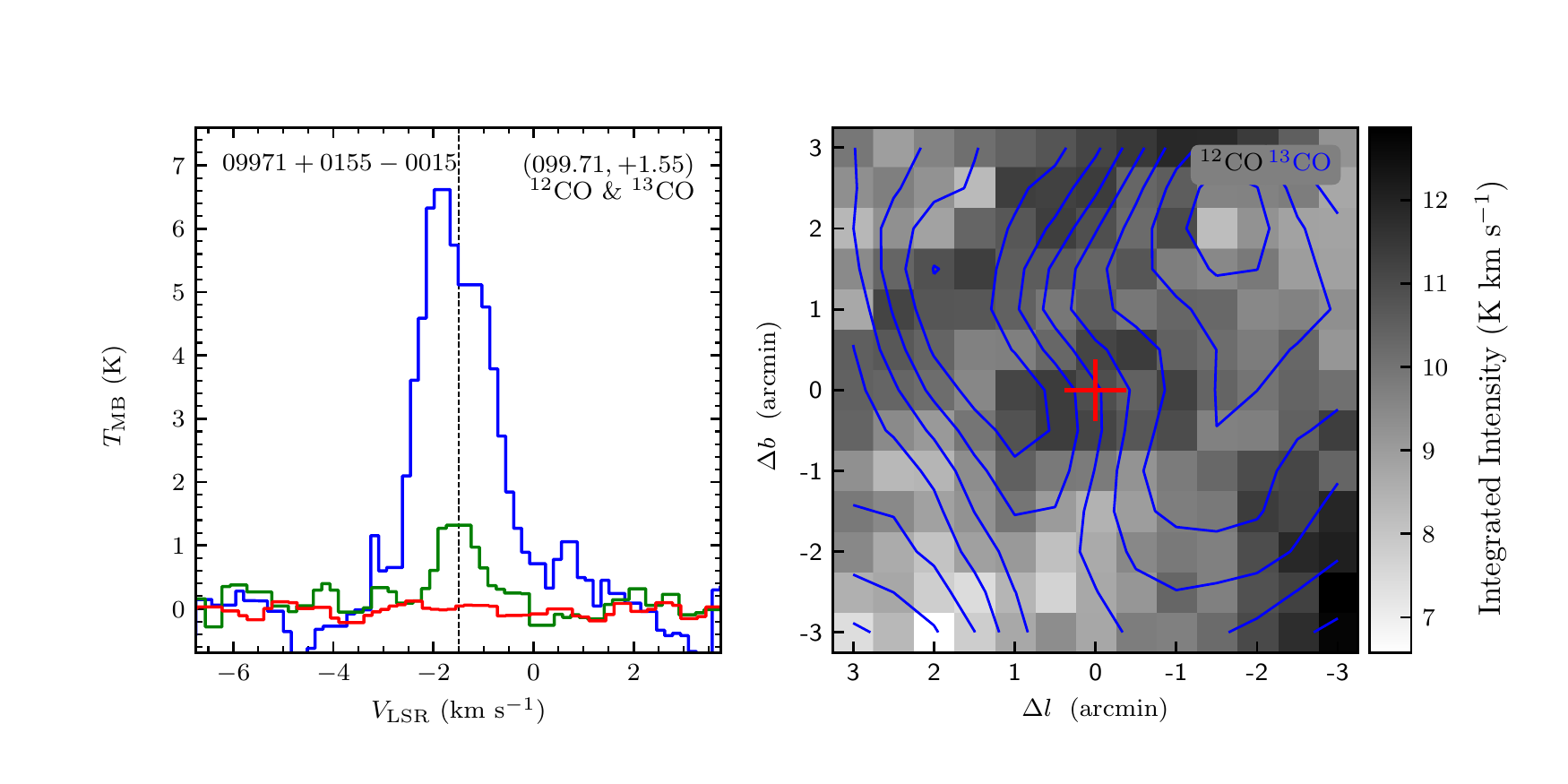}
\includegraphics[width=9.0cm,angle=0]{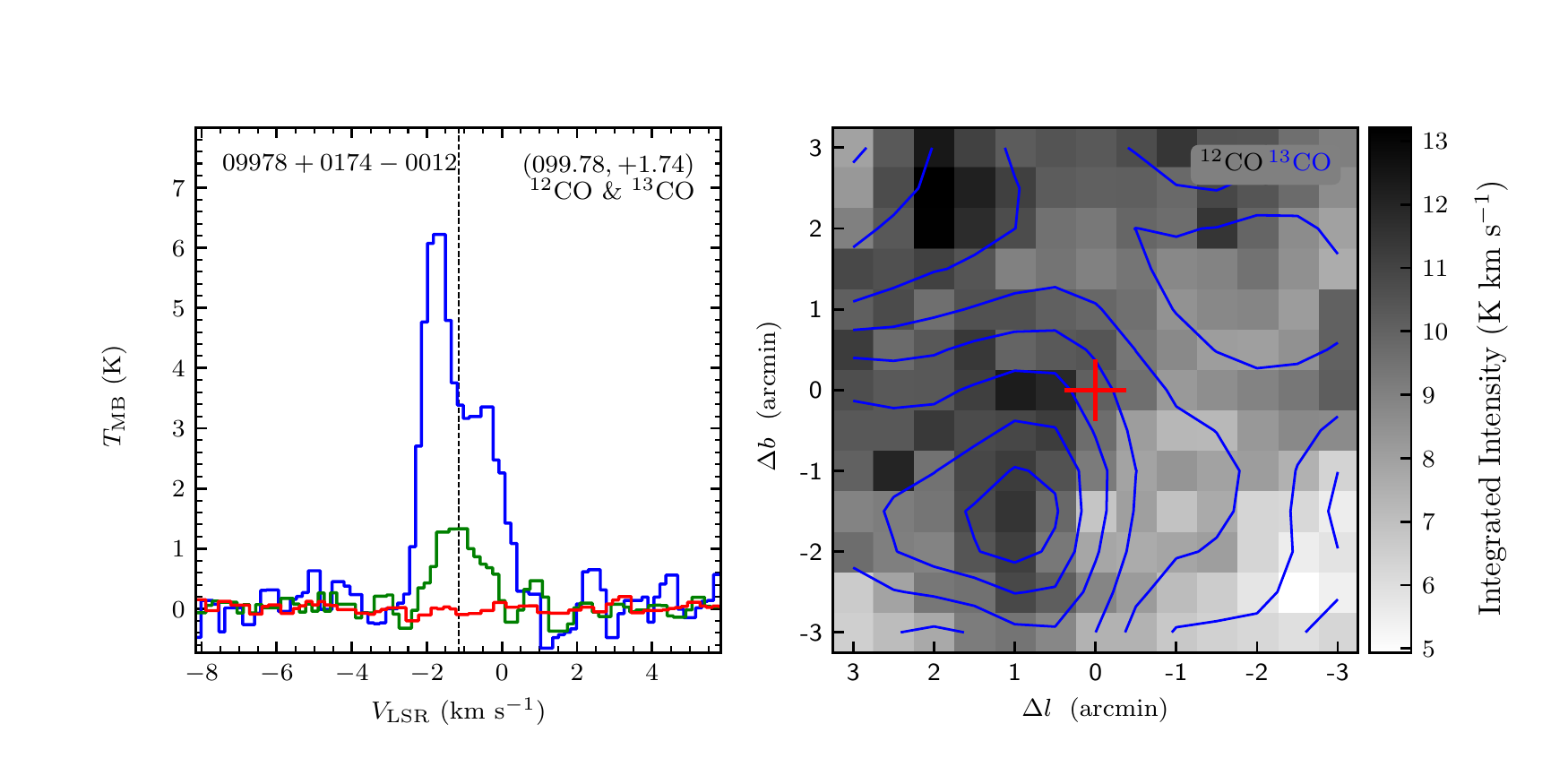}
\vspace{-0.5cm}

\includegraphics[width=9.0cm,angle=0]{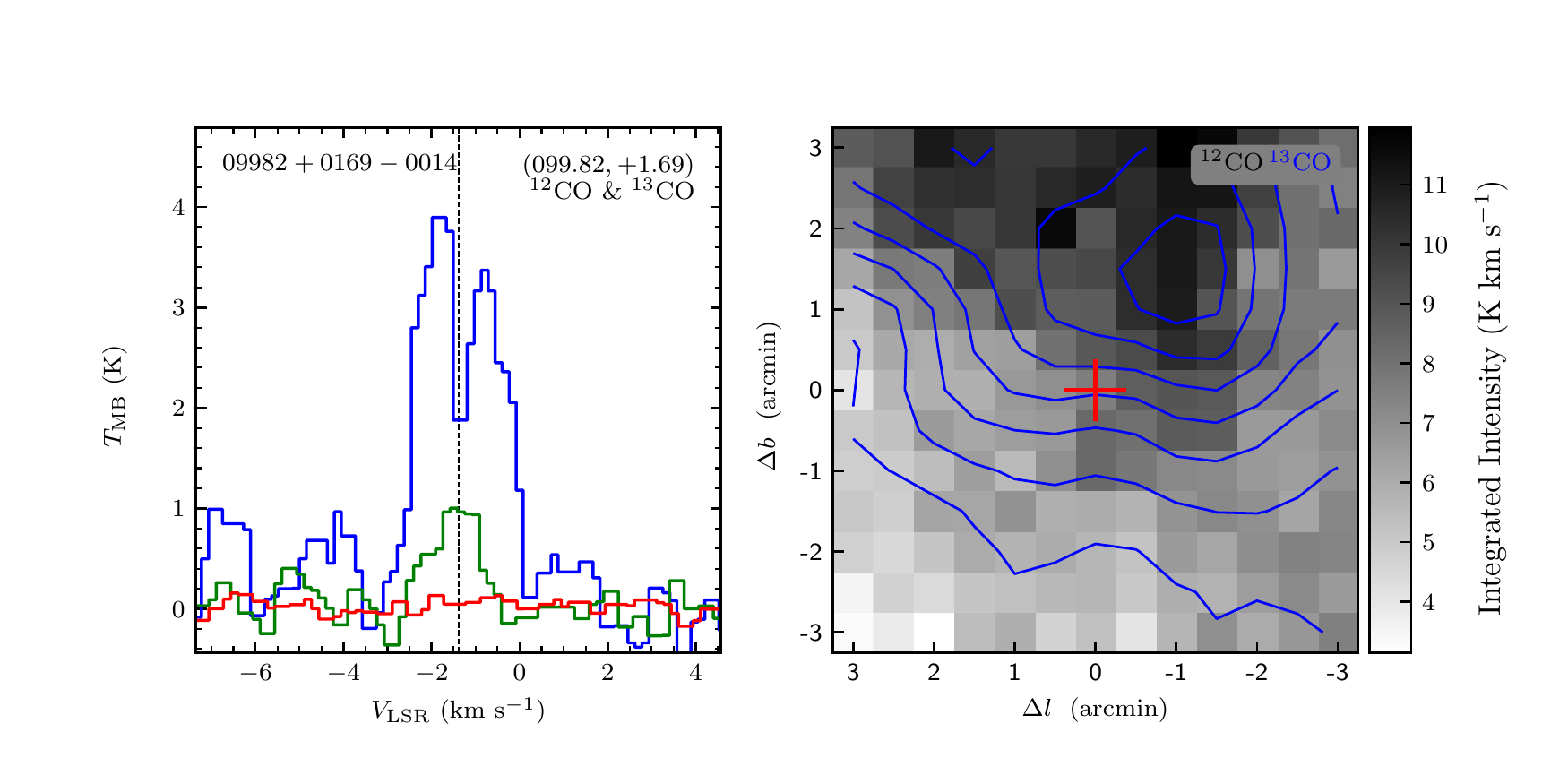}
\includegraphics[width=9.0cm,angle=0]{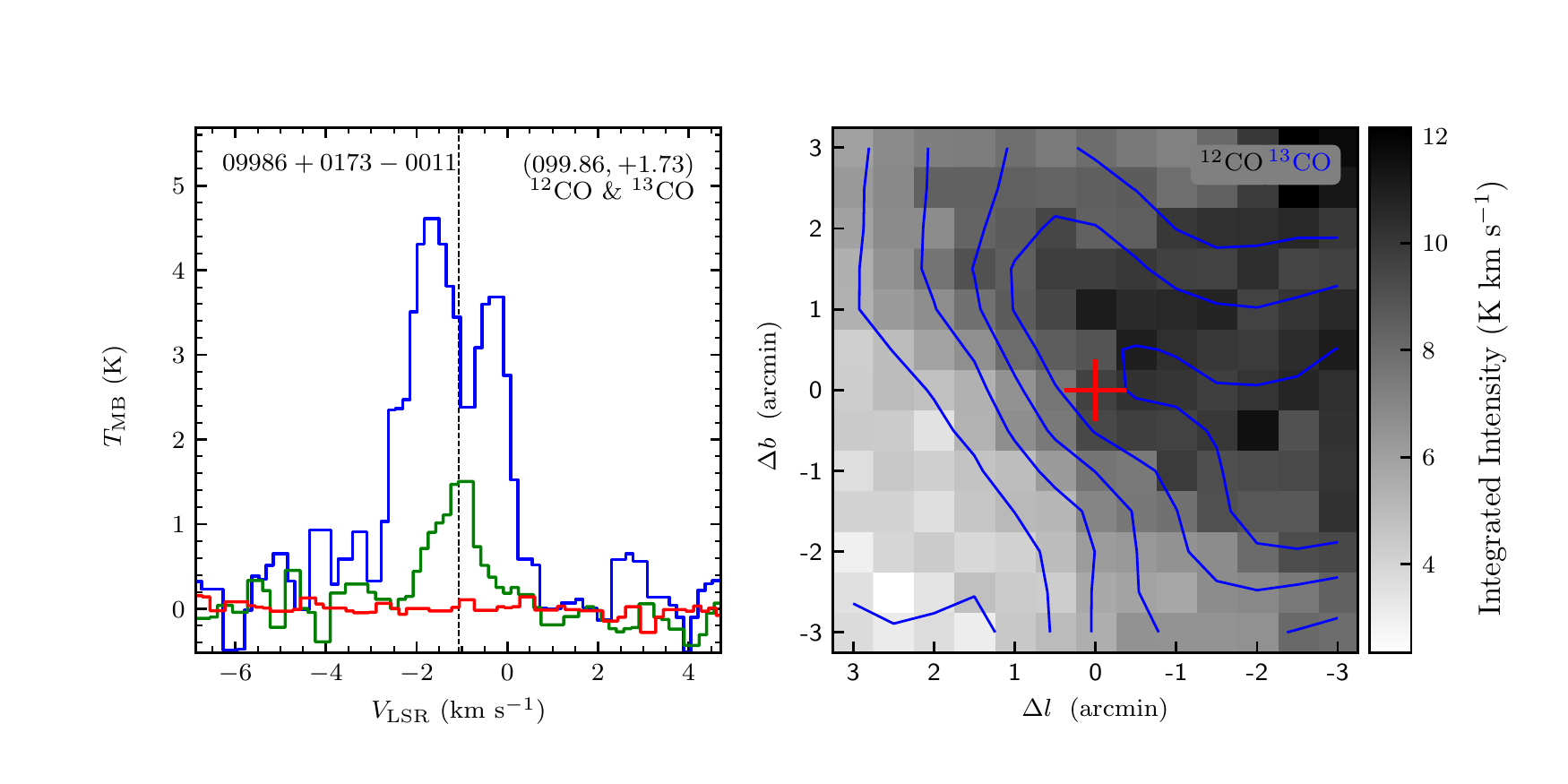}
\vspace{-0.5cm}

\includegraphics[width=9.0cm,angle=0]{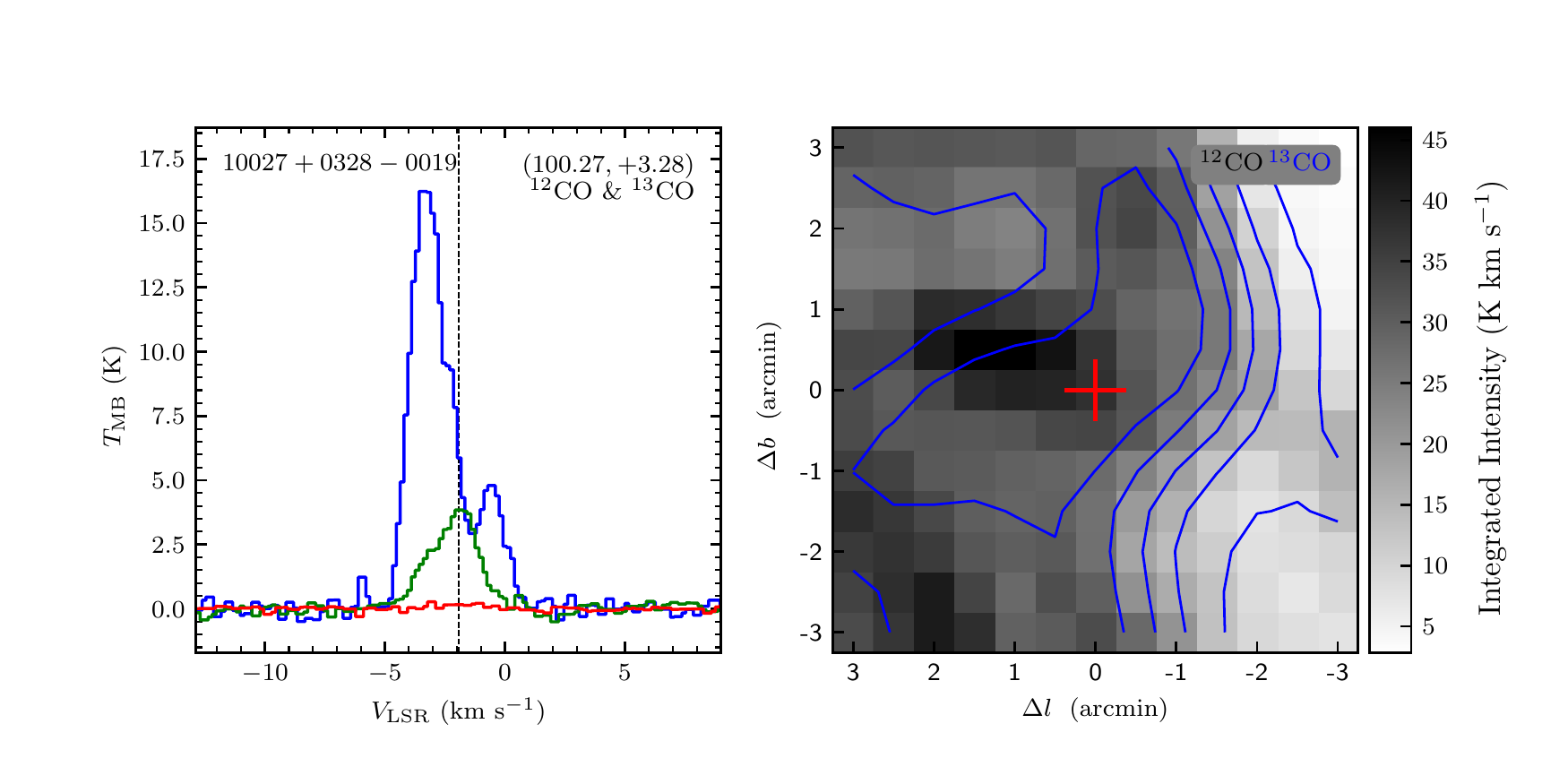}
\includegraphics[width=9.0cm,angle=0]{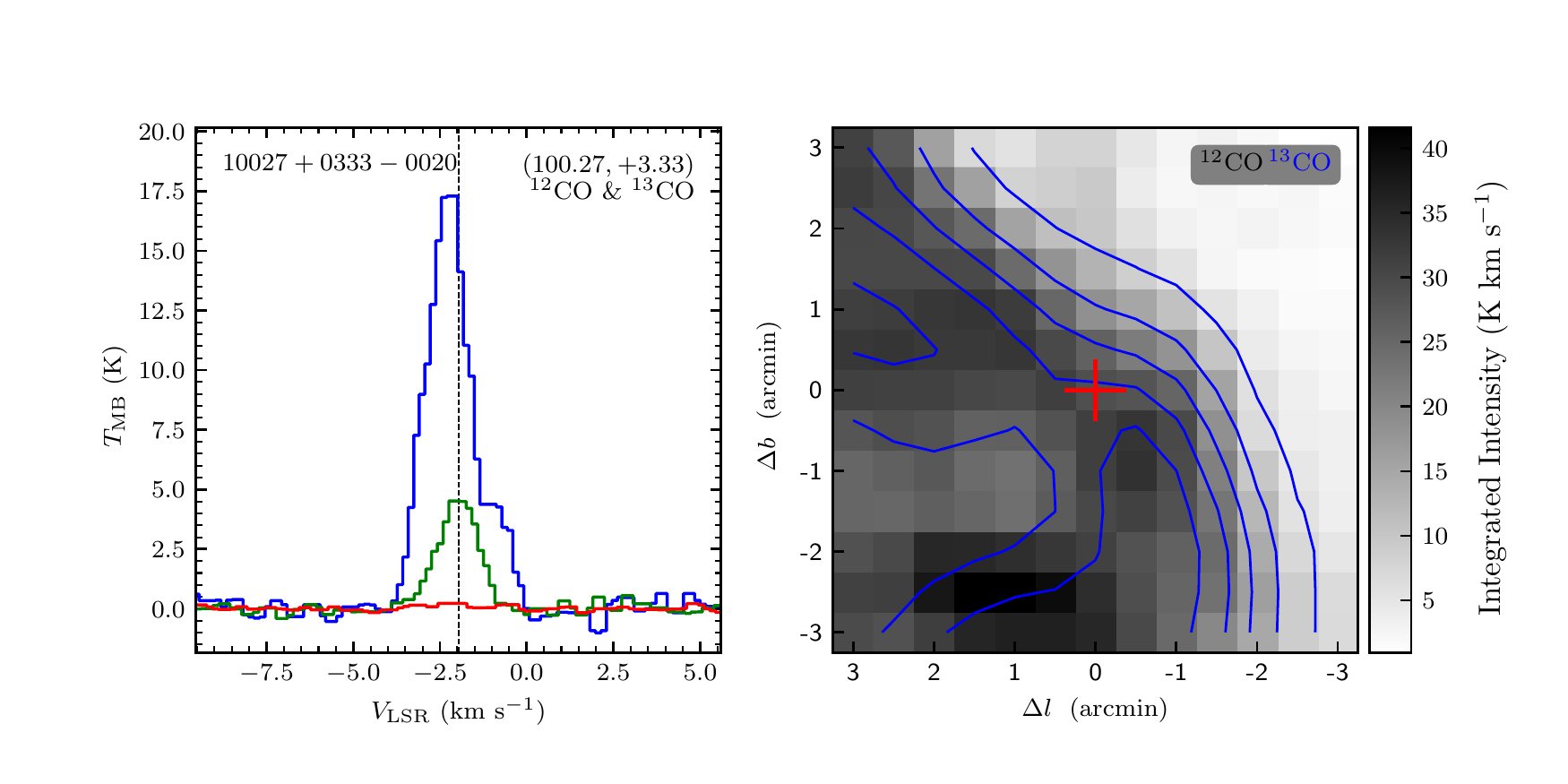}
\vspace{-0.5cm}

\includegraphics[width=9.0cm,angle=0]{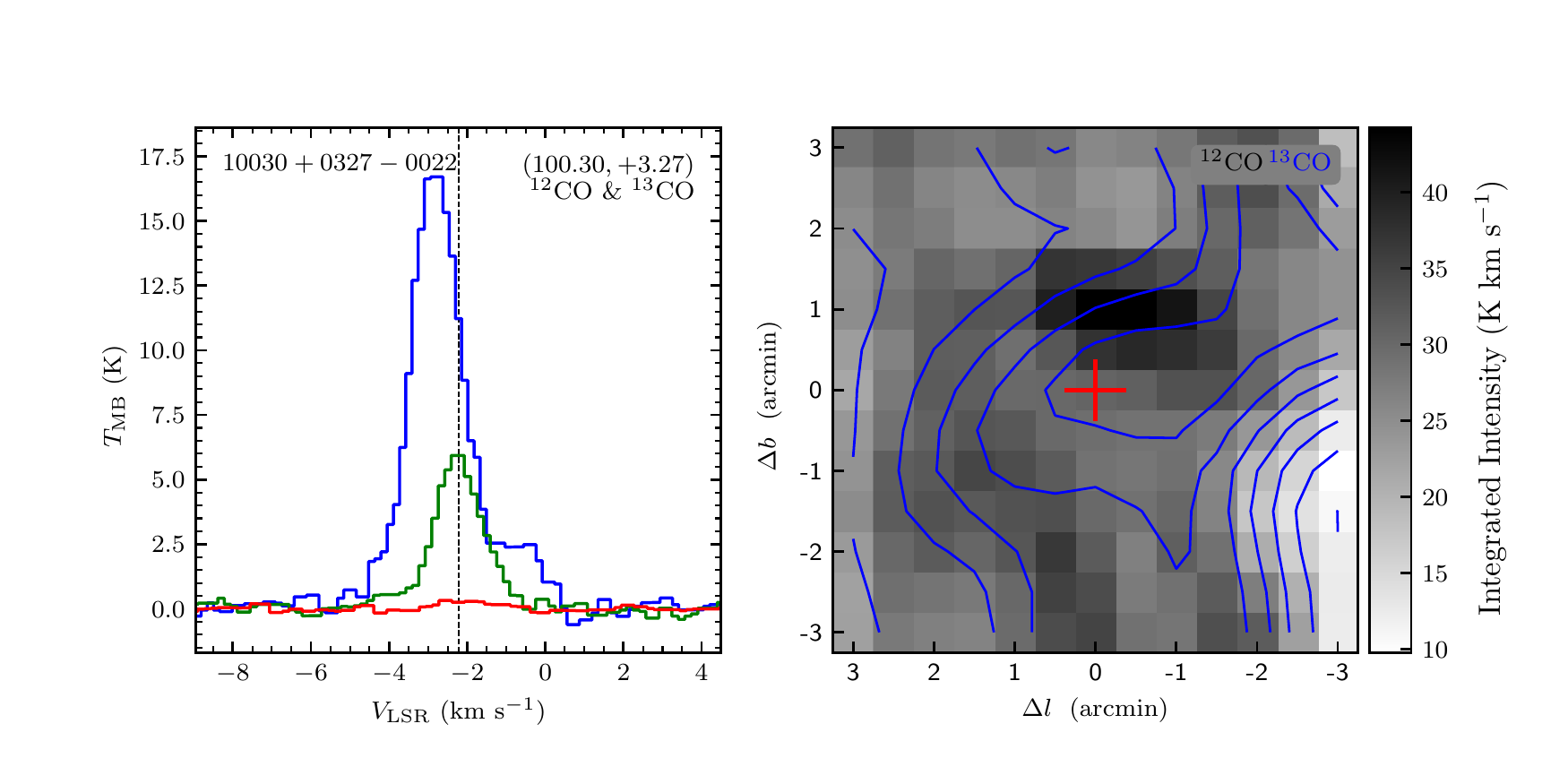}
\includegraphics[width=9.0cm,angle=0]{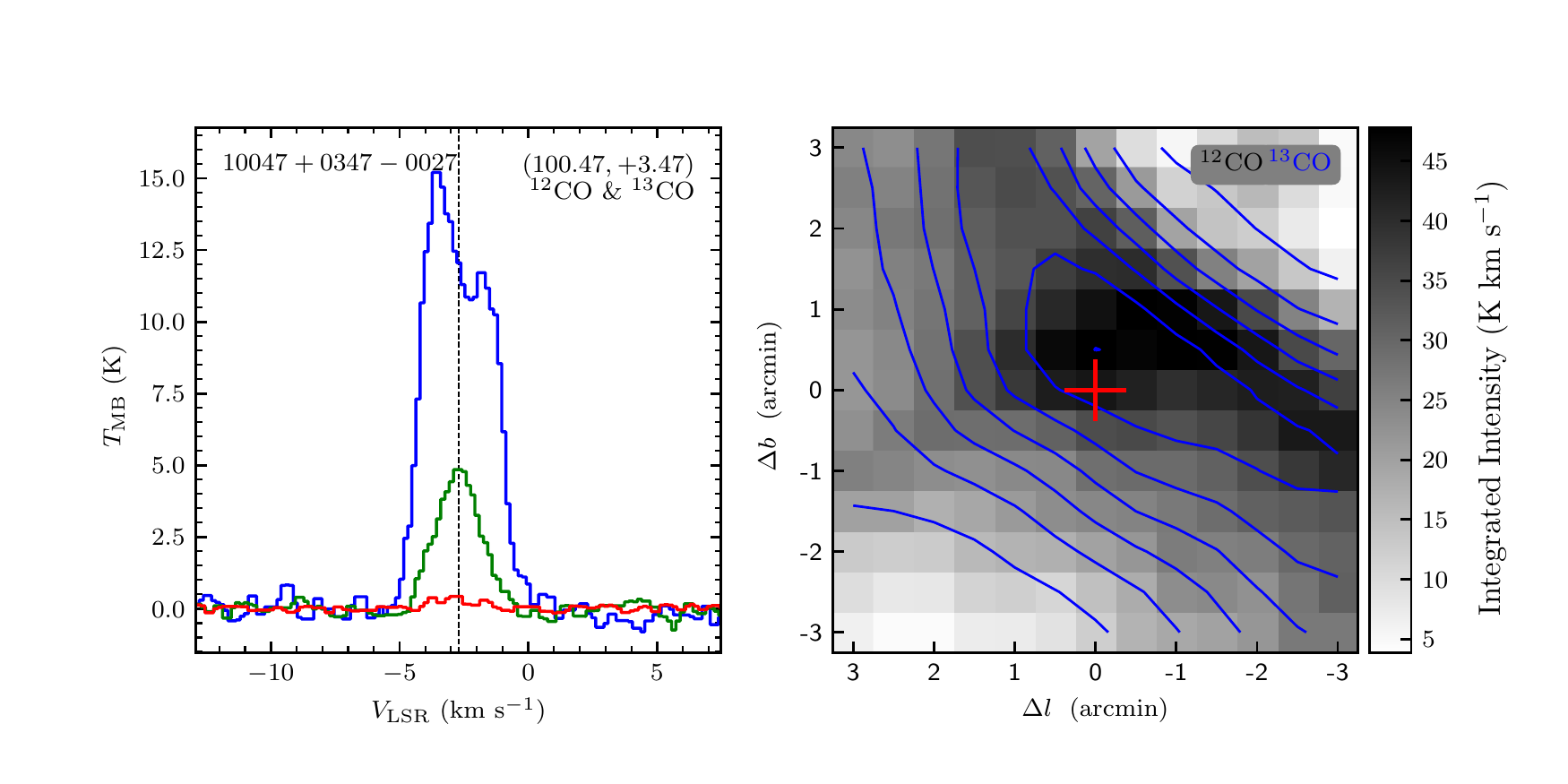}
\end{figure}
\clearpage

\begin{figure}
\includegraphics[width=9.0cm,angle=0]{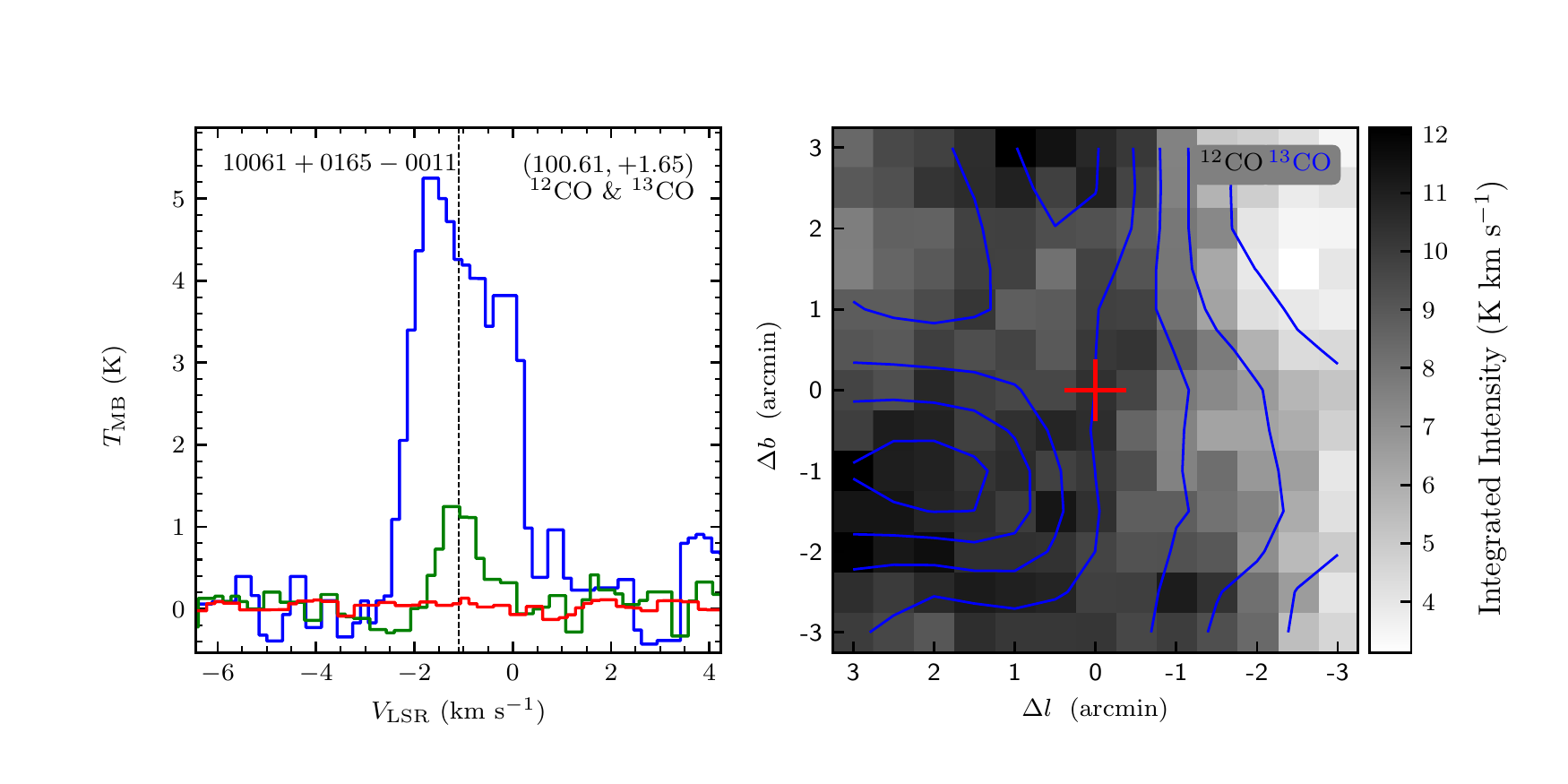}
\includegraphics[width=9.0cm,angle=0]{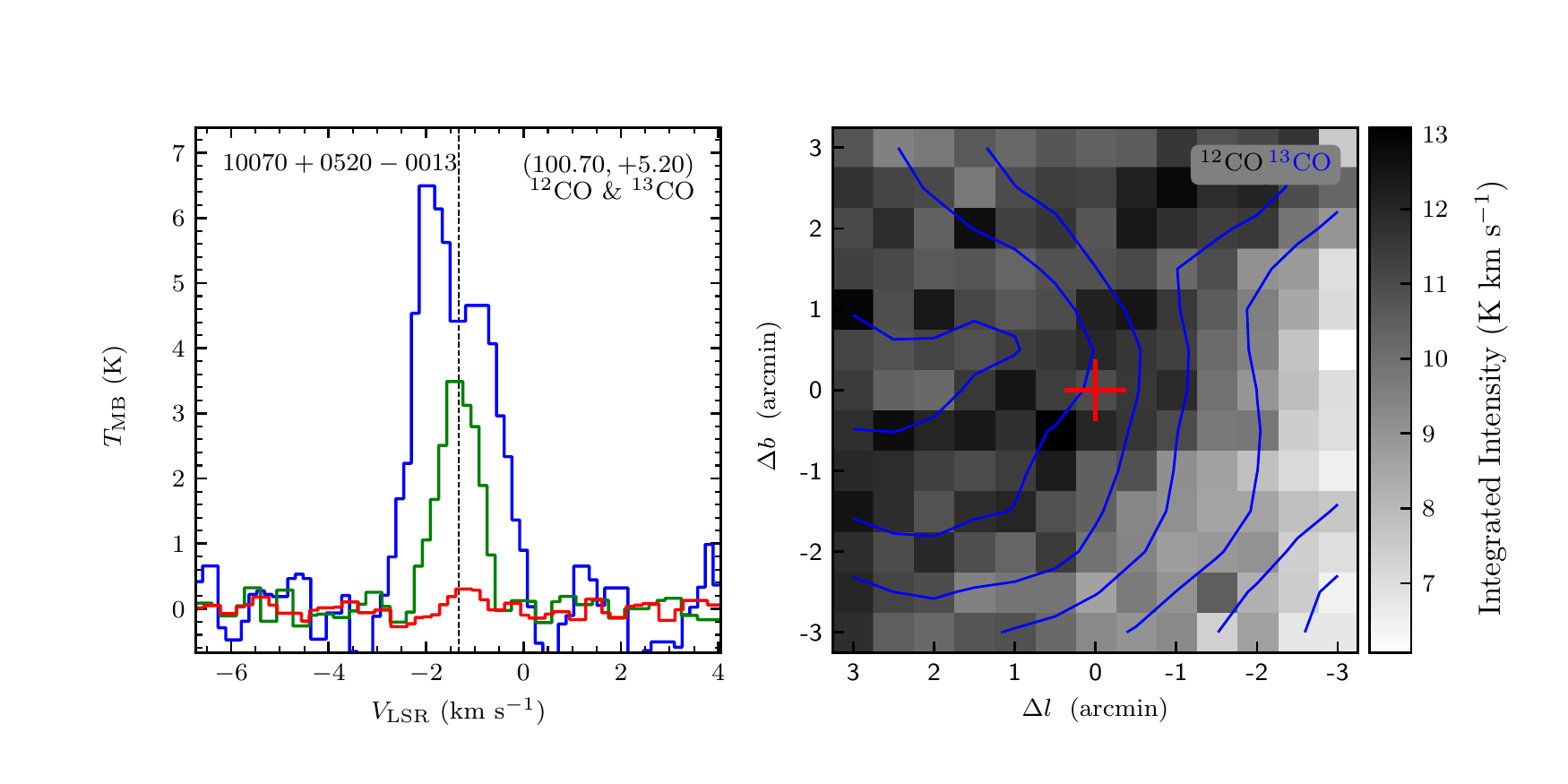}
\vspace{-0.5cm}

\includegraphics[width=9.0cm,angle=0]{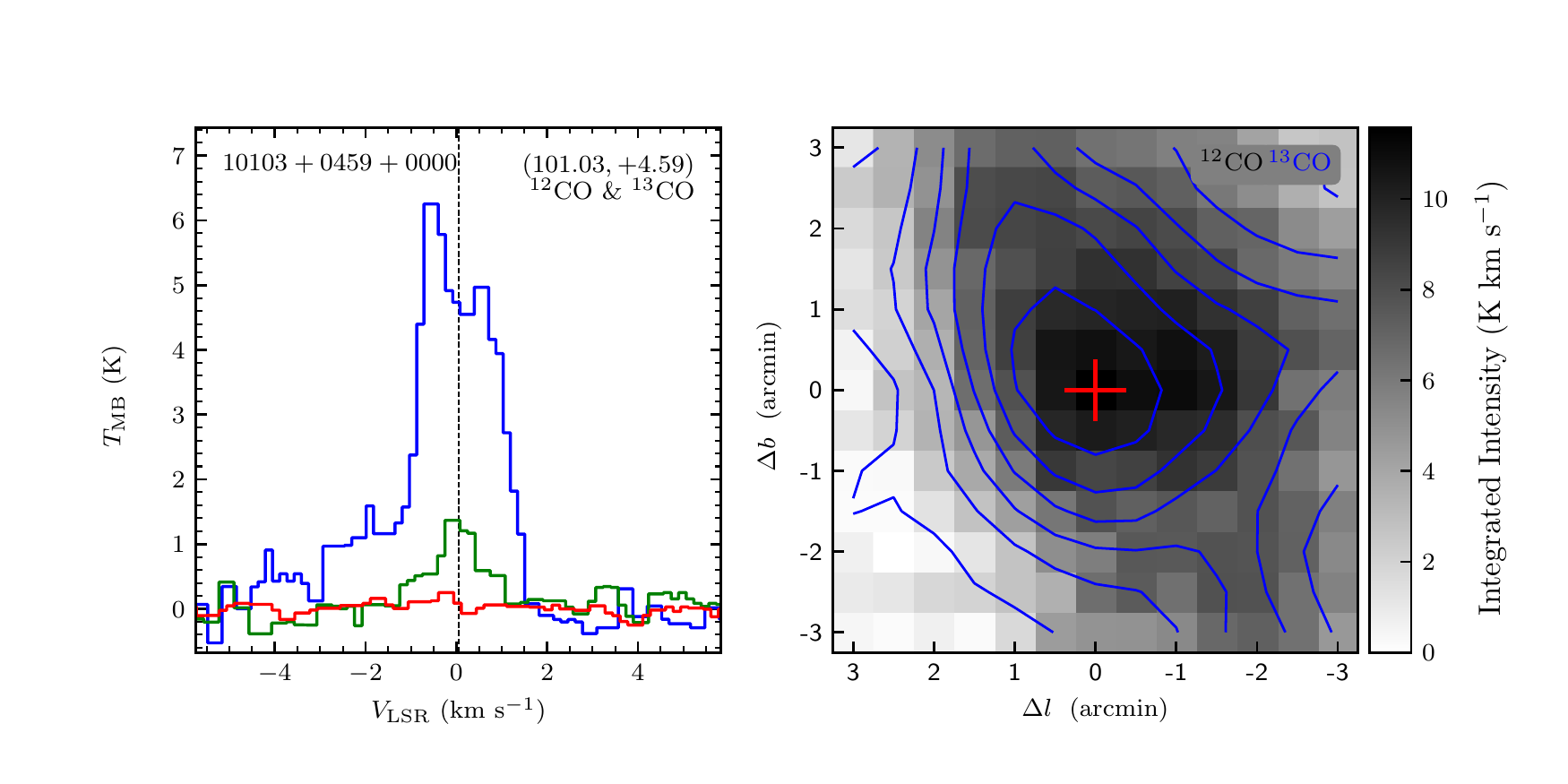}
\includegraphics[width=9.0cm,angle=0]{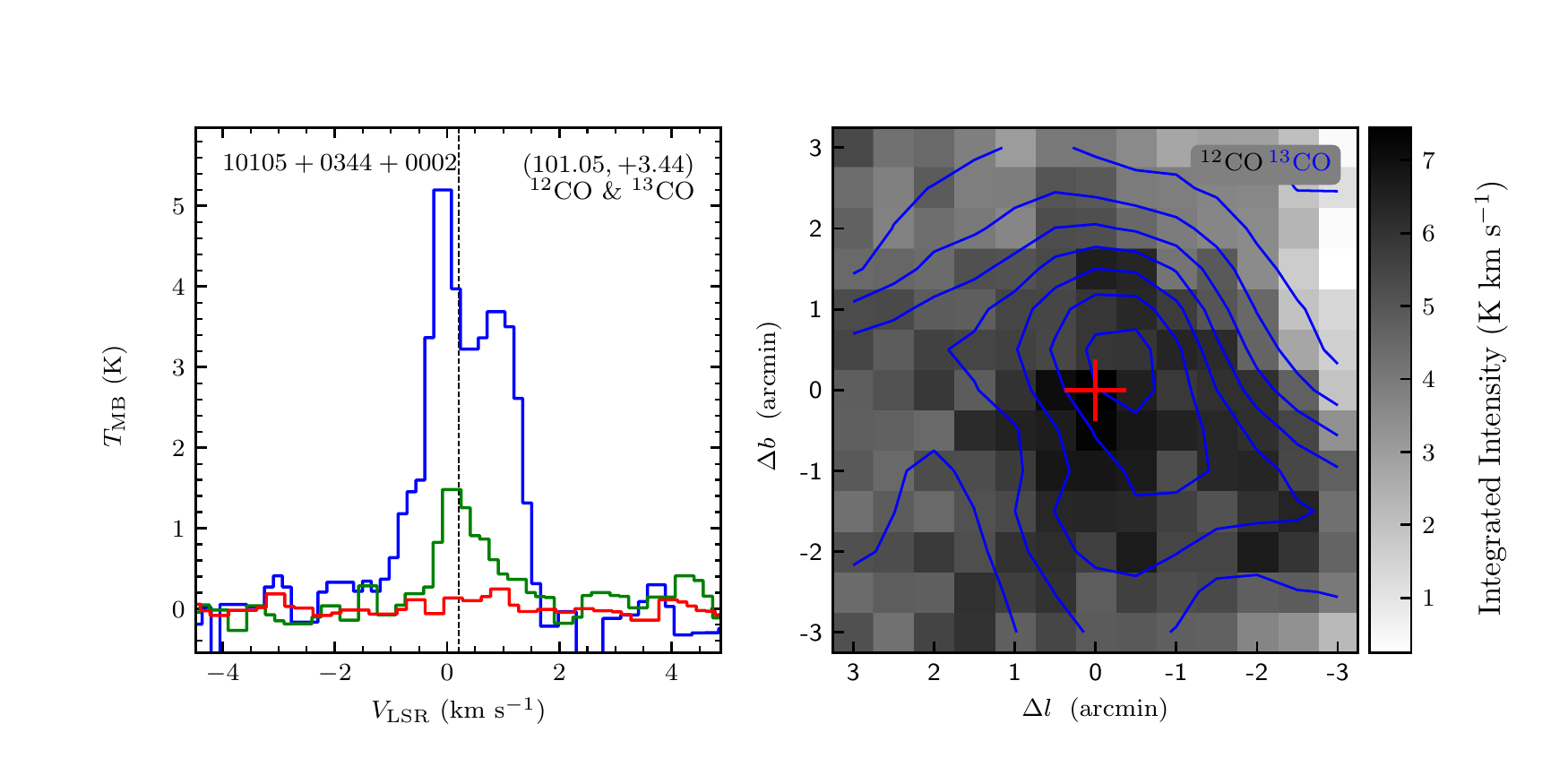}
\vspace{-0.5cm}

\includegraphics[width=9.0cm,angle=0]{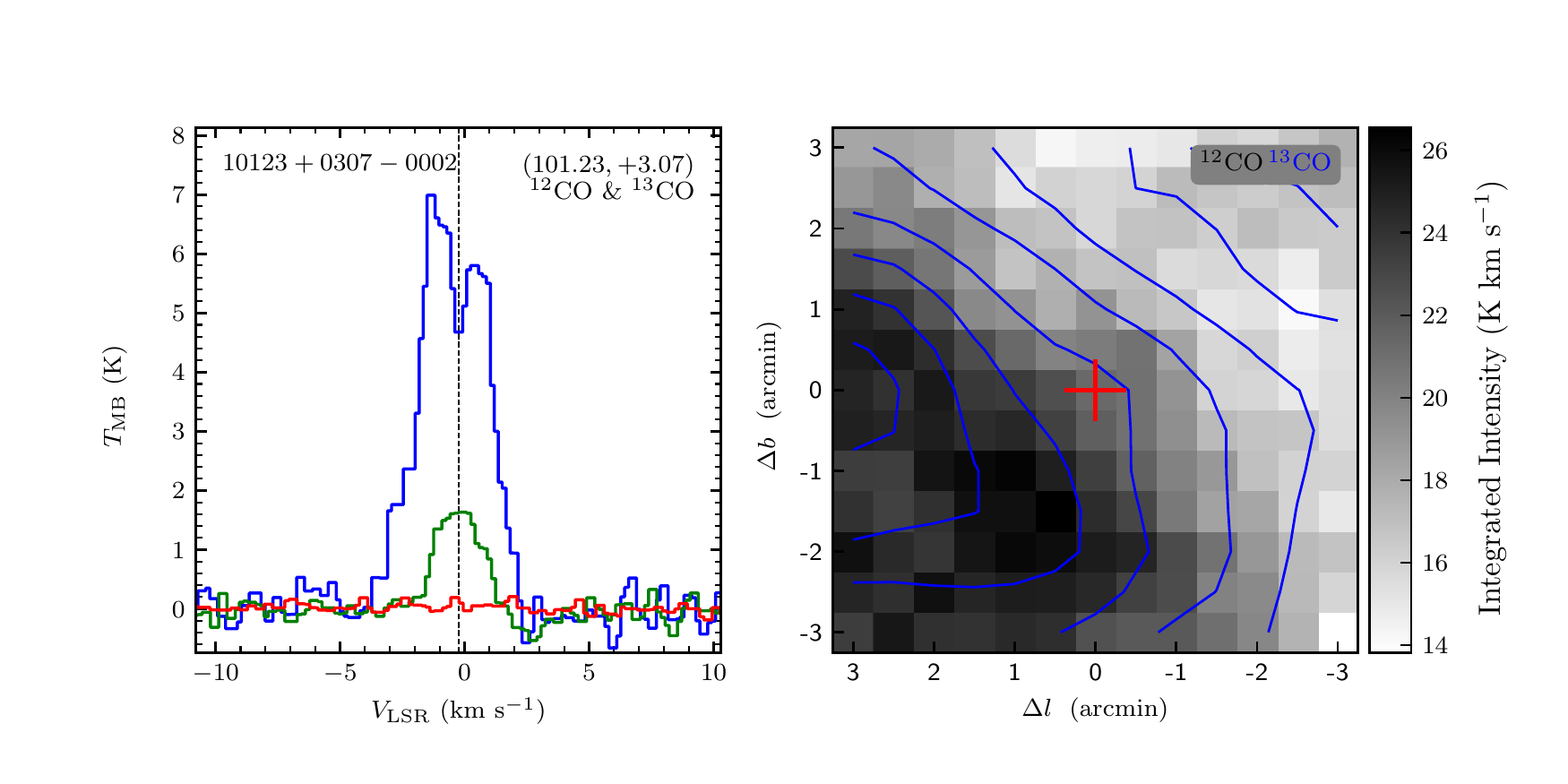}
\includegraphics[width=9.0cm,angle=0]{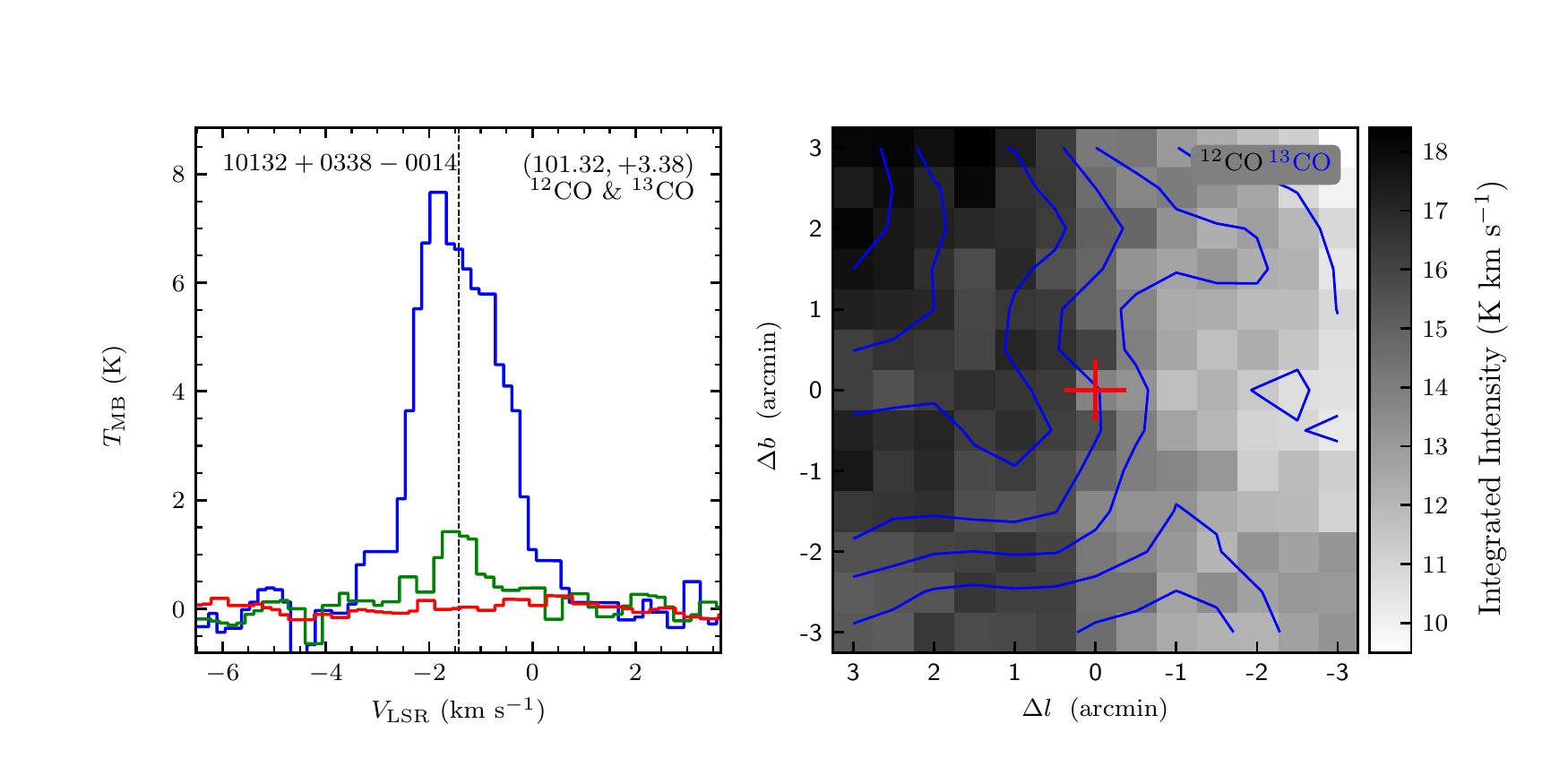}
\vspace{-0.5cm}

\includegraphics[width=9.0cm,angle=0]{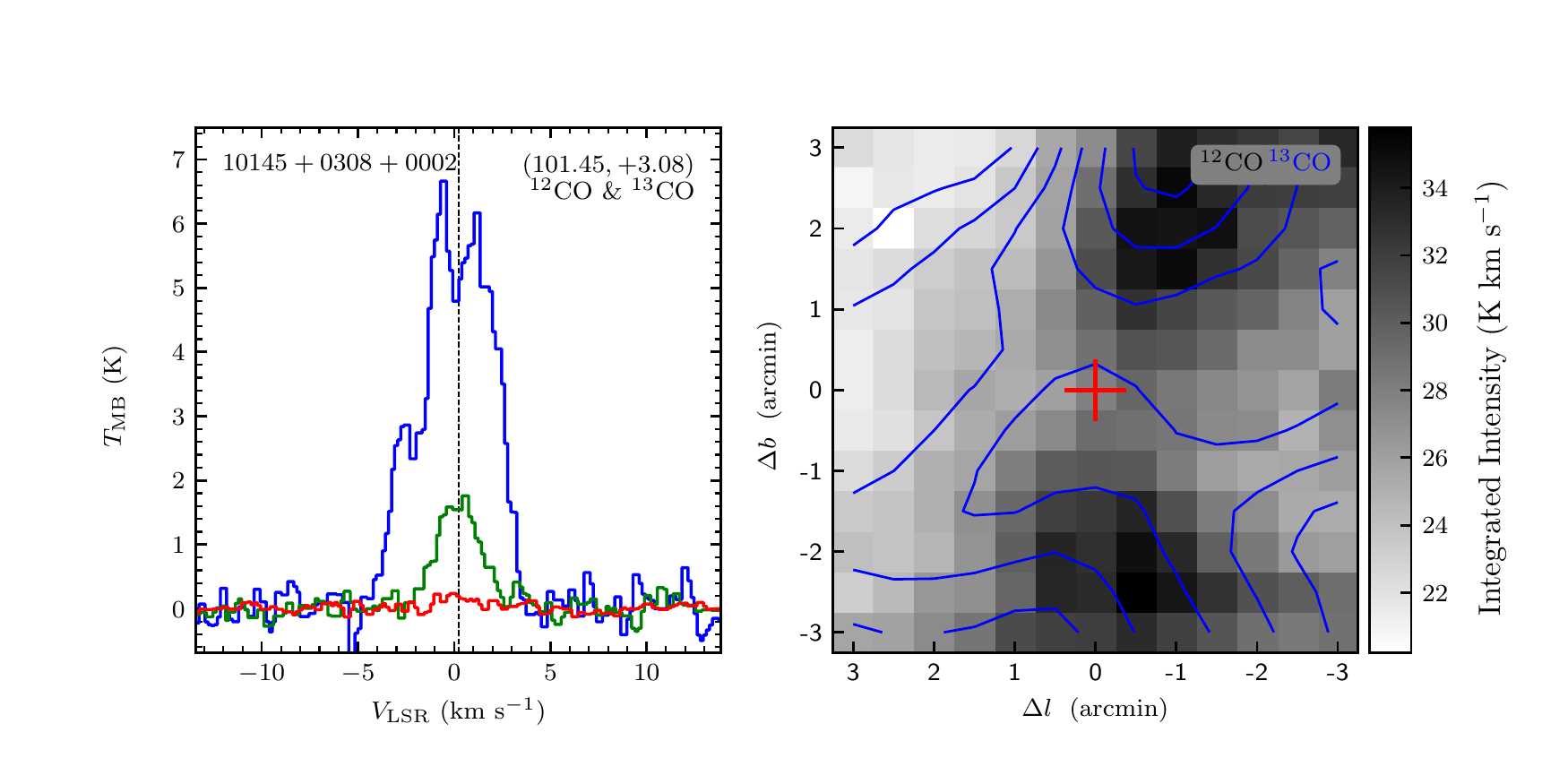}
\includegraphics[width=9.0cm,angle=0]{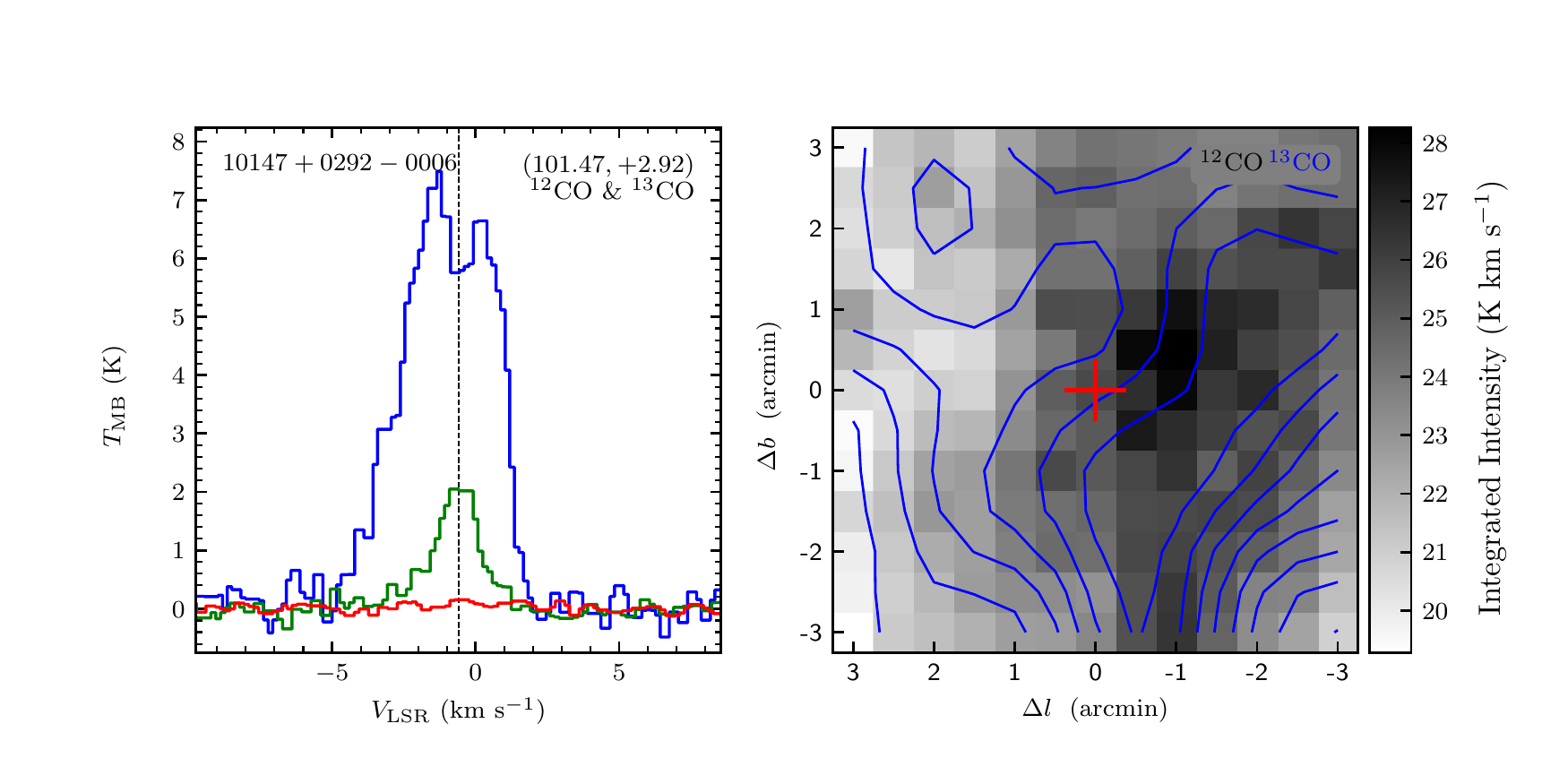}
\vspace{-0.5cm}

\includegraphics[width=9.0cm,angle=0]{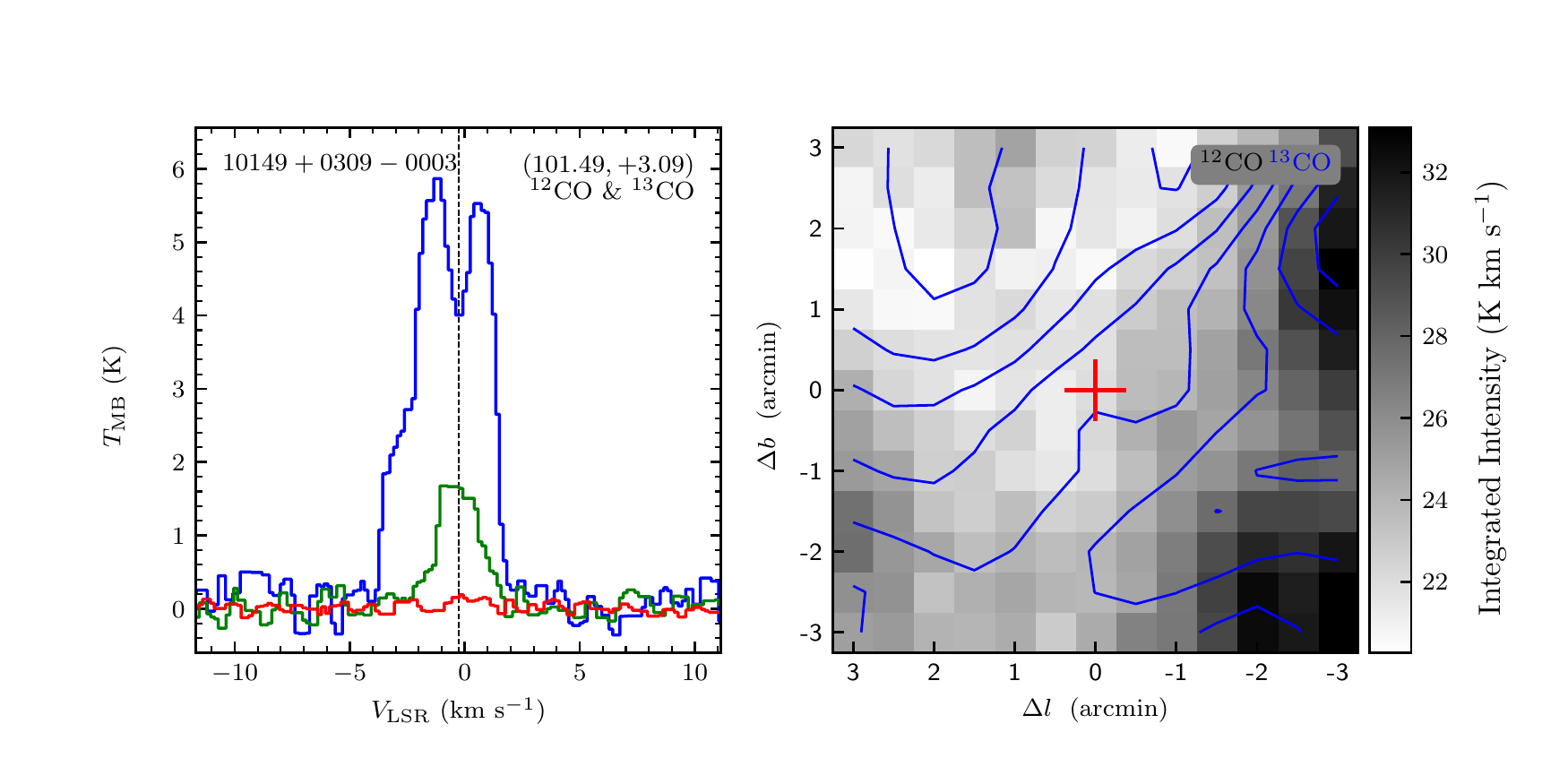}
\includegraphics[width=9.0cm,angle=0]{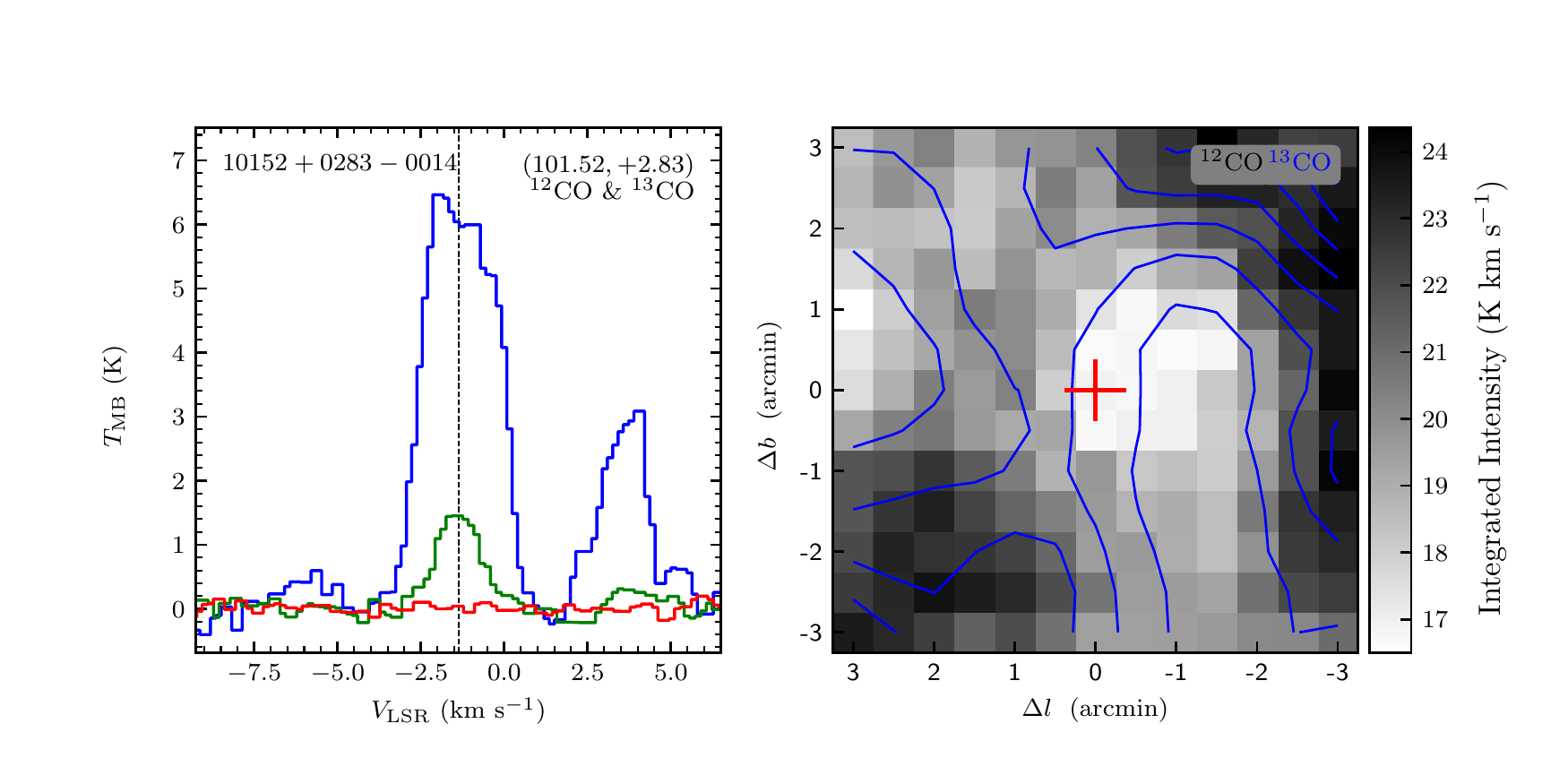}
\end{figure}
\clearpage

\begin{figure}
\includegraphics[width=9.0cm,angle=0]{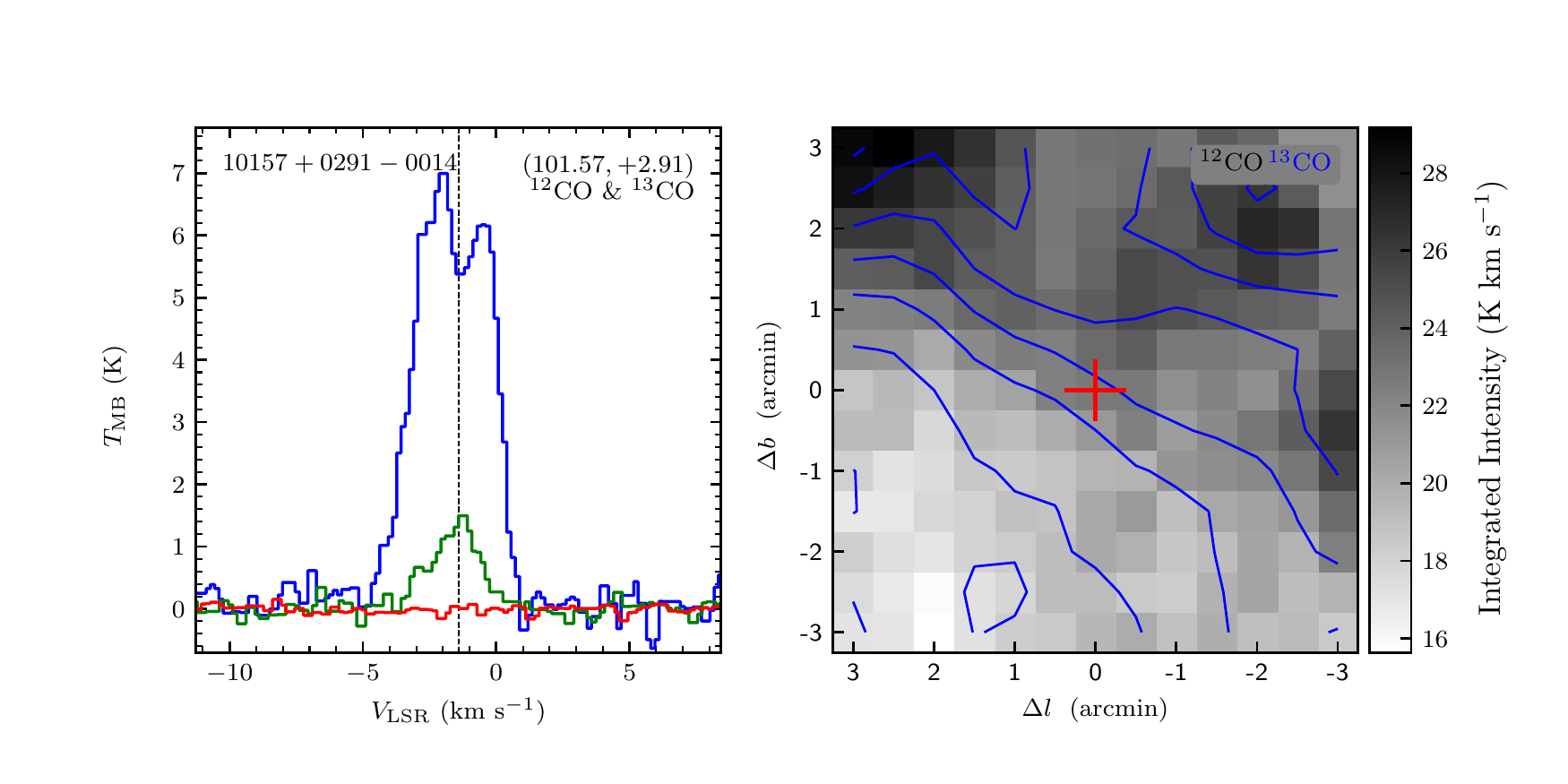}
\includegraphics[width=9.0cm,angle=0]{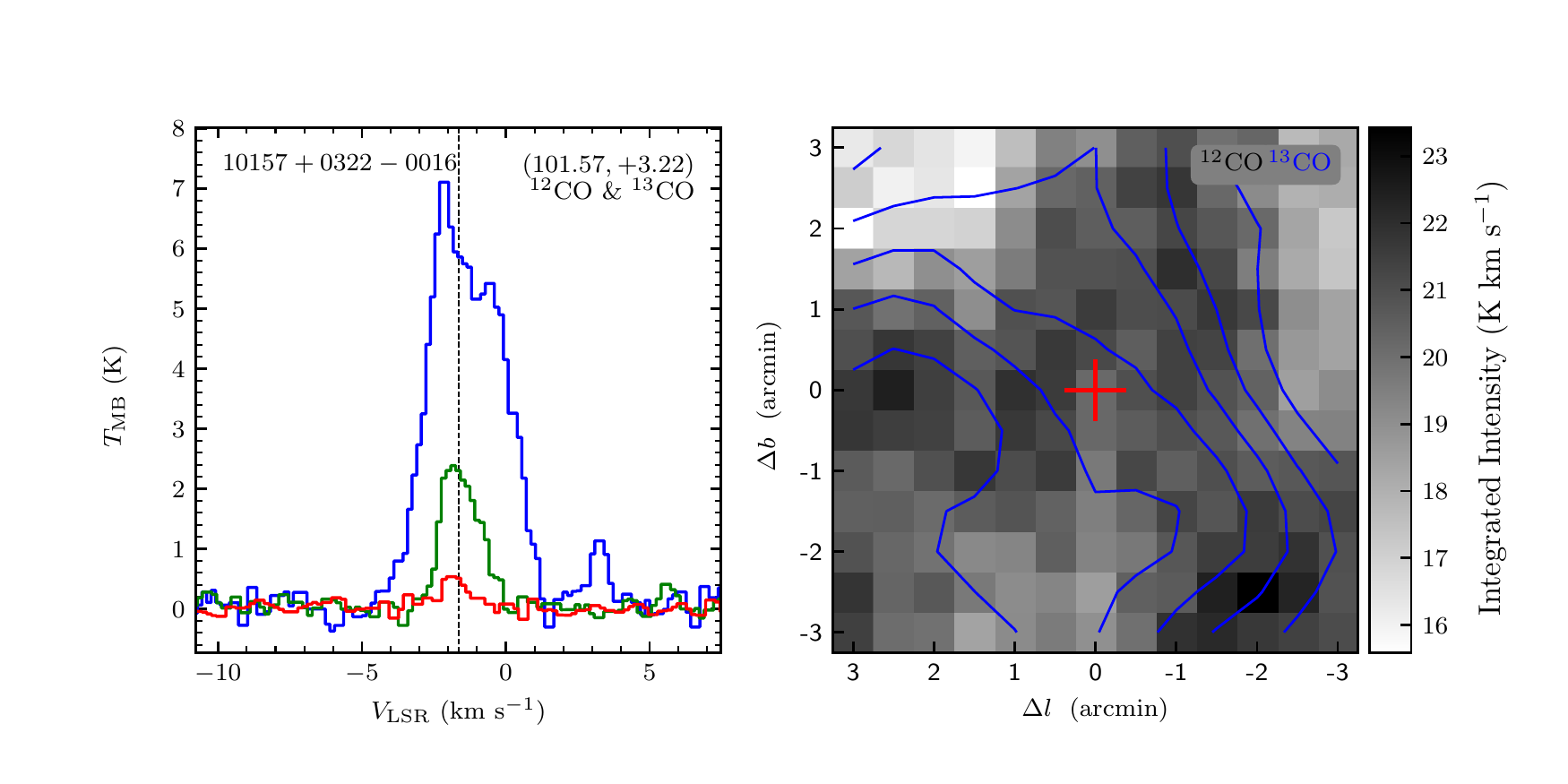}
\vspace{-0.5cm}

\includegraphics[width=9.0cm,angle=0]{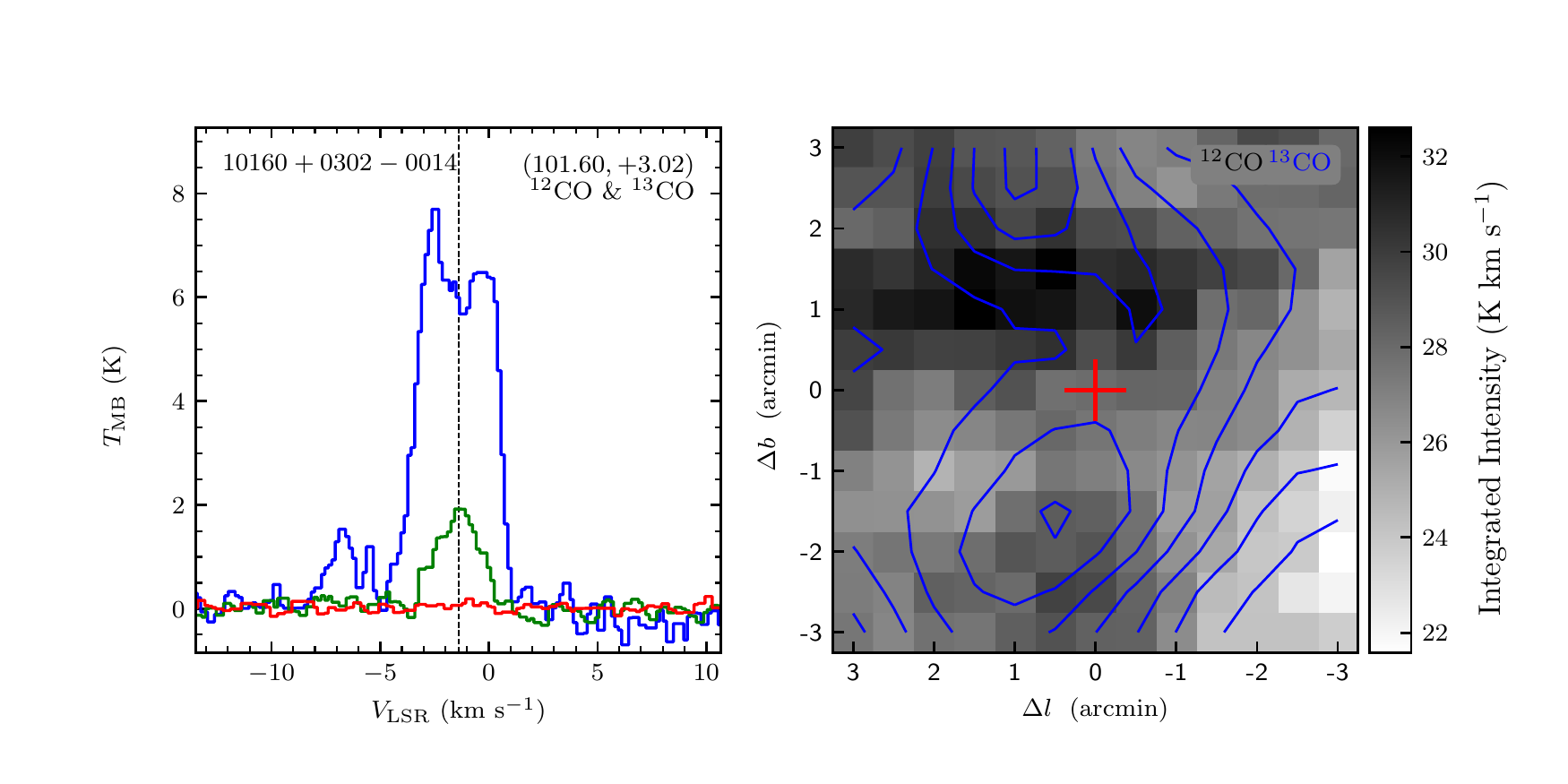}
\includegraphics[width=9.0cm,angle=0]{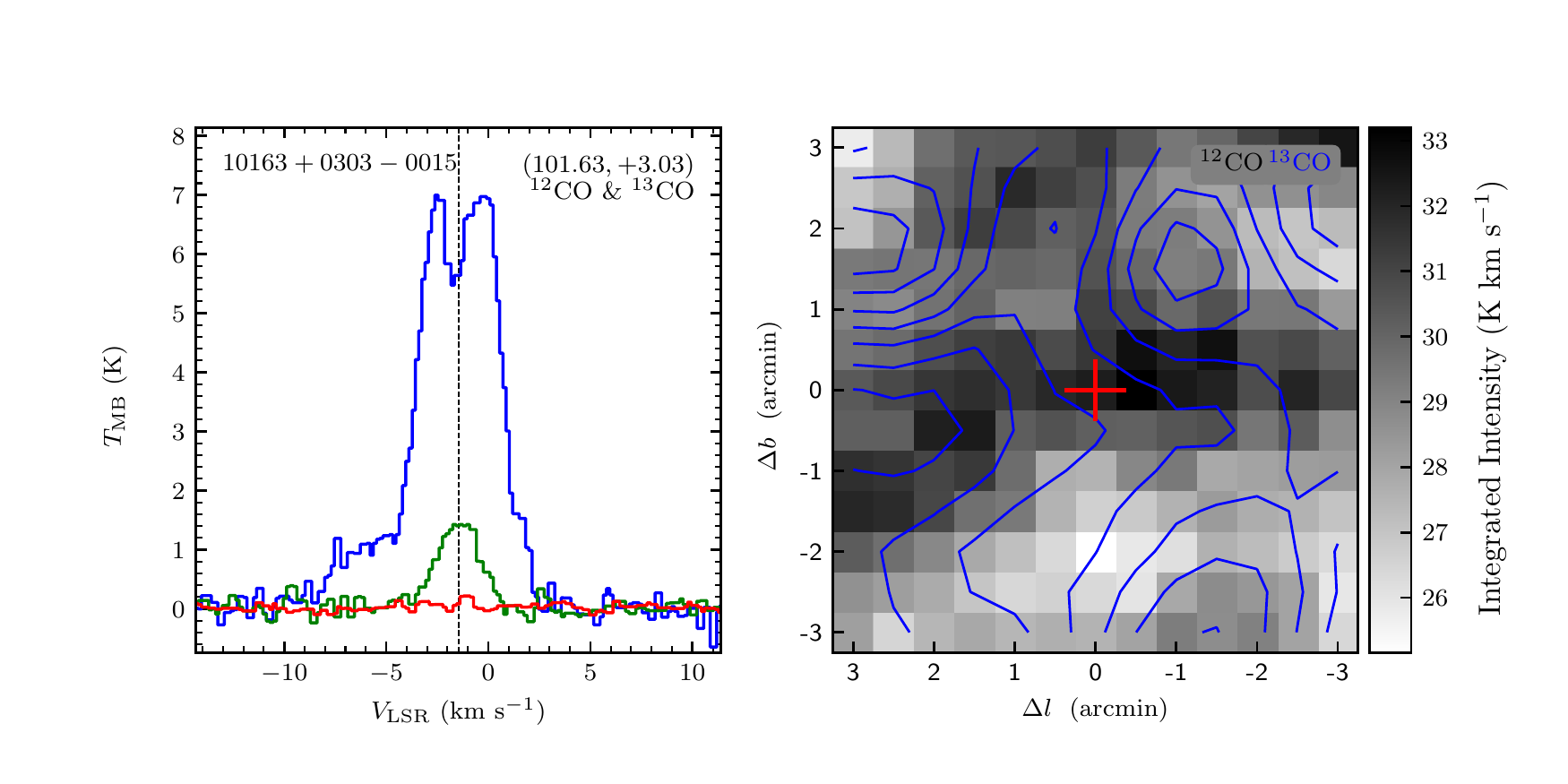}
\vspace{-0.5cm}

\includegraphics[width=9.0cm,angle=0]{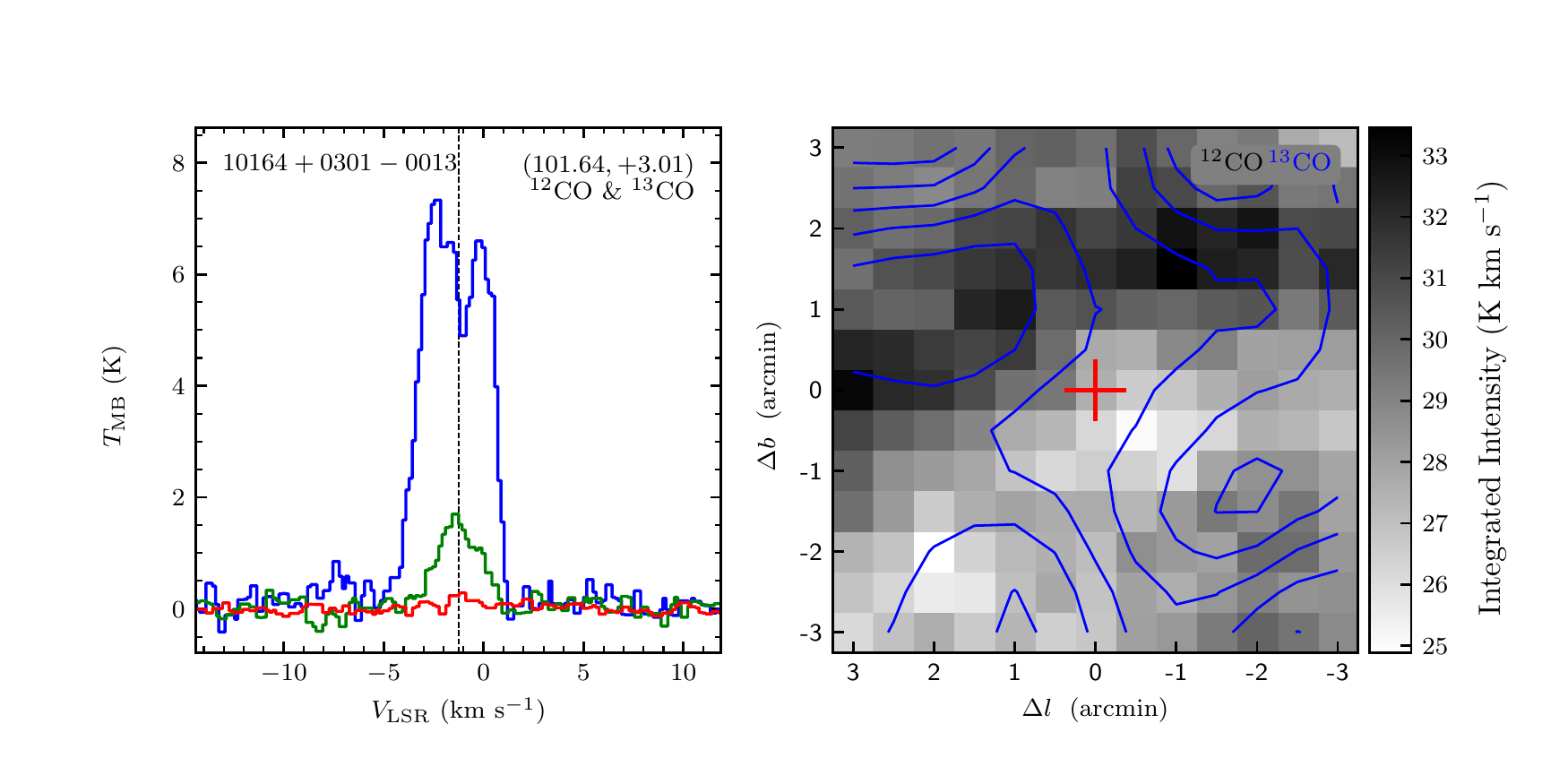}
\includegraphics[width=9.0cm,angle=0]{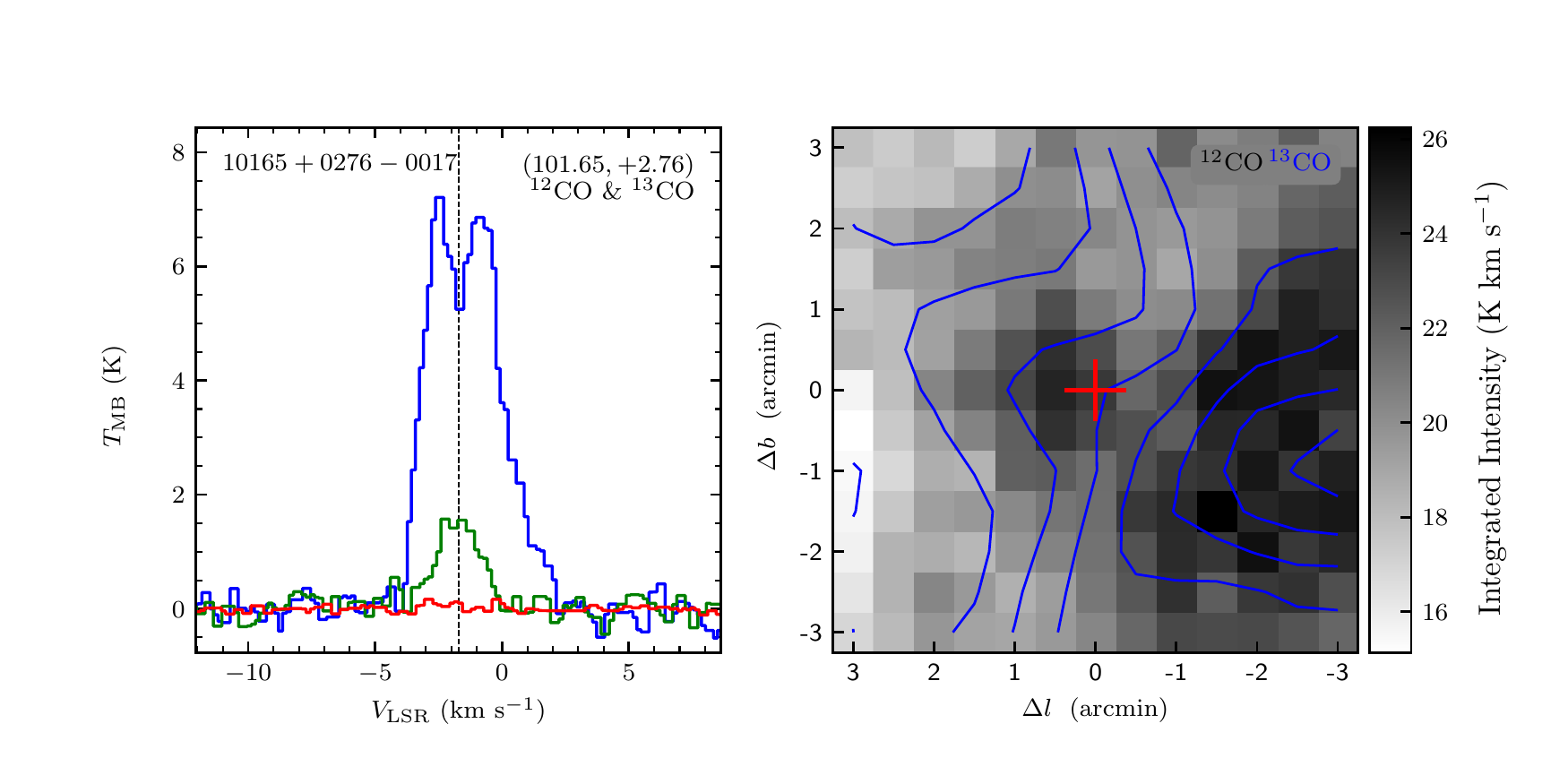}
\vspace{-0.5cm}

\includegraphics[width=9.0cm,angle=0]{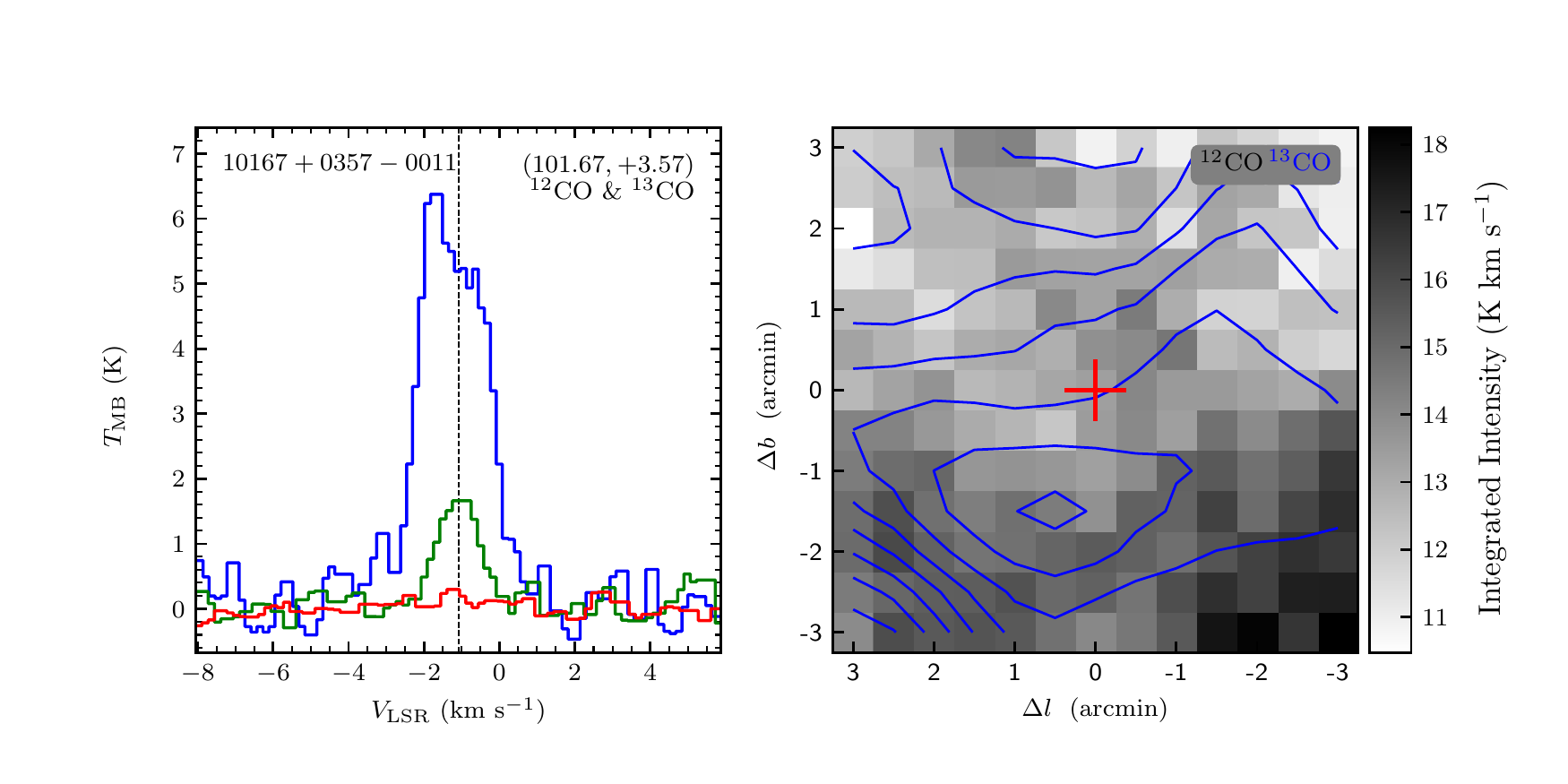}
\includegraphics[width=9.0cm,angle=0]{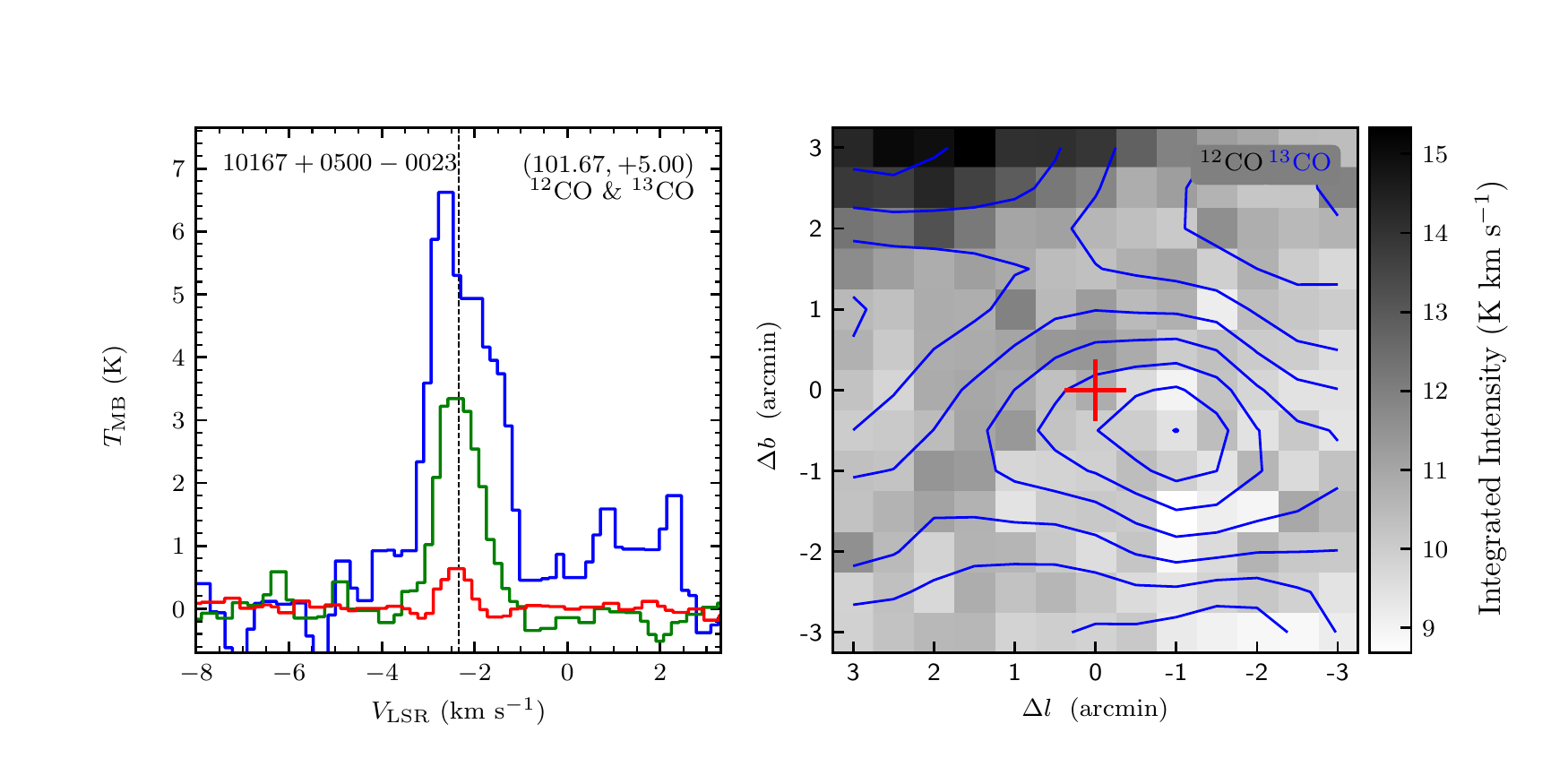}
\vspace{-0.5cm}

\includegraphics[width=9.0cm,angle=0]{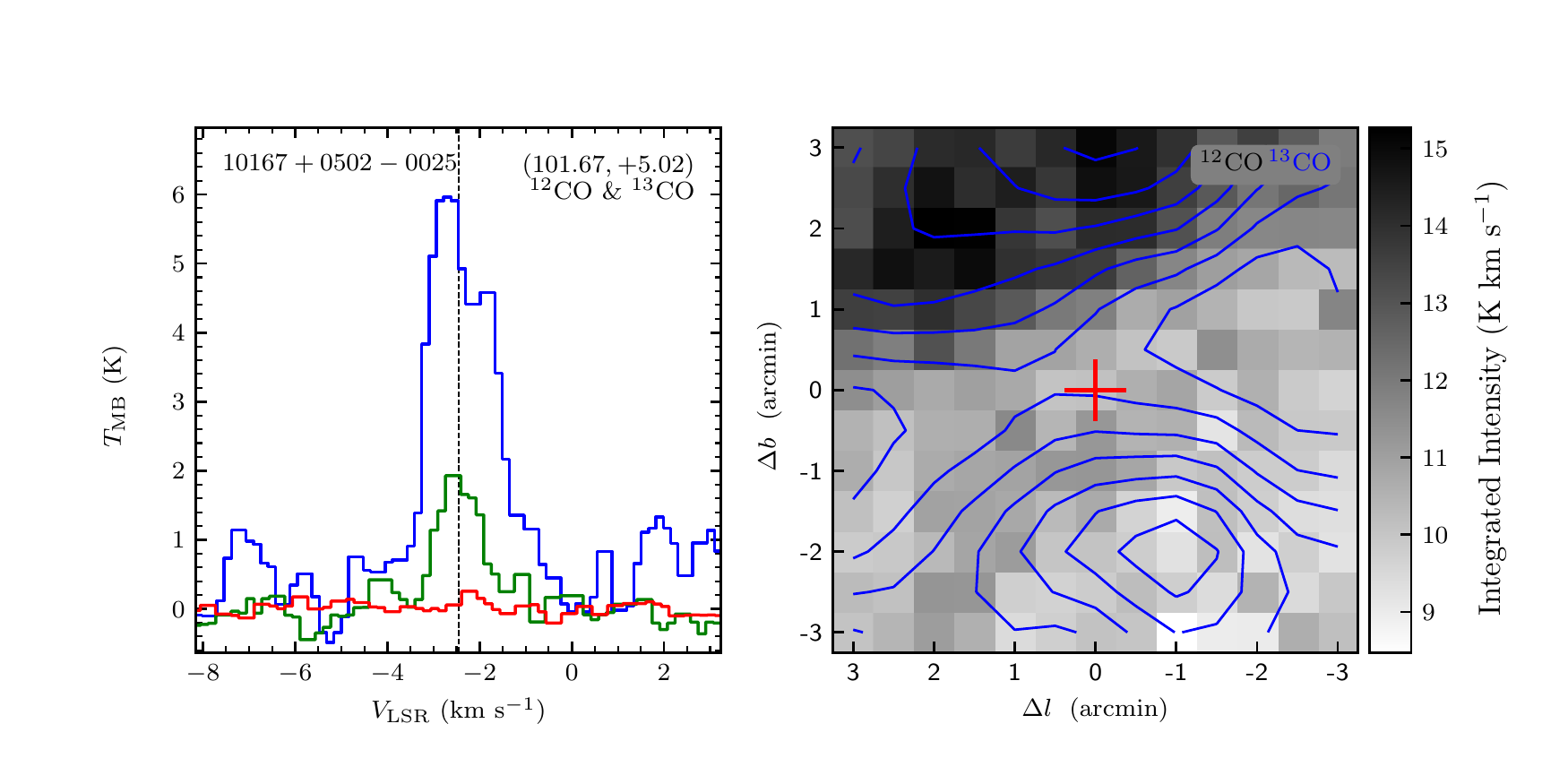}
\includegraphics[width=9.0cm,angle=0]{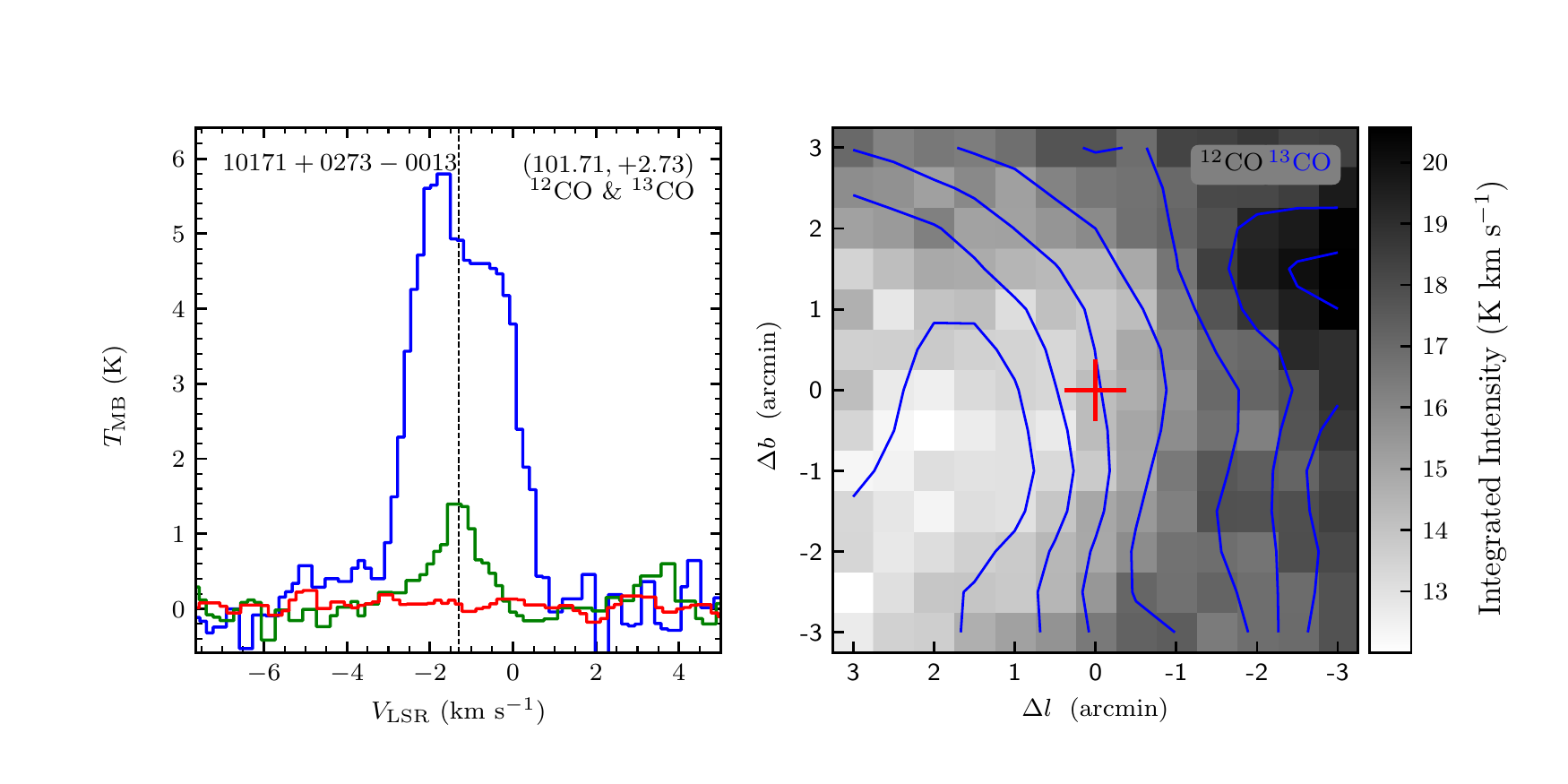}
\end{figure}
\clearpage

\begin{figure}
\includegraphics[width=9.0cm,angle=0]{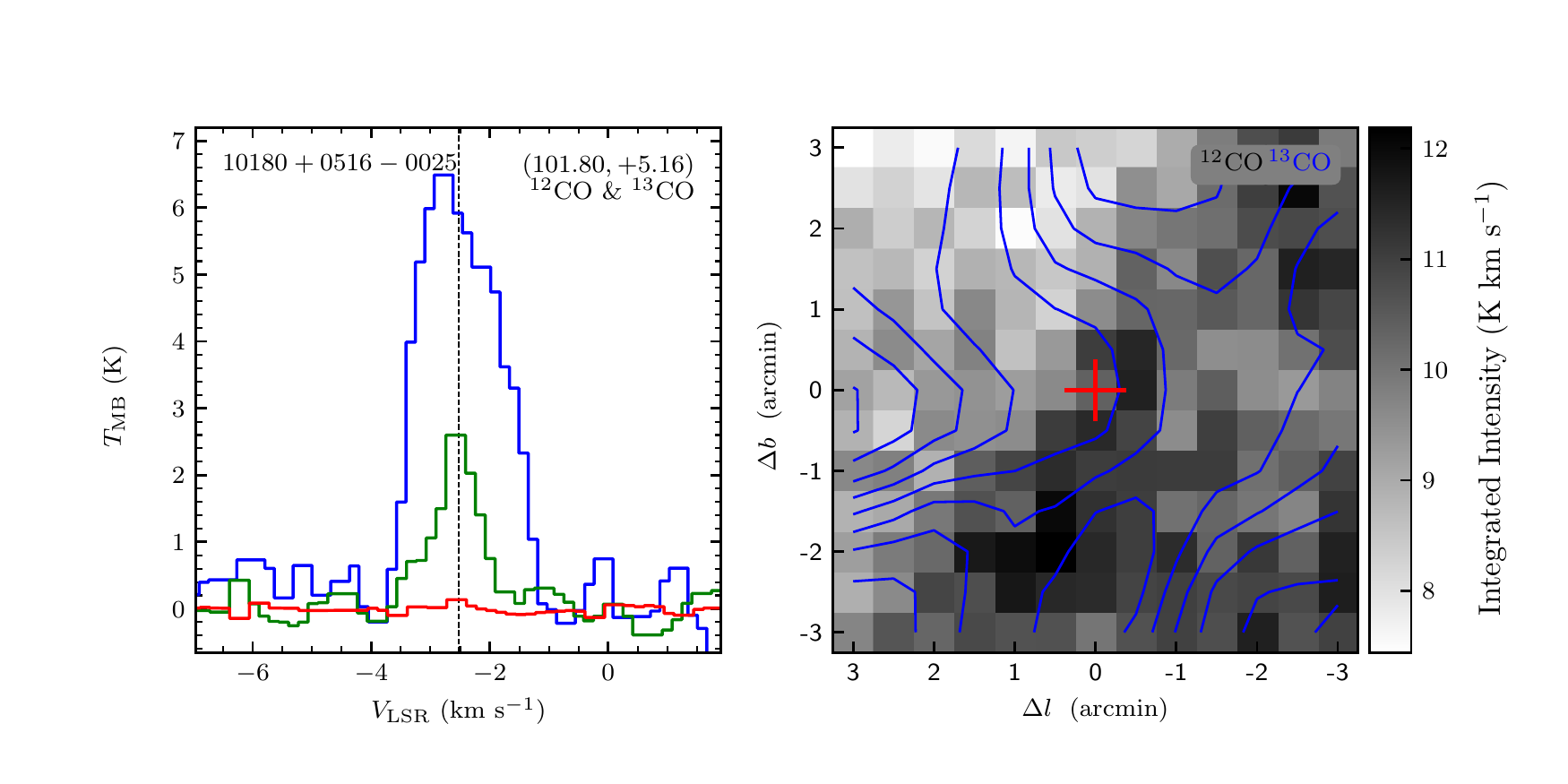}
\includegraphics[width=9.0cm,angle=0]{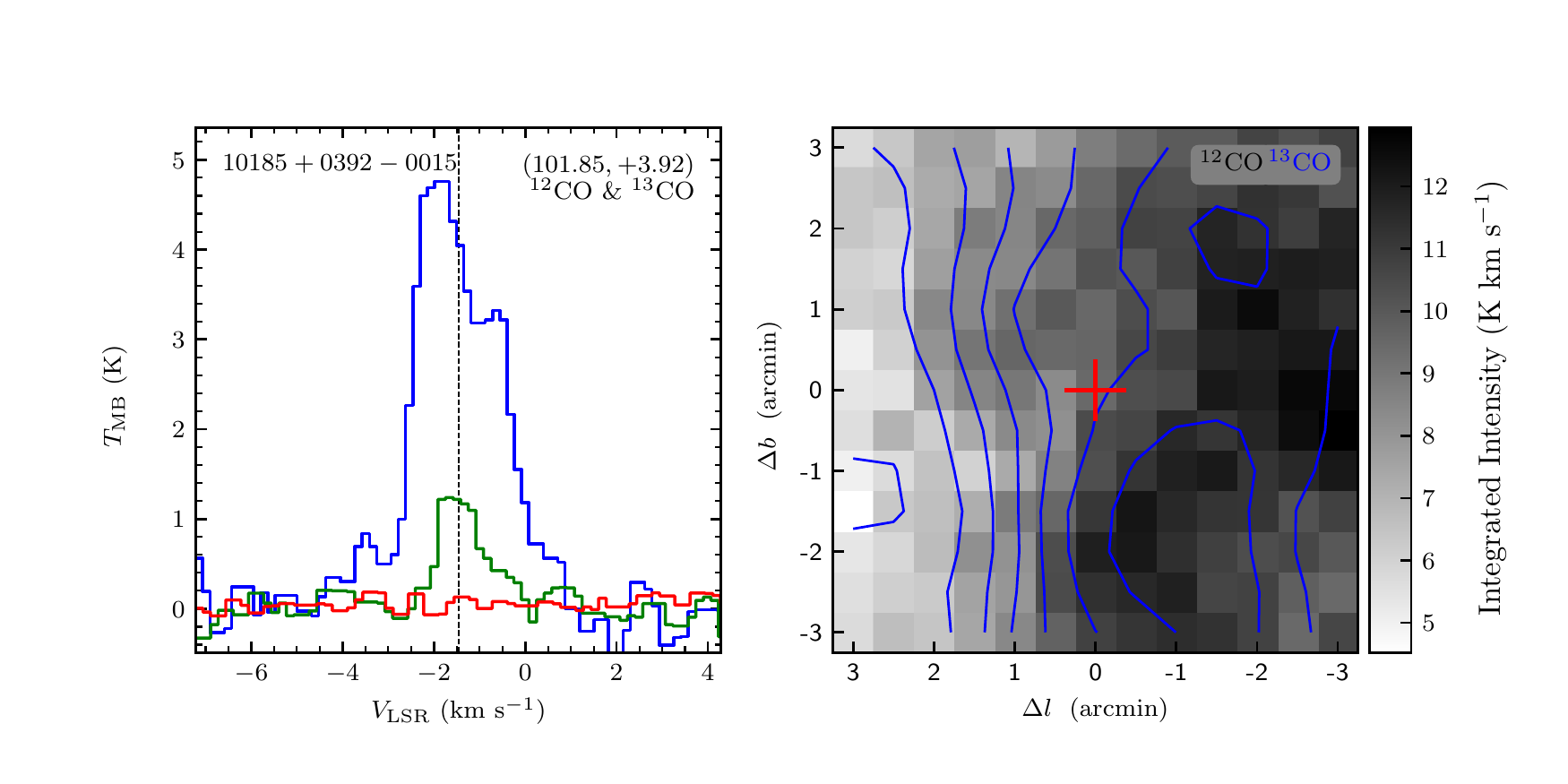}
\vspace{-0.5cm}

\includegraphics[width=9.0cm,angle=0]{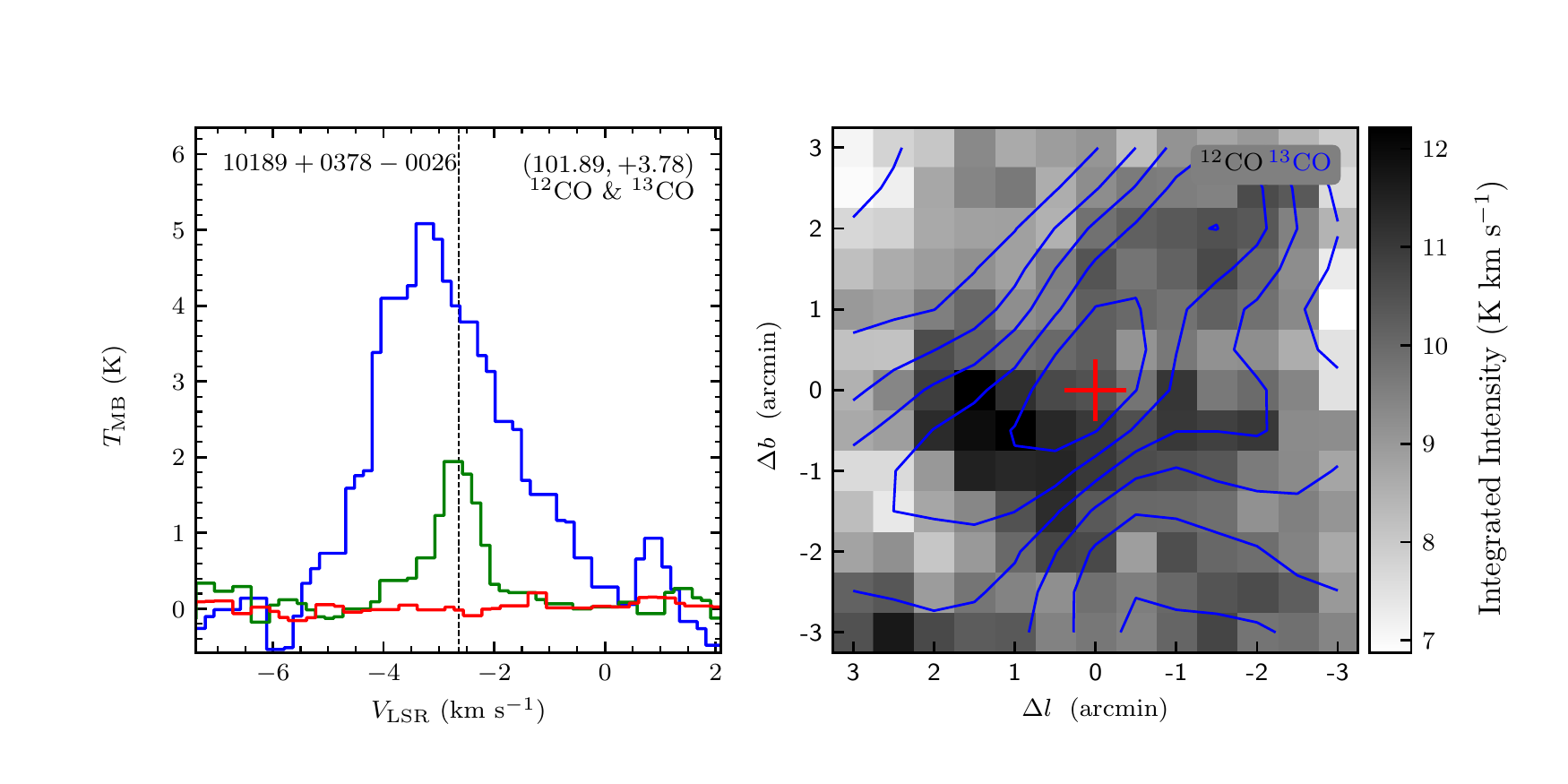}
\includegraphics[width=9.0cm,angle=0]{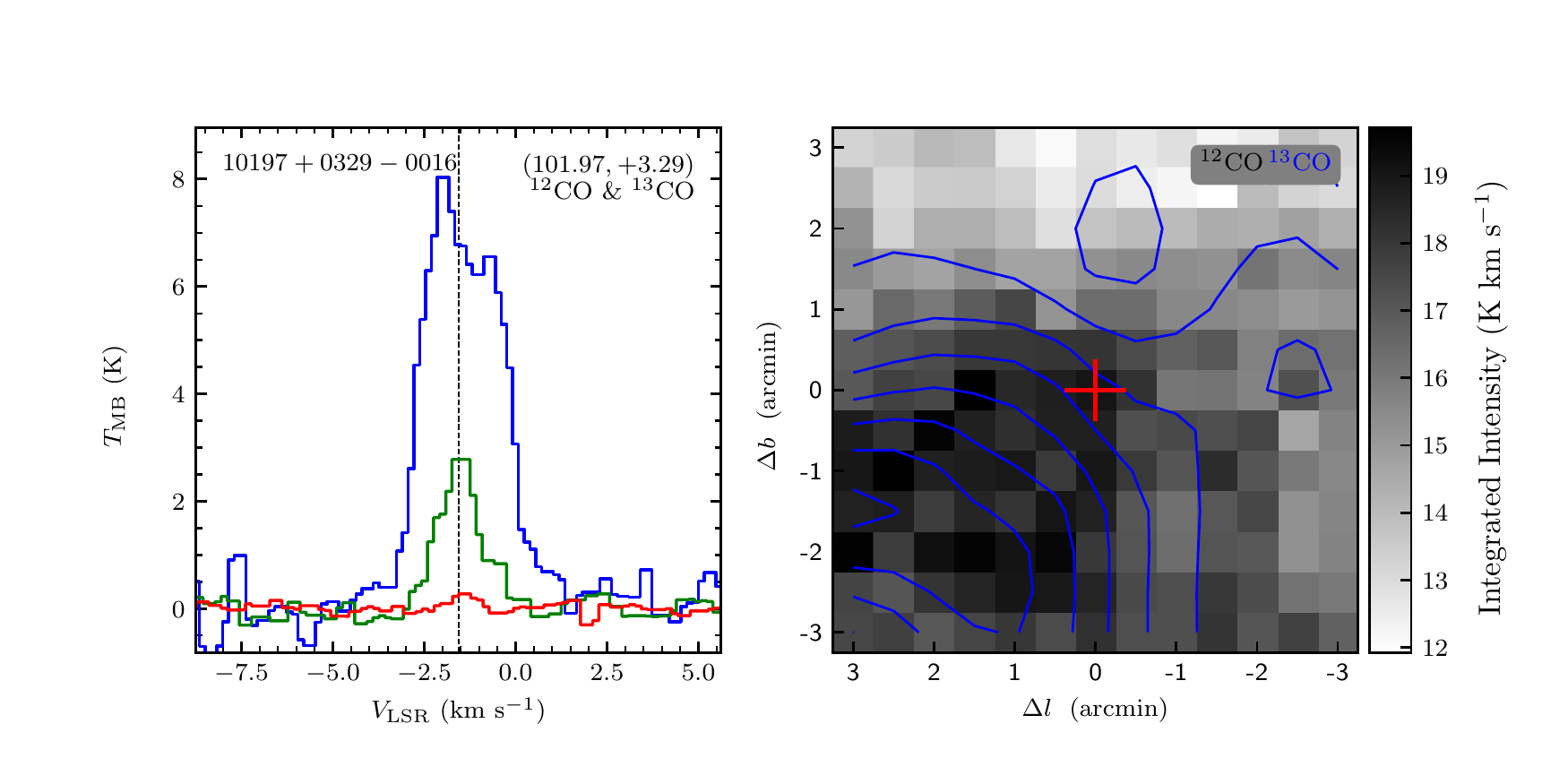}
\vspace{-0.5cm}

\includegraphics[width=9.0cm,angle=0]{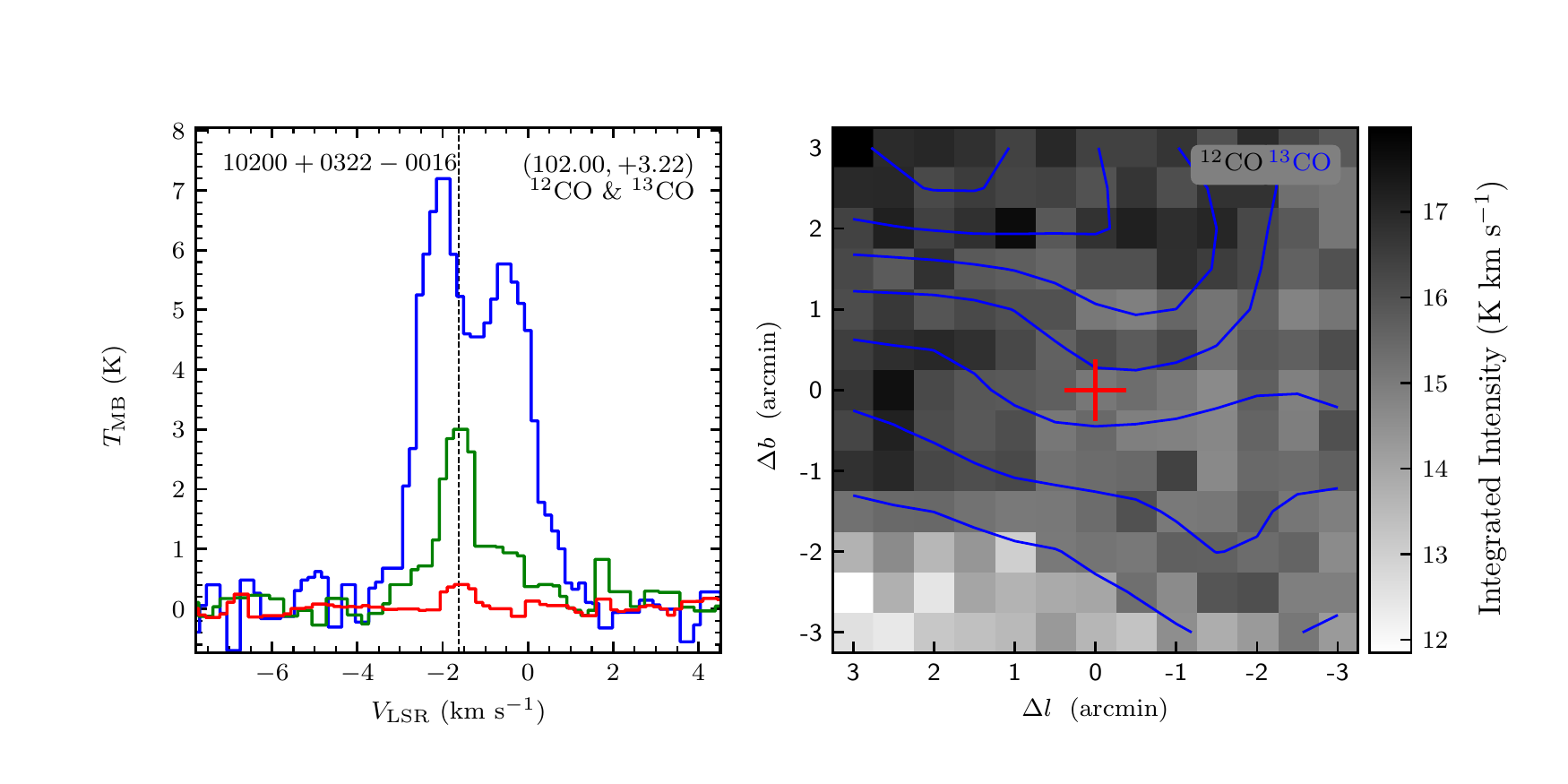}
\includegraphics[width=9.0cm,angle=0]{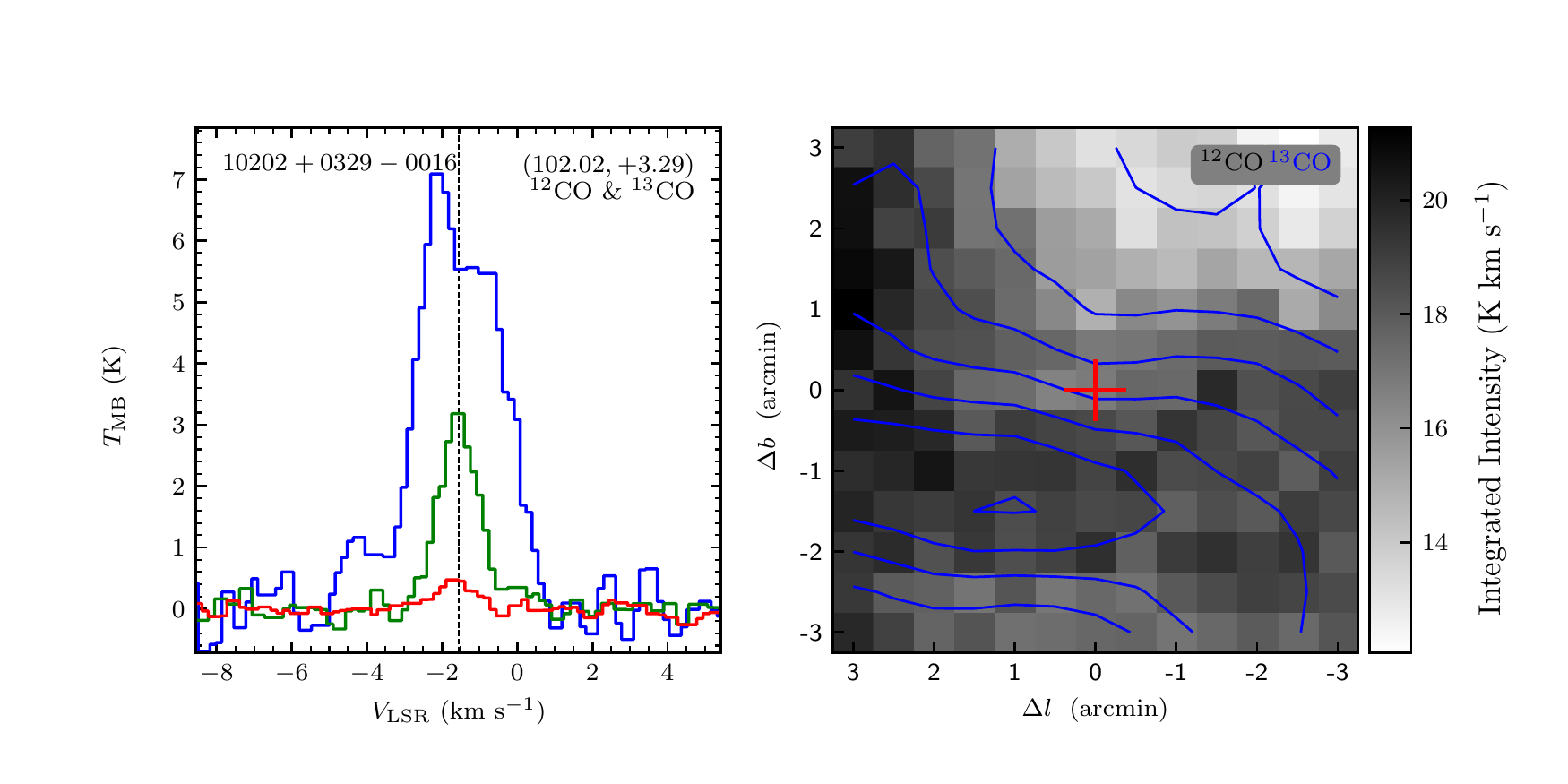}
\vspace{-0.5cm}

\includegraphics[width=9.0cm,angle=0]{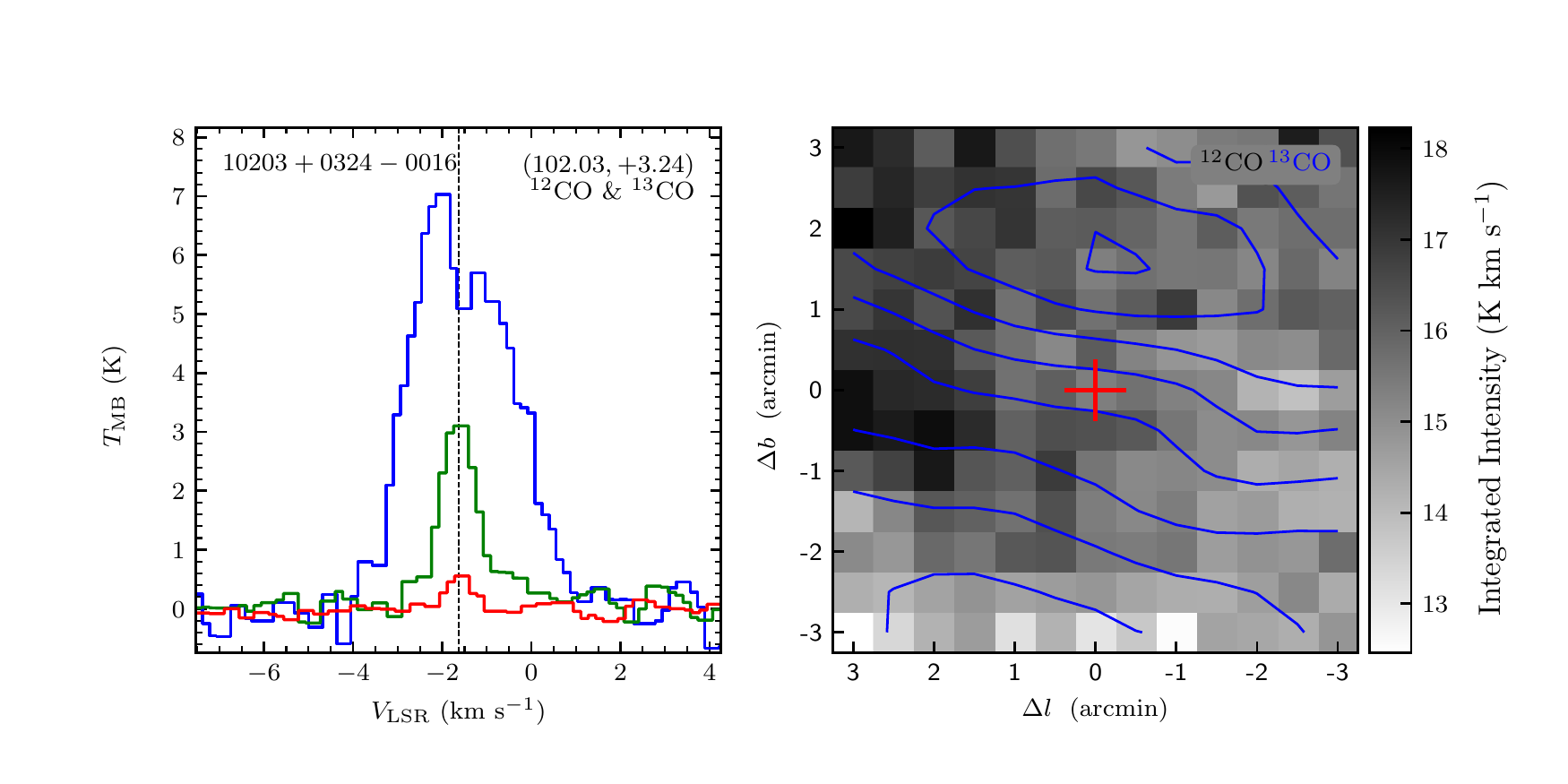}
\includegraphics[width=9.0cm,angle=0]{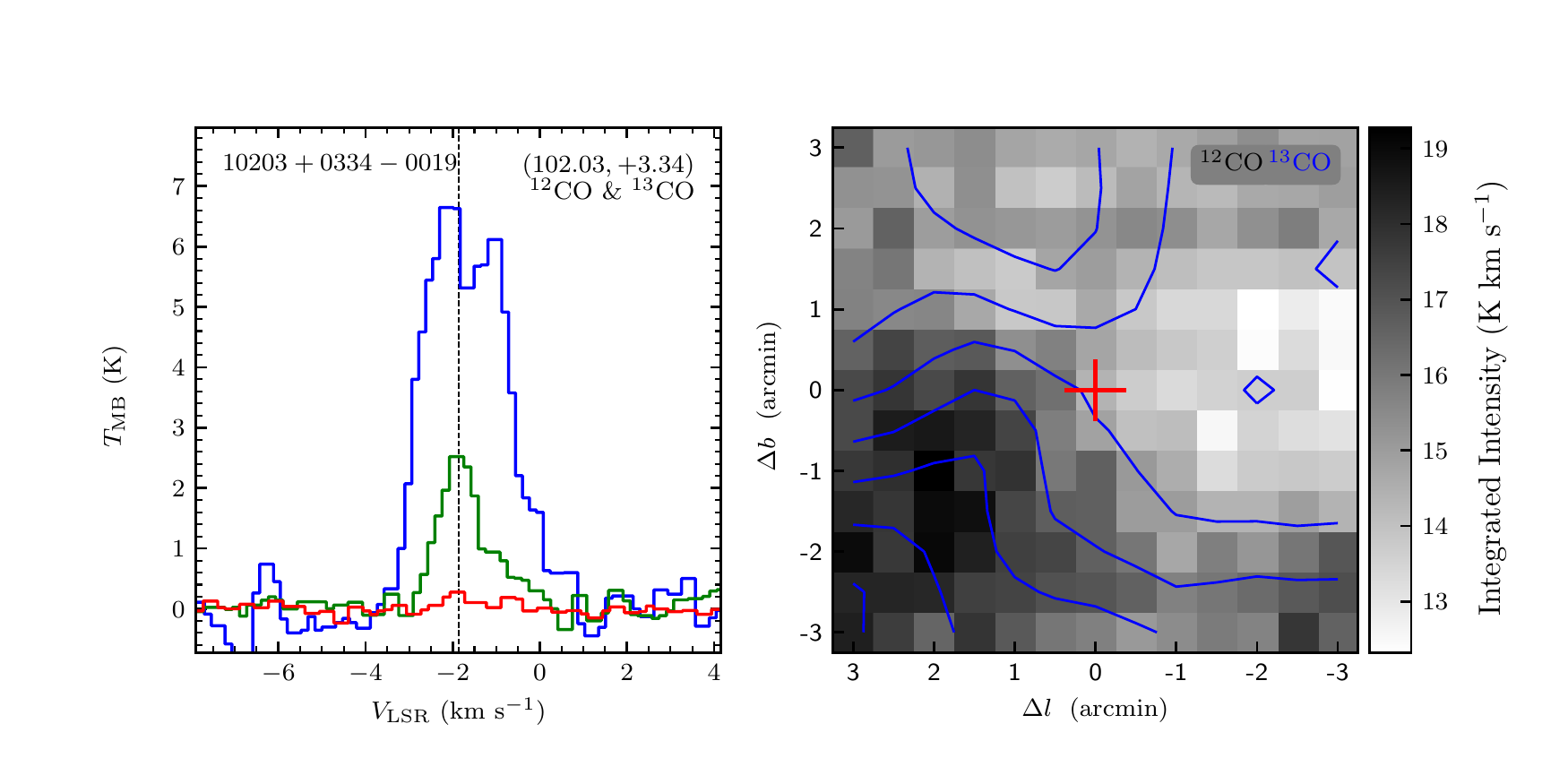}
\vspace{-0.5cm}

\includegraphics[width=9.0cm,angle=0]{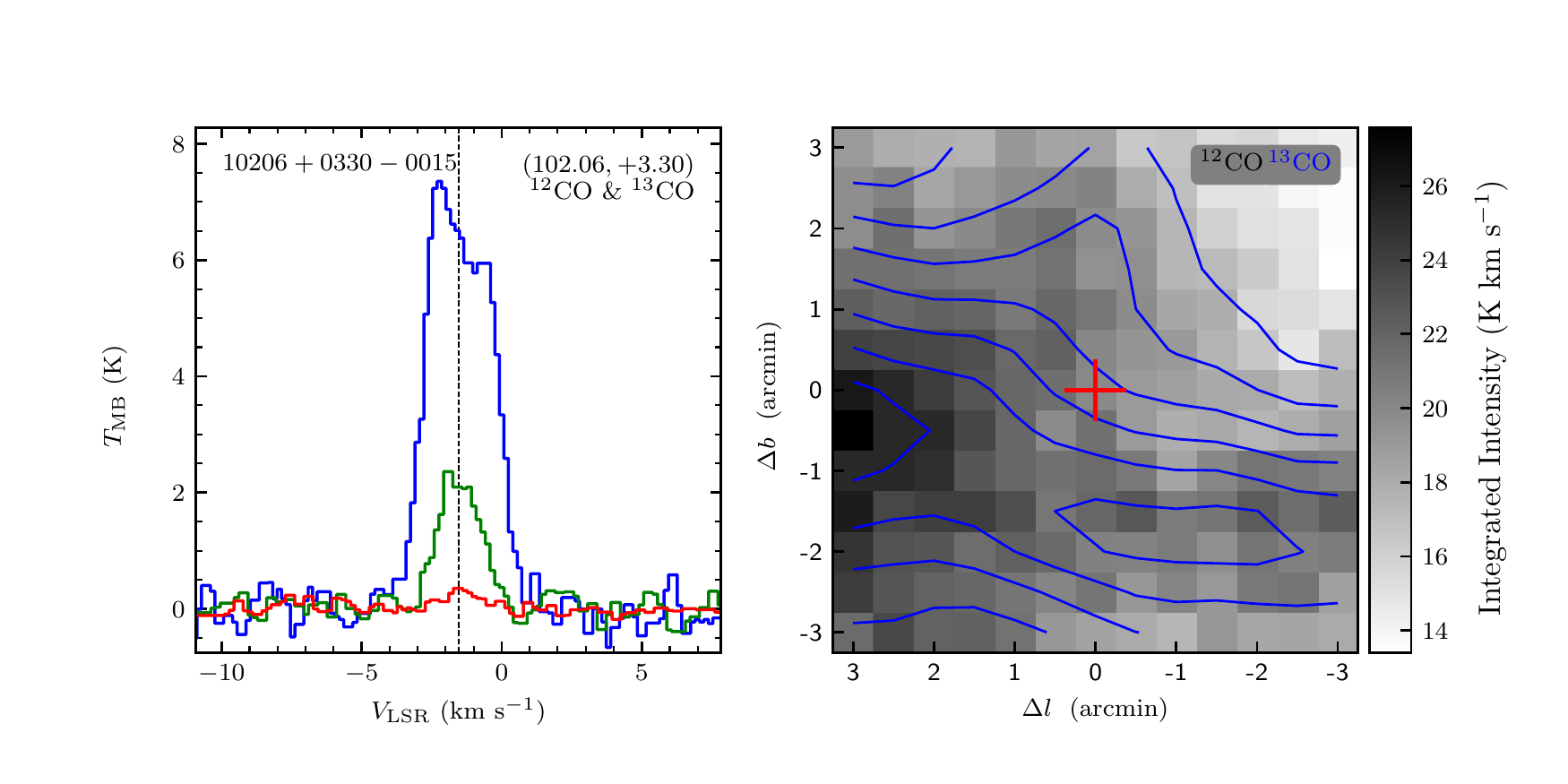}
\includegraphics[width=9.0cm,angle=0]{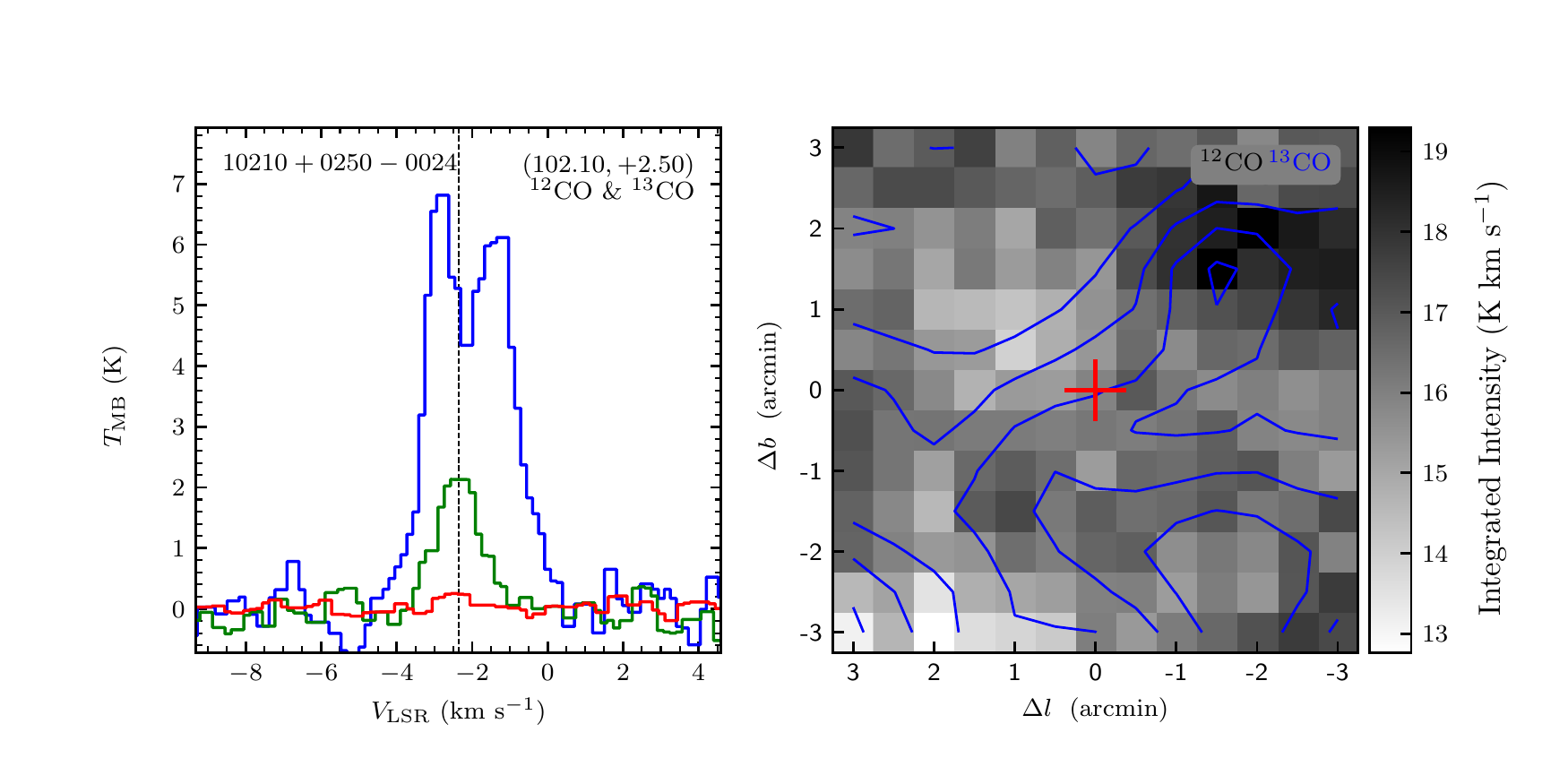}
\end{figure}
\clearpage

\begin{figure}
\includegraphics[width=9.0cm,angle=0]{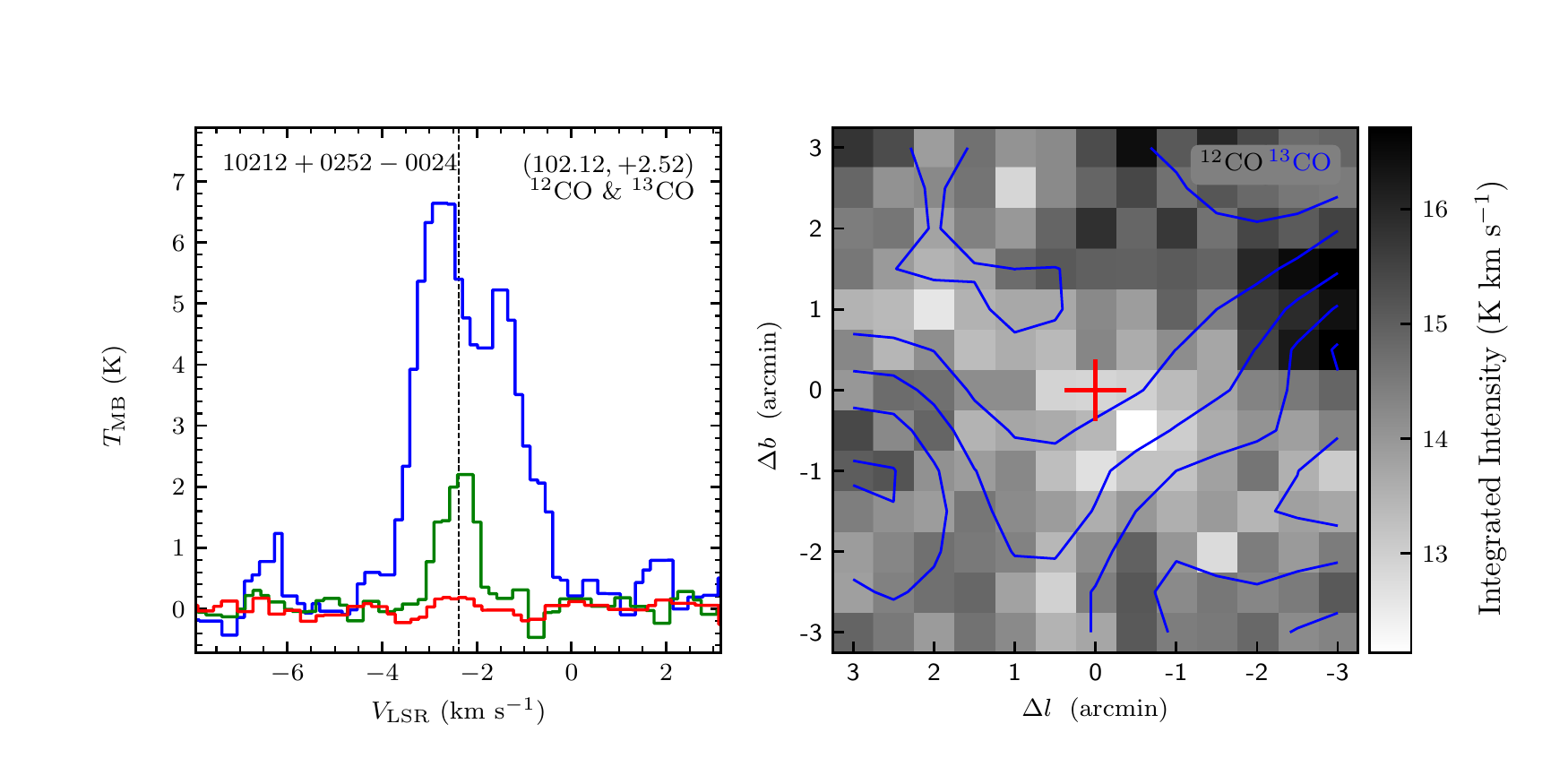}
\includegraphics[width=9.0cm,angle=0]{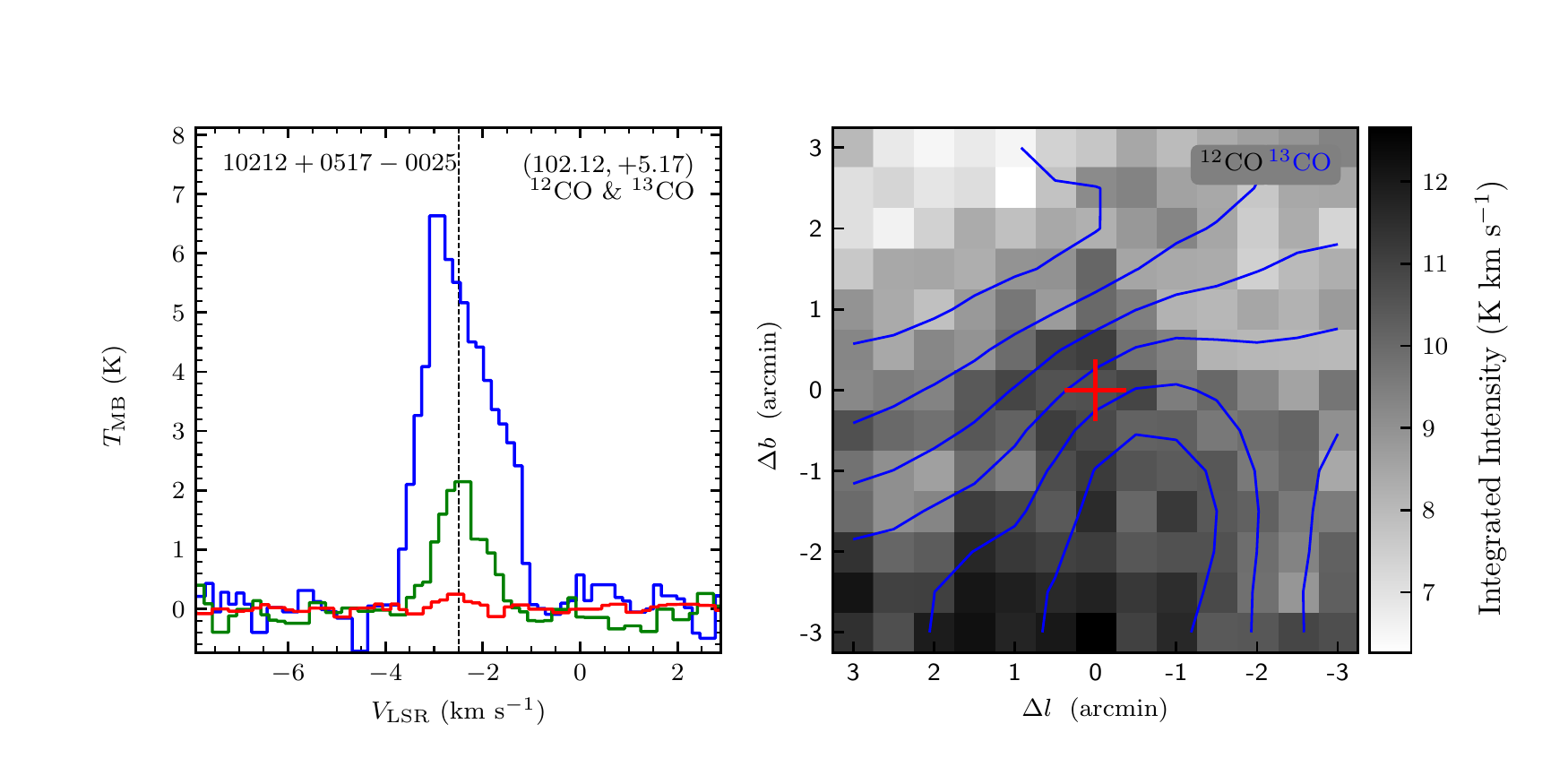}
\vspace{-0.5cm}

\includegraphics[width=9.0cm,angle=0]{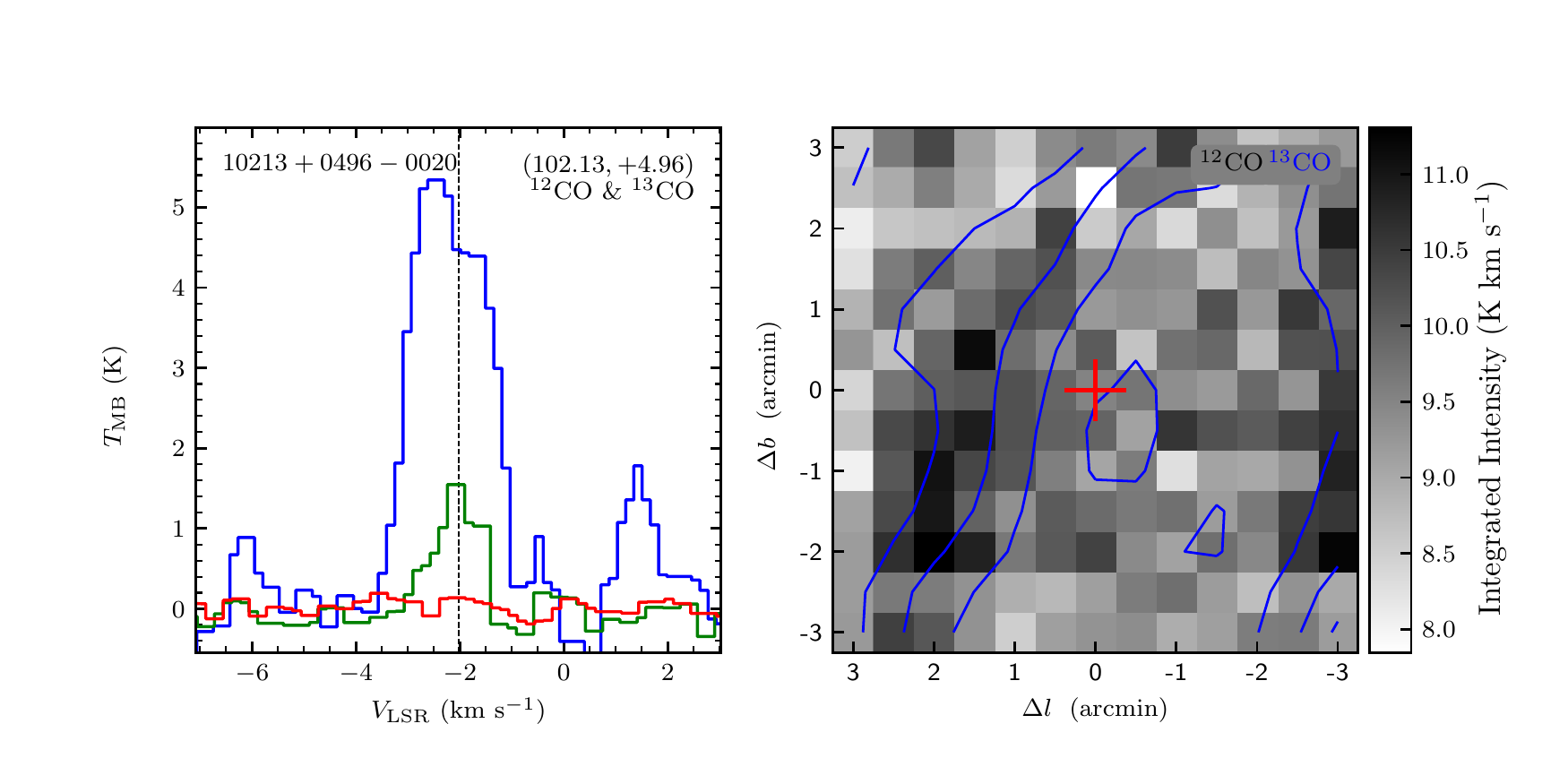}
\includegraphics[width=9.0cm,angle=0]{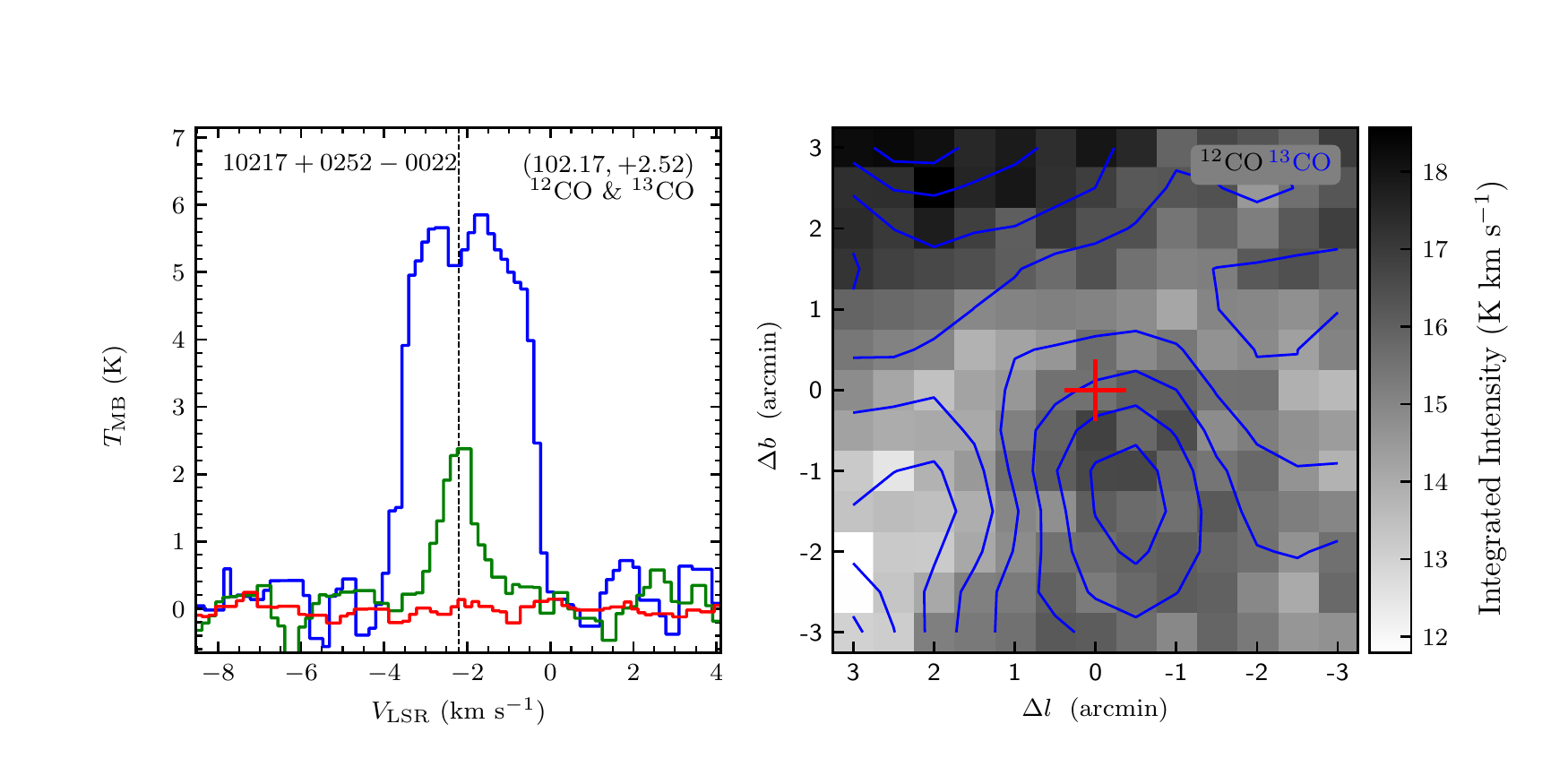}
\vspace{-0.5cm}

\includegraphics[width=9.0cm,angle=0]{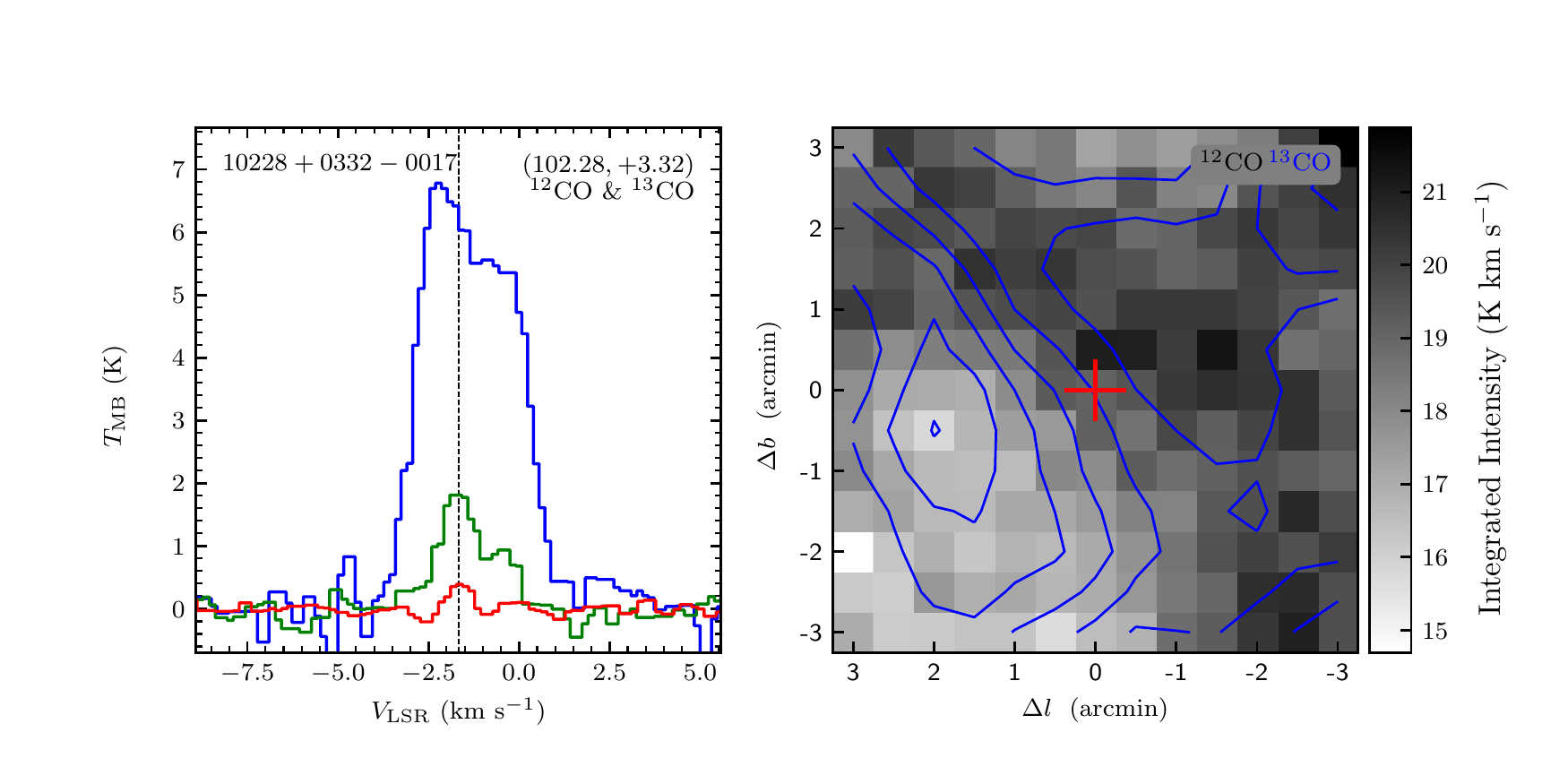}
\includegraphics[width=9.0cm,angle=0]{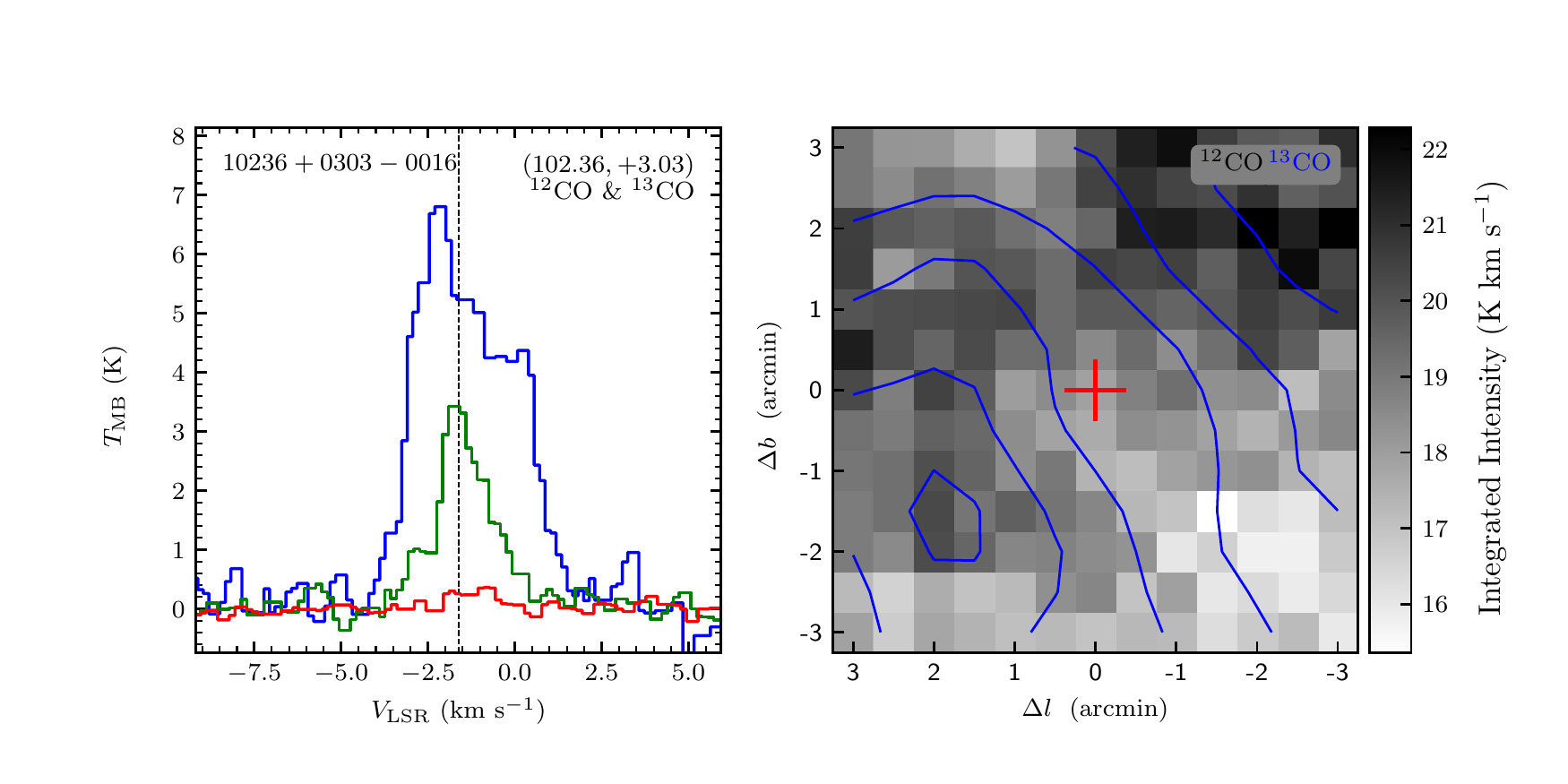}
\vspace{-0.5cm}

\includegraphics[width=9.0cm,angle=0]{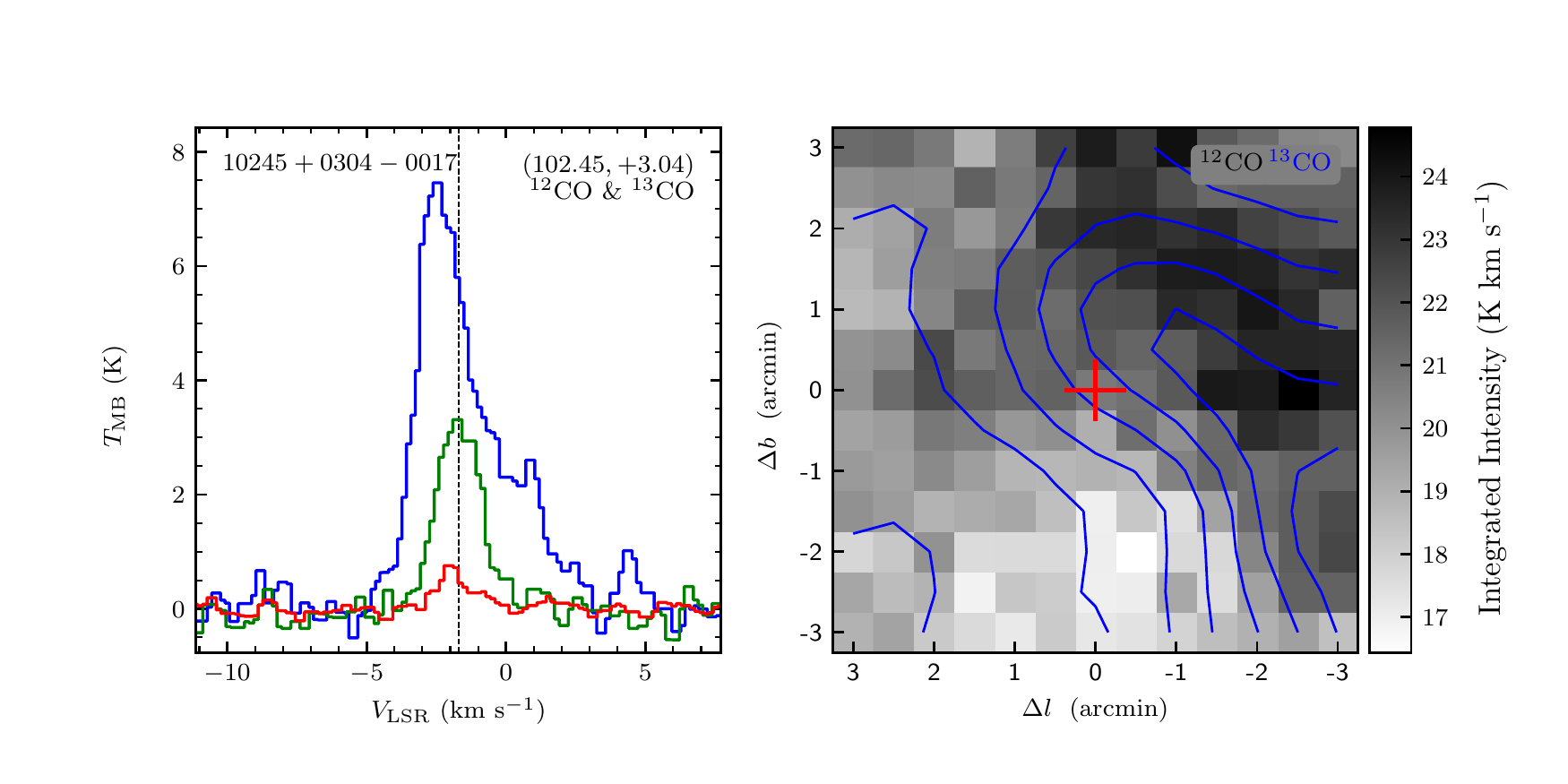}
\includegraphics[width=9.0cm,angle=0]{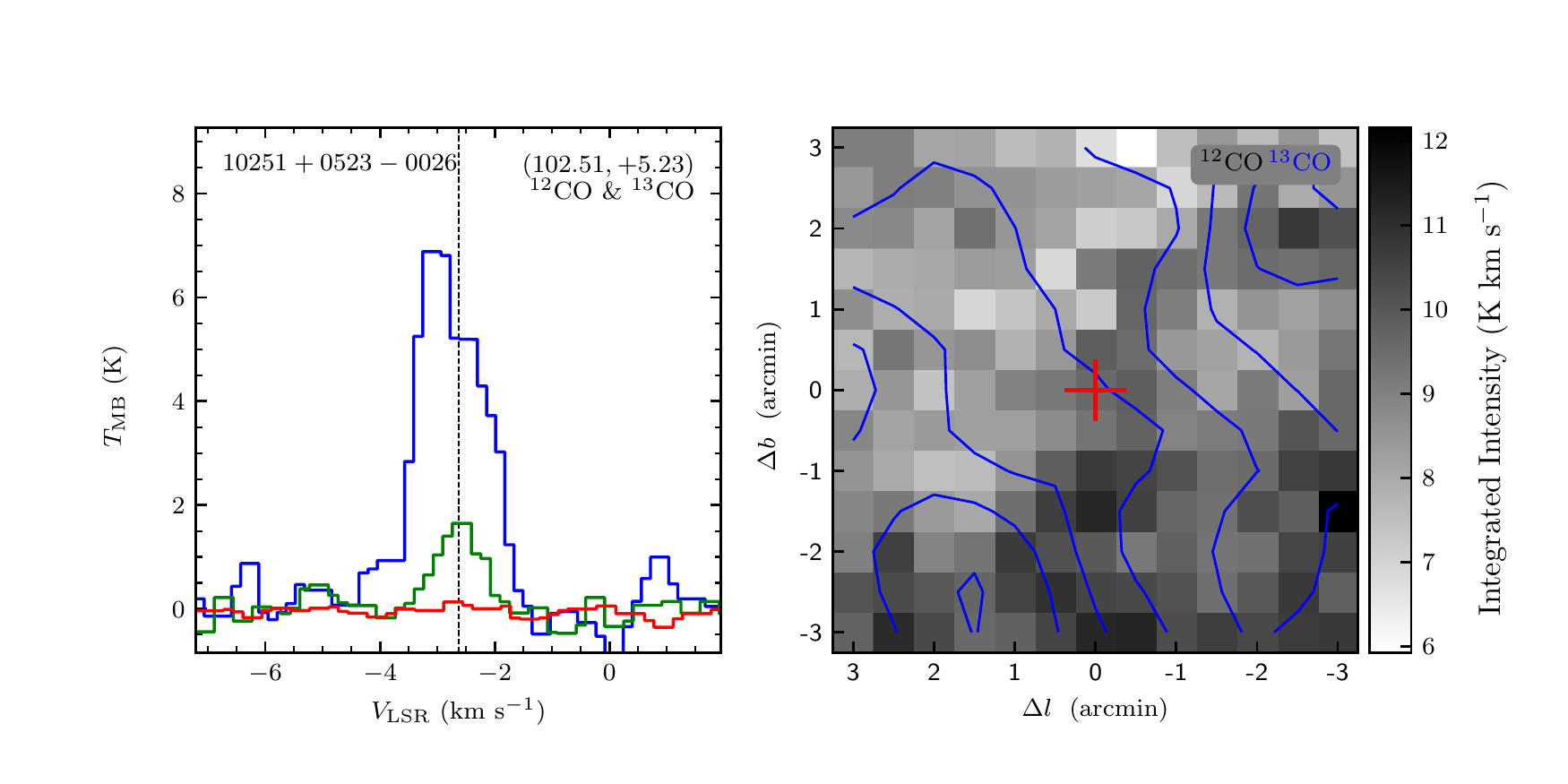}
\vspace{-0.5cm}

\includegraphics[width=9.0cm,angle=0]{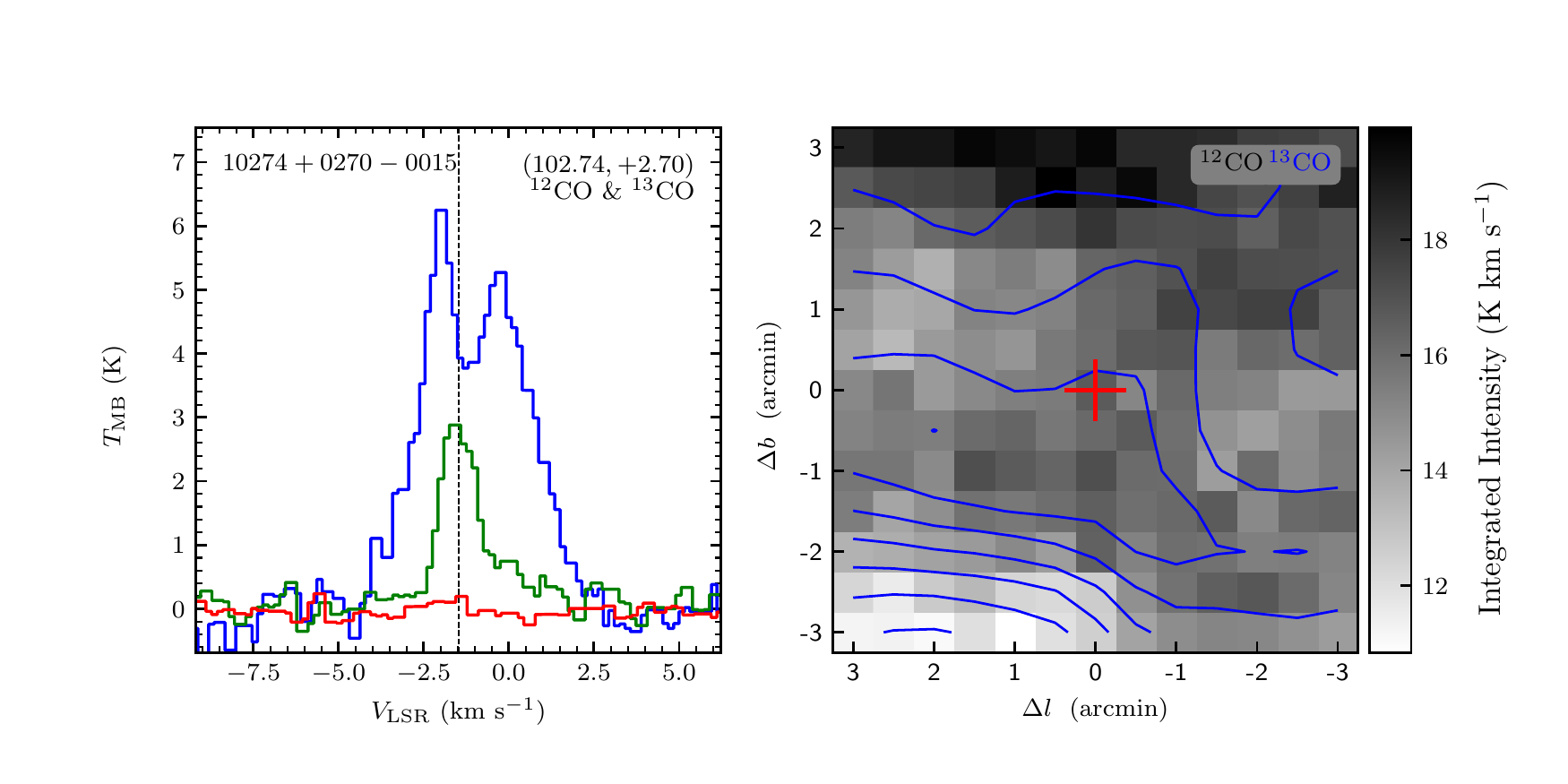}
\includegraphics[width=9.0cm,angle=0]{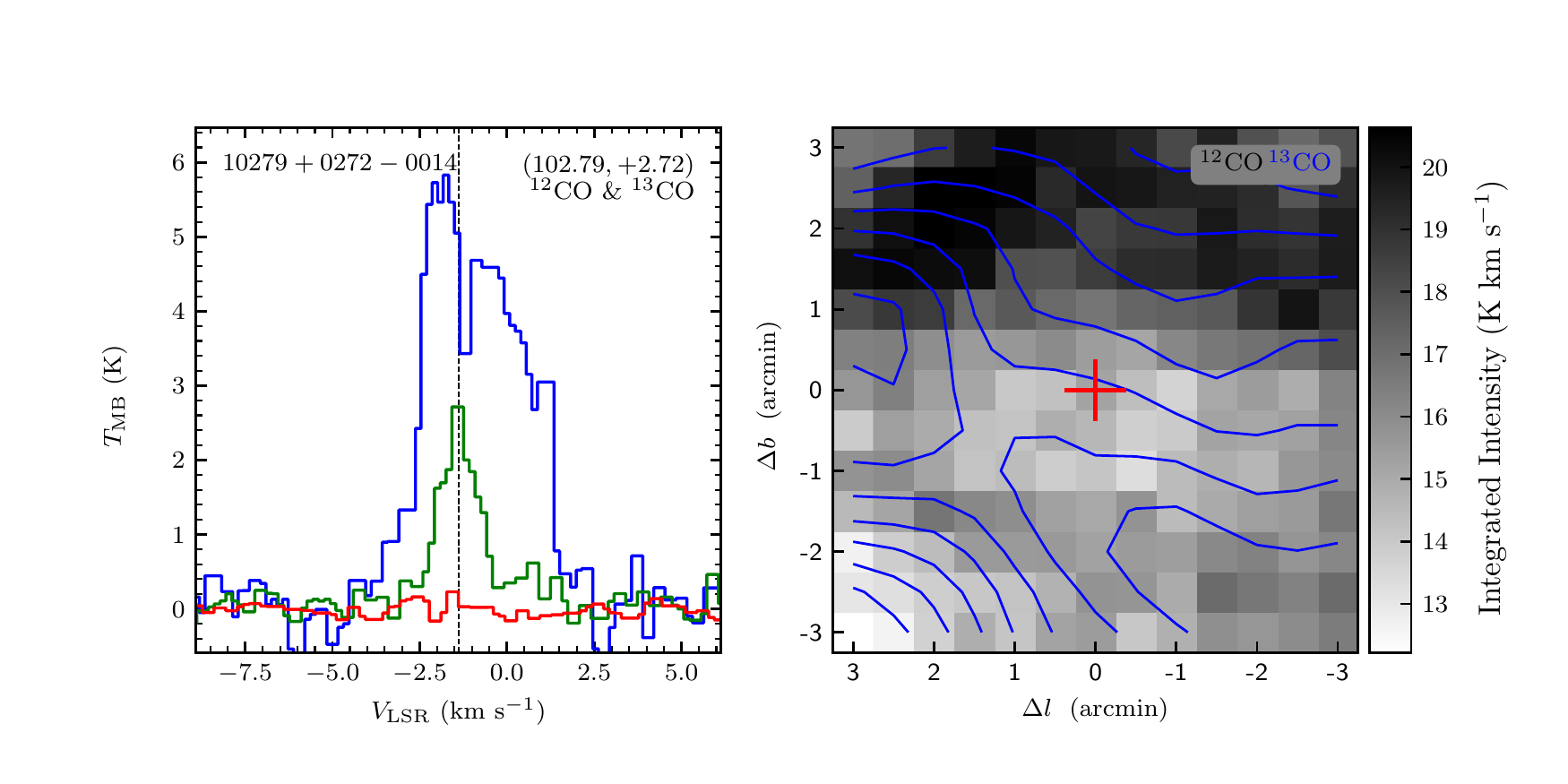}
\end{figure}
\clearpage

\begin{figure}
\includegraphics[width=9.0cm,angle=0]{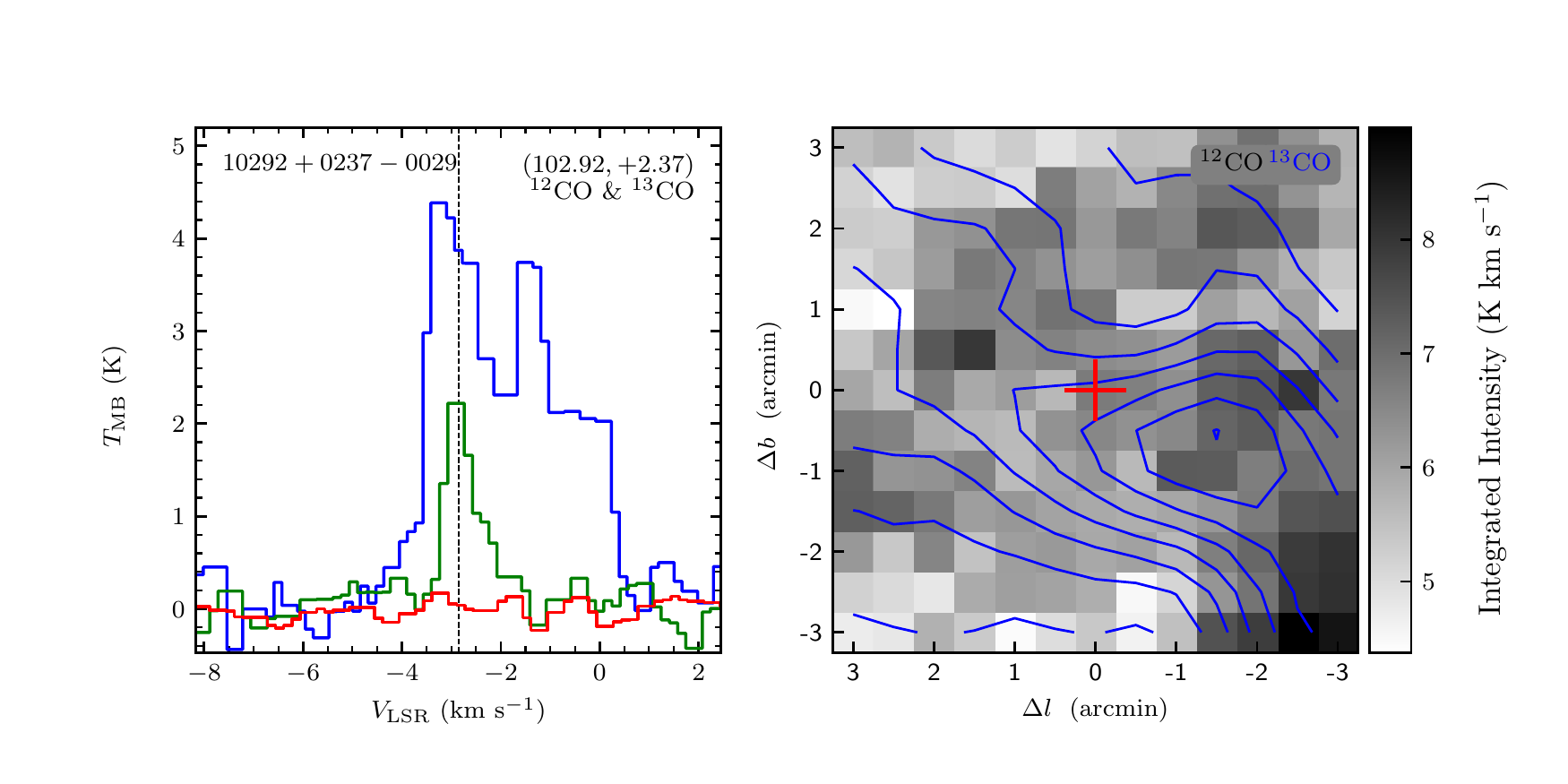}
\includegraphics[width=9.0cm,angle=0]{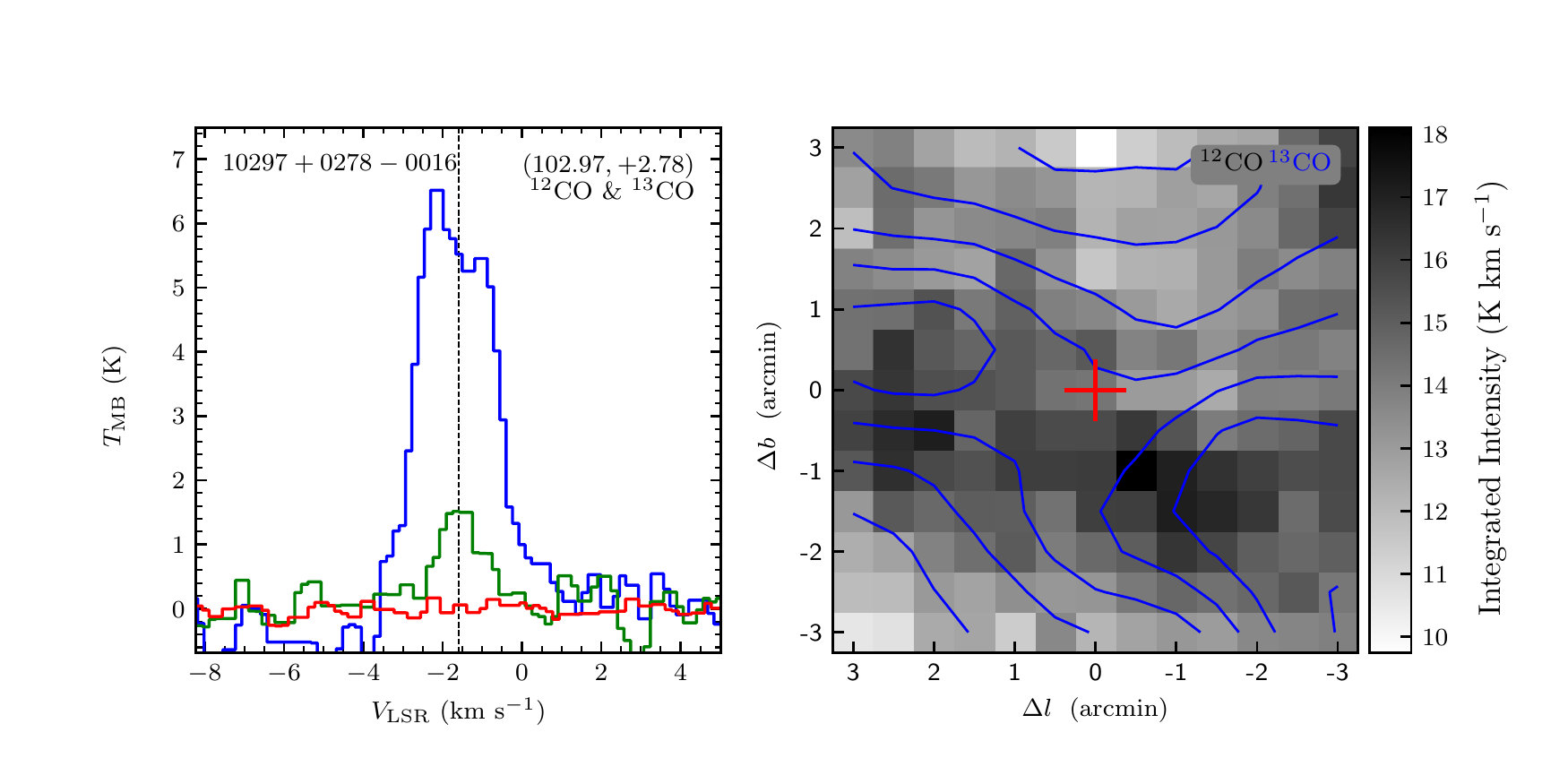}
\vspace{-0.5cm}

\includegraphics[width=9.0cm,angle=0]{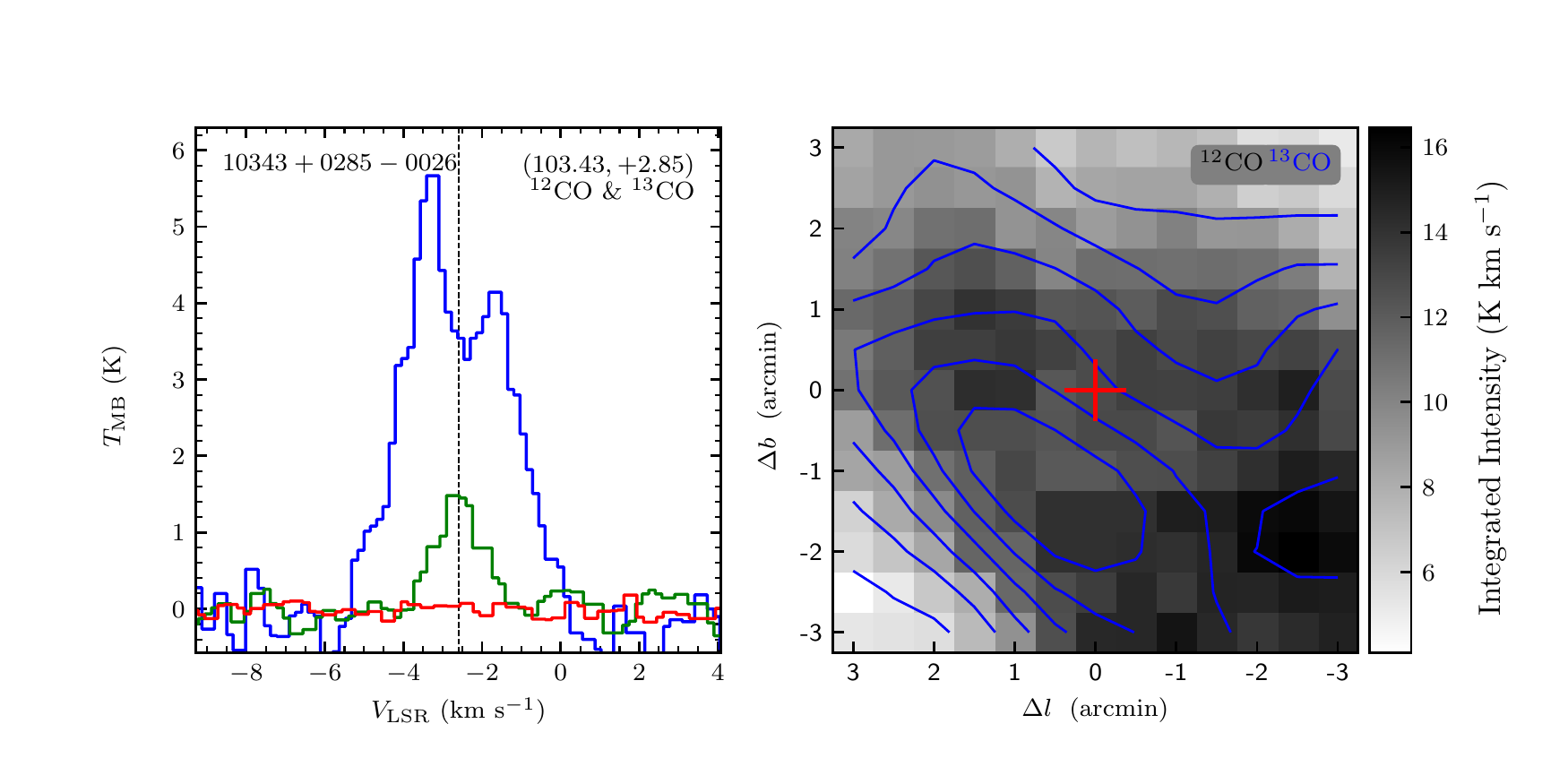}
\includegraphics[width=9.0cm,angle=0]{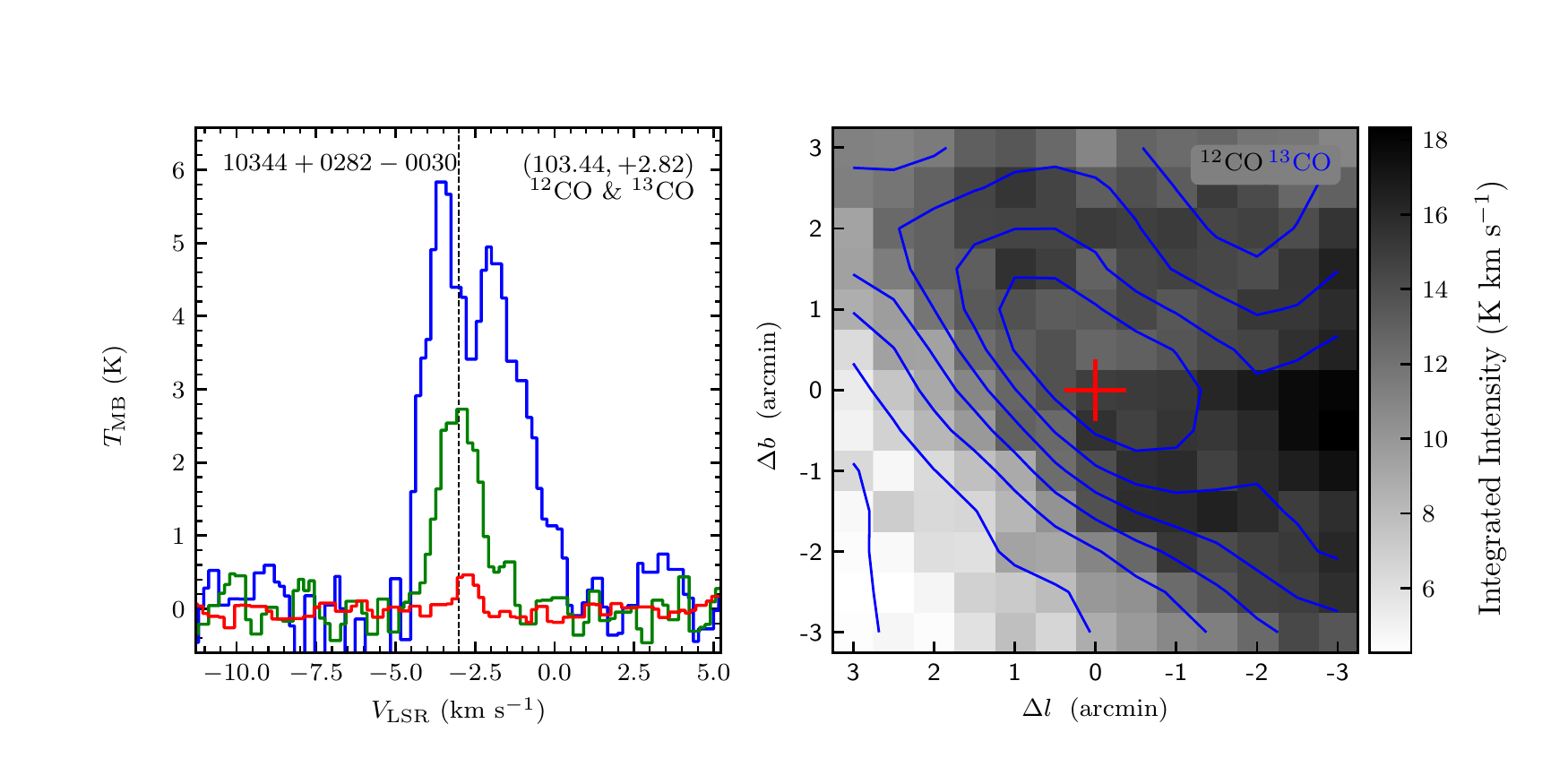}
\vspace{-0.5cm}

\includegraphics[width=9.0cm,angle=0]{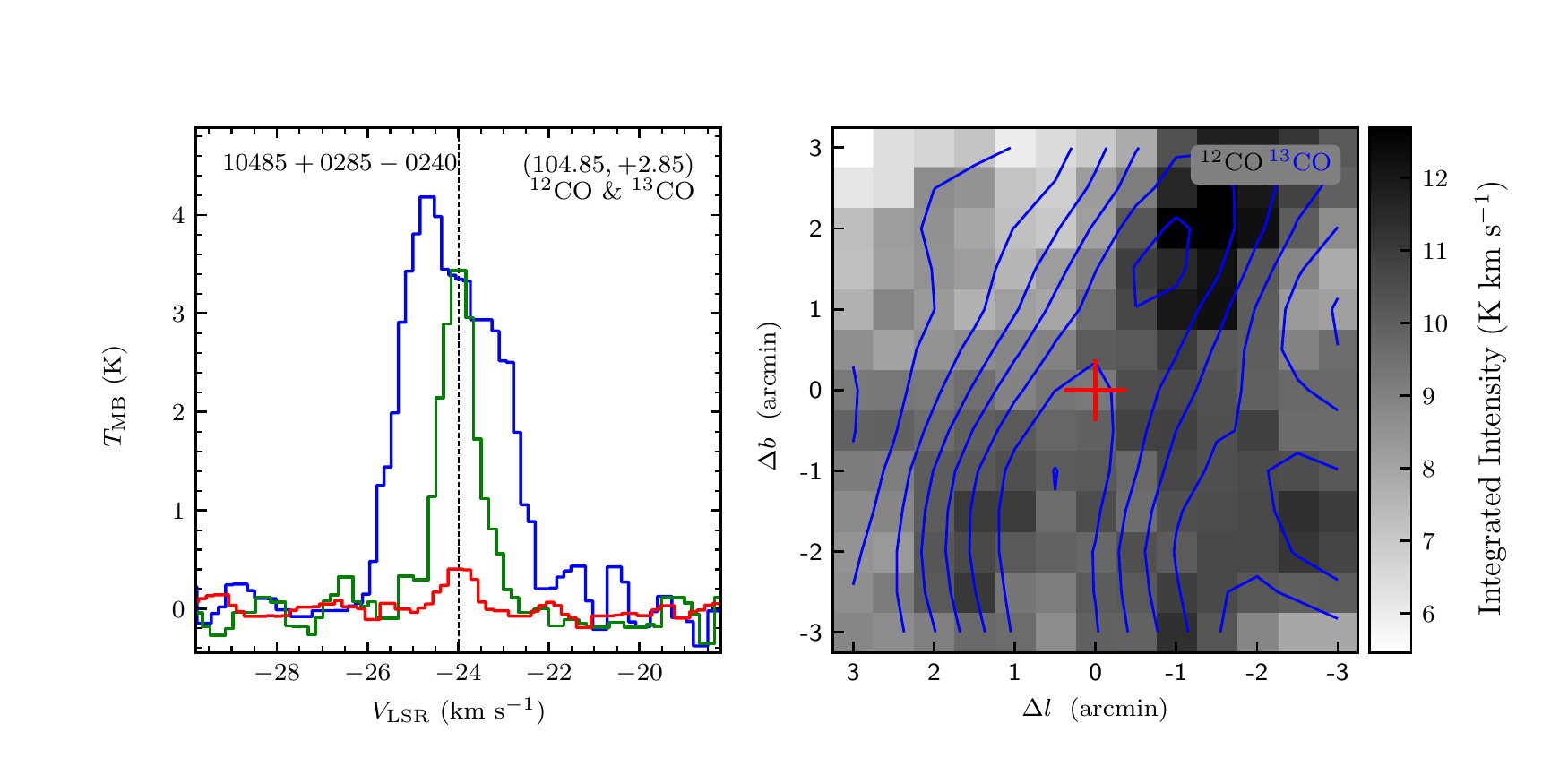}
\includegraphics[width=9.0cm,angle=0]{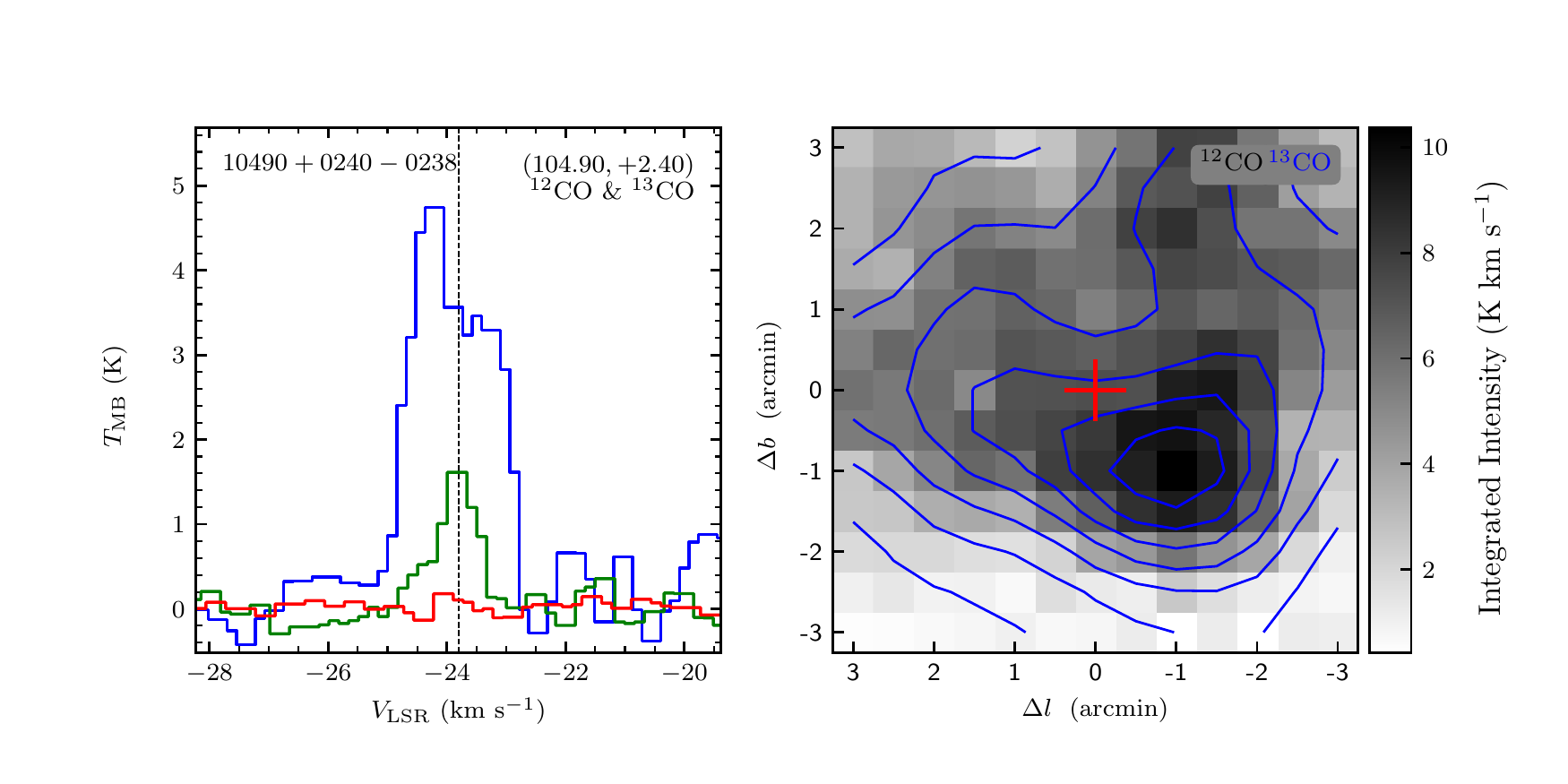}
\vspace{-0.5cm}

\includegraphics[width=9.0cm,angle=0]{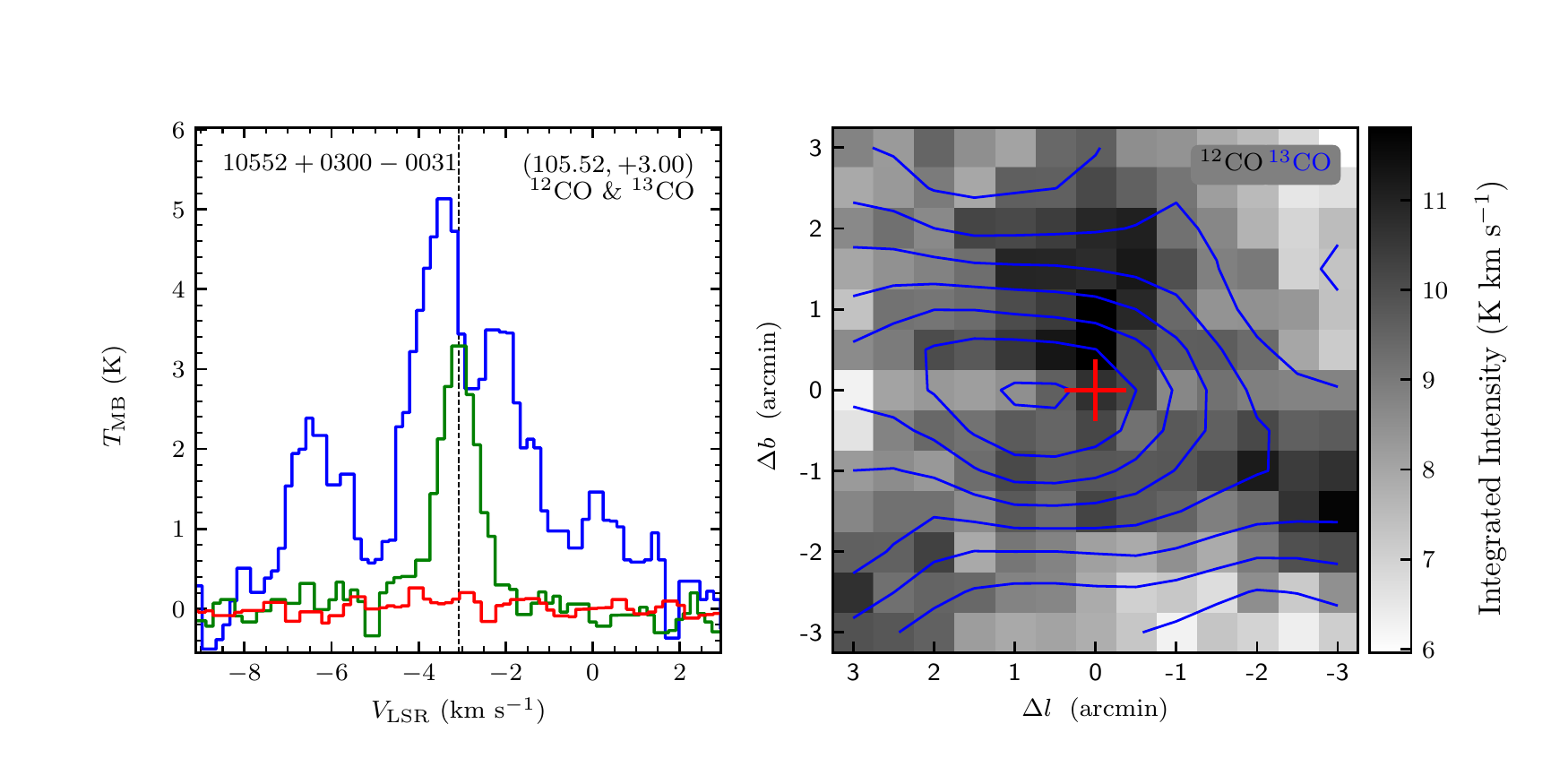}
\includegraphics[width=9.0cm,angle=0]{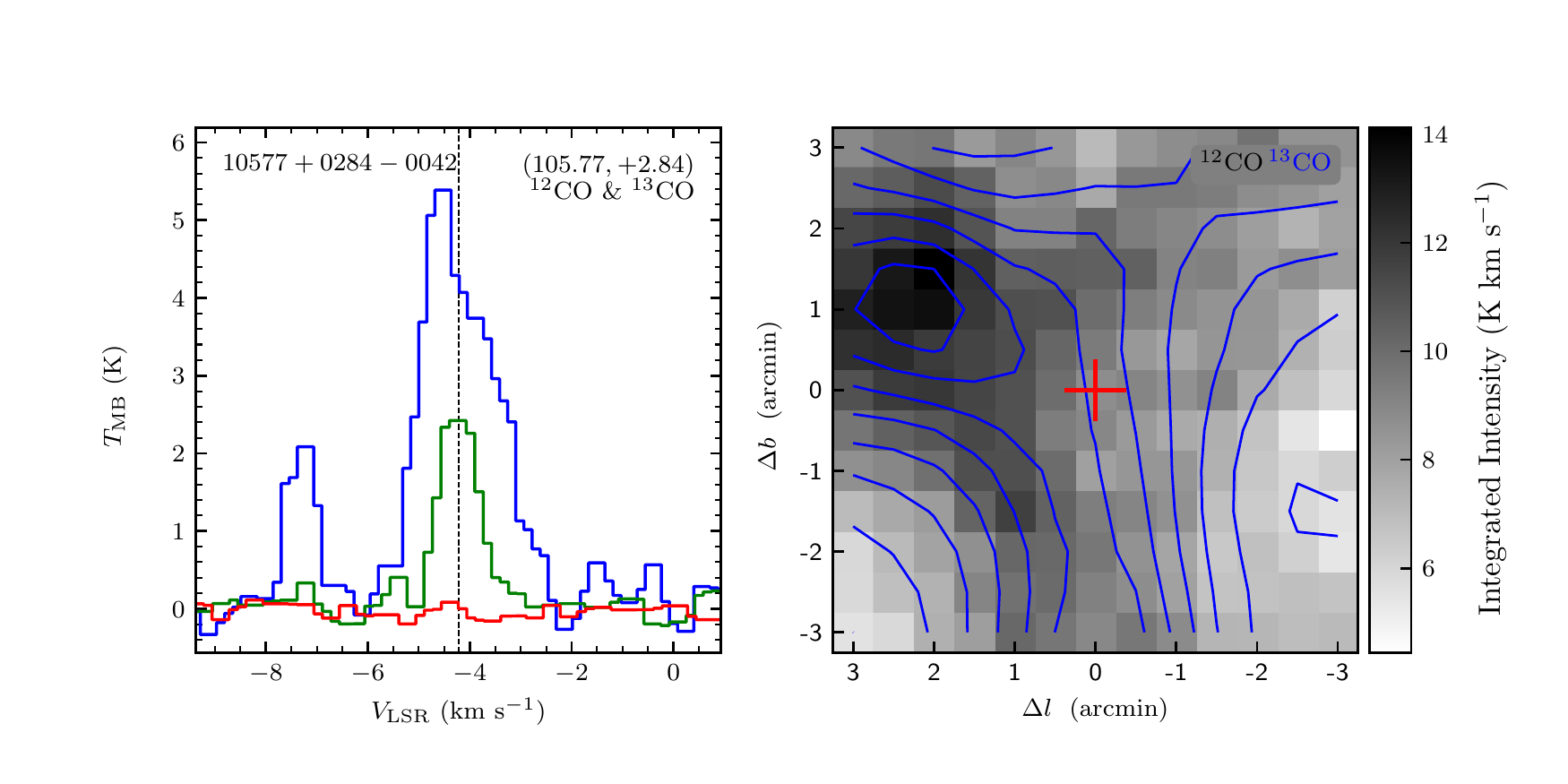}
\vspace{-0.5cm}

\includegraphics[width=9.0cm,angle=0]{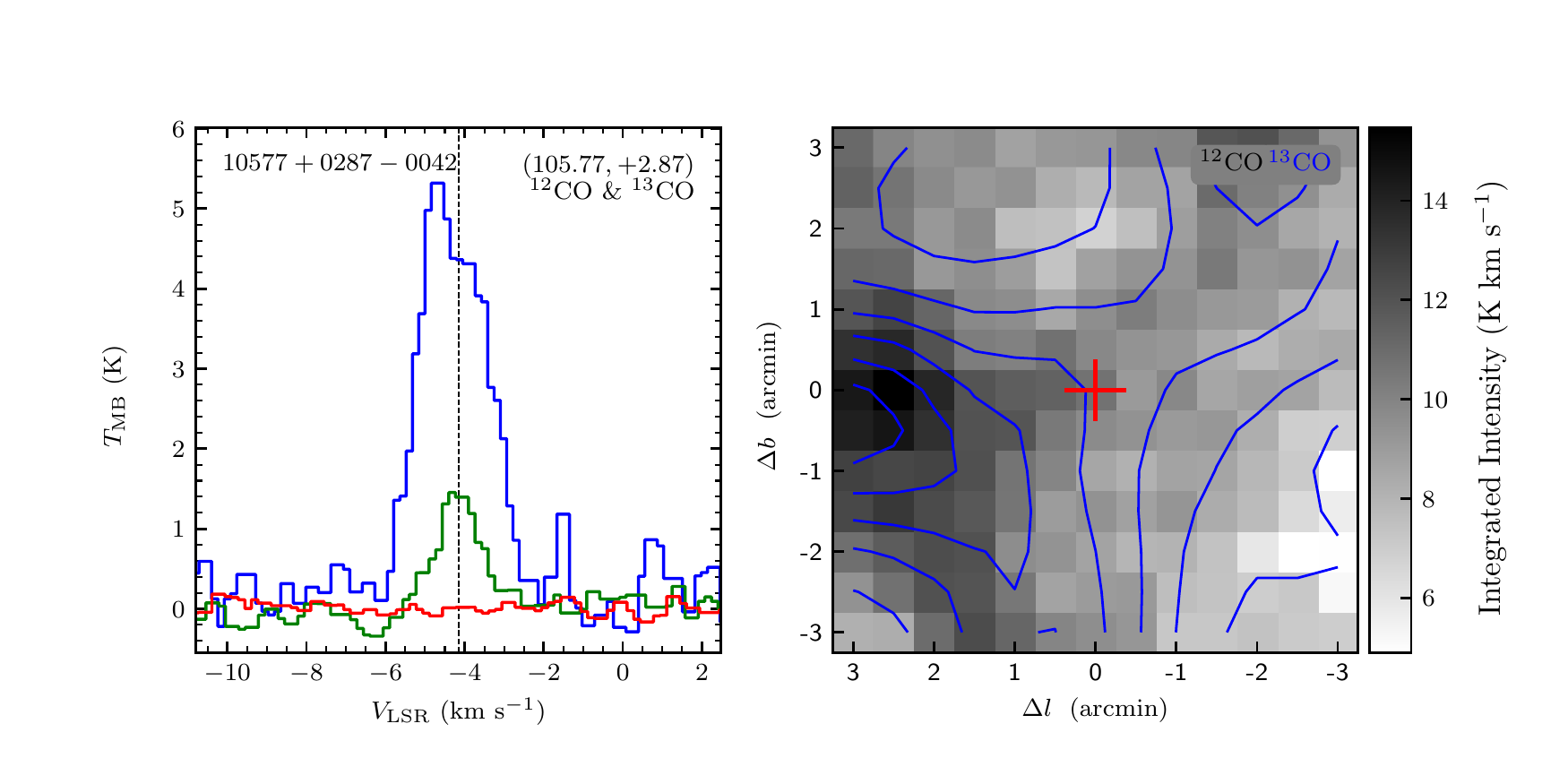}
\includegraphics[width=9.0cm,angle=0]{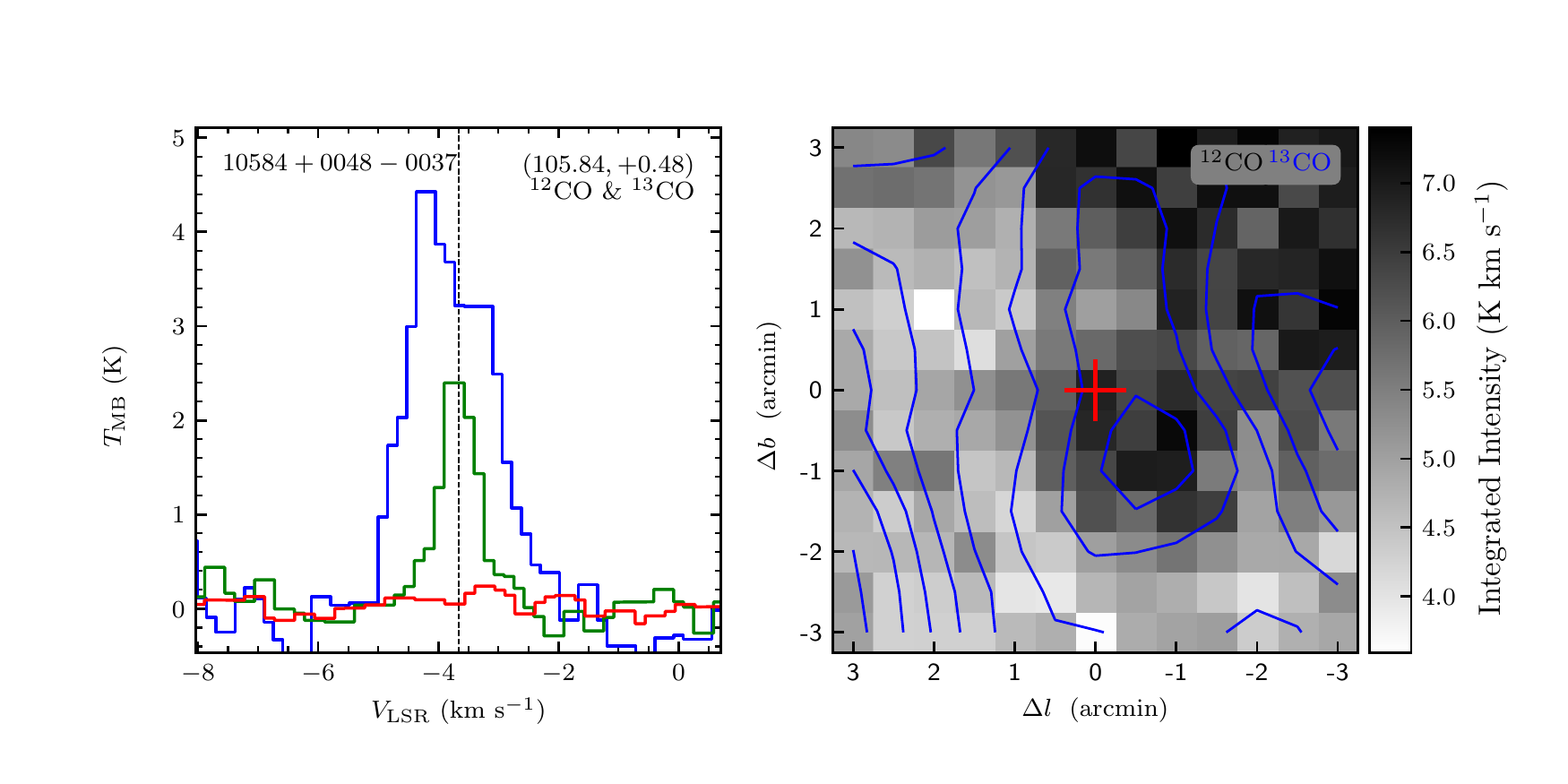}
\end{figure}
\clearpage

\begin{figure}
\includegraphics[width=9.0cm,angle=0]{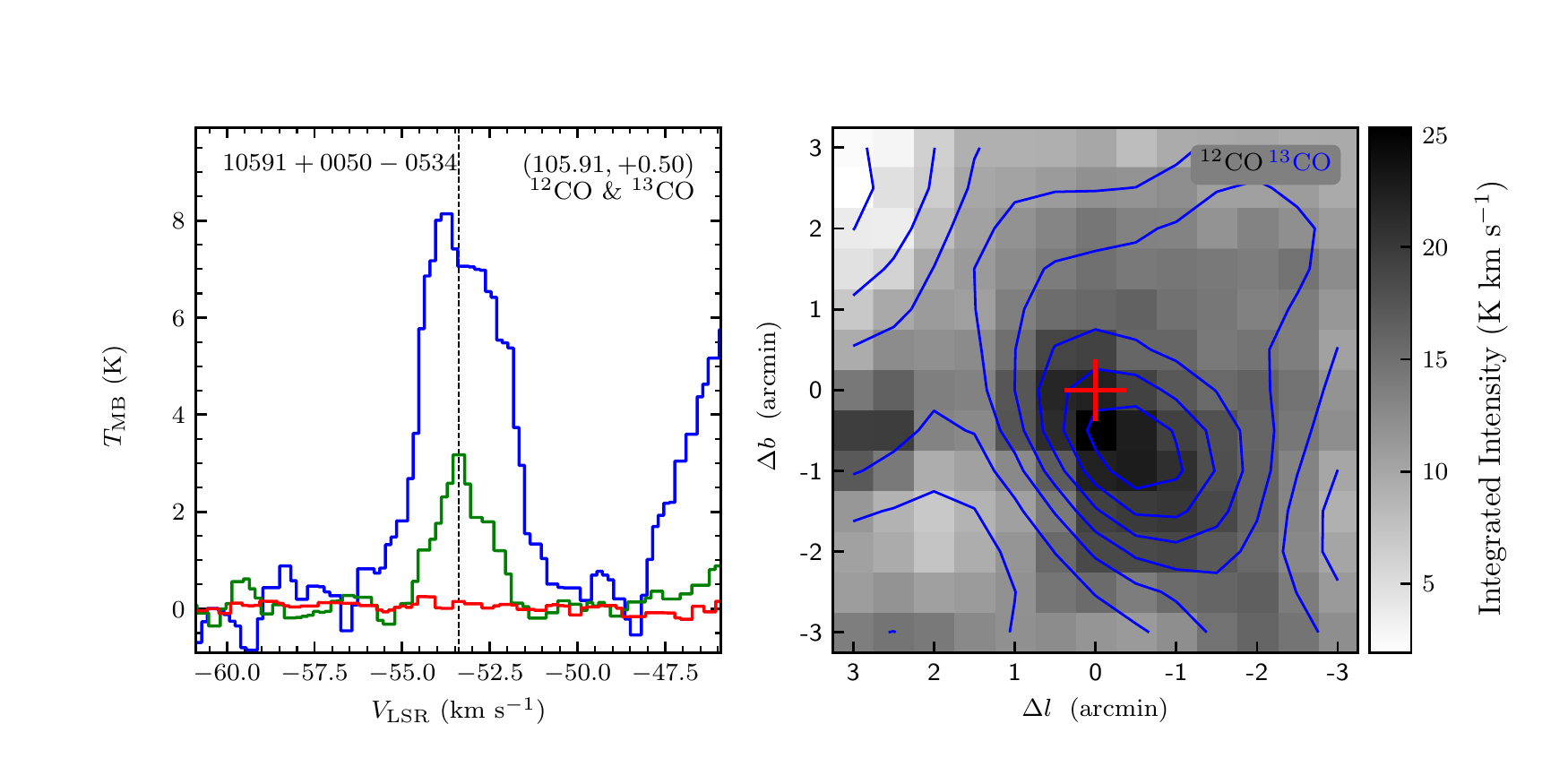}
\includegraphics[width=9.0cm,angle=0]{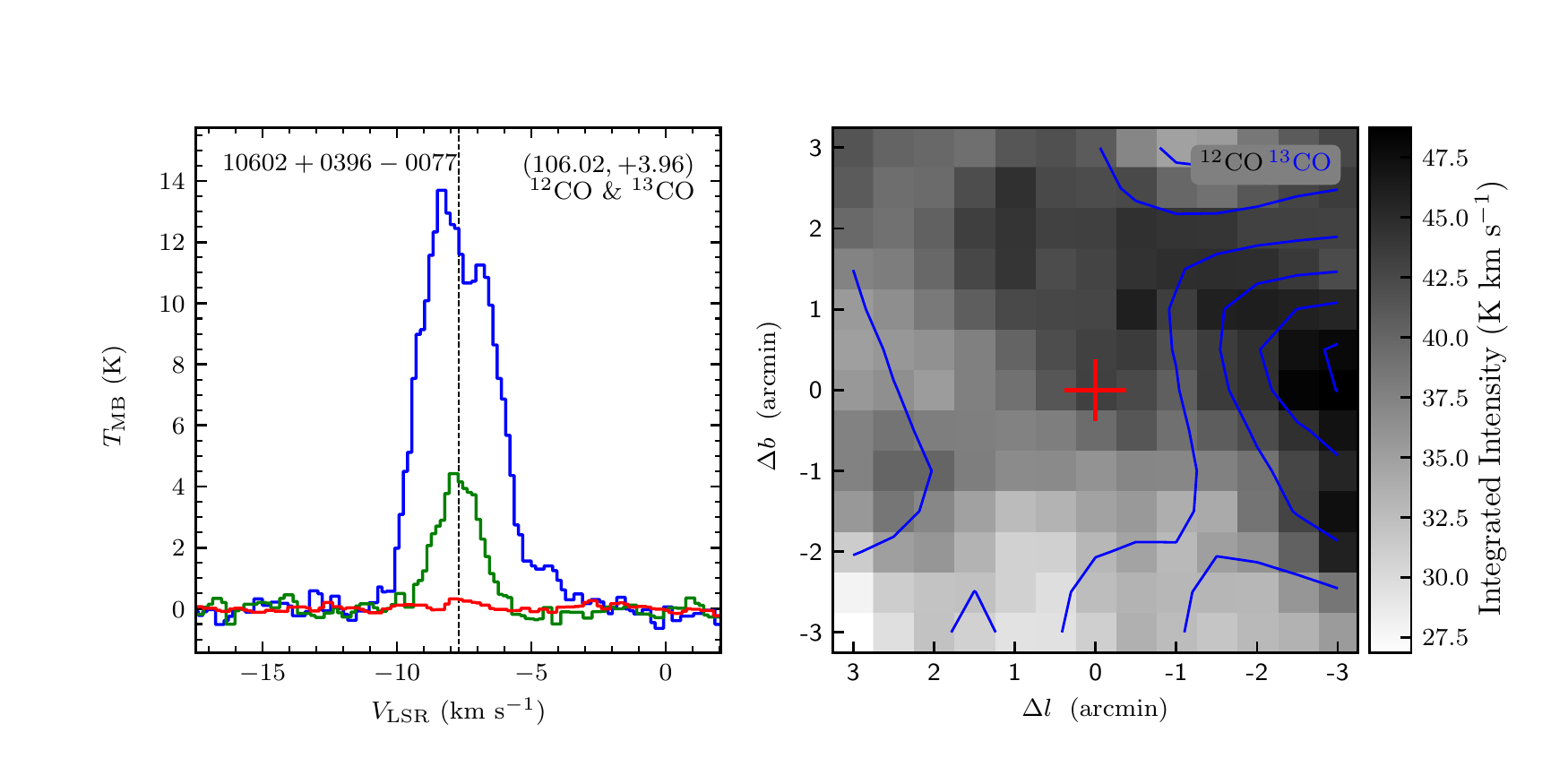}
\vspace{-0.5cm}

\includegraphics[width=9.0cm,angle=0]{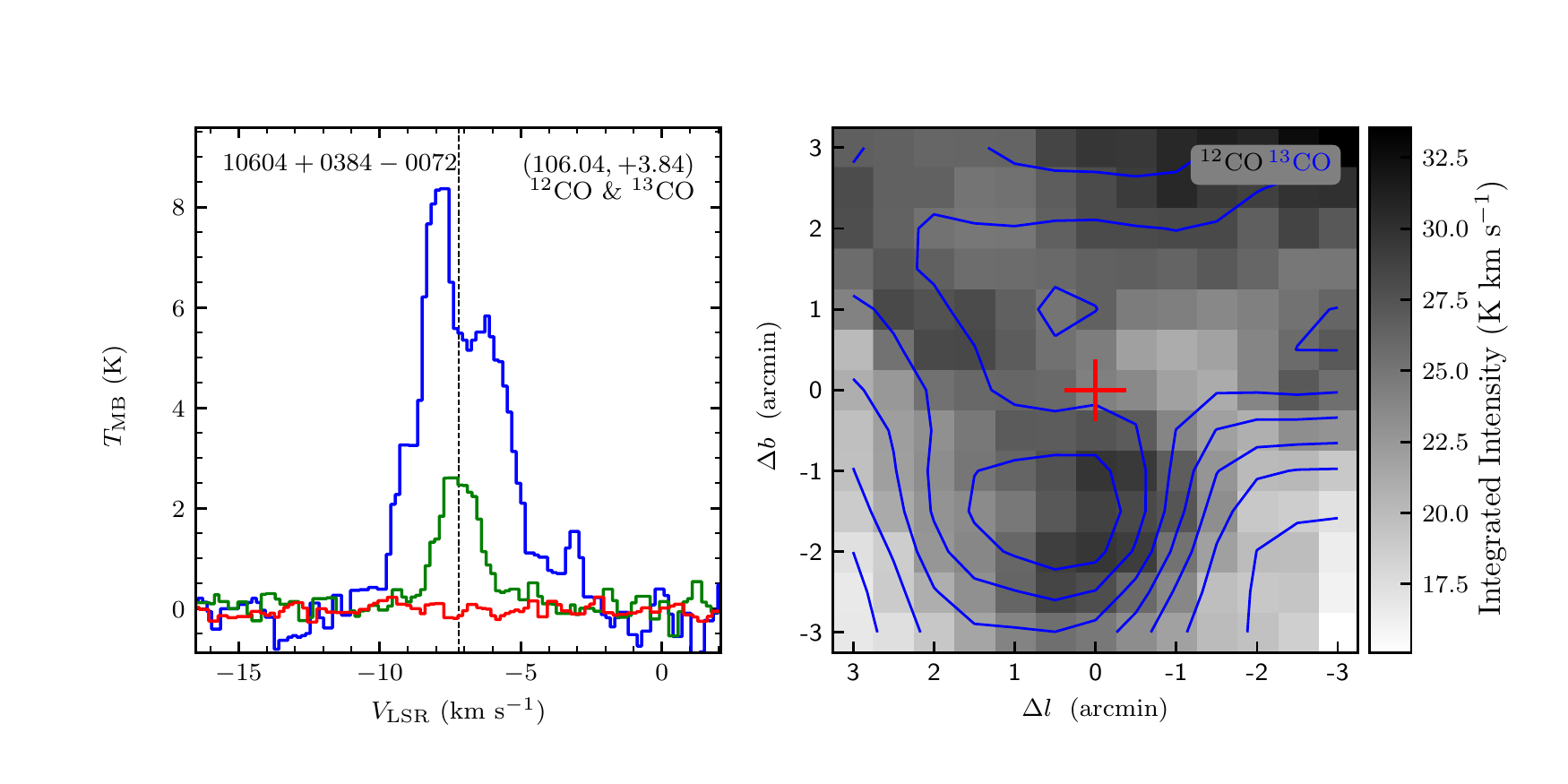}
\includegraphics[width=9.0cm,angle=0]{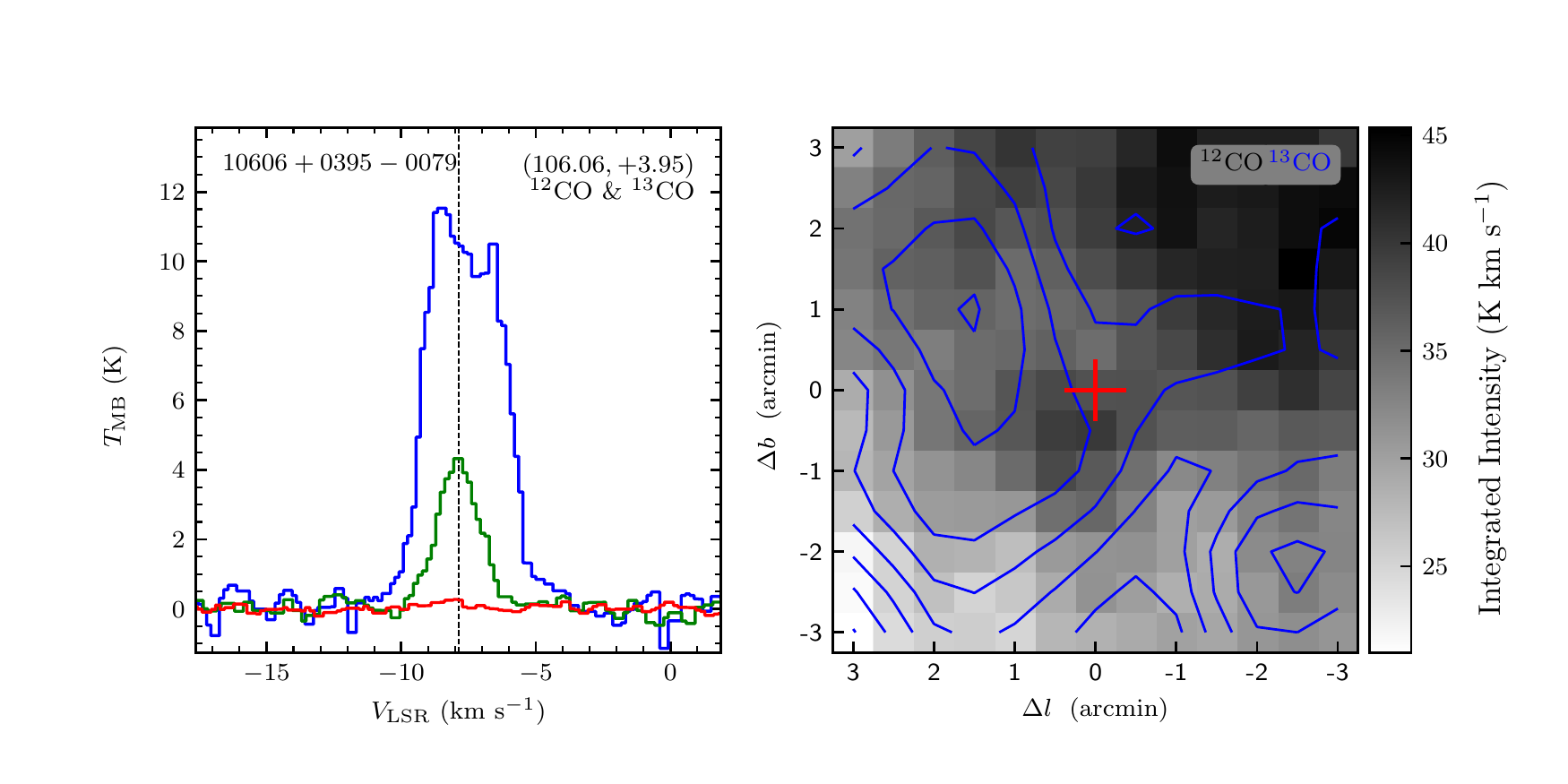}
\vspace{-0.5cm}

\includegraphics[width=9.0cm,angle=0]{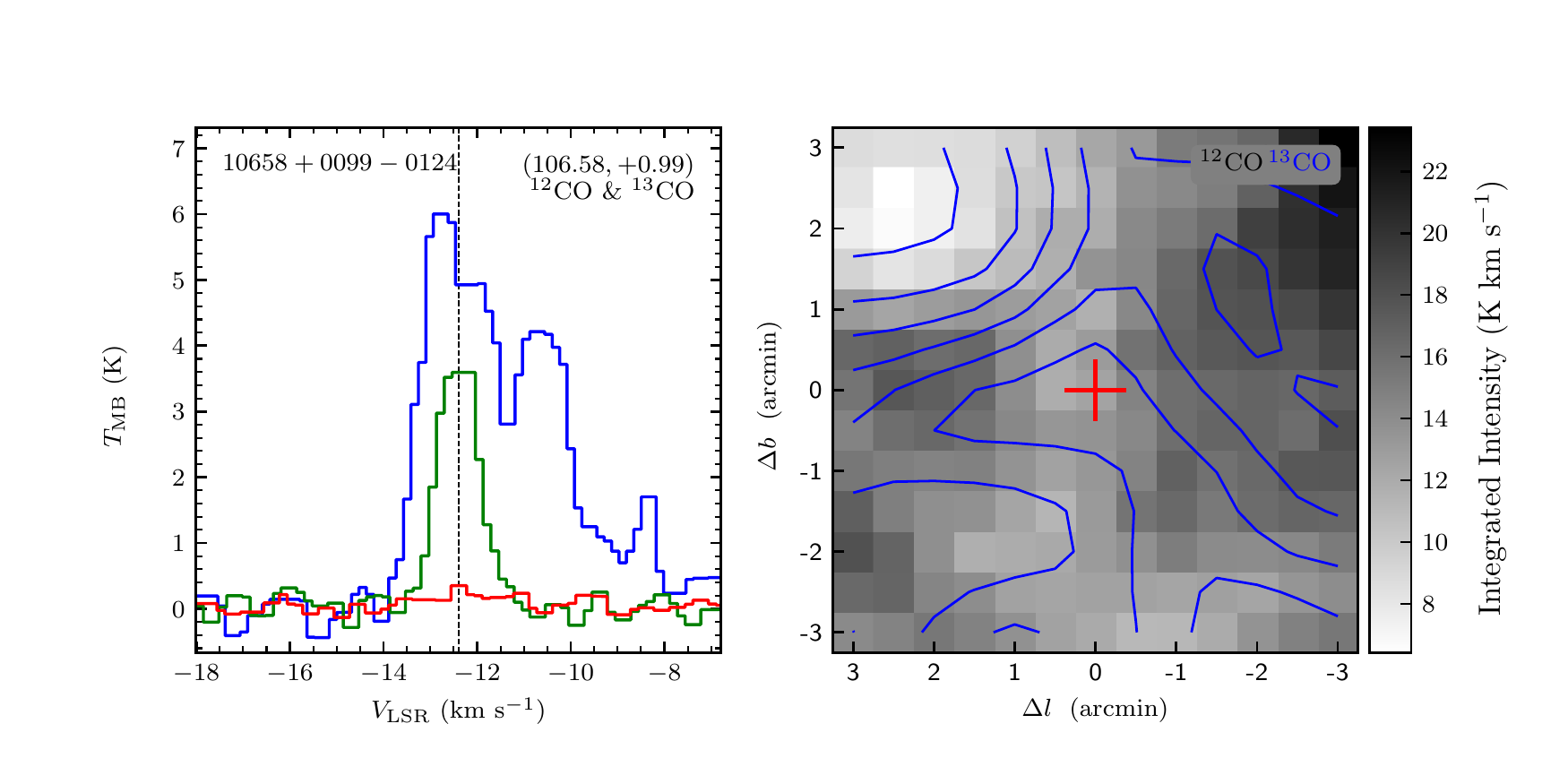}
\includegraphics[width=9.0cm,angle=0]{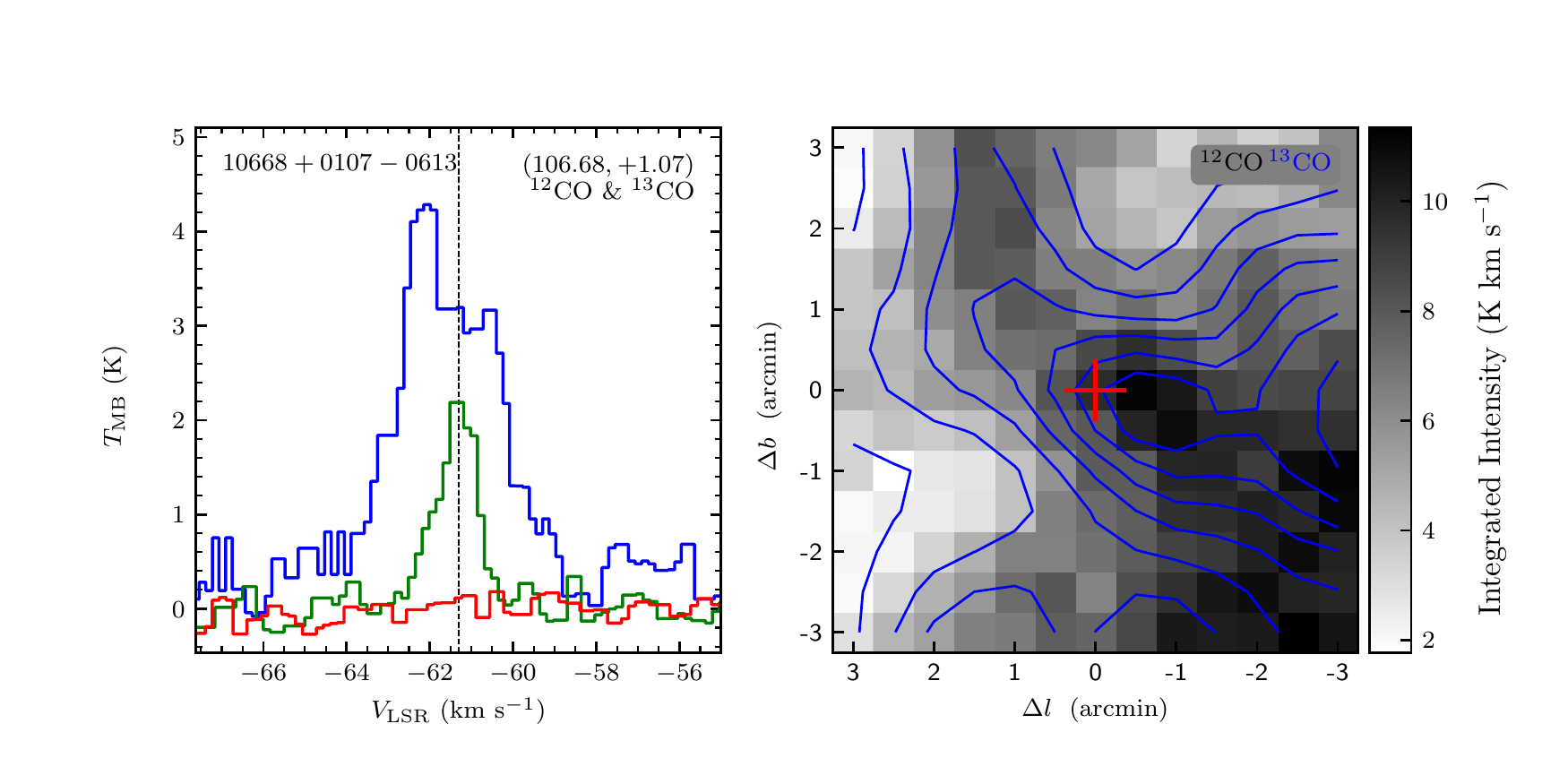}
\vspace{-0.5cm}

\includegraphics[width=9.0cm,angle=0]{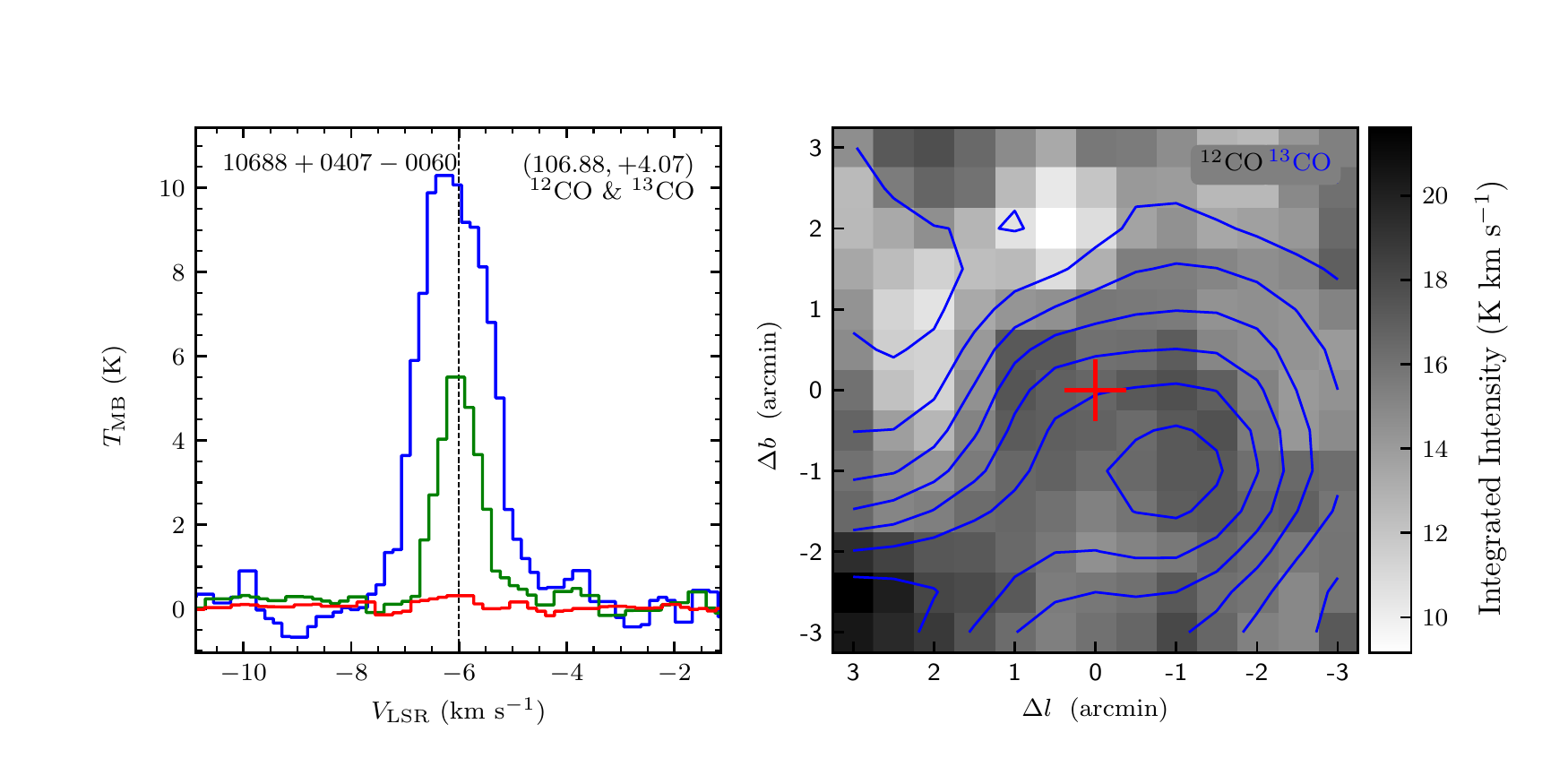}
\includegraphics[width=9.0cm,angle=0]{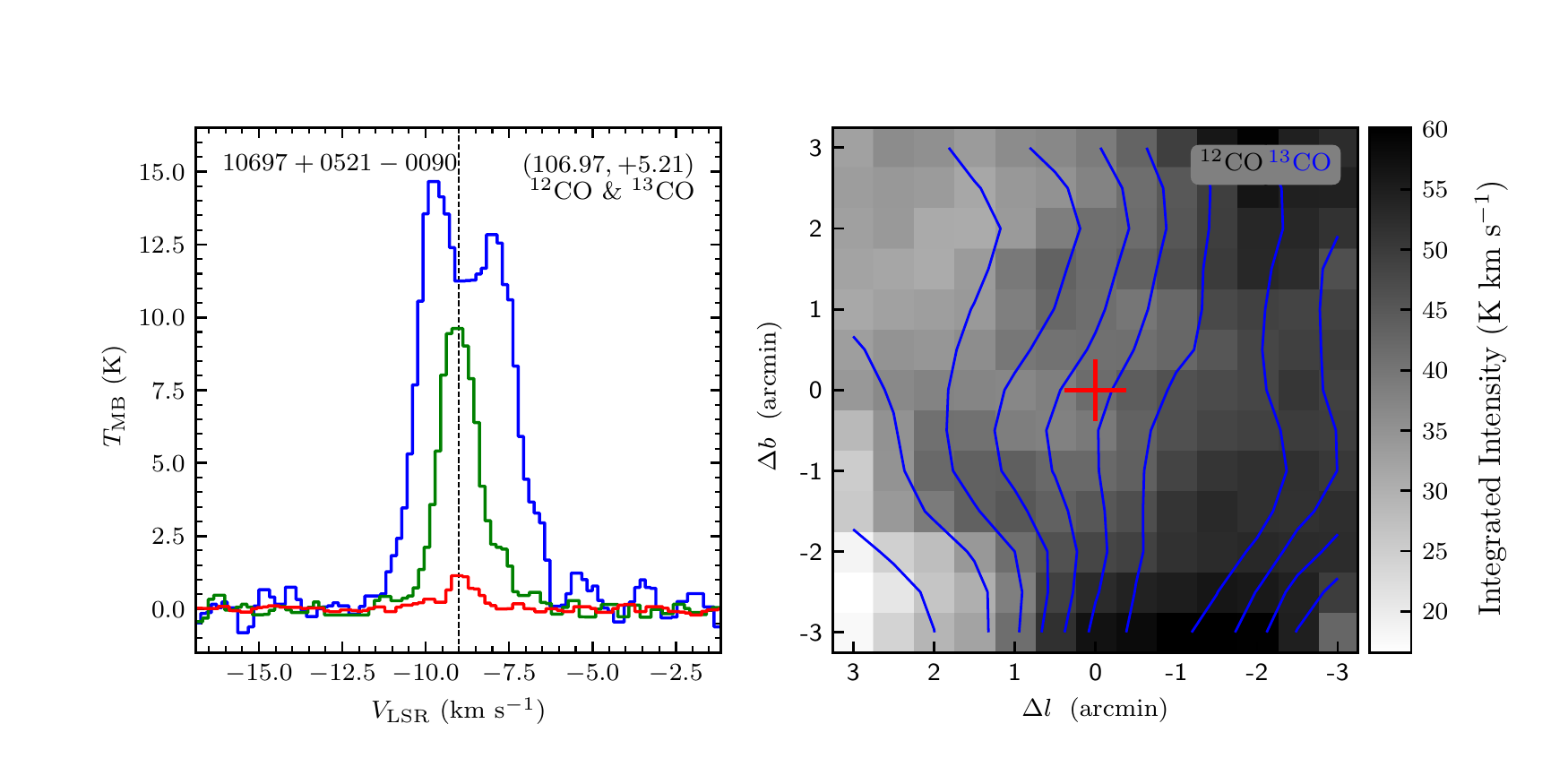}
\vspace{-0.5cm}

\includegraphics[width=9.0cm,angle=0]{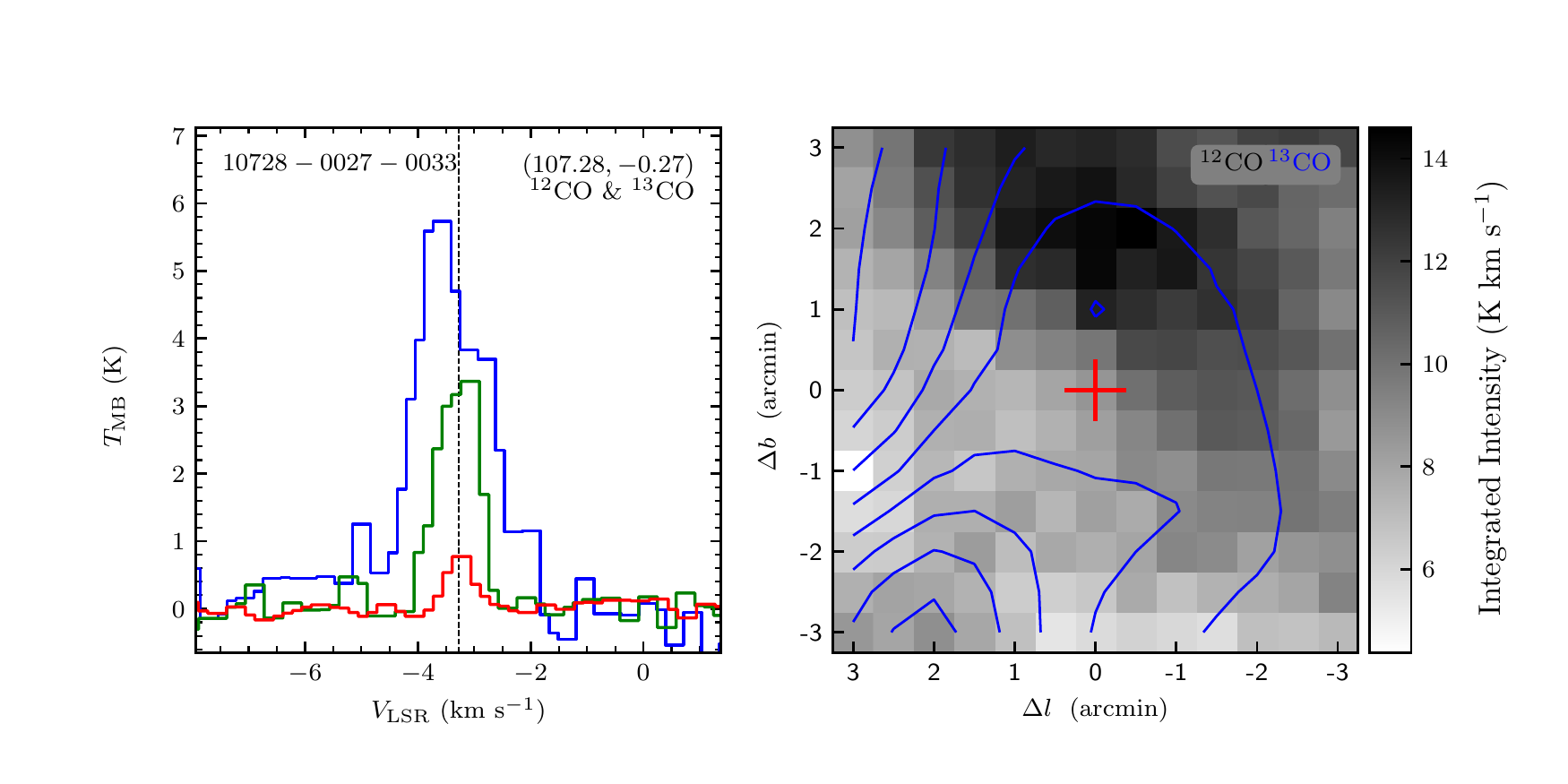}
\includegraphics[width=9.0cm,angle=0]{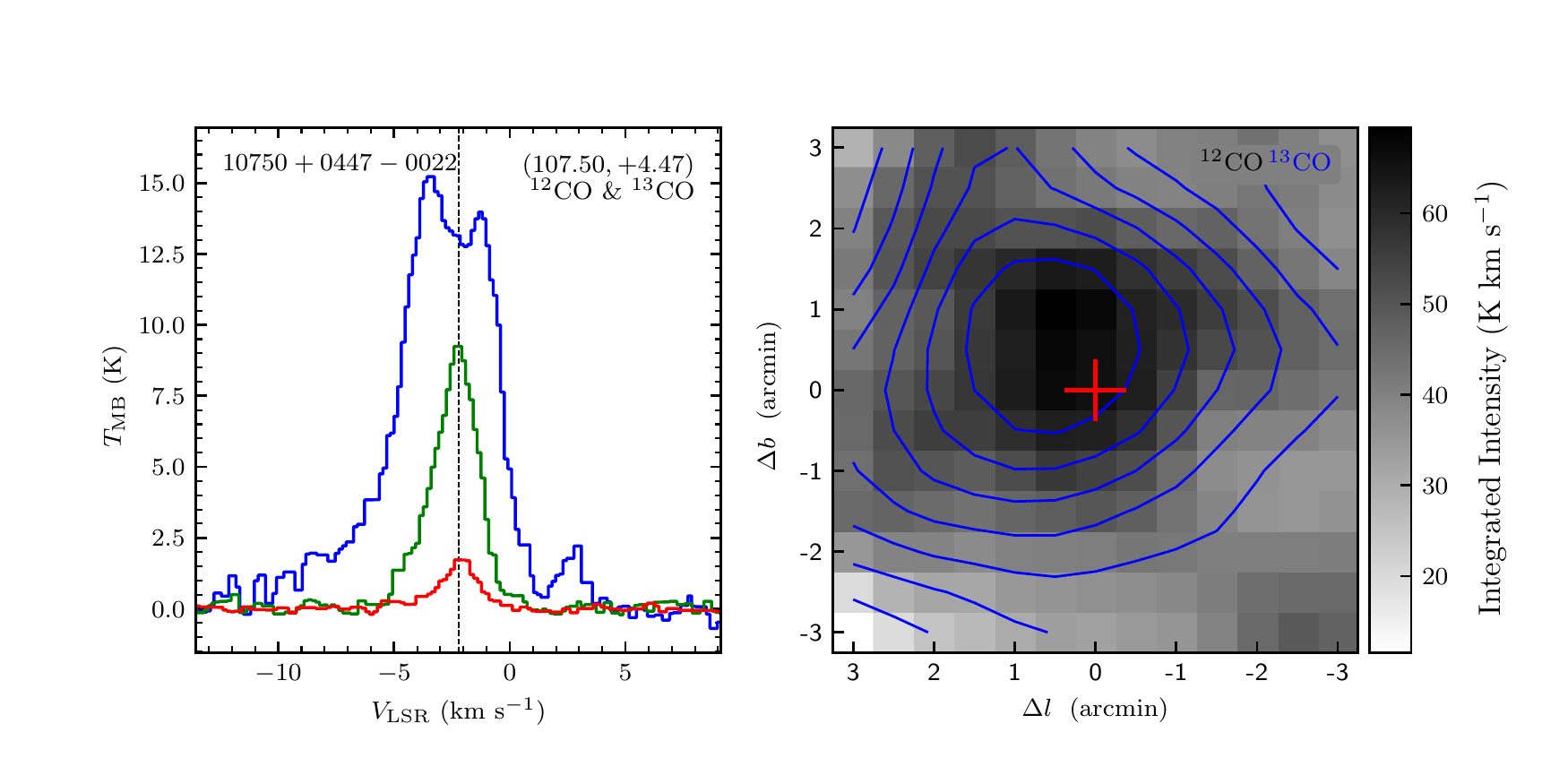}
\end{figure}
\clearpage

\begin{figure}
\includegraphics[width=9.0cm,angle=0]{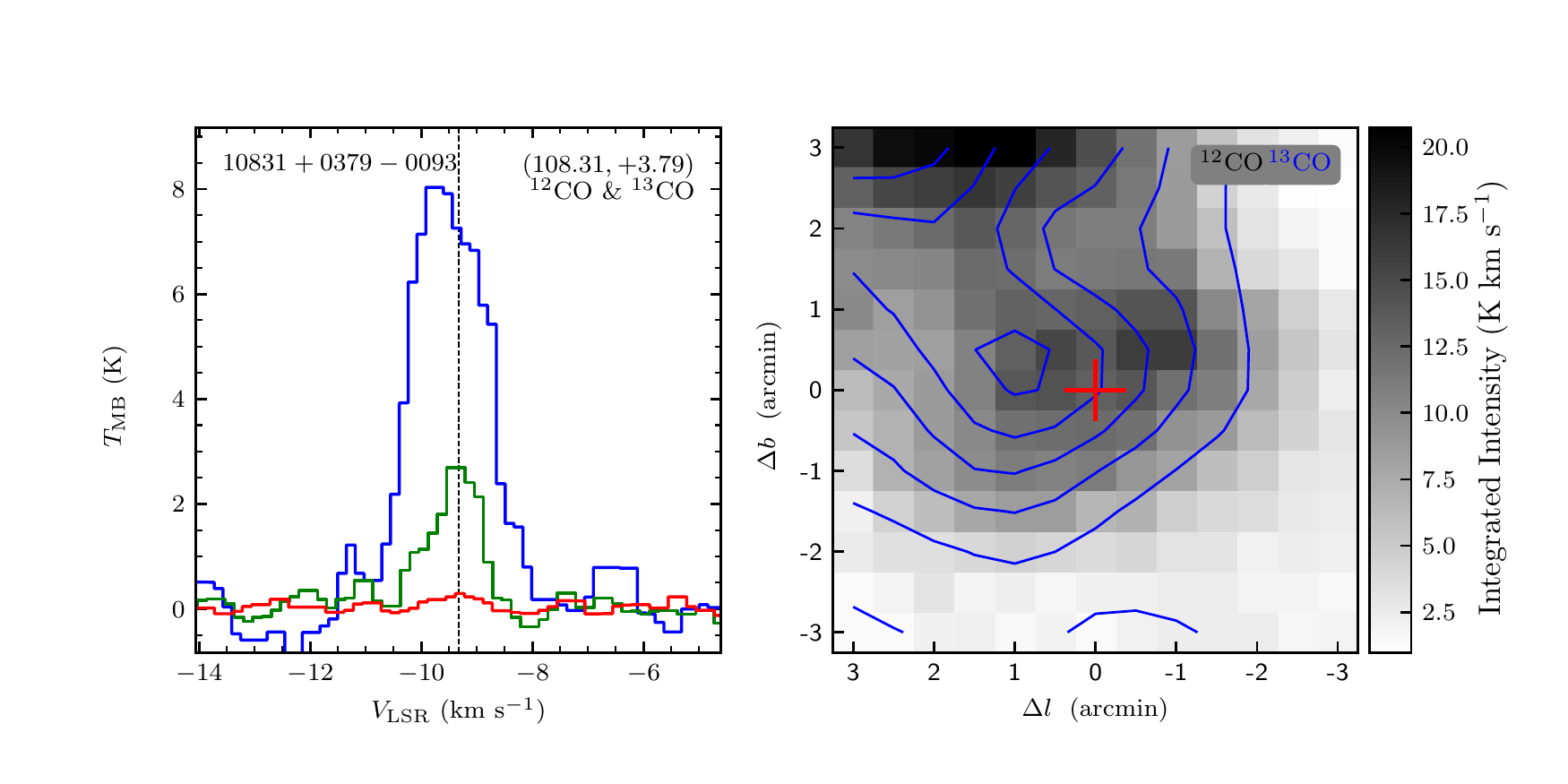}
\includegraphics[width=9.0cm,angle=0]{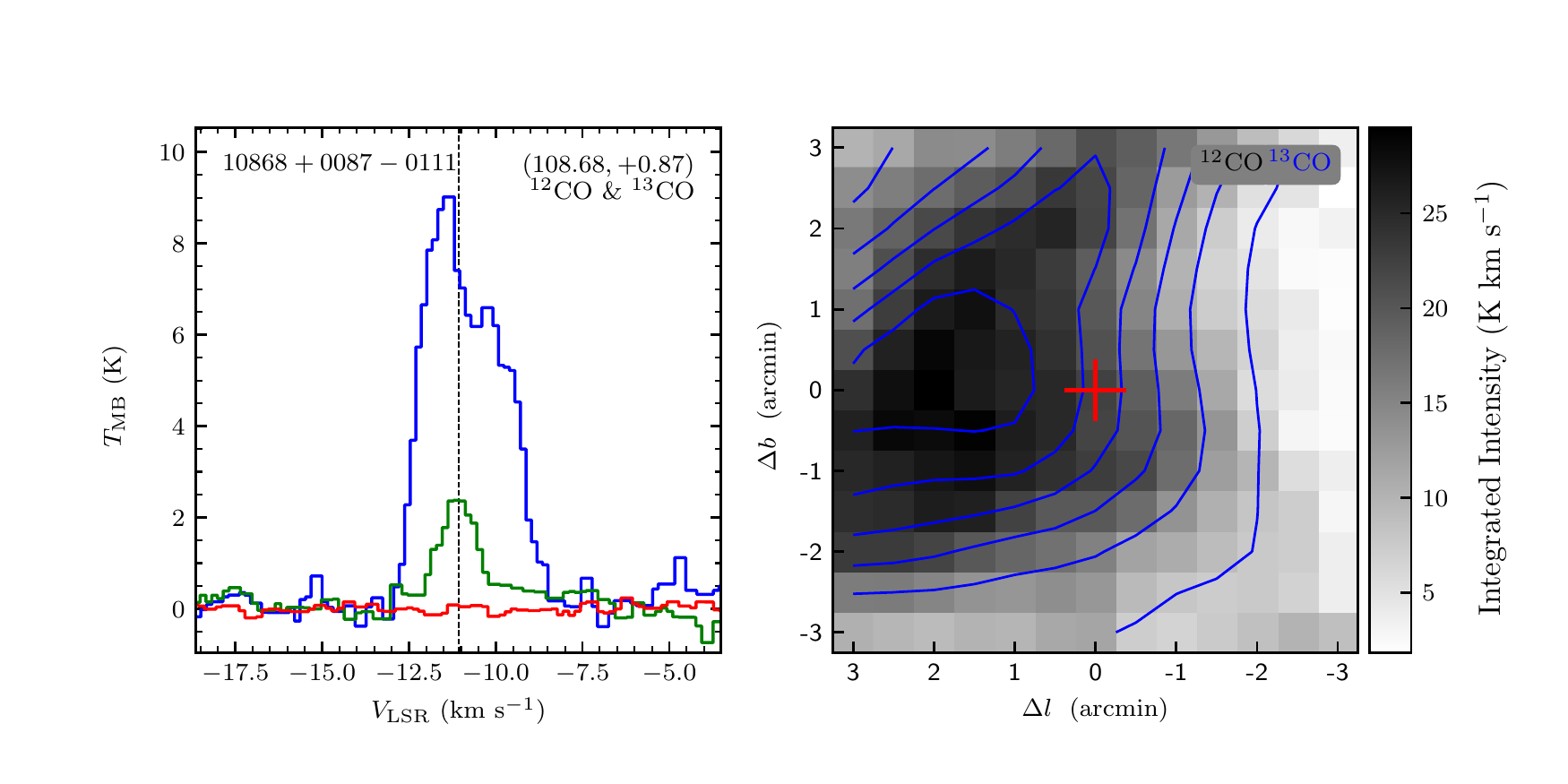}
\vspace{-0.5cm}

\includegraphics[width=9.0cm,angle=0]{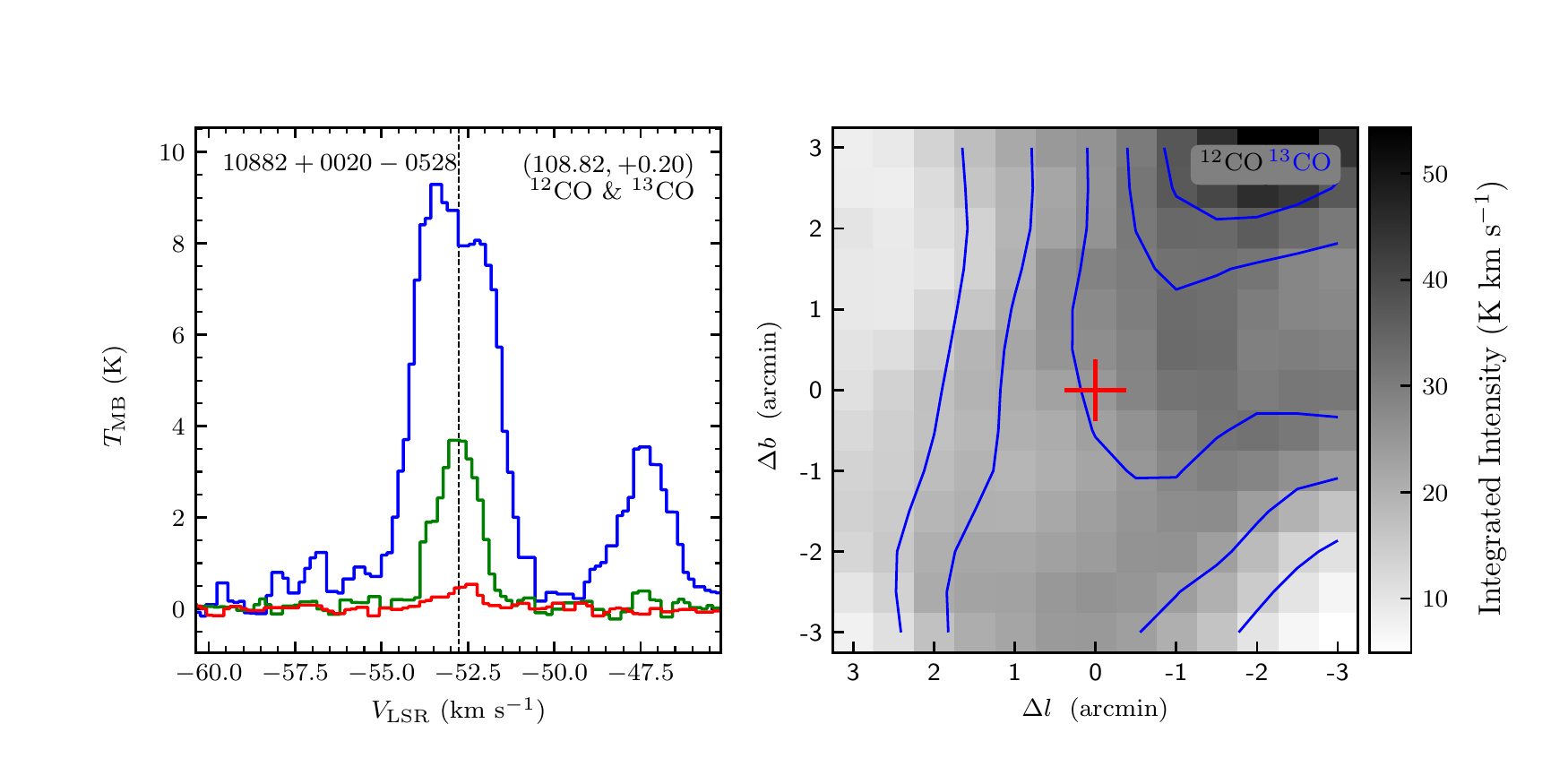}
\includegraphics[width=9.0cm,angle=0]{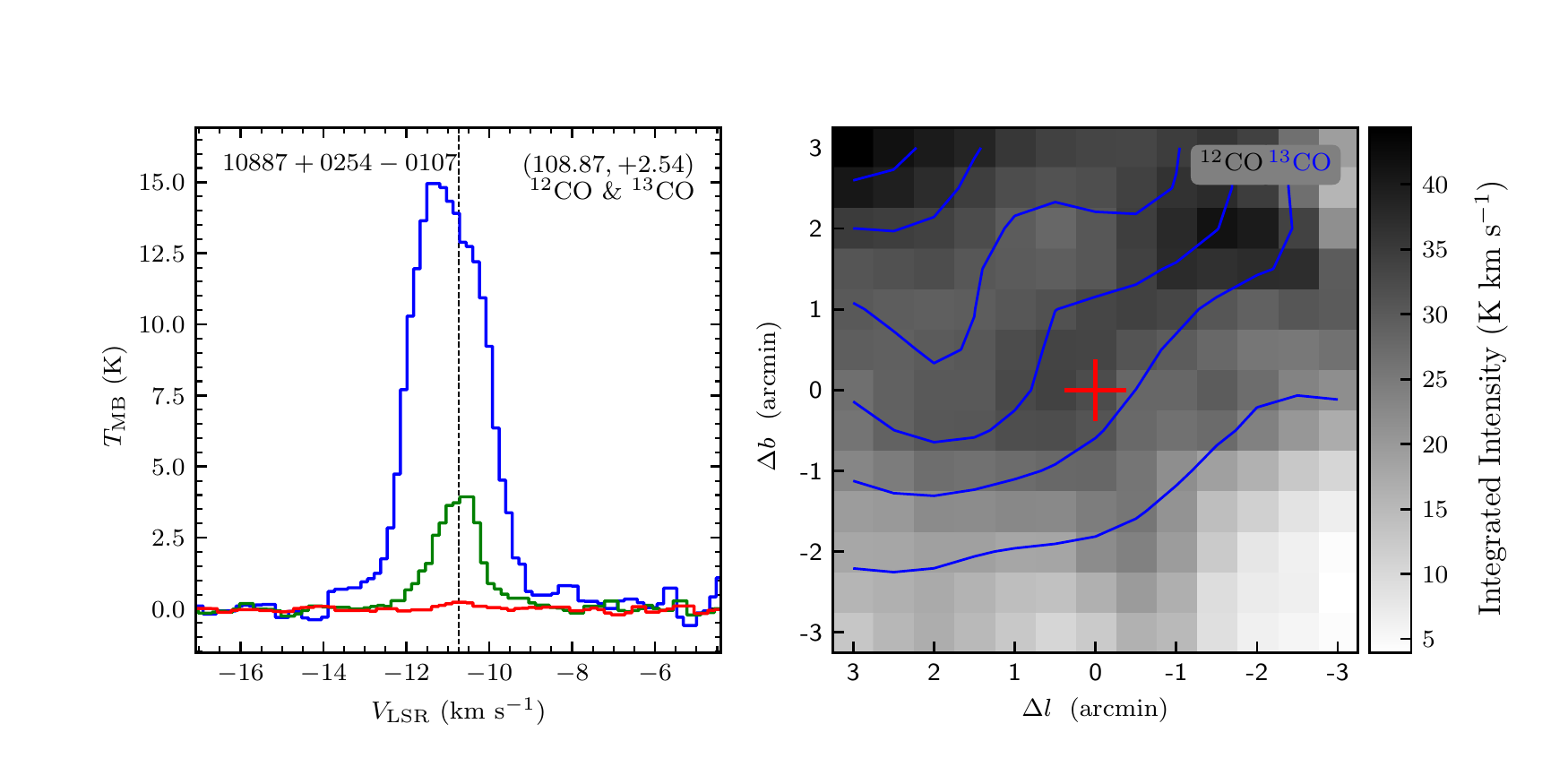}
\vspace{-0.5cm}

\includegraphics[width=9.0cm,angle=0]{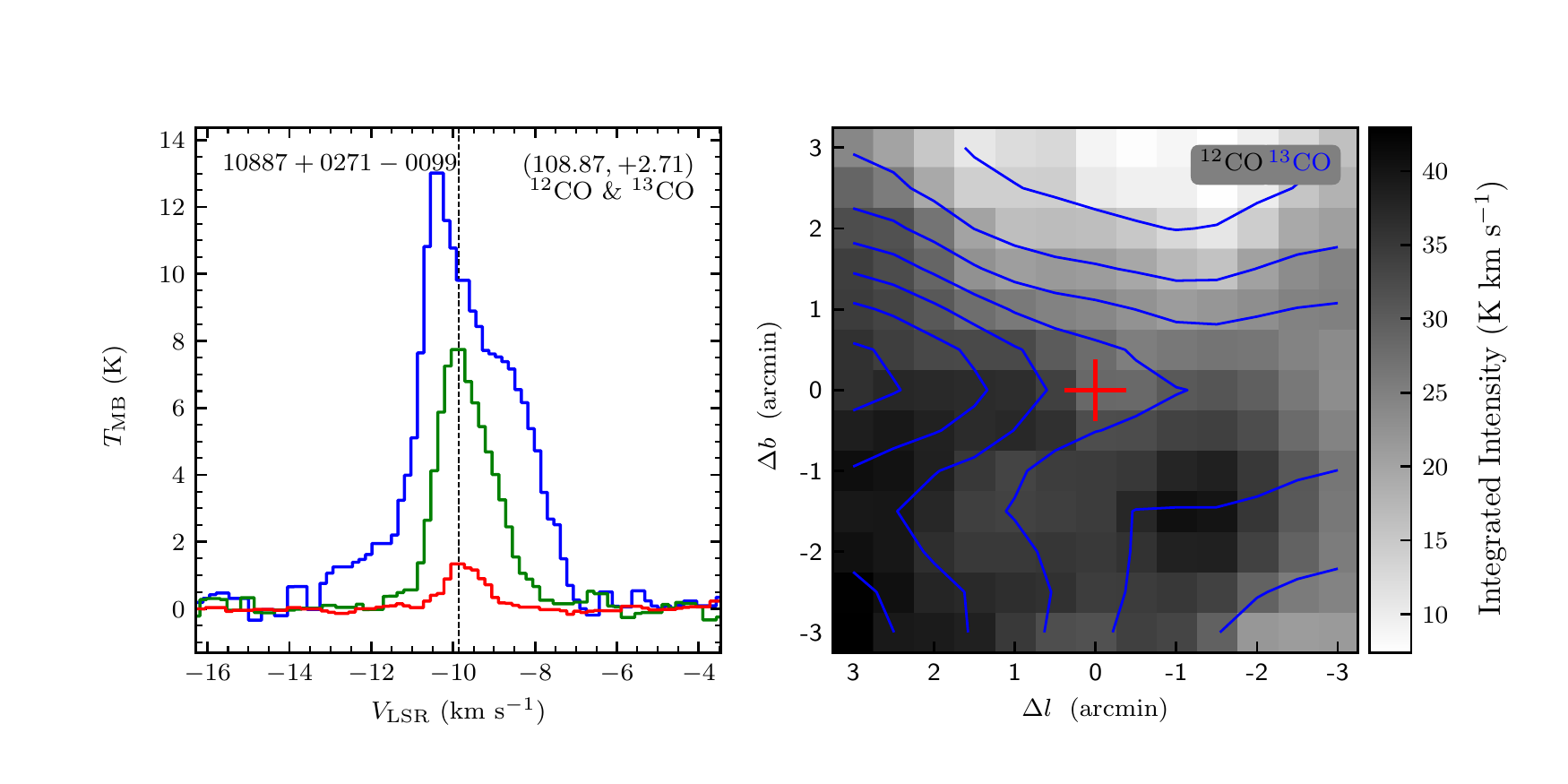}
\includegraphics[width=9.0cm,angle=0]{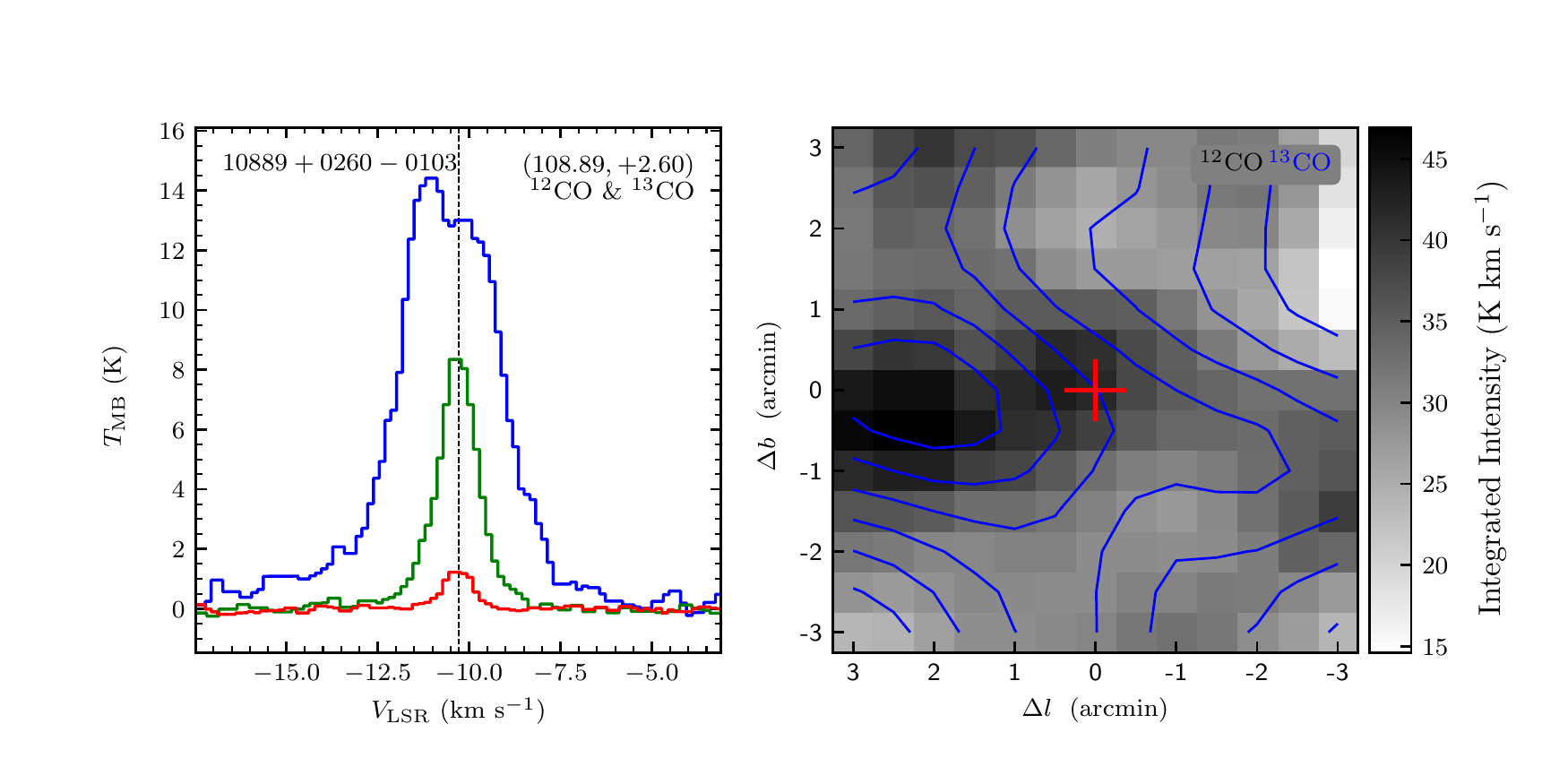}
\vspace{-0.5cm}

\includegraphics[width=9.0cm,angle=0]{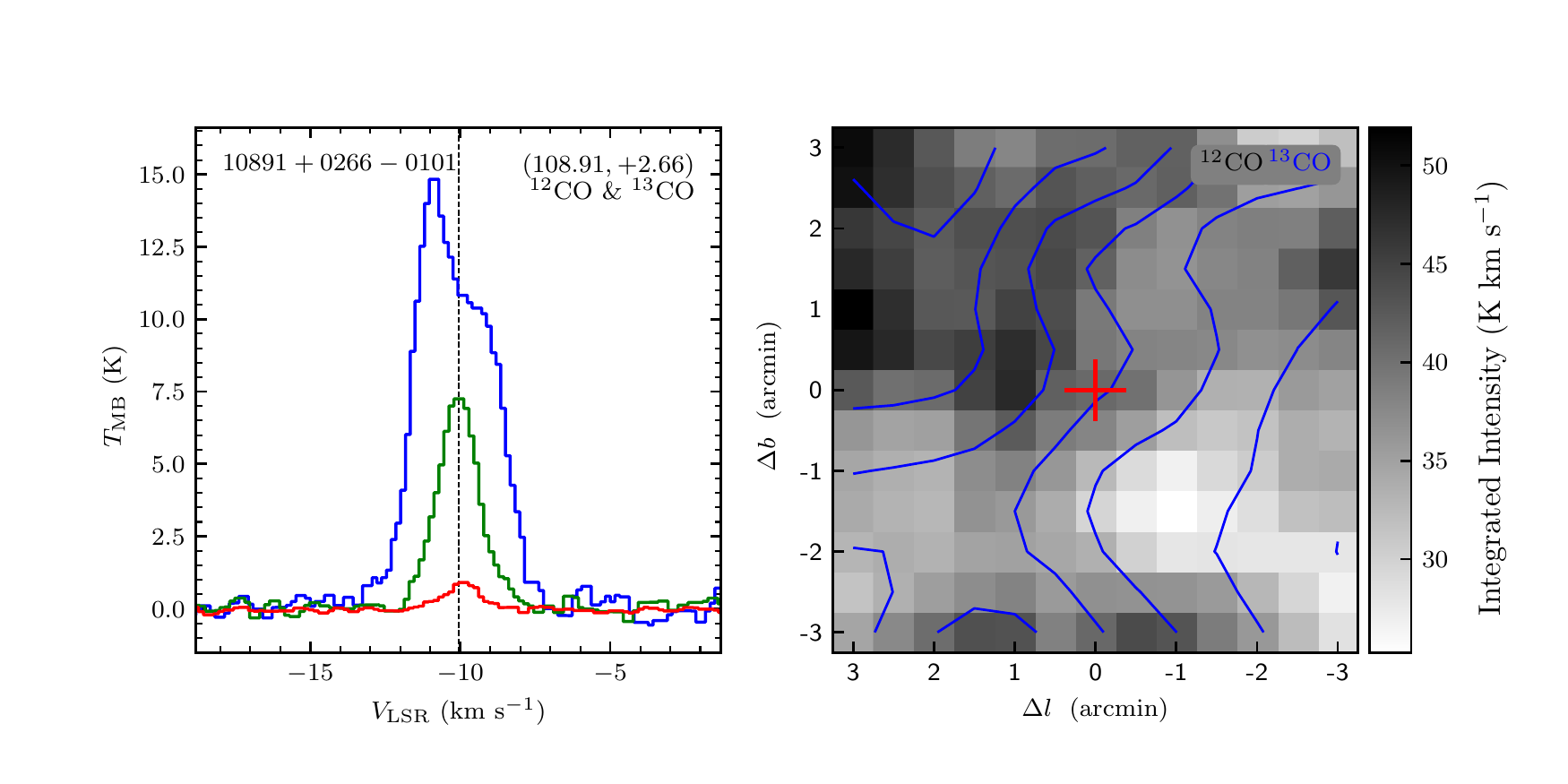}
\includegraphics[width=9.0cm,angle=0]{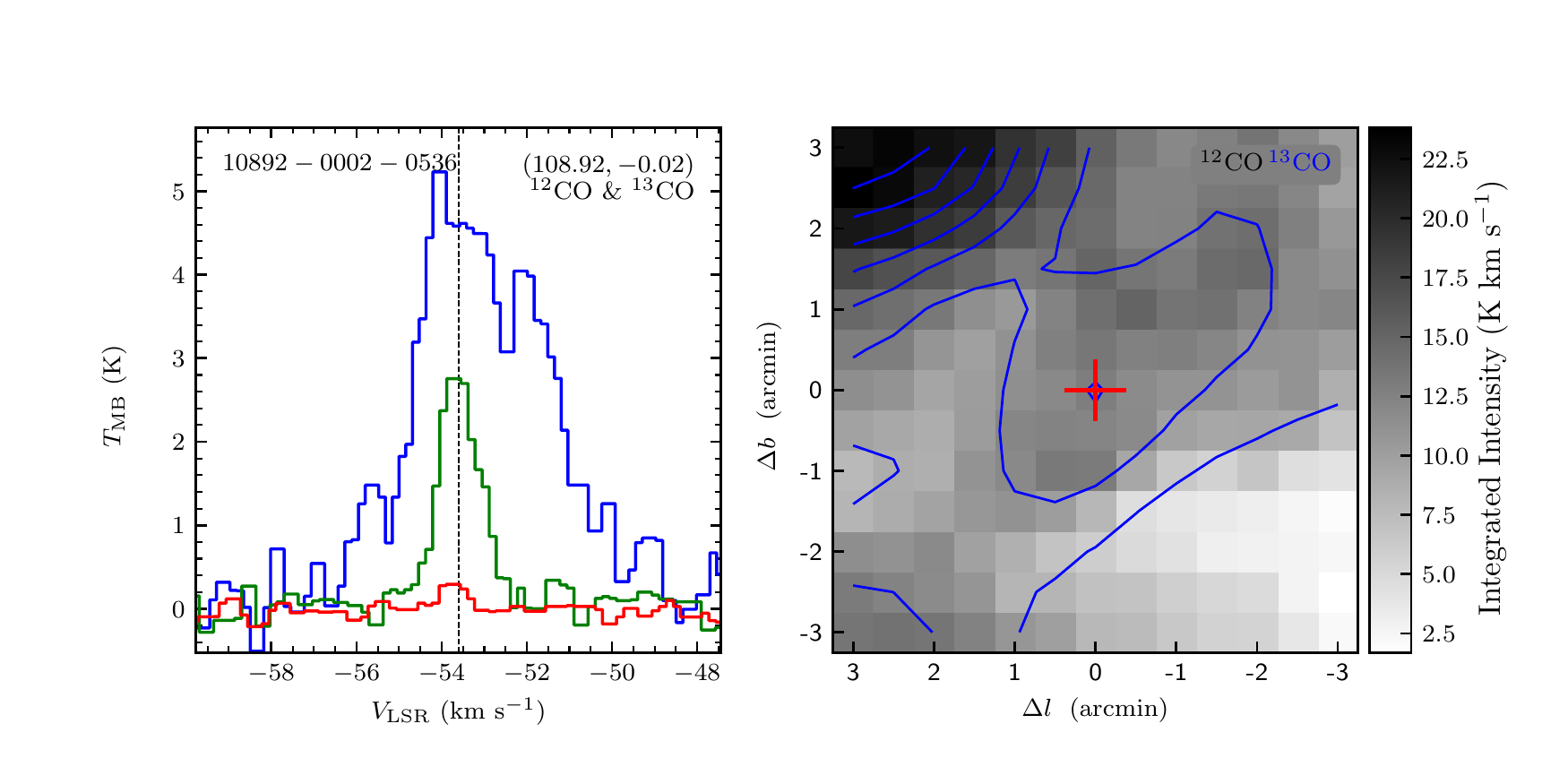}
\vspace{-0.5cm}

\includegraphics[width=9.0cm,angle=0]{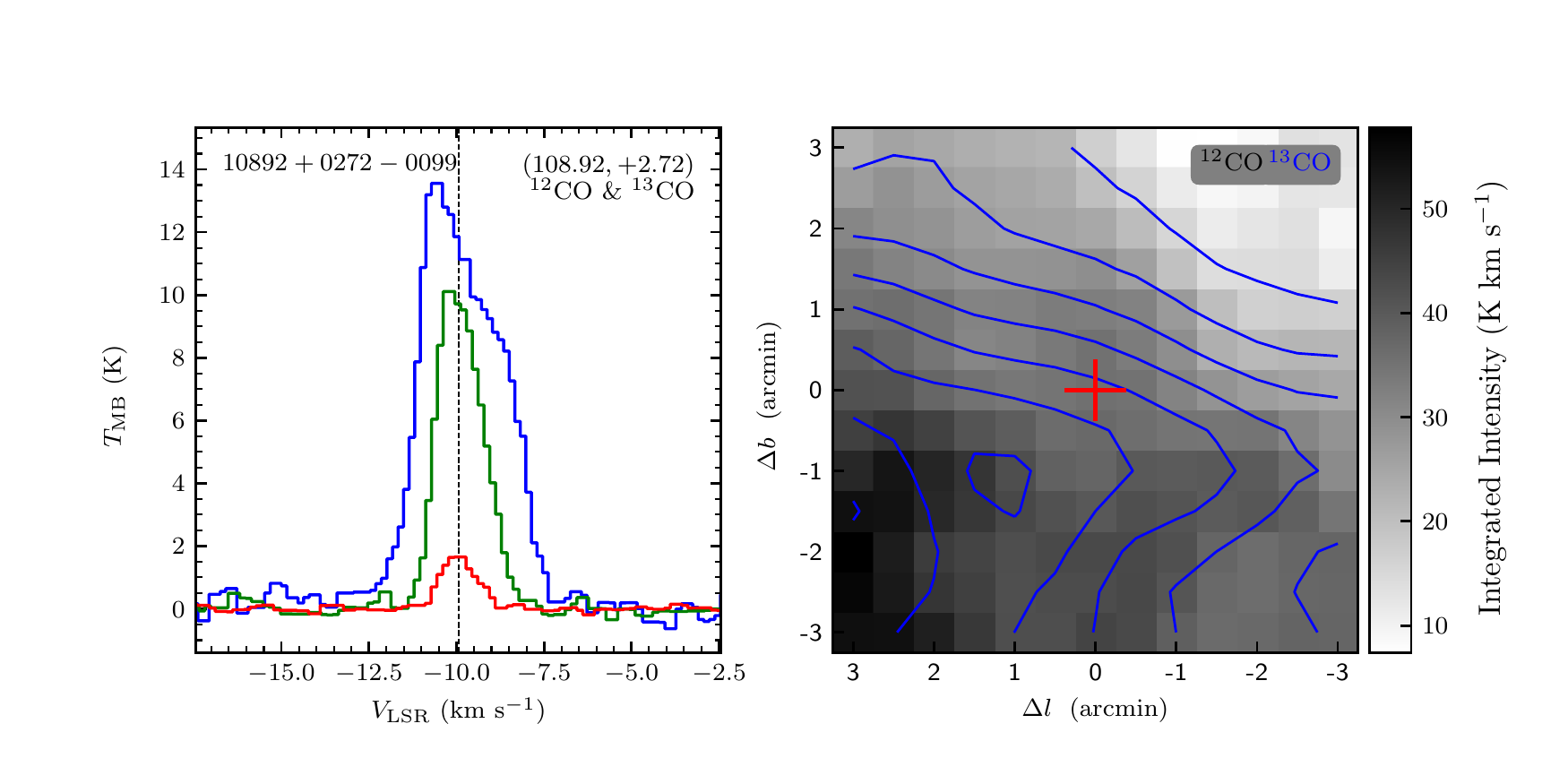}
\includegraphics[width=9.0cm,angle=0]{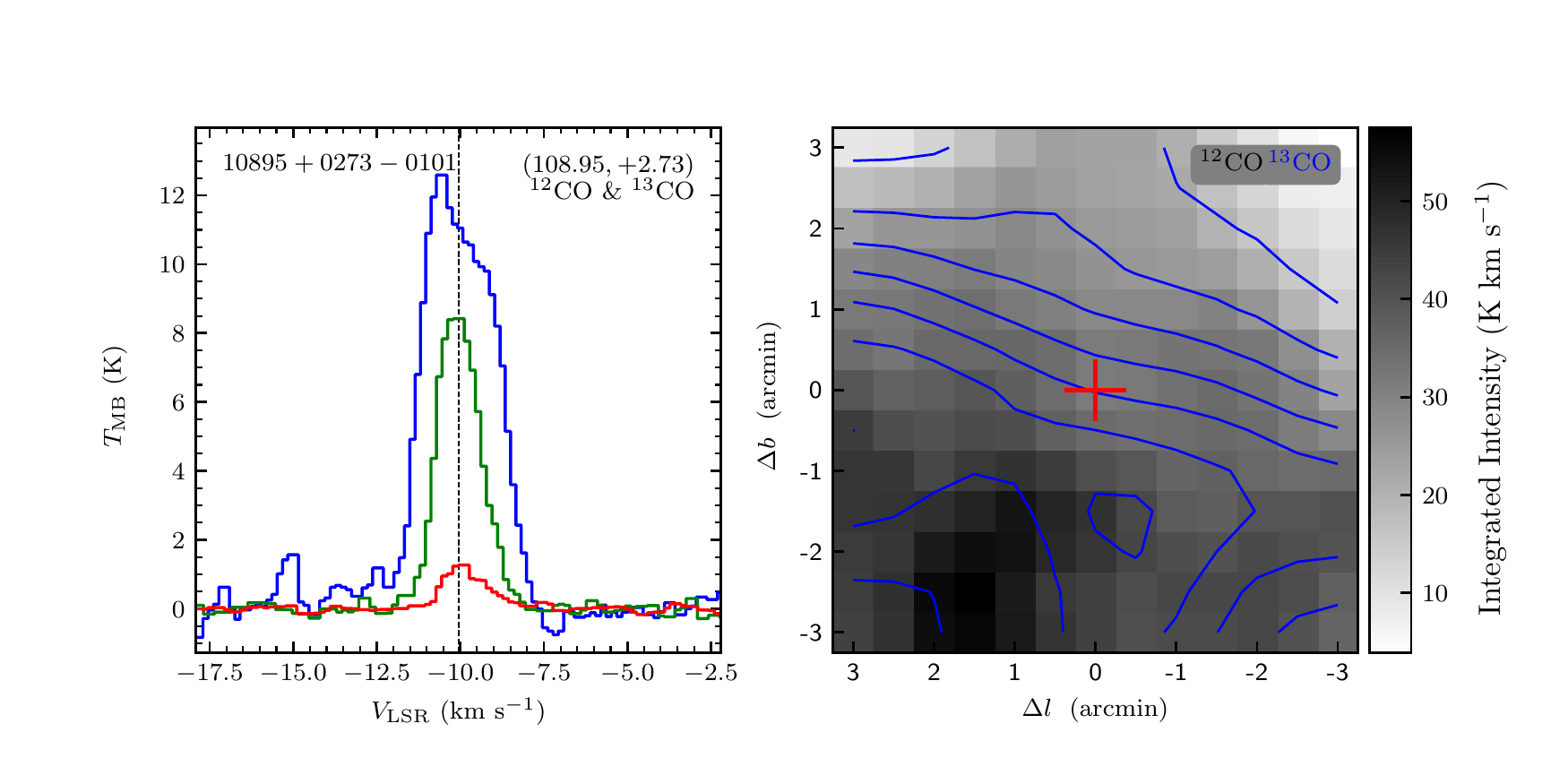}
\end{figure}
\clearpage

\begin{figure}
\includegraphics[width=9.0cm,angle=0]{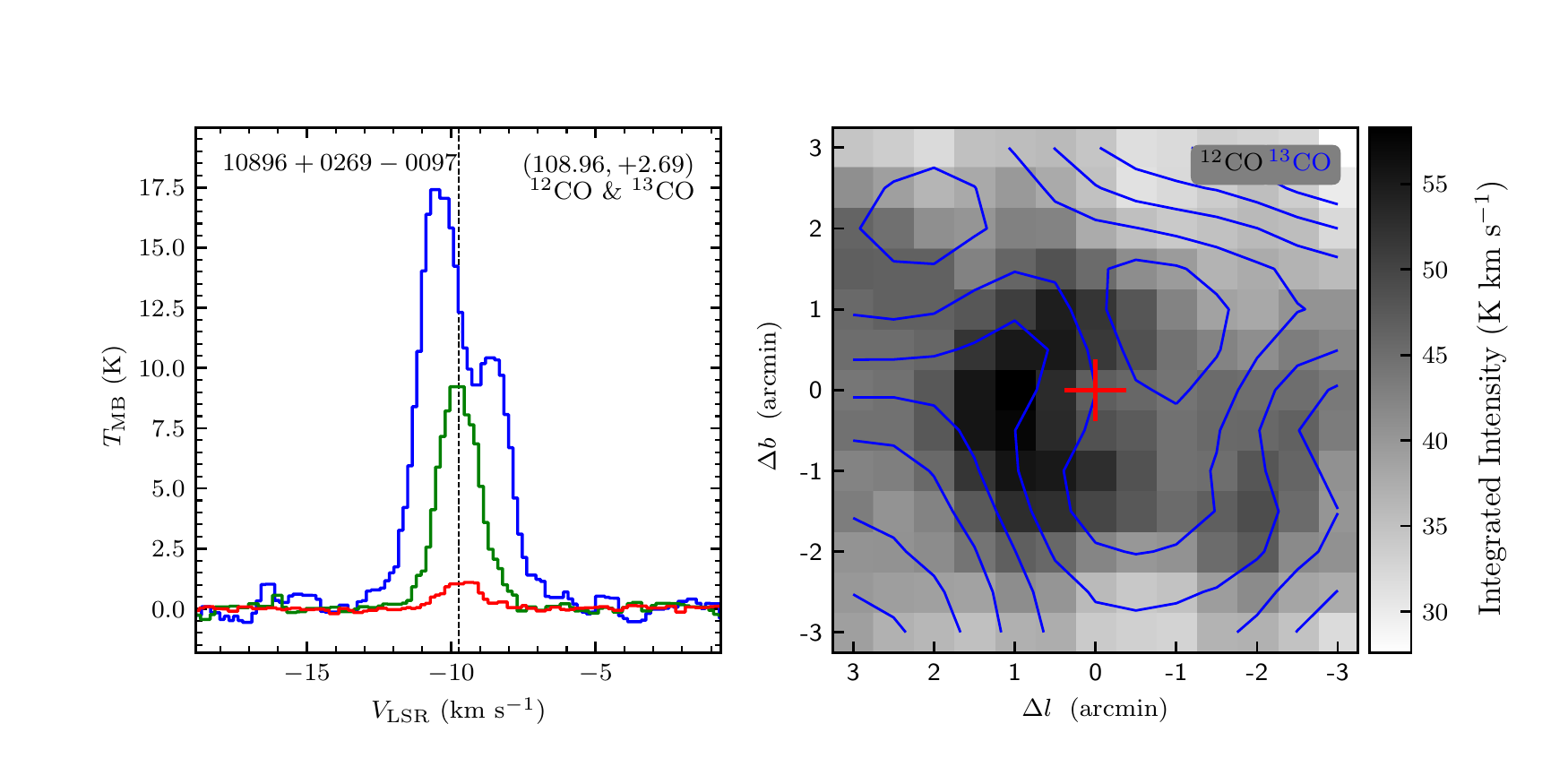}
\includegraphics[width=9.0cm,angle=0]{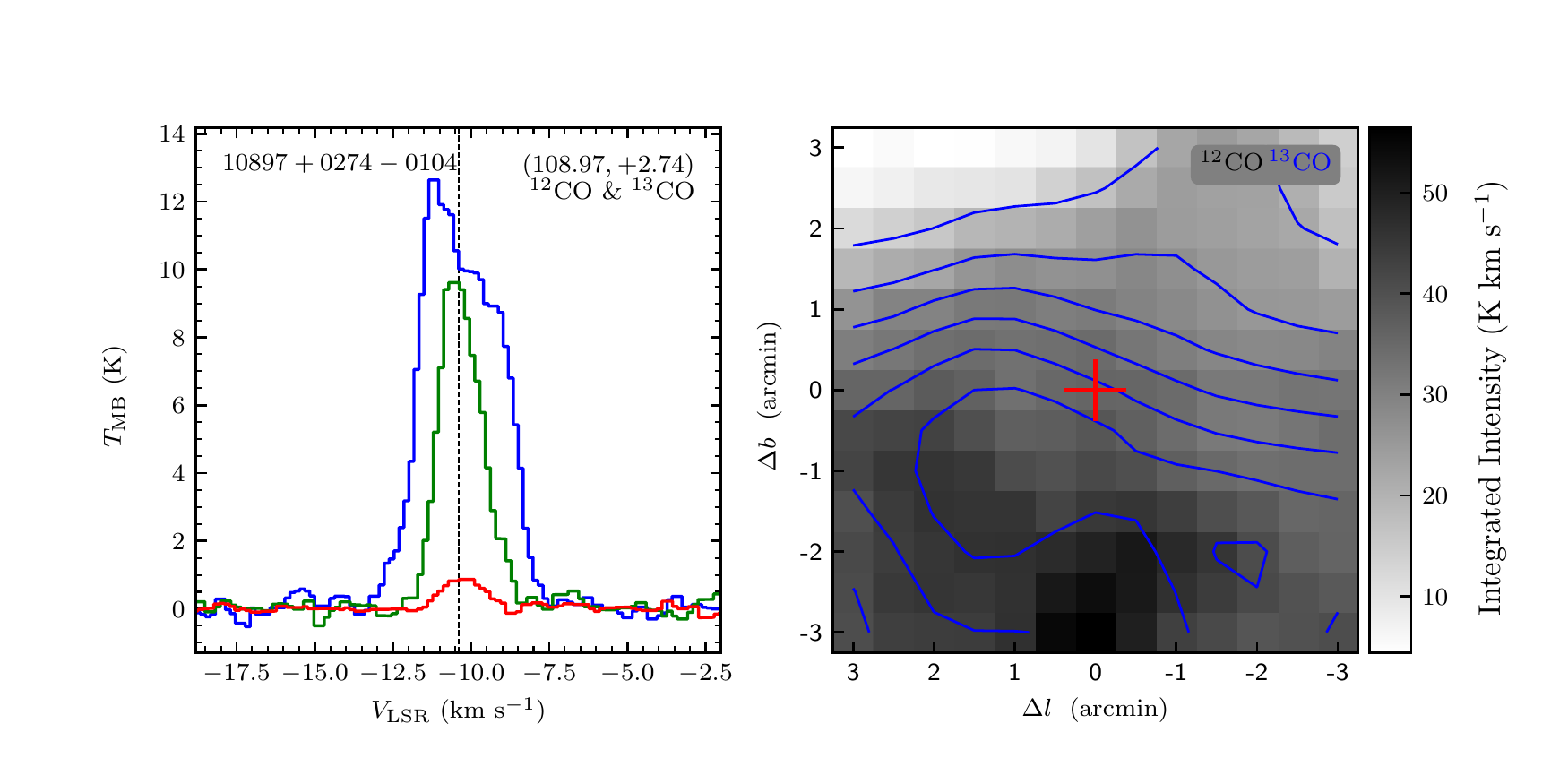}
\vspace{-0.5cm}

\includegraphics[width=9.0cm,angle=0]{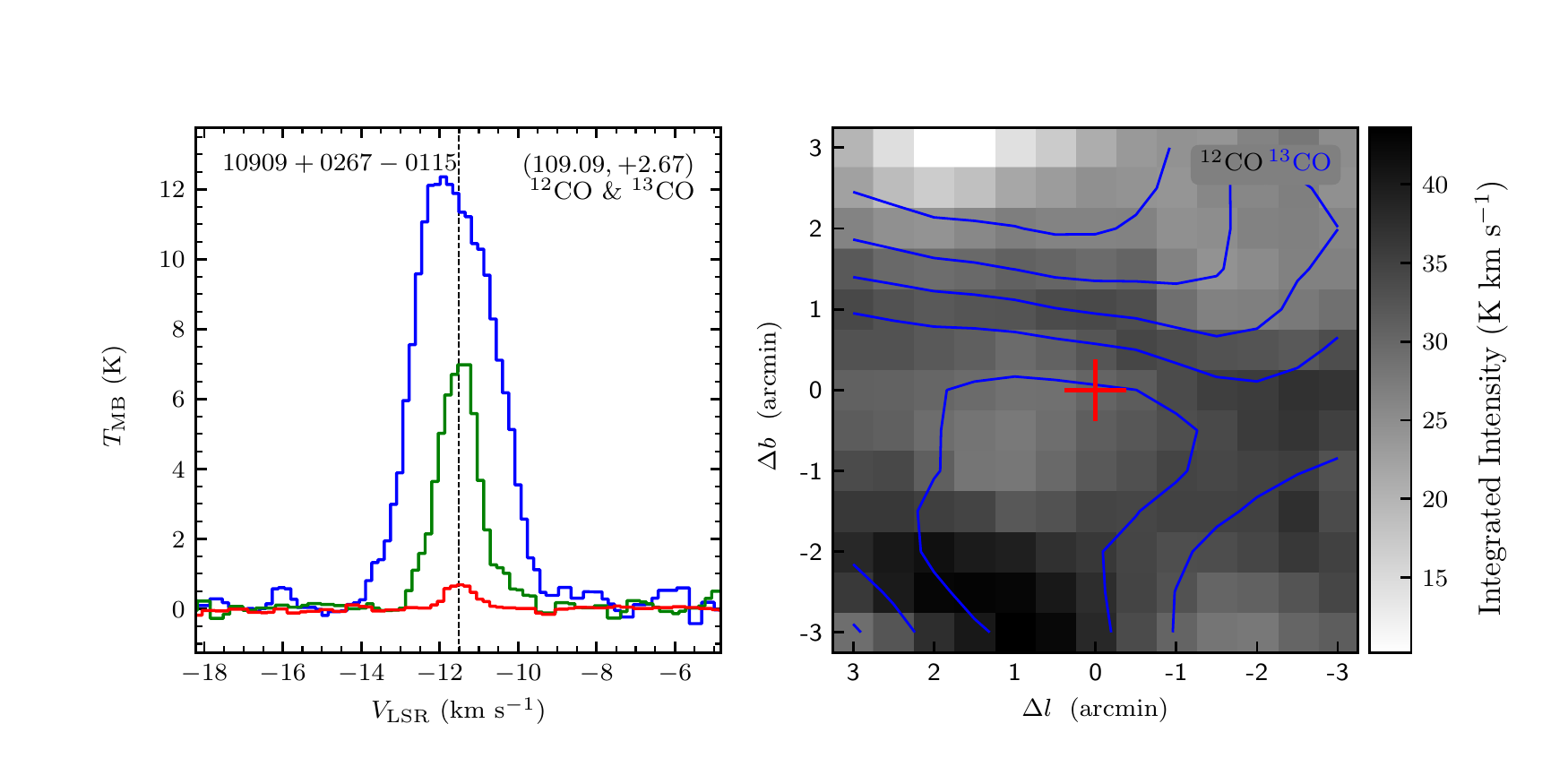}
\includegraphics[width=9.0cm,angle=0]{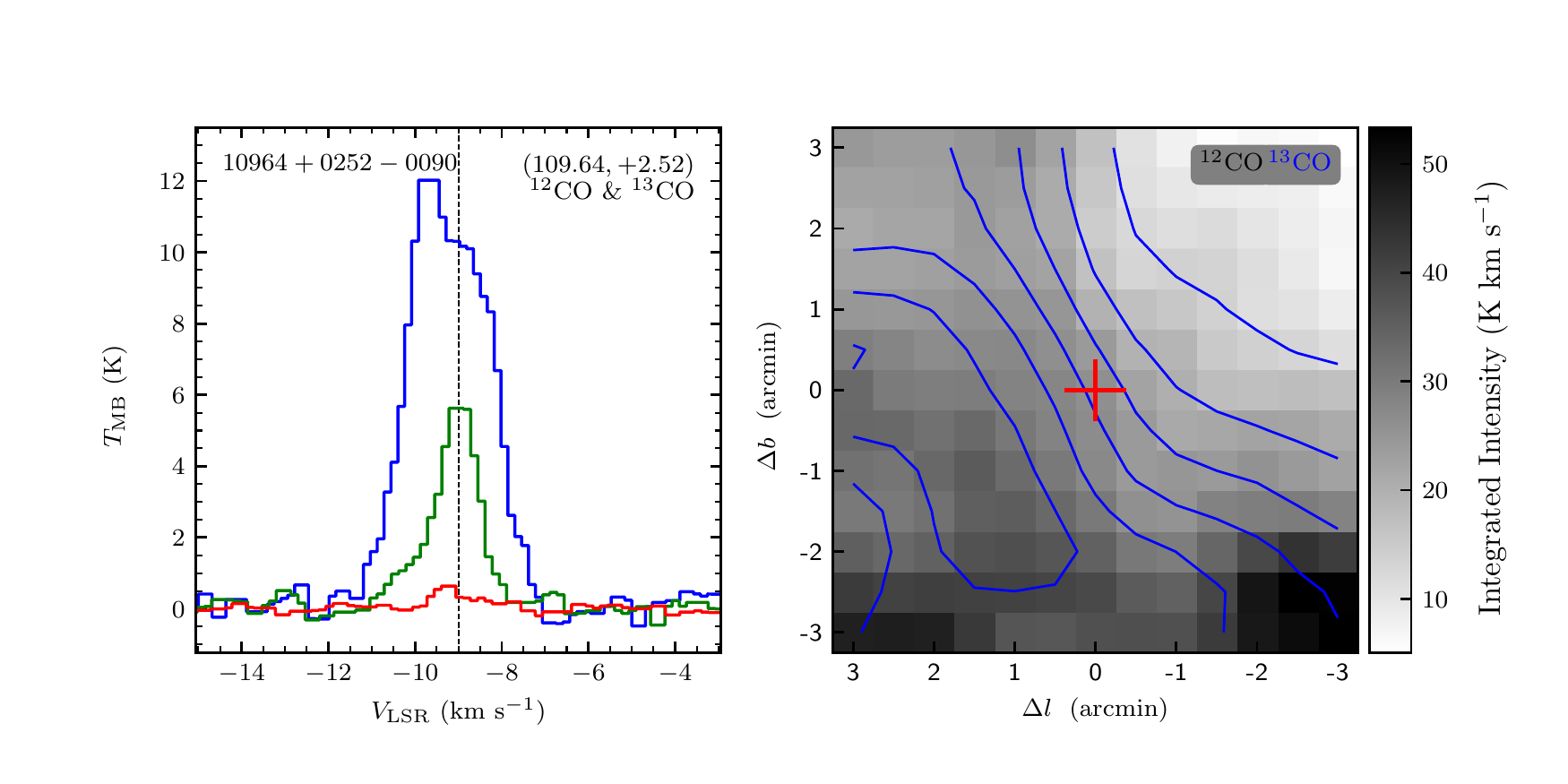}
\vspace{-0.5cm}

\includegraphics[width=9.0cm,angle=0]{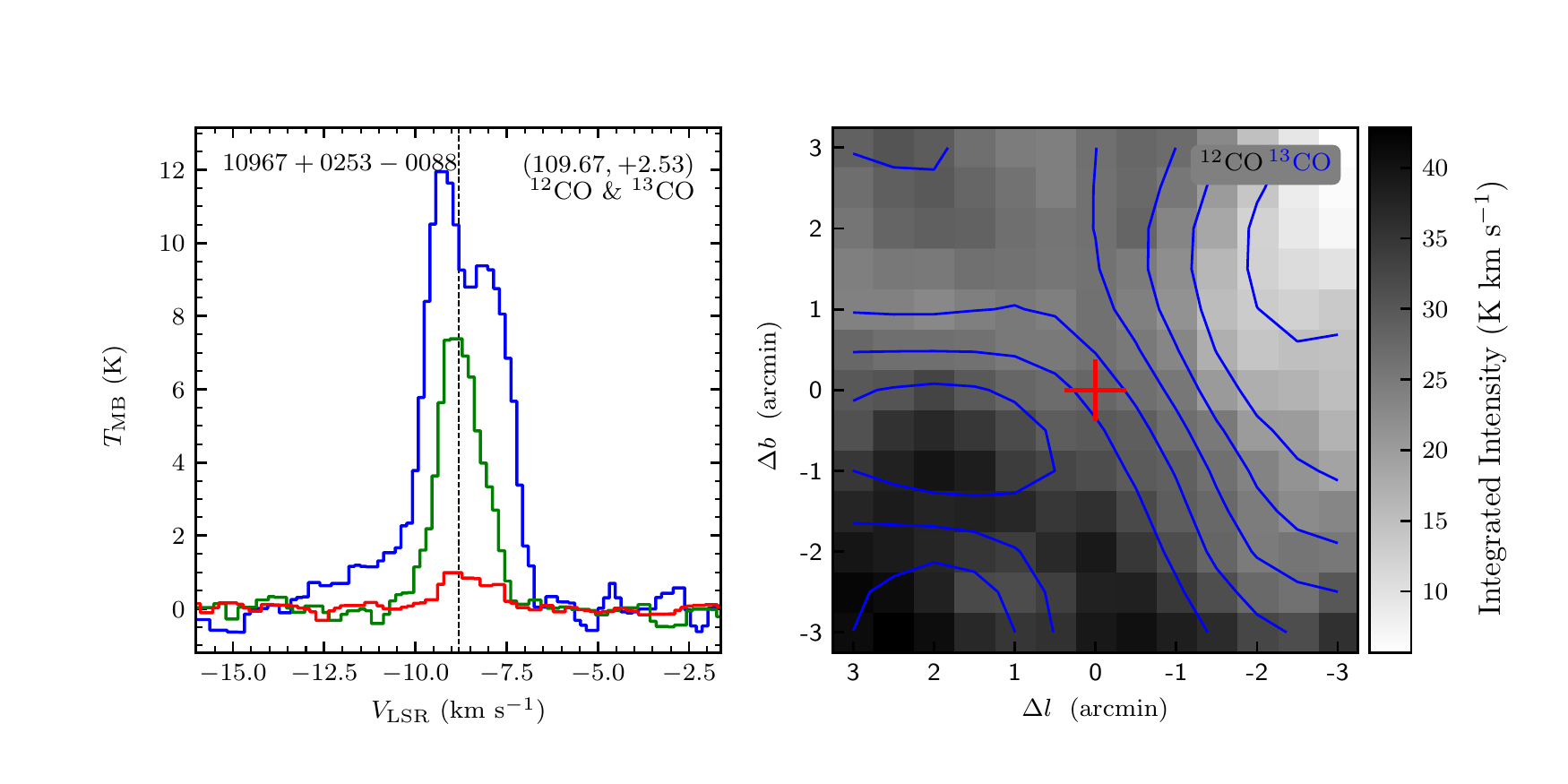}
\includegraphics[width=9.0cm,angle=0]{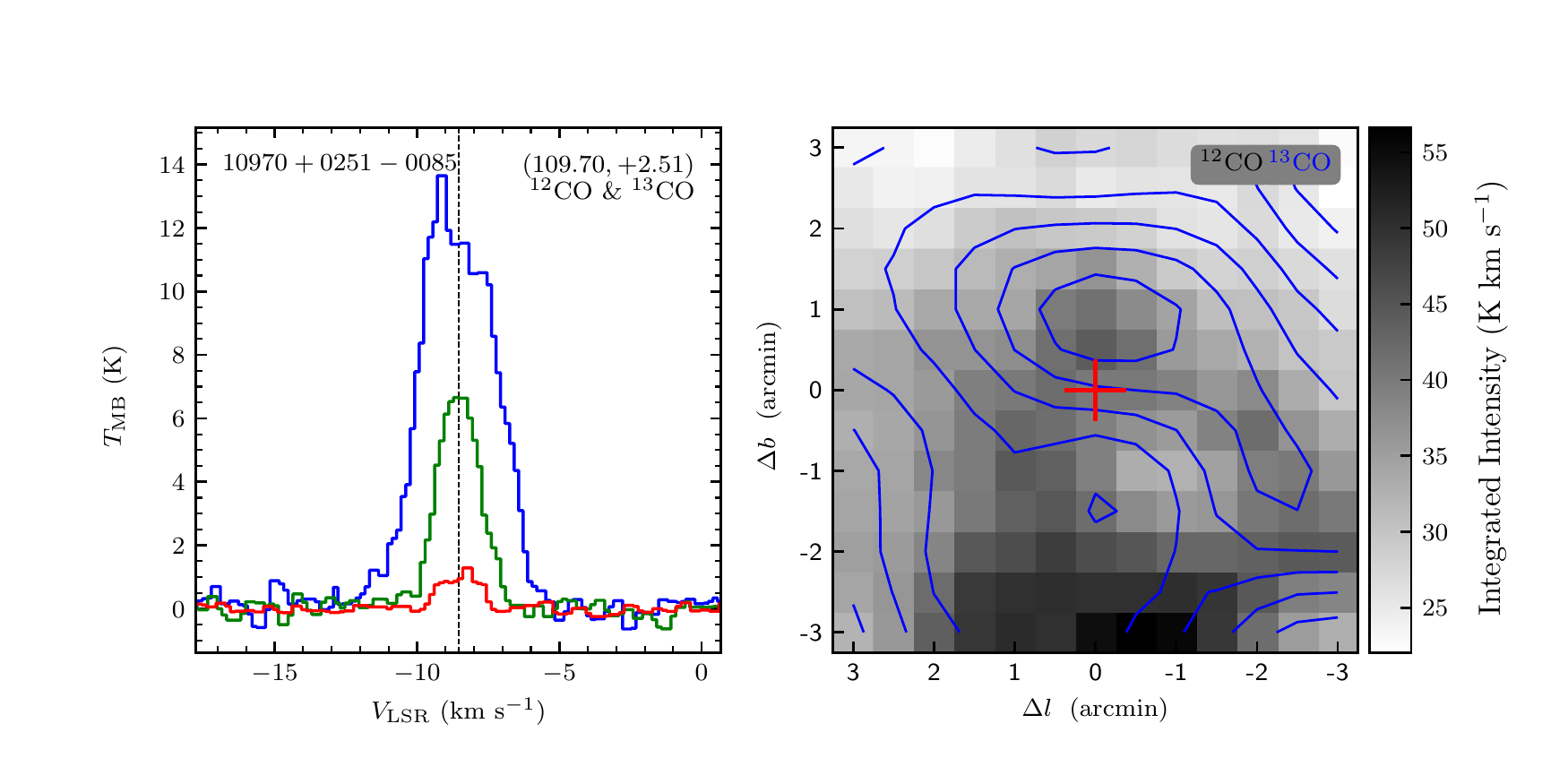}
\vspace{-0.5cm}

\includegraphics[width=9.0cm,angle=0]{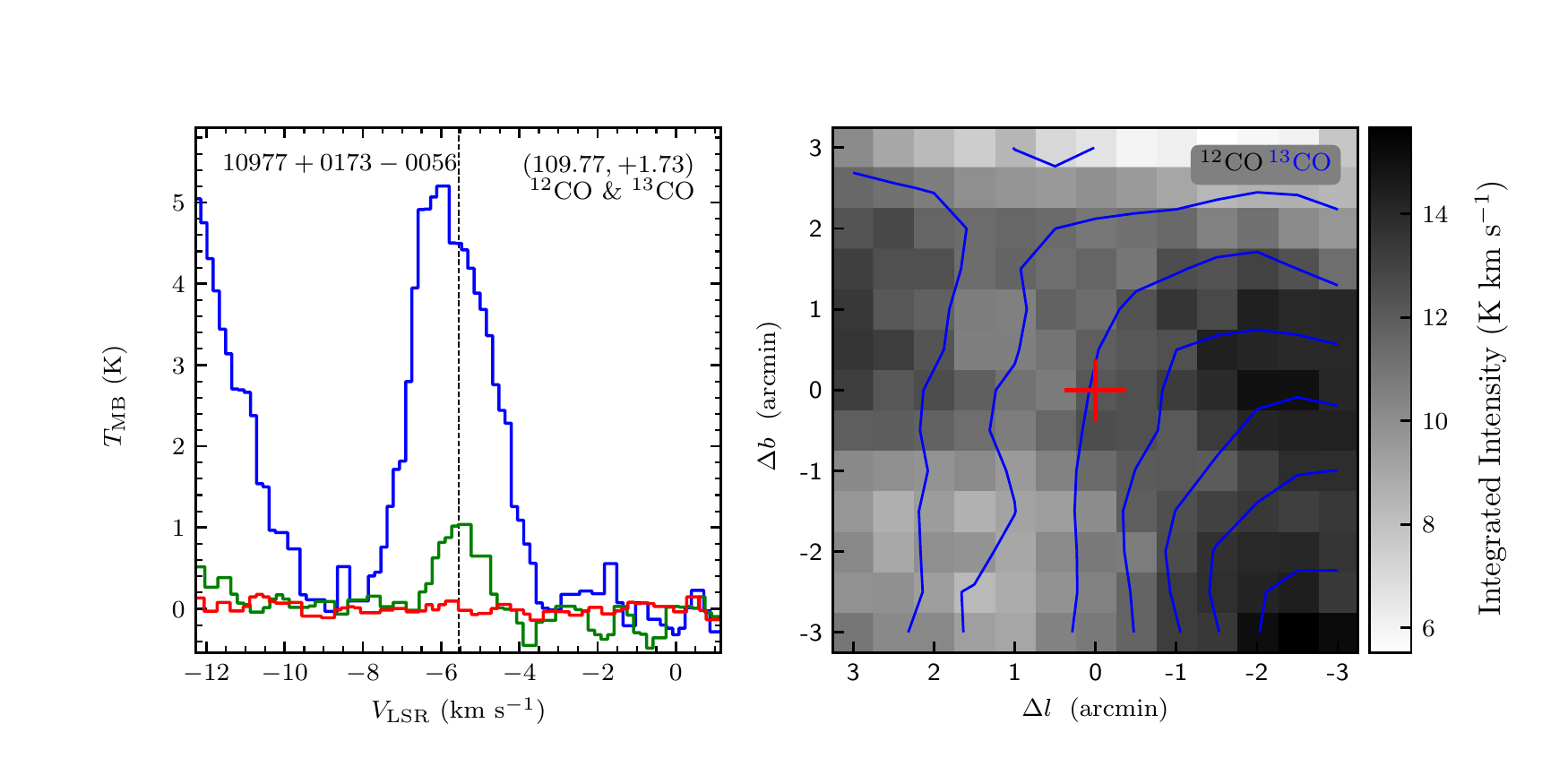}
\includegraphics[width=9.0cm,angle=0]{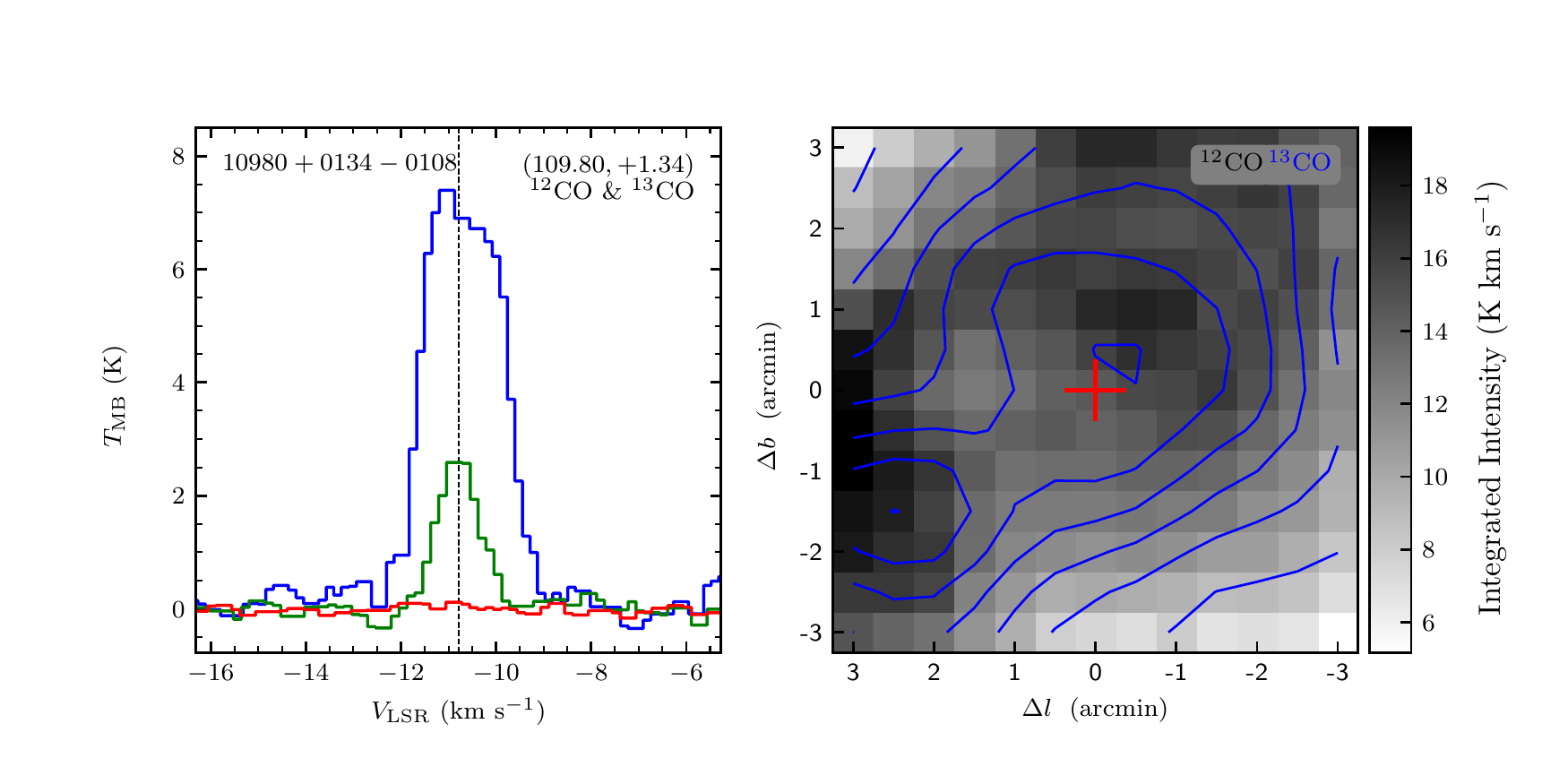}
\vspace{-0.5cm}

\includegraphics[width=9.0cm,angle=0]{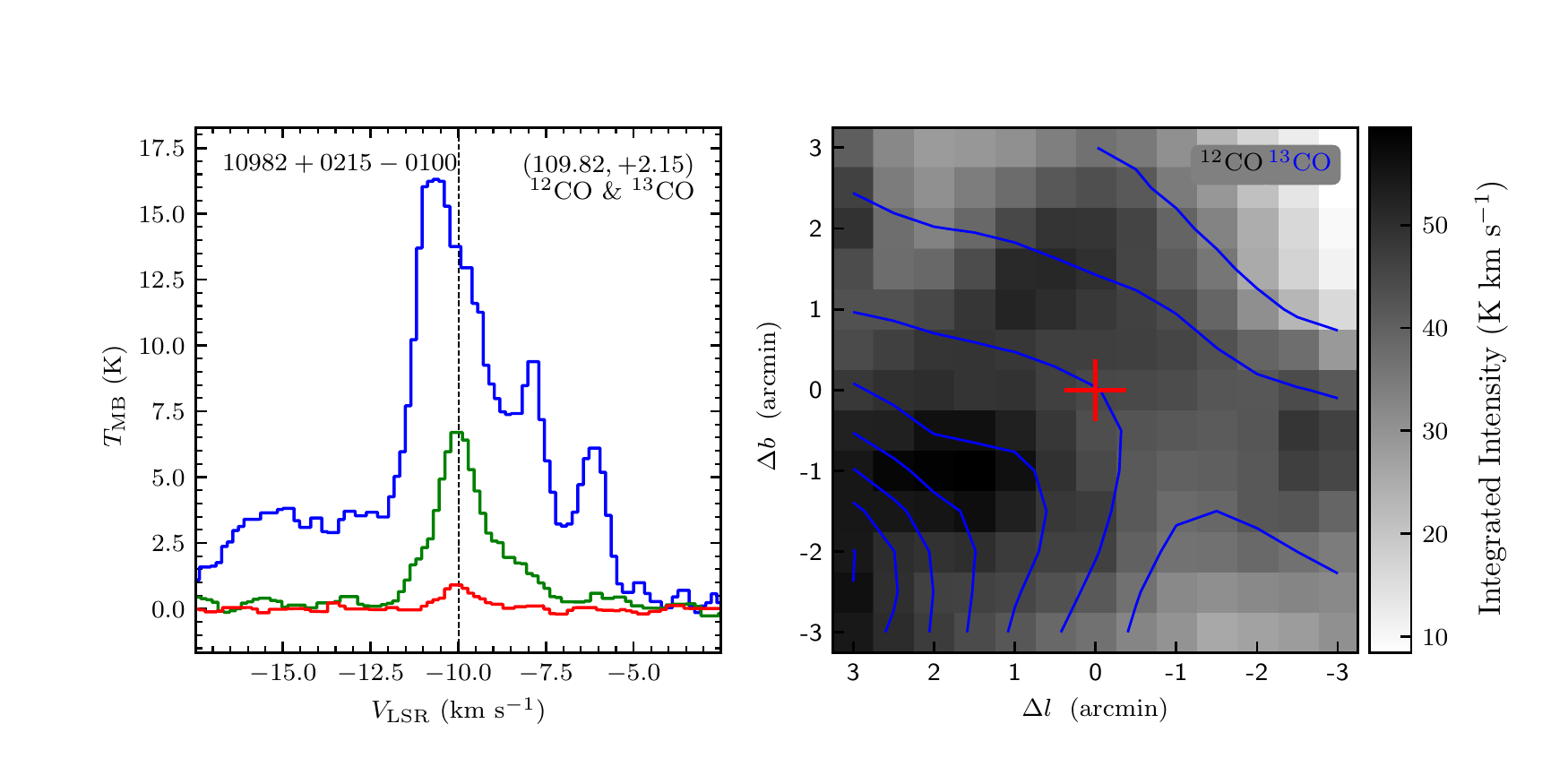}
\includegraphics[width=9.0cm,angle=0]{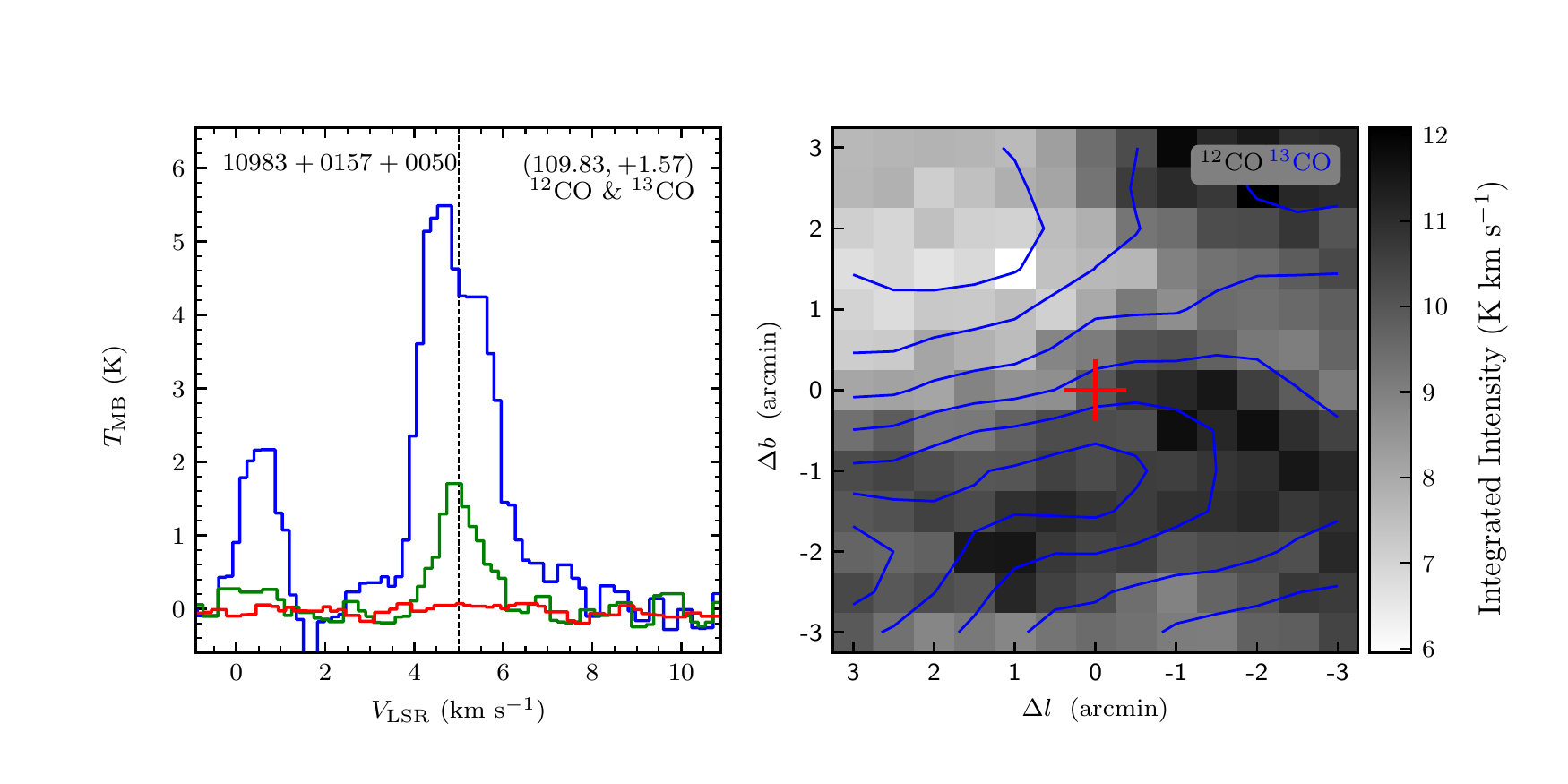}
\end{figure}
\clearpage

\begin{figure}
\includegraphics[width=9.0cm,angle=0]{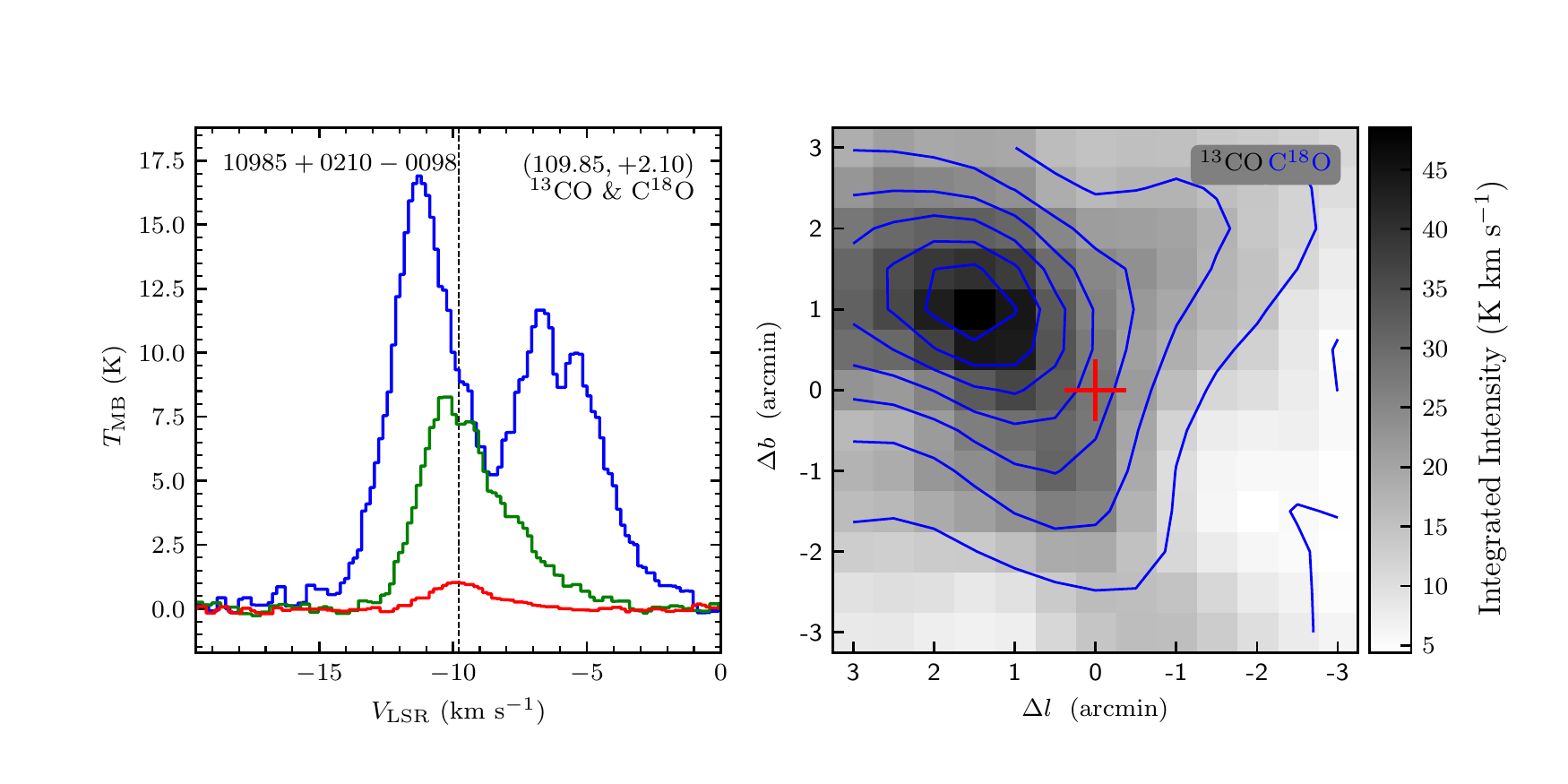}
\includegraphics[width=9.0cm,angle=0]{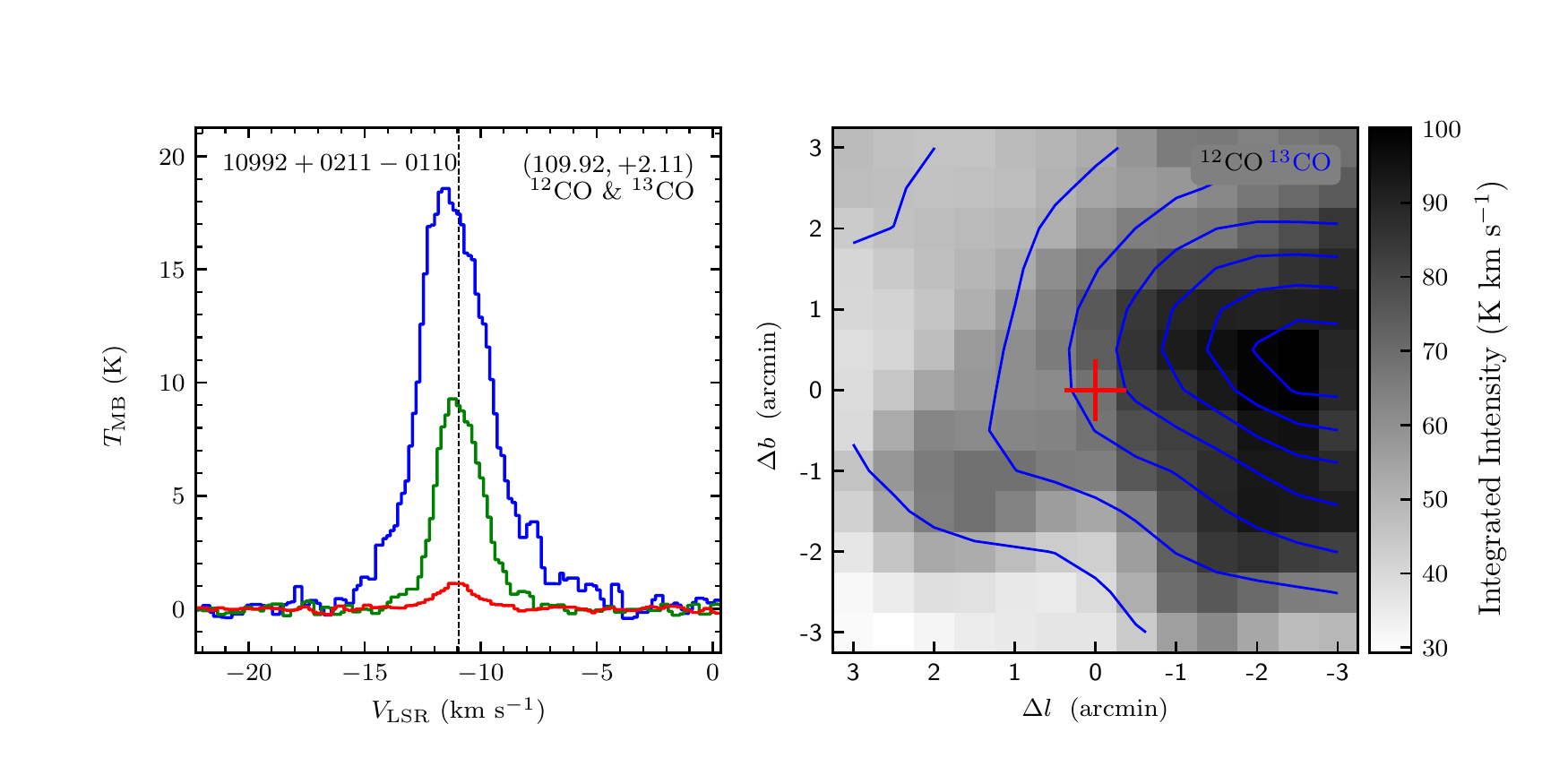}
\vspace{-0.5cm}

\includegraphics[width=9.0cm,angle=0]{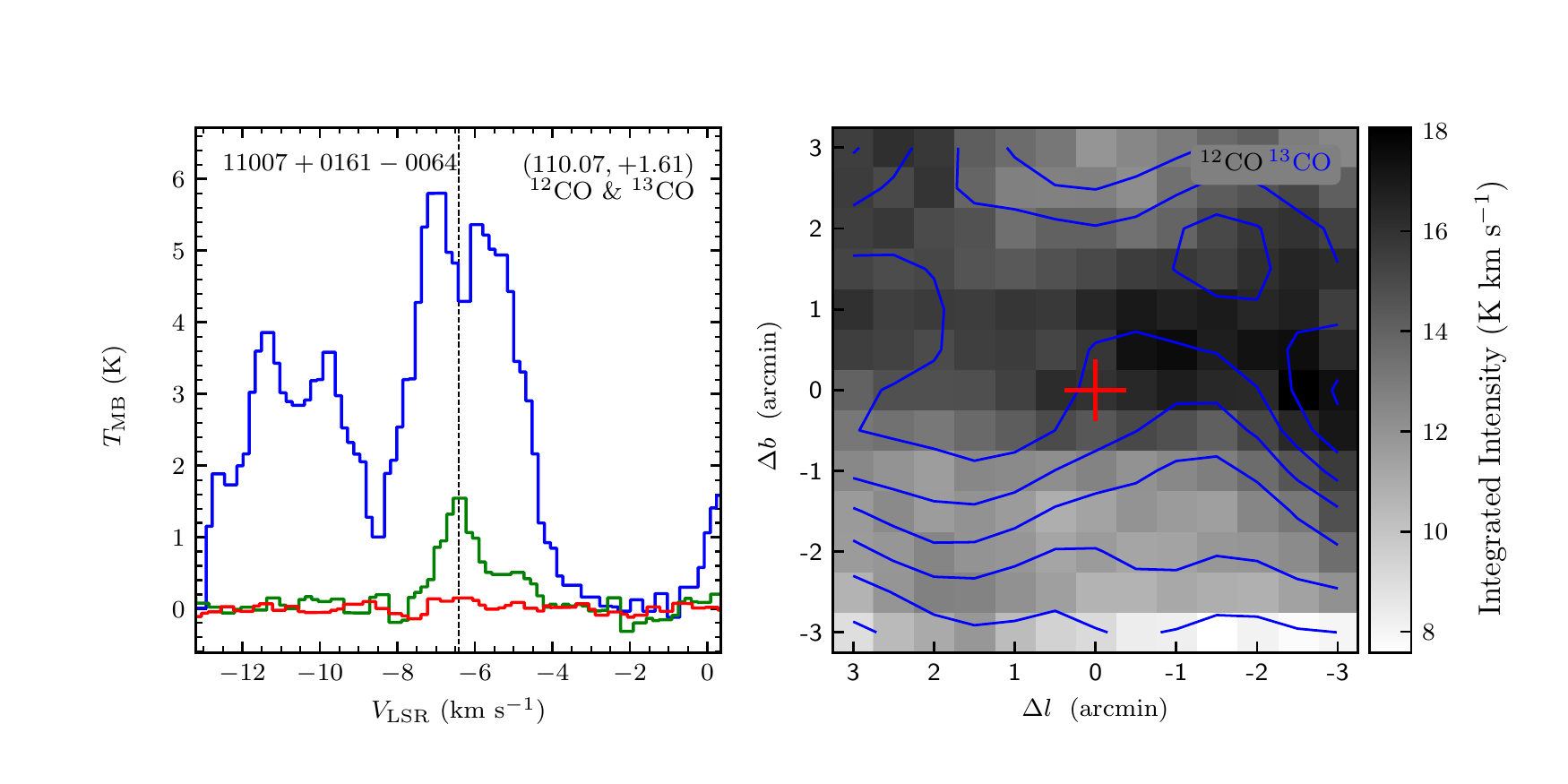}
\includegraphics[width=9.0cm,angle=0]{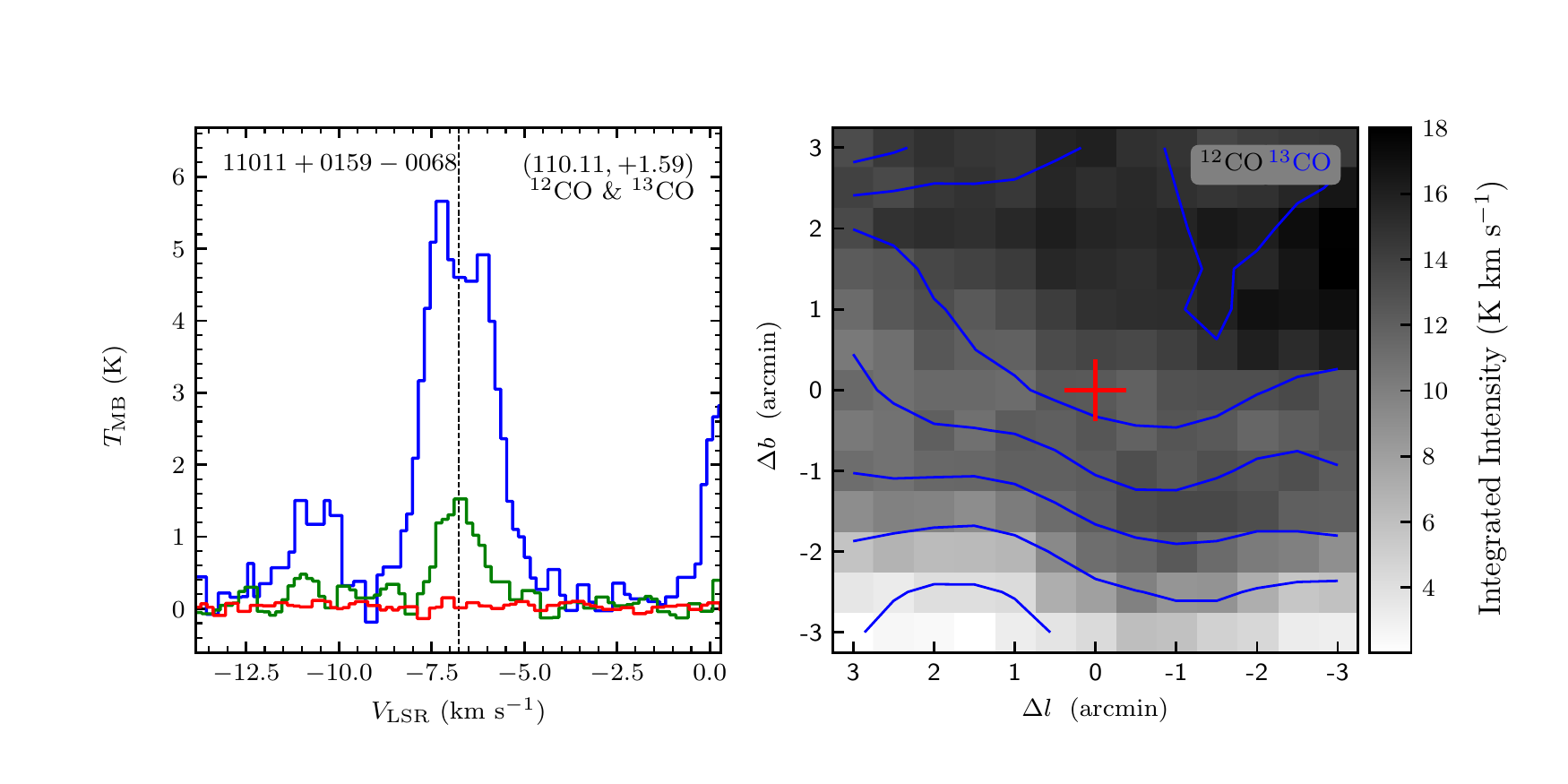}
\vspace{-0.5cm}

\includegraphics[width=9.0cm,angle=0]{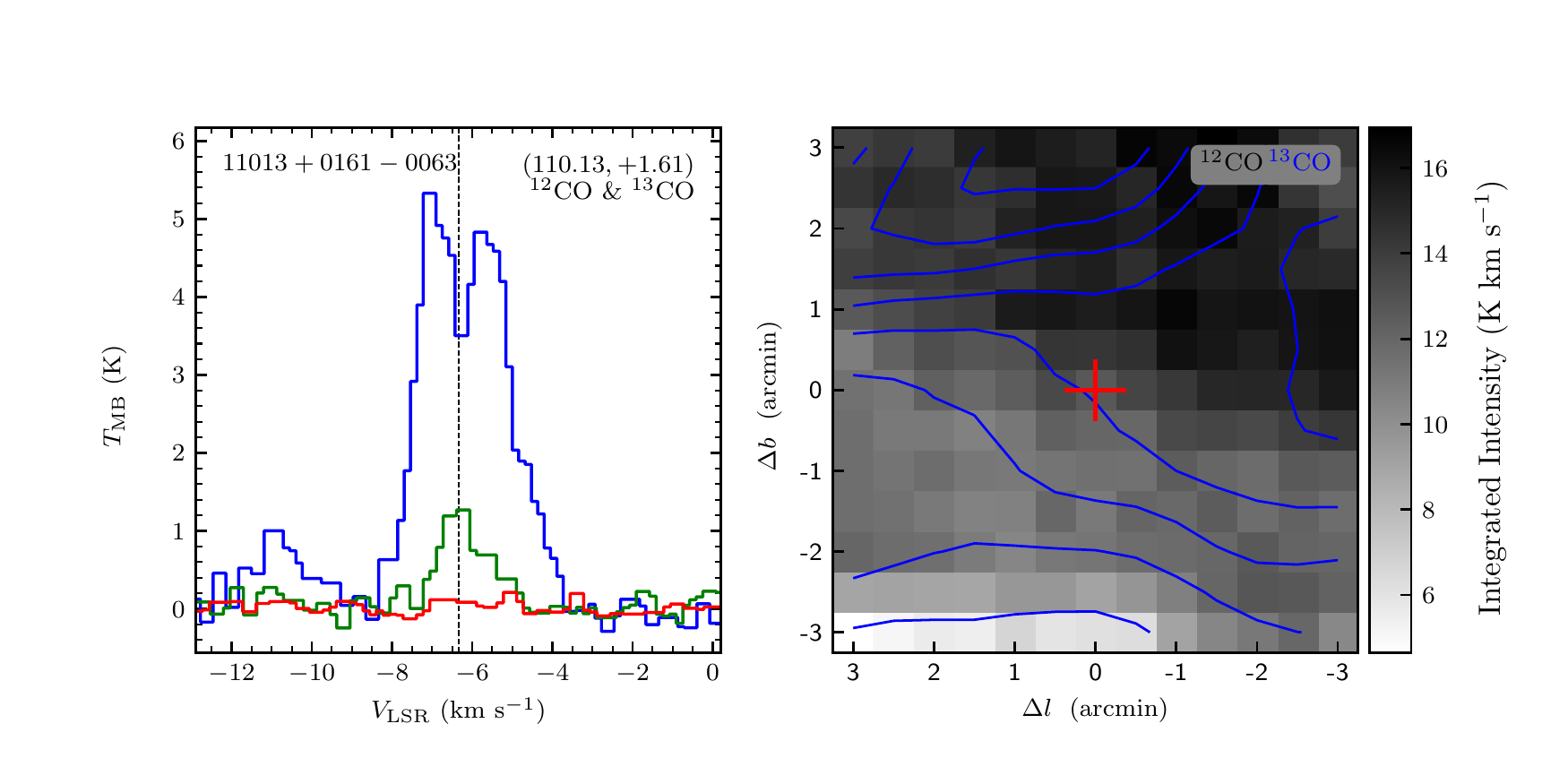}
\includegraphics[width=9.0cm,angle=0]{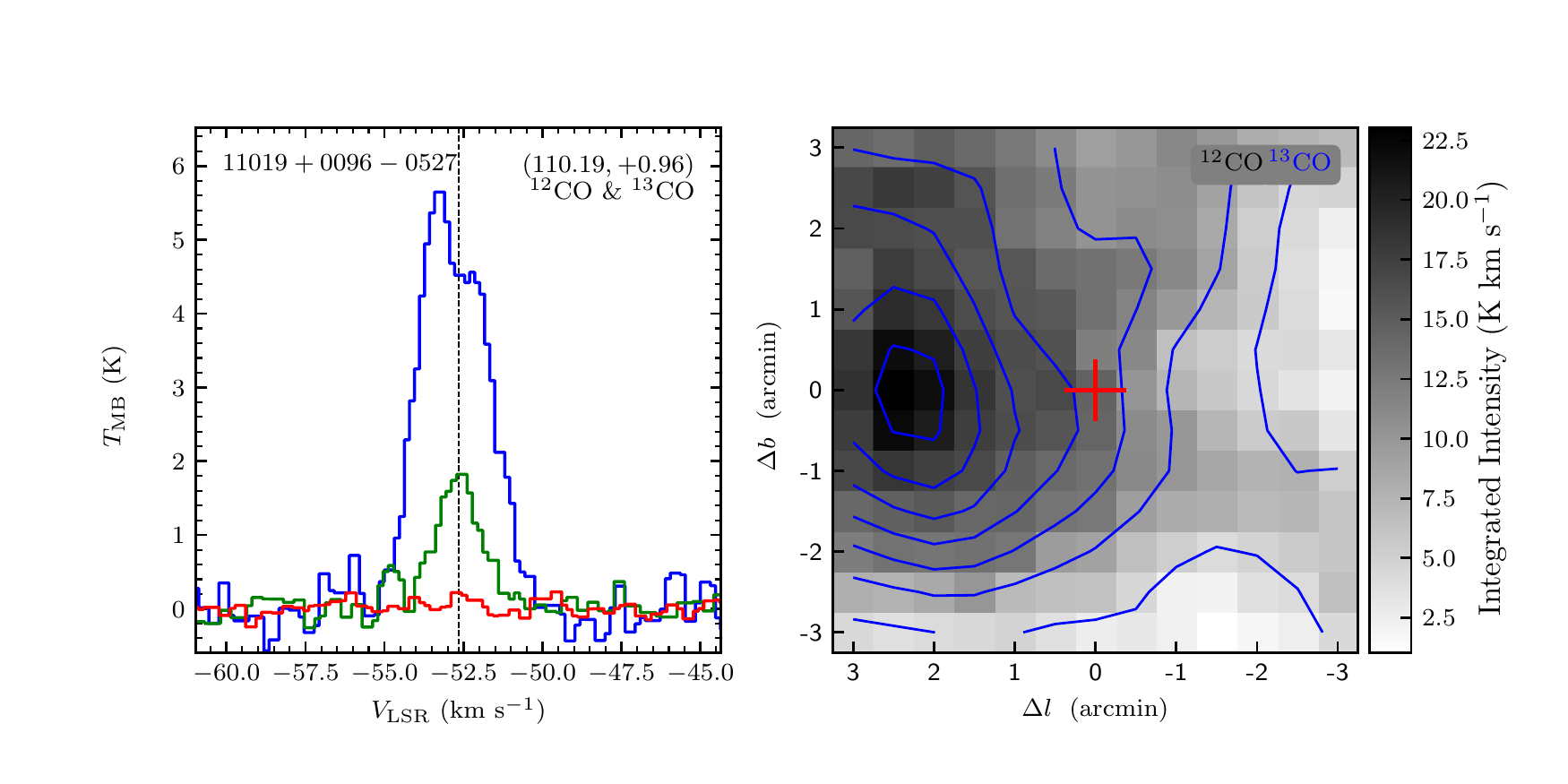}
\vspace{-0.5cm}

\includegraphics[width=9.0cm,angle=0]{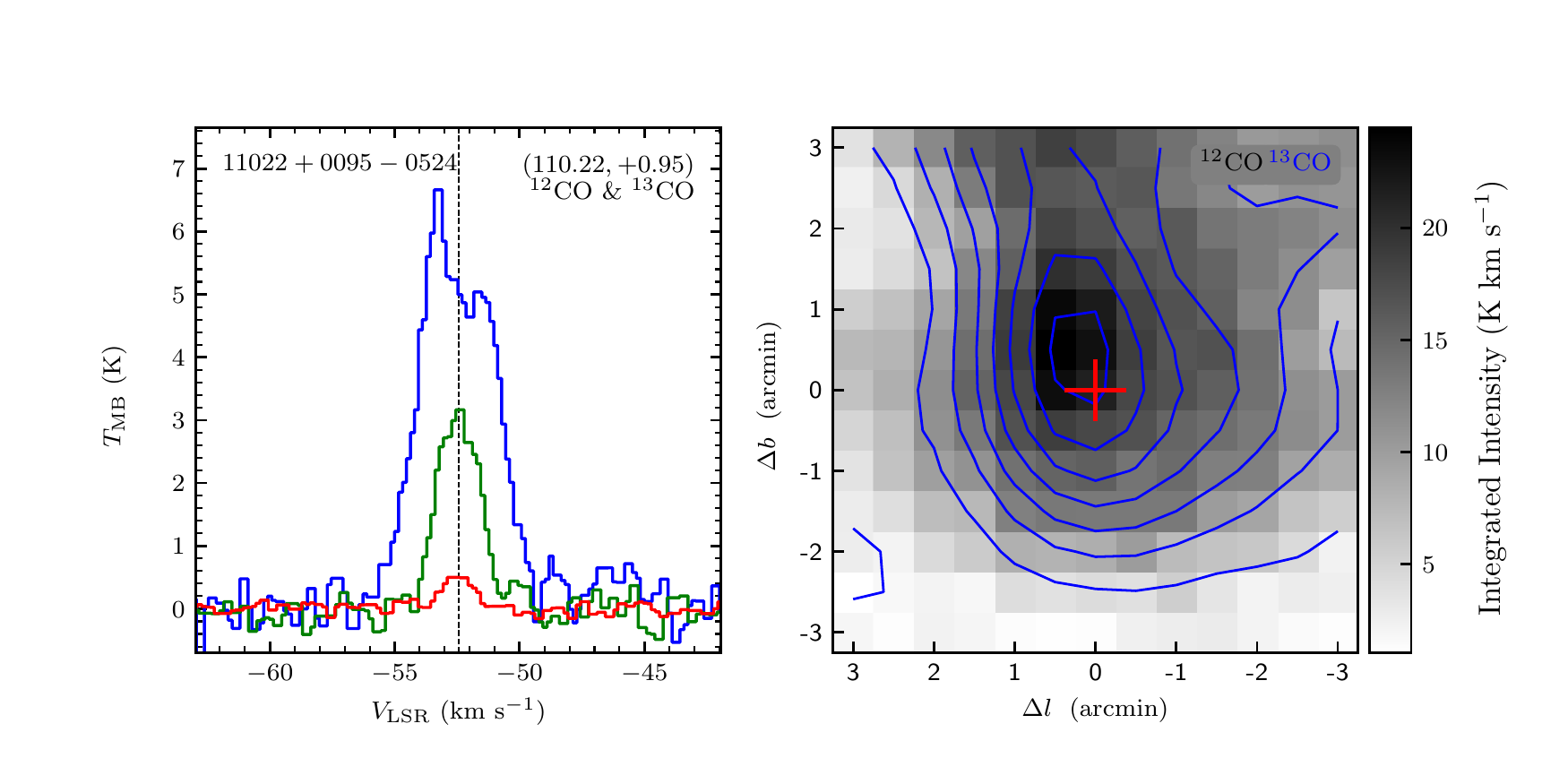}
\includegraphics[width=9.0cm,angle=0]{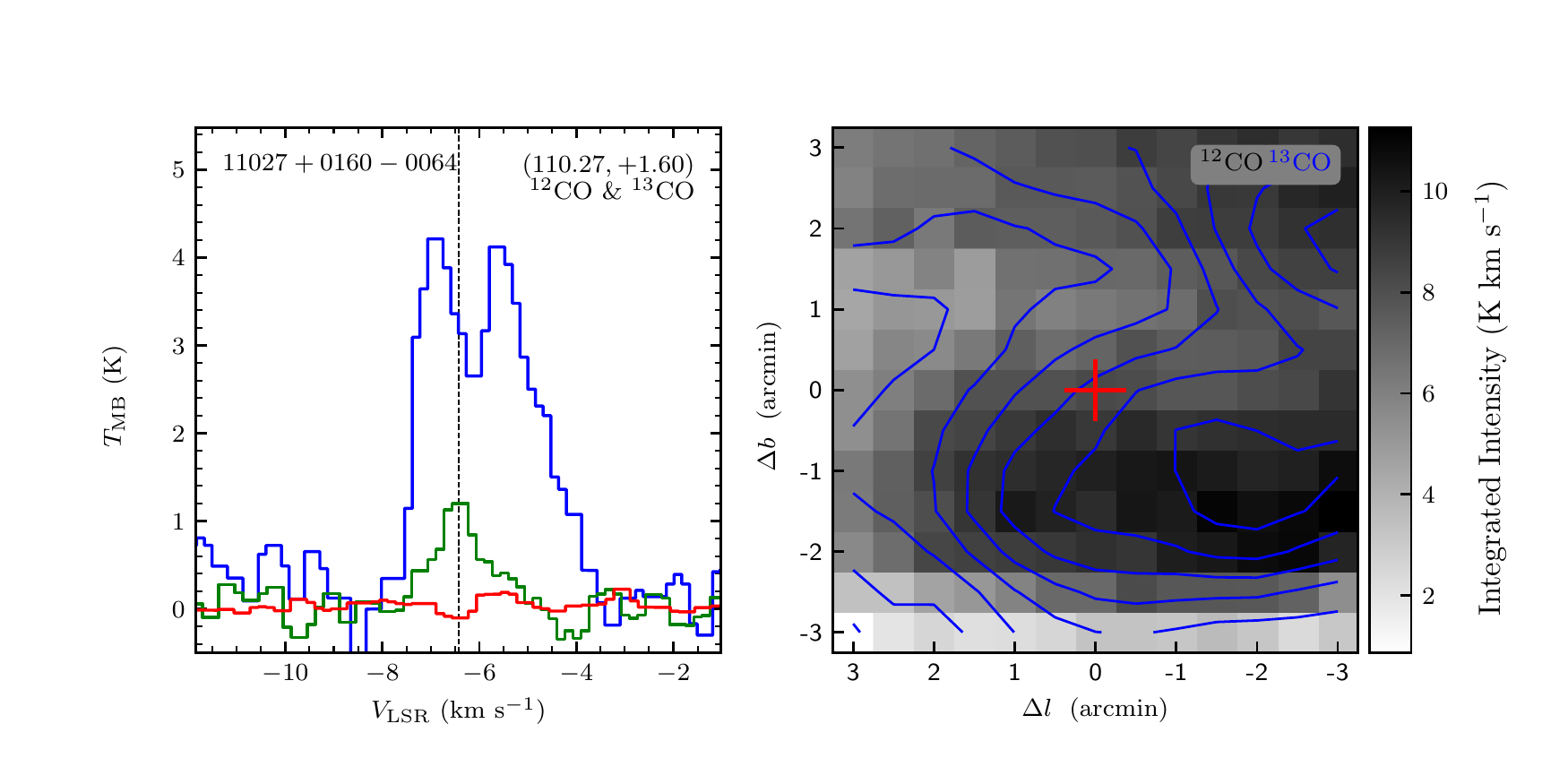}
\vspace{-0.5cm}

\includegraphics[width=9.0cm,angle=0]{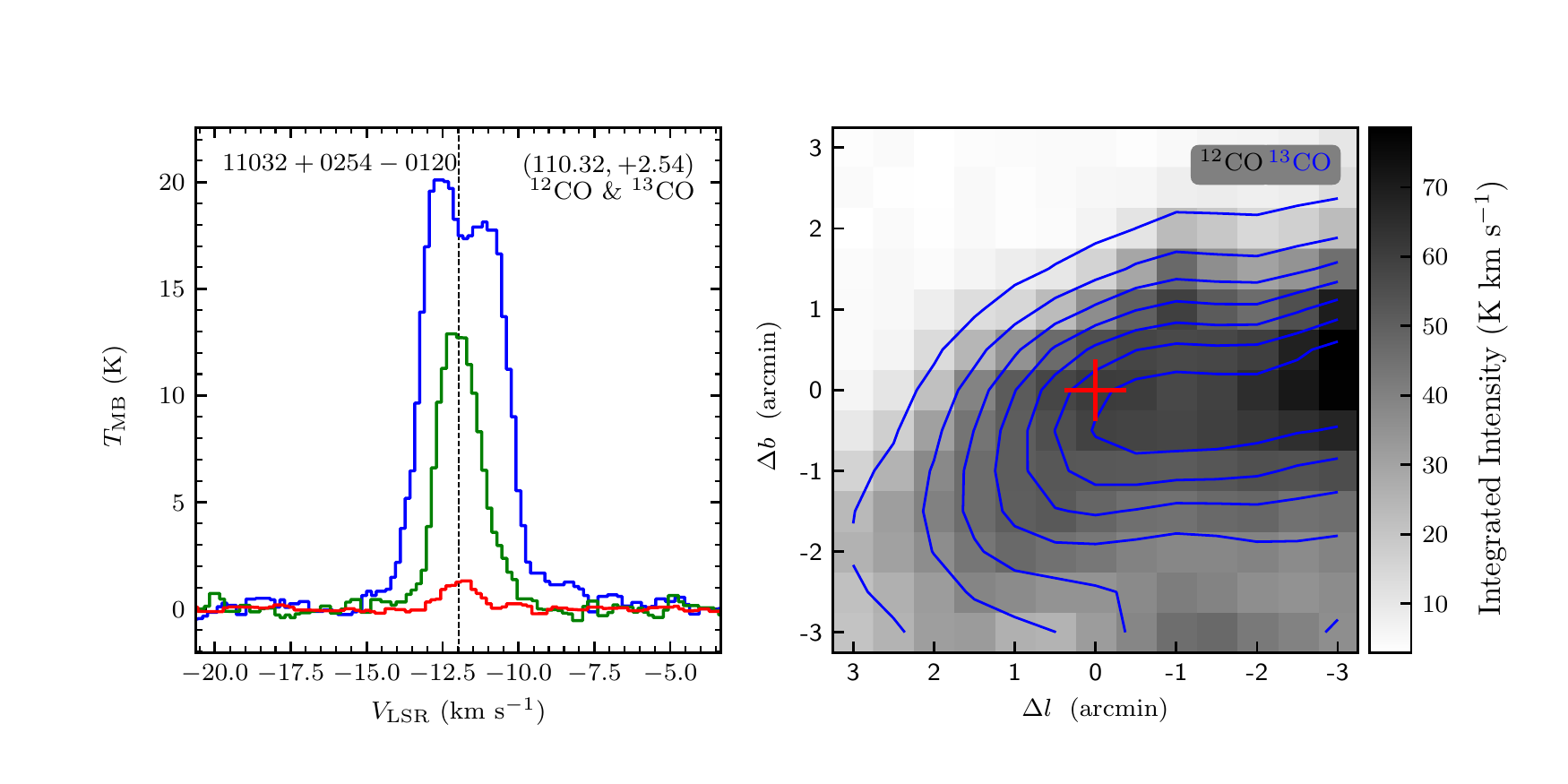}
\includegraphics[width=9.0cm,angle=0]{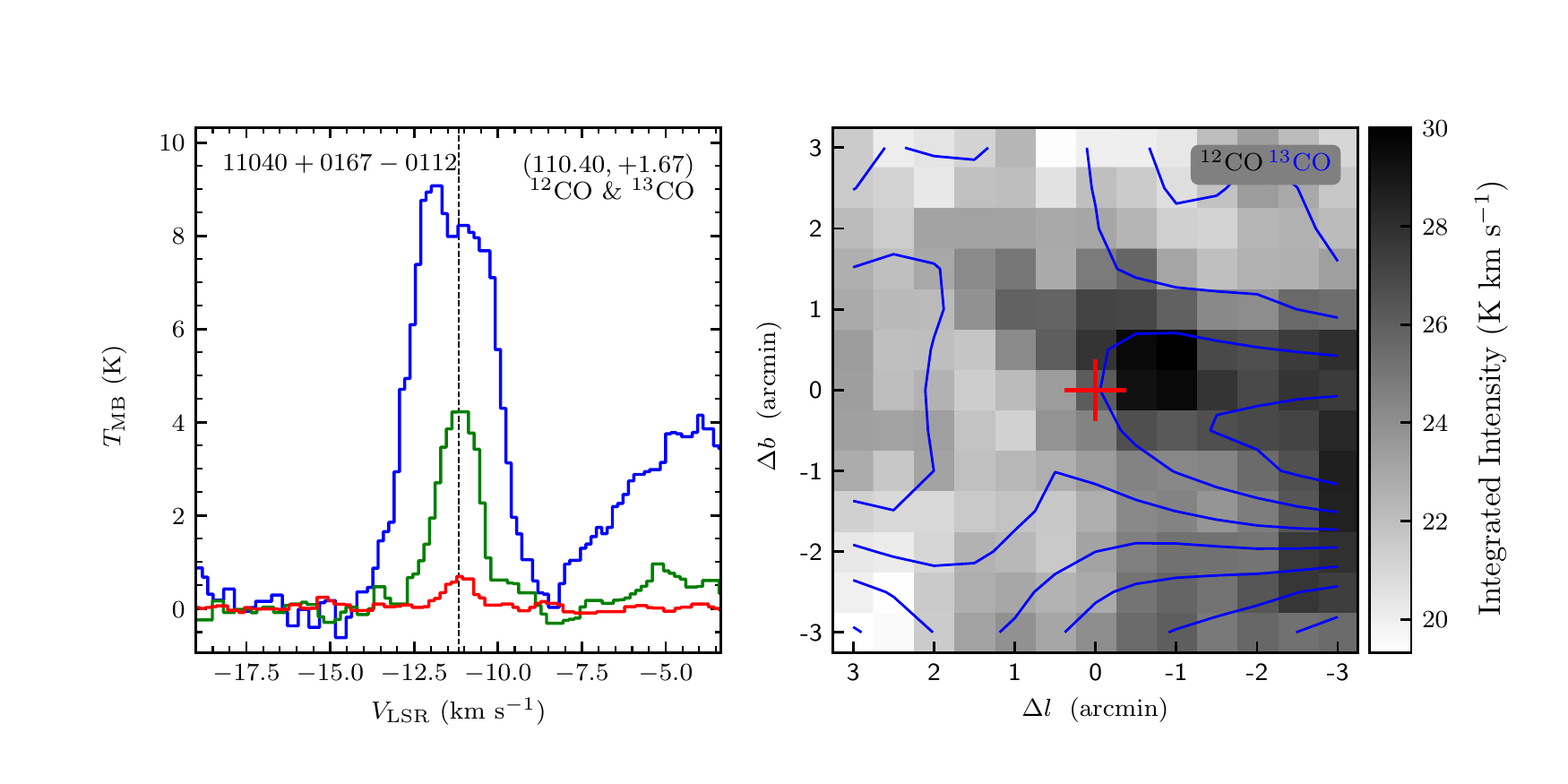}
\end{figure}
\clearpage

\begin{figure}
\includegraphics[width=9.0cm,angle=0]{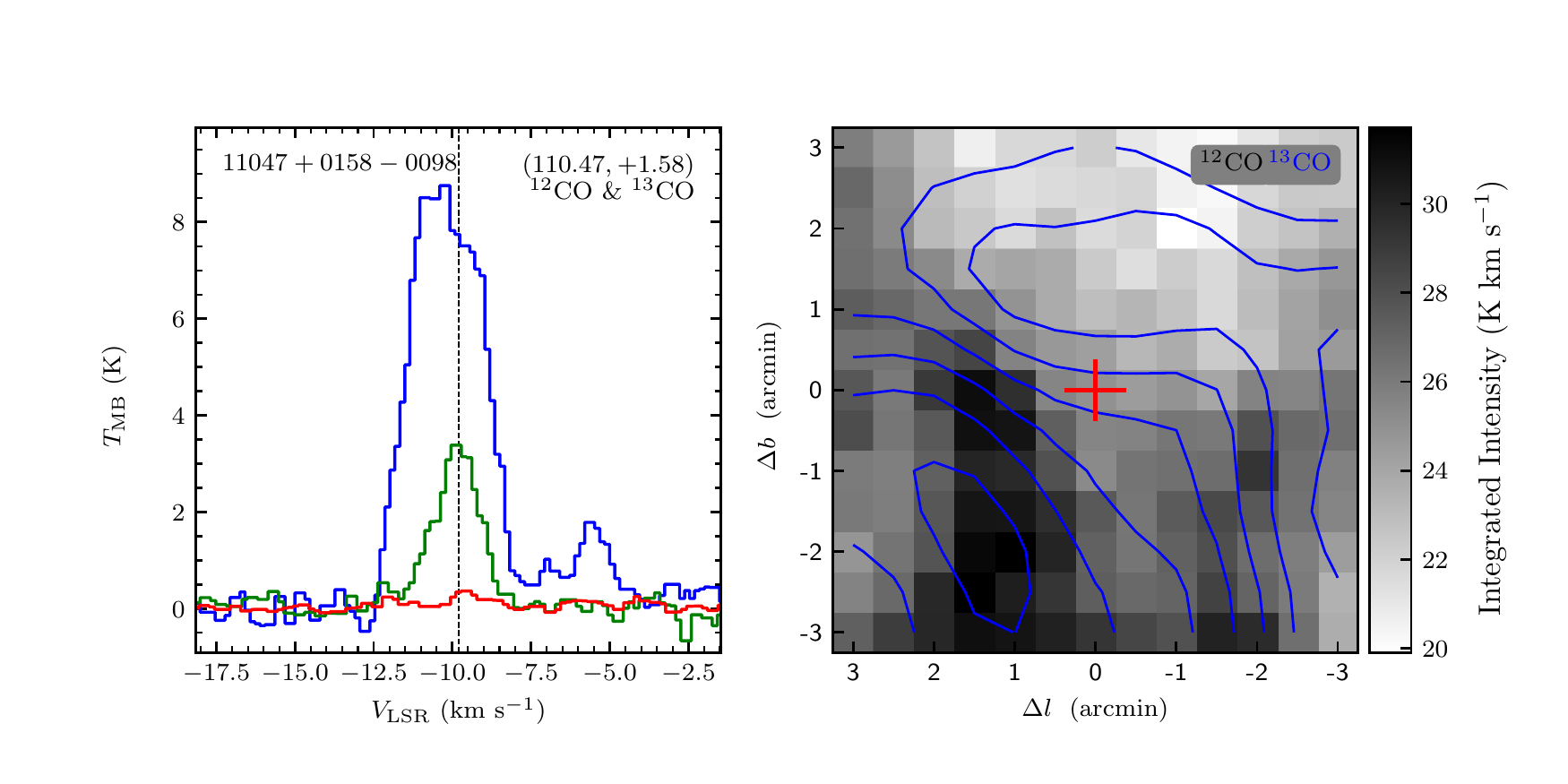}
\includegraphics[width=9.0cm,angle=0]{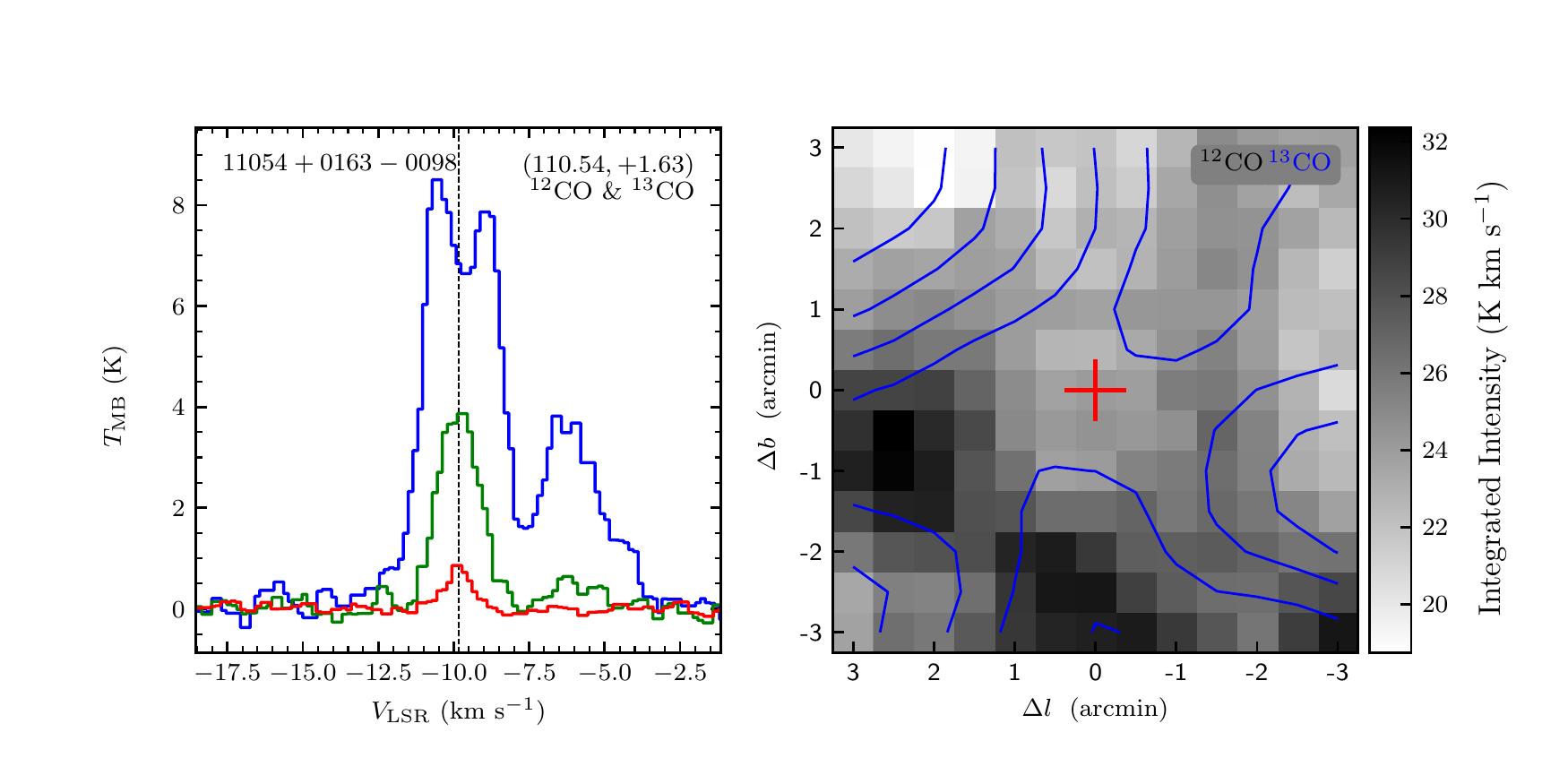}
\vspace{-0.5cm}

\includegraphics[width=9.0cm,angle=0]{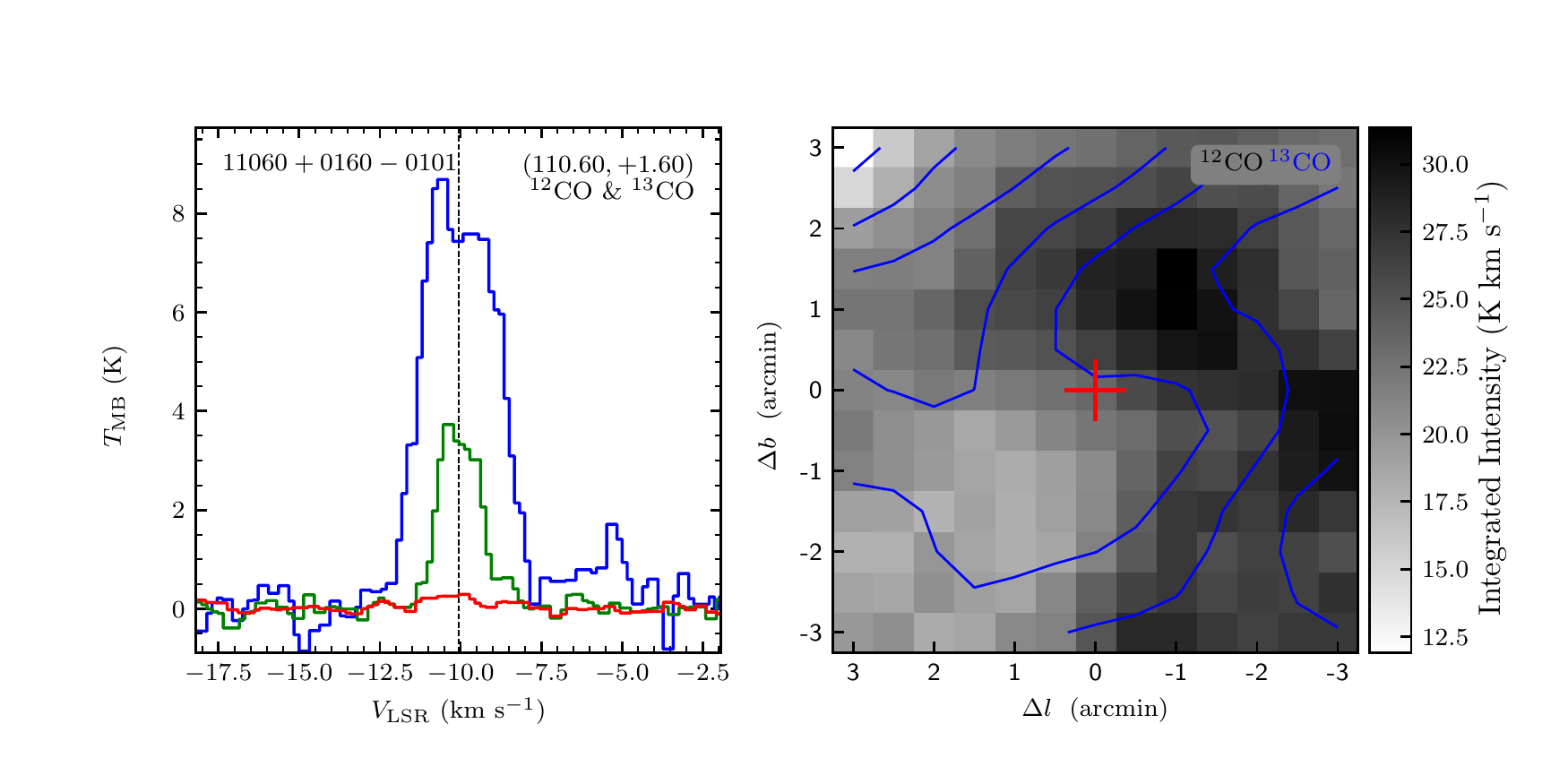}
\includegraphics[width=9.0cm,angle=0]{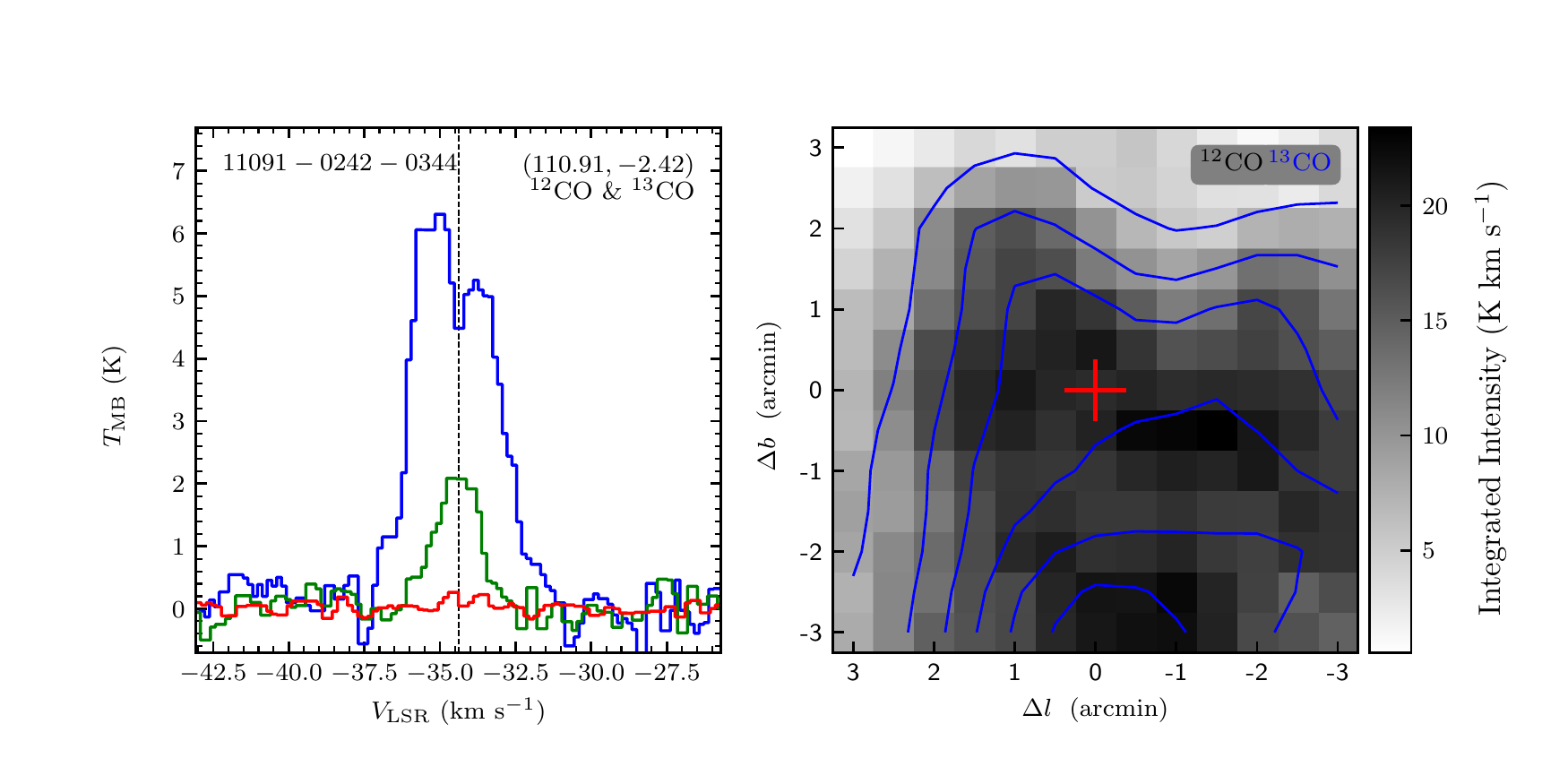}
\vspace{-0.5cm}

\includegraphics[width=9.0cm,angle=0]{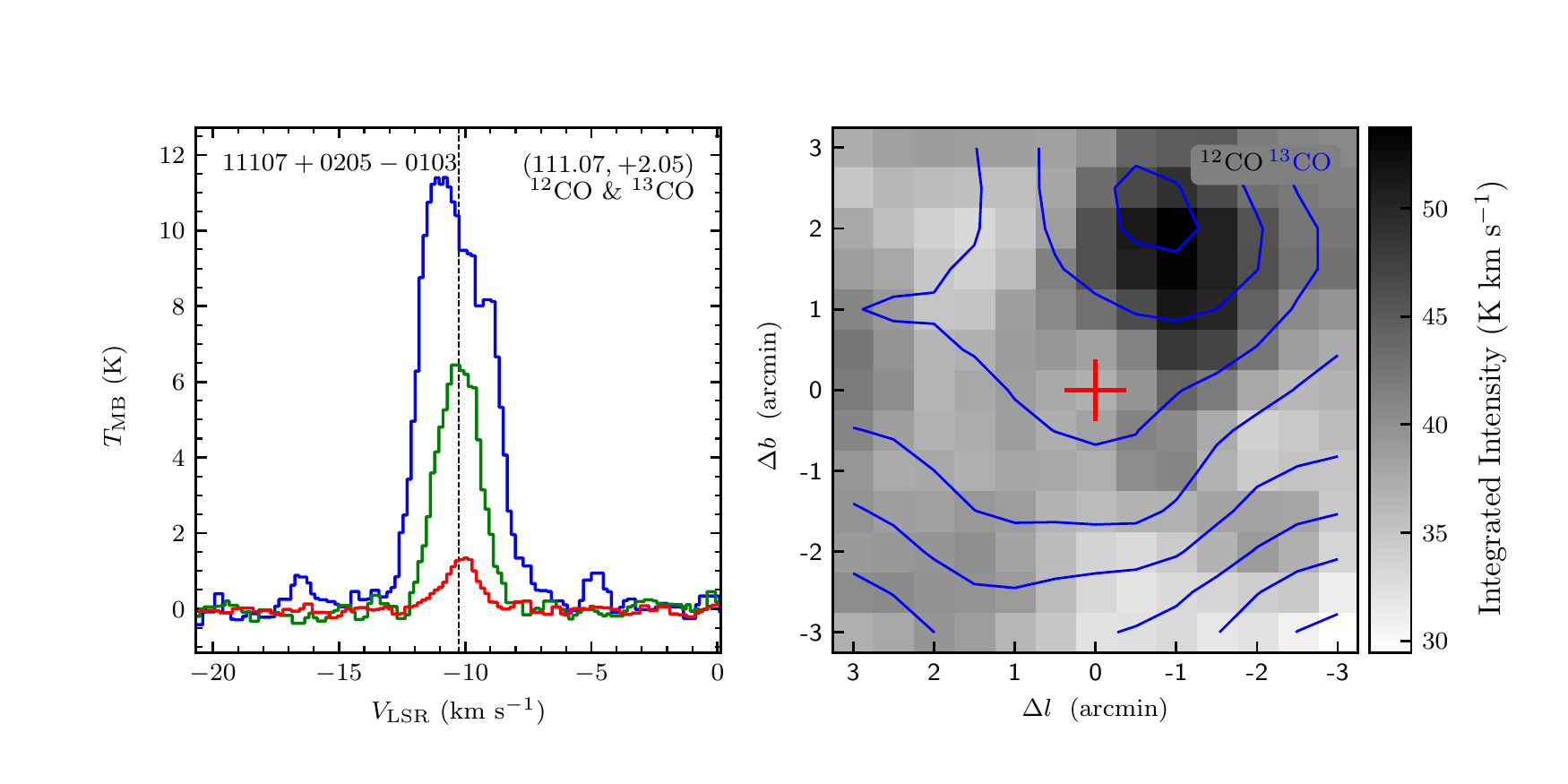}
\includegraphics[width=9.0cm,angle=0]{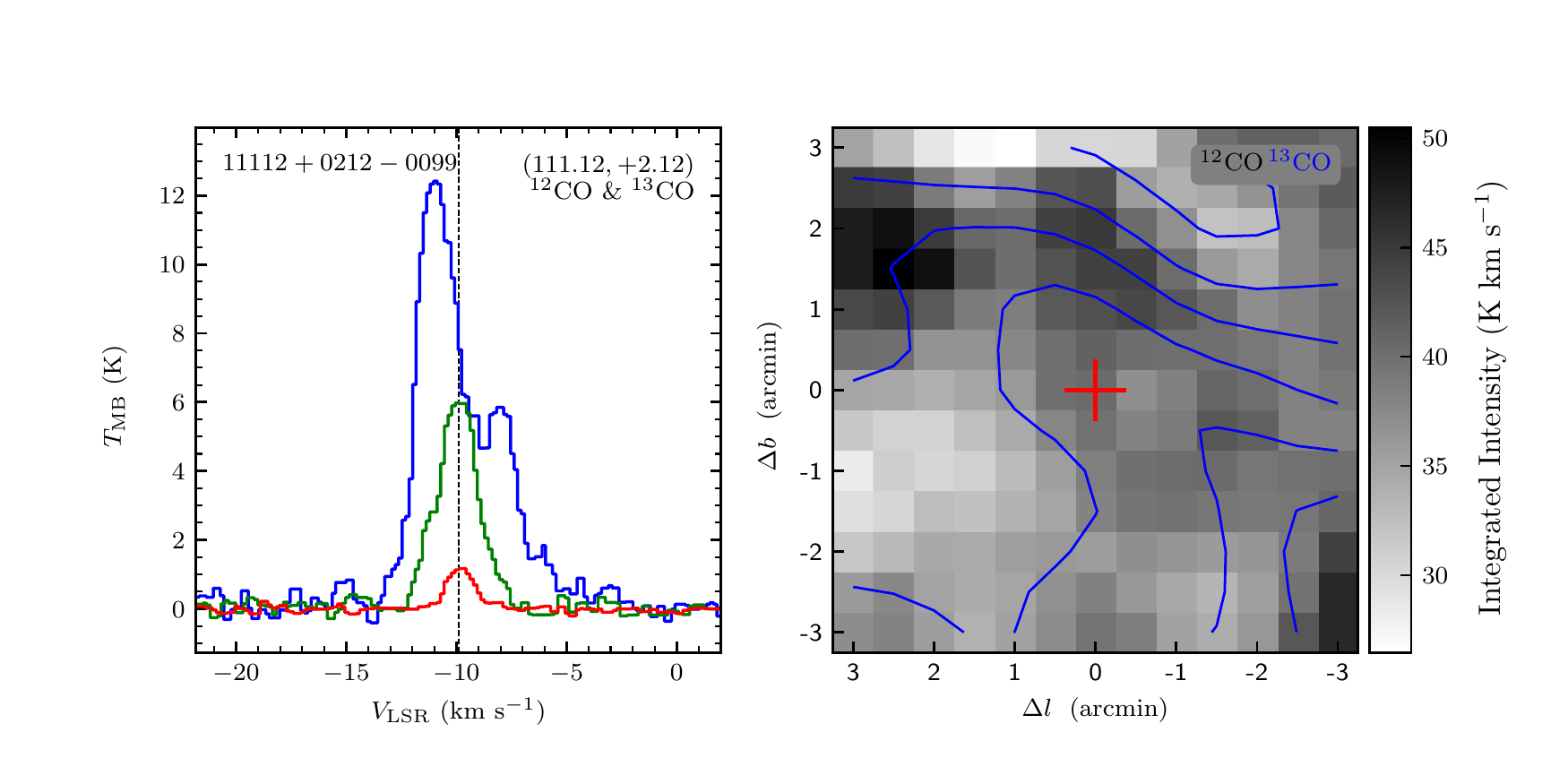}
\vspace{-0.5cm}

\includegraphics[width=9.0cm,angle=0]{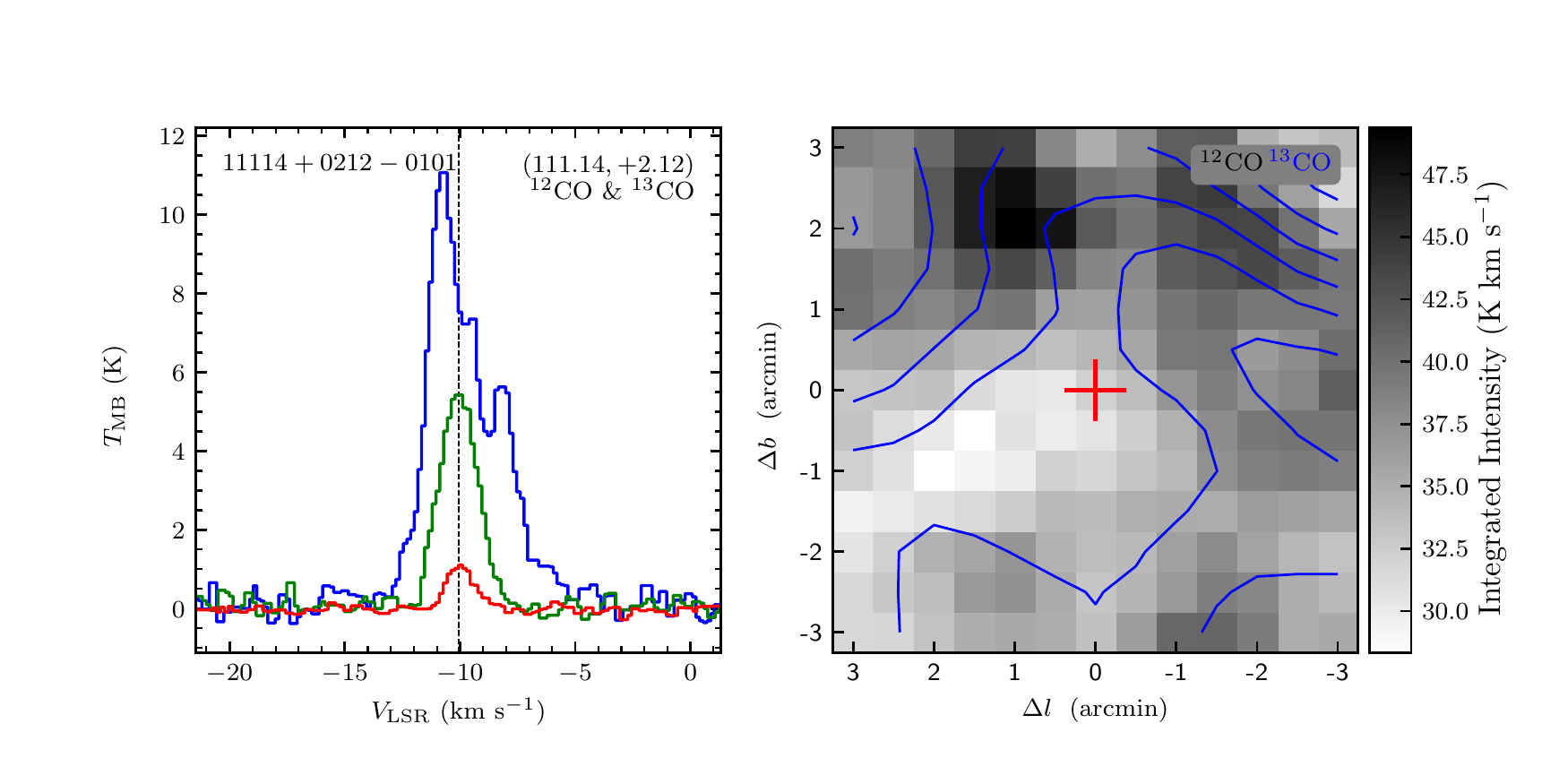}
\includegraphics[width=9.0cm,angle=0]{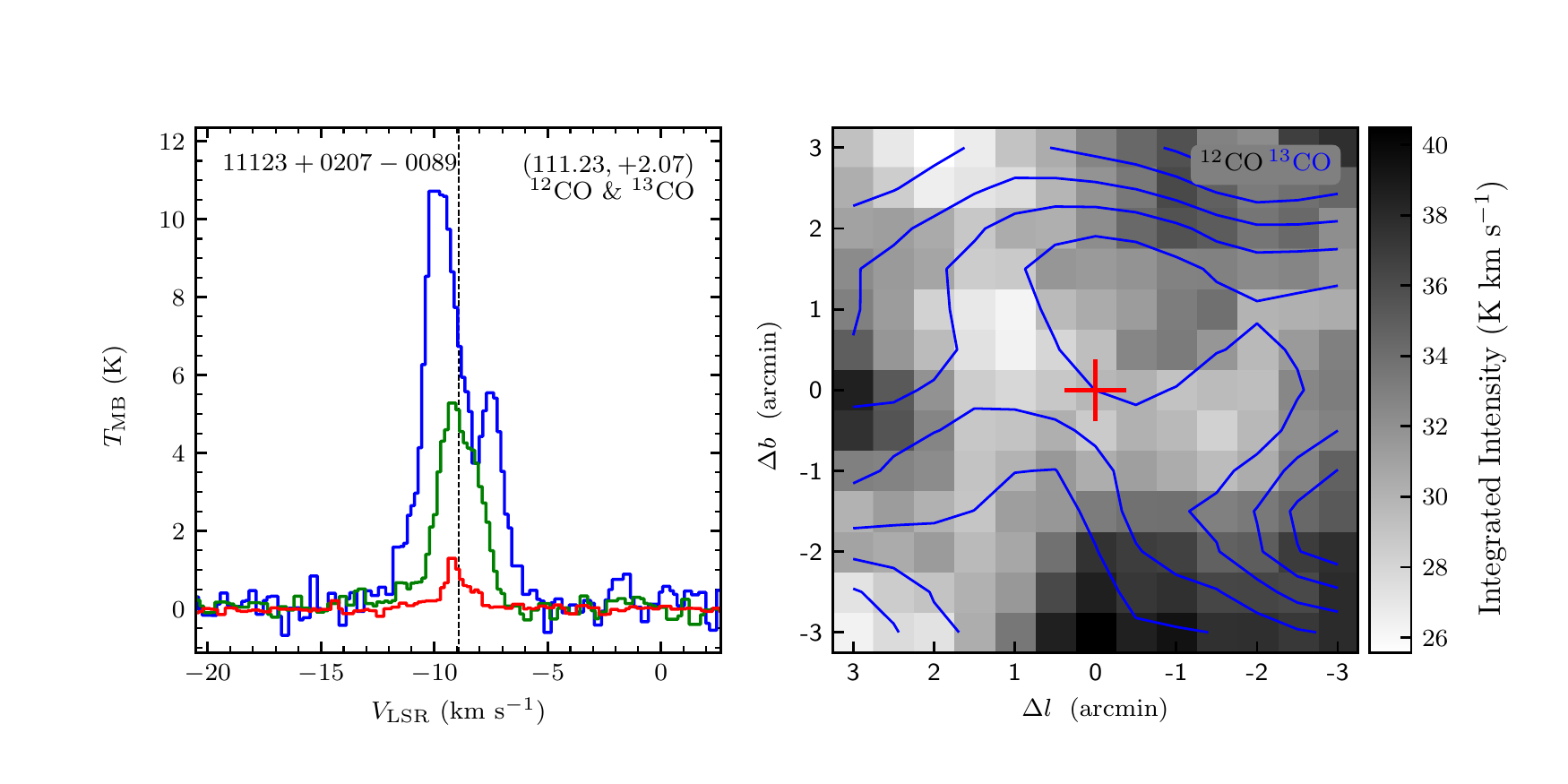}
\vspace{-0.5cm}

\includegraphics[width=9.0cm,angle=0]{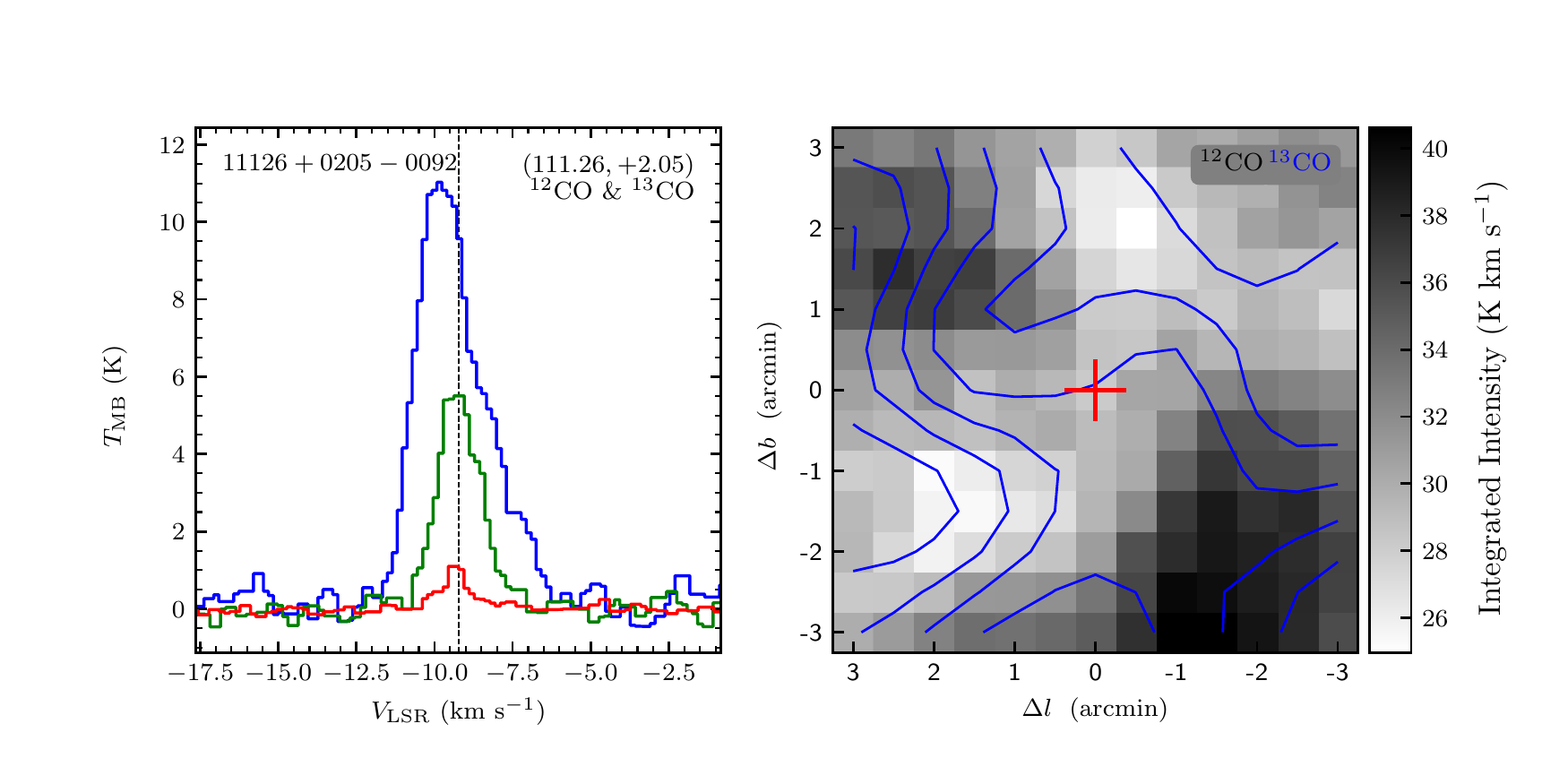}
\includegraphics[width=9.0cm,angle=0]{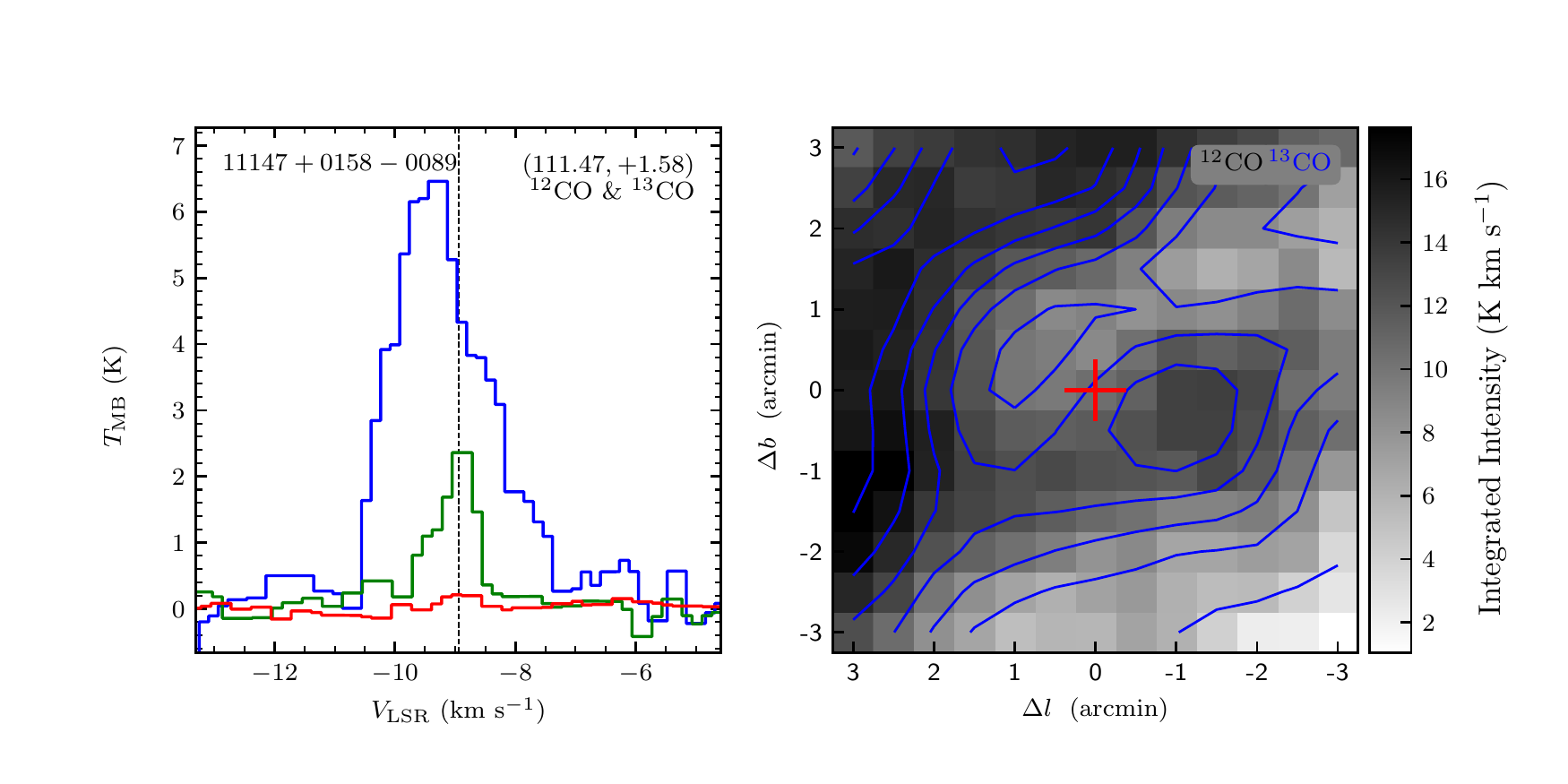}
\end{figure}
\clearpage

\begin{figure}
\includegraphics[width=9.0cm,angle=0]{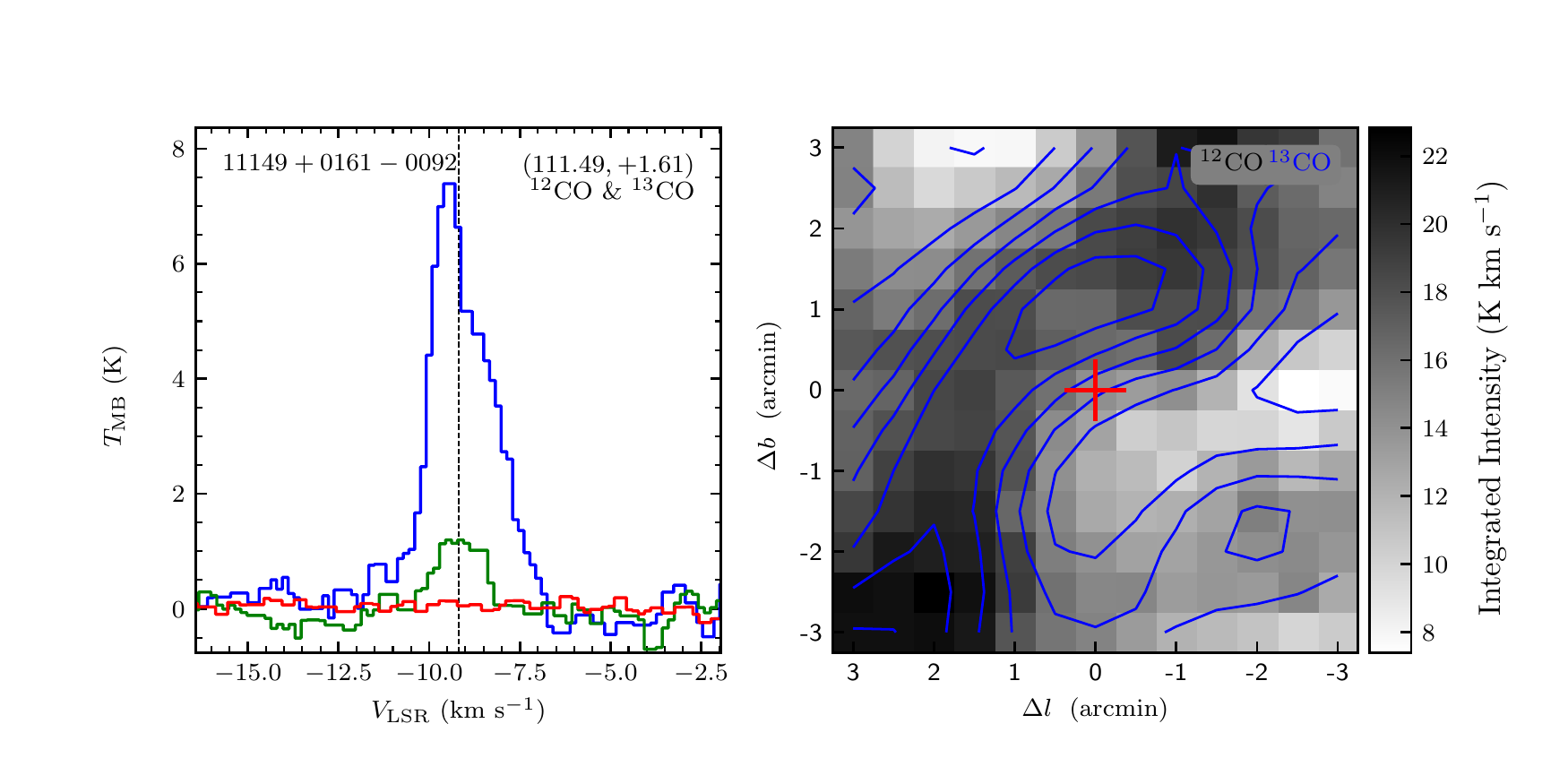}
\includegraphics[width=9.0cm,angle=0]{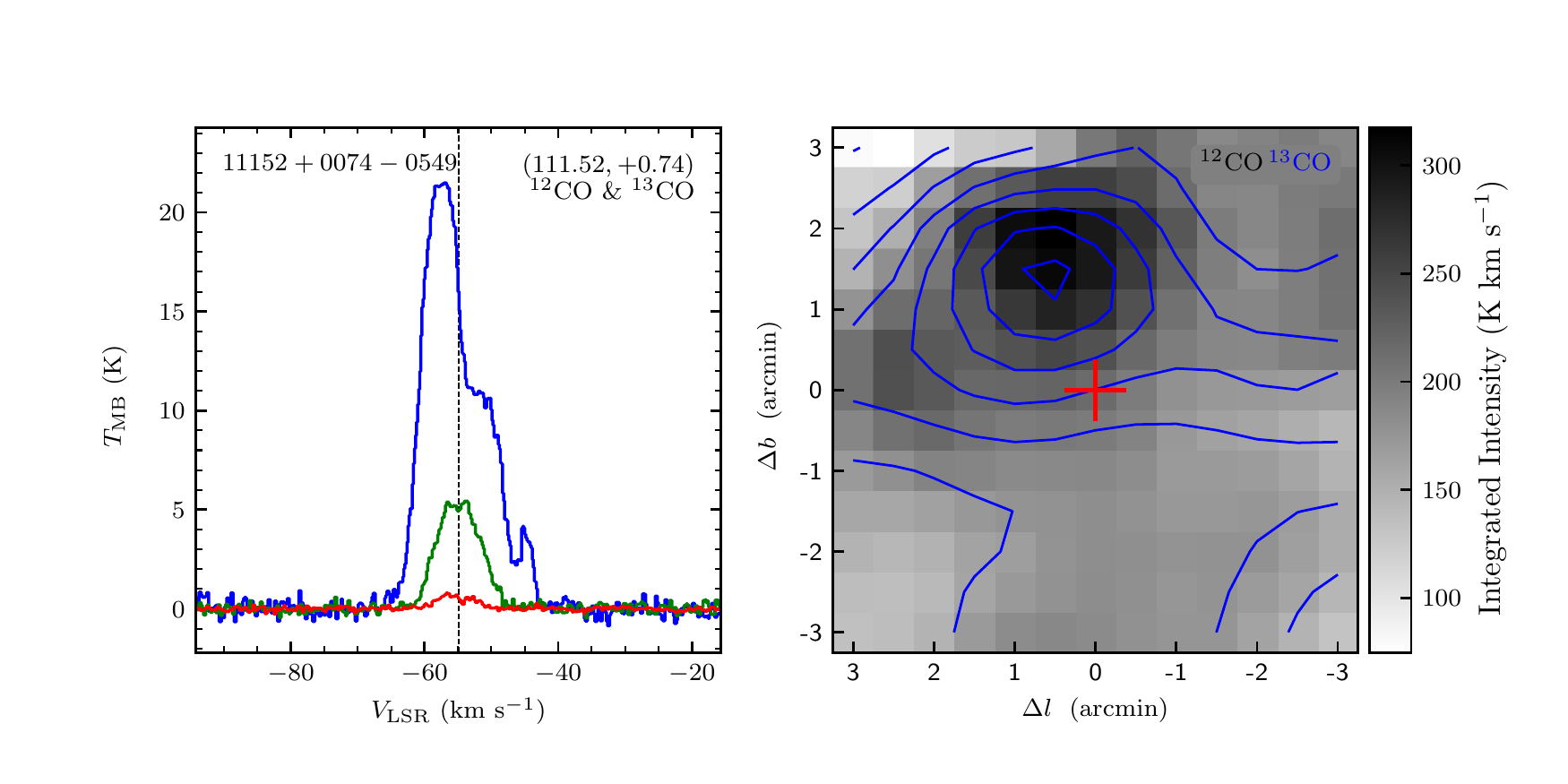}
\vspace{-0.5cm}

\includegraphics[width=9.0cm,angle=0]{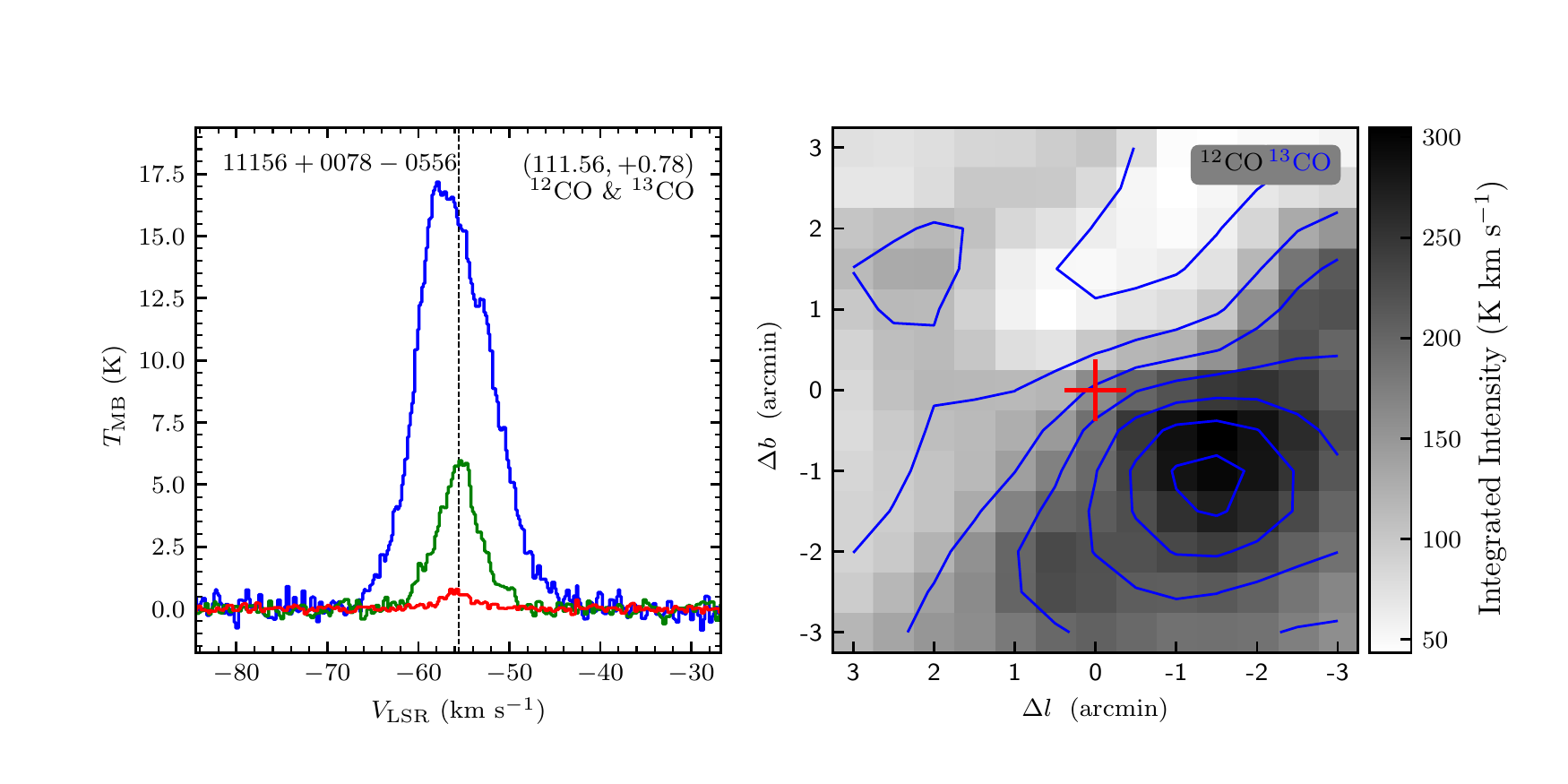}
\includegraphics[width=9.0cm,angle=0]{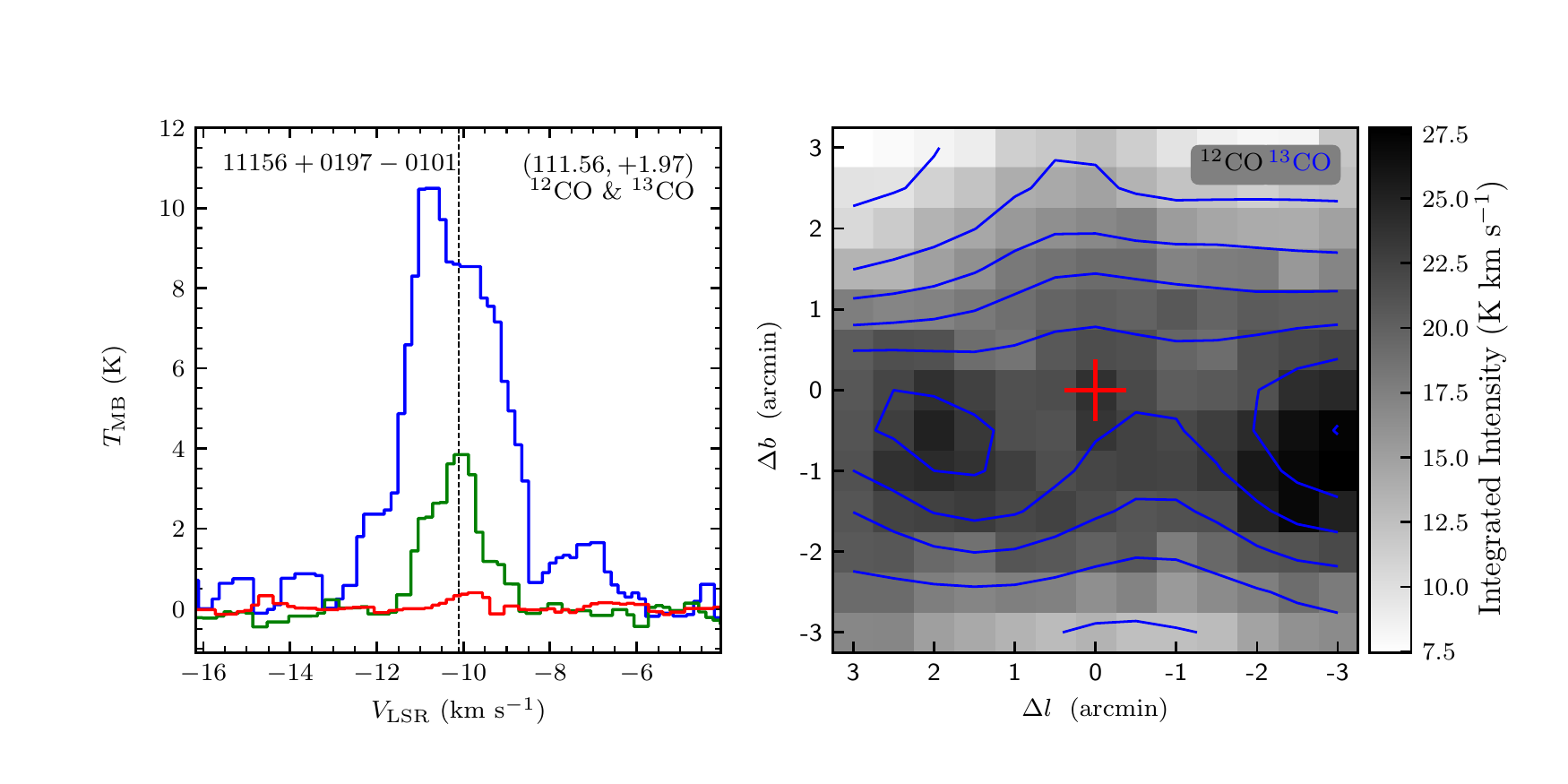}
\vspace{-0.5cm}

\includegraphics[width=9.0cm,angle=0]{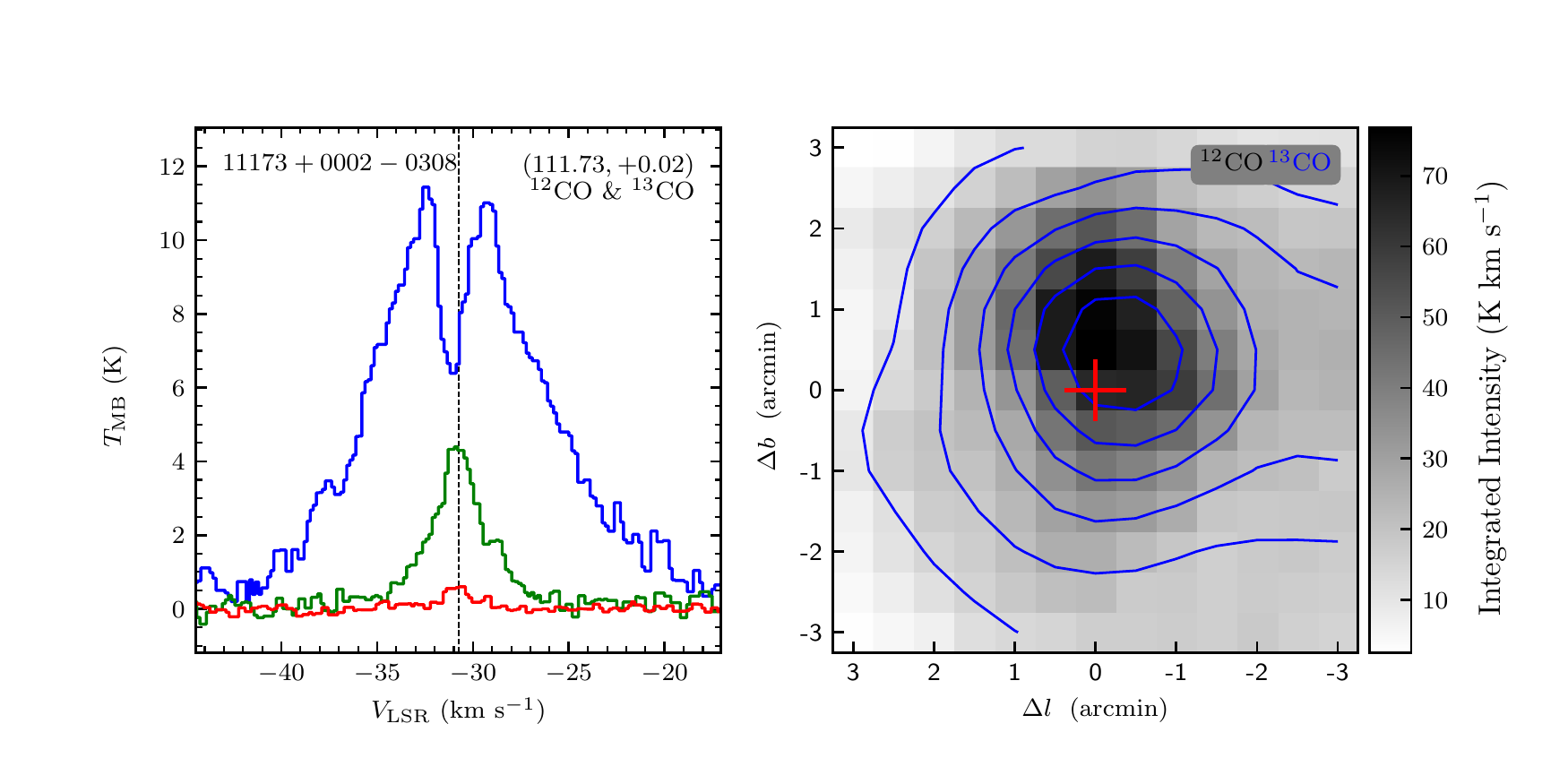}
\includegraphics[width=9.0cm,angle=0]{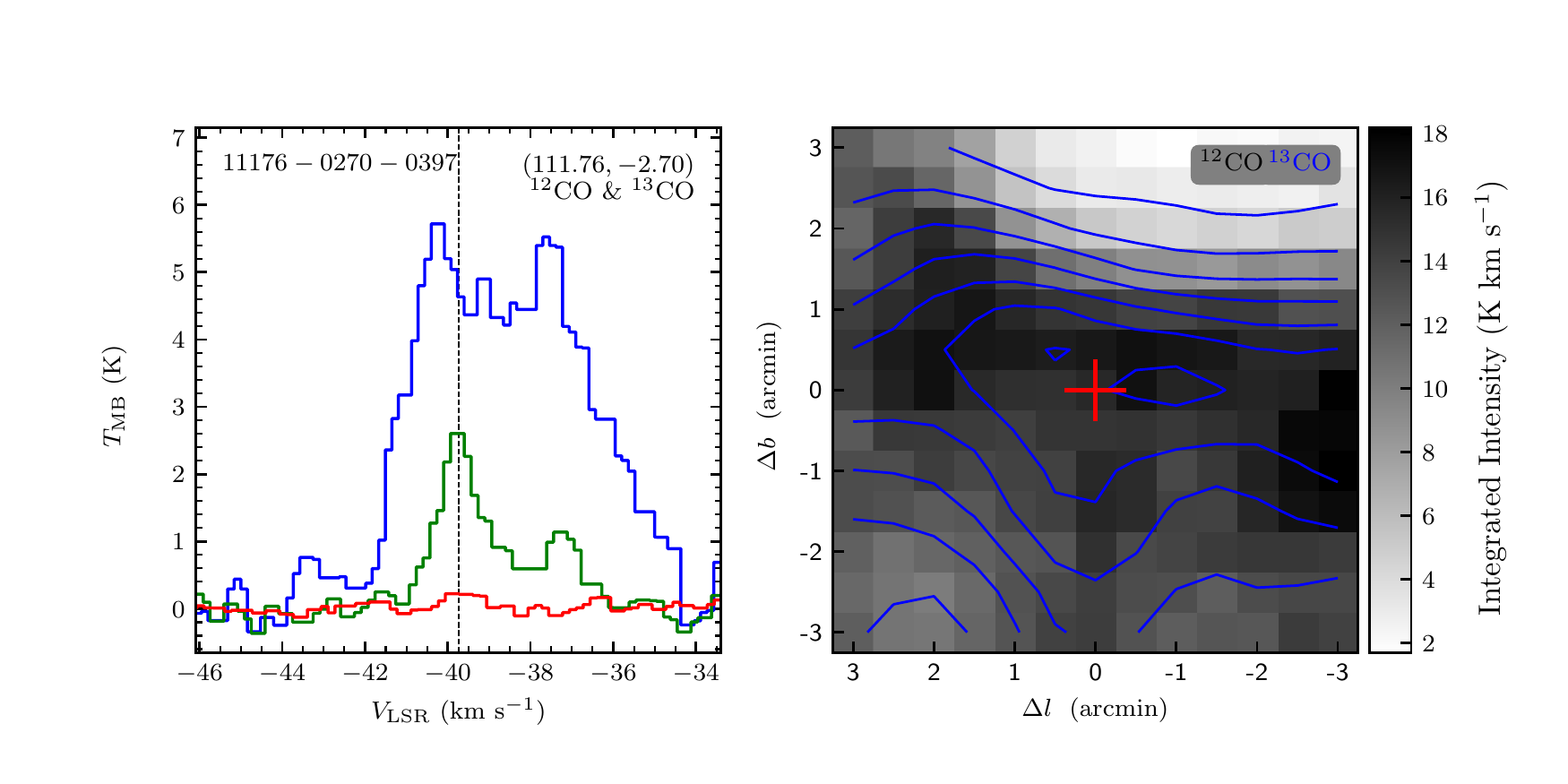}
\vspace{-0.5cm}

\includegraphics[width=9.0cm,angle=0]{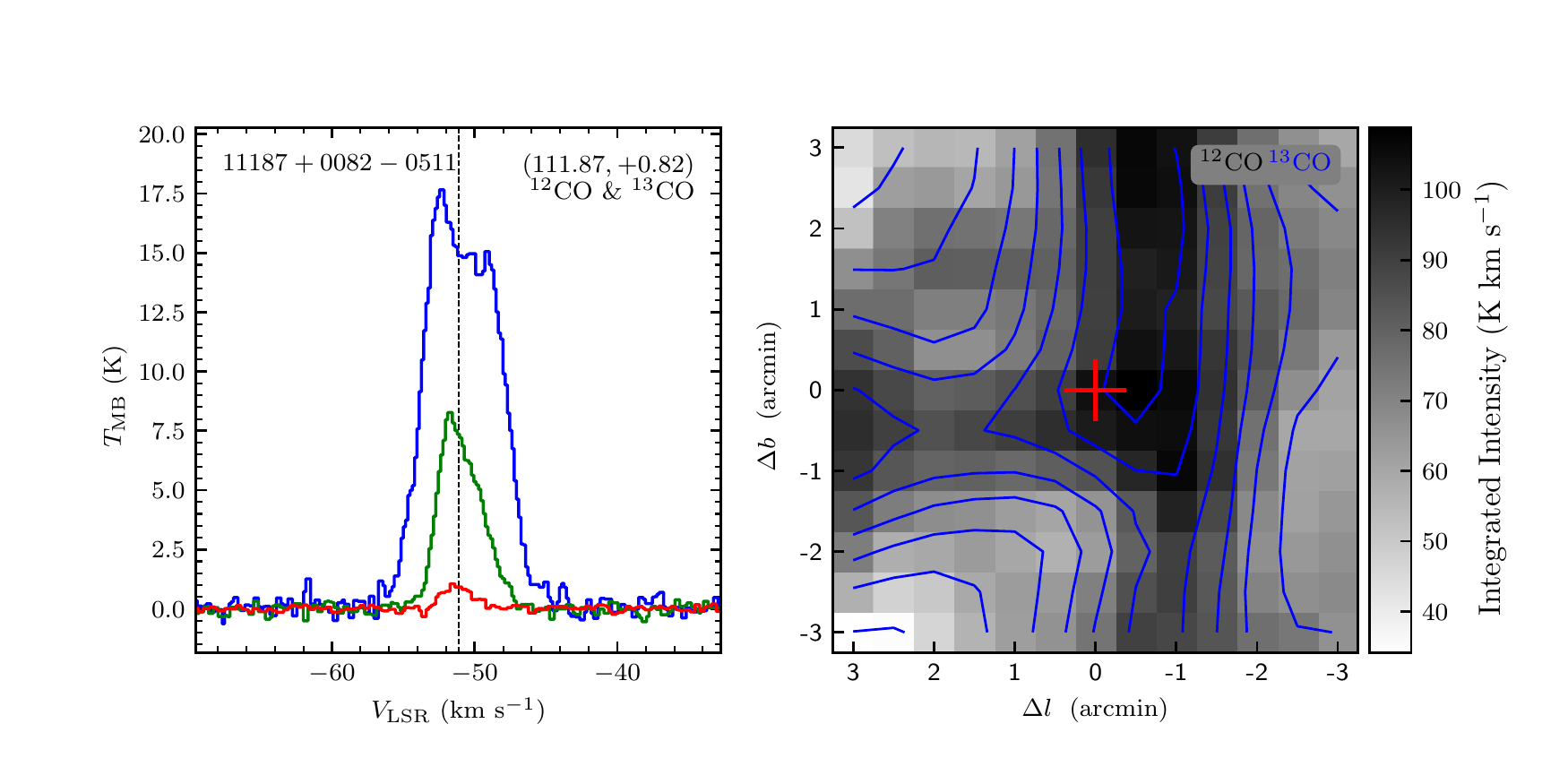}
\includegraphics[width=9.0cm,angle=0]{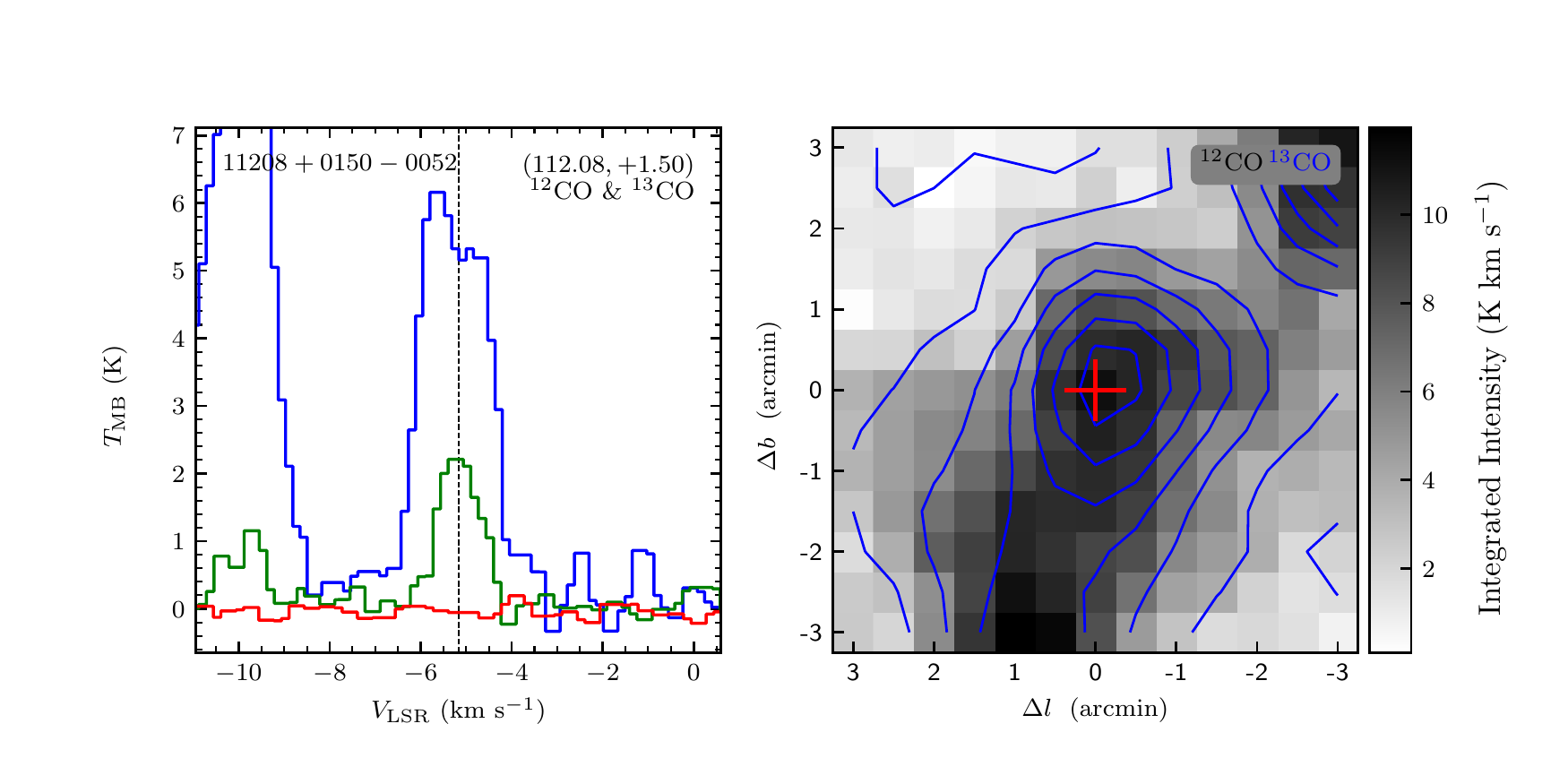}
\vspace{-0.5cm}

\includegraphics[width=9.0cm,angle=0]{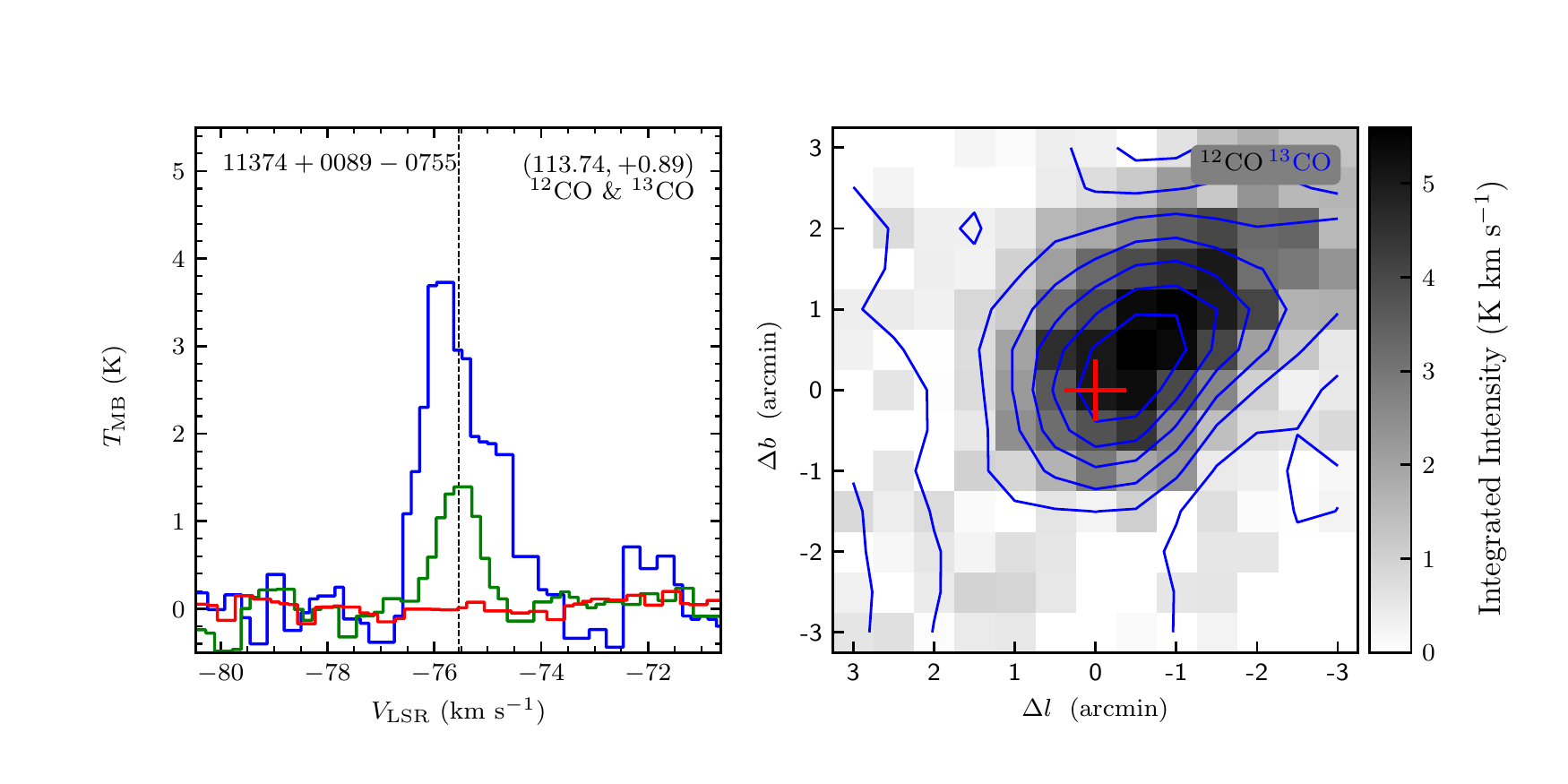}
\includegraphics[width=9.0cm,angle=0]{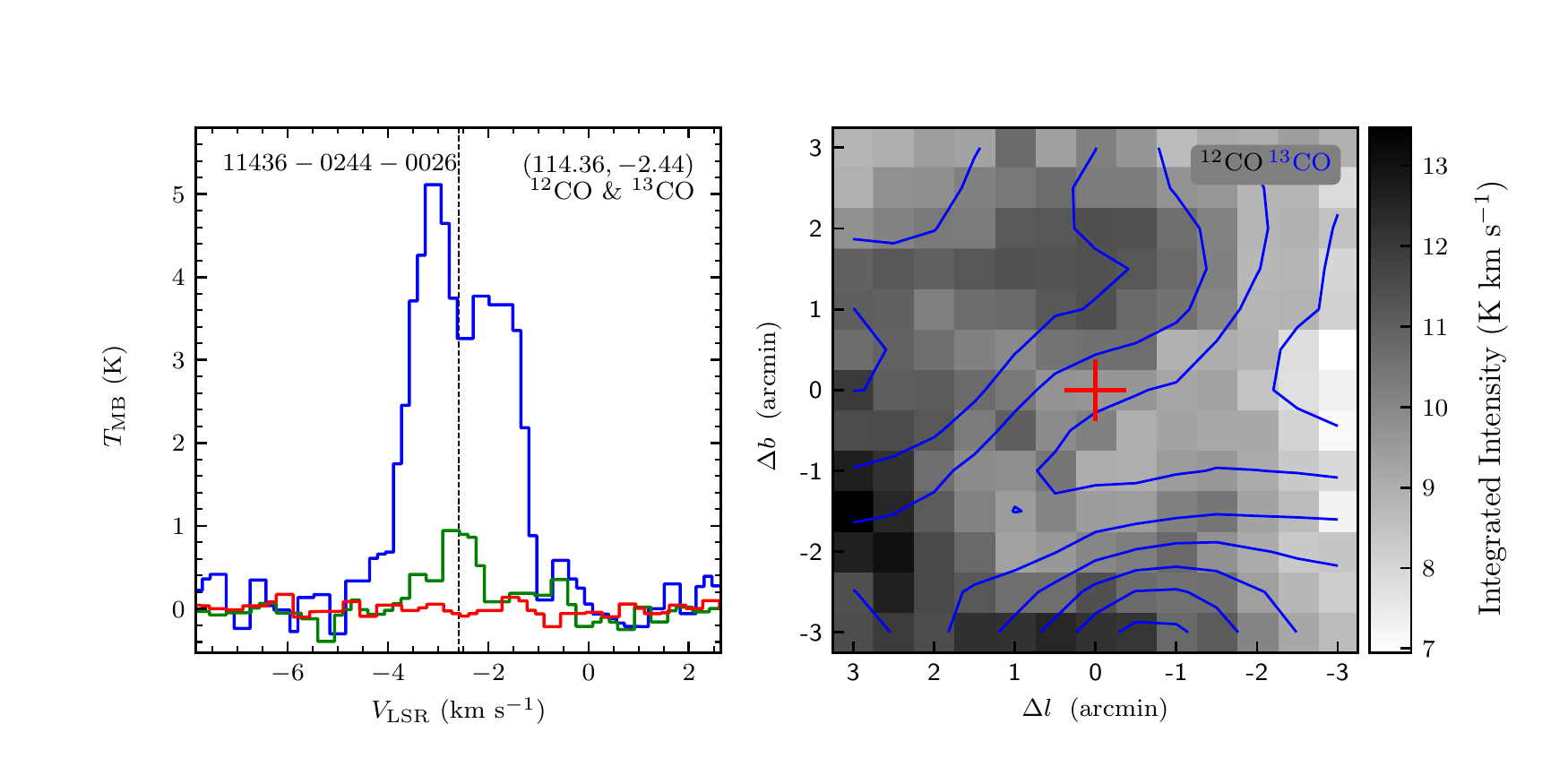}
\end{figure}
\clearpage

\begin{figure}
\includegraphics[width=9.0cm,angle=0]{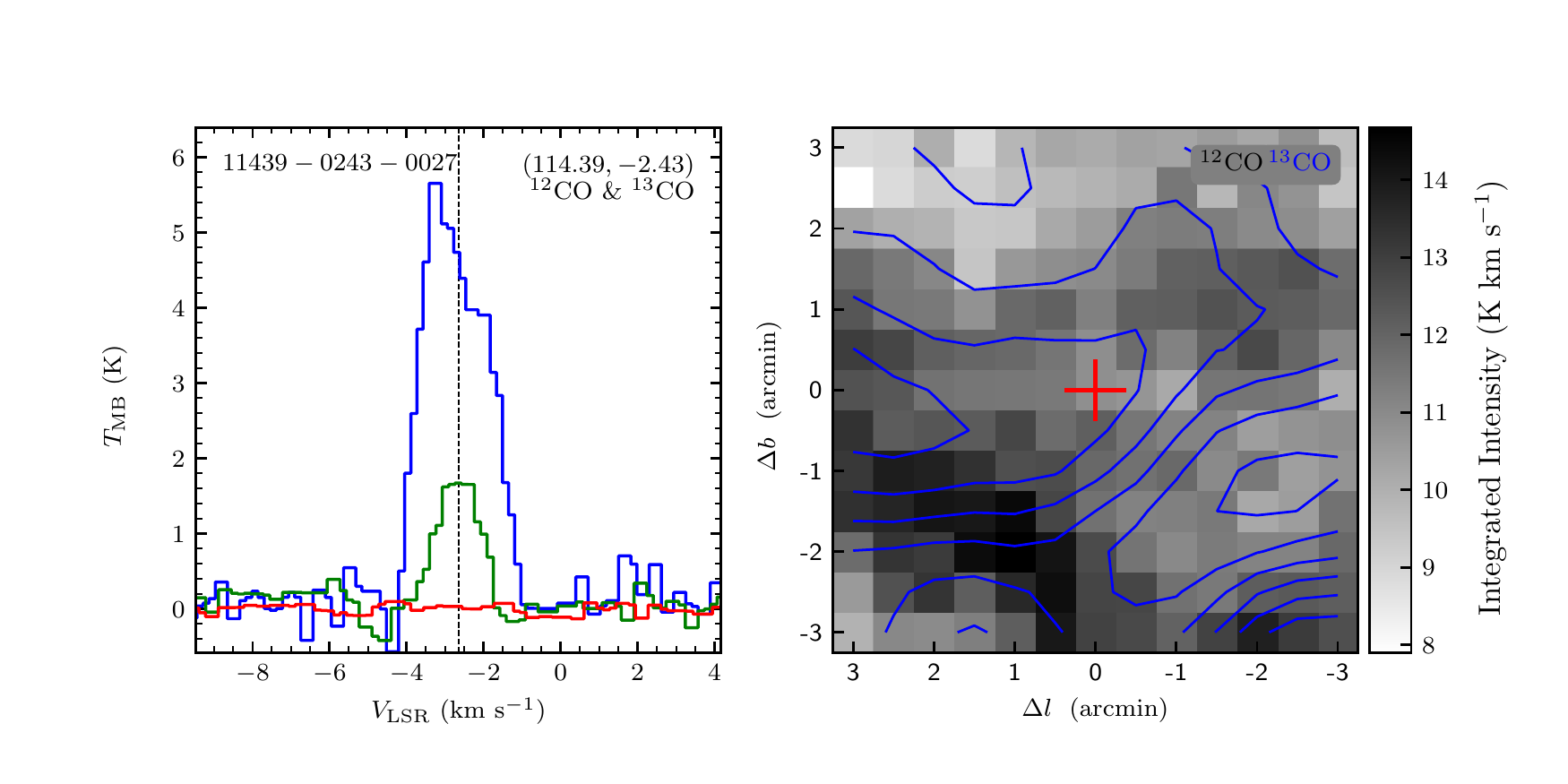}
\includegraphics[width=9.0cm,angle=0]{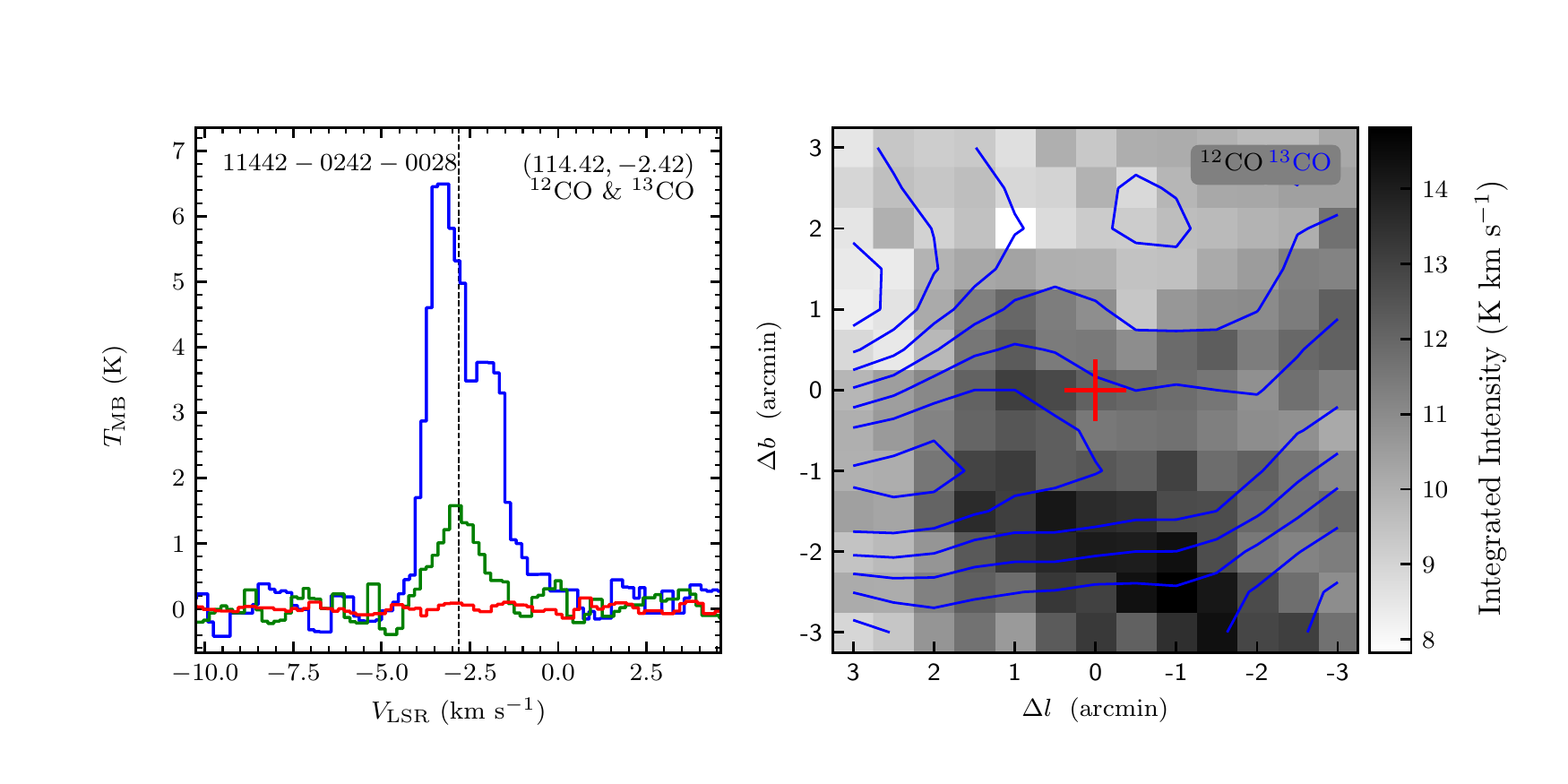}
\vspace{-0.5cm}

\includegraphics[width=9.0cm,angle=0]{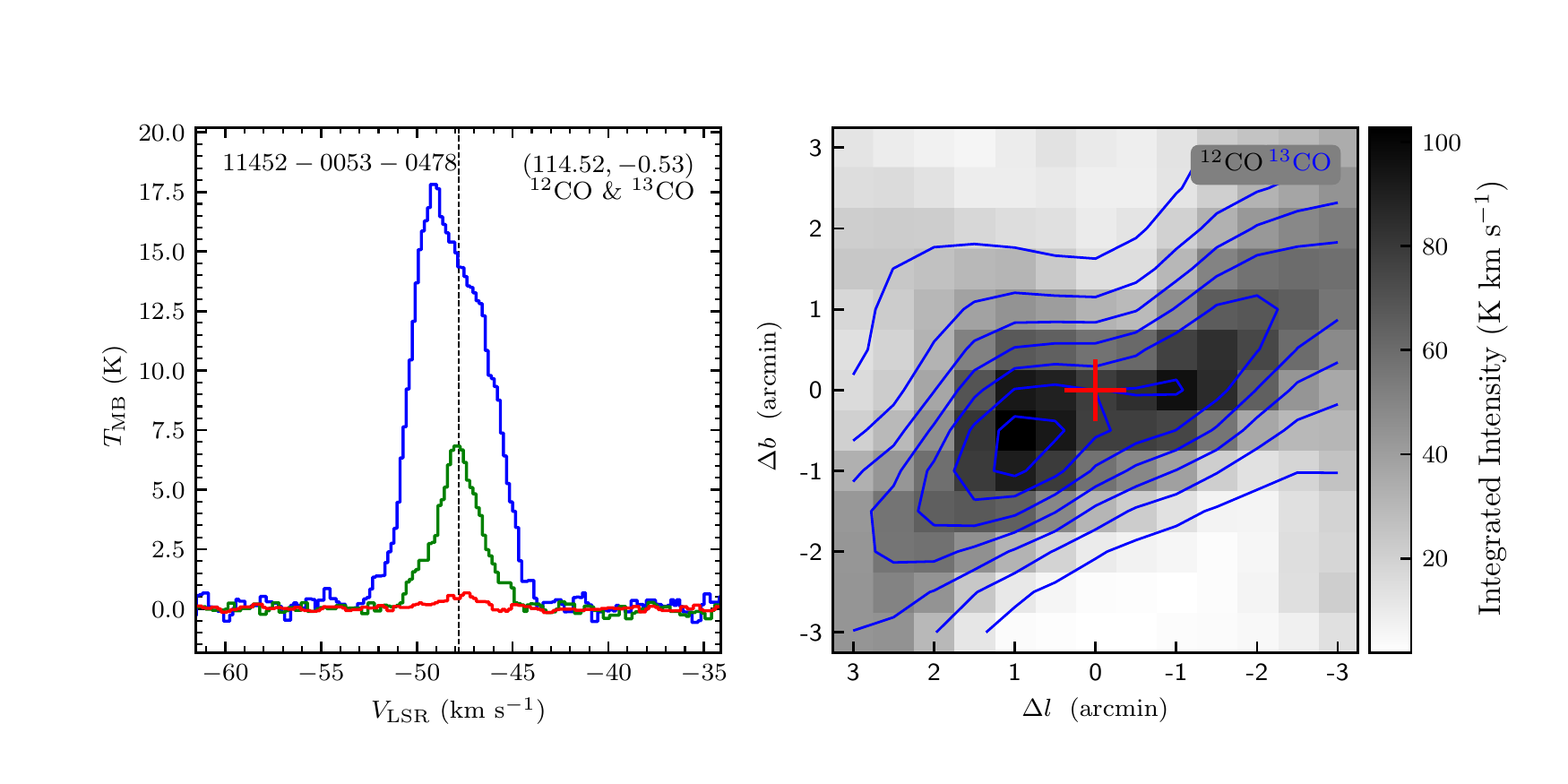}
\includegraphics[width=9.0cm,angle=0]{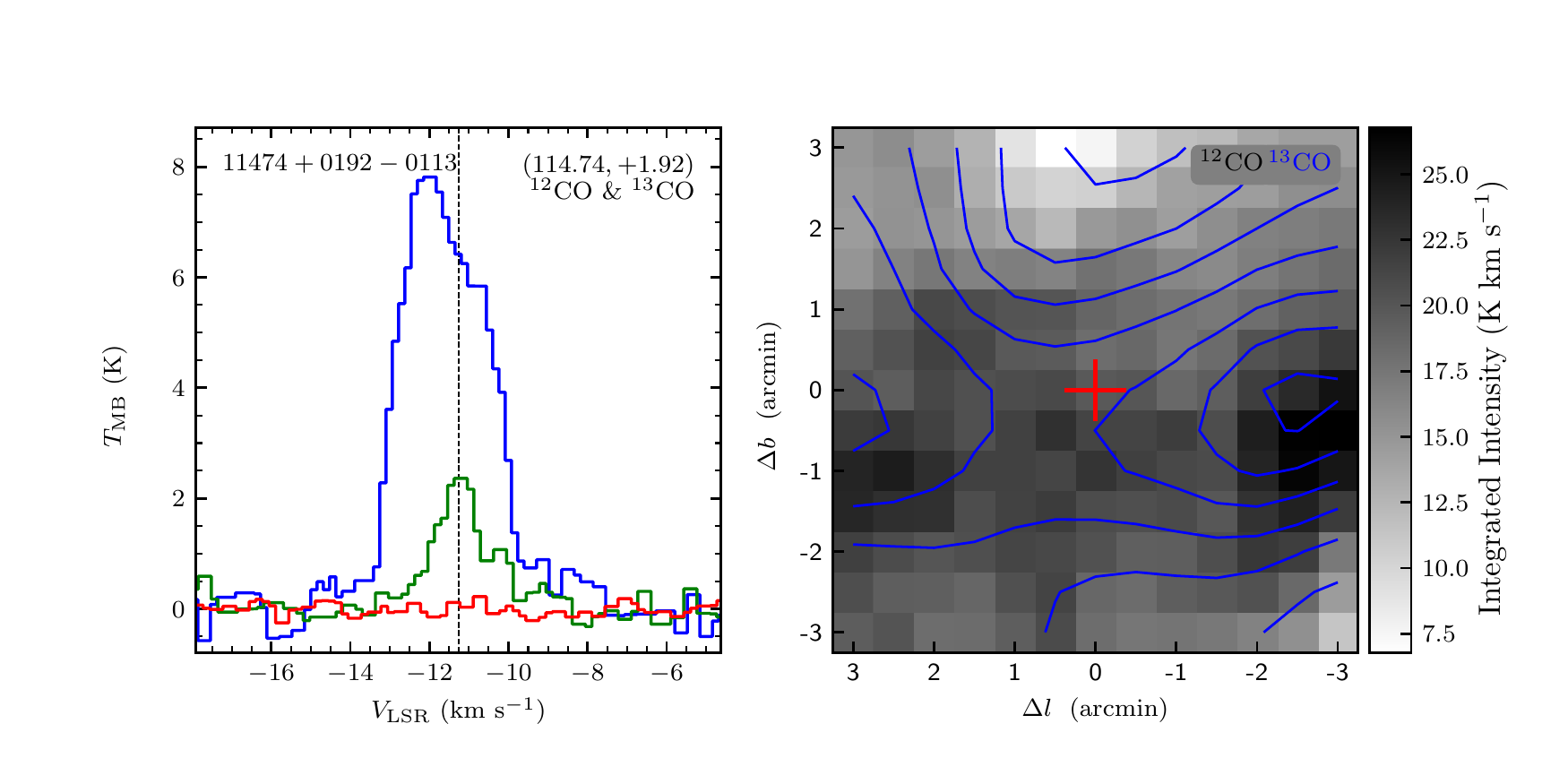}
\vspace{-0.5cm}

\includegraphics[width=9.0cm,angle=0]{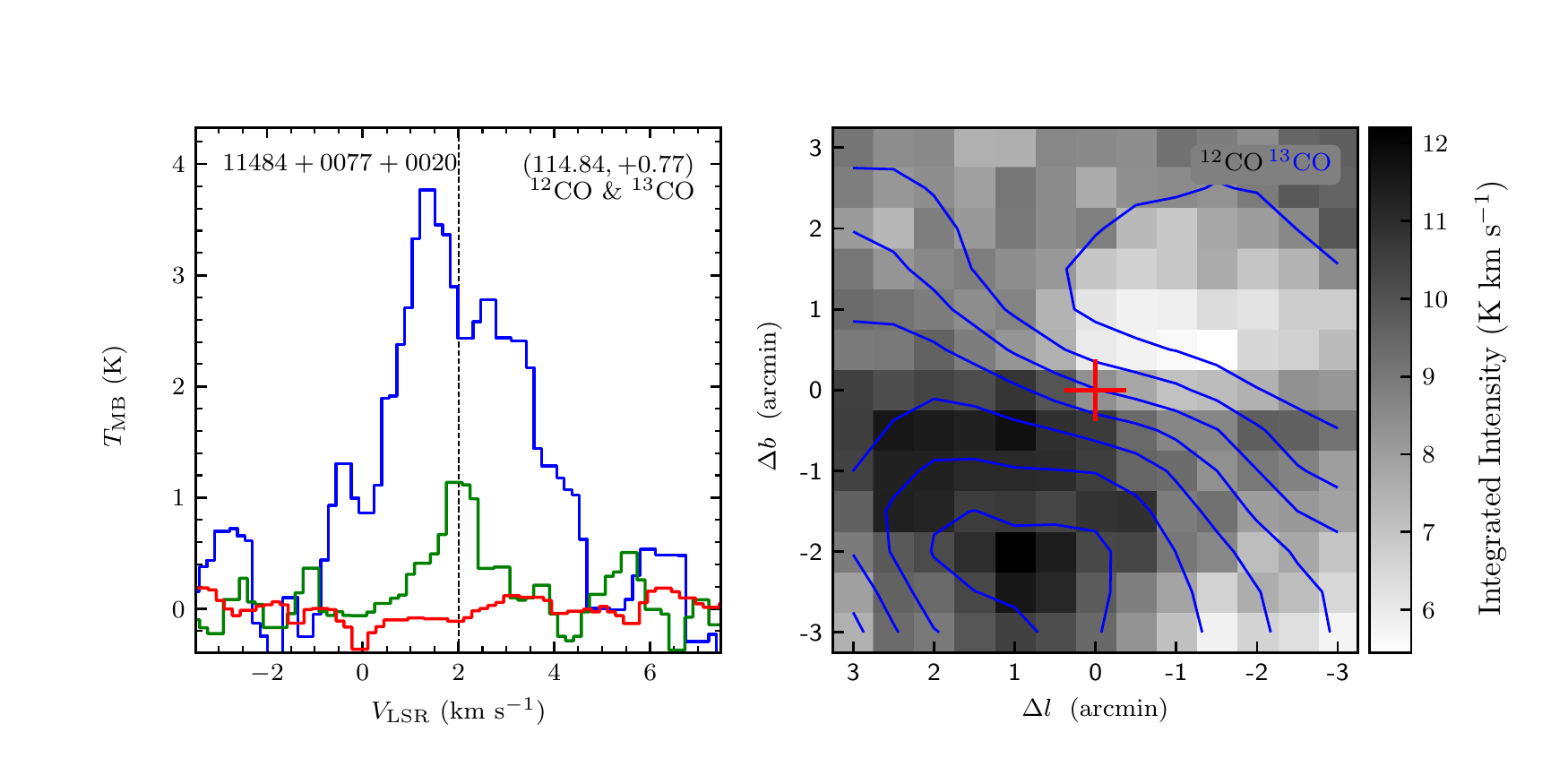}
\includegraphics[width=9.0cm,angle=0]{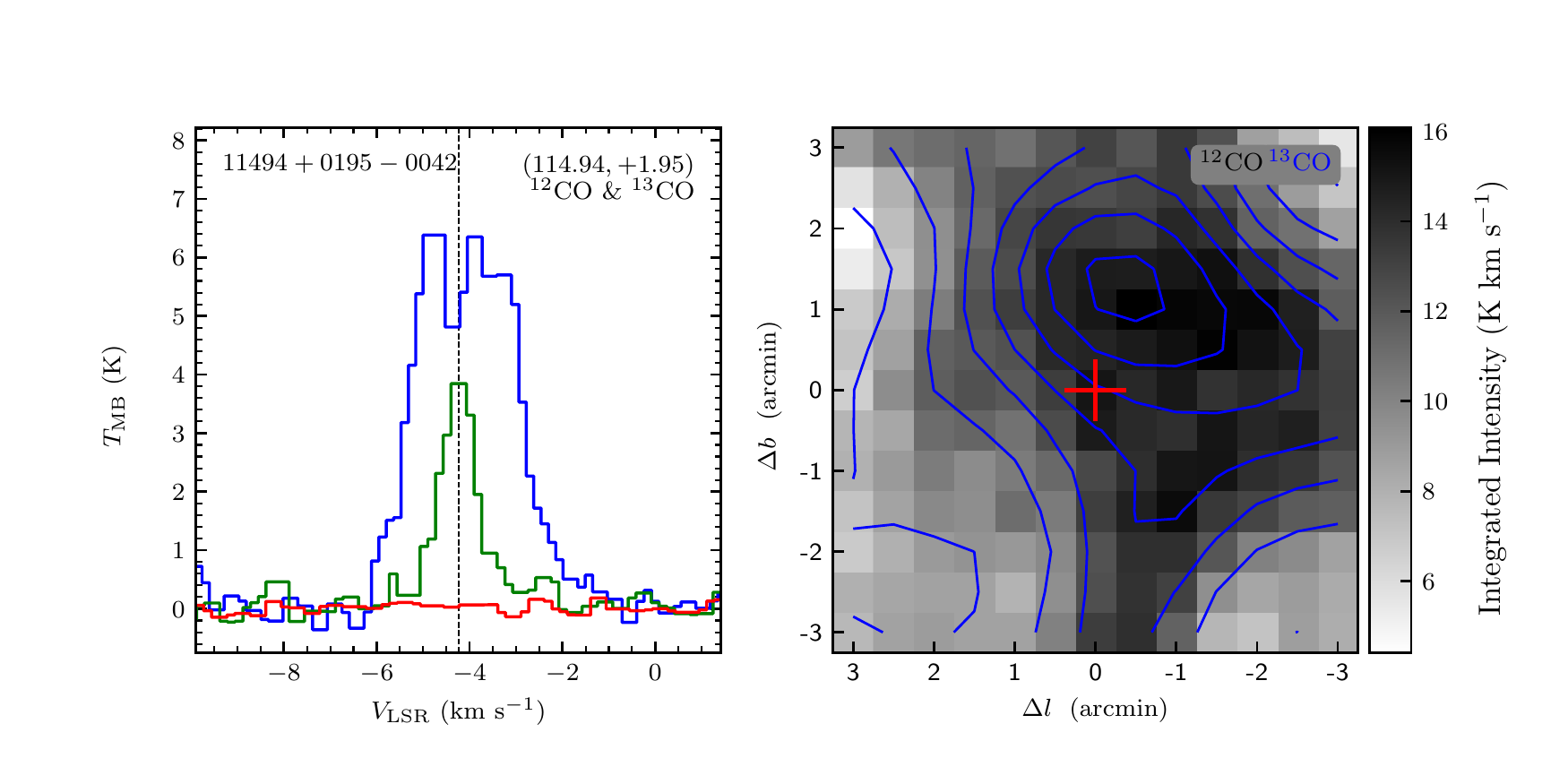}
\vspace{-0.5cm}

\includegraphics[width=9.0cm,angle=0]{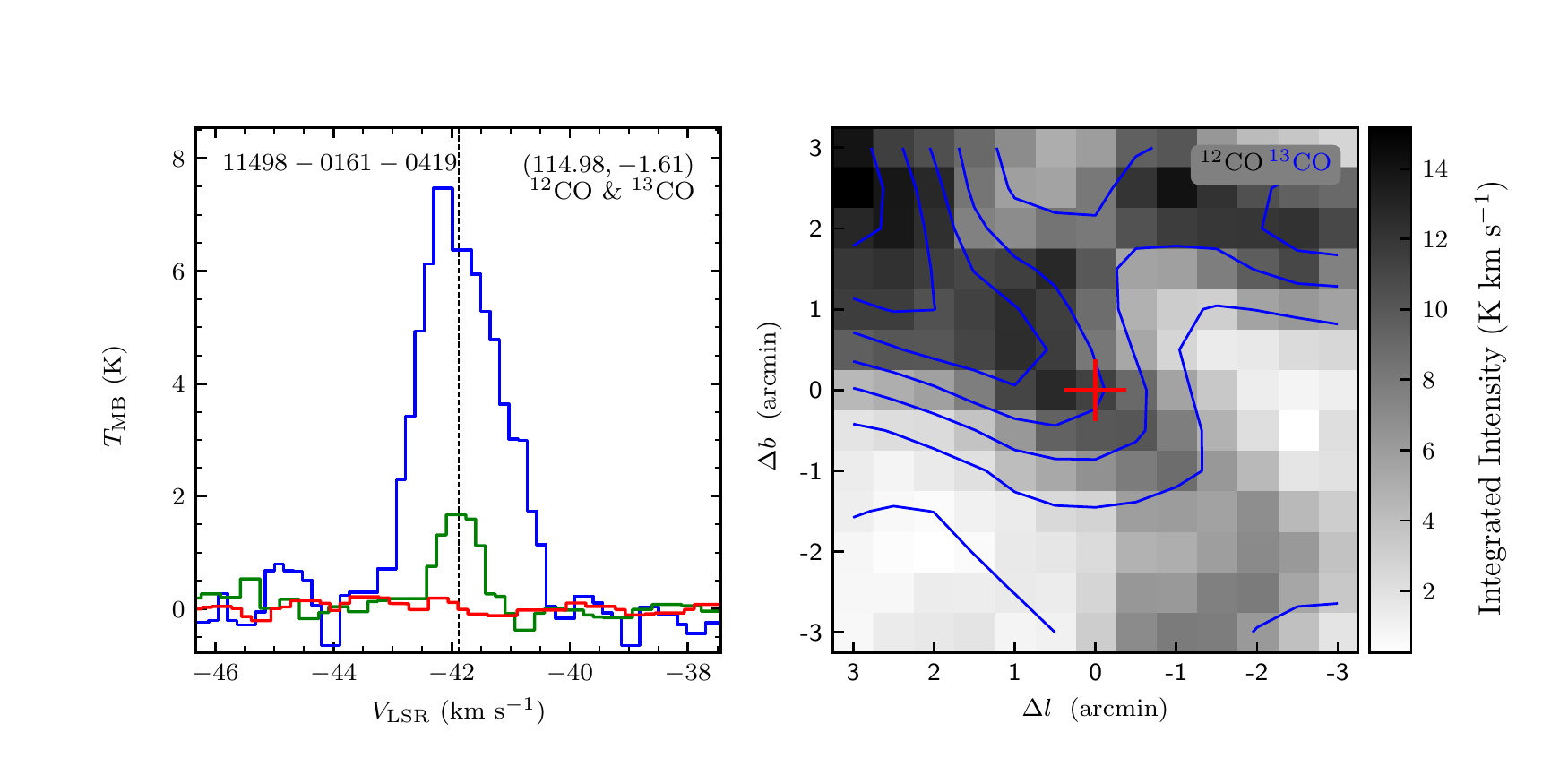}
\includegraphics[width=9.0cm,angle=0]{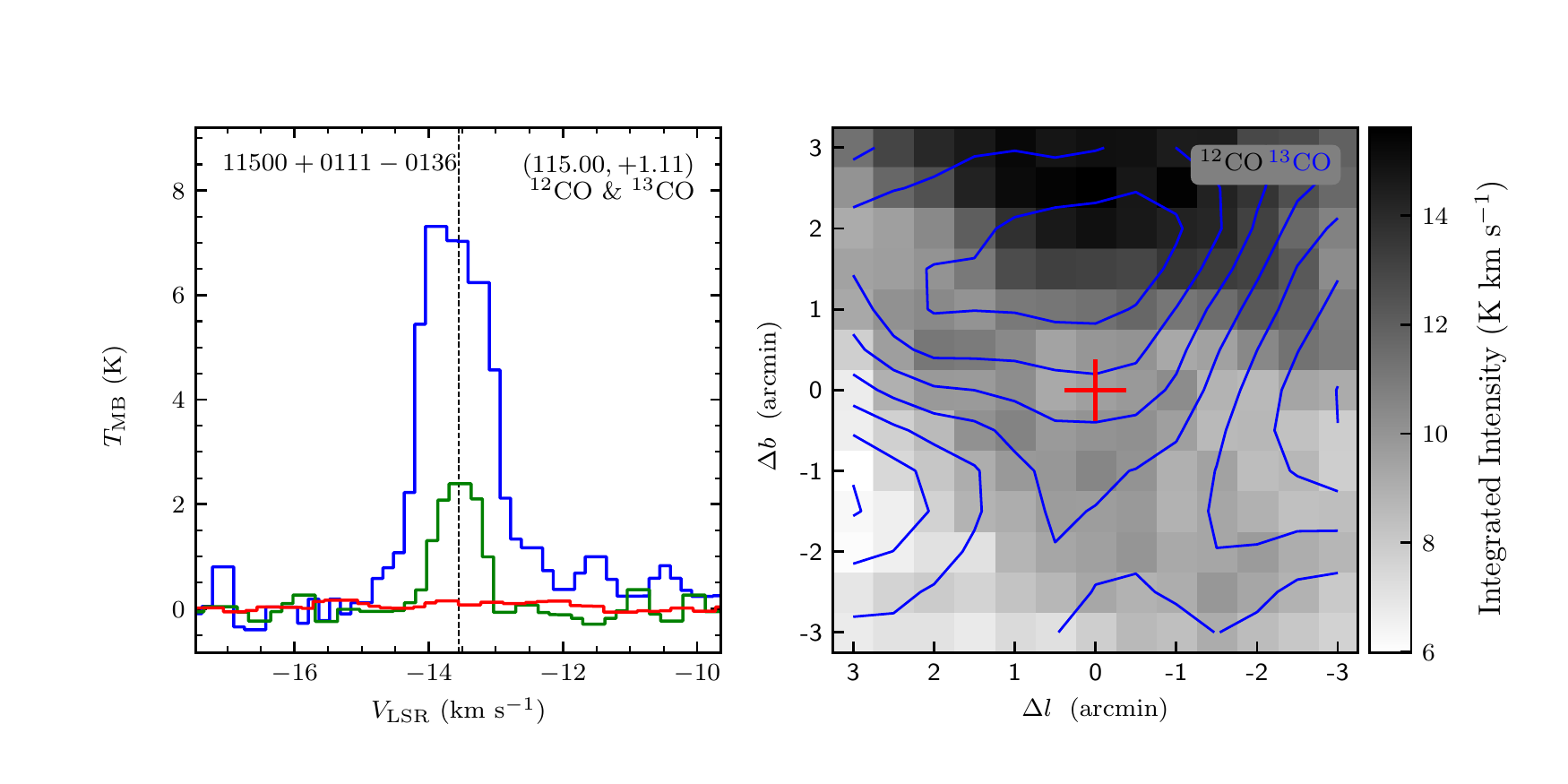}
\vspace{-0.5cm}

\includegraphics[width=9.0cm,angle=0]{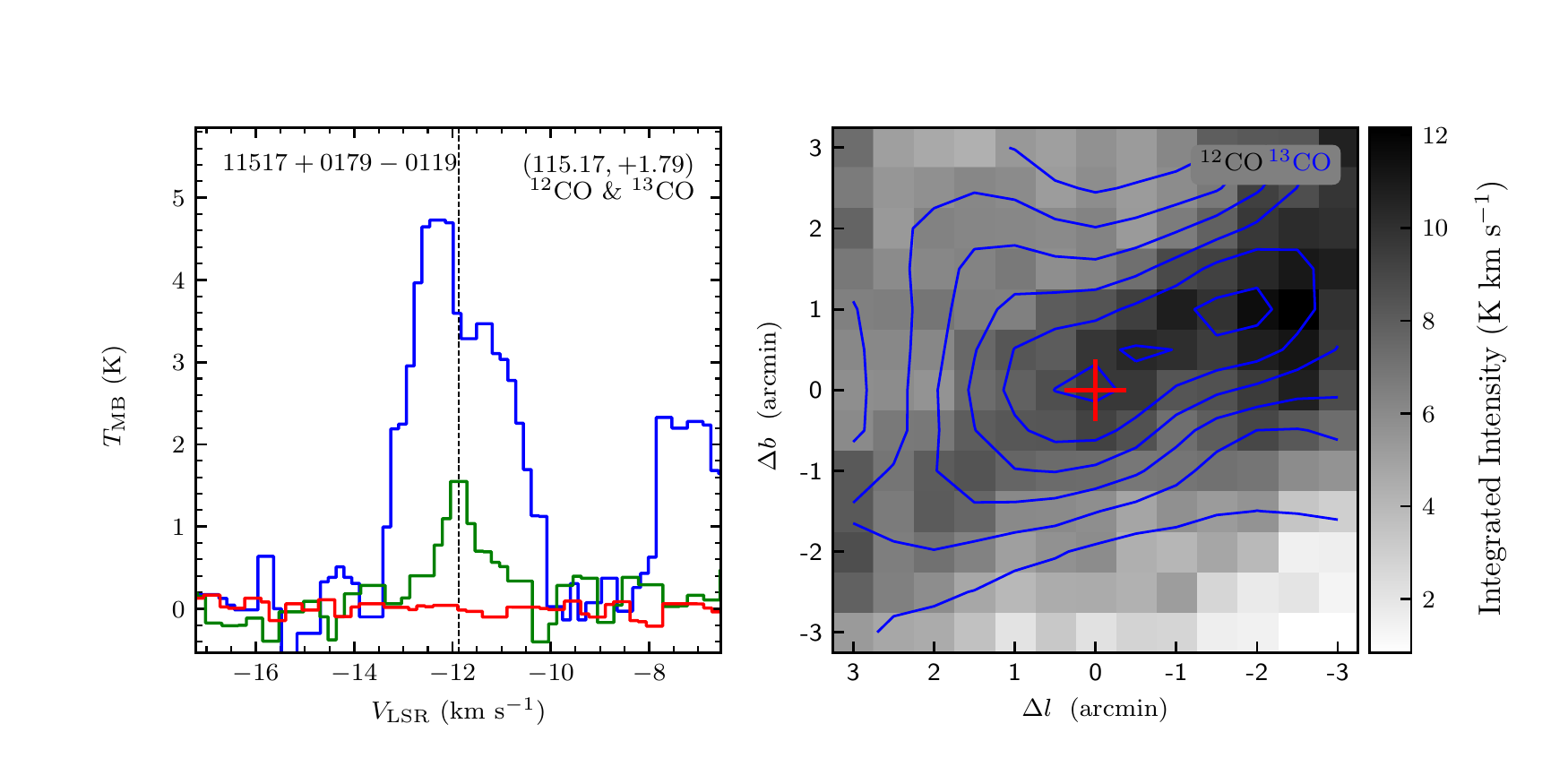}
\includegraphics[width=9.0cm,angle=0]{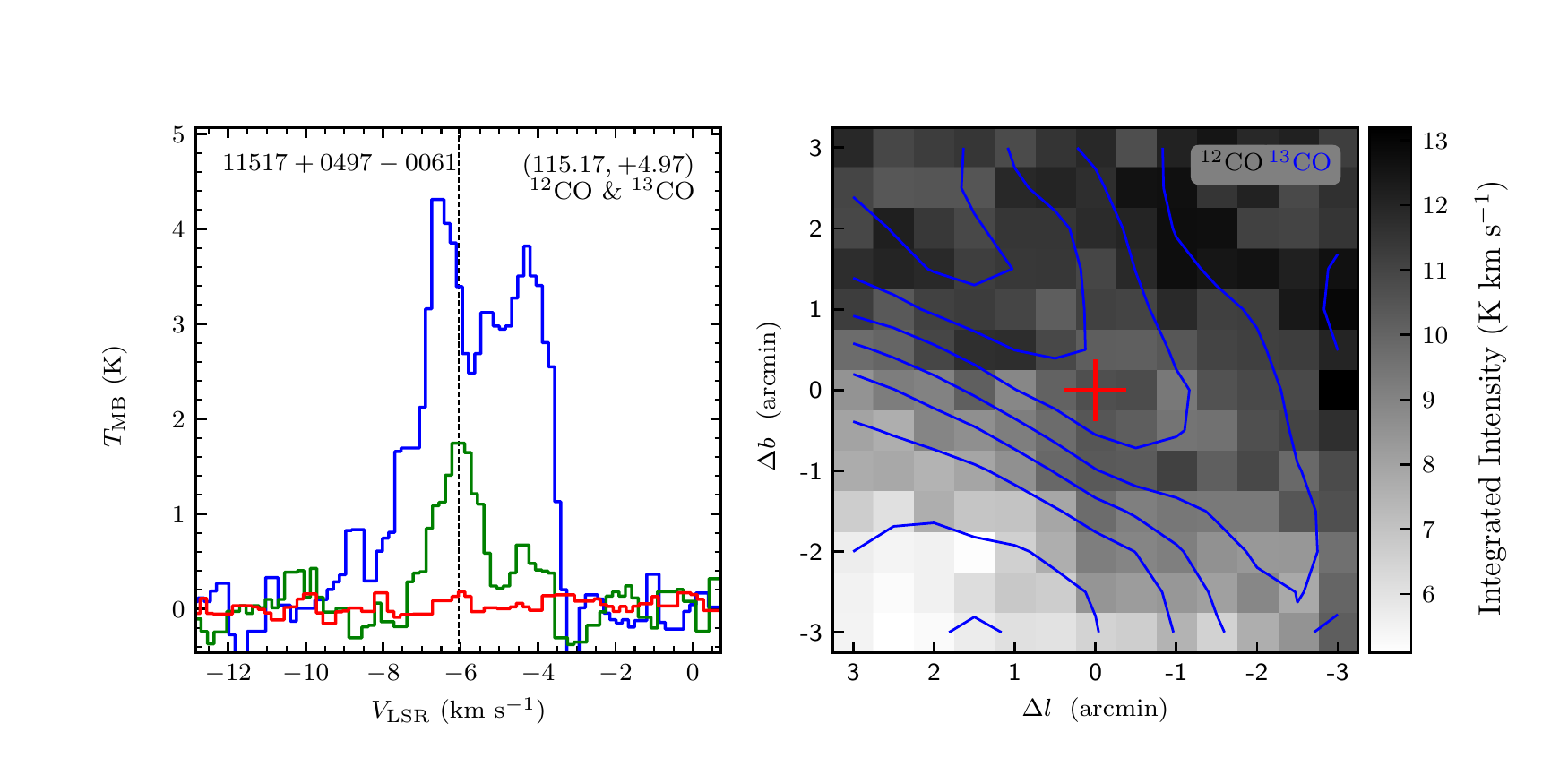}
\end{figure}
\clearpage

\begin{figure}
\includegraphics[width=9.0cm,angle=0]{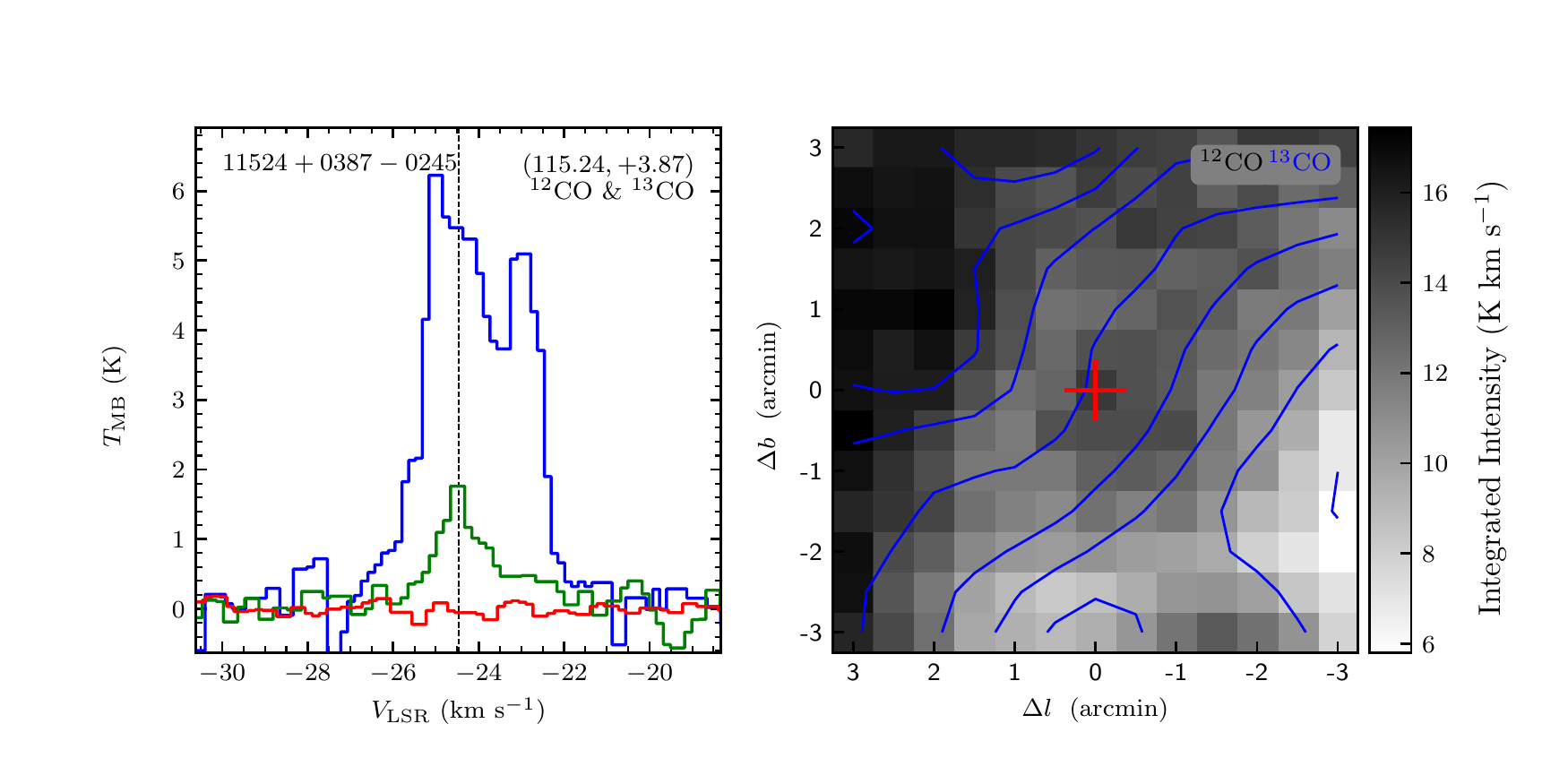}
\includegraphics[width=9.0cm,angle=0]{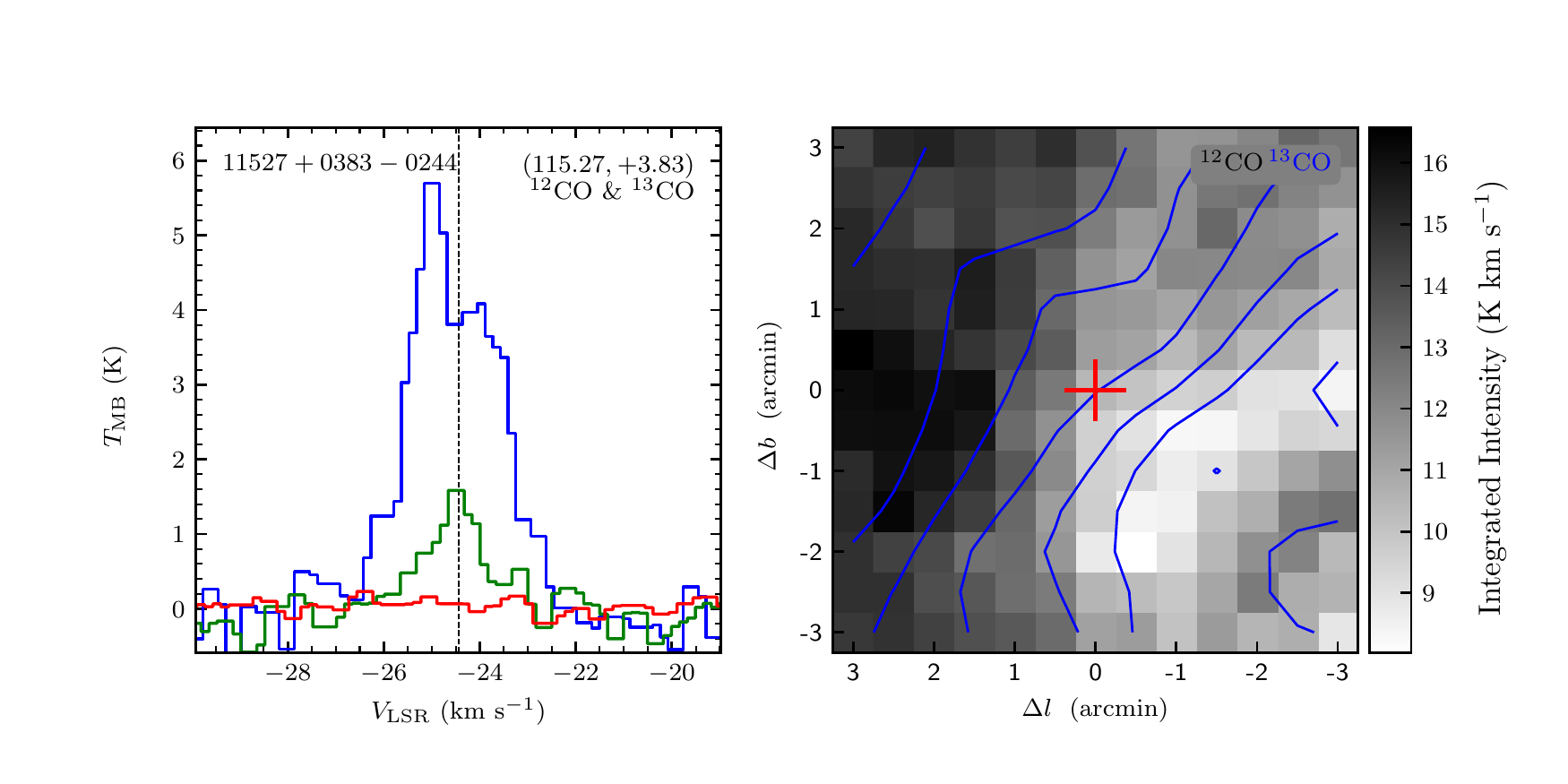}
\vspace{-0.5cm}

\includegraphics[width=9.0cm,angle=0]{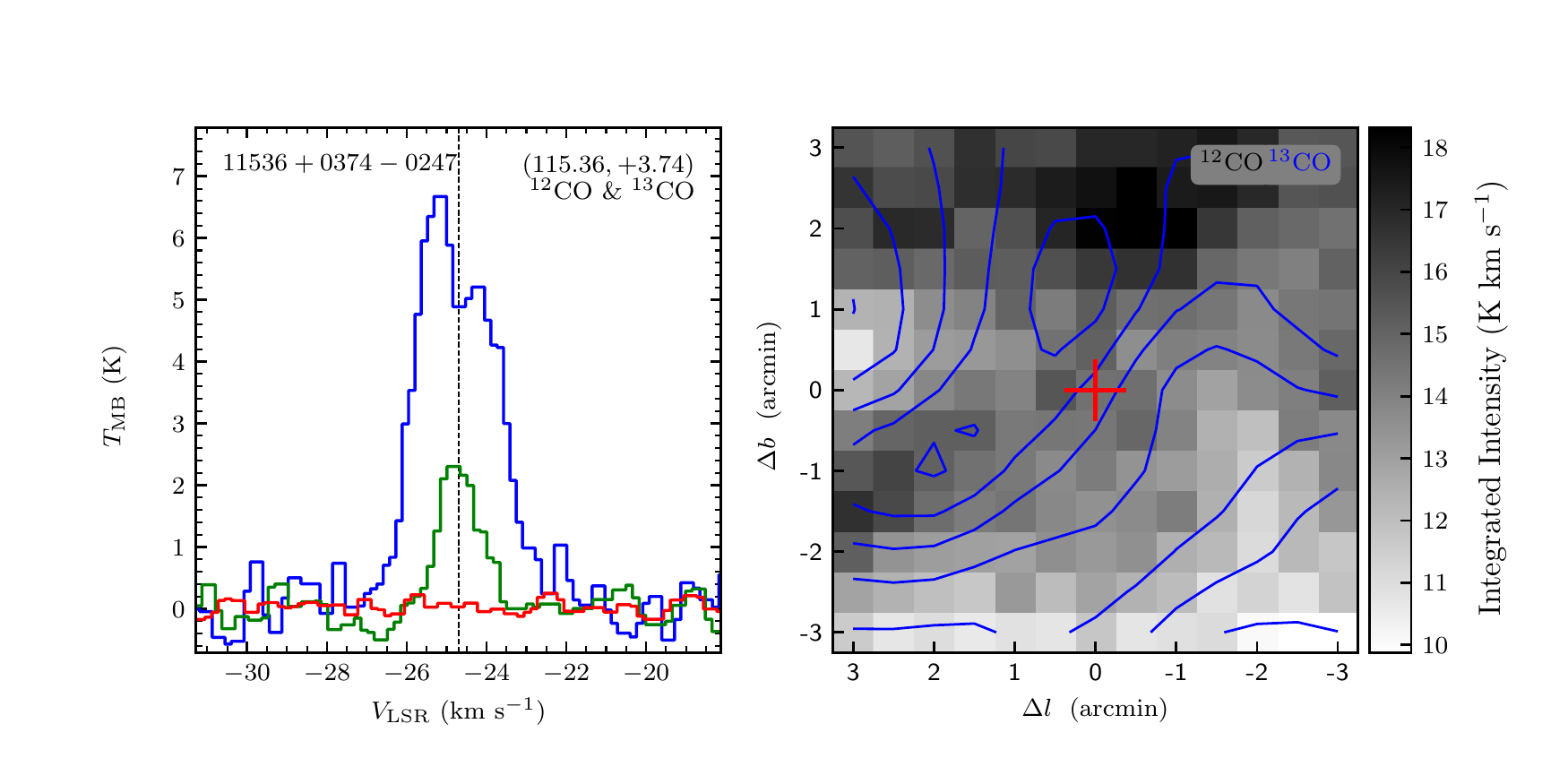}
\includegraphics[width=9.0cm,angle=0]{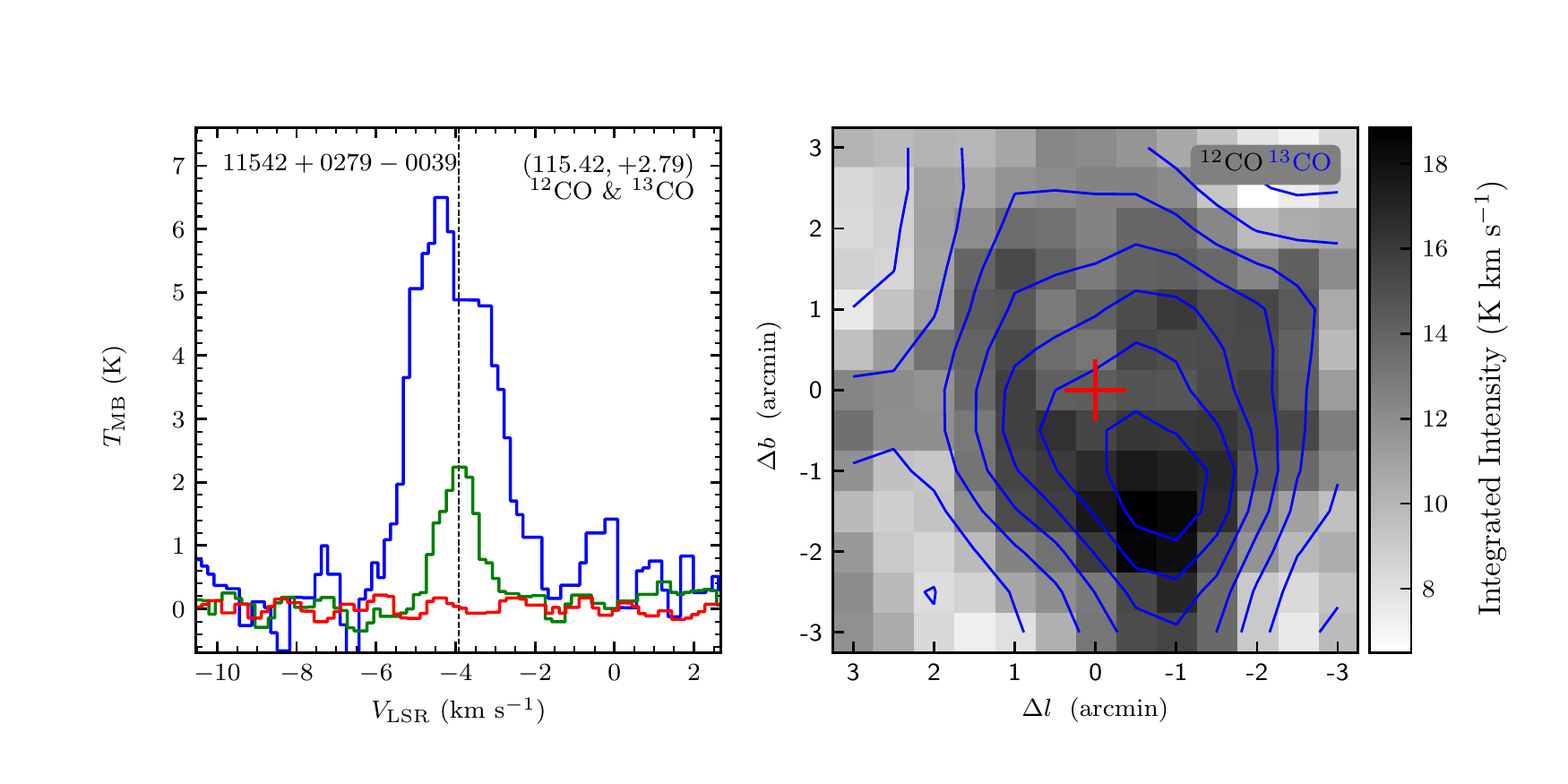}
\vspace{-0.5cm}

\includegraphics[width=9.0cm,angle=0]{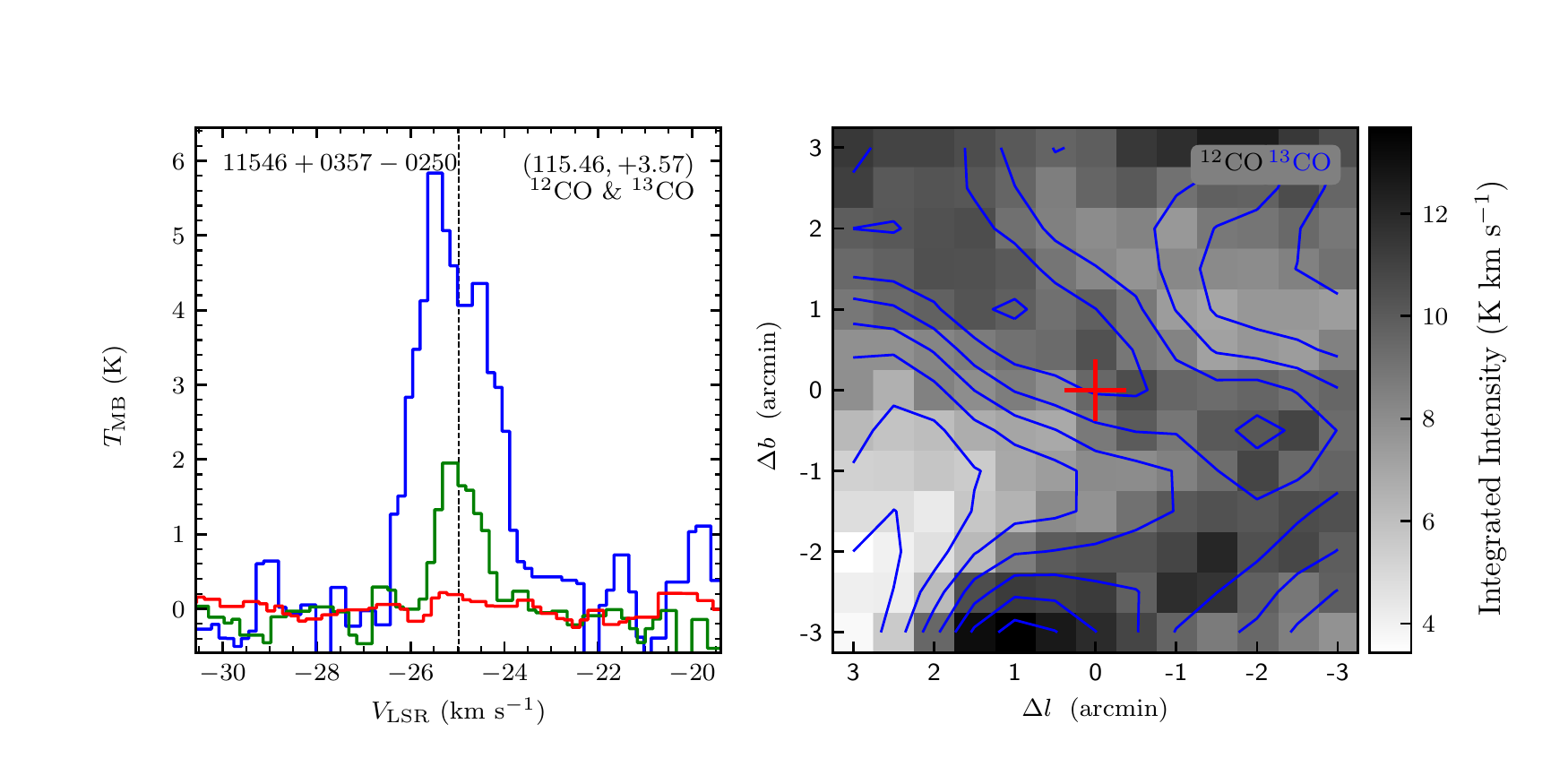}
\includegraphics[width=9.0cm,angle=0]{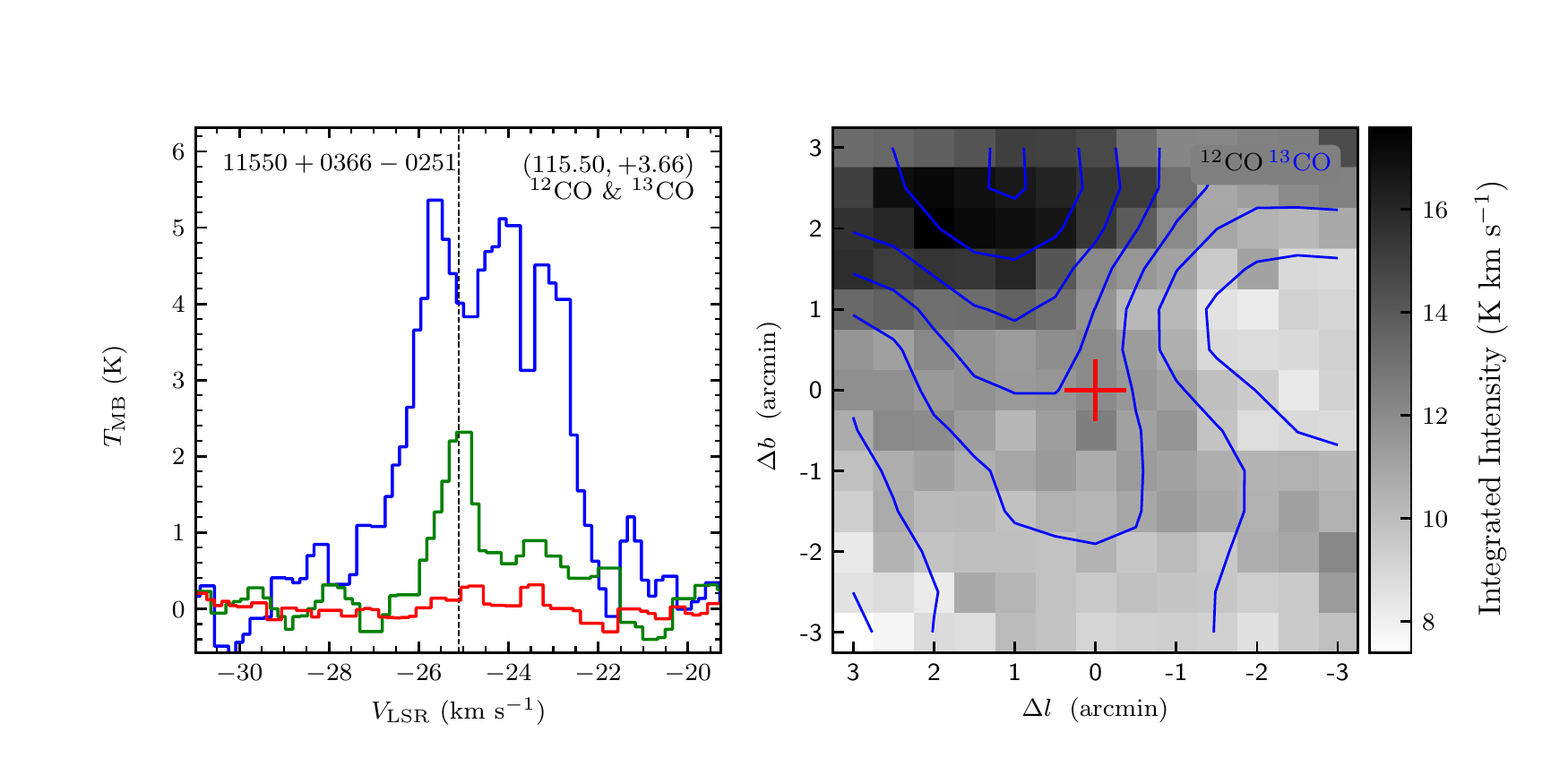}
\vspace{-0.5cm}

\includegraphics[width=9.0cm,angle=0]{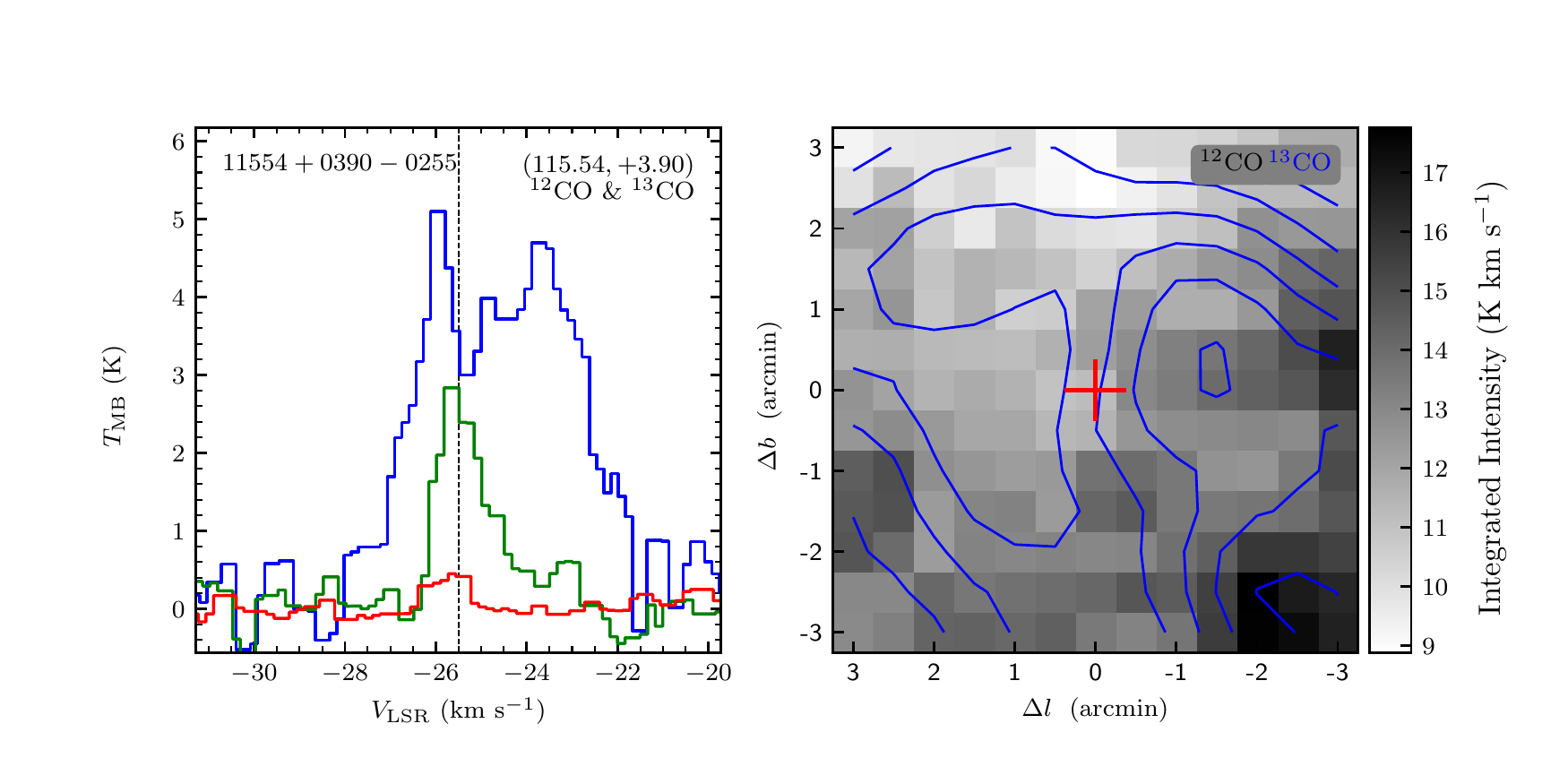}
\includegraphics[width=9.0cm,angle=0]{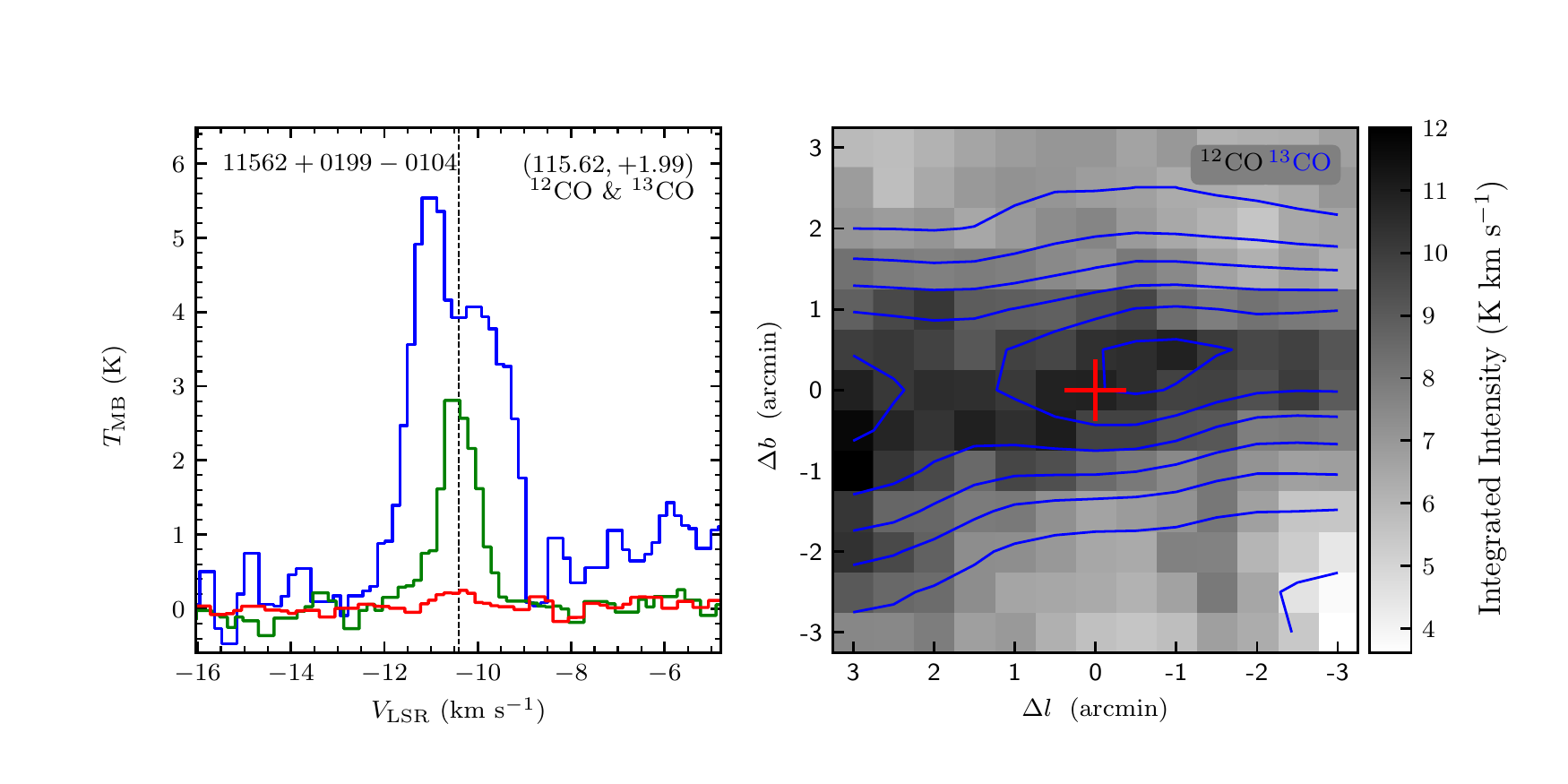}
\vspace{-0.5cm}

\includegraphics[width=9.0cm,angle=0]{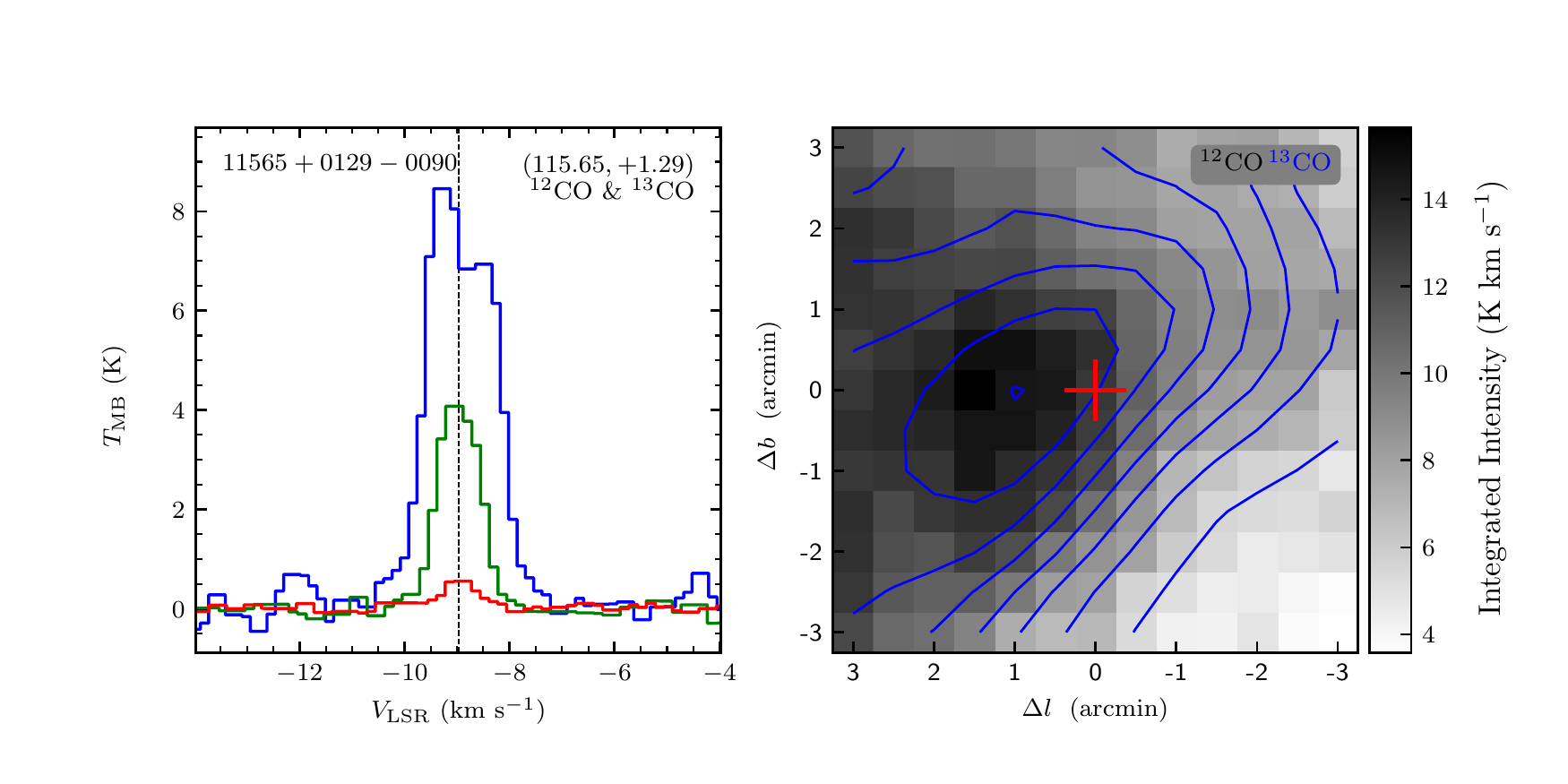}
\includegraphics[width=9.0cm,angle=0]{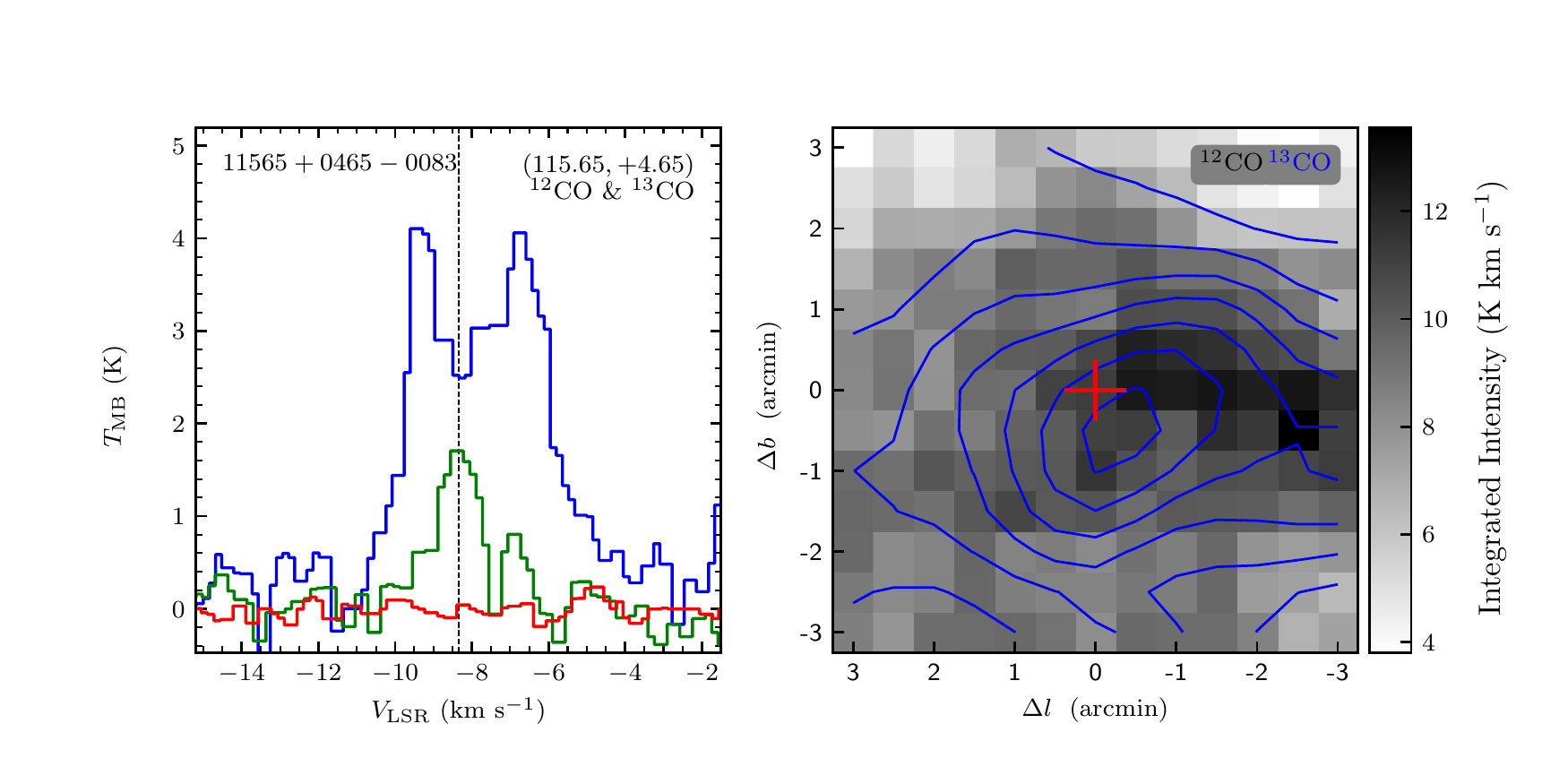}
\end{figure}
\clearpage

\begin{figure}
\includegraphics[width=9.0cm,angle=0]{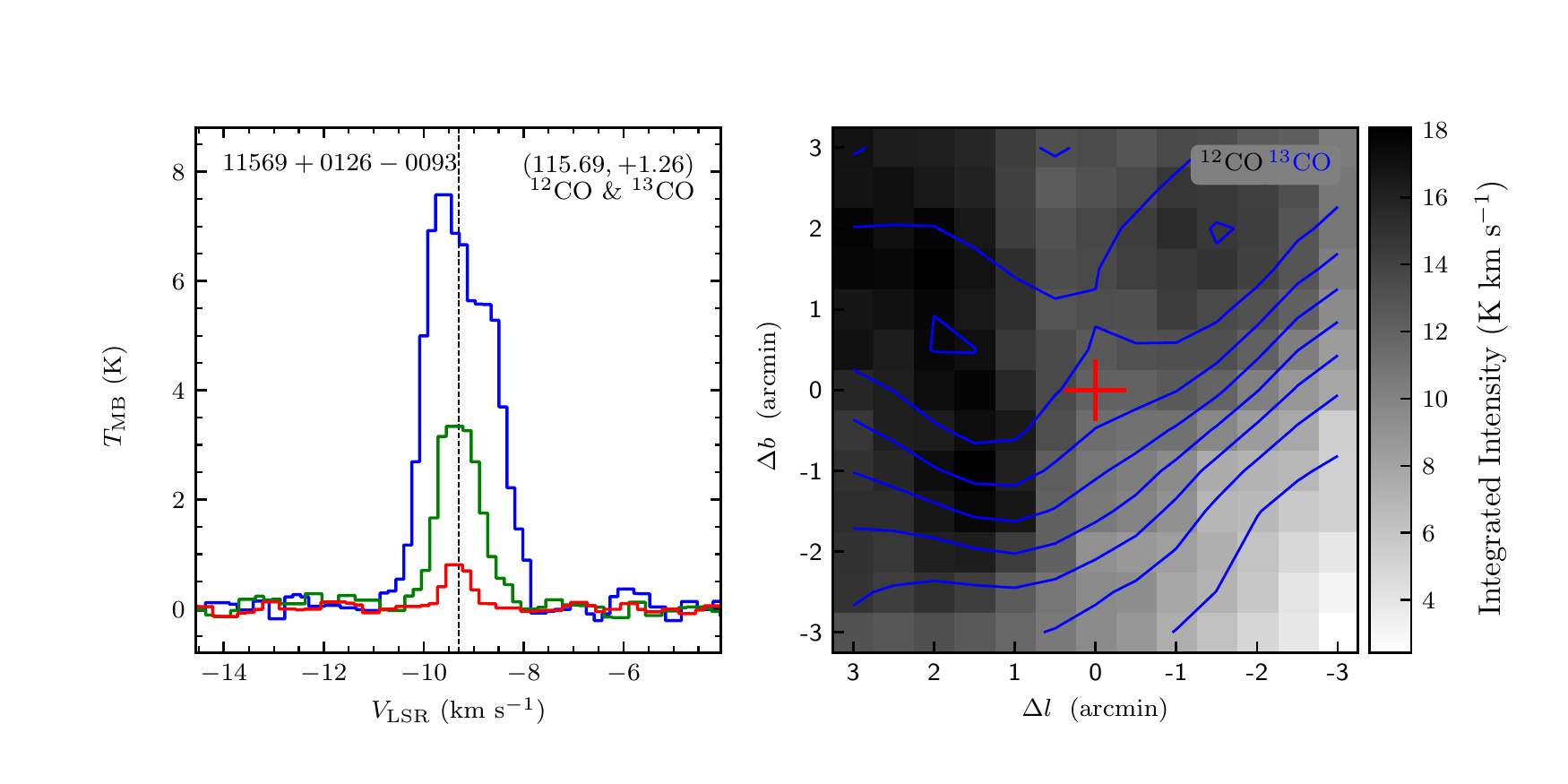}
\includegraphics[width=9.0cm,angle=0]{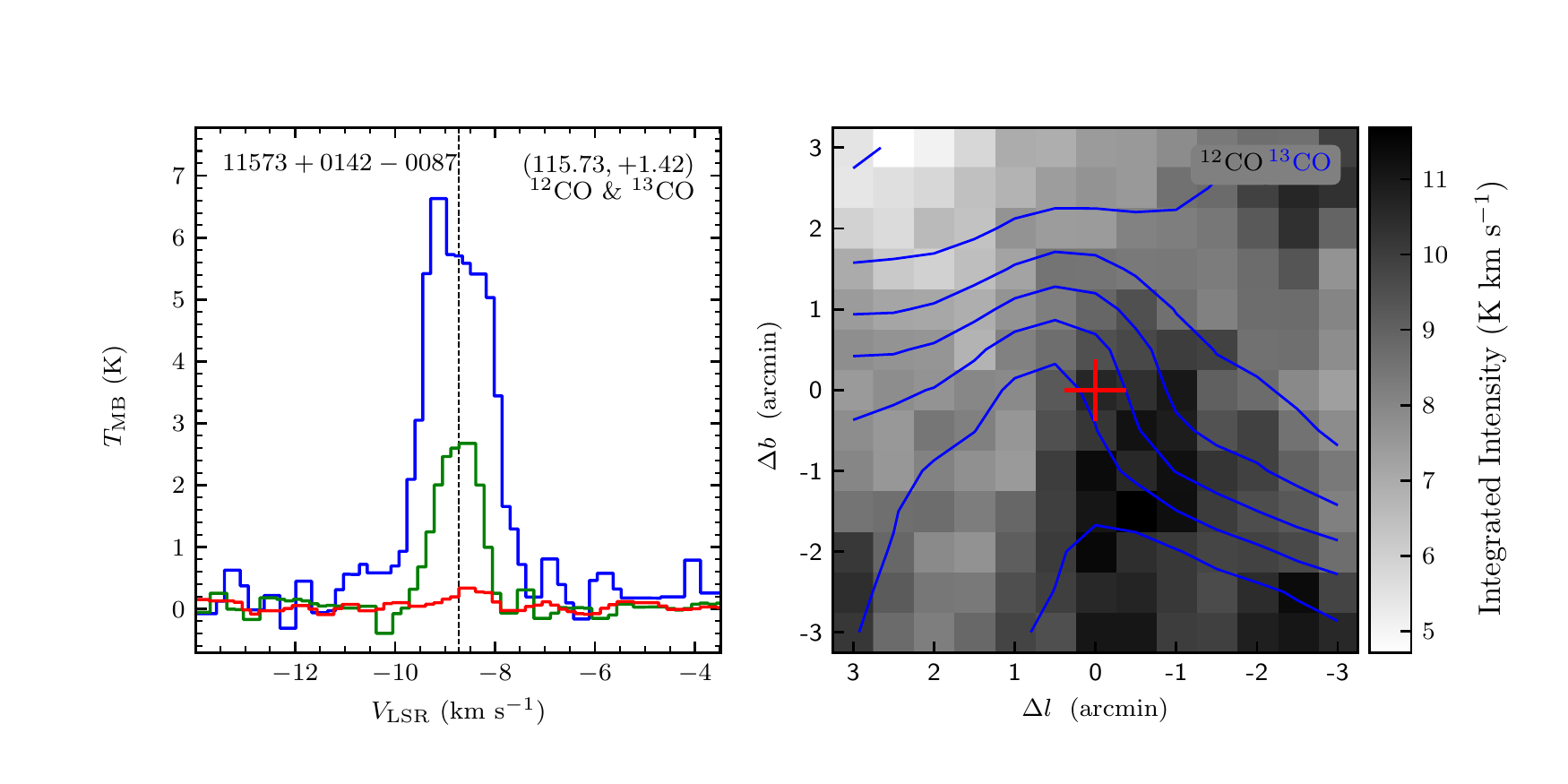}
\vspace{-0.5cm}

\includegraphics[width=9.0cm,angle=0]{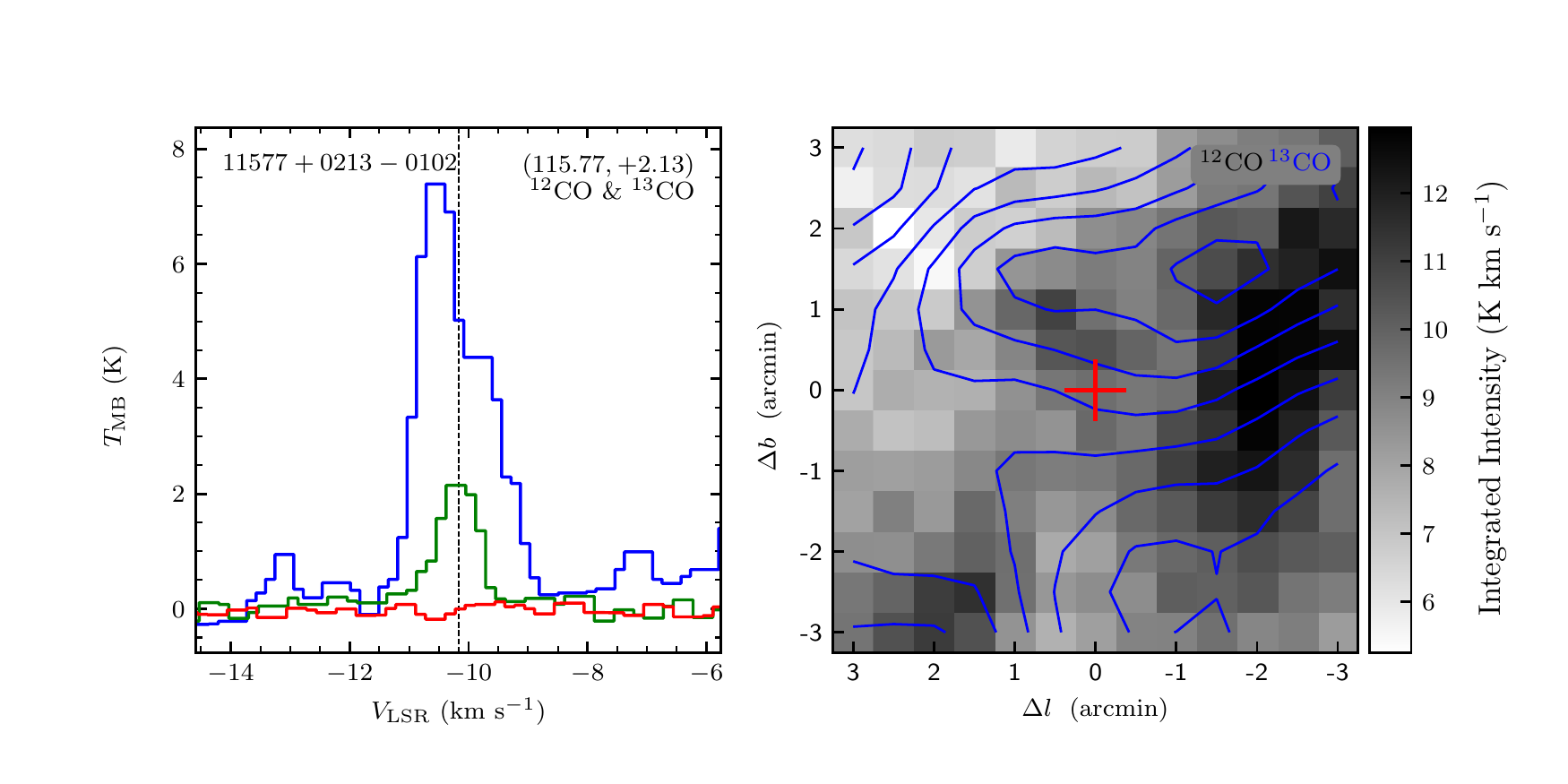}
\includegraphics[width=9.0cm,angle=0]{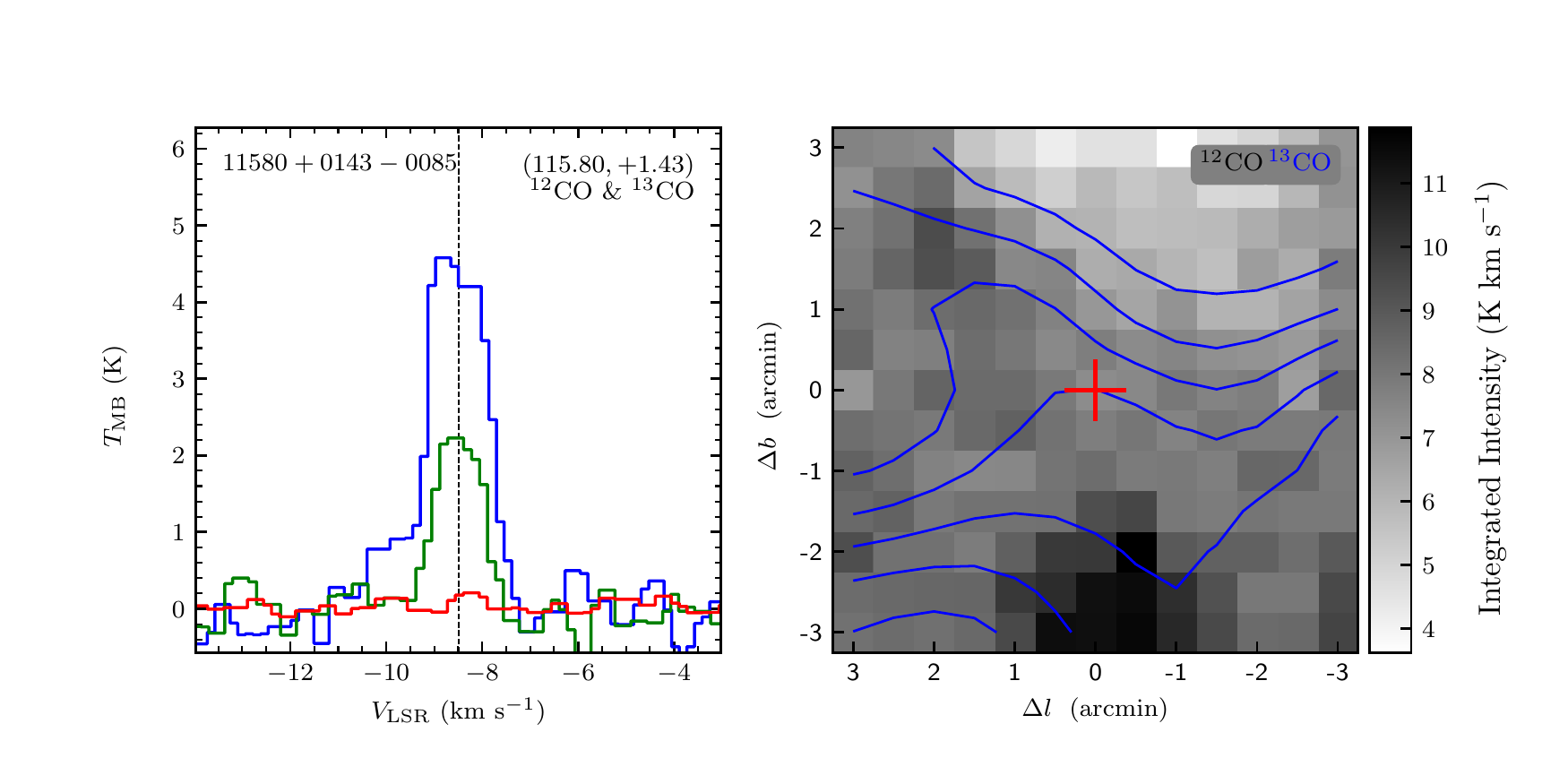}
\vspace{-0.5cm}

\includegraphics[width=9.0cm,angle=0]{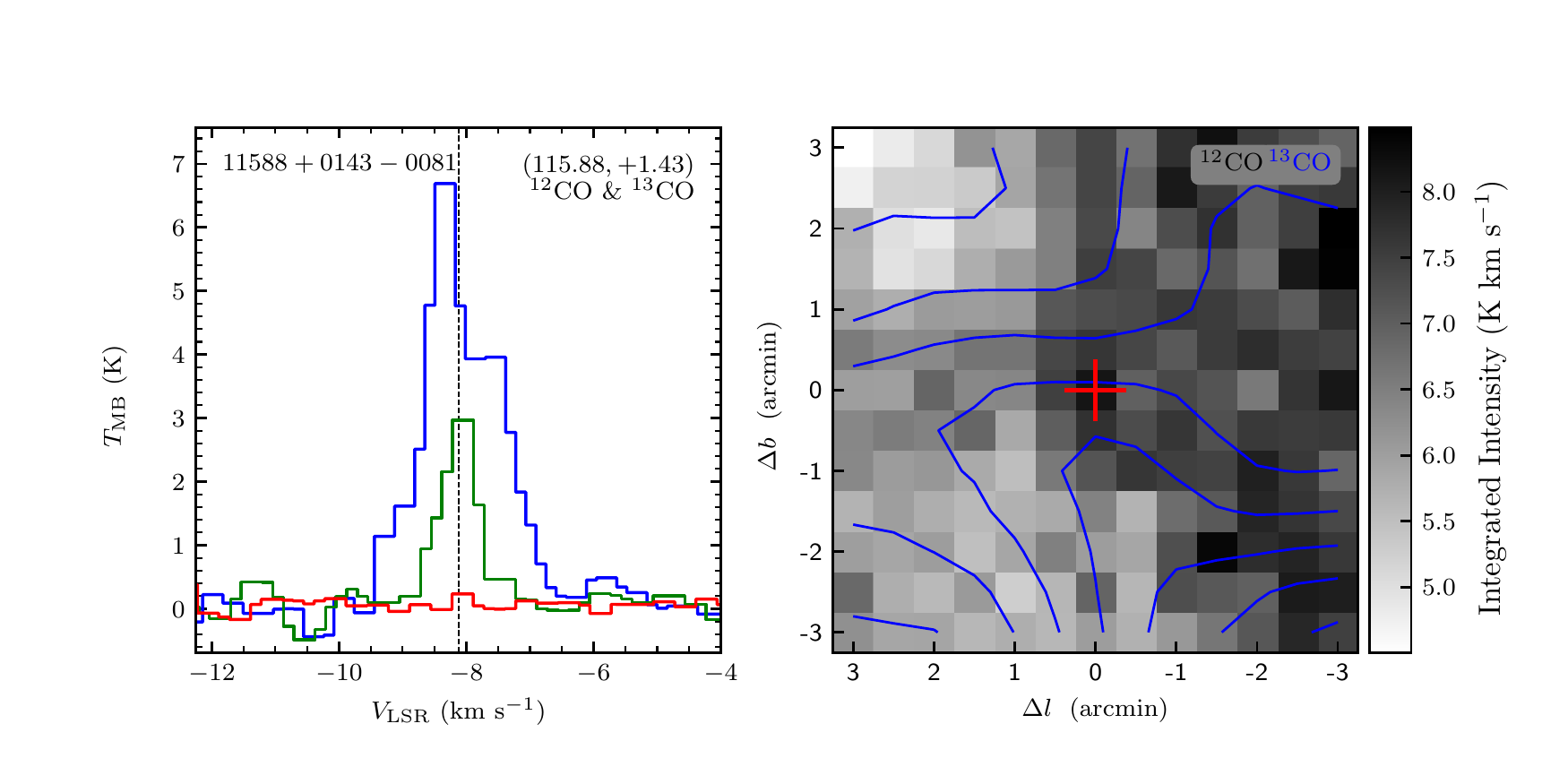}
\includegraphics[width=9.0cm,angle=0]{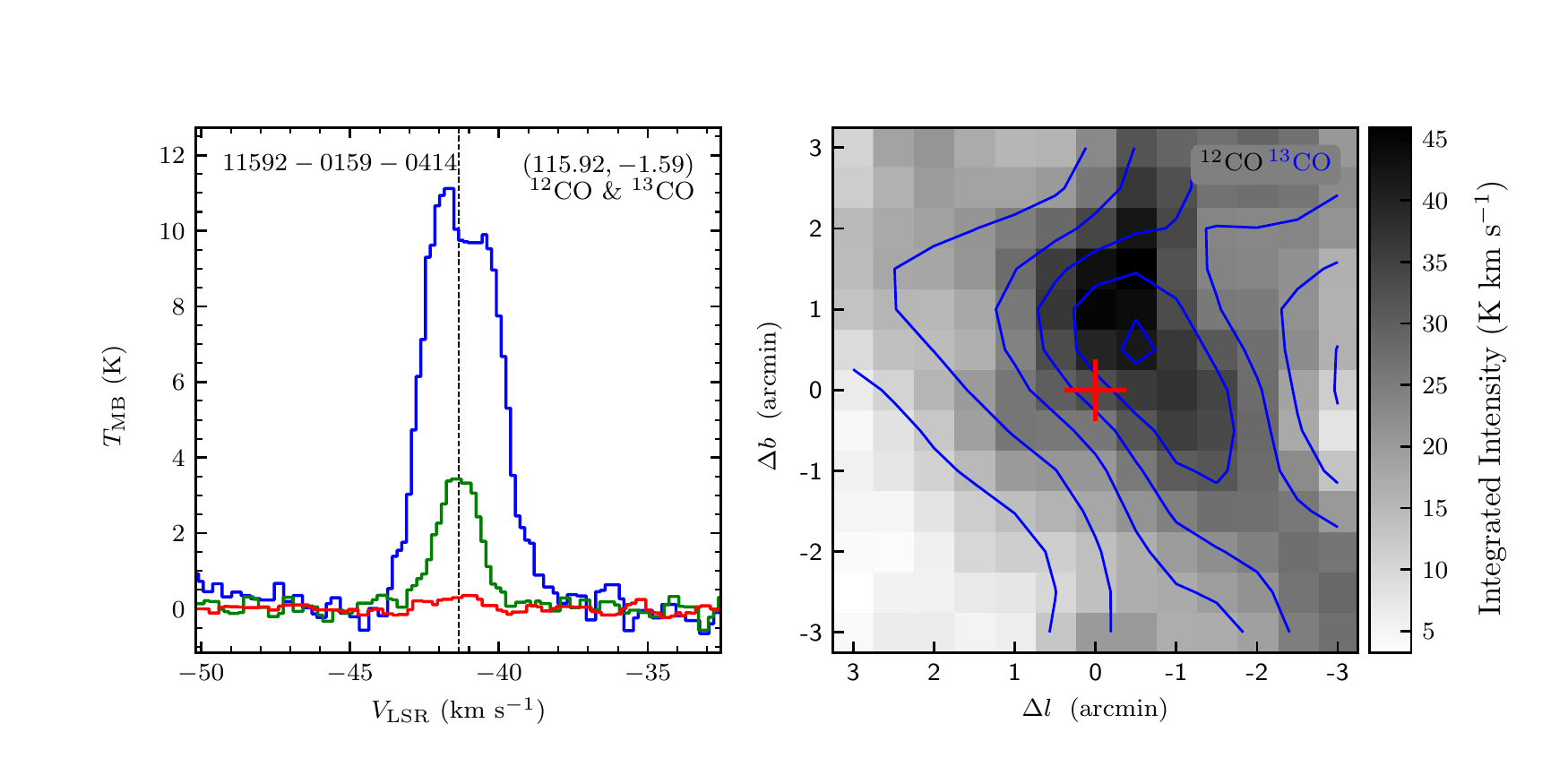}
\vspace{-0.5cm}

\includegraphics[width=9.0cm,angle=0]{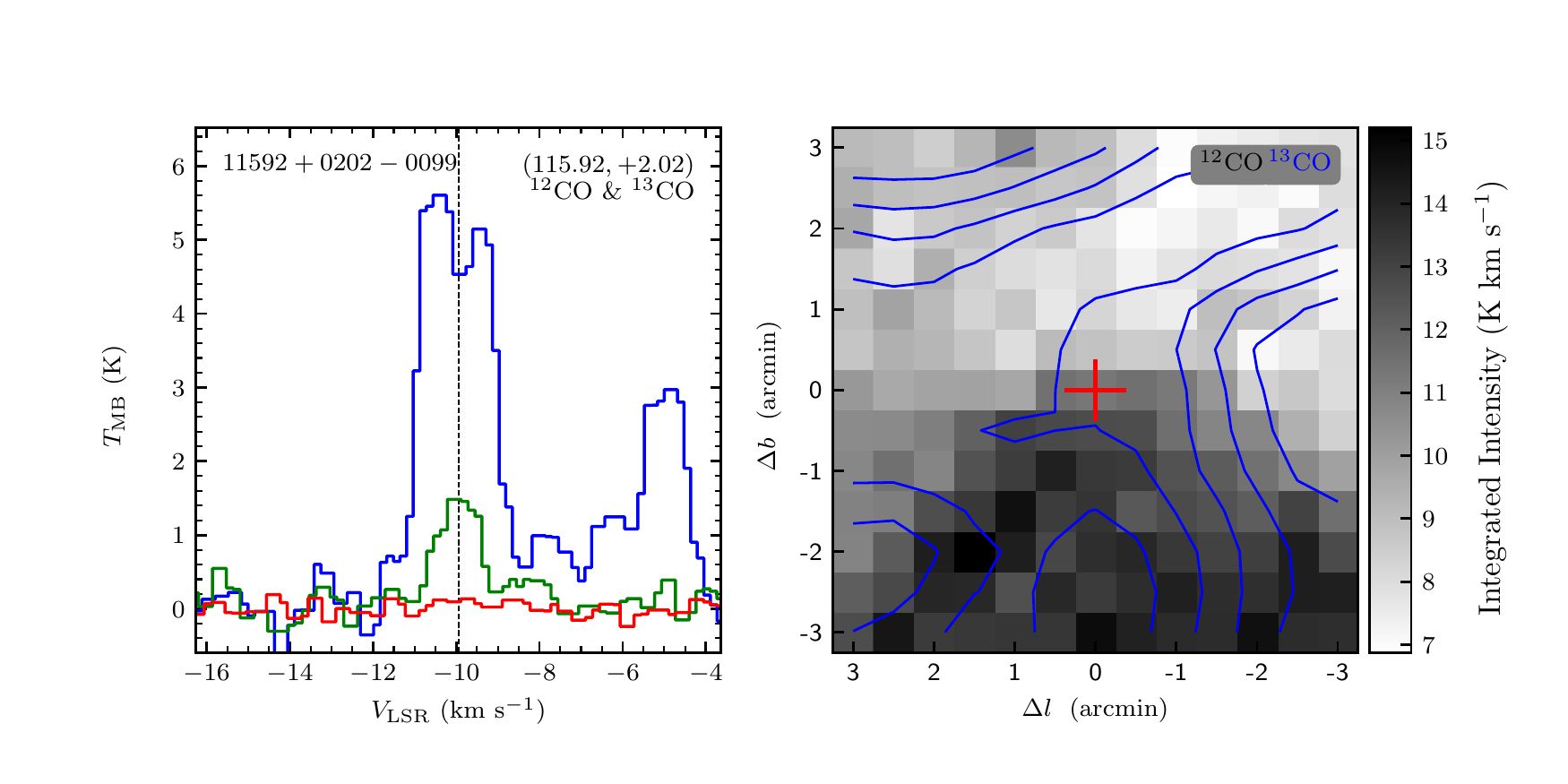}
\includegraphics[width=9.0cm,angle=0]{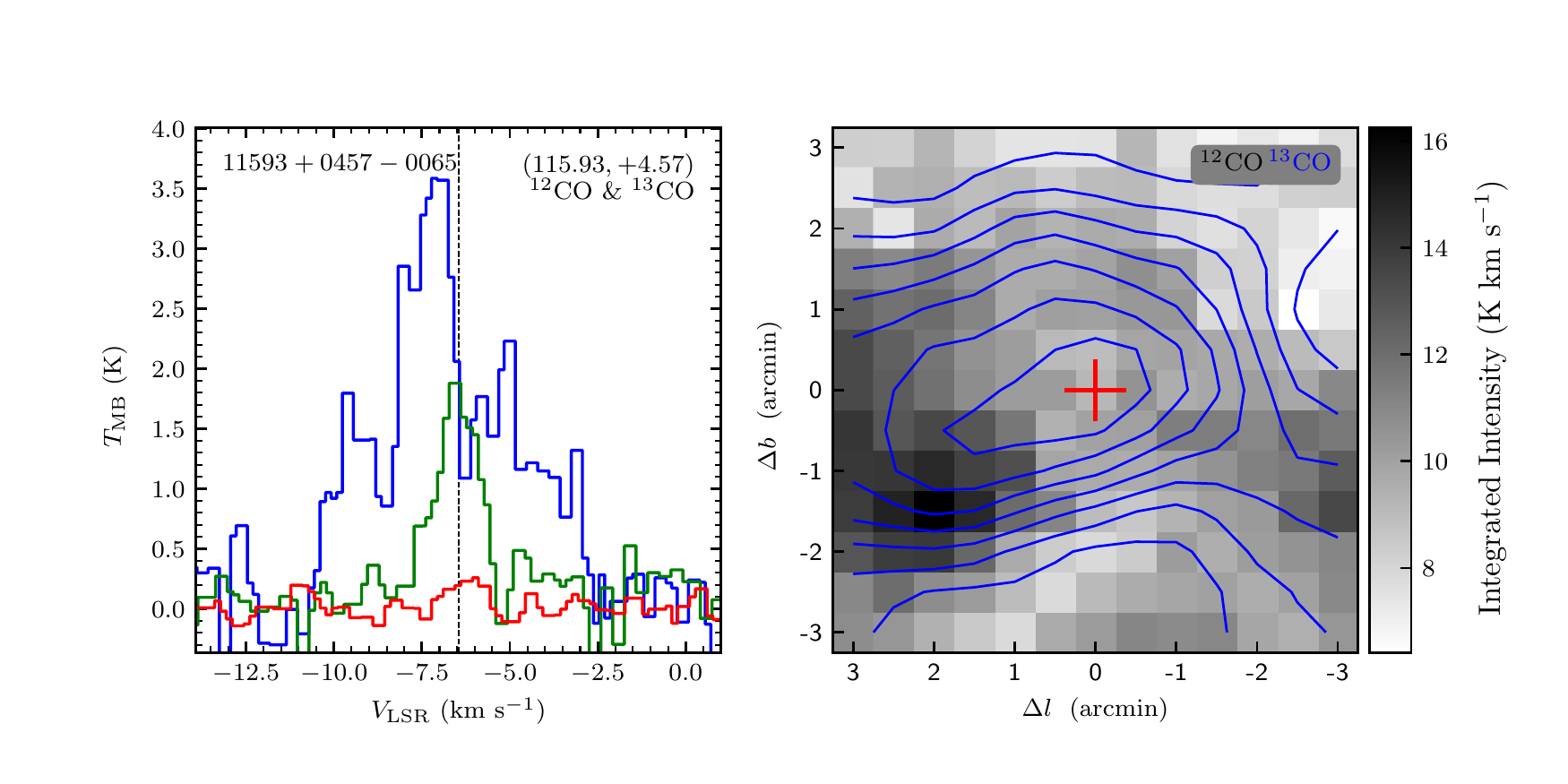}
\vspace{-0.5cm}

\includegraphics[width=9.0cm,angle=0]{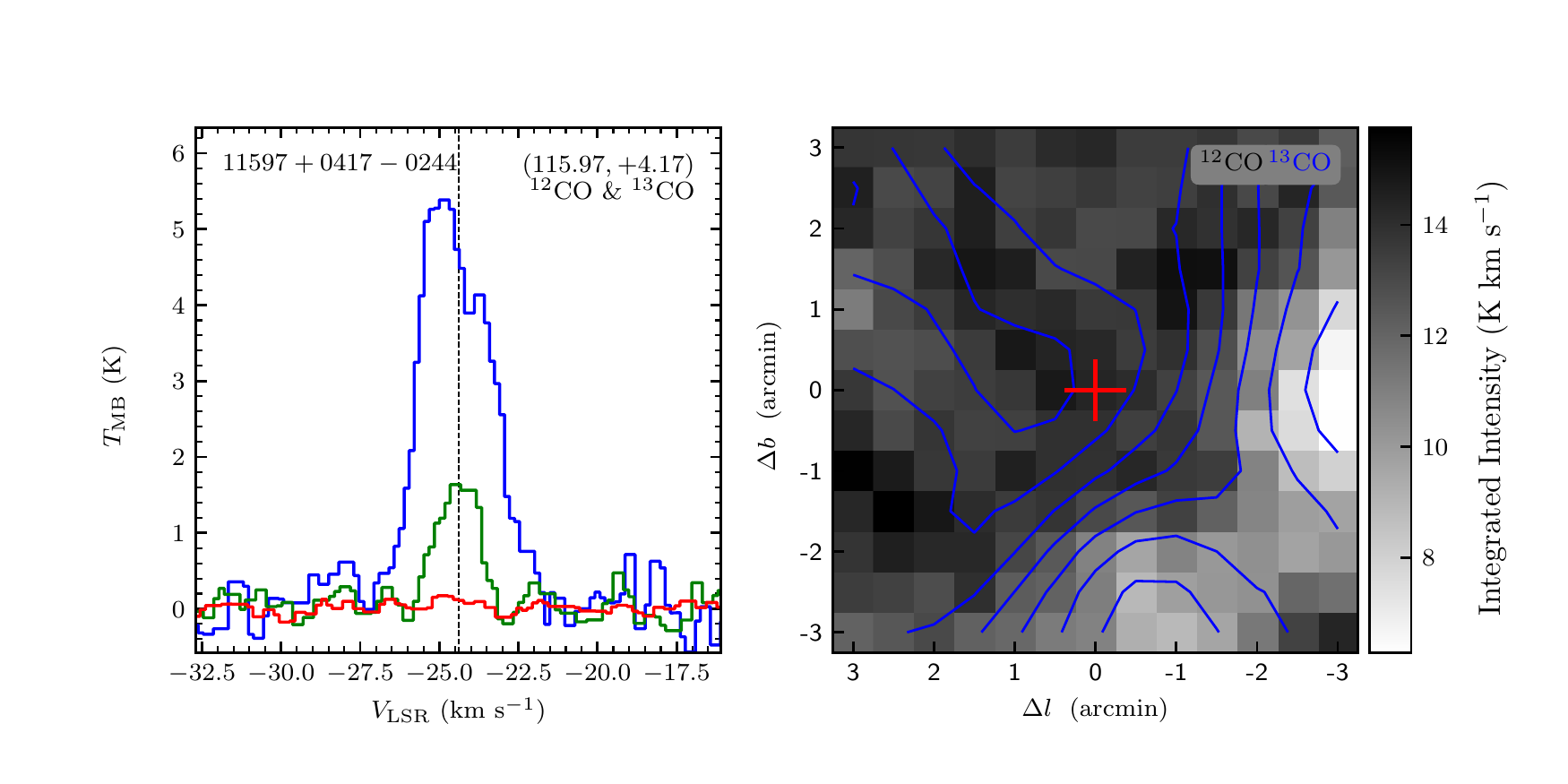}
\includegraphics[width=9.0cm,angle=0]{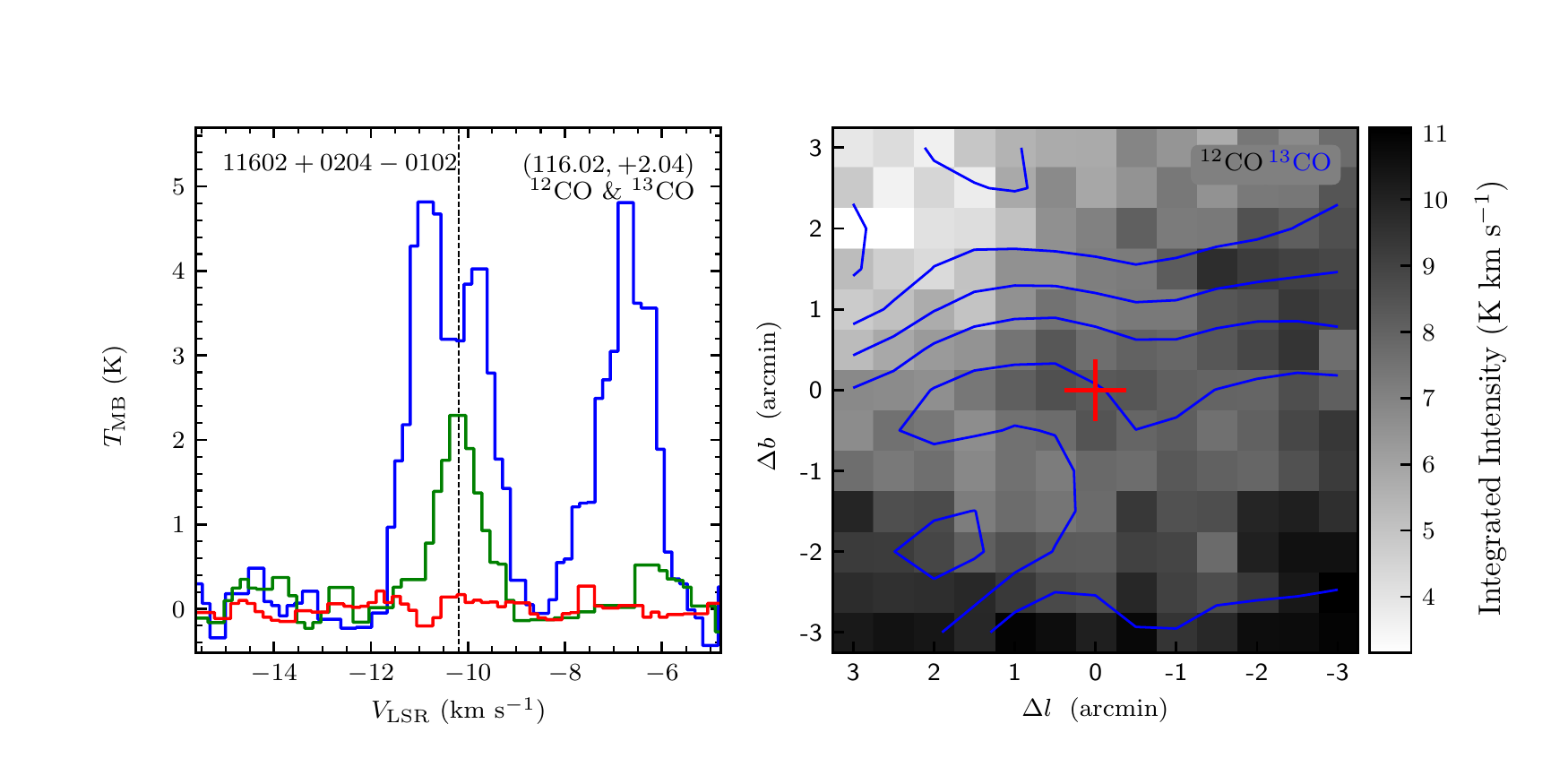}
\end{figure}
\clearpage

\begin{figure}
\includegraphics[width=9.0cm,angle=0]{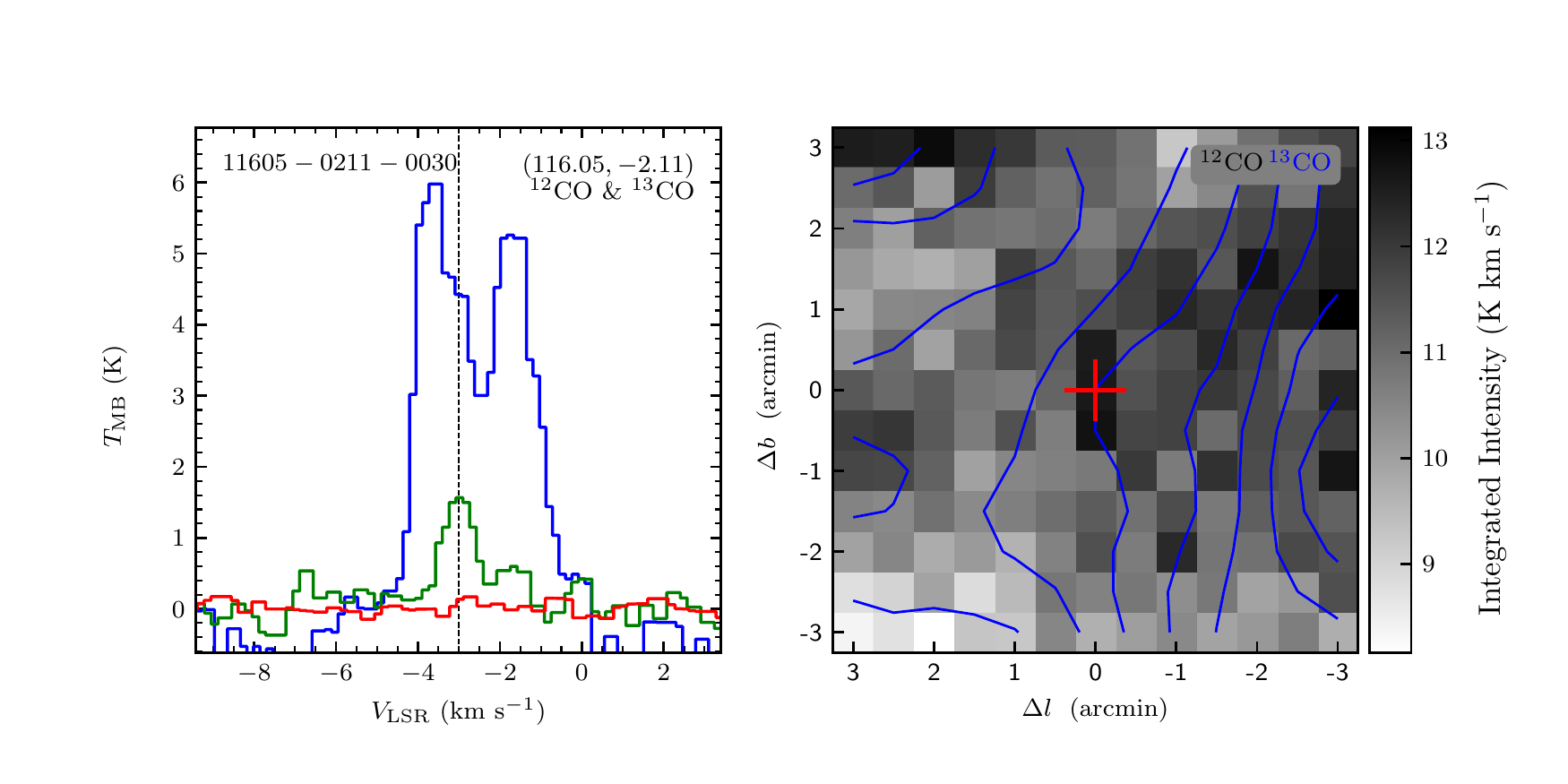}
\includegraphics[width=9.0cm,angle=0]{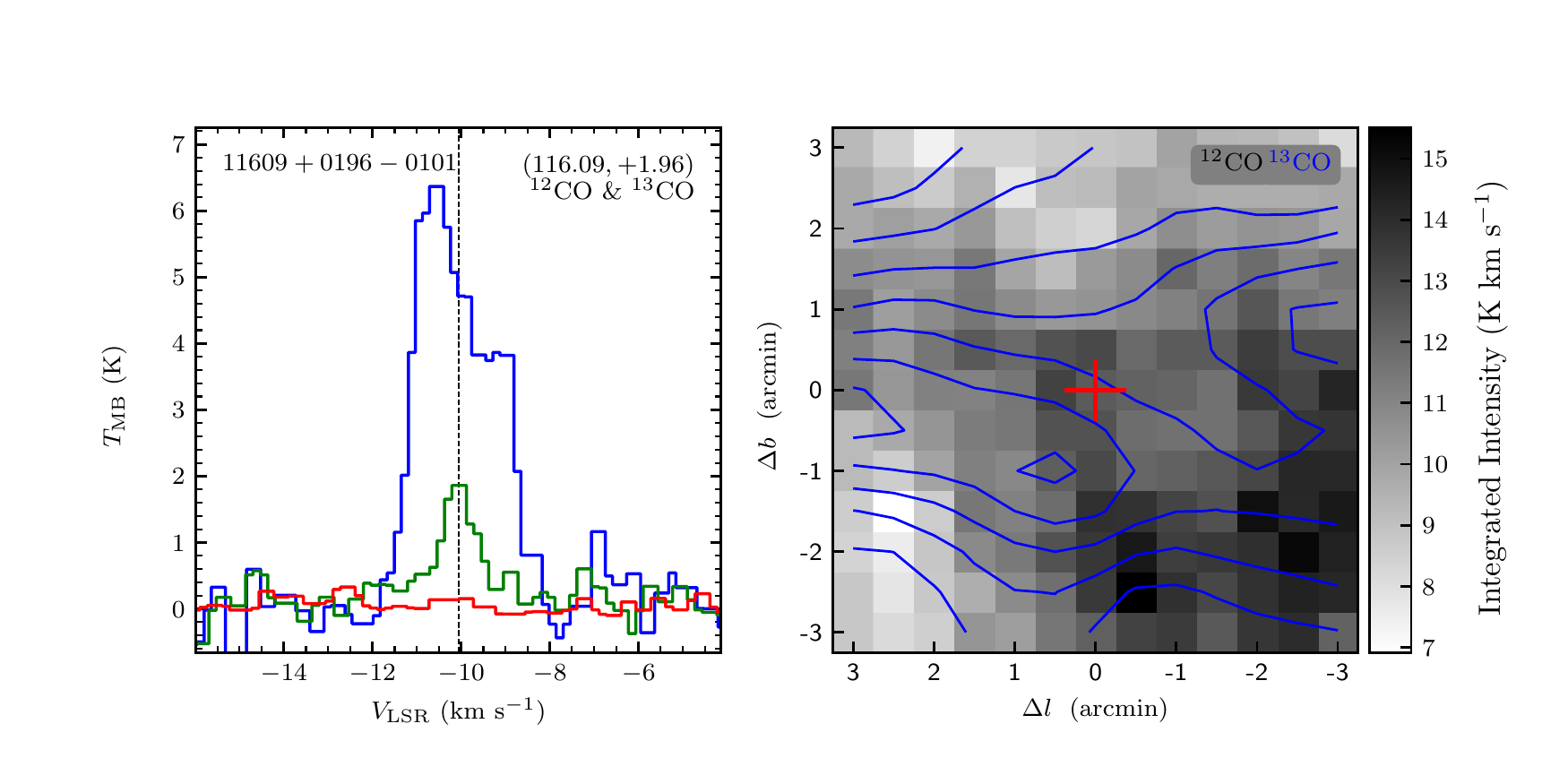}
\vspace{-0.5cm}

\includegraphics[width=9.0cm,angle=0]{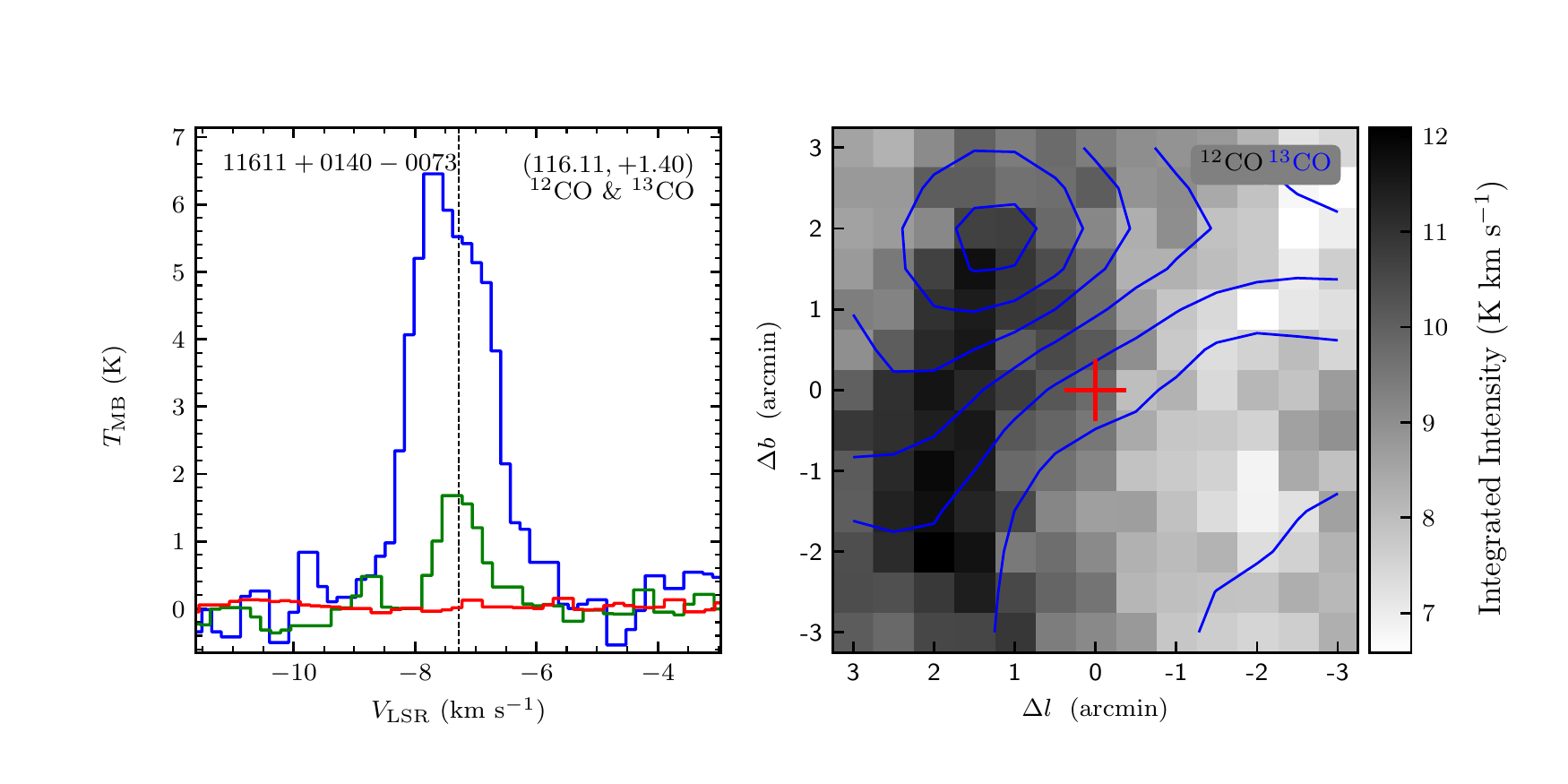}
\includegraphics[width=9.0cm,angle=0]{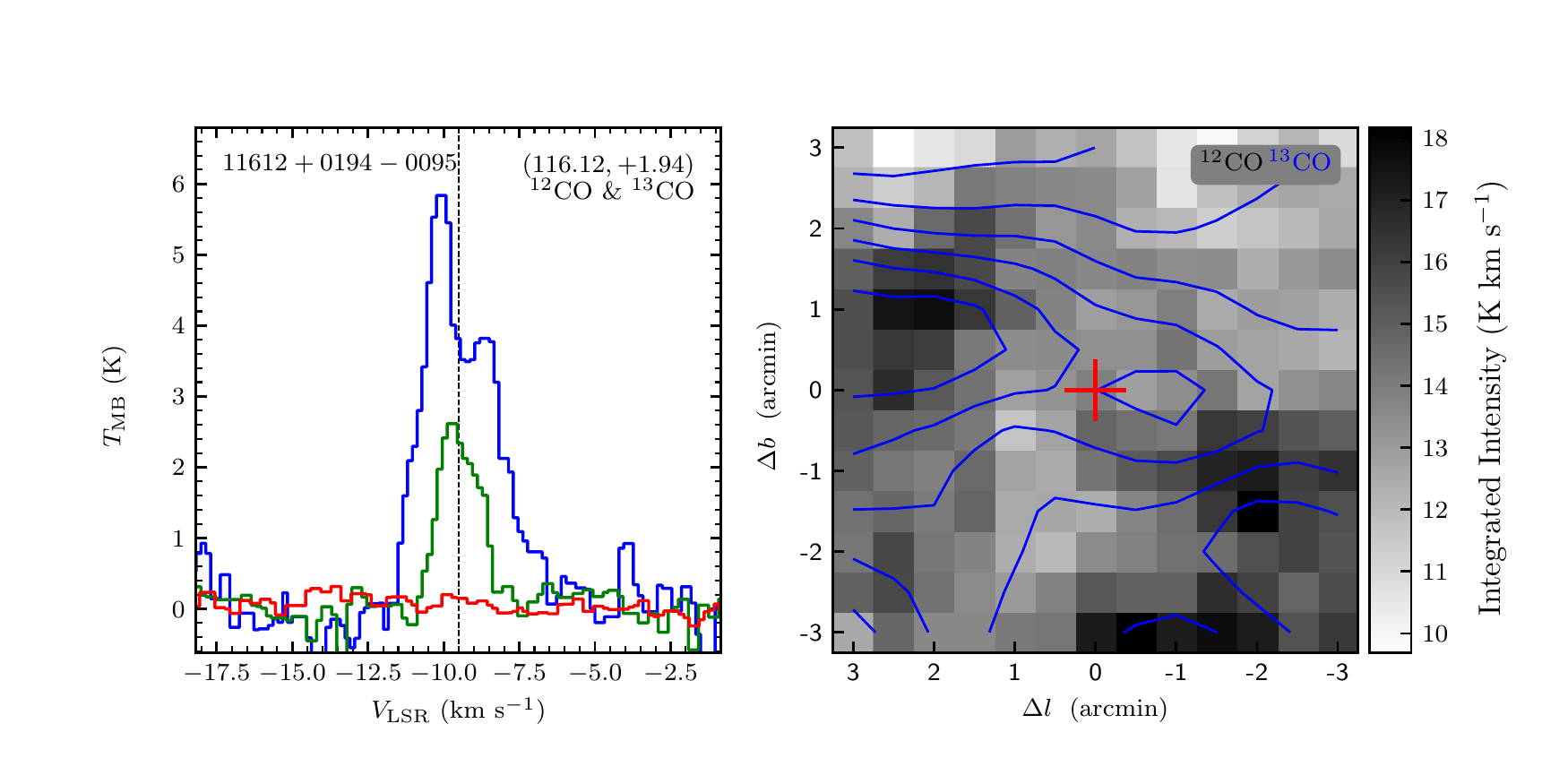}
\vspace{-0.5cm}

\includegraphics[width=9.0cm,angle=0]{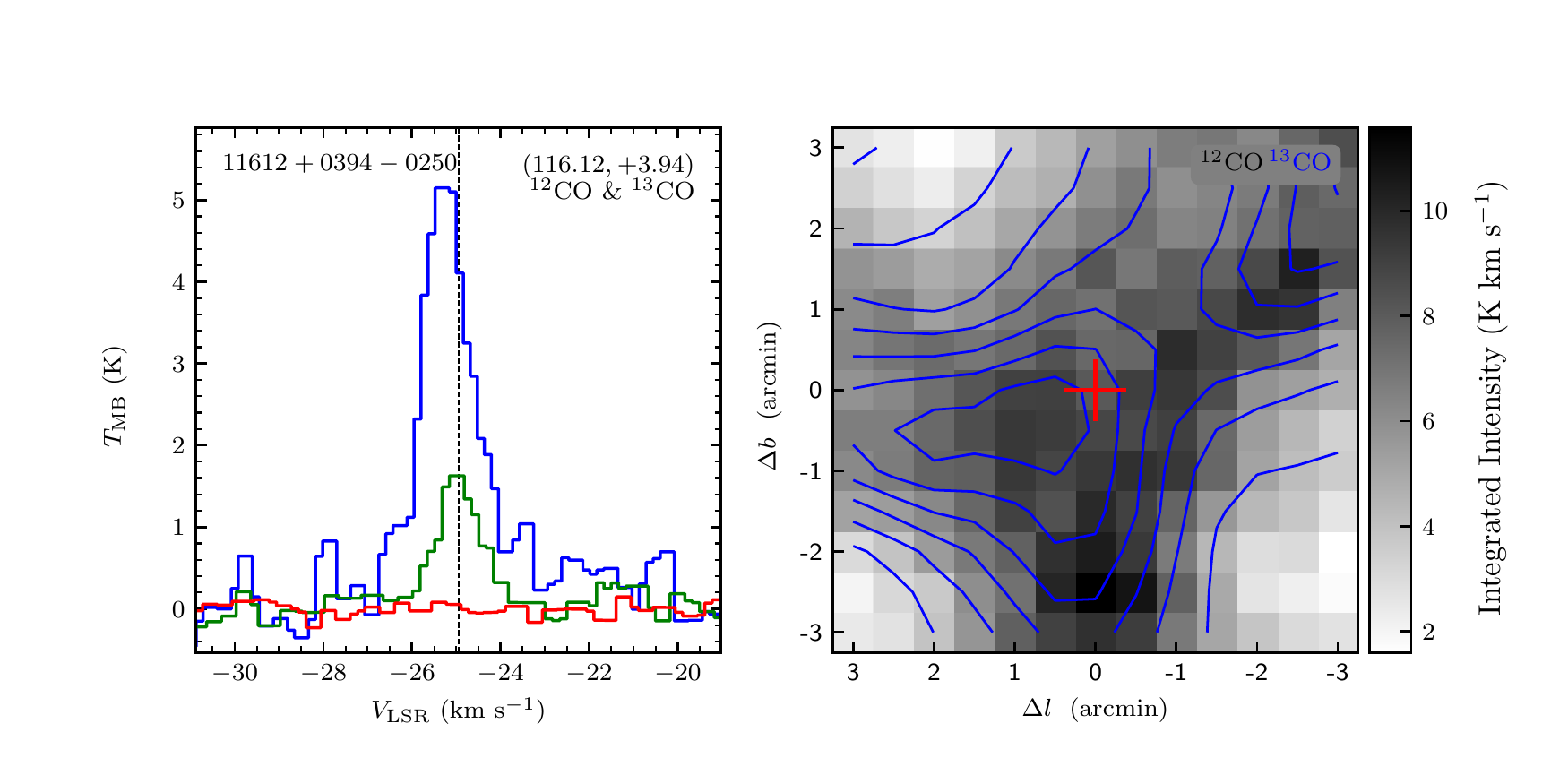}
\includegraphics[width=9.0cm,angle=0]{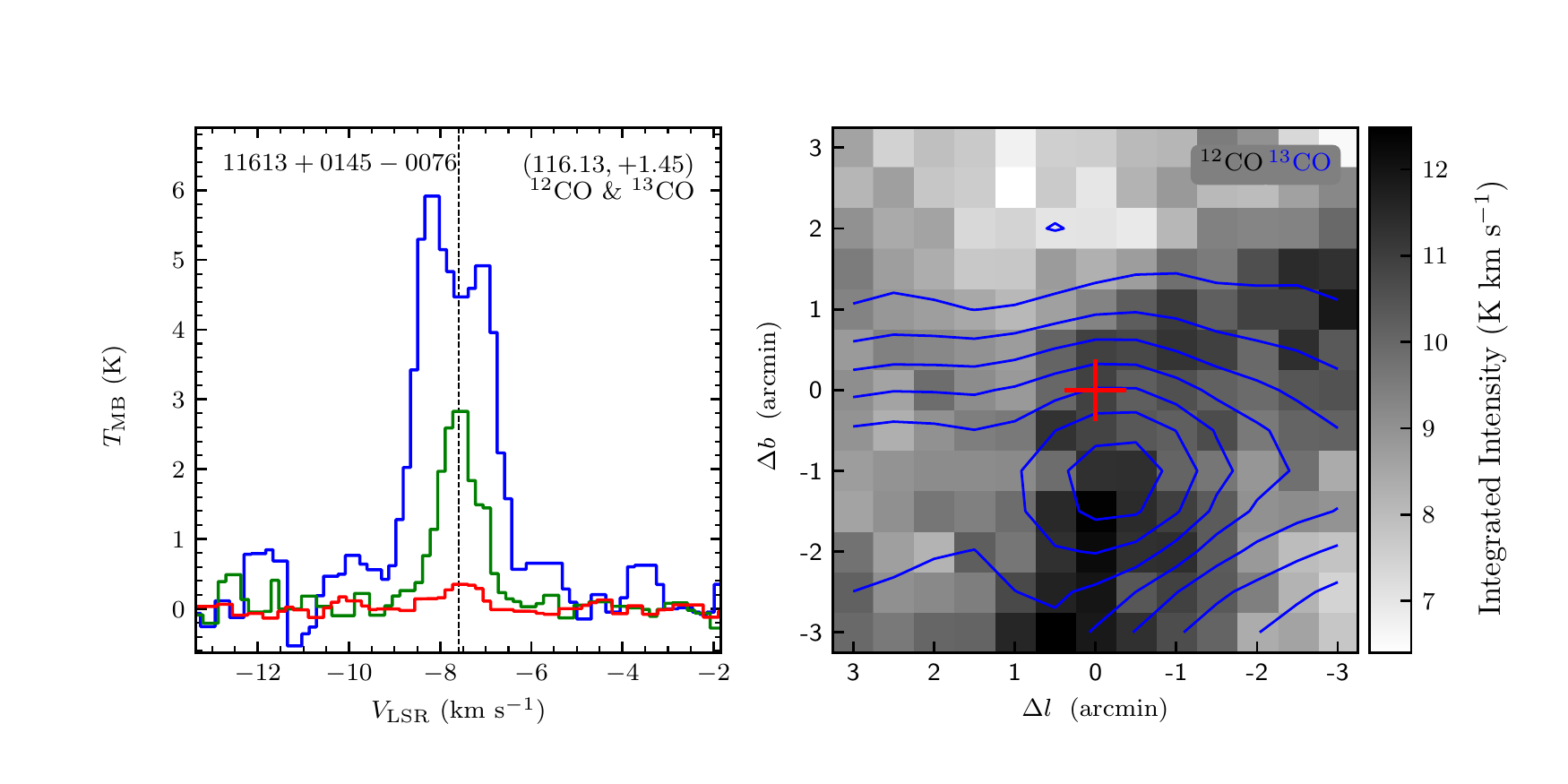}
\vspace{-0.5cm}

\includegraphics[width=9.0cm,angle=0]{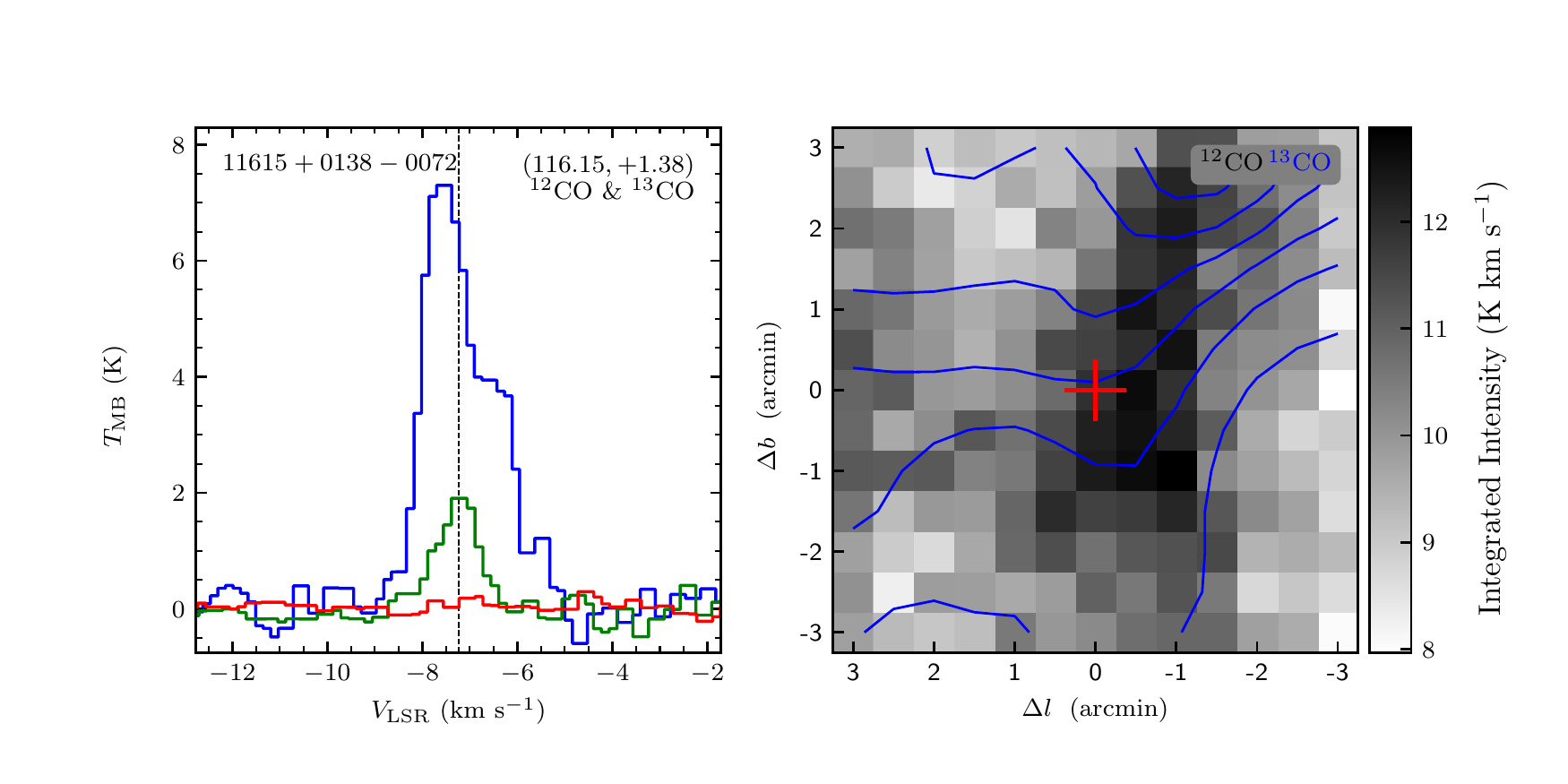}
\includegraphics[width=9.0cm,angle=0]{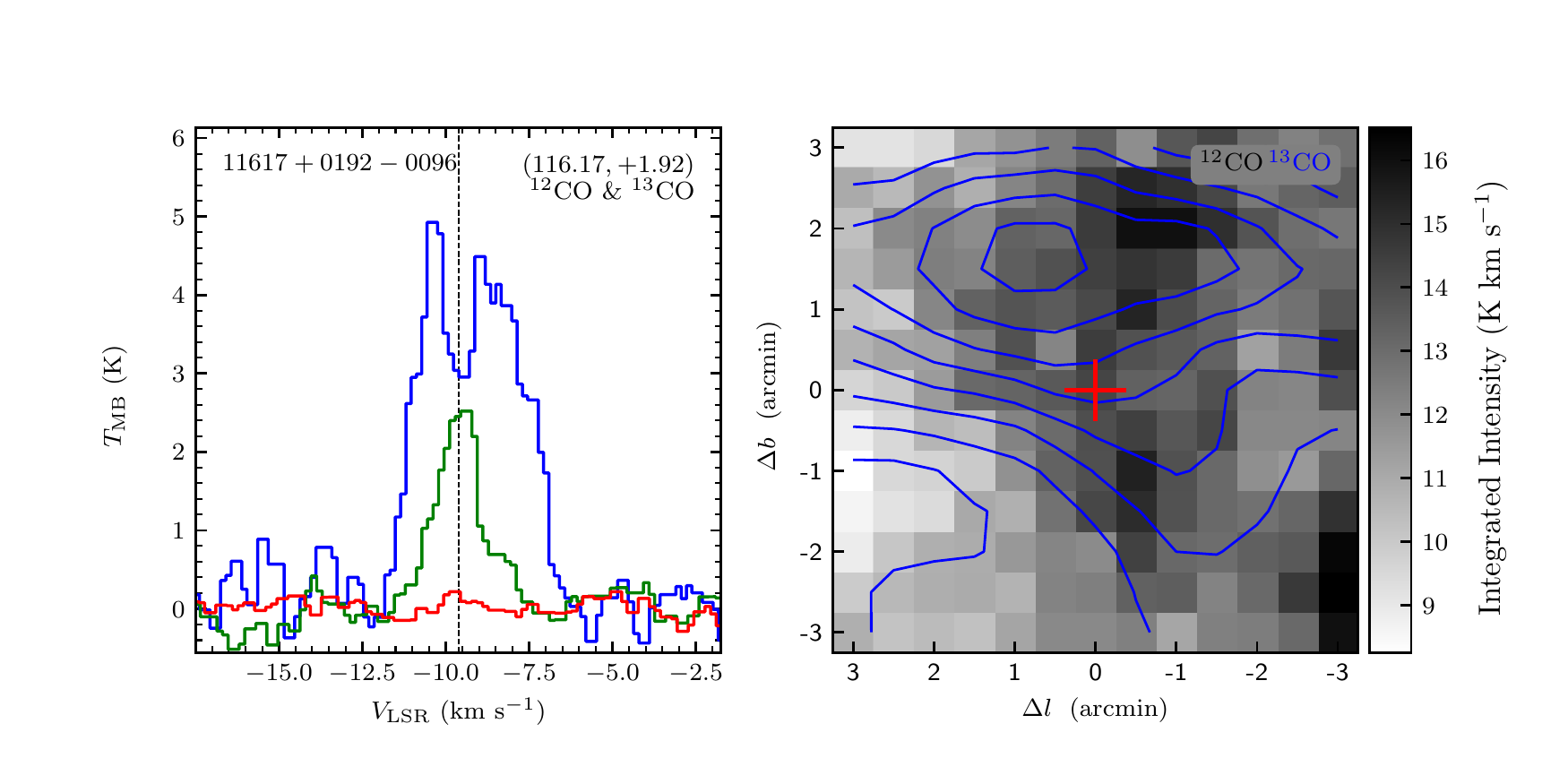}
\vspace{-0.5cm}

\includegraphics[width=9.0cm,angle=0]{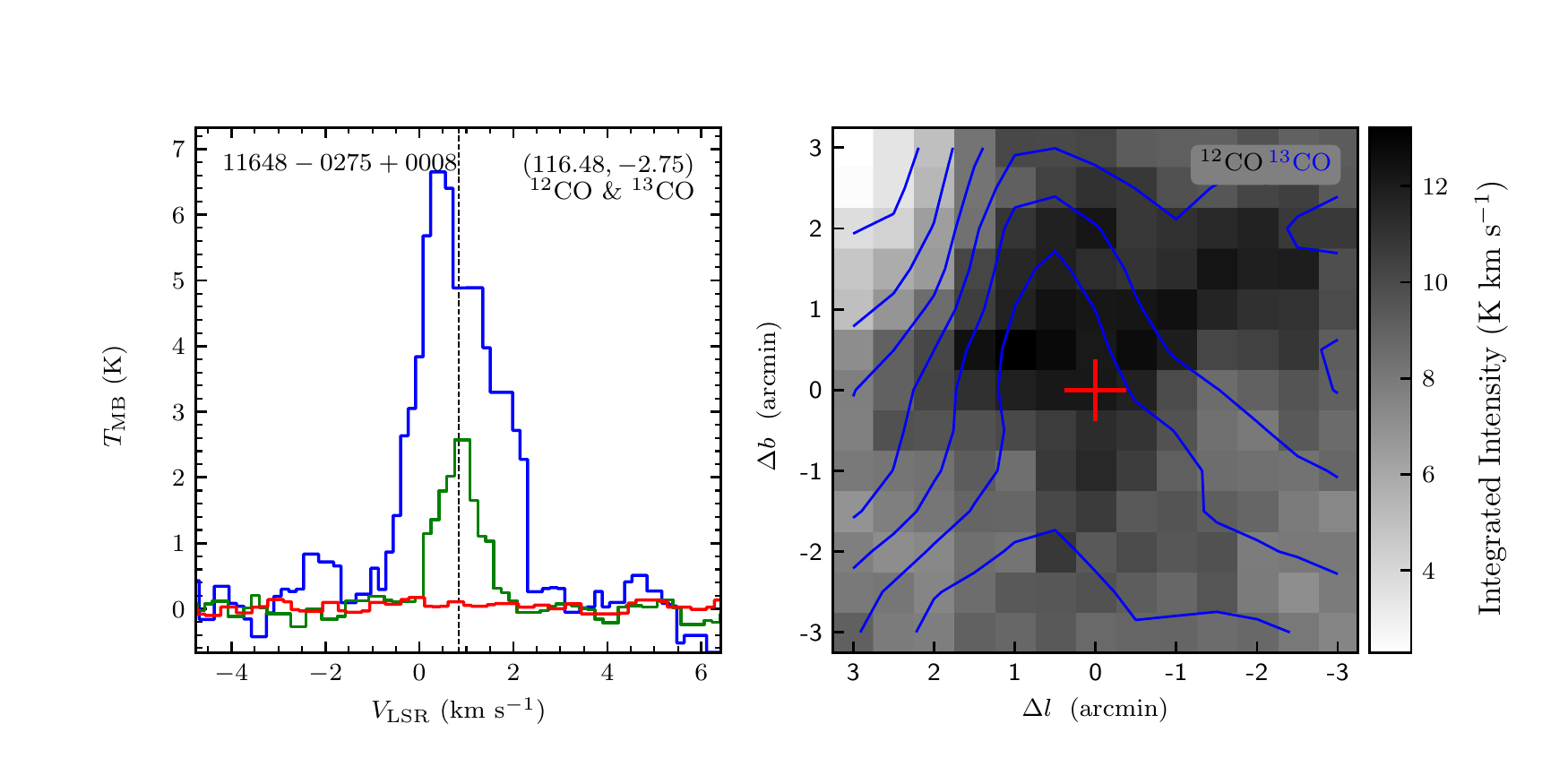}
\includegraphics[width=9.0cm,angle=0]{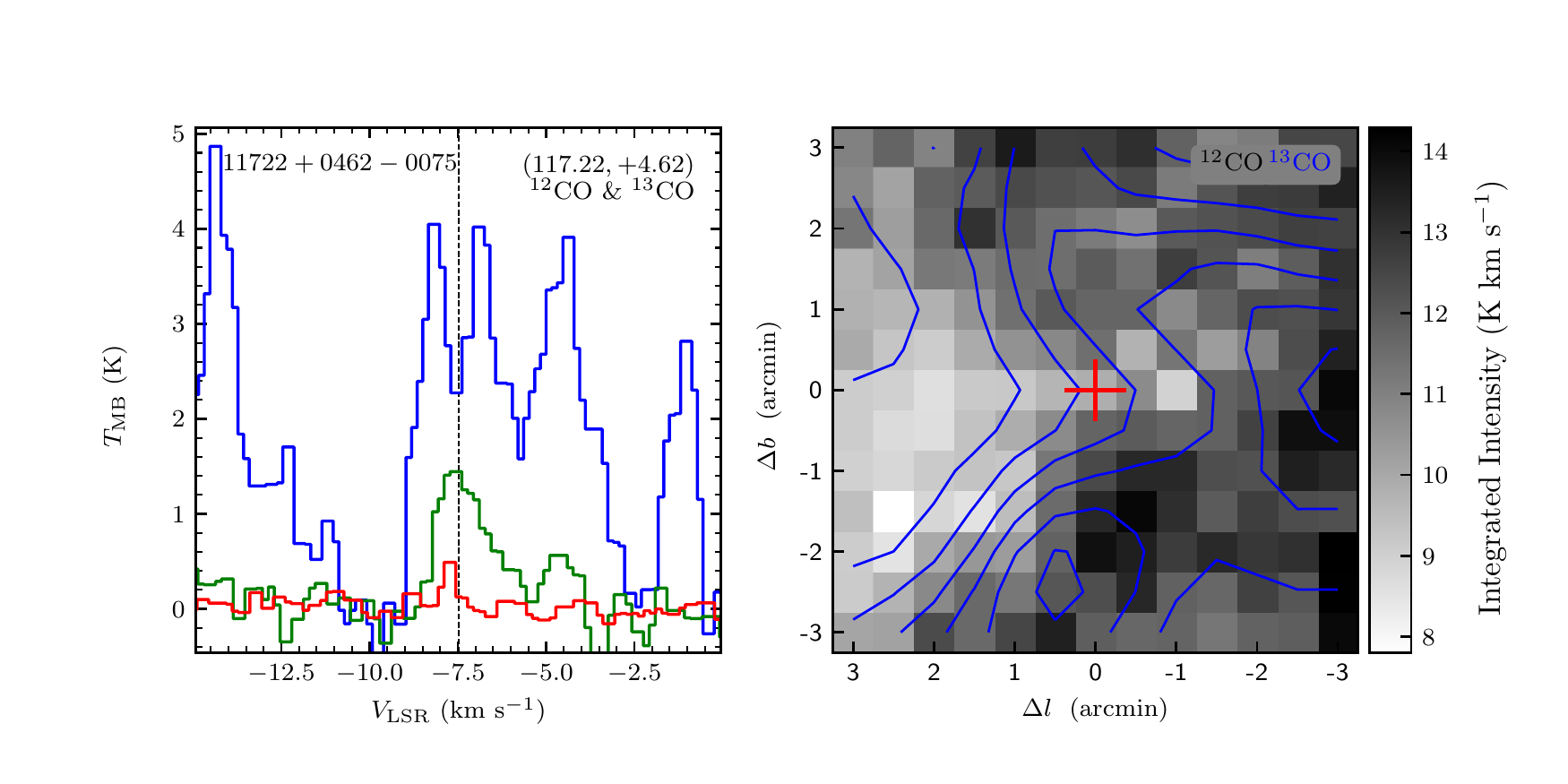}
\end{figure}
\clearpage

\begin{figure}
\includegraphics[width=9.0cm,angle=0]{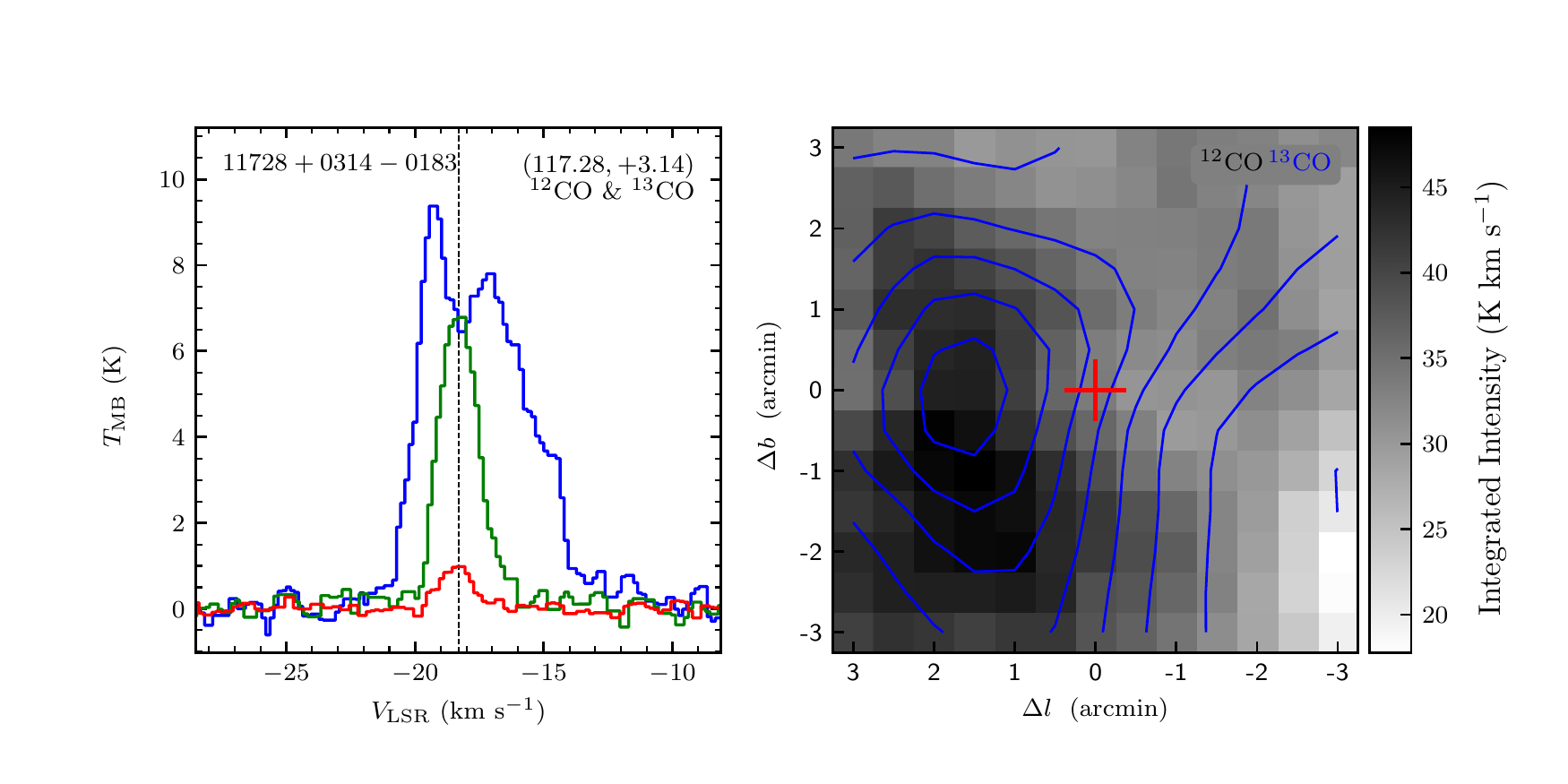}
\includegraphics[width=9.0cm,angle=0]{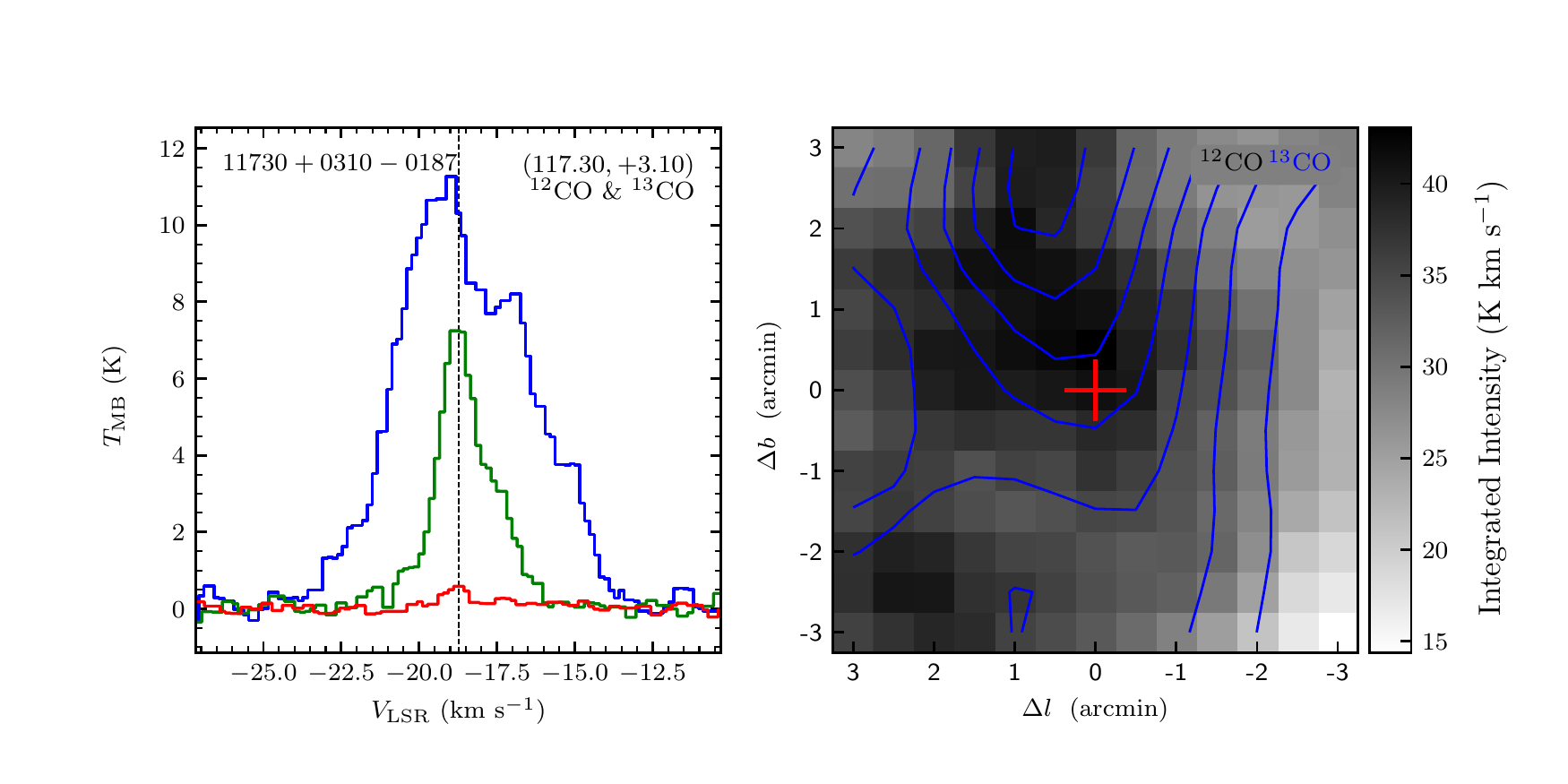}
\vspace{-0.5cm}

\includegraphics[width=9.0cm,angle=0]{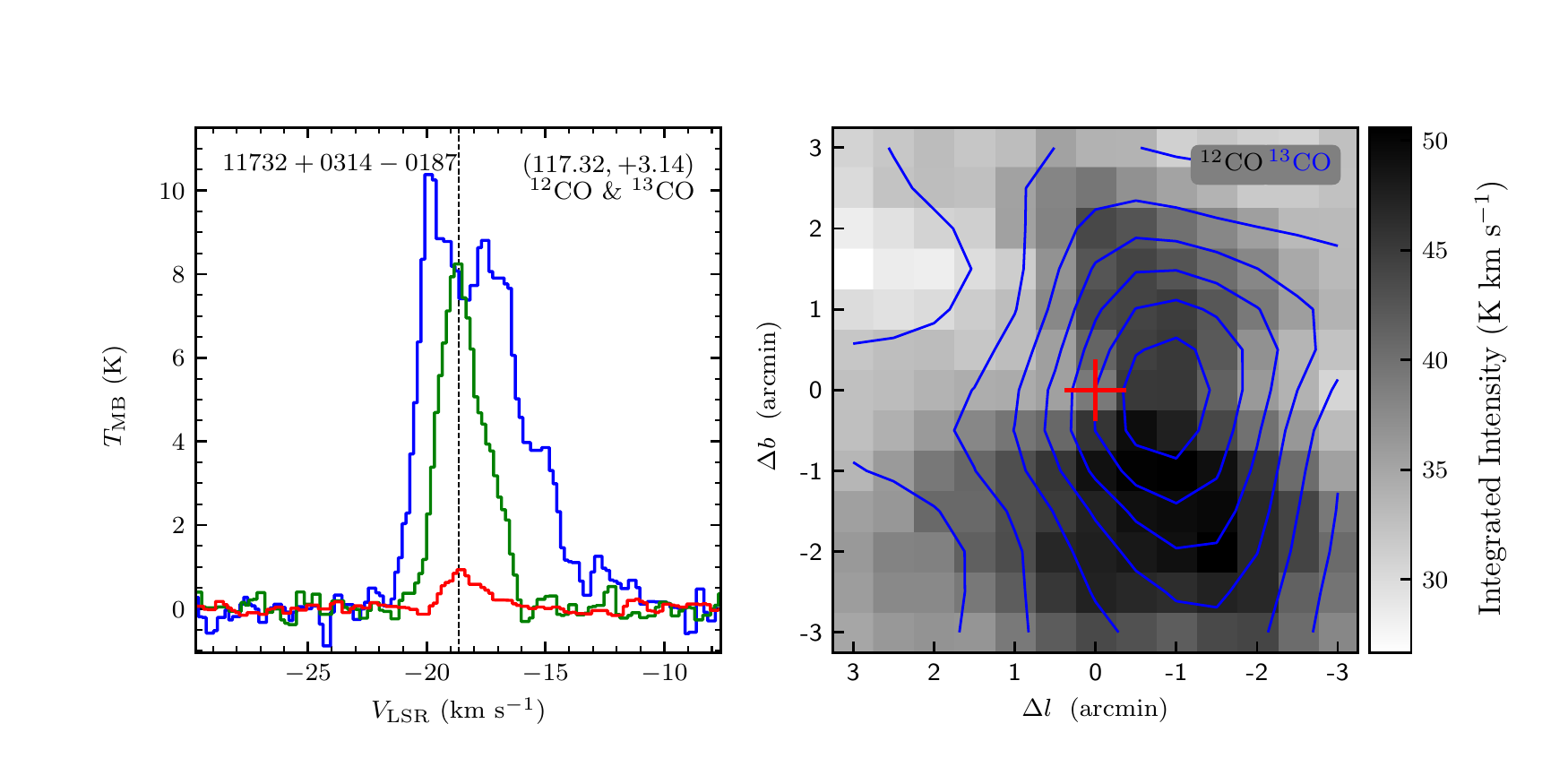}
\includegraphics[width=9.0cm,angle=0]{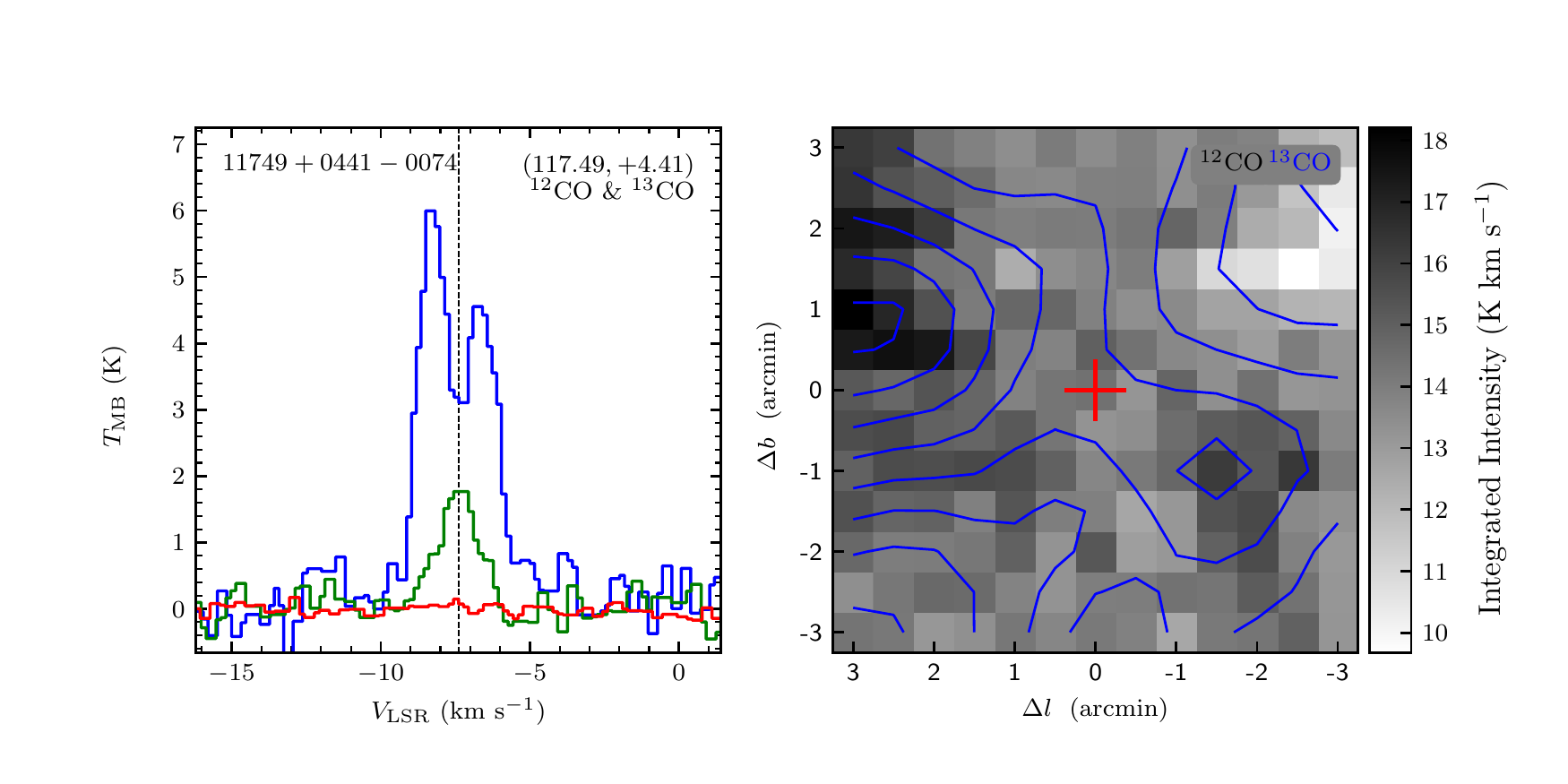}
\vspace{-0.5cm}

\includegraphics[width=9.0cm,angle=0]{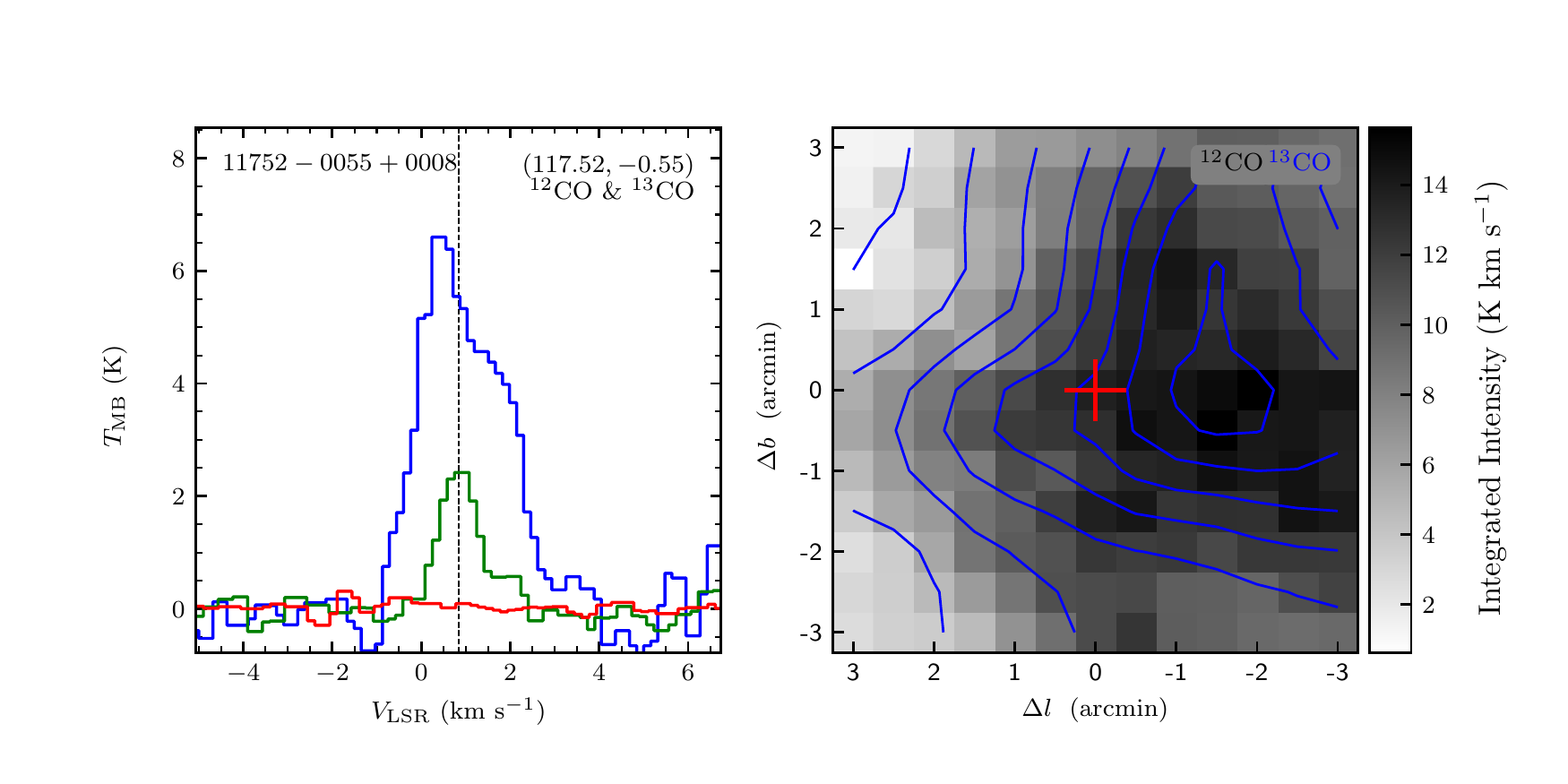}
\includegraphics[width=9.0cm,angle=0]{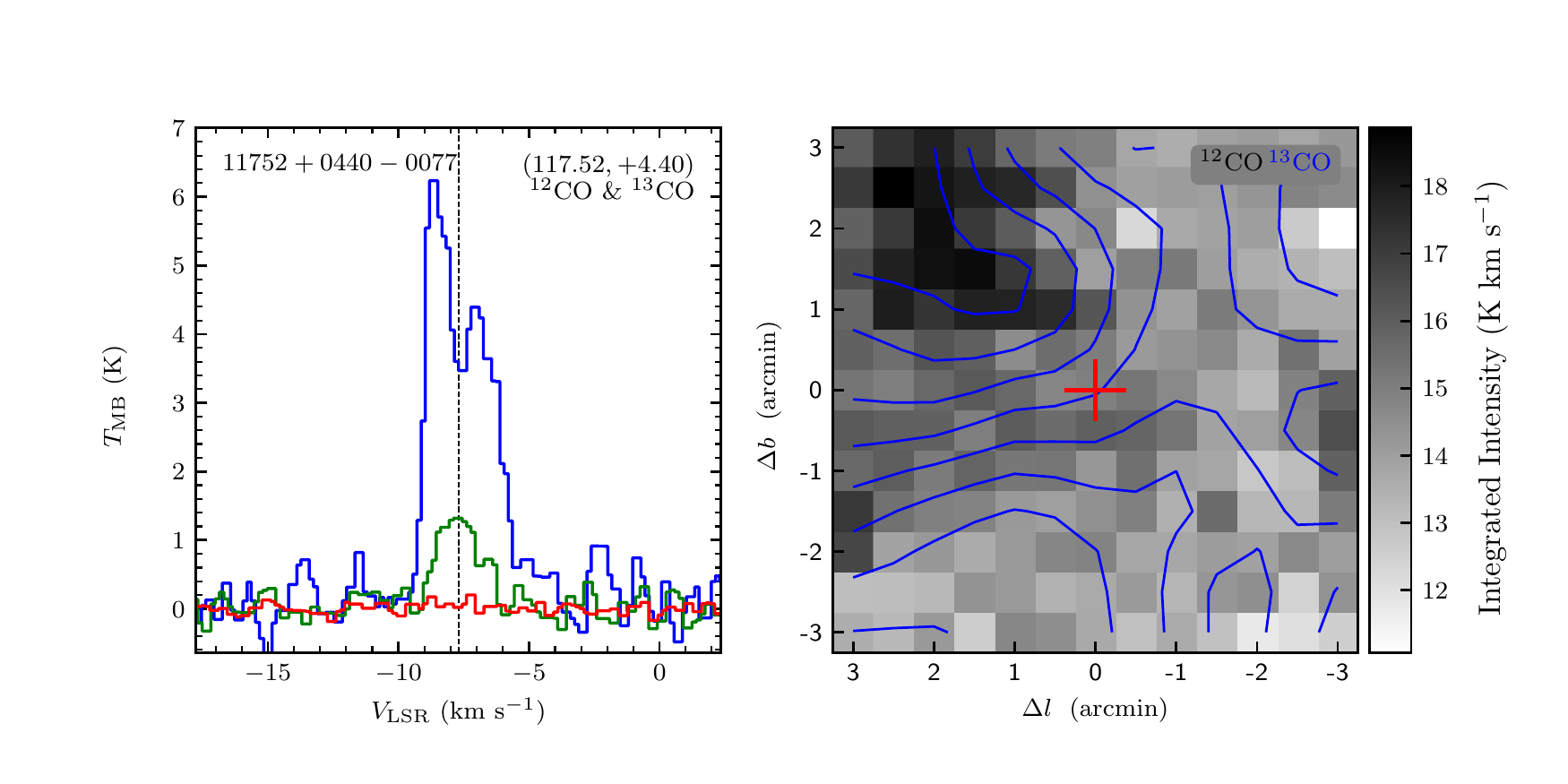}
\vspace{-0.5cm}

\includegraphics[width=9.0cm,angle=0]{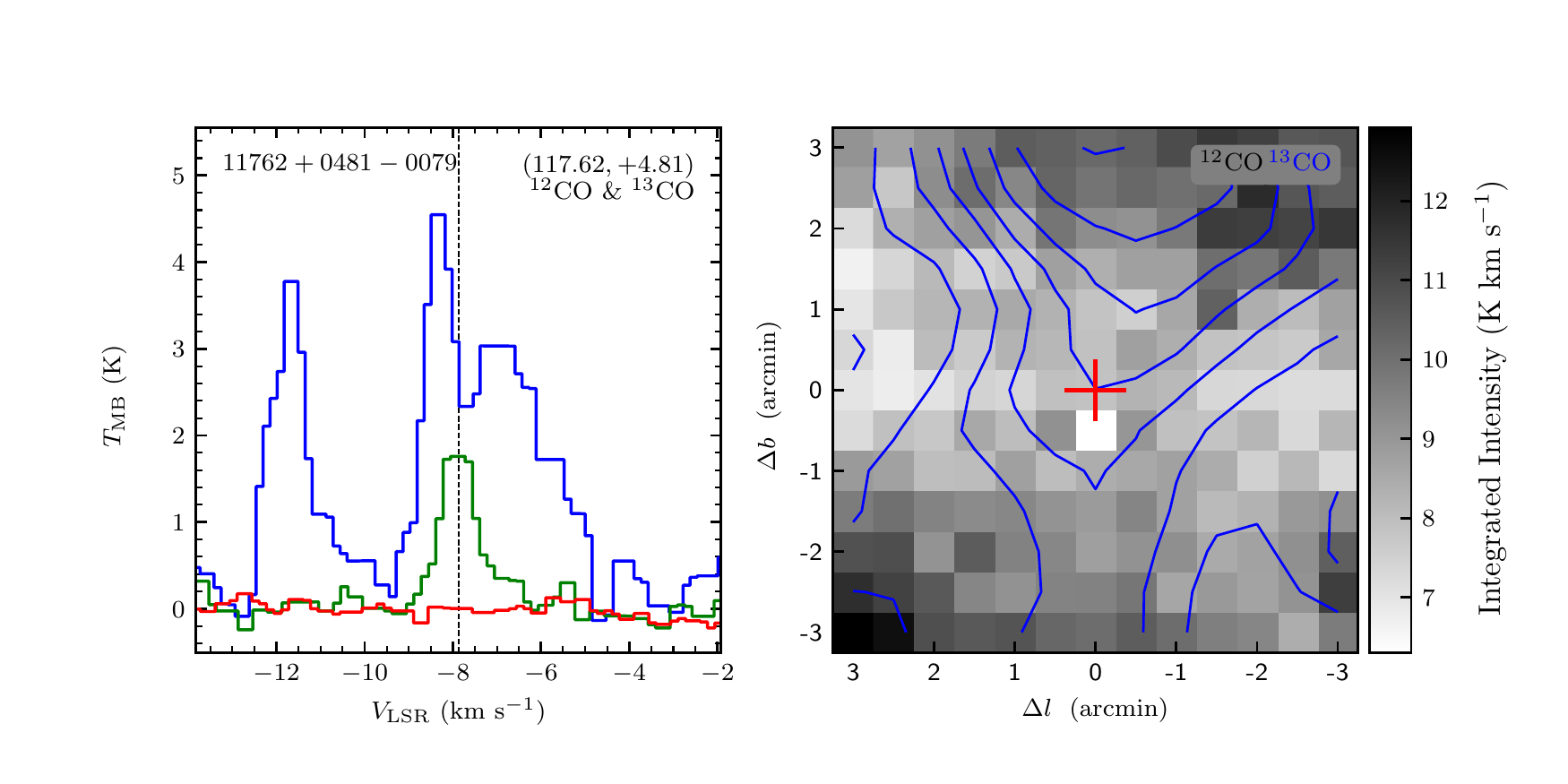}
\includegraphics[width=9.0cm,angle=0]{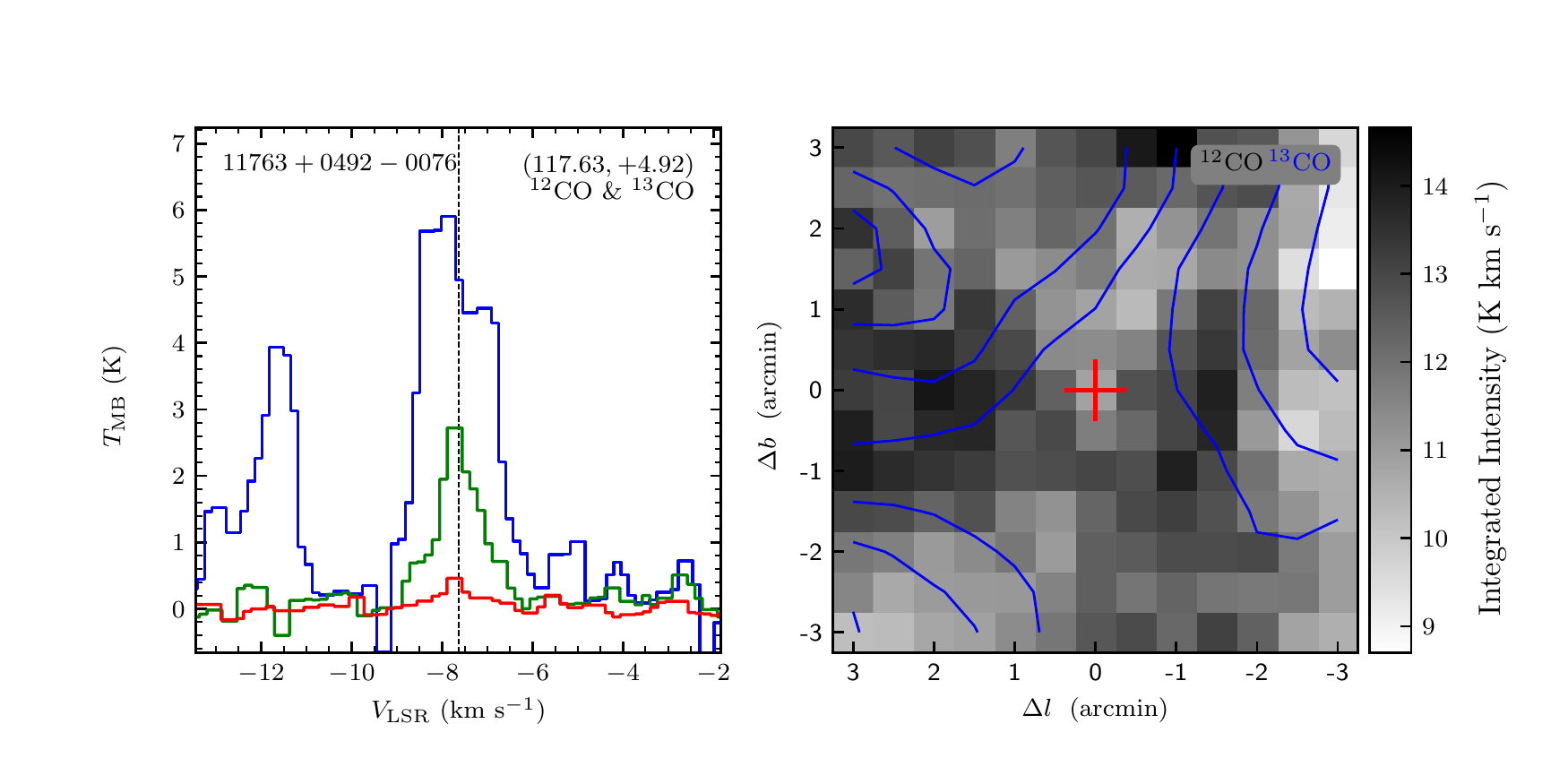}
\vspace{-0.5cm}

\includegraphics[width=9.0cm,angle=0]{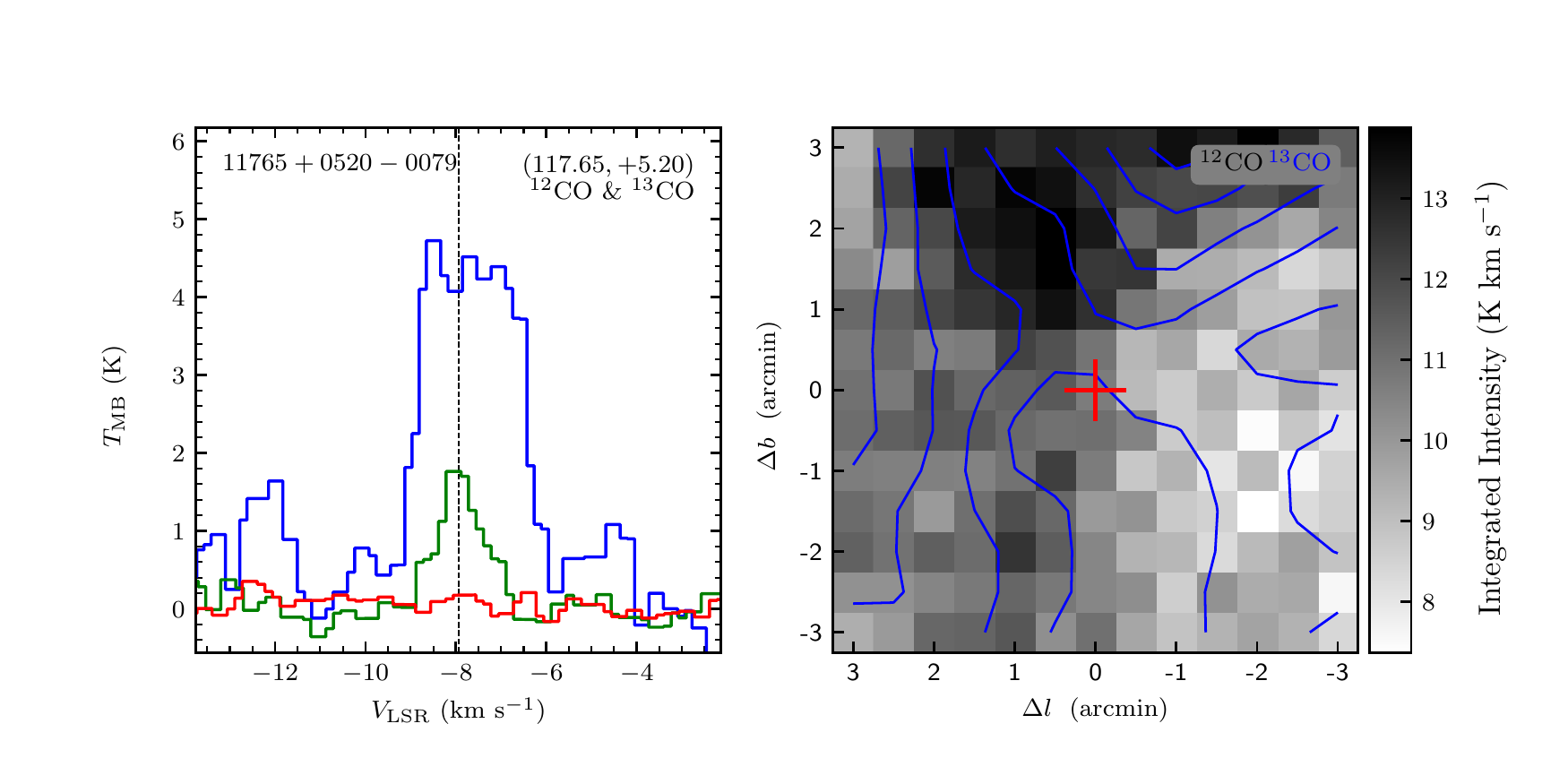}
\includegraphics[width=9.0cm,angle=0]{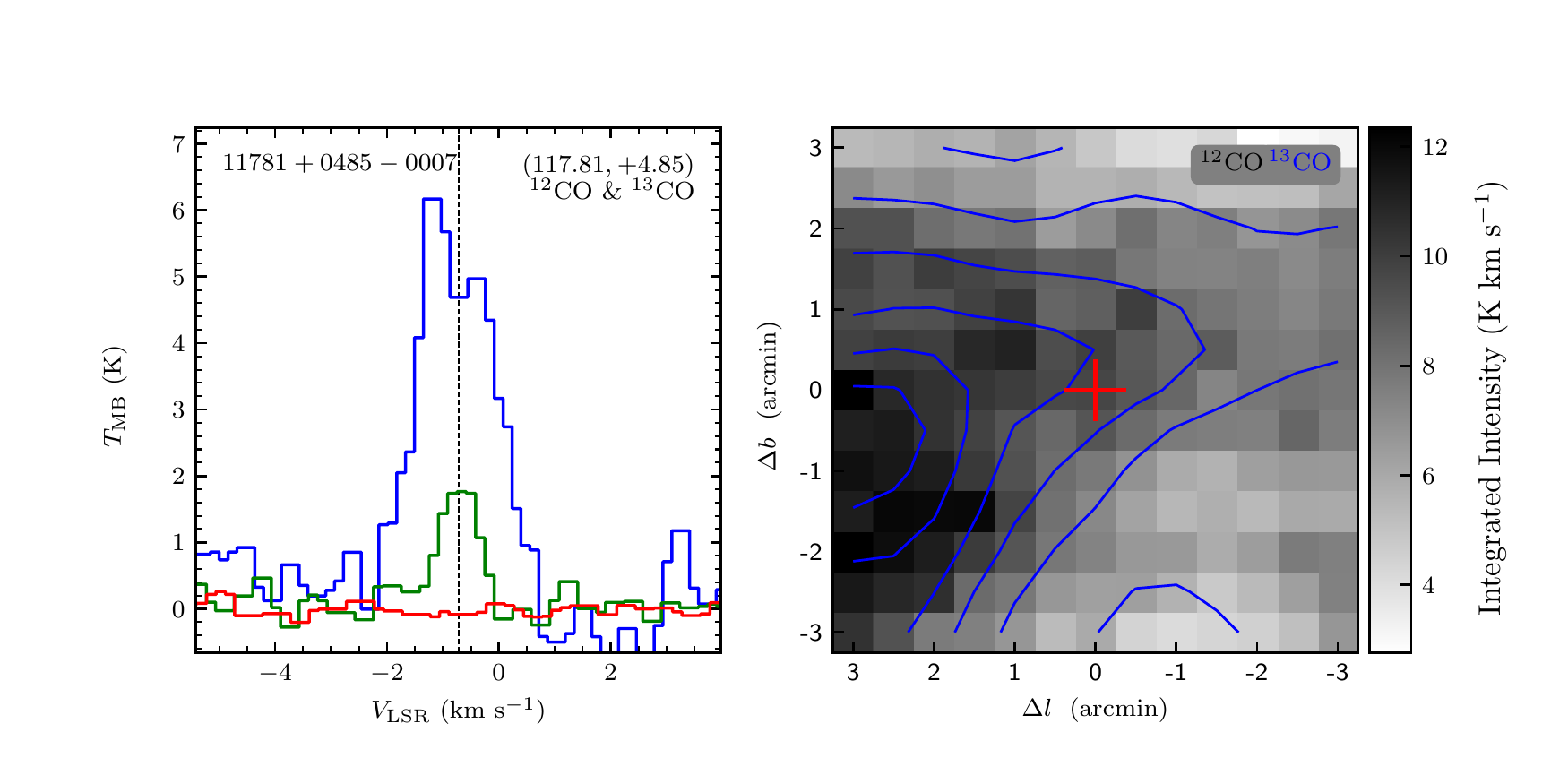}
\end{figure}
\clearpage

\begin{figure}
\includegraphics[width=9.0cm,angle=0]{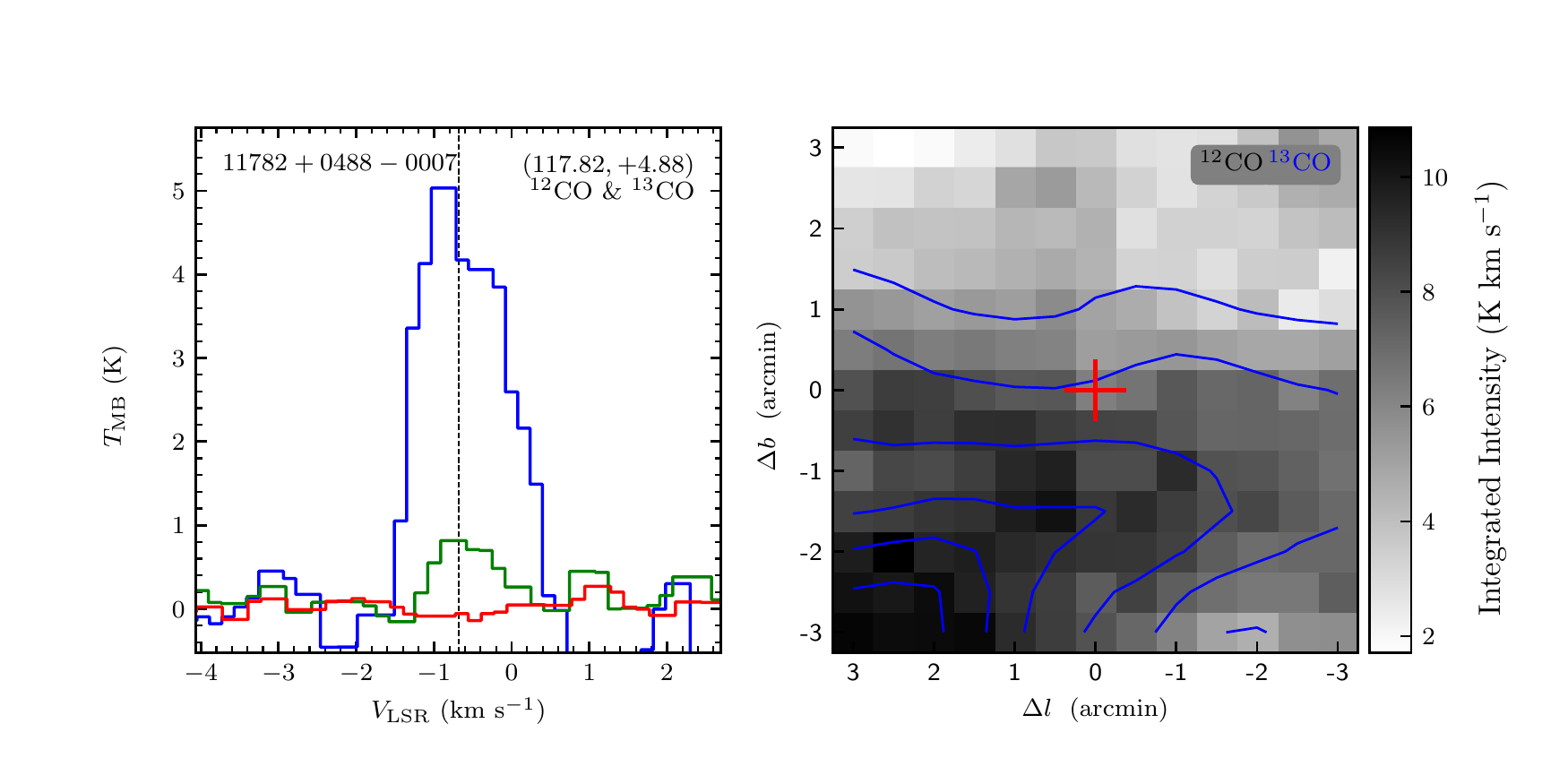}
\includegraphics[width=9.0cm,angle=0]{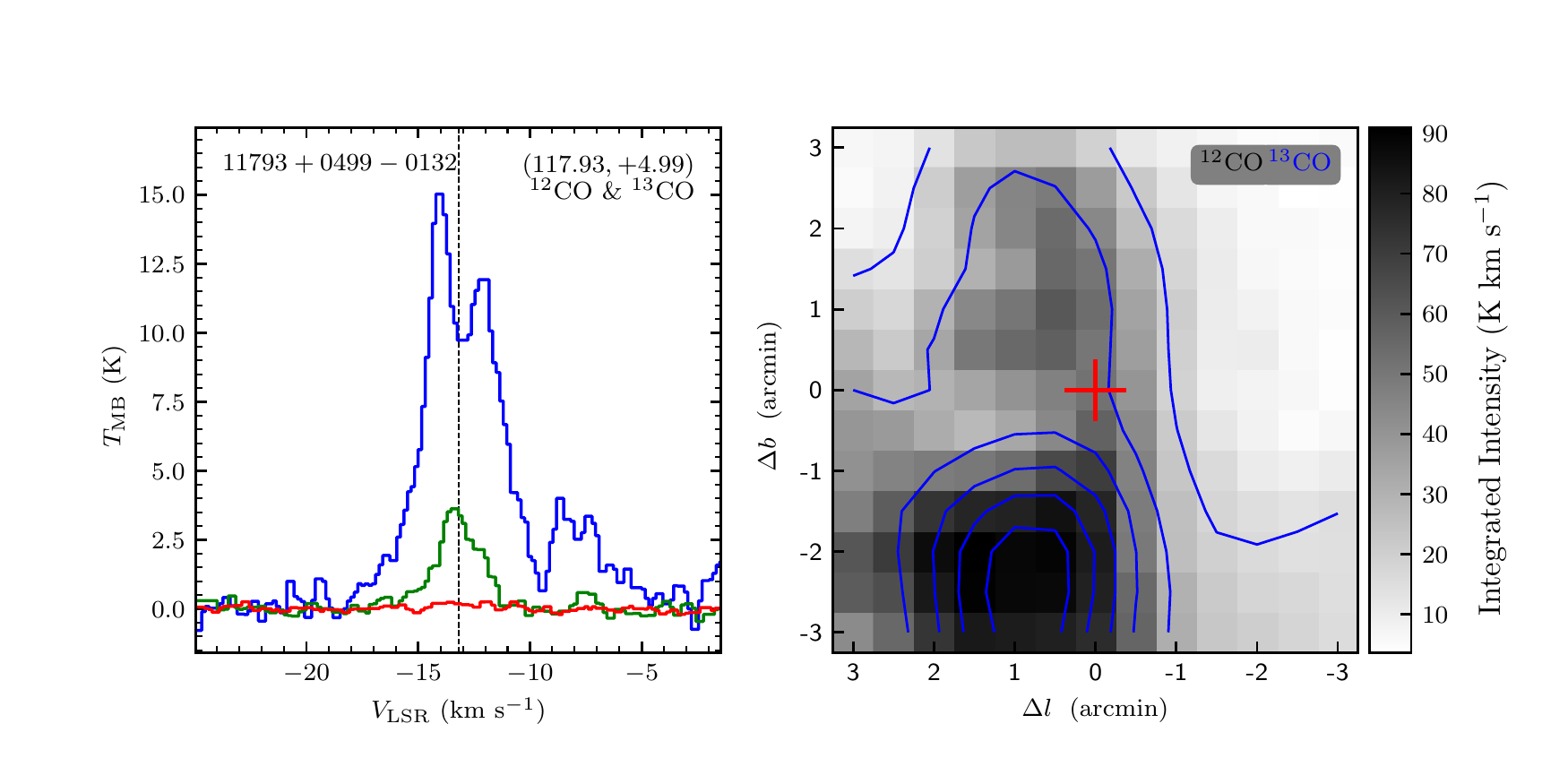}
\vspace{-0.5cm}

\includegraphics[width=9.0cm,angle=0]{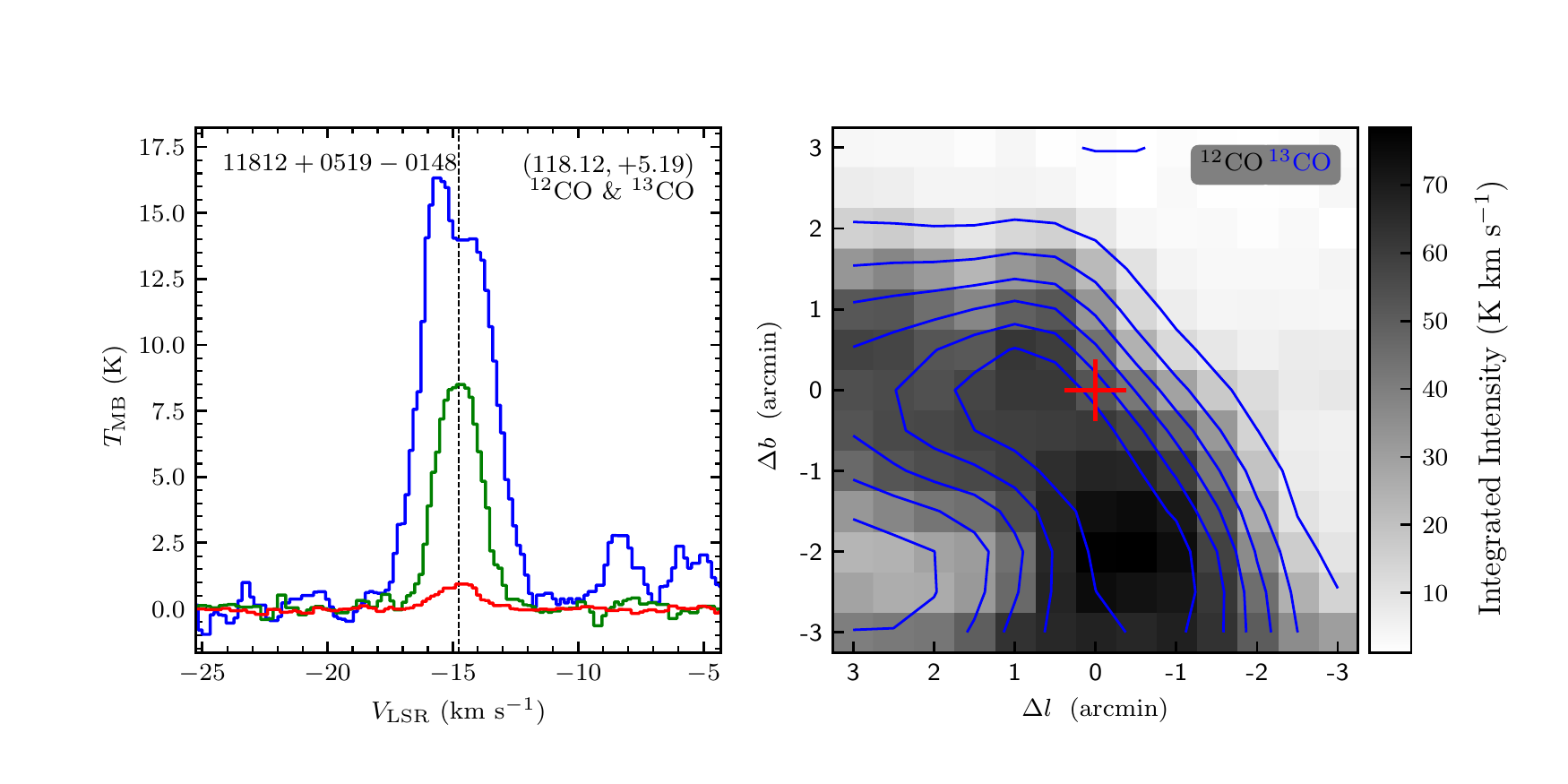}
\includegraphics[width=9.0cm,angle=0]{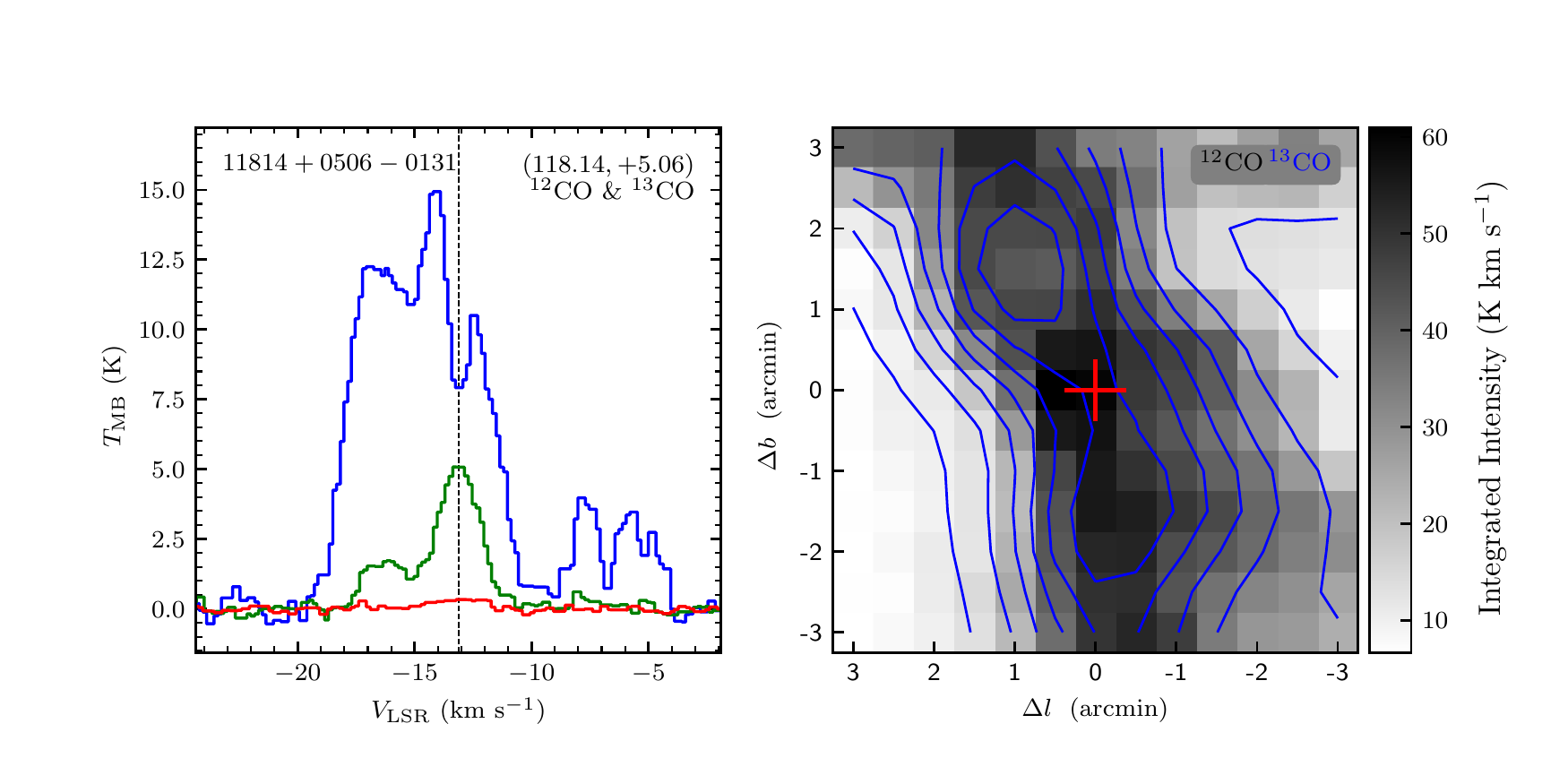}
\vspace{-0.5cm}

\includegraphics[width=9.0cm,angle=0]{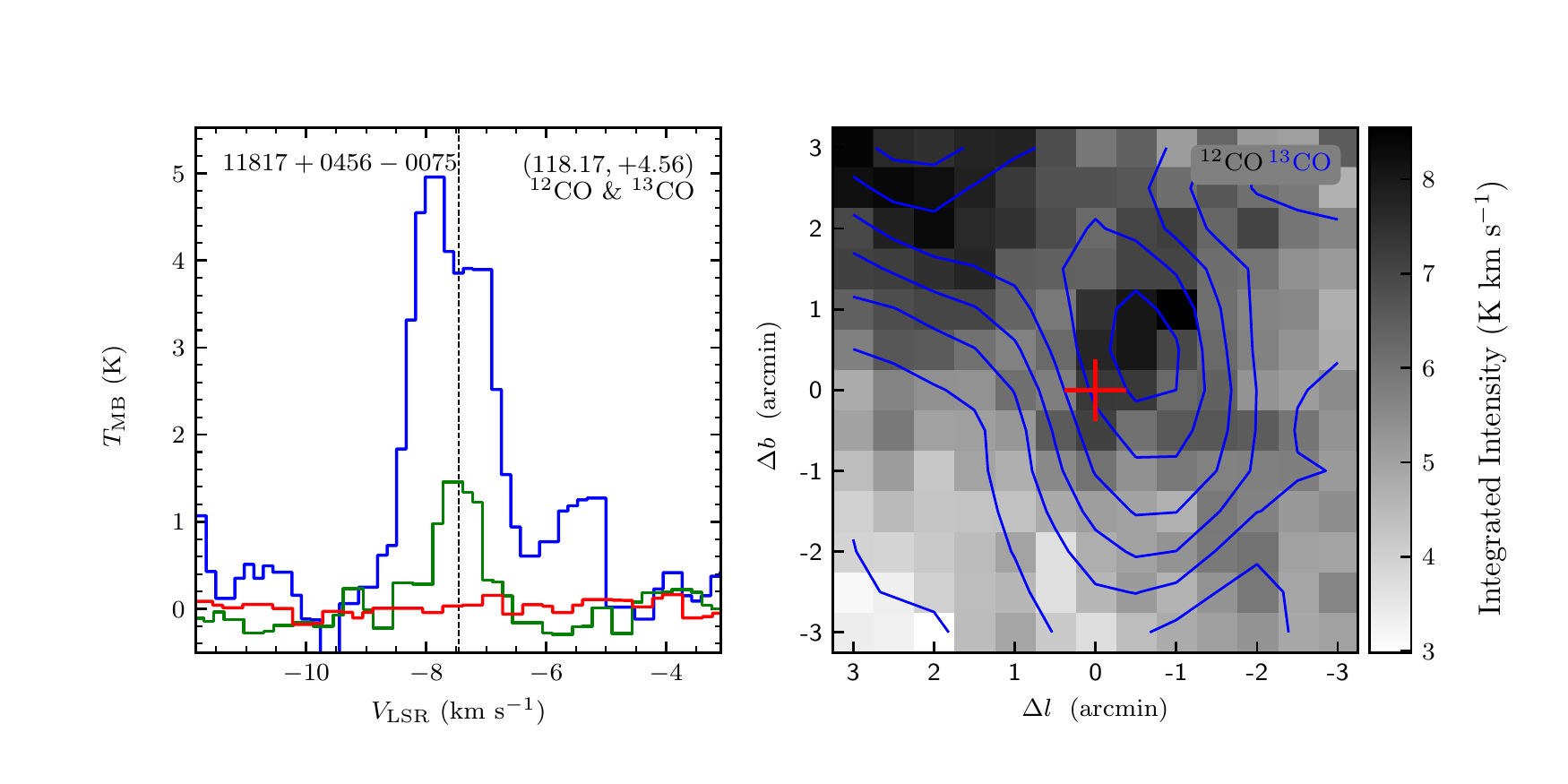}
\includegraphics[width=9.0cm,angle=0]{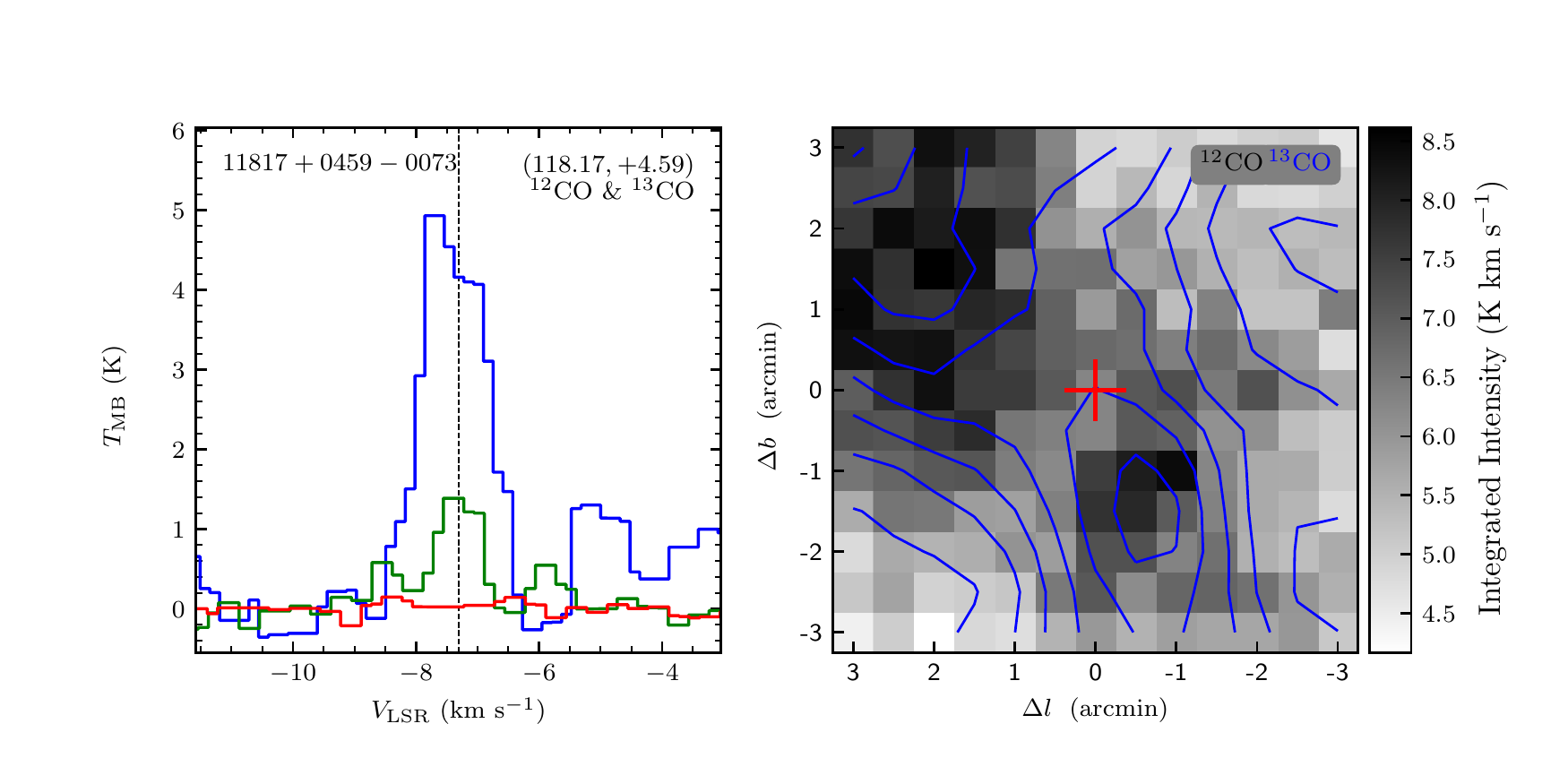}
\vspace{-0.5cm}

\includegraphics[width=9.0cm,angle=0]{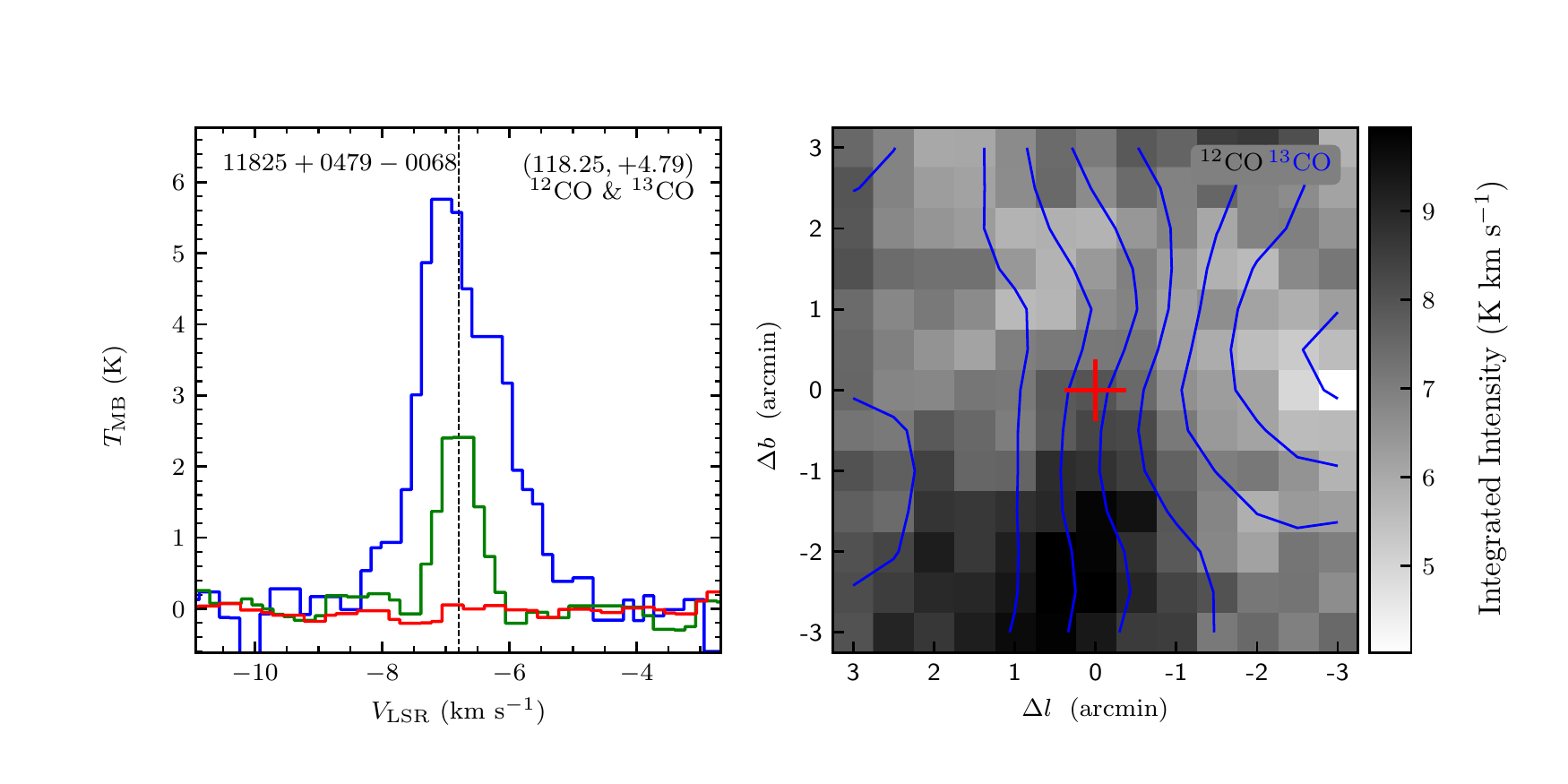}
\includegraphics[width=9.0cm,angle=0]{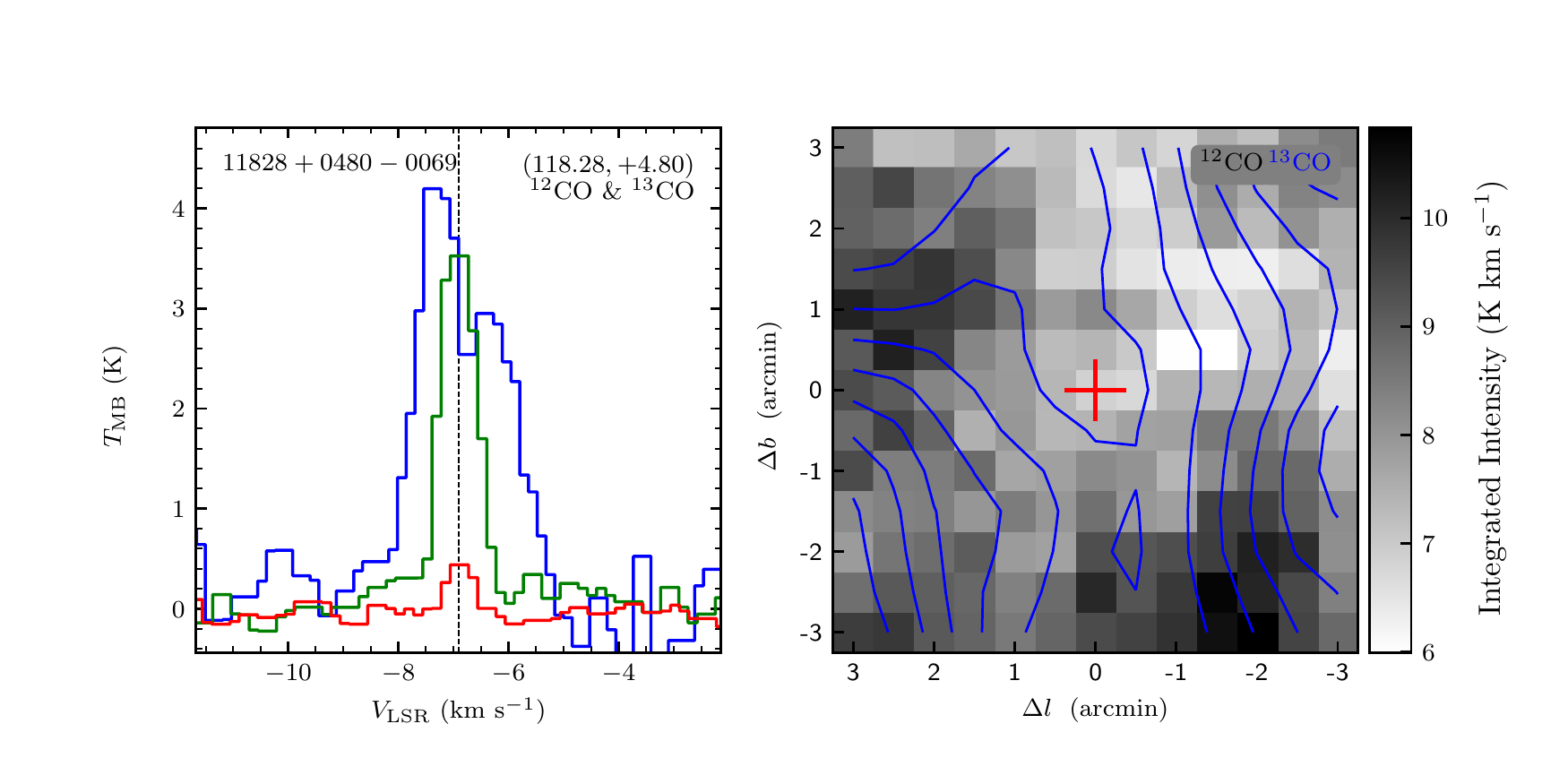}
\vspace{-0.5cm}

\includegraphics[width=9.0cm,angle=0]{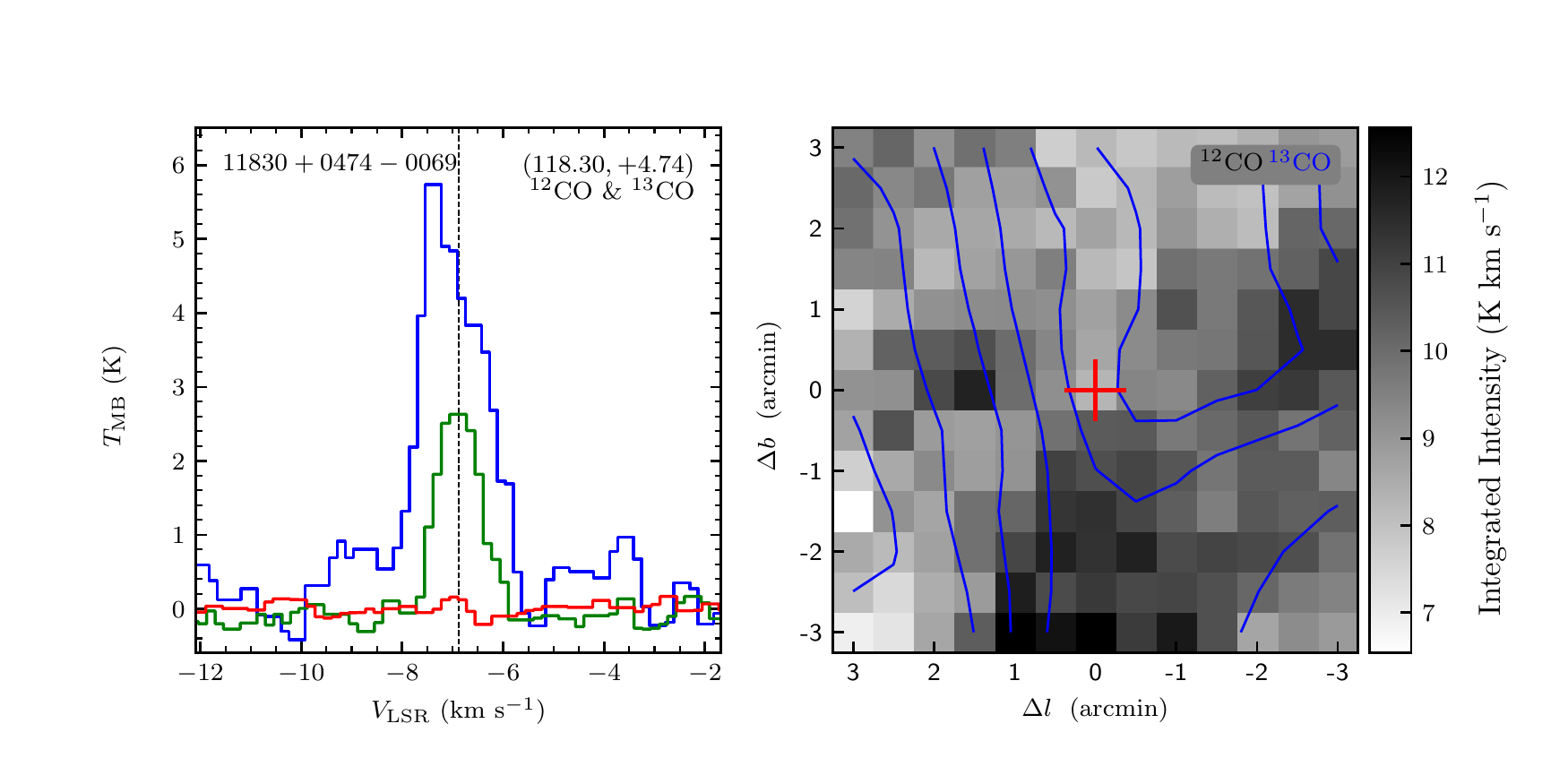}
\includegraphics[width=9.0cm,angle=0]{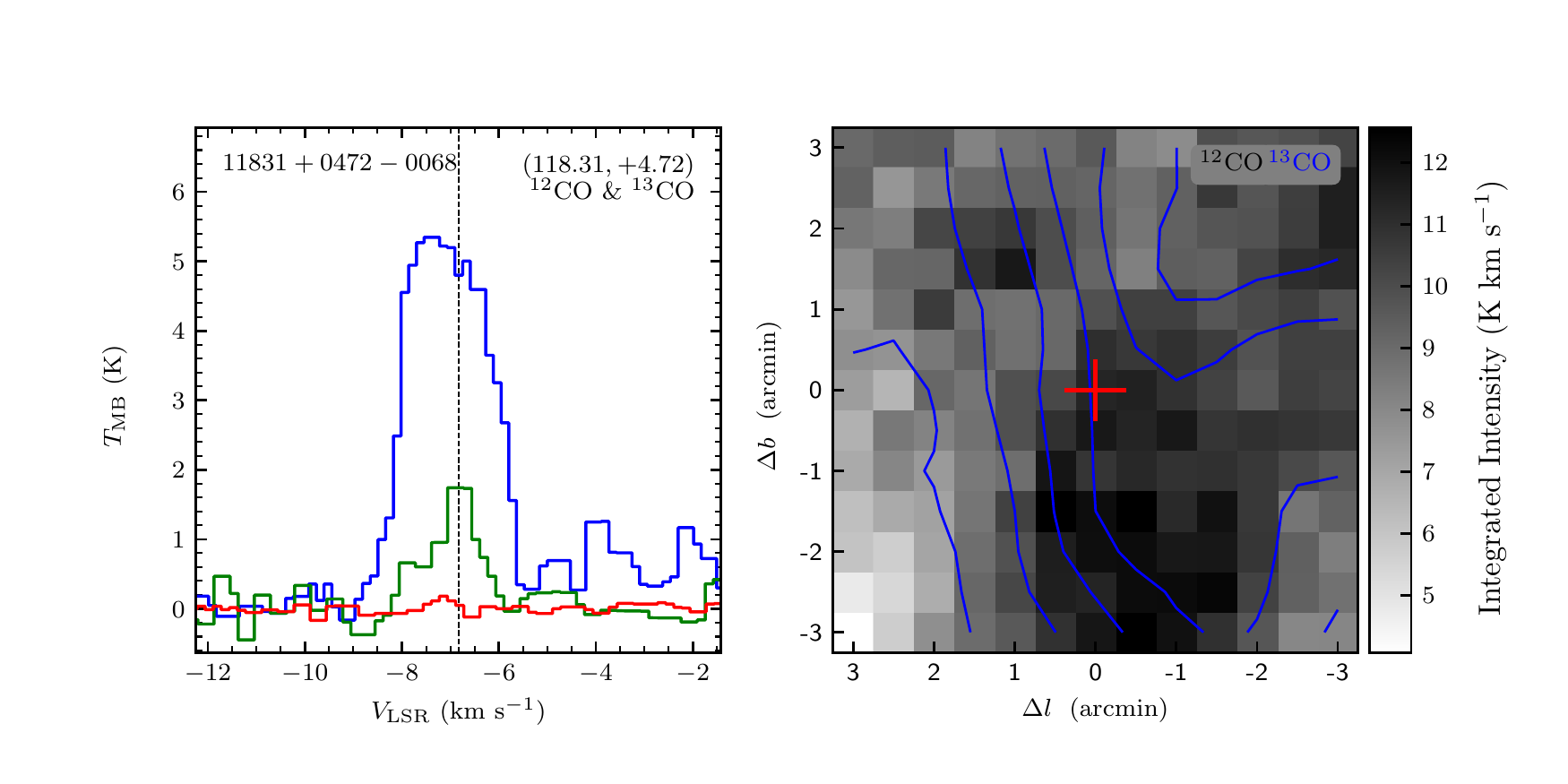}
\end{figure}
\clearpage

\begin{figure}
\includegraphics[width=9.0cm,angle=0]{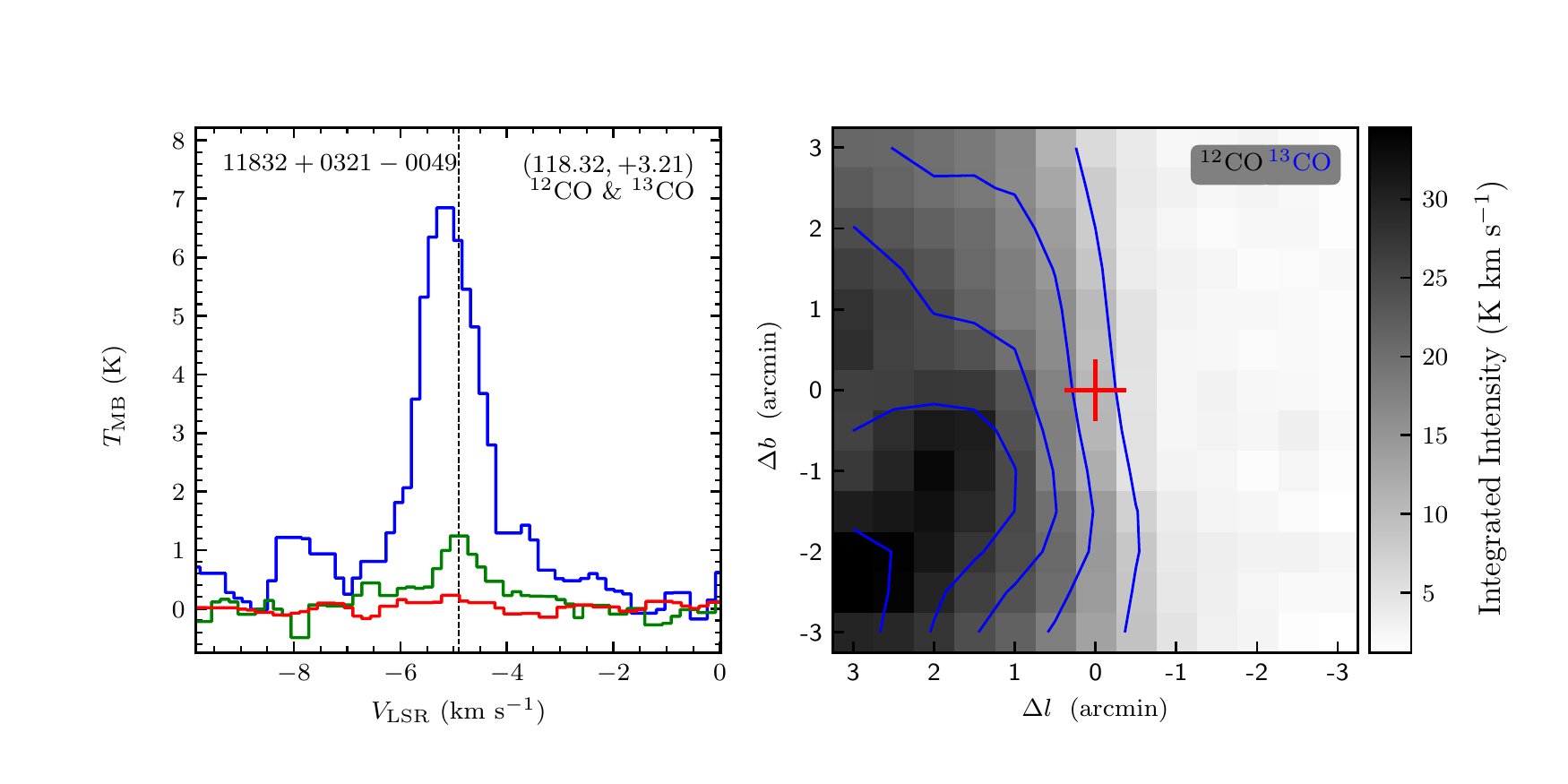}
\includegraphics[width=9.0cm,angle=0]{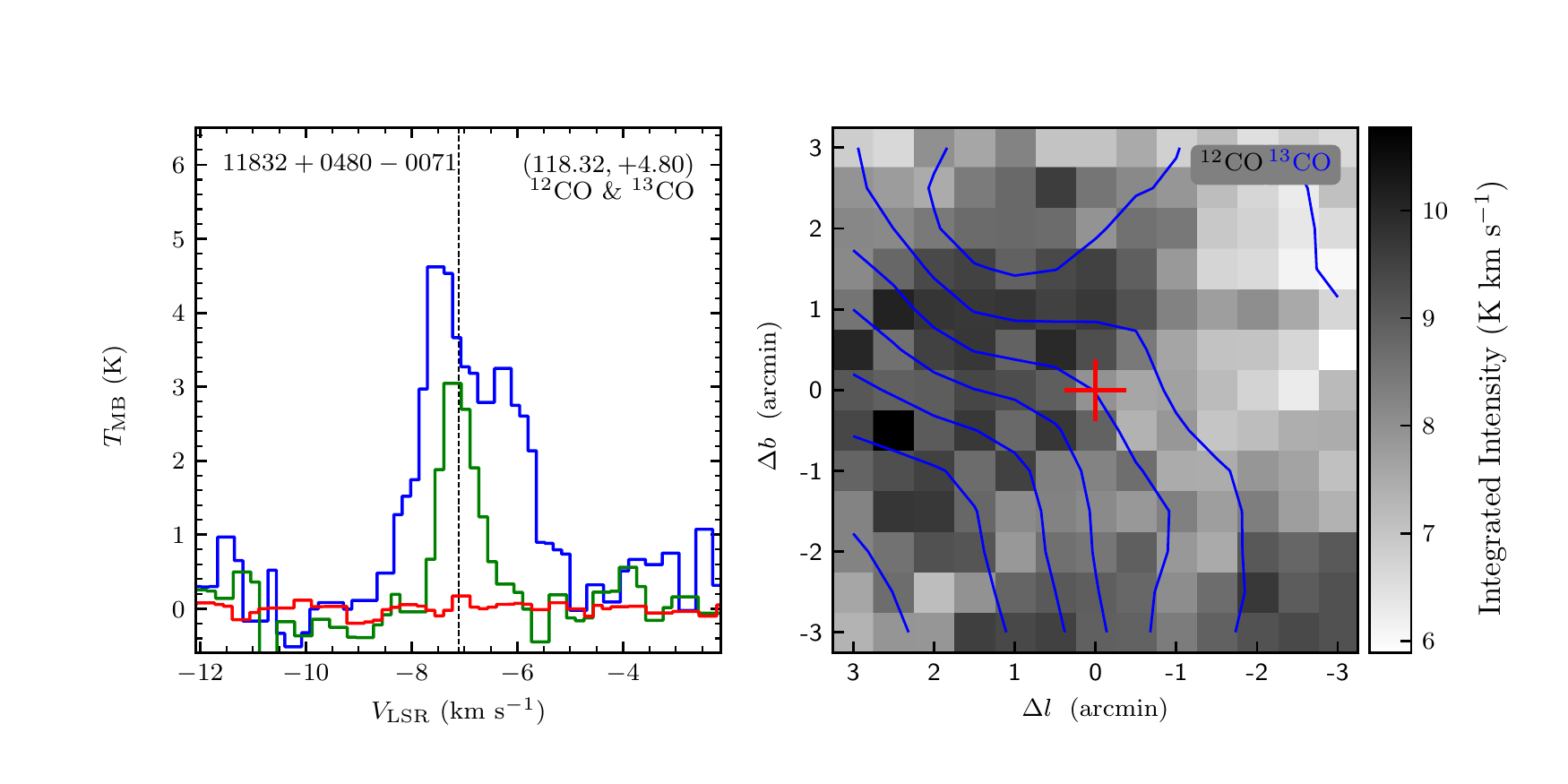}
\vspace{-0.5cm}

\includegraphics[width=9.0cm,angle=0]{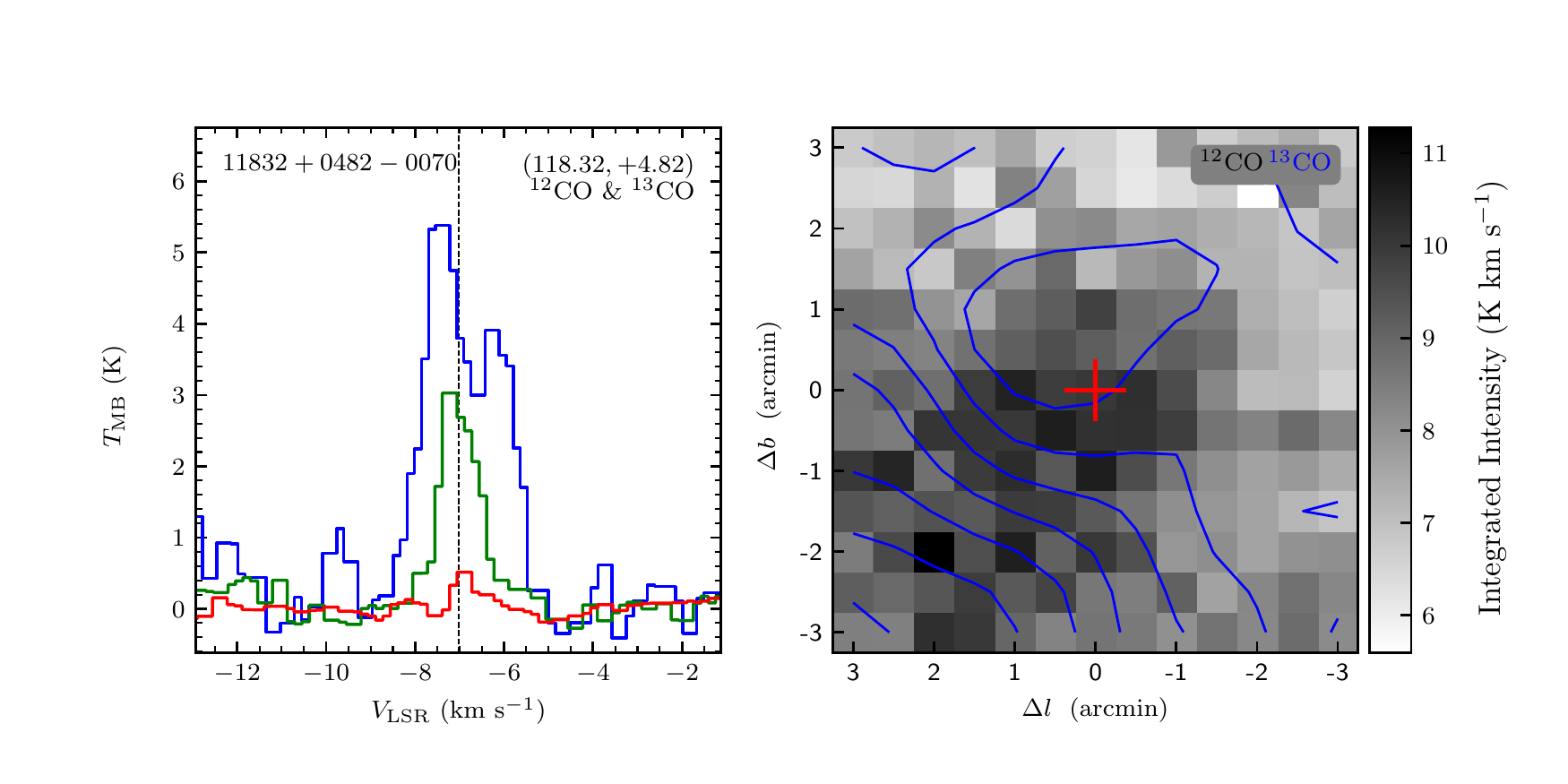}
\includegraphics[width=9.0cm,angle=0]{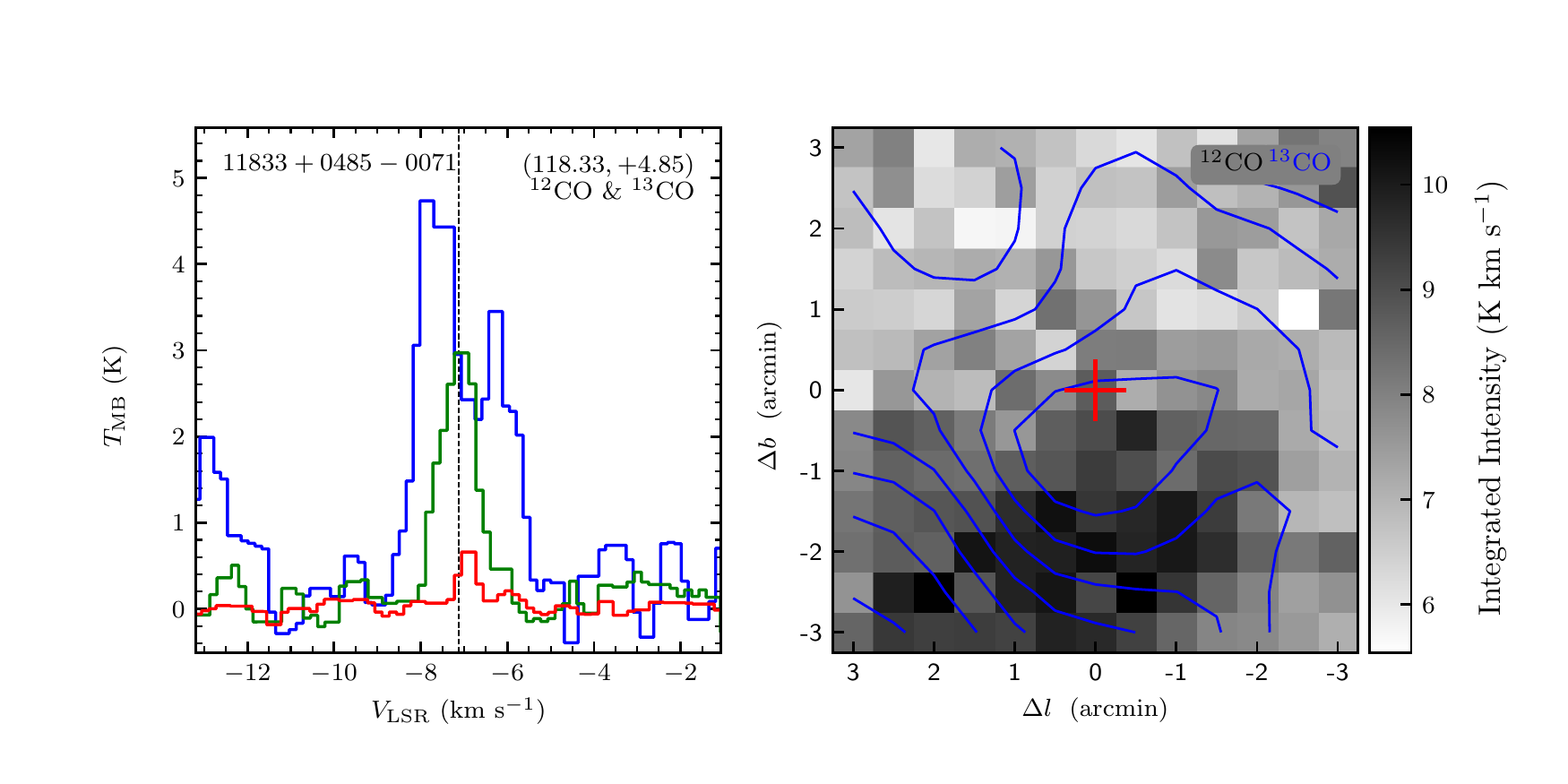}
\vspace{-0.5cm}

\includegraphics[width=9.0cm,angle=0]{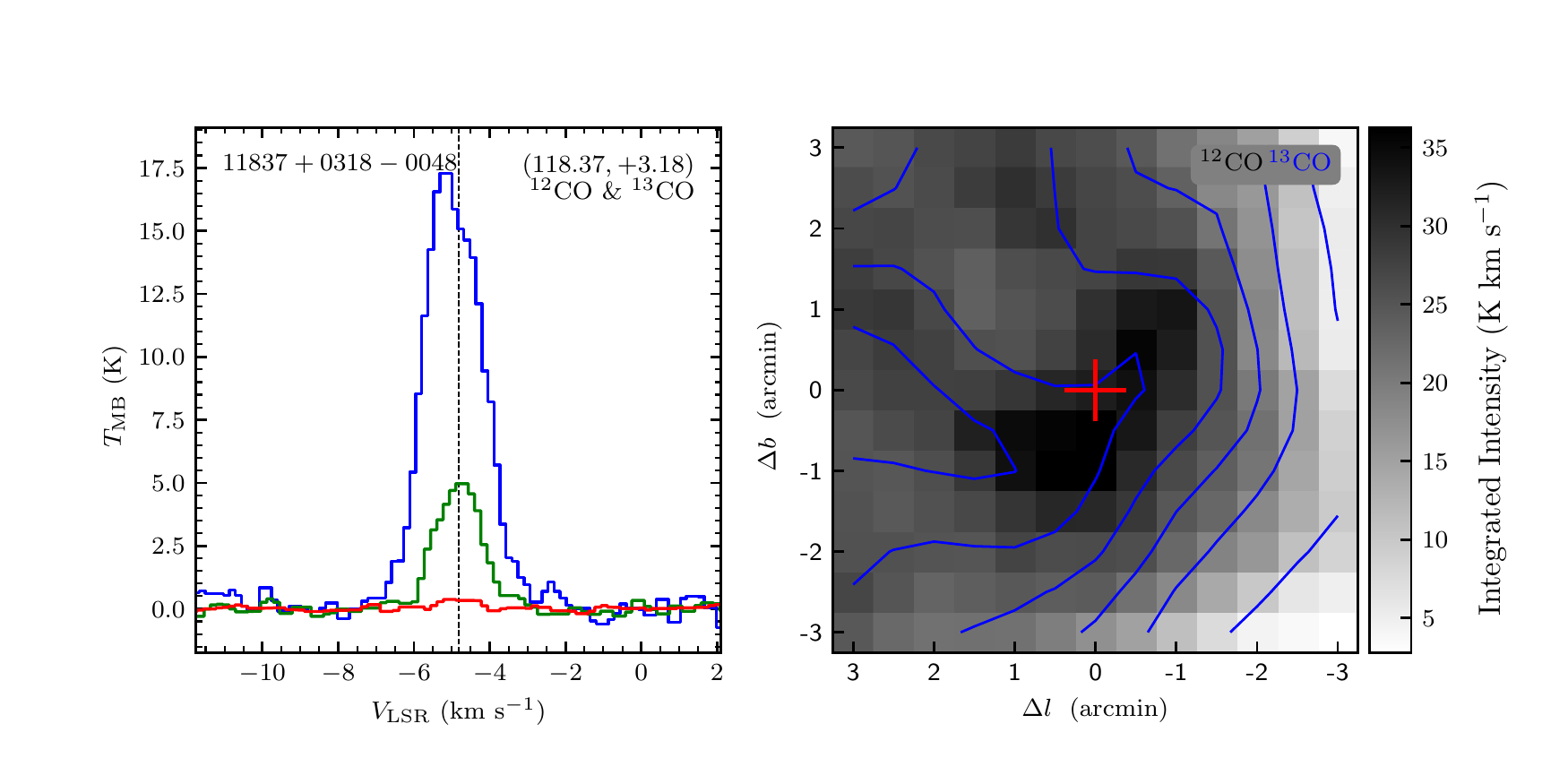}
\includegraphics[width=9.0cm,angle=0]{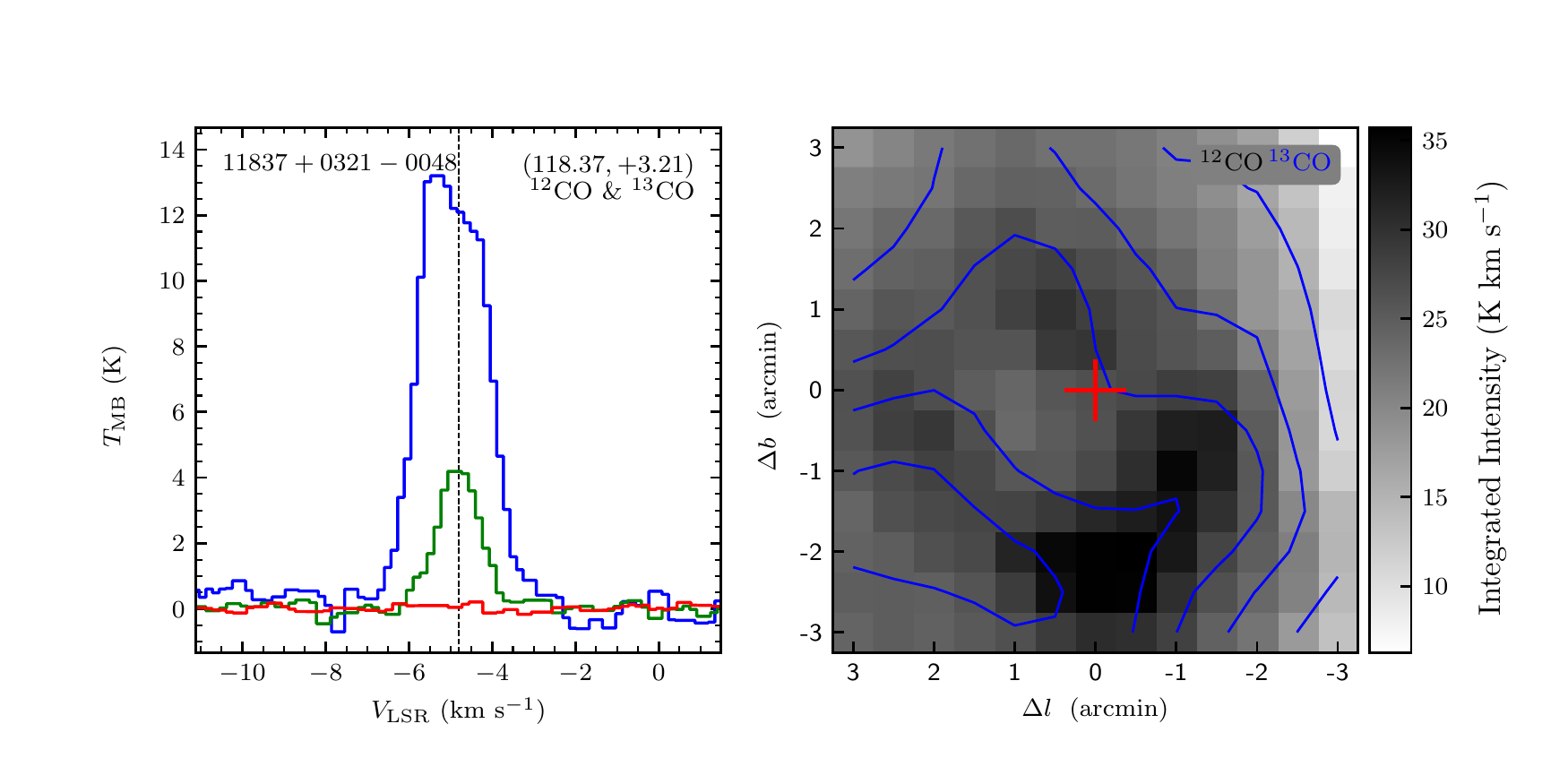}
\vspace{-0.5cm}

\includegraphics[width=9.0cm,angle=0]{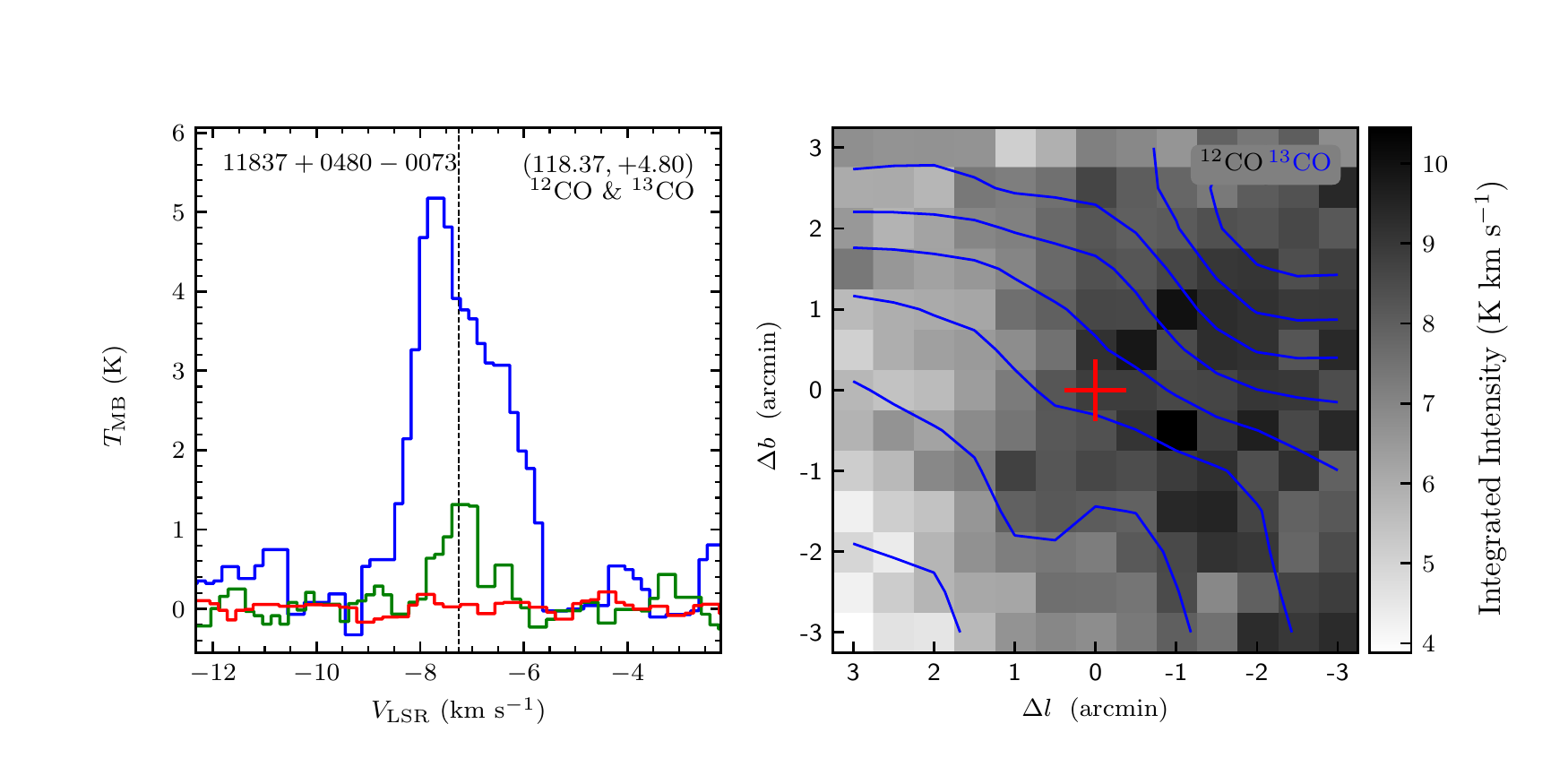}
\includegraphics[width=9.0cm,angle=0]{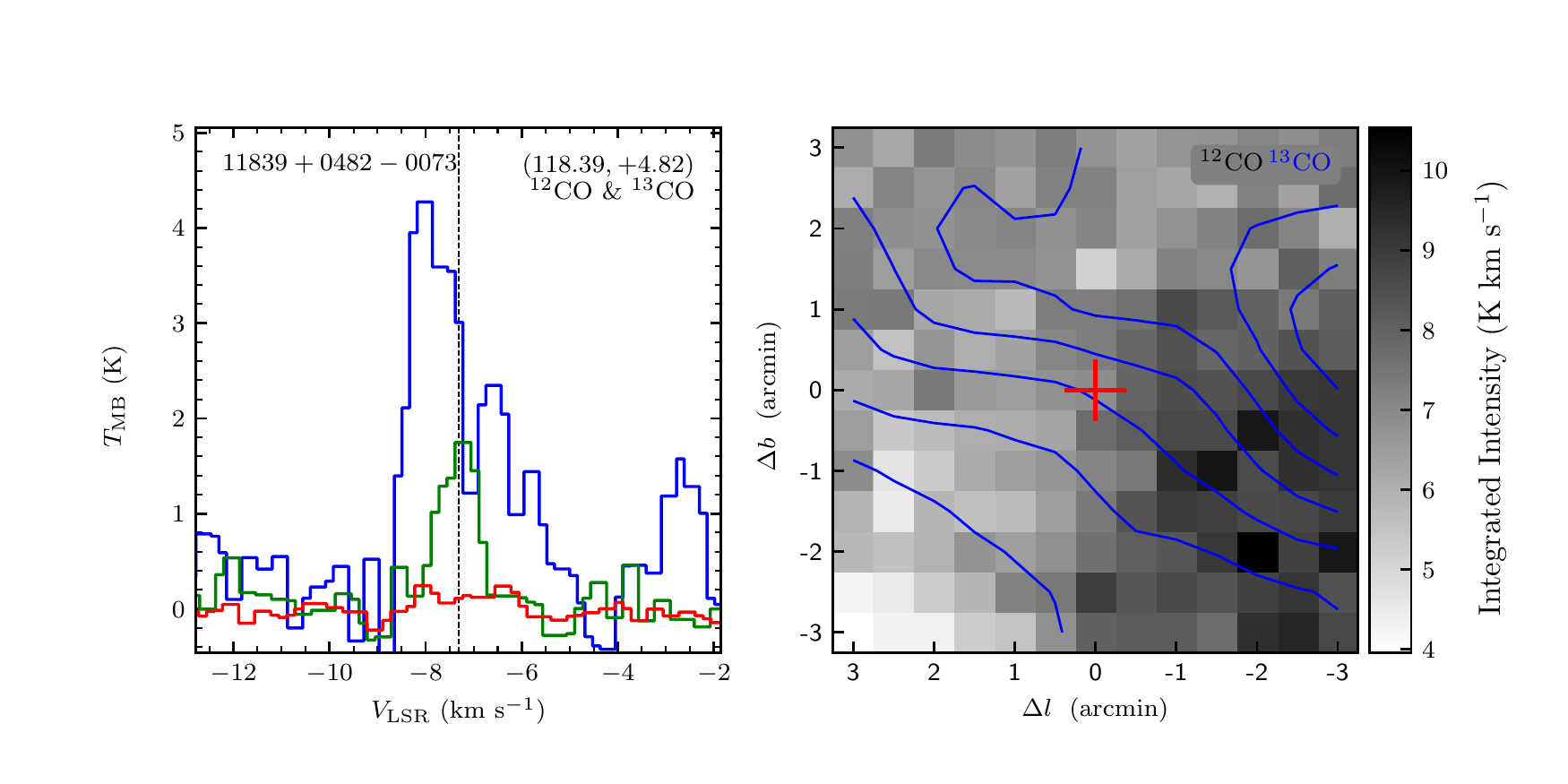}
\vspace{-0.5cm}

\includegraphics[width=9.0cm,angle=0]{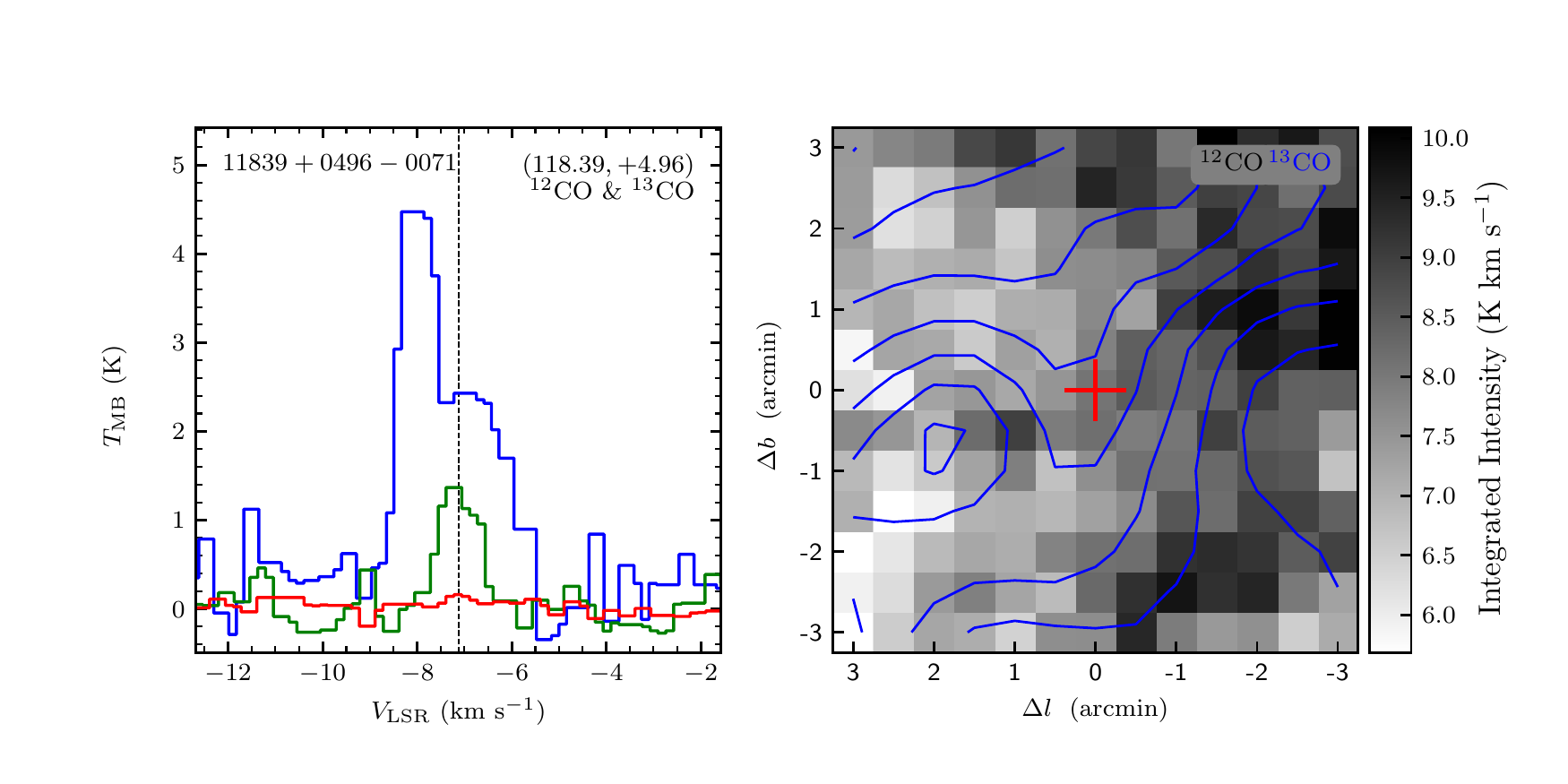}
\includegraphics[width=9.0cm,angle=0]{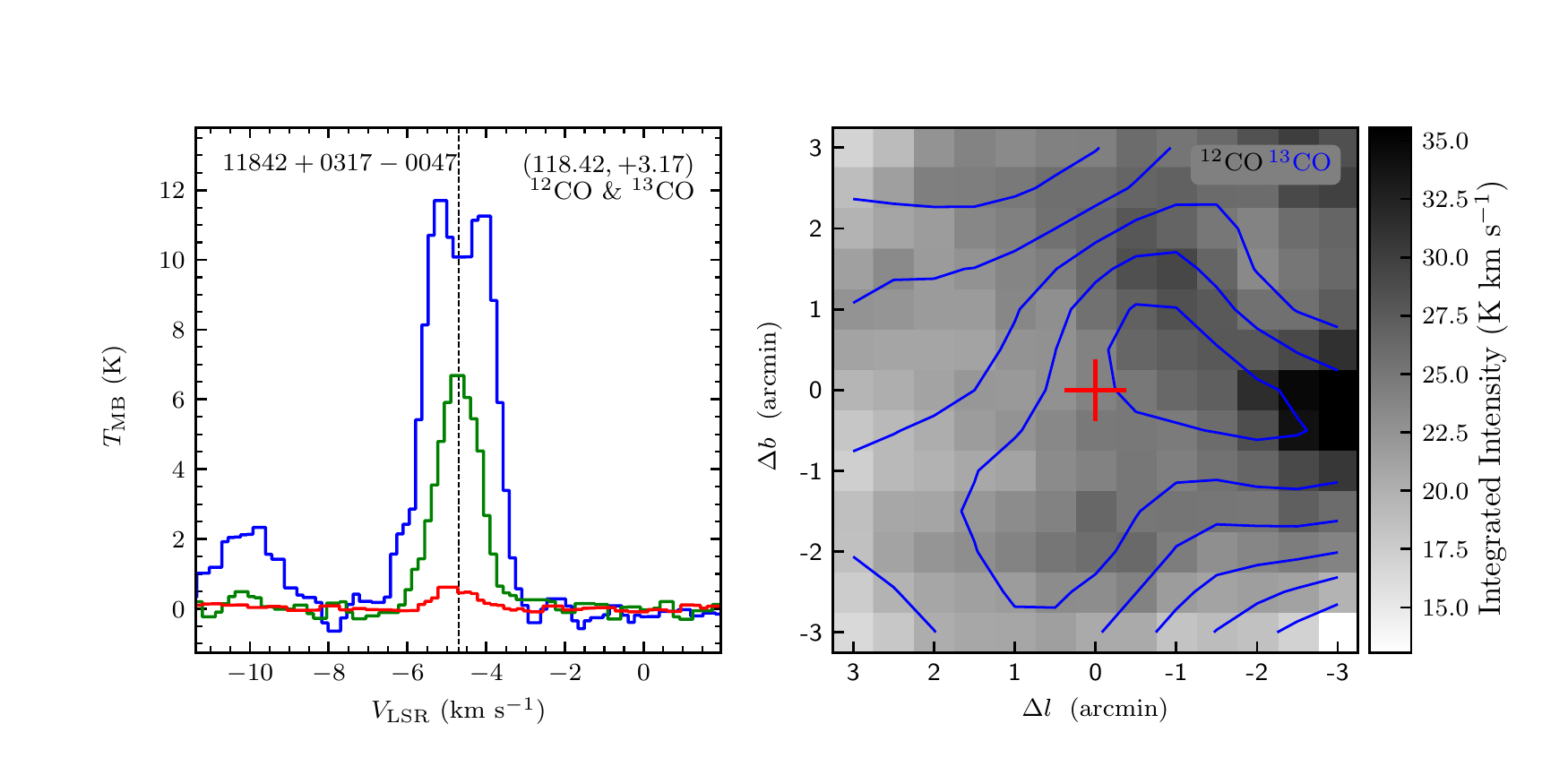}
\end{figure}
\clearpage

\begin{figure}
\includegraphics[width=9.0cm,angle=0]{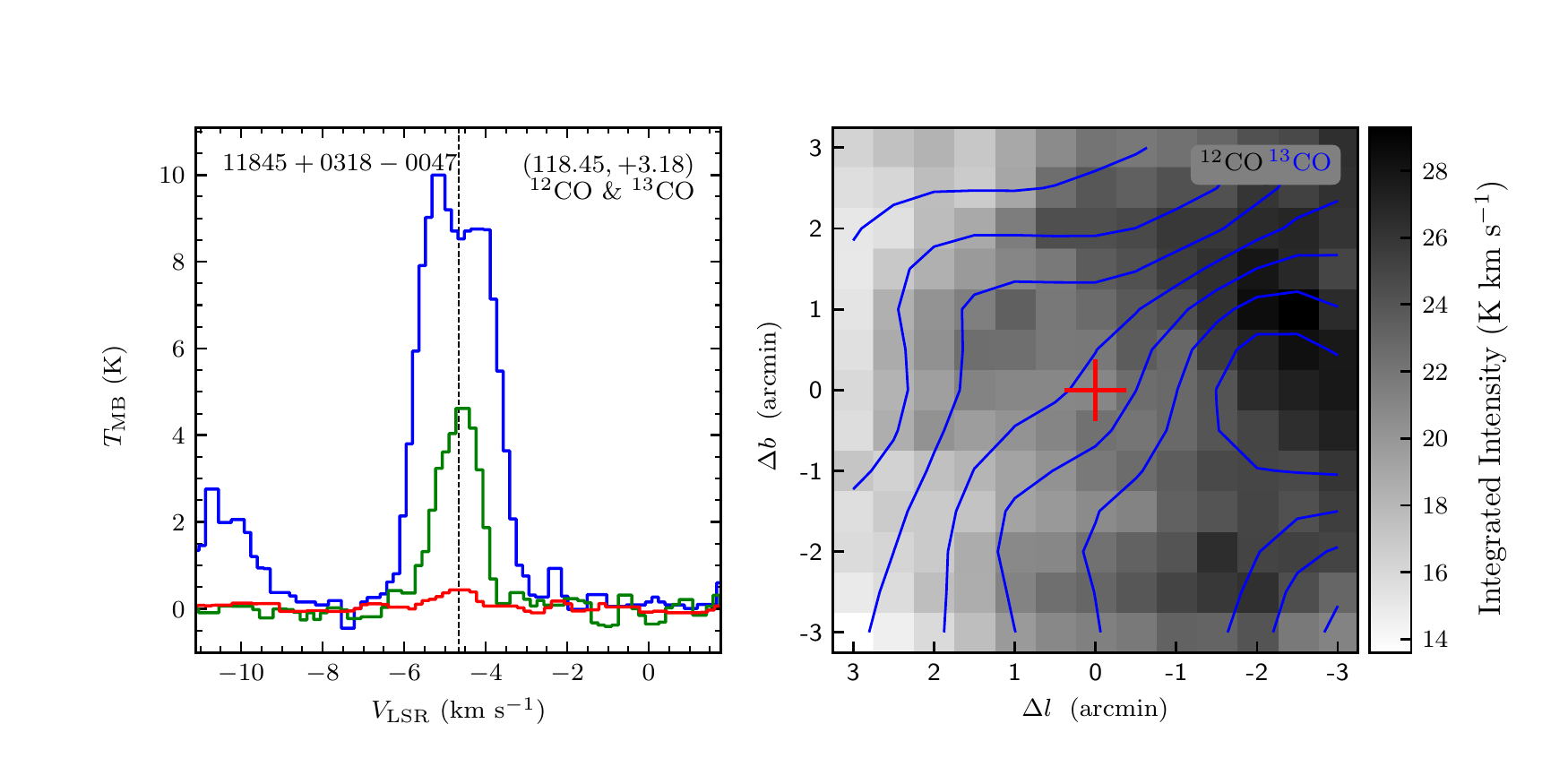}
\includegraphics[width=9.0cm,angle=0]{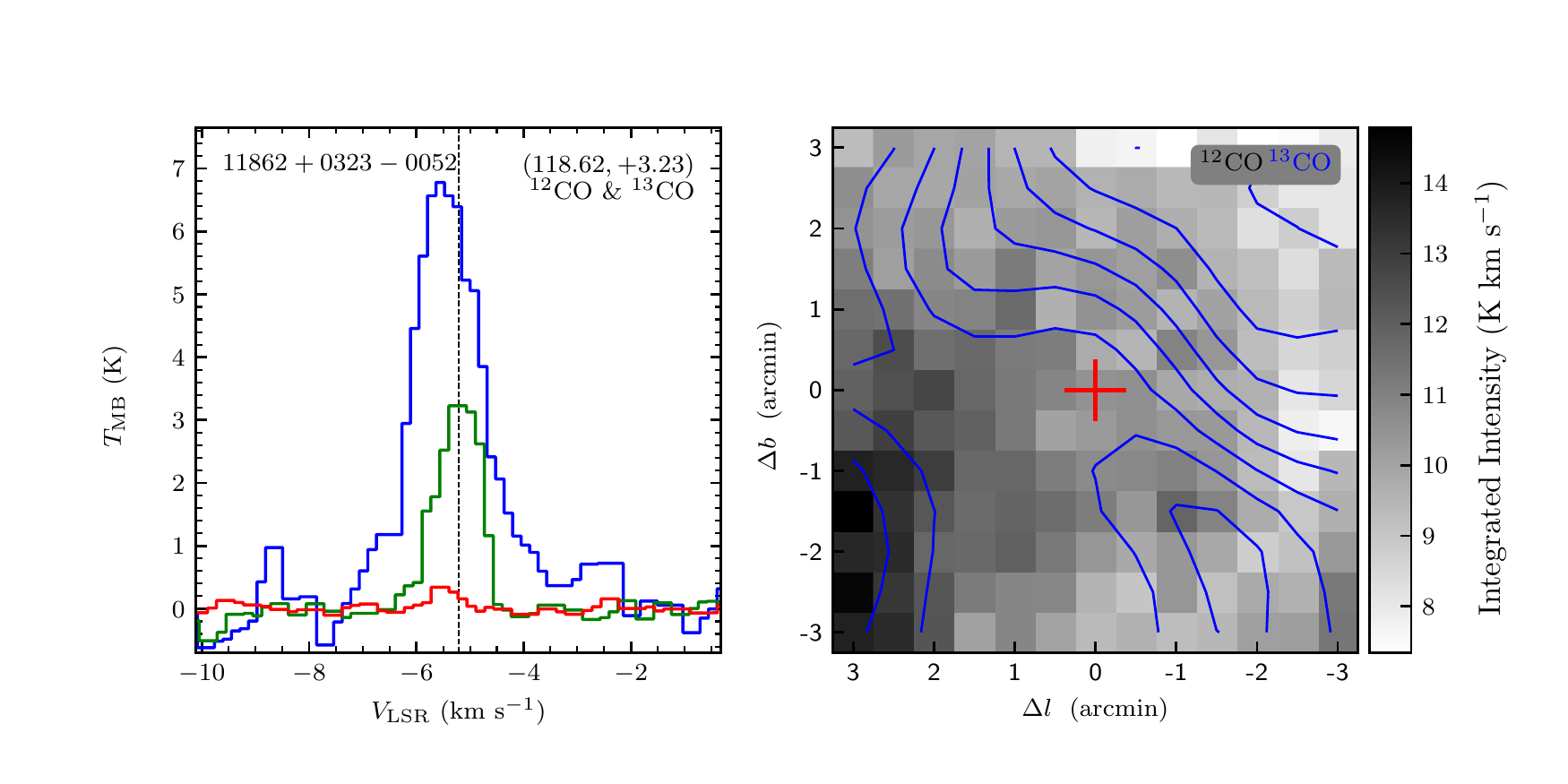}
\vspace{-0.5cm}

\includegraphics[width=9.0cm,angle=0]{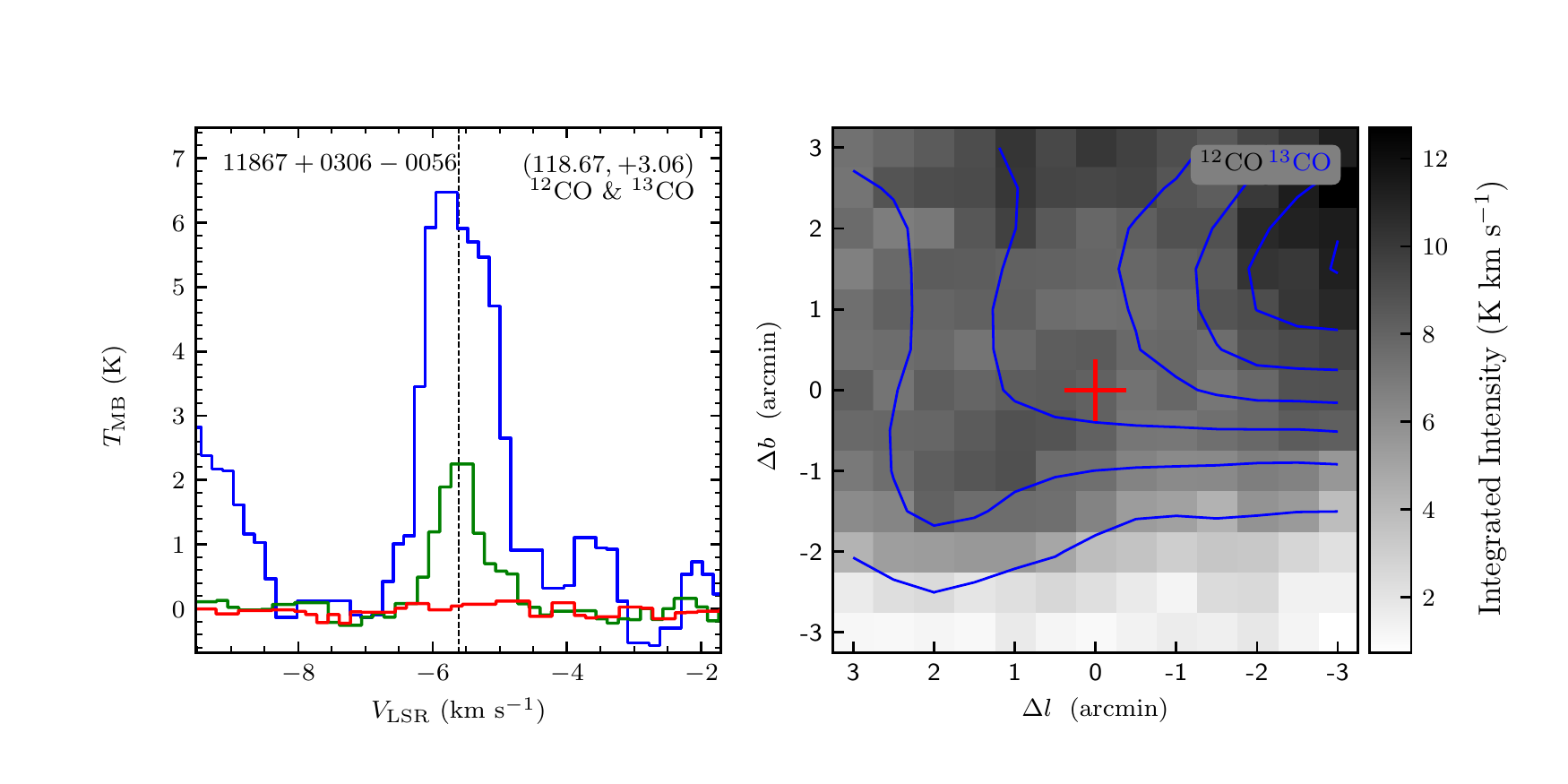}
\includegraphics[width=9.0cm,angle=0]{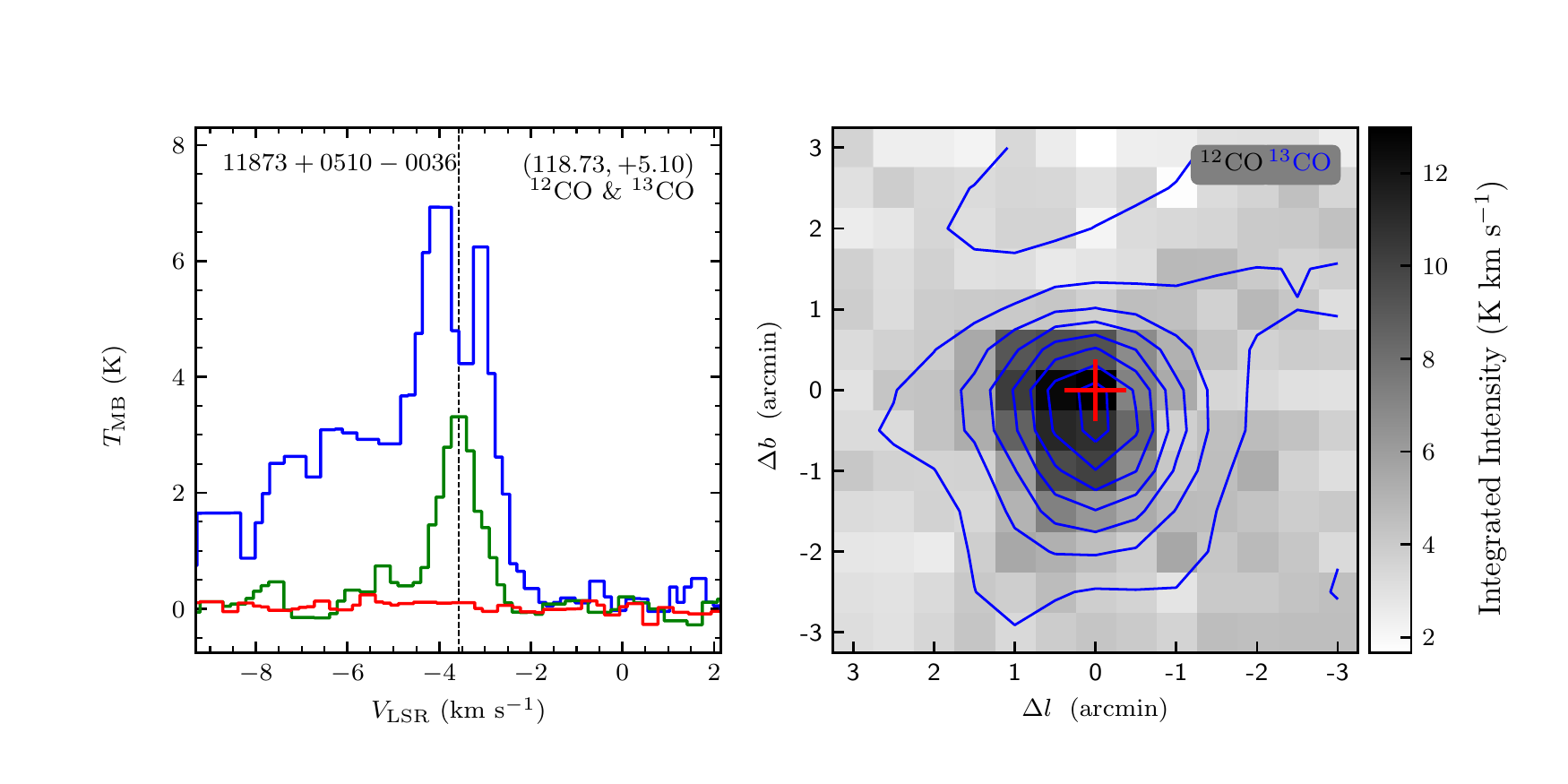}
\vspace{-0.5cm}

\includegraphics[width=9.0cm,angle=0]{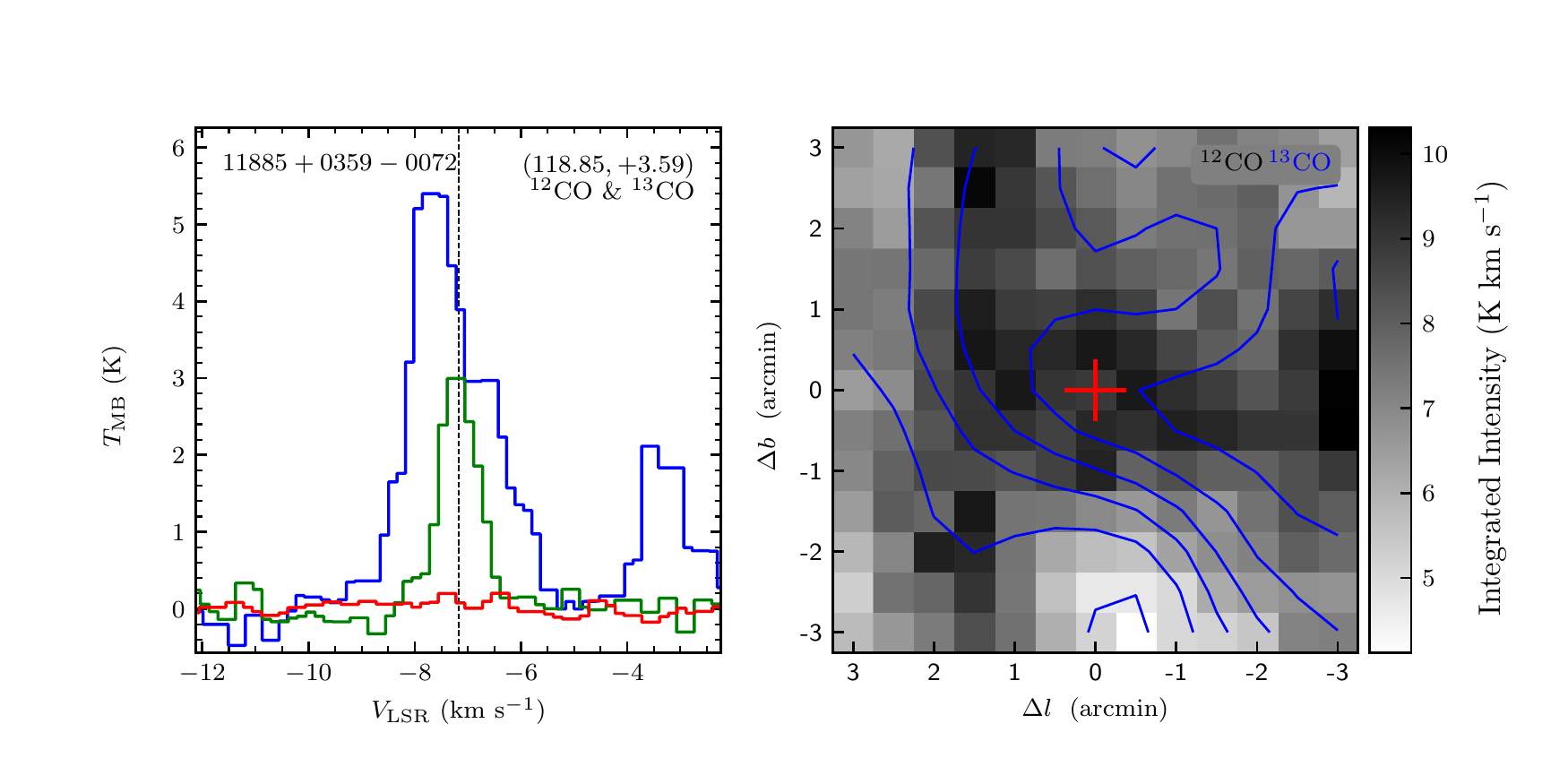}
\includegraphics[width=9.0cm,angle=0]{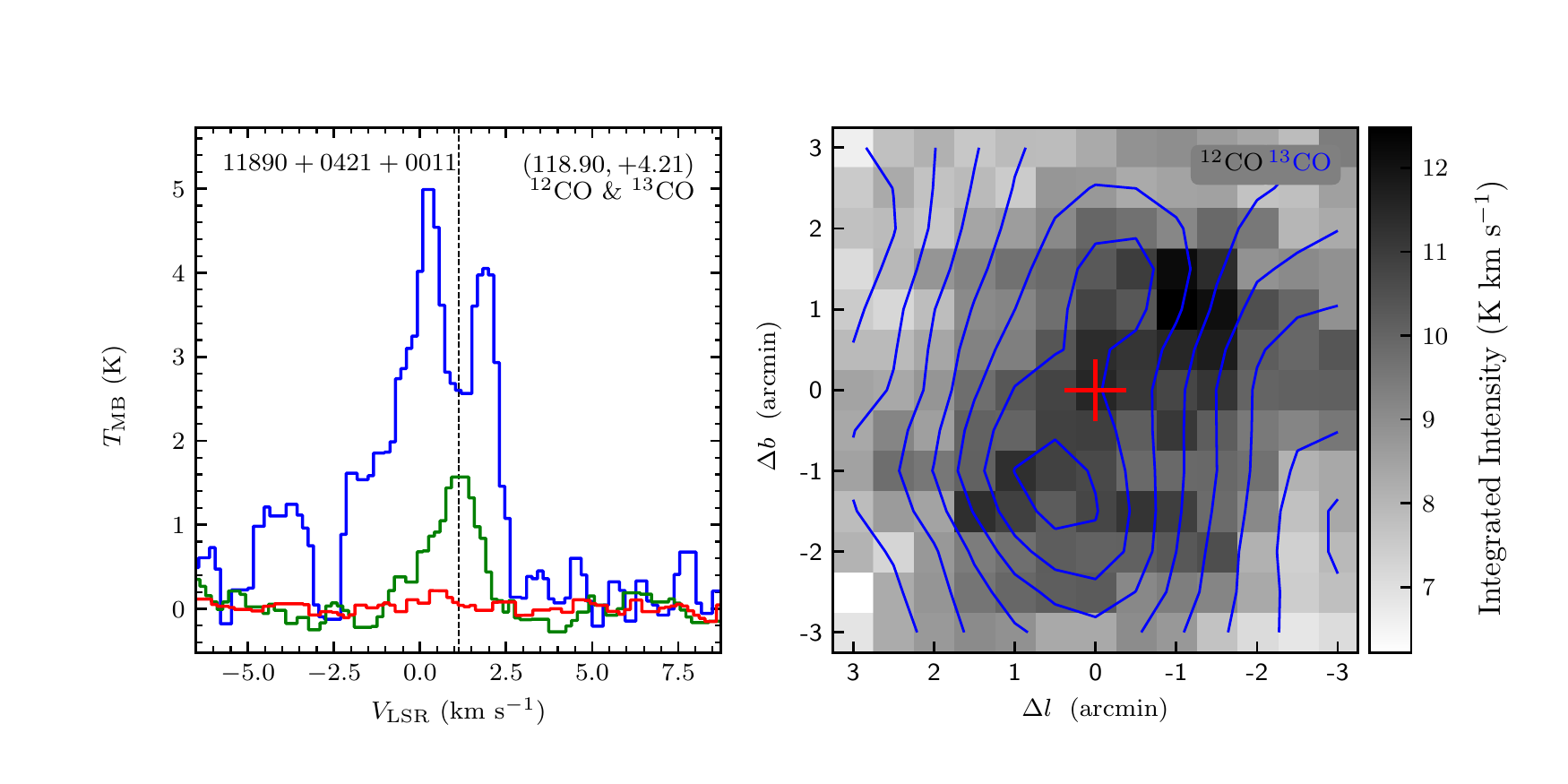}
\vspace{-0.5cm}

\includegraphics[width=9.0cm,angle=0]{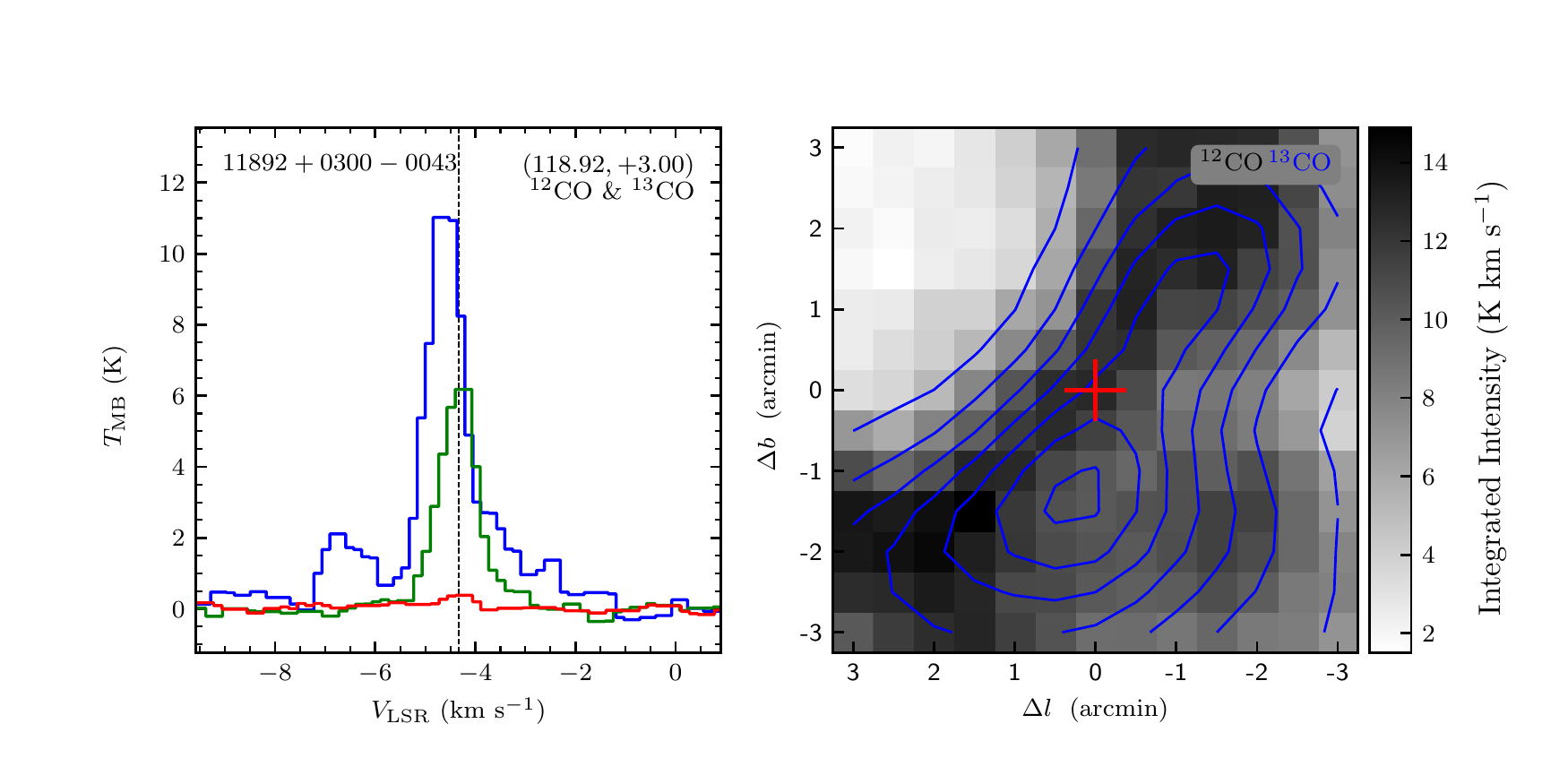}
\includegraphics[width=9.0cm,angle=0]{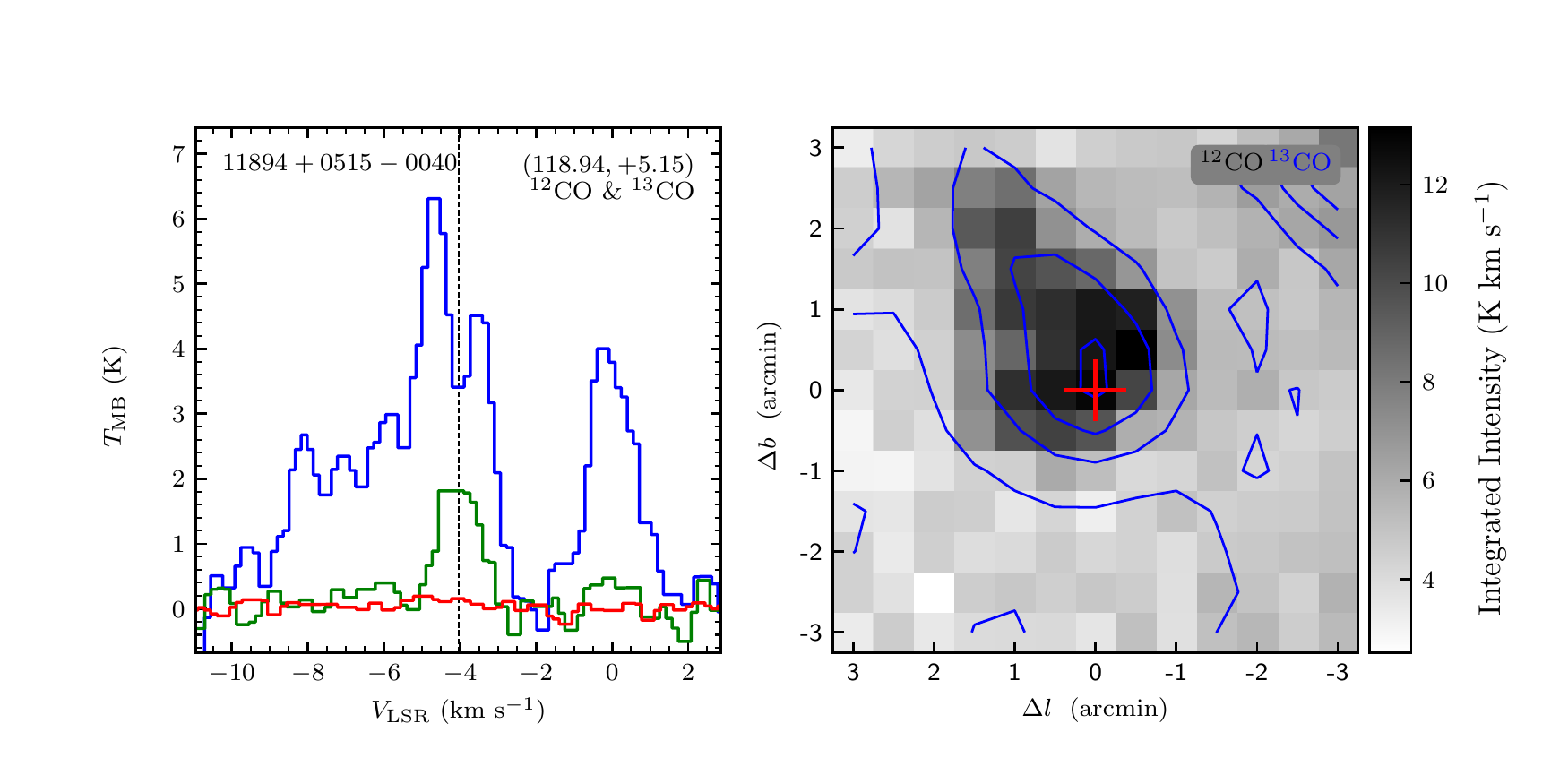}
\vspace{-0.5cm}

\includegraphics[width=9.0cm,angle=0]{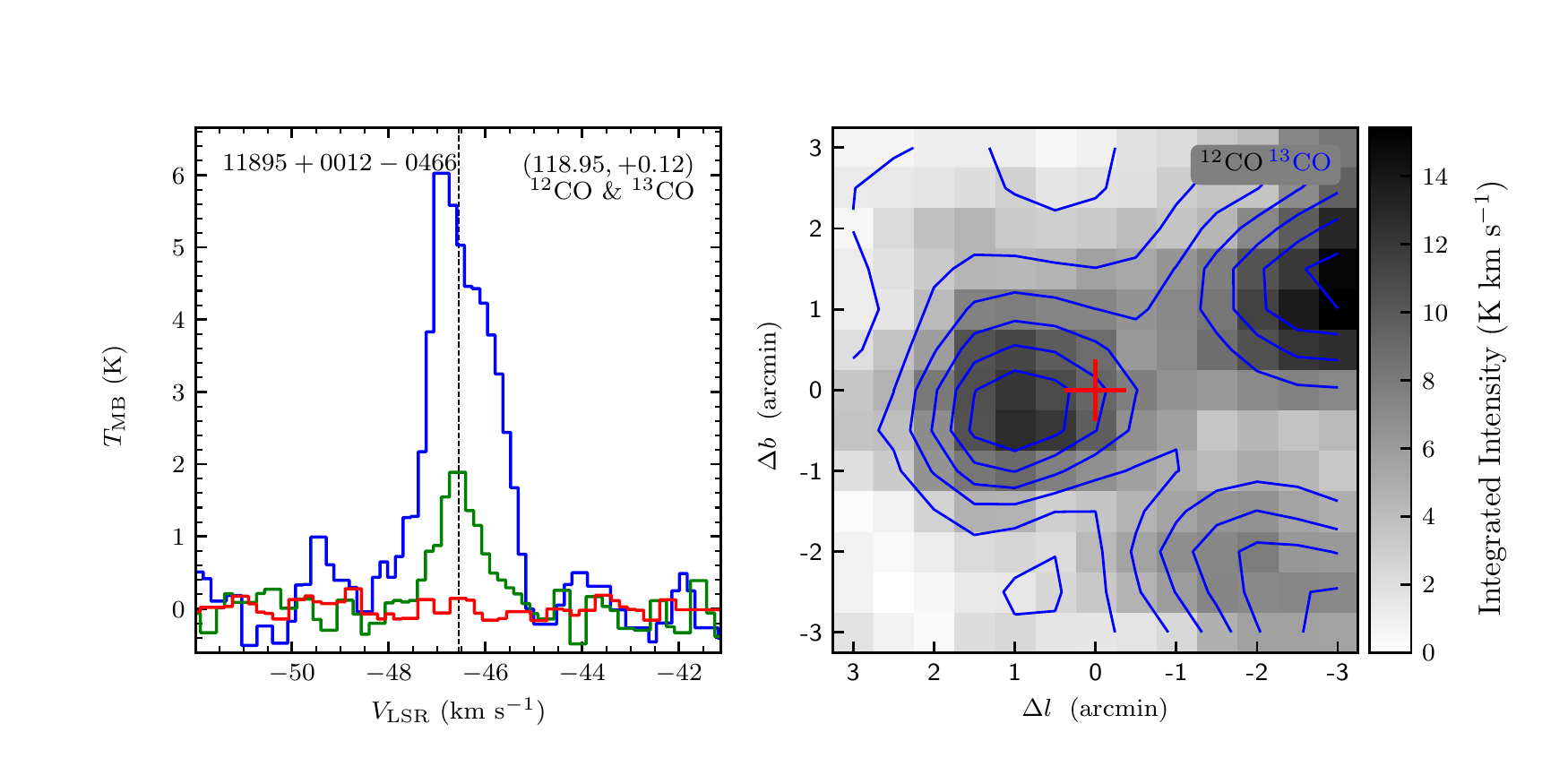}
\includegraphics[width=9.0cm,angle=0]{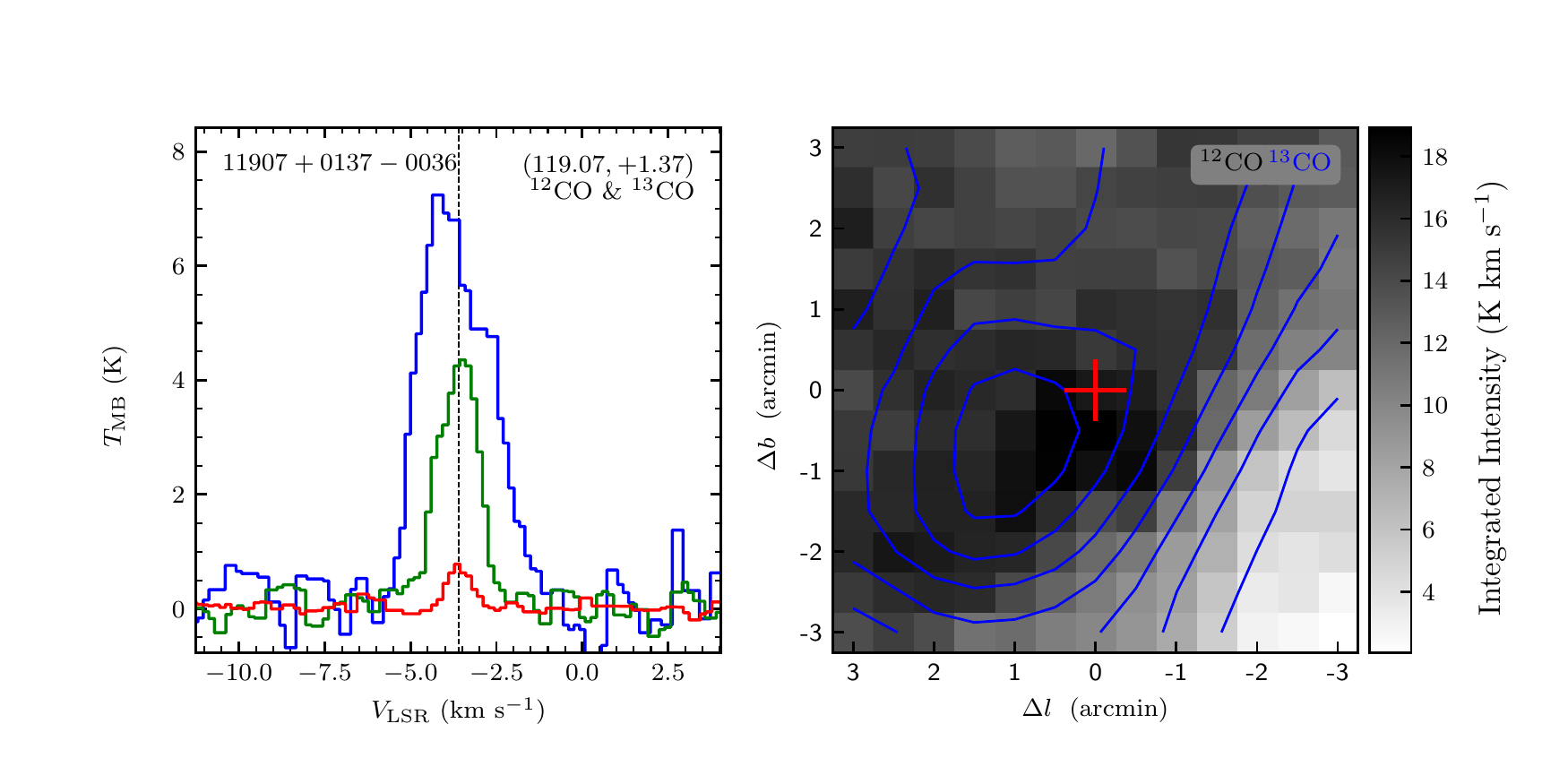}
\end{figure}
\clearpage

\begin{figure}
\includegraphics[width=9.0cm,angle=0]{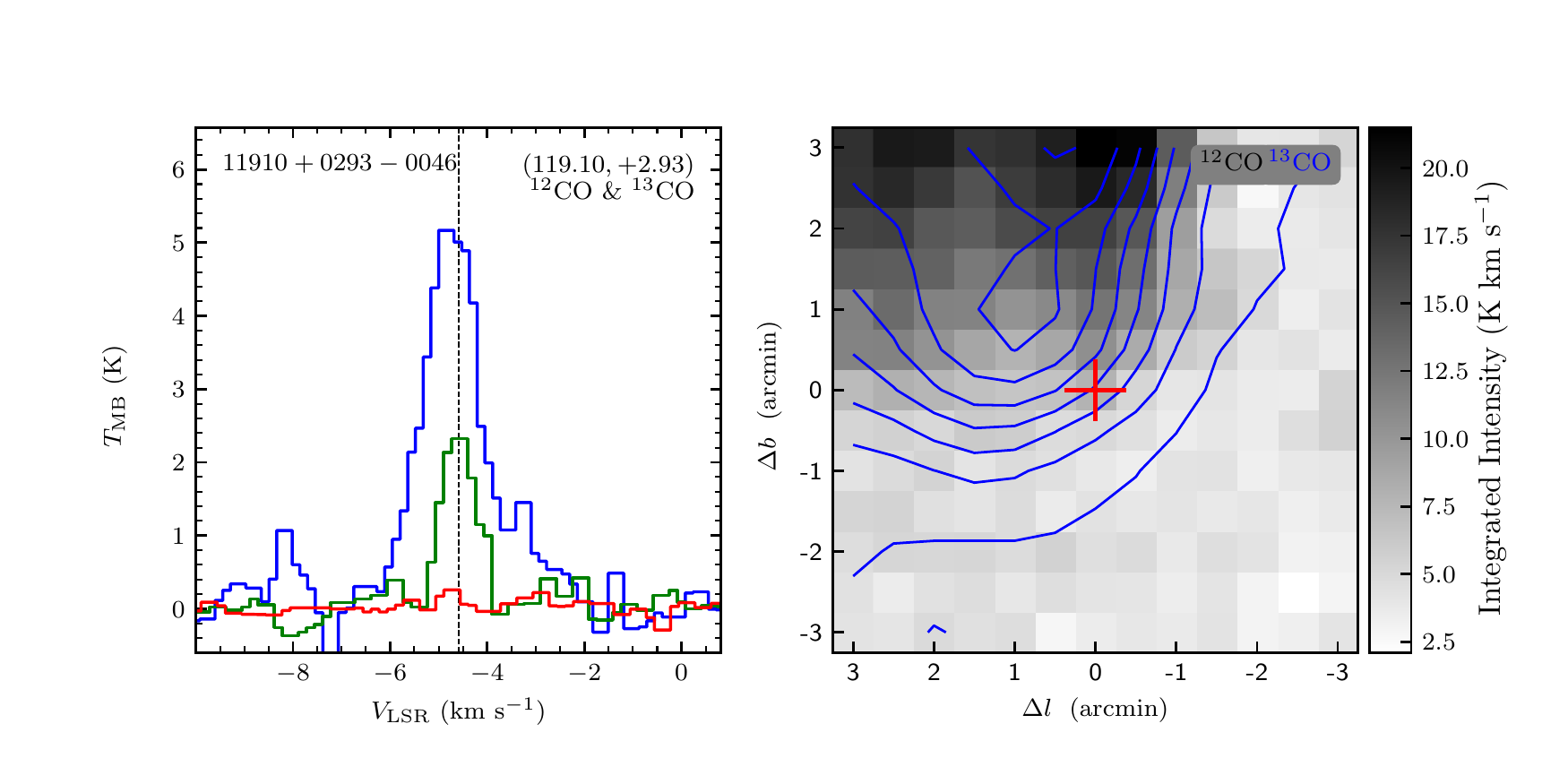}
\includegraphics[width=9.0cm,angle=0]{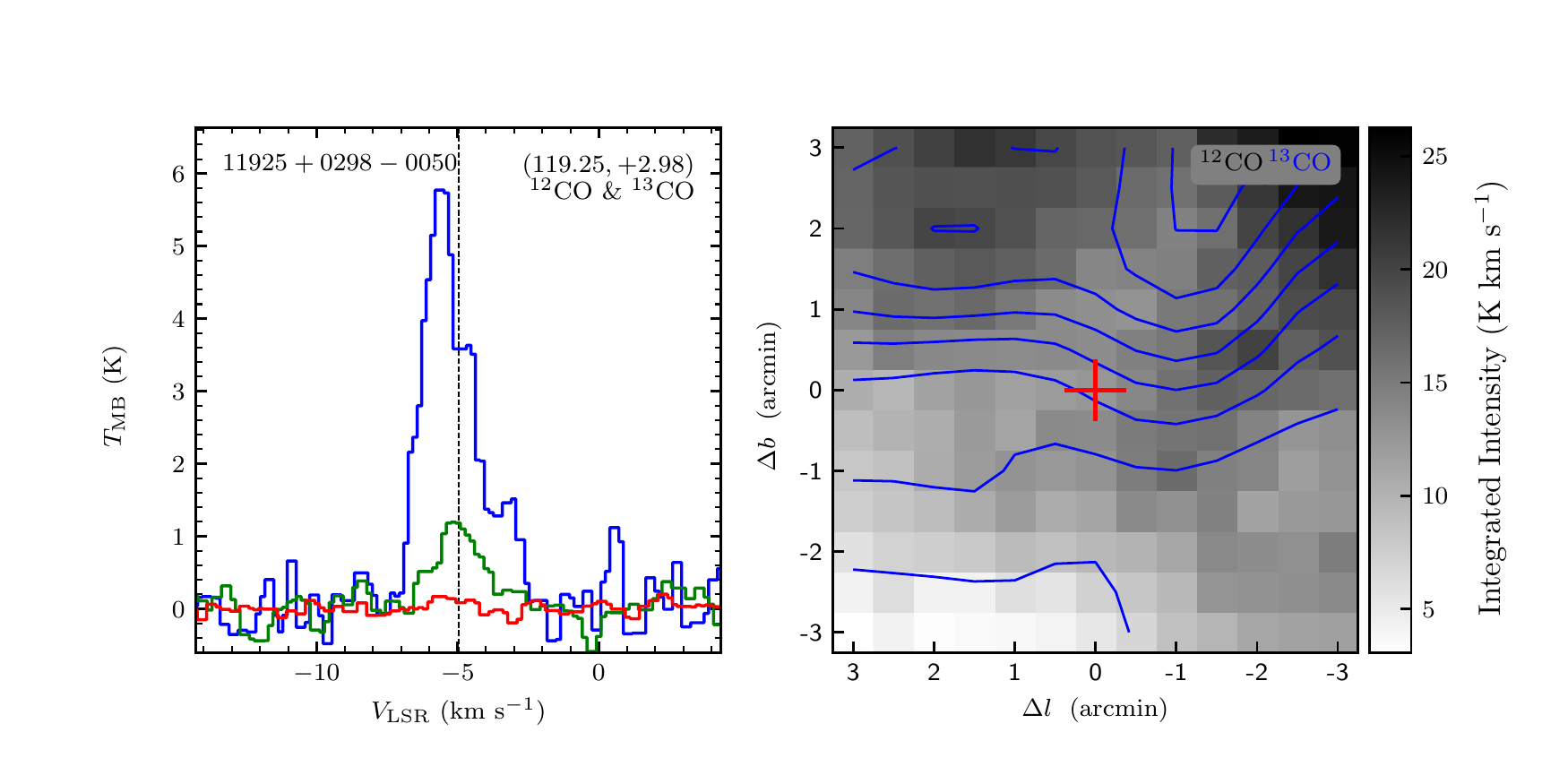}
\vspace{-0.5cm}

\includegraphics[width=9.0cm,angle=0]{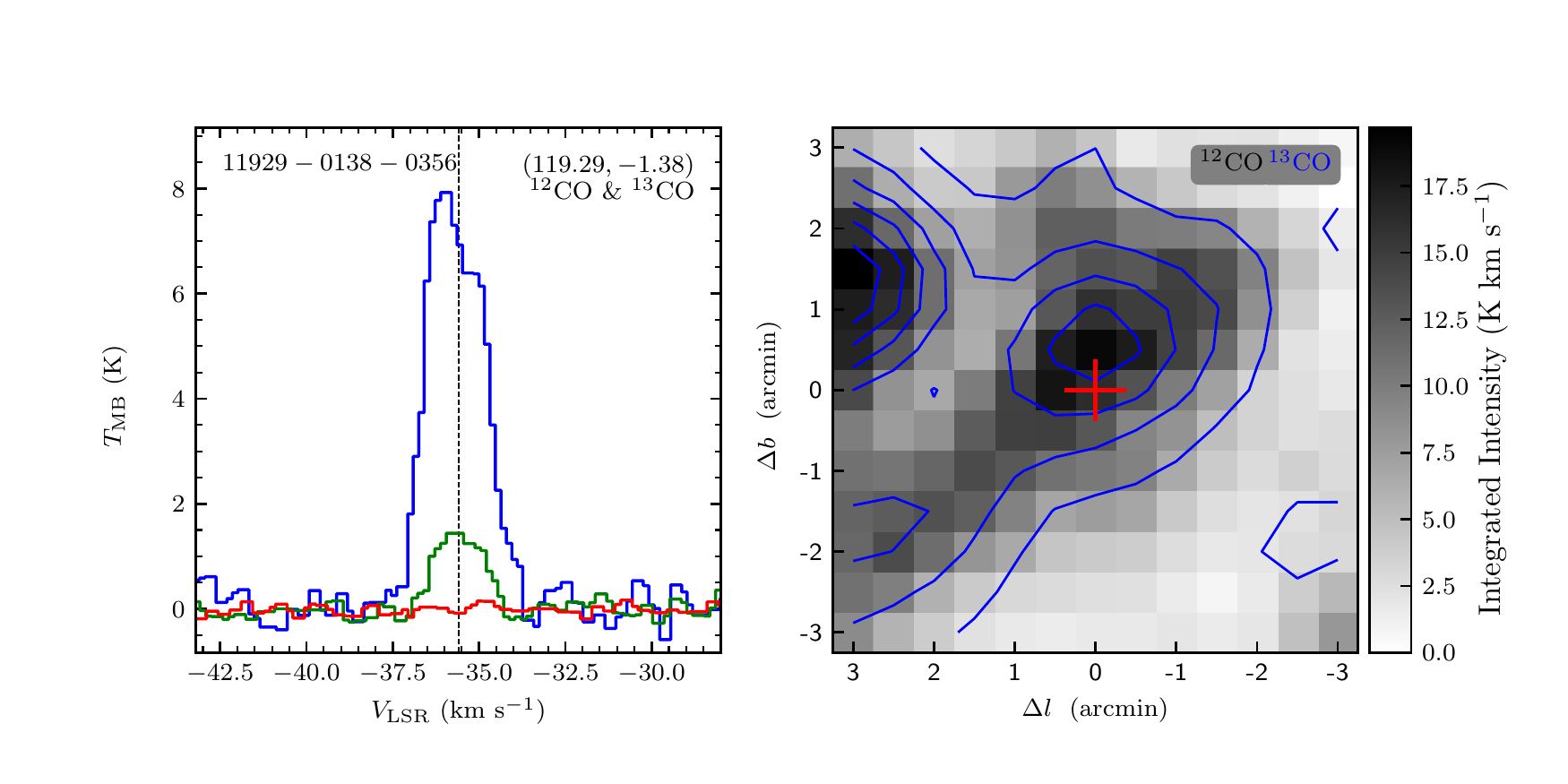}
\includegraphics[width=9.0cm,angle=0]{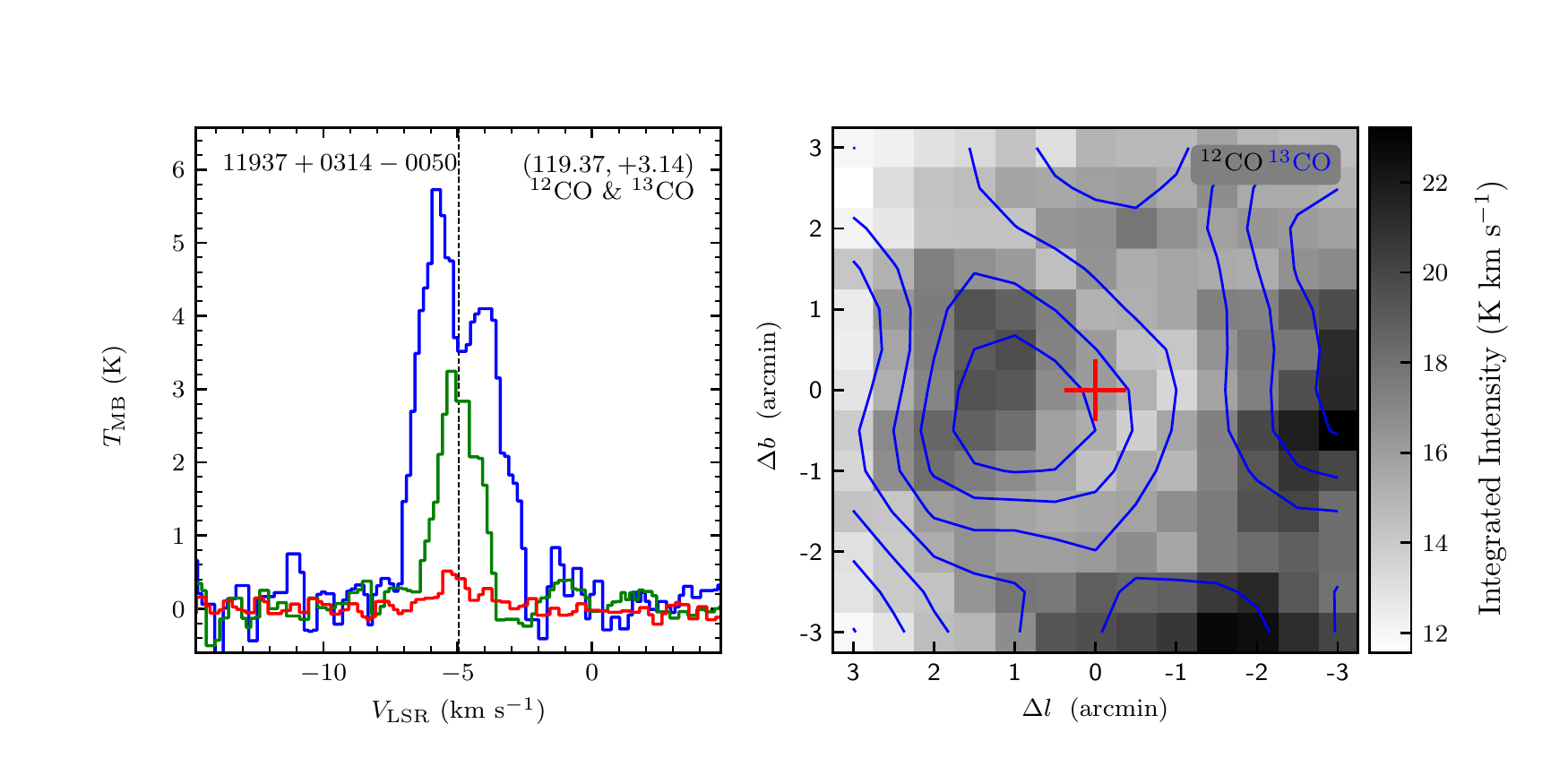}
\vspace{-0.5cm}

\includegraphics[width=9.0cm,angle=0]{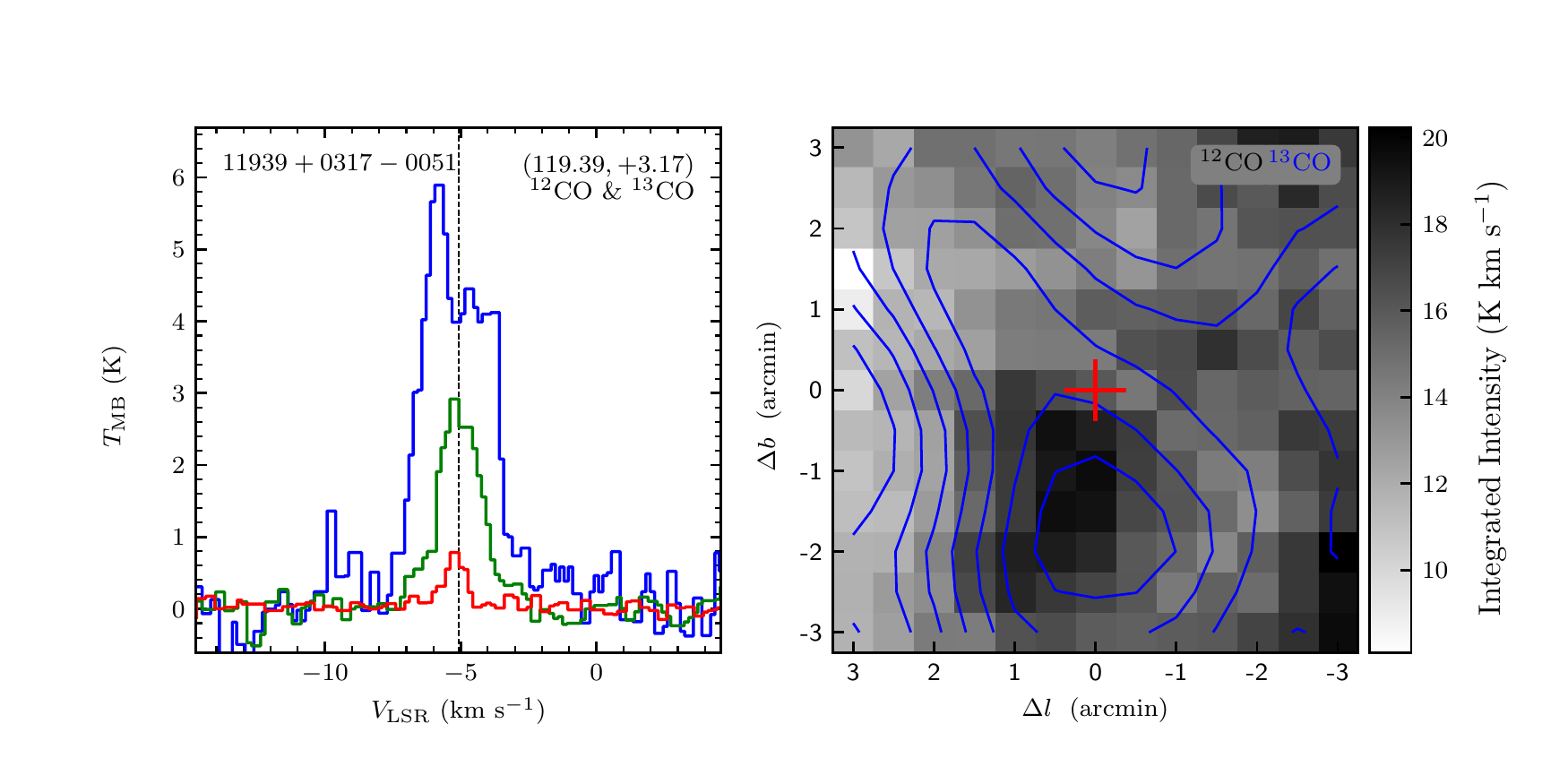}
\includegraphics[width=9.0cm,angle=0]{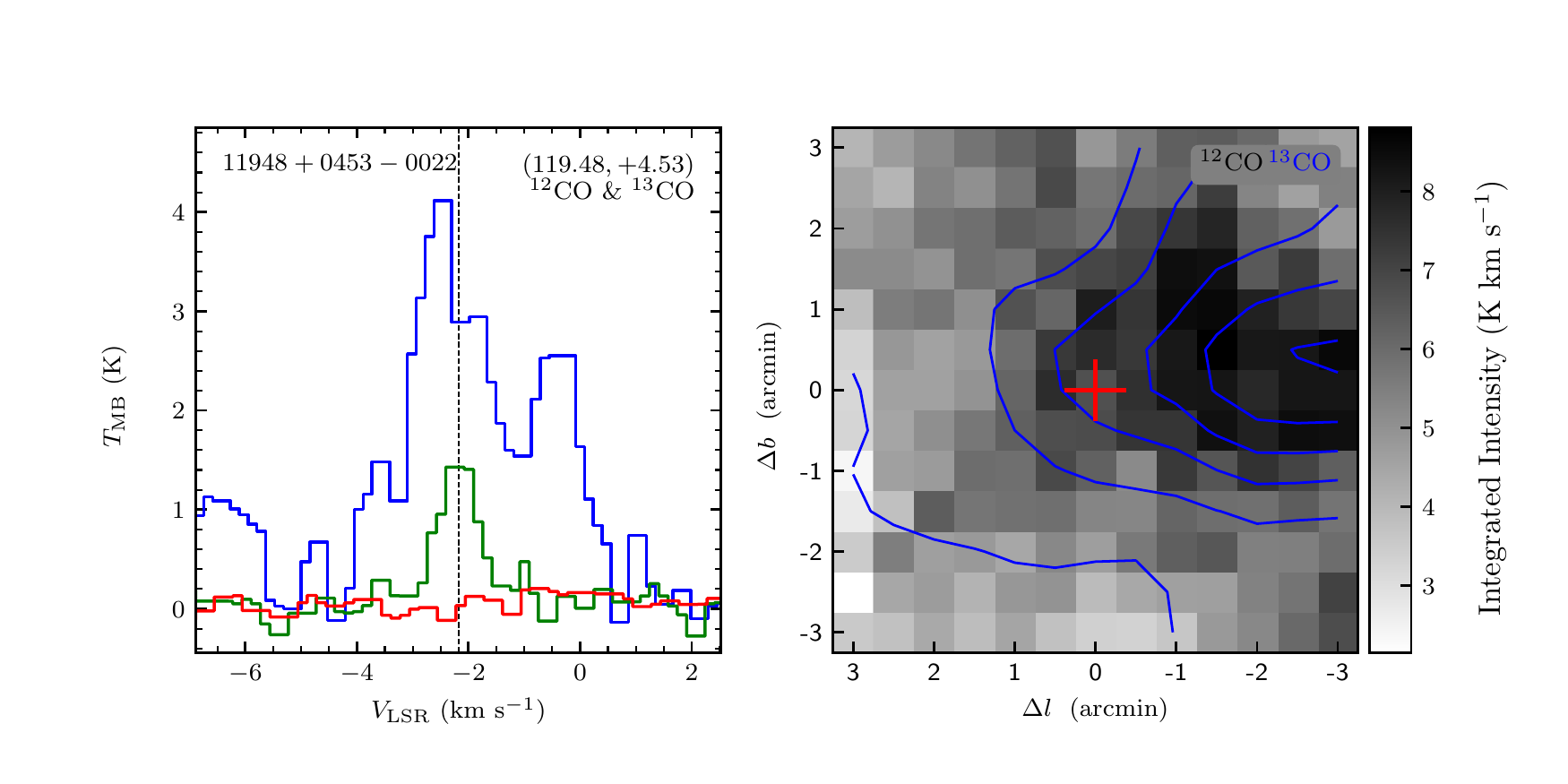}
\vspace{-0.5cm}

\includegraphics[width=9.0cm,angle=0]{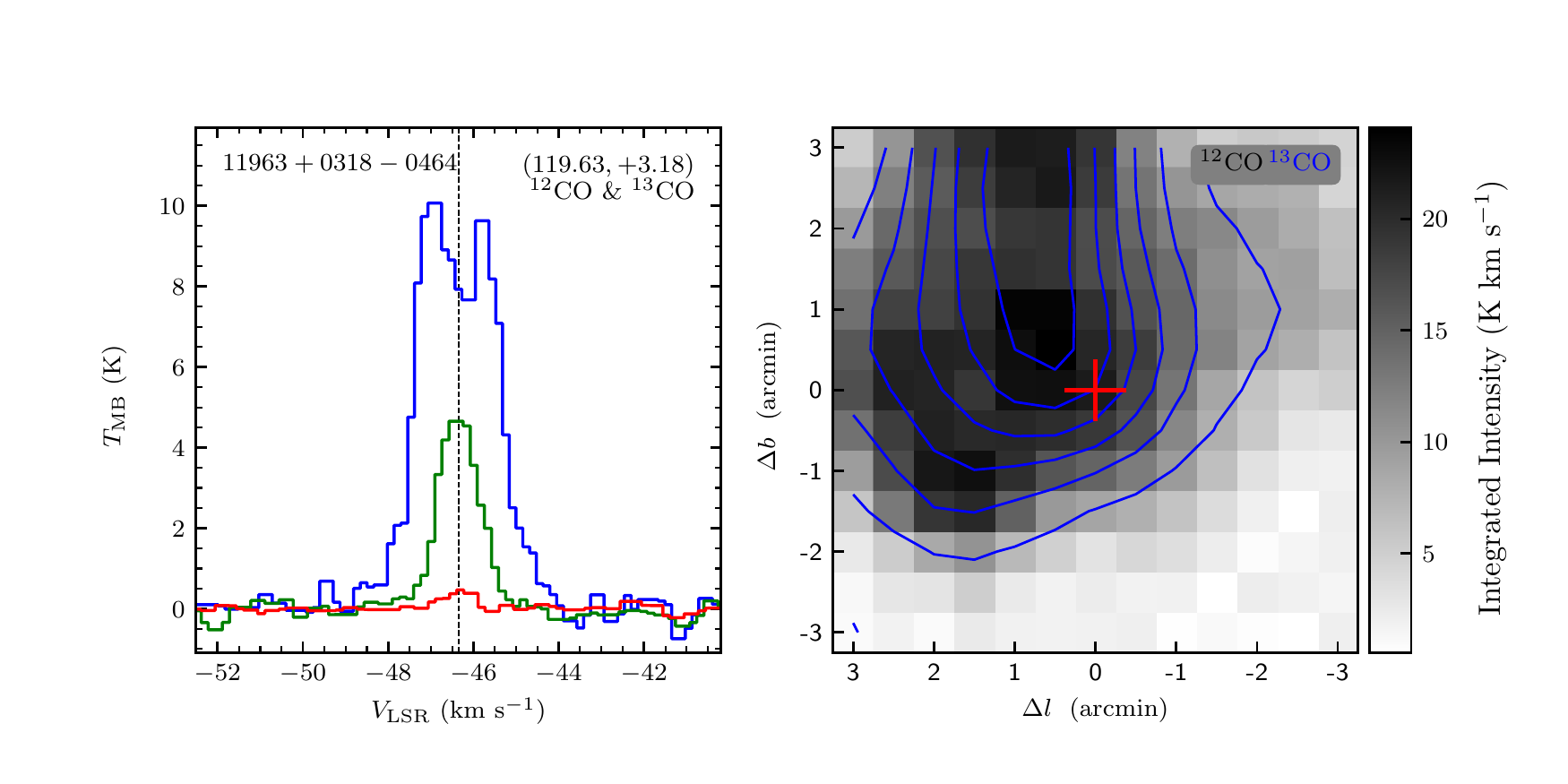}
\includegraphics[width=9.0cm,angle=0]{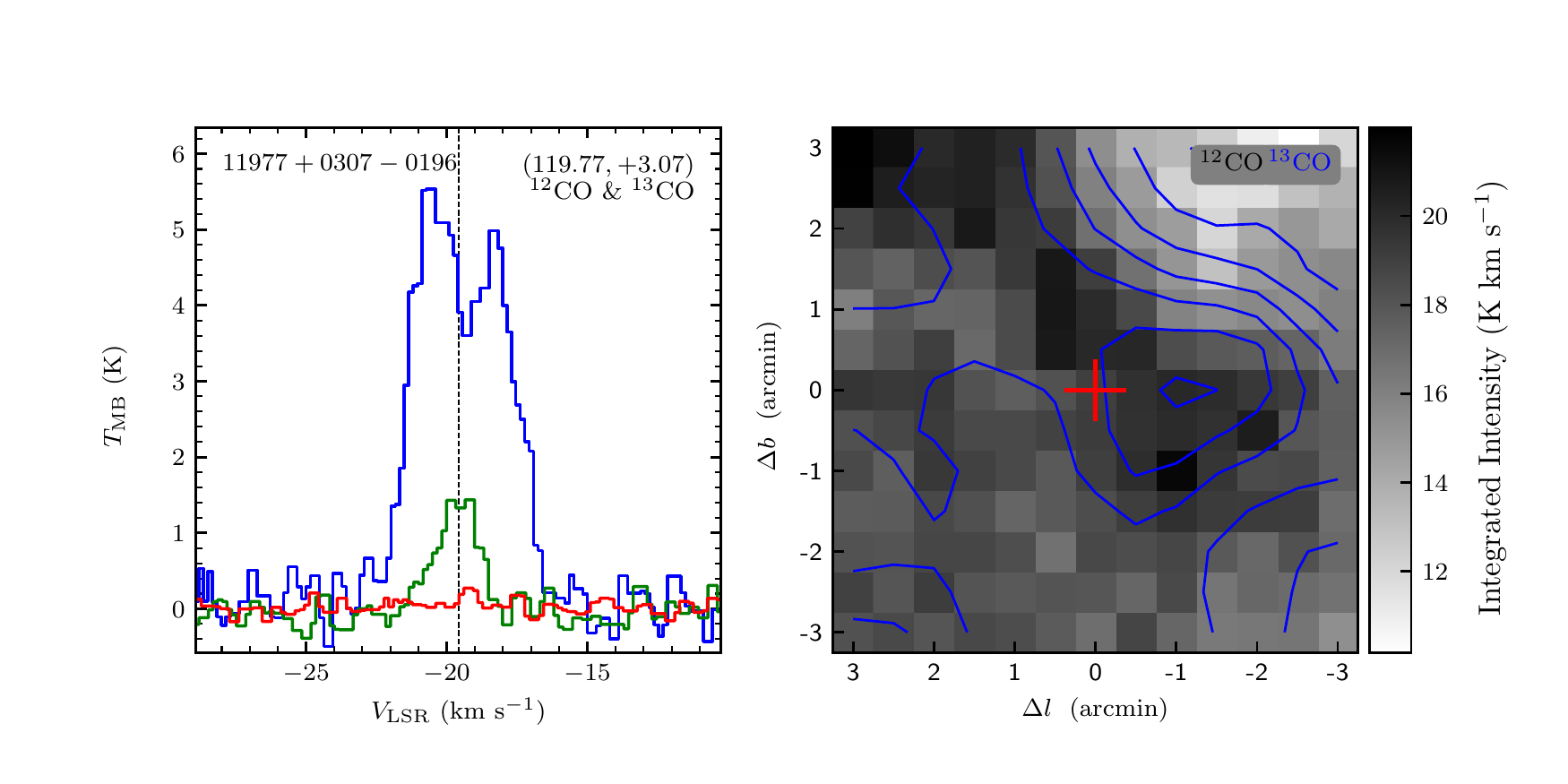}
\vspace{-0.5cm}

\includegraphics[width=9.0cm,angle=0]{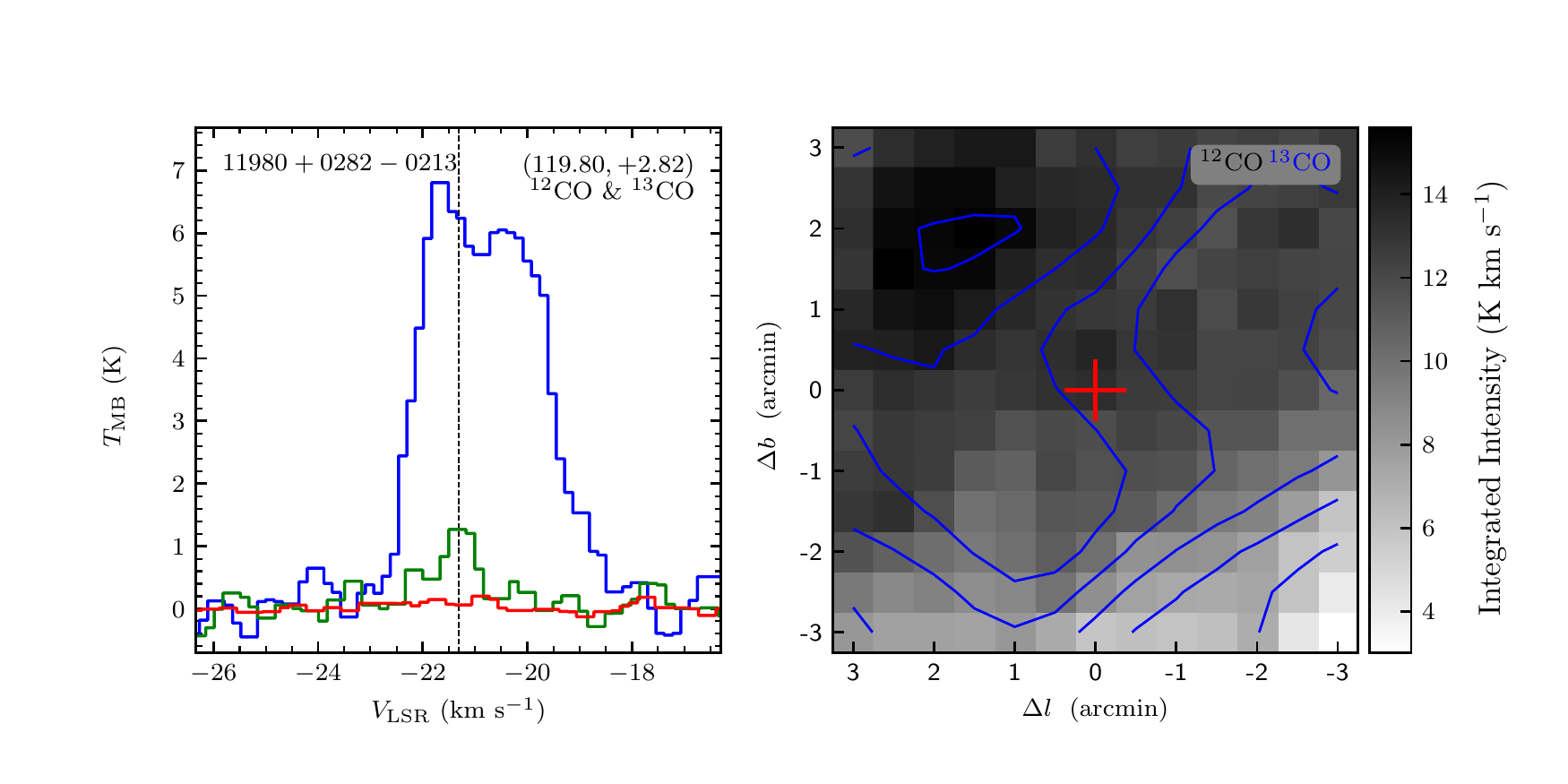}
\includegraphics[width=9.0cm,angle=0]{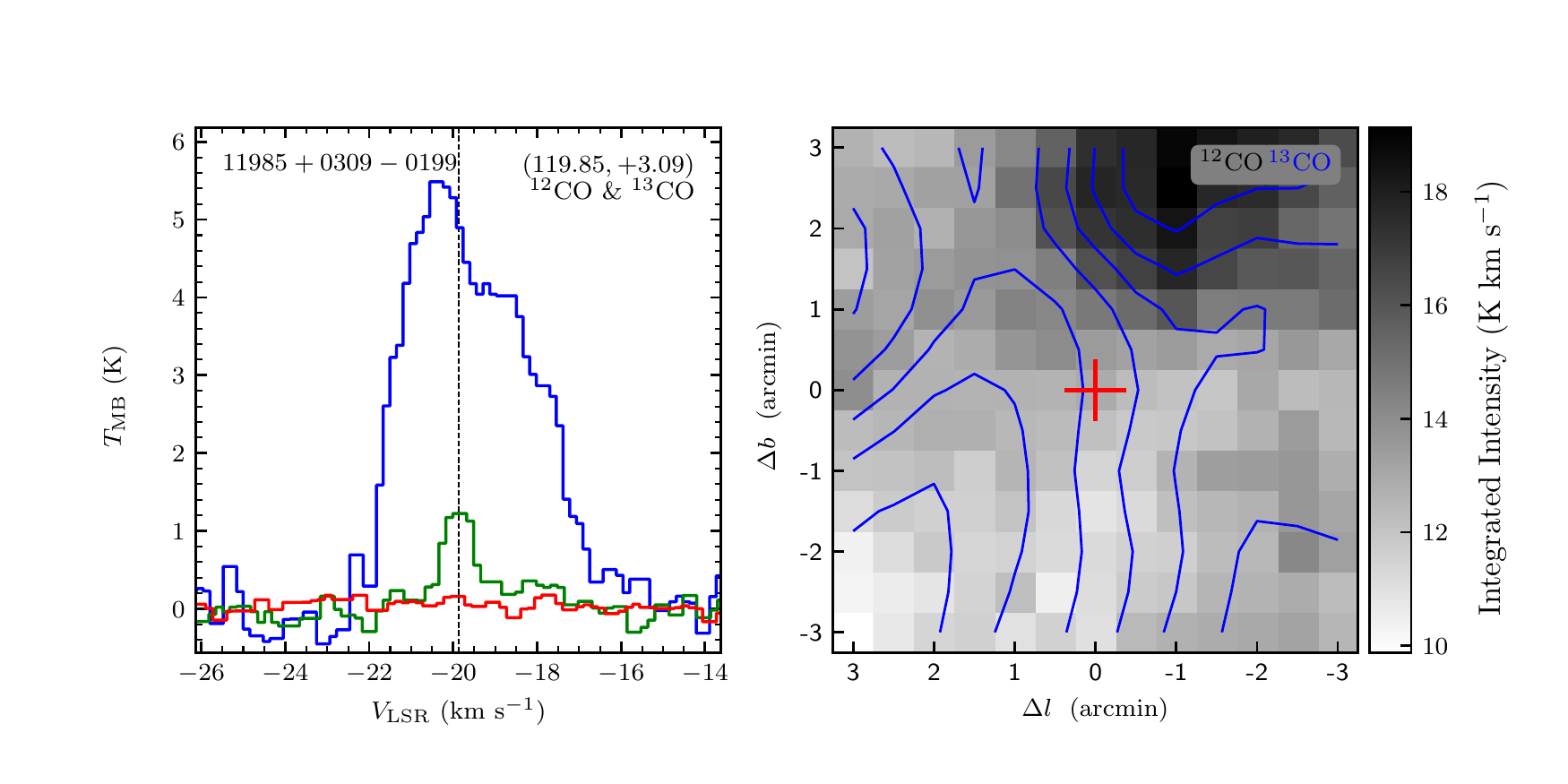}
\end{figure}
\clearpage

\begin{figure}
\includegraphics[width=9.0cm,angle=0]{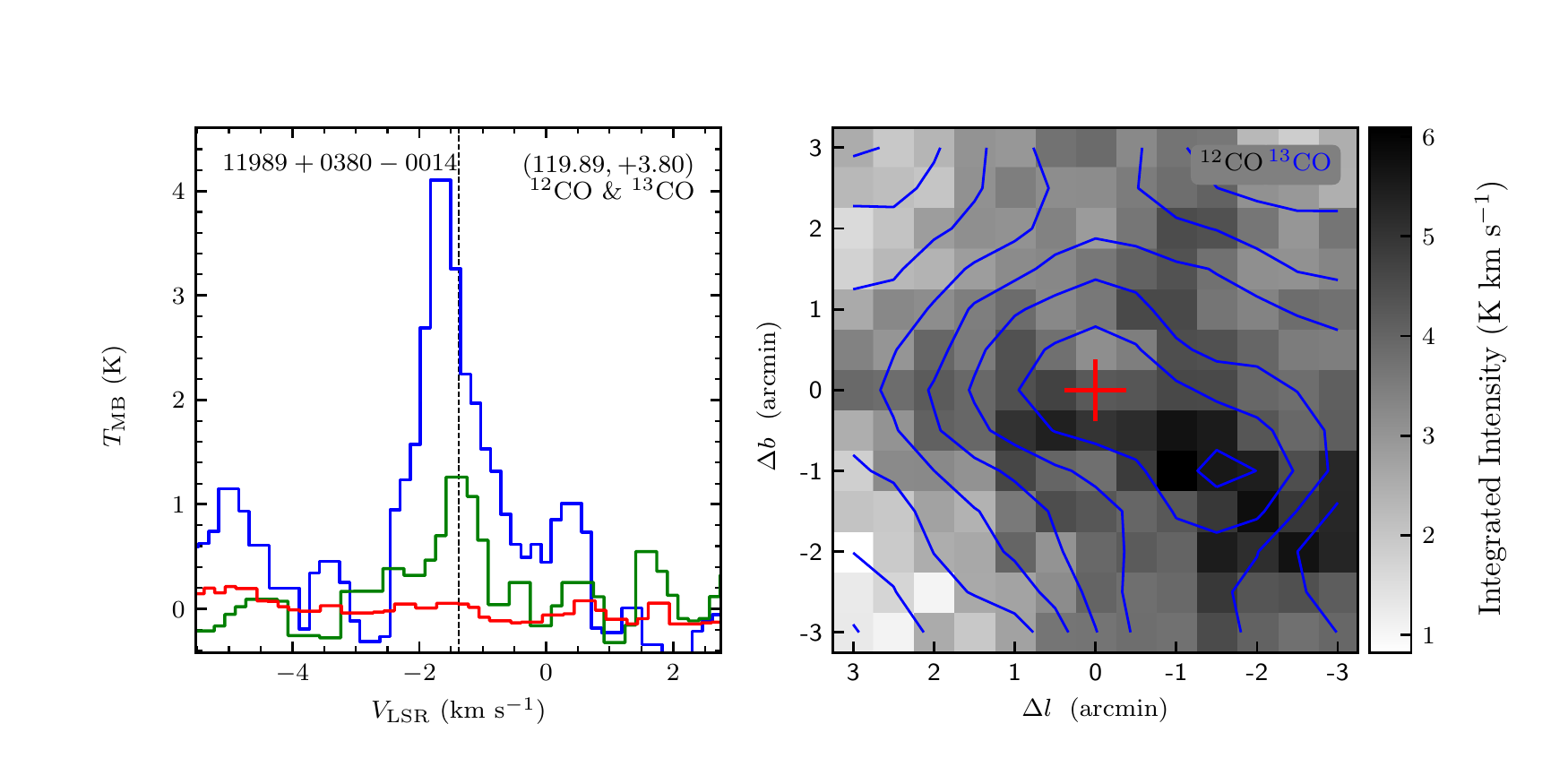}
\includegraphics[width=9.0cm,angle=0]{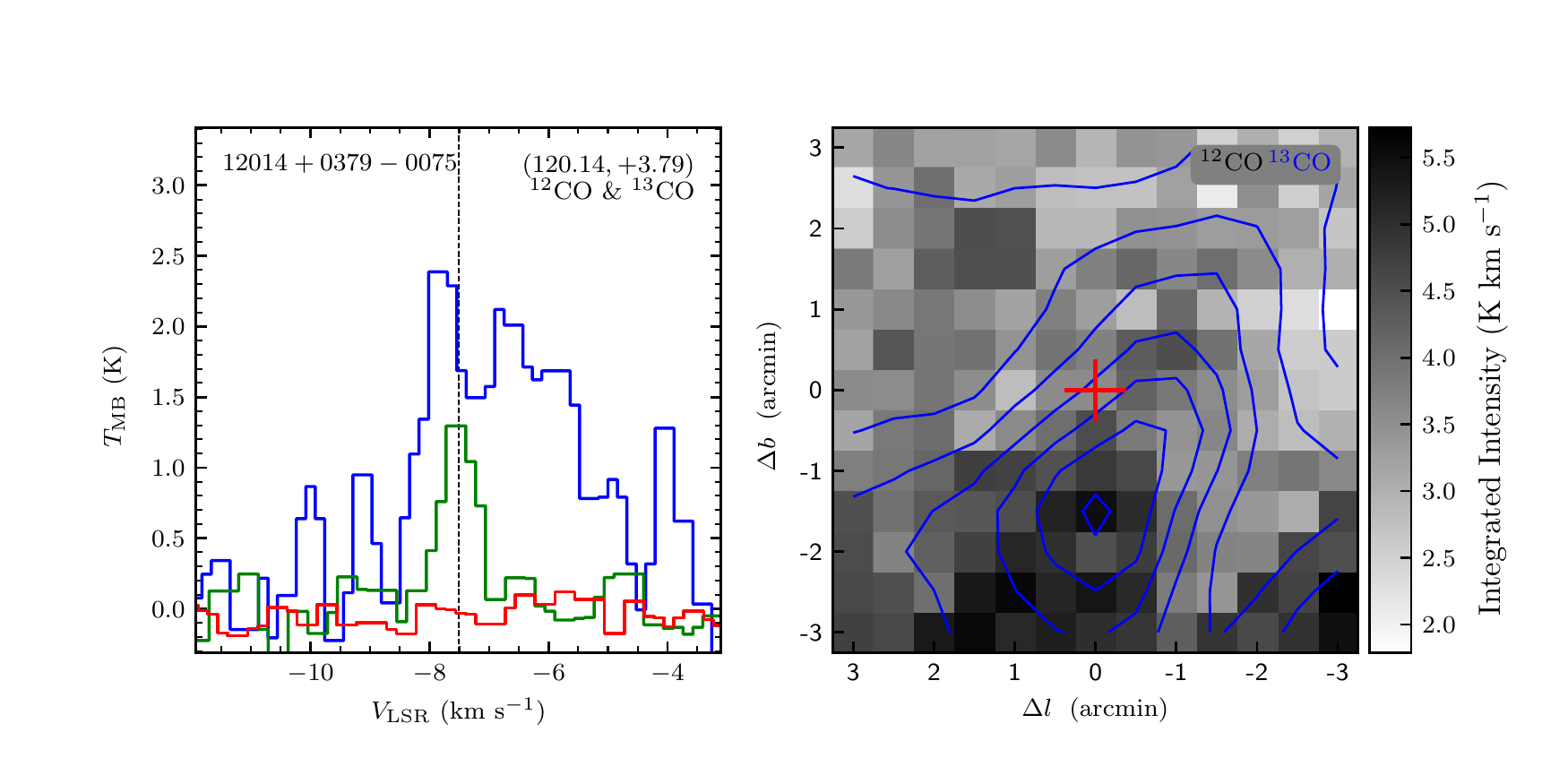}
\vspace{-0.5cm}

\includegraphics[width=9.0cm,angle=0]{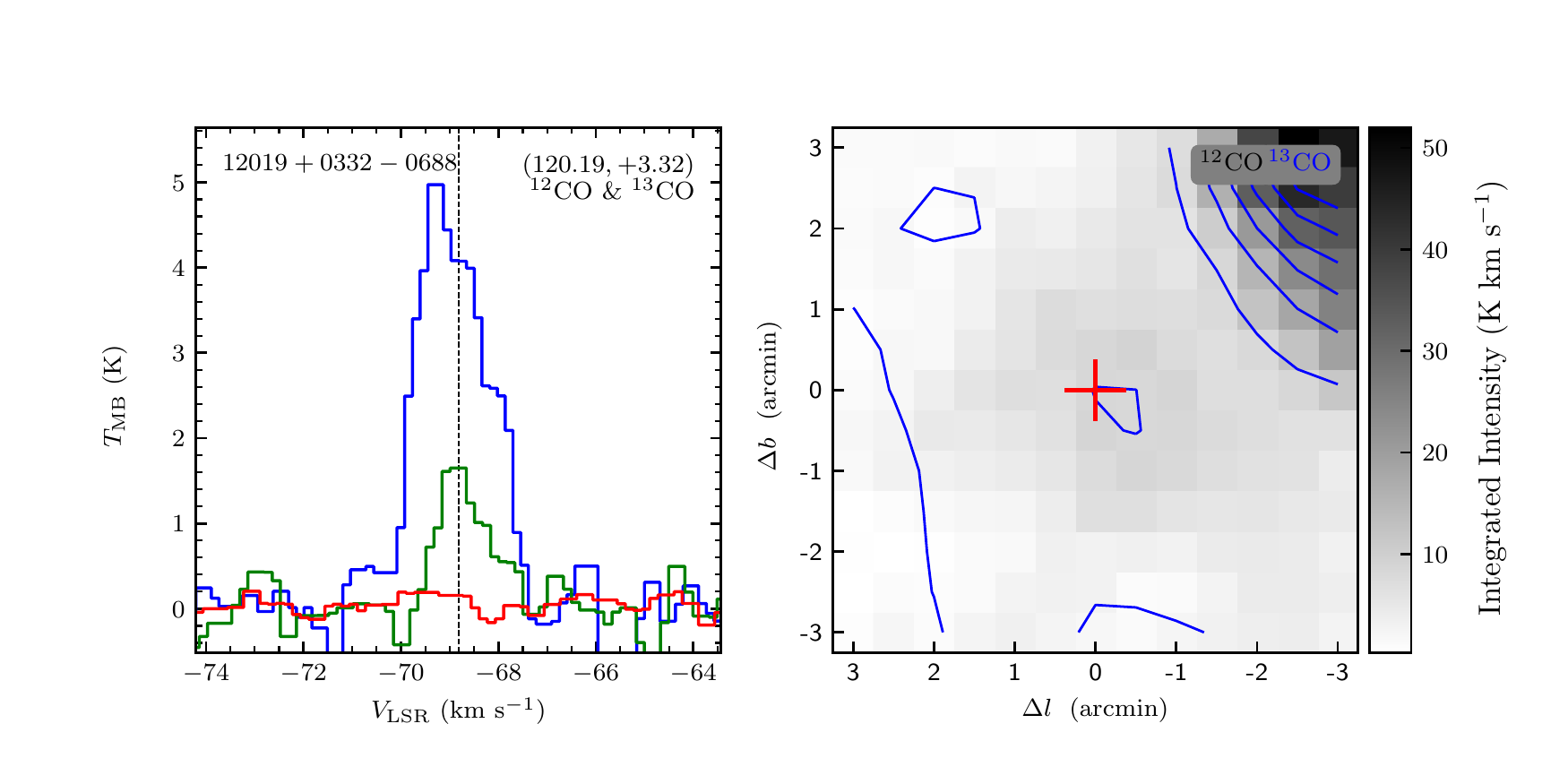}
\includegraphics[width=9.0cm,angle=0]{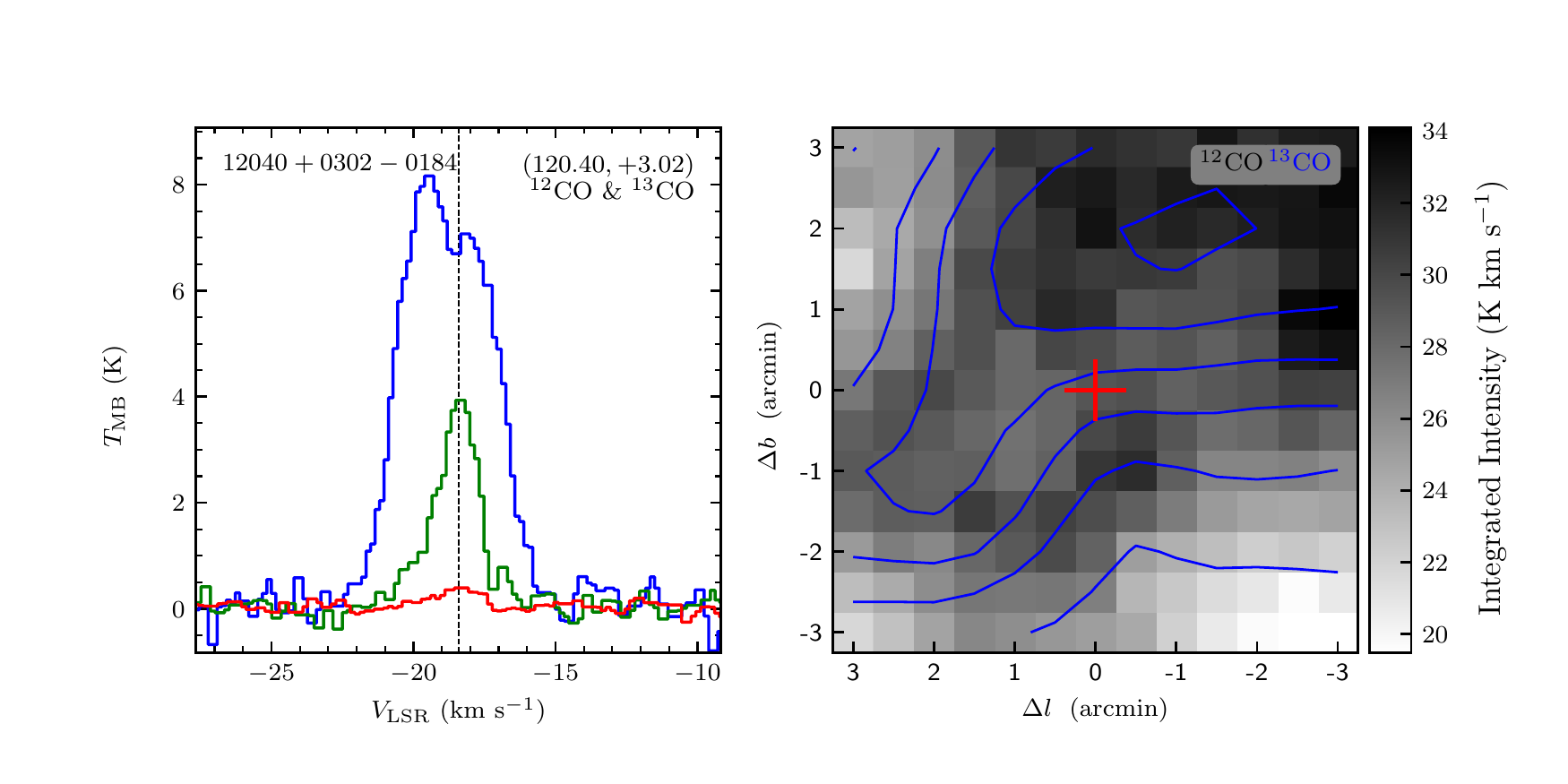}
\vspace{-0.5cm}

\includegraphics[width=9.0cm,angle=0]{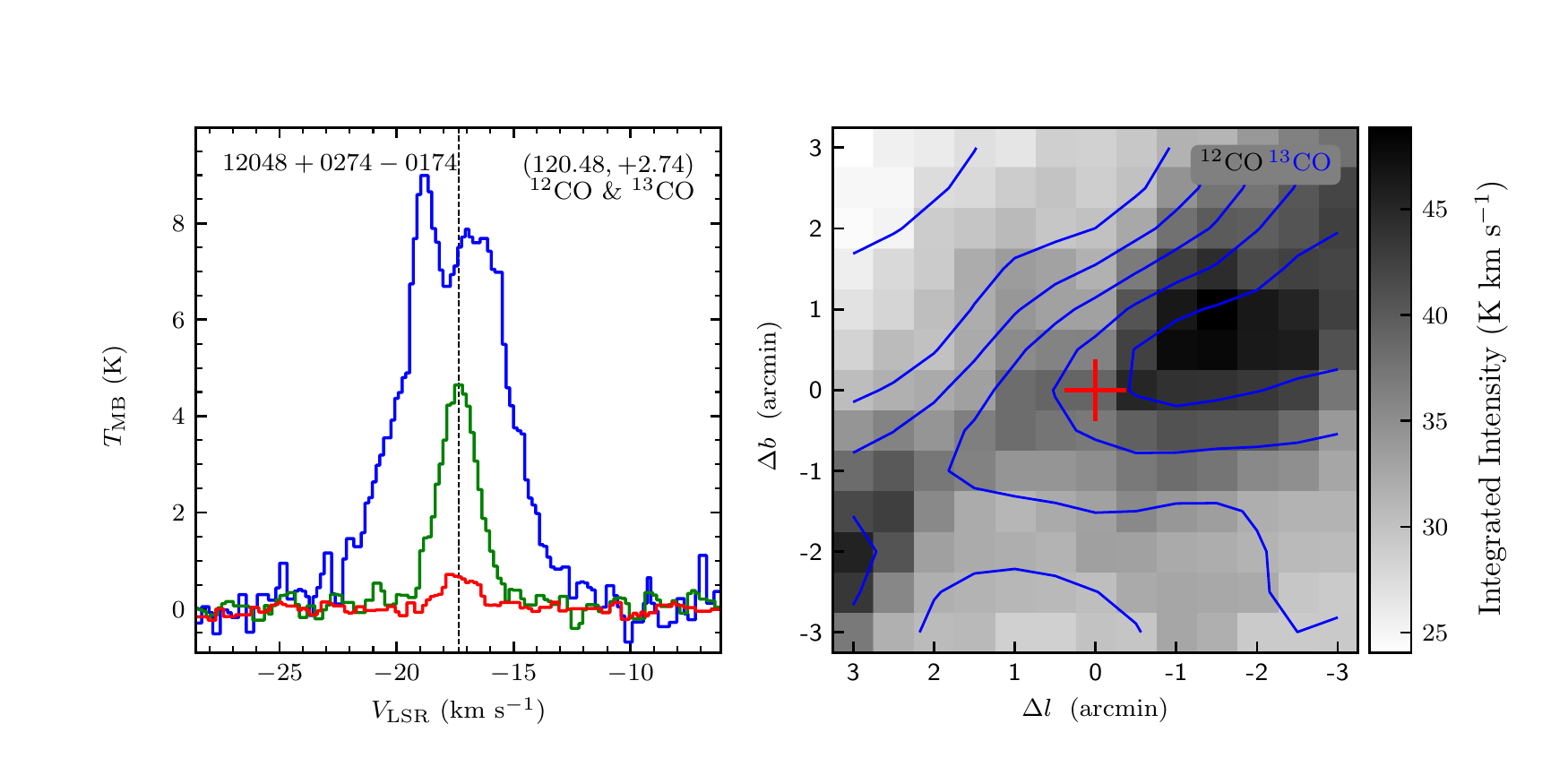}
\includegraphics[width=9.0cm,angle=0]{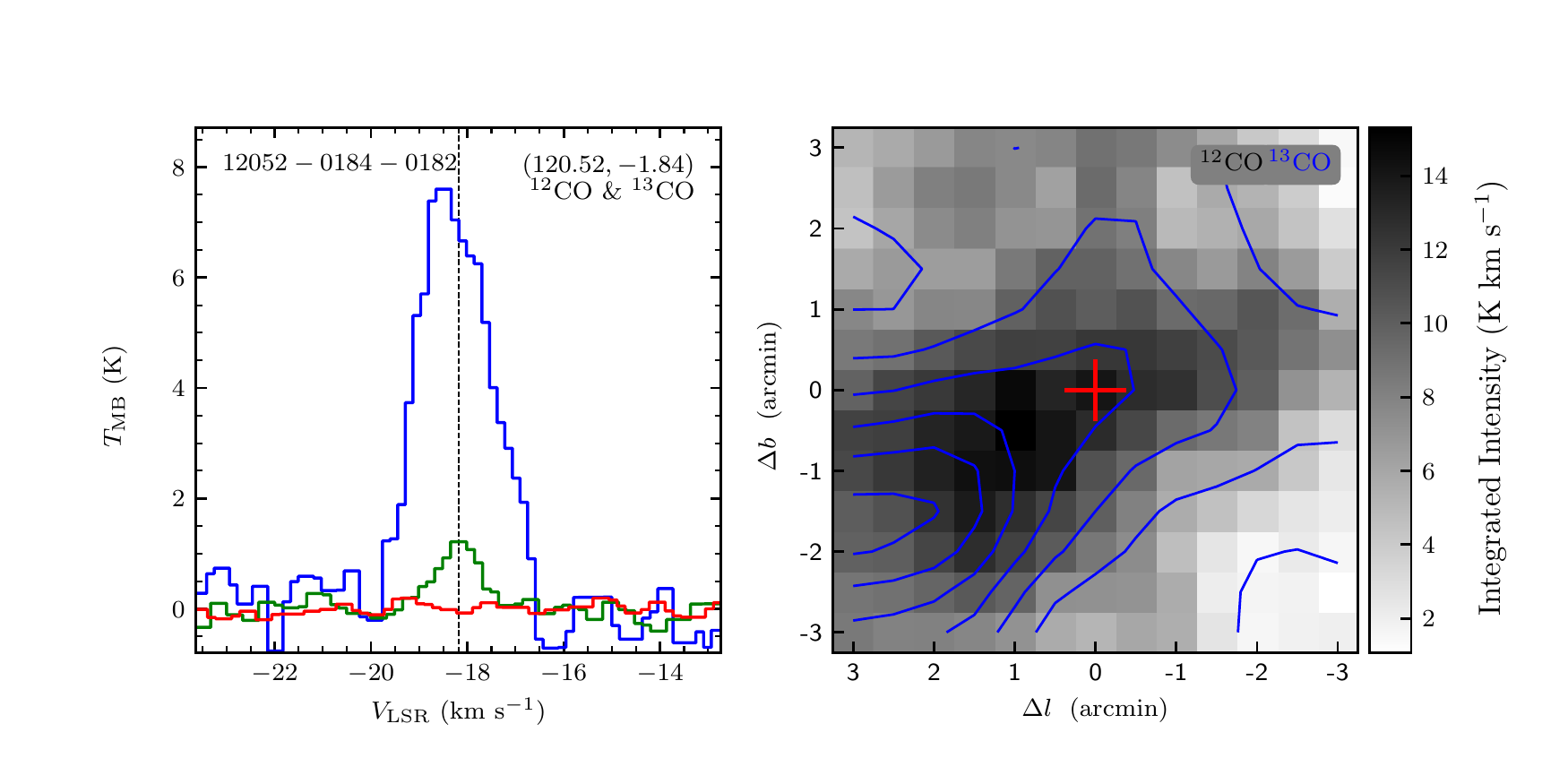}
\vspace{-0.5cm}

\includegraphics[width=9.0cm,angle=0]{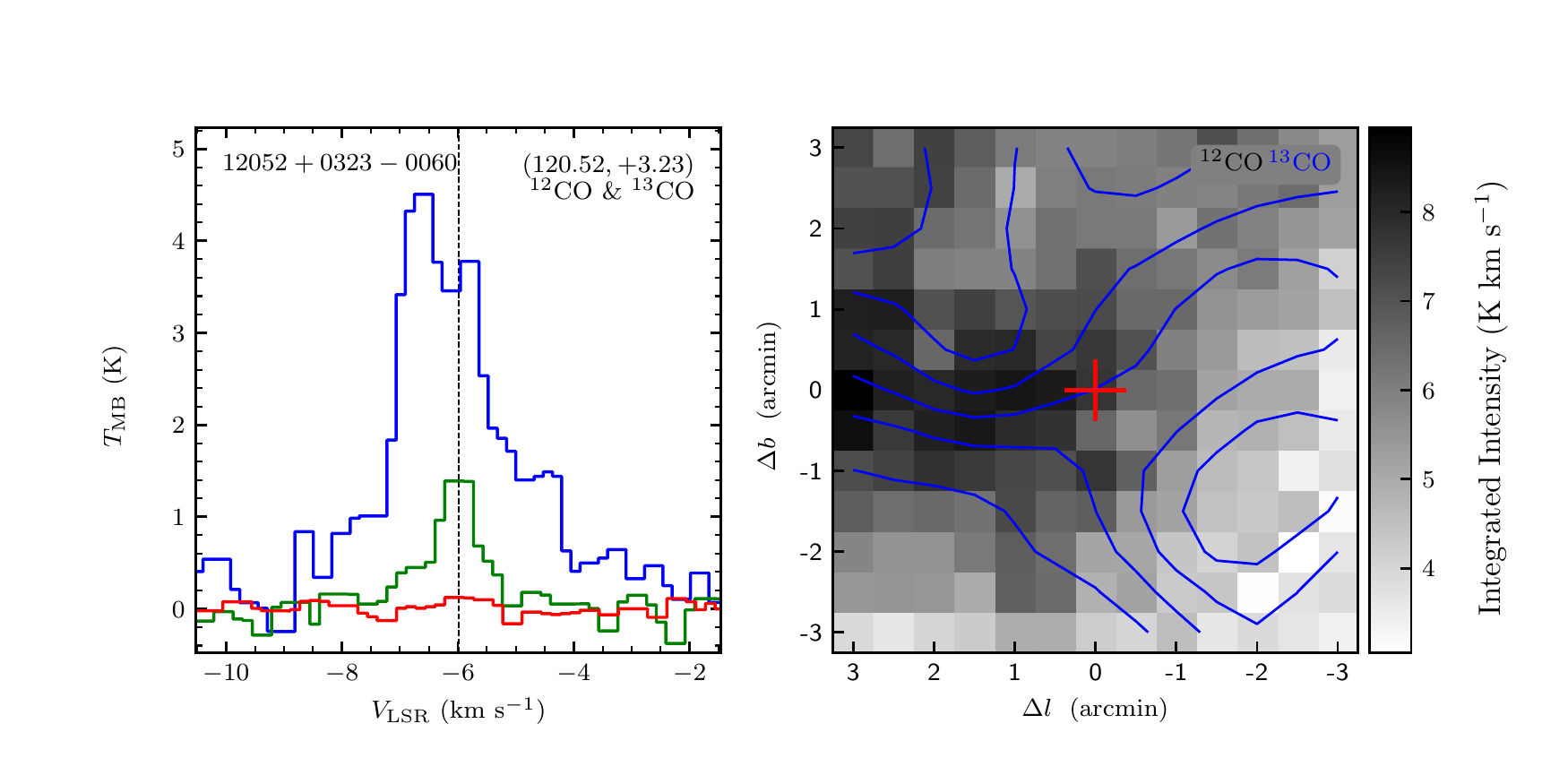}
\includegraphics[width=9.0cm,angle=0]{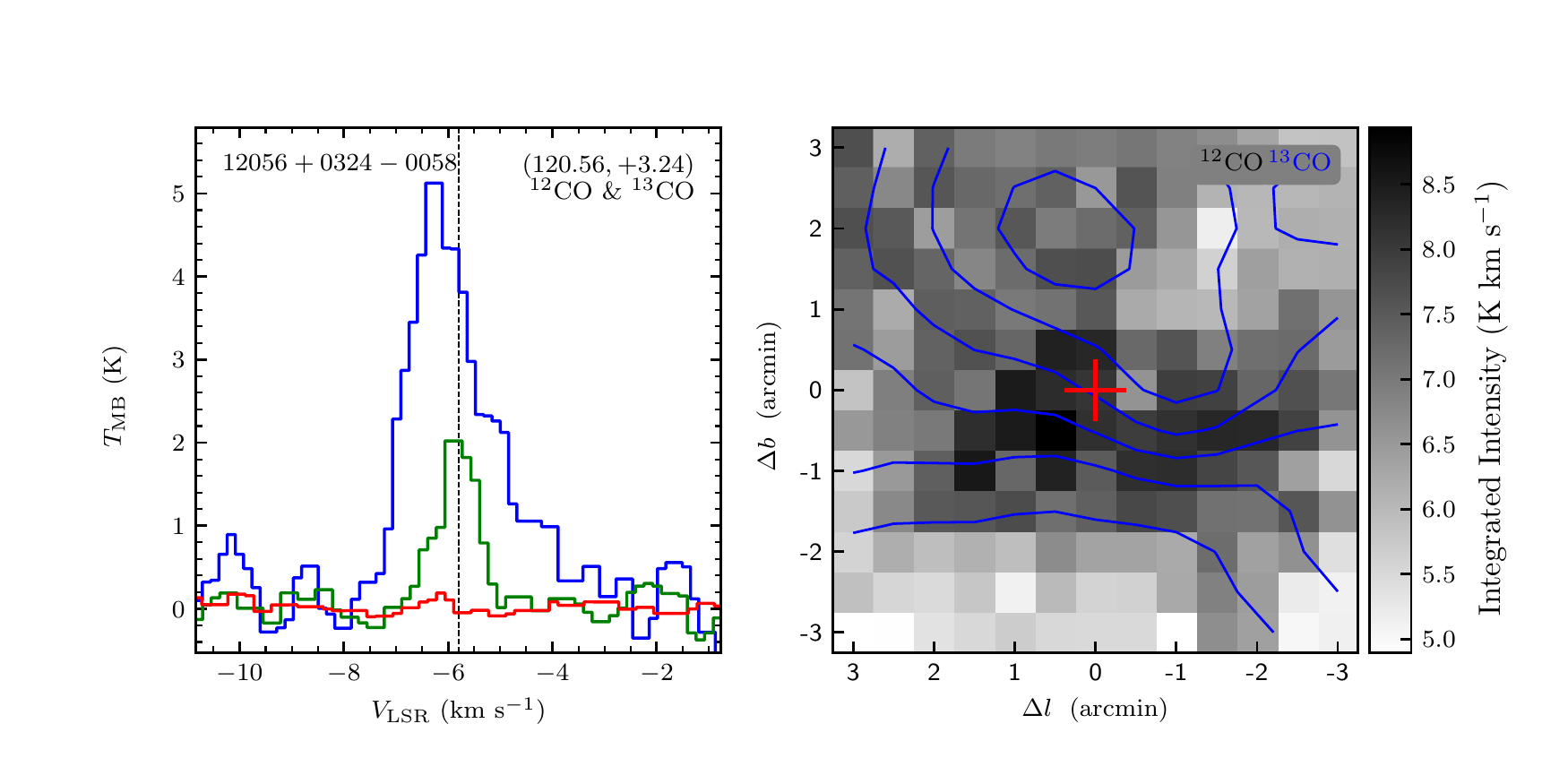}
\vspace{-0.5cm}

\includegraphics[width=9.0cm,angle=0]{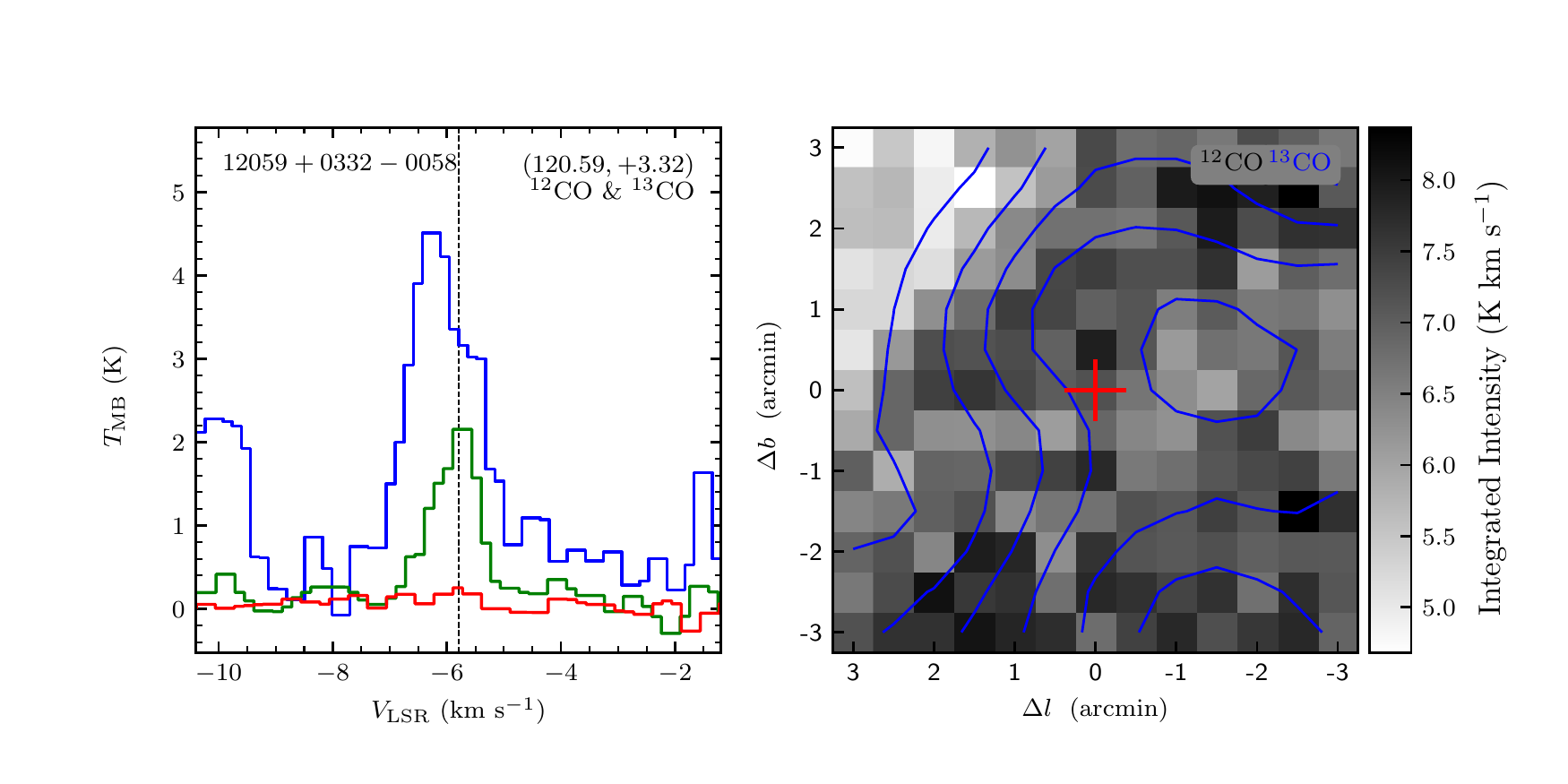}
\includegraphics[width=9.0cm,angle=0]{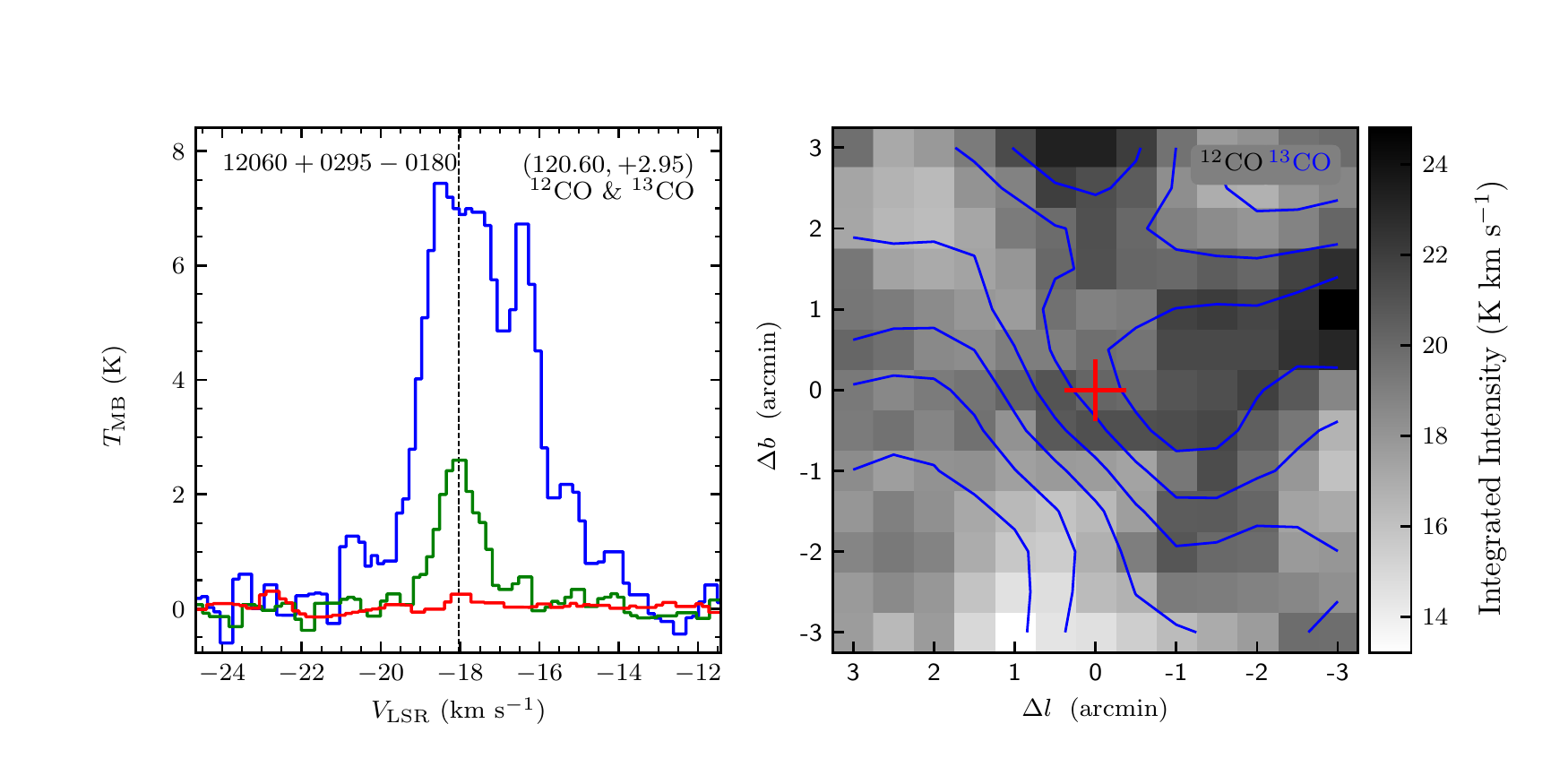}
\end{figure}
\clearpage

\begin{figure}
\includegraphics[width=9.0cm,angle=0]{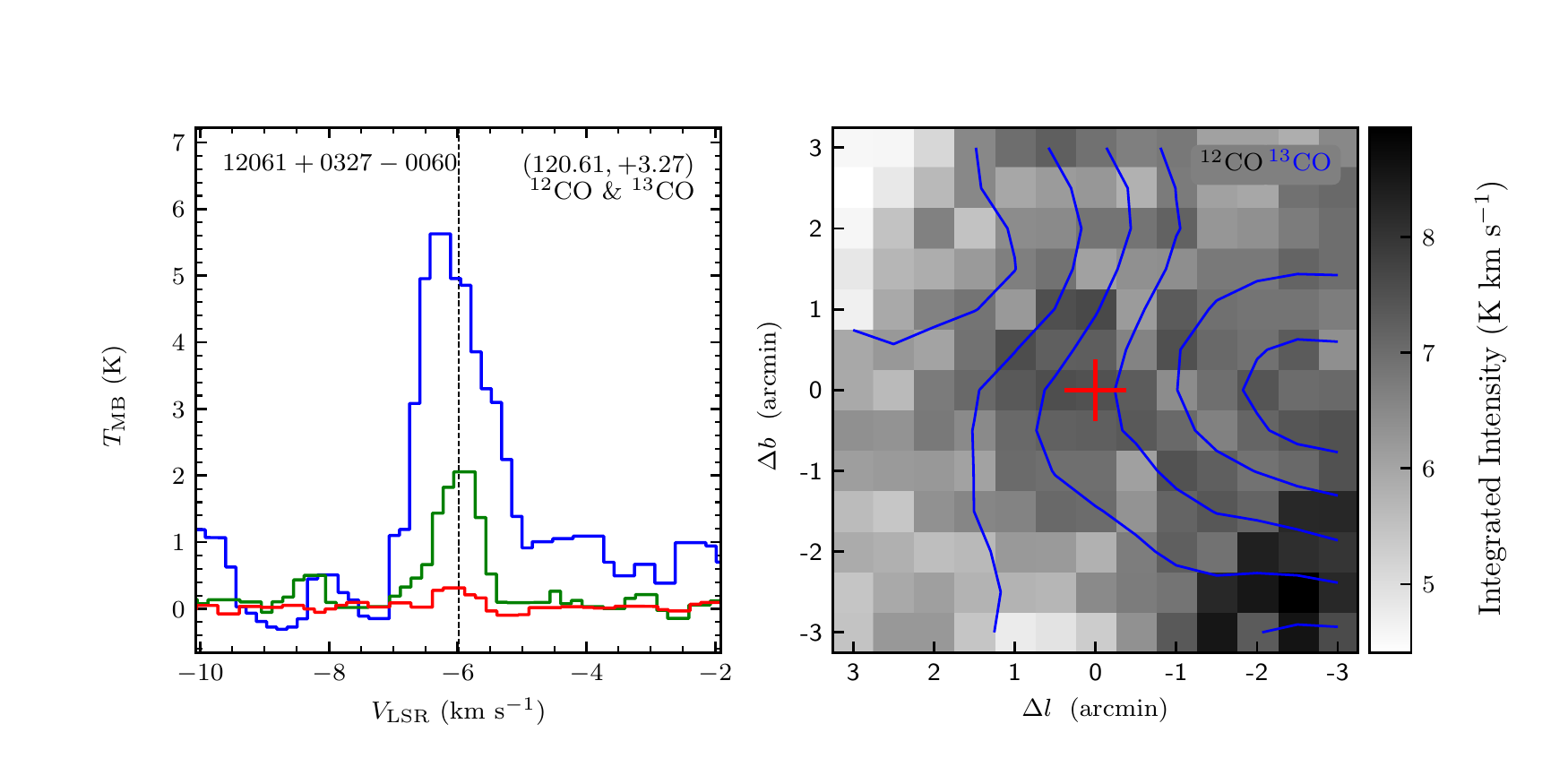}
\includegraphics[width=9.0cm,angle=0]{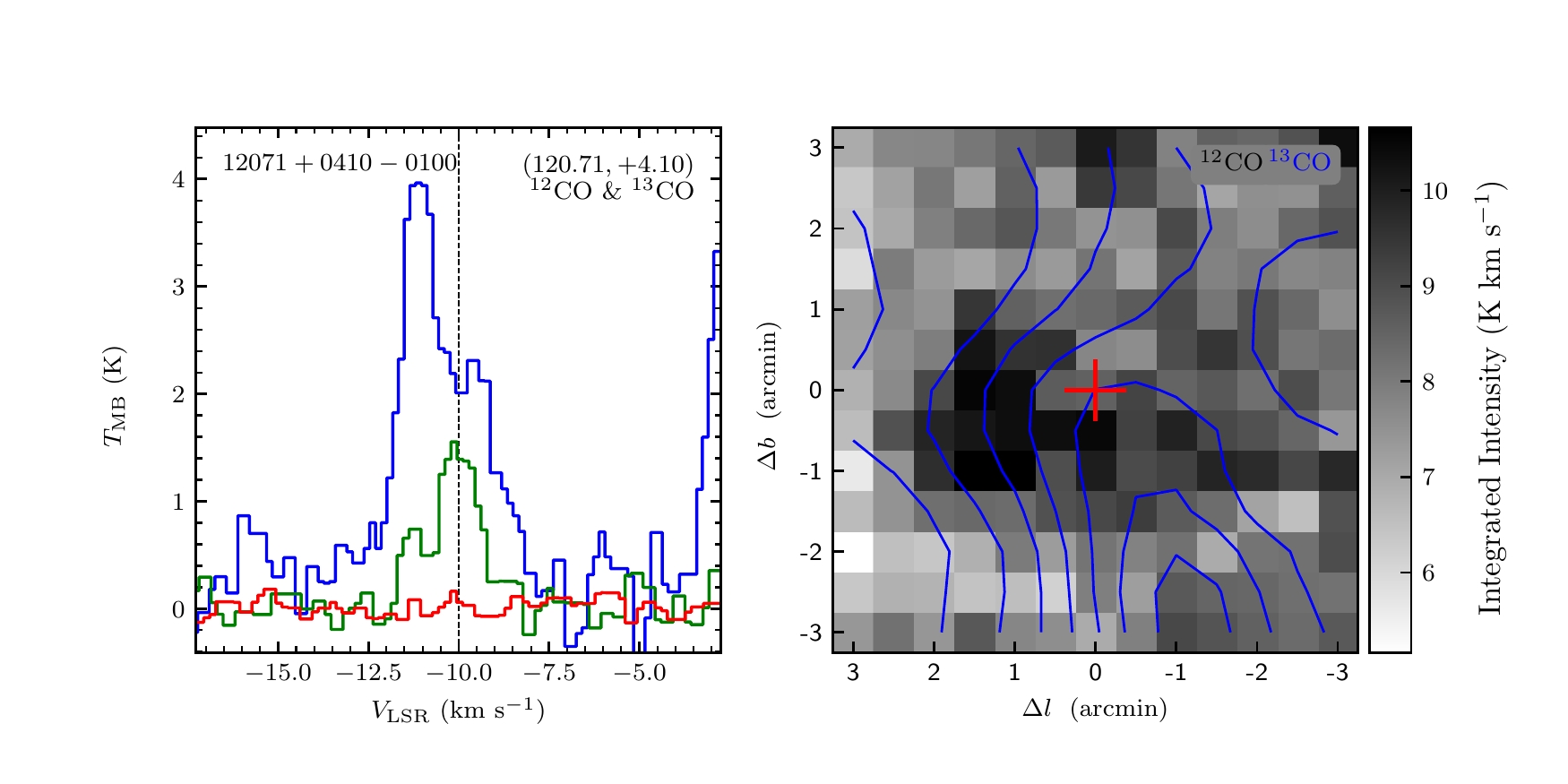}
\vspace{-0.5cm}

\includegraphics[width=9.0cm,angle=0]{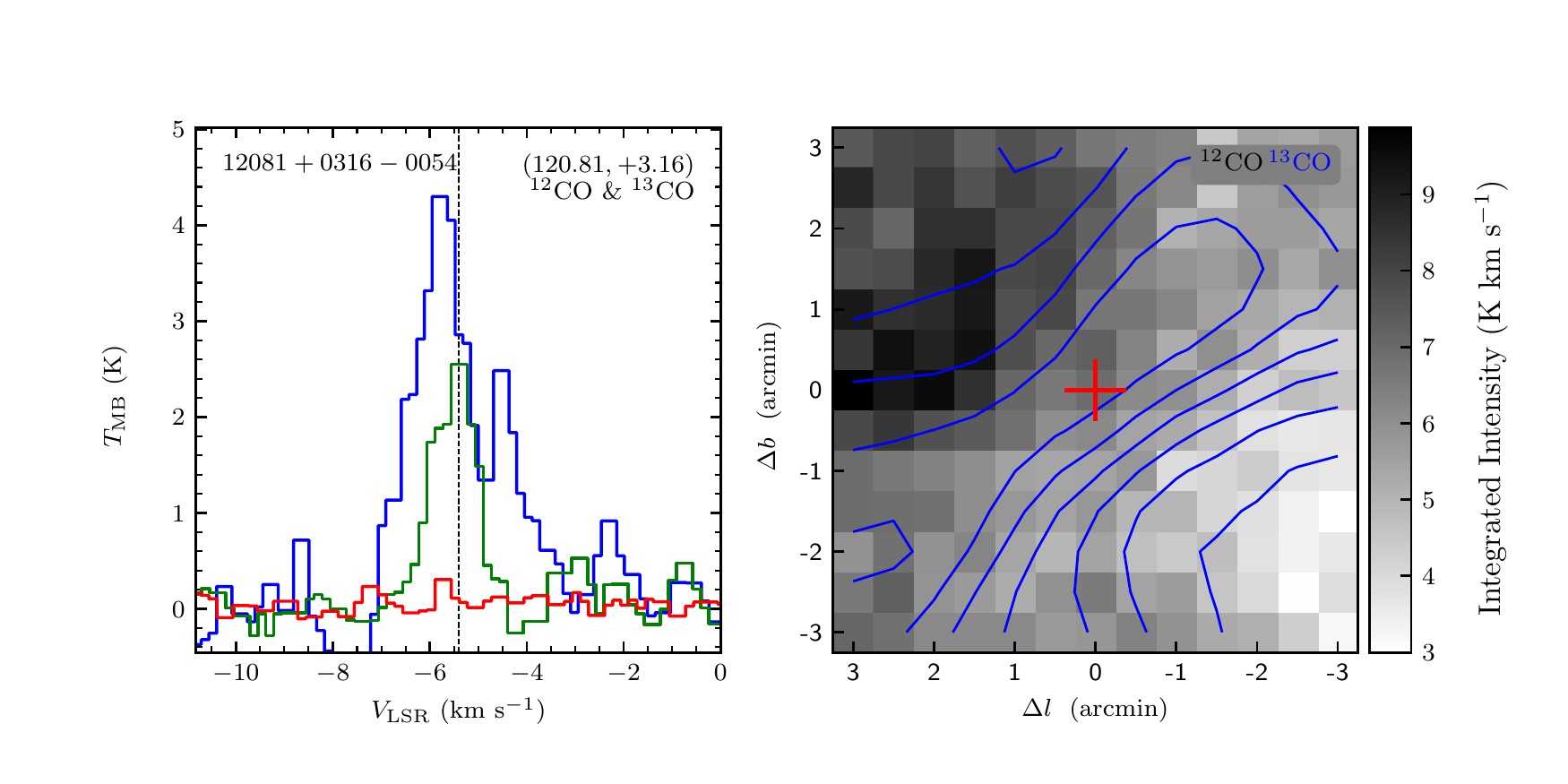}
\includegraphics[width=9.0cm,angle=0]{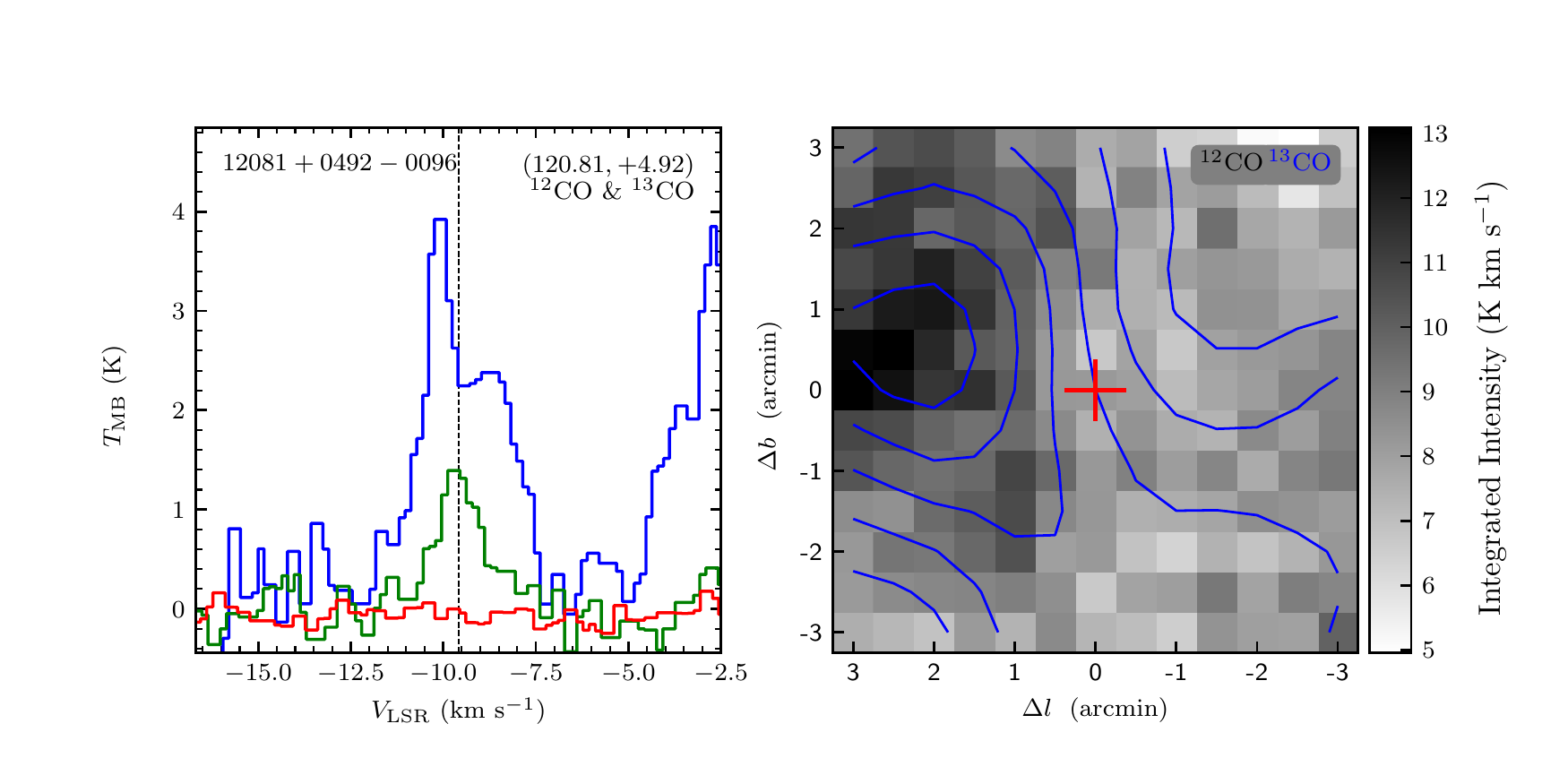}
\vspace{-0.5cm}

\includegraphics[width=9.0cm,angle=0]{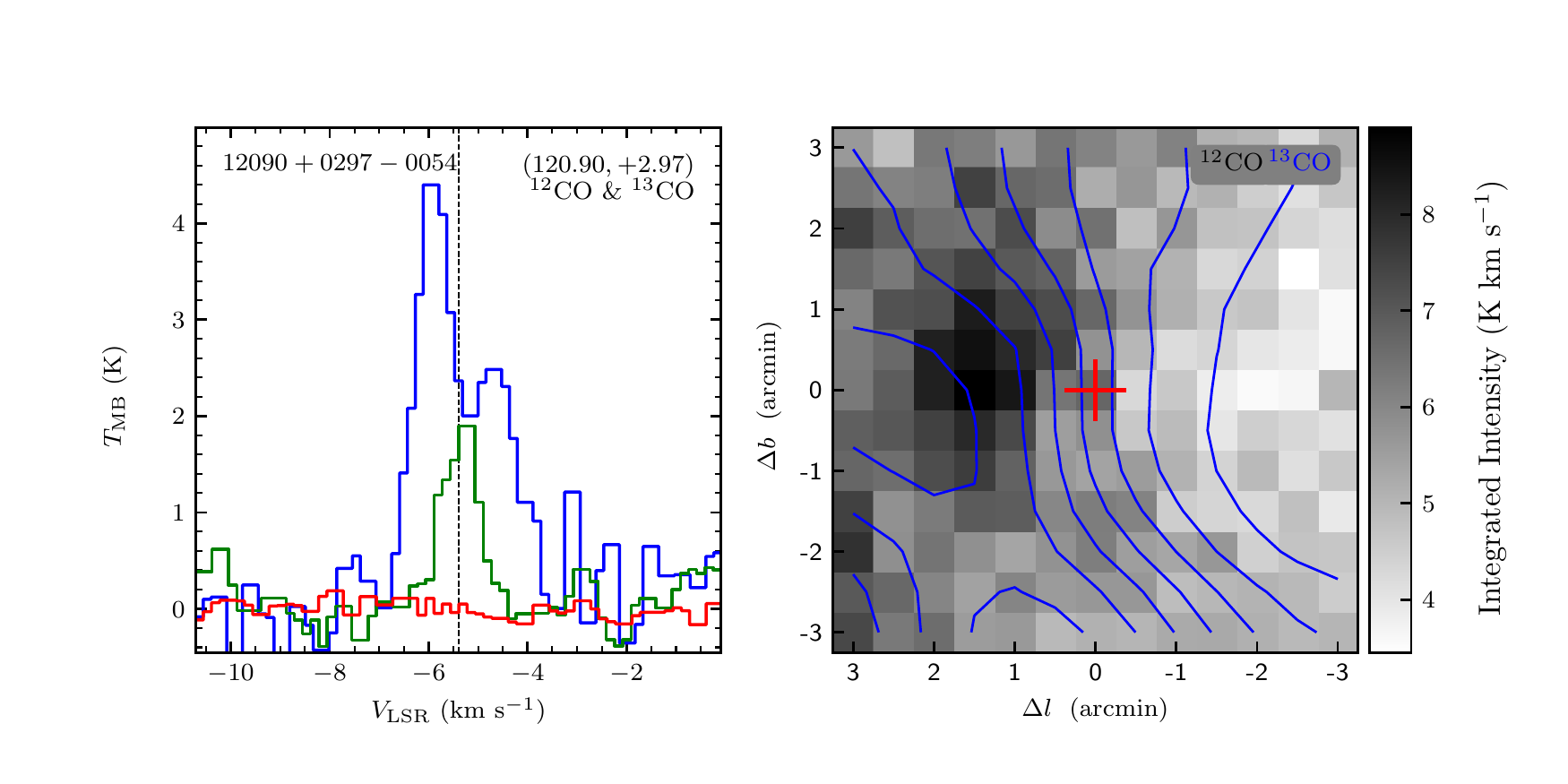}
\includegraphics[width=9.0cm,angle=0]{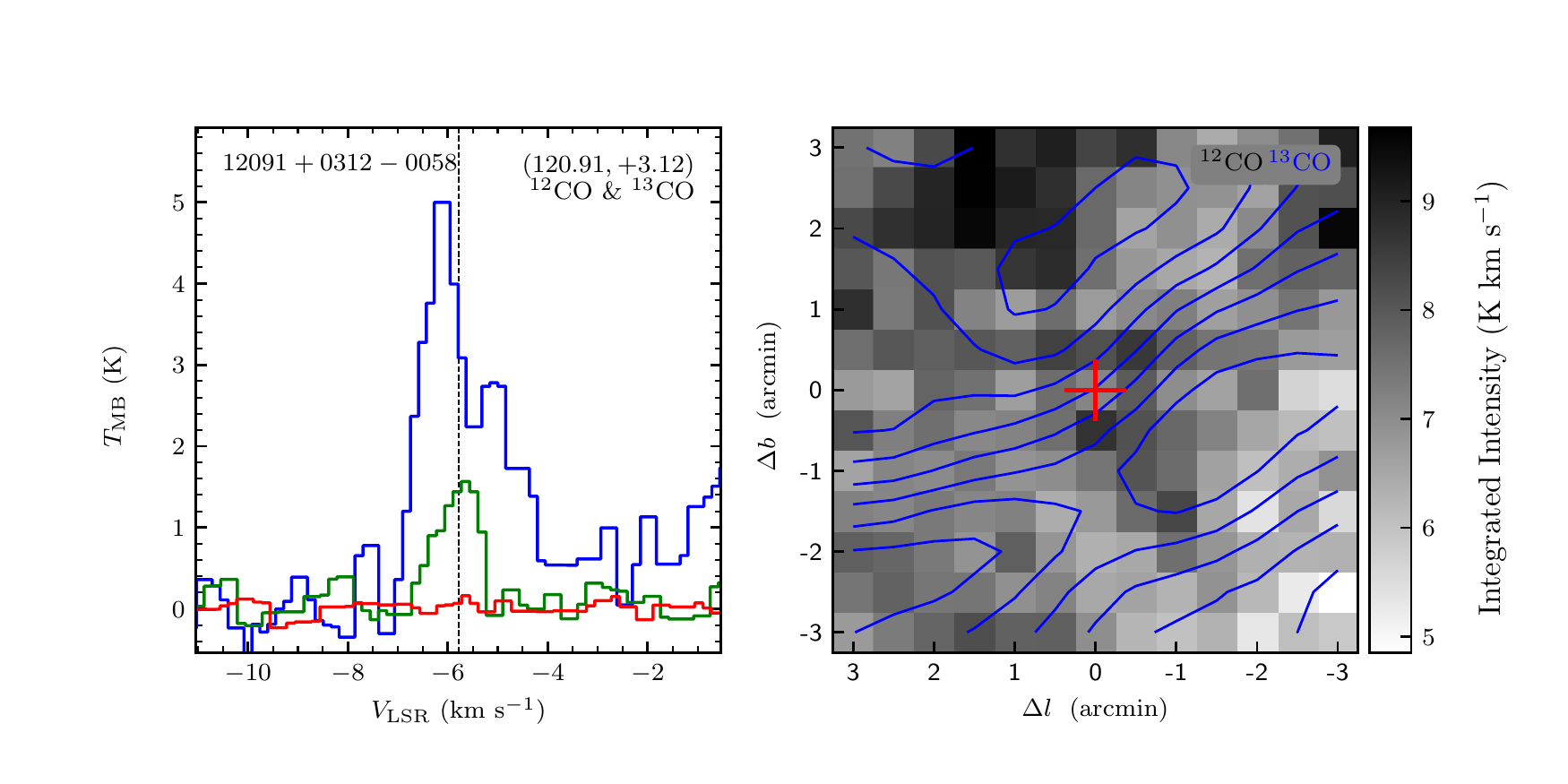}
\vspace{-0.5cm}

\includegraphics[width=9.0cm,angle=0]{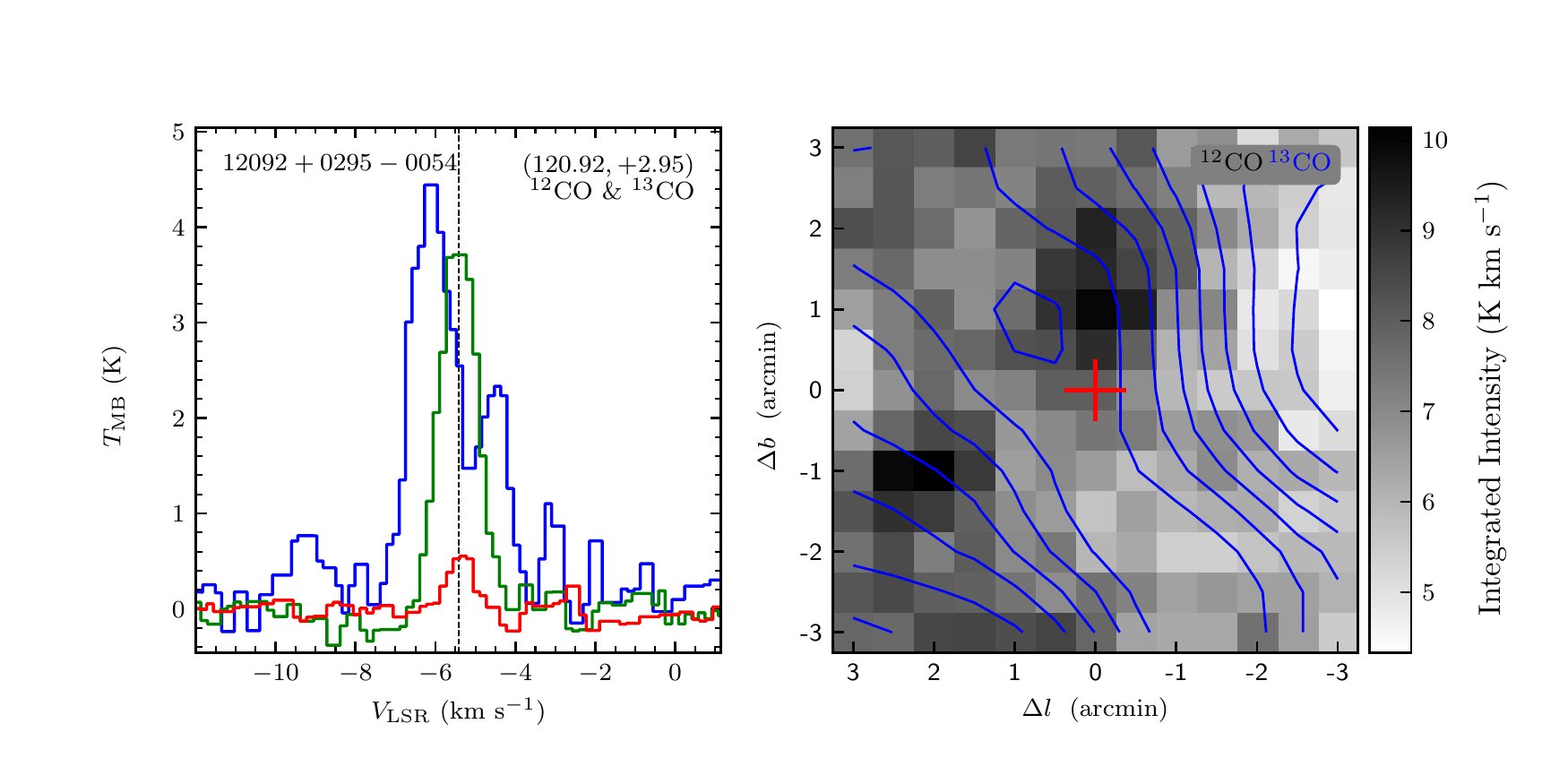}
\includegraphics[width=9.0cm,angle=0]{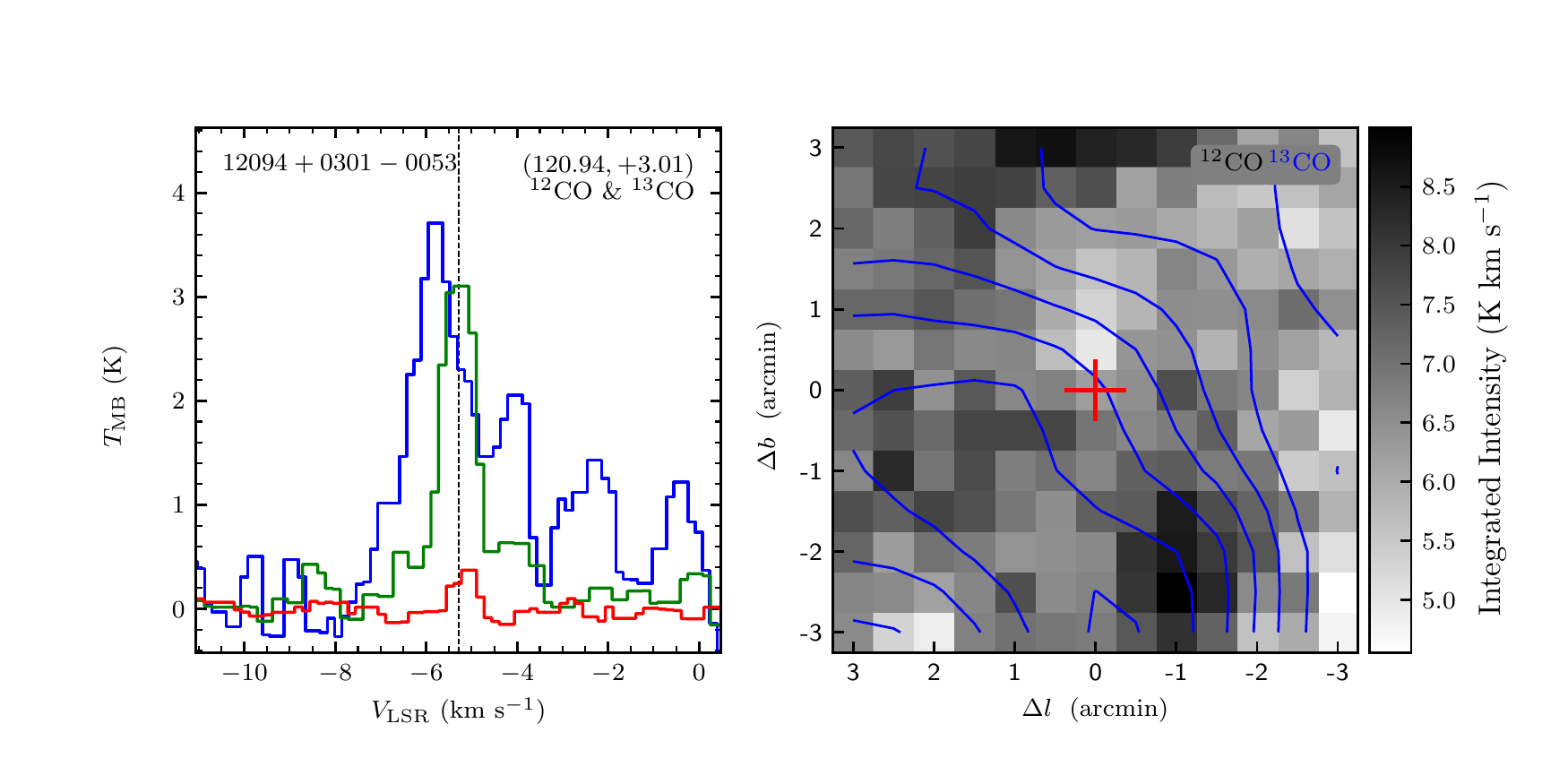}
\vspace{-0.5cm}

\includegraphics[width=9.0cm,angle=0]{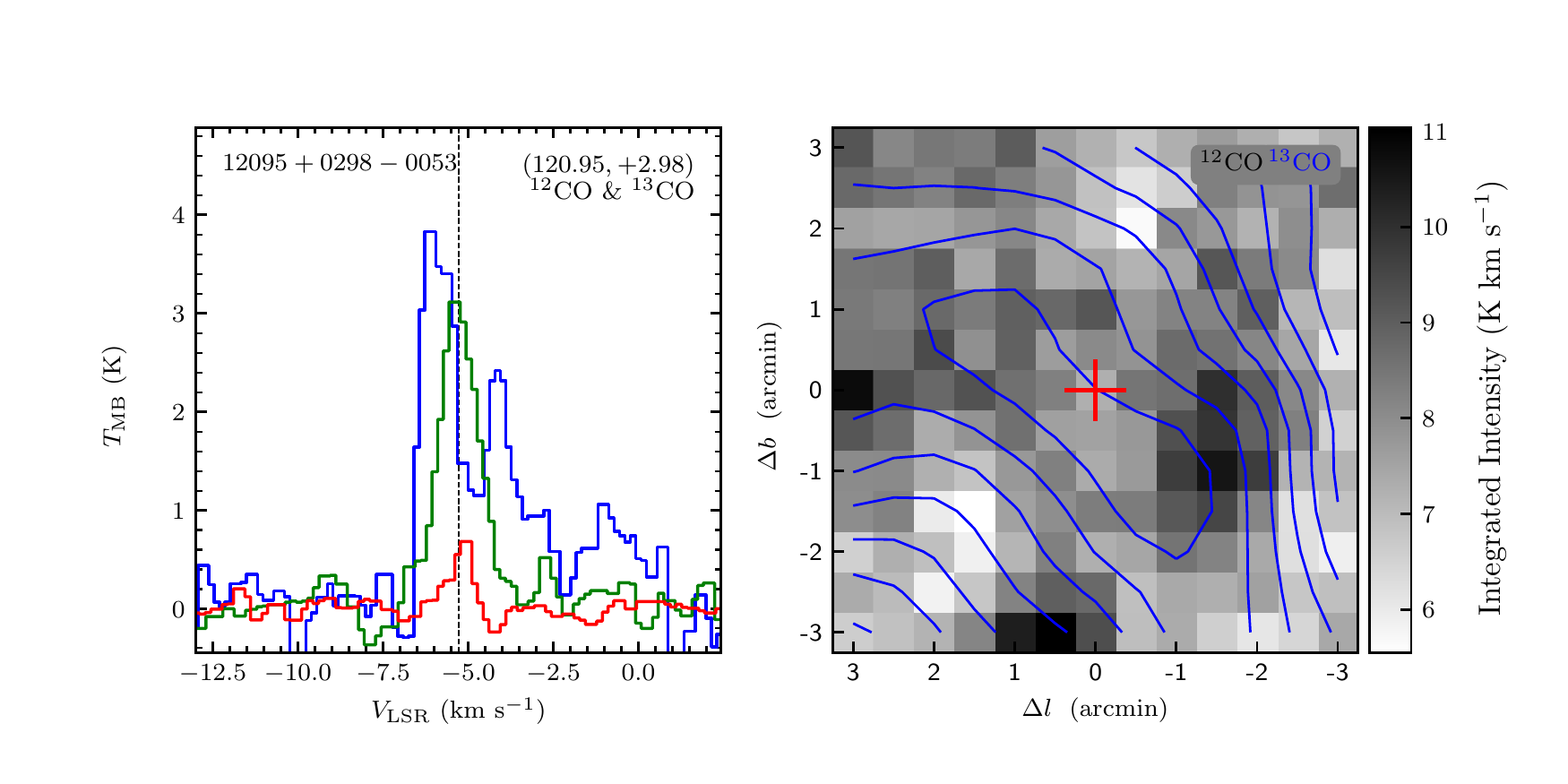}
\includegraphics[width=9.0cm,angle=0]{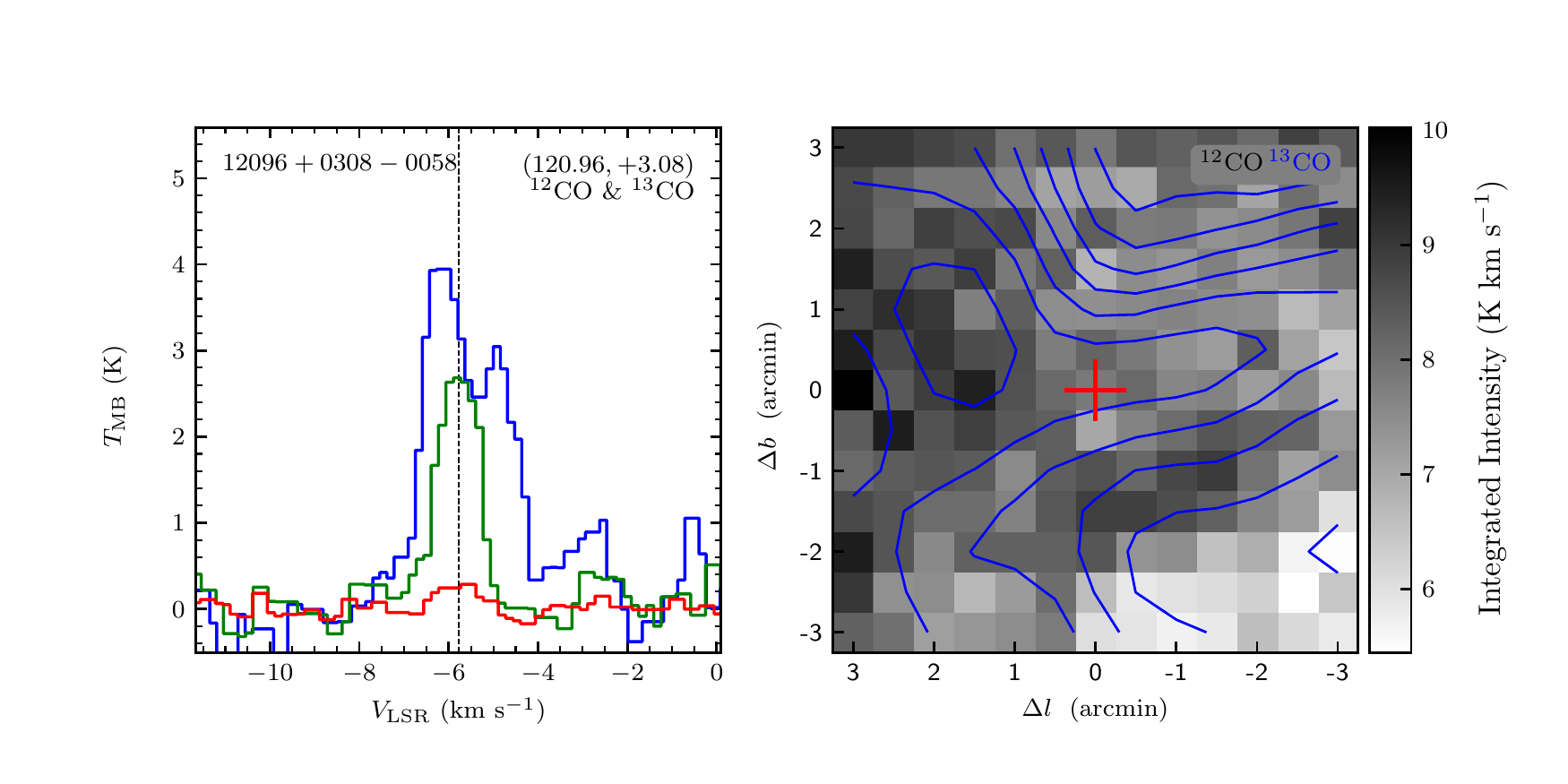}
\end{figure}
\clearpage

\begin{figure}
\includegraphics[width=9.0cm,angle=0]{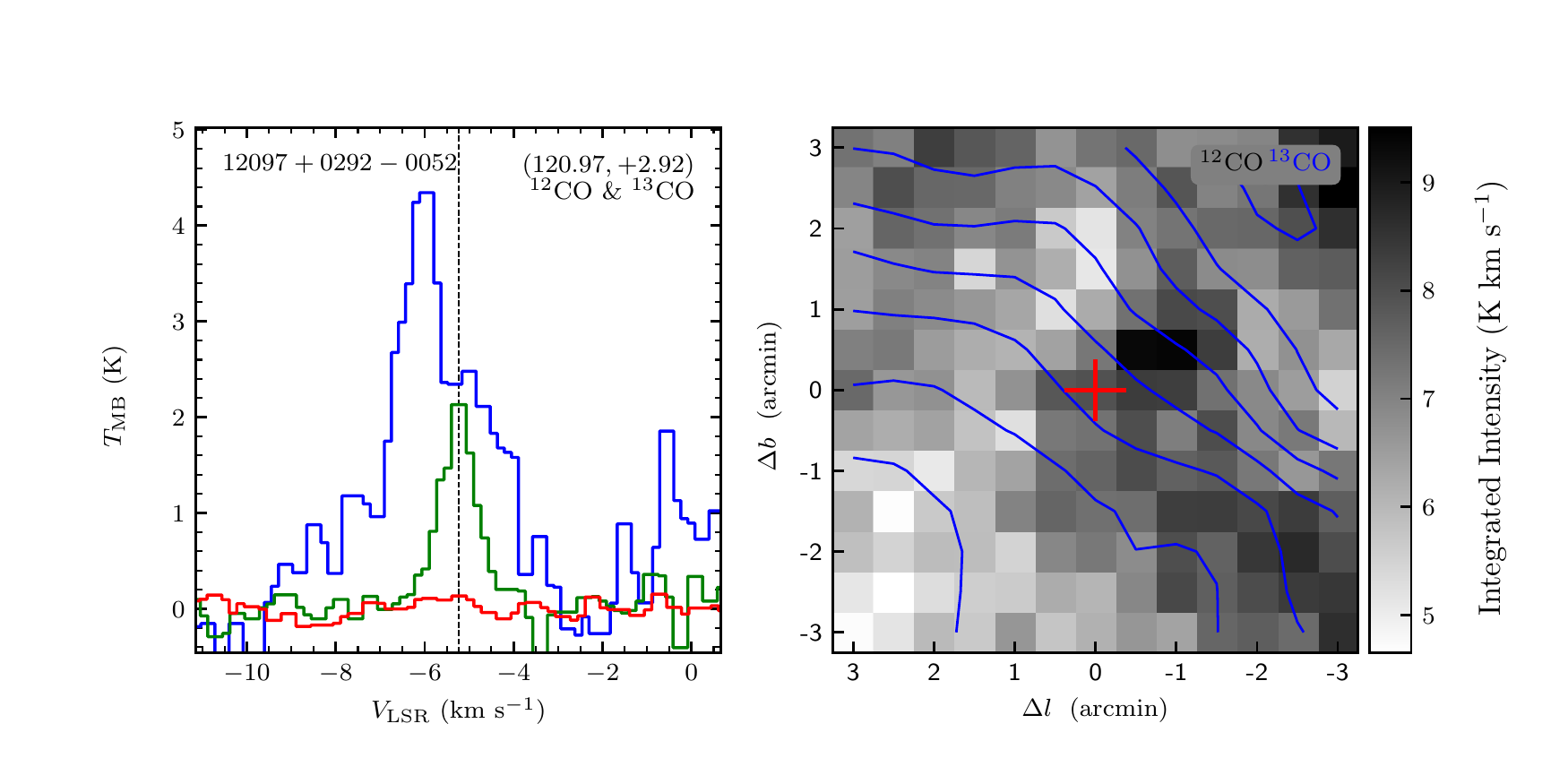}
\includegraphics[width=9.0cm,angle=0]{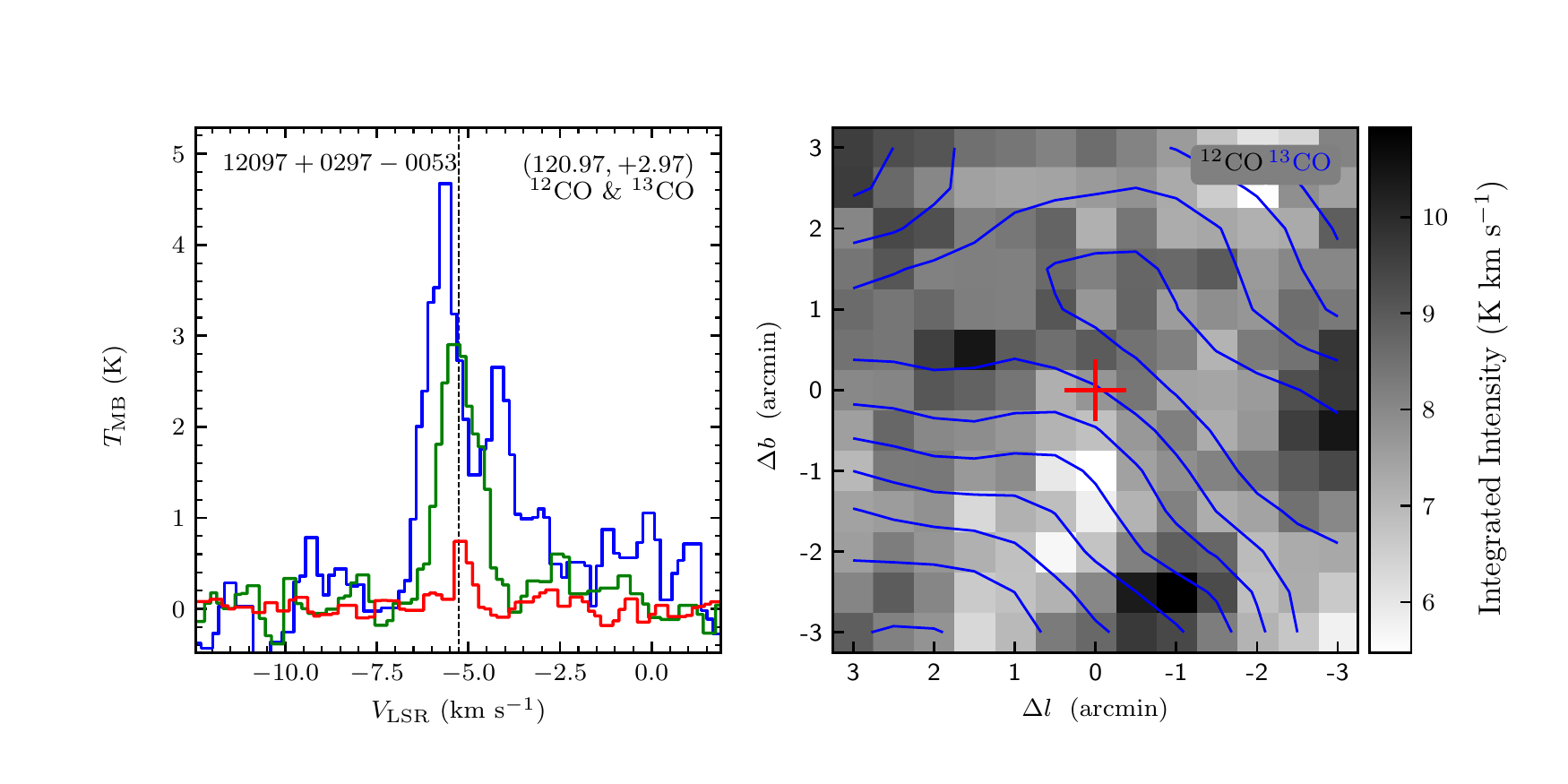}
\vspace{-0.5cm}

\includegraphics[width=9.0cm,angle=0]{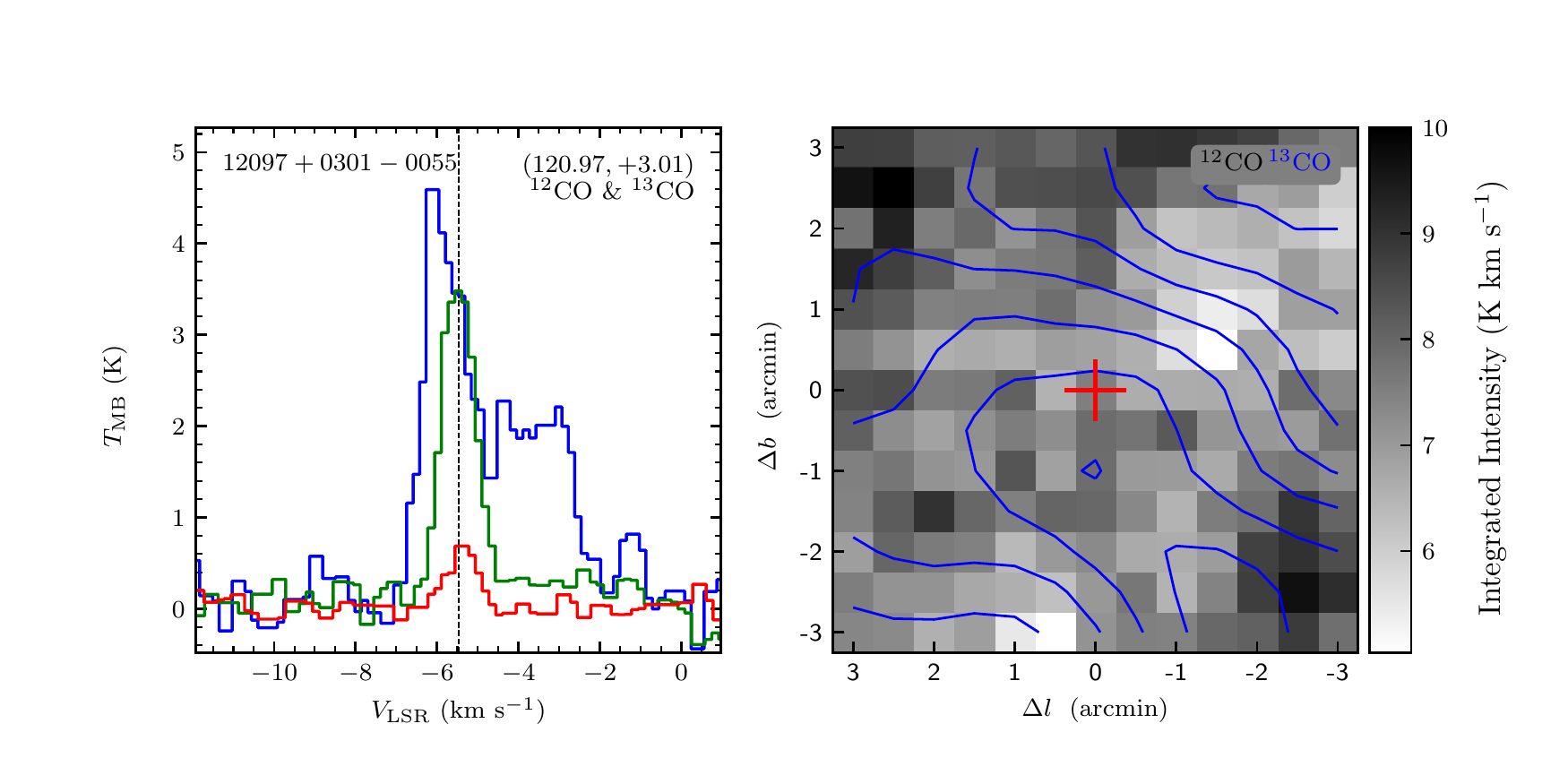}
\includegraphics[width=9.0cm,angle=0]{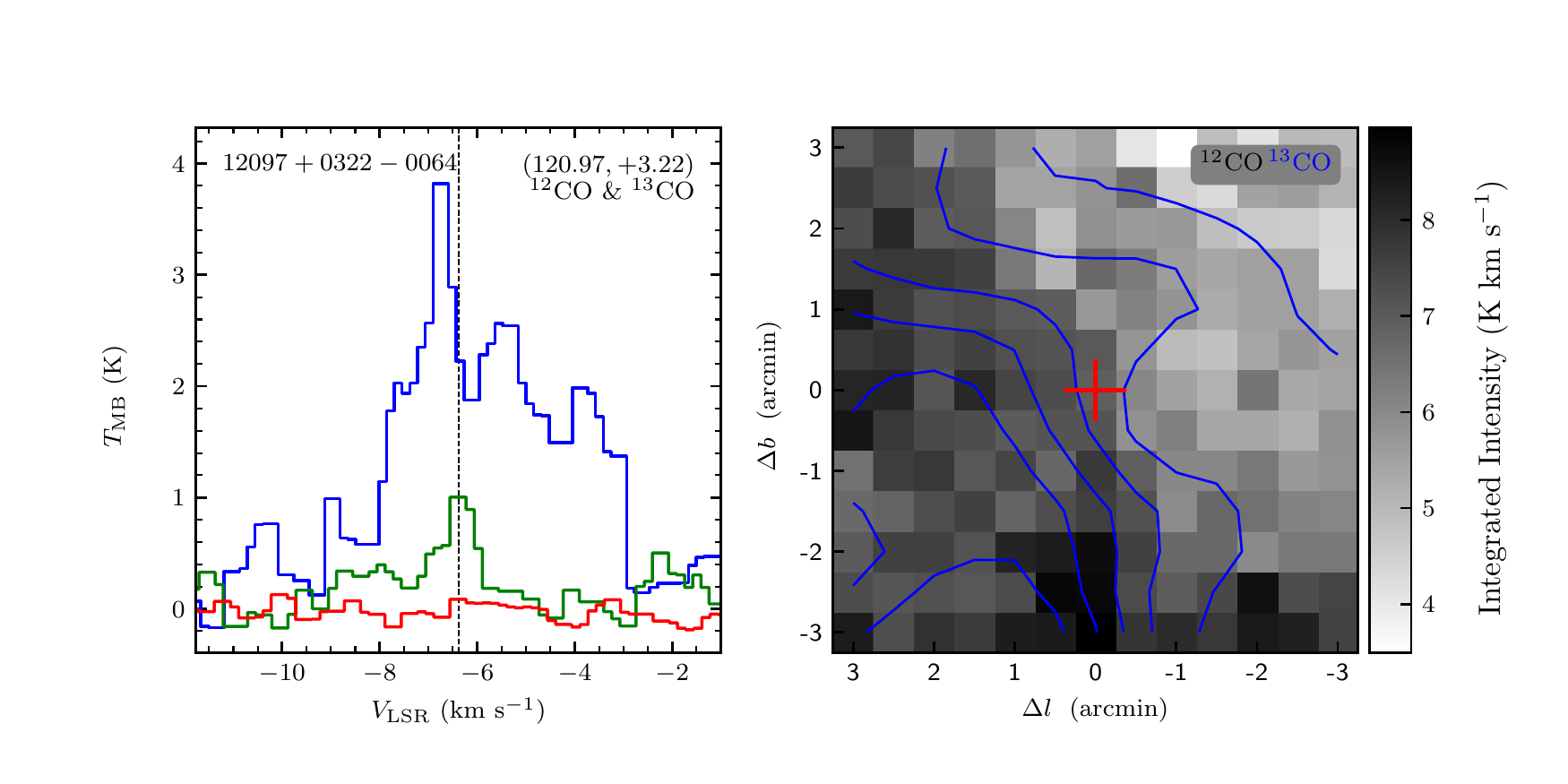}
\vspace{-0.5cm}

\includegraphics[width=9.0cm,angle=0]{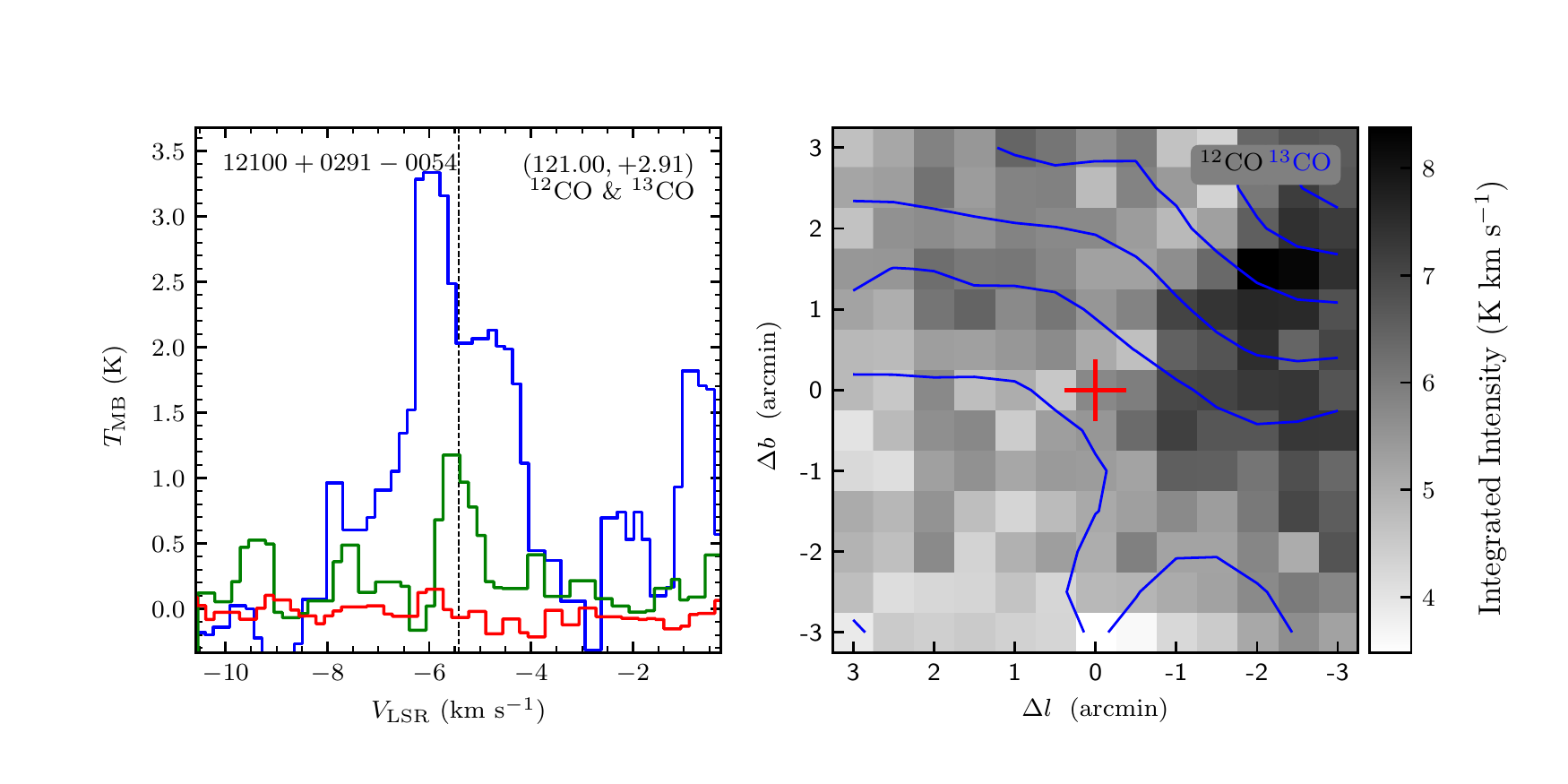}
\includegraphics[width=9.0cm,angle=0]{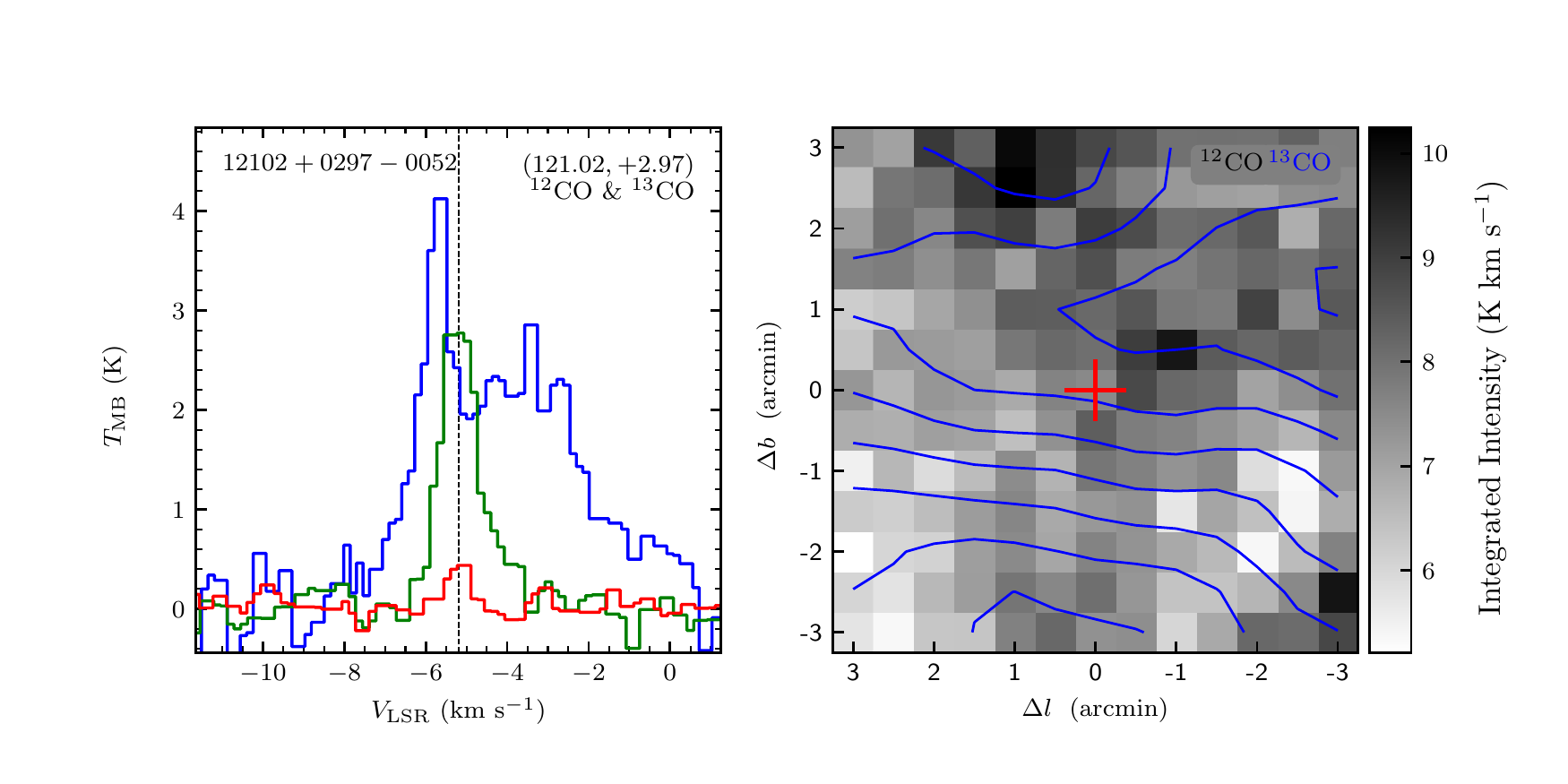}
\vspace{-0.5cm}

\includegraphics[width=9.0cm,angle=0]{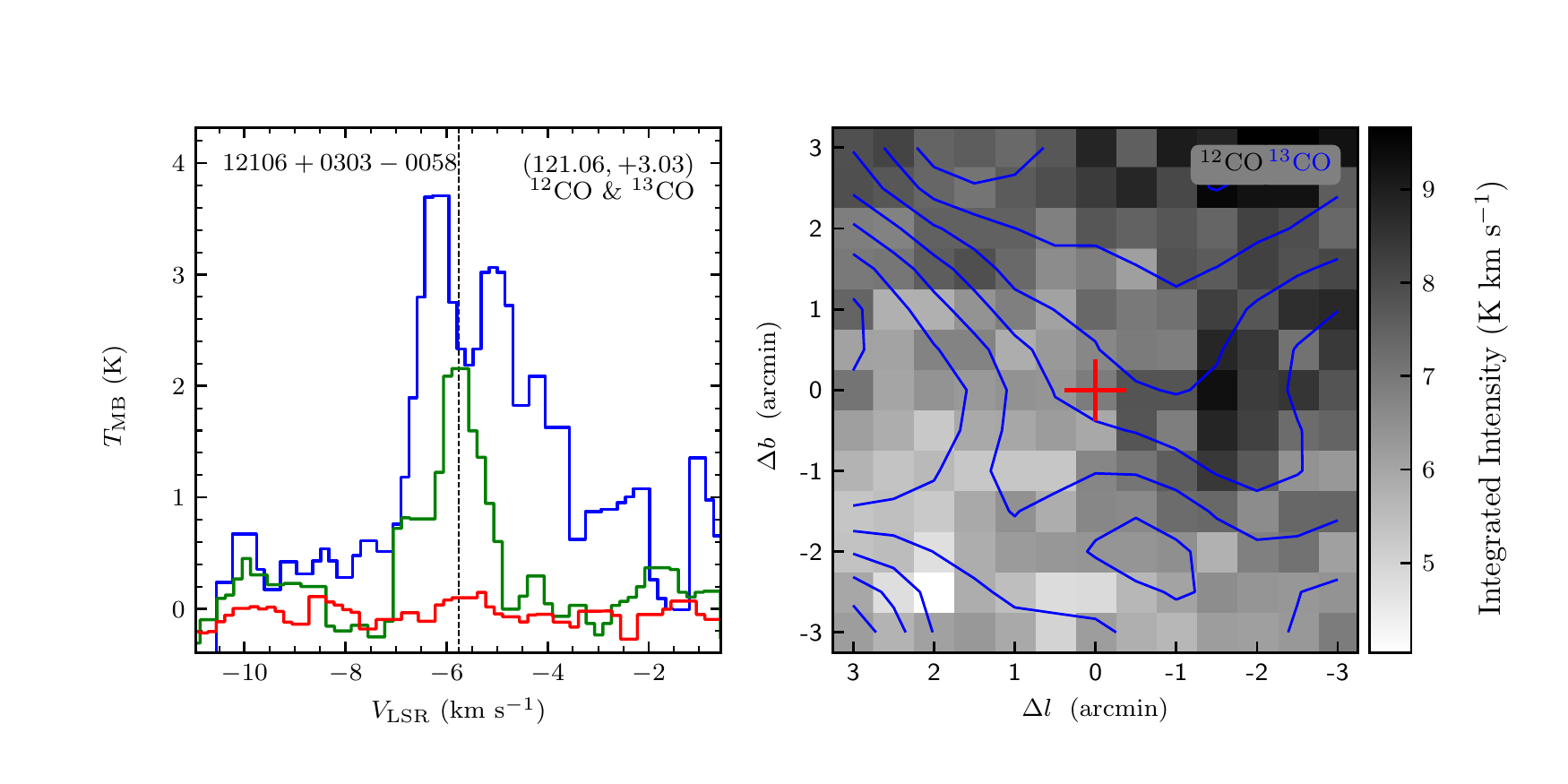}
\includegraphics[width=9.0cm,angle=0]{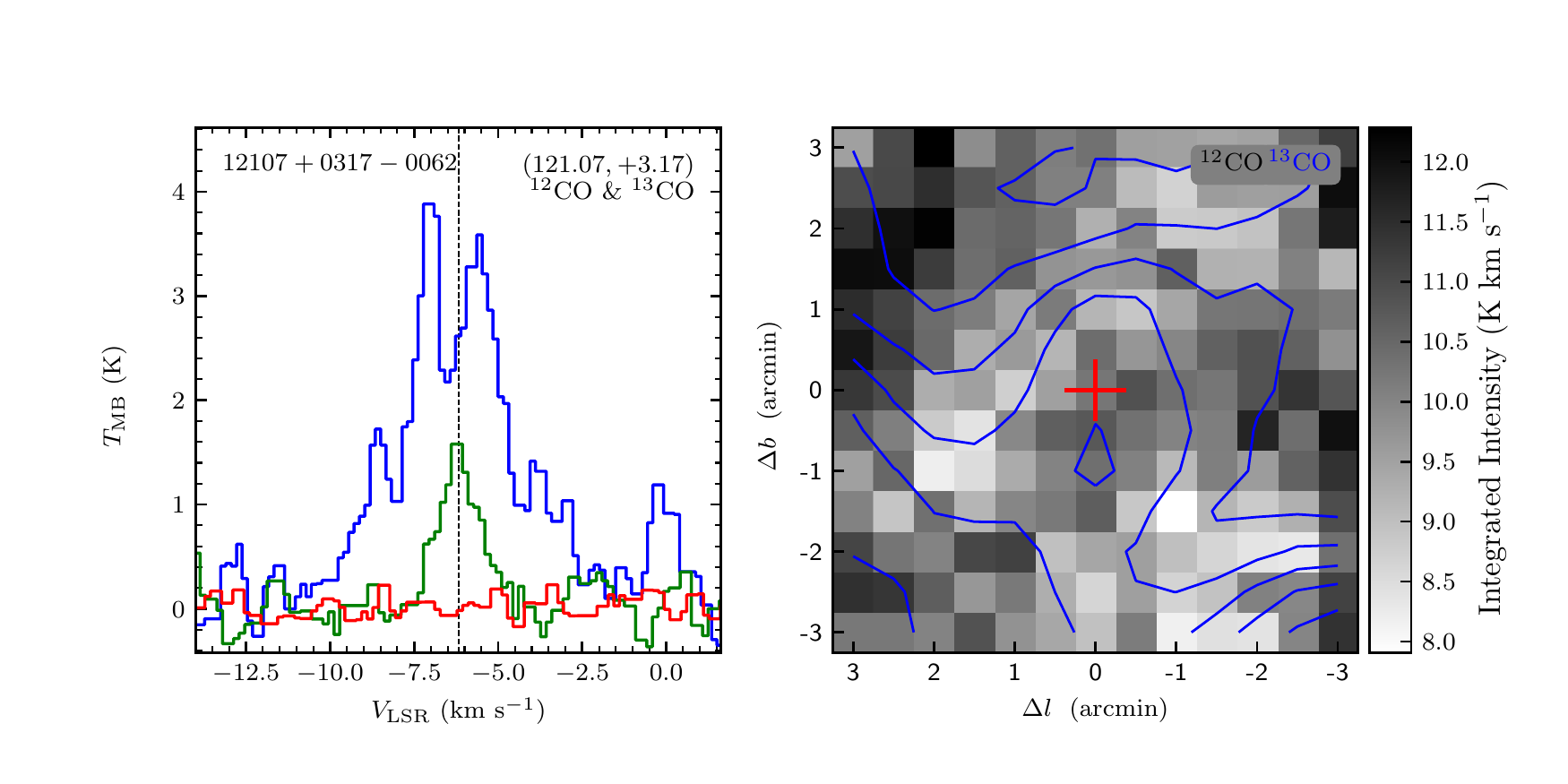}
\vspace{-0.5cm}

\includegraphics[width=9.0cm,angle=0]{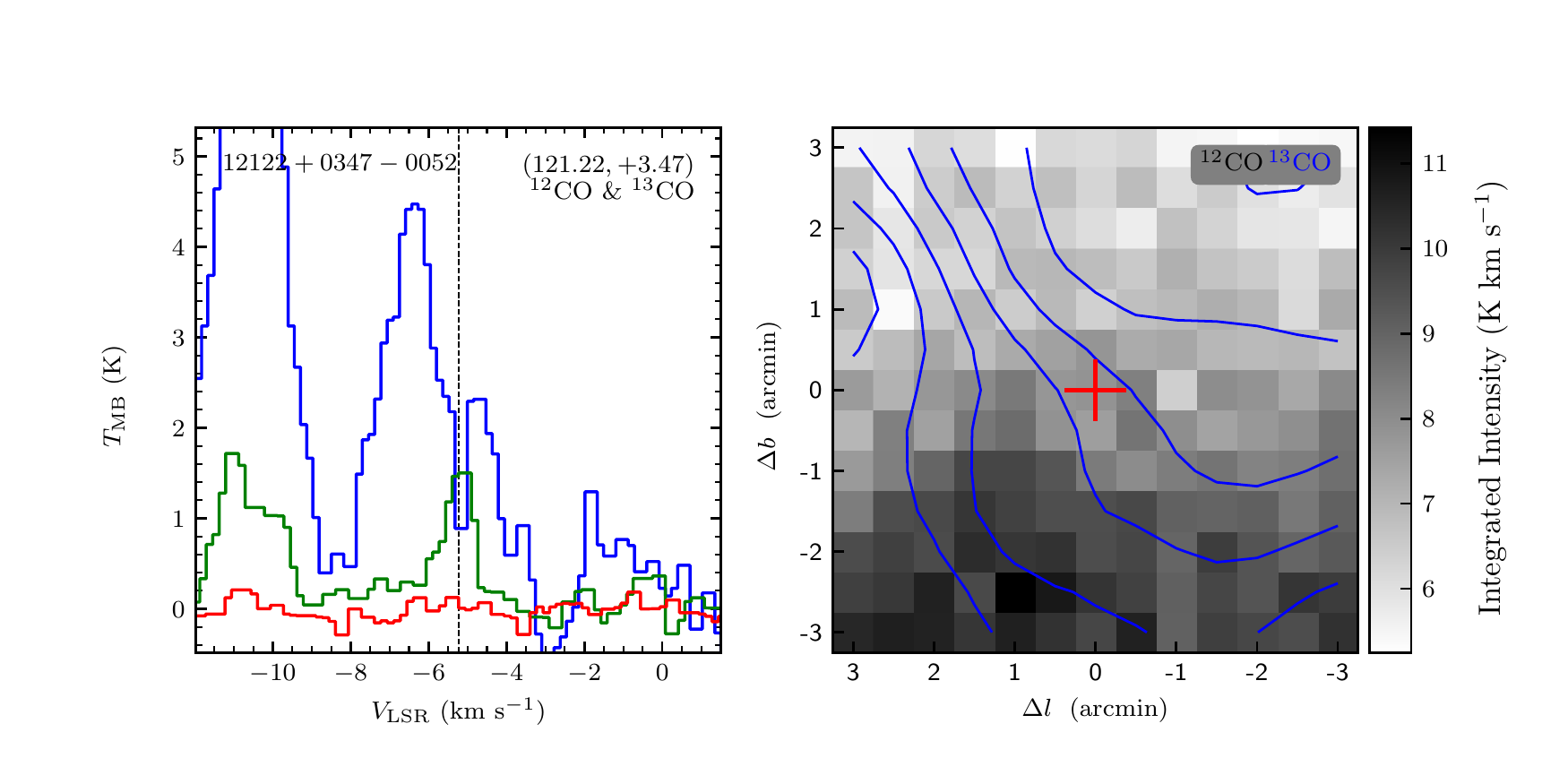}
\includegraphics[width=9.0cm,angle=0]{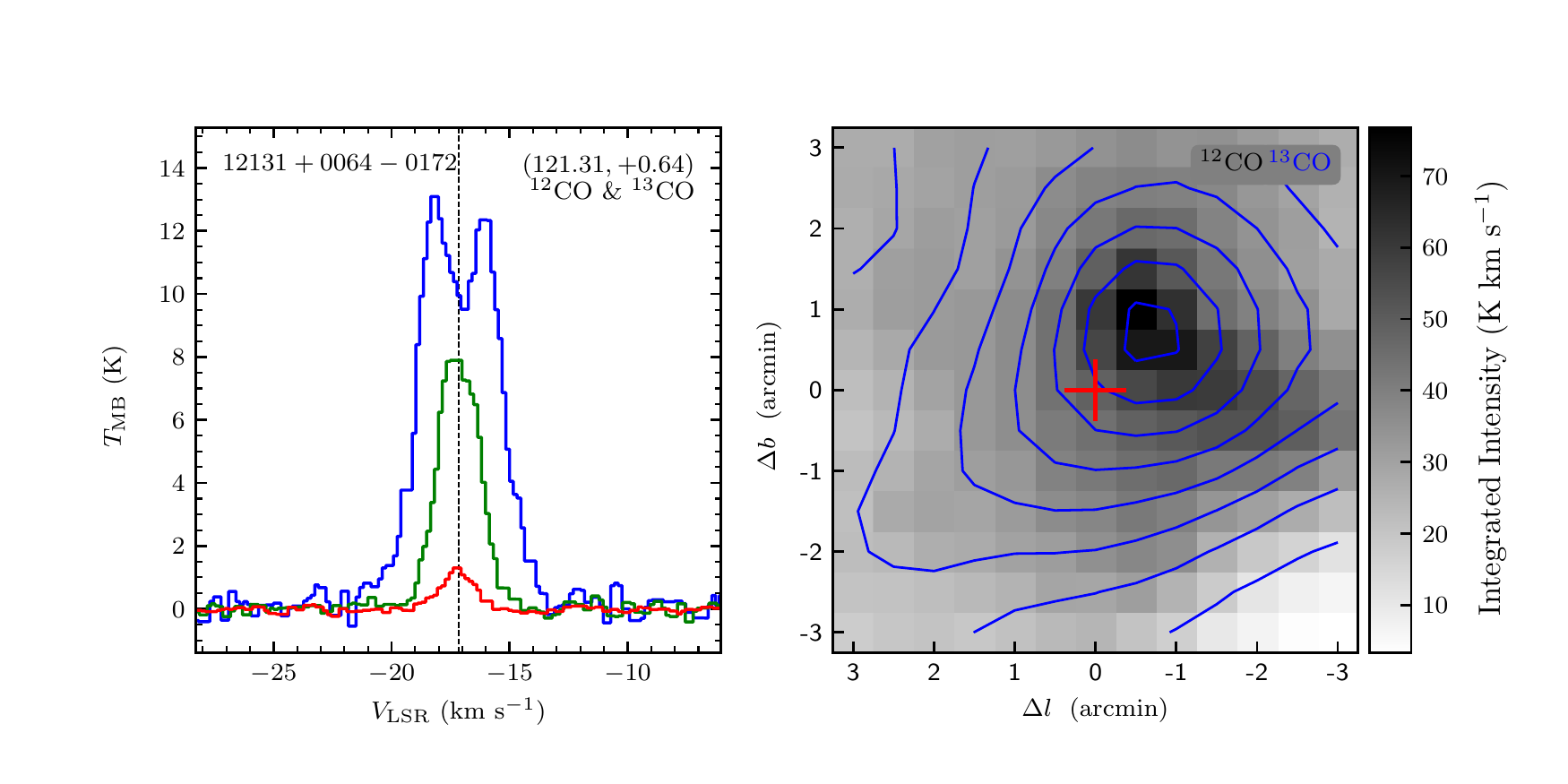}
\end{figure}
\clearpage

\begin{figure}
\includegraphics[width=9.0cm,angle=0]{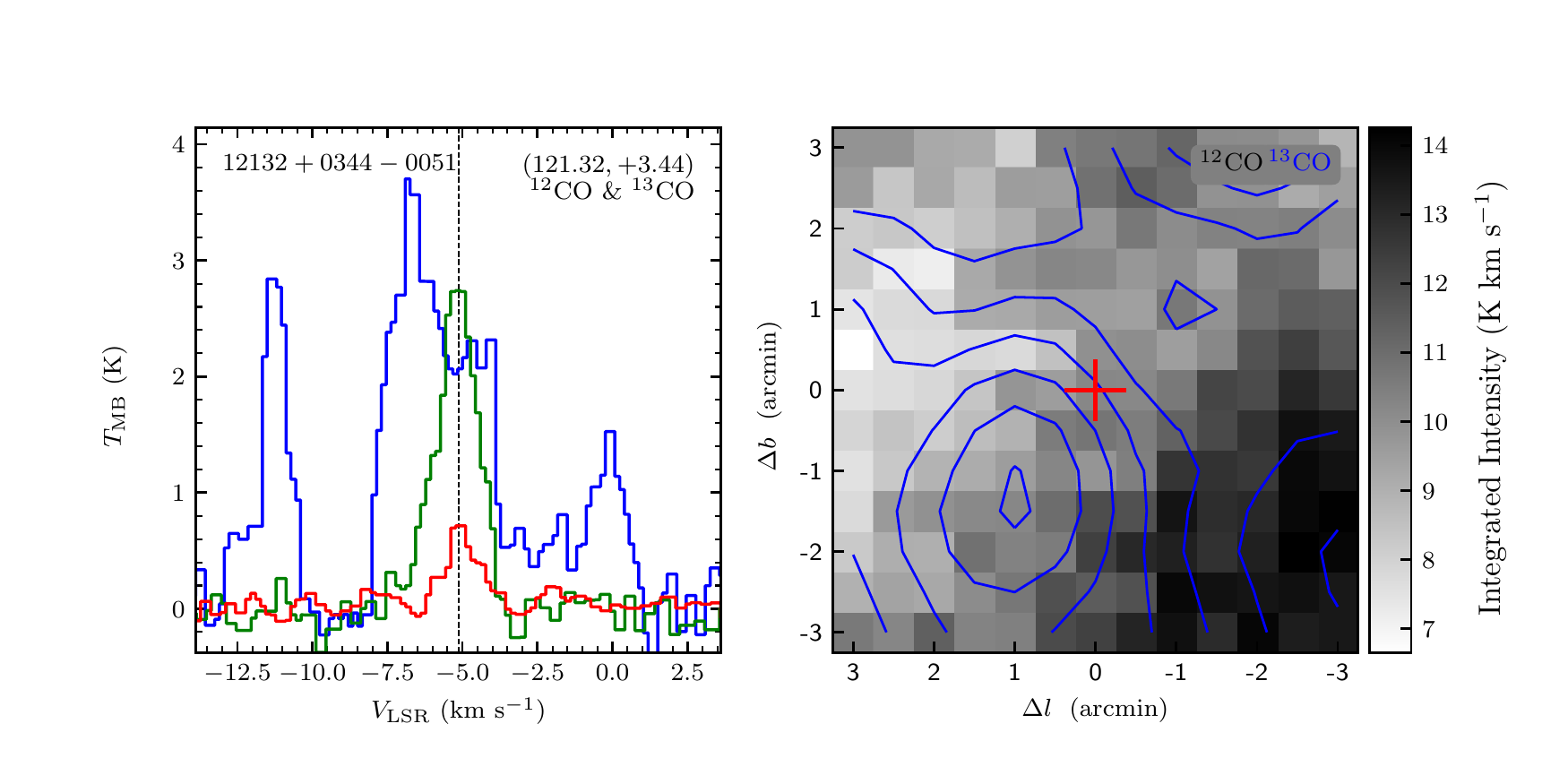}
\includegraphics[width=9.0cm,angle=0]{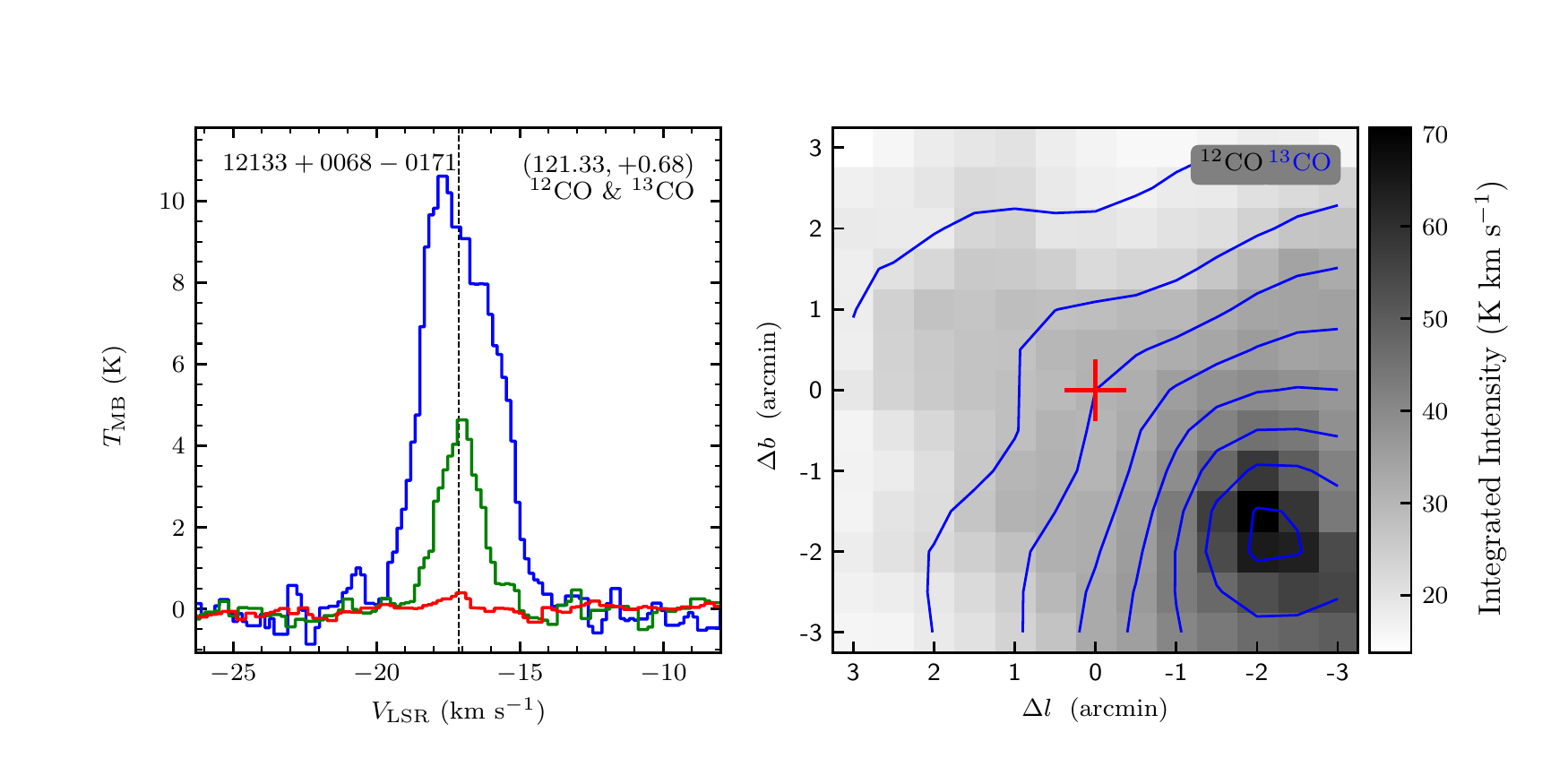}
\vspace{-0.5cm}

\includegraphics[width=9.0cm,angle=0]{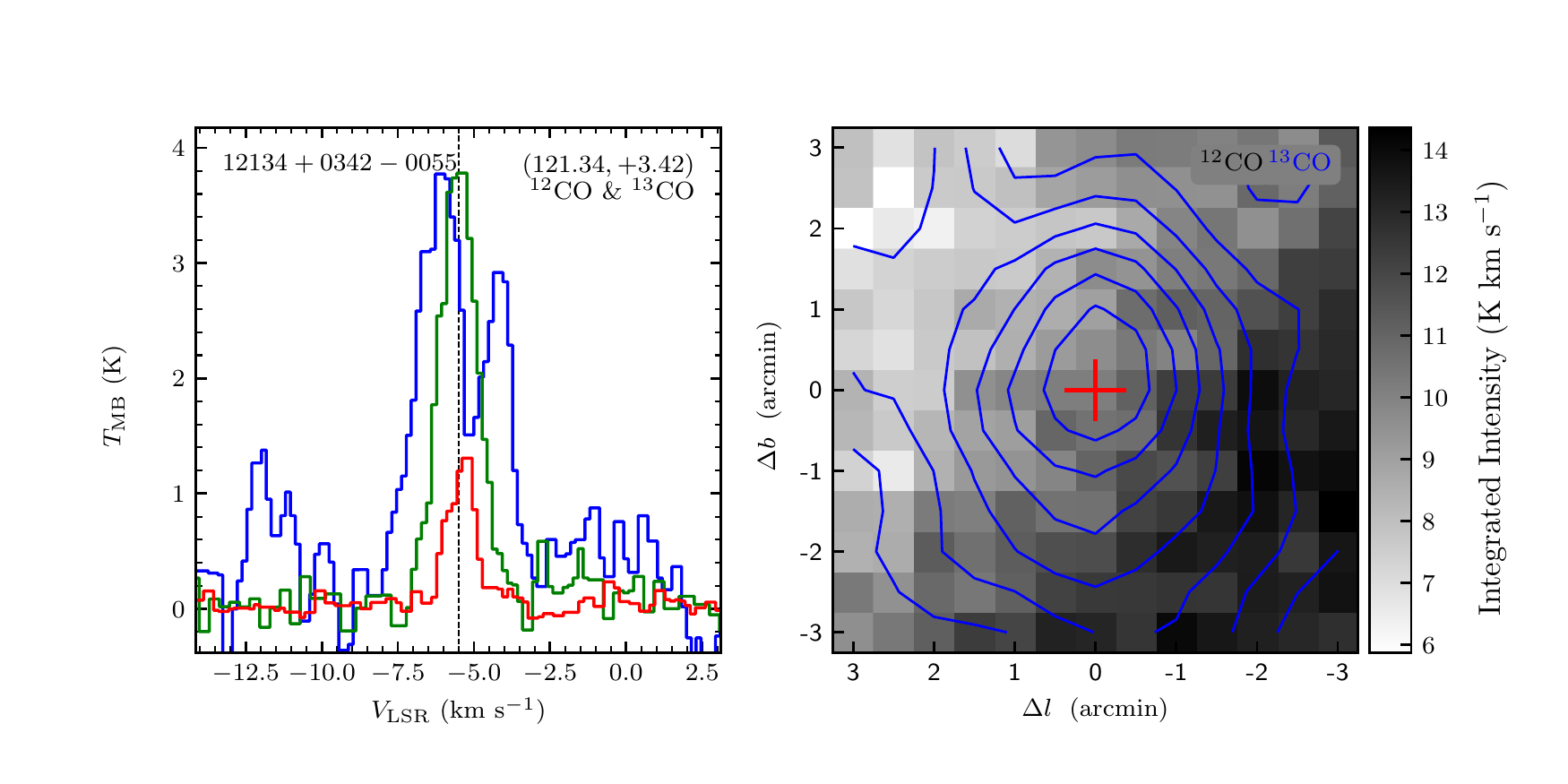}
\includegraphics[width=9.0cm,angle=0]{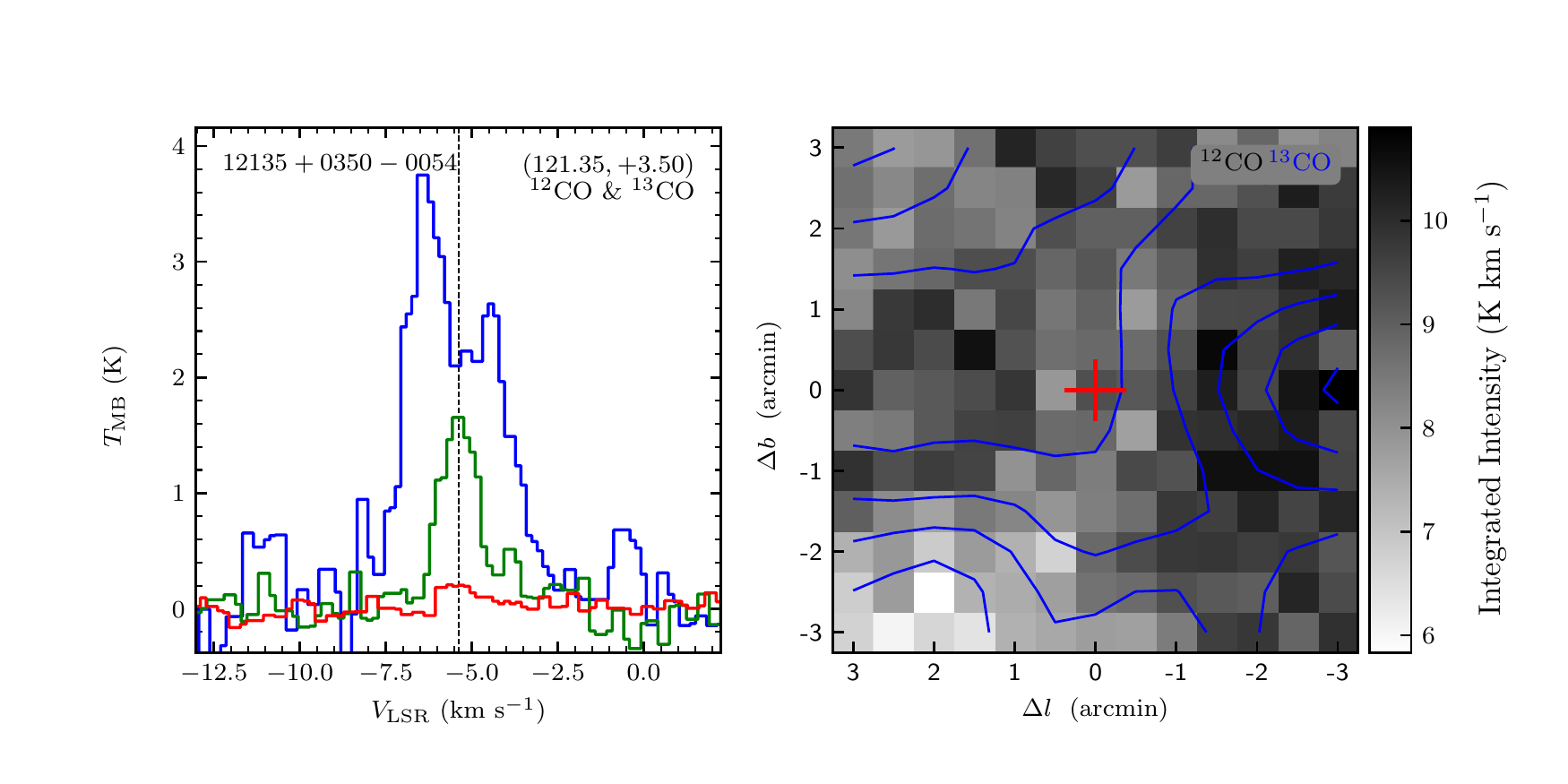}
\vspace{-0.5cm}

\includegraphics[width=9.0cm,angle=0]{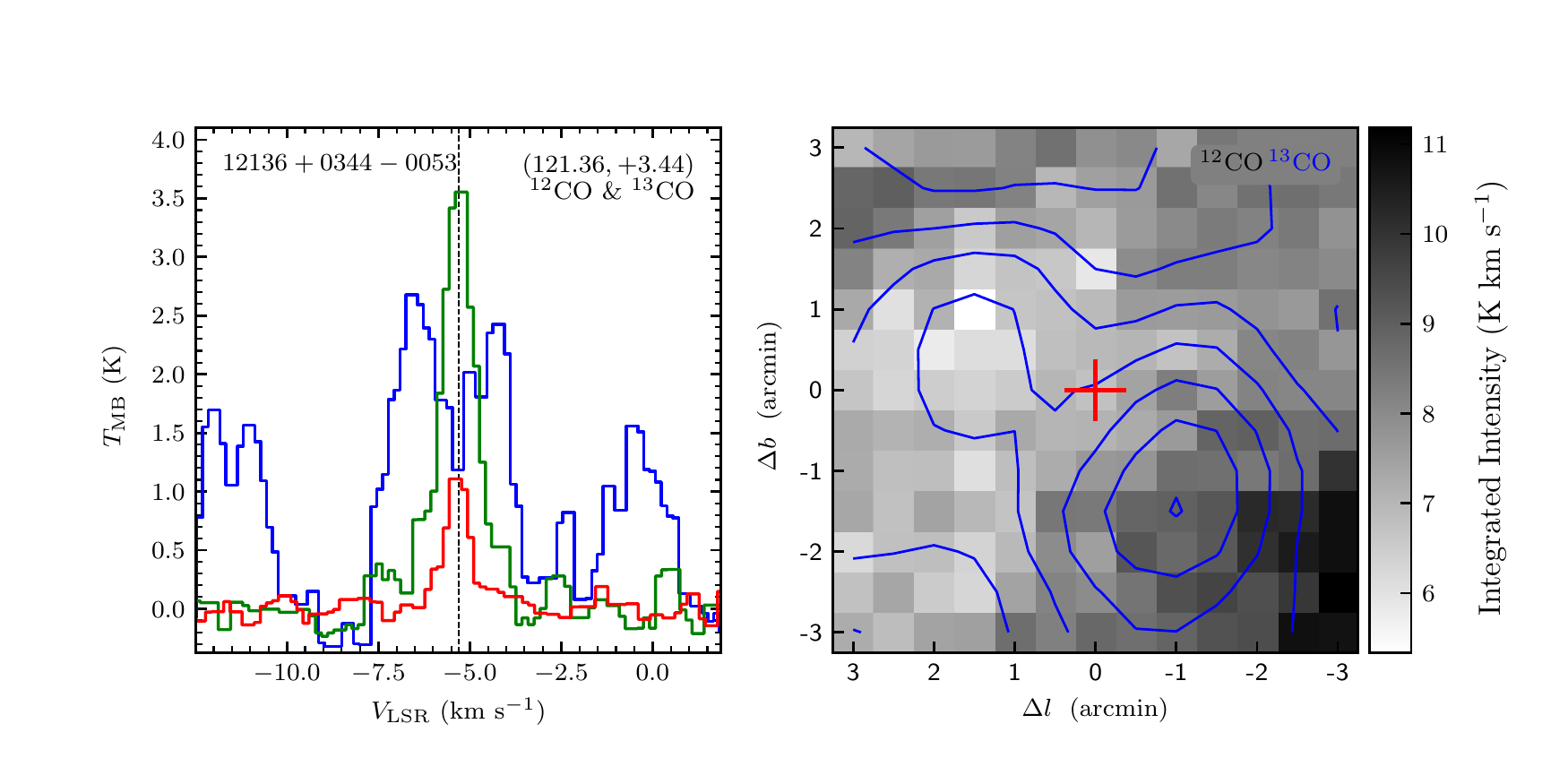}
\includegraphics[width=9.0cm,angle=0]{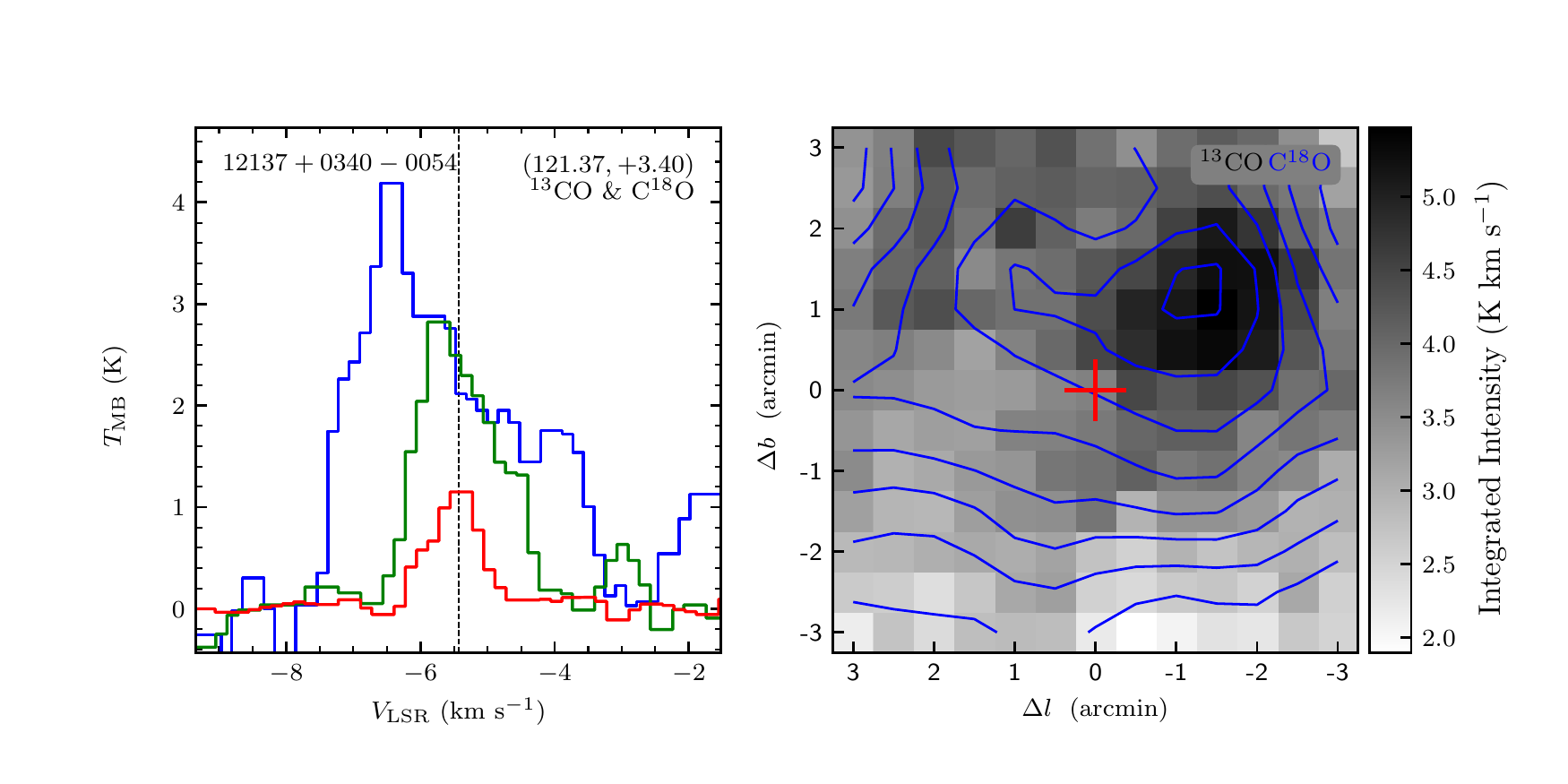}
\vspace{-0.5cm}

\includegraphics[width=9.0cm,angle=0]{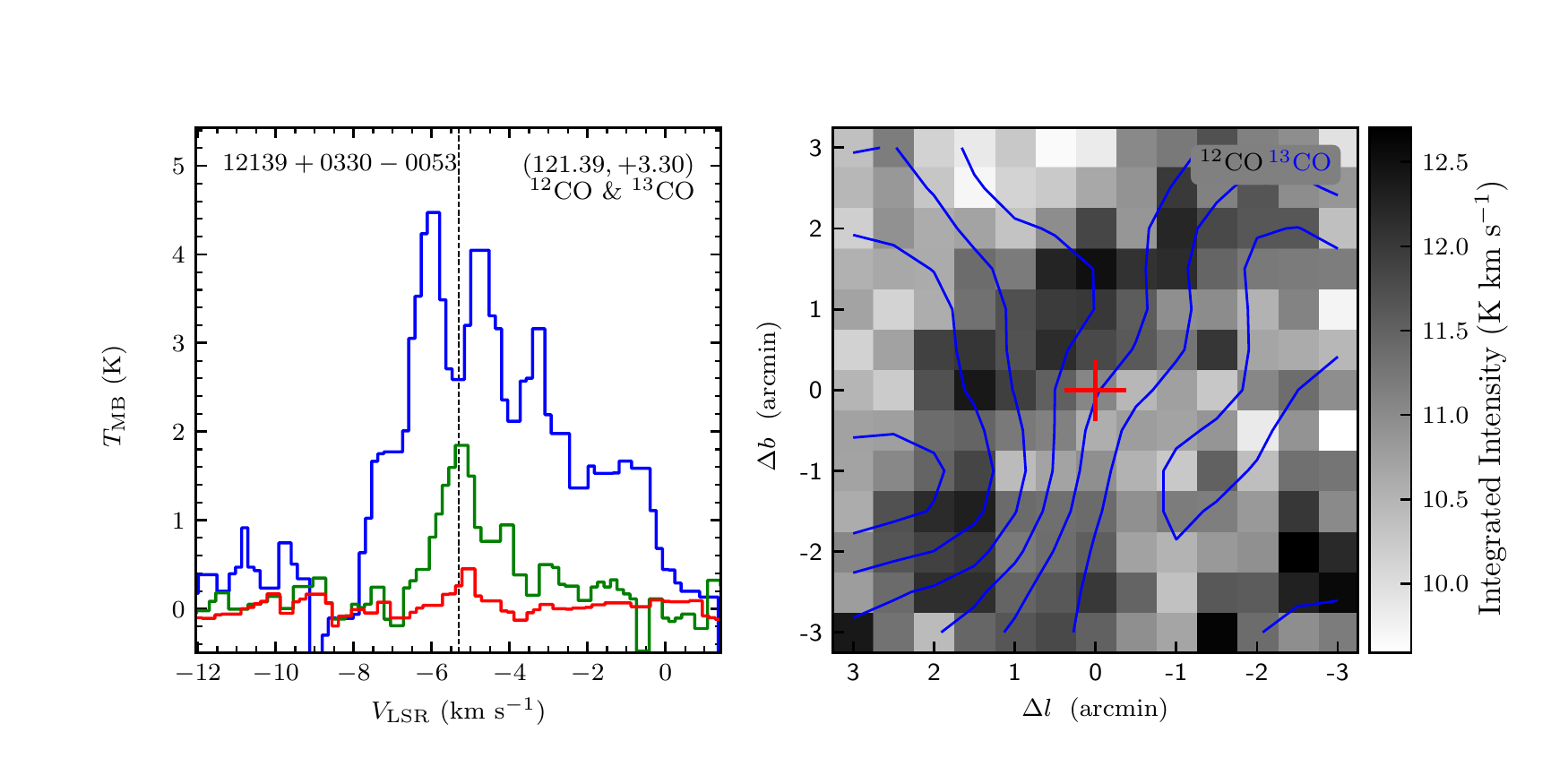}
\includegraphics[width=9.0cm,angle=0]{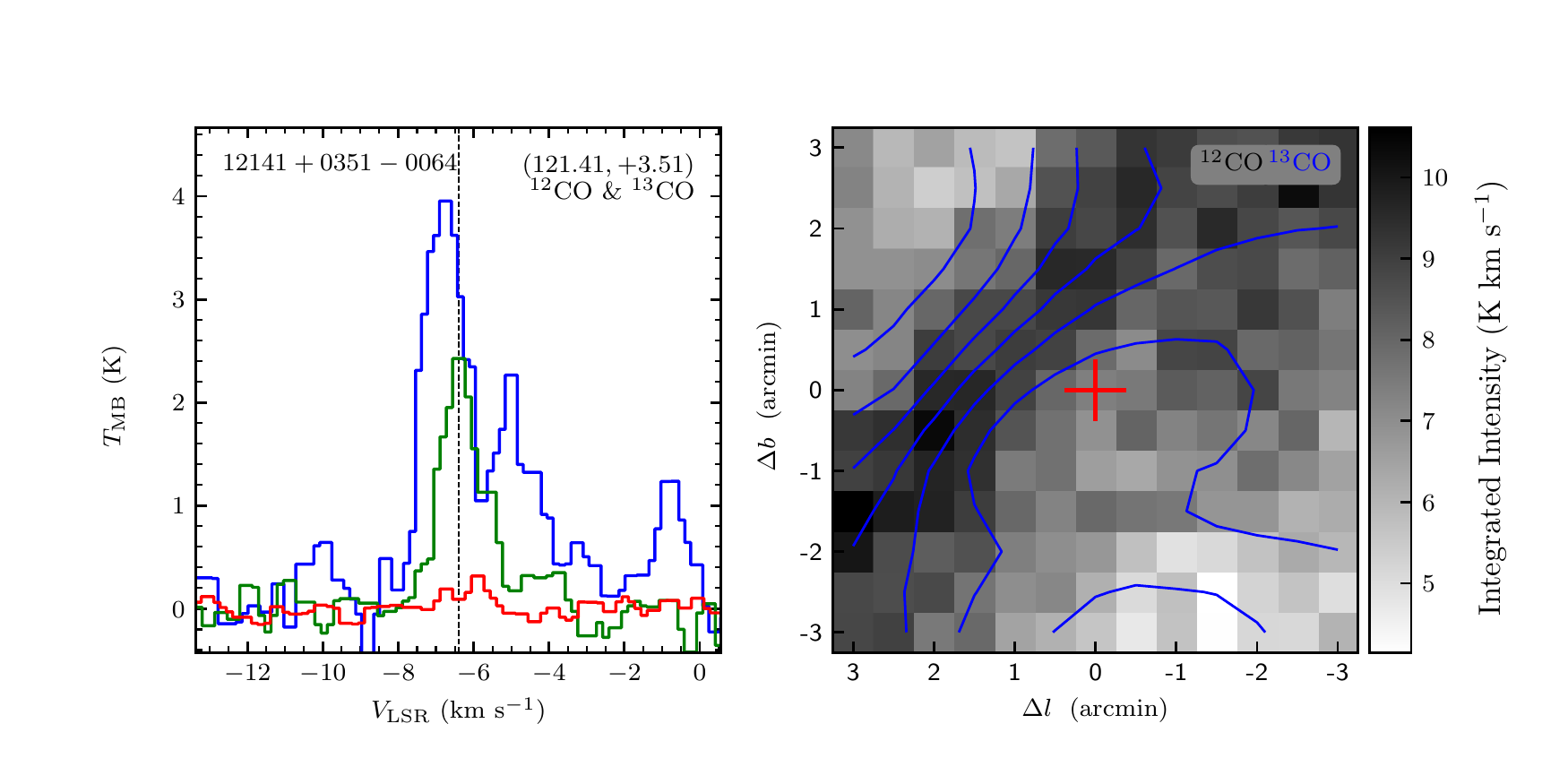}
\vspace{-0.5cm}

\includegraphics[width=9.0cm,angle=0]{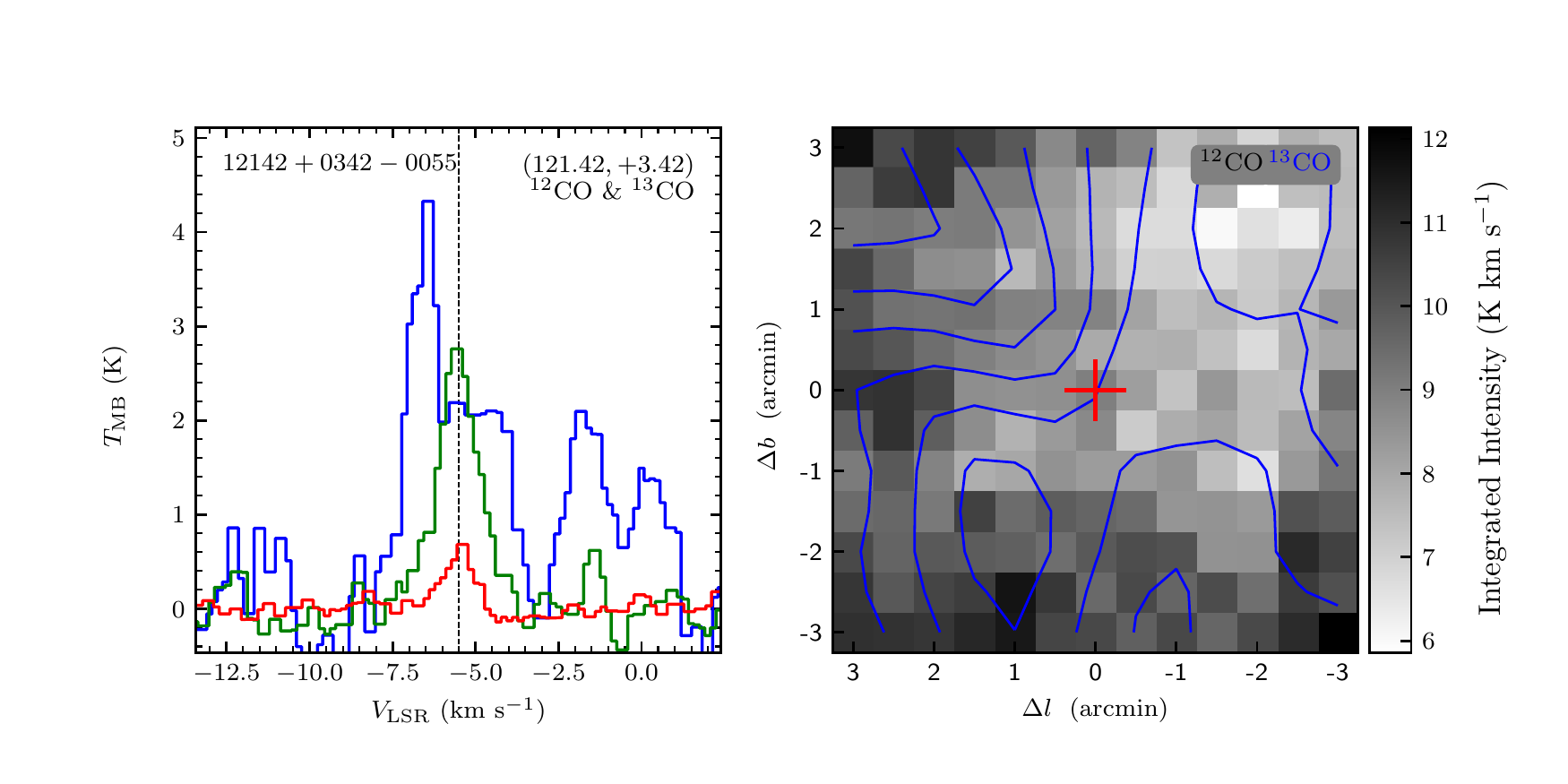}
\includegraphics[width=9.0cm,angle=0]{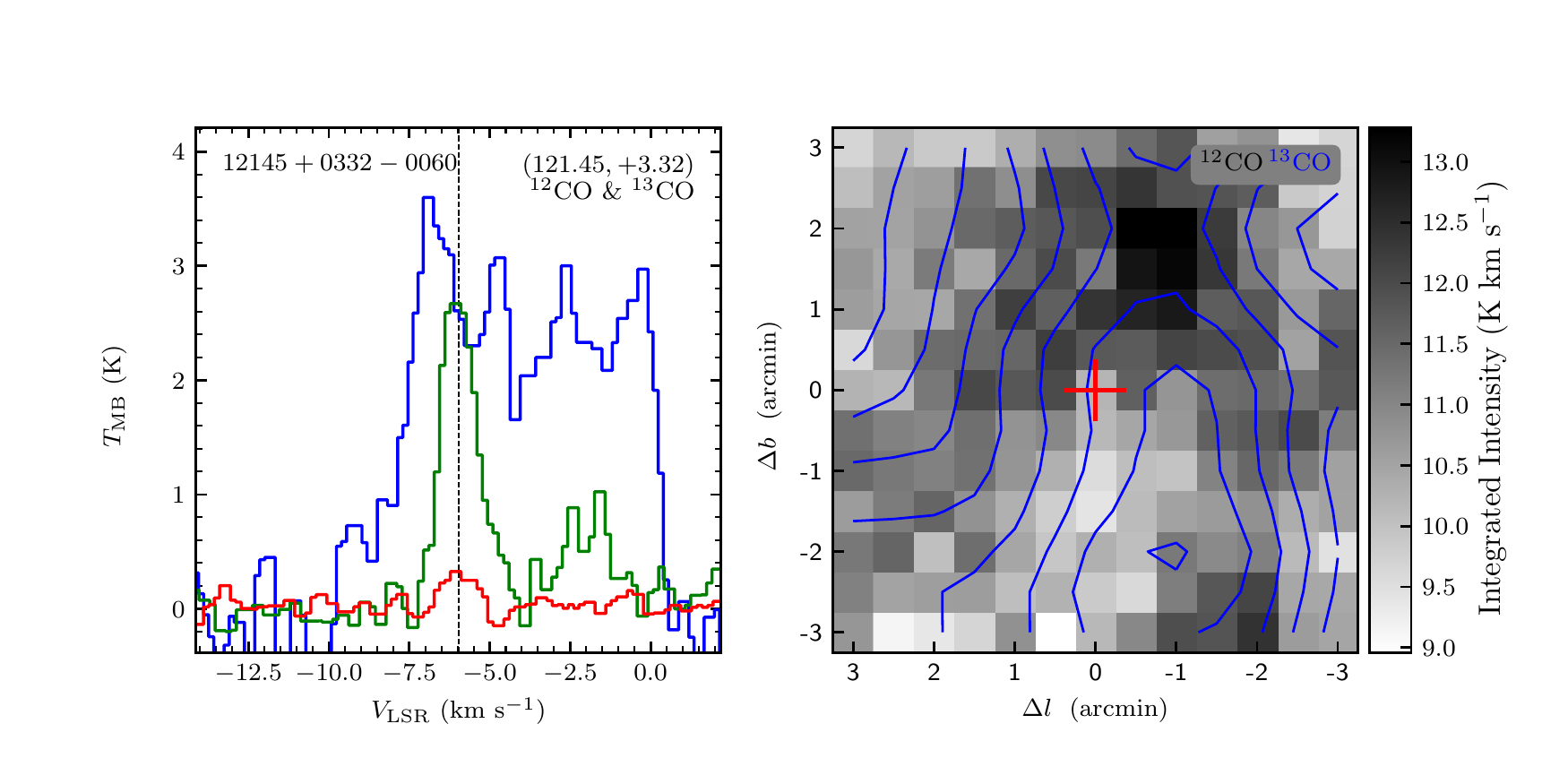}
\end{figure}
\clearpage

\begin{figure}
\includegraphics[width=9.0cm,angle=0]{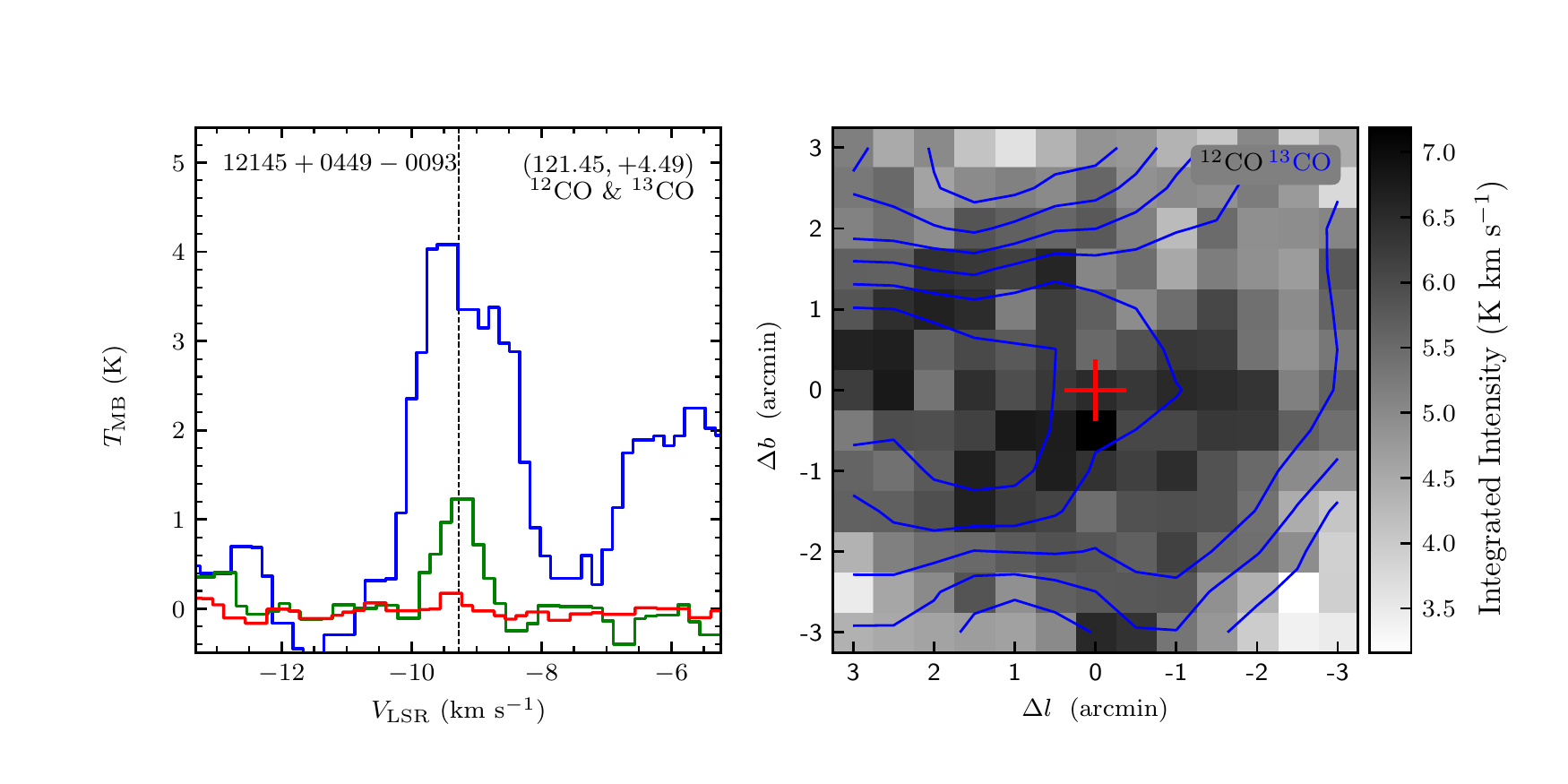}
\includegraphics[width=9.0cm,angle=0]{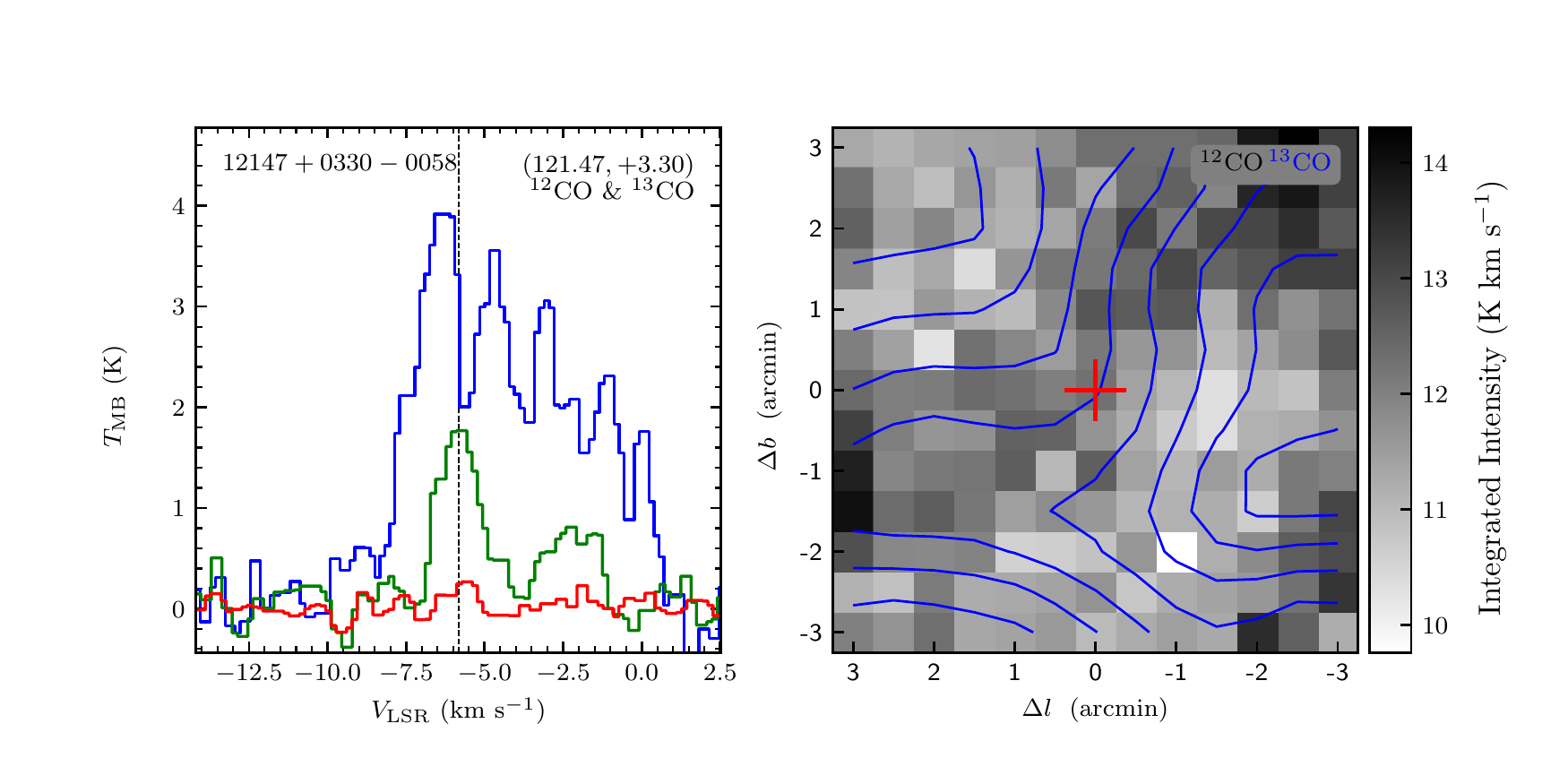}
\vspace{-0.5cm}

\includegraphics[width=9.0cm,angle=0]{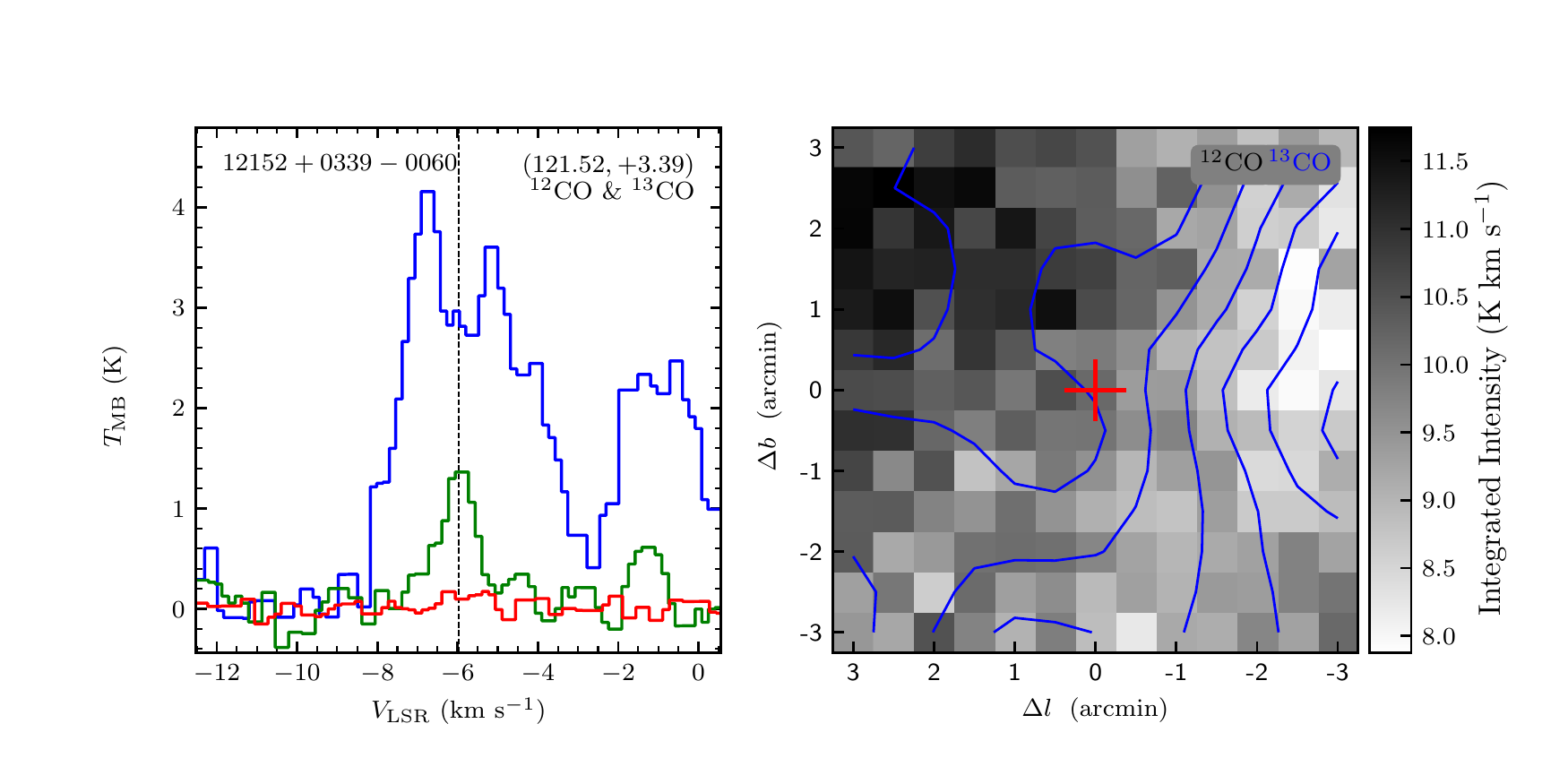}
\includegraphics[width=9.0cm,angle=0]{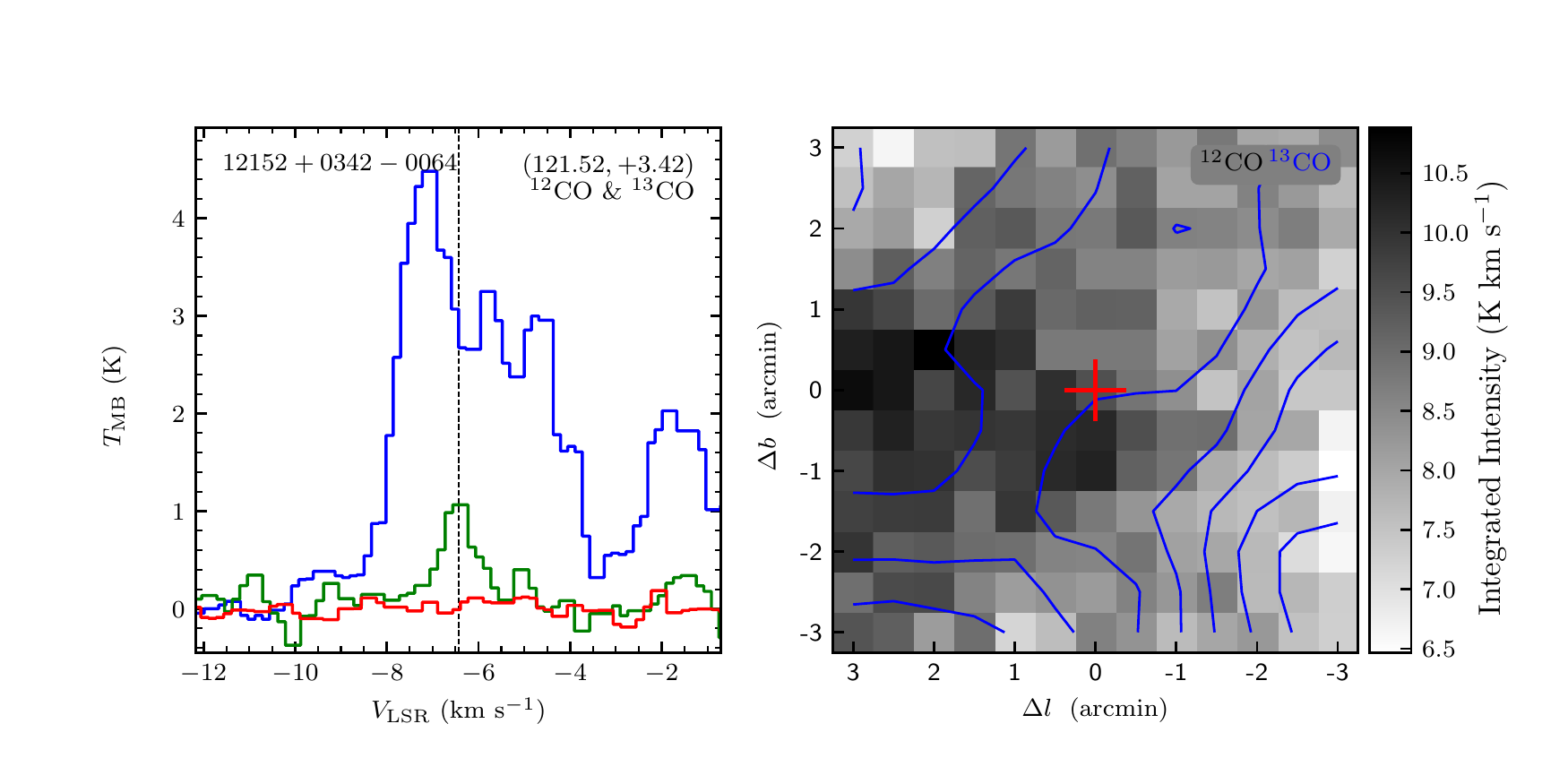}
\vspace{-0.5cm}

\includegraphics[width=9.0cm,angle=0]{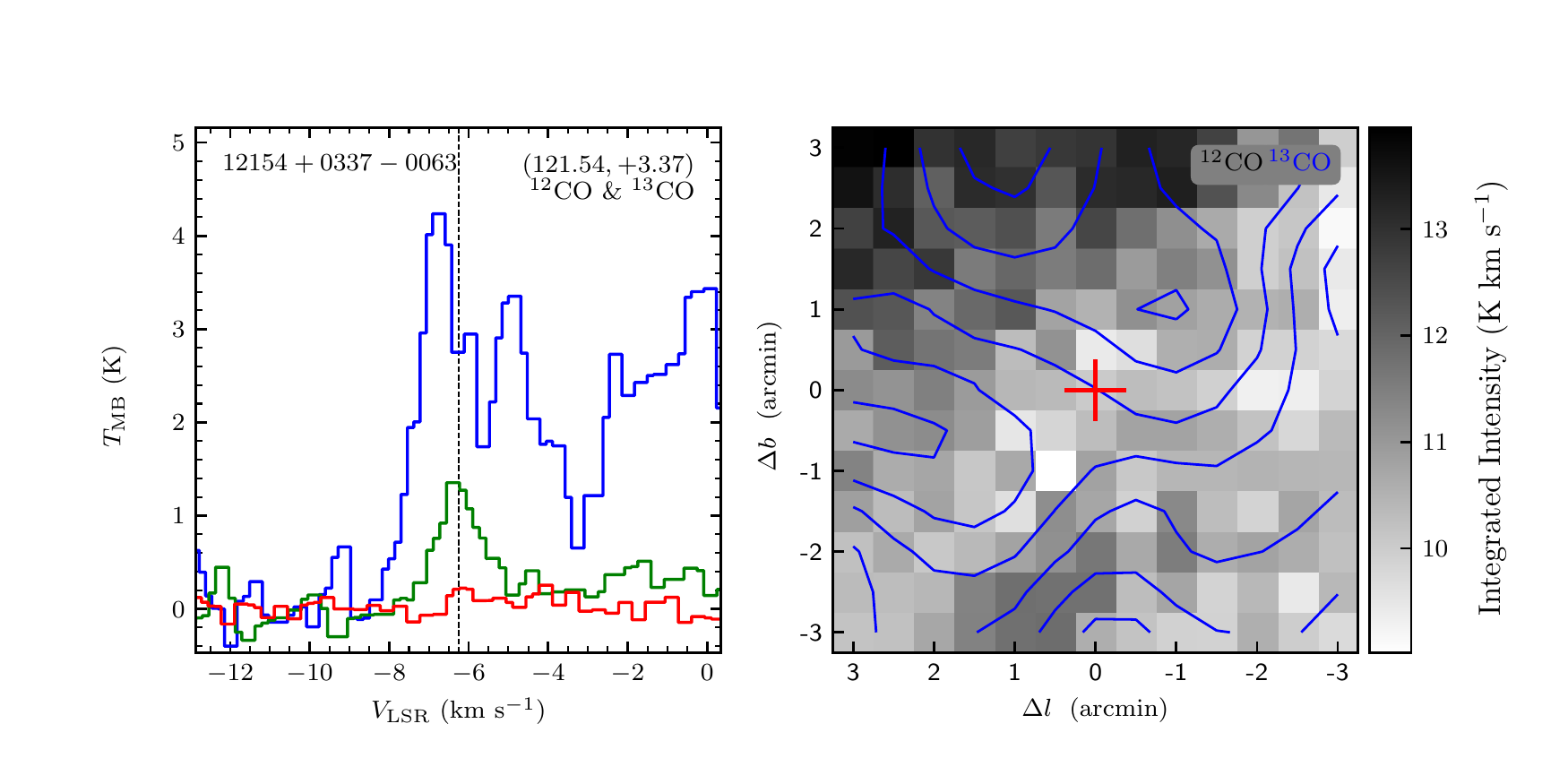}
\includegraphics[width=9.0cm,angle=0]{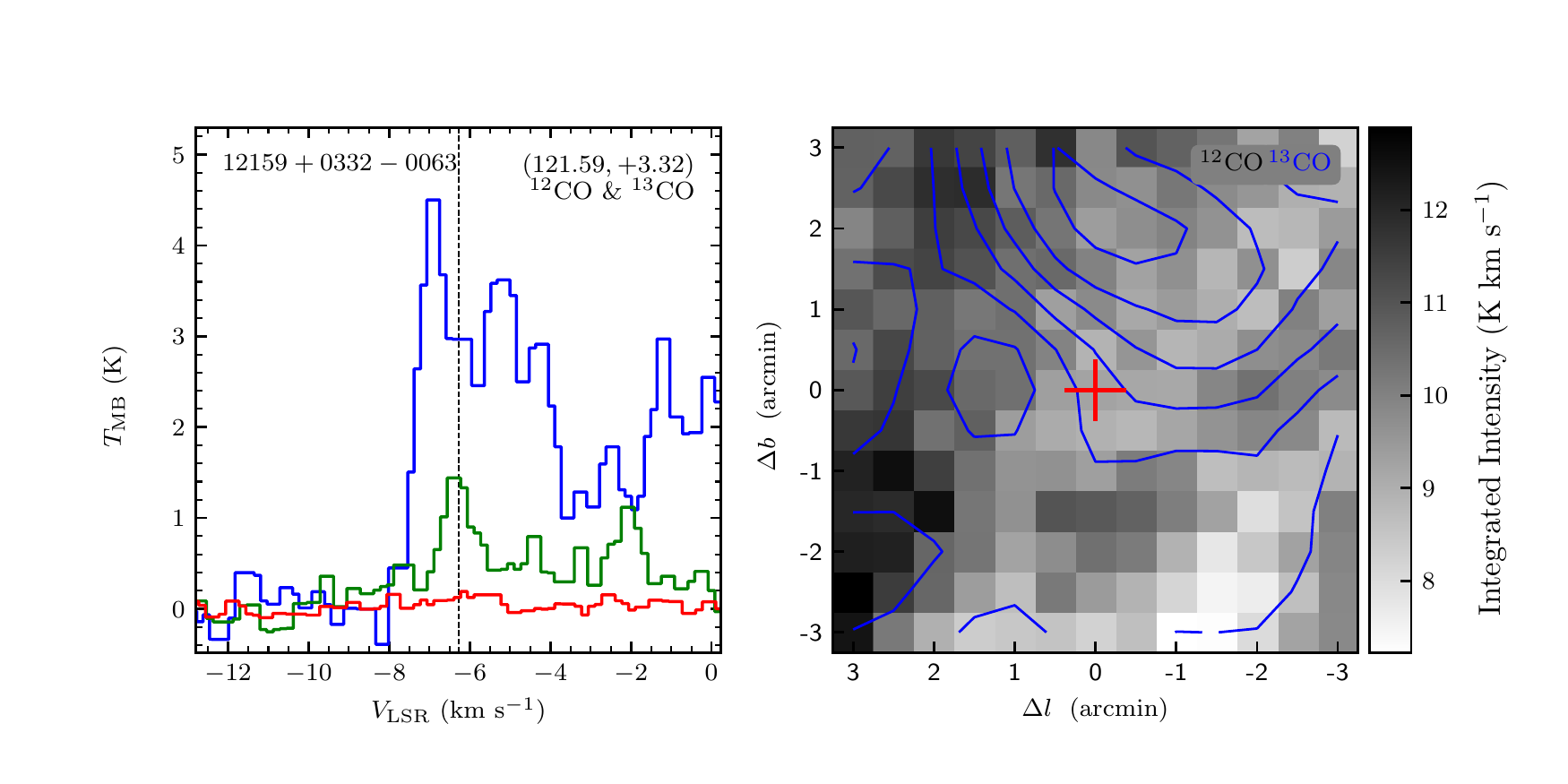}
\vspace{-0.5cm}

\includegraphics[width=9.0cm,angle=0]{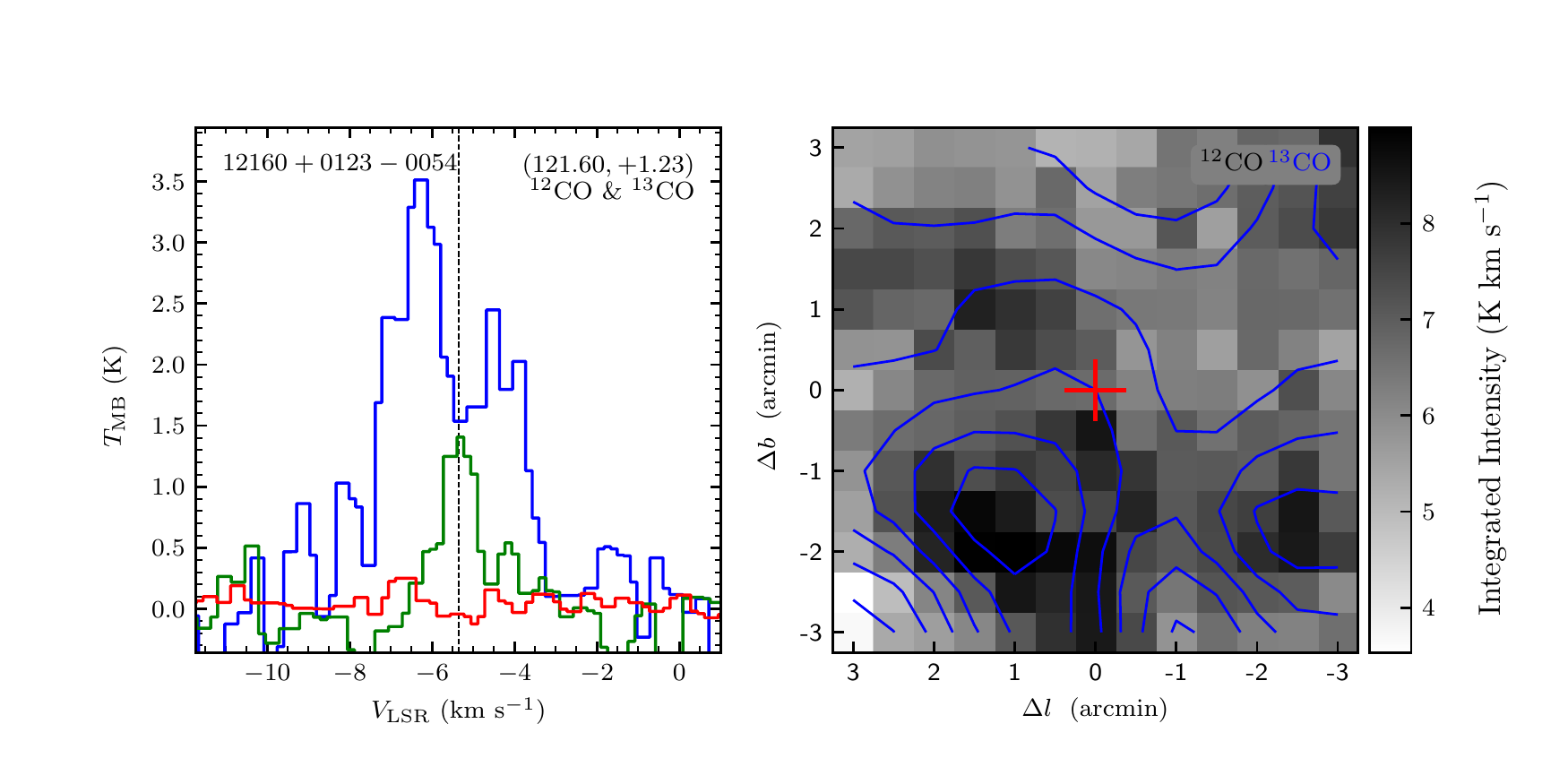}
\includegraphics[width=9.0cm,angle=0]{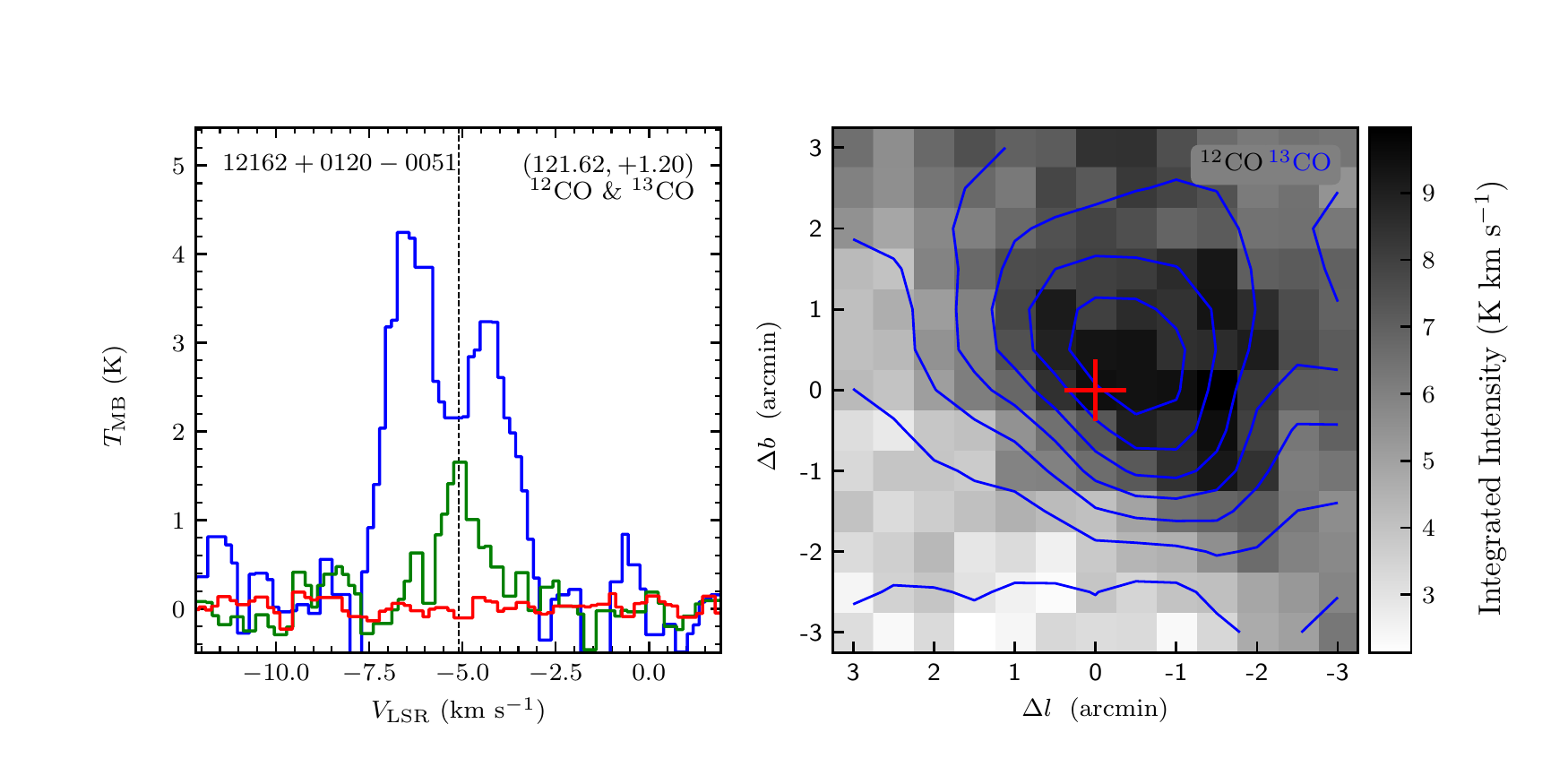}
\vspace{-0.5cm}

\includegraphics[width=9.0cm,angle=0]{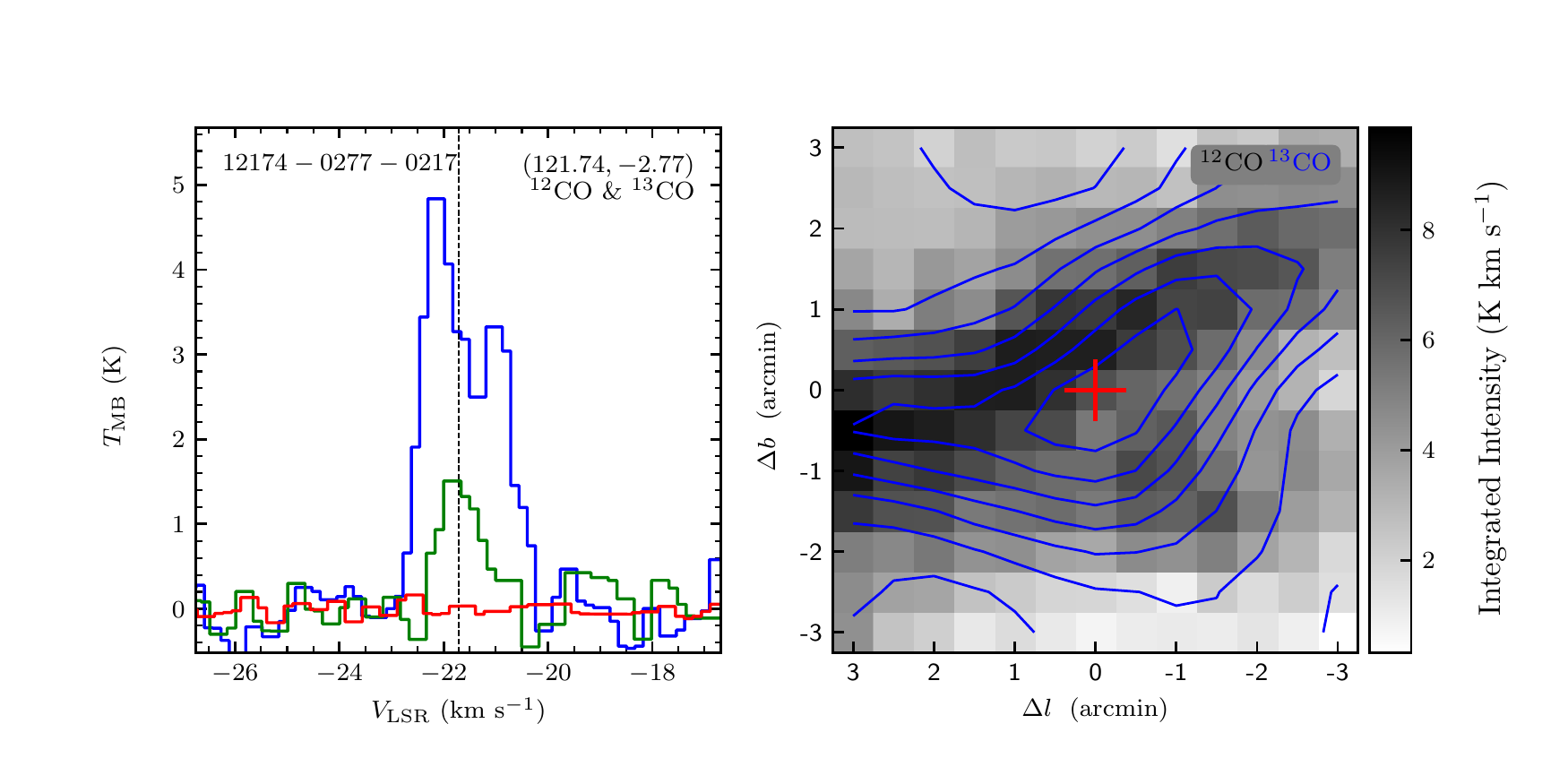}
\includegraphics[width=9.0cm,angle=0]{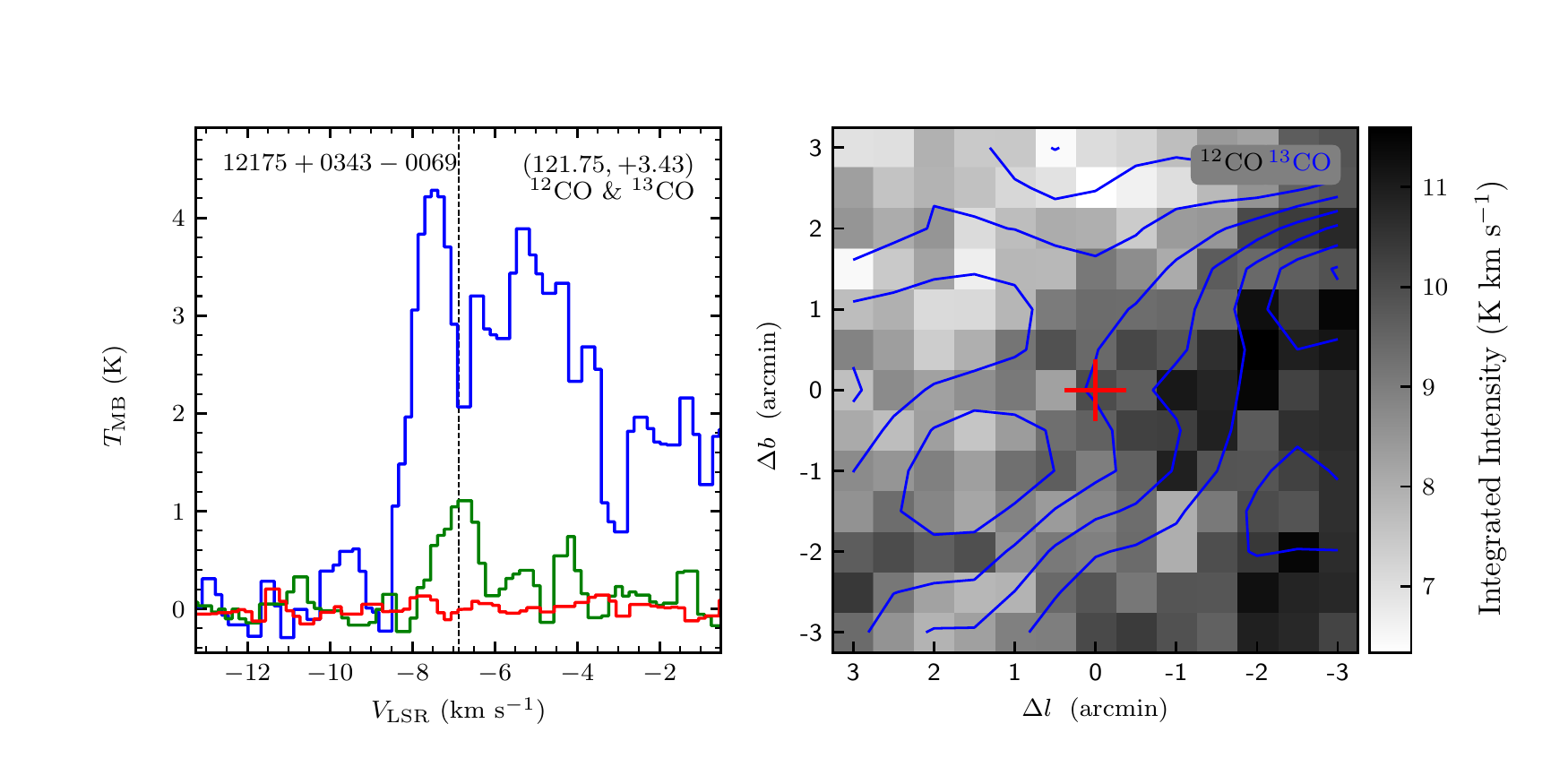}
\end{figure}
\clearpage

\begin{figure}
\includegraphics[width=9.0cm,angle=0]{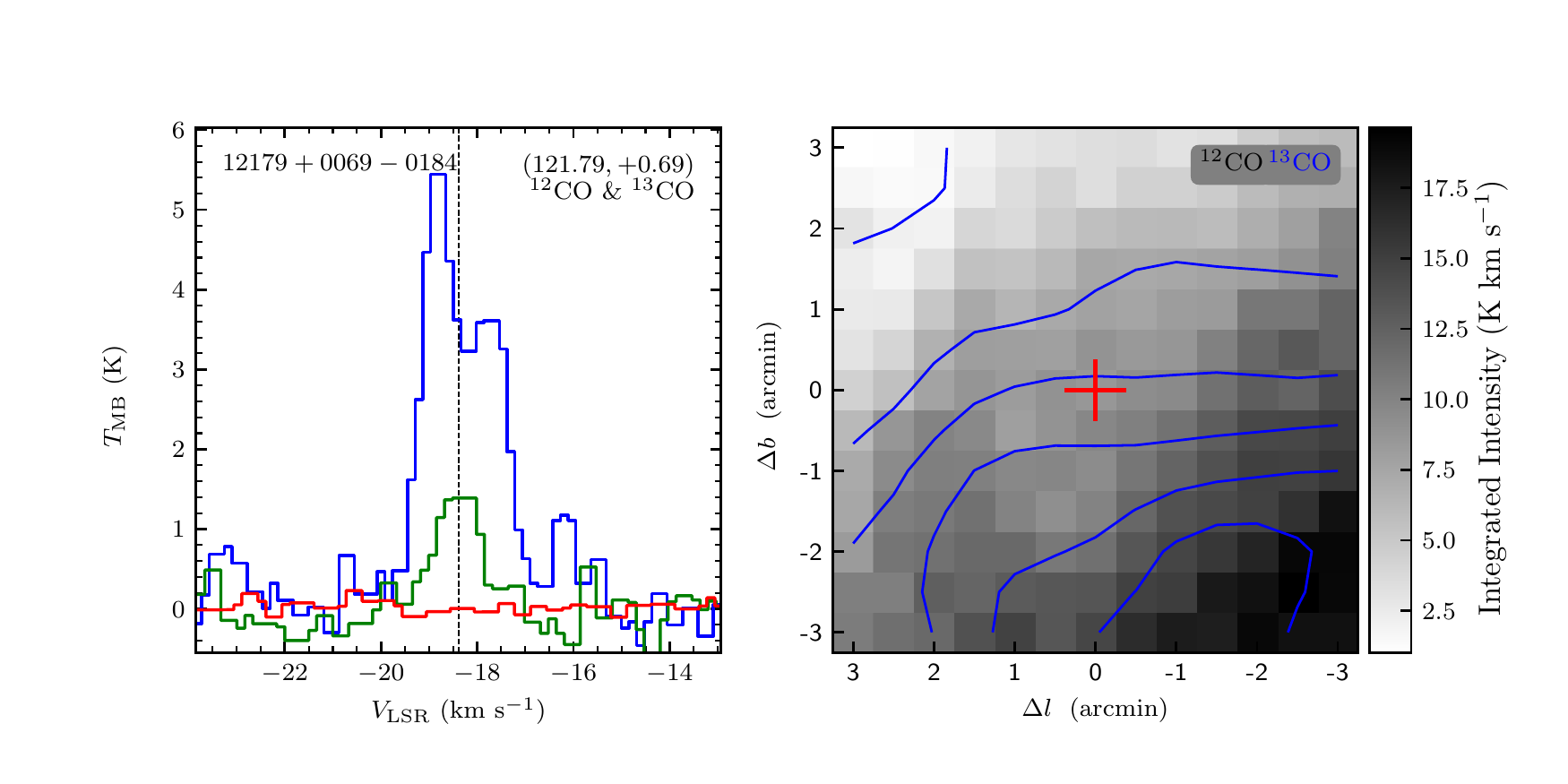}
\includegraphics[width=9.0cm,angle=0]{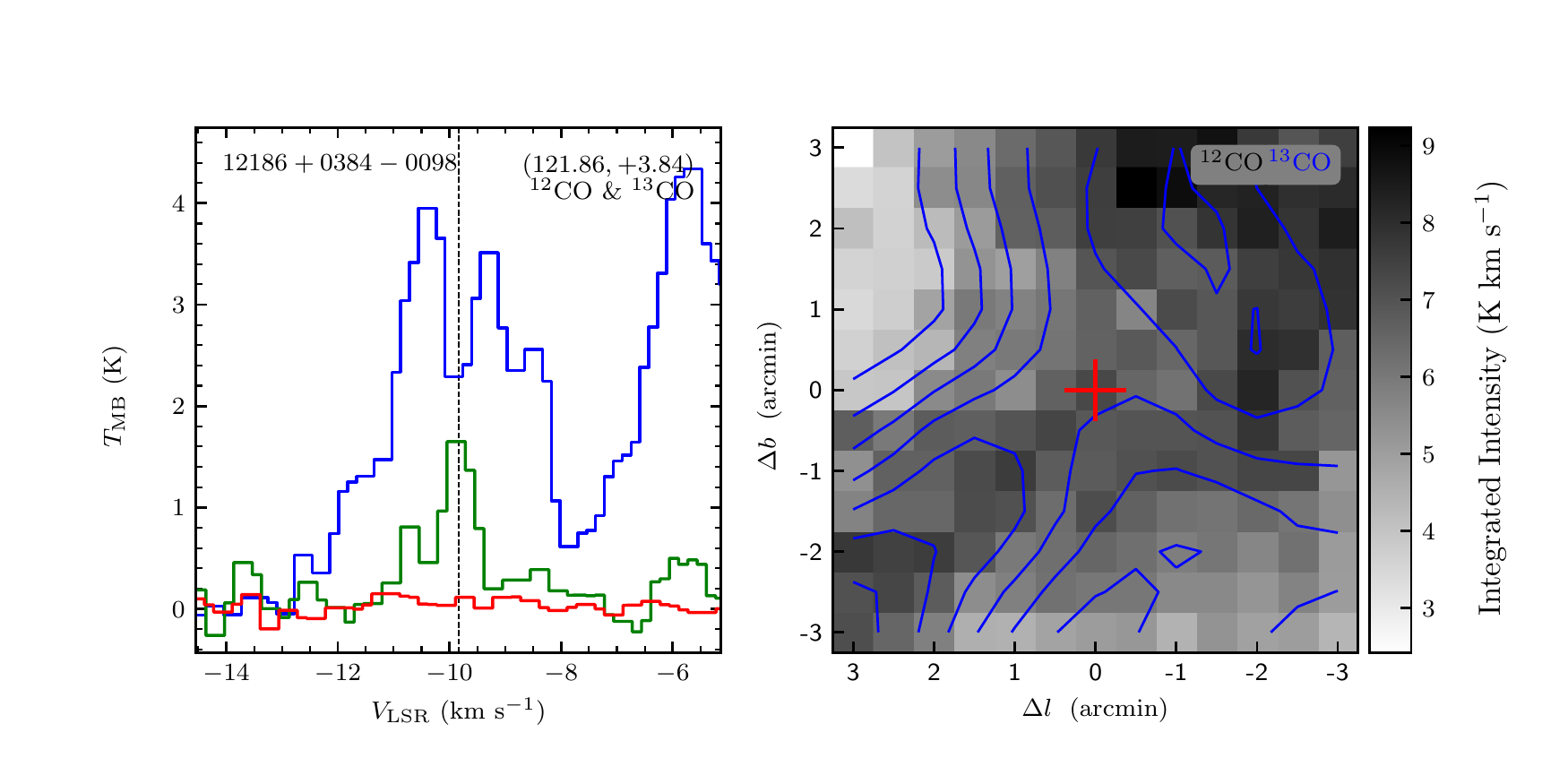}
\vspace{-0.5cm}

\includegraphics[width=9.0cm,angle=0]{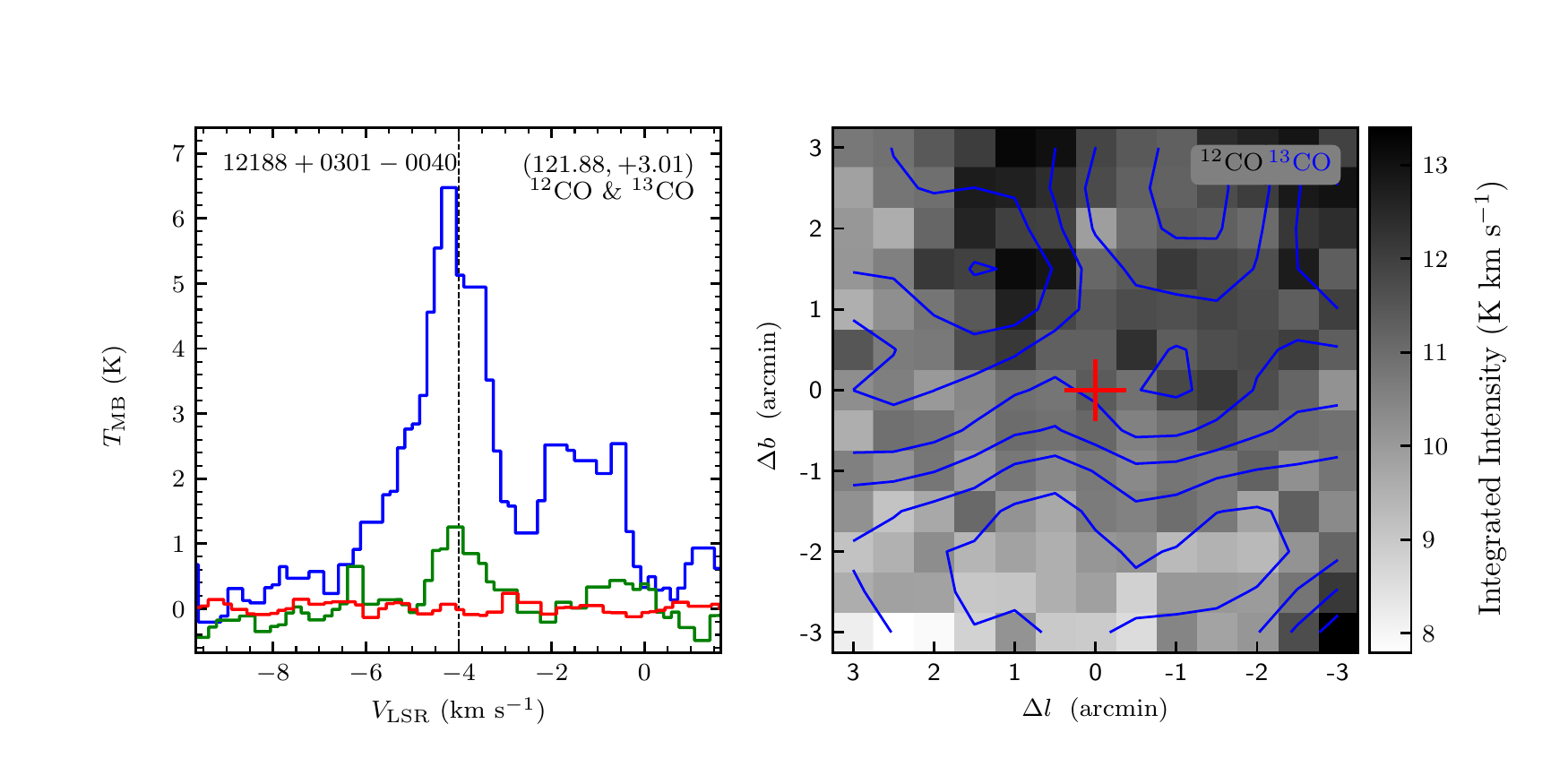}
\includegraphics[width=9.0cm,angle=0]{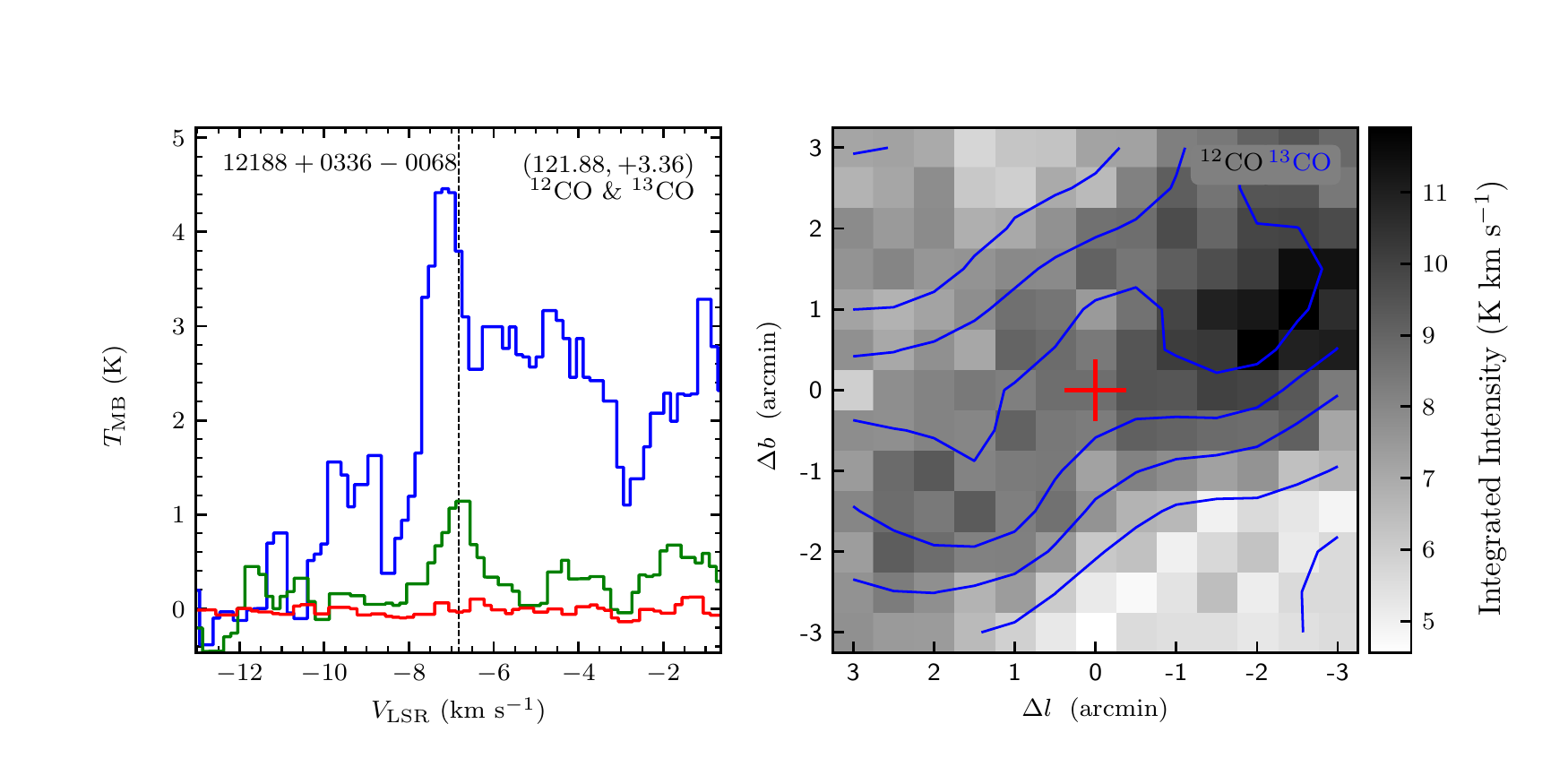}
\vspace{-0.5cm}

\includegraphics[width=9.0cm,angle=0]{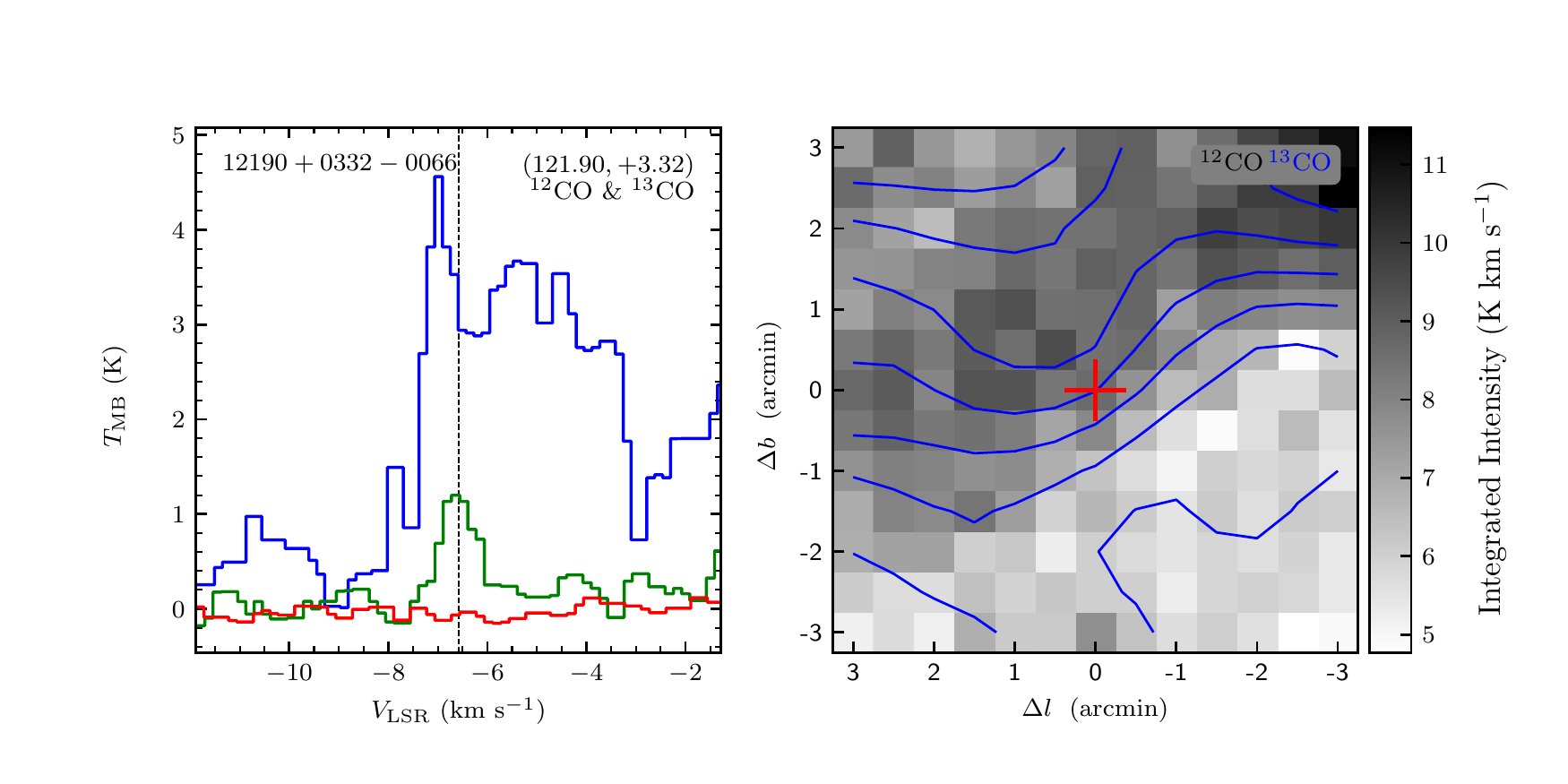}
\includegraphics[width=9.0cm,angle=0]{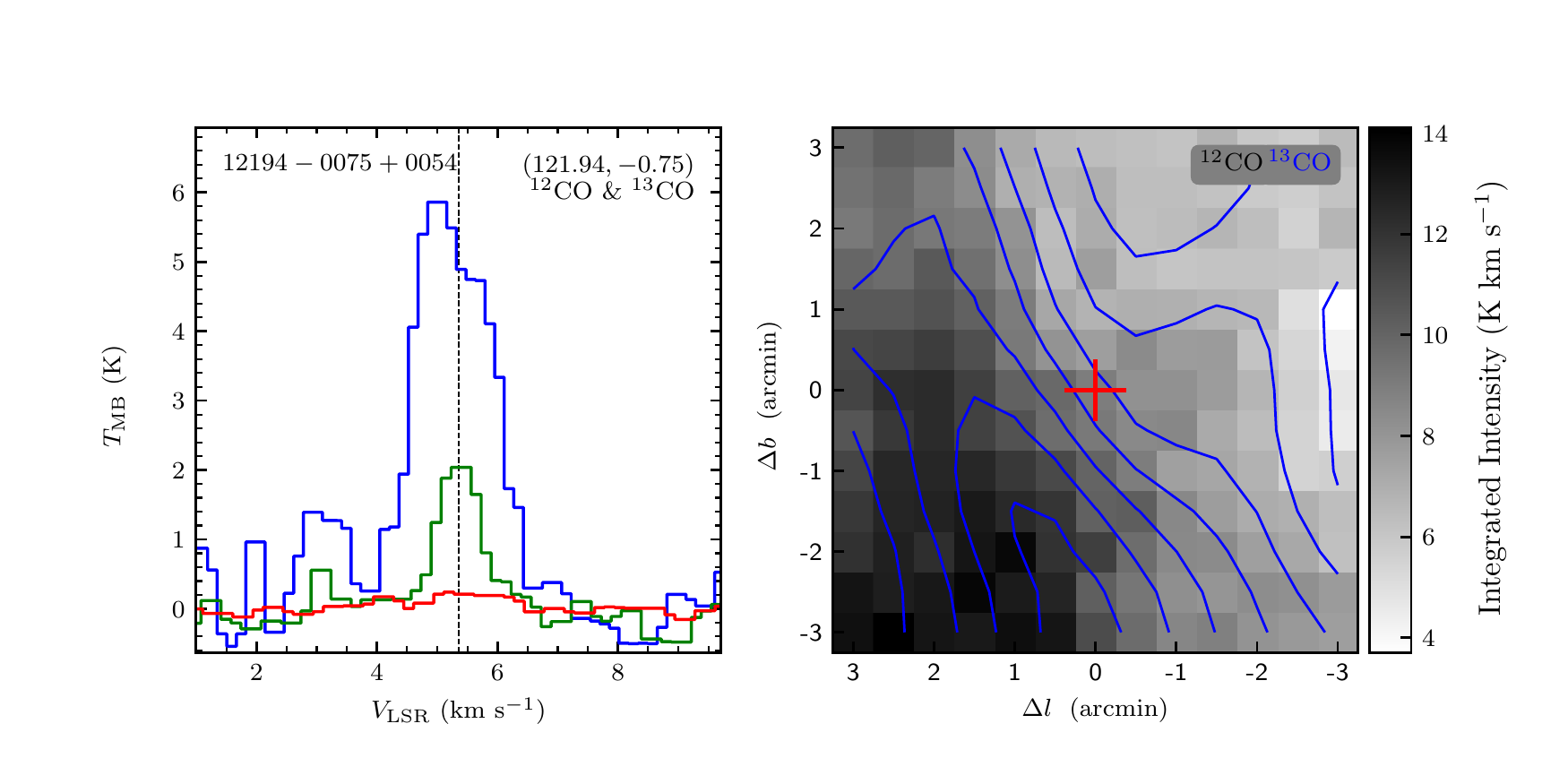}
\vspace{-0.5cm}

\includegraphics[width=9.0cm,angle=0]{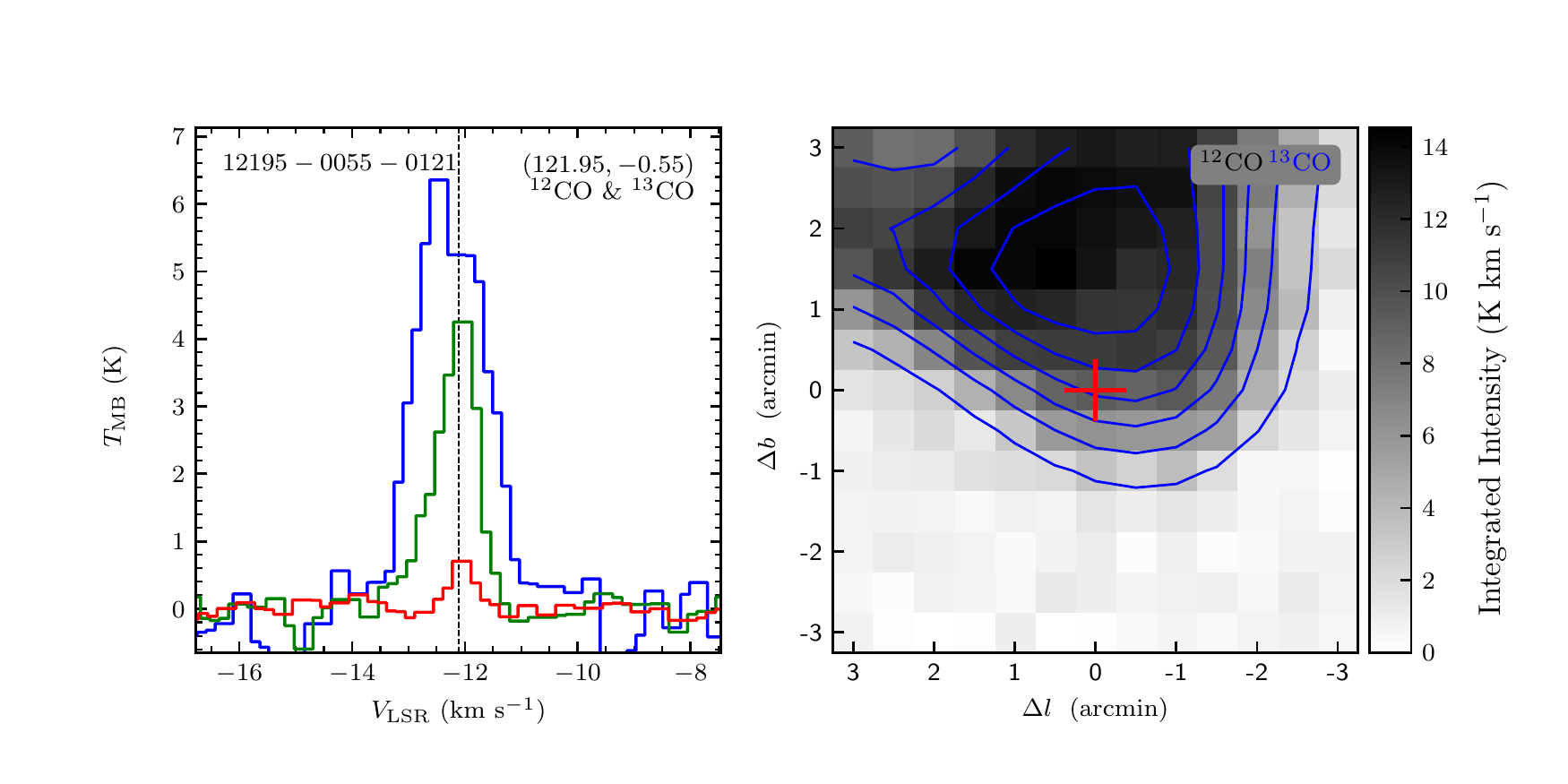}
\includegraphics[width=9.0cm,angle=0]{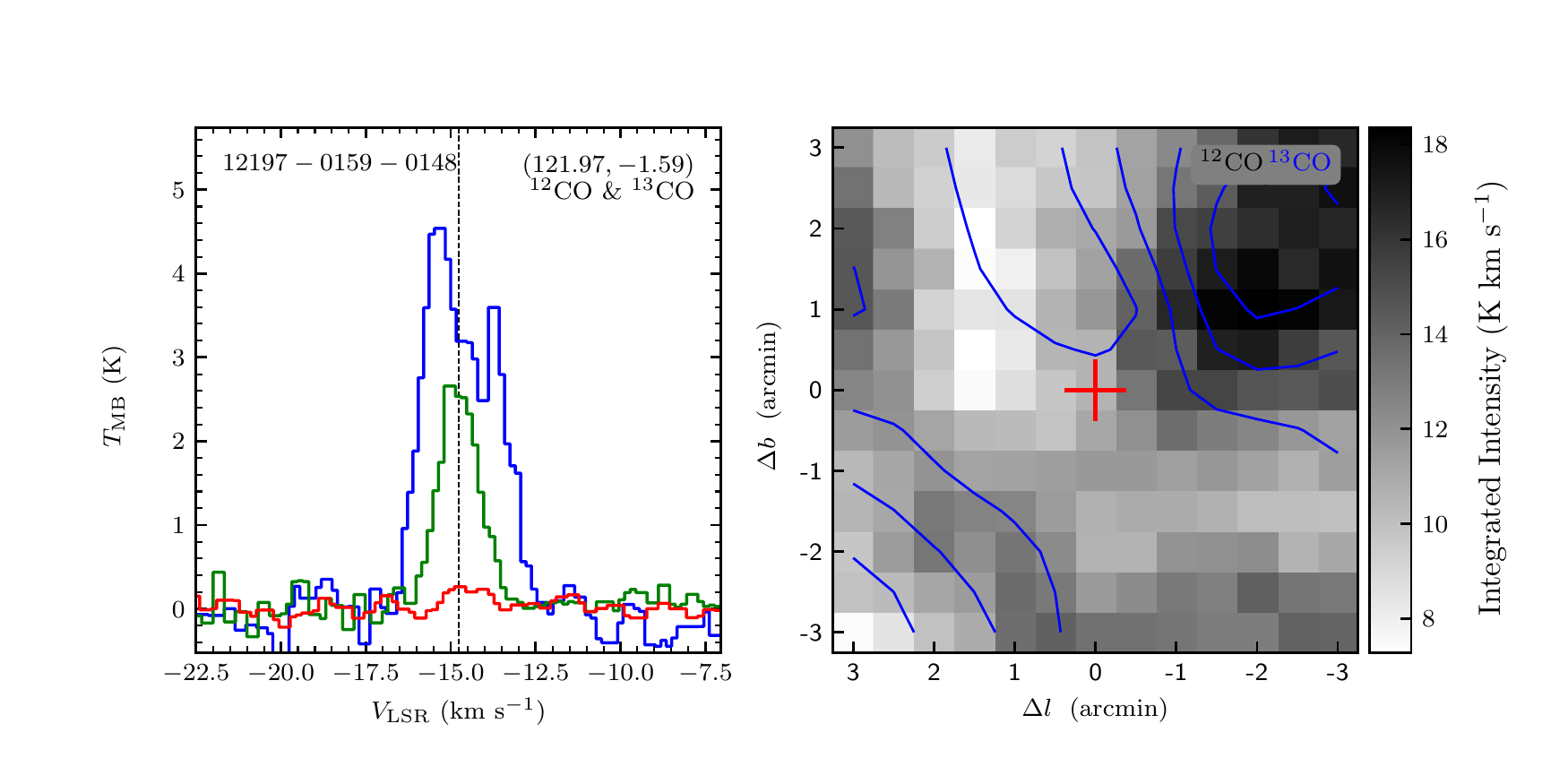}
\vspace{-0.5cm}

\includegraphics[width=9.0cm,angle=0]{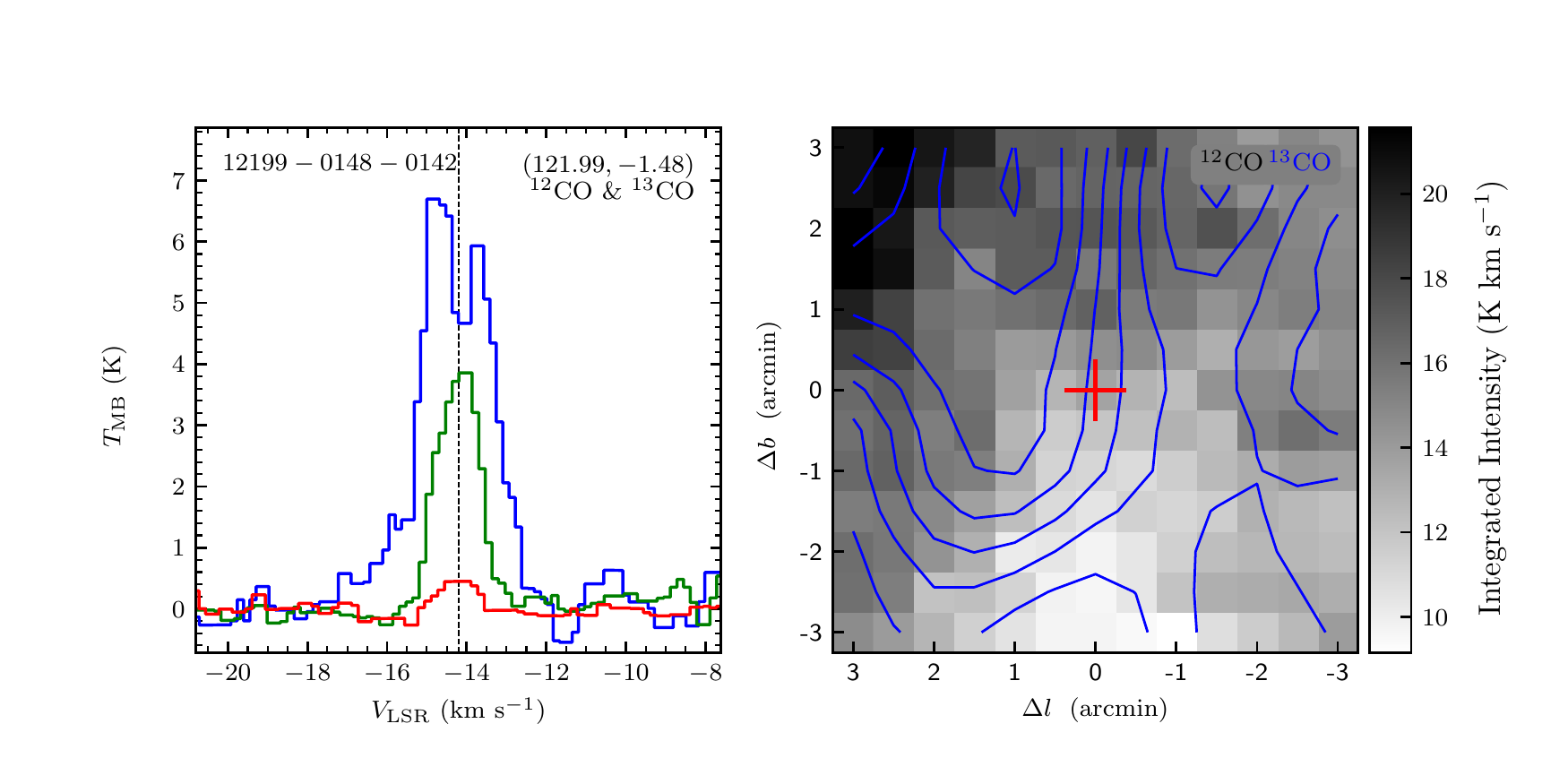}
\includegraphics[width=9.0cm,angle=0]{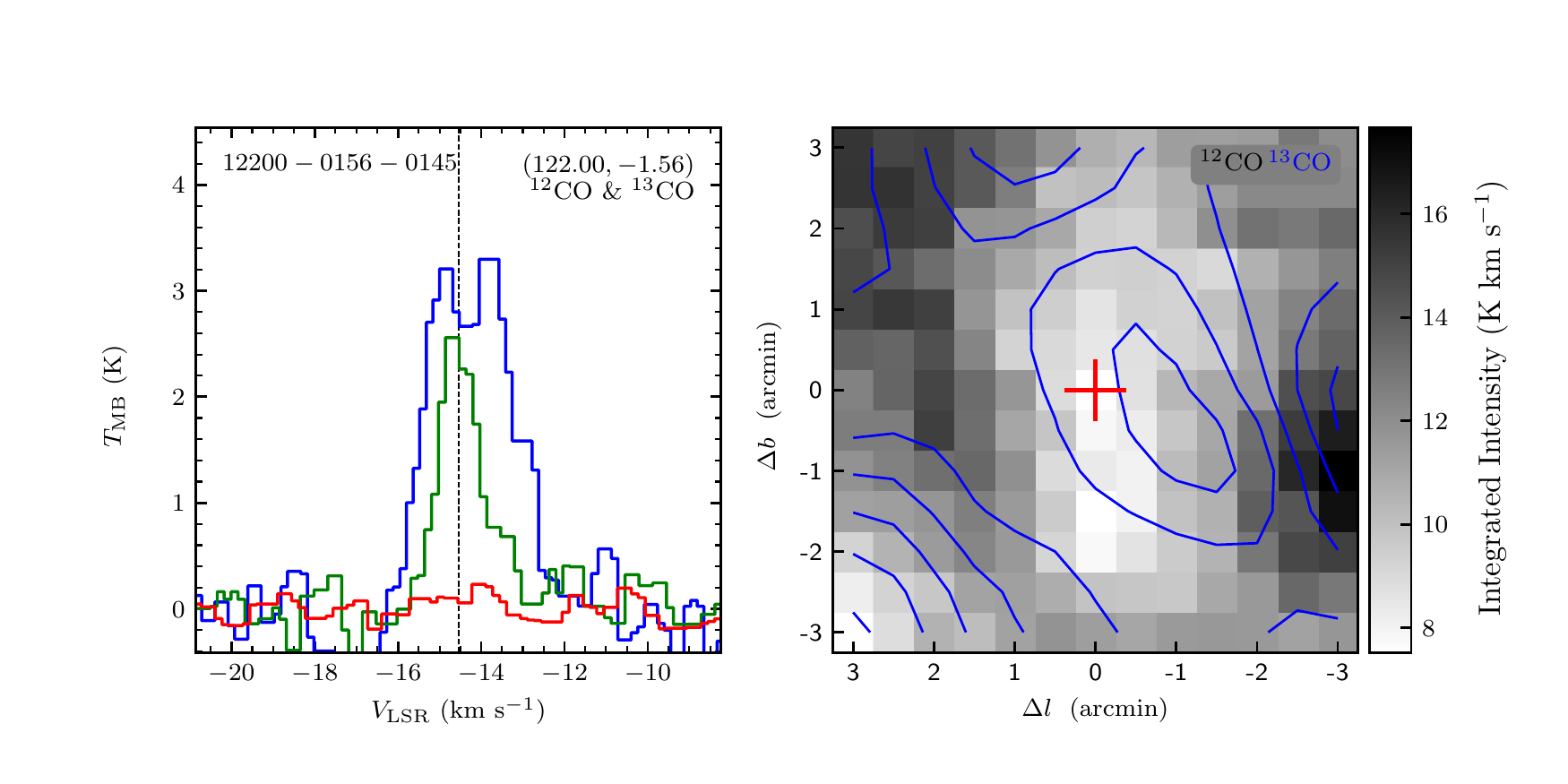}
\end{figure}
\clearpage

\begin{figure}
\includegraphics[width=9.0cm,angle=0]{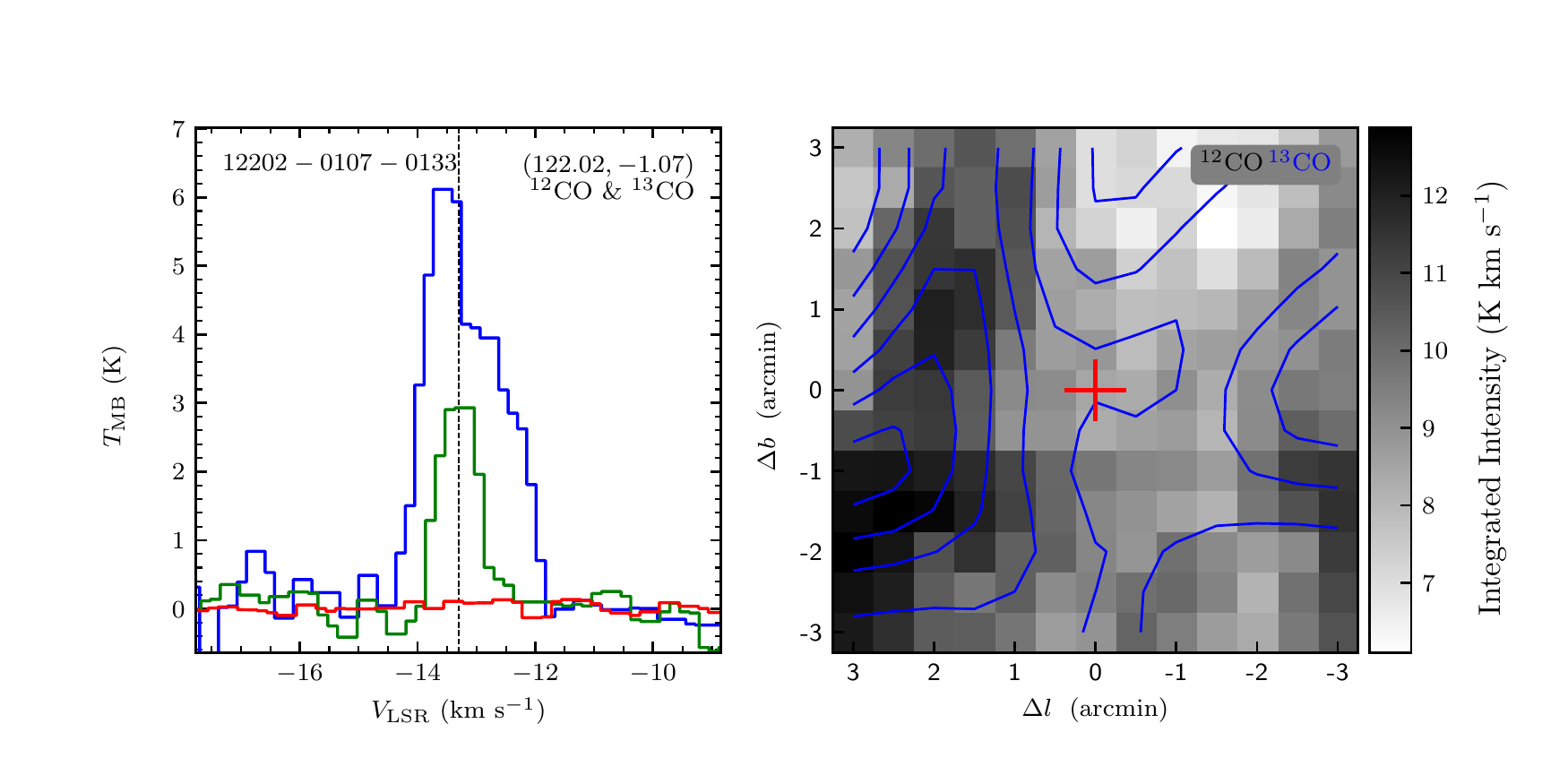}
\includegraphics[width=9.0cm,angle=0]{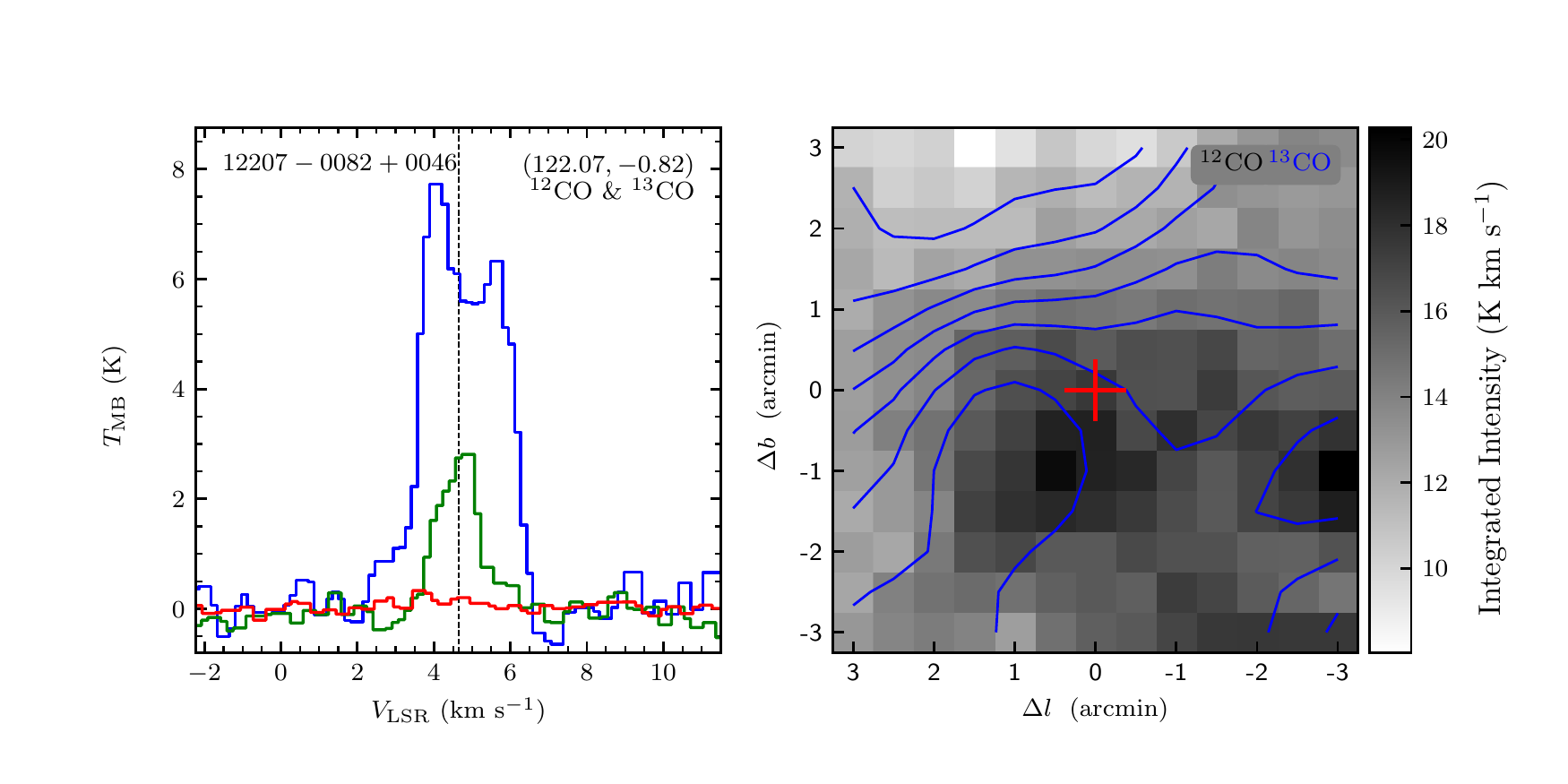}
\vspace{-0.5cm}

\includegraphics[width=9.0cm,angle=0]{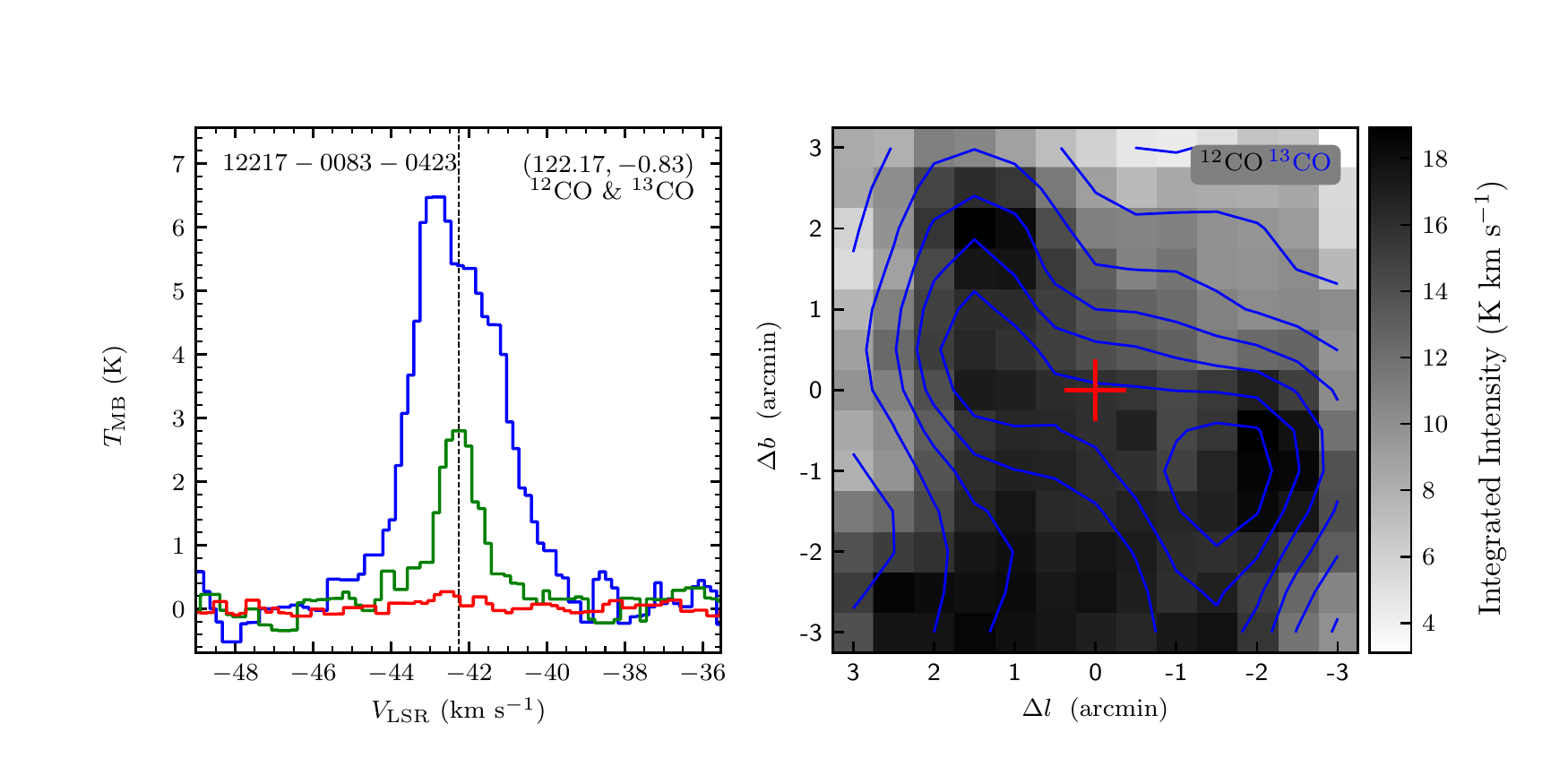}
\includegraphics[width=9.0cm,angle=0]{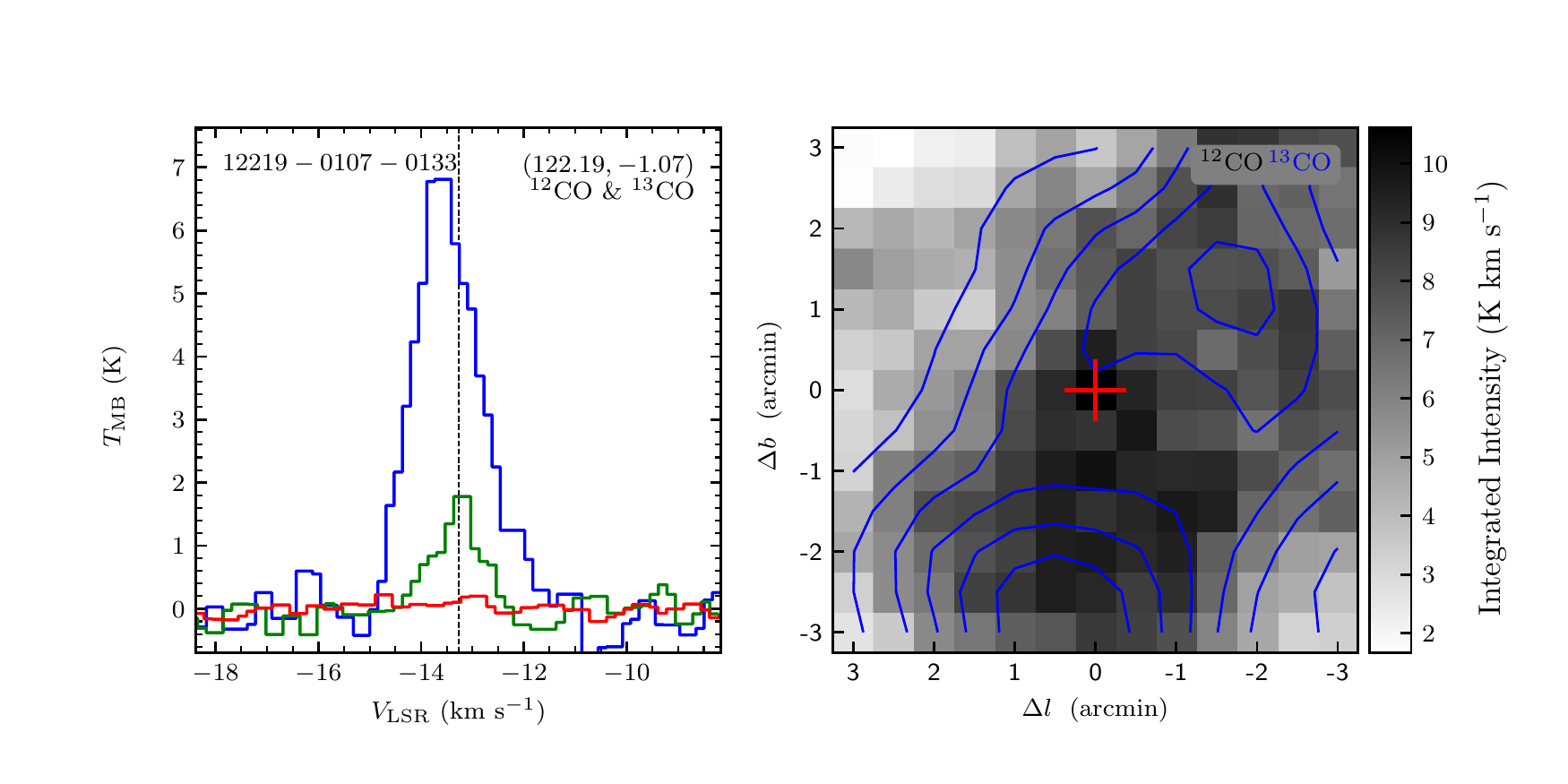}
\vspace{-0.5cm}

\includegraphics[width=9.0cm,angle=0]{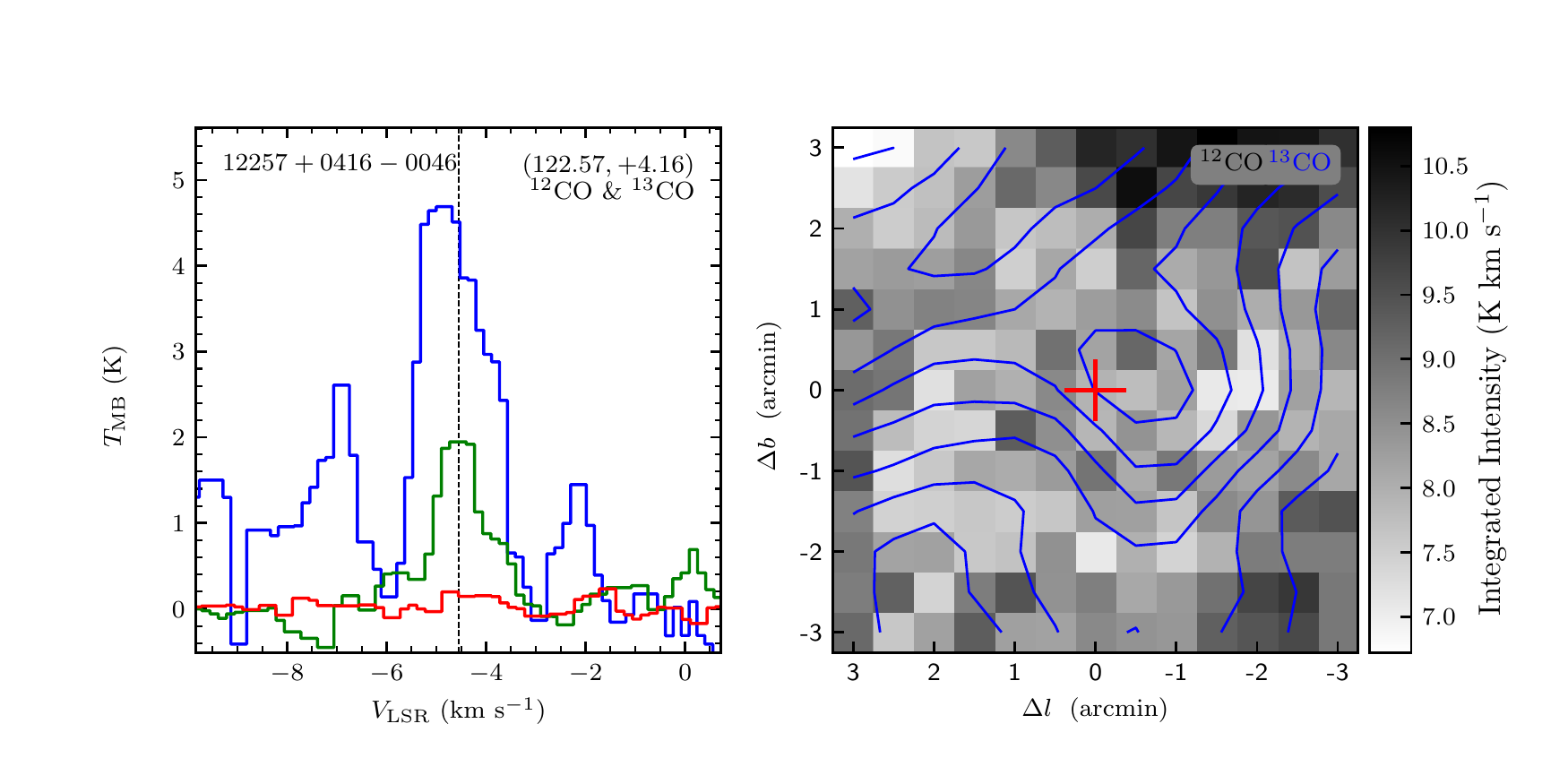}
\includegraphics[width=9.0cm,angle=0]{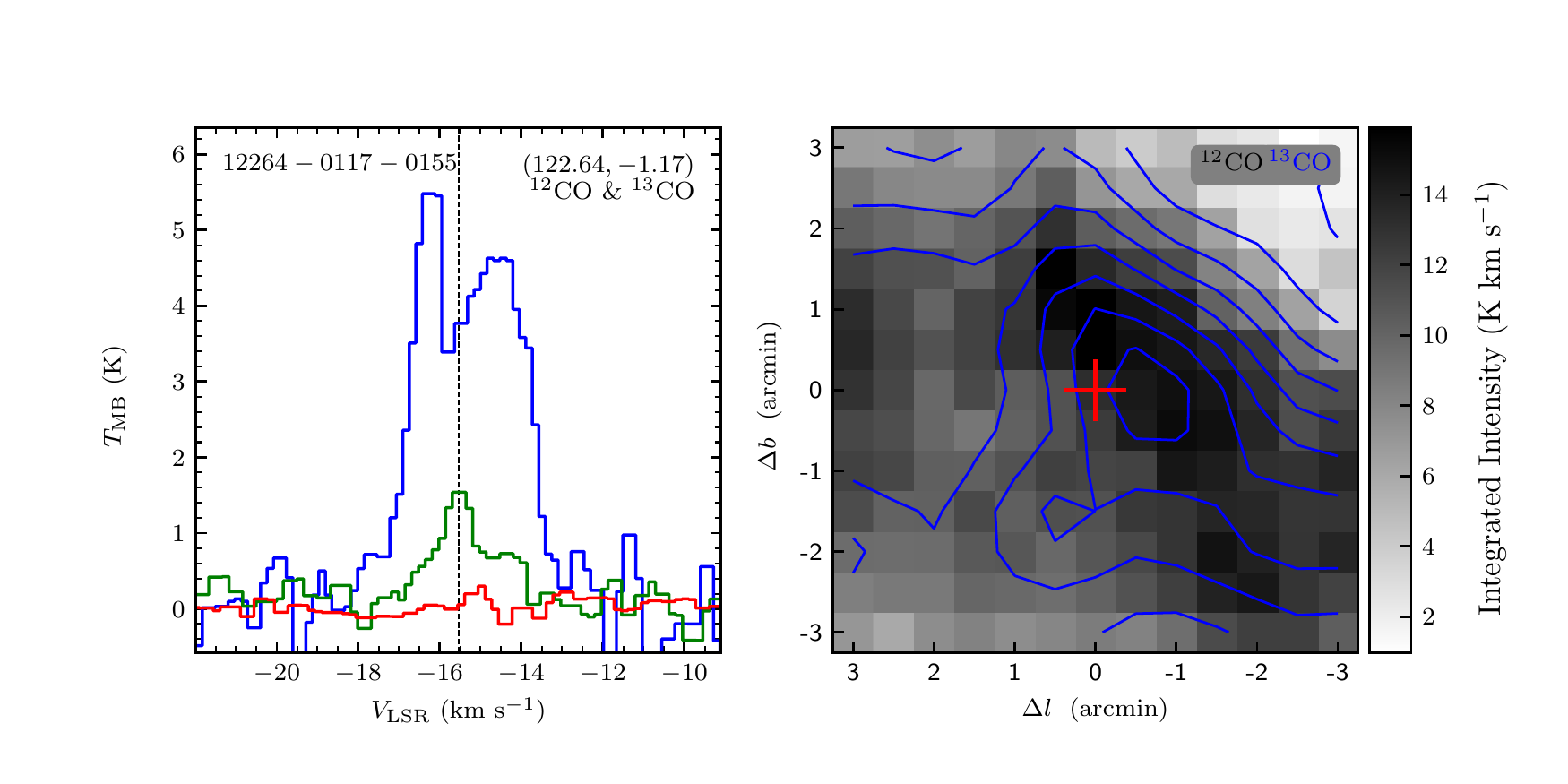}
\vspace{-0.5cm}

\includegraphics[width=9.0cm,angle=0]{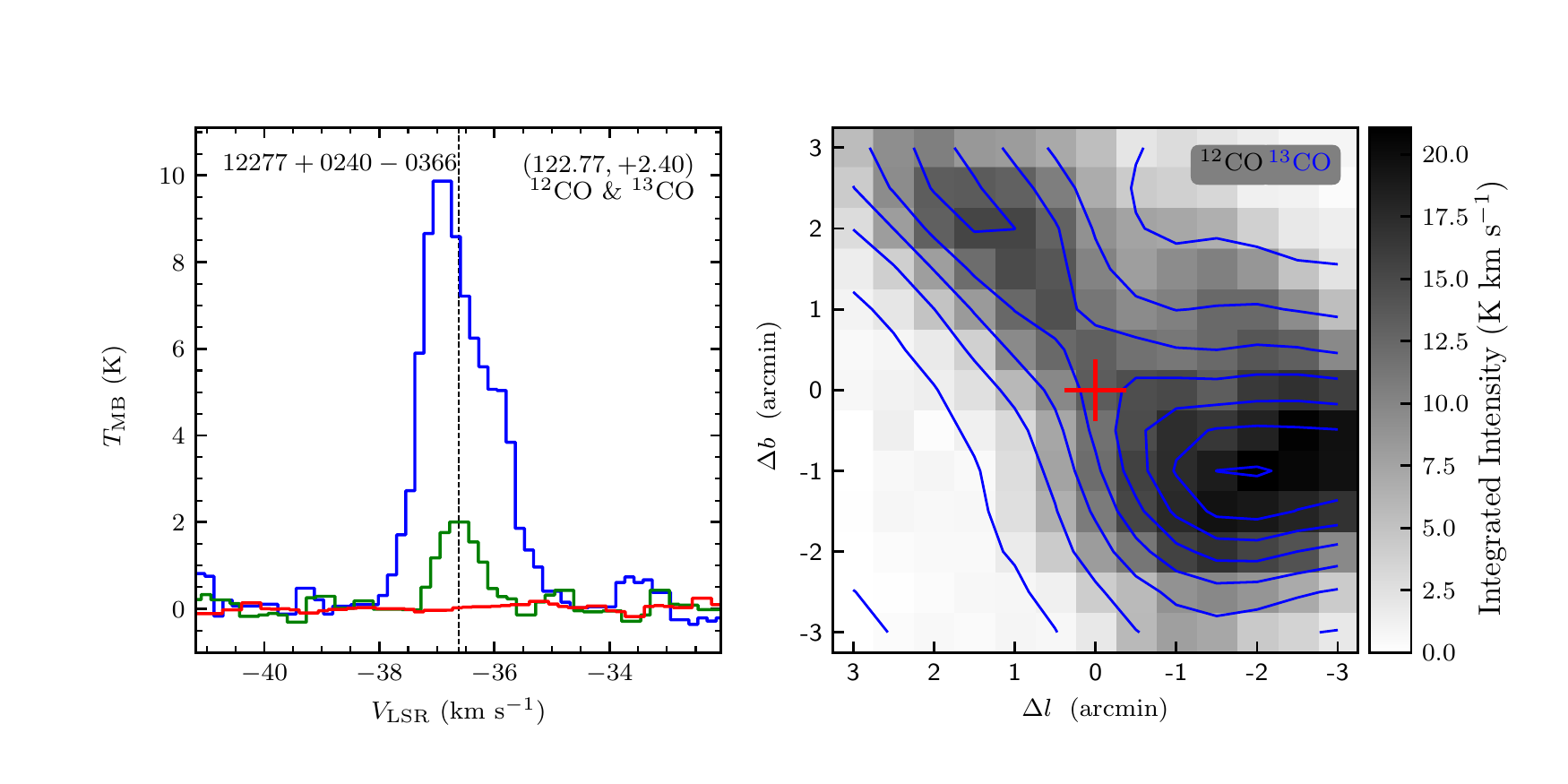}
\includegraphics[width=9.0cm,angle=0]{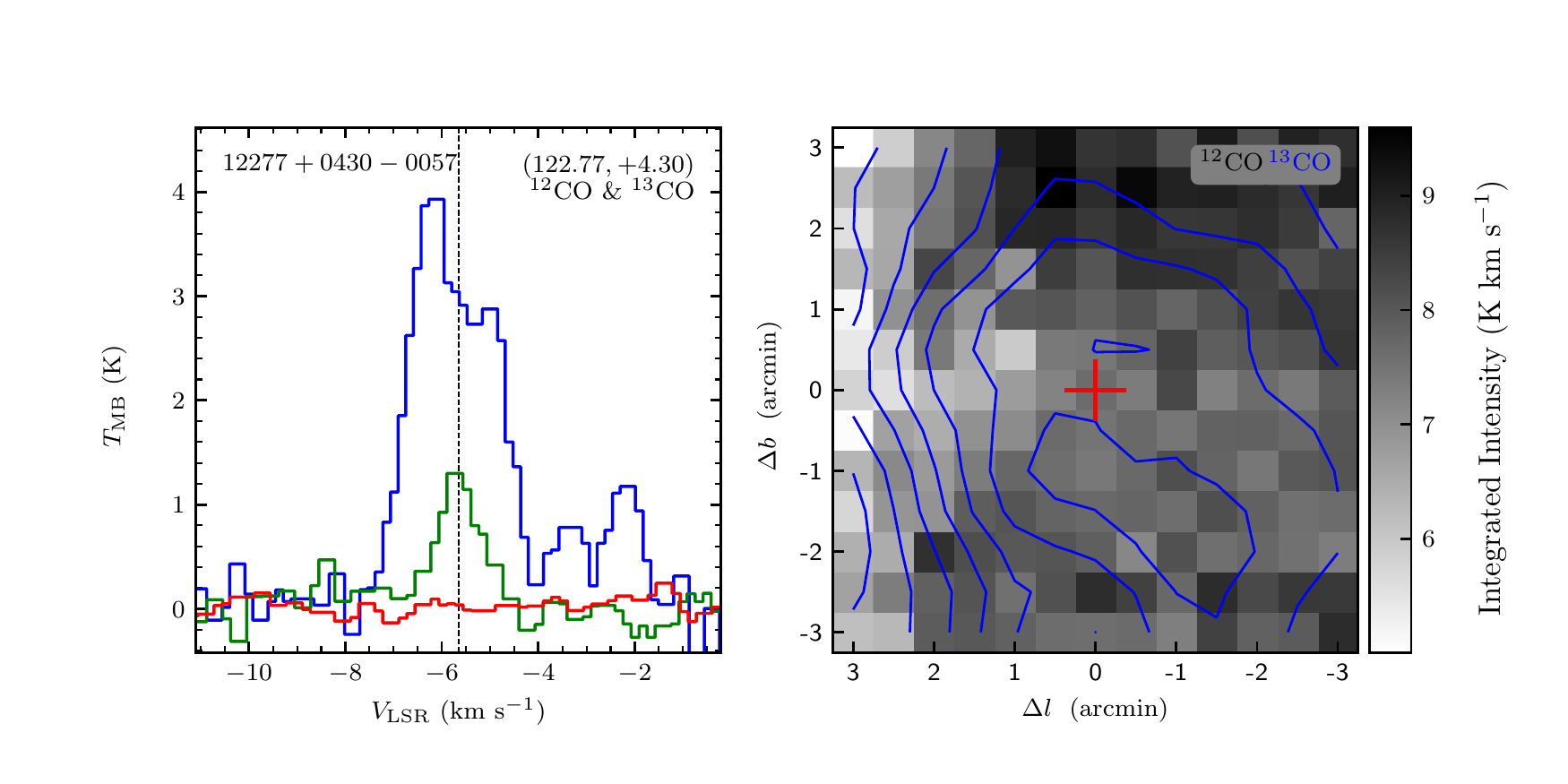}
\vspace{-0.5cm}

\includegraphics[width=9.0cm,angle=0]{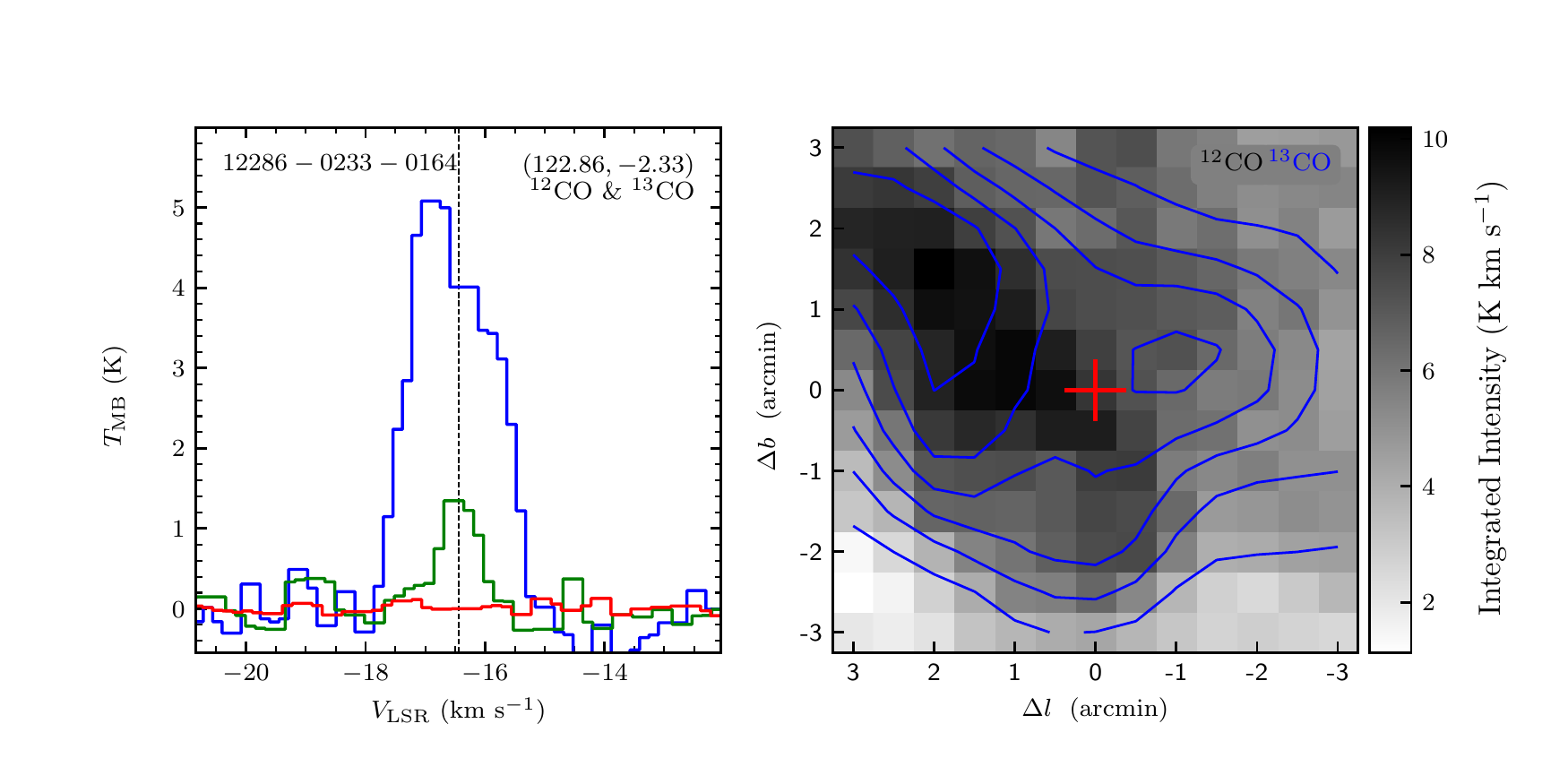}
\includegraphics[width=9.0cm,angle=0]{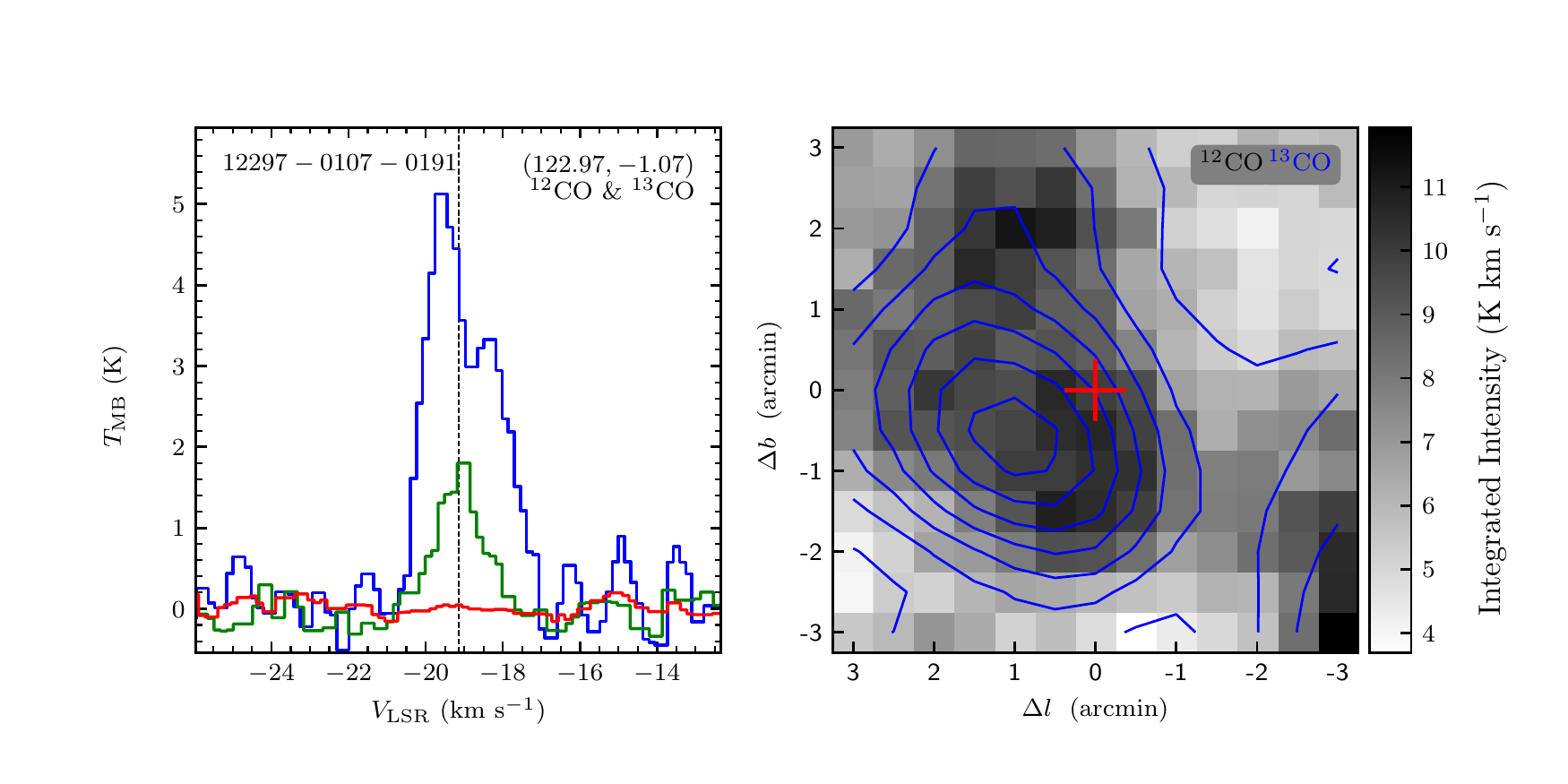}
\end{figure}
\clearpage

\begin{figure}
\includegraphics[width=9.0cm,angle=0]{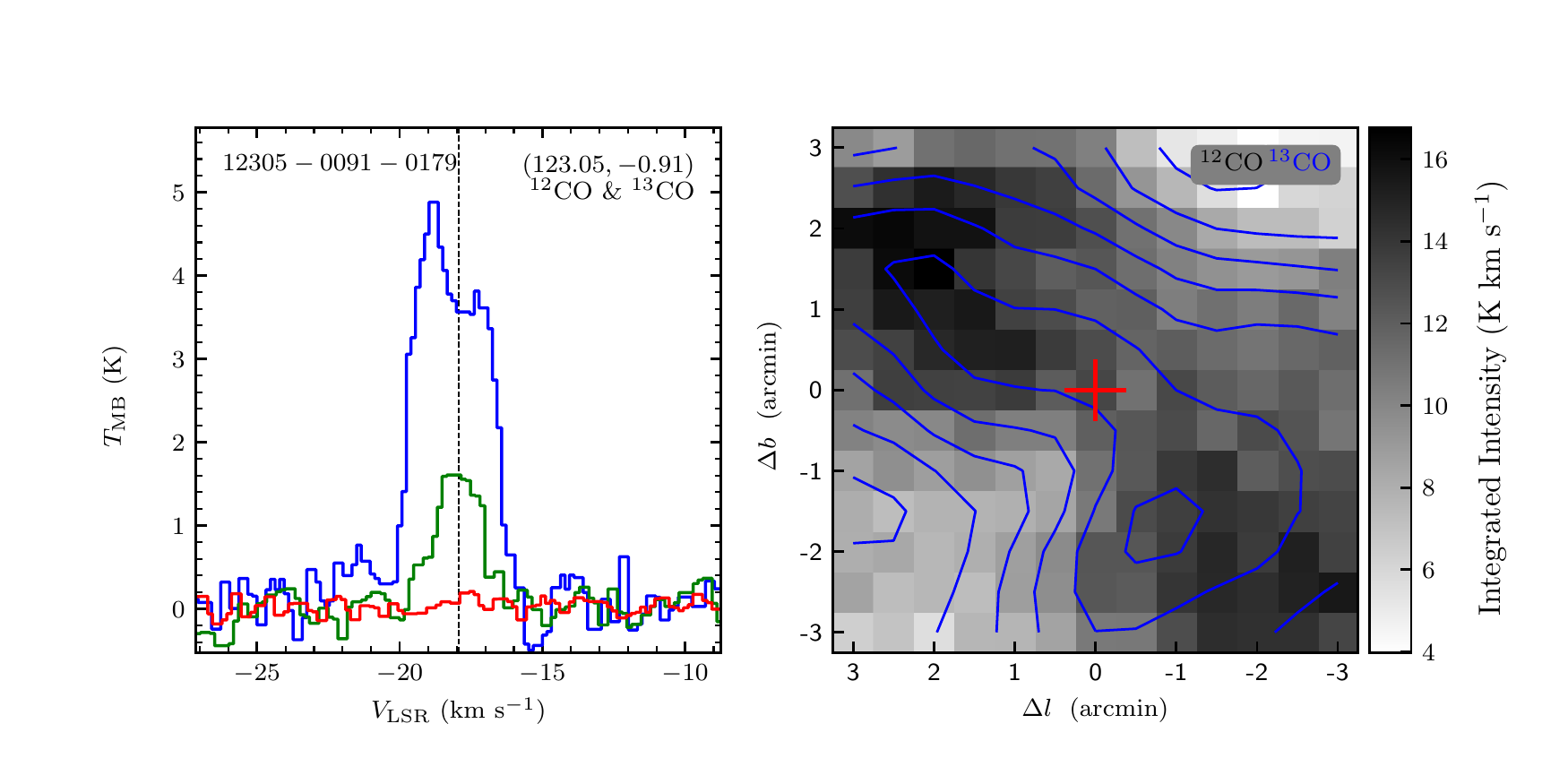}
\includegraphics[width=9.0cm,angle=0]{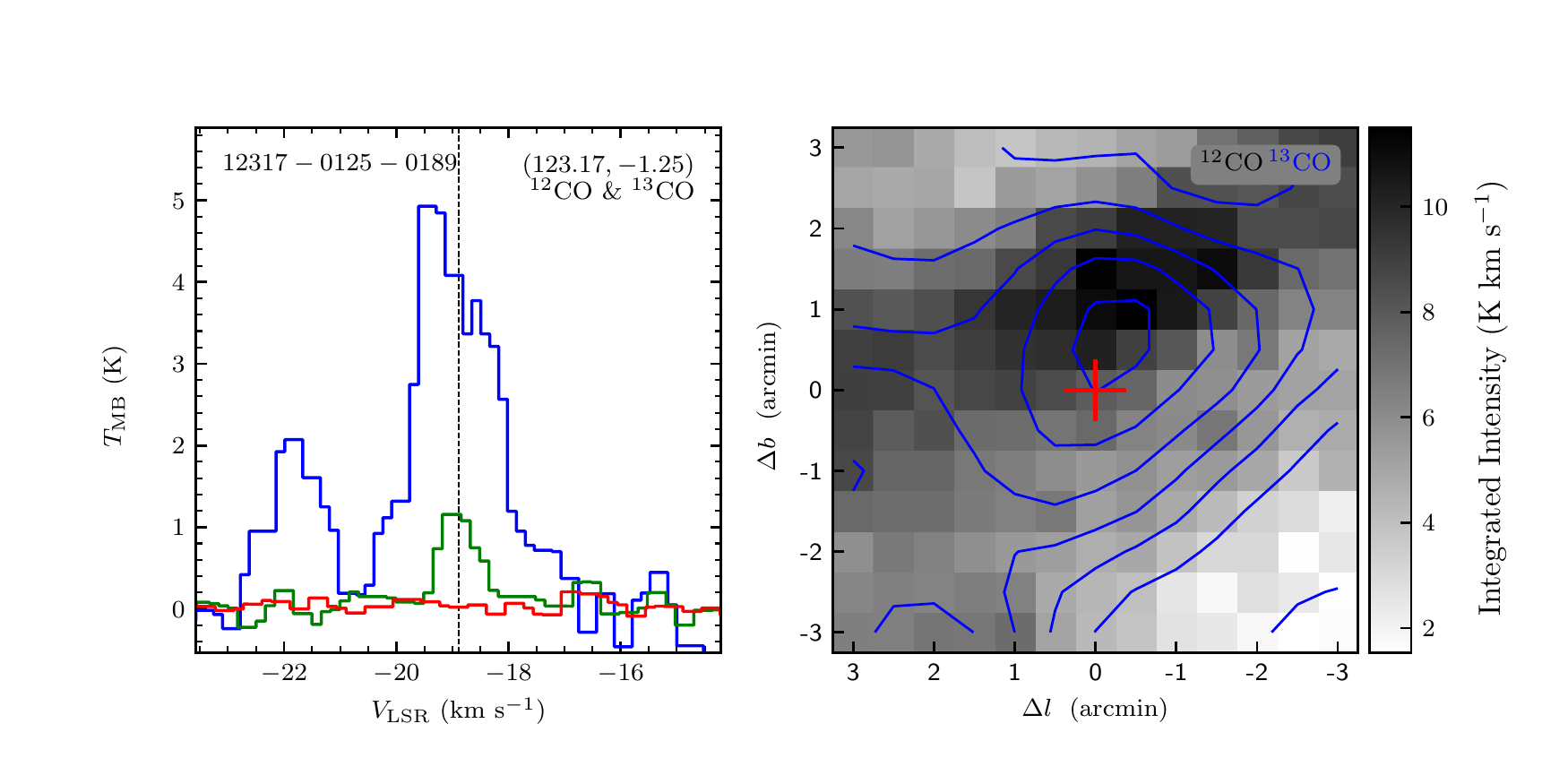}
\vspace{-0.5cm}

\includegraphics[width=9.0cm,angle=0]{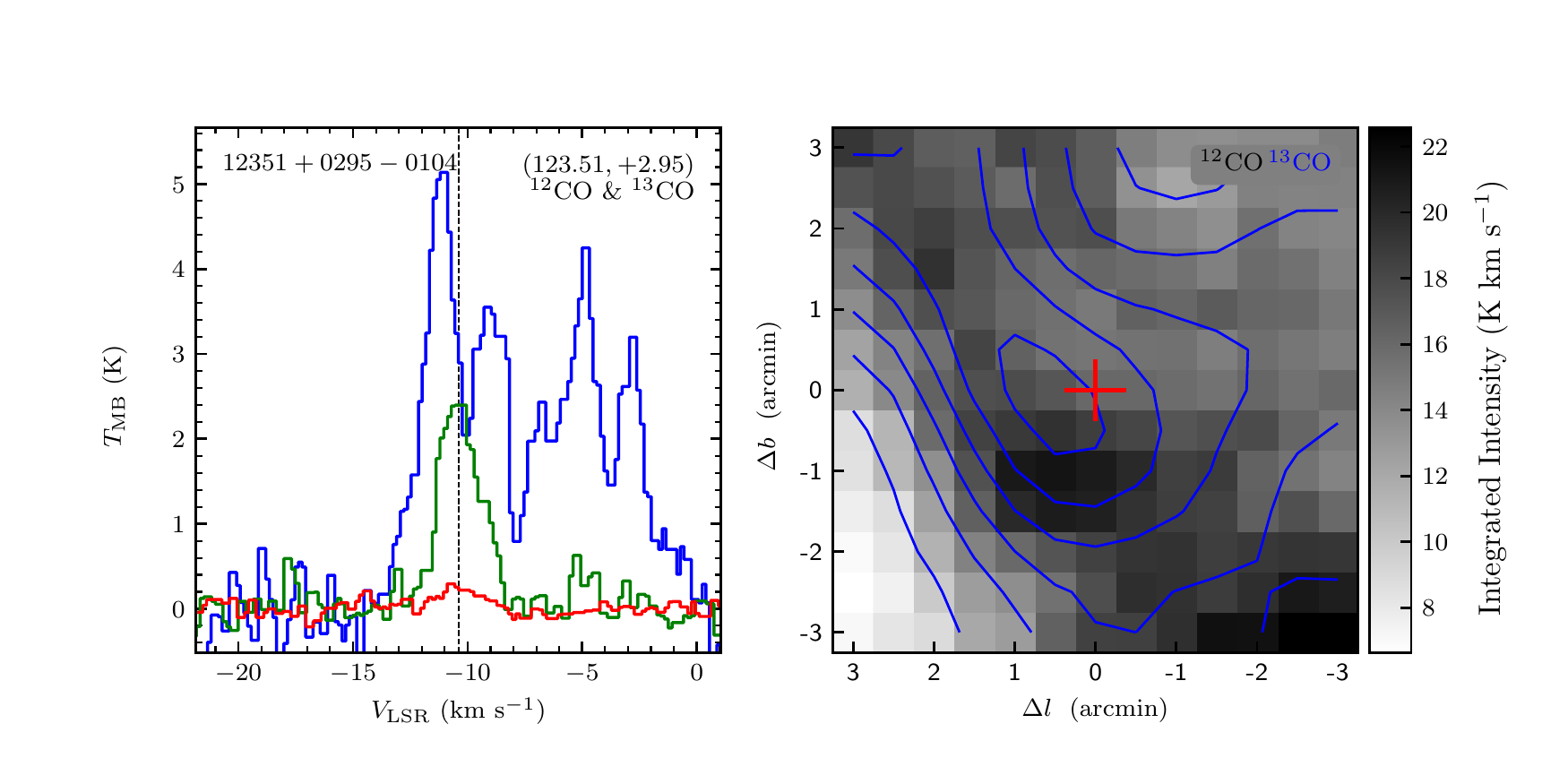}
\includegraphics[width=9.0cm,angle=0]{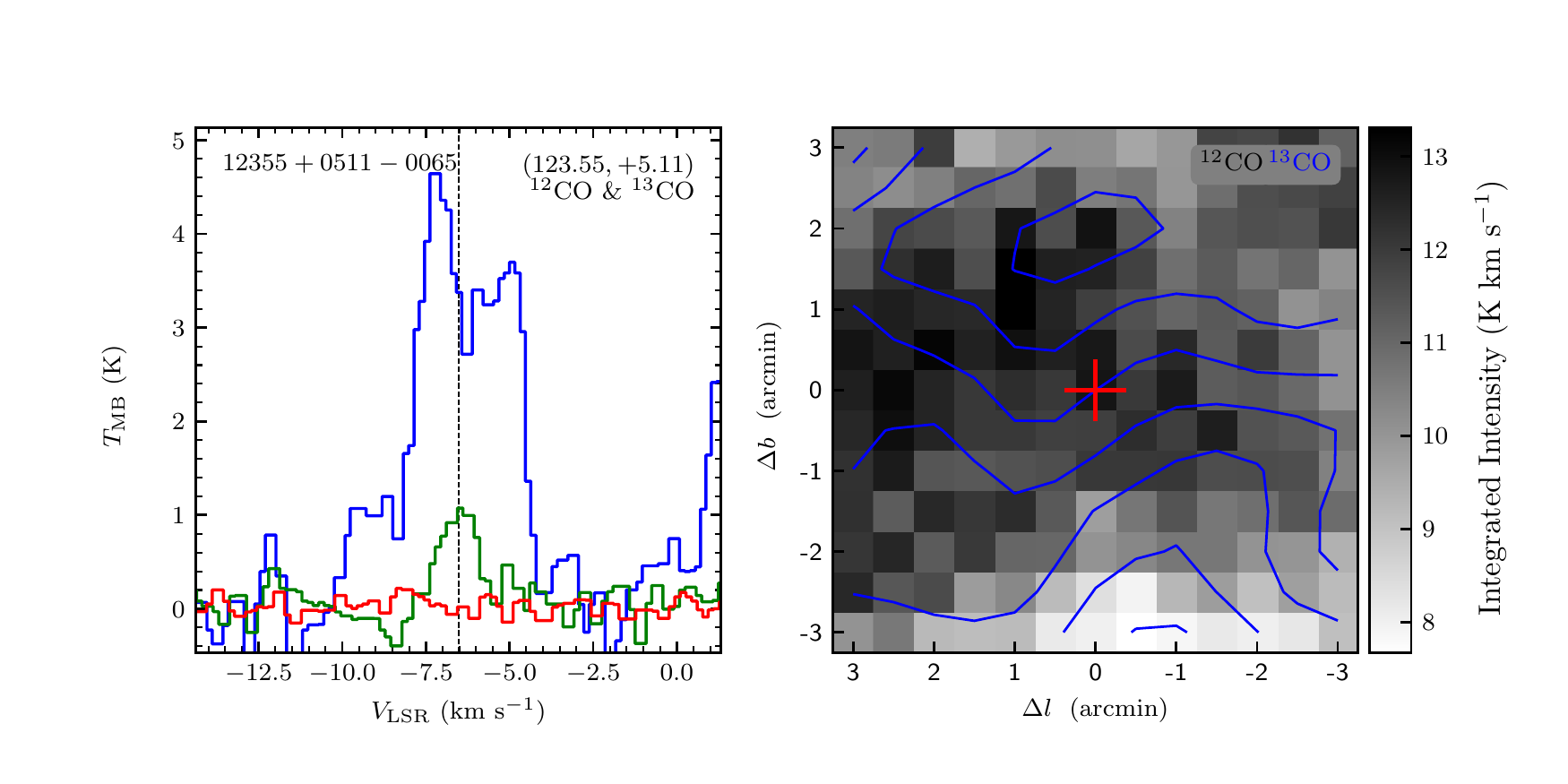}
\vspace{-0.5cm}

\includegraphics[width=9.0cm,angle=0]{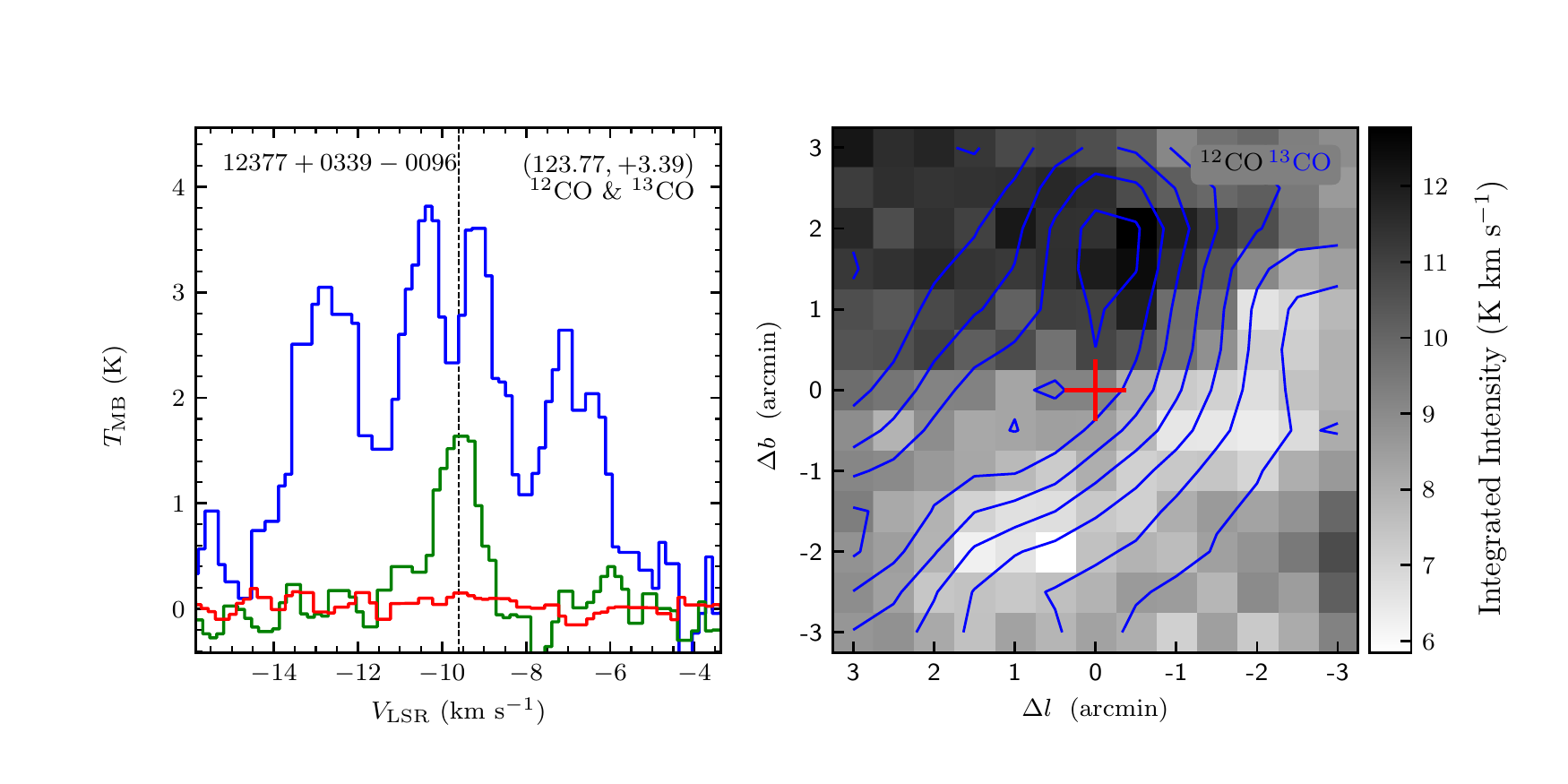}
\includegraphics[width=9.0cm,angle=0]{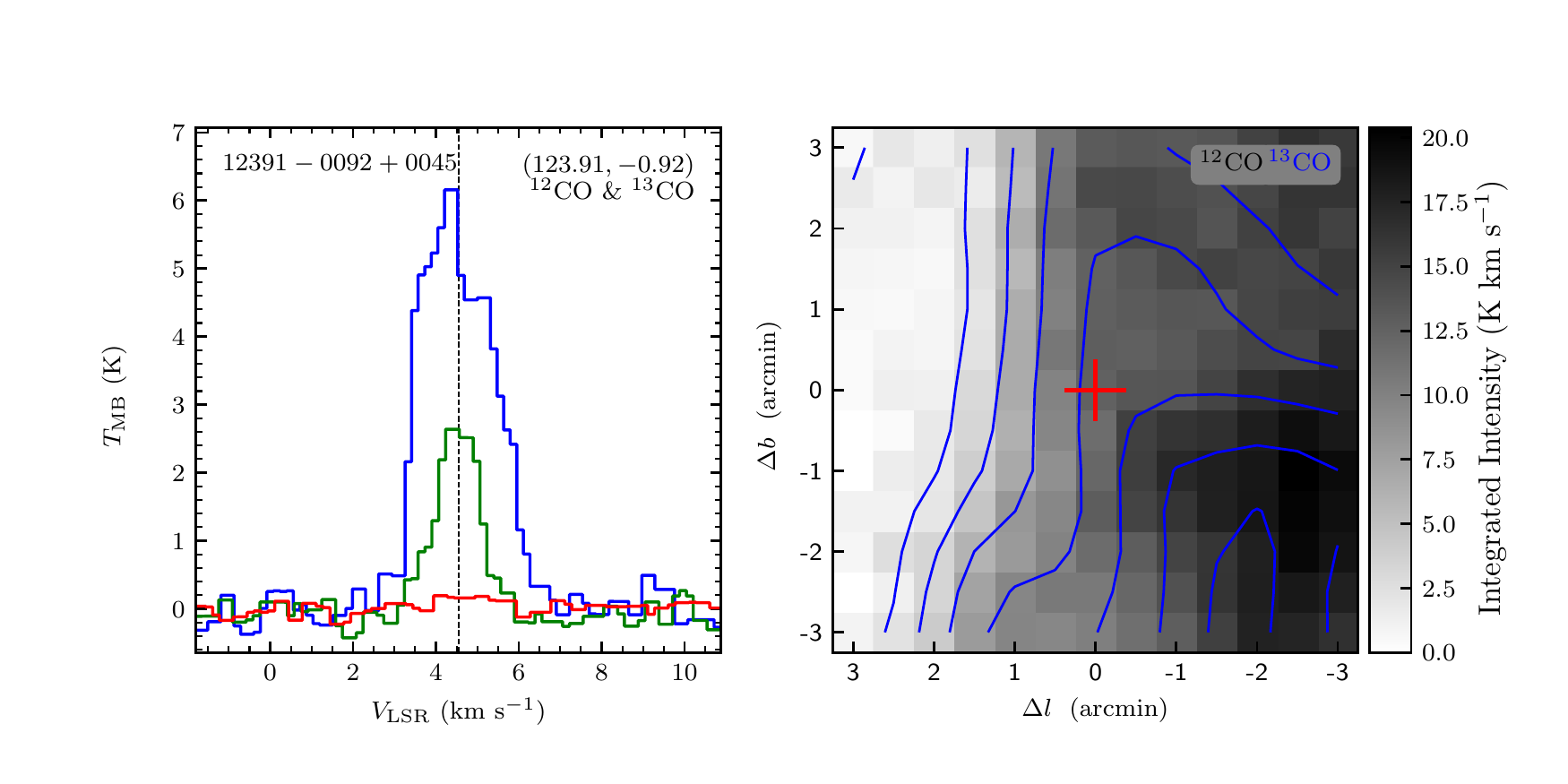}
\vspace{-0.5cm}

\includegraphics[width=9.0cm,angle=0]{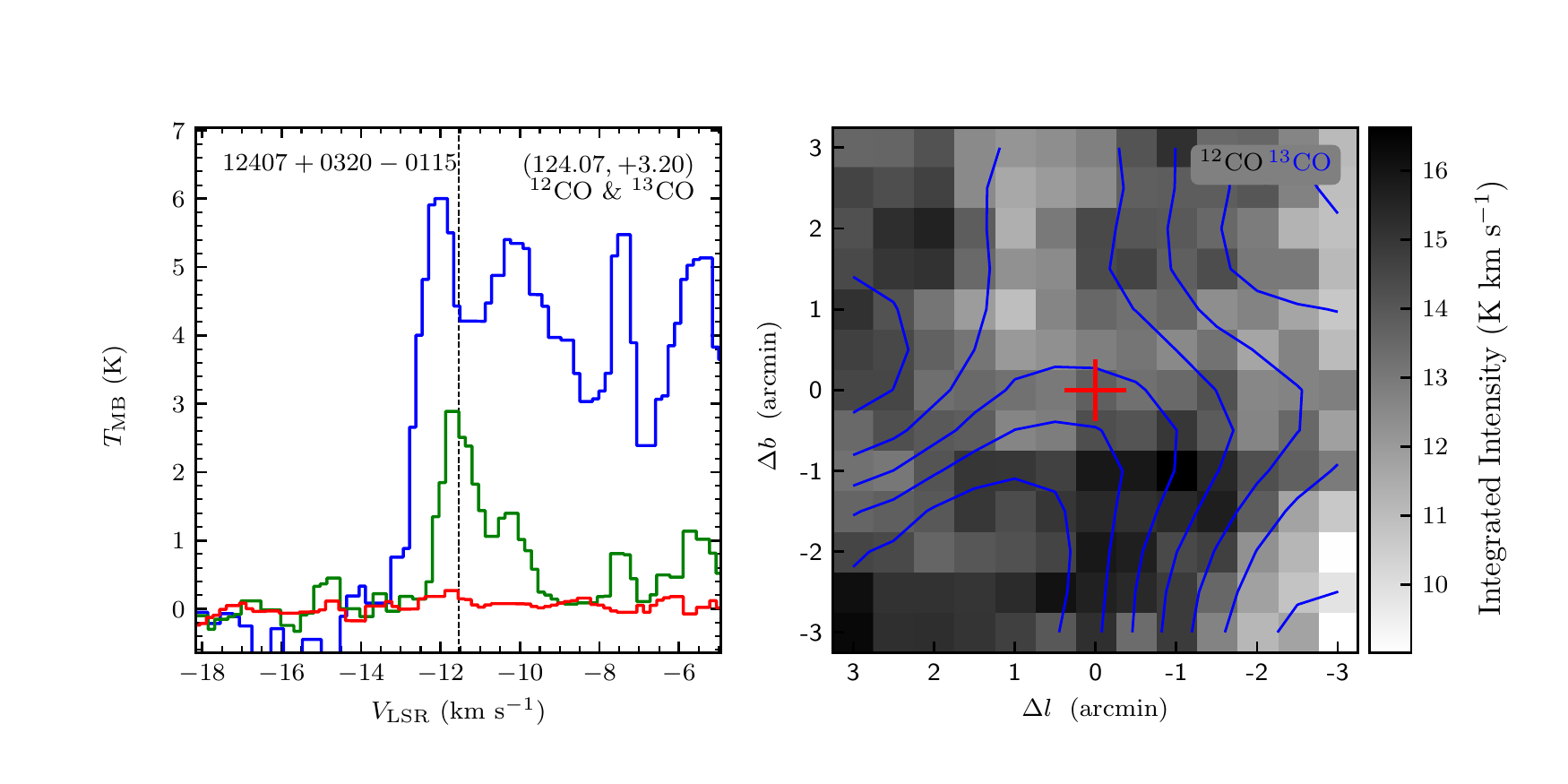}
\includegraphics[width=9.0cm,angle=0]{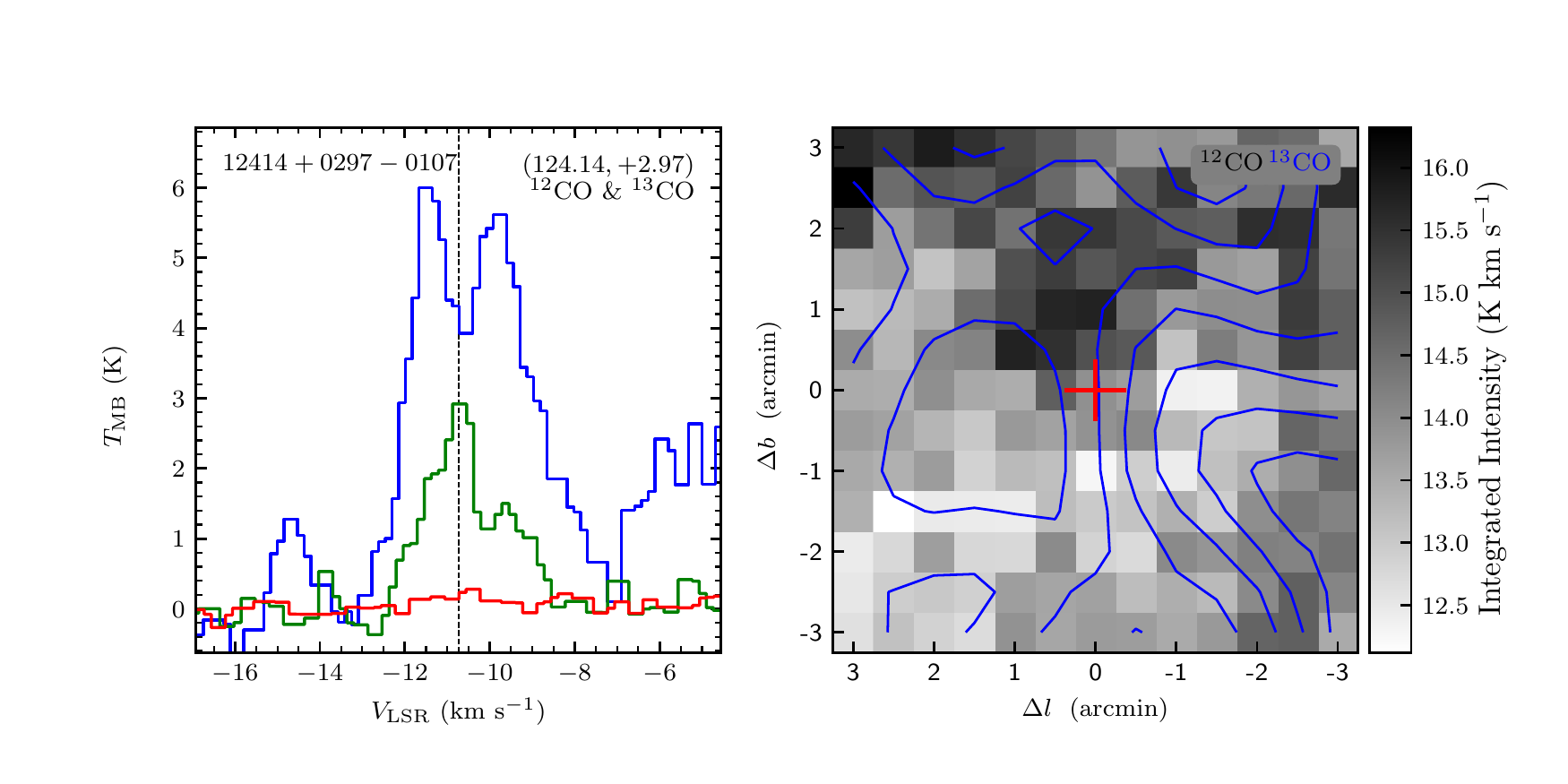}
\vspace{-0.5cm}

\includegraphics[width=9.0cm,angle=0]{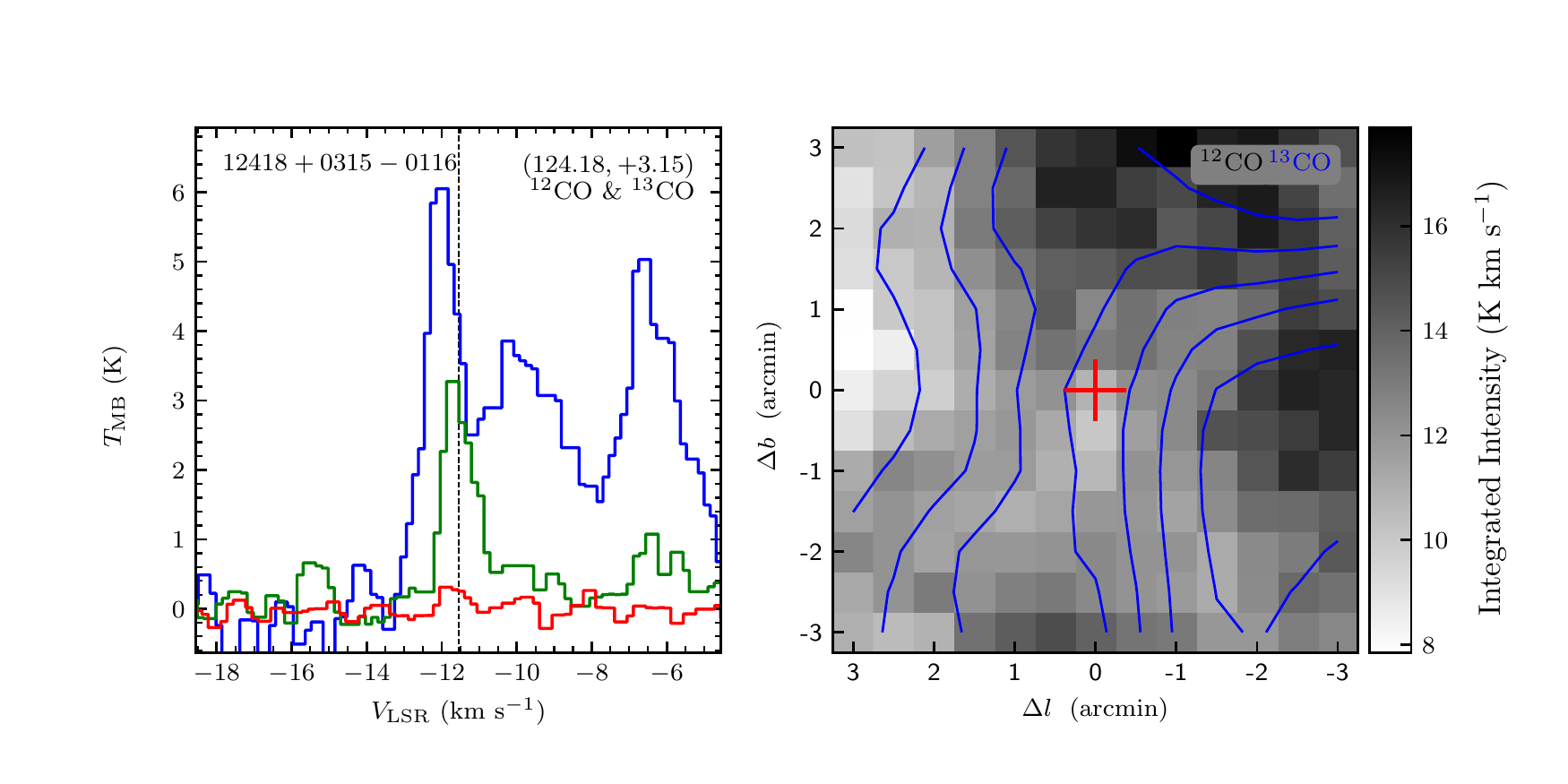}
\includegraphics[width=9.0cm,angle=0]{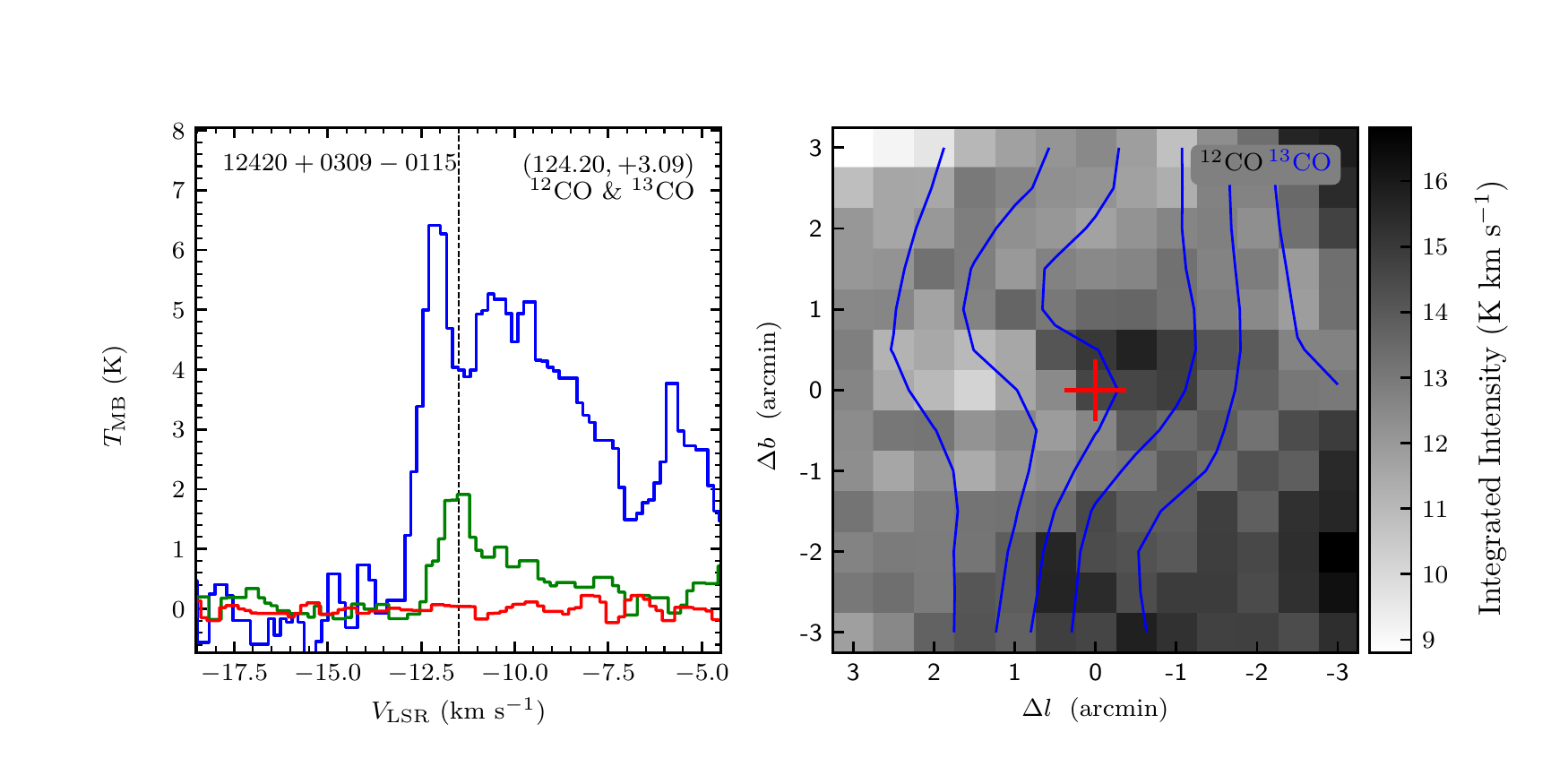}
\end{figure}
\clearpage

\begin{figure}
\includegraphics[width=9.0cm,angle=0]{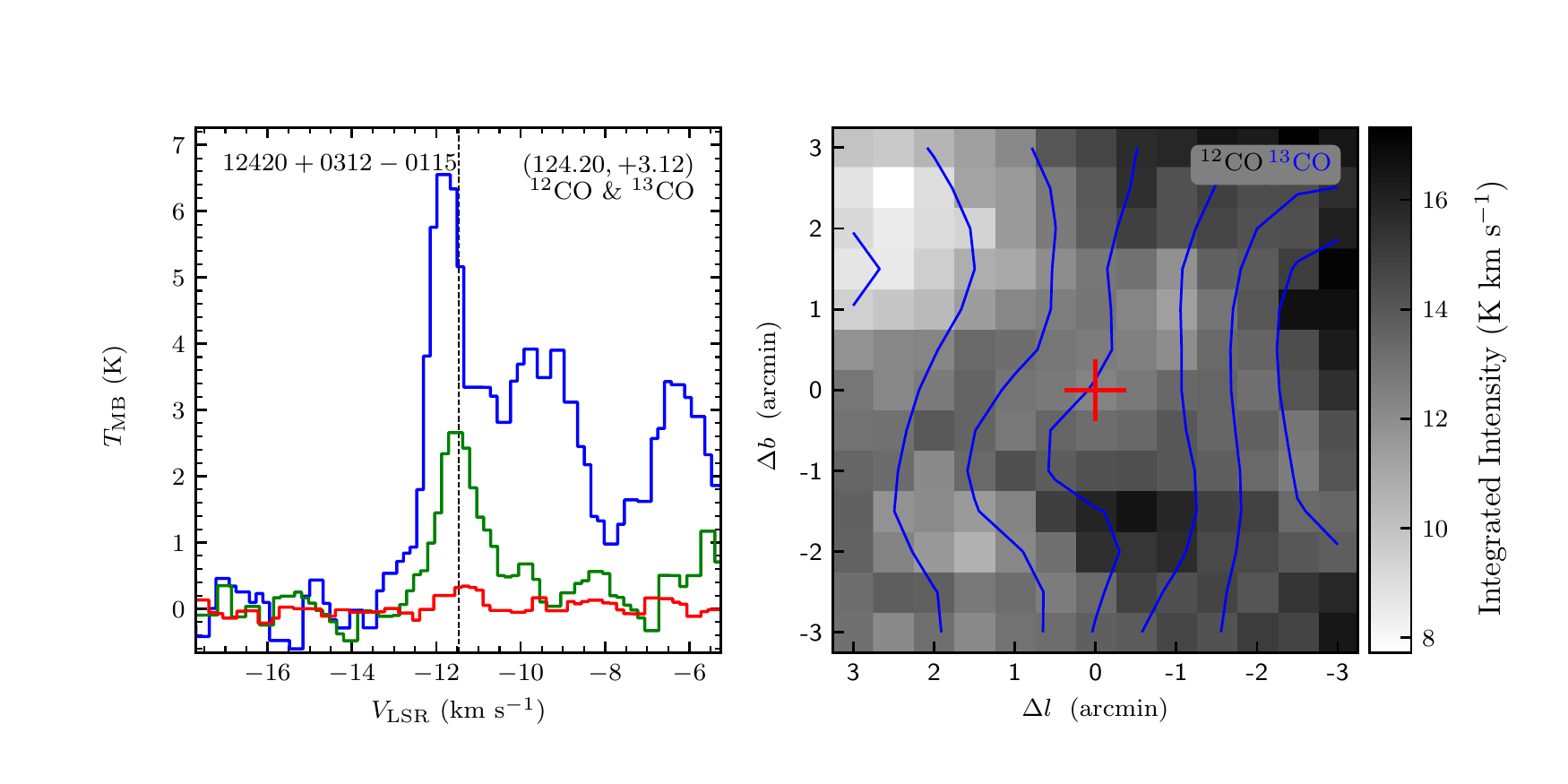}
\includegraphics[width=9.0cm,angle=0]{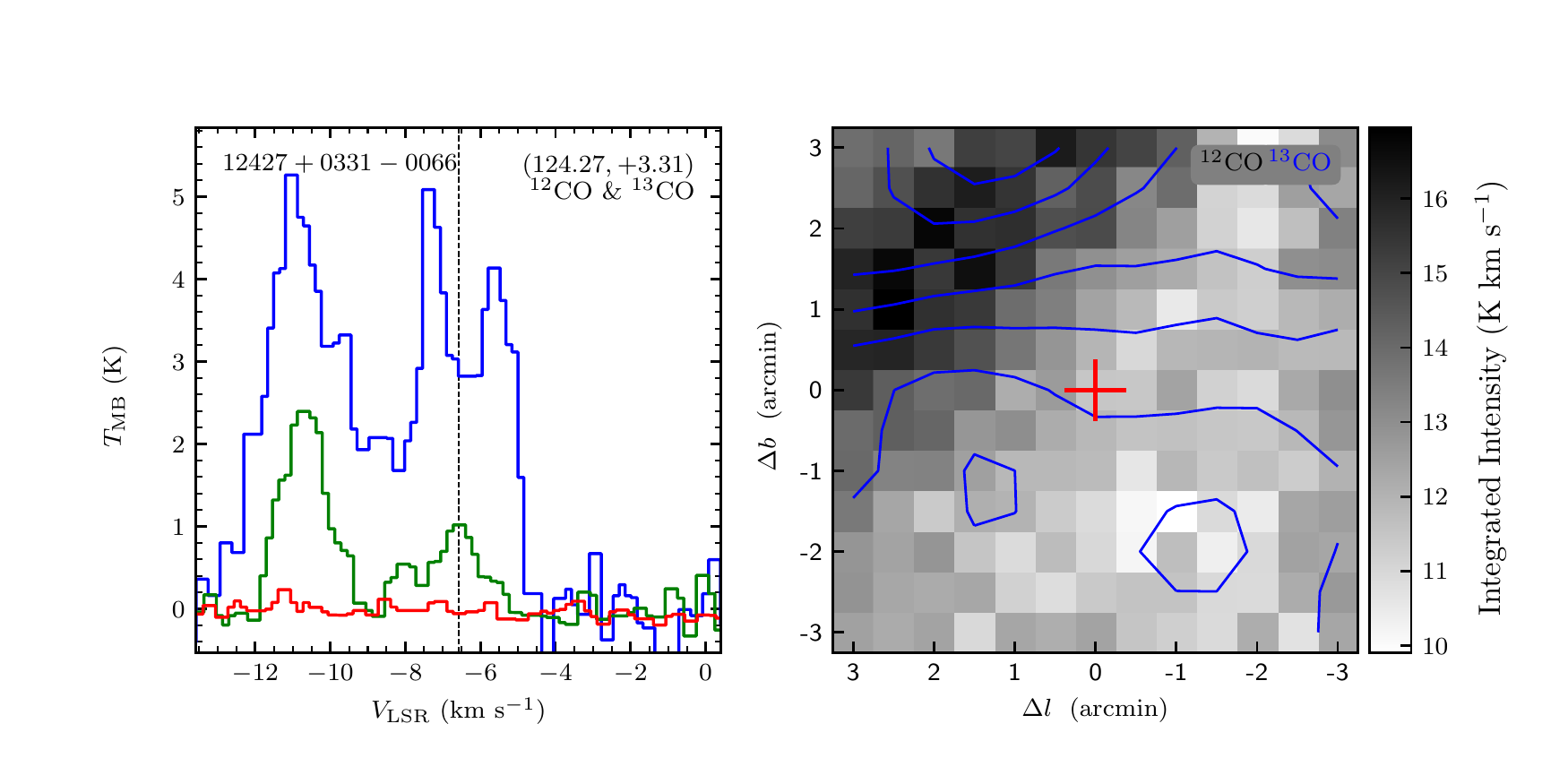}
\vspace{-0.5cm}

\includegraphics[width=9.0cm,angle=0]{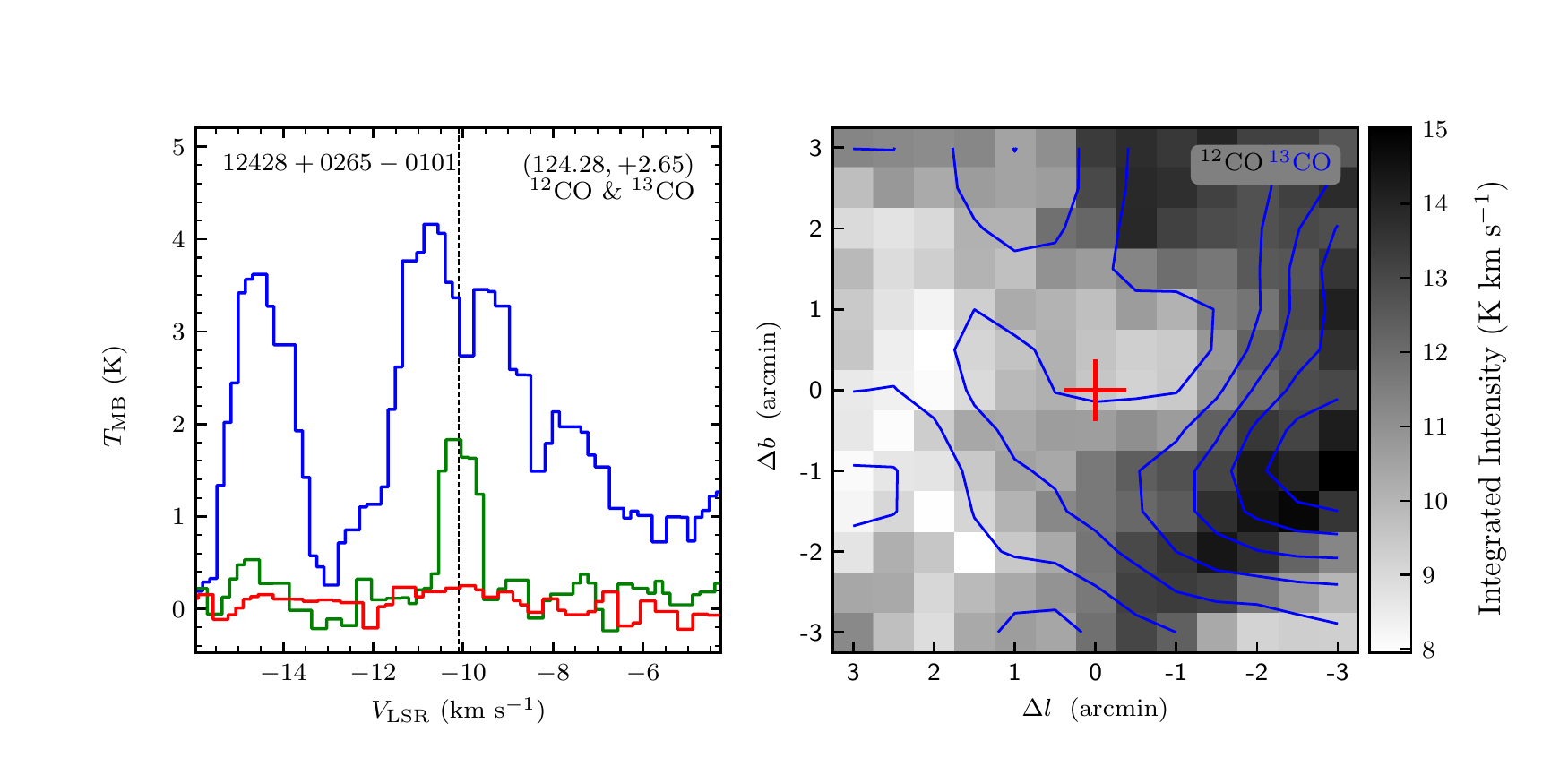}
\includegraphics[width=9.0cm,angle=0]{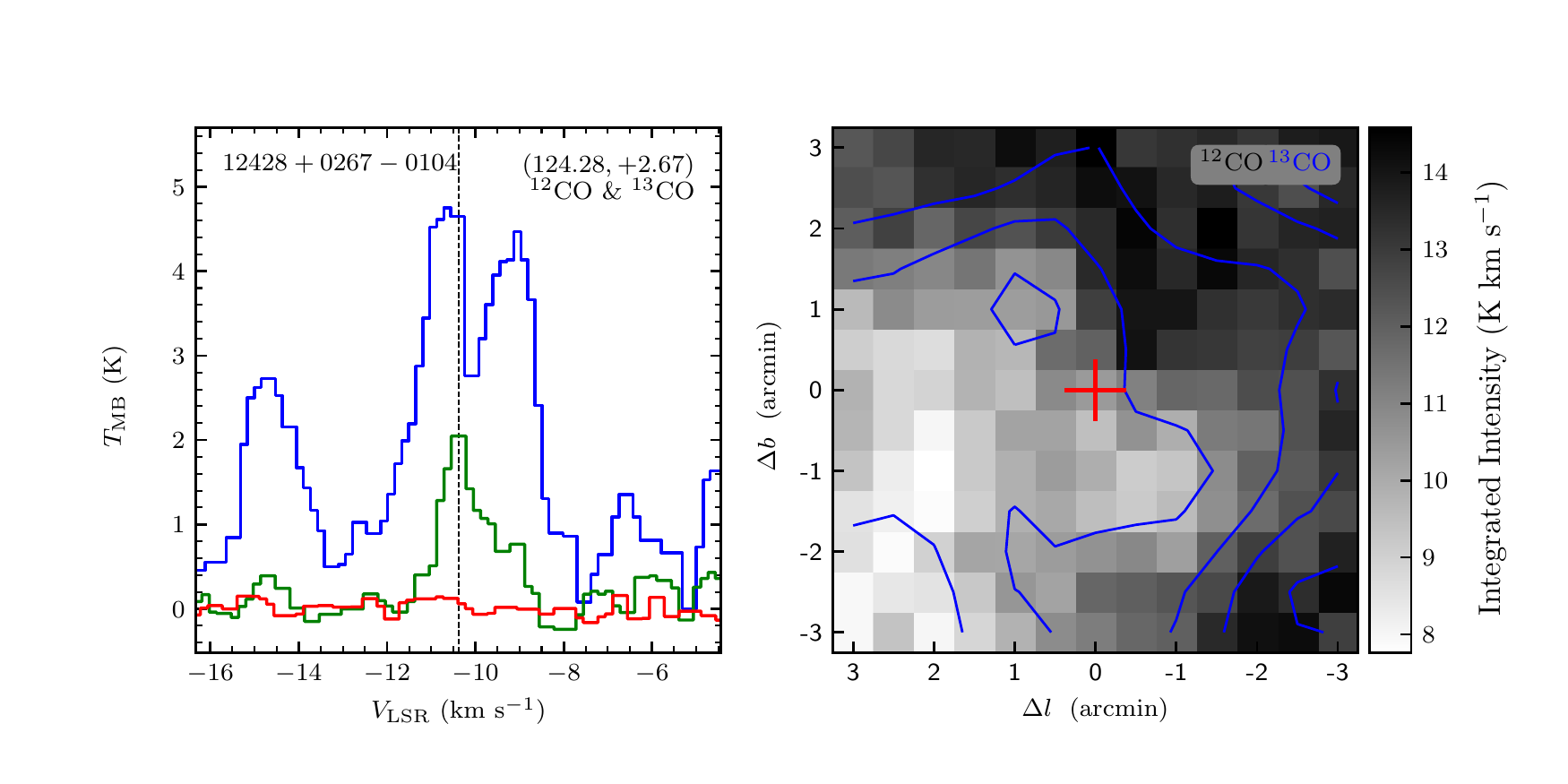}
\vspace{-0.5cm}

\includegraphics[width=9.0cm,angle=0]{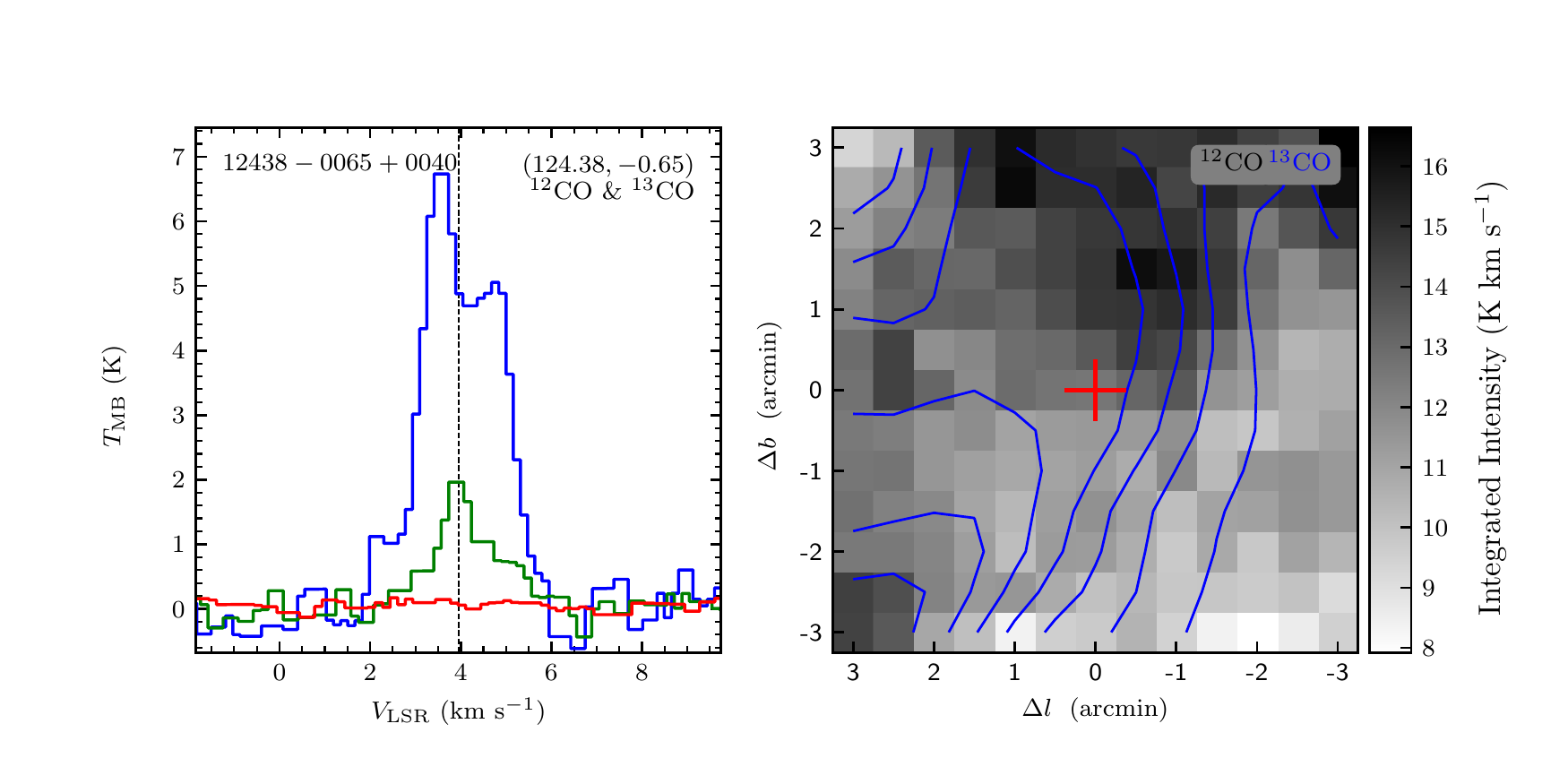}
\includegraphics[width=9.0cm,angle=0]{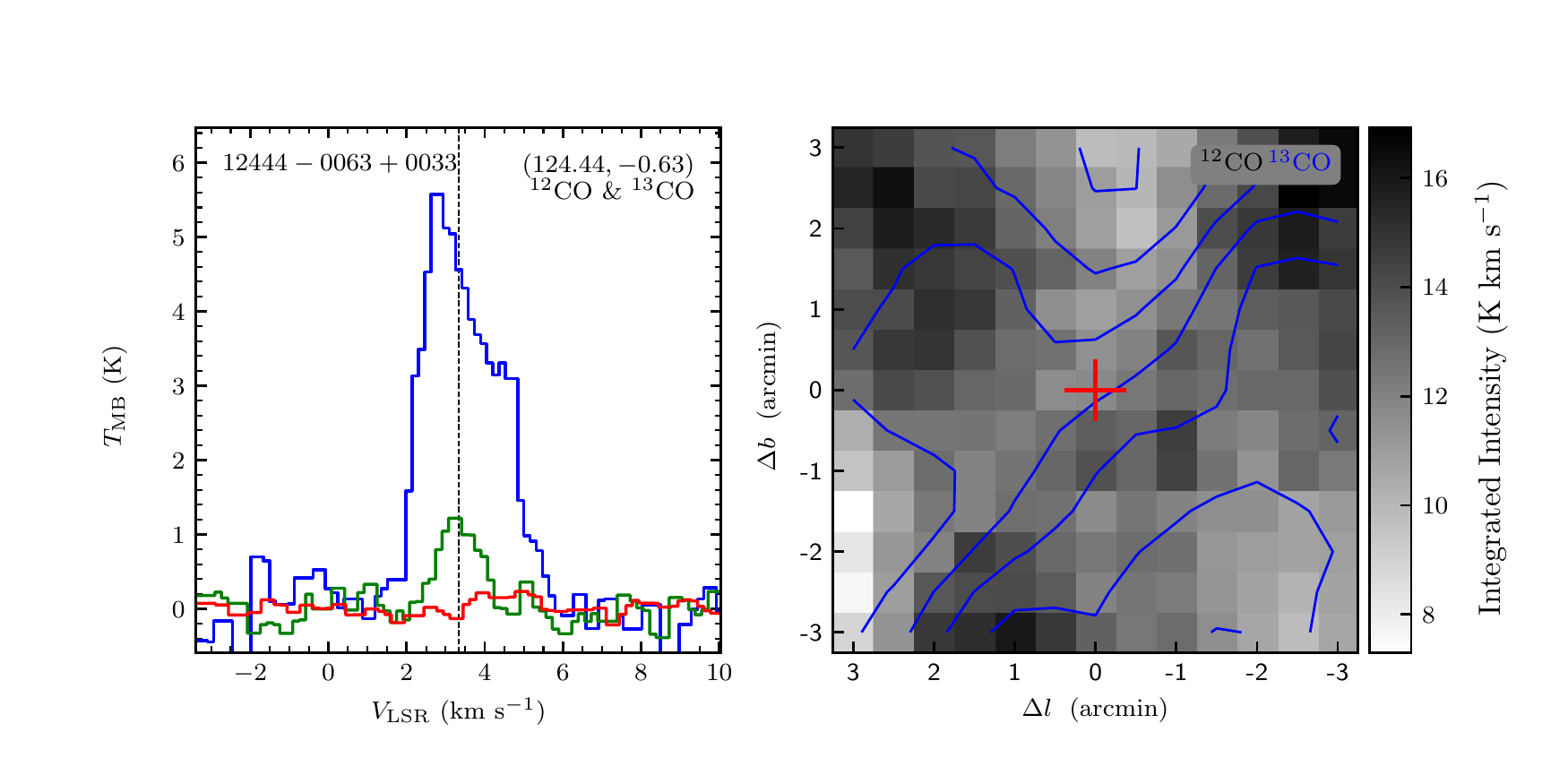}
\vspace{-0.5cm}

\includegraphics[width=9.0cm,angle=0]{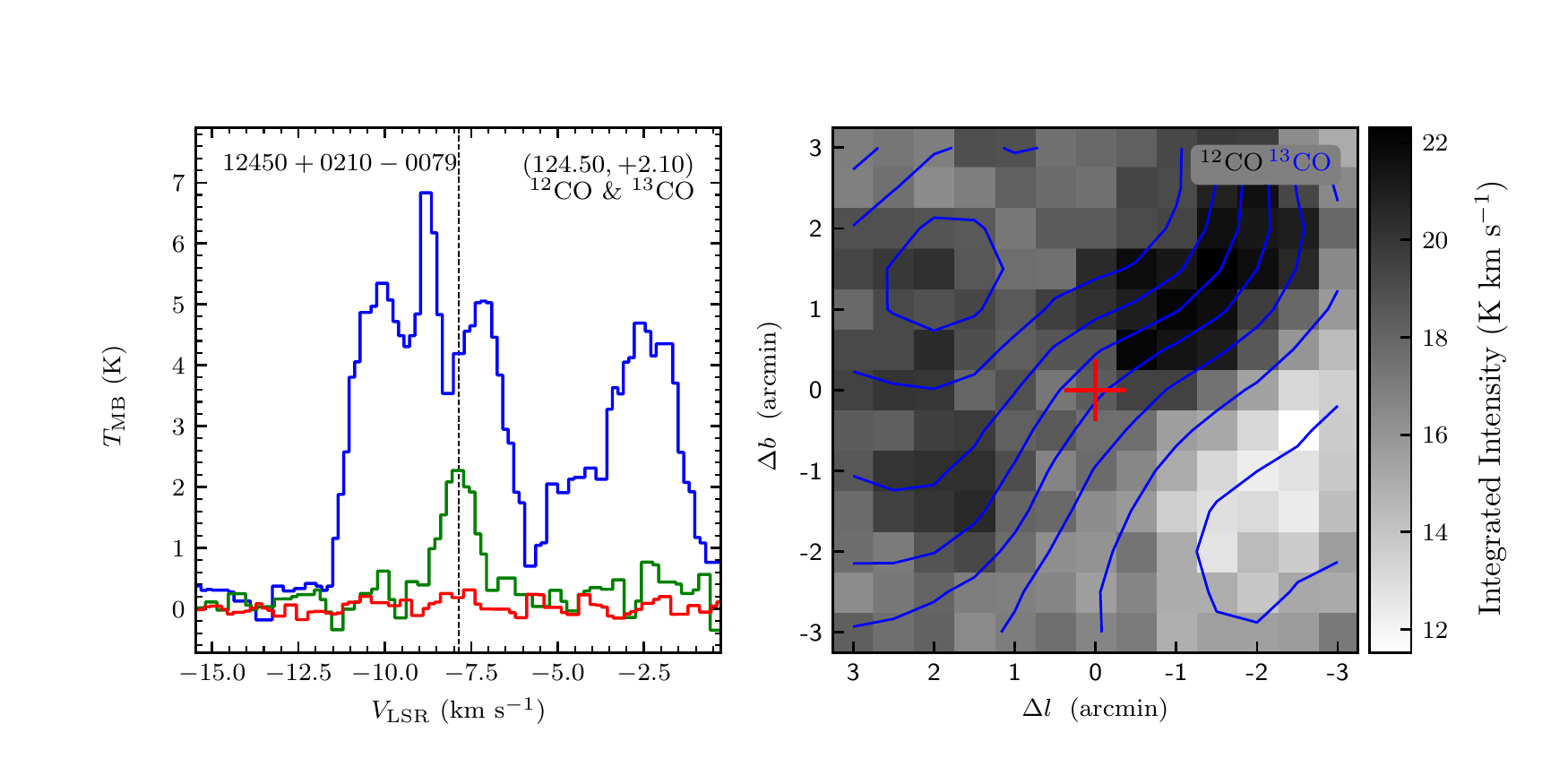}
\includegraphics[width=9.0cm,angle=0]{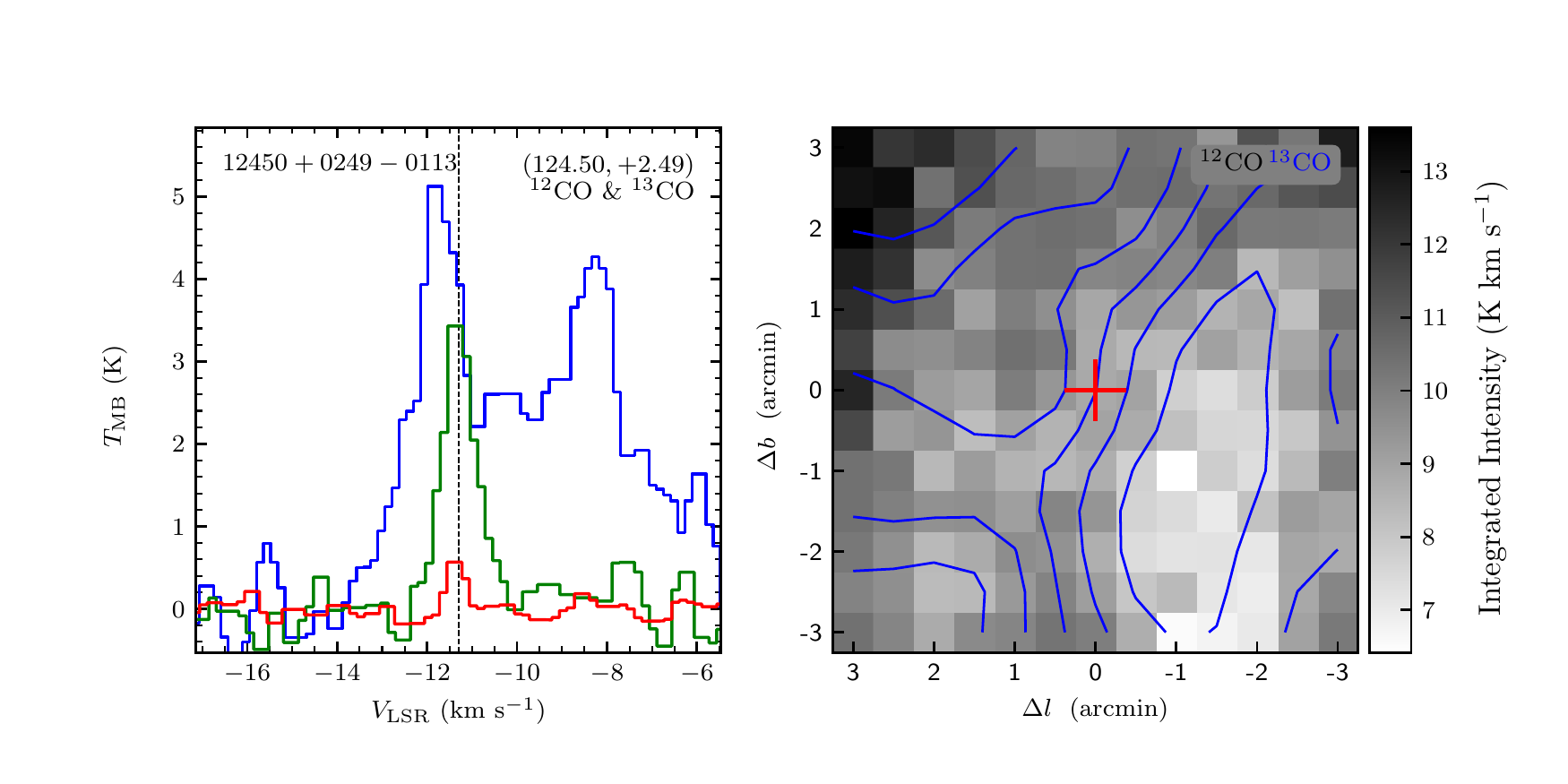}
\vspace{-0.5cm}

\includegraphics[width=9.0cm,angle=0]{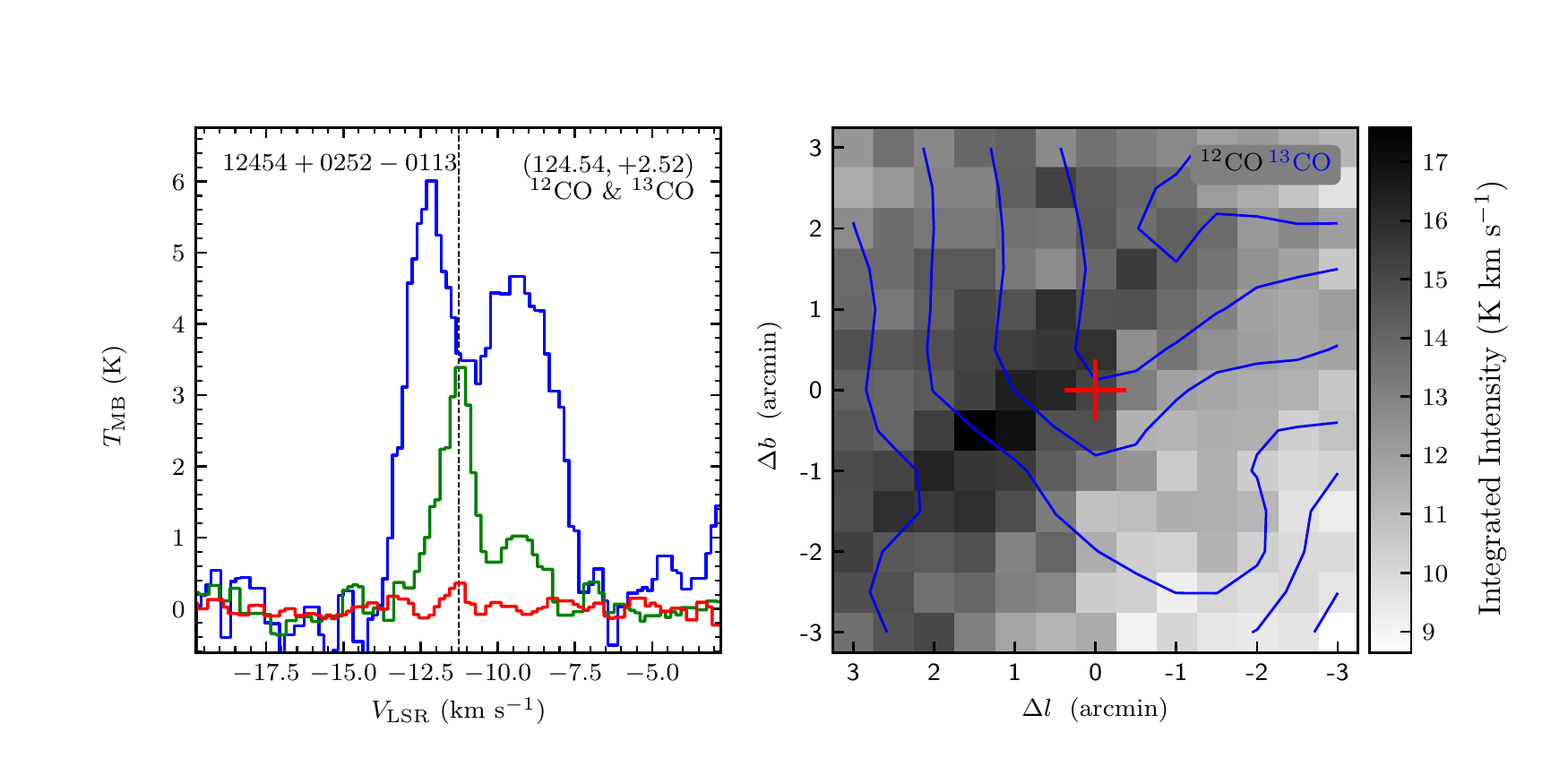}
\includegraphics[width=9.0cm,angle=0]{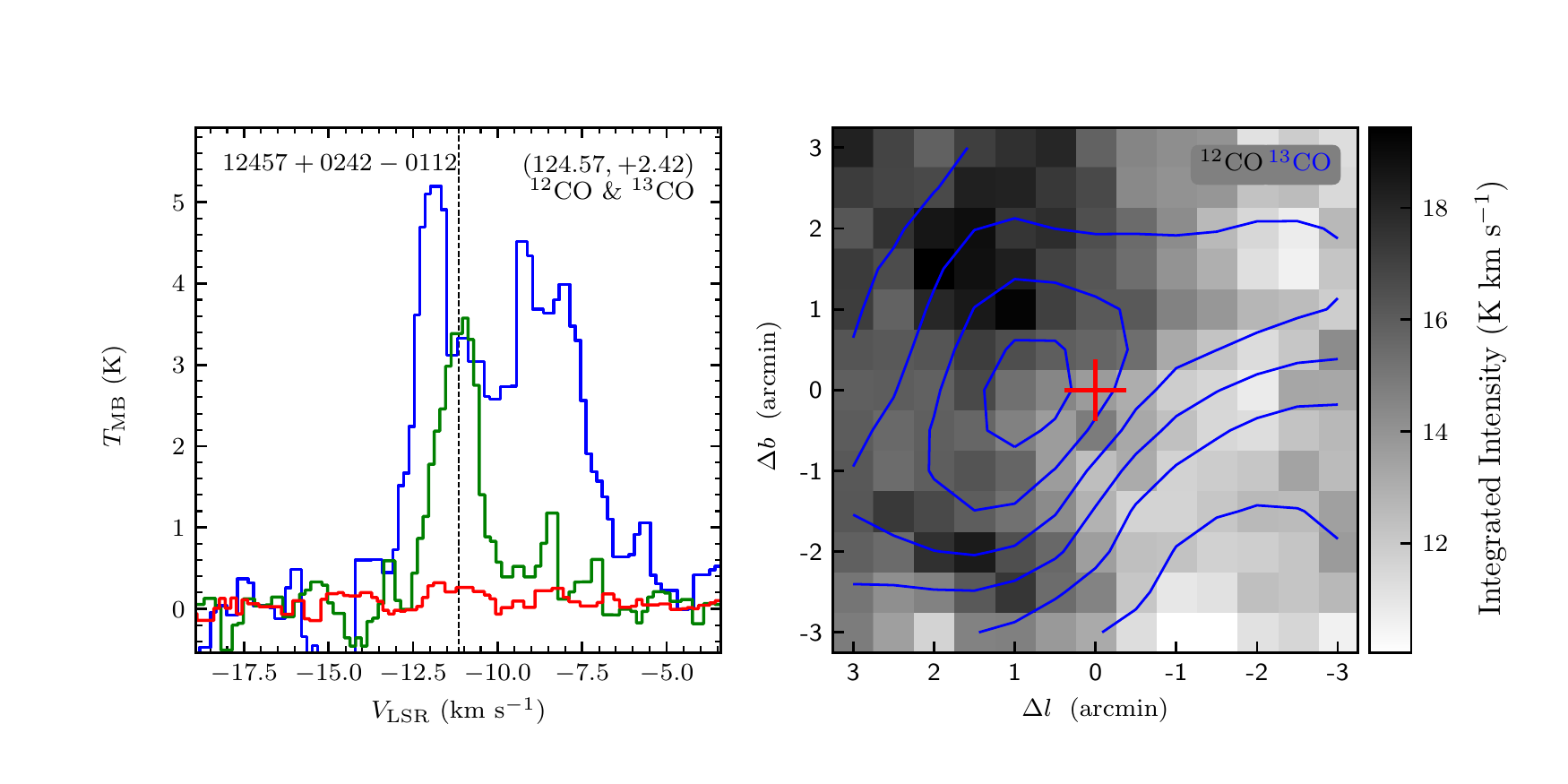}
\end{figure}
\clearpage

\begin{figure}
\includegraphics[width=9.0cm,angle=0]{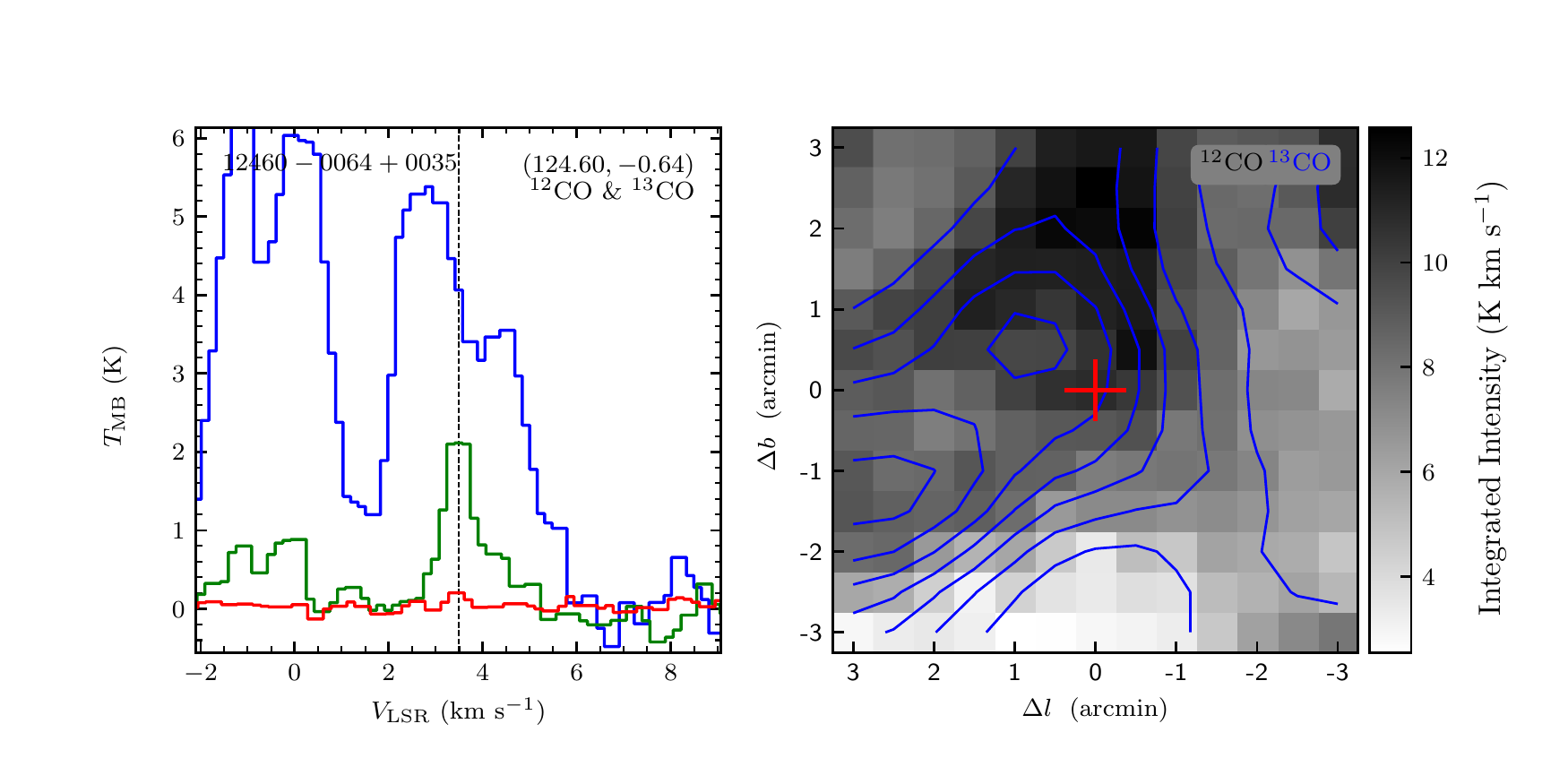}
\includegraphics[width=9.0cm,angle=0]{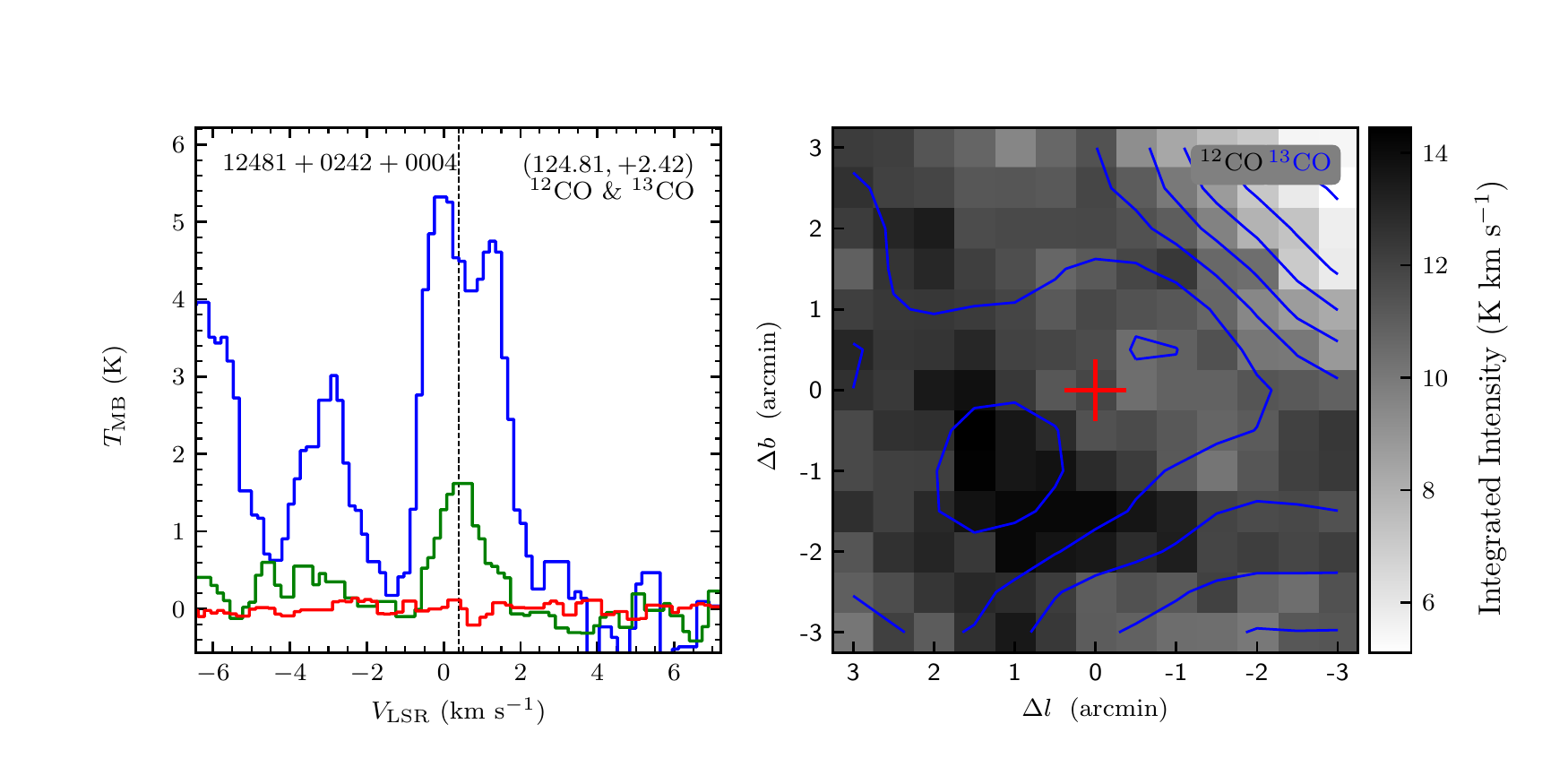}
\vspace{-0.5cm}

\includegraphics[width=9.0cm,angle=0]{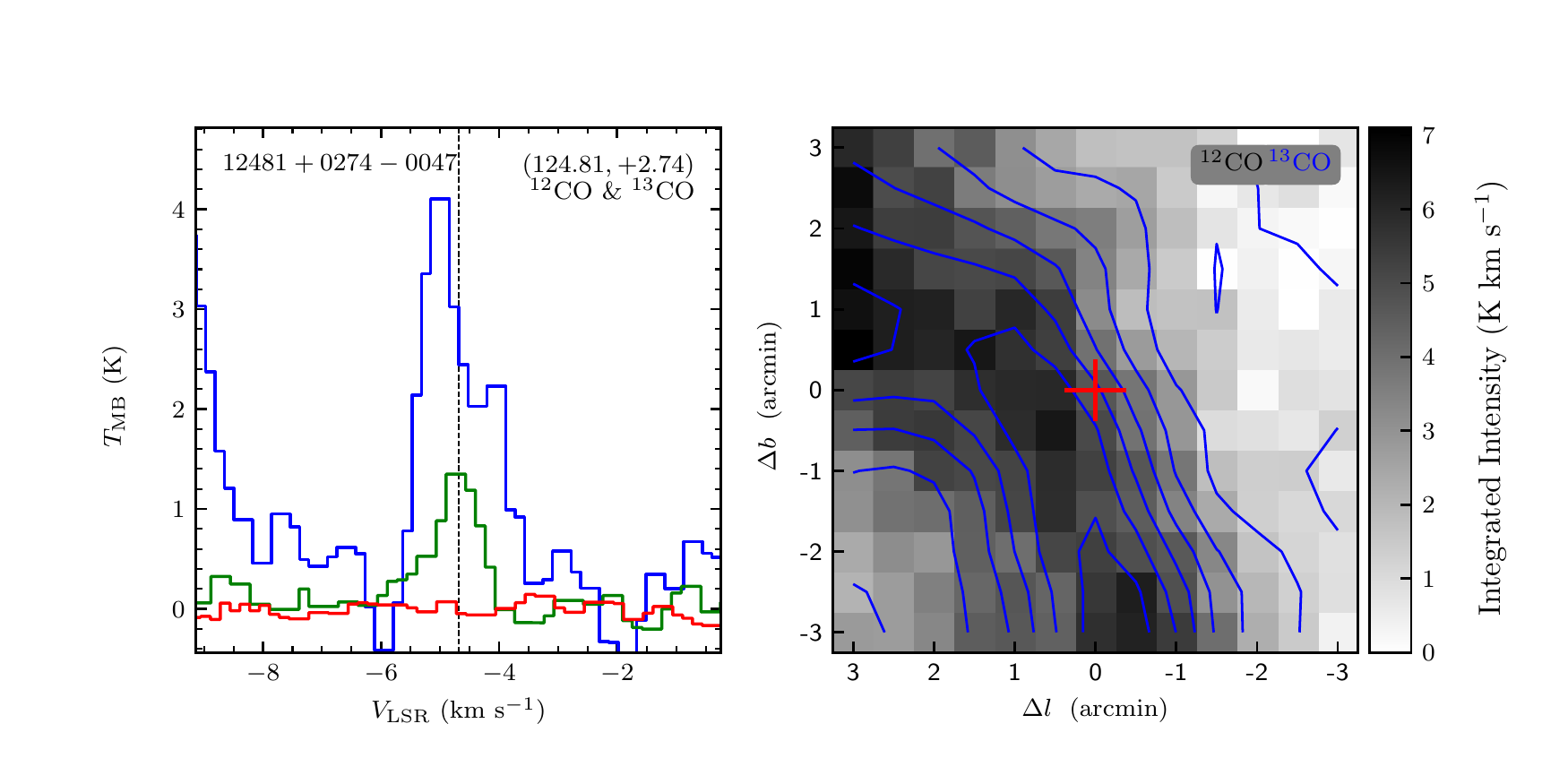}
\includegraphics[width=9.0cm,angle=0]{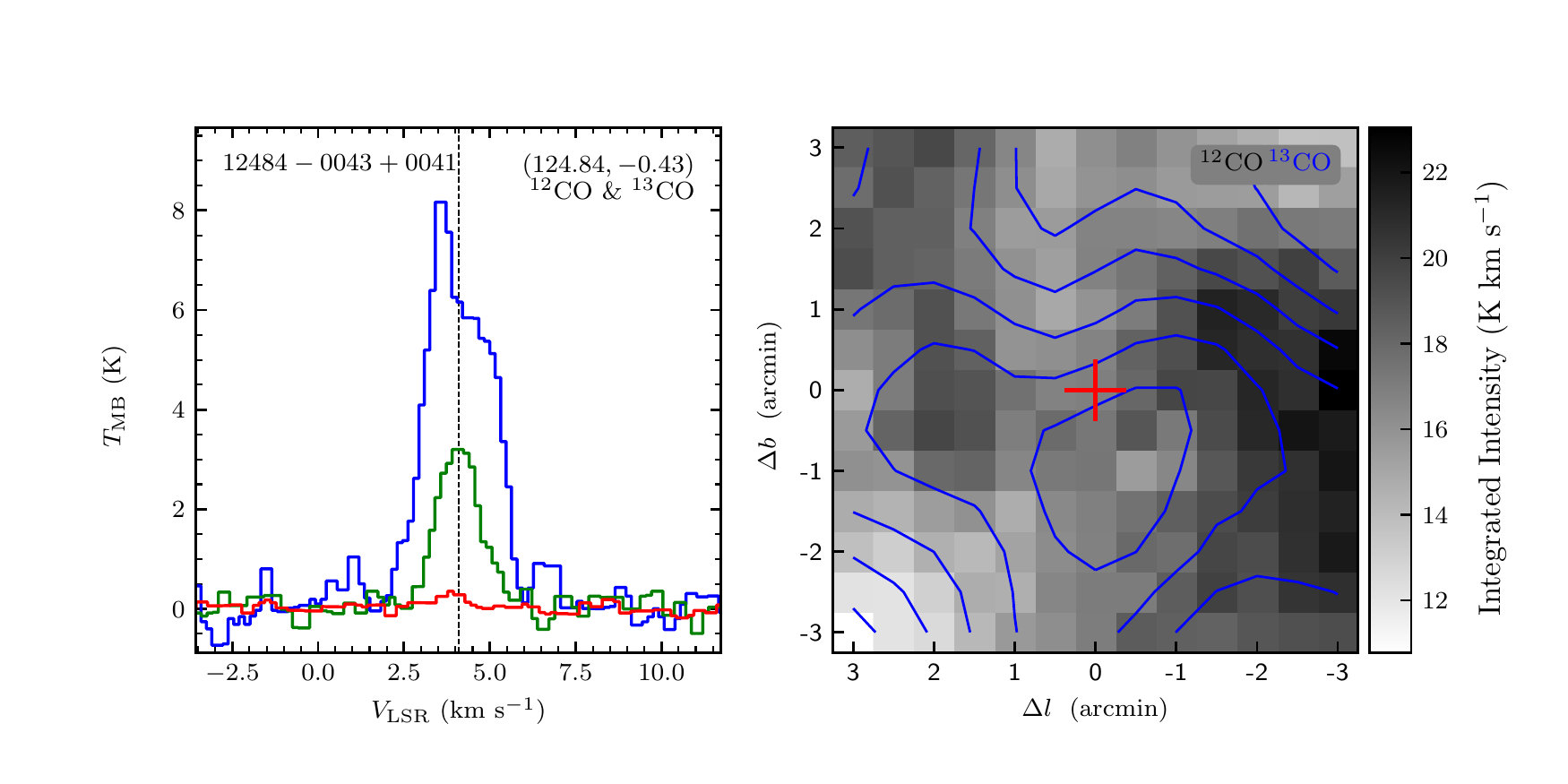}
\vspace{-0.5cm}

\includegraphics[width=9.0cm,angle=0]{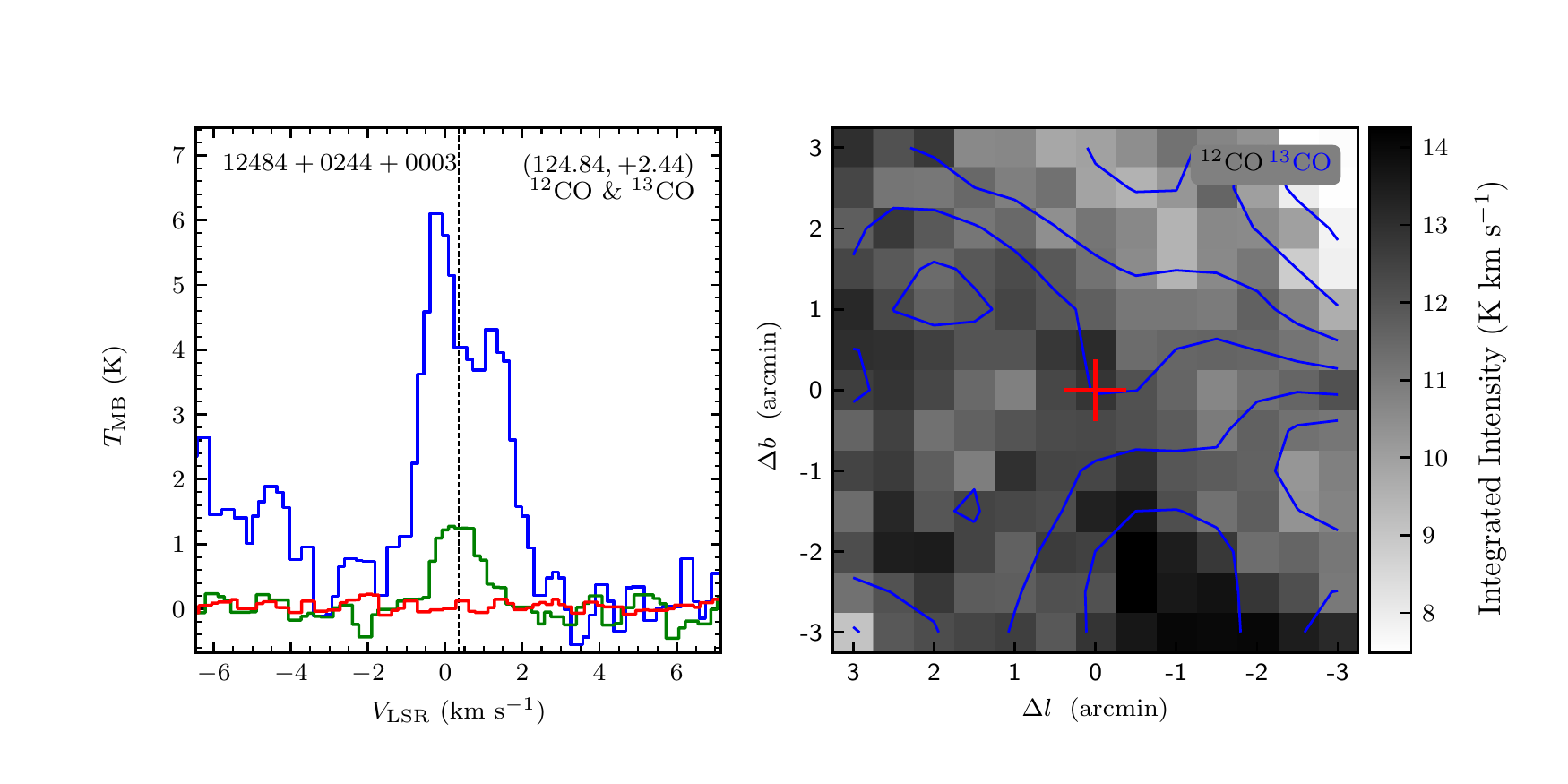}
\includegraphics[width=9.0cm,angle=0]{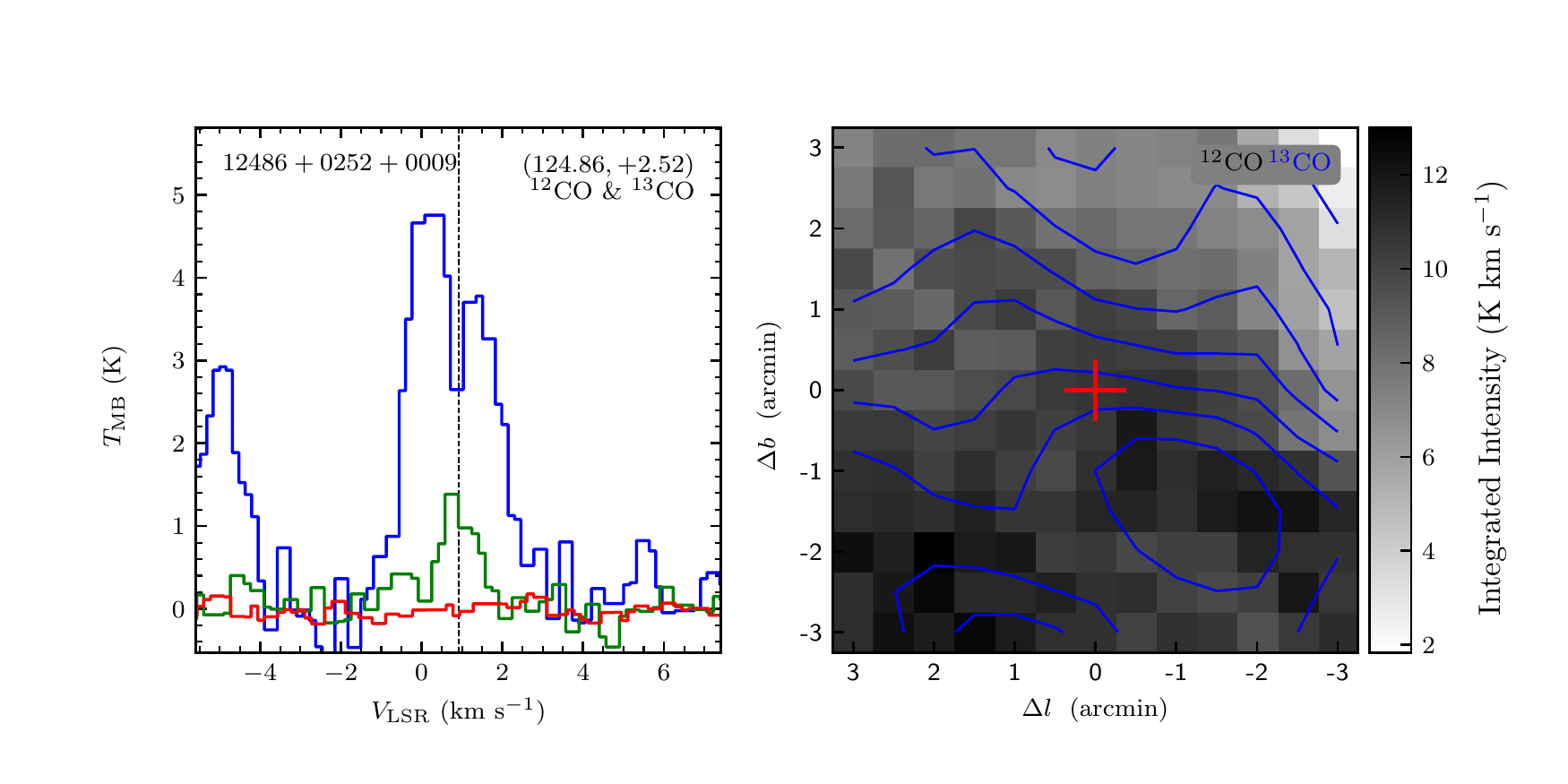}
\vspace{-0.5cm}

\includegraphics[width=9.0cm,angle=0]{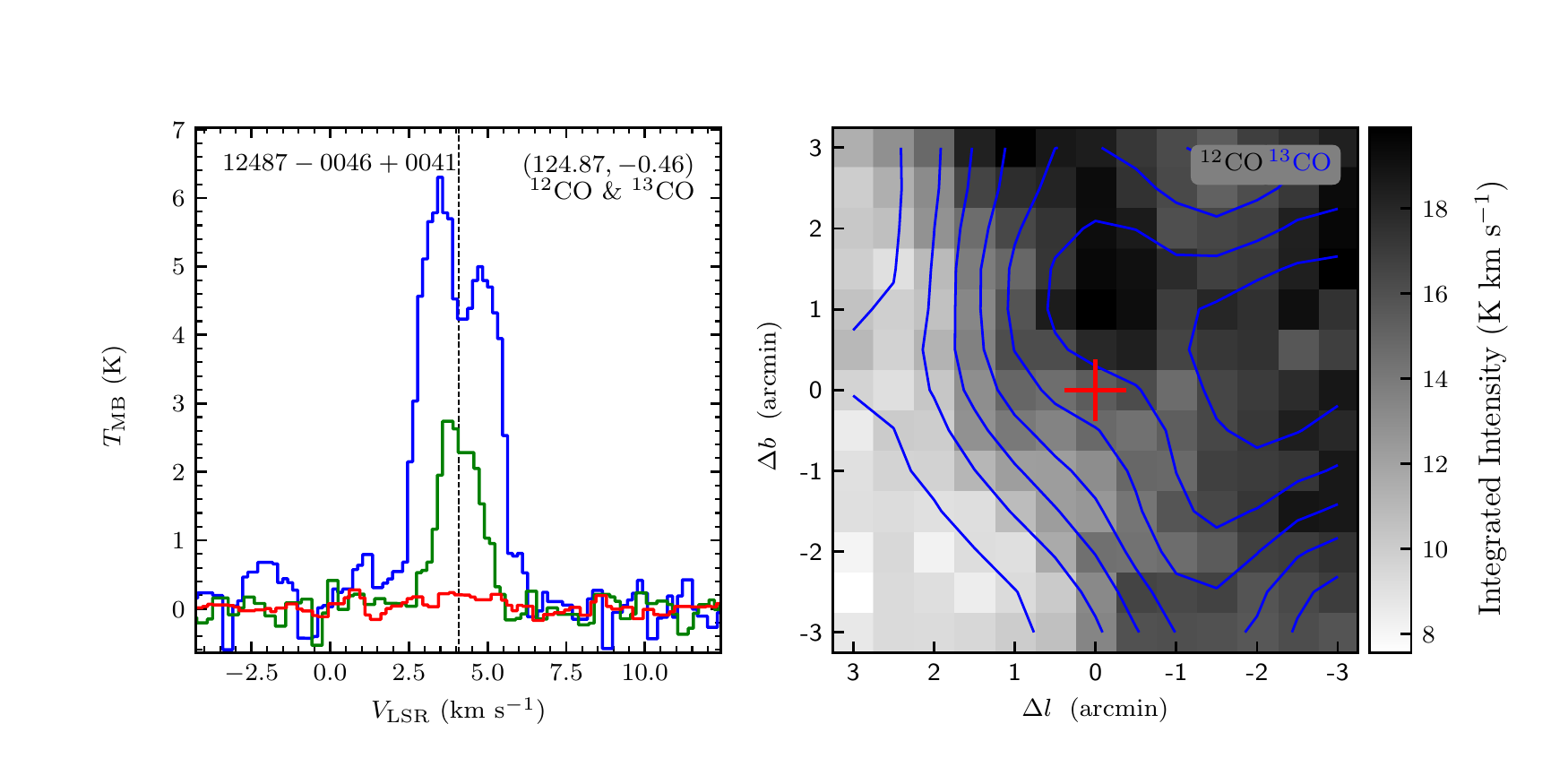}
\includegraphics[width=9.0cm,angle=0]{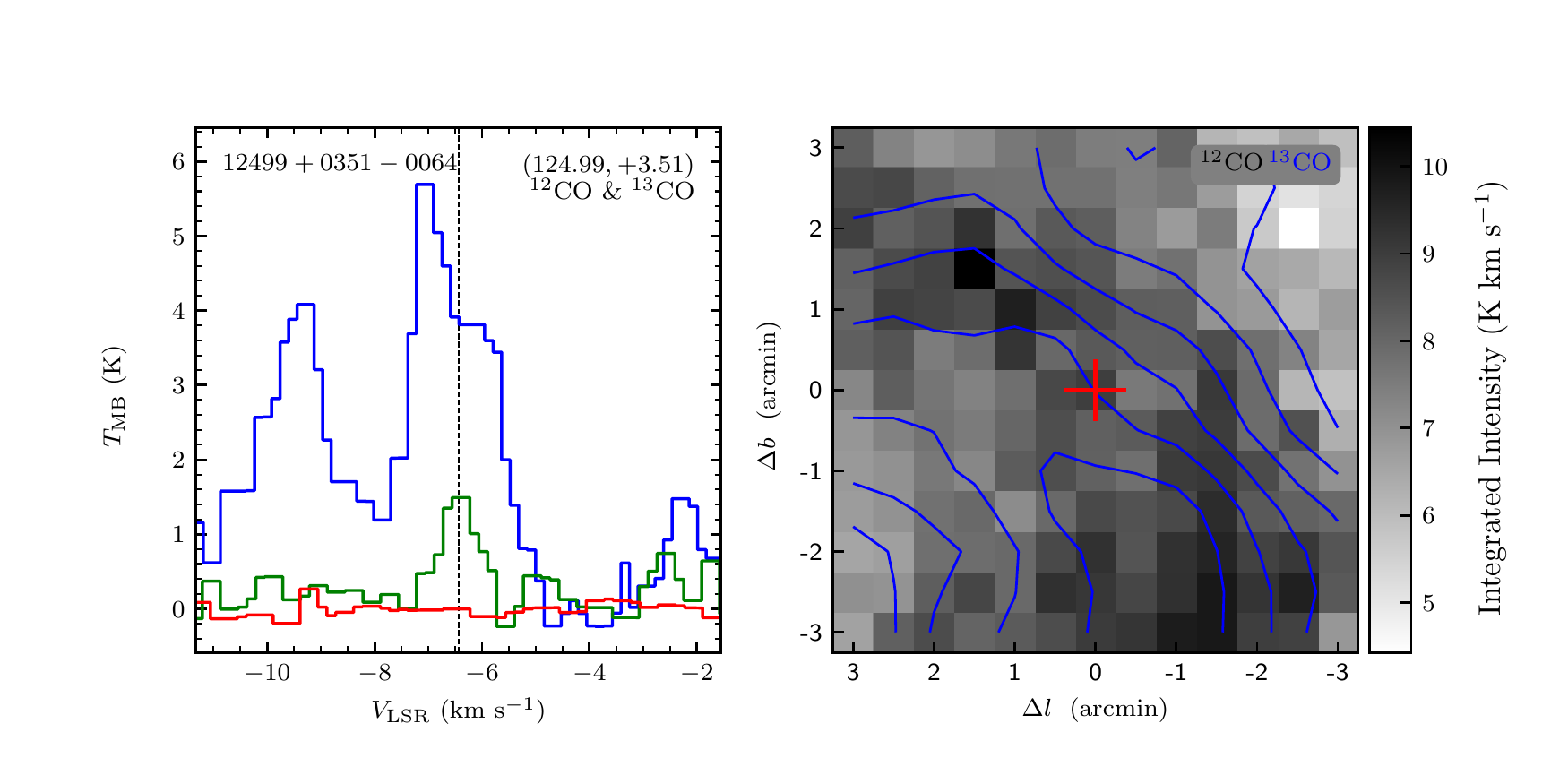}
\vspace{-0.5cm}

\includegraphics[width=9.0cm,angle=0]{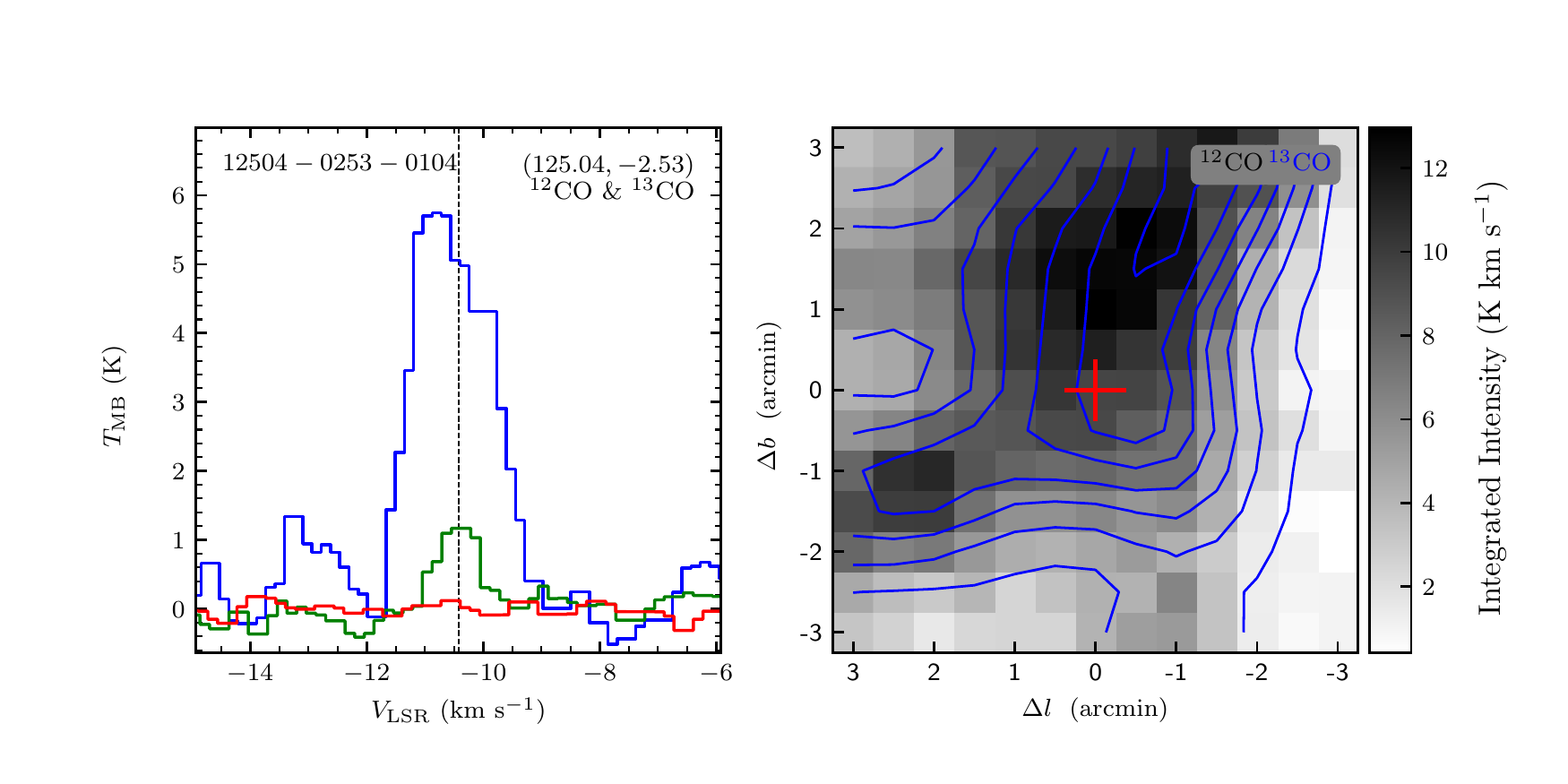}
\includegraphics[width=9.0cm,angle=0]{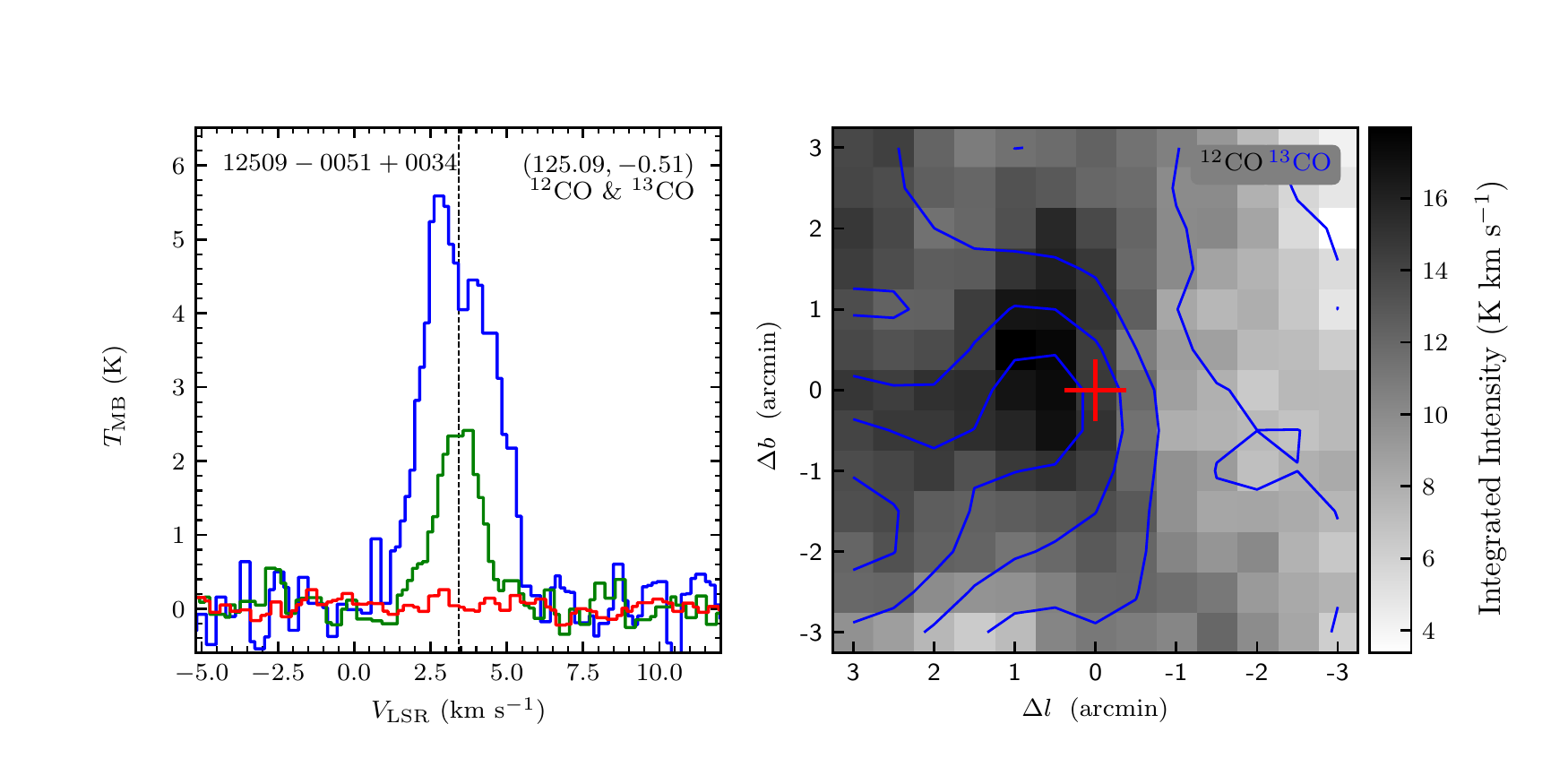}
\end{figure}
\clearpage

\begin{figure}
\includegraphics[width=9.0cm,angle=0]{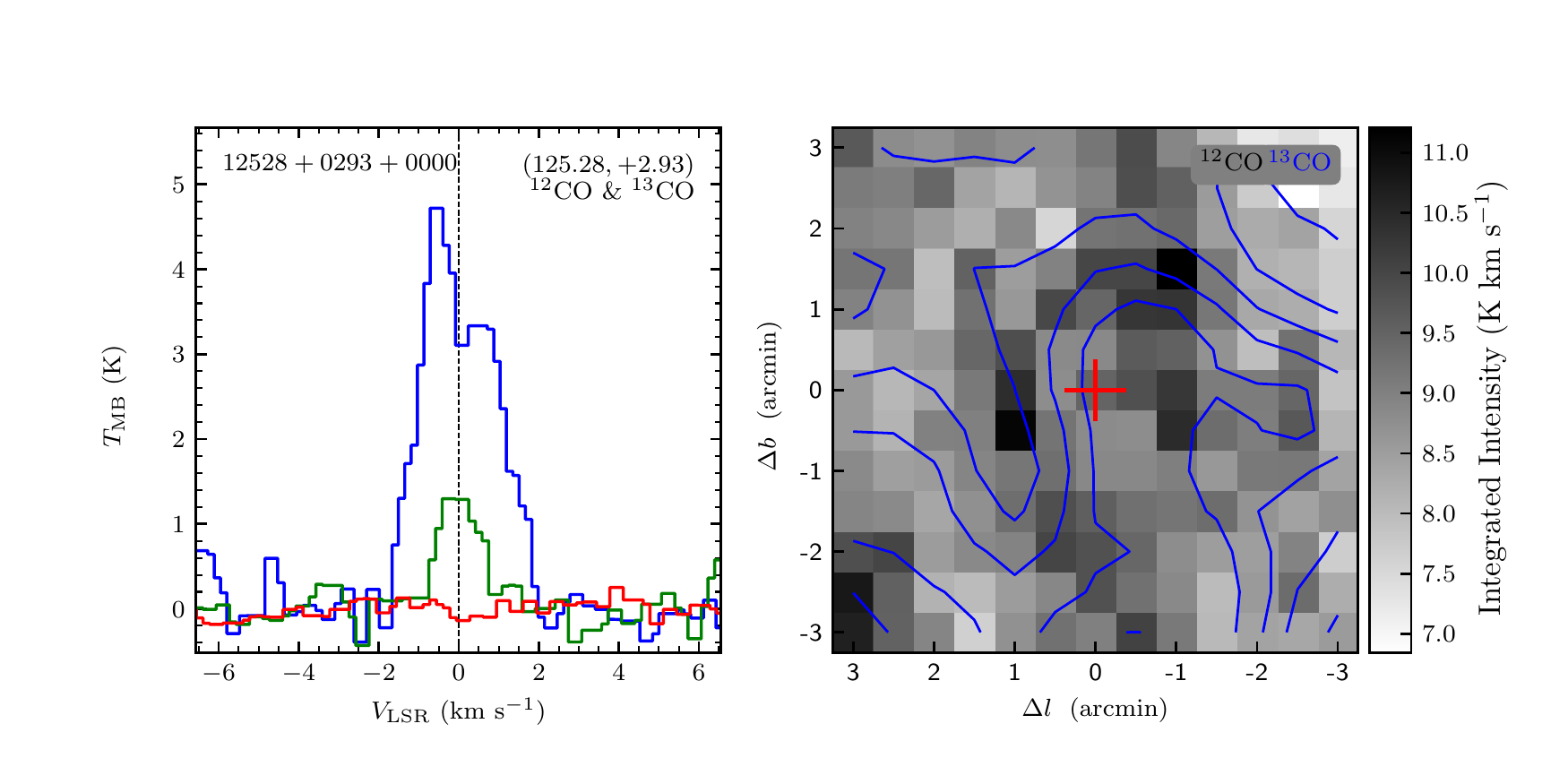}
\includegraphics[width=9.0cm,angle=0]{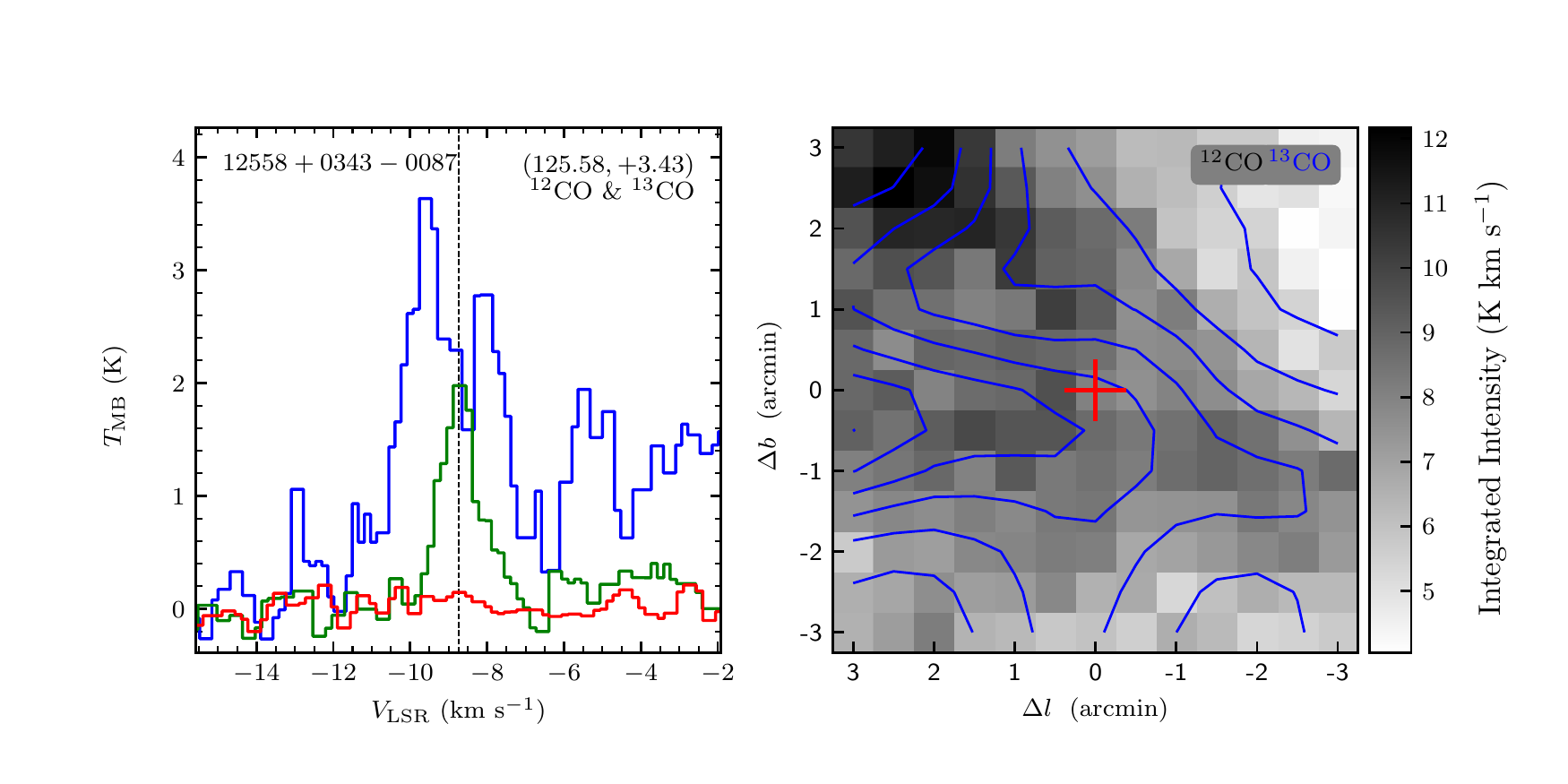}
\vspace{-0.5cm}

\includegraphics[width=9.0cm,angle=0]{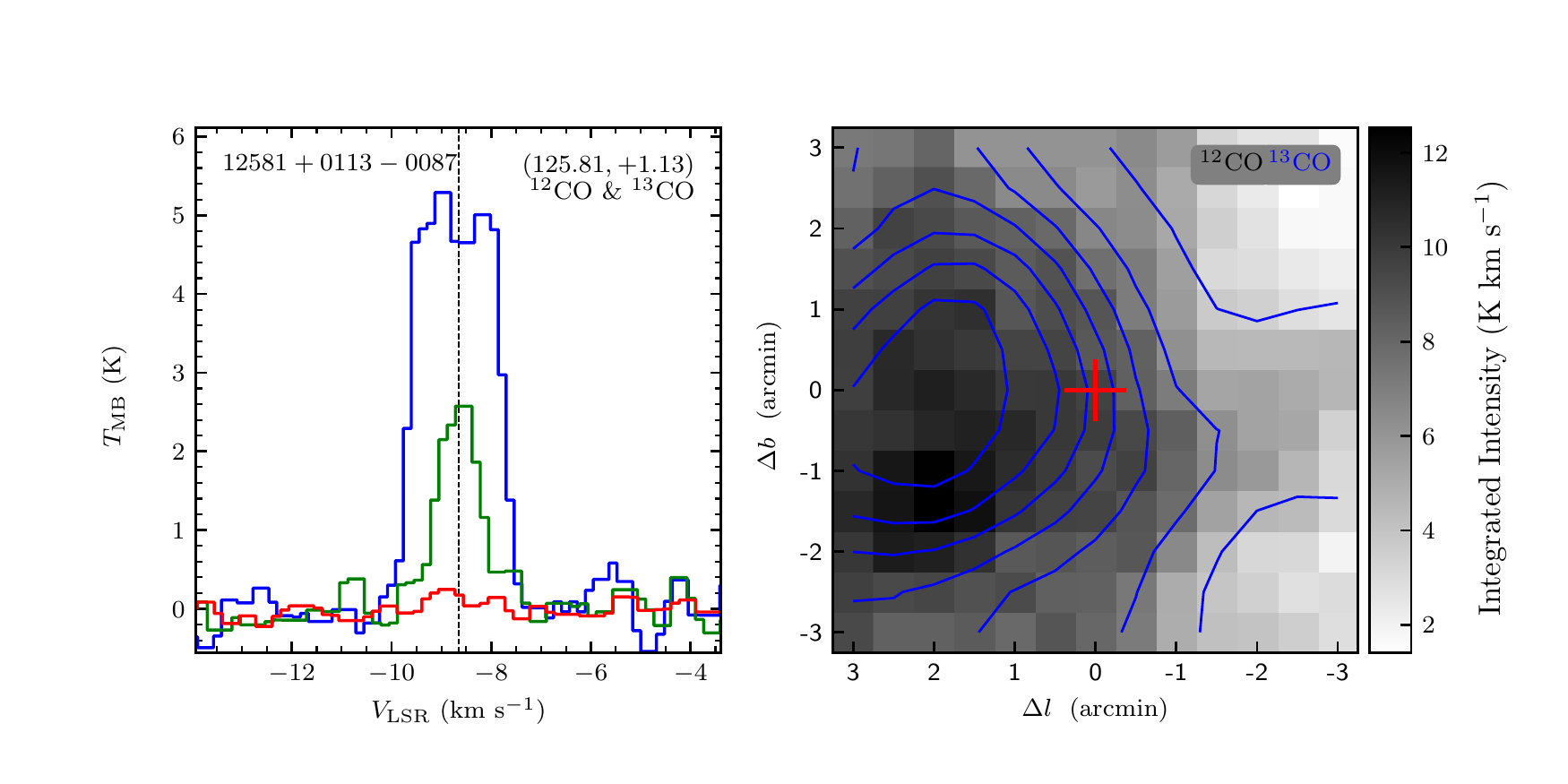}
\includegraphics[width=9.0cm,angle=0]{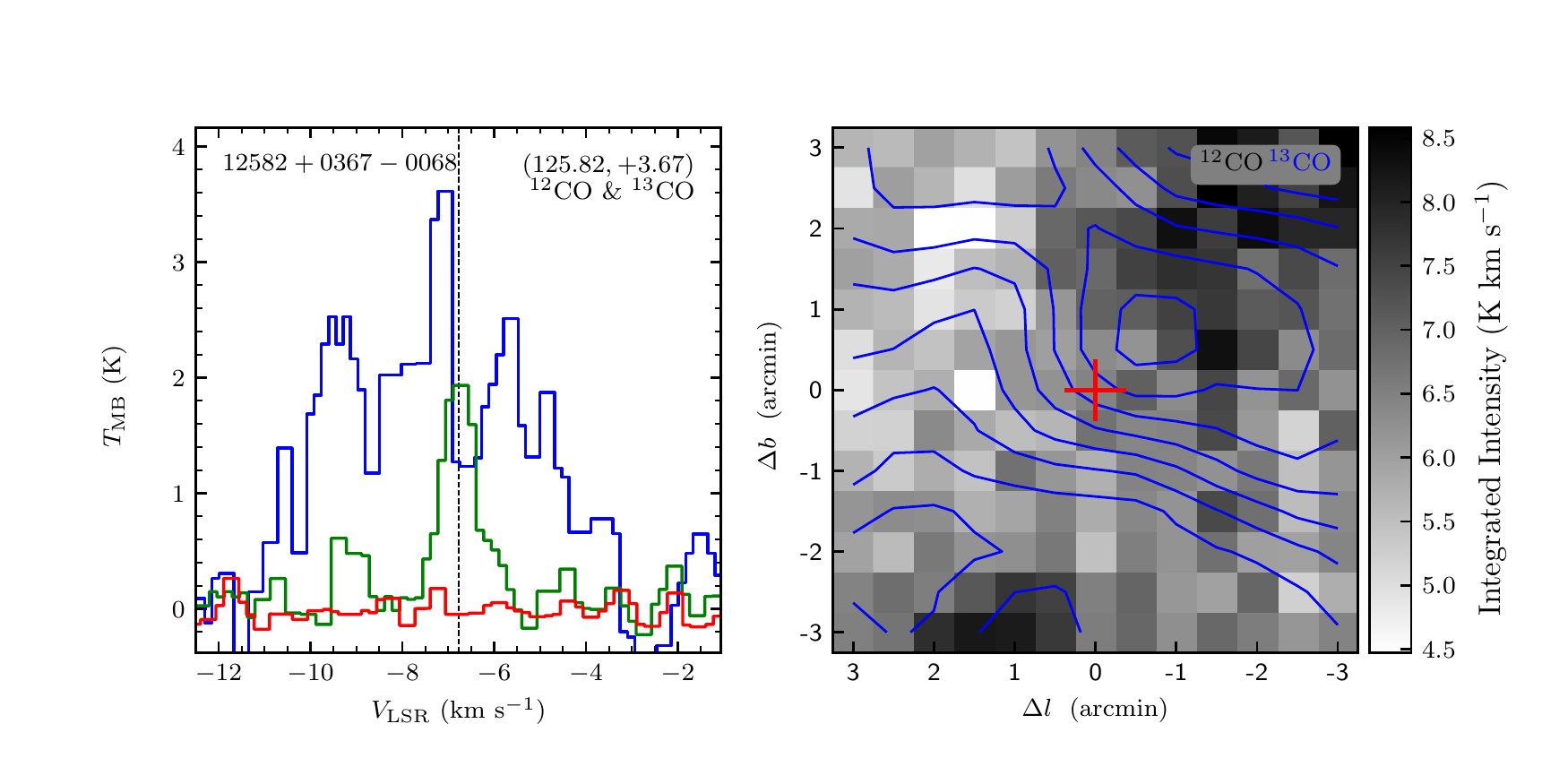}
\vspace{-0.5cm}

\includegraphics[width=9.0cm,angle=0]{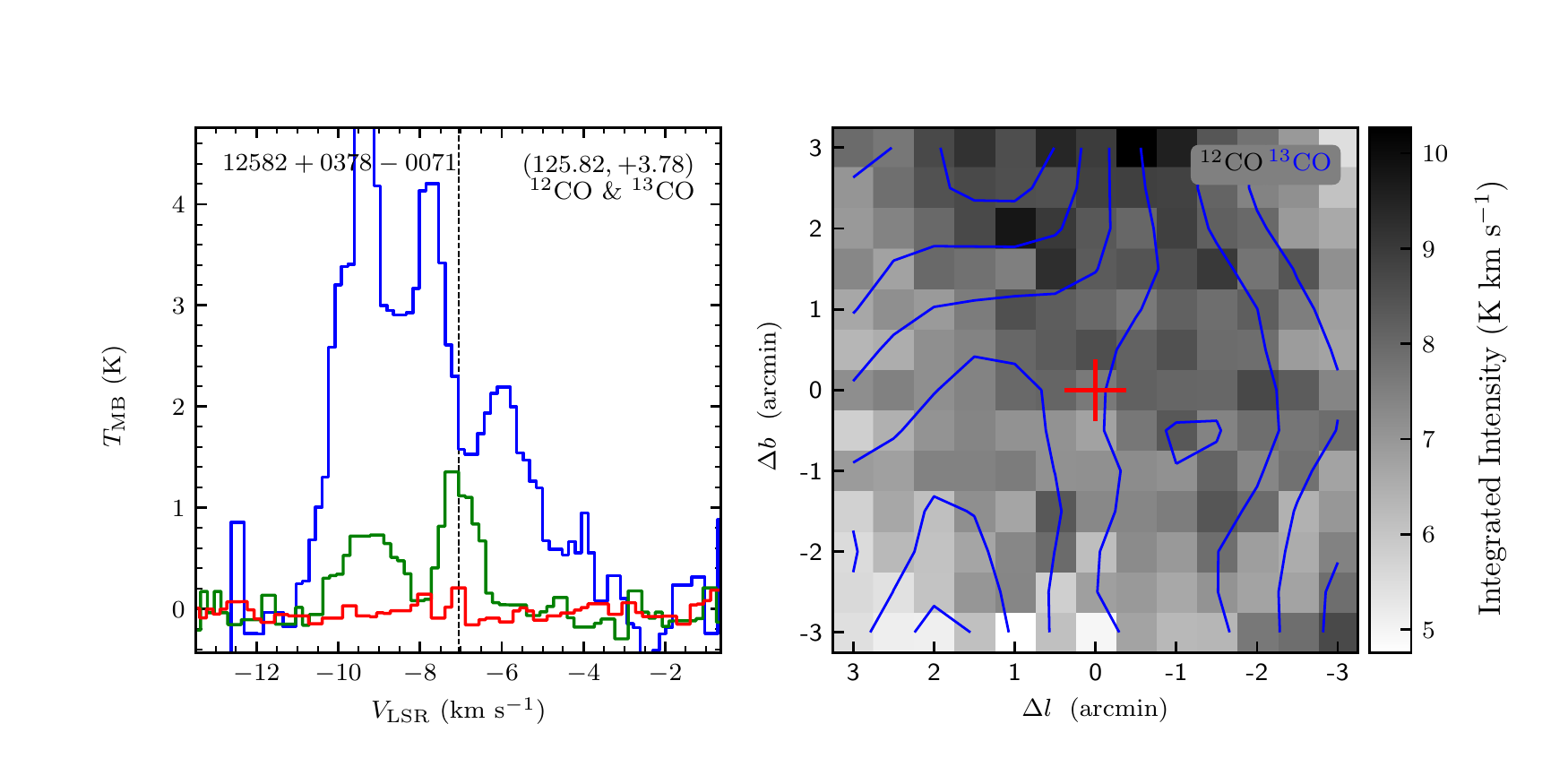}
\includegraphics[width=9.0cm,angle=0]{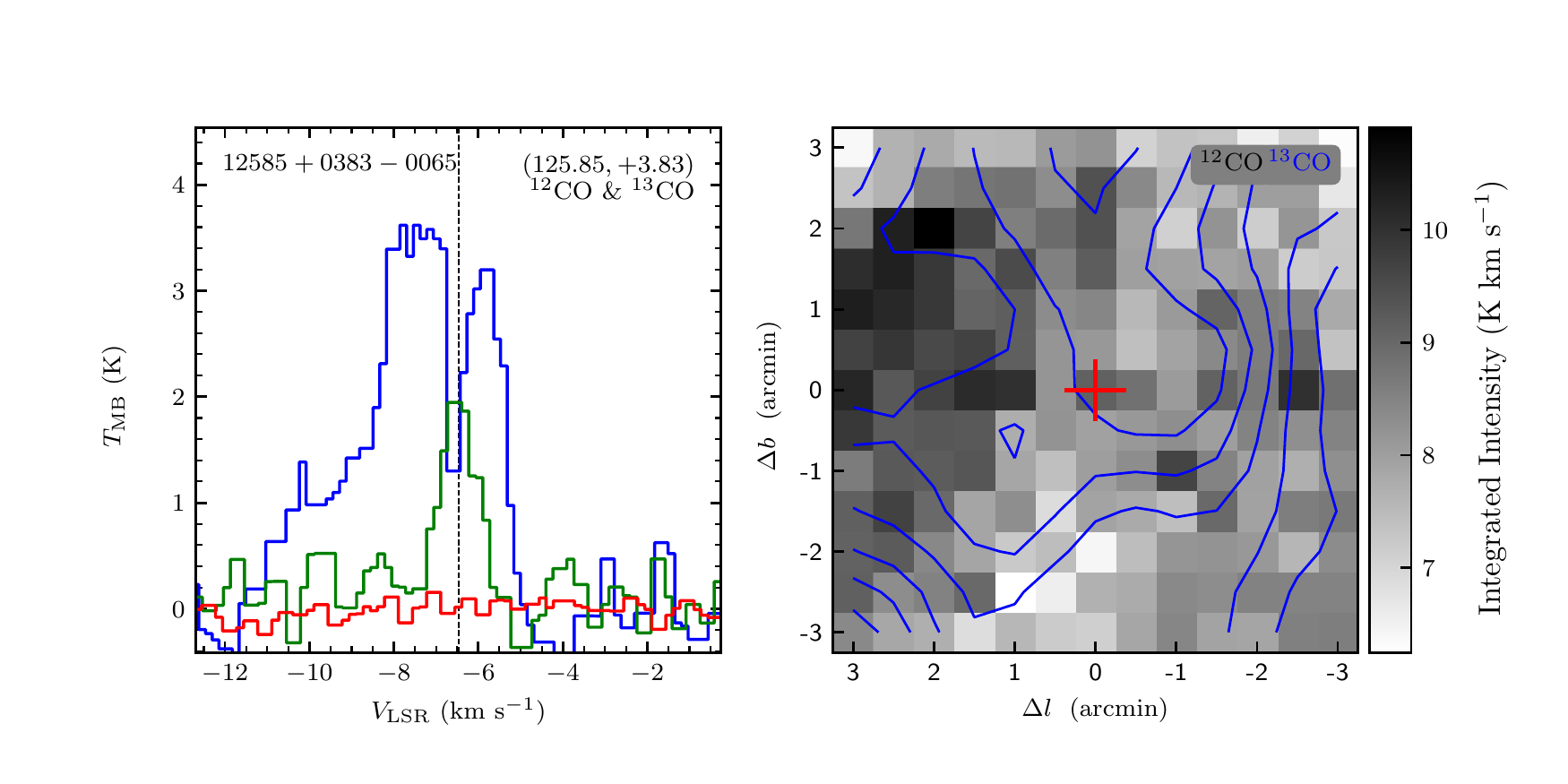}
\vspace{-0.5cm}

\includegraphics[width=9.0cm,angle=0]{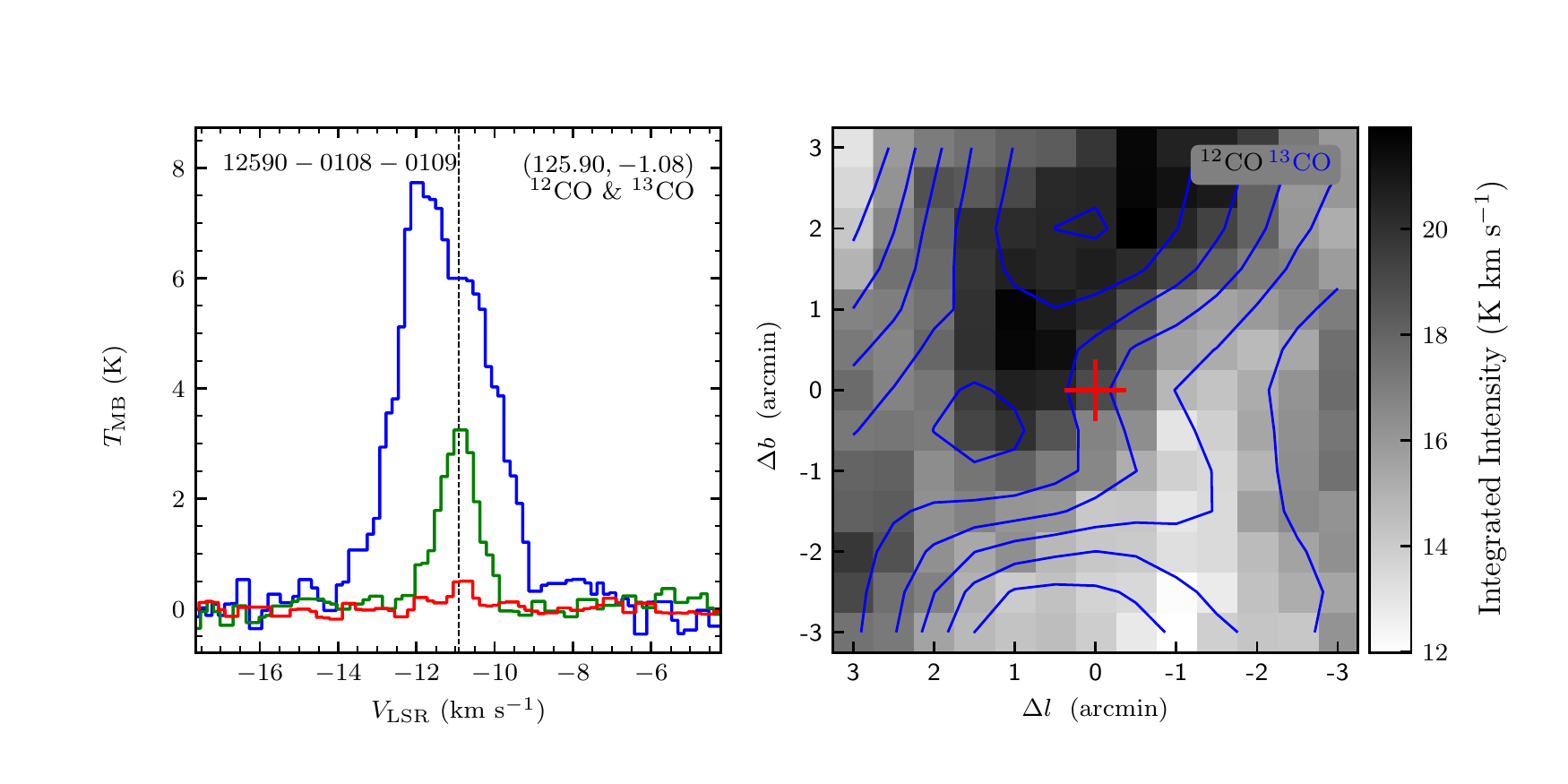}
\includegraphics[width=9.0cm,angle=0]{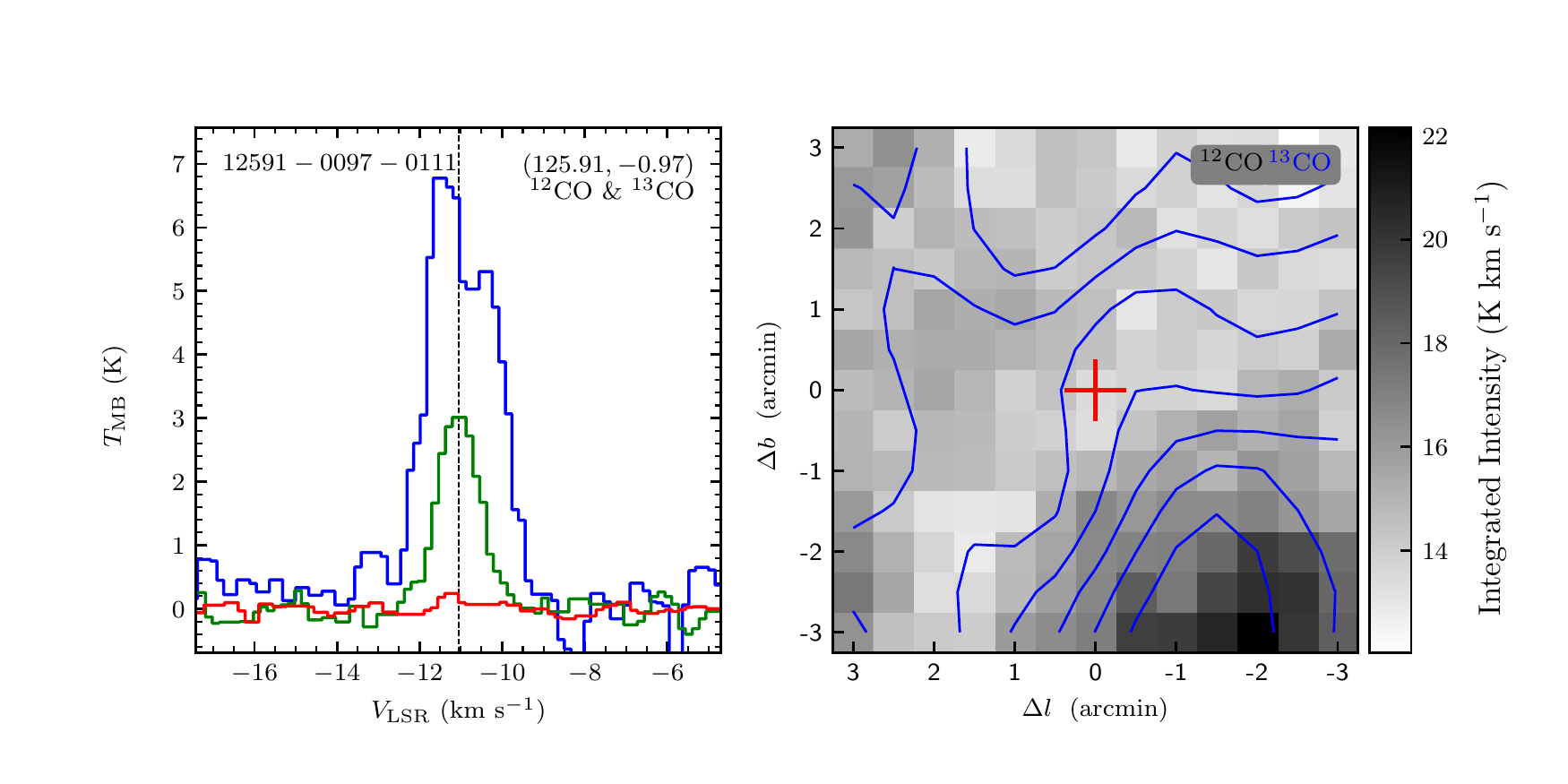}
\vspace{-0.5cm}

\includegraphics[width=9.0cm,angle=0]{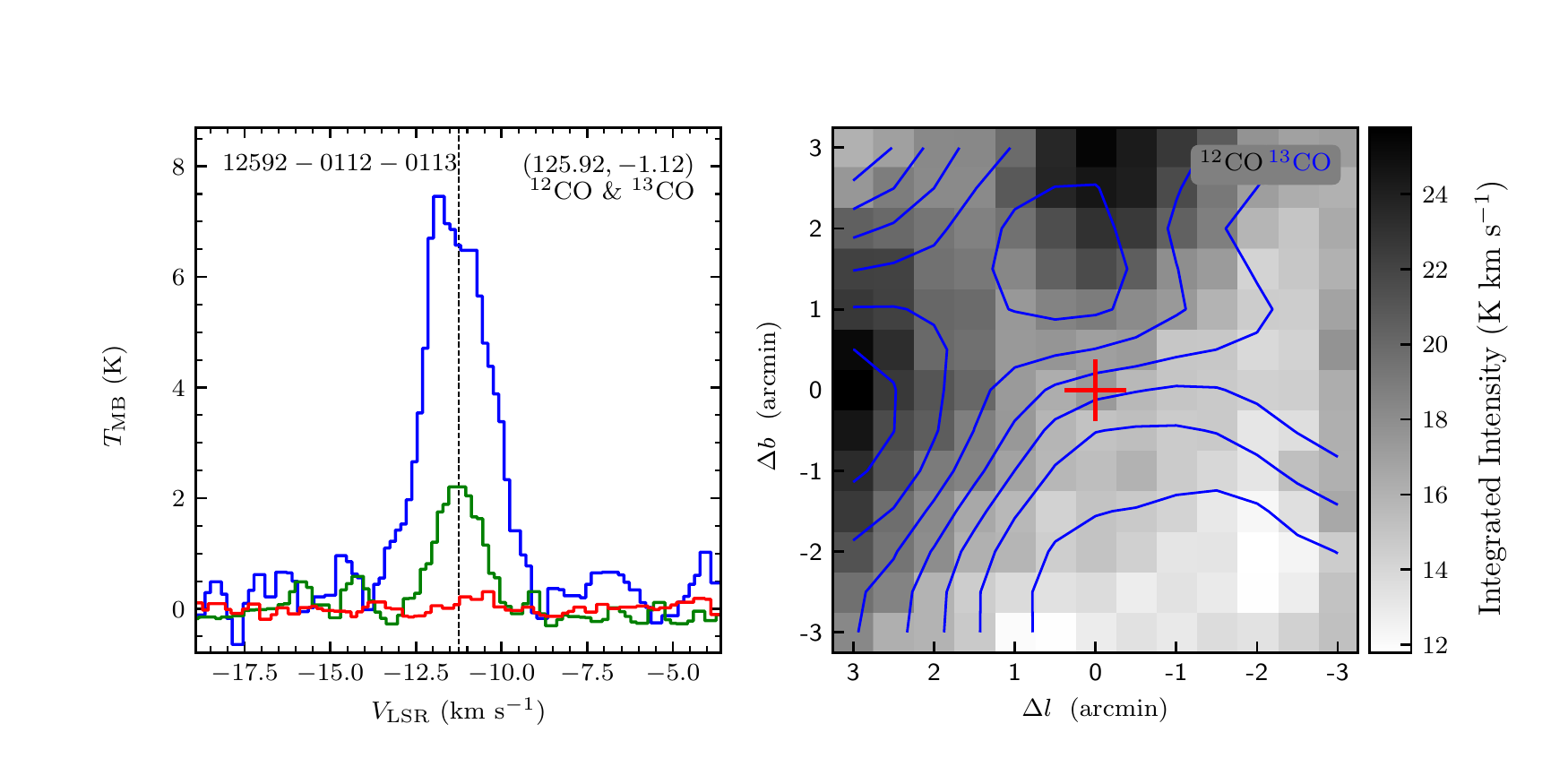}
\includegraphics[width=9.0cm,angle=0]{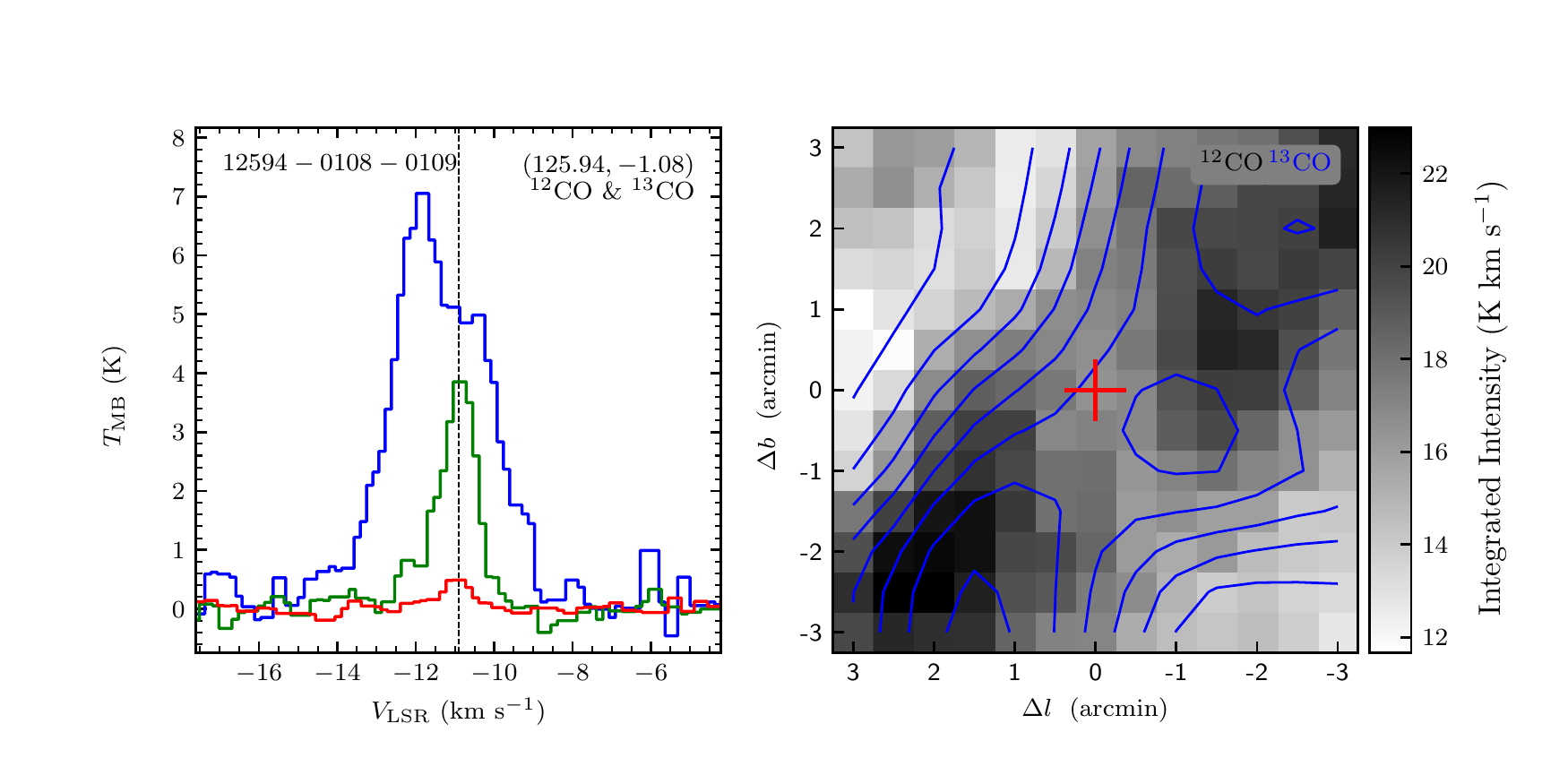}
\end{figure}
\clearpage

\begin{figure}
\includegraphics[width=9.0cm,angle=0]{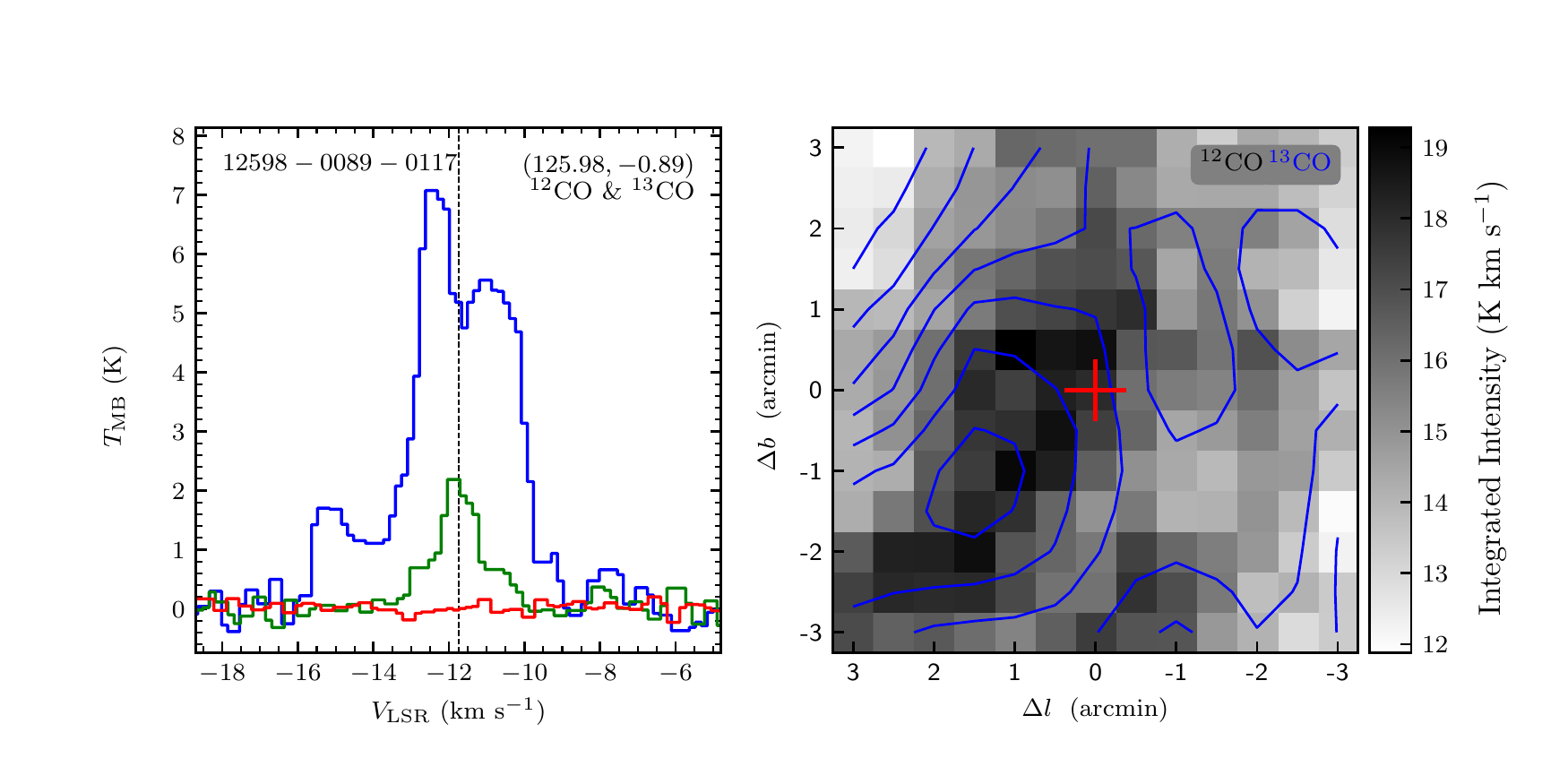}
\includegraphics[width=9.0cm,angle=0]{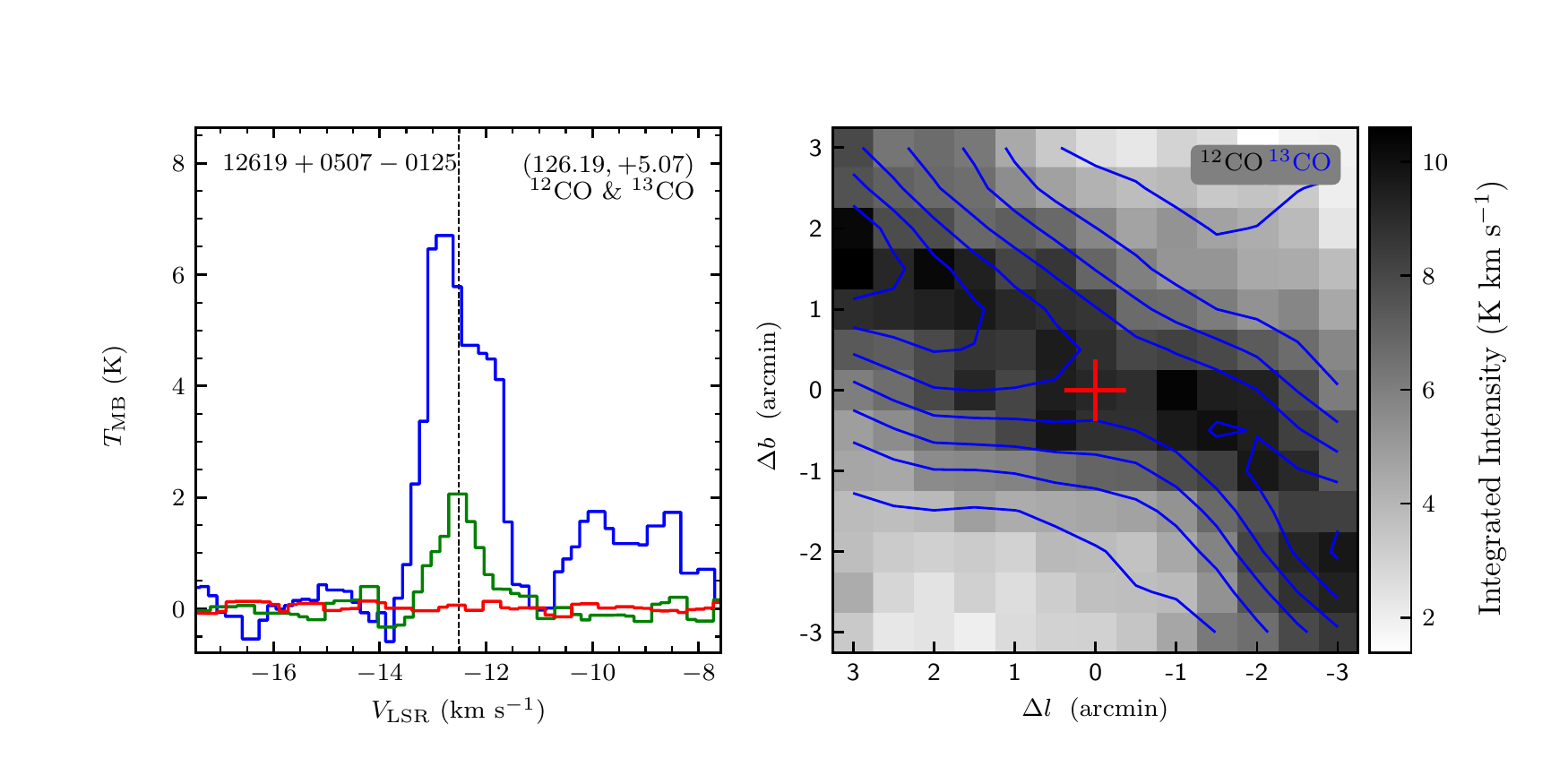}
\vspace{-0.5cm}

\includegraphics[width=9.0cm,angle=0]{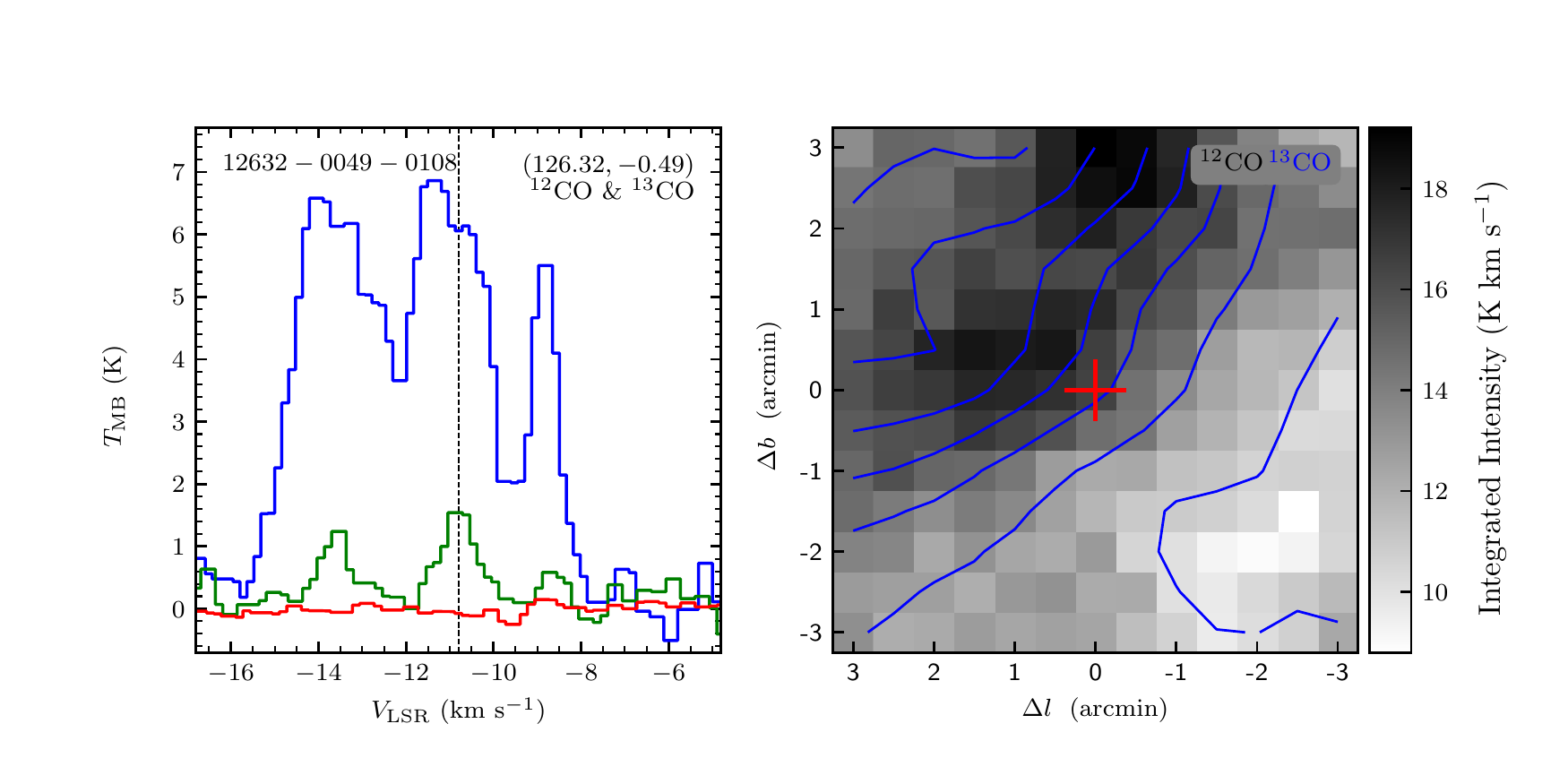}
\includegraphics[width=9.0cm,angle=0]{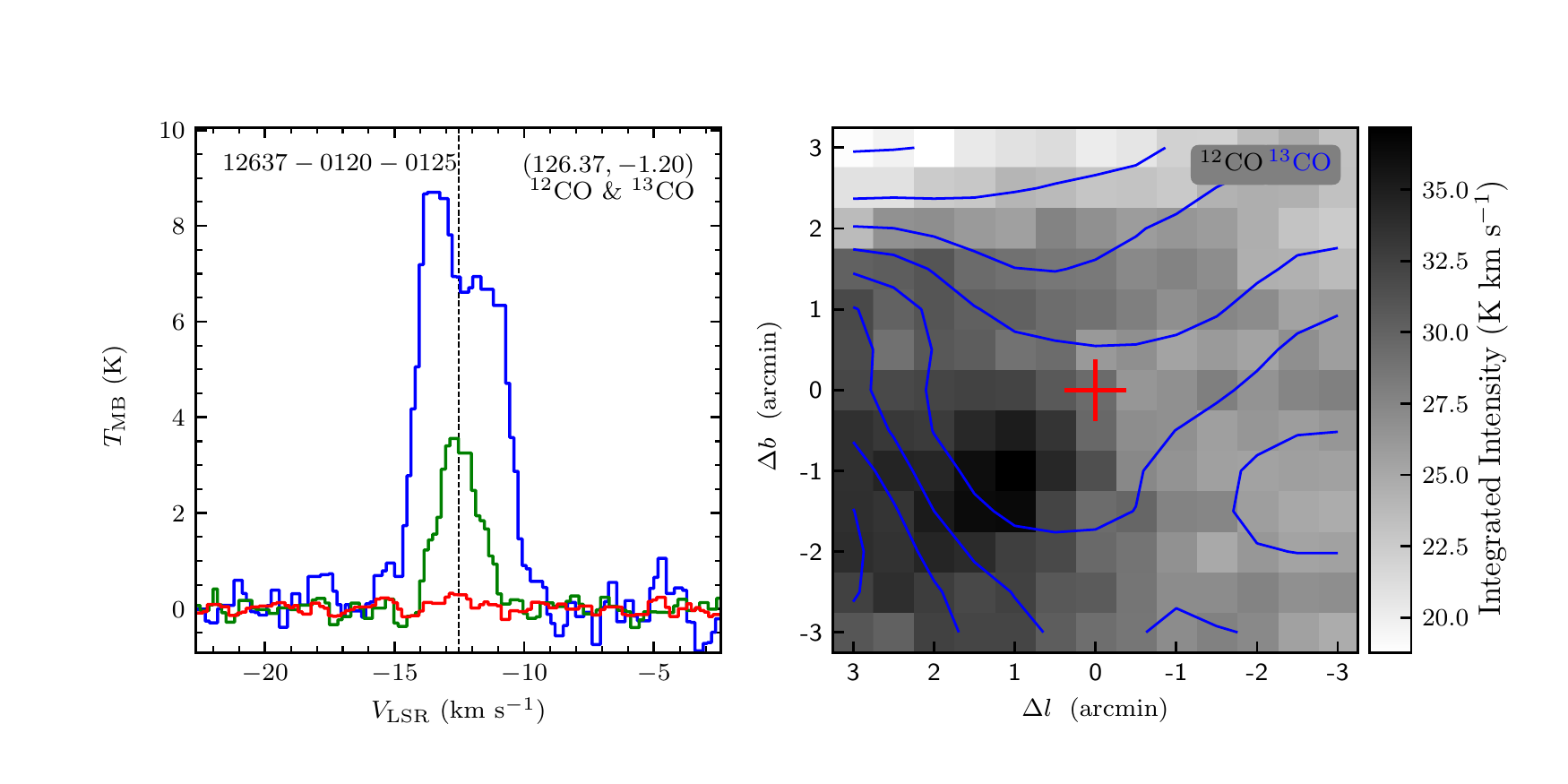}
\vspace{-0.5cm}

\includegraphics[width=9.0cm,angle=0]{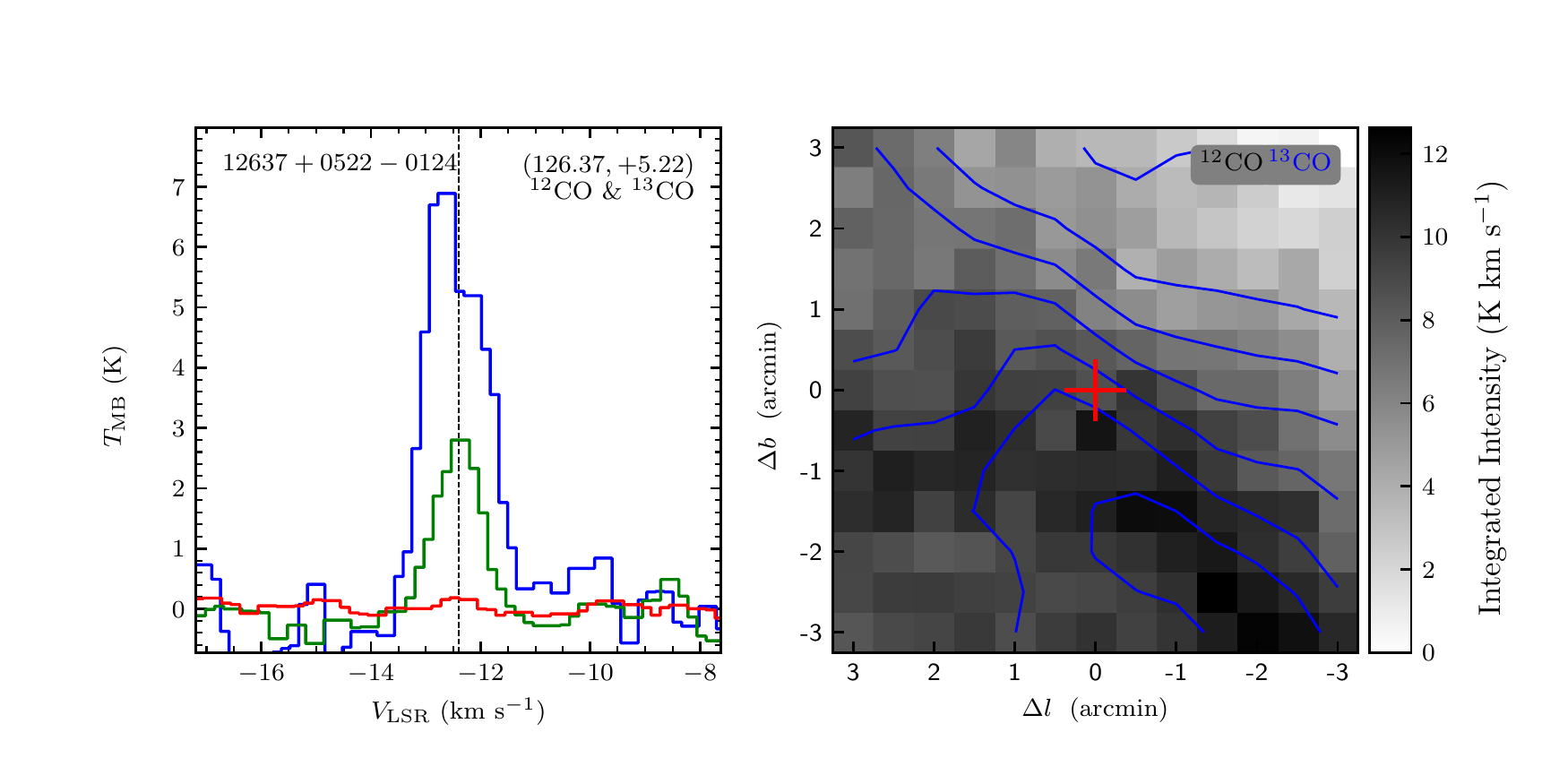}
\includegraphics[width=9.0cm,angle=0]{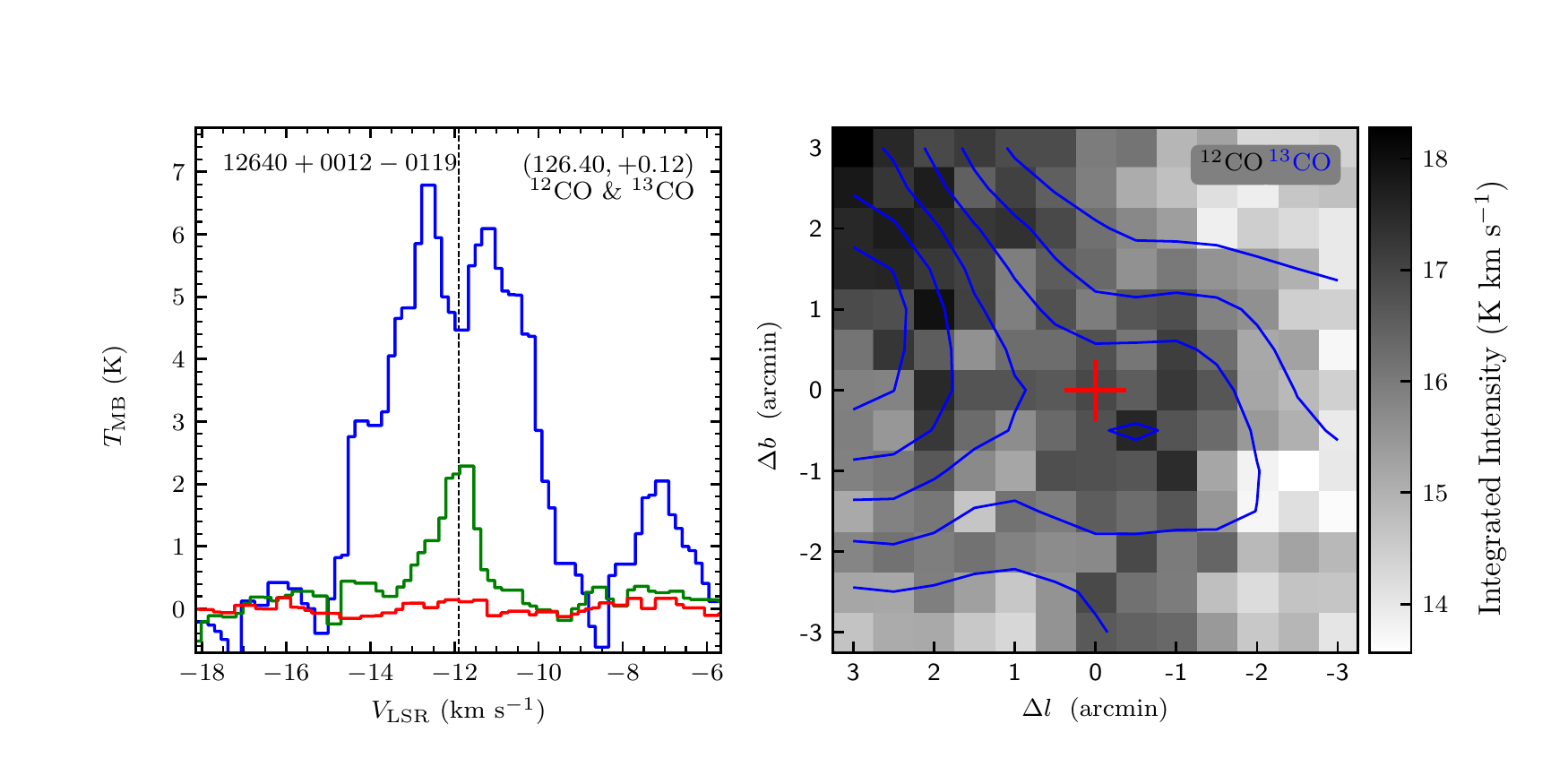}
\vspace{-0.5cm}

\includegraphics[width=9.0cm,angle=0]{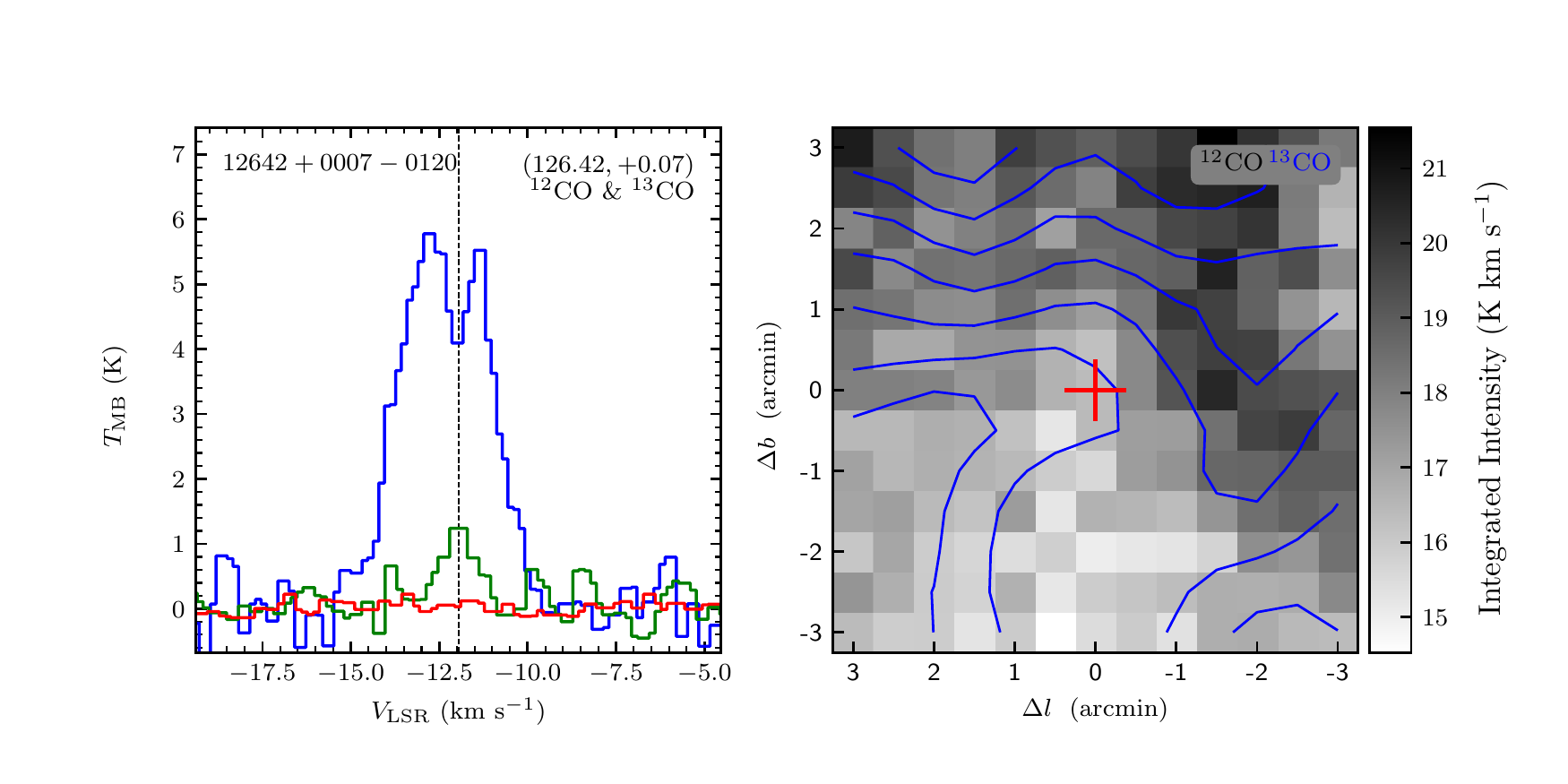}
\includegraphics[width=9.0cm,angle=0]{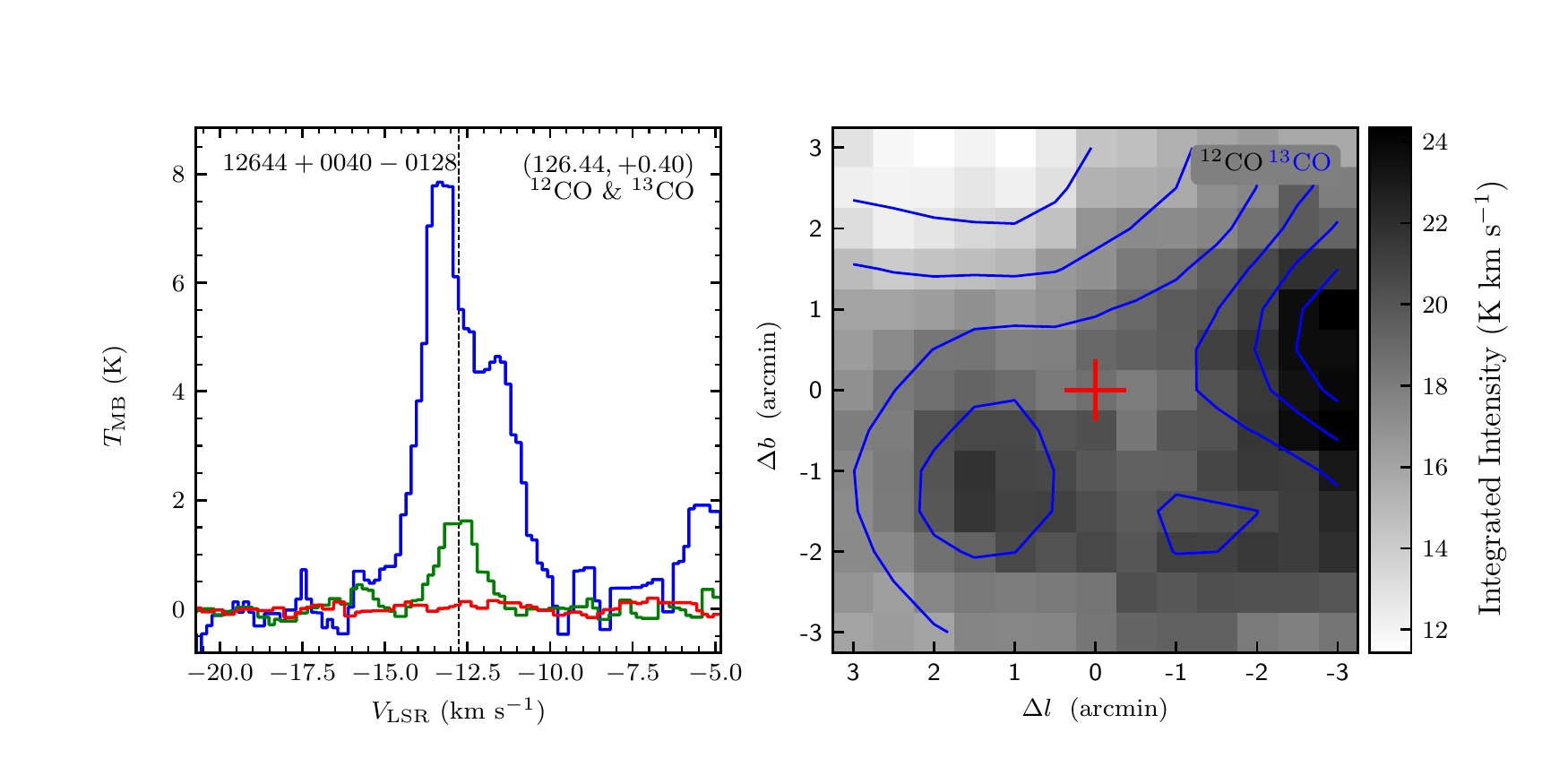}
\vspace{-0.5cm}

\includegraphics[width=9.0cm,angle=0]{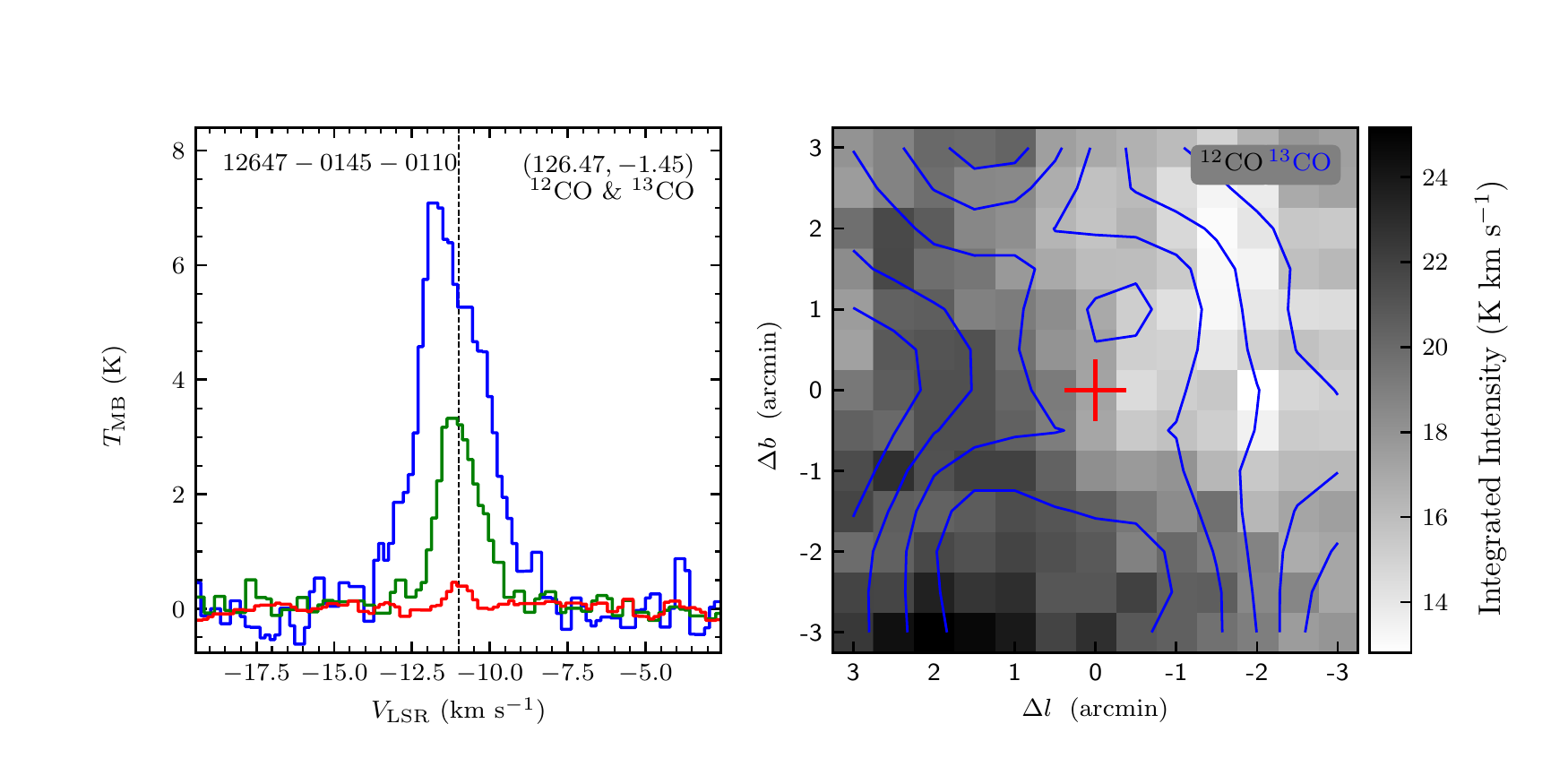}
\includegraphics[width=9.0cm,angle=0]{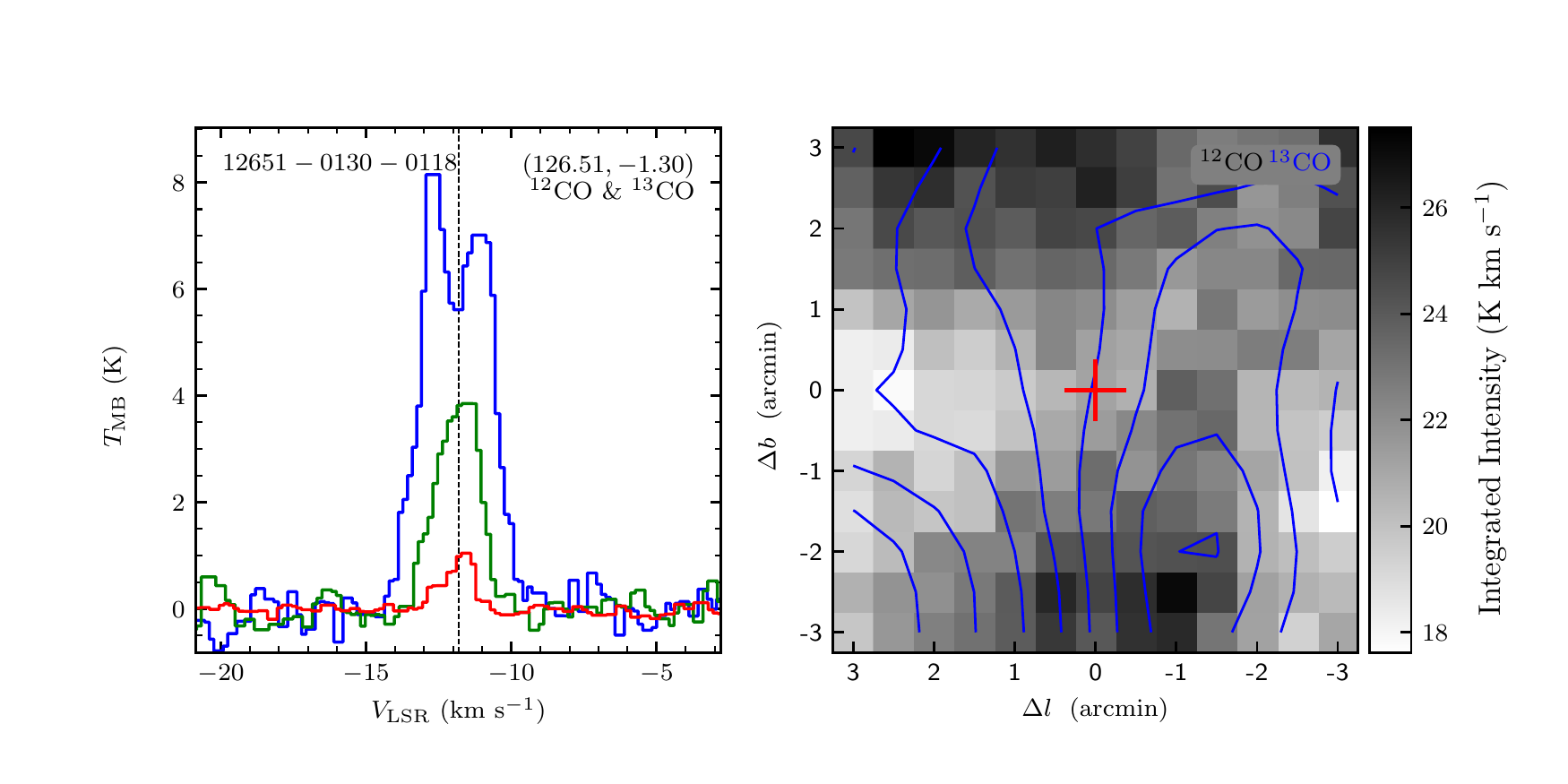}
\end{figure}
\clearpage

\begin{figure}
\includegraphics[width=9.0cm,angle=0]{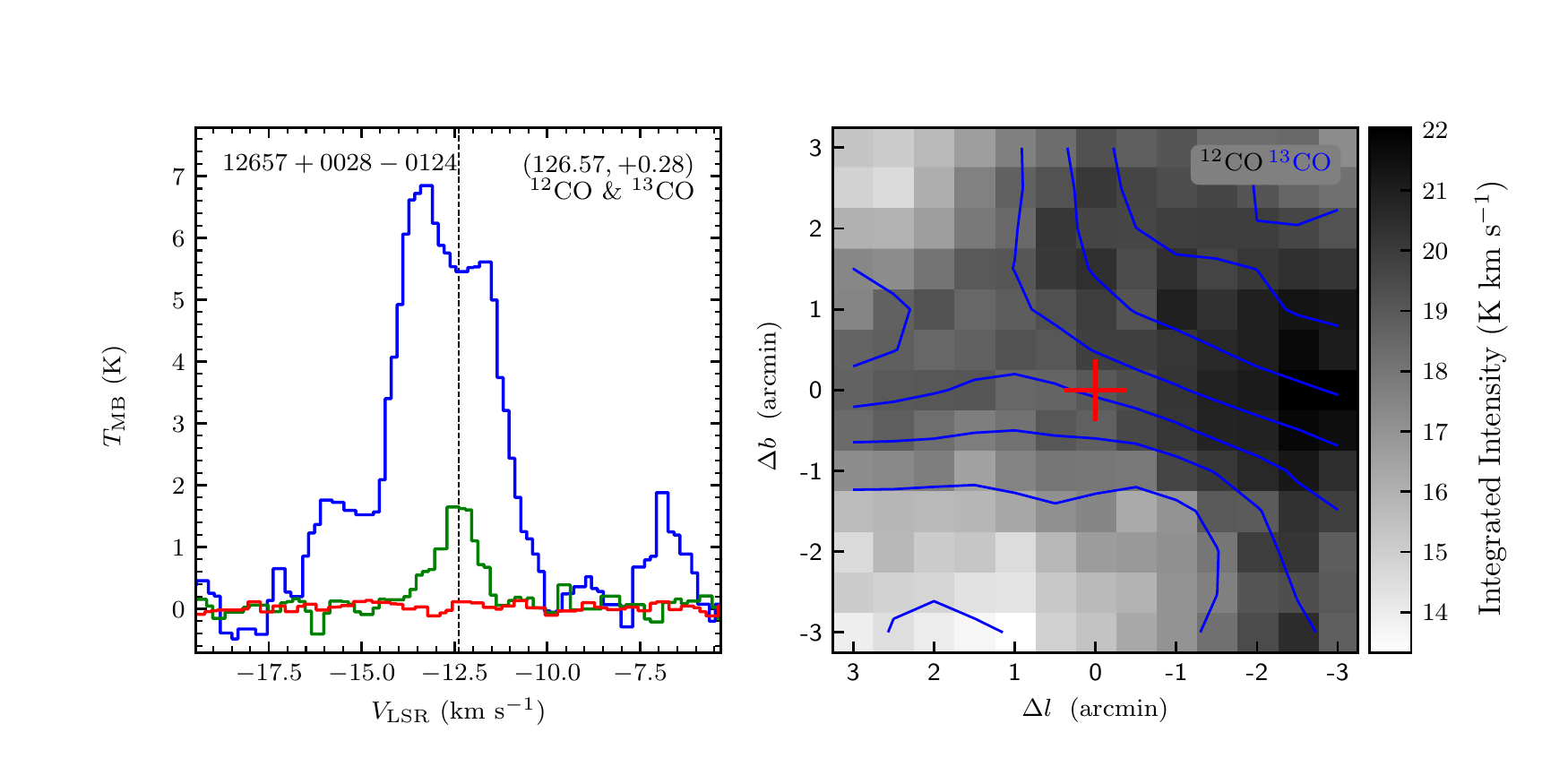}
\includegraphics[width=9.0cm,angle=0]{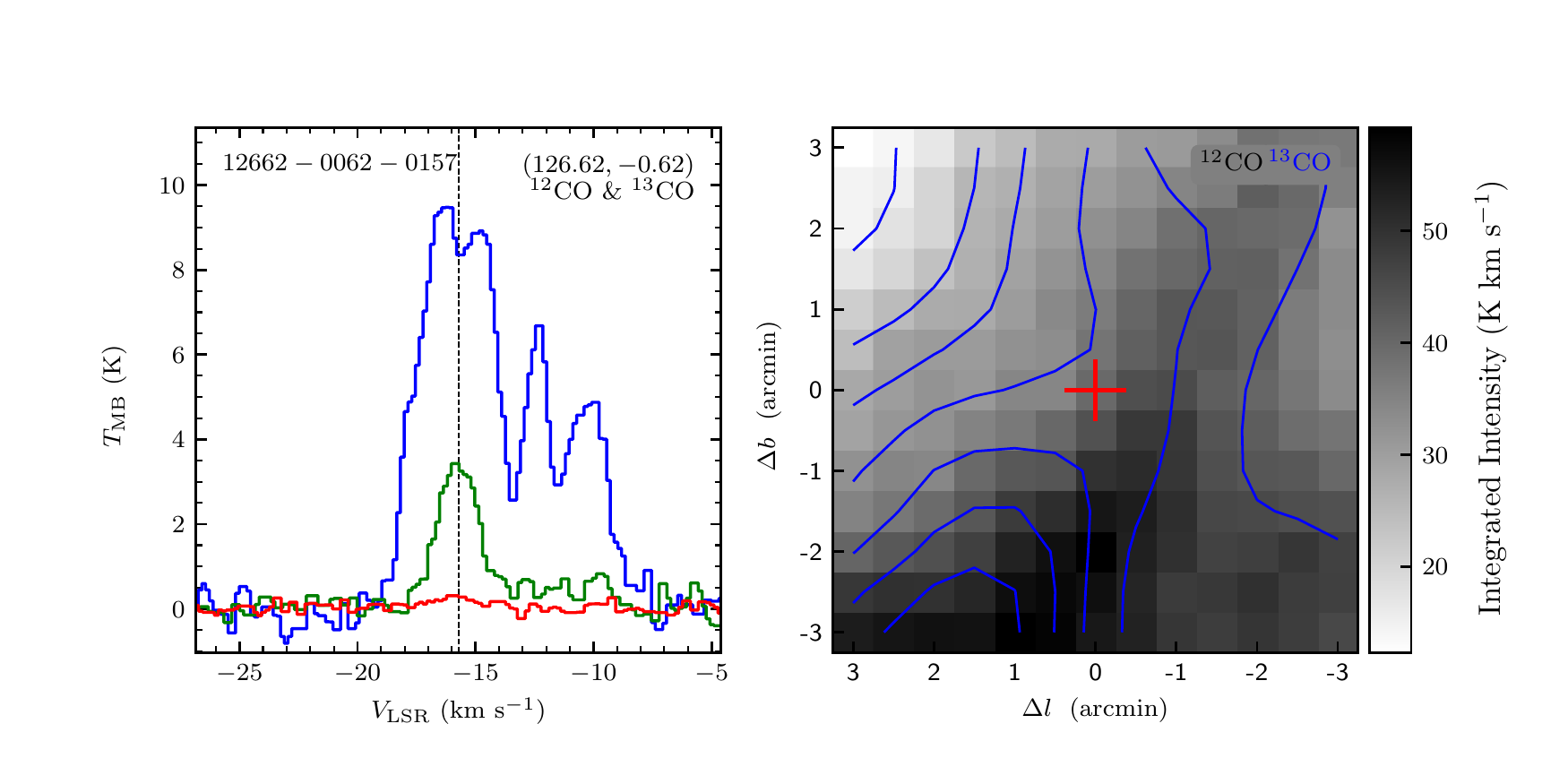}
\vspace{-0.5cm}

\includegraphics[width=9.0cm,angle=0]{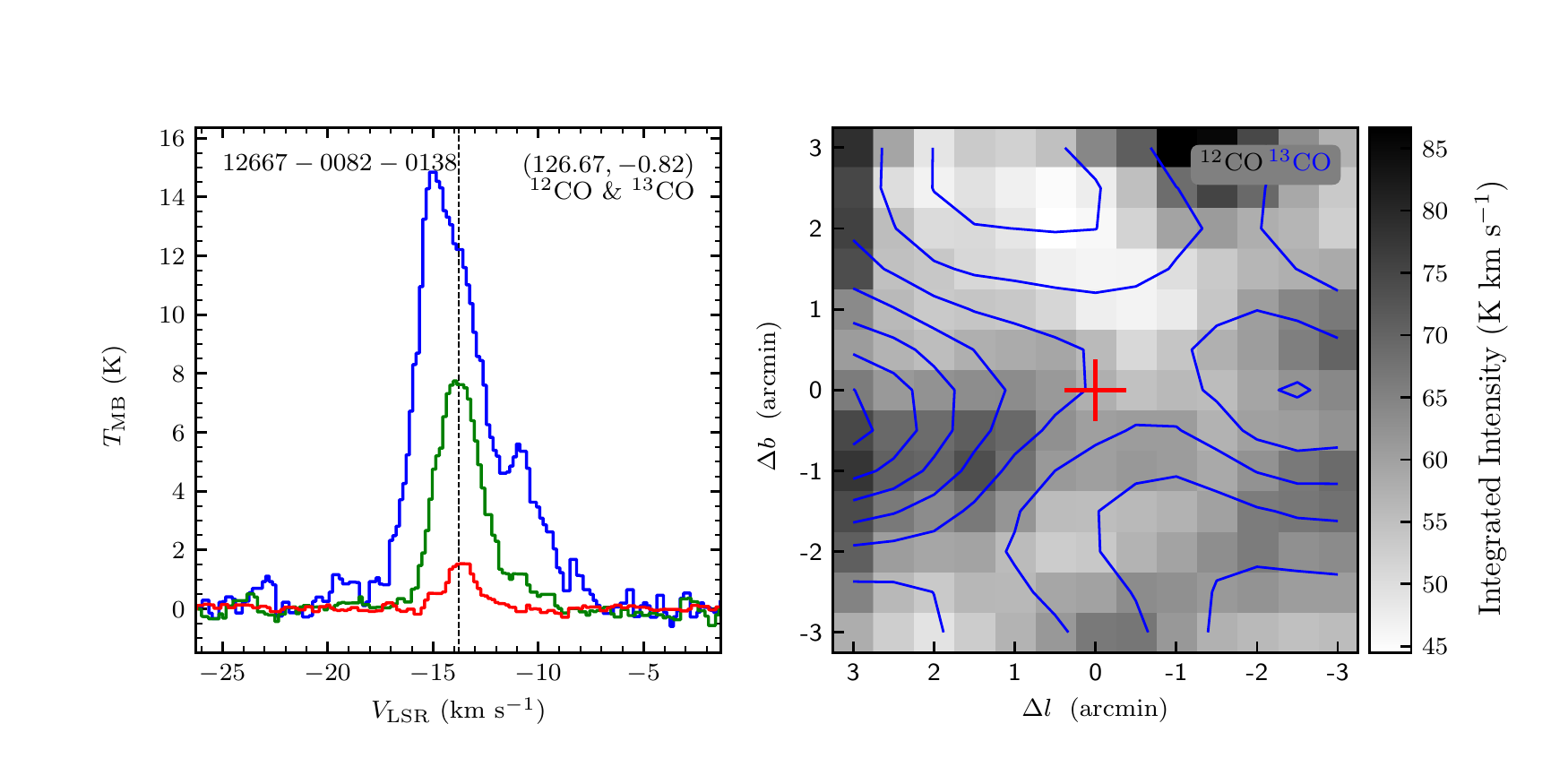}
\includegraphics[width=9.0cm,angle=0]{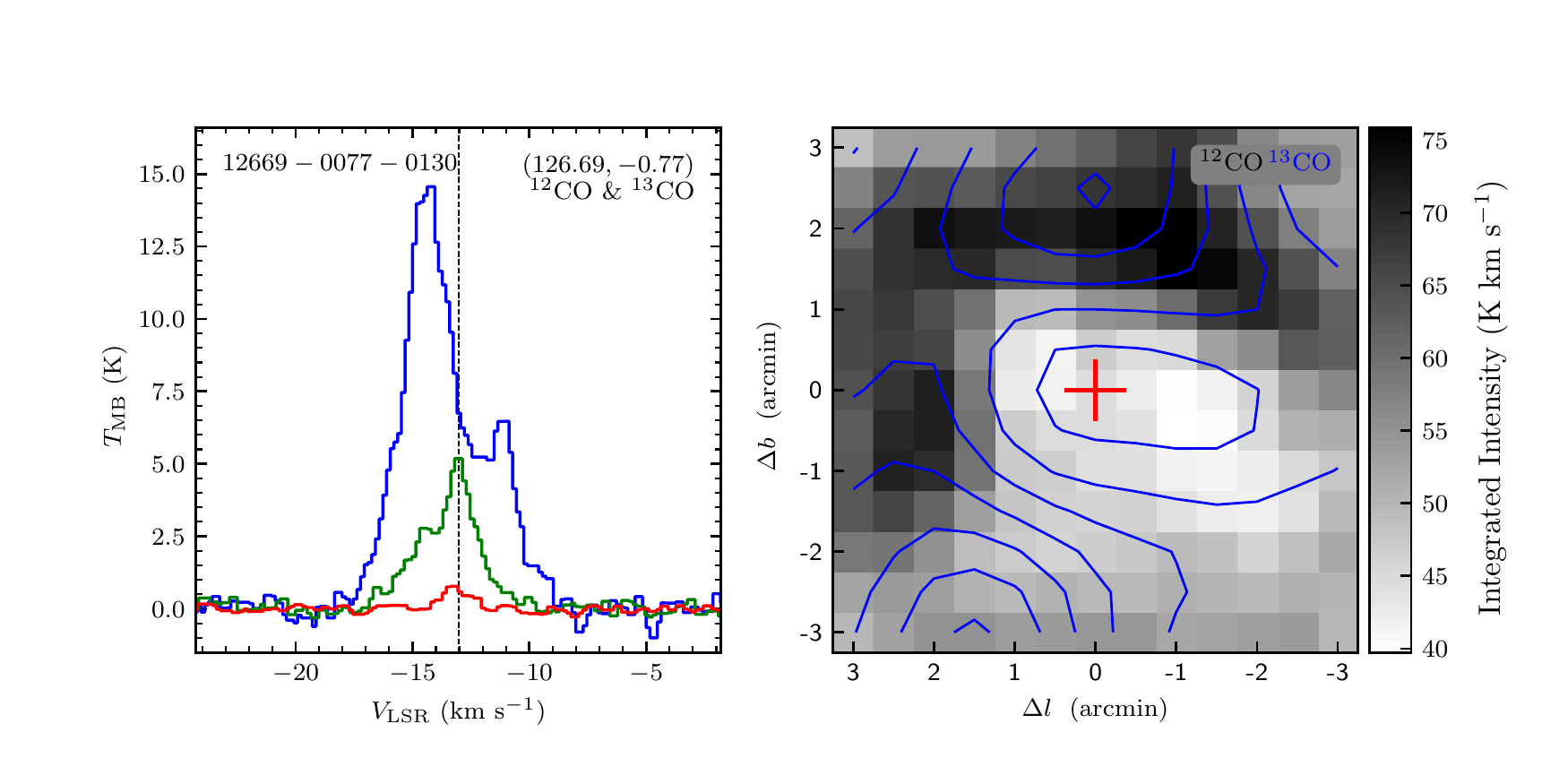}
\vspace{-0.5cm}

\includegraphics[width=9.0cm,angle=0]{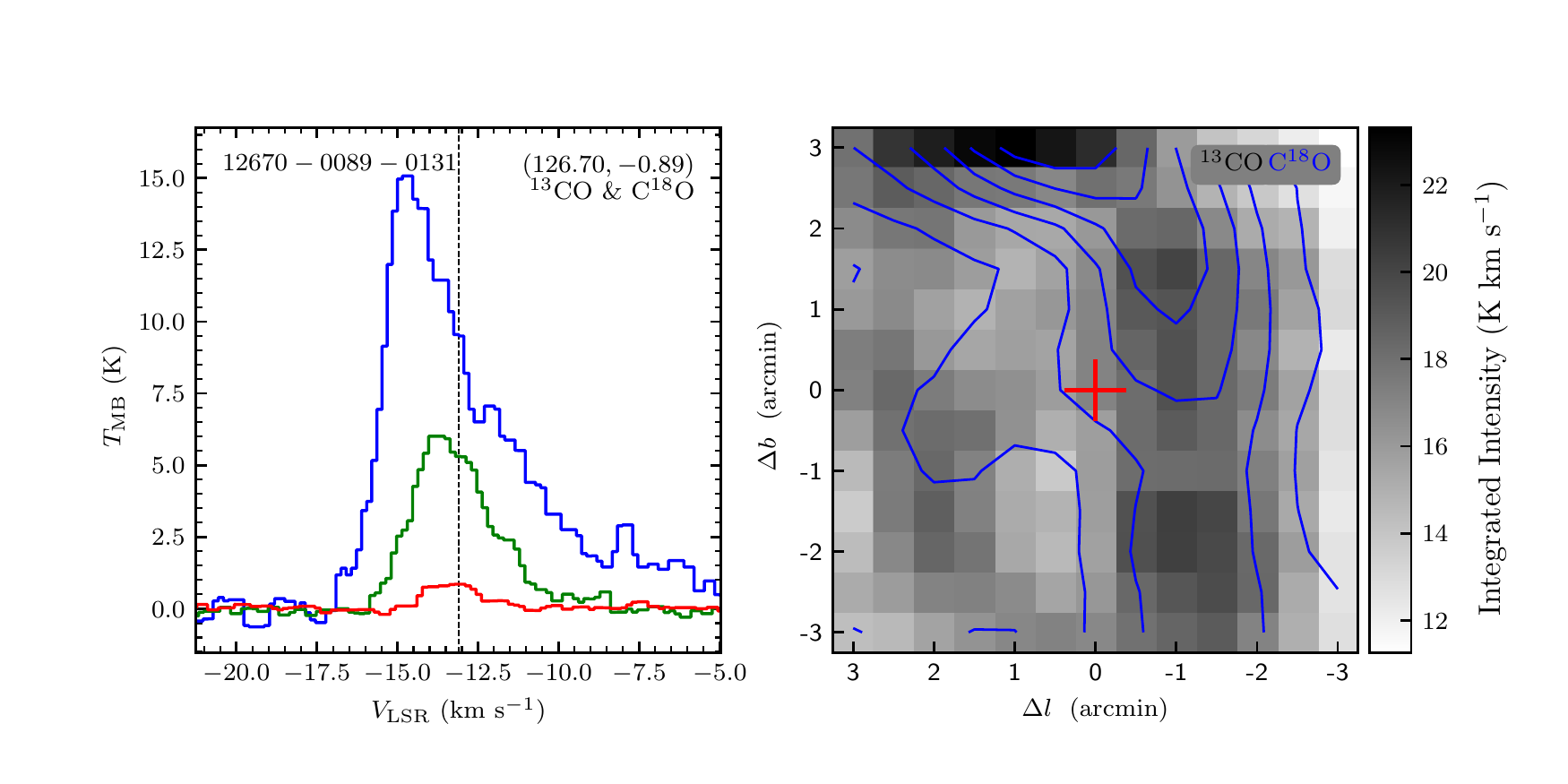}
\includegraphics[width=9.0cm,angle=0]{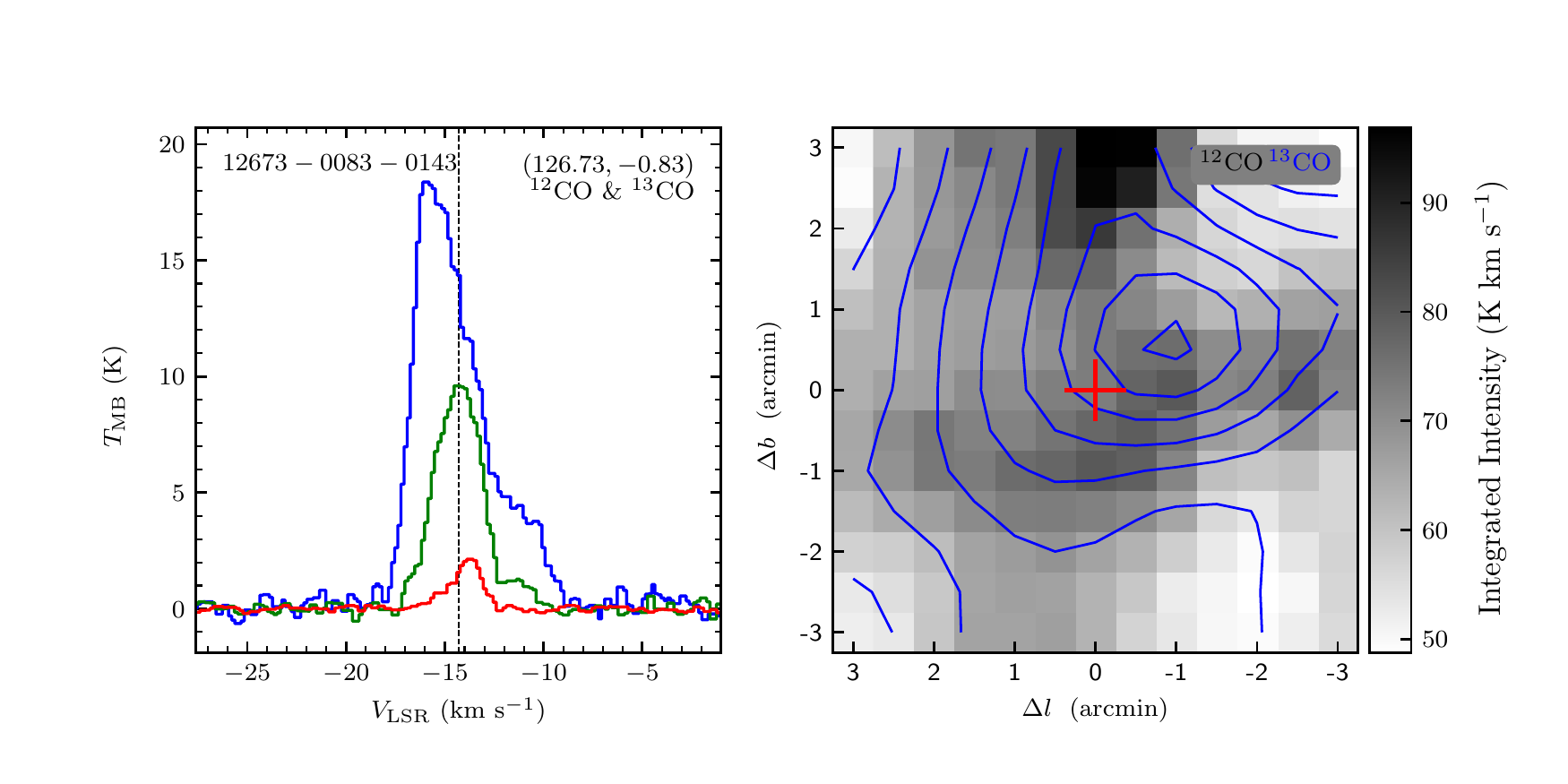}
\vspace{-0.5cm}

\includegraphics[width=9.0cm,angle=0]{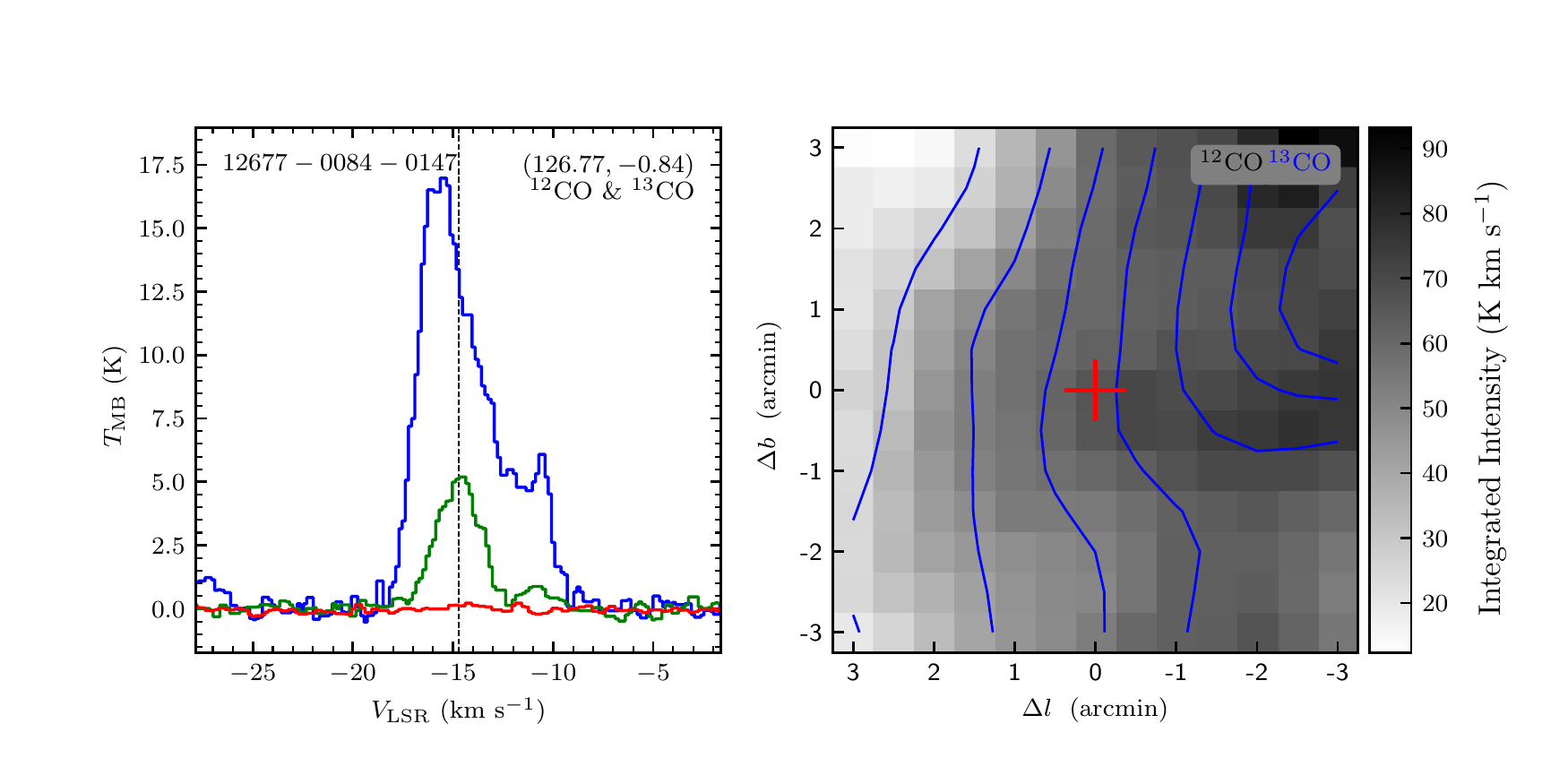}
\includegraphics[width=9.0cm,angle=0]{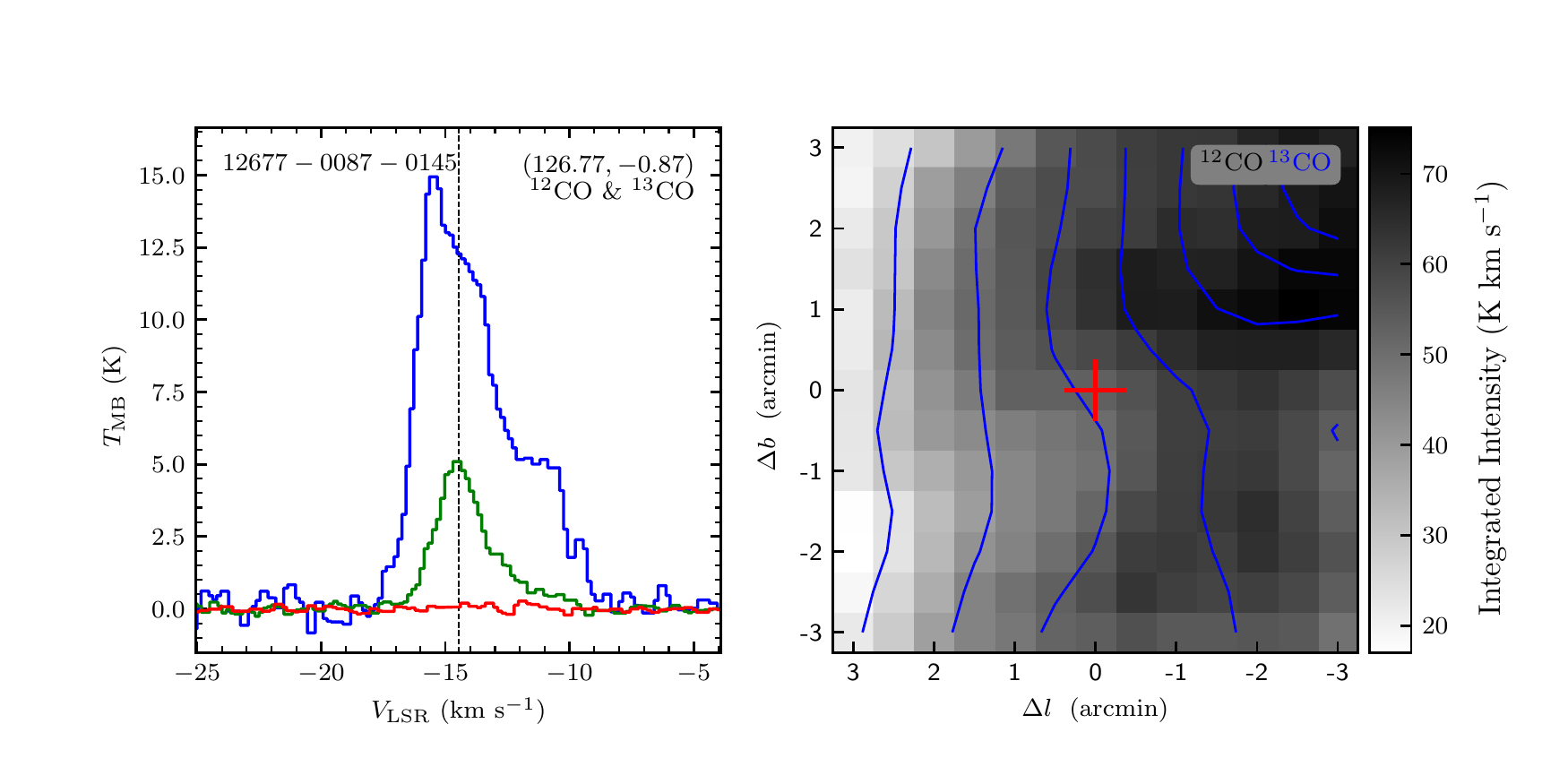}
\vspace{-0.5cm}

\includegraphics[width=9.0cm,angle=0]{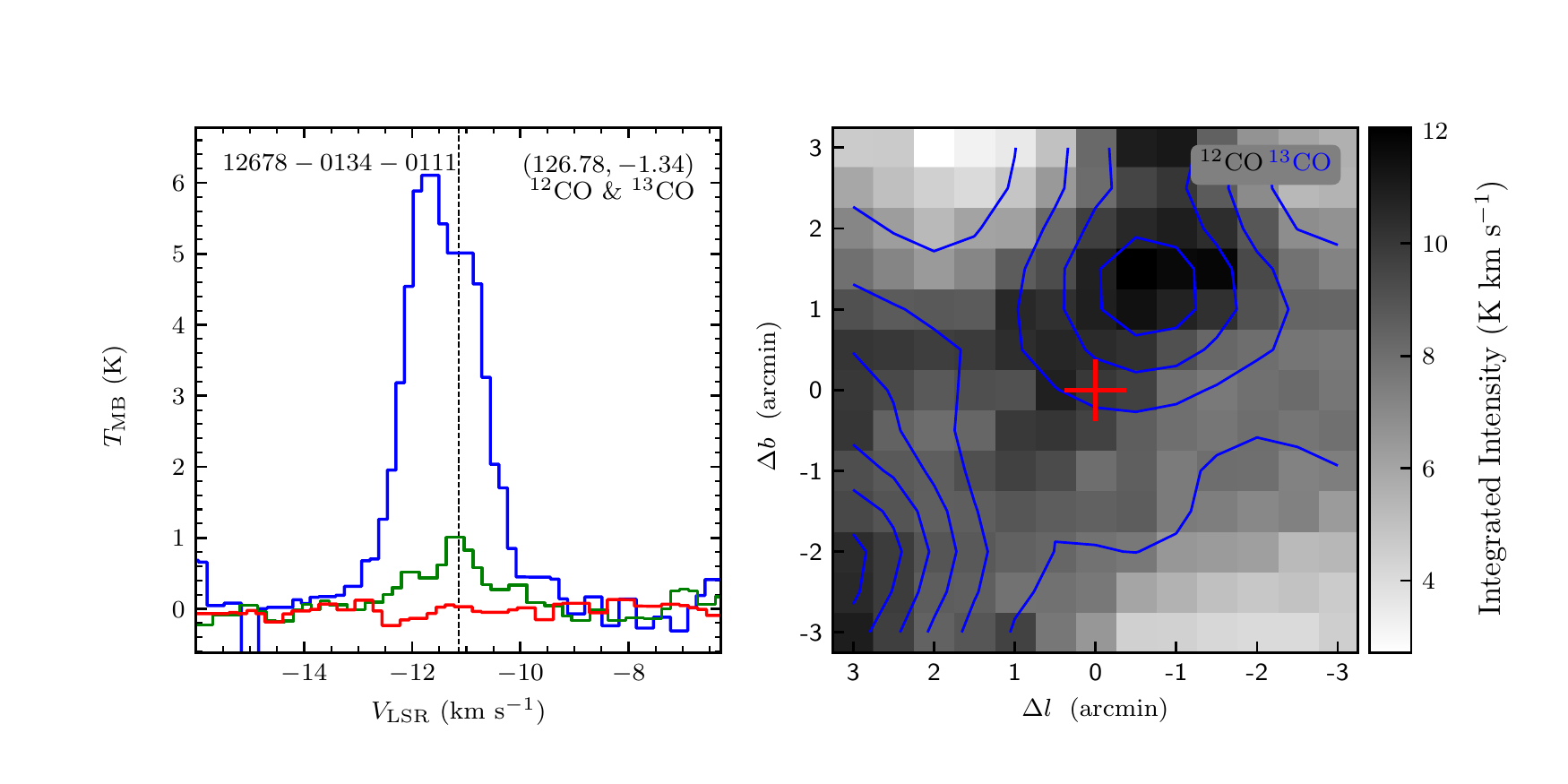}
\includegraphics[width=9.0cm,angle=0]{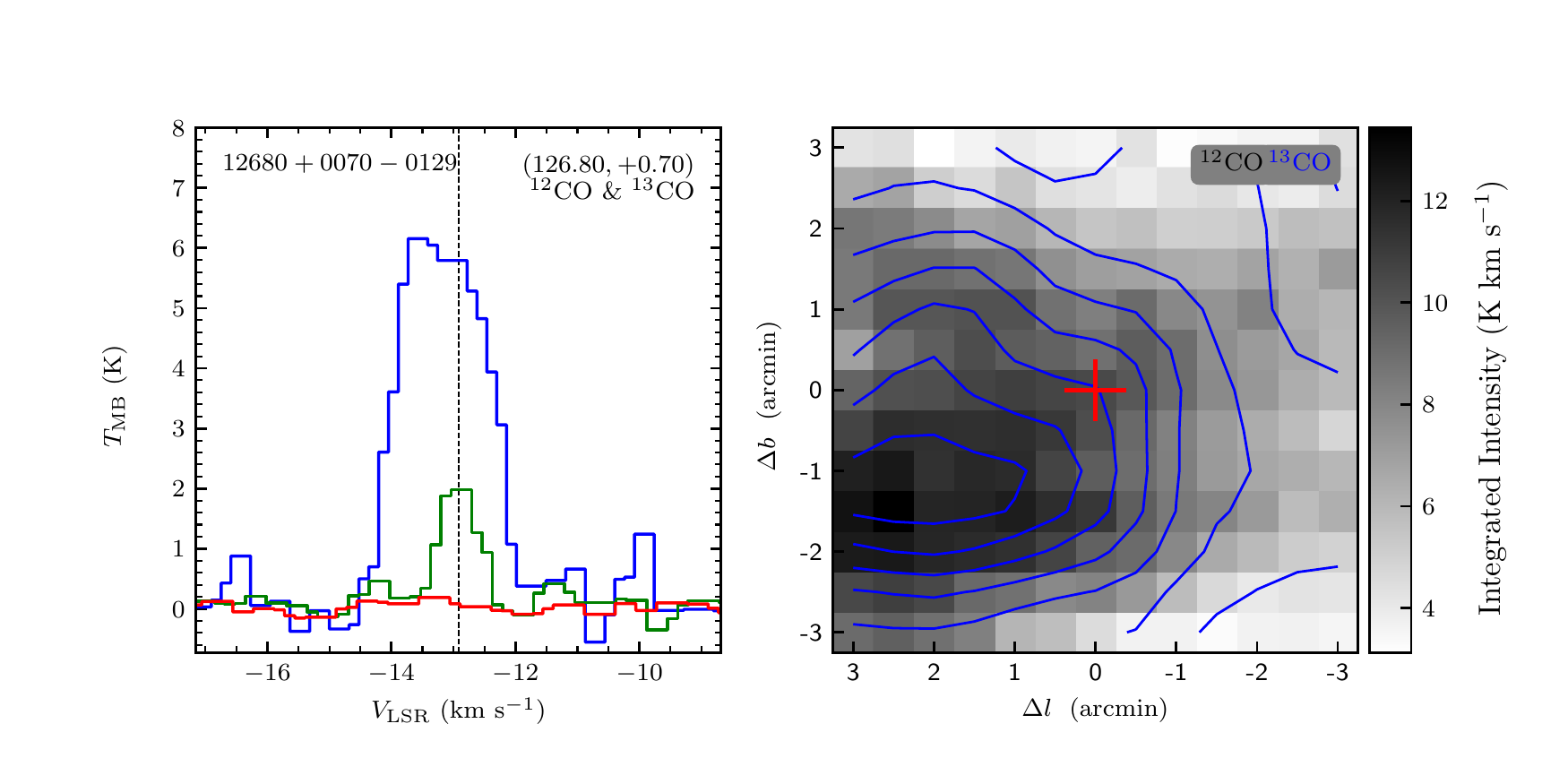}
\end{figure}
\clearpage

\begin{figure}
\includegraphics[width=9.0cm,angle=0]{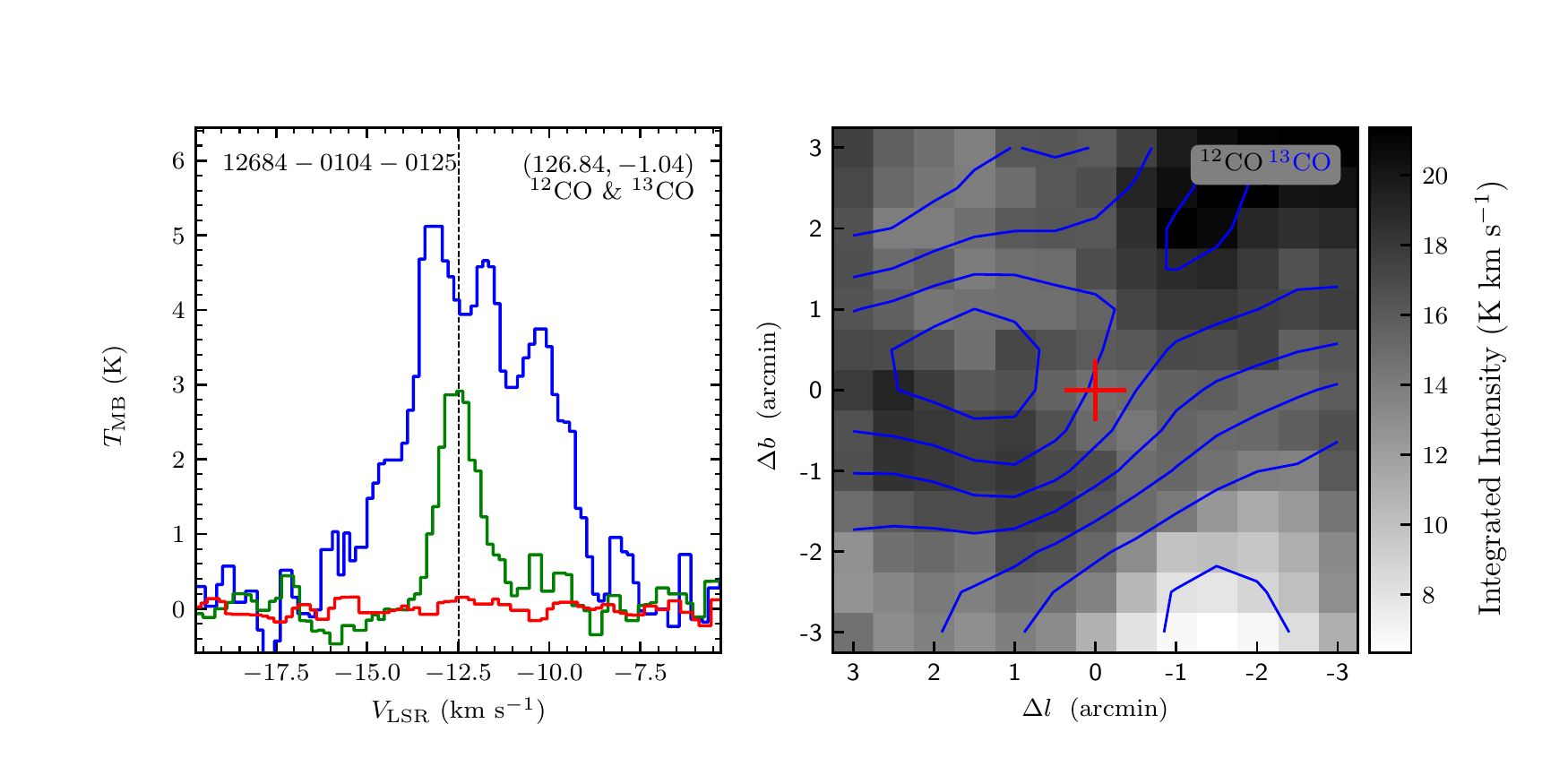}
\includegraphics[width=9.0cm,angle=0]{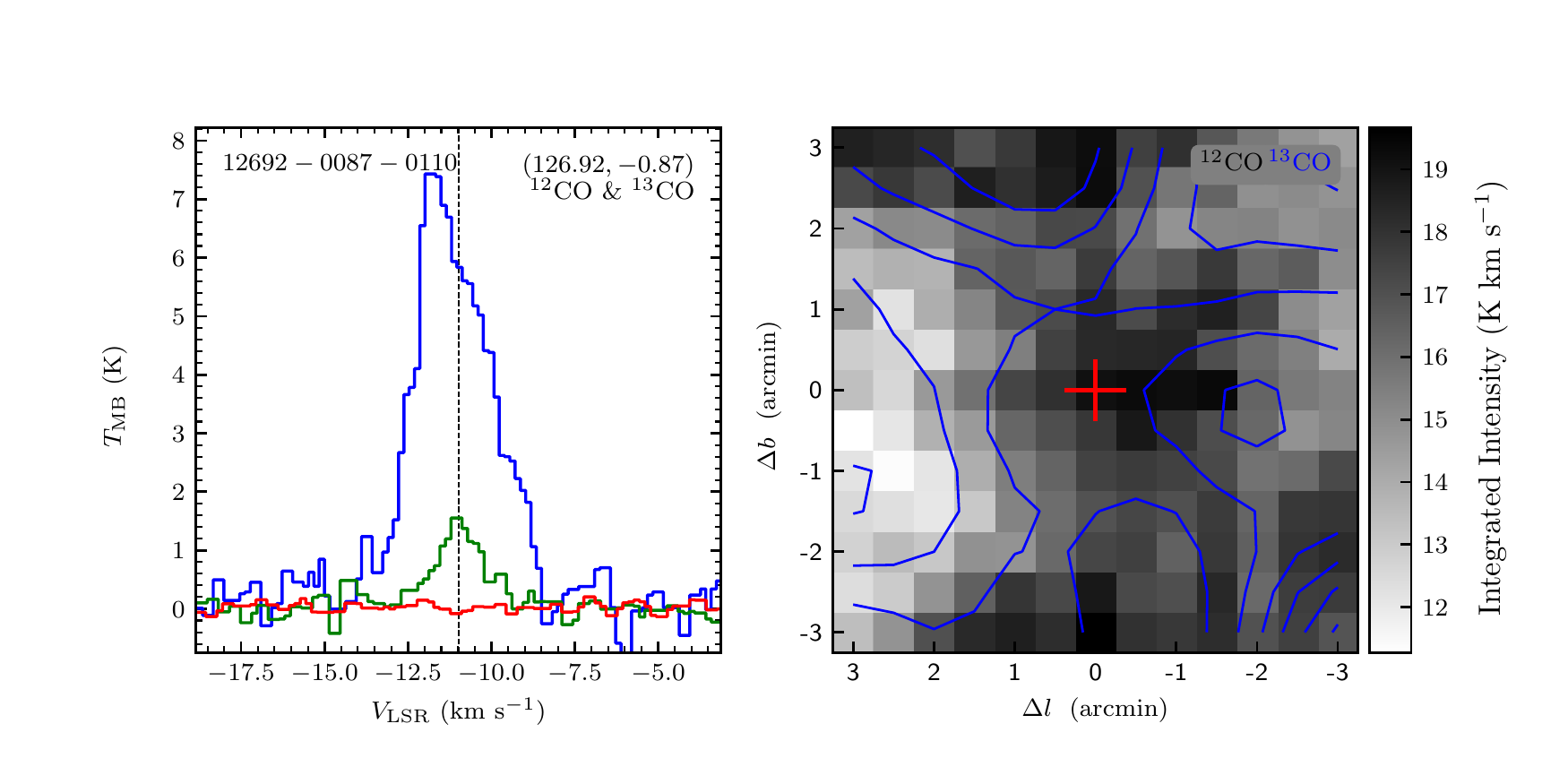}
\vspace{-0.5cm}

\includegraphics[width=9.0cm,angle=0]{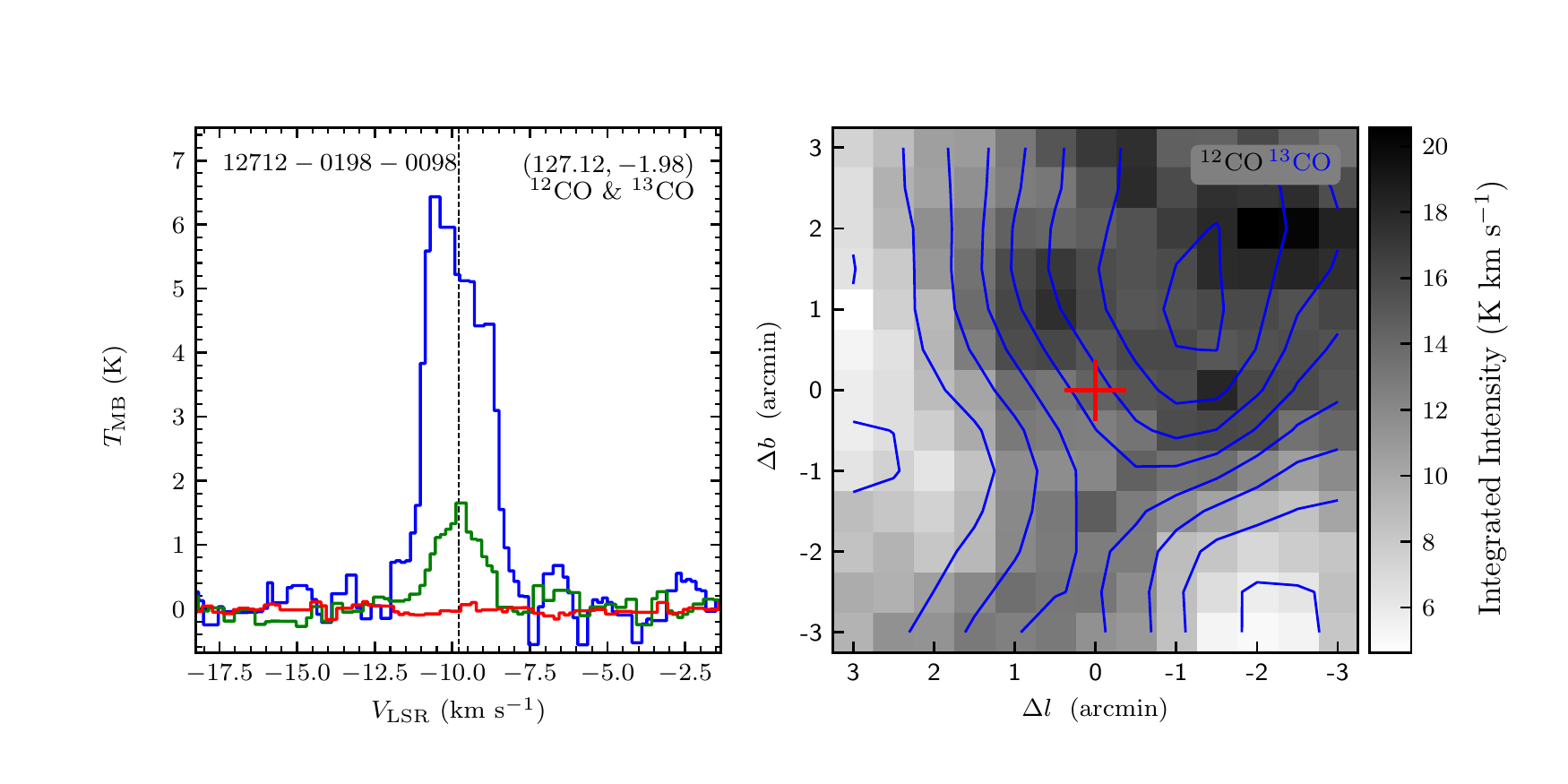}
\includegraphics[width=9.0cm,angle=0]{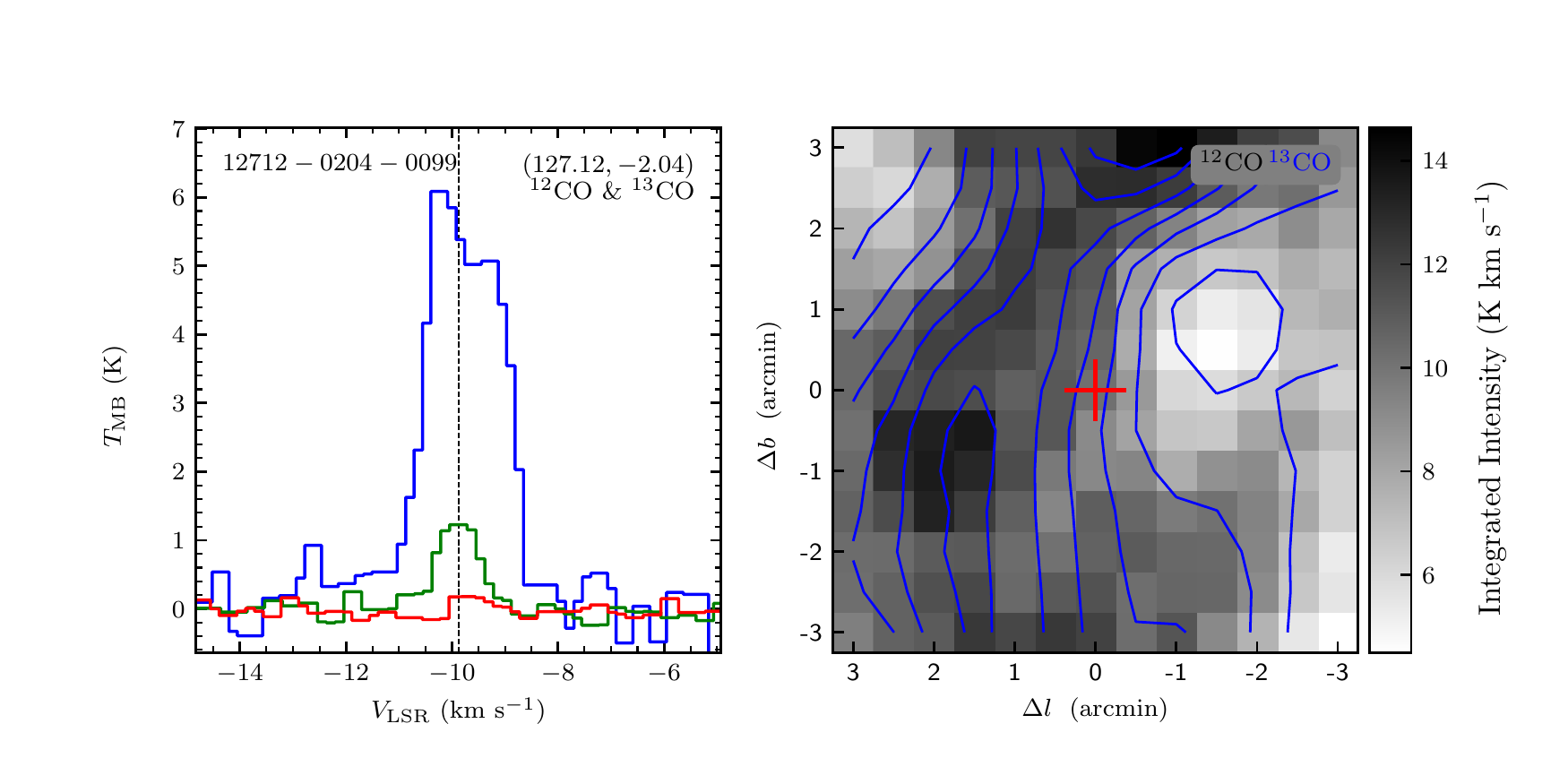}
\vspace{-0.5cm}

\includegraphics[width=9.0cm,angle=0]{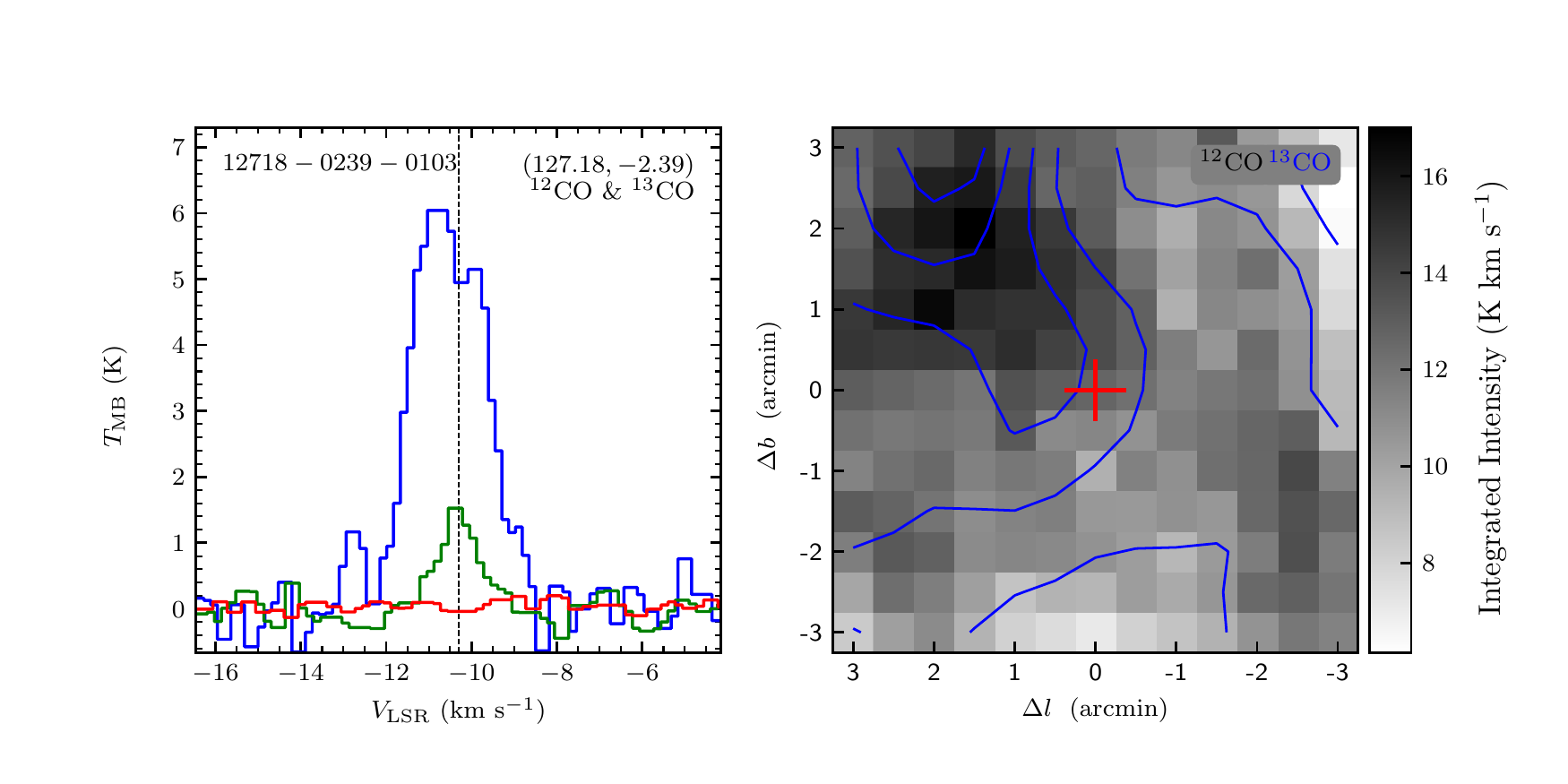}
\includegraphics[width=9.0cm,angle=0]{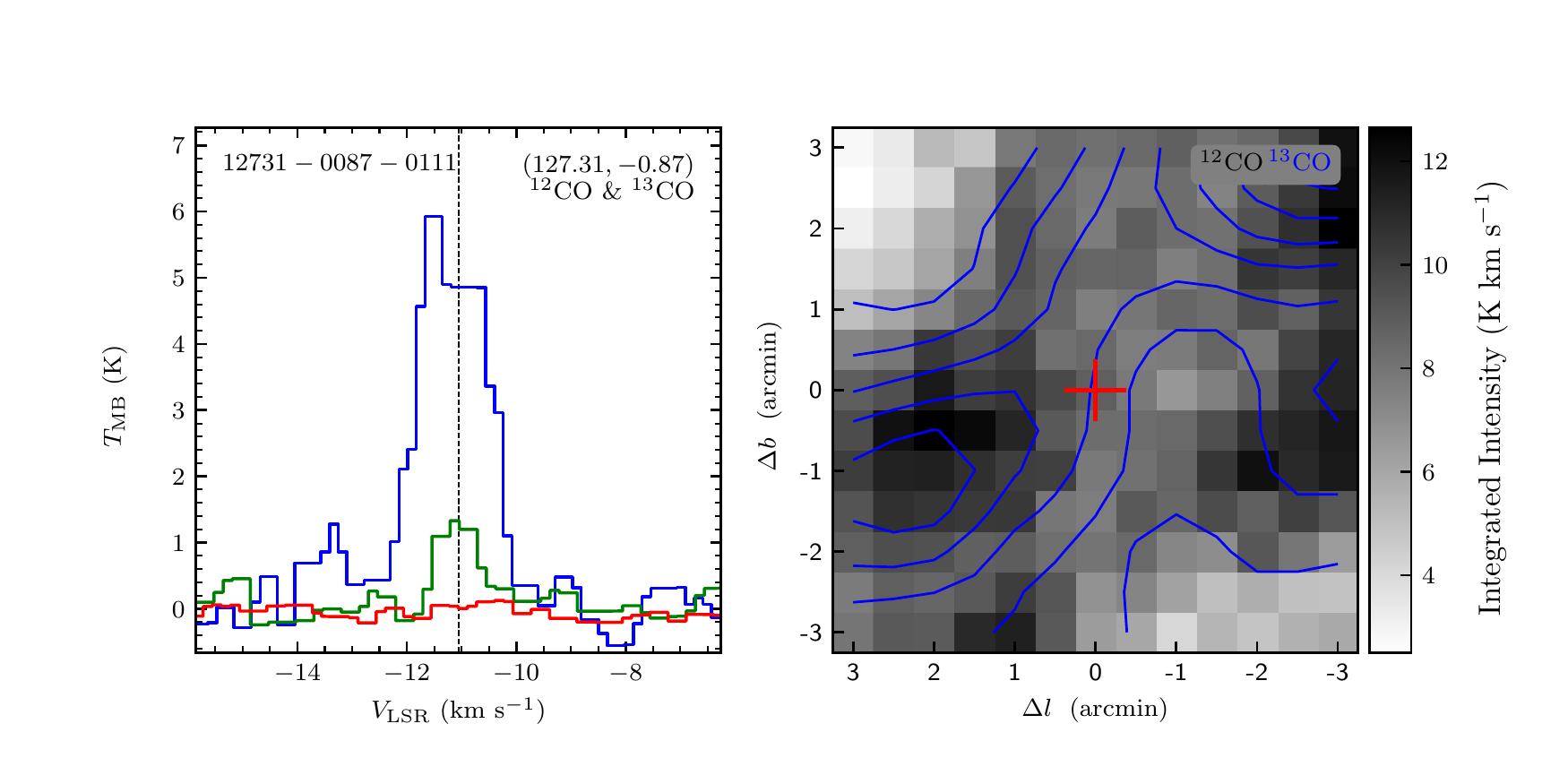}
\vspace{-0.5cm}

\includegraphics[width=9.0cm,angle=0]{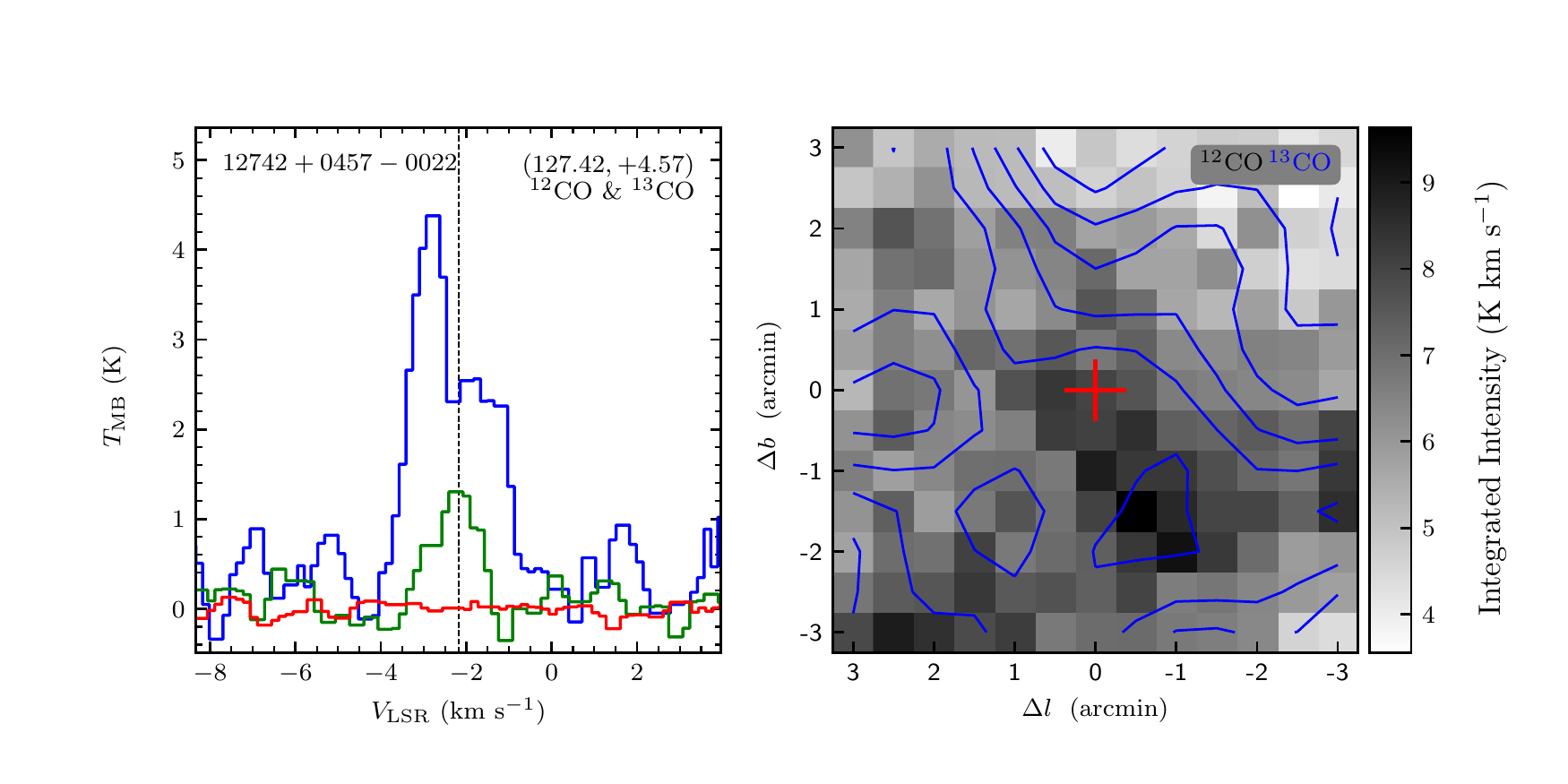}
\includegraphics[width=9.0cm,angle=0]{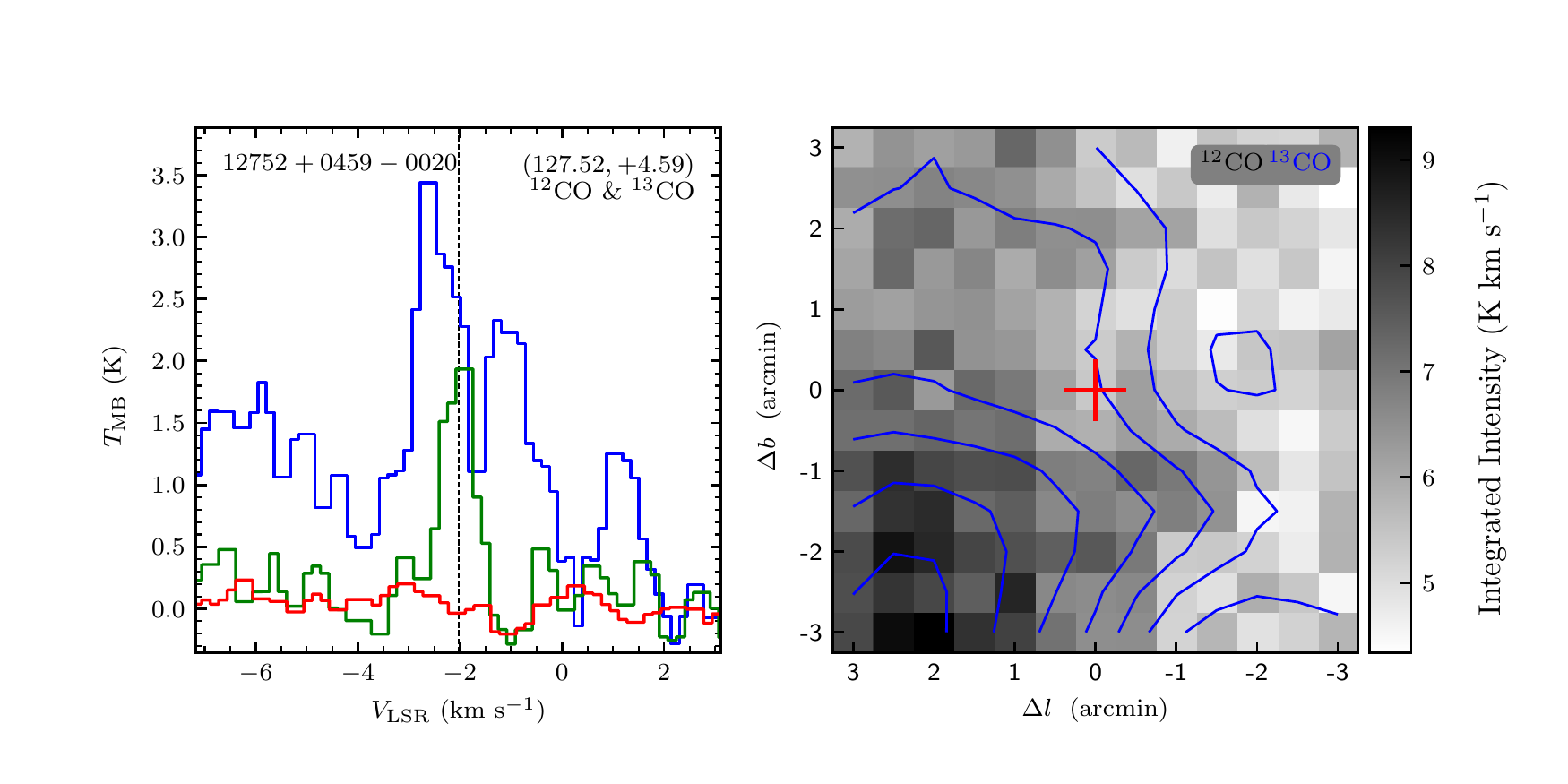}
\vspace{-0.5cm}

\includegraphics[width=9.0cm,angle=0]{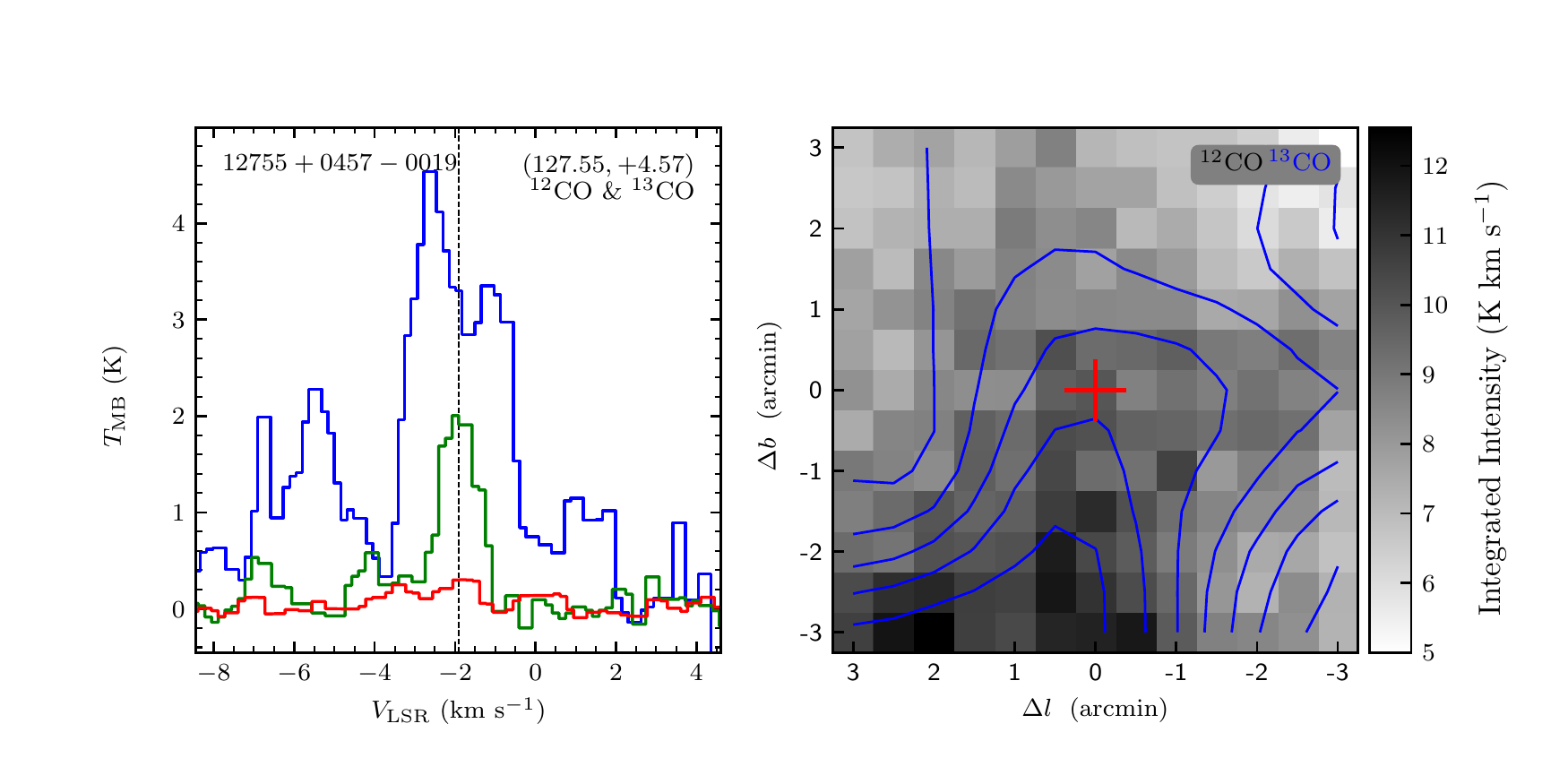}
\includegraphics[width=9.0cm,angle=0]{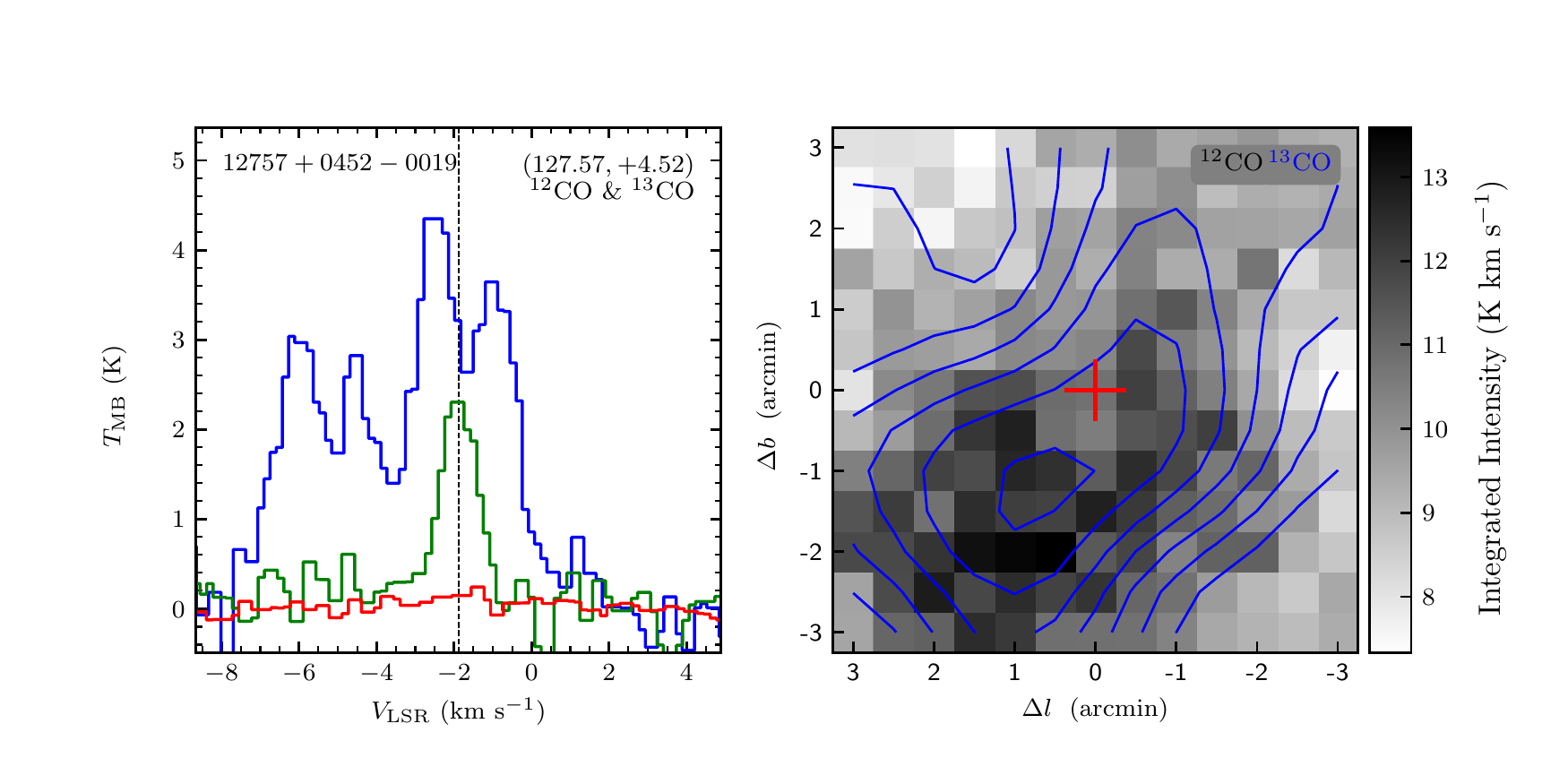}
\end{figure}
\clearpage

\begin{figure}
\includegraphics[width=9.0cm,angle=0]{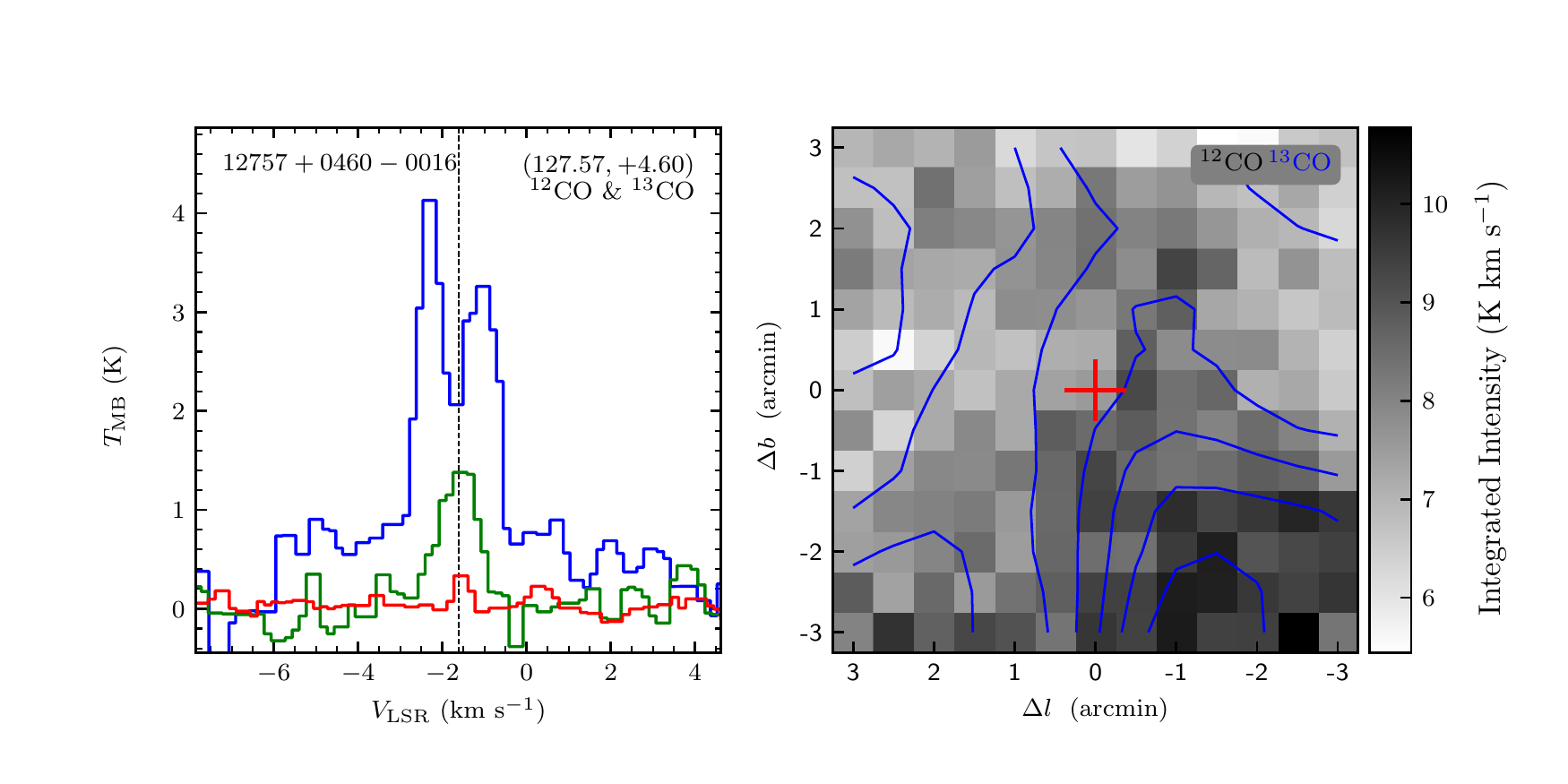}
\includegraphics[width=9.0cm,angle=0]{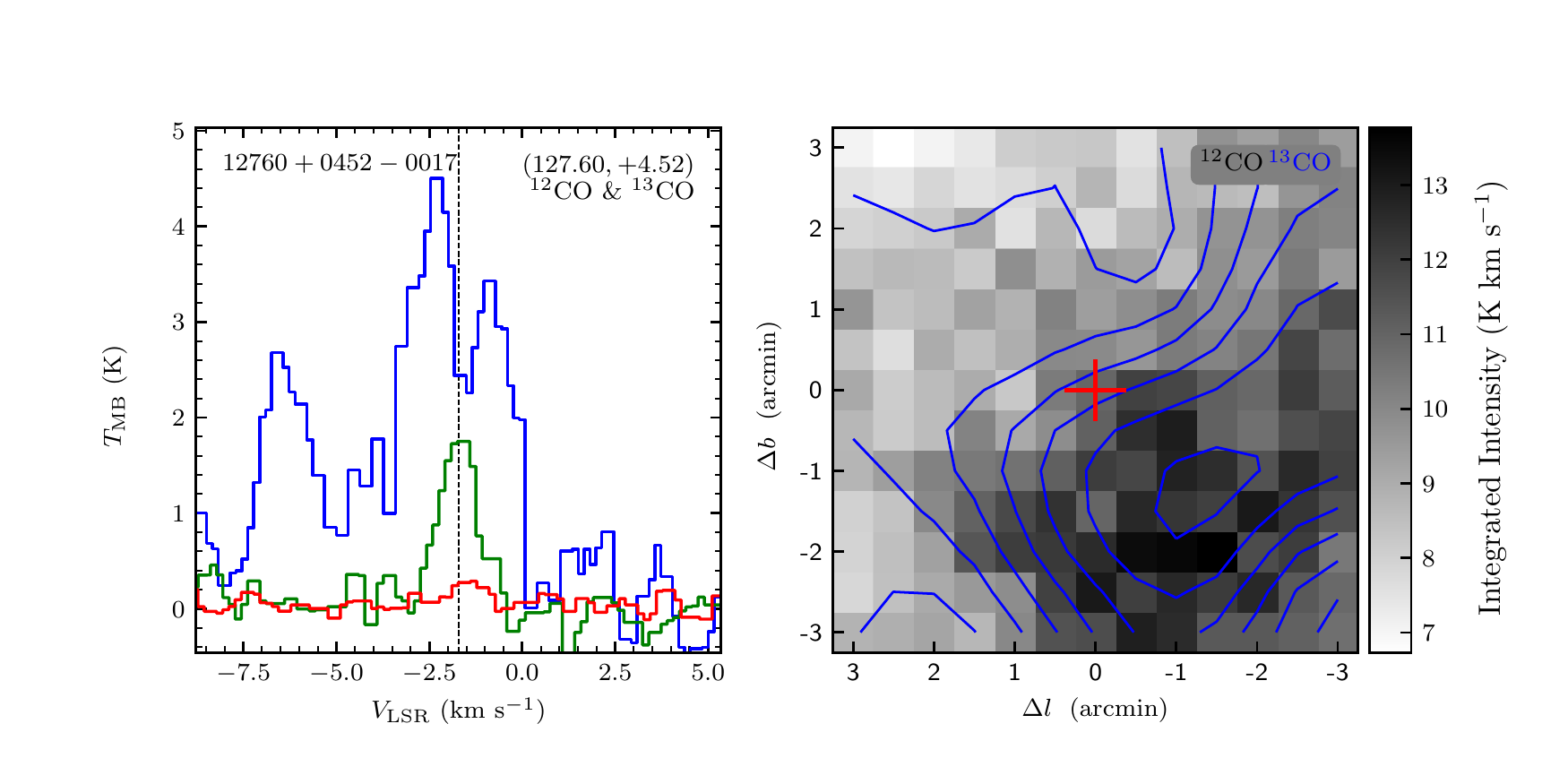}
\vspace{-0.5cm}

\includegraphics[width=9.0cm,angle=0]{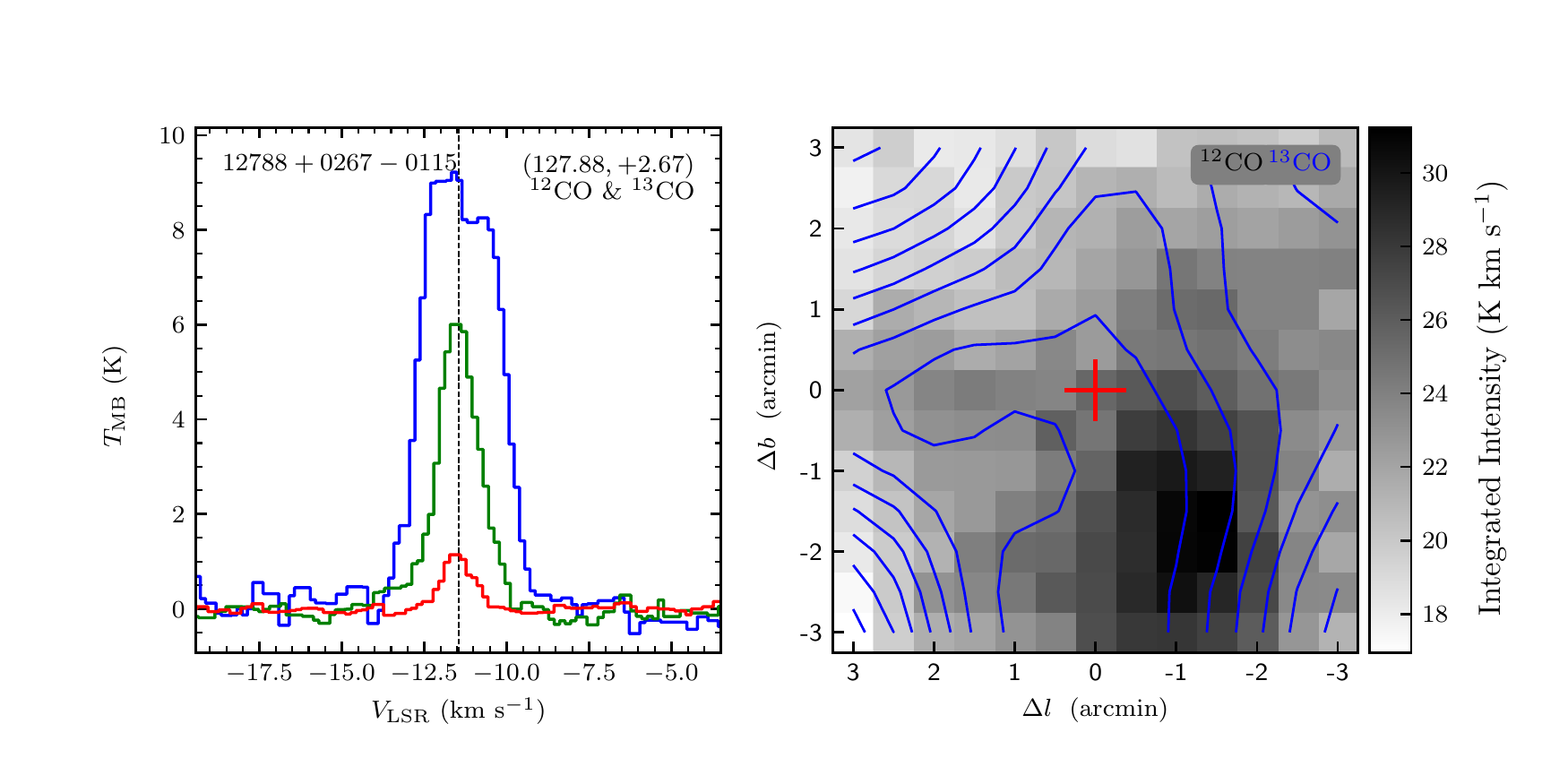}
\includegraphics[width=9.0cm,angle=0]{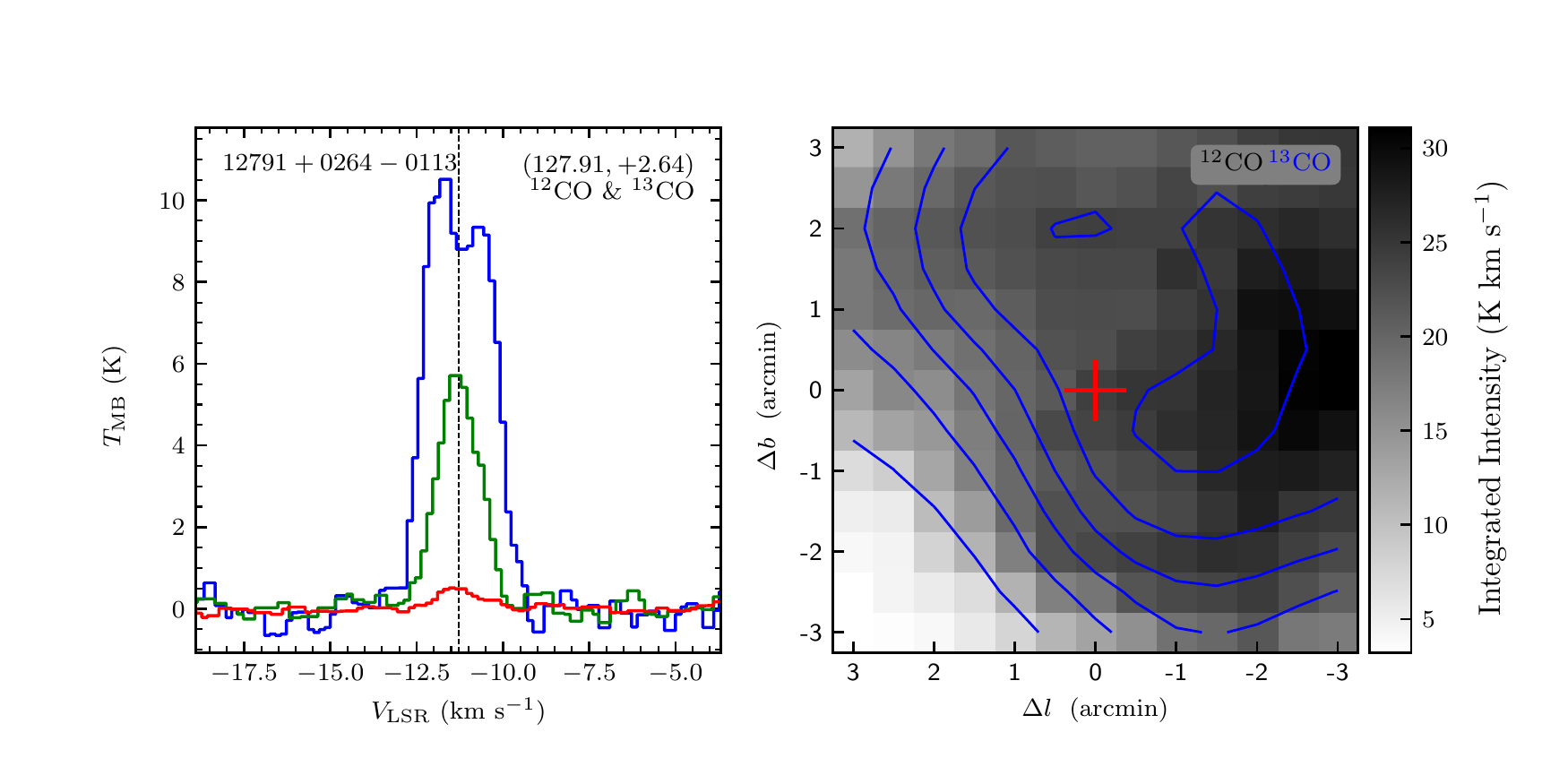}
\vspace{-0.5cm}

\includegraphics[width=9.0cm,angle=0]{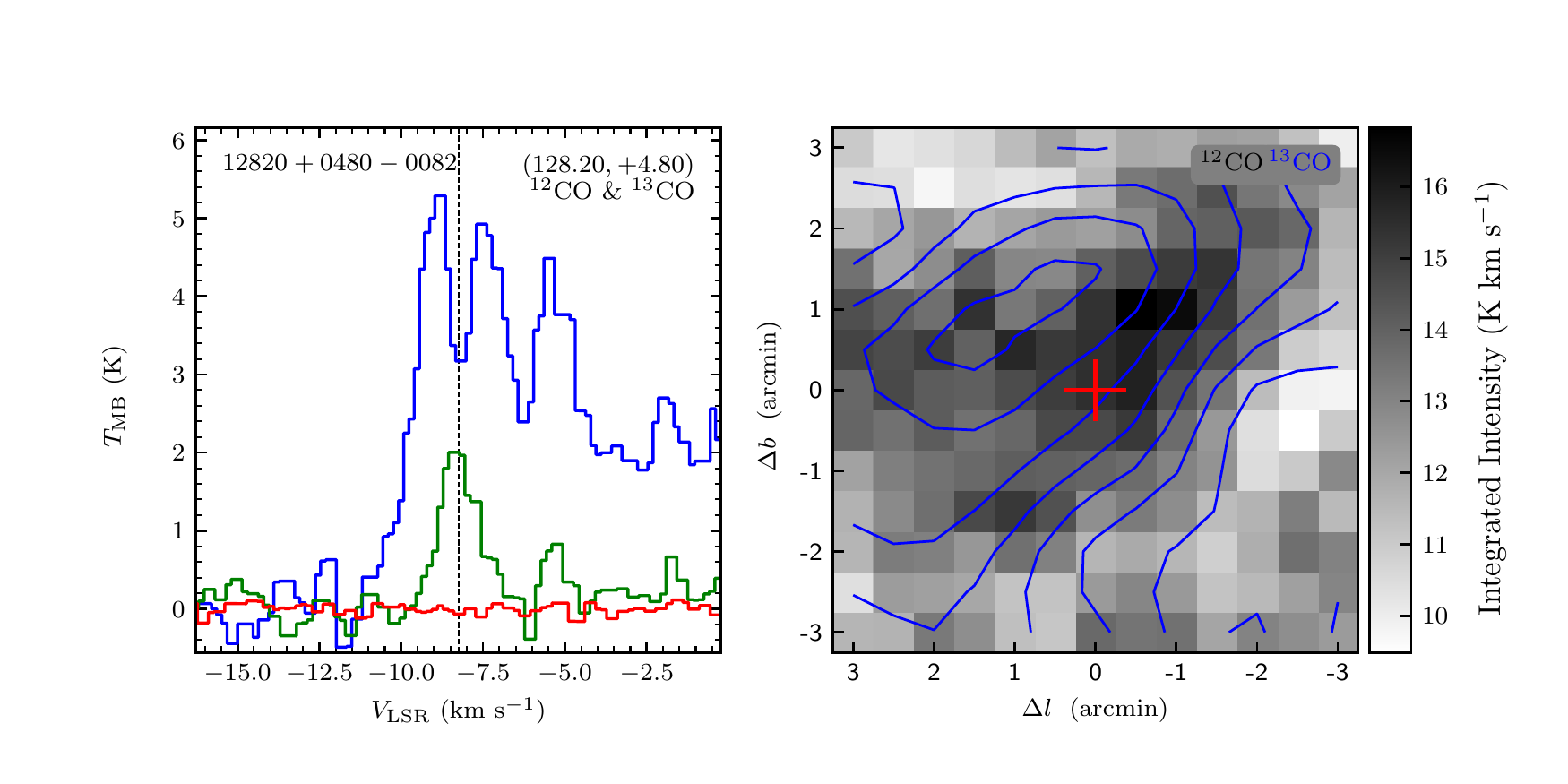}
\includegraphics[width=9.0cm,angle=0]{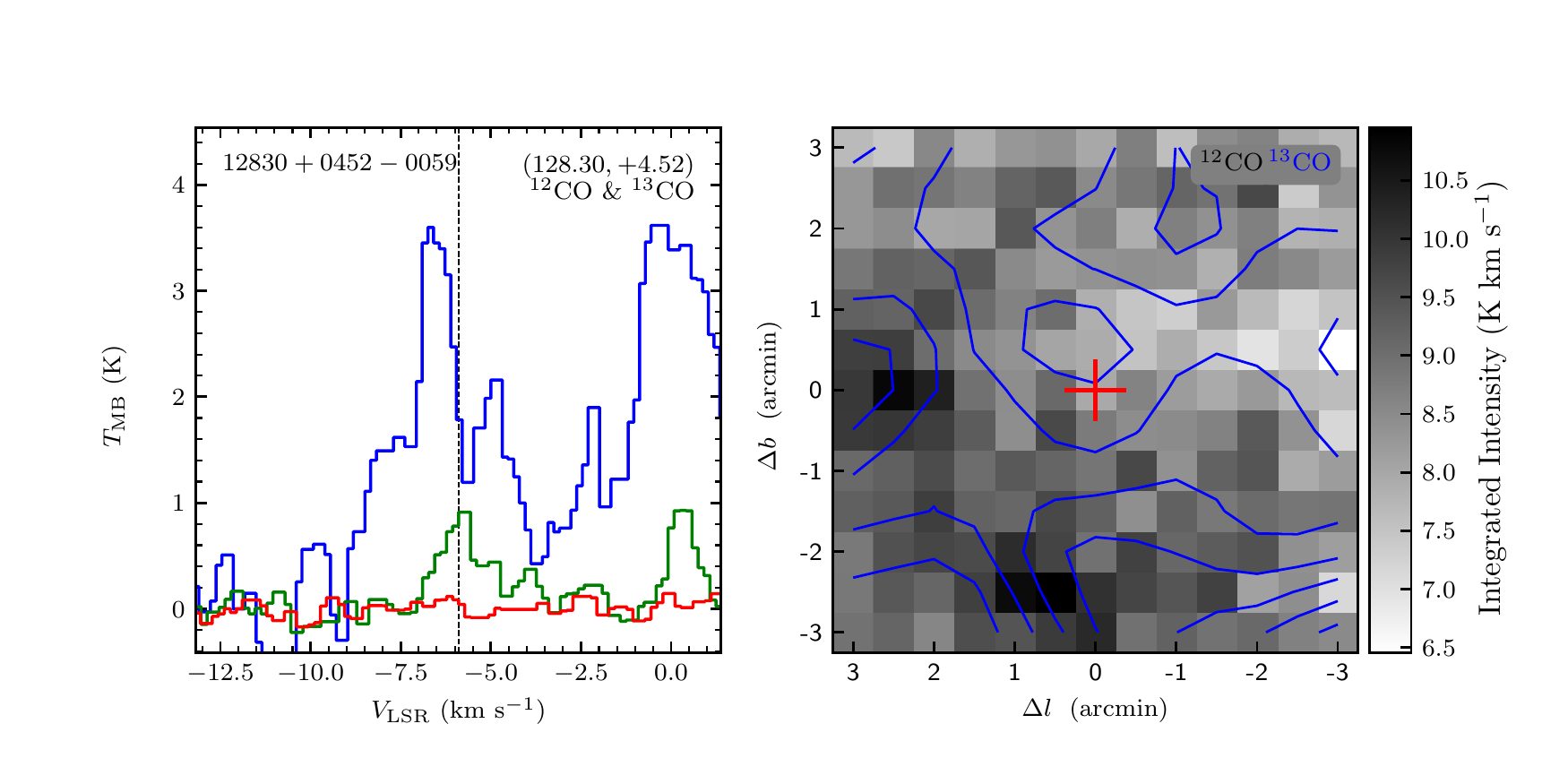}
\vspace{-0.5cm}

\includegraphics[width=9.0cm,angle=0]{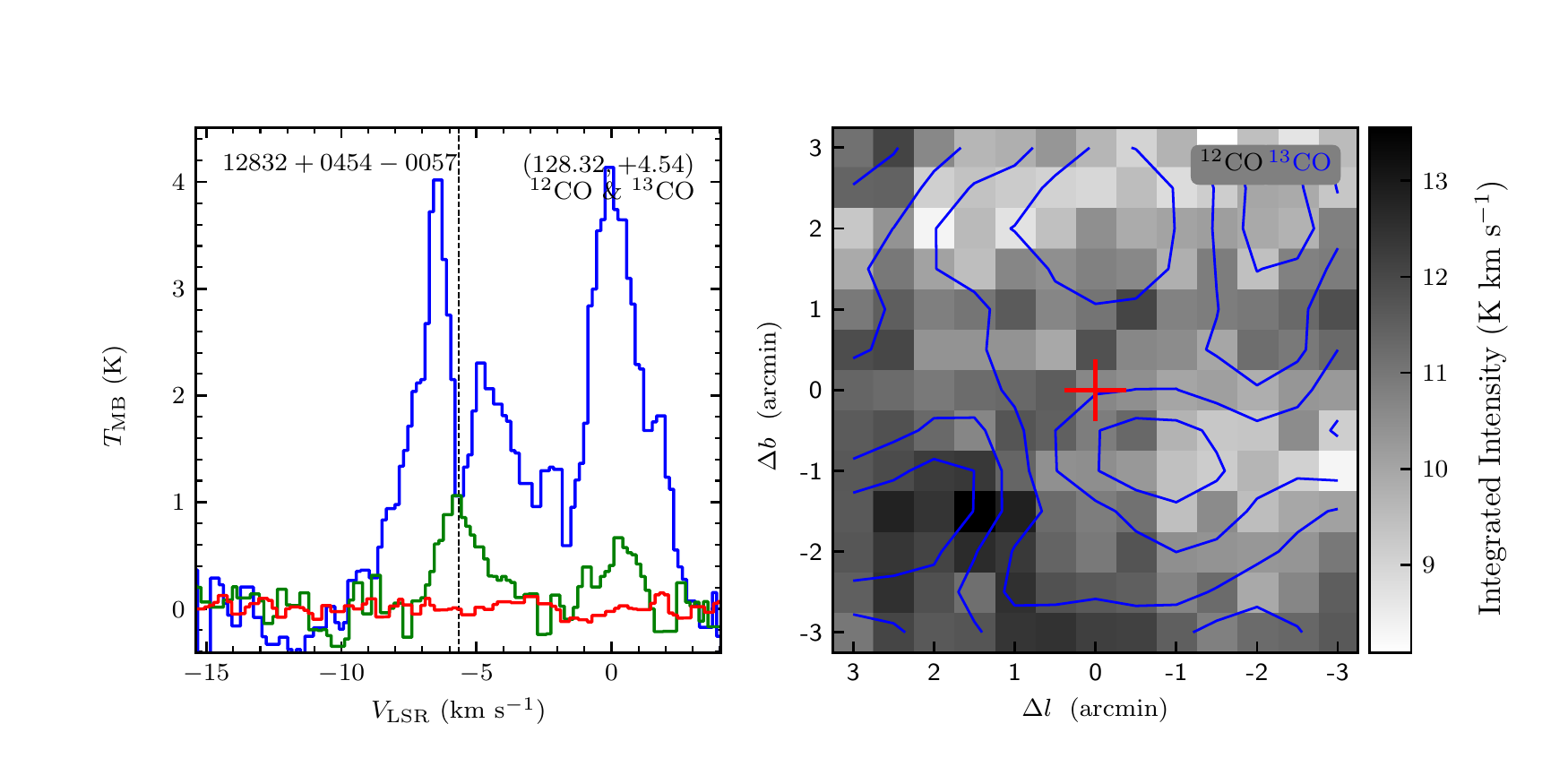}
\includegraphics[width=9.0cm,angle=0]{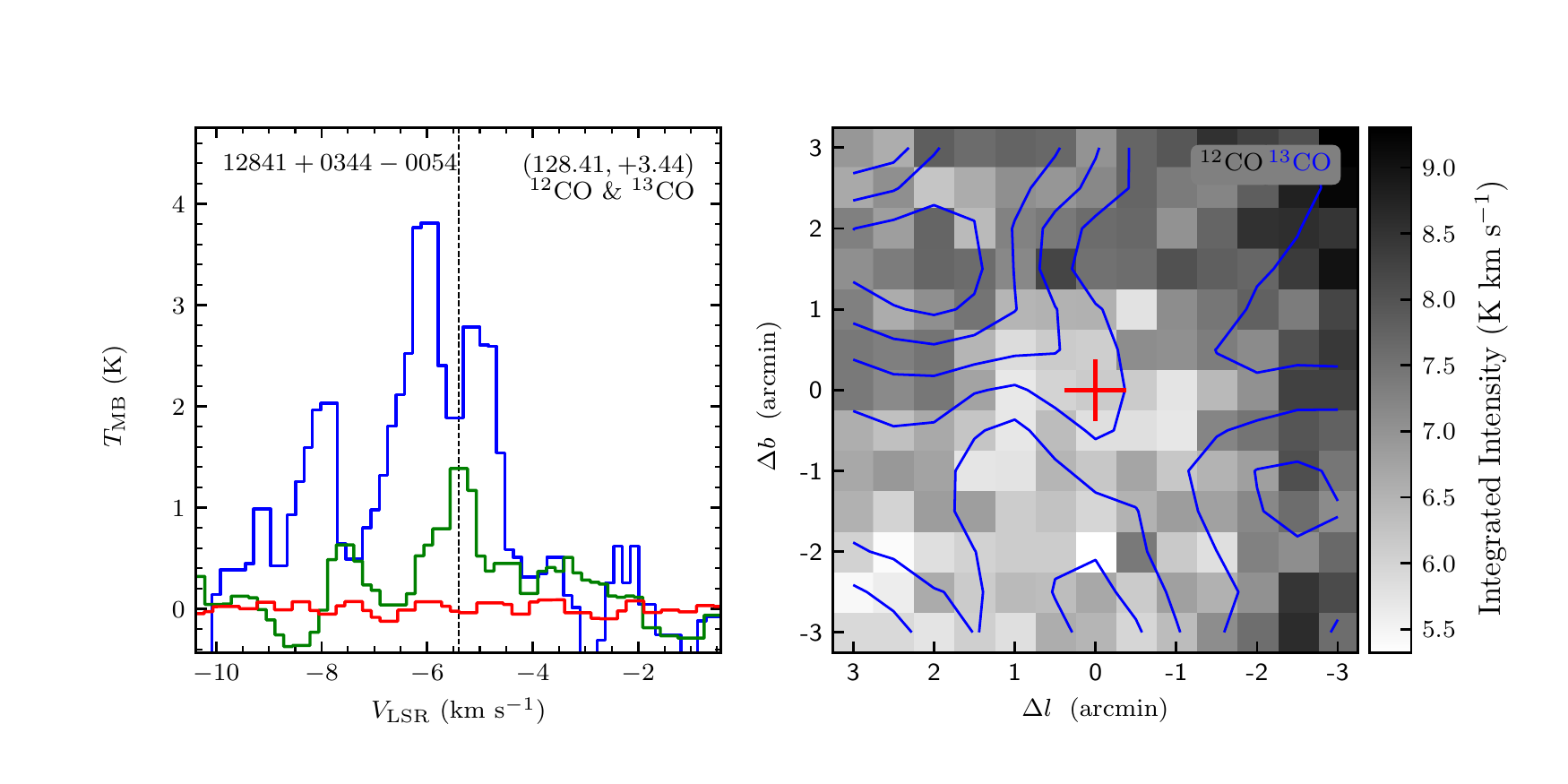}
\vspace{-0.5cm}

\includegraphics[width=9.0cm,angle=0]{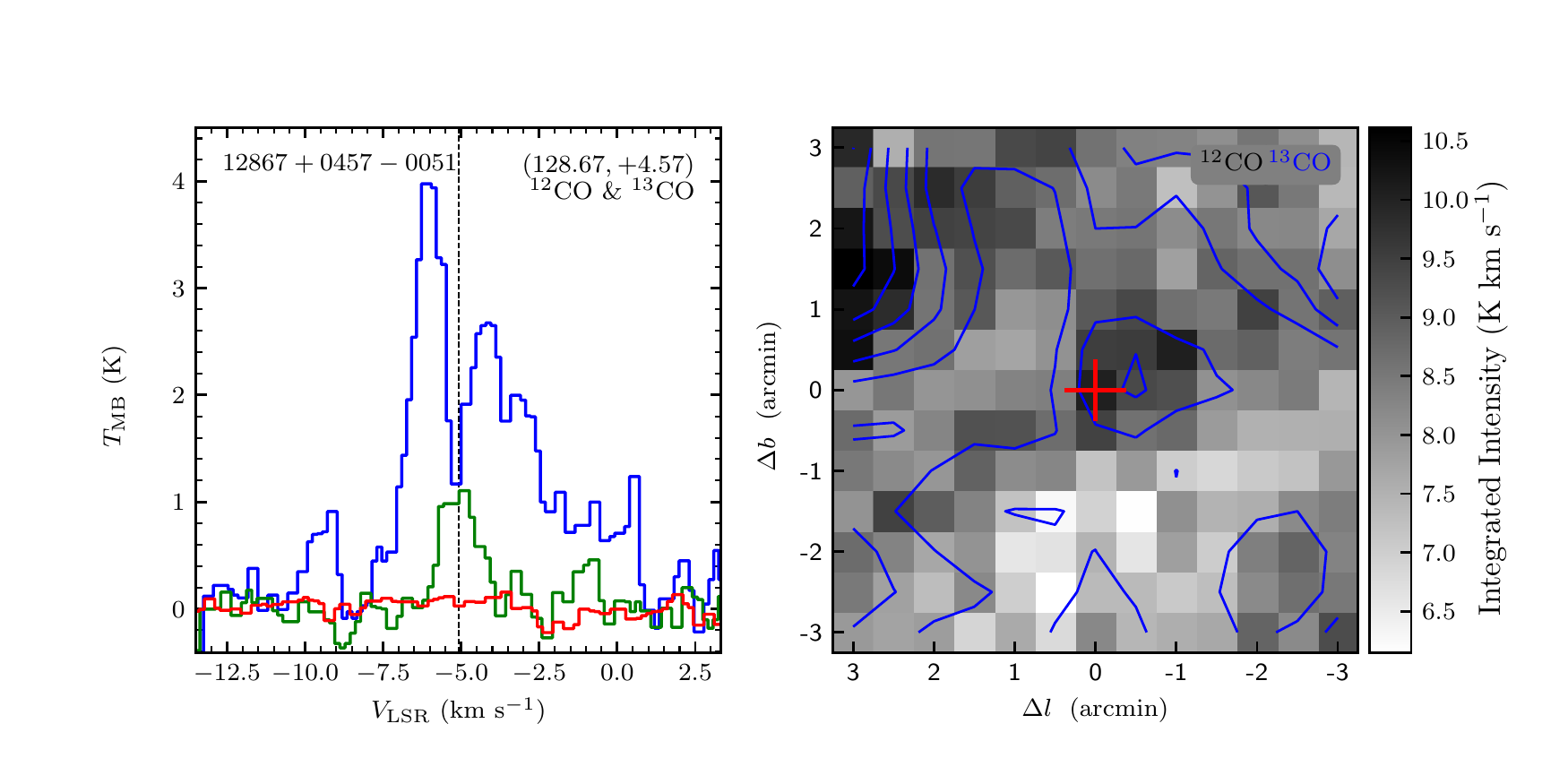}
\includegraphics[width=9.0cm,angle=0]{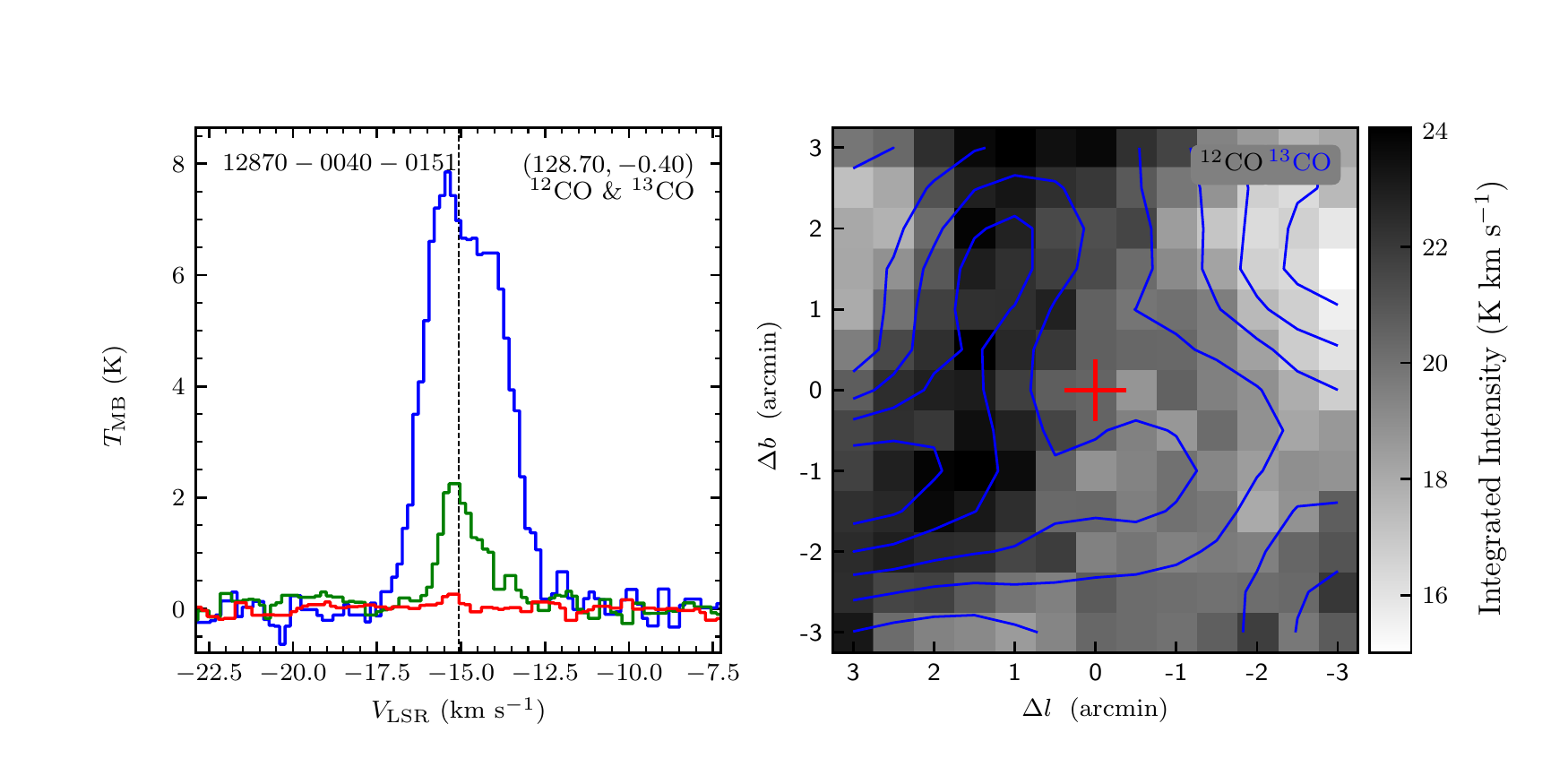}
\end{figure}
\clearpage

\begin{figure}
\includegraphics[width=9.0cm,angle=0]{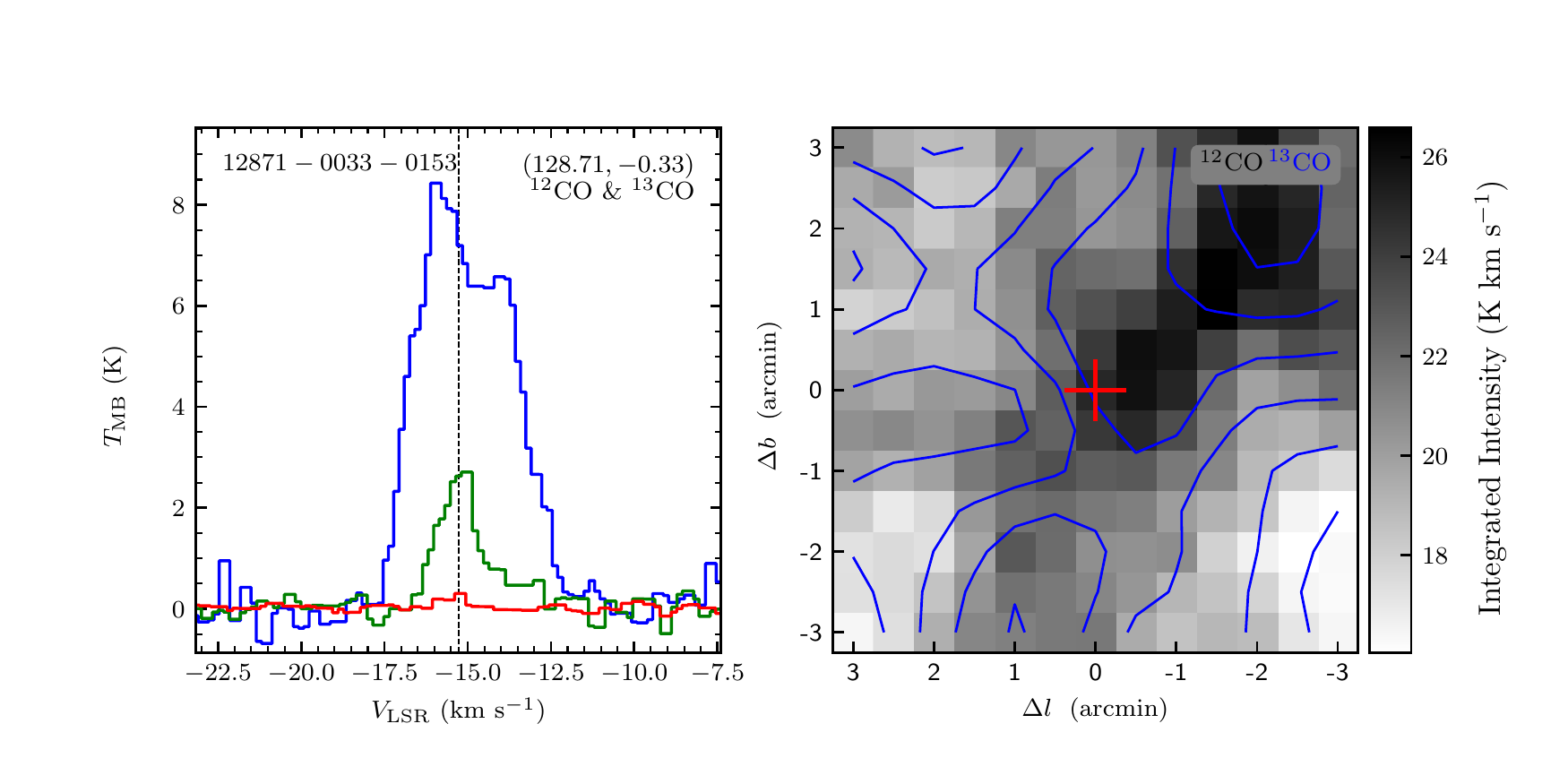}
\includegraphics[width=9.0cm,angle=0]{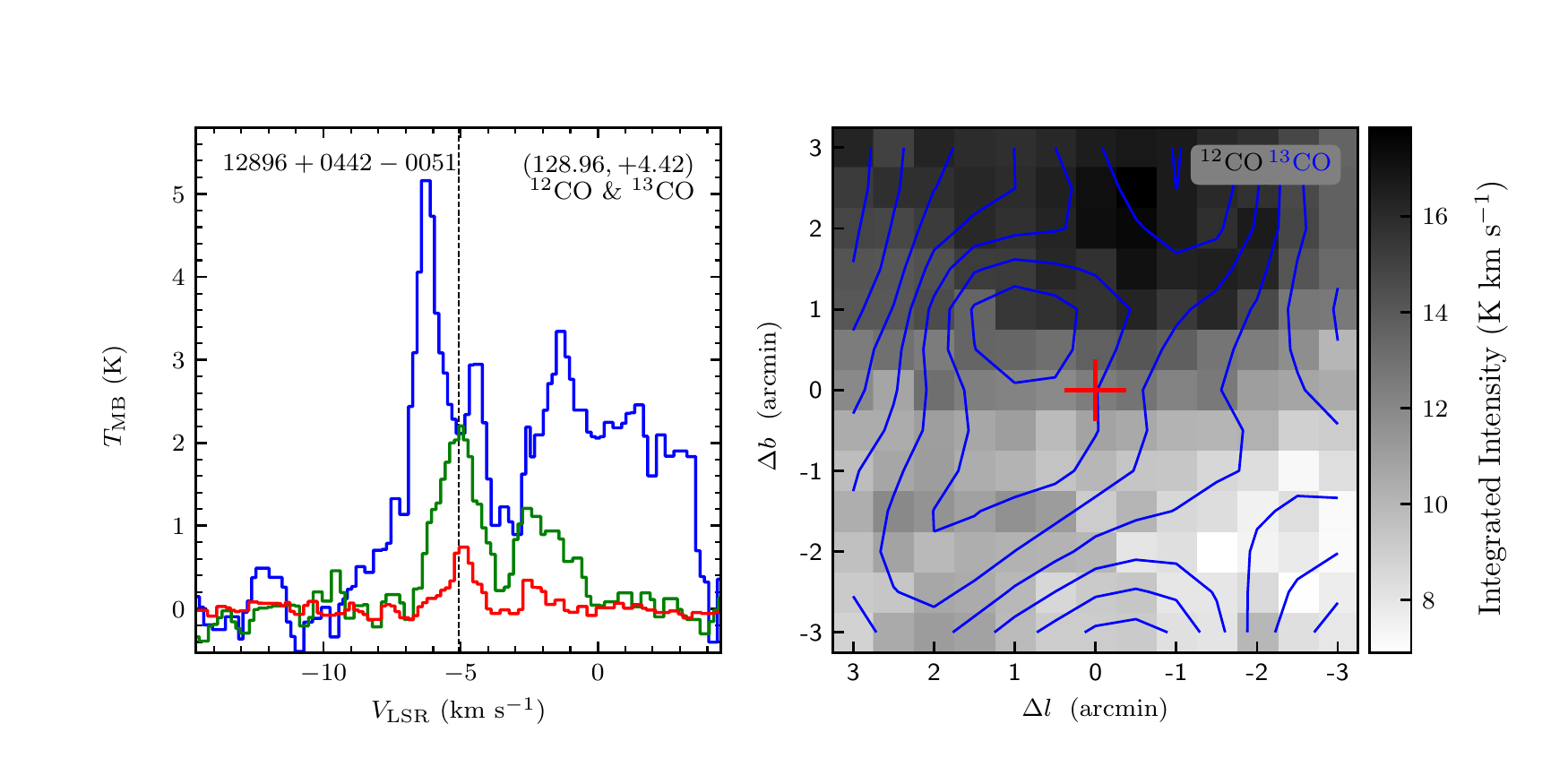}
\vspace{-0.5cm}

\includegraphics[width=9.0cm,angle=0]{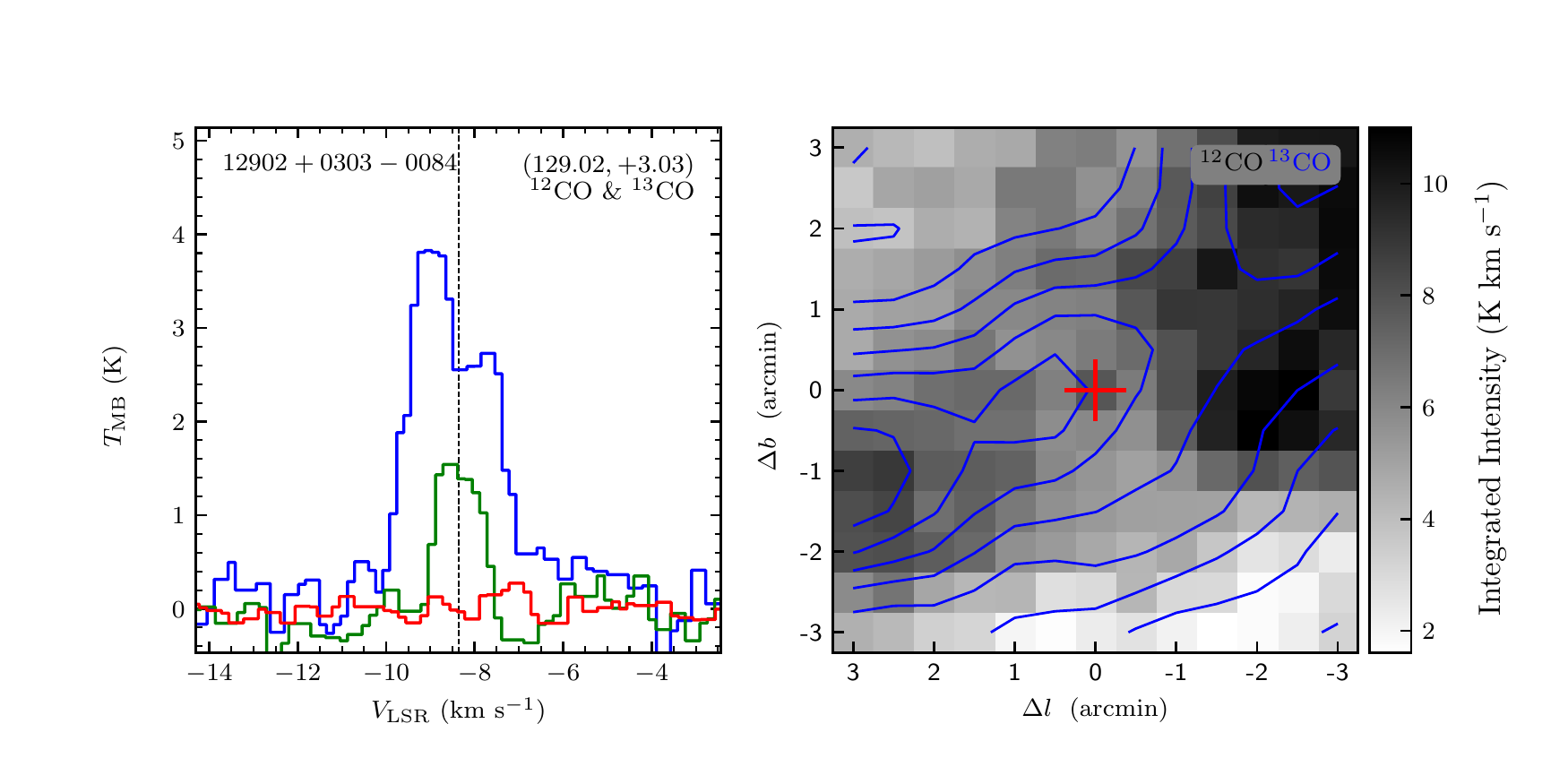}
\includegraphics[width=9.0cm,angle=0]{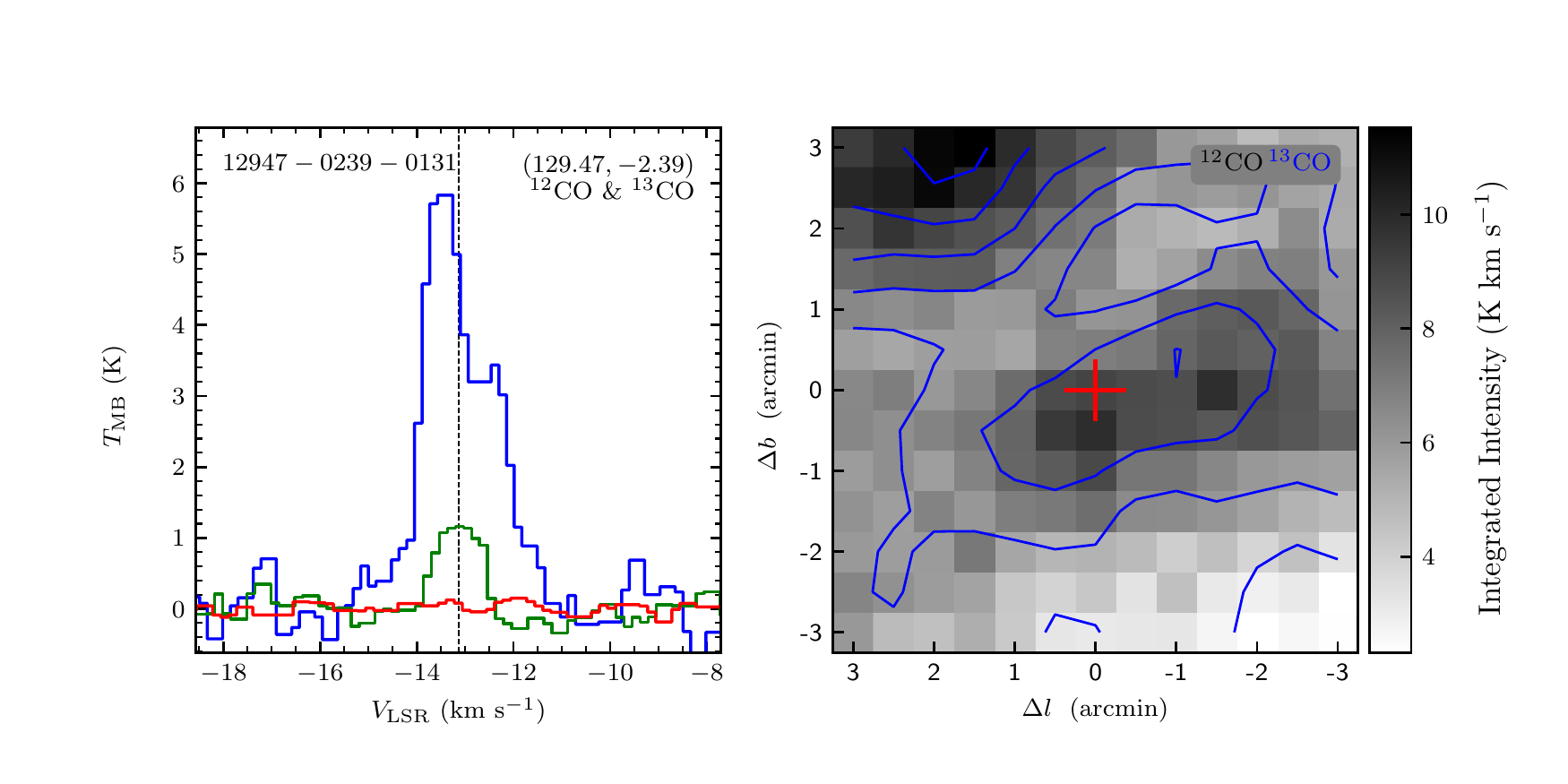}
\vspace{-0.5cm}

\includegraphics[width=9.0cm,angle=0]{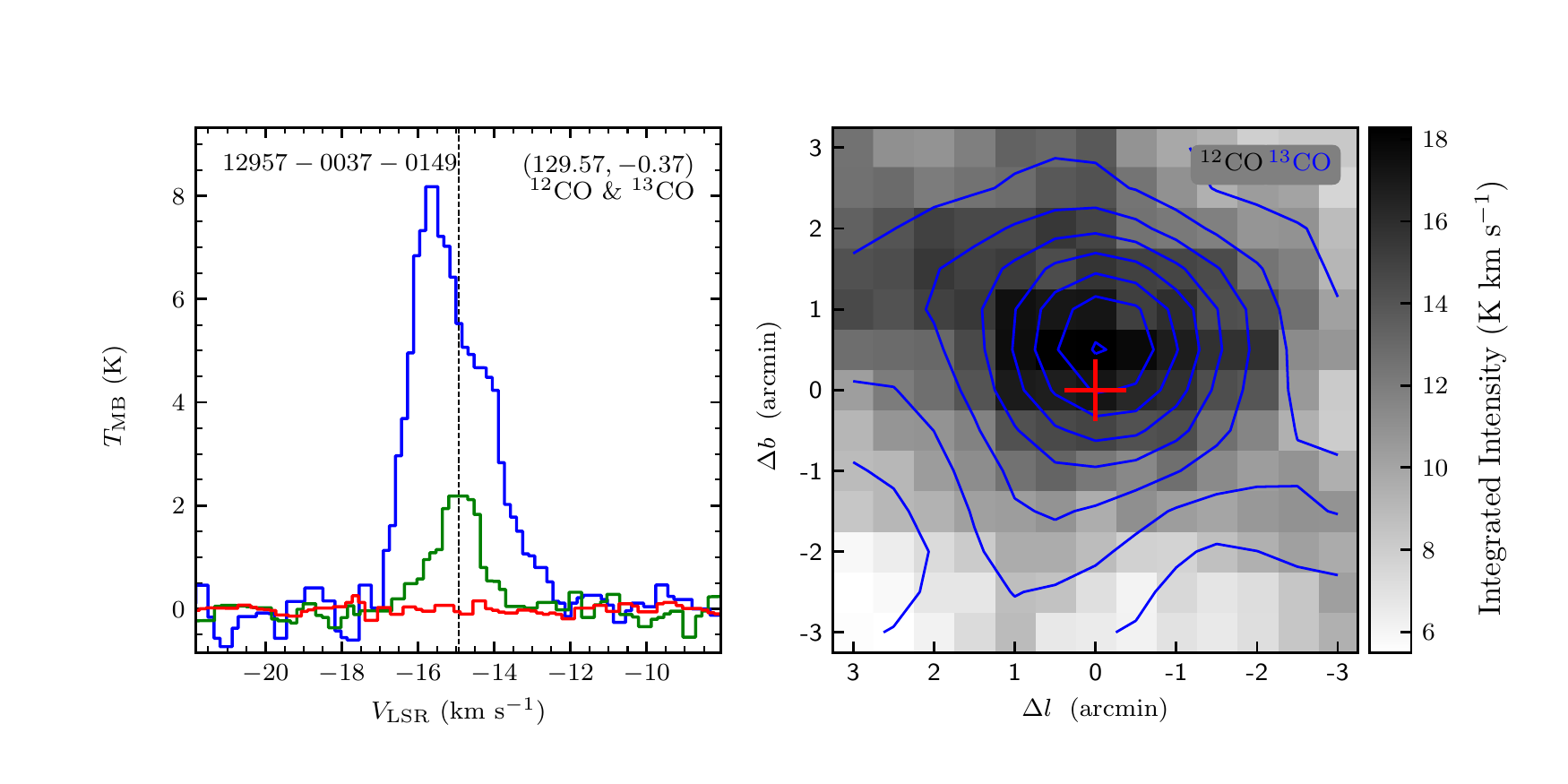}
\includegraphics[width=9.0cm,angle=0]{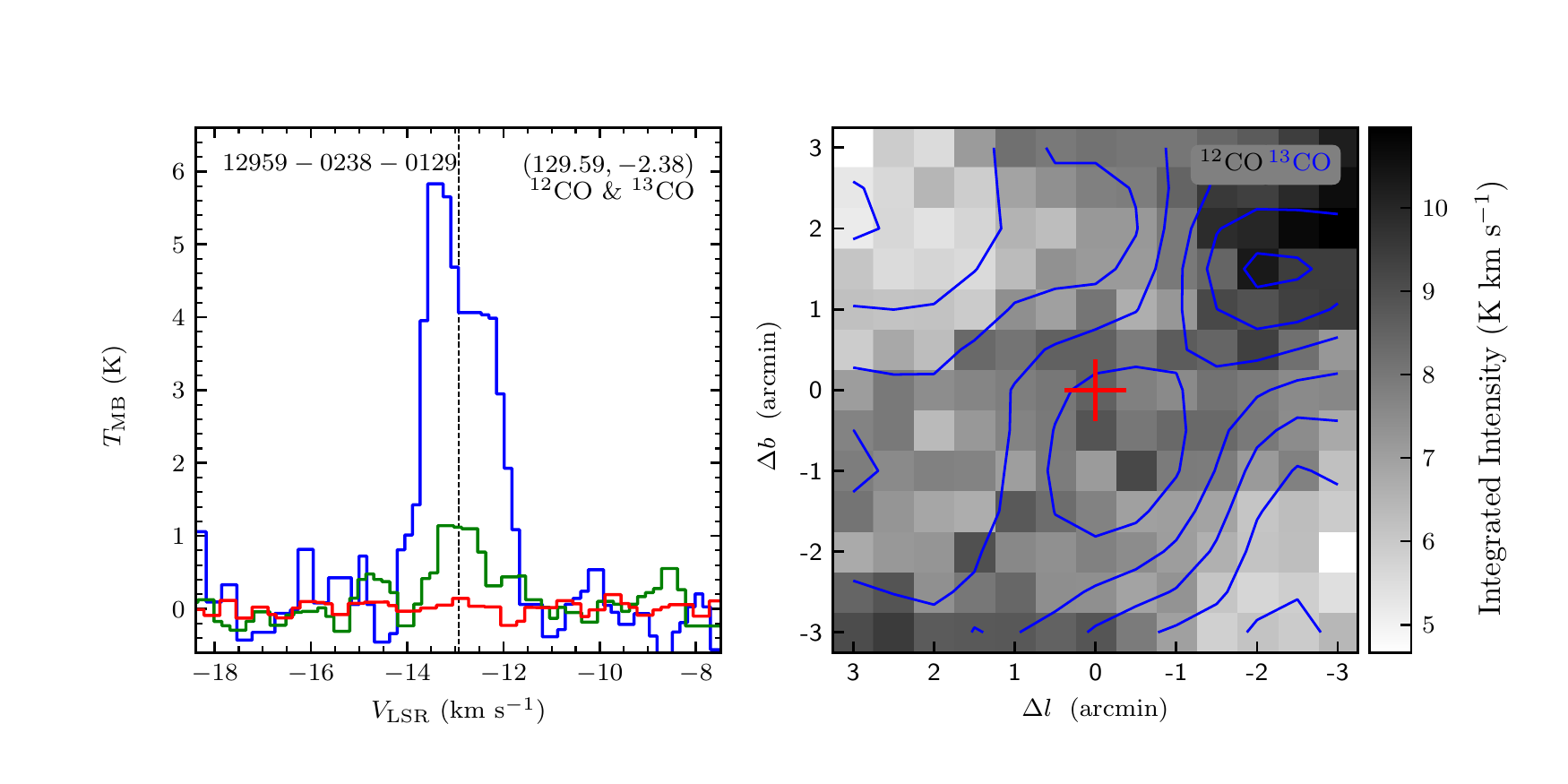}
\vspace{-0.5cm}

\includegraphics[width=9.0cm,angle=0]{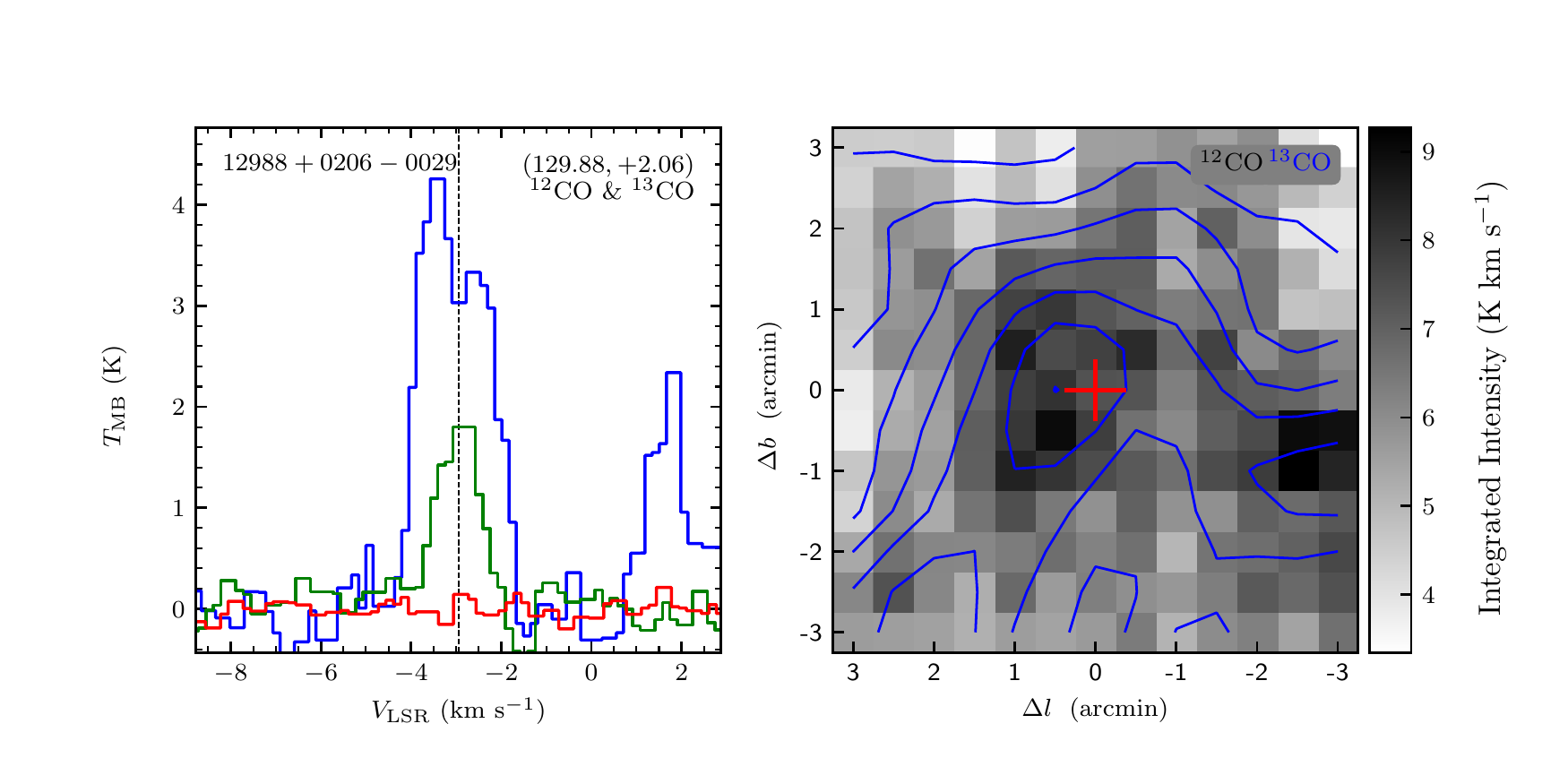}
\includegraphics[width=9.0cm,angle=0]{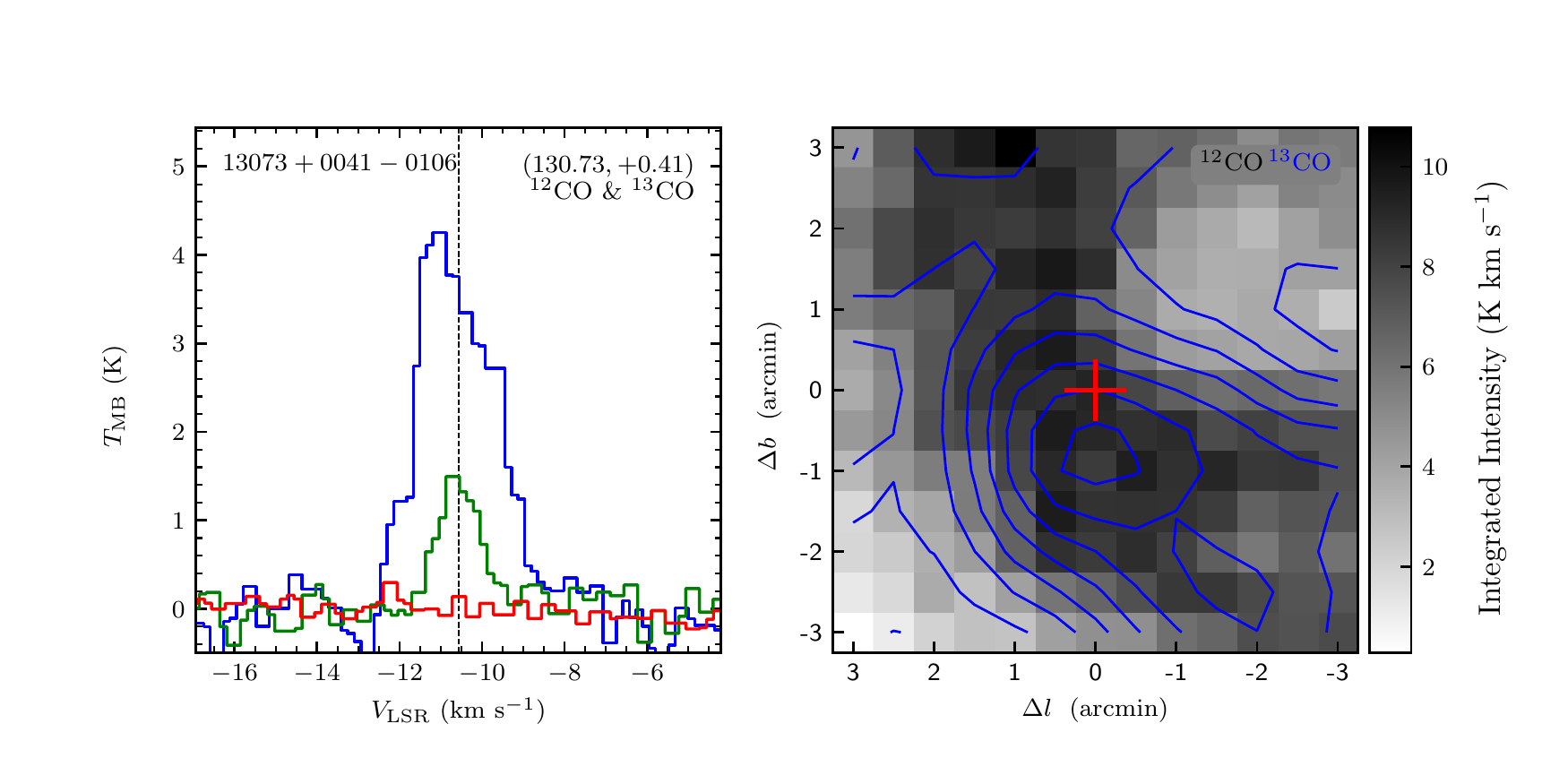}
\vspace{-0.5cm}

\includegraphics[width=9.0cm,angle=0]{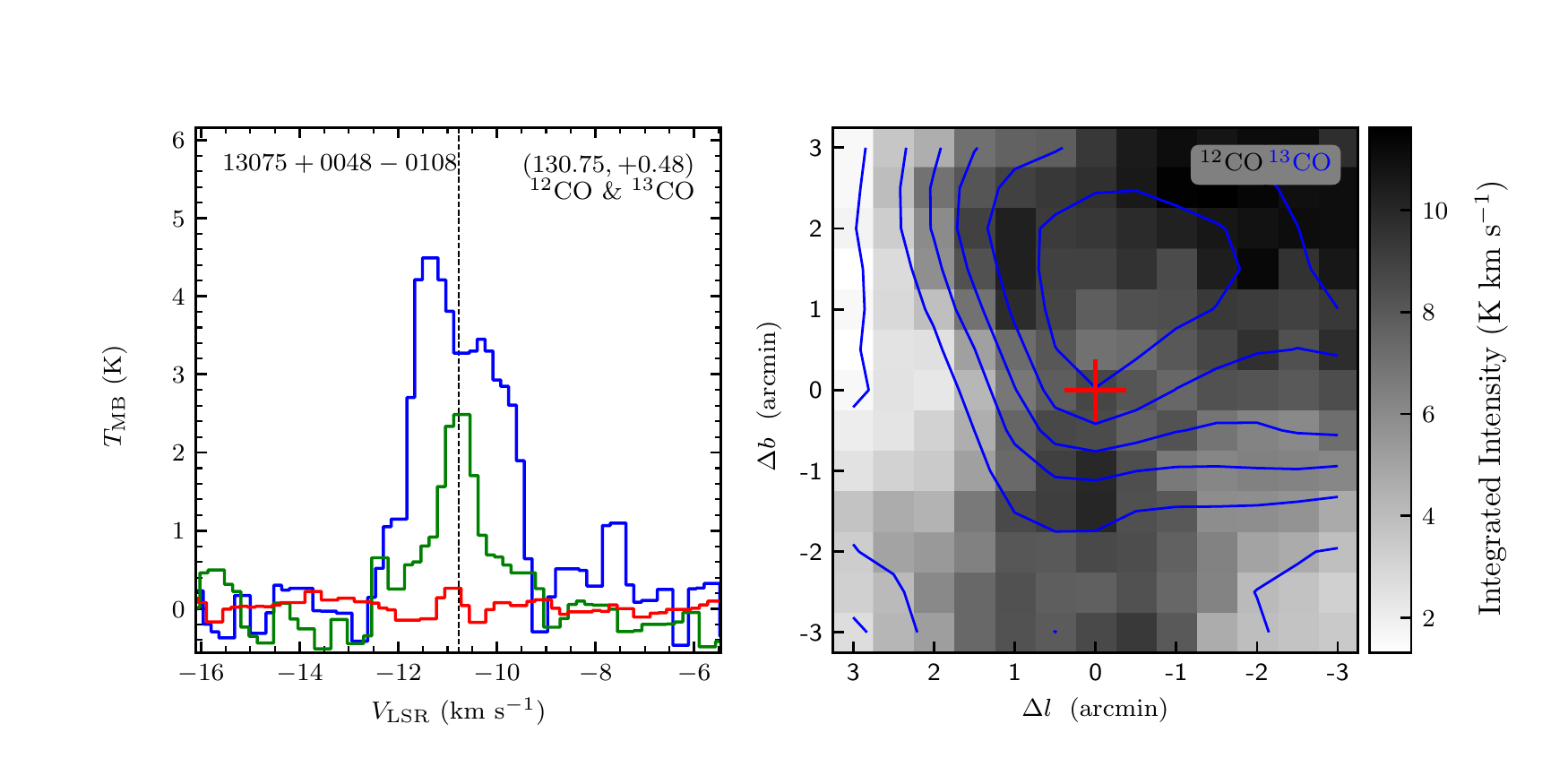}
\includegraphics[width=9.0cm,angle=0]{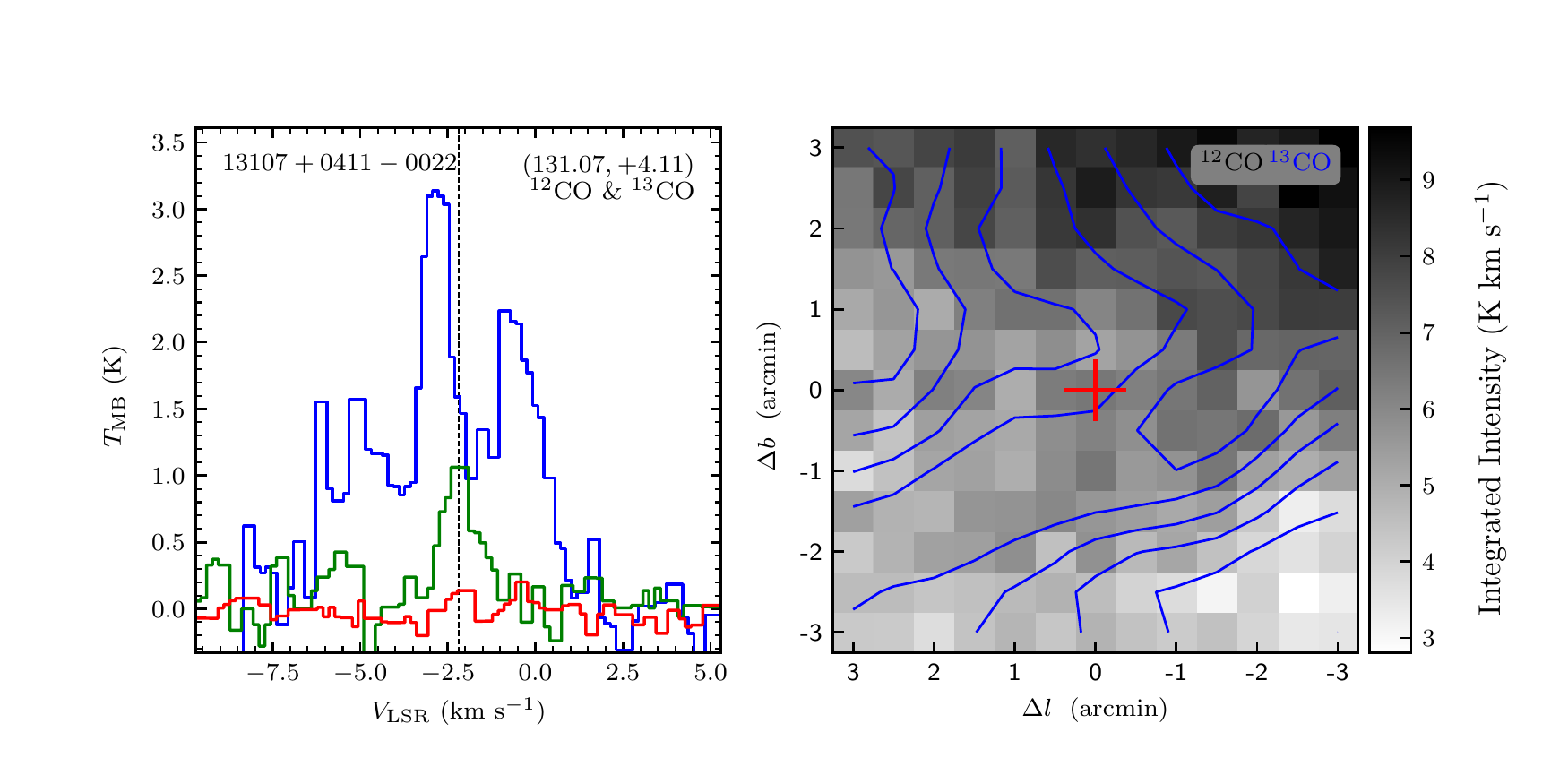}
\end{figure}
\clearpage

\begin{figure}
\includegraphics[width=9.0cm,angle=0]{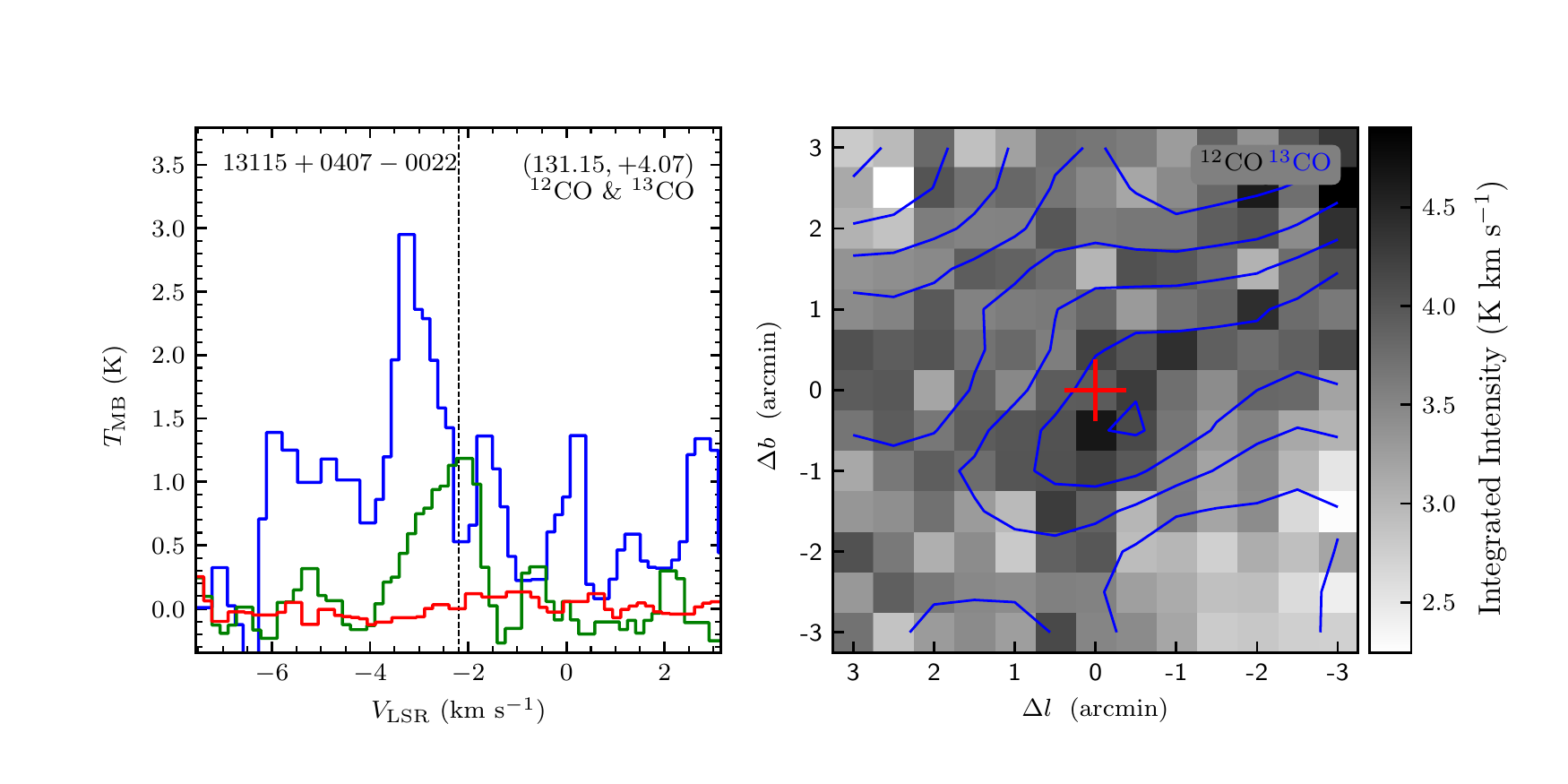}
\includegraphics[width=9.0cm,angle=0]{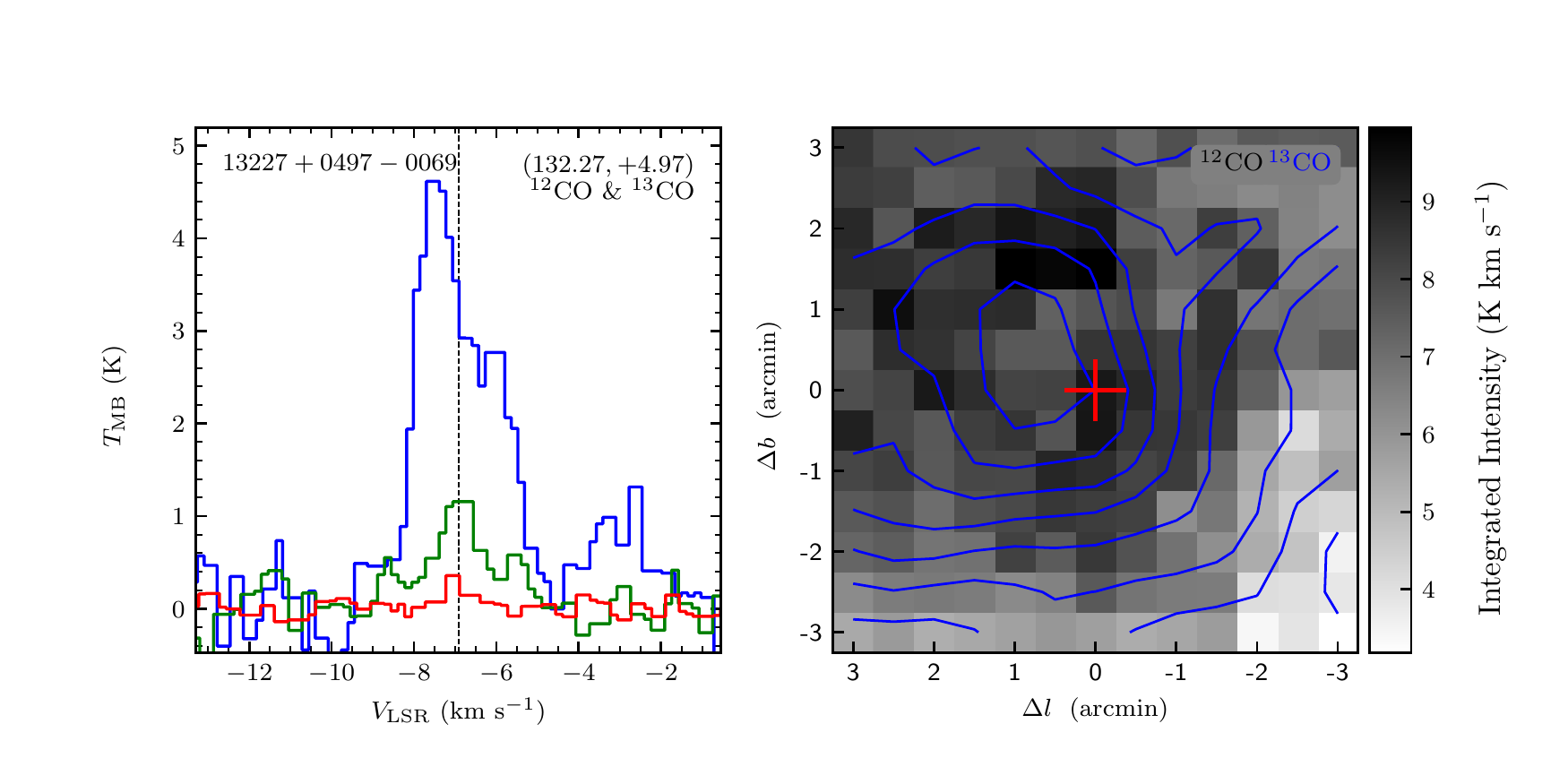}
\vspace{-0.5cm}

\includegraphics[width=9.0cm,angle=0]{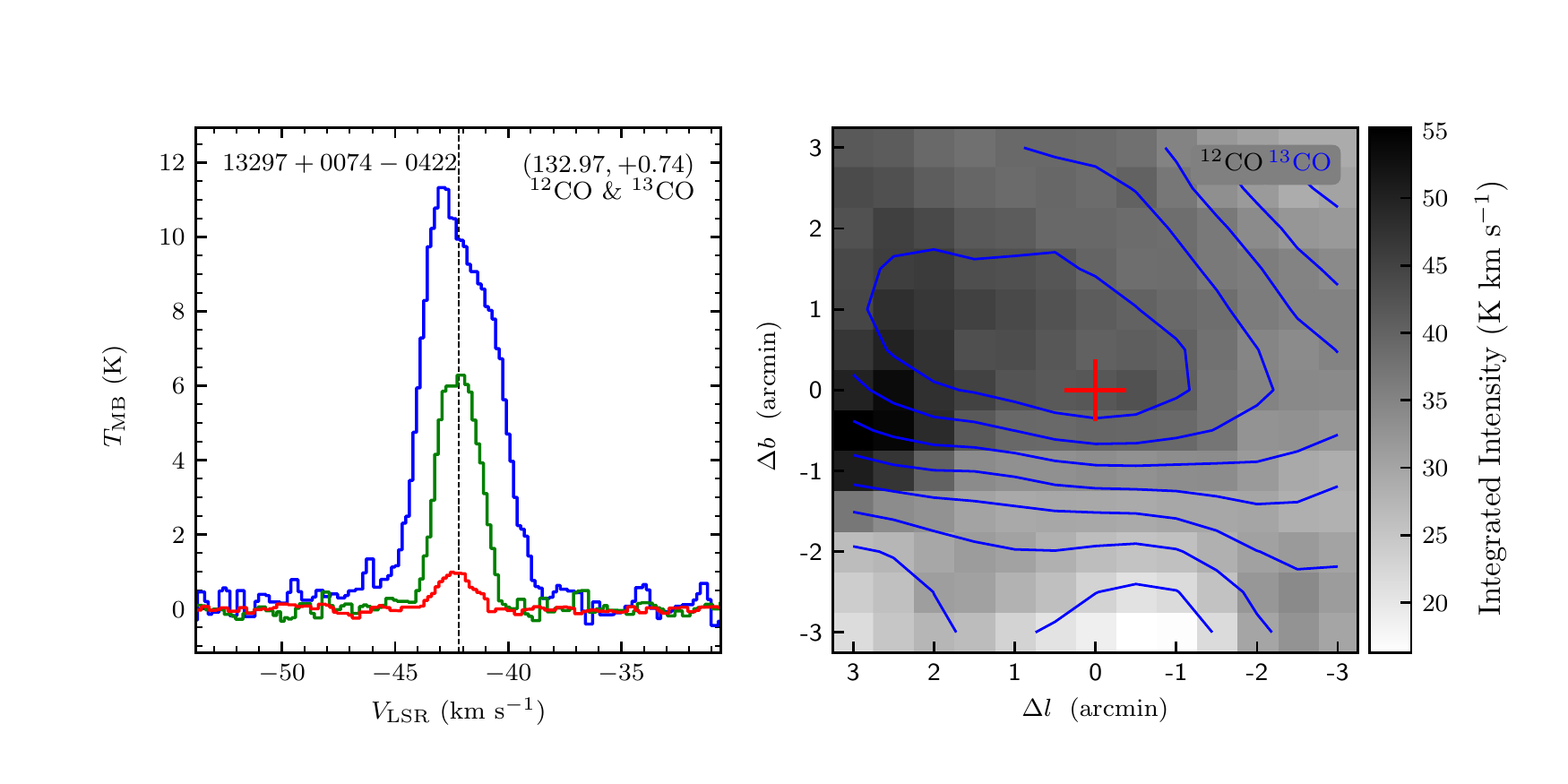}
\includegraphics[width=9.0cm,angle=0]{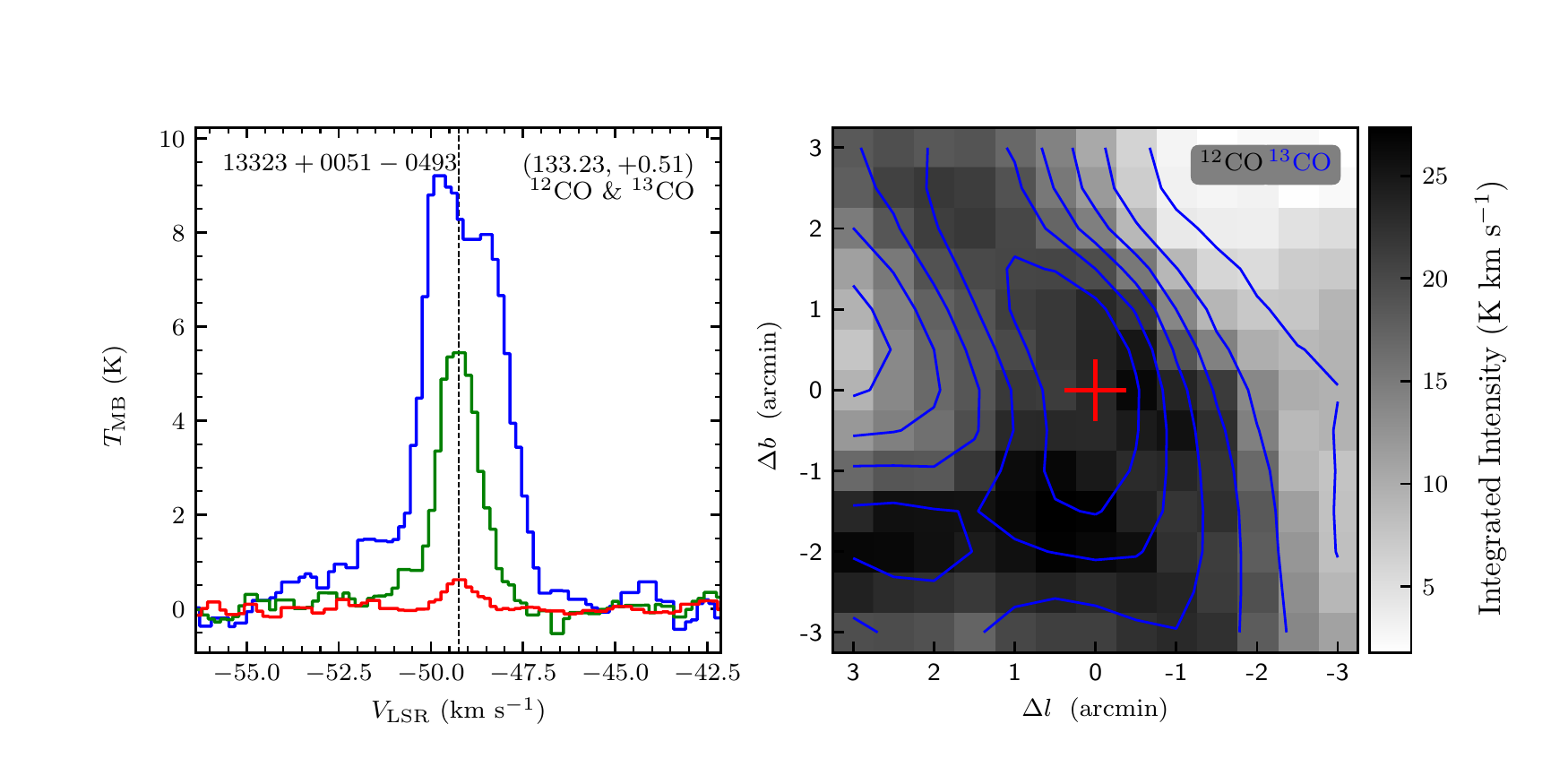}
\vspace{-0.5cm}

\includegraphics[width=9.0cm,angle=0]{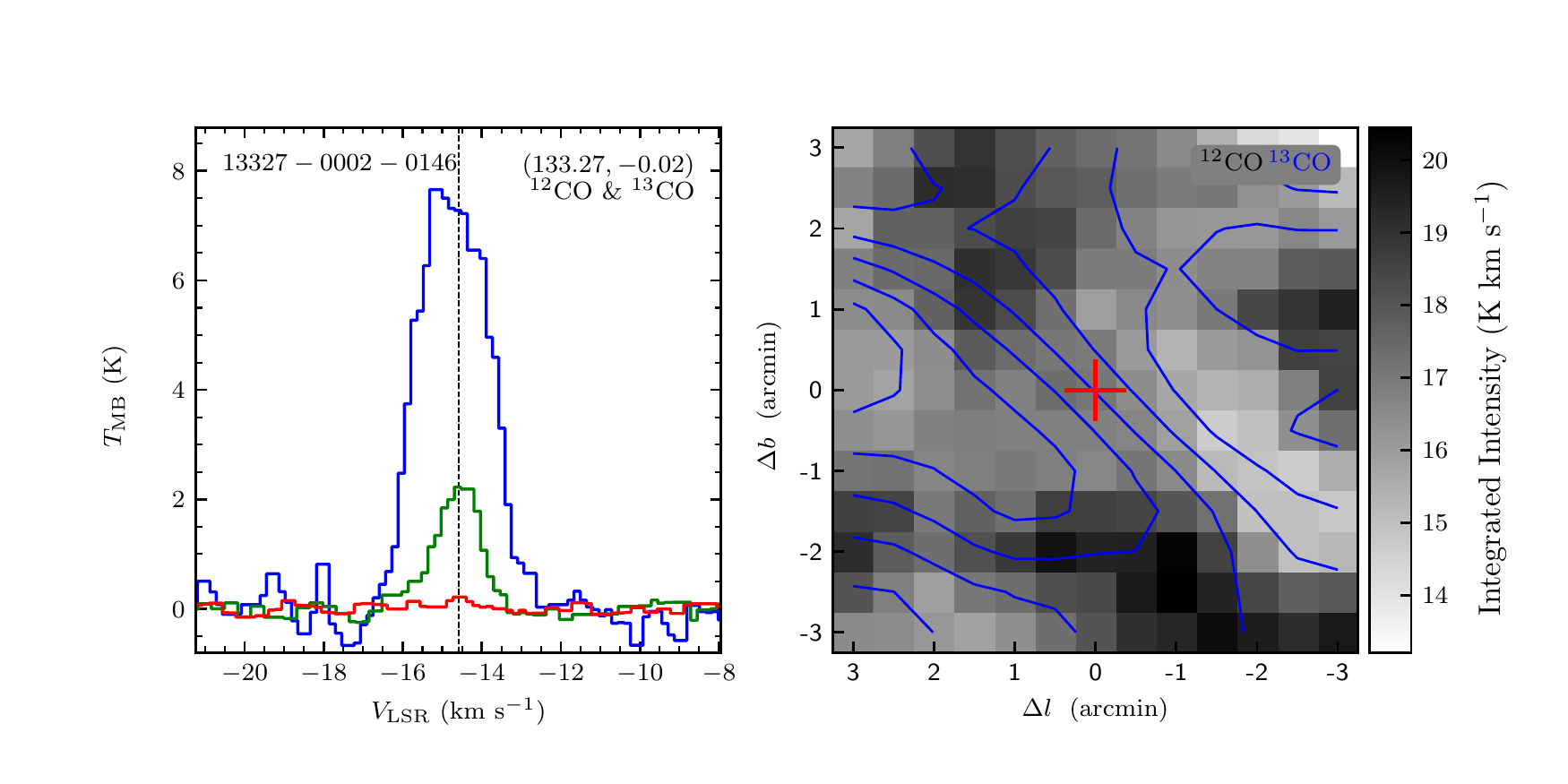}
\includegraphics[width=9.0cm,angle=0]{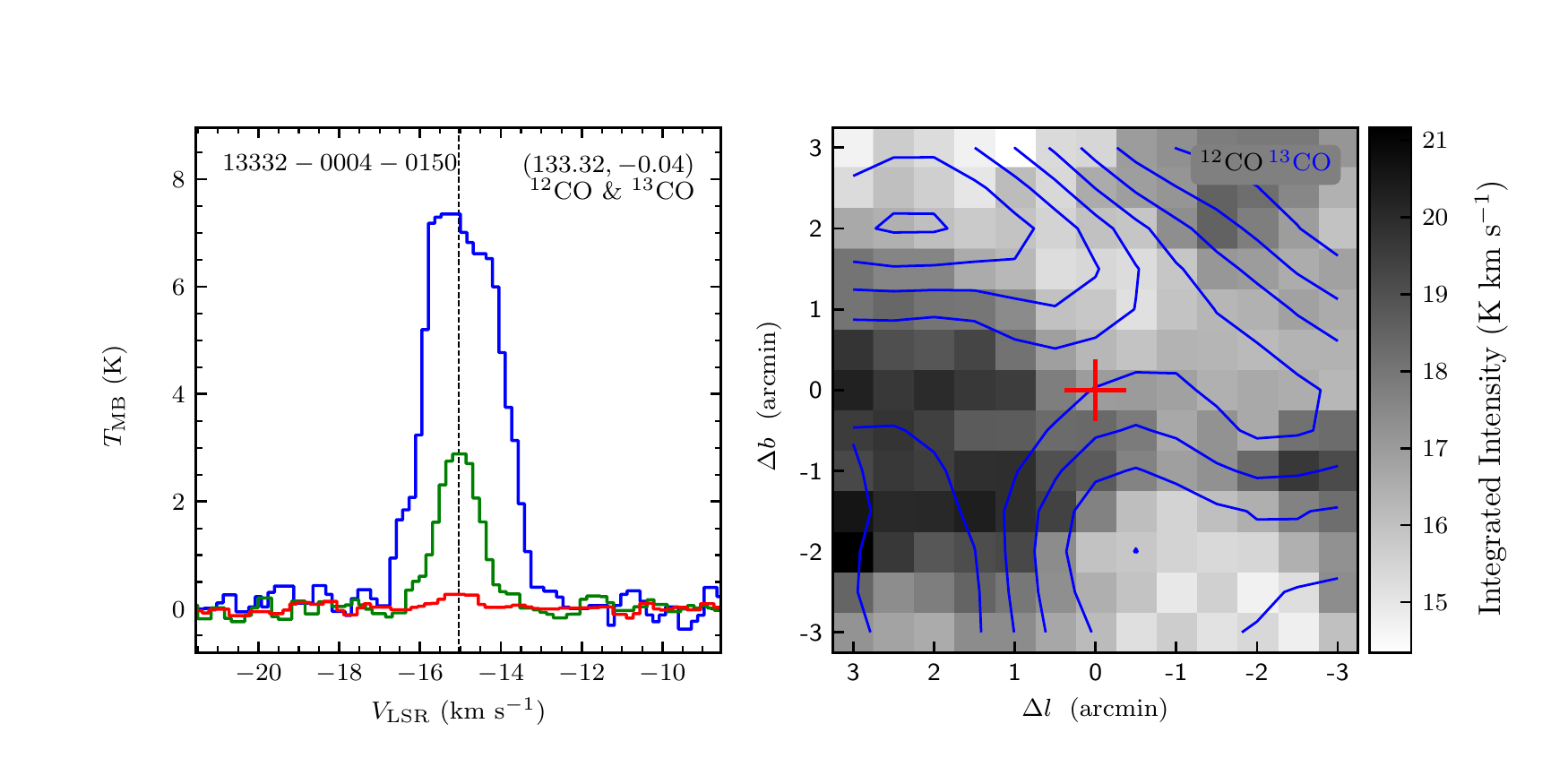}
\vspace{-0.5cm}

\includegraphics[width=9.0cm,angle=0]{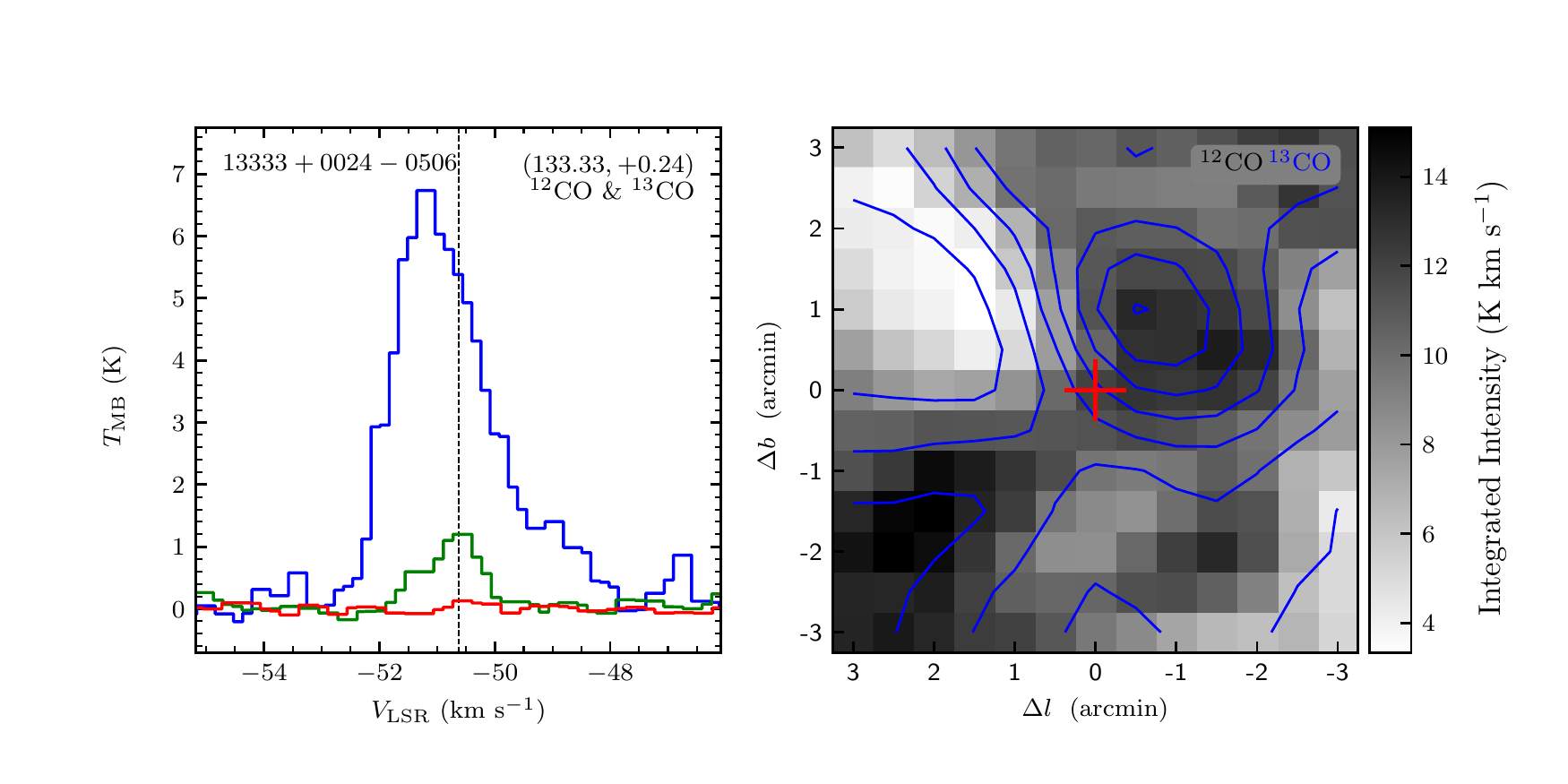}
\includegraphics[width=9.0cm,angle=0]{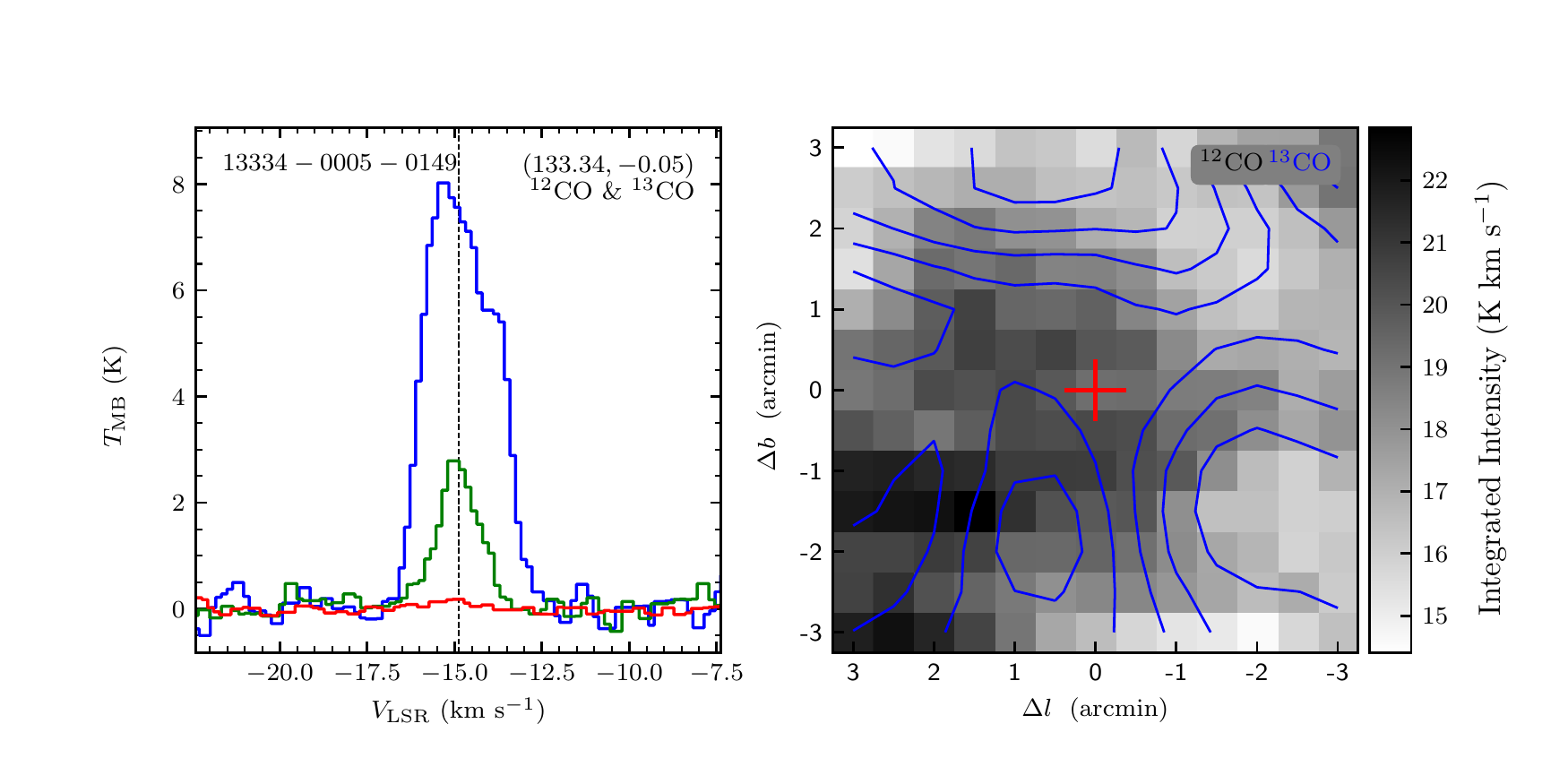}
\vspace{-0.5cm}

\includegraphics[width=9.0cm,angle=0]{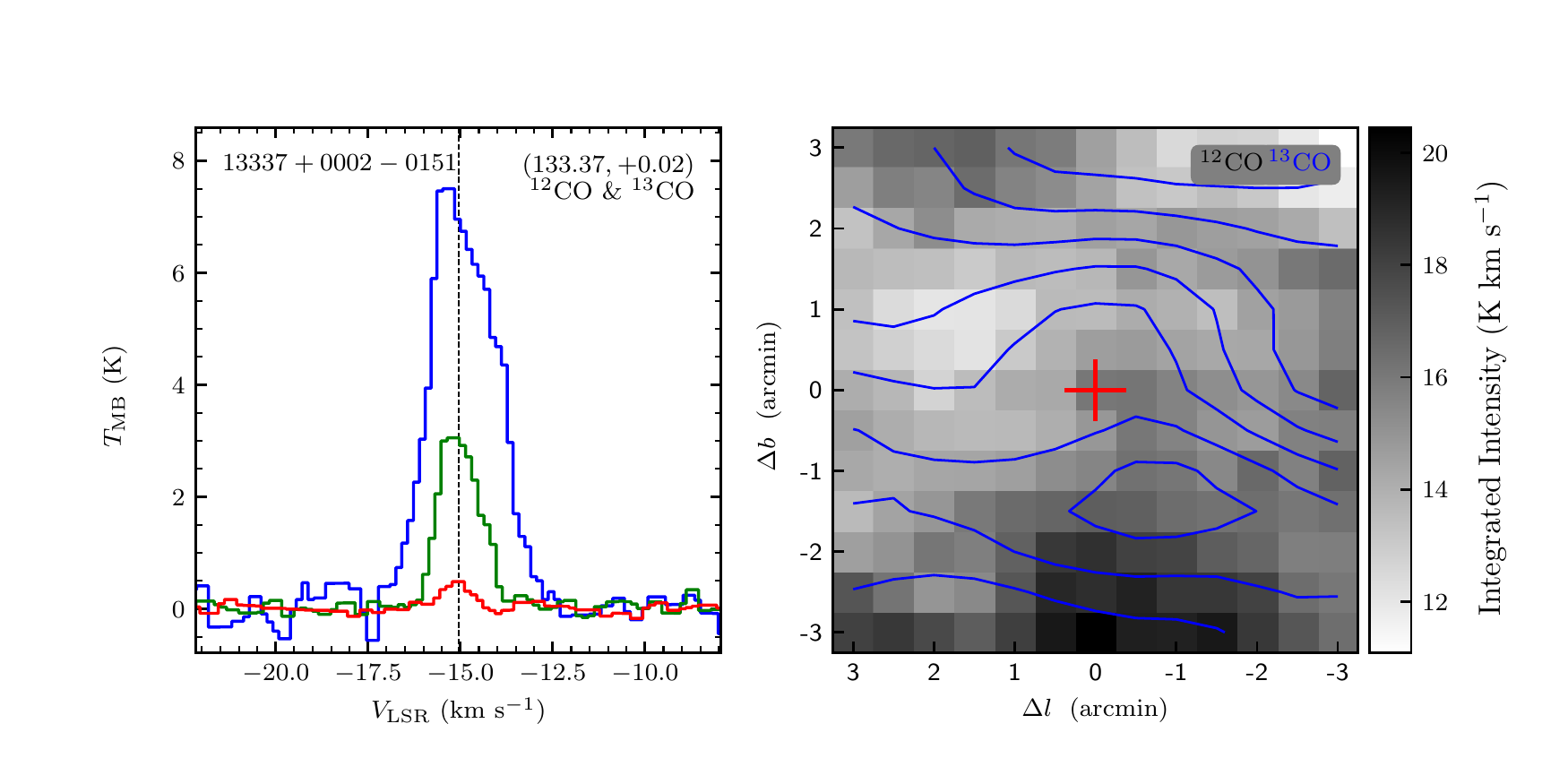}
\includegraphics[width=9.0cm,angle=0]{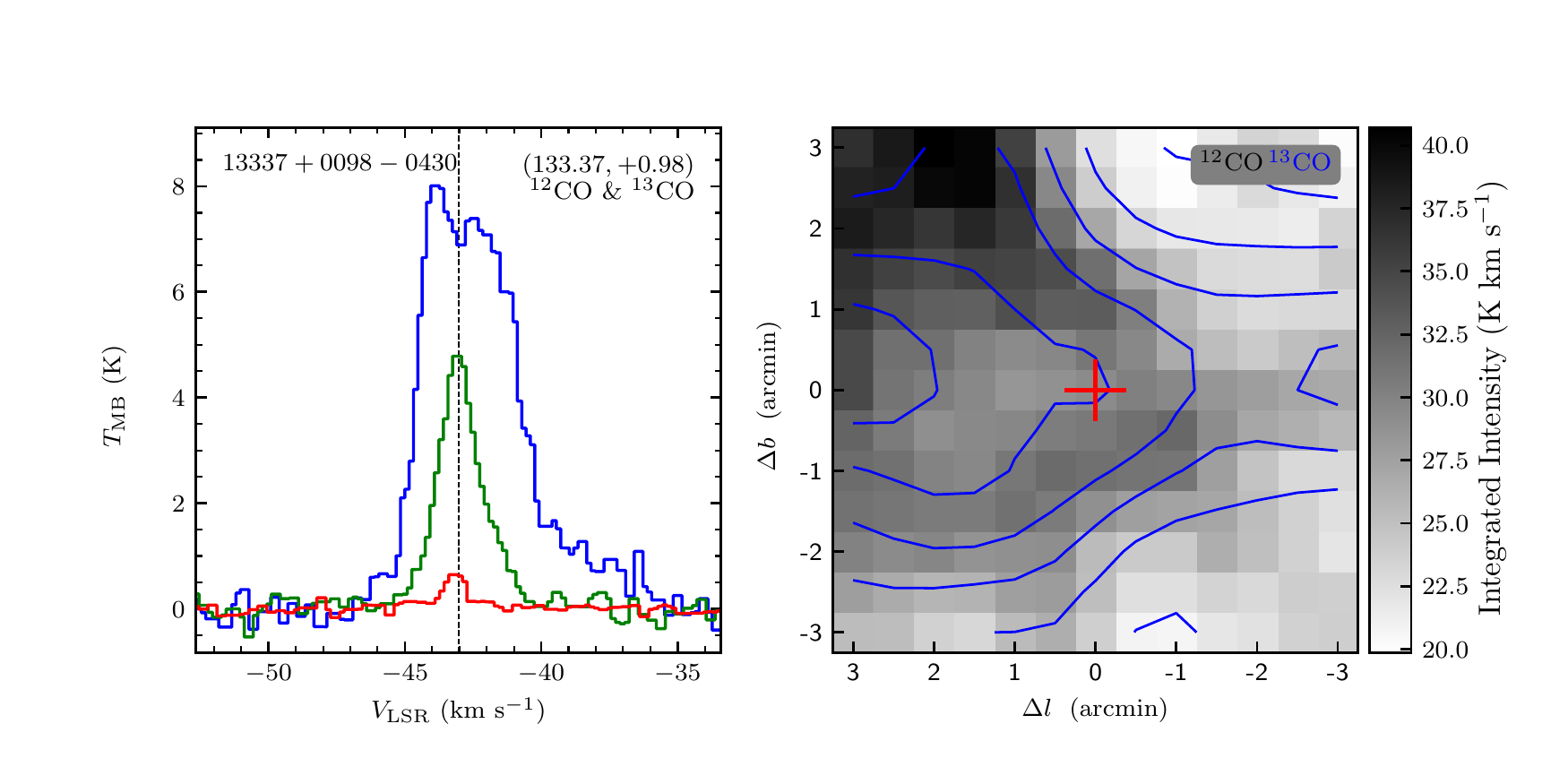}
\end{figure}
\clearpage

\begin{figure}
\includegraphics[width=9.0cm,angle=0]{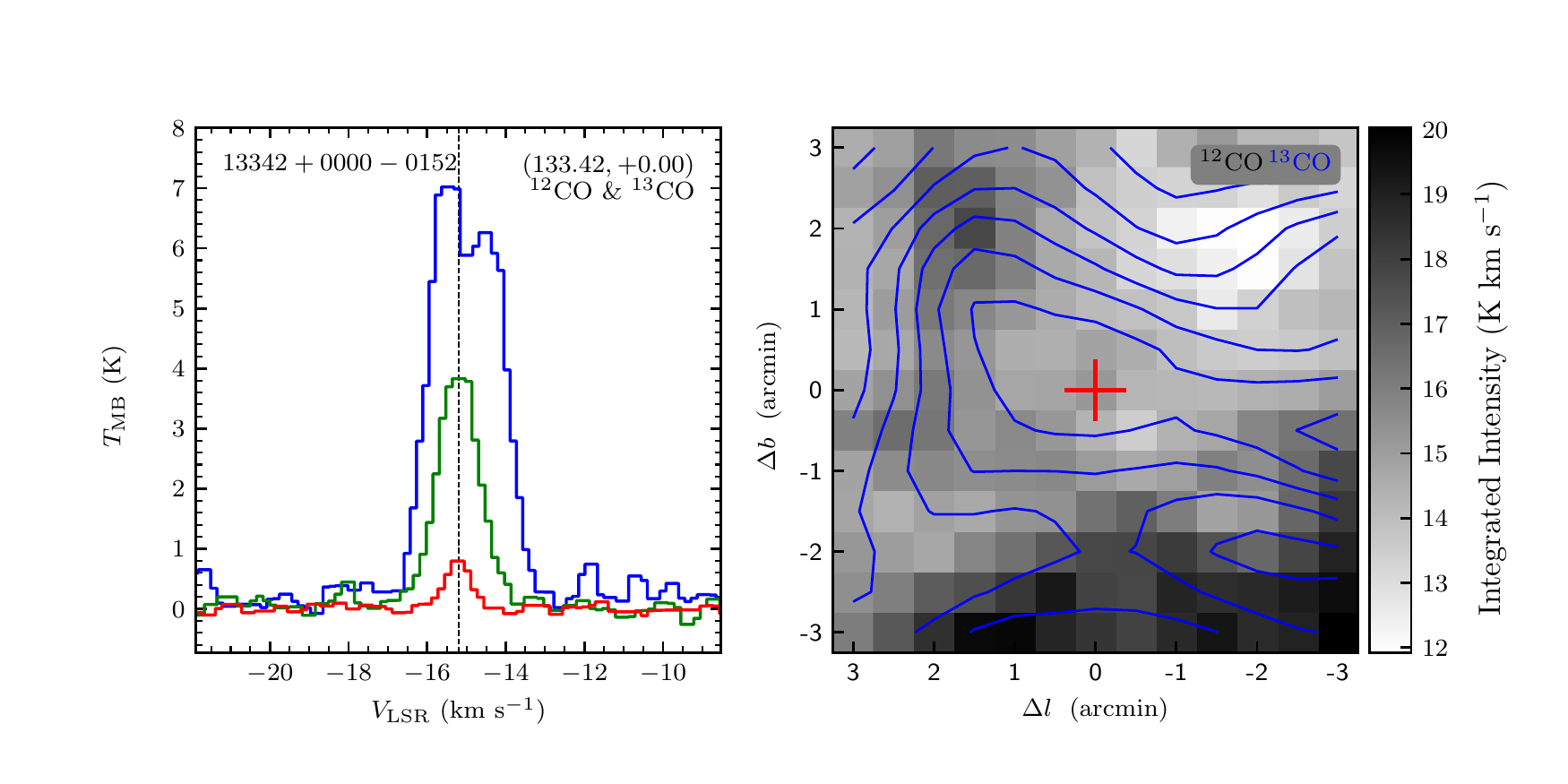}
\includegraphics[width=9.0cm,angle=0]{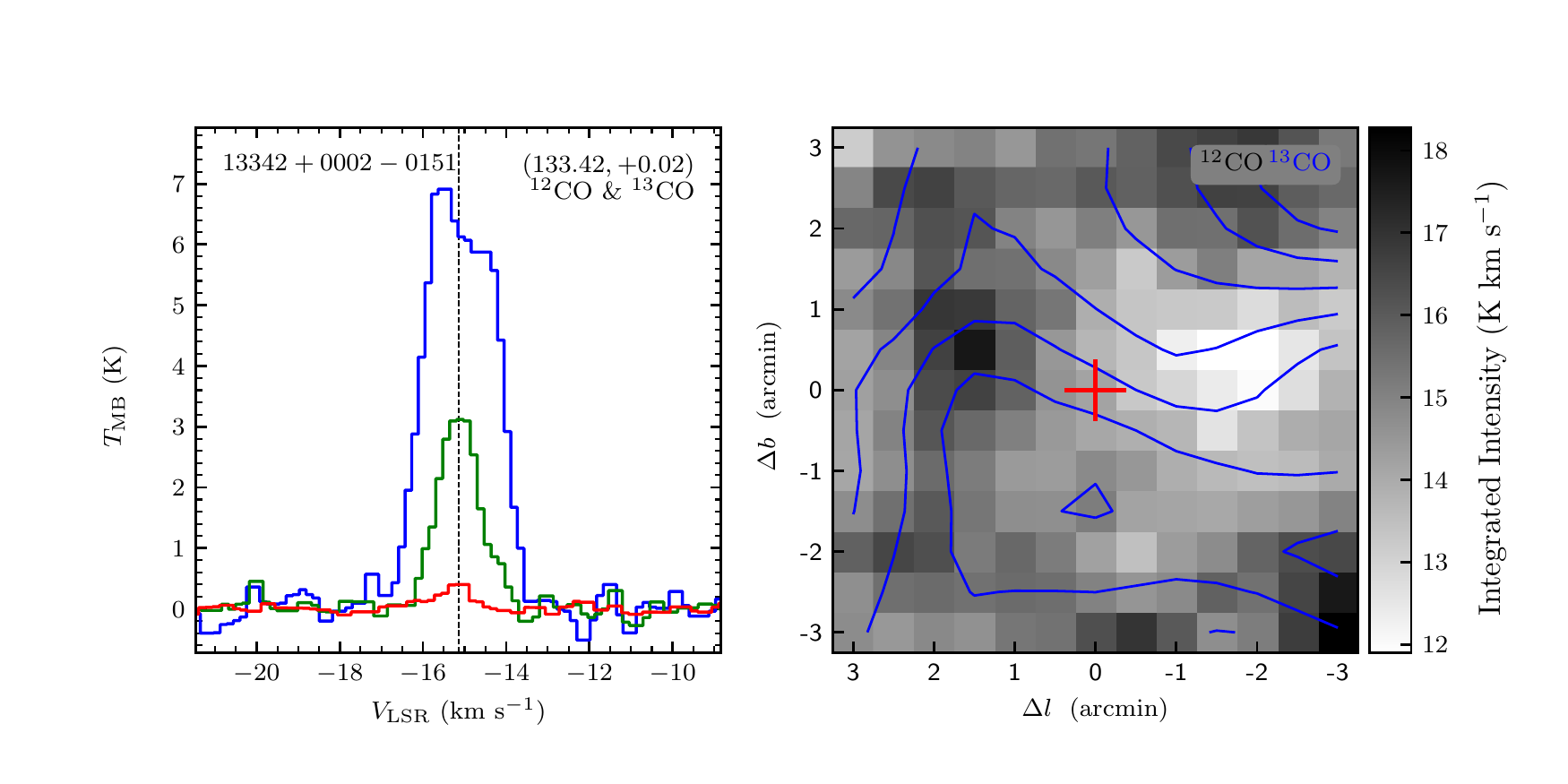}
\vspace{-0.5cm}

\includegraphics[width=9.0cm,angle=0]{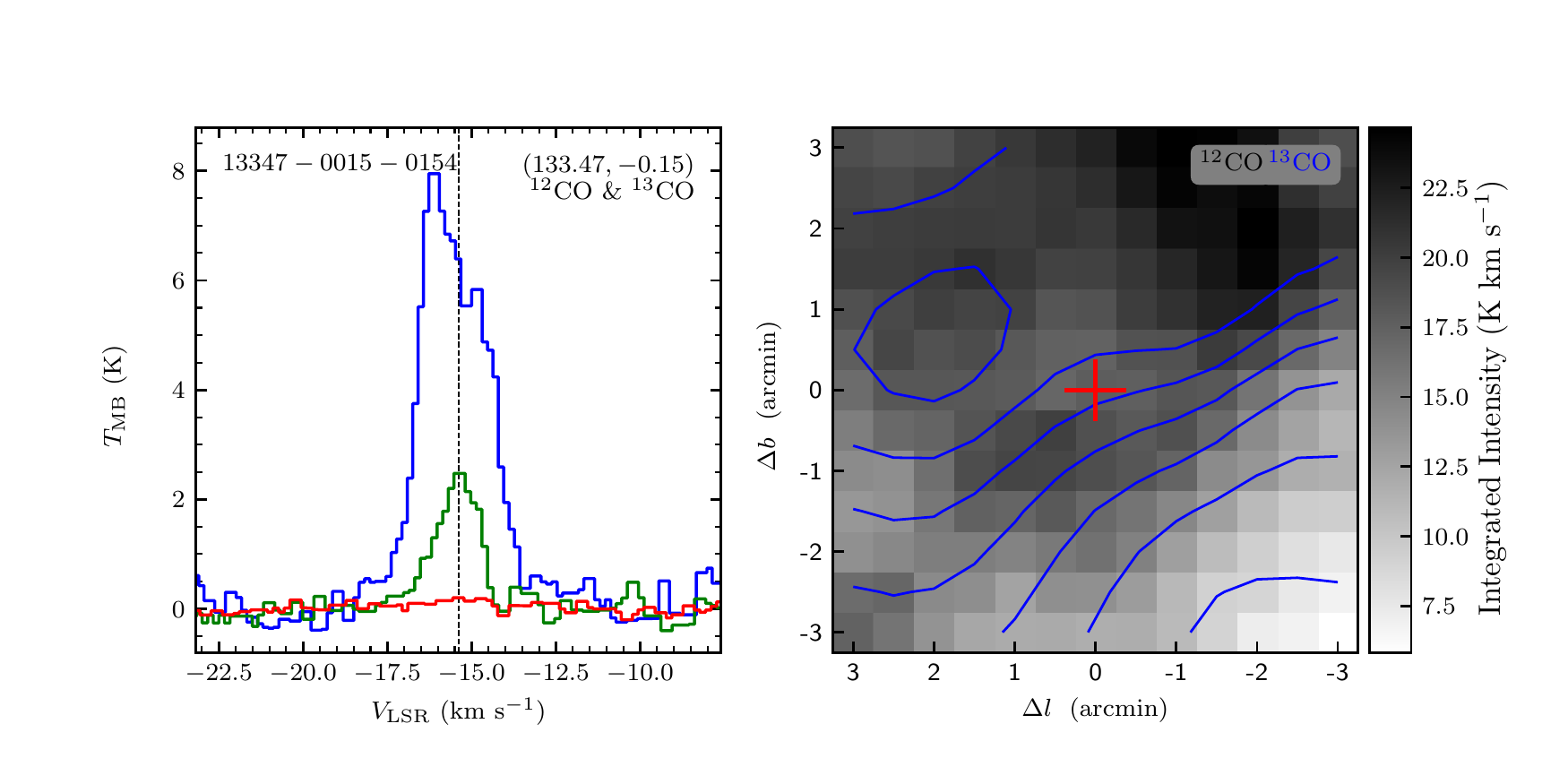}
\includegraphics[width=9.0cm,angle=0]{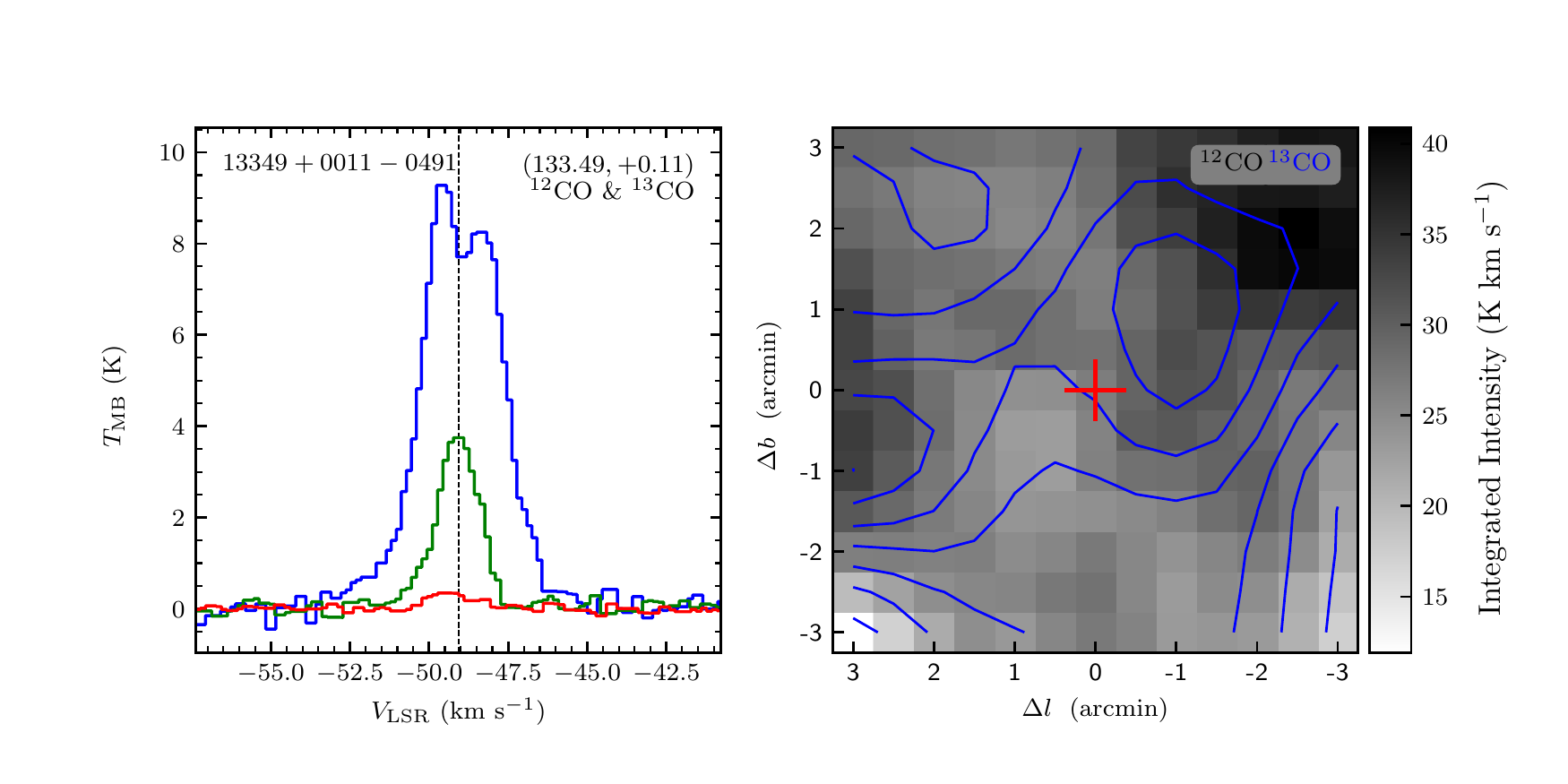}
\vspace{-0.5cm}

\includegraphics[width=9.0cm,angle=0]{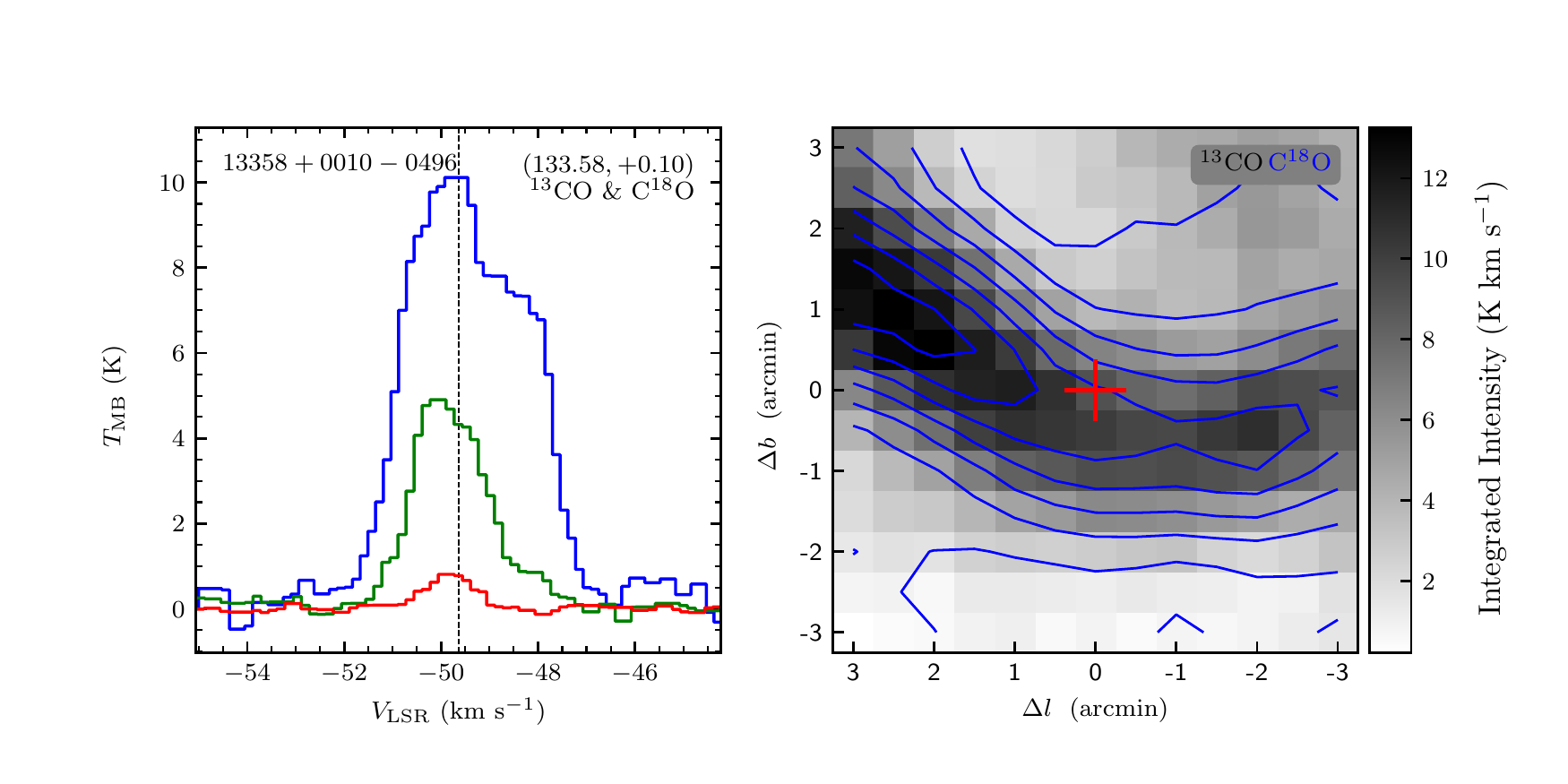}
\includegraphics[width=9.0cm,angle=0]{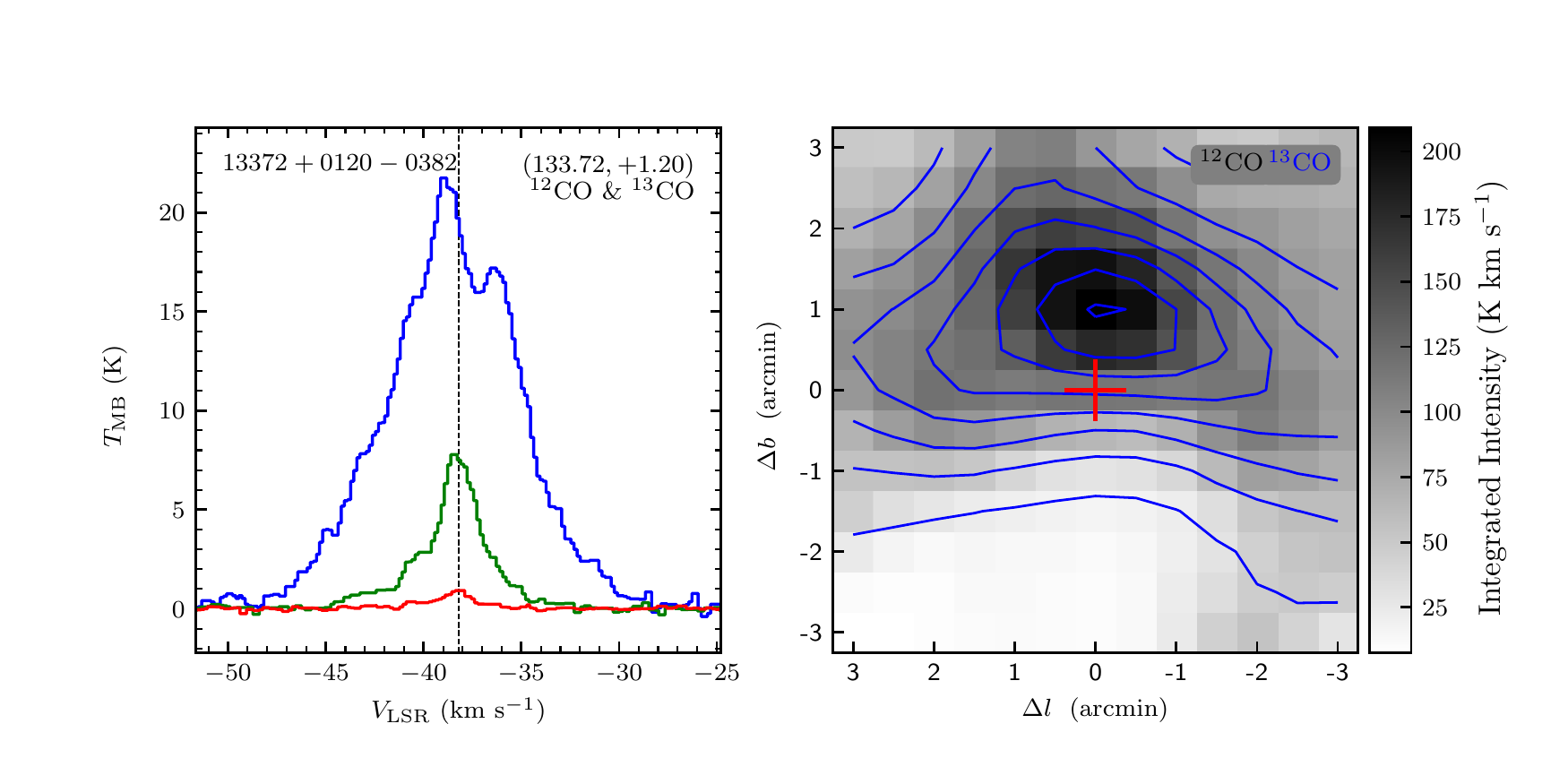}
\vspace{-0.5cm}

\includegraphics[width=9.0cm,angle=0]{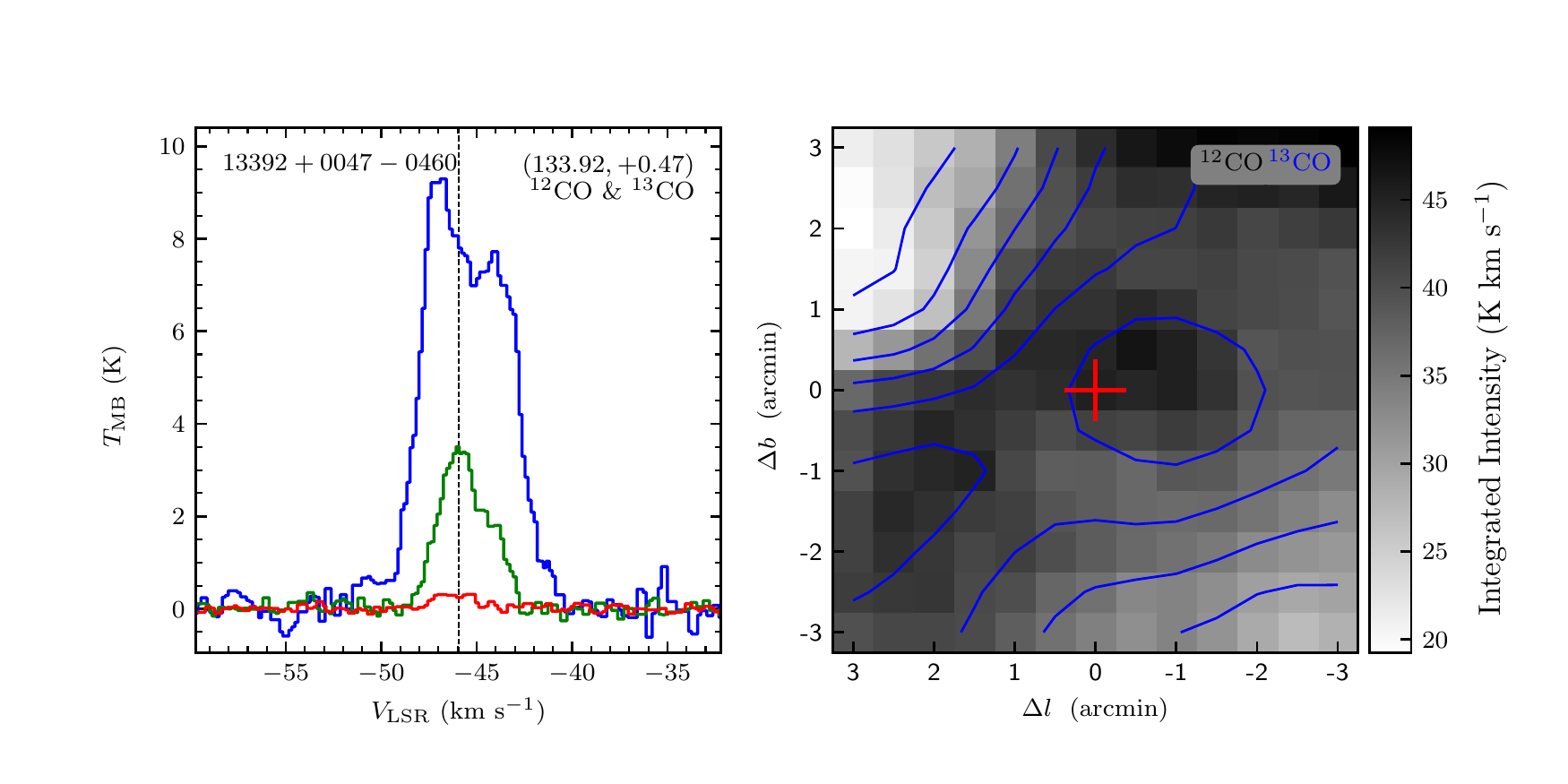}
\includegraphics[width=9.0cm,angle=0]{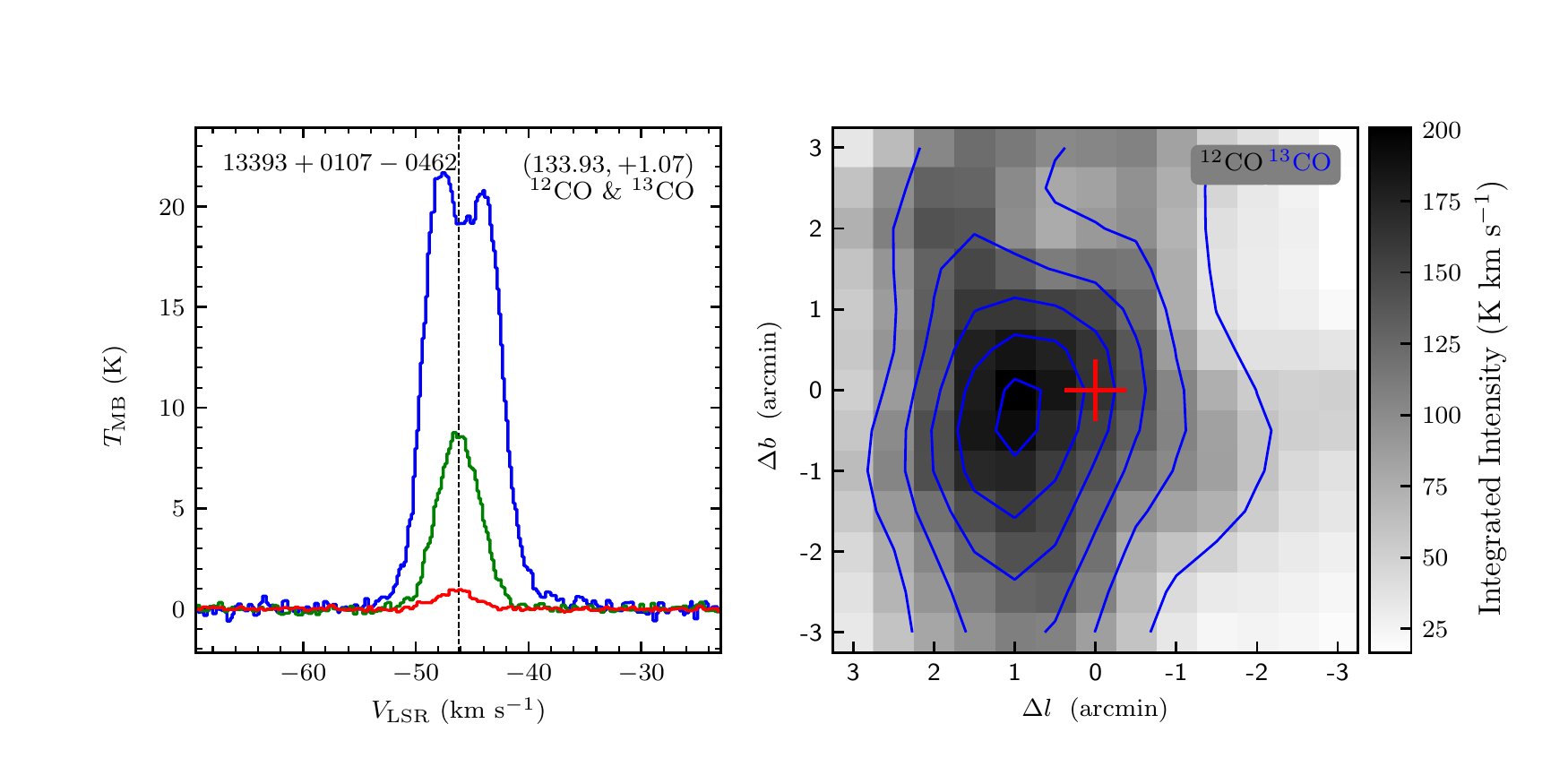}
\vspace{-0.5cm}

\includegraphics[width=9.0cm,angle=0]{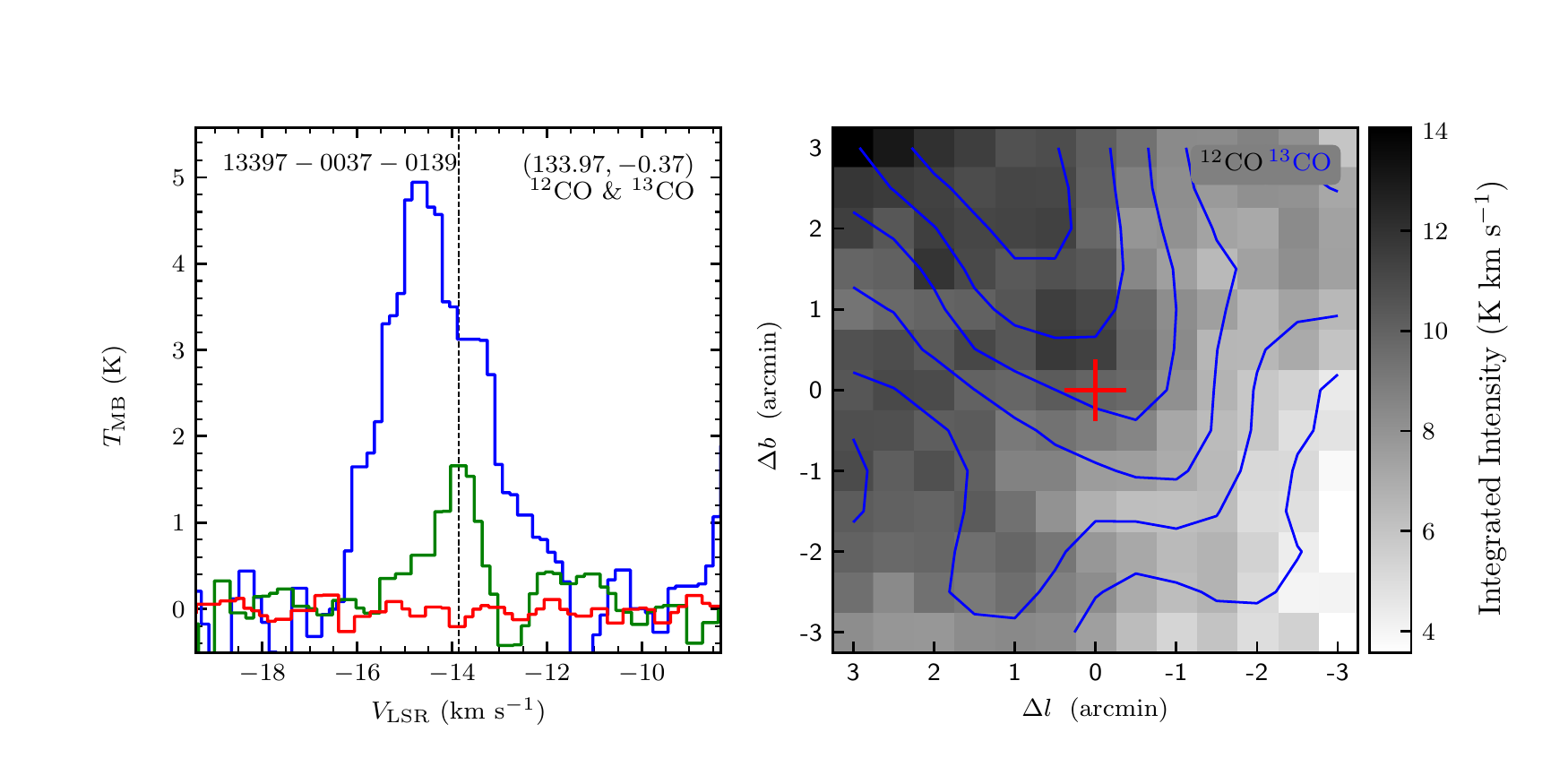}
\includegraphics[width=9.0cm,angle=0]{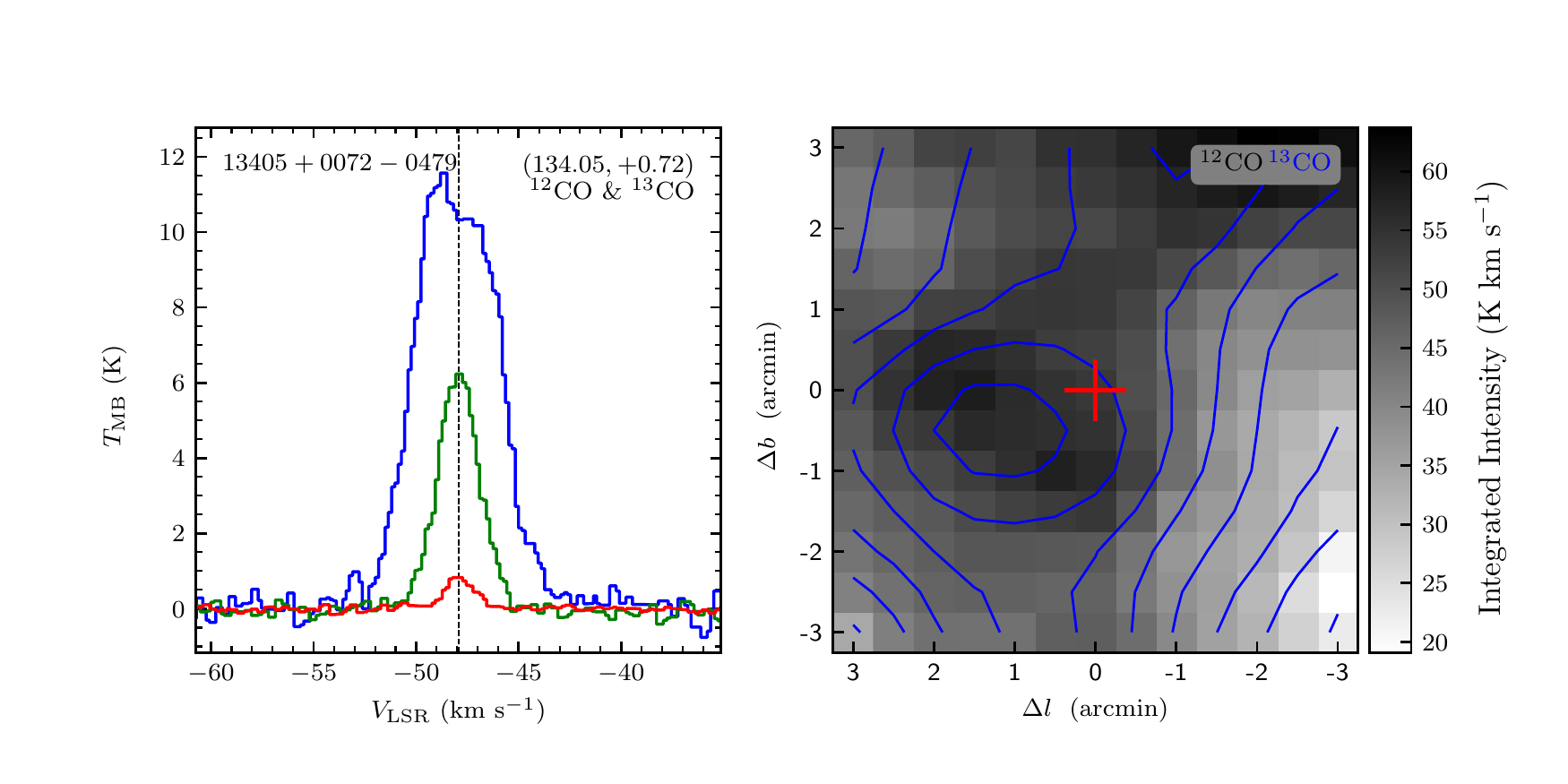}
\end{figure}
\clearpage

\begin{figure}
\includegraphics[width=9.0cm,angle=0]{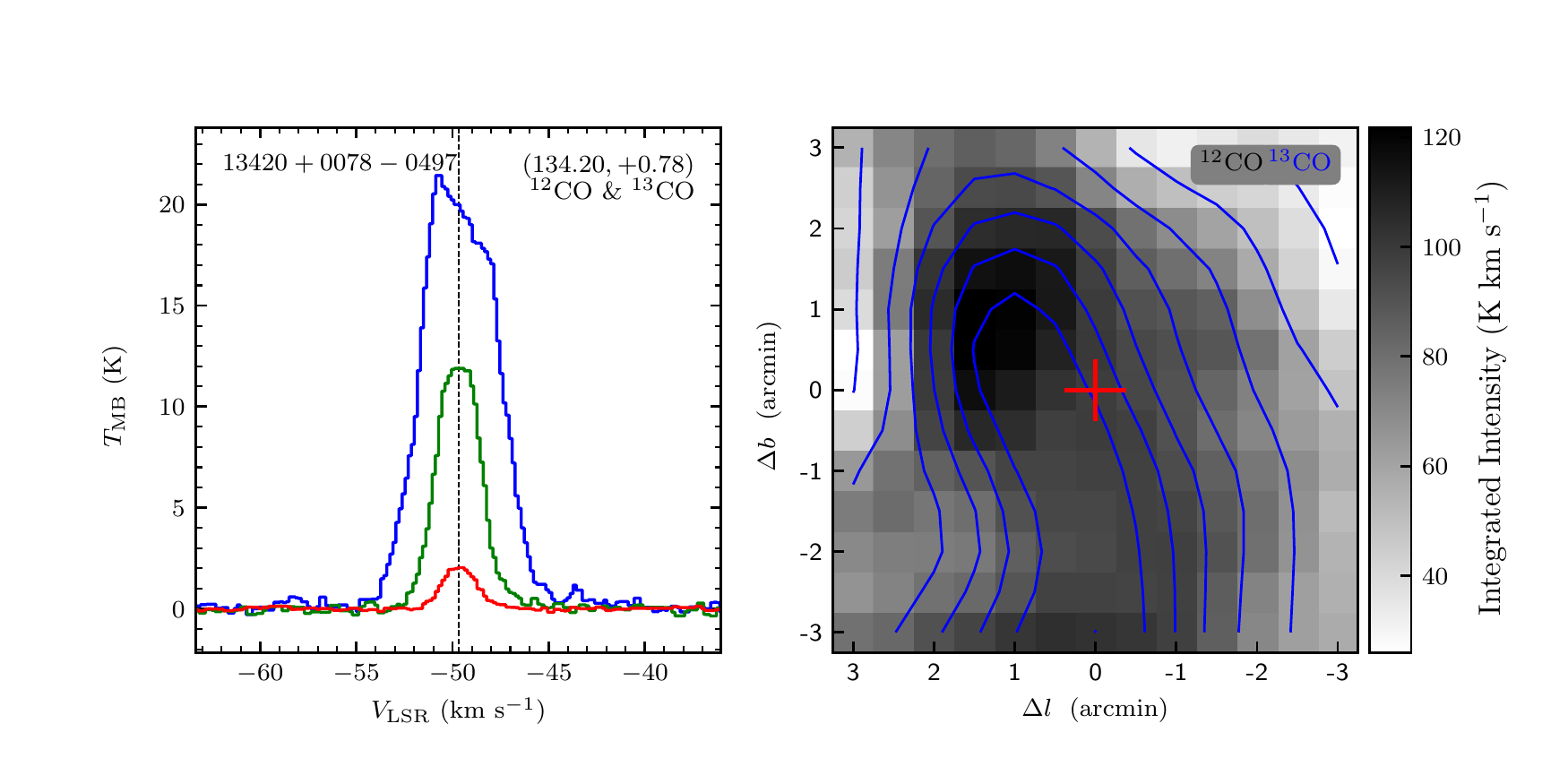}
\includegraphics[width=9.0cm,angle=0]{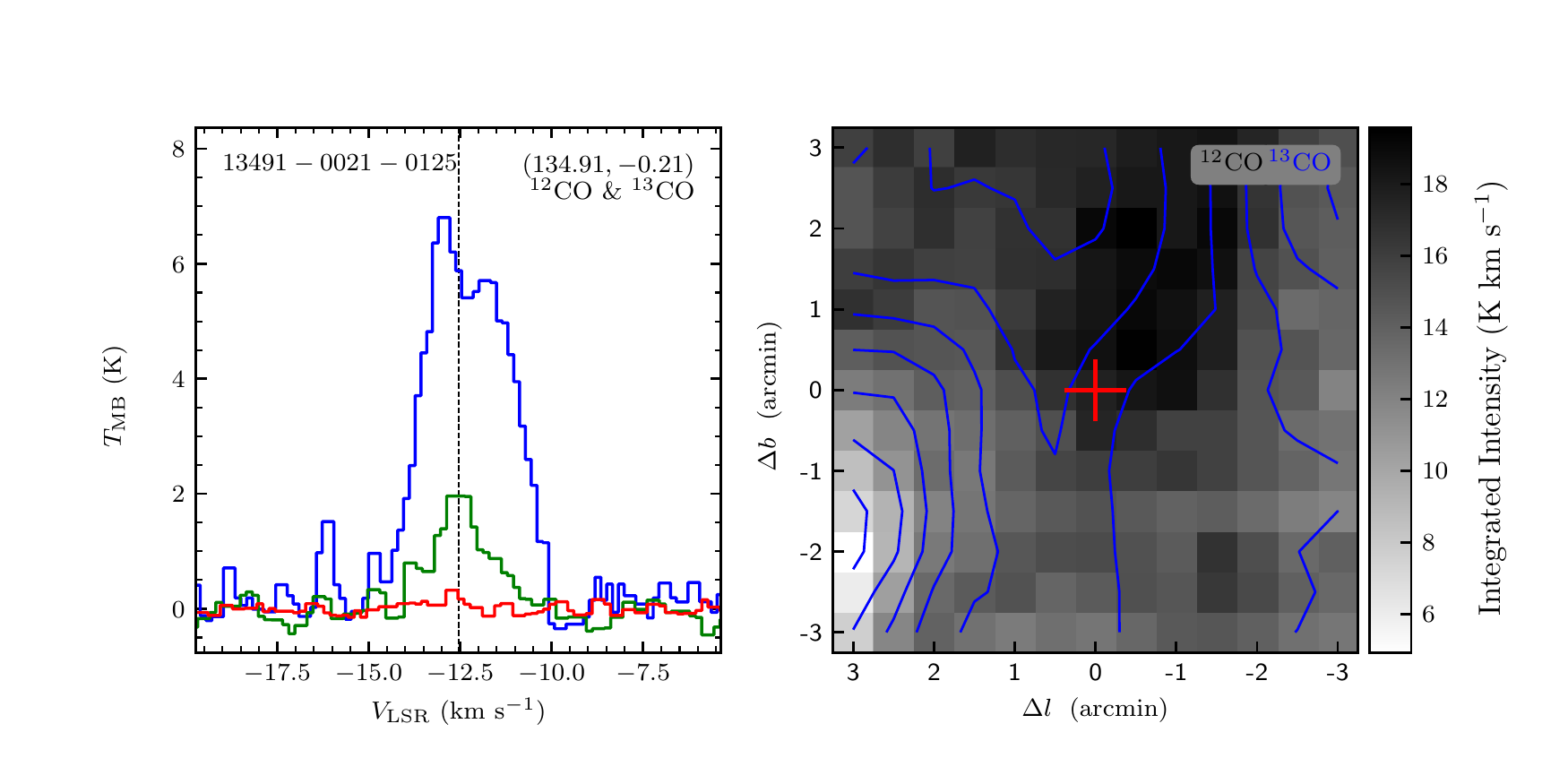}
\vspace{-0.5cm}

\includegraphics[width=9.0cm,angle=0]{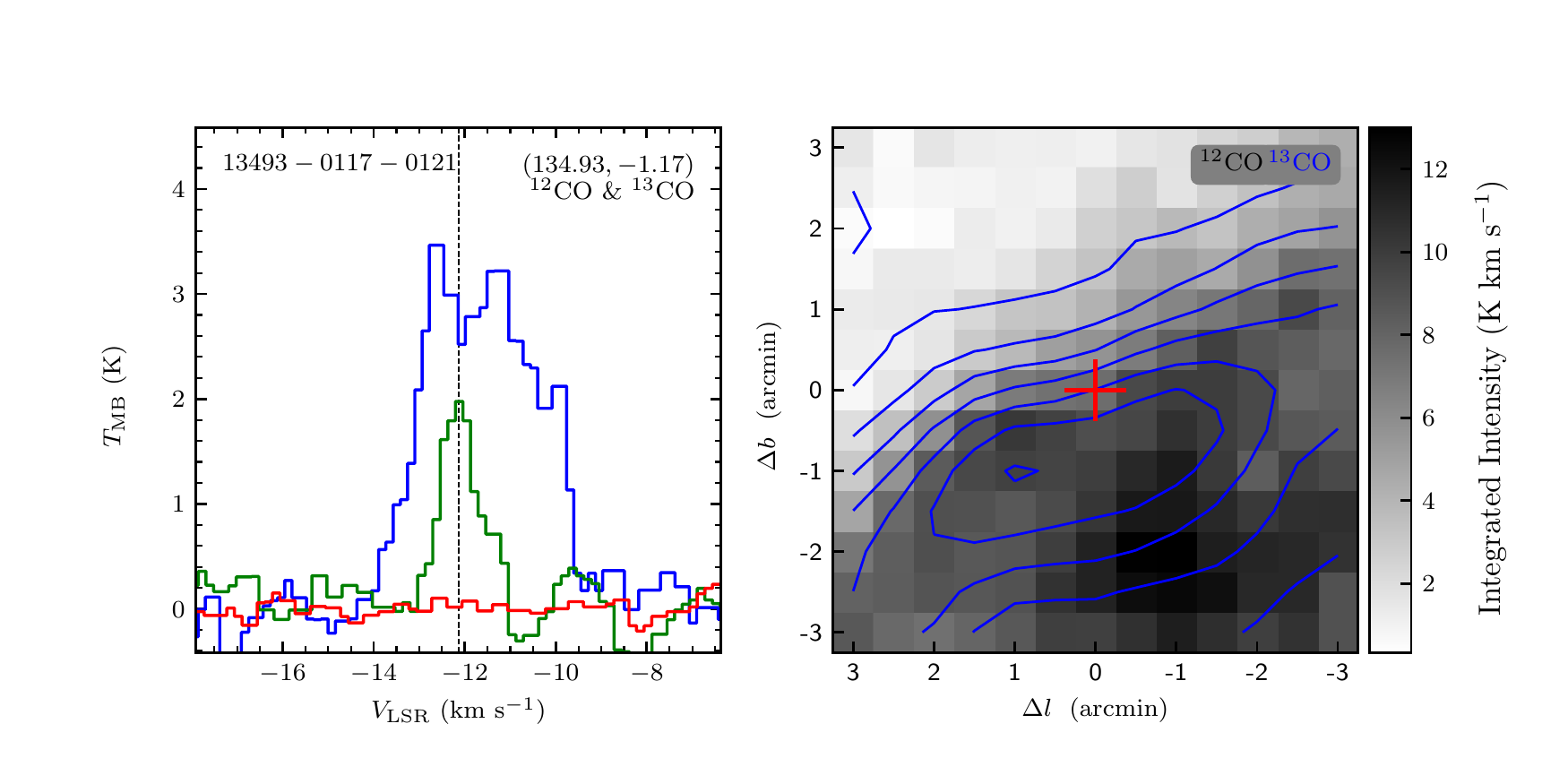}
\includegraphics[width=9.0cm,angle=0]{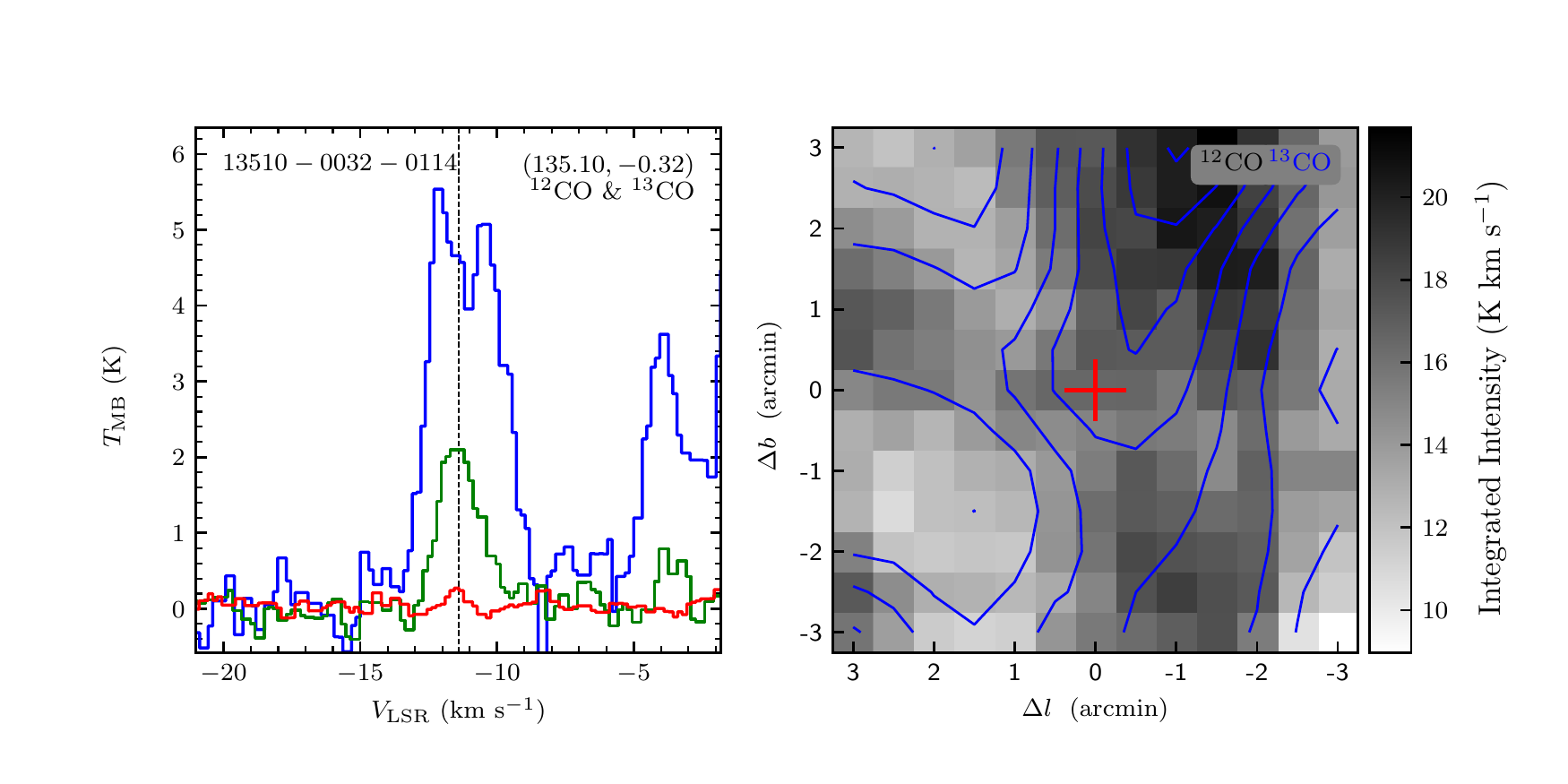}
\vspace{-0.5cm}

\includegraphics[width=9.0cm,angle=0]{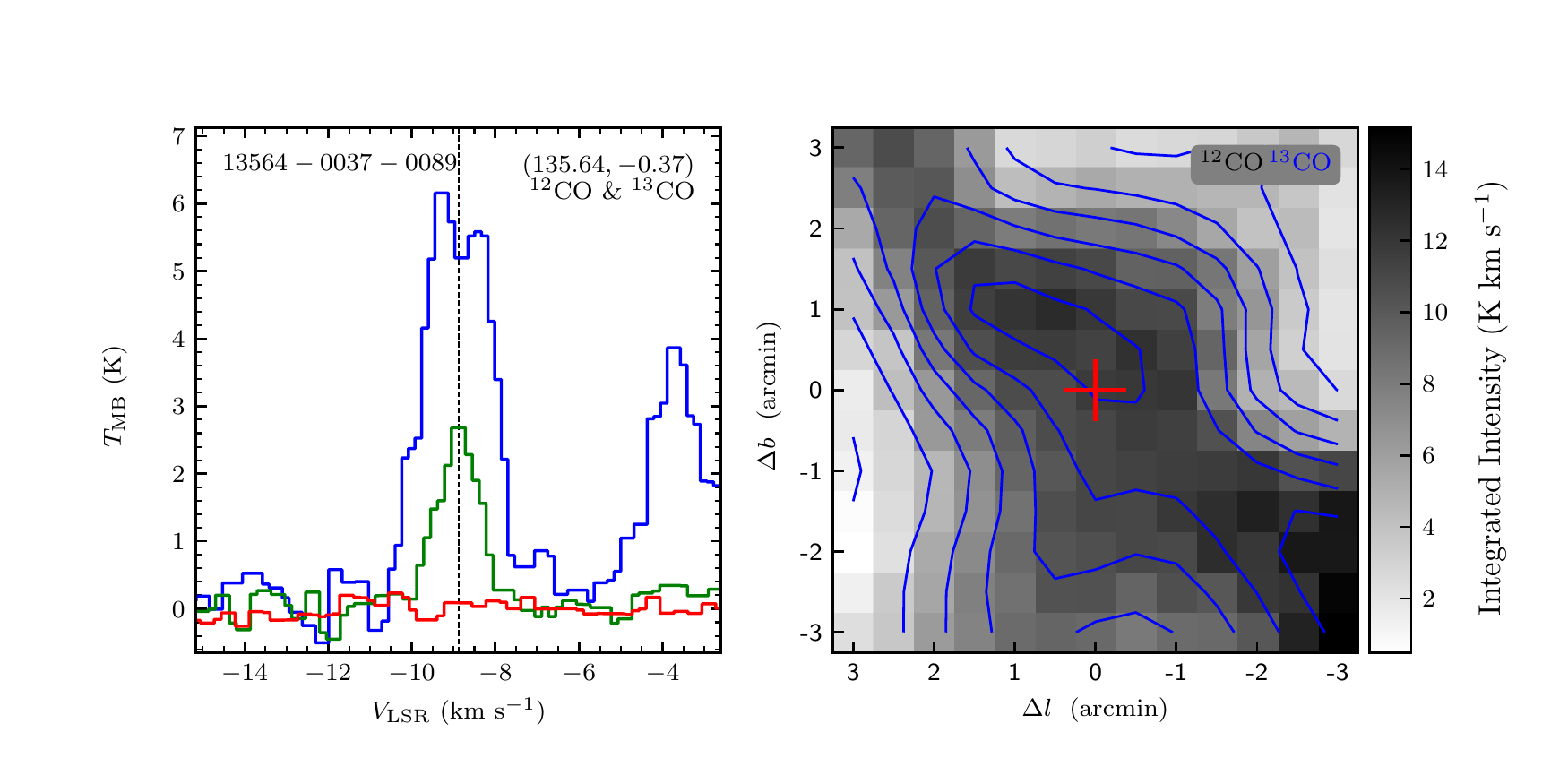}
\includegraphics[width=9.0cm,angle=0]{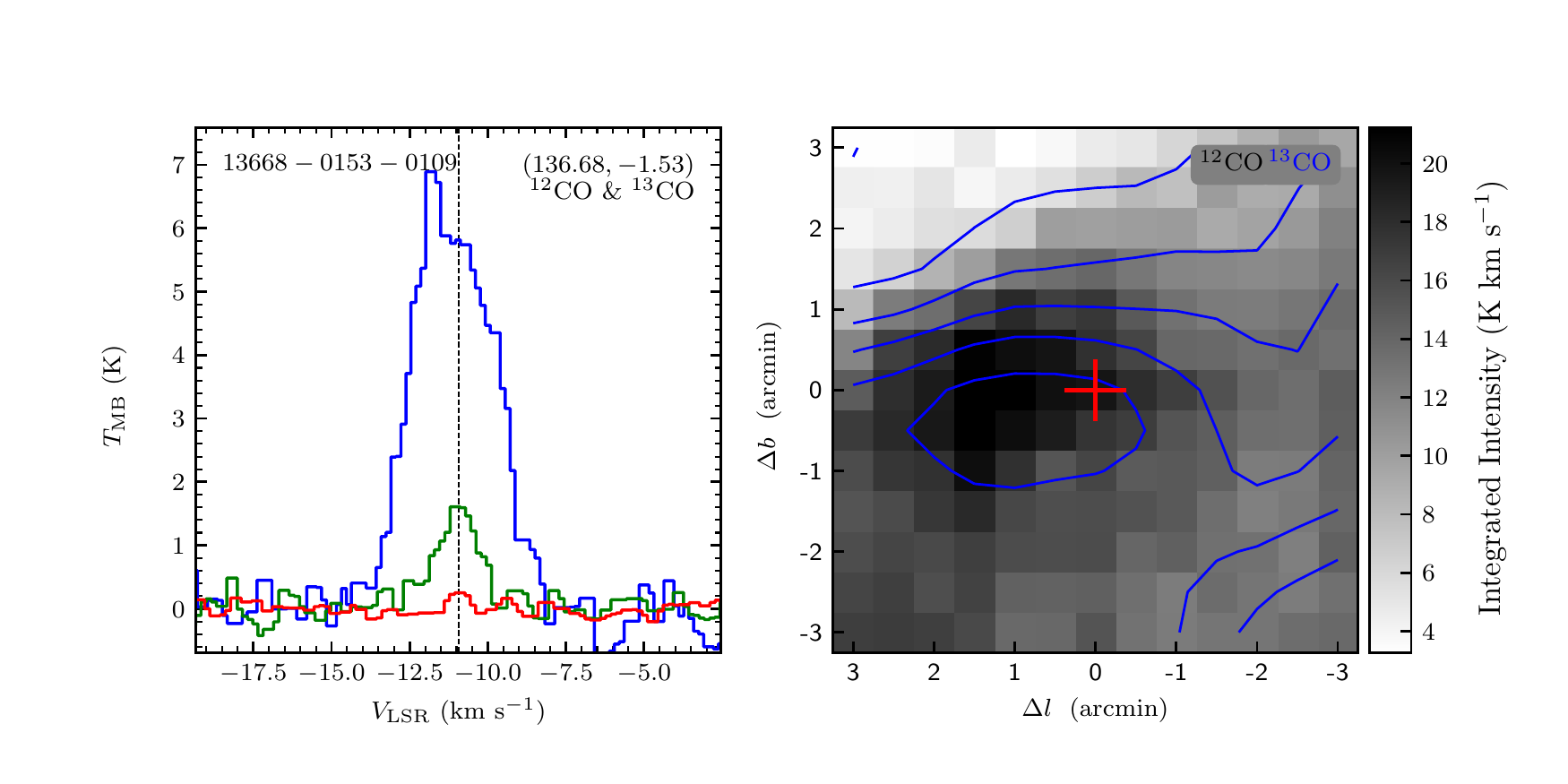}
\vspace{-0.5cm}

\includegraphics[width=9.0cm,angle=0]{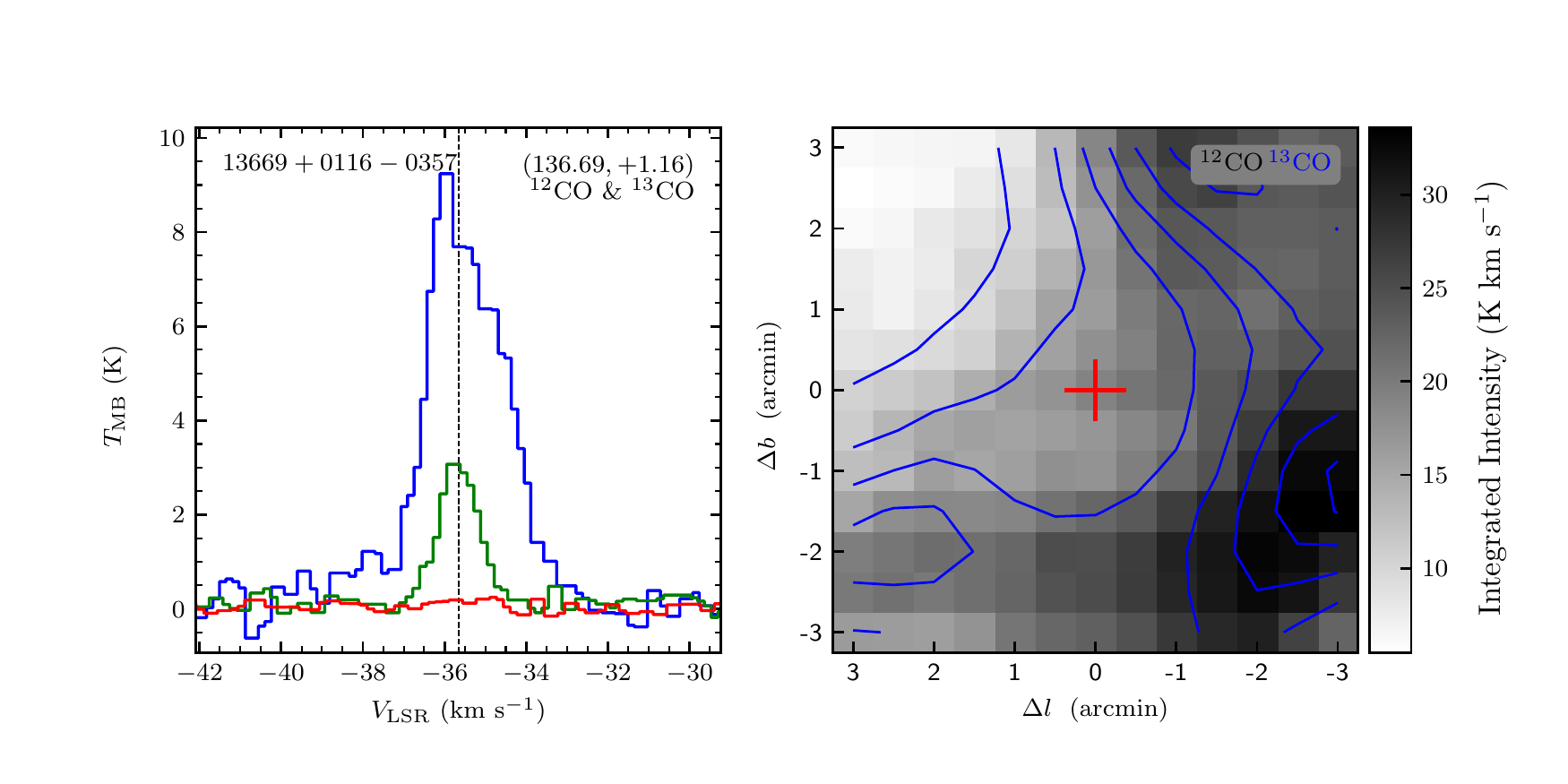}
\includegraphics[width=9.0cm,angle=0]{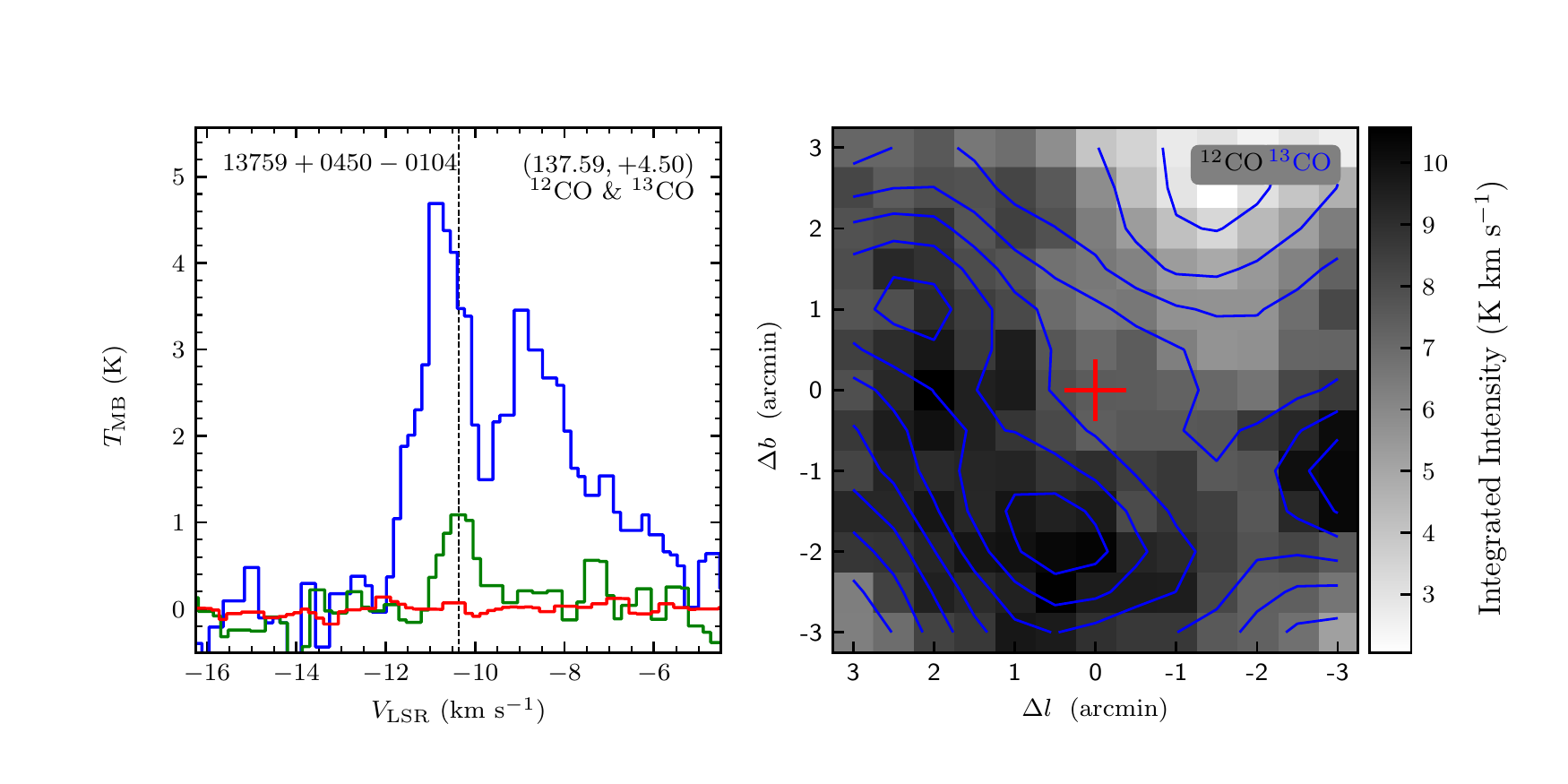}
\vspace{-0.5cm}

\includegraphics[width=9.0cm,angle=0]{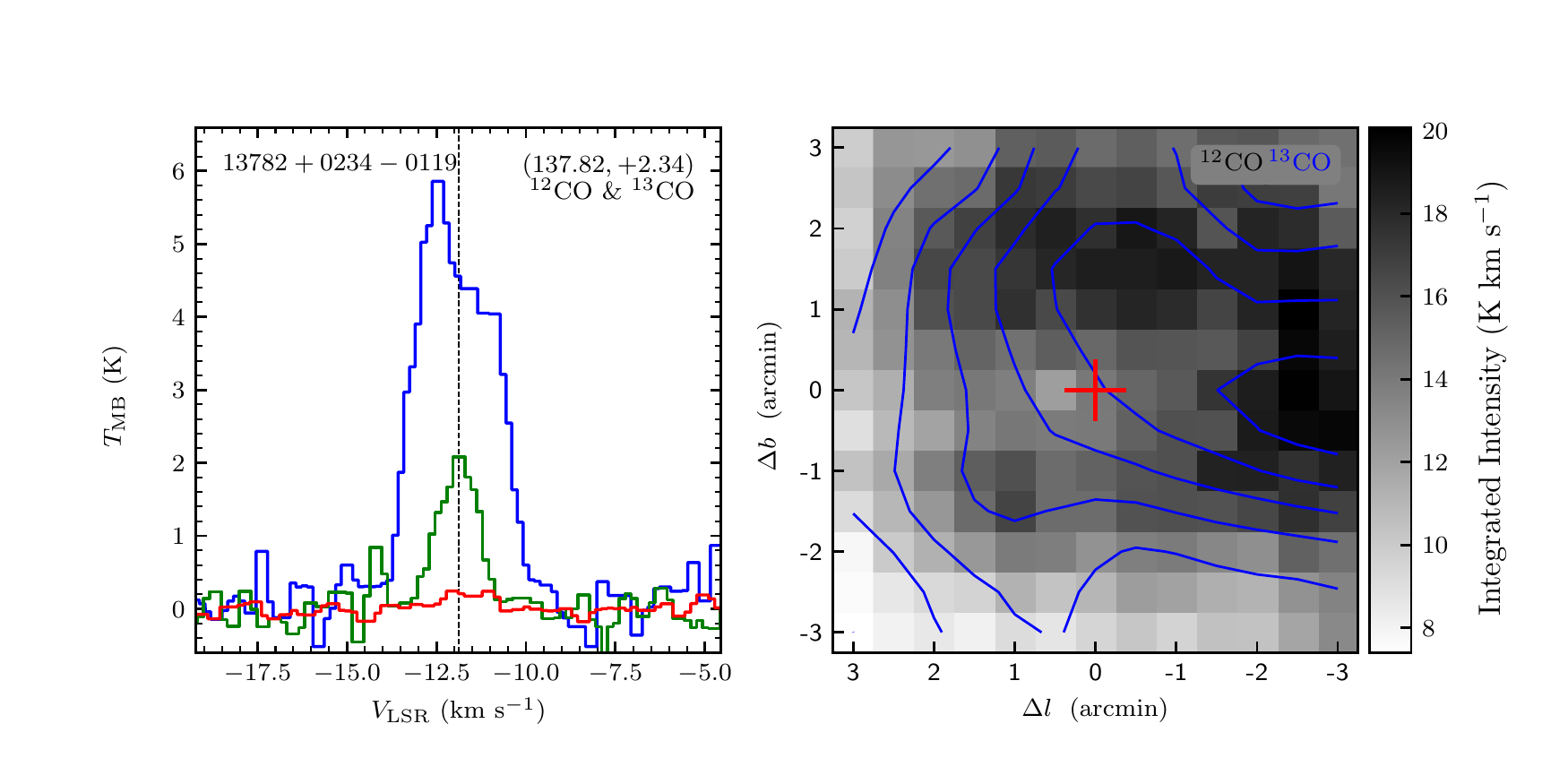}
\includegraphics[width=9.0cm,angle=0]{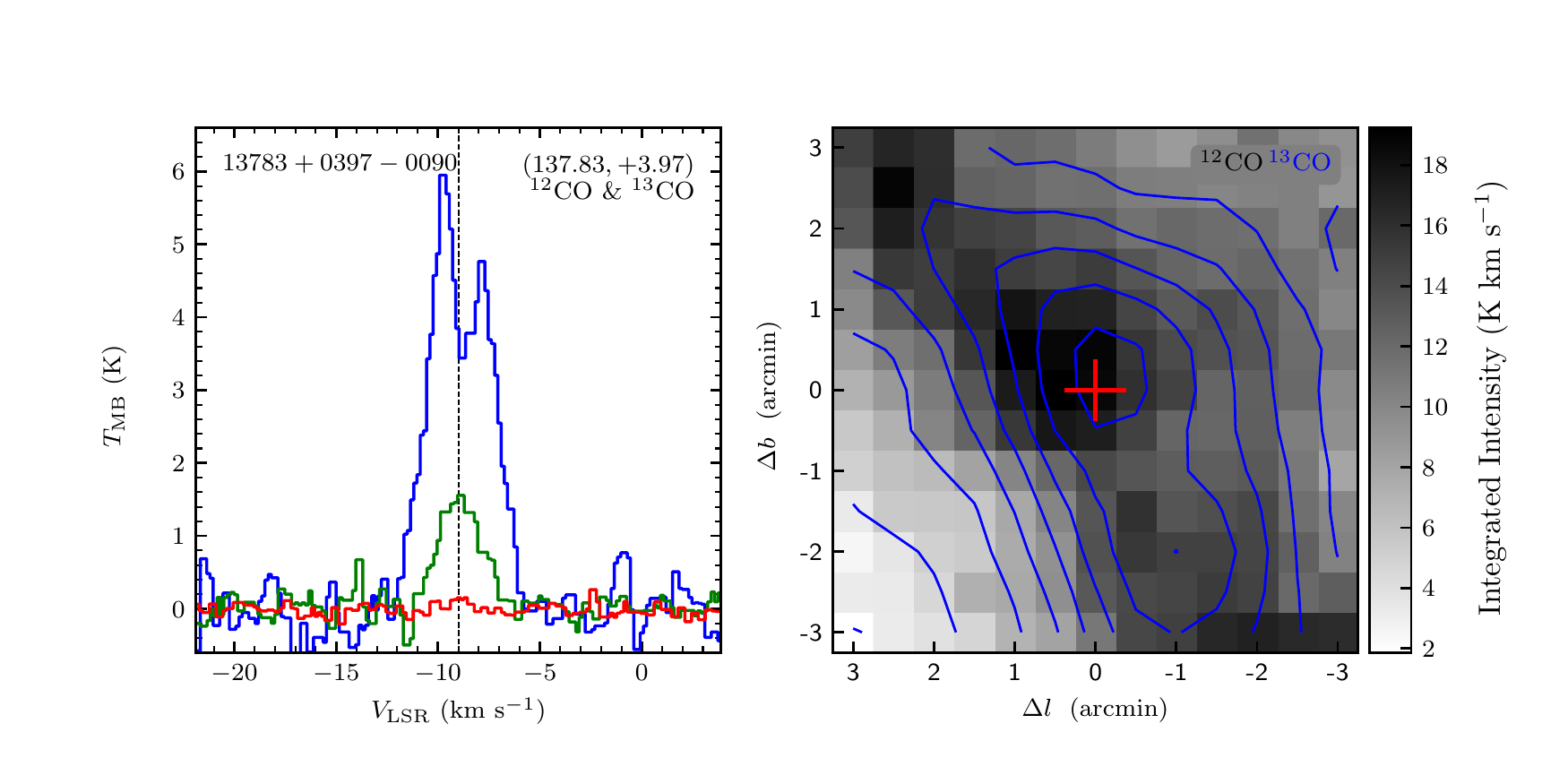}
\end{figure}
\clearpage

\begin{figure}
\includegraphics[width=9.0cm,angle=0]{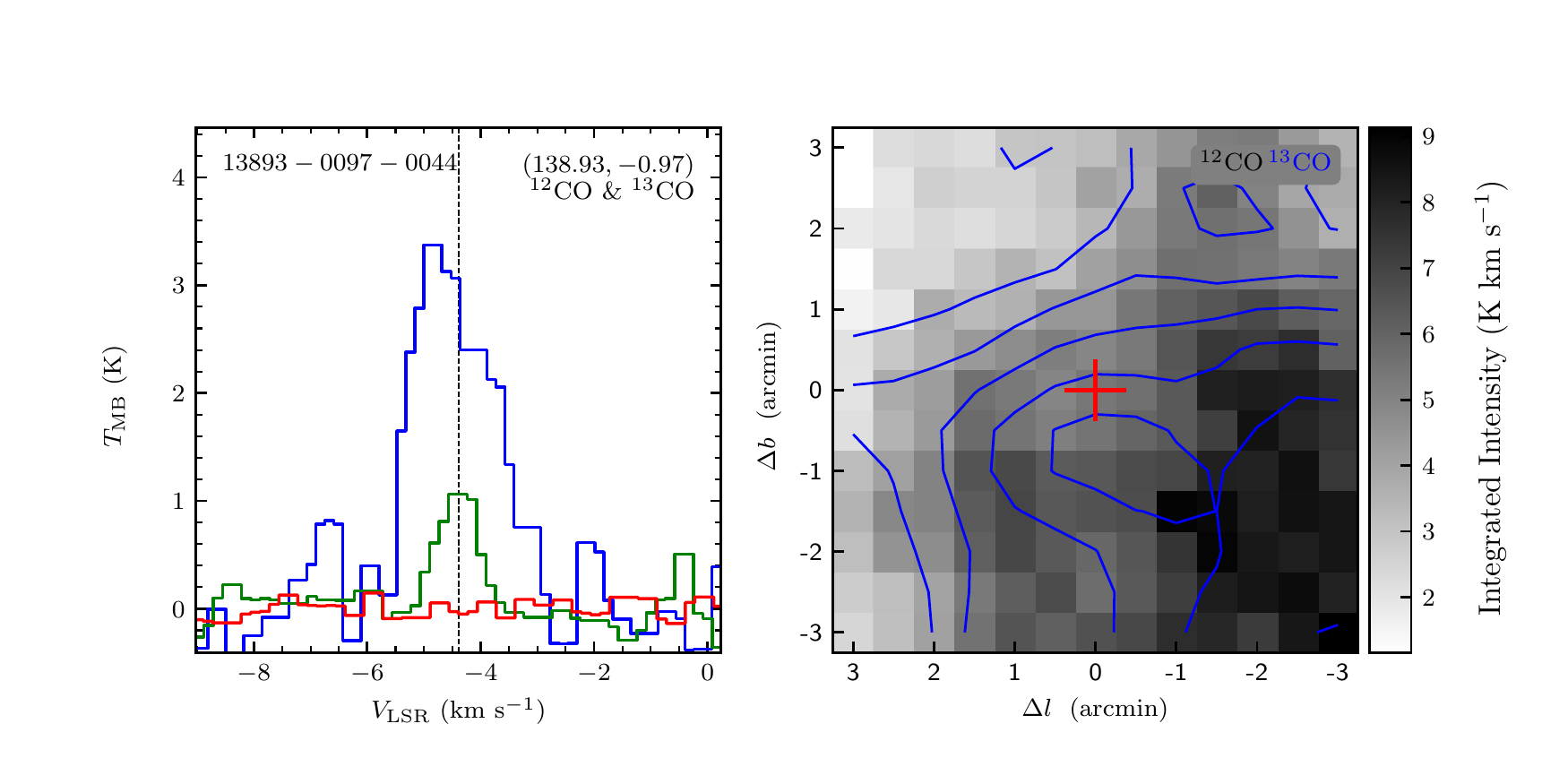}
\includegraphics[width=9.0cm,angle=0]{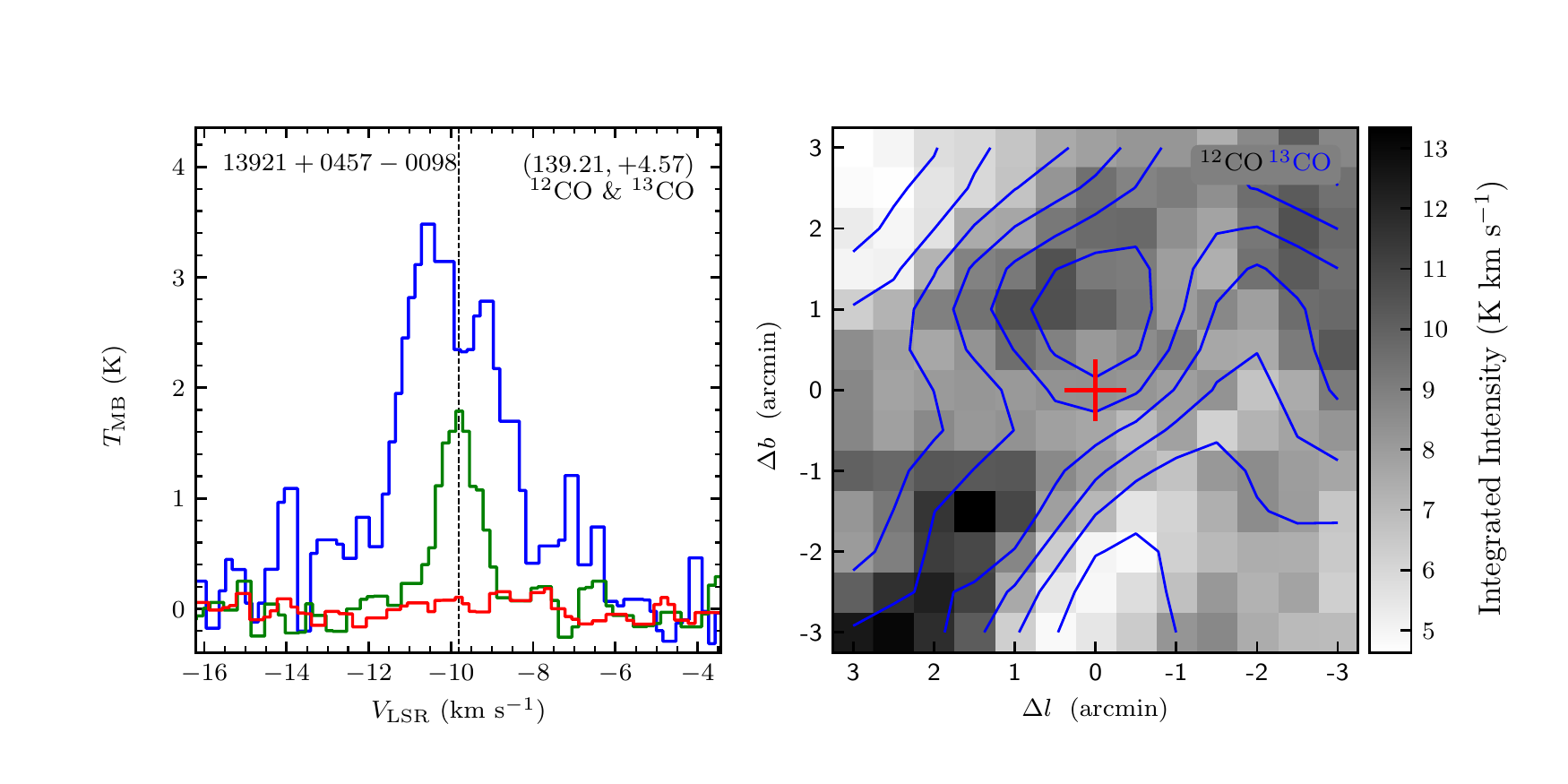}
\vspace{-0.5cm}

\includegraphics[width=9.0cm,angle=0]{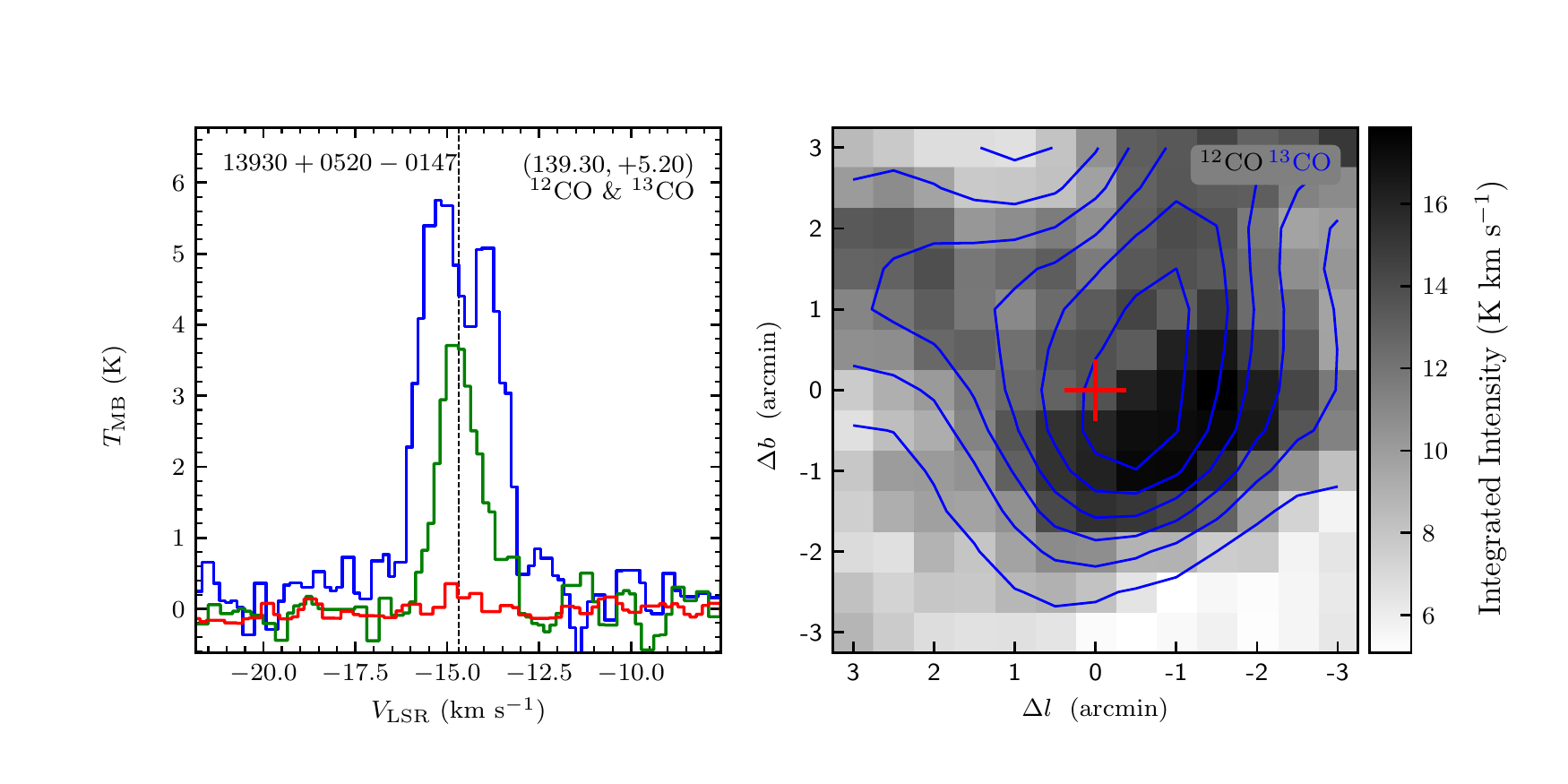}
\includegraphics[width=9.0cm,angle=0]{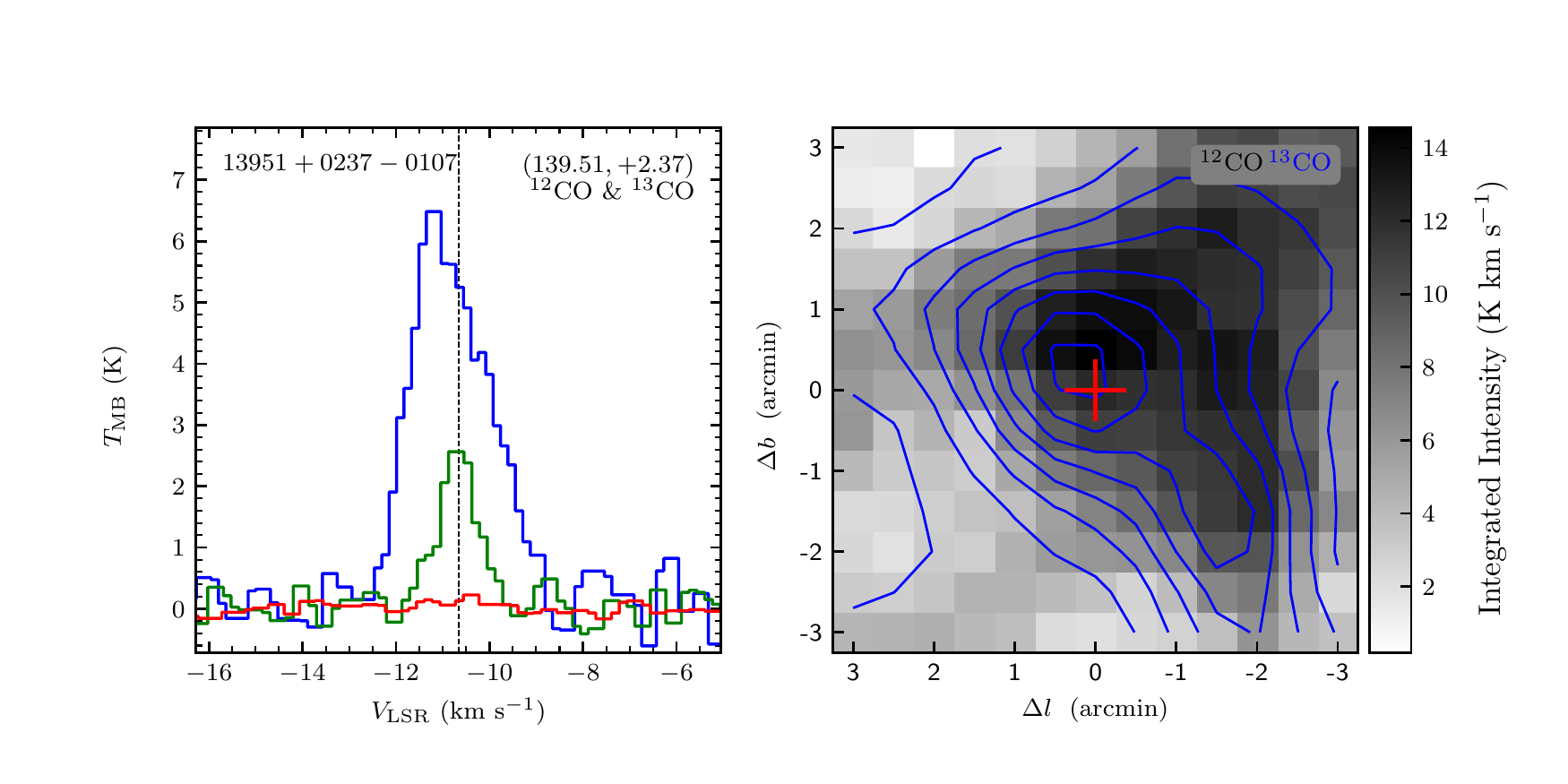}
\vspace{-0.5cm}

\includegraphics[width=9.0cm,angle=0]{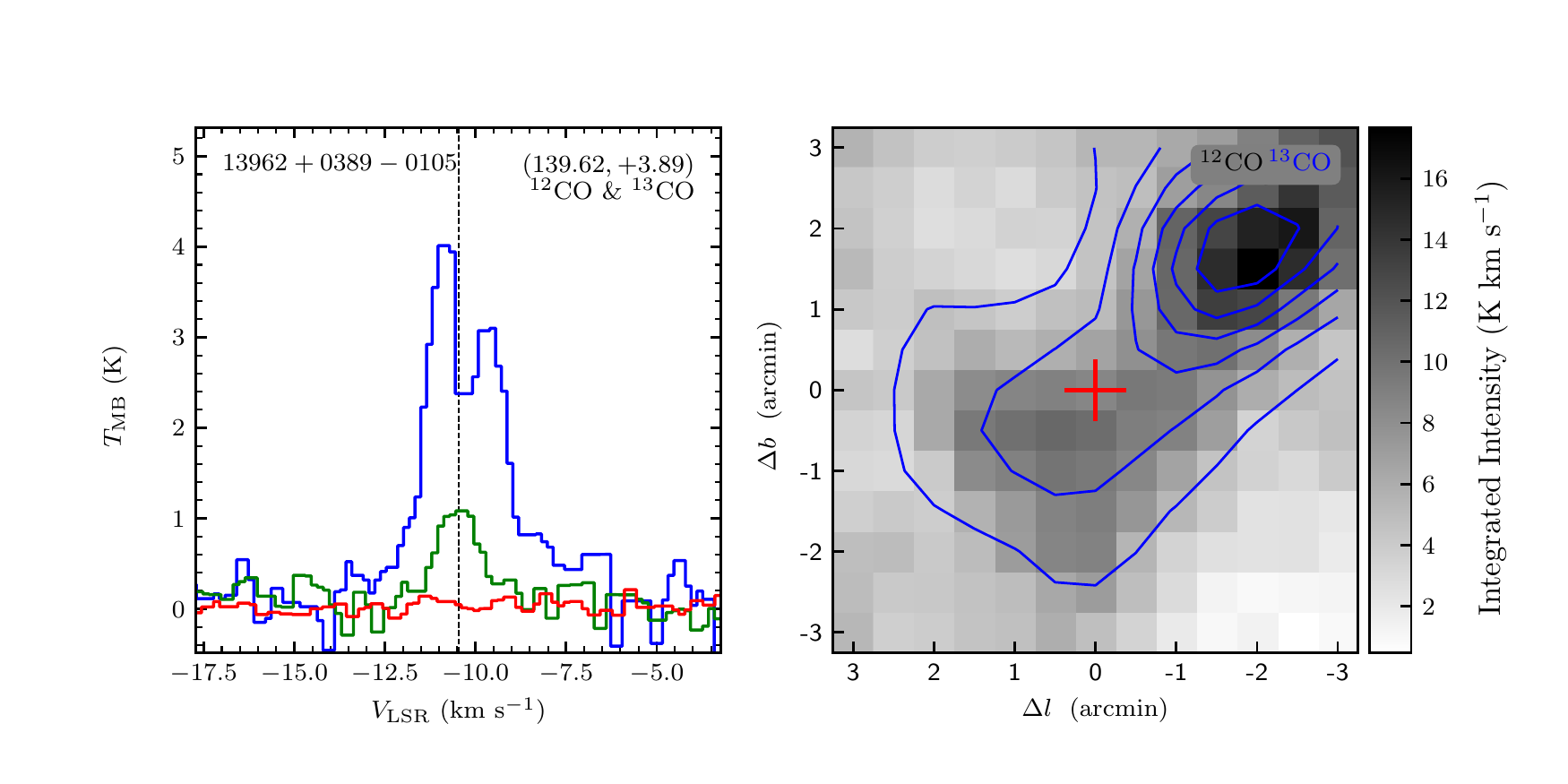}
\includegraphics[width=9.0cm,angle=0]{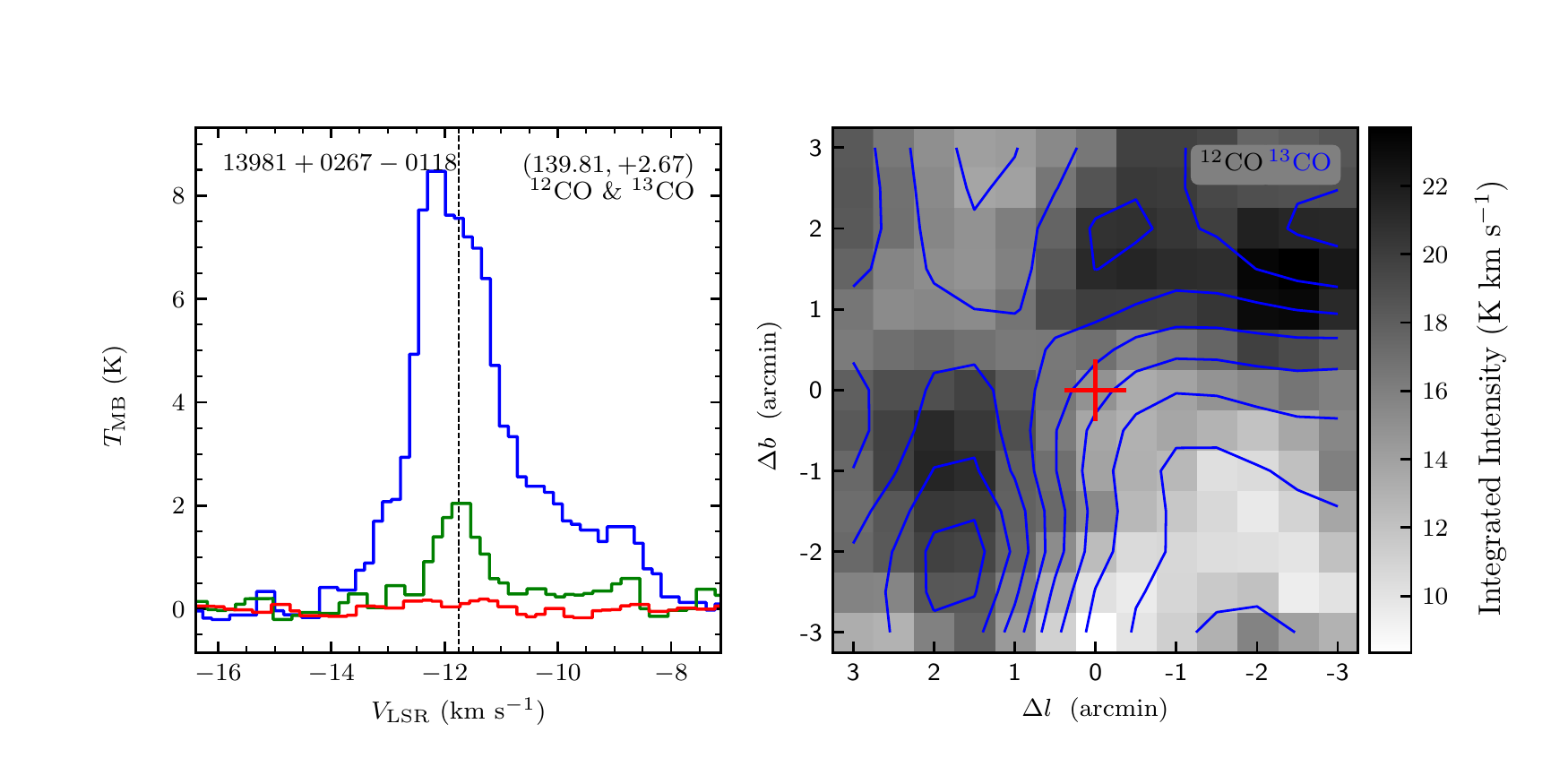}
\vspace{-0.5cm}

\includegraphics[width=9.0cm,angle=0]{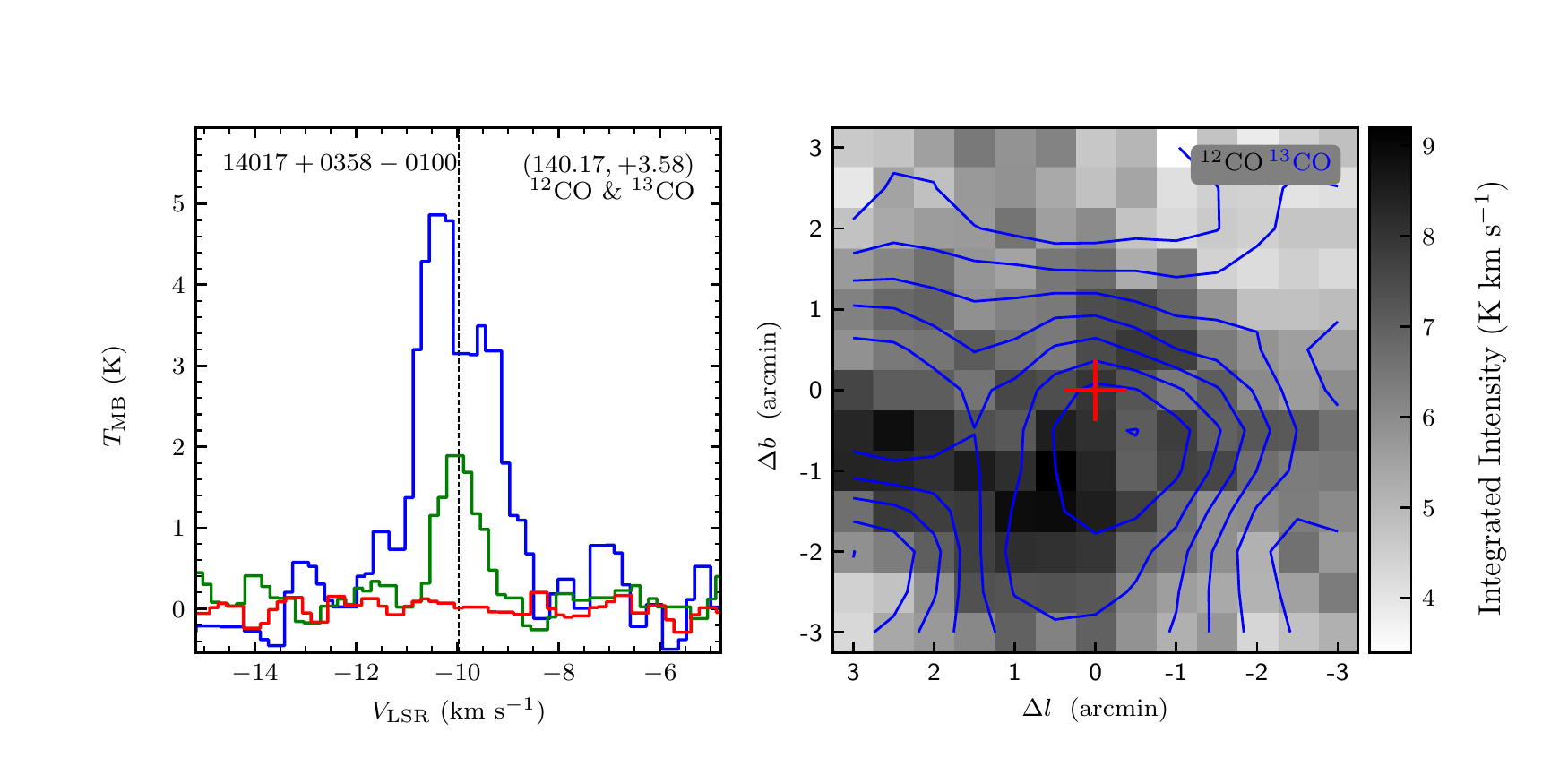}
\includegraphics[width=9.0cm,angle=0]{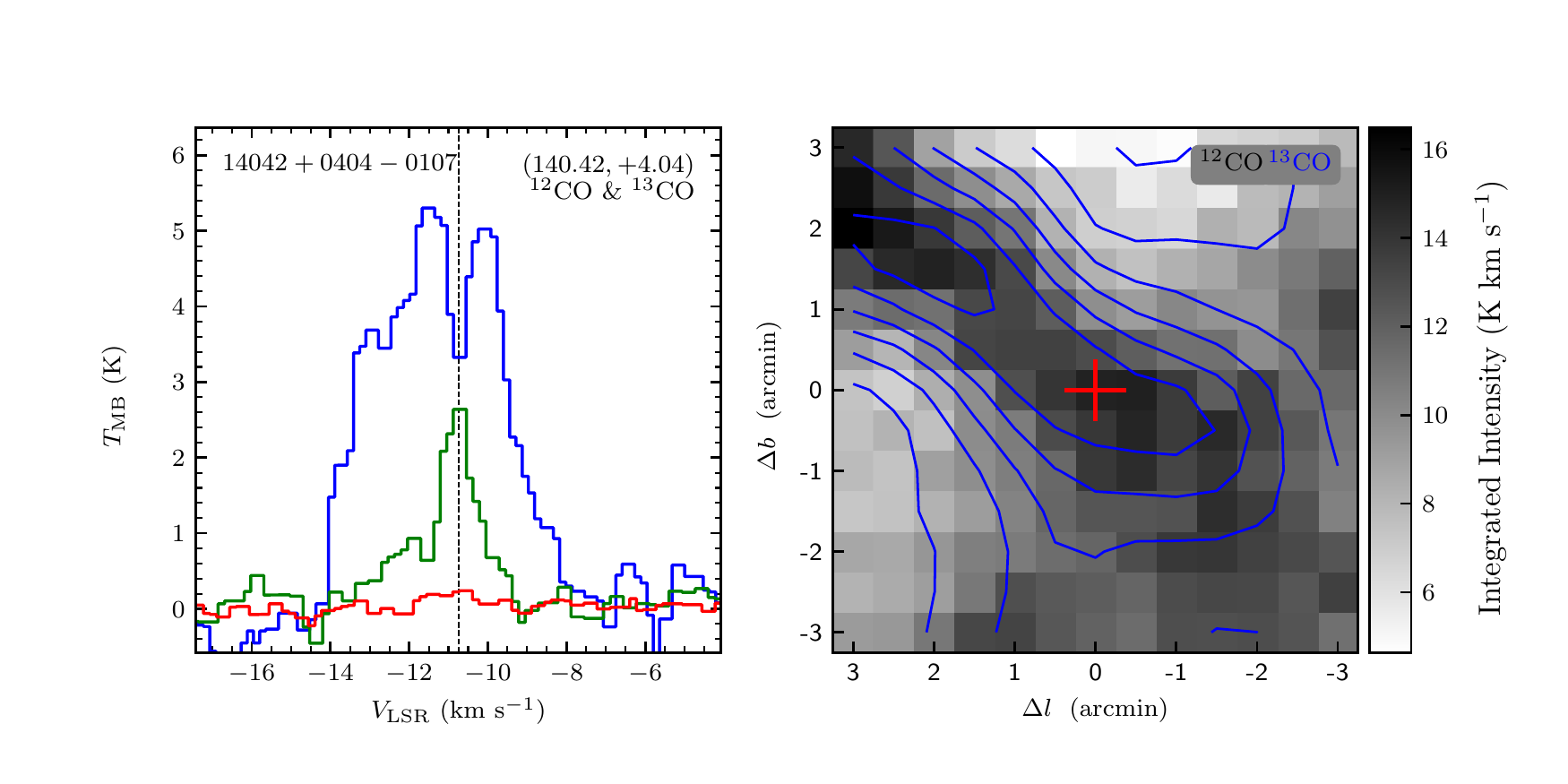}
\vspace{-0.5cm}

\includegraphics[width=9.0cm,angle=0]{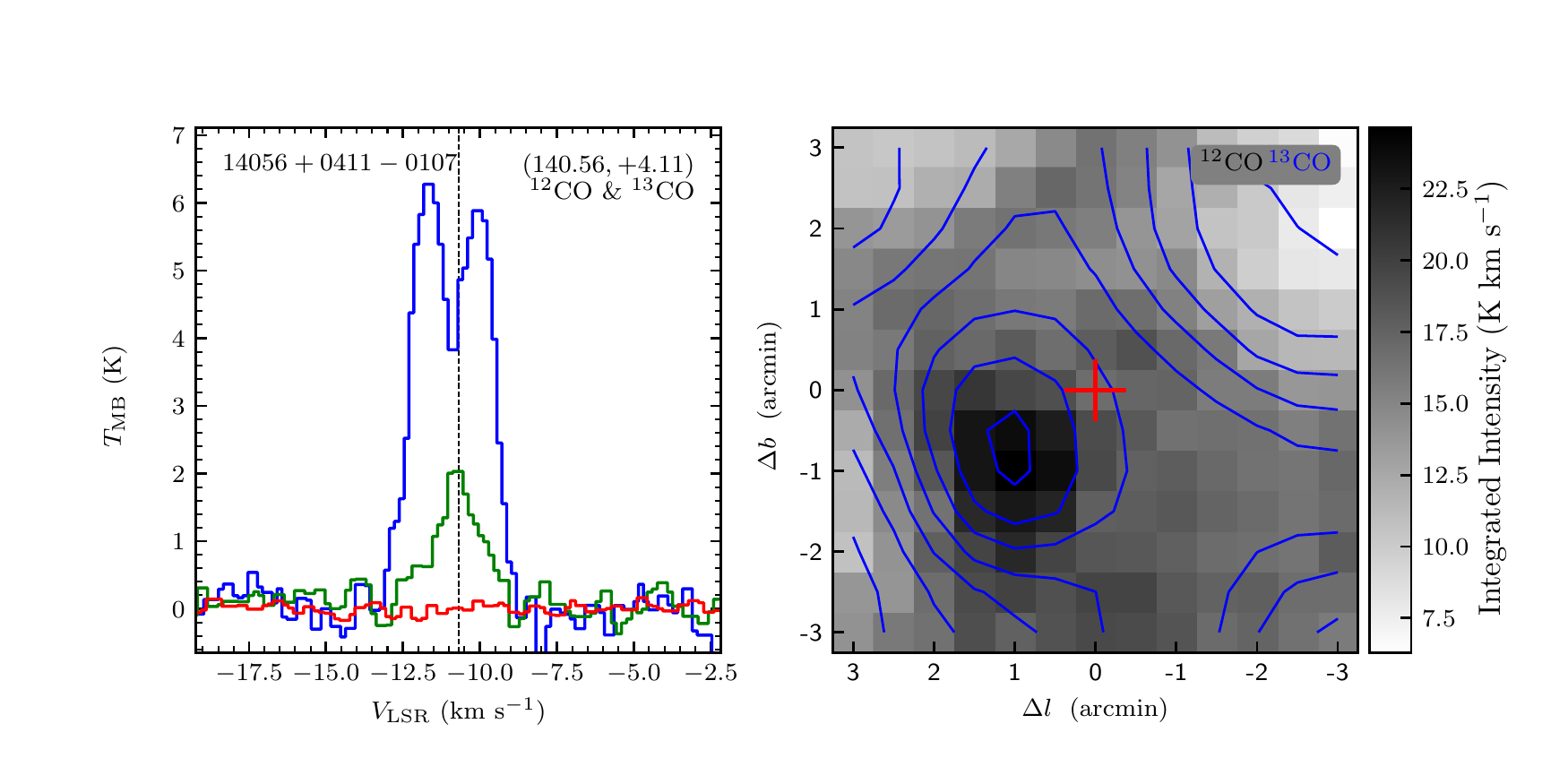}
\includegraphics[width=9.0cm,angle=0]{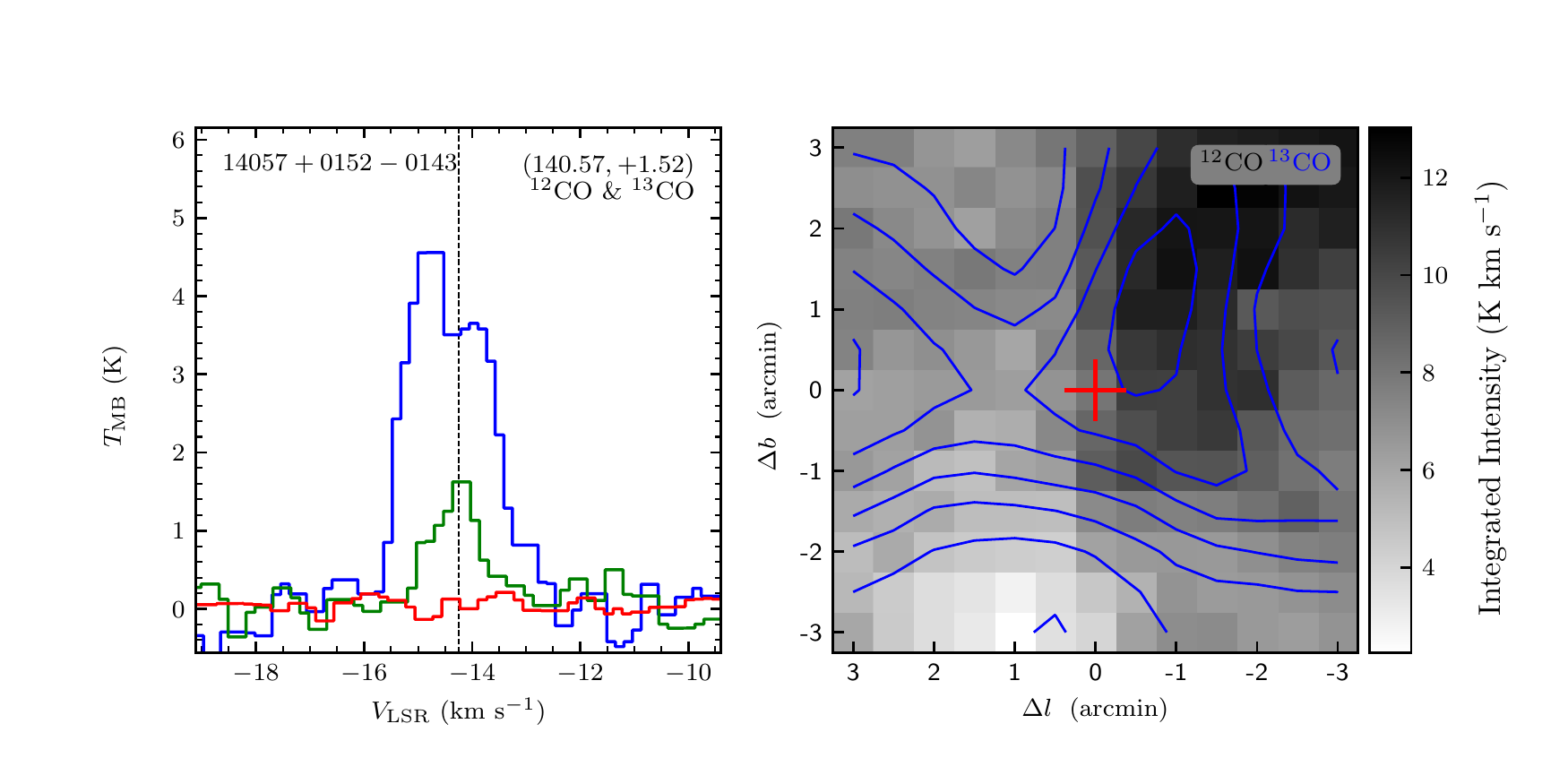}
\end{figure}
\clearpage

\begin{figure}
\includegraphics[width=9.0cm,angle=0]{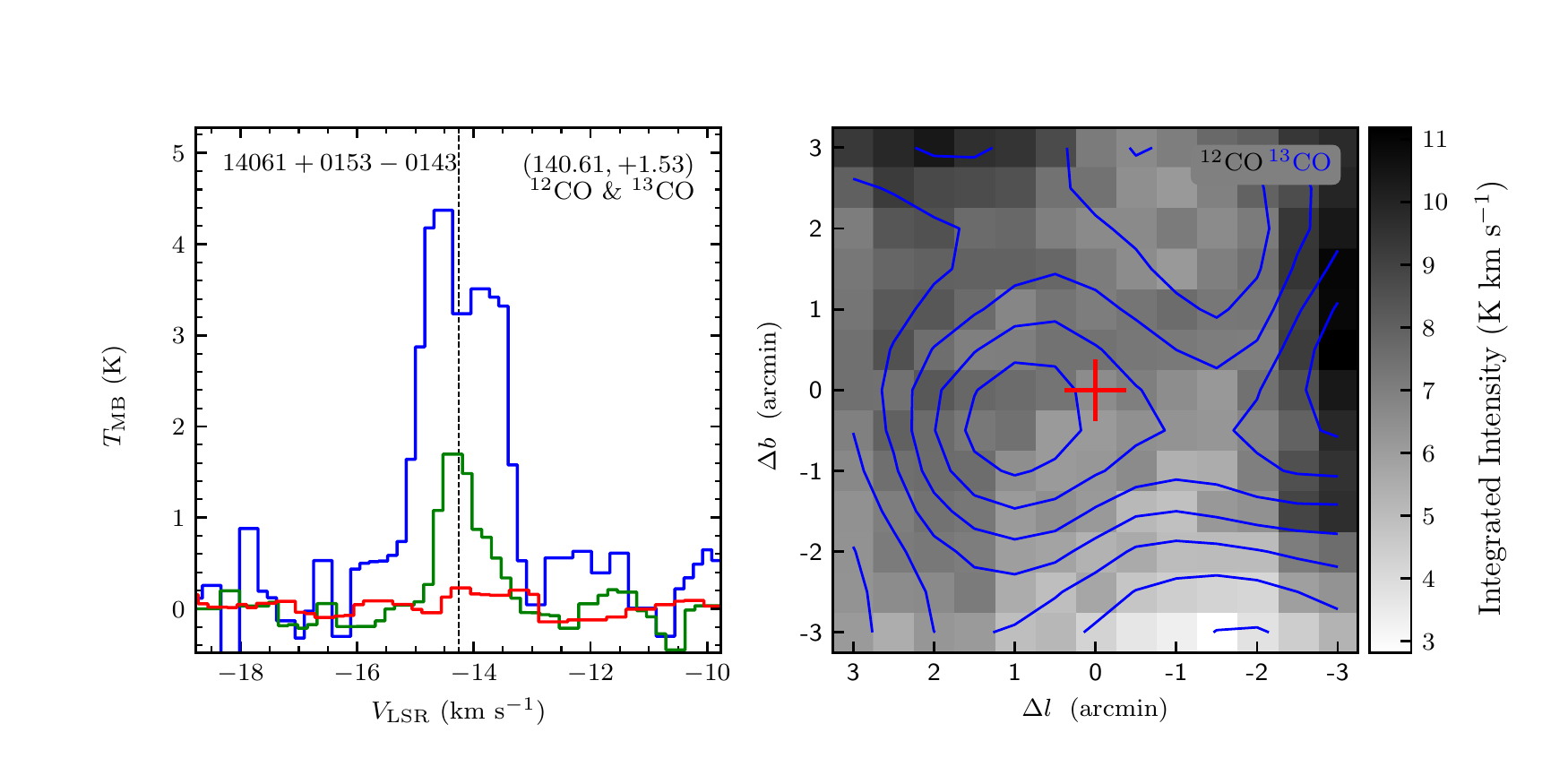}
\includegraphics[width=9.0cm,angle=0]{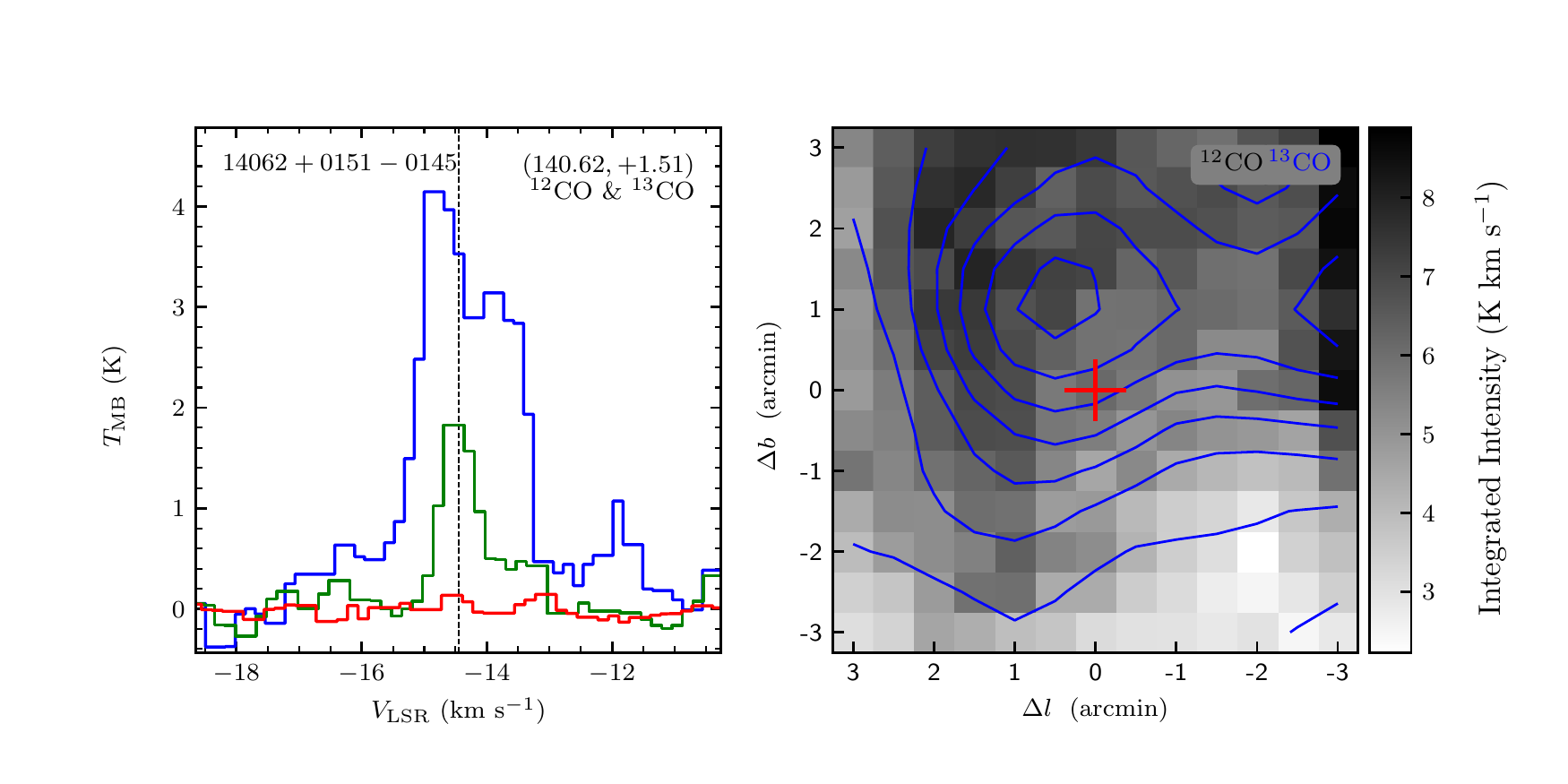}
\vspace{-0.5cm}

\includegraphics[width=9.0cm,angle=0]{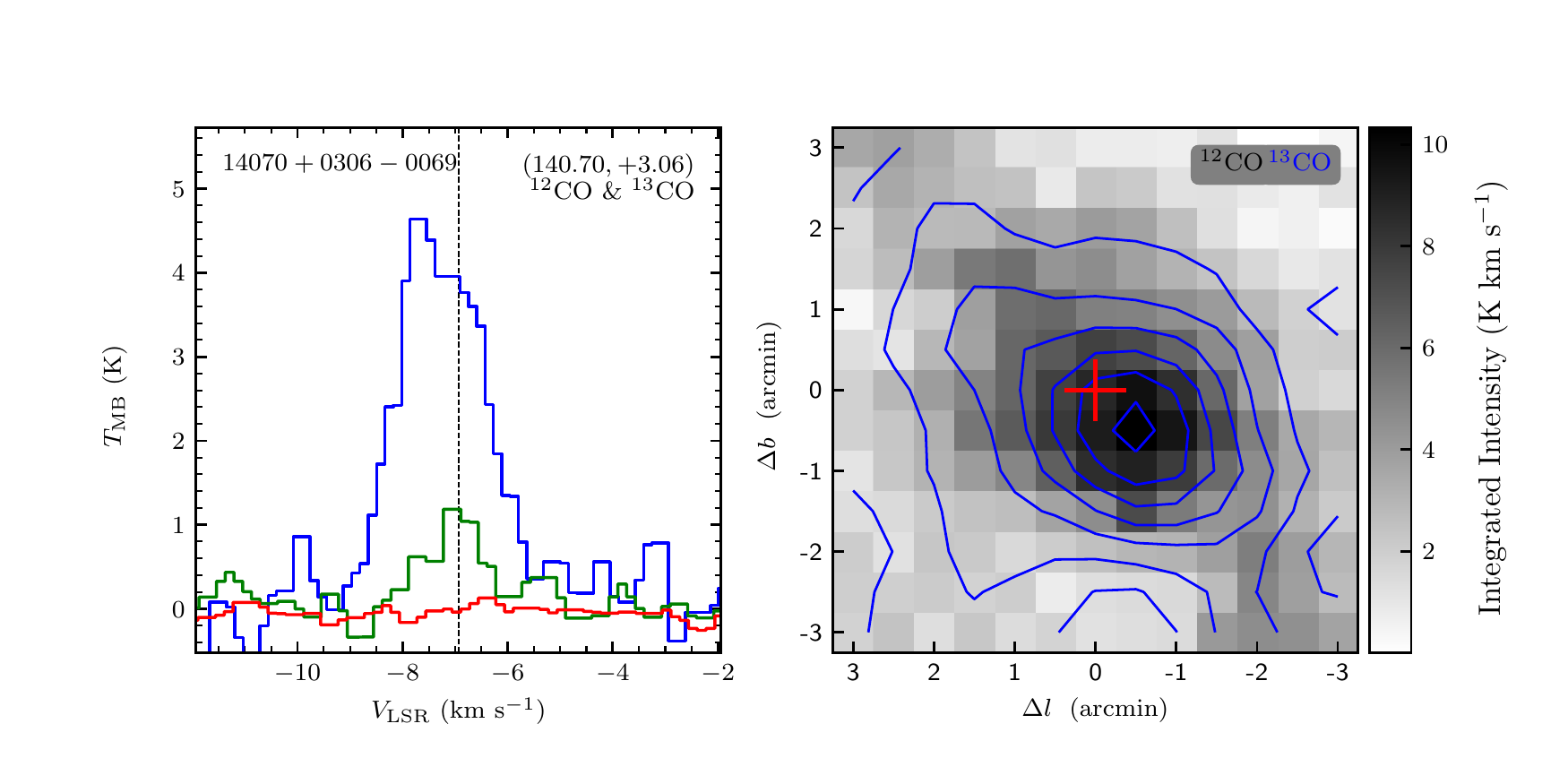}
\includegraphics[width=9.0cm,angle=0]{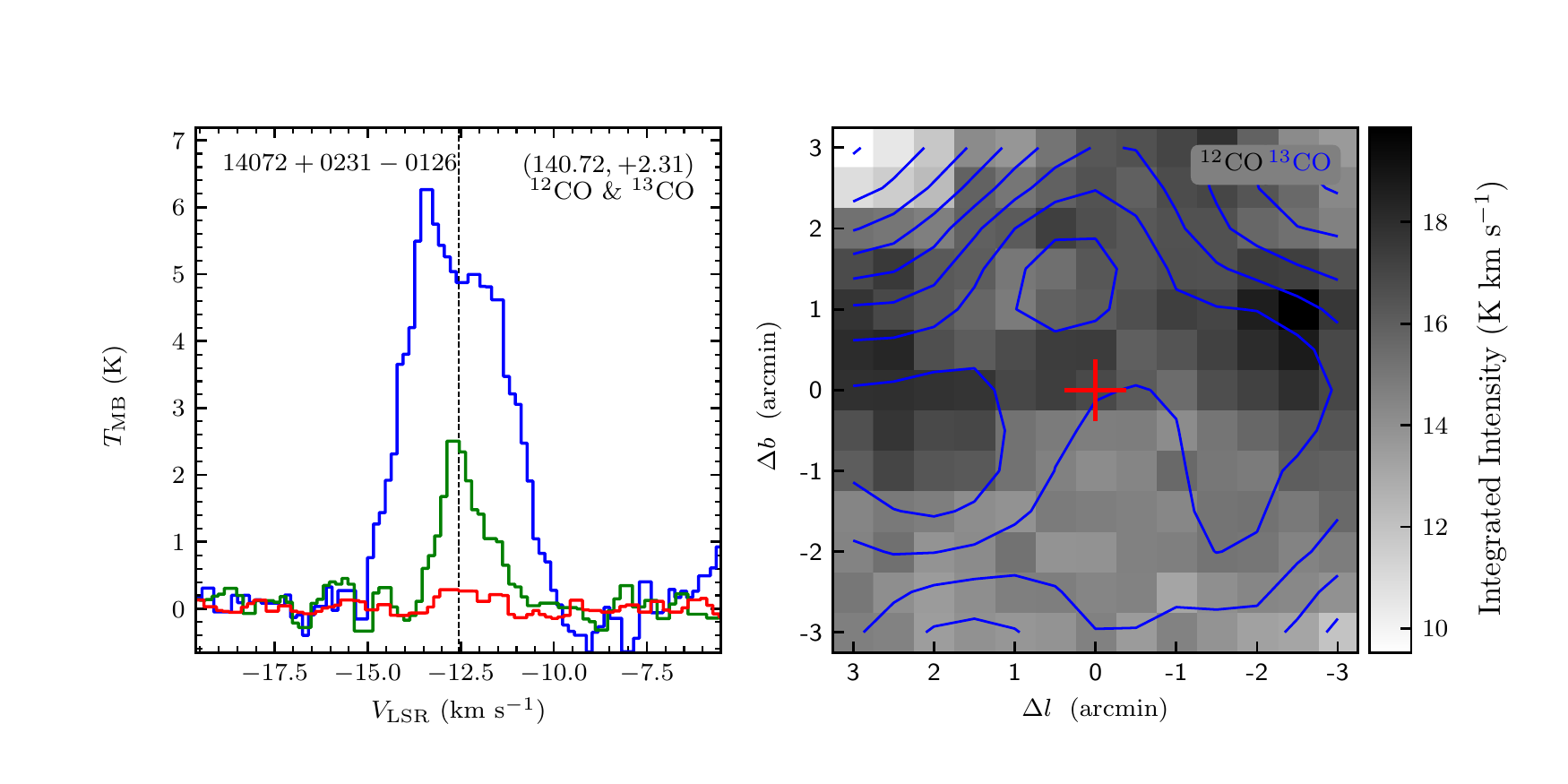}
\vspace{-0.5cm}

\includegraphics[width=9.0cm,angle=0]{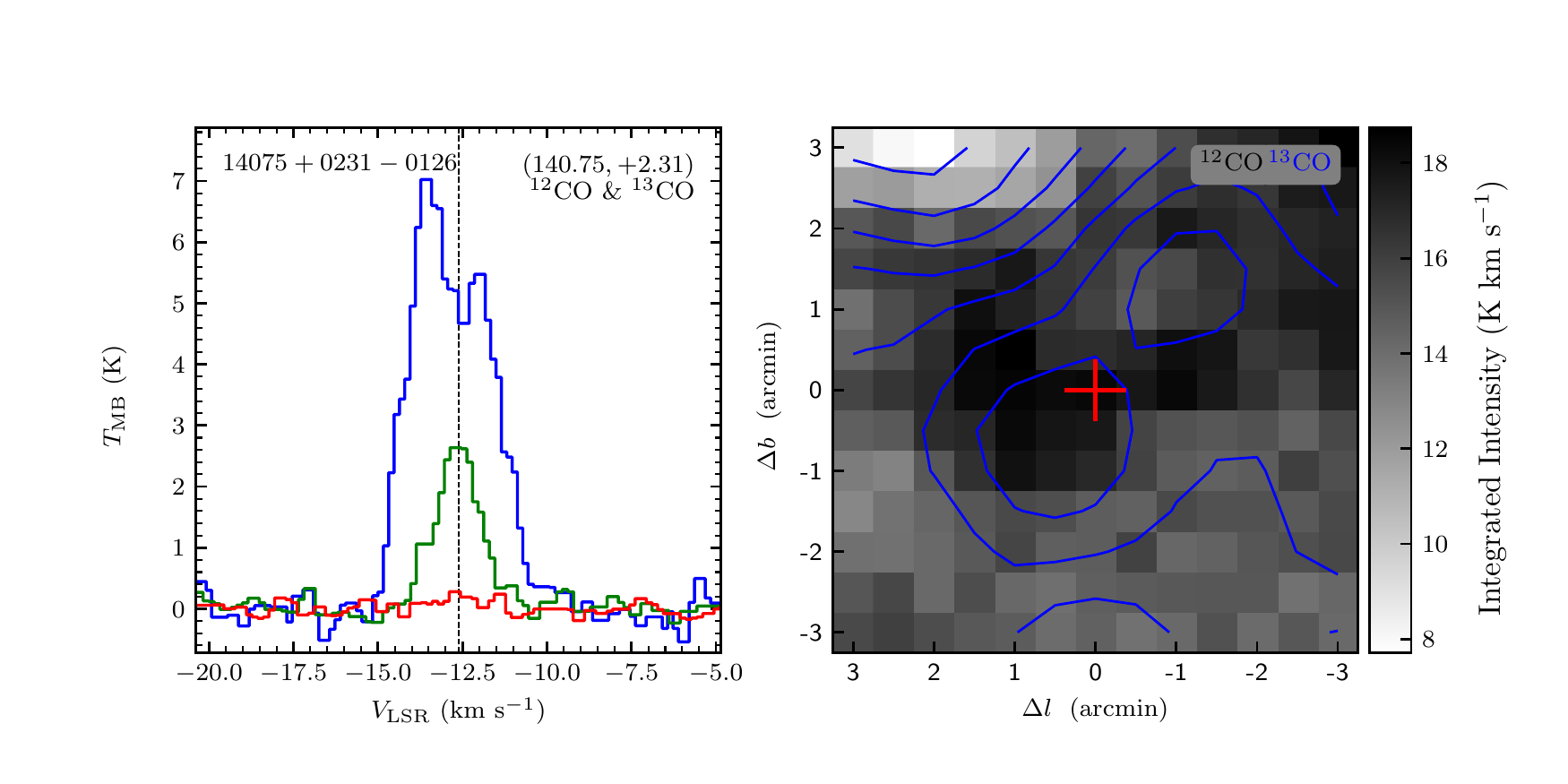}
\includegraphics[width=9.0cm,angle=0]{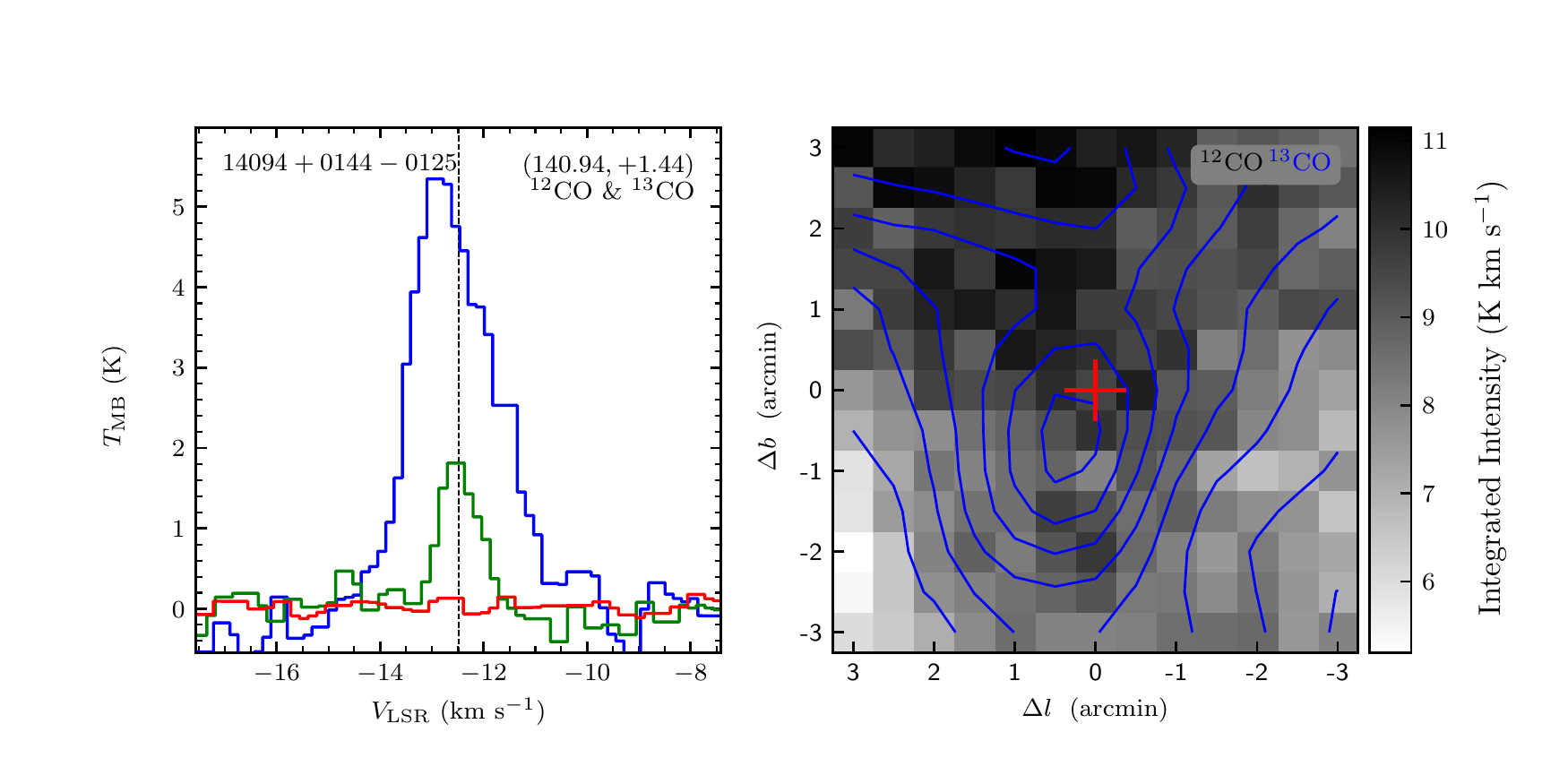}
\vspace{-0.5cm}

\includegraphics[width=9.0cm,angle=0]{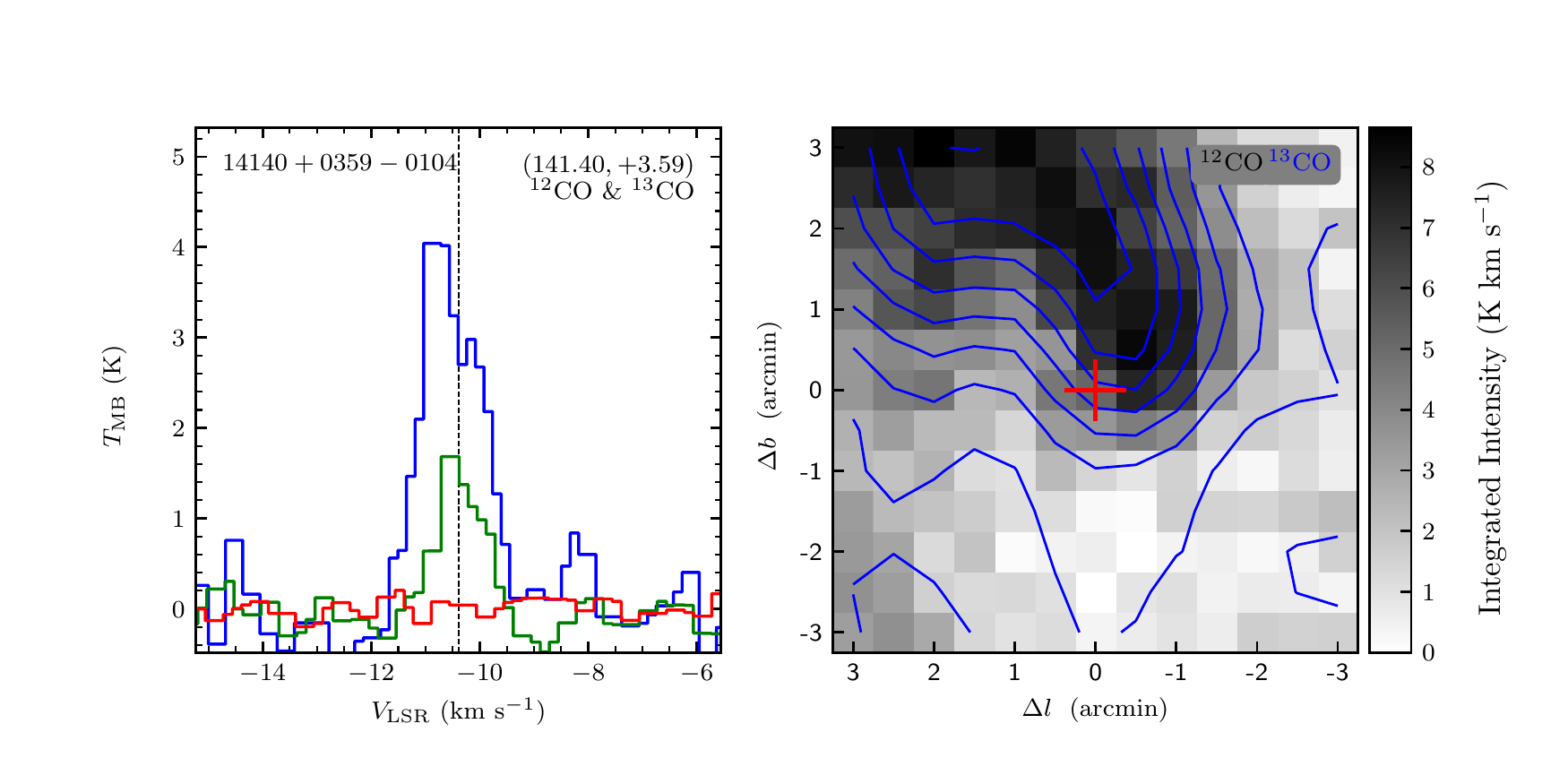}
\includegraphics[width=9.0cm,angle=0]{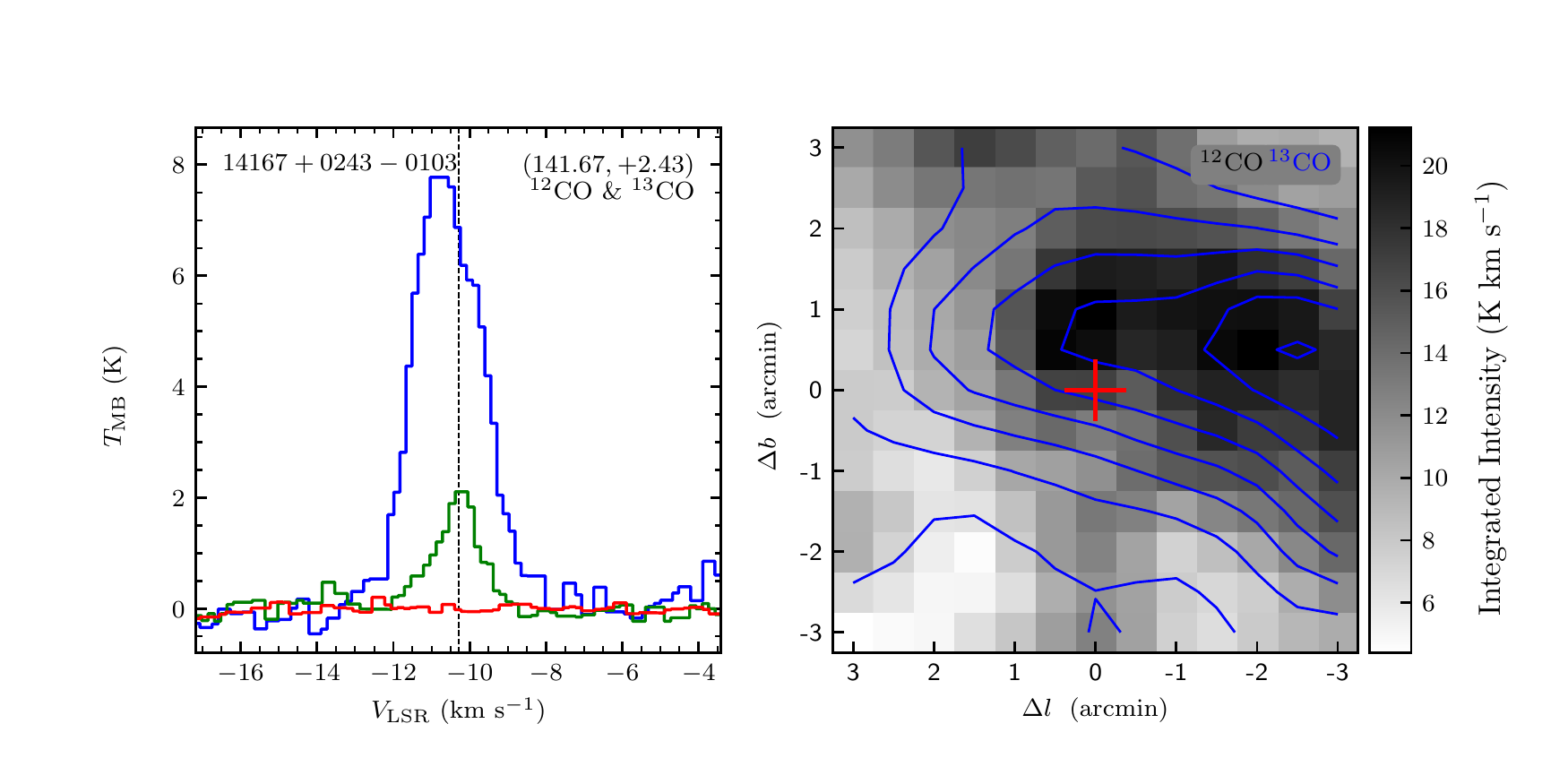}
\vspace{-0.5cm}

\includegraphics[width=9.0cm,angle=0]{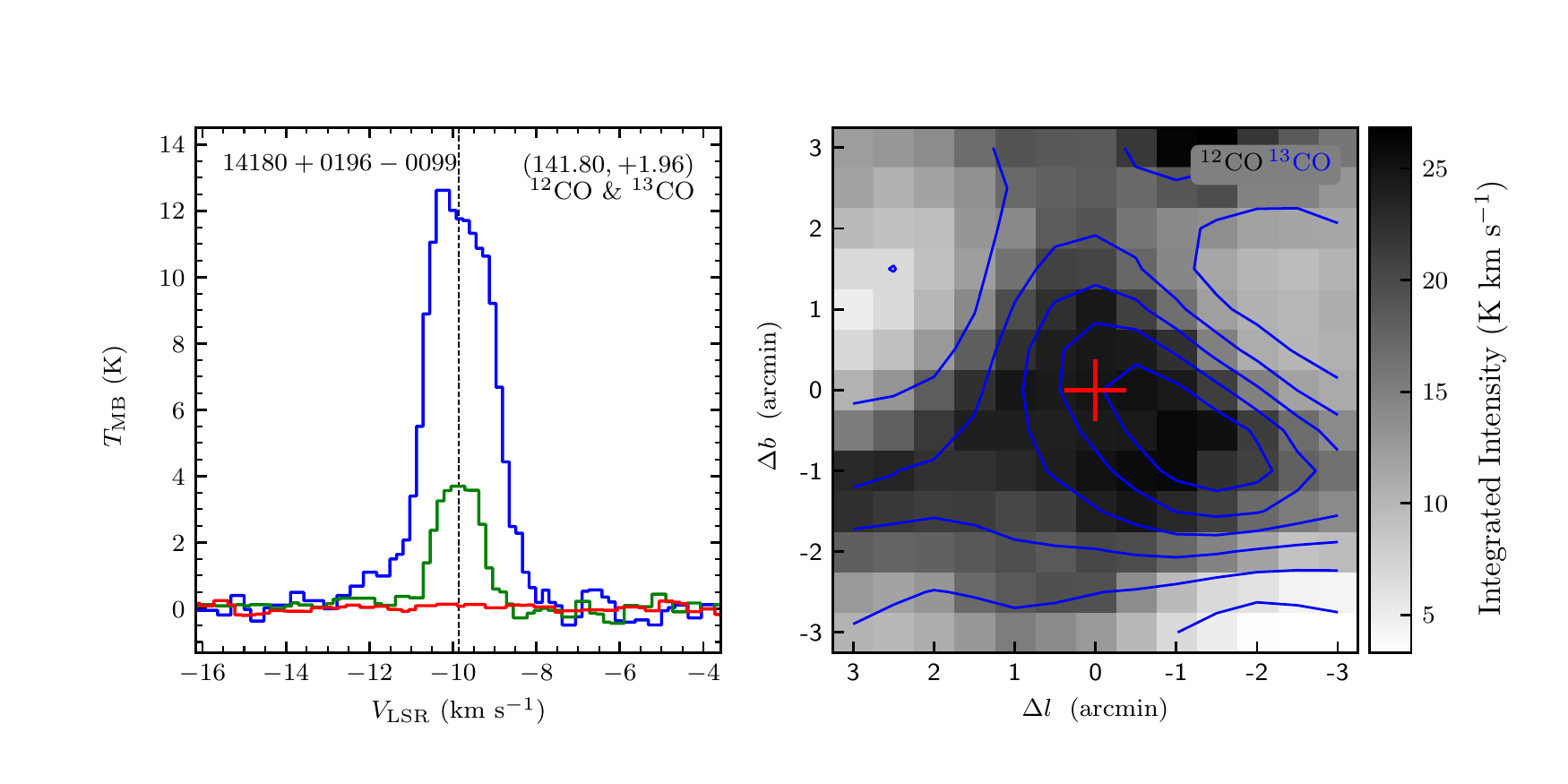}
\includegraphics[width=9.0cm,angle=0]{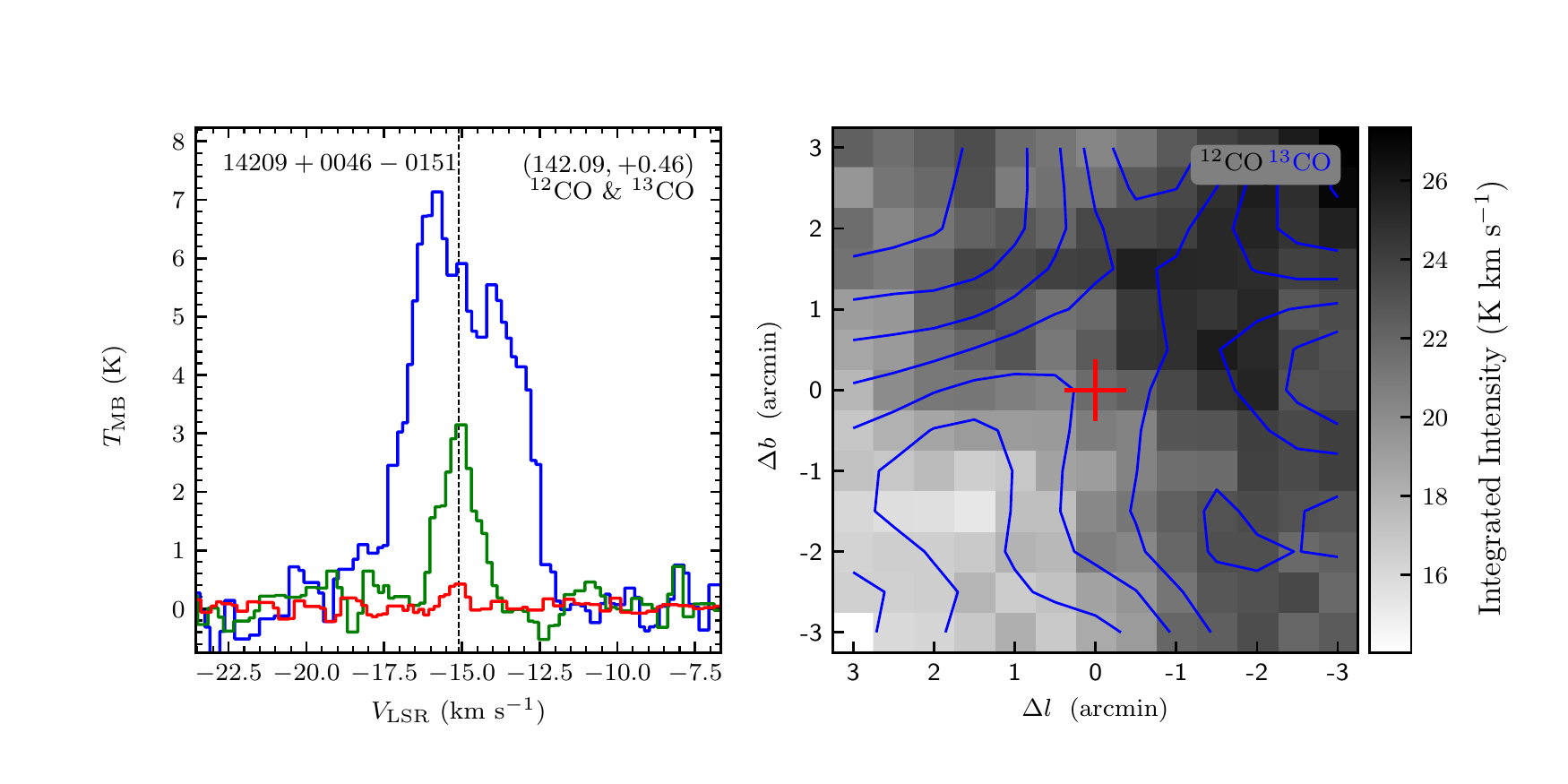}
\end{figure}
\clearpage

\begin{figure}
\includegraphics[width=9.0cm,angle=0]{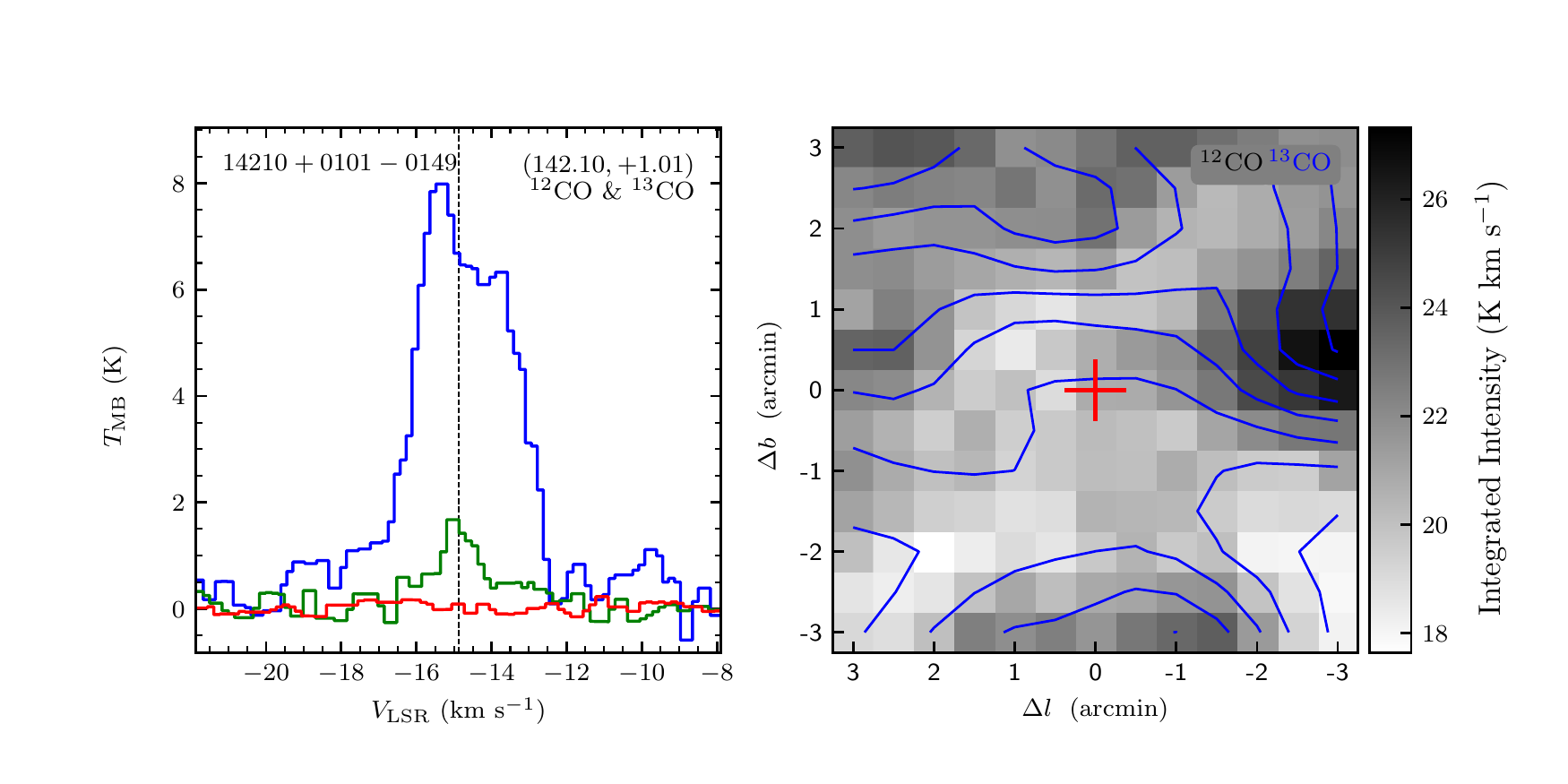}
\includegraphics[width=9.0cm,angle=0]{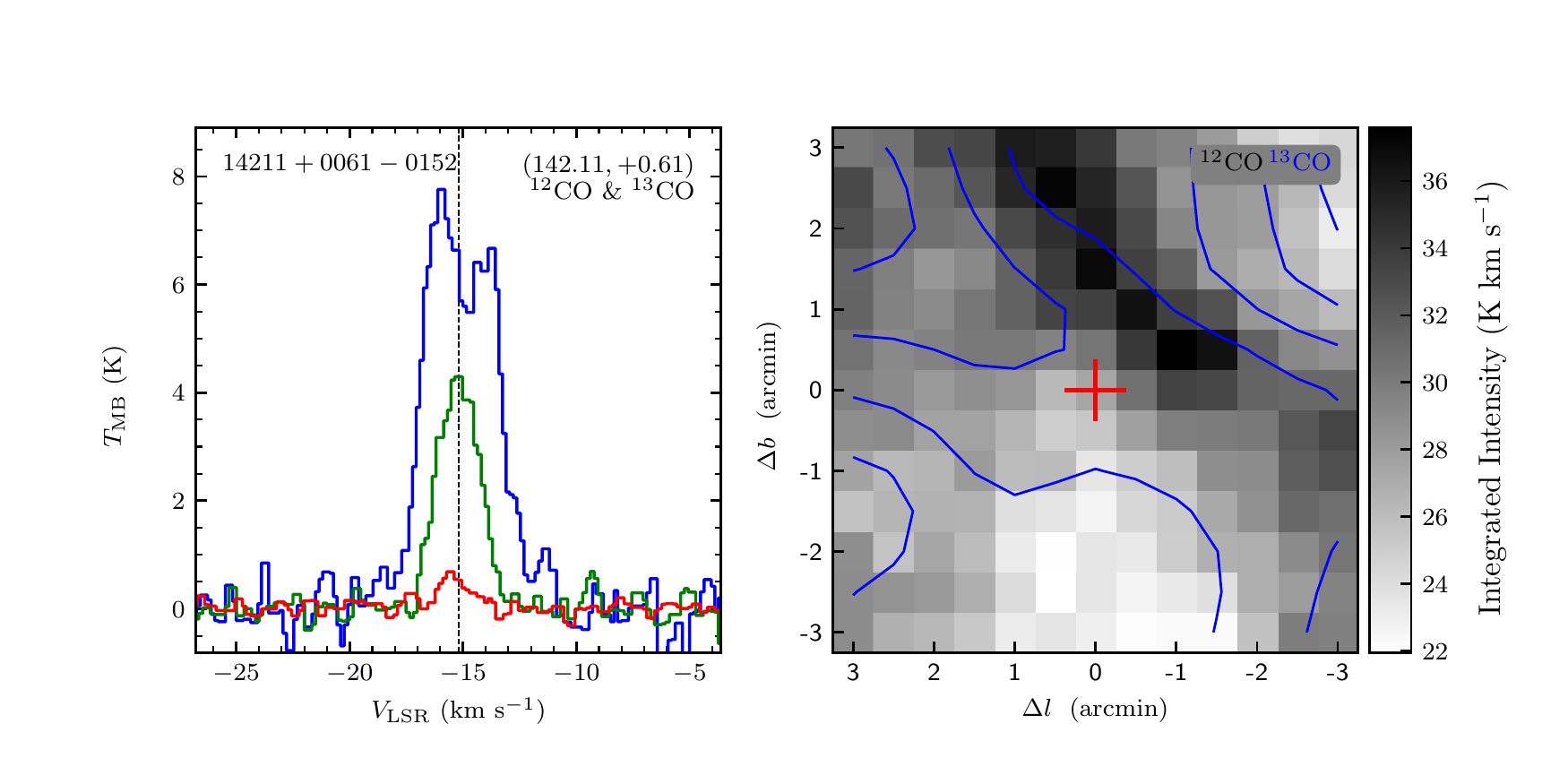}
\vspace{-0.5cm}

\includegraphics[width=9.0cm,angle=0]{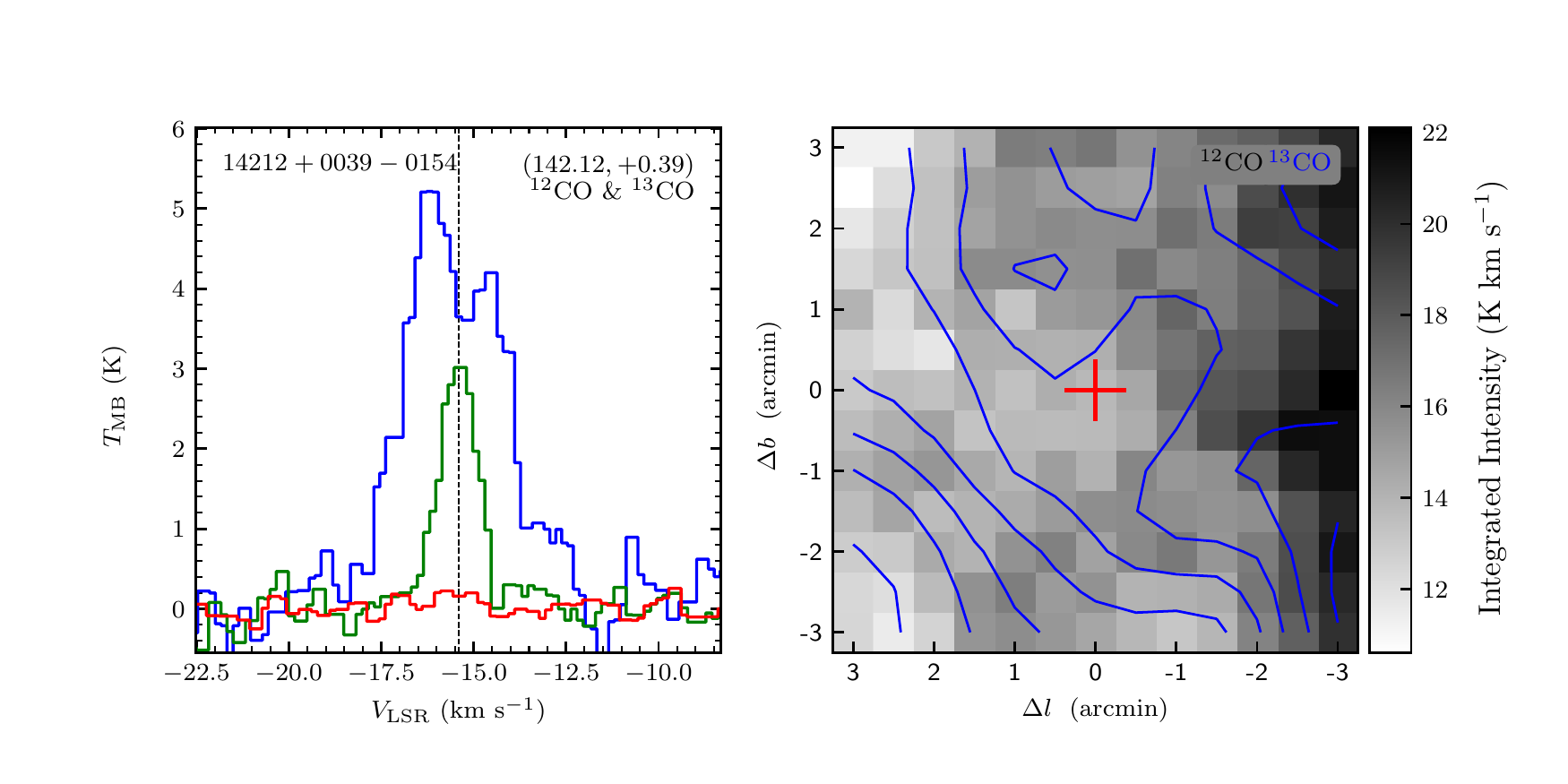}
\includegraphics[width=9.0cm,angle=0]{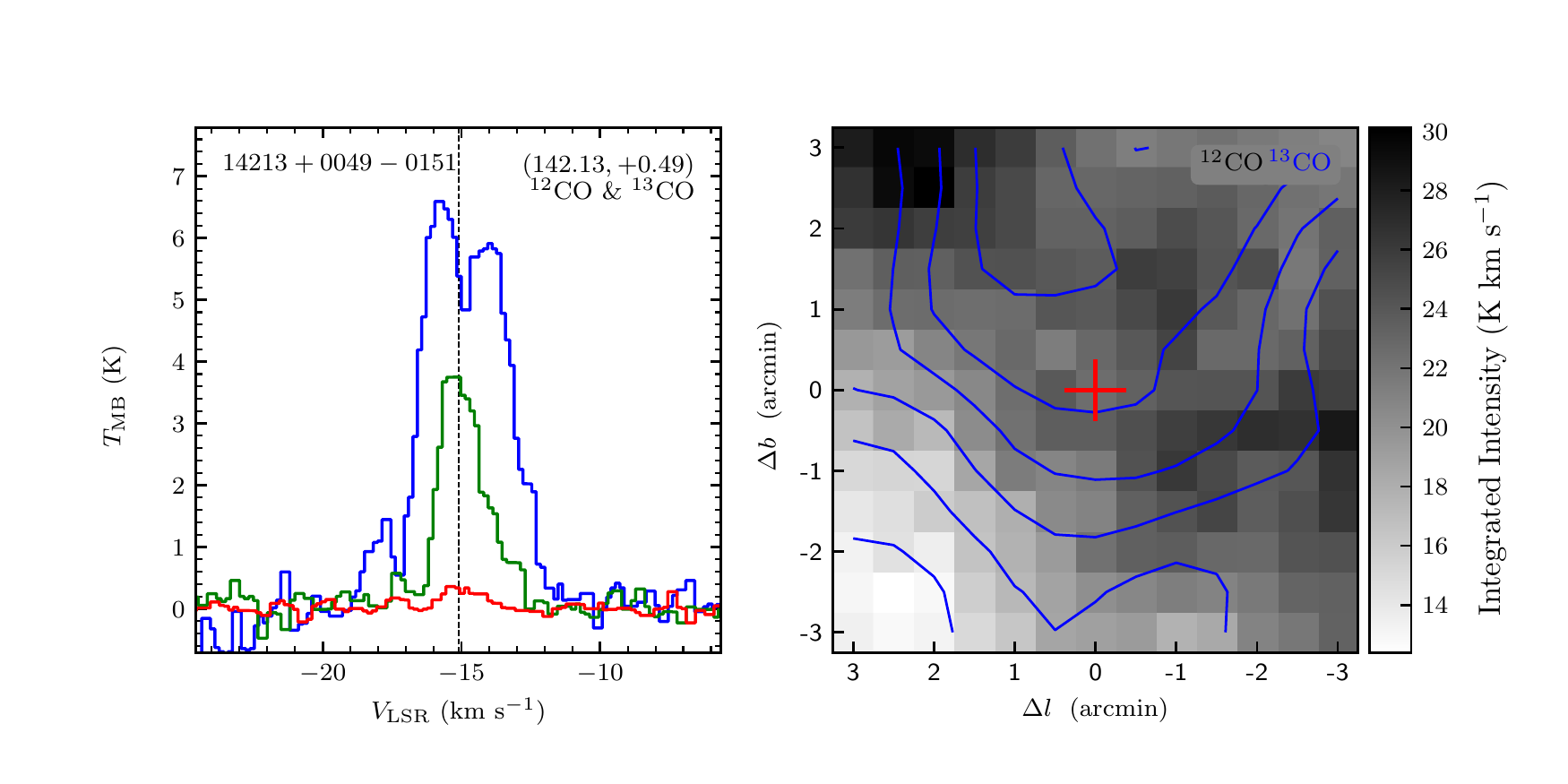}
\vspace{-0.5cm}

\includegraphics[width=9.0cm,angle=0]{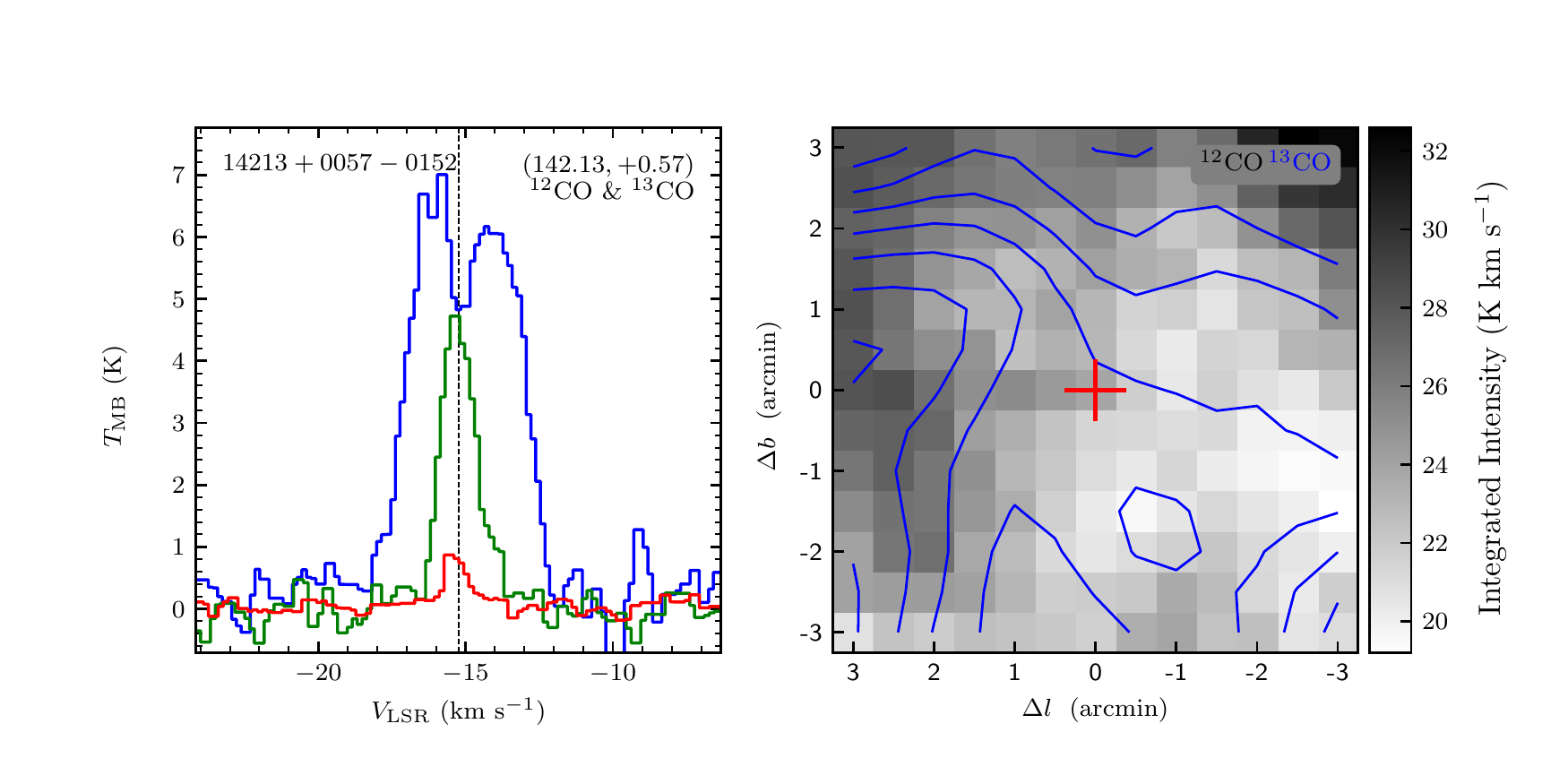}
\includegraphics[width=9.0cm,angle=0]{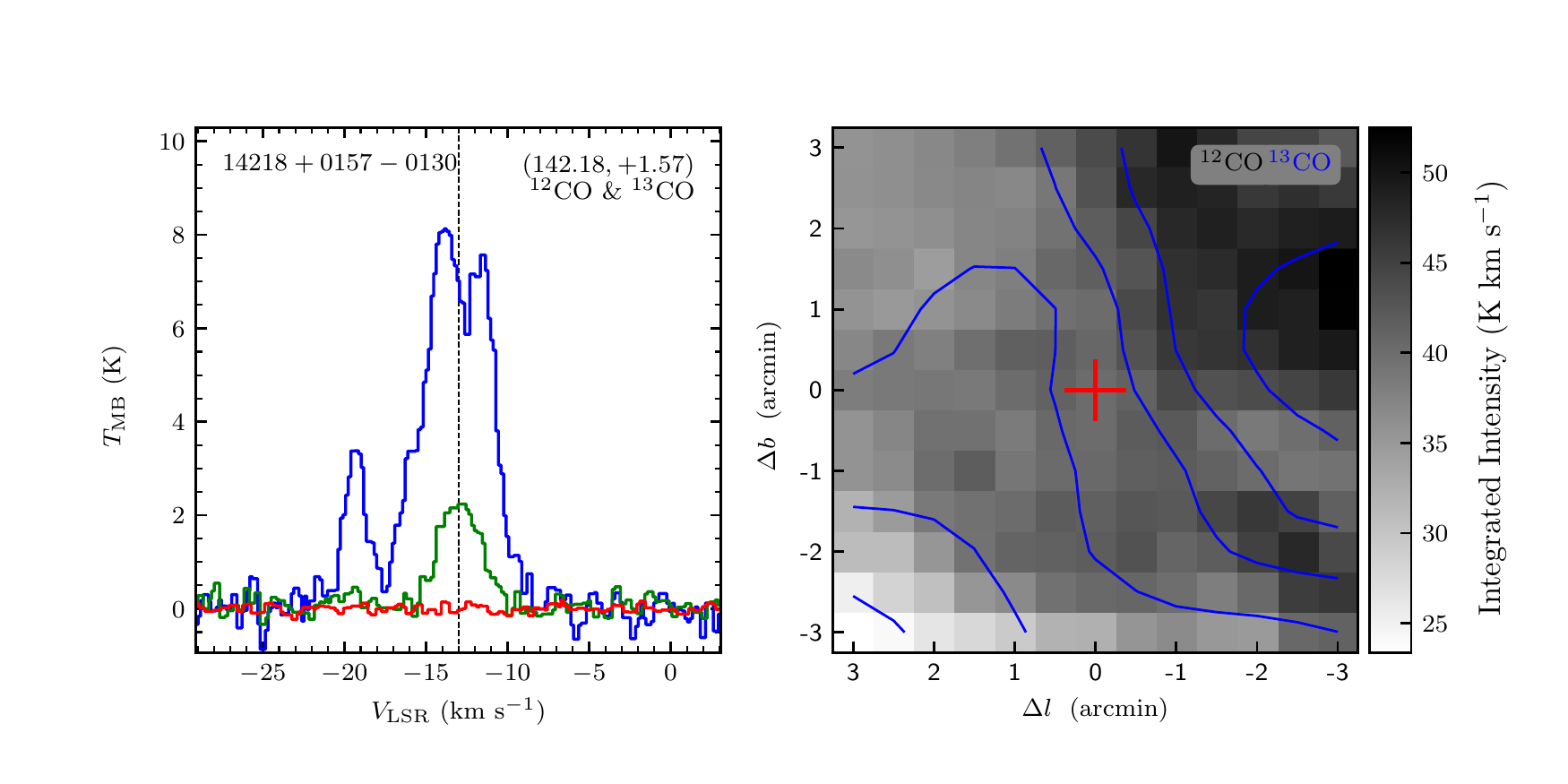}
\vspace{-0.5cm}

\includegraphics[width=9.0cm,angle=0]{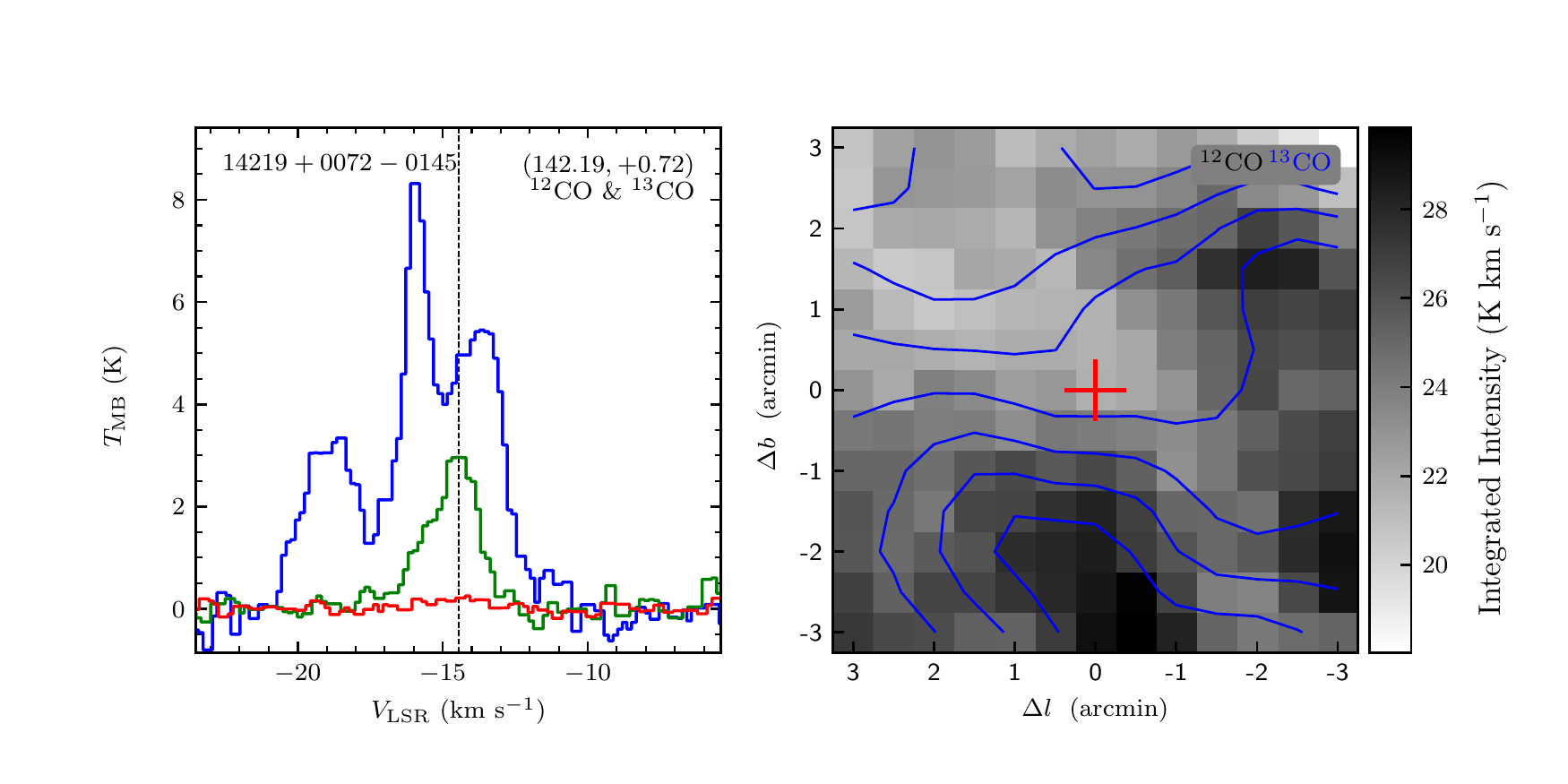}
\includegraphics[width=9.0cm,angle=0]{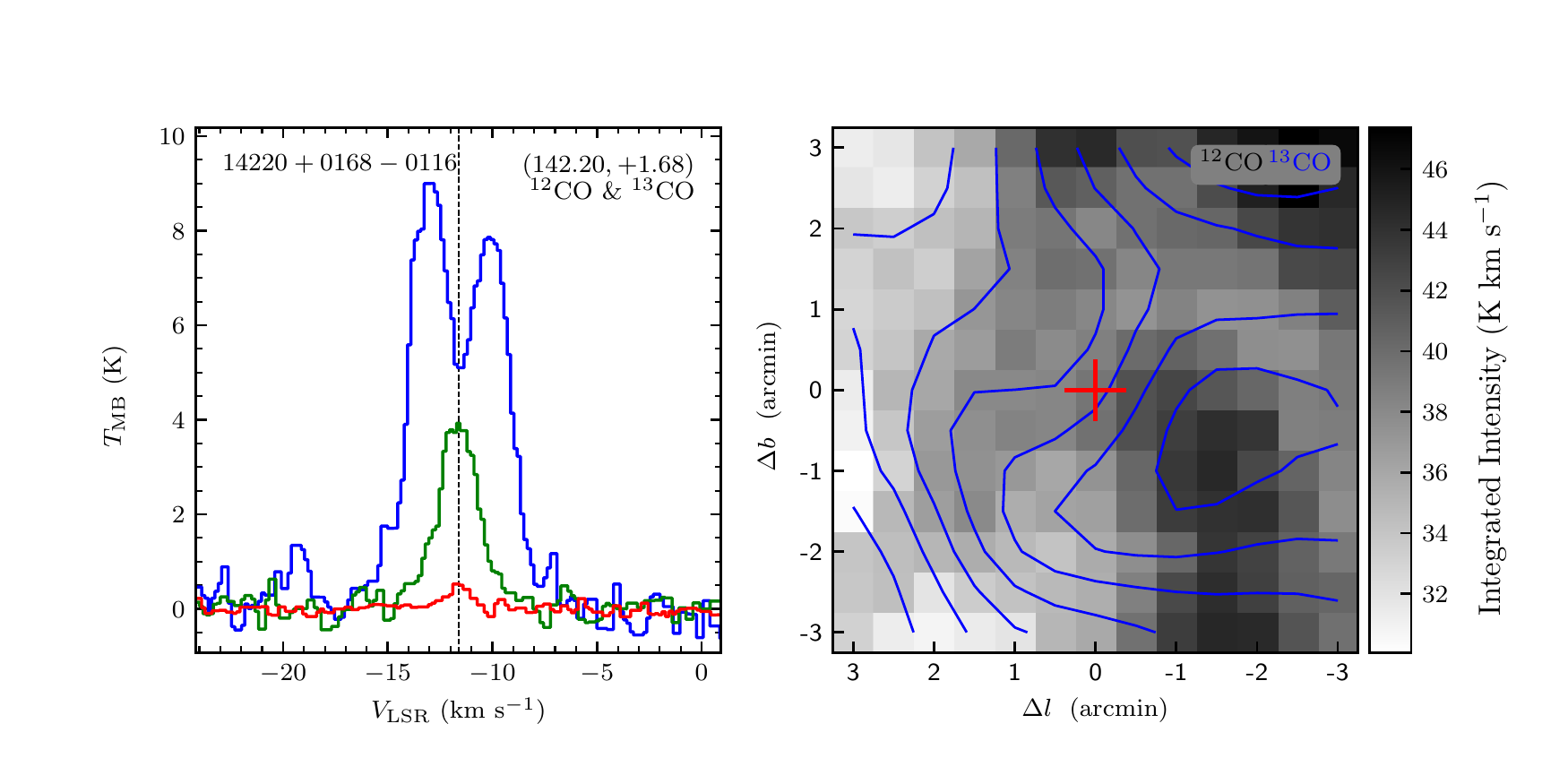}
\vspace{-0.5cm}

\includegraphics[width=9.0cm,angle=0]{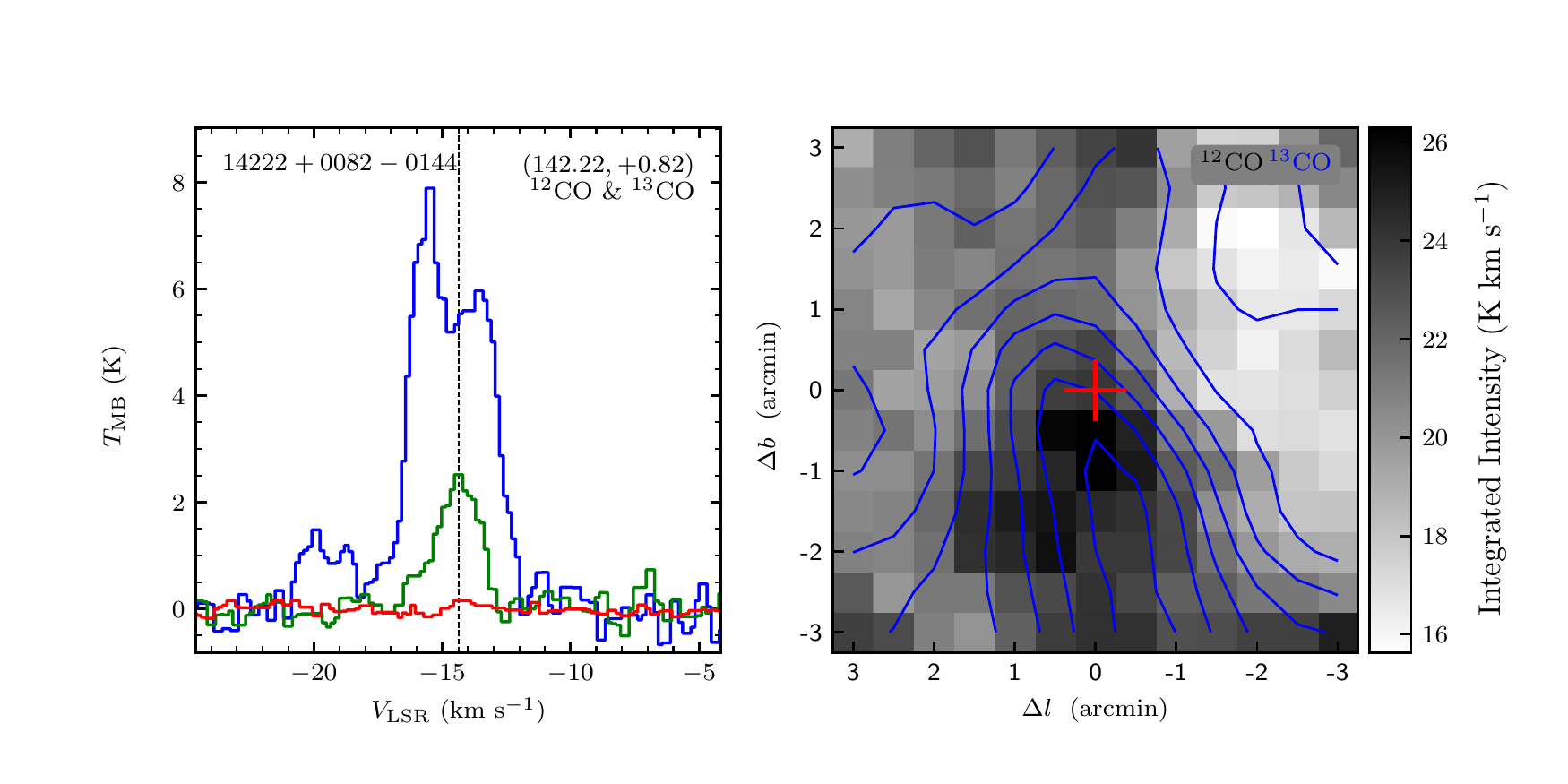}
\includegraphics[width=9.0cm,angle=0]{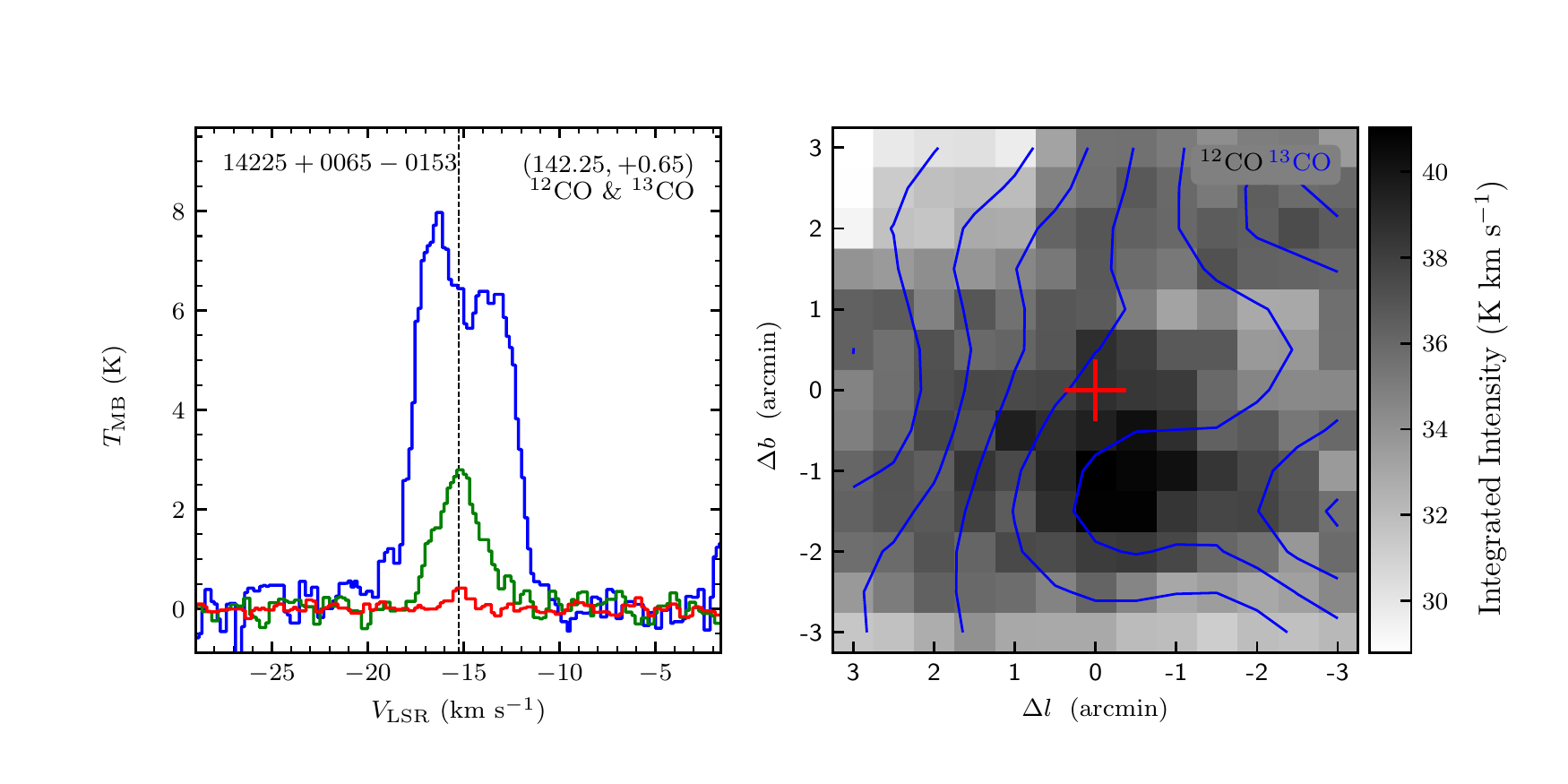}
\end{figure}
\clearpage

\begin{figure}
\includegraphics[width=9.0cm,angle=0]{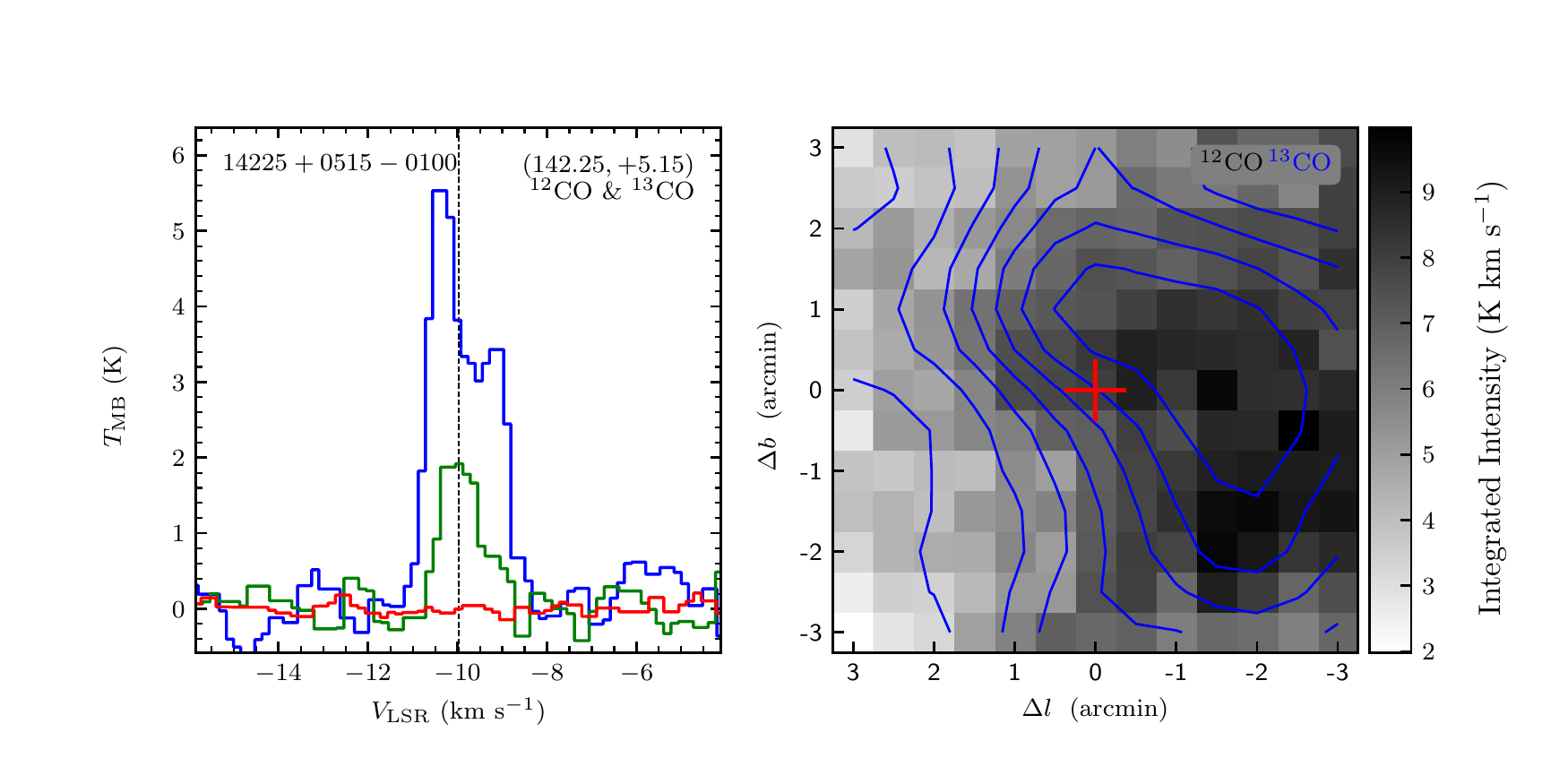}
\includegraphics[width=9.0cm,angle=0]{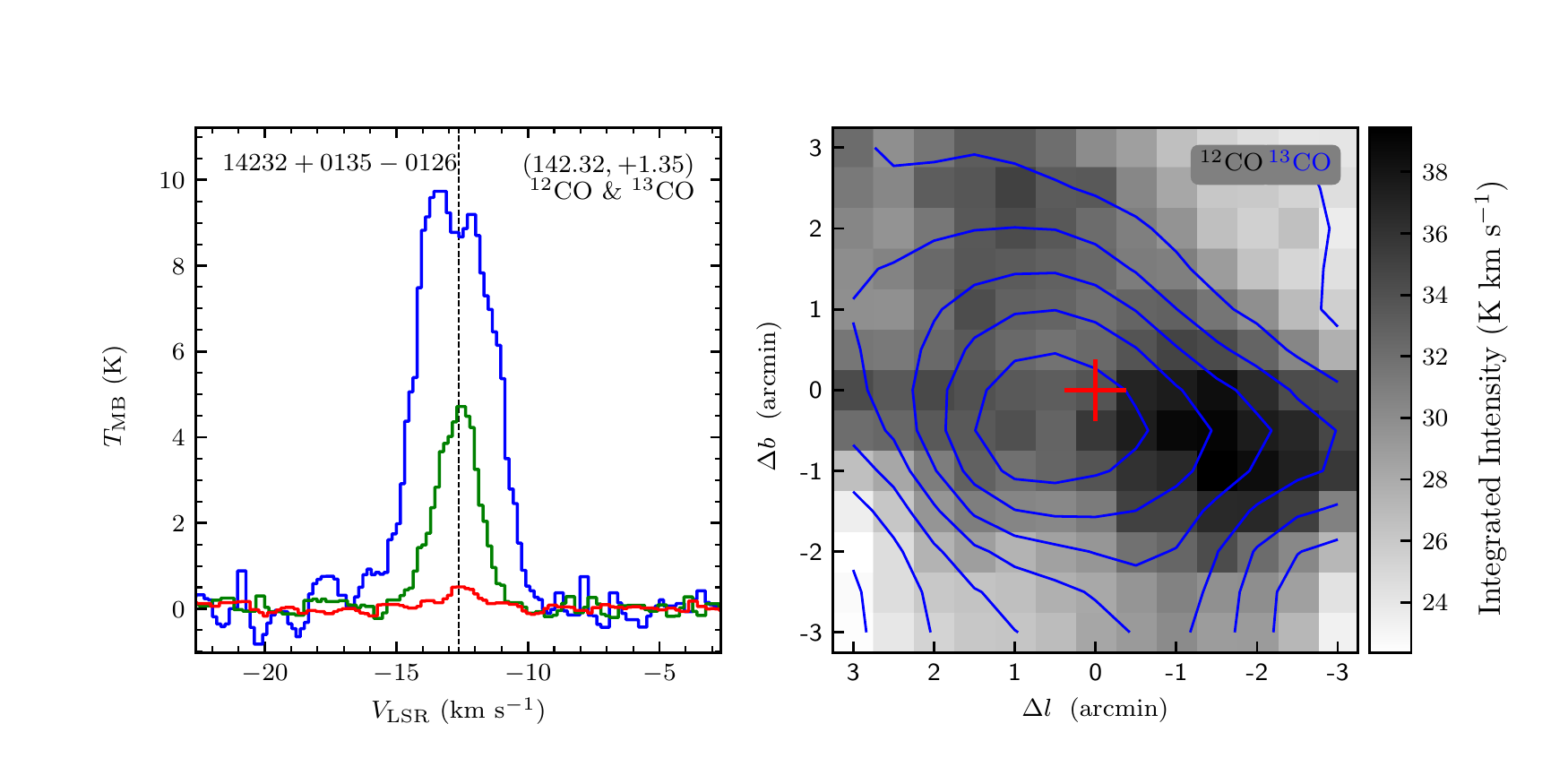}
\vspace{-0.5cm}

\includegraphics[width=9.0cm,angle=0]{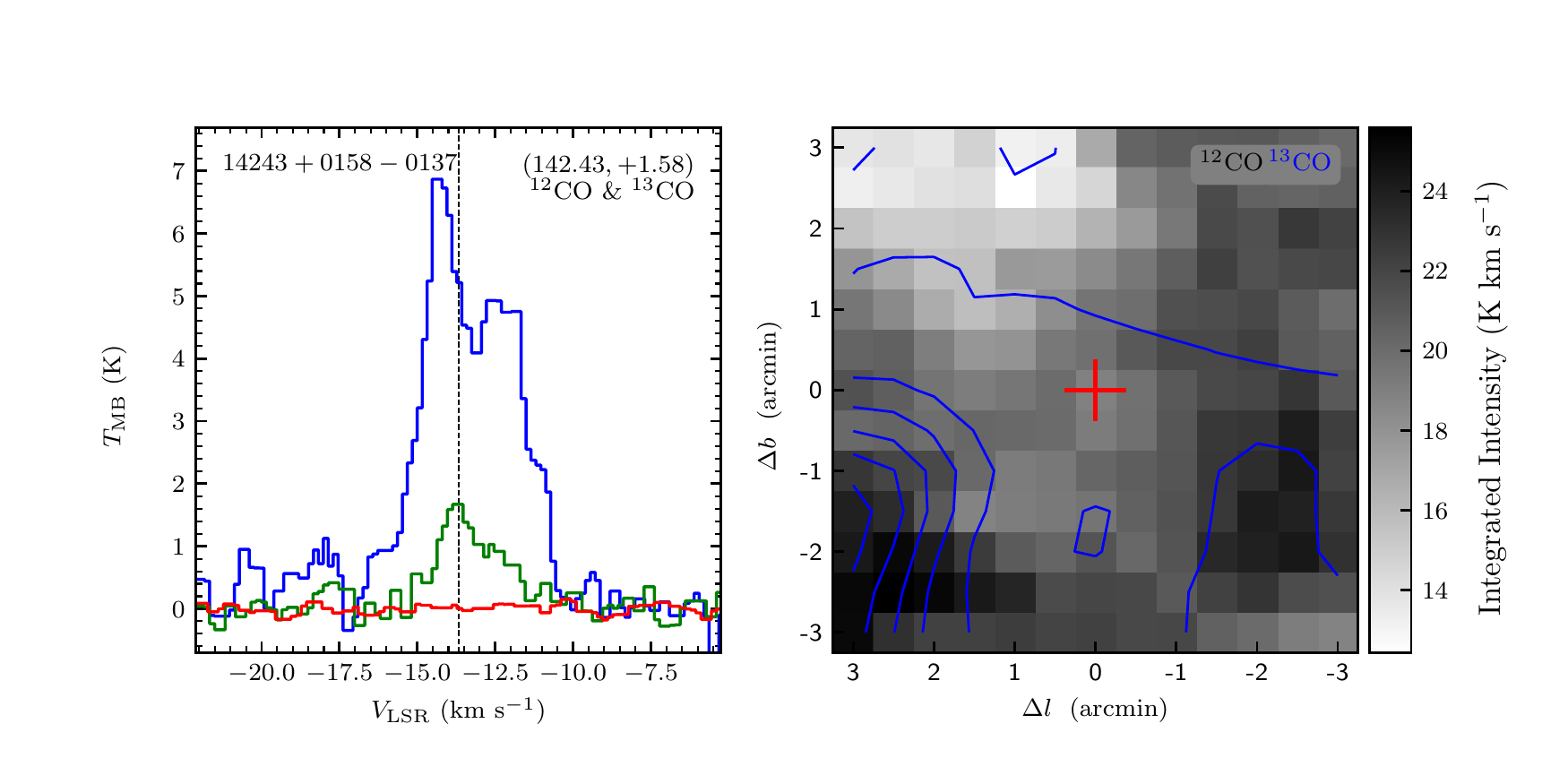}
\includegraphics[width=9.0cm,angle=0]{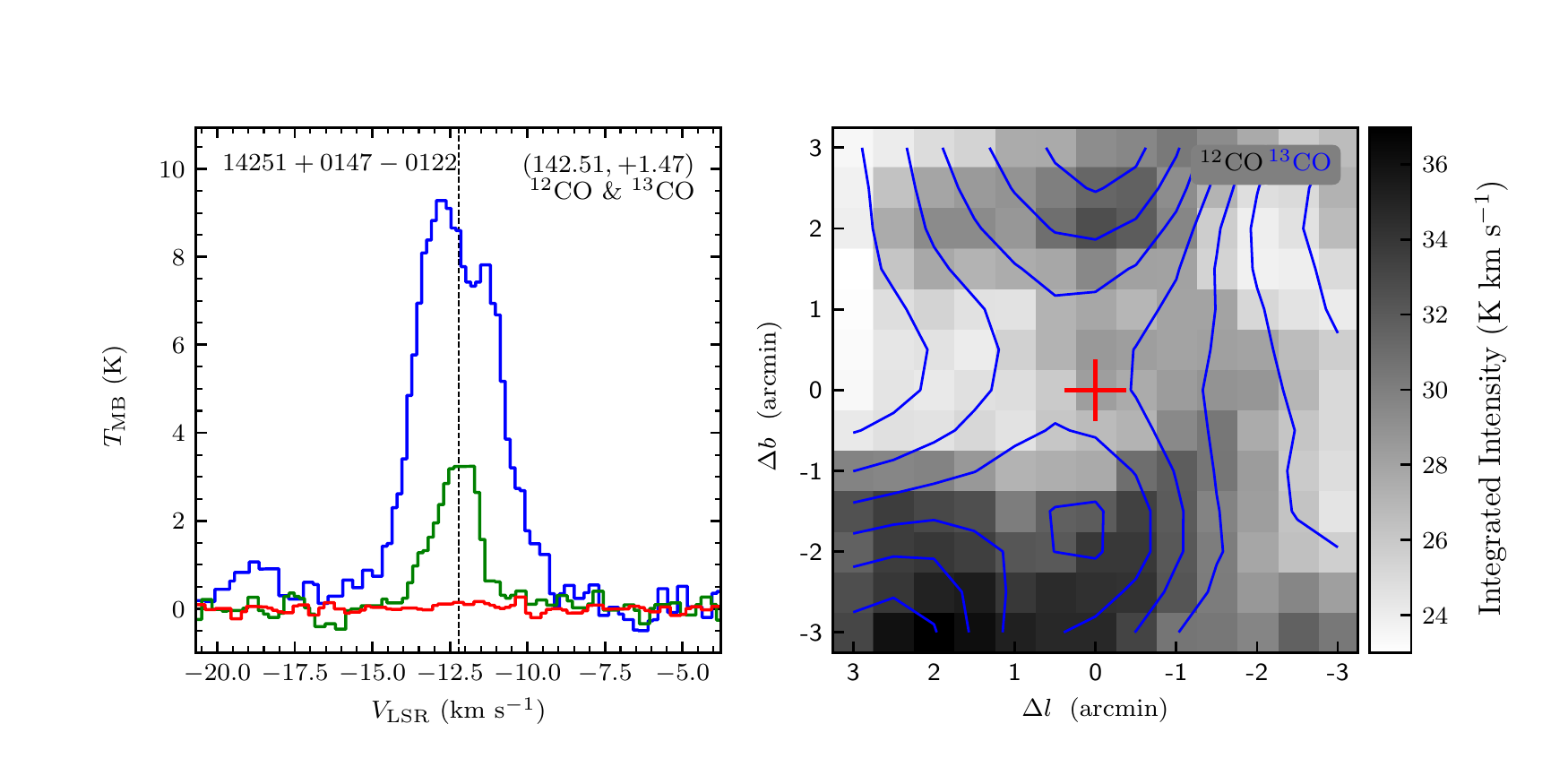}
\vspace{-0.5cm}

\includegraphics[width=9.0cm,angle=0]{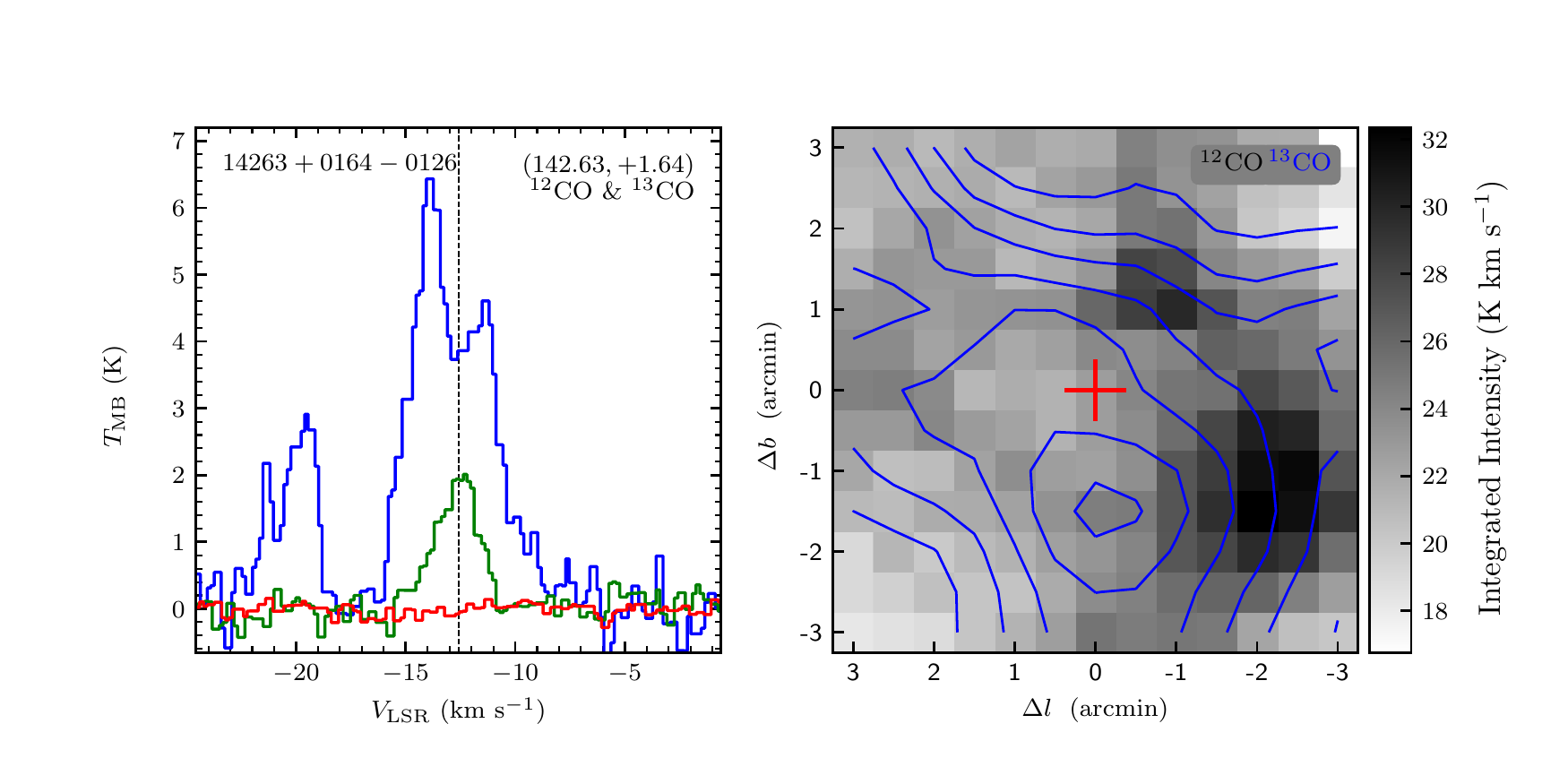}
\includegraphics[width=9.0cm,angle=0]{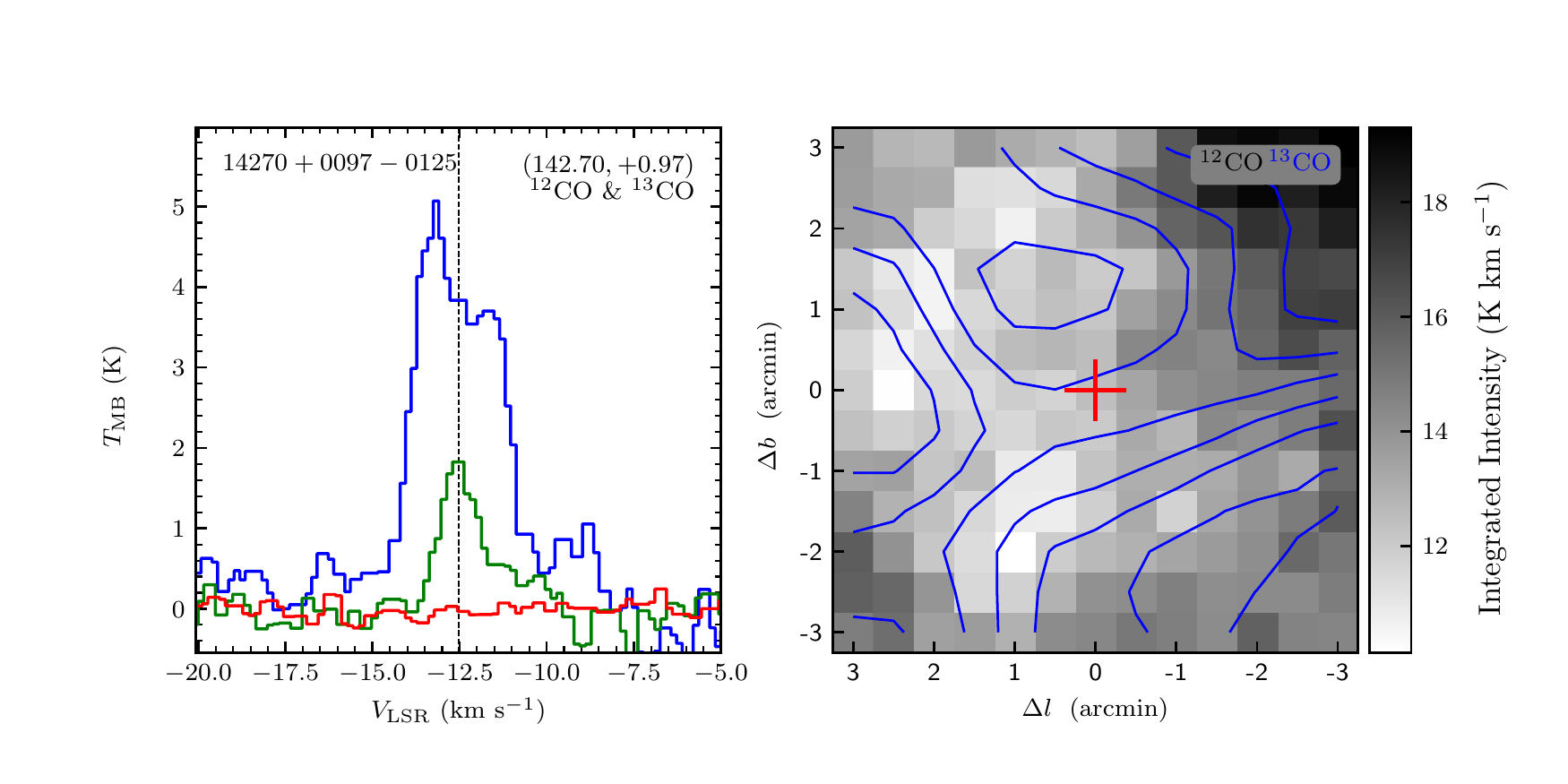}
\vspace{-0.5cm}

\includegraphics[width=9.0cm,angle=0]{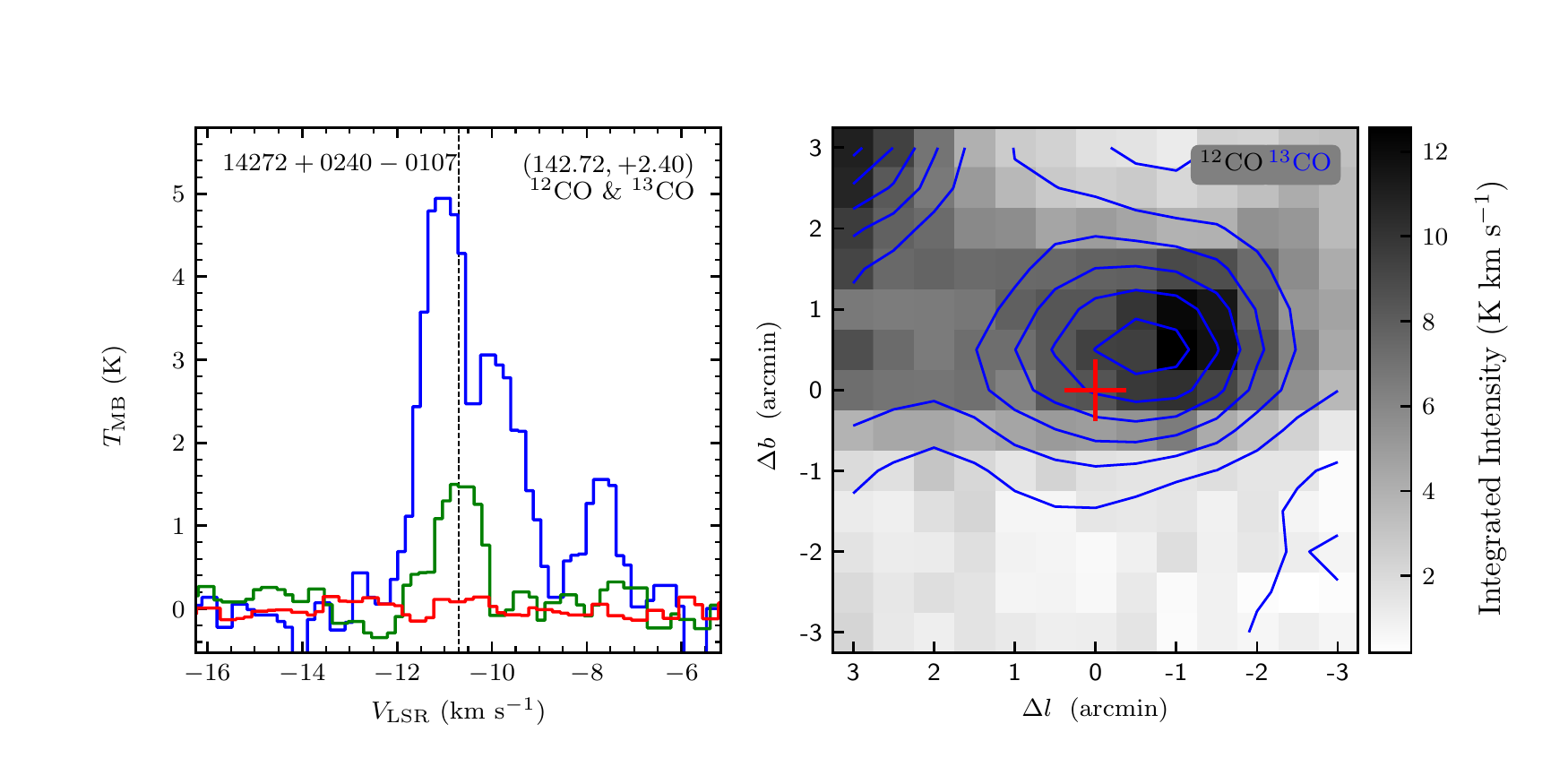}
\includegraphics[width=9.0cm,angle=0]{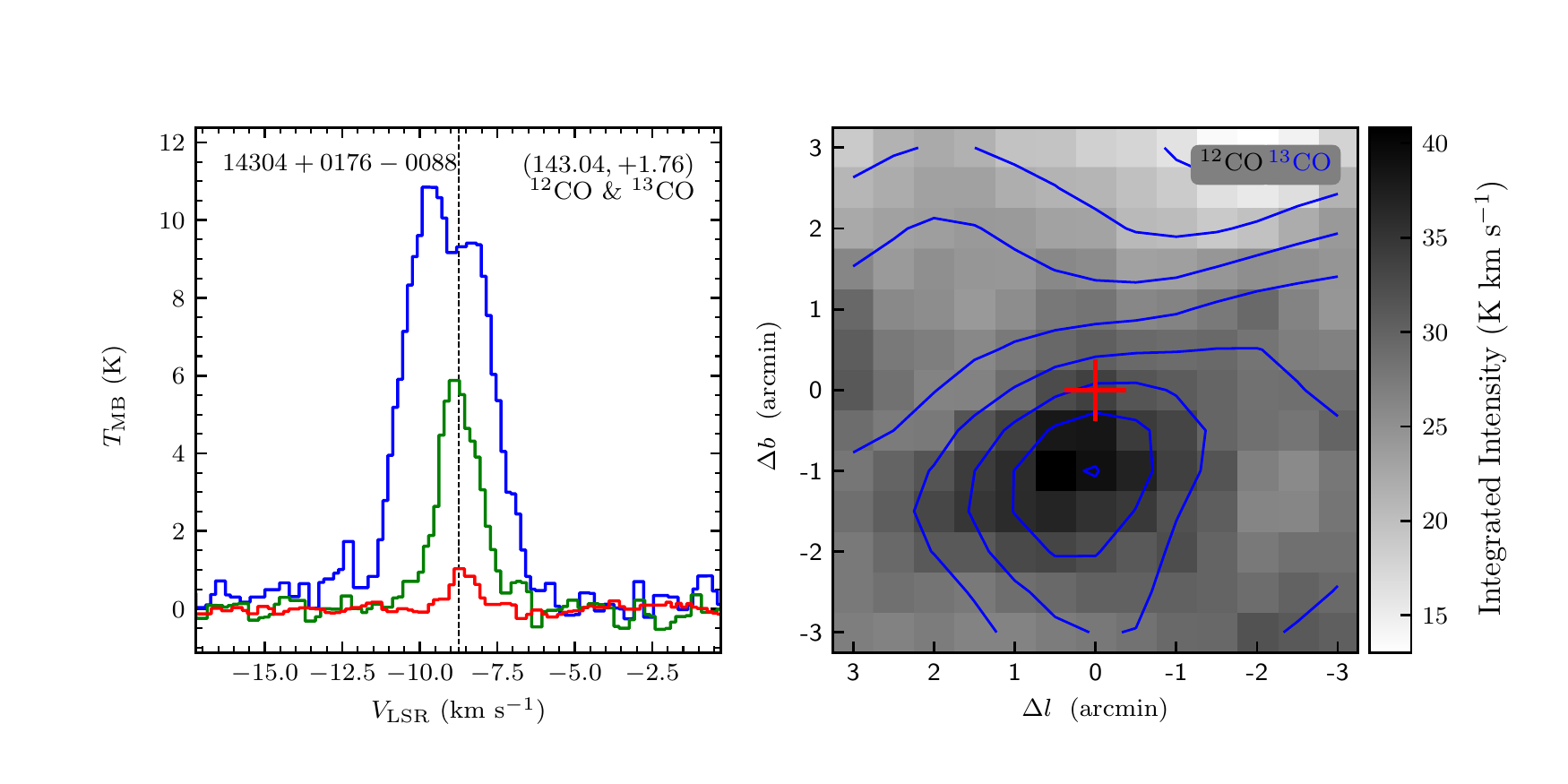}
\vspace{-0.5cm}

\includegraphics[width=9.0cm,angle=0]{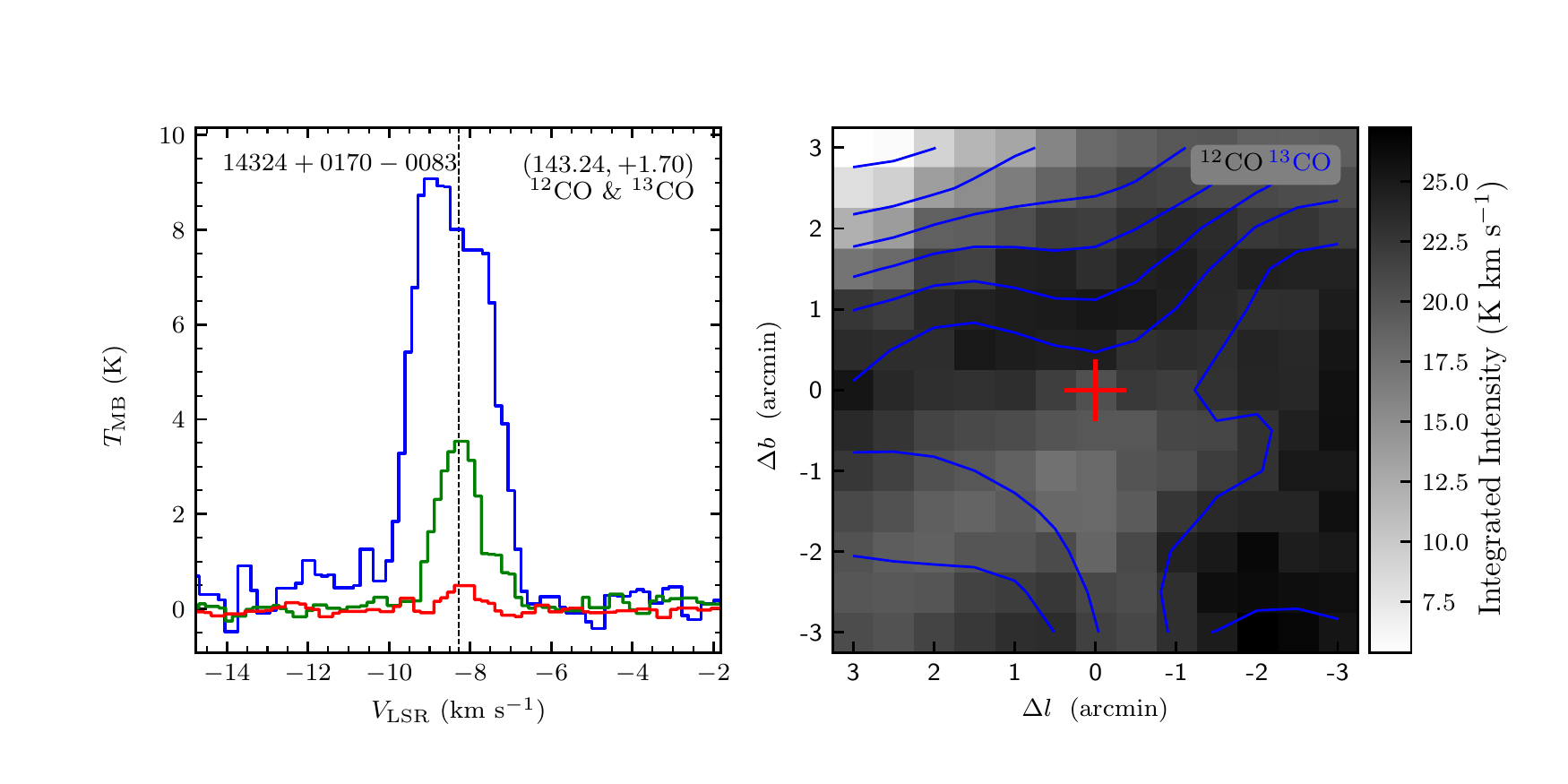}
\includegraphics[width=9.0cm,angle=0]{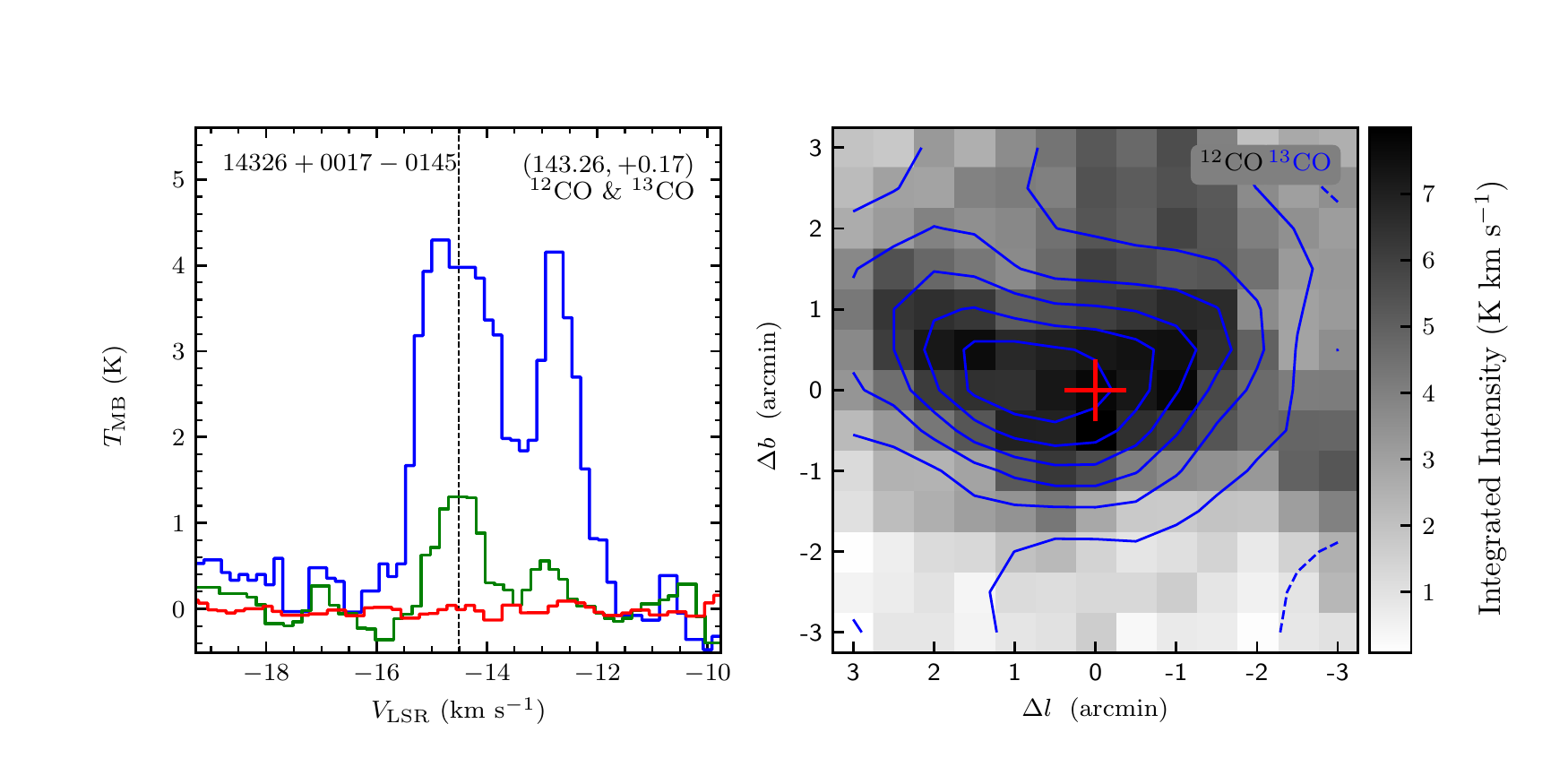}
\end{figure}
\clearpage

\begin{figure}
\includegraphics[width=9.0cm,angle=0]{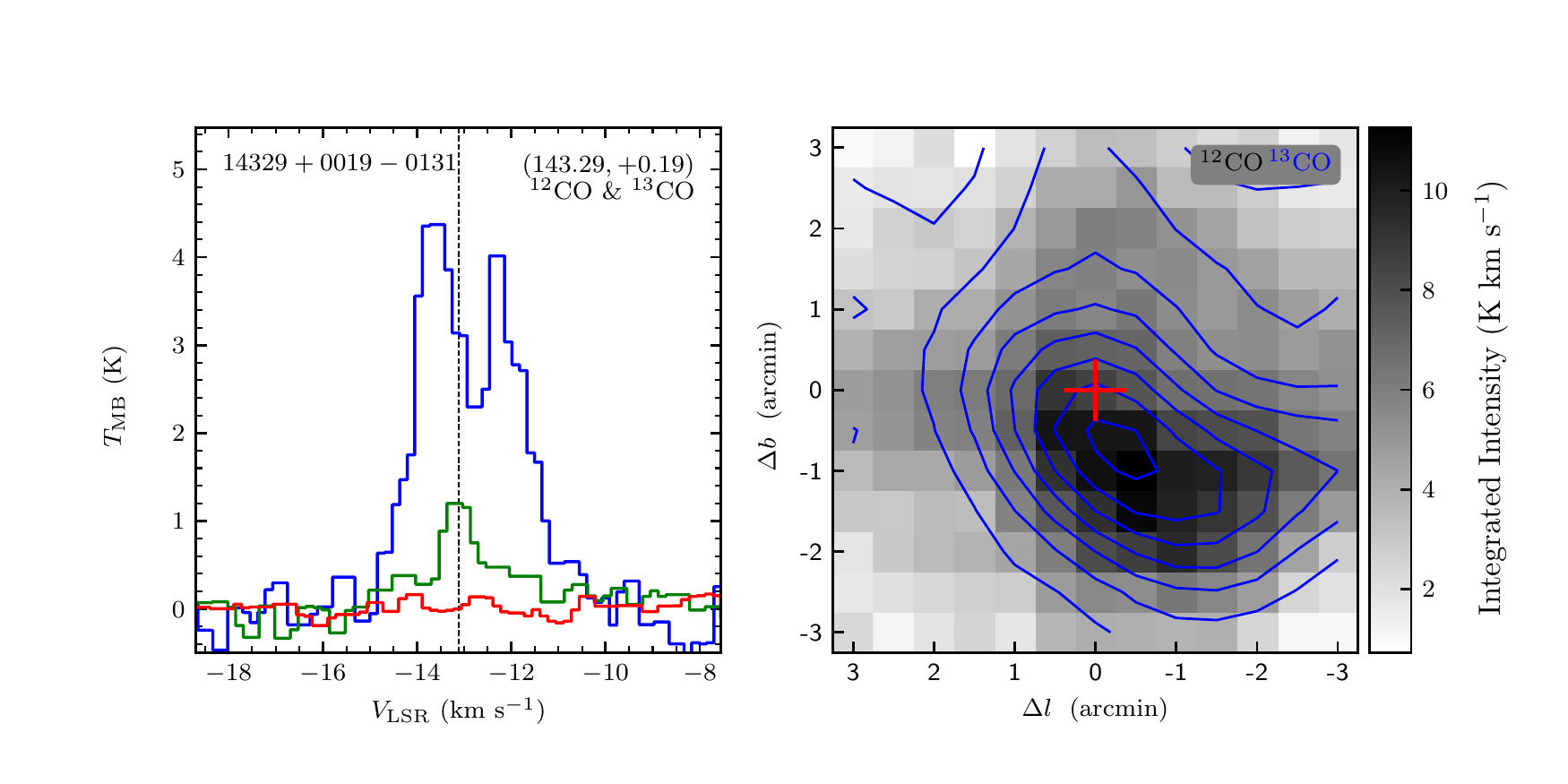}
\includegraphics[width=9.0cm,angle=0]{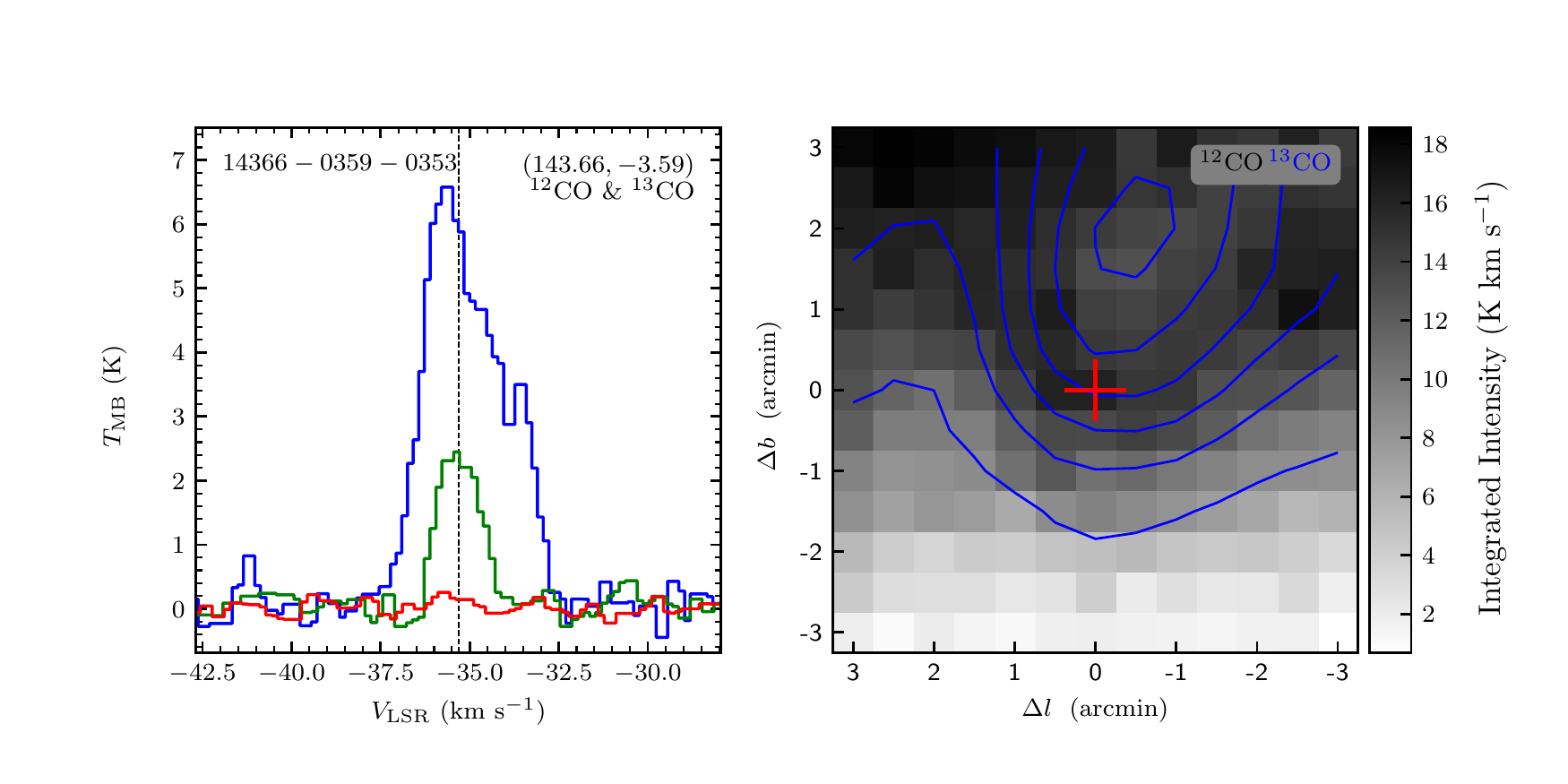}
\vspace{-0.5cm}

\includegraphics[width=9.0cm,angle=0]{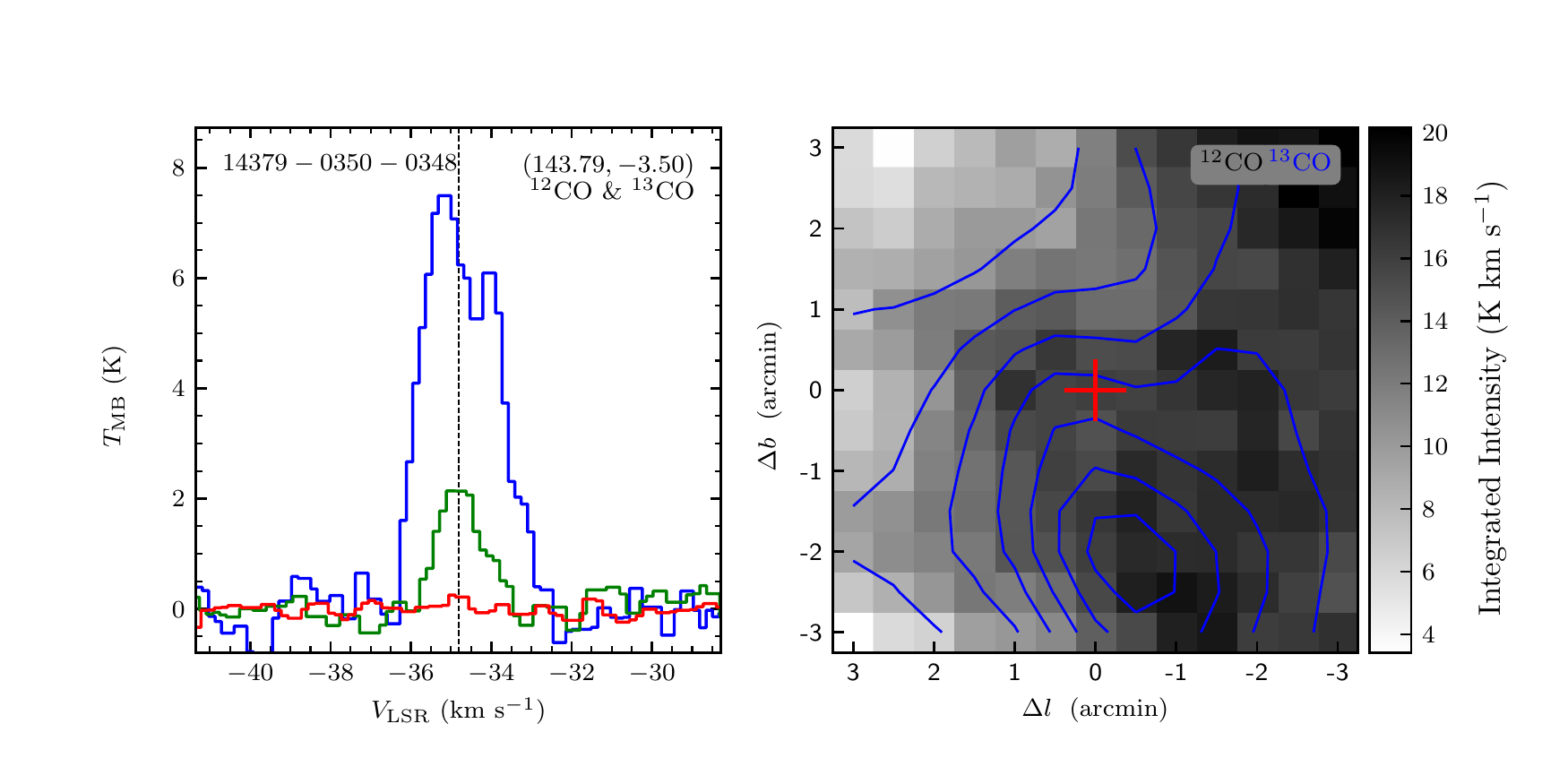}
\includegraphics[width=9.0cm,angle=0]{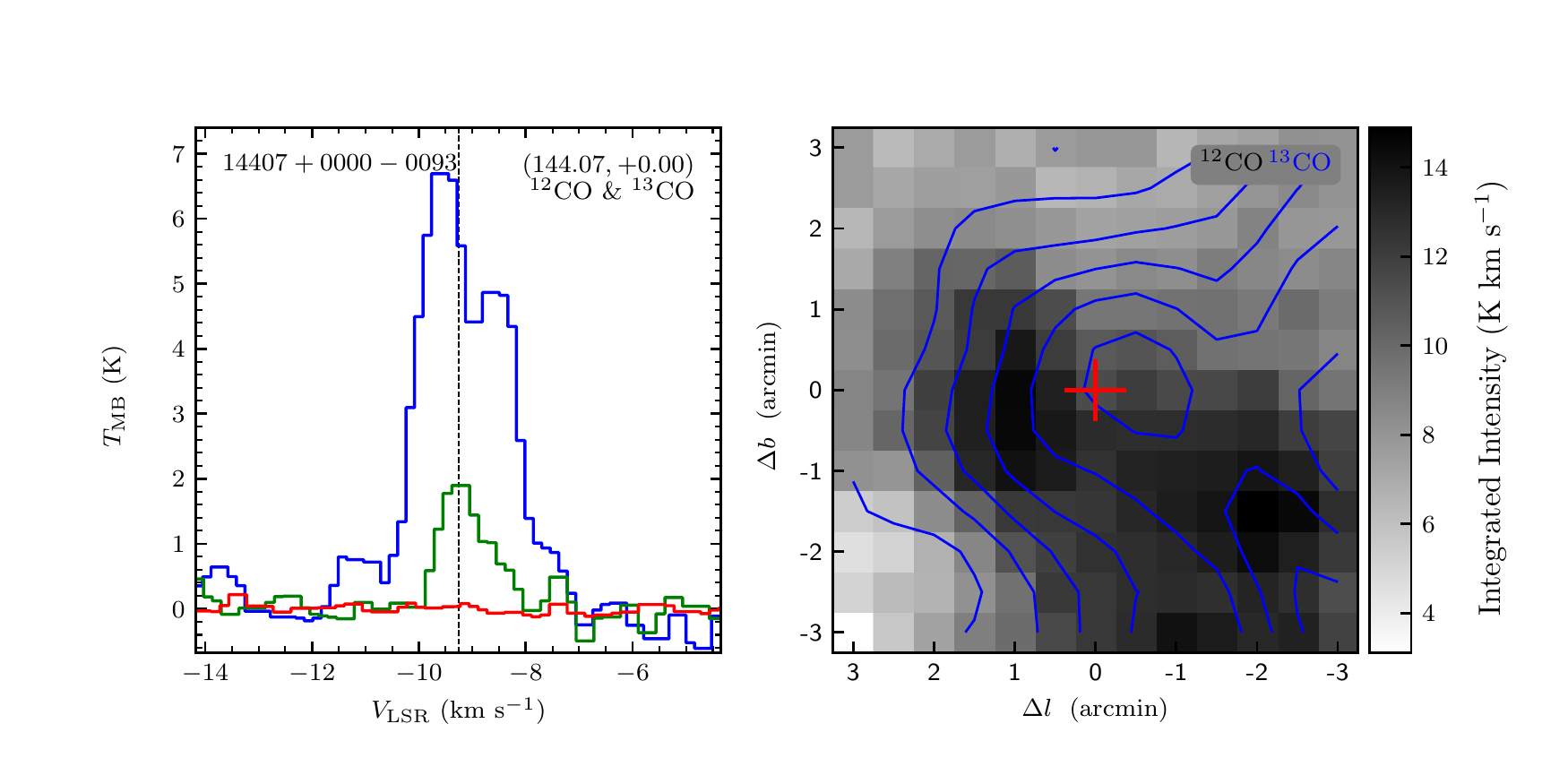}
\vspace{-0.5cm}

\includegraphics[width=9.0cm,angle=0]{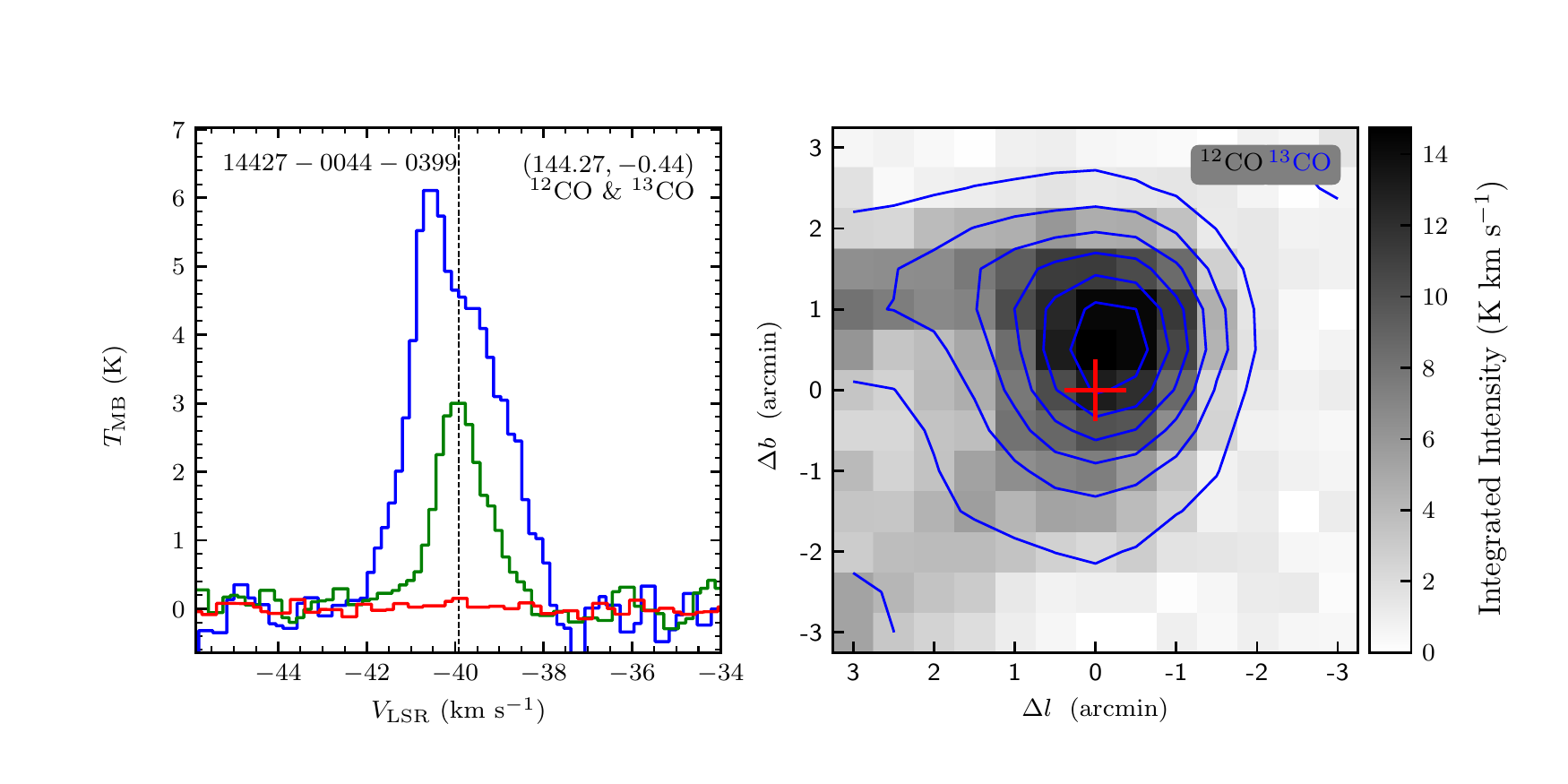}
\includegraphics[width=9.0cm,angle=0]{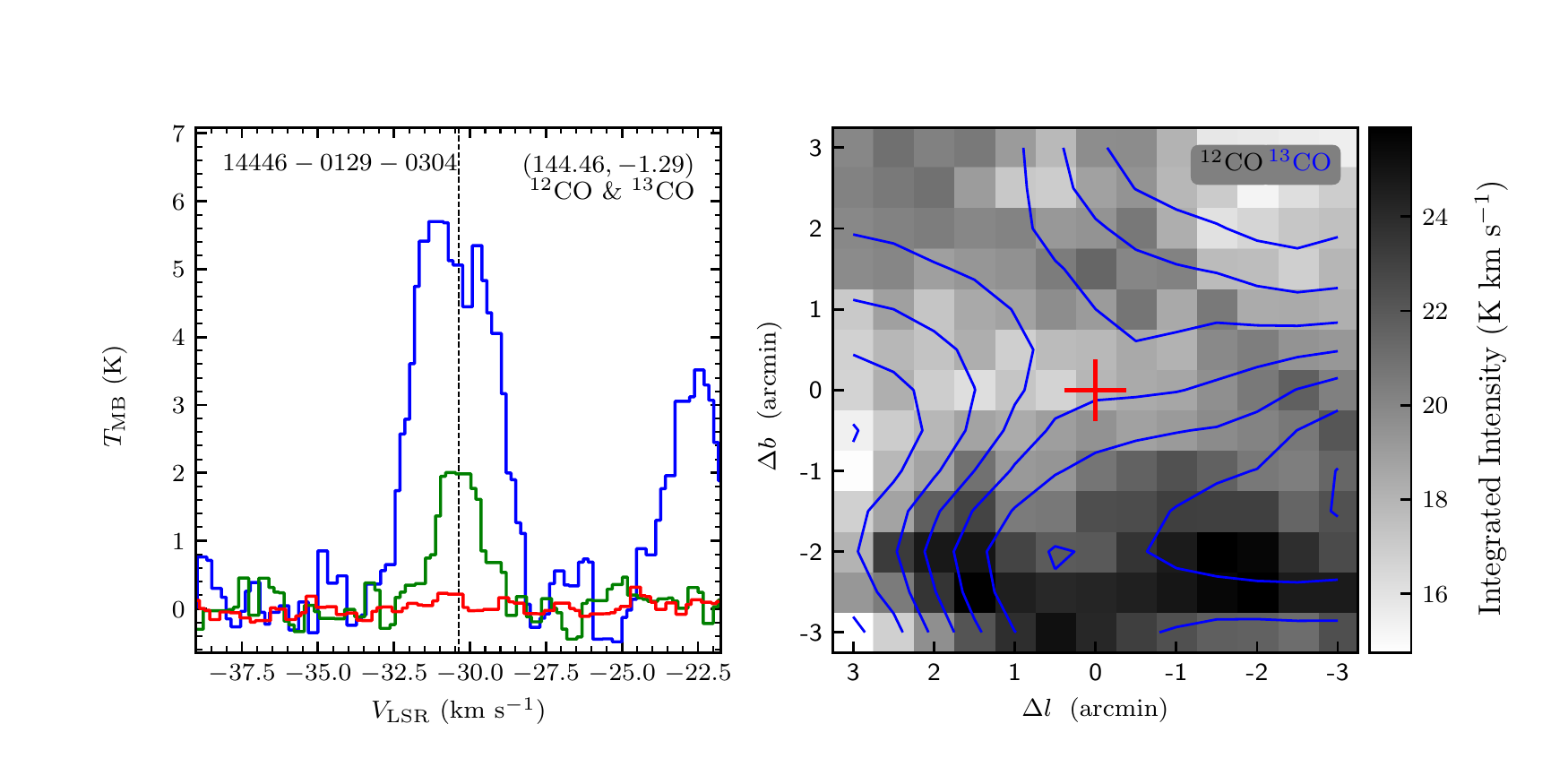}
\vspace{-0.5cm}

\includegraphics[width=9.0cm,angle=0]{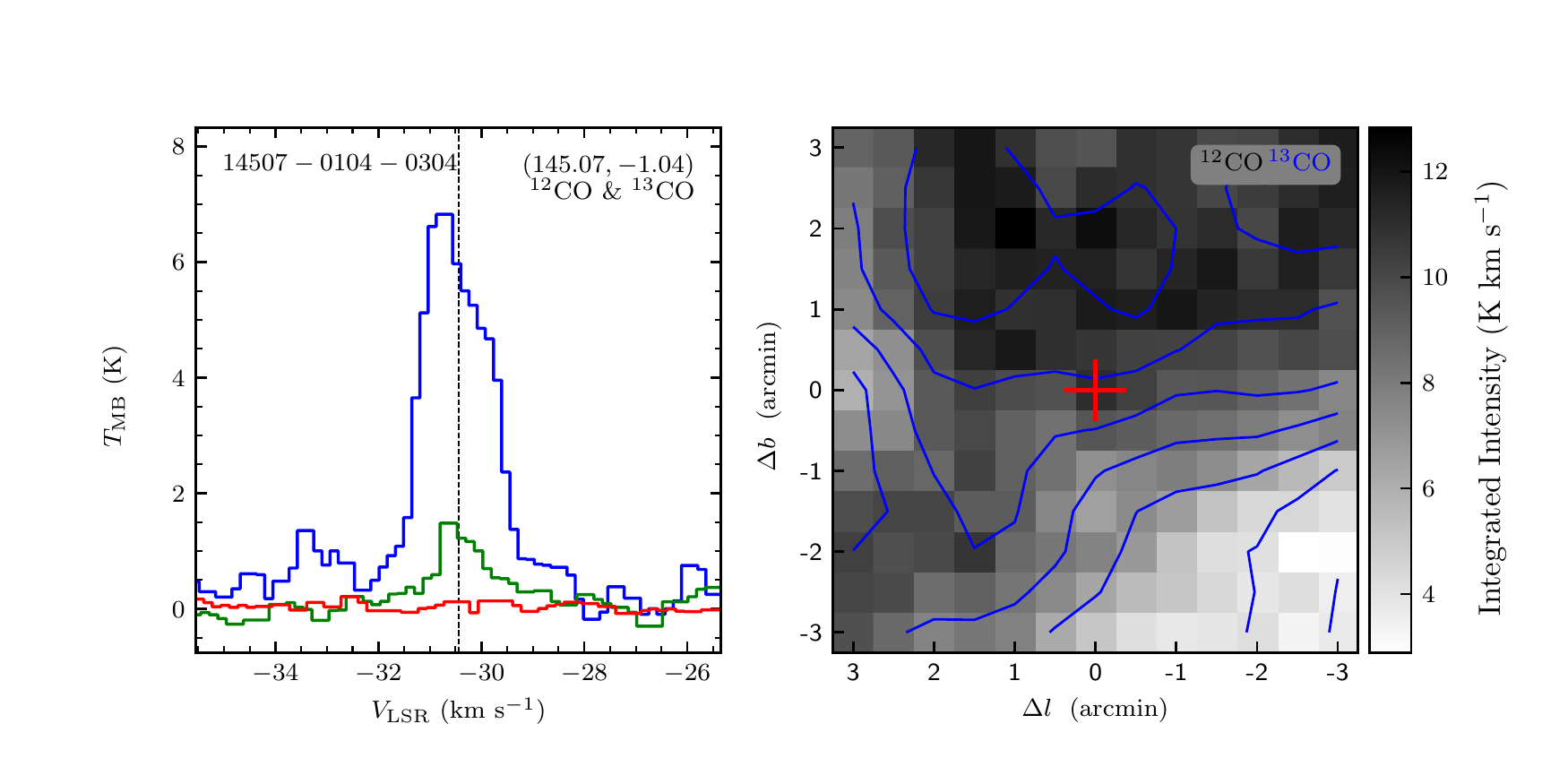}
\includegraphics[width=9.0cm,angle=0]{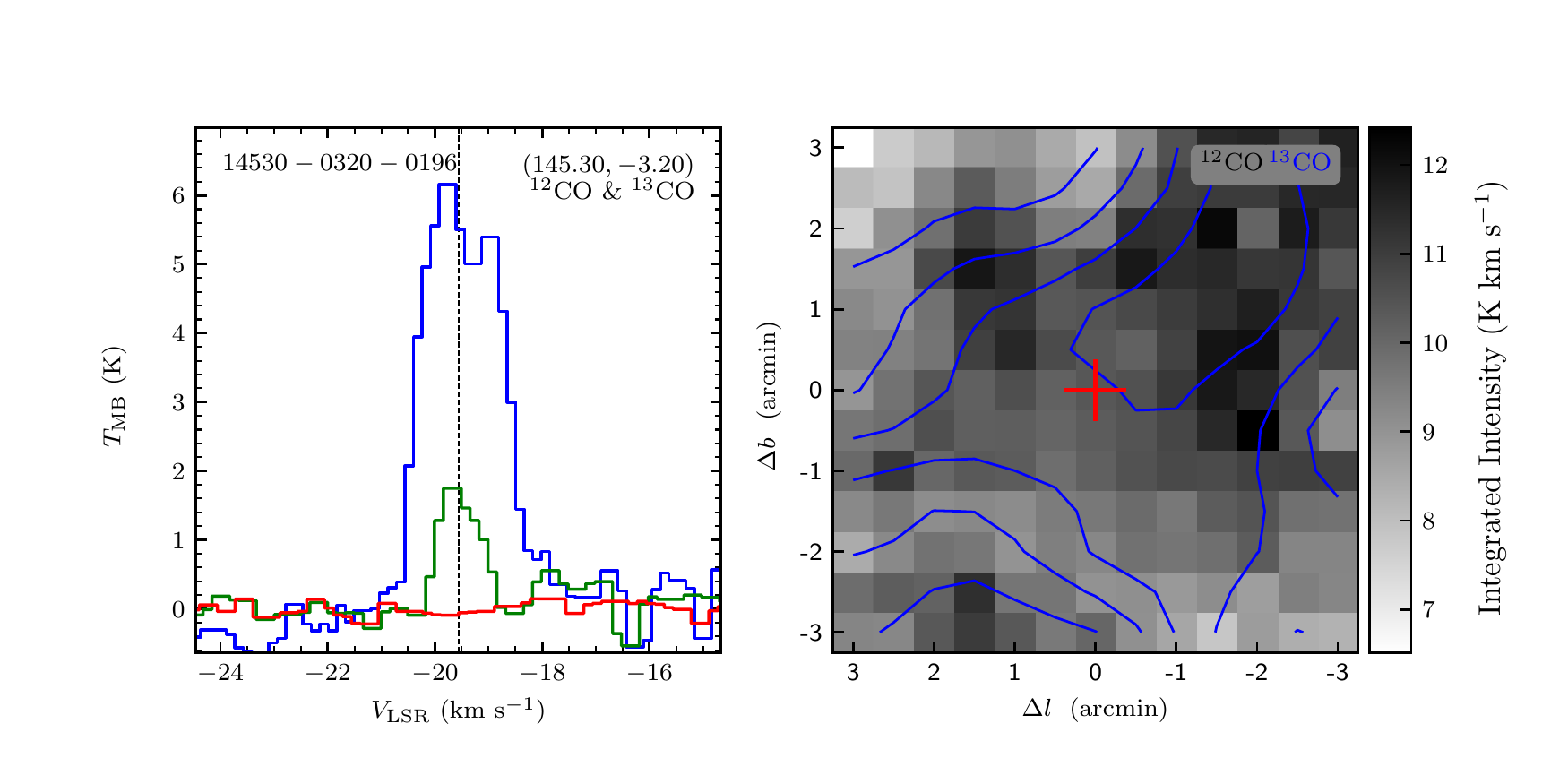}
\vspace{-0.5cm}

\includegraphics[width=9.0cm,angle=0]{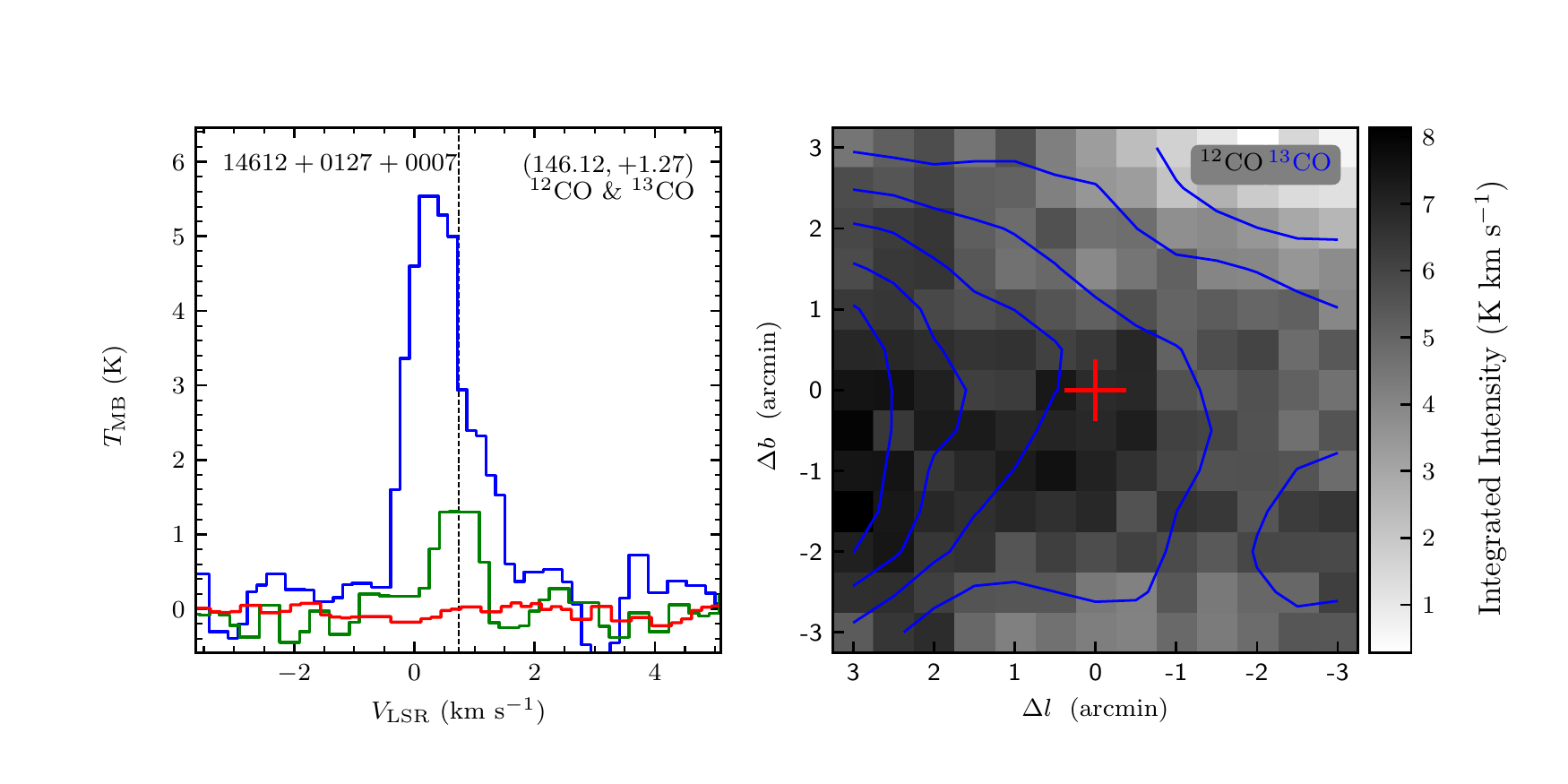}
\includegraphics[width=9.0cm,angle=0]{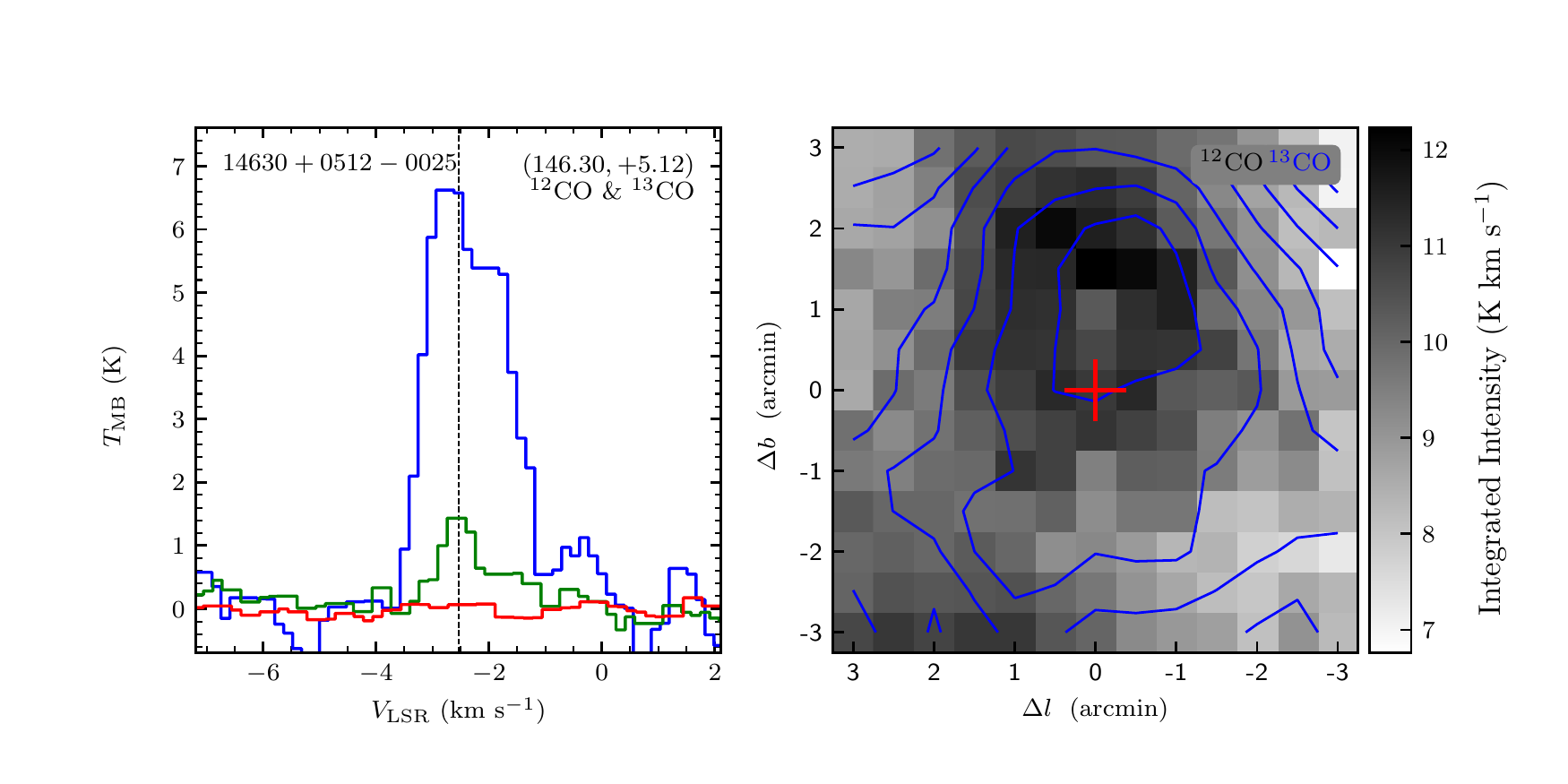}
\end{figure}
\clearpage

\begin{figure}
\includegraphics[width=9.0cm,angle=0]{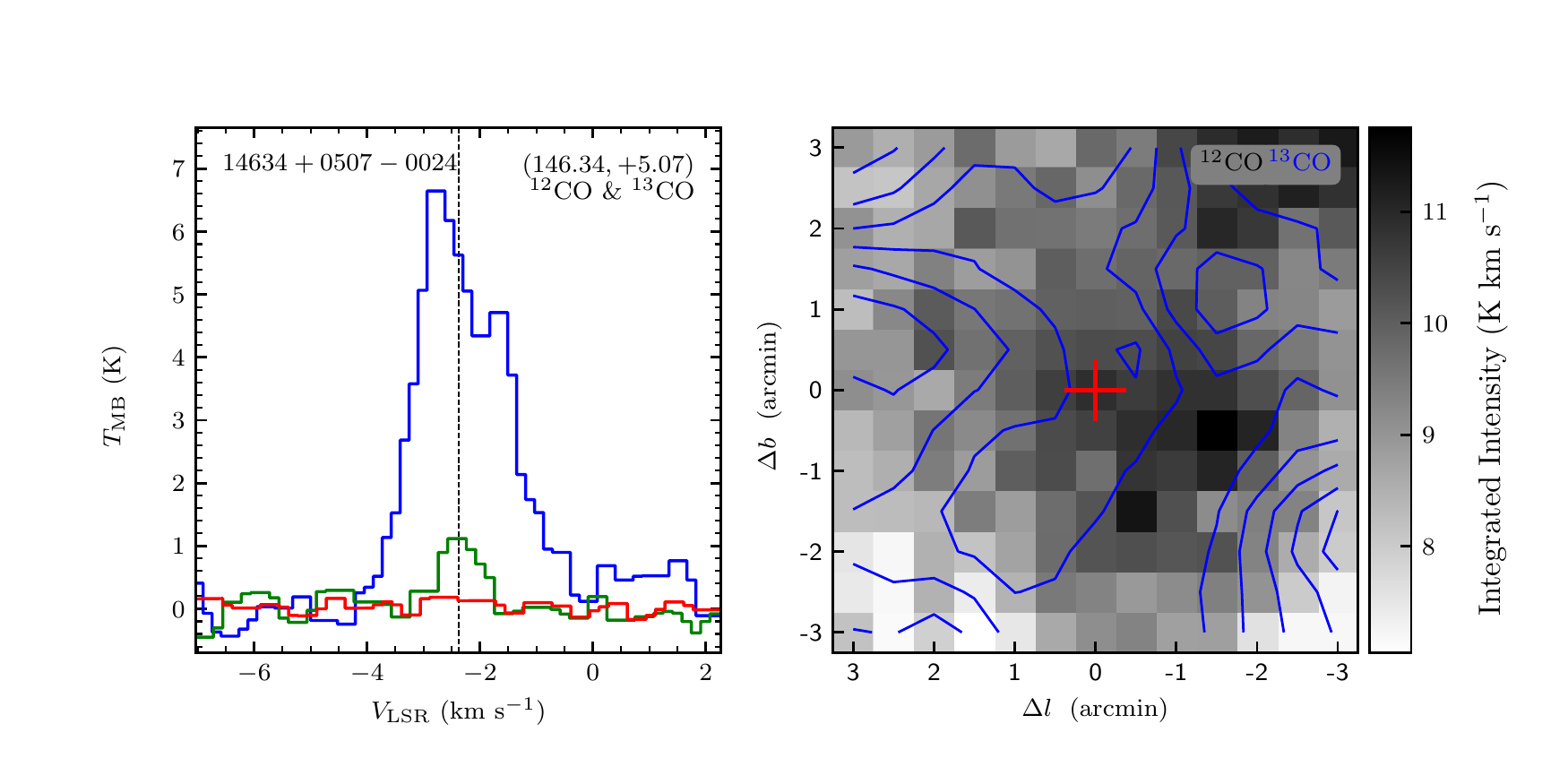}
\includegraphics[width=9.0cm,angle=0]{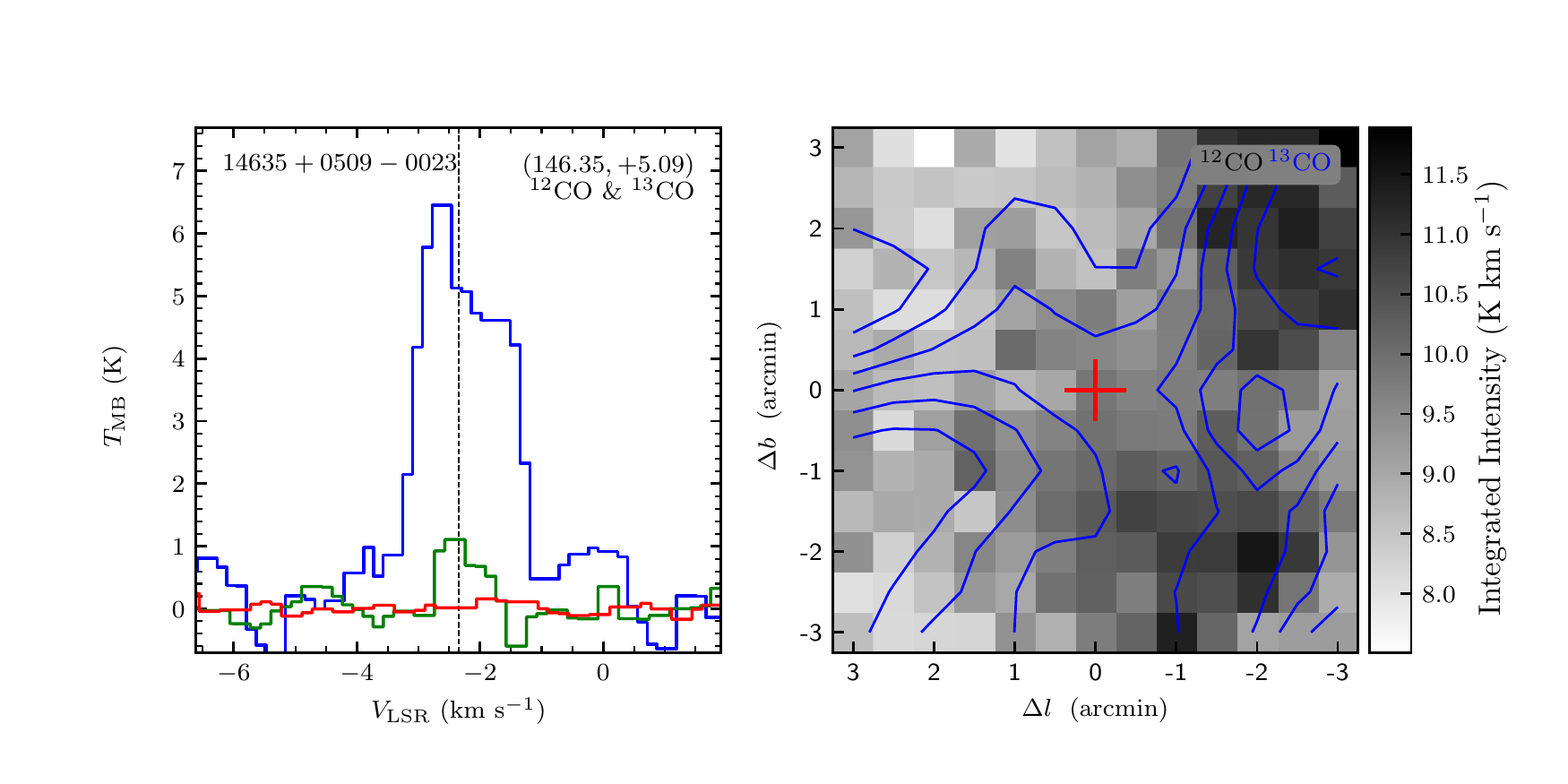}
\vspace{-0.5cm}

\includegraphics[width=9.0cm,angle=0]{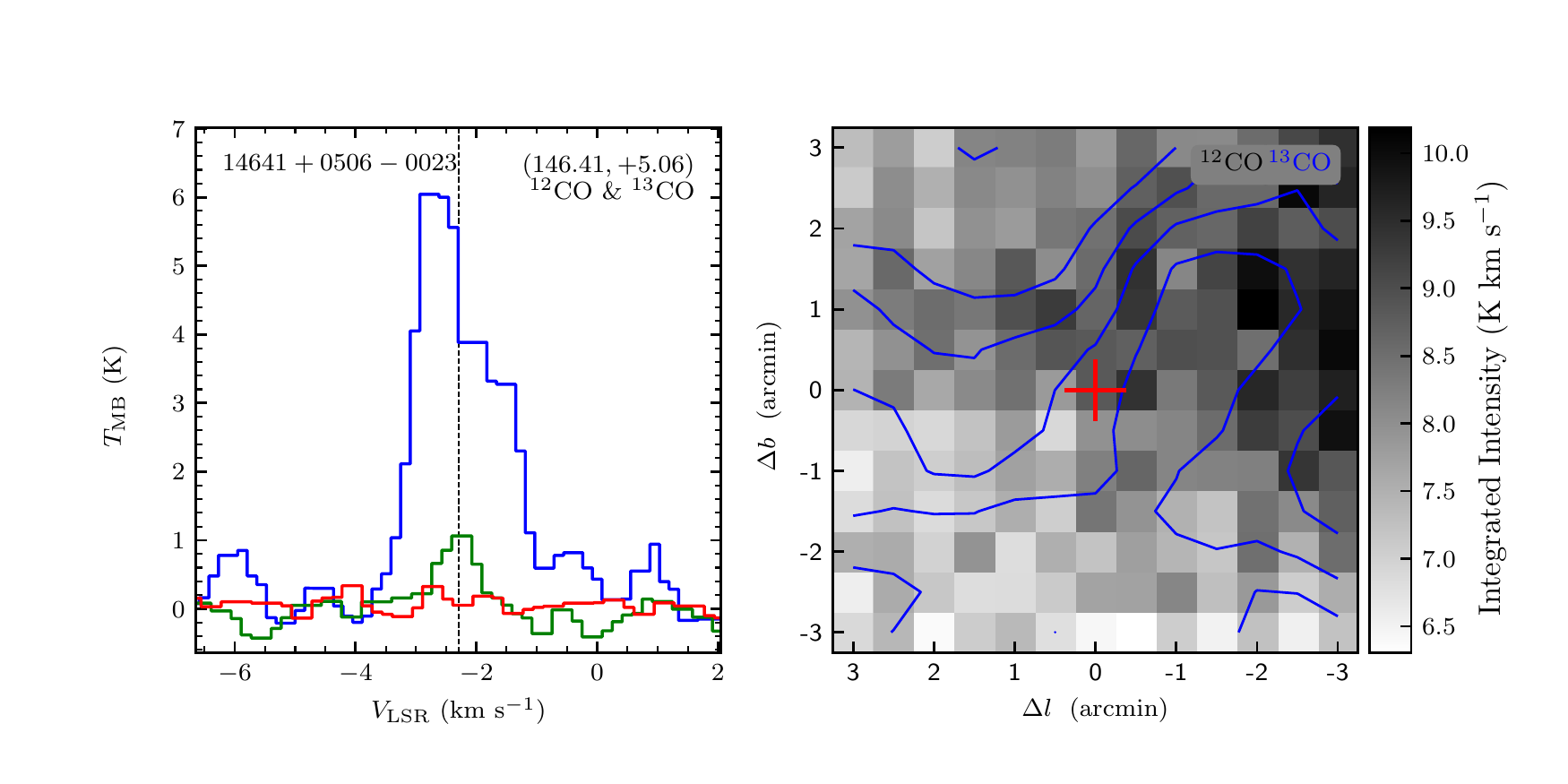}
\includegraphics[width=9.0cm,angle=0]{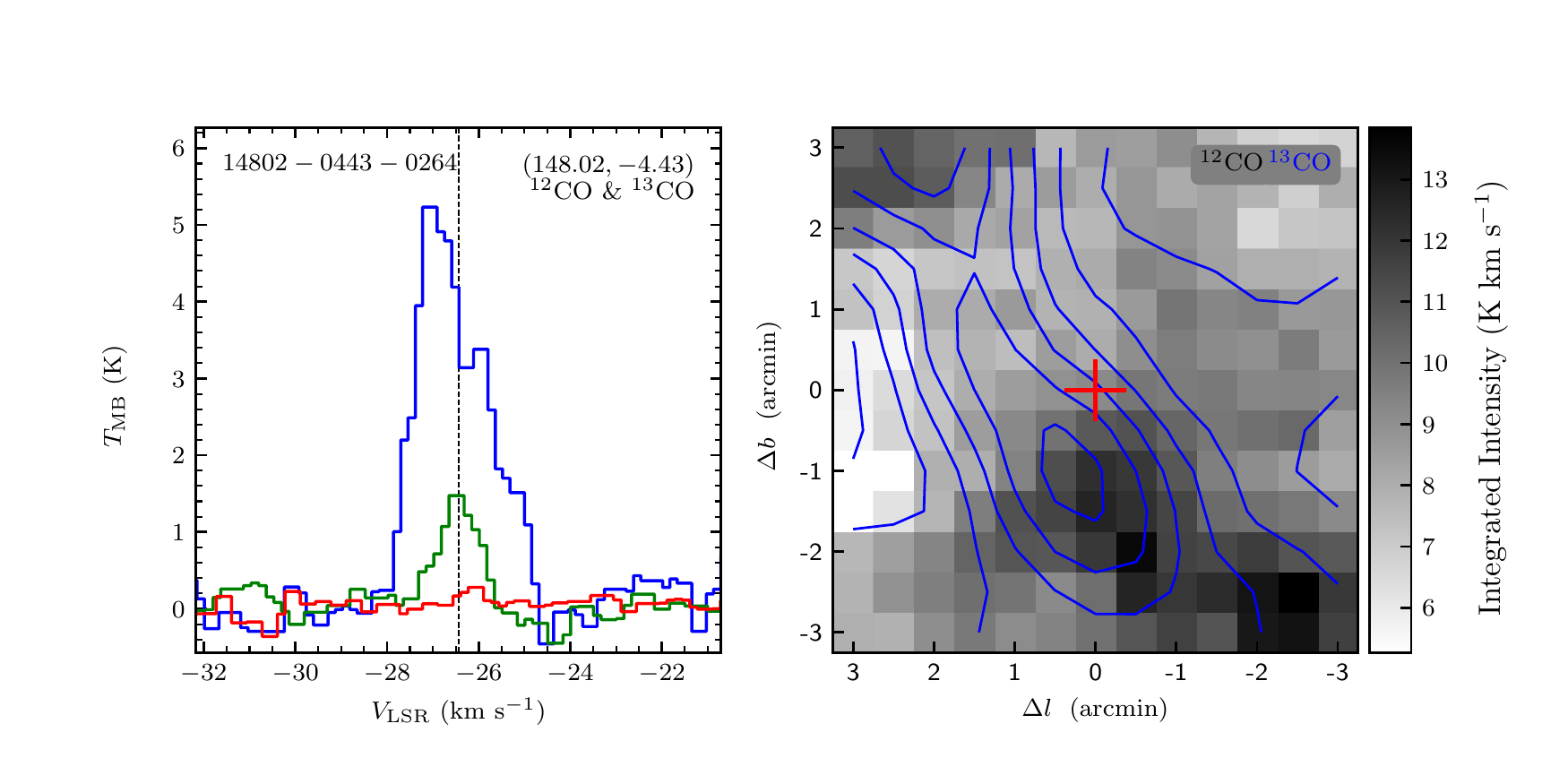}
\vspace{-0.5cm}

\includegraphics[width=9.0cm,angle=0]{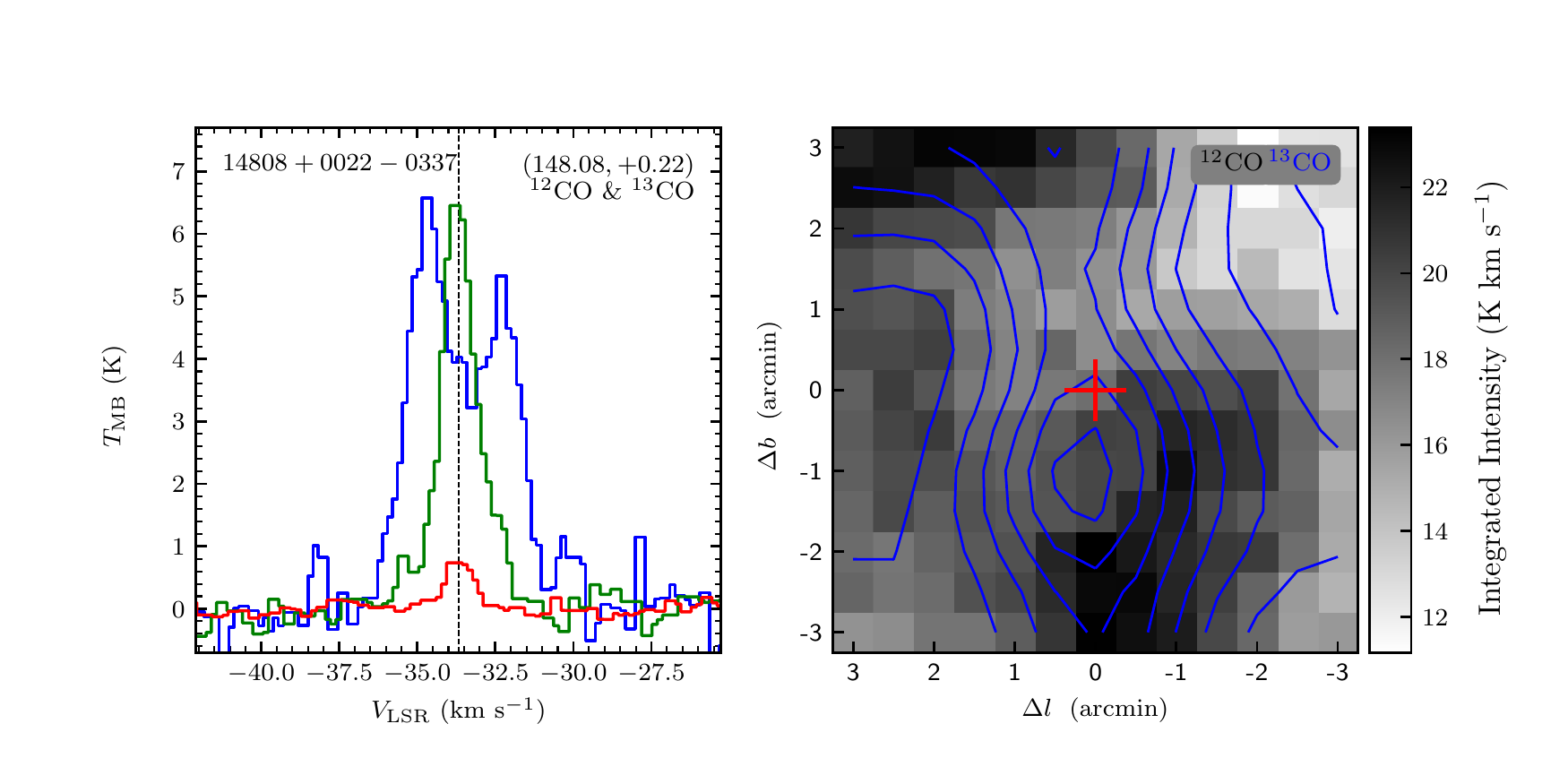}
\includegraphics[width=9.0cm,angle=0]{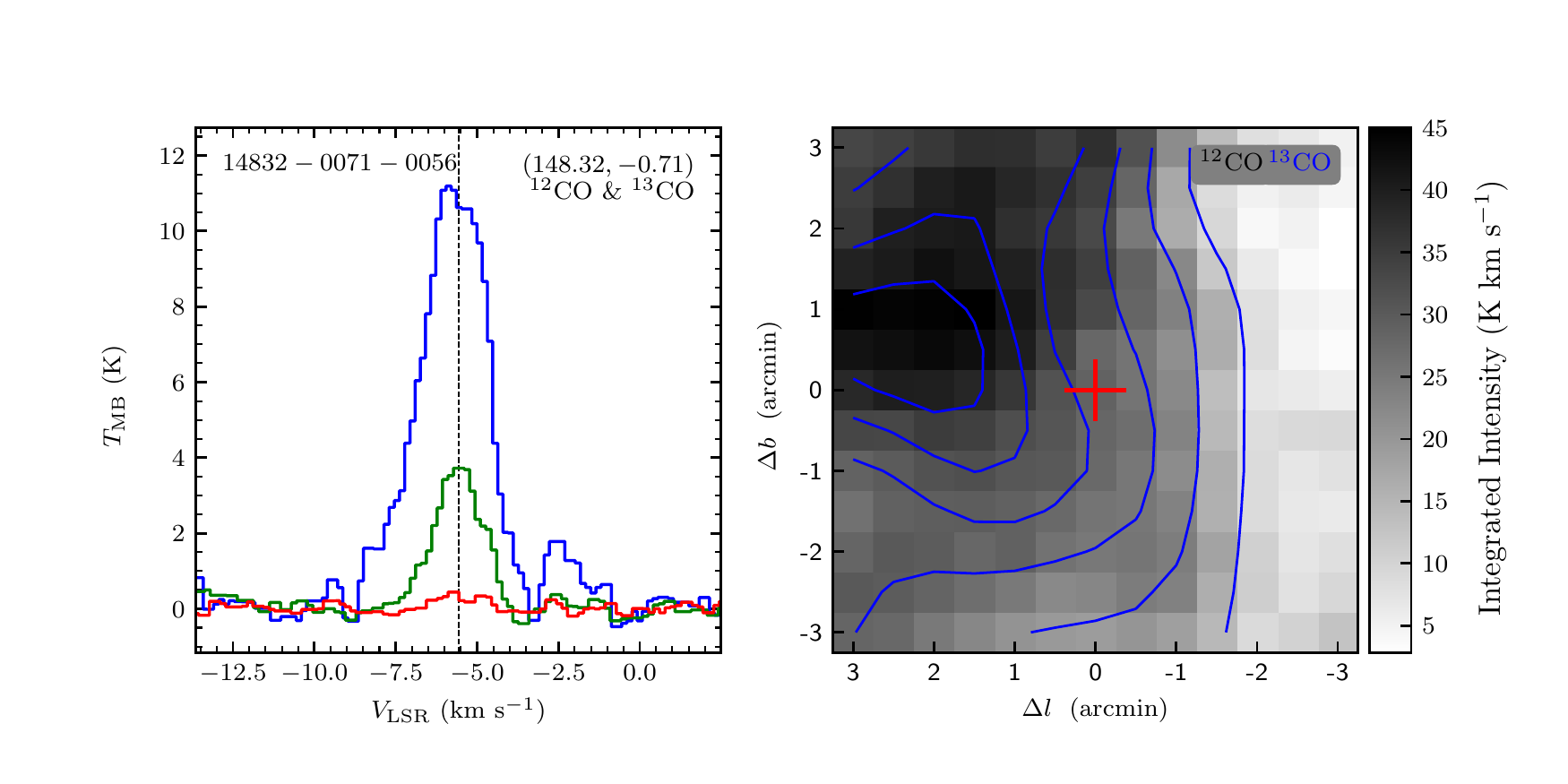}
\vspace{-0.5cm}

\includegraphics[width=9.0cm,angle=0]{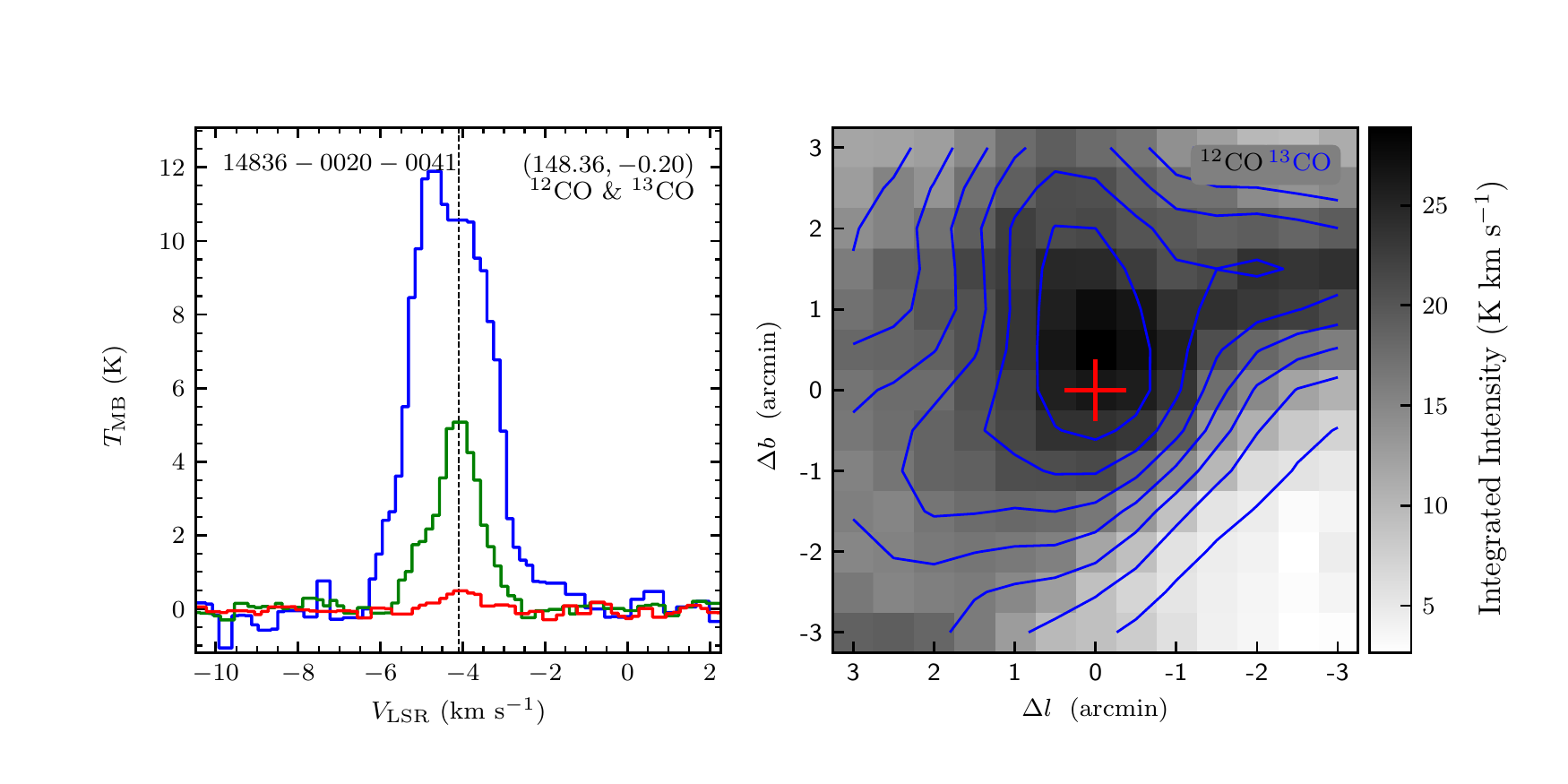}
\includegraphics[width=9.0cm,angle=0]{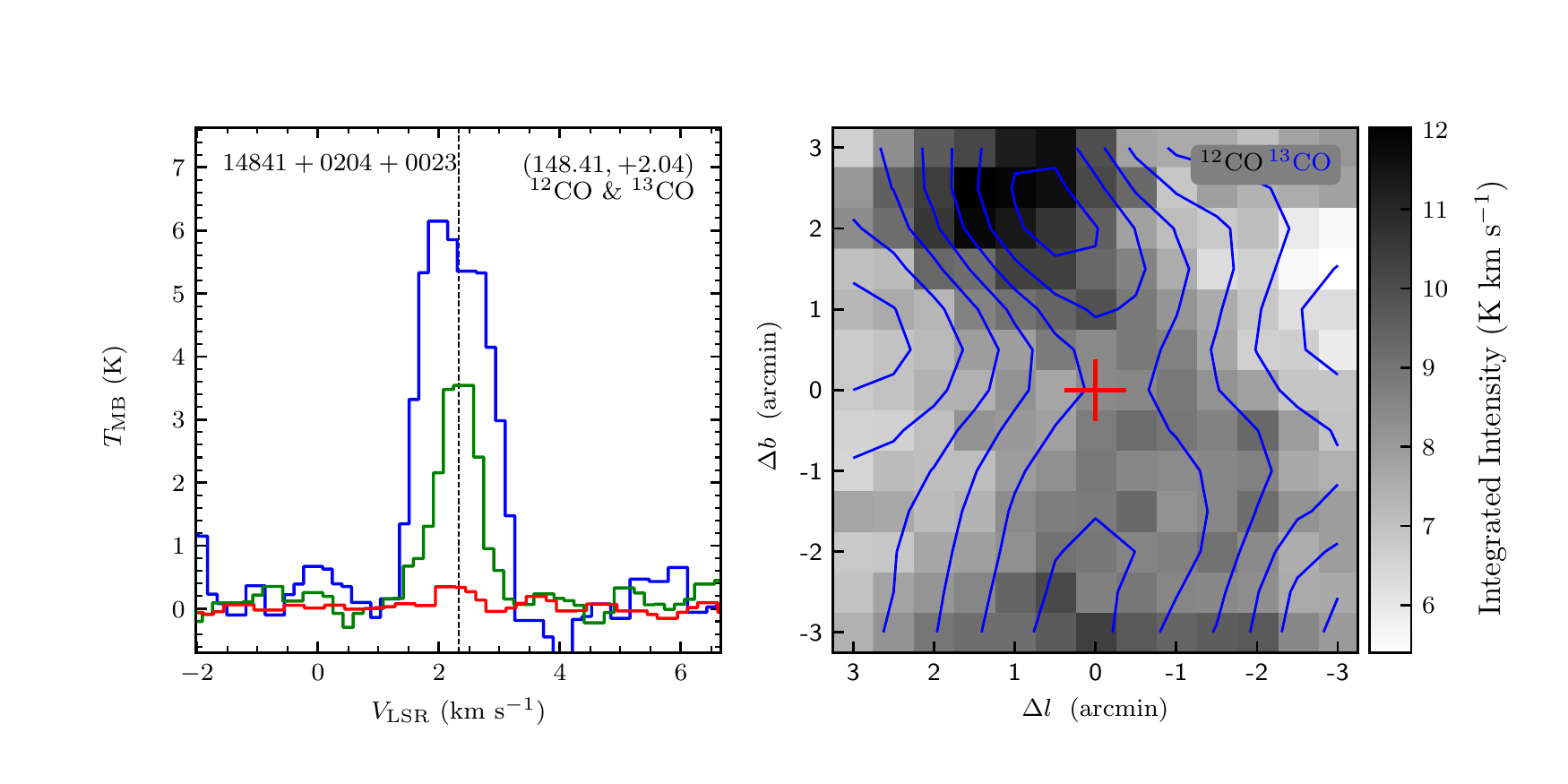}
\vspace{-0.5cm}

\includegraphics[width=9.0cm,angle=0]{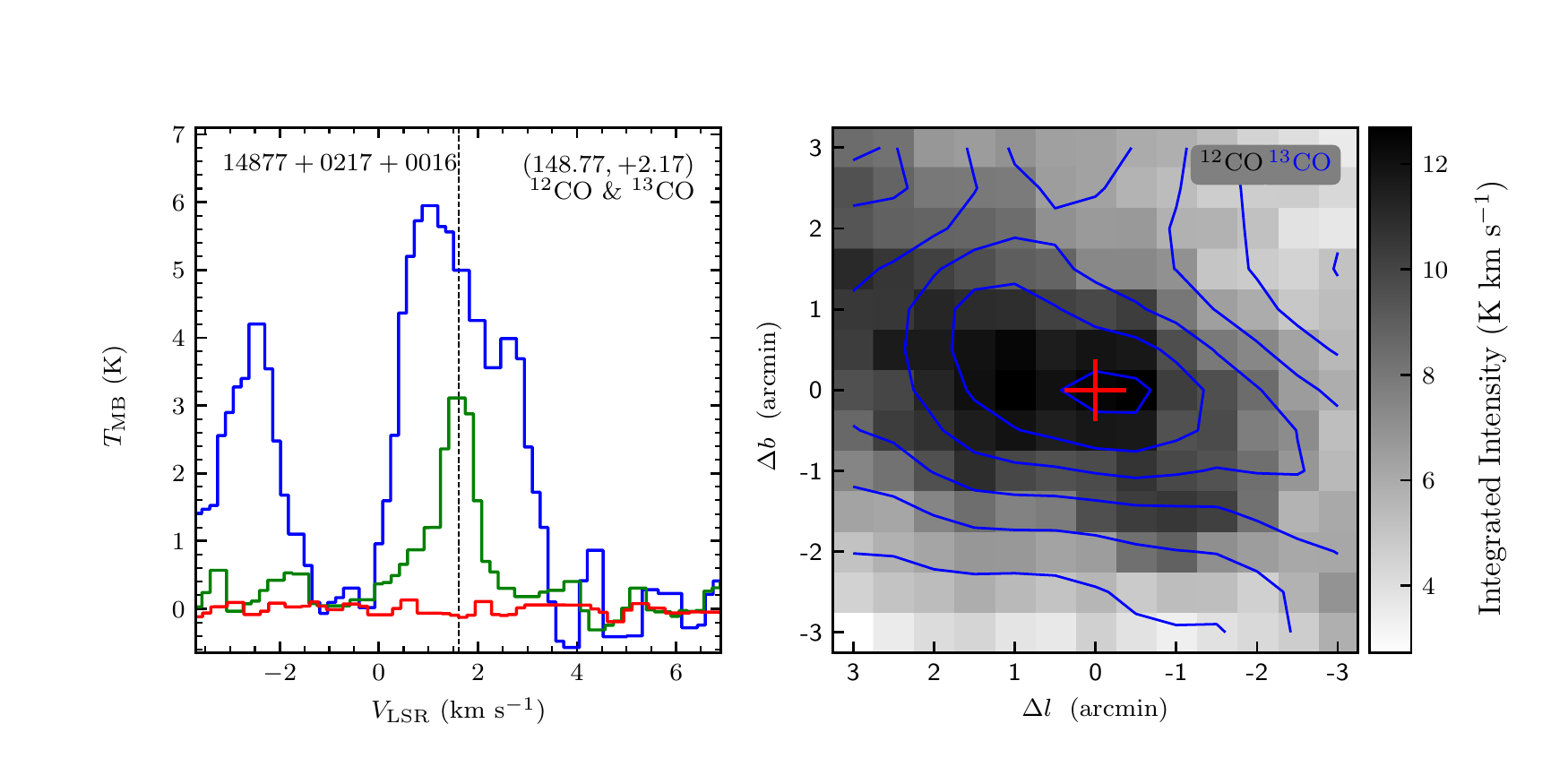}
\includegraphics[width=9.0cm,angle=0]{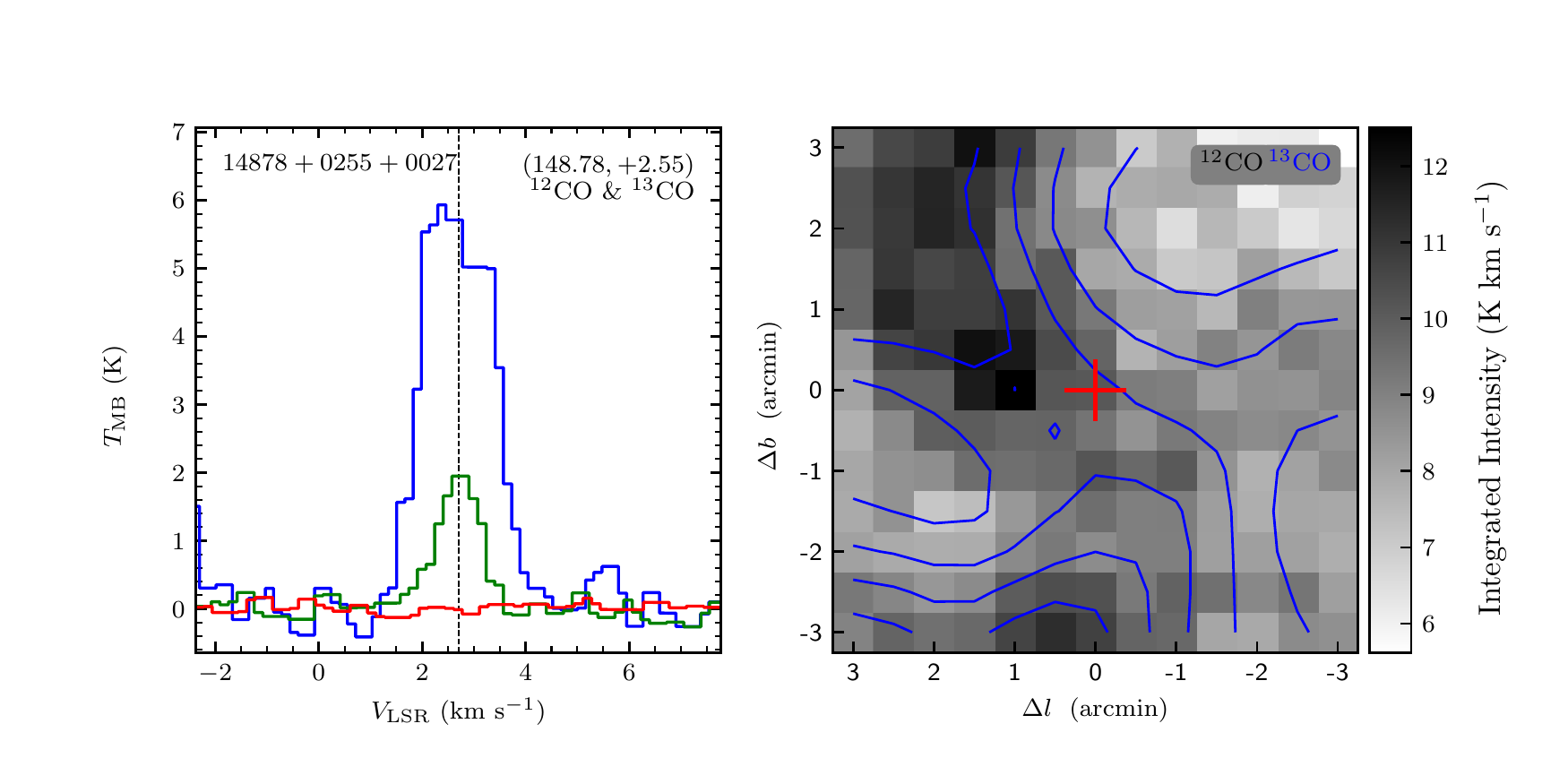}
\end{figure}
\clearpage

\begin{figure}
\includegraphics[width=9.0cm,angle=0]{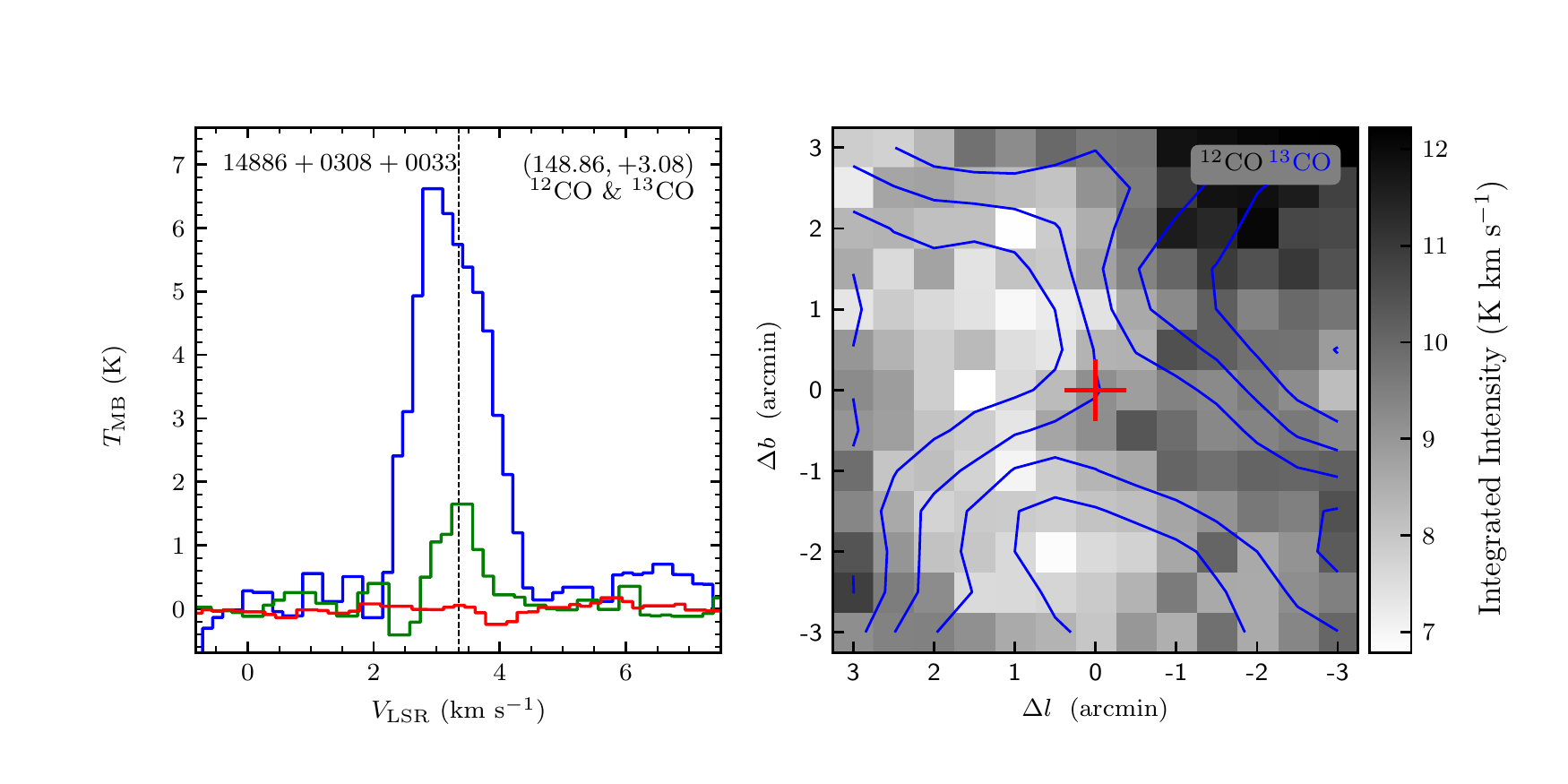}
\includegraphics[width=9.0cm,angle=0]{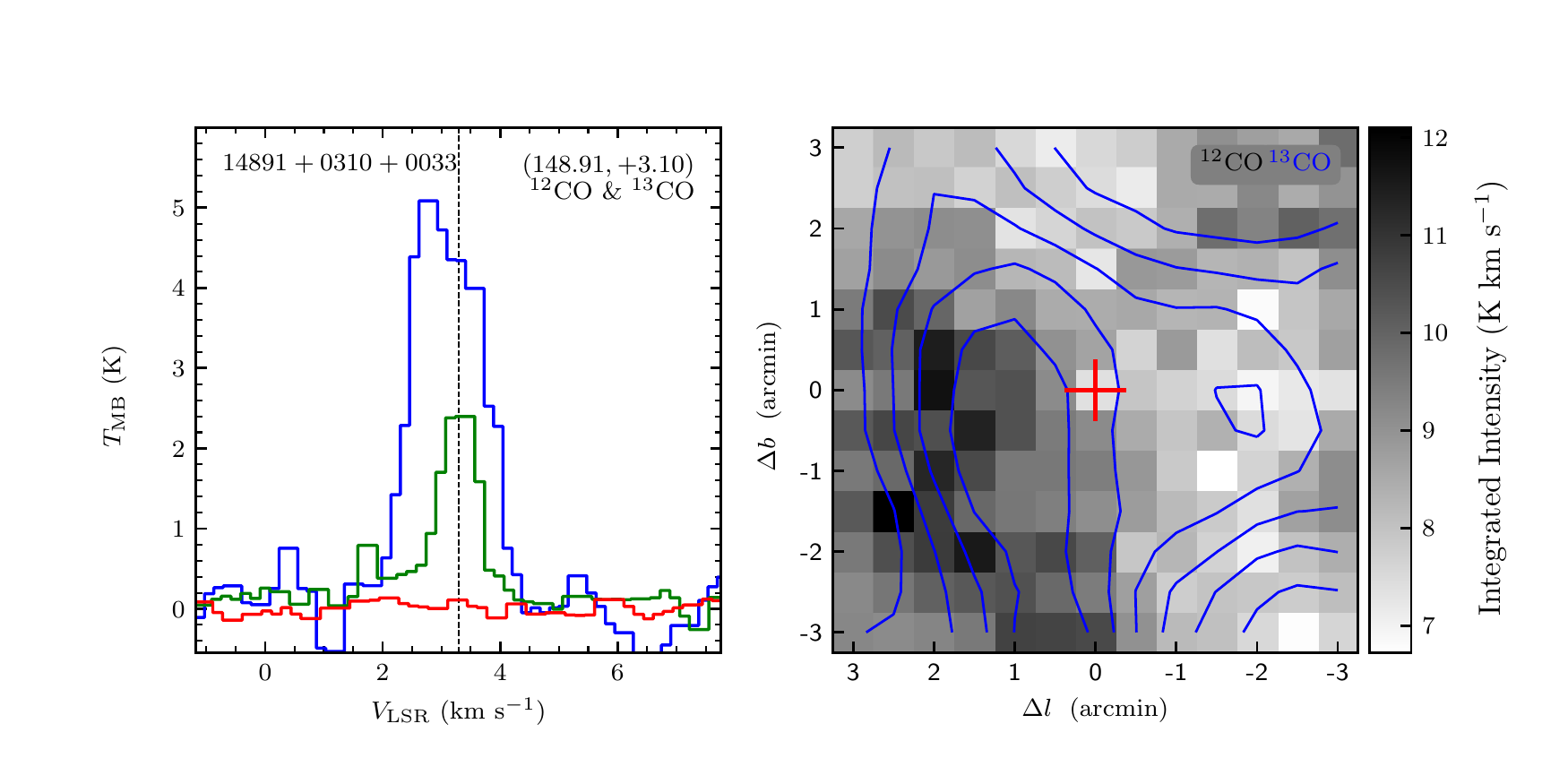}
\vspace{-0.5cm}

\includegraphics[width=9.0cm,angle=0]{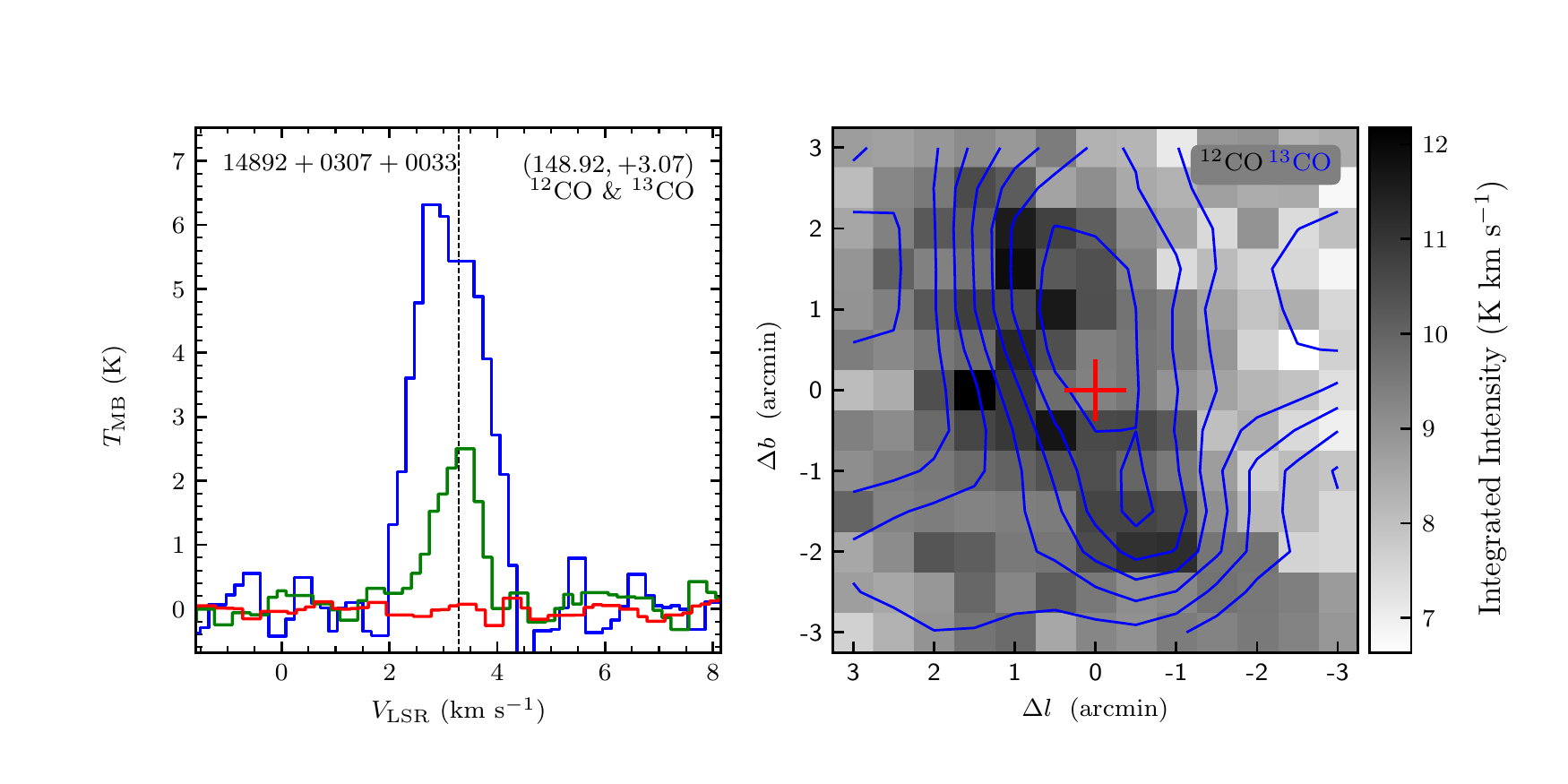}
\includegraphics[width=9.0cm,angle=0]{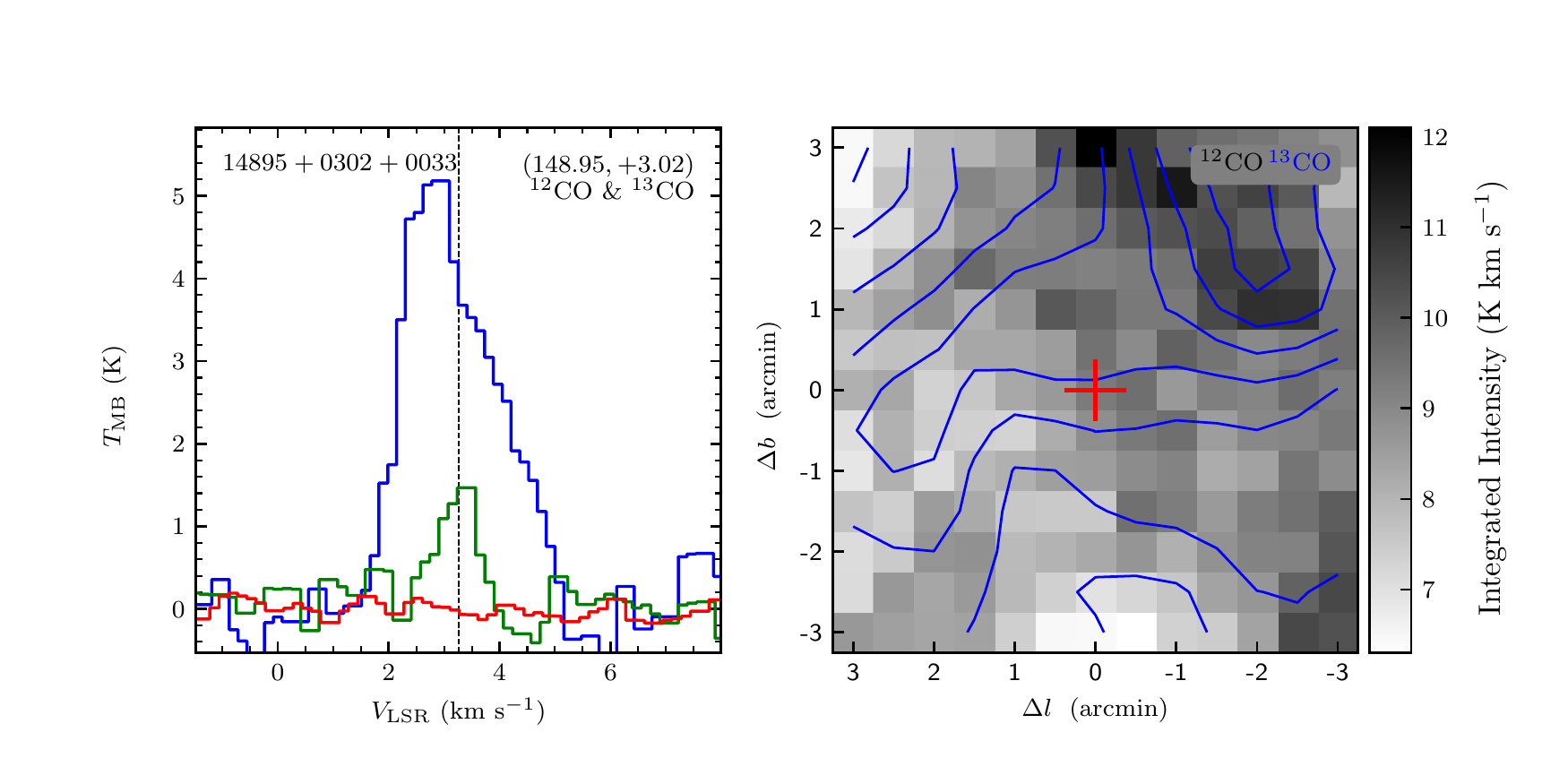}
\vspace{-0.5cm}

\includegraphics[width=9.0cm,angle=0]{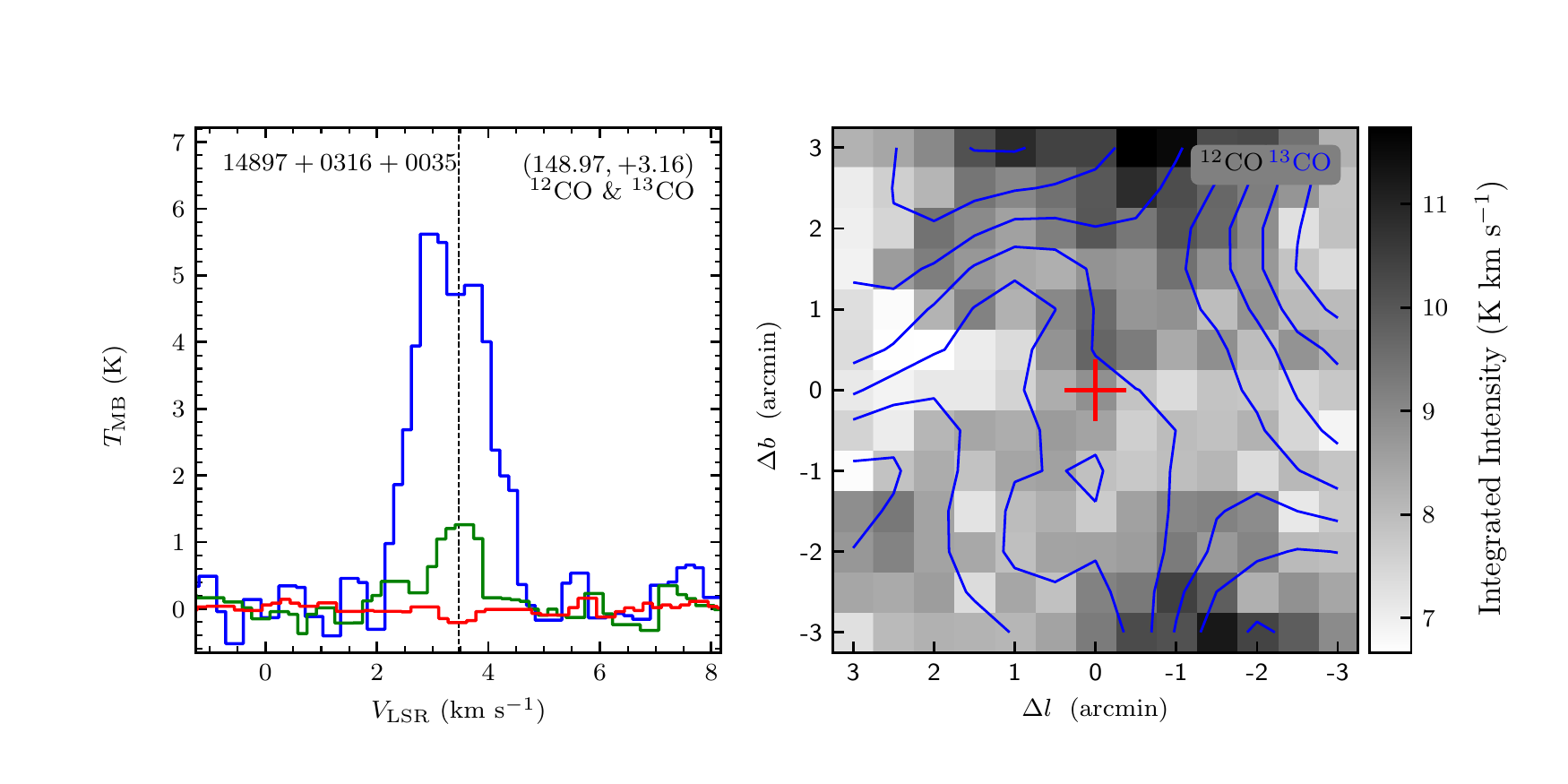}
\includegraphics[width=9.0cm,angle=0]{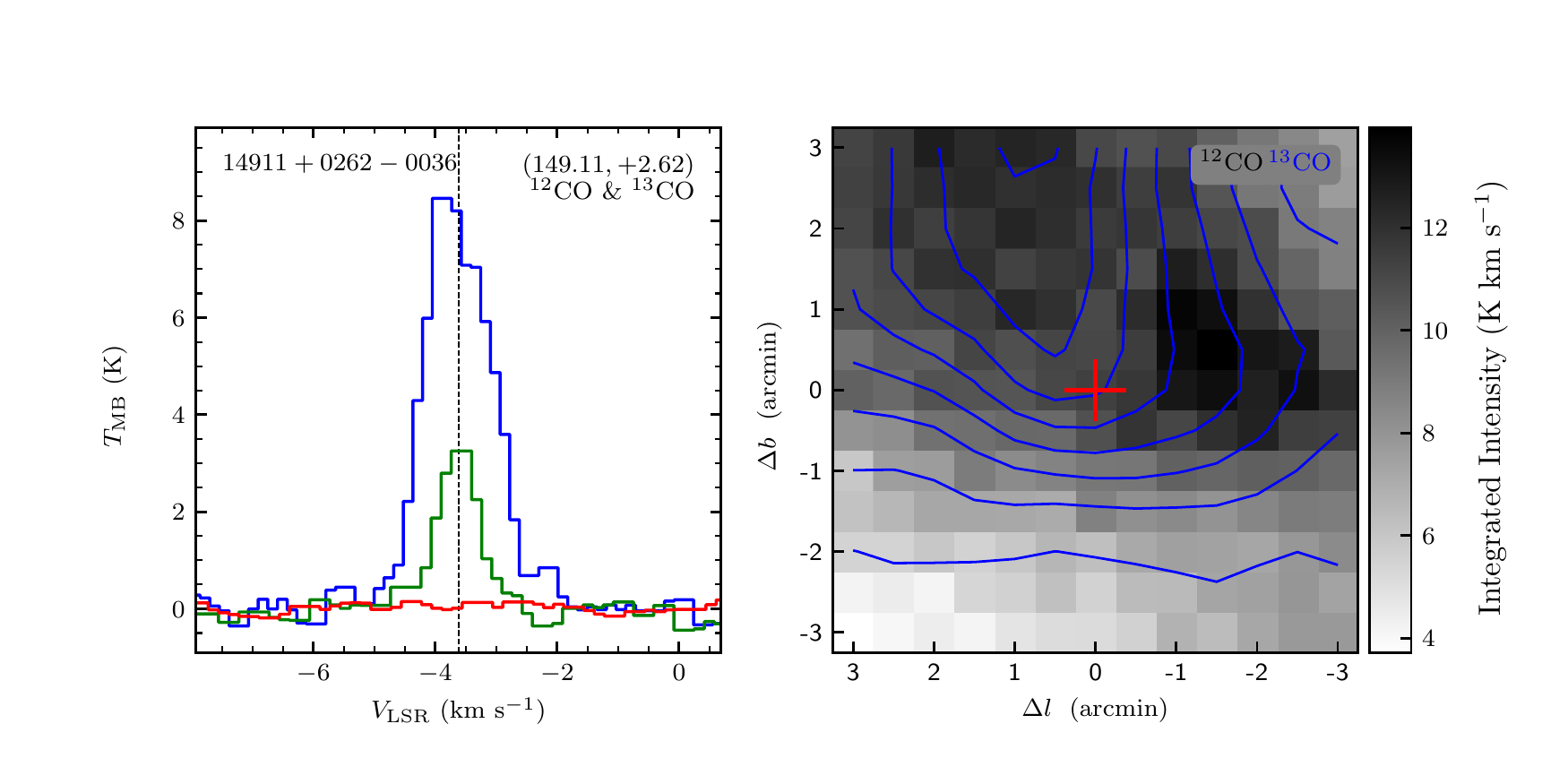}
\vspace{-0.5cm}

\includegraphics[width=9.0cm,angle=0]{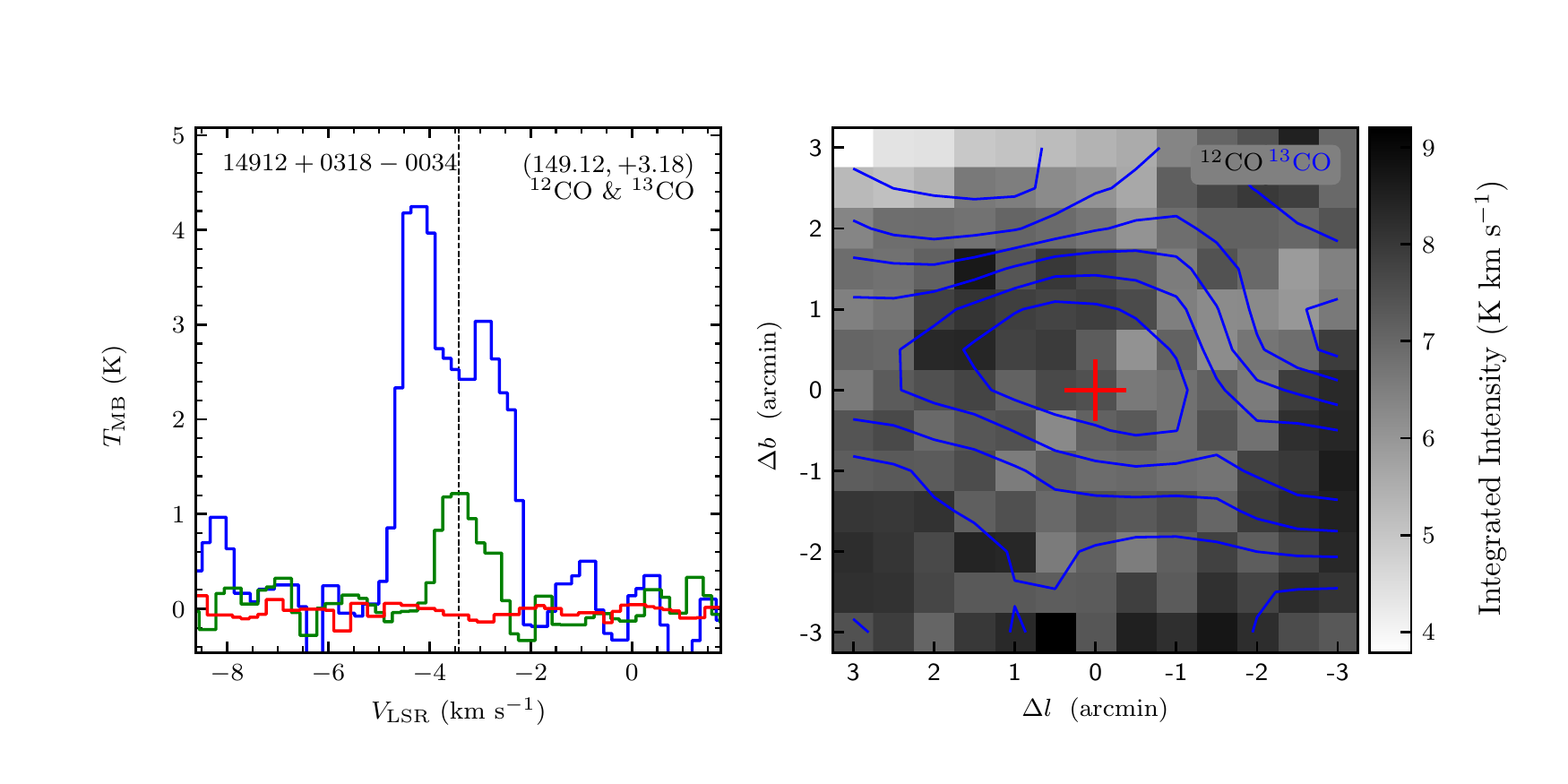}
\includegraphics[width=9.0cm,angle=0]{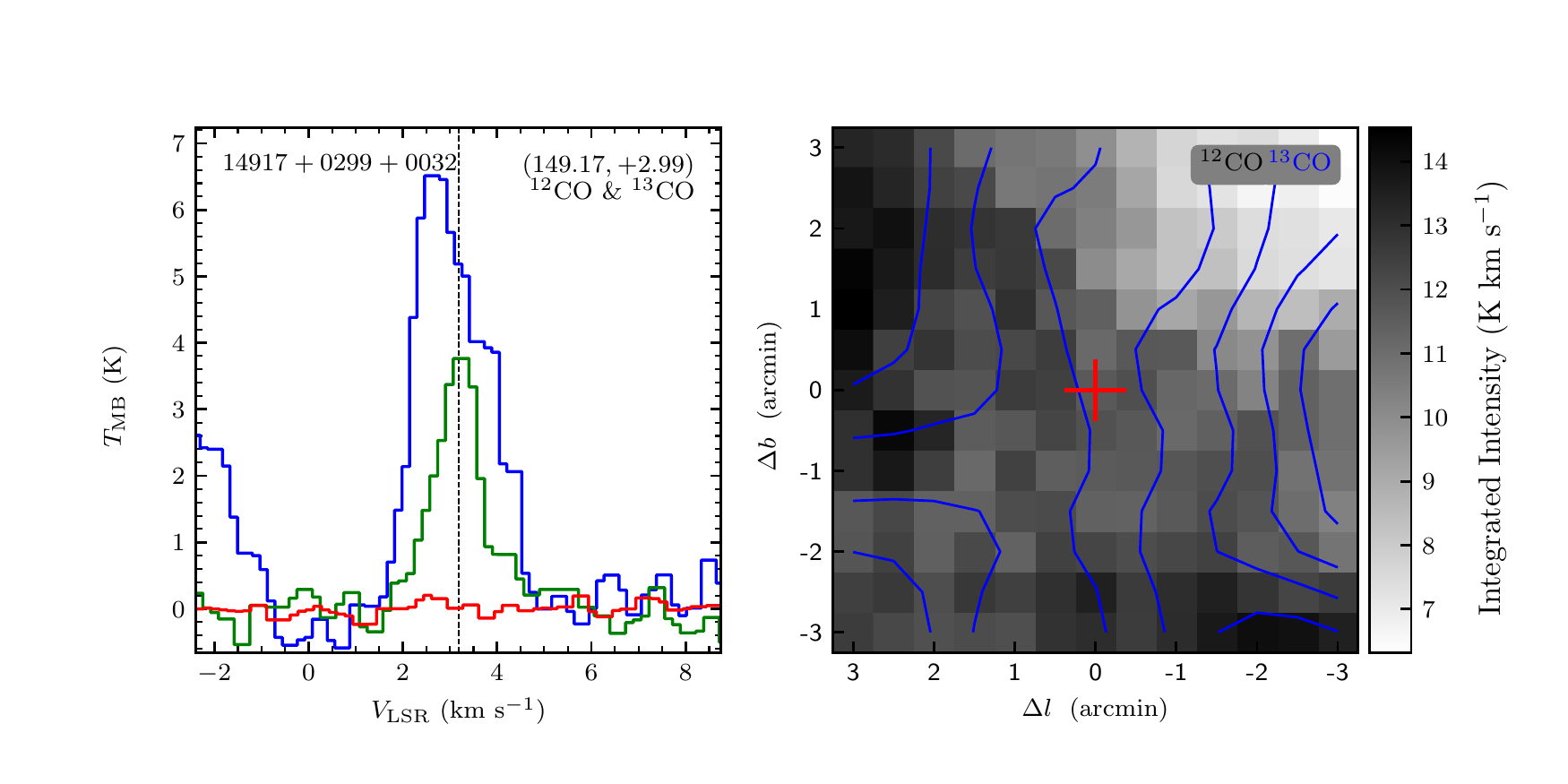}
\vspace{-0.5cm}

\includegraphics[width=9.0cm,angle=0]{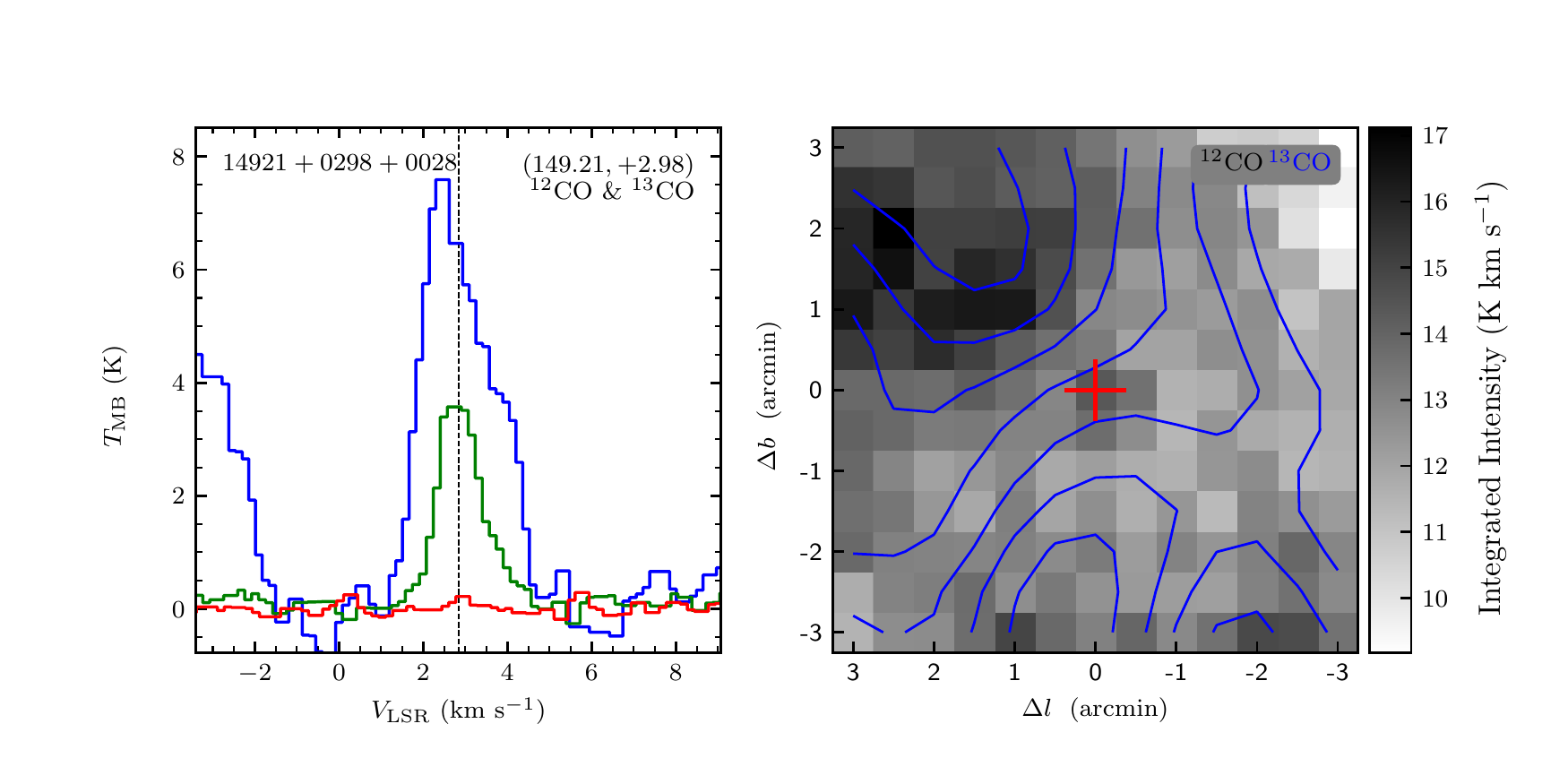}
\includegraphics[width=9.0cm,angle=0]{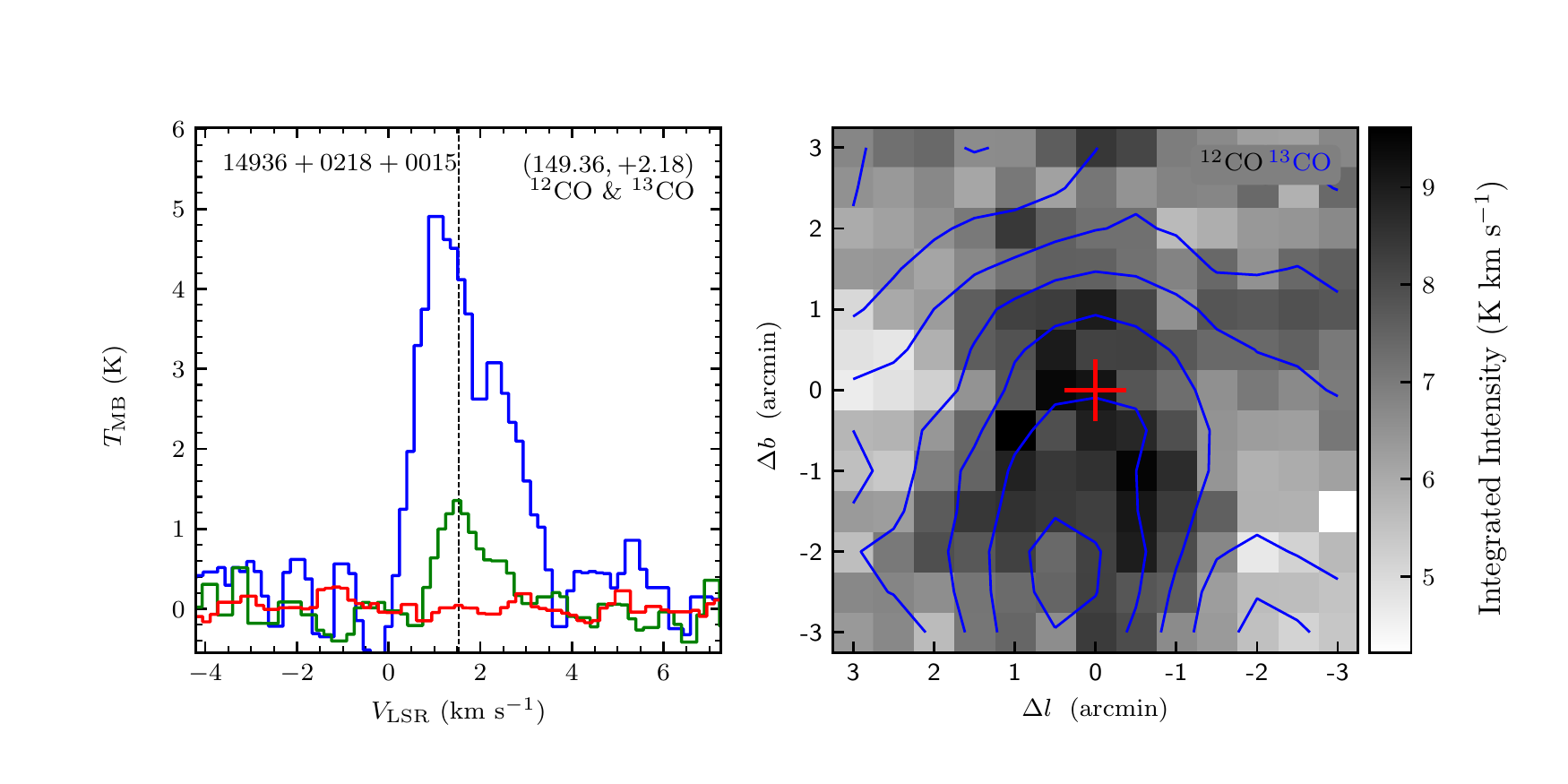}
\end{figure}
\clearpage

\begin{figure}
\includegraphics[width=9.0cm,angle=0]{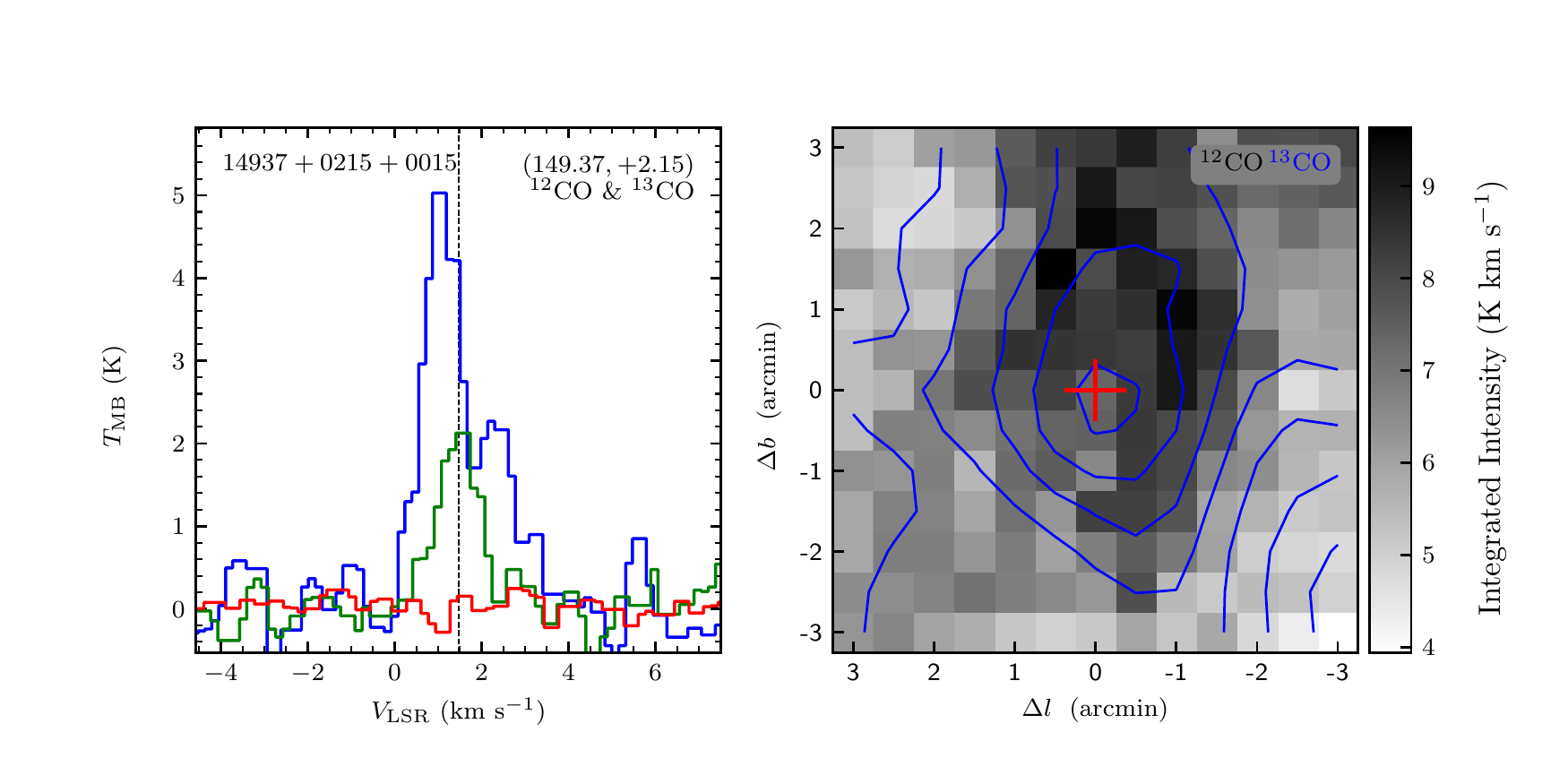}
\includegraphics[width=9.0cm,angle=0]{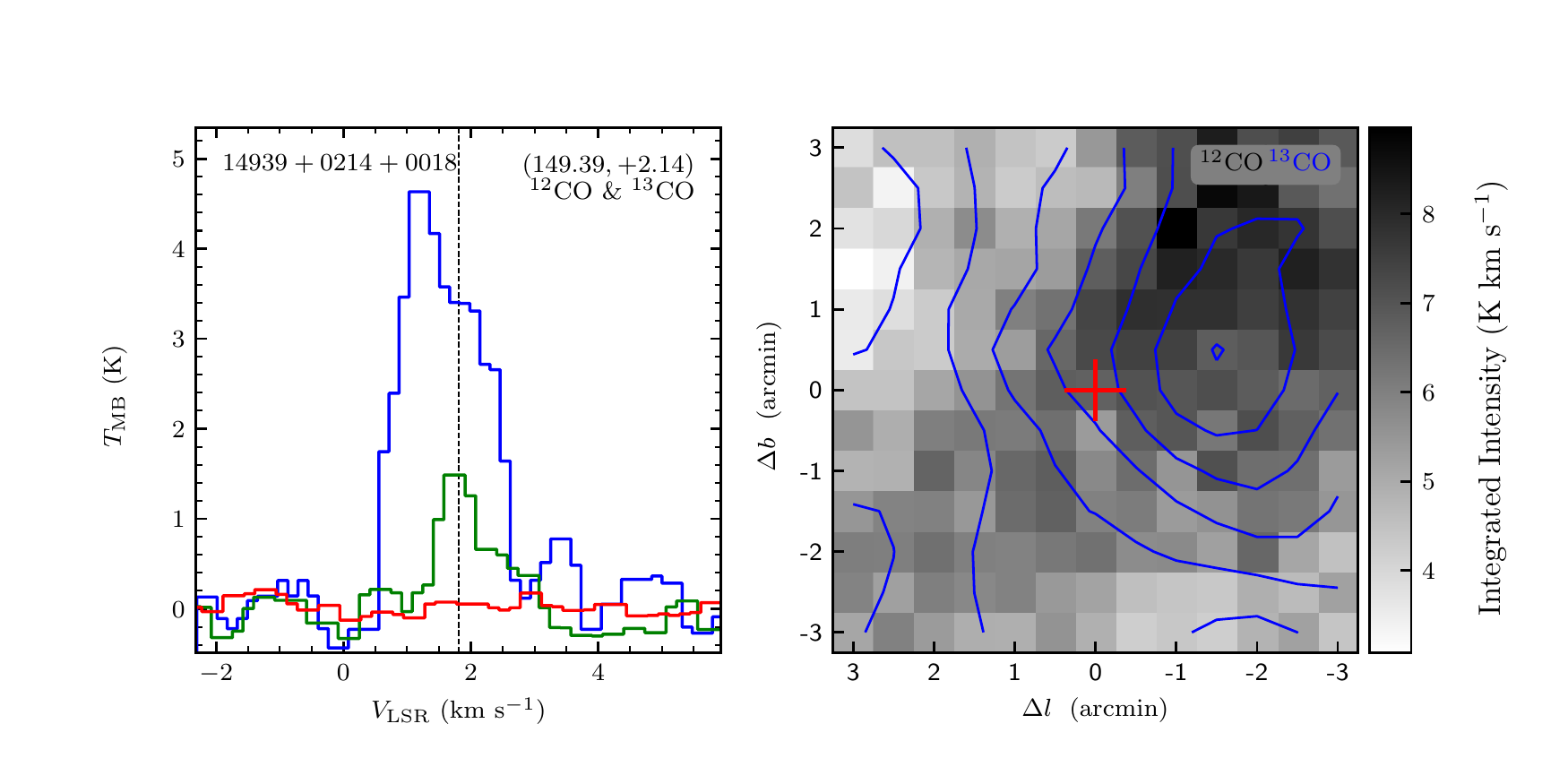}
\vspace{-0.5cm}

\includegraphics[width=9.0cm,angle=0]{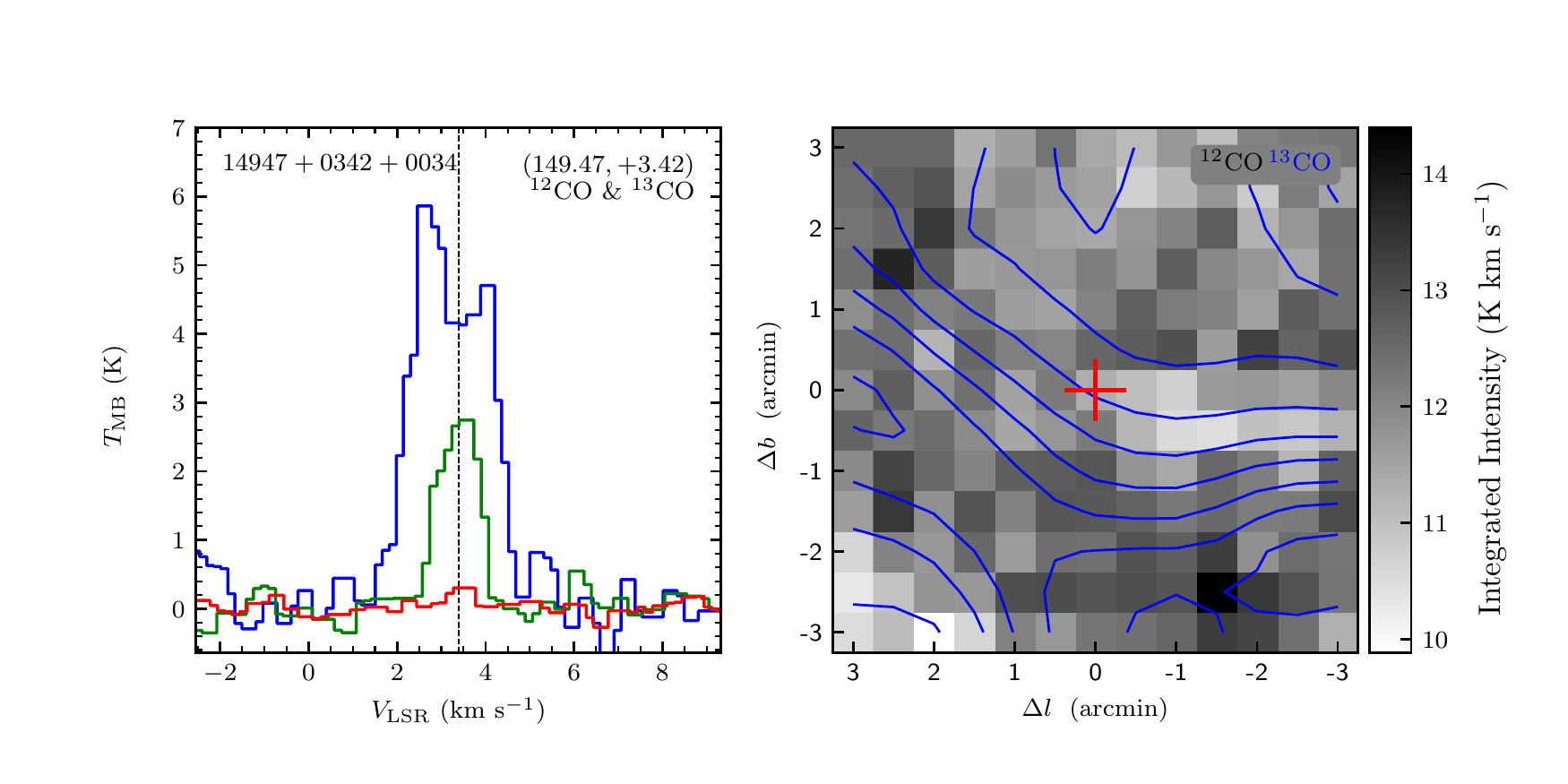}
\includegraphics[width=9.0cm,angle=0]{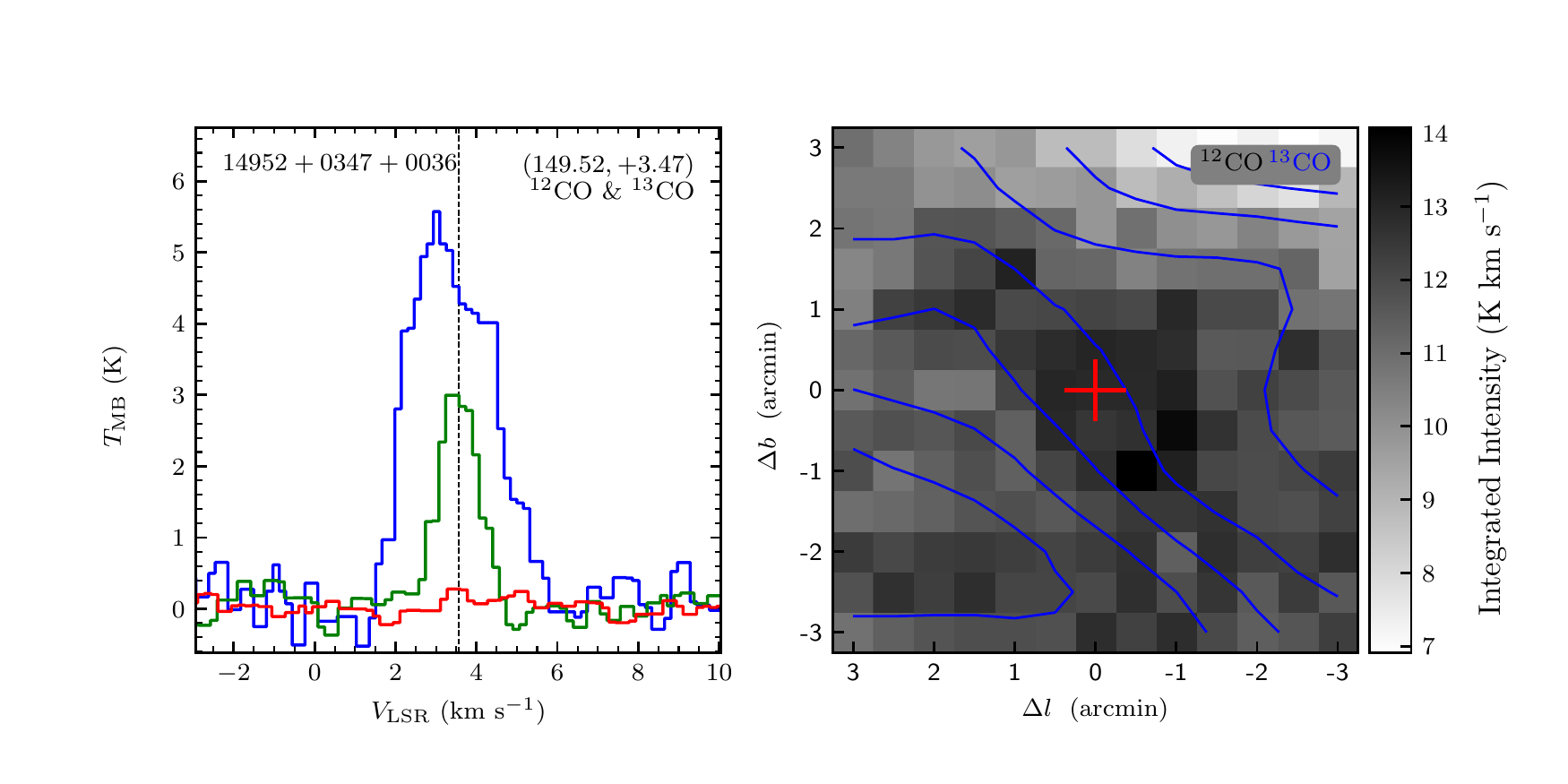}
\vspace{-0.5cm}

\includegraphics[width=9.0cm,angle=0]{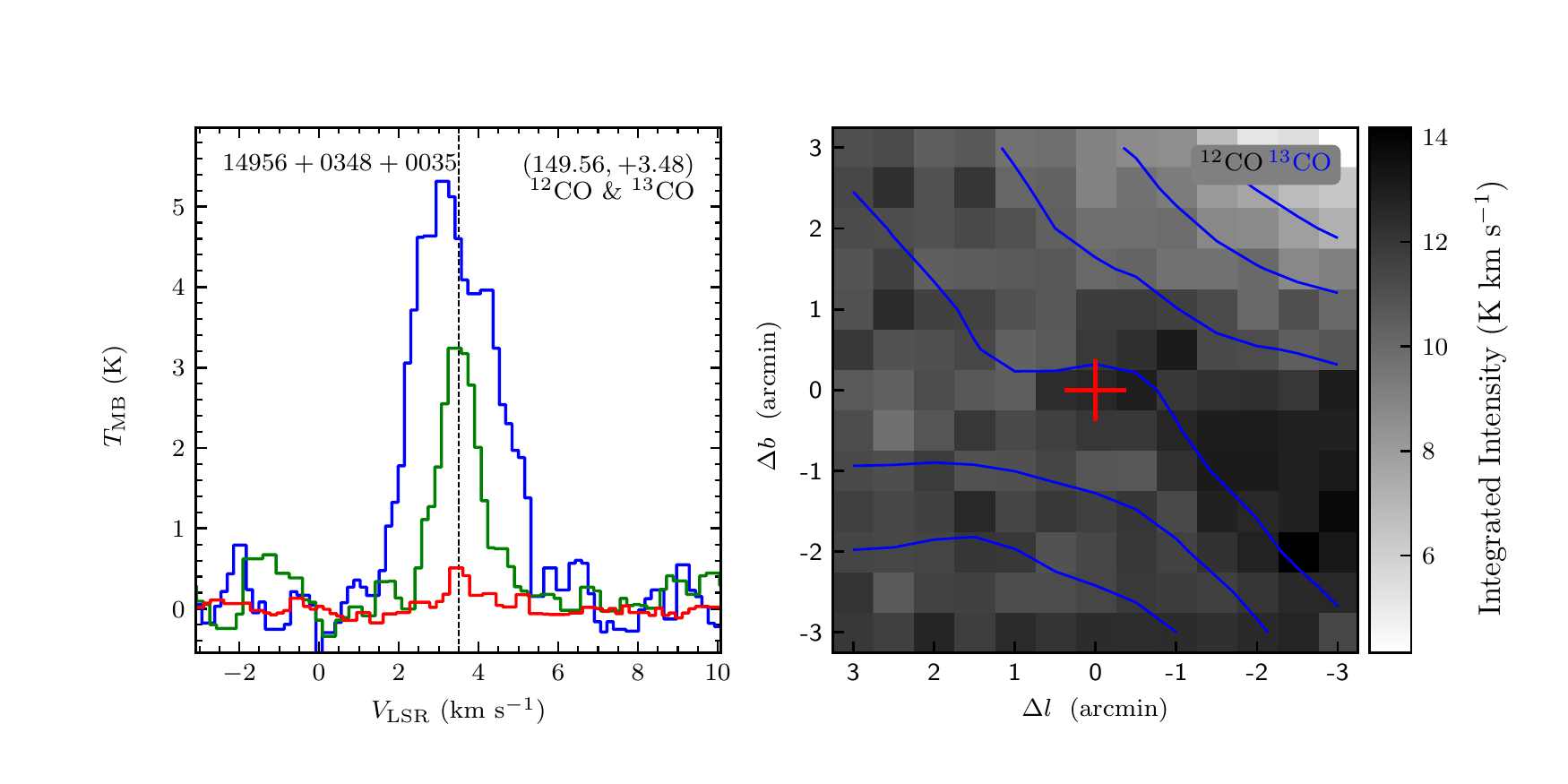}
\includegraphics[width=9.0cm,angle=0]{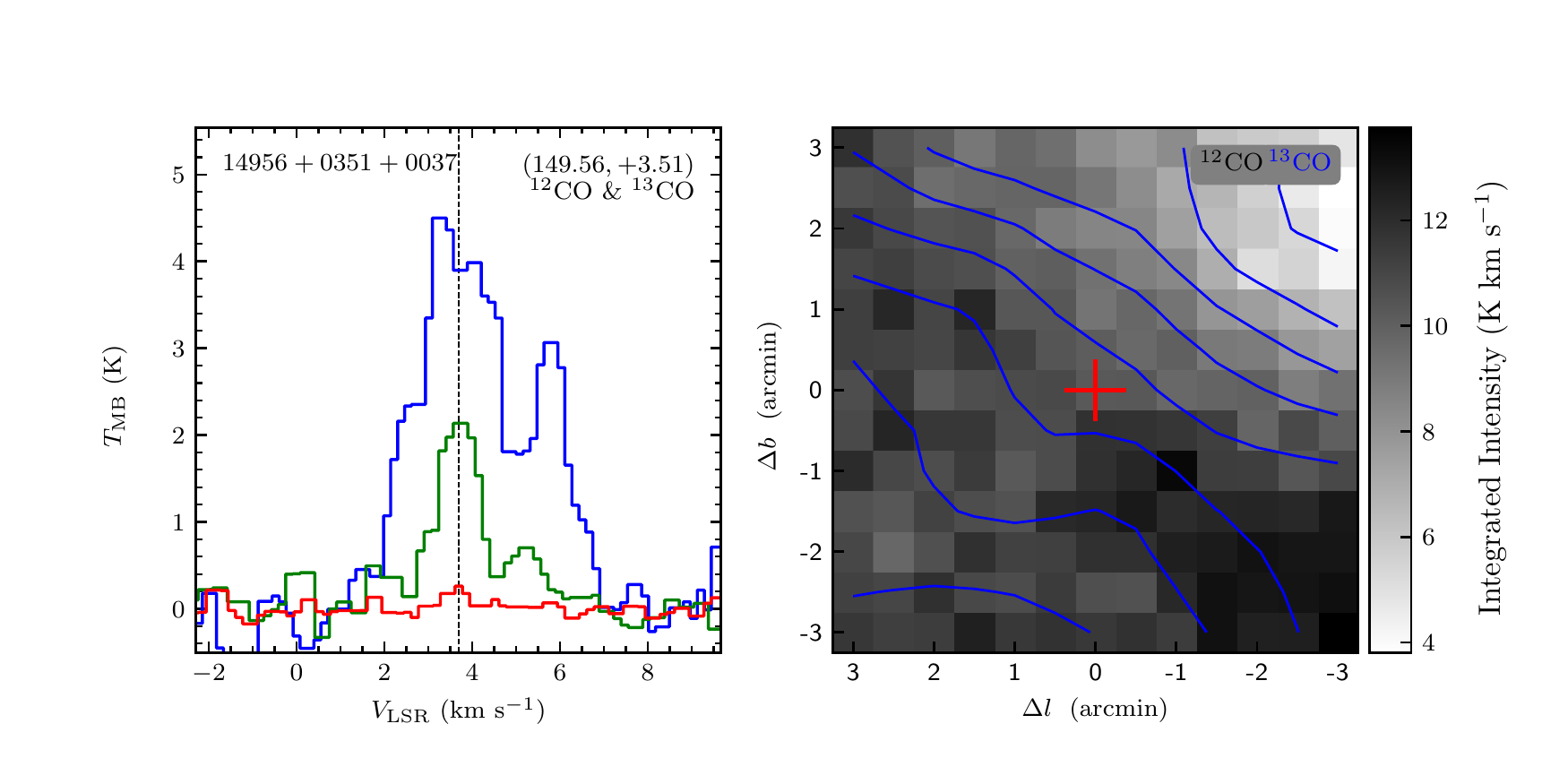}
\vspace{-0.5cm}

\includegraphics[width=9.0cm,angle=0]{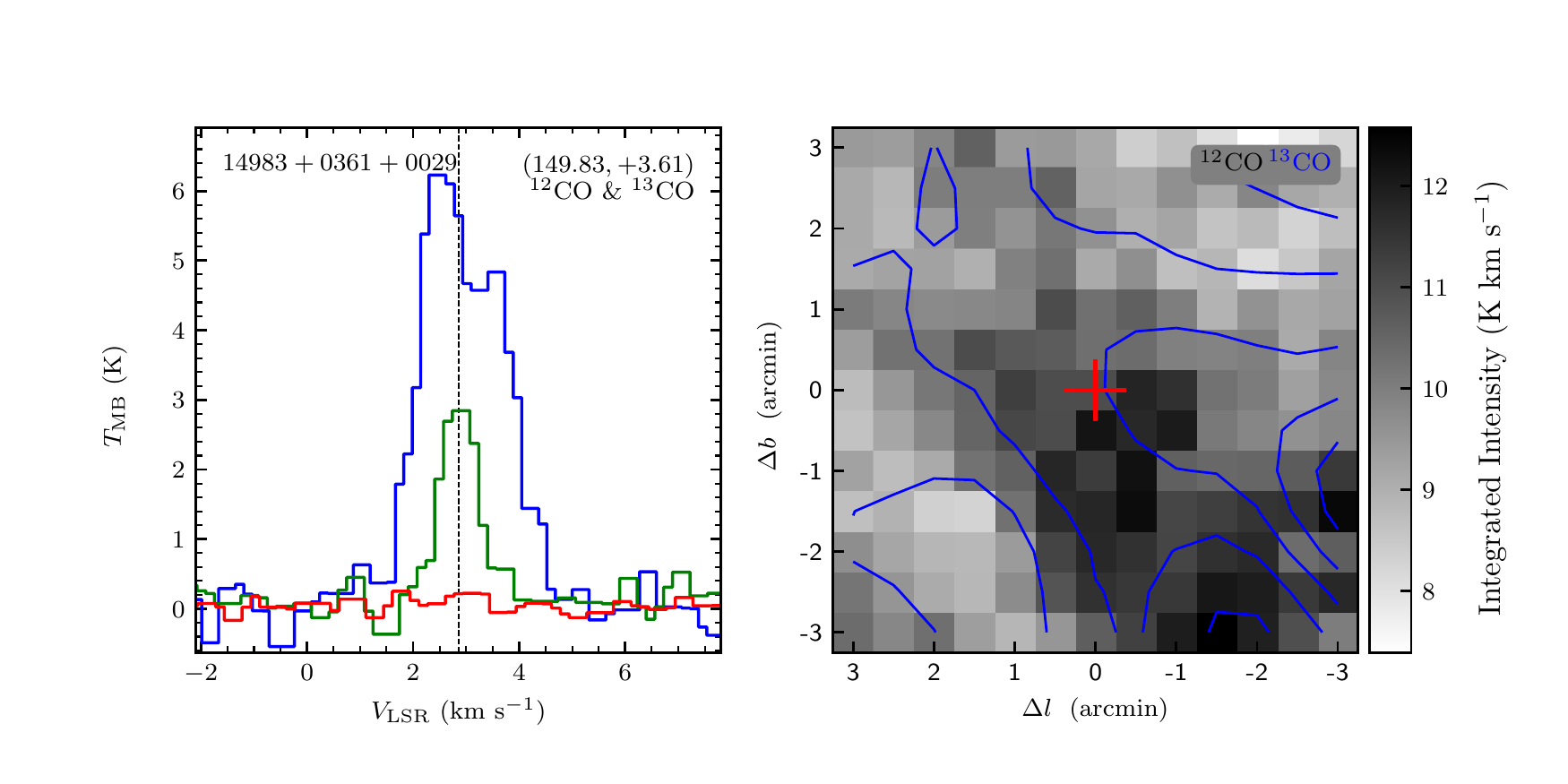}
\includegraphics[width=9.0cm,angle=0]{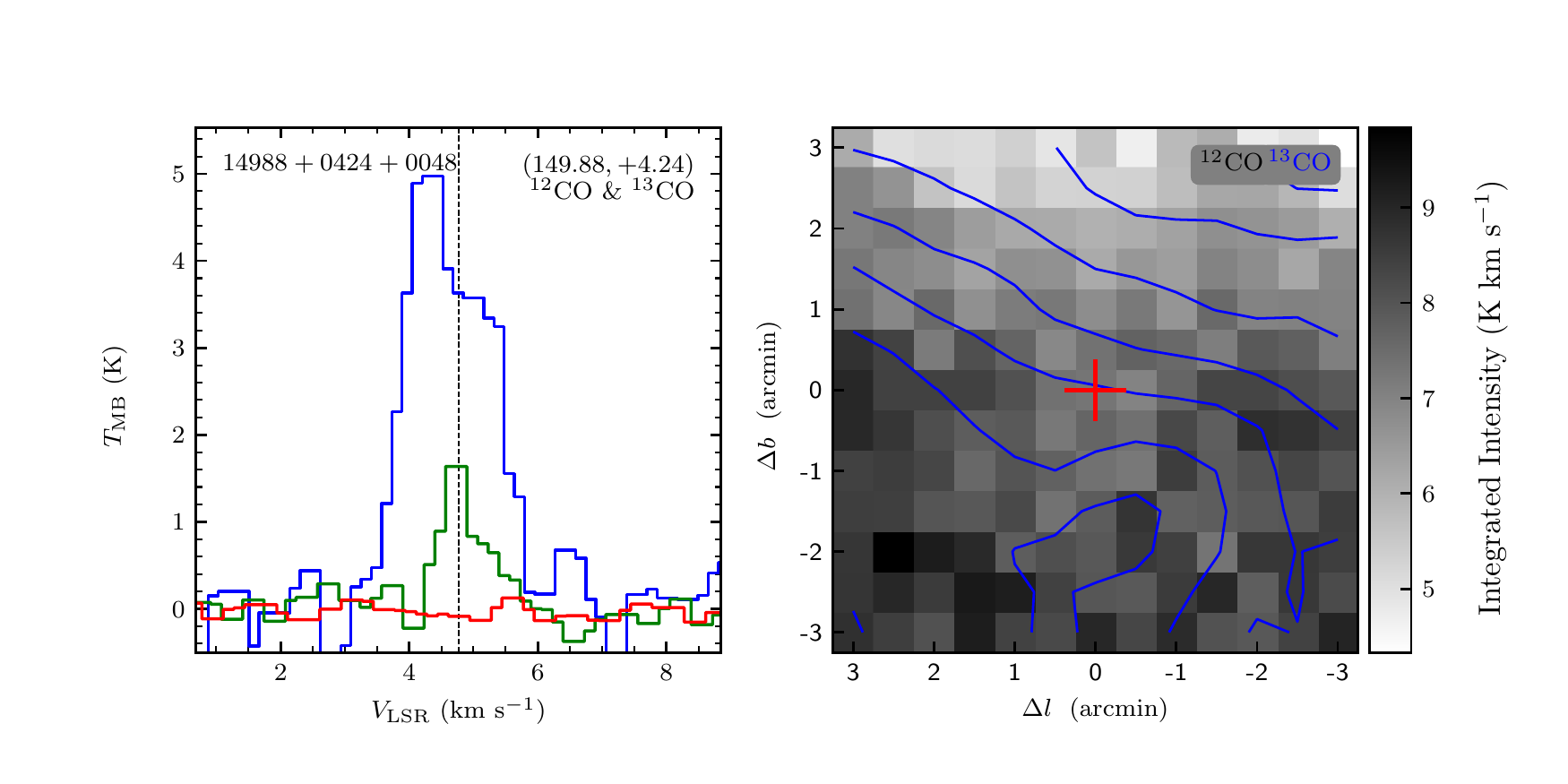}
\vspace{-0.5cm}

\includegraphics[width=9.0cm,angle=0]{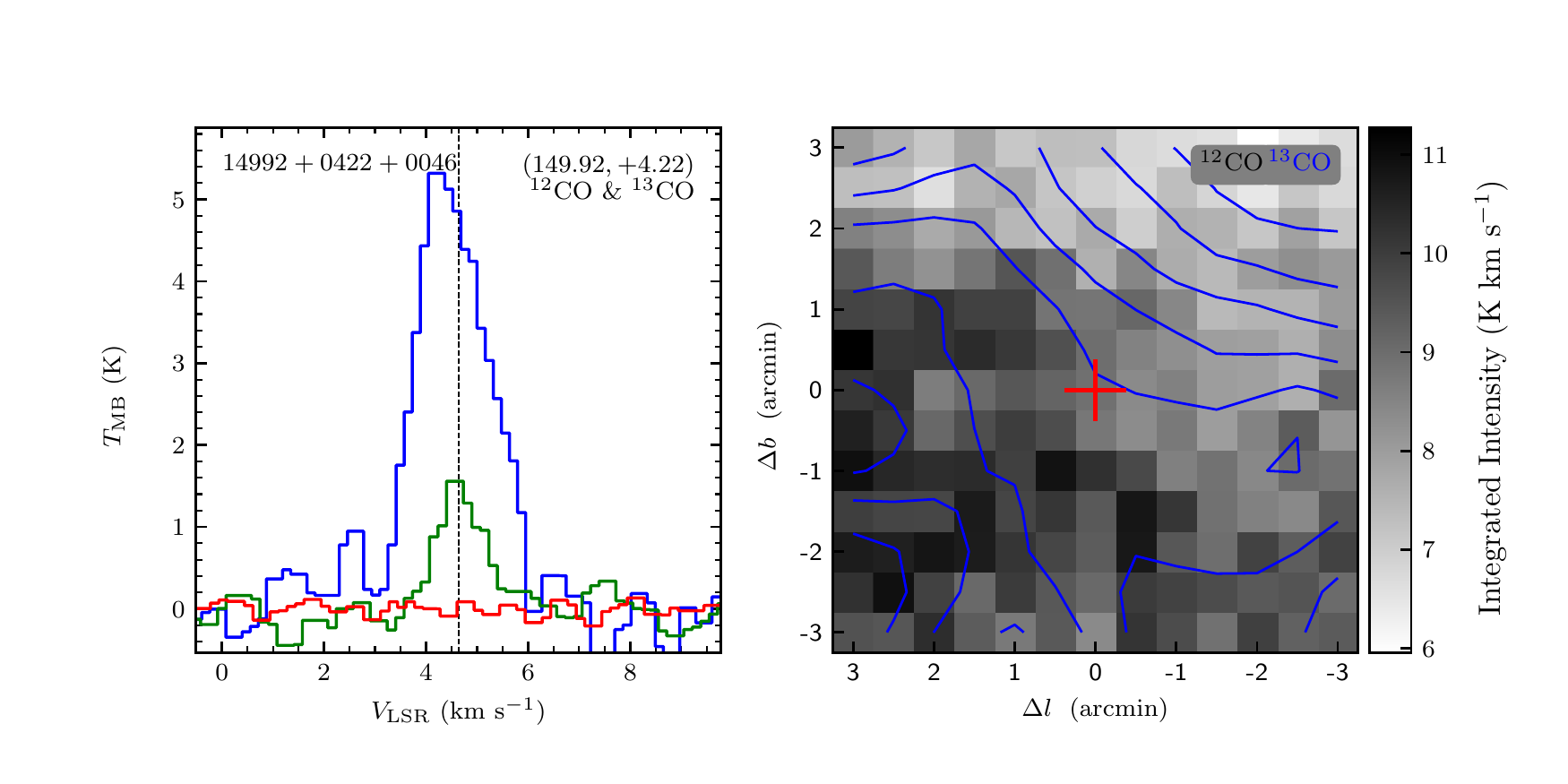}
\includegraphics[width=9.0cm,angle=0]{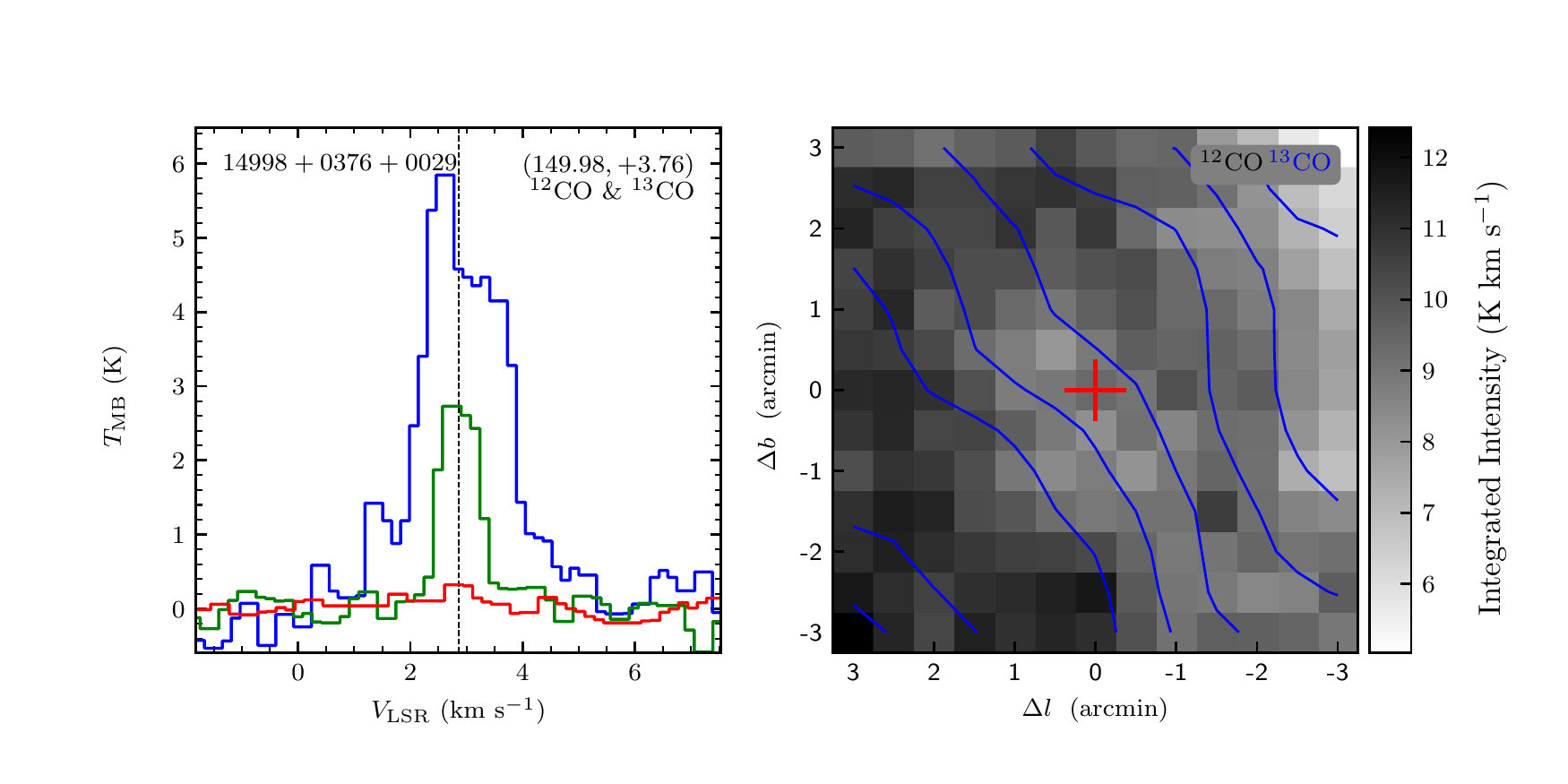}
\end{figure}
\clearpage

\begin{figure}
\includegraphics[width=9.0cm,angle=0]{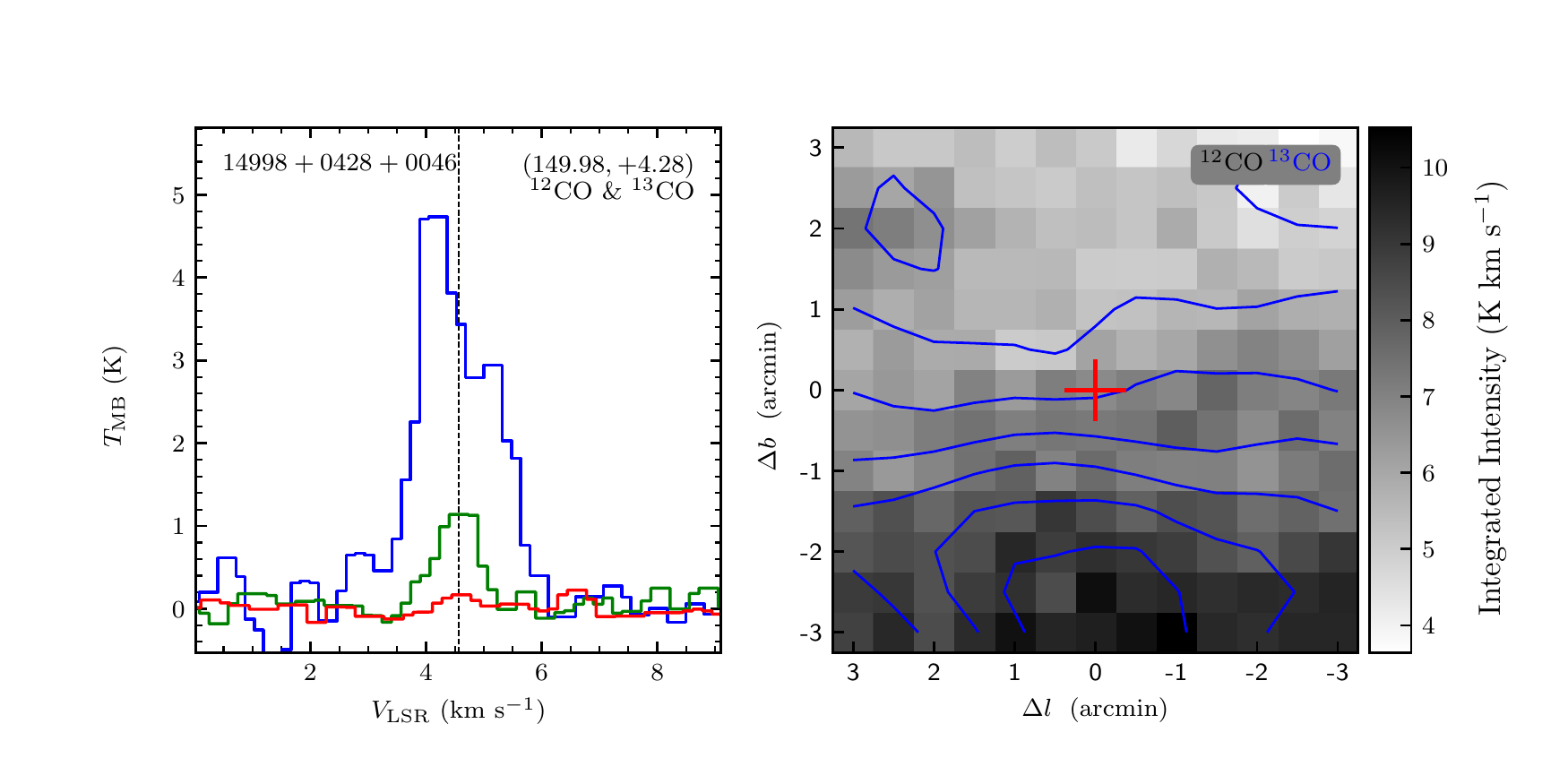}
\includegraphics[width=9.0cm,angle=0]{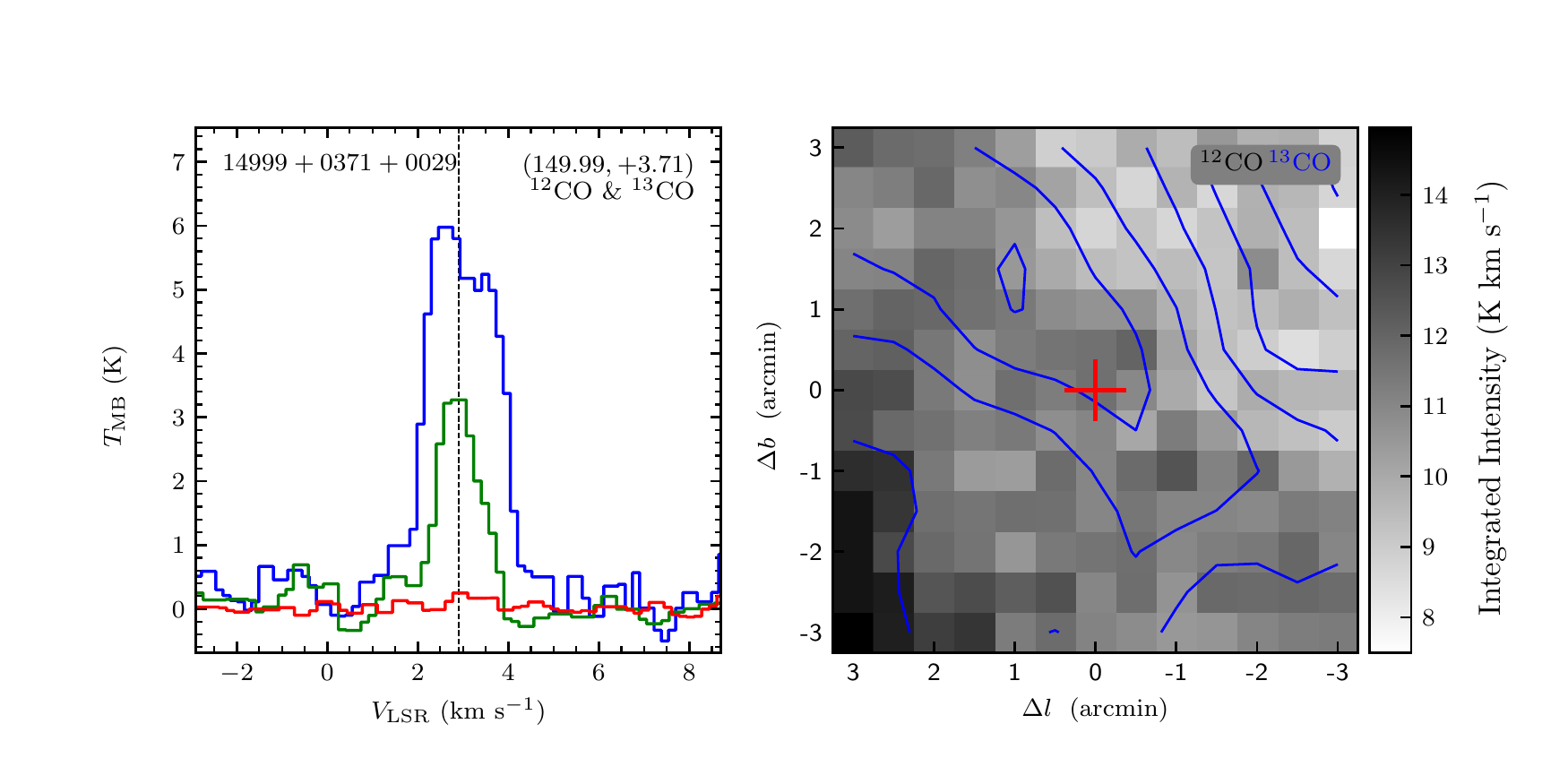}
\vspace{-0.5cm}

\includegraphics[width=9.0cm,angle=0]{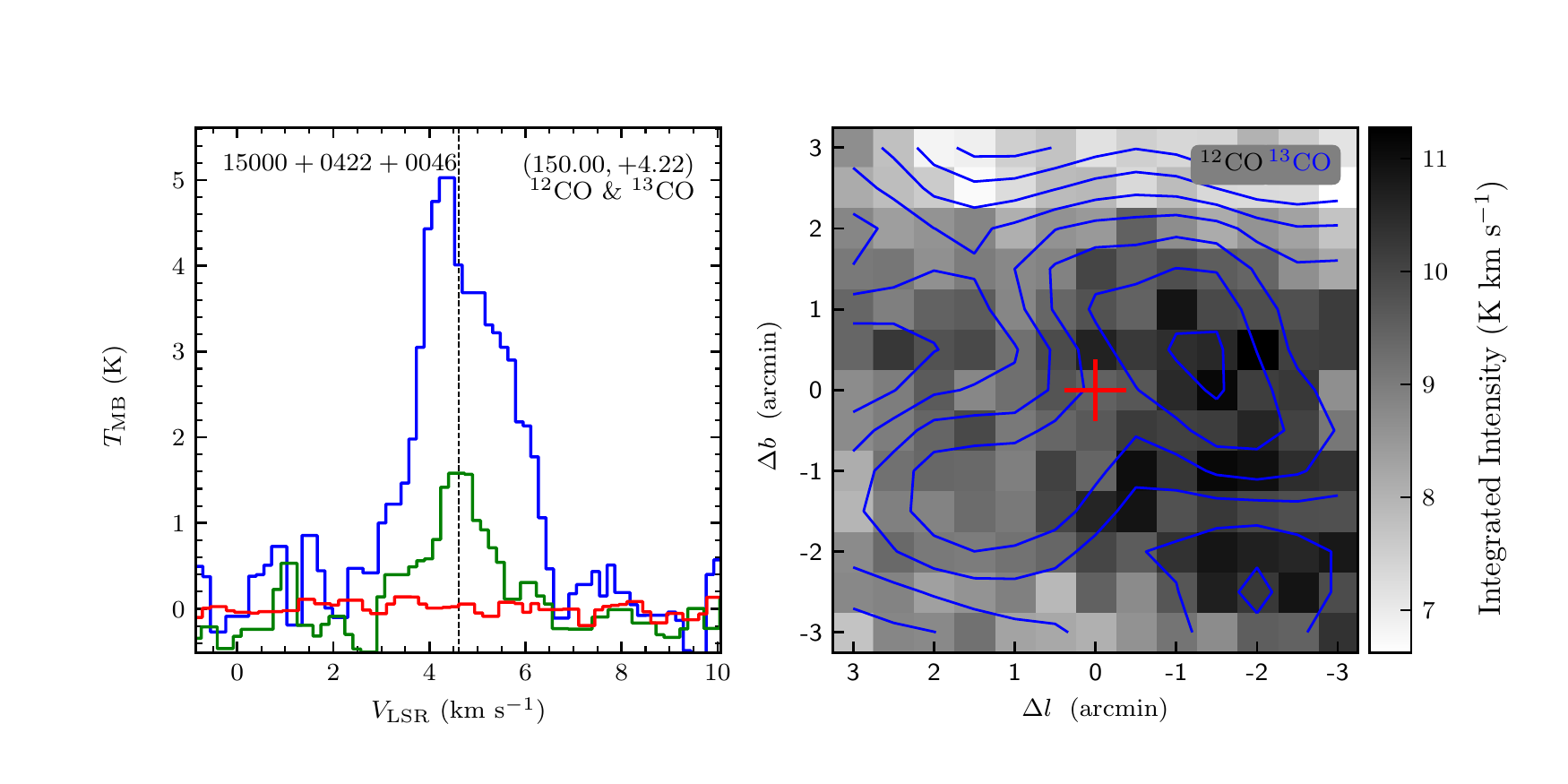}
\includegraphics[width=9.0cm,angle=0]{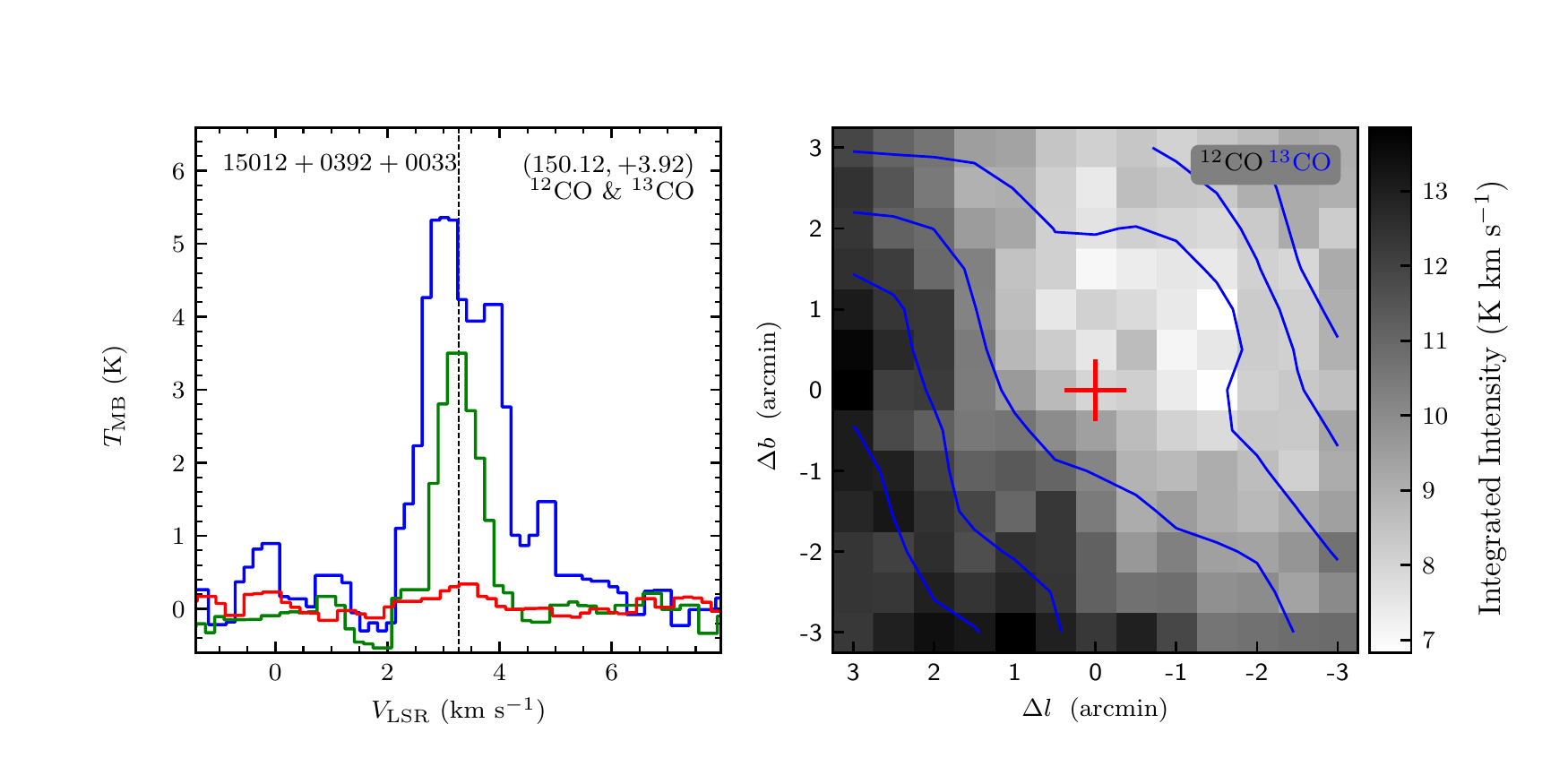}
\vspace{-0.5cm}

\includegraphics[width=9.0cm,angle=0]{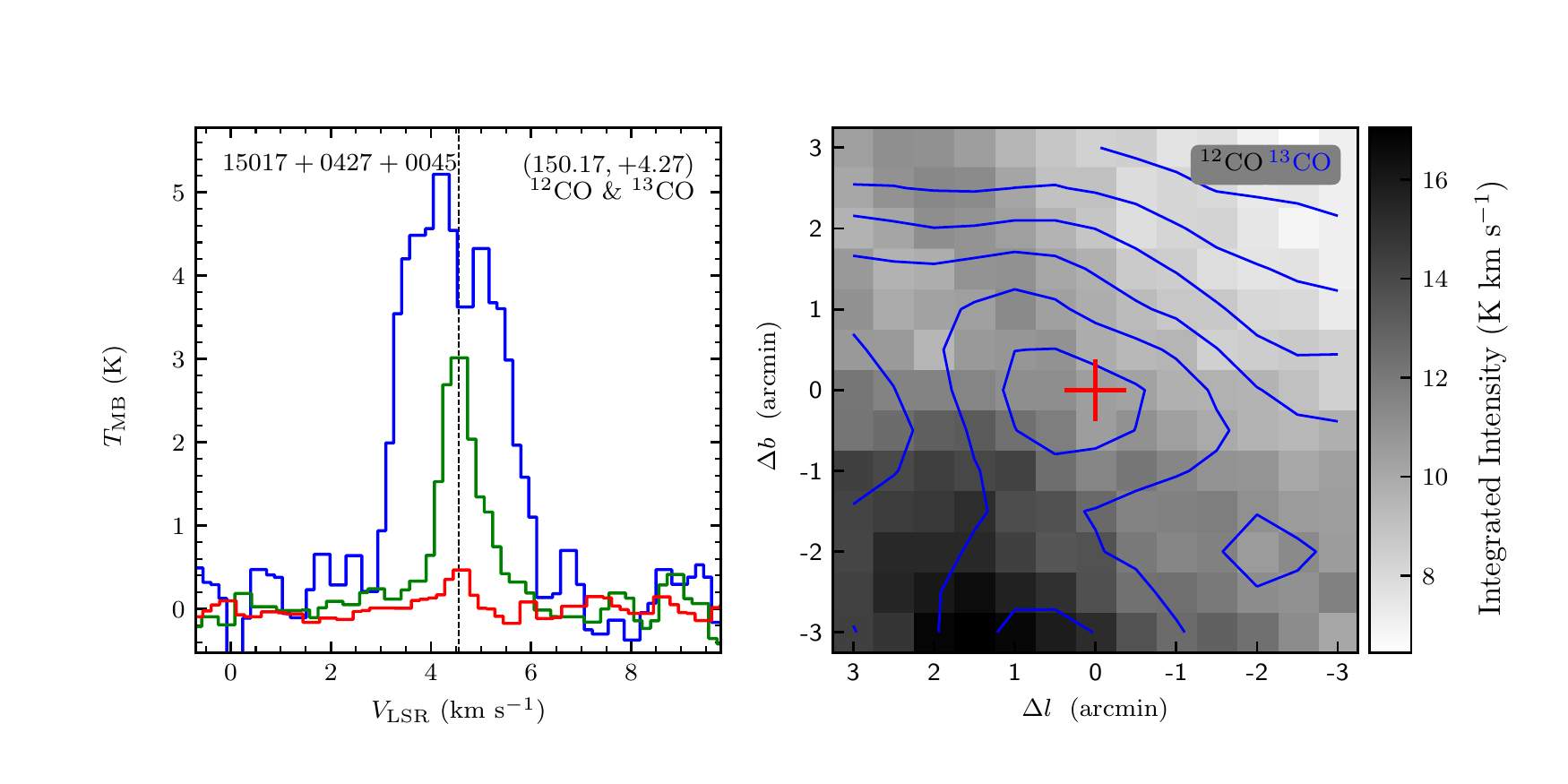}
\includegraphics[width=9.0cm,angle=0]{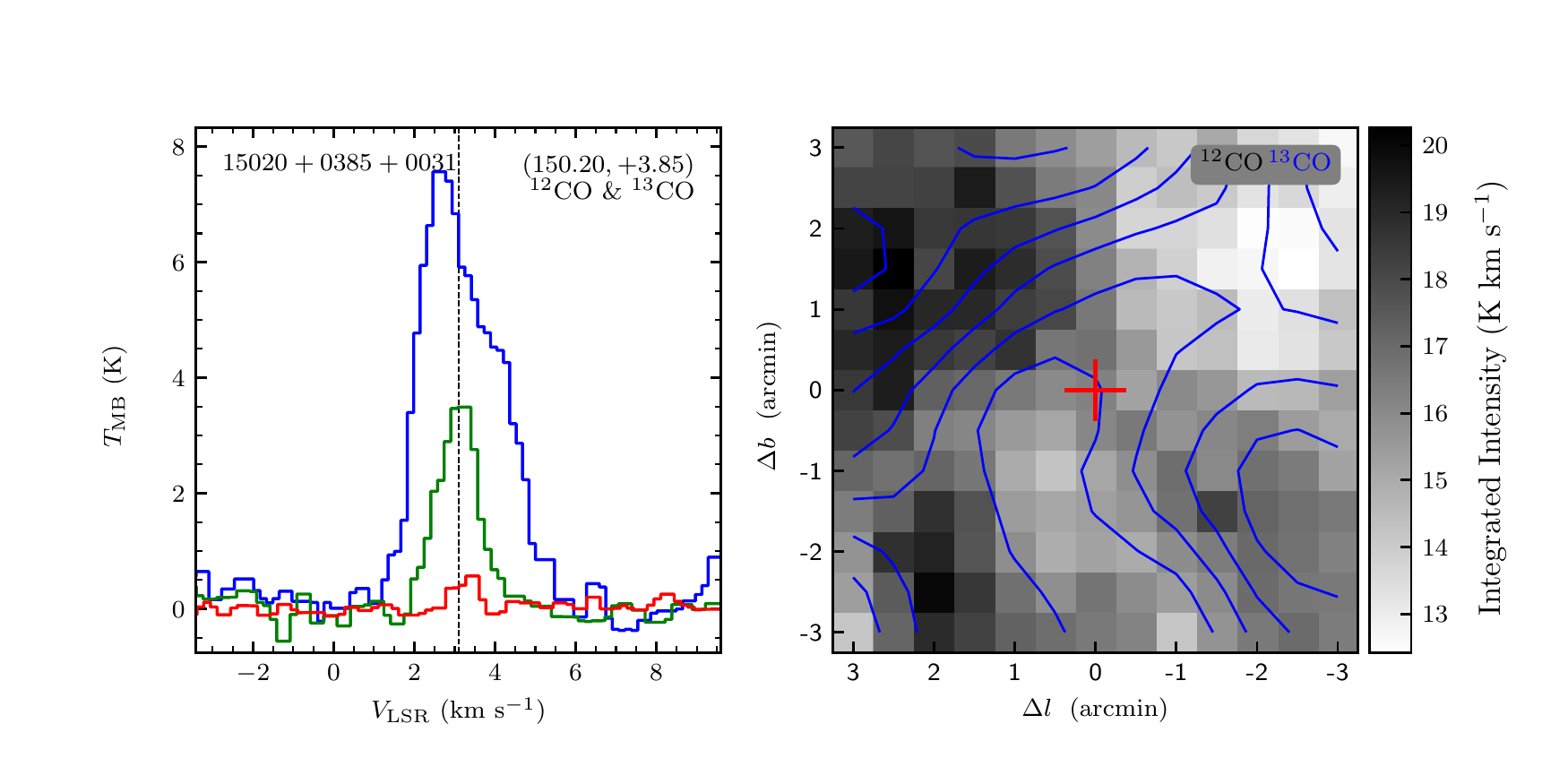}
\vspace{-0.5cm}

\includegraphics[width=9.0cm,angle=0]{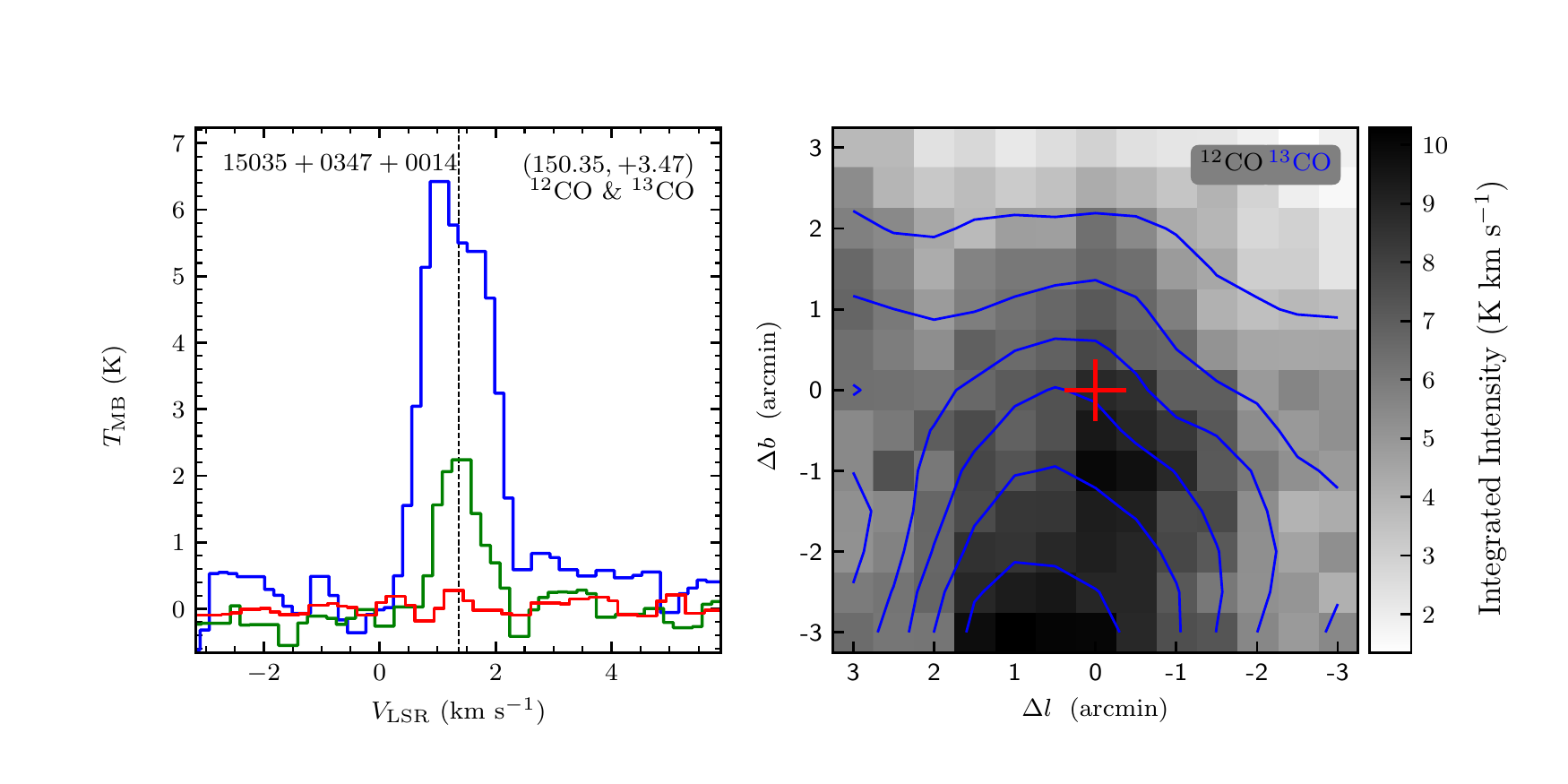}
\includegraphics[width=9.0cm,angle=0]{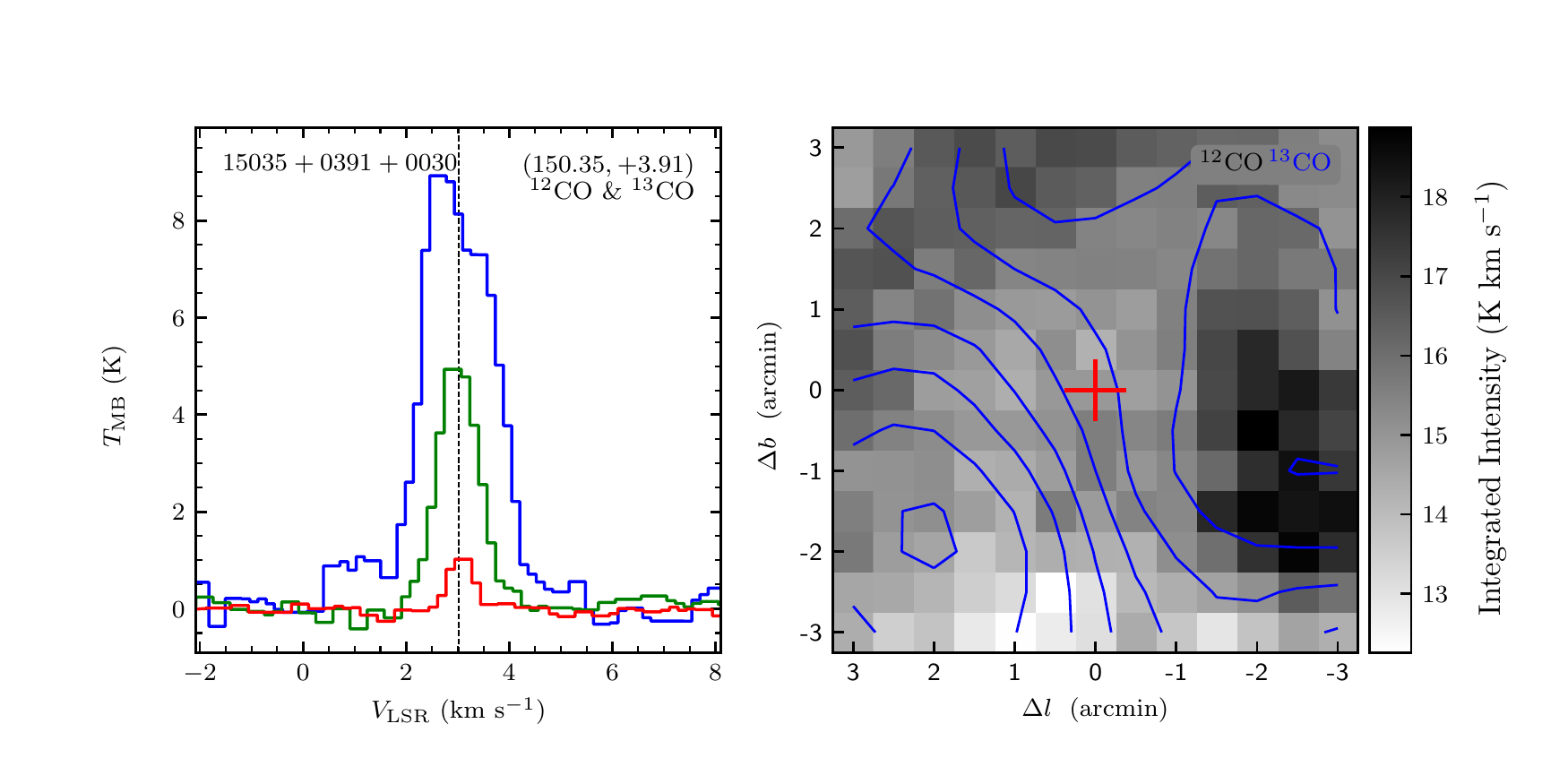}
\vspace{-0.5cm}

\includegraphics[width=9.0cm,angle=0]{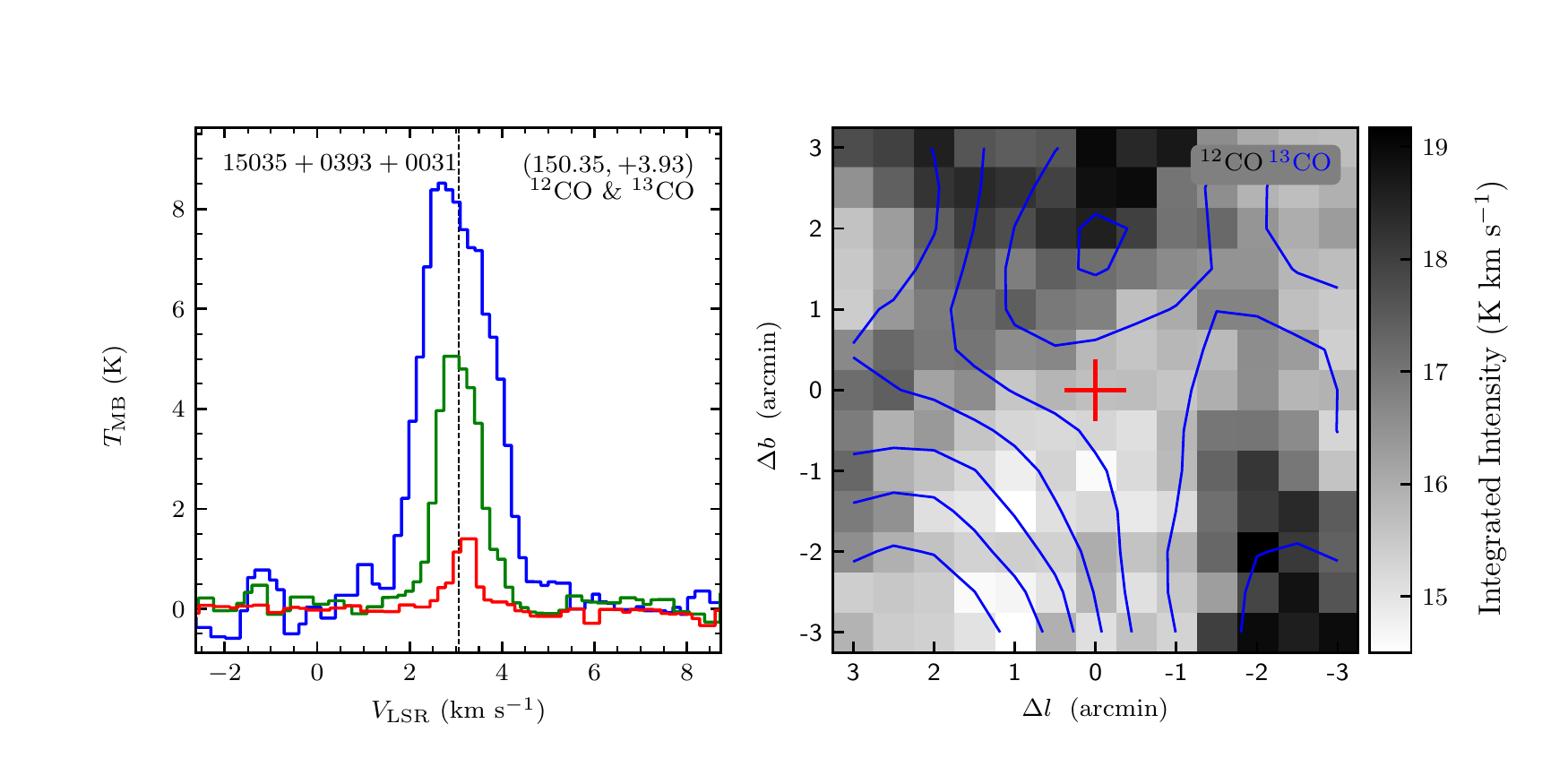}
\includegraphics[width=9.0cm,angle=0]{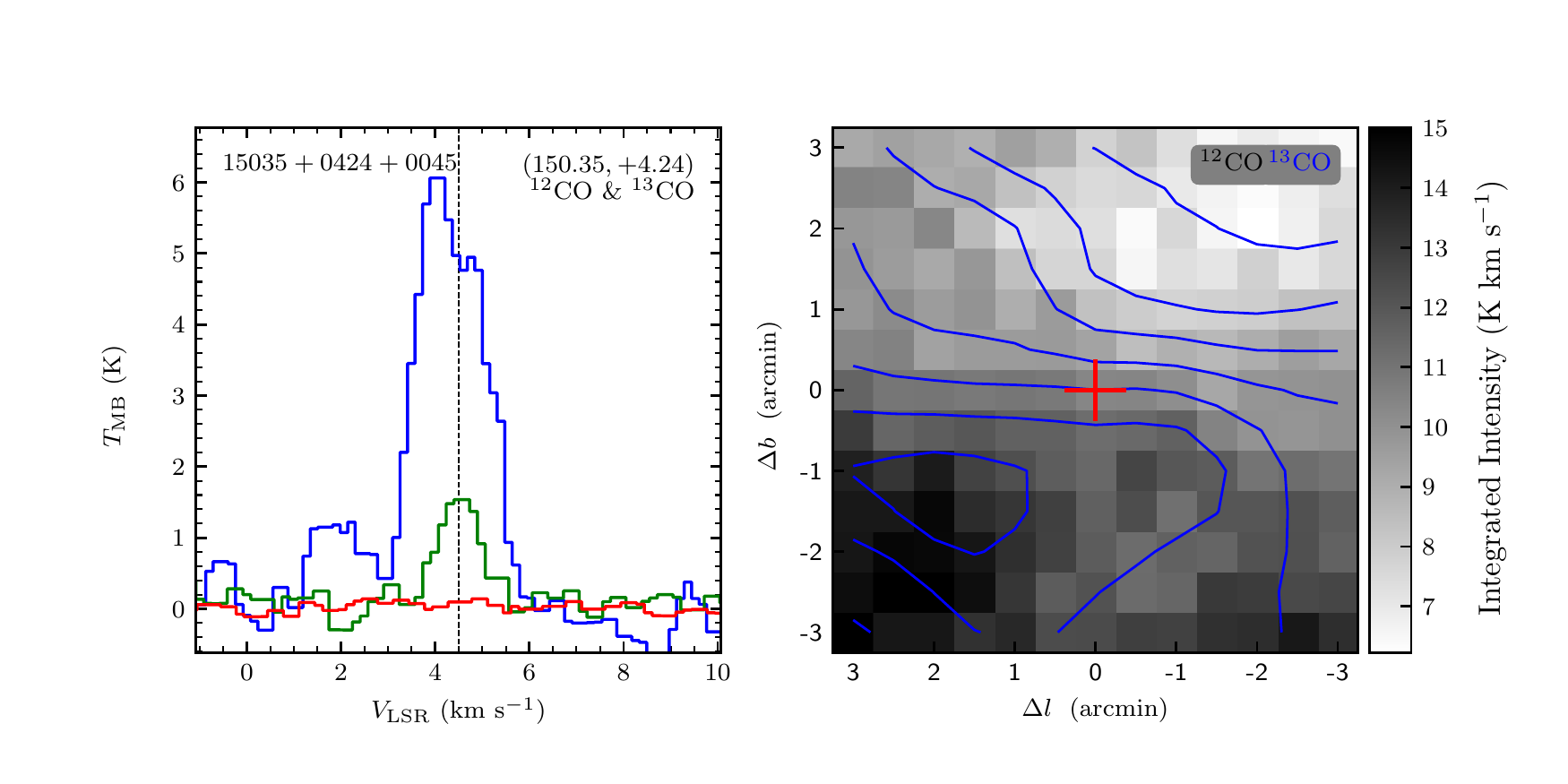}
\end{figure}
\clearpage

\begin{figure}
\includegraphics[width=9.0cm,angle=0]{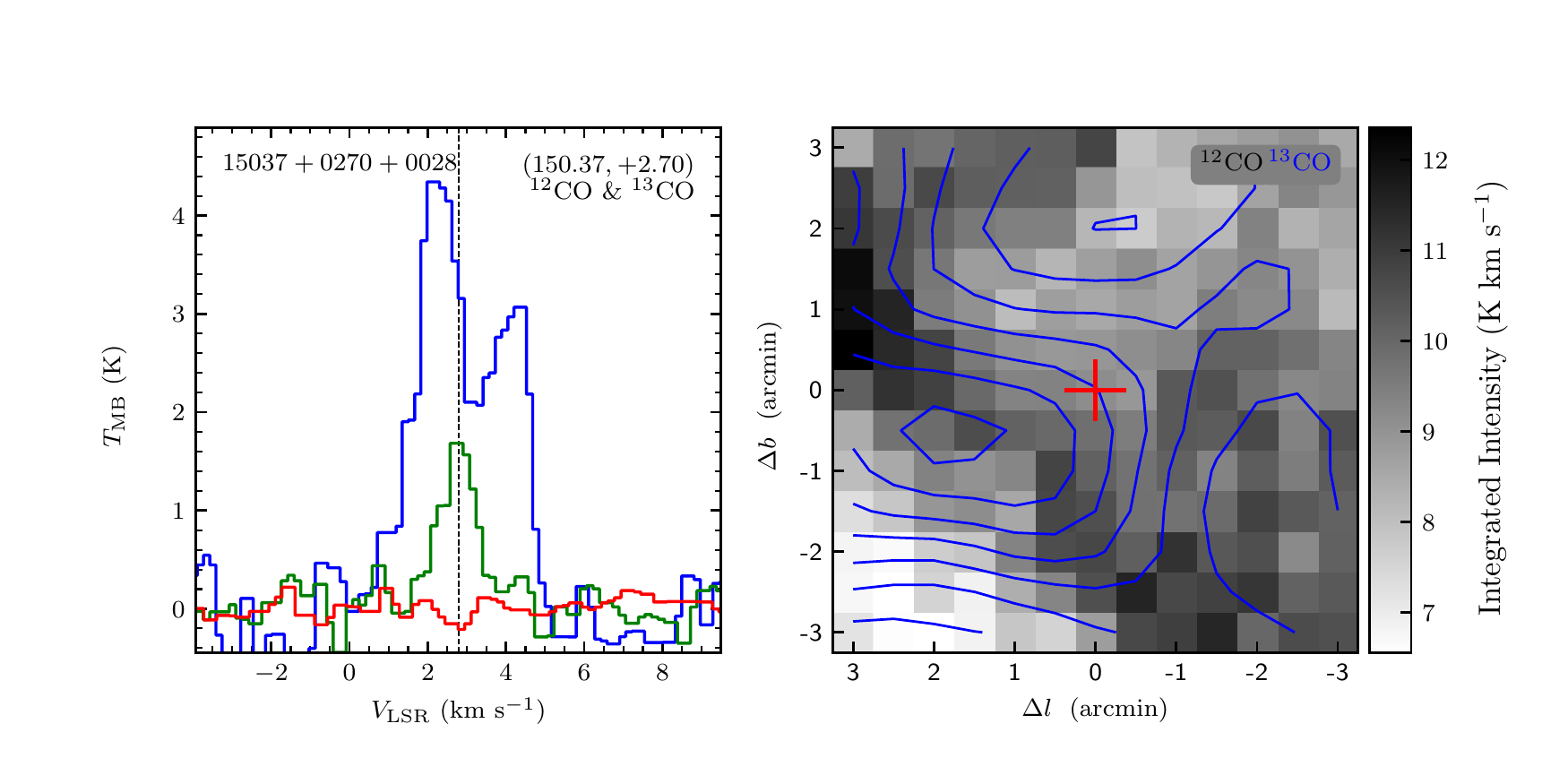}
\includegraphics[width=9.0cm,angle=0]{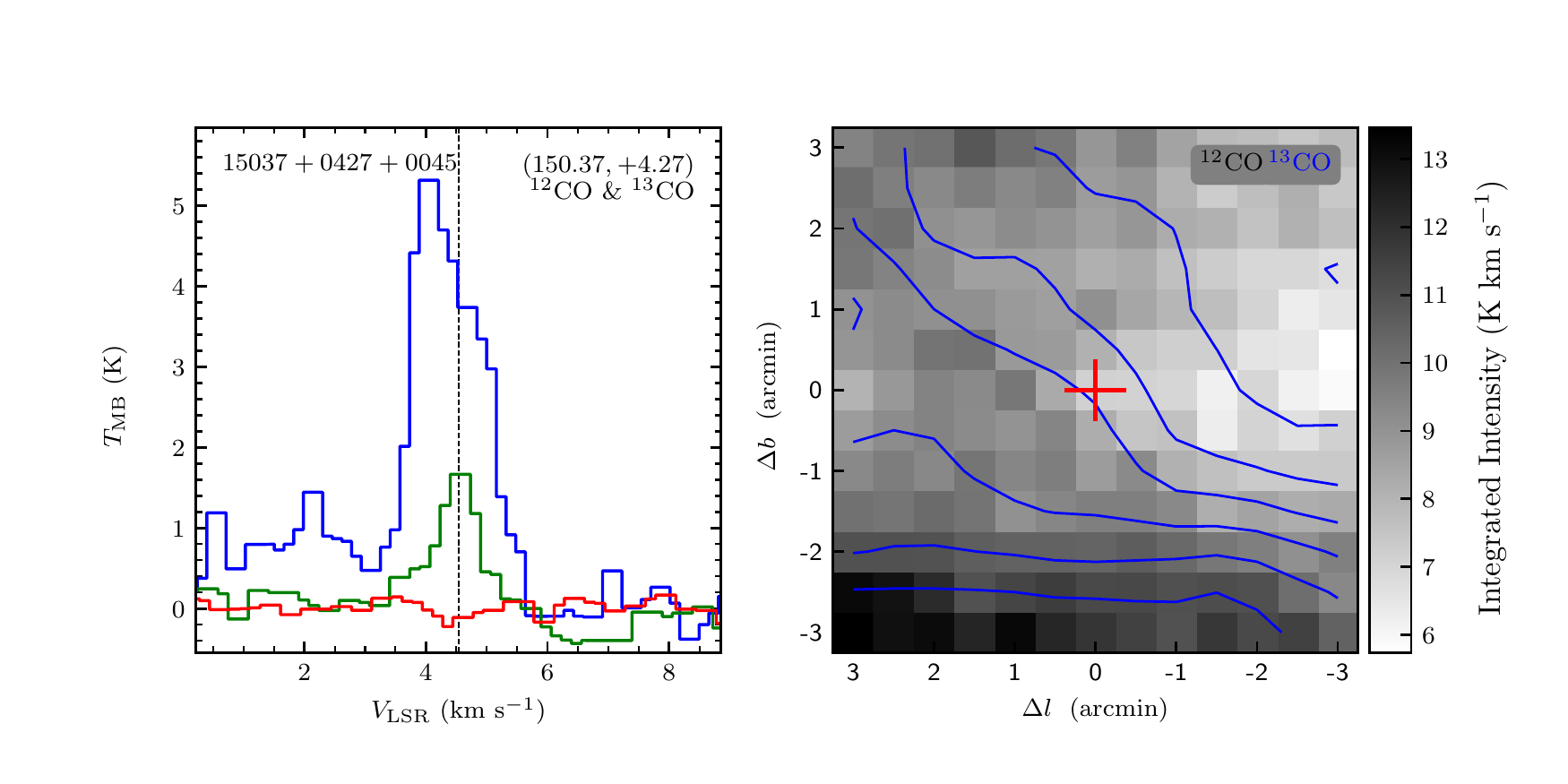}
\vspace{-0.5cm}

\includegraphics[width=9.0cm,angle=0]{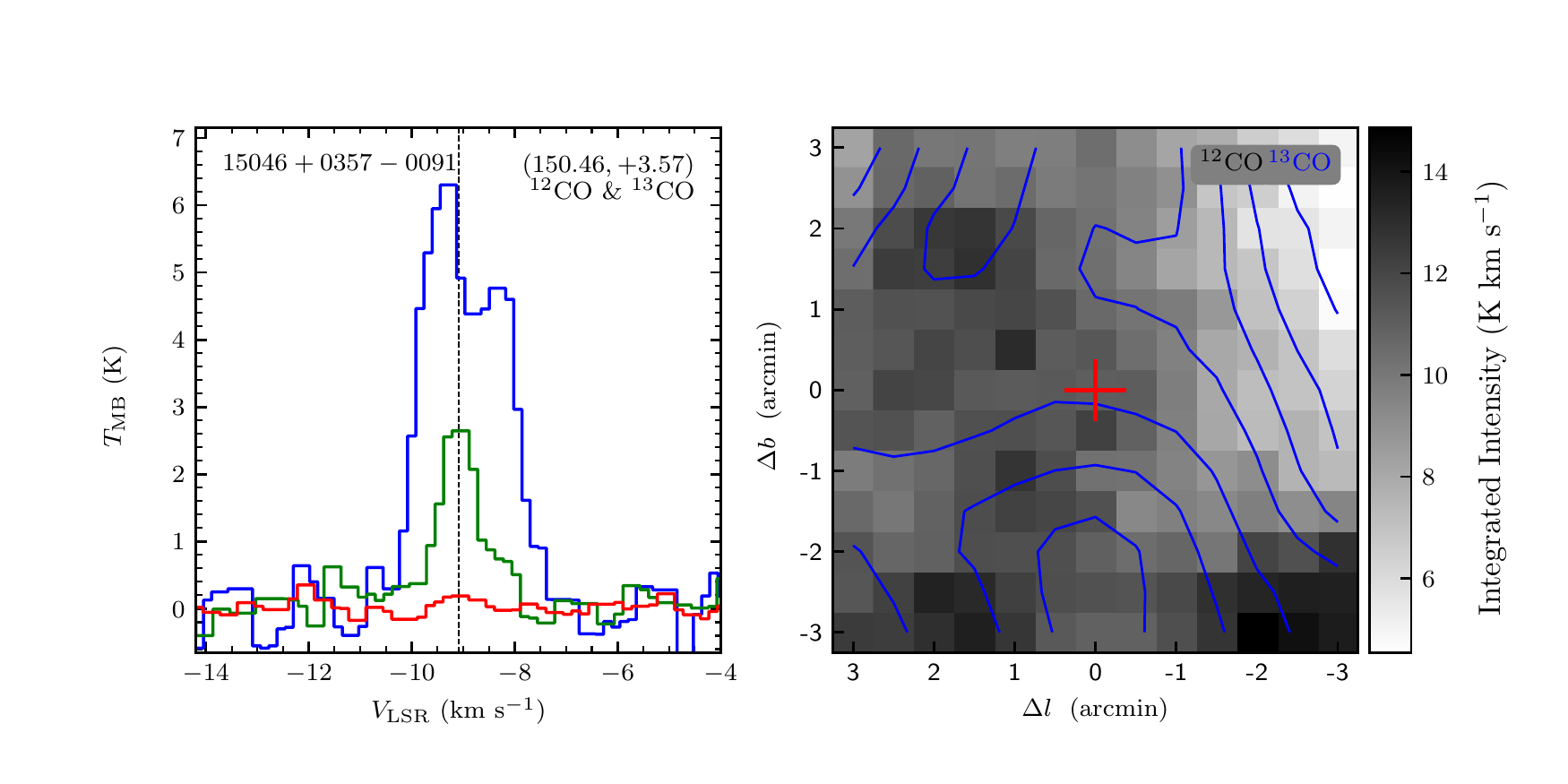}
\includegraphics[width=9.0cm,angle=0]{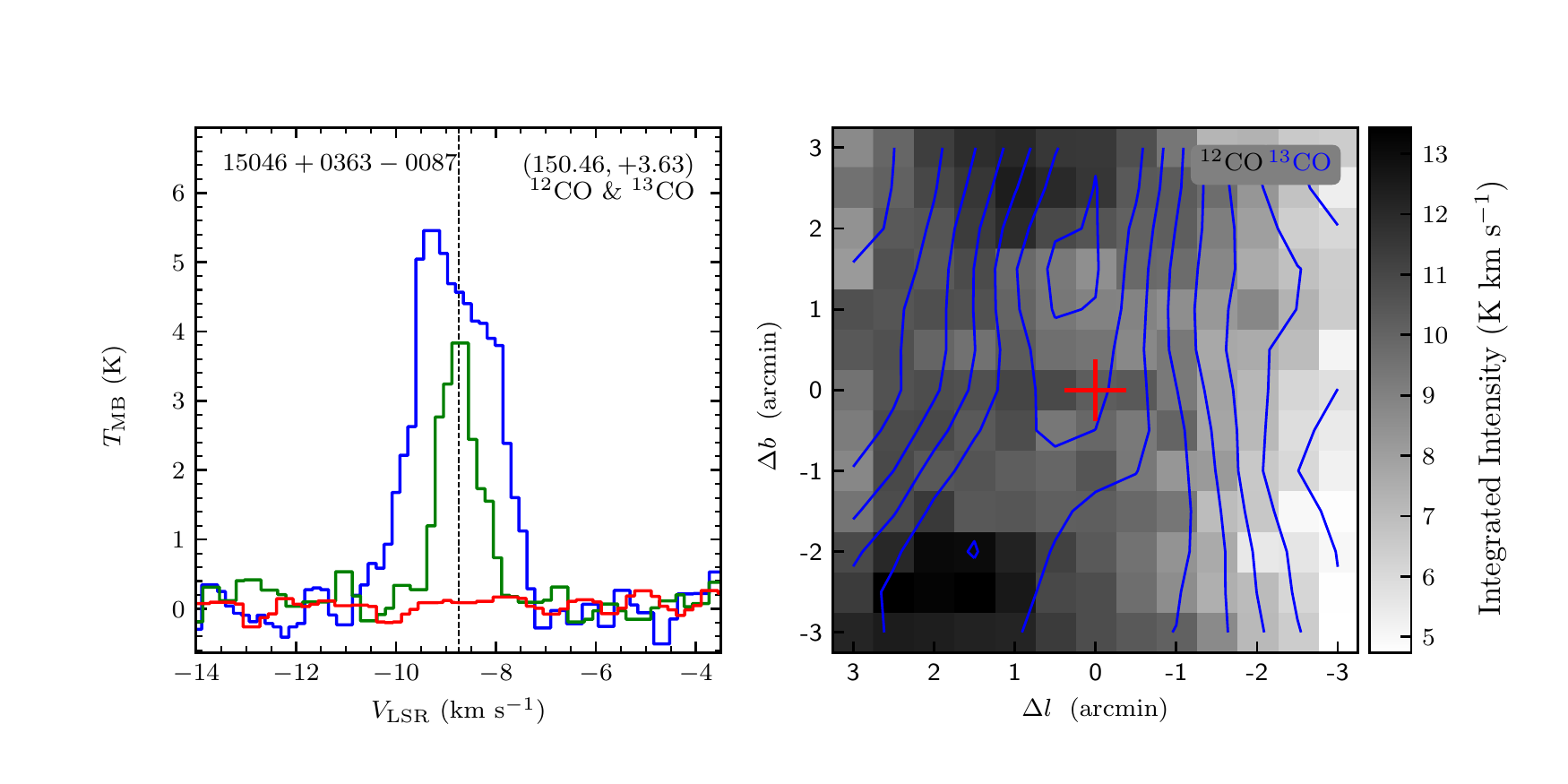}
\vspace{-0.5cm}

\includegraphics[width=9.0cm,angle=0]{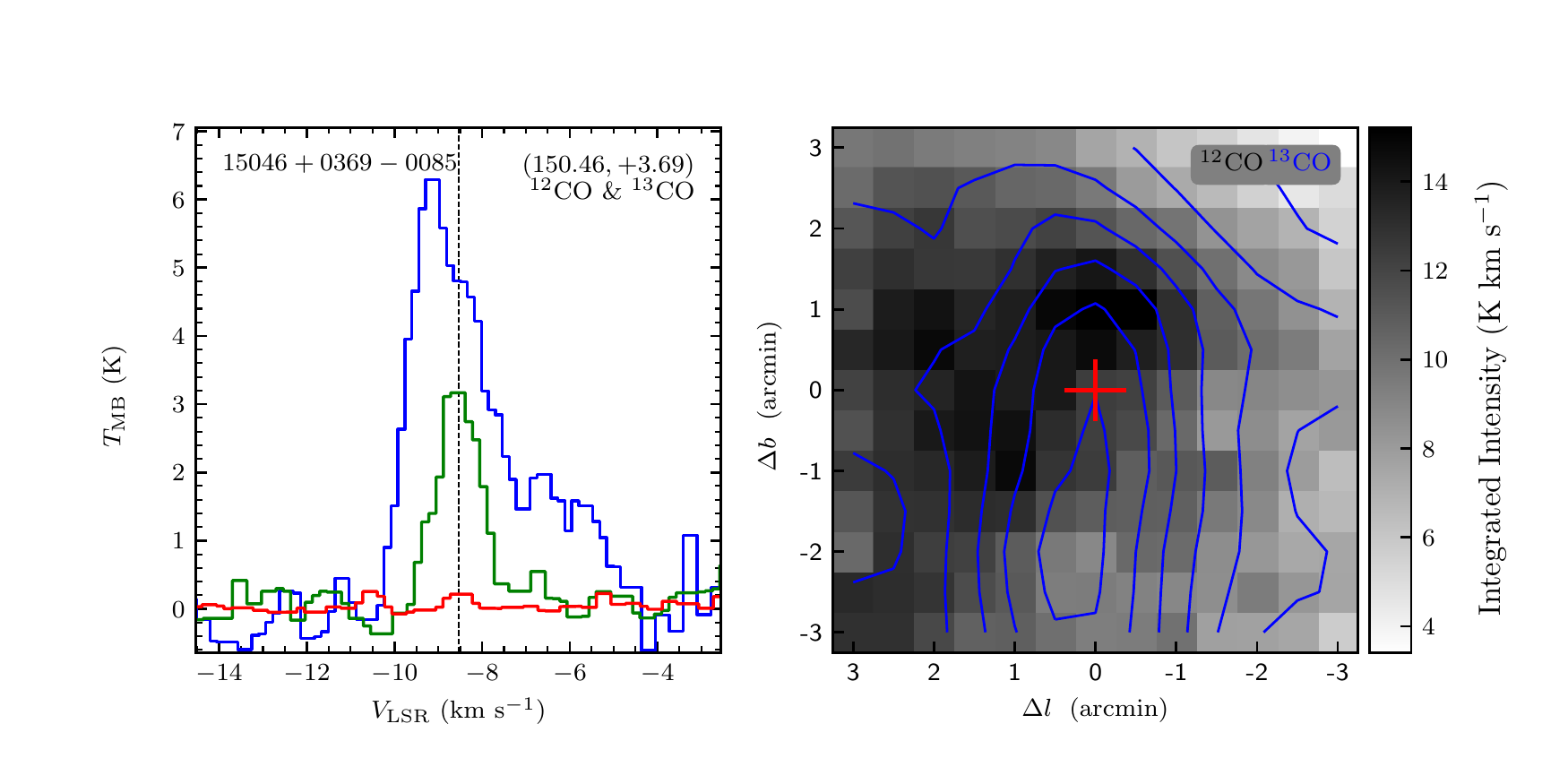}
\includegraphics[width=9.0cm,angle=0]{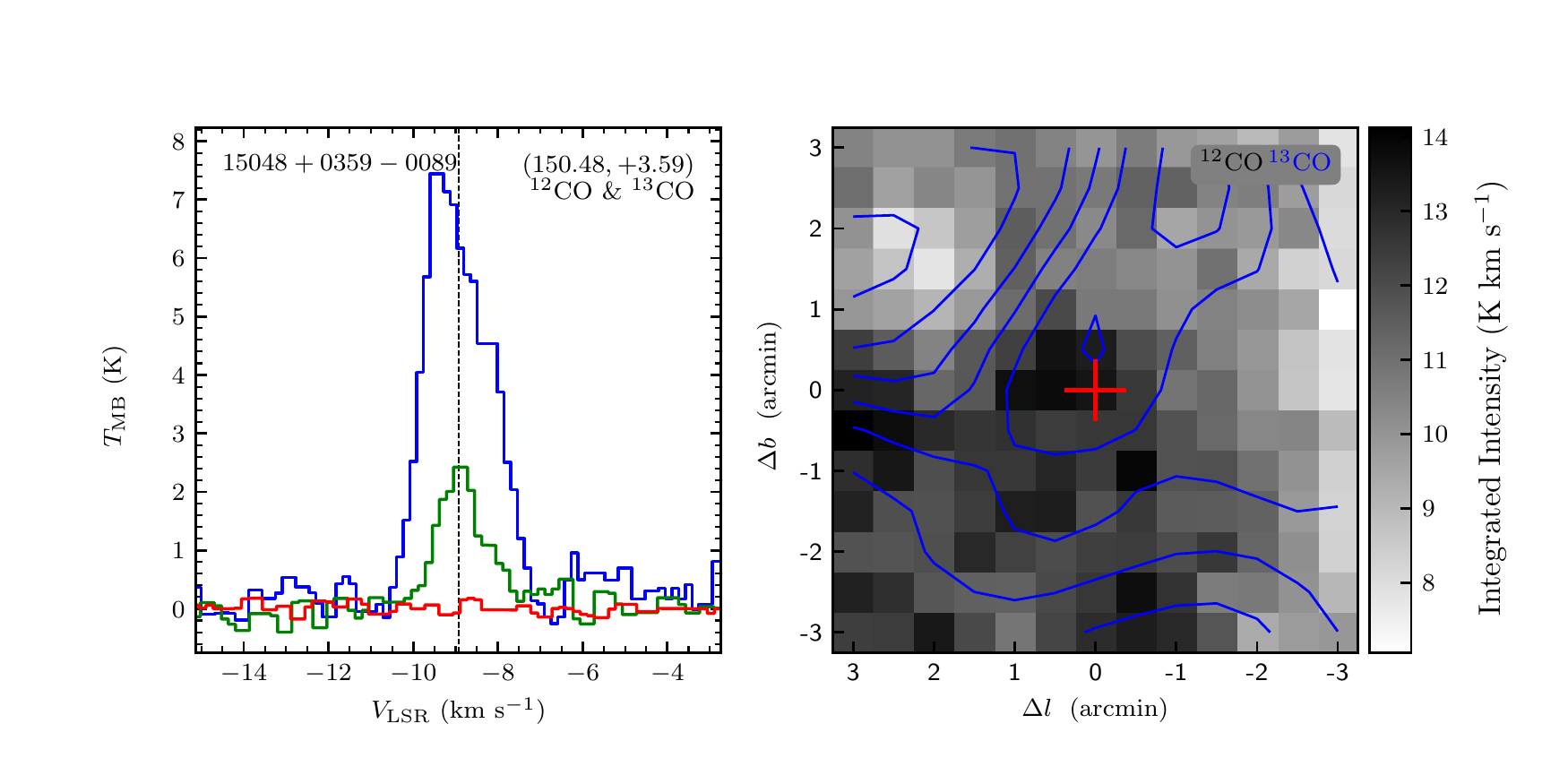}
\vspace{-0.5cm}

\includegraphics[width=9.0cm,angle=0]{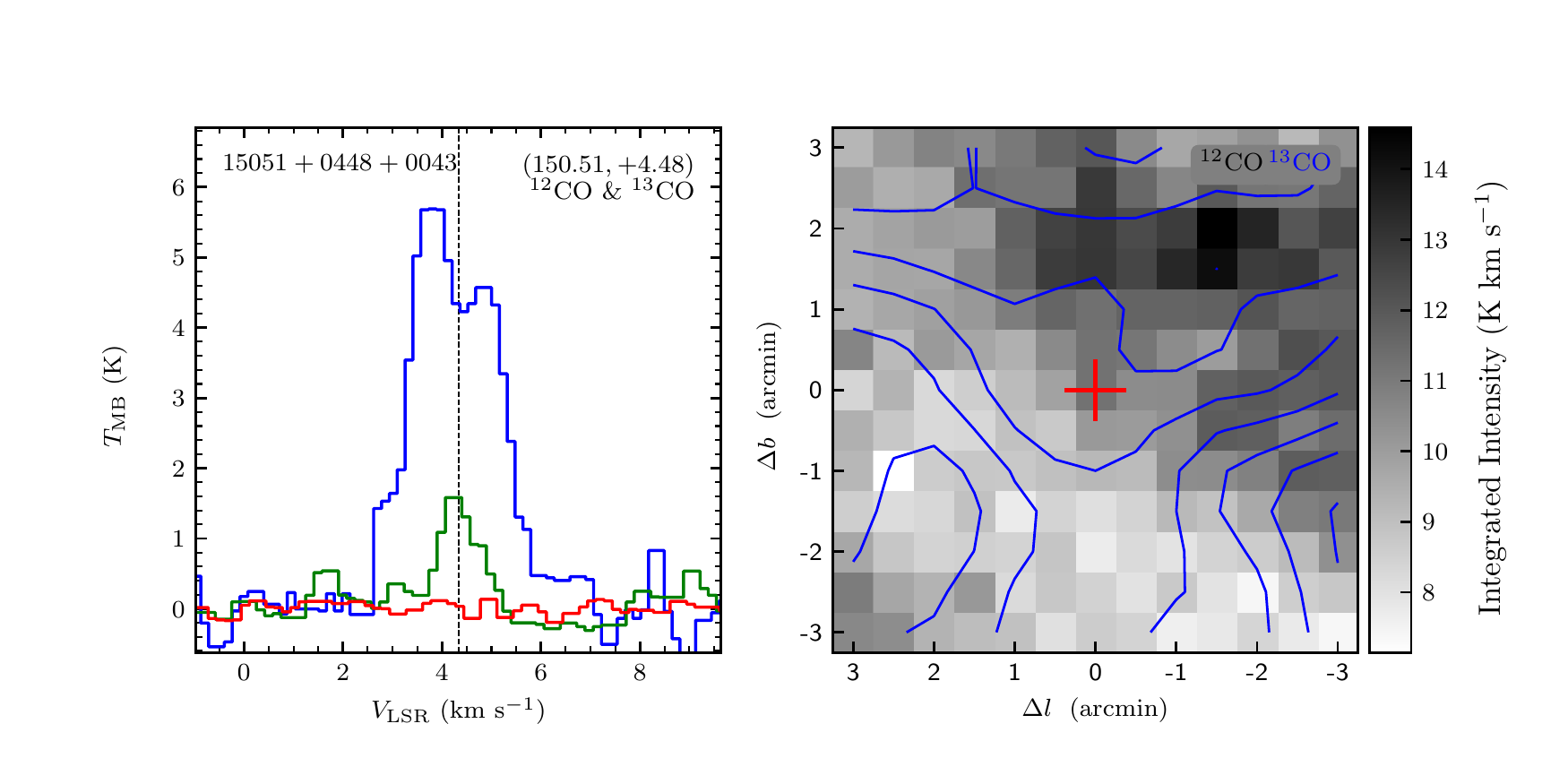}
\includegraphics[width=9.0cm,angle=0]{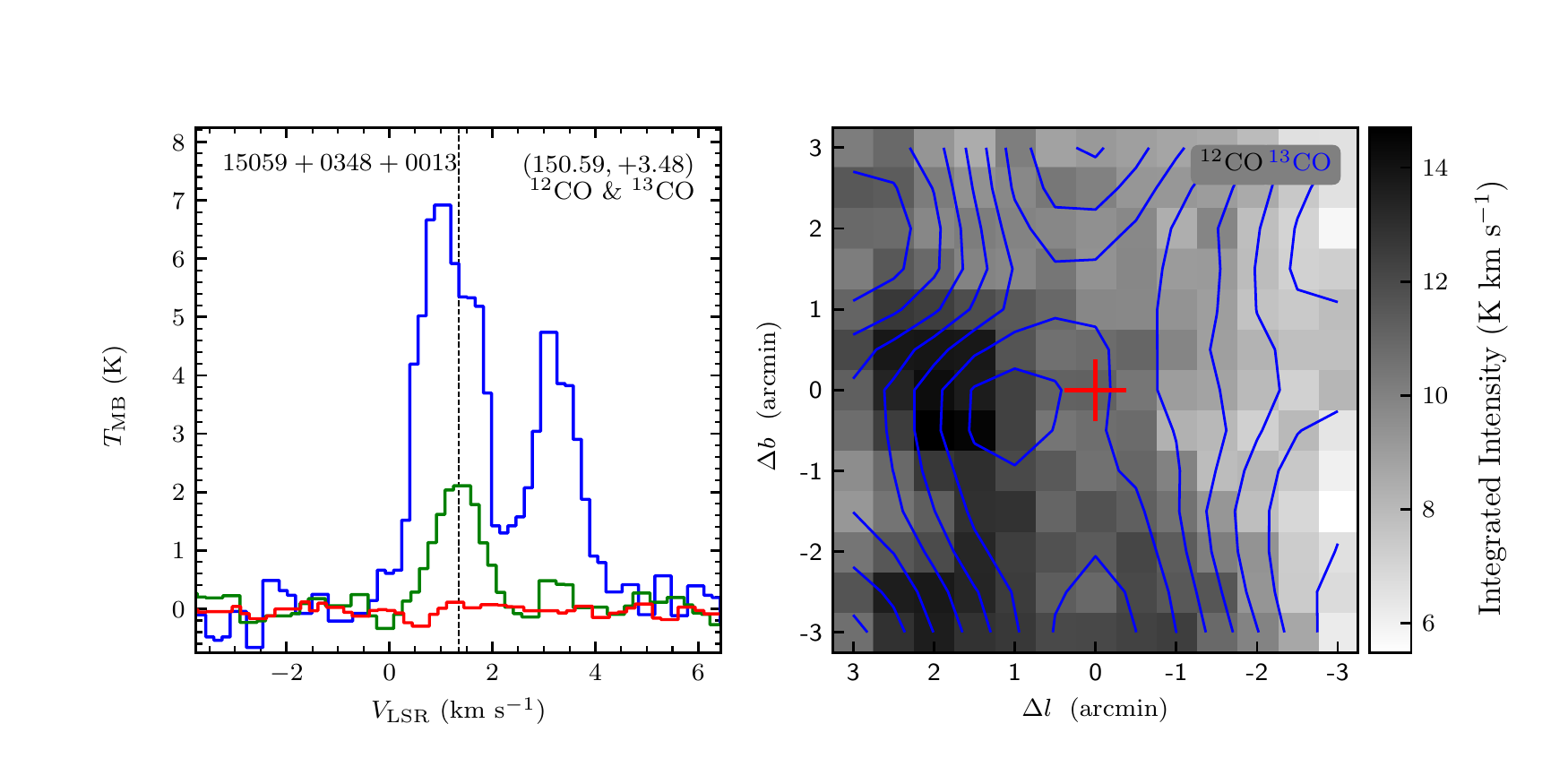}
\vspace{-0.5cm}

\includegraphics[width=9.0cm,angle=0]{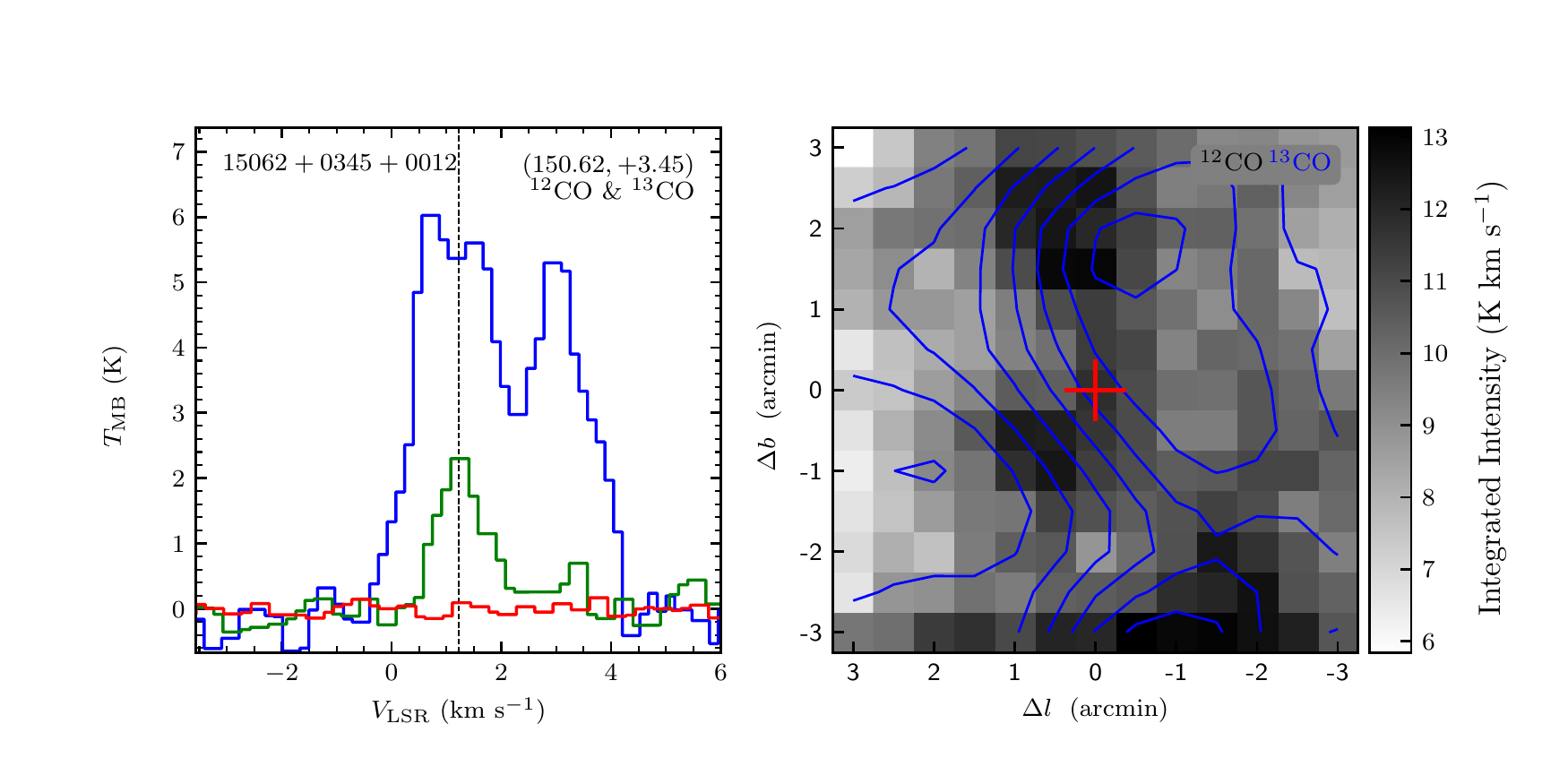}
\includegraphics[width=9.0cm,angle=0]{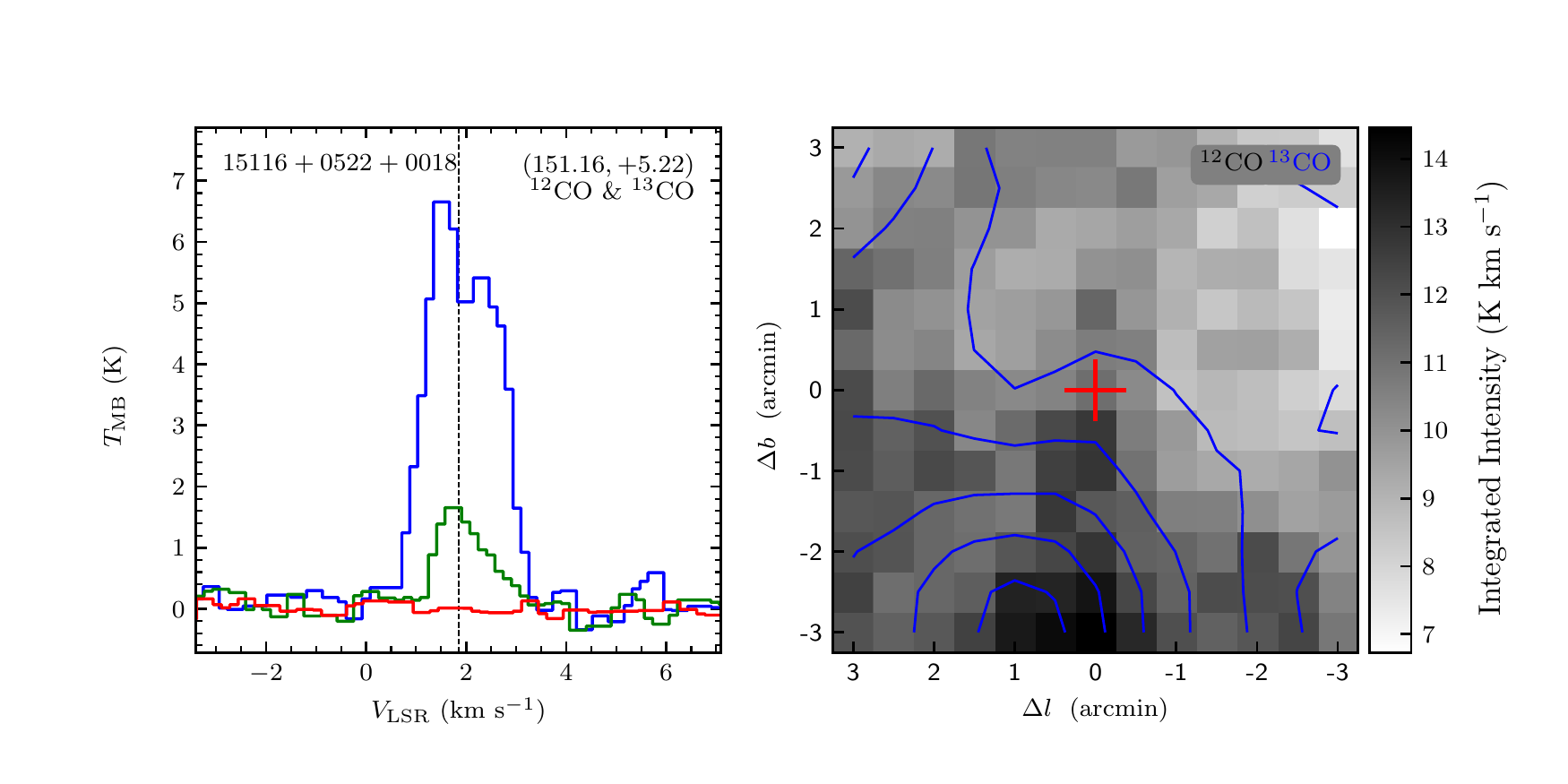}
\end{figure}
\clearpage

\begin{figure}
\includegraphics[width=9.0cm,angle=0]{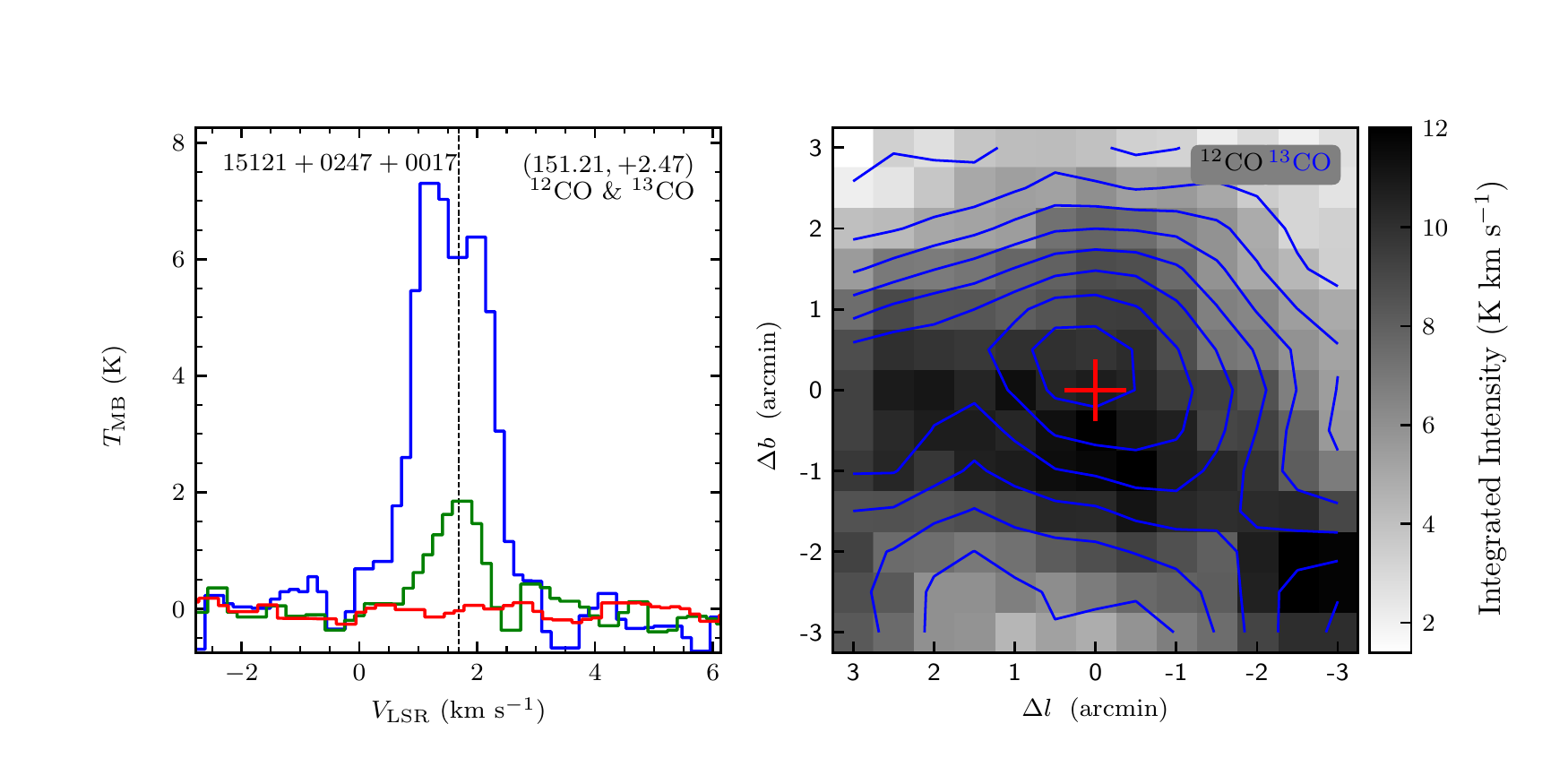}
\includegraphics[width=9.0cm,angle=0]{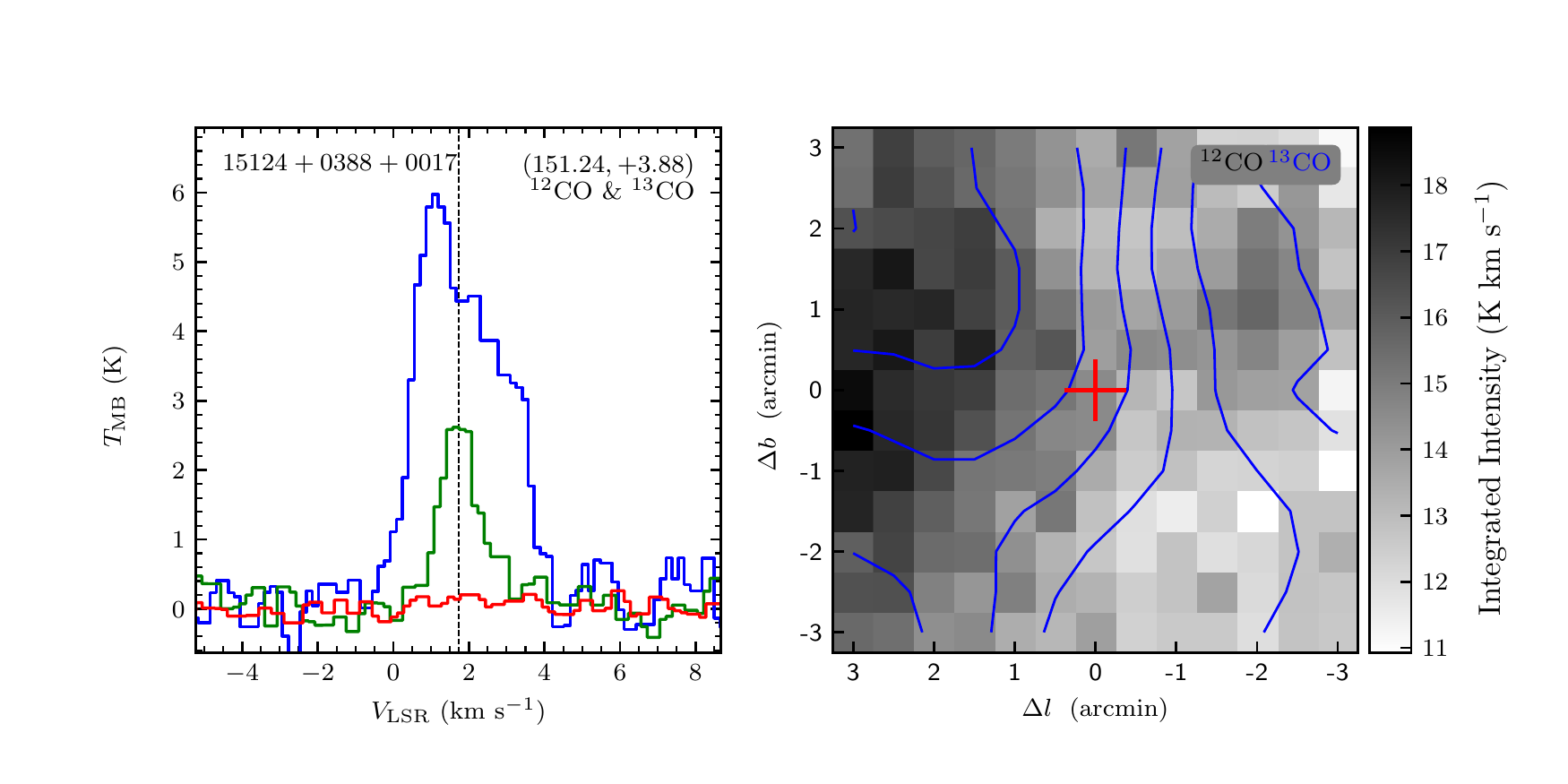}
\vspace{-0.5cm}

\includegraphics[width=9.0cm,angle=0]{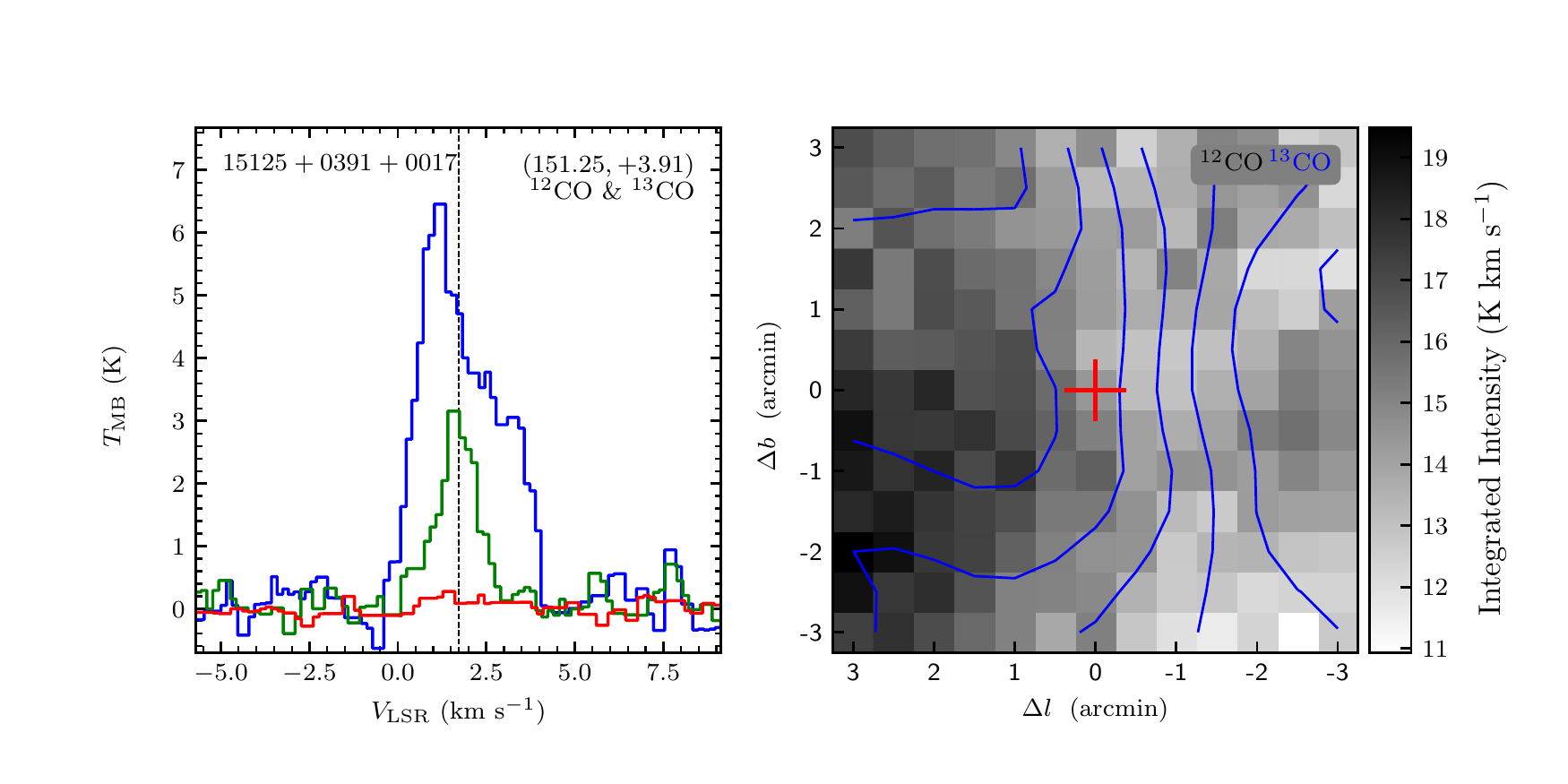}
\includegraphics[width=9.0cm,angle=0]{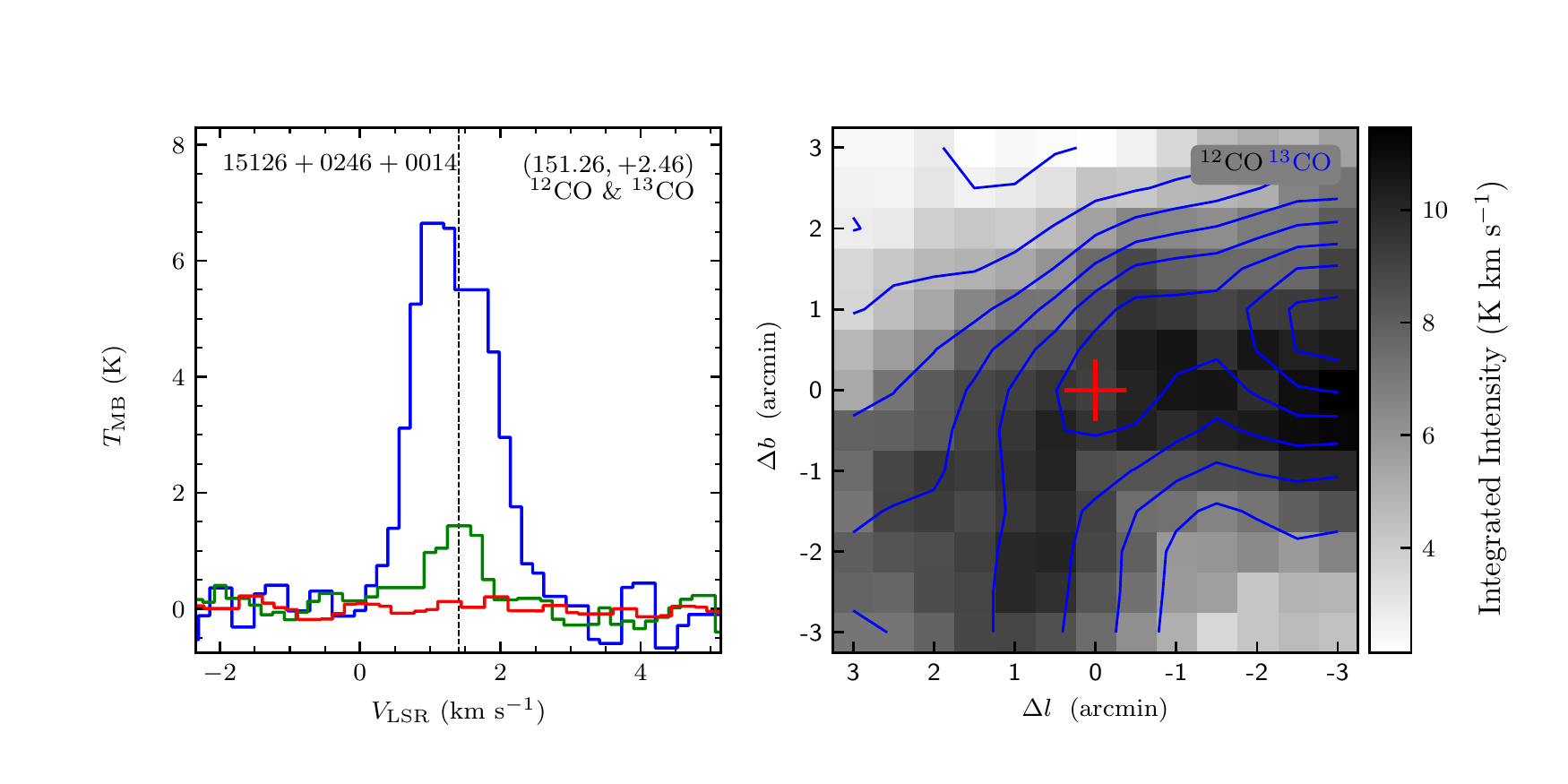}
\vspace{-0.5cm}

\includegraphics[width=9.0cm,angle=0]{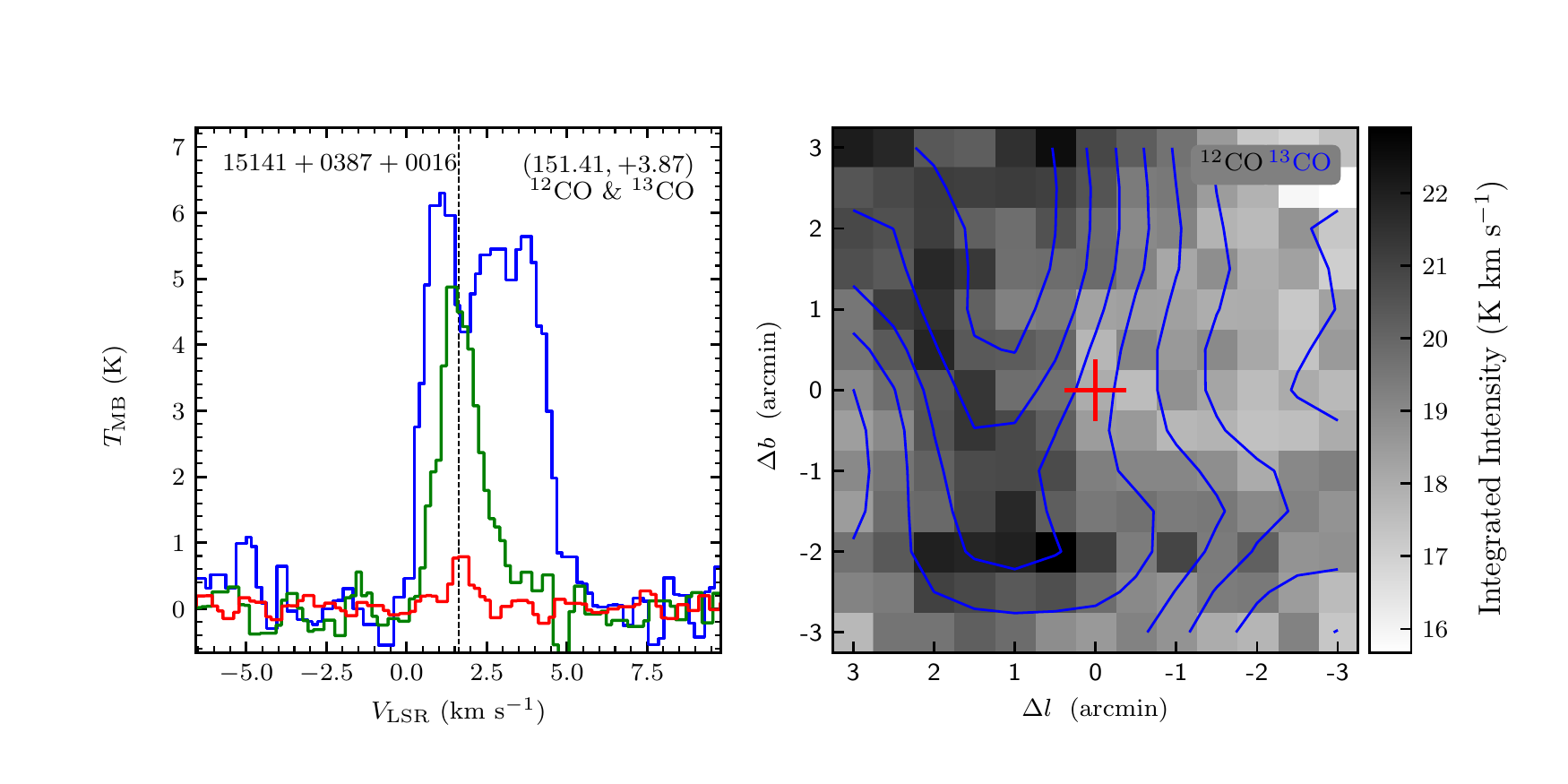}
\includegraphics[width=9.0cm,angle=0]{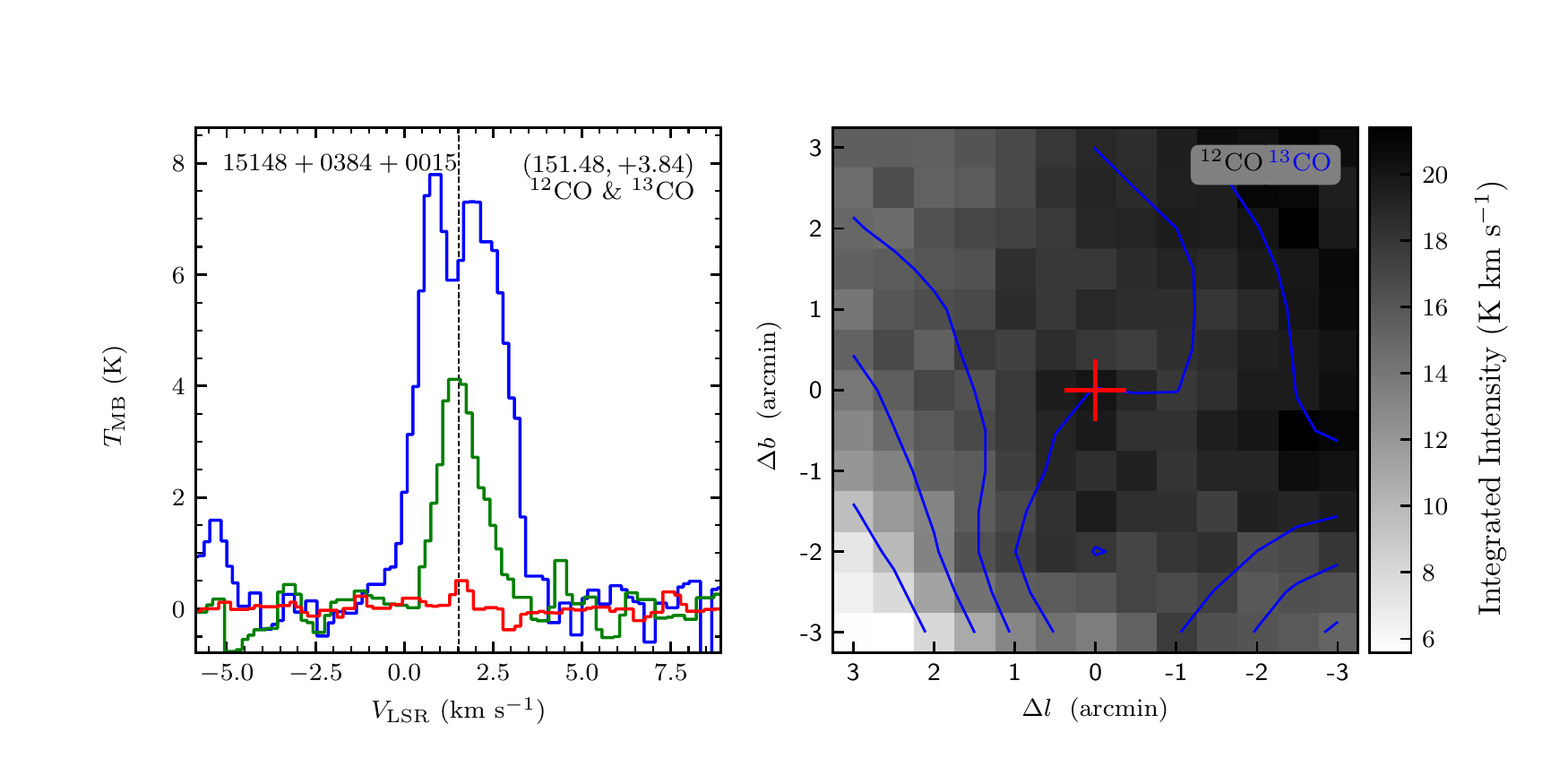}
\vspace{-0.5cm}

\includegraphics[width=9.0cm,angle=0]{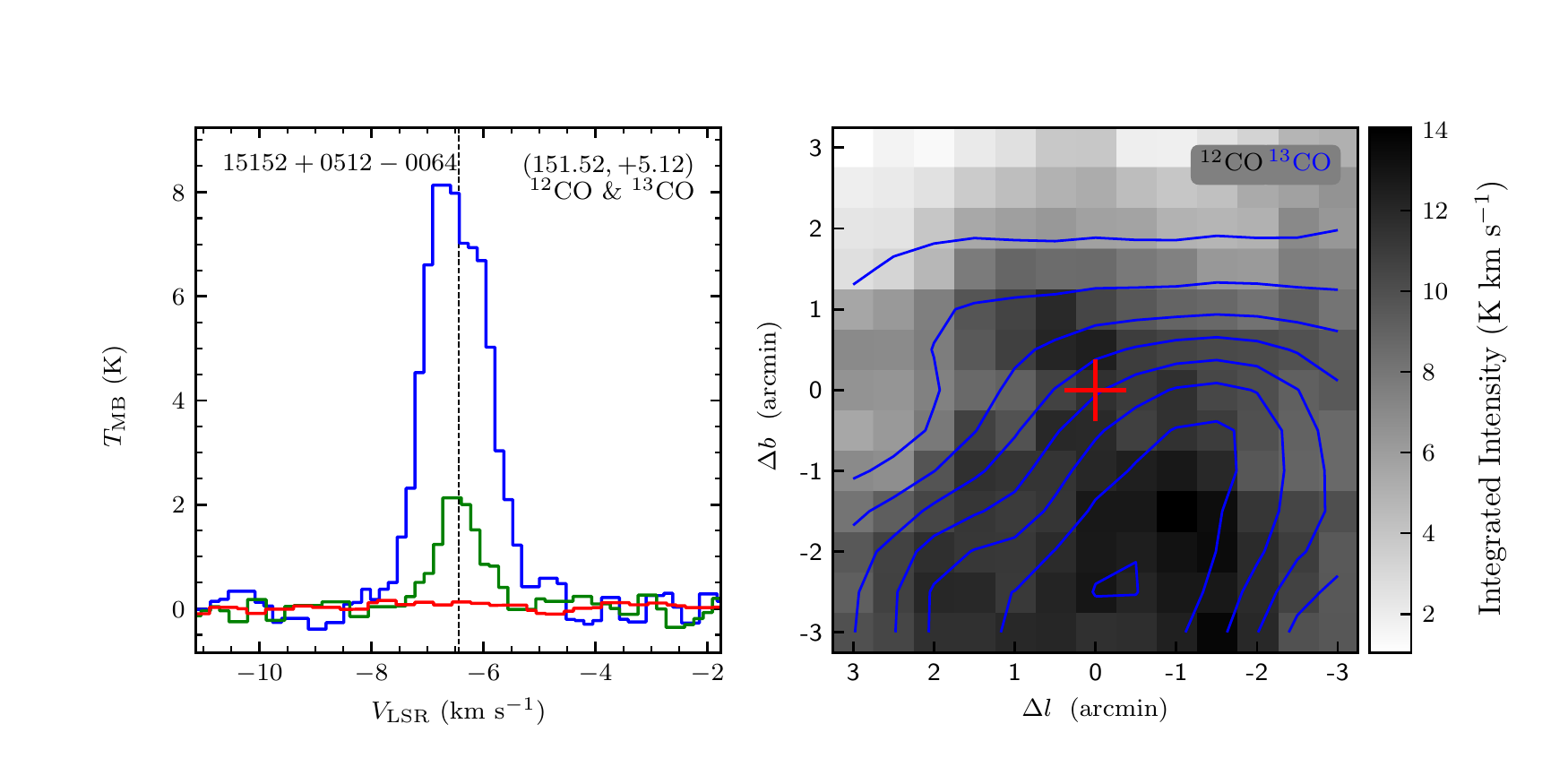}
\includegraphics[width=9.0cm,angle=0]{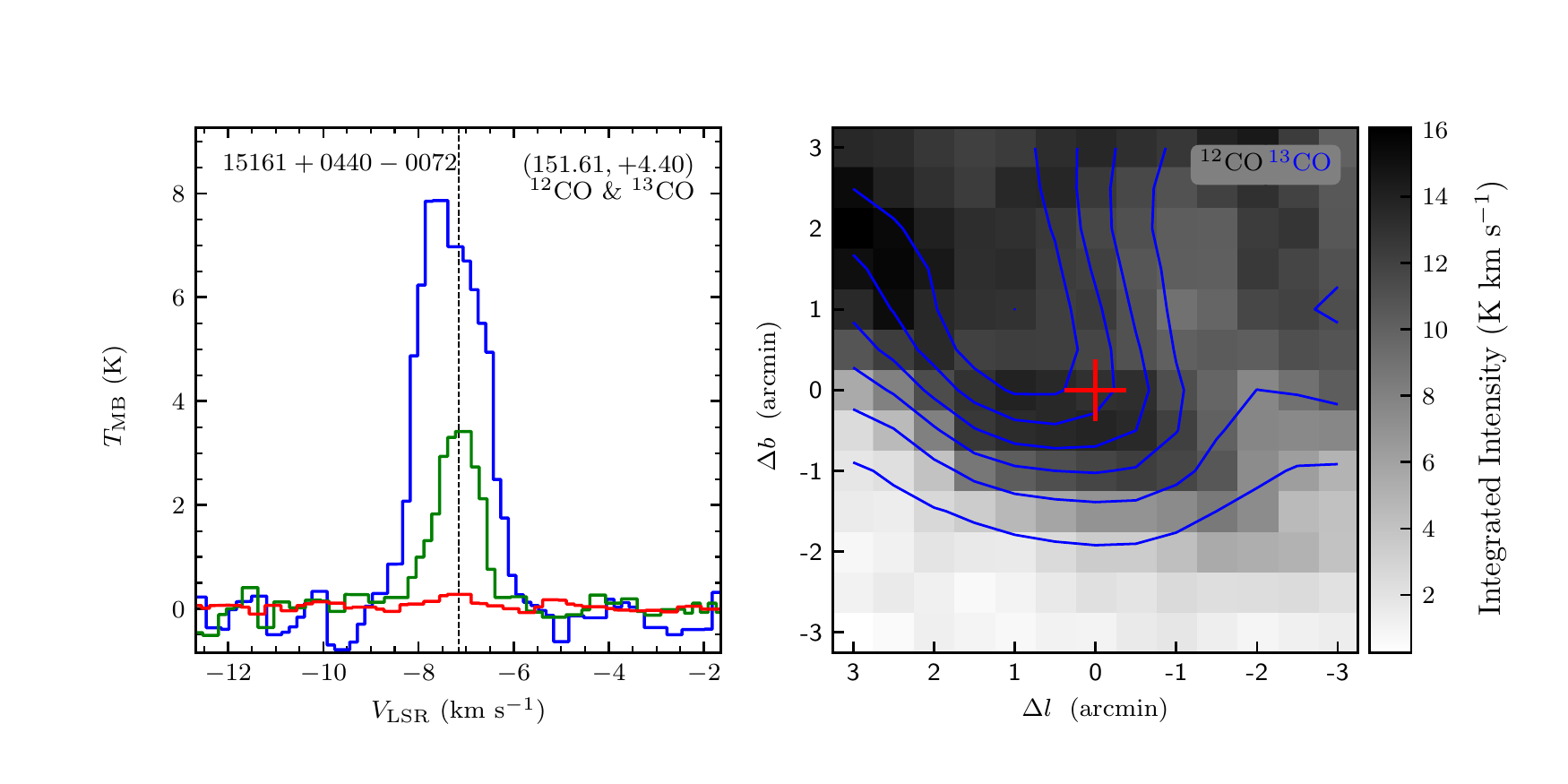}
\vspace{-0.5cm}

\includegraphics[width=9.0cm,angle=0]{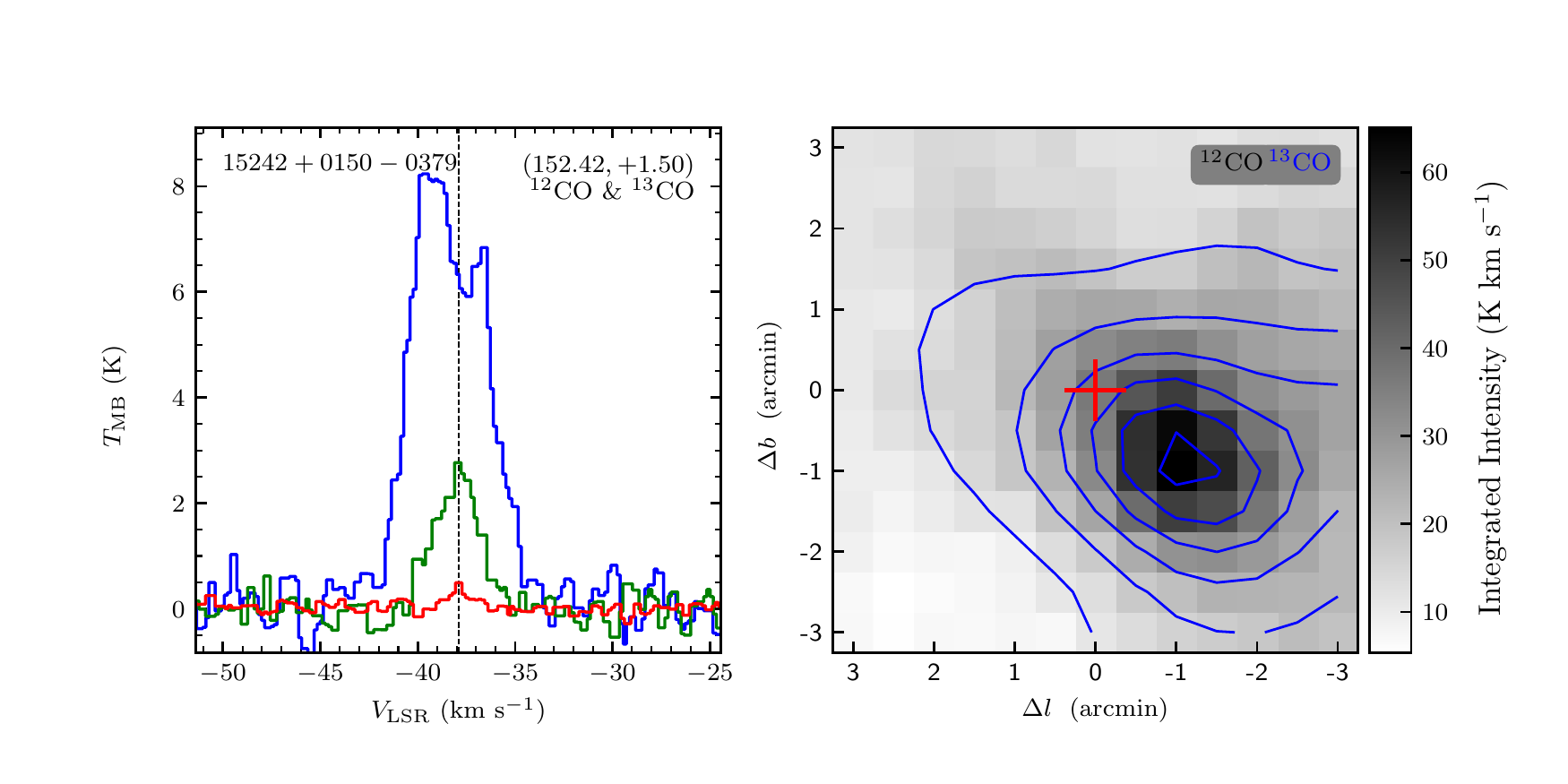}
\includegraphics[width=9.0cm,angle=0]{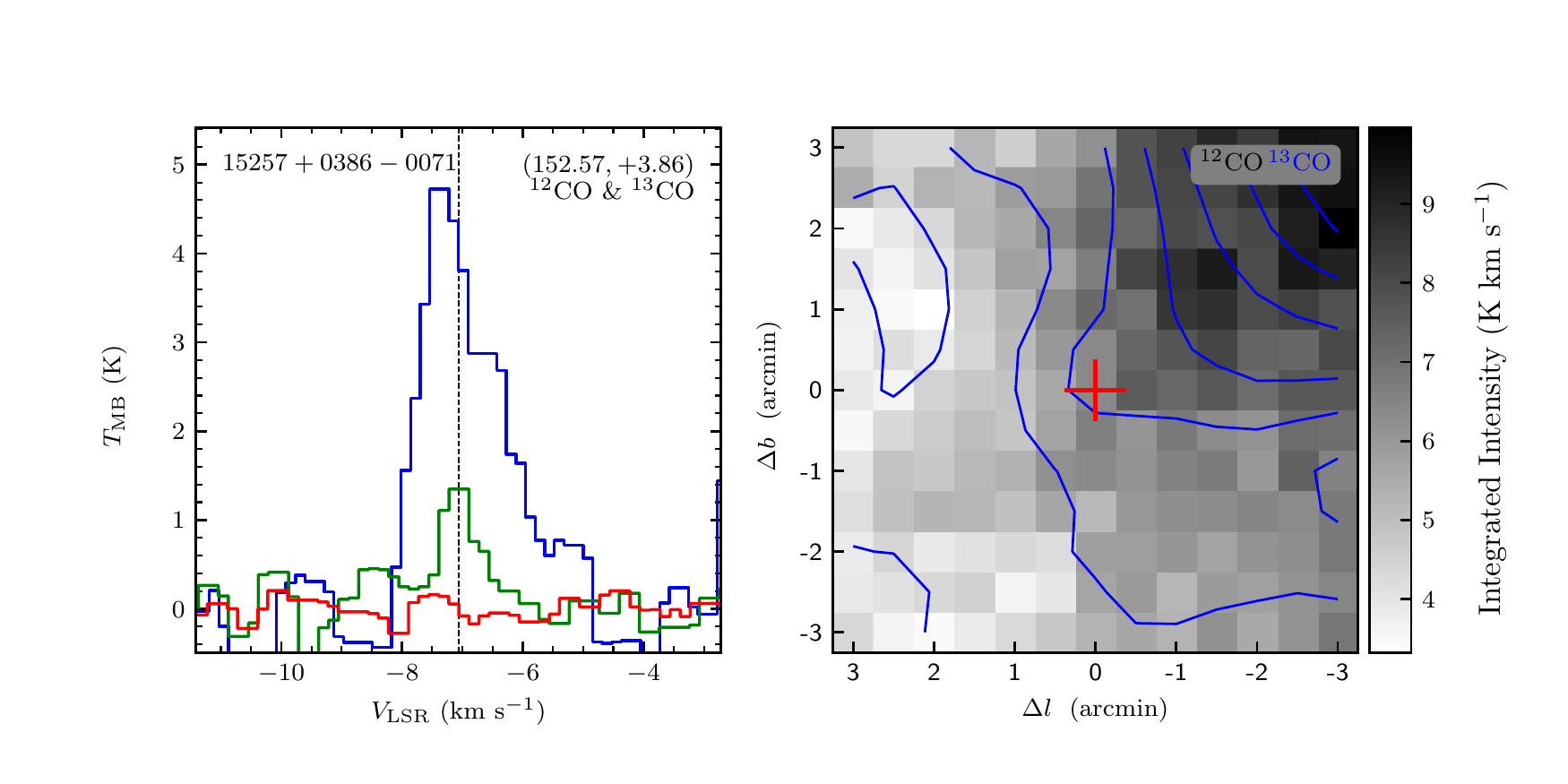}
\end{figure}
\clearpage

\begin{figure}
\includegraphics[width=9.0cm,angle=0]{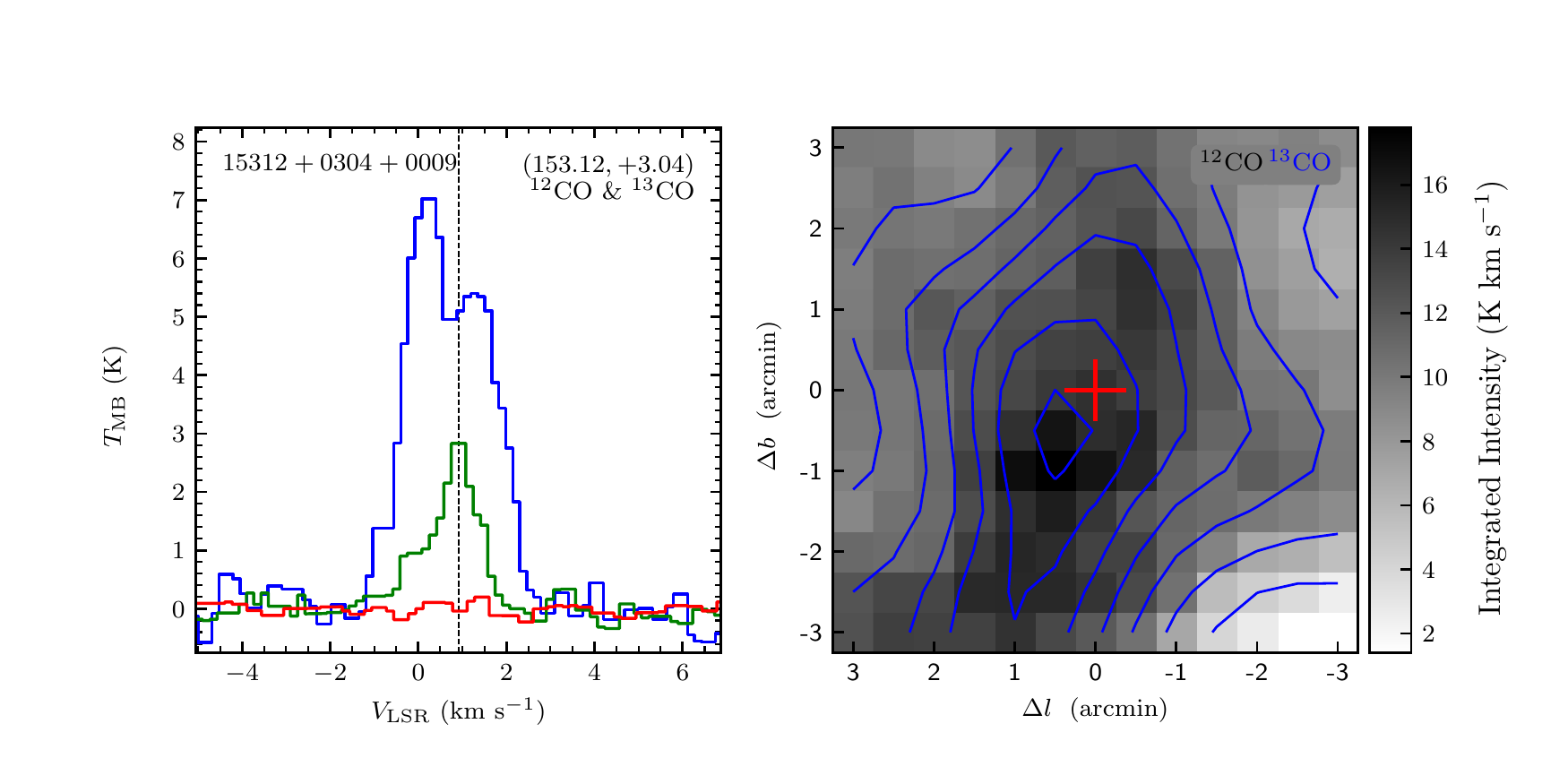}
\includegraphics[width=9.0cm,angle=0]{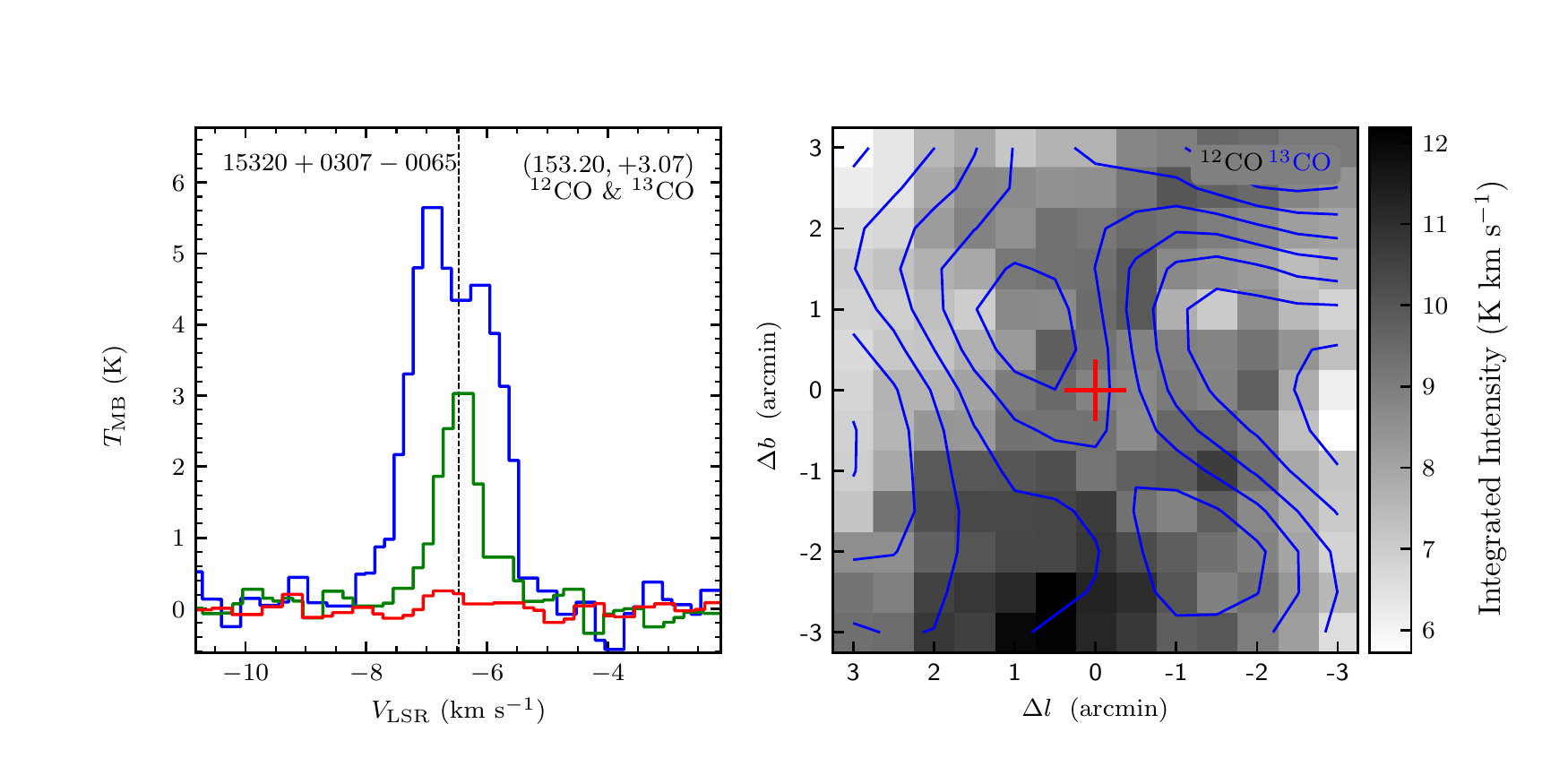}
\vspace{-0.5cm}

\includegraphics[width=9.0cm,angle=0]{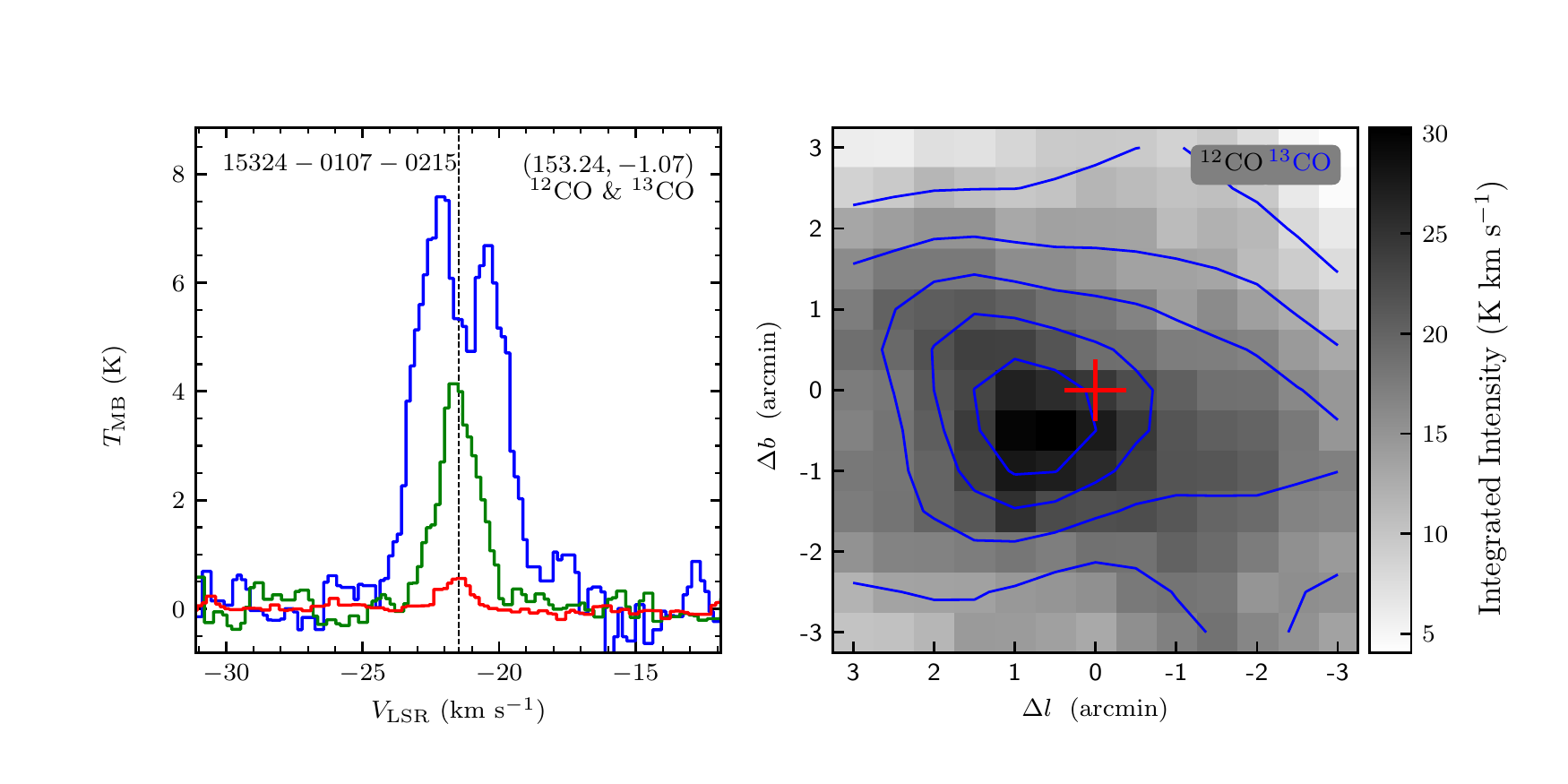}
\includegraphics[width=9.0cm,angle=0]{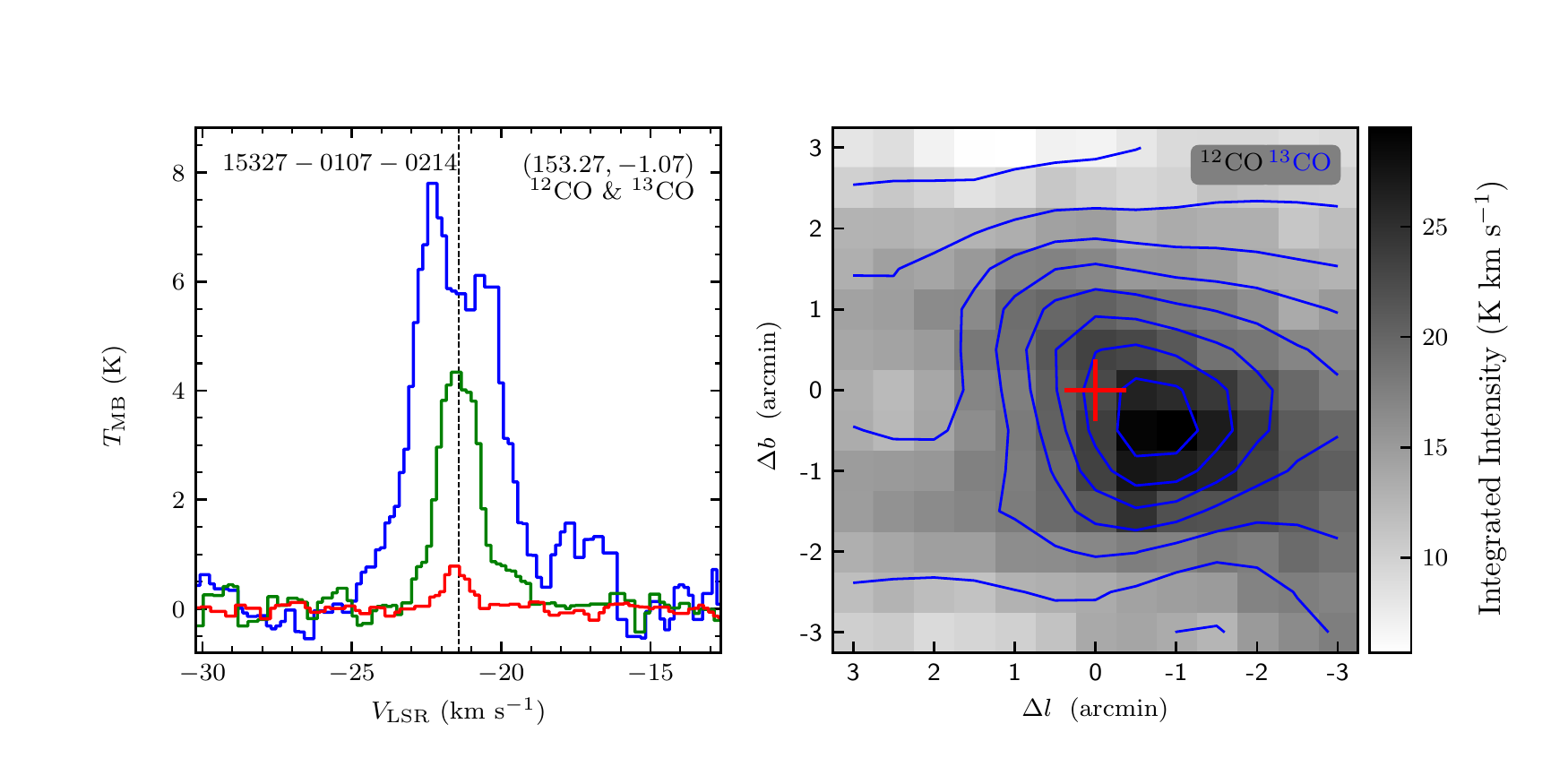}
\vspace{-0.5cm}

\includegraphics[width=9.0cm,angle=0]{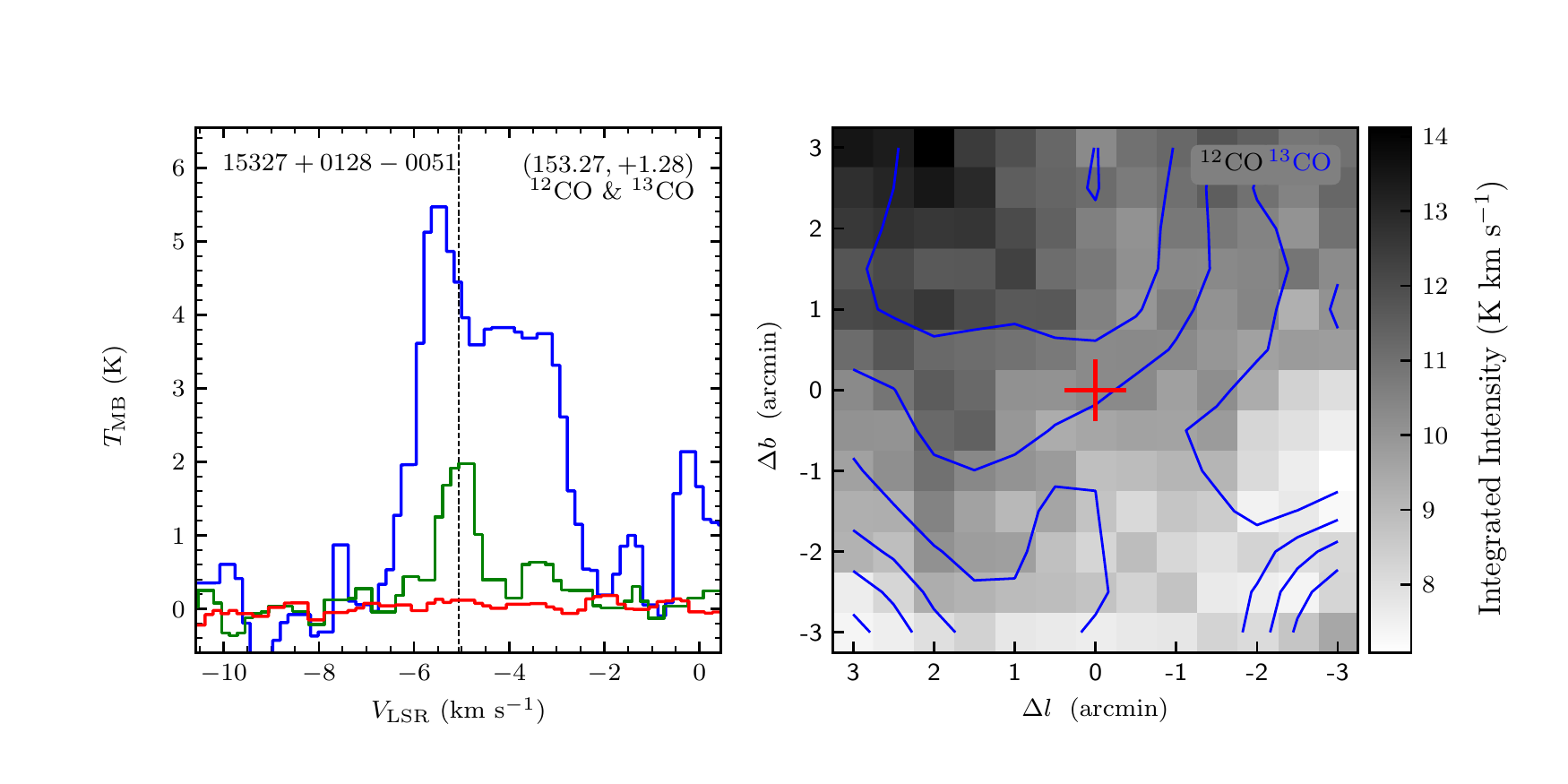}
\includegraphics[width=9.0cm,angle=0]{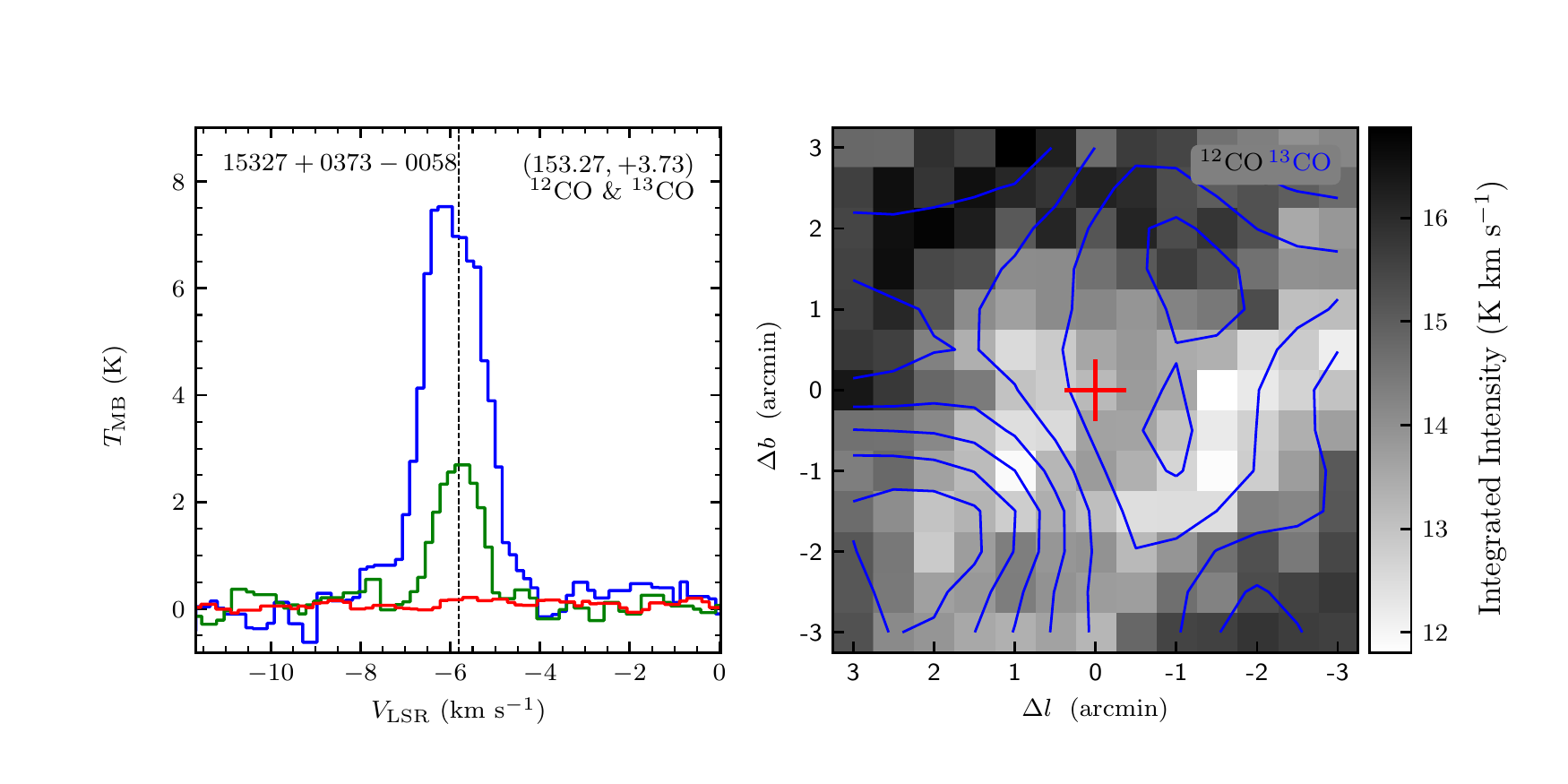}
\vspace{-0.5cm}

\includegraphics[width=9.0cm,angle=0]{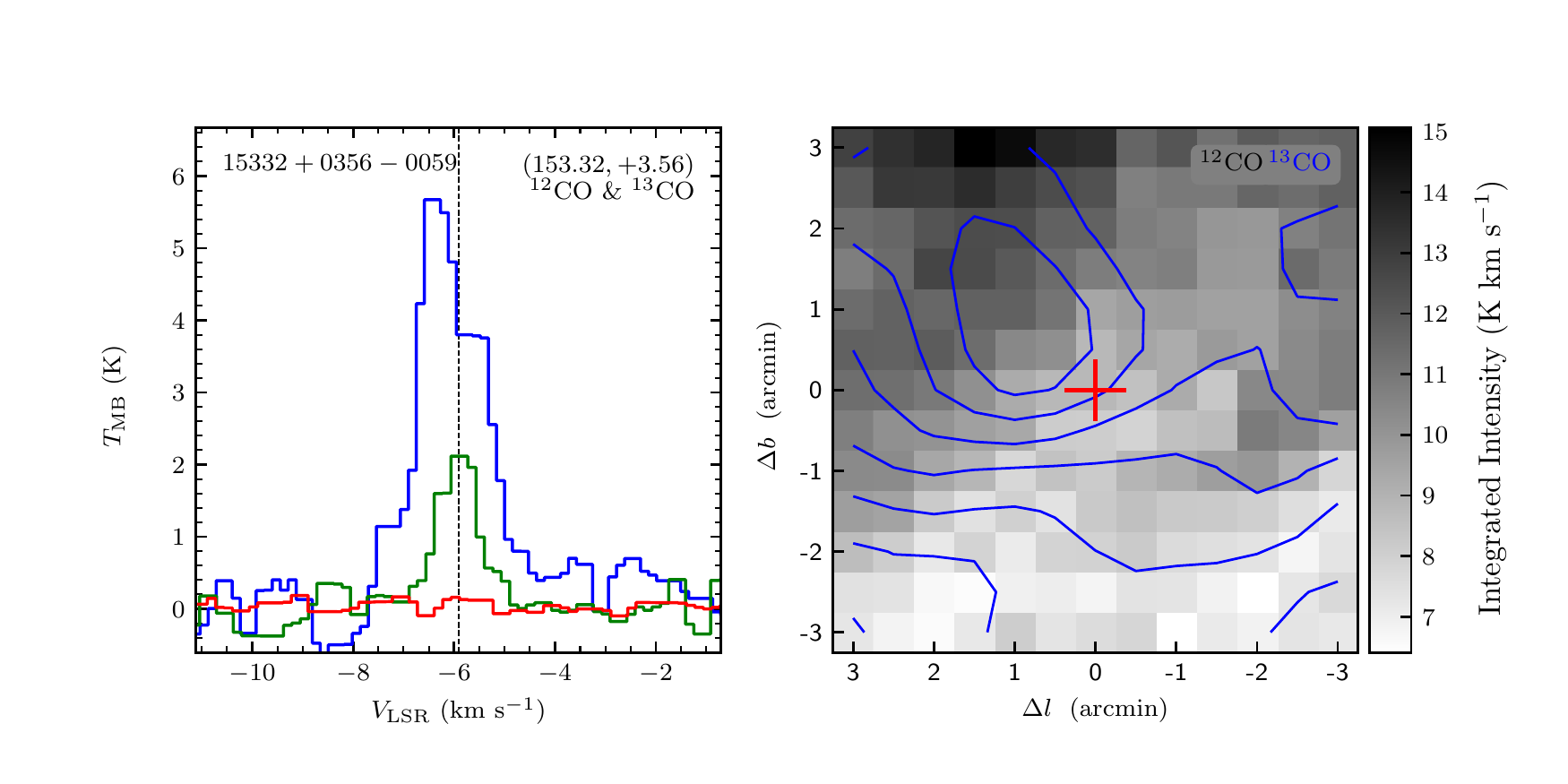}
\includegraphics[width=9.0cm,angle=0]{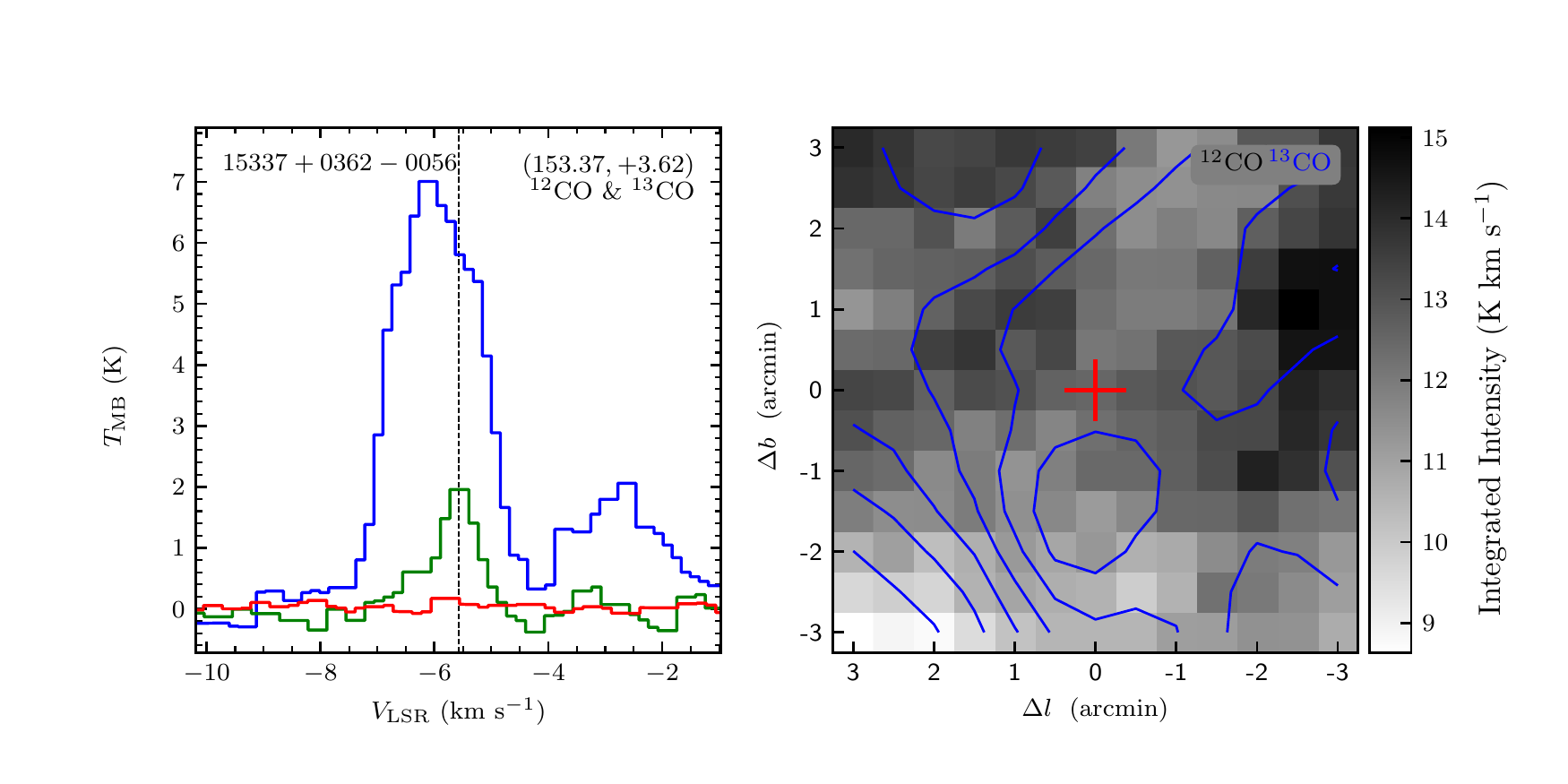}
\vspace{-0.5cm}

\includegraphics[width=9.0cm,angle=0]{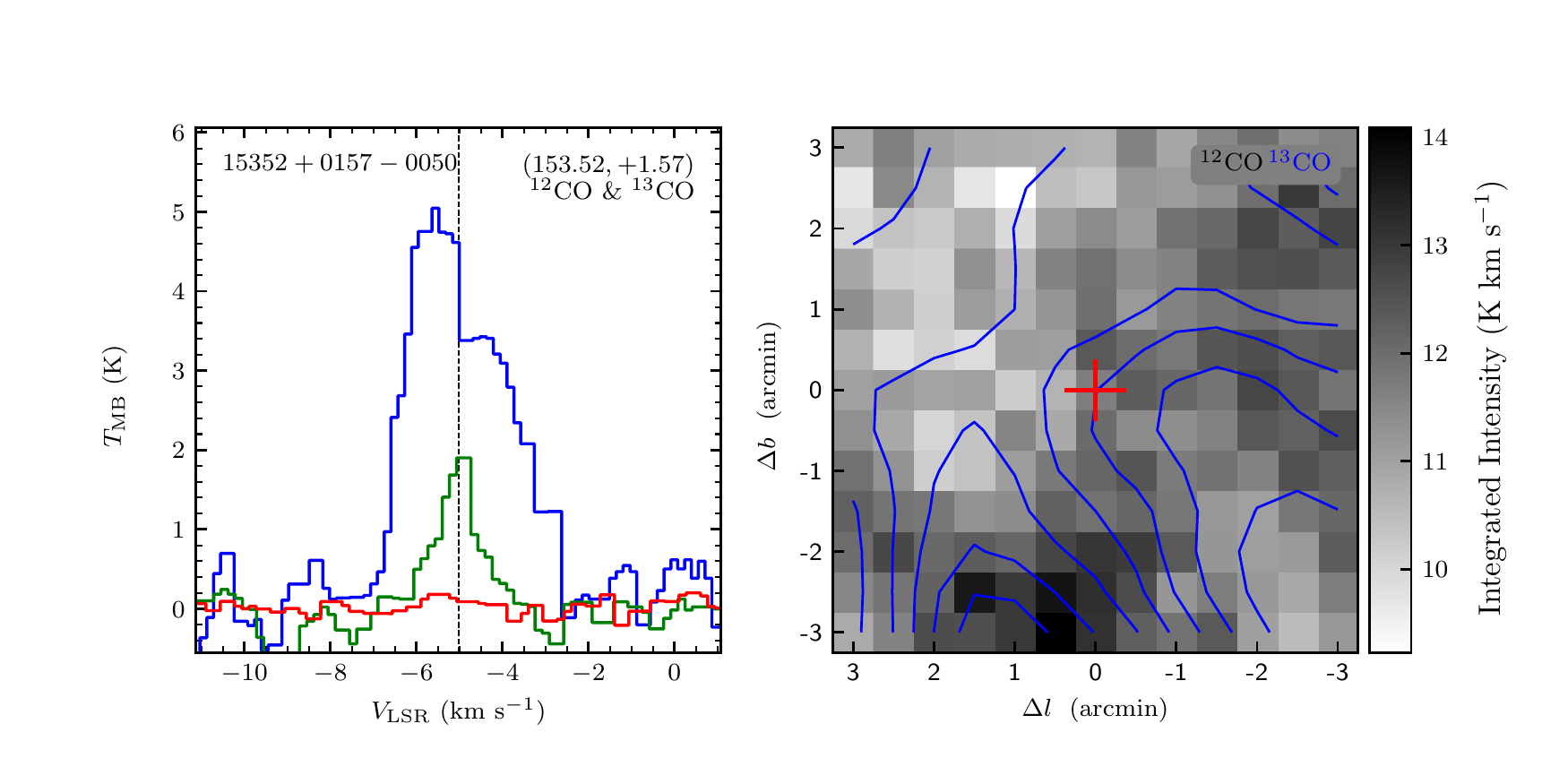}
\includegraphics[width=9.0cm,angle=0]{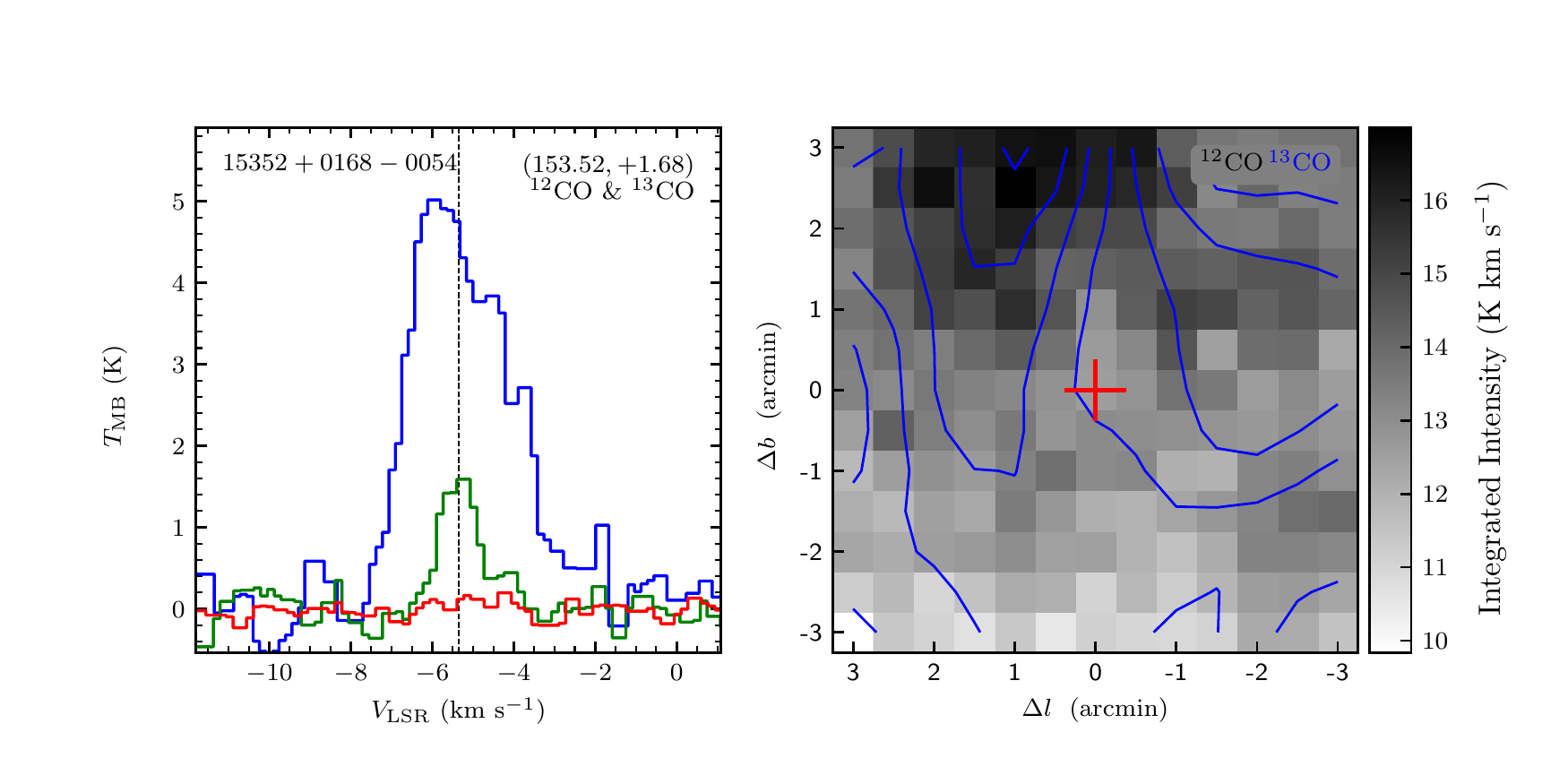}
\end{figure}
\clearpage

\begin{figure}
\includegraphics[width=9.0cm,angle=0]{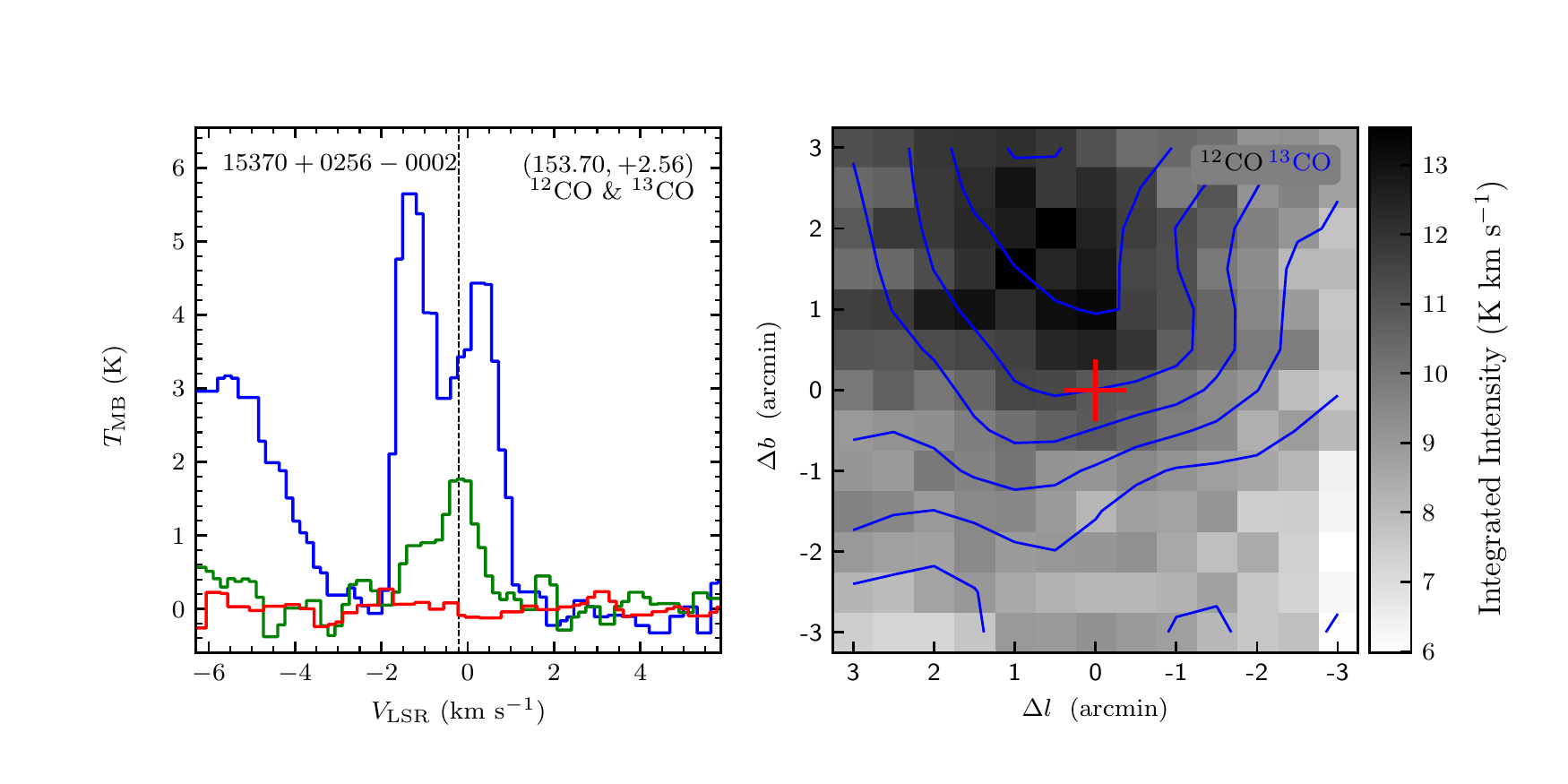}
\includegraphics[width=9.0cm,angle=0]{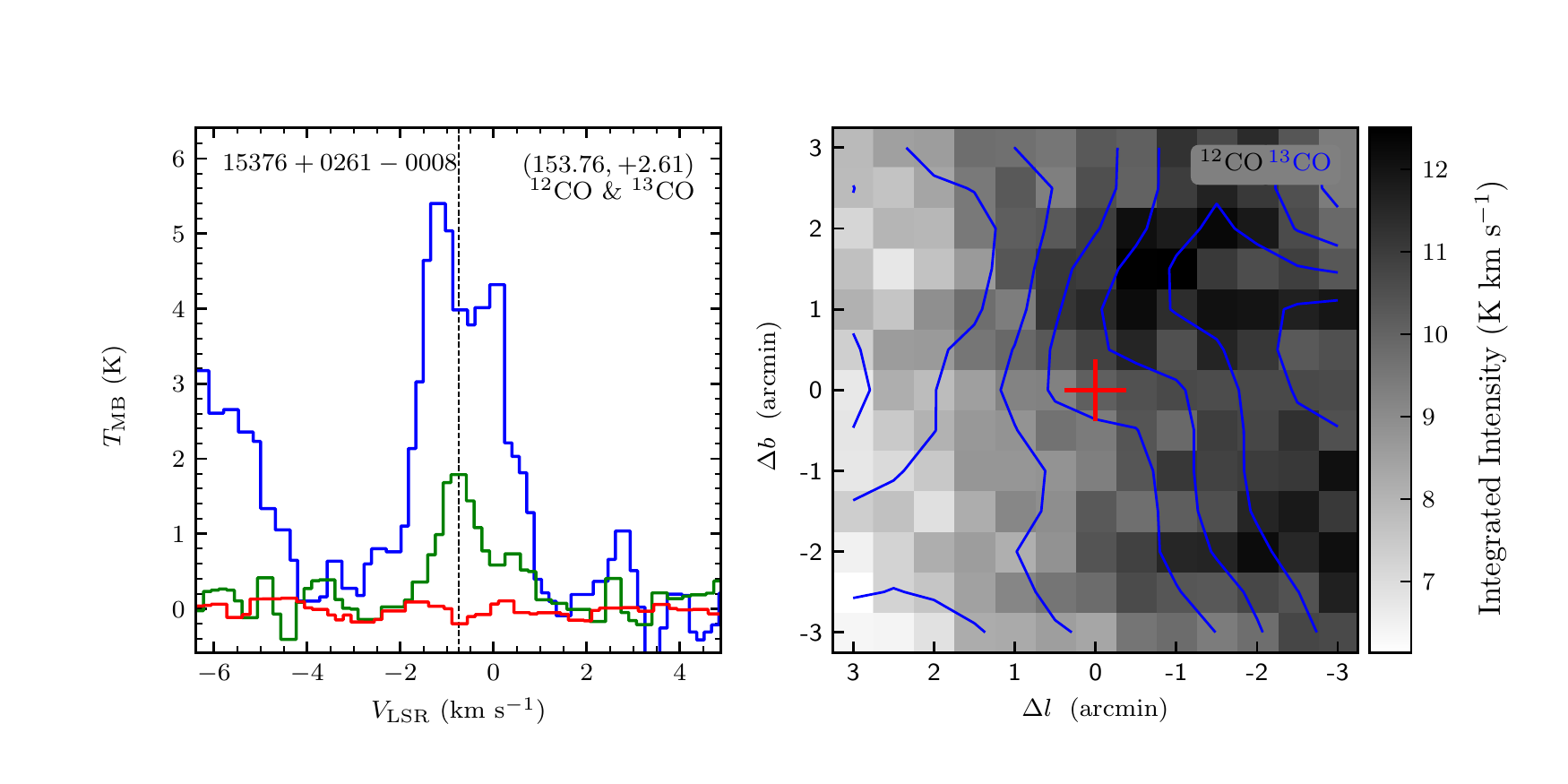}
\vspace{-0.5cm}

\includegraphics[width=9.0cm,angle=0]{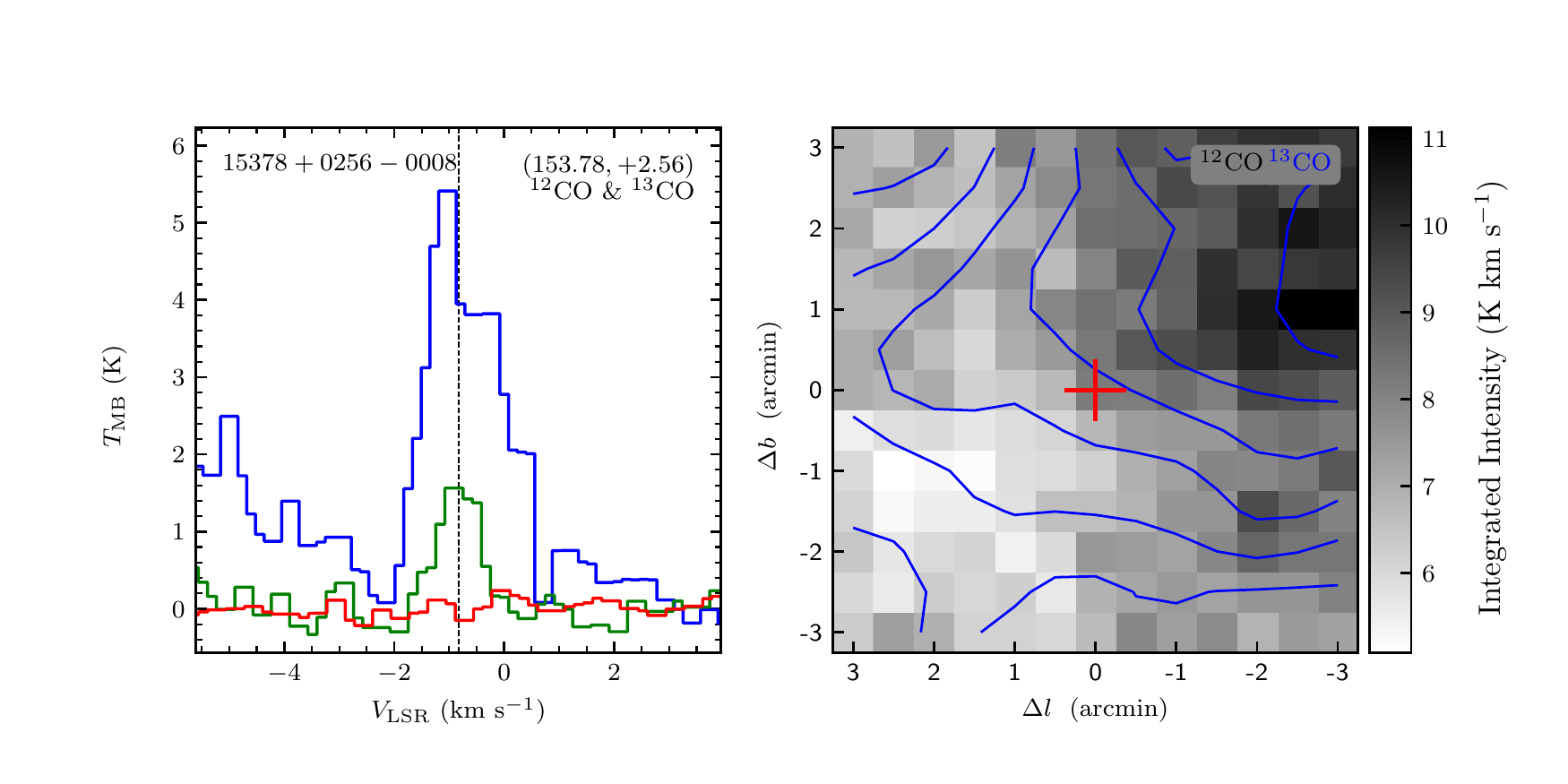}
\includegraphics[width=9.0cm,angle=0]{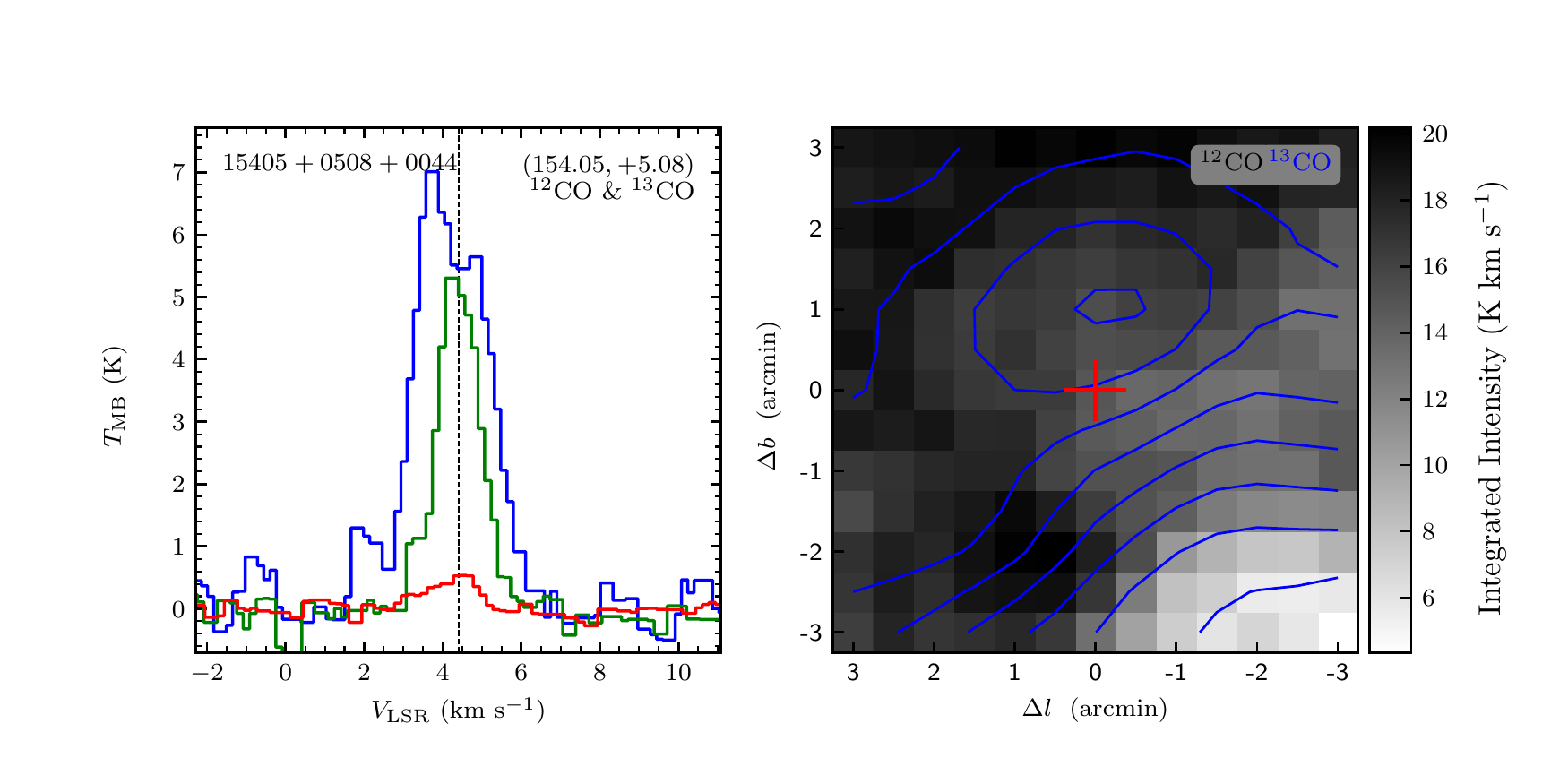}
\vspace{-0.5cm}

\includegraphics[width=9.0cm,angle=0]{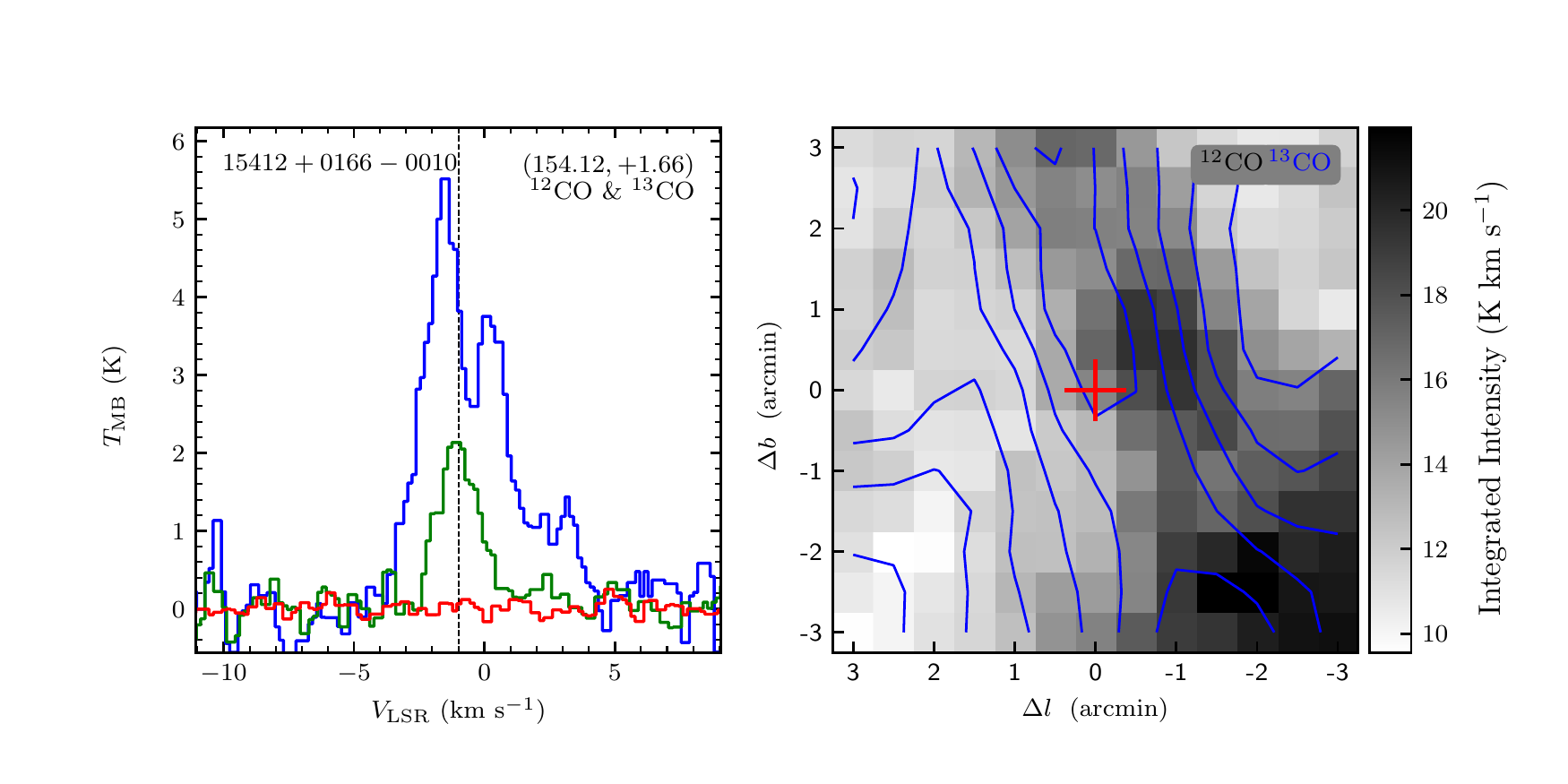}
\includegraphics[width=9.0cm,angle=0]{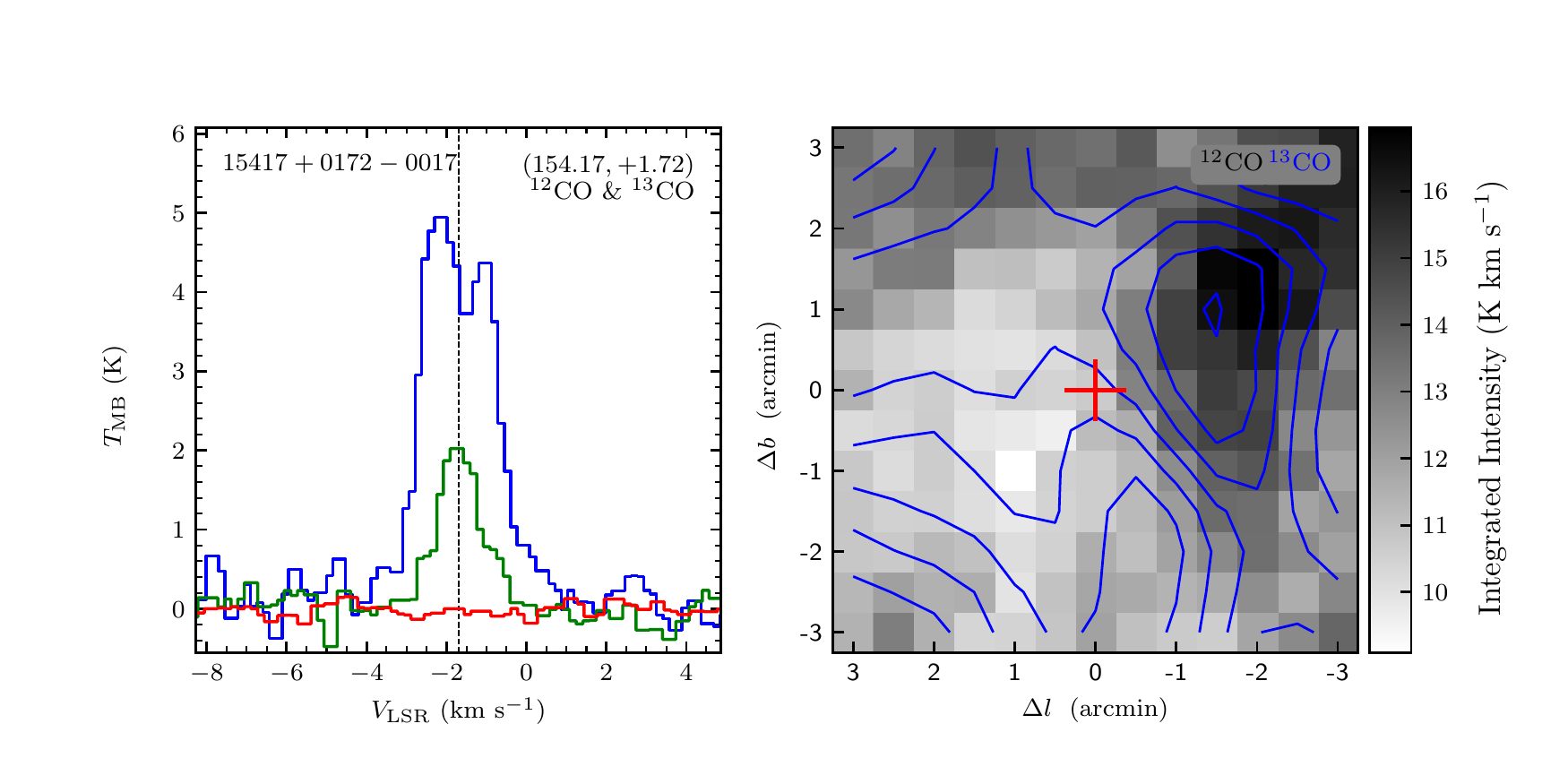}
\vspace{-0.5cm}

\includegraphics[width=9.0cm,angle=0]{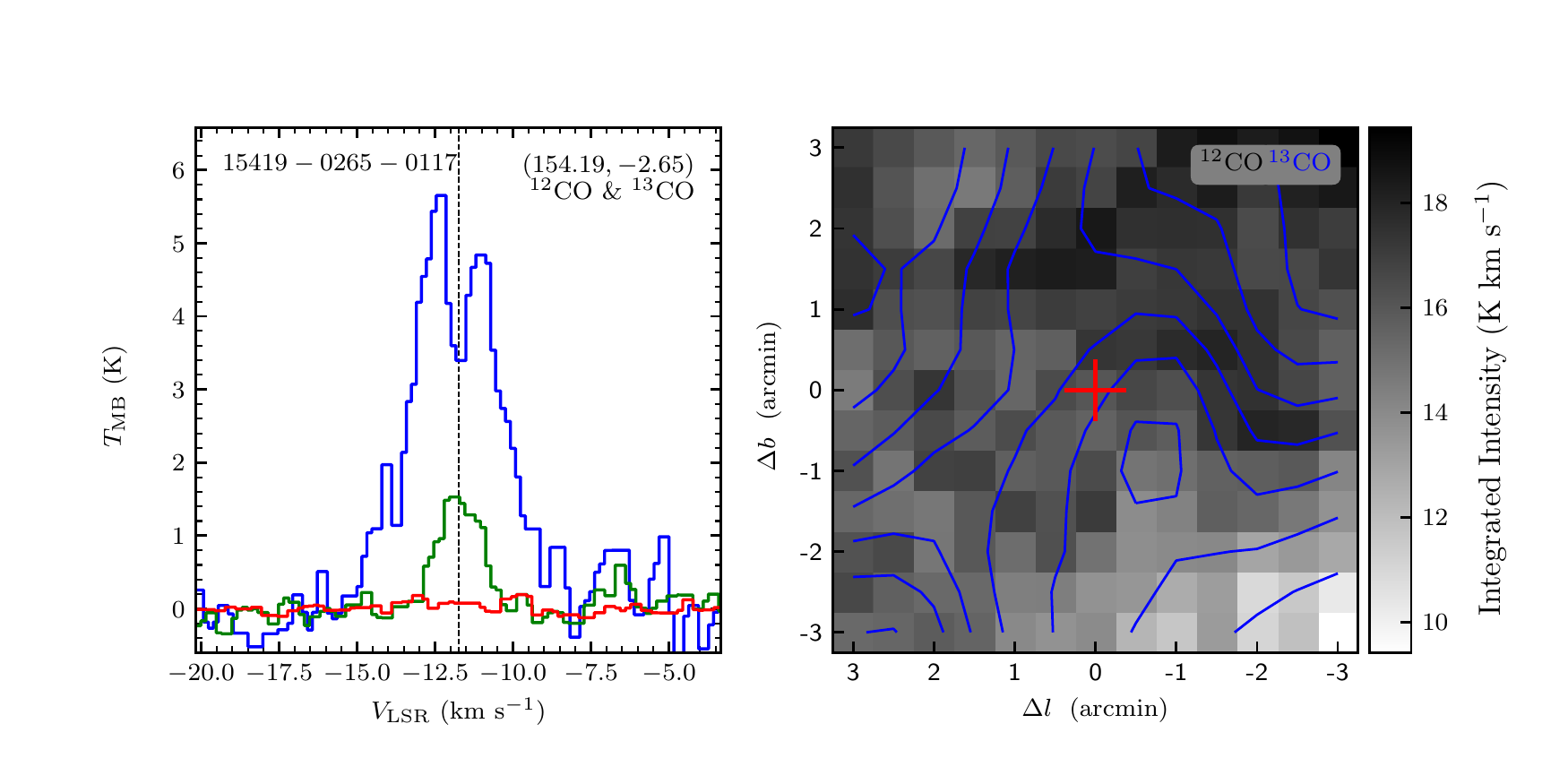}
\includegraphics[width=9.0cm,angle=0]{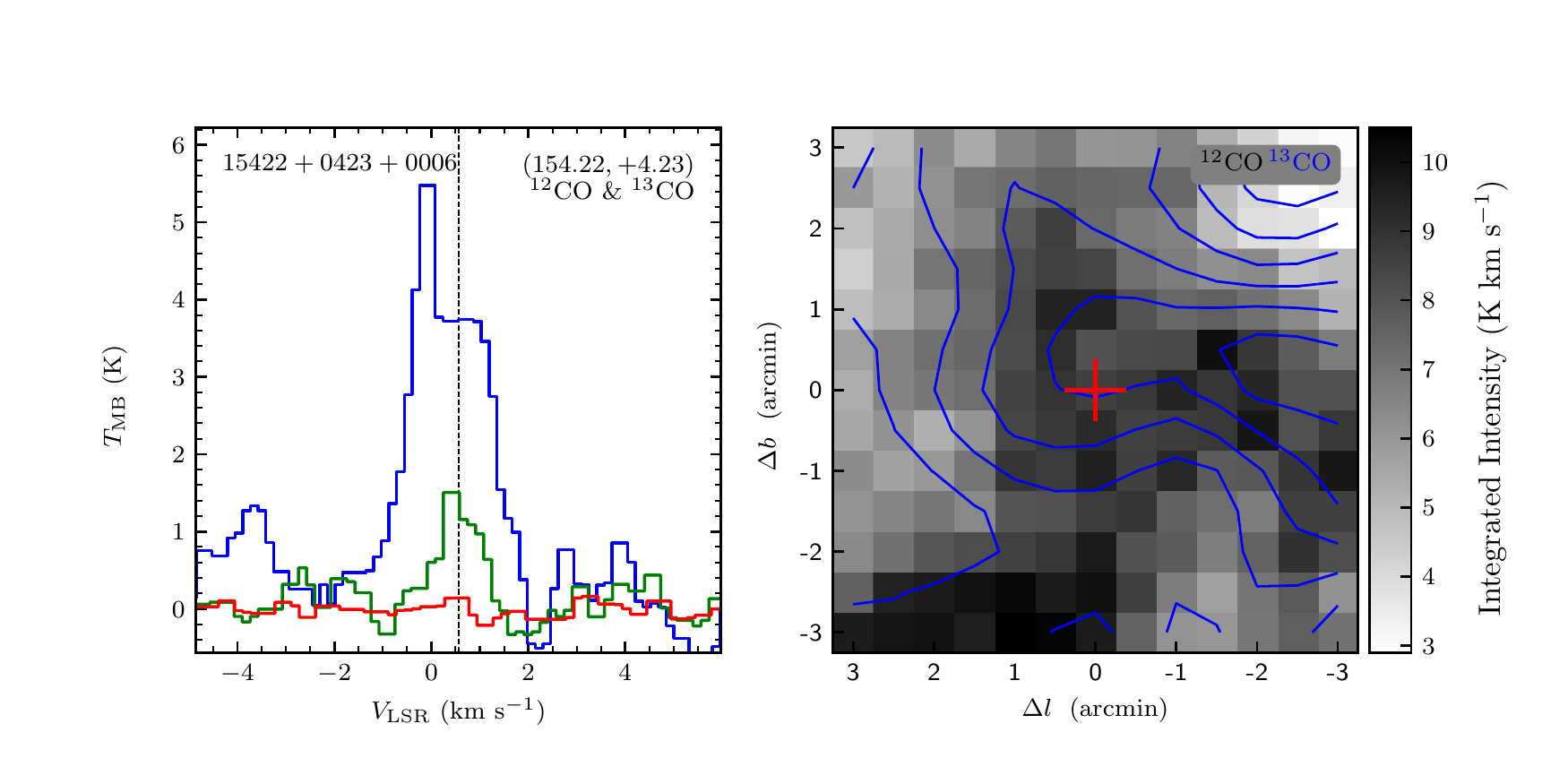}
\vspace{-0.5cm}

\includegraphics[width=9.0cm,angle=0]{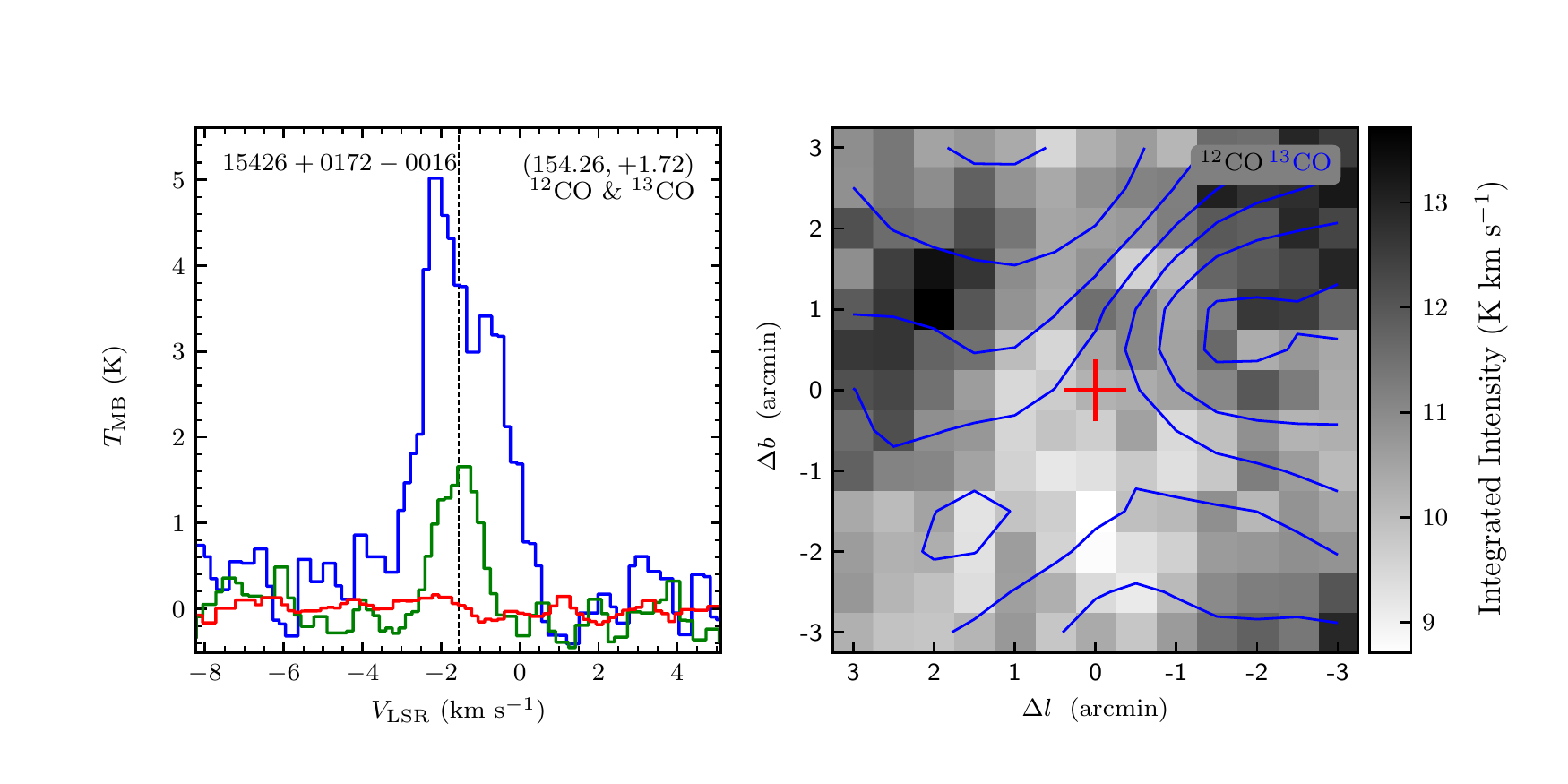}
\includegraphics[width=9.0cm,angle=0]{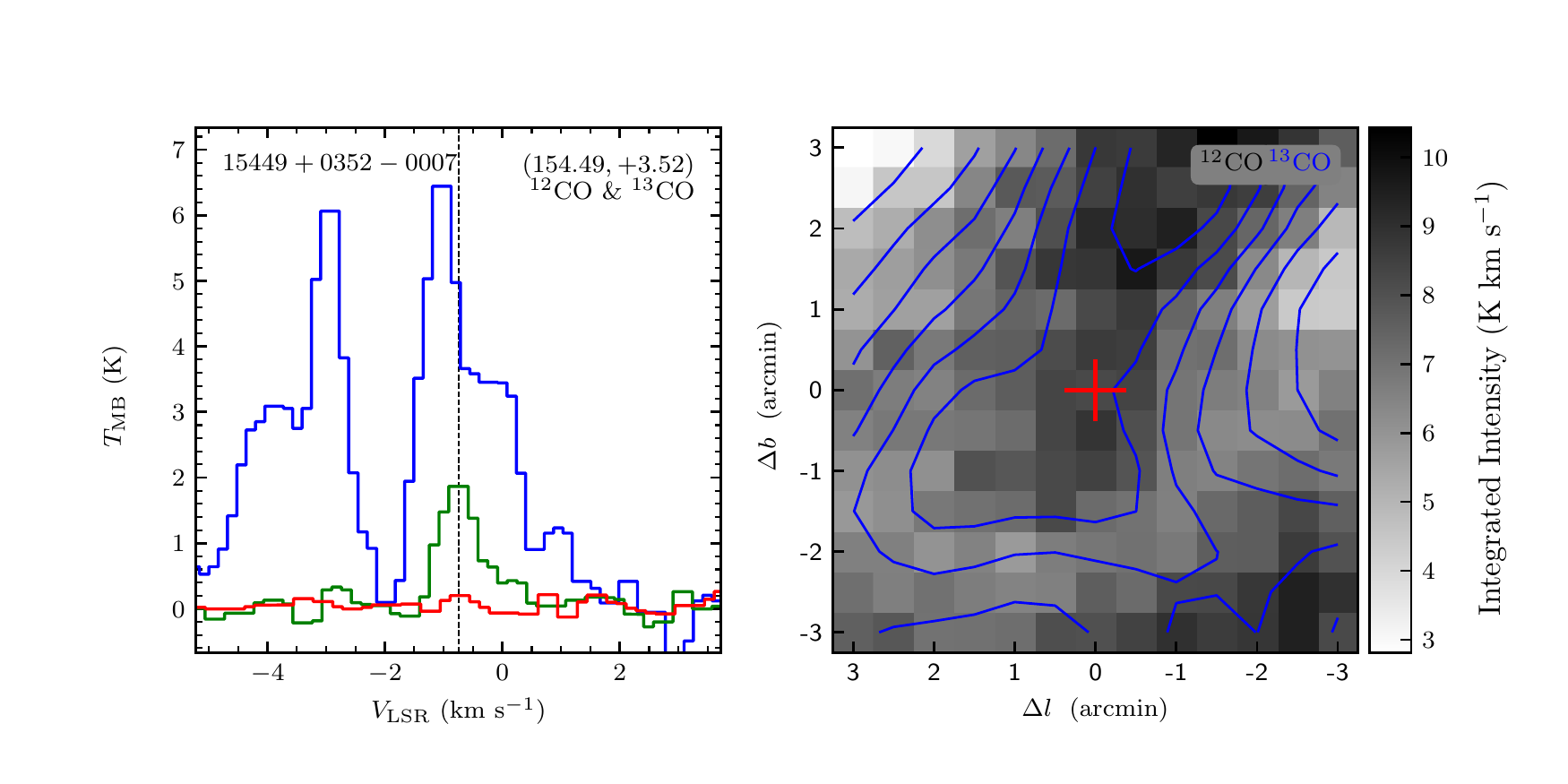}
\end{figure}
\clearpage

\begin{figure}
\includegraphics[width=9.0cm,angle=0]{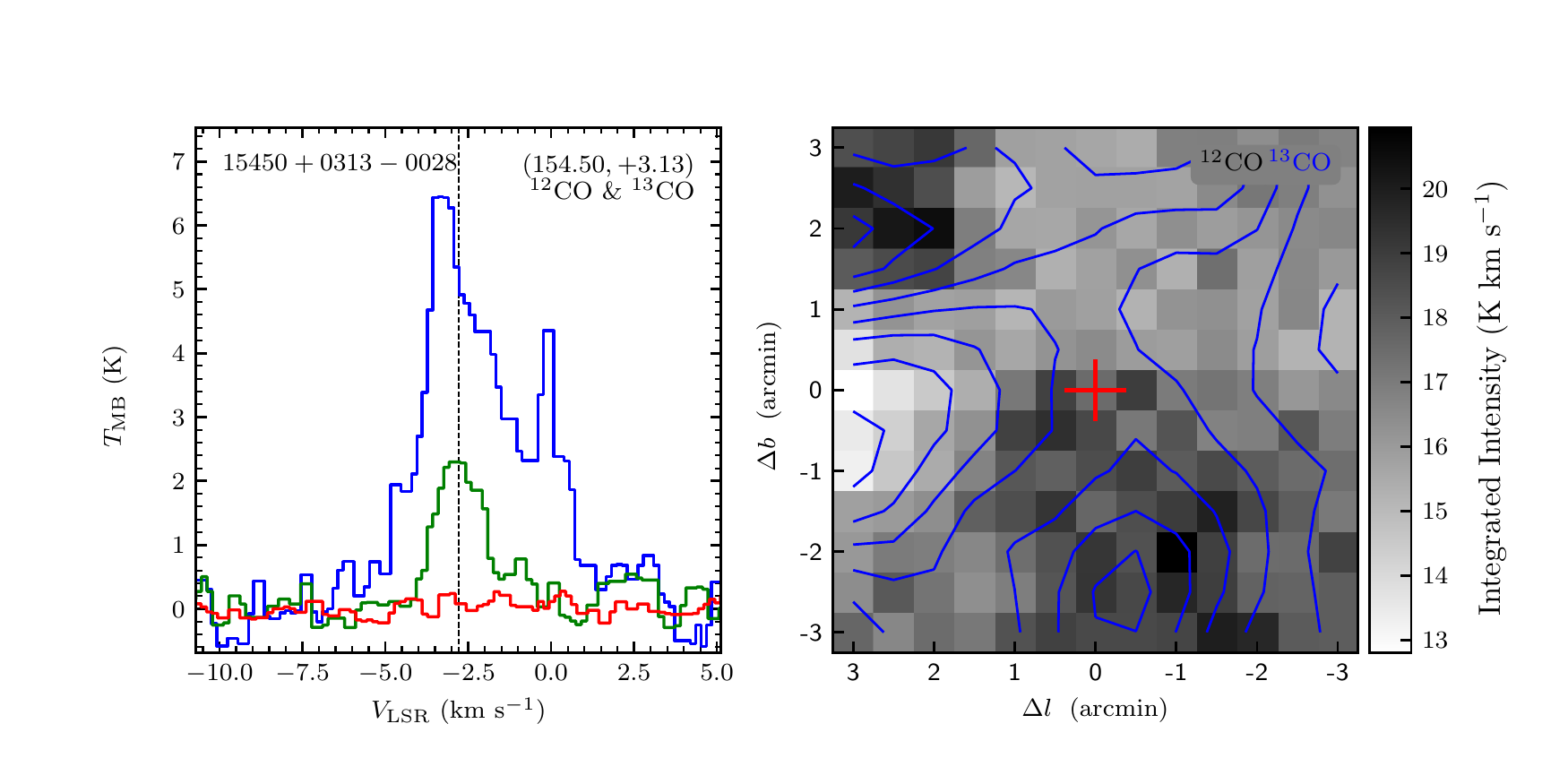}
\includegraphics[width=9.0cm,angle=0]{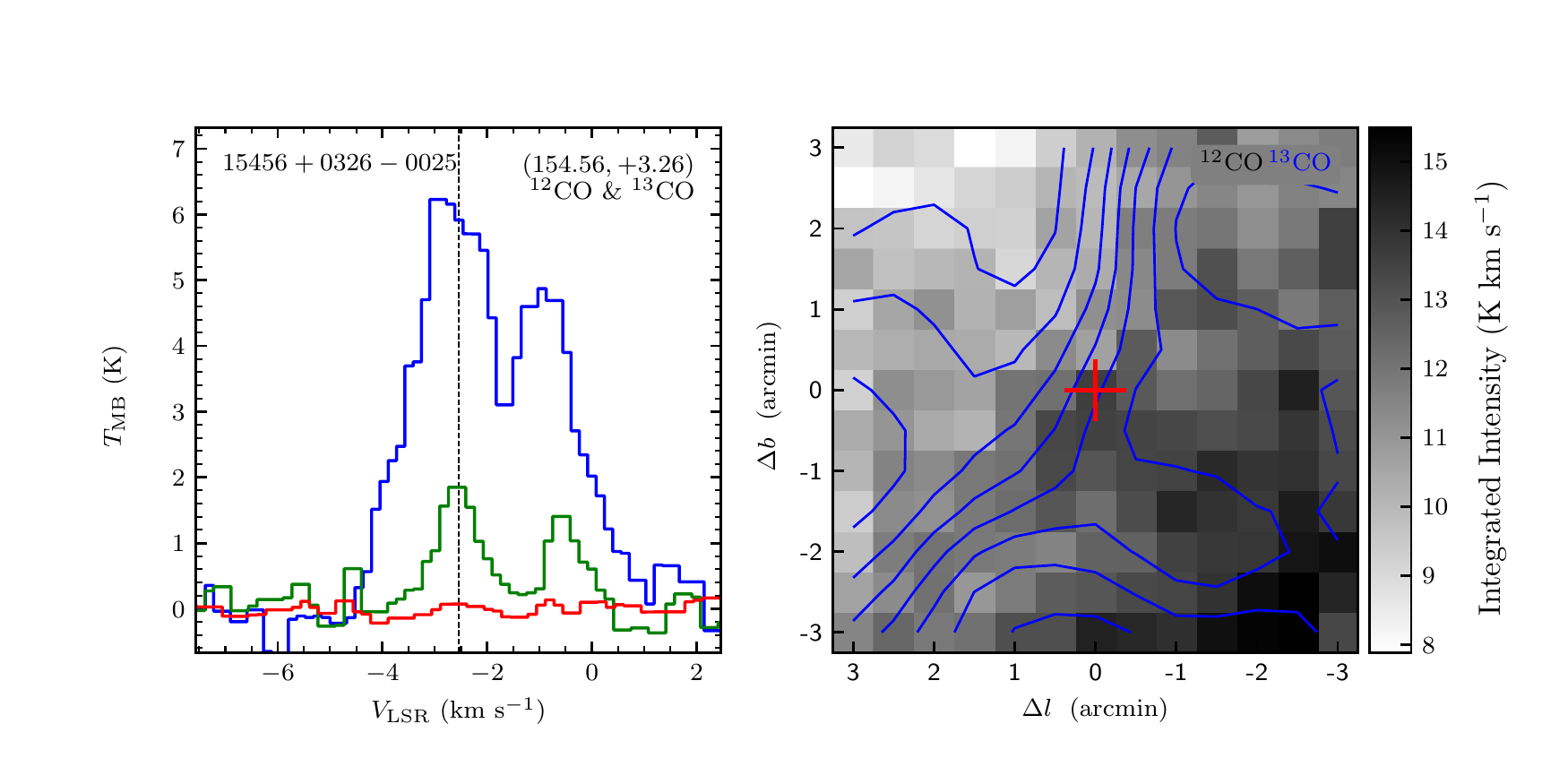}
\vspace{-0.5cm}

\includegraphics[width=9.0cm,angle=0]{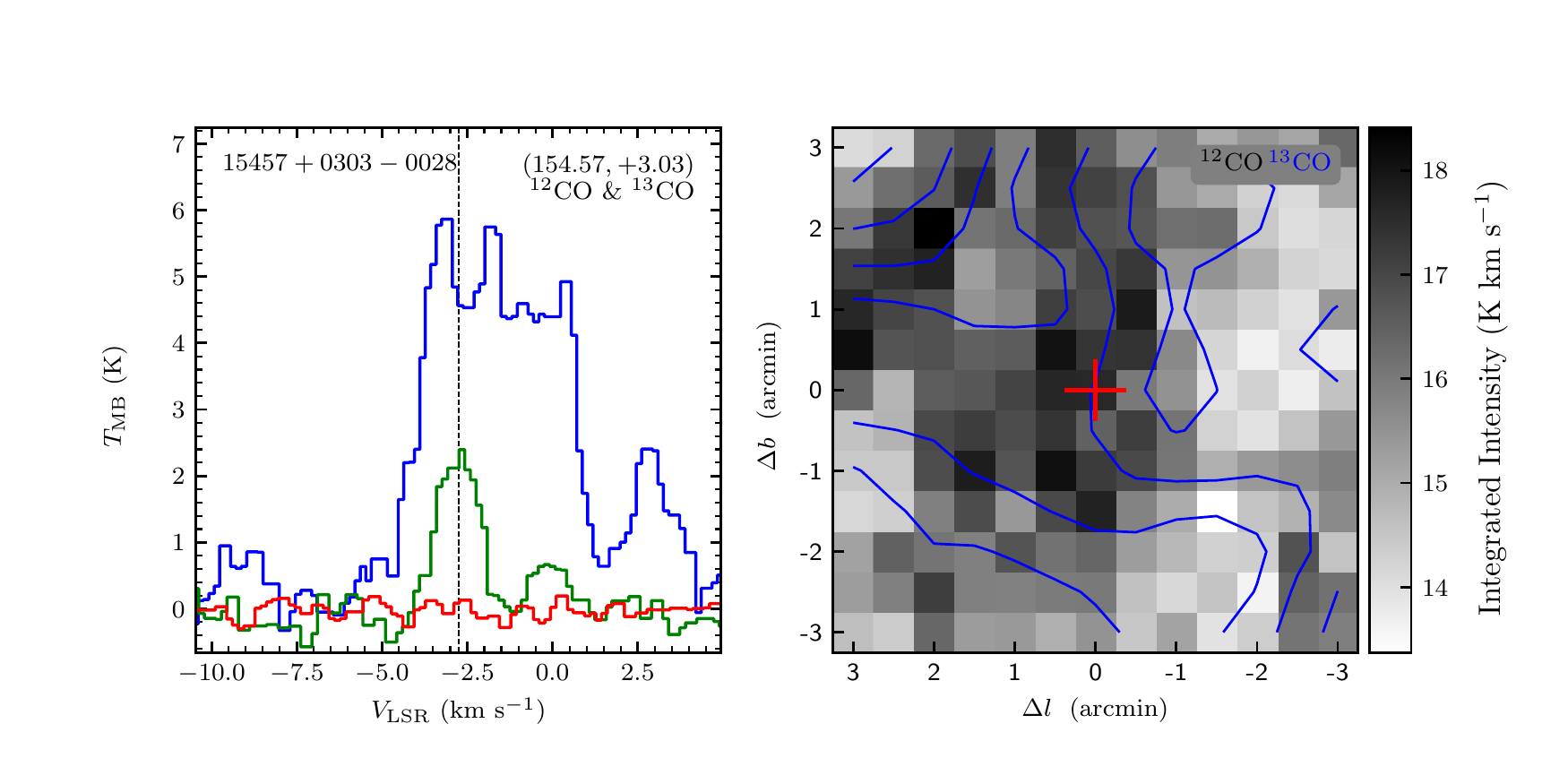}
\includegraphics[width=9.0cm,angle=0]{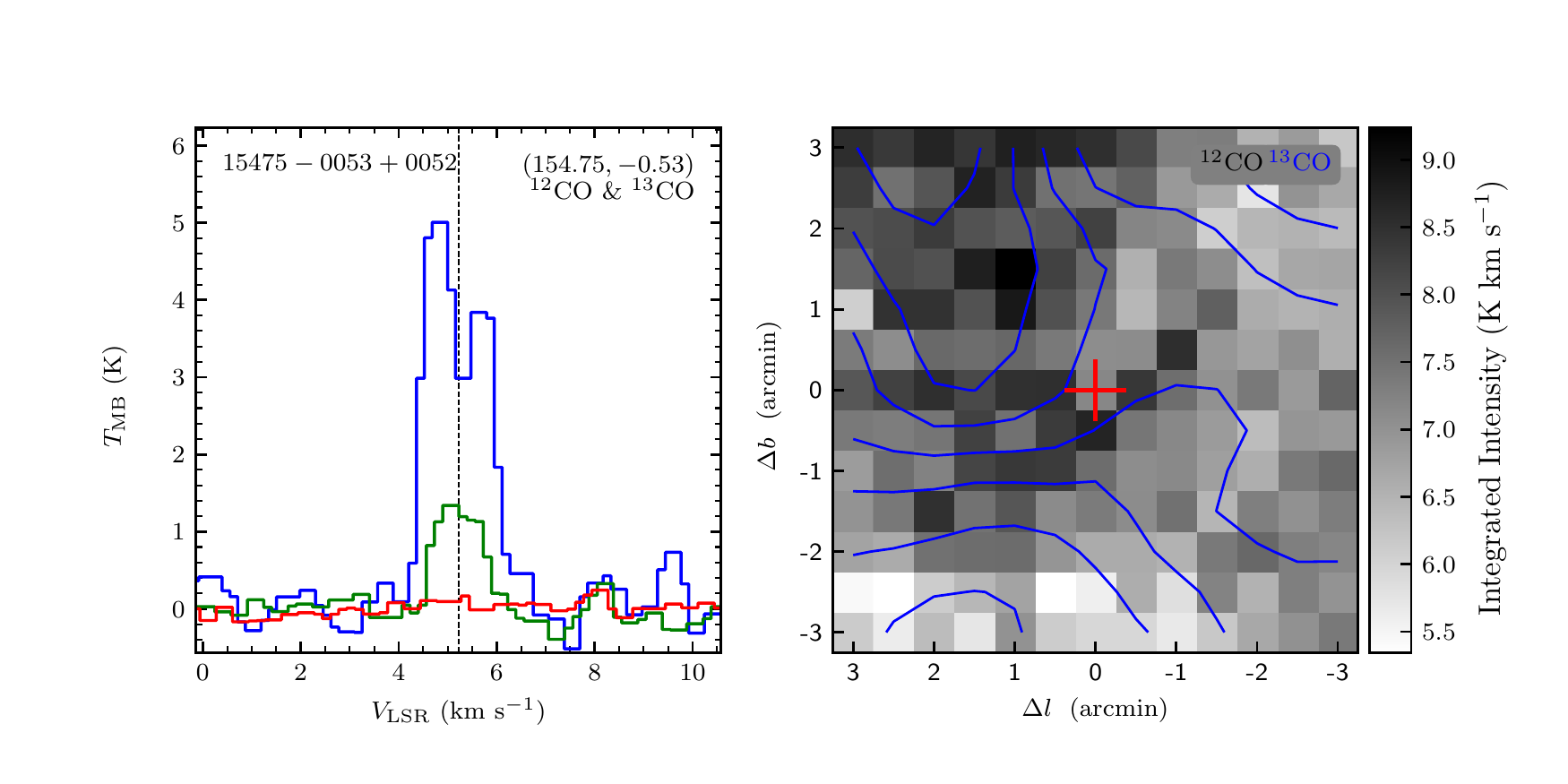}
\vspace{-0.5cm}

\includegraphics[width=9.0cm,angle=0]{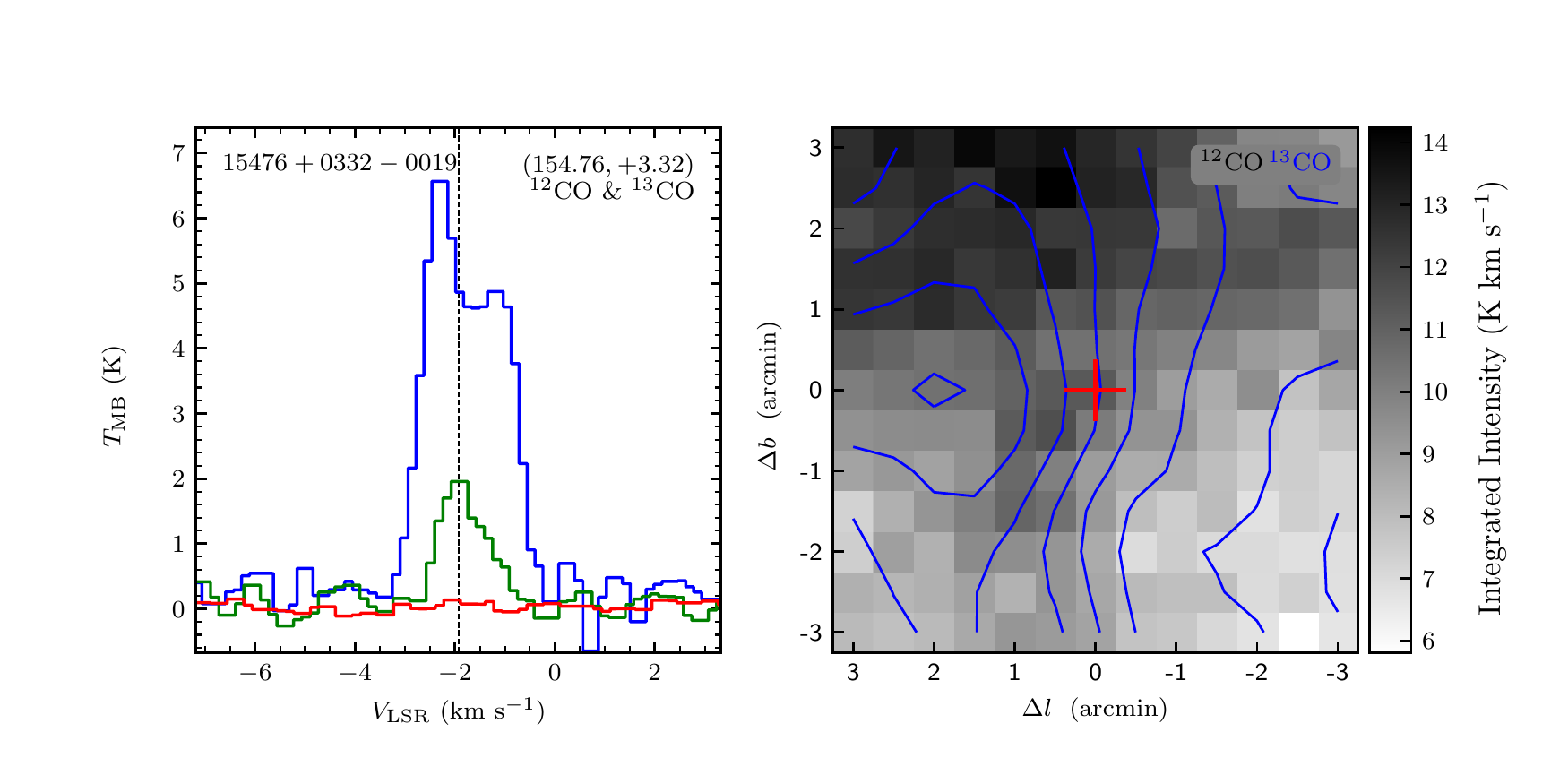}
\includegraphics[width=9.0cm,angle=0]{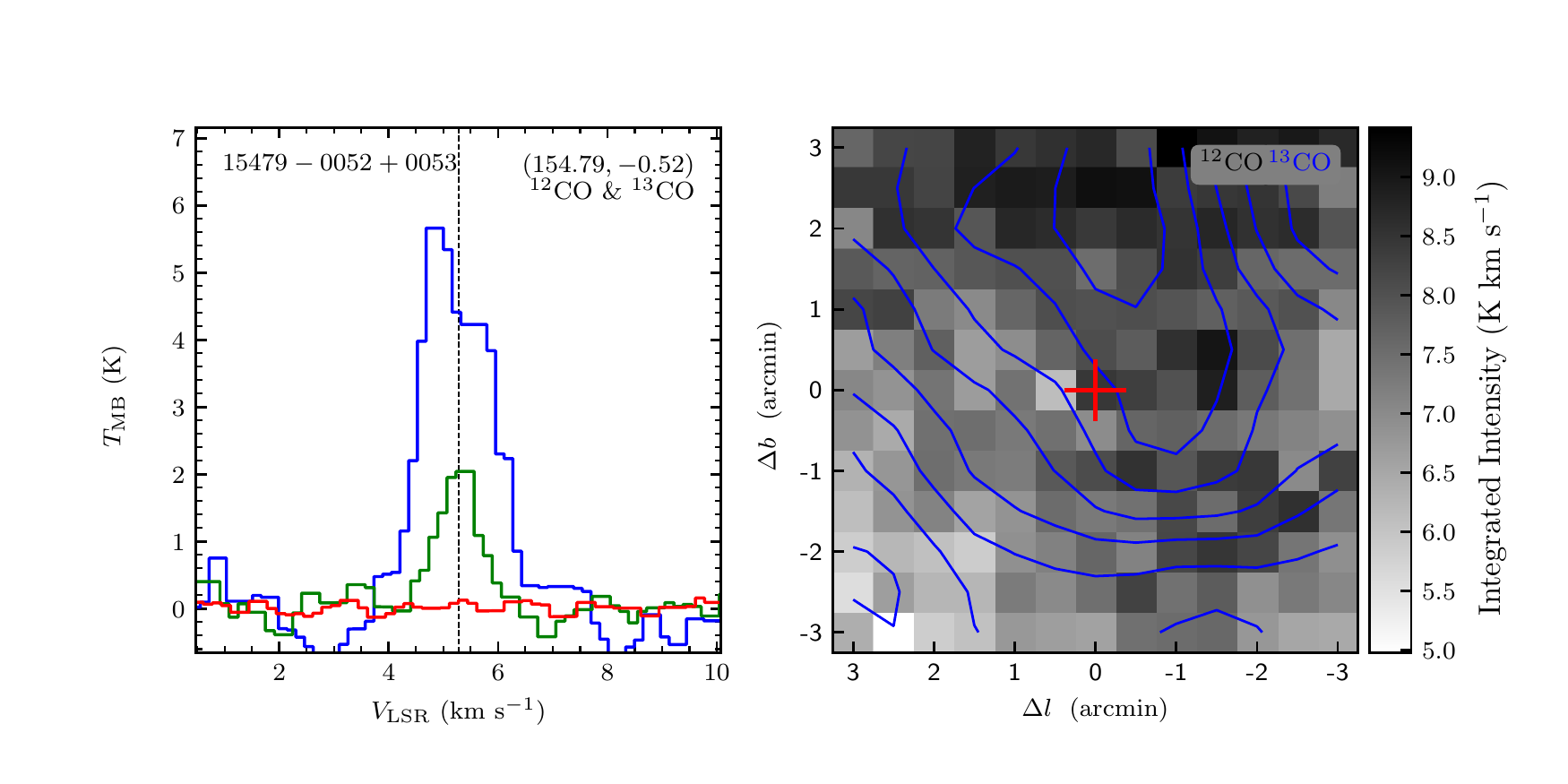}
\vspace{-0.5cm}

\includegraphics[width=9.0cm,angle=0]{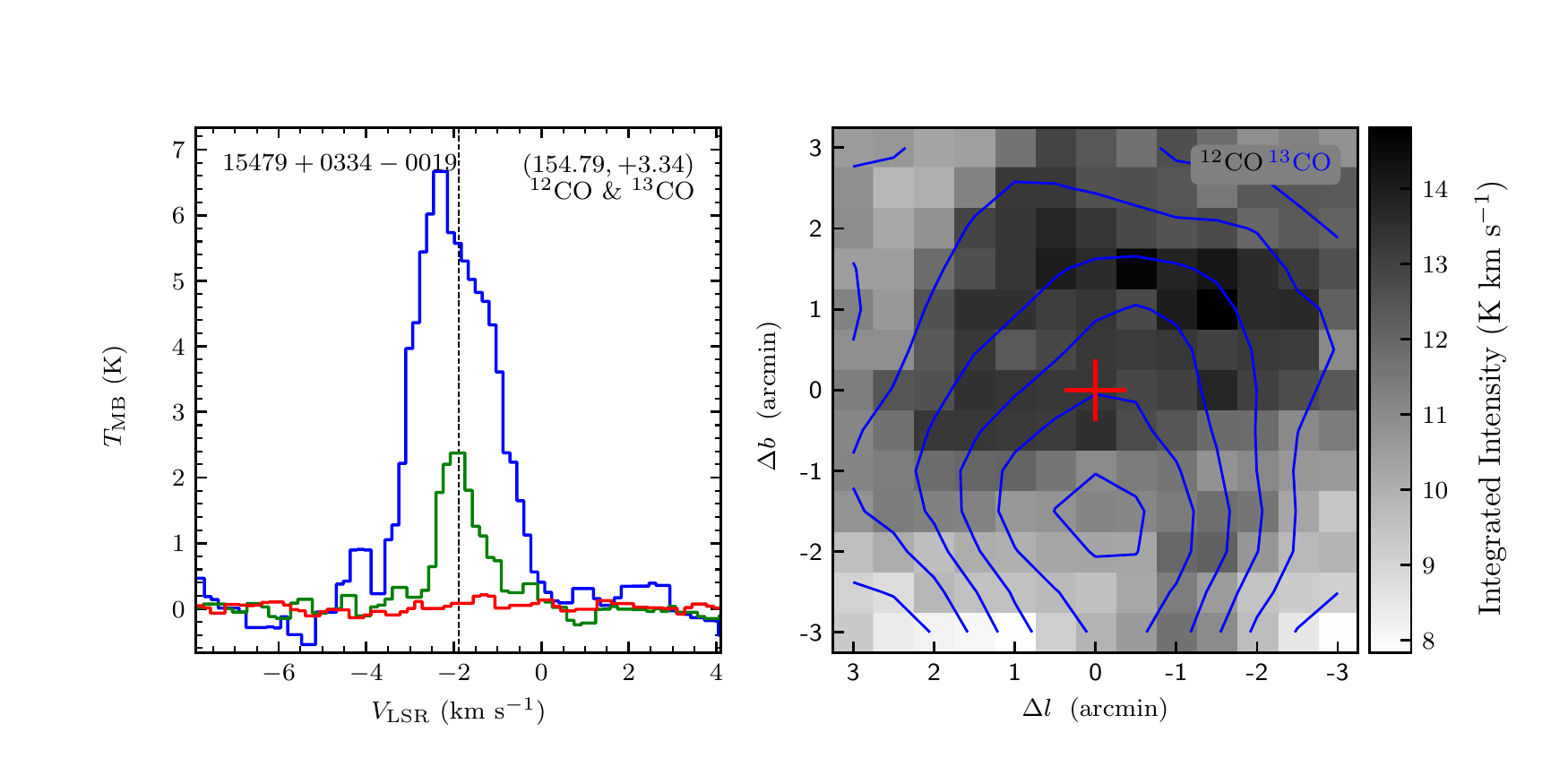}
\includegraphics[width=9.0cm,angle=0]{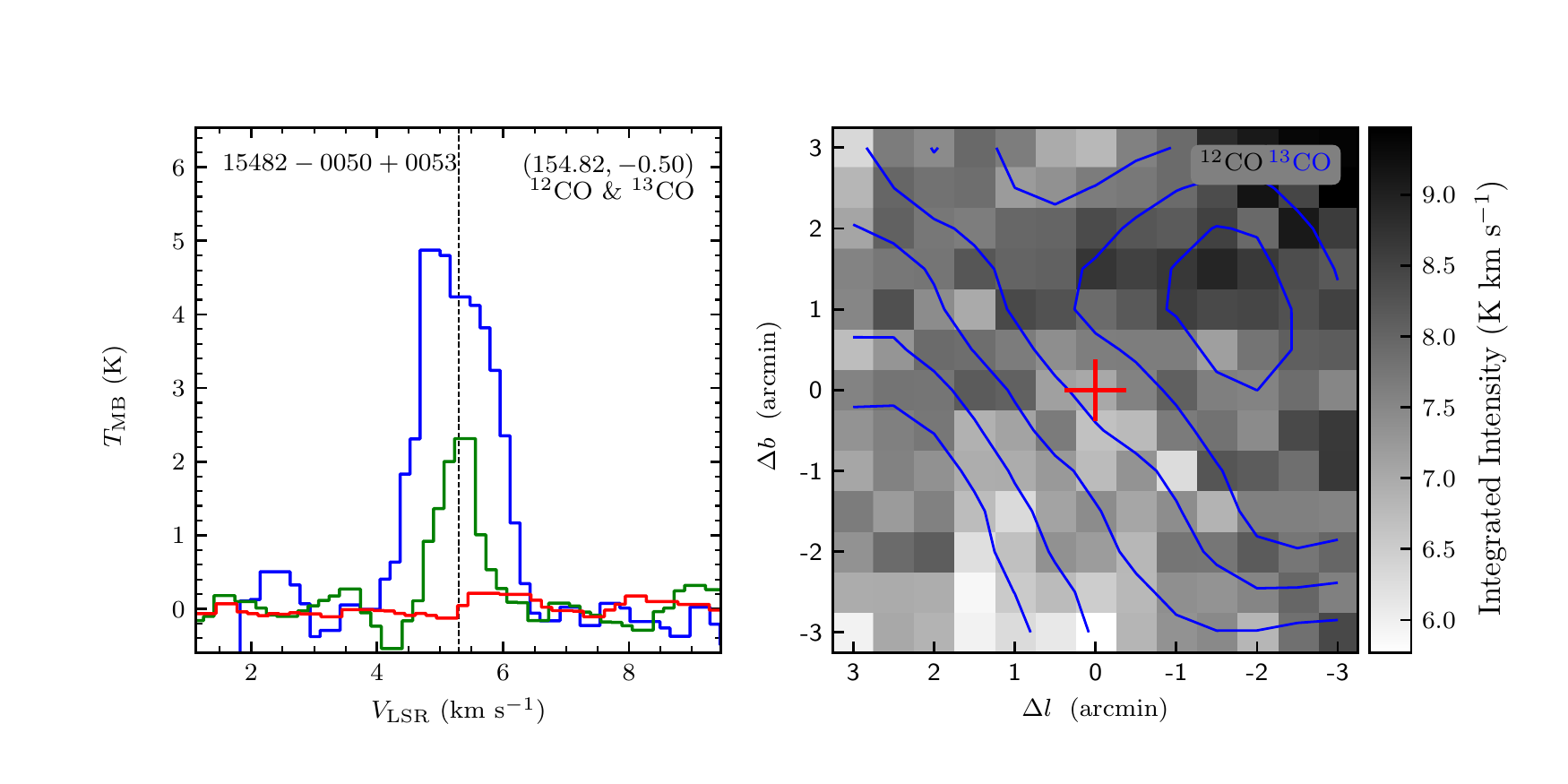}
\vspace{-0.5cm}

\includegraphics[width=9.0cm,angle=0]{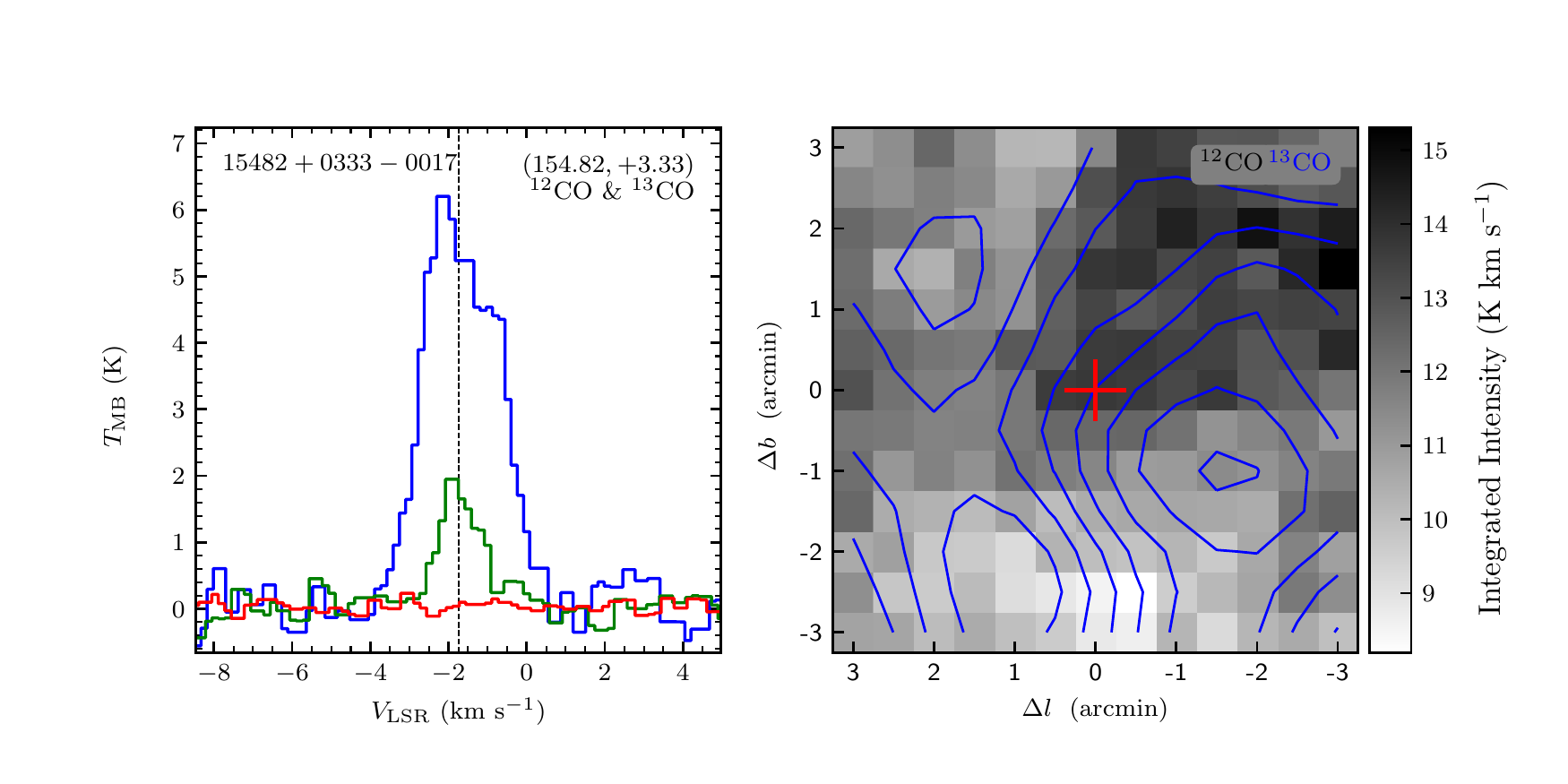}
\includegraphics[width=9.0cm,angle=0]{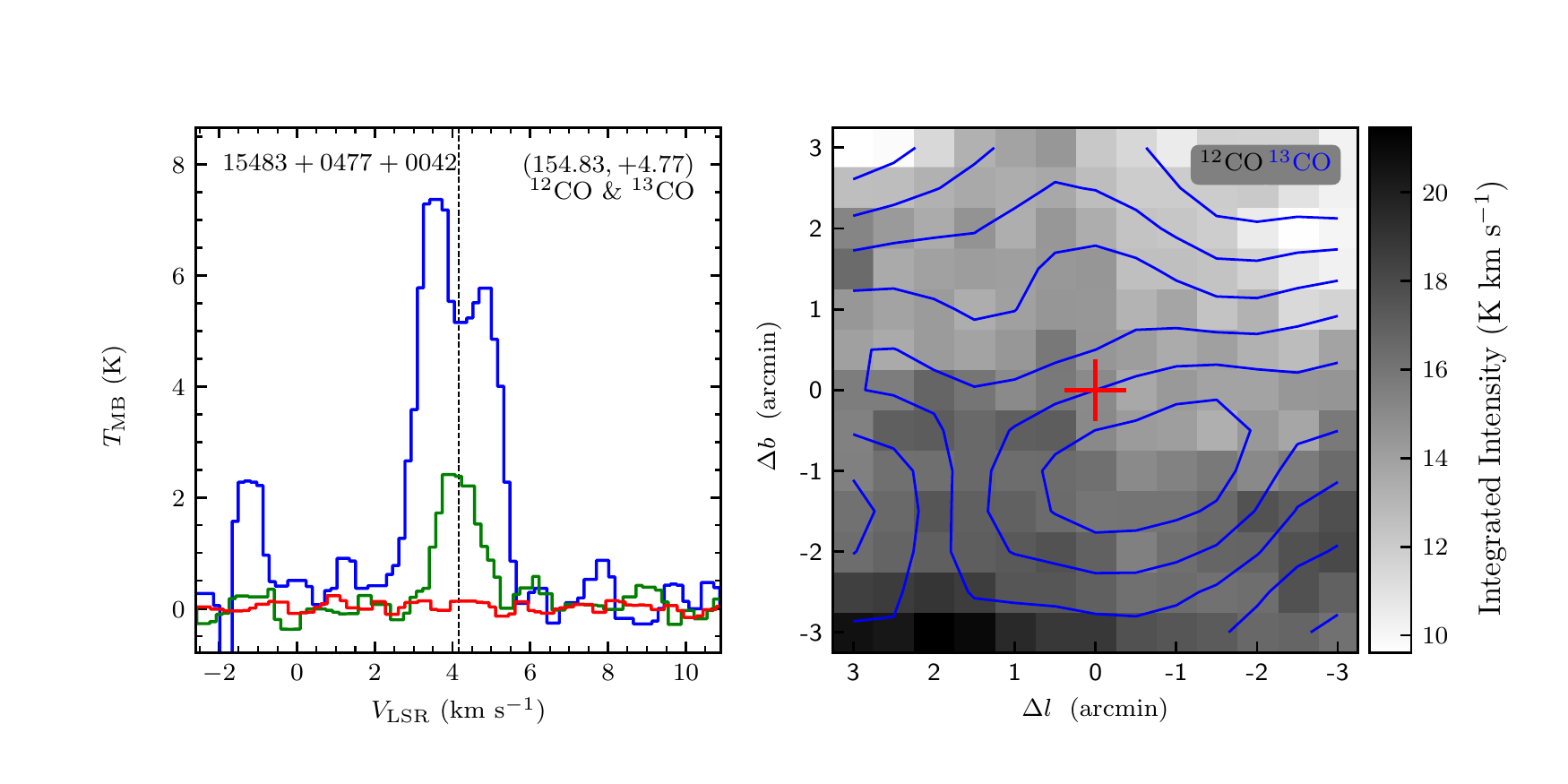}
\end{figure}
\clearpage

\begin{figure}
\includegraphics[width=9.0cm,angle=0]{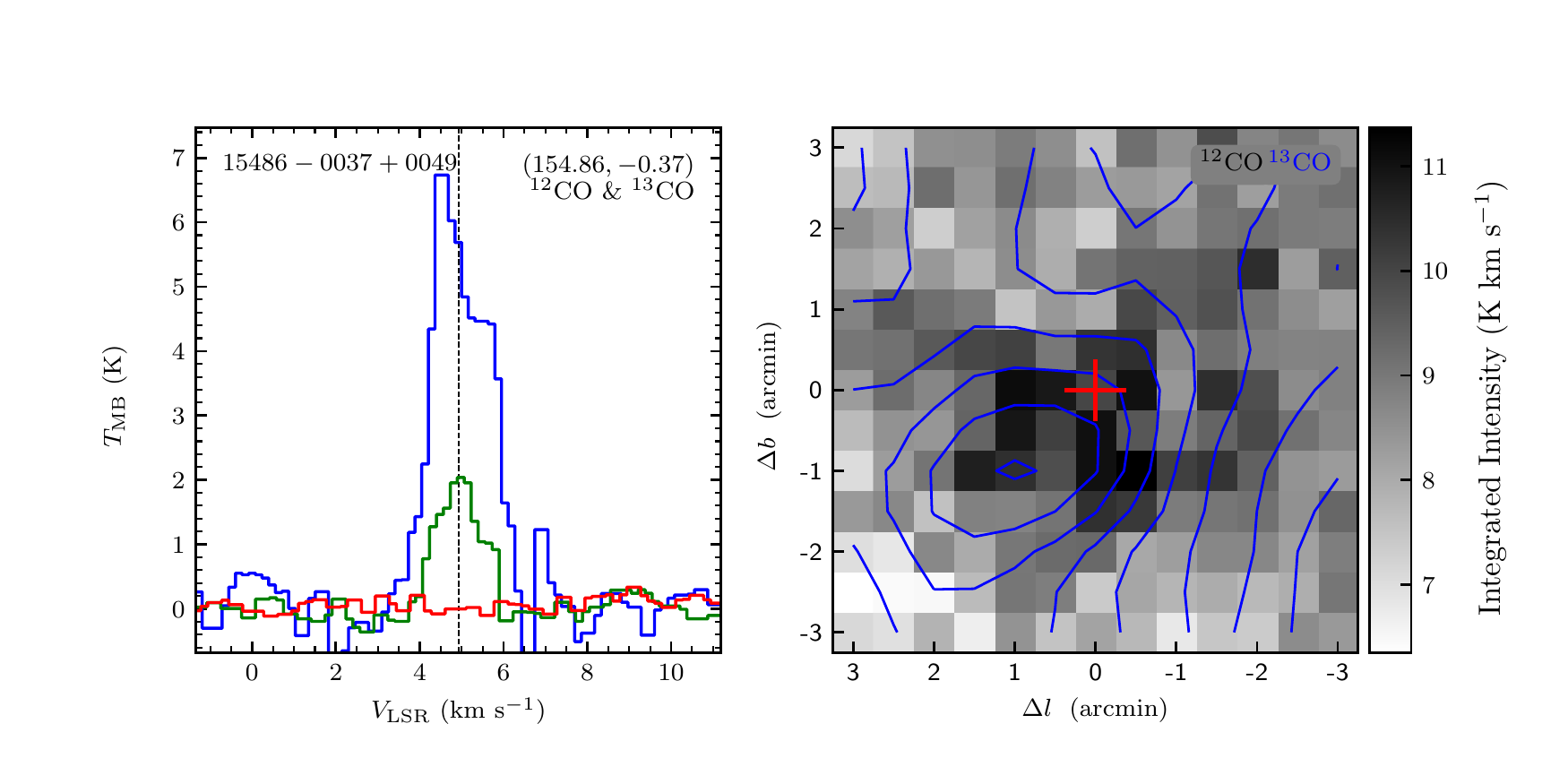}
\includegraphics[width=9.0cm,angle=0]{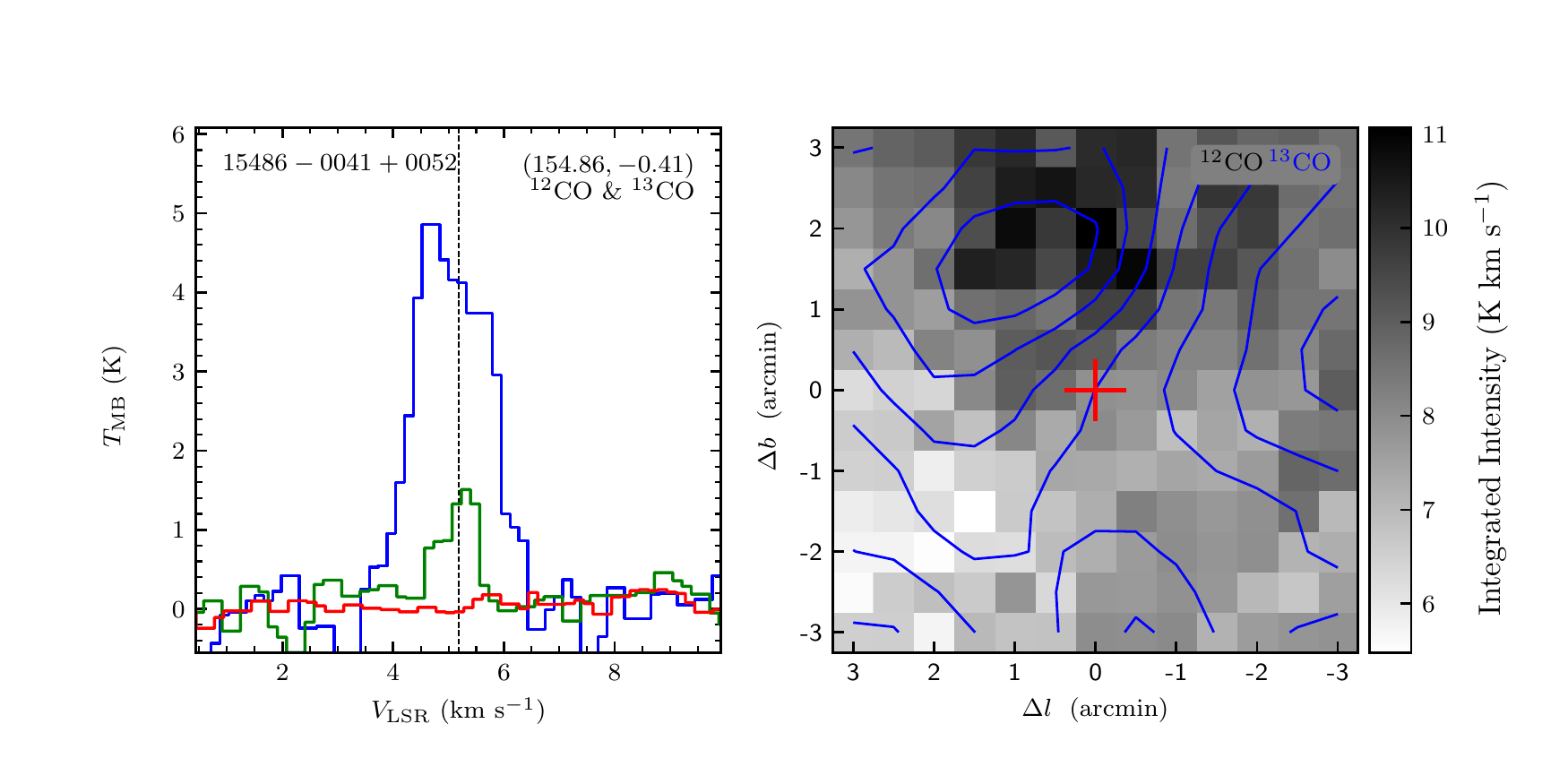}
\vspace{-0.5cm}

\includegraphics[width=9.0cm,angle=0]{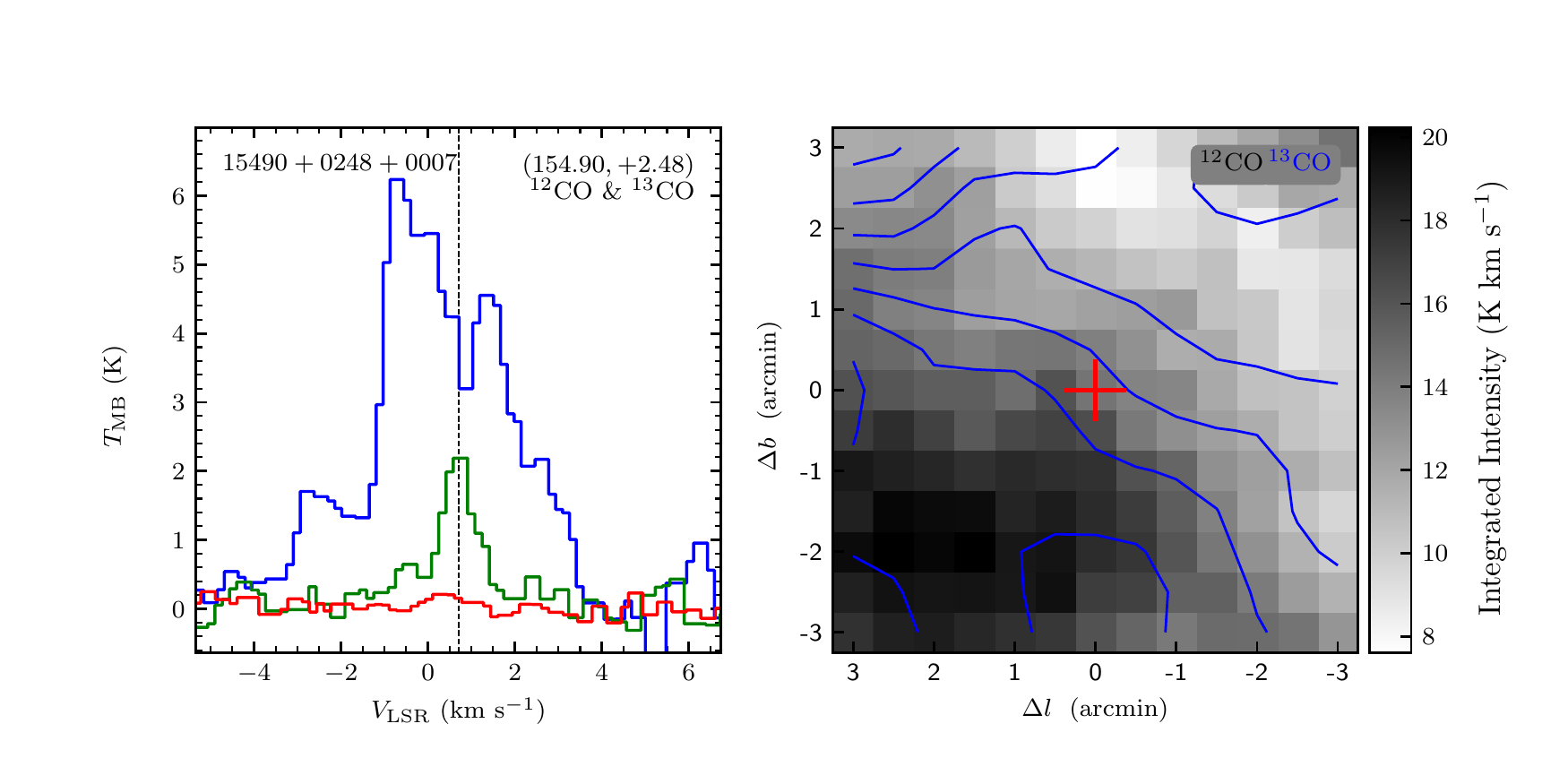}
\includegraphics[width=9.0cm,angle=0]{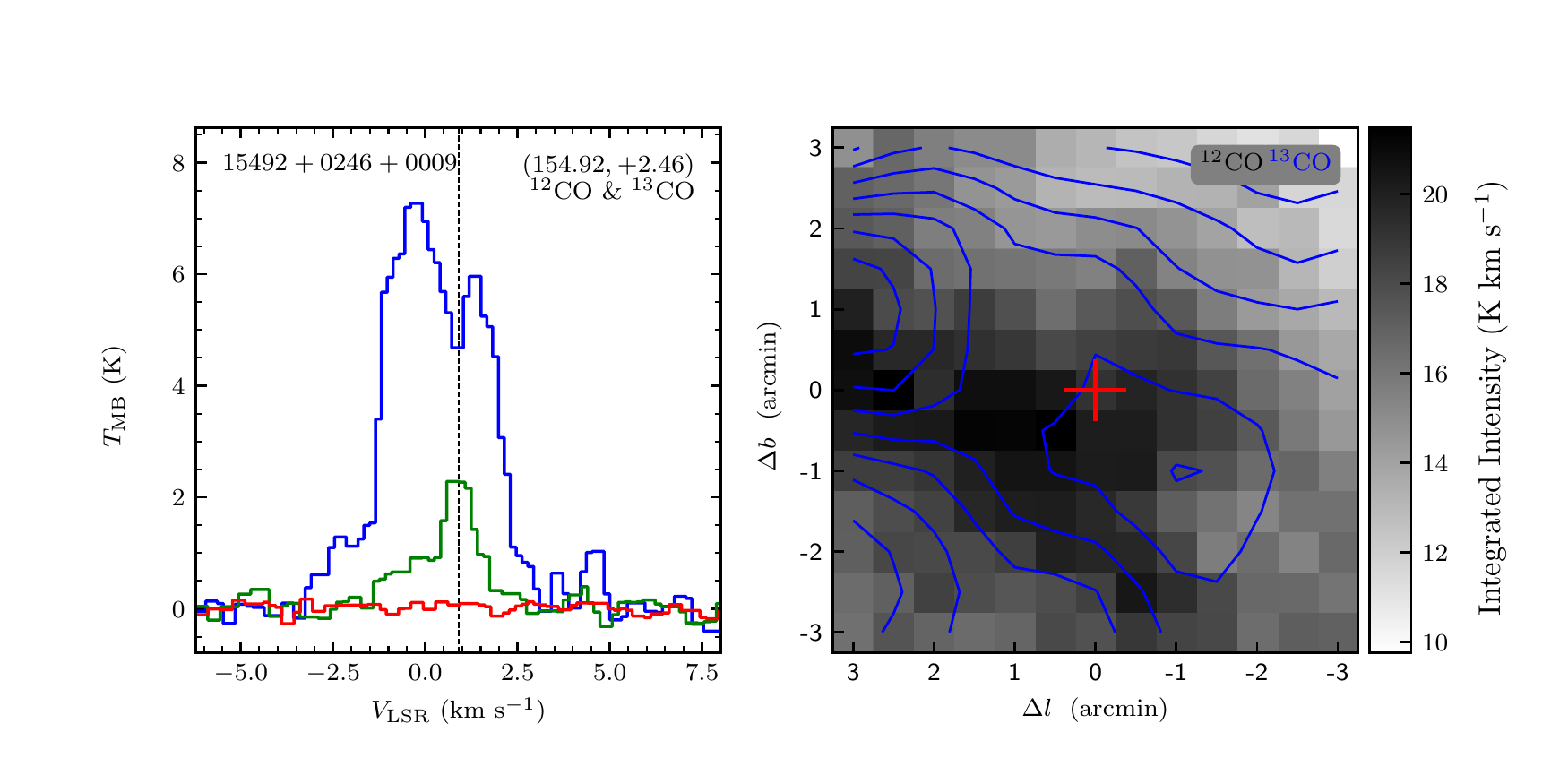}
\vspace{-0.5cm}

\includegraphics[width=9.0cm,angle=0]{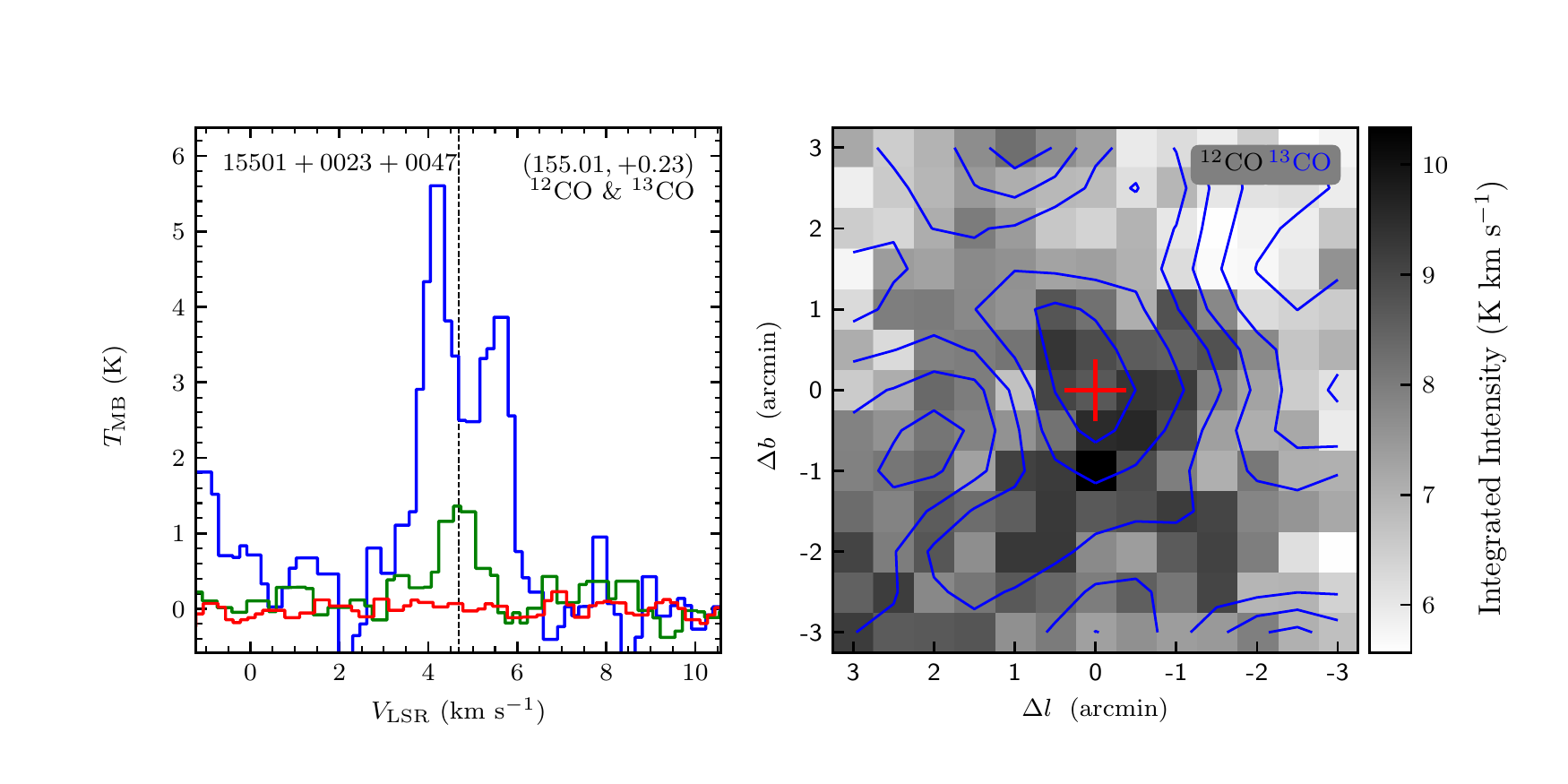}
\includegraphics[width=9.0cm,angle=0]{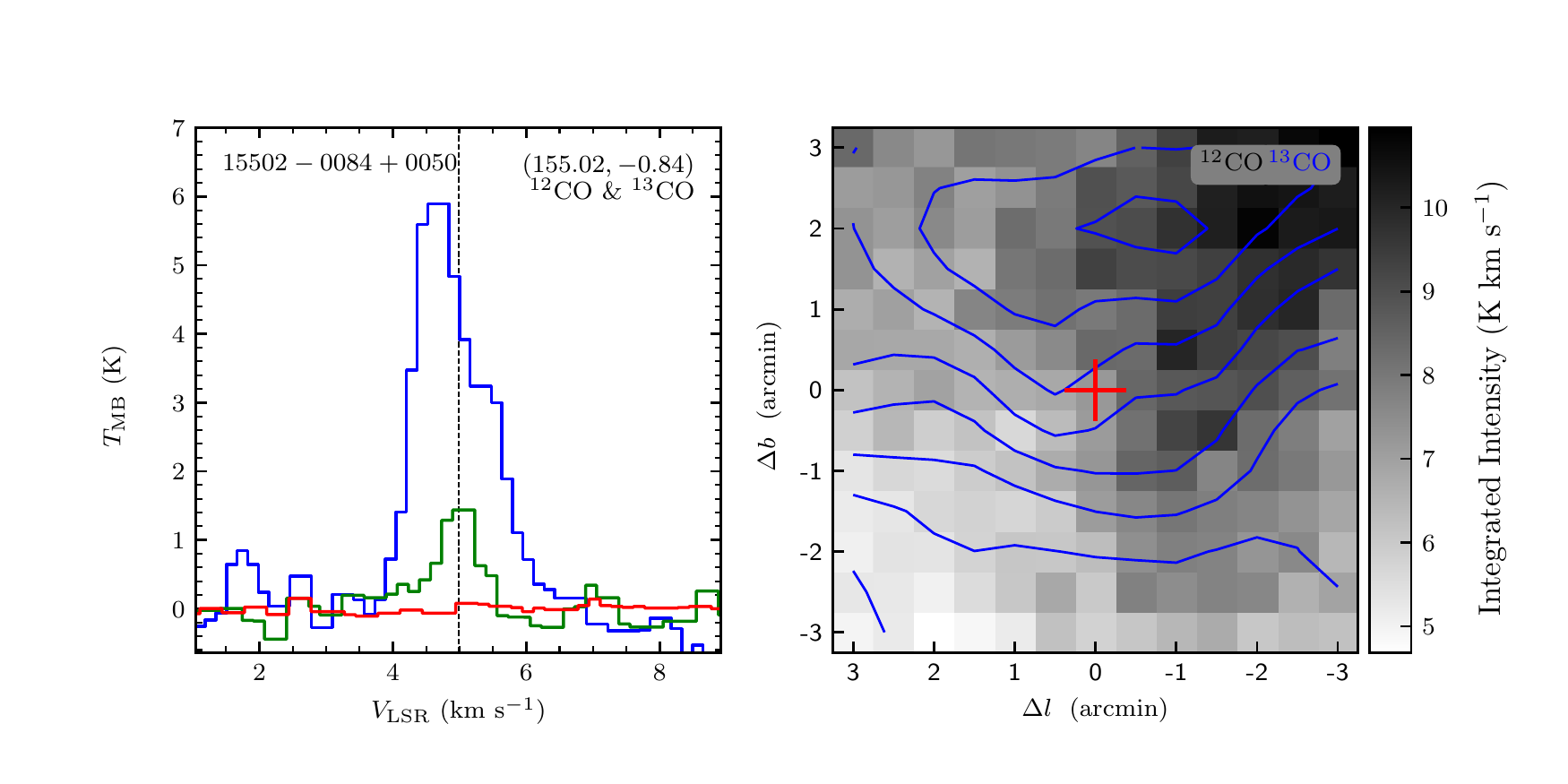}
\vspace{-0.5cm}

\includegraphics[width=9.0cm,angle=0]{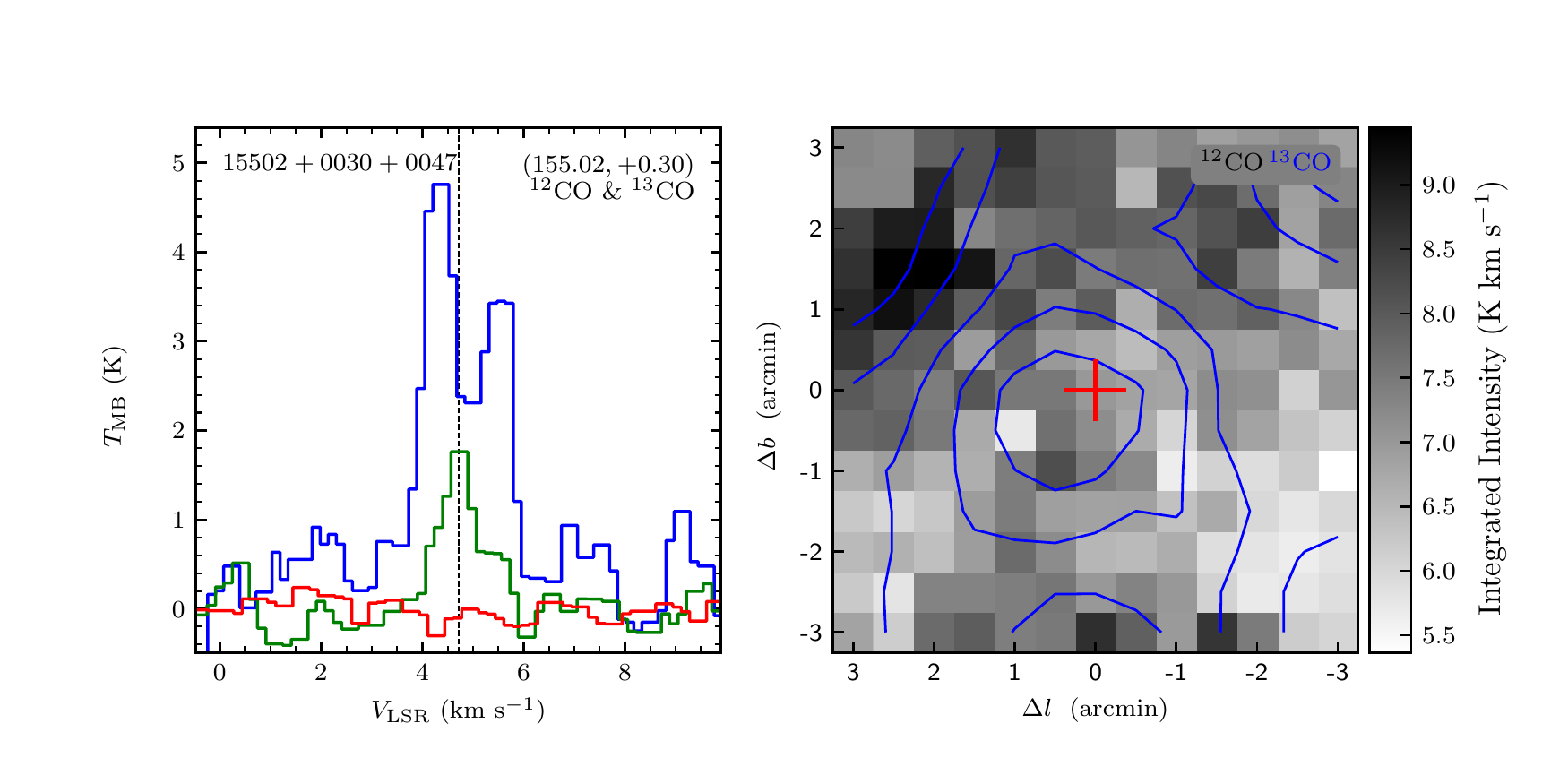}
\includegraphics[width=9.0cm,angle=0]{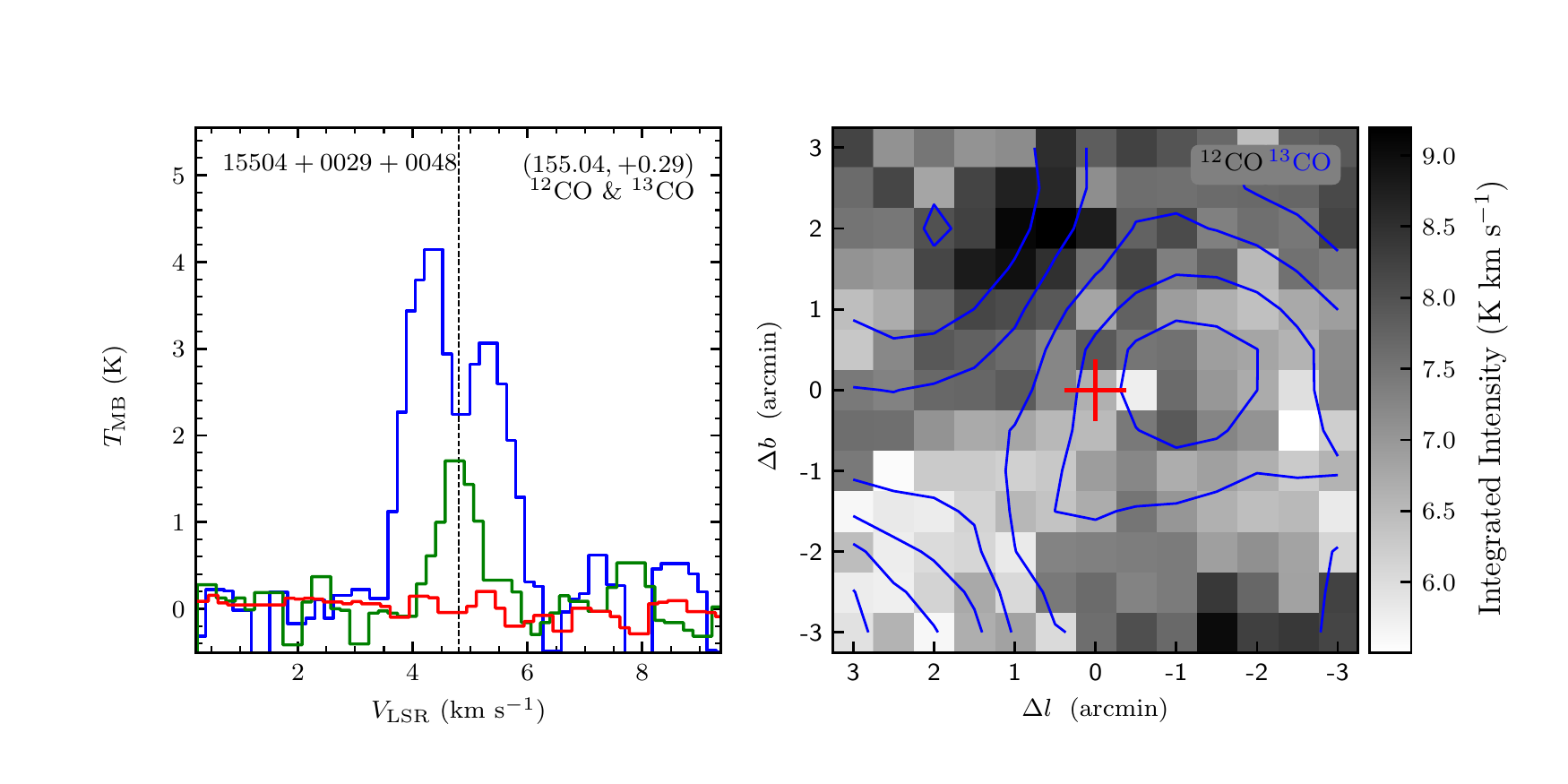}
\vspace{-0.5cm}

\includegraphics[width=9.0cm,angle=0]{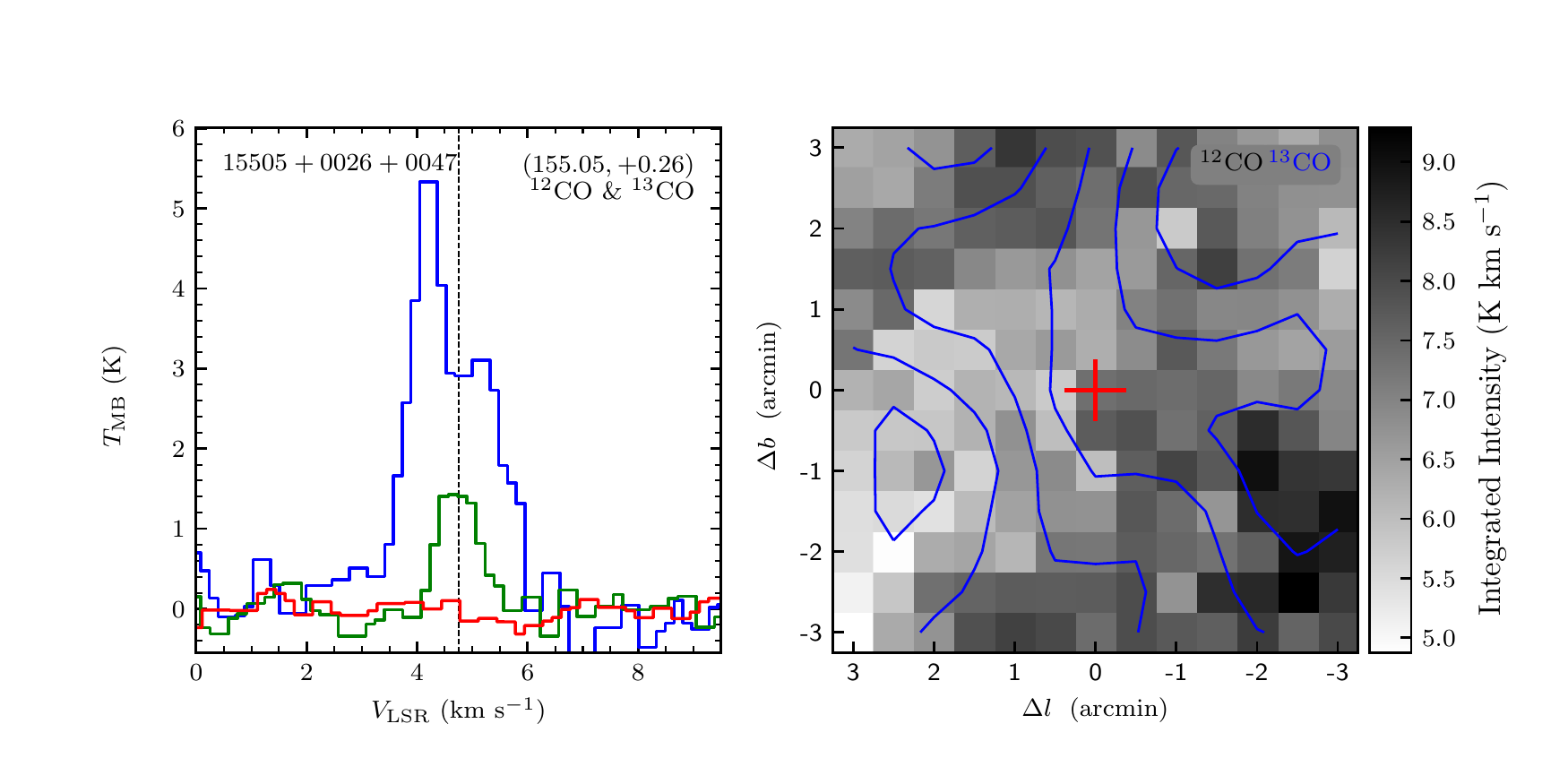}
\includegraphics[width=9.0cm,angle=0]{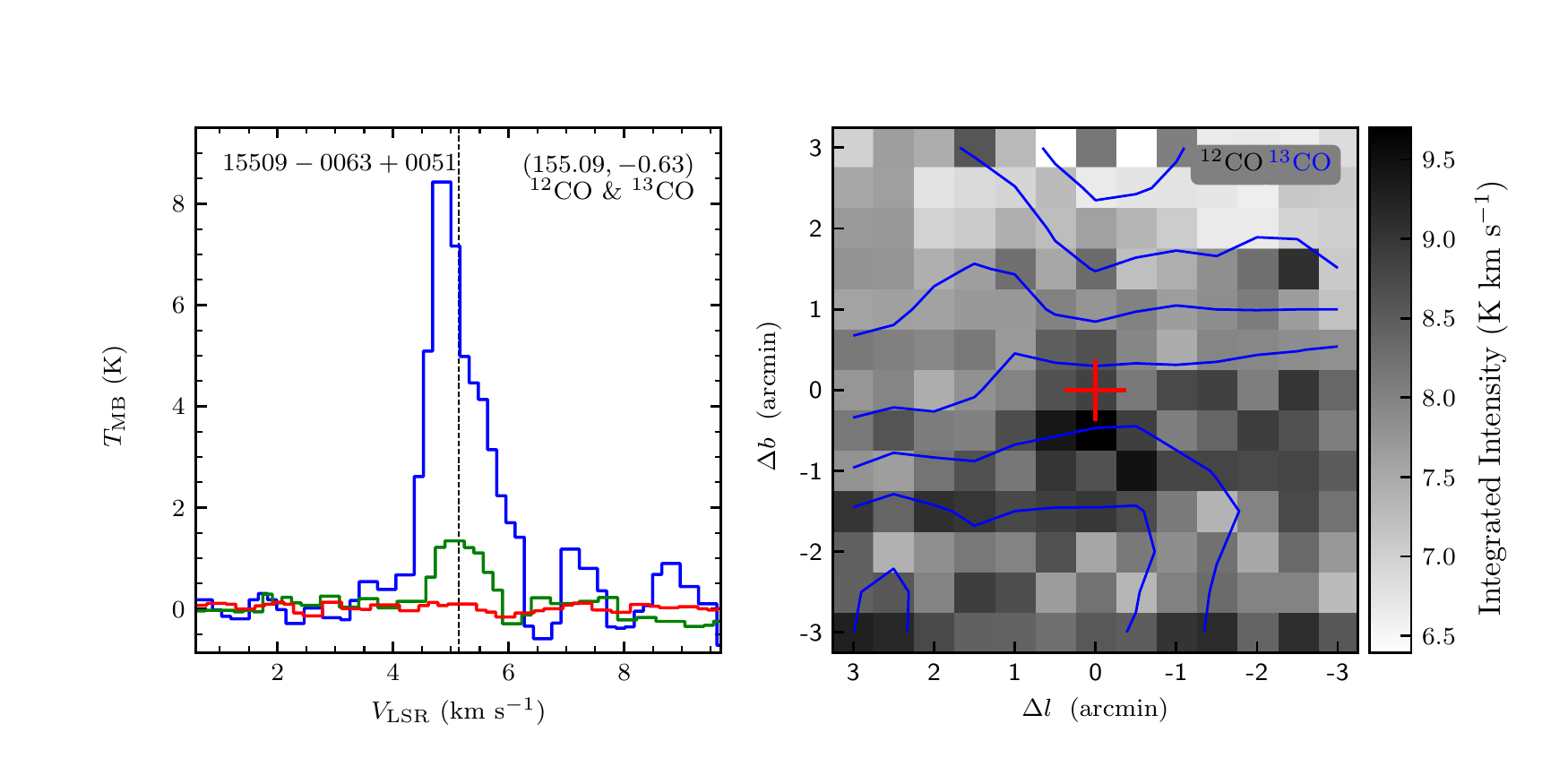}
\end{figure}
\clearpage

\begin{figure}
\includegraphics[width=9.0cm,angle=0]{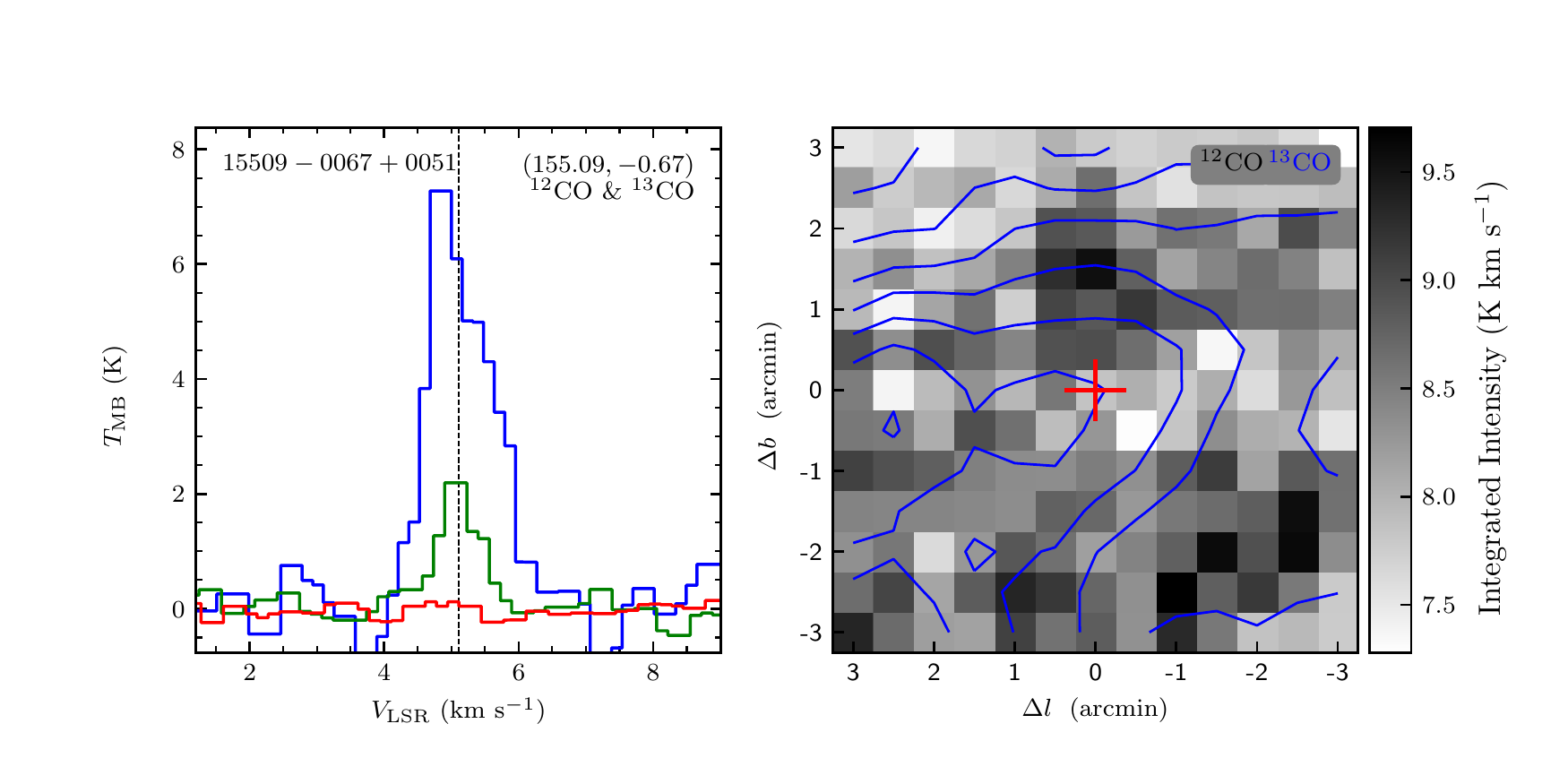}
\includegraphics[width=9.0cm,angle=0]{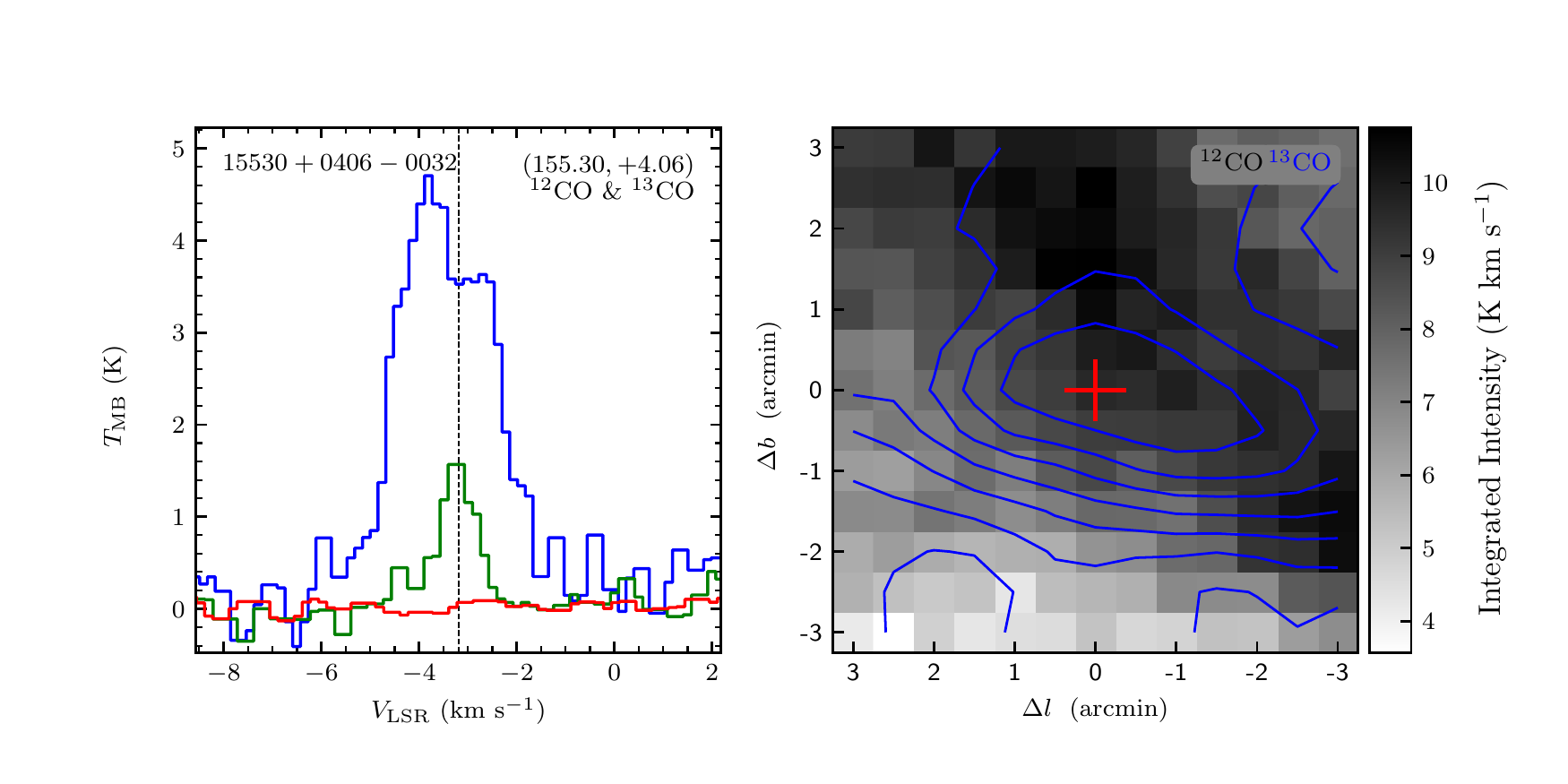}
\vspace{-0.5cm}

\includegraphics[width=9.0cm,angle=0]{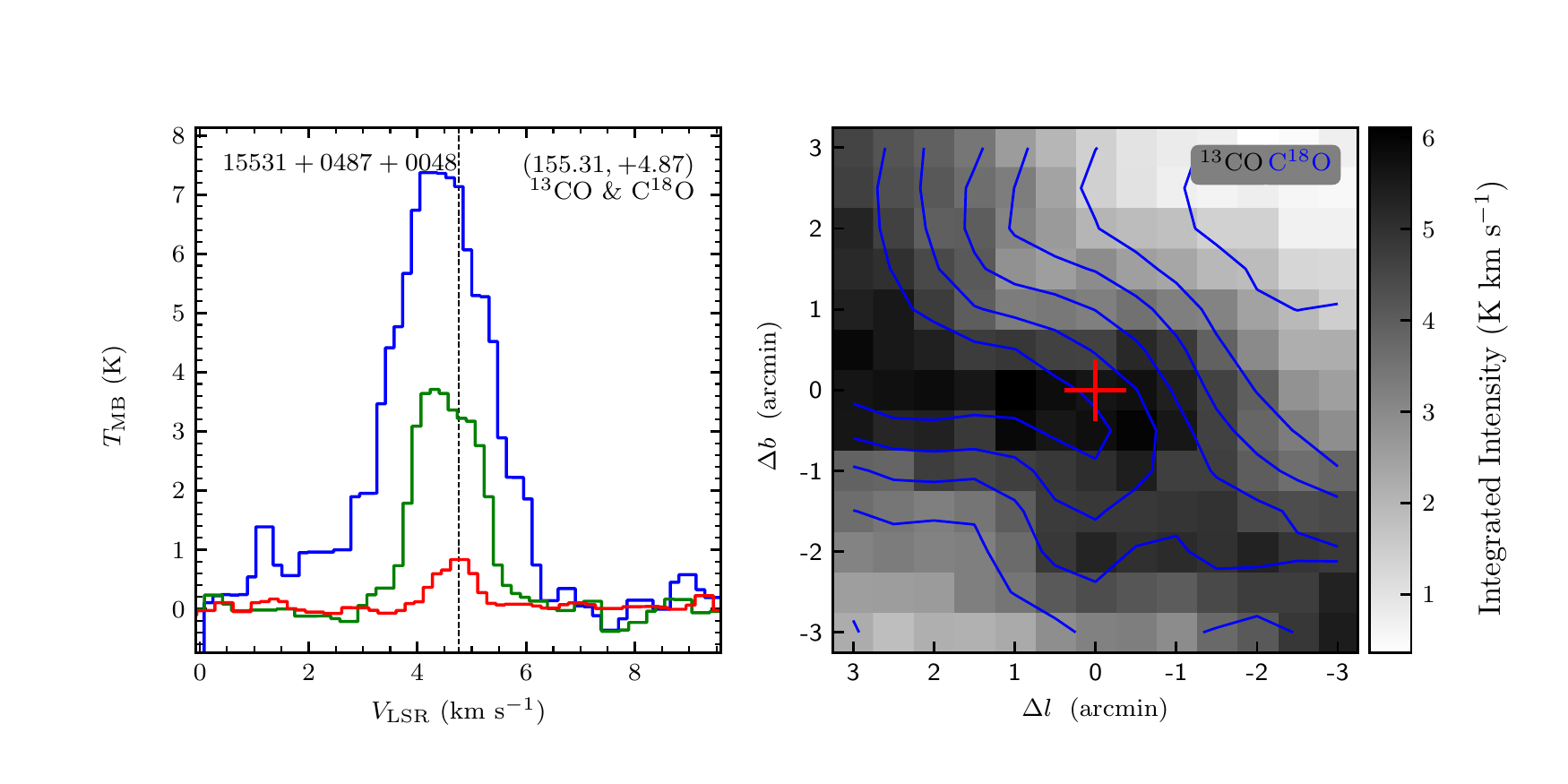}
\includegraphics[width=9.0cm,angle=0]{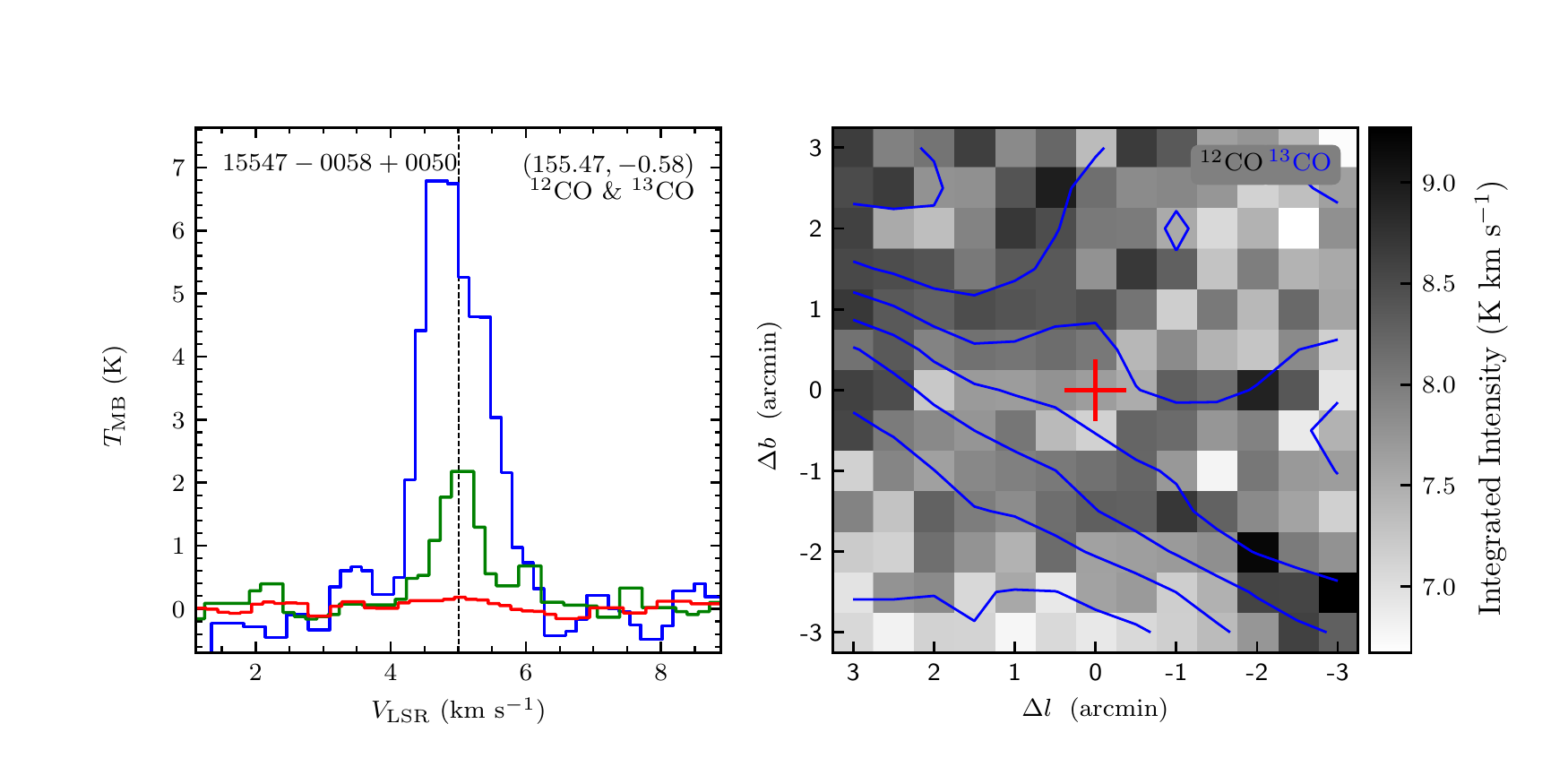}
\vspace{-0.5cm}

\includegraphics[width=9.0cm,angle=0]{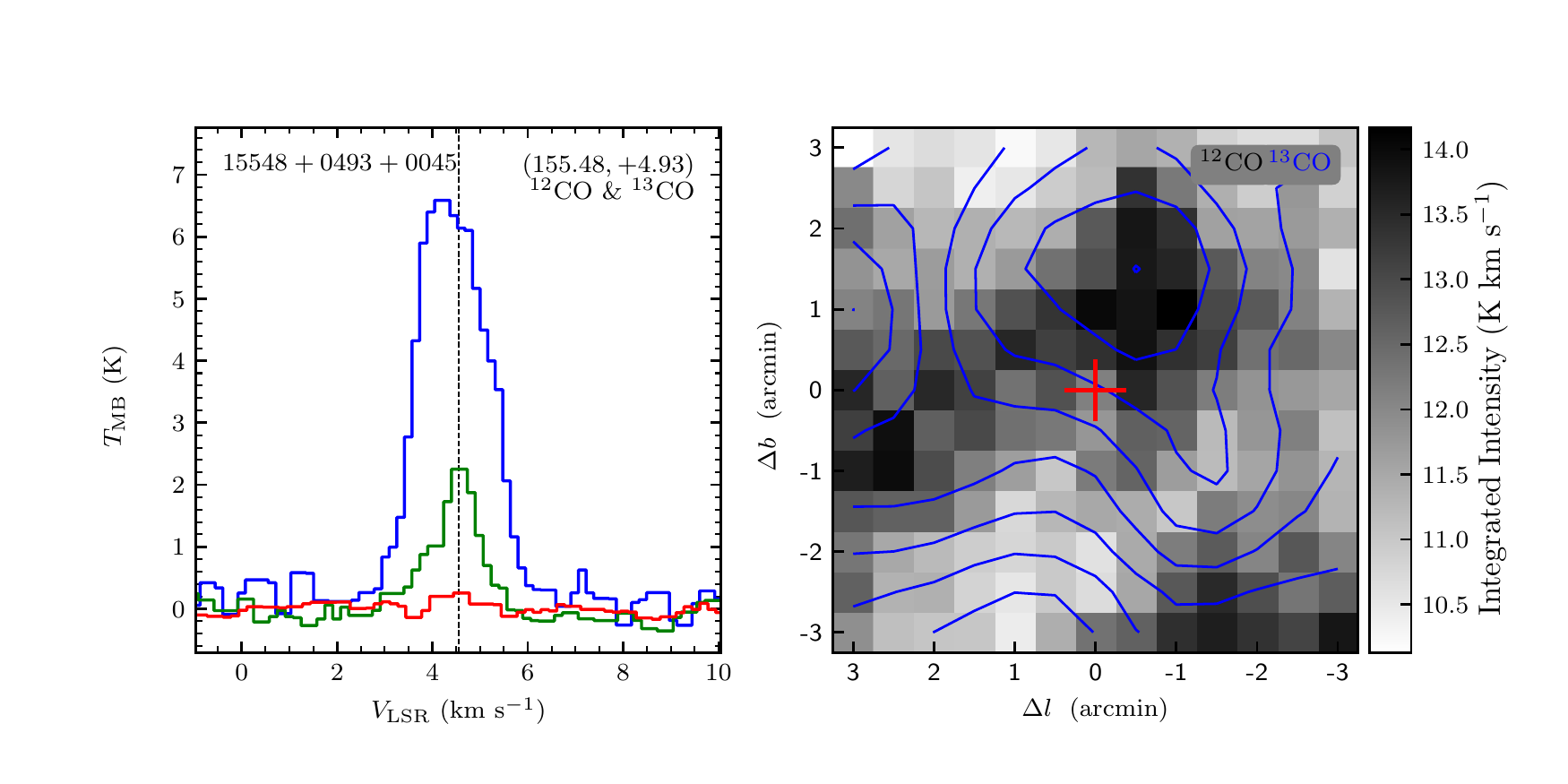}
\includegraphics[width=9.0cm,angle=0]{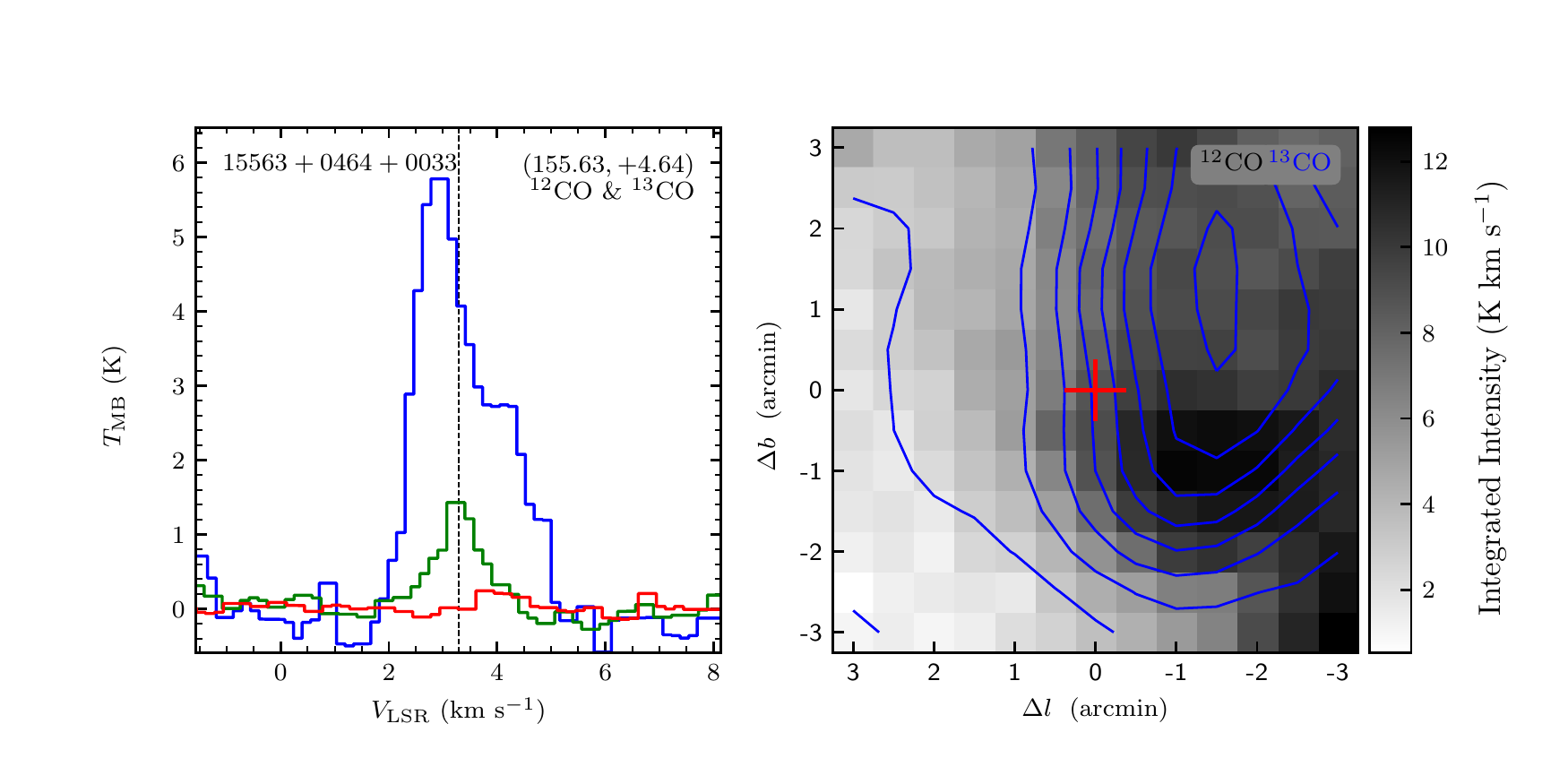}
\vspace{-0.5cm}

\includegraphics[width=9.0cm,angle=0]{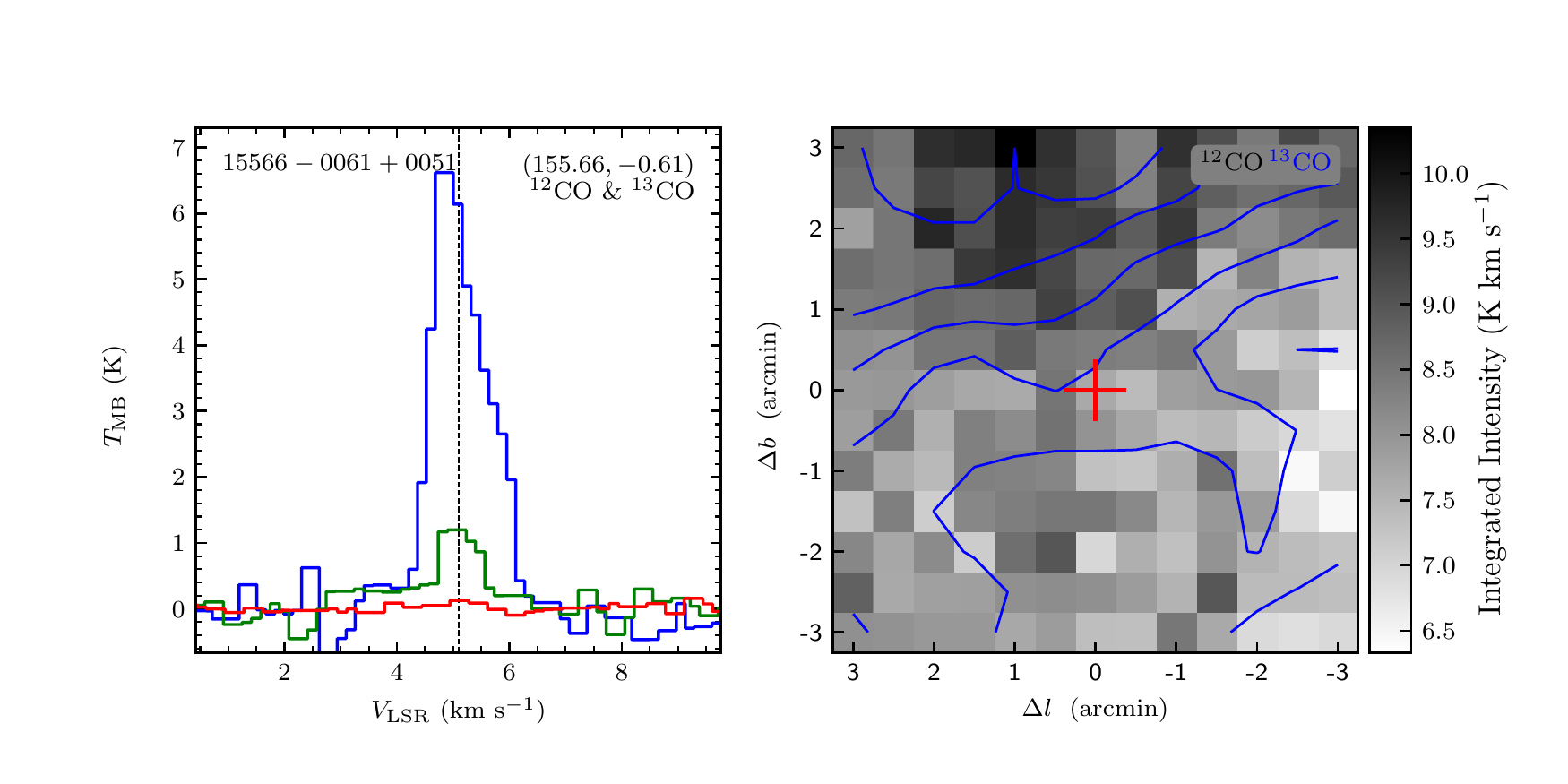}
\includegraphics[width=9.0cm,angle=0]{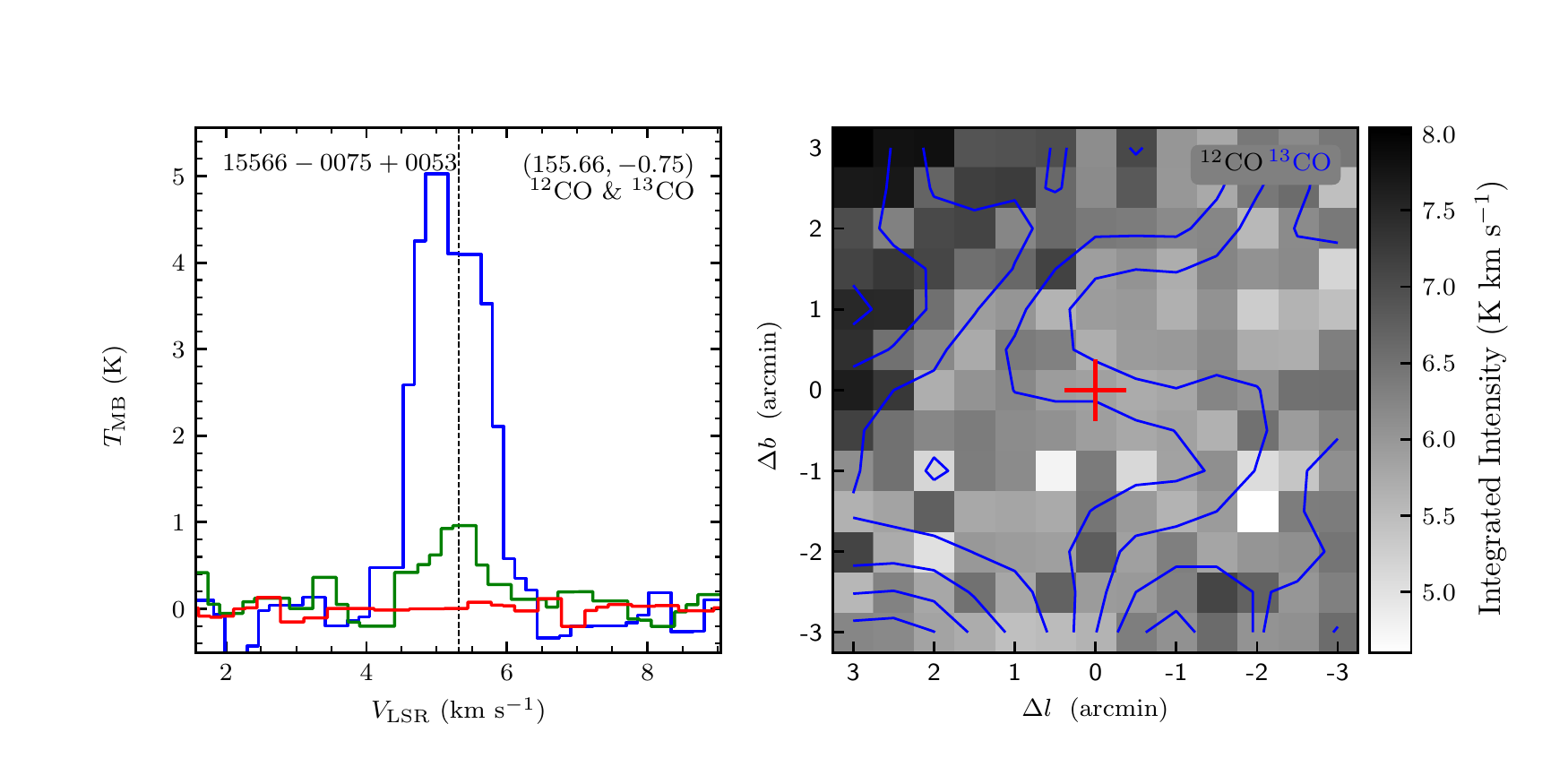}
\vspace{-0.5cm}

\includegraphics[width=9.0cm,angle=0]{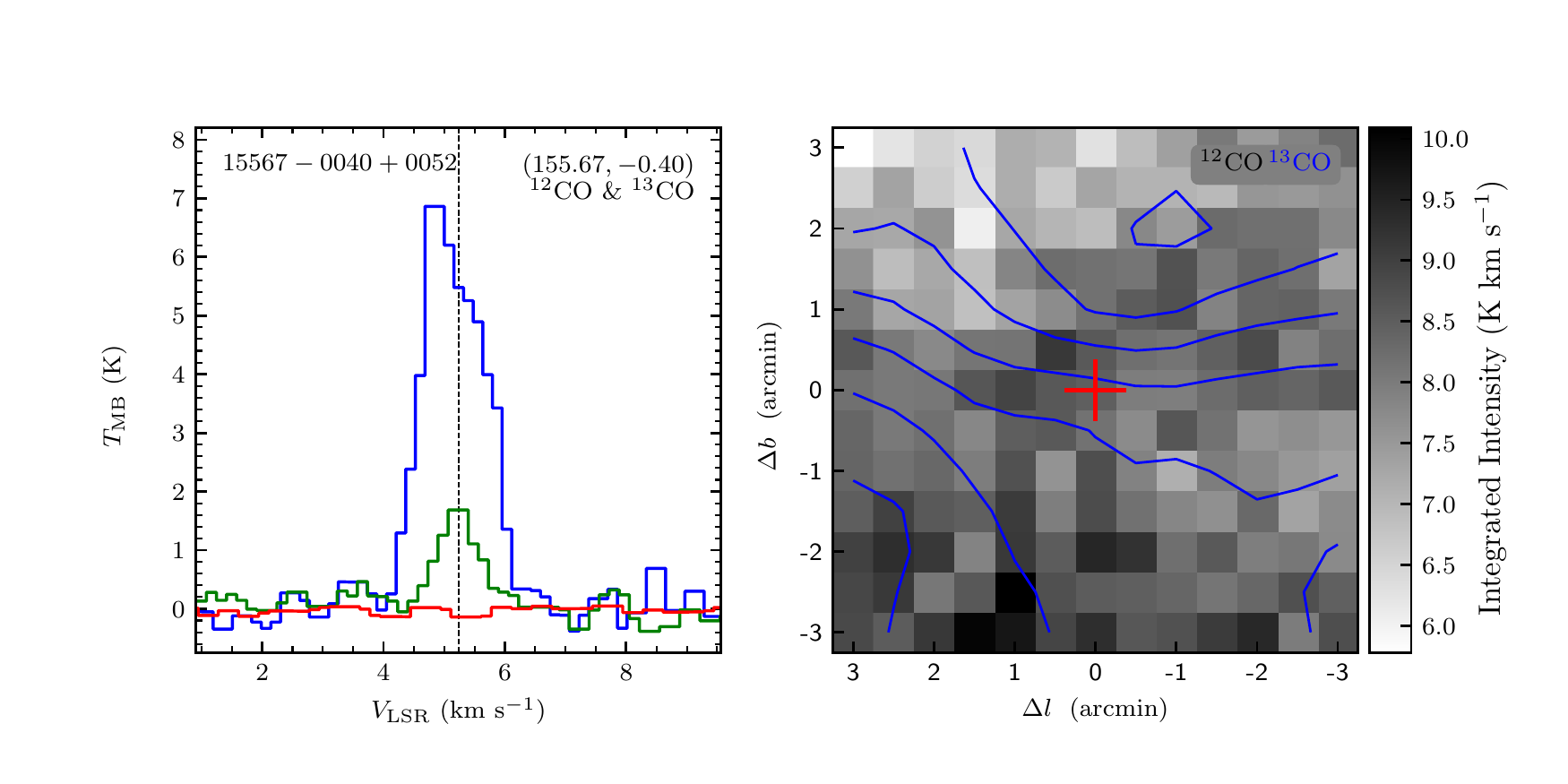}
\includegraphics[width=9.0cm,angle=0]{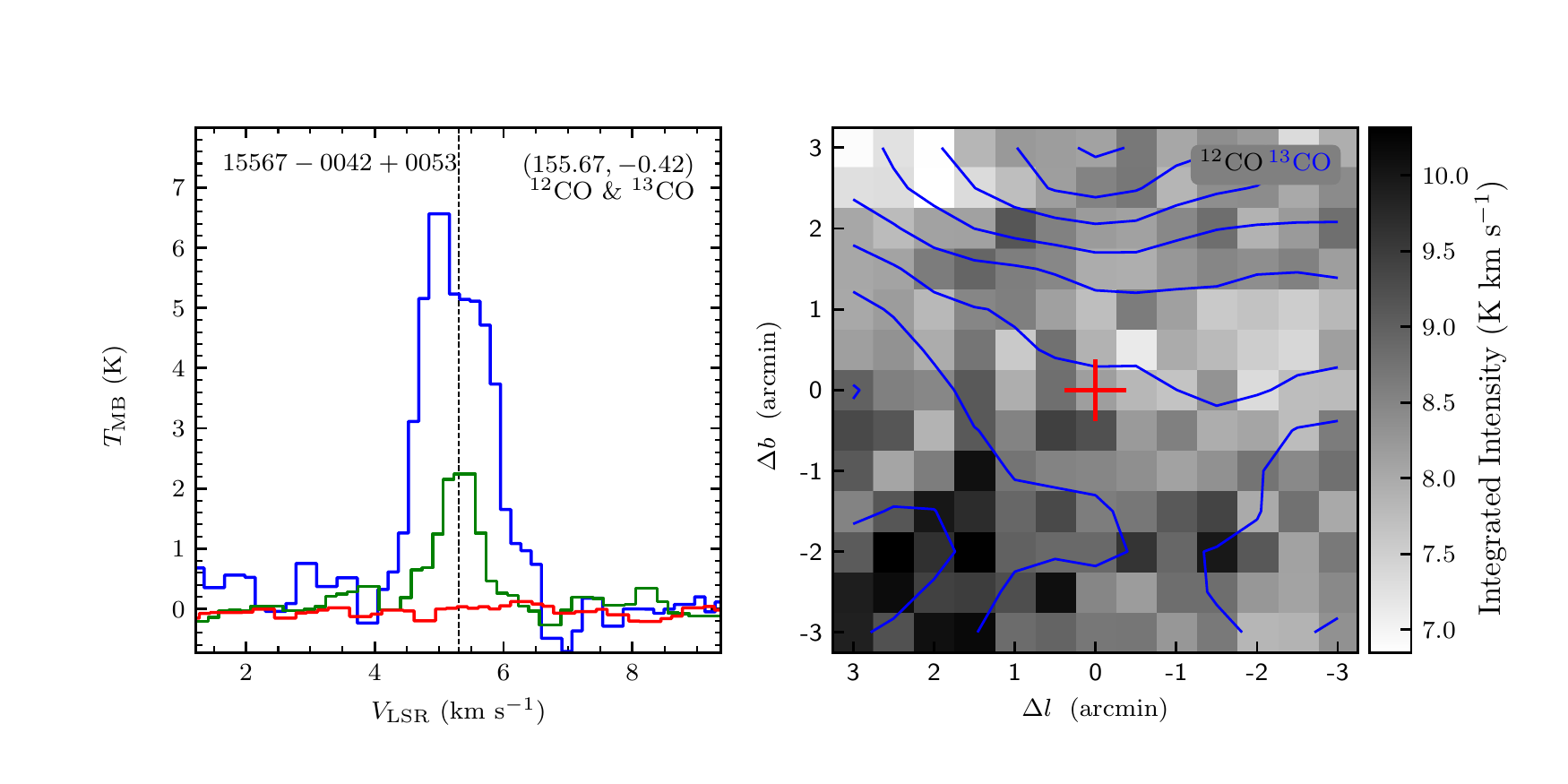}
\end{figure}
\clearpage

\begin{figure}
\includegraphics[width=9.0cm,angle=0]{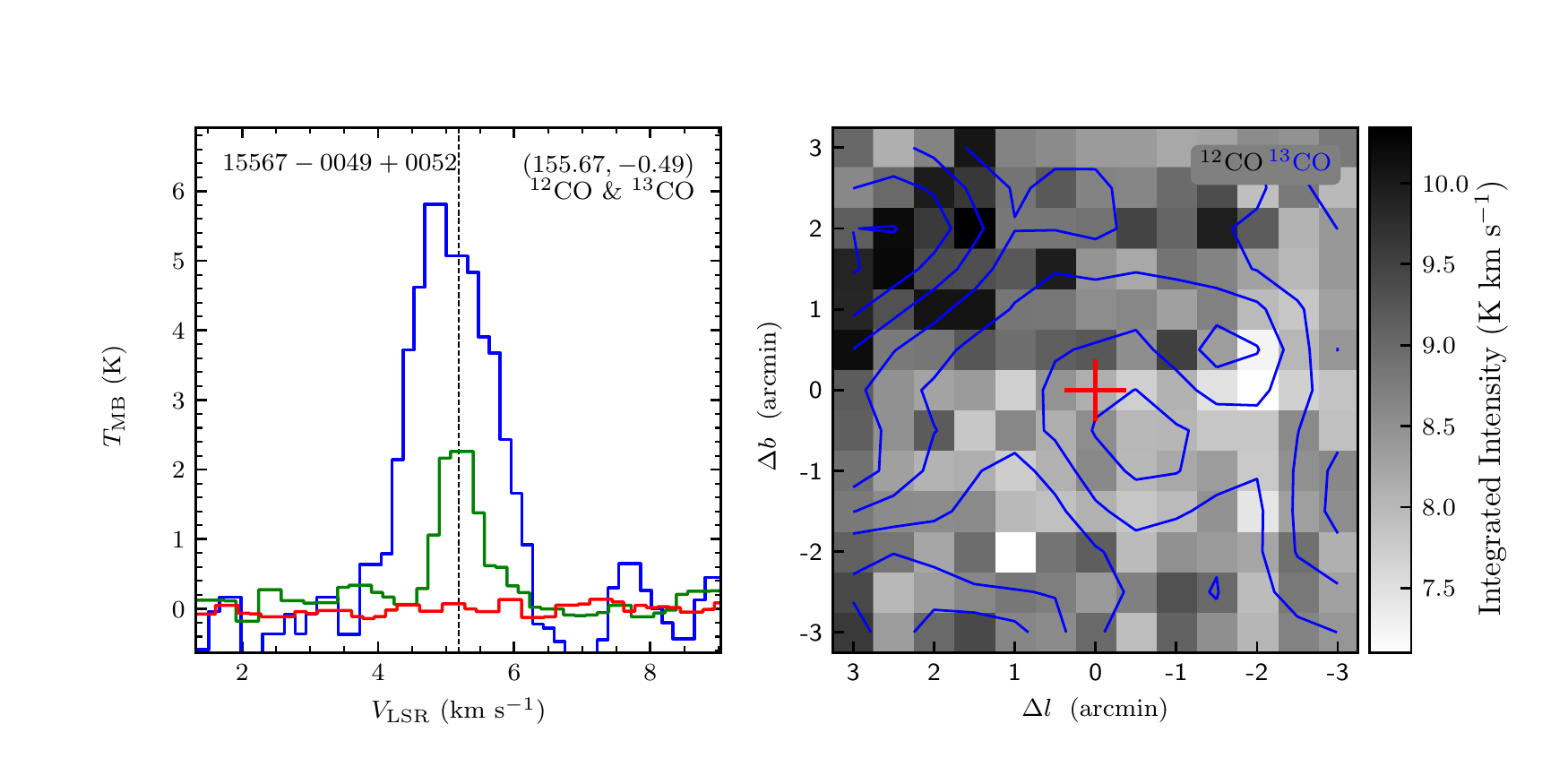}
\includegraphics[width=9.0cm,angle=0]{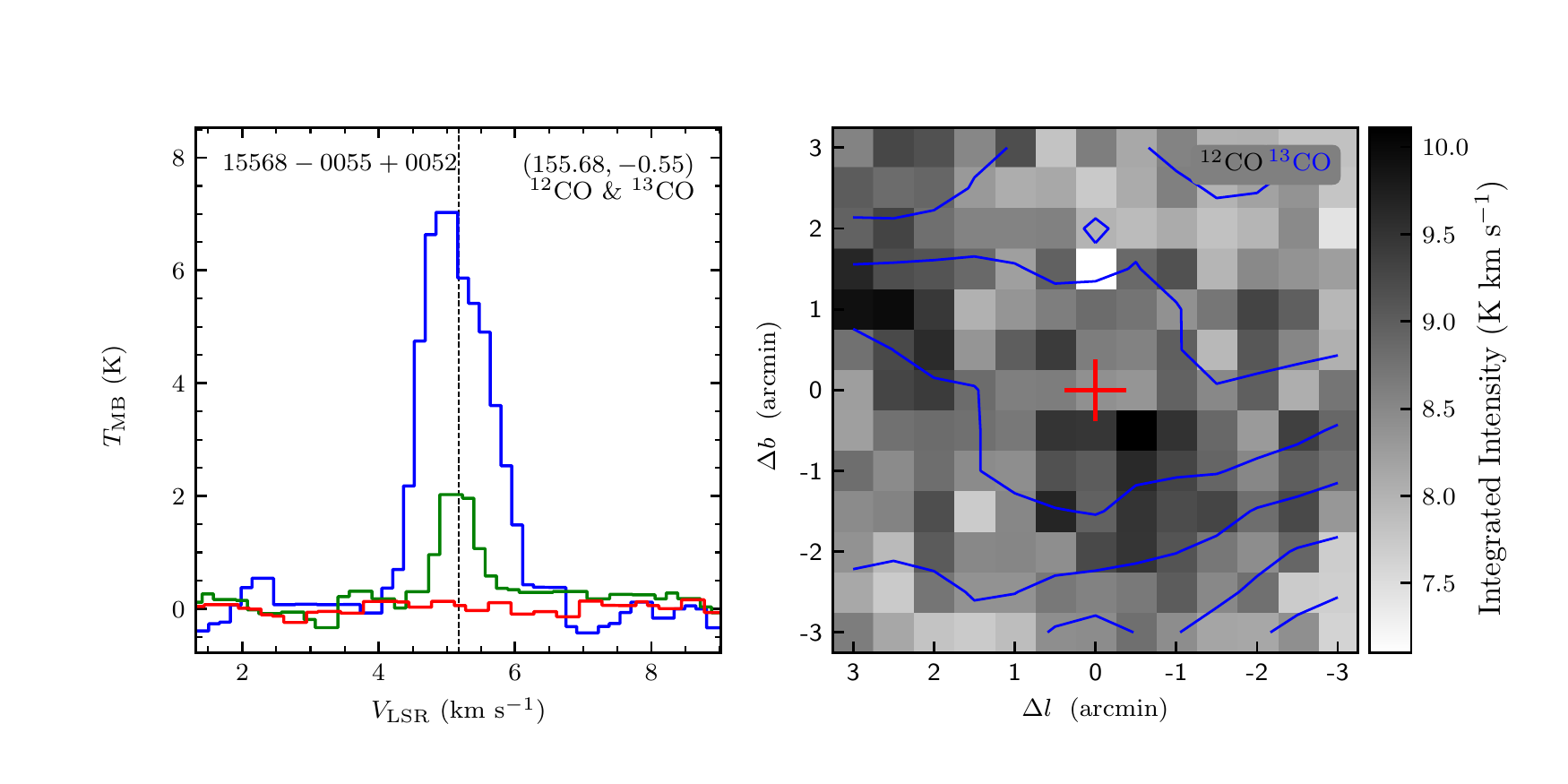}
\vspace{-0.5cm}

\includegraphics[width=9.0cm,angle=0]{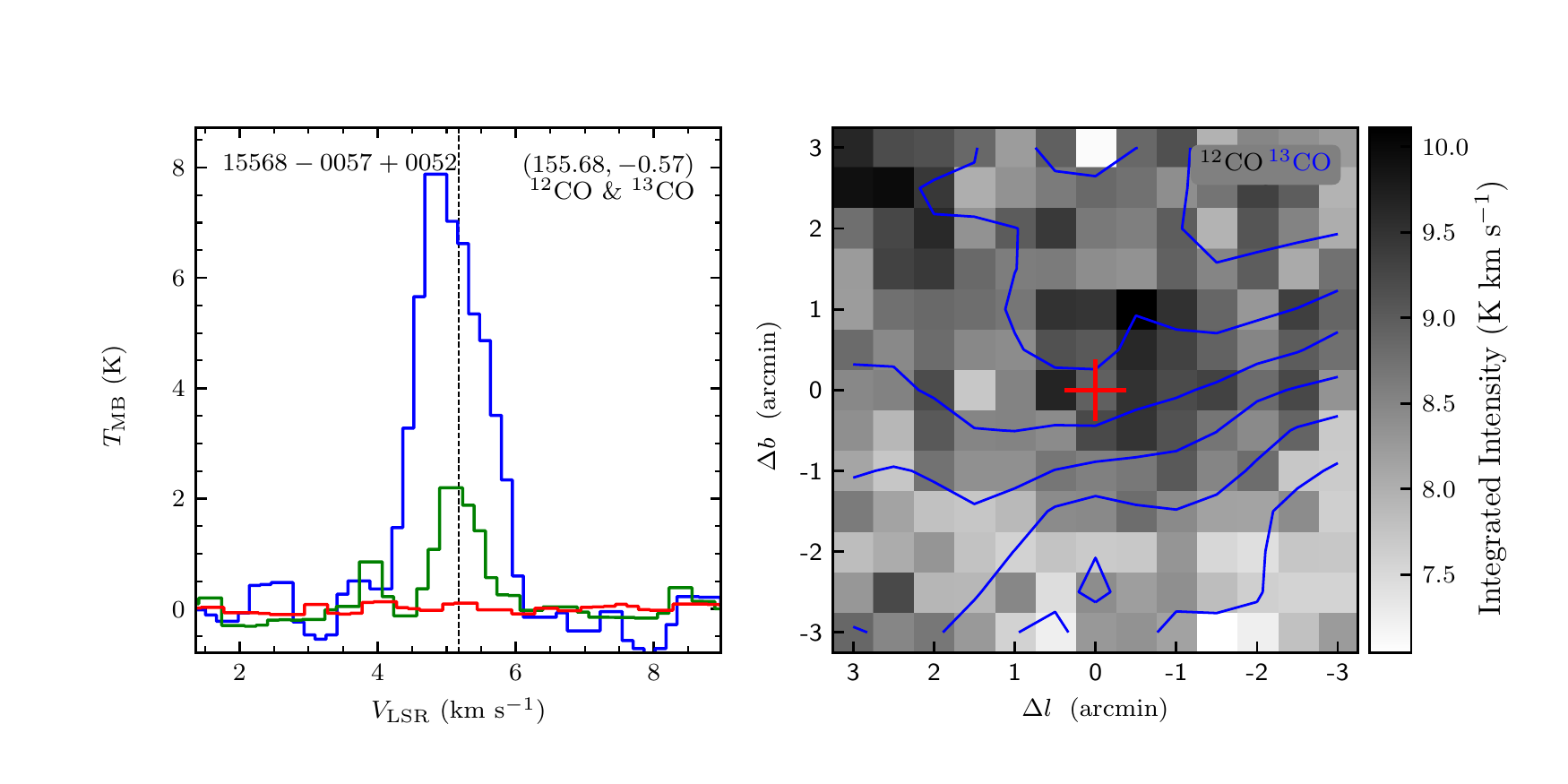}
\includegraphics[width=9.0cm,angle=0]{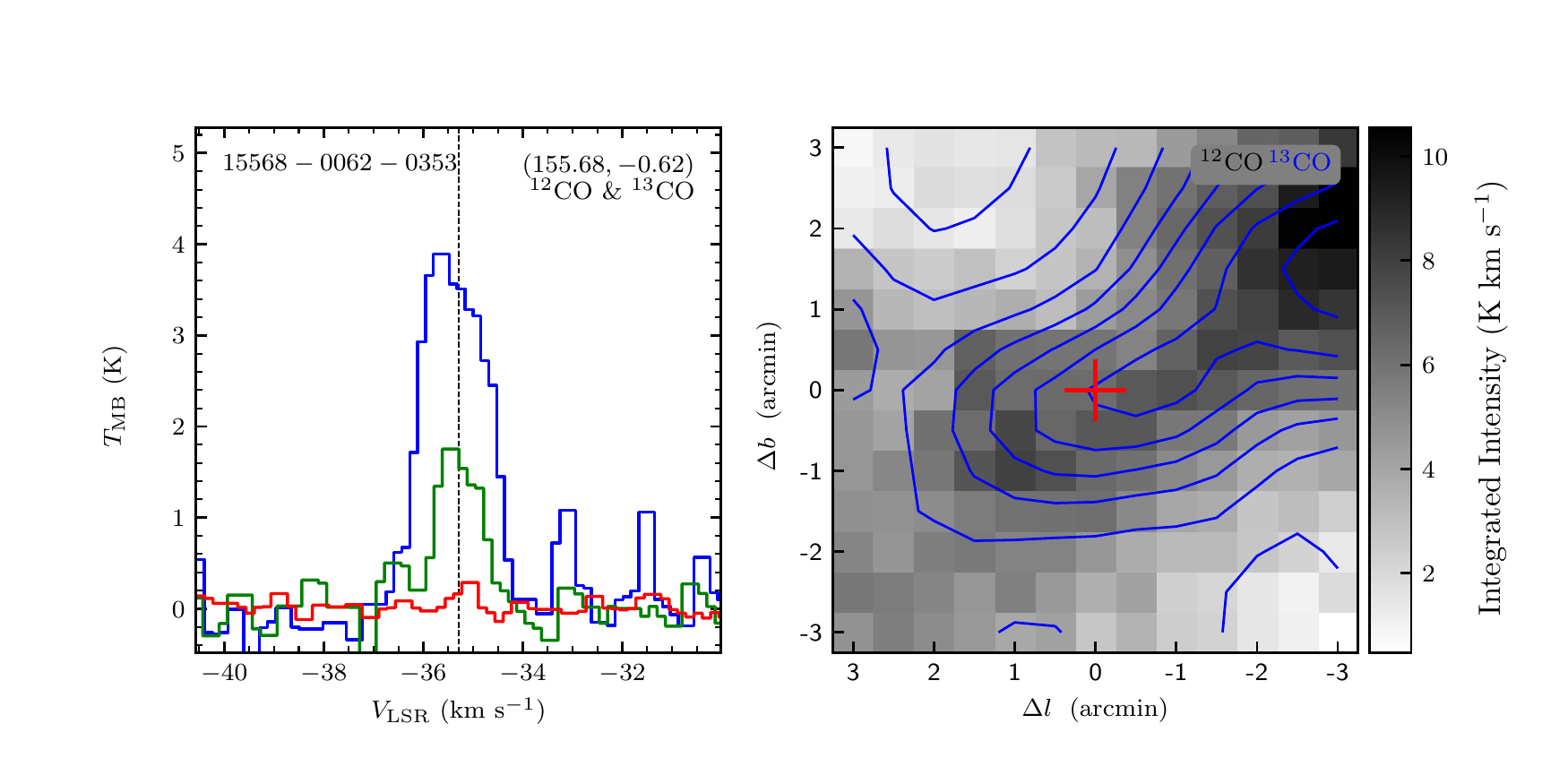}
\vspace{-0.5cm}

\includegraphics[width=9.0cm,angle=0]{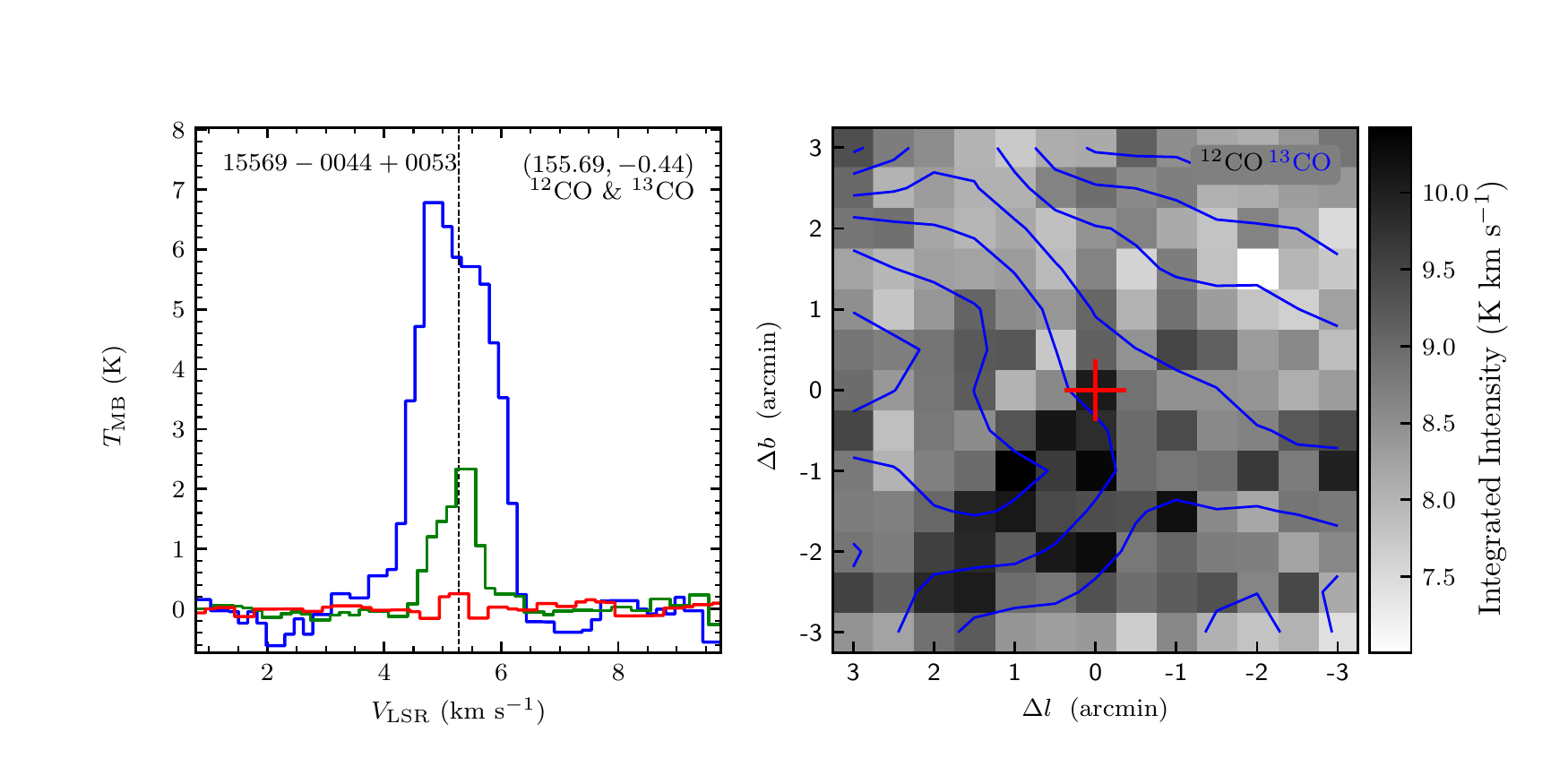}
\includegraphics[width=9.0cm,angle=0]{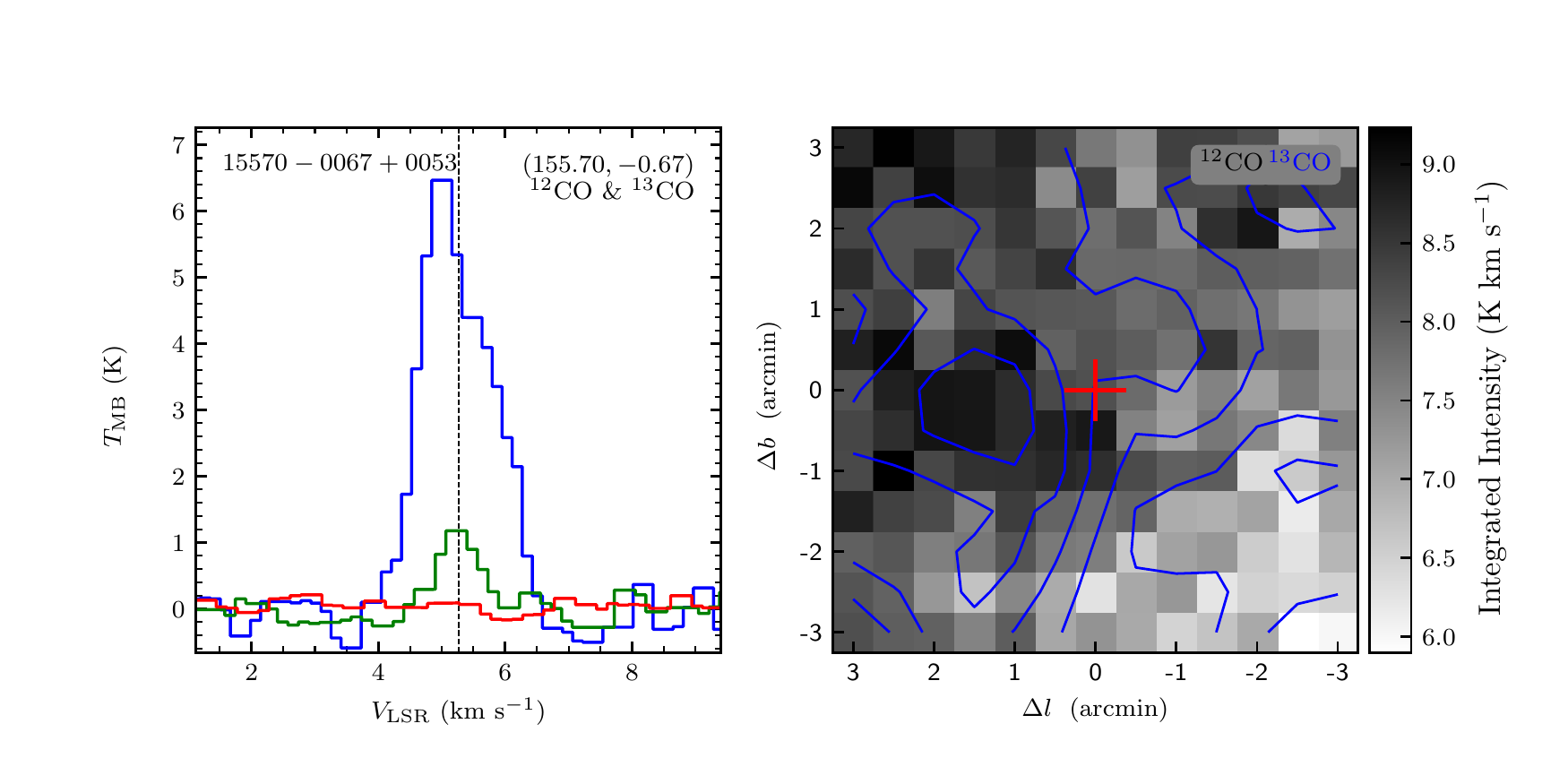}
\vspace{-0.5cm}

\includegraphics[width=9.0cm,angle=0]{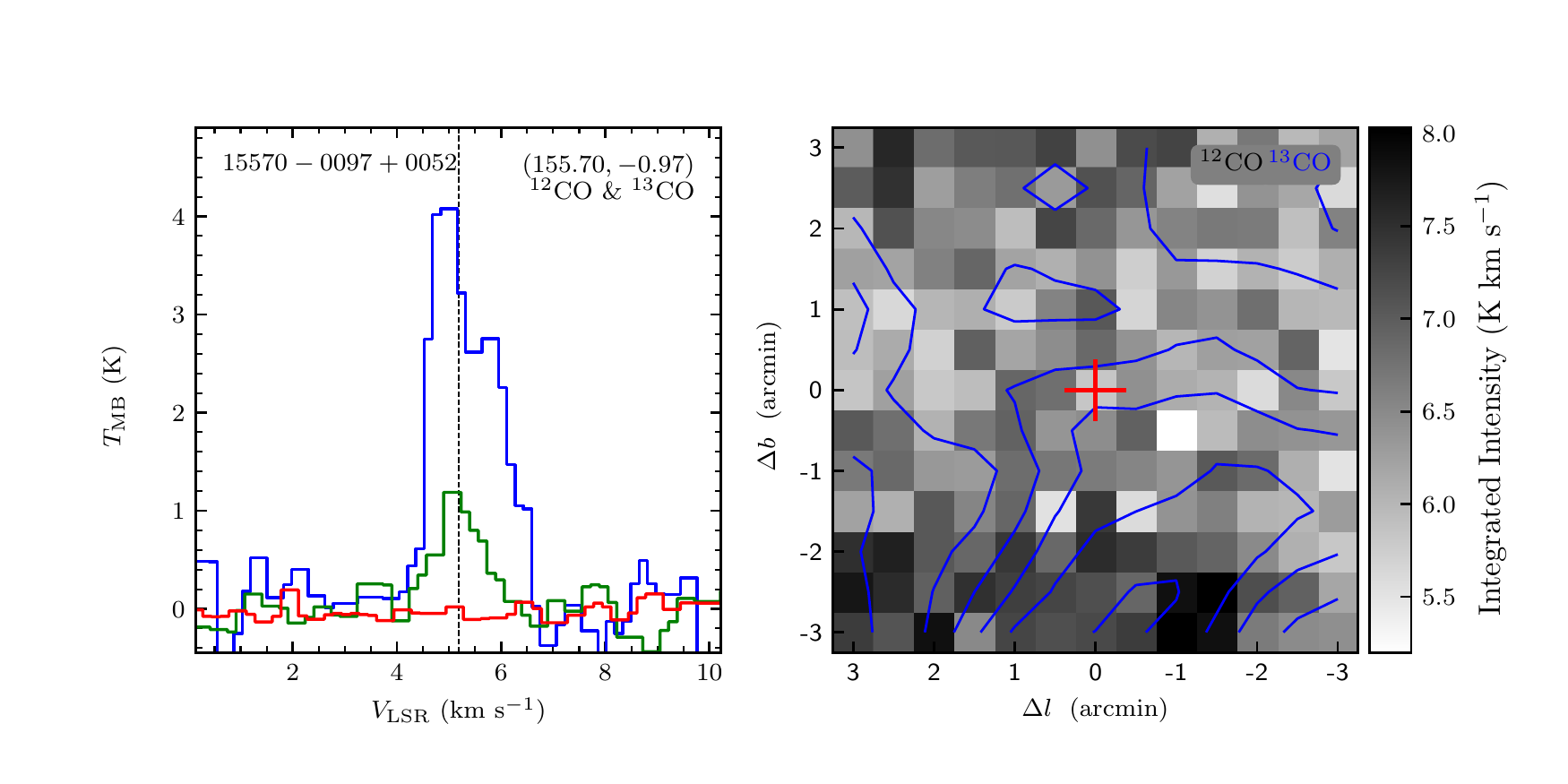}
\includegraphics[width=9.0cm,angle=0]{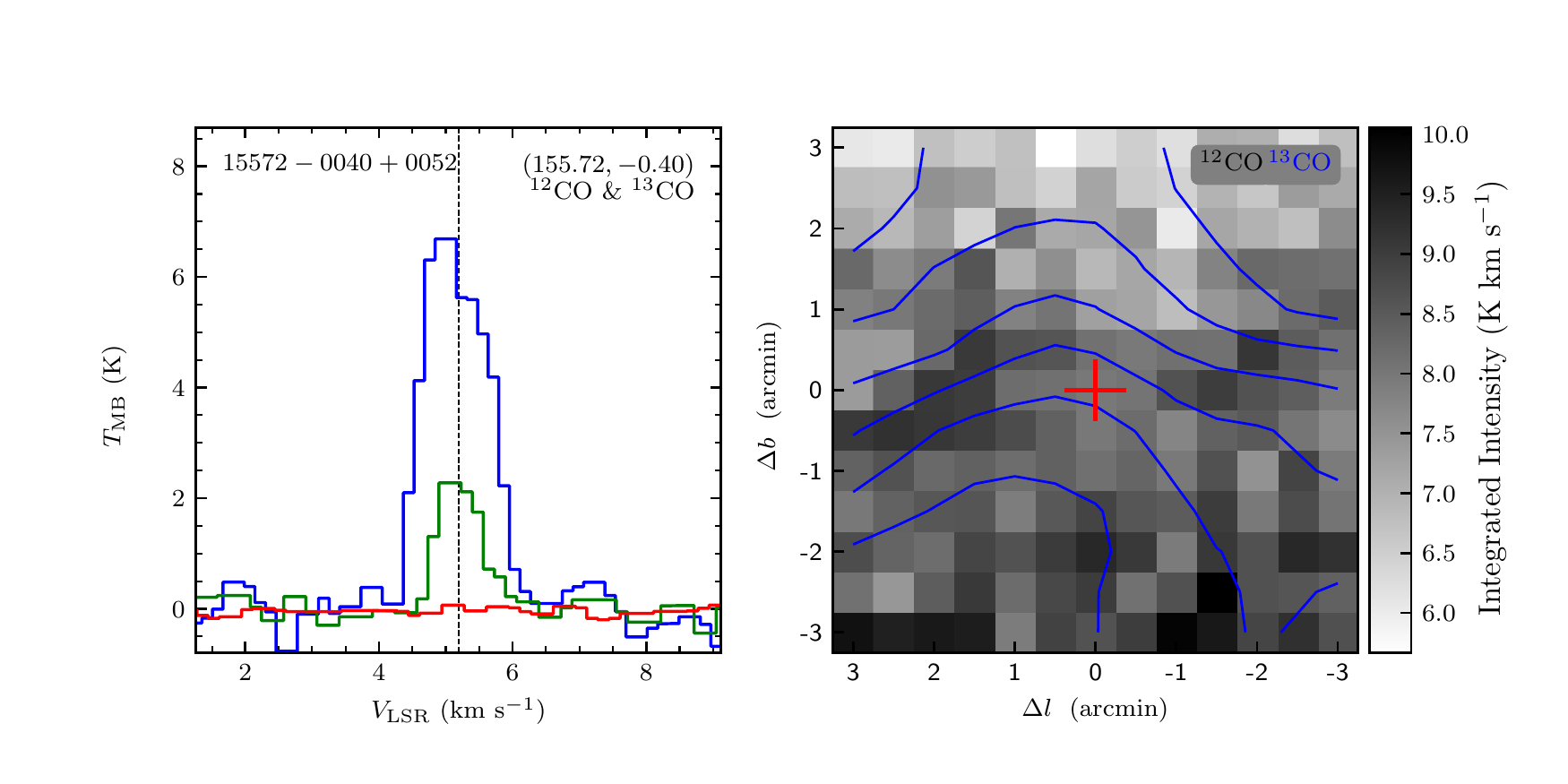}
\vspace{-0.5cm}

\includegraphics[width=9.0cm,angle=0]{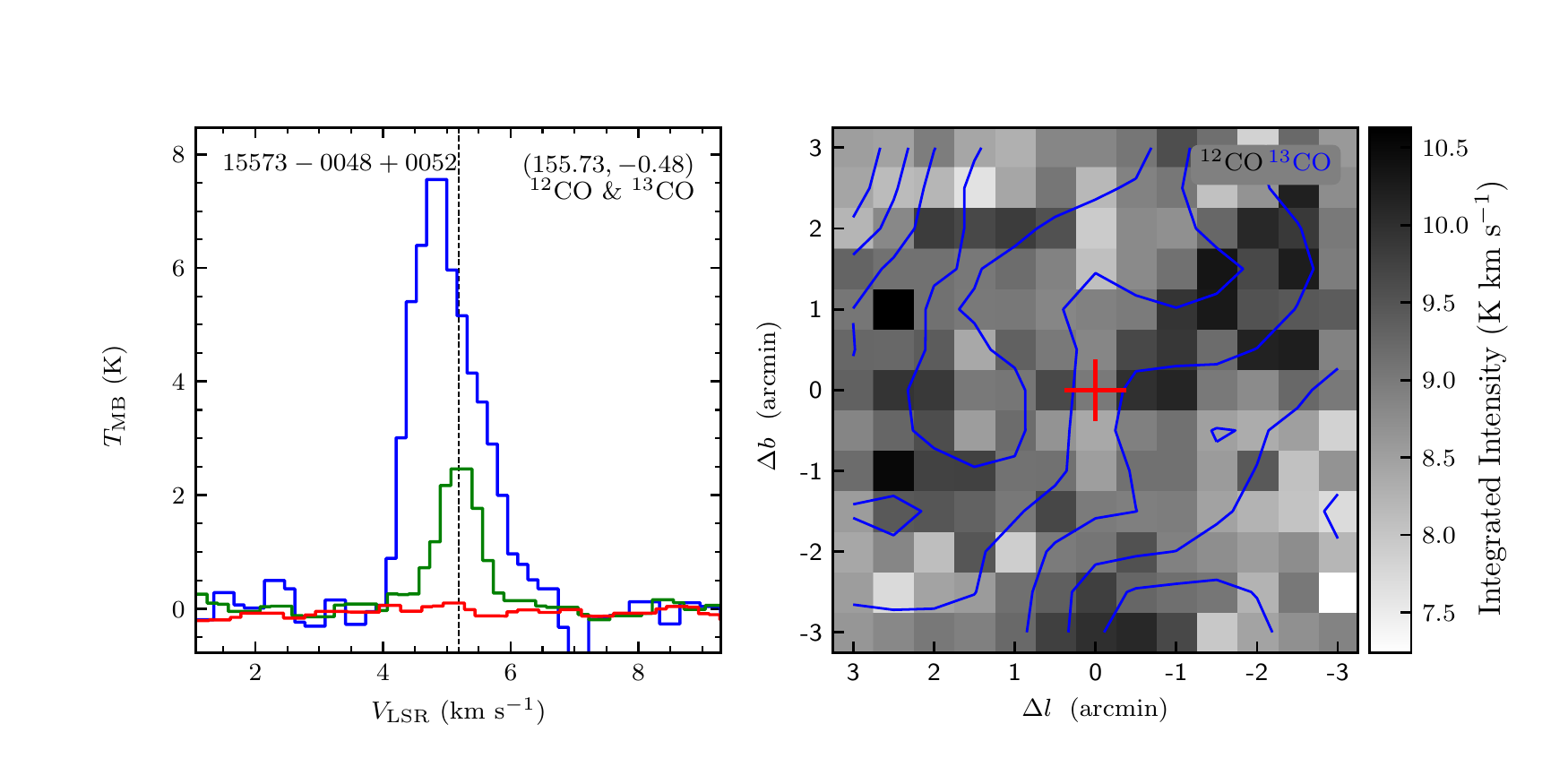}
\includegraphics[width=9.0cm,angle=0]{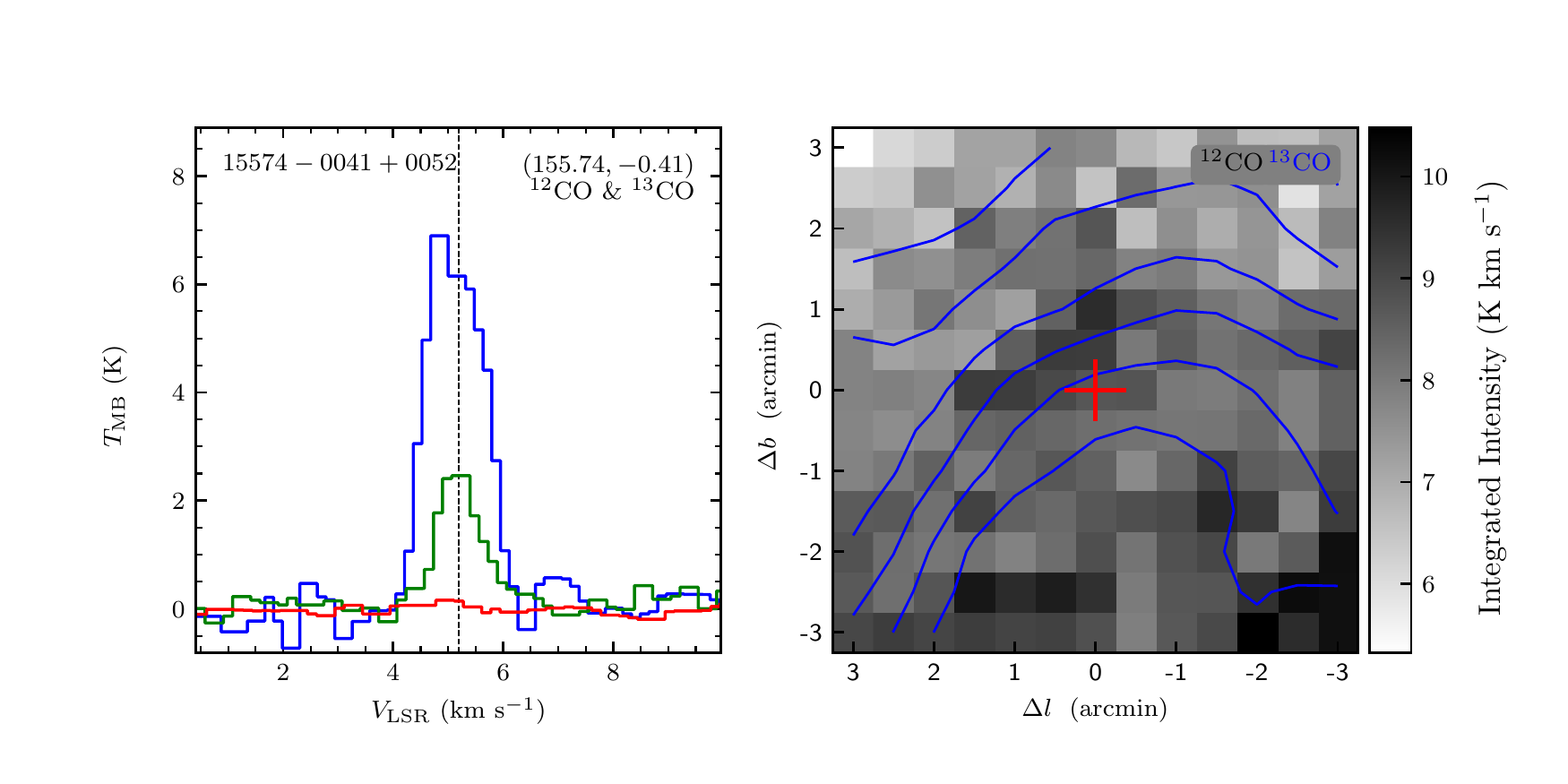}
\end{figure}
\clearpage

\begin{figure}
\includegraphics[width=9.0cm,angle=0]{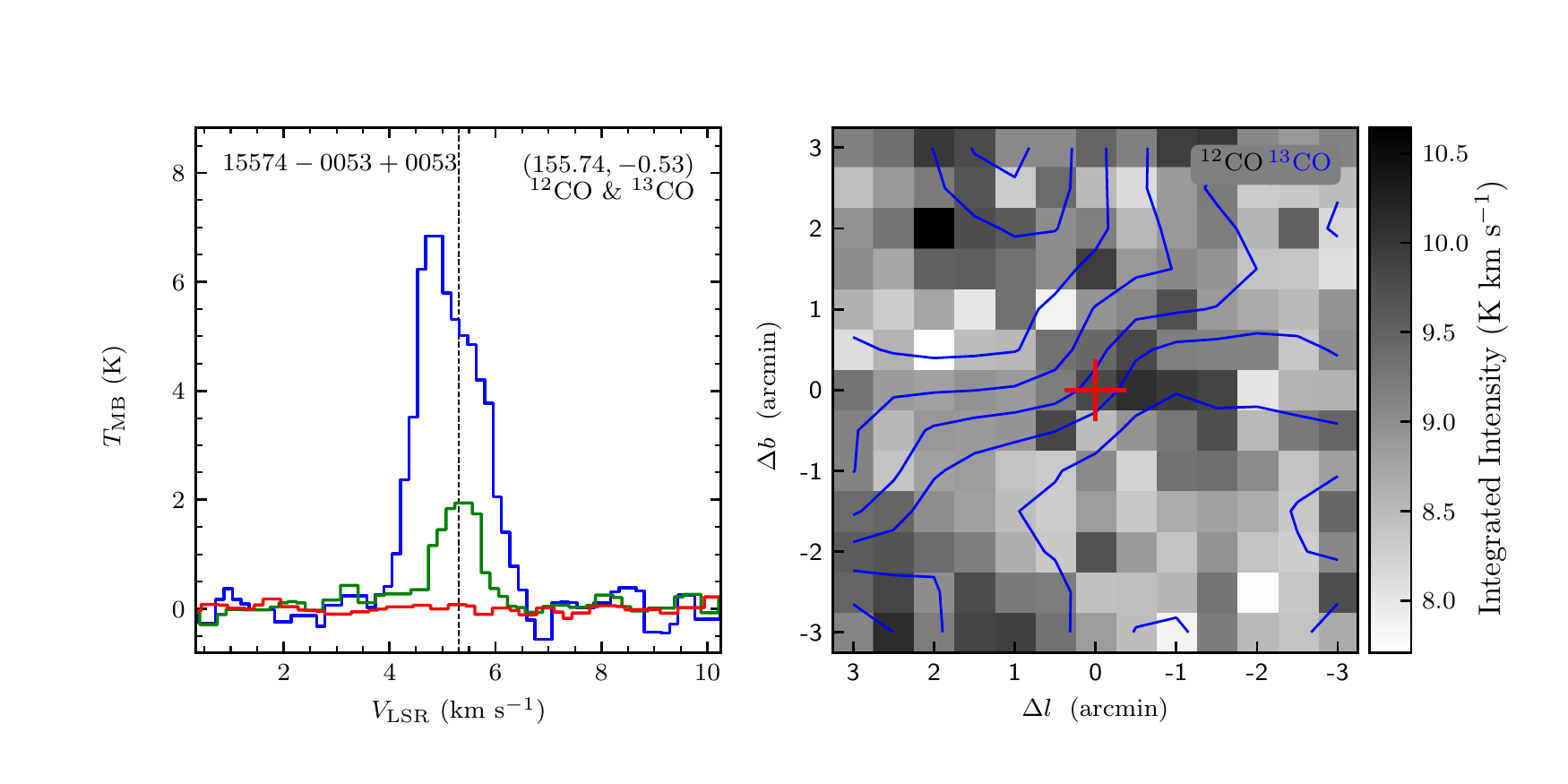}
\includegraphics[width=9.0cm,angle=0]{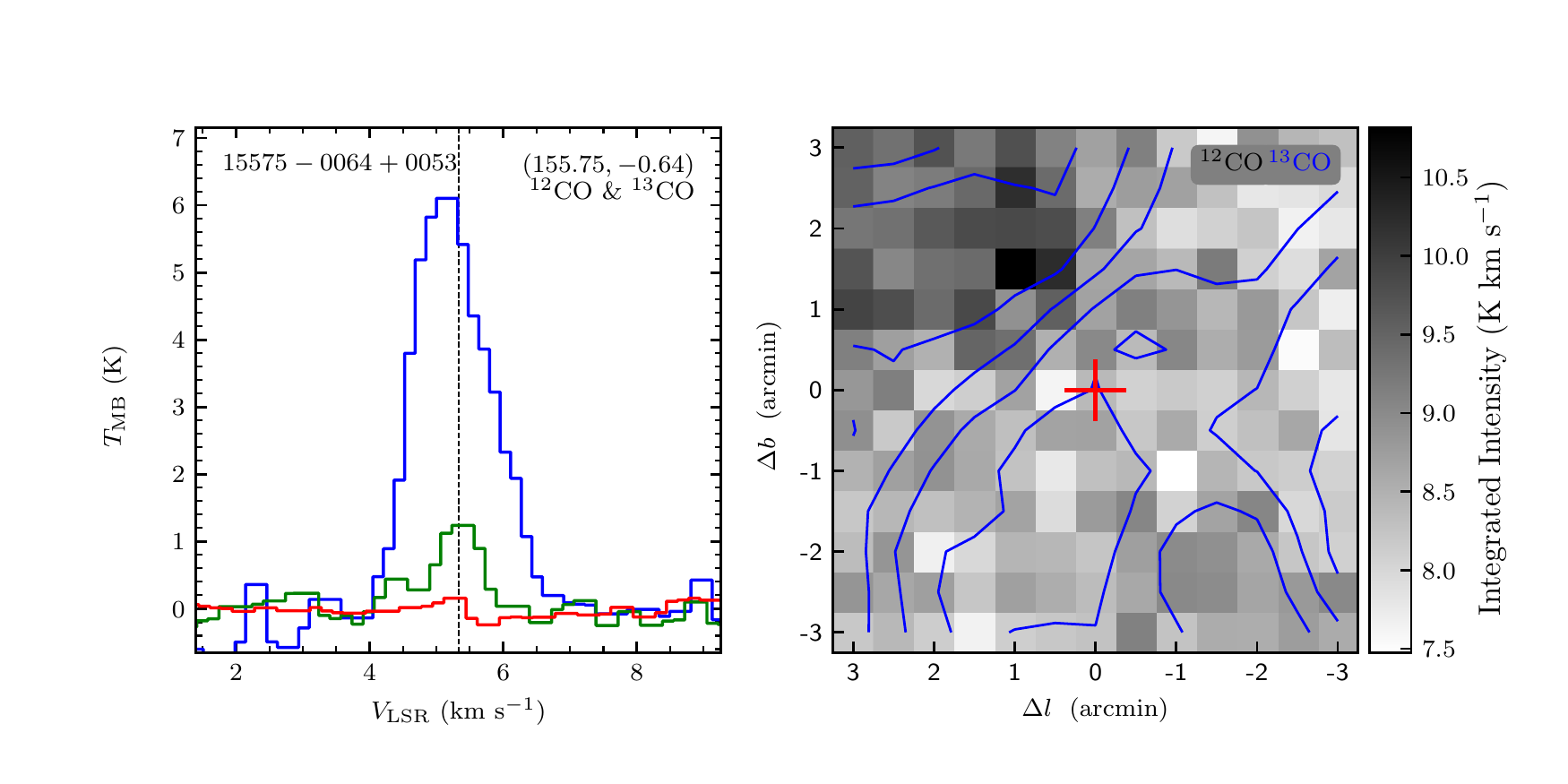}
\vspace{-0.5cm}

\includegraphics[width=9.0cm,angle=0]{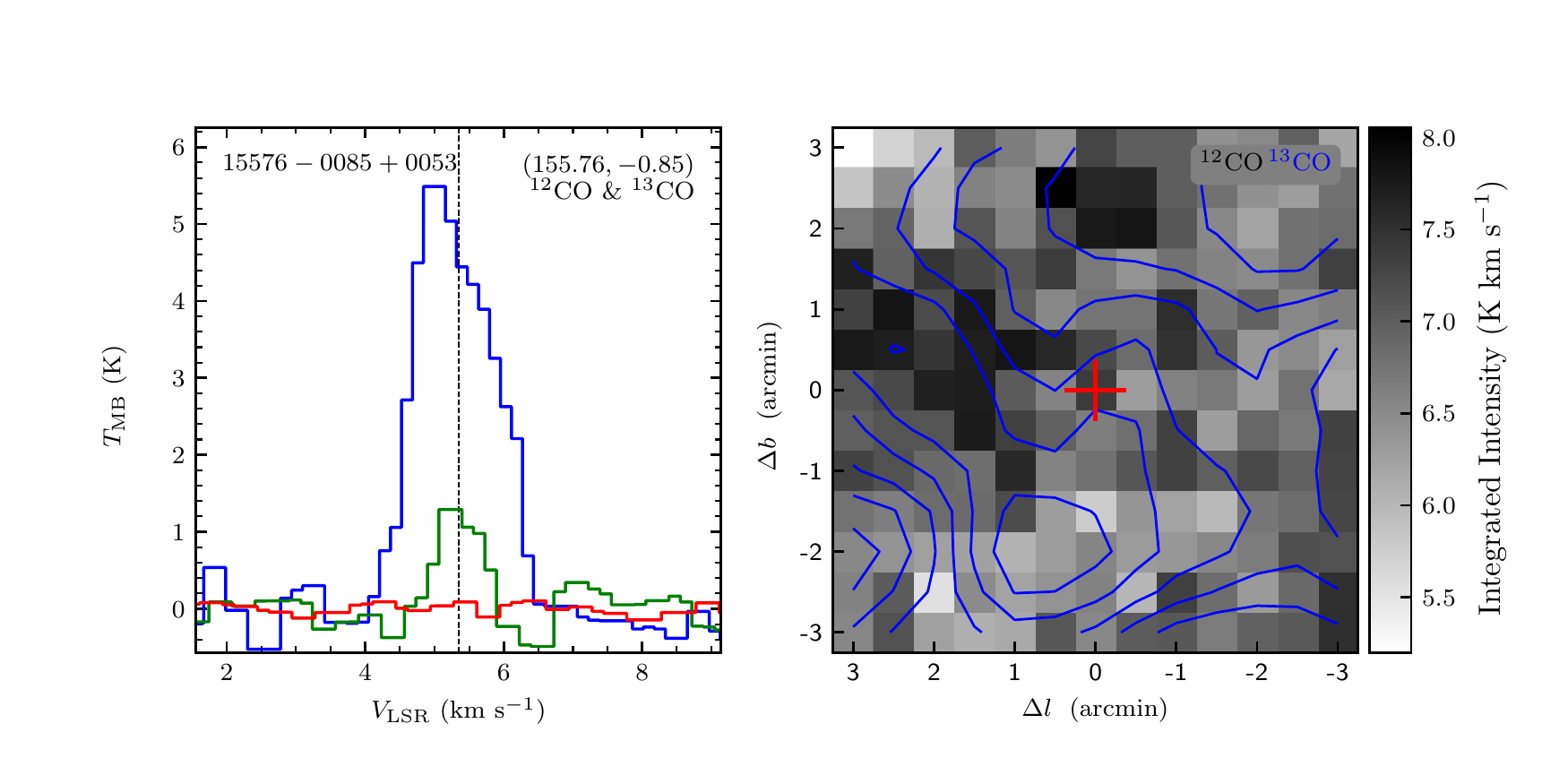}
\includegraphics[width=9.0cm,angle=0]{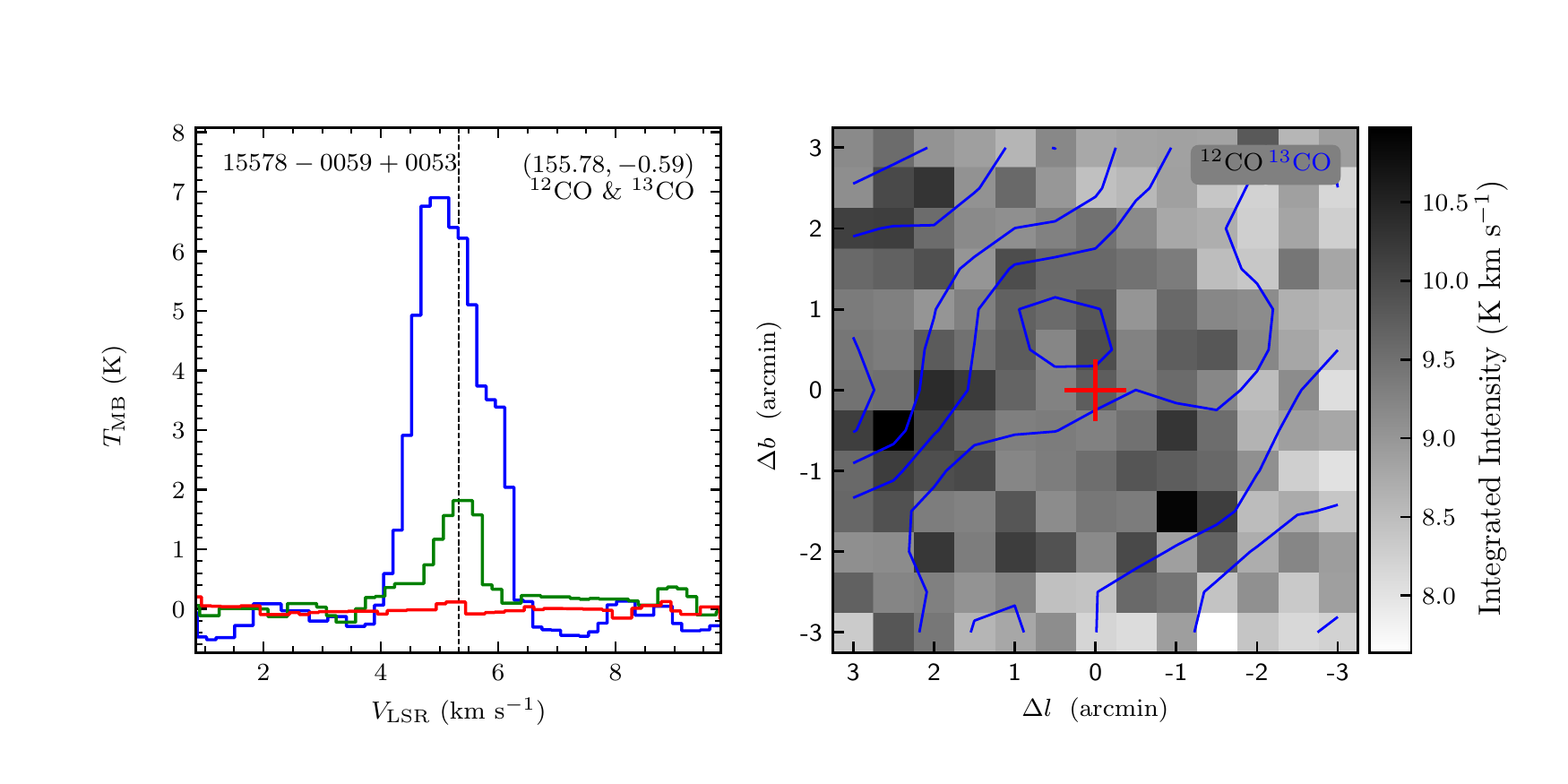}
\vspace{-0.5cm}

\includegraphics[width=9.0cm,angle=0]{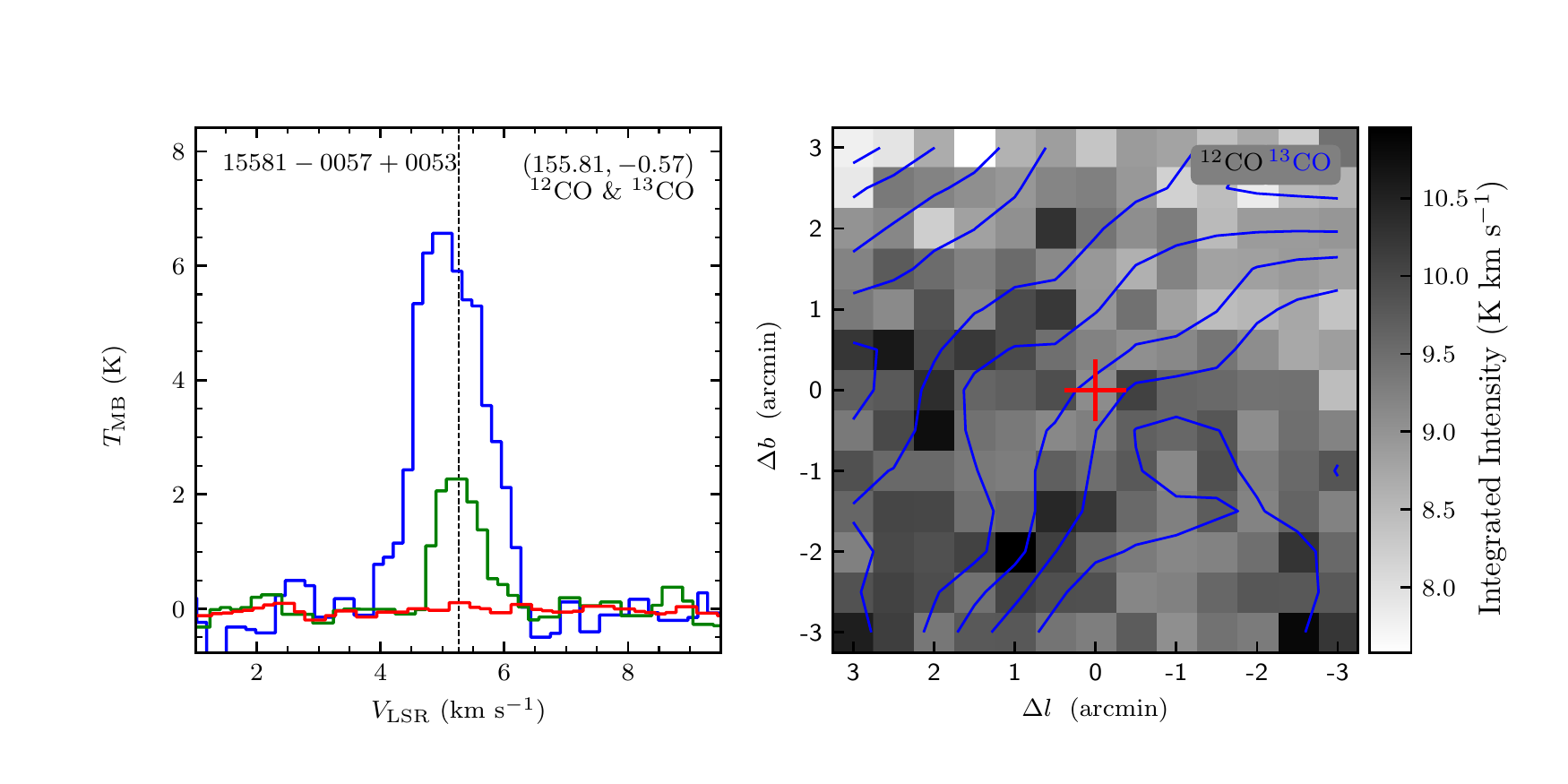}
\includegraphics[width=9.0cm,angle=0]{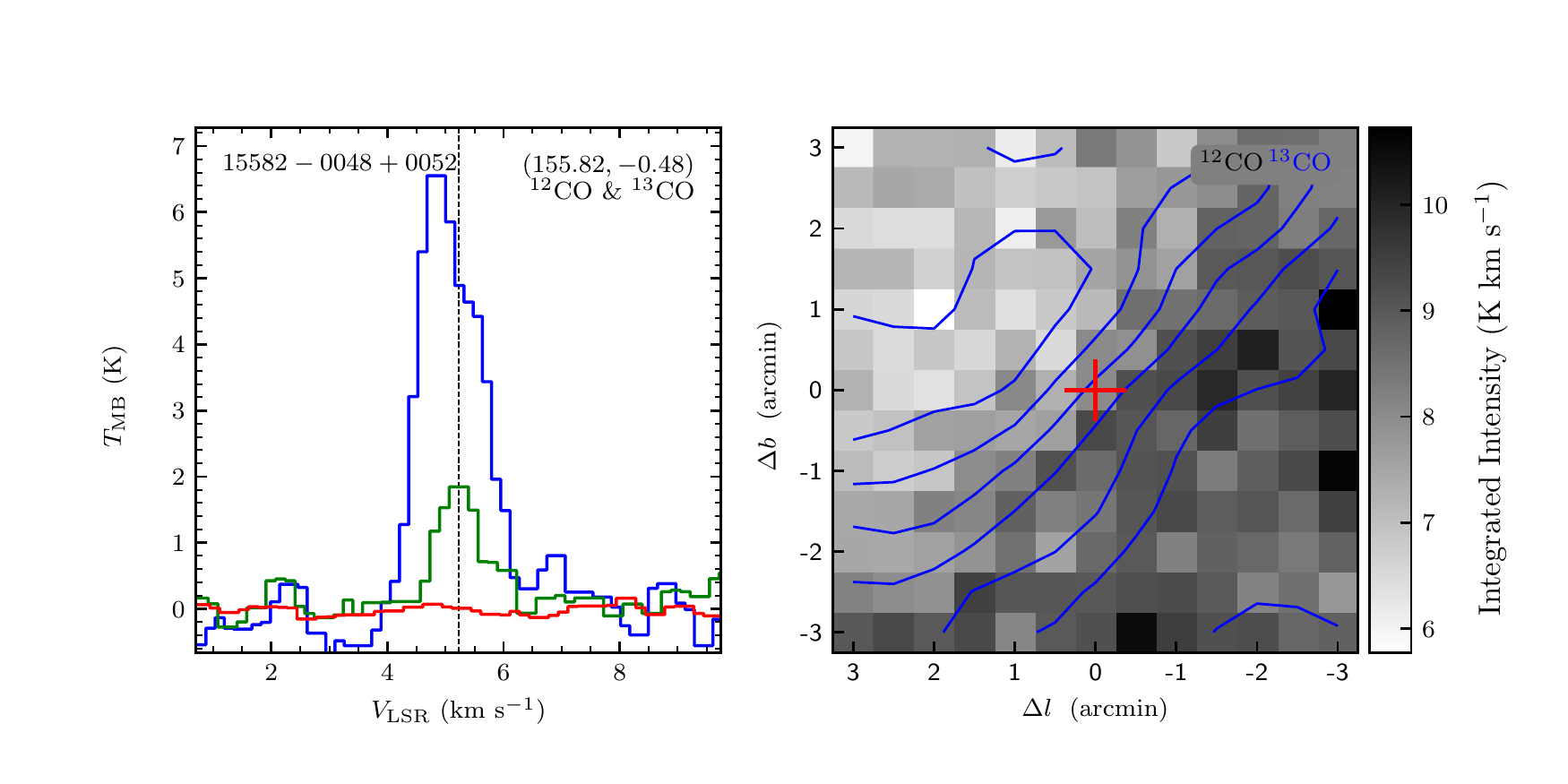}
\vspace{-0.5cm}

\includegraphics[width=9.0cm,angle=0]{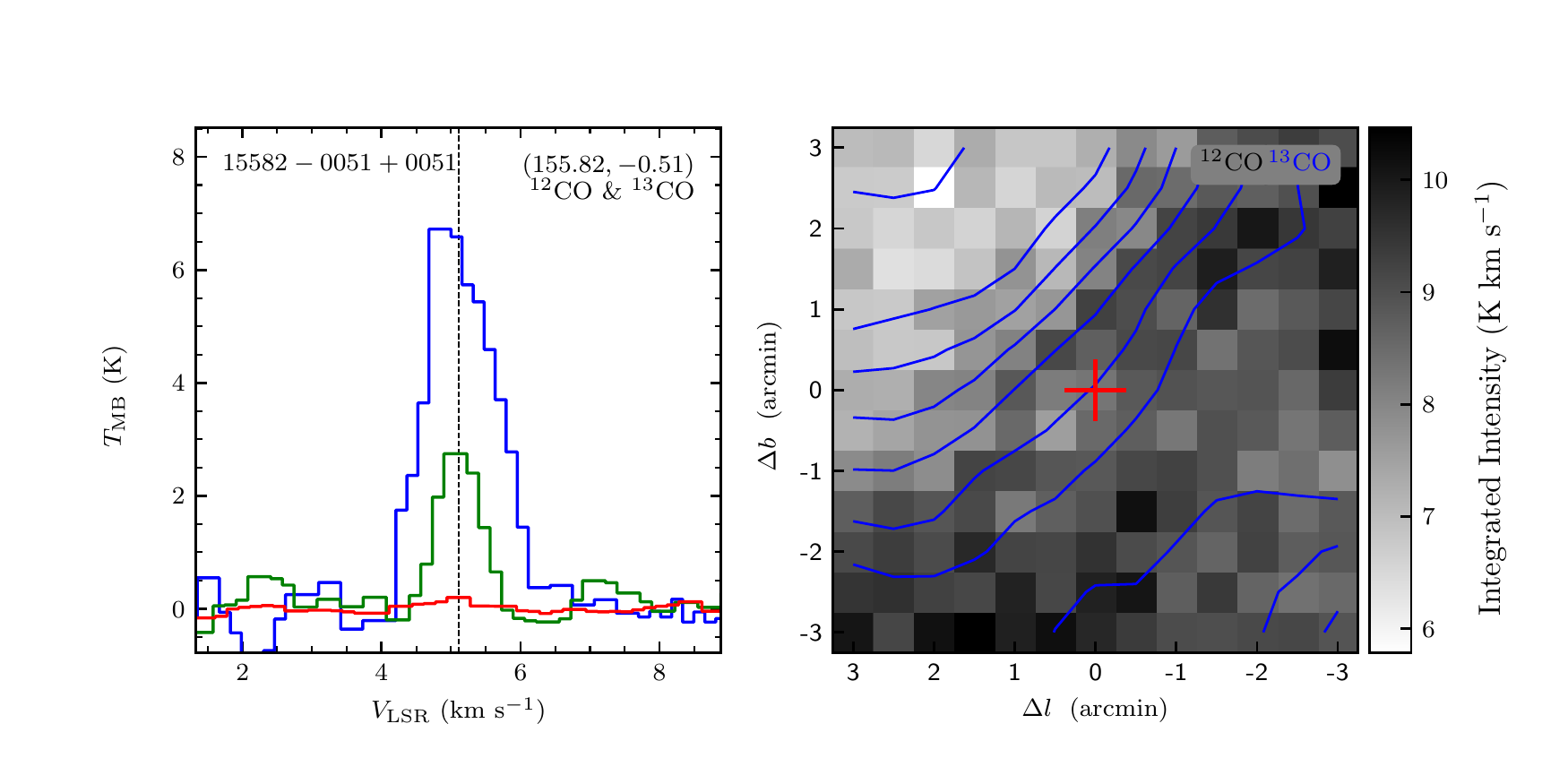}
\includegraphics[width=9.0cm,angle=0]{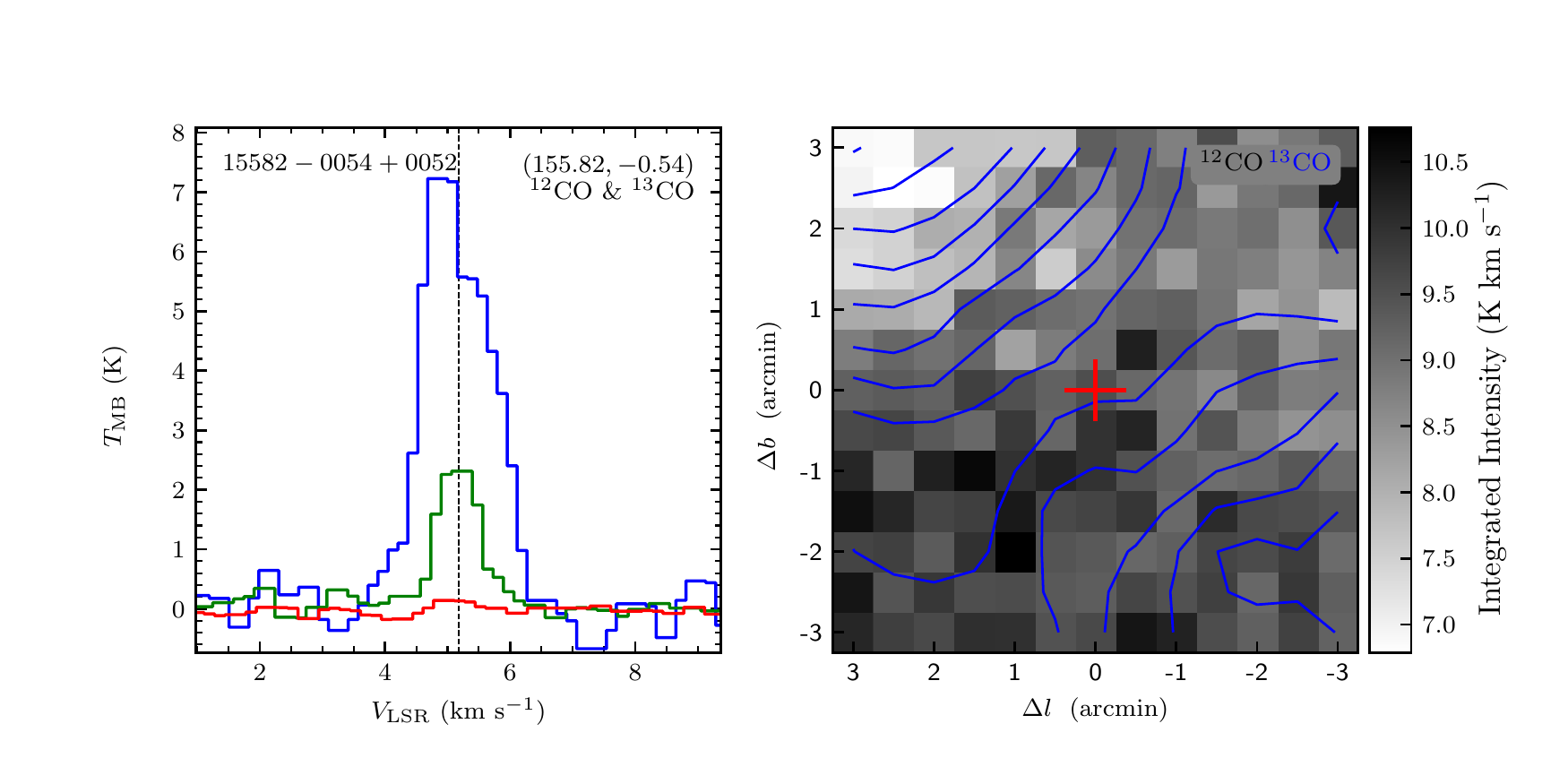}
\vspace{-0.5cm}

\includegraphics[width=9.0cm,angle=0]{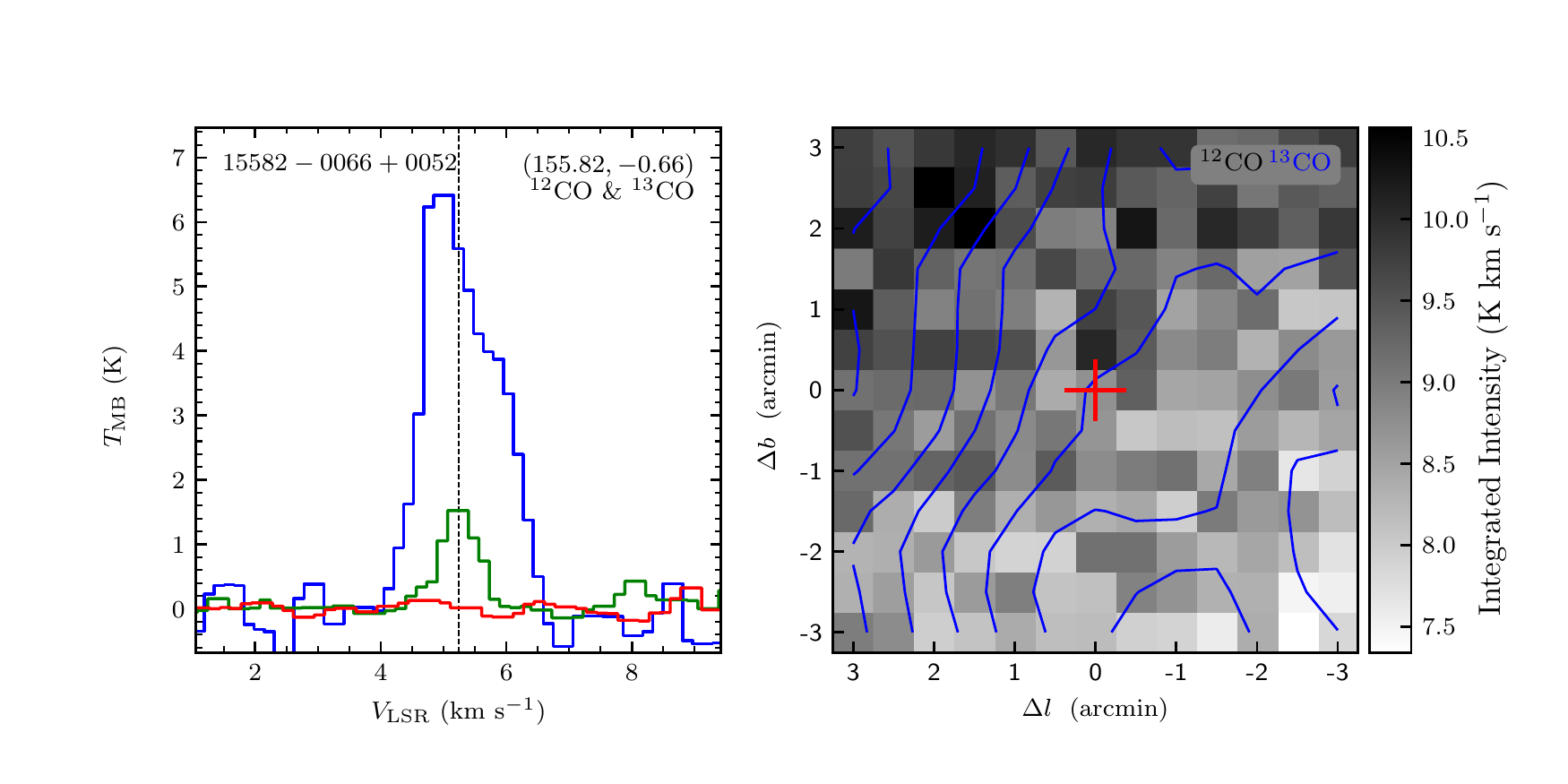}
\includegraphics[width=9.0cm,angle=0]{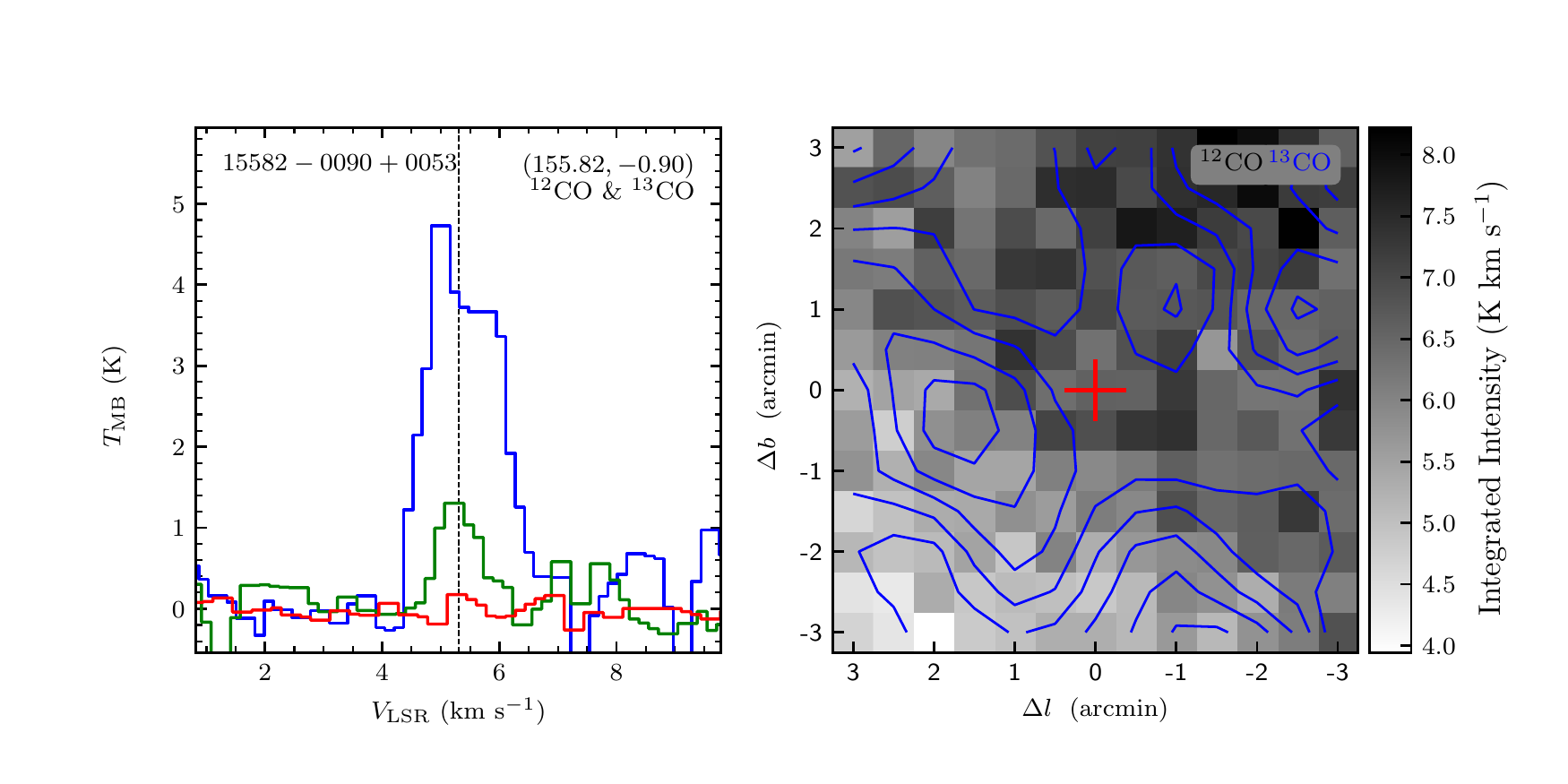}
\end{figure}
\clearpage

\begin{figure}
\includegraphics[width=9.0cm,angle=0]{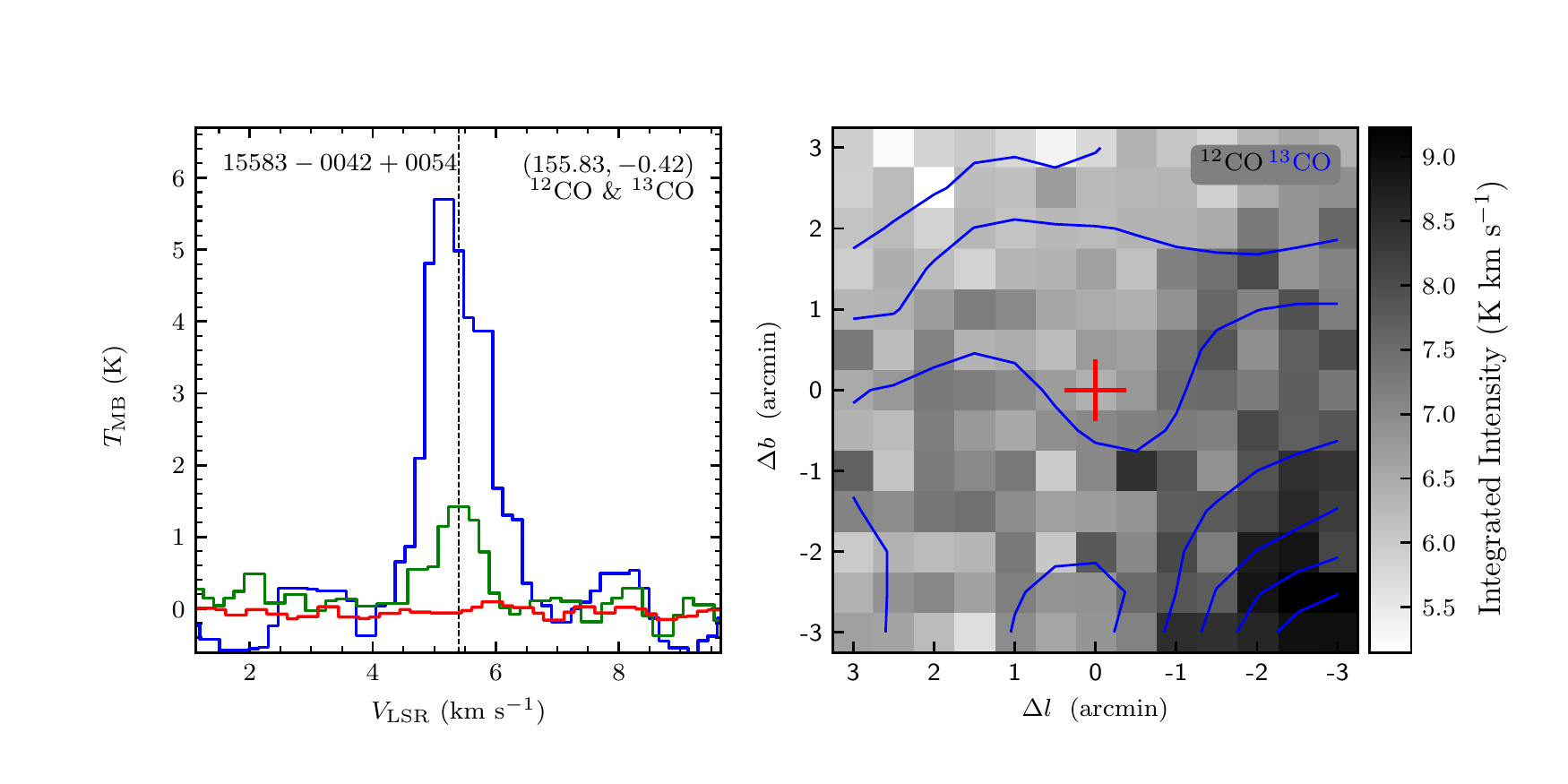}
\includegraphics[width=9.0cm,angle=0]{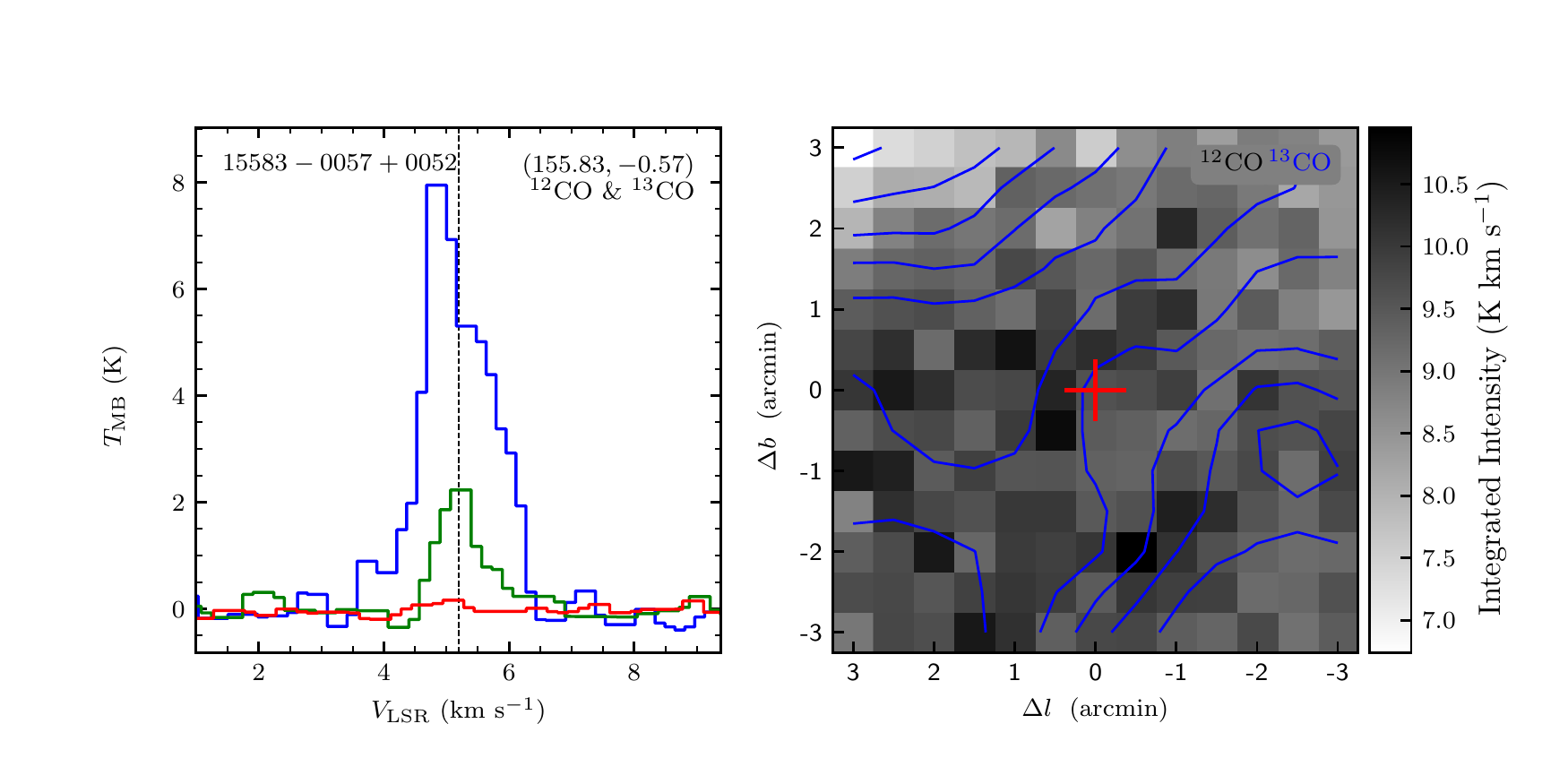}
\vspace{-0.5cm}

\includegraphics[width=9.0cm,angle=0]{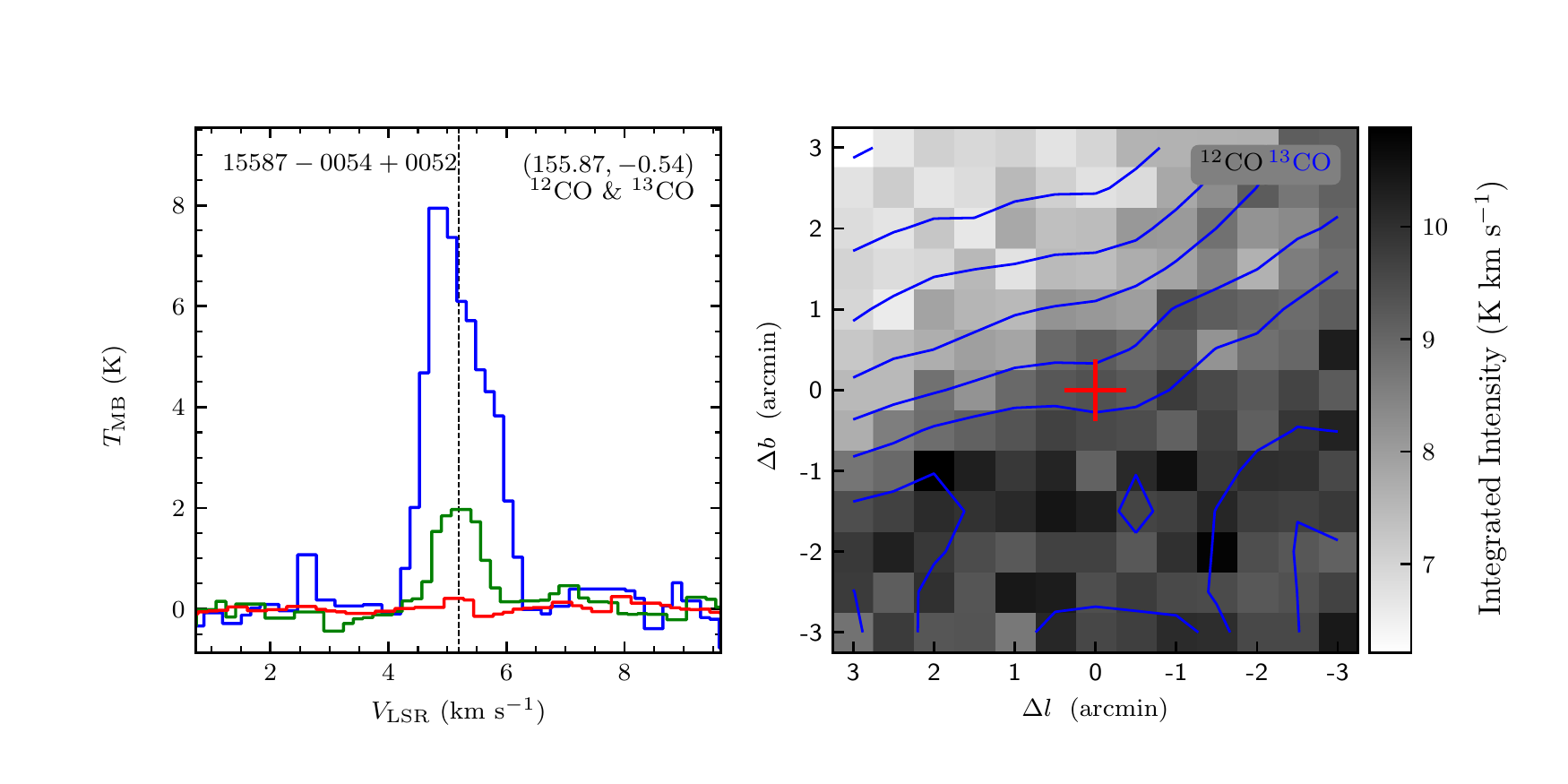}
\includegraphics[width=9.0cm,angle=0]{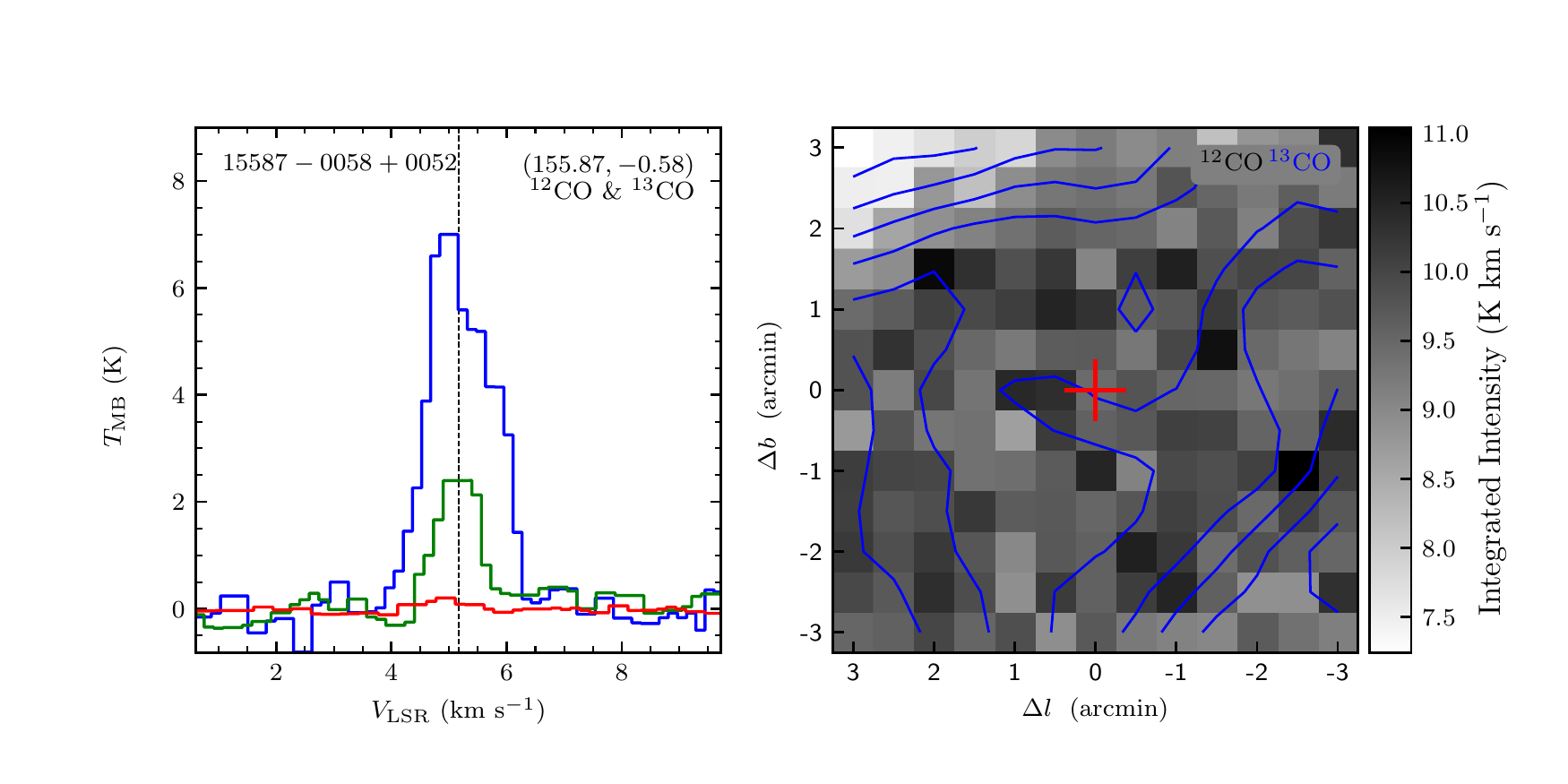}
\vspace{-0.5cm}

\includegraphics[width=9.0cm,angle=0]{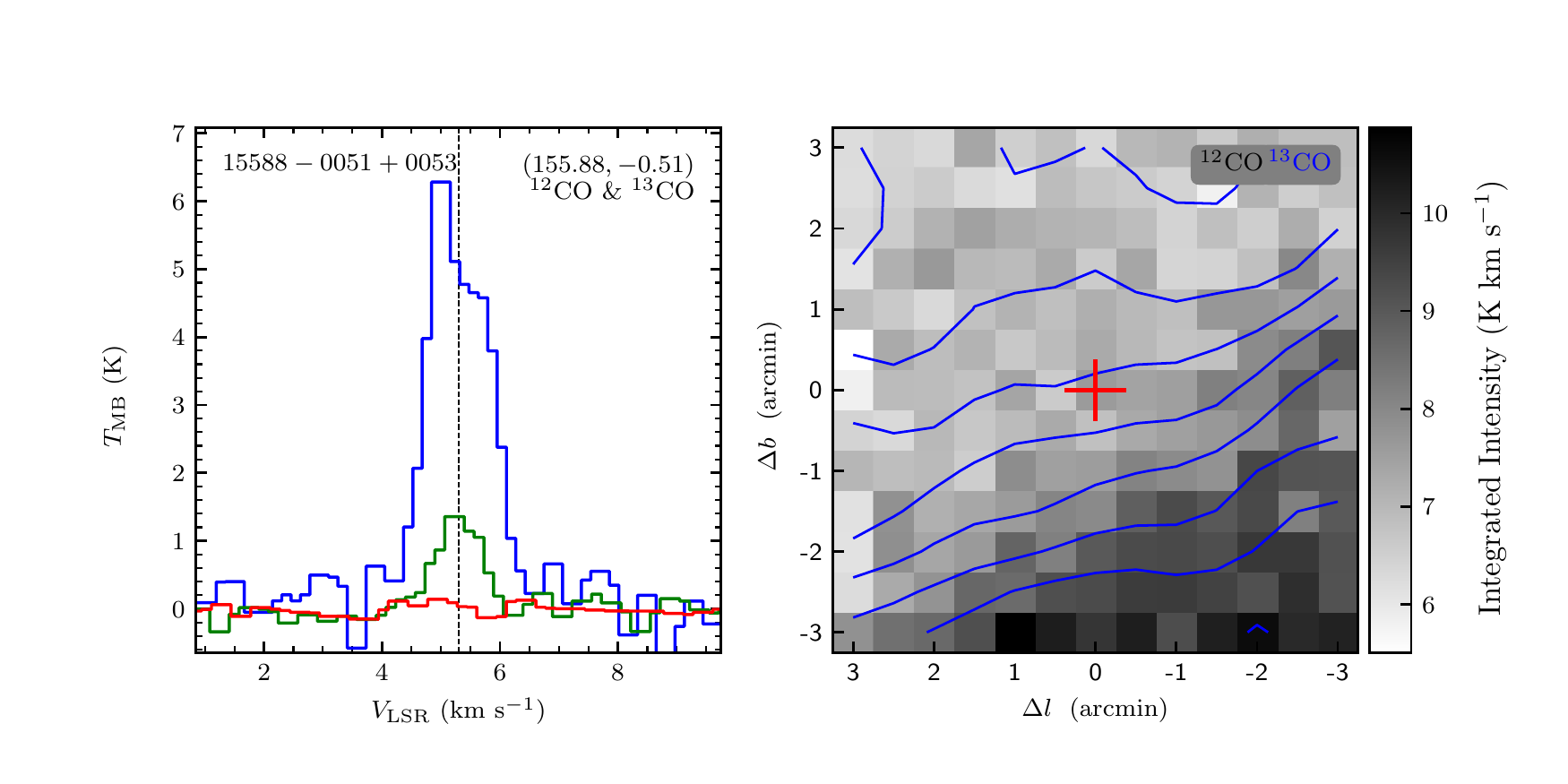}
\includegraphics[width=9.0cm,angle=0]{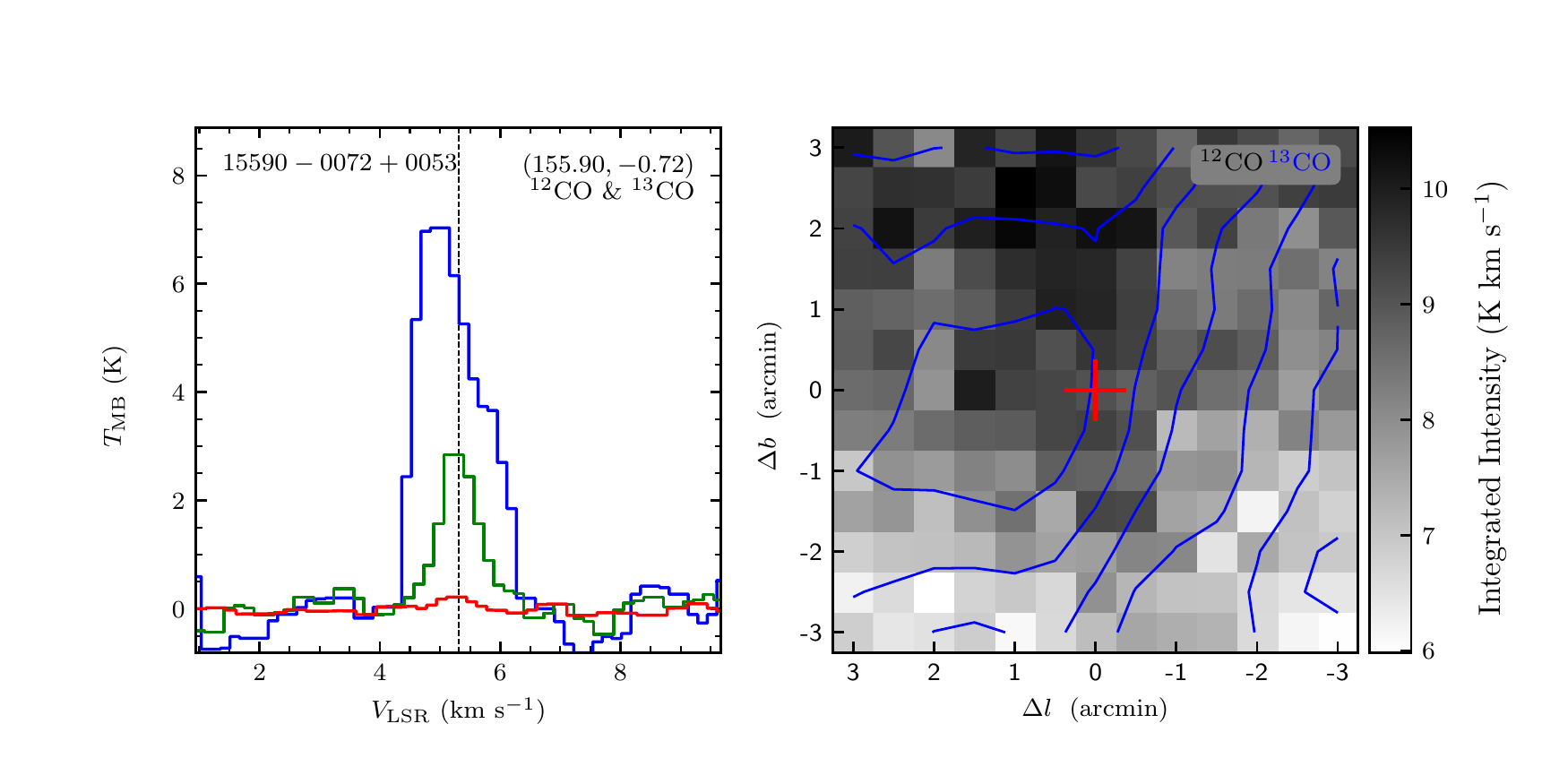}
\vspace{-0.5cm}

\includegraphics[width=9.0cm,angle=0]{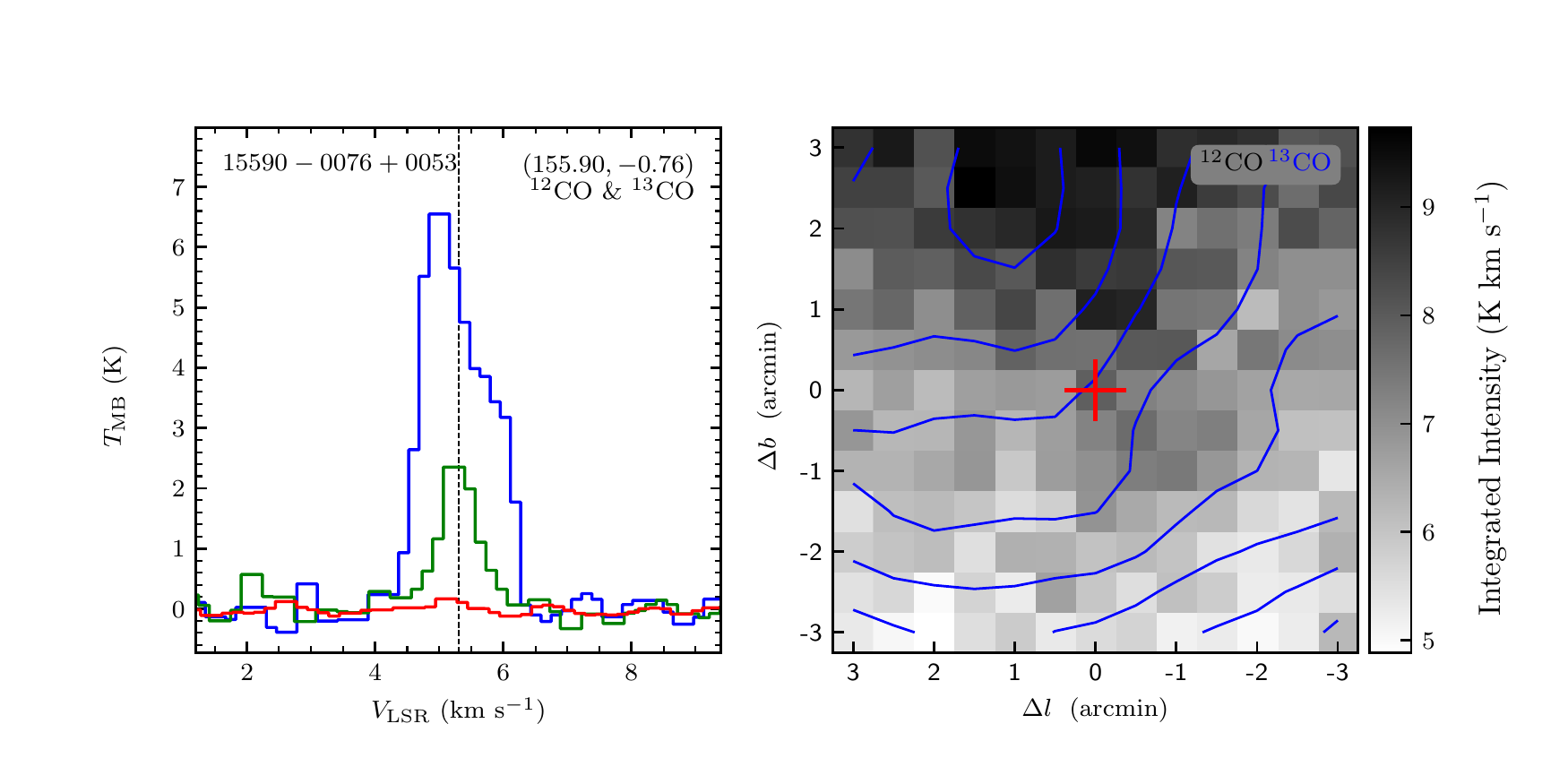}
\includegraphics[width=9.0cm,angle=0]{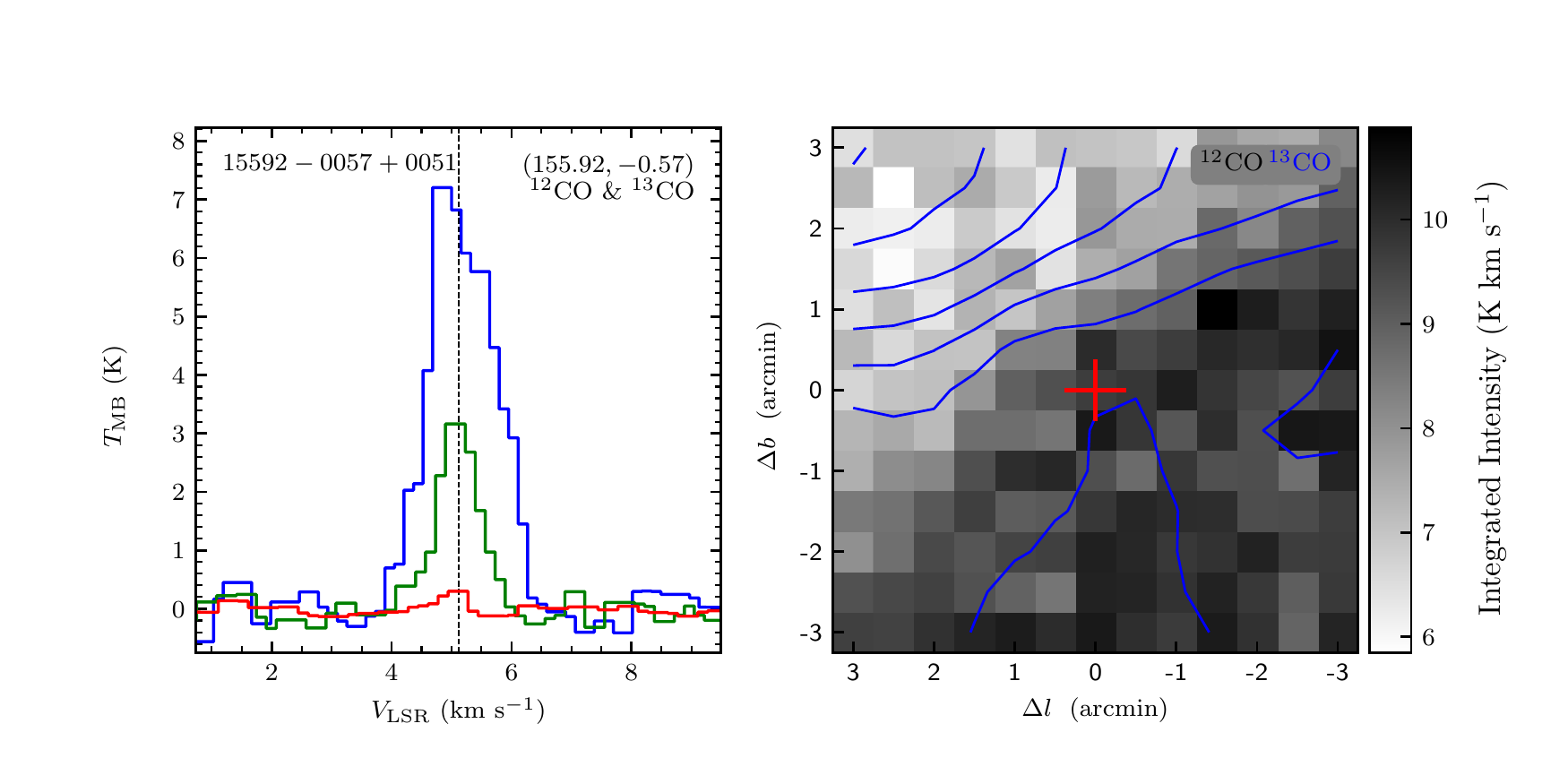}
\vspace{-0.5cm}

\includegraphics[width=9.0cm,angle=0]{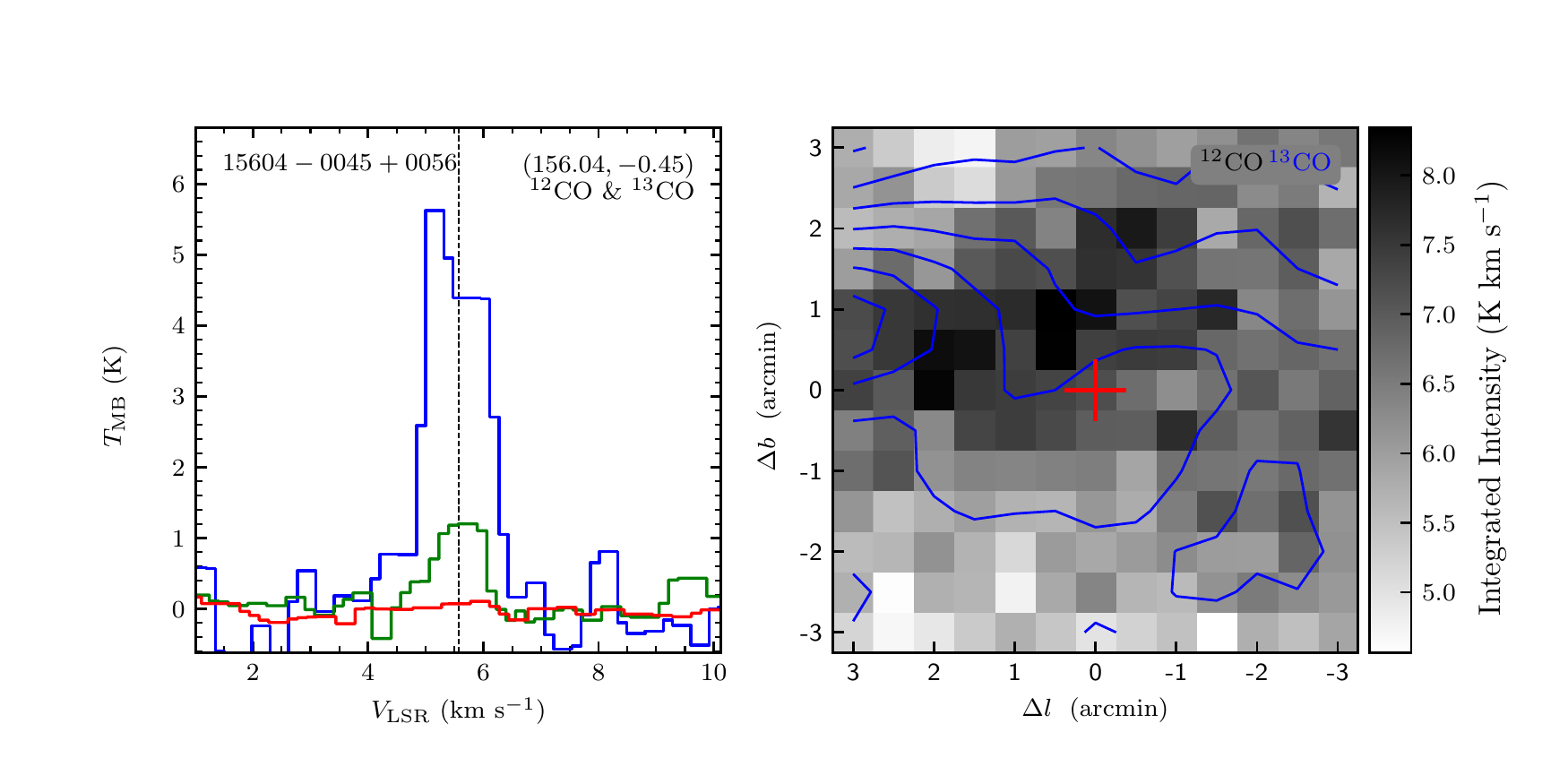}
\includegraphics[width=9.0cm,angle=0]{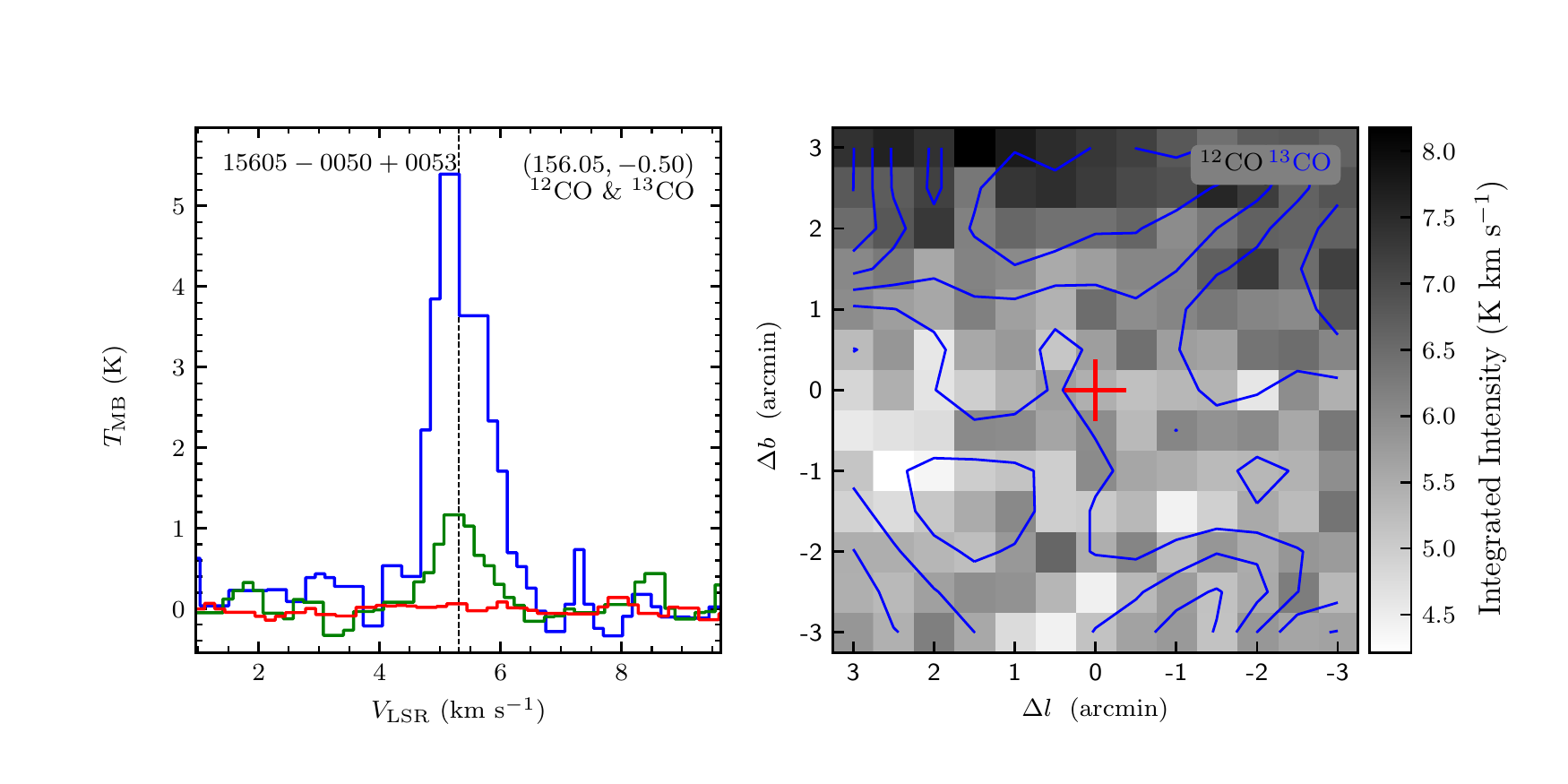}
\end{figure}
\clearpage

\begin{figure}
\includegraphics[width=9.0cm,angle=0]{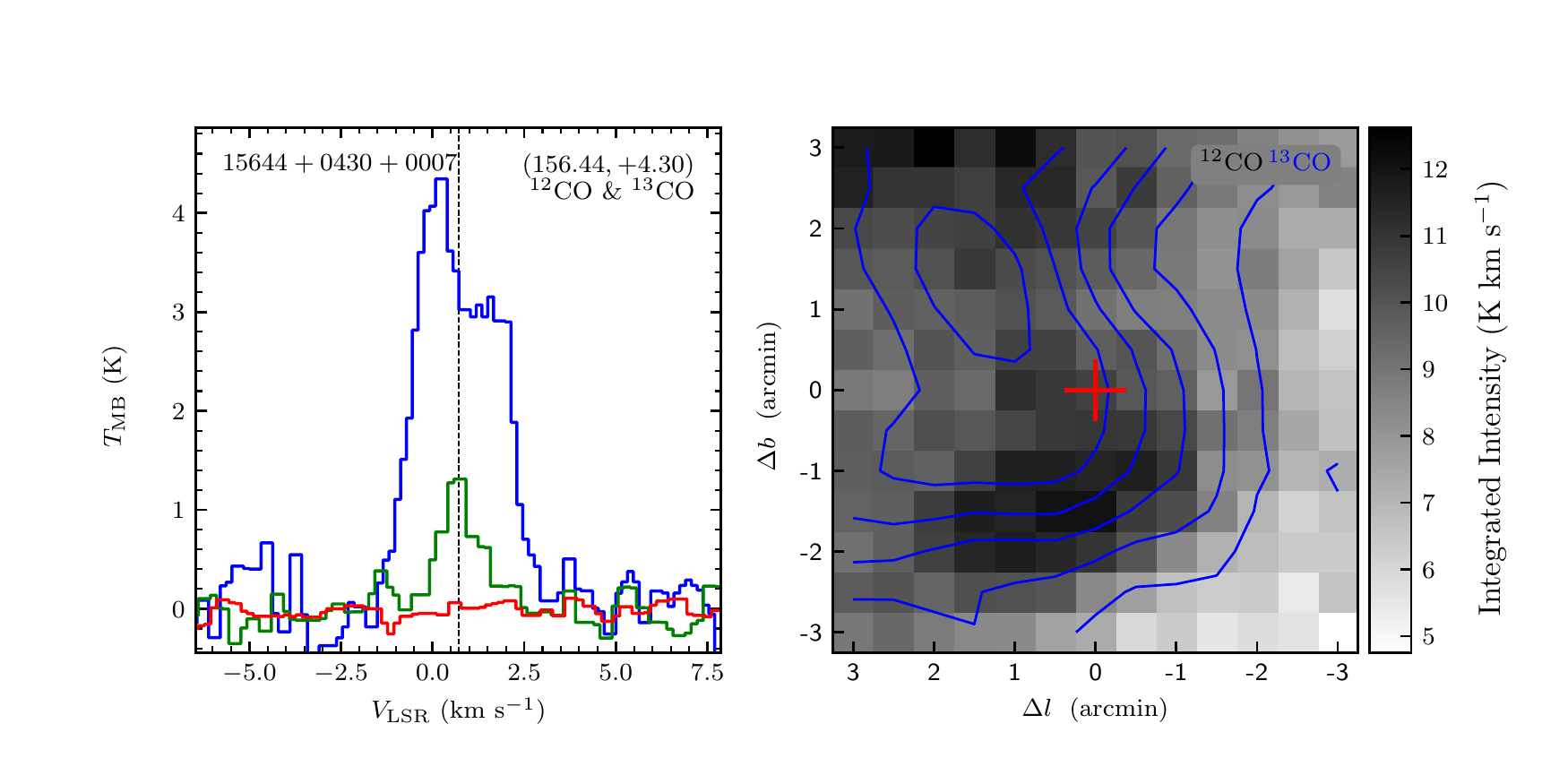}
\includegraphics[width=9.0cm,angle=0]{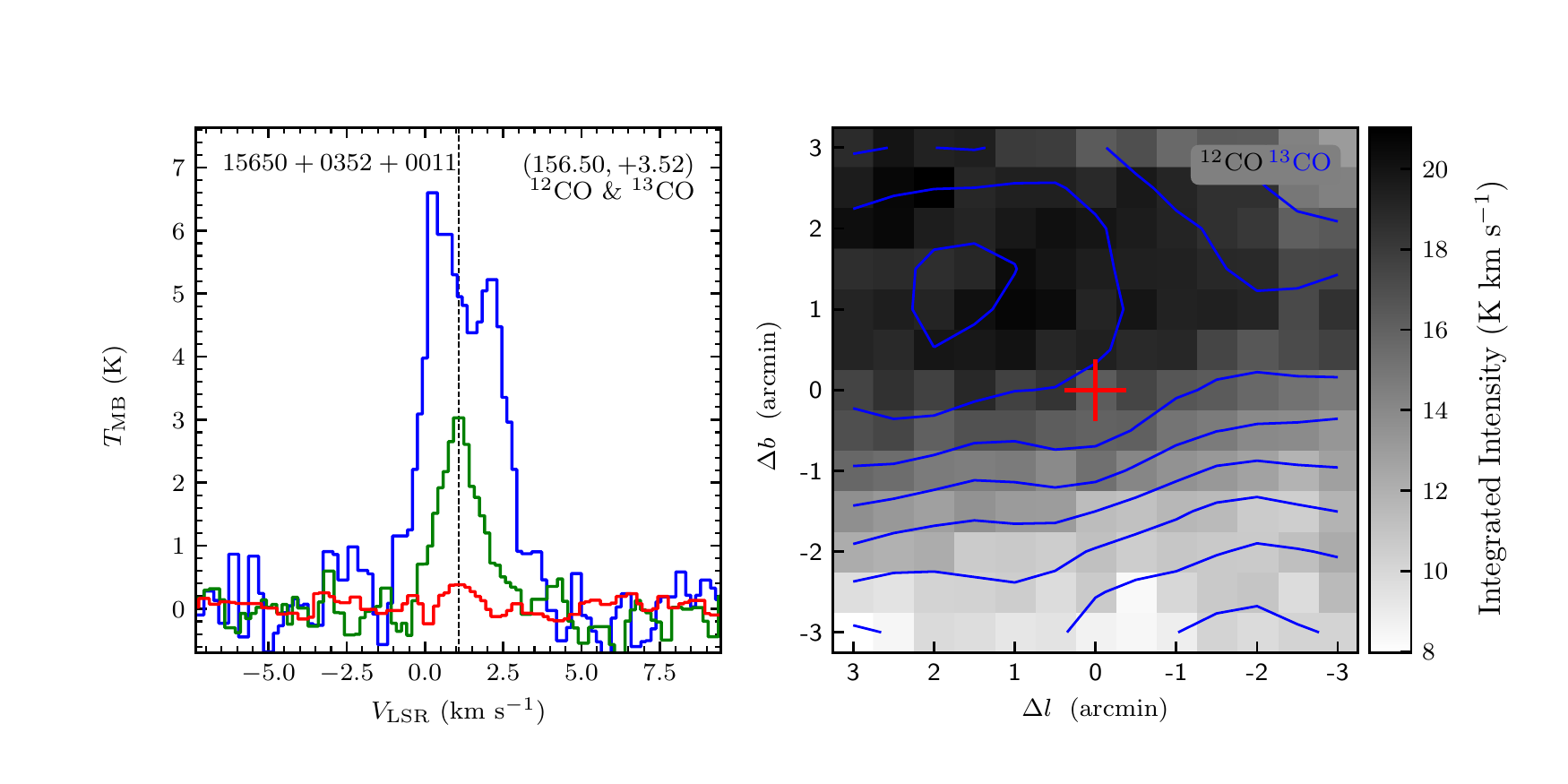}
\vspace{-0.5cm}

\includegraphics[width=9.0cm,angle=0]{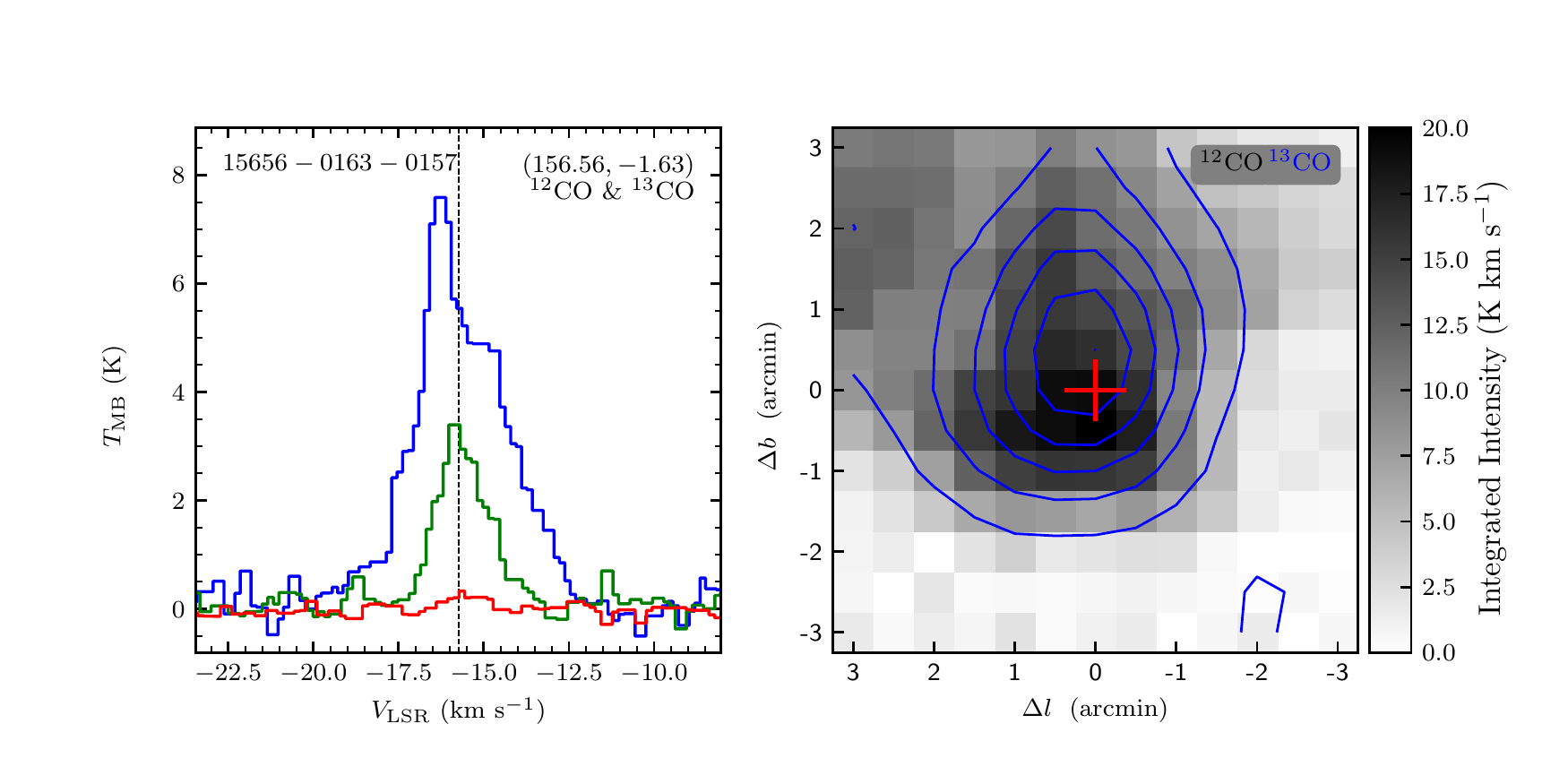}
\includegraphics[width=9.0cm,angle=0]{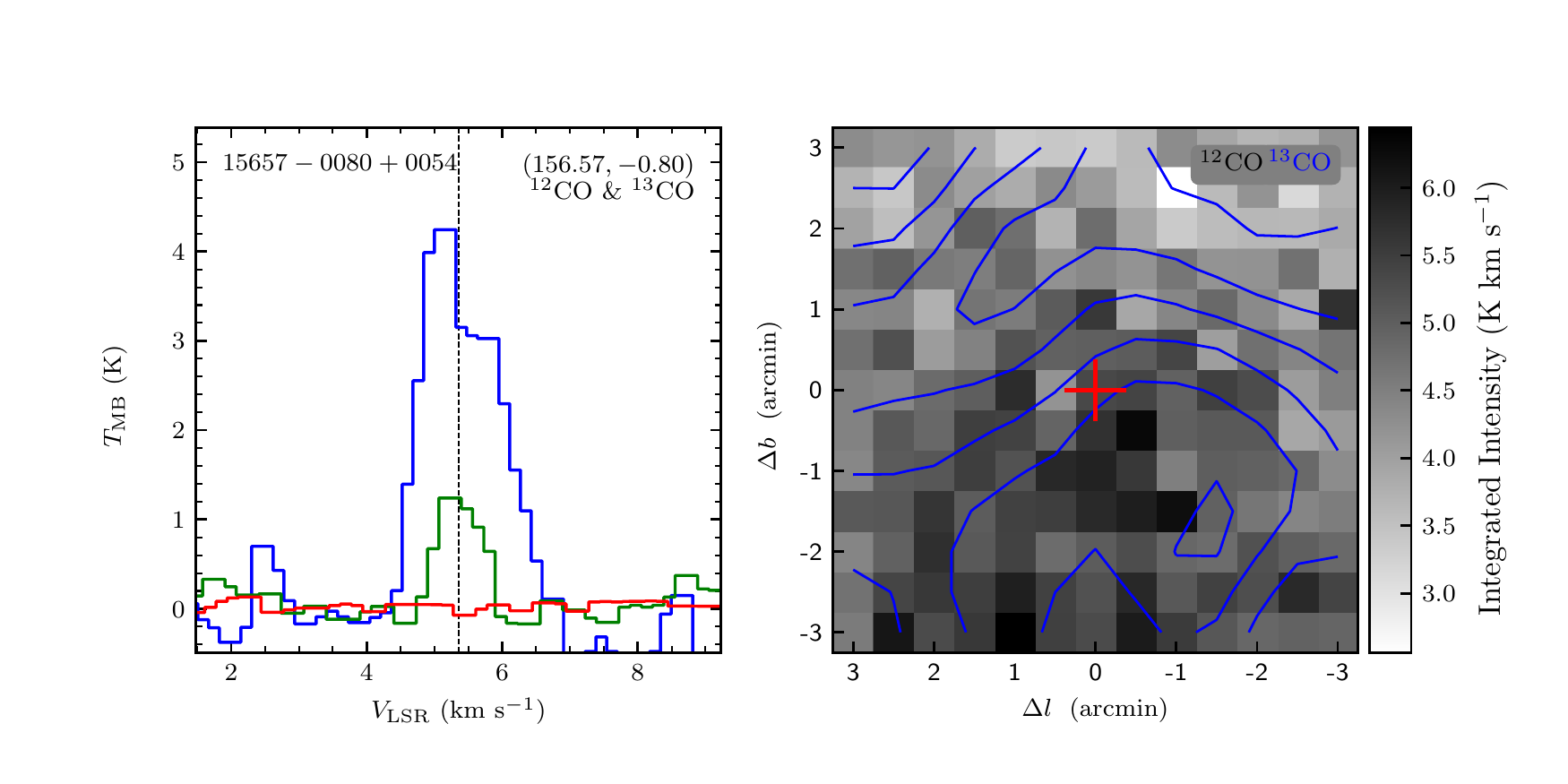}
\vspace{-0.5cm}

\includegraphics[width=9.0cm,angle=0]{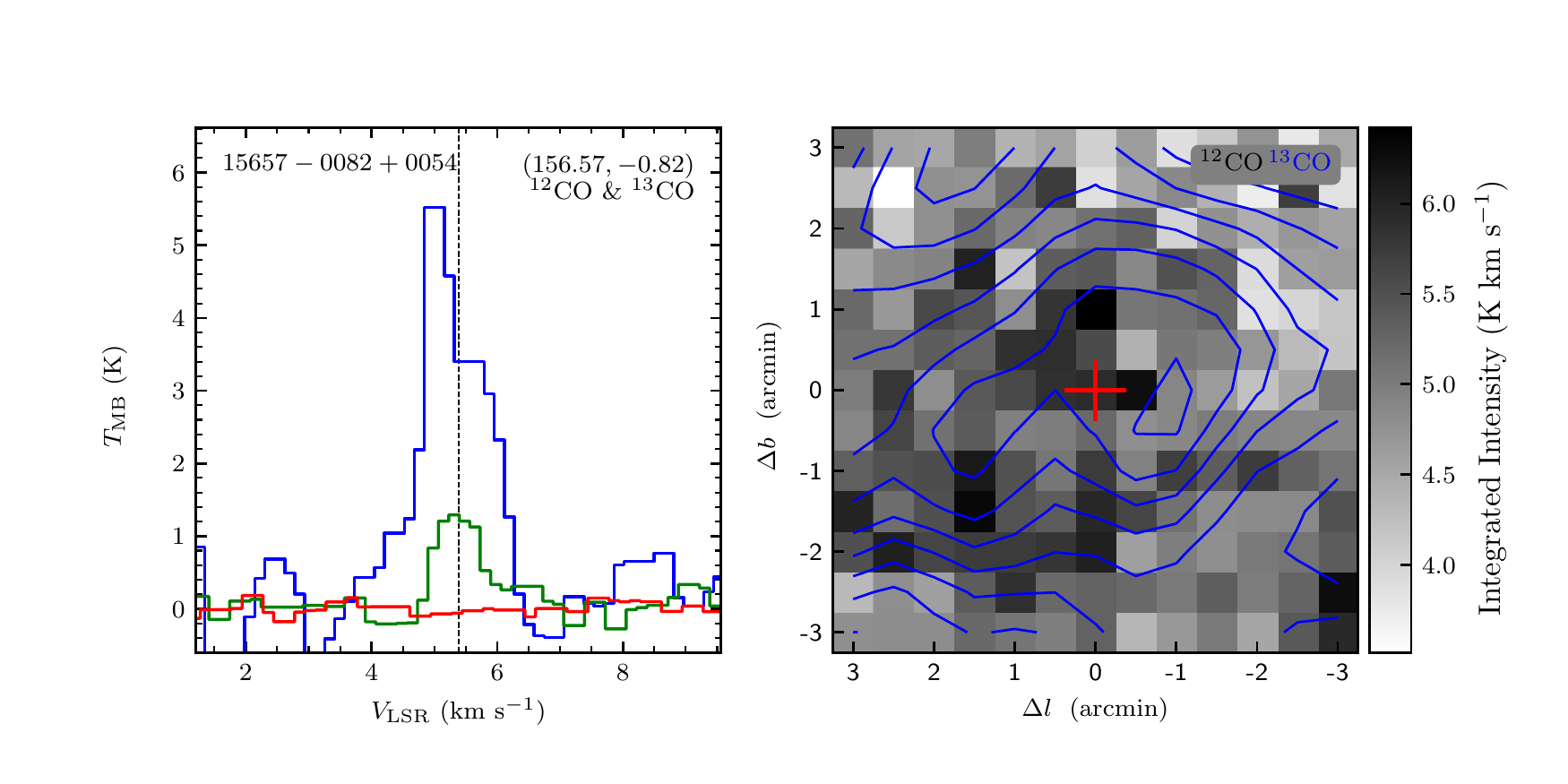}
\includegraphics[width=9.0cm,angle=0]{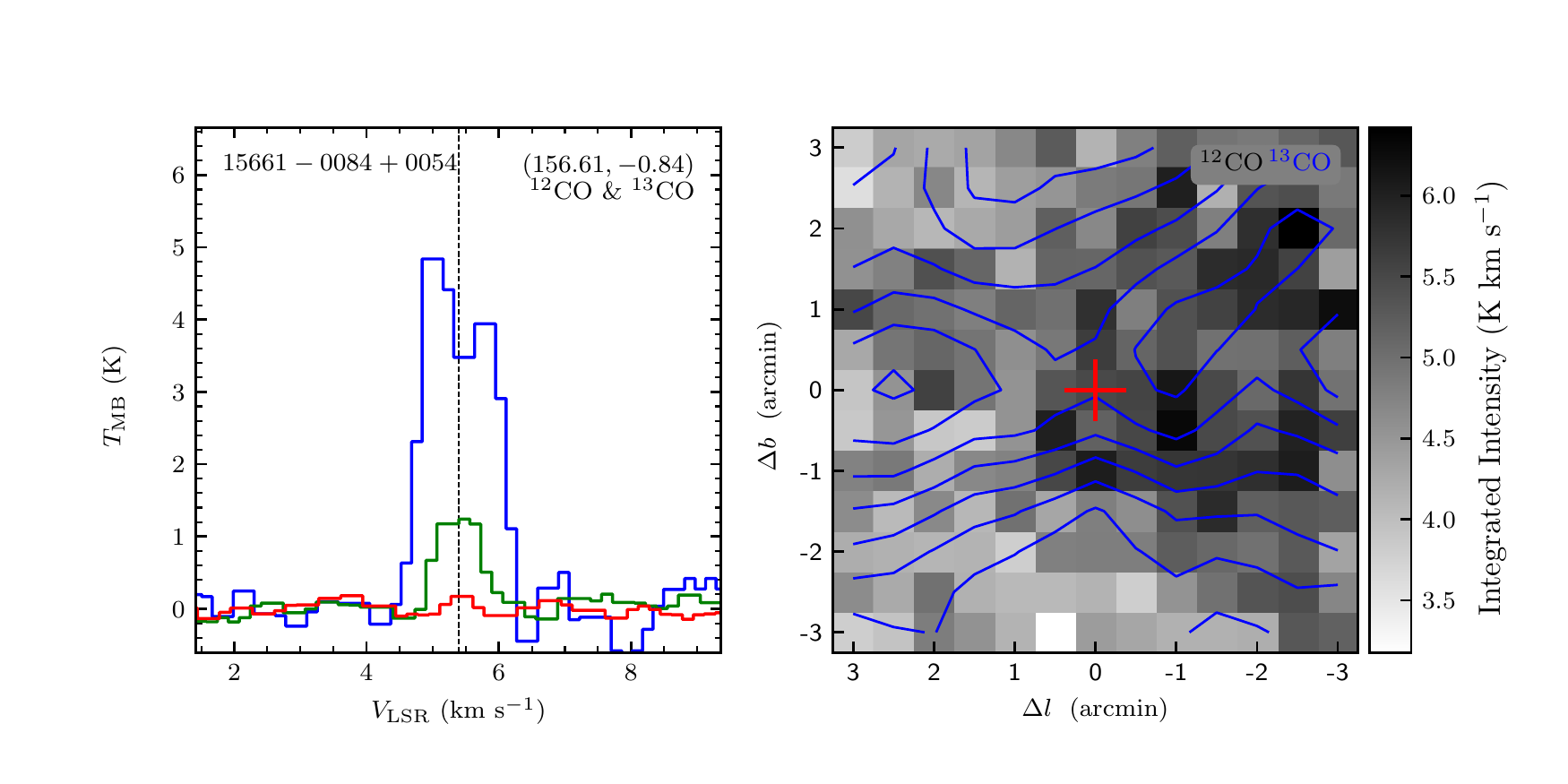}
\vspace{-0.5cm}

\includegraphics[width=9.0cm,angle=0]{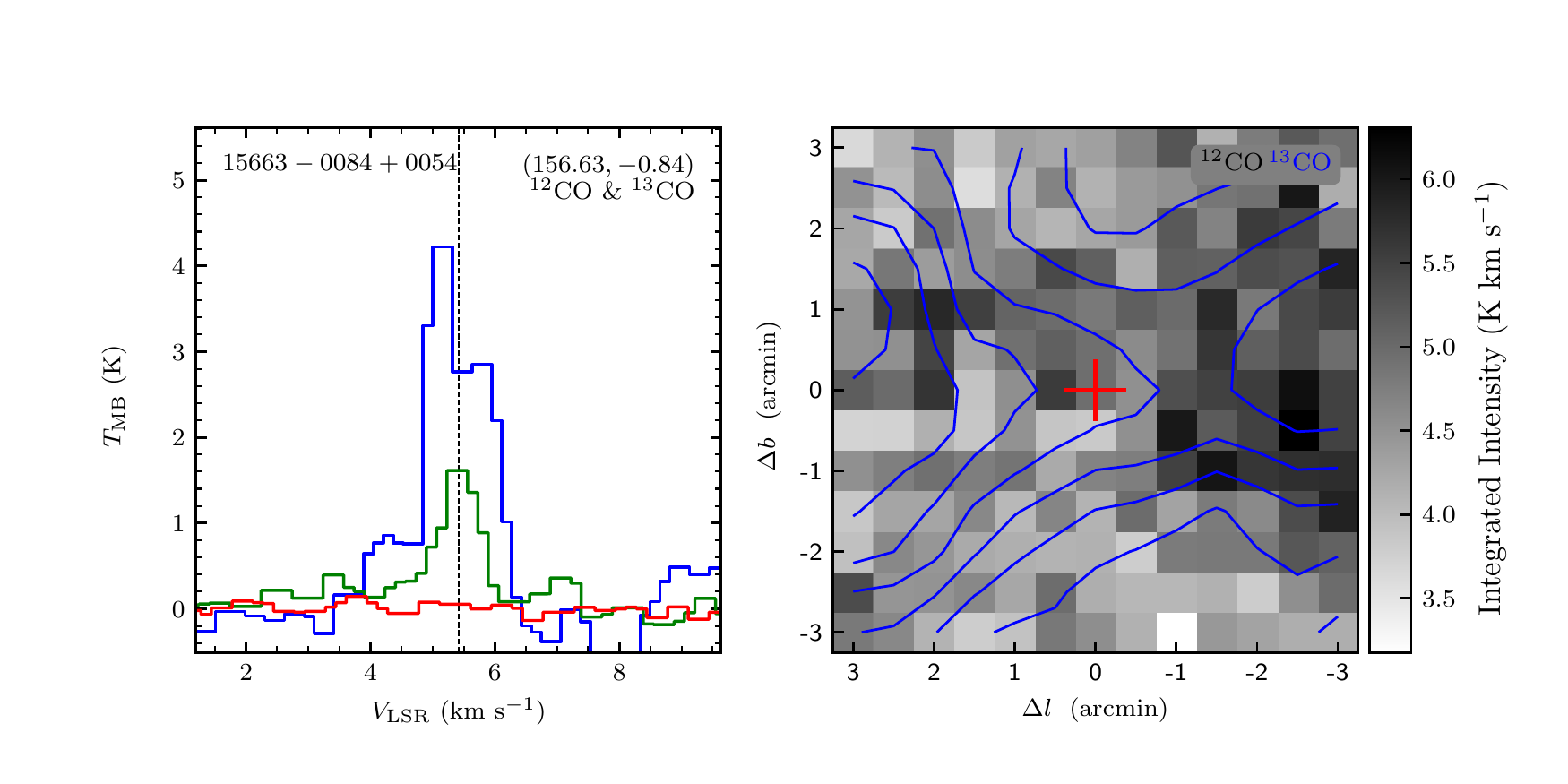}
\includegraphics[width=9.0cm,angle=0]{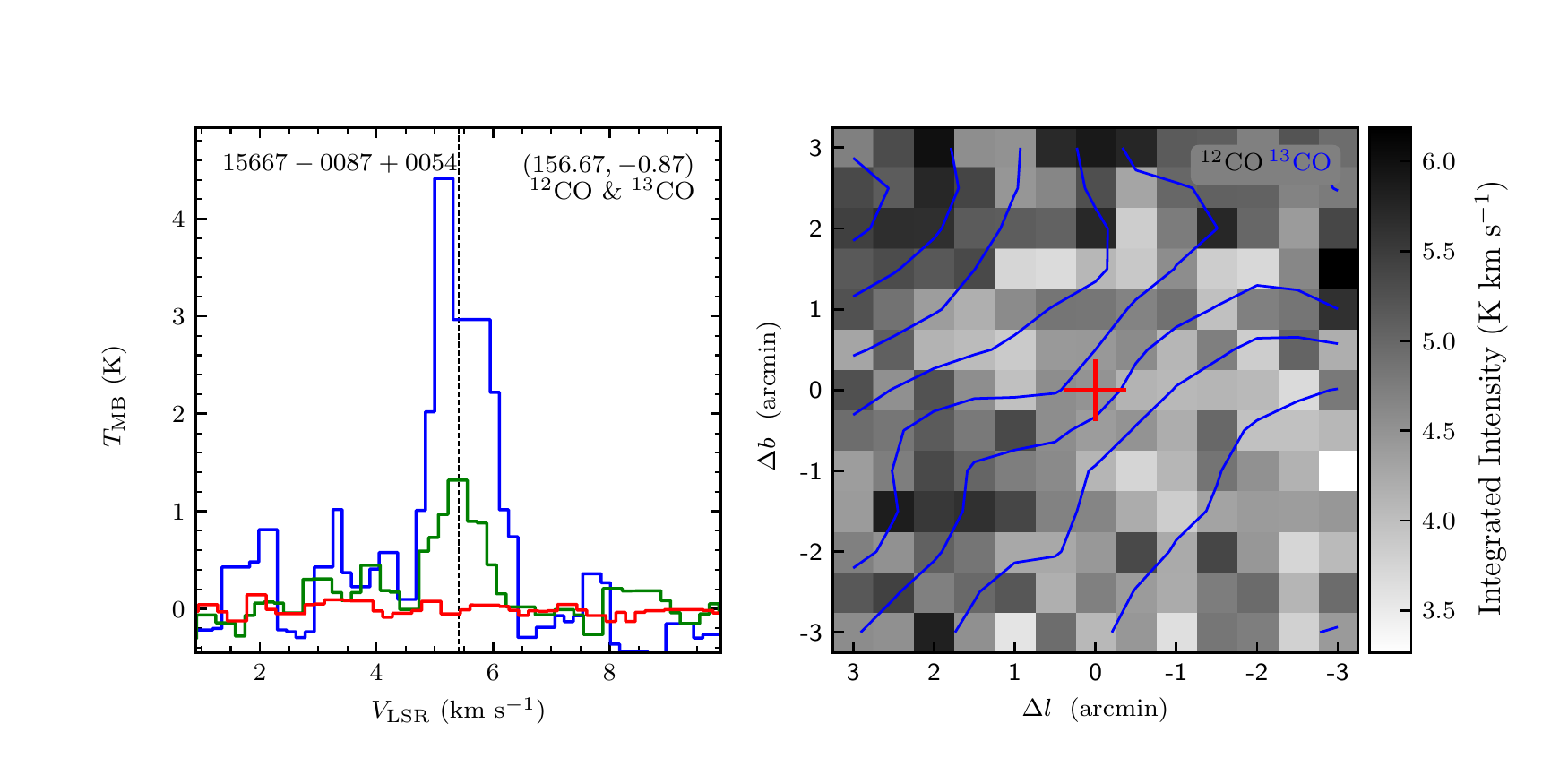}
\vspace{-0.5cm}

\includegraphics[width=9.0cm,angle=0]{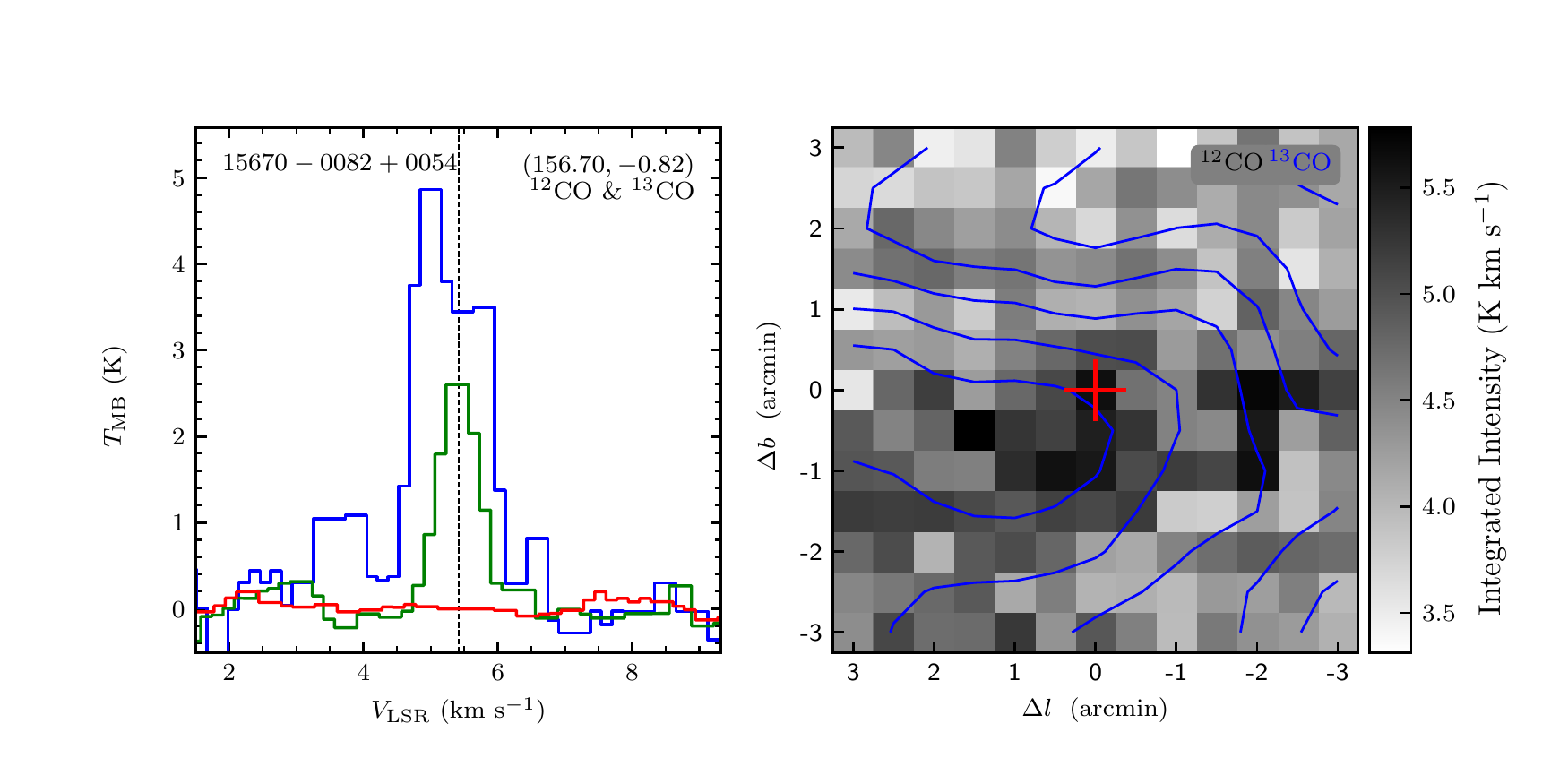}
\includegraphics[width=9.0cm,angle=0]{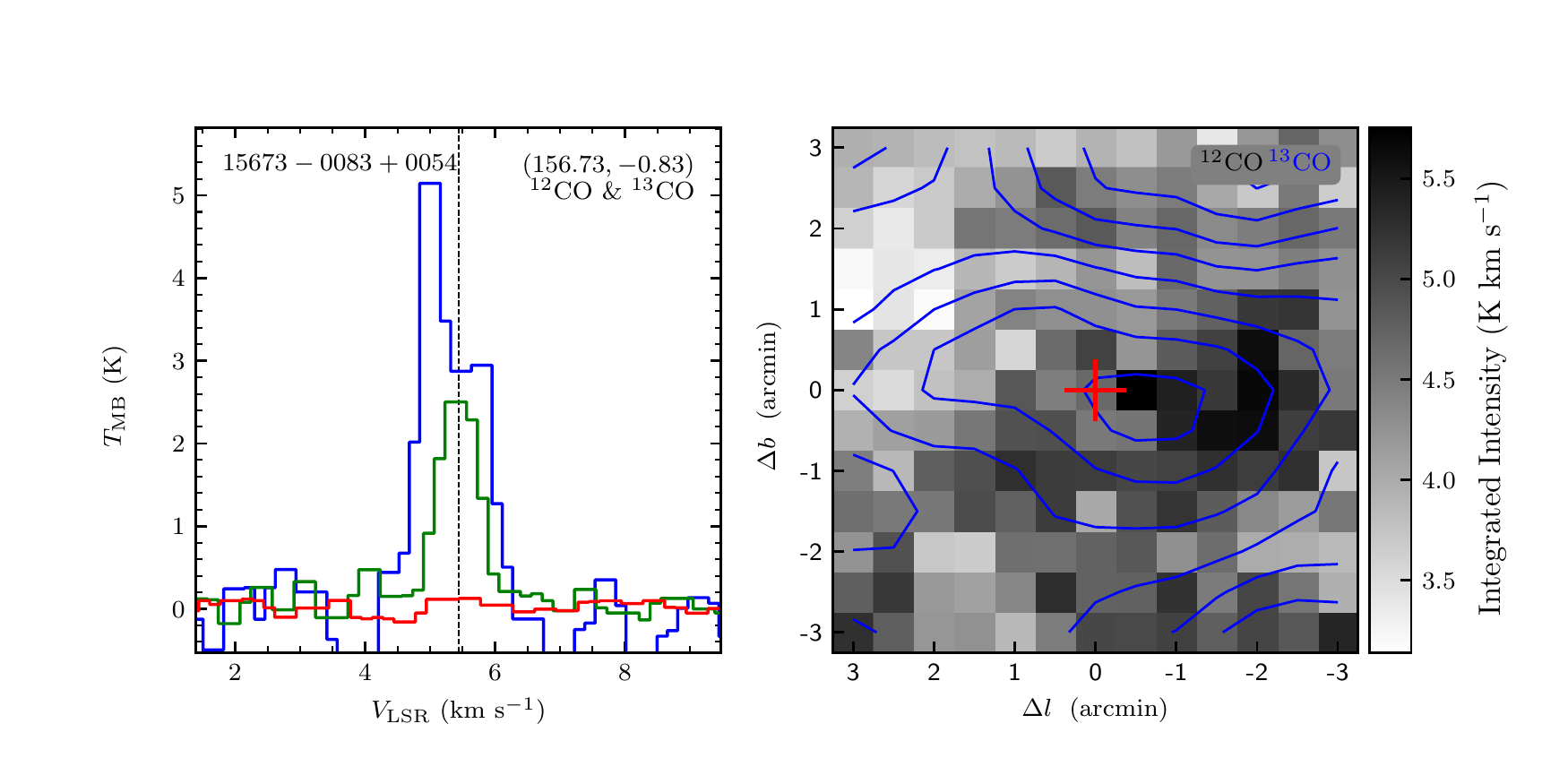}
\end{figure}
\clearpage

\begin{figure}
\includegraphics[width=9.0cm,angle=0]{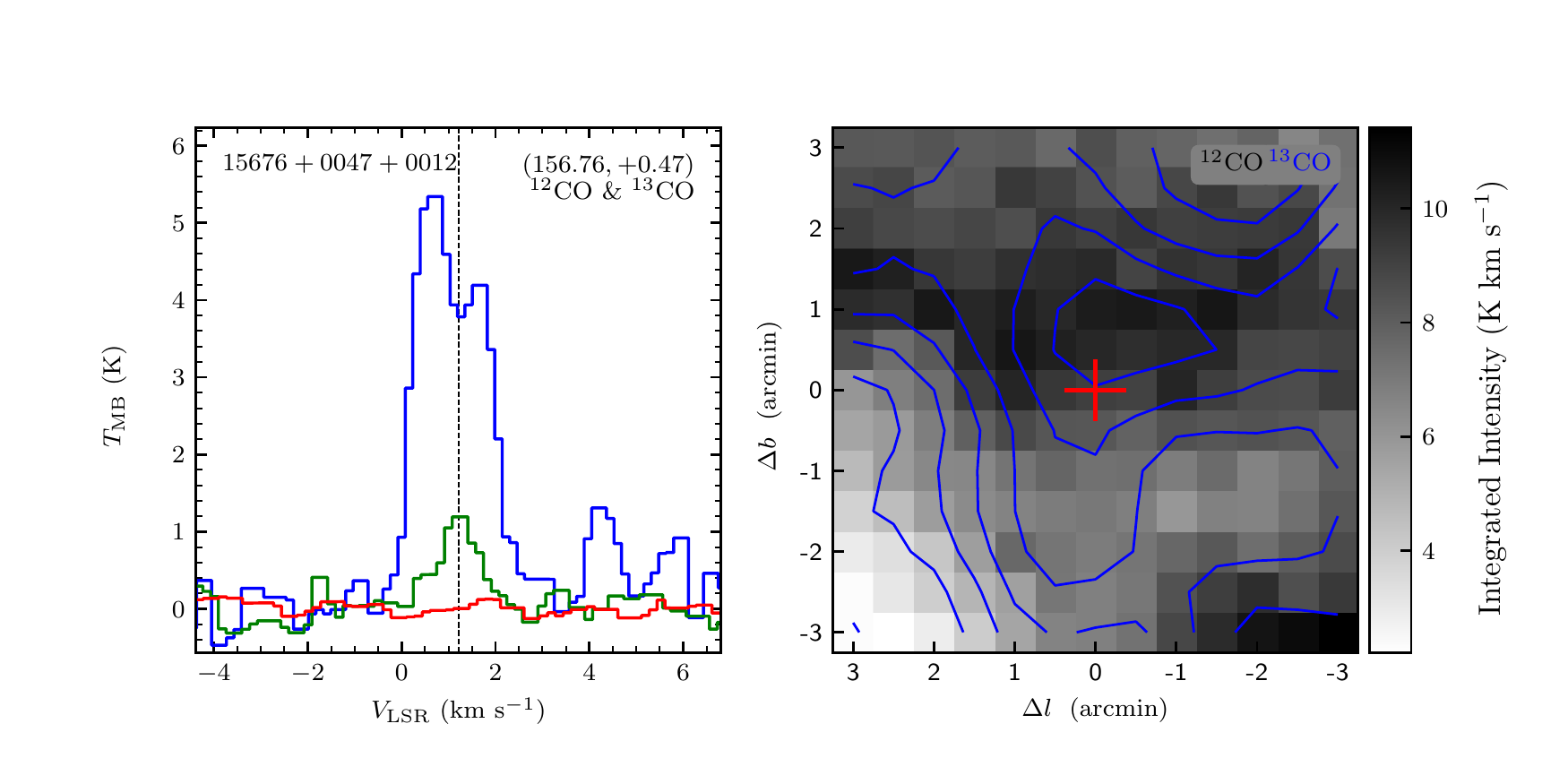}
\includegraphics[width=9.0cm,angle=0]{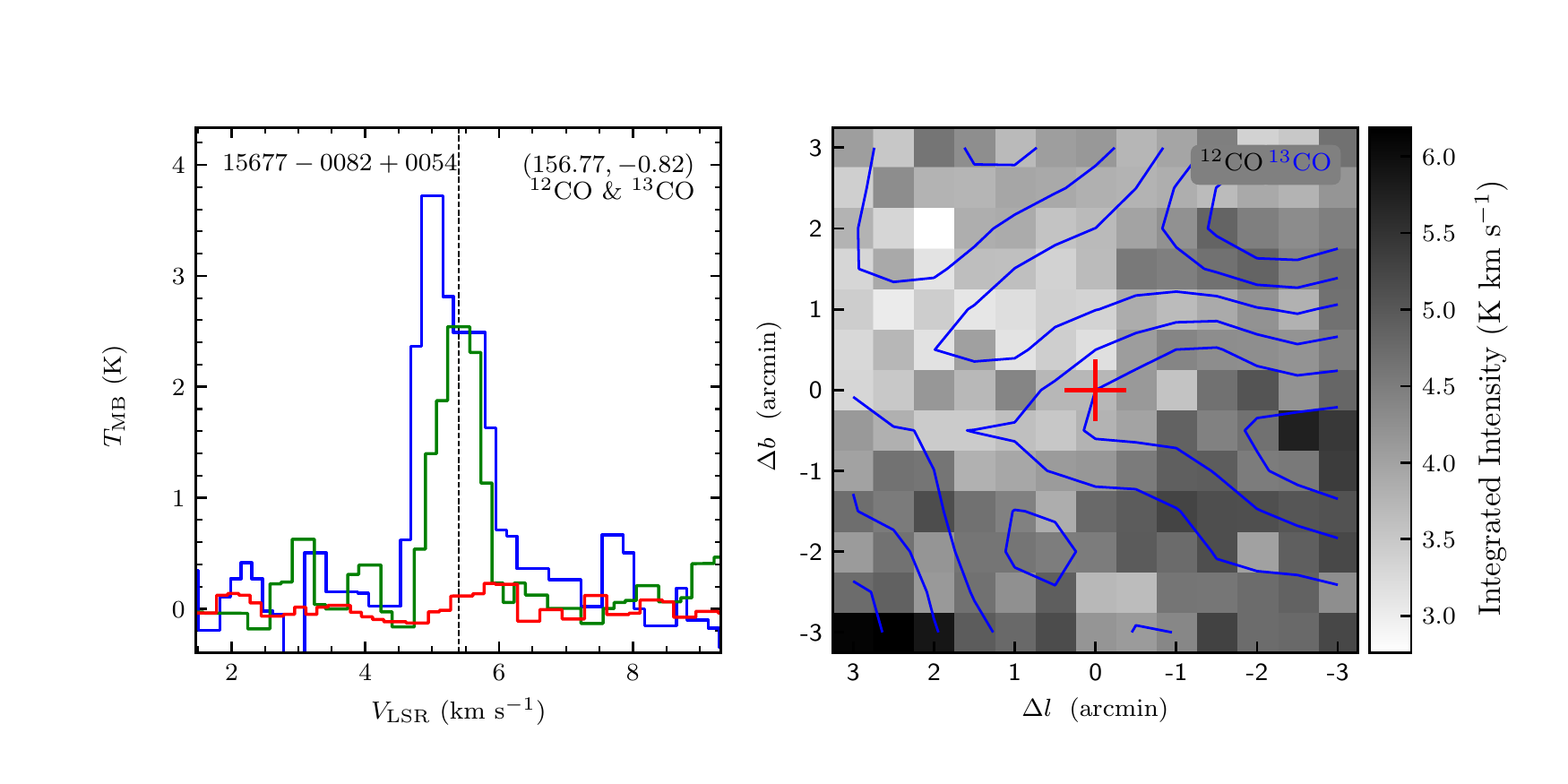}
\vspace{-0.5cm}

\includegraphics[width=9.0cm,angle=0]{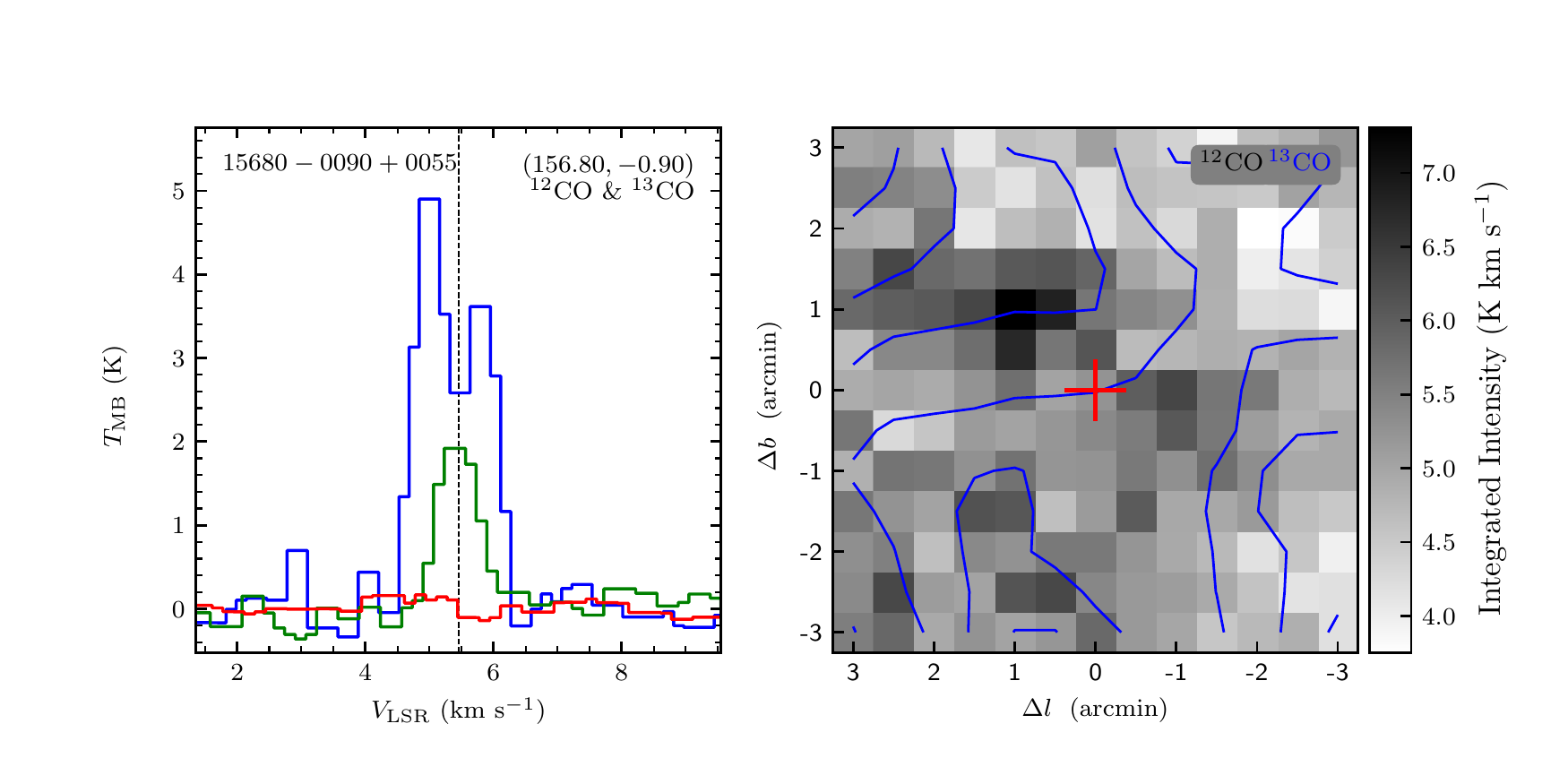}
\includegraphics[width=9.0cm,angle=0]{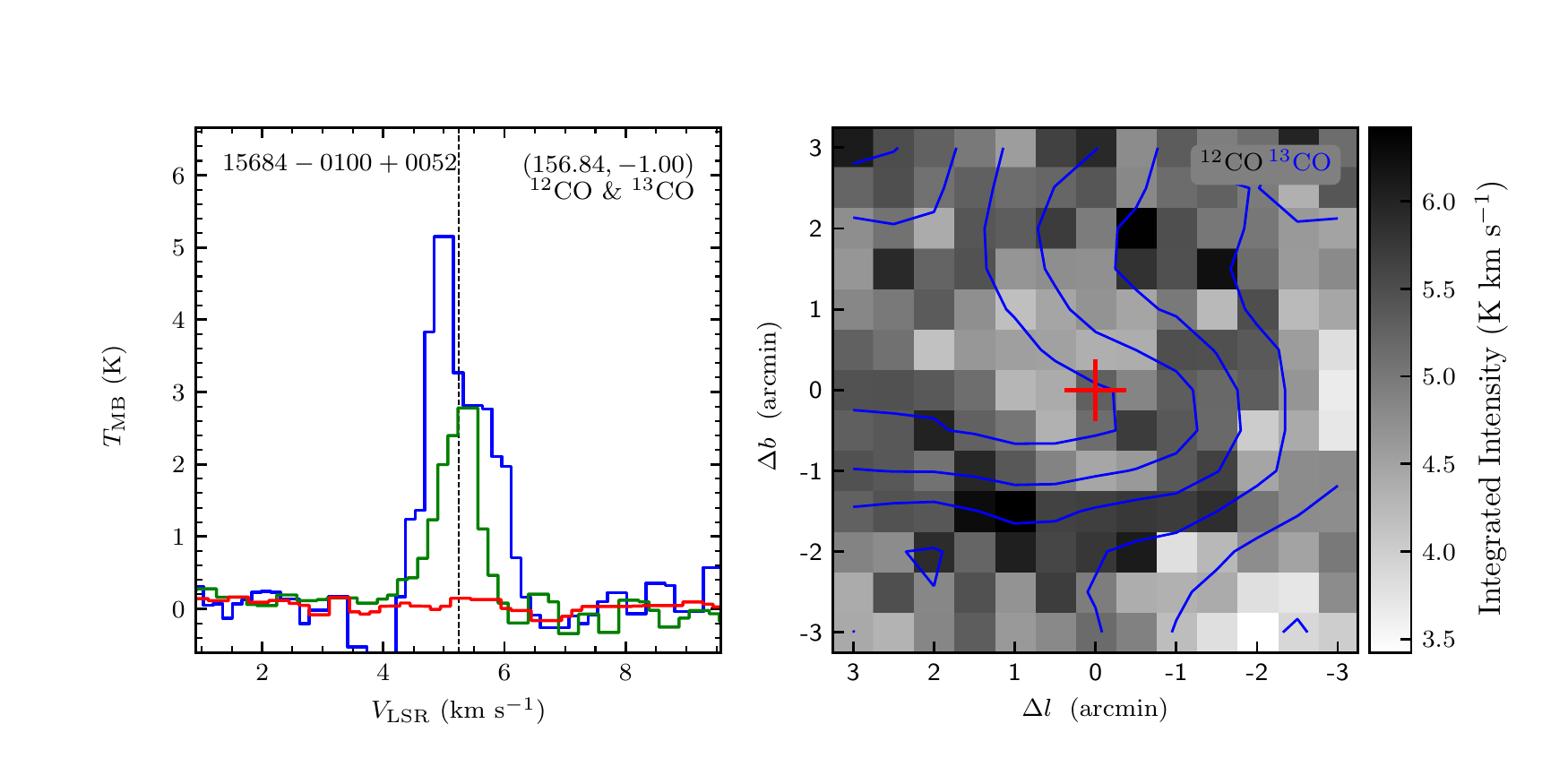}
\vspace{-0.5cm}

\includegraphics[width=9.0cm,angle=0]{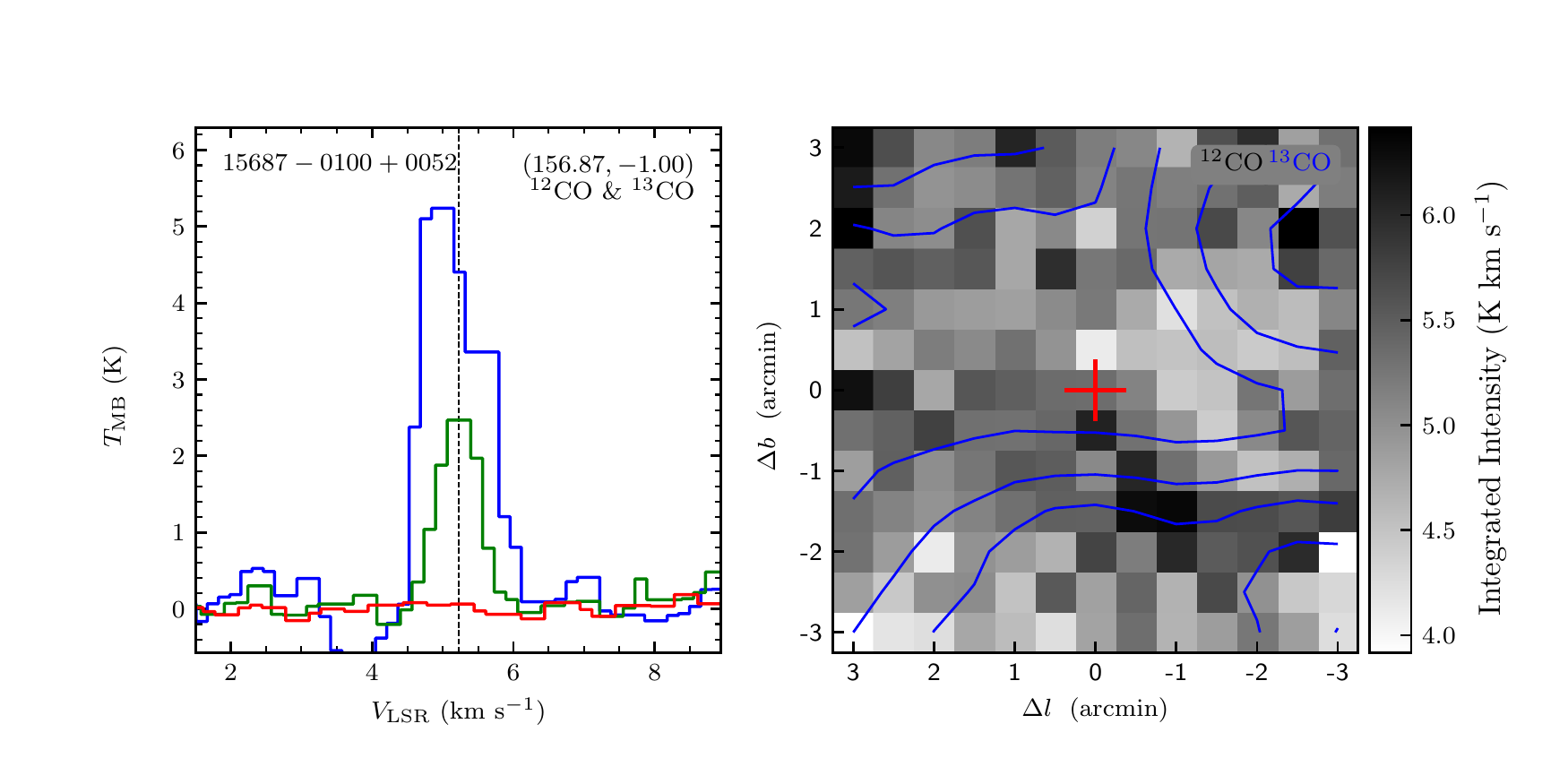}
\includegraphics[width=9.0cm,angle=0]{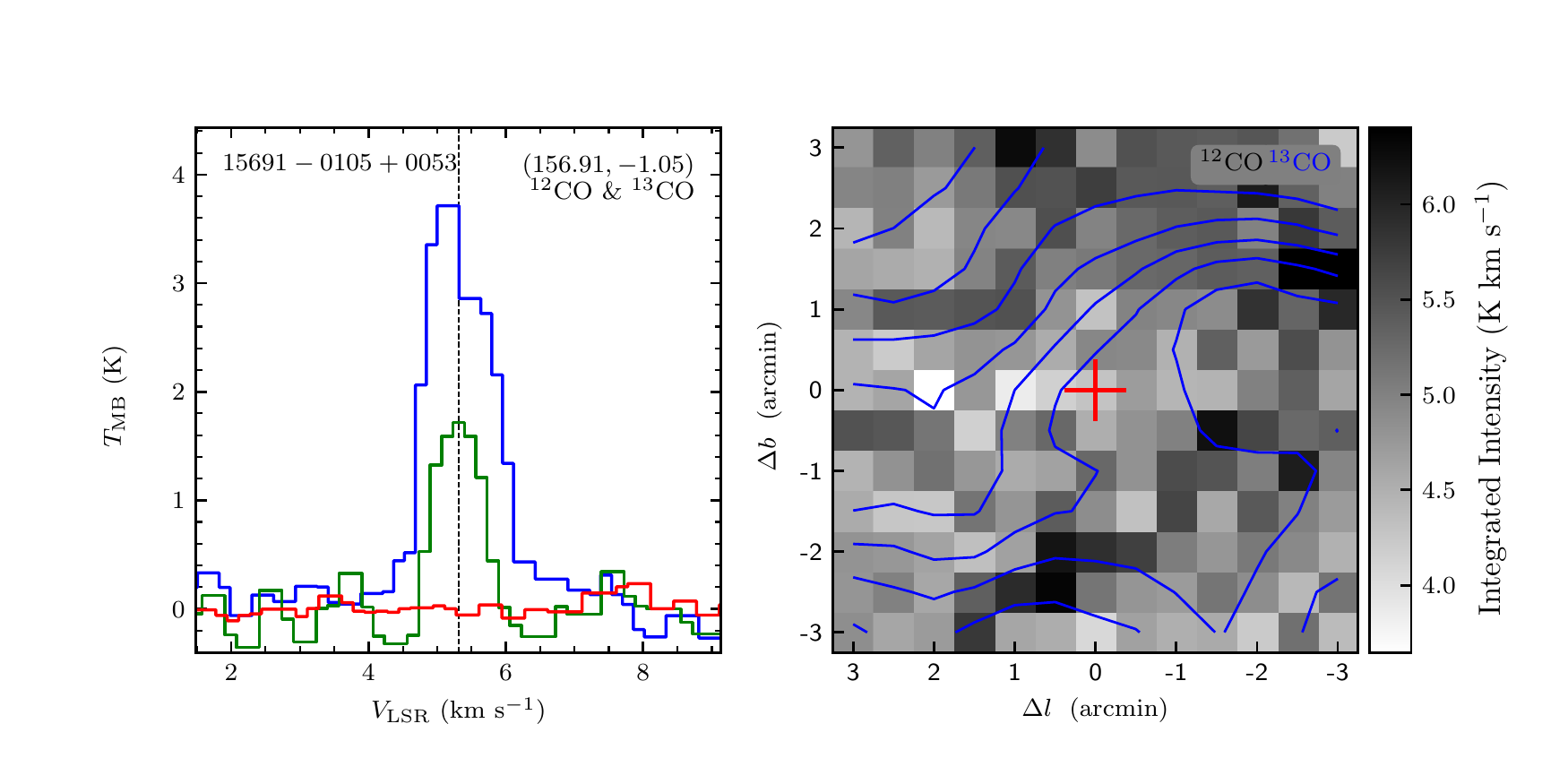}
\vspace{-0.5cm}

\includegraphics[width=9.0cm,angle=0]{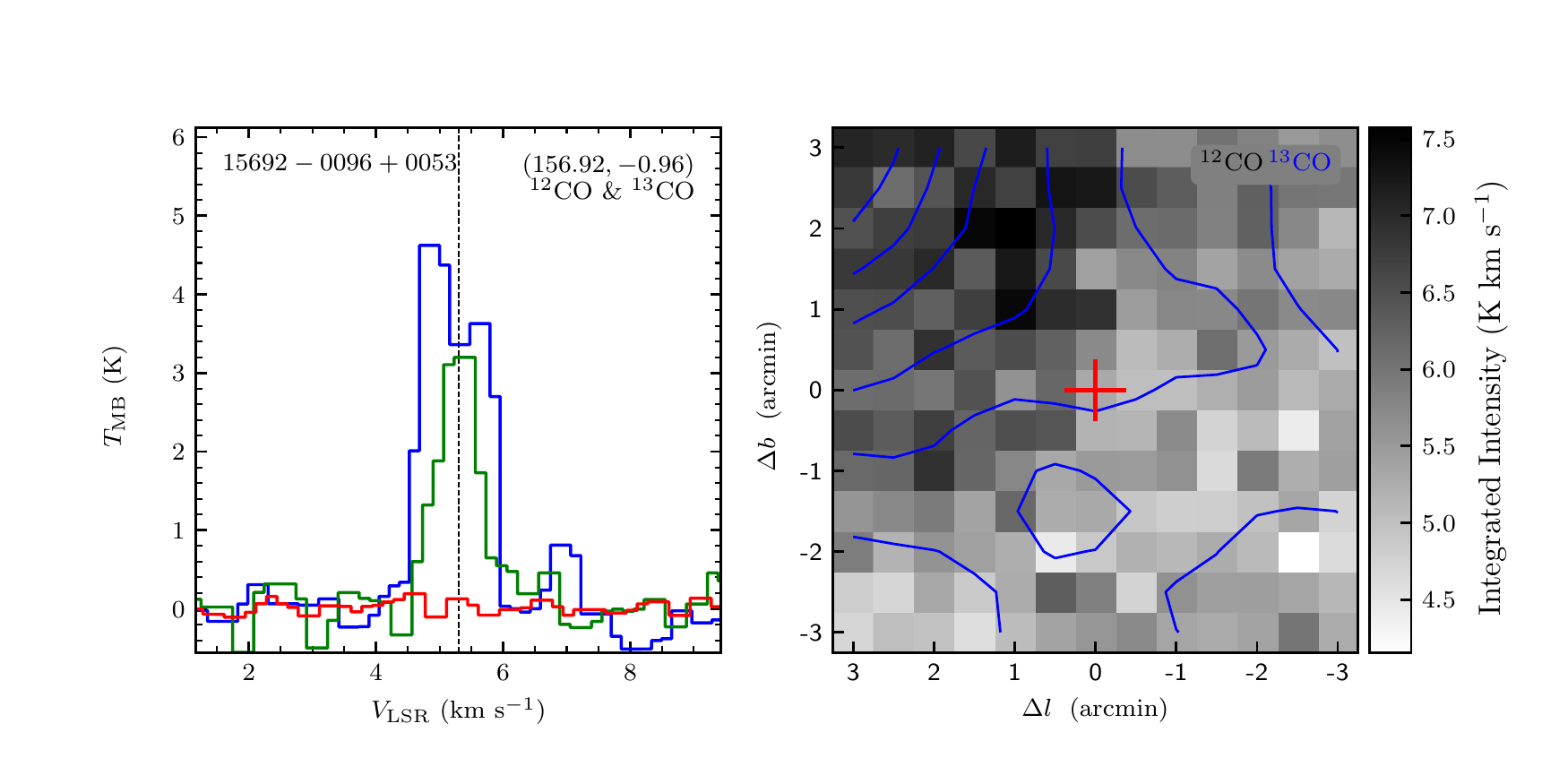}
\includegraphics[width=9.0cm,angle=0]{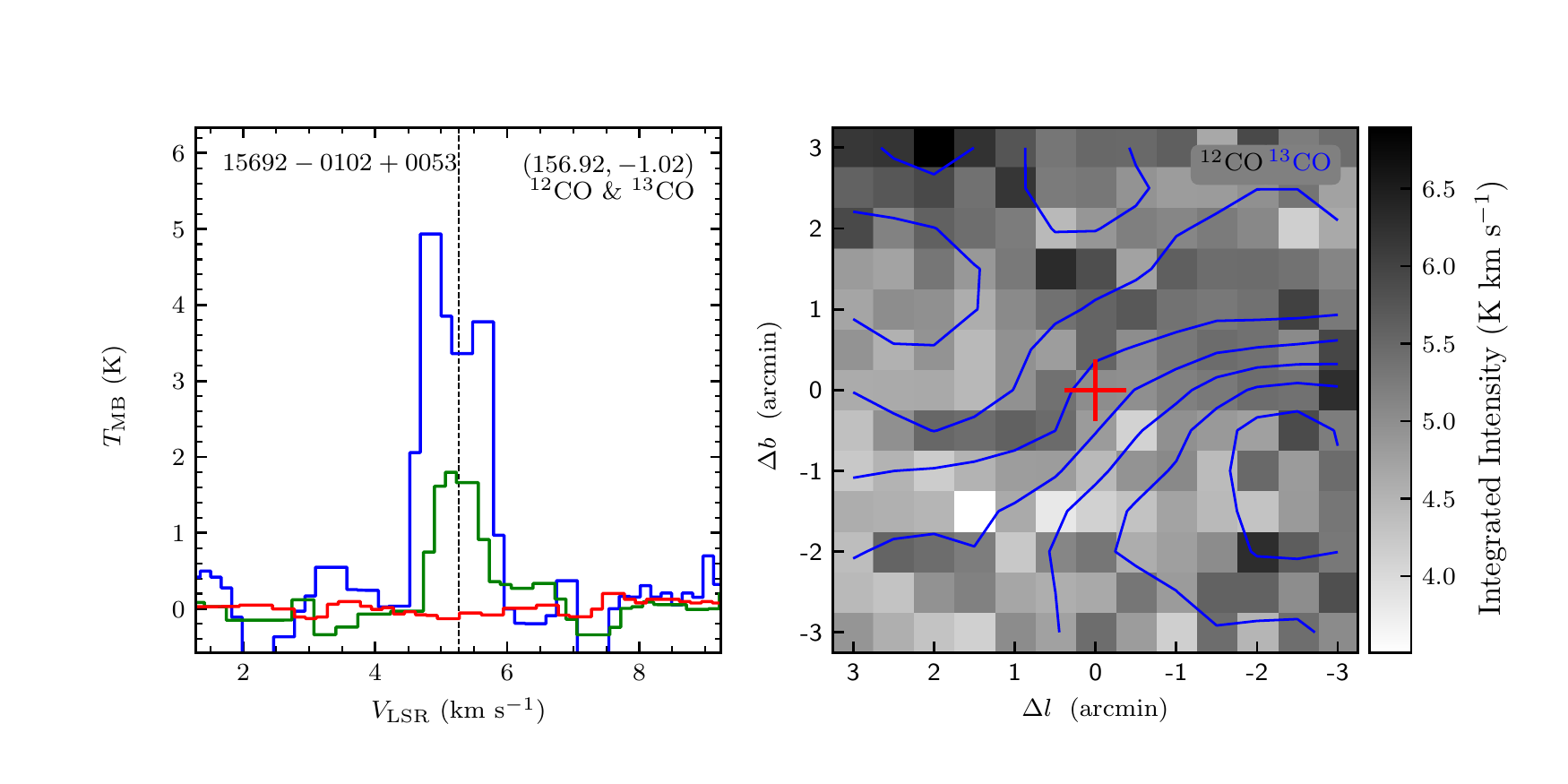}
\vspace{-0.5cm}

\includegraphics[width=9.0cm,angle=0]{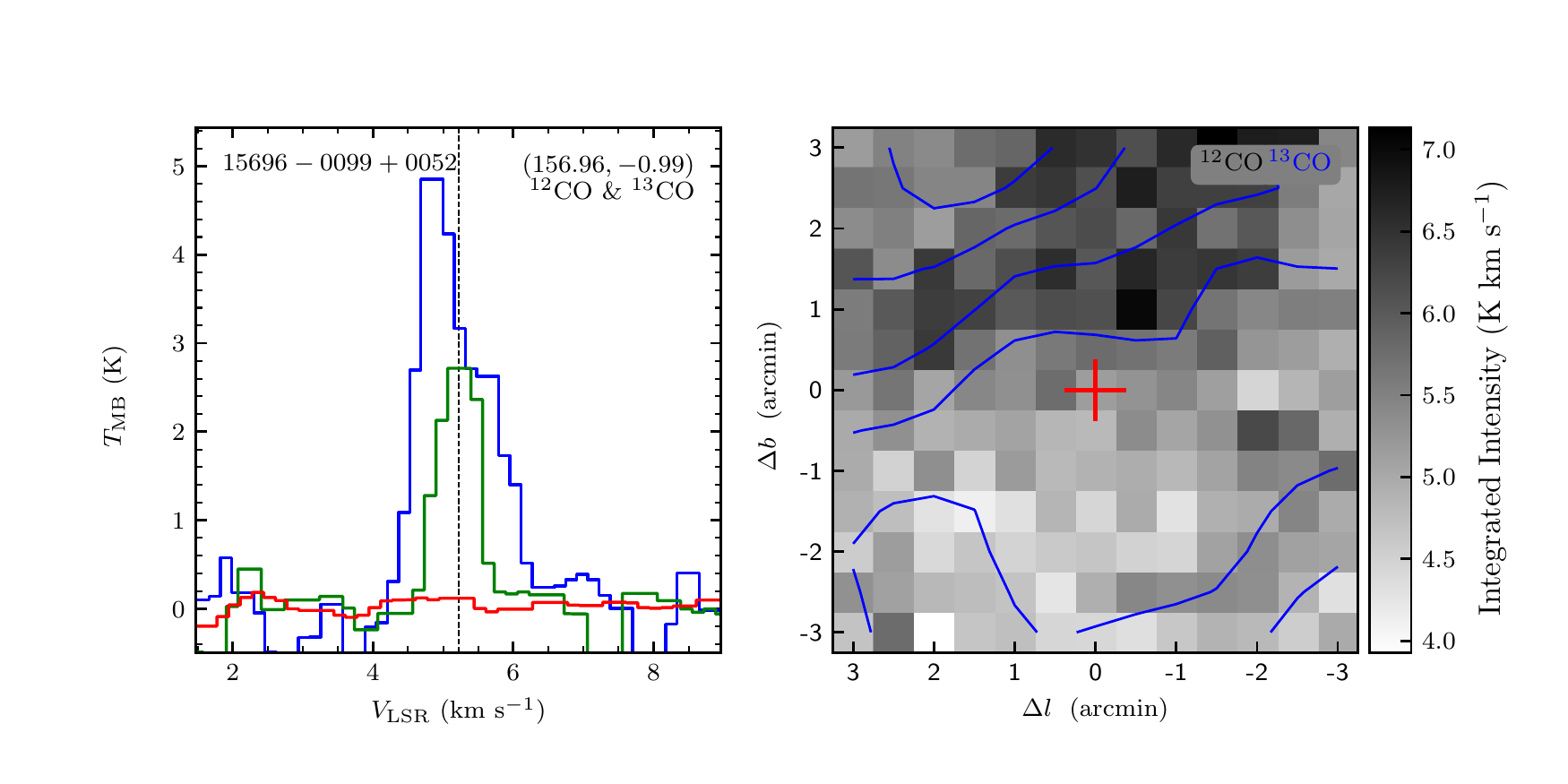}
\includegraphics[width=9.0cm,angle=0]{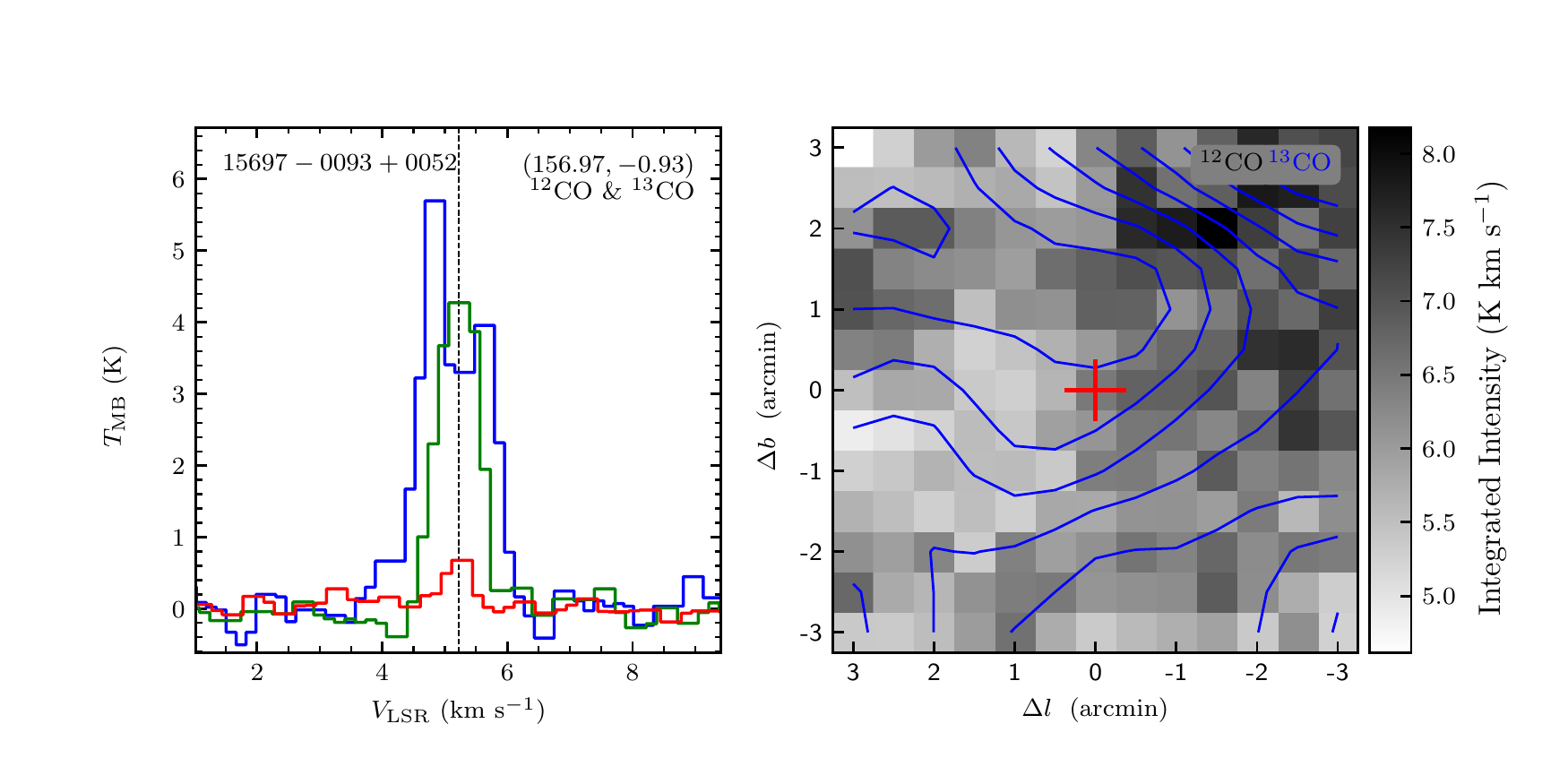}
\end{figure}
\clearpage

\begin{figure}
\includegraphics[width=9.0cm,angle=0]{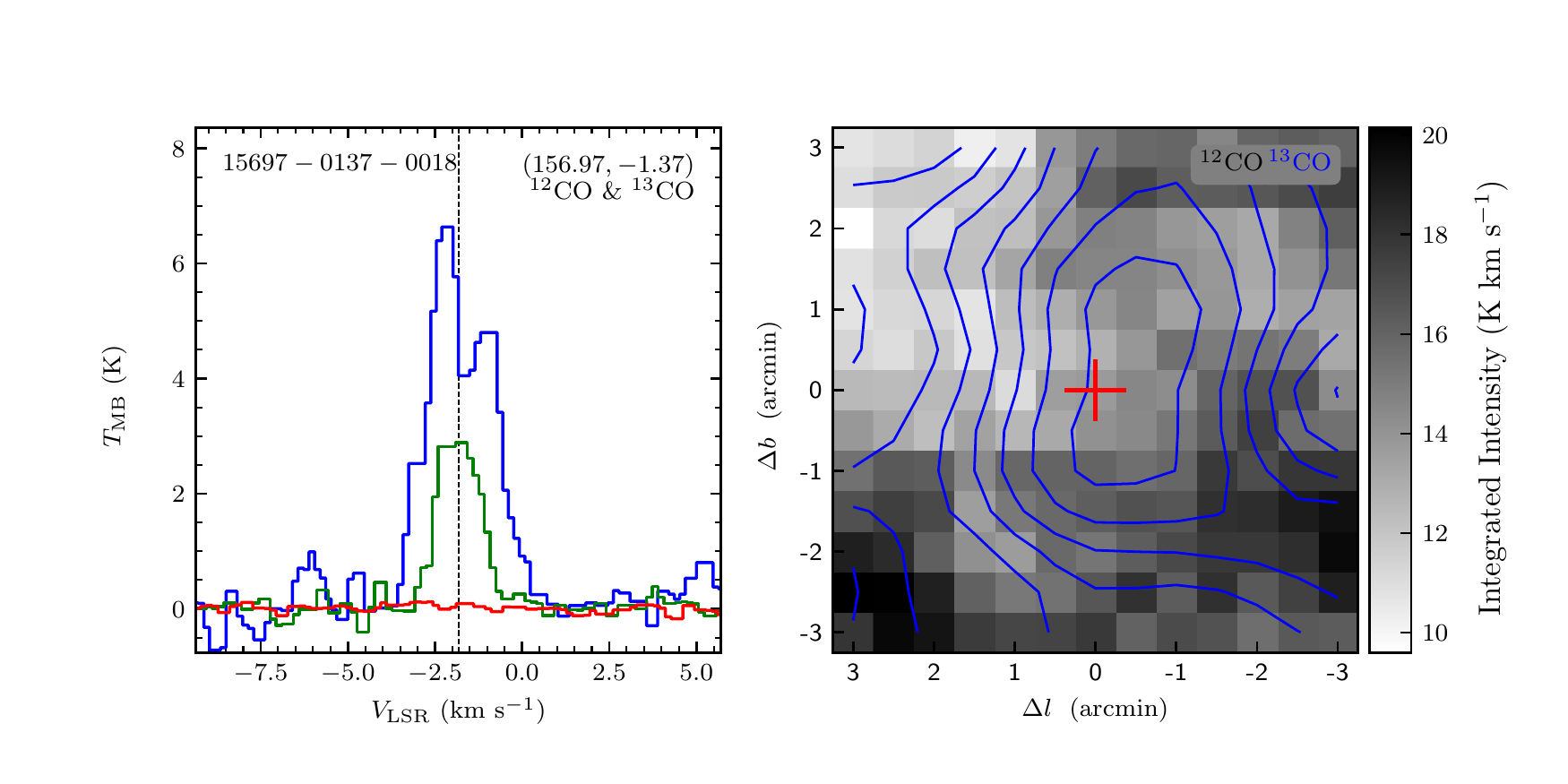}
\includegraphics[width=9.0cm,angle=0]{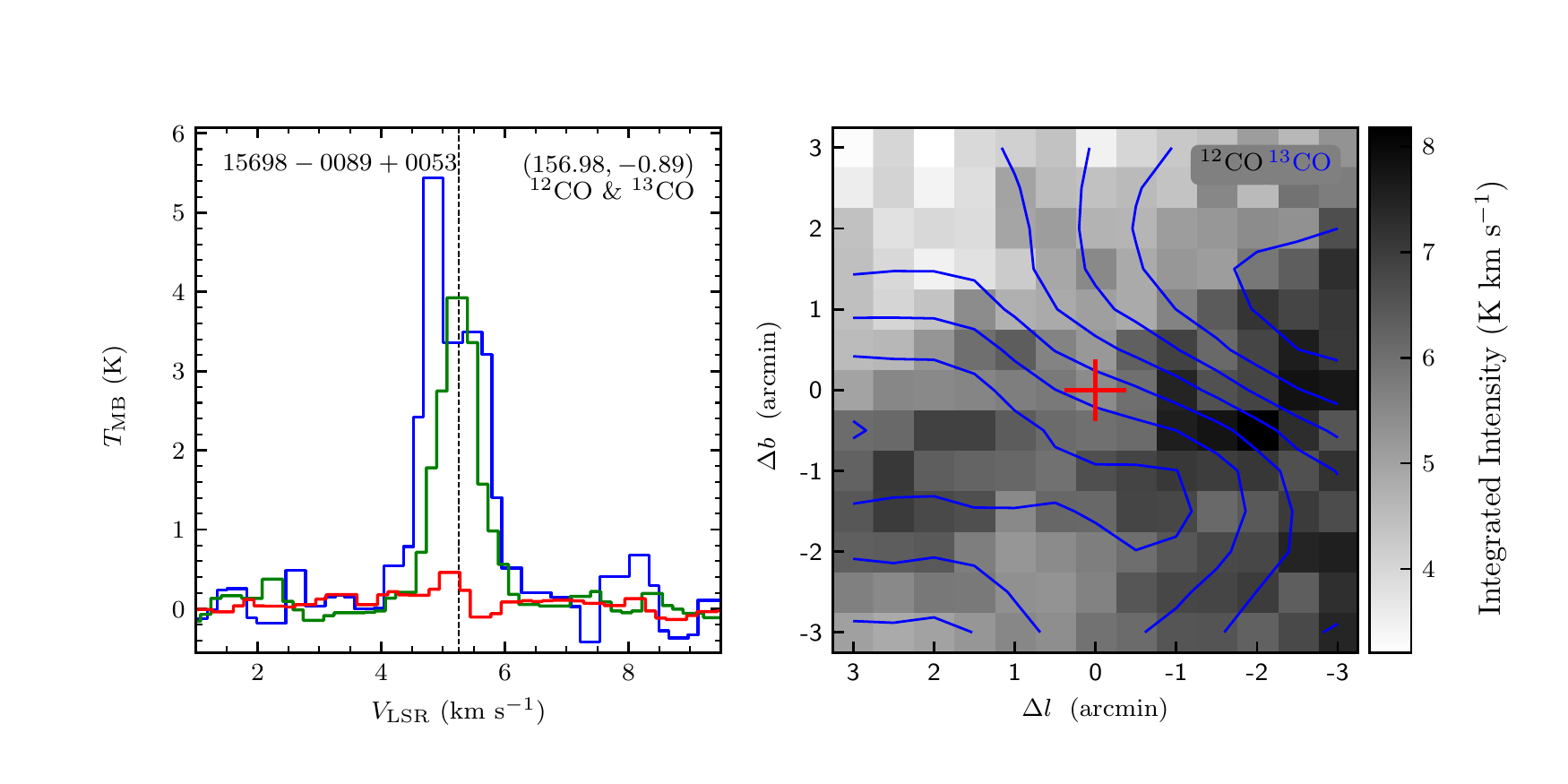}
\vspace{-0.5cm}

\includegraphics[width=9.0cm,angle=0]{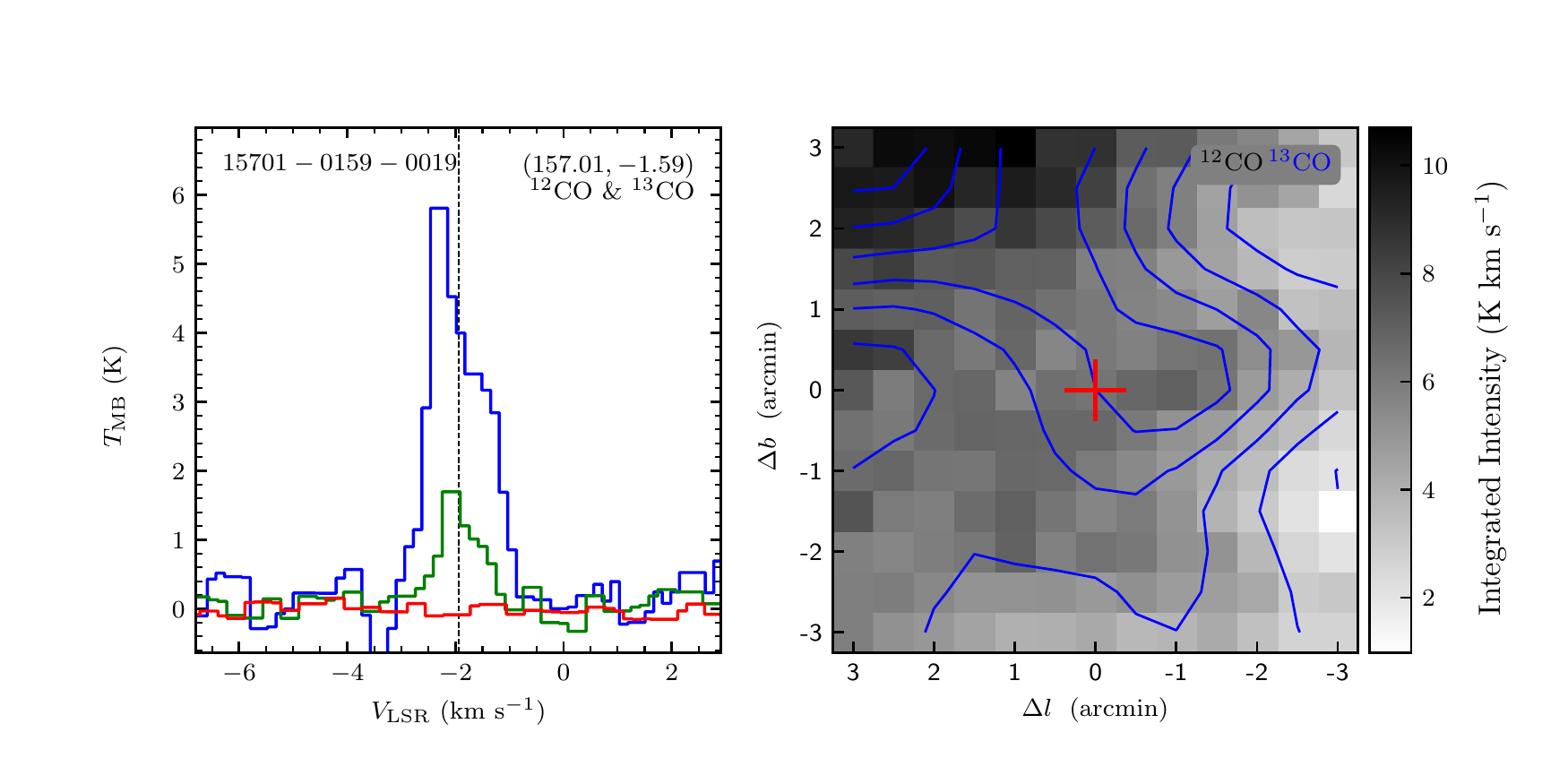}
\includegraphics[width=9.0cm,angle=0]{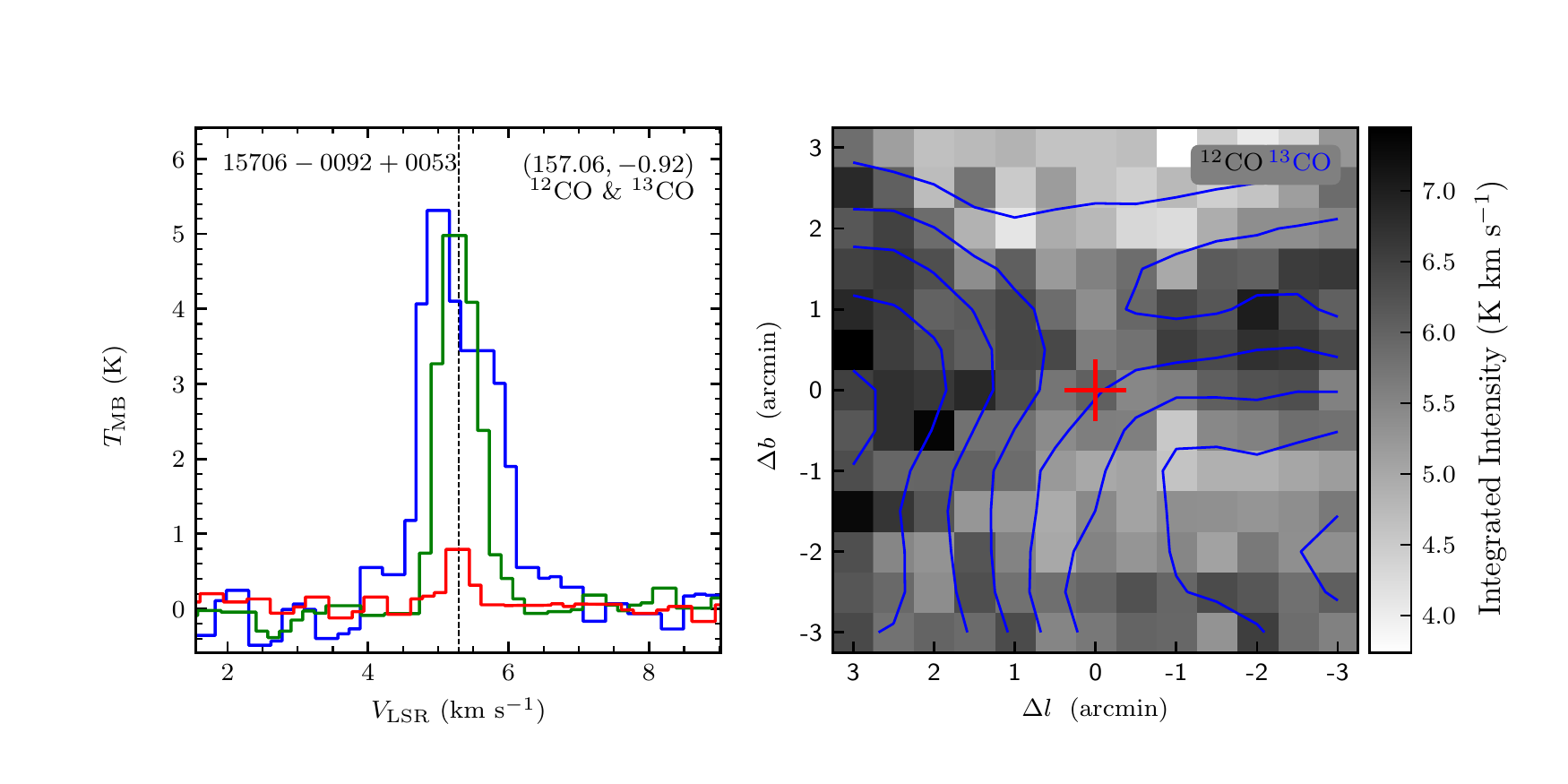}
\vspace{-0.5cm}

\includegraphics[width=9.0cm,angle=0]{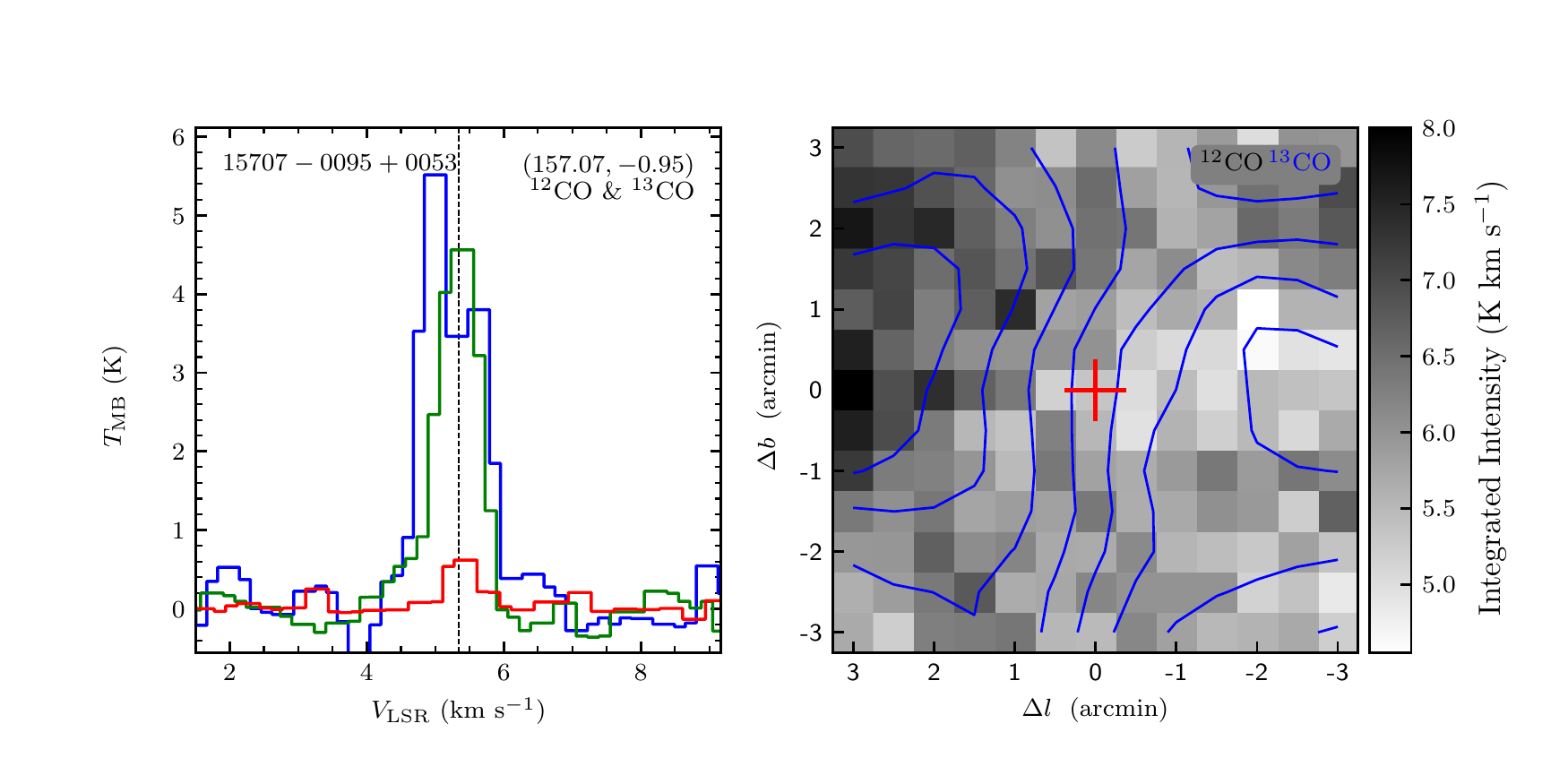}
\includegraphics[width=9.0cm,angle=0]{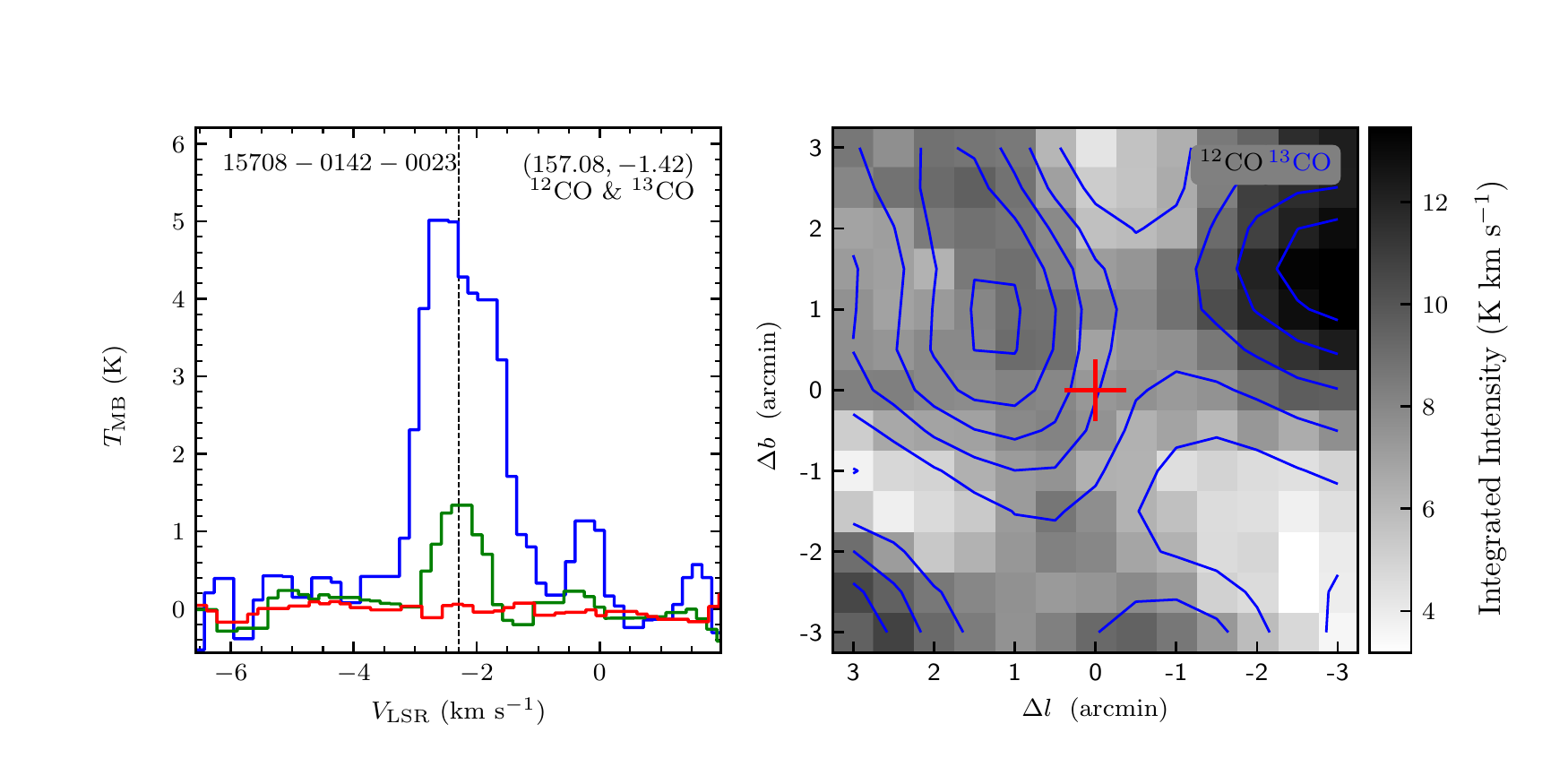}
\vspace{-0.5cm}

\includegraphics[width=9.0cm,angle=0]{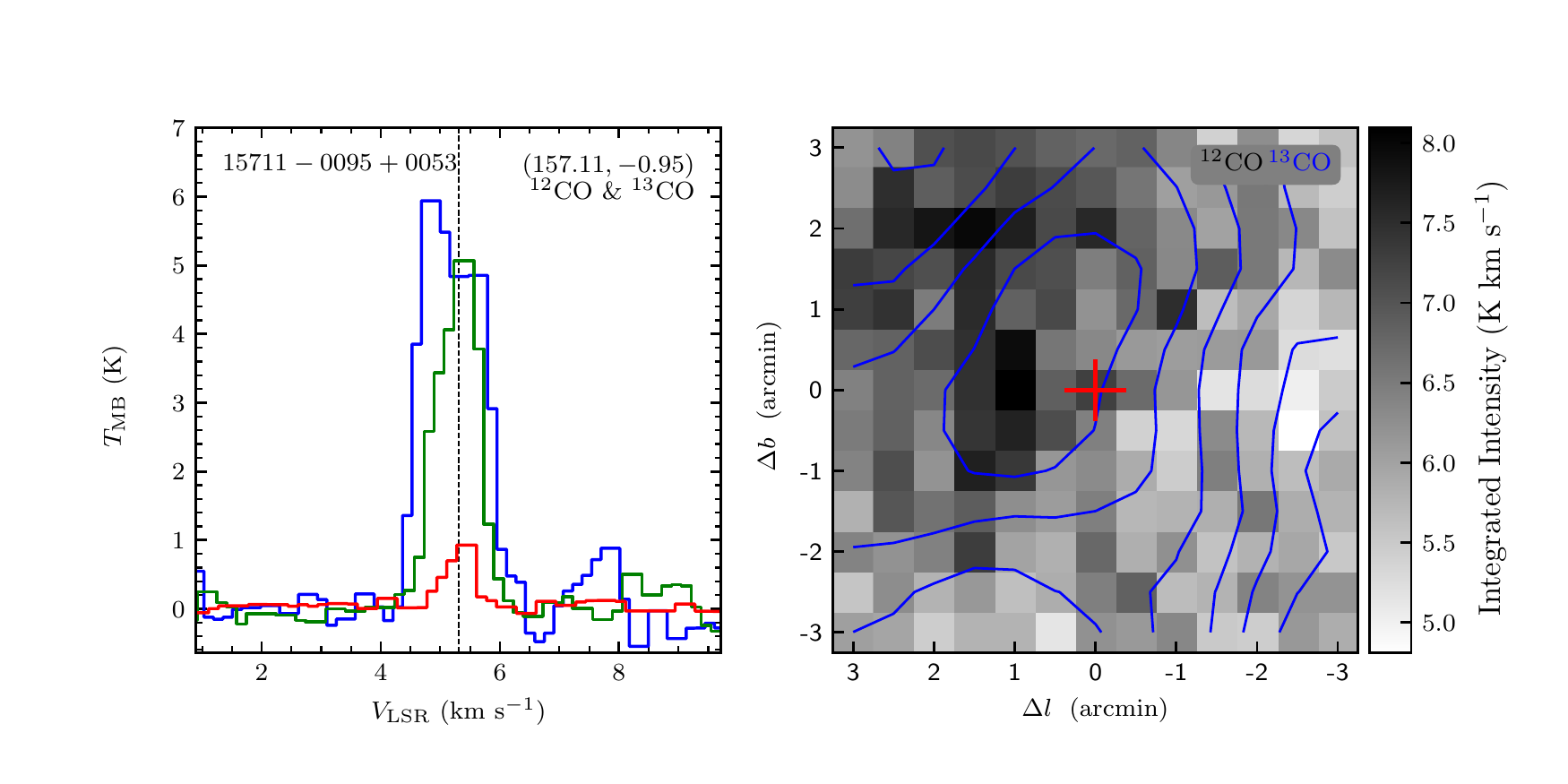}
\includegraphics[width=9.0cm,angle=0]{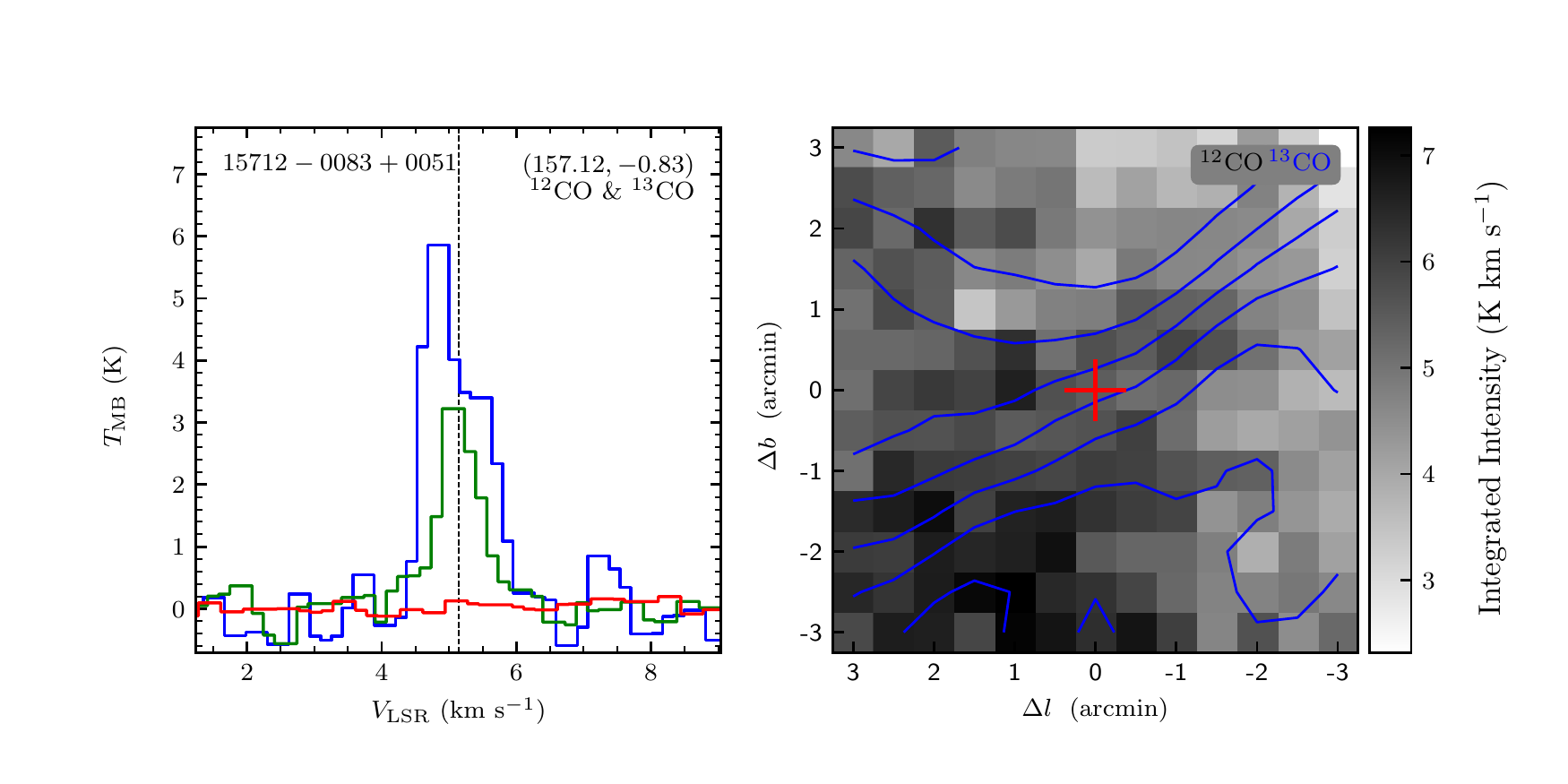}
\vspace{-0.5cm}

\includegraphics[width=9.0cm,angle=0]{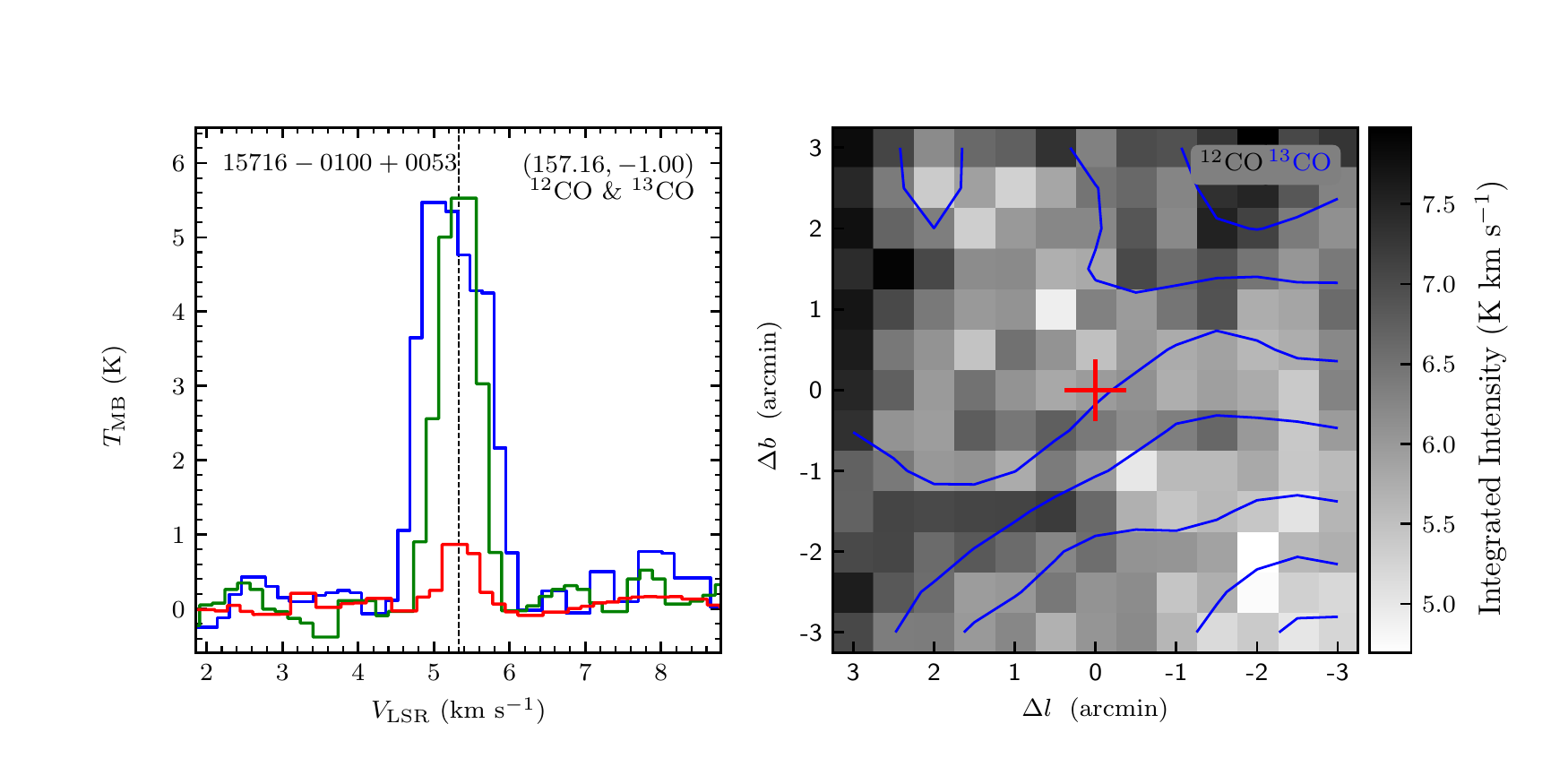}
\includegraphics[width=9.0cm,angle=0]{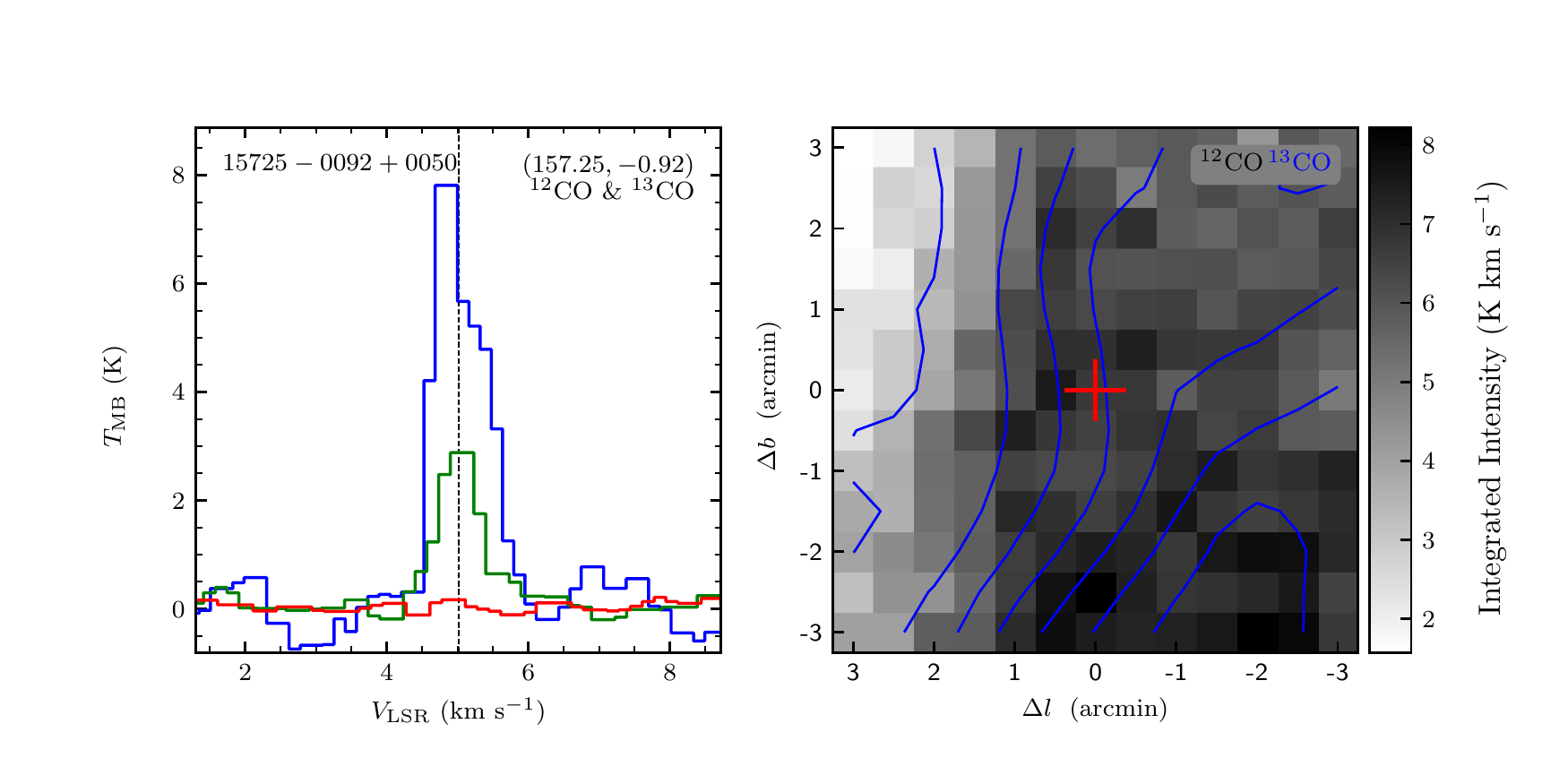}
\end{figure}
\clearpage

\begin{figure}
\includegraphics[width=9.0cm,angle=0]{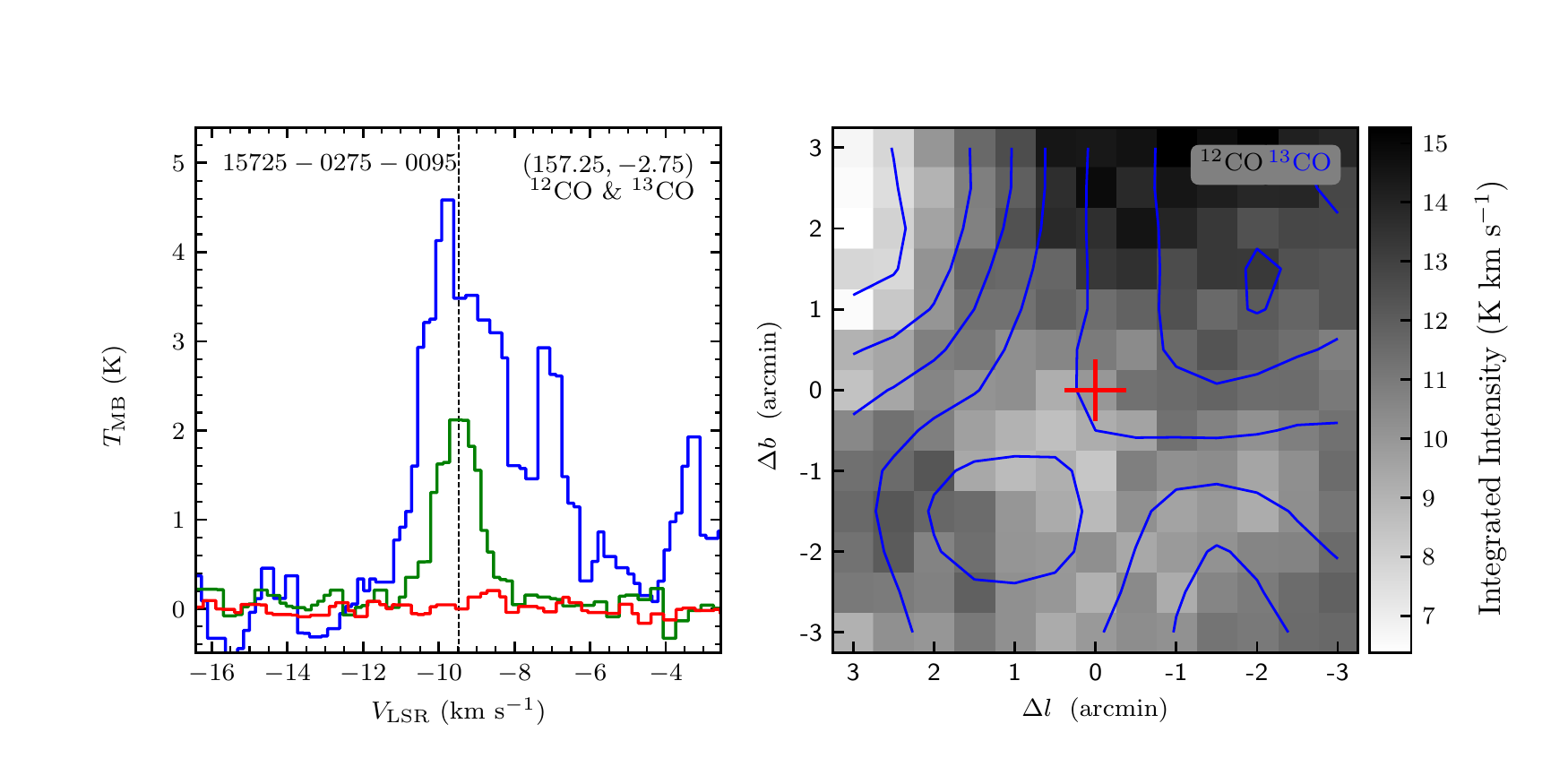}
\includegraphics[width=9.0cm,angle=0]{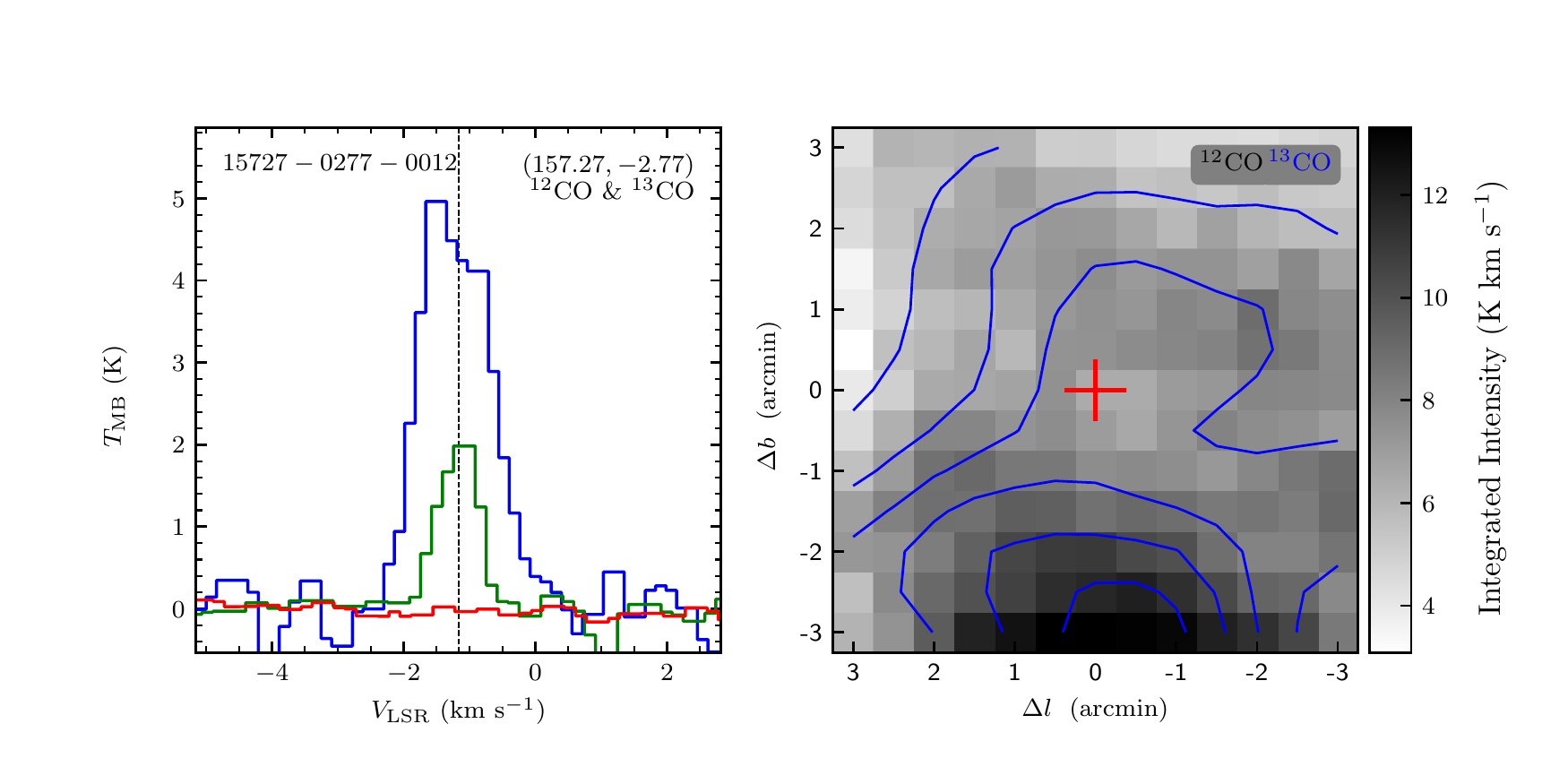}
\vspace{-0.5cm}

\includegraphics[width=9.0cm,angle=0]{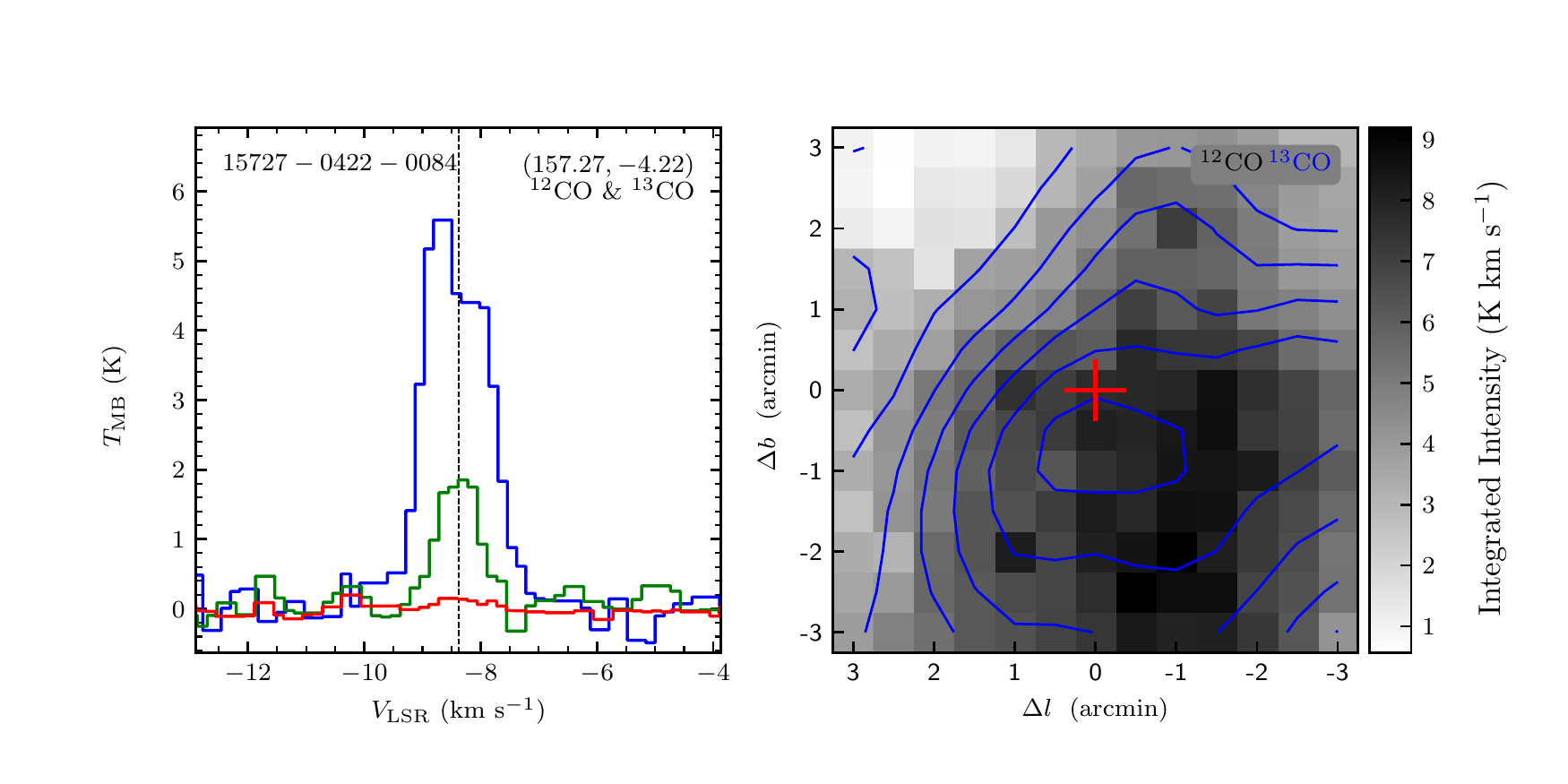}
\includegraphics[width=9.0cm,angle=0]{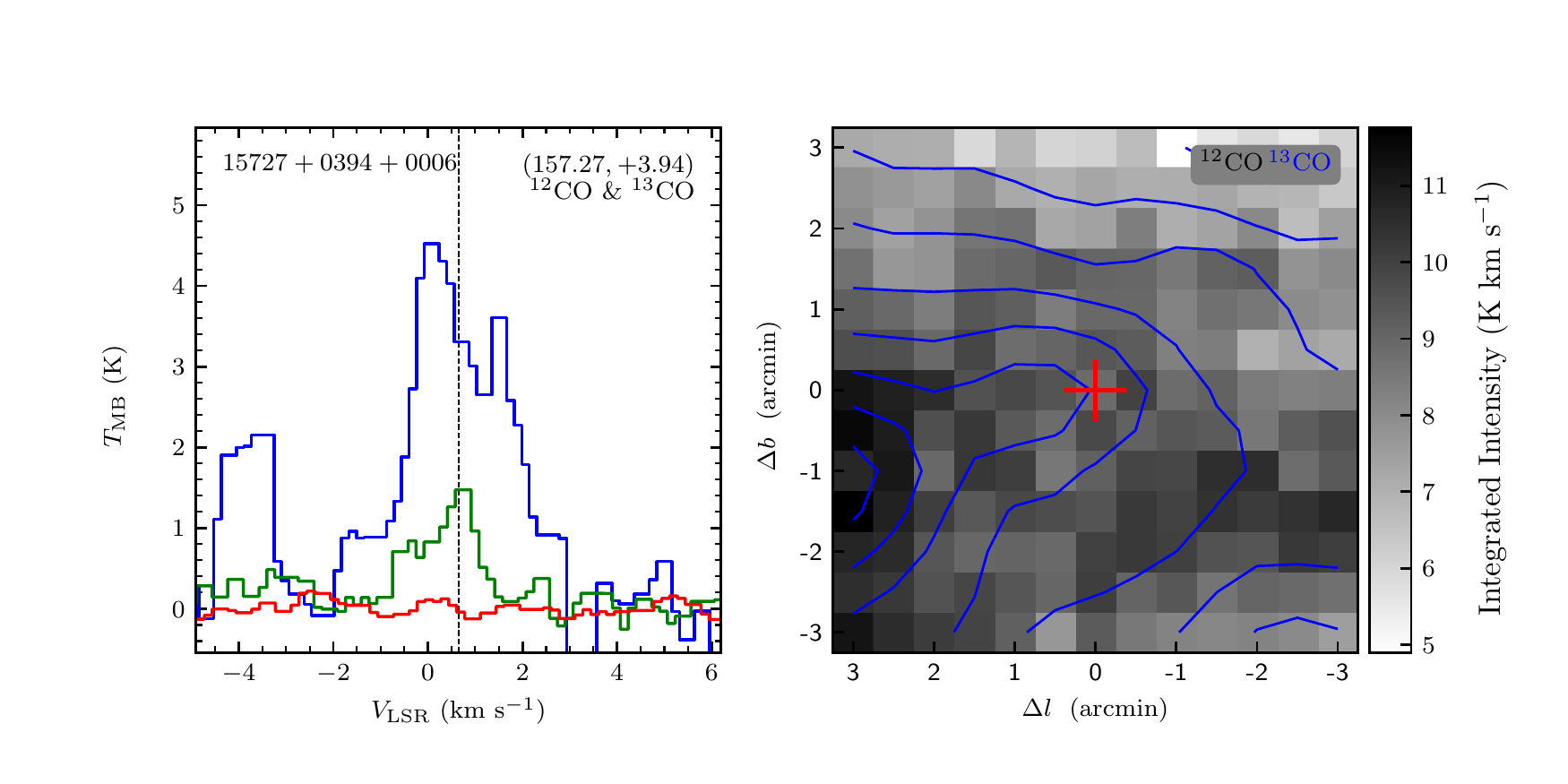}
\vspace{-0.5cm}

\includegraphics[width=9.0cm,angle=0]{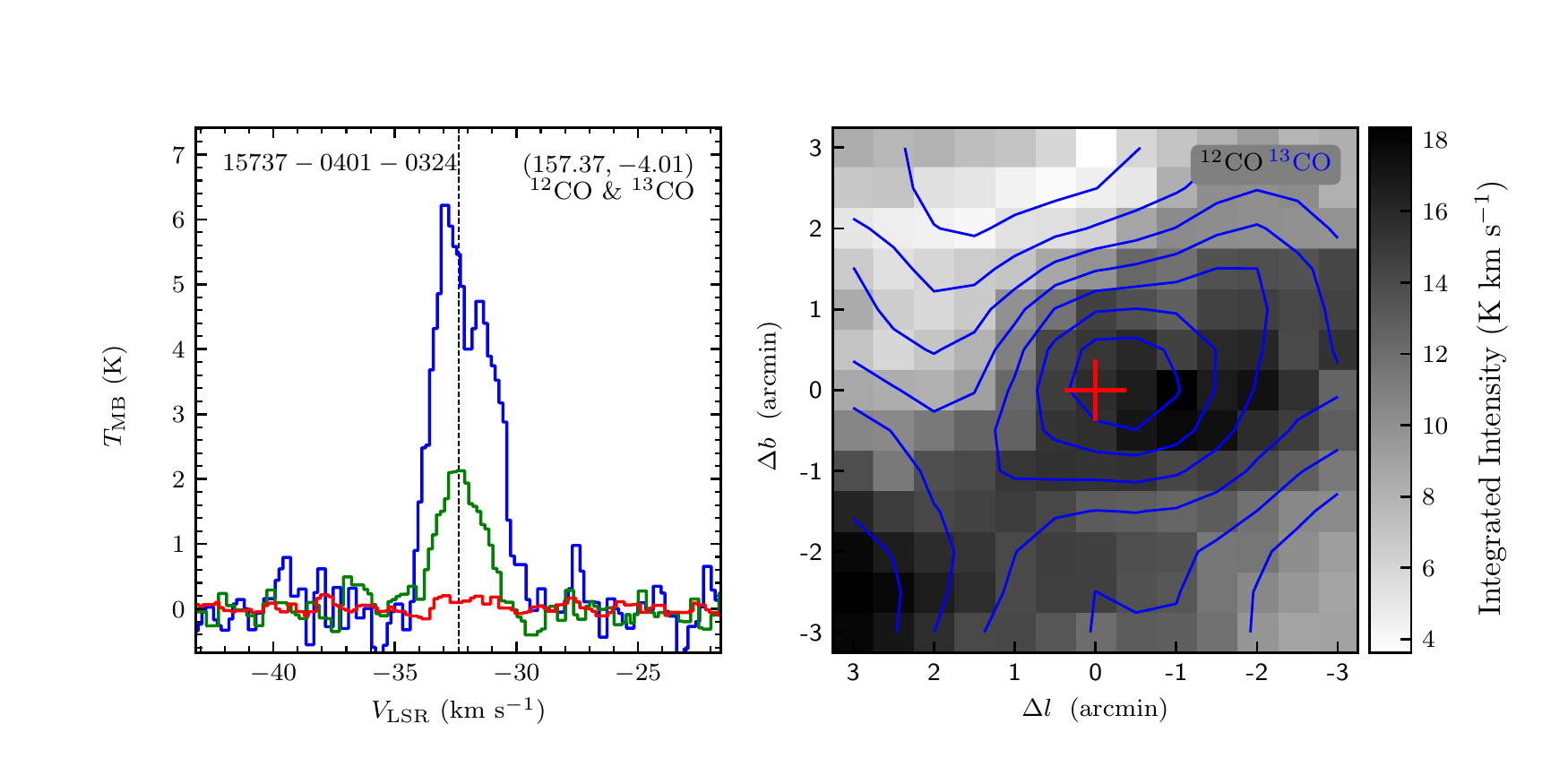}
\includegraphics[width=9.0cm,angle=0]{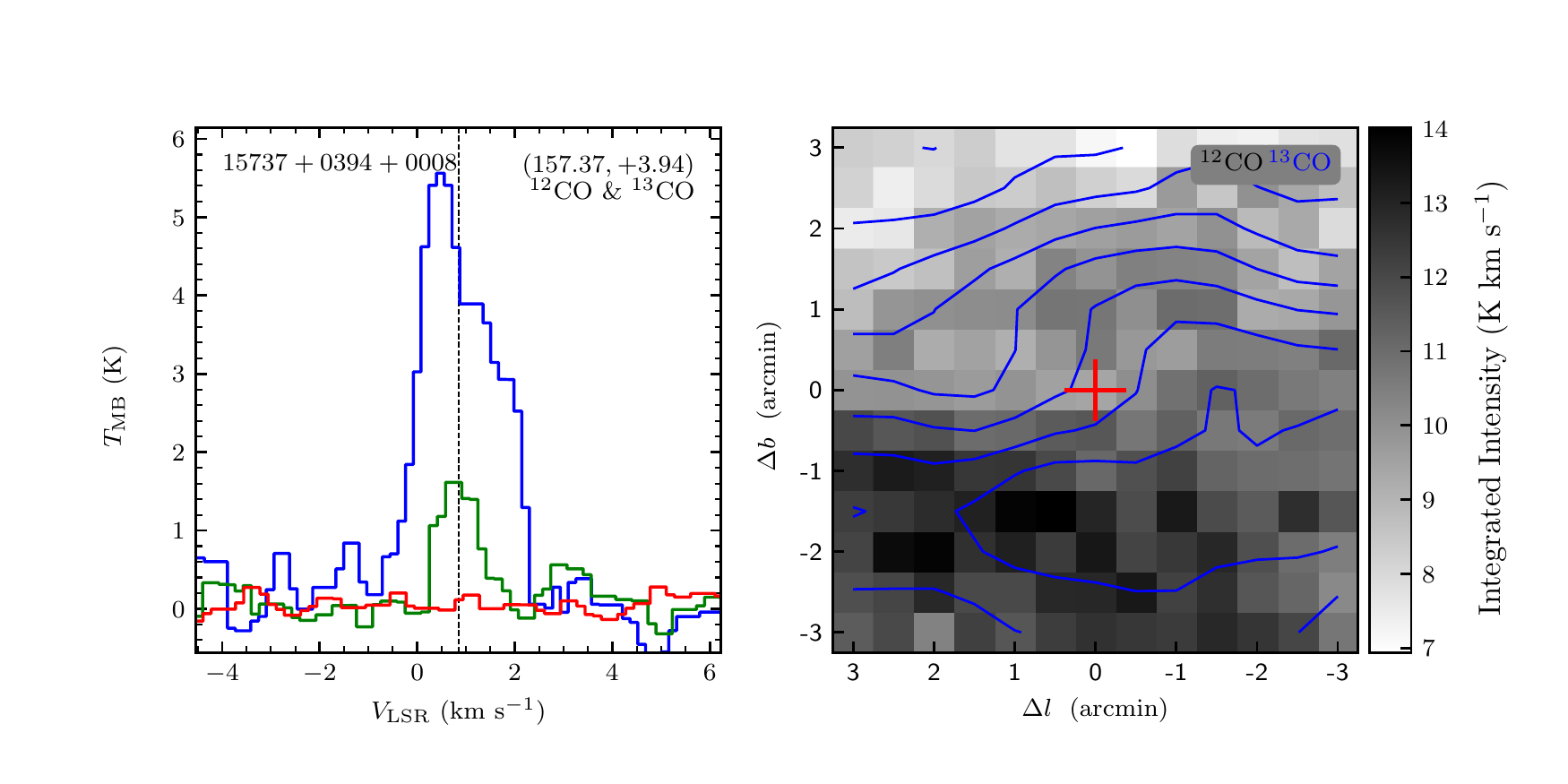}
\vspace{-0.5cm}

\includegraphics[width=9.0cm,angle=0]{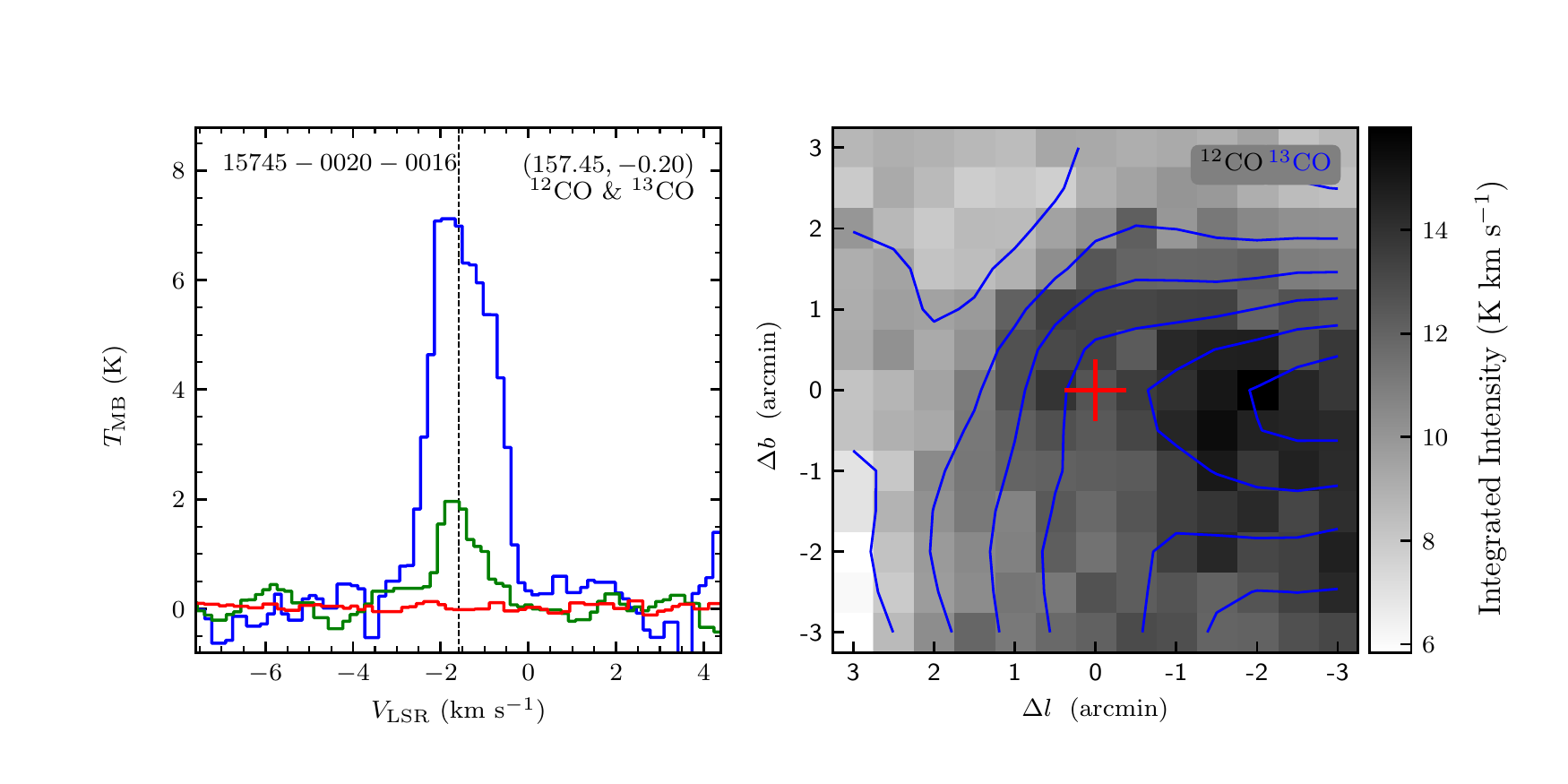}
\includegraphics[width=9.0cm,angle=0]{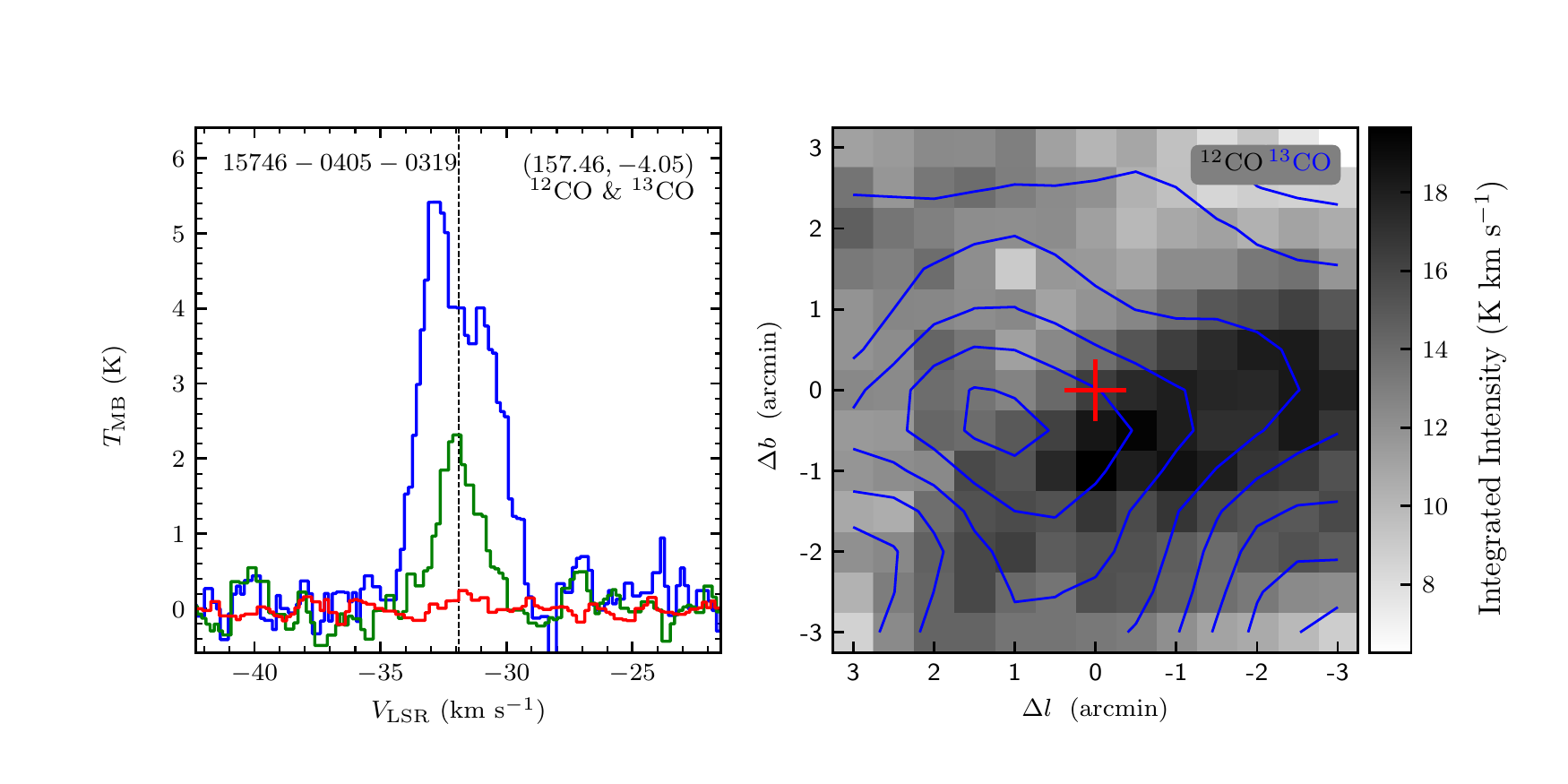}
\vspace{-0.5cm}

\includegraphics[width=9.0cm,angle=0]{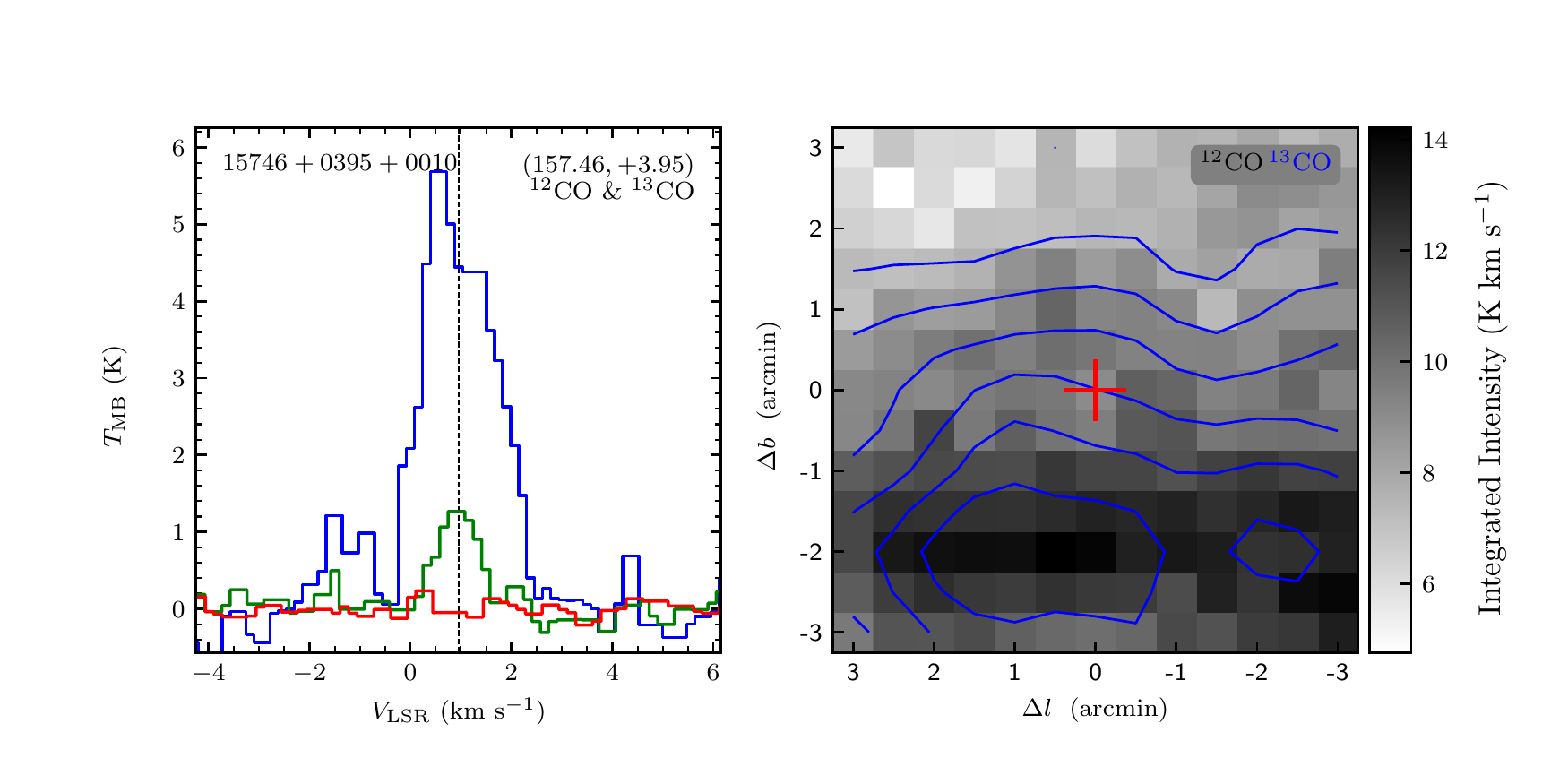}
\includegraphics[width=9.0cm,angle=0]{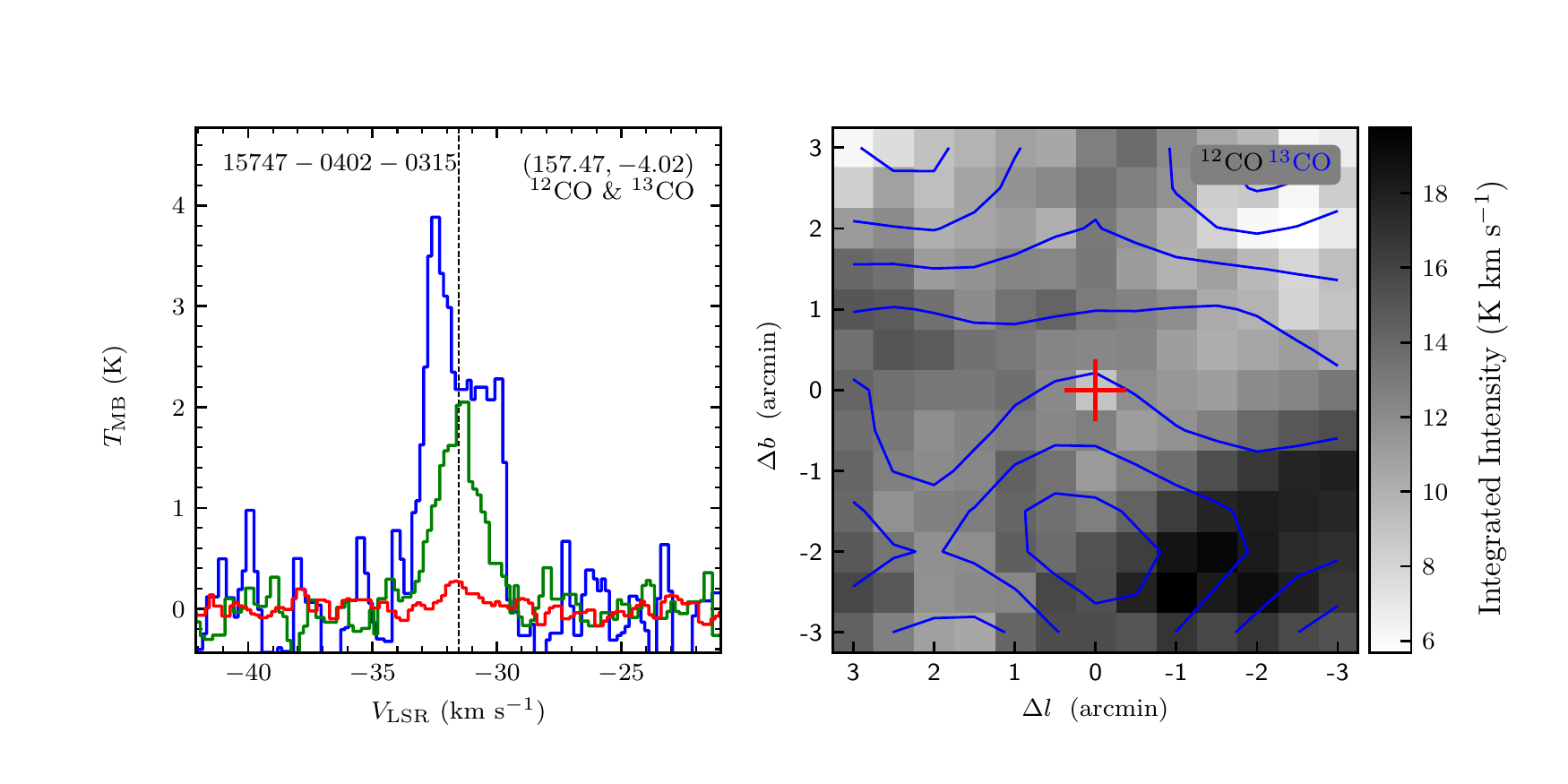}
\end{figure}
\clearpage

\begin{figure}
\includegraphics[width=9.0cm,angle=0]{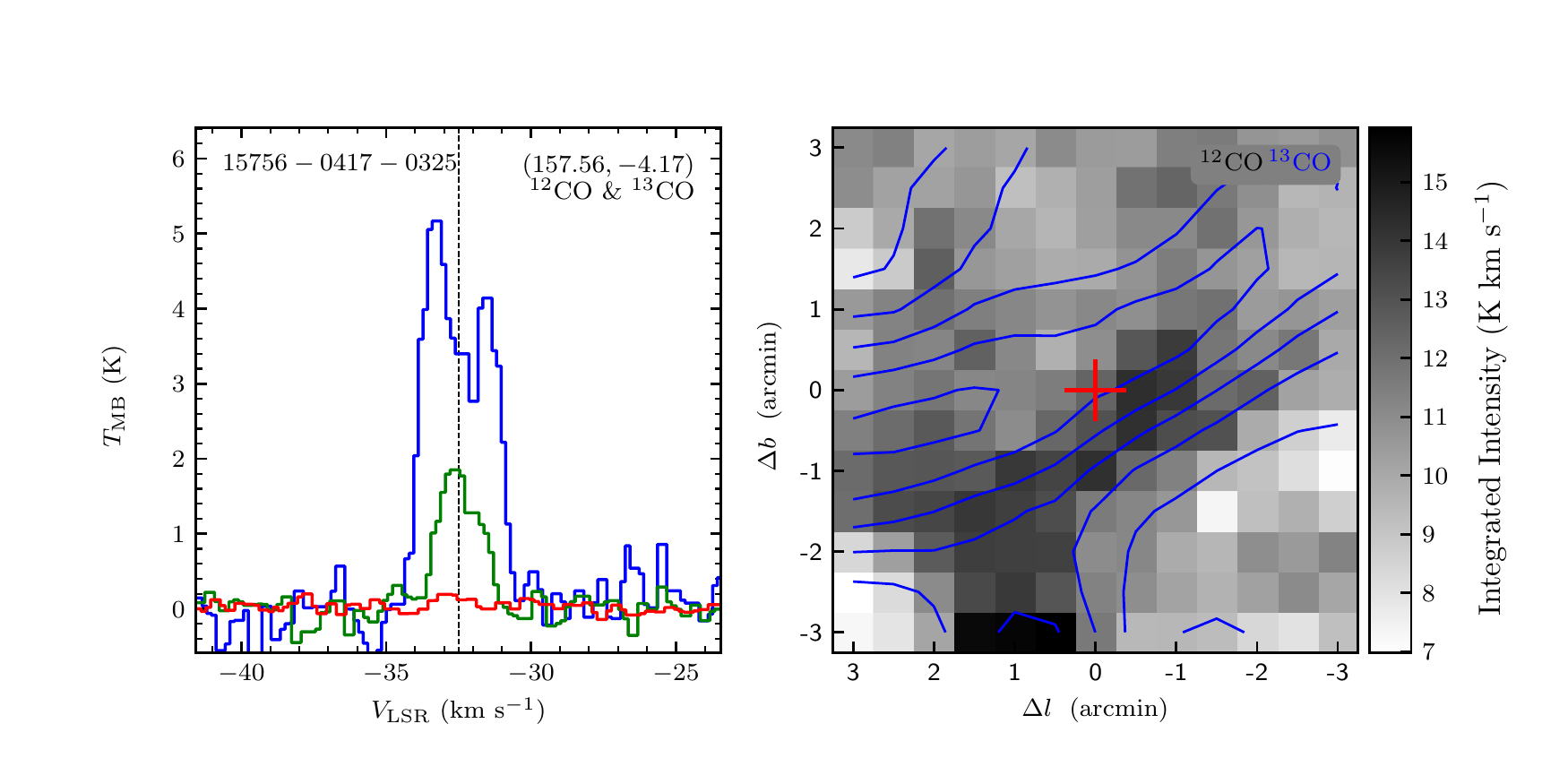}
\includegraphics[width=9.0cm,angle=0]{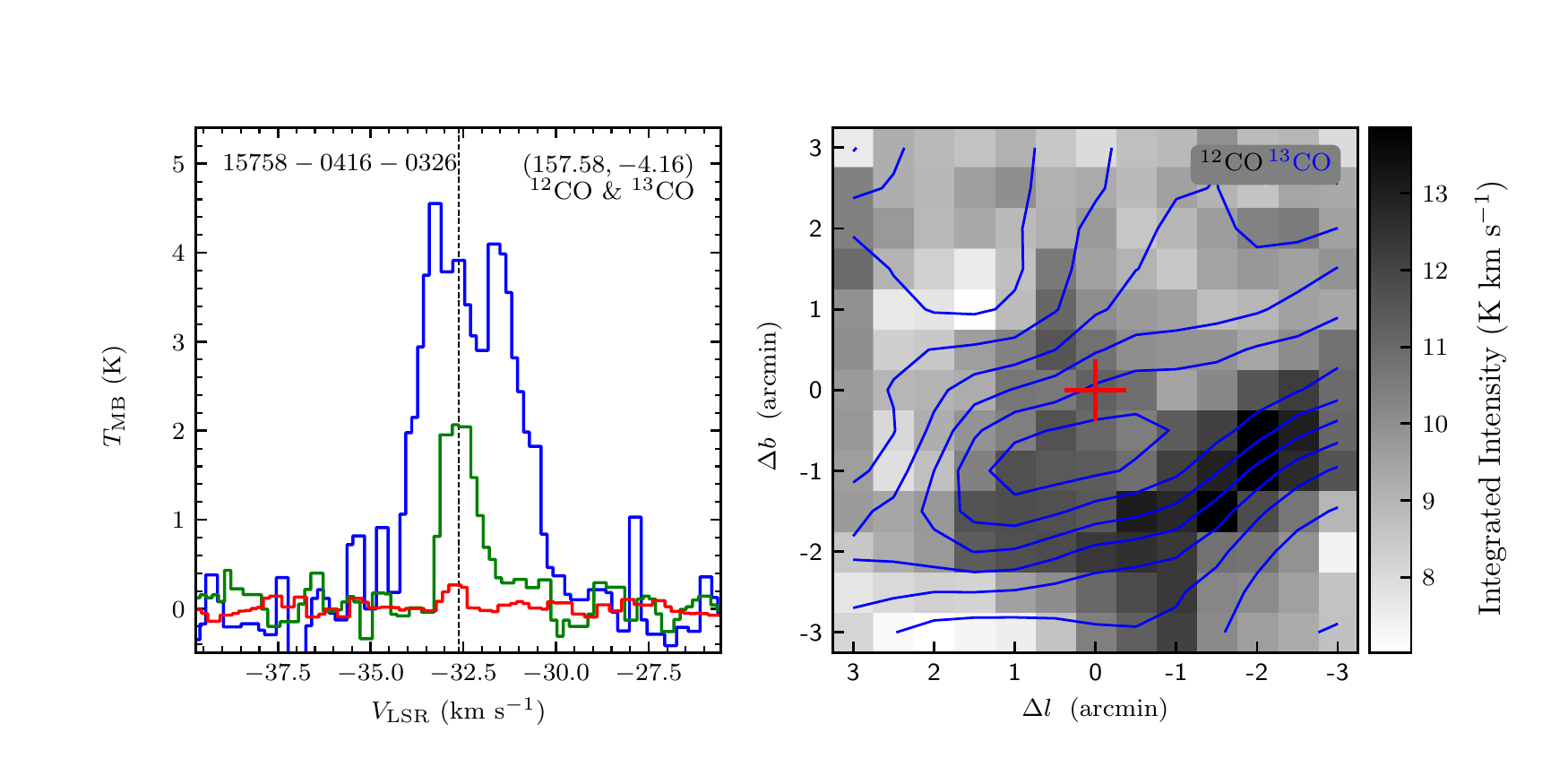}
\vspace{-0.5cm}

\includegraphics[width=9.0cm,angle=0]{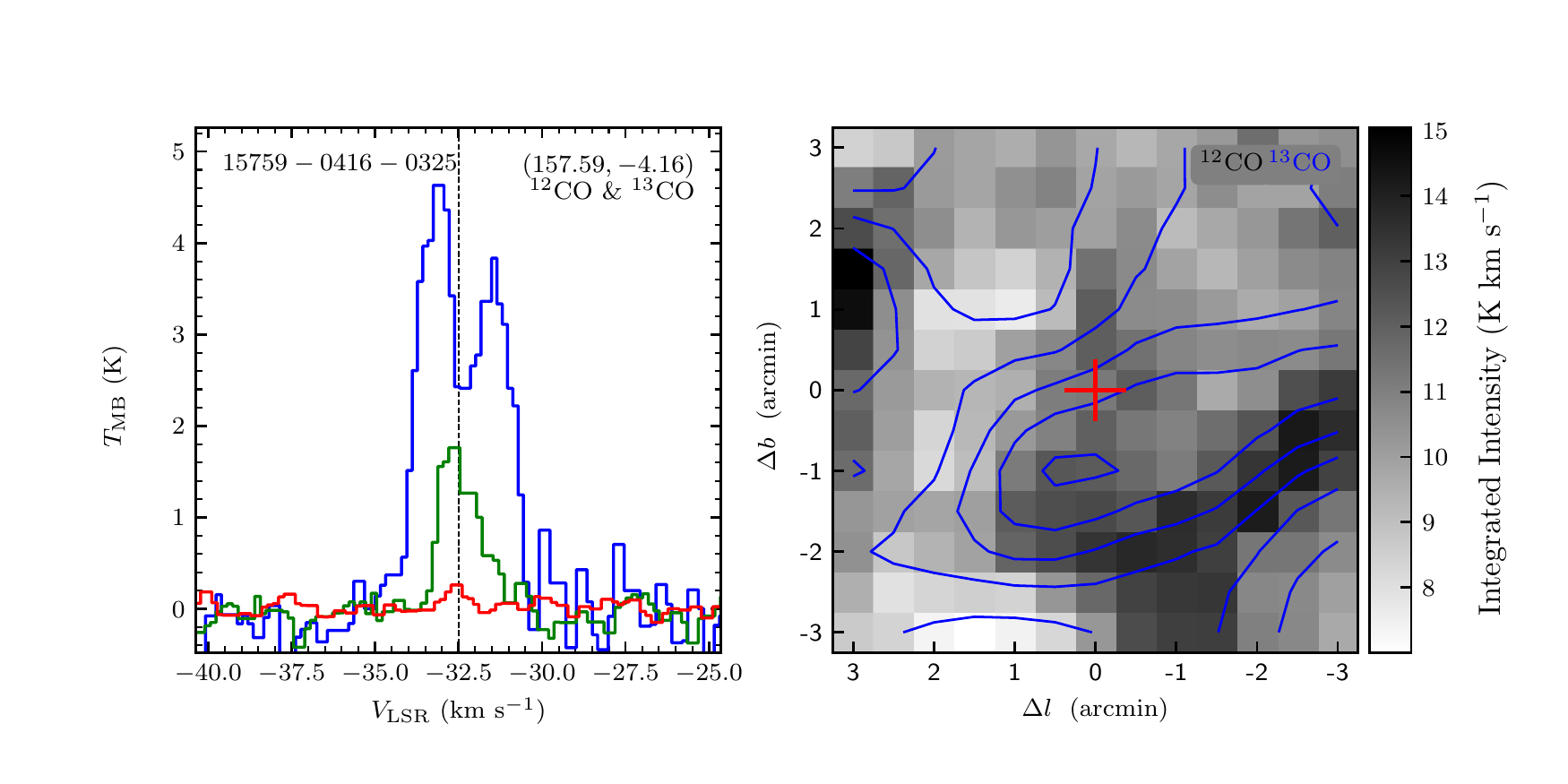}
\includegraphics[width=9.0cm,angle=0]{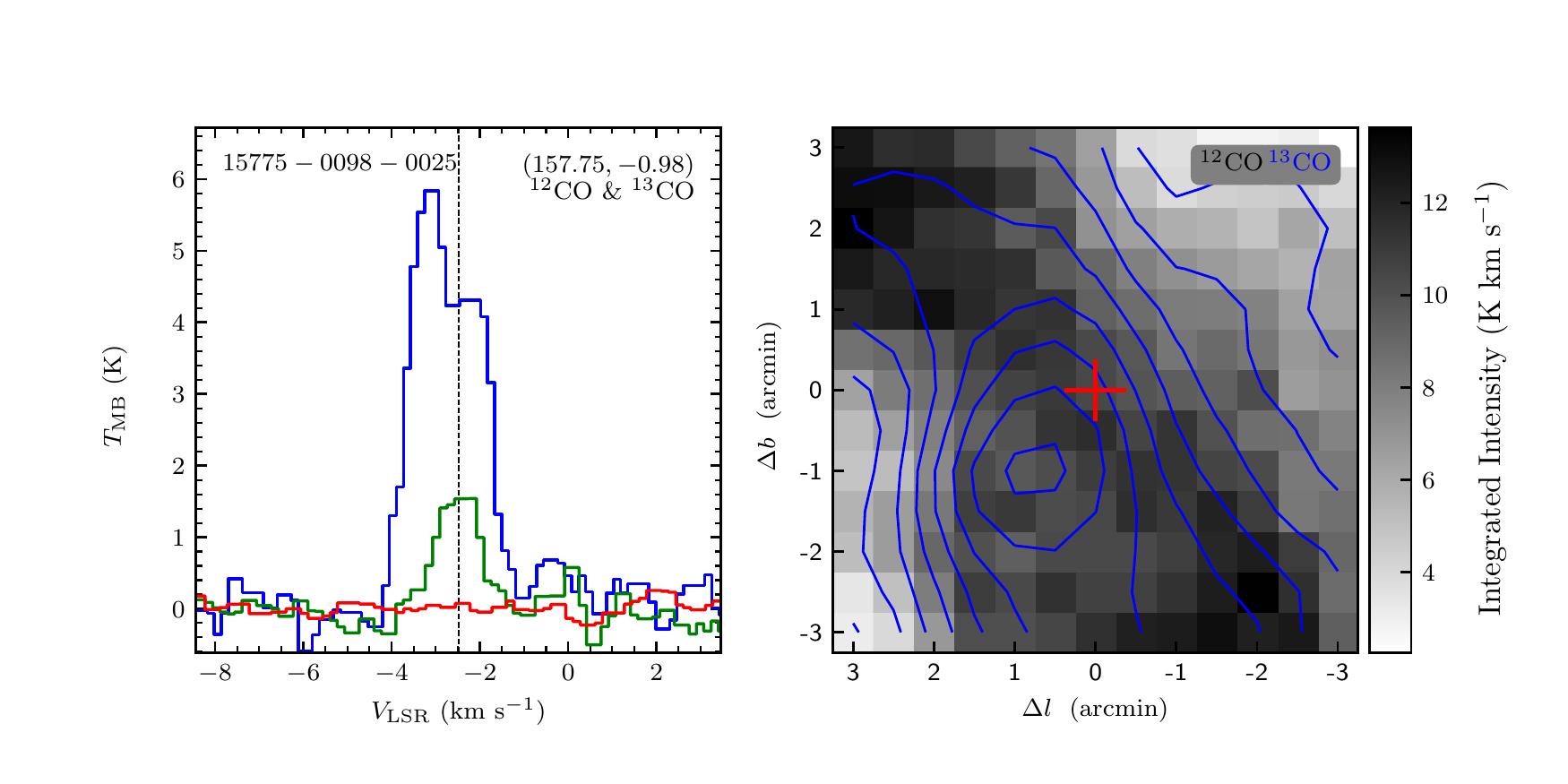}
\vspace{-0.5cm}

\includegraphics[width=9.0cm,angle=0]{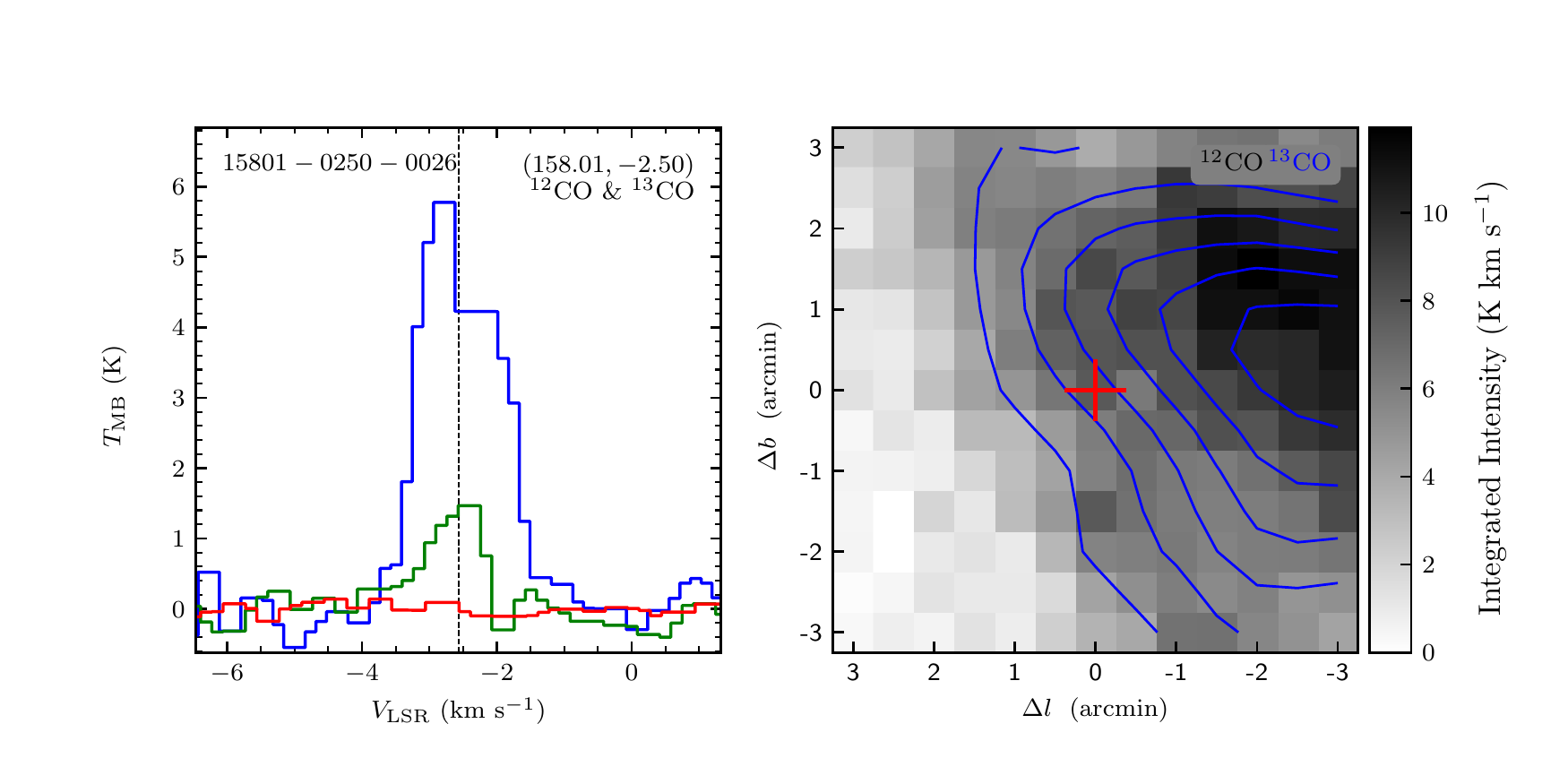}
\includegraphics[width=9.0cm,angle=0]{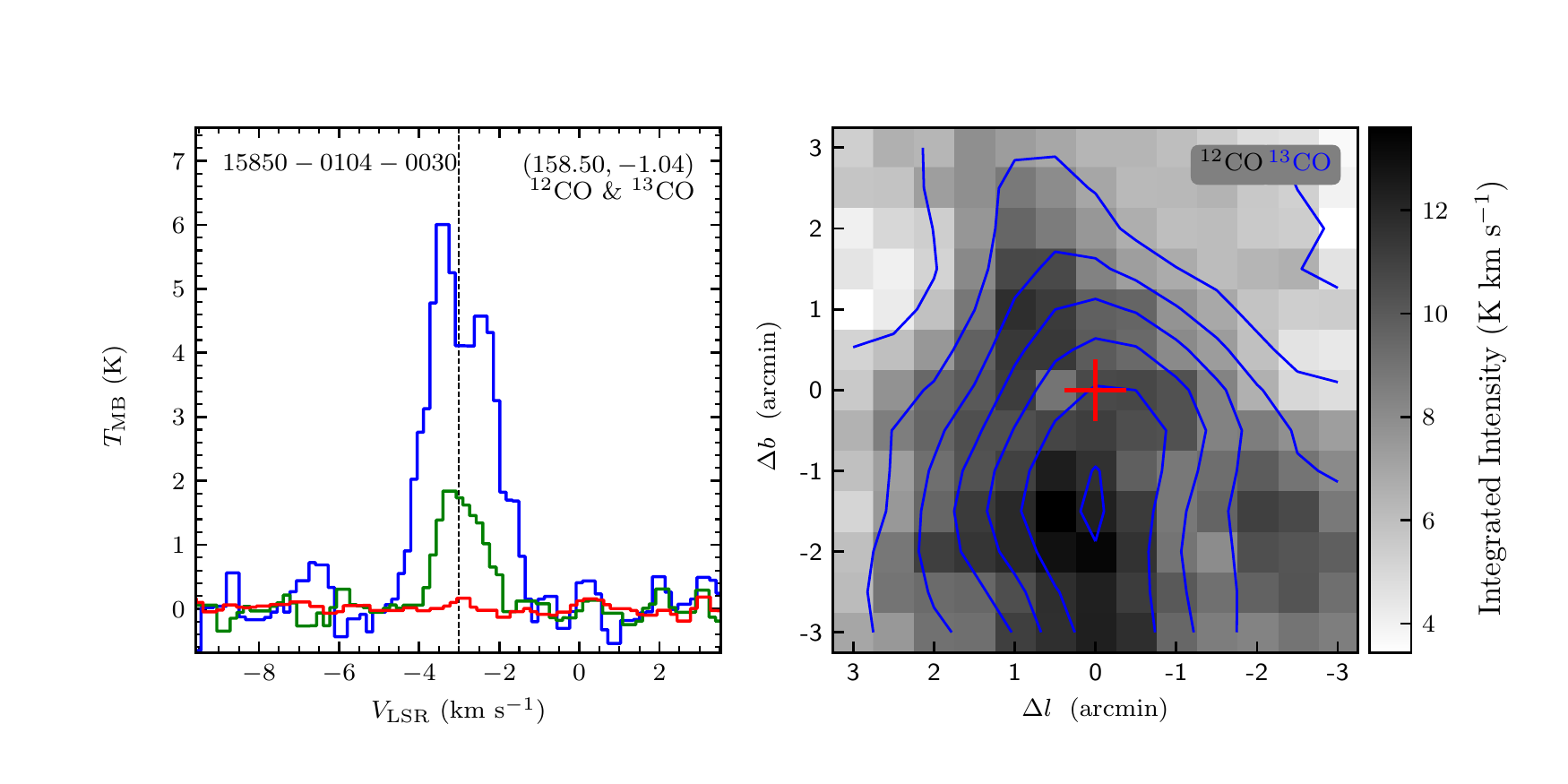}
\vspace{-0.5cm}

\includegraphics[width=9.0cm,angle=0]{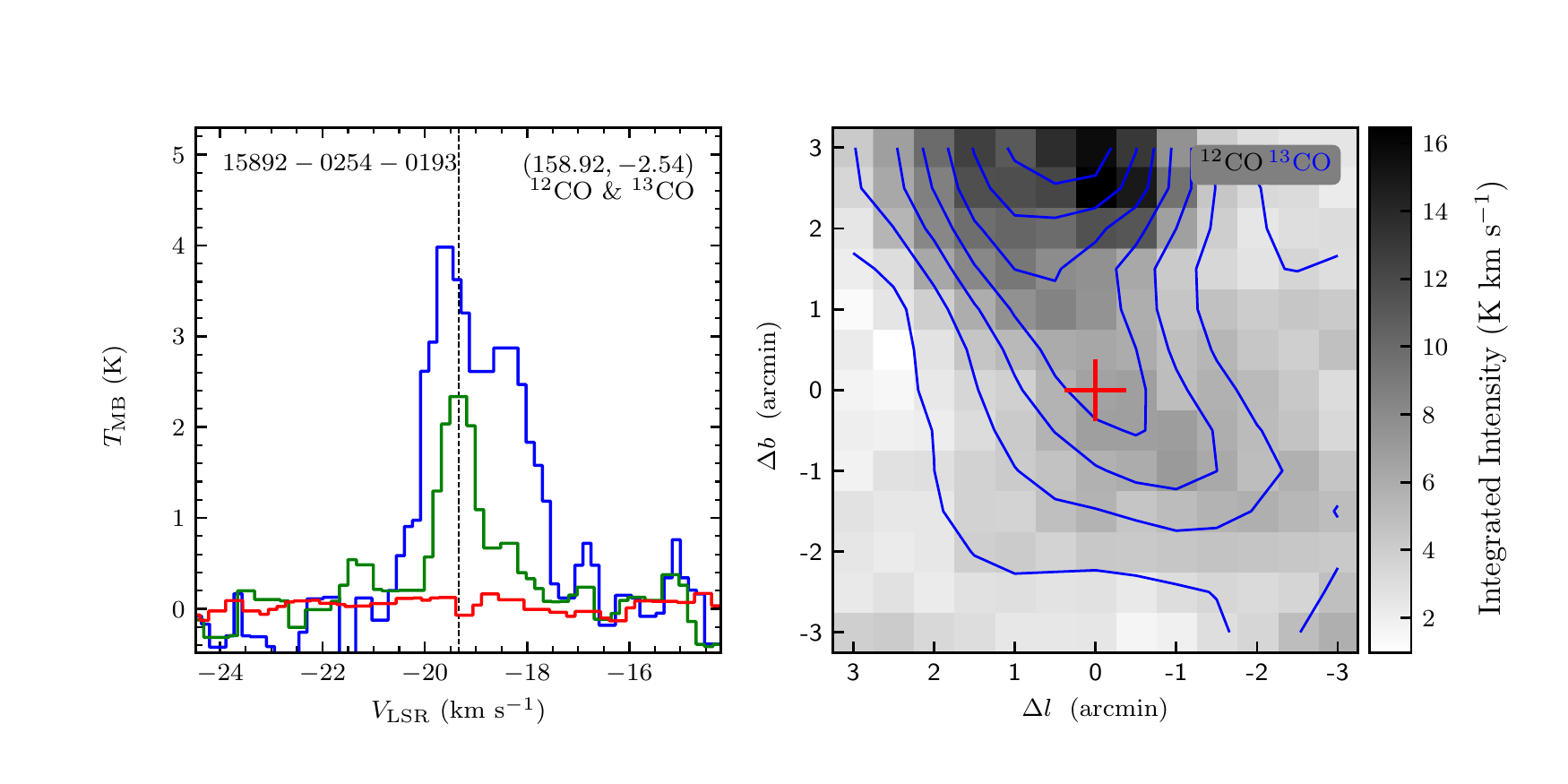}
\includegraphics[width=9.0cm,angle=0]{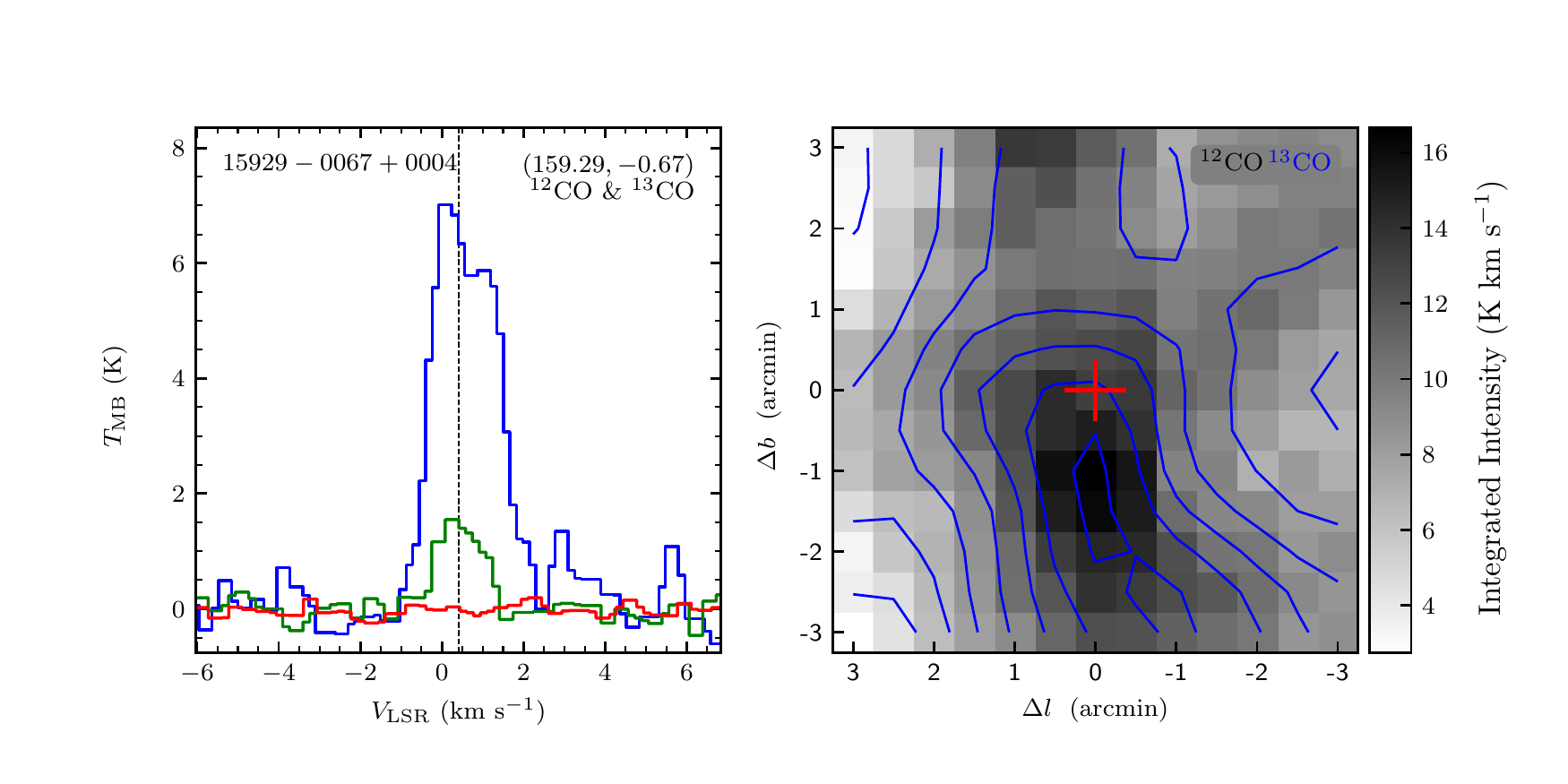}
\vspace{-0.5cm}

\includegraphics[width=9.0cm,angle=0]{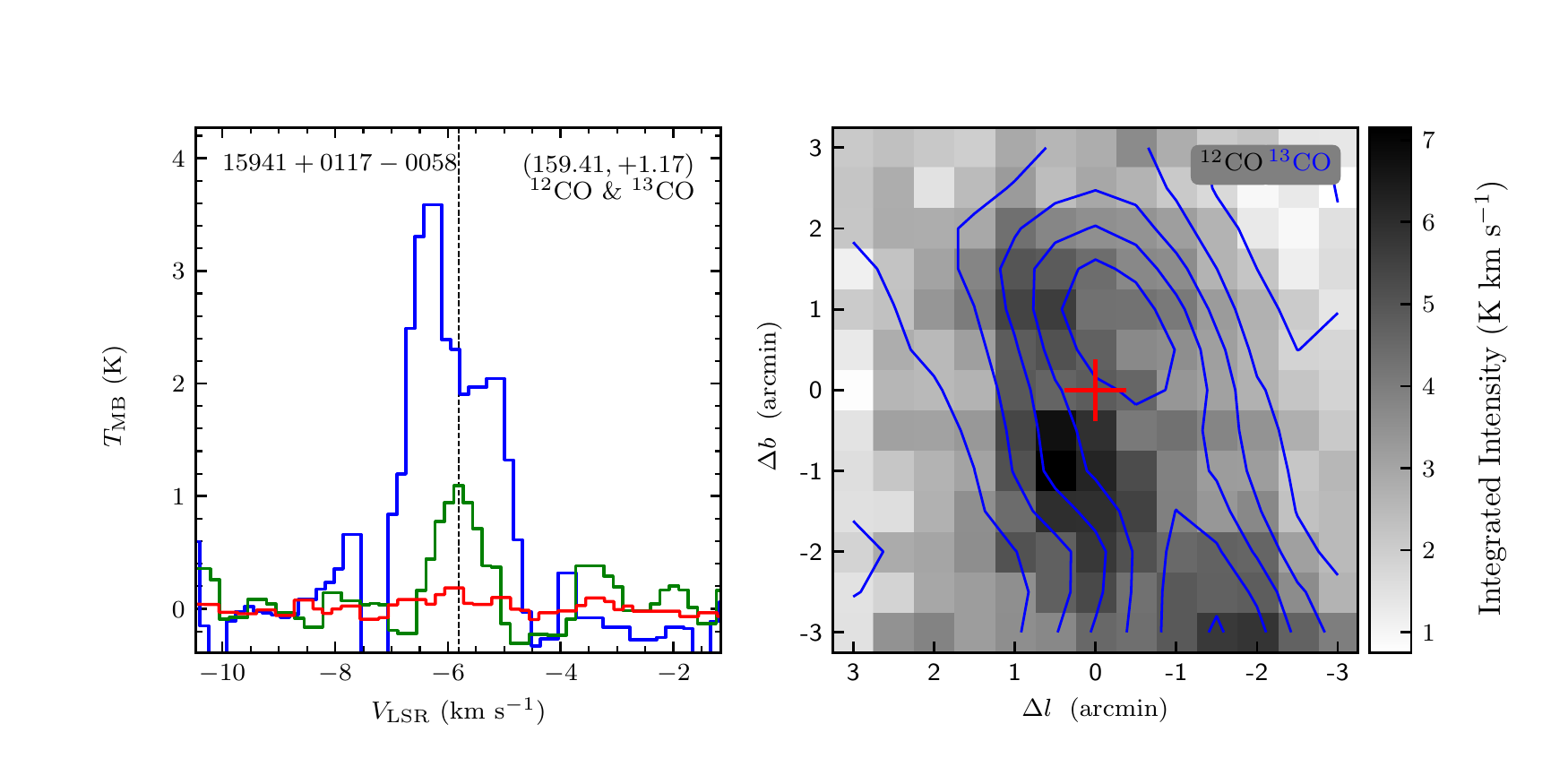}
\includegraphics[width=9.0cm,angle=0]{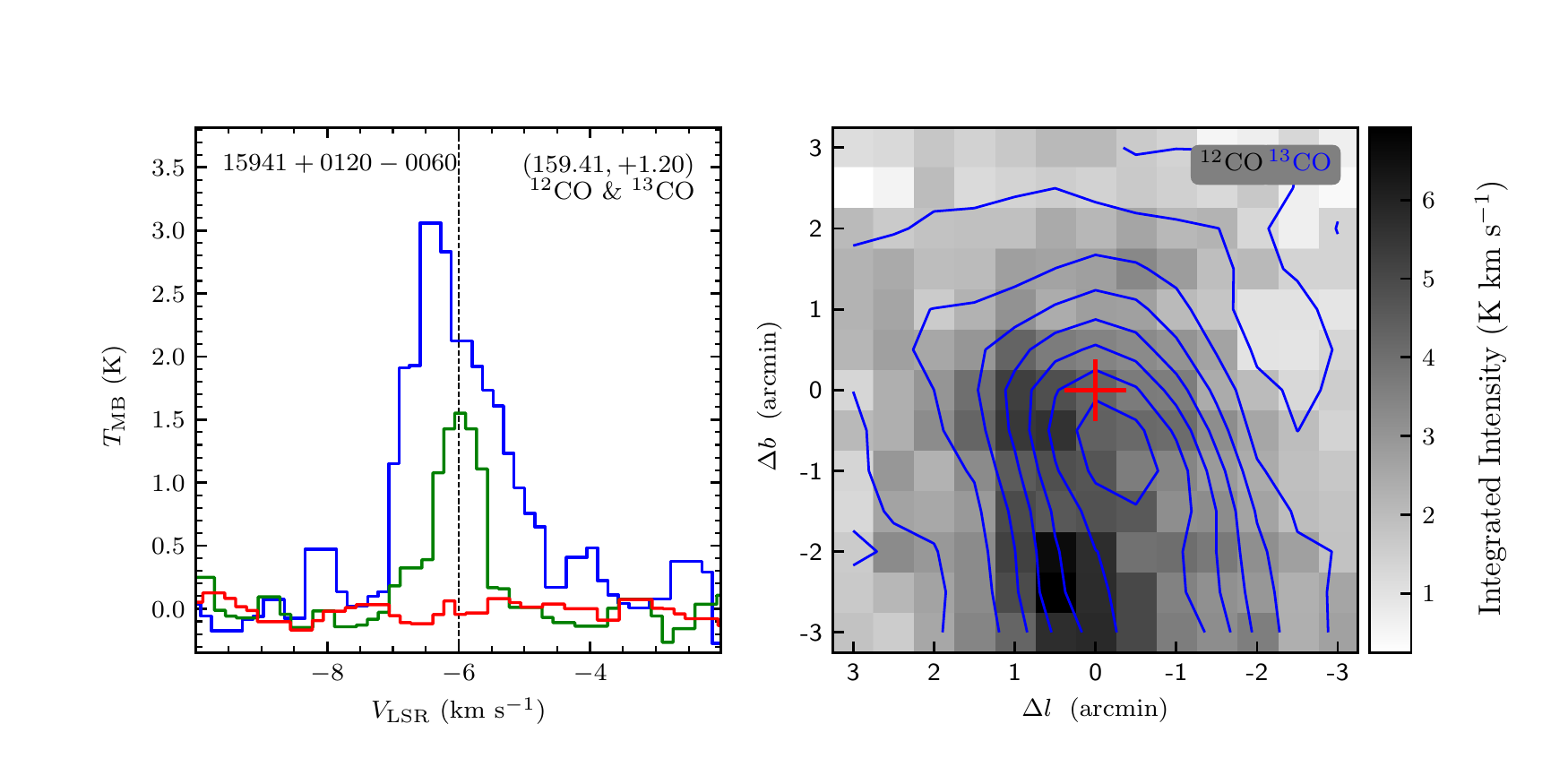}
\end{figure}
\clearpage

\begin{figure}
\includegraphics[width=9.0cm,angle=0]{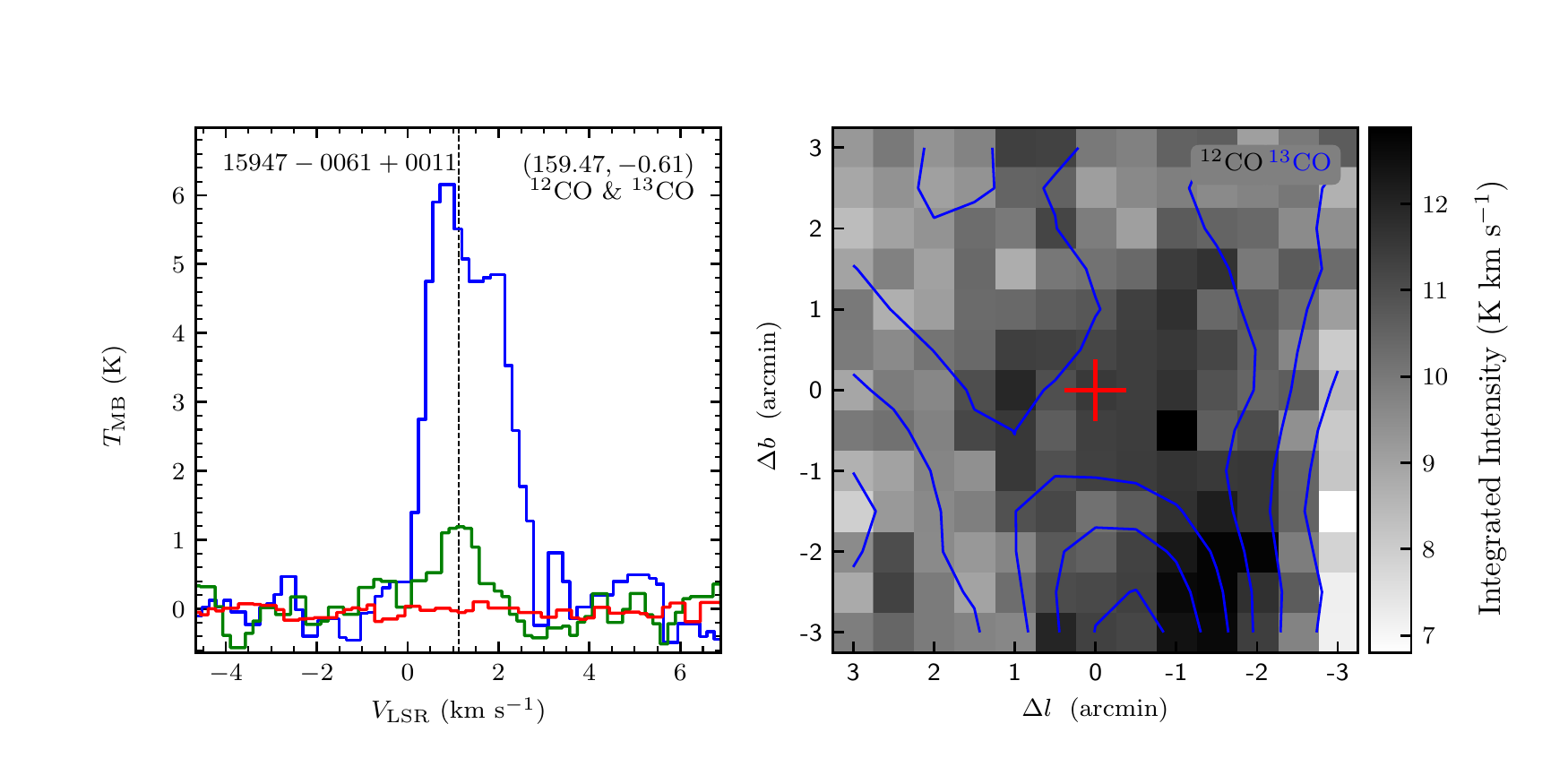}
\includegraphics[width=9.0cm,angle=0]{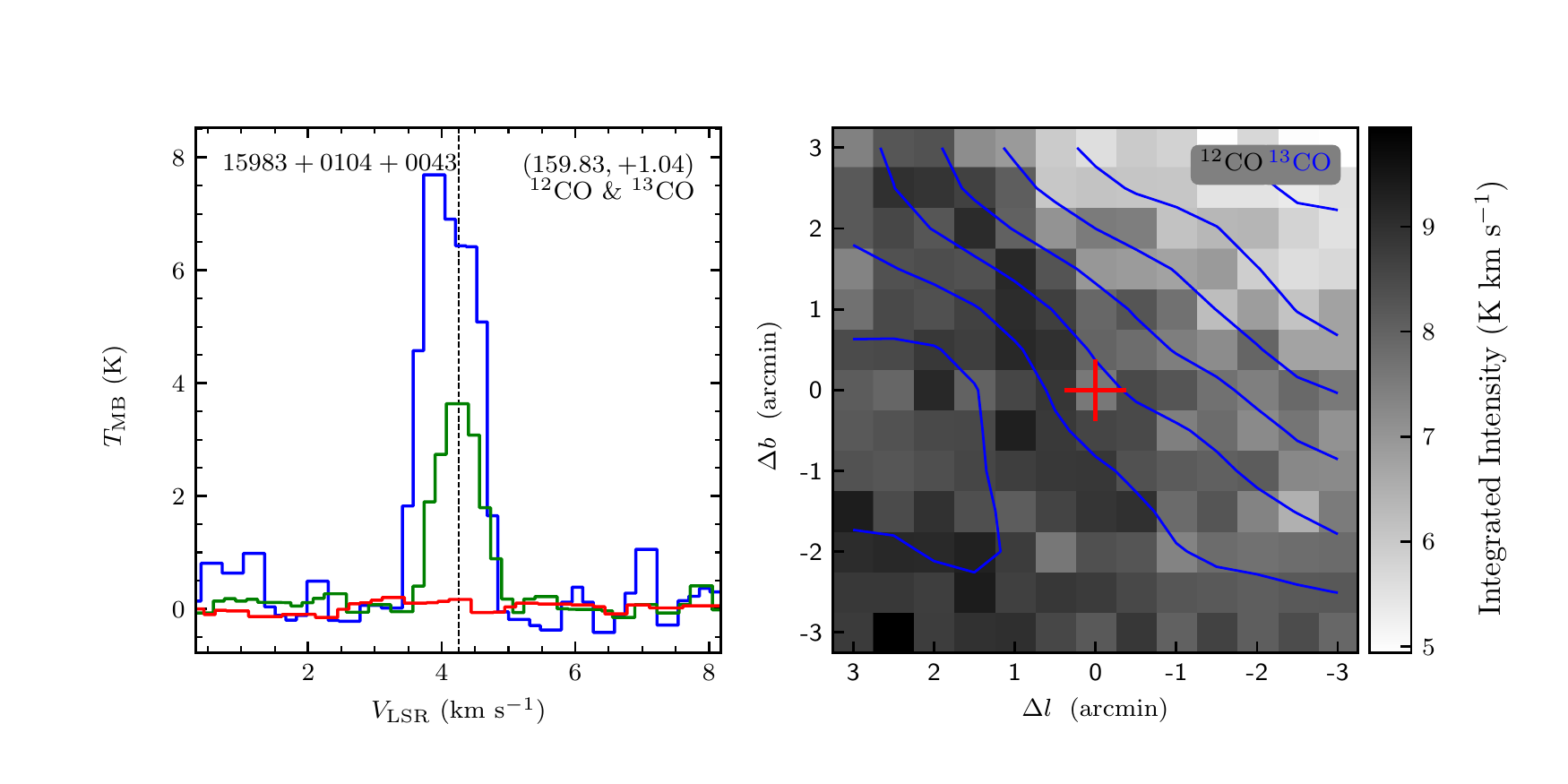}
\vspace{-0.5cm}

\includegraphics[width=9.0cm,angle=0]{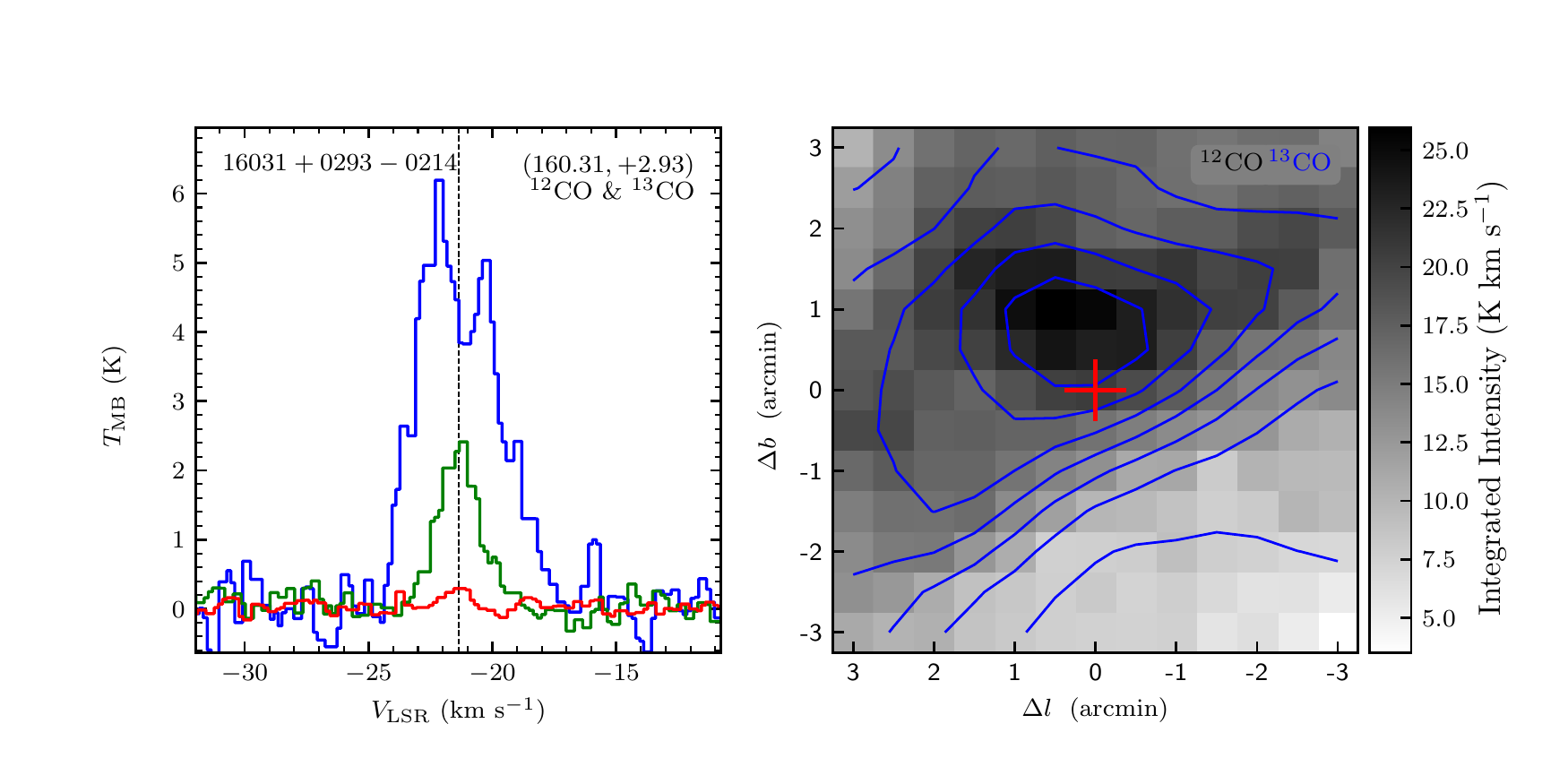}
\includegraphics[width=9.0cm,angle=0]{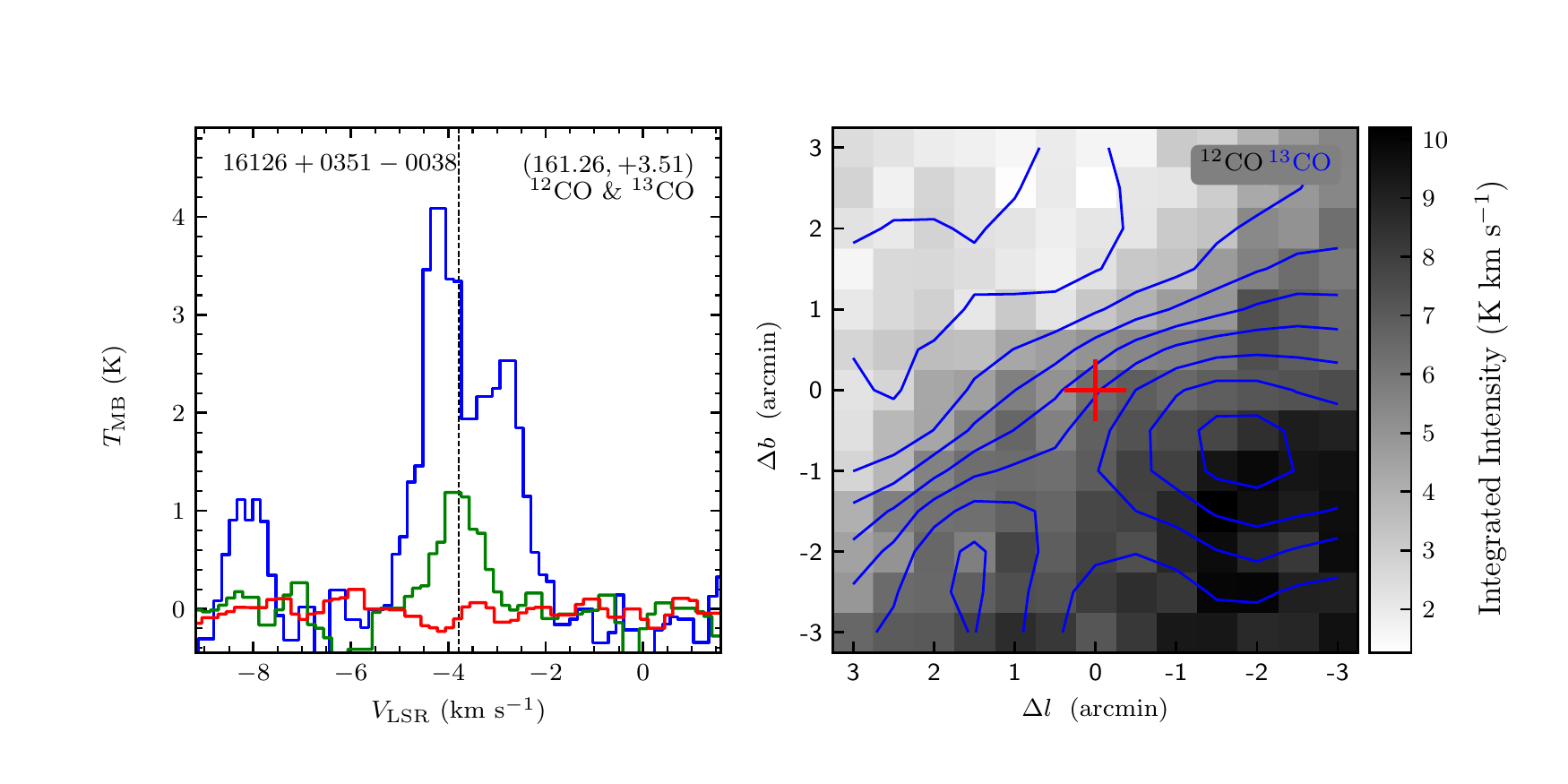}
\vspace{-0.5cm}

\includegraphics[width=9.0cm,angle=0]{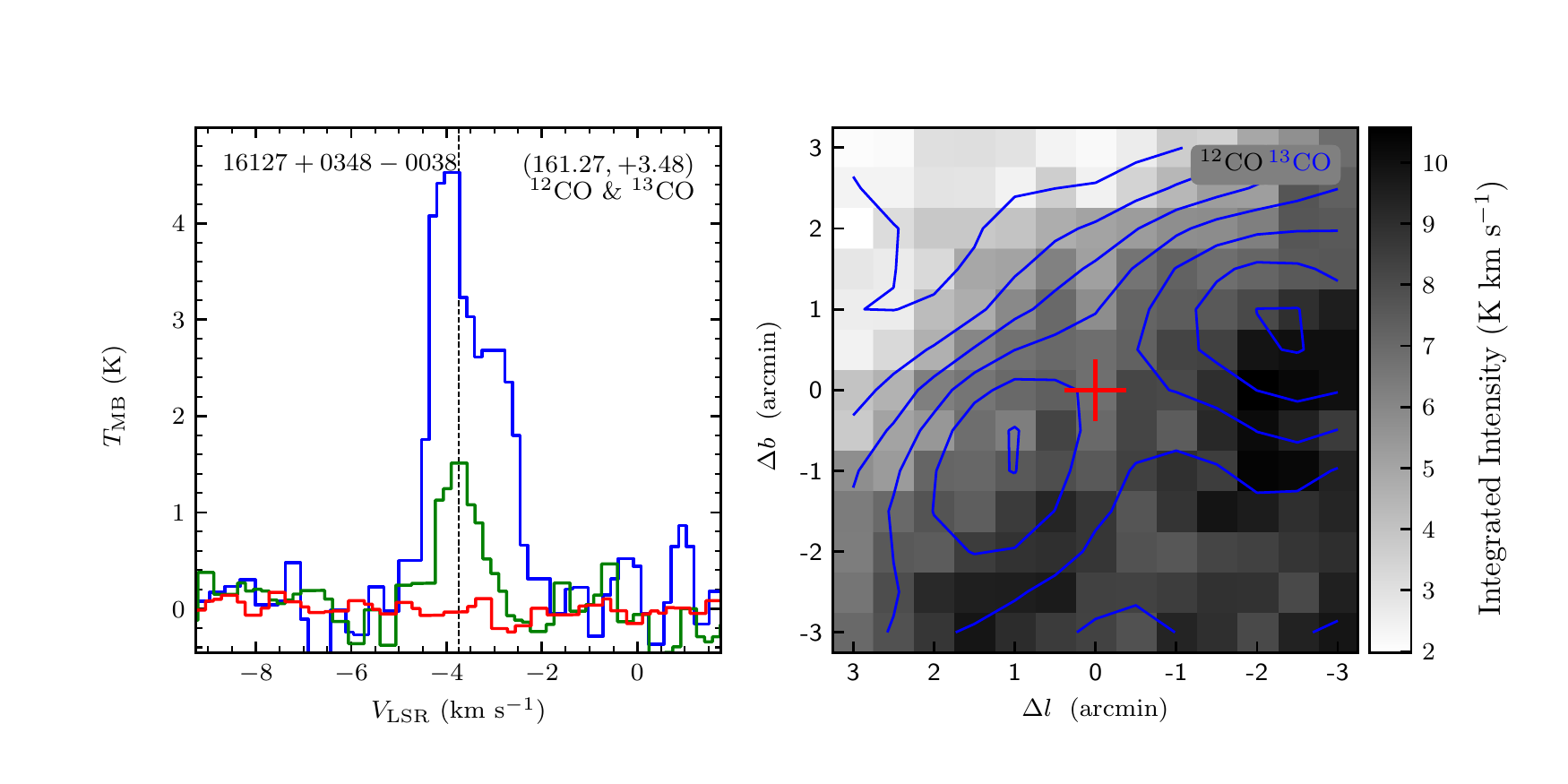}
\includegraphics[width=9.0cm,angle=0]{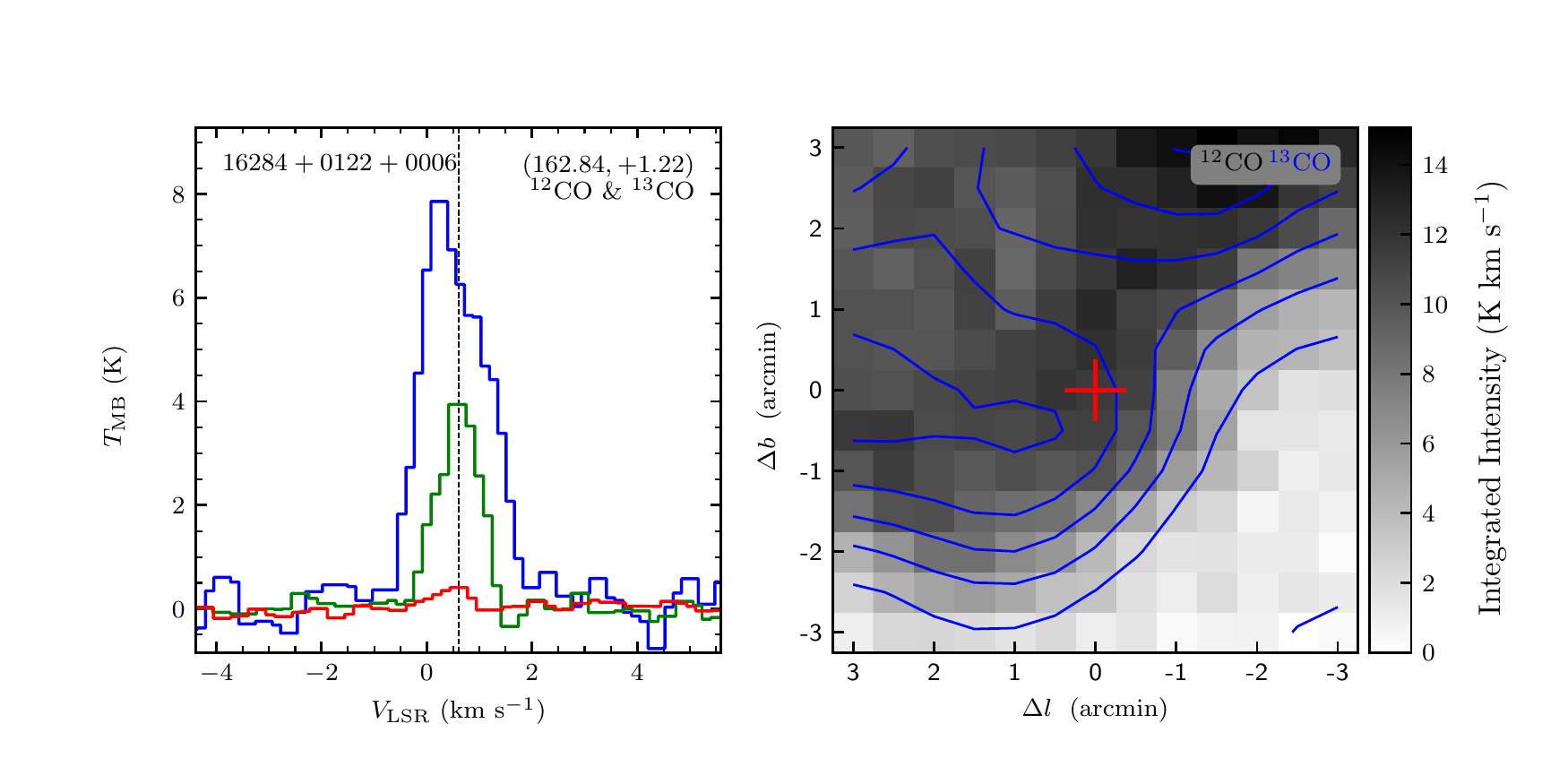}
\vspace{-0.5cm}

\includegraphics[width=9.0cm,angle=0]{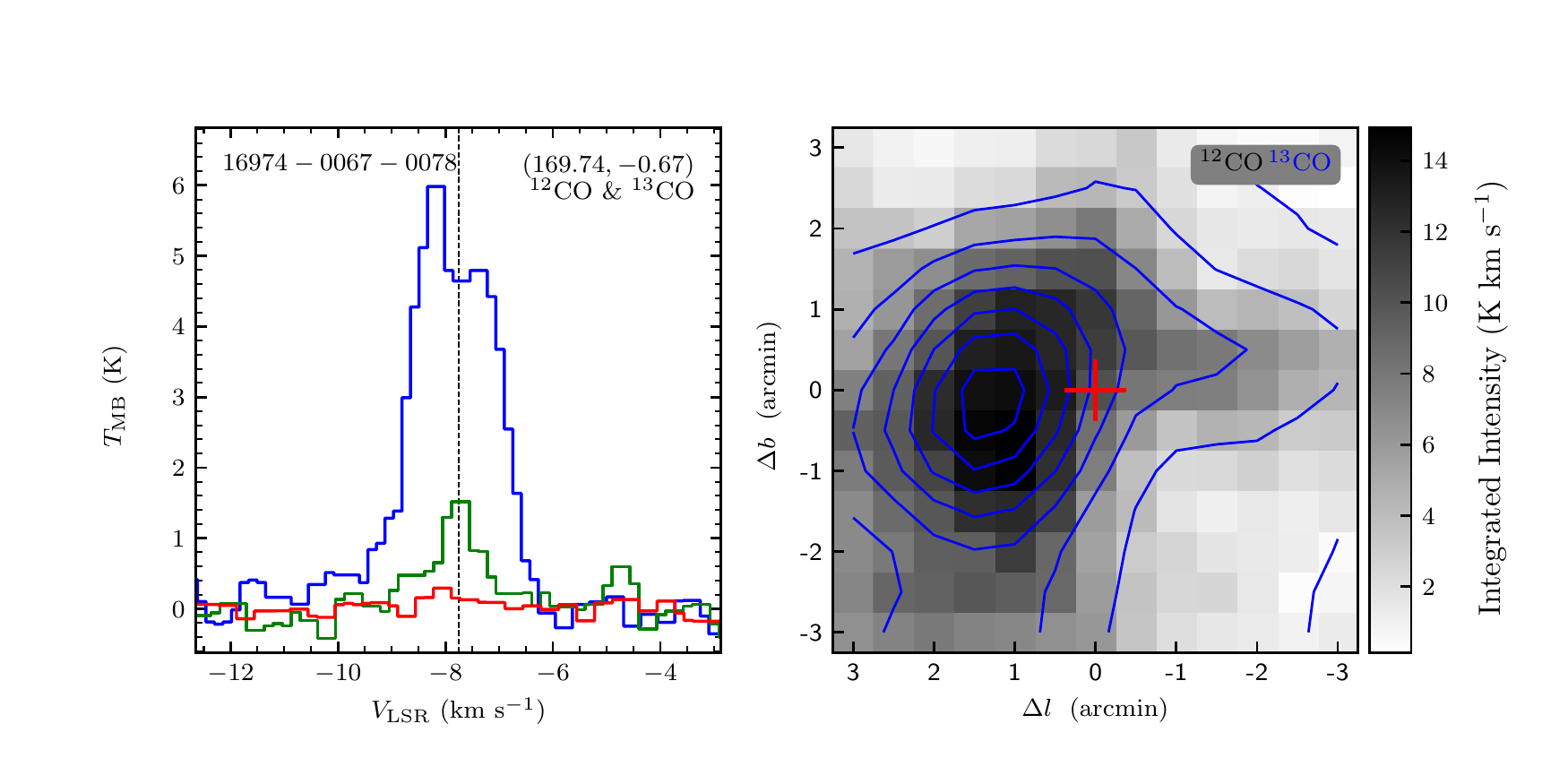}
\includegraphics[width=9.0cm,angle=0]{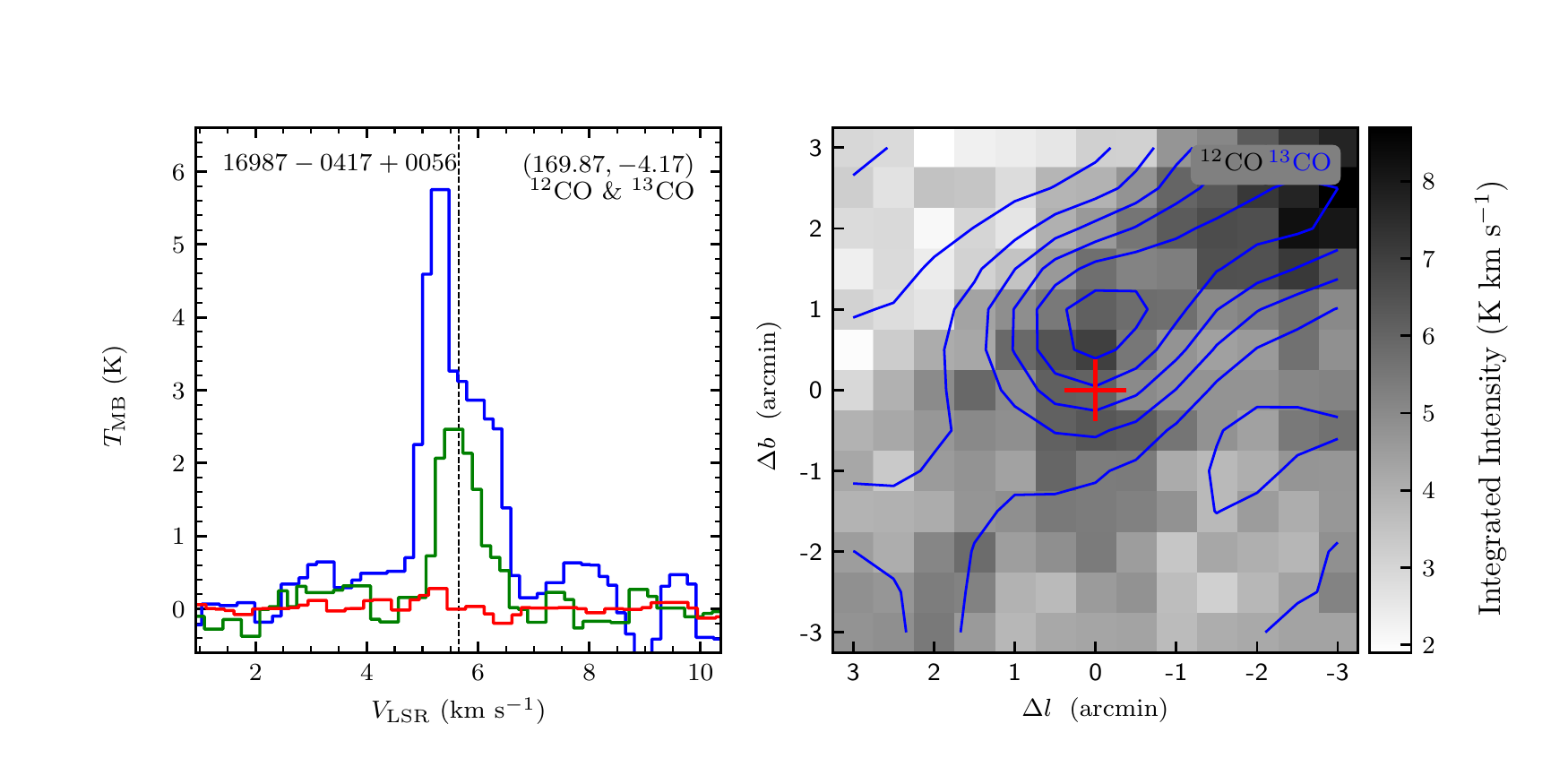}
\vspace{-0.5cm}

\includegraphics[width=9.0cm,angle=0]{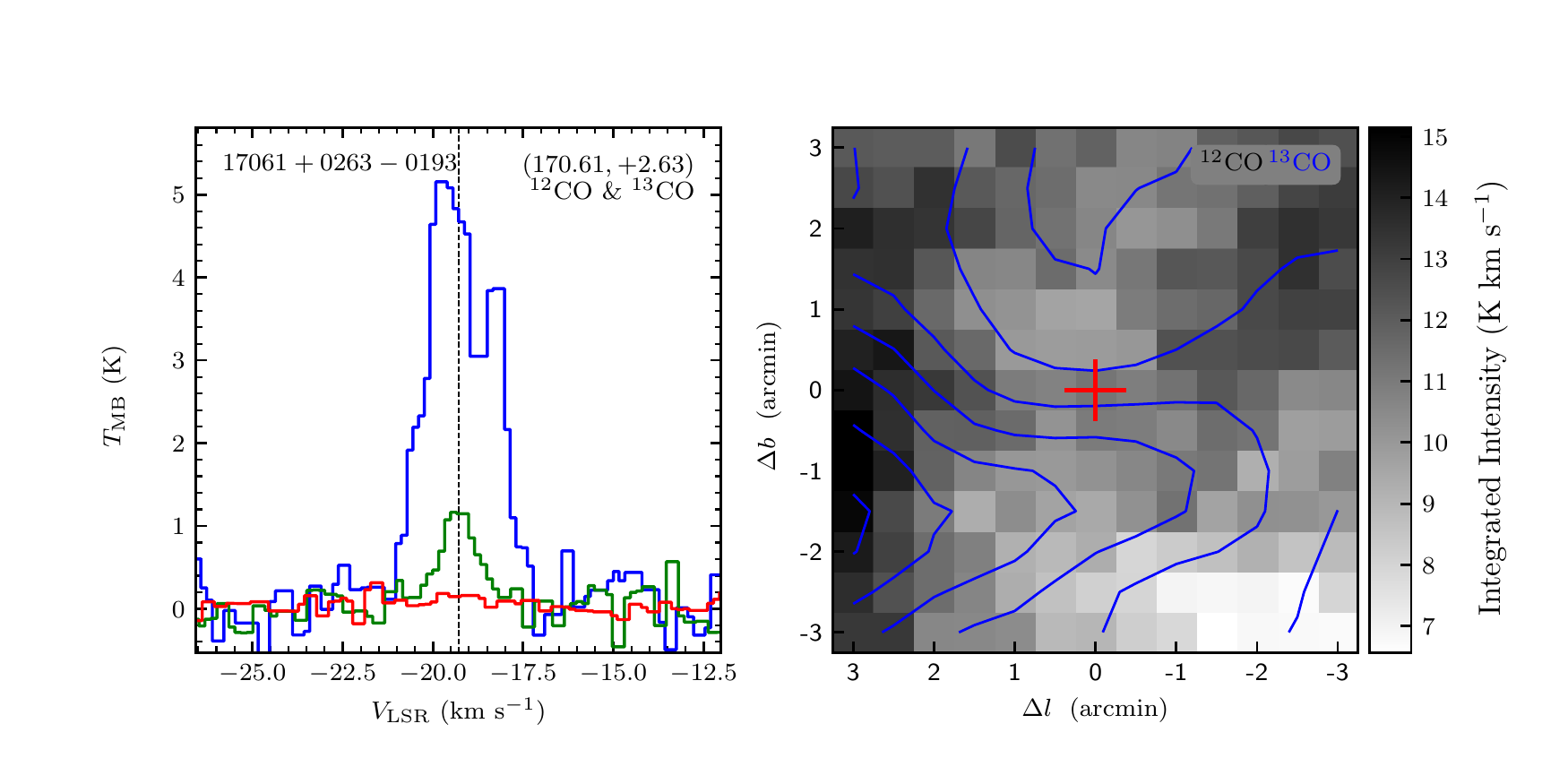}
\includegraphics[width=9.0cm,angle=0]{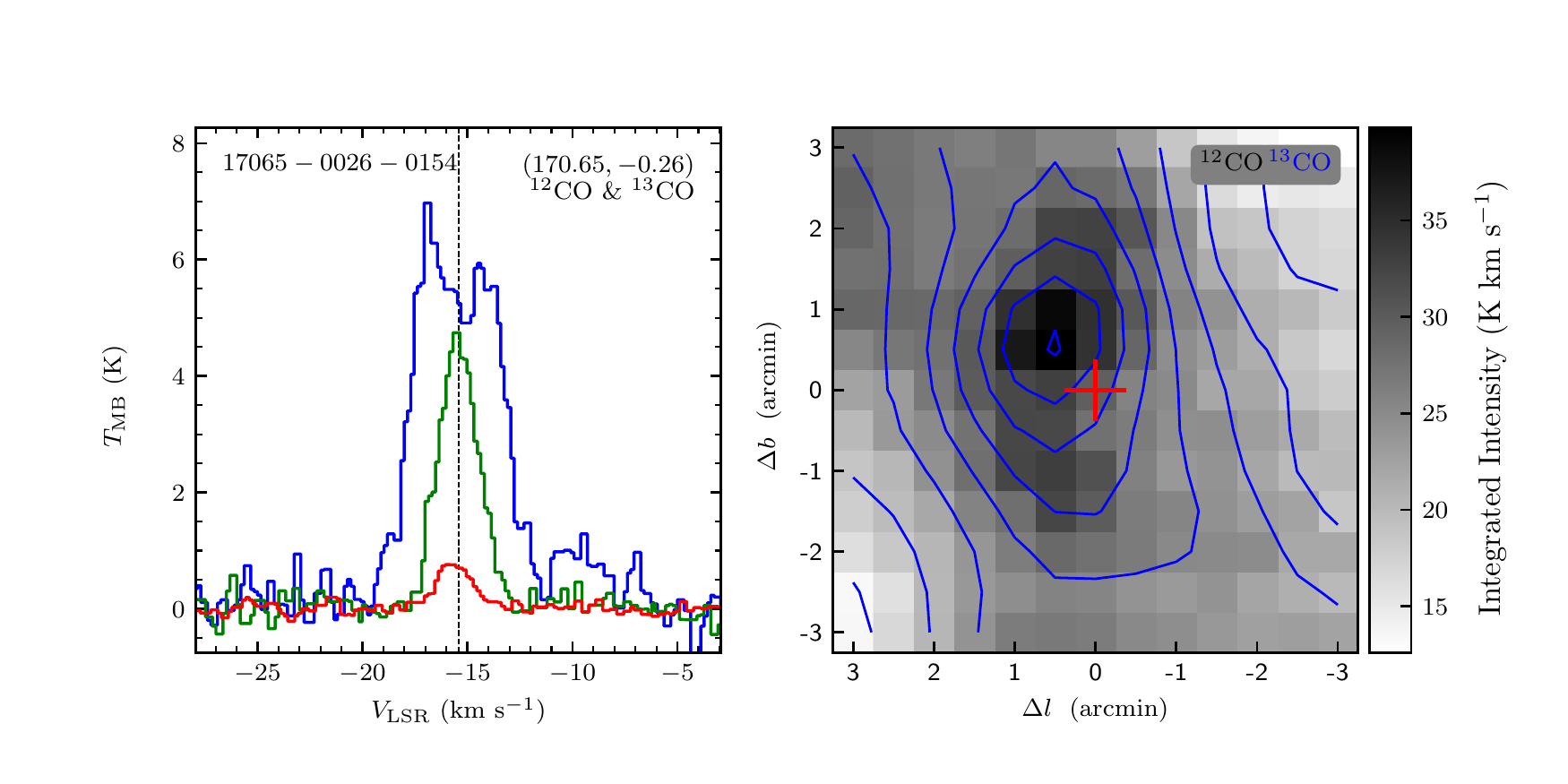}
\end{figure}
\clearpage

\begin{figure}
\includegraphics[width=9.0cm,angle=0]{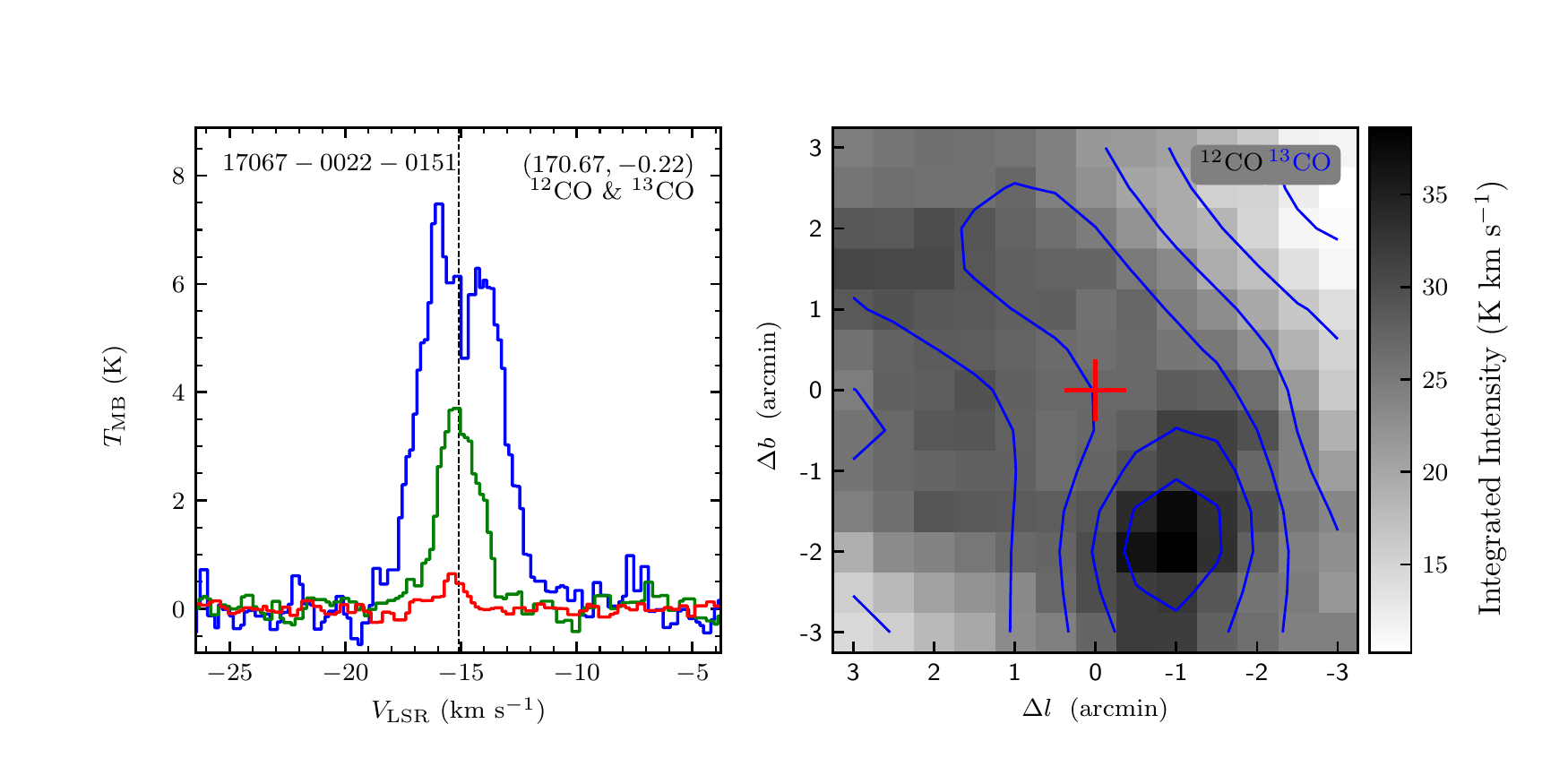}
\includegraphics[width=9.0cm,angle=0]{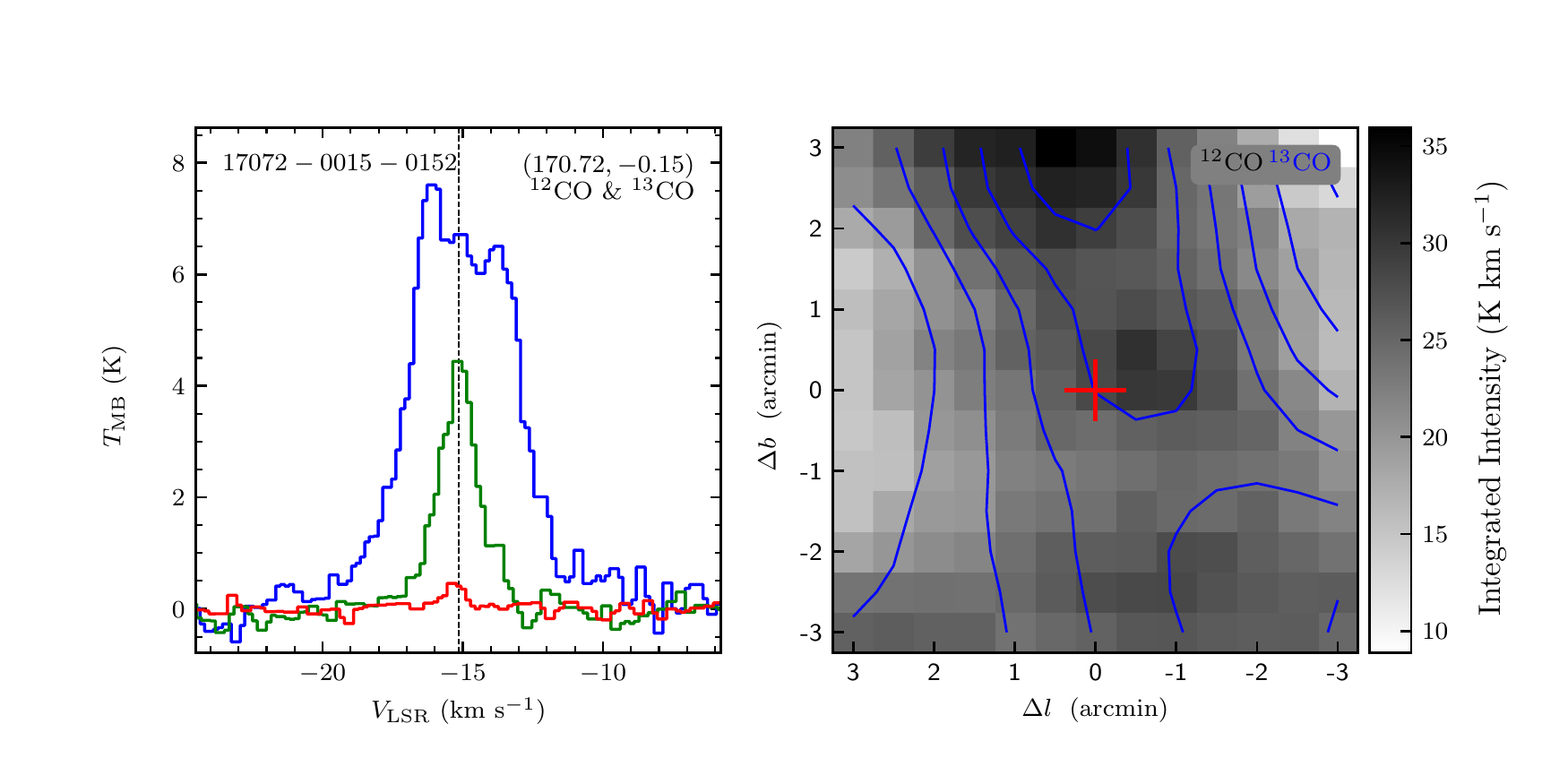}
\vspace{-0.5cm}

\includegraphics[width=9.0cm,angle=0]{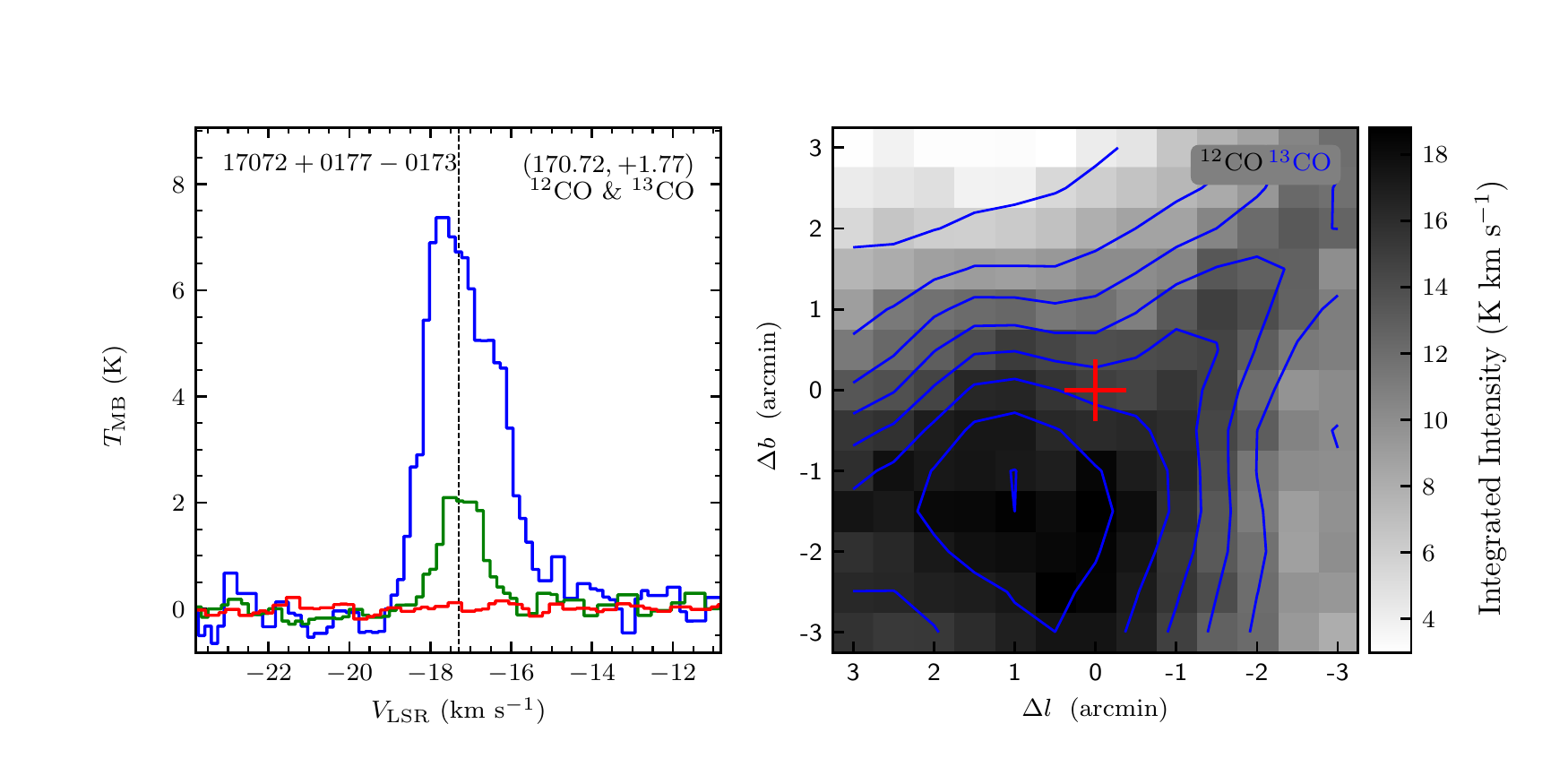}
\includegraphics[width=9.0cm,angle=0]{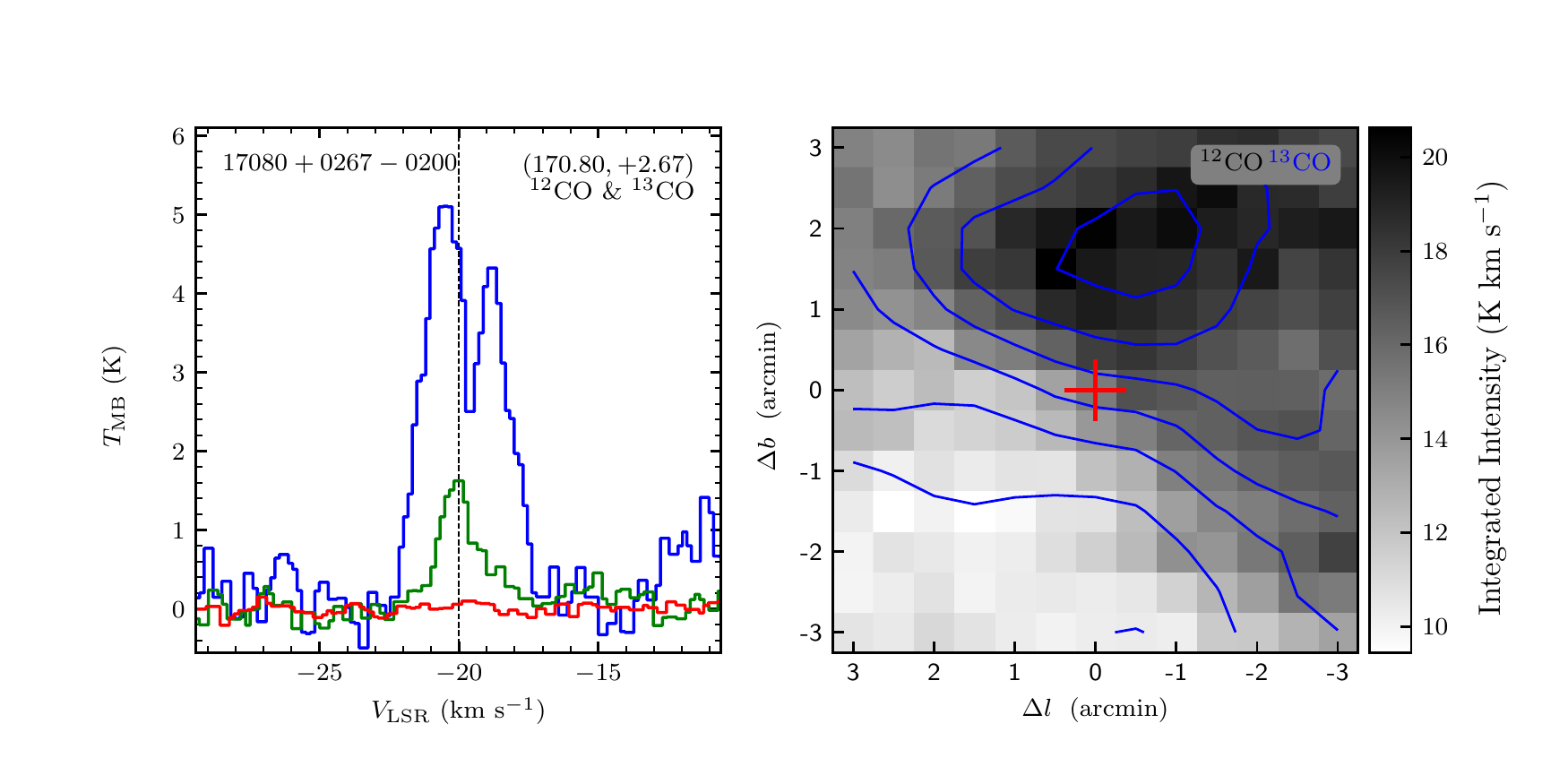}
\vspace{-0.5cm}

\includegraphics[width=9.0cm,angle=0]{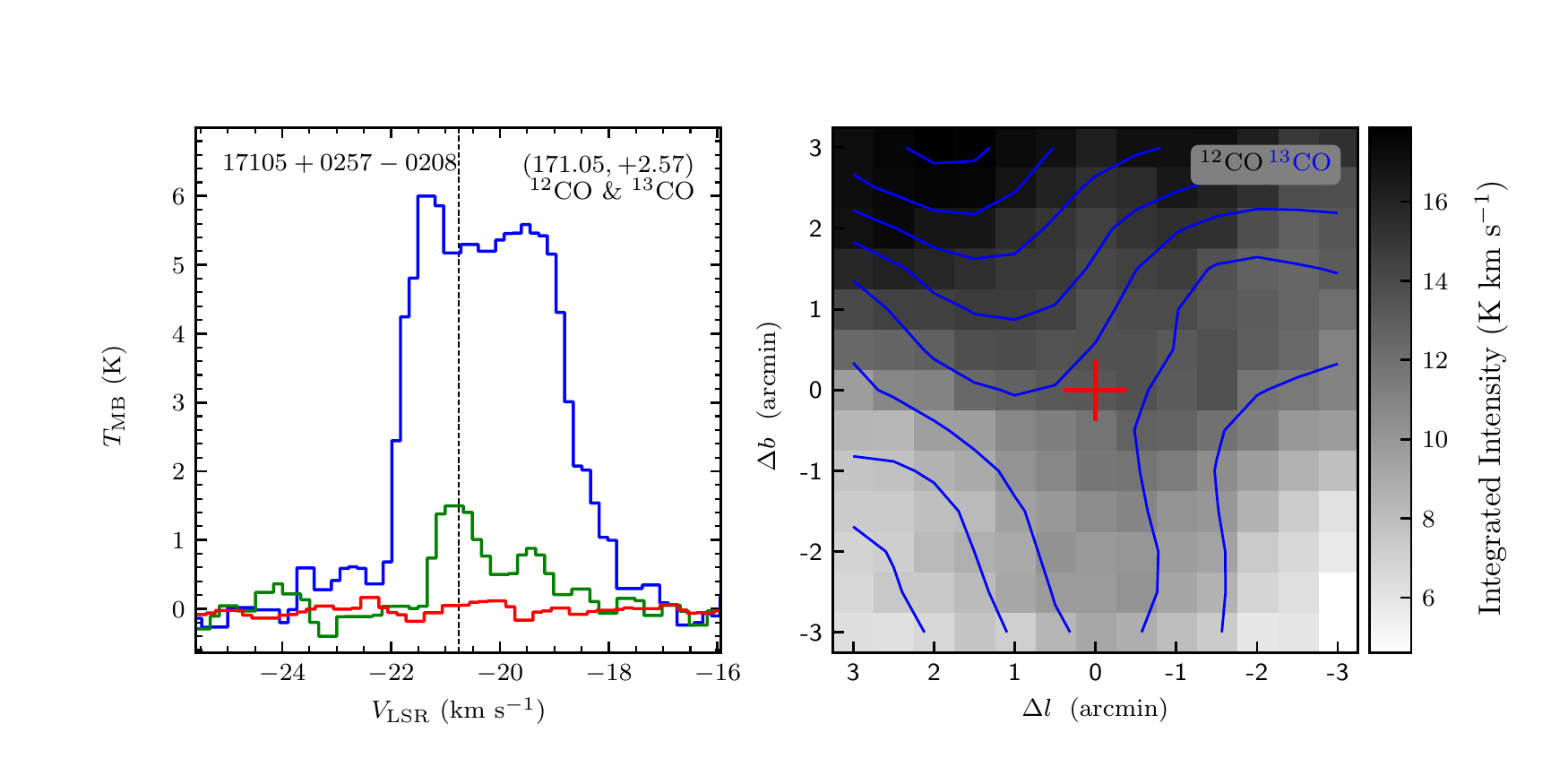}
\includegraphics[width=9.0cm,angle=0]{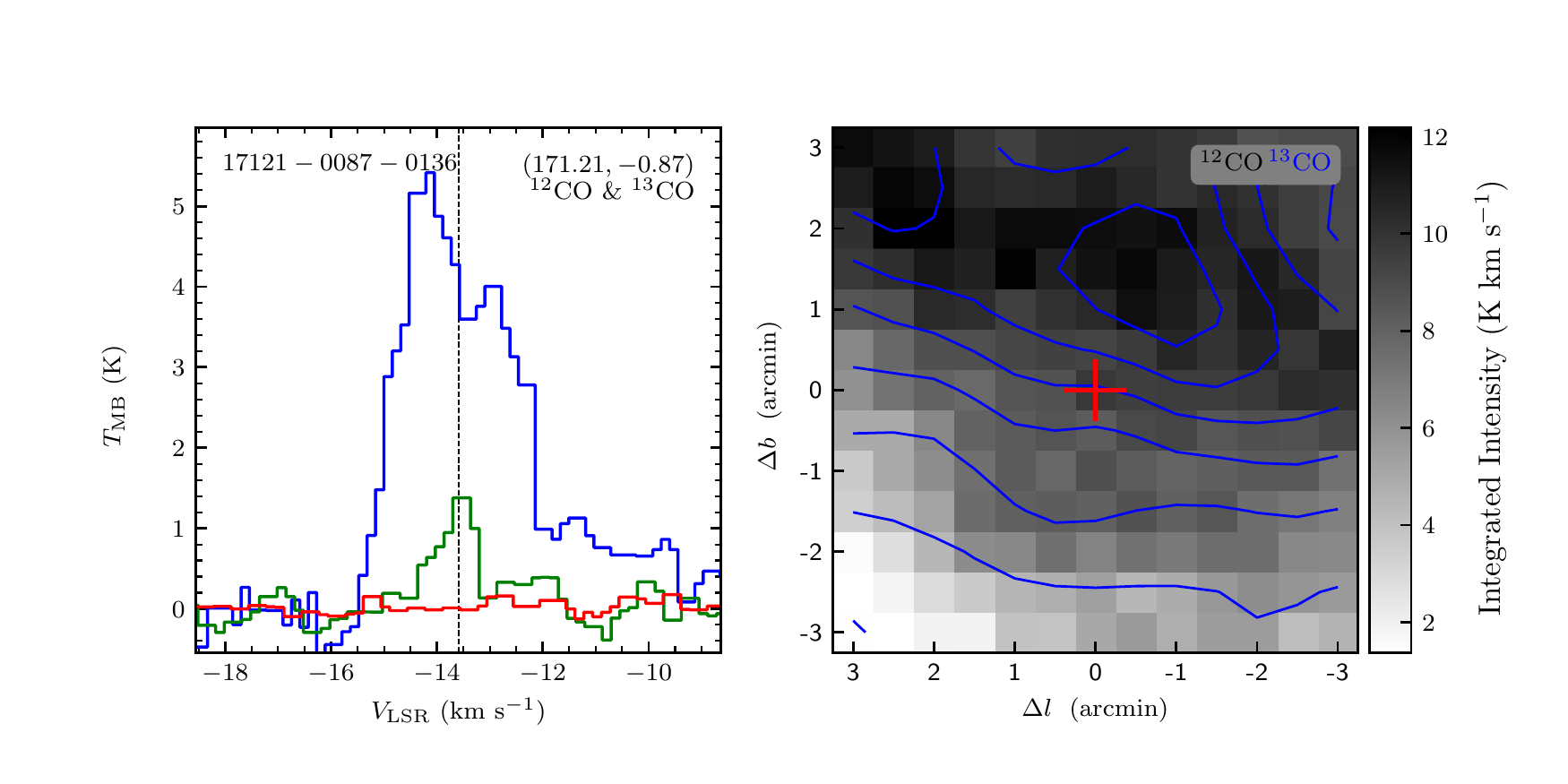}
\vspace{-0.5cm}

\includegraphics[width=9.0cm,angle=0]{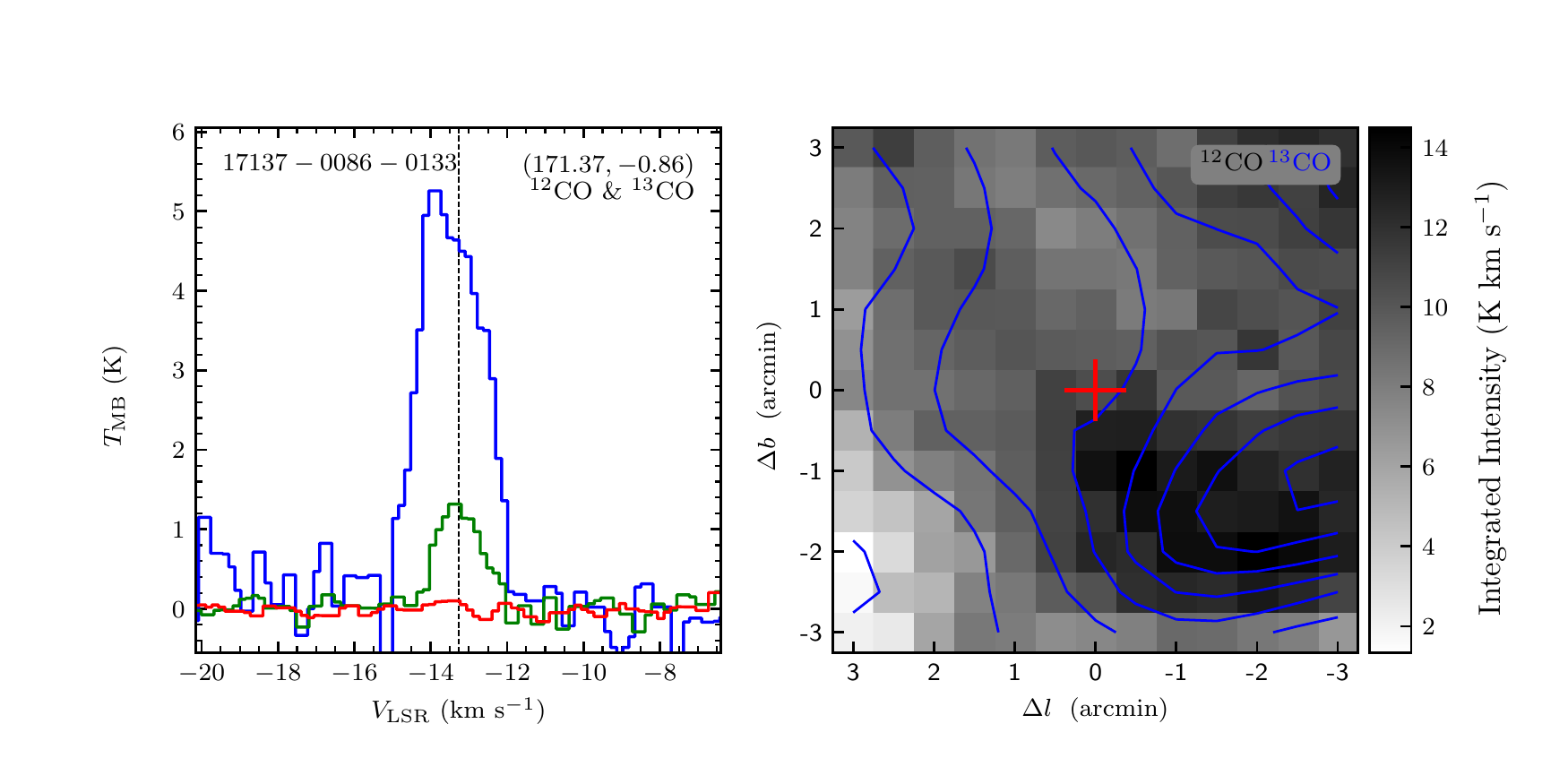}
\includegraphics[width=9.0cm,angle=0]{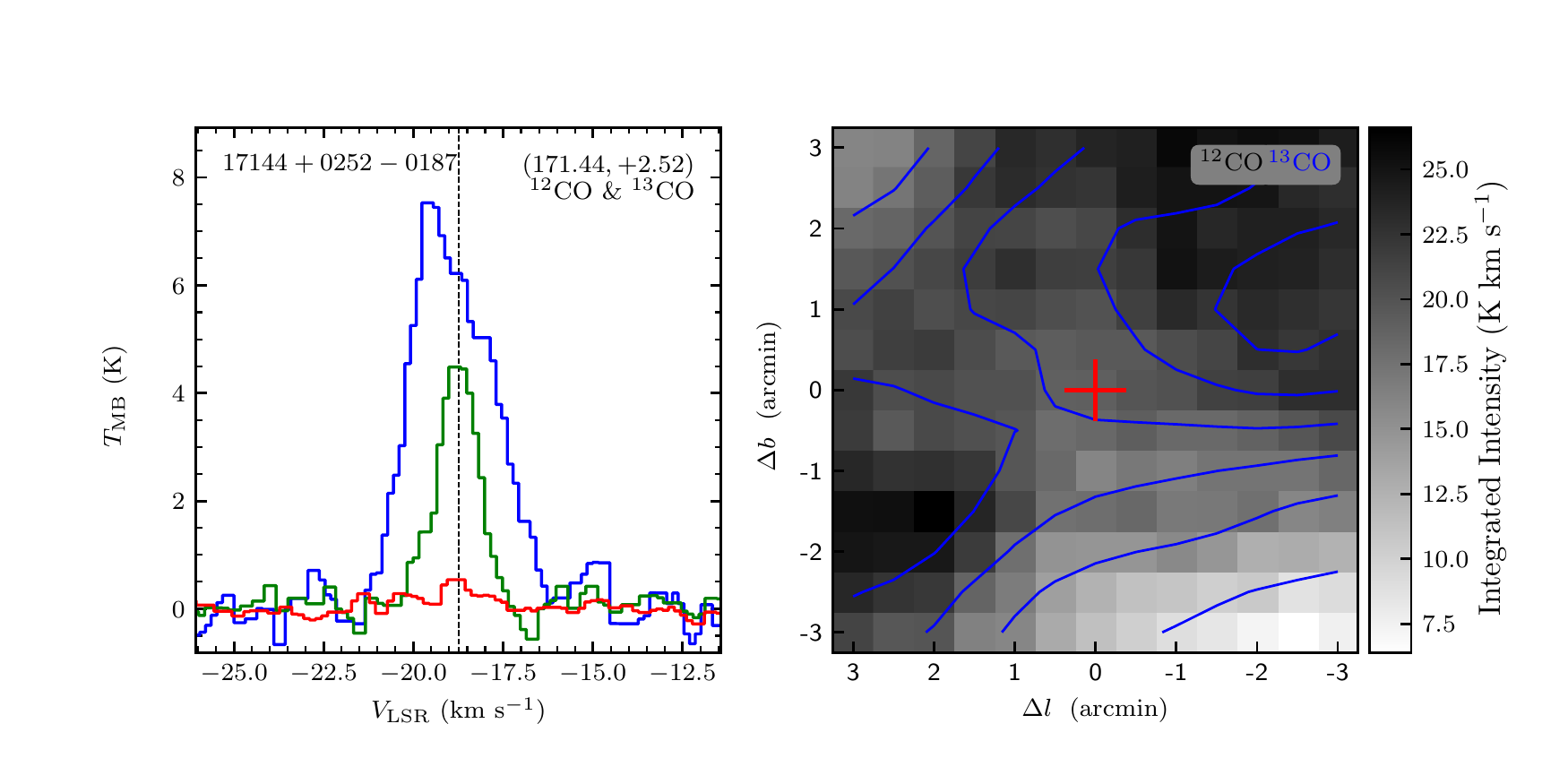}
\vspace{-0.5cm}

\includegraphics[width=9.0cm,angle=0]{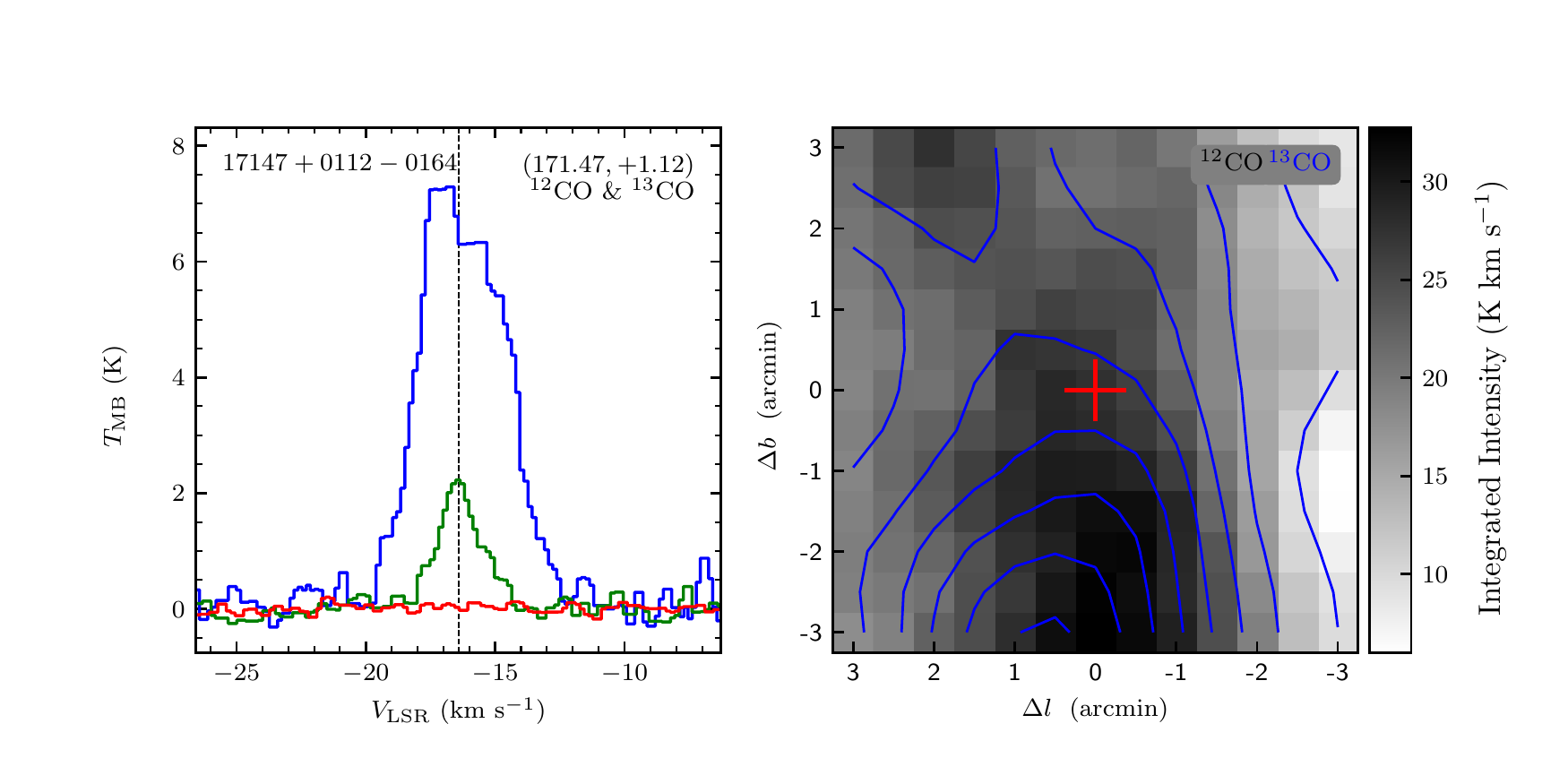}
\includegraphics[width=9.0cm,angle=0]{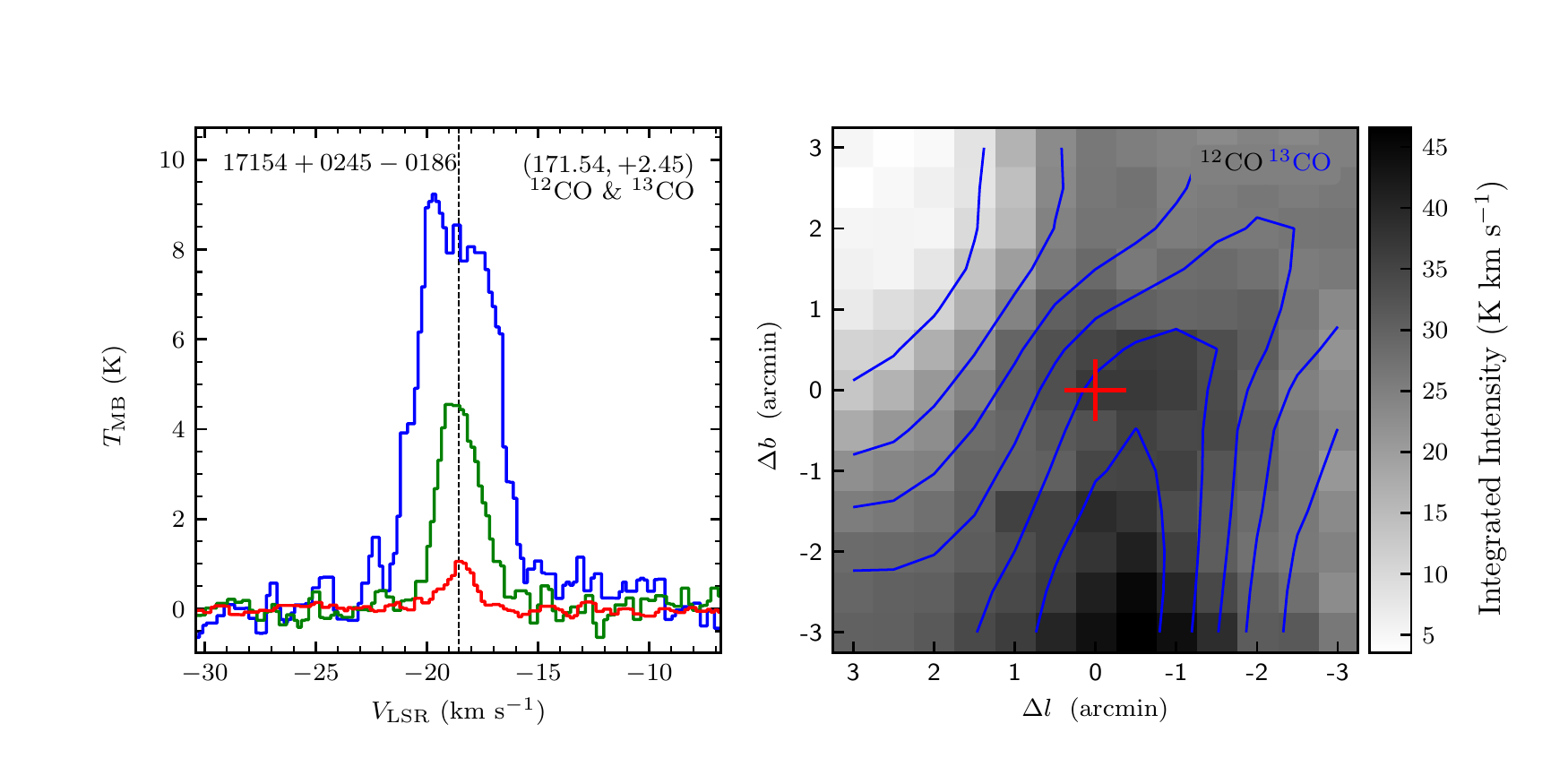}
\end{figure}
\clearpage

\begin{figure}
\includegraphics[width=9.0cm,angle=0]{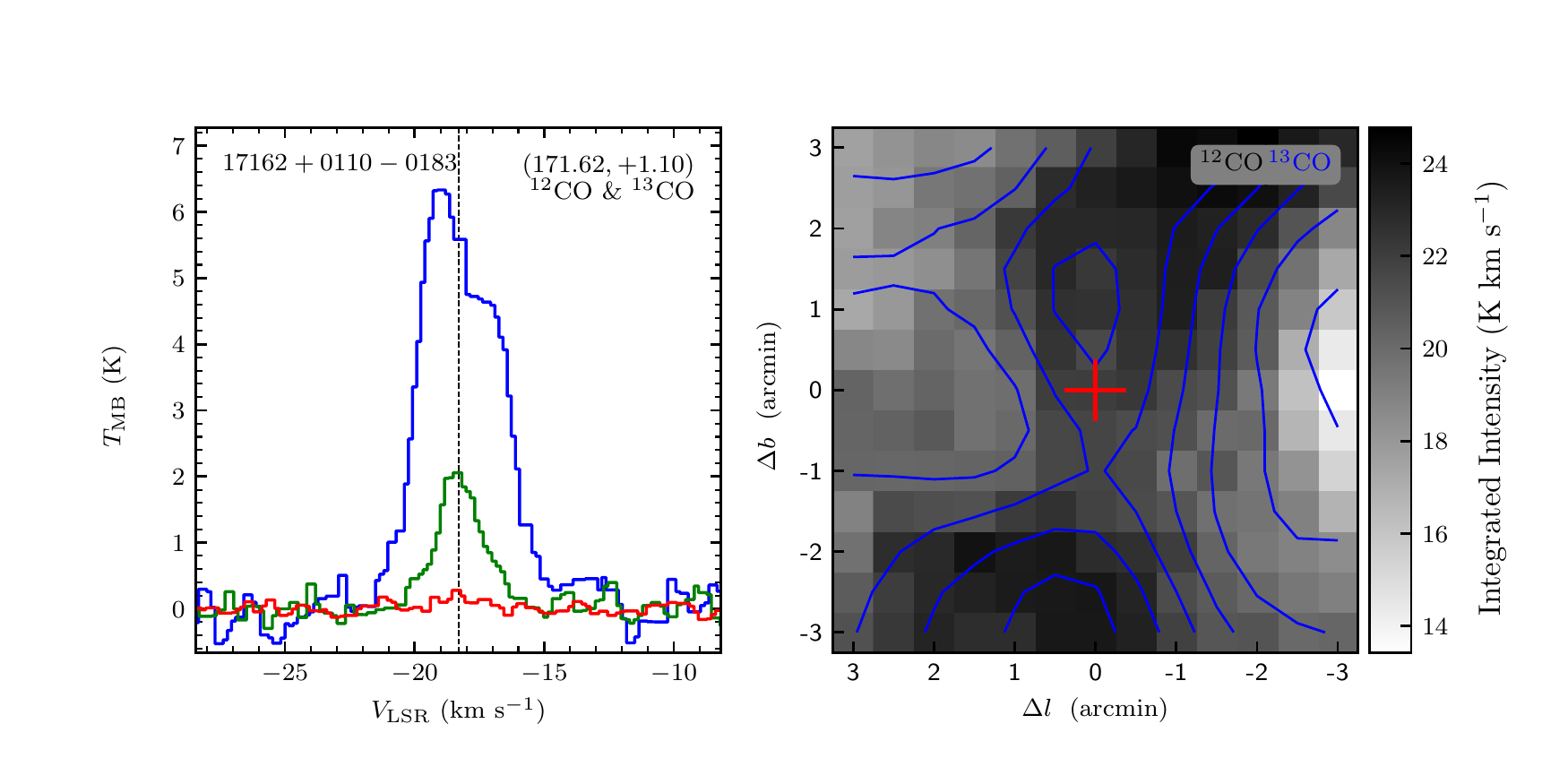}
\includegraphics[width=9.0cm,angle=0]{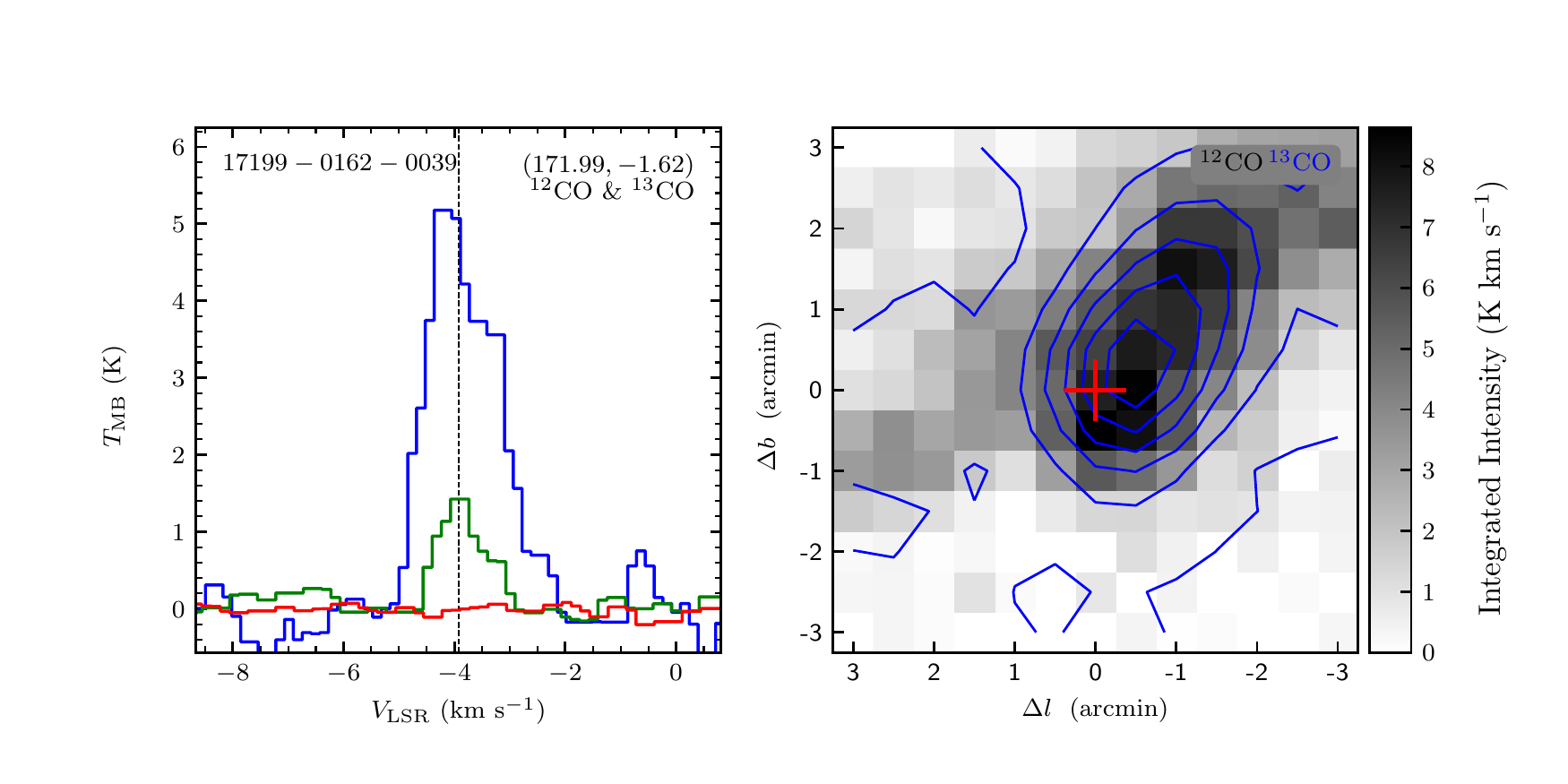}
\vspace{-0.5cm}

\includegraphics[width=9.0cm,angle=0]{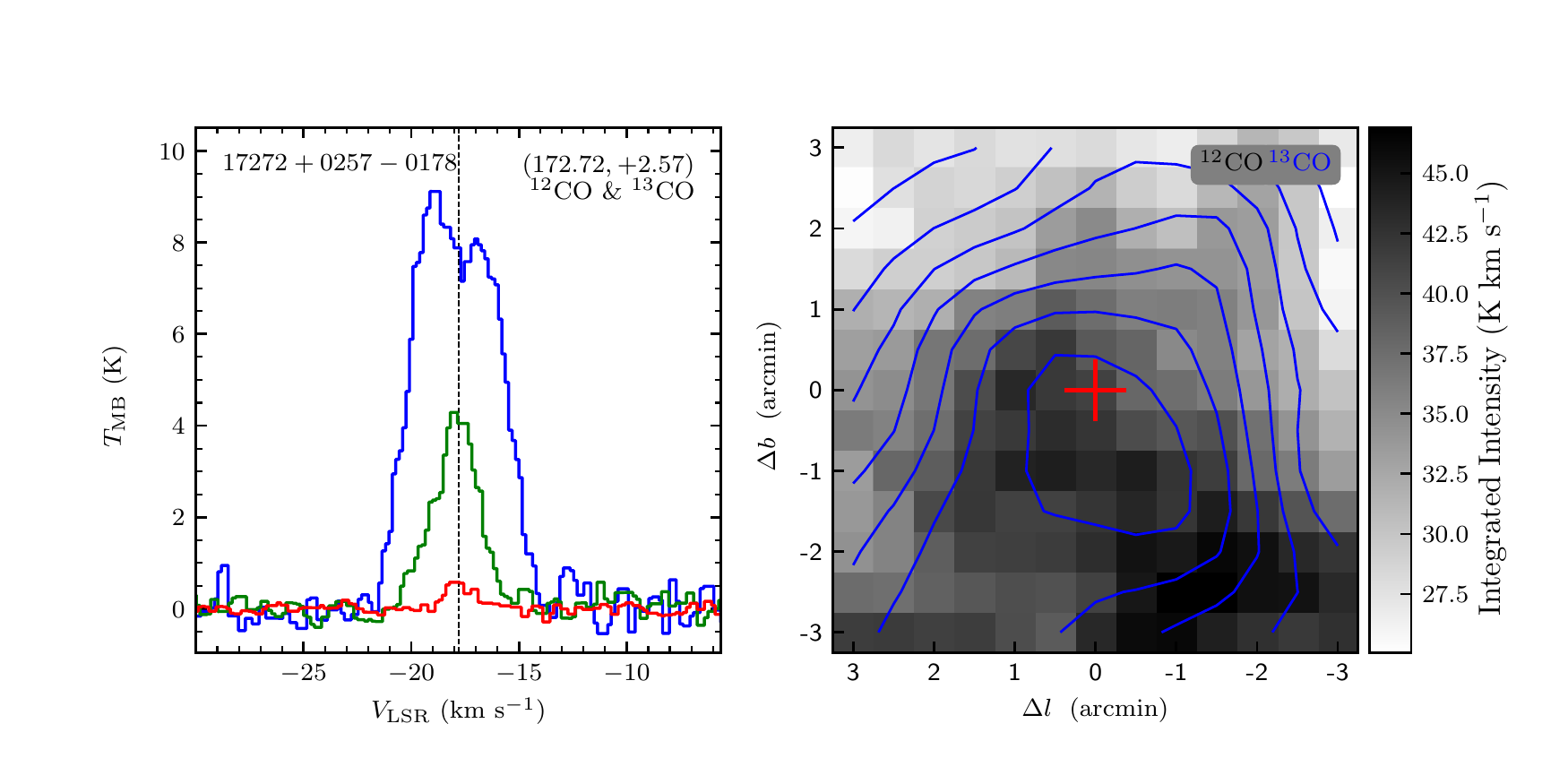}
\includegraphics[width=9.0cm,angle=0]{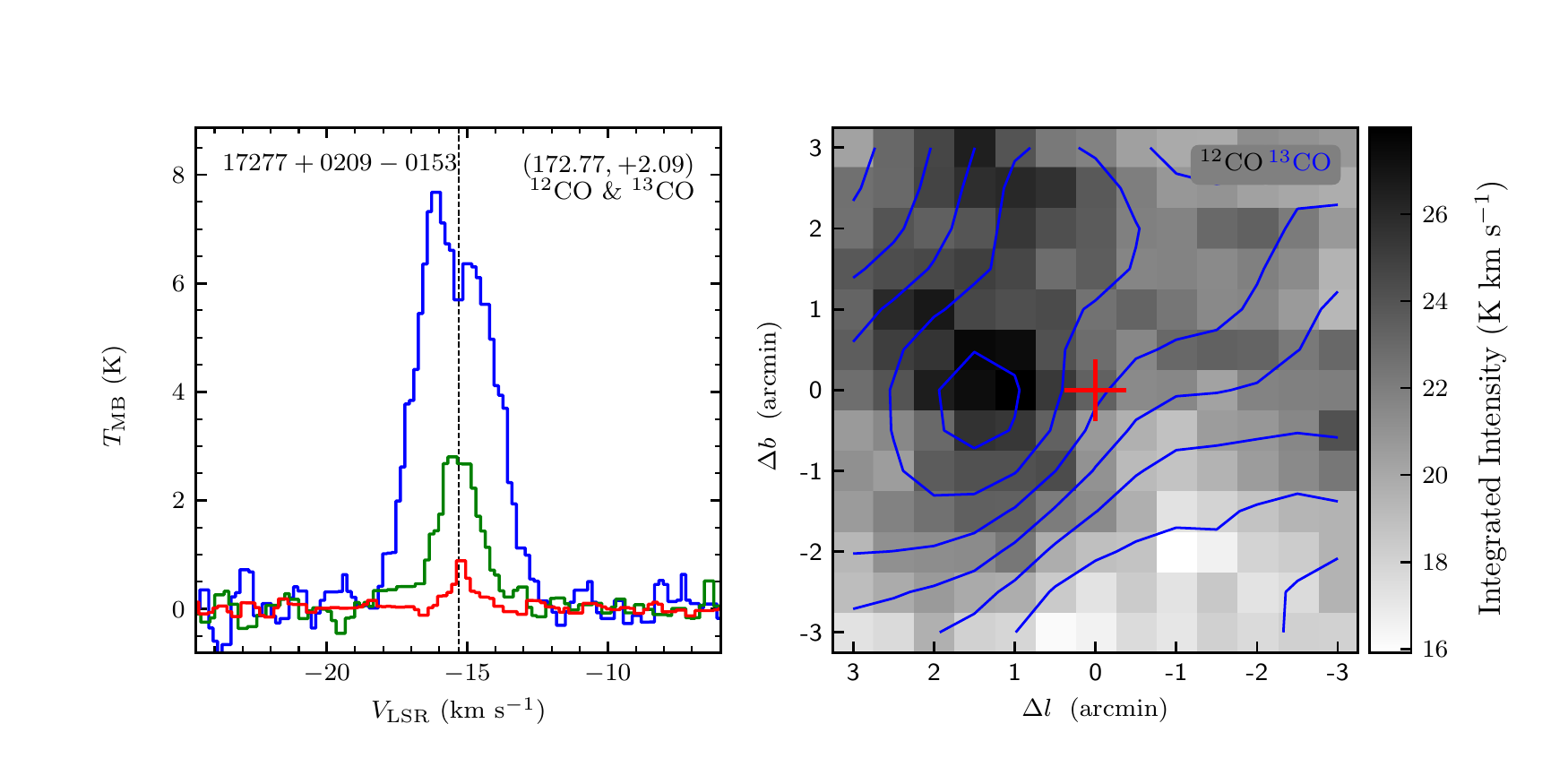}
\vspace{-0.5cm}

\includegraphics[width=9.0cm,angle=0]{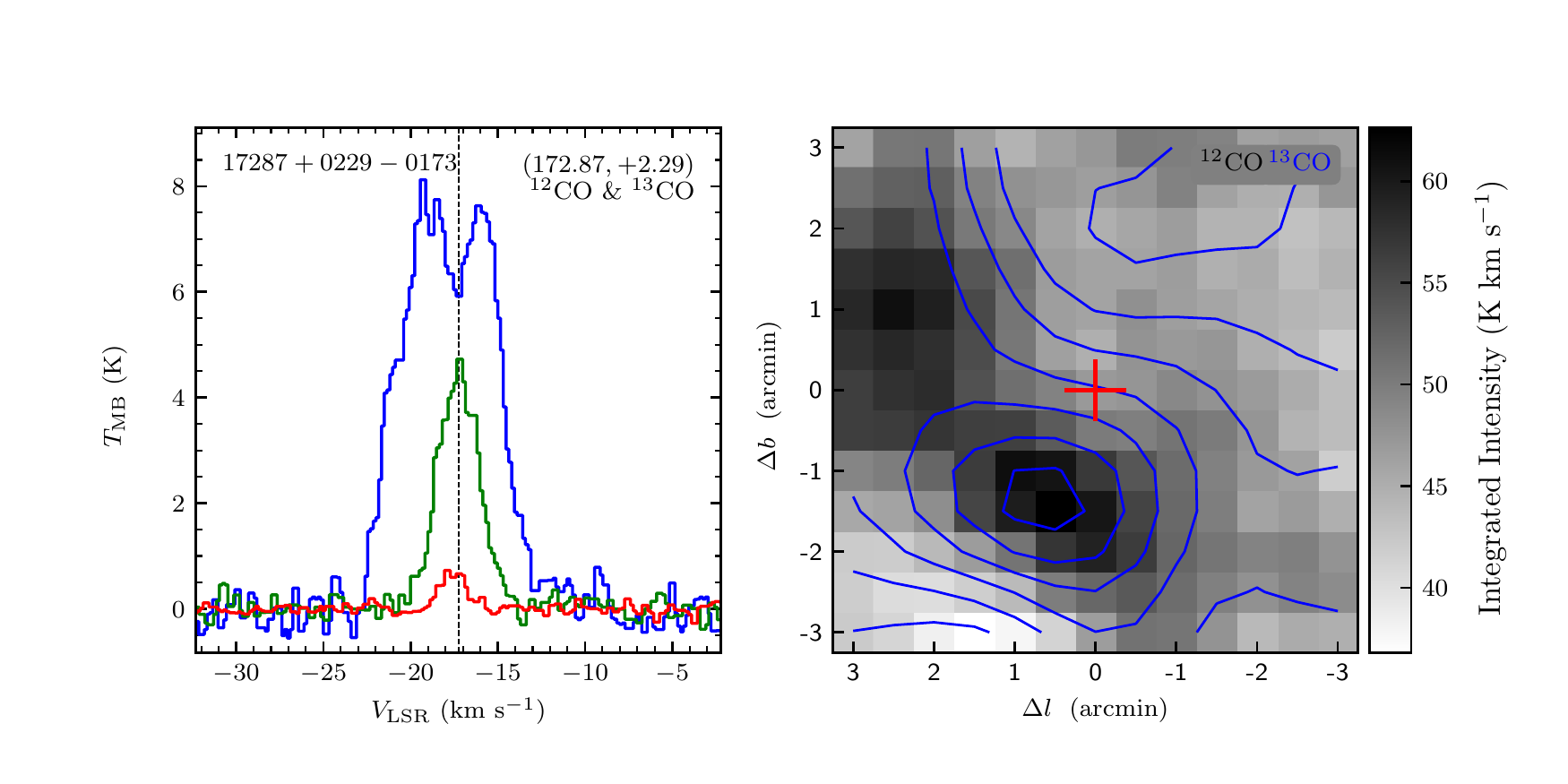}
\includegraphics[width=9.0cm,angle=0]{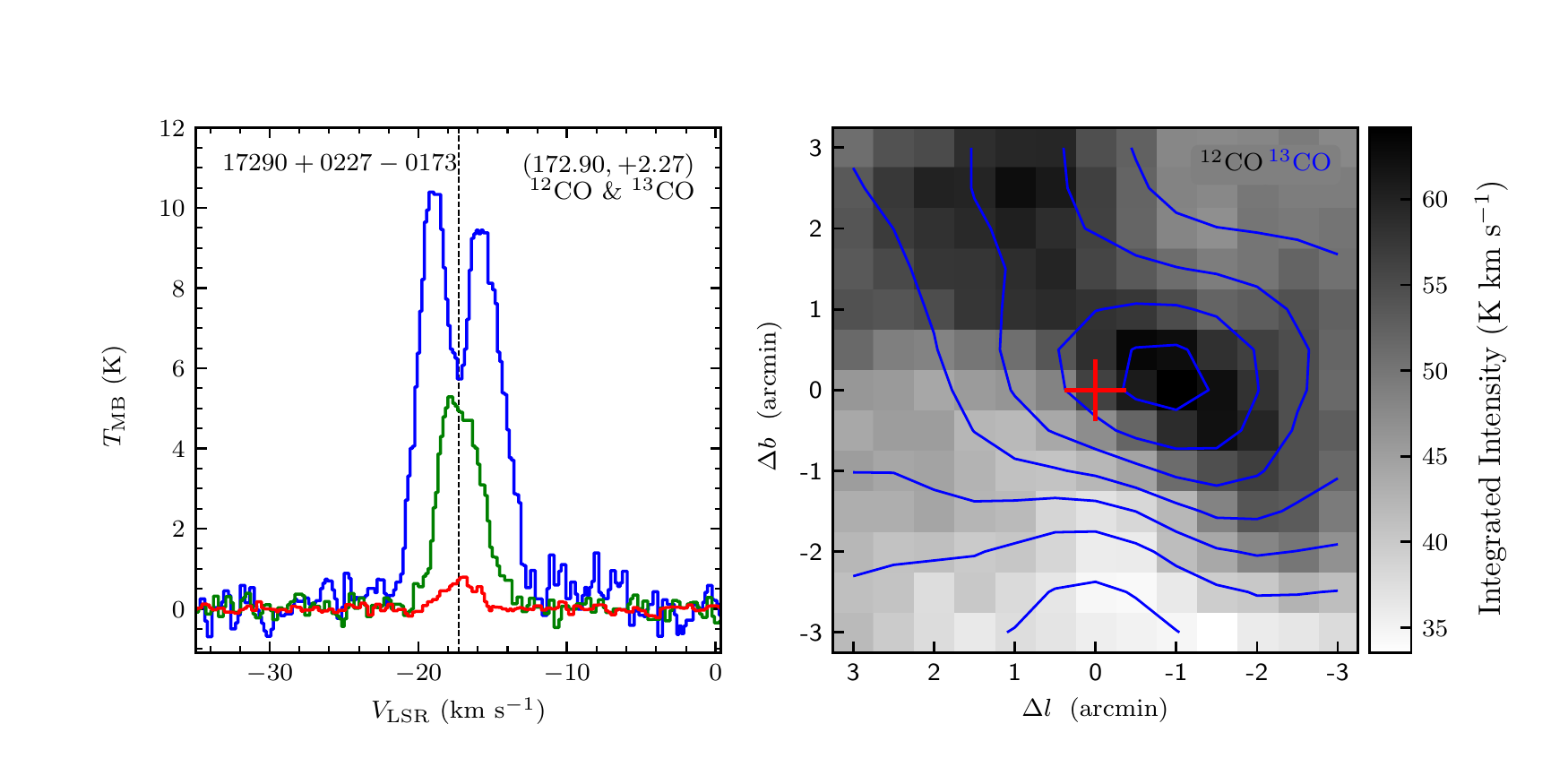}
\vspace{-0.5cm}

\includegraphics[width=9.0cm,angle=0]{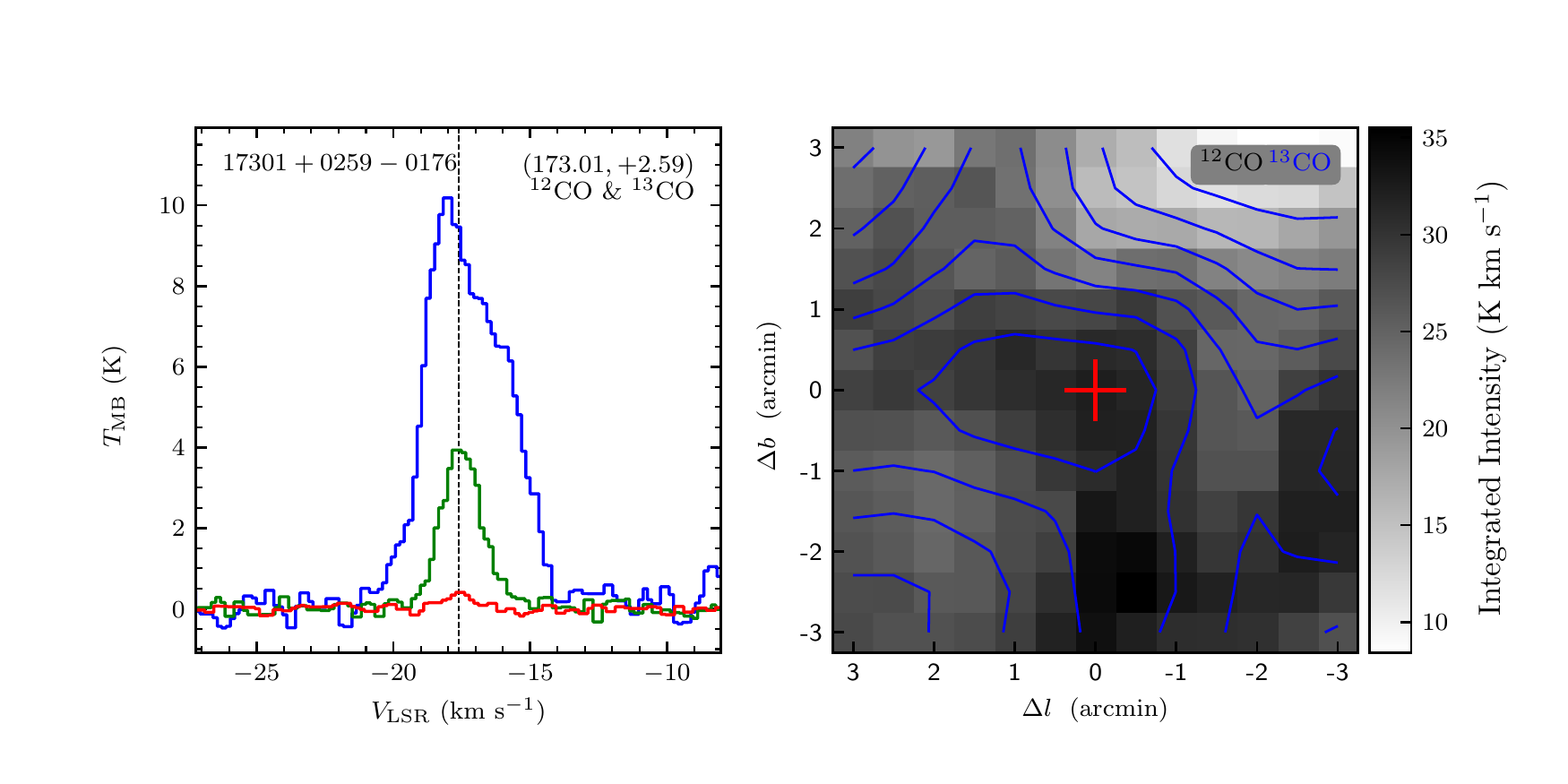}
\includegraphics[width=9.0cm,angle=0]{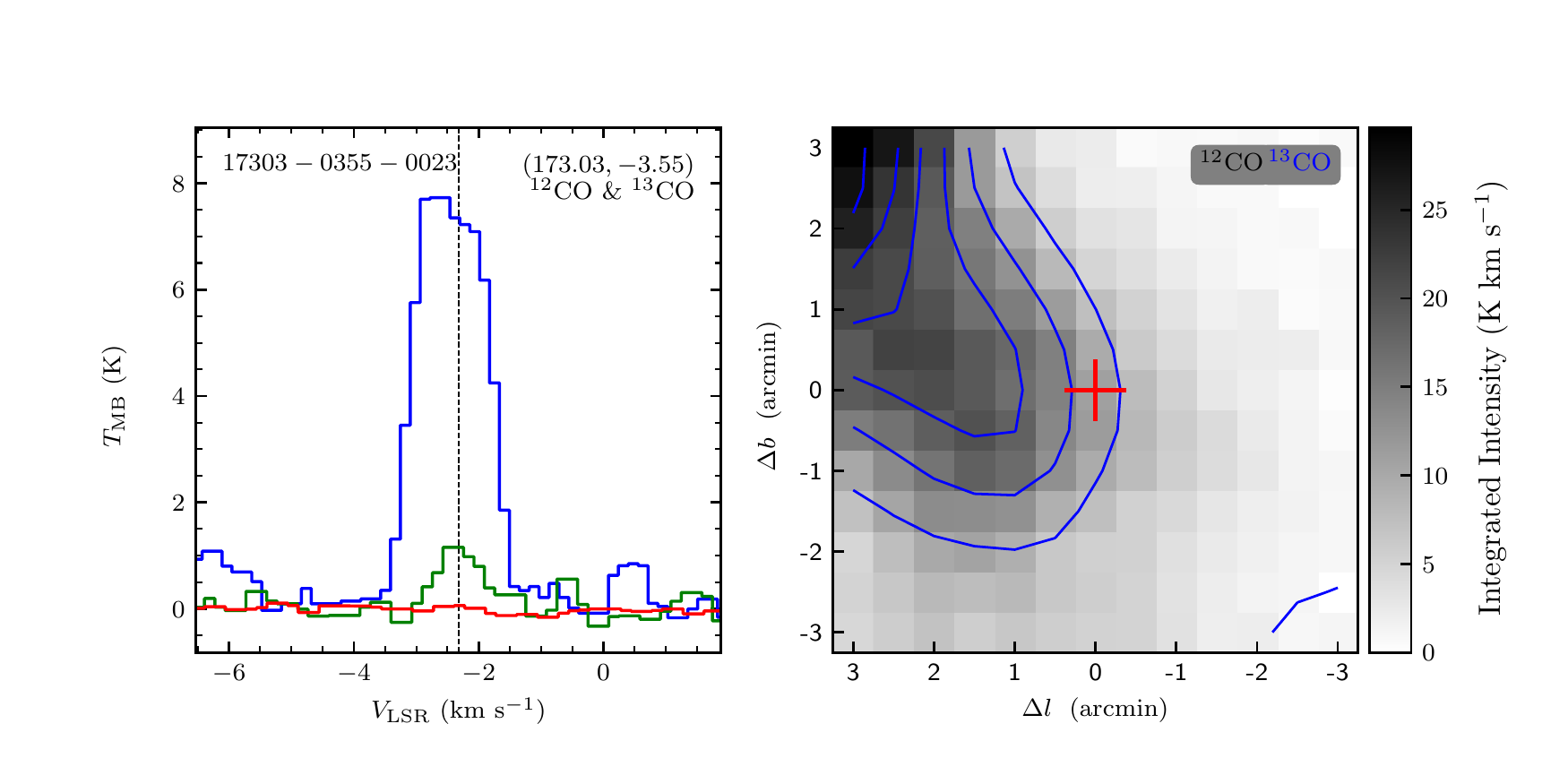}
\vspace{-0.5cm}

\includegraphics[width=9.0cm,angle=0]{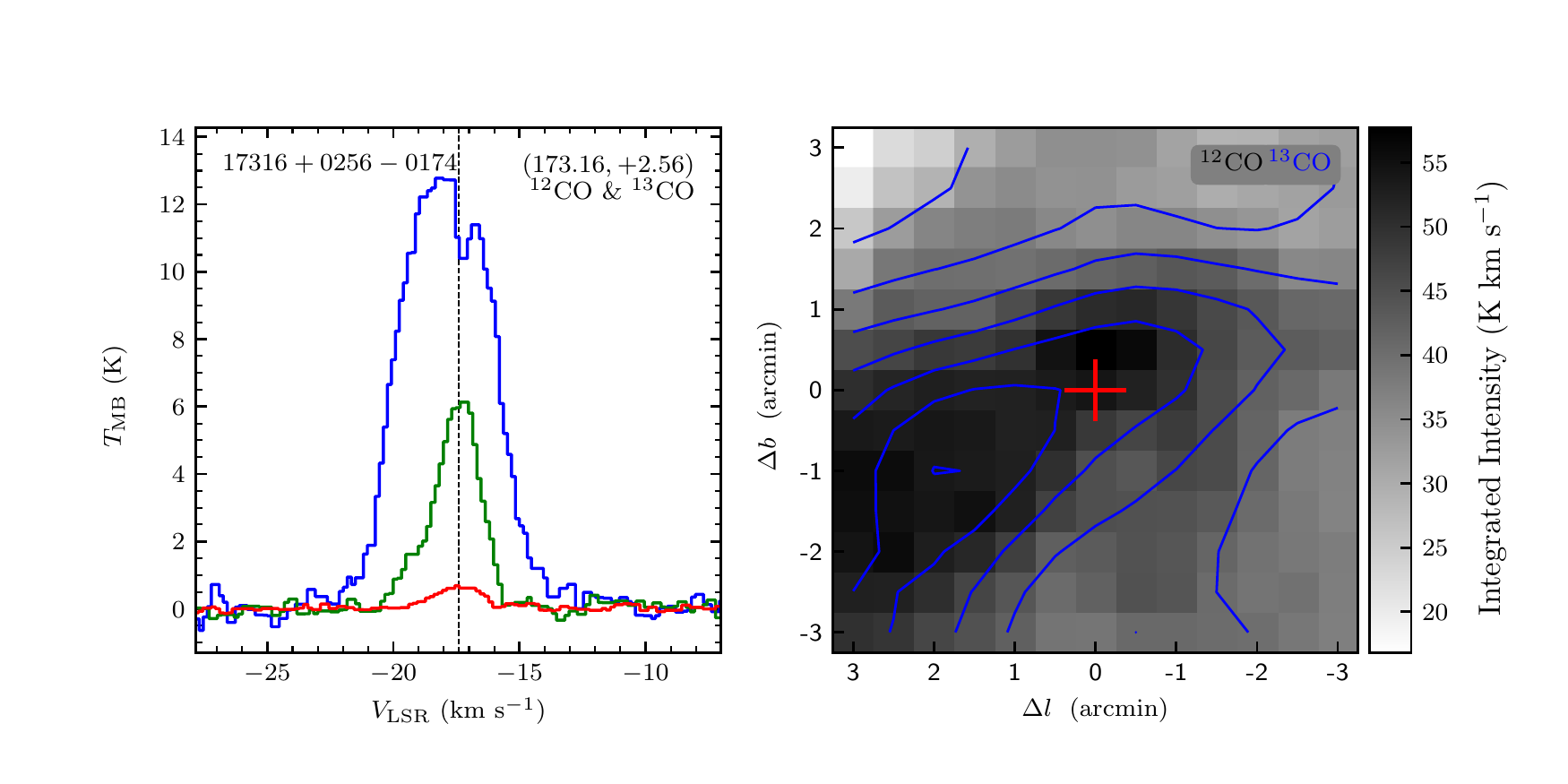}
\includegraphics[width=9.0cm,angle=0]{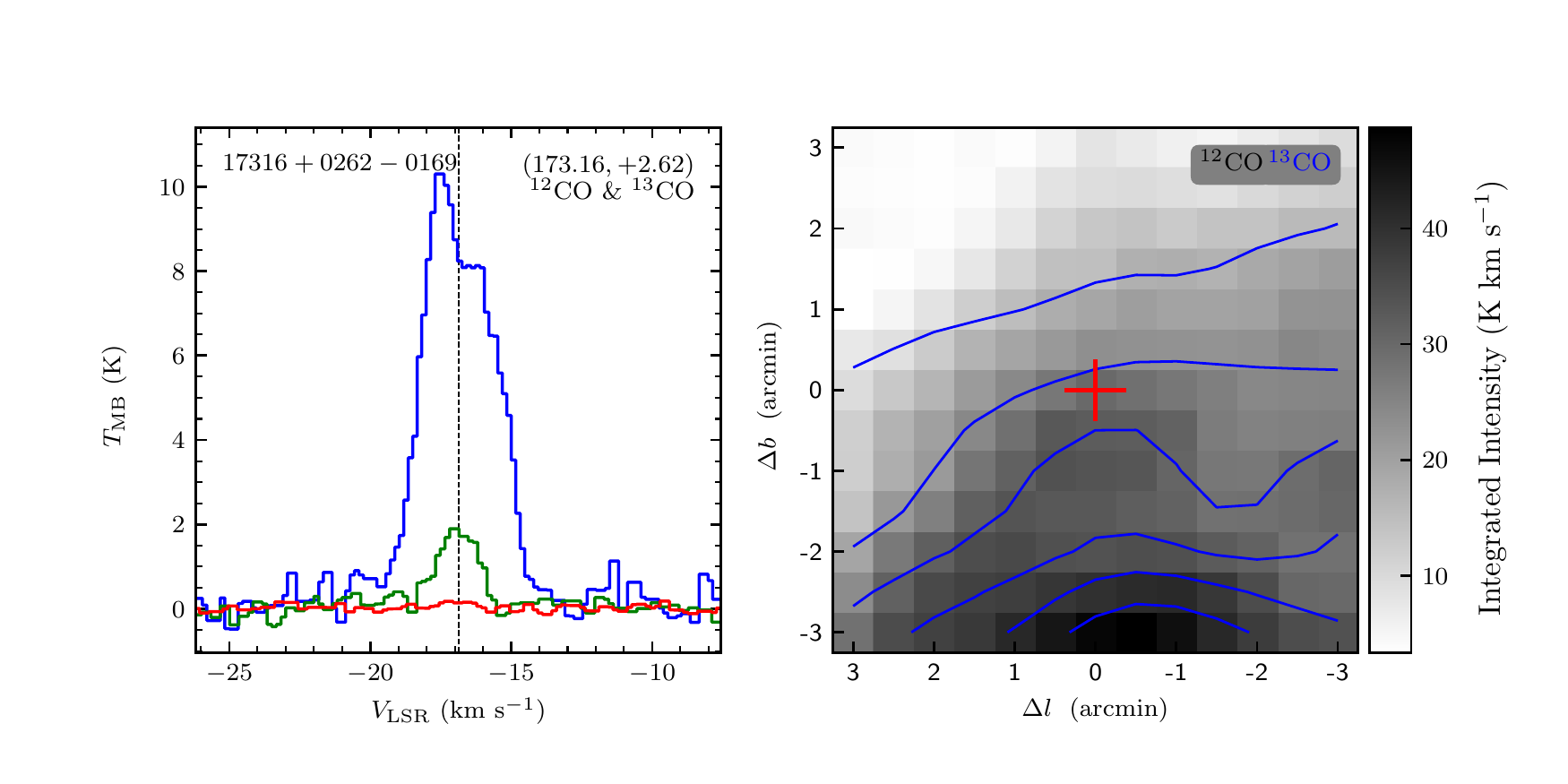}
\end{figure}
\clearpage

\begin{figure}
\includegraphics[width=9.0cm,angle=0]{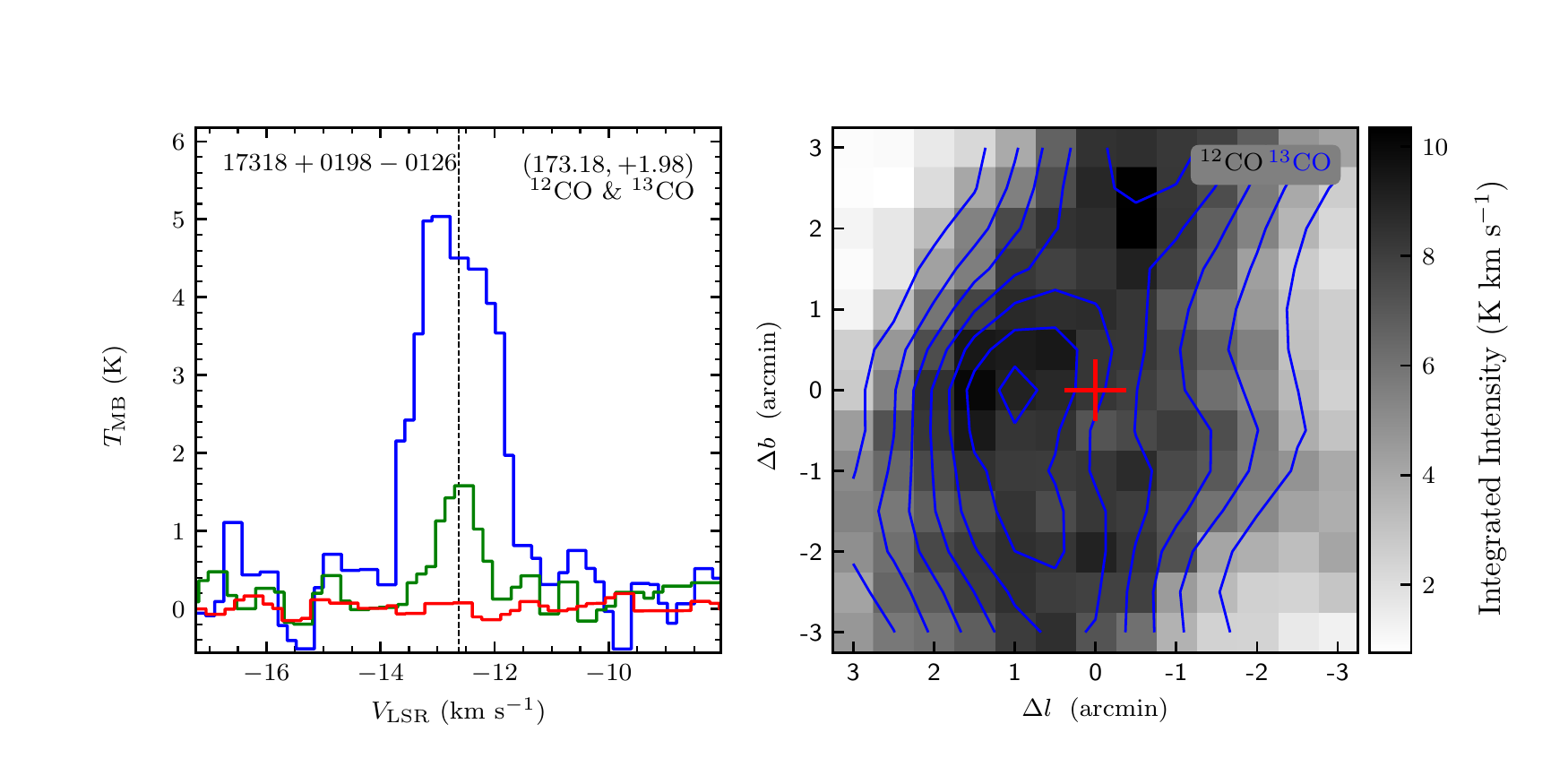}
\includegraphics[width=9.0cm,angle=0]{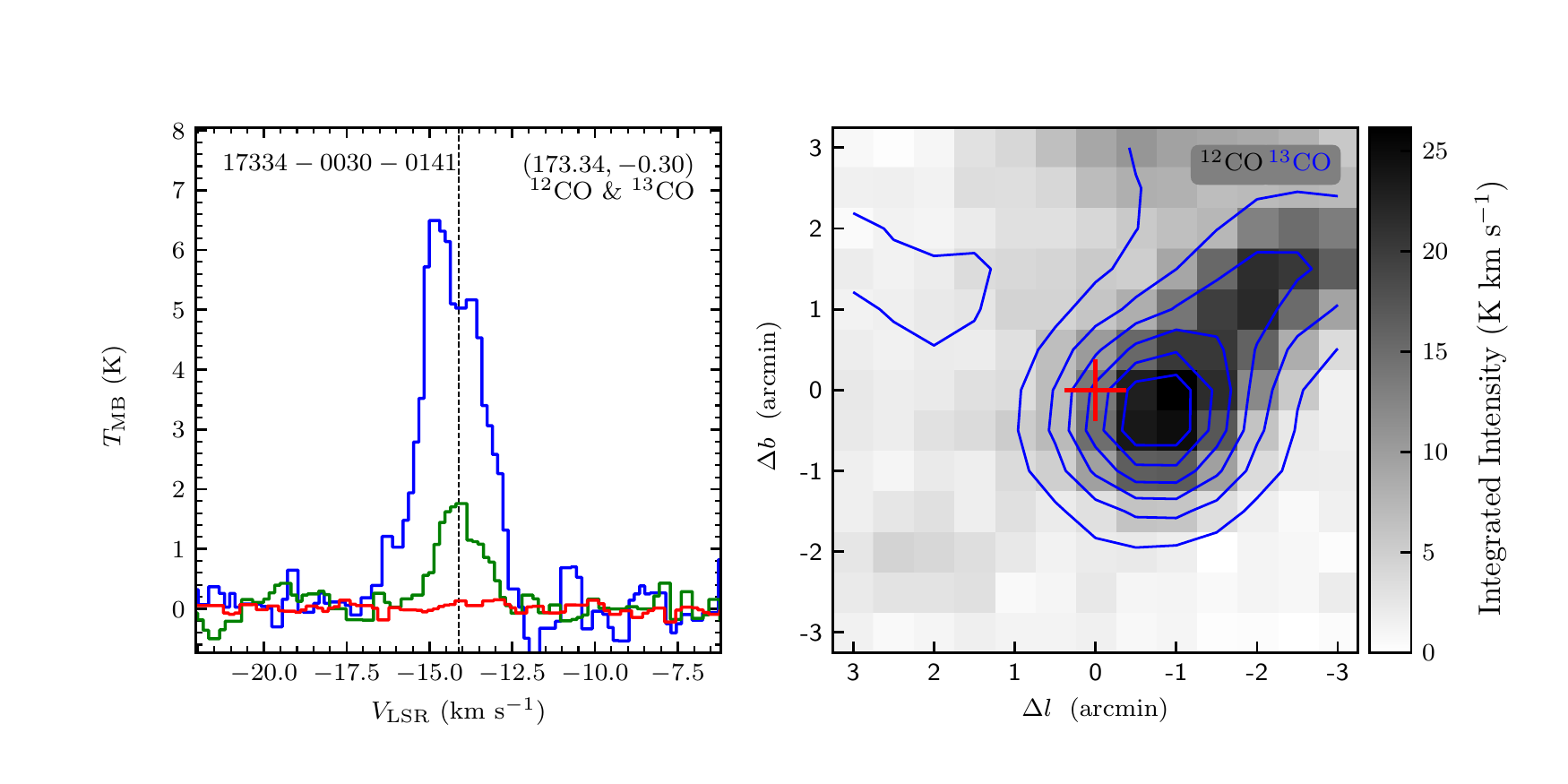}
\vspace{-0.5cm}

\includegraphics[width=9.0cm,angle=0]{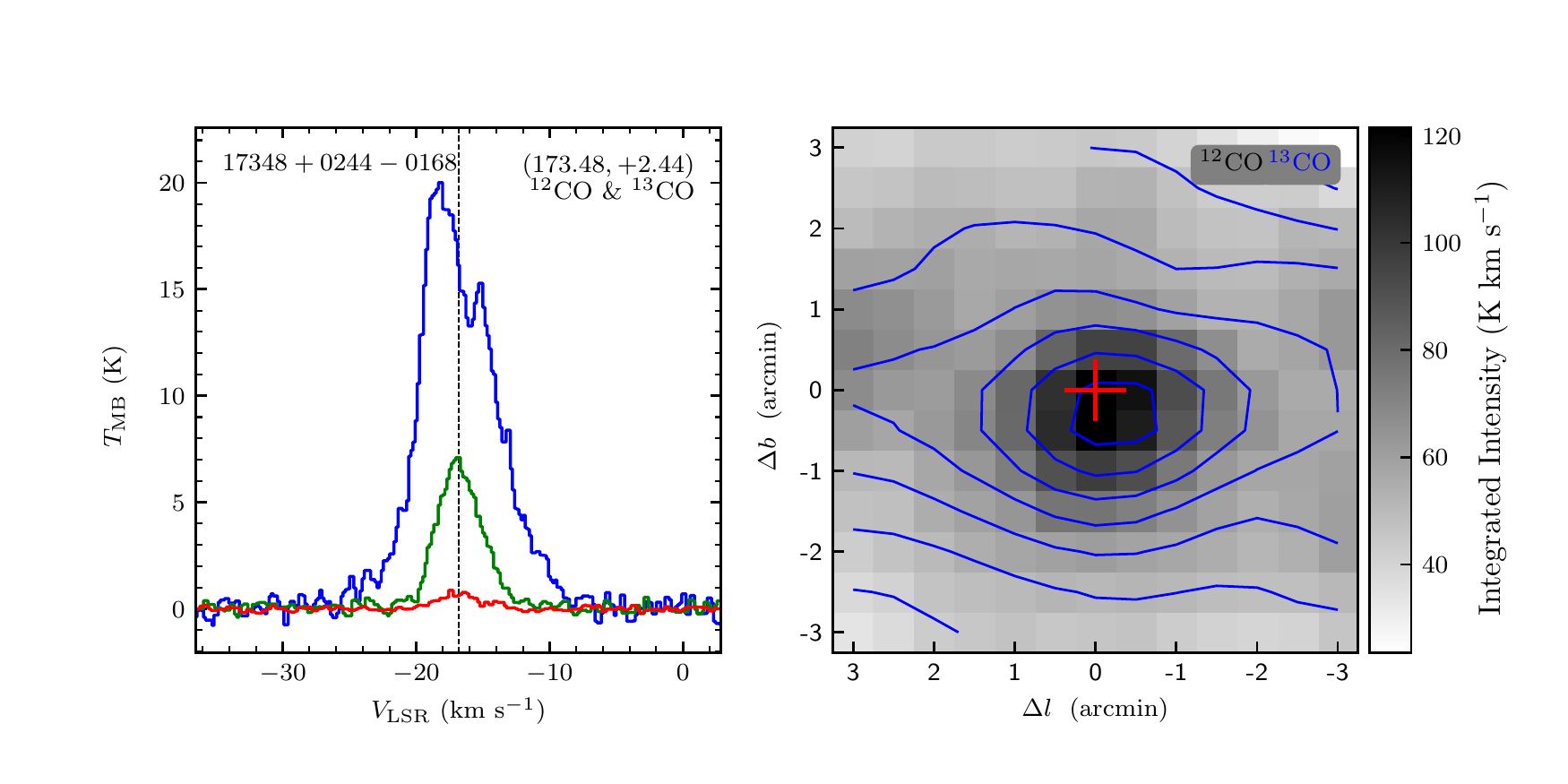}
\includegraphics[width=9.0cm,angle=0]{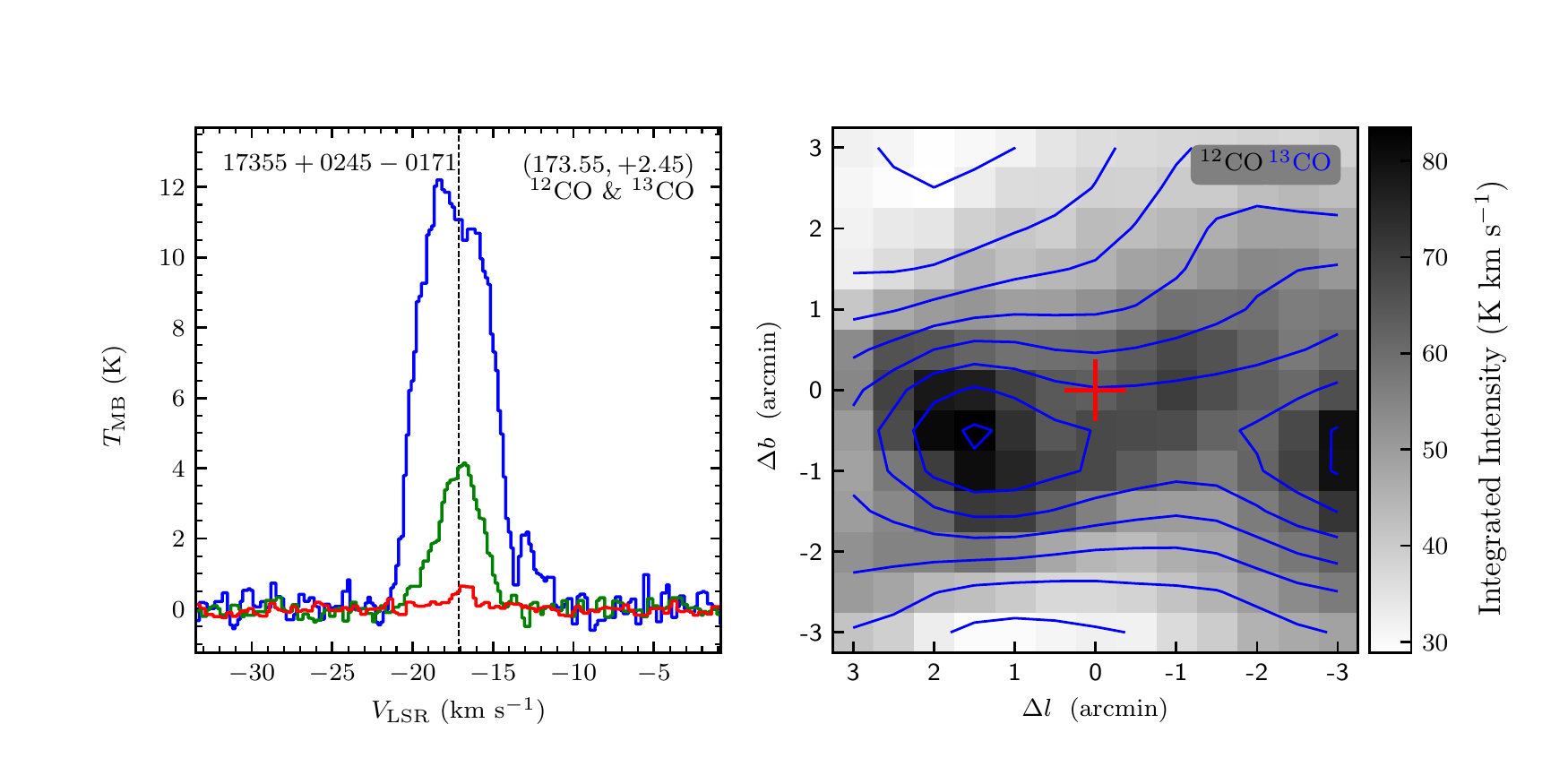}
\vspace{-0.5cm}

\includegraphics[width=9.0cm,angle=0]{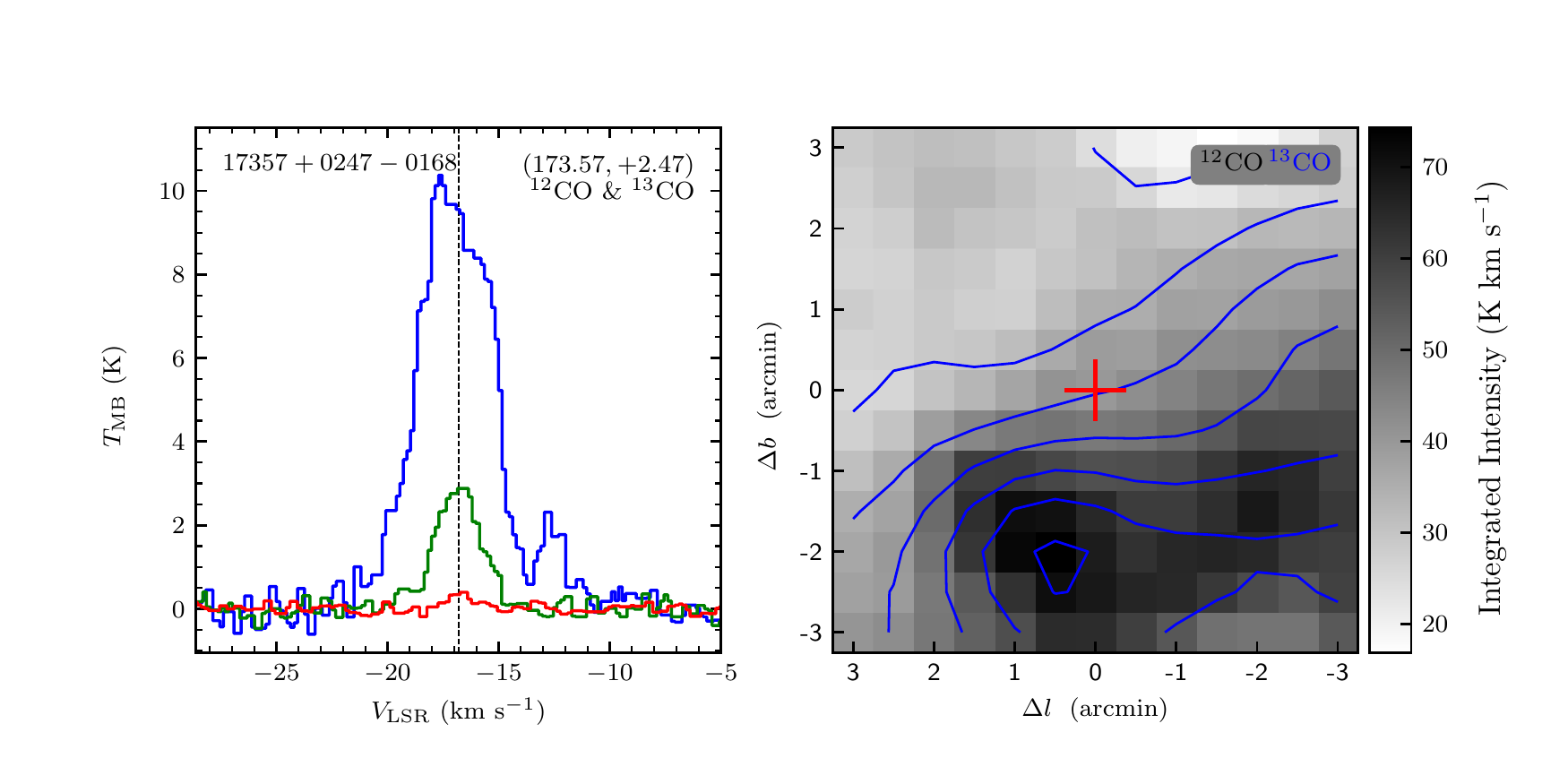}
\includegraphics[width=9.0cm,angle=0]{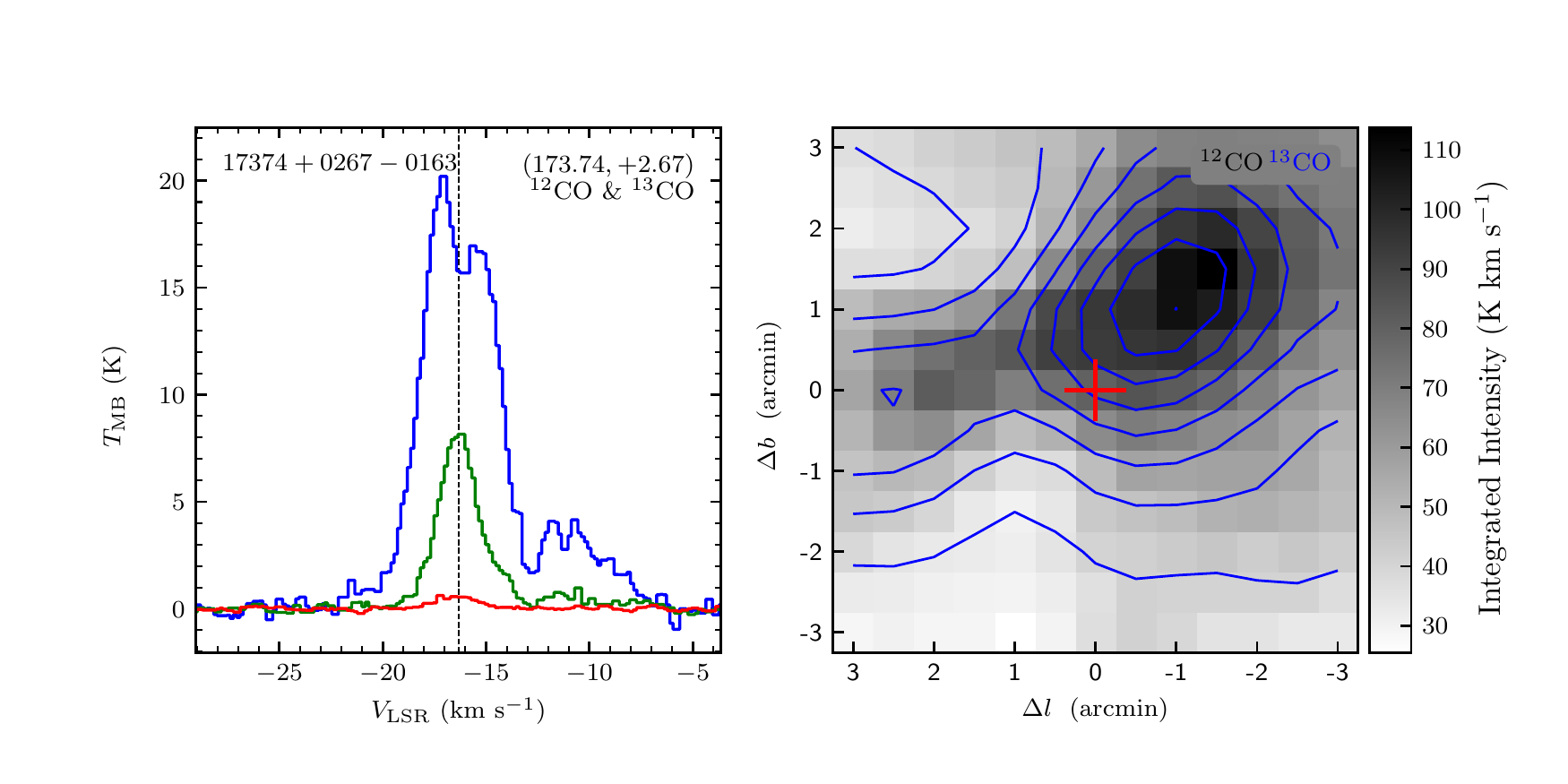}
\vspace{-0.5cm}

\includegraphics[width=9.0cm,angle=0]{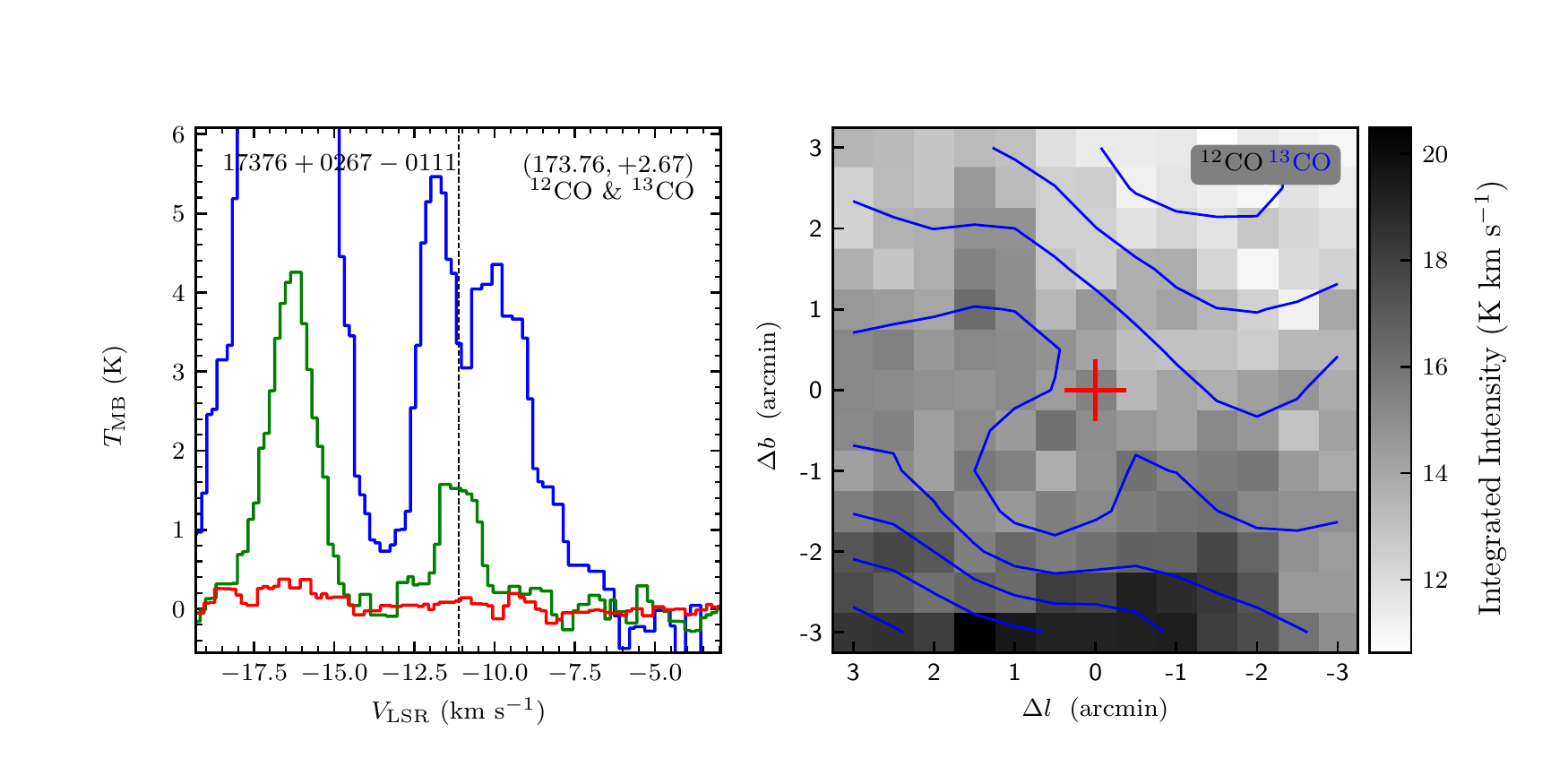}
\includegraphics[width=9.0cm,angle=0]{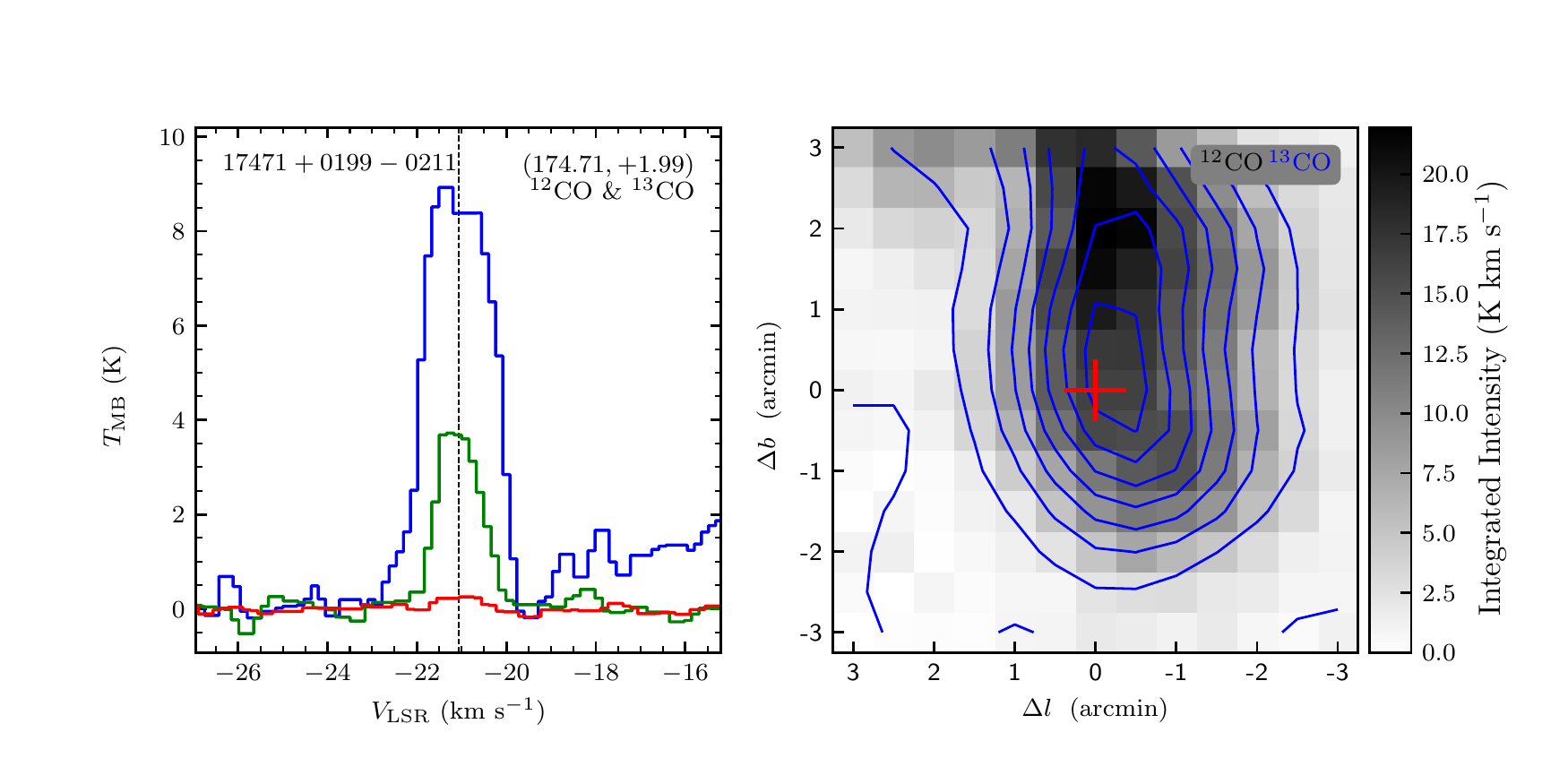}
\vspace{-0.5cm}

\includegraphics[width=9.0cm,angle=0]{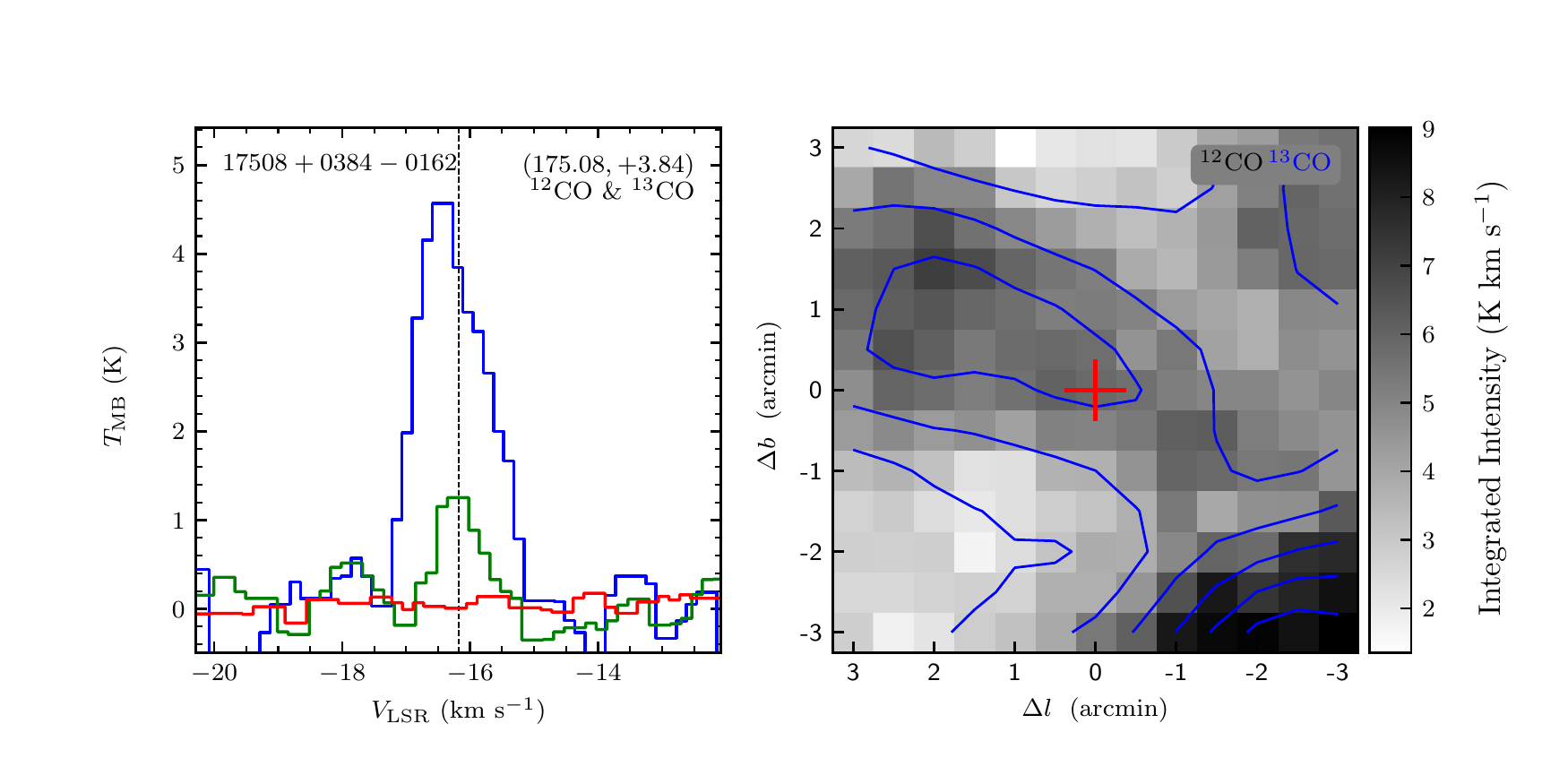}
\includegraphics[width=9.0cm,angle=0]{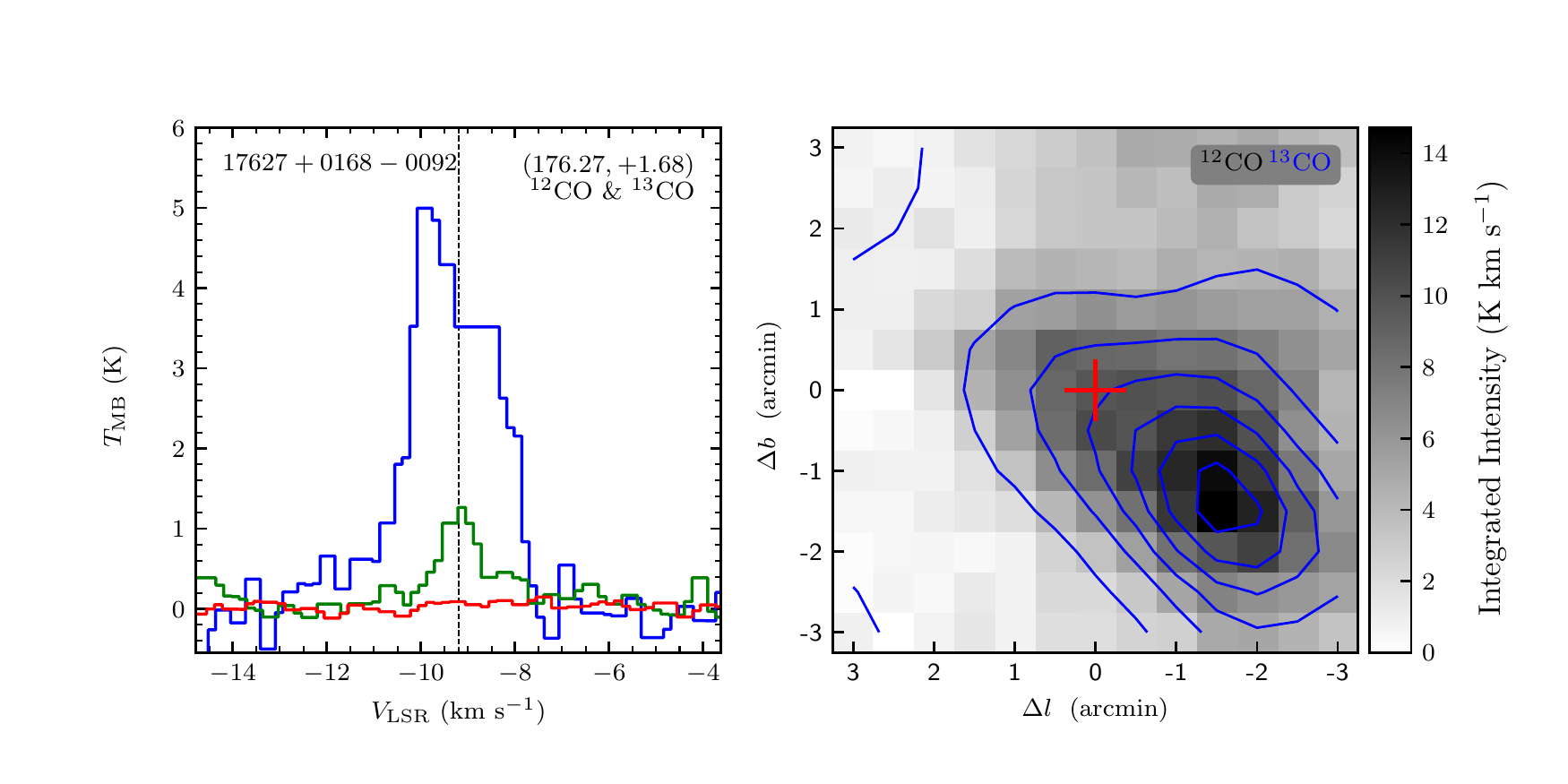}
\end{figure}
\clearpage

\begin{figure}
\includegraphics[width=9.0cm,angle=0]{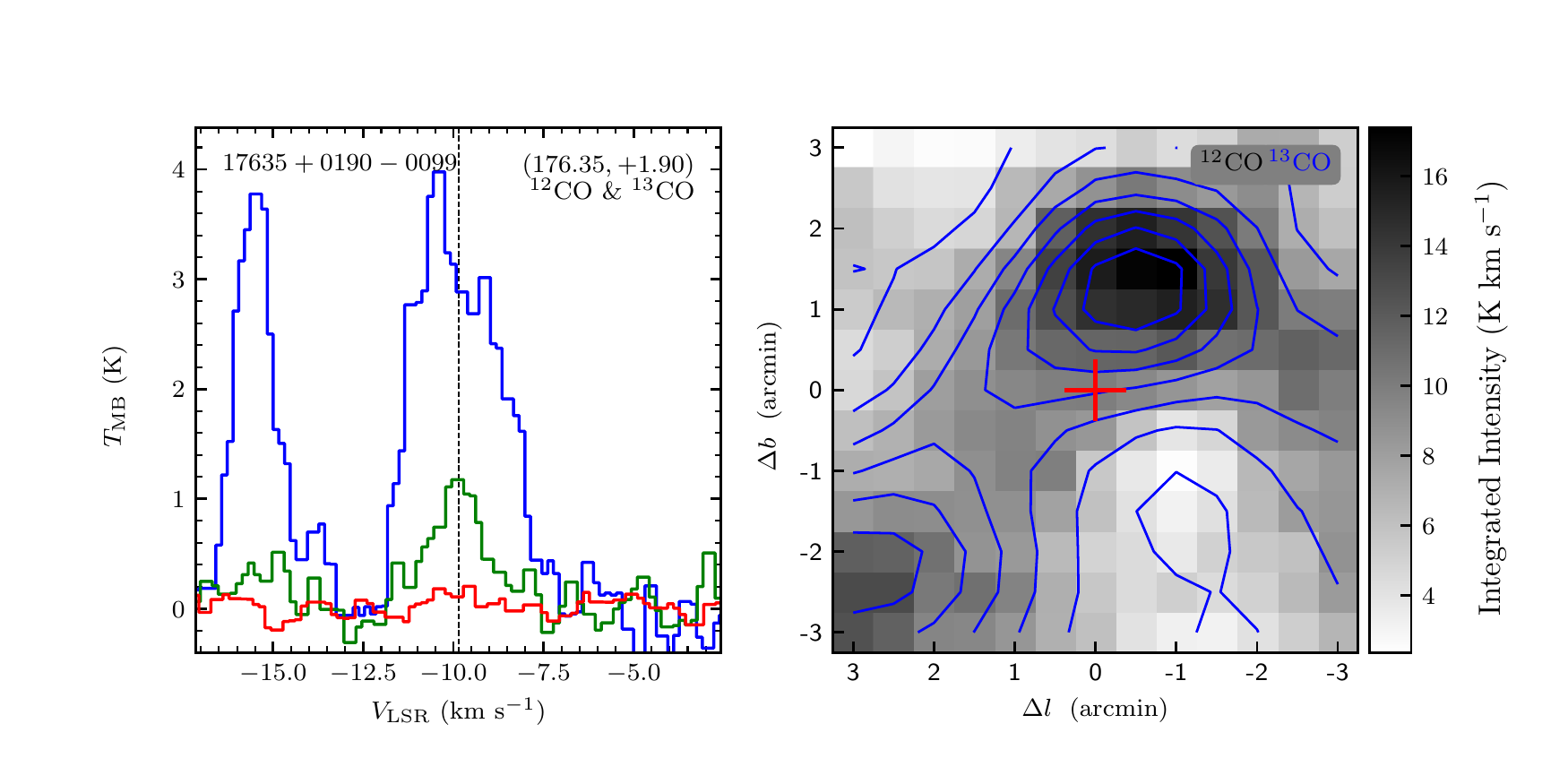}
\includegraphics[width=9.0cm,angle=0]{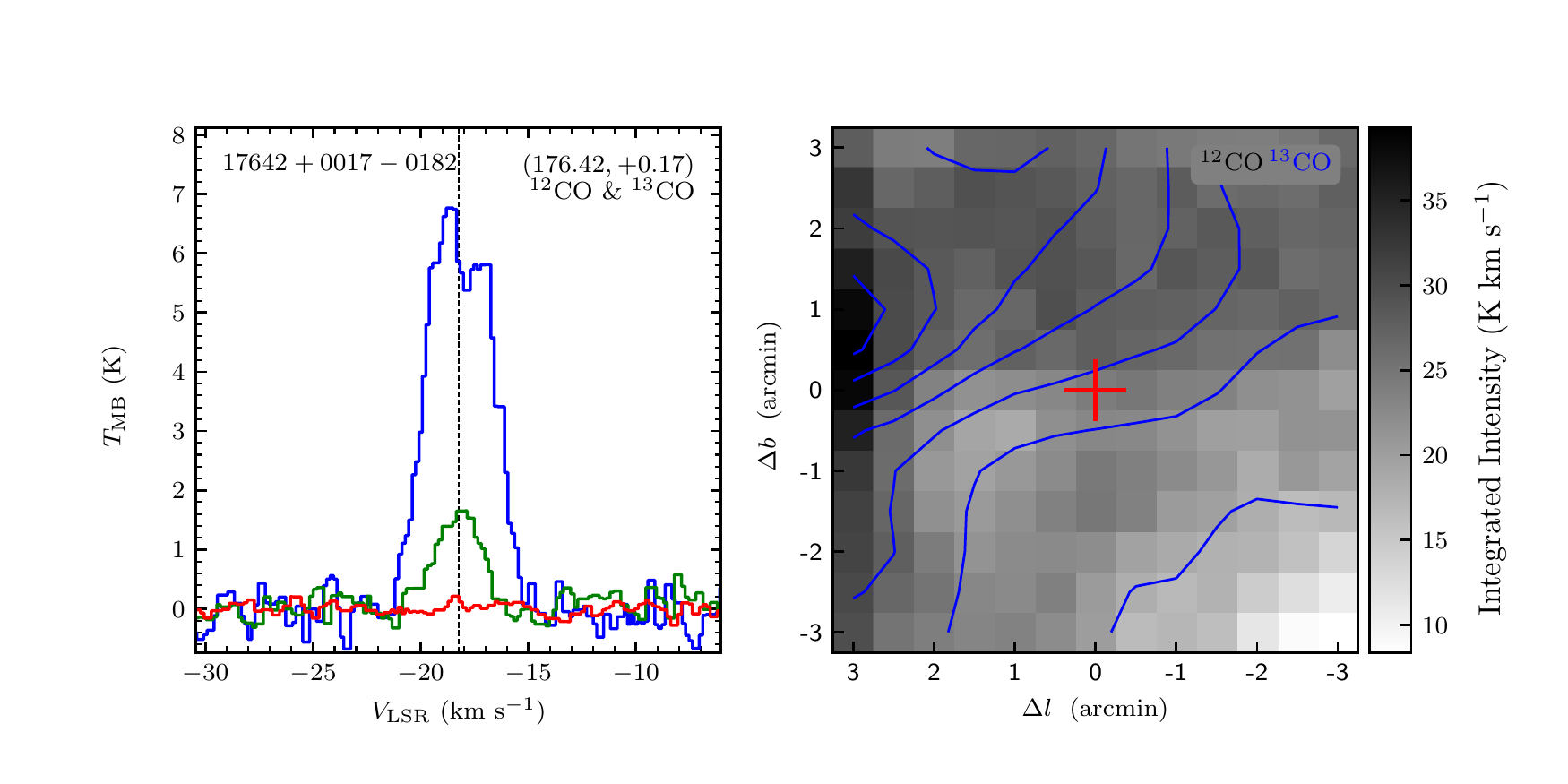}
\vspace{-0.5cm}

\includegraphics[width=9.0cm,angle=0]{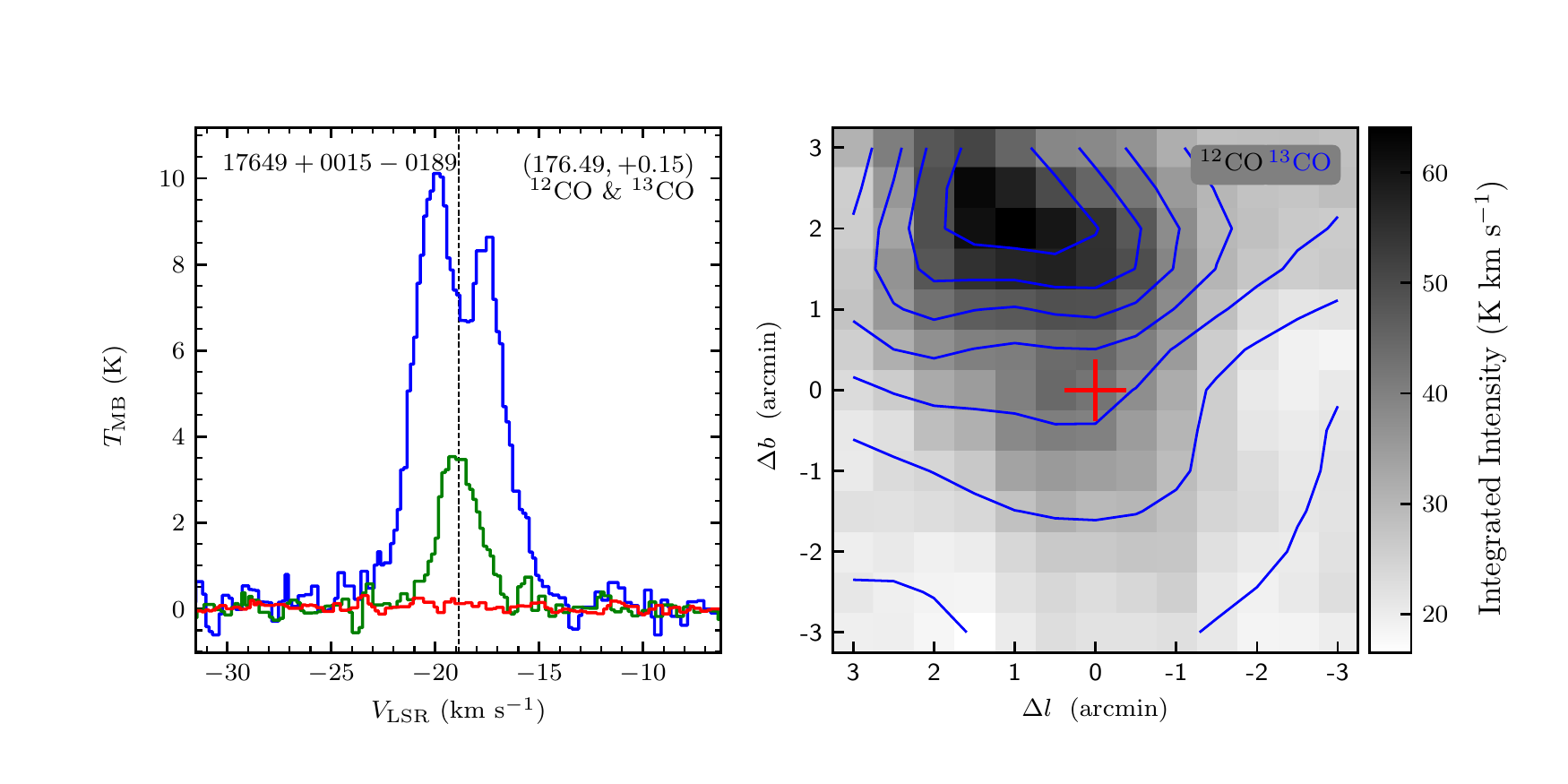}
\includegraphics[width=9.0cm,angle=0]{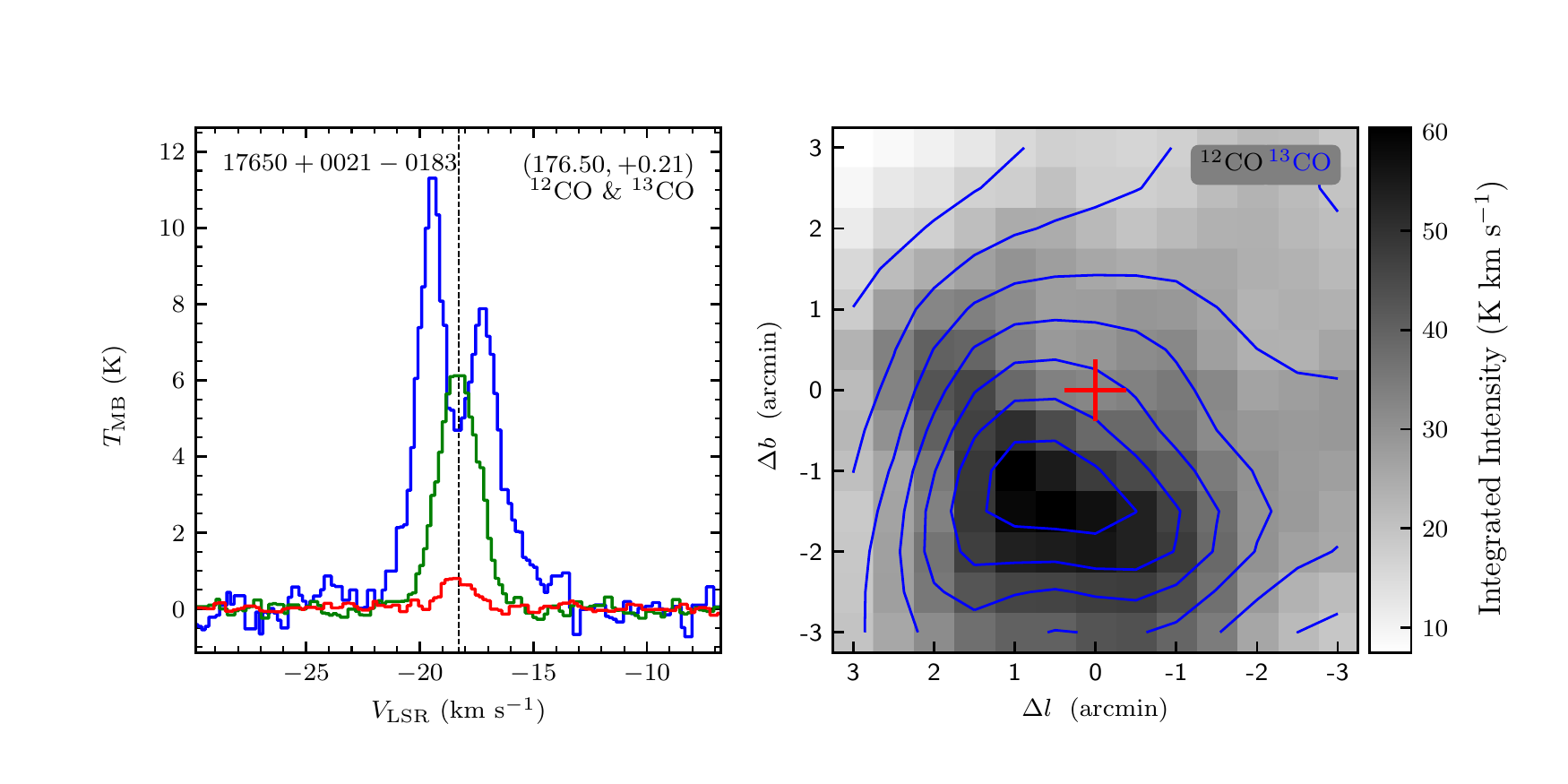}
\vspace{-0.5cm}

\includegraphics[width=9.0cm,angle=0]{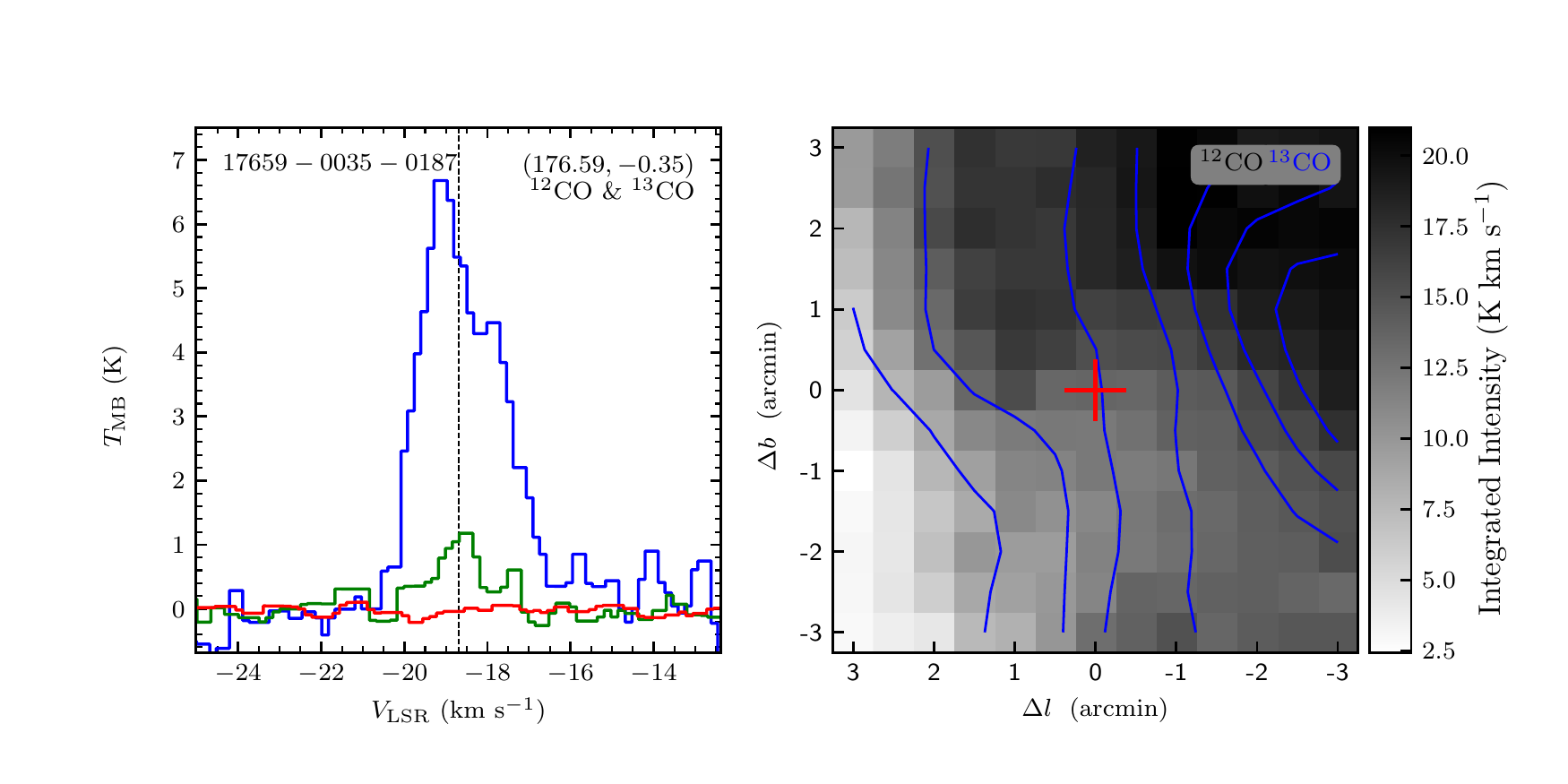}
\includegraphics[width=9.0cm,angle=0]{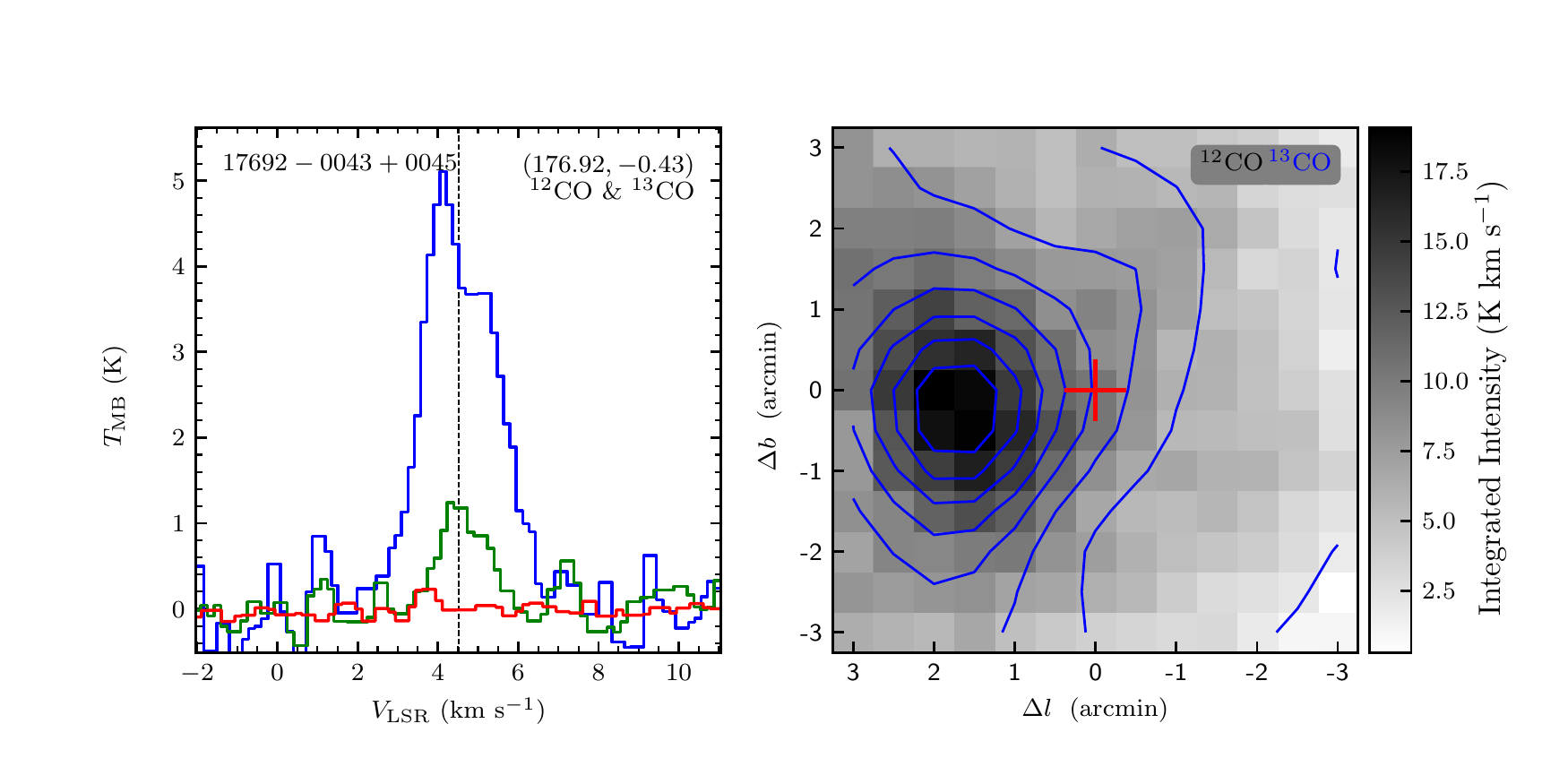}
\vspace{-0.5cm}

\includegraphics[width=9.0cm,angle=0]{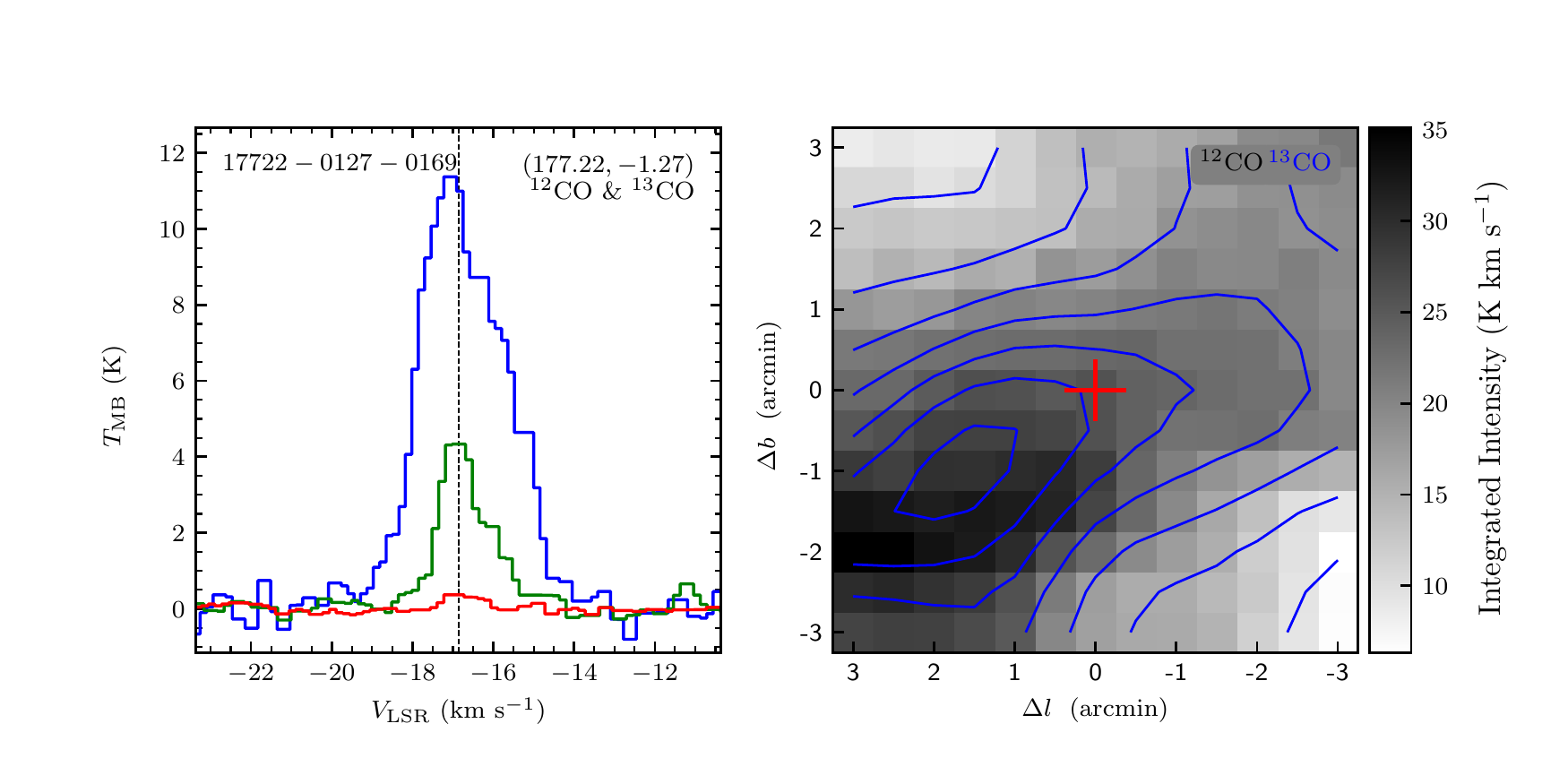}
\includegraphics[width=9.0cm,angle=0]{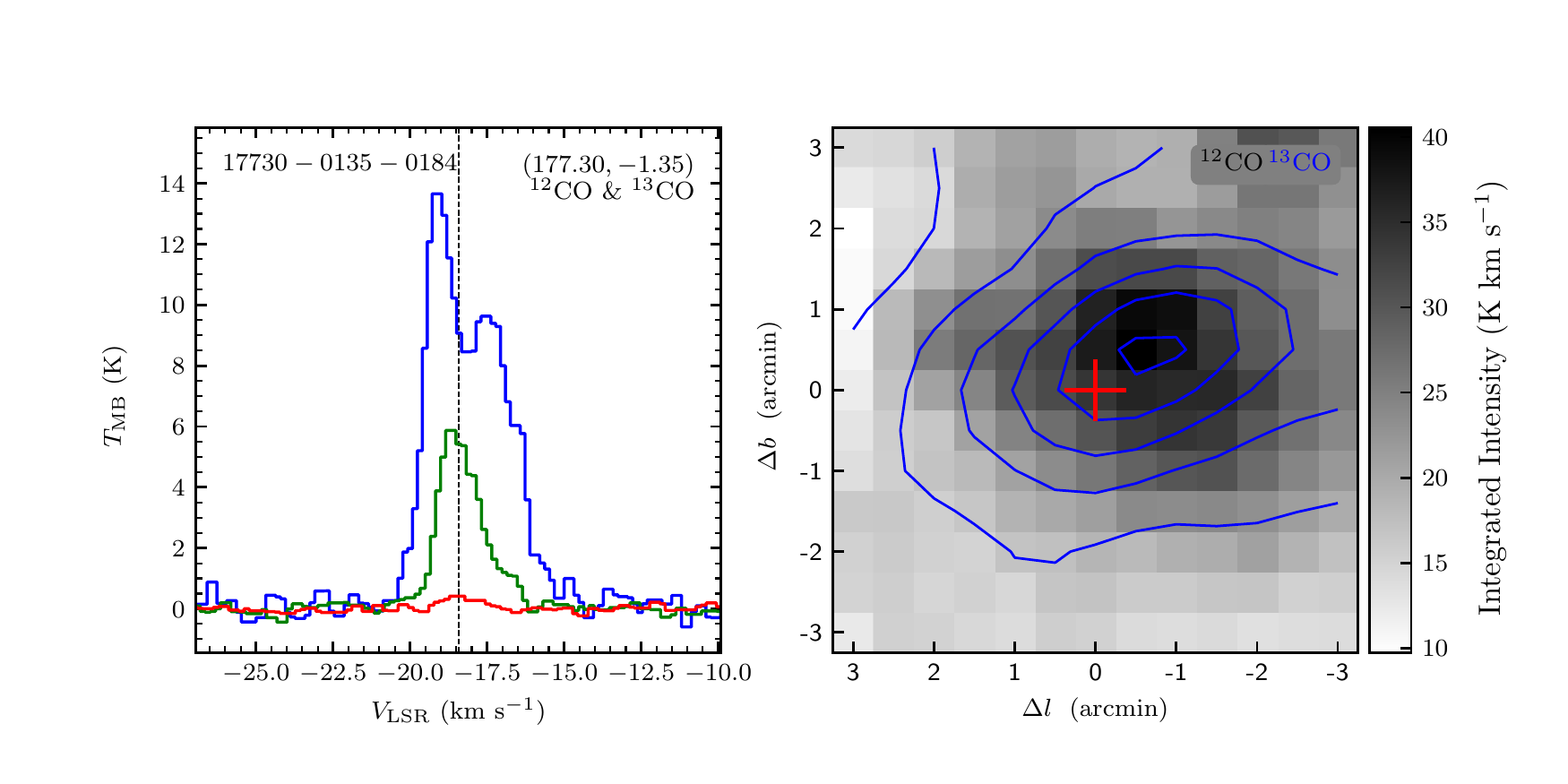}
\vspace{-0.5cm}

\includegraphics[width=9.0cm,angle=0]{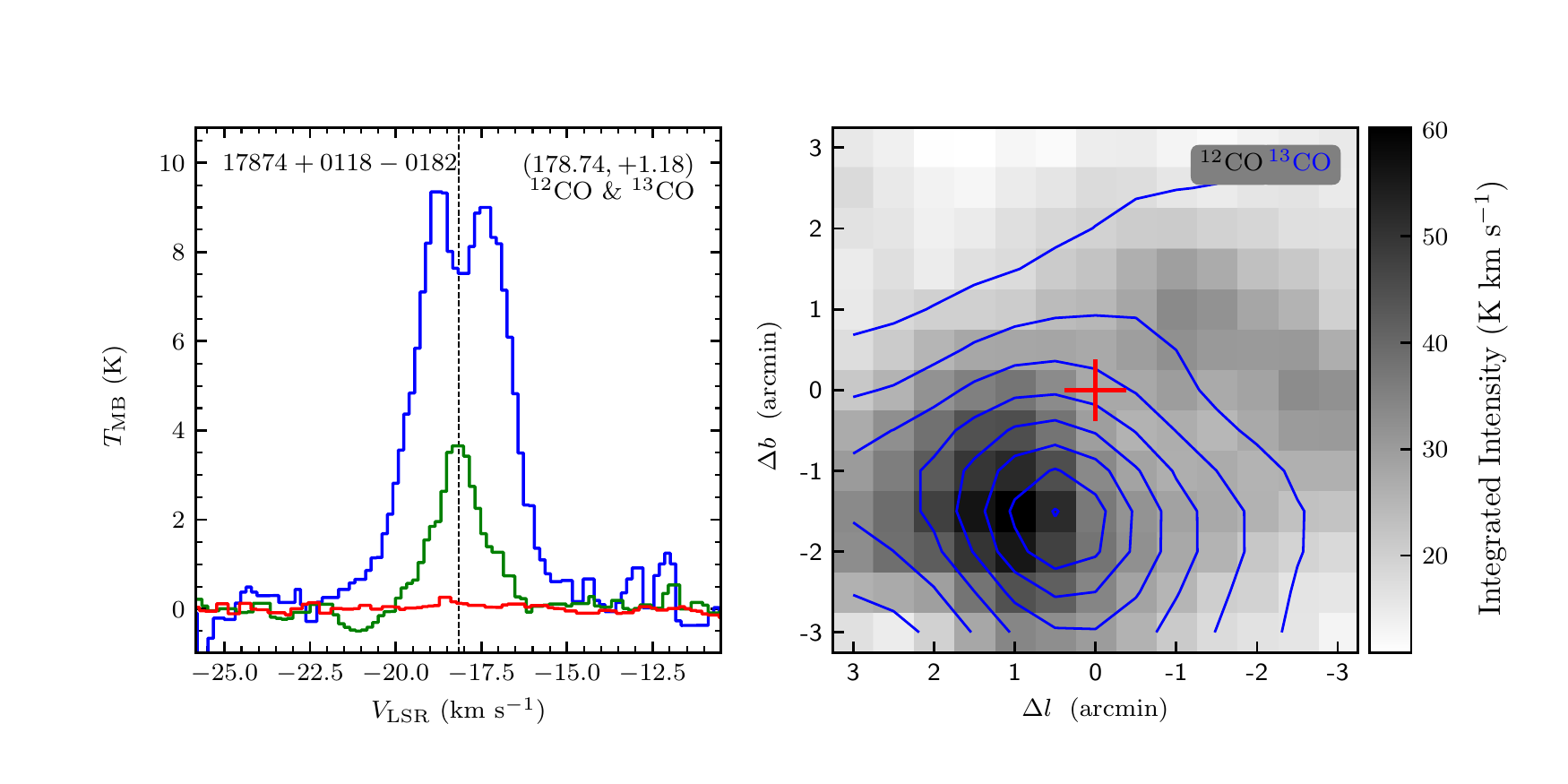}
\includegraphics[width=9.0cm,angle=0]{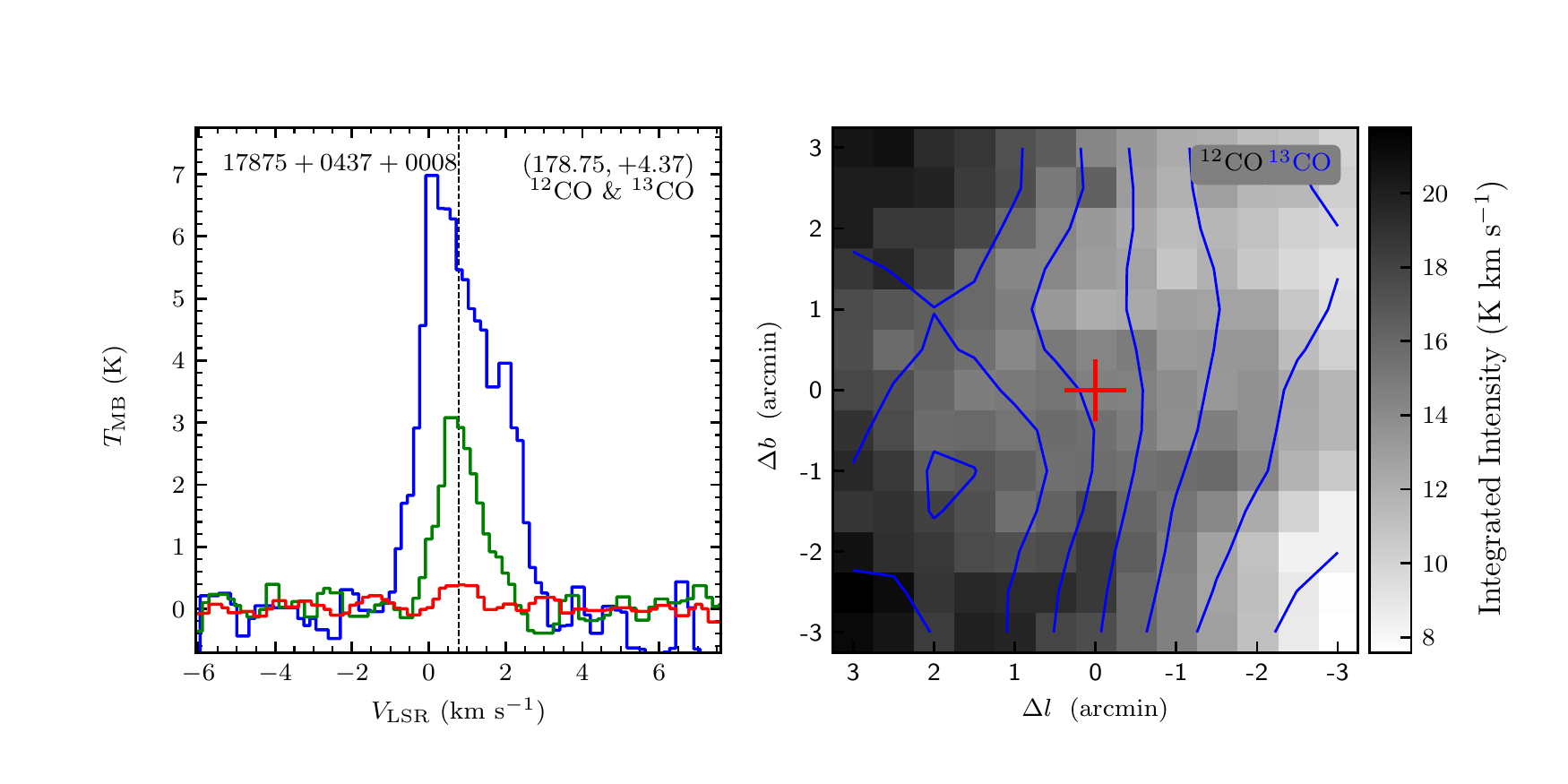}
\end{figure}
\clearpage

\begin{figure}
\includegraphics[width=9.0cm,angle=0]{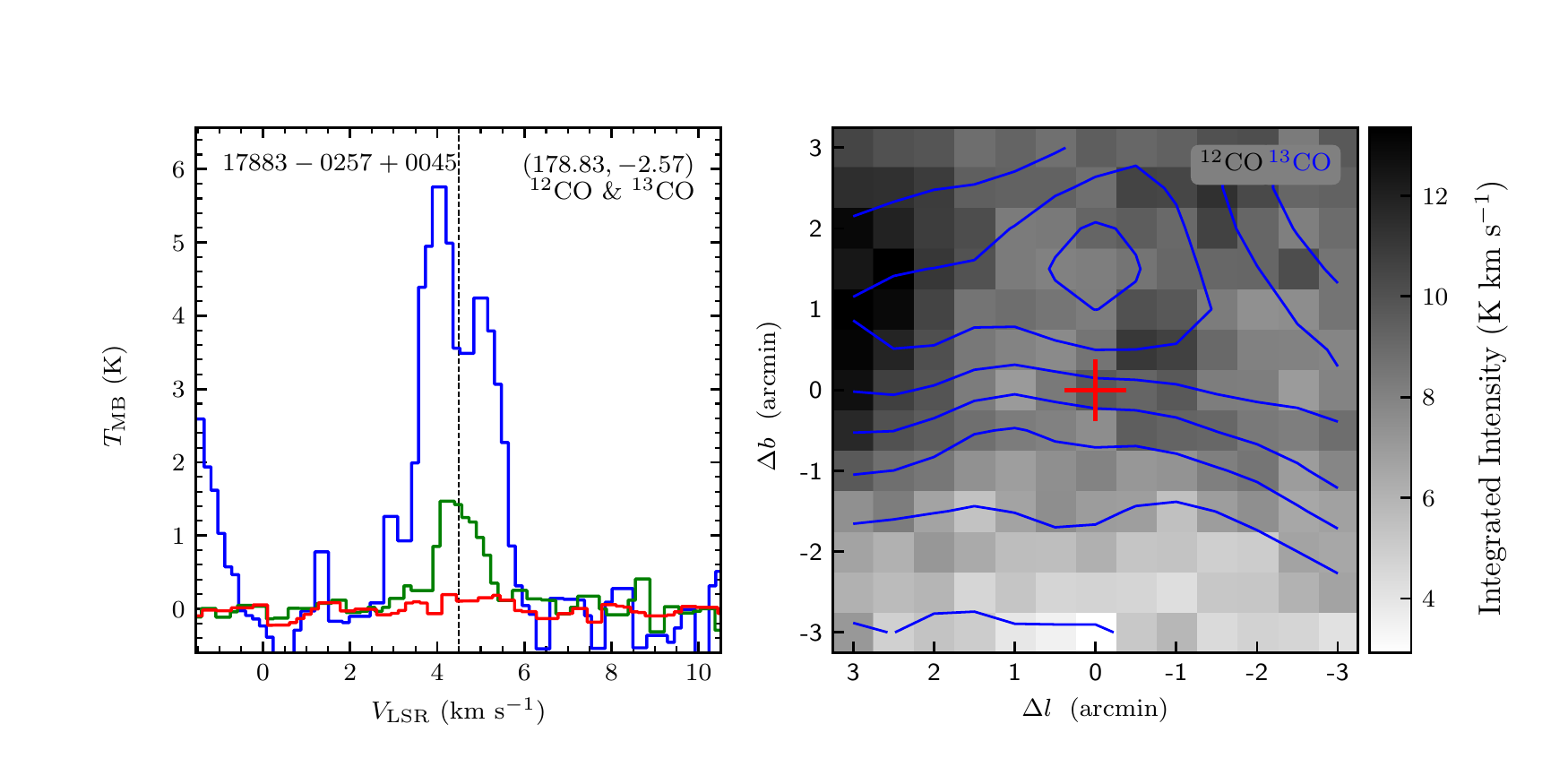}
\includegraphics[width=9.0cm,angle=0]{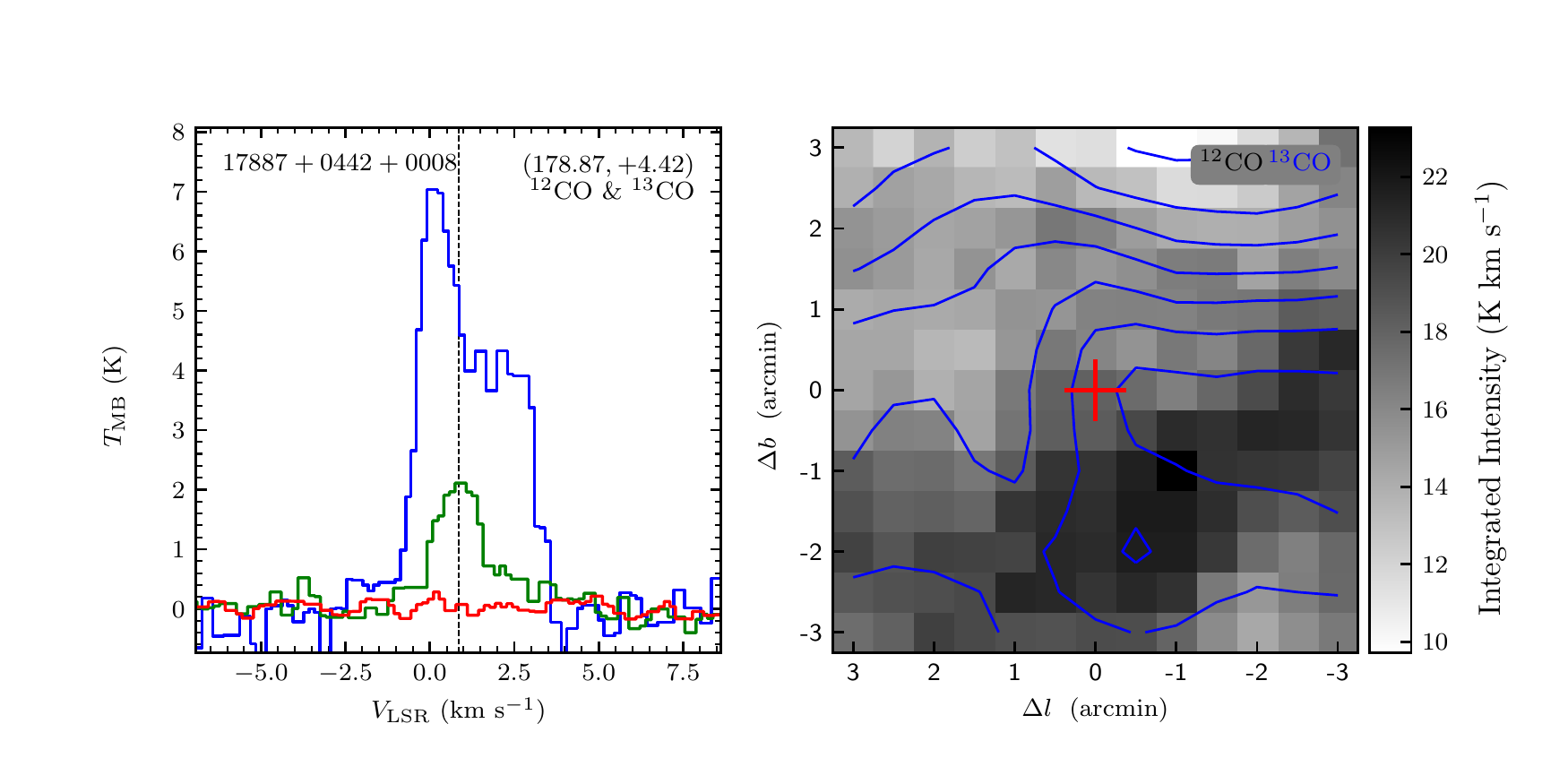}
\vspace{-0.5cm}

\includegraphics[width=9.0cm,angle=0]{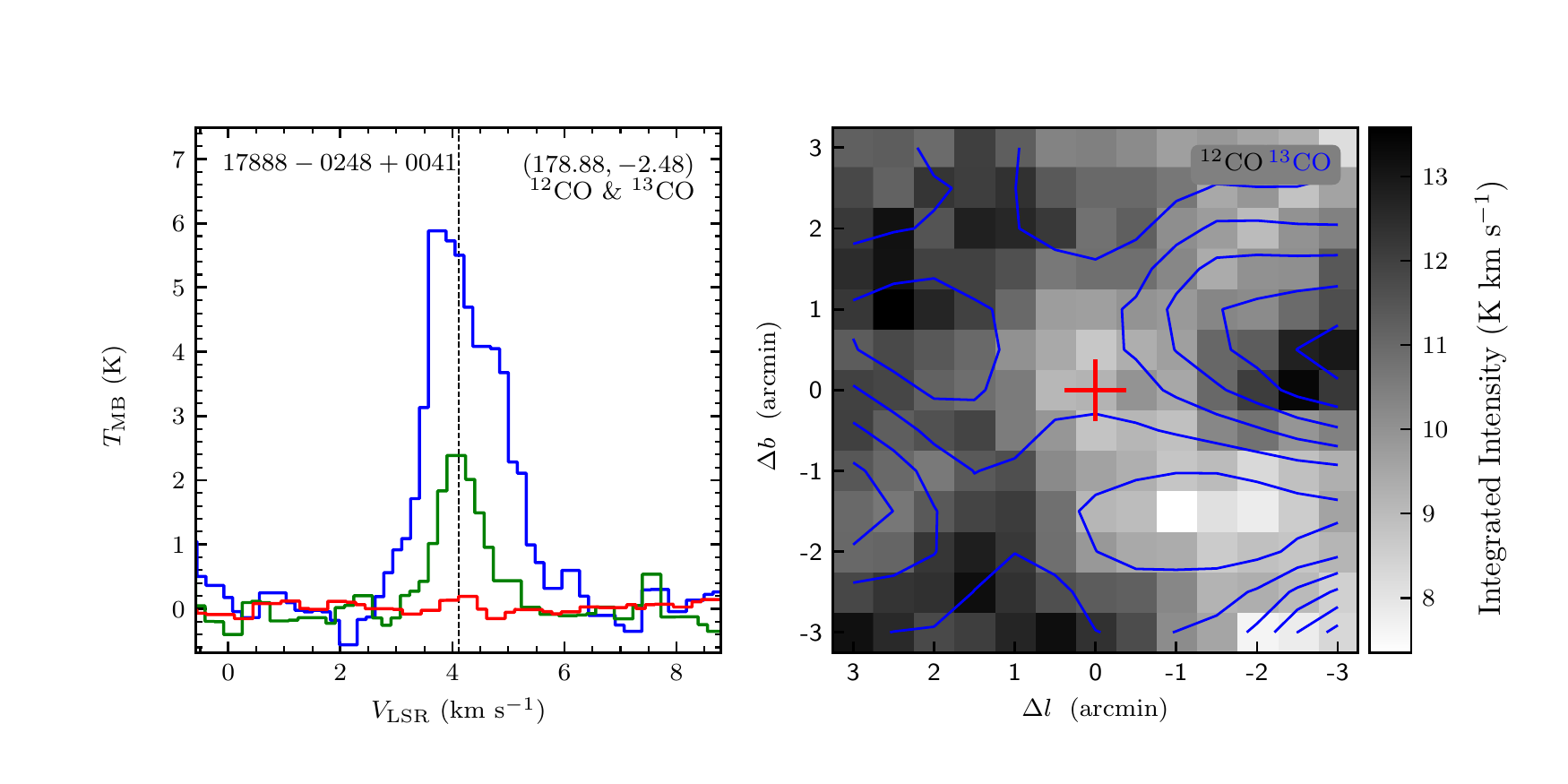}
\includegraphics[width=9.0cm,angle=0]{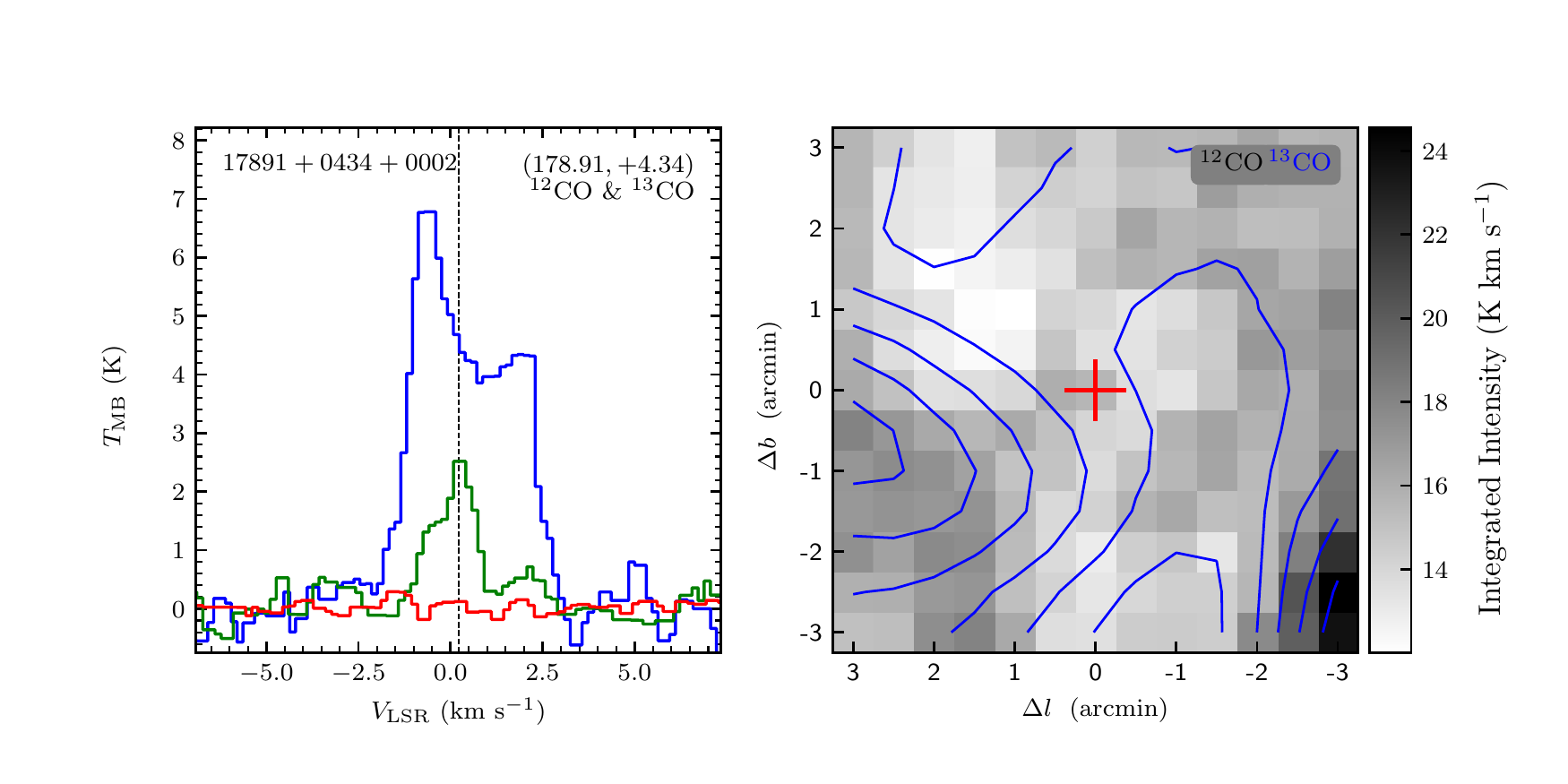}
\vspace{-0.5cm}

\includegraphics[width=9.0cm,angle=0]{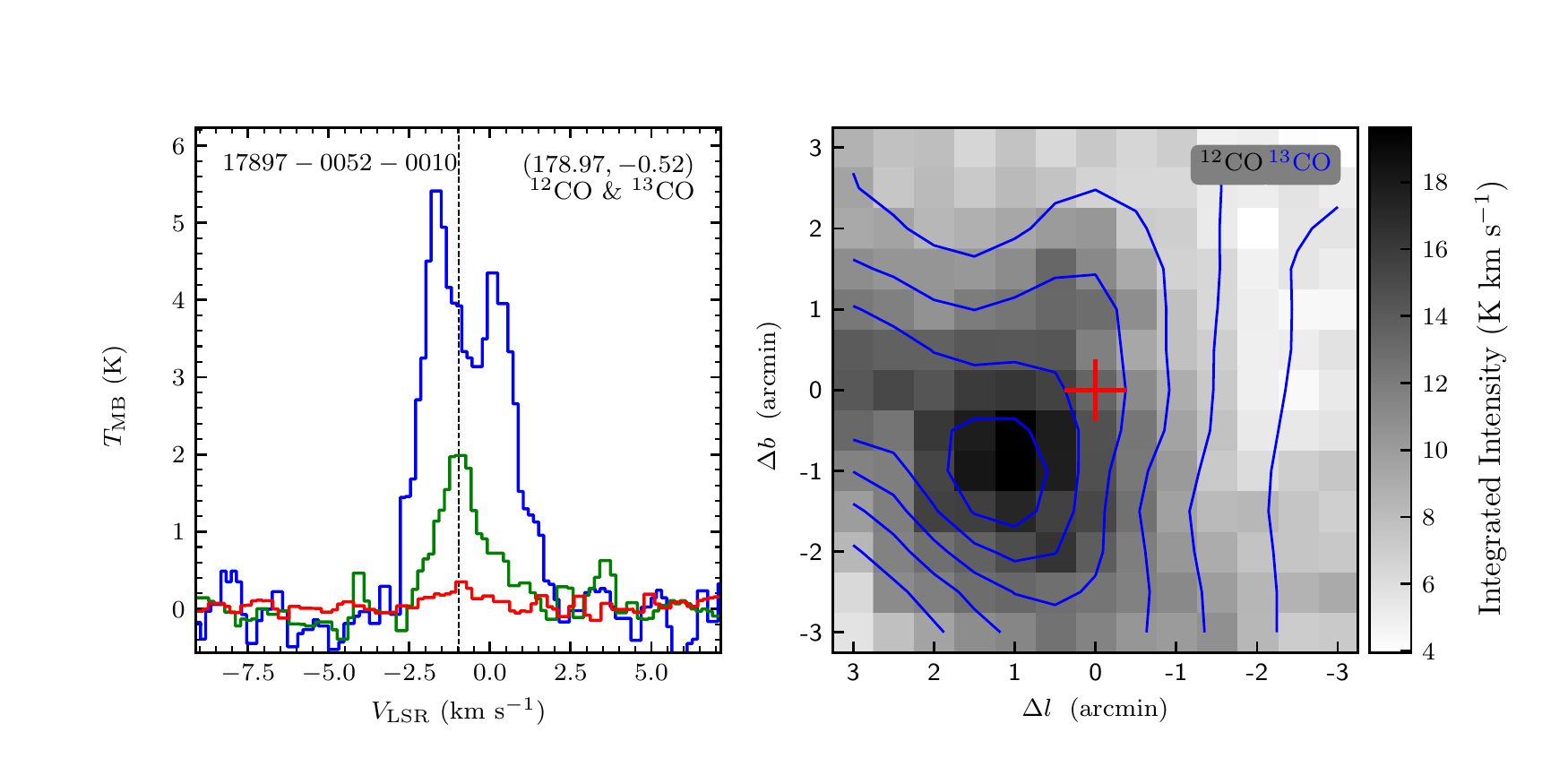}
\includegraphics[width=9.0cm,angle=0]{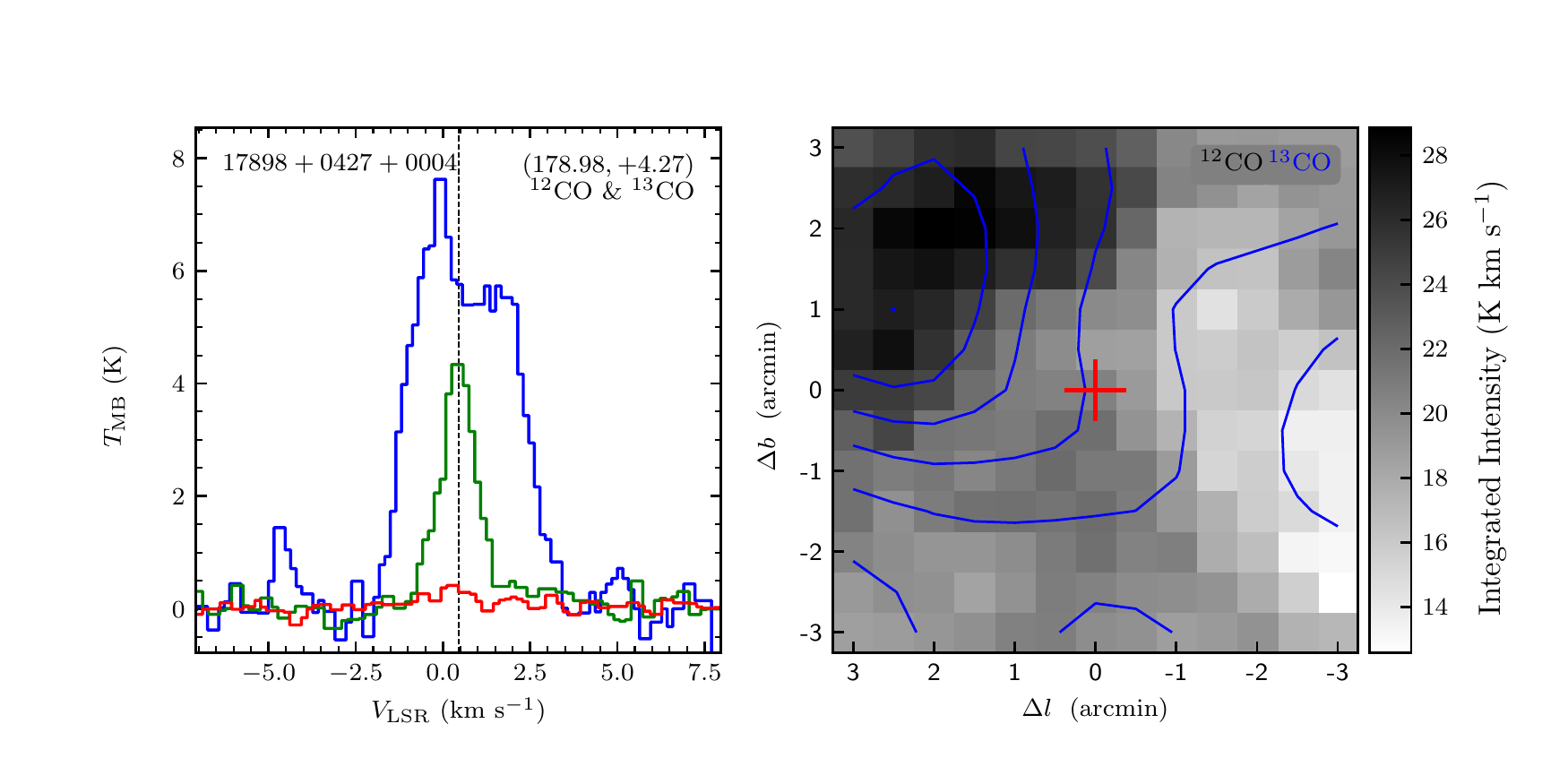}
\vspace{-0.5cm}

\includegraphics[width=9.0cm,angle=0]{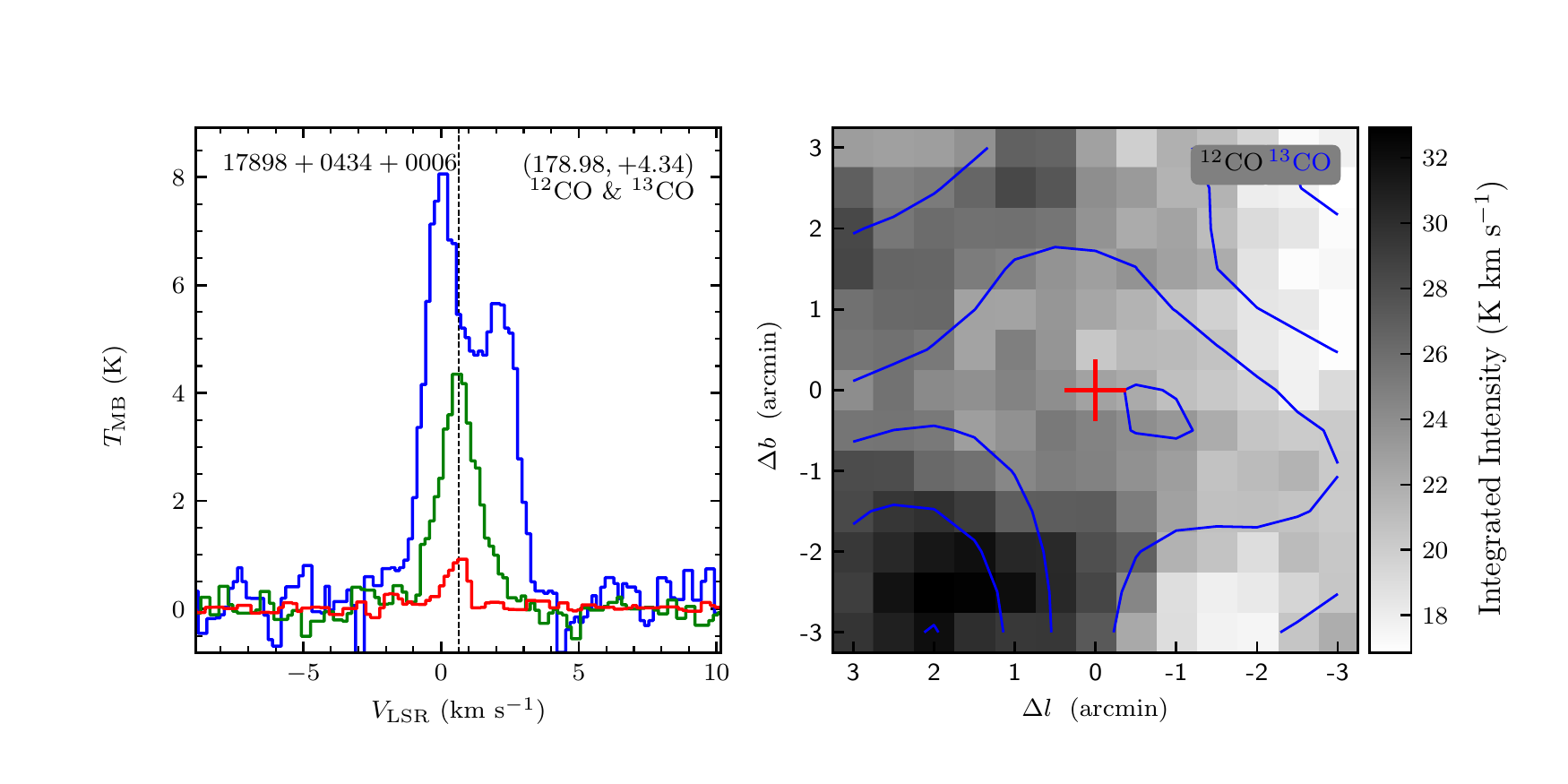}
\includegraphics[width=9.0cm,angle=0]{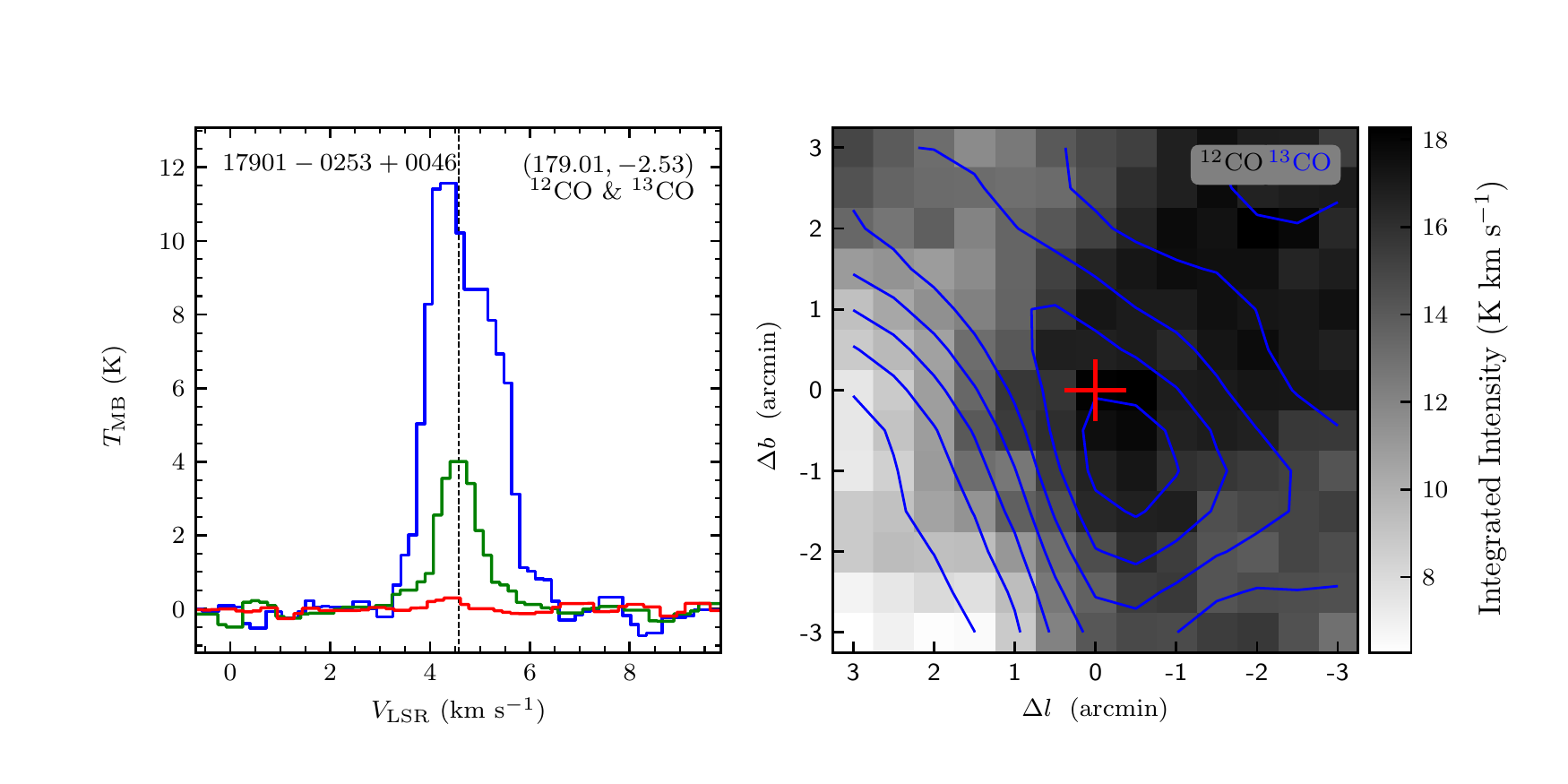}
\vspace{-0.5cm}

\includegraphics[width=9.0cm,angle=0]{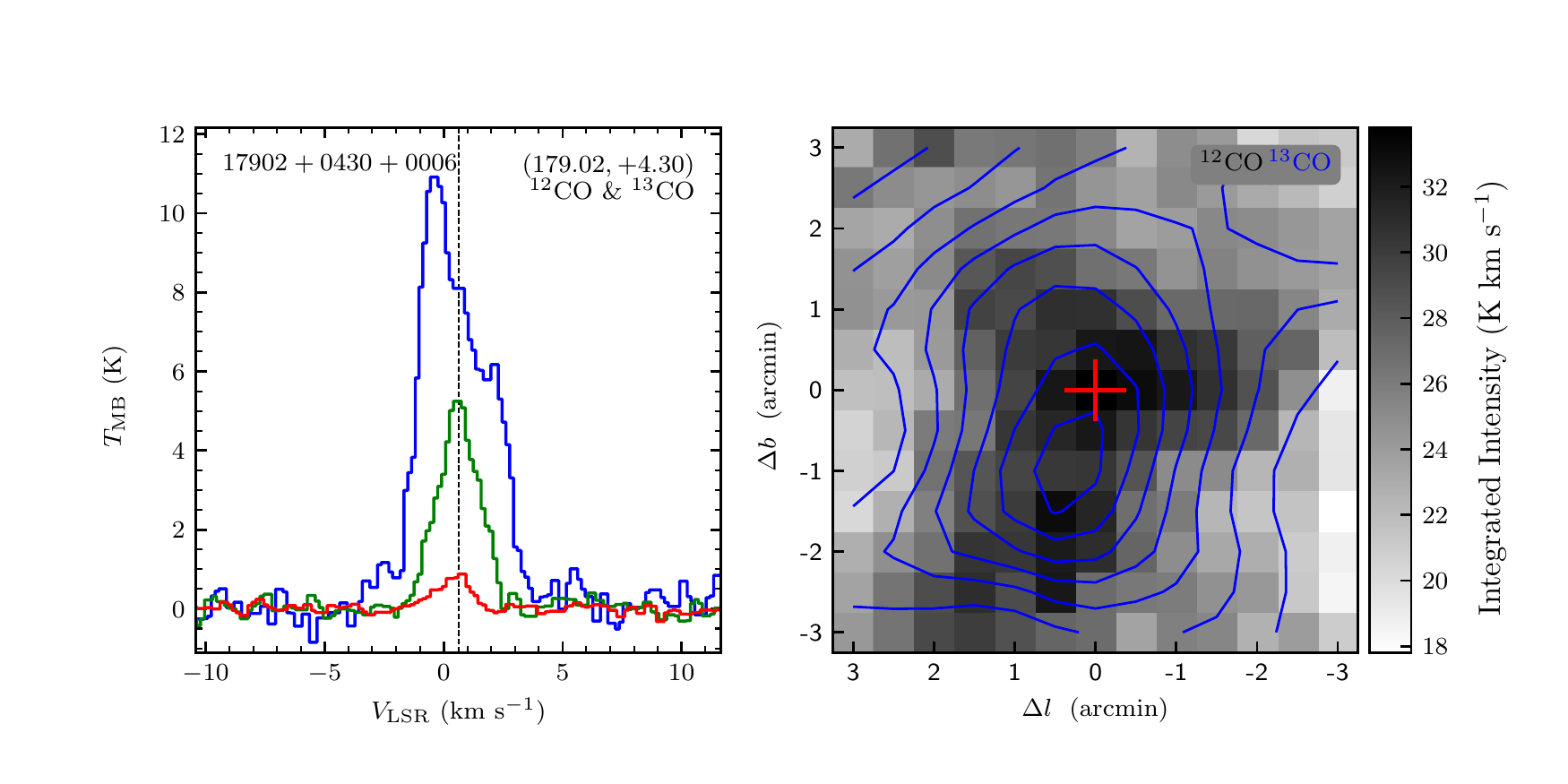}
\includegraphics[width=9.0cm,angle=0]{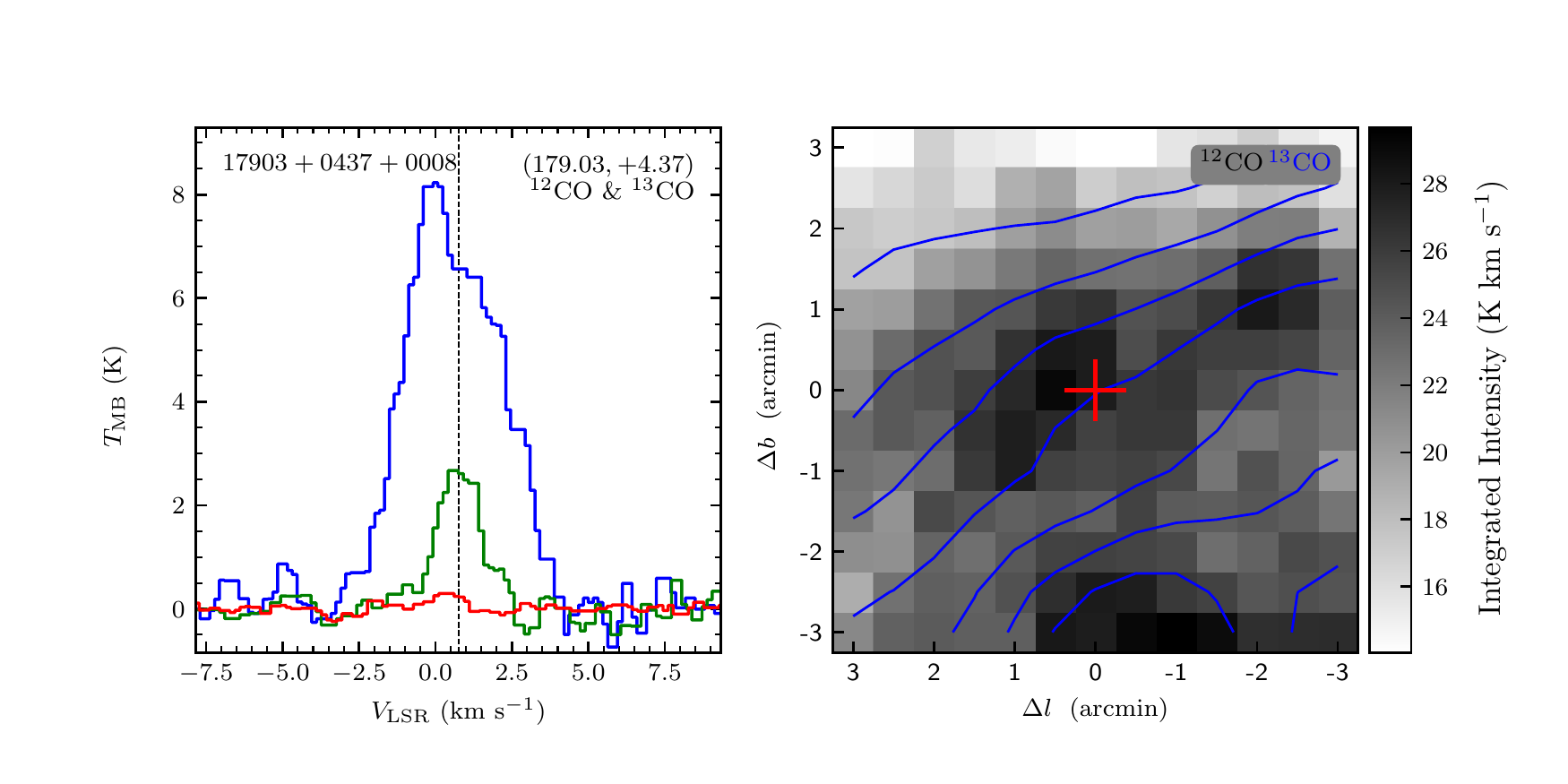}
\end{figure}
\clearpage

\begin{figure}
\includegraphics[width=9.0cm,angle=0]{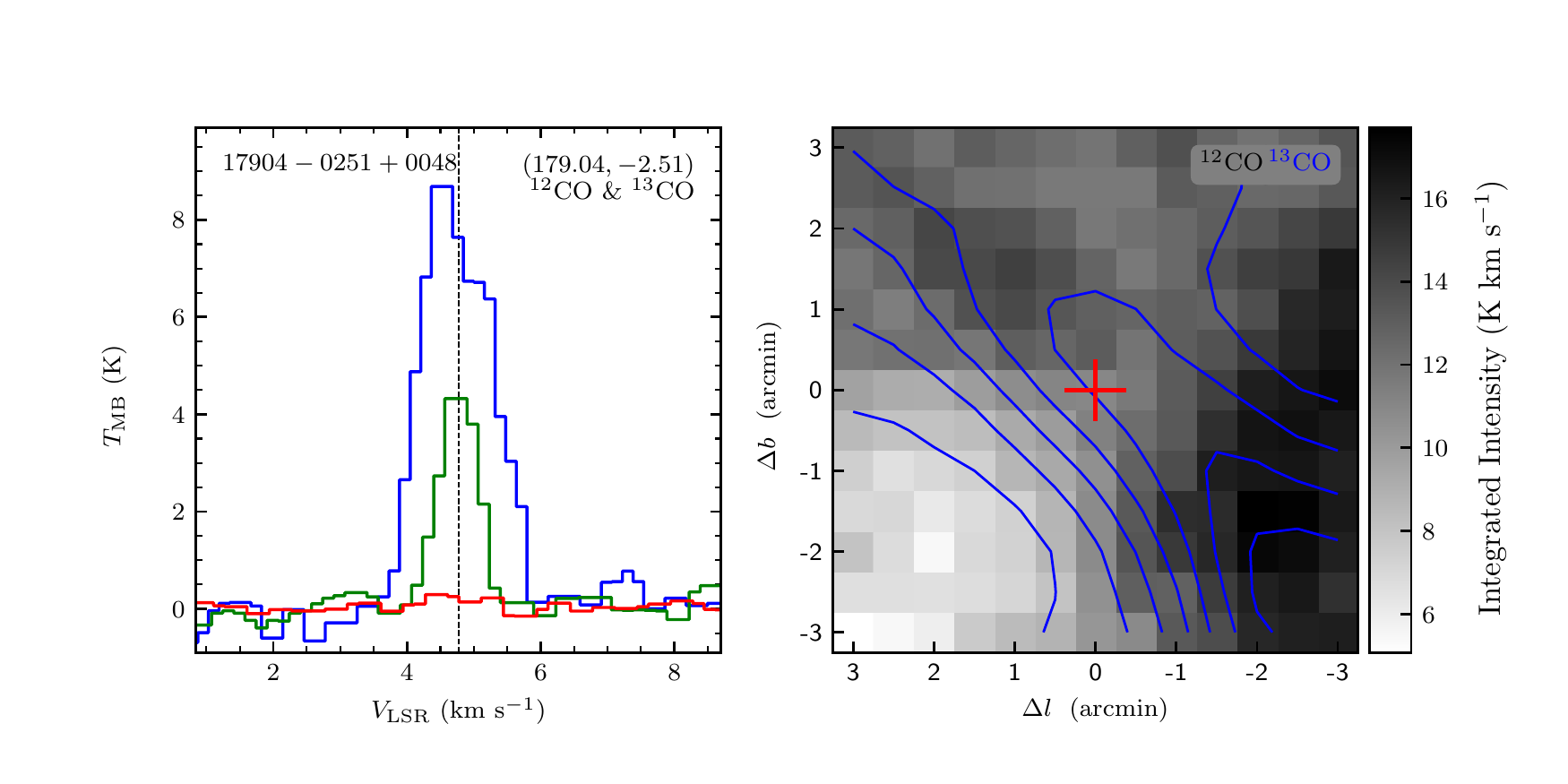}
\includegraphics[width=9.0cm,angle=0]{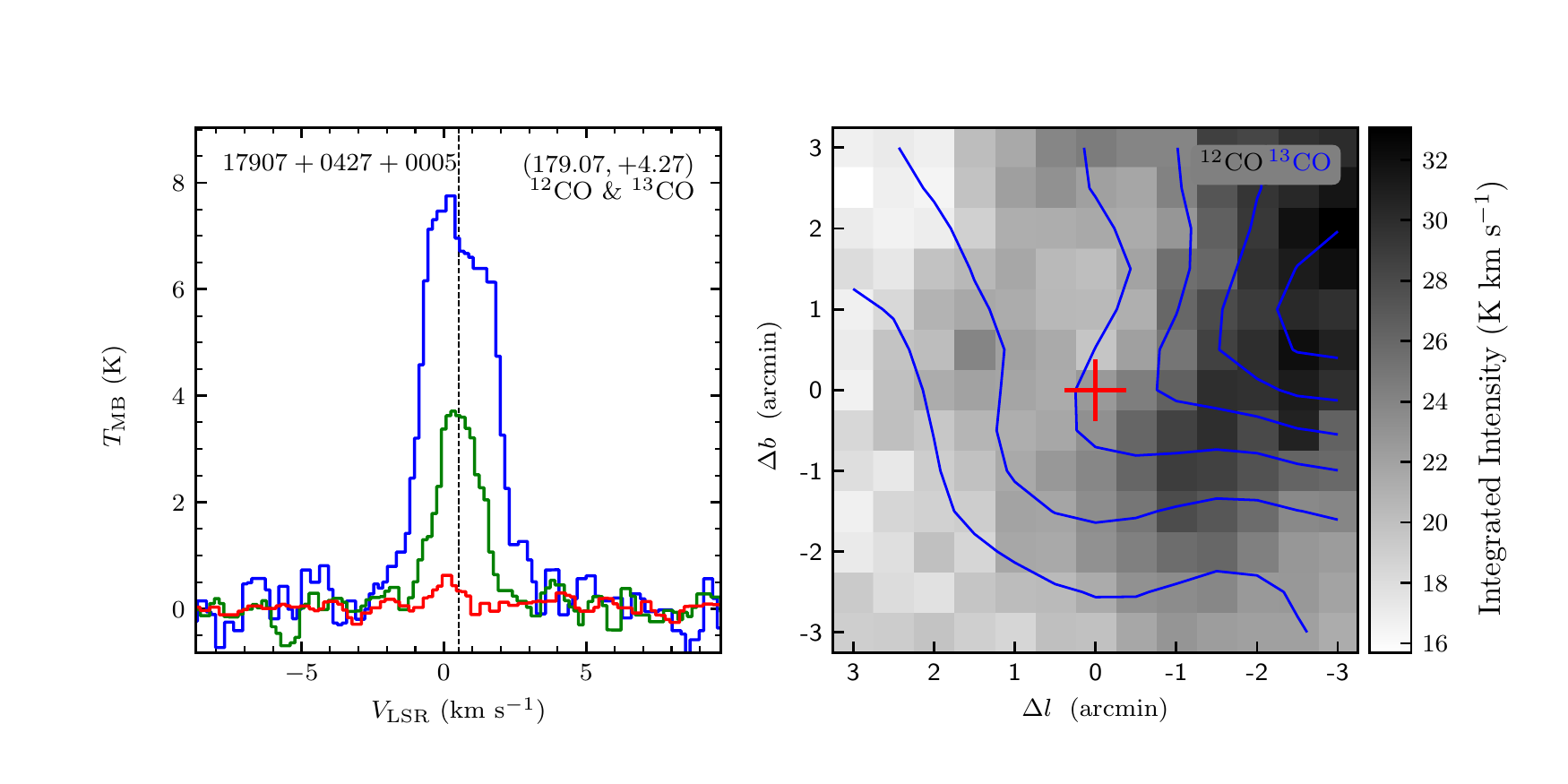}
\vspace{-0.5cm}

\includegraphics[width=9.0cm,angle=0]{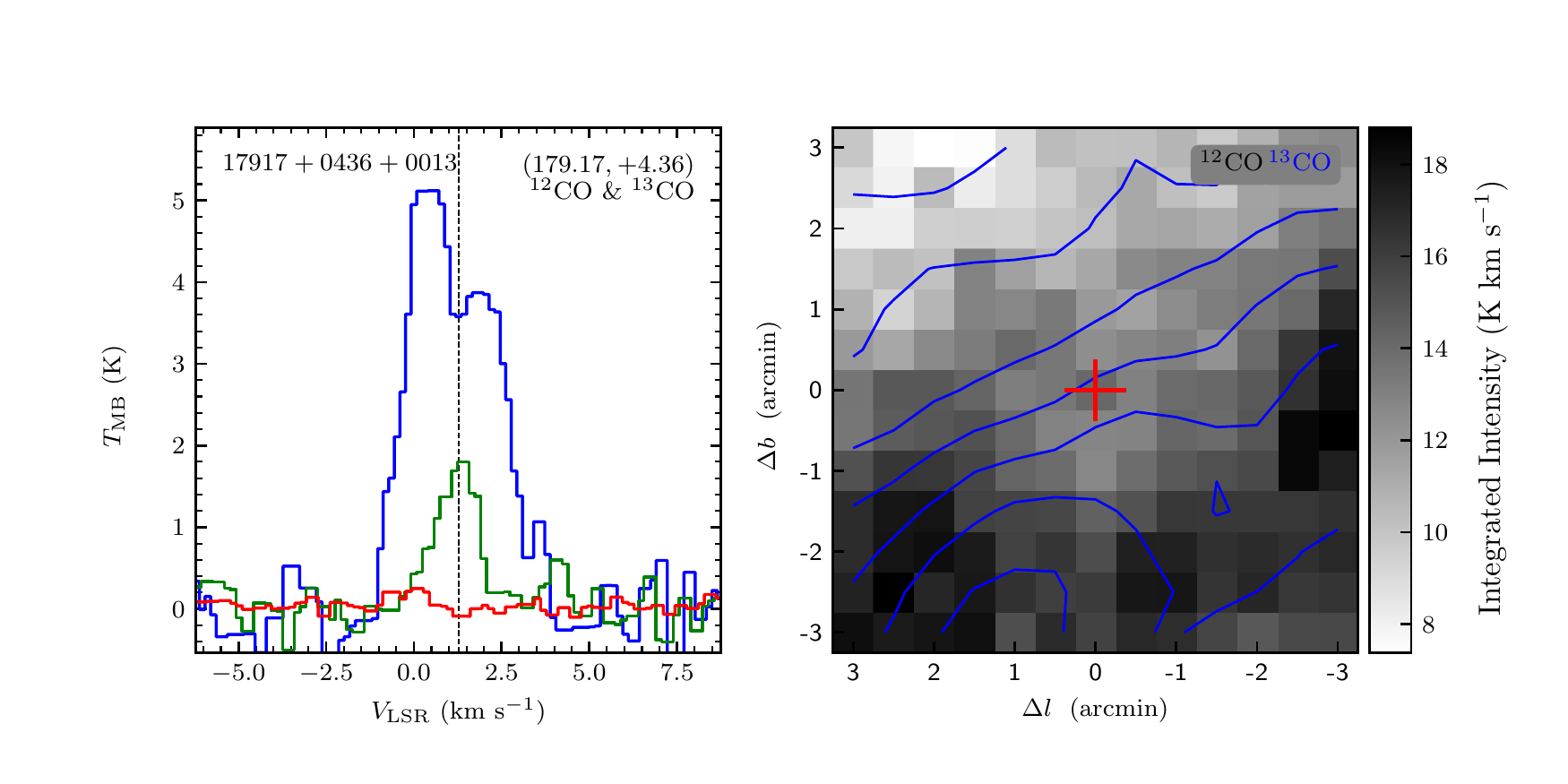}
\includegraphics[width=9.0cm,angle=0]{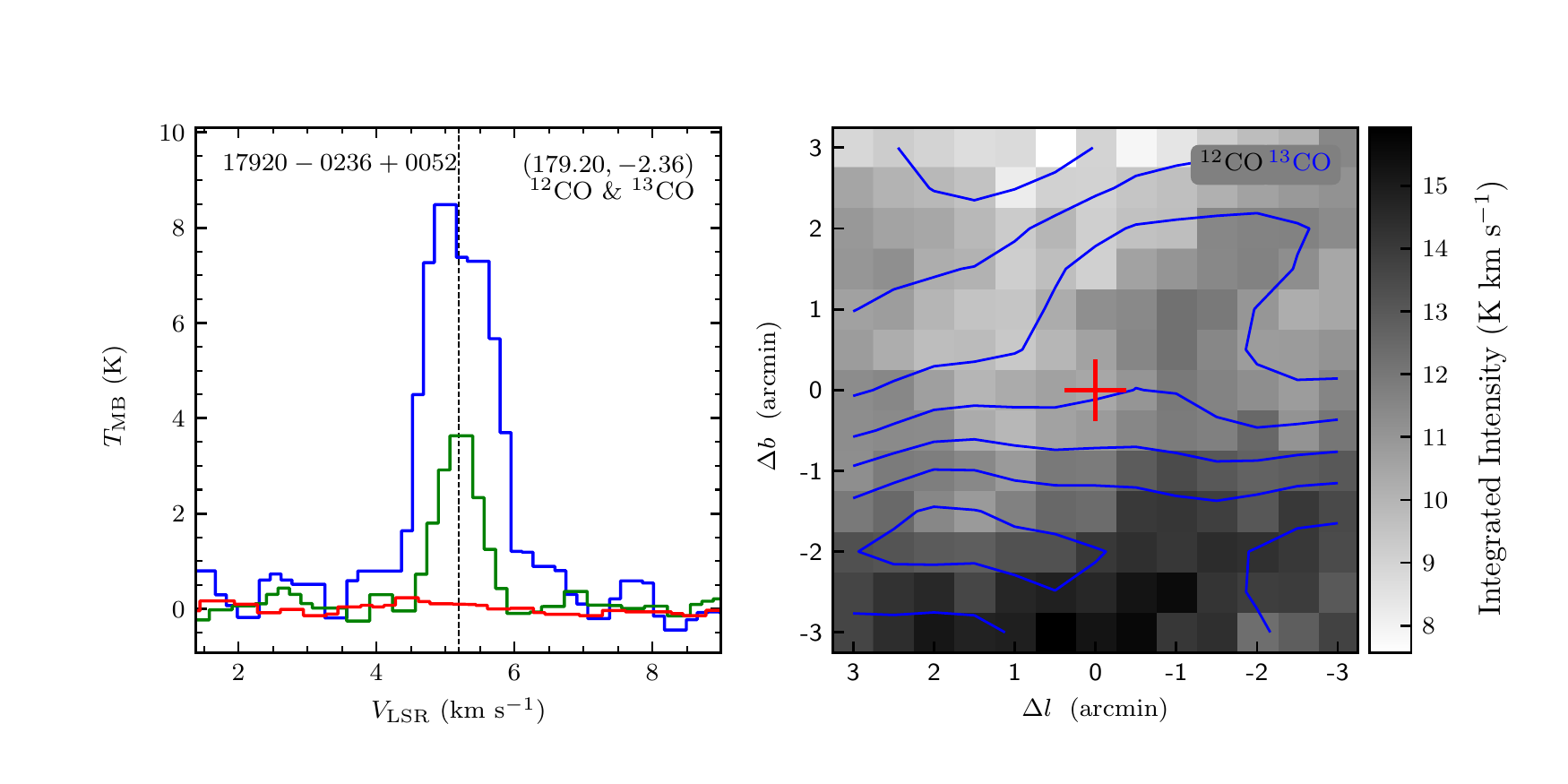}
\vspace{-0.5cm}

\includegraphics[width=9.0cm,angle=0]{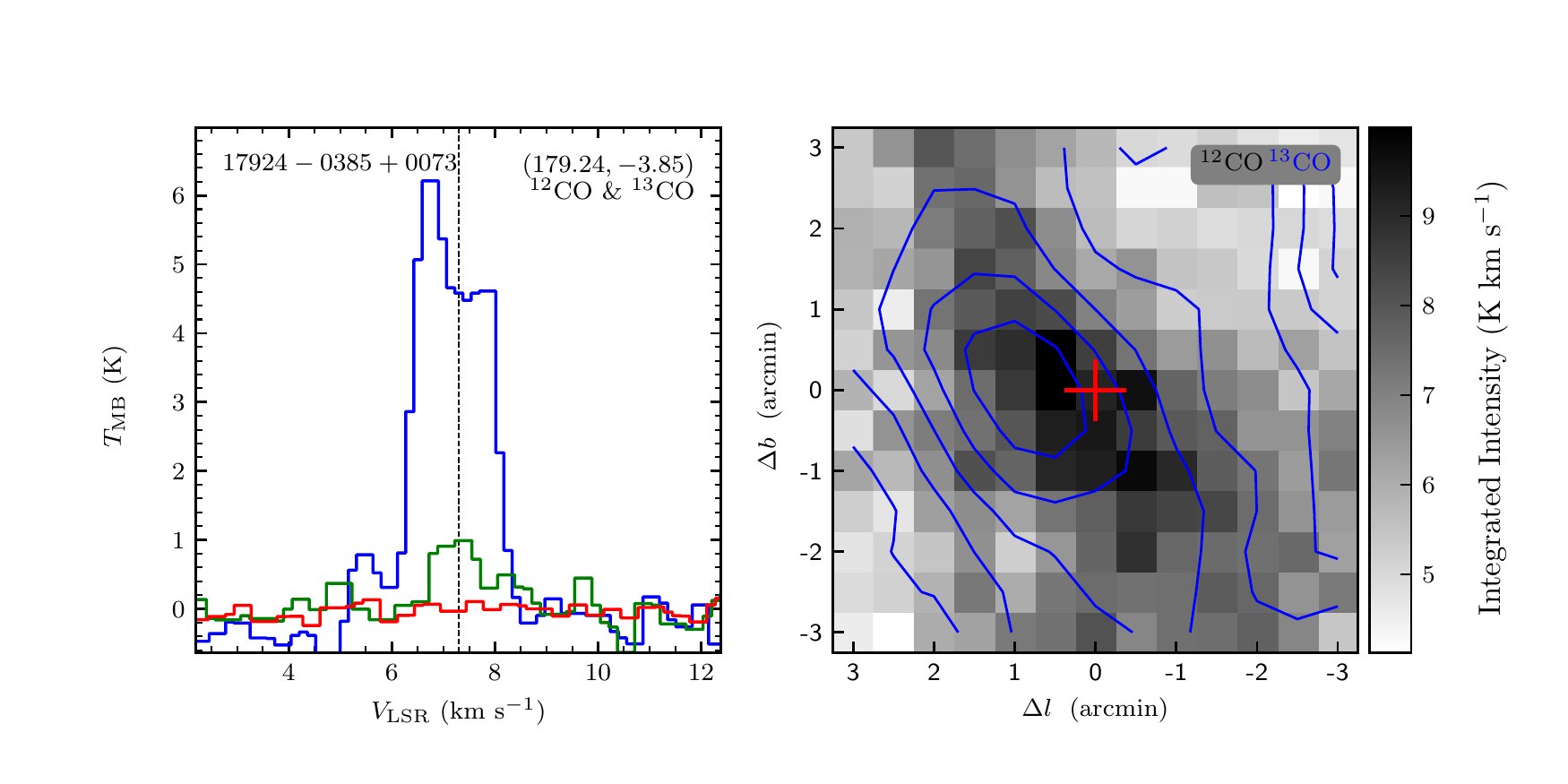}
\includegraphics[width=9.0cm,angle=0]{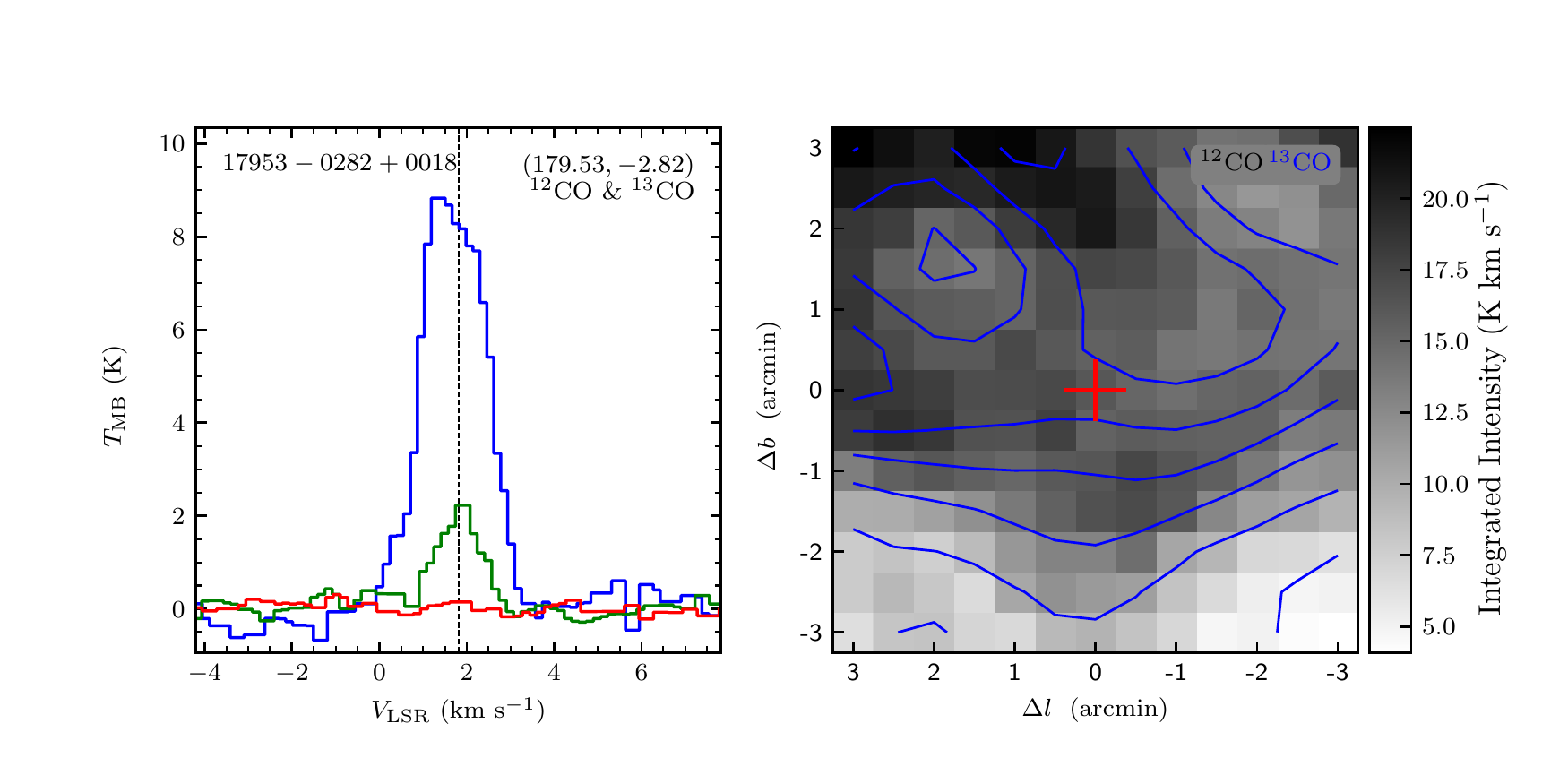}
\vspace{-0.5cm}

\includegraphics[width=9.0cm,angle=0]{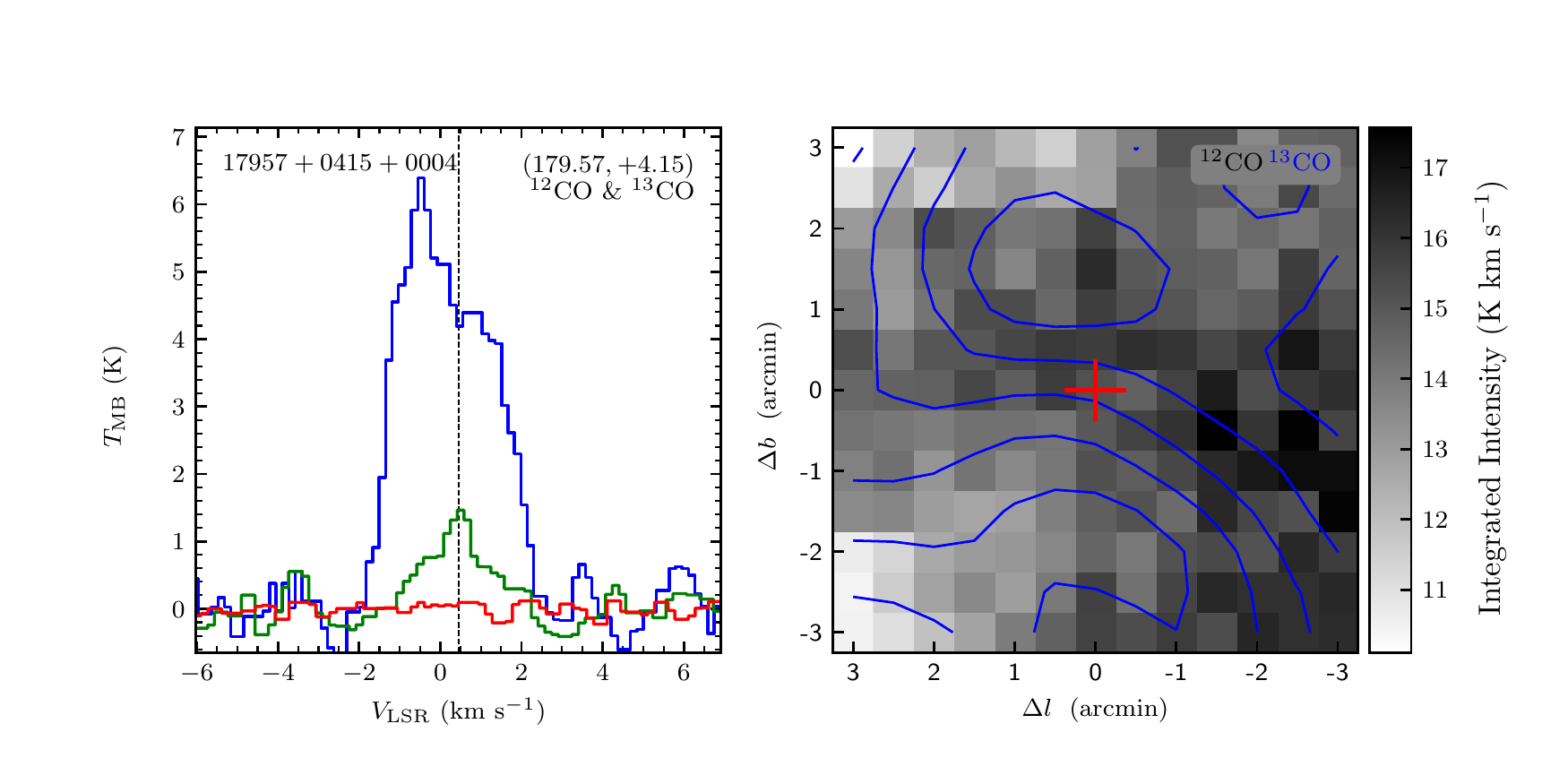}
\includegraphics[width=9.0cm,angle=0]{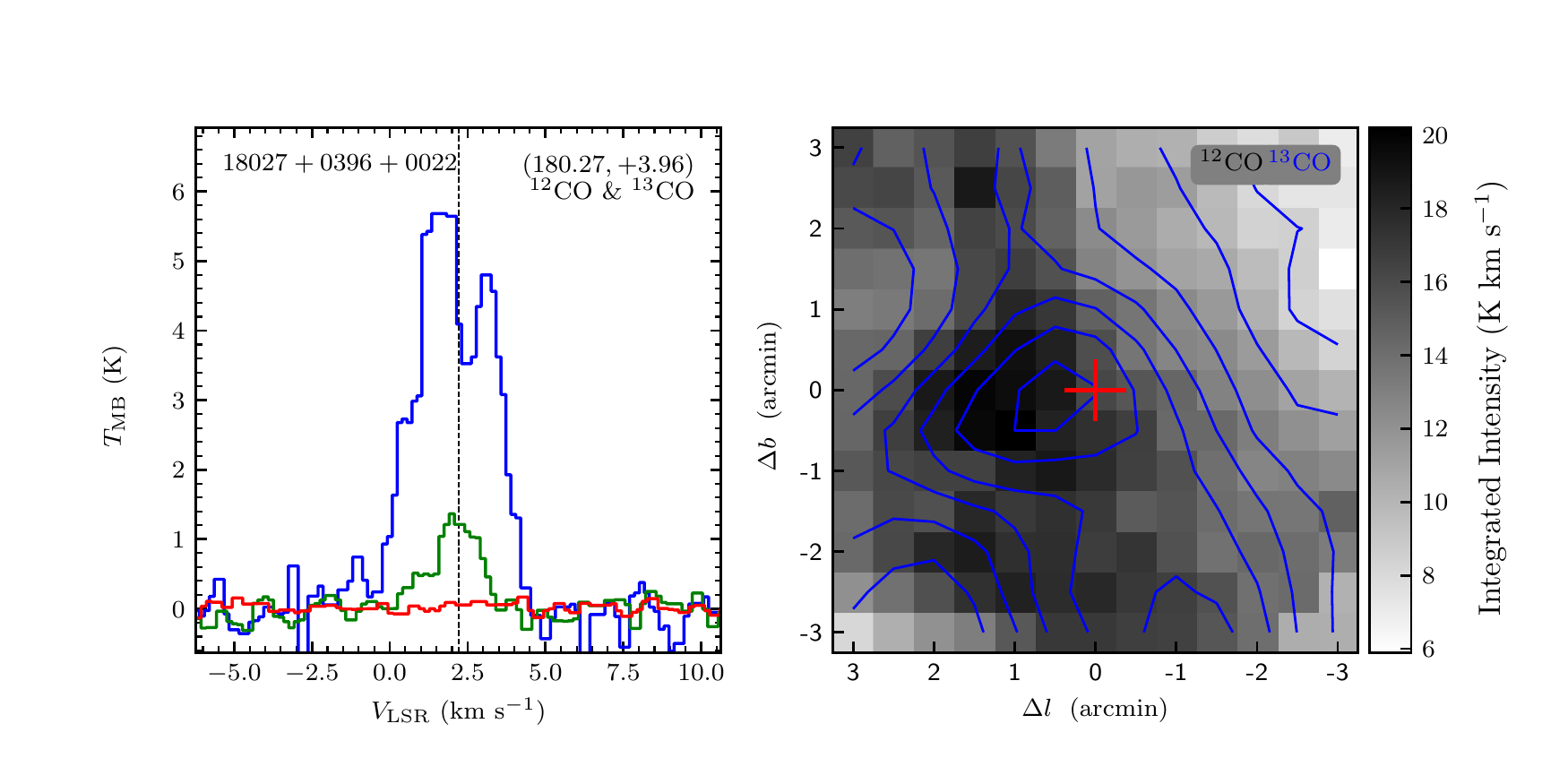}
\vspace{-0.5cm}

\includegraphics[width=9.0cm,angle=0]{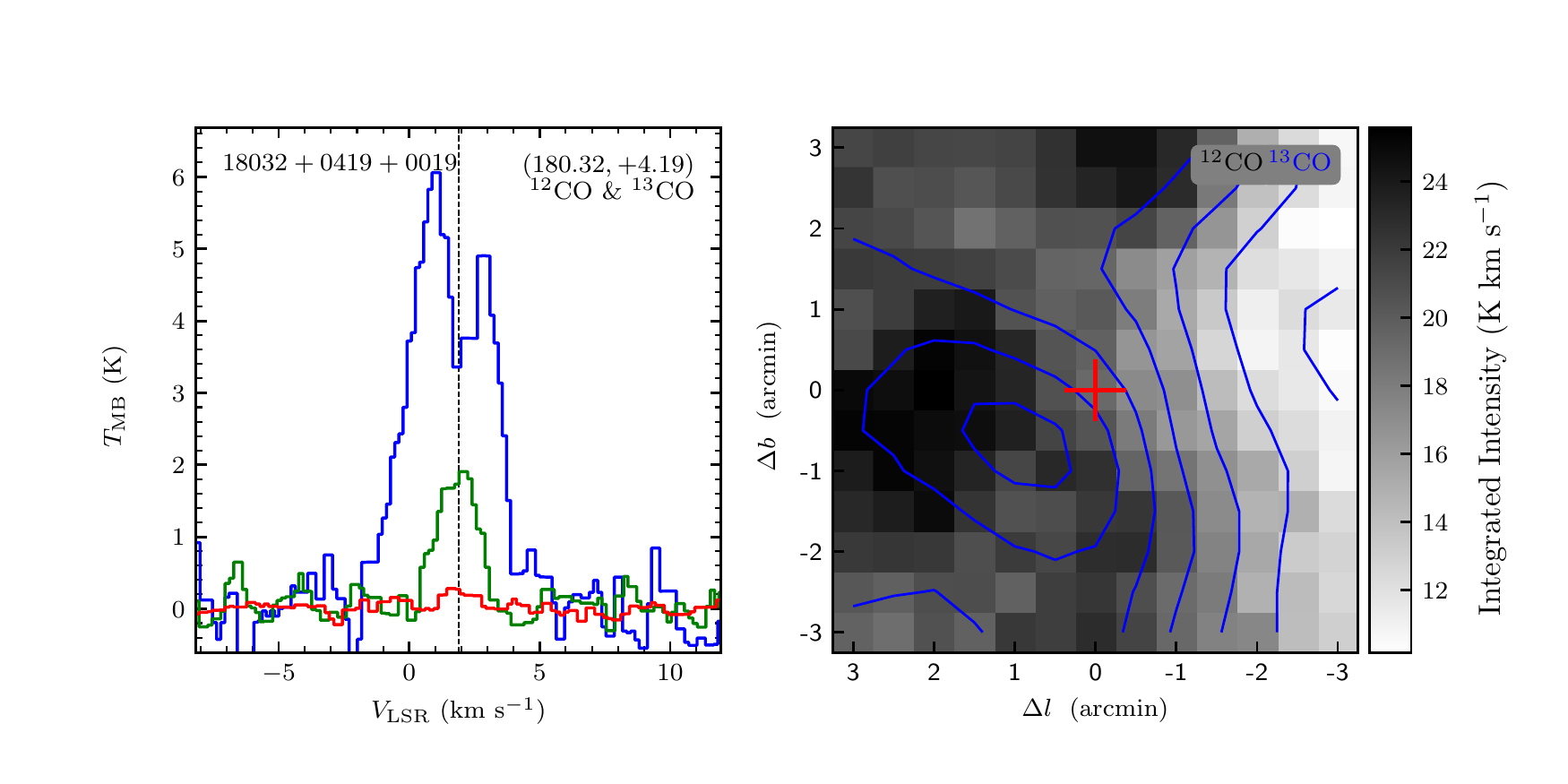}
\includegraphics[width=9.0cm,angle=0]{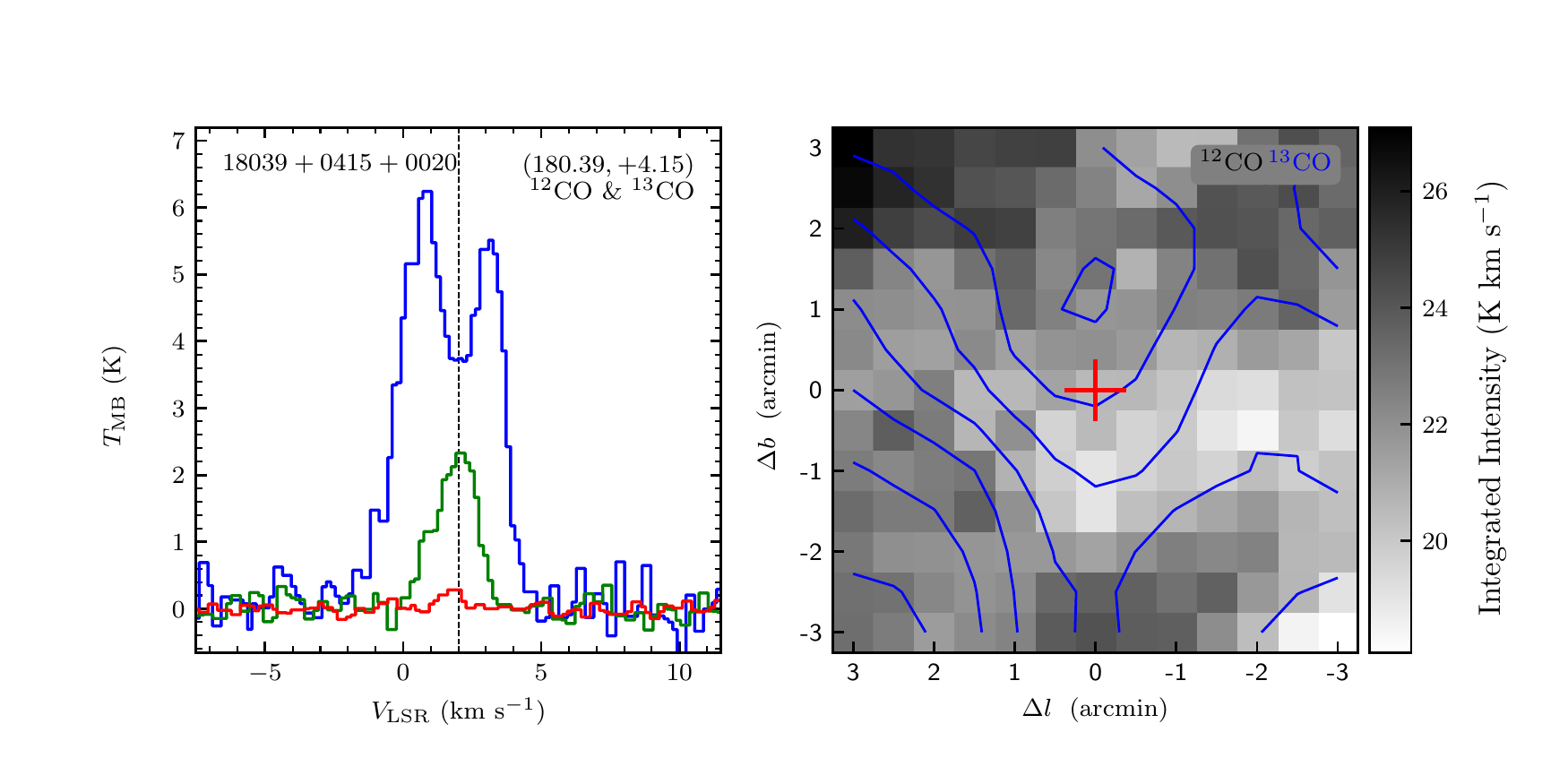}
\end{figure}
\clearpage

\begin{figure}
\includegraphics[width=9.0cm,angle=0]{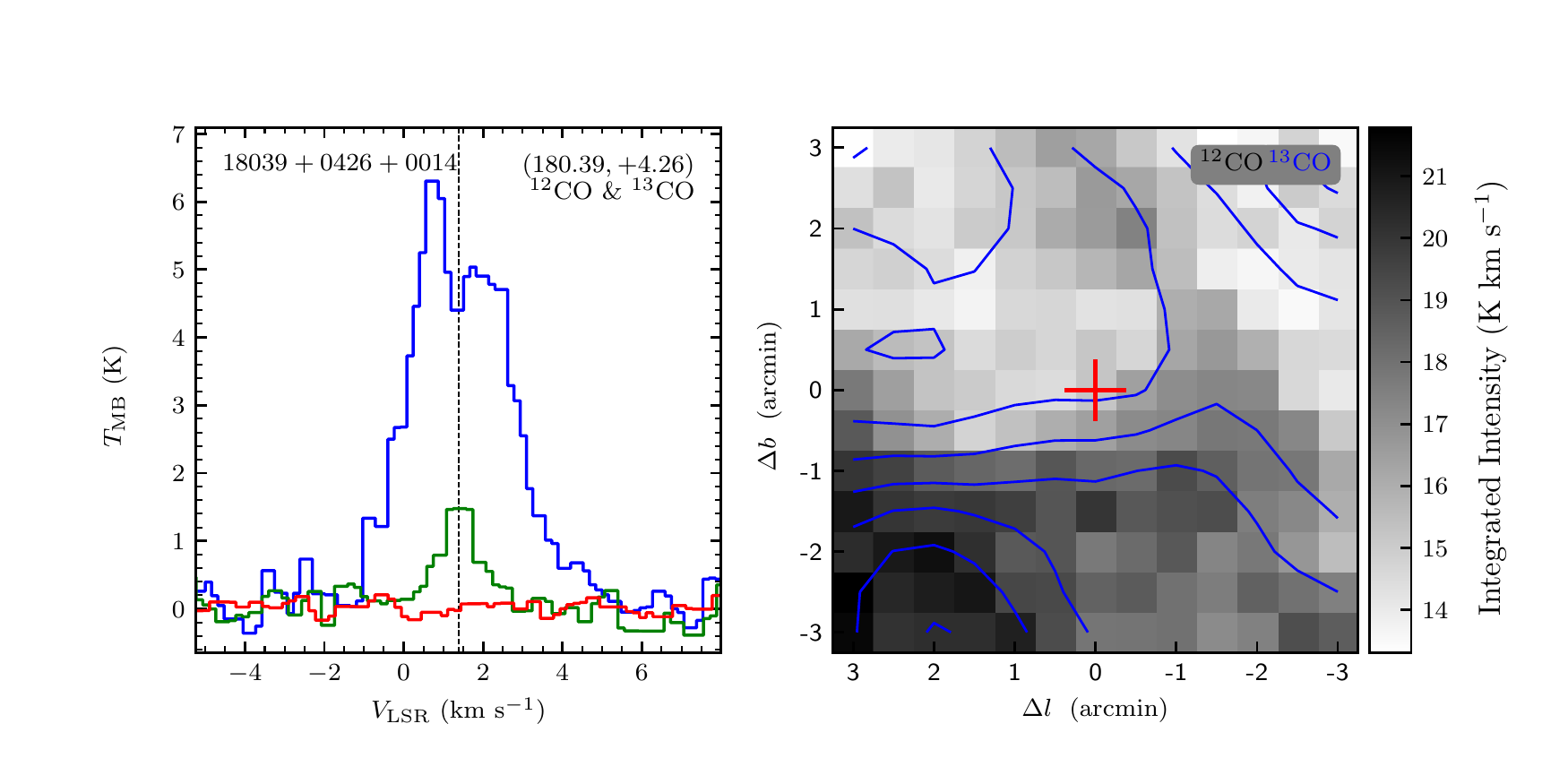}
\includegraphics[width=9.0cm,angle=0]{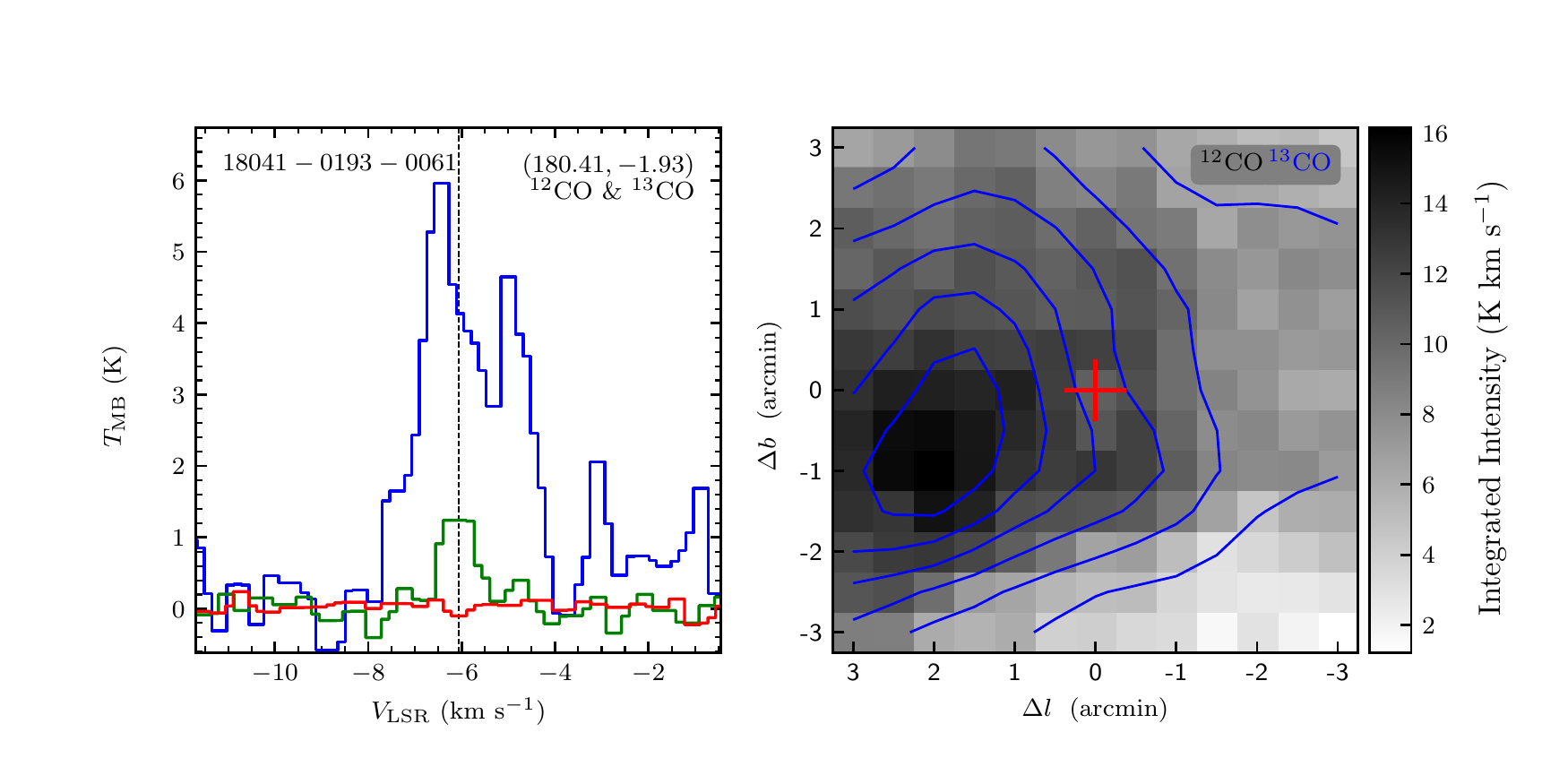}
\vspace{-0.5cm}

\includegraphics[width=9.0cm,angle=0]{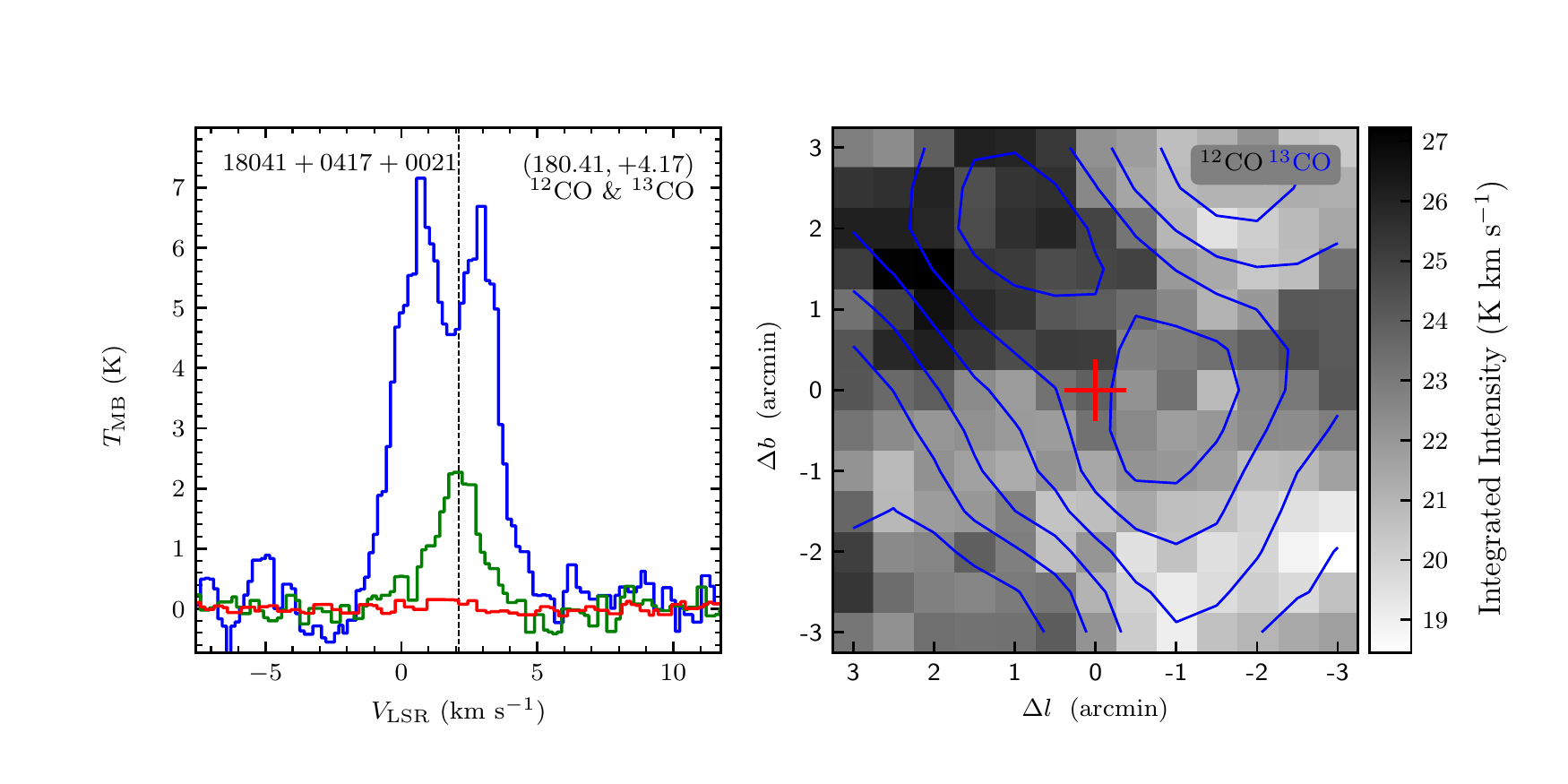}
\includegraphics[width=9.0cm,angle=0]{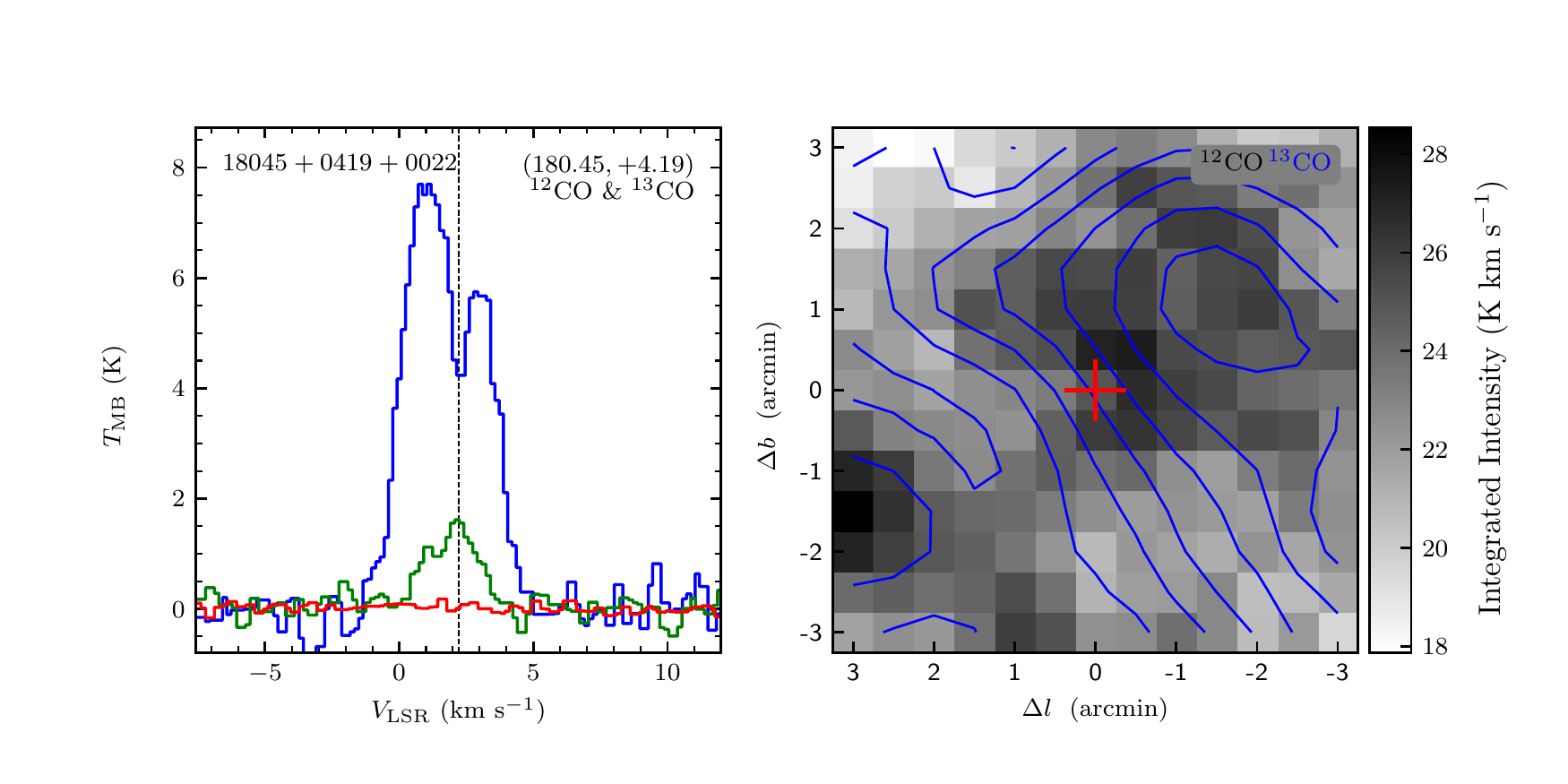}
\vspace{-0.5cm}

\includegraphics[width=9.0cm,angle=0]{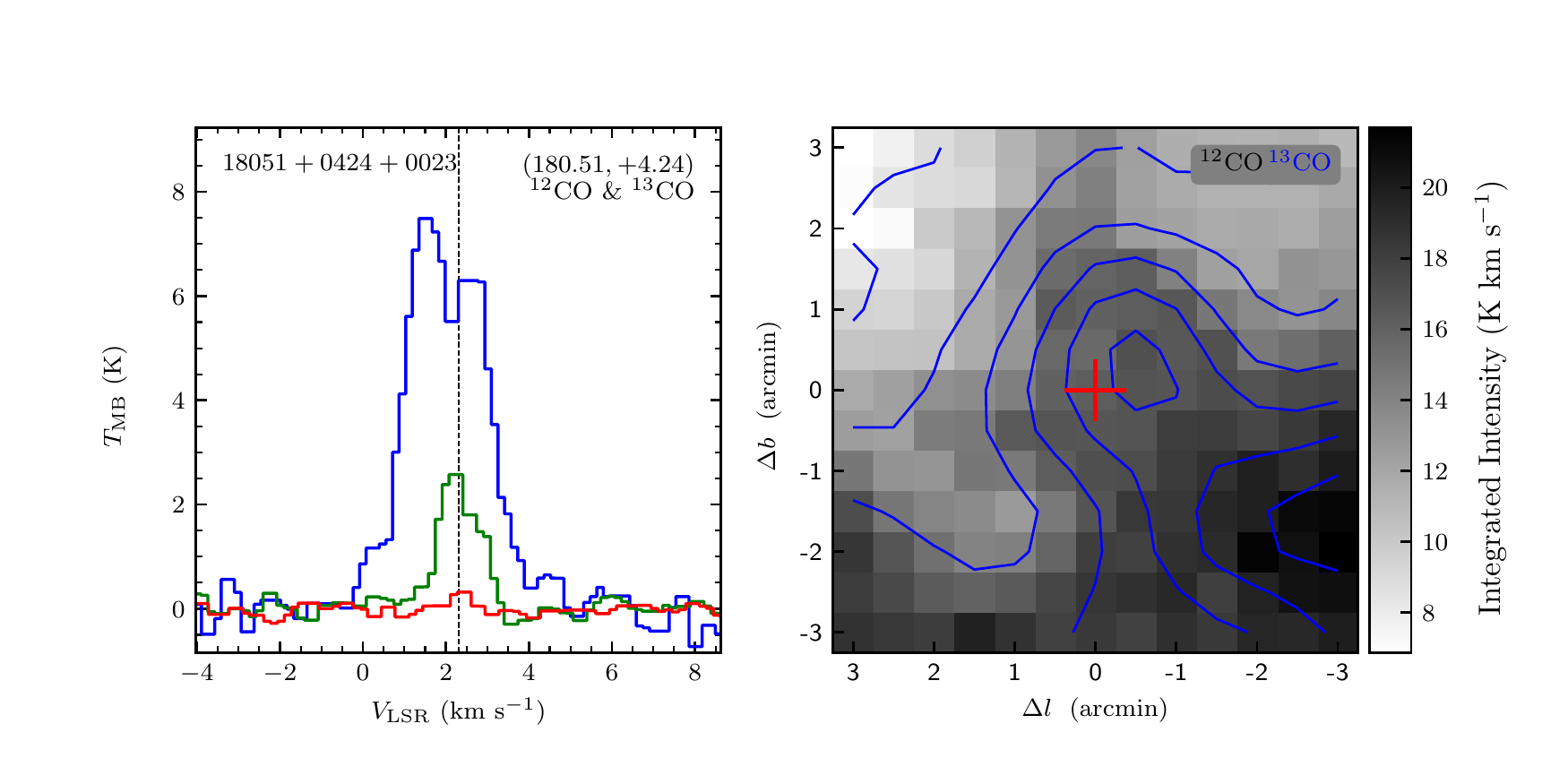}
\includegraphics[width=9.0cm,angle=0]{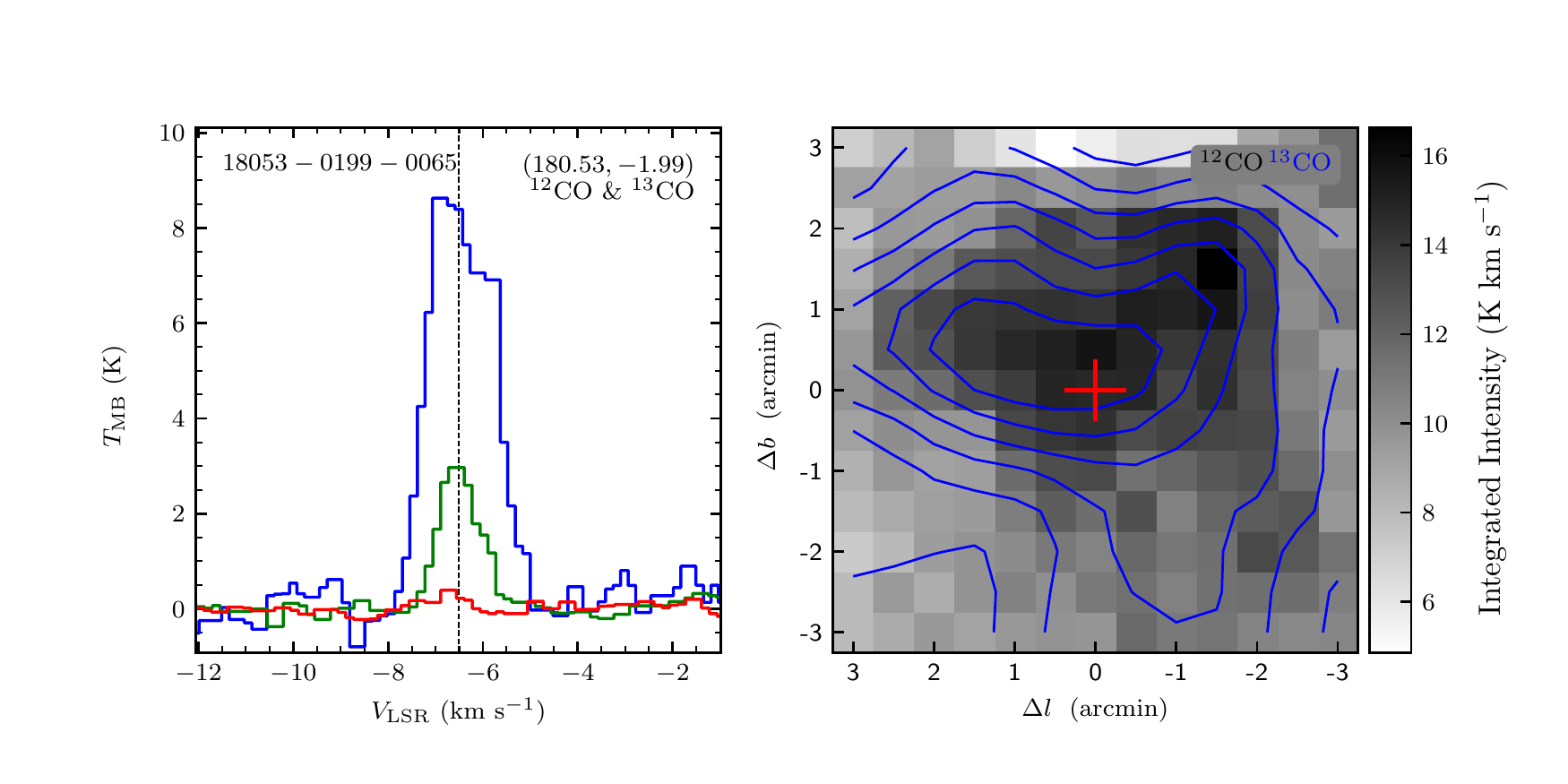}
\vspace{-0.5cm}

\includegraphics[width=9.0cm,angle=0]{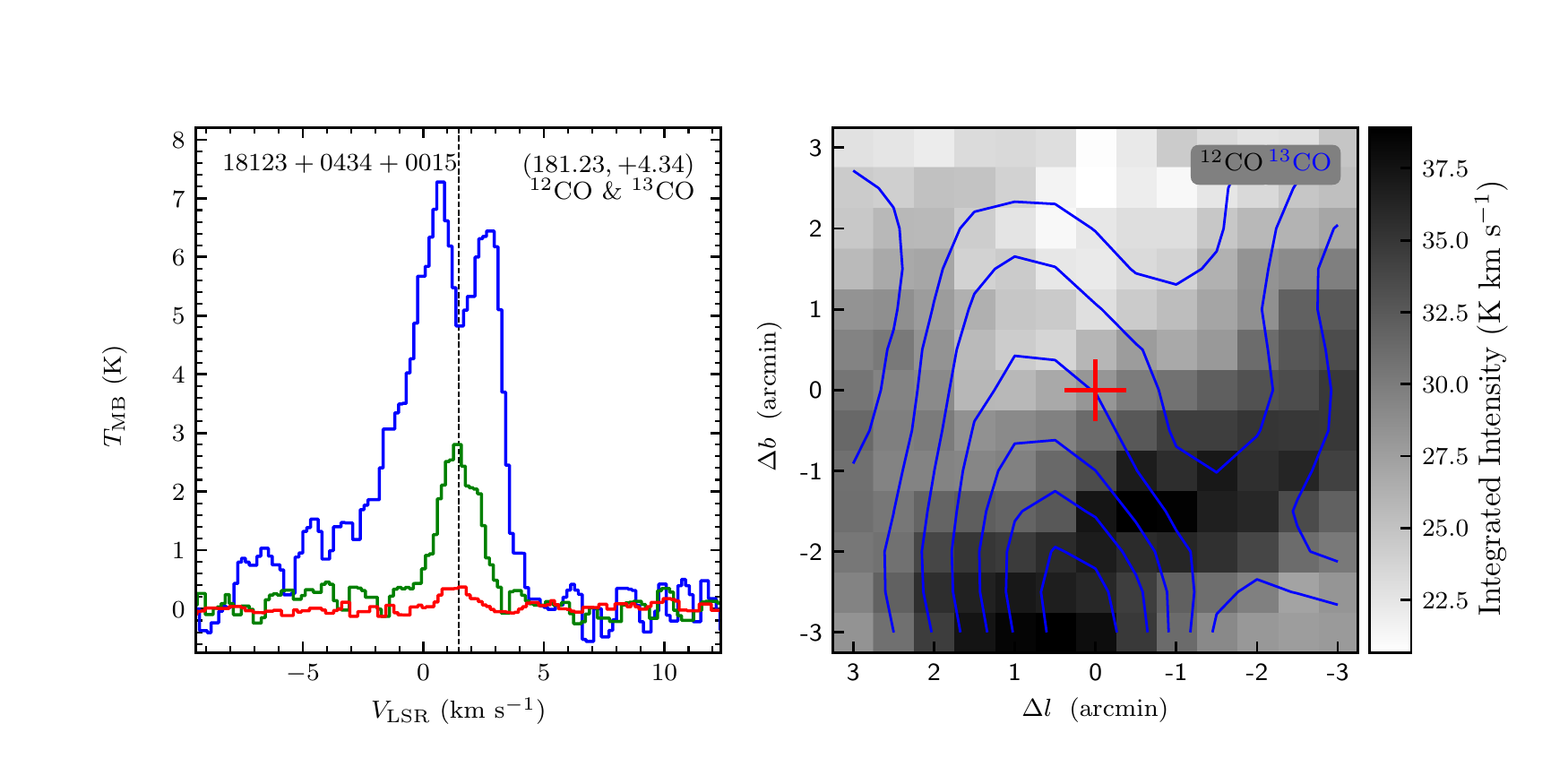}
\includegraphics[width=9.0cm,angle=0]{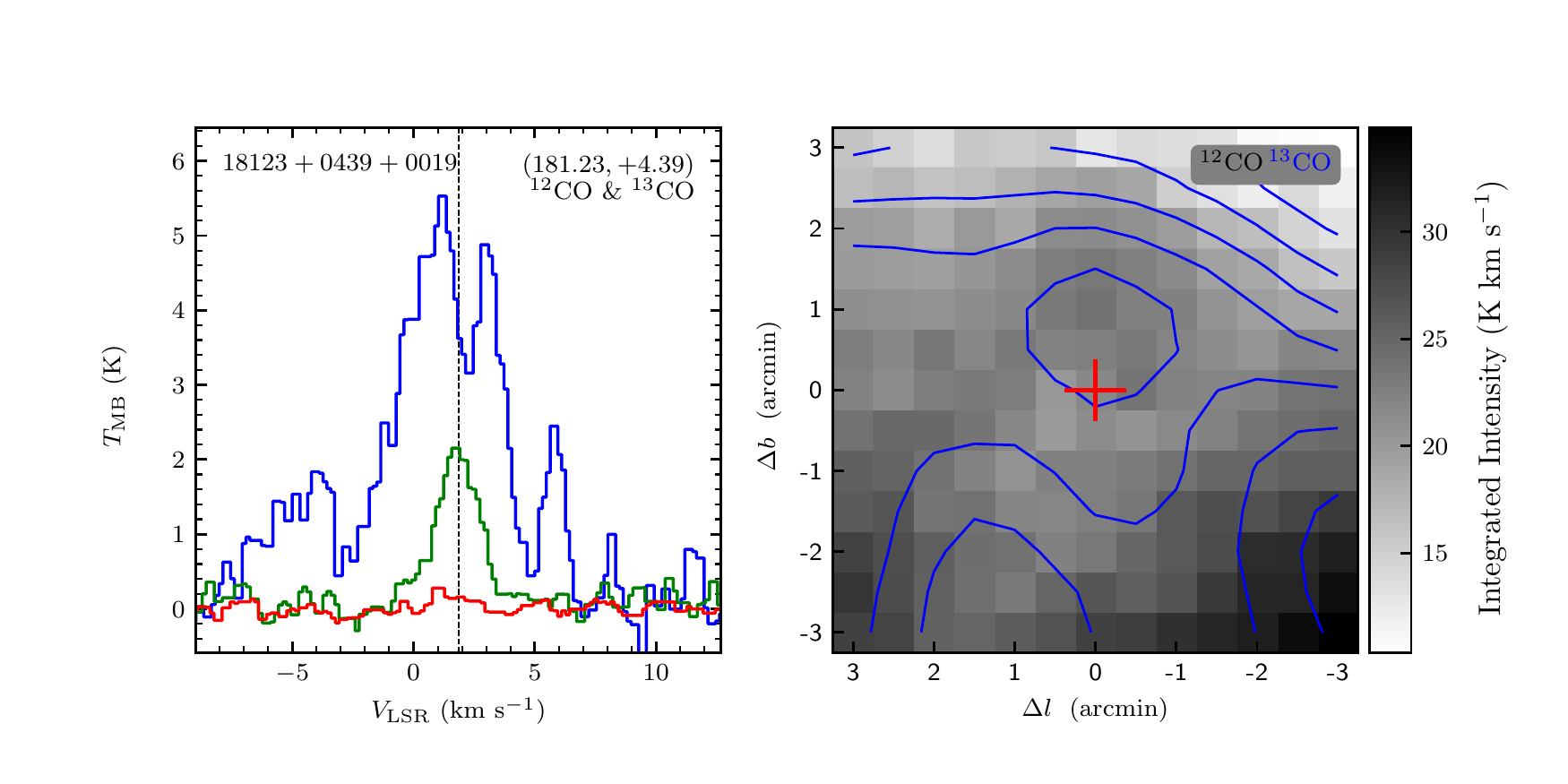}
\vspace{-0.5cm}

\includegraphics[width=9.0cm,angle=0]{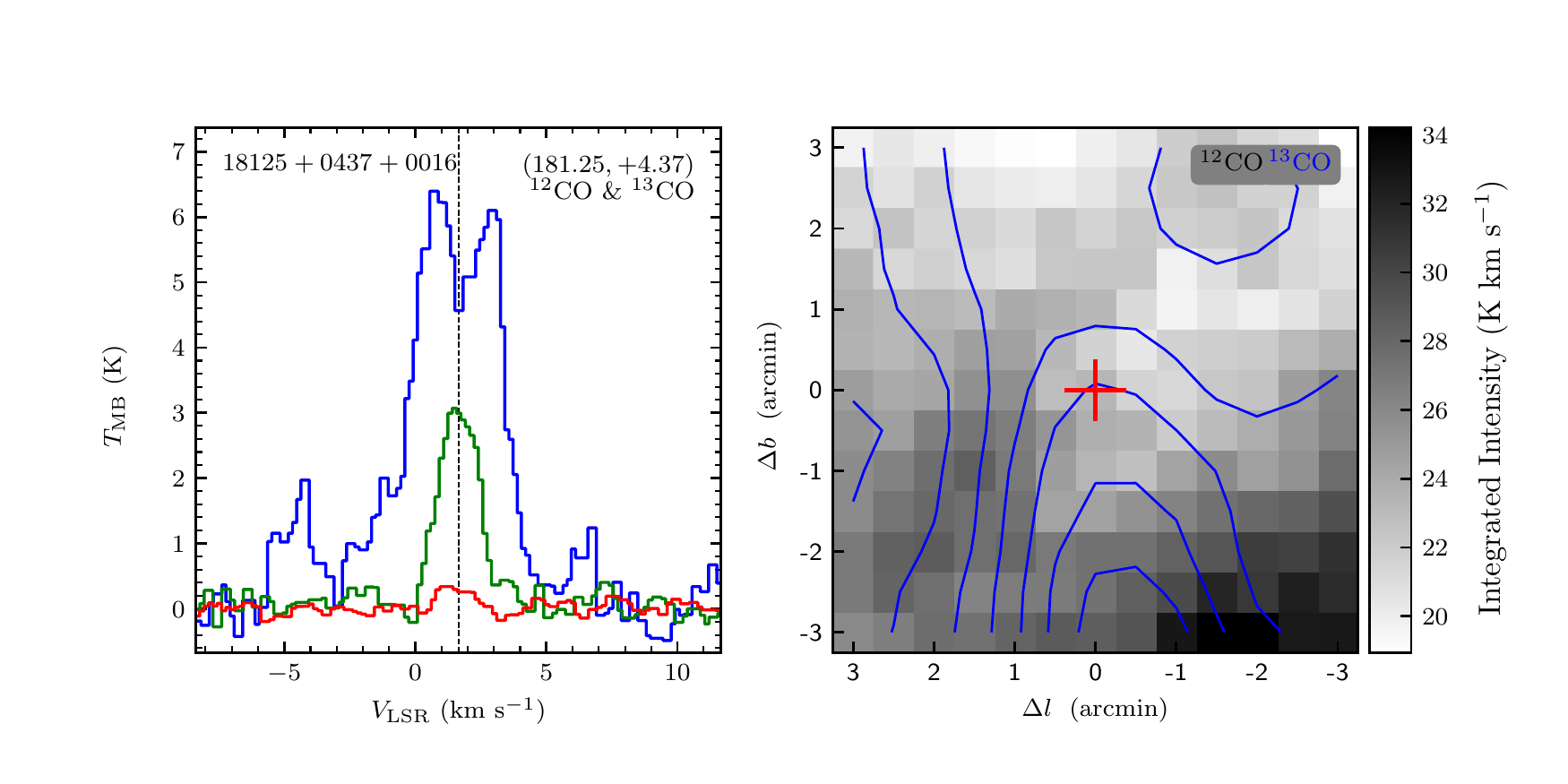}
\includegraphics[width=9.0cm,angle=0]{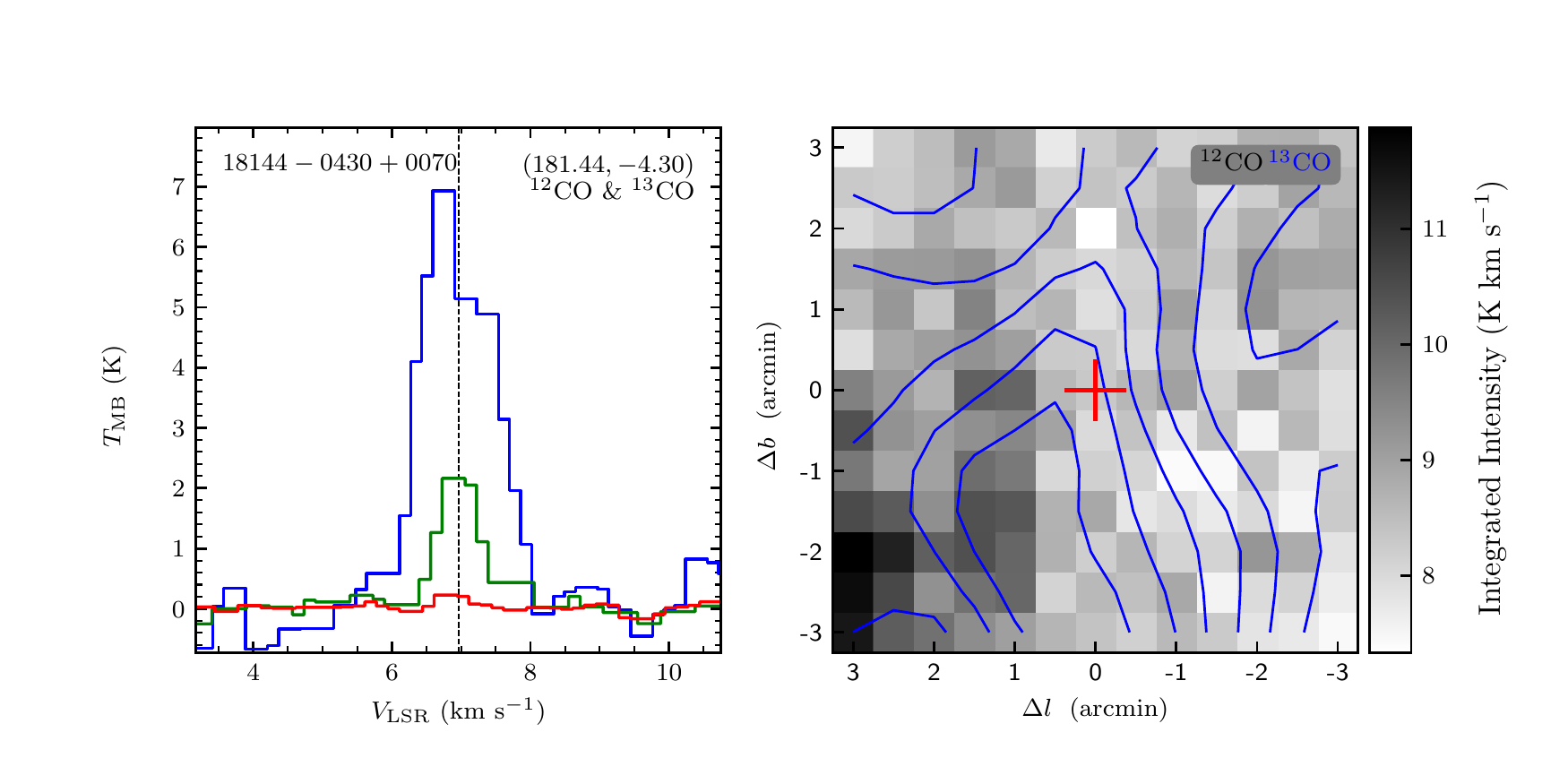}
\end{figure}
\clearpage

\begin{figure}
\includegraphics[width=9.0cm,angle=0]{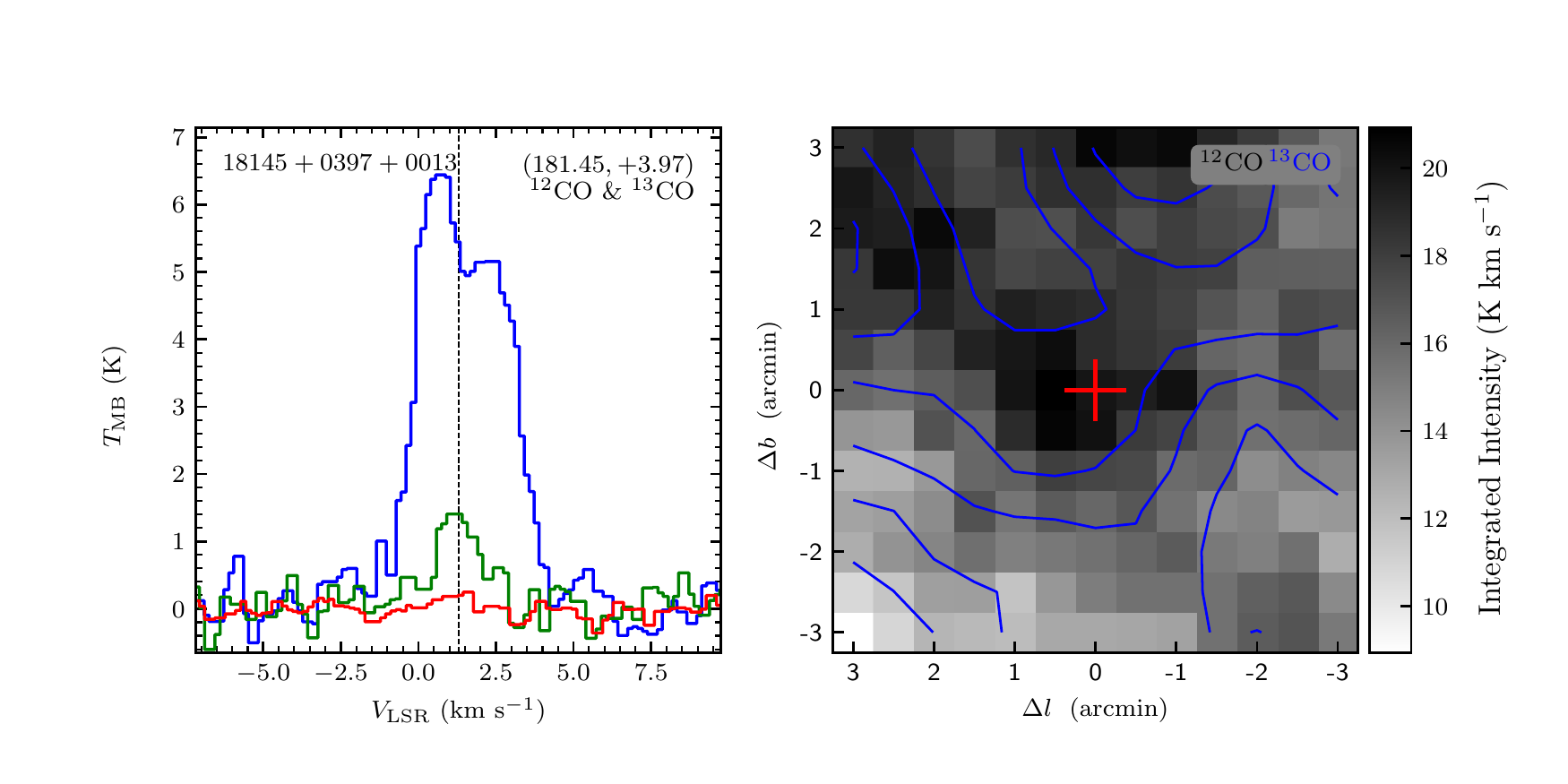}
\includegraphics[width=9.0cm,angle=0]{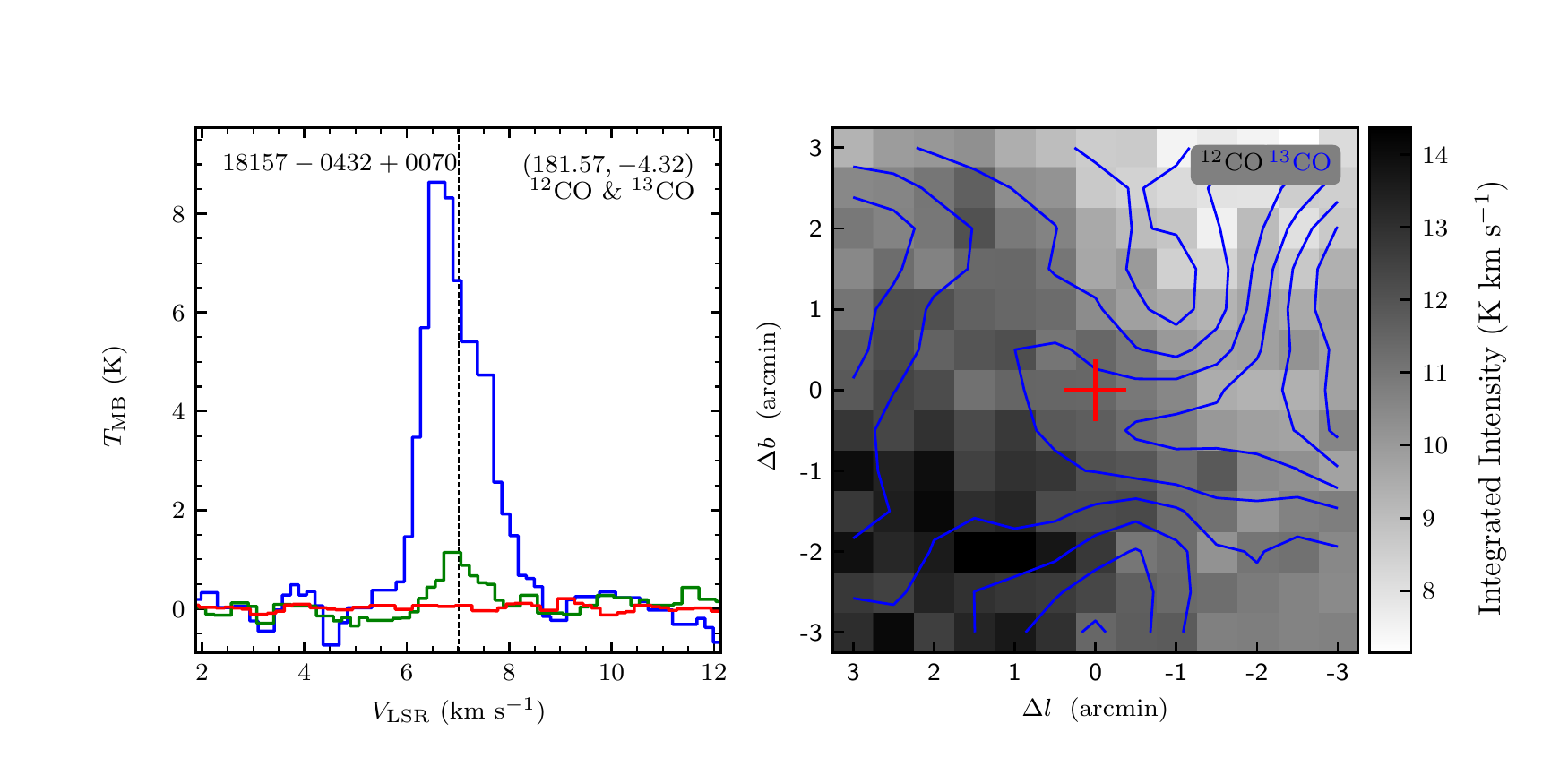}
\vspace{-0.5cm}

\includegraphics[width=9.0cm,angle=0]{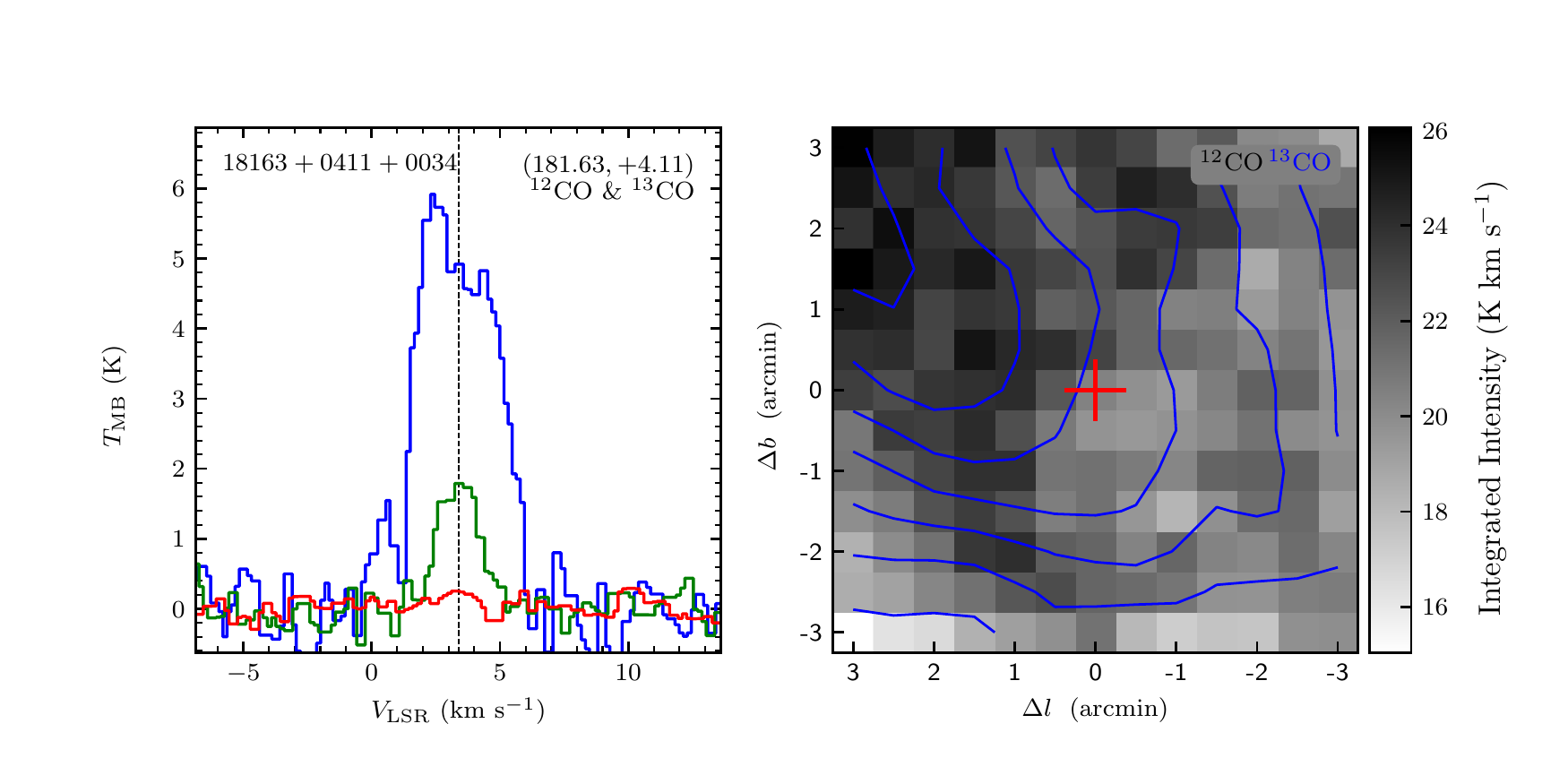}
\includegraphics[width=9.0cm,angle=0]{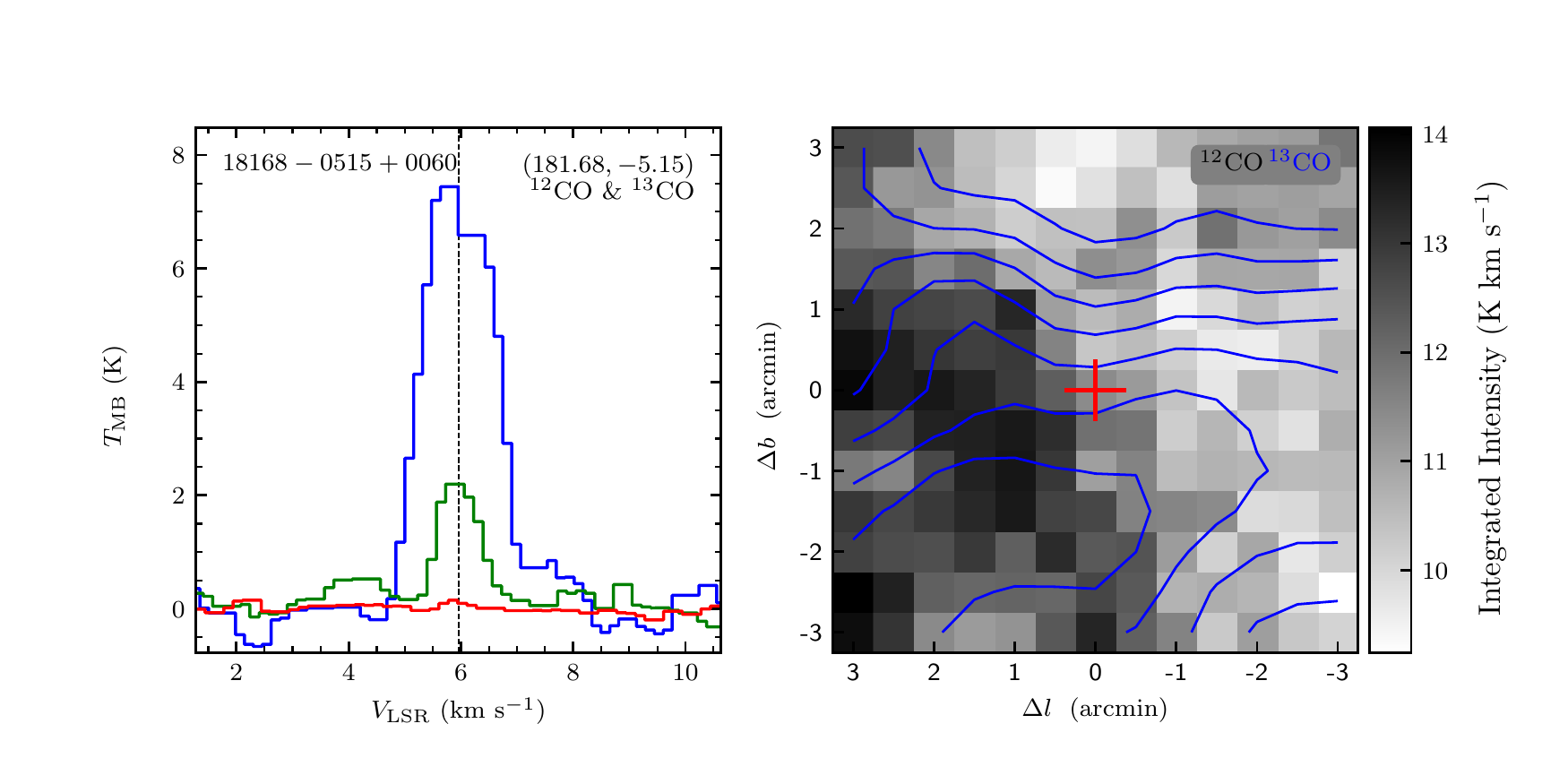}
\vspace{-0.5cm}

\includegraphics[width=9.0cm,angle=0]{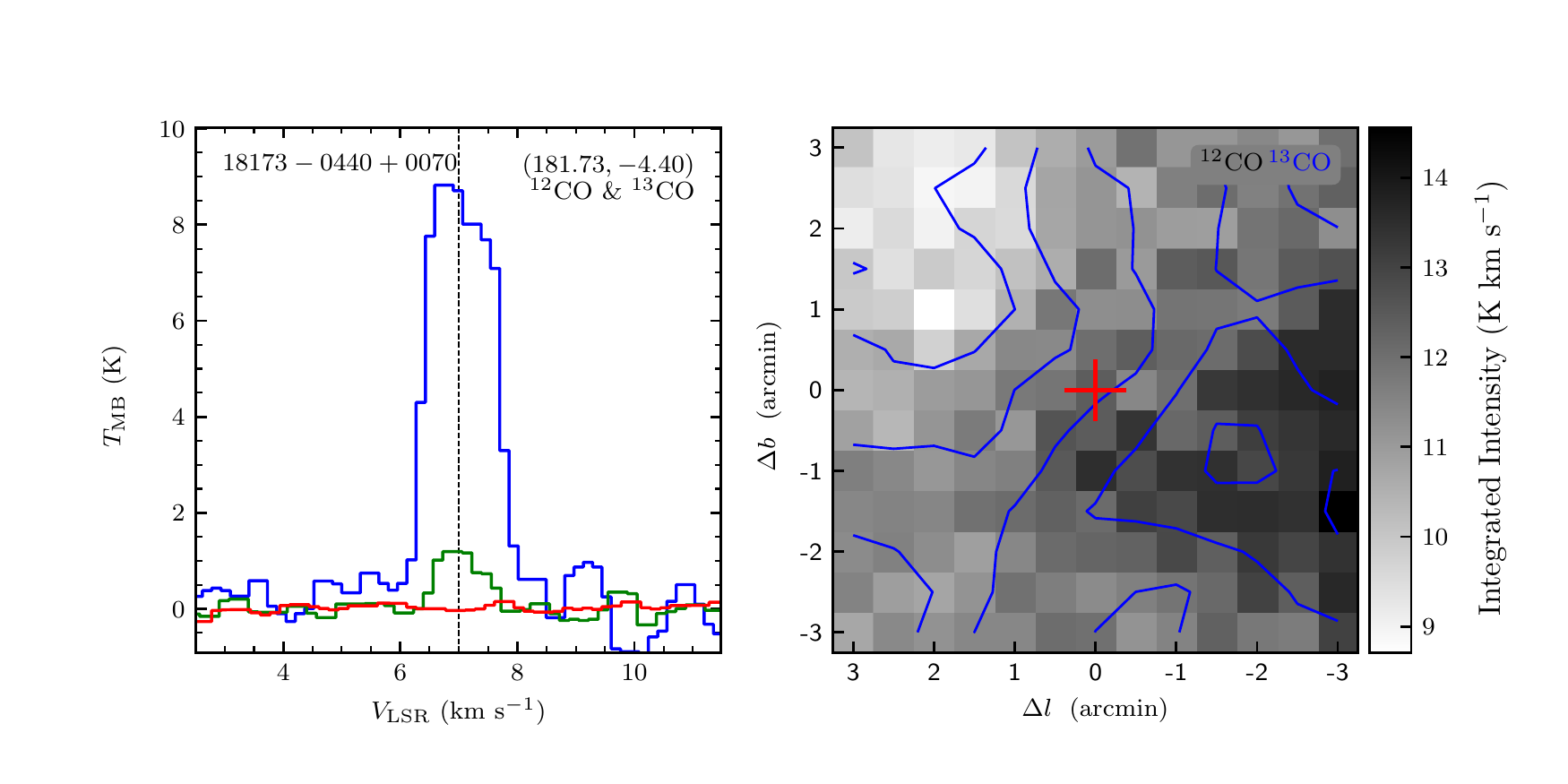}
\includegraphics[width=9.0cm,angle=0]{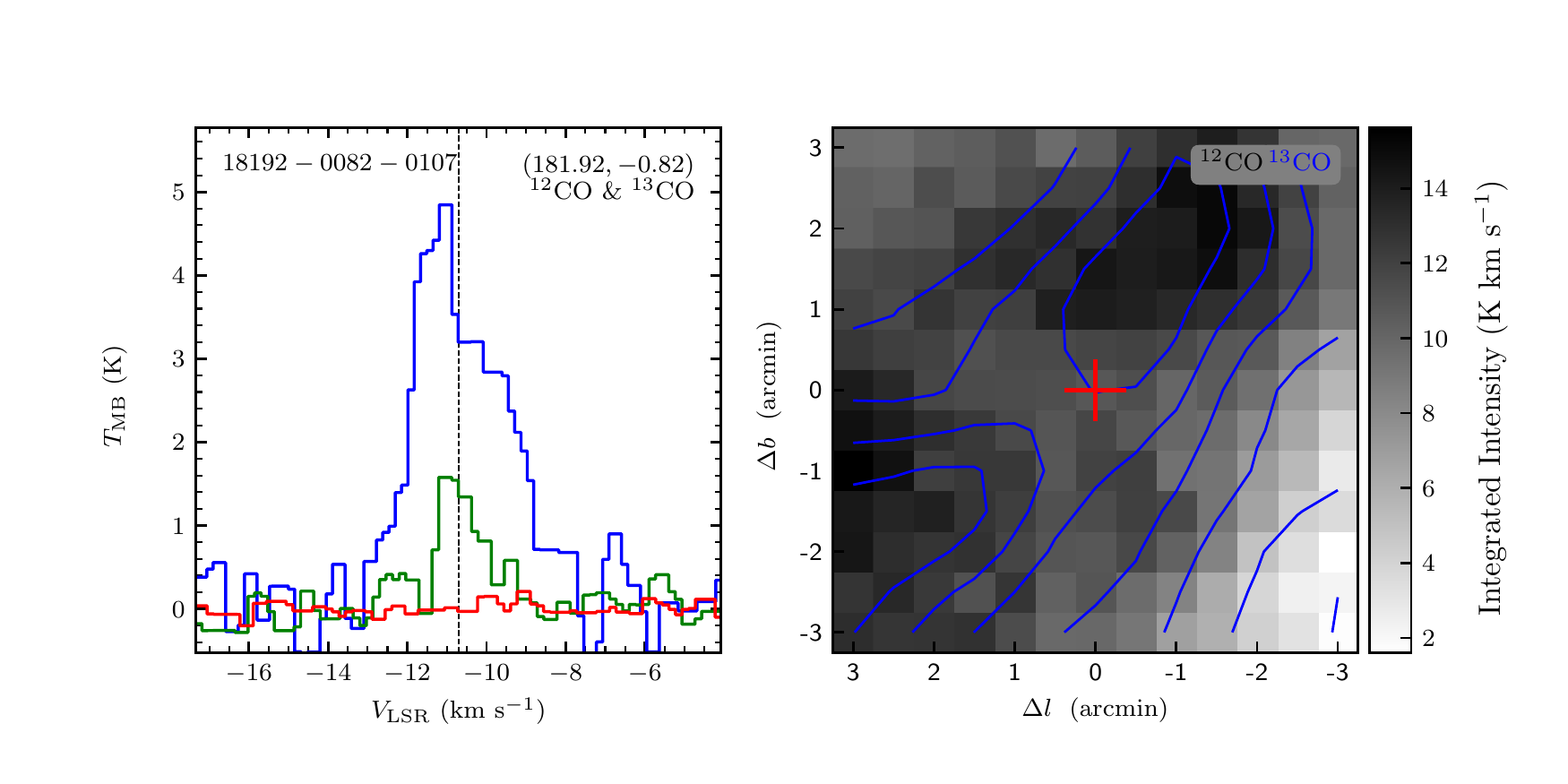}
\vspace{-0.5cm}

\includegraphics[width=9.0cm,angle=0]{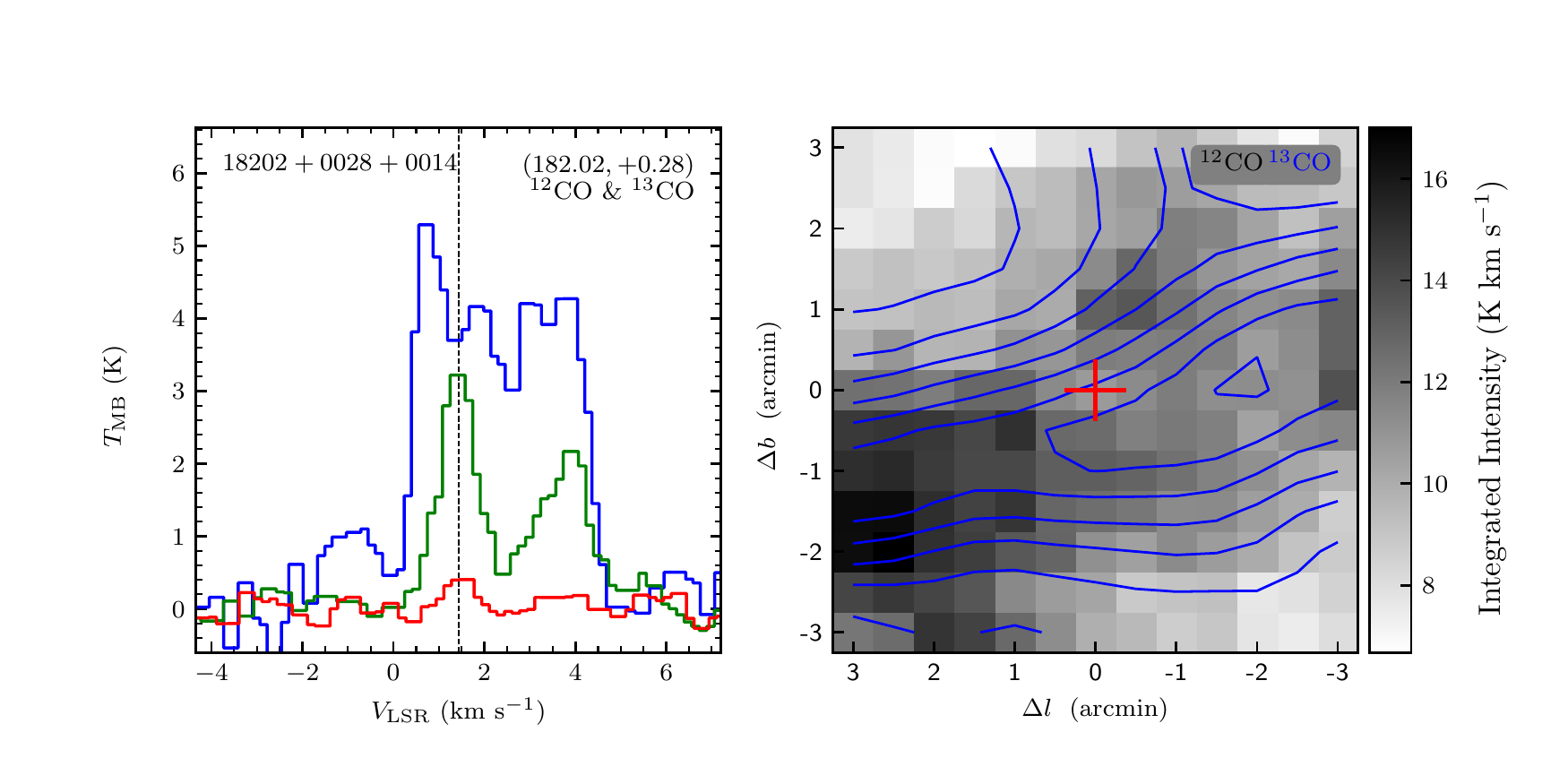}
\includegraphics[width=9.0cm,angle=0]{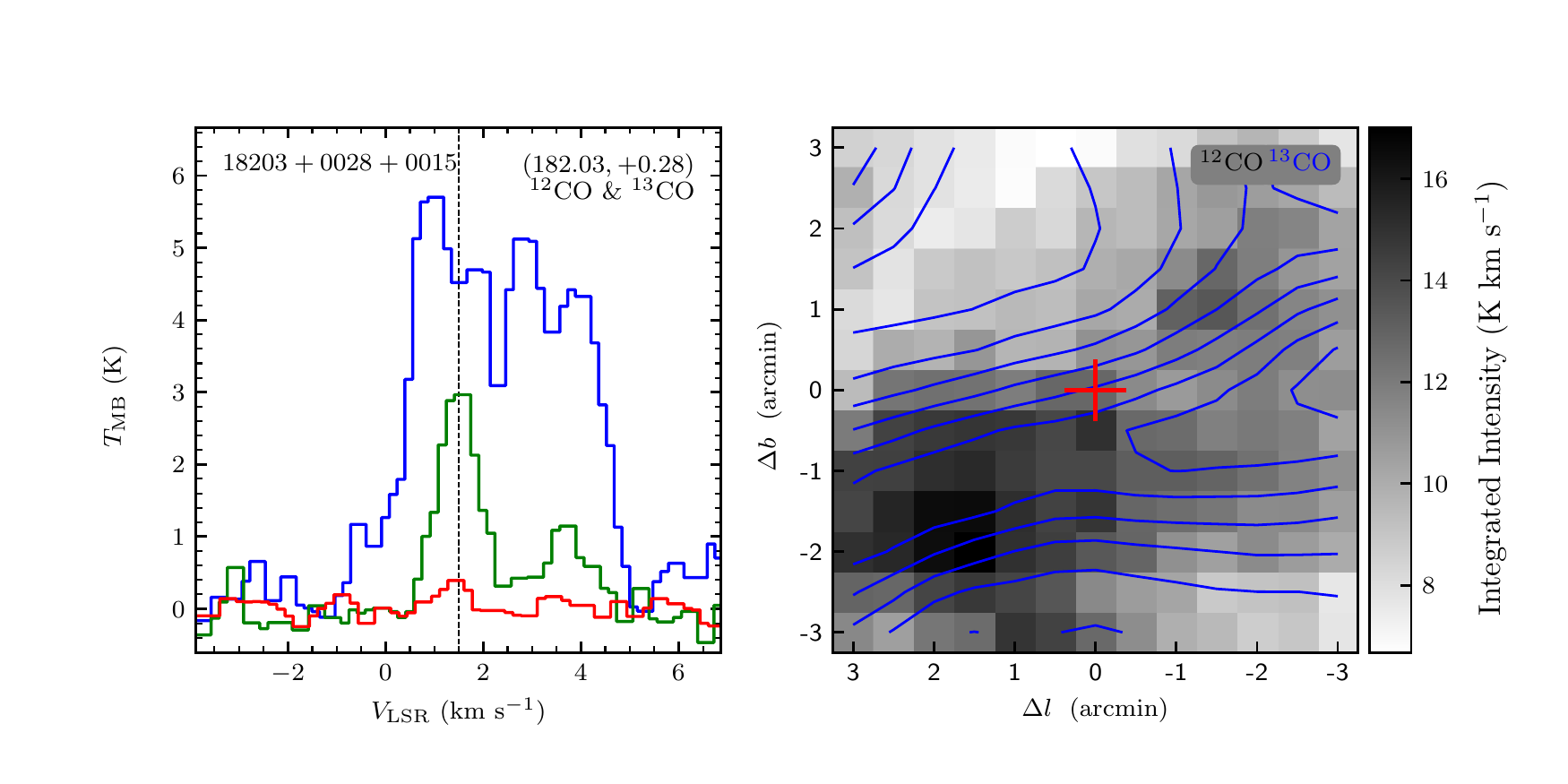}
\vspace{-0.5cm}

\includegraphics[width=9.0cm,angle=0]{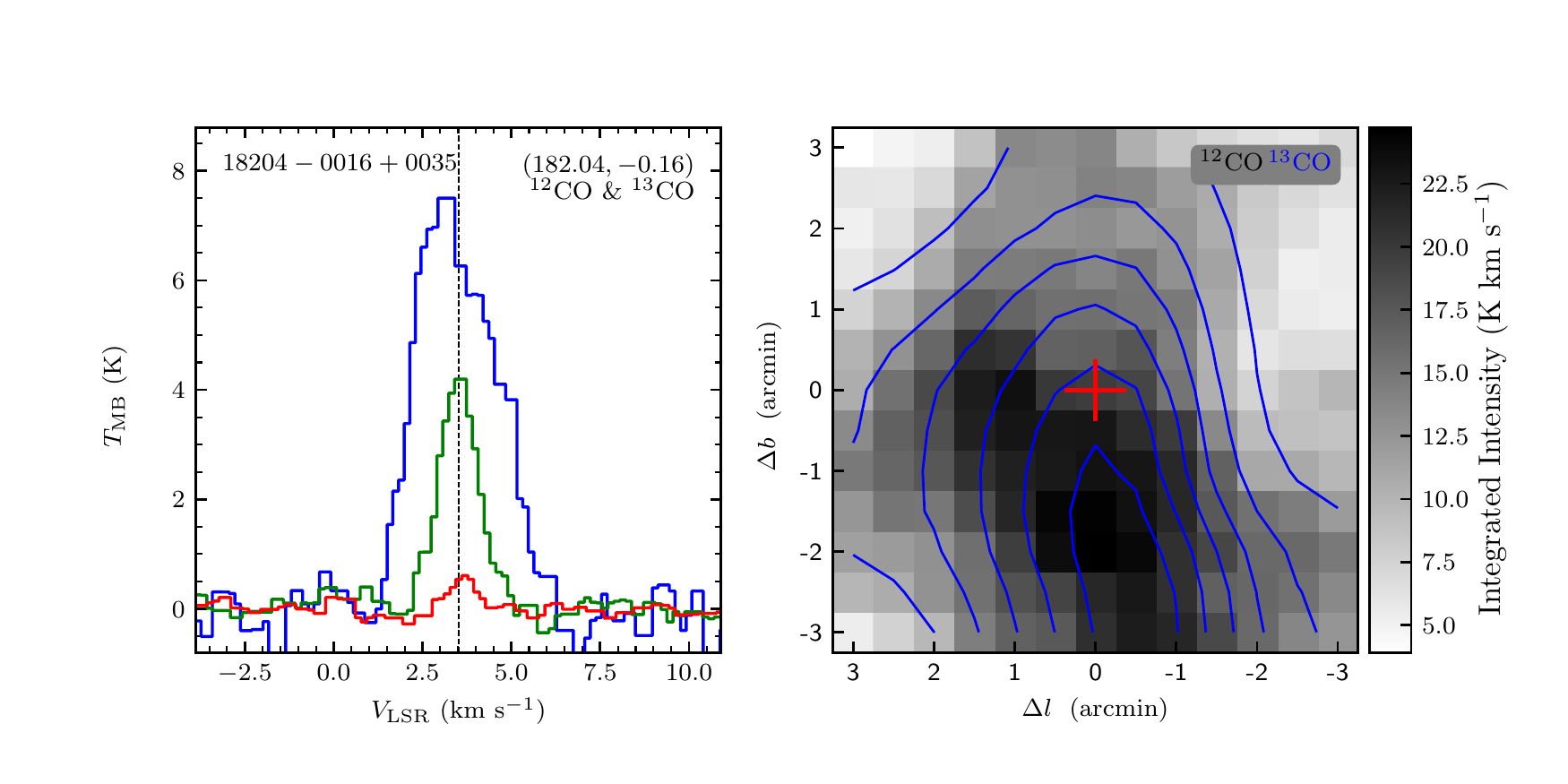}
\includegraphics[width=9.0cm,angle=0]{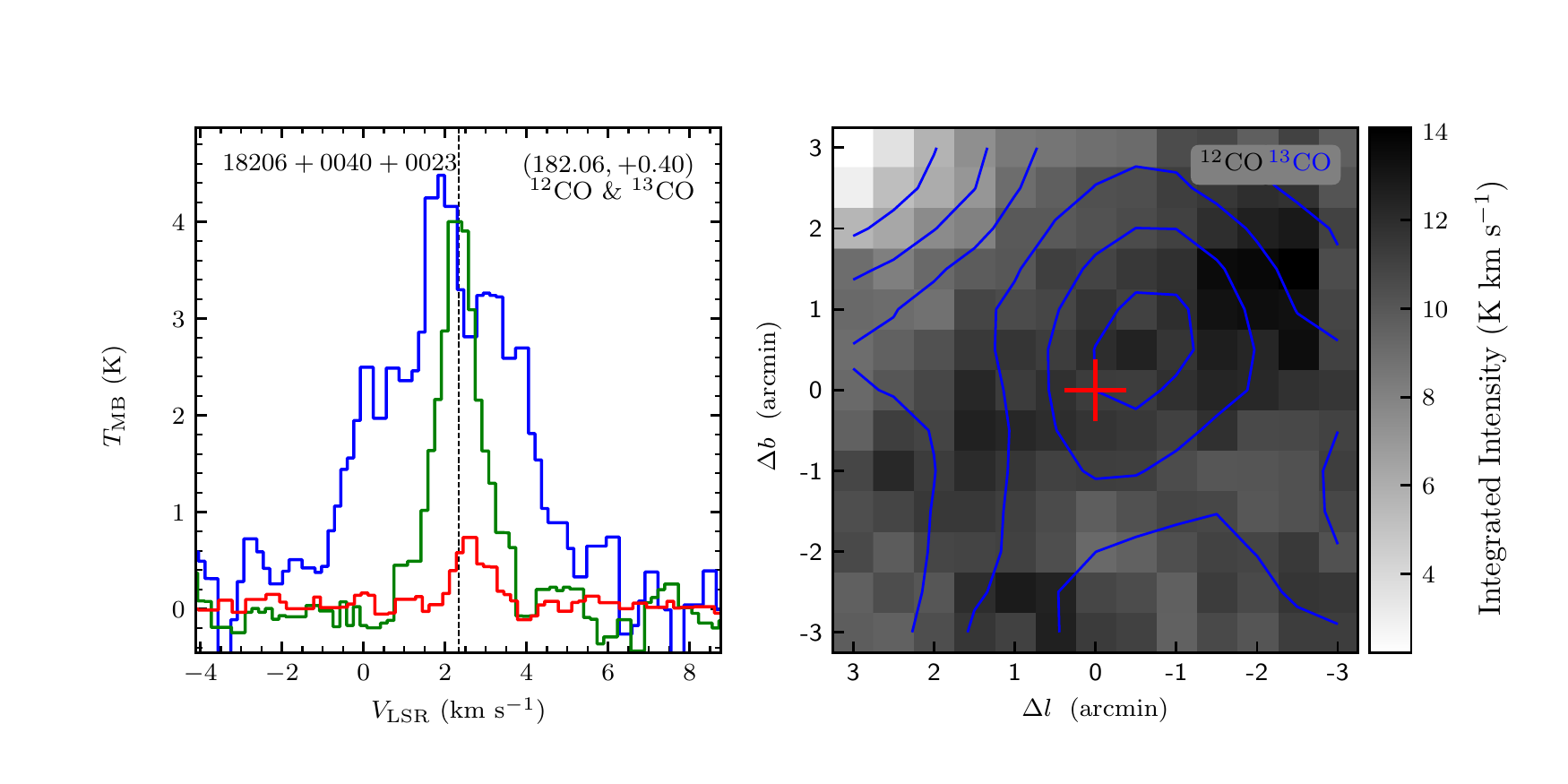}
\end{figure}
\clearpage

\begin{figure}
\includegraphics[width=9.0cm,angle=0]{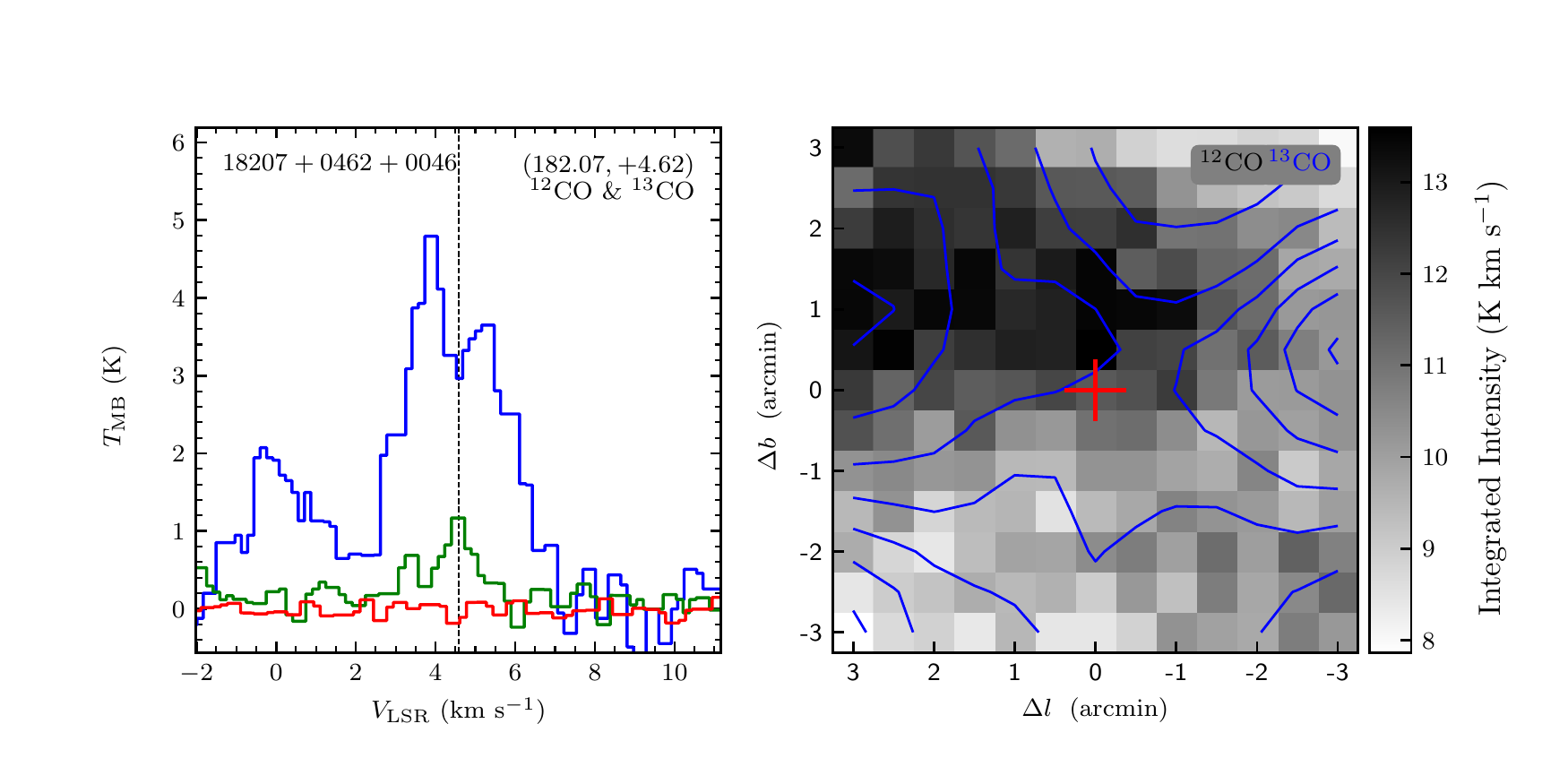}
\includegraphics[width=9.0cm,angle=0]{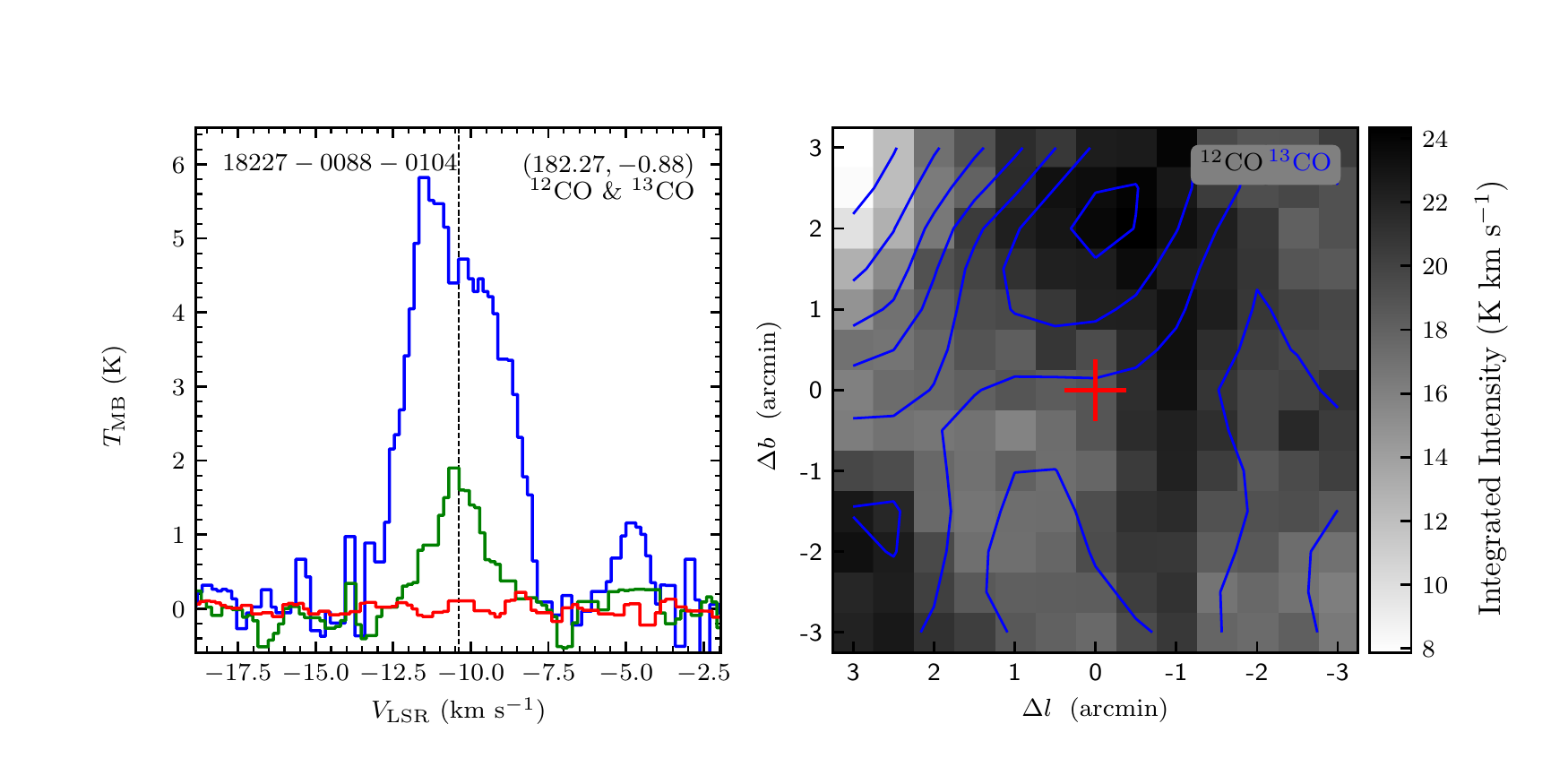}
\vspace{-0.5cm}

\includegraphics[width=9.0cm,angle=0]{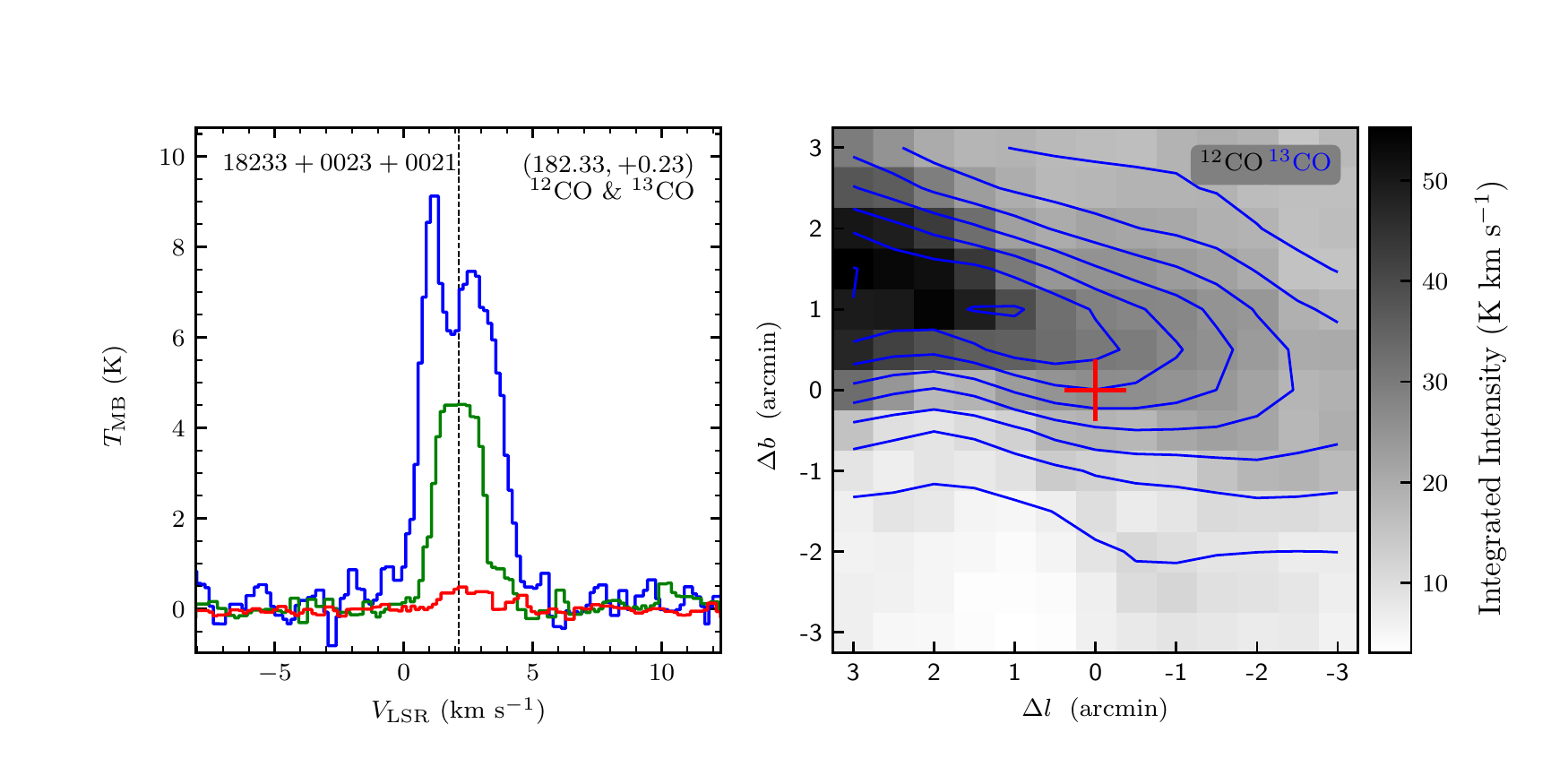}
\includegraphics[width=9.0cm,angle=0]{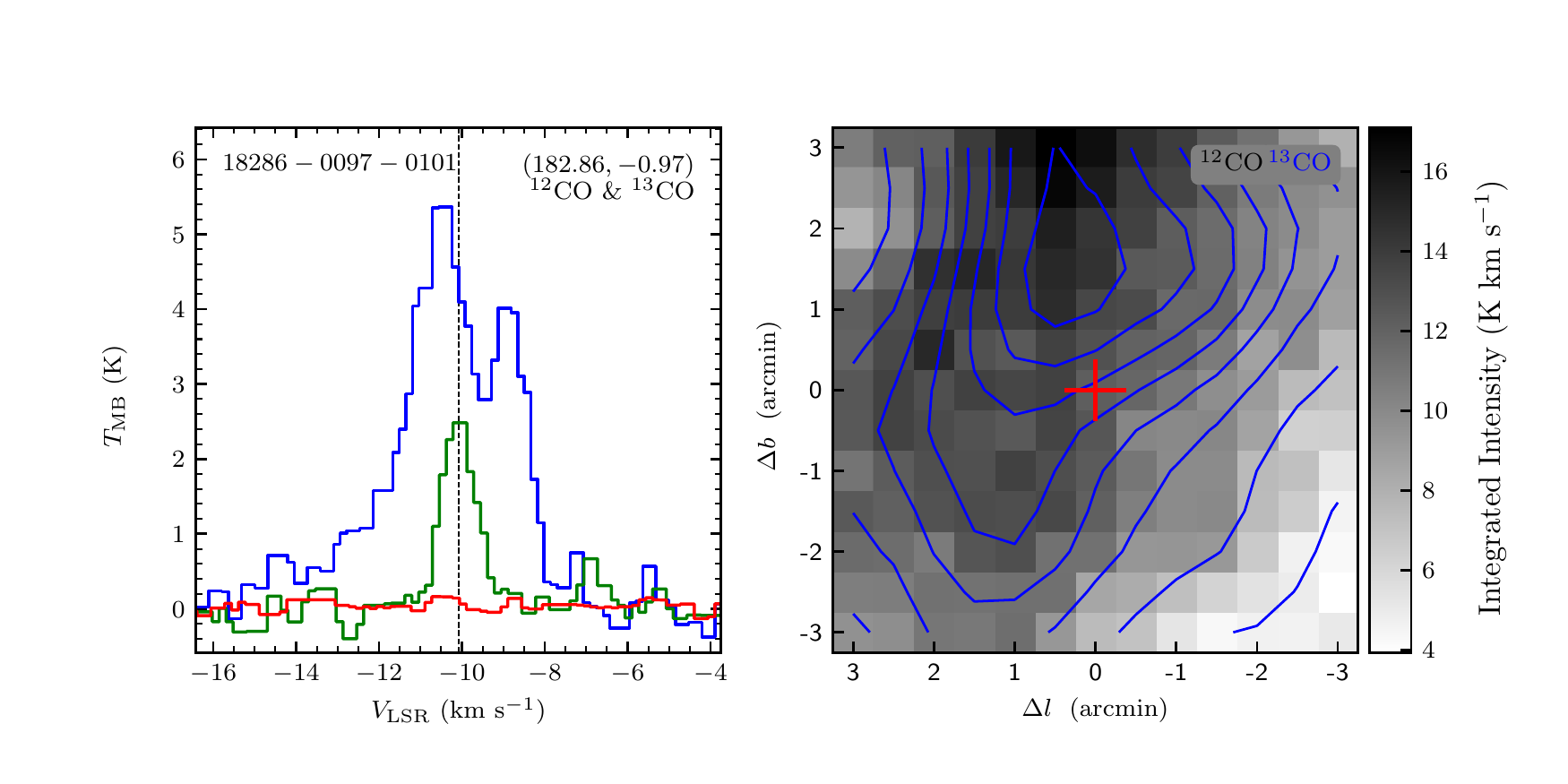}
\vspace{-0.5cm}

\includegraphics[width=9.0cm,angle=0]{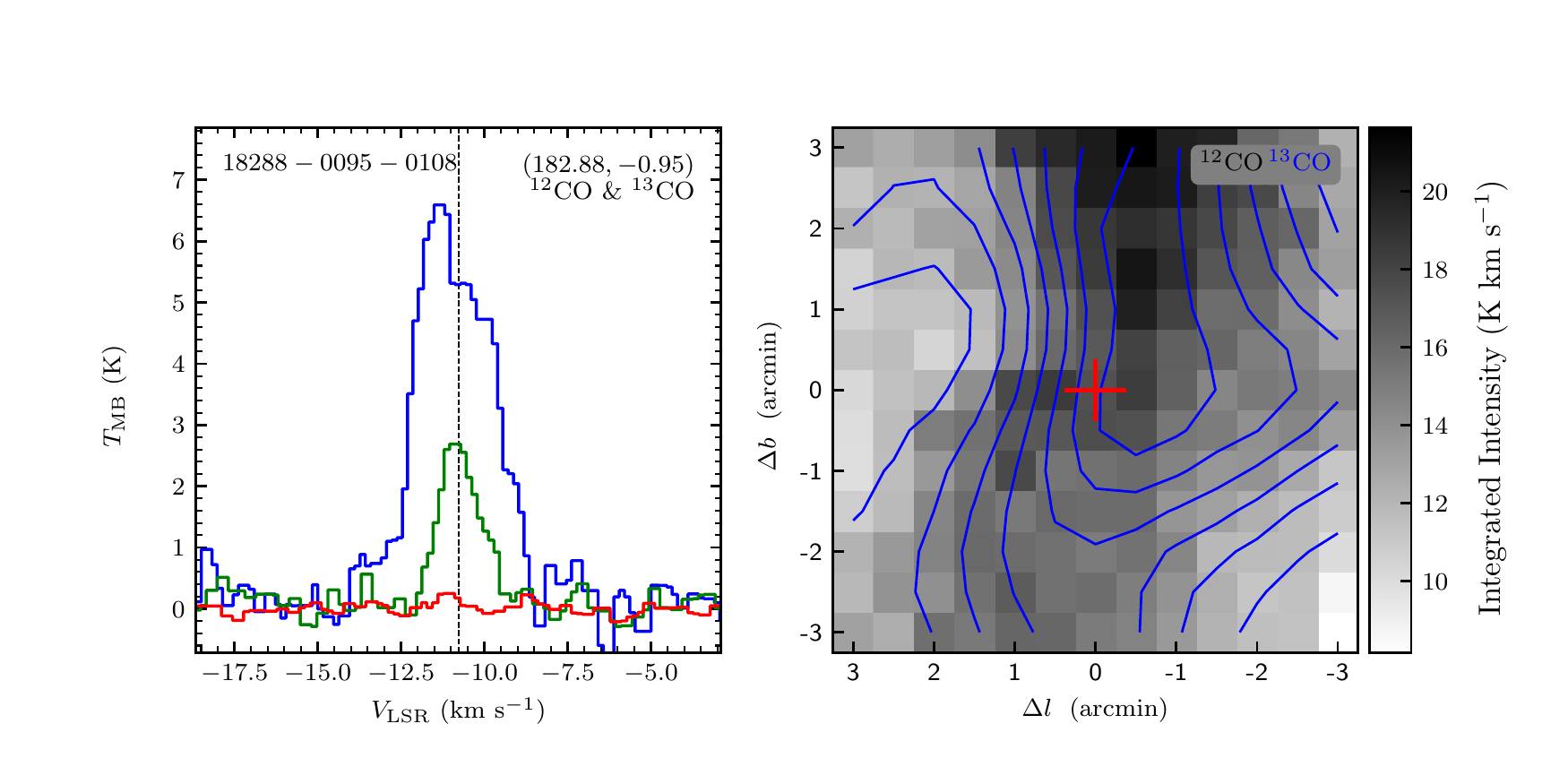}
\includegraphics[width=9.0cm,angle=0]{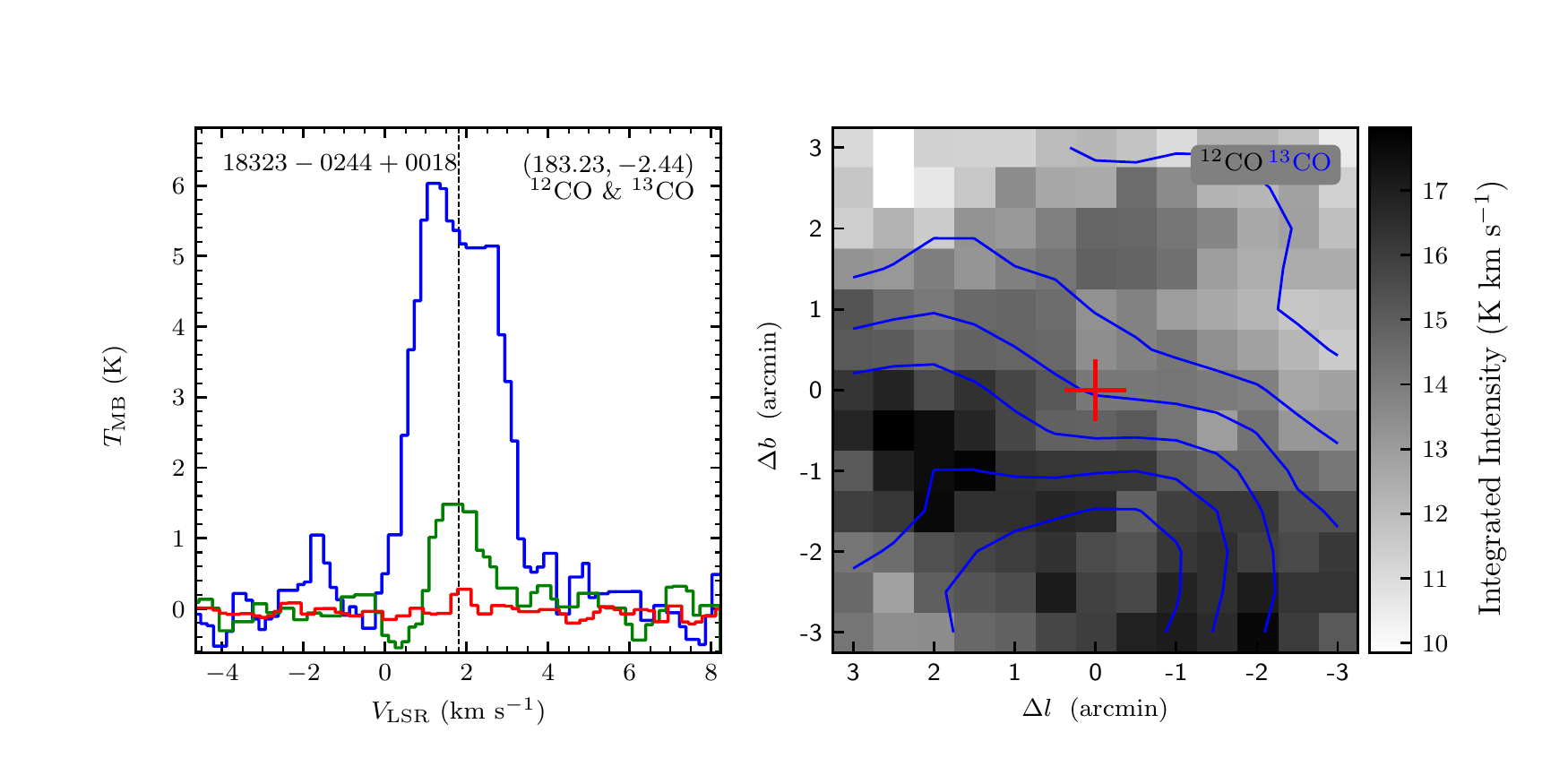}
\vspace{-0.5cm}

\includegraphics[width=9.0cm,angle=0]{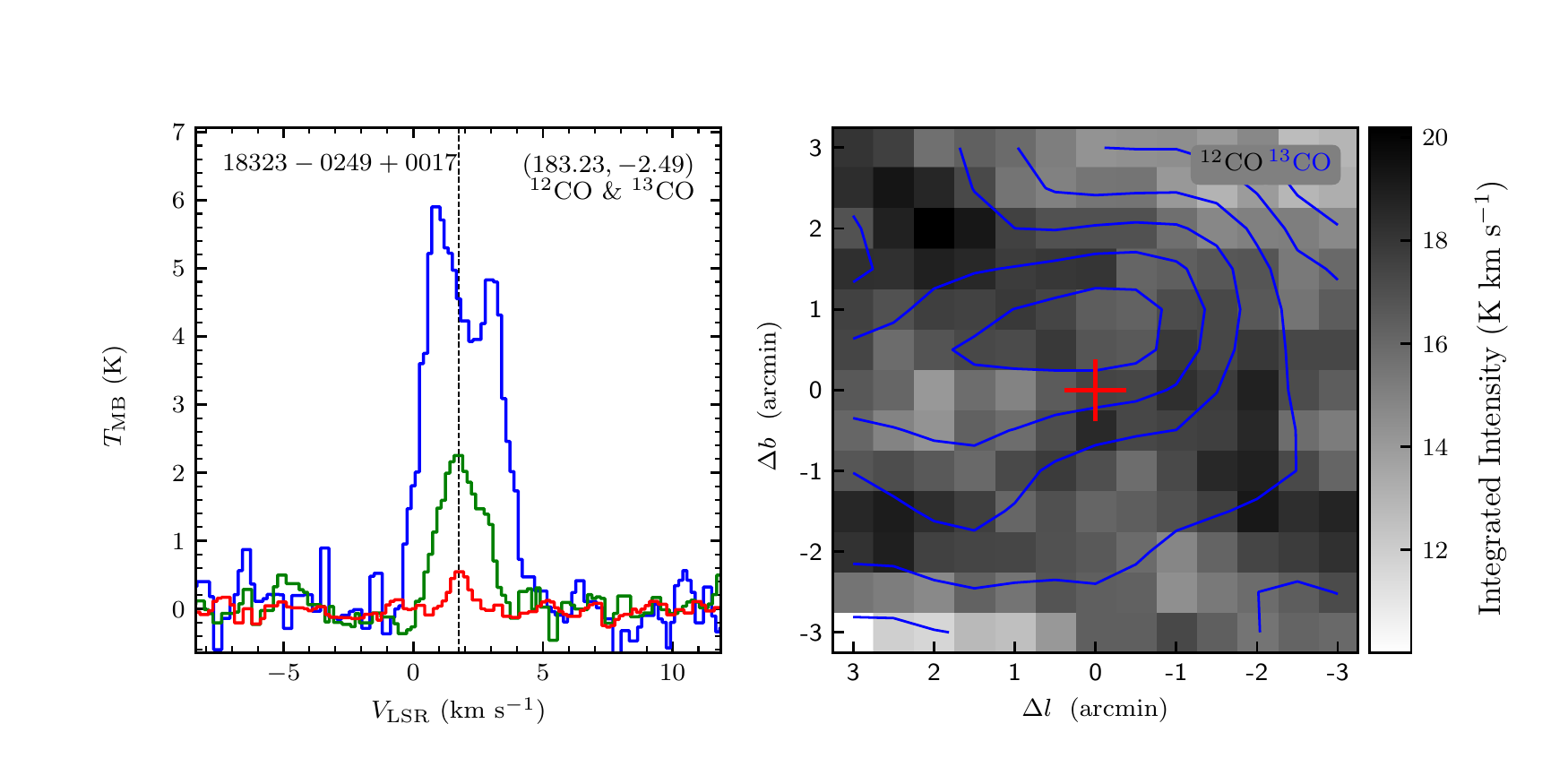}
\includegraphics[width=9.0cm,angle=0]{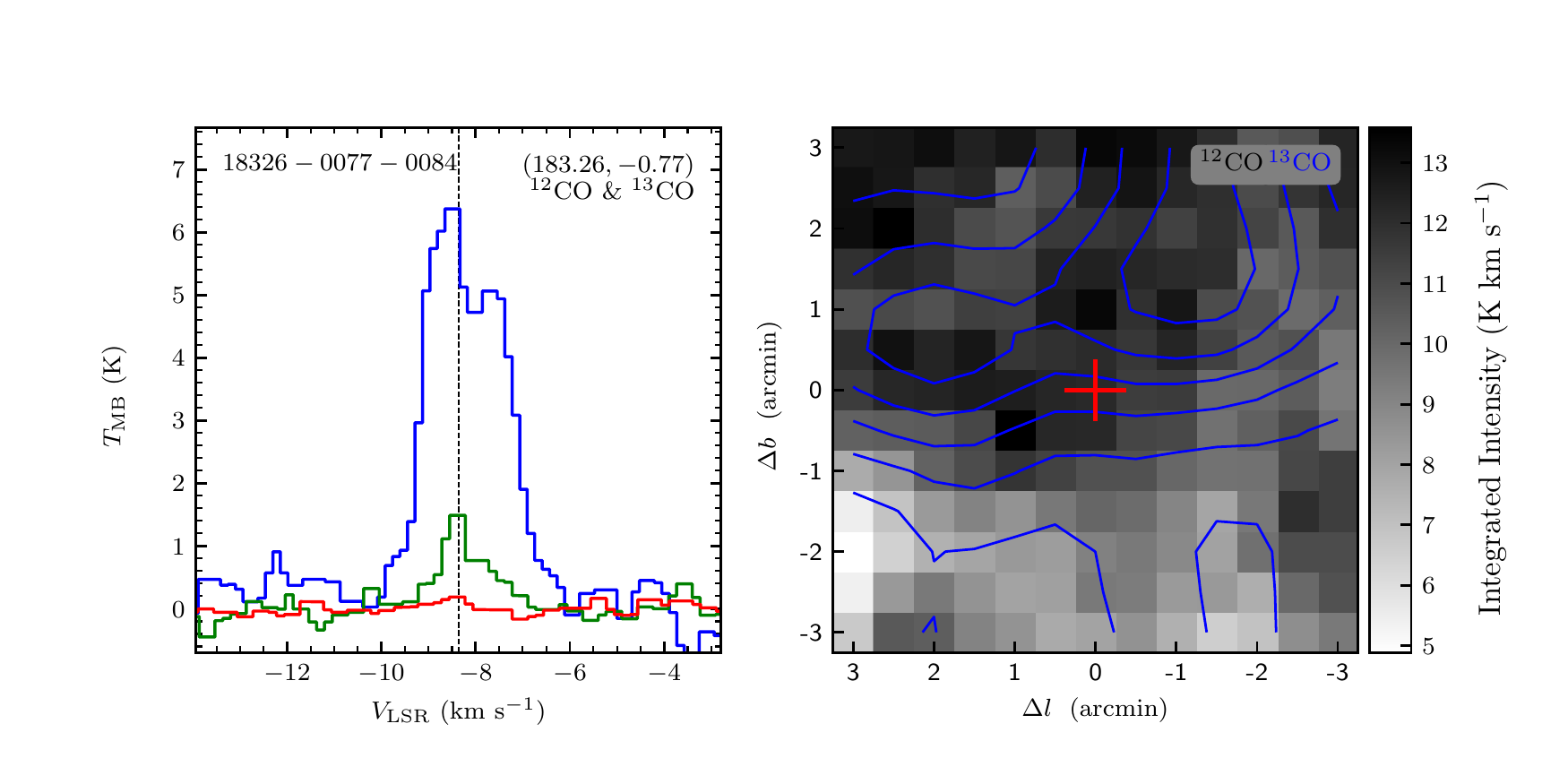}
\vspace{-0.5cm}

\includegraphics[width=9.0cm,angle=0]{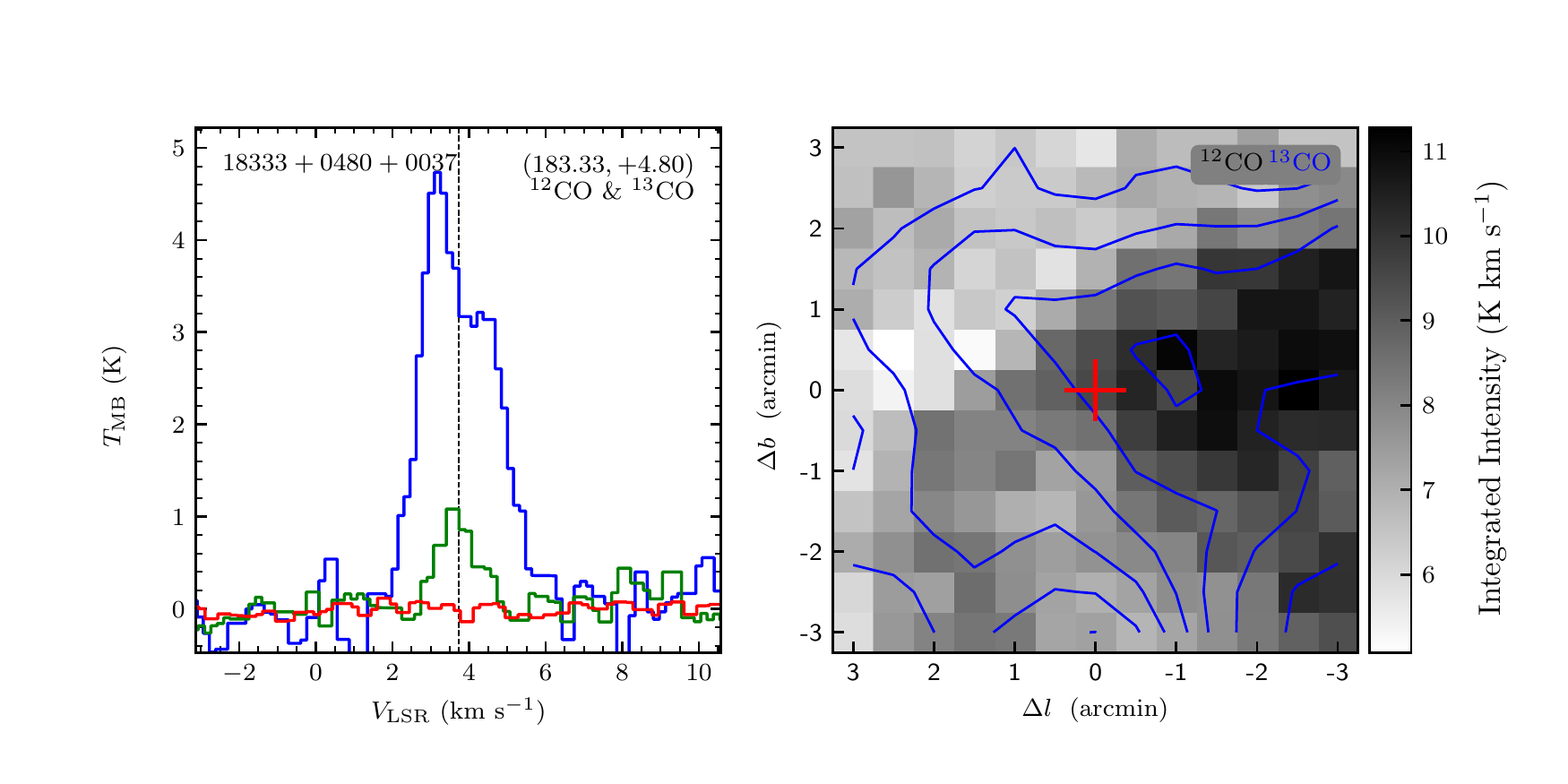}
\includegraphics[width=9.0cm,angle=0]{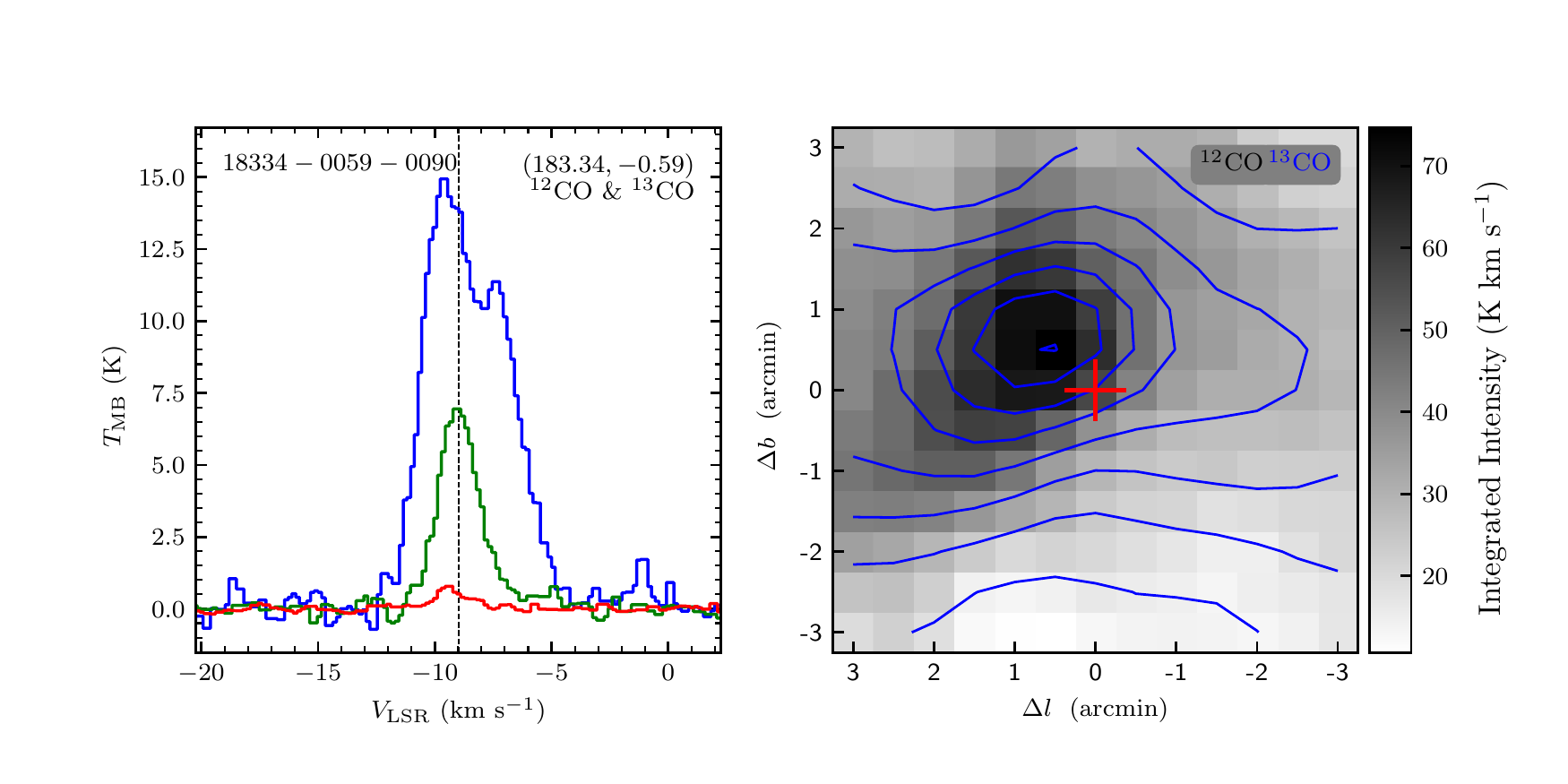}
\end{figure}
\clearpage

\begin{figure}
\includegraphics[width=9.0cm,angle=0]{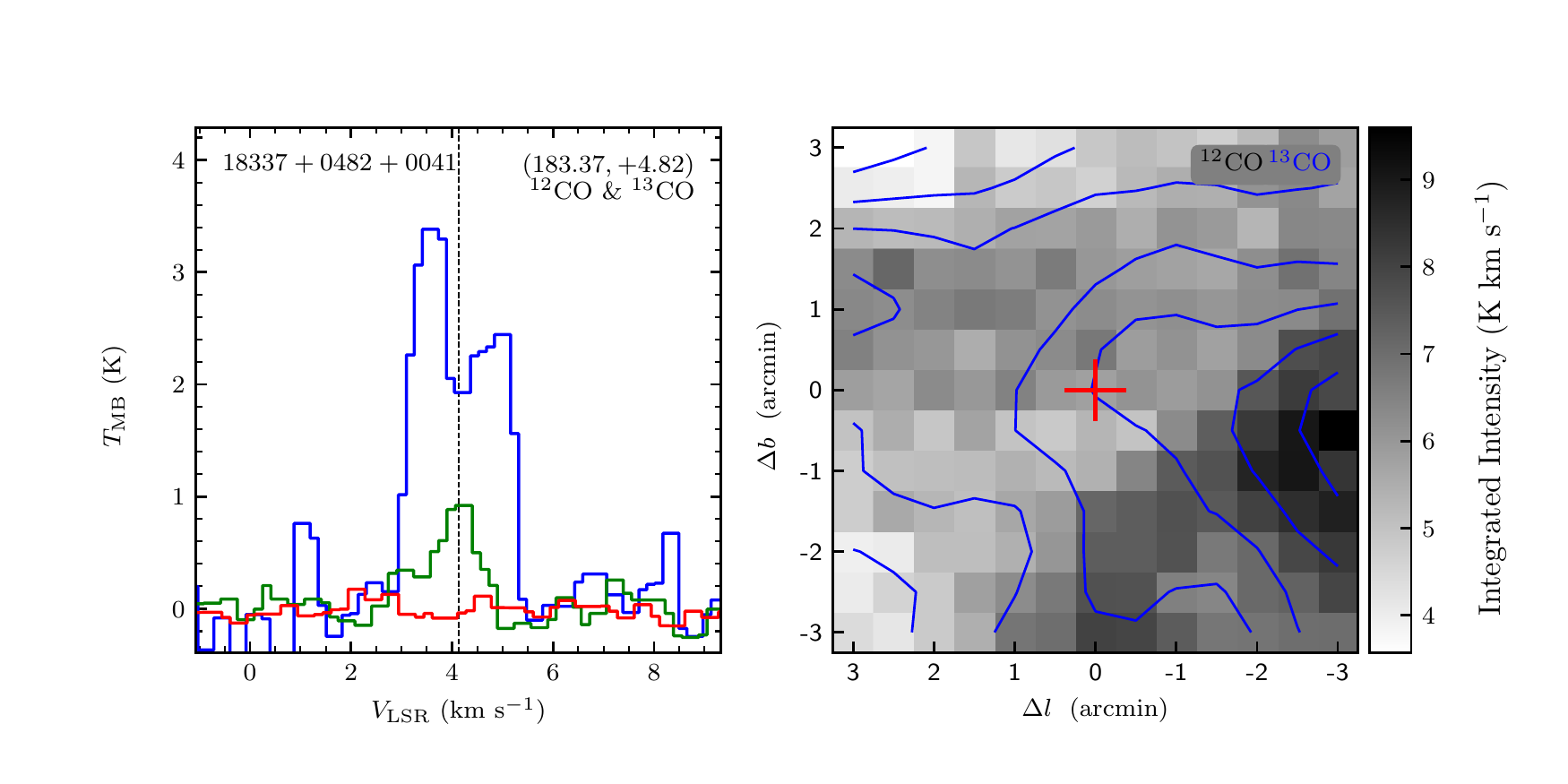}
\includegraphics[width=9.0cm,angle=0]{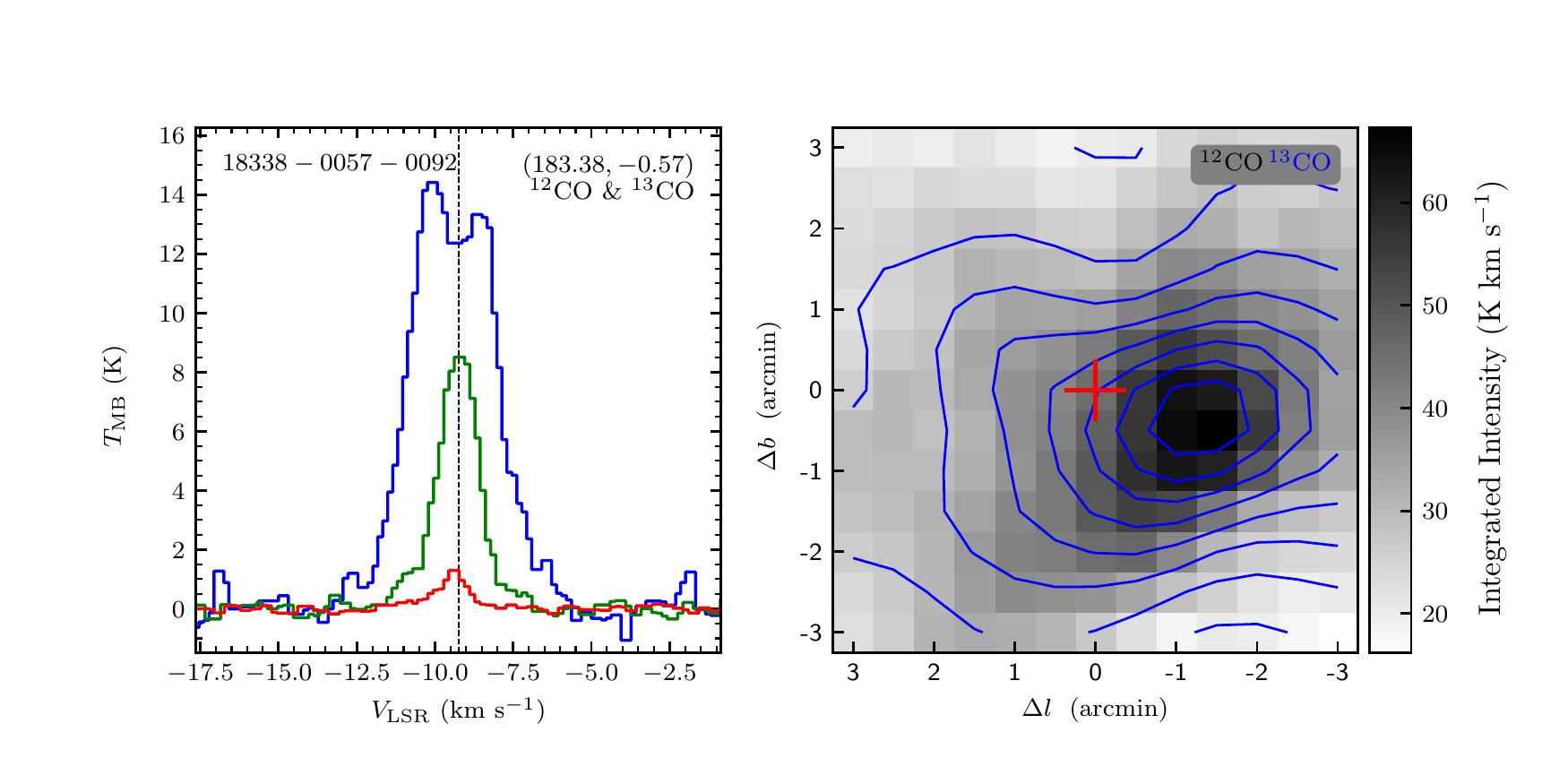}
\vspace{-0.5cm}

\includegraphics[width=9.0cm,angle=0]{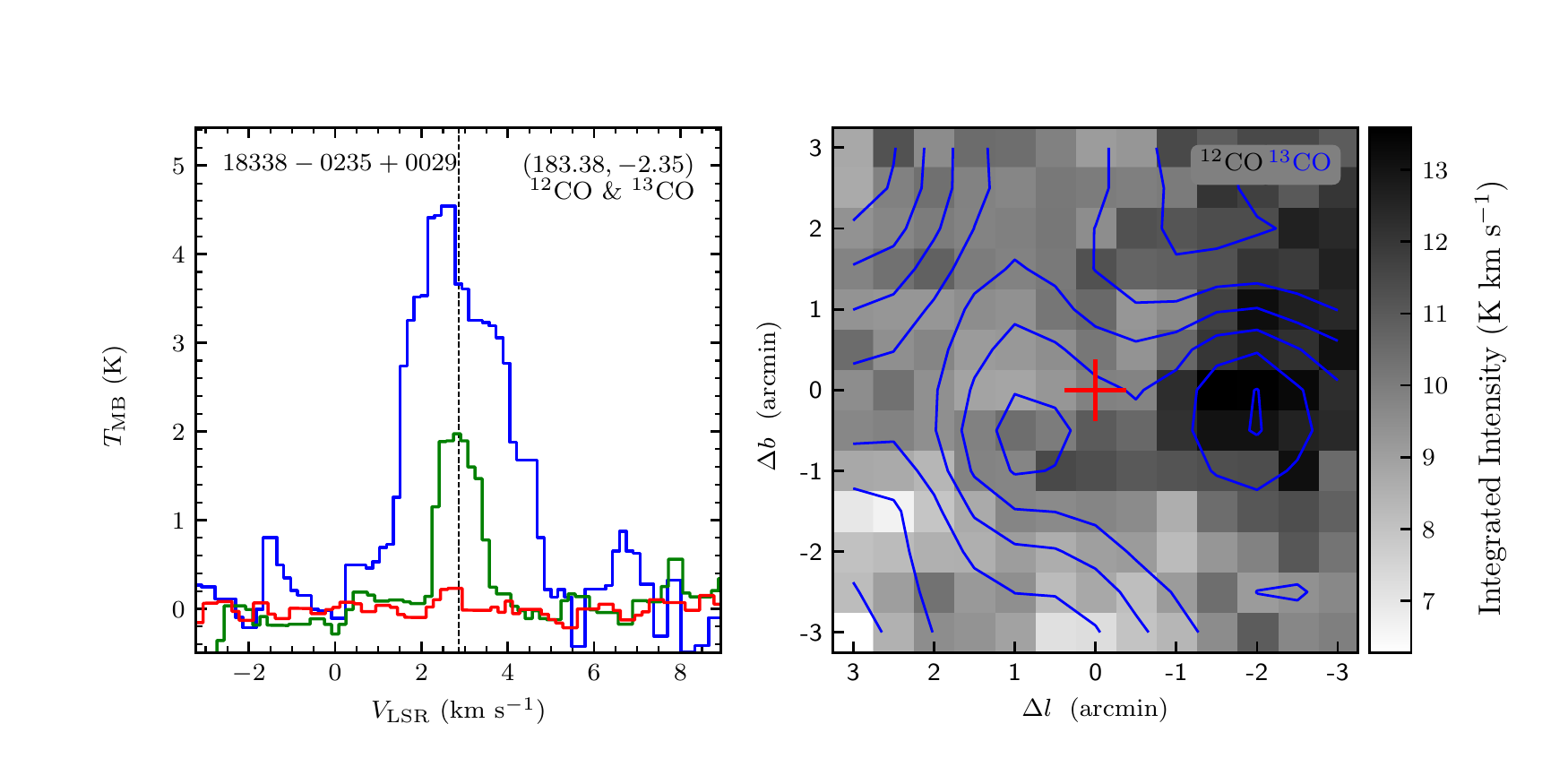}
\includegraphics[width=9.0cm,angle=0]{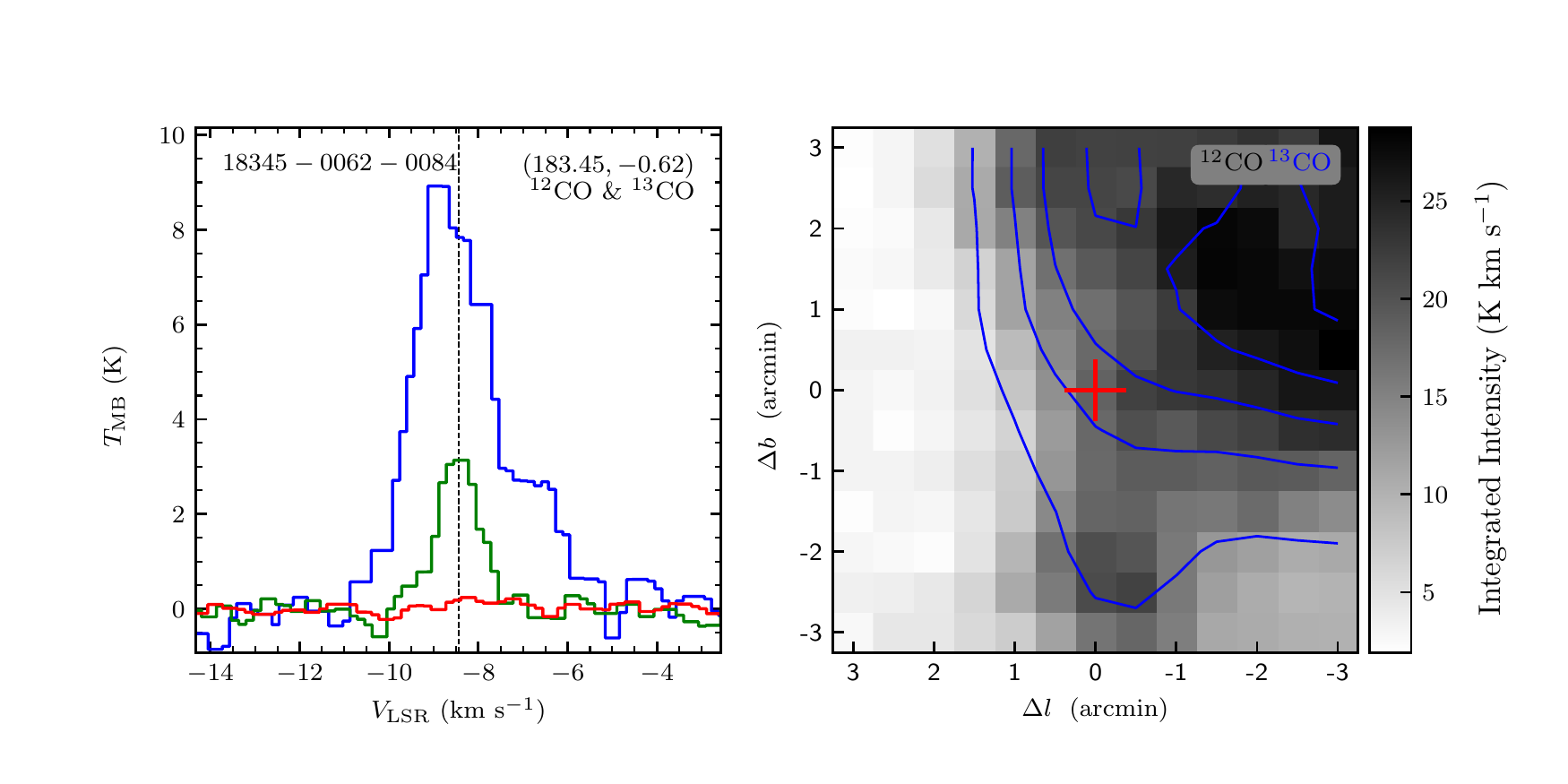}
\vspace{-0.5cm}

\includegraphics[width=9.0cm,angle=0]{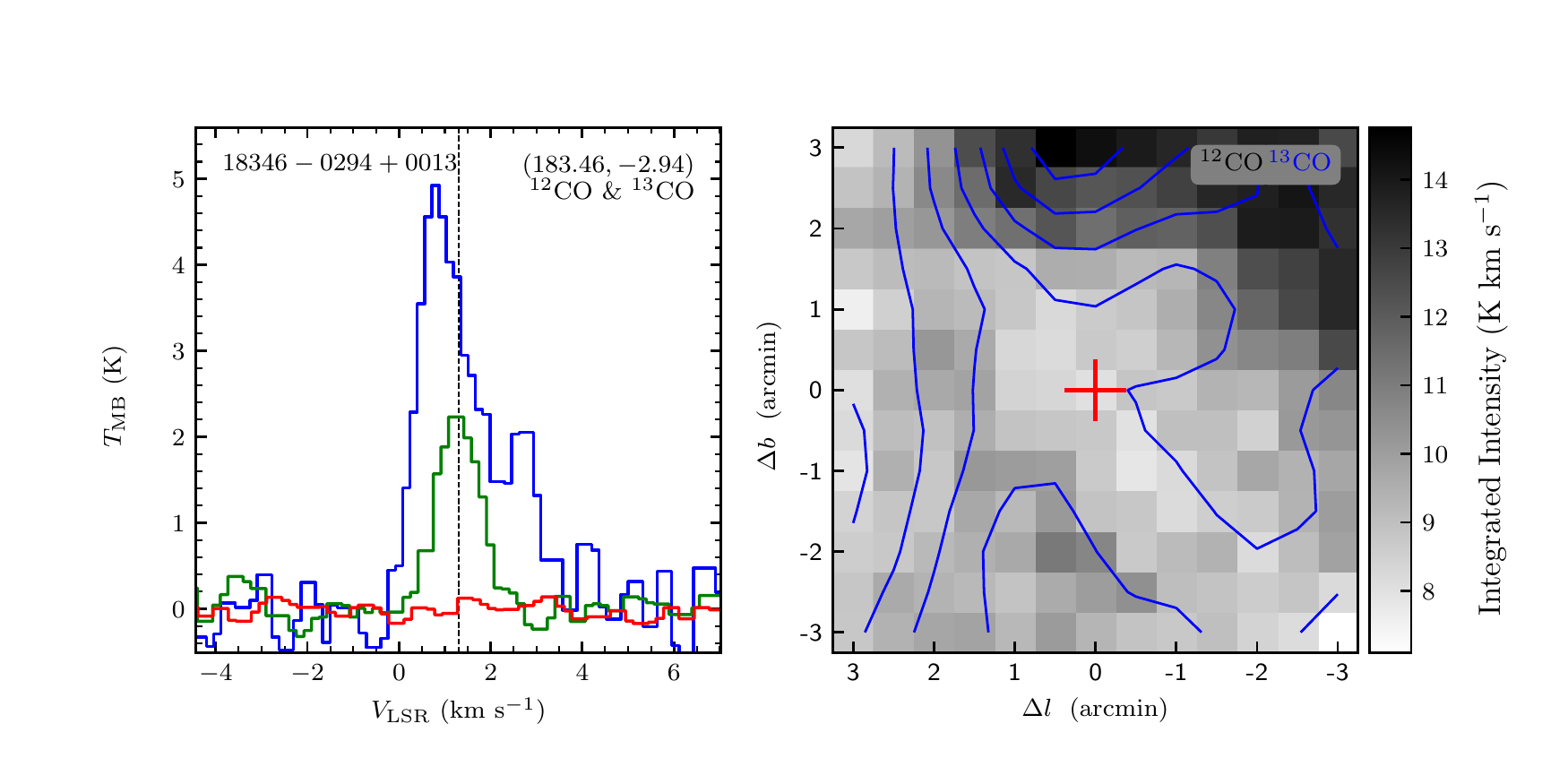}
\includegraphics[width=9.0cm,angle=0]{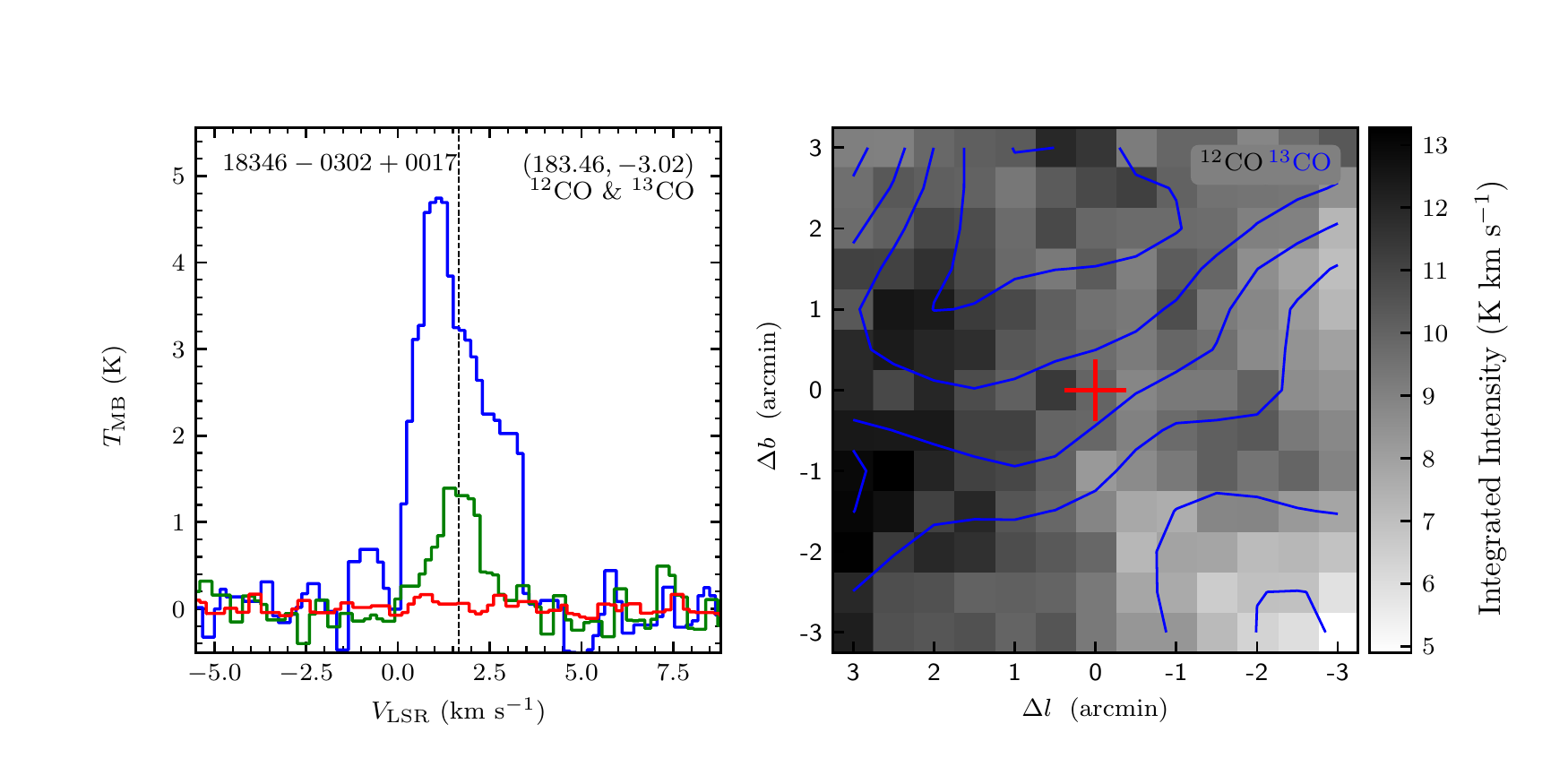}
\vspace{-0.5cm}

\includegraphics[width=9.0cm,angle=0]{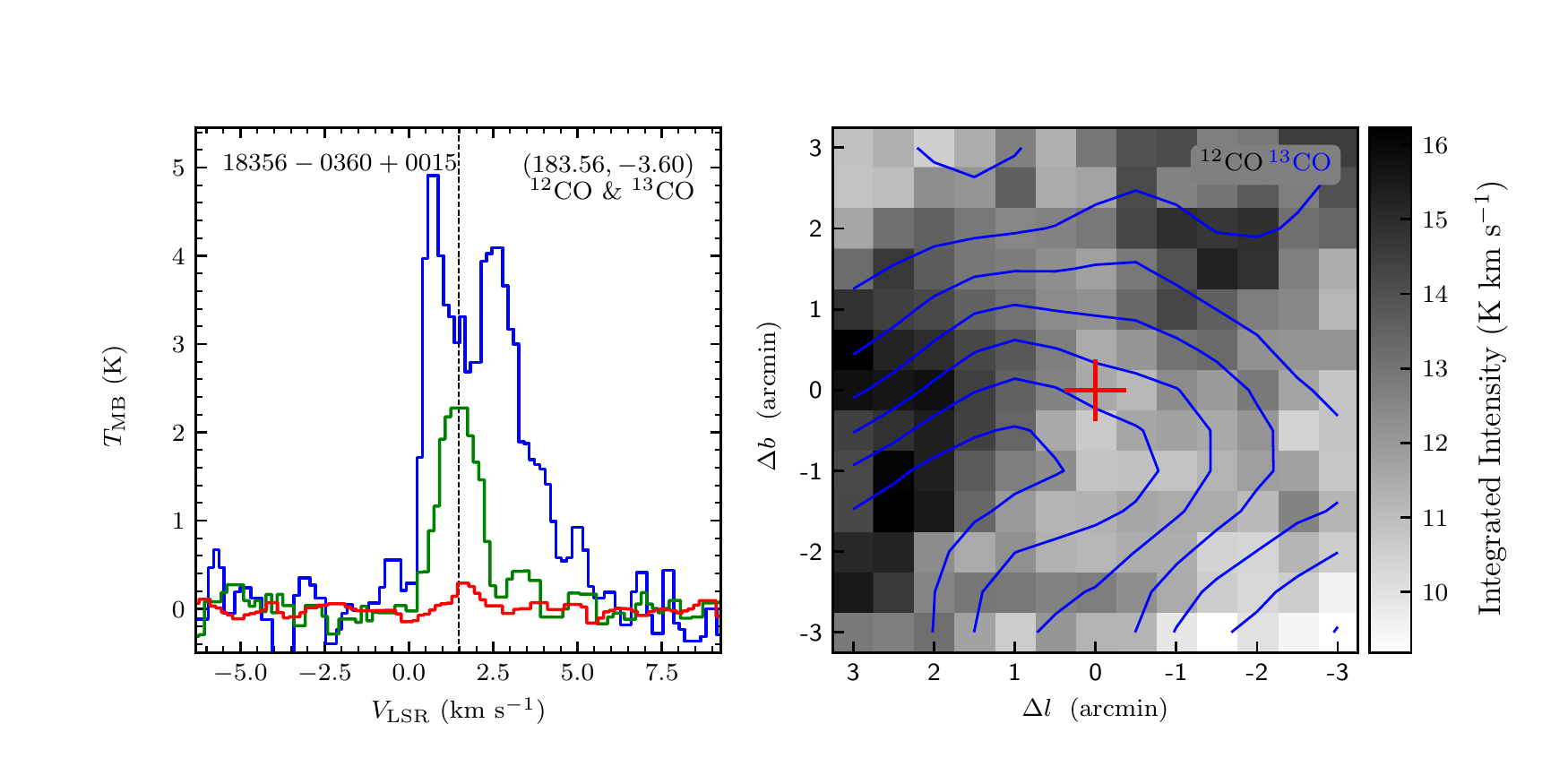}
\includegraphics[width=9.0cm,angle=0]{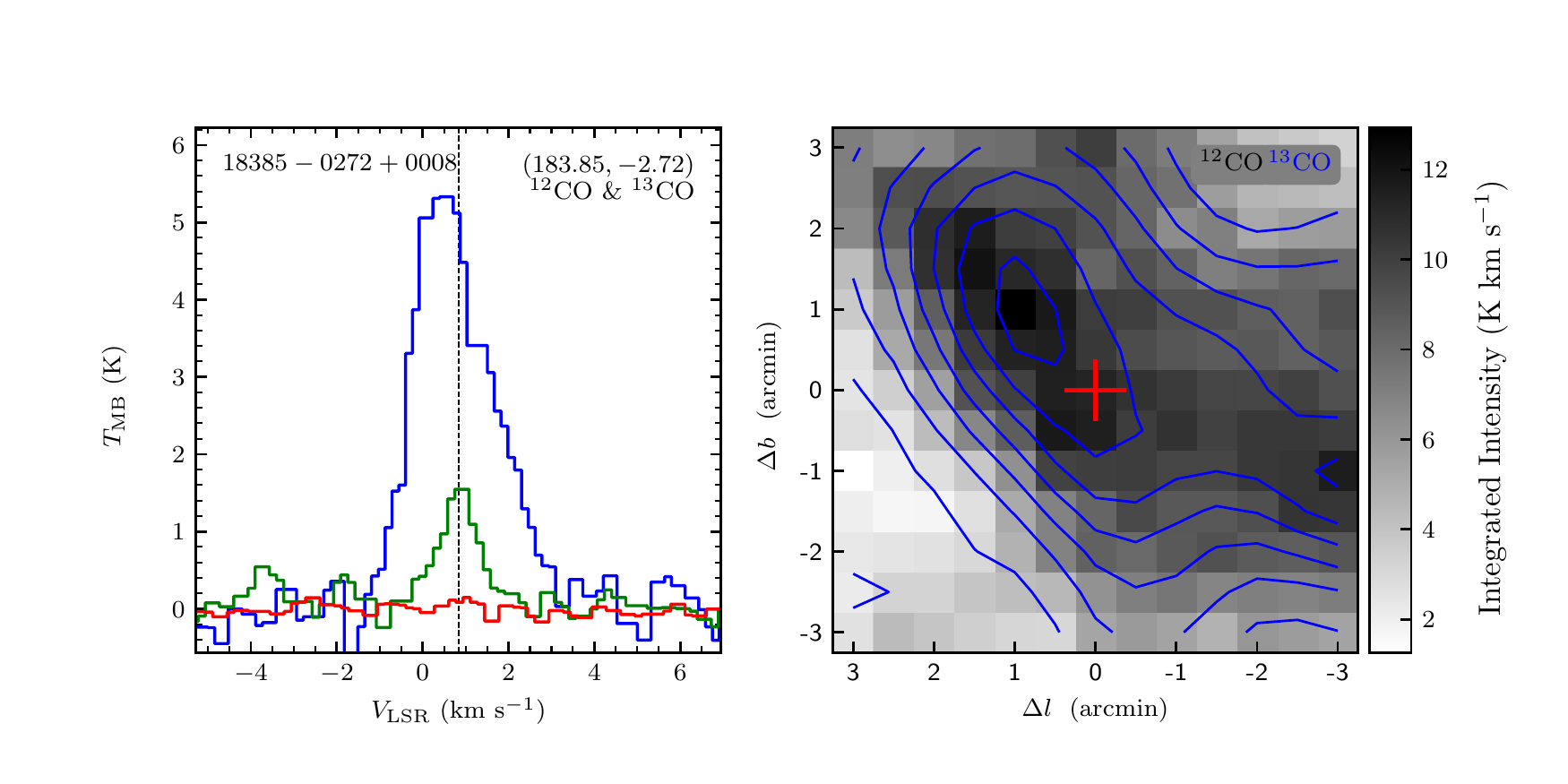}
\vspace{-0.5cm}

\includegraphics[width=9.0cm,angle=0]{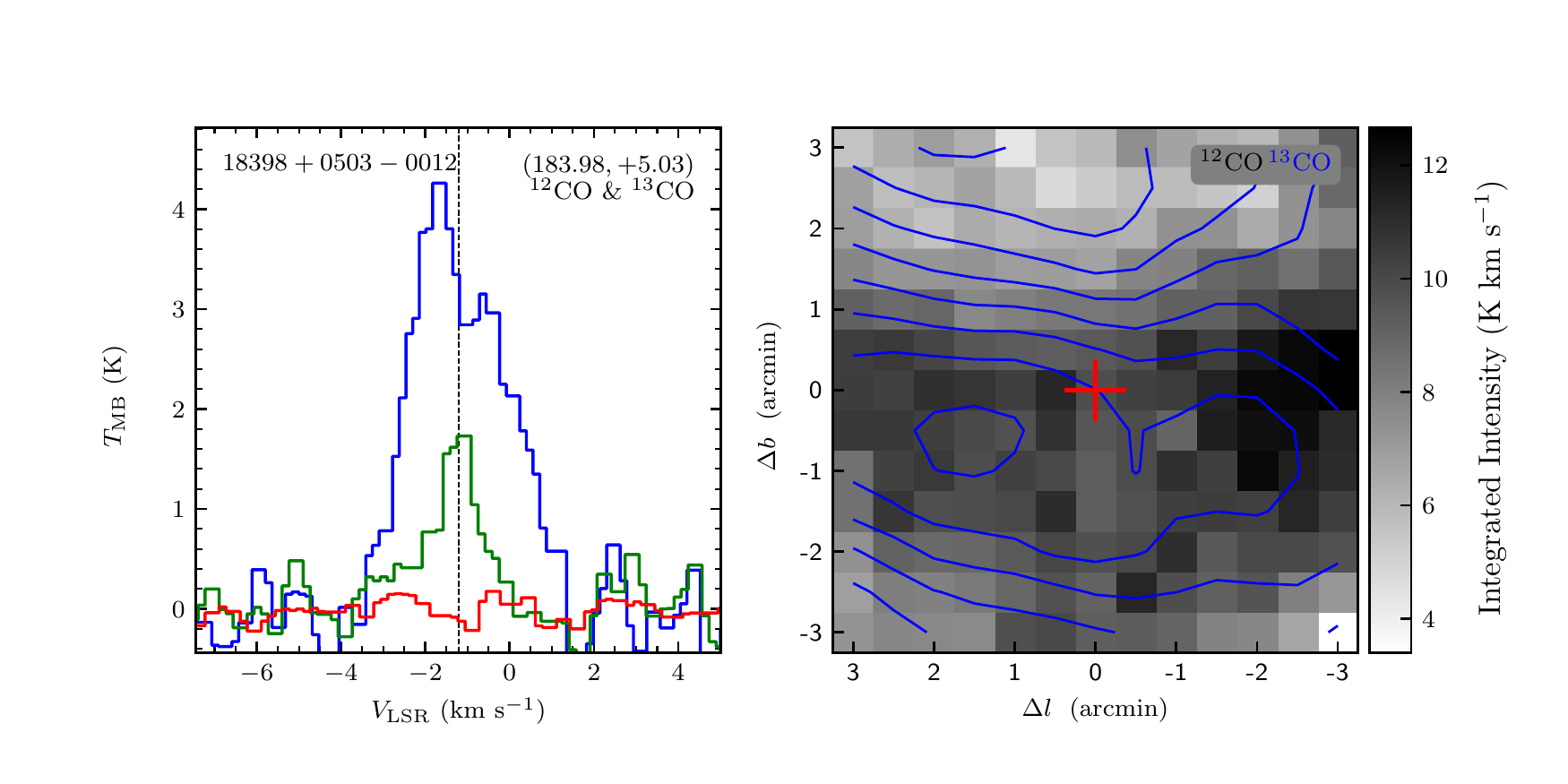}
\includegraphics[width=9.0cm,angle=0]{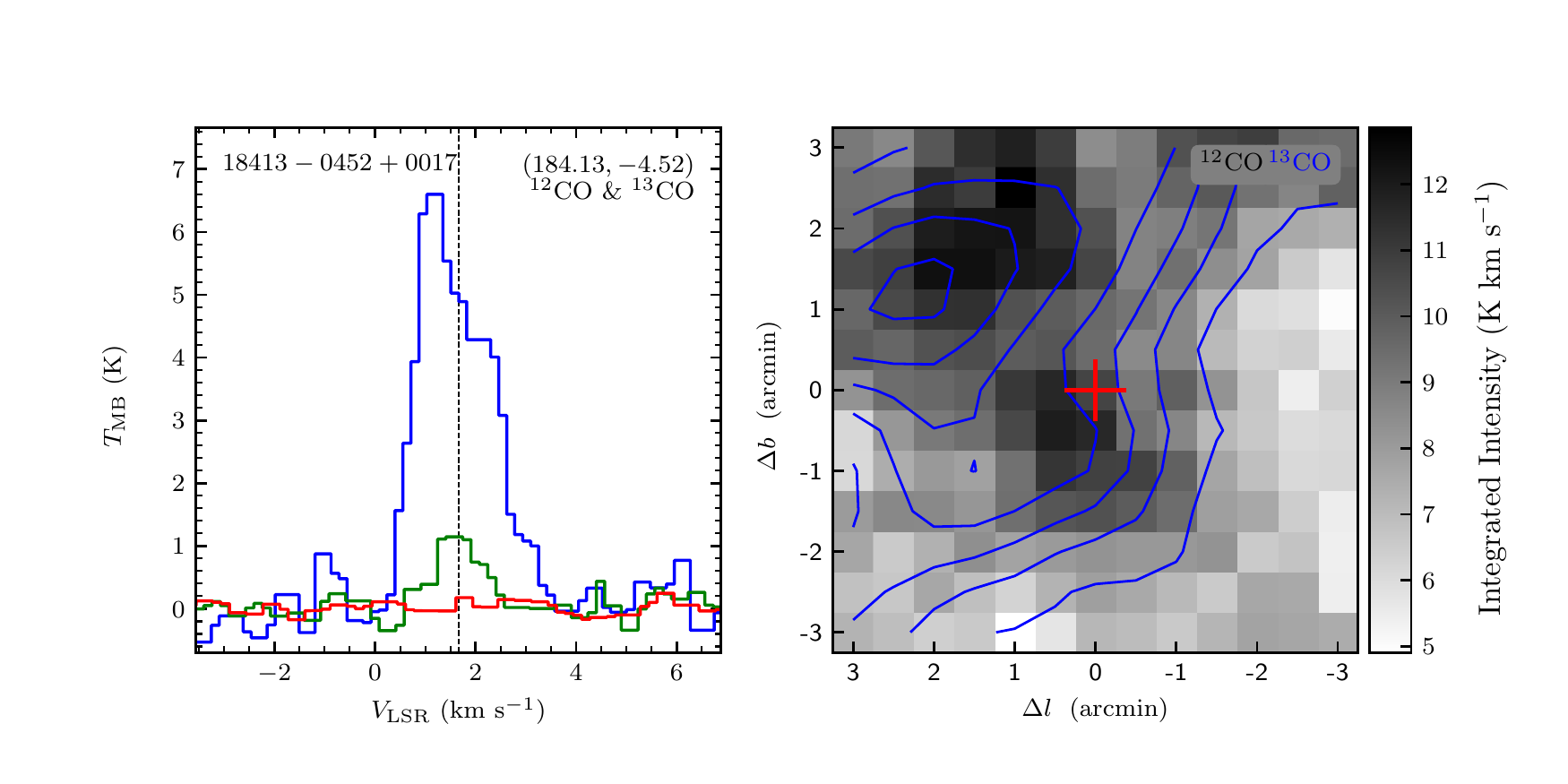}
\end{figure}
\clearpage

\begin{figure}
\includegraphics[width=9.0cm,angle=0]{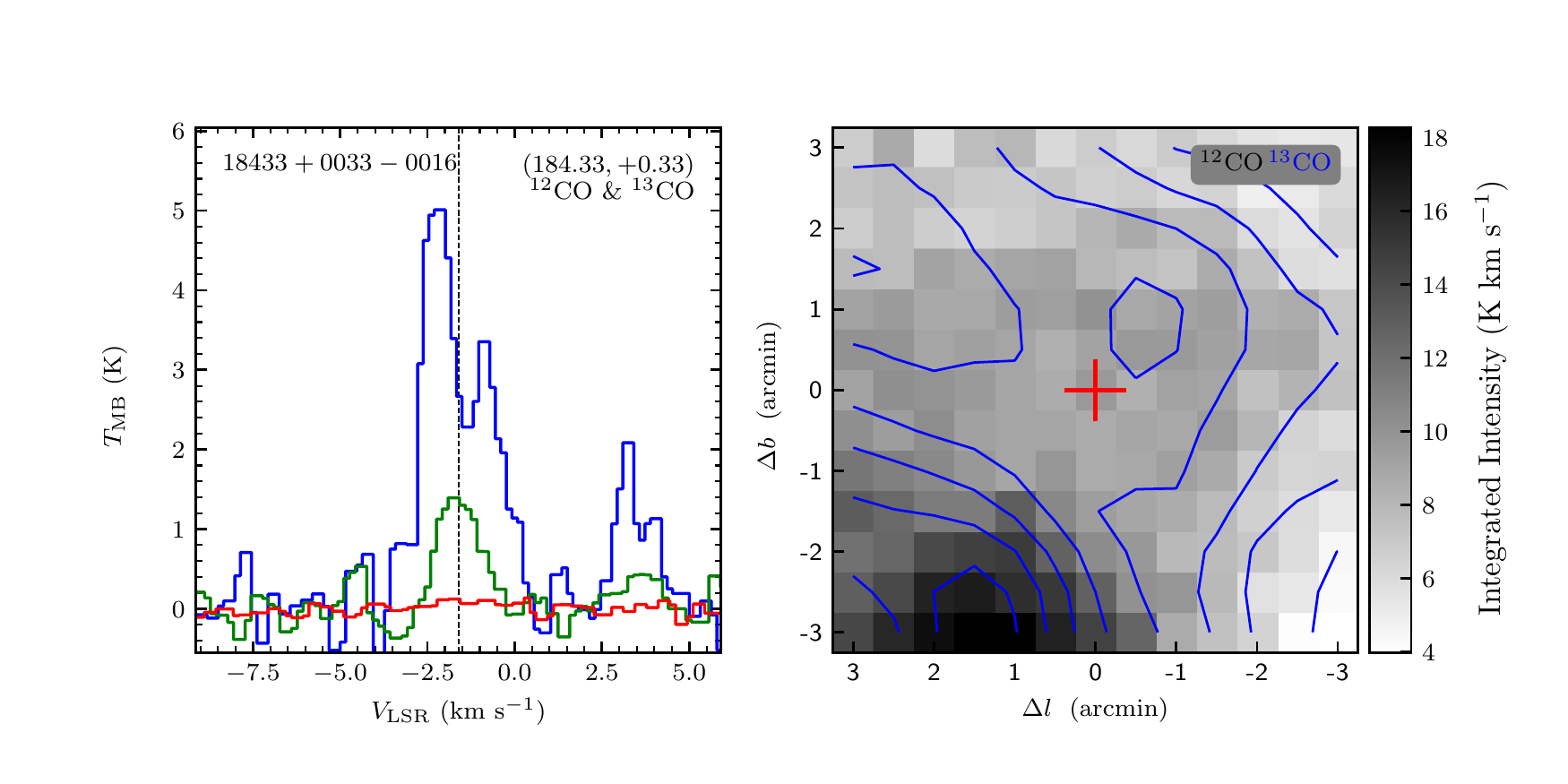}
\includegraphics[width=9.0cm,angle=0]{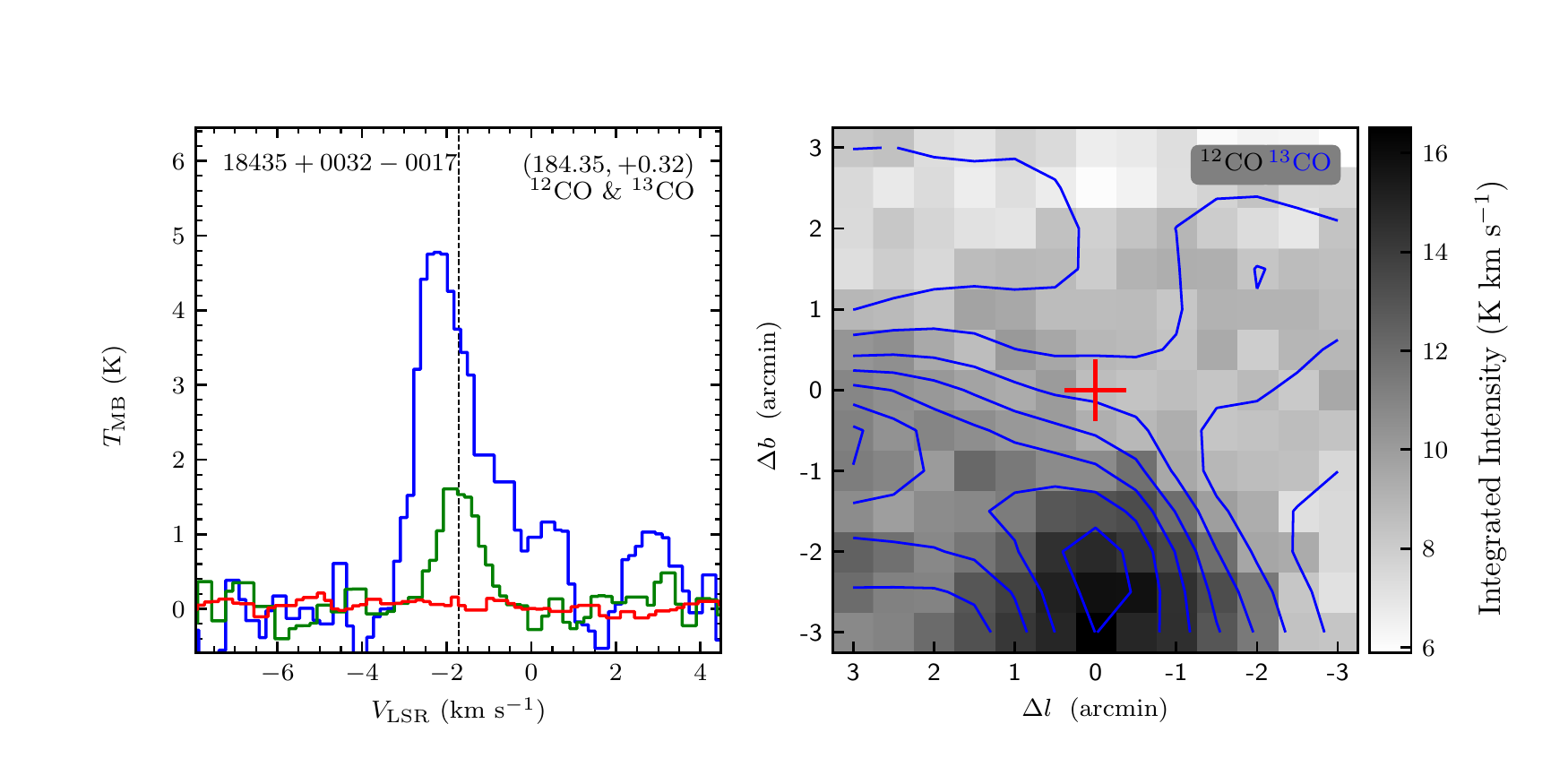}
\vspace{-0.5cm}

\includegraphics[width=9.0cm,angle=0]{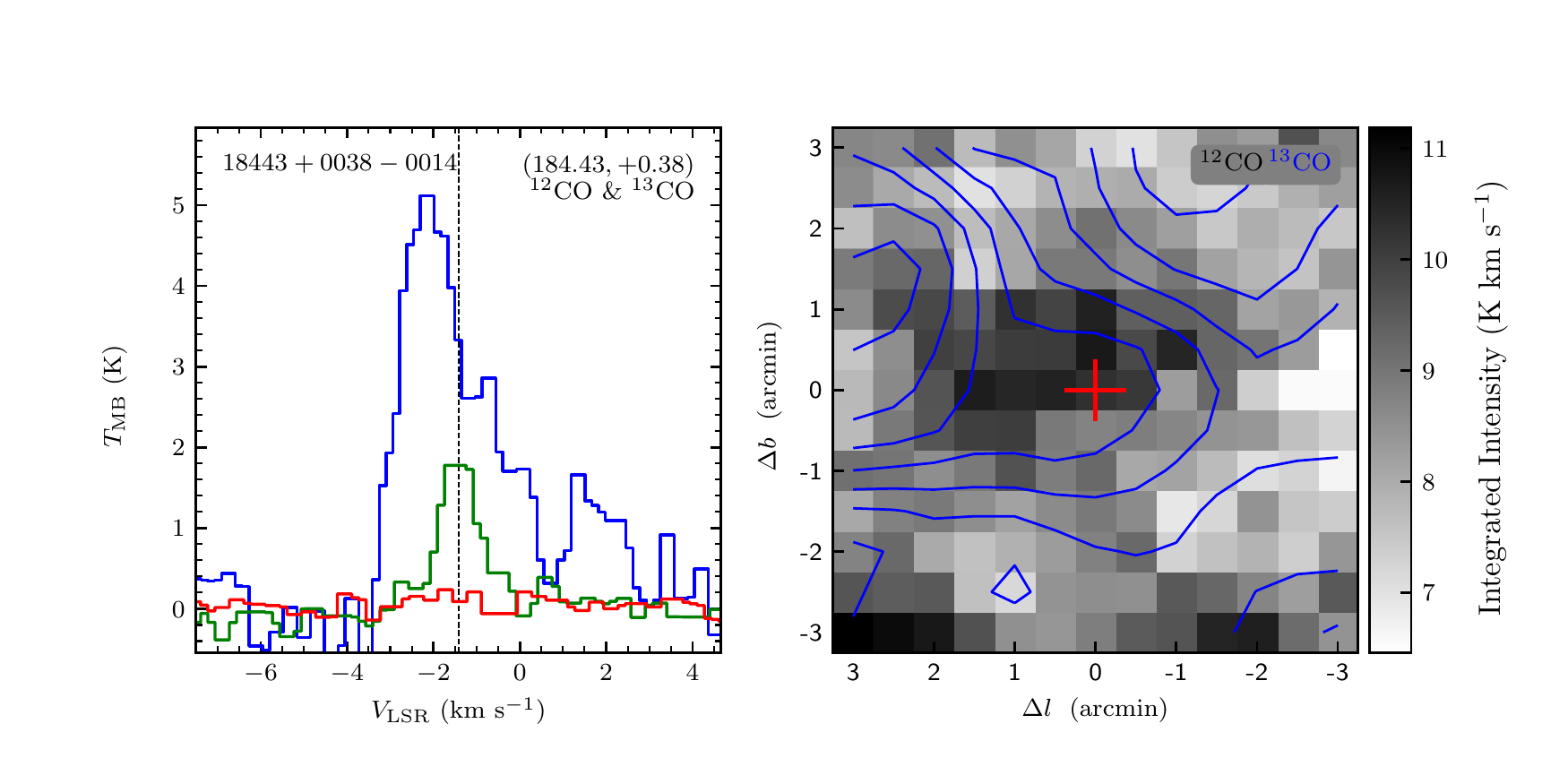}
\includegraphics[width=9.0cm,angle=0]{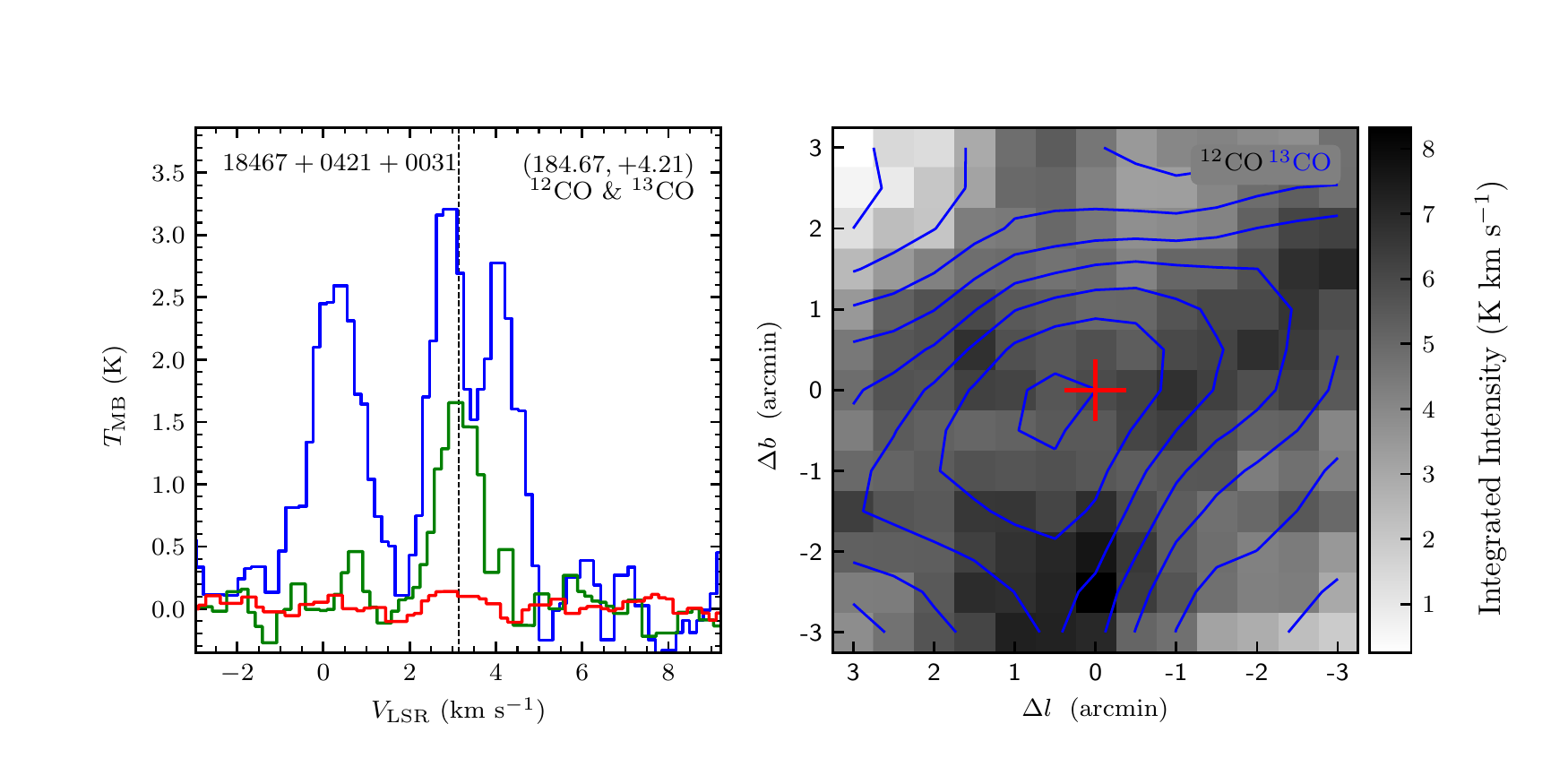}
\vspace{-0.5cm}

\includegraphics[width=9.0cm,angle=0]{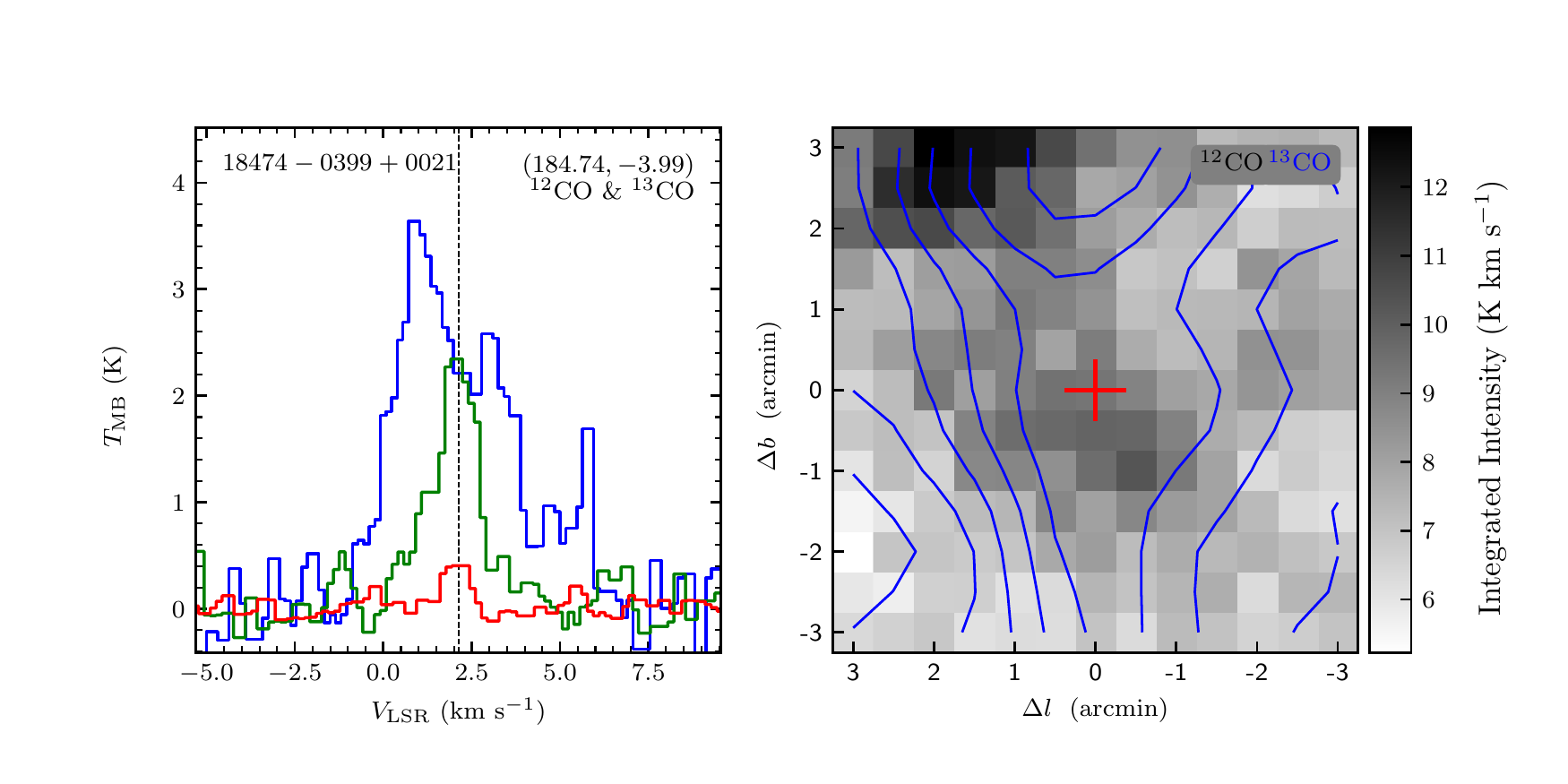}
\includegraphics[width=9.0cm,angle=0]{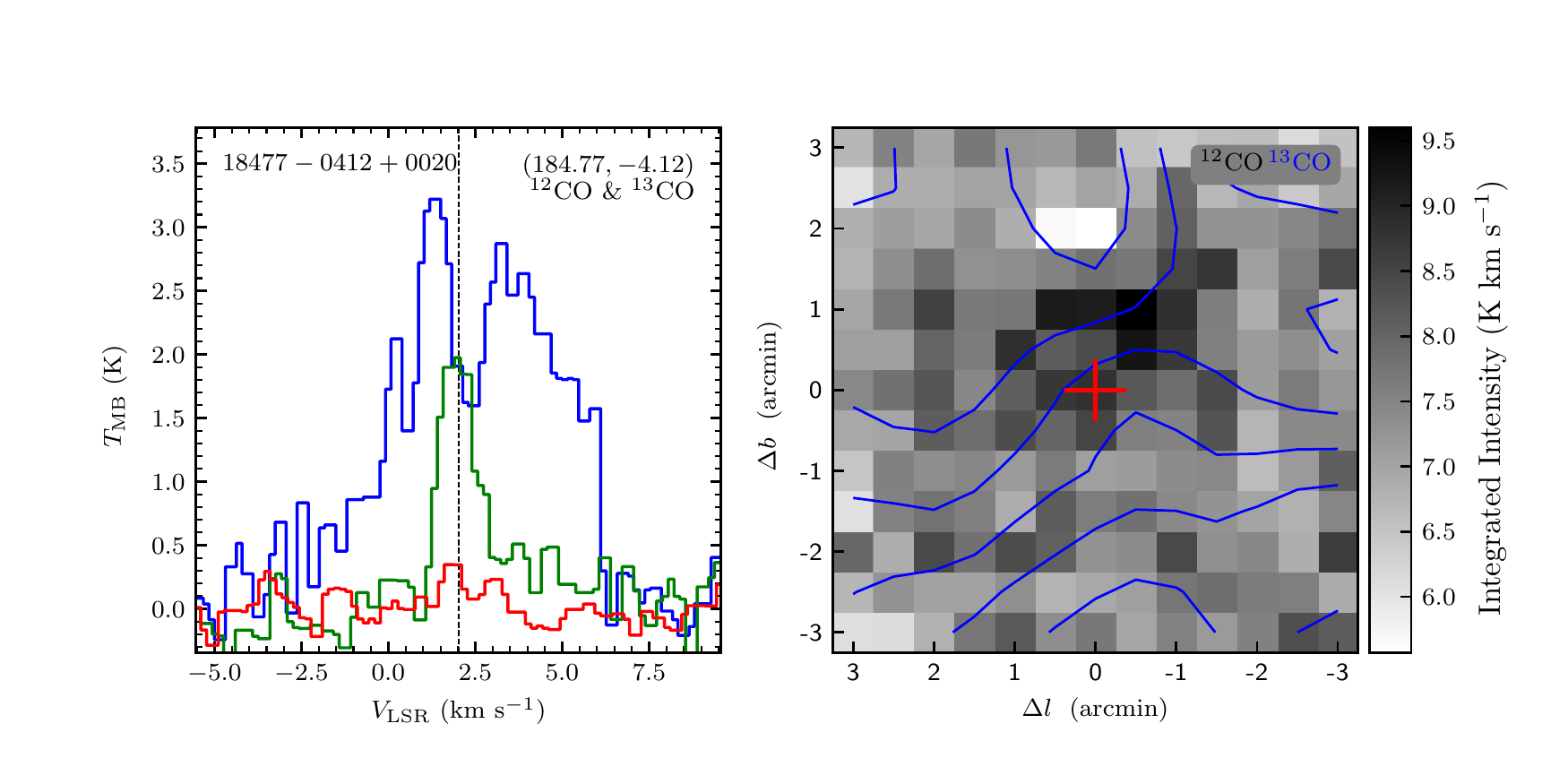}
\vspace{-0.5cm}

\includegraphics[width=9.0cm,angle=0]{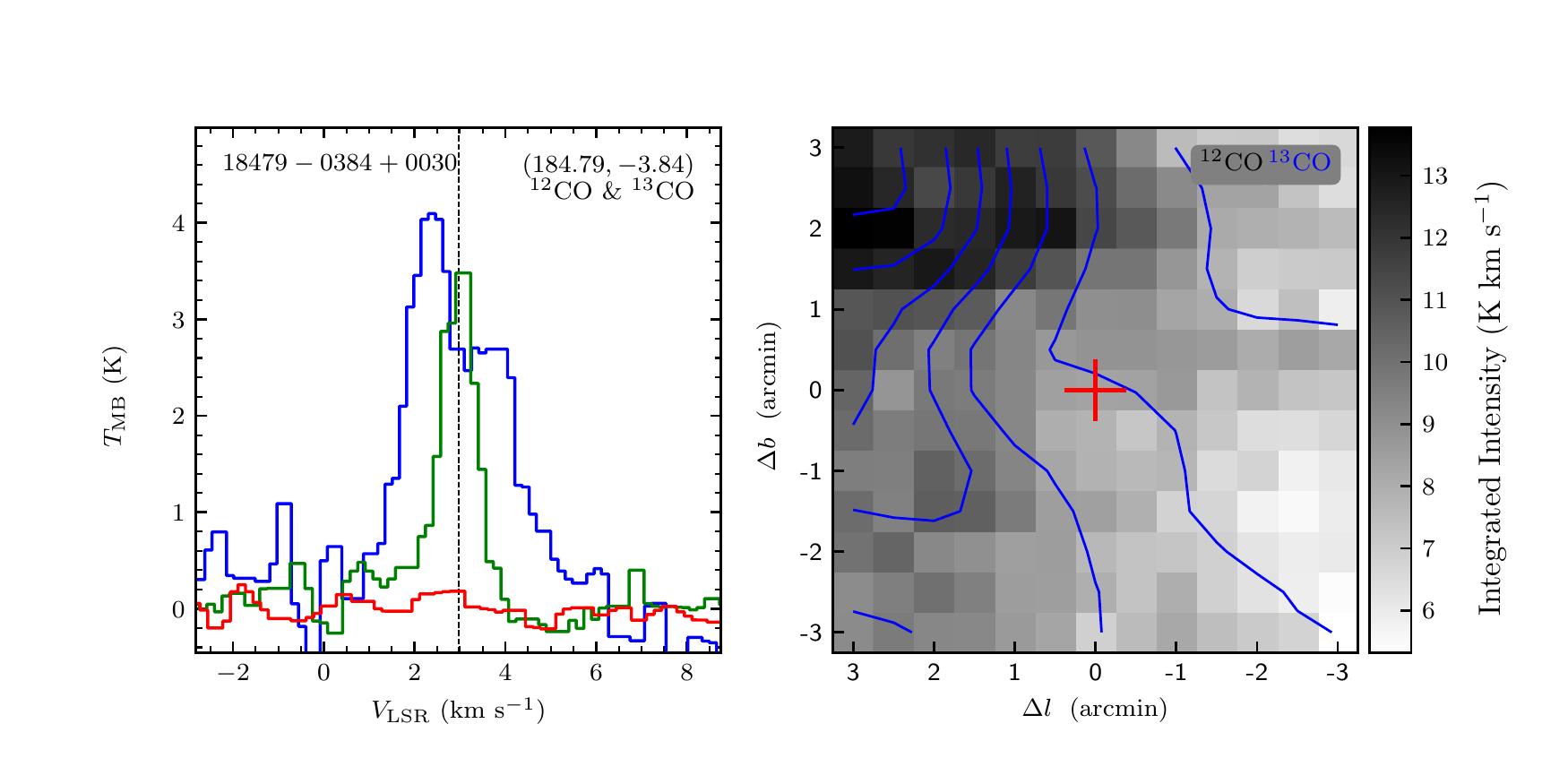}
\includegraphics[width=9.0cm,angle=0]{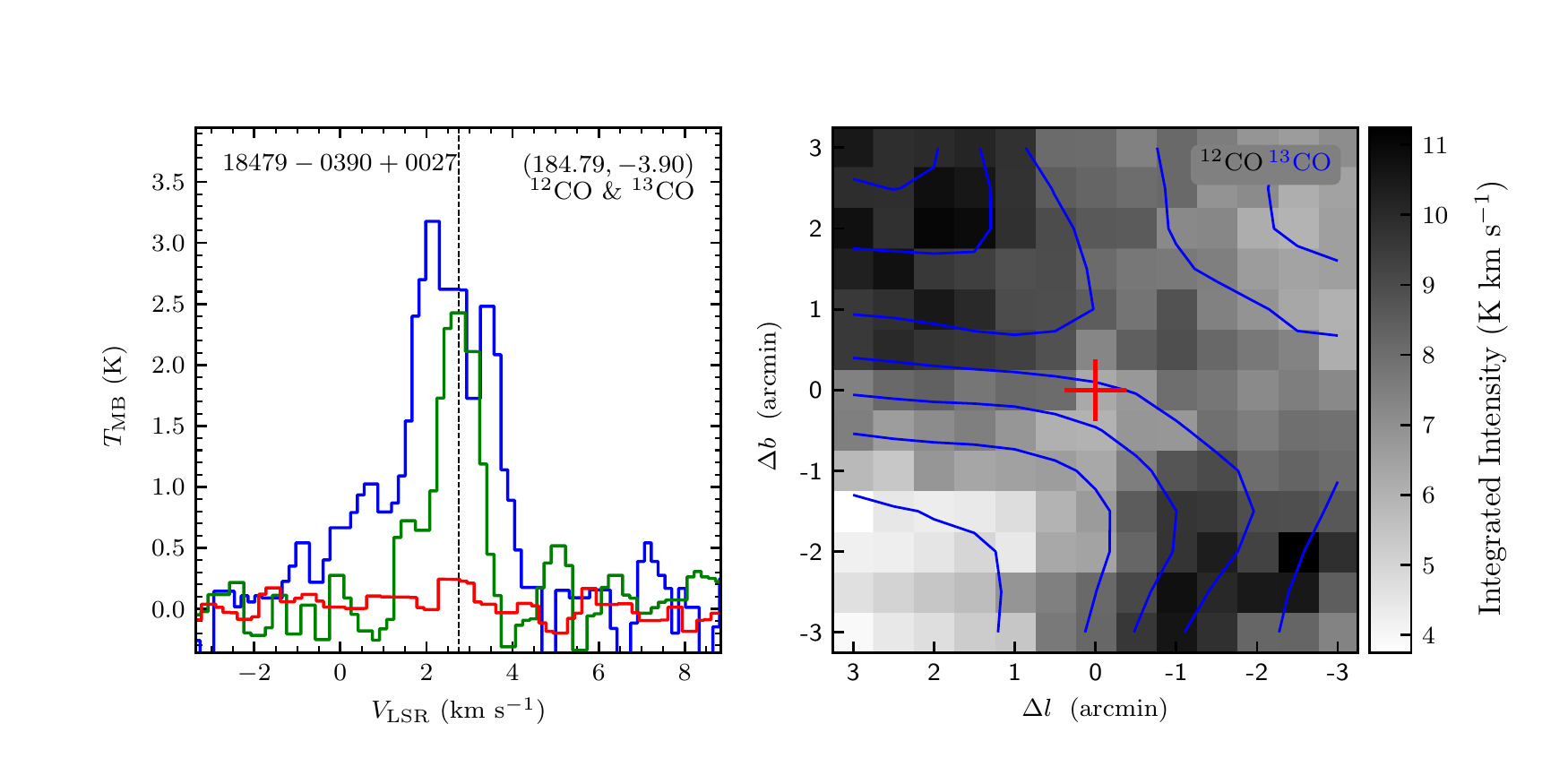}
\vspace{-0.5cm}

\includegraphics[width=9.0cm,angle=0]{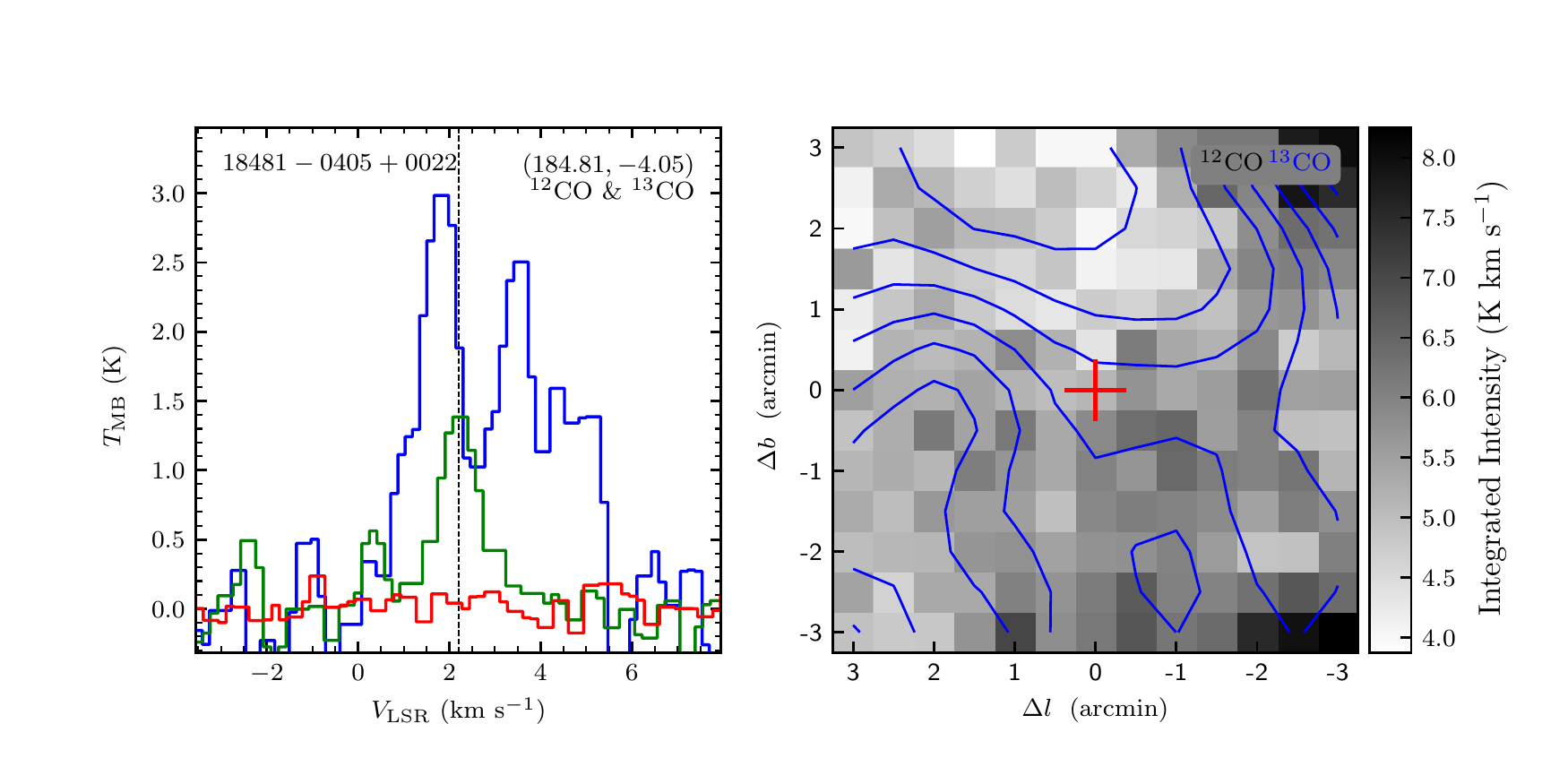}
\includegraphics[width=9.0cm,angle=0]{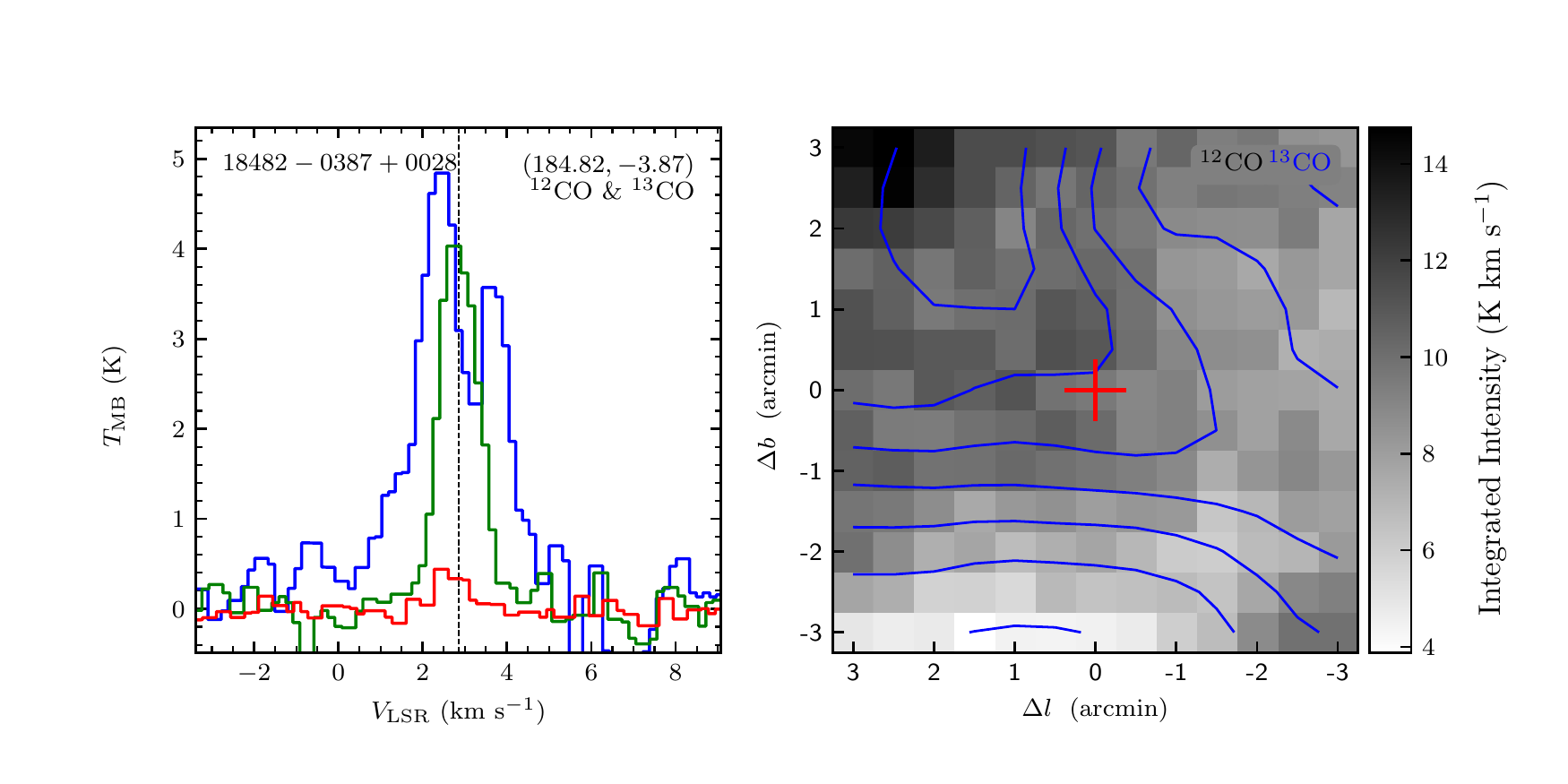}
\end{figure}
\clearpage

\begin{figure}
\includegraphics[width=9.0cm,angle=0]{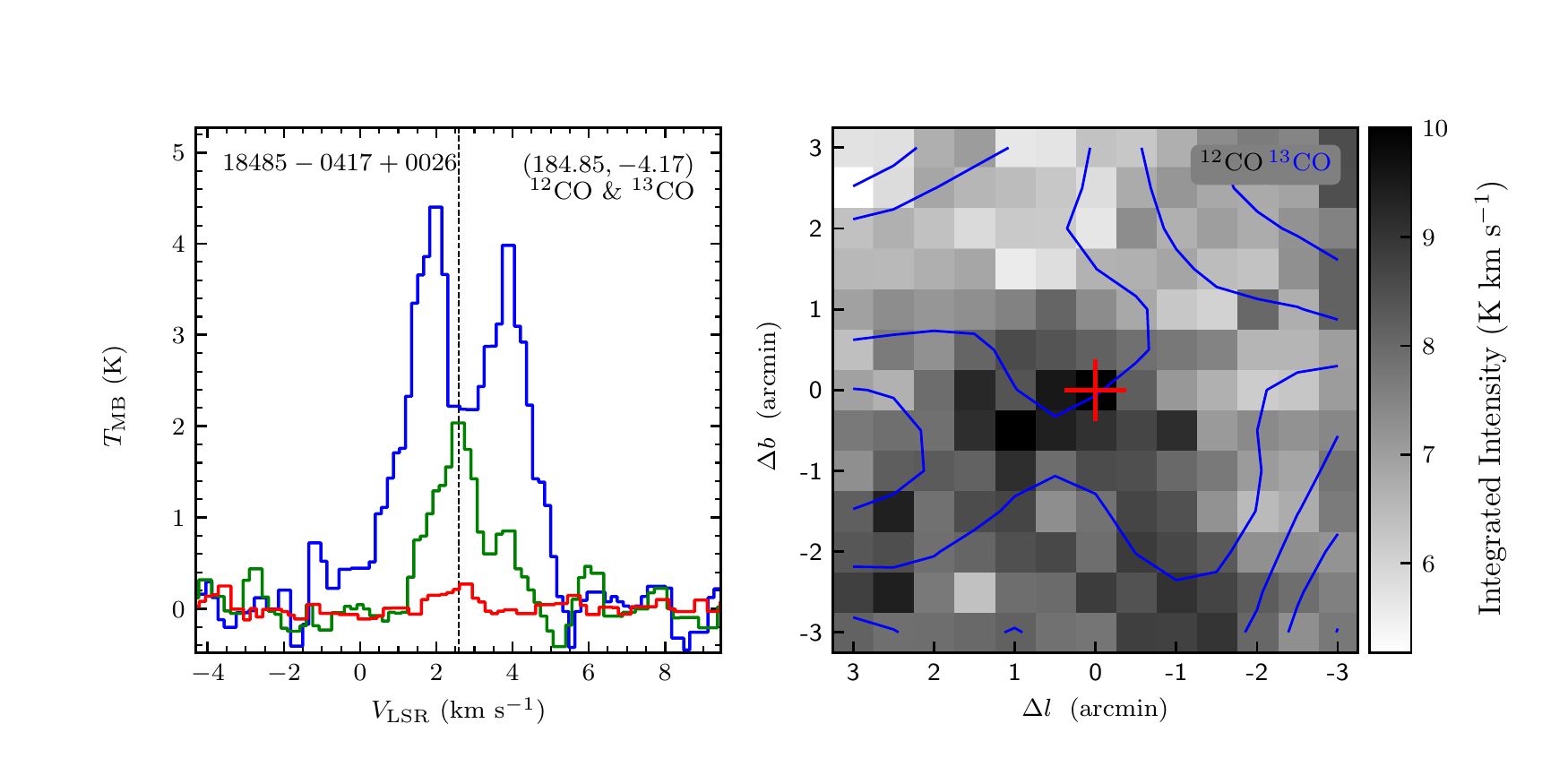}
\includegraphics[width=9.0cm,angle=0]{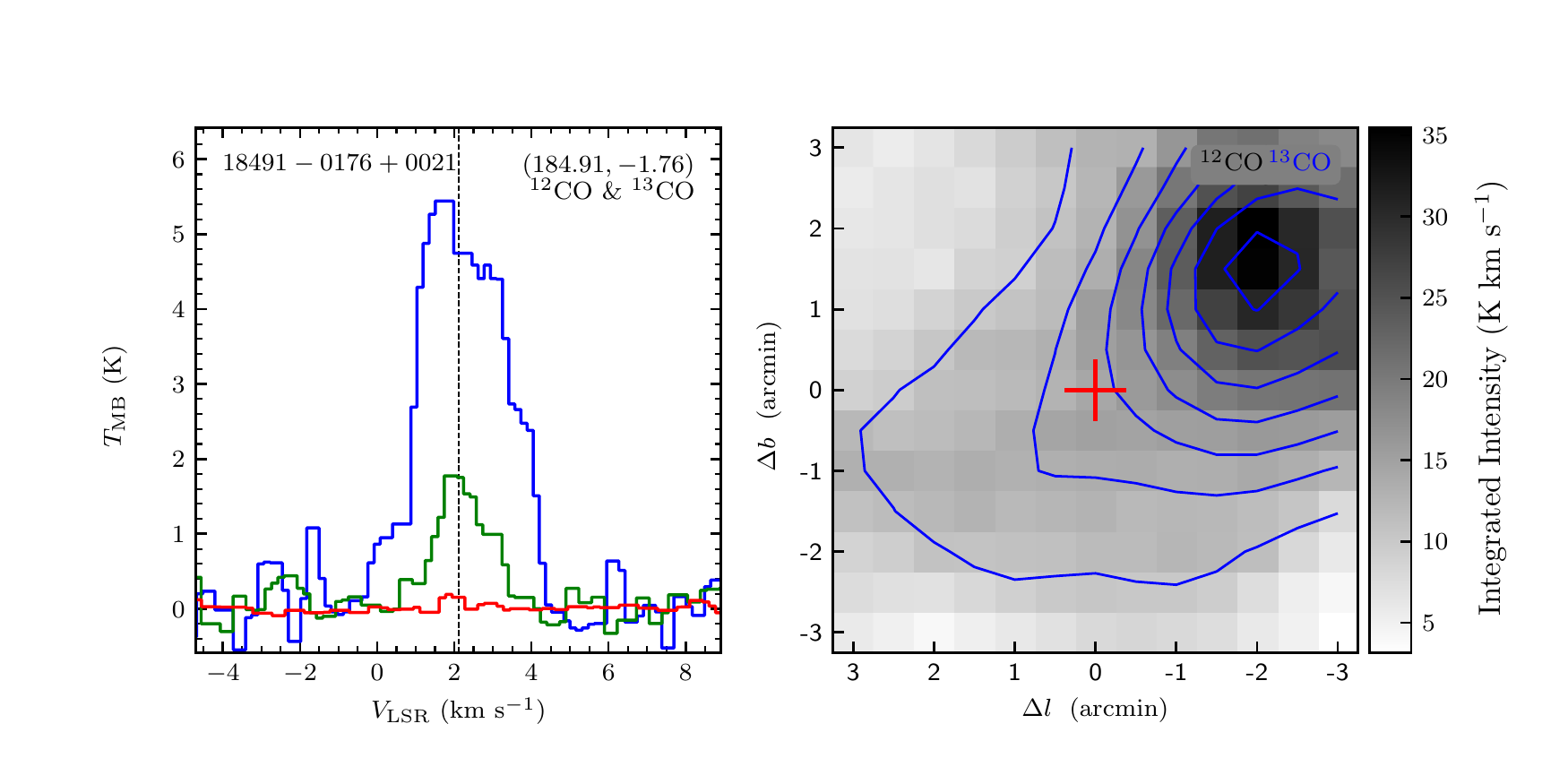}
\vspace{-0.5cm}

\includegraphics[width=9.0cm,angle=0]{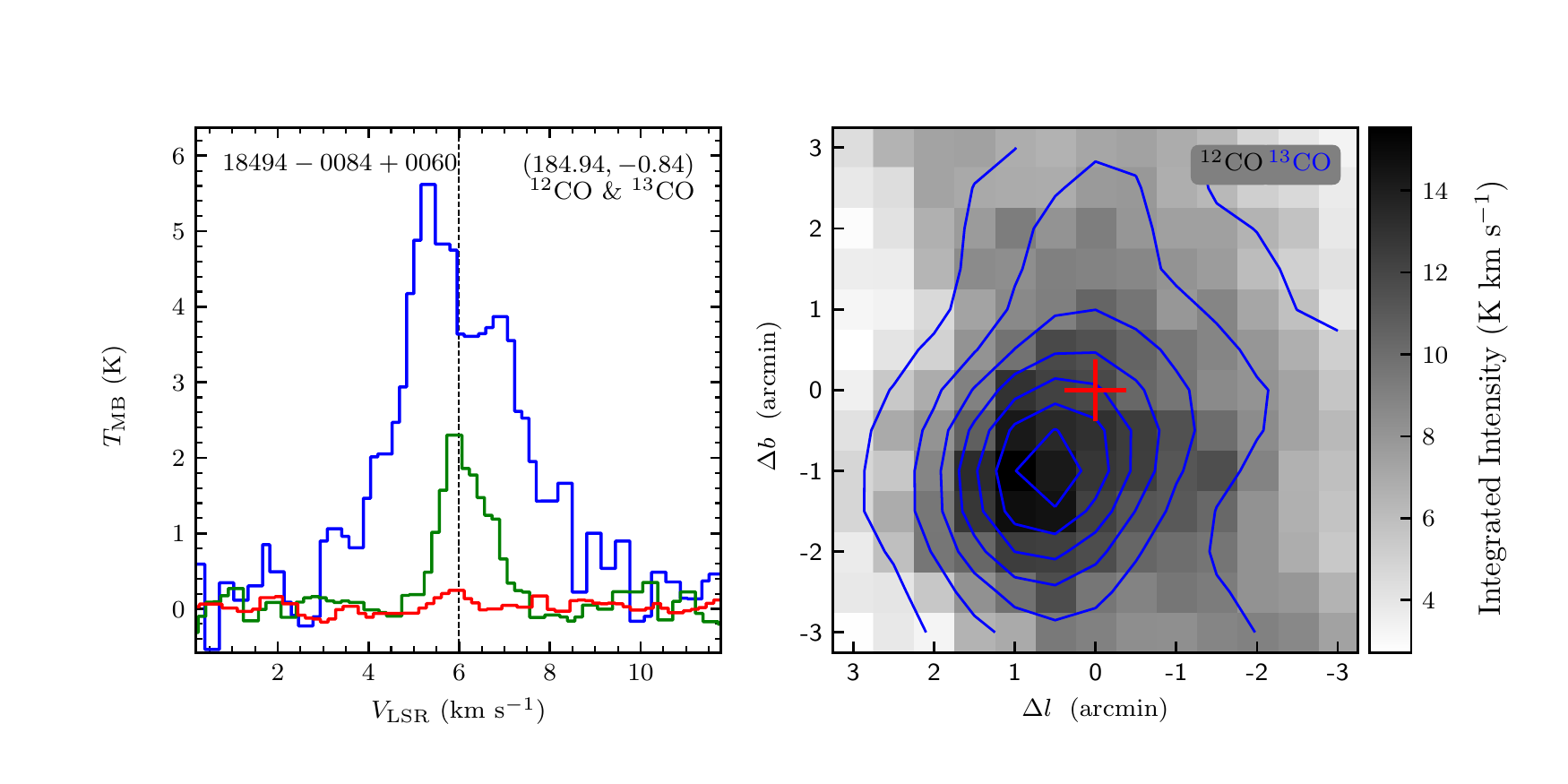}
\includegraphics[width=9.0cm,angle=0]{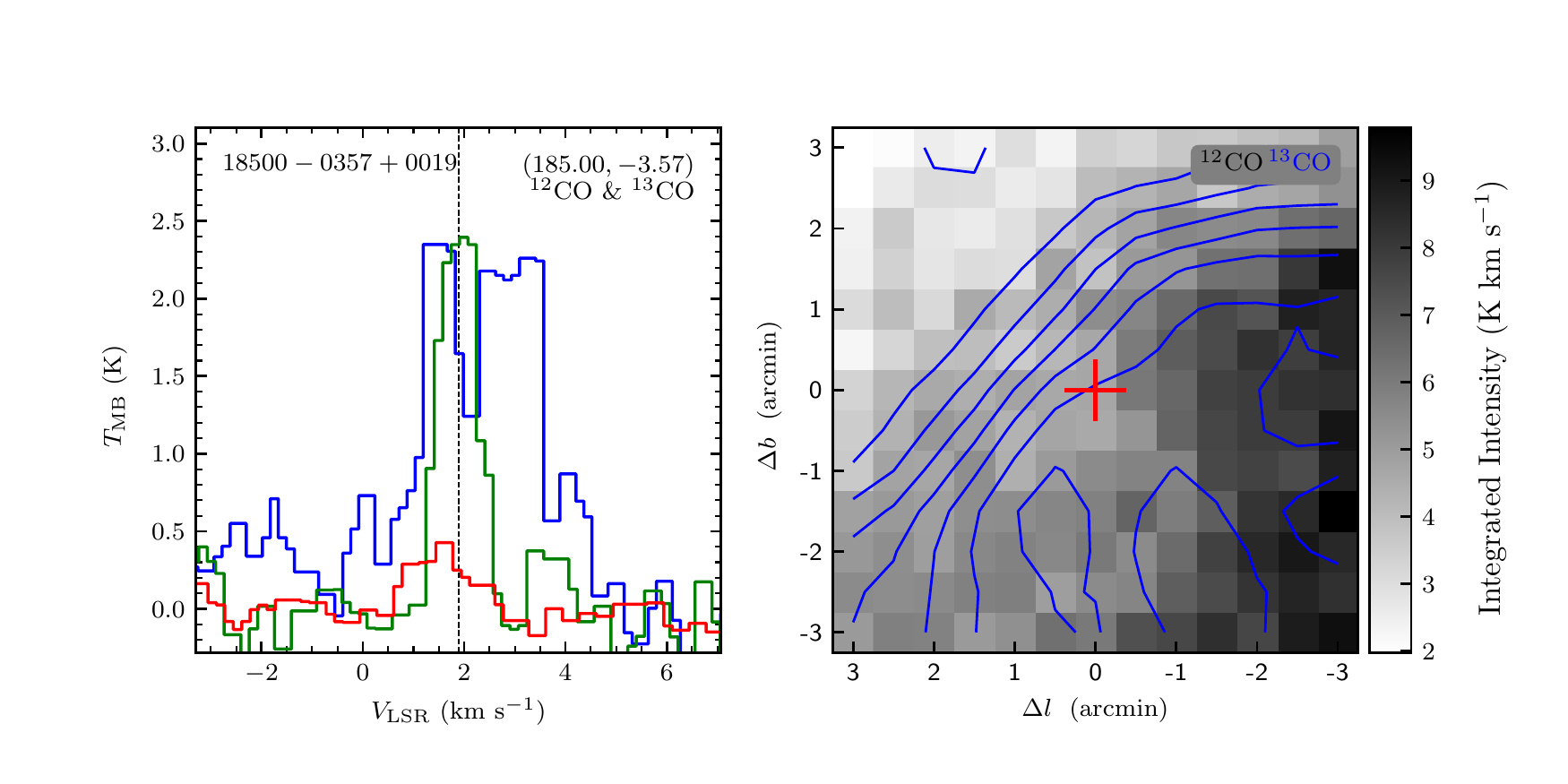}
\vspace{-0.5cm}

\includegraphics[width=9.0cm,angle=0]{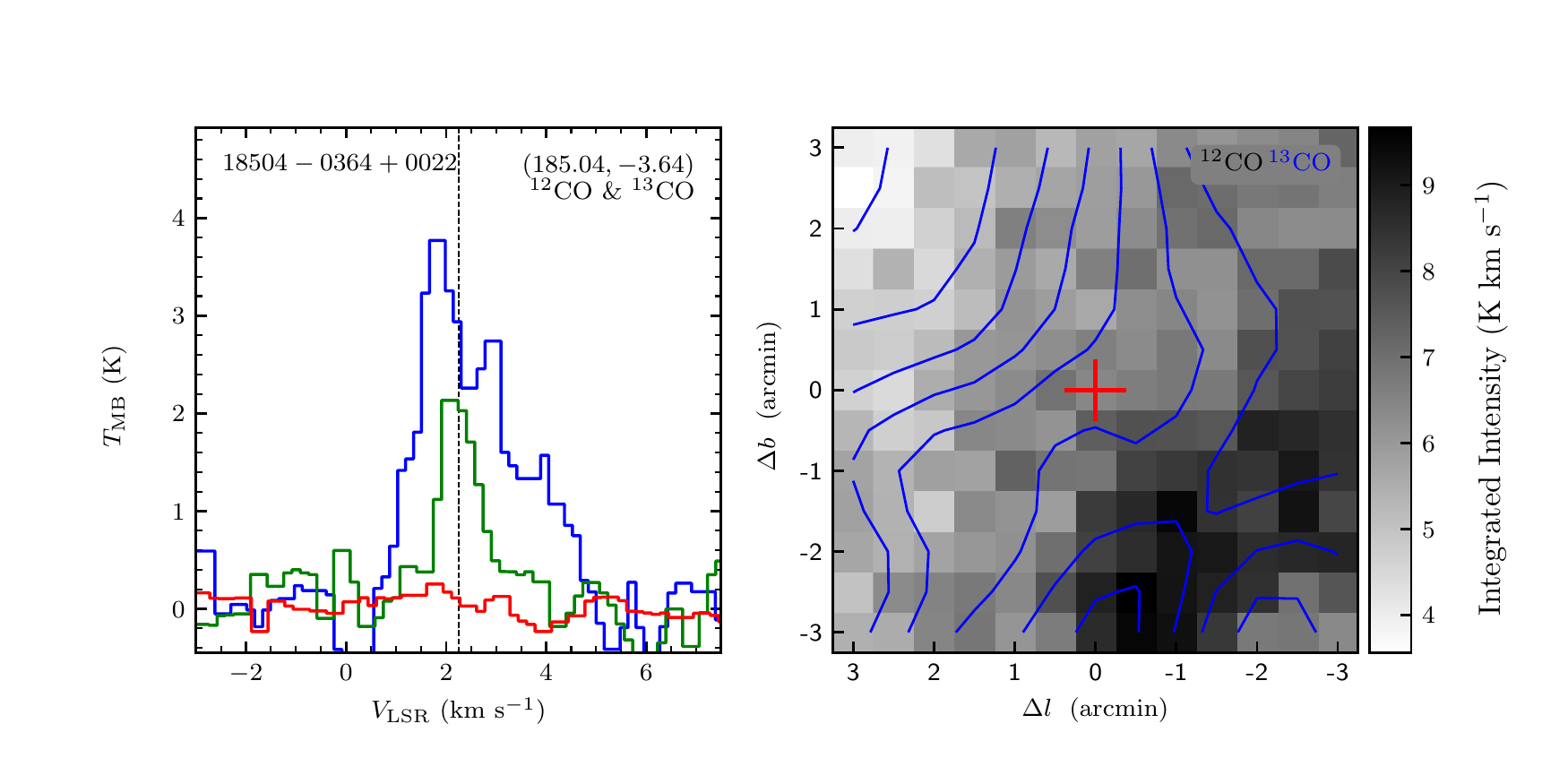}
\includegraphics[width=9.0cm,angle=0]{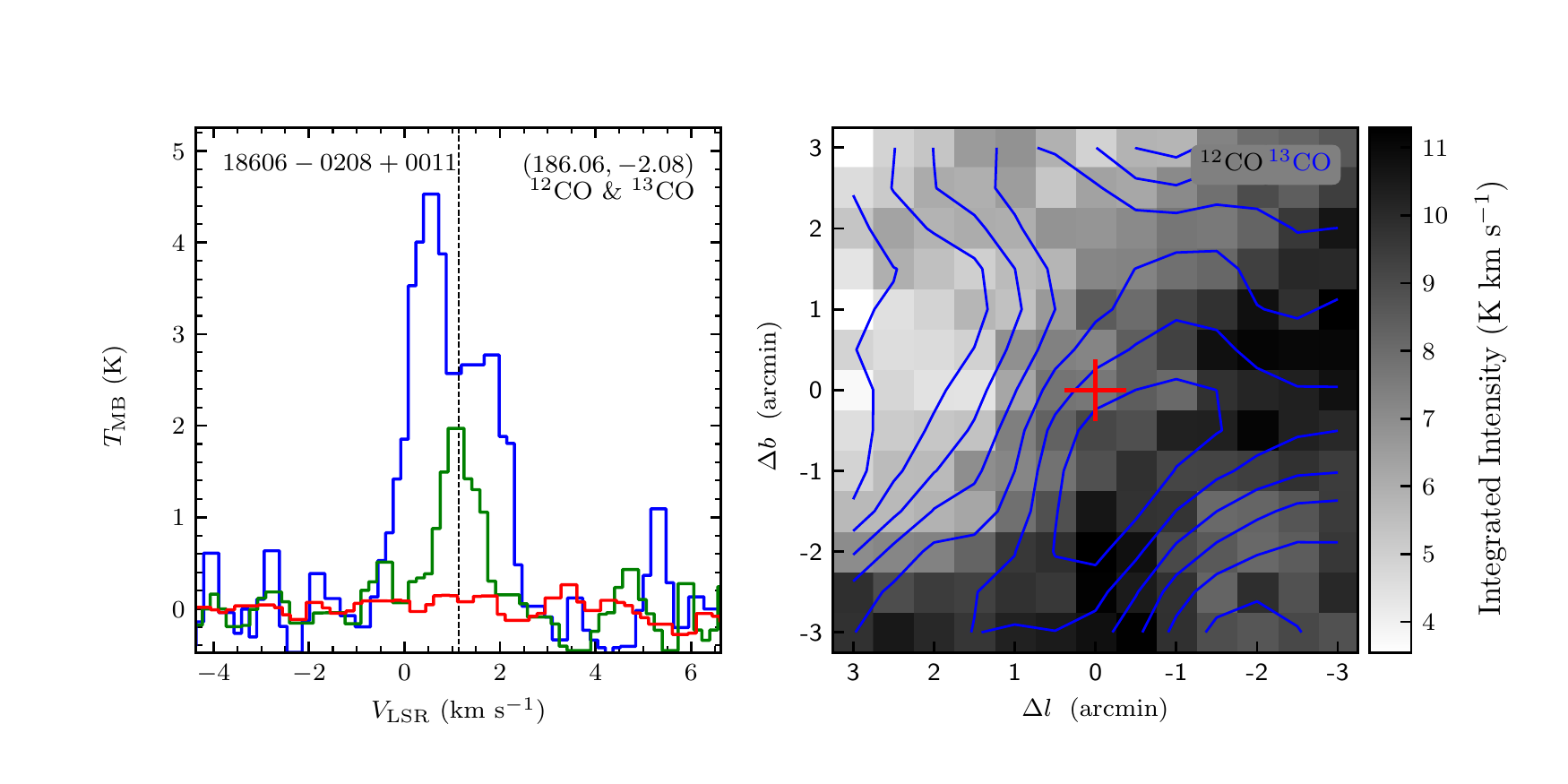}
\vspace{-0.5cm}

\includegraphics[width=9.0cm,angle=0]{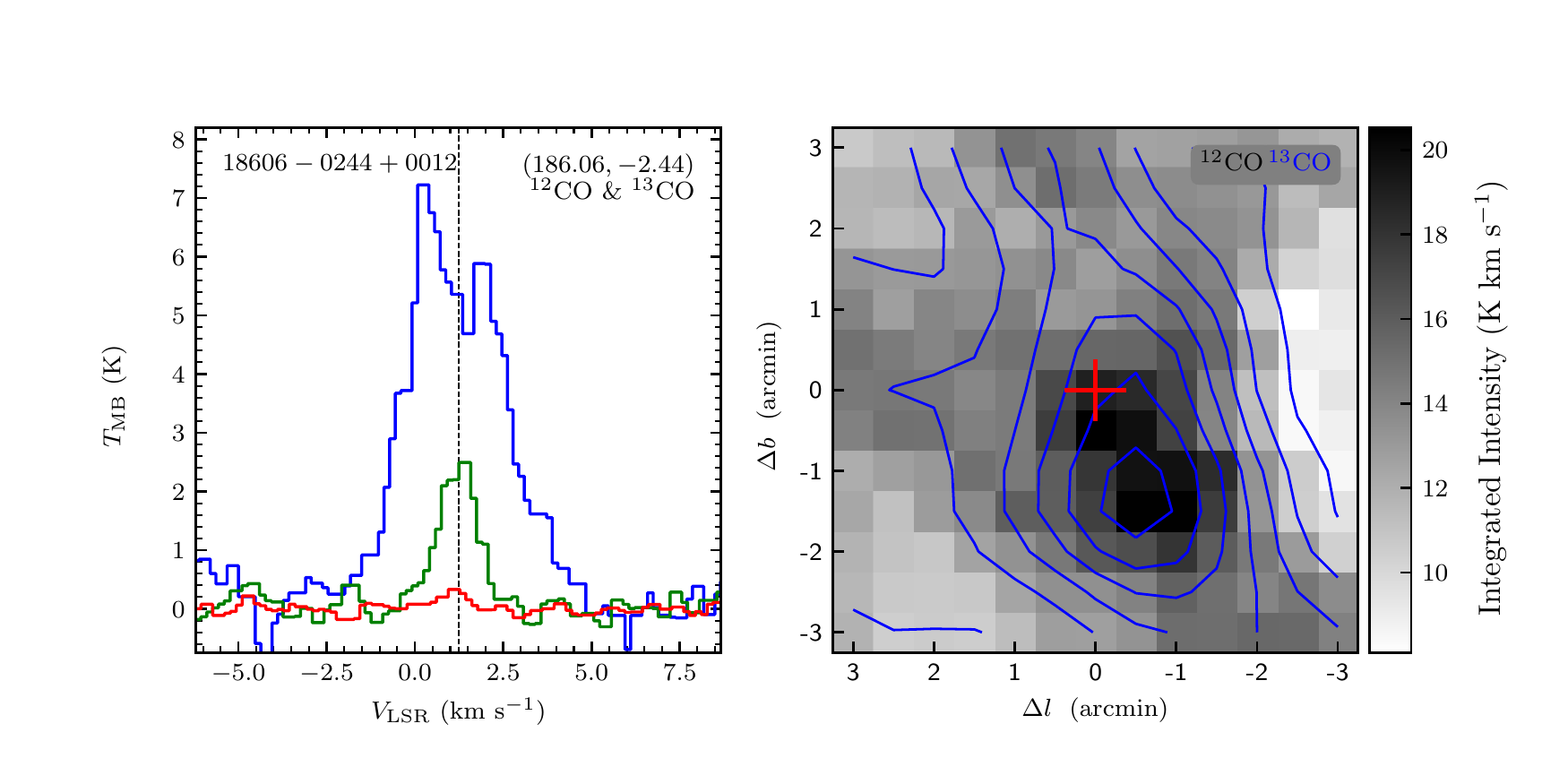}
\includegraphics[width=9.0cm,angle=0]{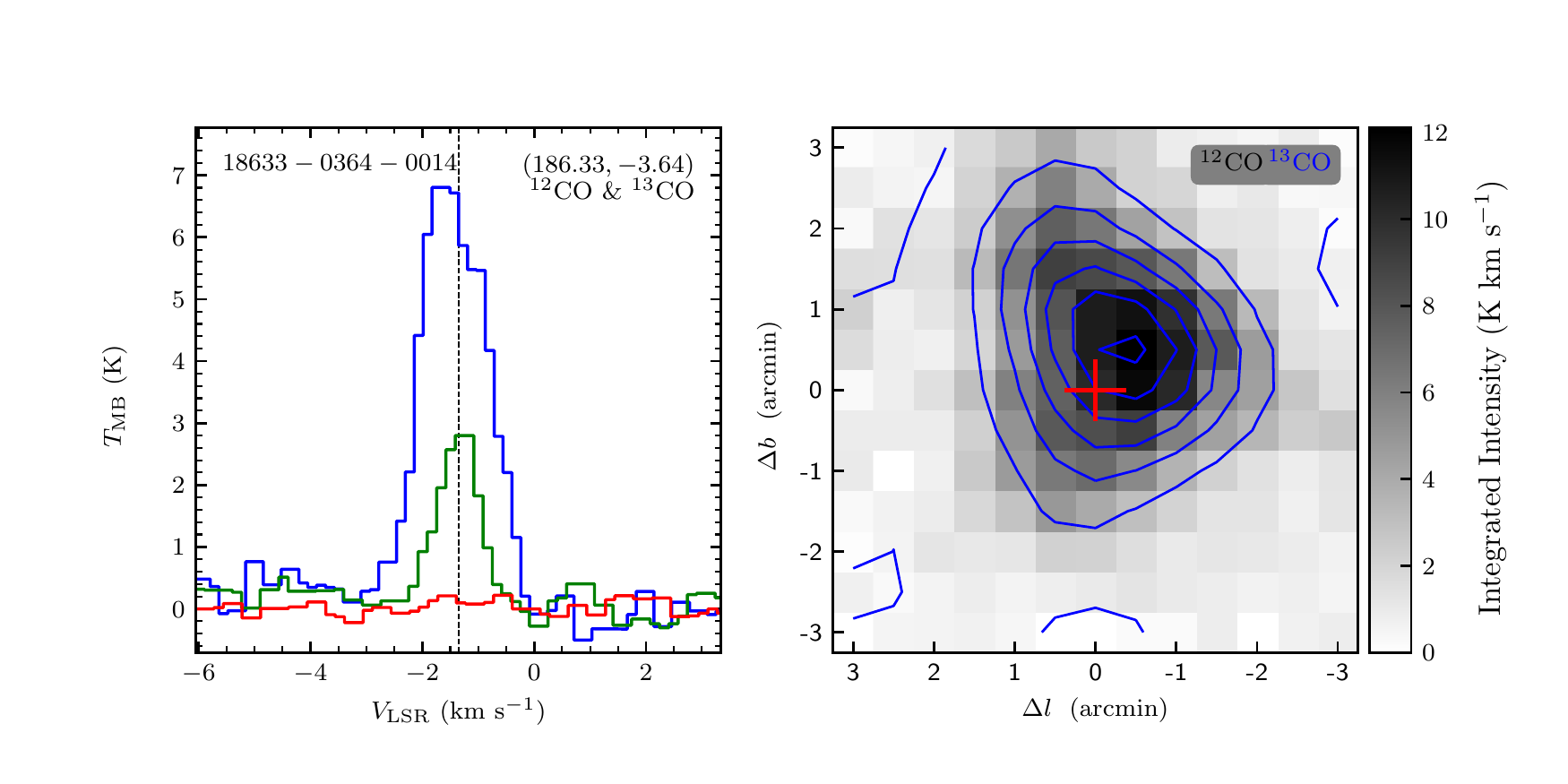}
\vspace{-0.5cm}

\includegraphics[width=9.0cm,angle=0]{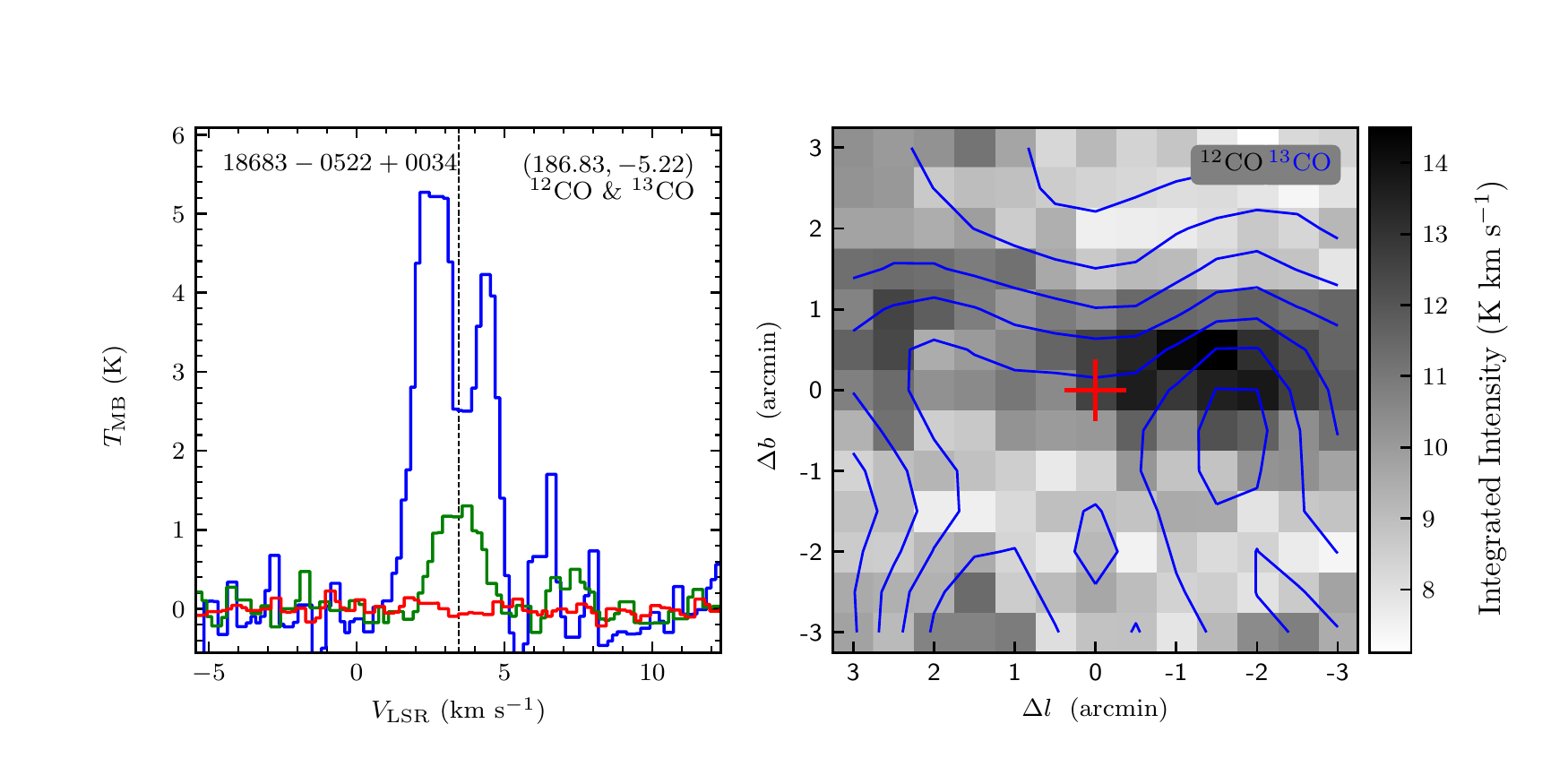}
\includegraphics[width=9.0cm,angle=0]{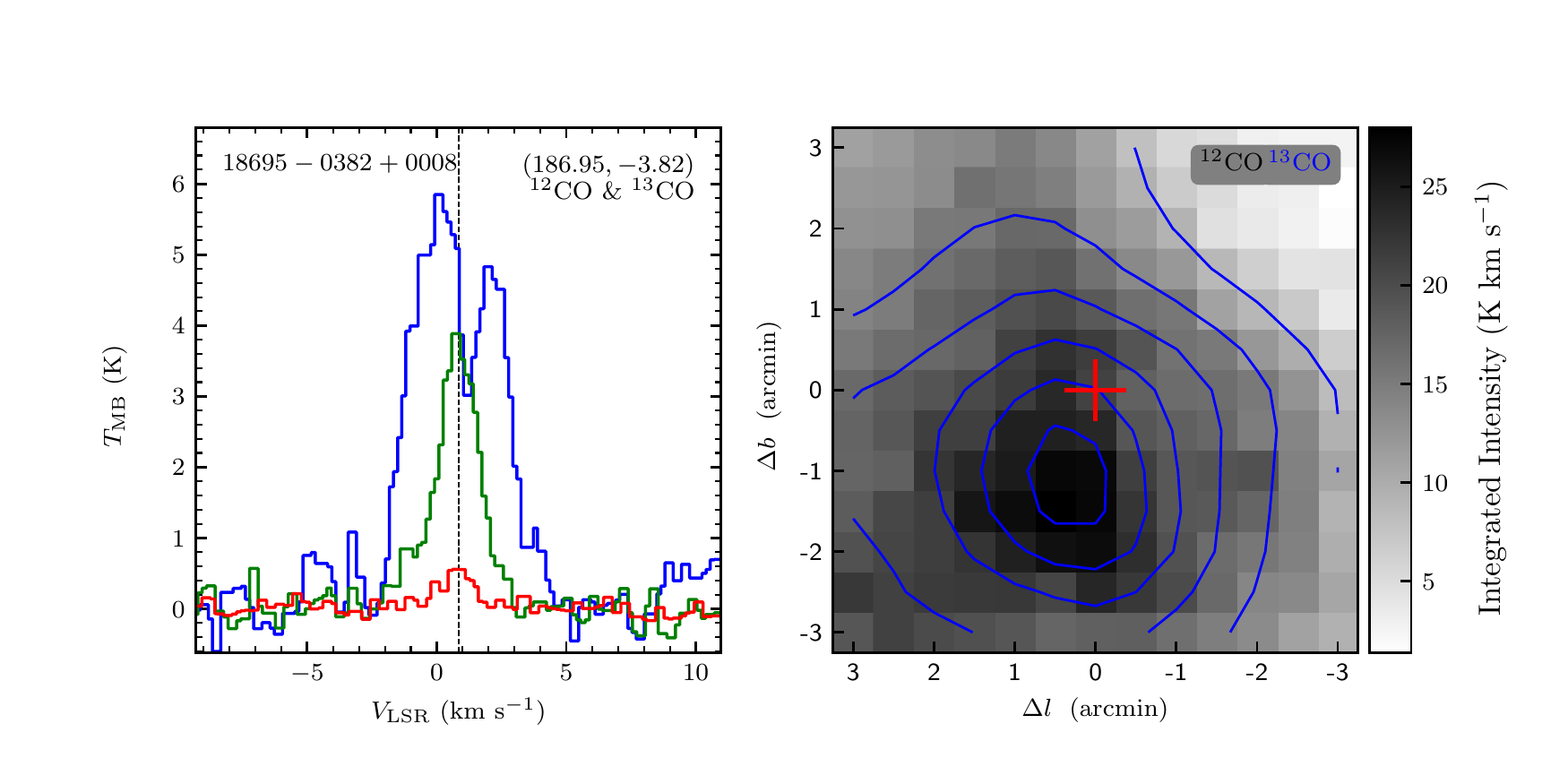}
\end{figure}
\clearpage

\begin{figure}
\includegraphics[width=9.0cm,angle=0]{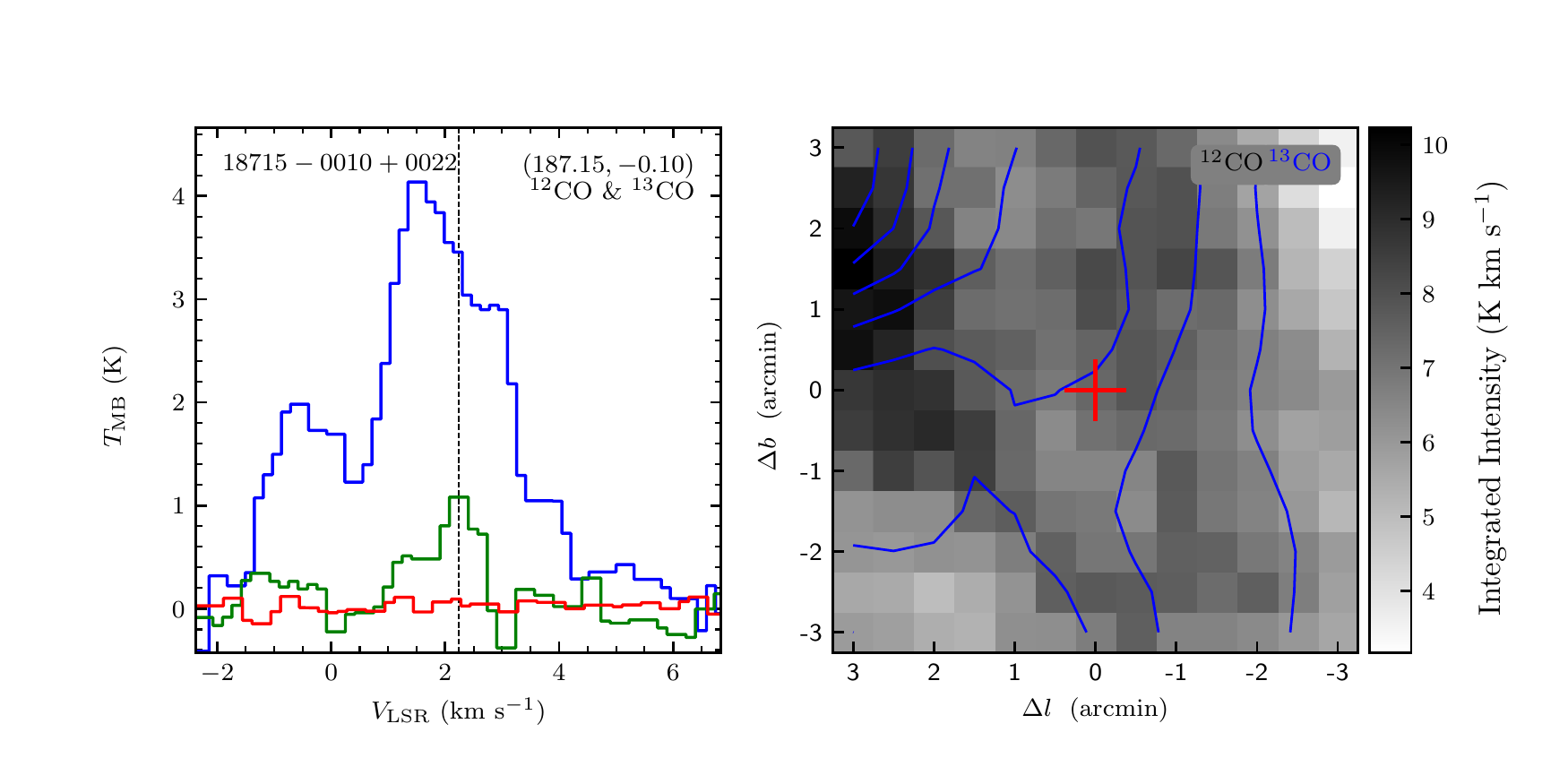}
\includegraphics[width=9.0cm,angle=0]{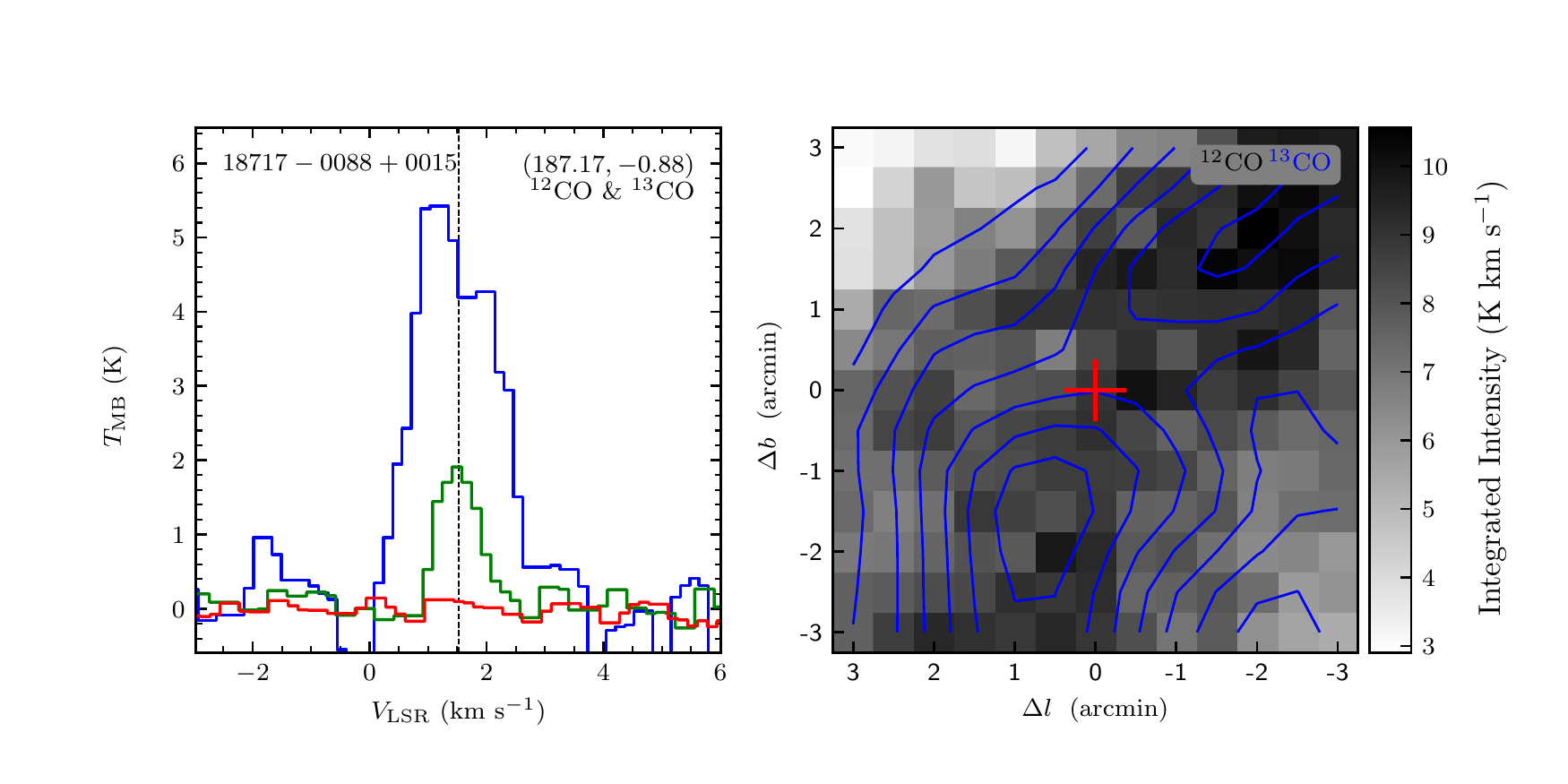}
\vspace{-0.5cm}

\includegraphics[width=9.0cm,angle=0]{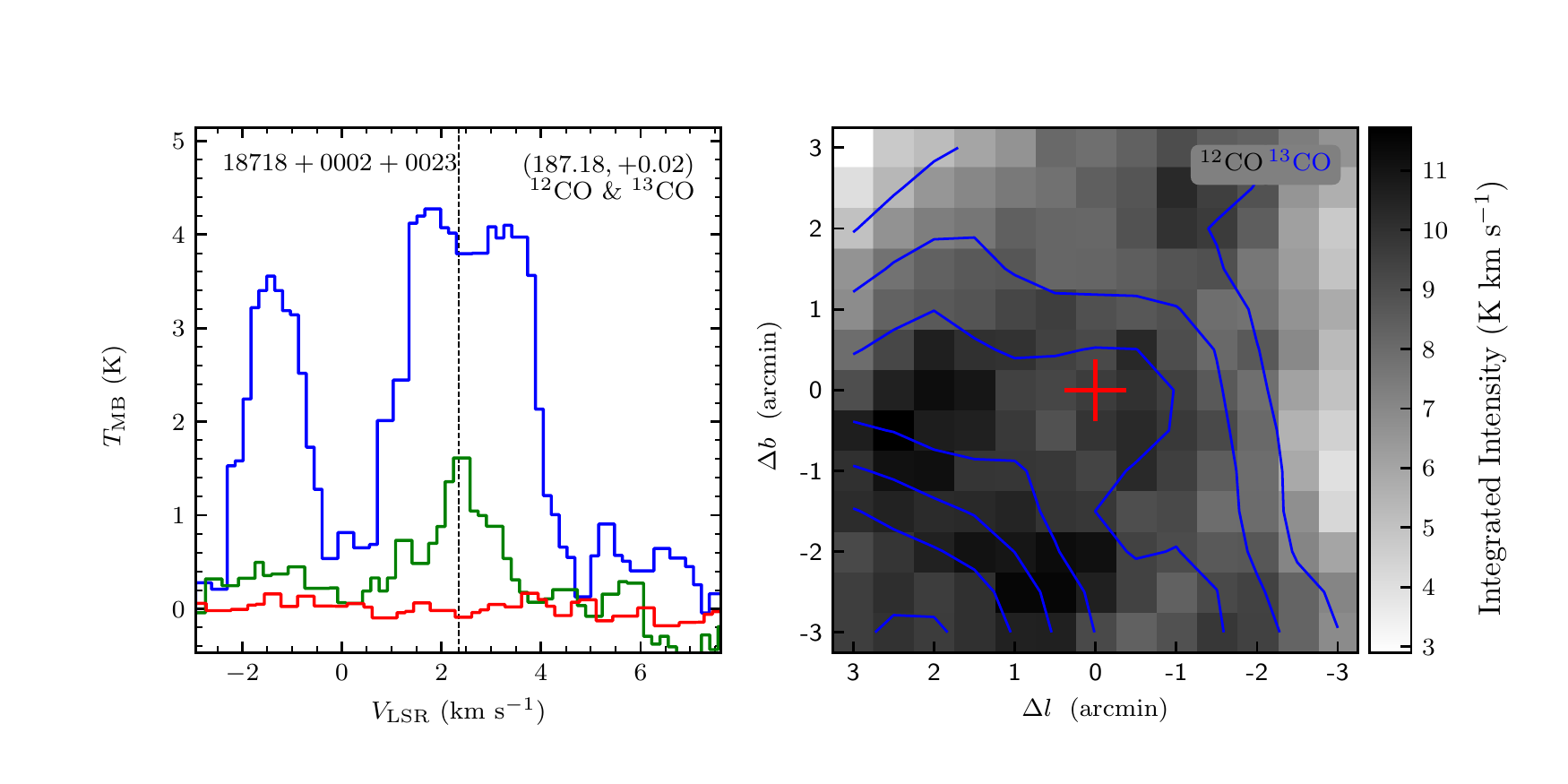}
\includegraphics[width=9.0cm,angle=0]{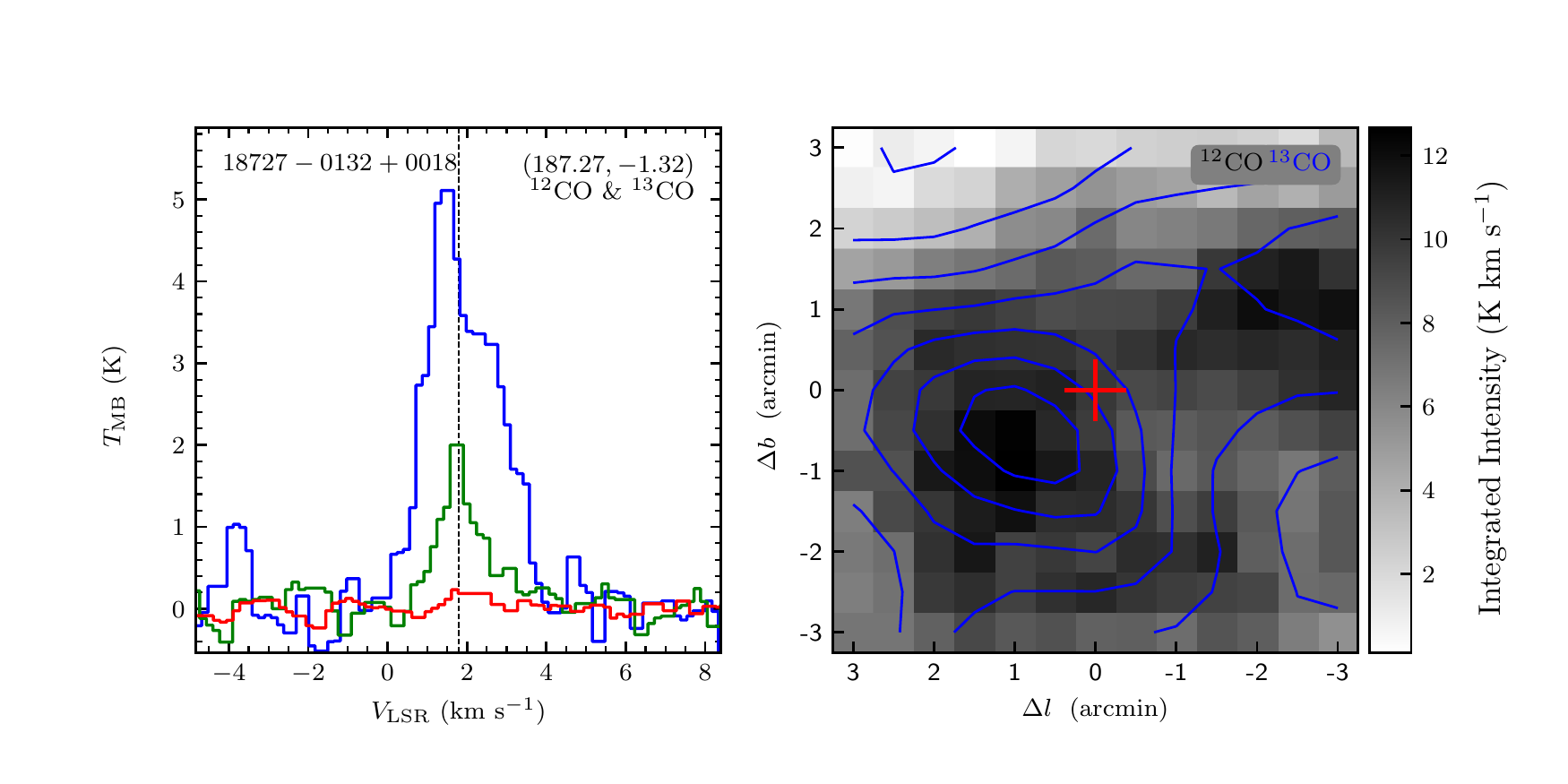}
\vspace{-0.5cm}

\includegraphics[width=9.0cm,angle=0]{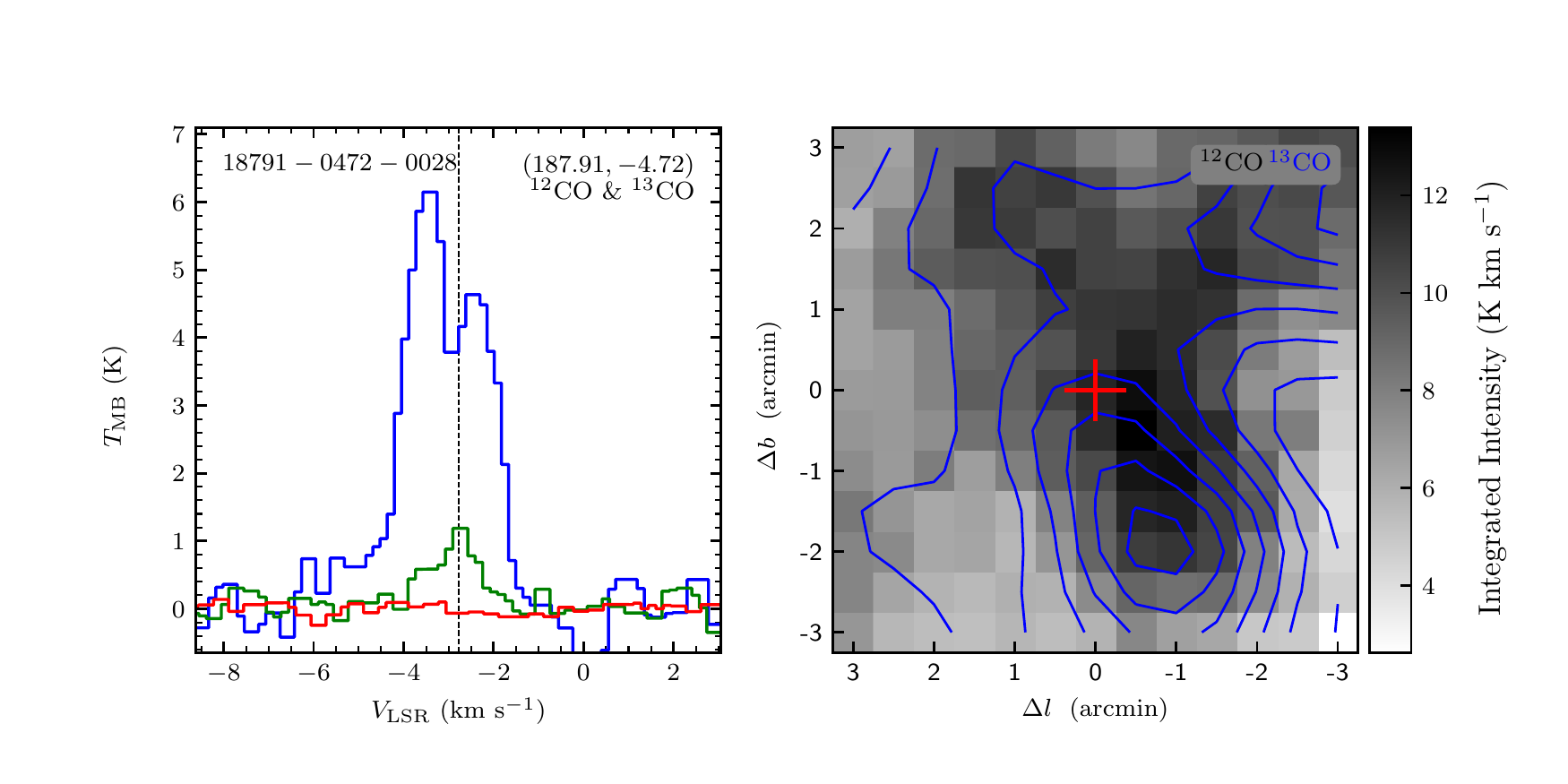}
\includegraphics[width=9.0cm,angle=0]{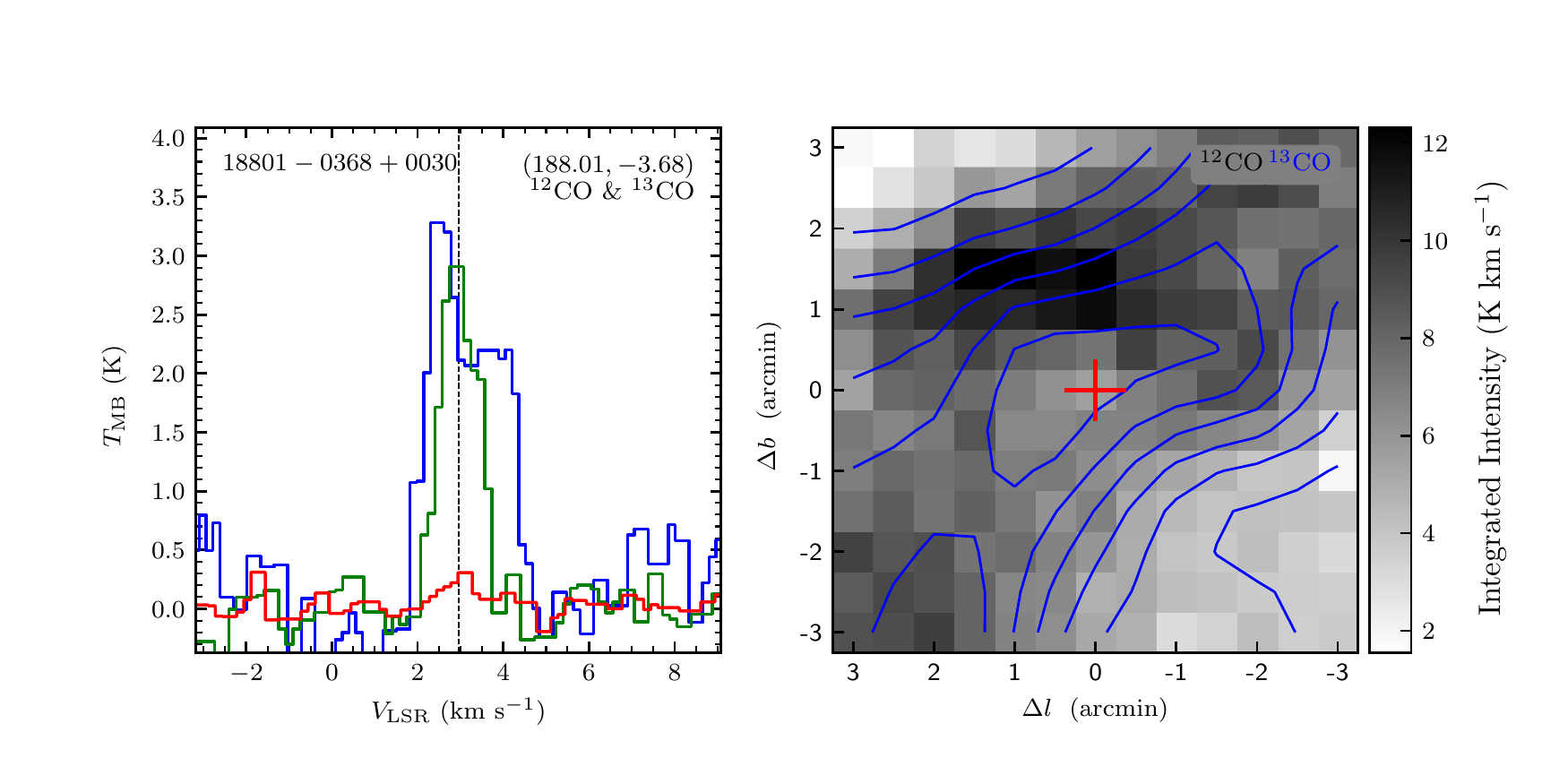}
\vspace{-0.5cm}

\includegraphics[width=9.0cm,angle=0]{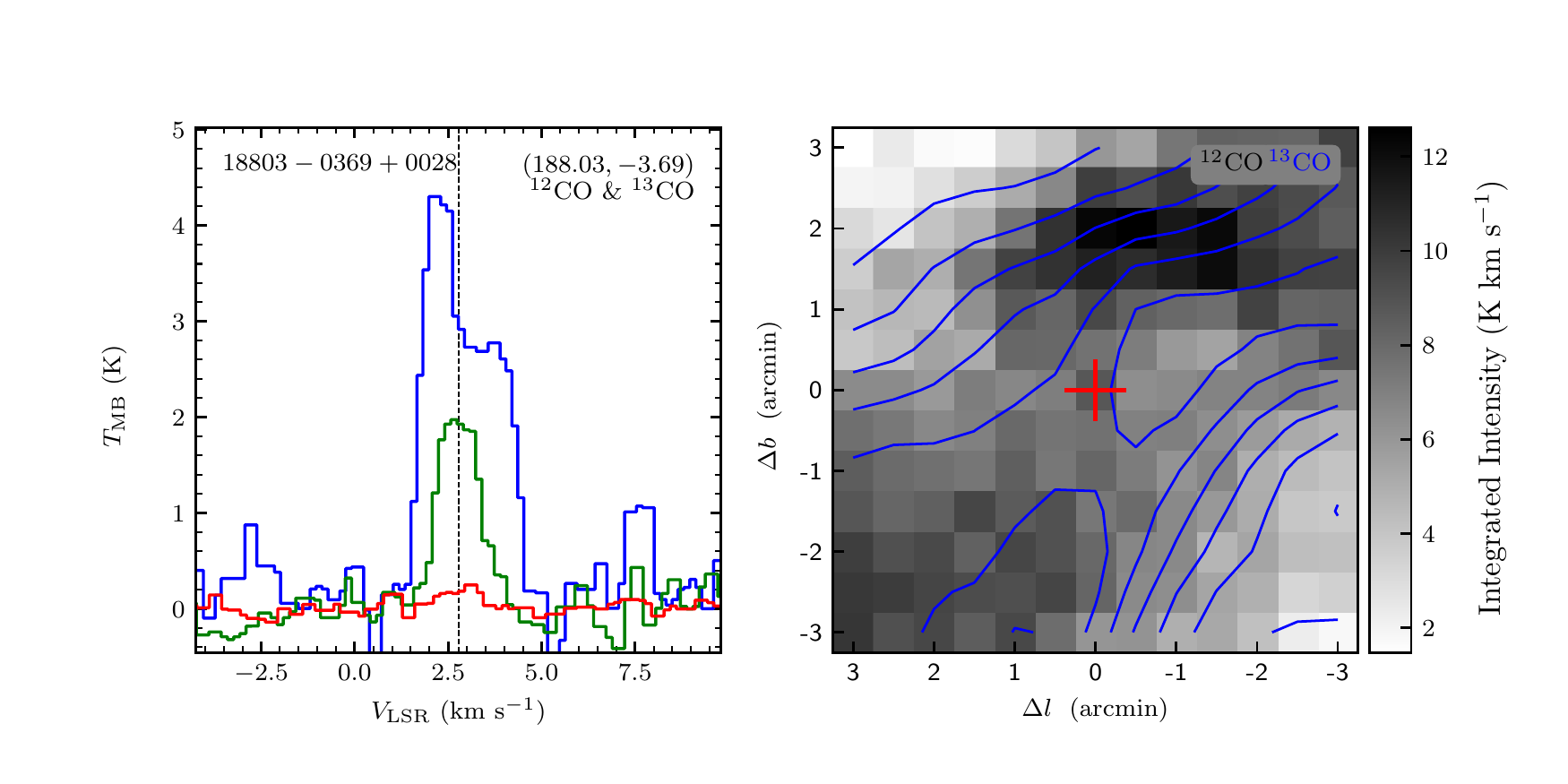}
\includegraphics[width=9.0cm,angle=0]{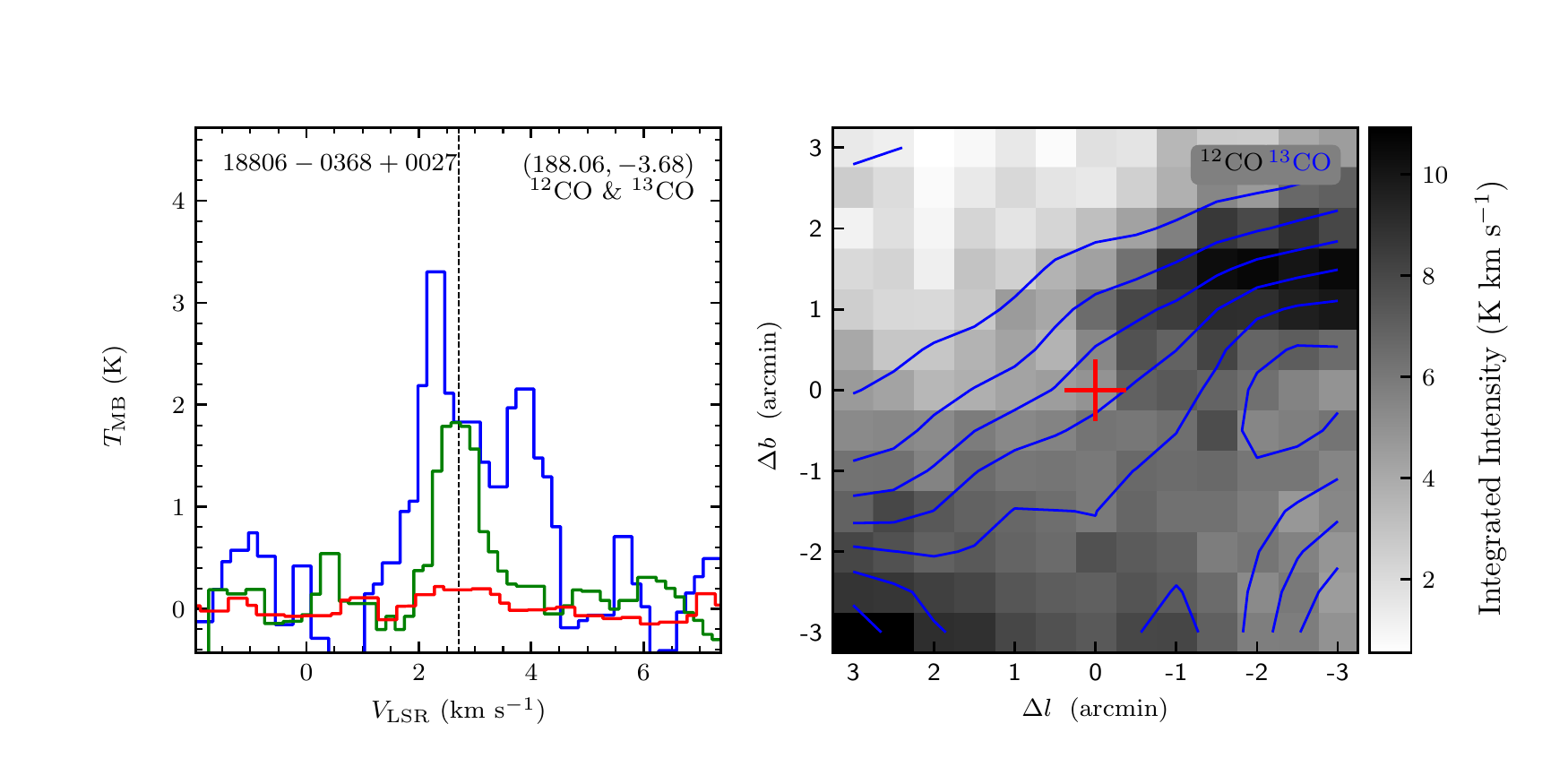}
\vspace{-0.5cm}

\includegraphics[width=9.0cm,angle=0]{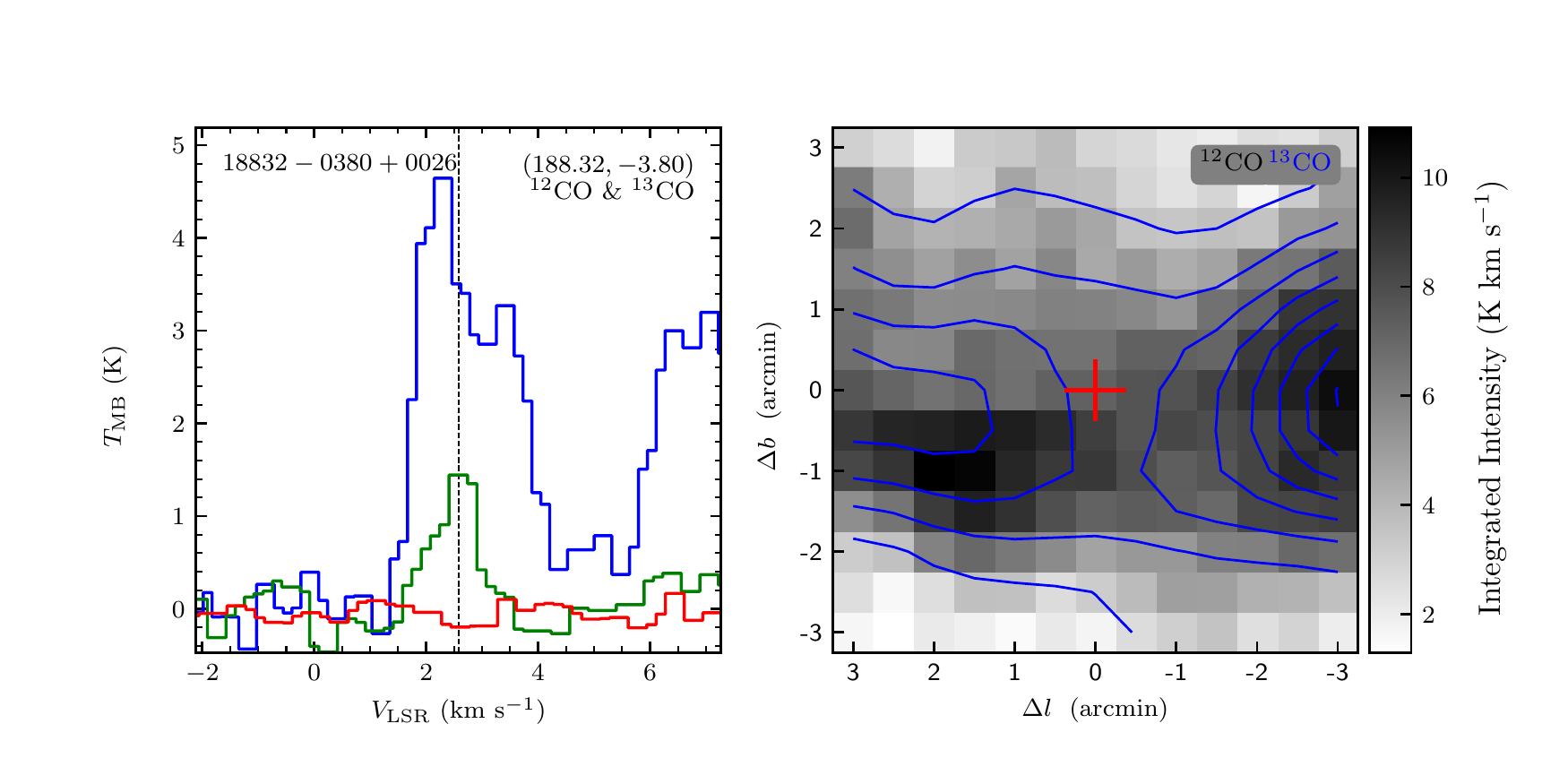}
\includegraphics[width=9.0cm,angle=0]{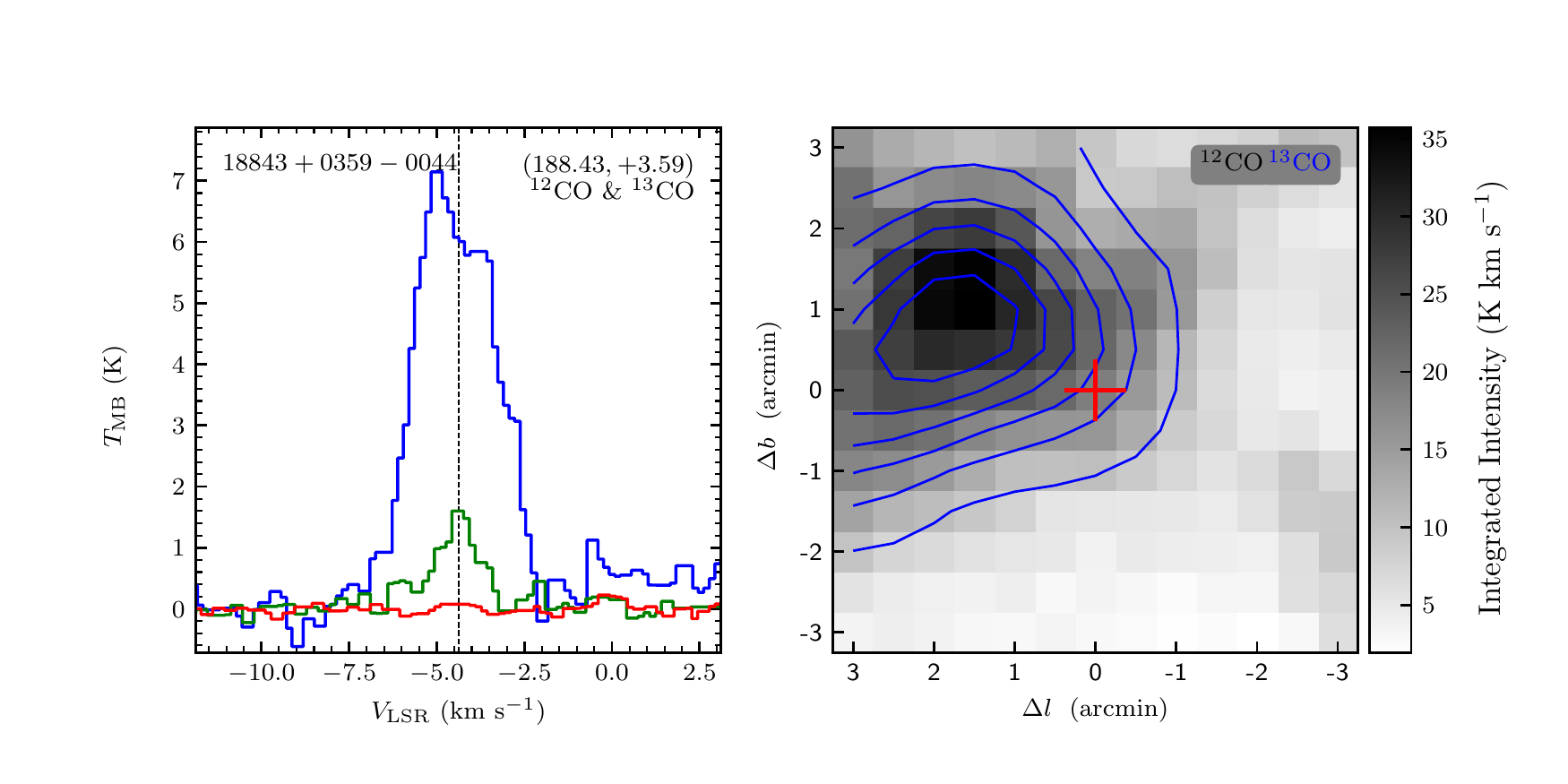}
\end{figure}
\clearpage

\begin{figure}
\includegraphics[width=9.0cm,angle=0]{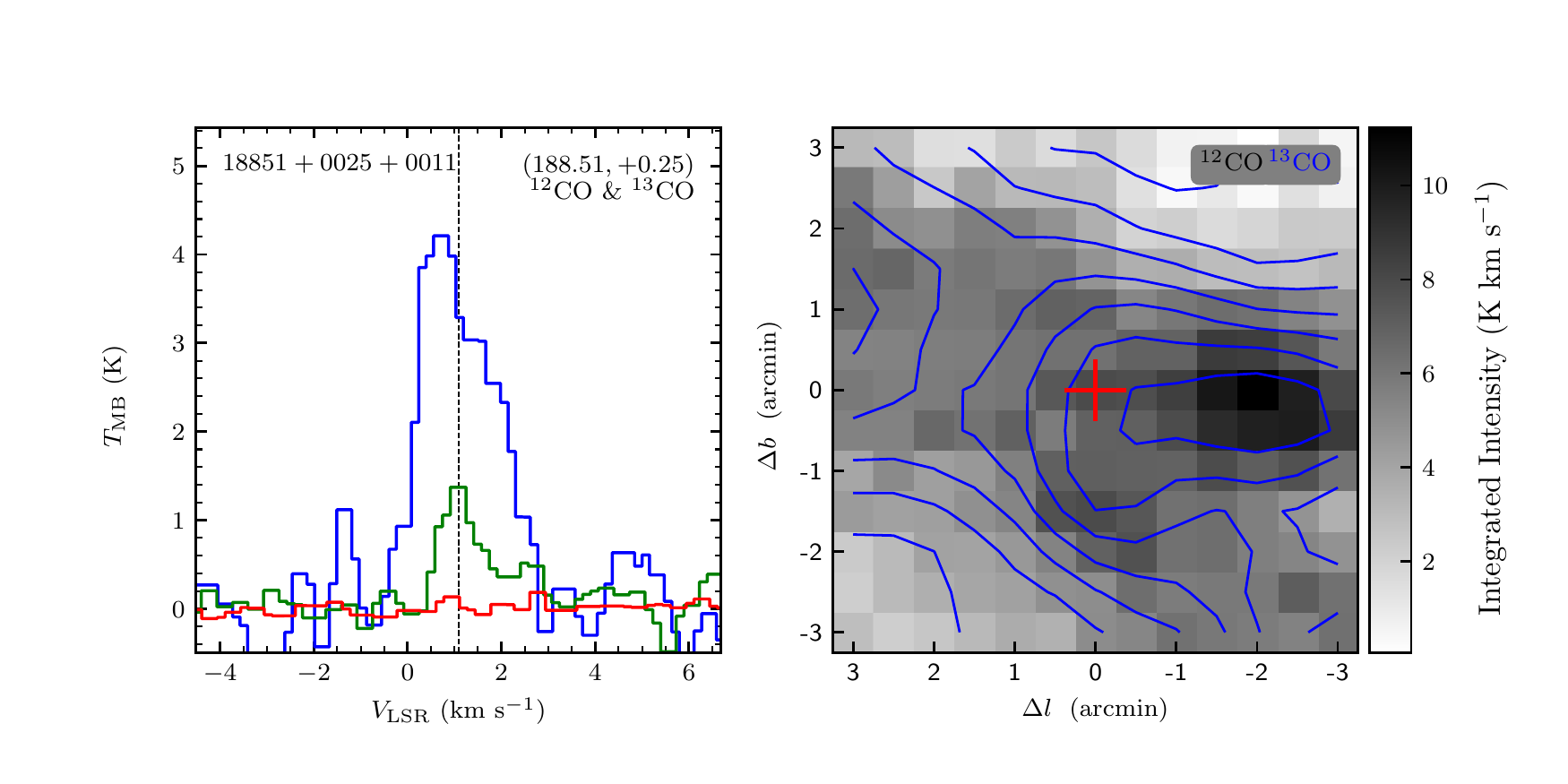}
\includegraphics[width=9.0cm,angle=0]{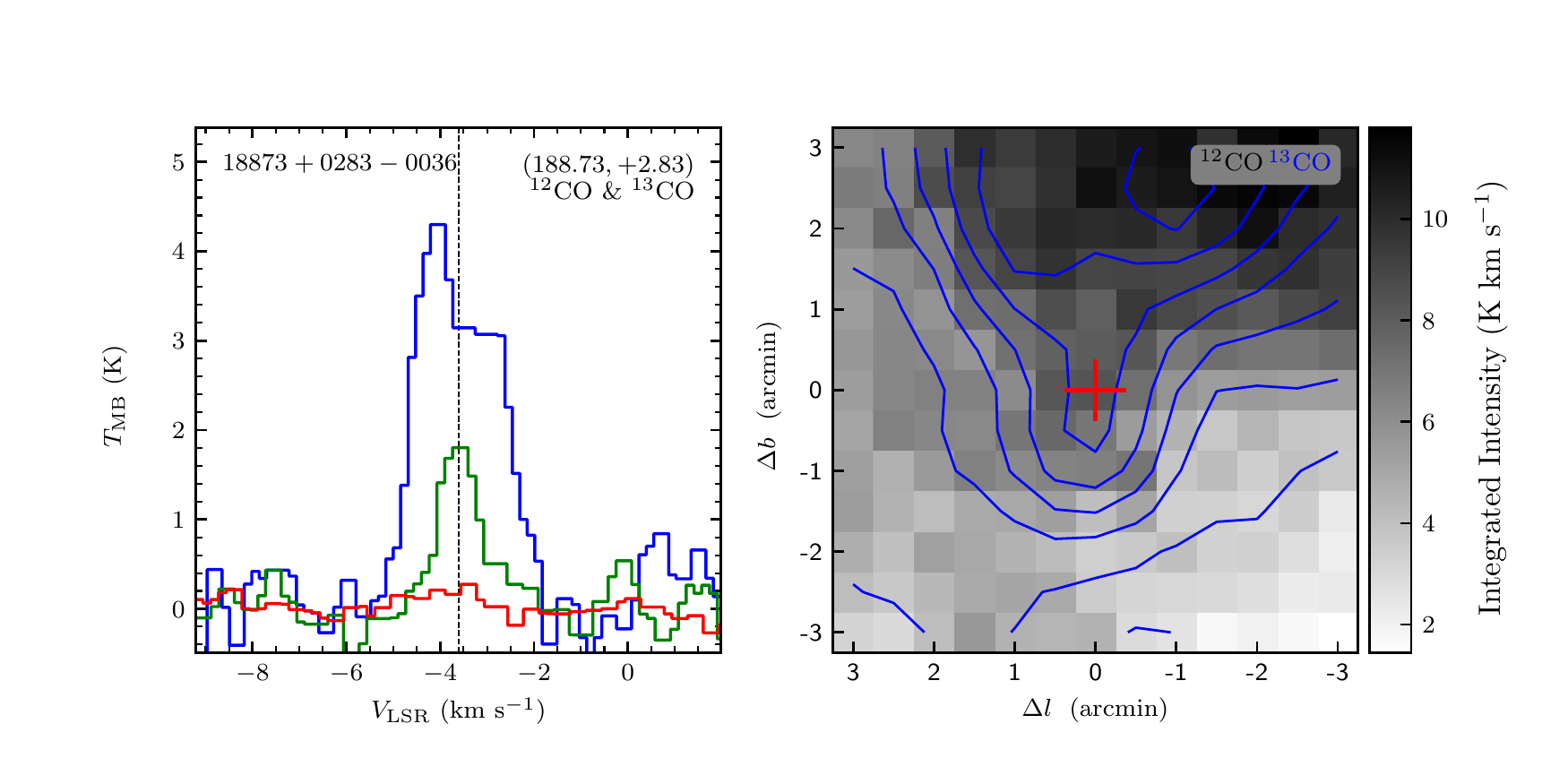}
\vspace{-0.5cm}

\includegraphics[width=9.0cm,angle=0]{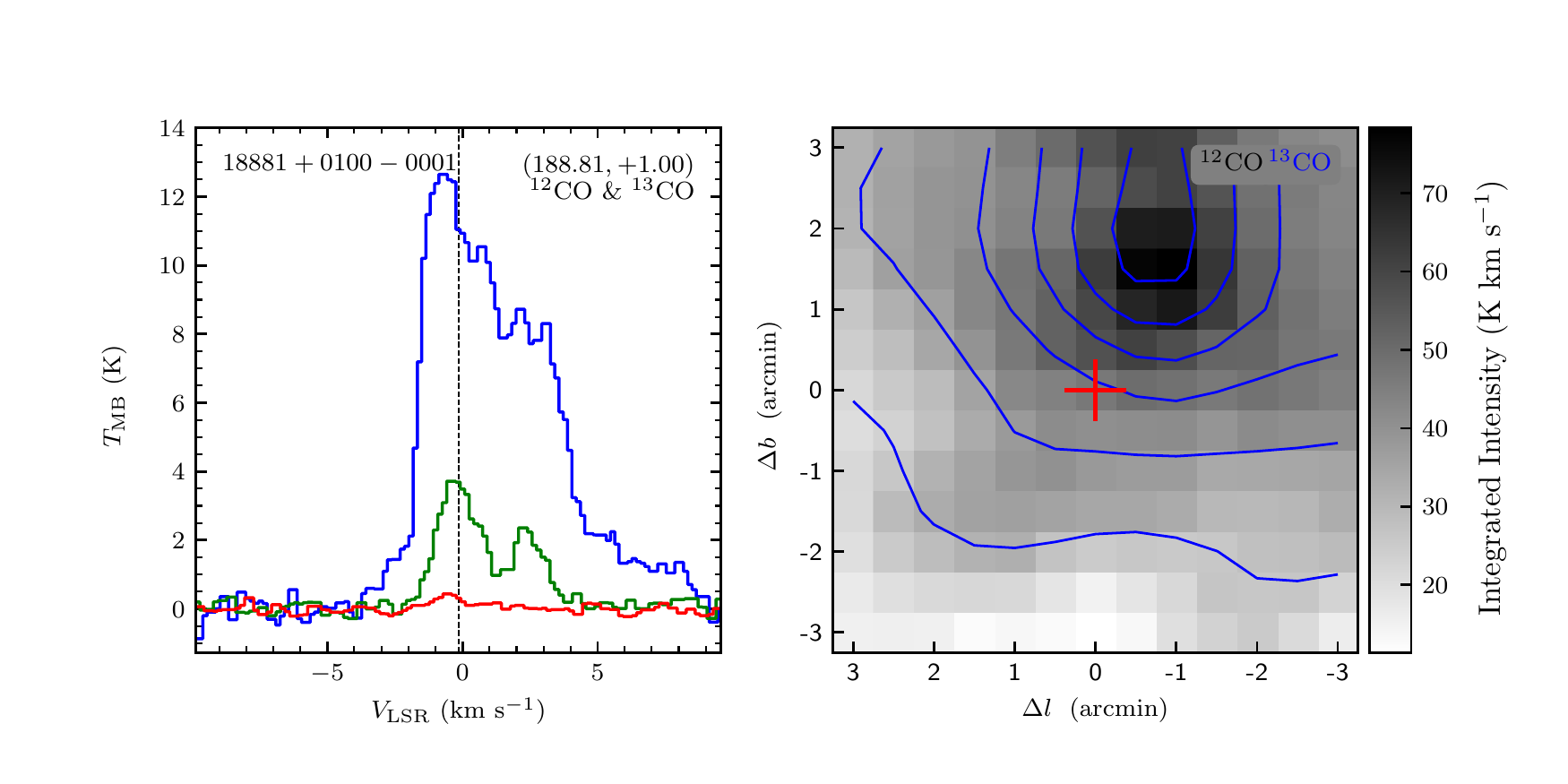}
\includegraphics[width=9.0cm,angle=0]{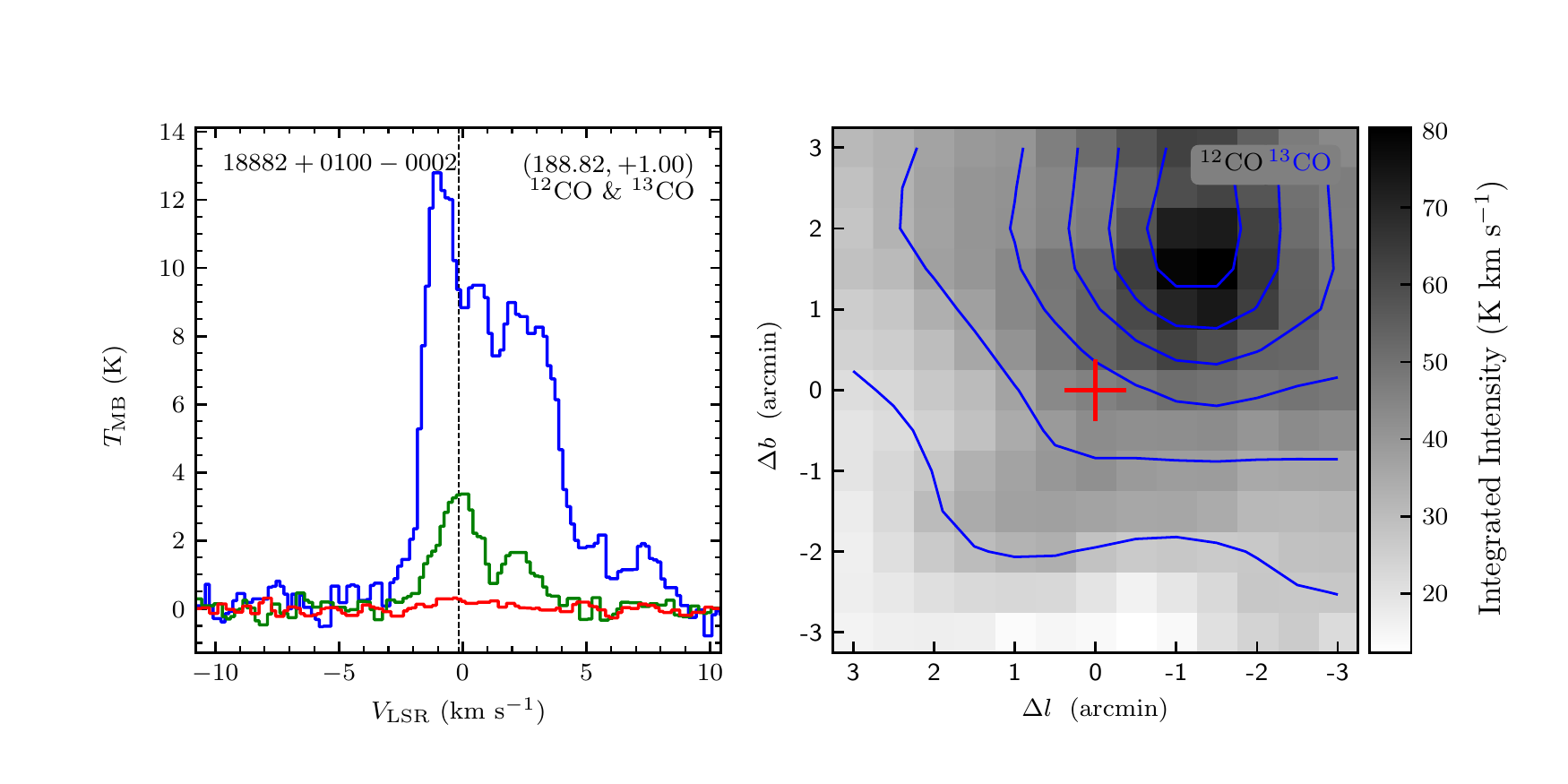}
\vspace{-0.5cm}

\includegraphics[width=9.0cm,angle=0]{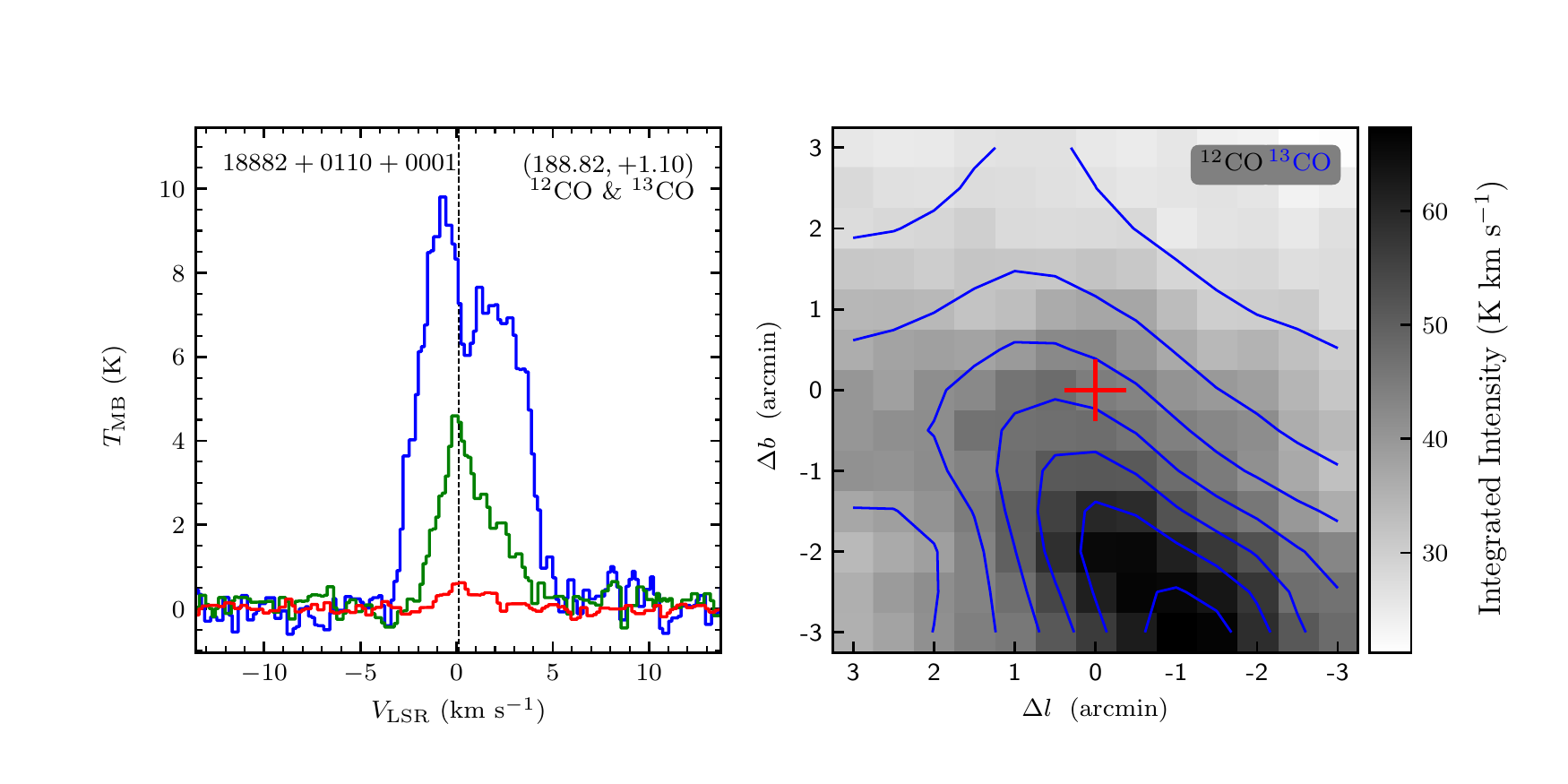}
\includegraphics[width=9.0cm,angle=0]{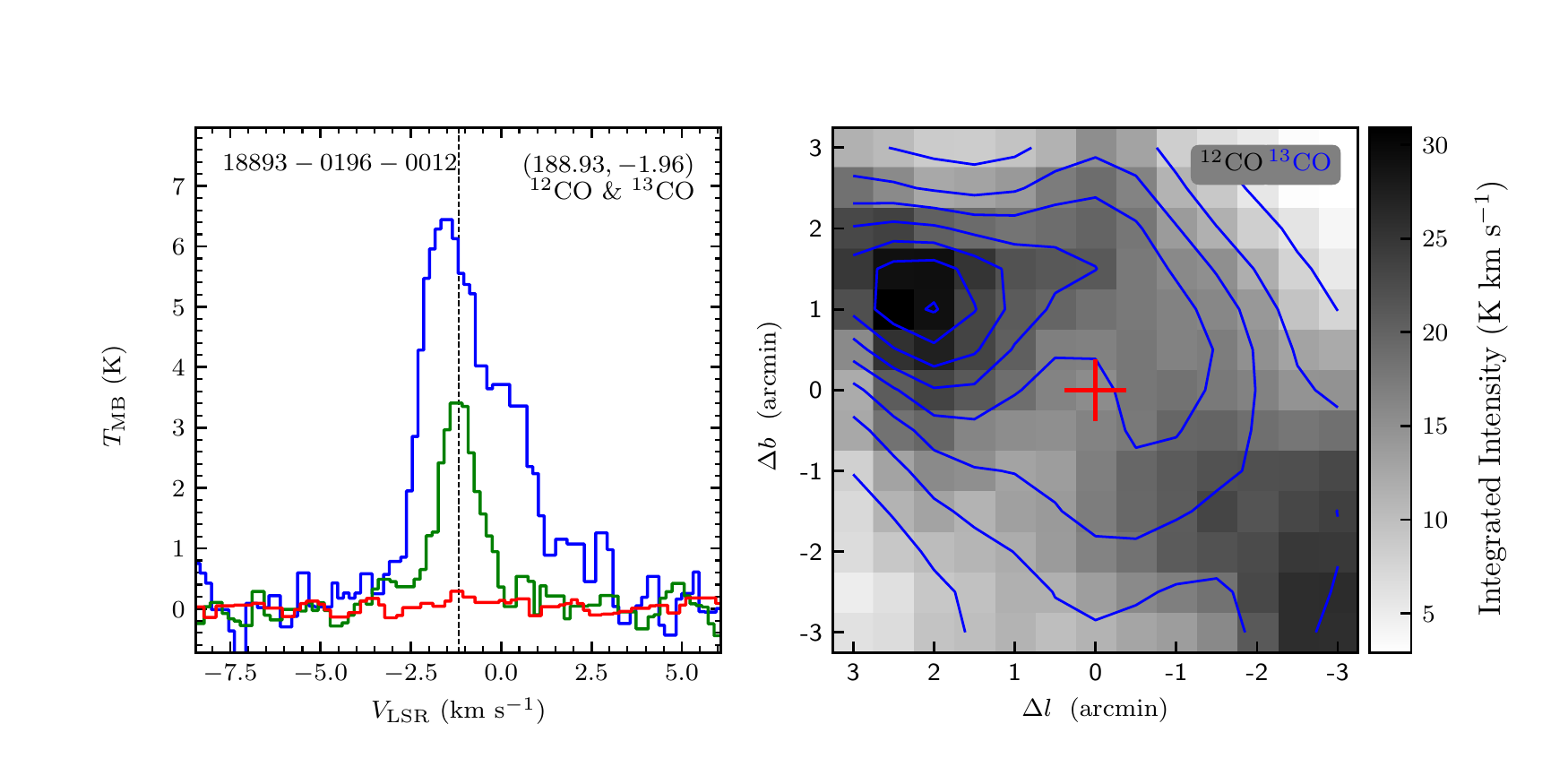}
\vspace{-0.5cm}

\includegraphics[width=9.0cm,angle=0]{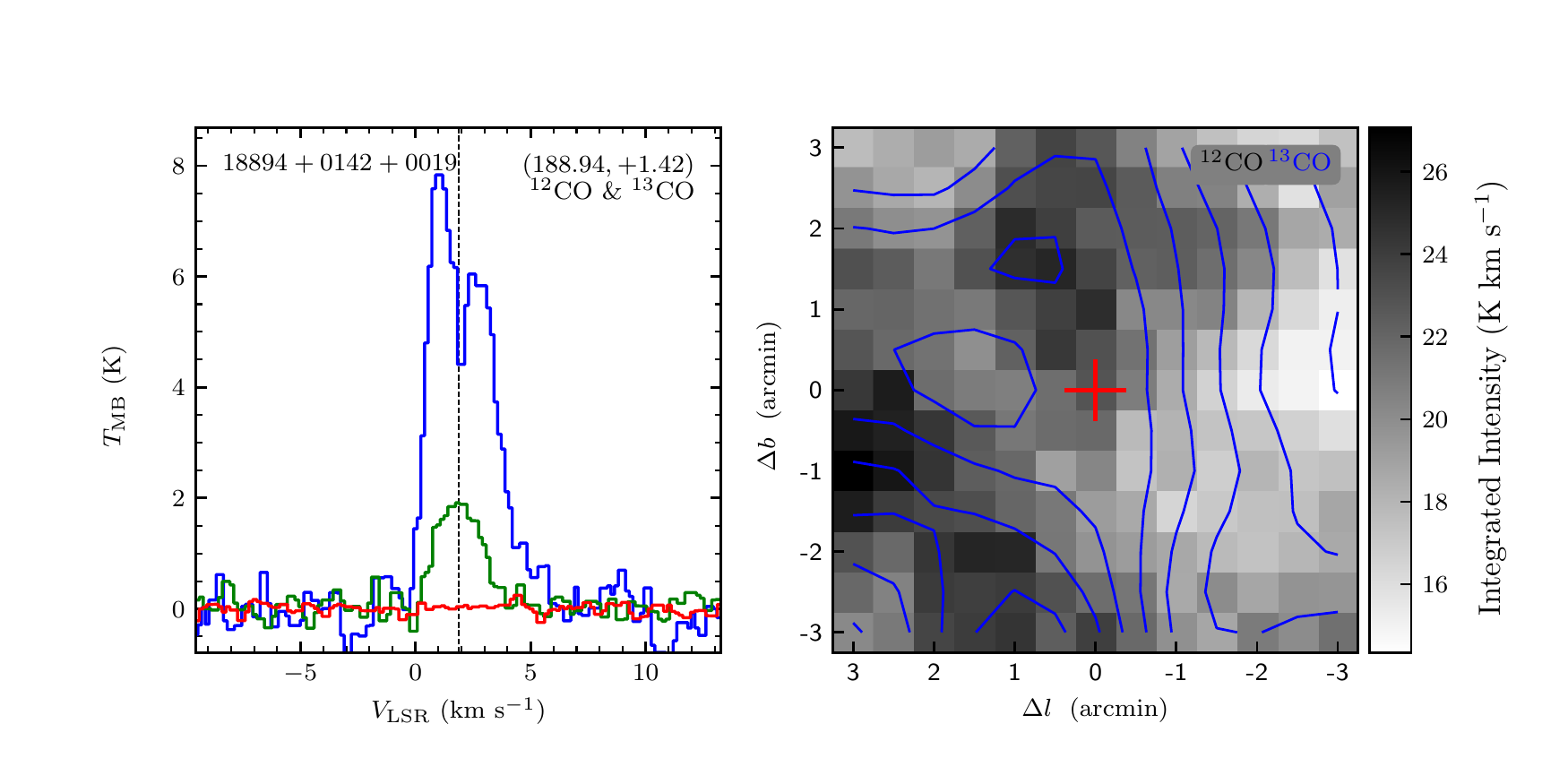}
\includegraphics[width=9.0cm,angle=0]{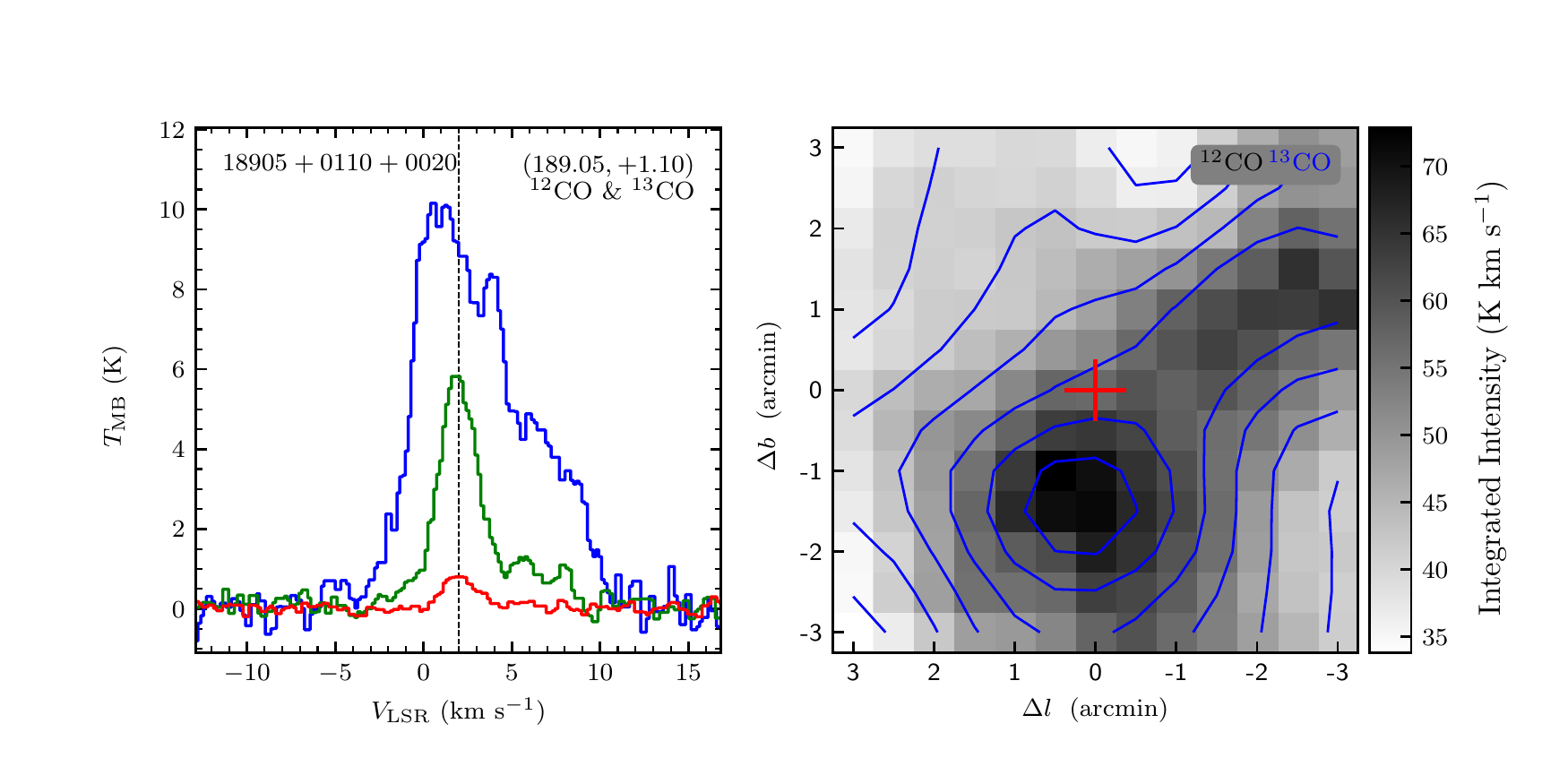}
\vspace{-0.5cm}

\includegraphics[width=9.0cm,angle=0]{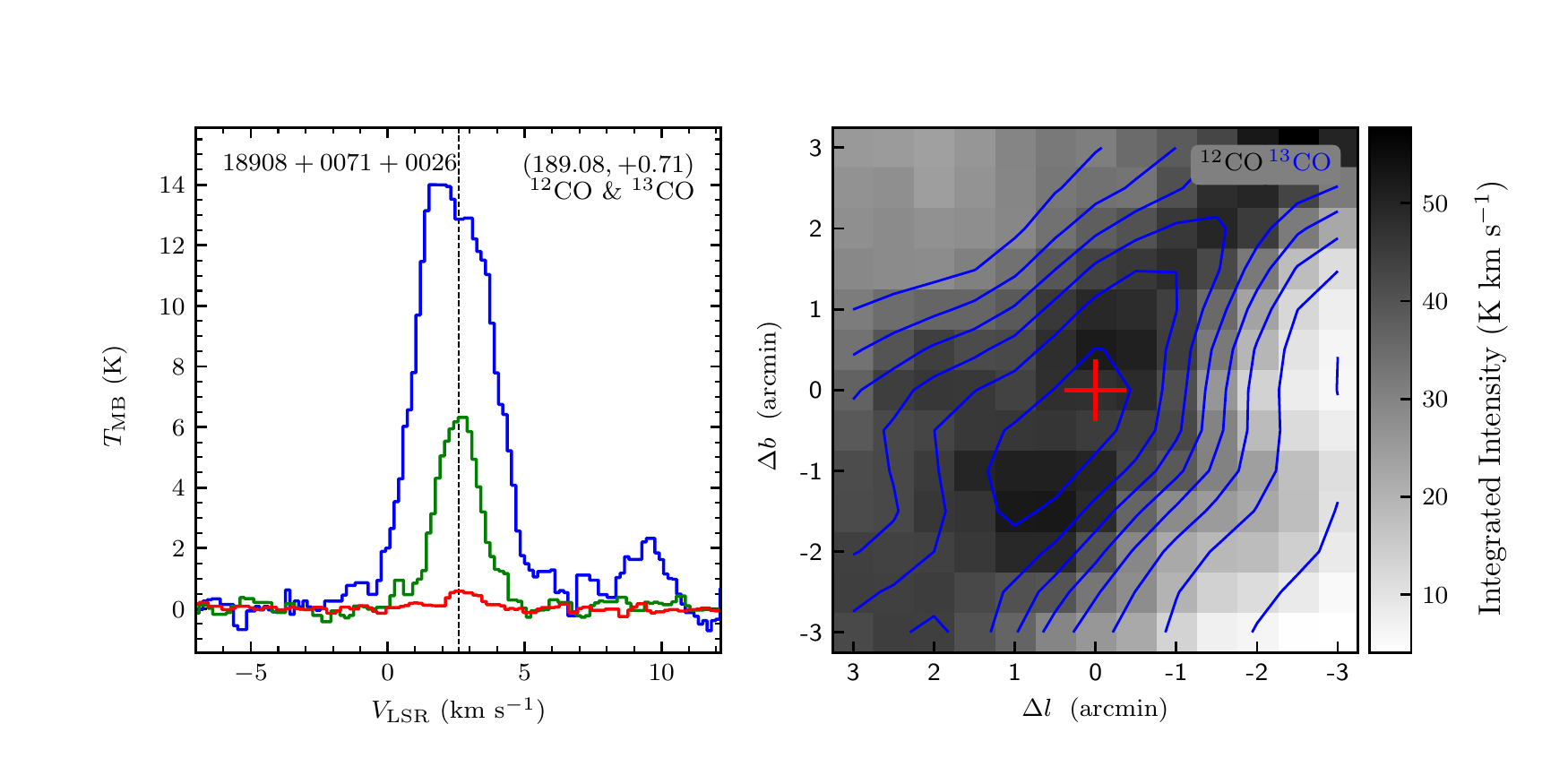}
\includegraphics[width=9.0cm,angle=0]{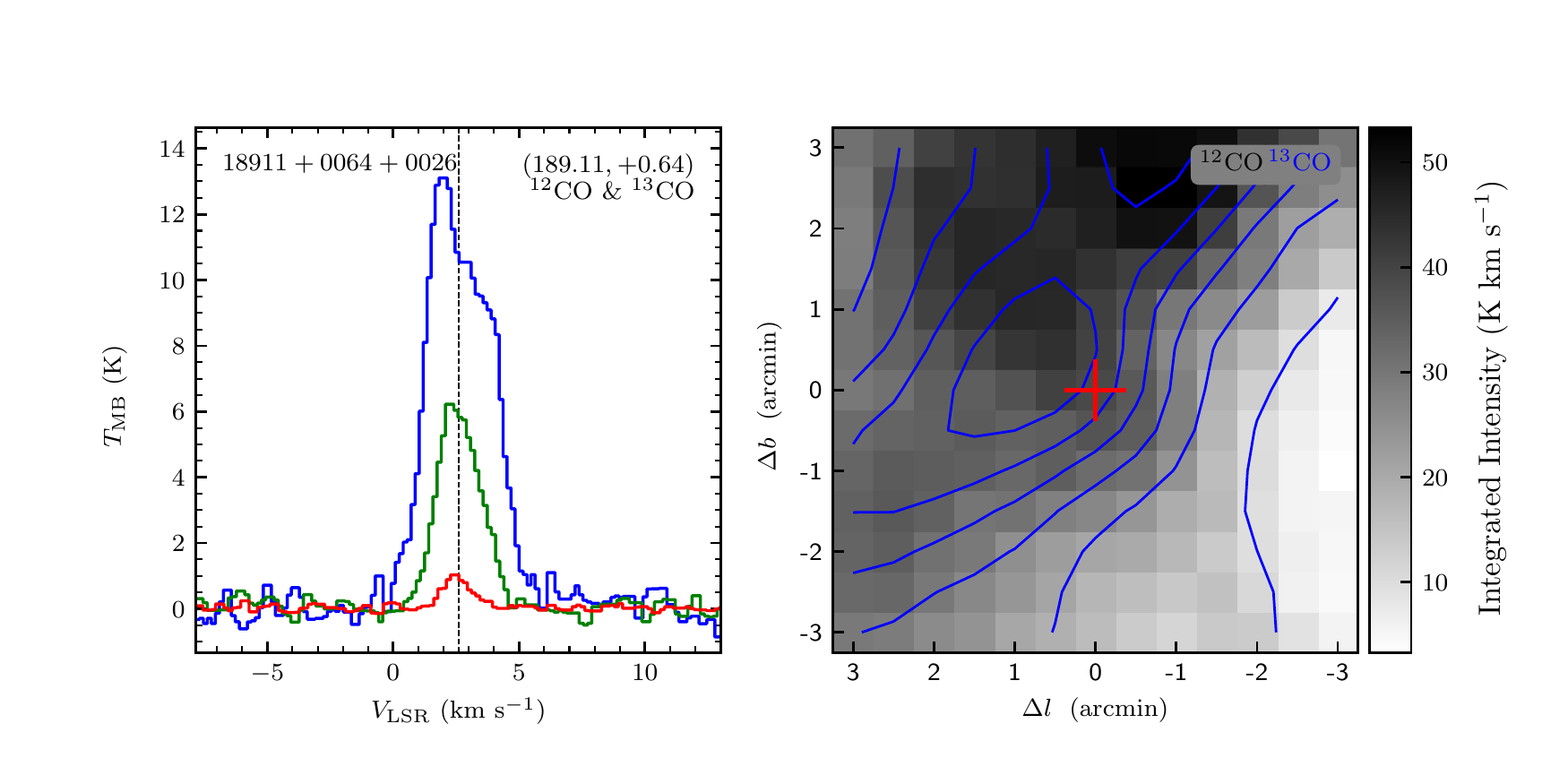}
\end{figure}
\clearpage

\begin{figure}
\includegraphics[width=9.0cm,angle=0]{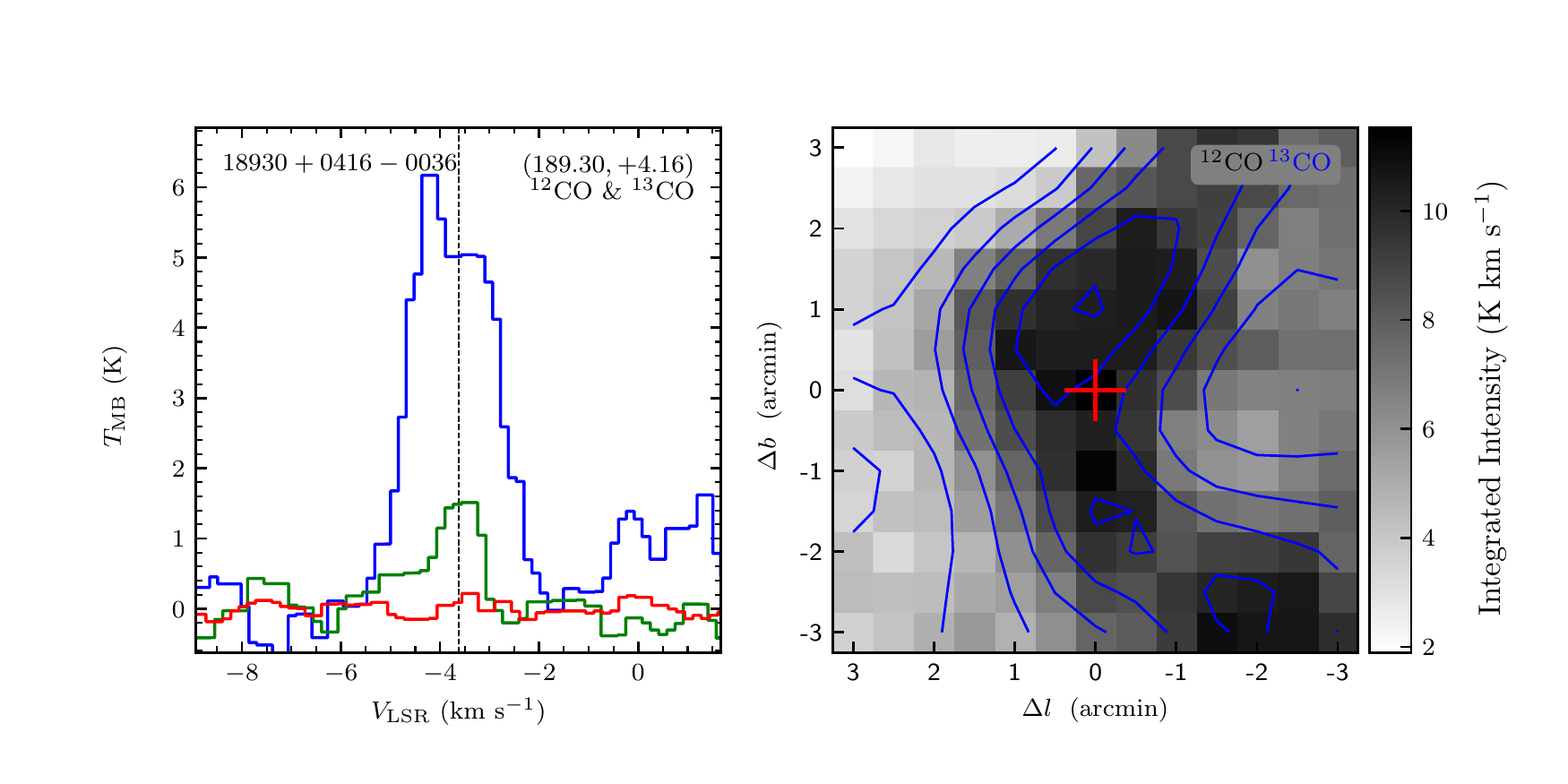}
\includegraphics[width=9.0cm,angle=0]{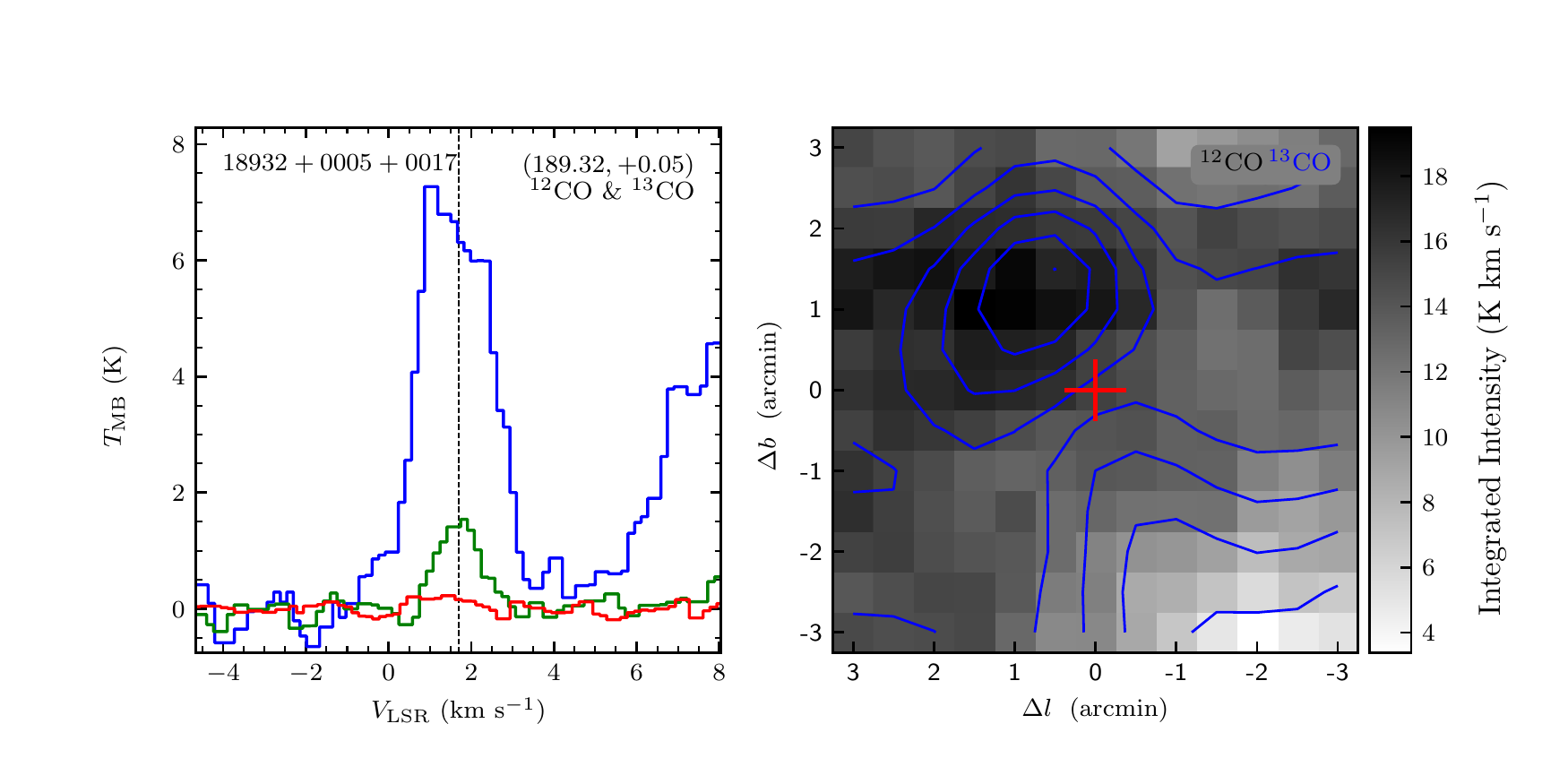}
\vspace{-0.5cm}

\includegraphics[width=9.0cm,angle=0]{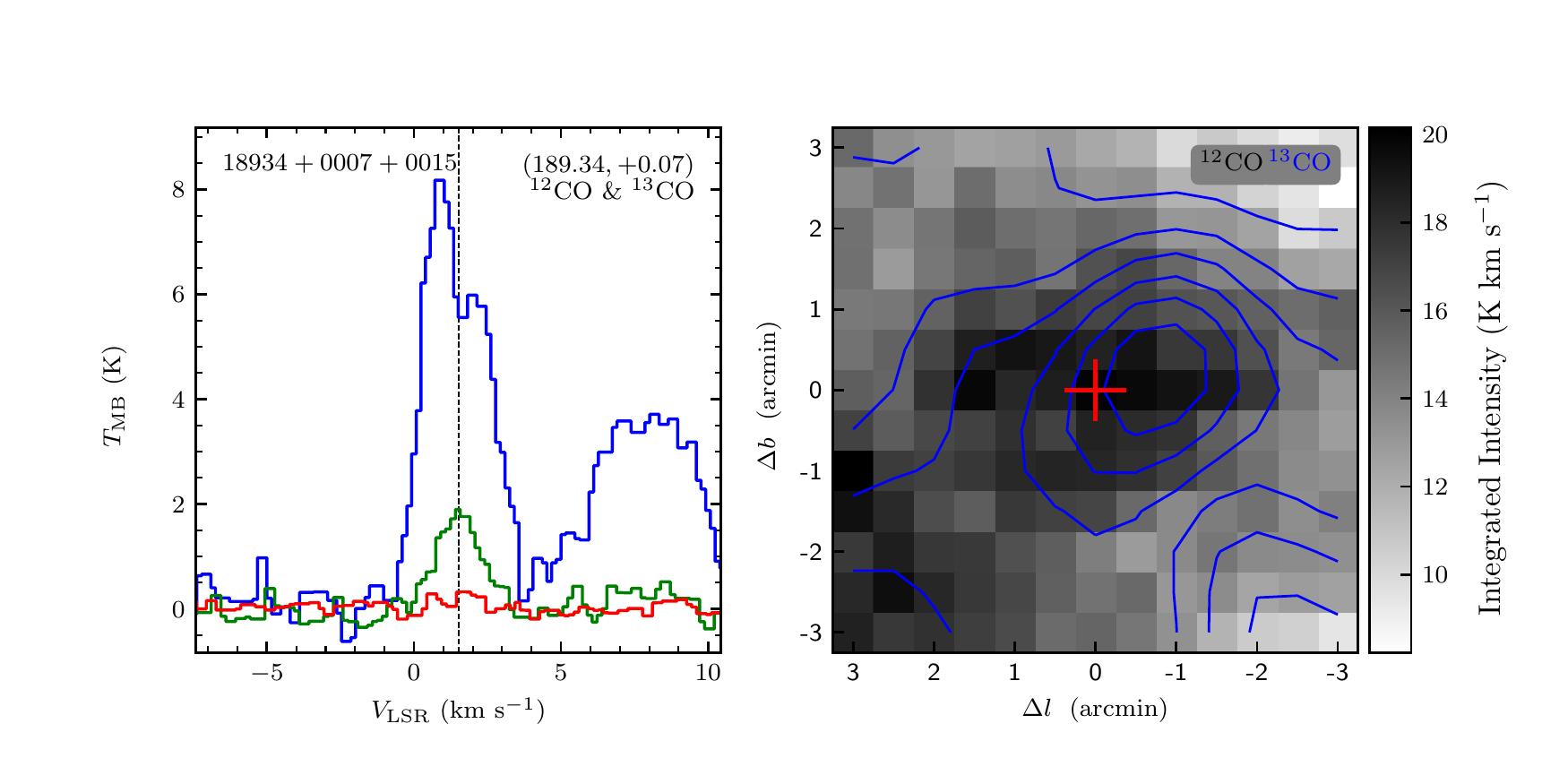}
\includegraphics[width=9.0cm,angle=0]{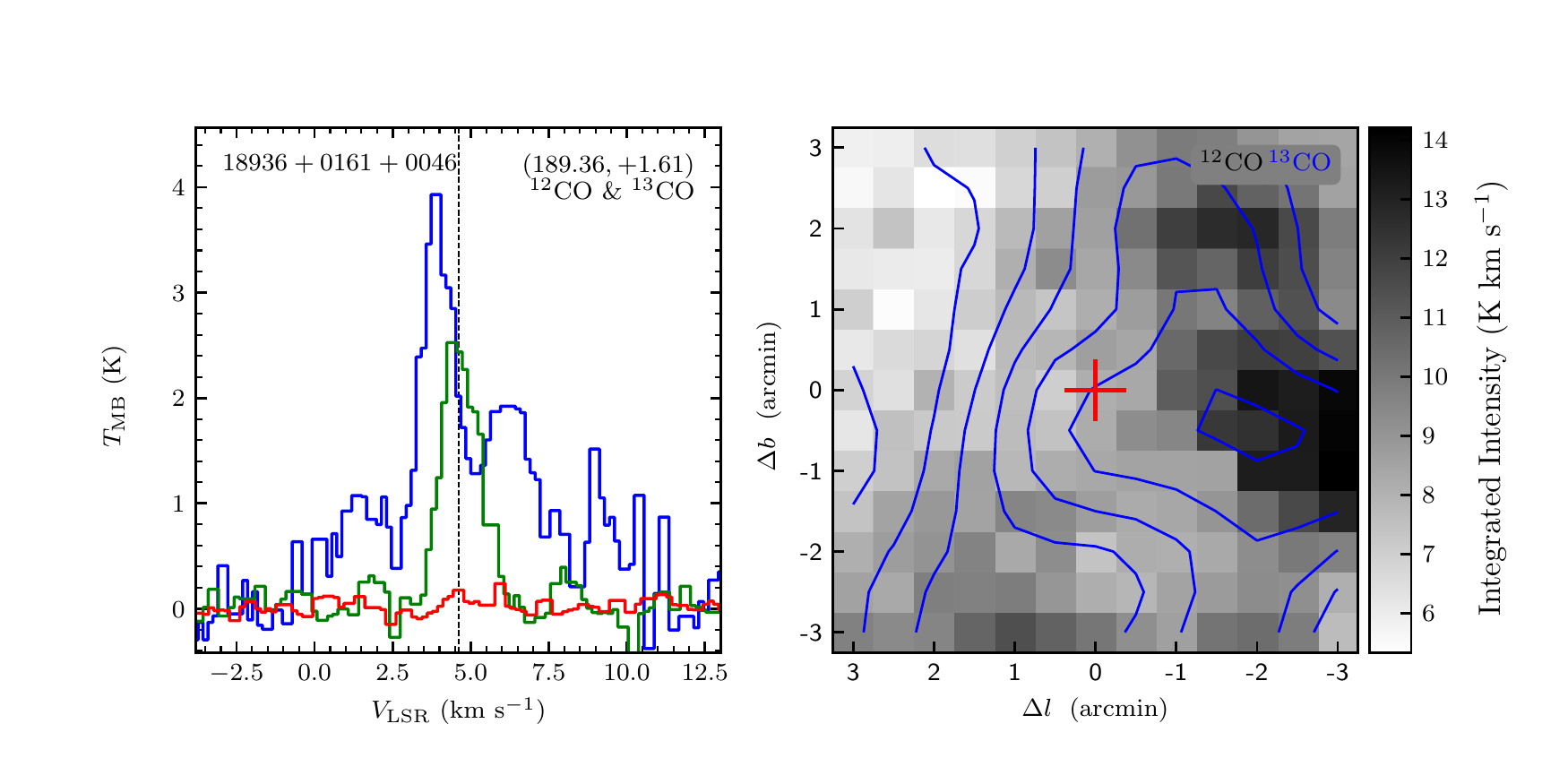}
\vspace{-0.5cm}

\includegraphics[width=9.0cm,angle=0]{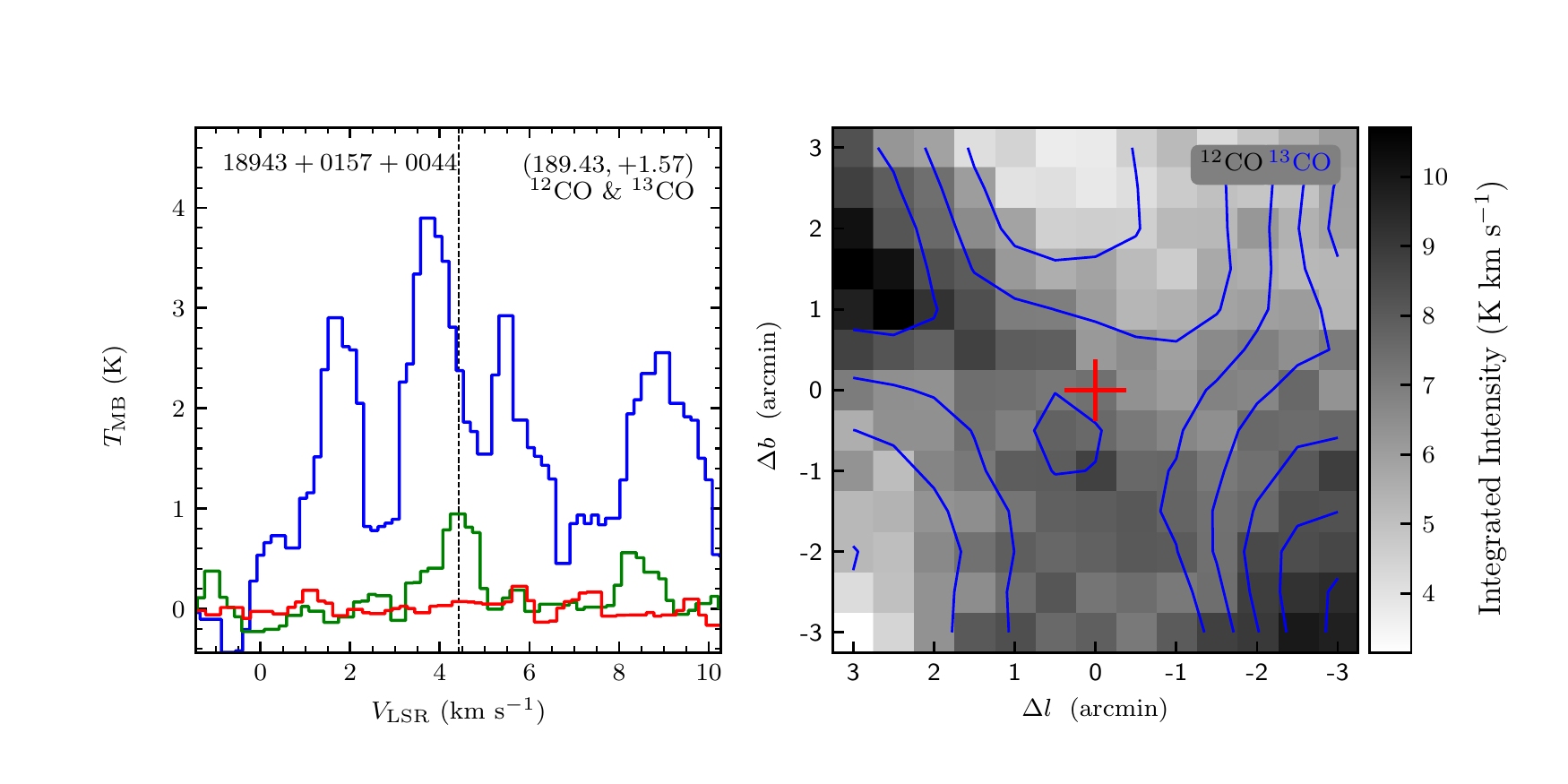}
\includegraphics[width=9.0cm,angle=0]{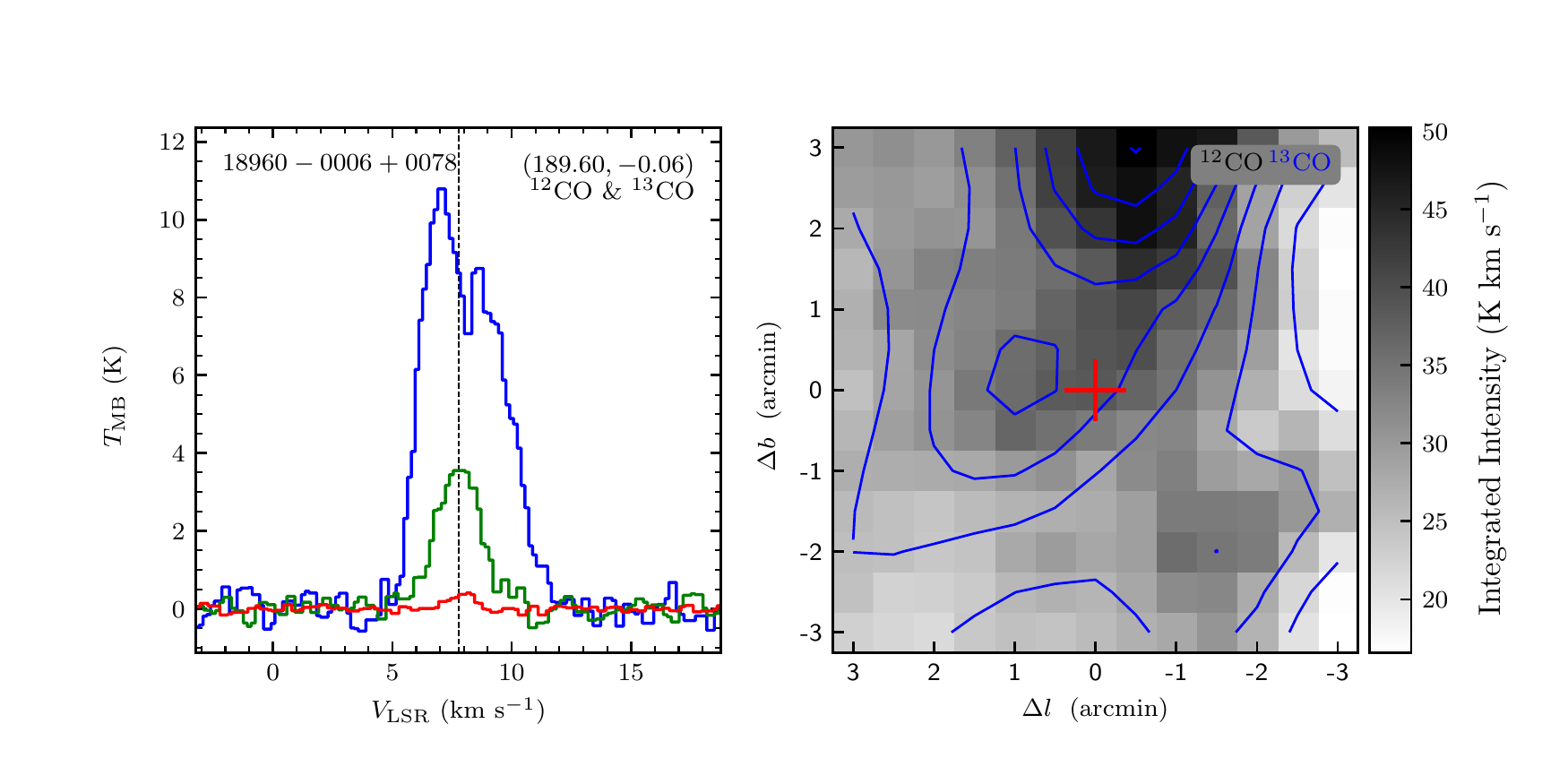}
\vspace{-0.5cm}

\includegraphics[width=9.0cm,angle=0]{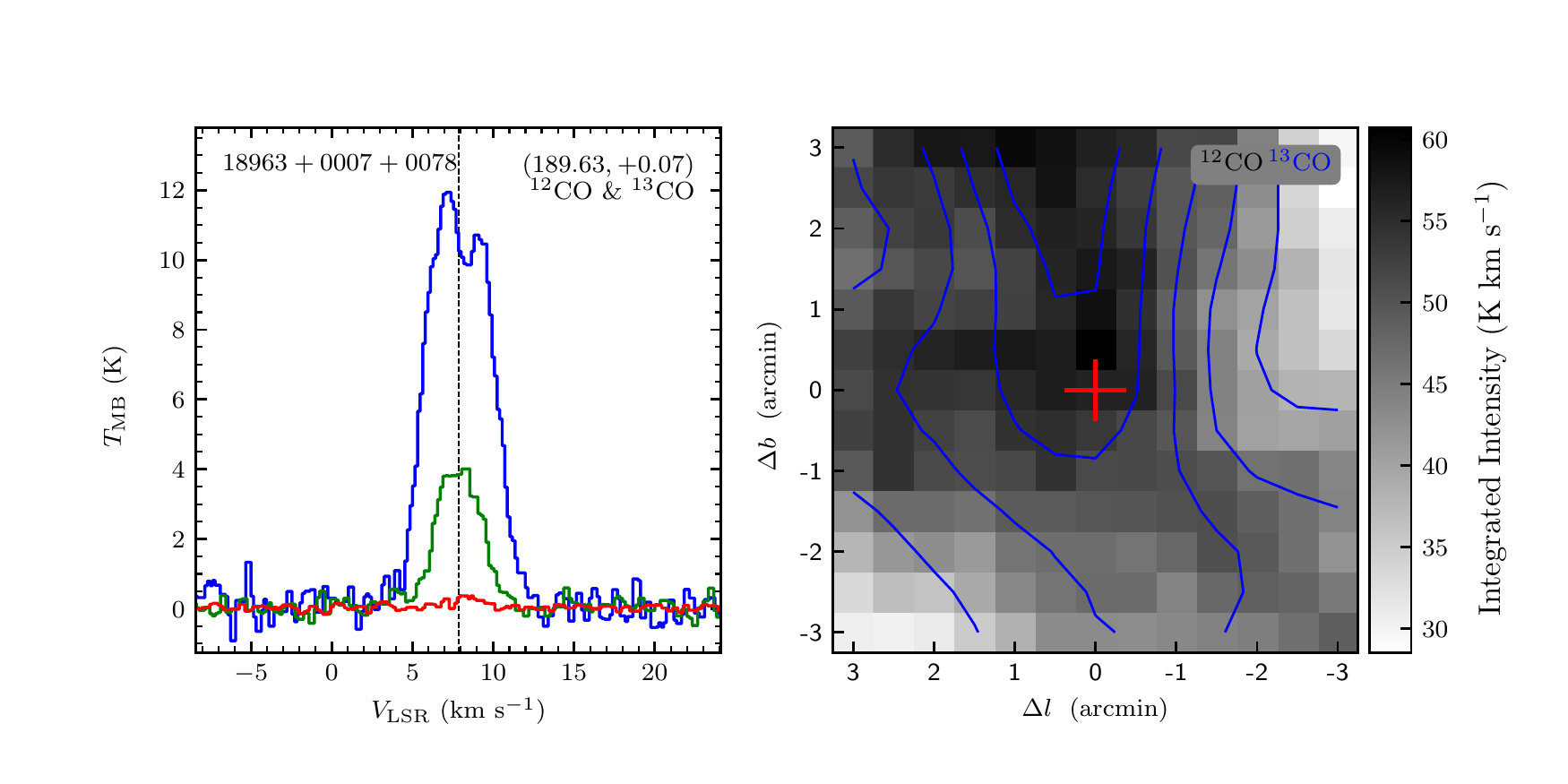}
\includegraphics[width=9.0cm,angle=0]{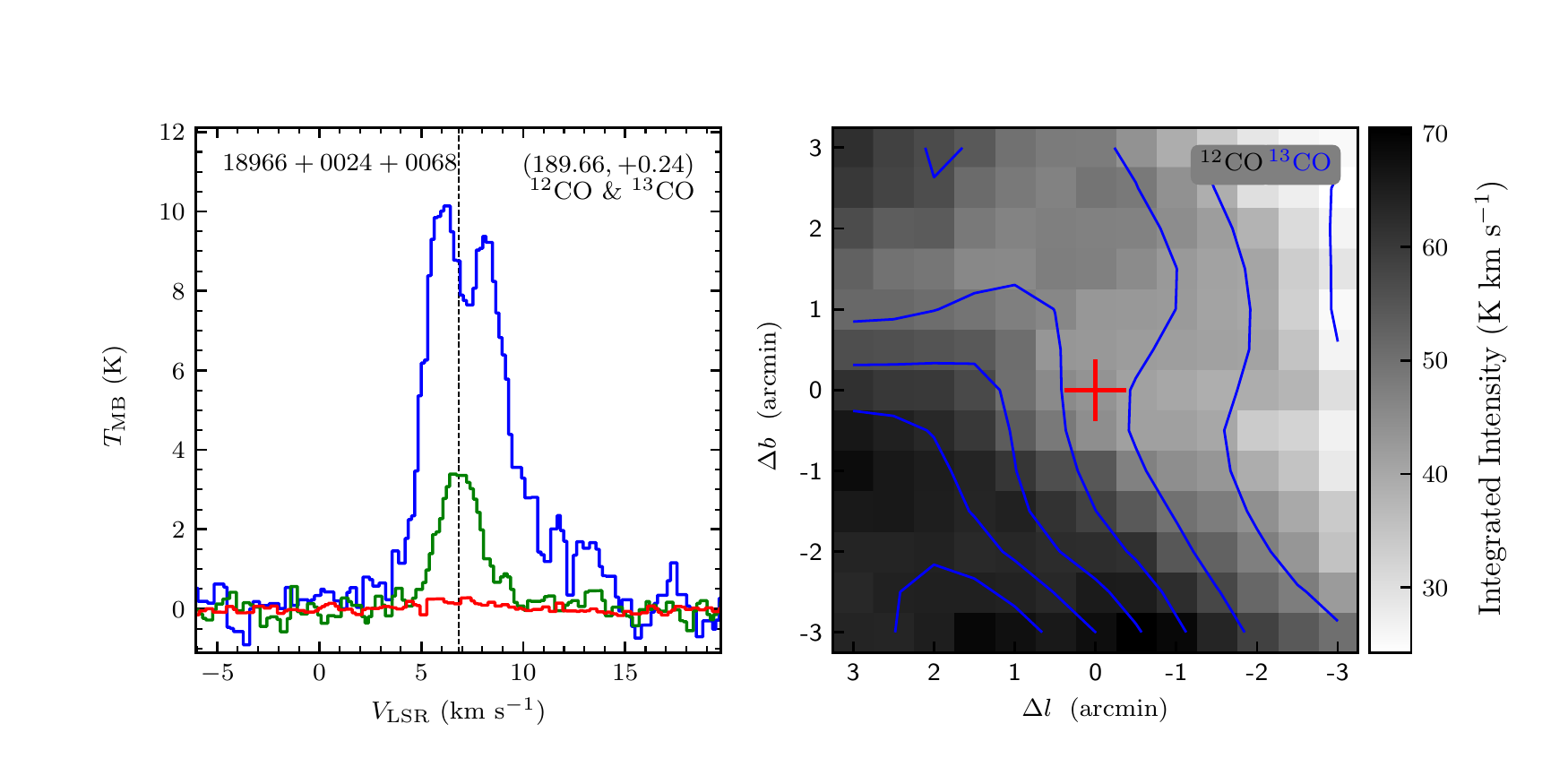}
\vspace{-0.5cm}

\includegraphics[width=9.0cm,angle=0]{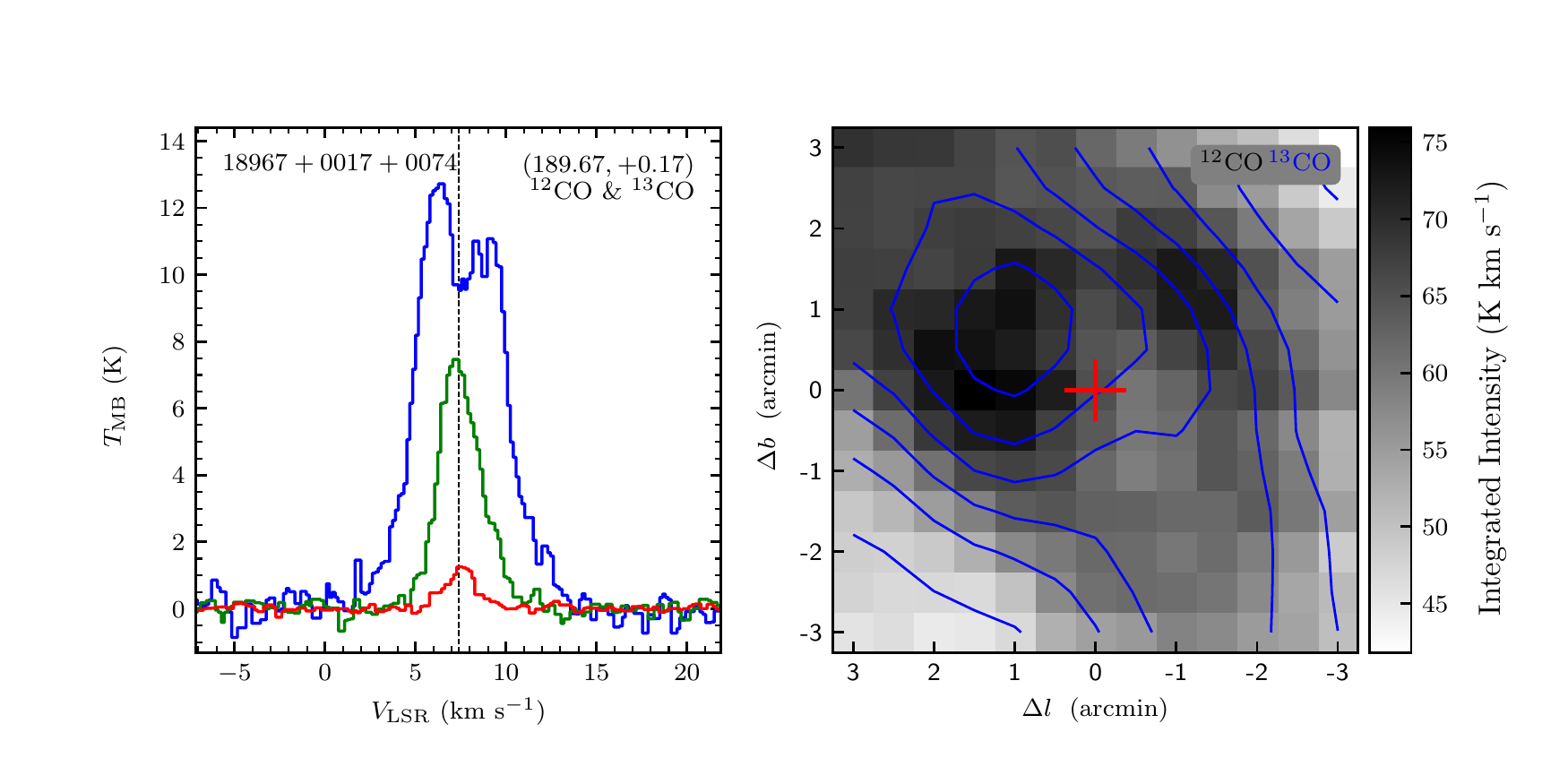}
\includegraphics[width=9.0cm,angle=0]{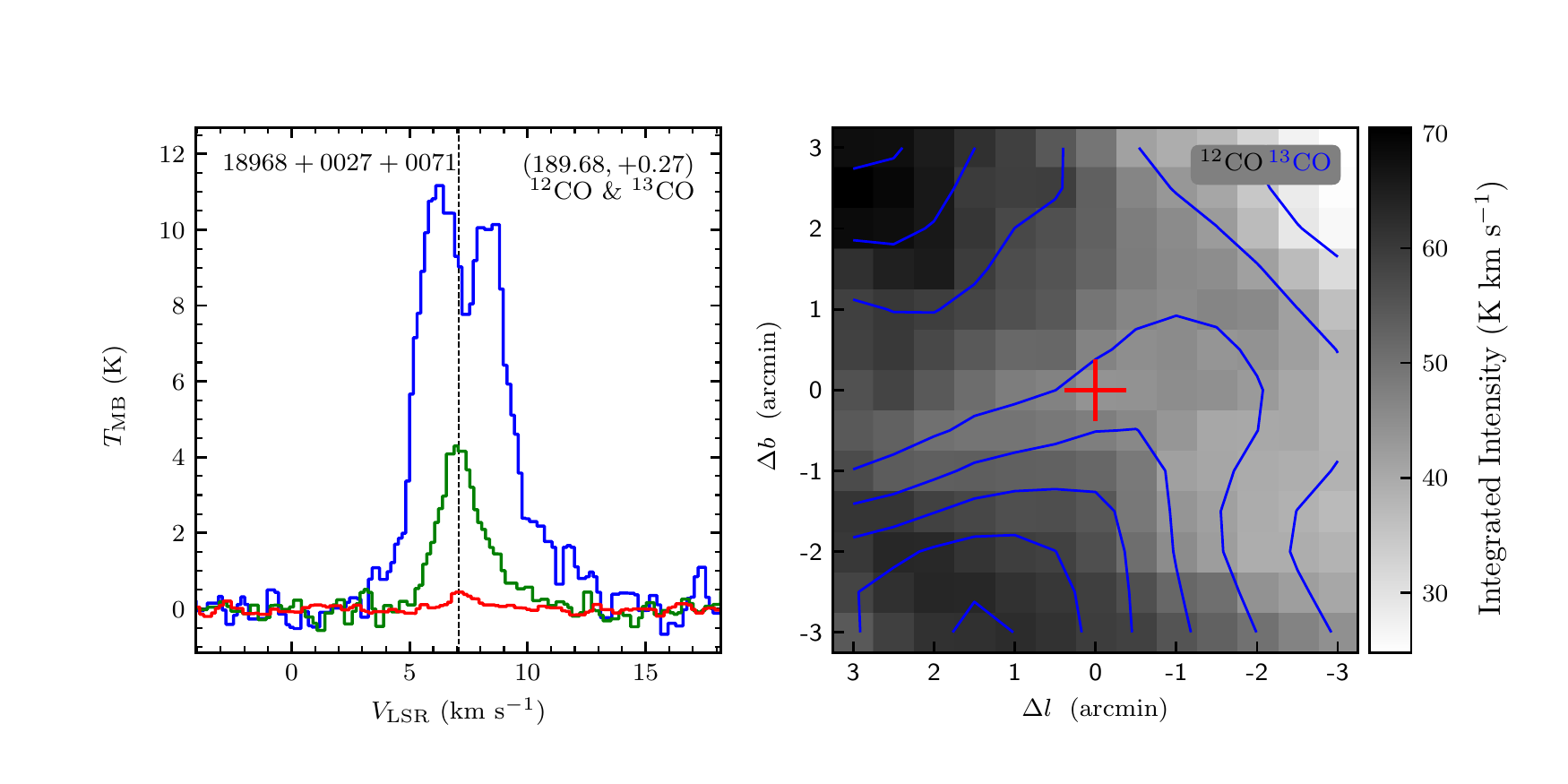}
\end{figure}
\clearpage

\begin{figure}
\includegraphics[width=9.0cm,angle=0]{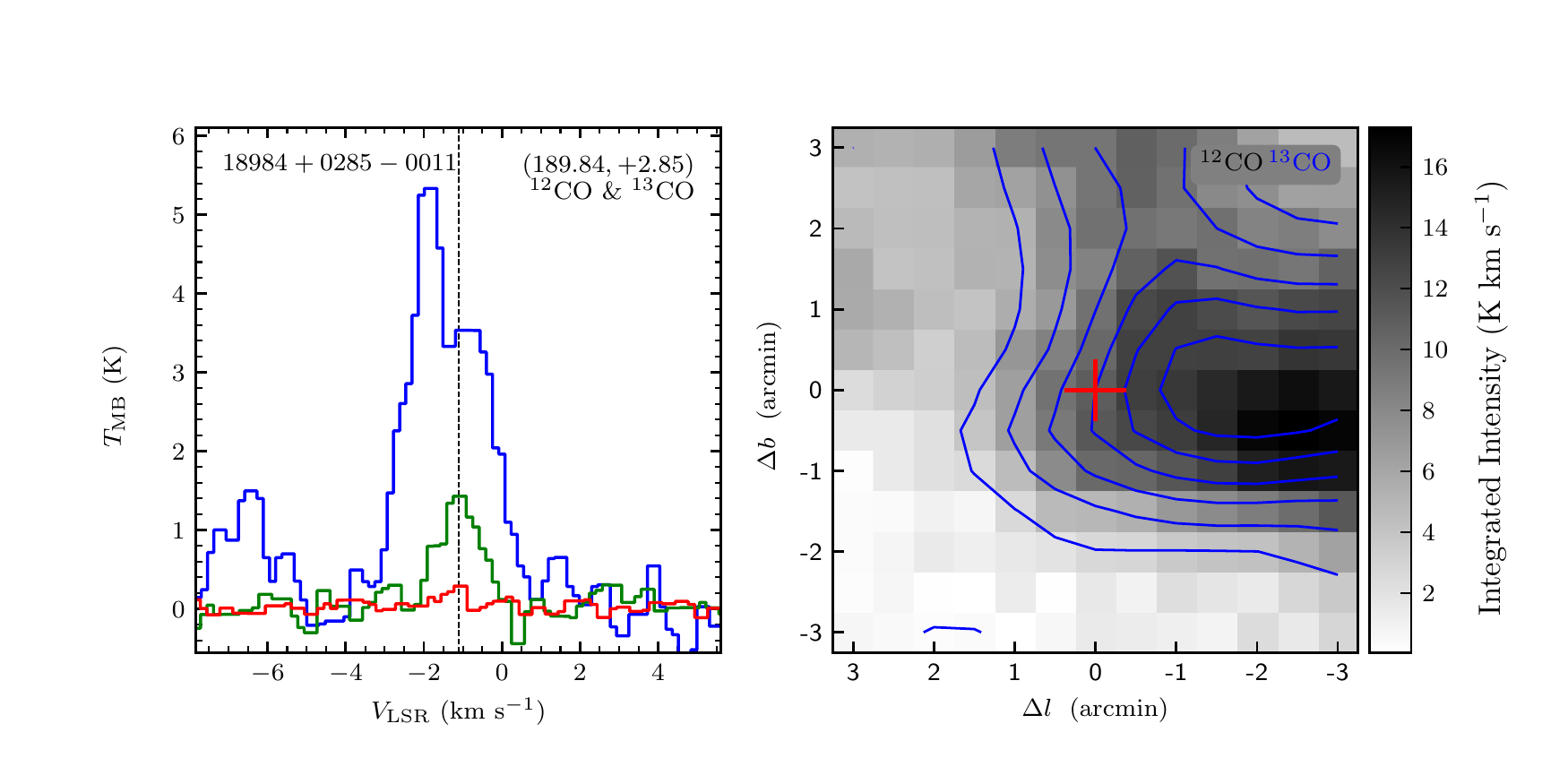}
\includegraphics[width=9.0cm,angle=0]{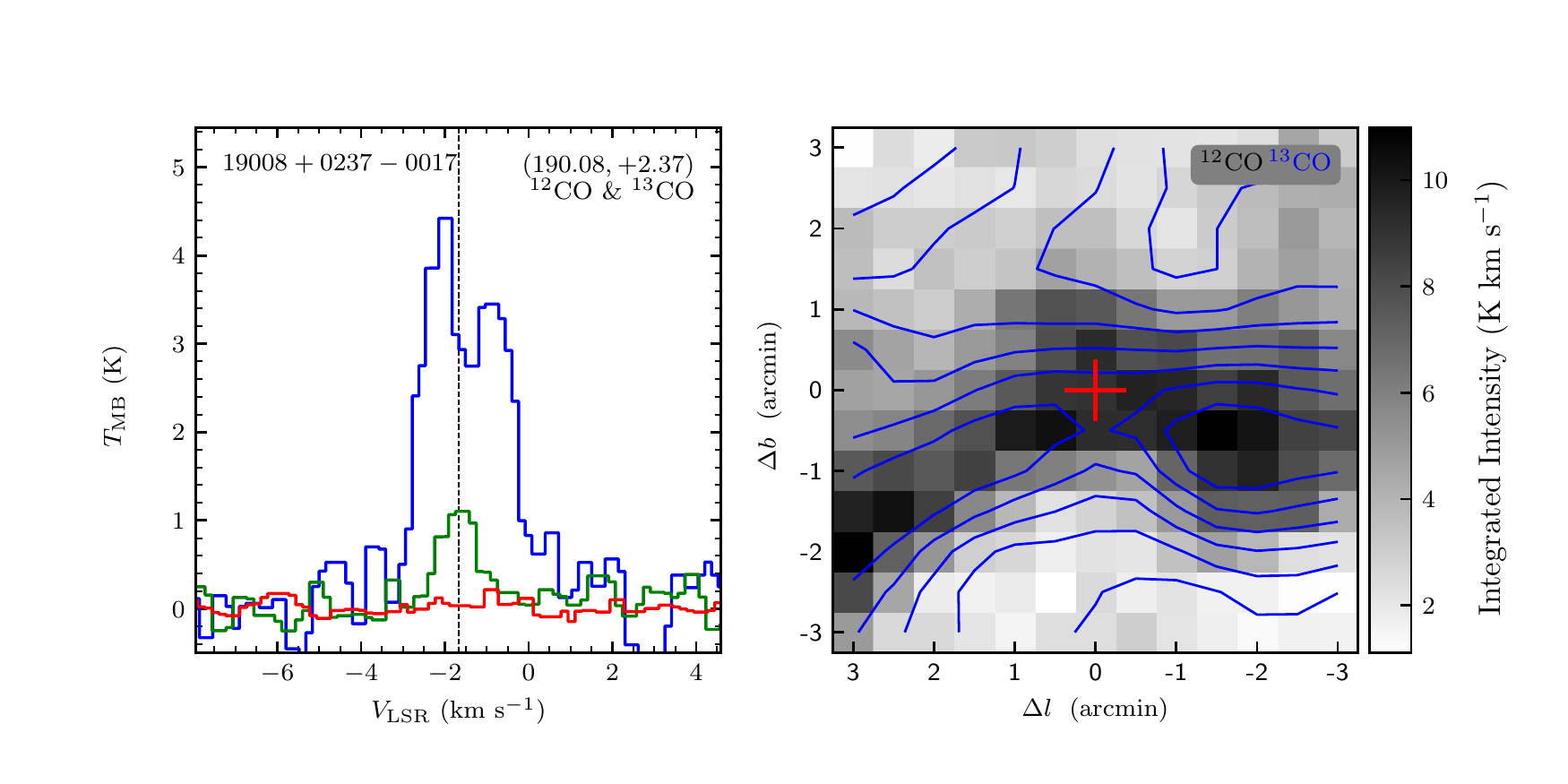}
\vspace{-0.5cm}

\includegraphics[width=9.0cm,angle=0]{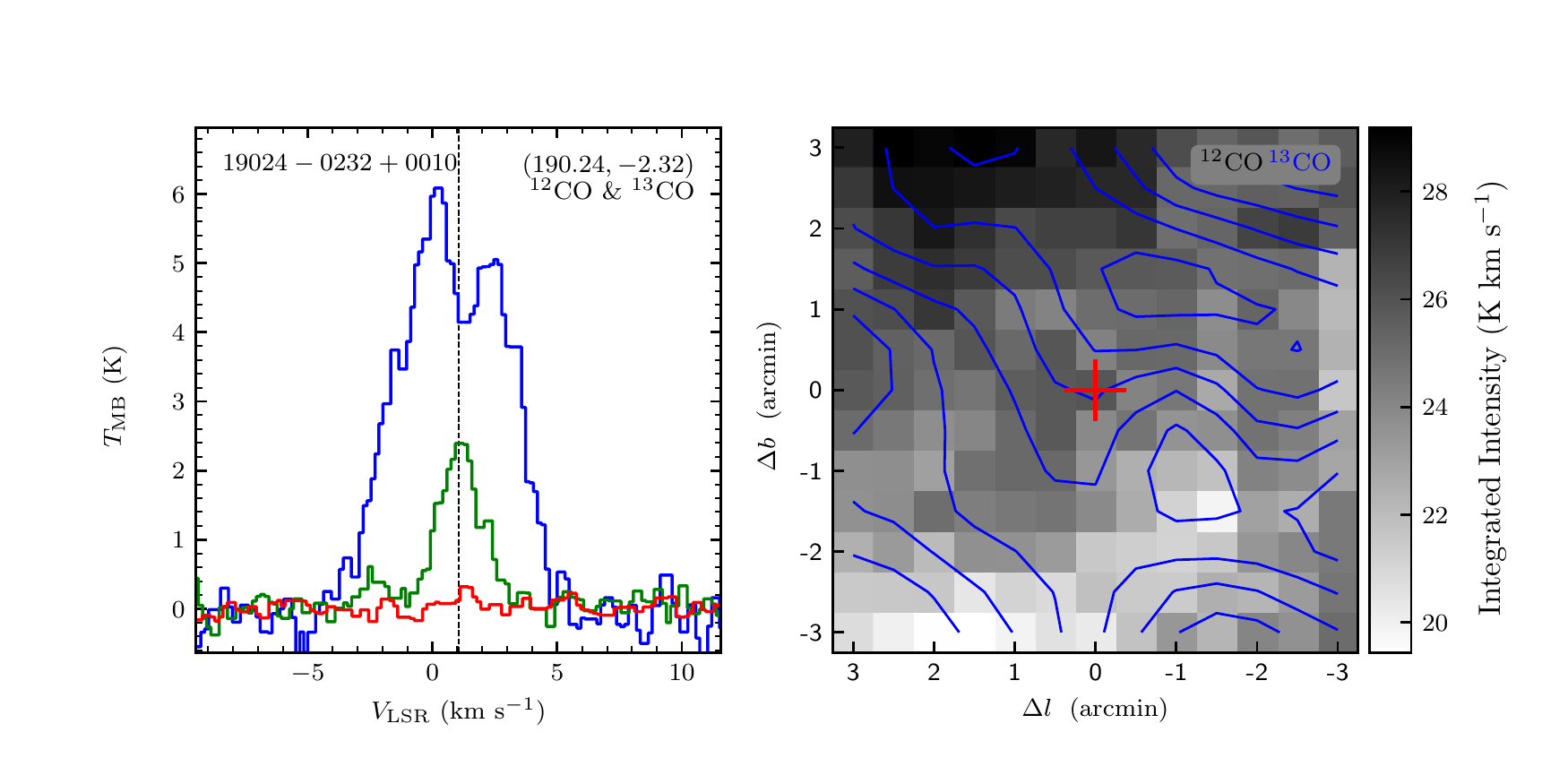}
\includegraphics[width=9.0cm,angle=0]{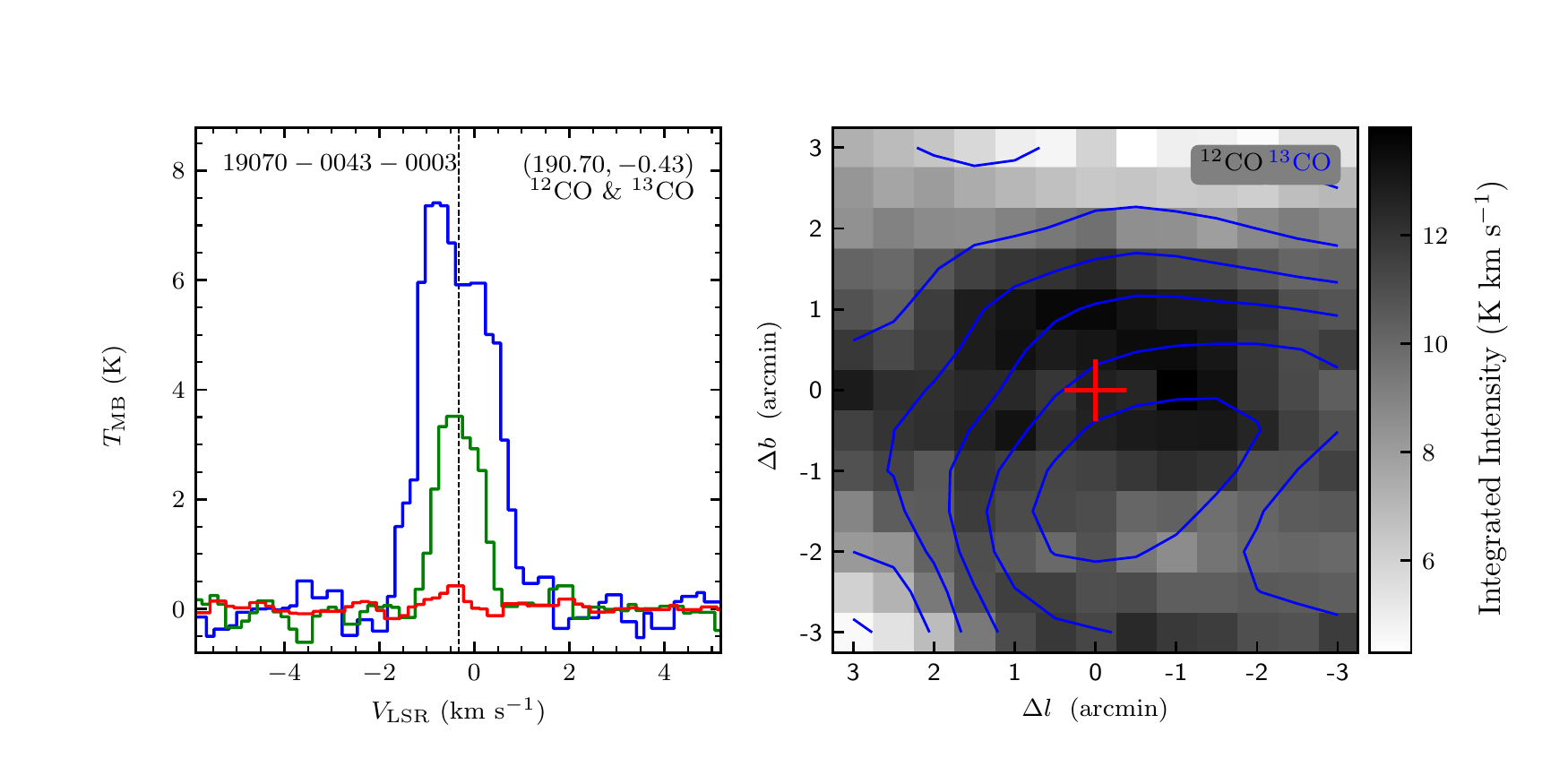}
\vspace{-0.5cm}

\includegraphics[width=9.0cm,angle=0]{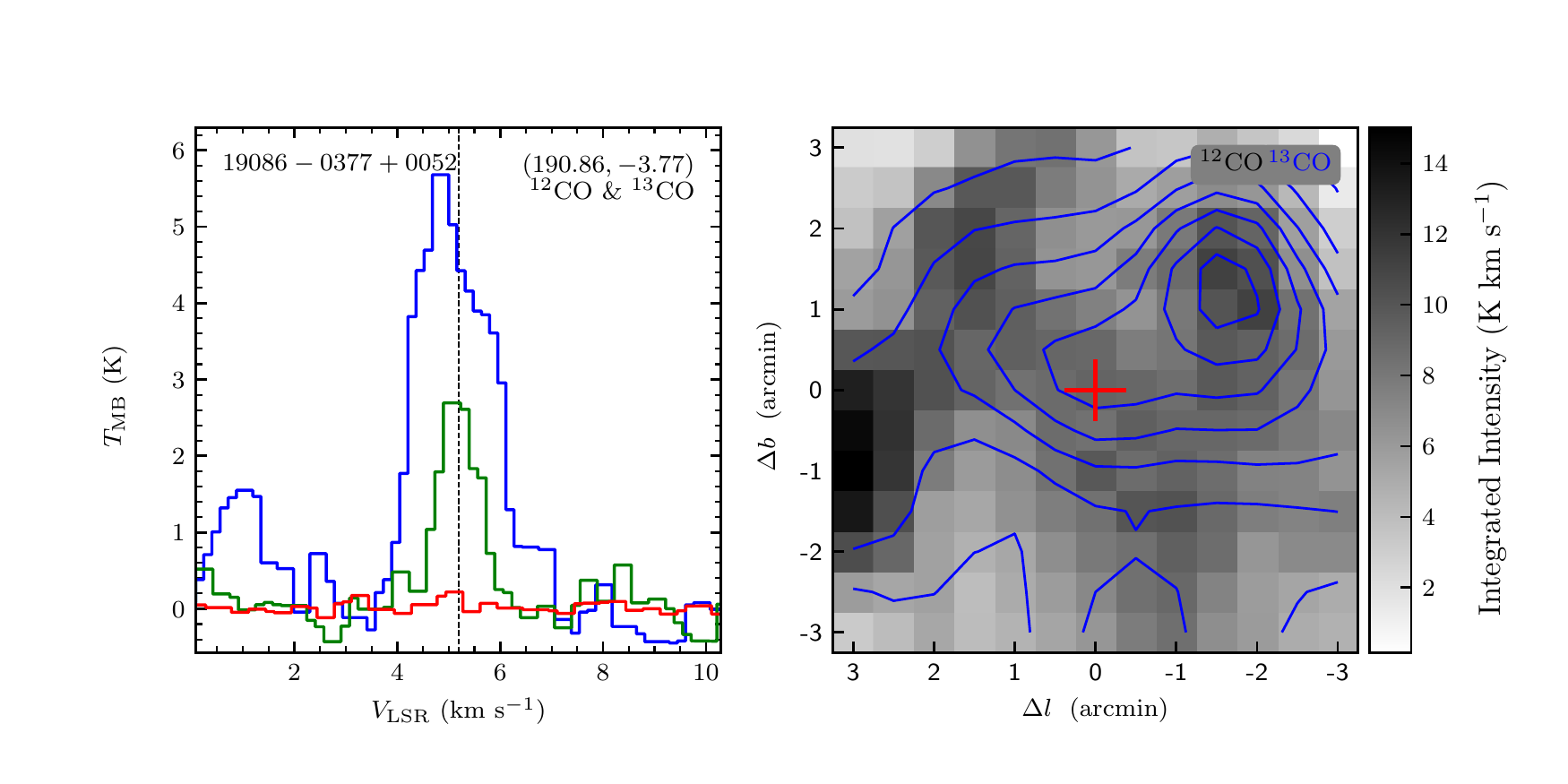}
\includegraphics[width=9.0cm,angle=0]{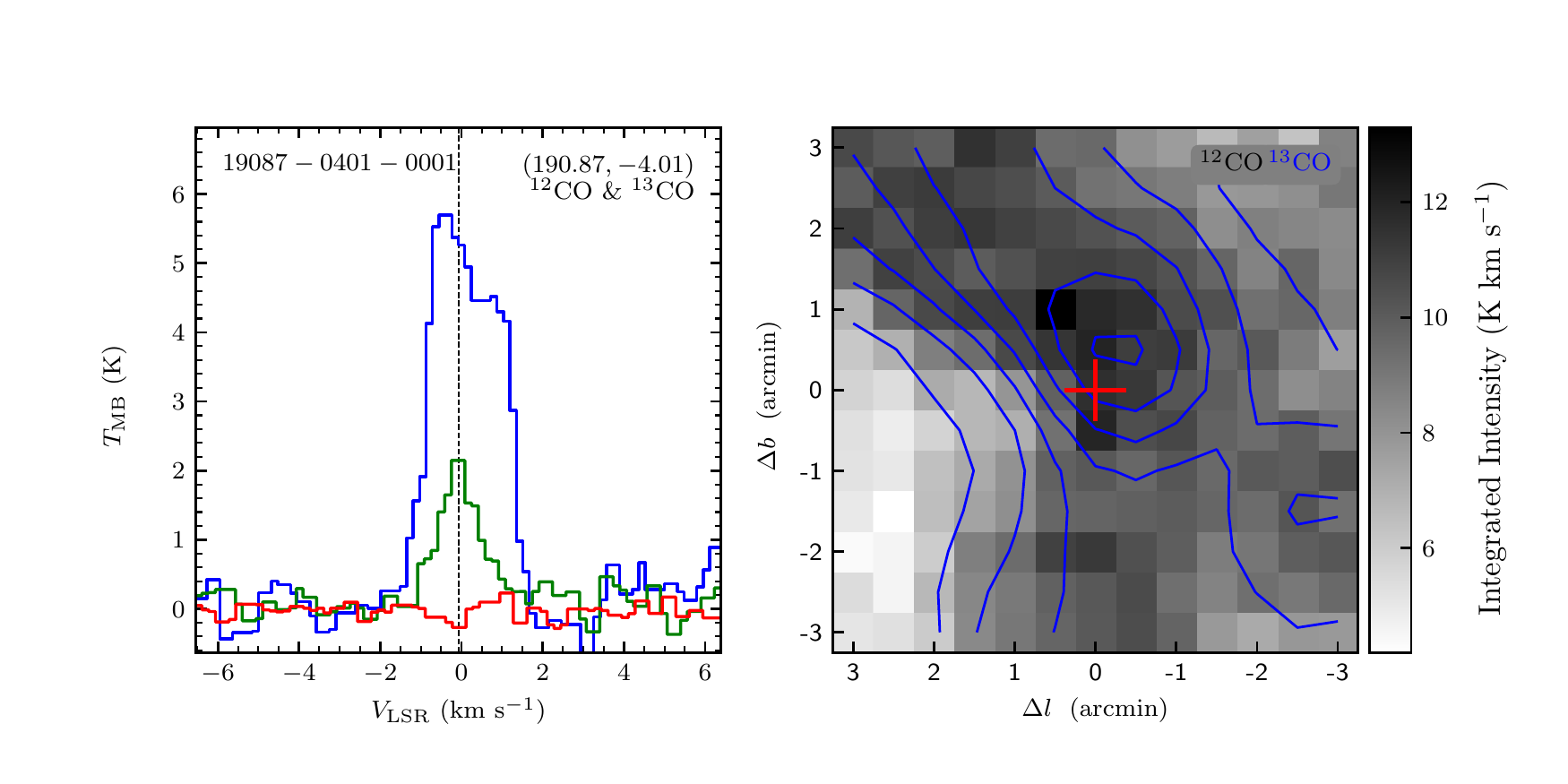}
\vspace{-0.5cm}

\includegraphics[width=9.0cm,angle=0]{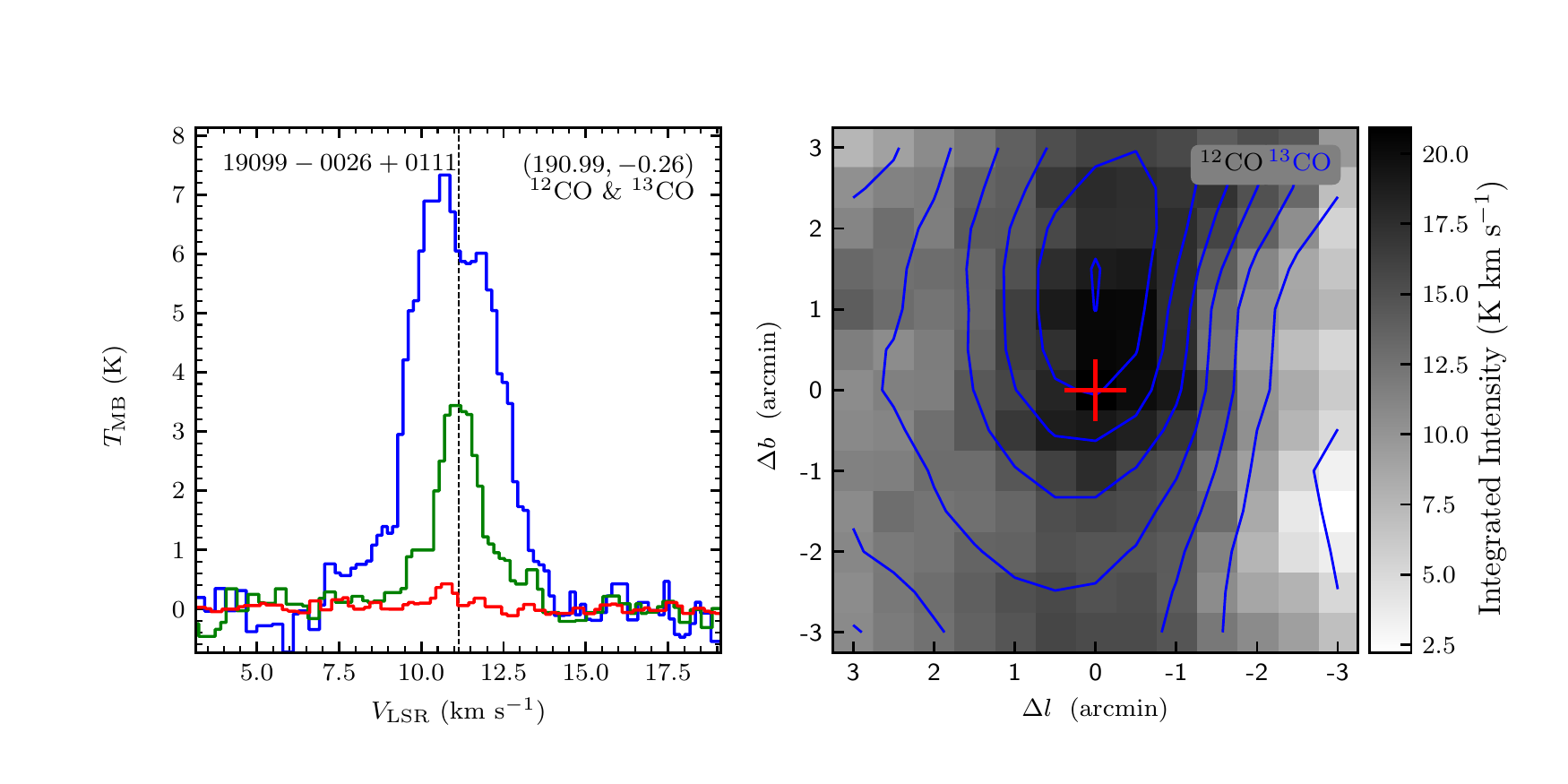}
\includegraphics[width=9.0cm,angle=0]{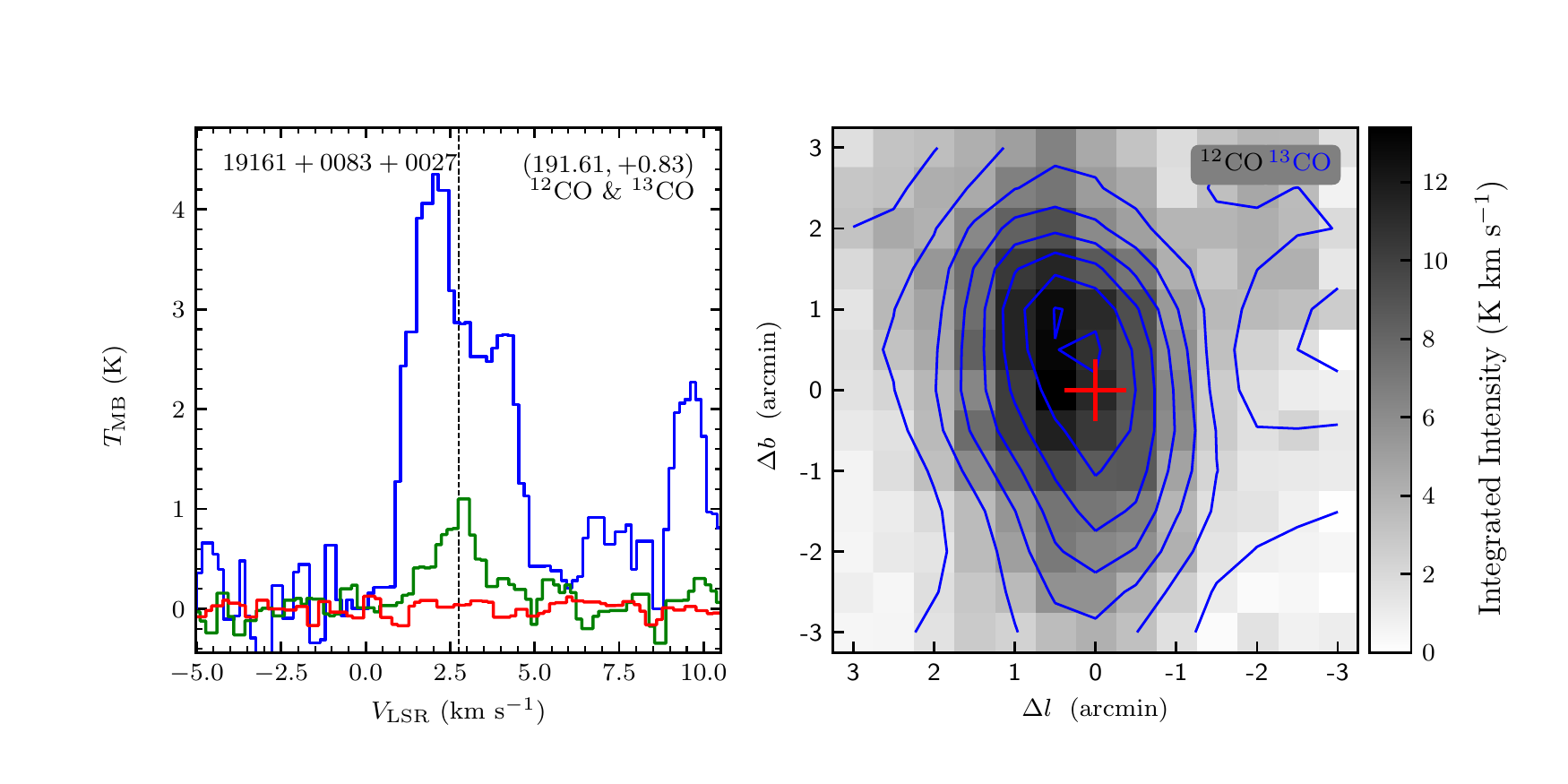}
\vspace{-0.5cm}

\includegraphics[width=9.0cm,angle=0]{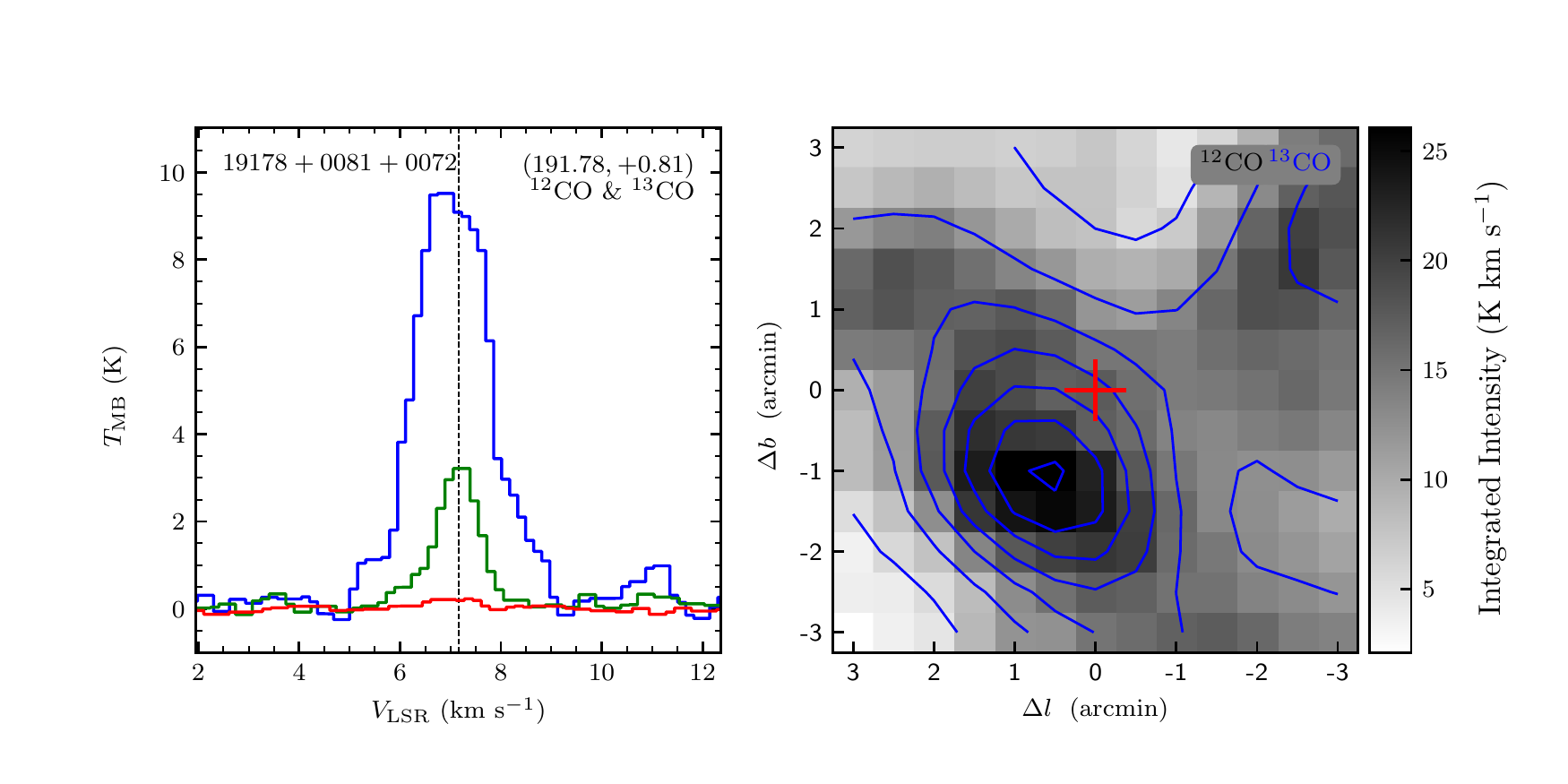}
\includegraphics[width=9.0cm,angle=0]{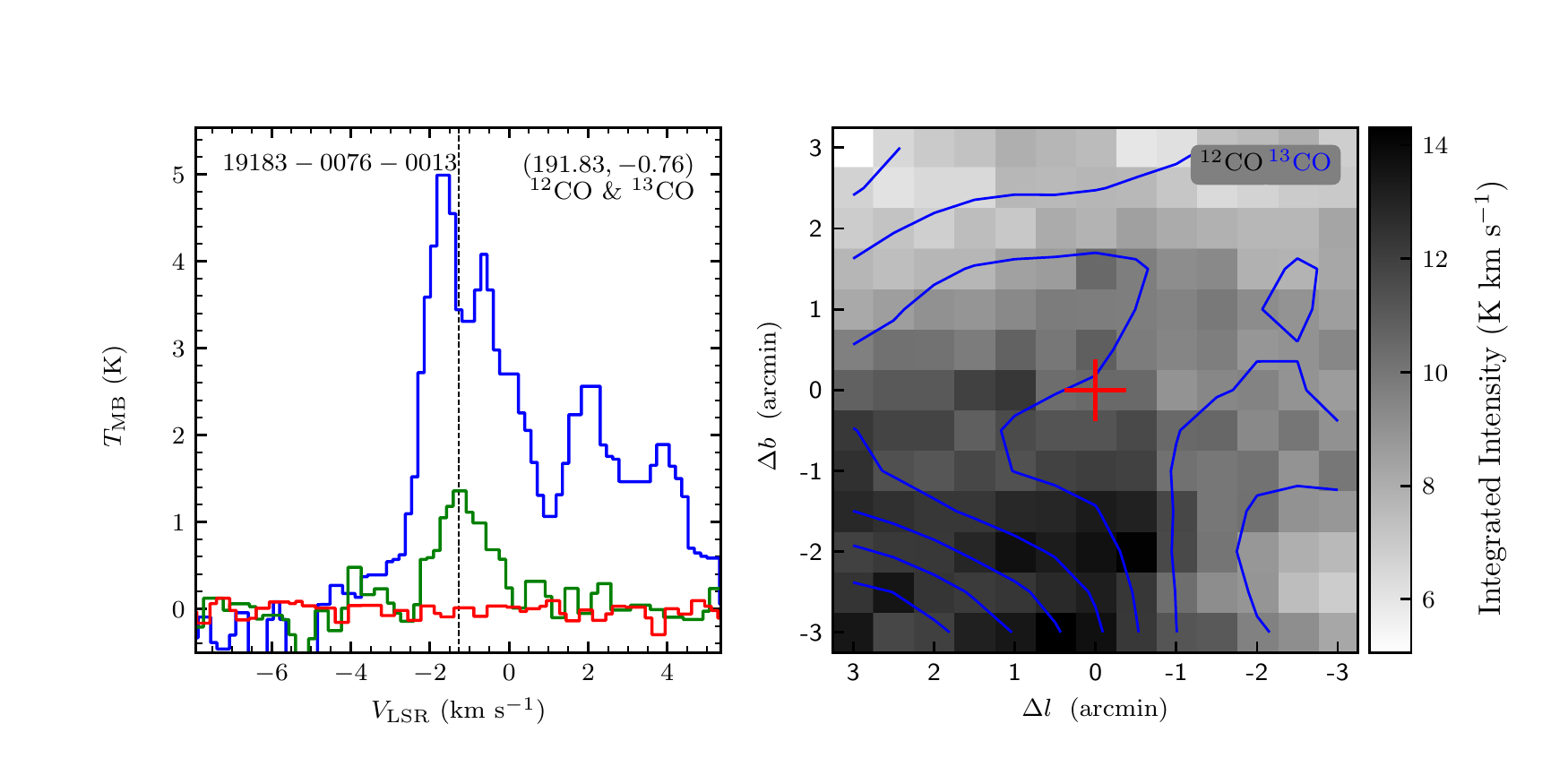}
\end{figure}
\clearpage

\begin{figure}
\includegraphics[width=9.0cm,angle=0]{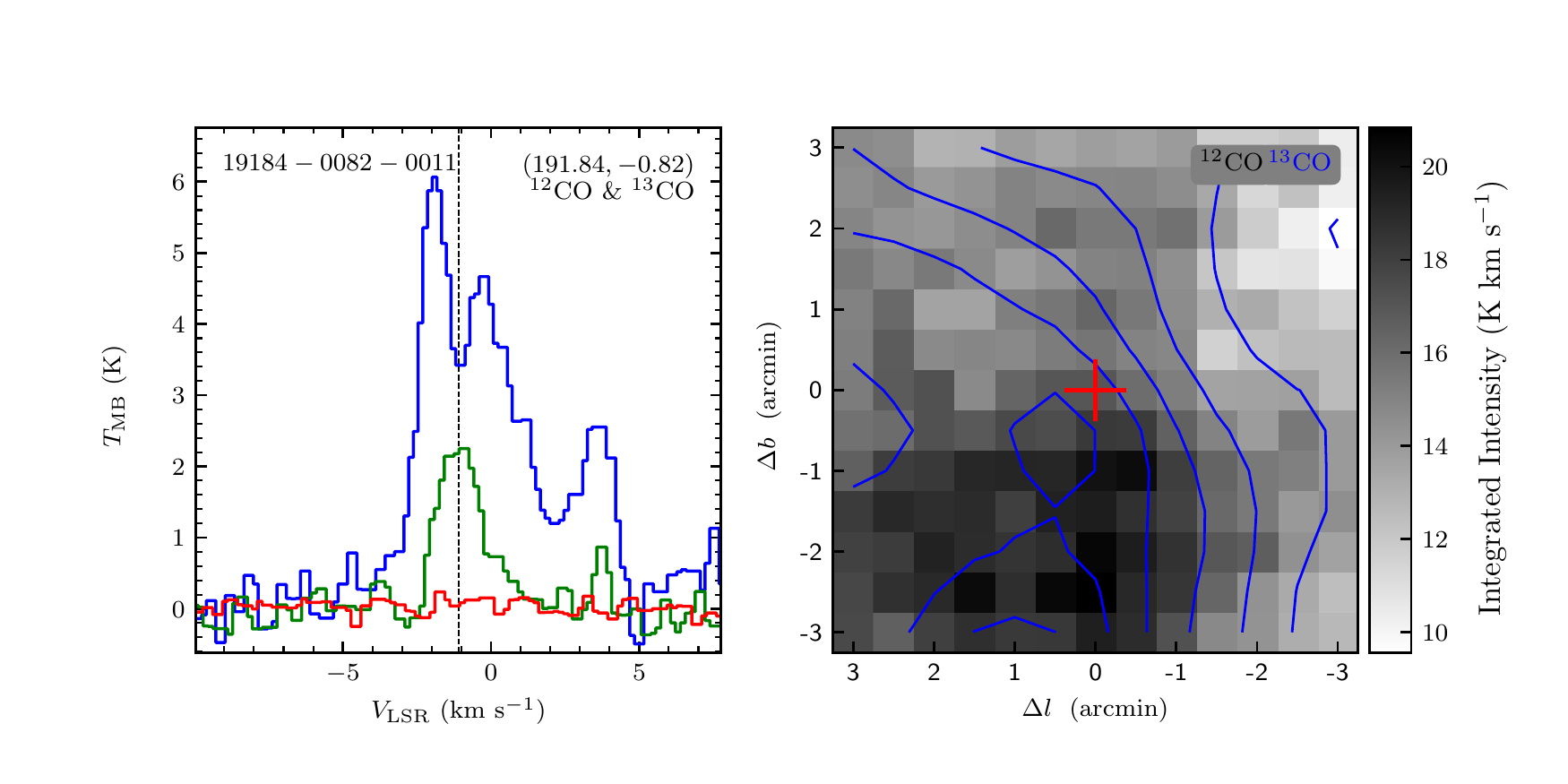}
\includegraphics[width=9.0cm,angle=0]{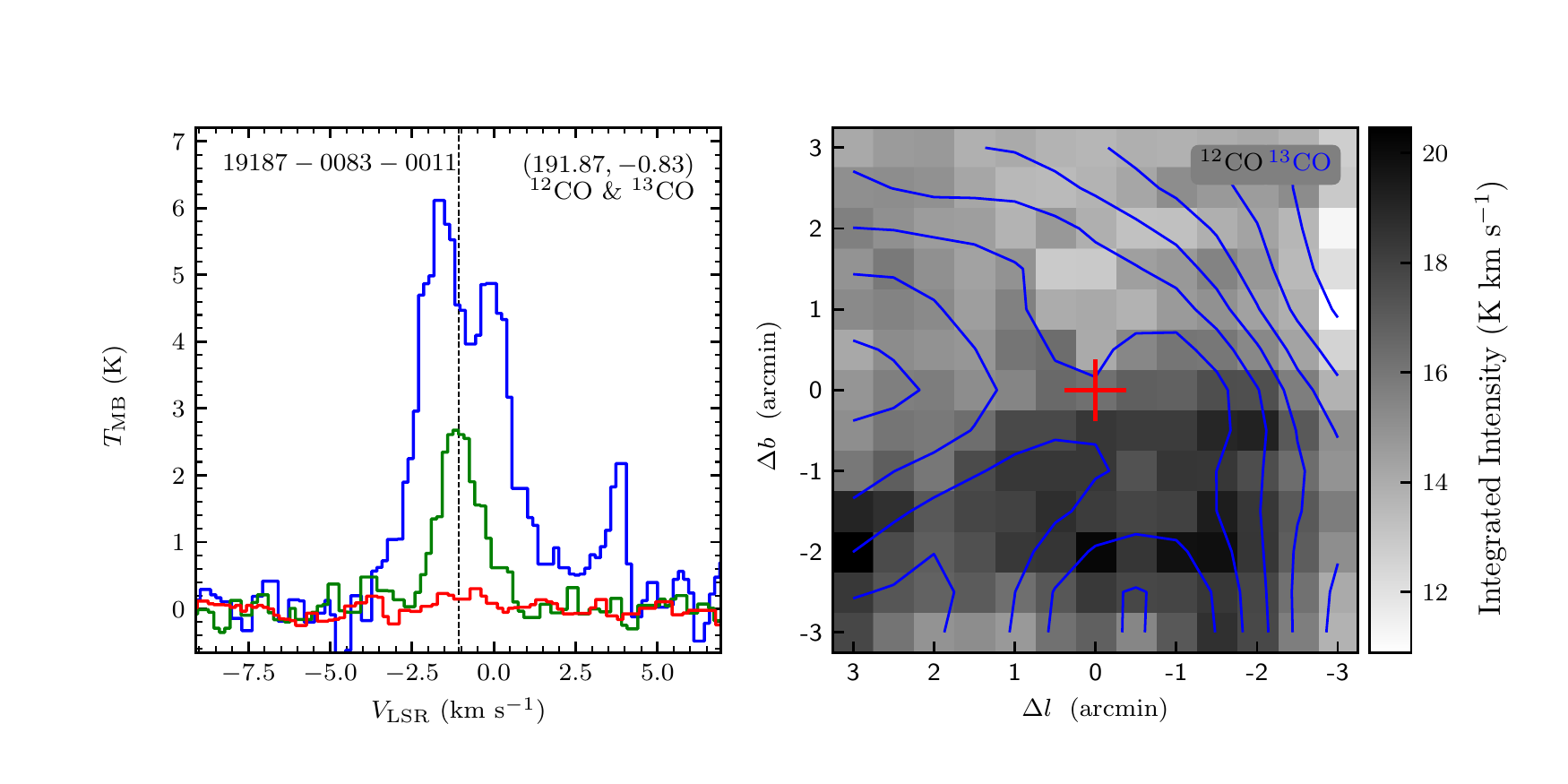}
\vspace{-0.5cm}

\includegraphics[width=9.0cm,angle=0]{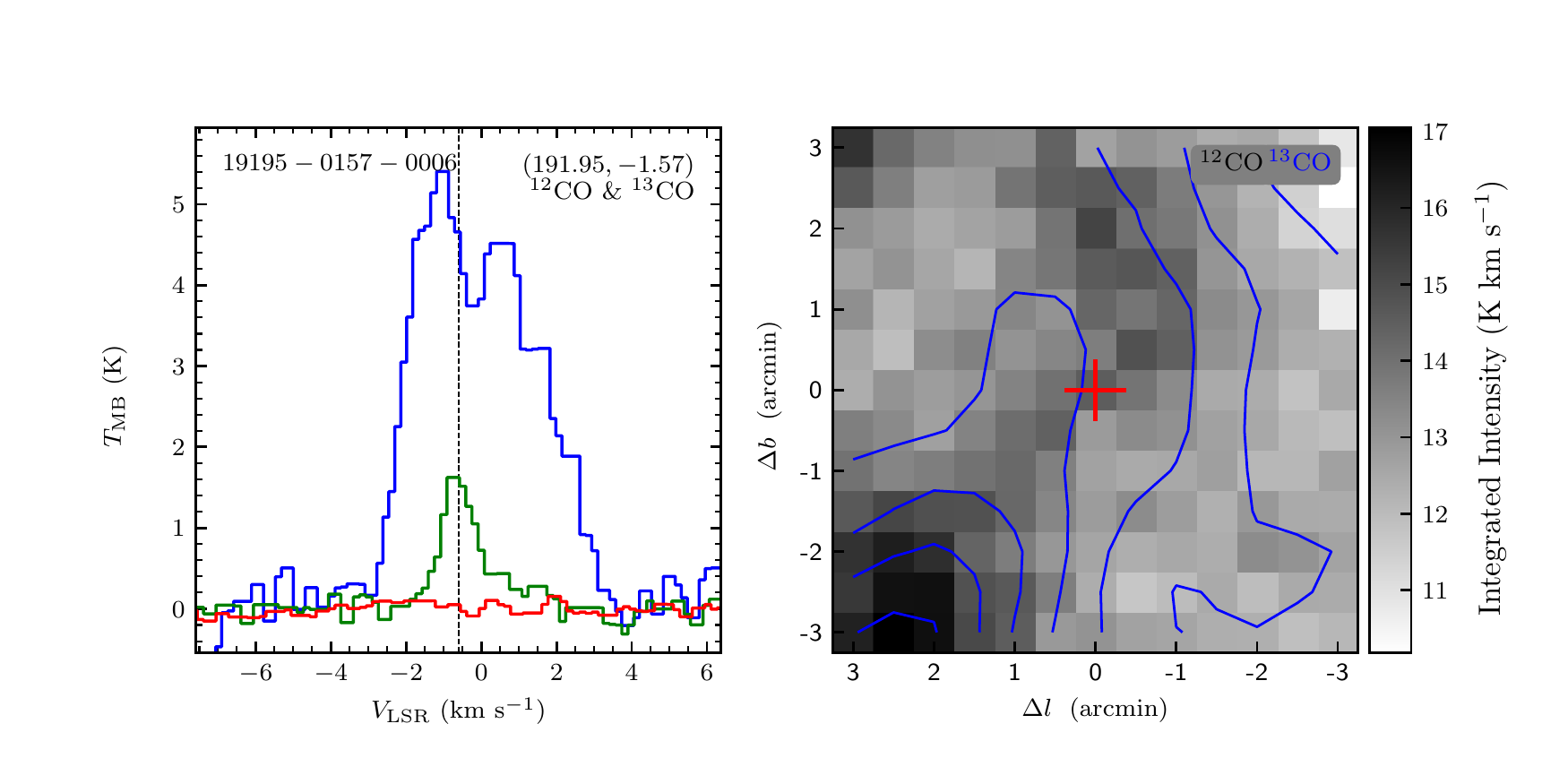}
\includegraphics[width=9.0cm,angle=0]{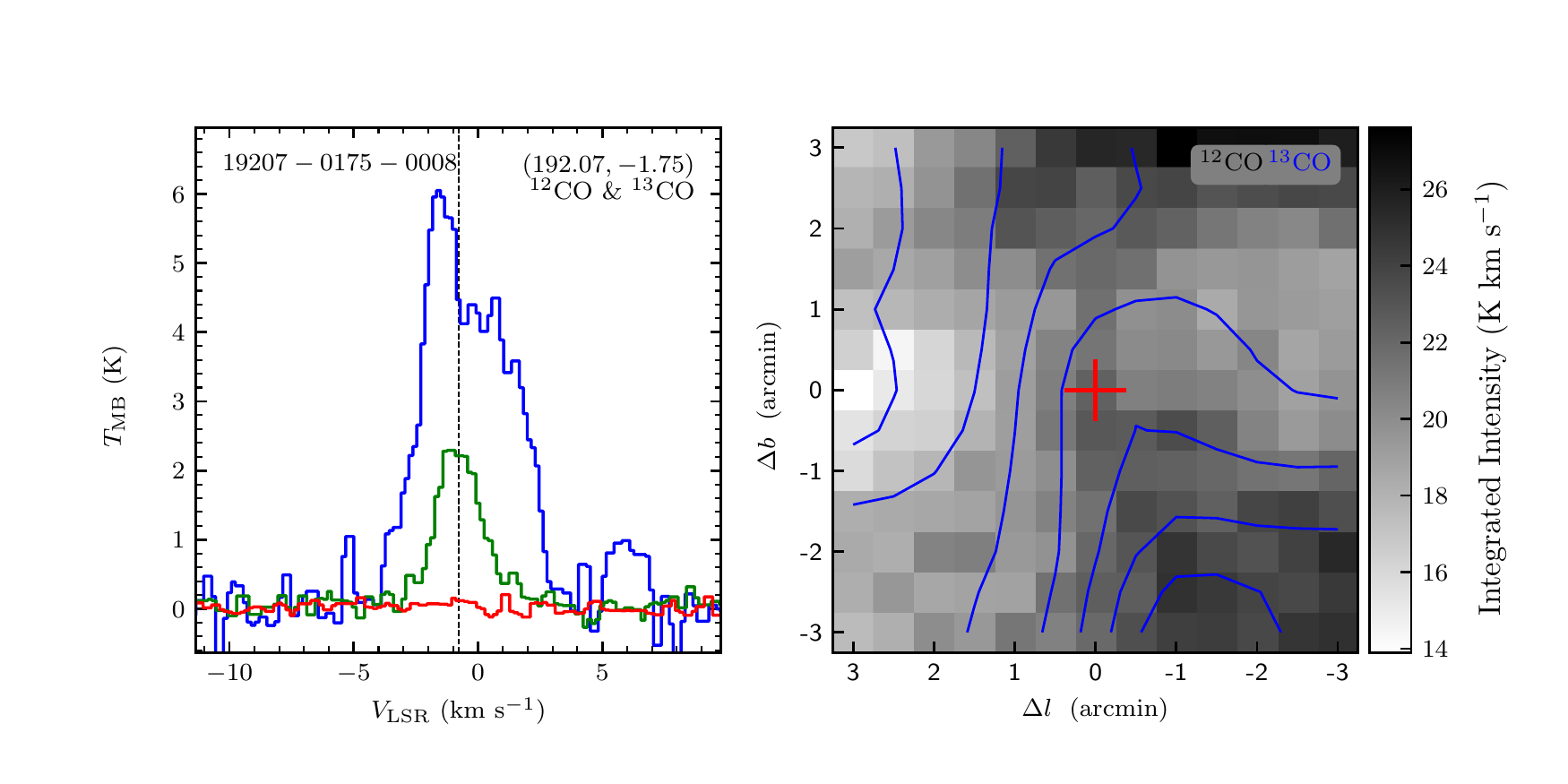}
\vspace{-0.5cm}

\includegraphics[width=9.0cm,angle=0]{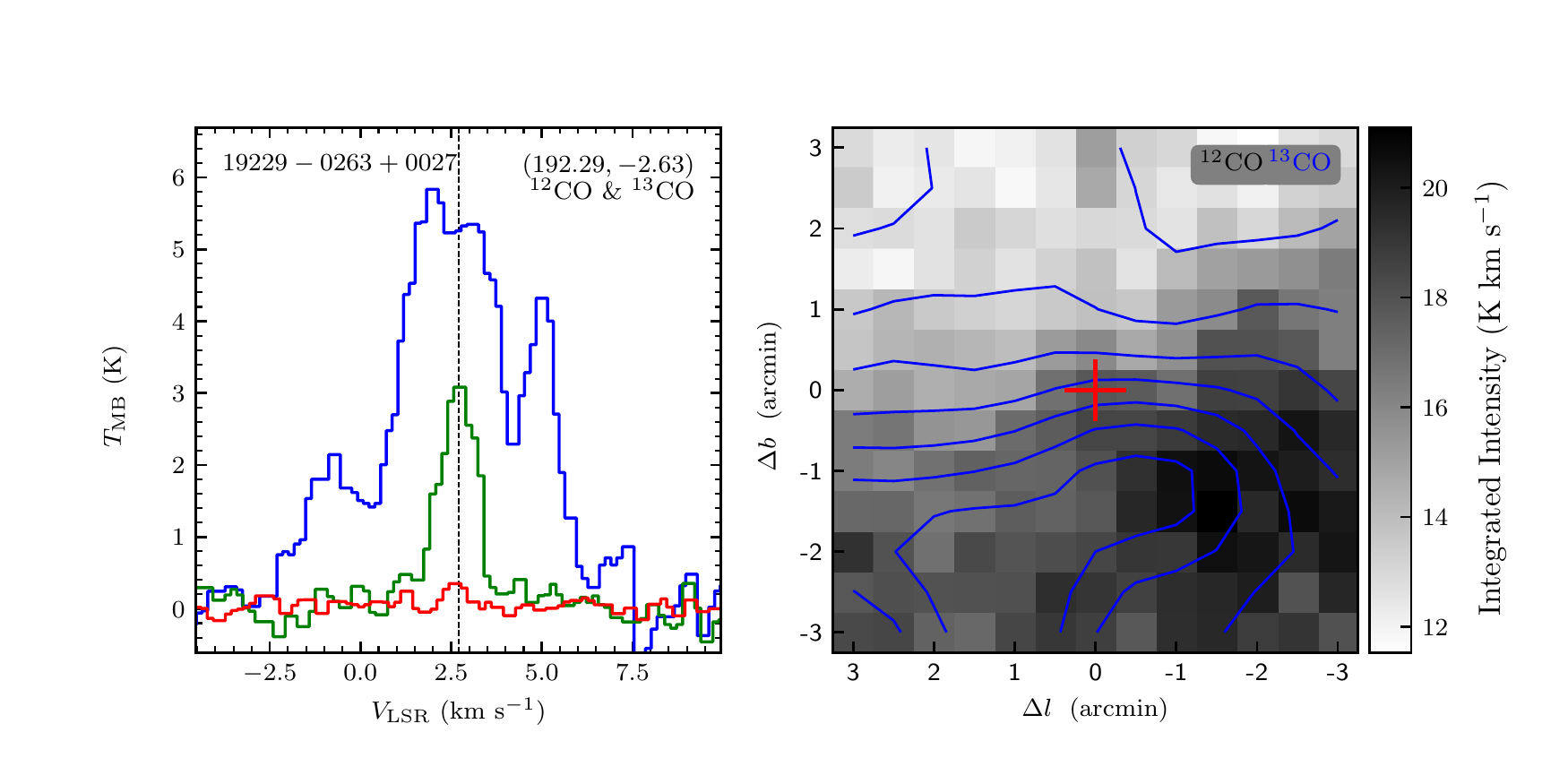}
\includegraphics[width=9.0cm,angle=0]{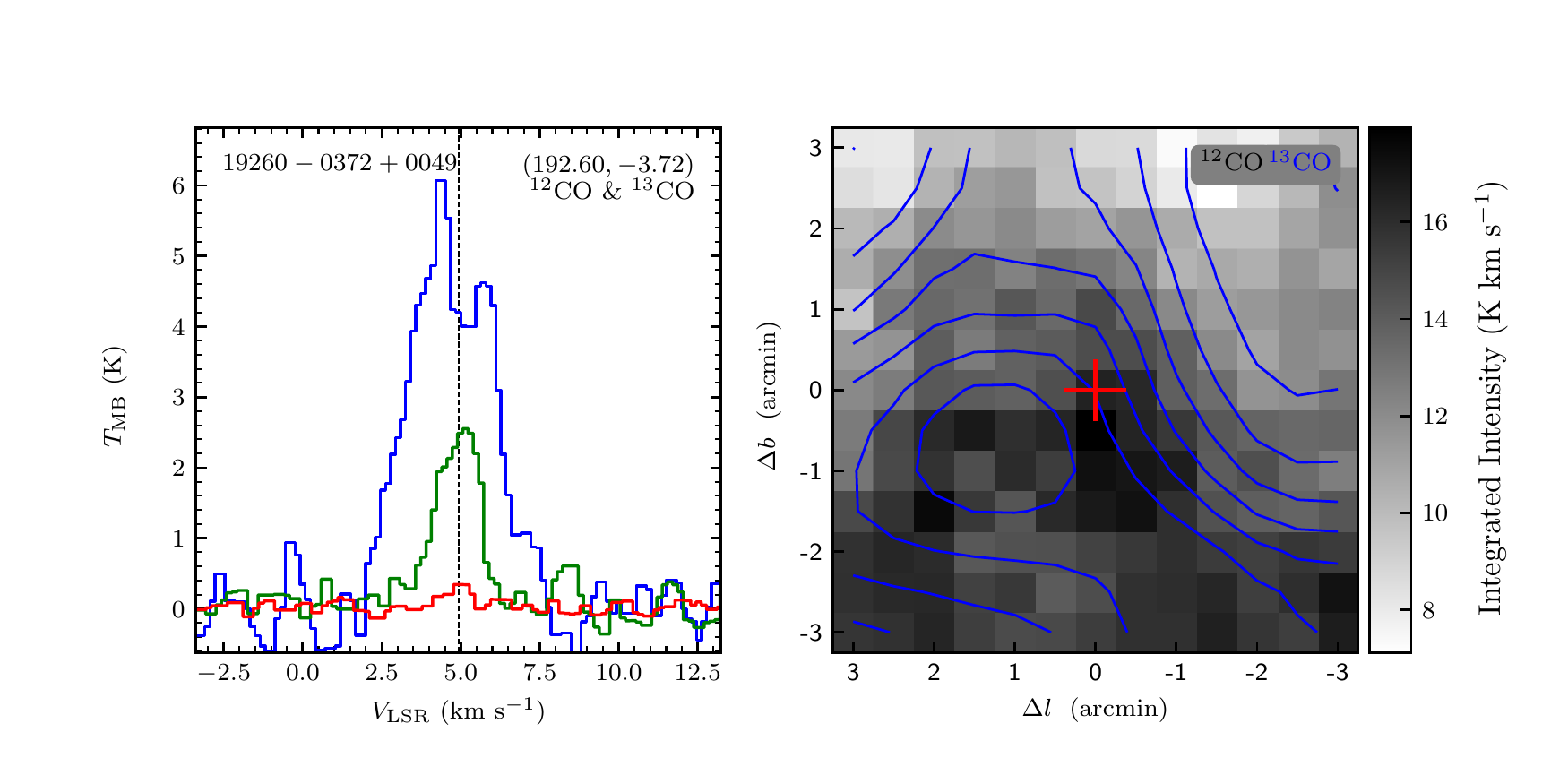}
\vspace{-0.5cm}

\includegraphics[width=9.0cm,angle=0]{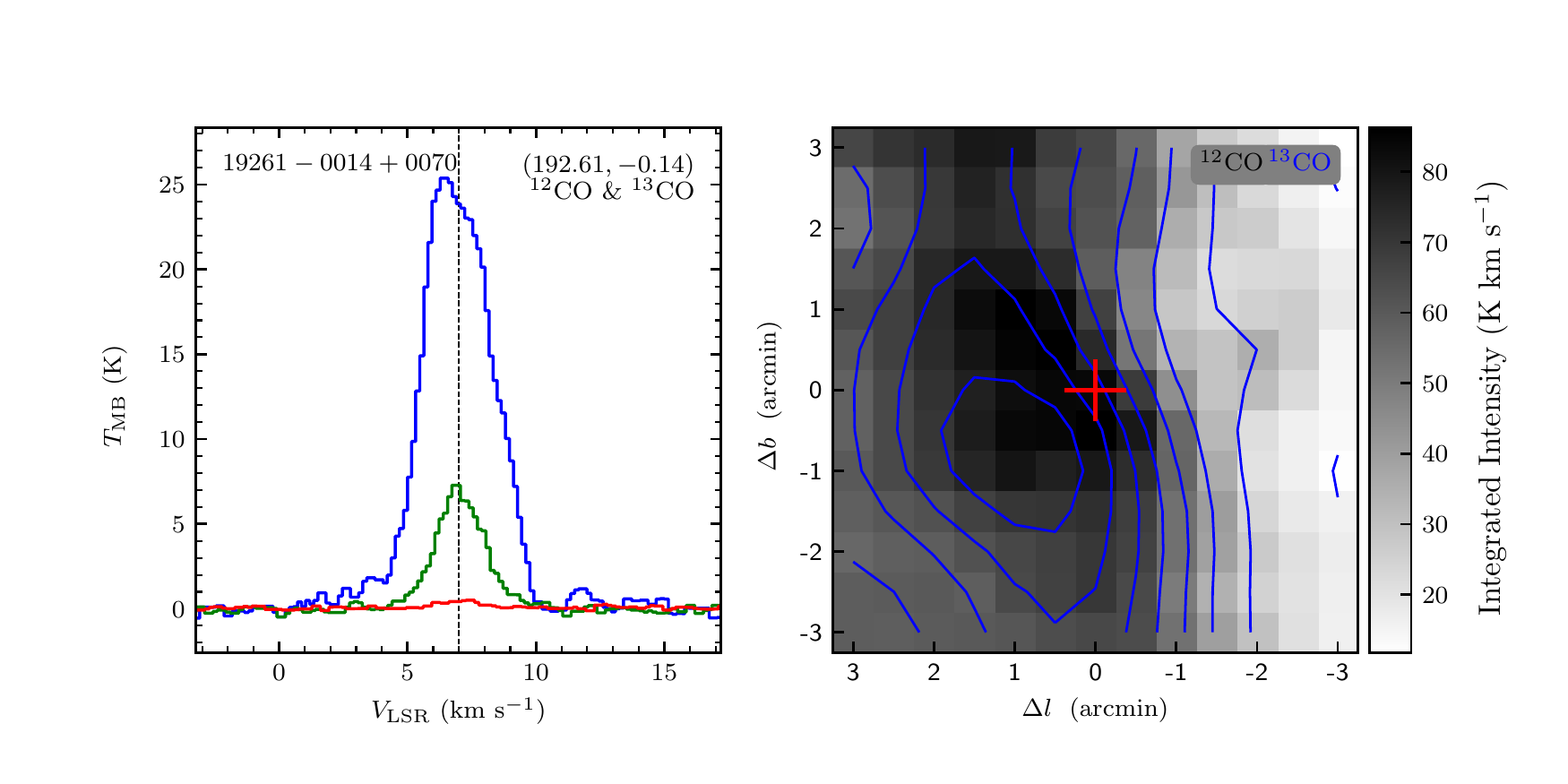}
\includegraphics[width=9.0cm,angle=0]{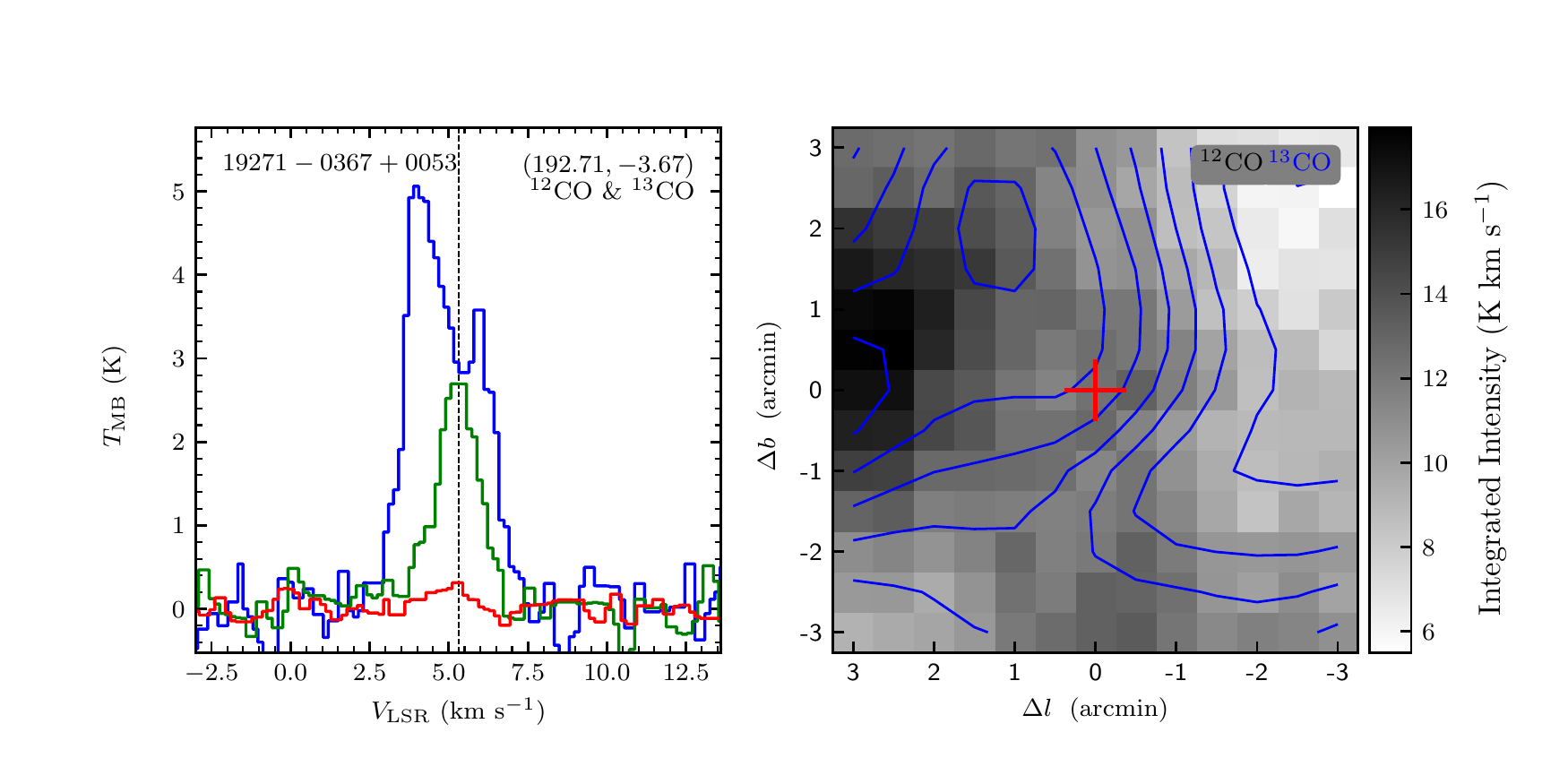}
\vspace{-0.5cm}

\includegraphics[width=9.0cm,angle=0]{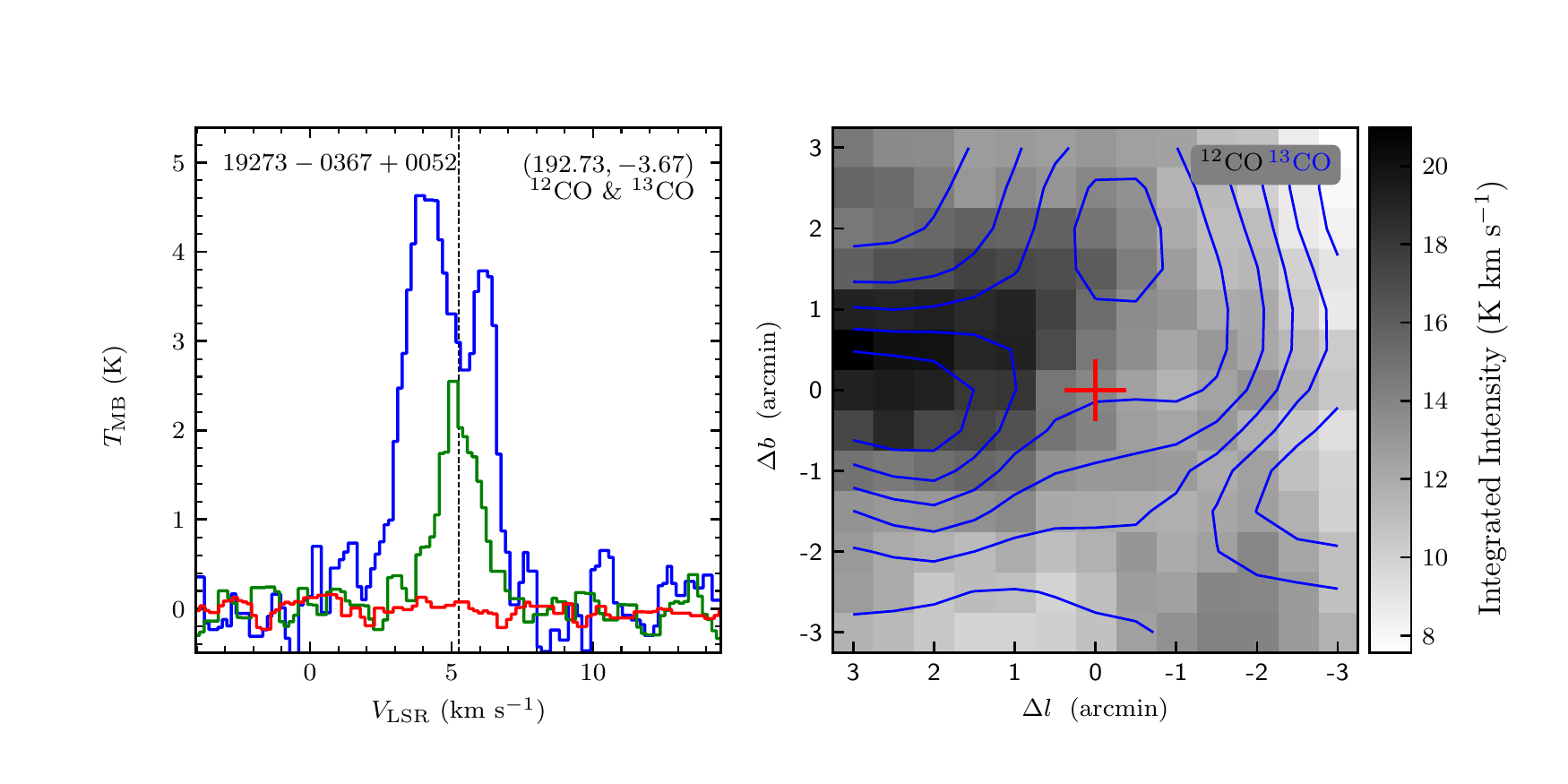}
\includegraphics[width=9.0cm,angle=0]{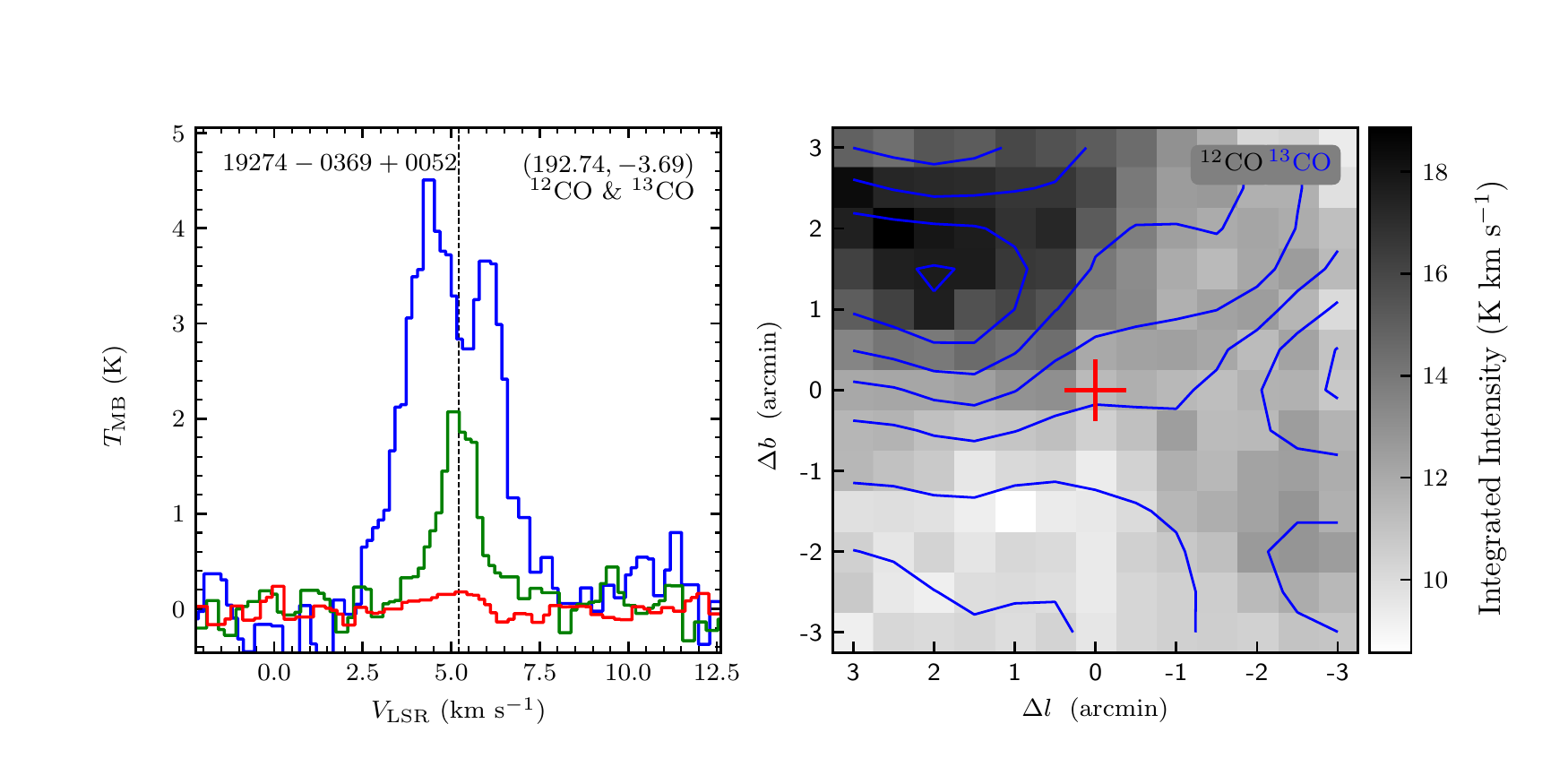}
\end{figure}
\clearpage

\begin{figure}
\includegraphics[width=9.0cm,angle=0]{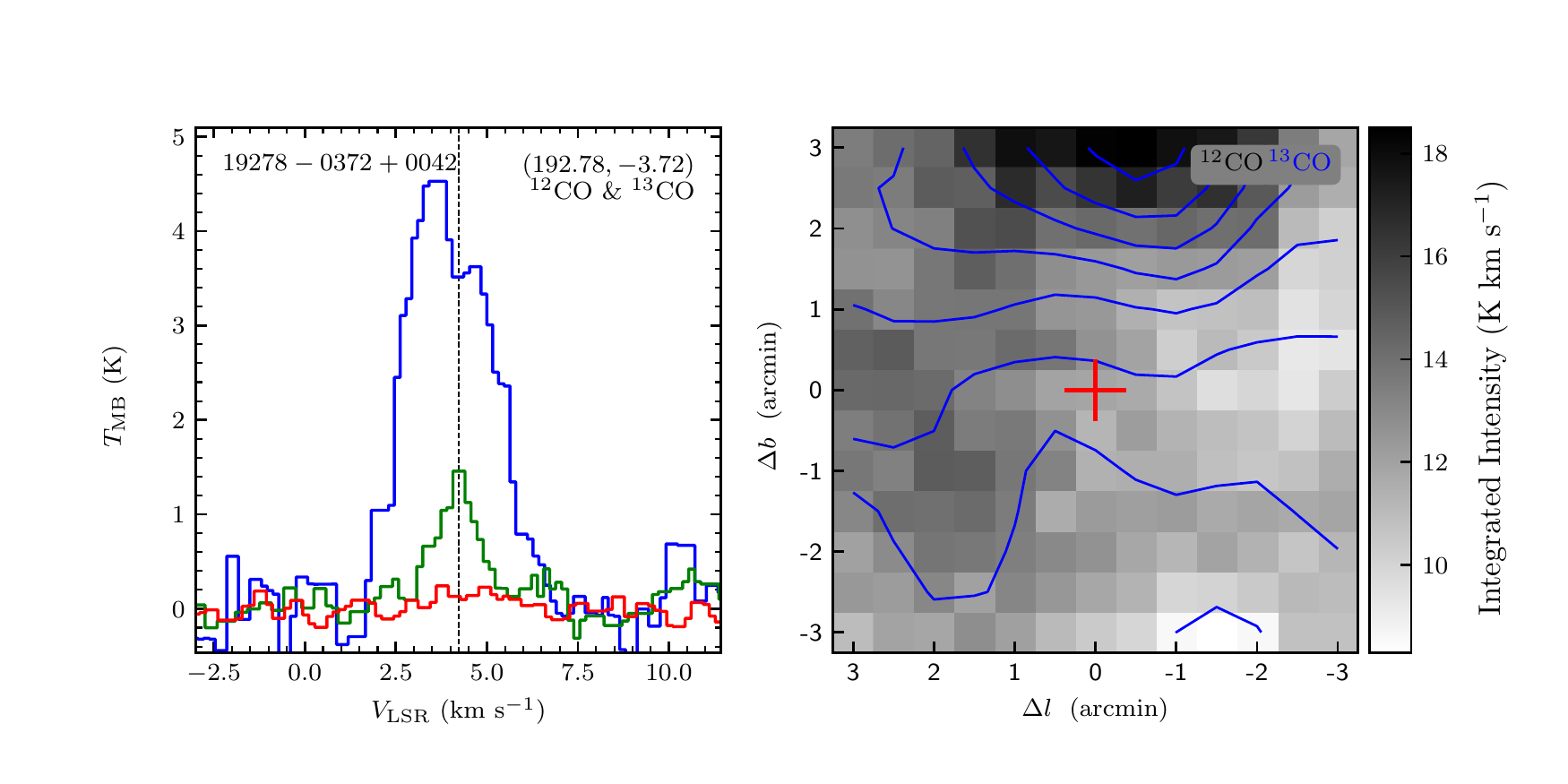}
\includegraphics[width=9.0cm,angle=0]{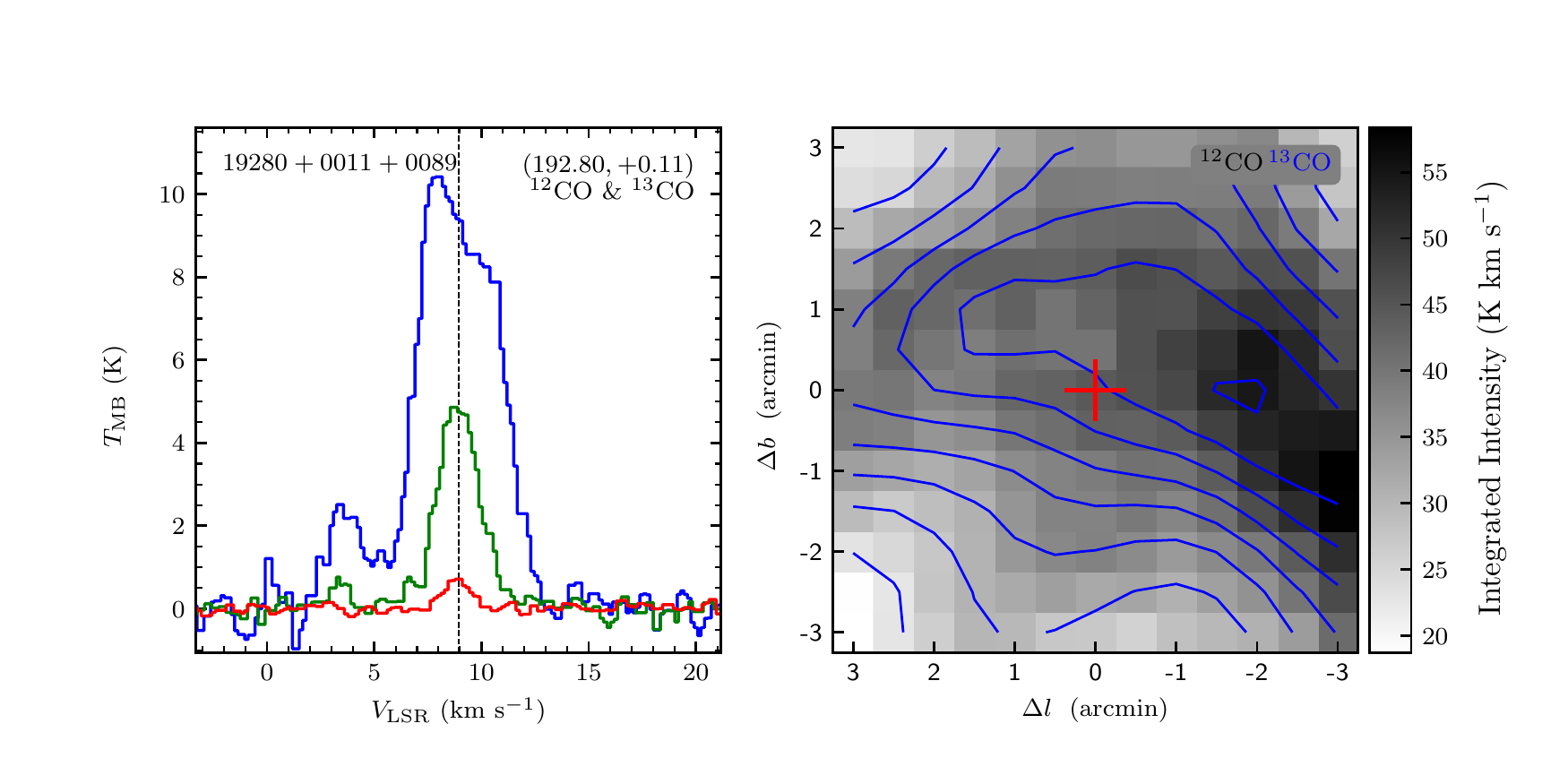}
\vspace{-0.5cm}

\includegraphics[width=9.0cm,angle=0]{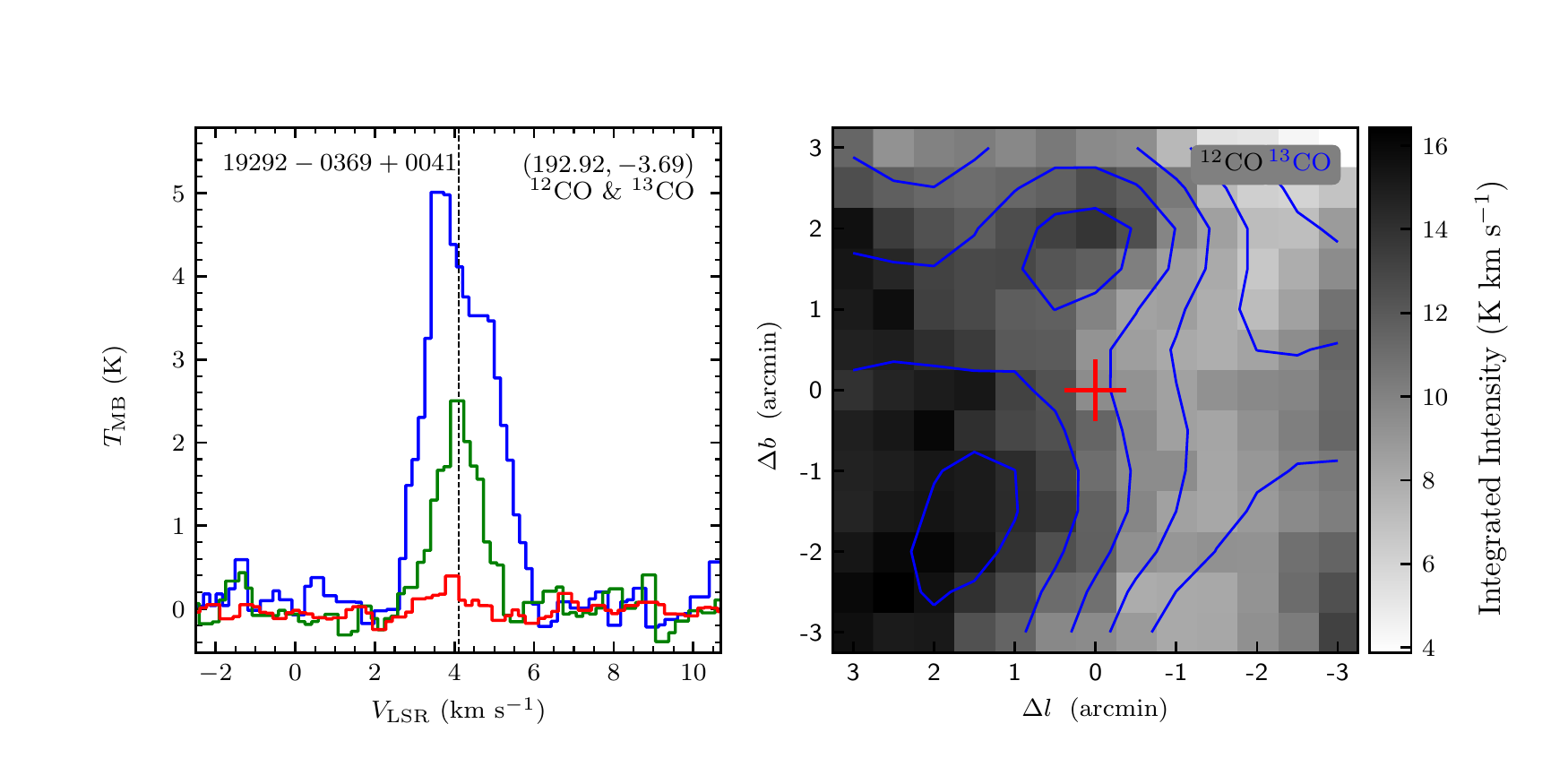}
\includegraphics[width=9.0cm,angle=0]{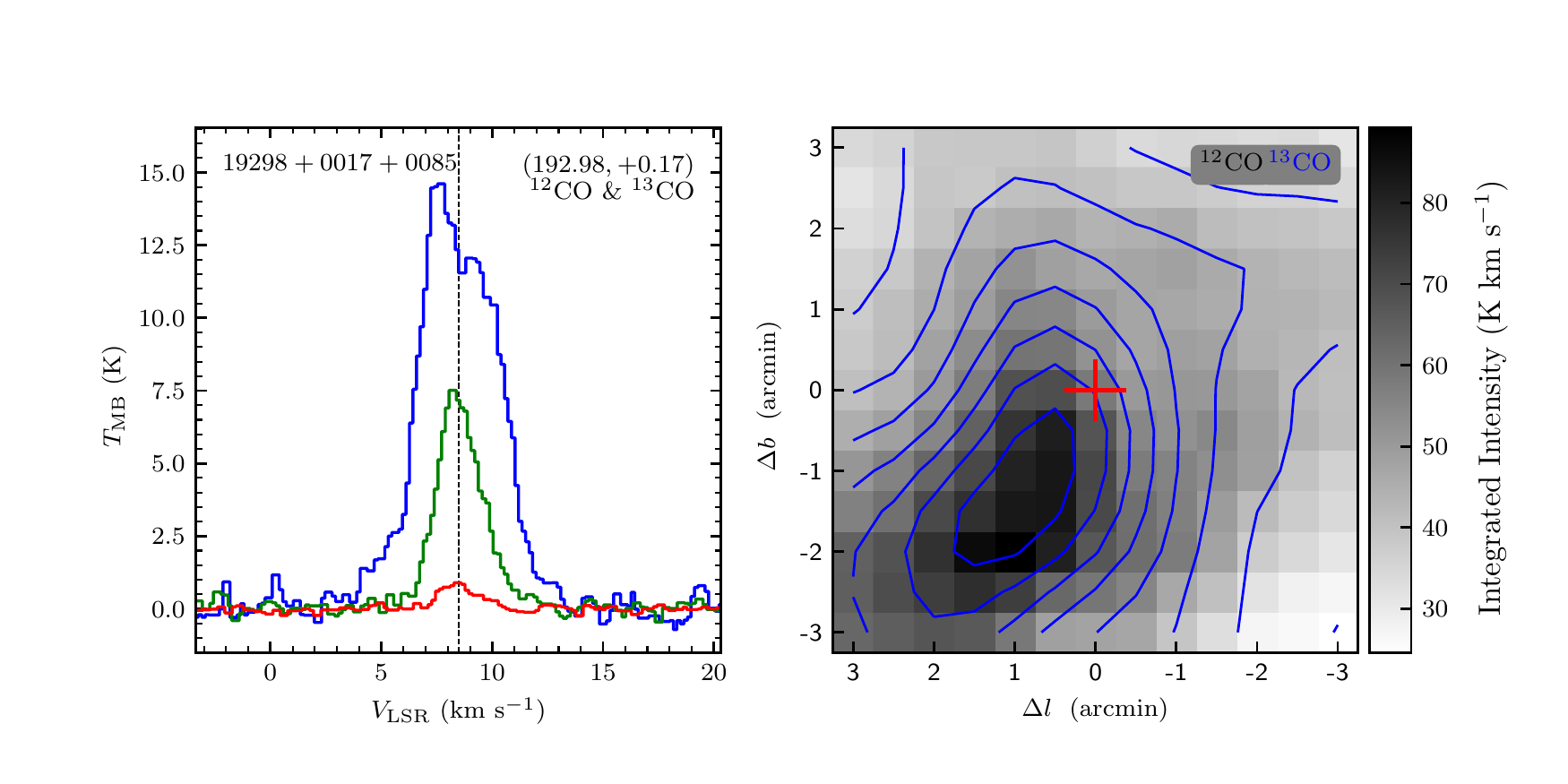}
\vspace{-0.5cm}

\includegraphics[width=9.0cm,angle=0]{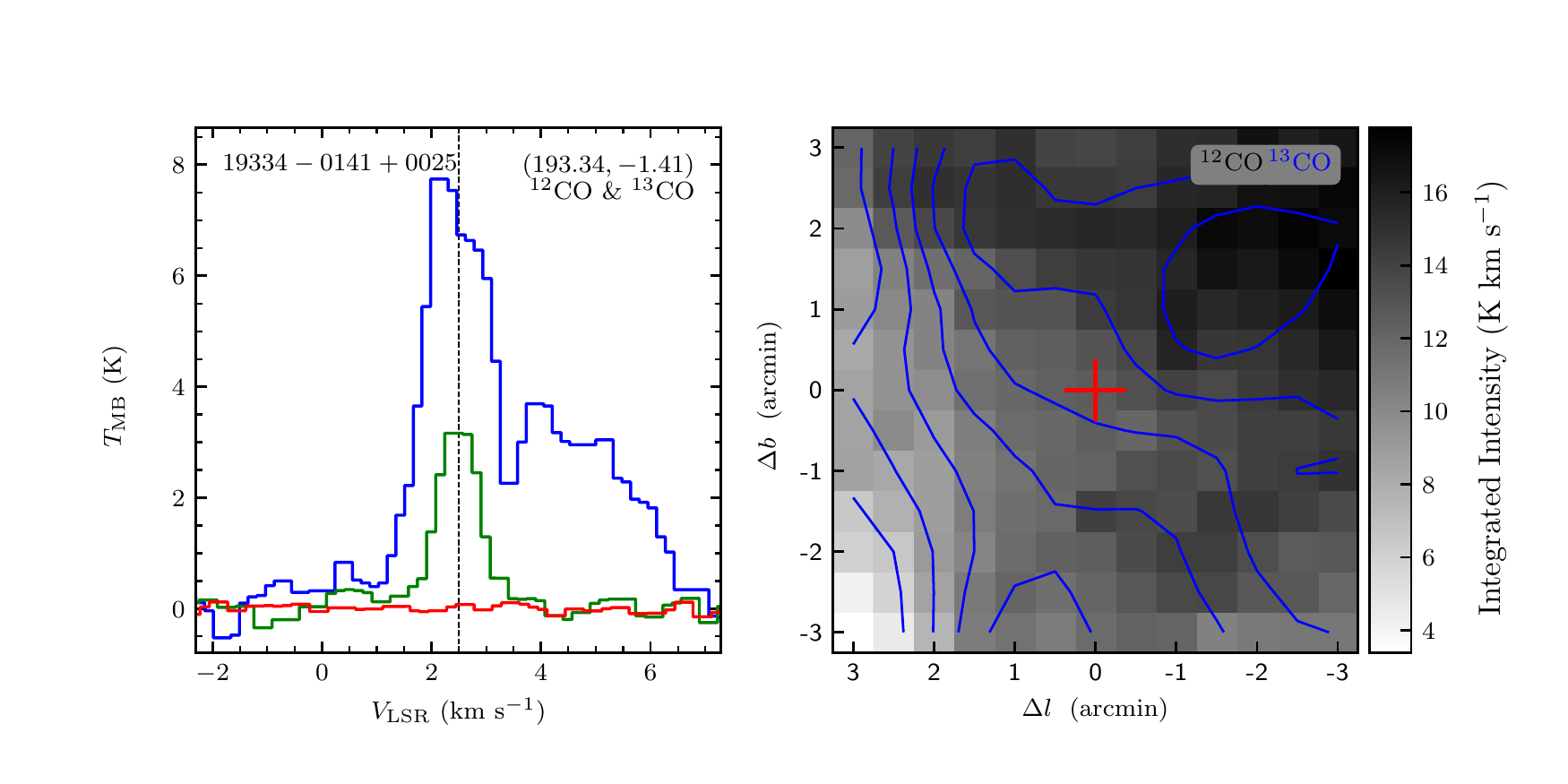}
\includegraphics[width=9.0cm,angle=0]{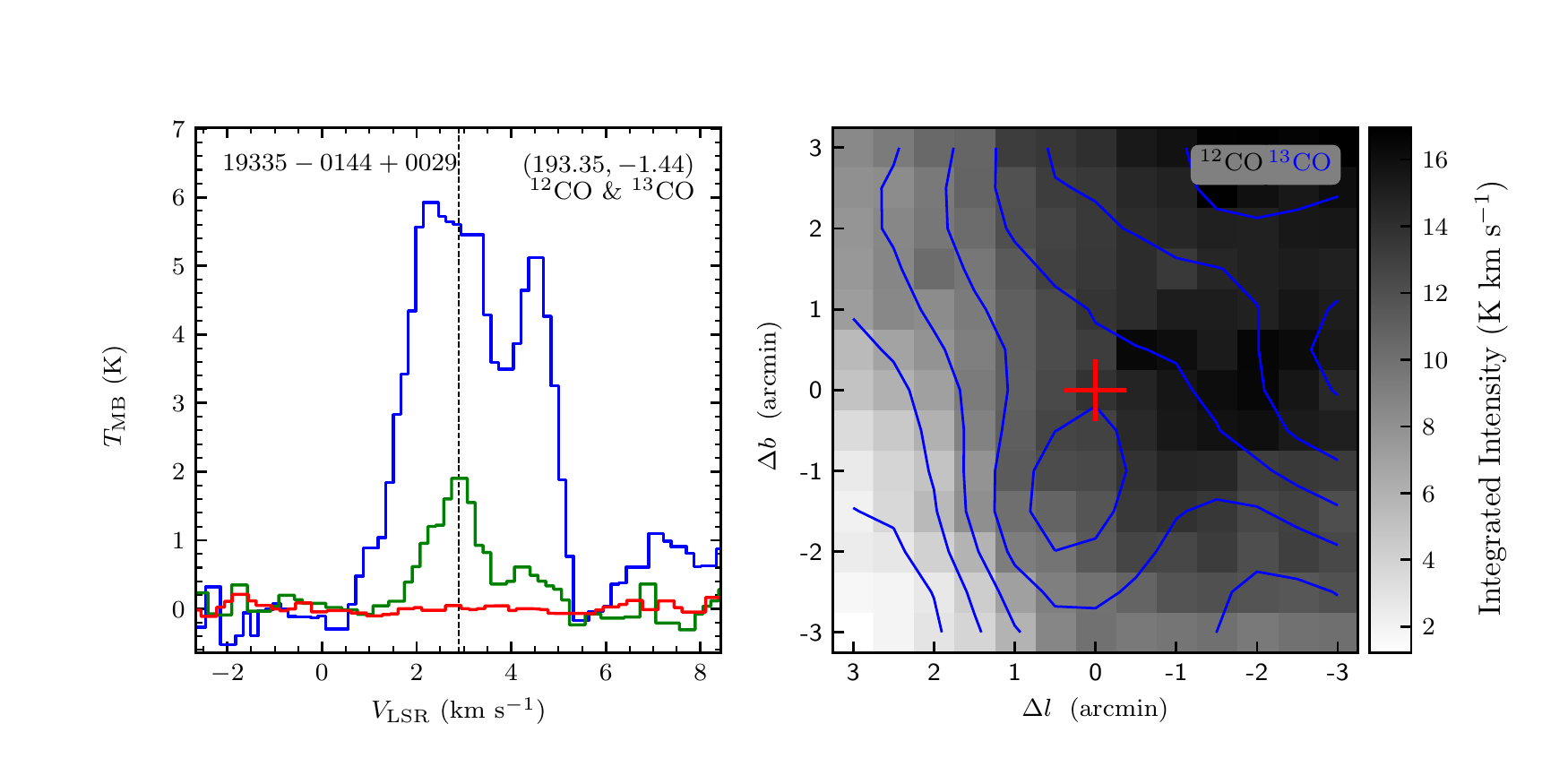}
\vspace{-0.5cm}

\includegraphics[width=9.0cm,angle=0]{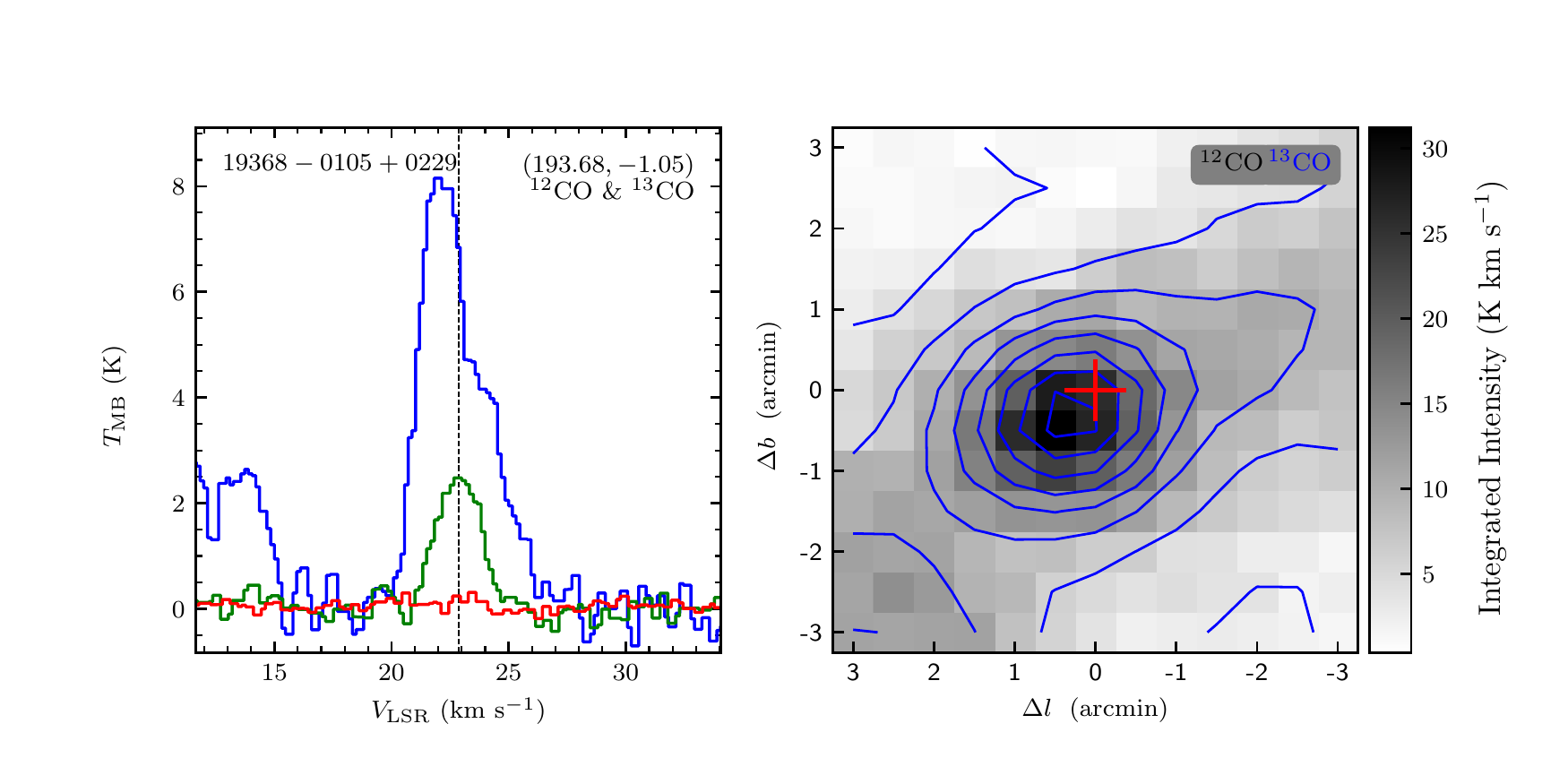}
\includegraphics[width=9.0cm,angle=0]{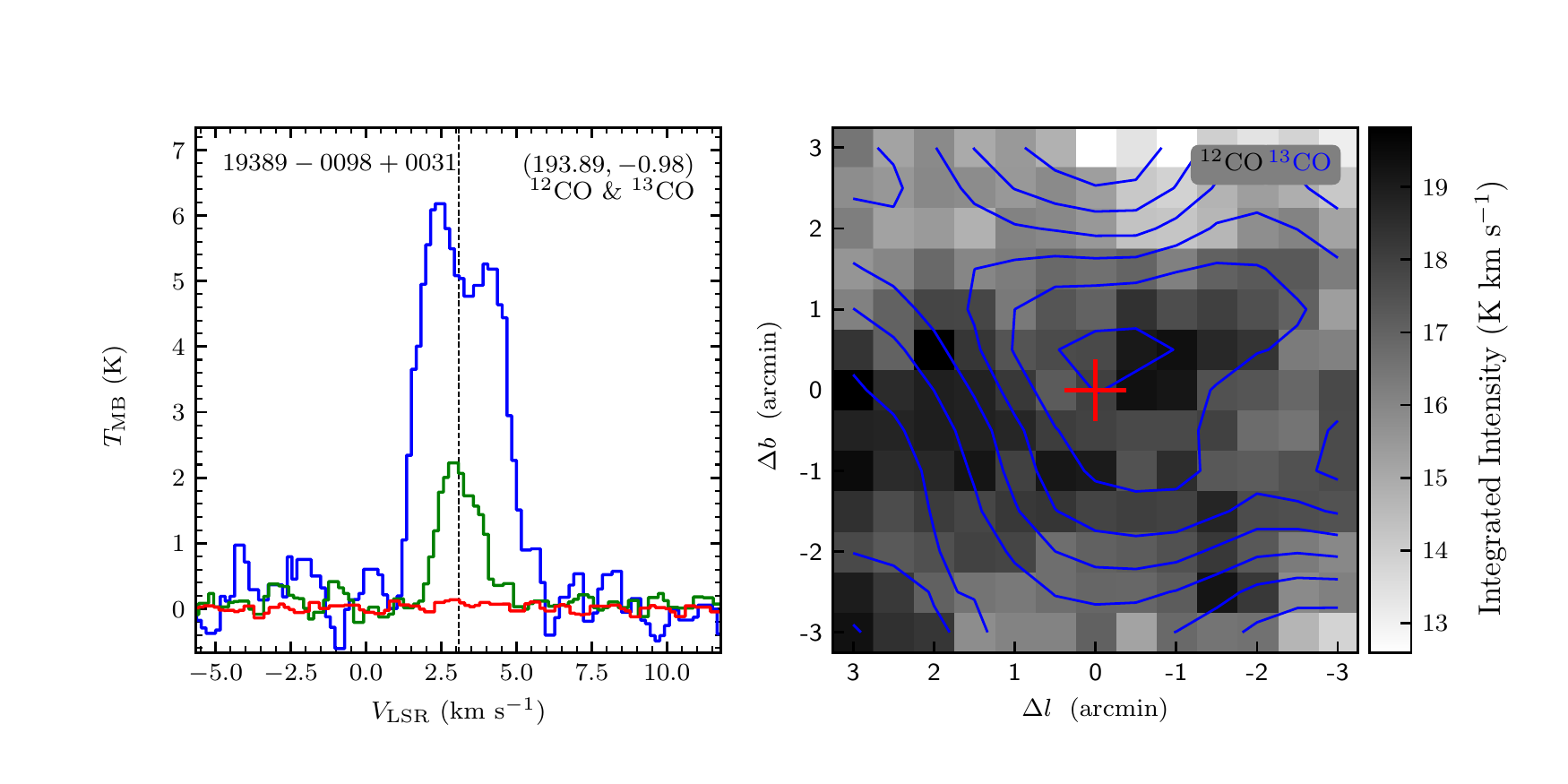}
\vspace{-0.5cm}

\includegraphics[width=9.0cm,angle=0]{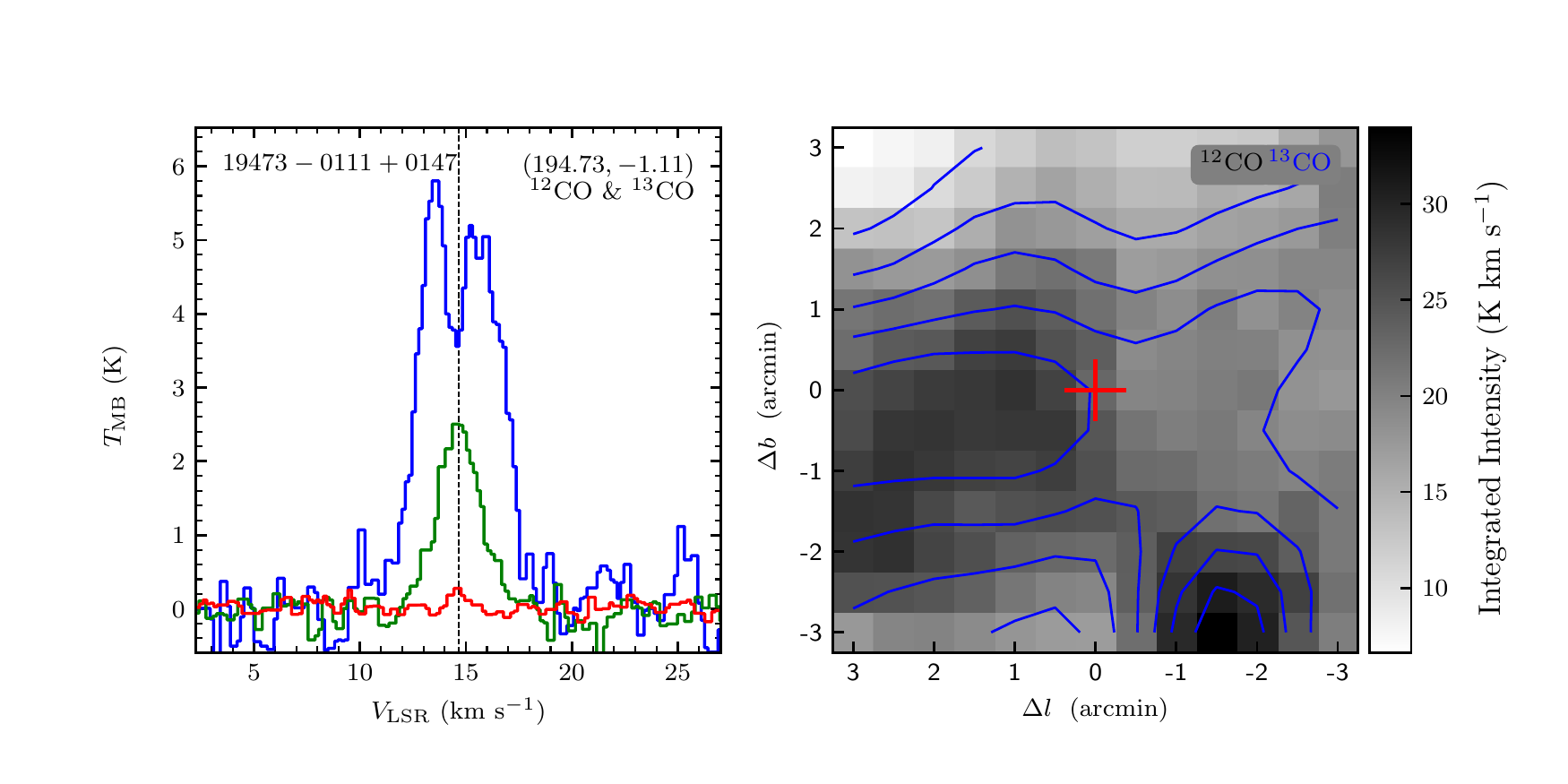}
\includegraphics[width=9.0cm,angle=0]{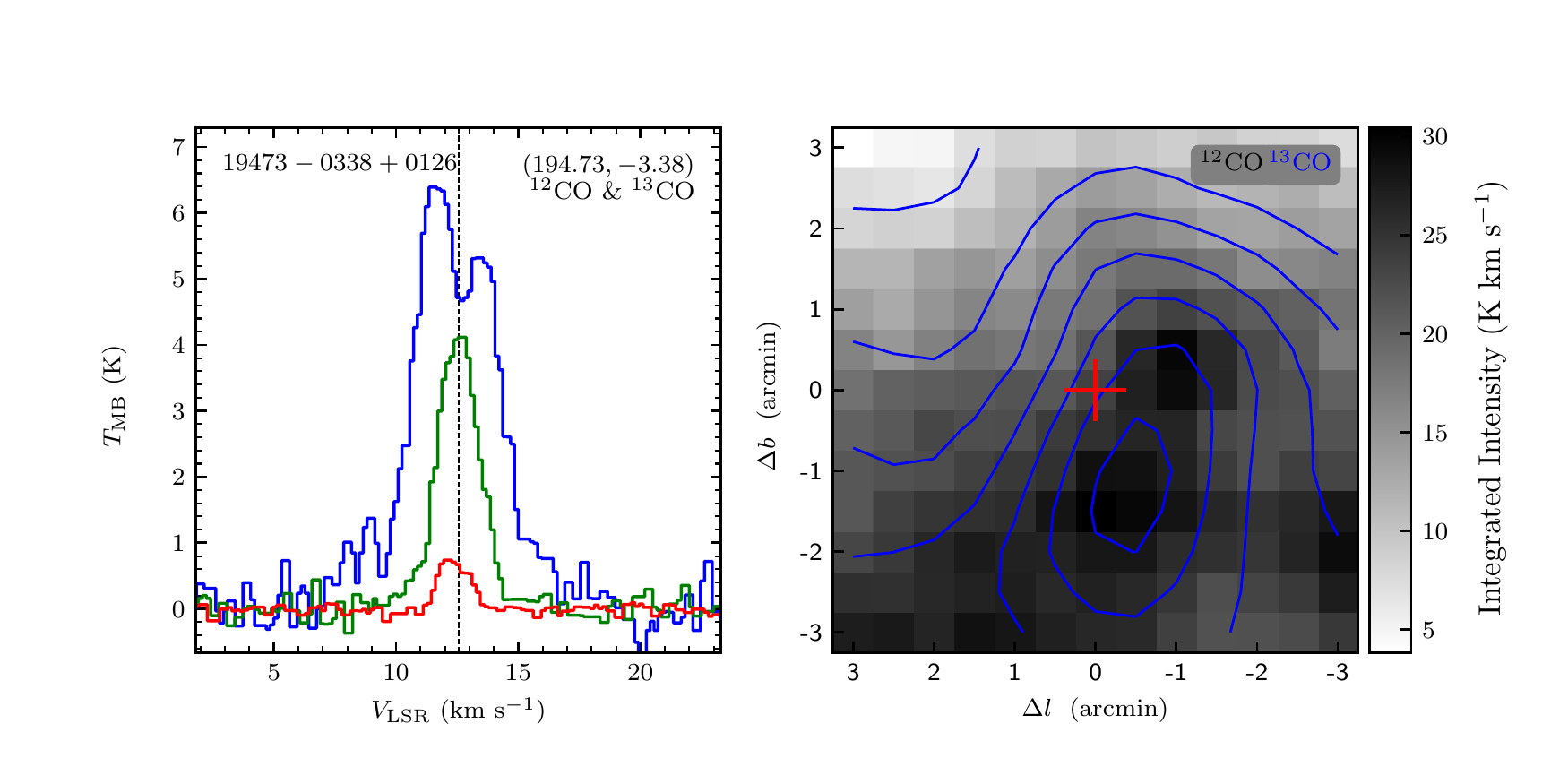}
\end{figure}
\clearpage

\begin{figure}
\includegraphics[width=9.0cm,angle=0]{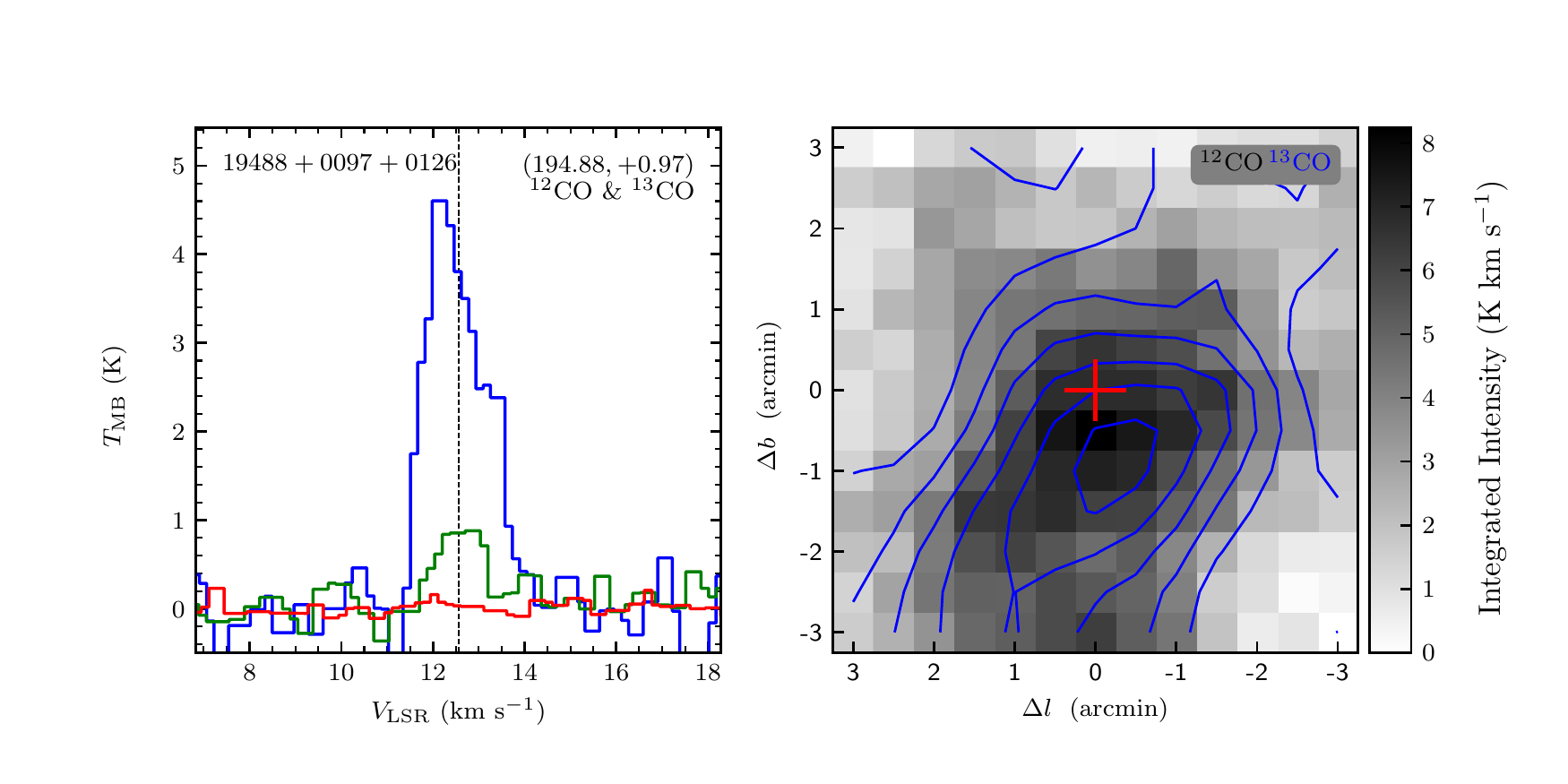}
\includegraphics[width=9.0cm,angle=0]{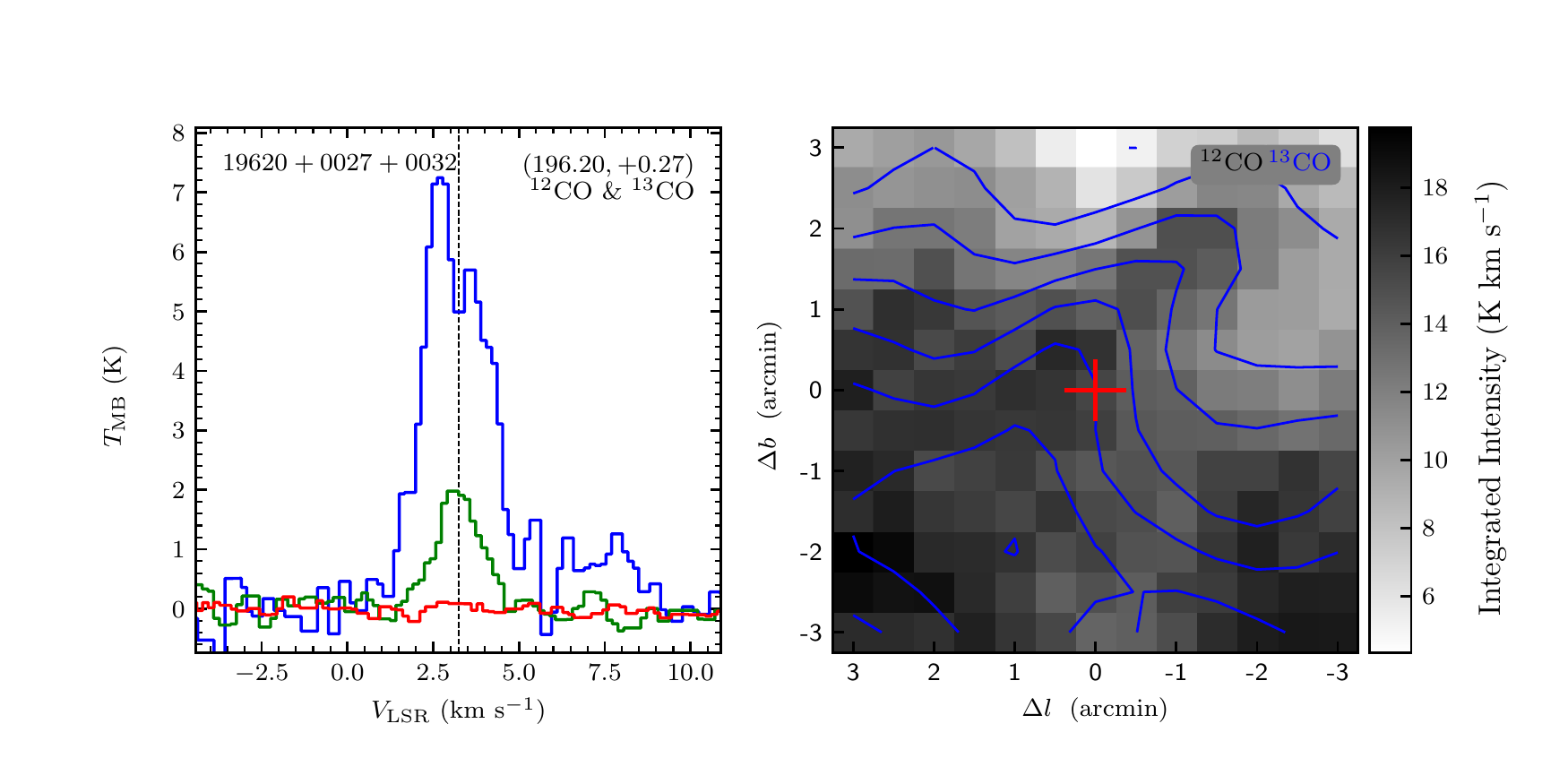}
\vspace{-0.5cm}

\includegraphics[width=9.0cm,angle=0]{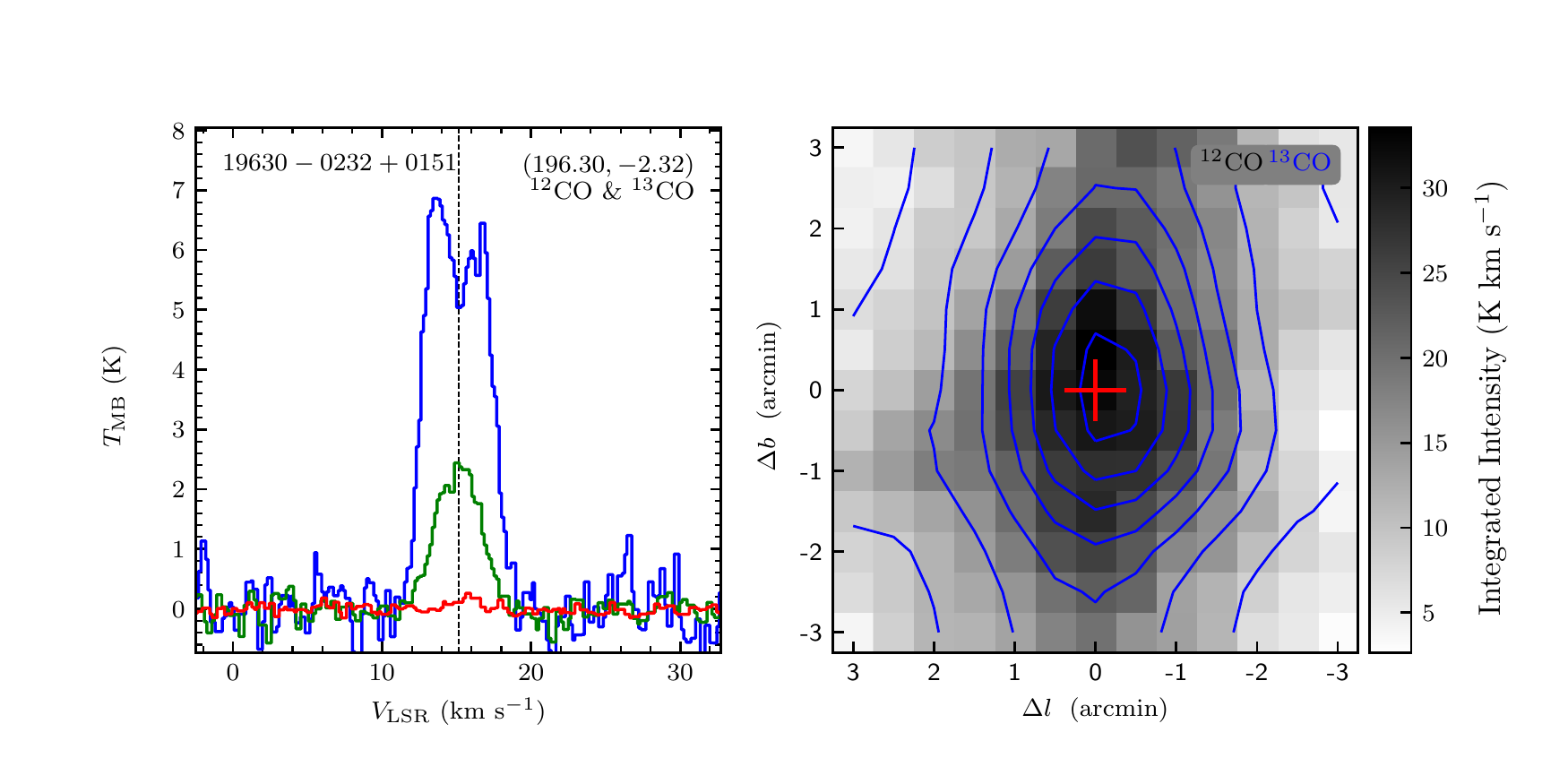}
\includegraphics[width=9.0cm,angle=0]{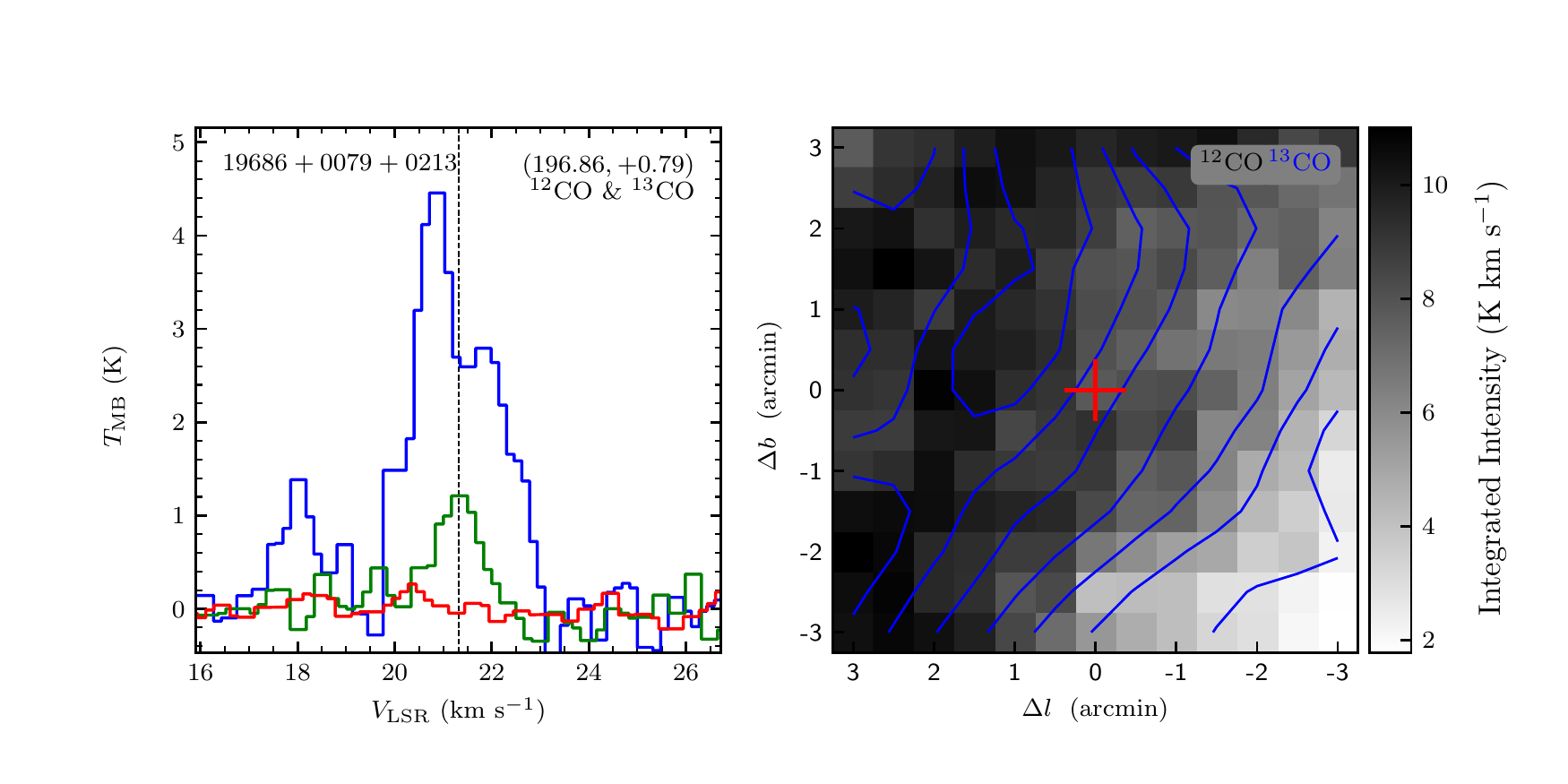}
\vspace{-0.5cm}

\includegraphics[width=9.0cm,angle=0]{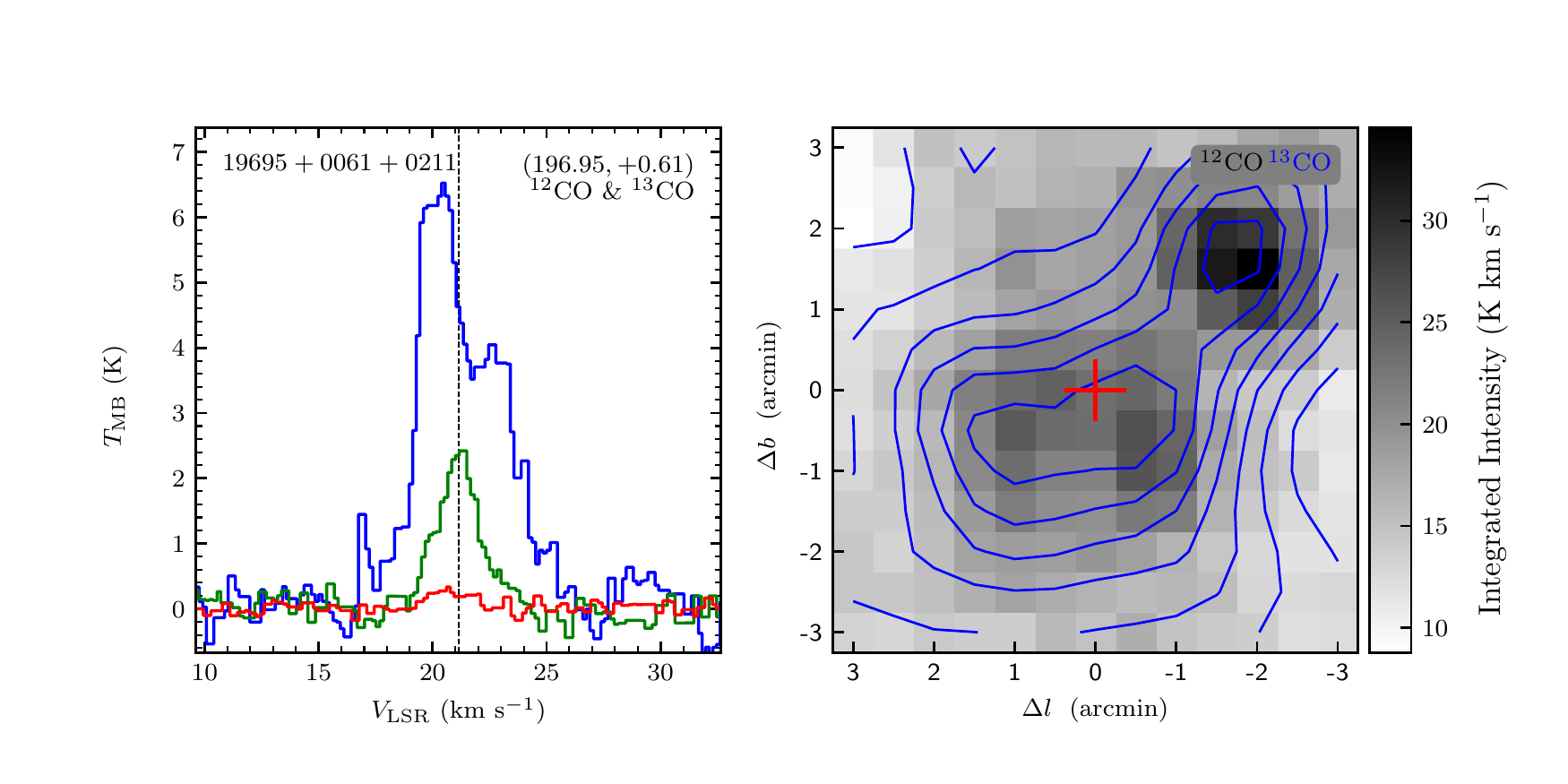}
\includegraphics[width=9.0cm,angle=0]{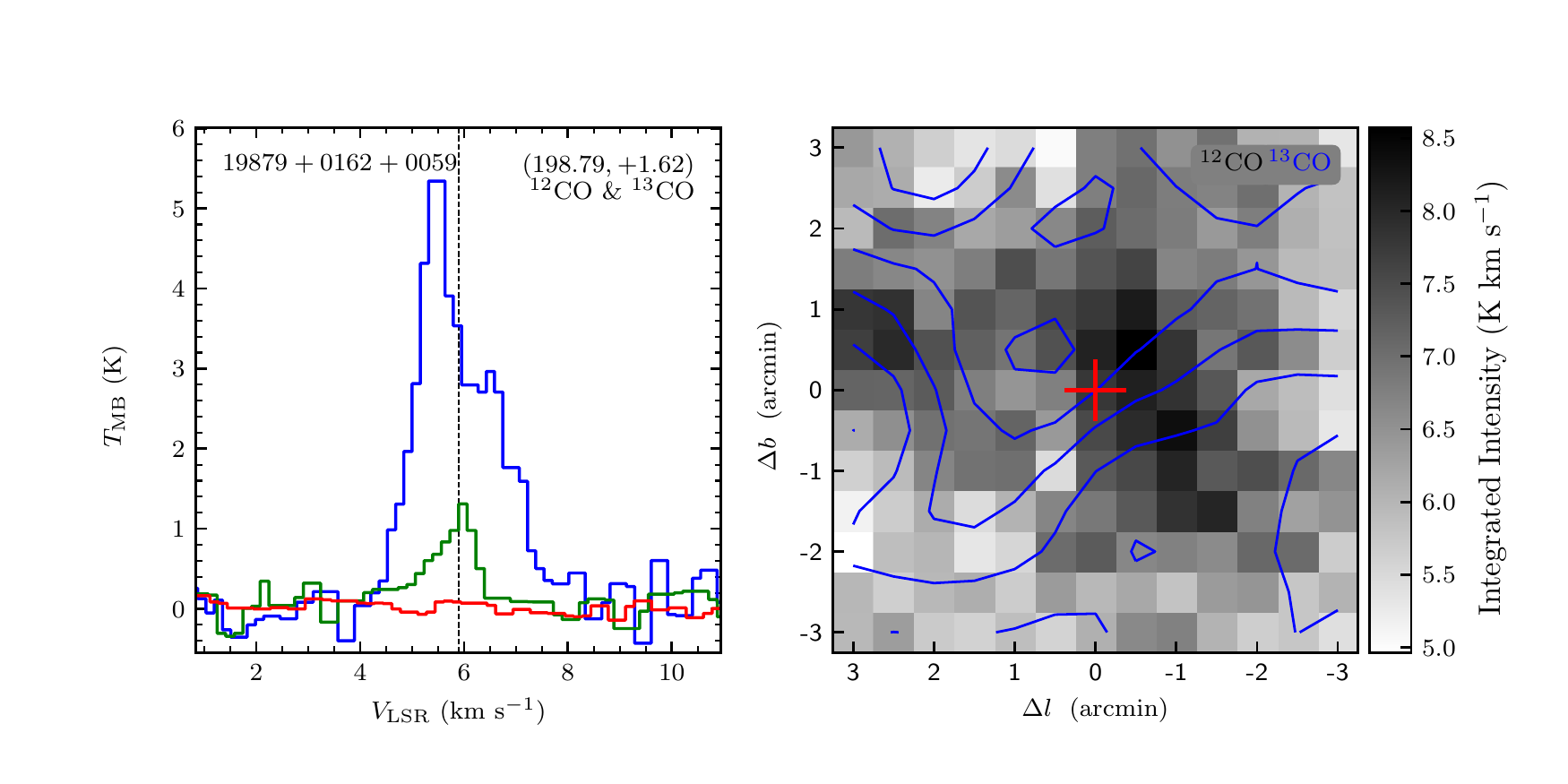}
\vspace{-0.5cm}

\includegraphics[width=9.0cm,angle=0]{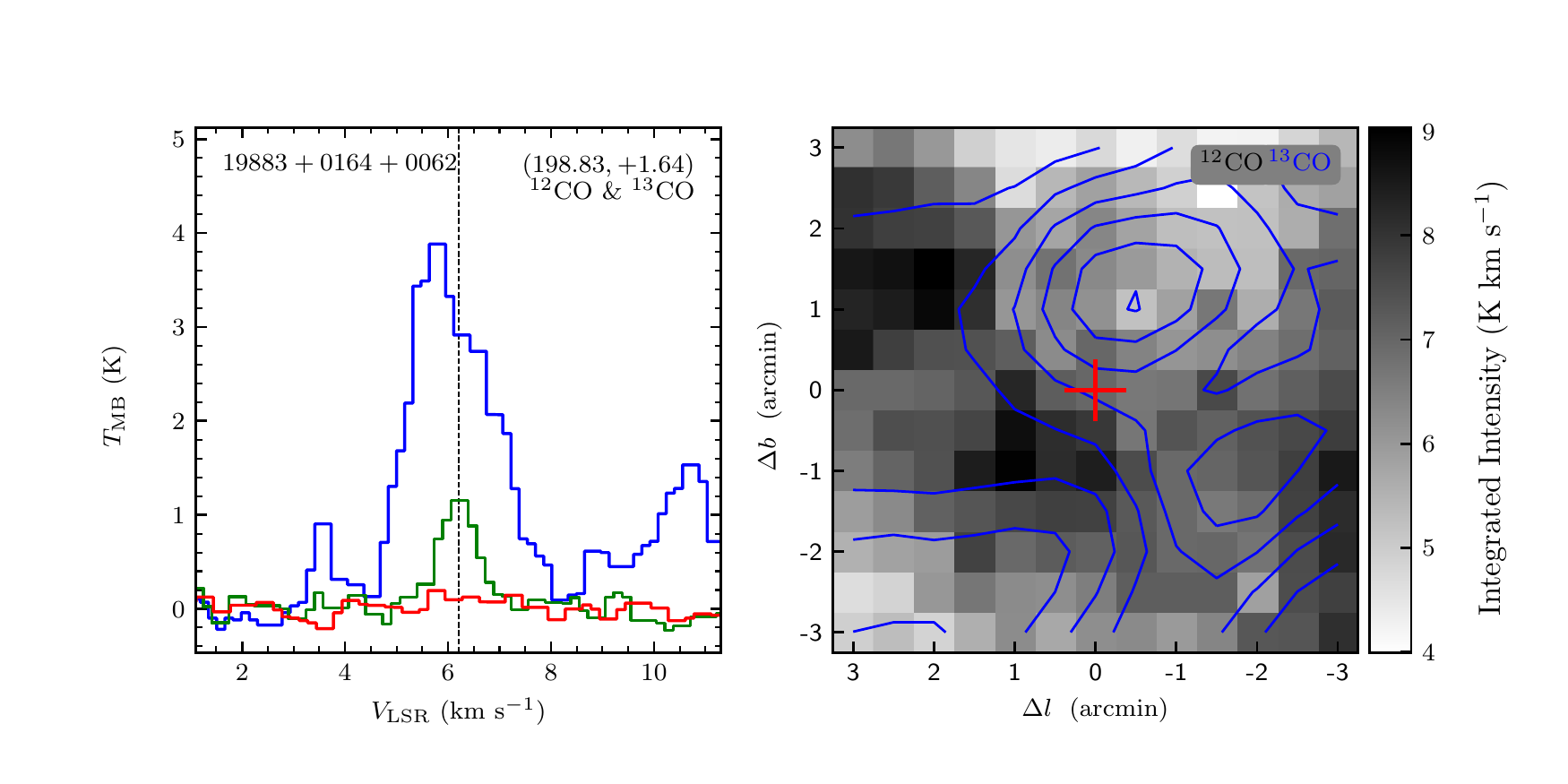}
\includegraphics[width=9.0cm,angle=0]{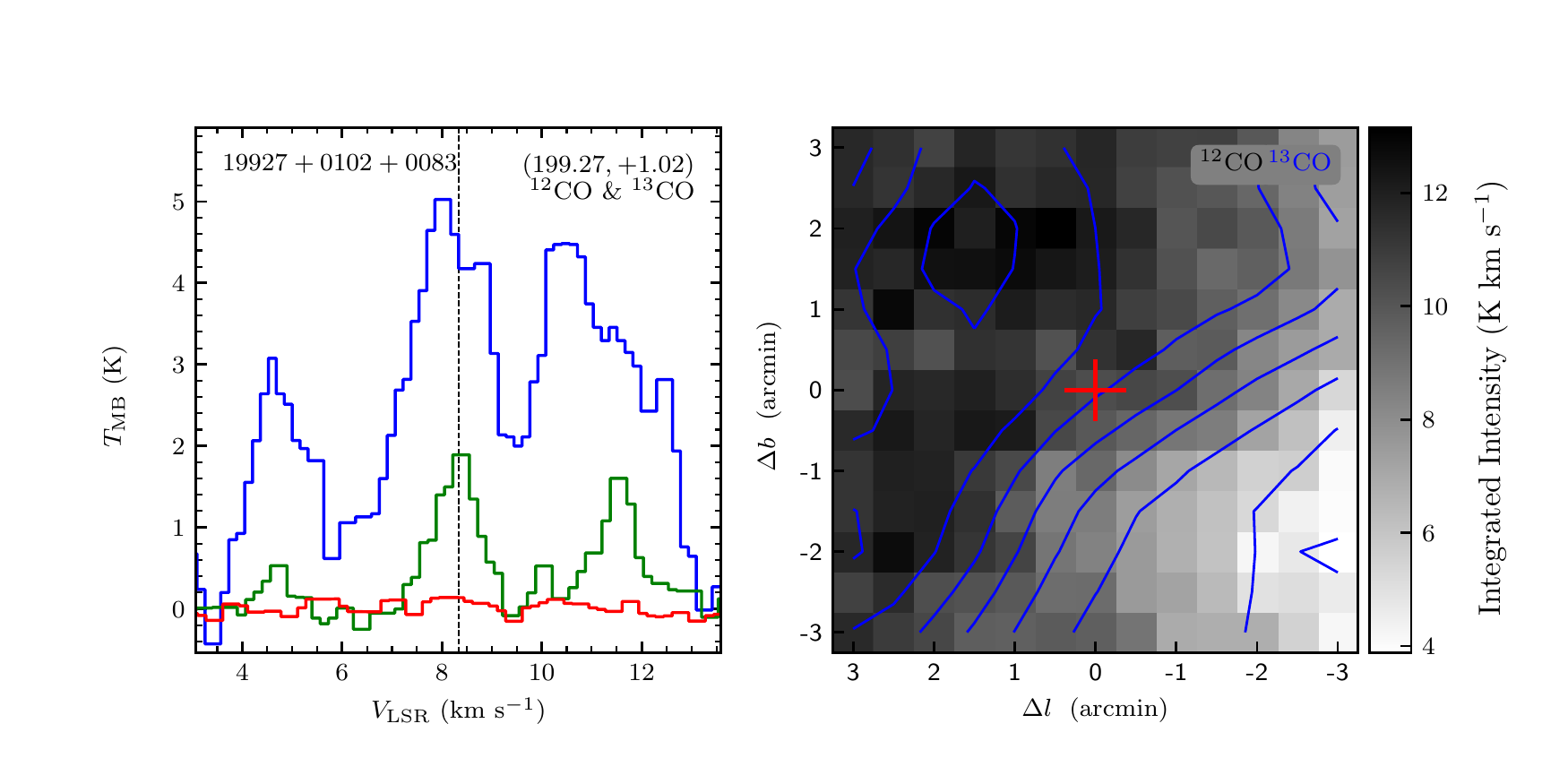}
\vspace{-0.5cm}

\includegraphics[width=9.0cm,angle=0]{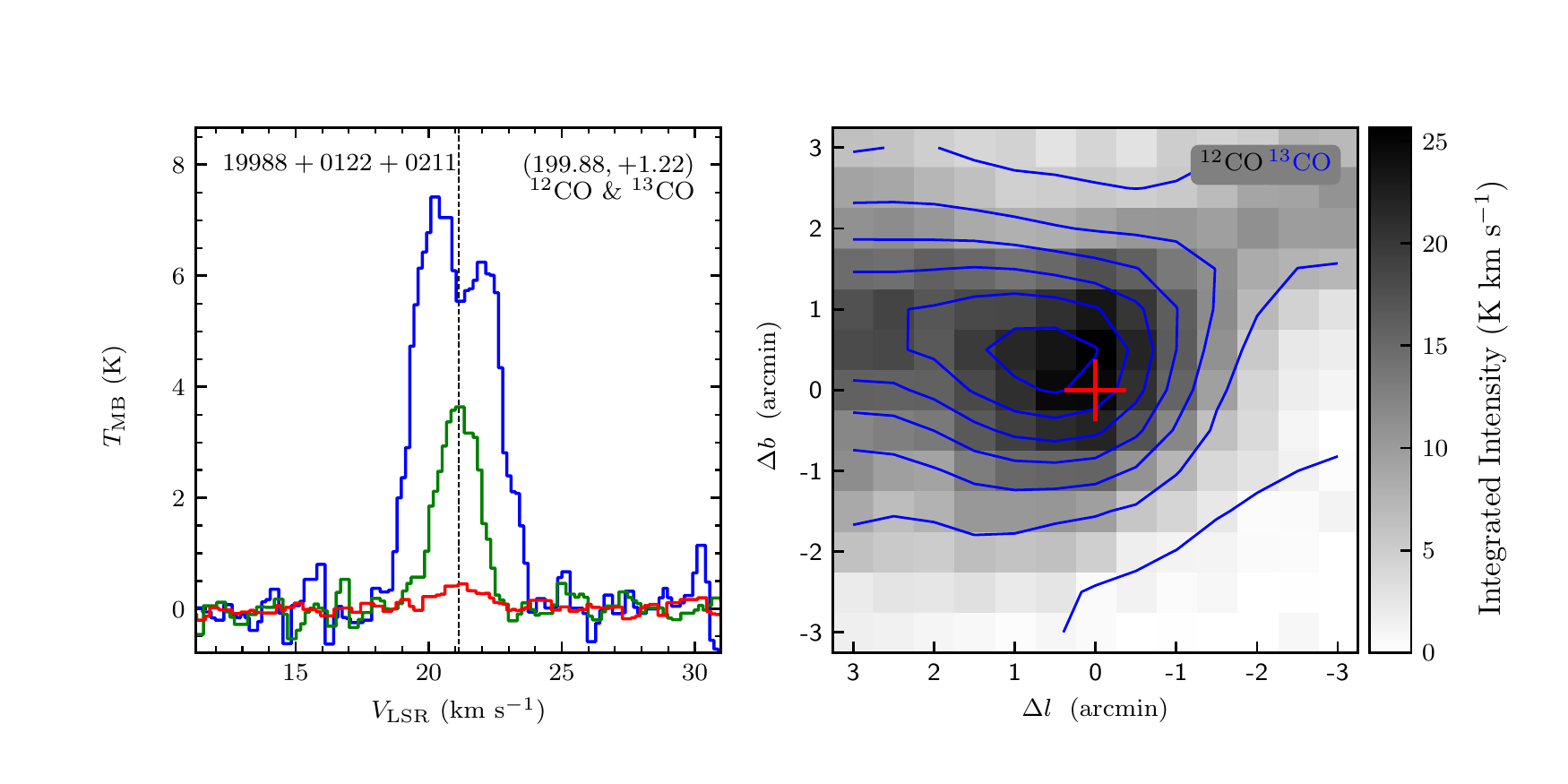}
\includegraphics[width=9.0cm,angle=0]{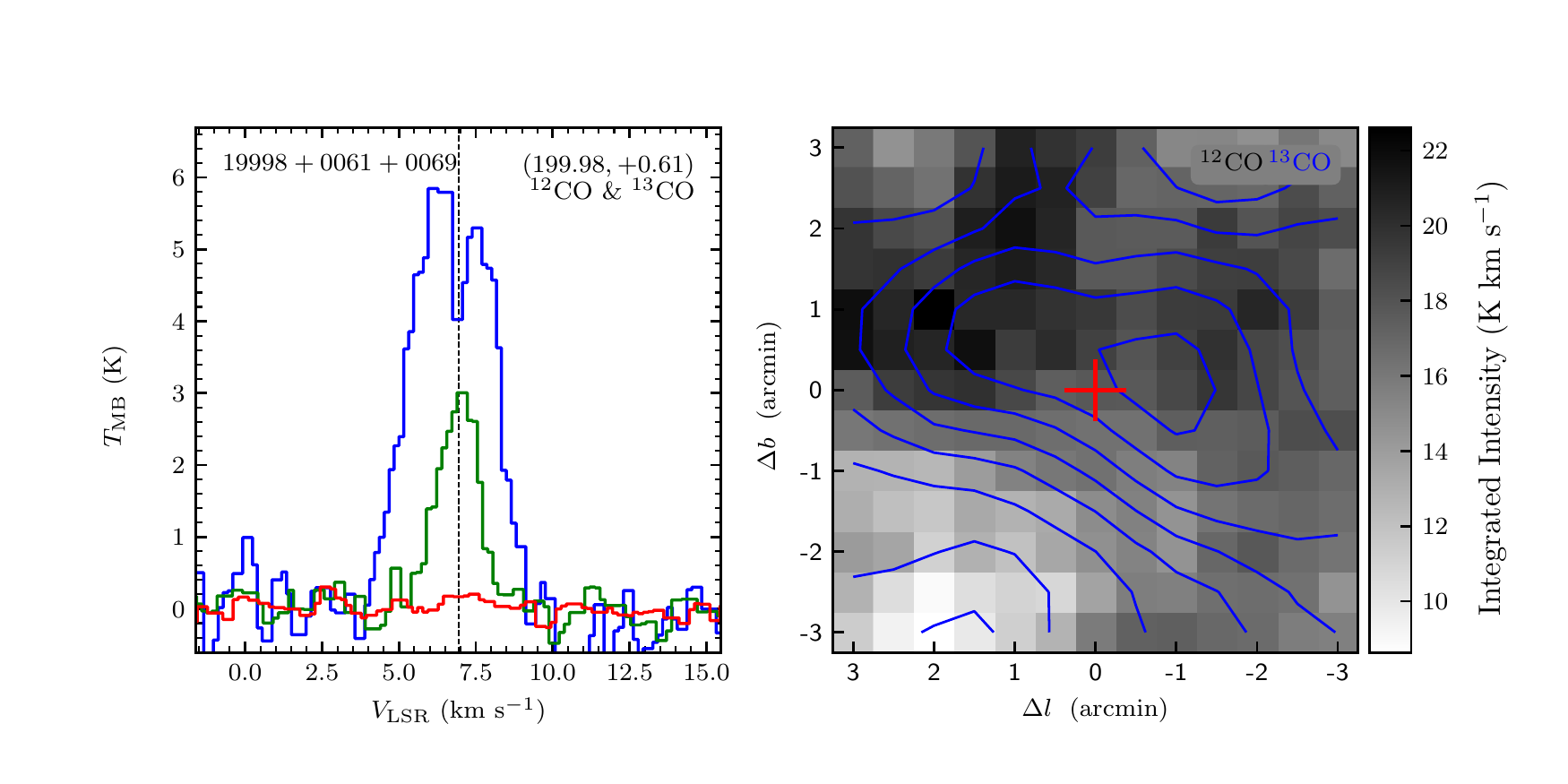}
\end{figure}
\clearpage

\begin{figure}
\includegraphics[width=9.0cm,angle=0]{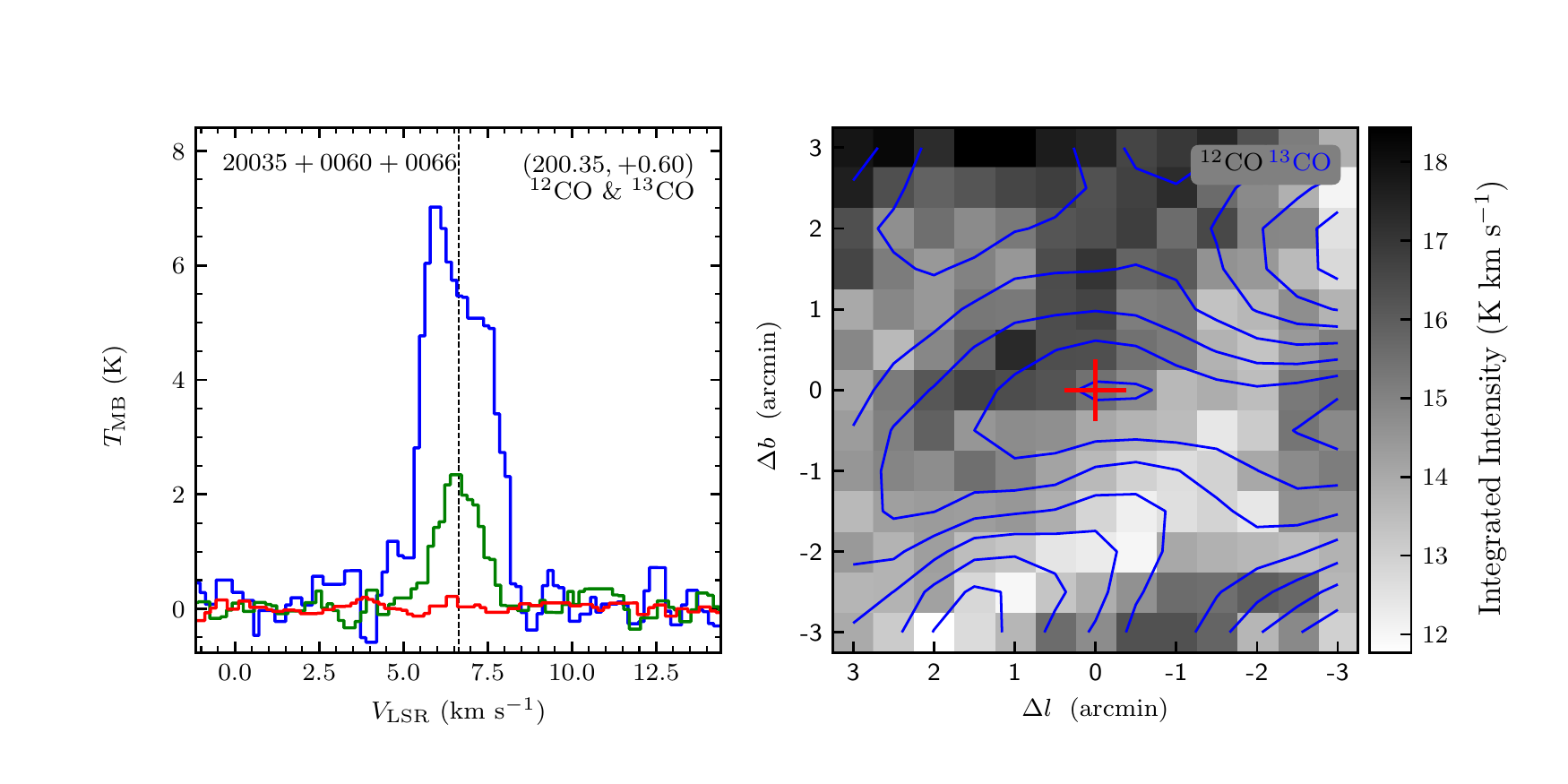}
\includegraphics[width=9.0cm,angle=0]{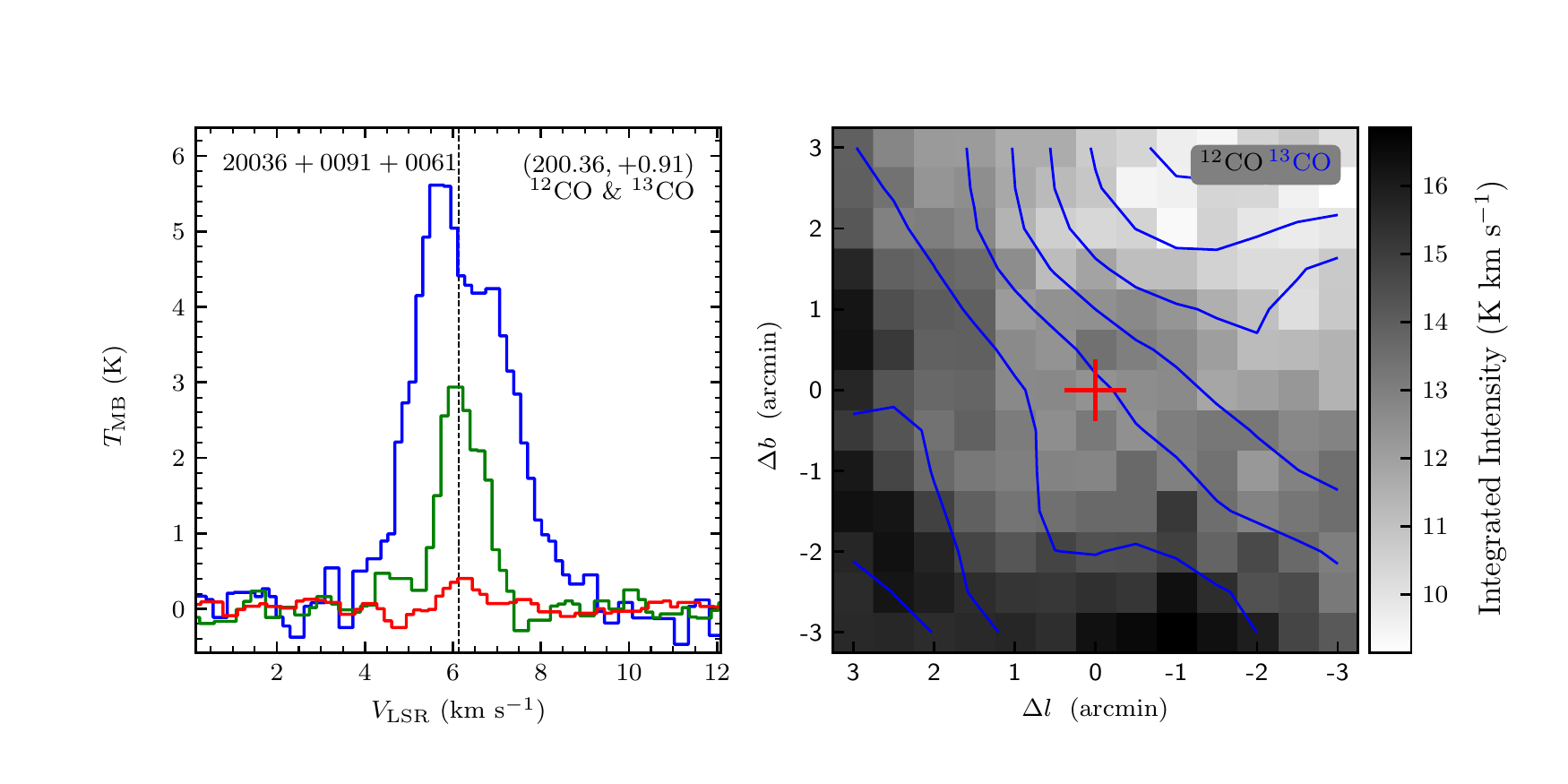}
\vspace{-0.5cm}

\includegraphics[width=9.0cm,angle=0]{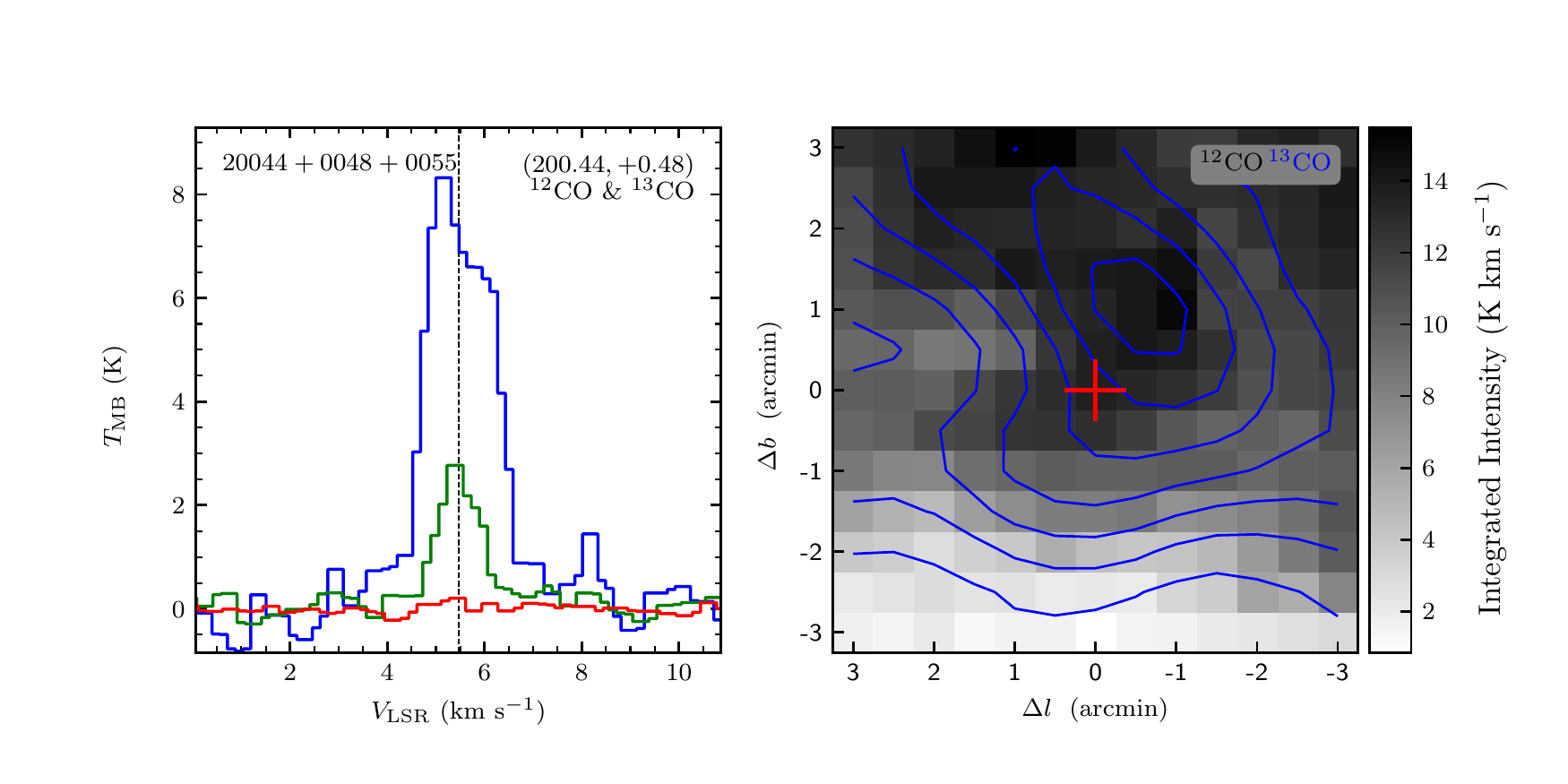}
\includegraphics[width=9.0cm,angle=0]{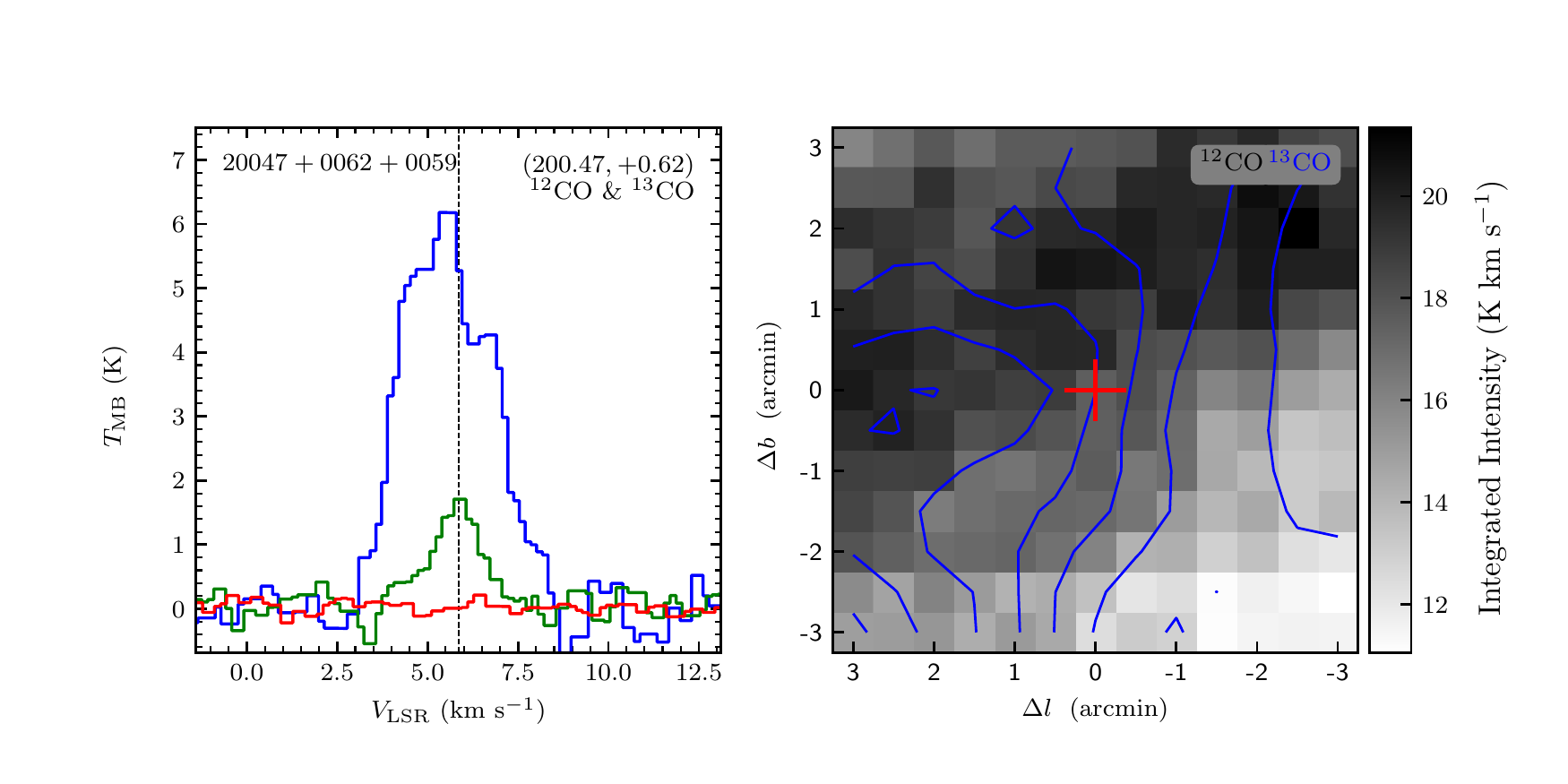}
\vspace{-0.5cm}

\includegraphics[width=9.0cm,angle=0]{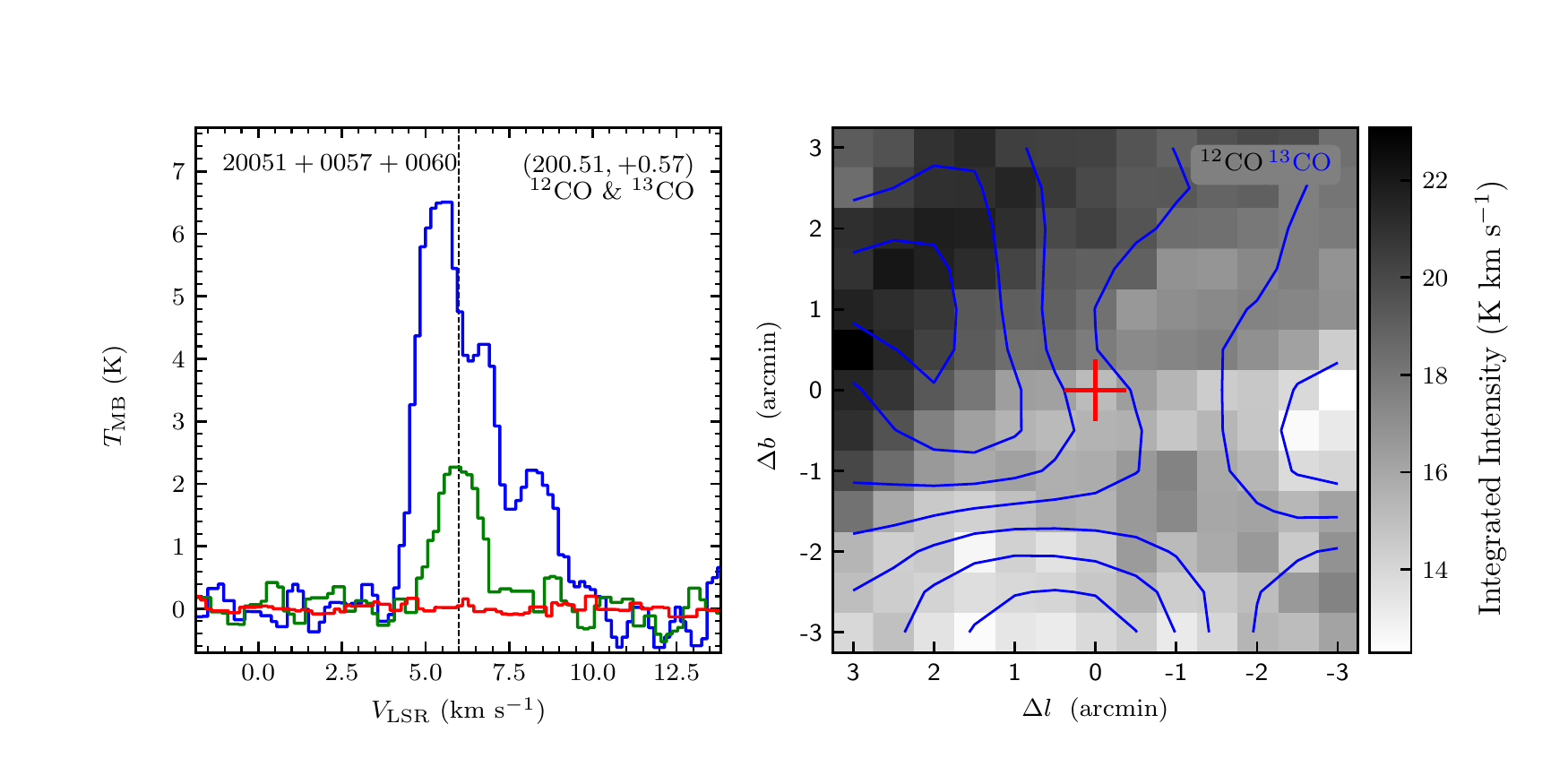}
\includegraphics[width=9.0cm,angle=0]{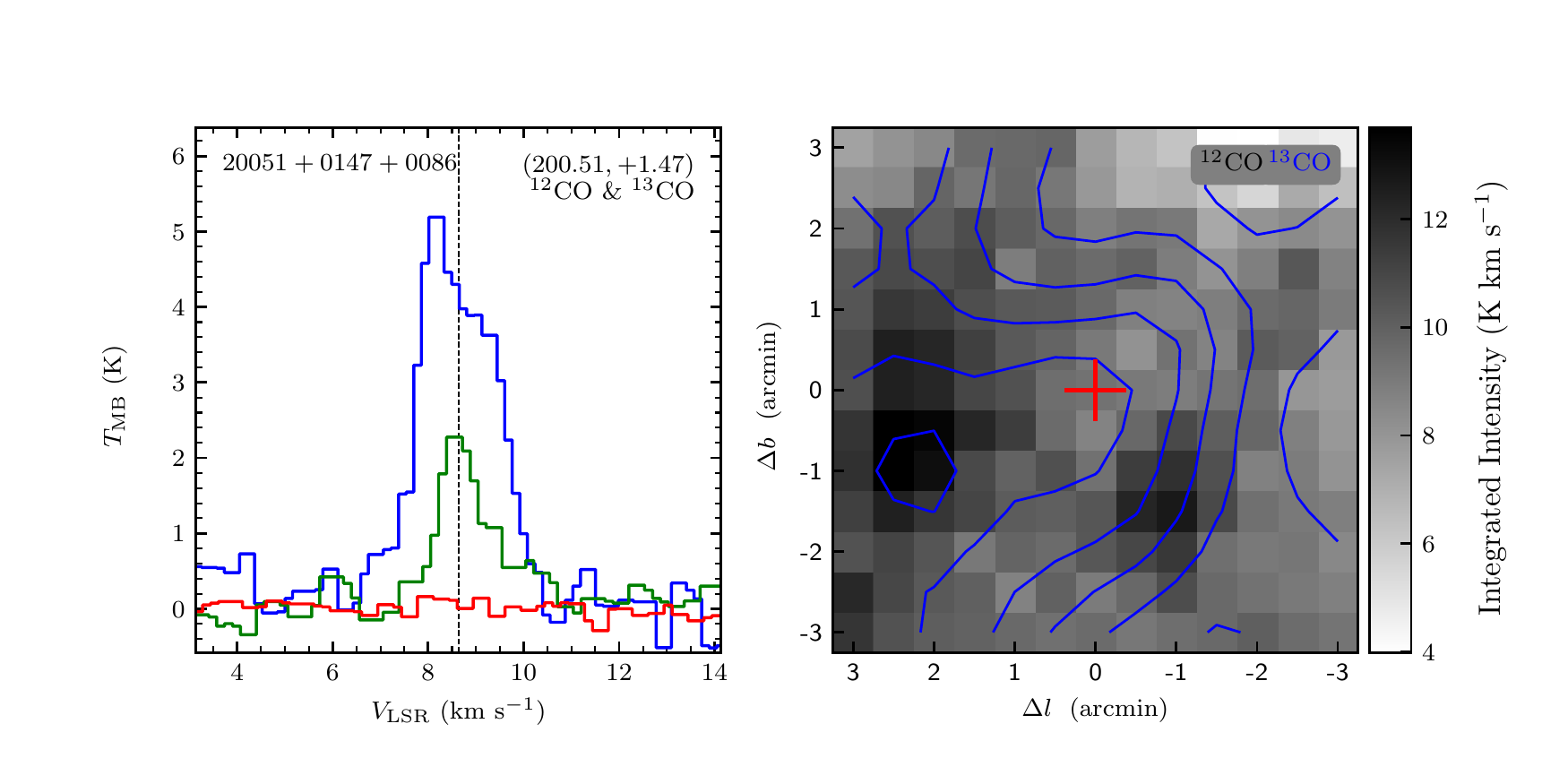}
\vspace{-0.5cm}

\includegraphics[width=9.0cm,angle=0]{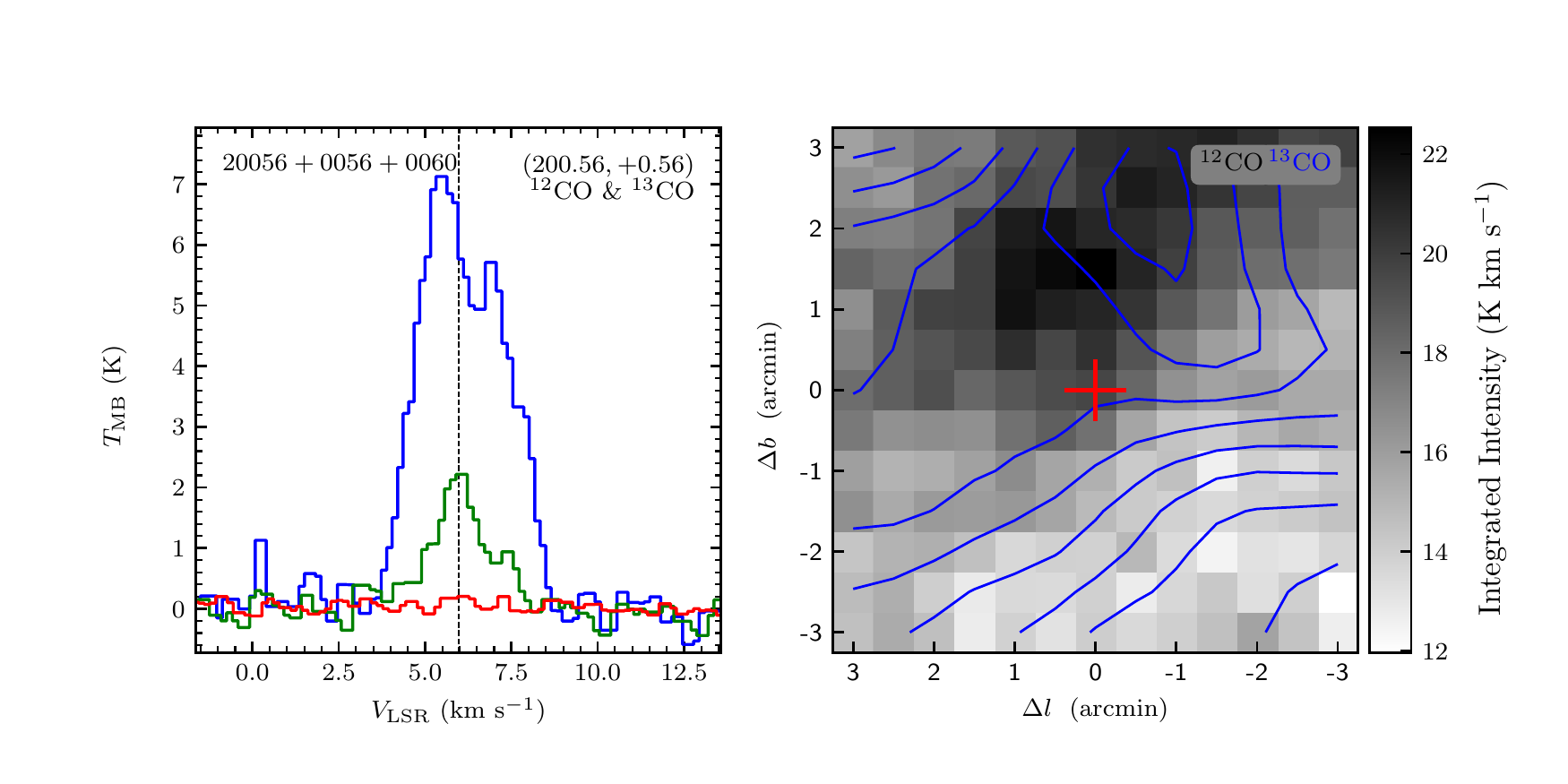}
\includegraphics[width=9.0cm,angle=0]{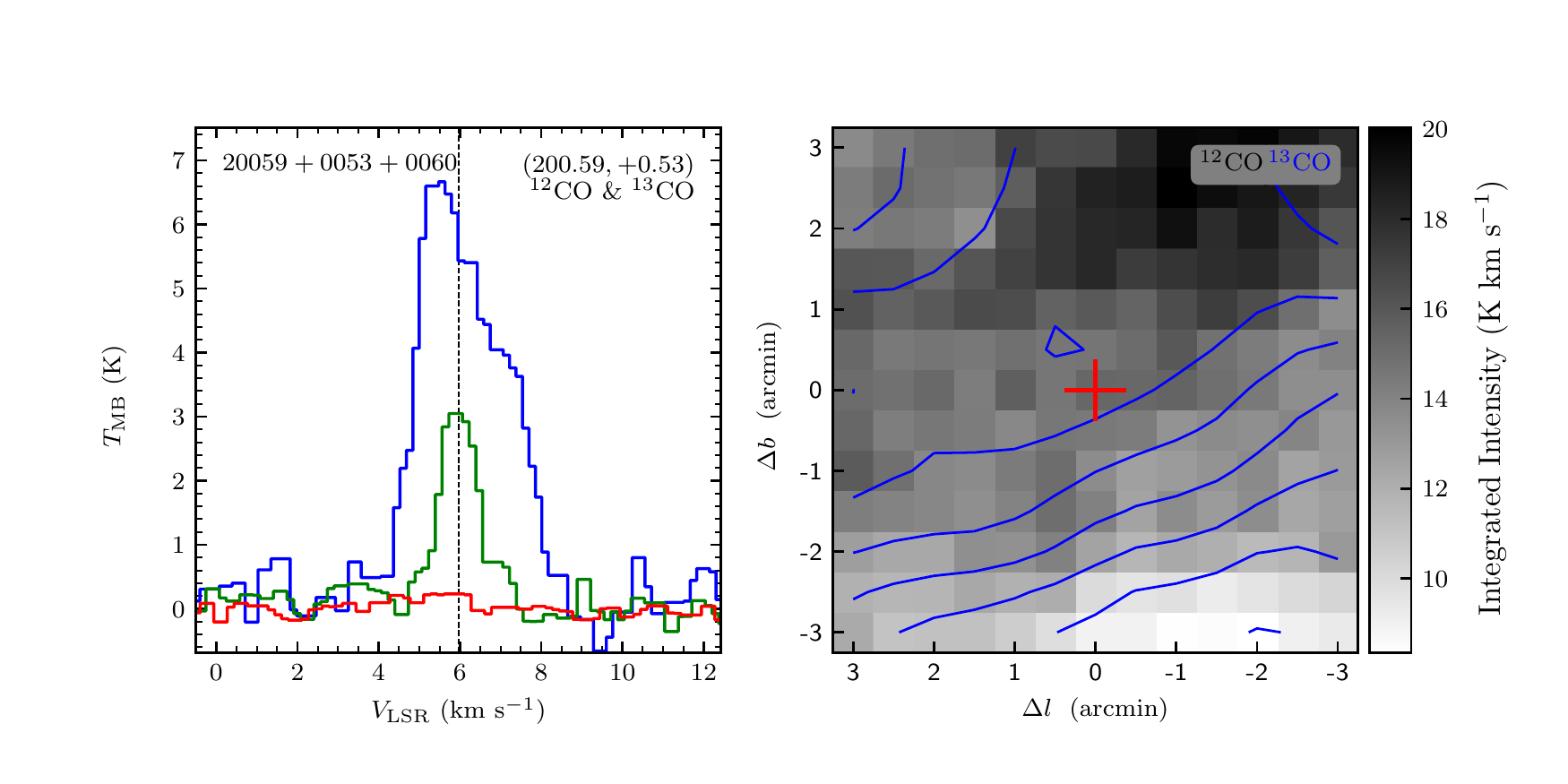}
\vspace{-0.5cm}

\includegraphics[width=9.0cm,angle=0]{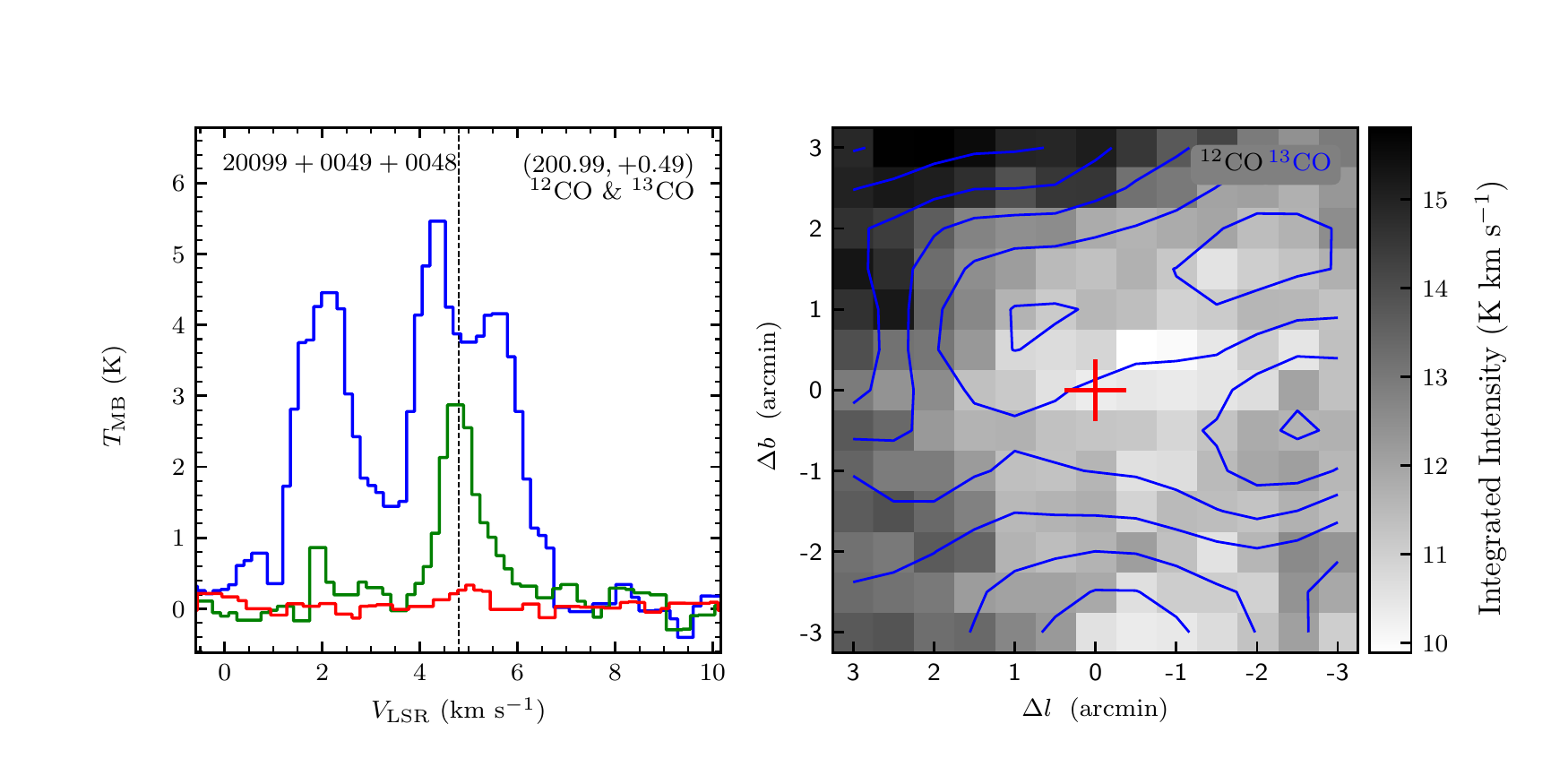}
\includegraphics[width=9.0cm,angle=0]{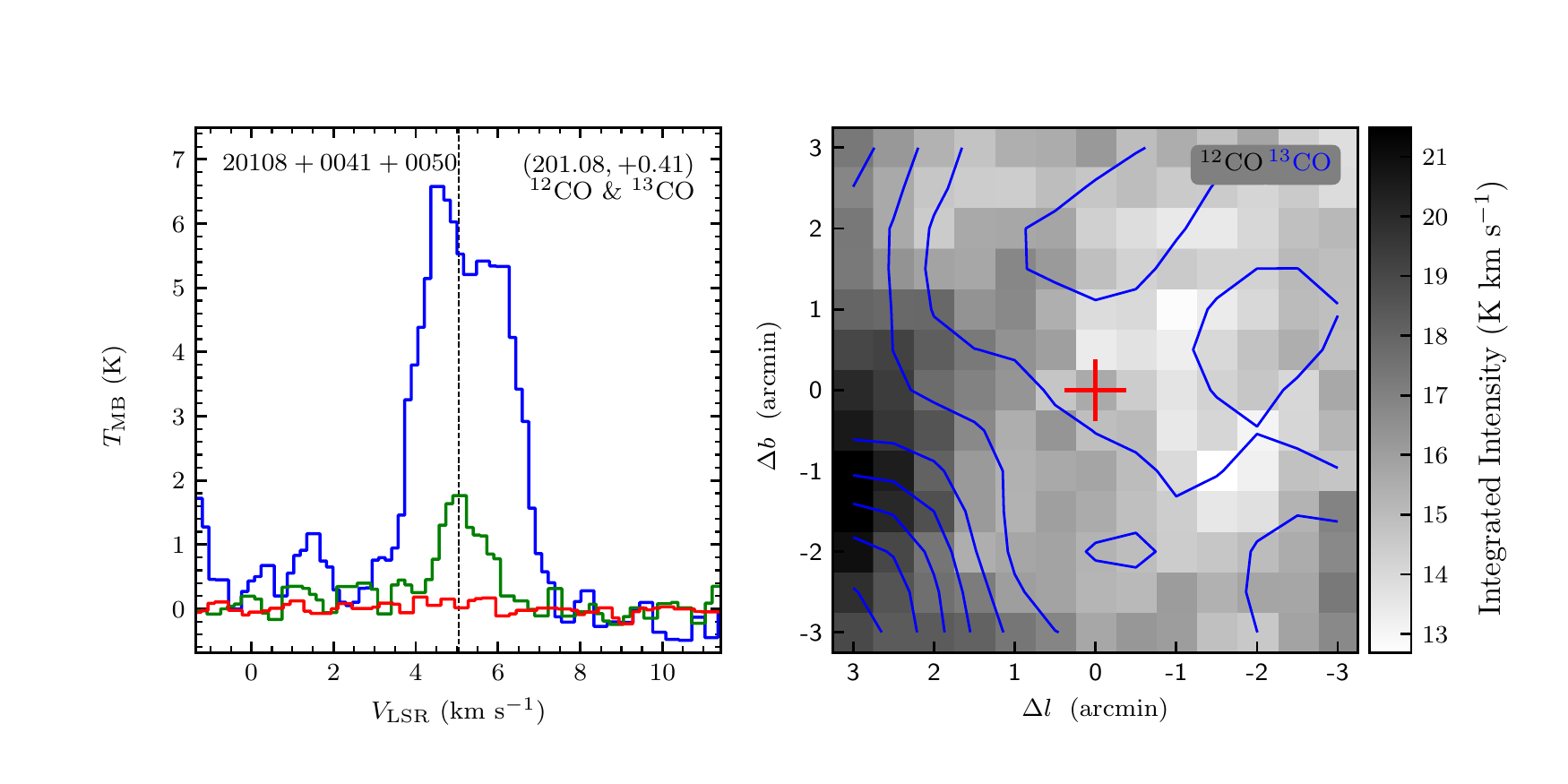}
\end{figure}
\clearpage

\begin{figure}
\includegraphics[width=9.0cm,angle=0]{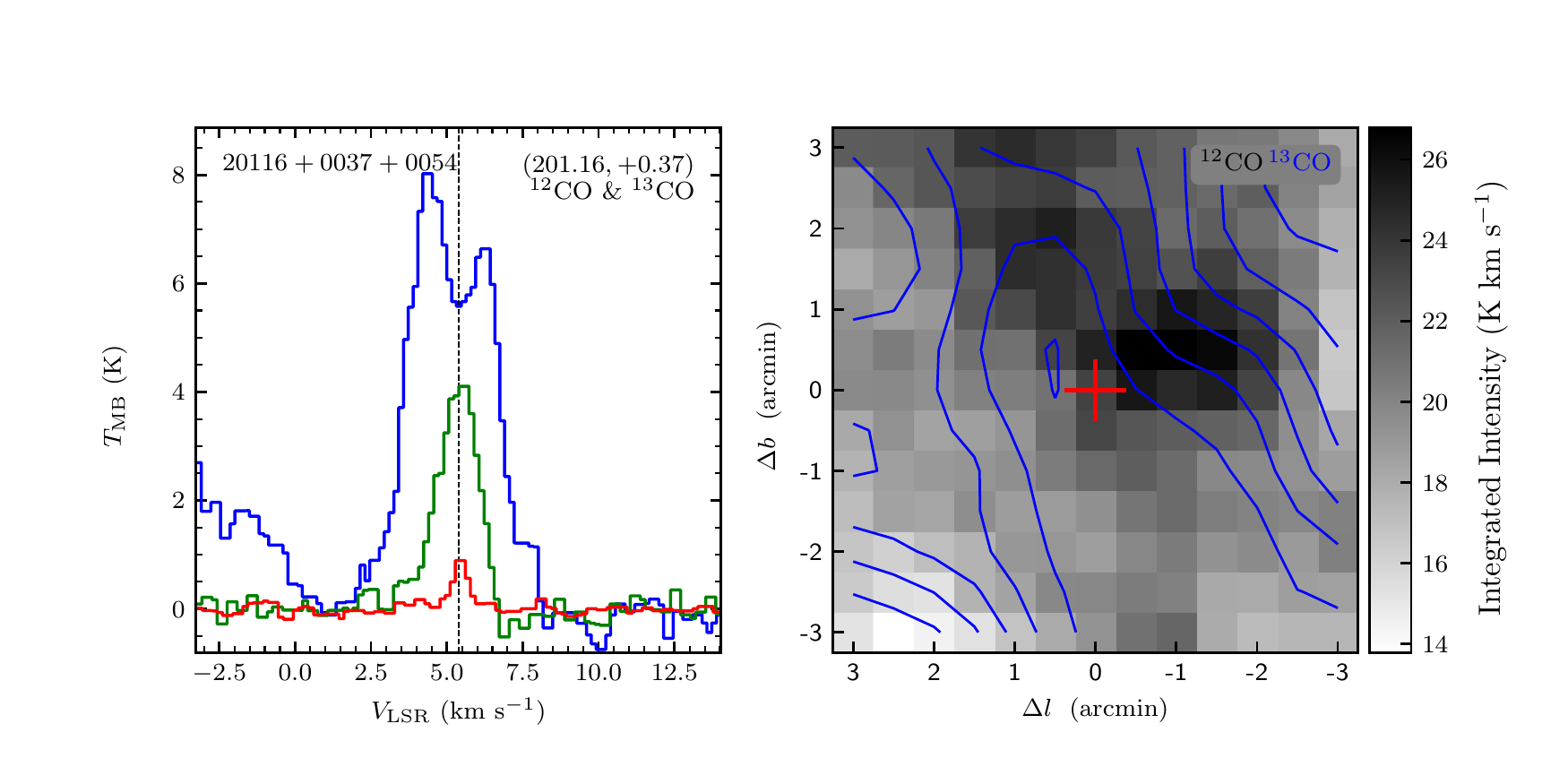}
\includegraphics[width=9.0cm,angle=0]{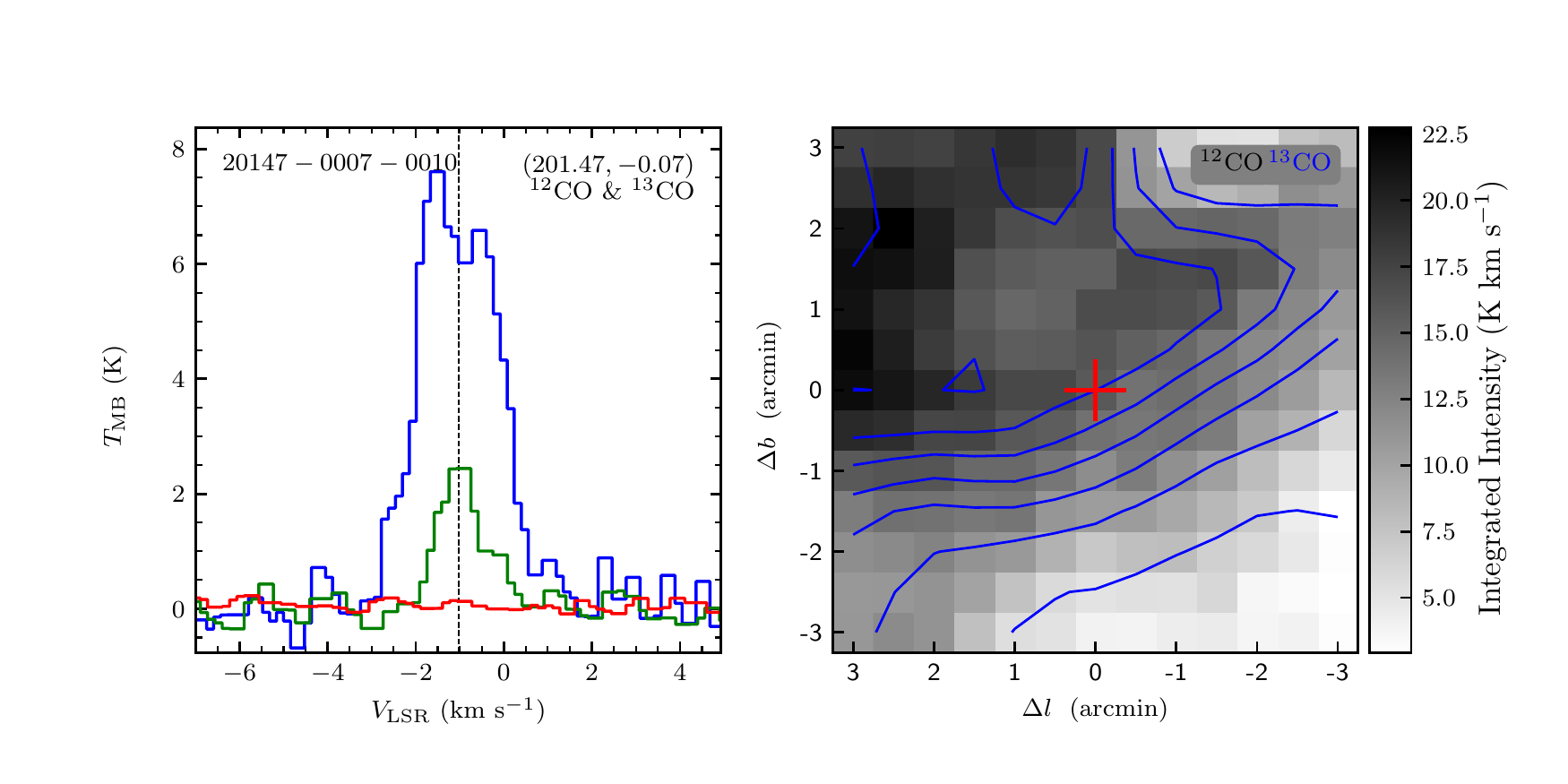}
\vspace{-0.5cm}

\includegraphics[width=9.0cm,angle=0]{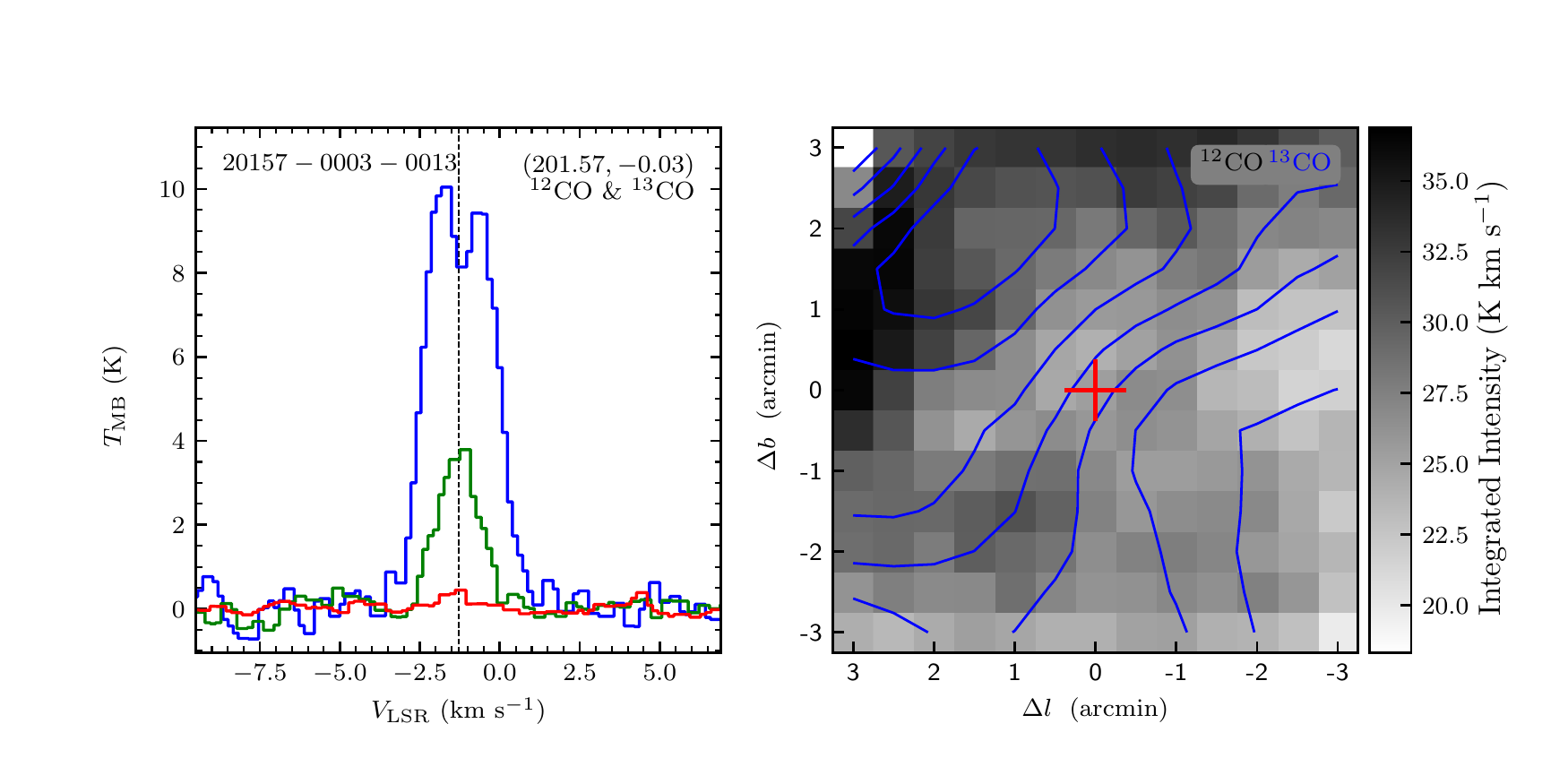}
\includegraphics[width=9.0cm,angle=0]{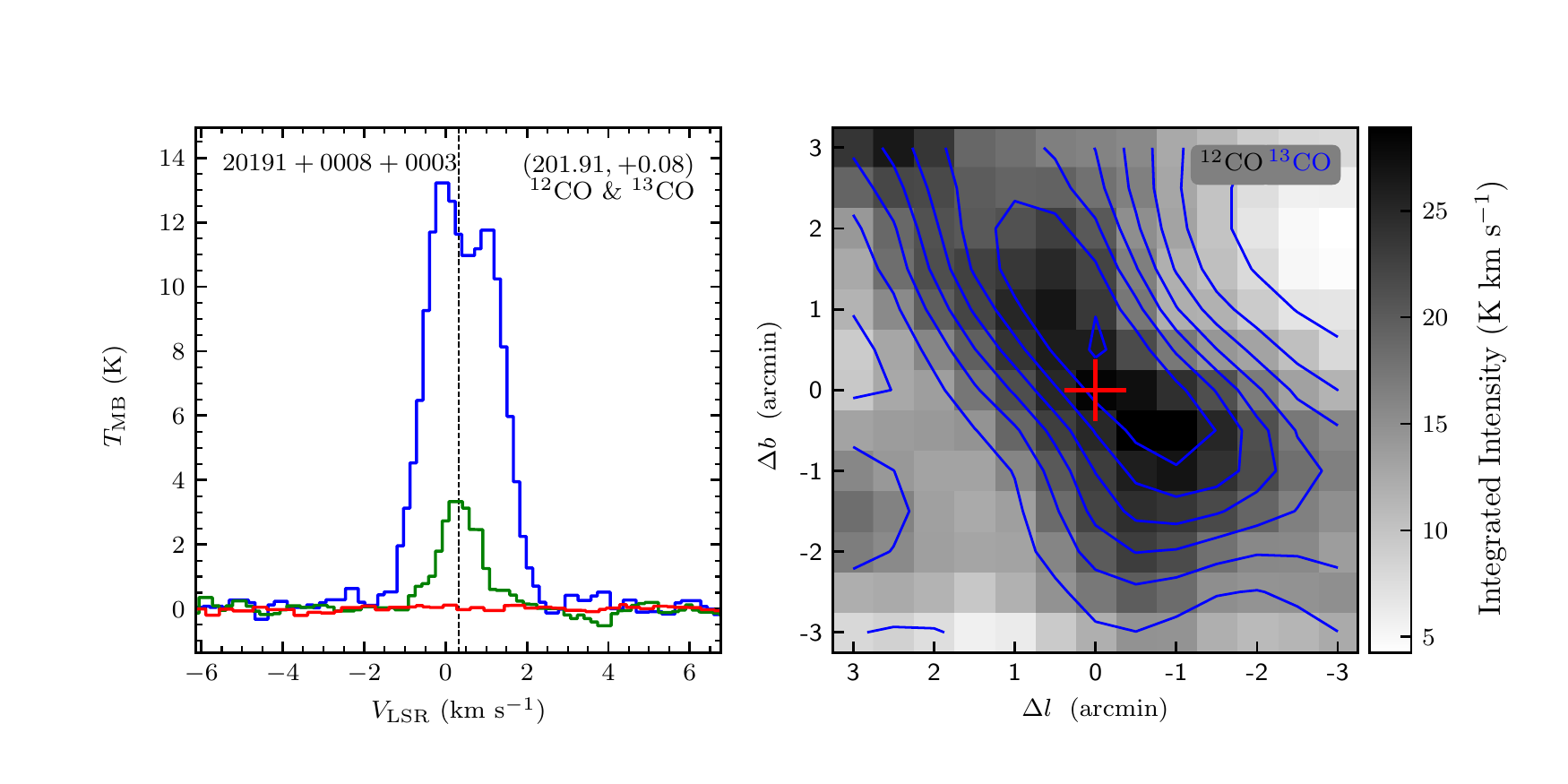}
\vspace{-0.5cm}

\includegraphics[width=9.0cm,angle=0]{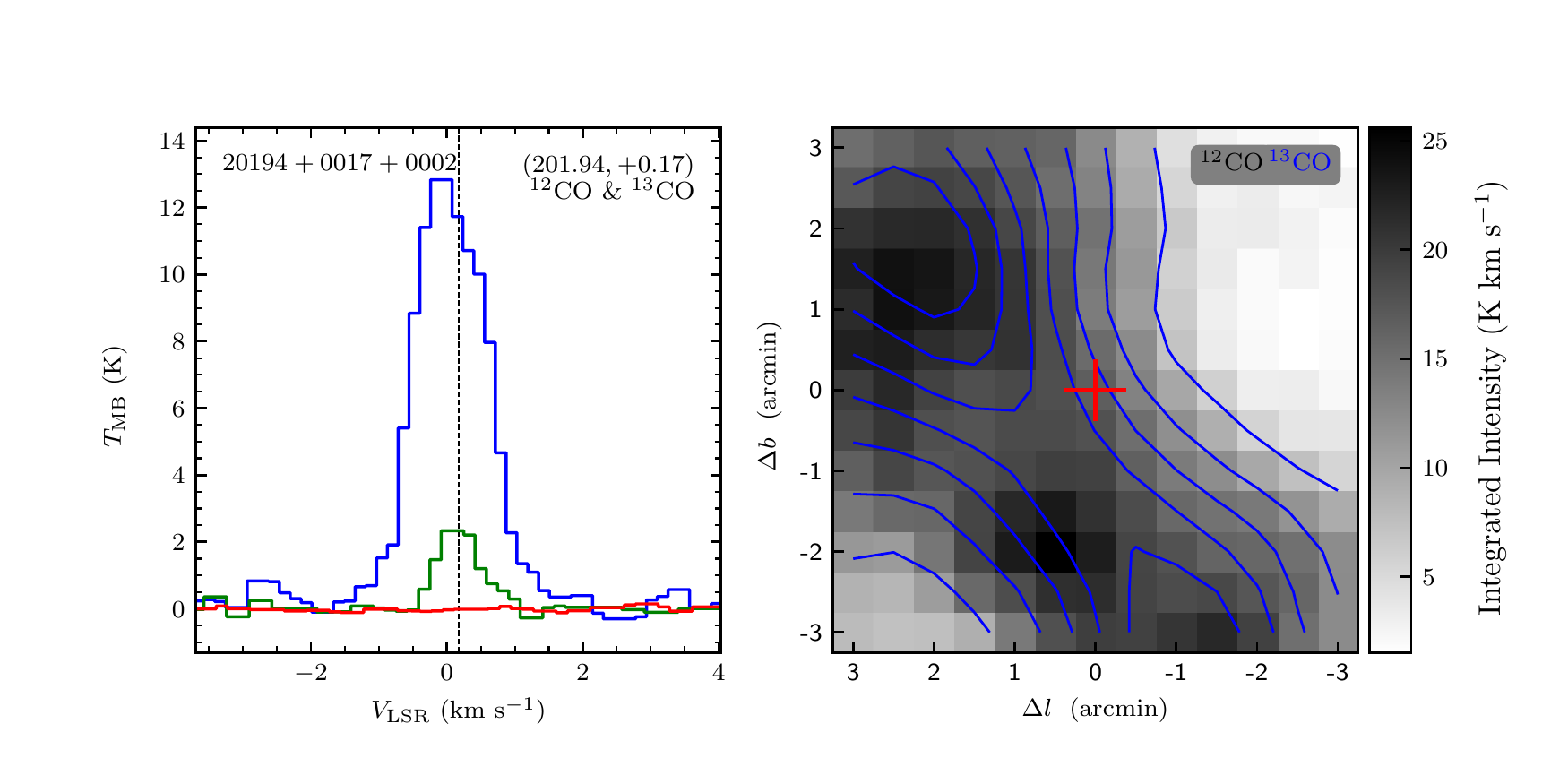}
\includegraphics[width=9.0cm,angle=0]{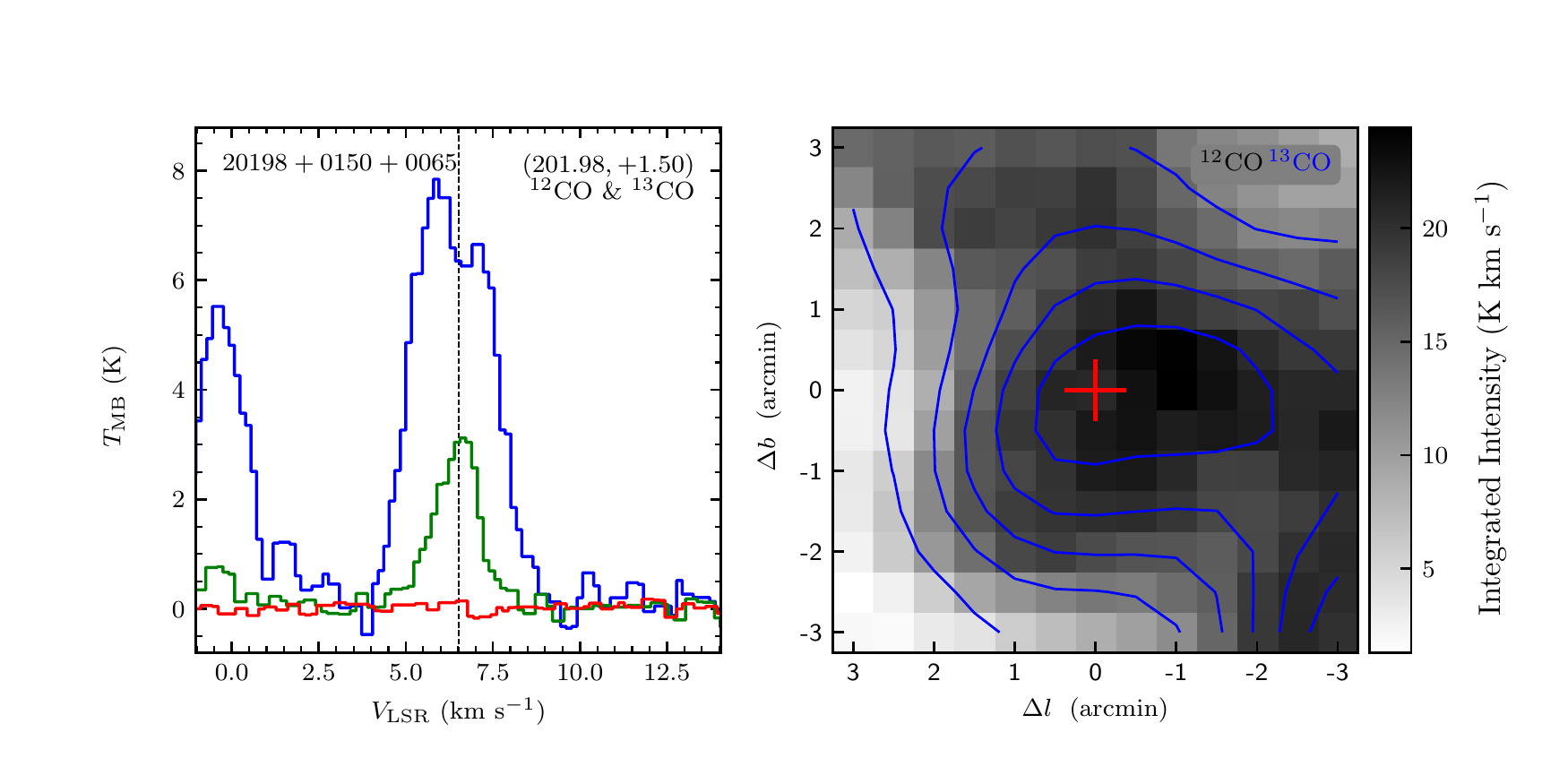}
\vspace{-0.5cm}

\includegraphics[width=9.0cm,angle=0]{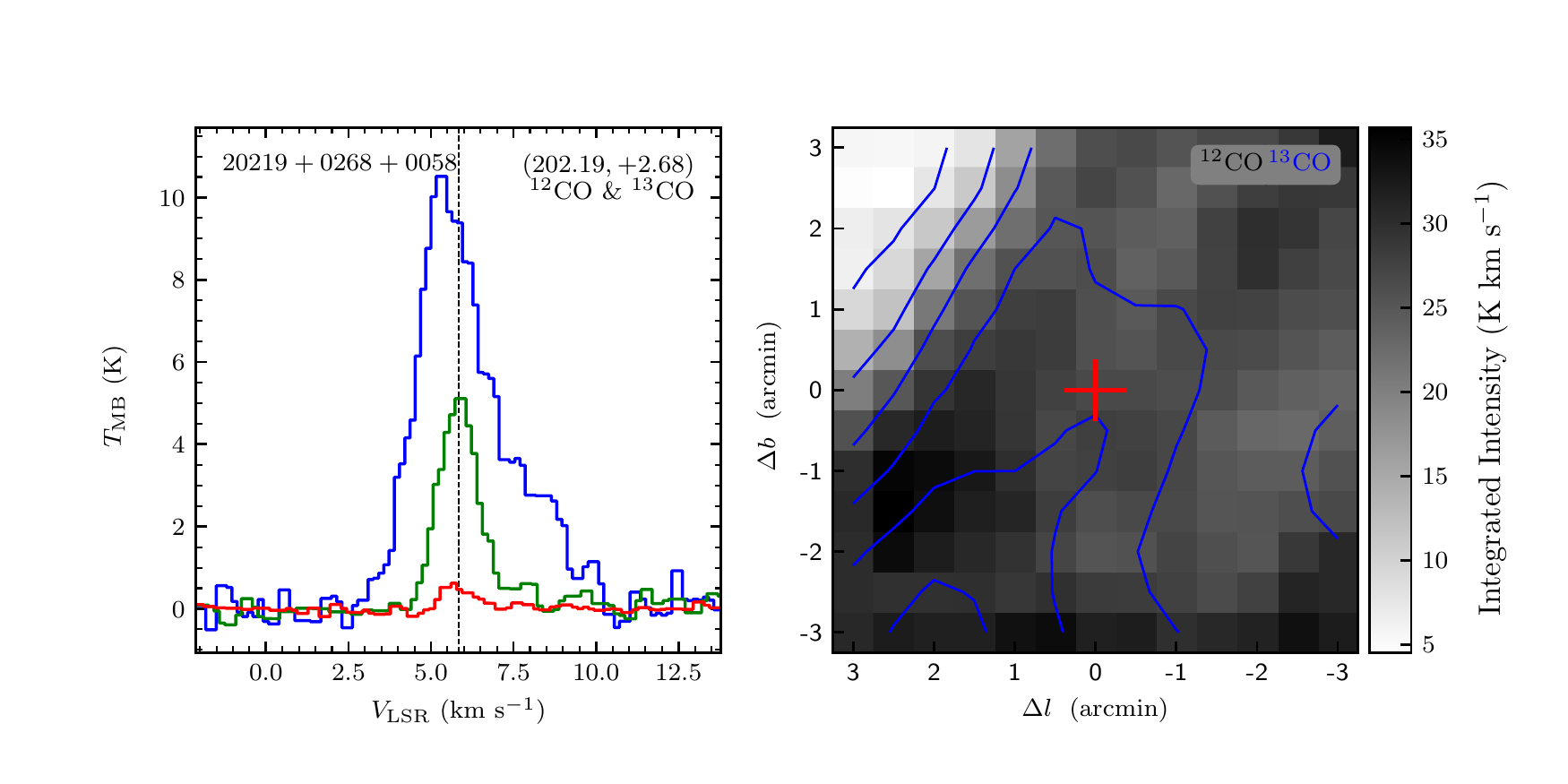}
\includegraphics[width=9.0cm,angle=0]{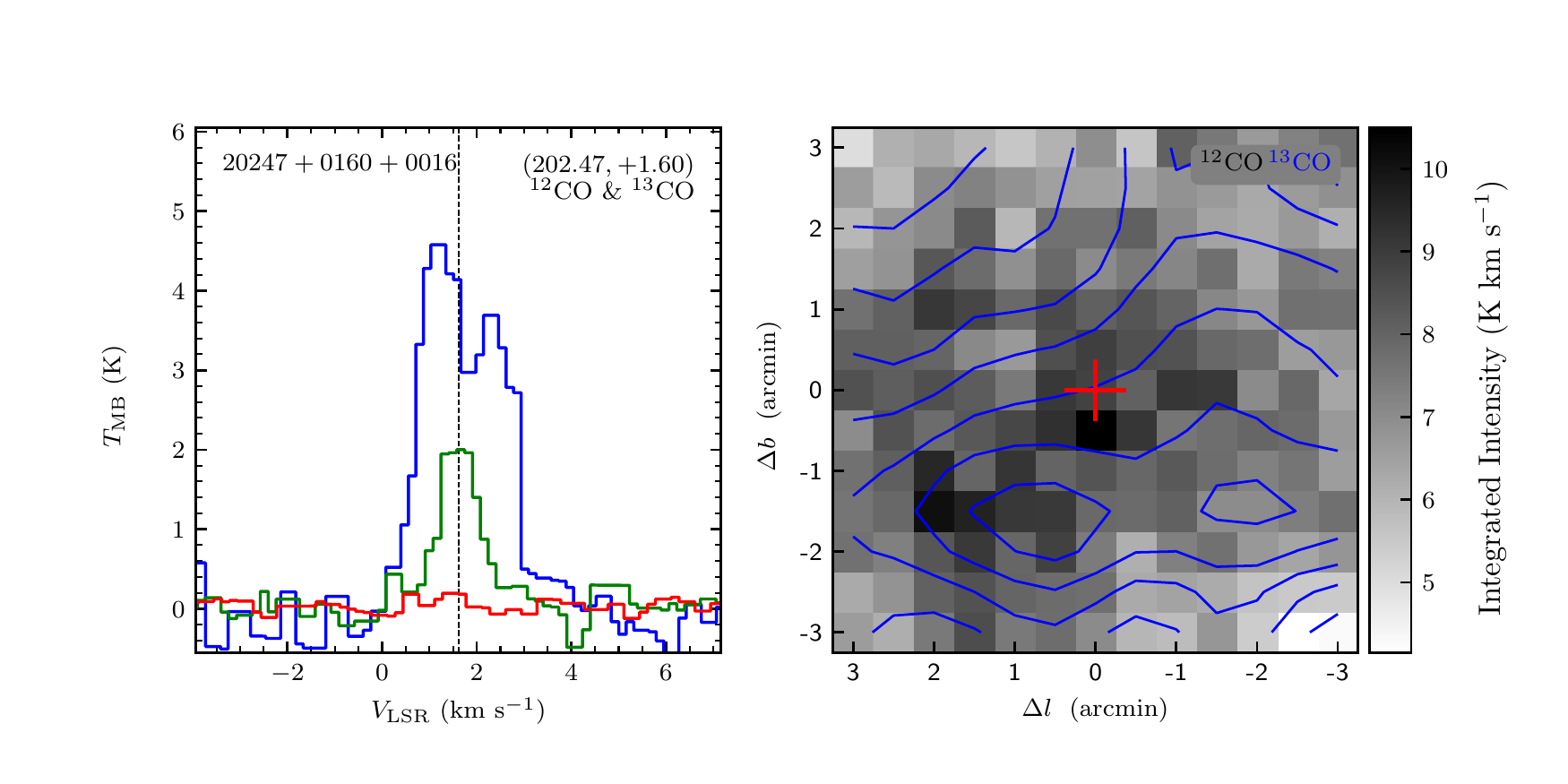}
\vspace{-0.5cm}

\includegraphics[width=9.0cm,angle=0]{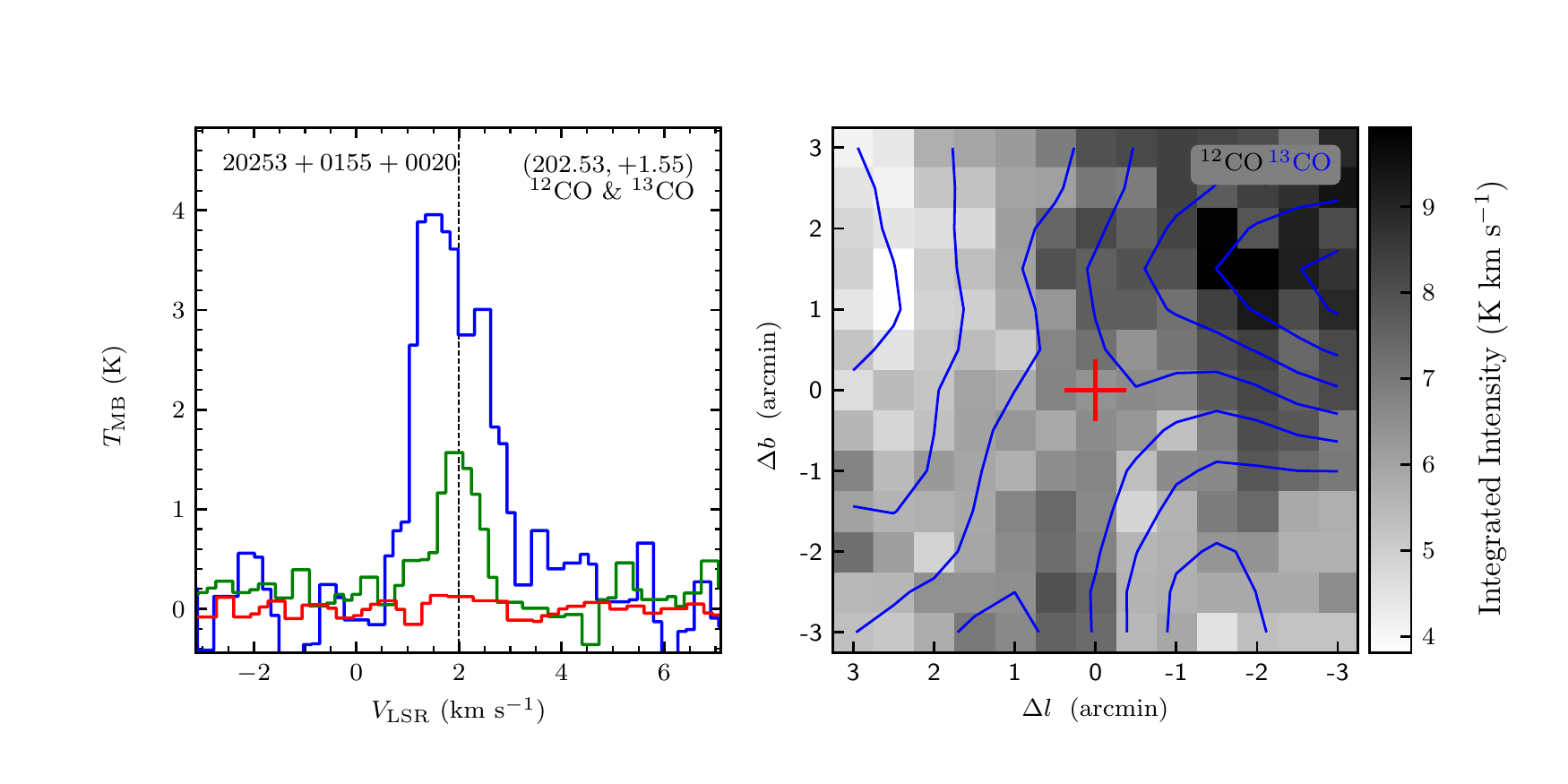}
\includegraphics[width=9.0cm,angle=0]{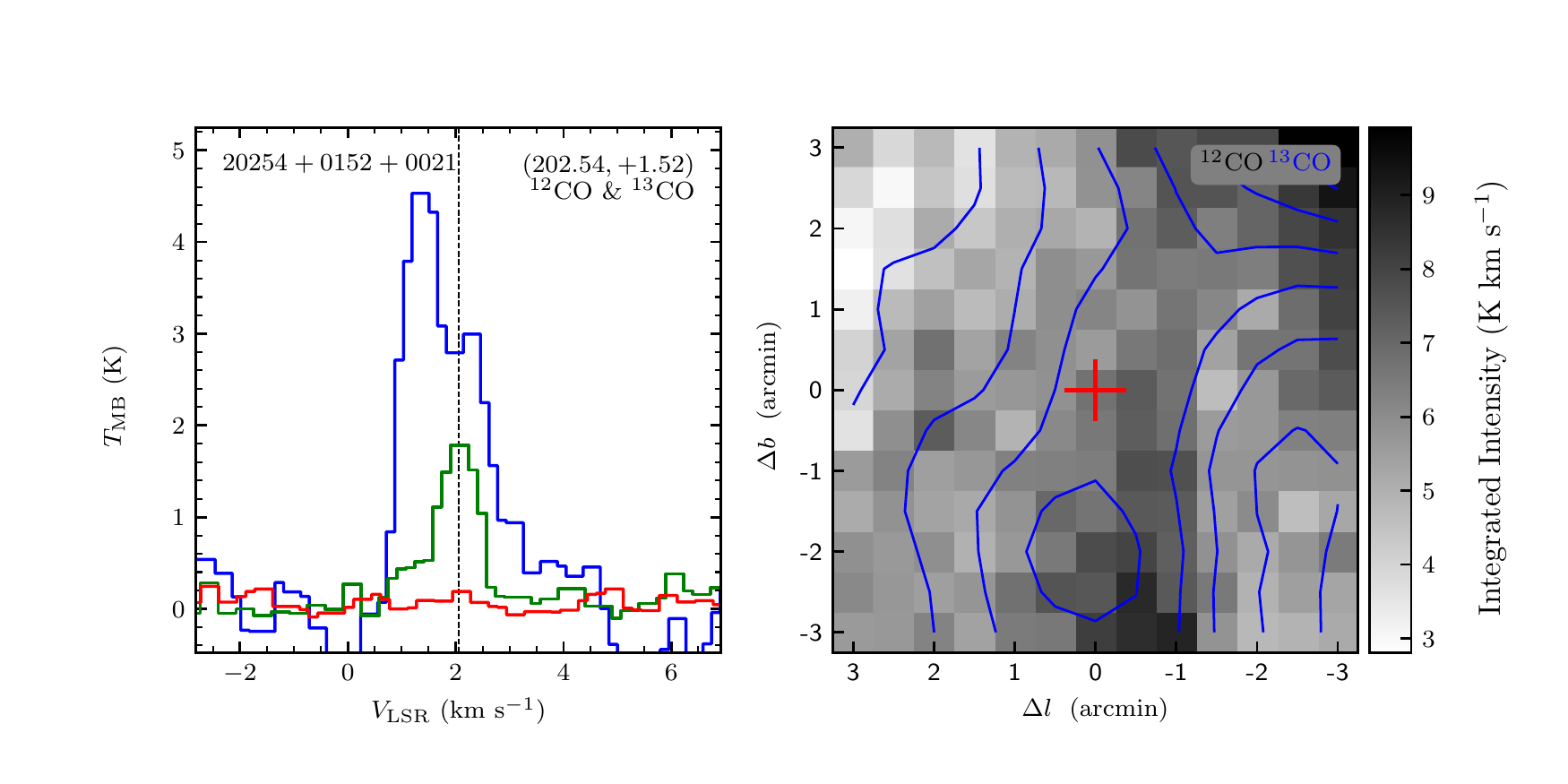}
\end{figure}
\clearpage

\begin{figure}
\includegraphics[width=9.0cm,angle=0]{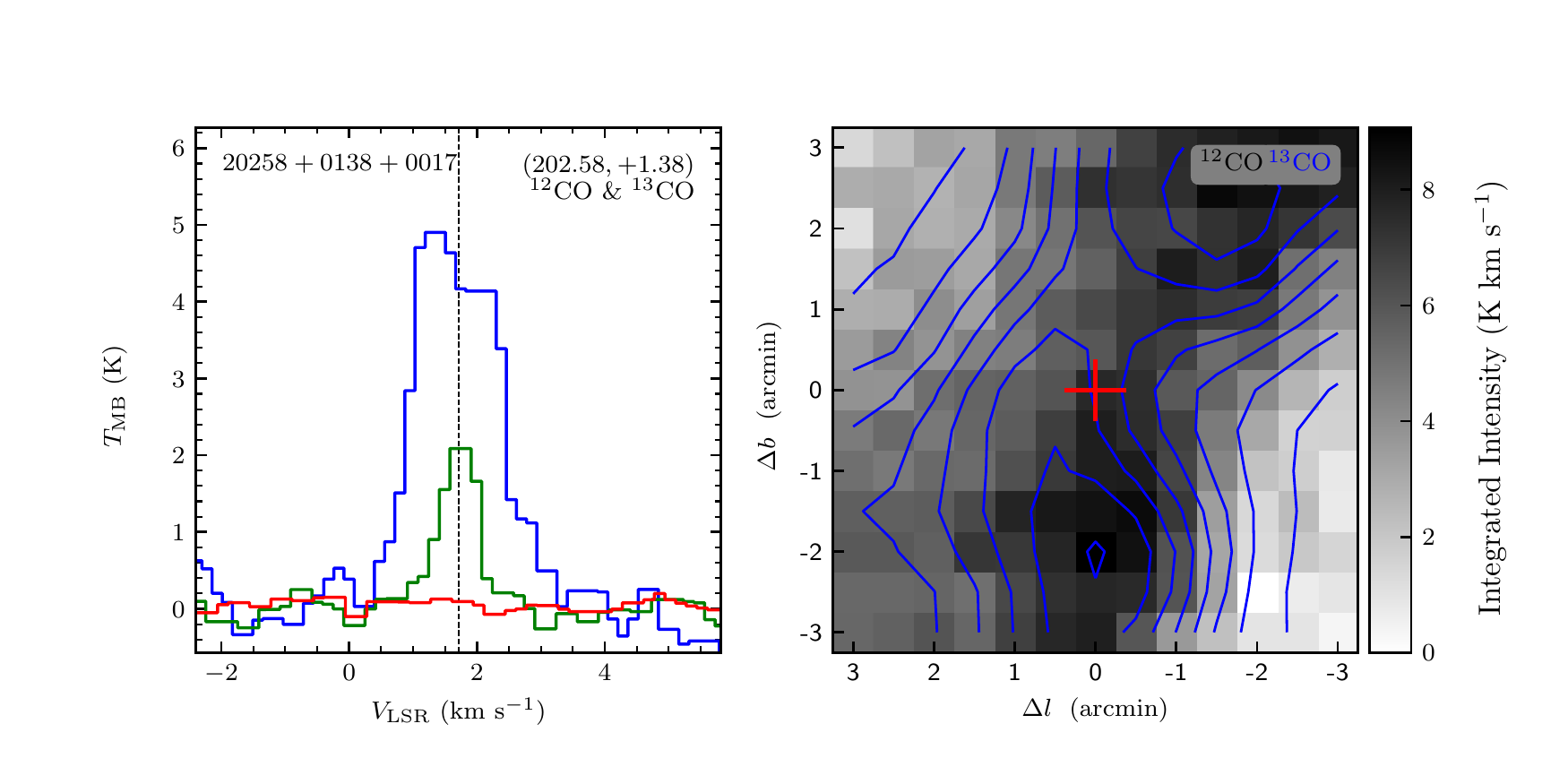}
\includegraphics[width=9.0cm,angle=0]{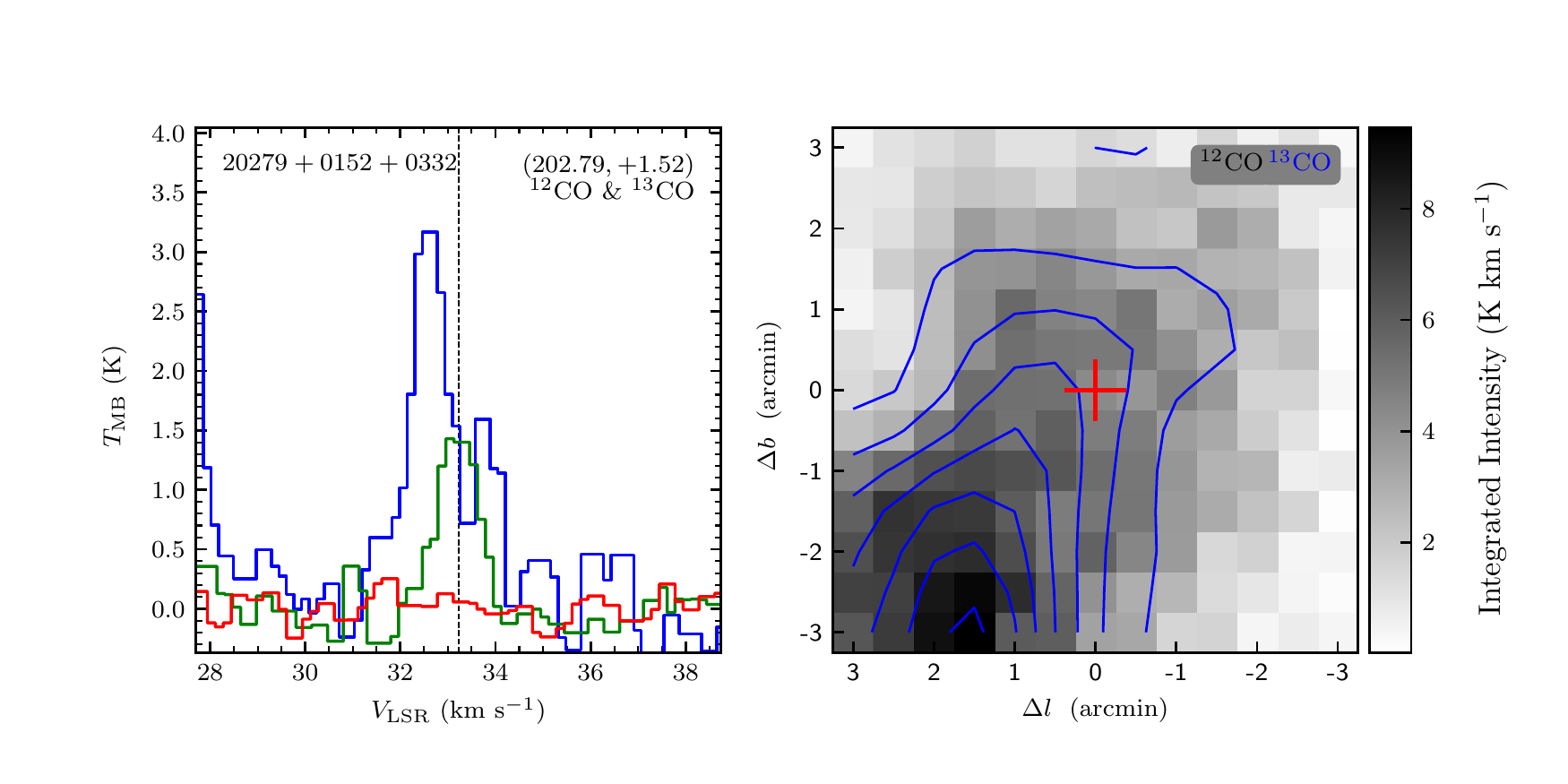}
\vspace{-0.5cm}

\includegraphics[width=9.0cm,angle=0]{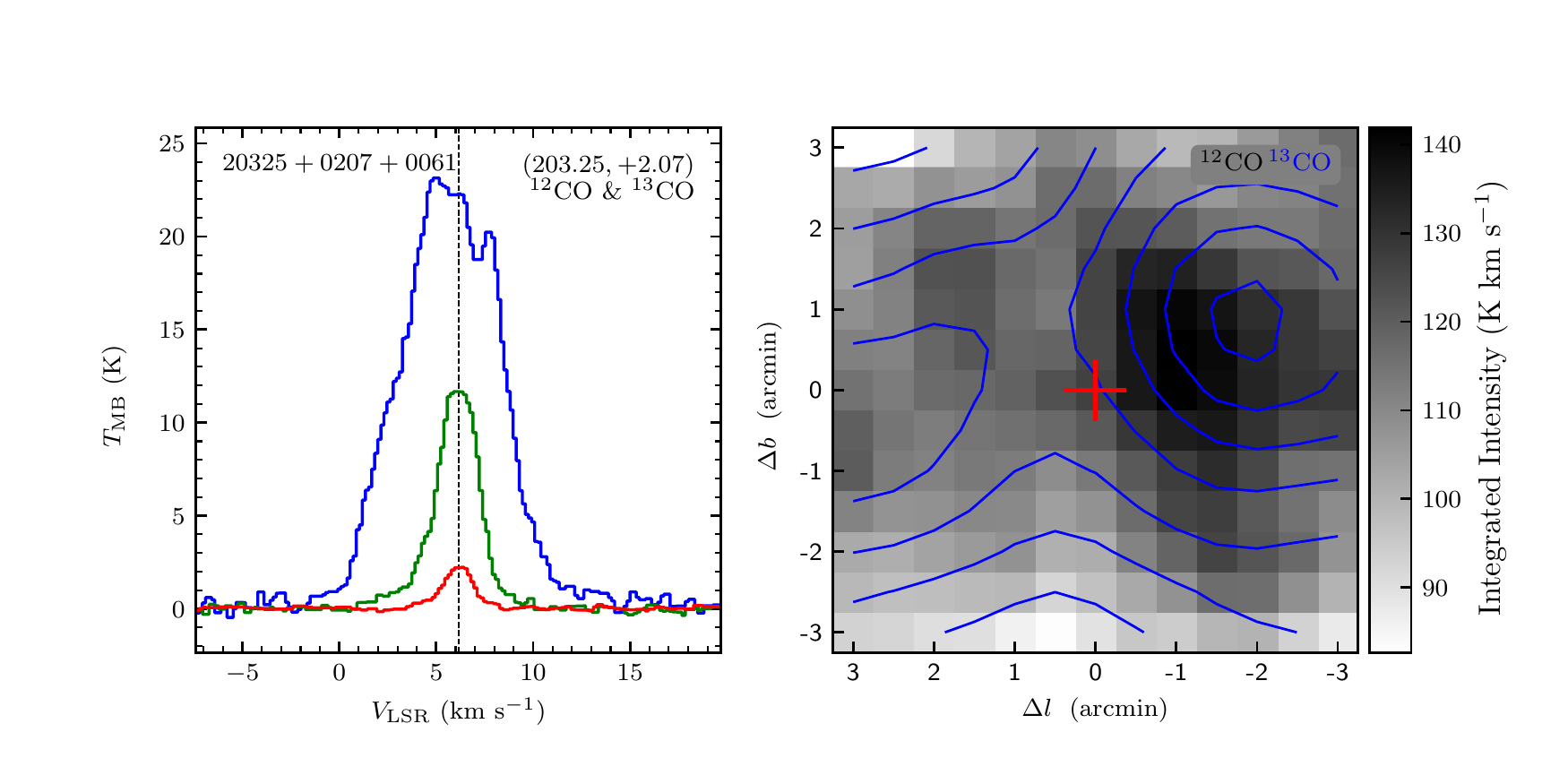}
\includegraphics[width=9.0cm,angle=0]{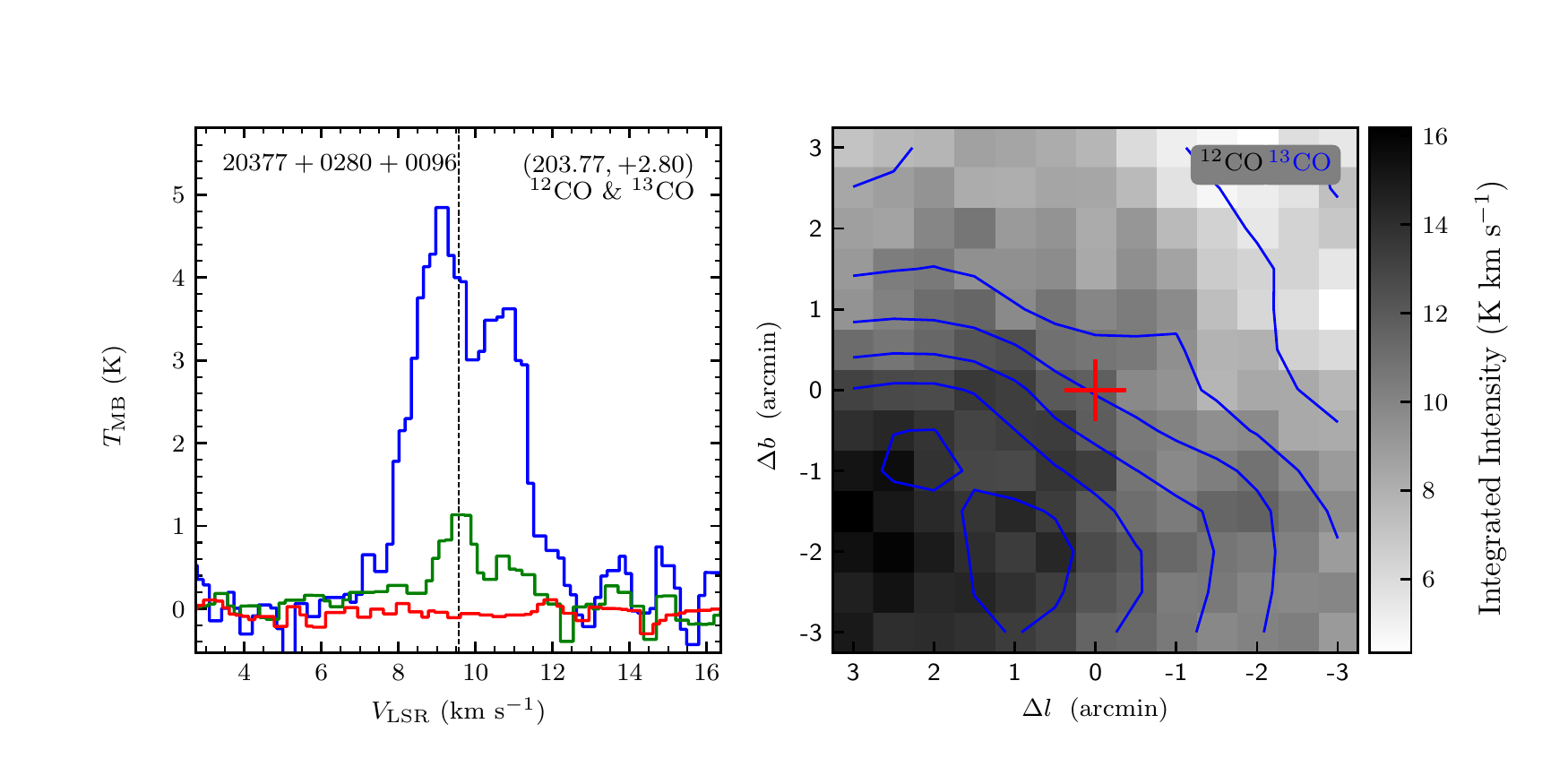}
\vspace{-0.5cm}

\includegraphics[width=9.0cm,angle=0]{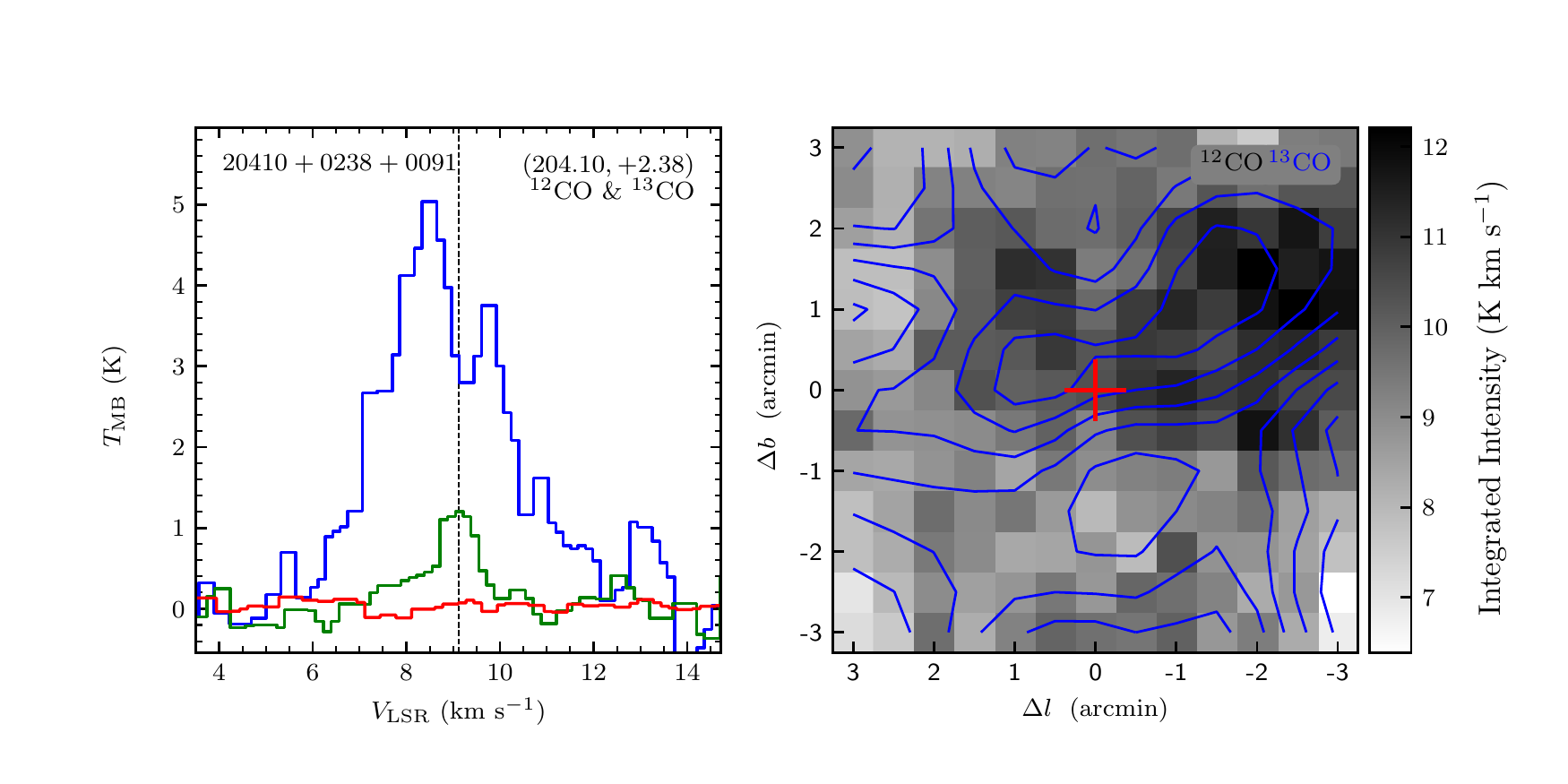}
\includegraphics[width=9.0cm,angle=0]{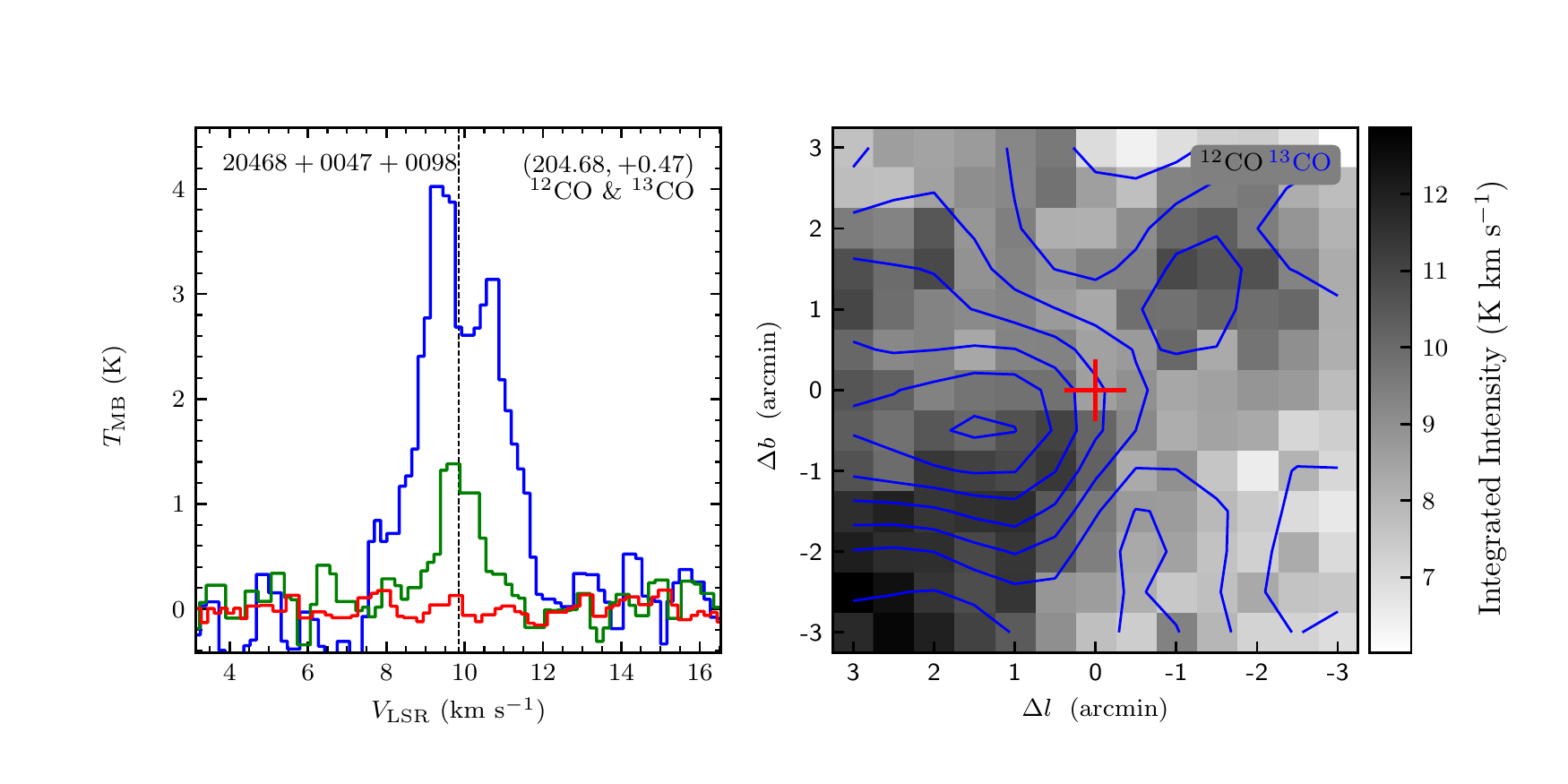}
\vspace{-0.5cm}

\includegraphics[width=9.0cm,angle=0]{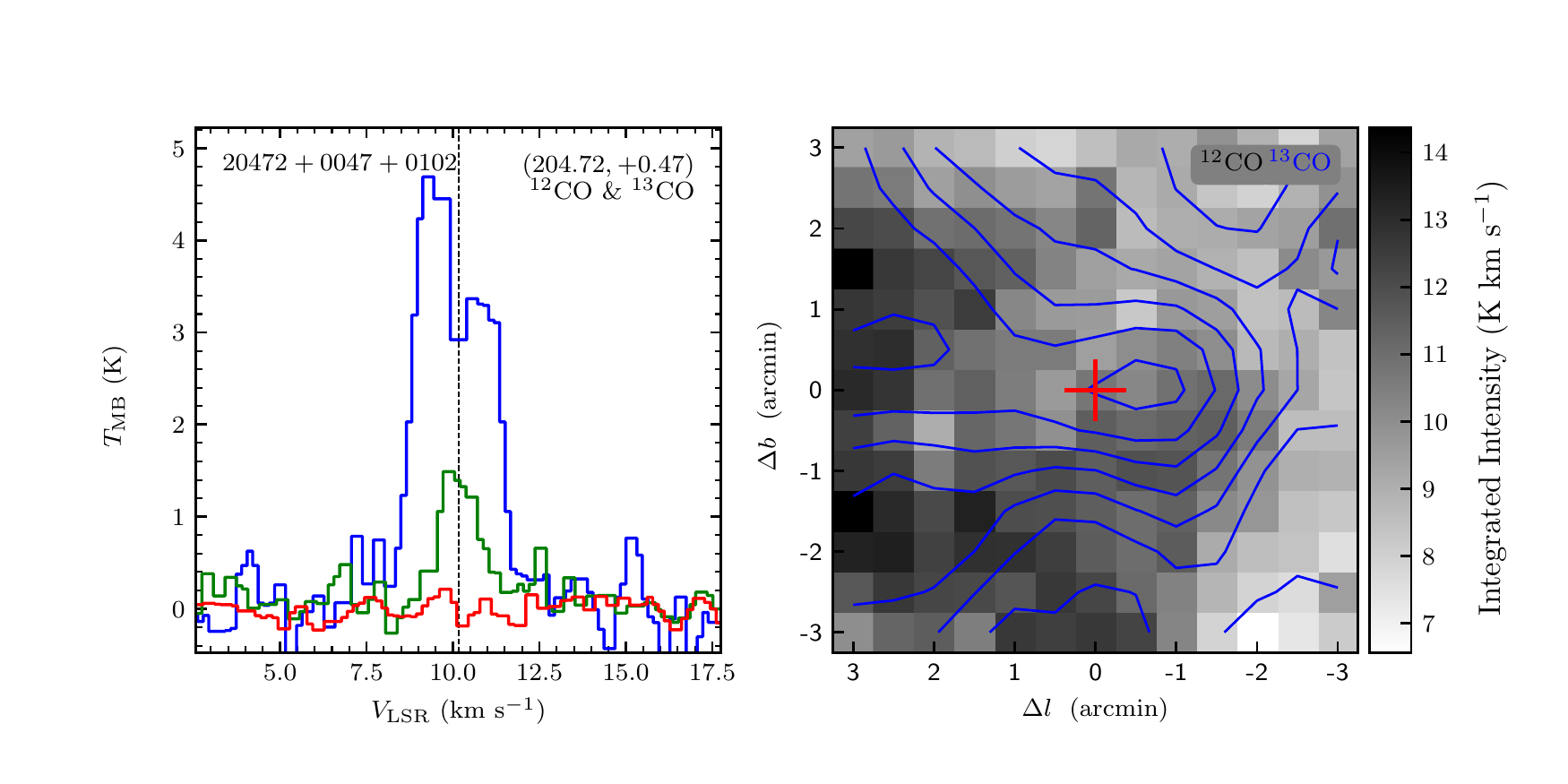}
\includegraphics[width=9.0cm,angle=0]{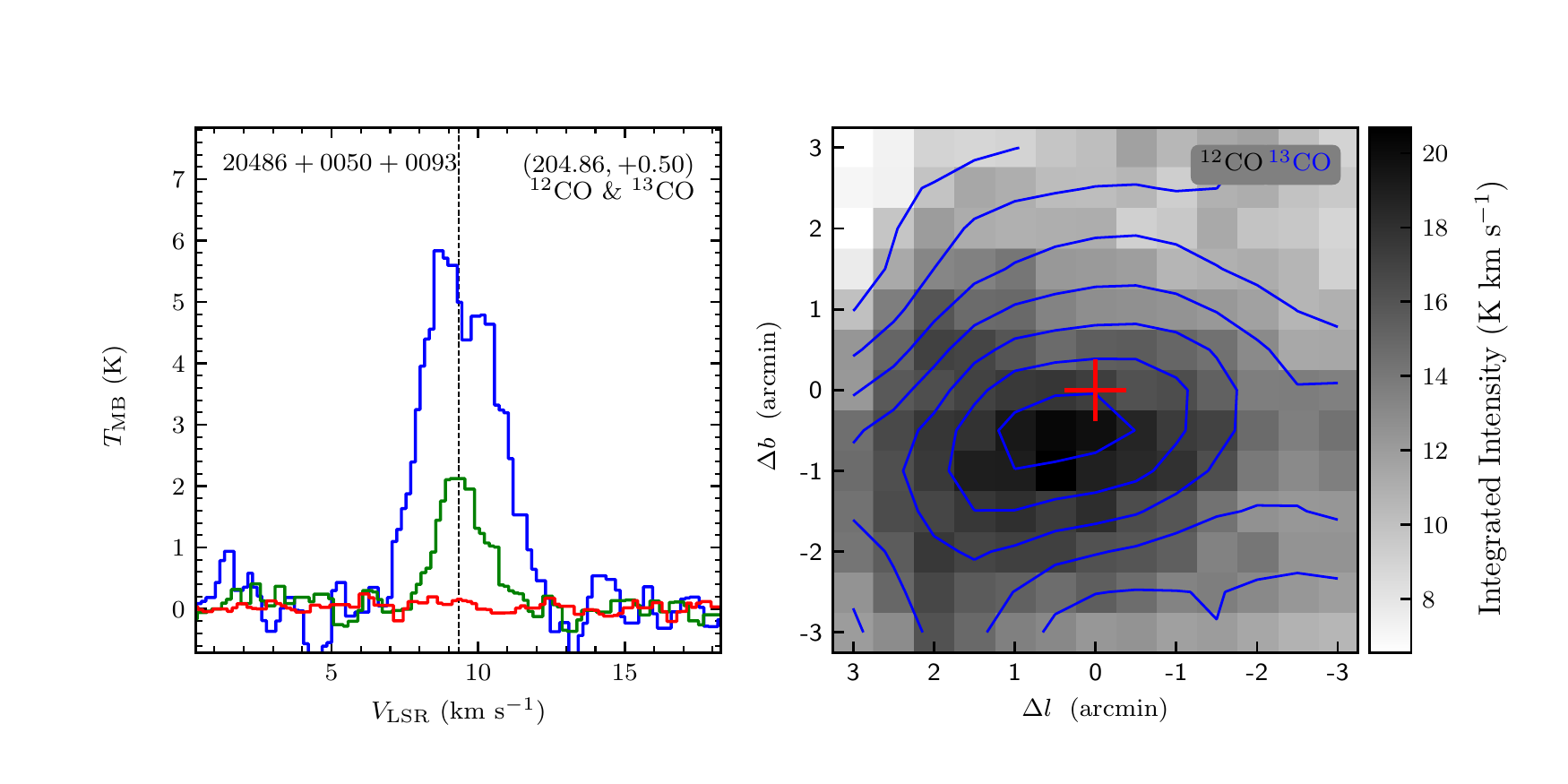}
\vspace{-0.5cm}

\includegraphics[width=9.0cm,angle=0]{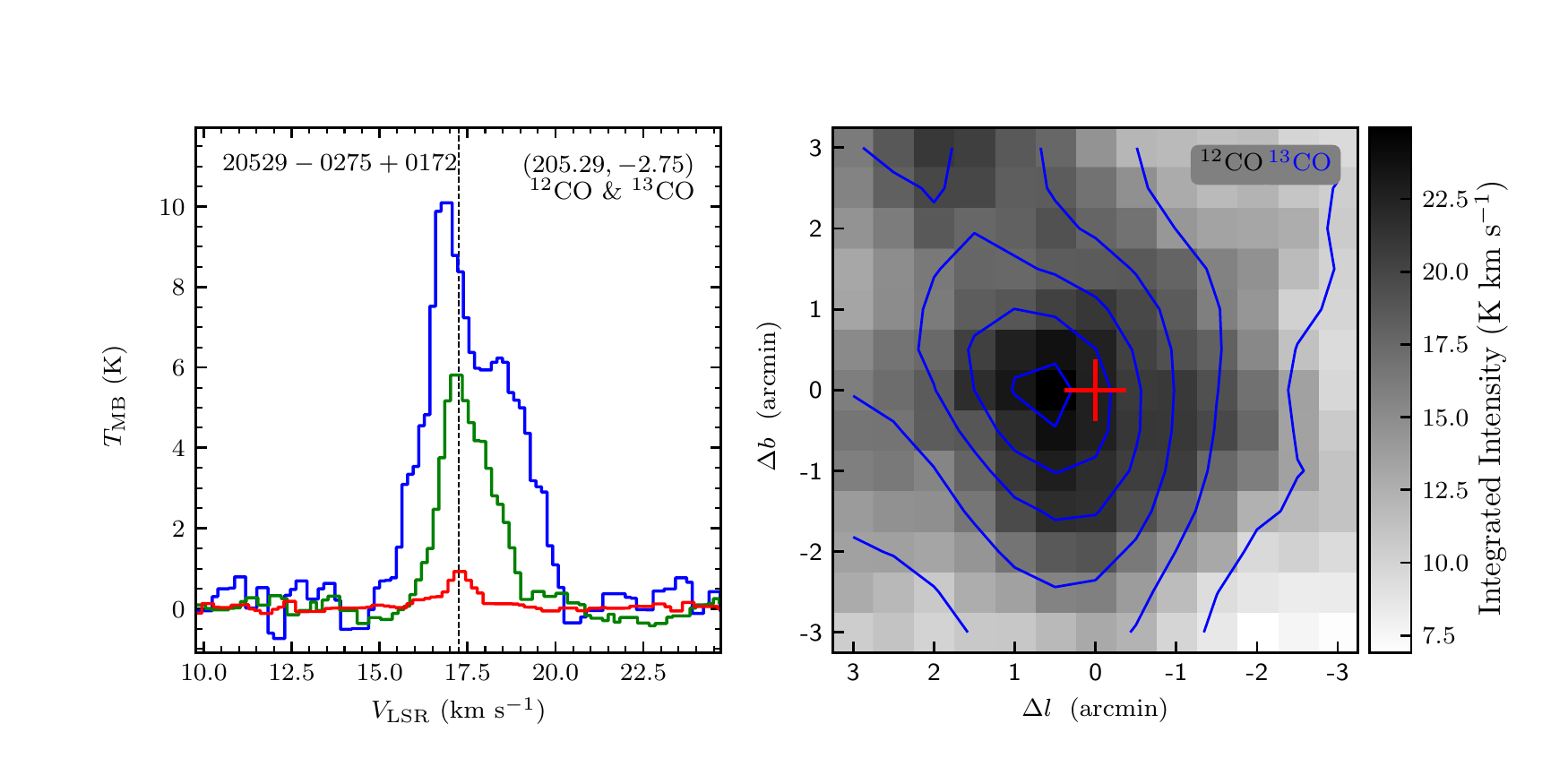}
\includegraphics[width=9.0cm,angle=0]{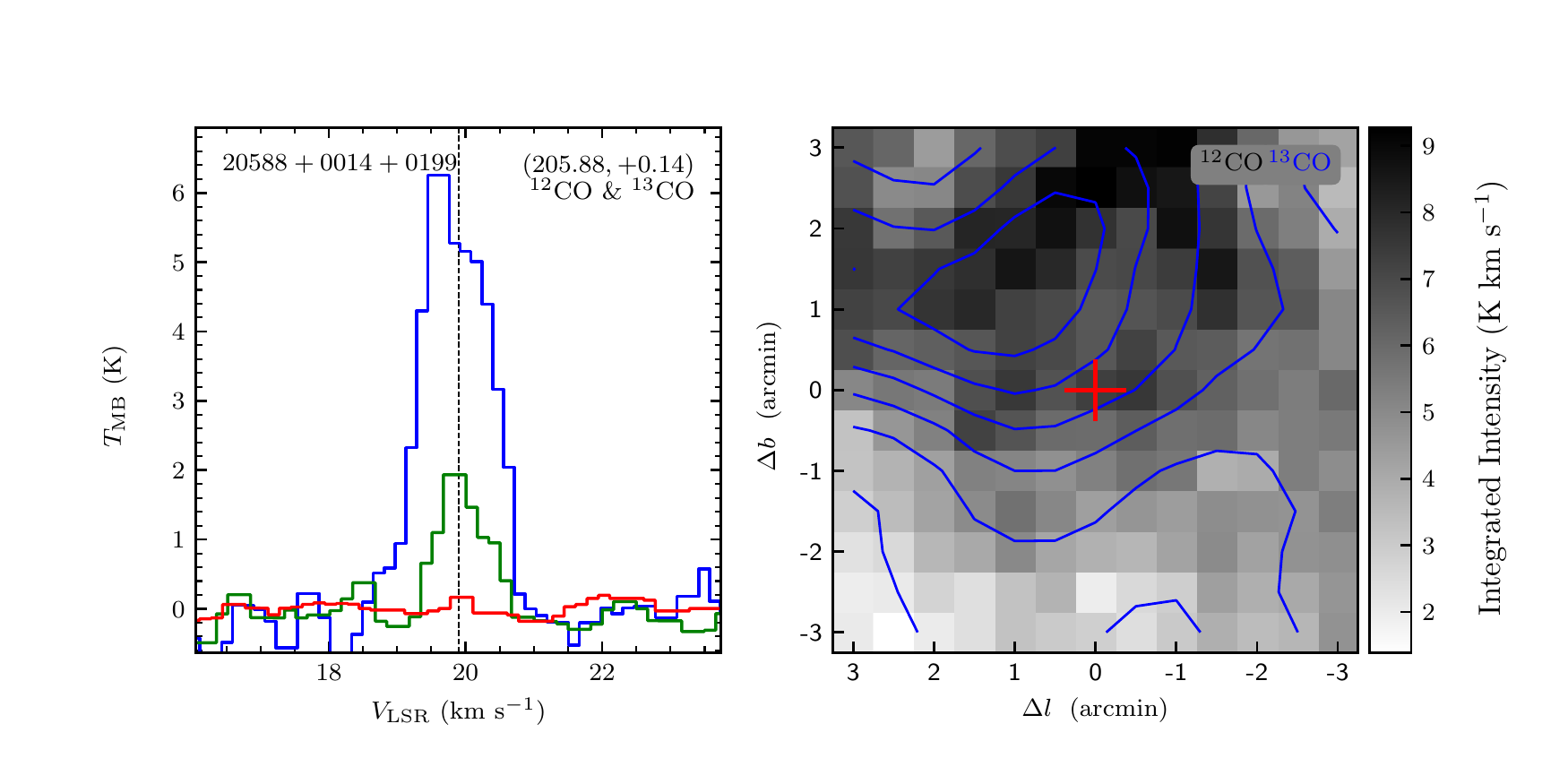}
\end{figure}
\clearpage

\begin{figure}
\includegraphics[width=9.0cm,angle=0]{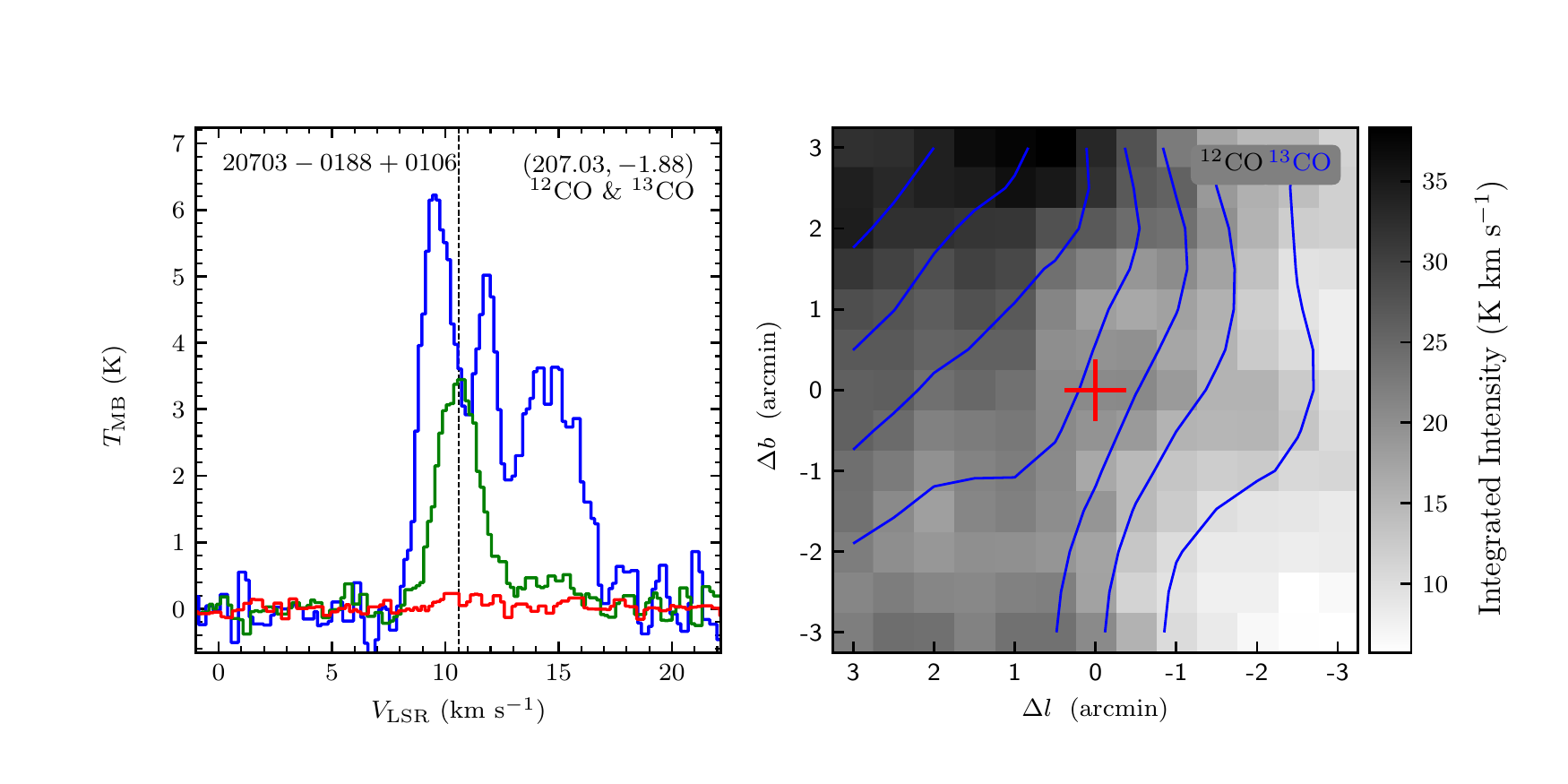}
\includegraphics[width=9.0cm,angle=0]{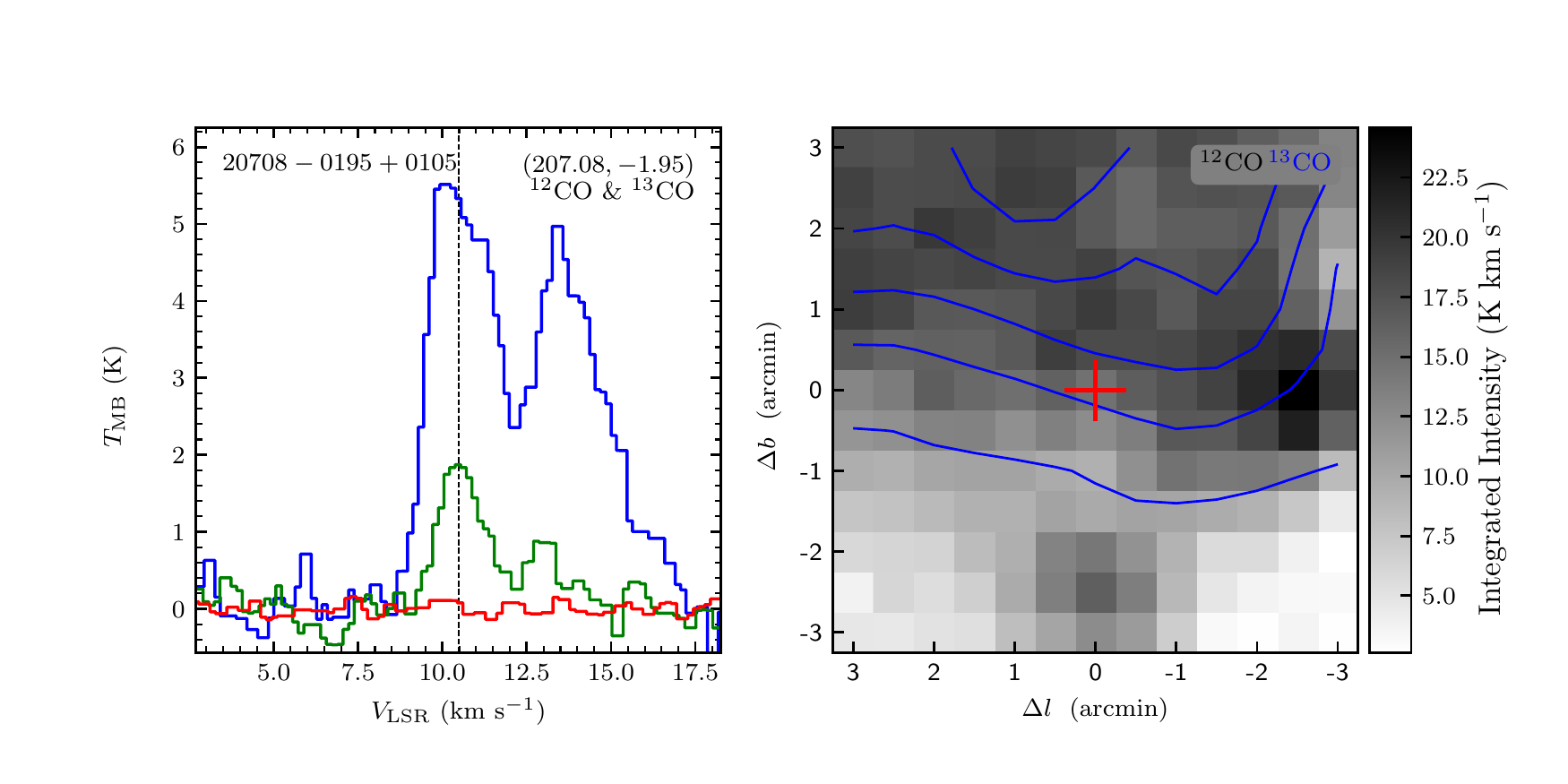}
\vspace{-0.5cm}

\includegraphics[width=9.0cm,angle=0]{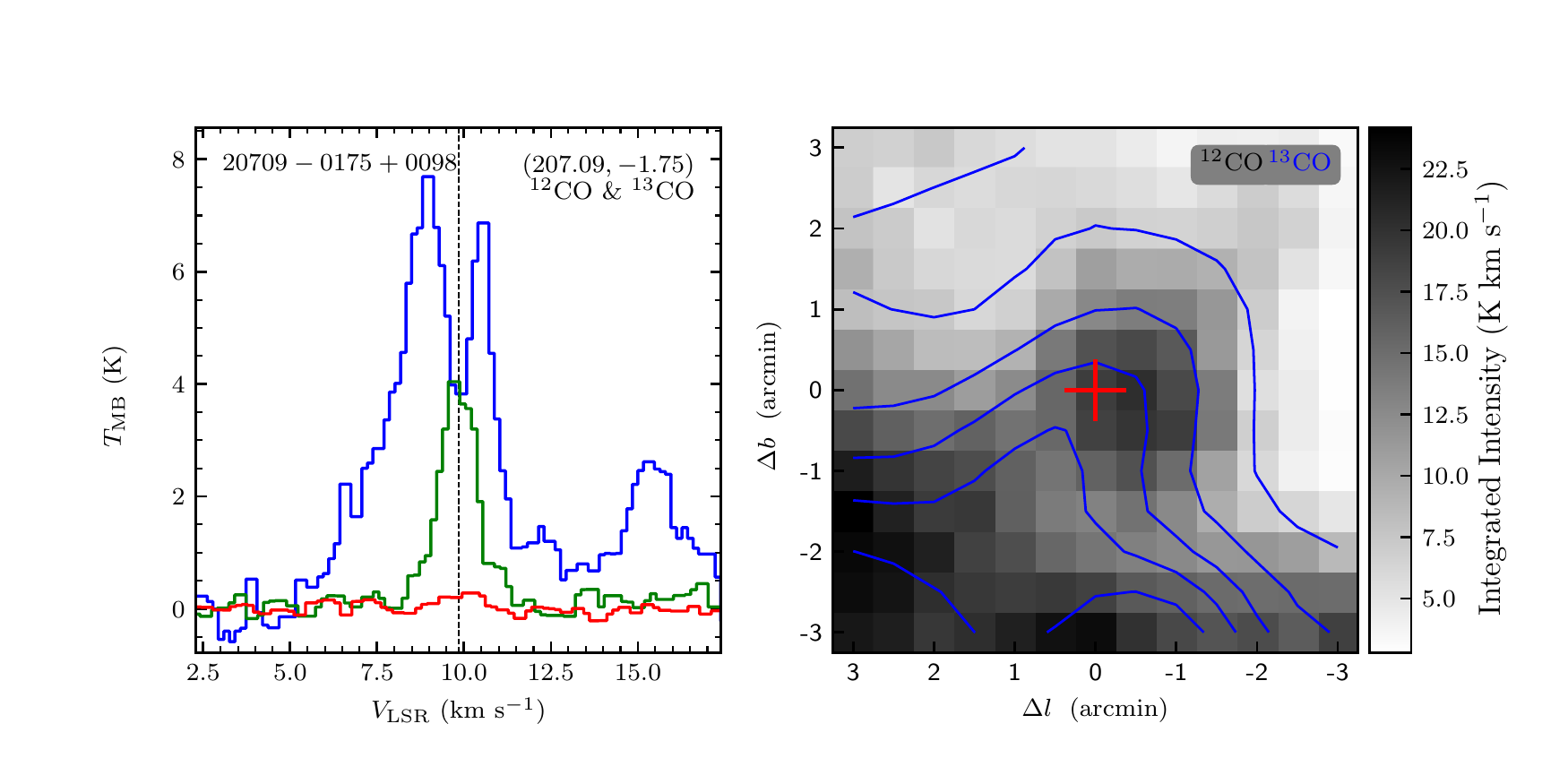}
\includegraphics[width=9.0cm,angle=0]{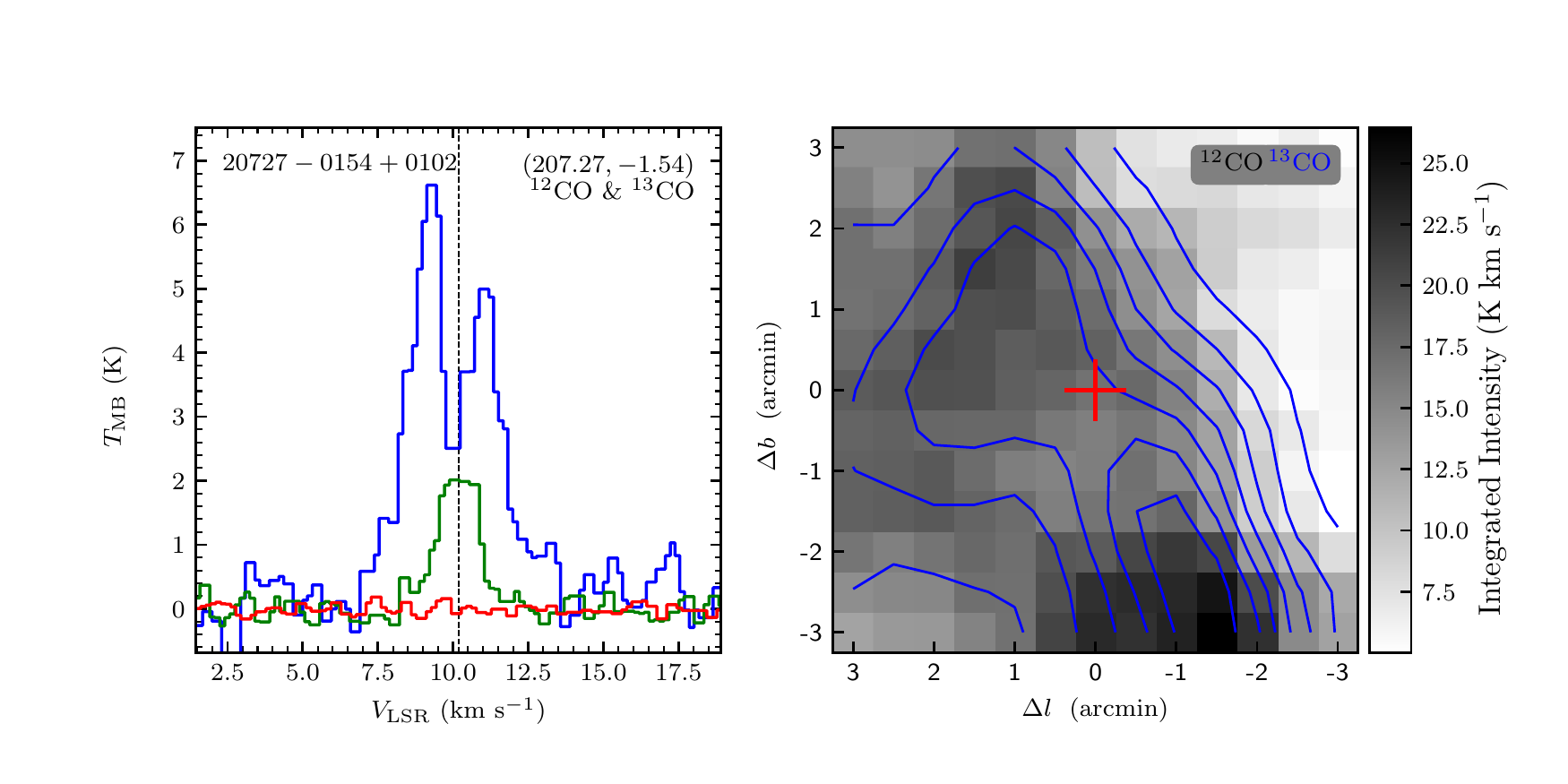}
\vspace{-0.5cm}

\includegraphics[width=9.0cm,angle=0]{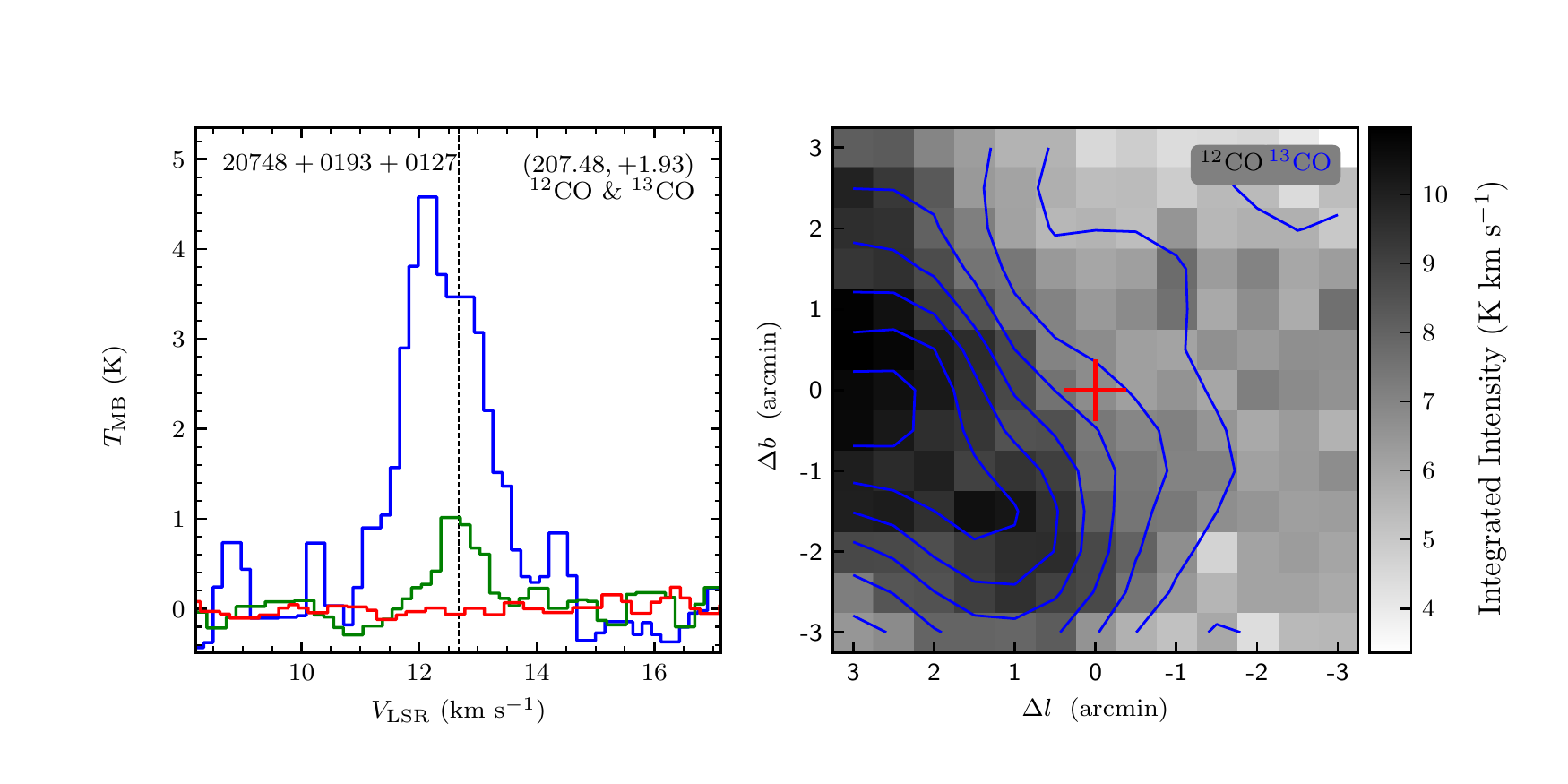}
\includegraphics[width=9.0cm,angle=0]{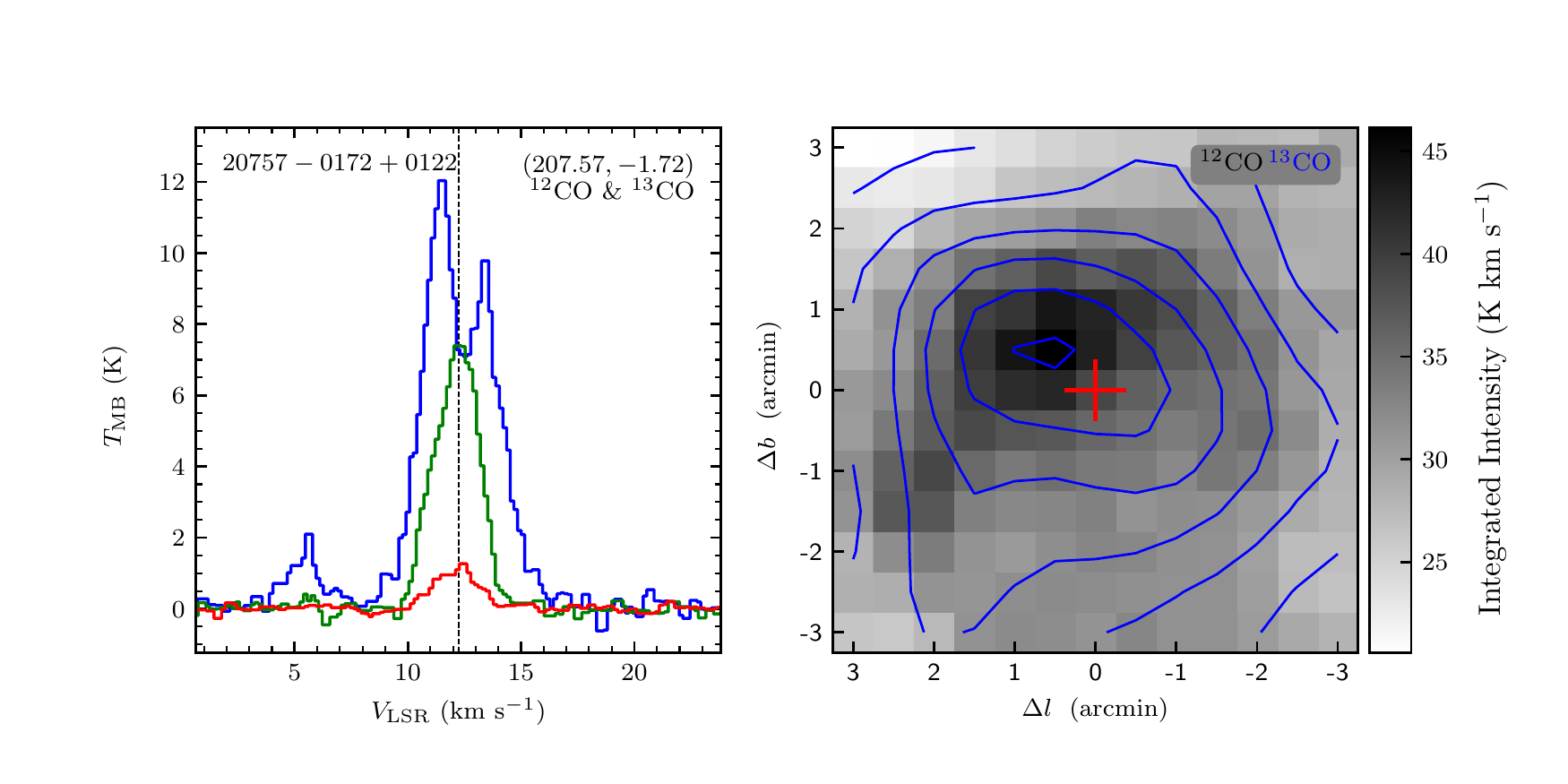}
\vspace{-0.5cm}

\includegraphics[width=9.0cm,angle=0]{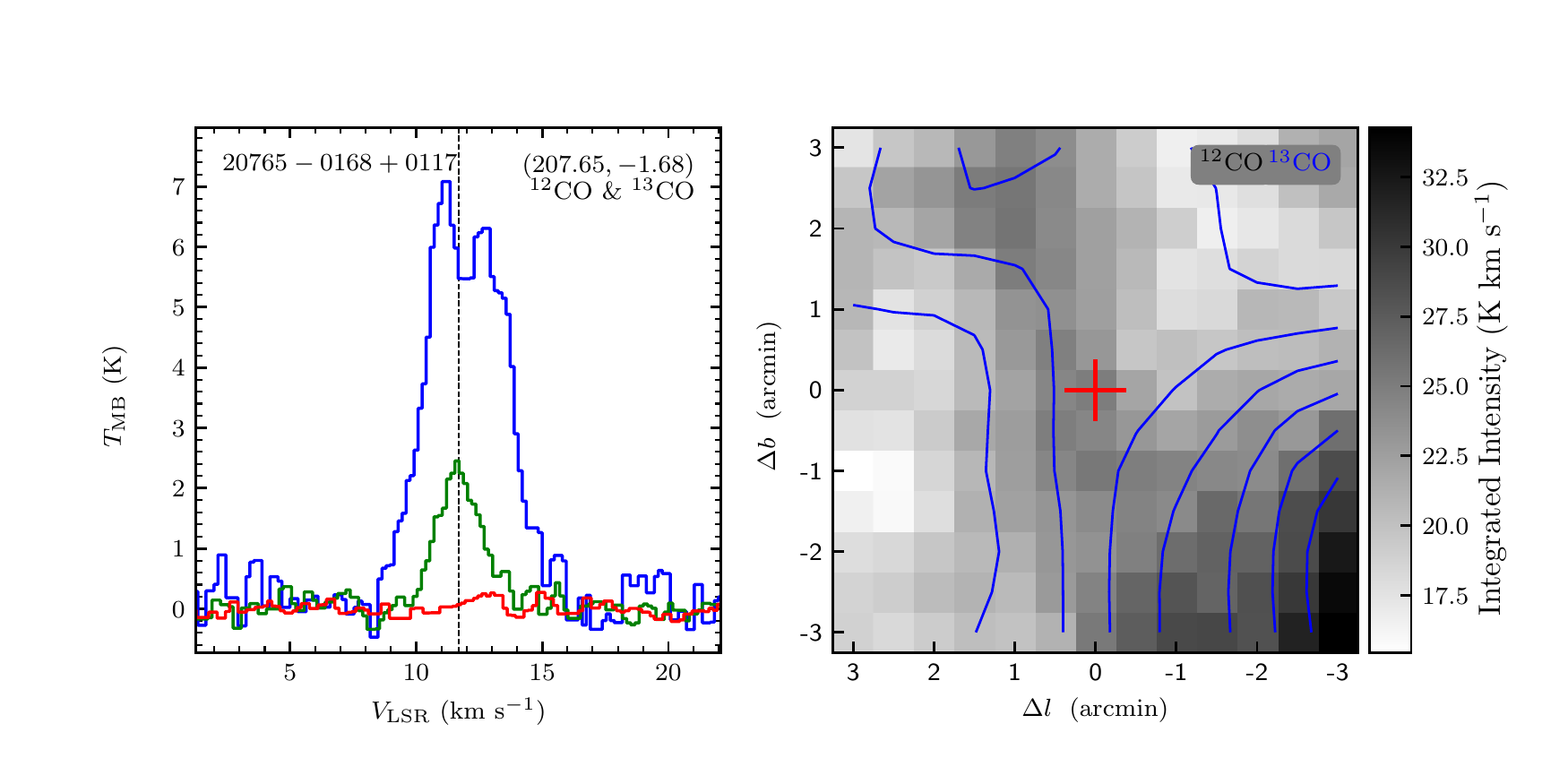}
\includegraphics[width=9.0cm,angle=0]{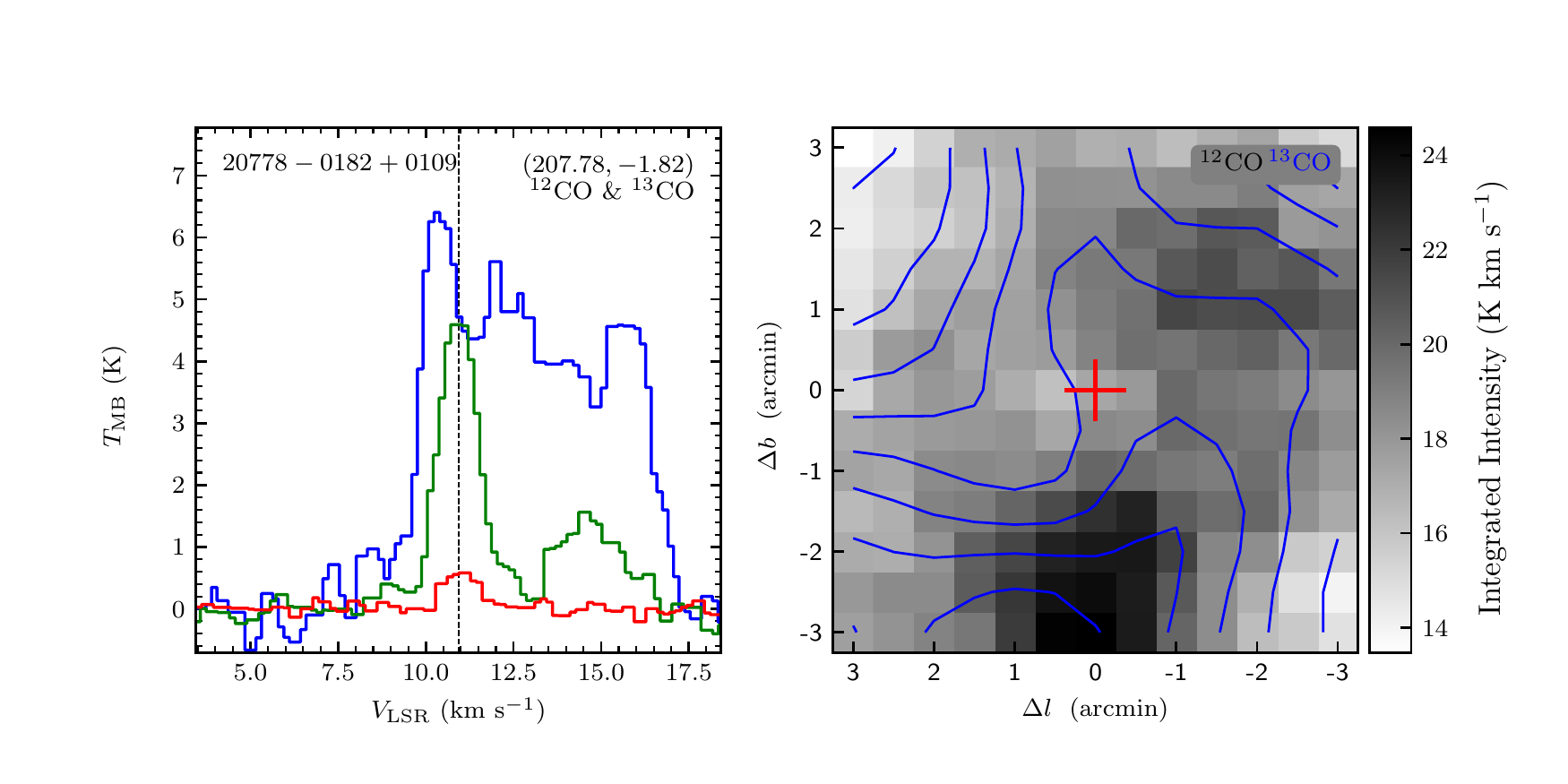}
\vspace{-0.5cm}

\includegraphics[width=9.0cm,angle=0]{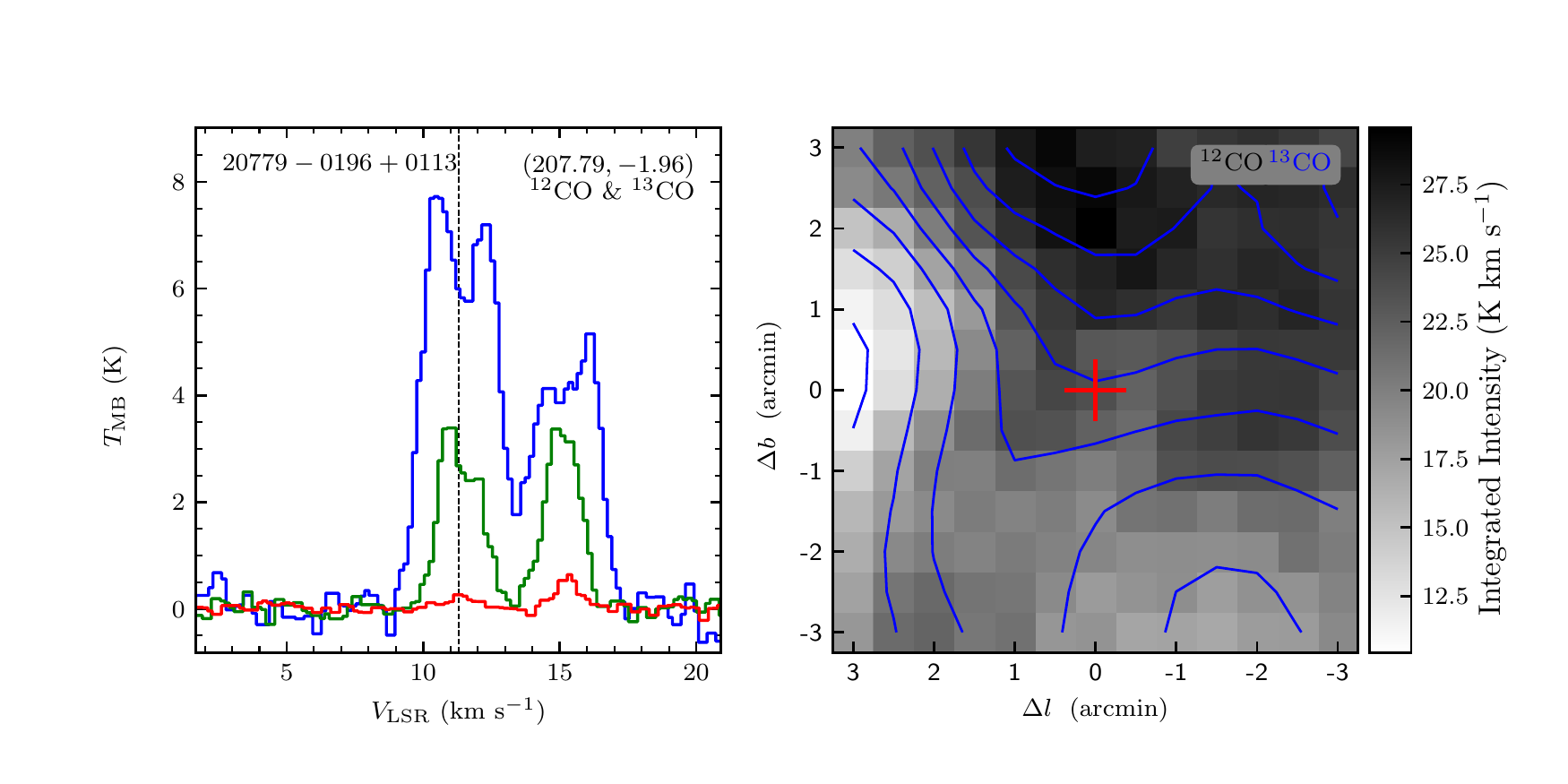}
\includegraphics[width=9.0cm,angle=0]{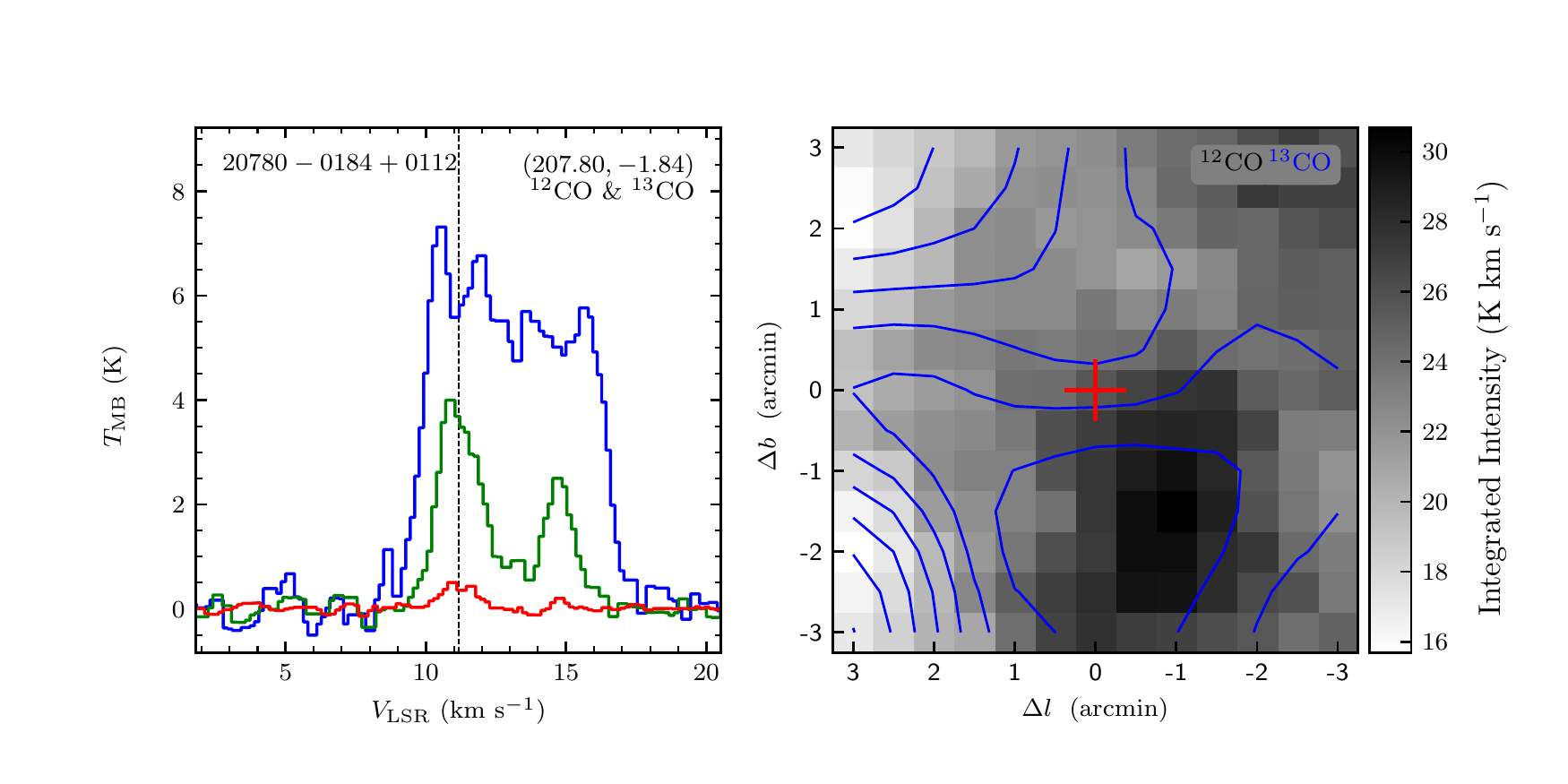}
\end{figure}
\clearpage

\begin{figure}
\includegraphics[width=9.0cm,angle=0]{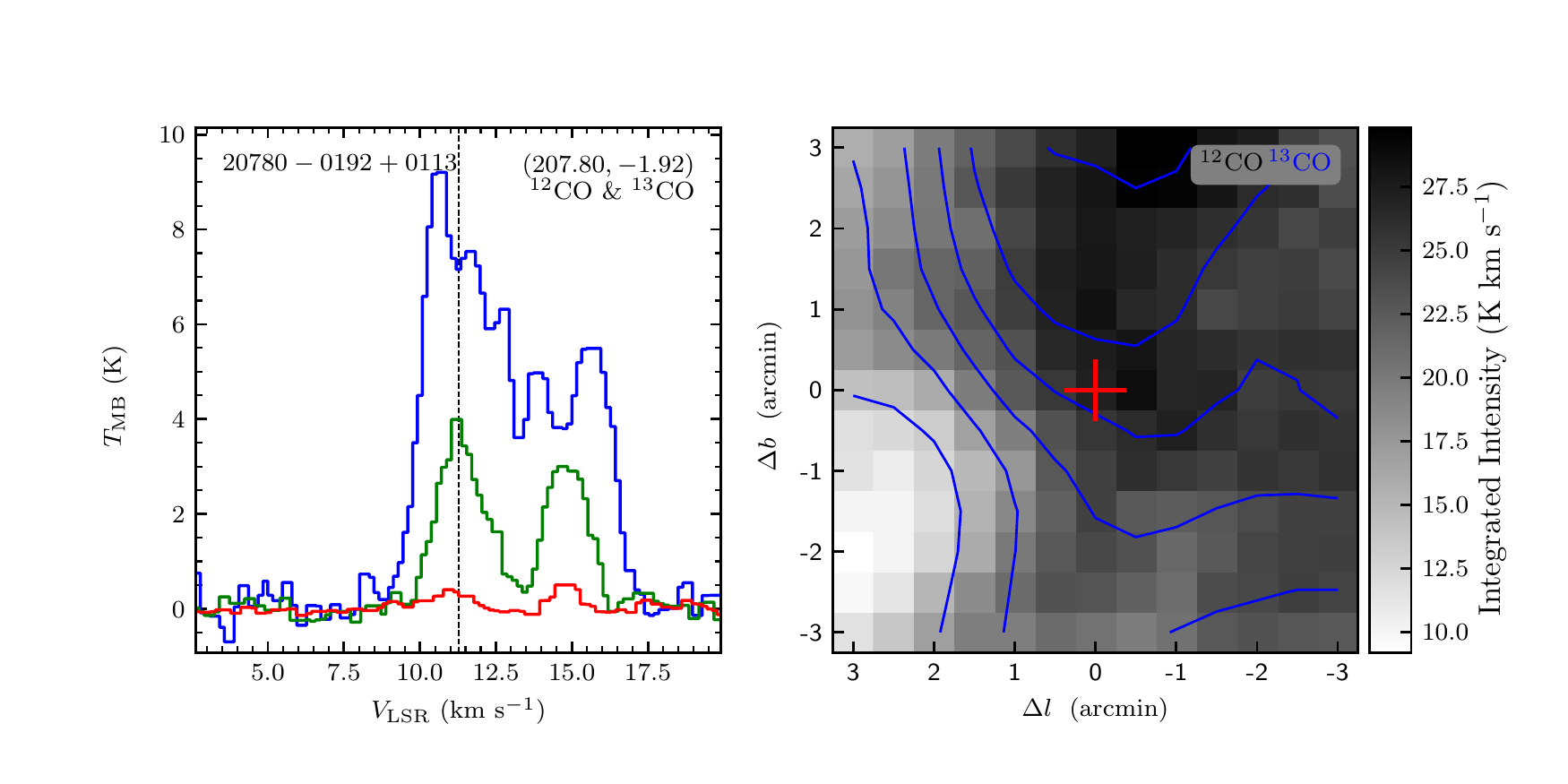}
\includegraphics[width=9.0cm,angle=0]{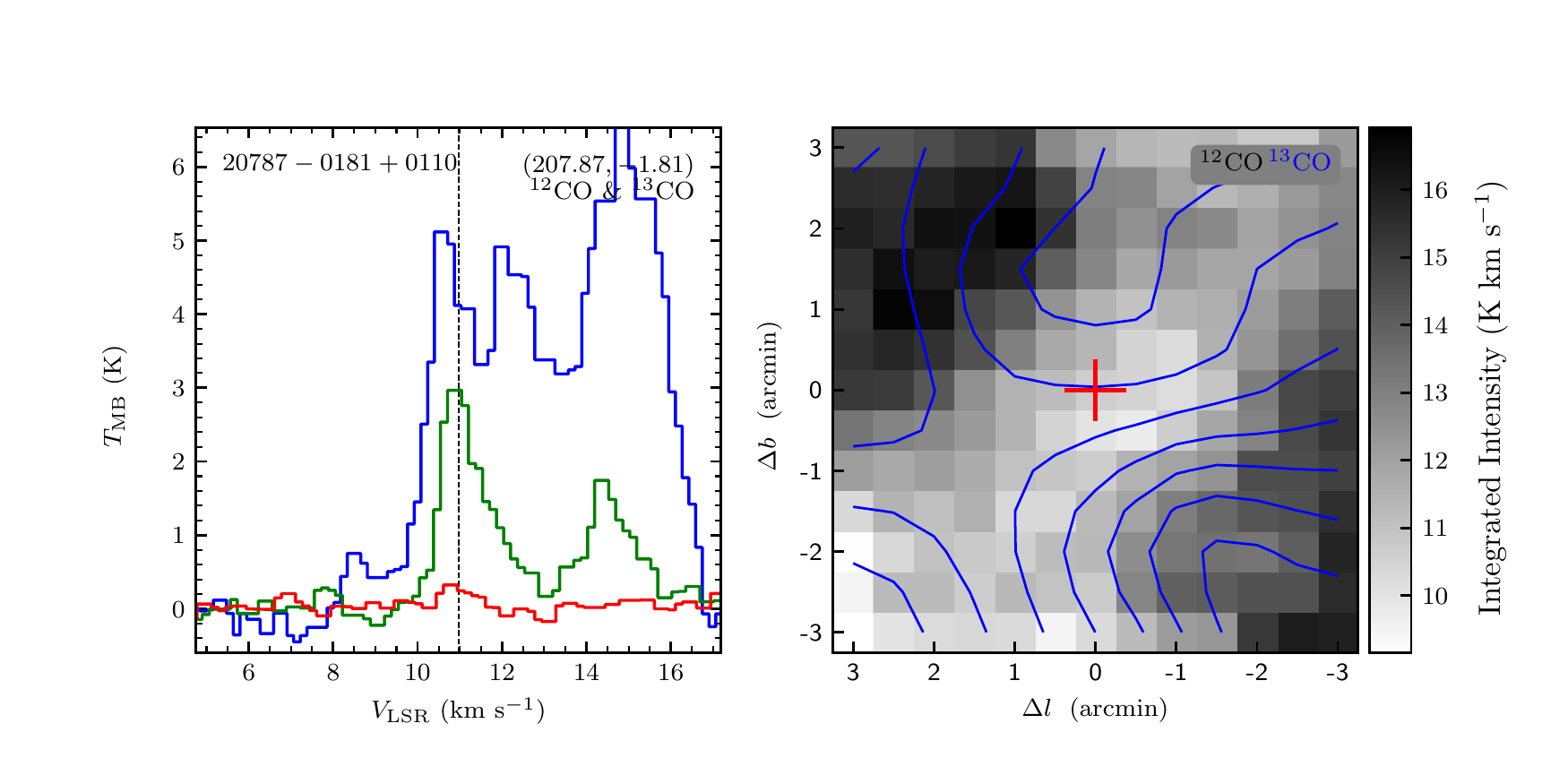}
\vspace{-0.5cm}

\includegraphics[width=9.0cm,angle=0]{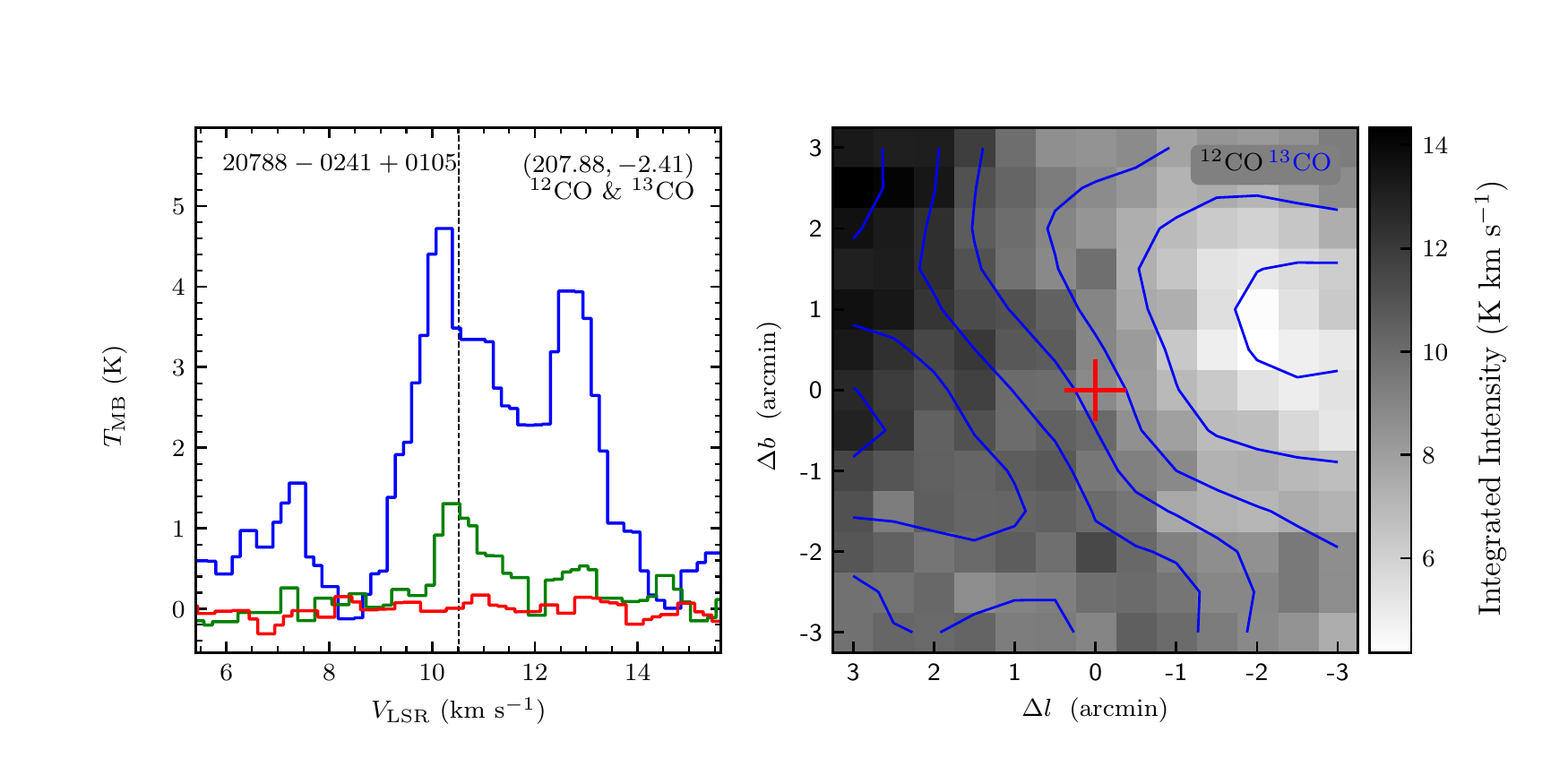}
\includegraphics[width=9.0cm,angle=0]{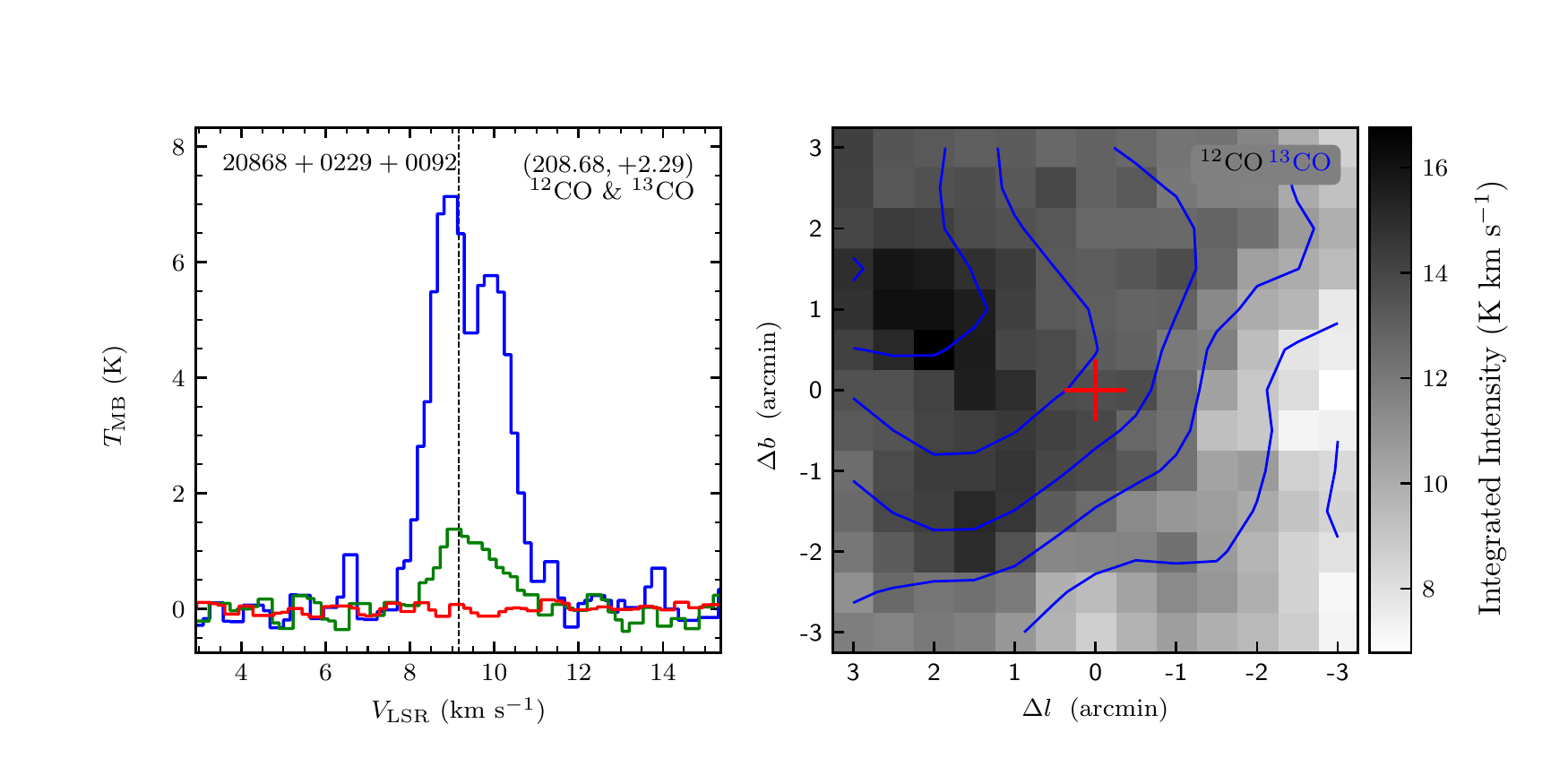}
\vspace{-0.5cm}

\includegraphics[width=9.0cm,angle=0]{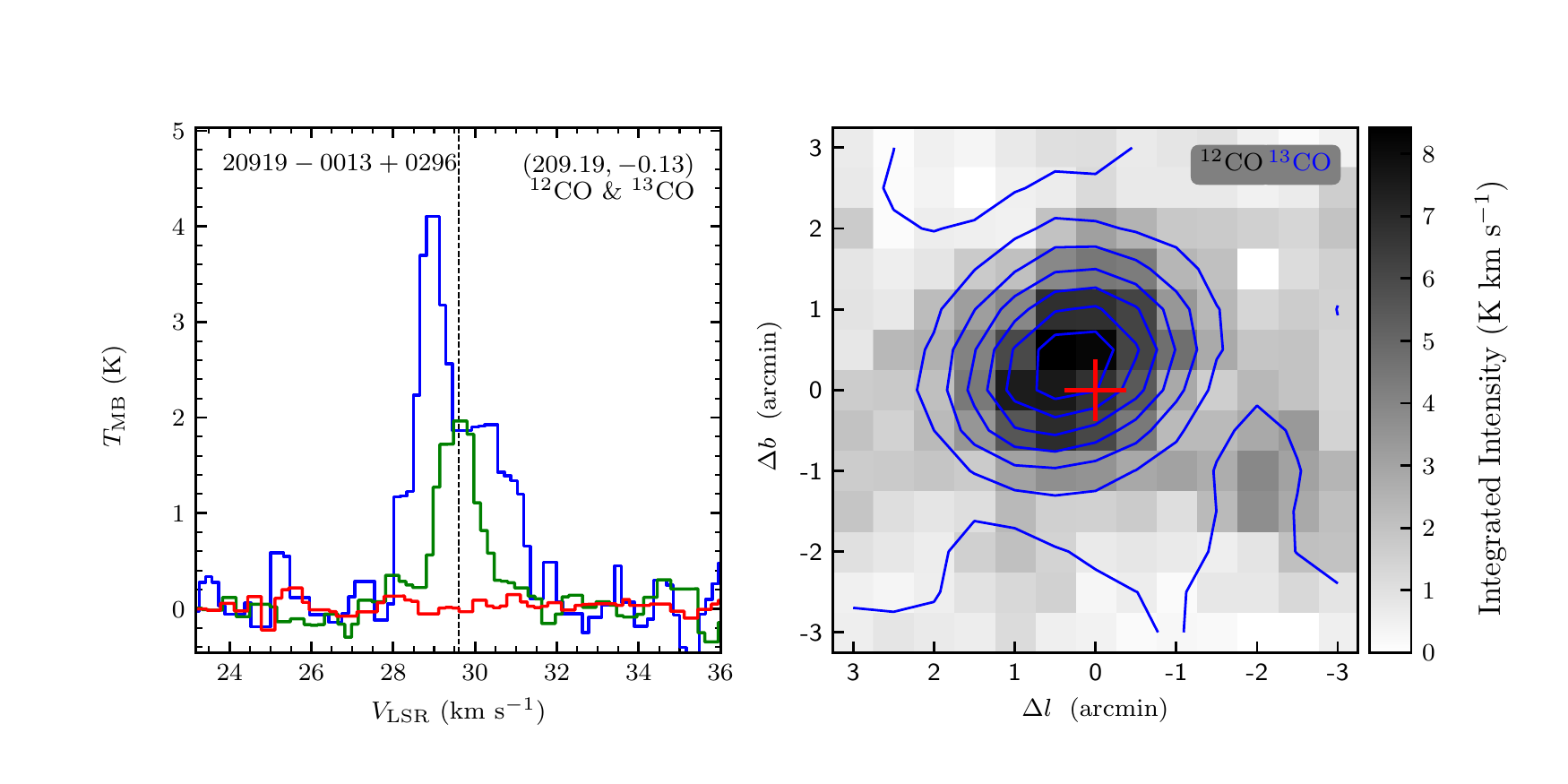}
\includegraphics[width=9.0cm,angle=0]{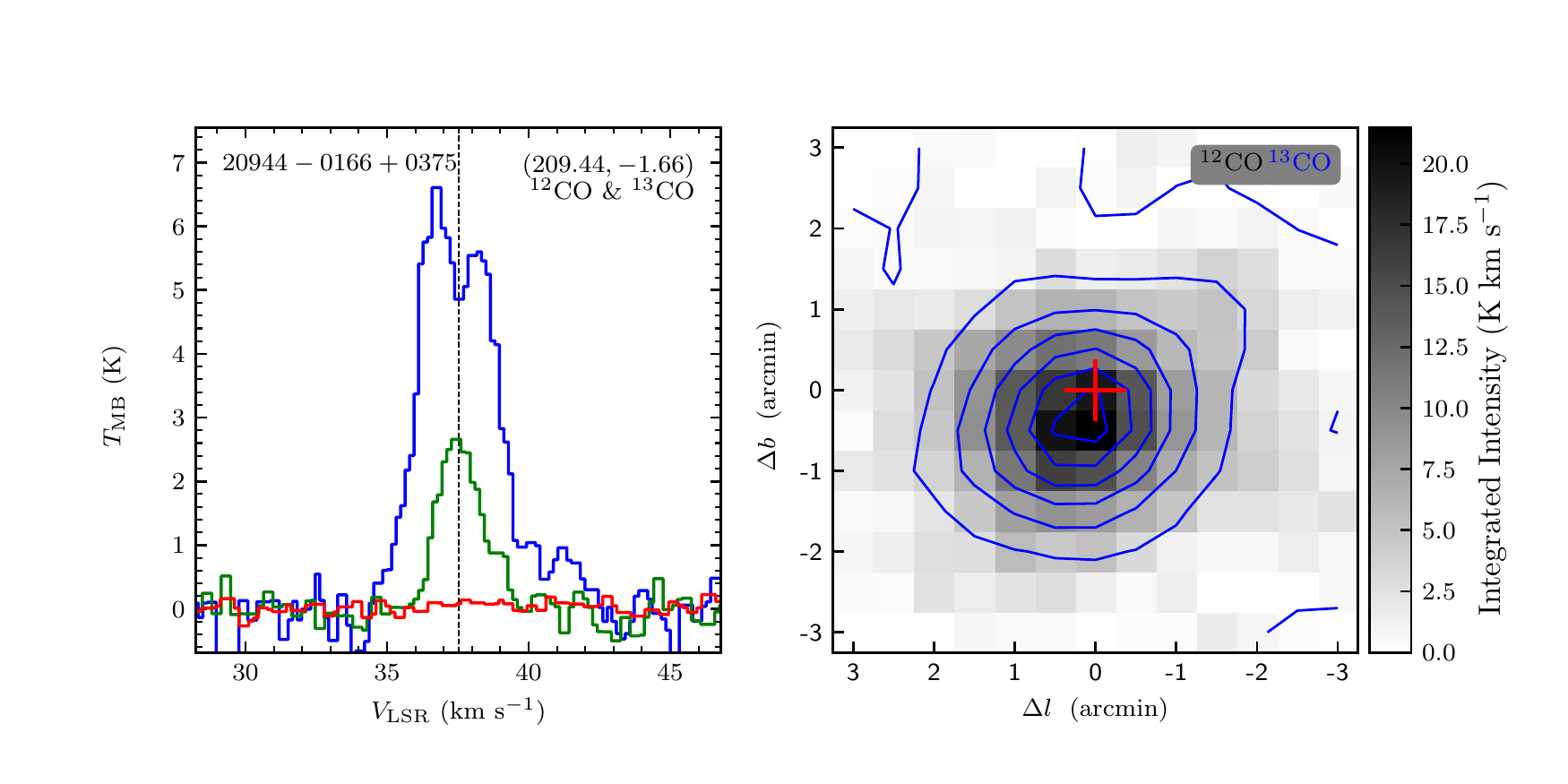}
\vspace{-0.5cm}

\includegraphics[width=9.0cm,angle=0]{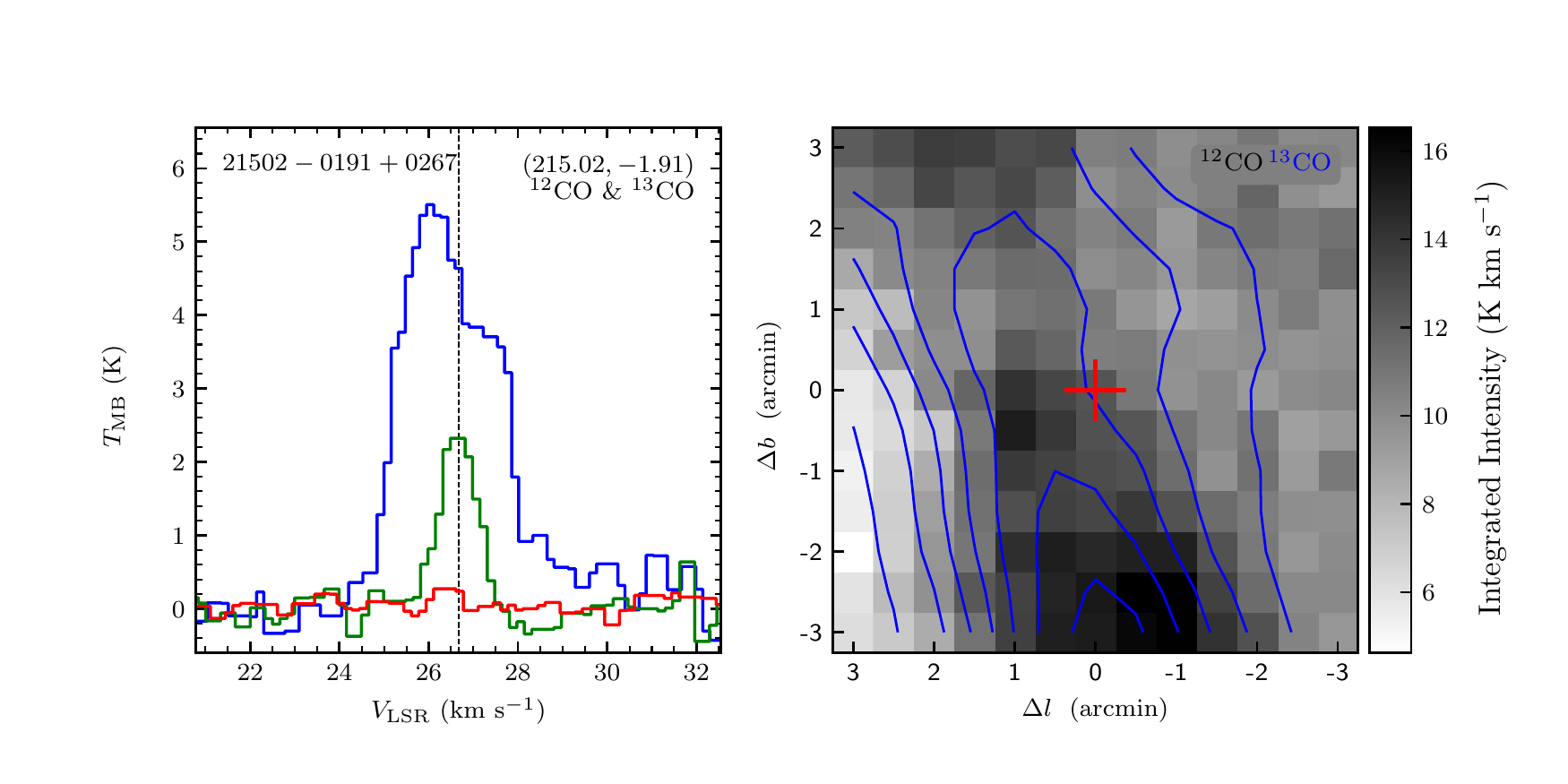}
\includegraphics[width=9.0cm,angle=0]{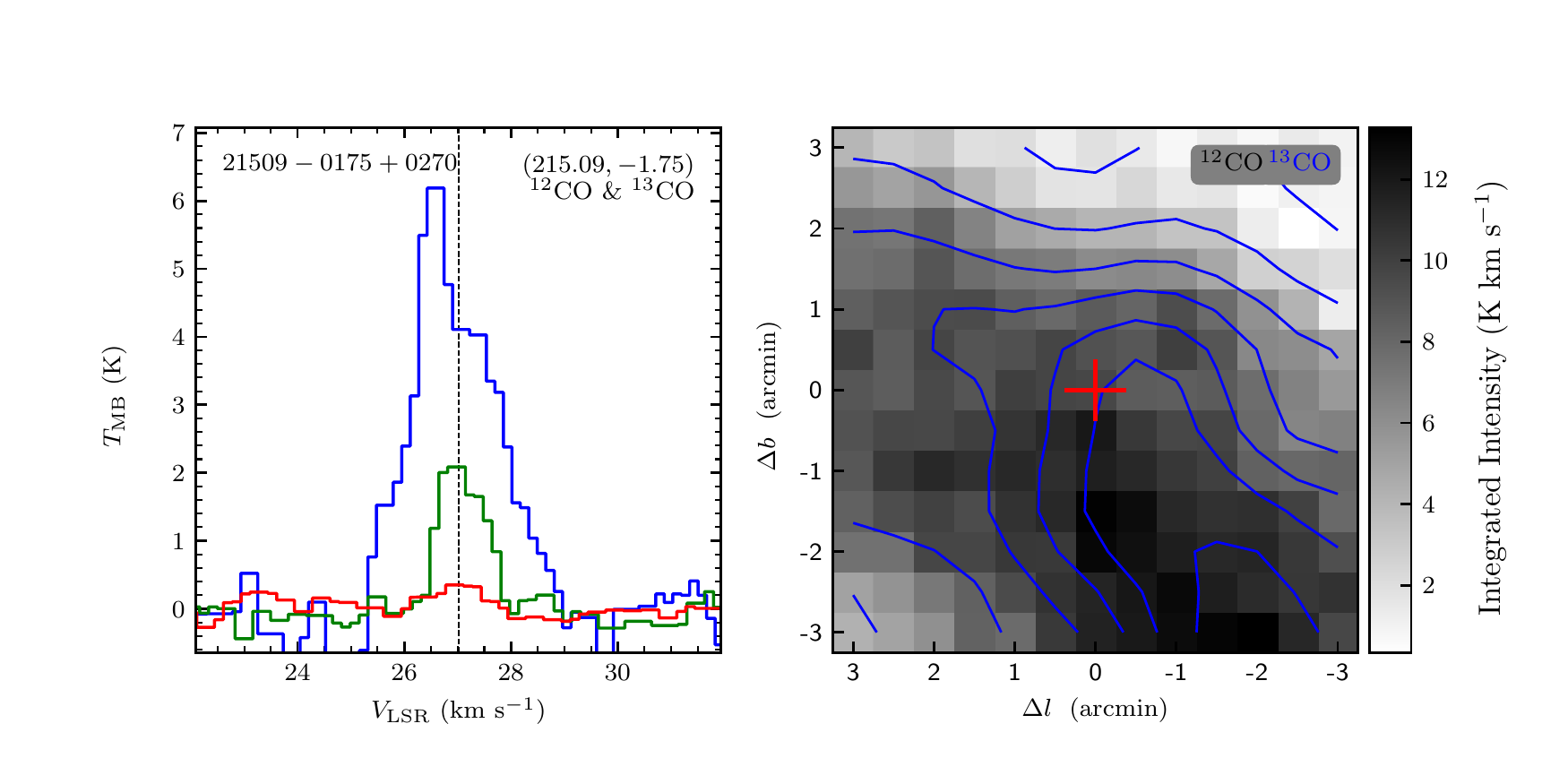}
\vspace{-0.5cm}

\includegraphics[width=9.0cm,angle=0]{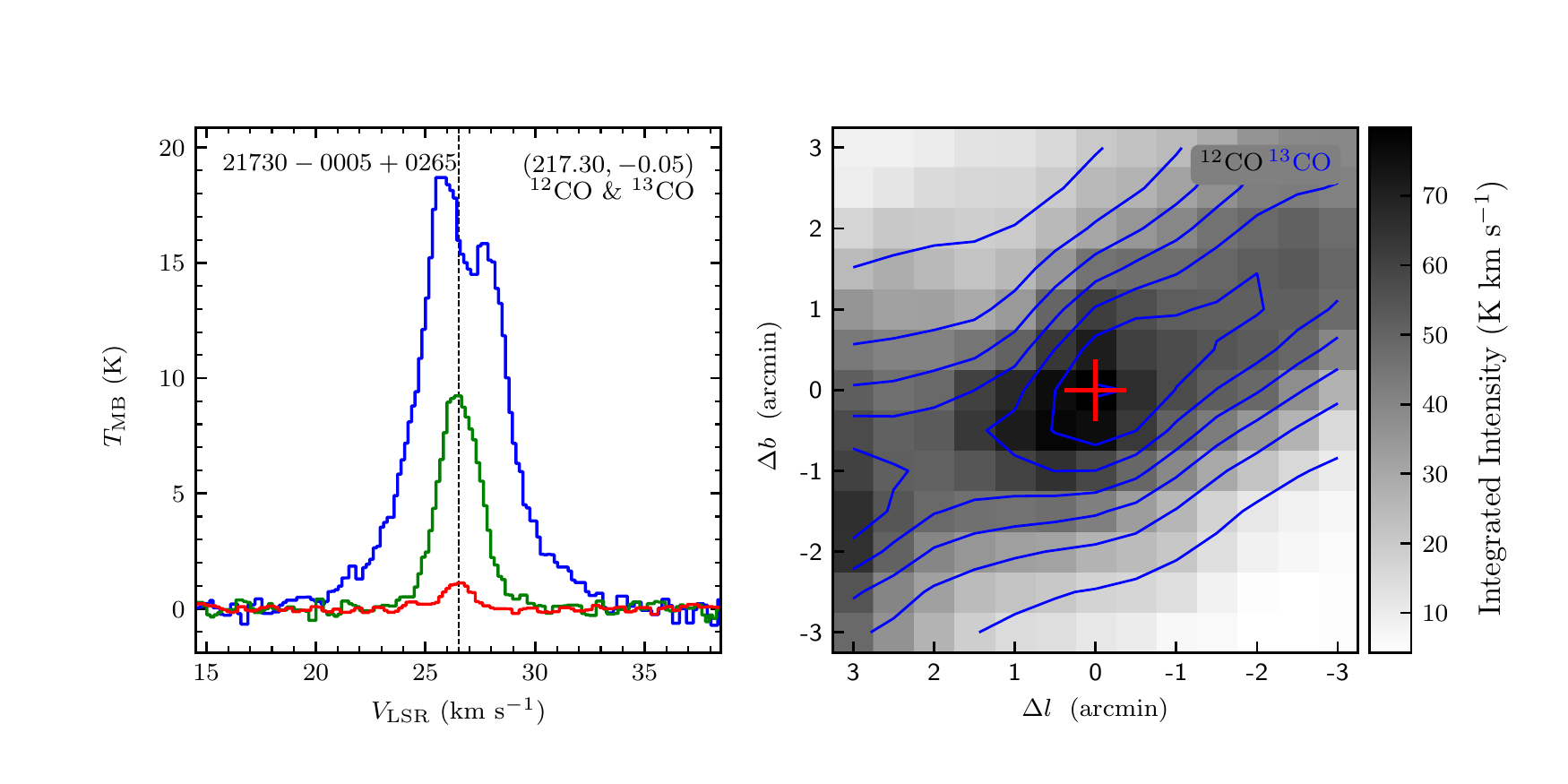}
\includegraphics[width=9.0cm,angle=0]{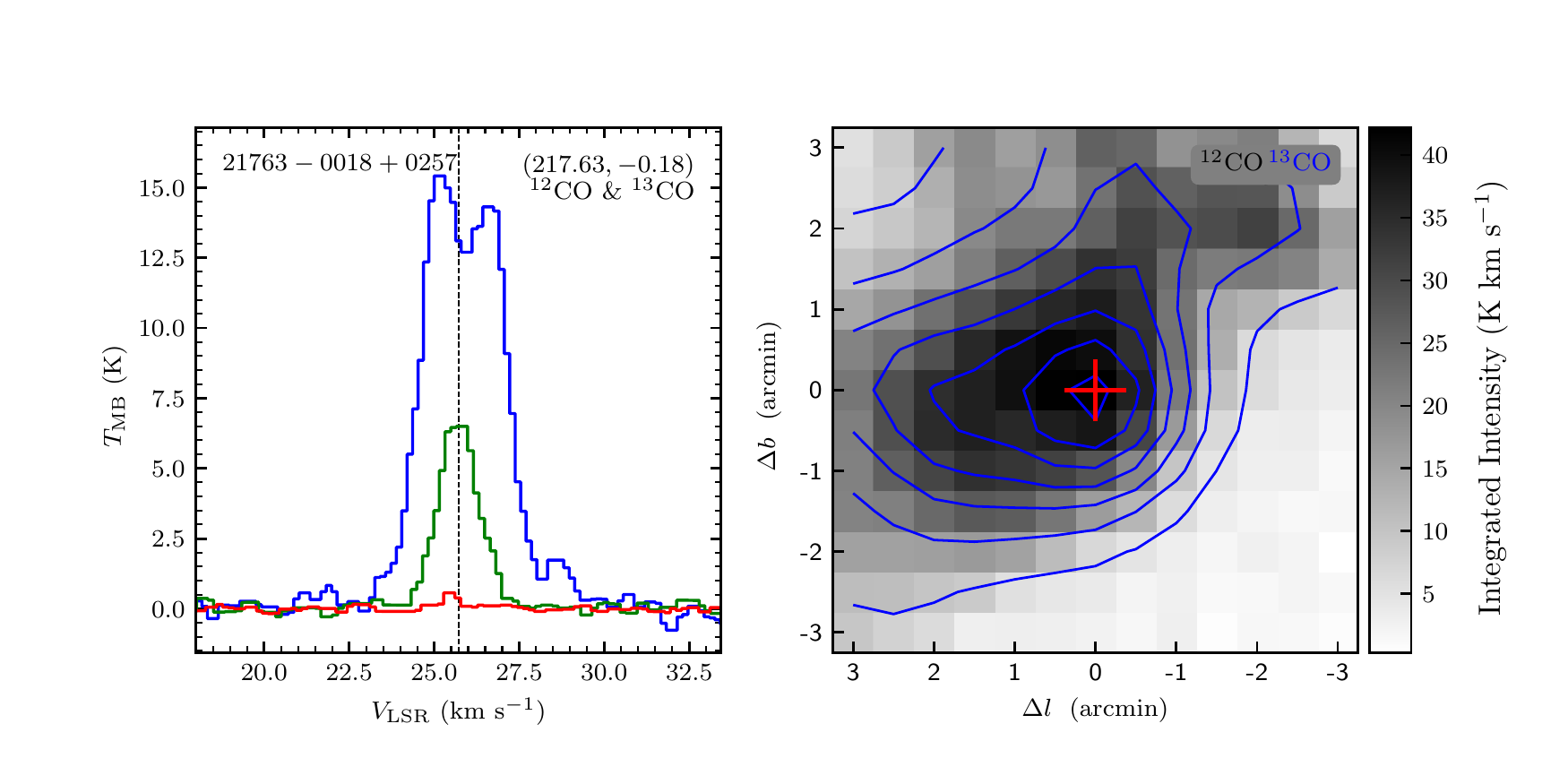}
\end{figure}
\clearpage

\begin{figure}
\includegraphics[width=9.0cm,angle=0]{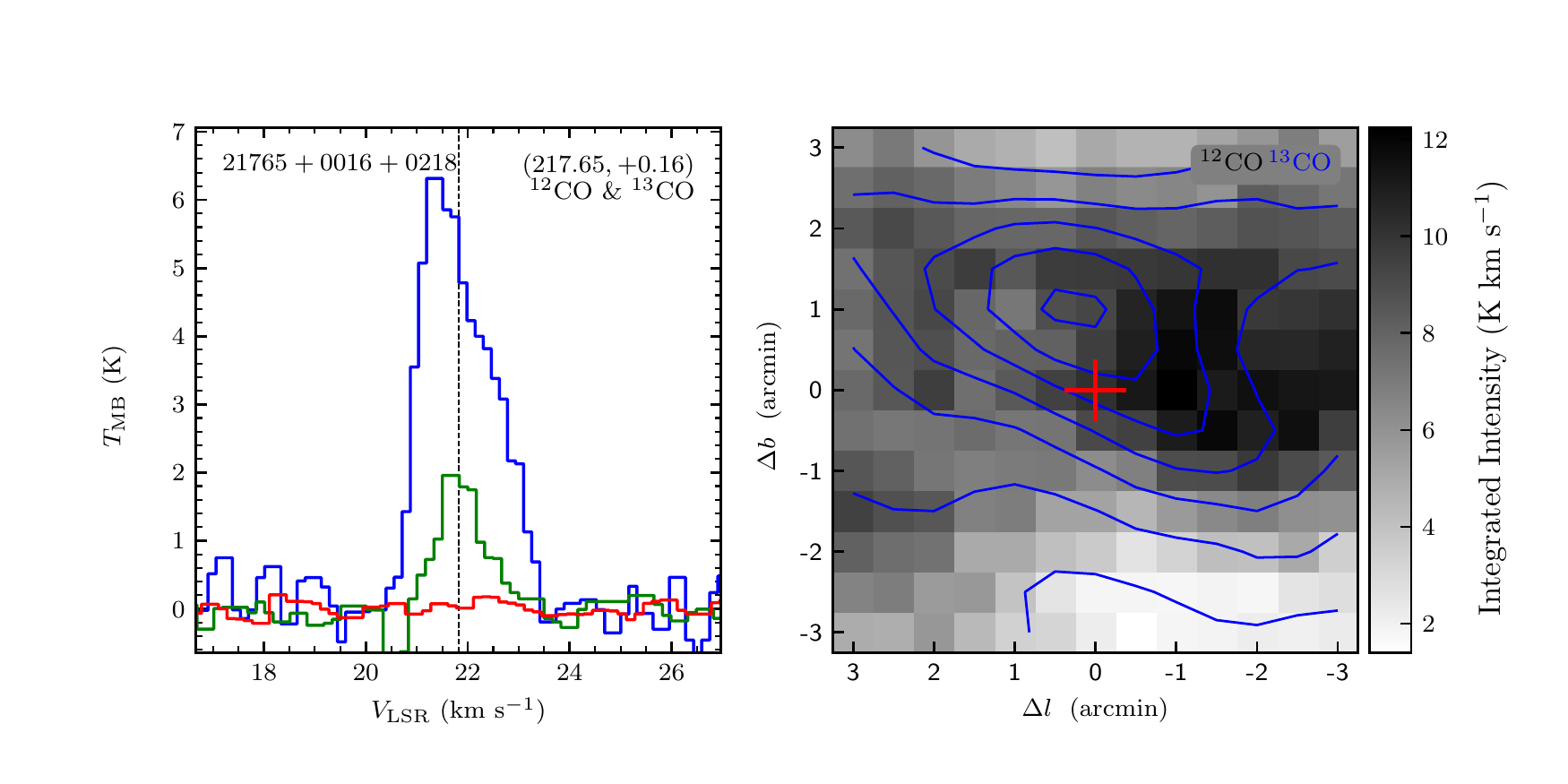}
\includegraphics[width=9.0cm,angle=0]{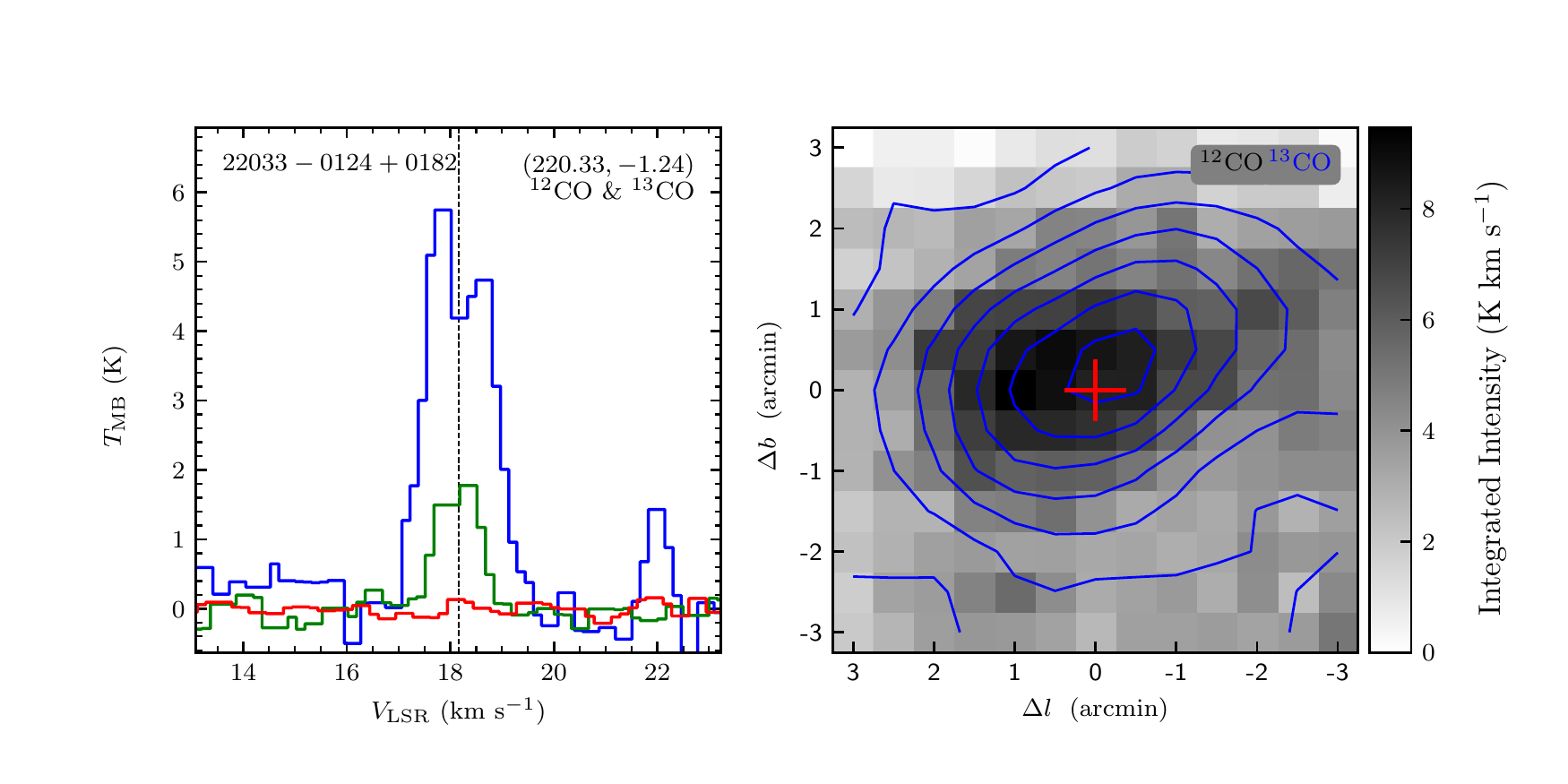}
\vspace{-0.5cm}

\includegraphics[width=9.0cm,angle=0]{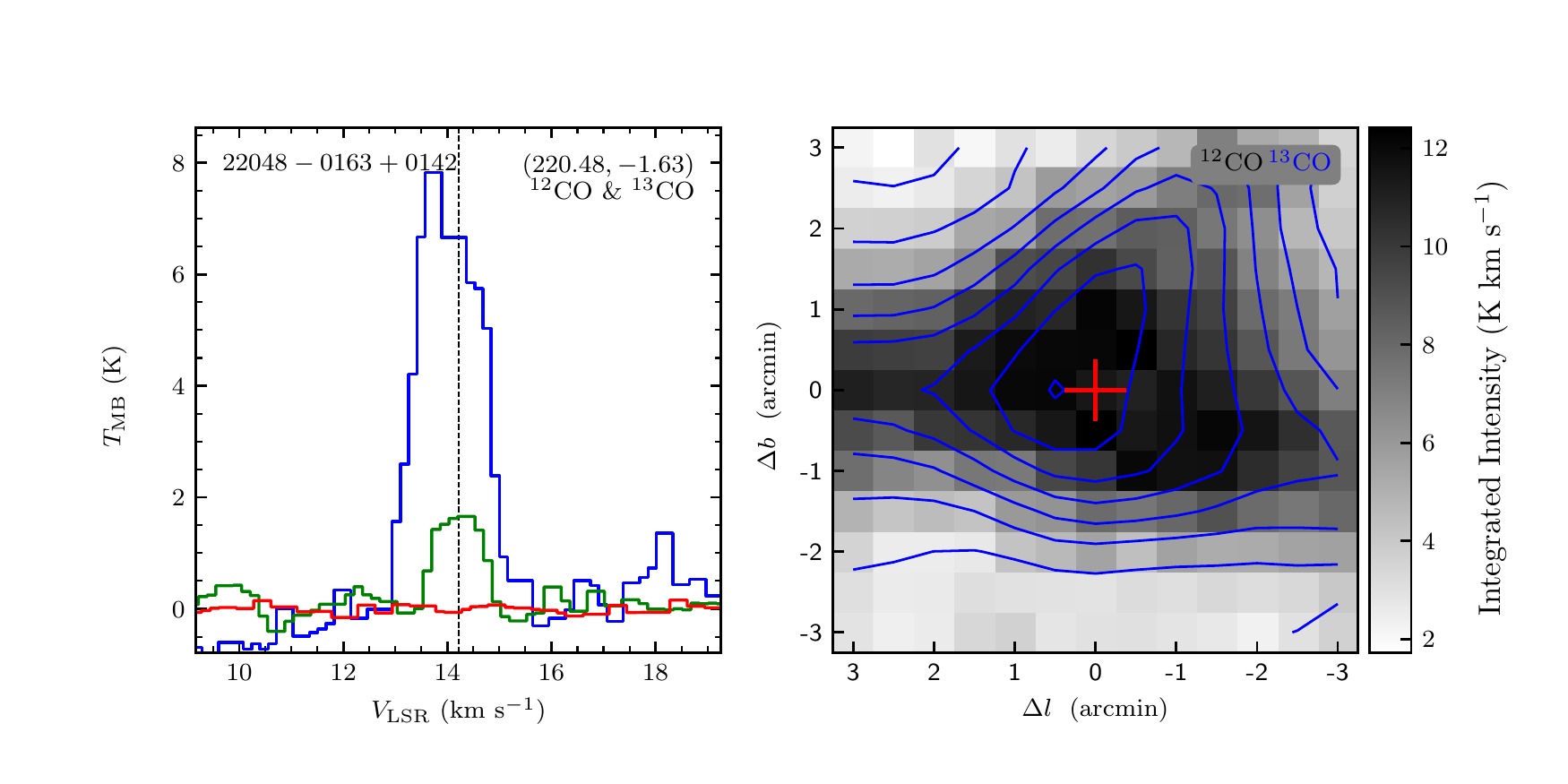}
\includegraphics[width=9.0cm,angle=0]{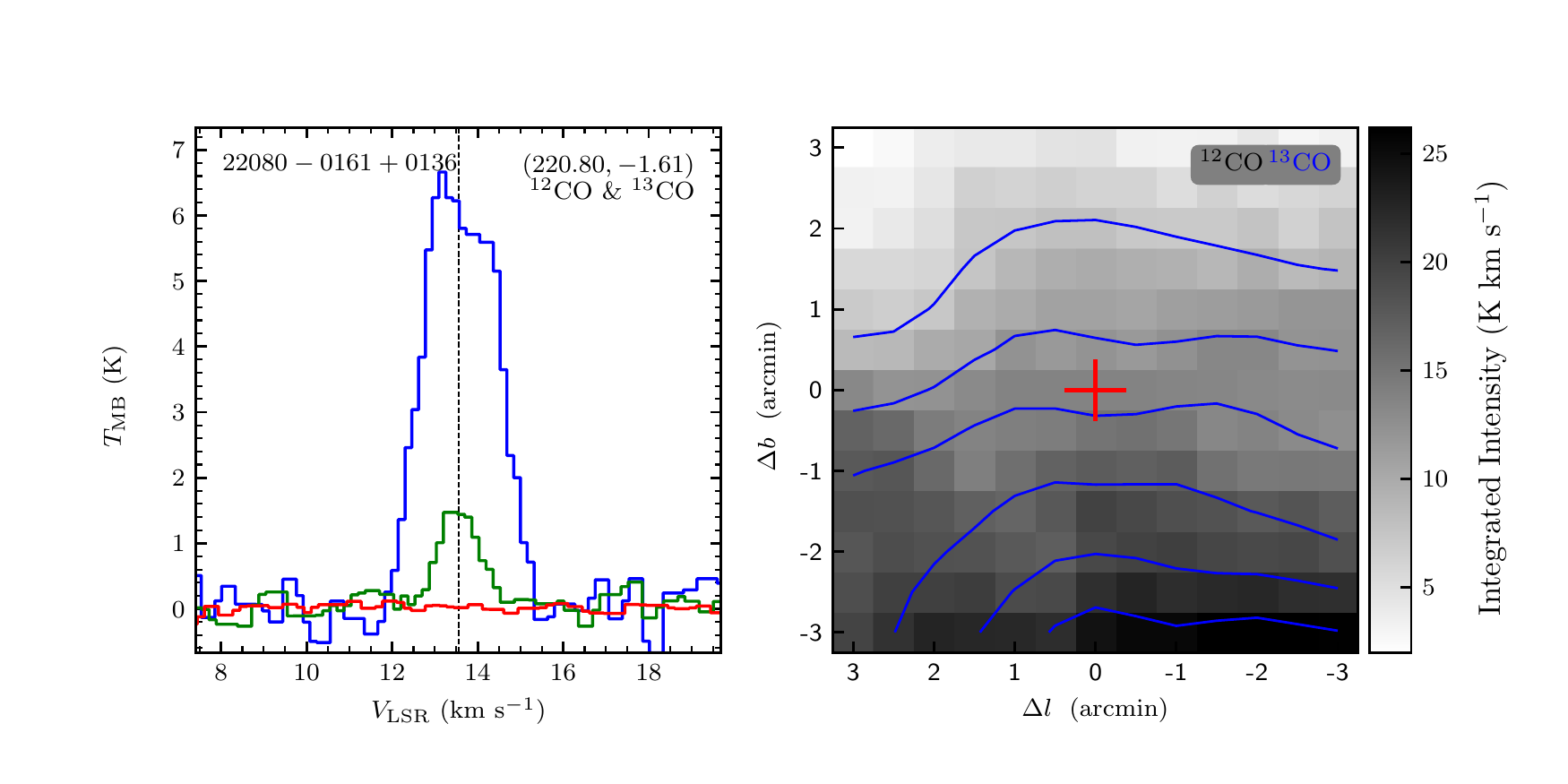}
\vspace{-0.5cm}

\includegraphics[width=9.0cm,angle=0]{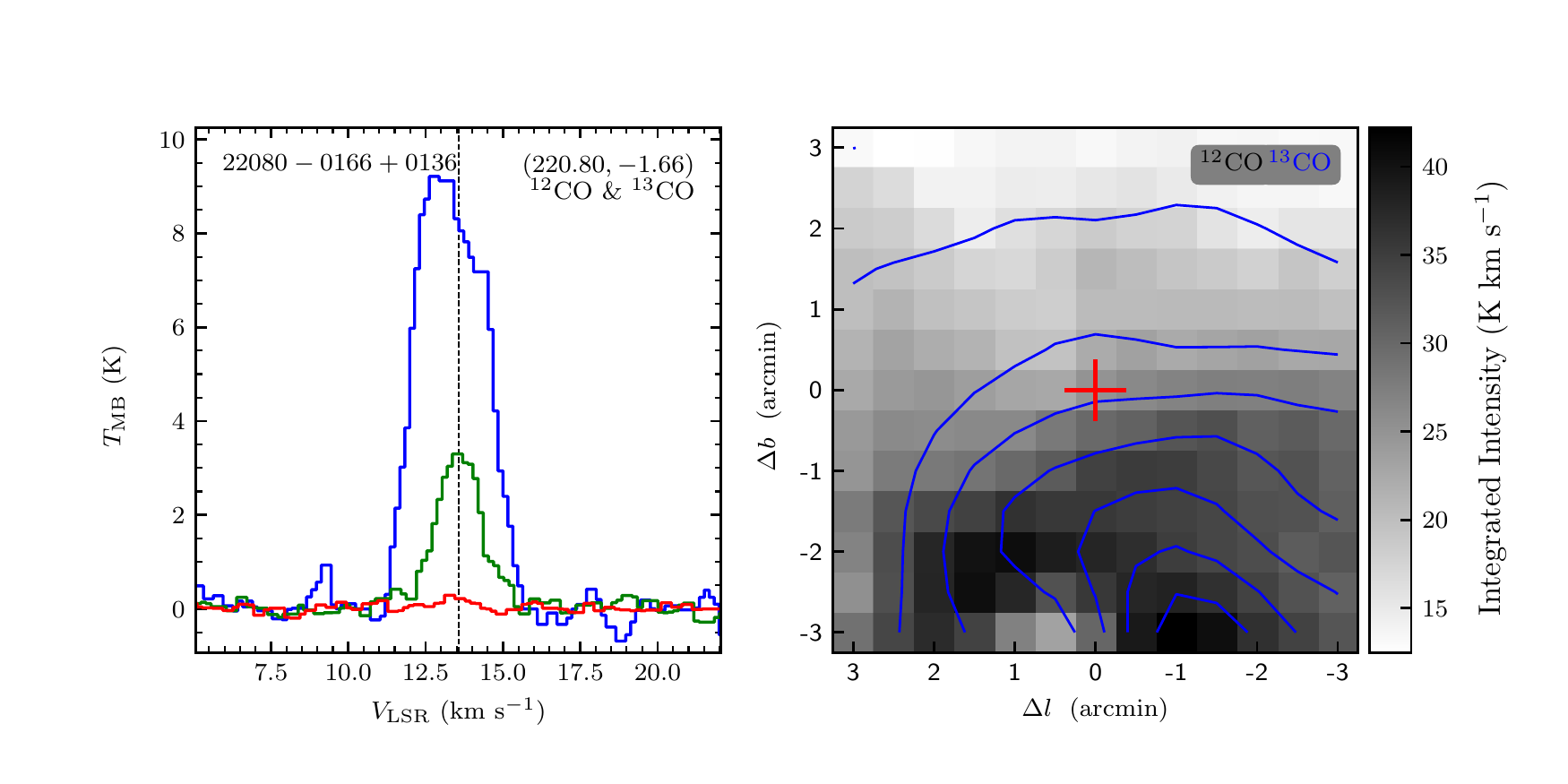}
\includegraphics[width=9.0cm,angle=0]{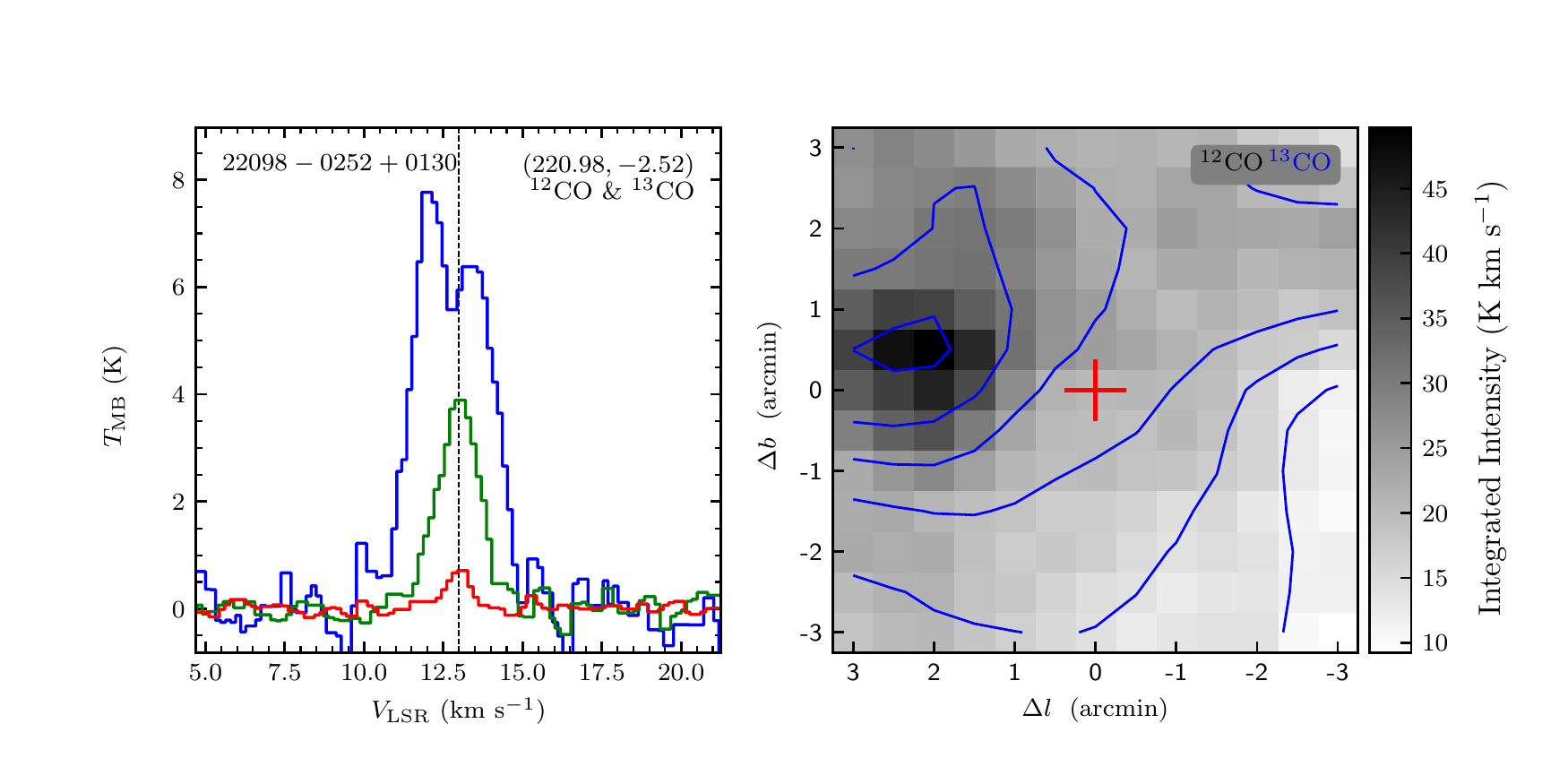}
\vspace{-0.5cm}

\includegraphics[width=9.0cm,angle=0]{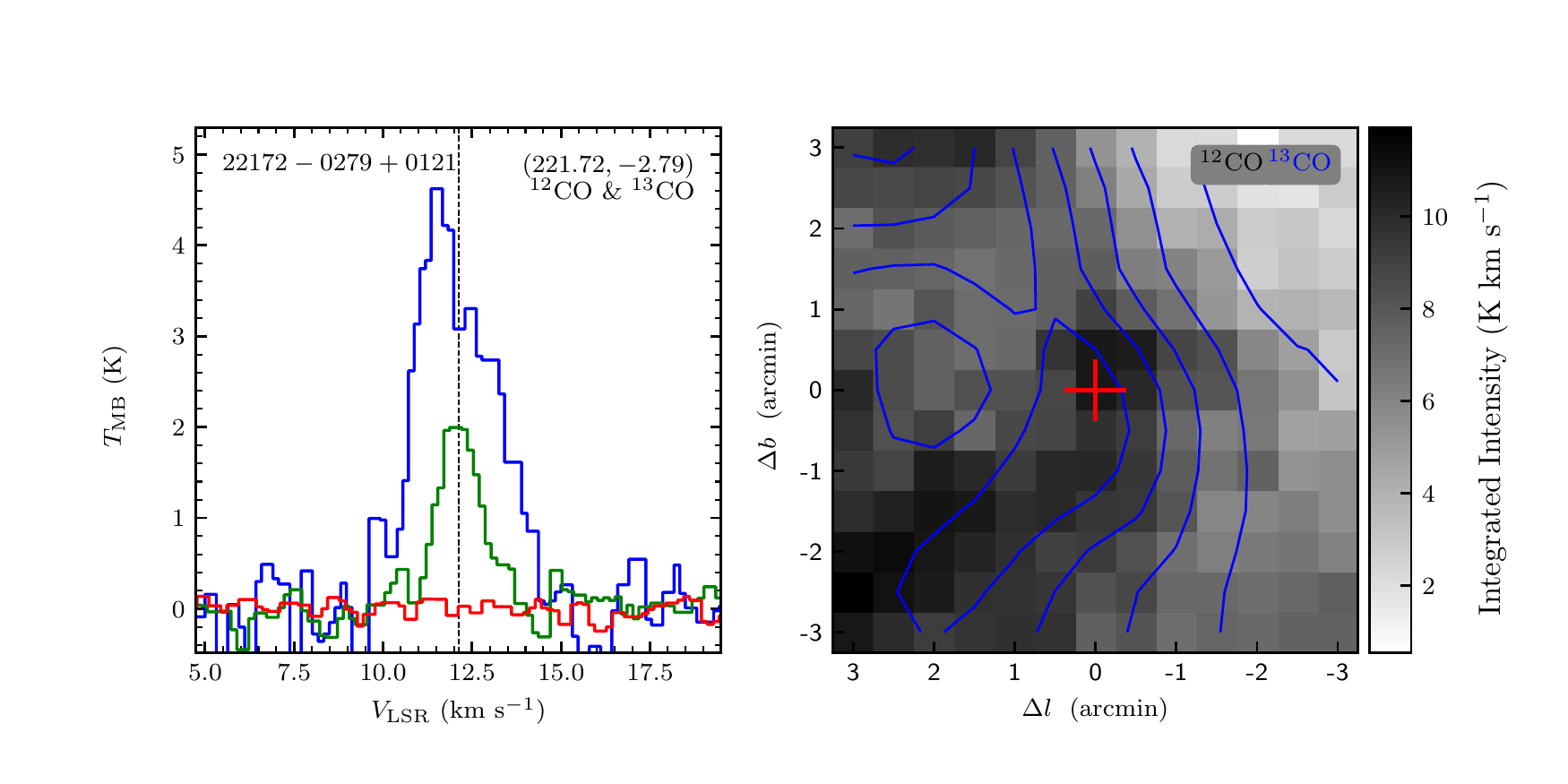}
\includegraphics[width=9.0cm,angle=0]{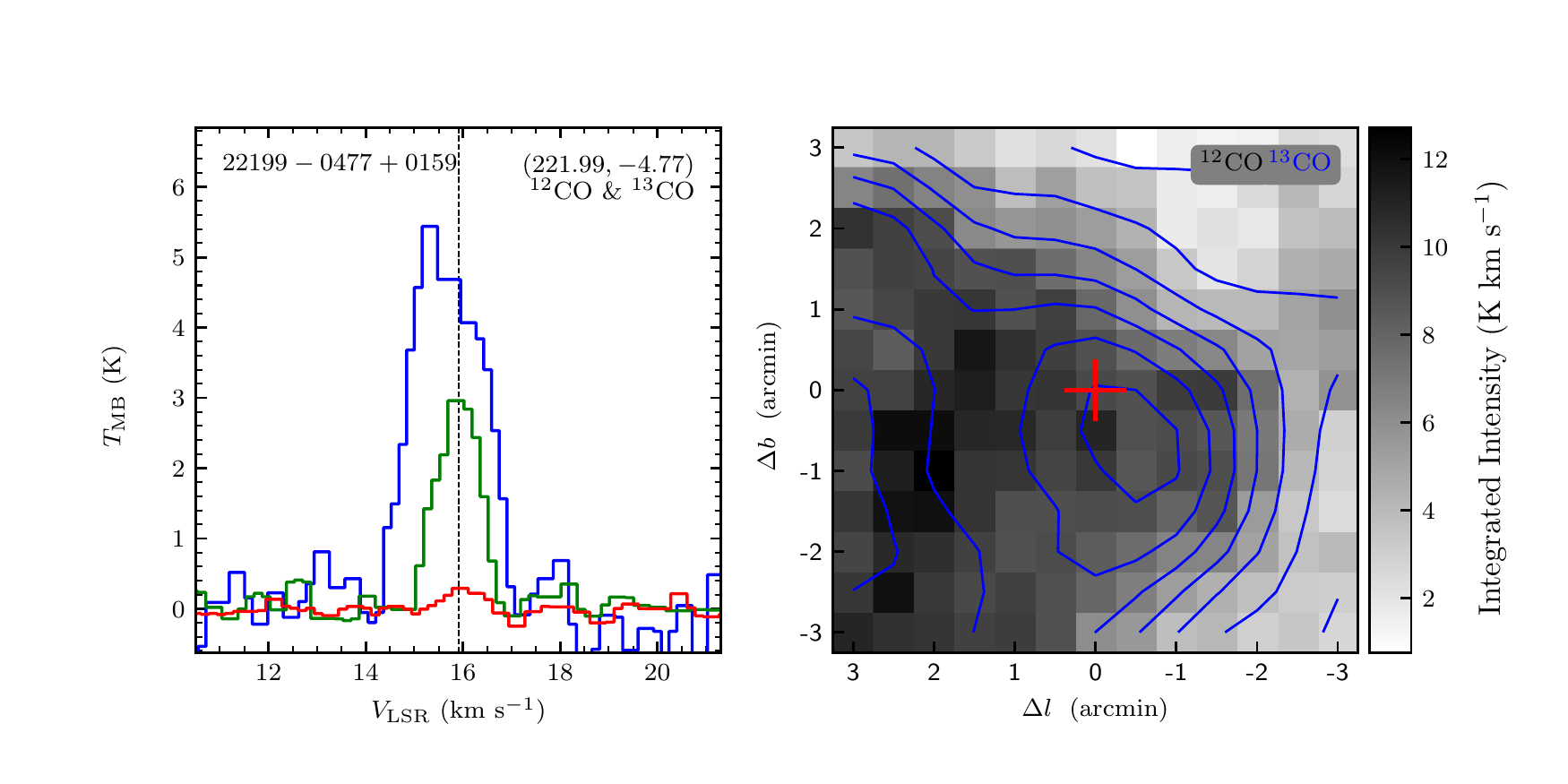}
\vspace{-0.5cm}

\includegraphics[width=9.0cm,angle=0]{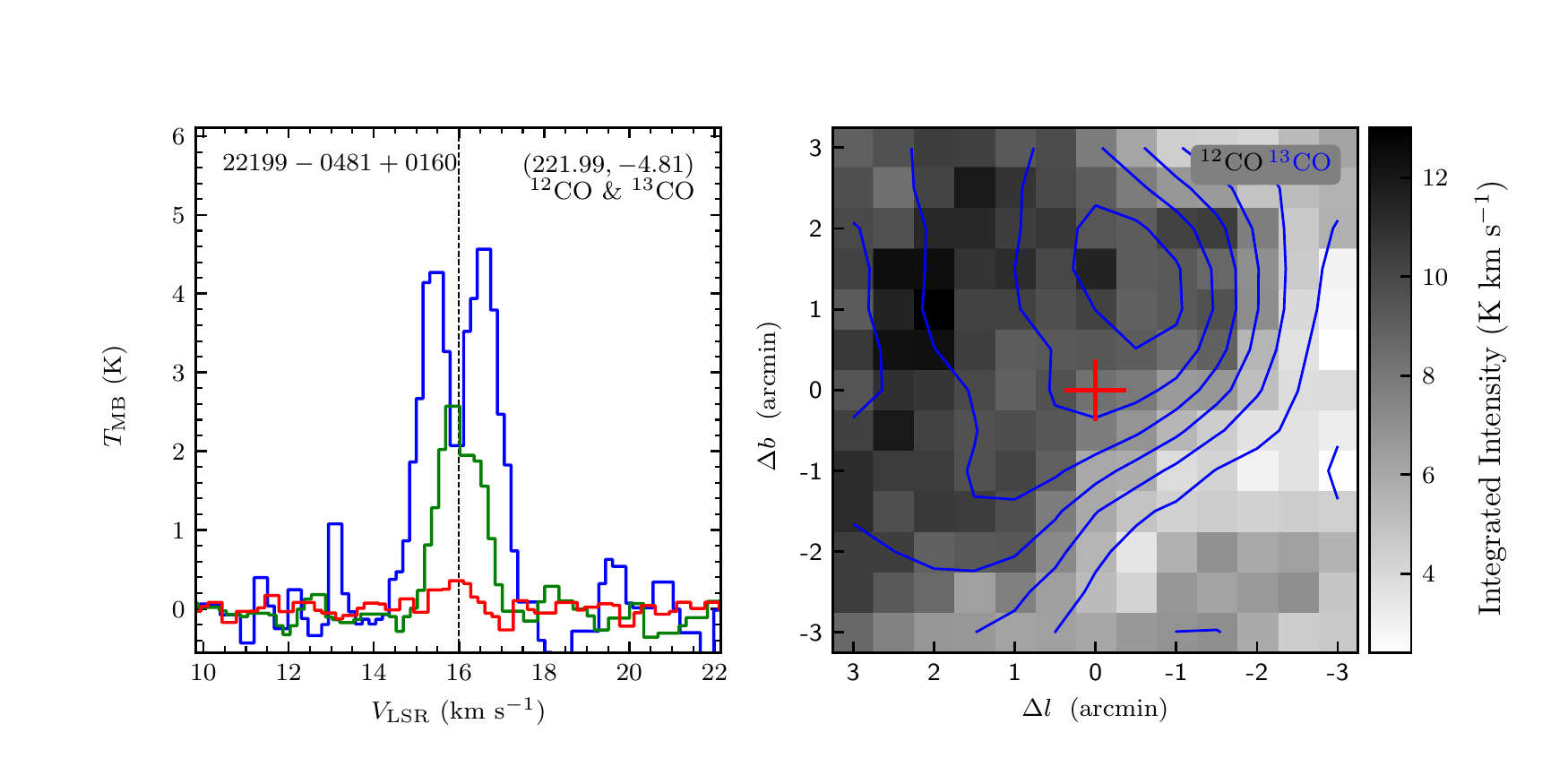}
\includegraphics[width=9.0cm,angle=0]{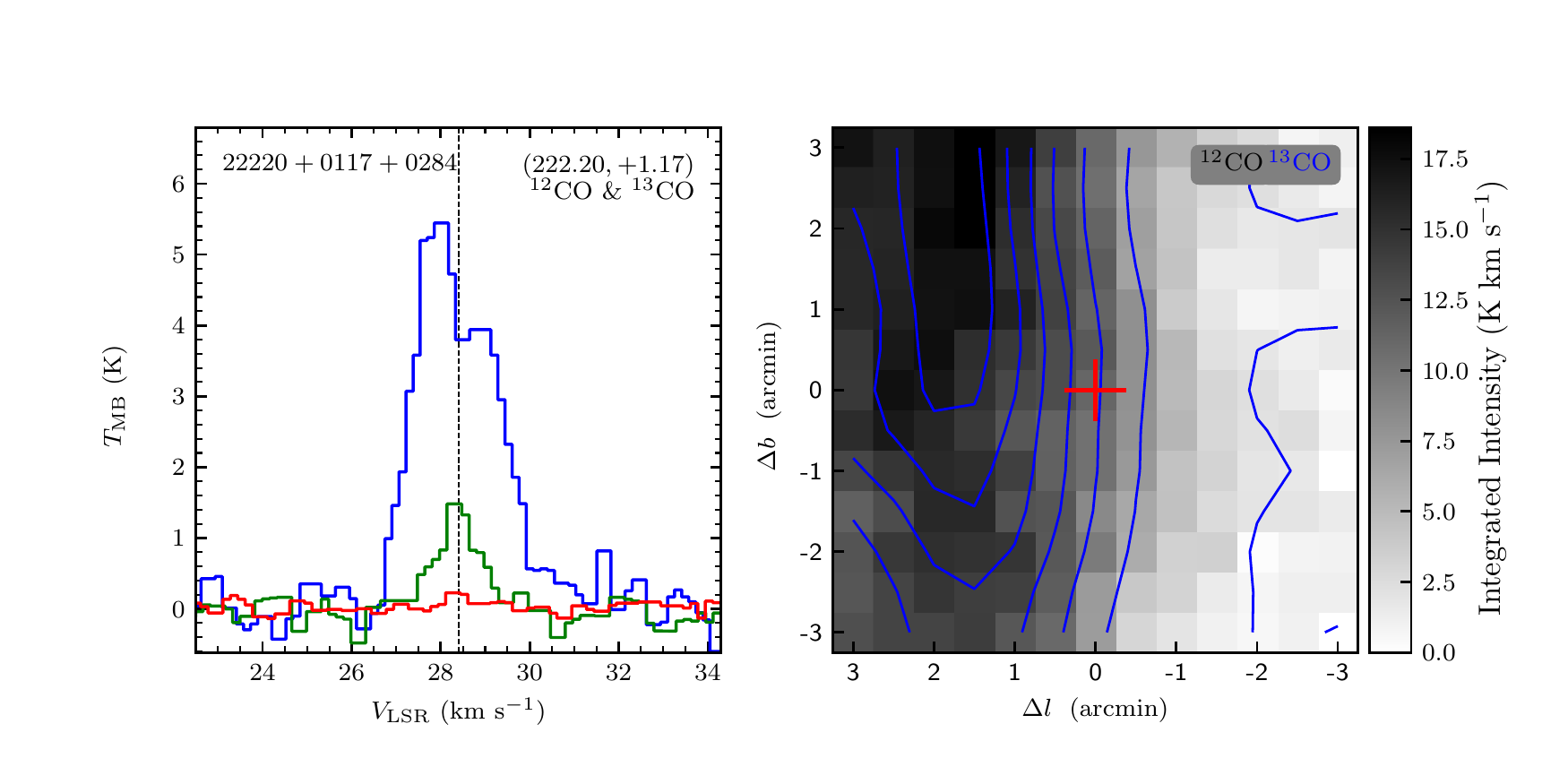}
\end{figure}
\clearpage

\begin{figure}
\includegraphics[width=9.0cm,angle=0]{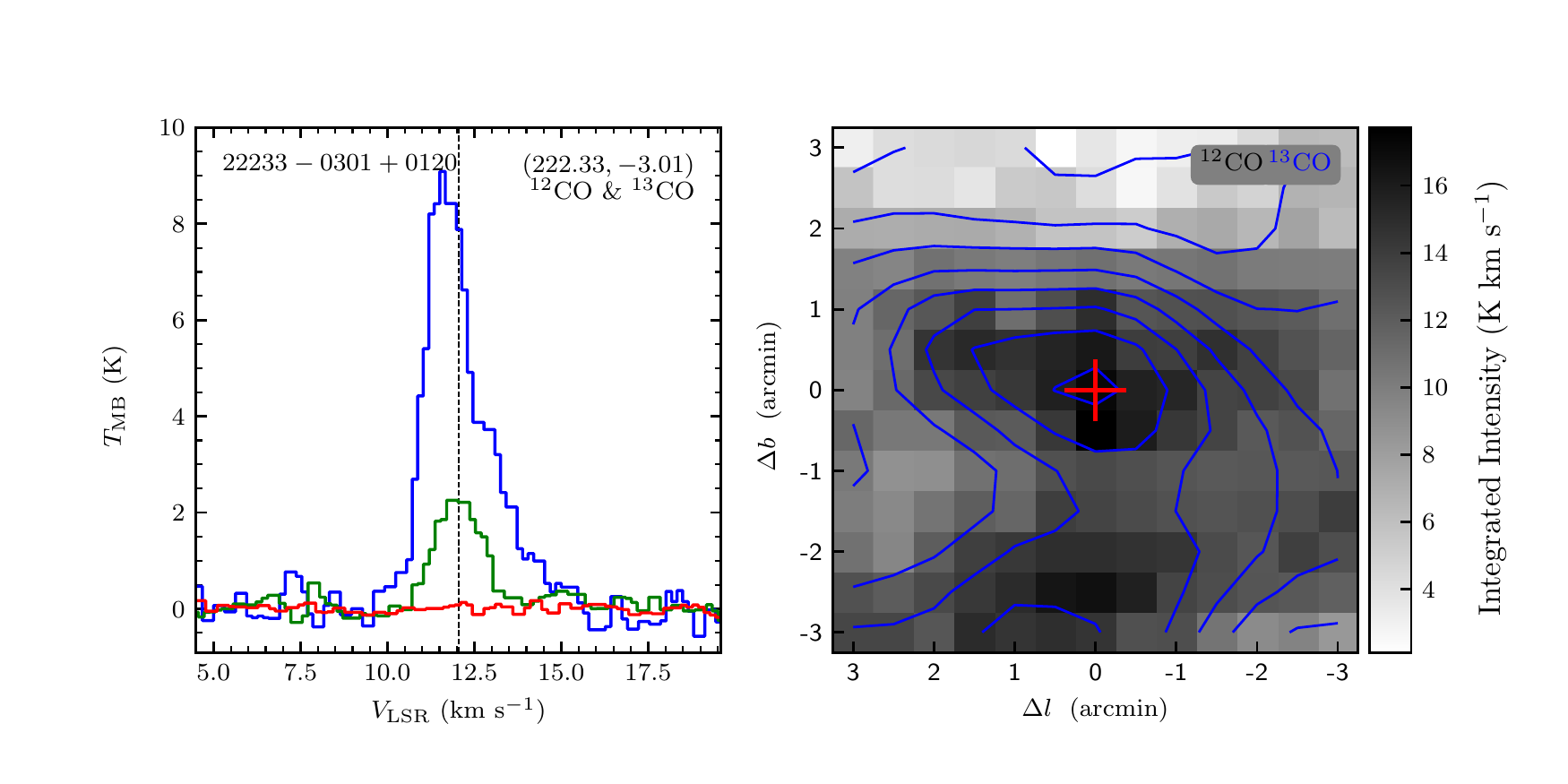}
\includegraphics[width=9.0cm,angle=0]{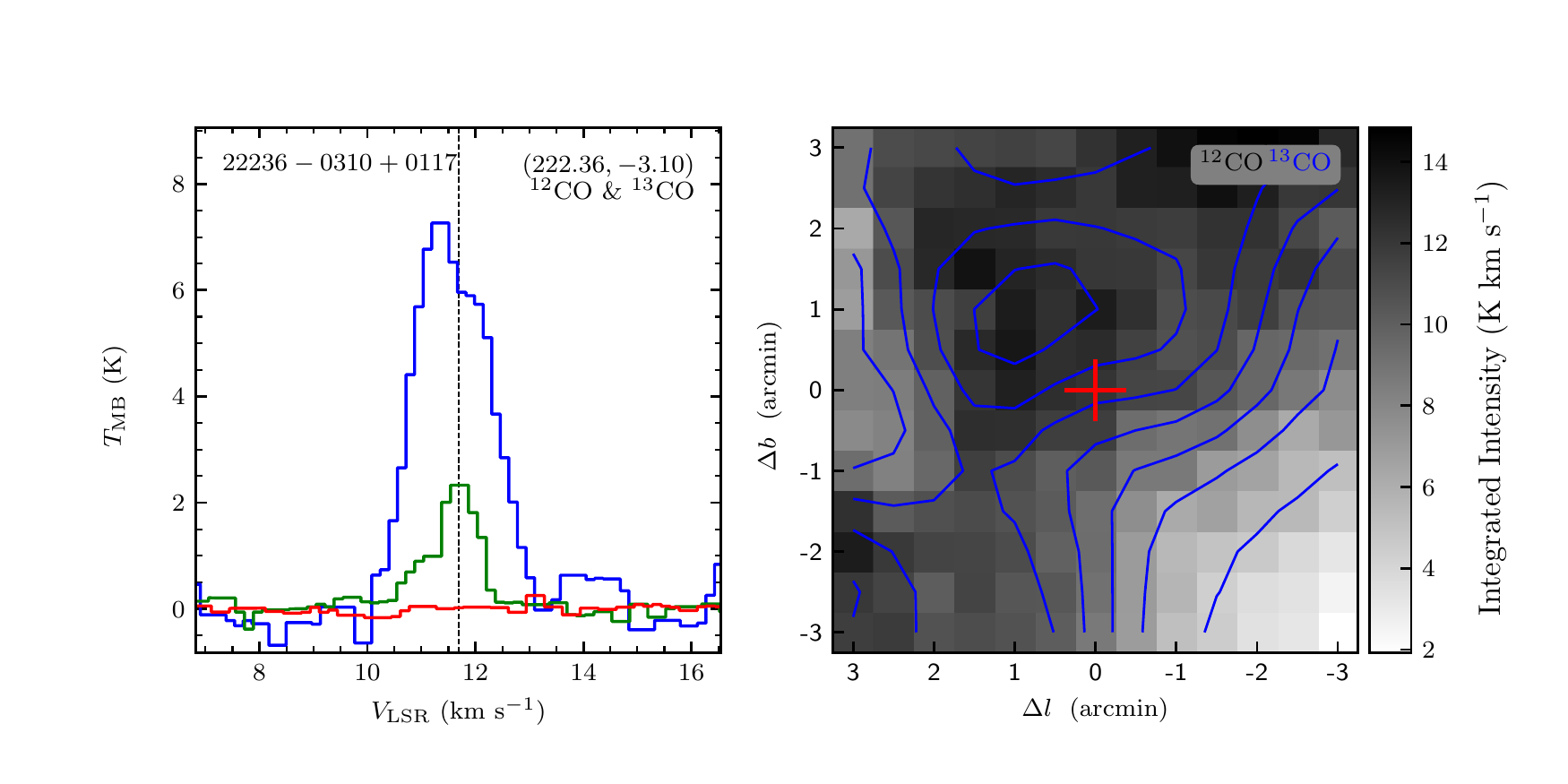}
\vspace{-0.5cm}

\includegraphics[width=9.0cm,angle=0]{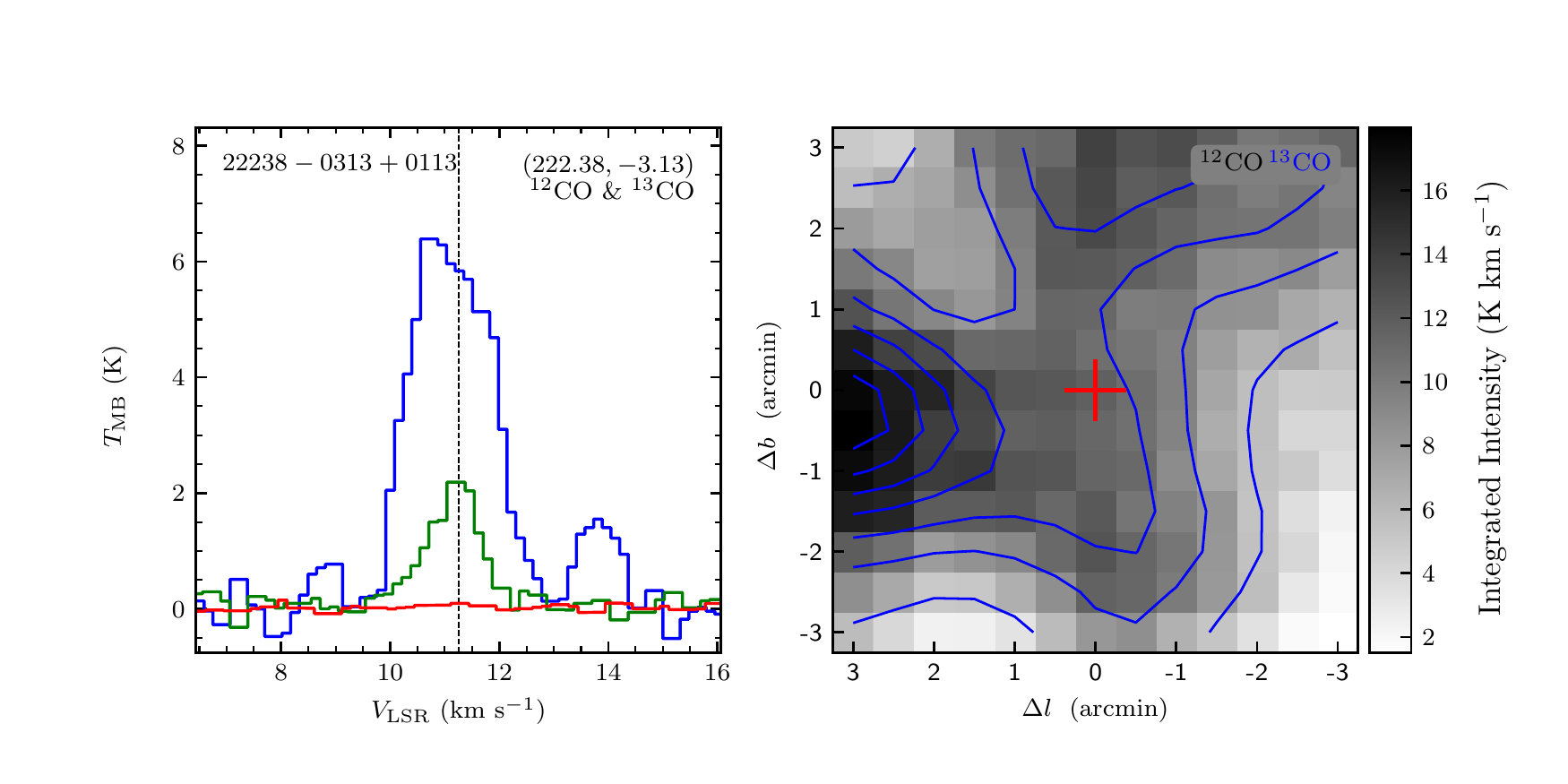}
\includegraphics[width=9.0cm,angle=0]{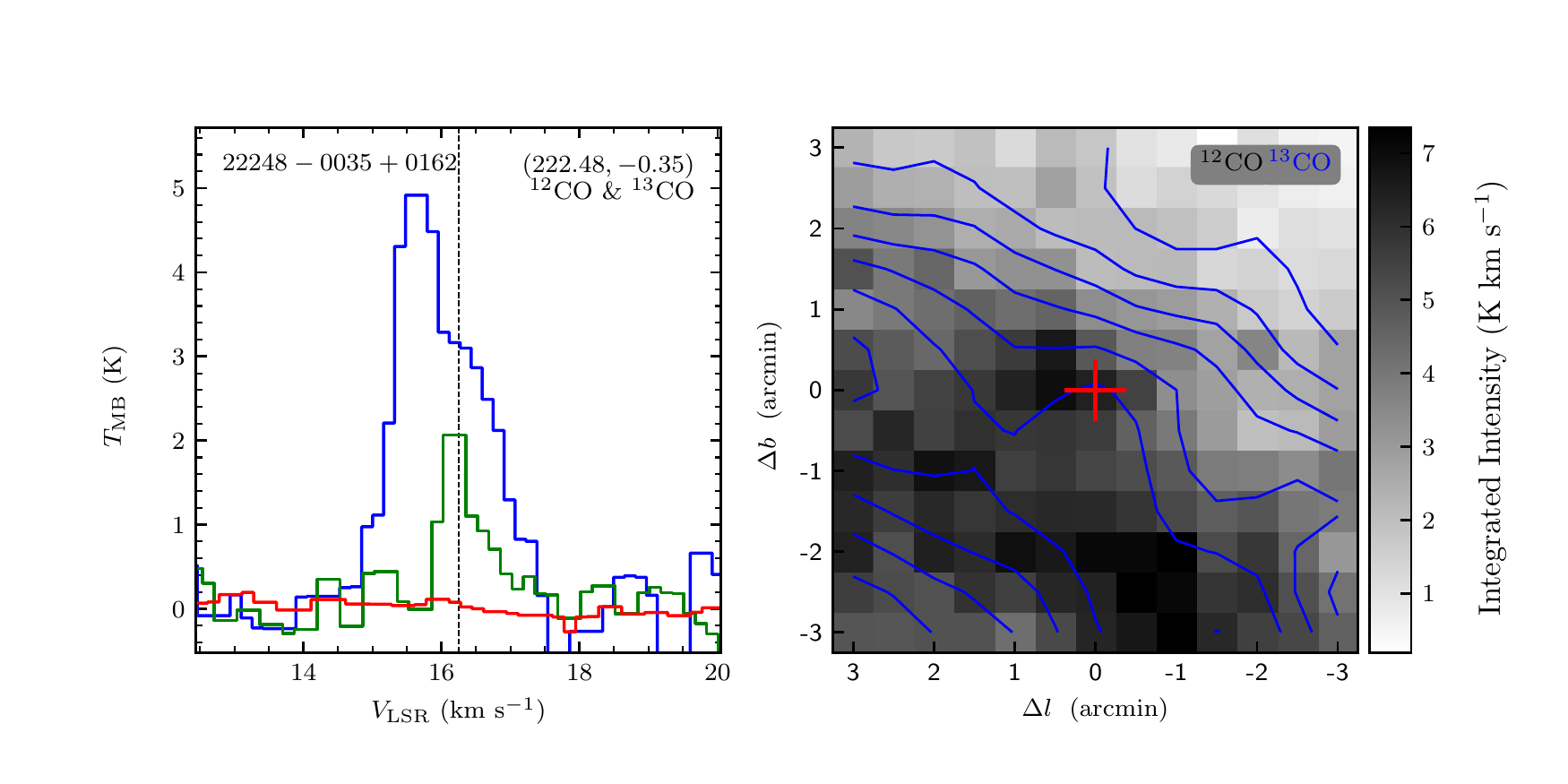}
\vspace{-0.5cm}

\includegraphics[width=9.0cm,angle=0]{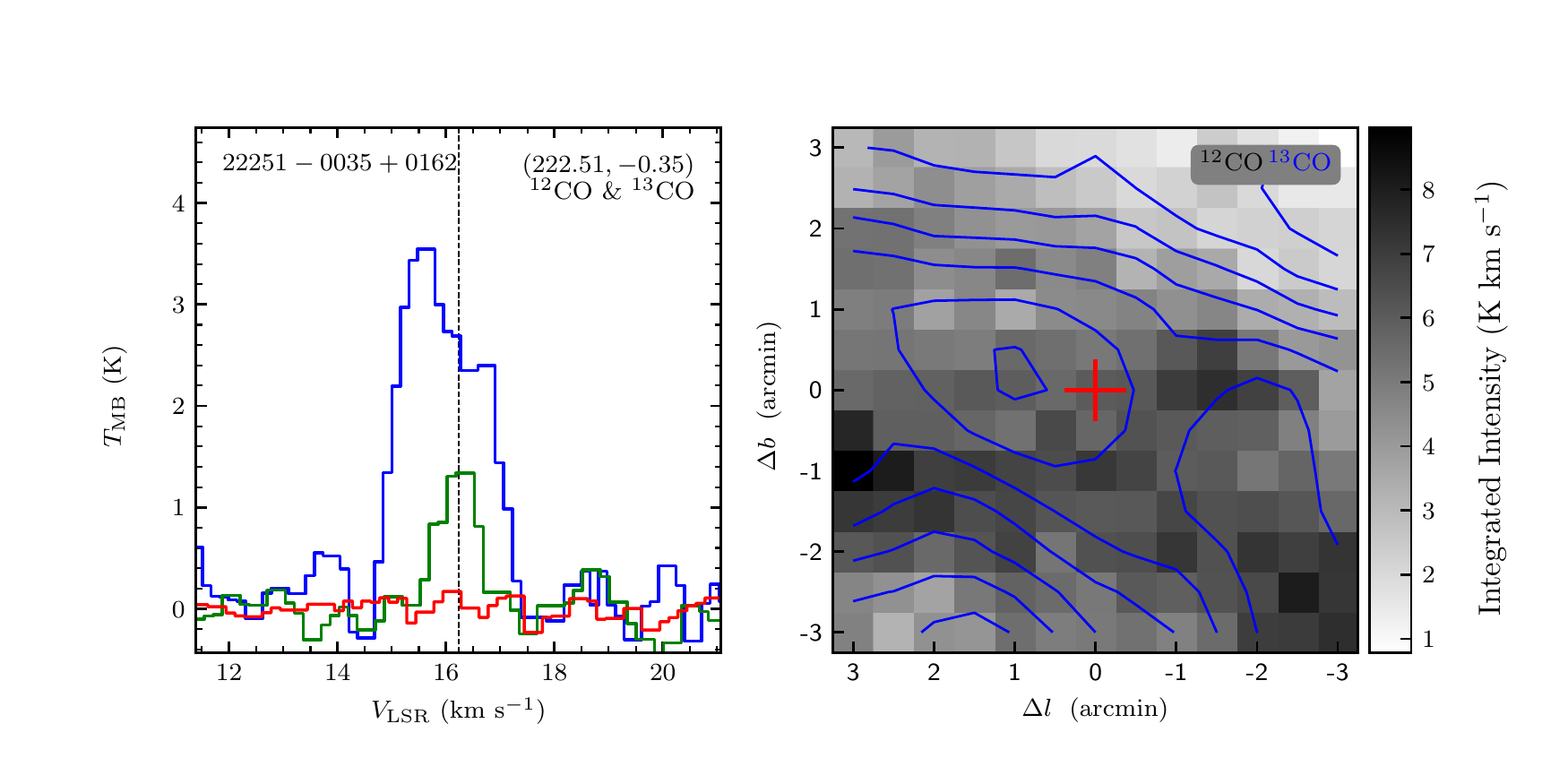}
\includegraphics[width=9.0cm,angle=0]{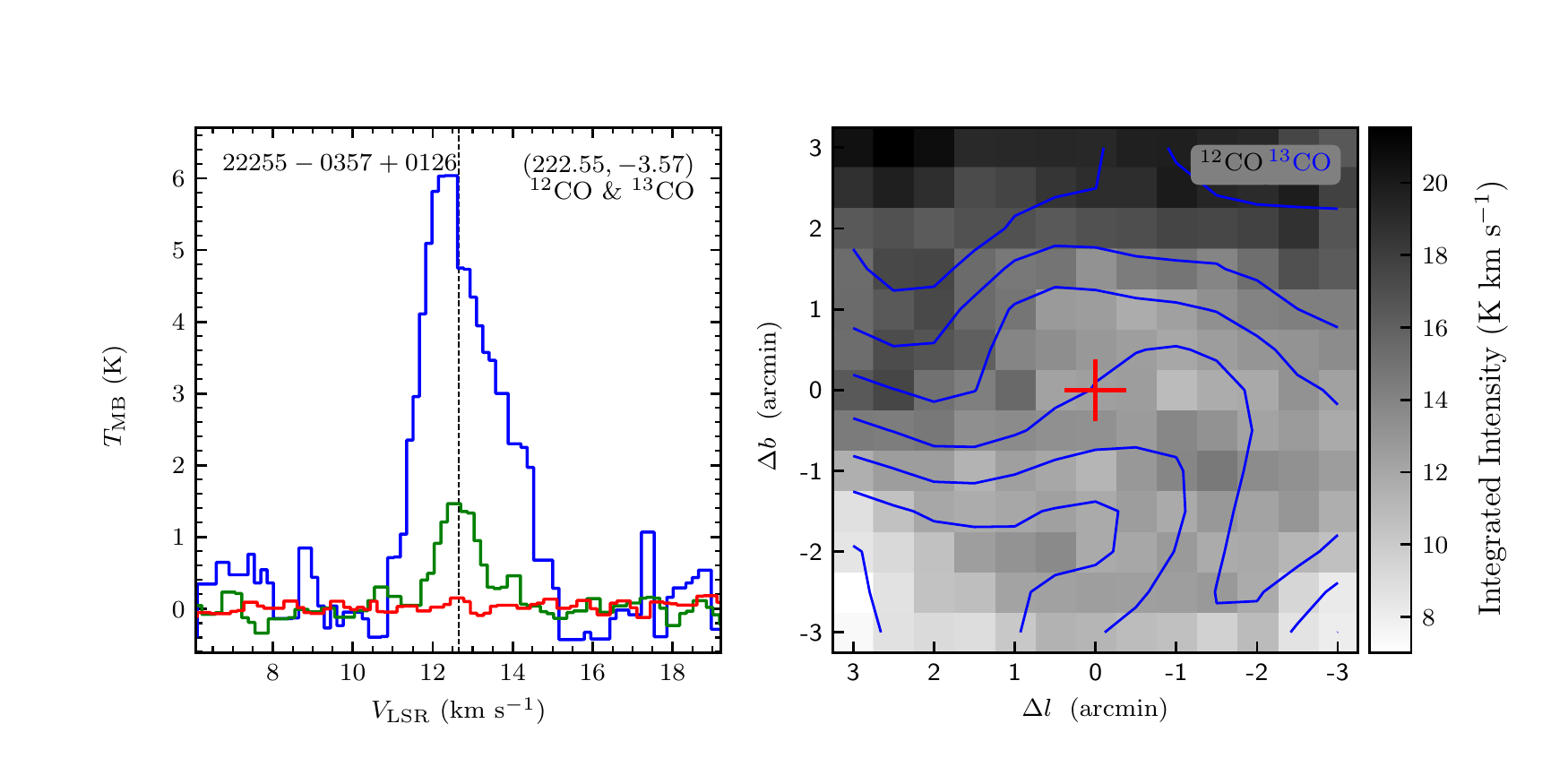}
\vspace{-0.5cm}

\includegraphics[width=9.0cm,angle=0]{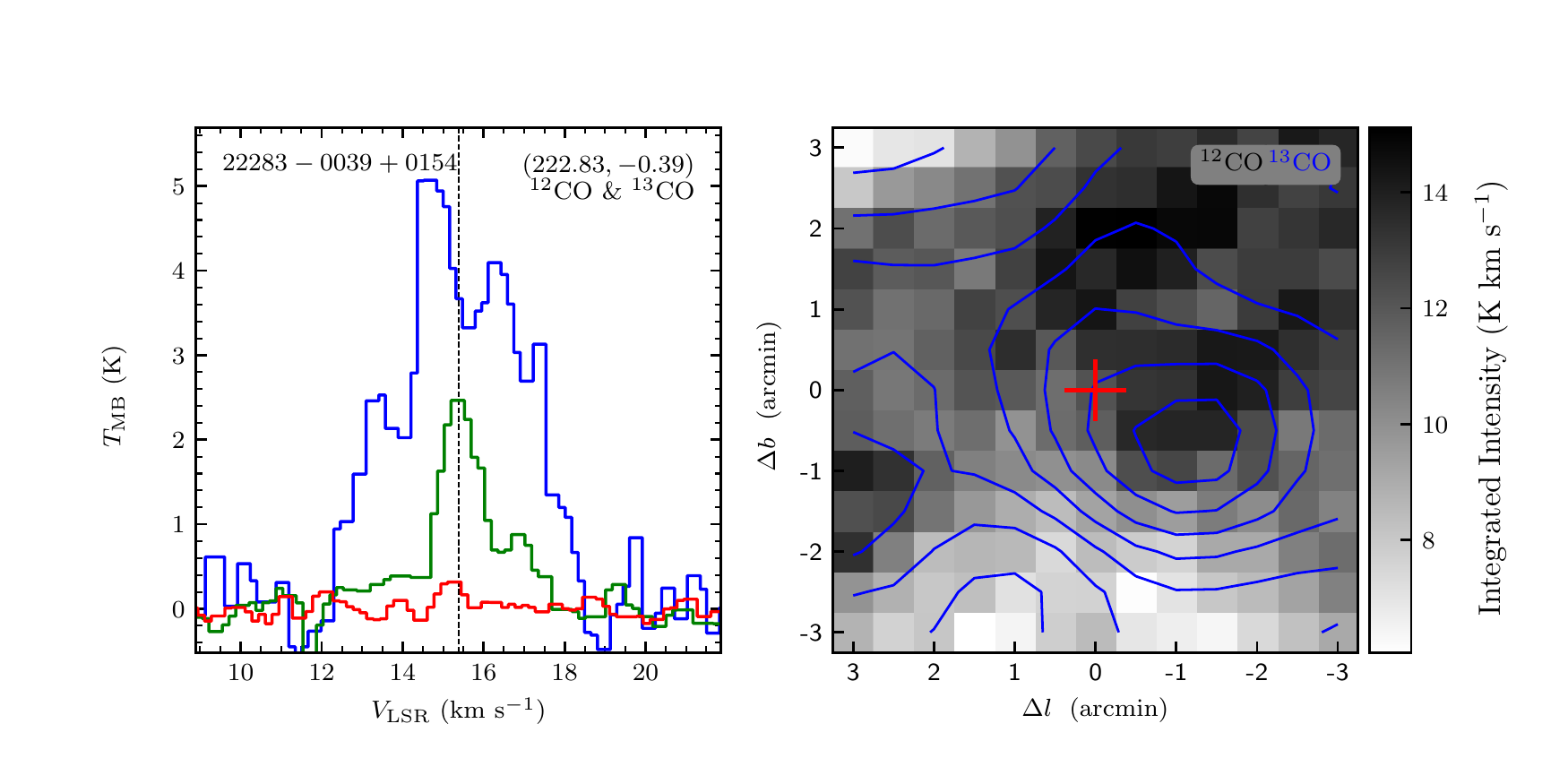}
\includegraphics[width=9.0cm,angle=0]{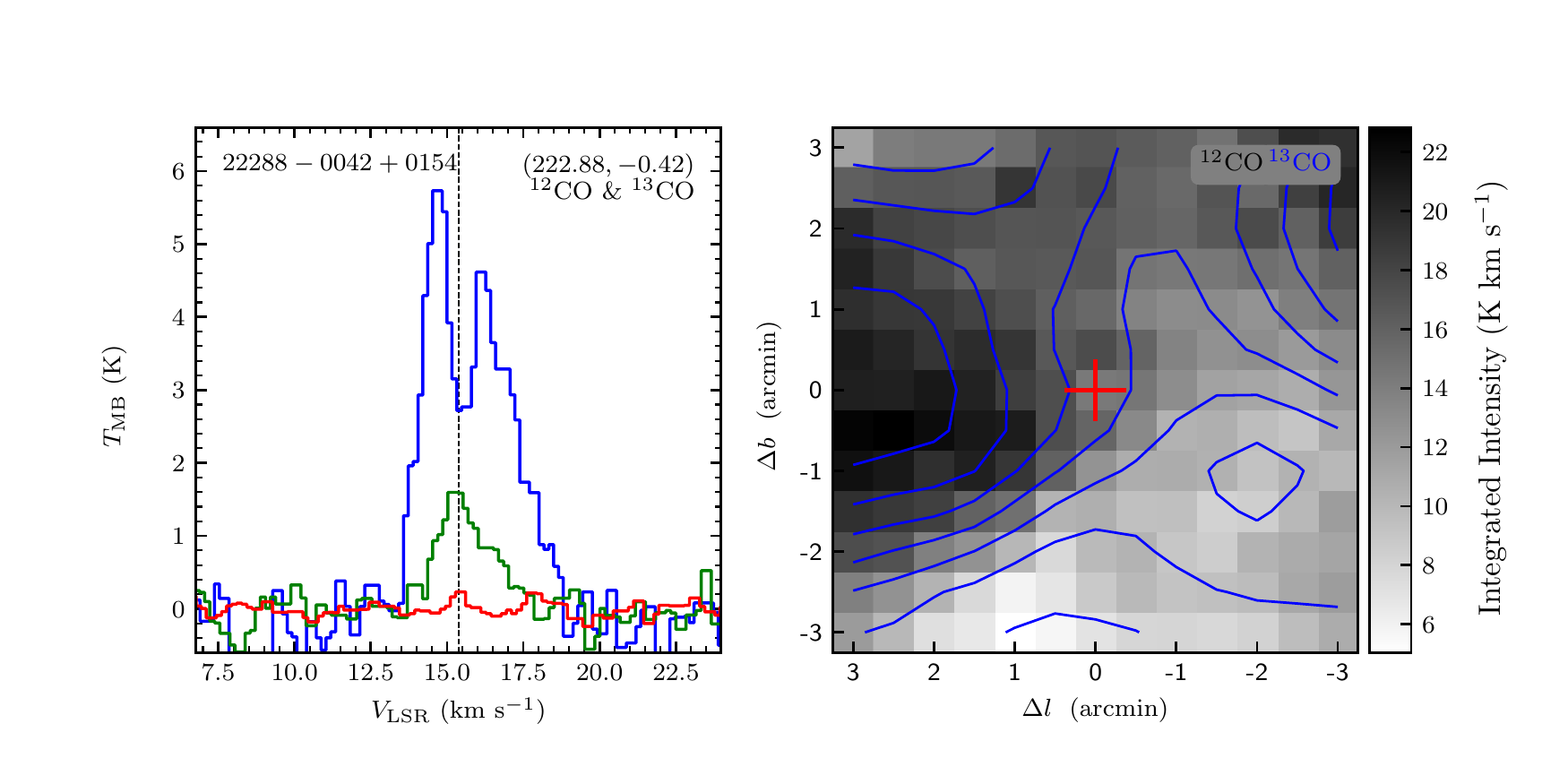}
\vspace{-0.5cm}

\includegraphics[width=9.0cm,angle=0]{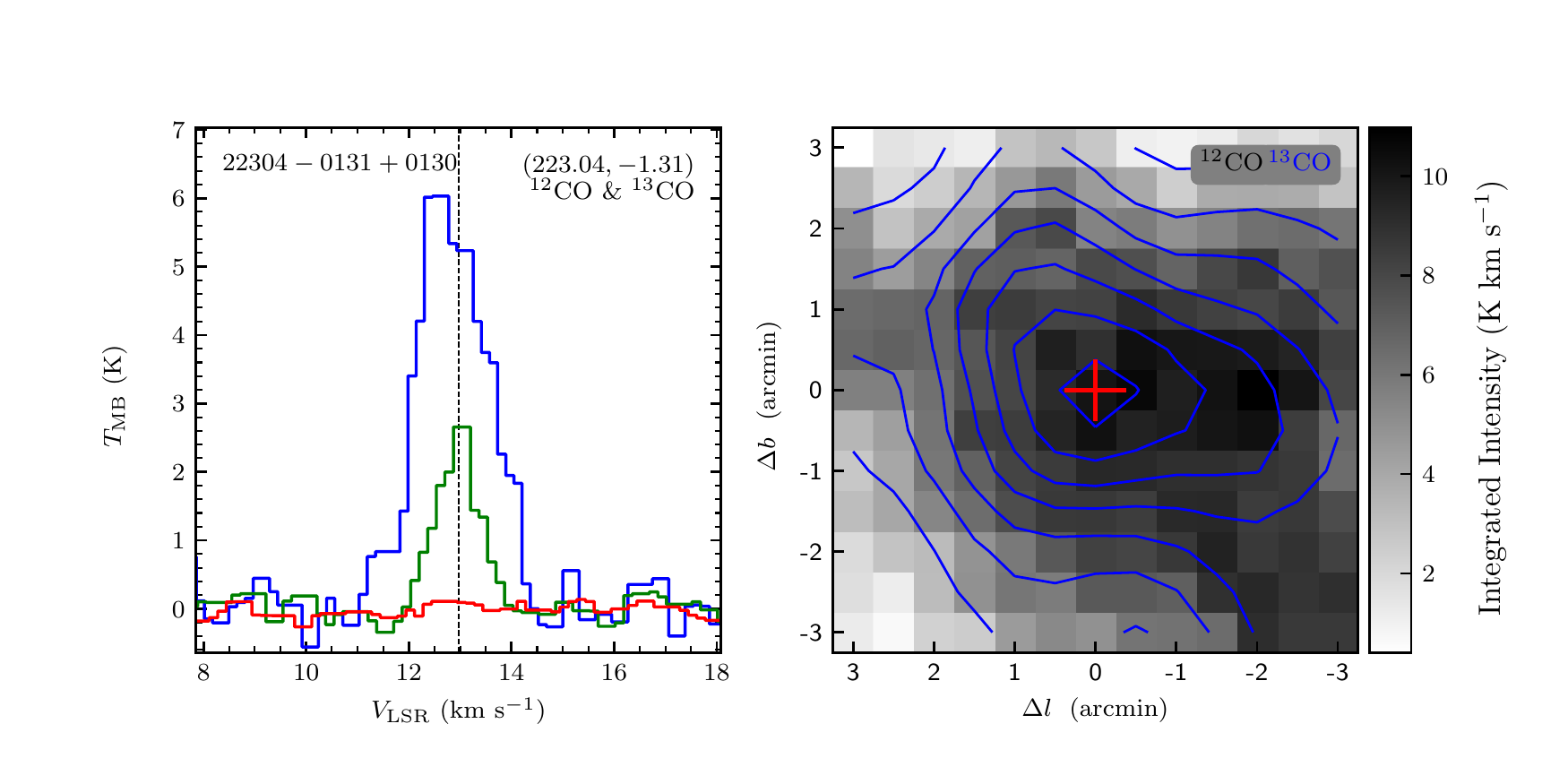}
\includegraphics[width=9.0cm,angle=0]{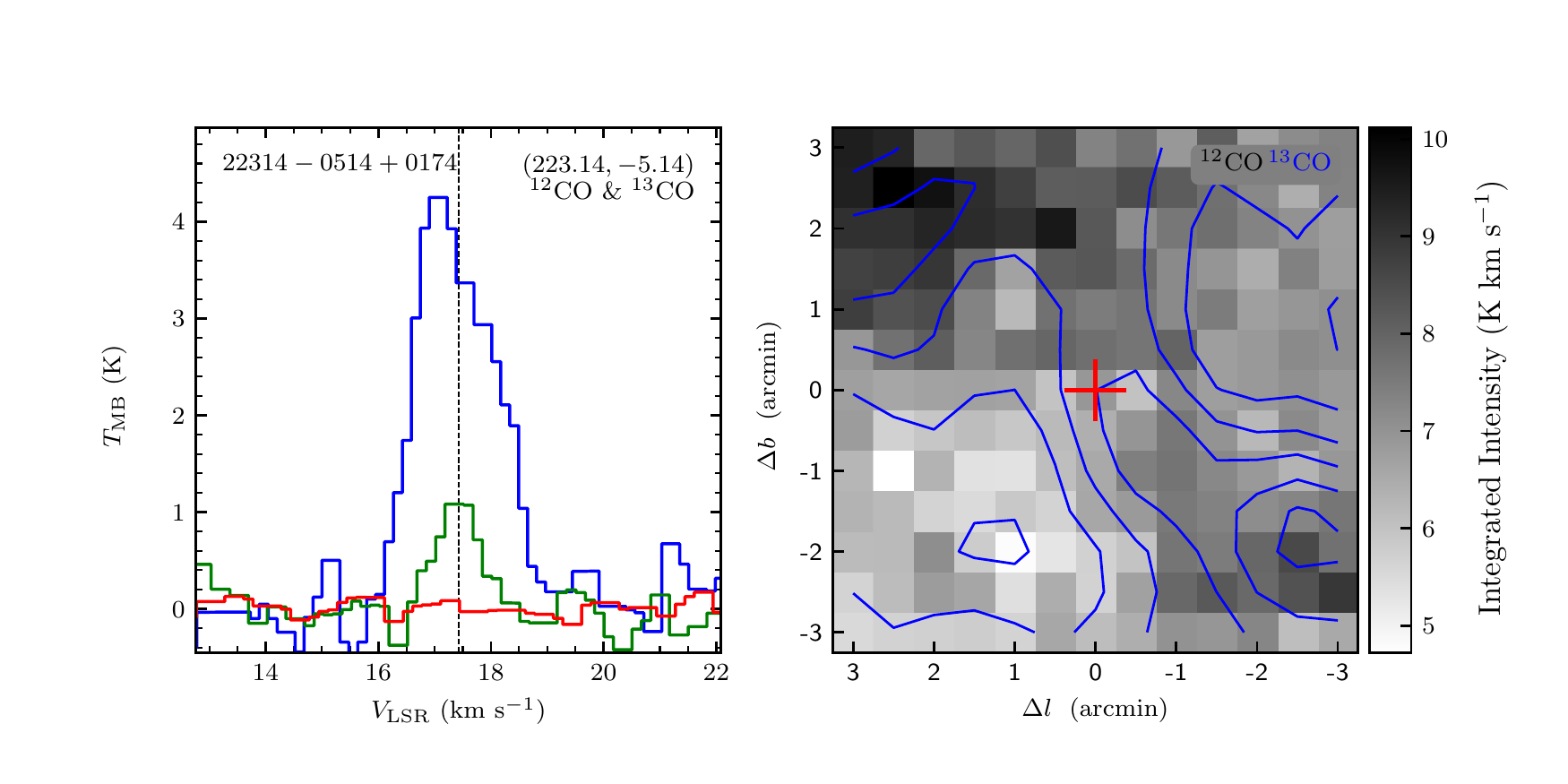}
\end{figure}
\clearpage

\begin{figure}
\includegraphics[width=9.0cm,angle=0]{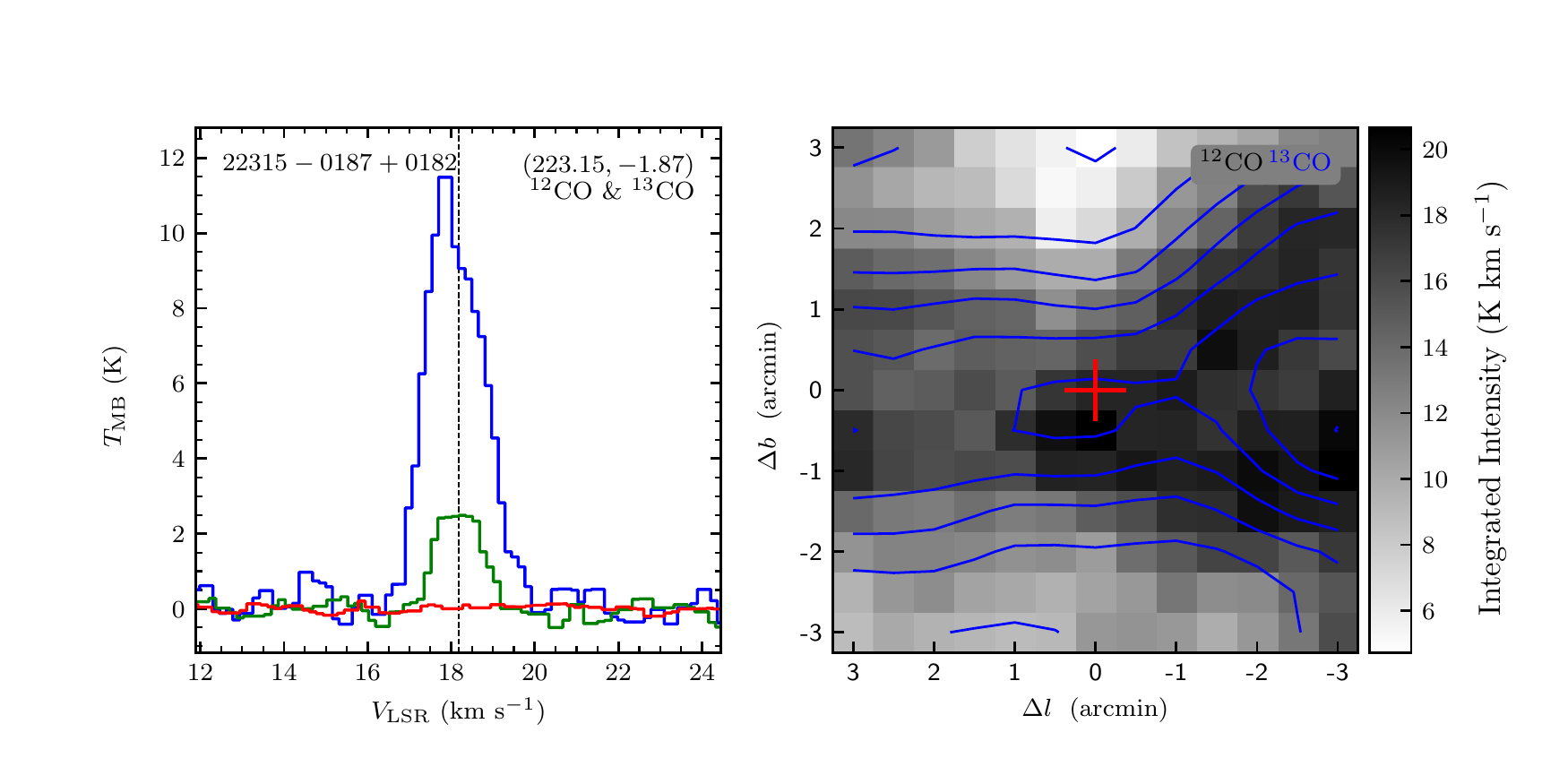}
\includegraphics[width=9.0cm,angle=0]{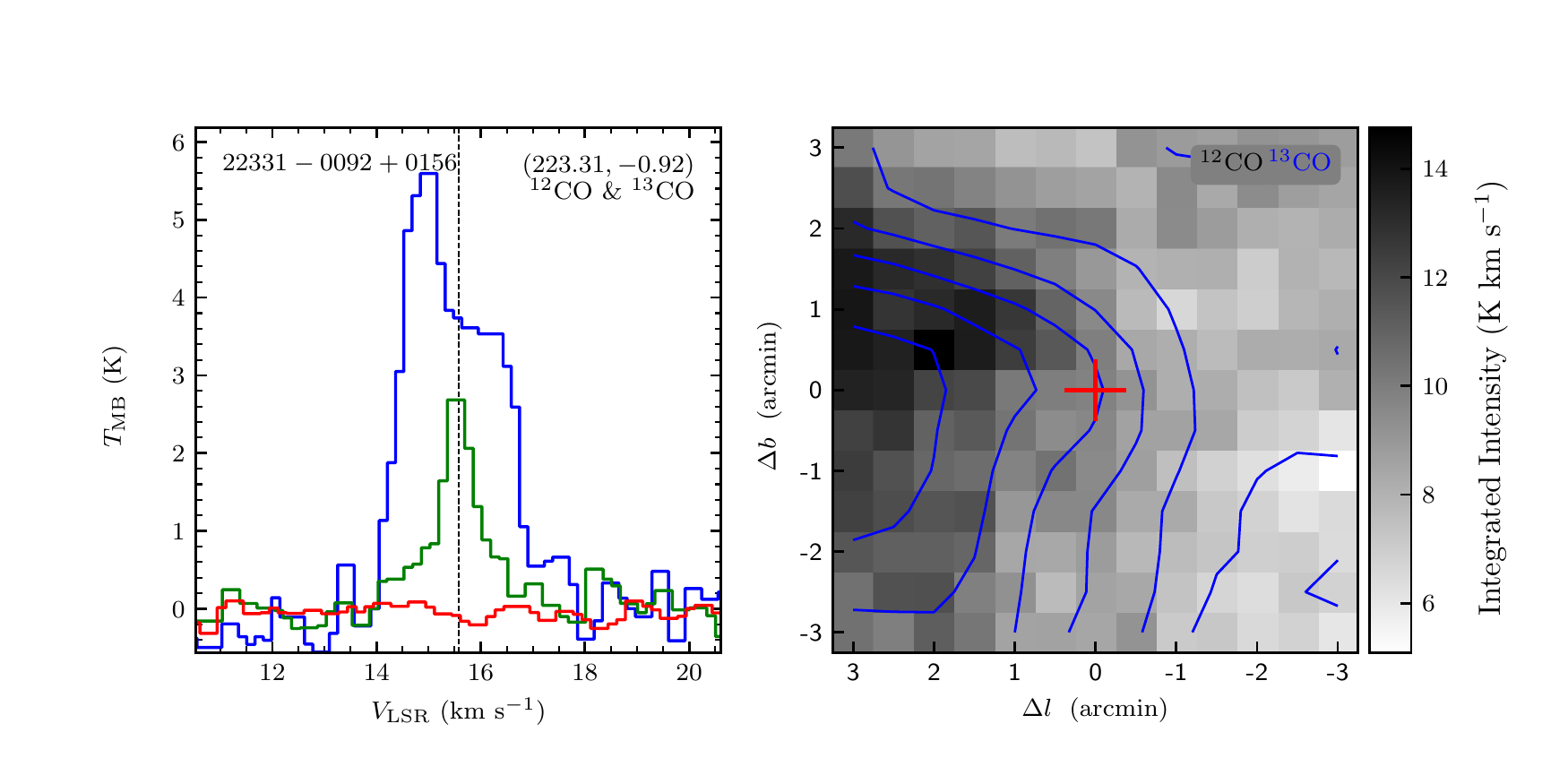}
\vspace{-0.5cm}

\includegraphics[width=9.0cm,angle=0]{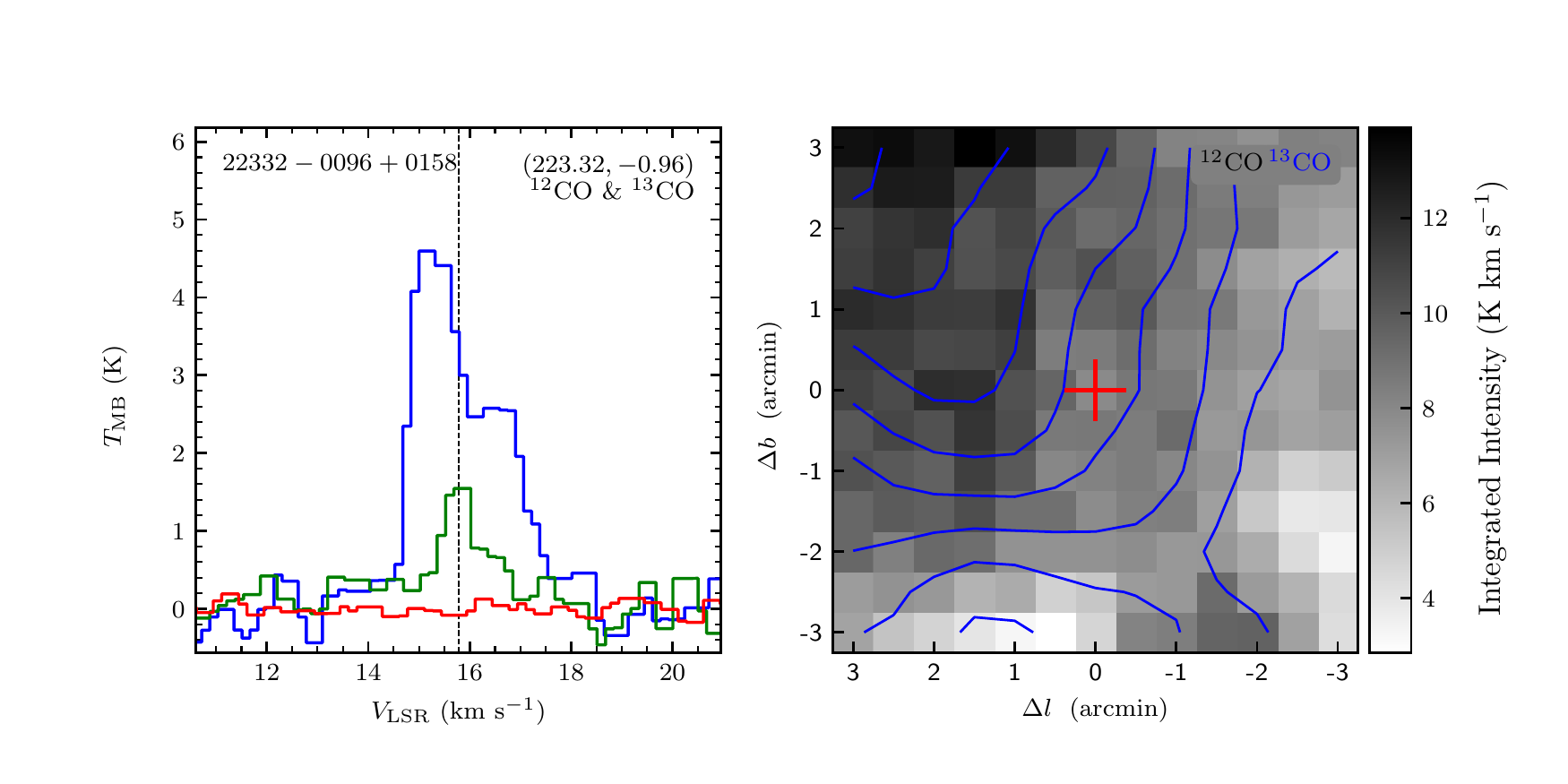}
\includegraphics[width=9.0cm,angle=0]{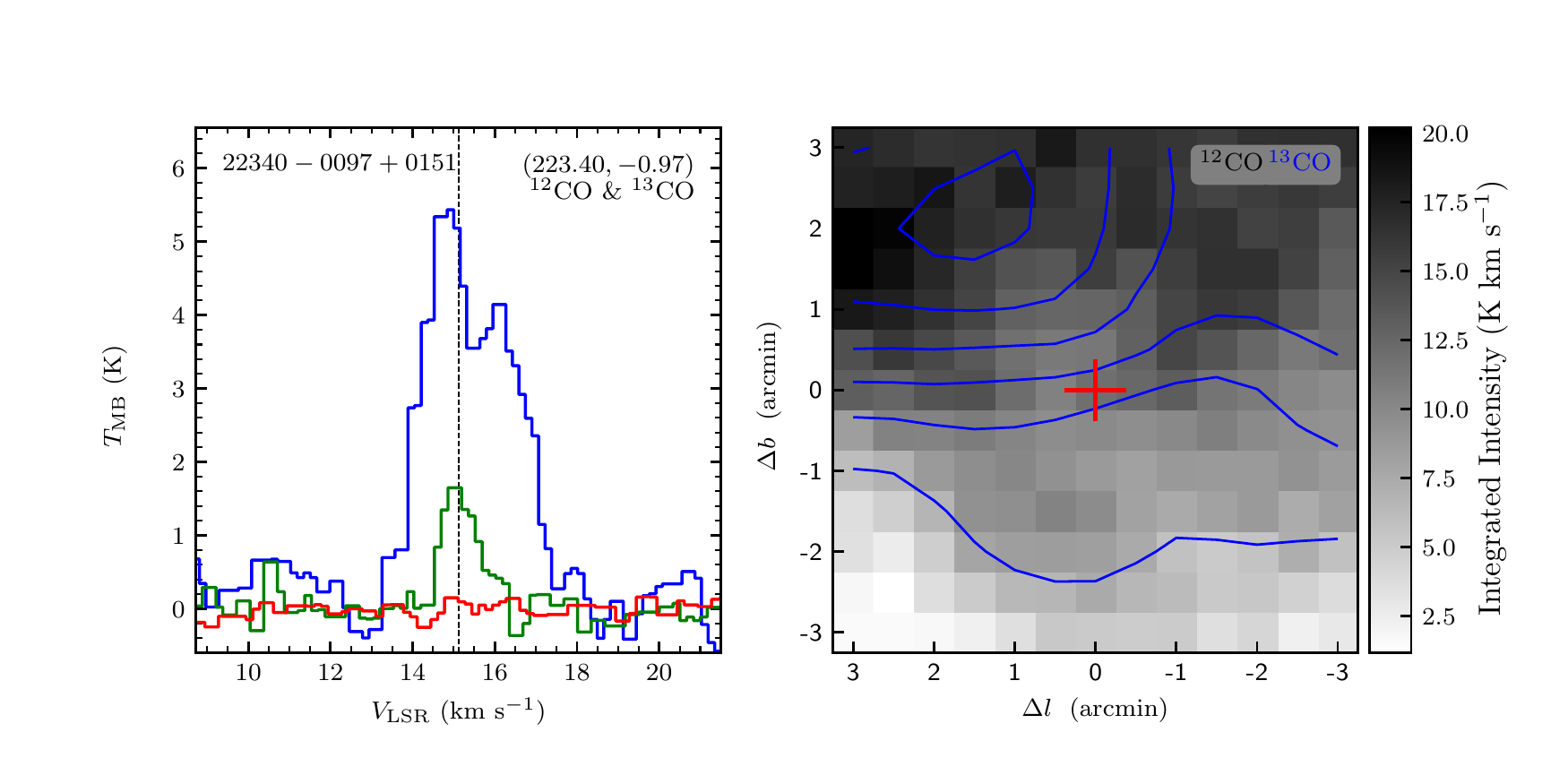}
\vspace{-0.5cm}

\includegraphics[width=9.0cm,angle=0]{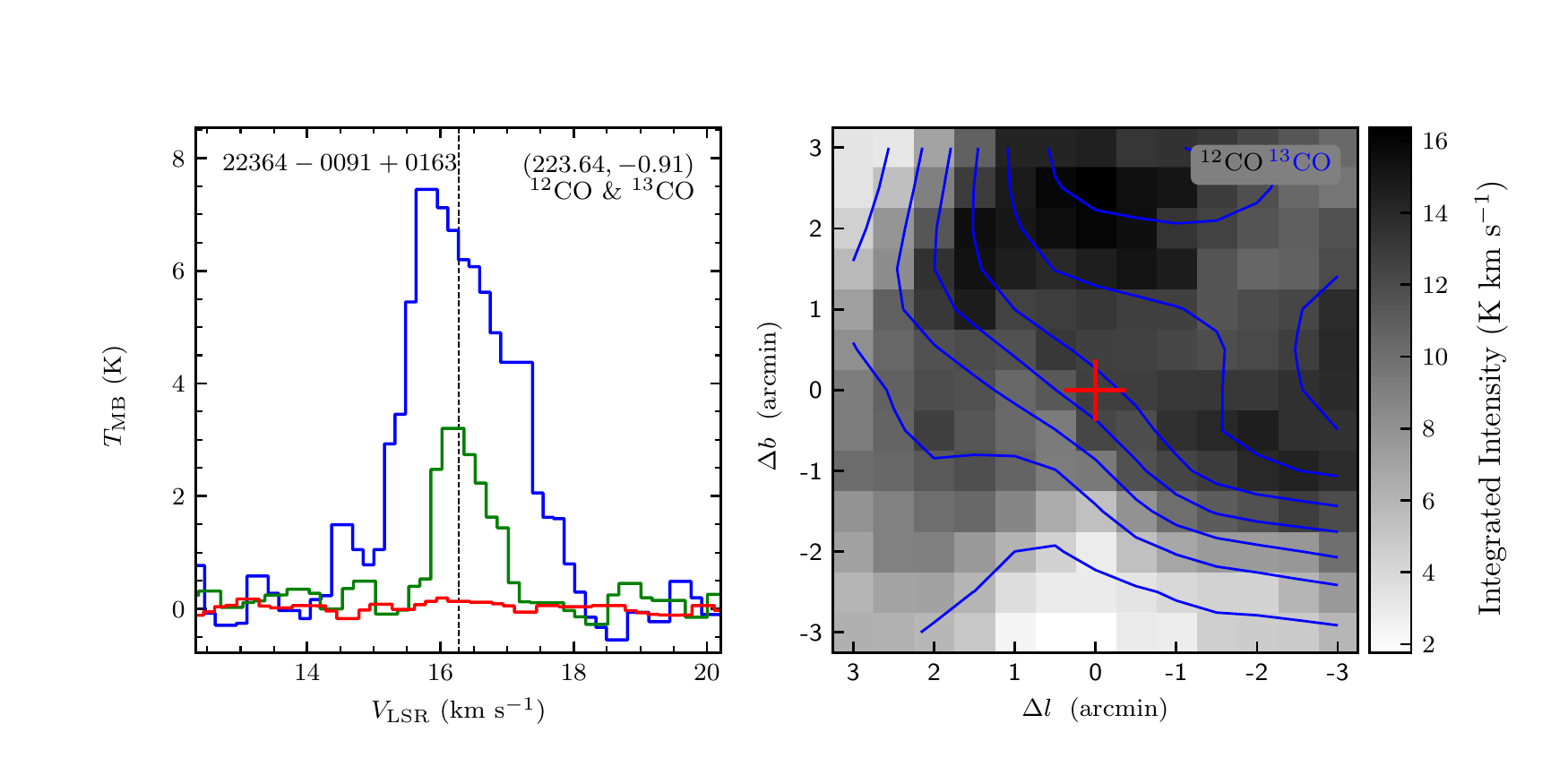}
\includegraphics[width=9.0cm,angle=0]{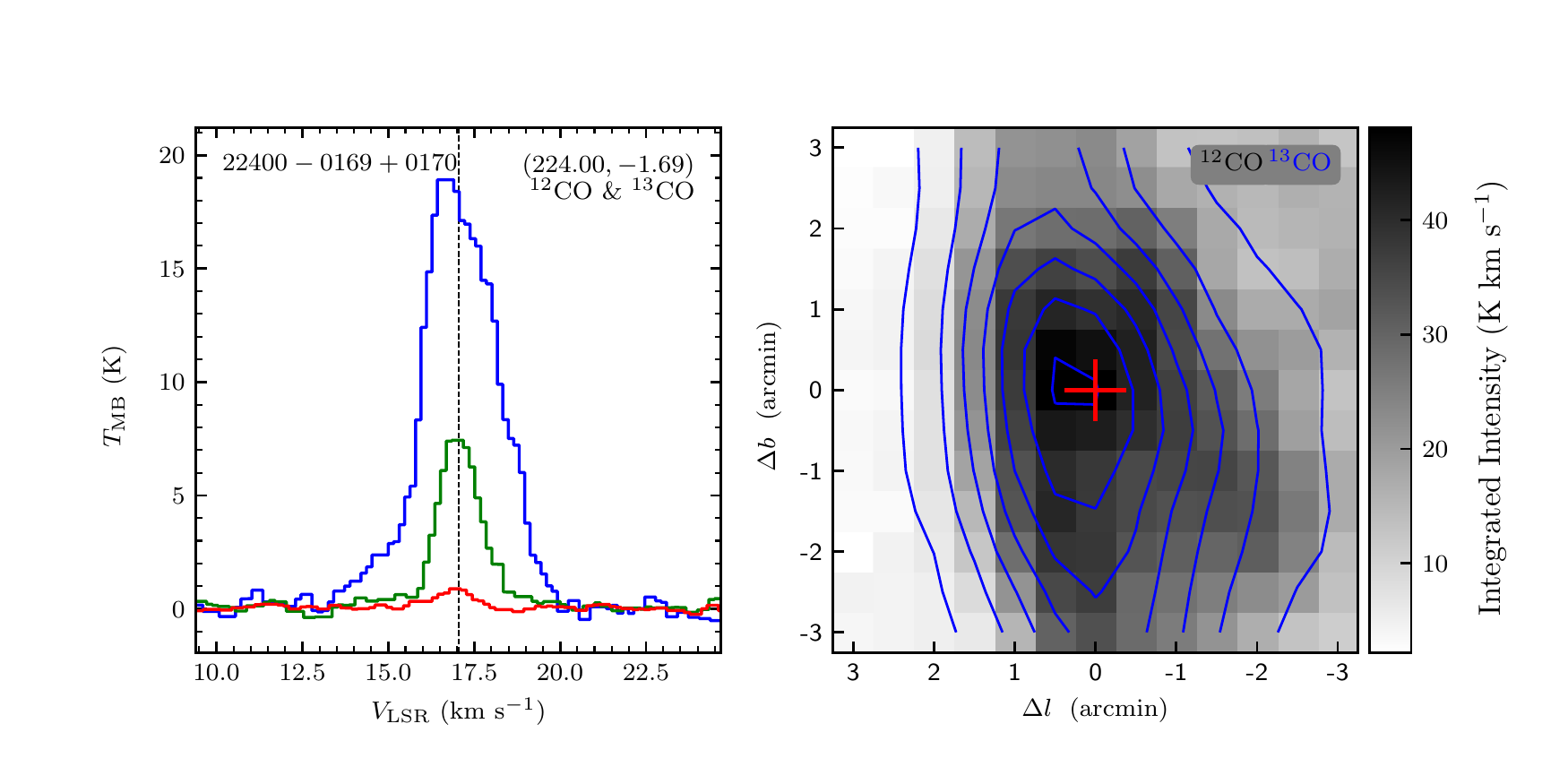}
\vspace{-0.5cm}

\includegraphics[width=9.0cm,angle=0]{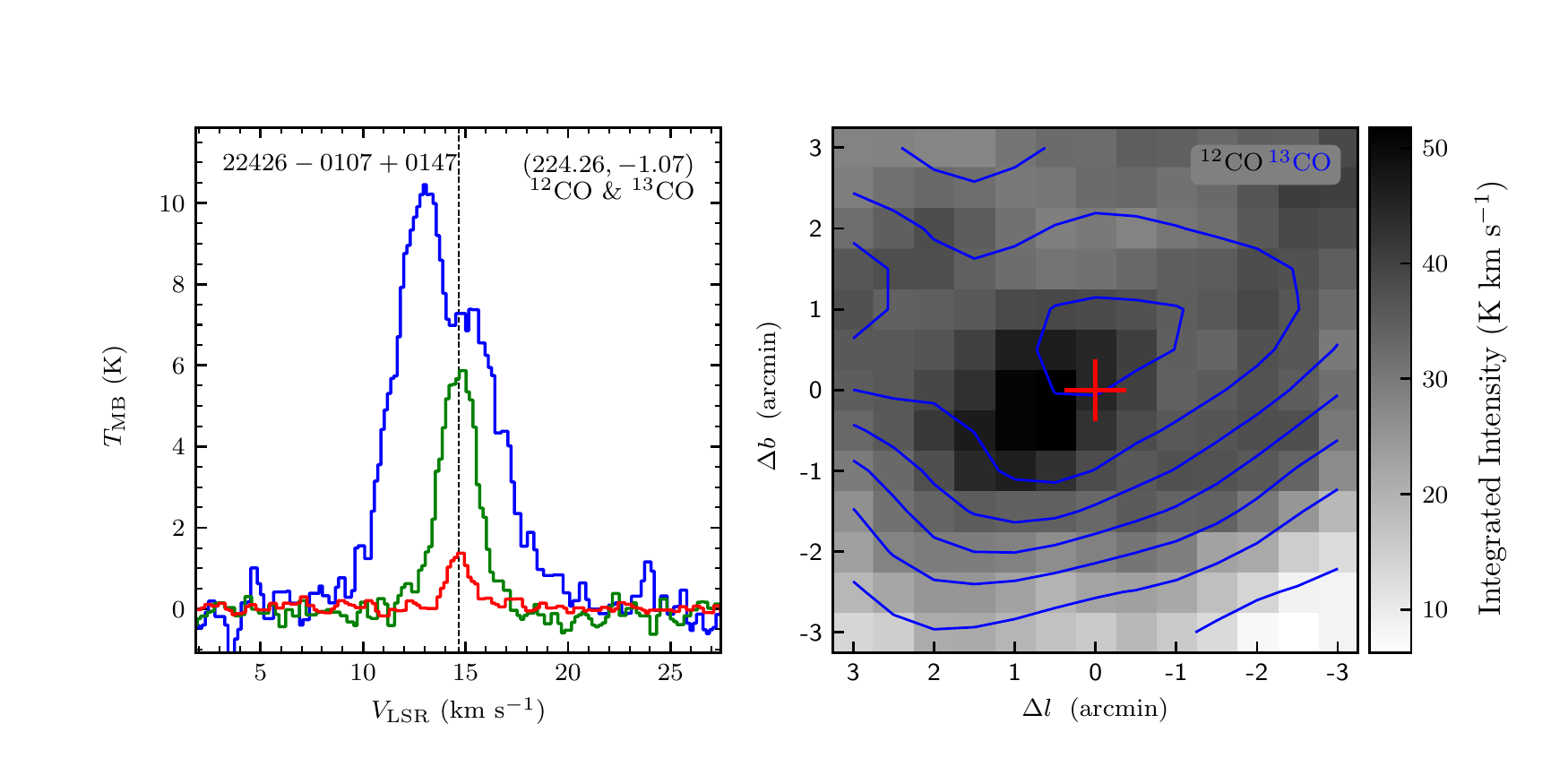}
\includegraphics[width=9.0cm,angle=0]{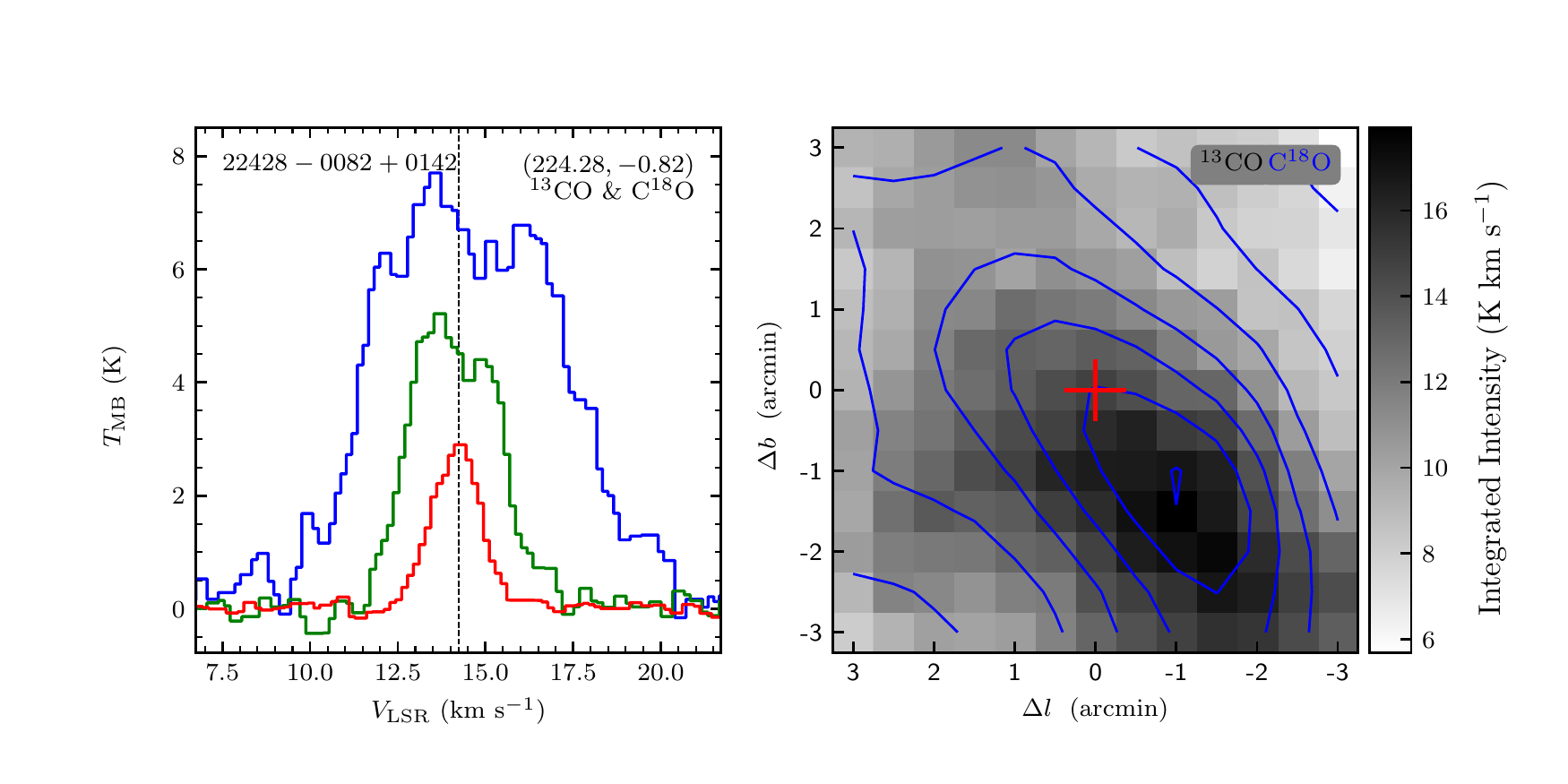}
\vspace{-0.5cm}

\includegraphics[width=9.0cm,angle=0]{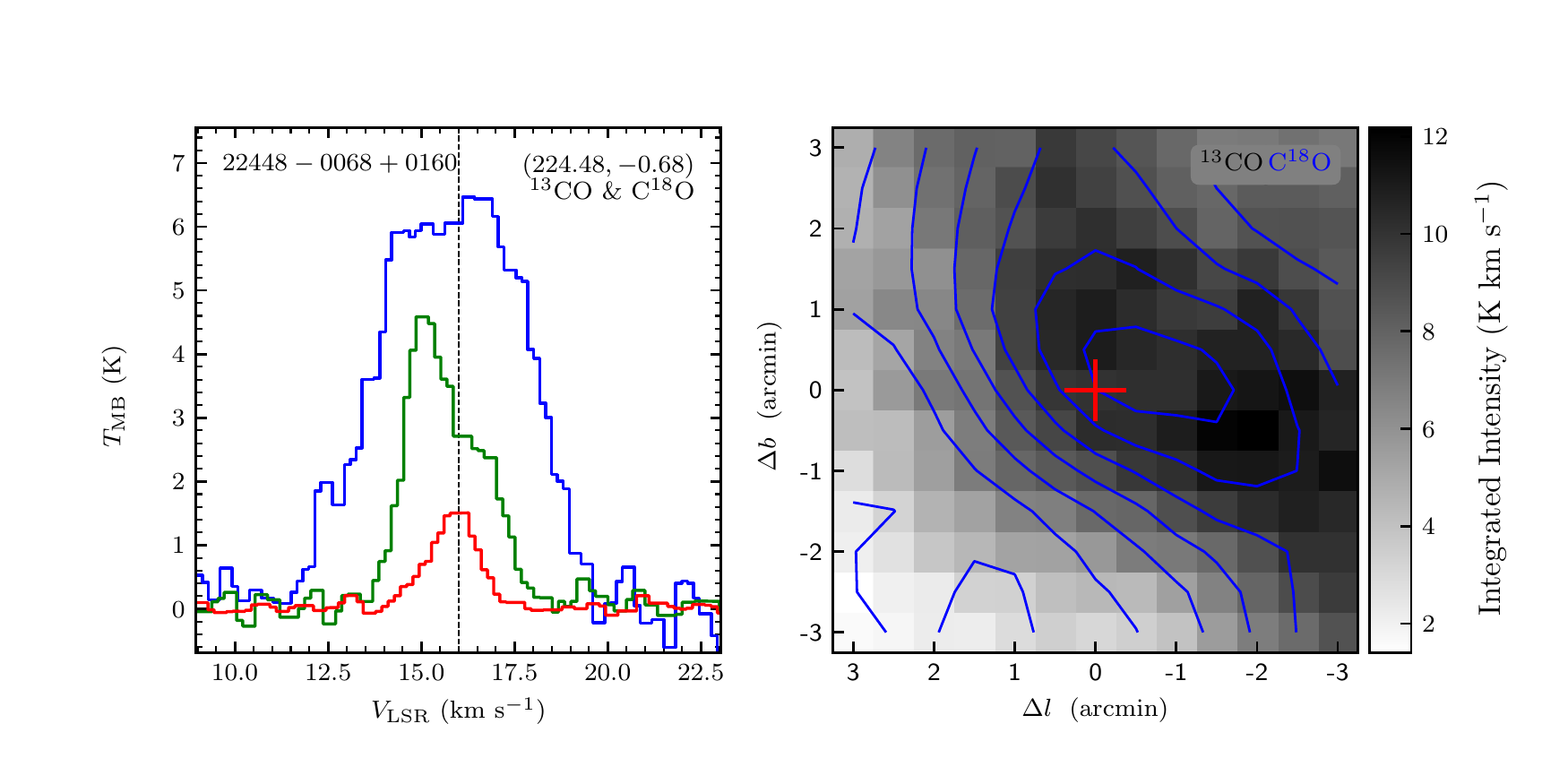}
\includegraphics[width=9.0cm,angle=0]{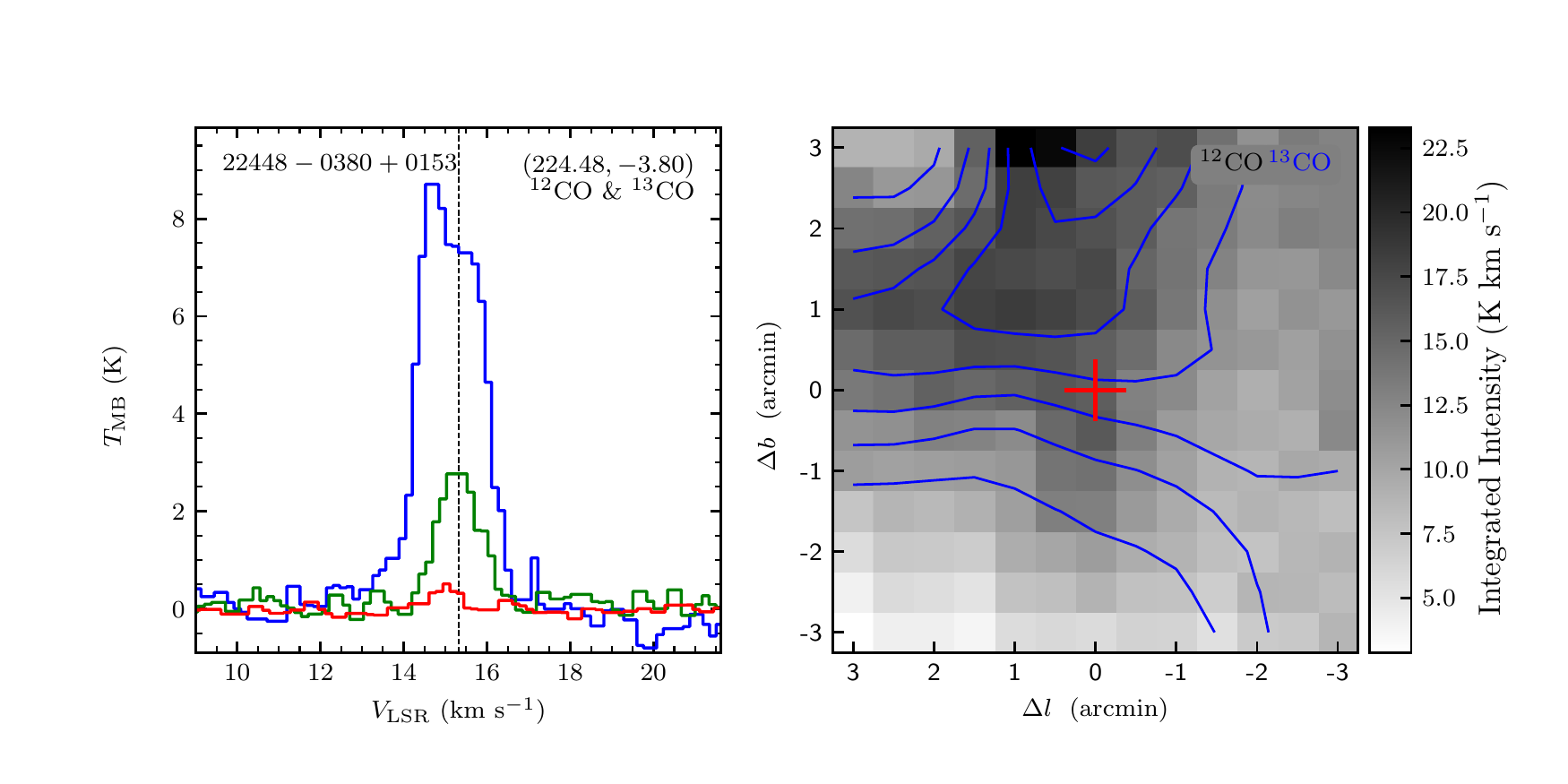}
\end{figure}
\clearpage

\begin{figure}
\includegraphics[width=9.0cm,angle=0]{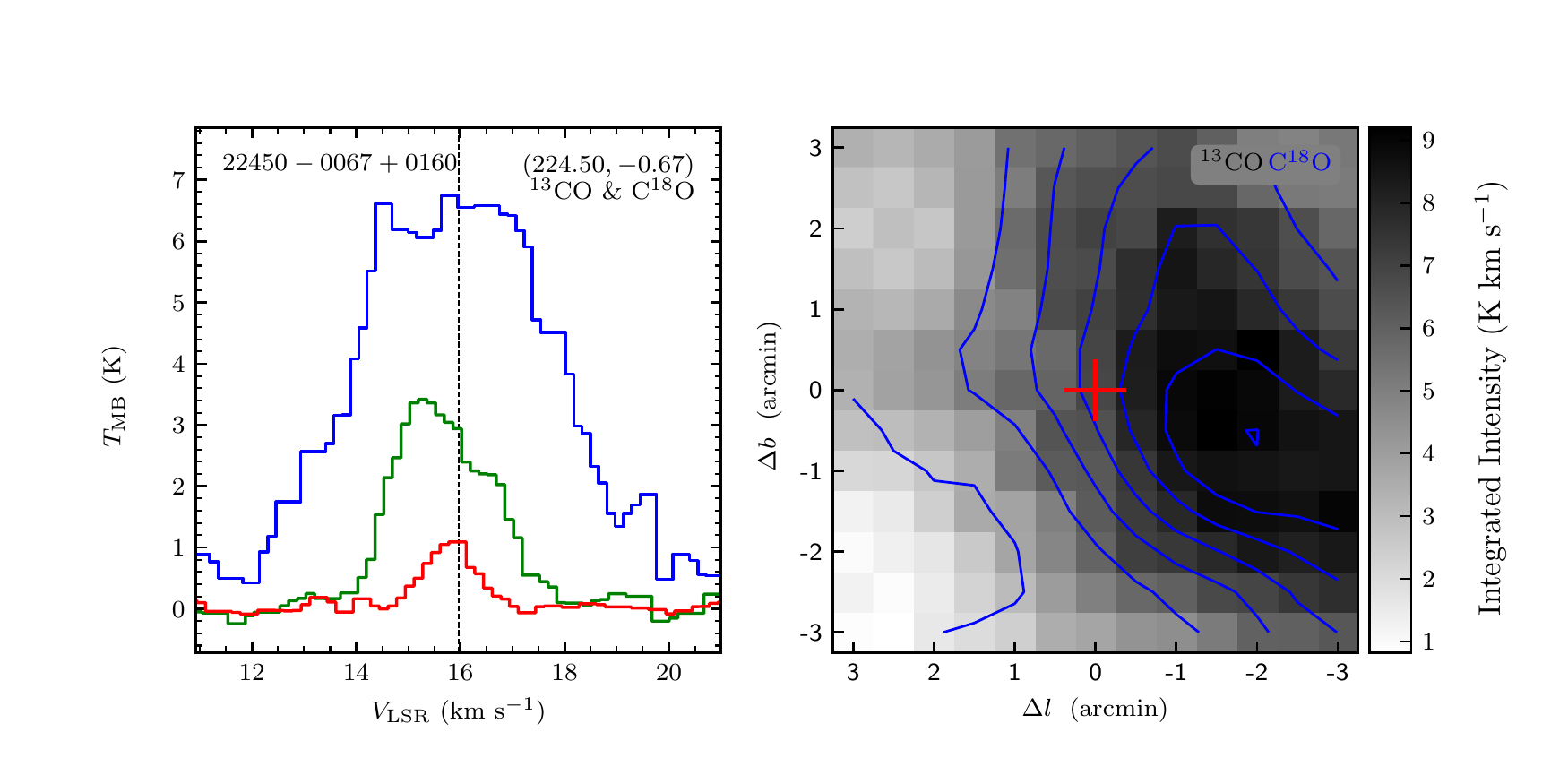}
\includegraphics[width=9.0cm,angle=0]{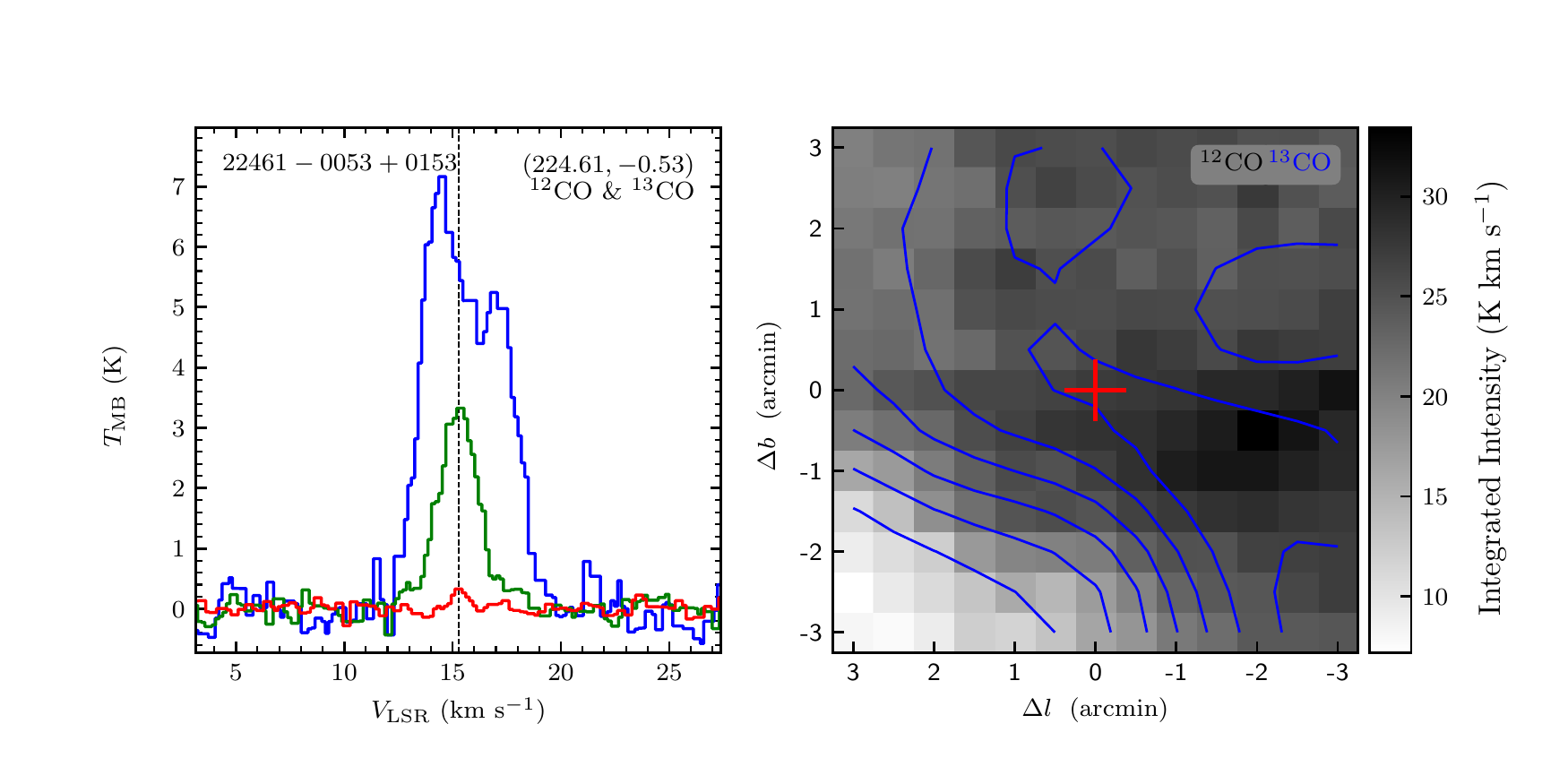}
\vspace{-0.5cm}

\includegraphics[width=9.0cm,angle=0]{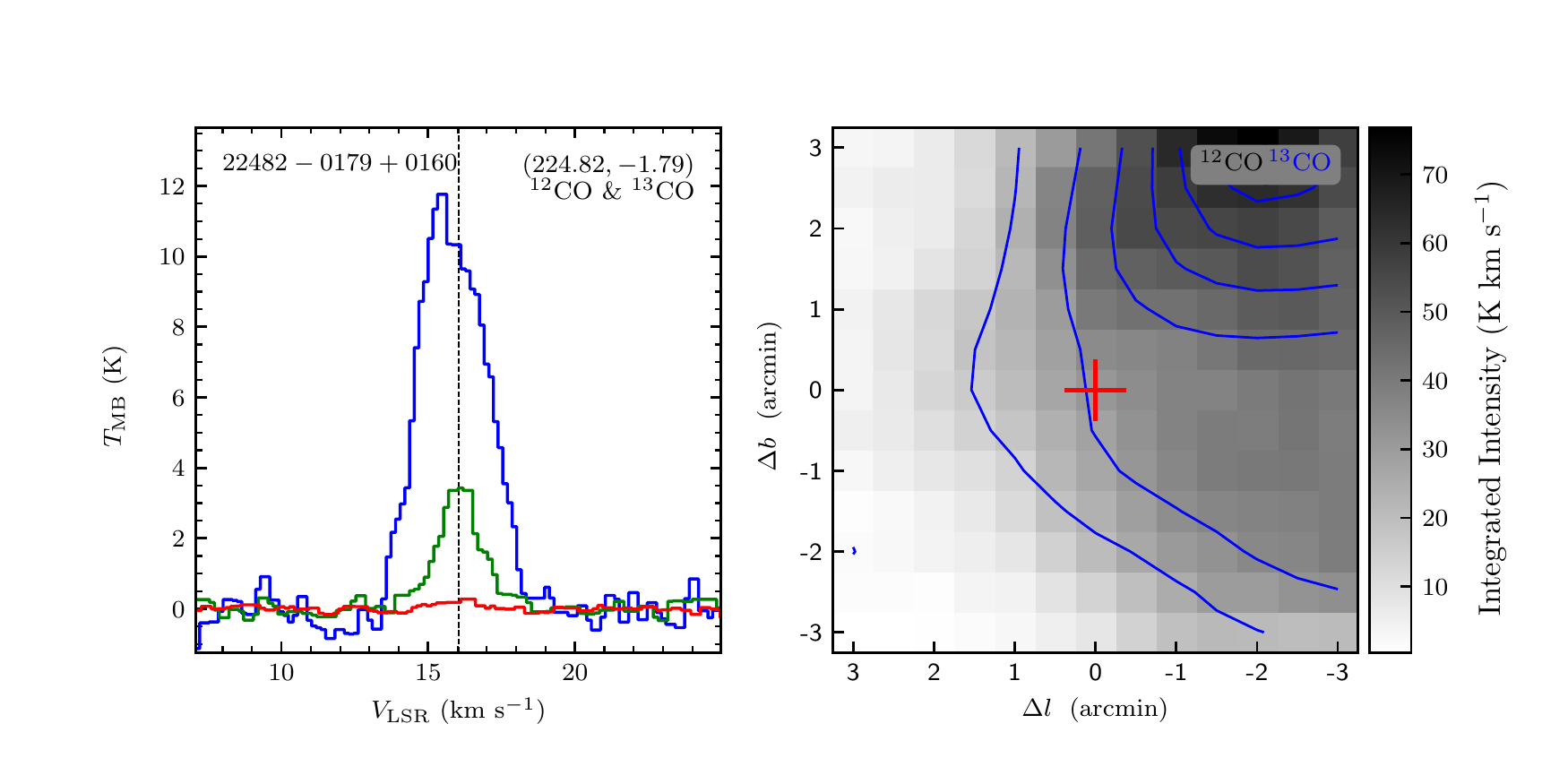}
\includegraphics[width=9.0cm,angle=0]{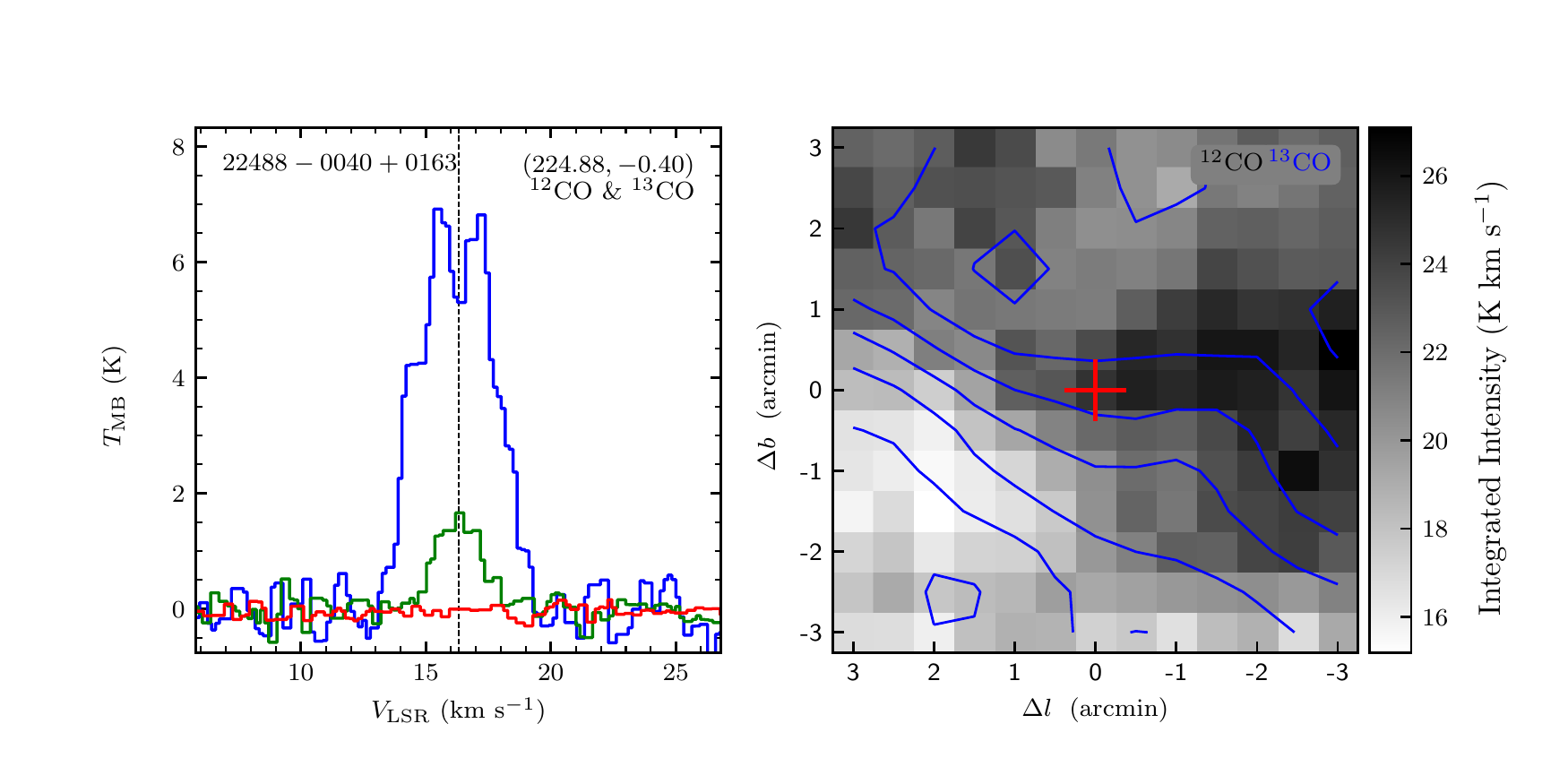}
\vspace{-0.5cm}

\includegraphics[width=9.0cm,angle=0]{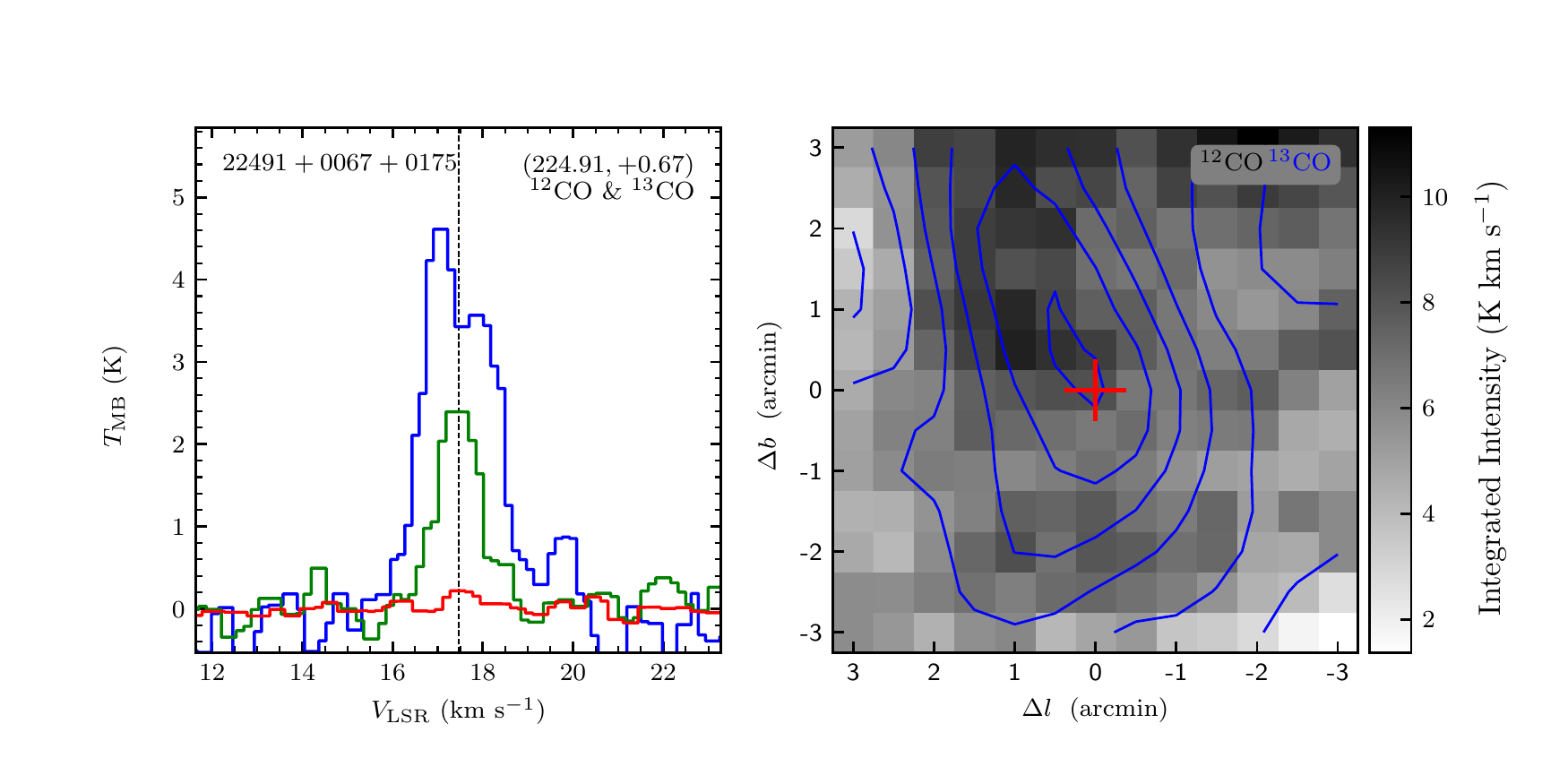}
\includegraphics[width=9.0cm,angle=0]{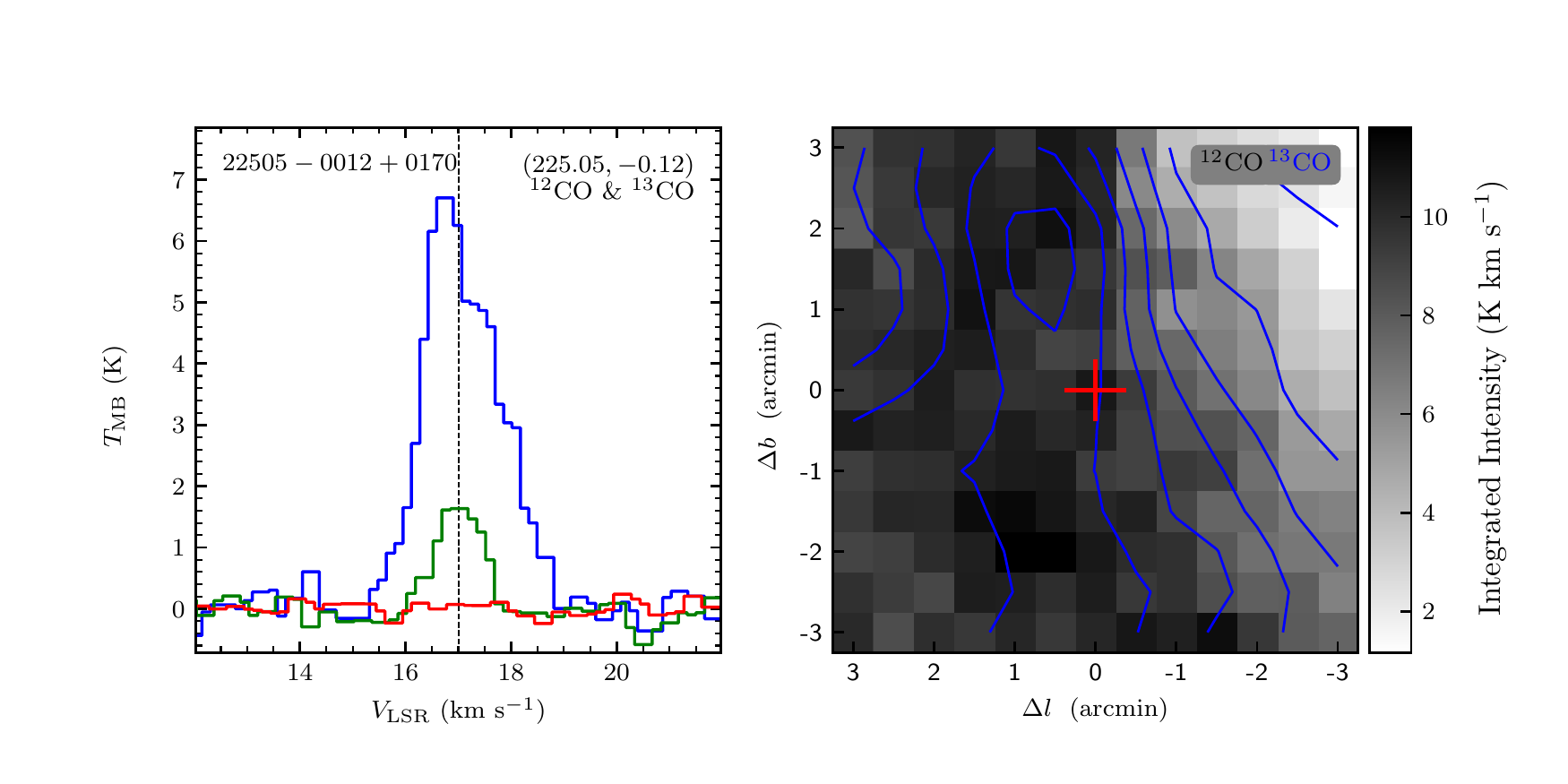}
\vspace{-0.5cm}

\includegraphics[width=9.0cm,angle=0]{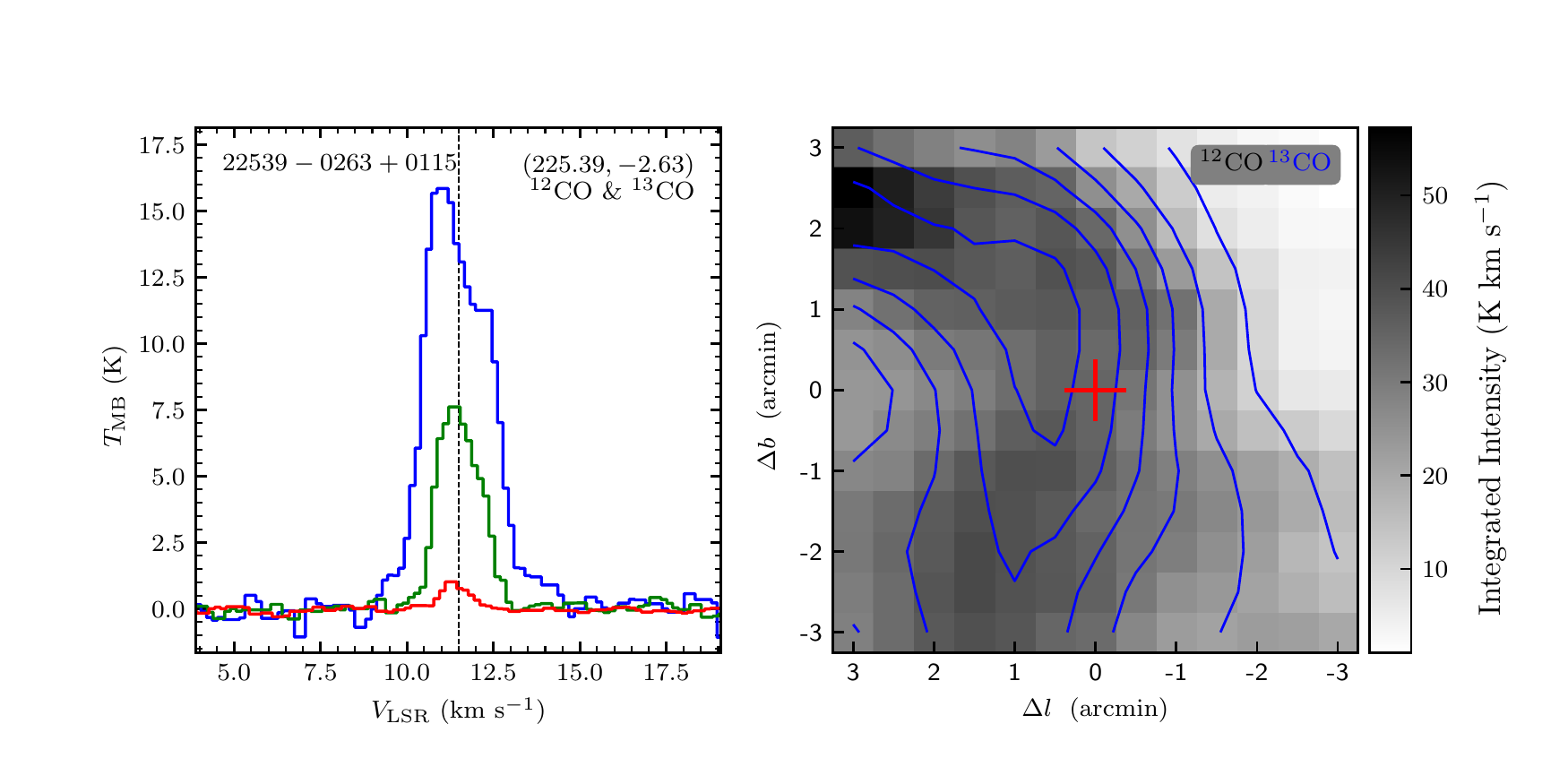}
\includegraphics[width=9.0cm,angle=0]{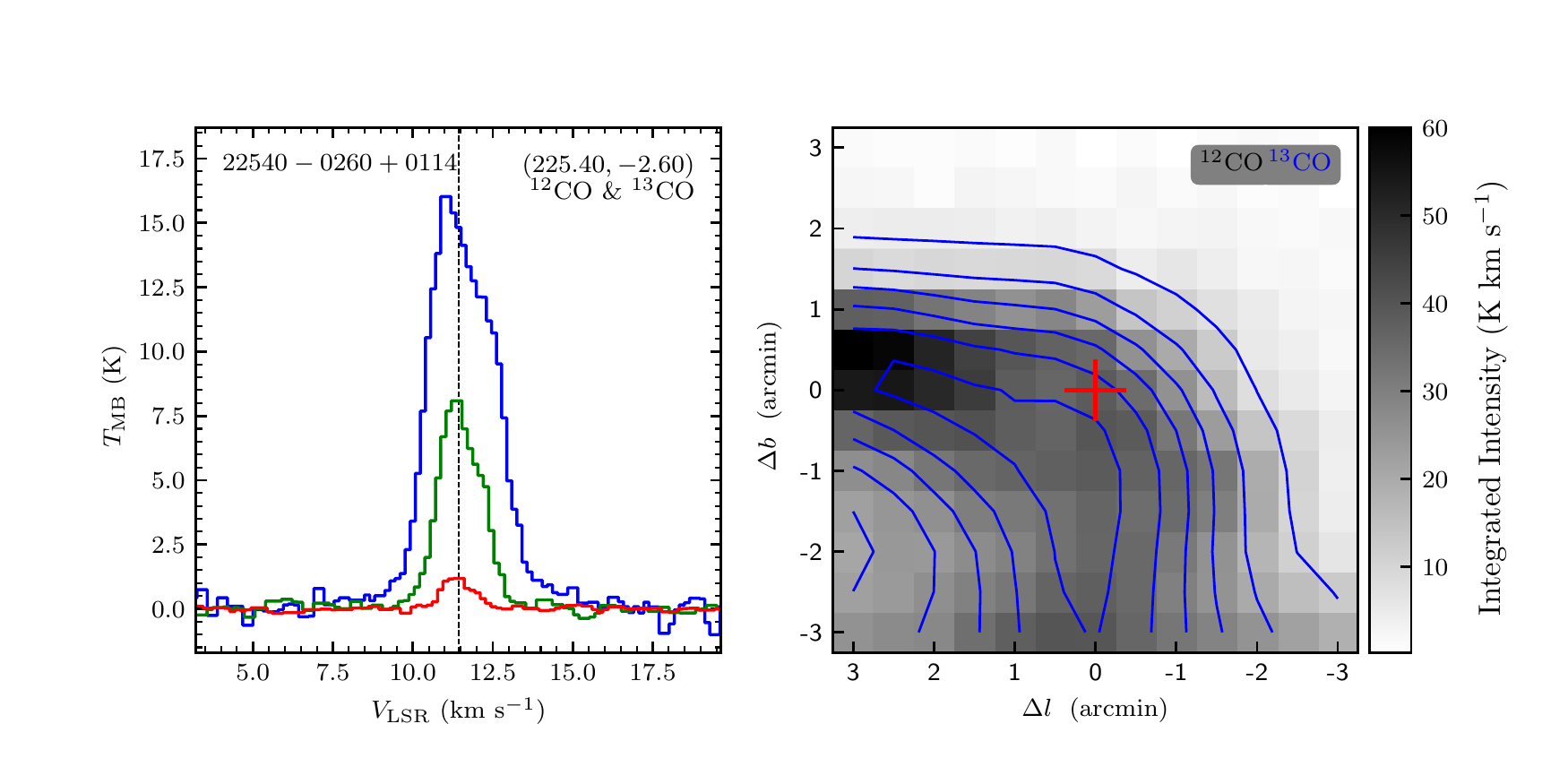}
\vspace{-0.5cm}

\includegraphics[width=9.0cm,angle=0]{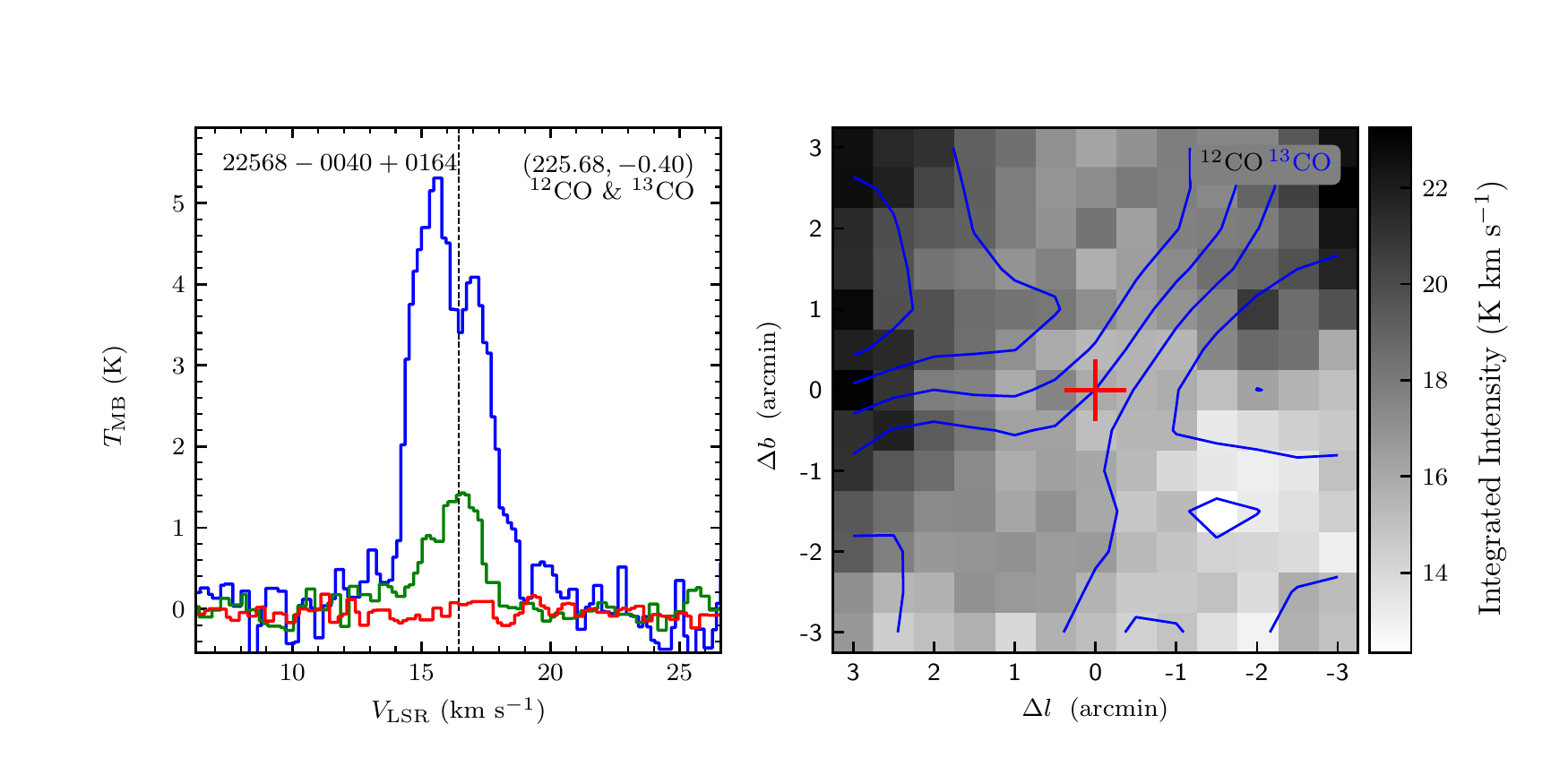}
\includegraphics[width=9.0cm,angle=0]{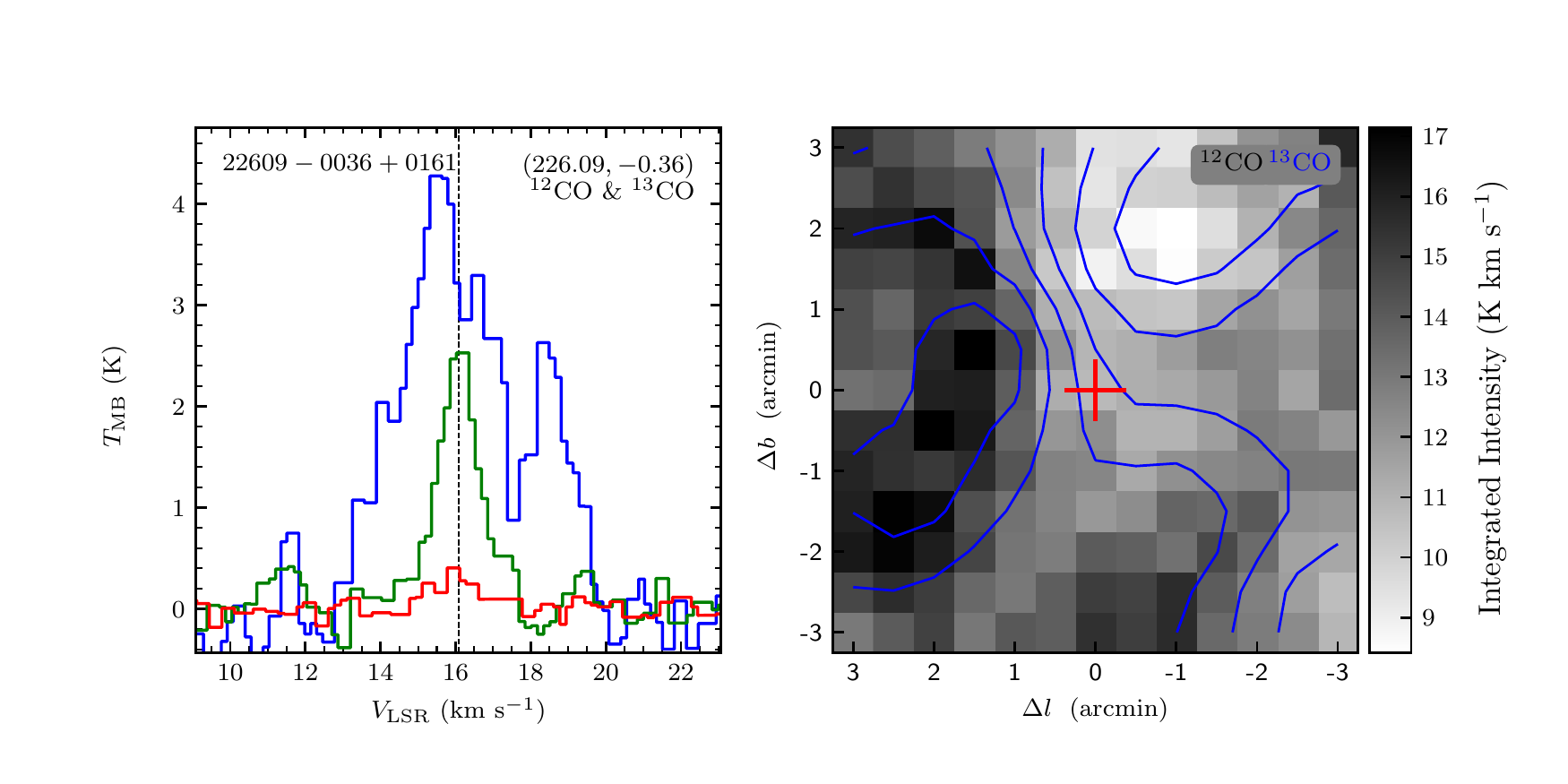}
\end{figure}
\clearpage

\begin{figure}
\includegraphics[width=9.0cm,angle=0]{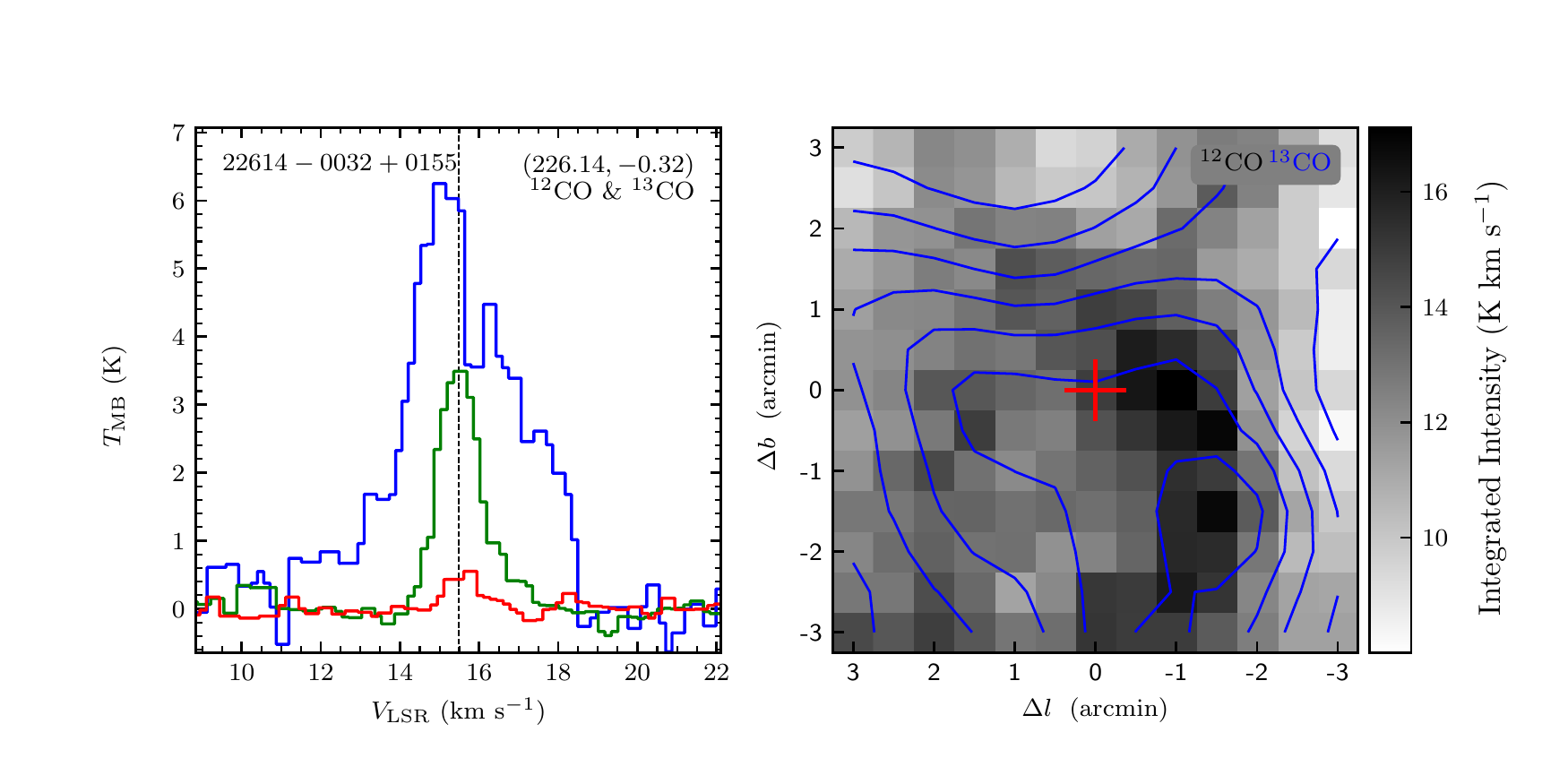}
\includegraphics[width=9.0cm,angle=0]{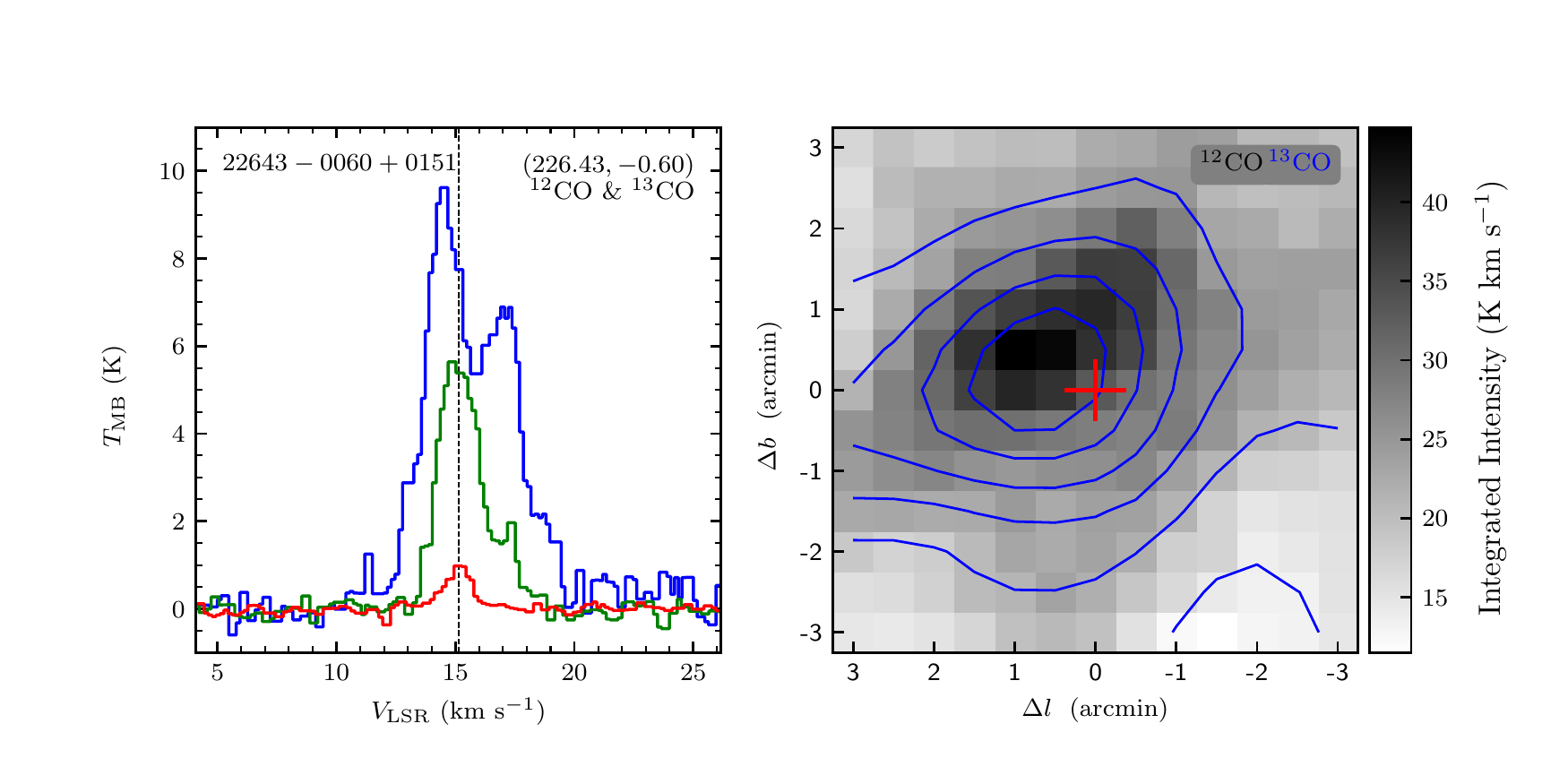}
\vspace{-0.5cm}

\includegraphics[width=9.0cm,angle=0]{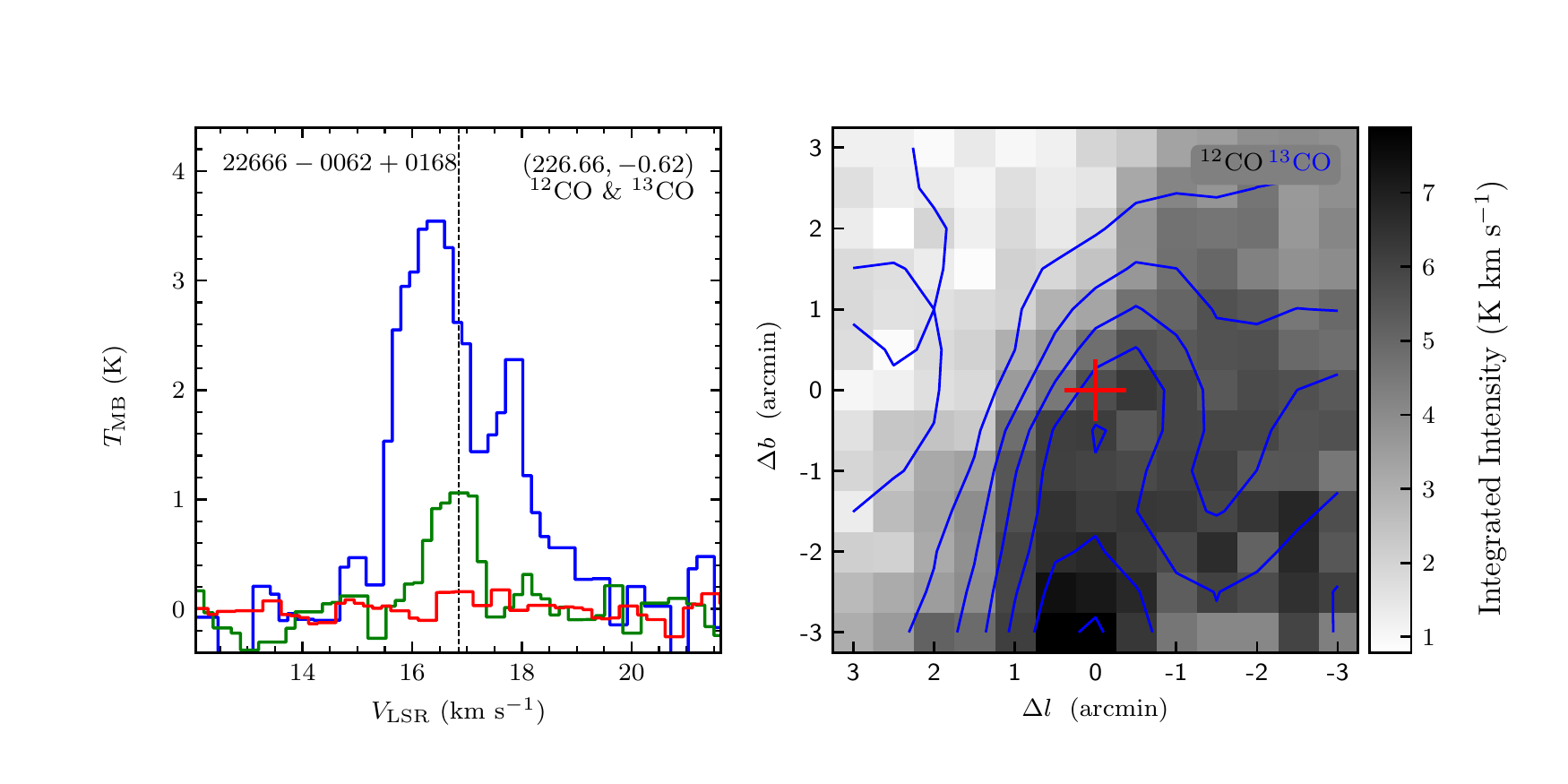}
\end{figure}

\bibliographystyle{raa}
\bibliography{reference.bib}
\end{document}